\documentclass[10pt]{article}
\usepackage[margin=1in]{geometry}
\usepackage{amsfonts,amsmath,amssymb}
\usepackage[none]{hyphenat}
\usepackage{fancyhdr}
\usepackage{graphicx}
\usepackage{bigints}
\usepackage{float}
\usepackage[nottoc,notlot,notlof]{tocbibind}

\usepackage[utf8]{inputenc}

\usepackage{afterpage}
\newcommand\myemptypage{
    \null
    \thispagestyle{empty}
    \addtocounter{page}{-1}
    \newpage
    }

\usepackage{amssymb,amsmath,tikz}
\usepackage{pgfplots} 
\usepackage{amsthm}
\usepackage{mathrsfs}
\usepackage{pifont}
\usepackage{slashed}
\usepackage{mathtools}
\usepackage{cancel}
\usepackage{makeidx}

\PassOptionsToPackage{linktocpage}{hyperref}

\showoutput
\DeclareMathOperator*{\supp}{supp}
\DeclareMathOperator*{\Tr}{Tr \,}
\DeclareMathOperator*{\Sp}{Spec \,}
\DeclareMathOperator*{\cl}{cl \,}
\DeclareMathOperator*{\intt}{int}
\DeclareMathOperator*{\Aut}{Aut}
\DeclareMathOperator*{\Ass}{Ass}
\DeclareMathOperator*{\rev}{rev}
\DeclareMathOperator*{\inv}{inv}
\DeclareMathOperator*{\vac}{vac}
\DeclareMathOperator*{\nonvac}{nonvac}
\DeclareMathOperator*{\unphys}{unphys}
\DeclareMathOperator*{\diag}{diag}

\DeclareMathOperator*{\Op}{Op}
\DeclareMathOperator*{\p}{p}
\DeclareMathOperator*{\x}{x}
\DeclareMathOperator*{\y}{y}
\DeclareMathOperator*{\q}{k}
\DeclareMathOperator*{\n}{n}

\DeclareMathOperator*{\dvol}{d \, vol}
\DeclareMathOperator*{\Dom}{Dom}
\DeclareMathOperator*{\dist}{dist}
\DeclareMathOperator*{\ind}{ind}
\DeclareMathOperator*{\Inv}{Inv}
\DeclareMathOperator*{\In}{In}

\renewcommand{\qedsymbol}{$\blacksquare$}

\newcommand{\tvect}[2]{%
  \ensuremath{\Bigl(\negthinspace\begin{smallmatrix}#1\\#2\end{smallmatrix}\Bigr)}}

\theoremstyle{plain}

\newtheorem*{twr*}{Theorem}
\newtheorem*{lem*}{Lemma}
\newtheorem{twr}{Theorem}
\newtheorem{lem}{Lemma}
\newtheorem{defin}{Definition}
\newtheorem*{defin*}{Definition}
\newtheorem{rem}{Remark}
\newtheorem*{rem*}{Remark}
\newtheorem{cor}{Corollary}
\newtheorem{cor*}{Corollary}
\newtheorem{ex}{Example}

\newtheorem*{notn*}{Notation}
\newtheorem*{wiener-ito*}{Wiener-It\^o-Segal Decomposition}
\newtheorem*{prop*}{Proposition}

\newcommand{\ud}{\mathrm{d}}
\DeclareMathAlphabet{\mathpzc}{OT1}{pzc}{m}{it}

\makeindex

\begin{document}

\begin{titlepage}
\begin{center}
\vspace*{1cm}
\large{\textbf{CAUSAL PERTURBATIVE QFT}}\\
\large{\textbf{AND SPACE-TIME GEOMETRY}}\\
\vspace*{2cm}
\small{JAROS{\L}AW WAWRZYCKI}\\[1mm]
\tiny{
Joint Institute for Nuclear Research, 141980 Dubna, Russia}\\[1mm]
\tiny{\today}\\
\vfill\begin{abstract}
This work is devoted to the causal perturbative Quantum Field Theory (QFT) due to Bogoliubov, including QED
and other realistic QFT. It is given the white noise formulation of this theory.
The white noise analysis and the Hida operators as the creation and annihilation operators
for free fields are used. The whole Bogoliubov method is unchanged. Causal axioms of such QFT make sense on any
globally causal space-times. 
Perturbative QFT with Hida operators is analysed on the flat Minkowski space-time and on the static
Einstein Universe (EU). 
On the flat Minkowski space-time this allowed us to go a step further in the analysis 
of the adiabatic limit problem, the existence of which has a much wider scope than 
in the theory based on  Wightman's operator valued distributions. The natural 
condition of existence of the adiabatic limit (as generalized integral kernel operators of white noise calculus) 
for higher-order contributions to the interacting fields, together with
the remaining assumptions of causal QFT (including the natural invariance conditions, e.g. gauge invariance), allowed 
us to reduce the freedom in choice of renormalization and imposed nontrivial 
conditions on the masses of elementary particles. 
Obtained results on the flat space-time are 
confirmed by the analogous results obtained for QFT on UE, where, moreover, the operators of 
the interacting fields are much more regular, so, e.g. a consequent treatment 
of the bound state problem becomes possible on EU. Our approach also enables 
the analysis of the infrared asymptotics of QED on flat Minkowski spacetime, 
which we also present.
\end{abstract}
\end{center}
\vfill
\end{titlepage}

\myemptypage

\tableofcontents
\thispagestyle{empty}
\clearpage

\myemptypage

\section{Introduction}\label{intro}

This work is concerned with the causal perturbative approach to Quantum Field Theories (QFT), initiated by 
St\"uckelberg, Bogoliubov and Shirkov \cite{Bogoliubov_Shirkov}, and developed mainly by
Epstein, Glaser \cite{Epstein-Glaser}, Blanchard, Seneor, Kazakov, Shirkov \cite{Kazakov_Shirkov}, 
Duch \cite{BlaSen} \cite{duch}, D\"utch, Krahe and Scharf and Fredenhagen \cite{DKS1}-\cite{DKS4}, \cite{DutFred}
and others.

In causal perturbative approach to QFT the infra-red-divergence (IR) problem is clearly separated from the ultra-violet-divergence
(UV) problem by using a space-time function $x \mapsto g(x)$  as coupling ``constant''. 
The UV-problem is essentially solved within this approach, \cite{Epstein-Glaser},   -- the origin of infinite counter terms of the renormalization scheme is well understood by now, i.e. using the counter terms (renormalization) 
is equivalent to the causal perturbative construction of the perturbative series due to Bogoliubov-Epstein-Glaser, 
improved by the Epstein-Glaser splitting of causal distributions \cite{Epstein-Glaser}, 
developed further for QED, and other physical theories with non-Abelian gauge mainly by
D\"utch, Krahe and Scharf, \cite{DKS1}-\cite{DKS4},
where no infinite counter terms appear but instead one uses recurrence rules for the construction of the chronological
product of fields regarded as operator-valued distributions and the Epstein-Glaser splitting. The renormalization scheme 
is now incorporated into the following recurrence causal rules for the chronological product \cite{Epstein-Glaser}, \cite{DKS1}-\cite{DKS4},
\cite{DutFred}, \cite{Scharf}: 
\begin{enumerate}

\item[1)]
causality, 

\item[2)]
symmetricity, 

\item[3)]
unitarity, 

\item[4)]
Poincar\'e covariance (more precisely: $T_4 \circledS SL(2, \mathbb{C})$-covariance) 

\item[5)]
Ward identities -- quantum version of gauge invariance (e. g. in case of QED),

\item[6)]
preservation of the Steinmann scaling degree, 
\end{enumerate}
part of the remaining freedom may be reduced by imposing the natural 
field equations for the interacting field (which is always possible for the standard gauge fields)
and the rest of the remaining freedom is pertinent to the St\"uckelberg-Petermann renormalization group. All the recurrence rules should be regarded as important physical laws which incorporate the whole content of the standard pragmatic approach including the renormalization scheme. 
 
The most important and still open problems are the following. 

\begin{enumerate}

\item[(a)] The problem of
existence of the adiabatic limit ($g \mapsto 1$ \emph{constant function over the whole space-time})
in each order separately. This is the IR-problem or the Adiabatic Limit Problem.

\item[(b)] 
The convergence of the formal perturbative series for interacting fields (with $g=1$), but compare \cite{Kazakov_Shirkov}. 

\end{enumerate}

In this work we give a positive solution to the Adiabatic Limit Problem for QED, i.e. the problem (a), 
and give a contribution to the problem (b) for QED, including QED and other realistic causal perturbative QFT on the Minkowski
space-time and other globally causal space-times with compact Cauchy surfaces and nonzero curvature. 
 
The method is based solely on substitution into the casual perturbative series the free fields of the theory 
which are constructed with the help of white noise calculus. The whole causal perturbative method of Bogliubov-Epstein-Glaser 
remains unchanged. The whole point in constructing the free fields within the white nose setup lies in the fact that
it  allows us to treat them equivalently as integral kernel operators with vector-valued kernels in the sense of 
Obata \cite{obataJFA},
and opens us to the effective theory of such operators worked out by the Japanese School of Hida which developed the ideas
of Berezin's integral kernel operators in the Fock space as well as Fock expansions for generalized operators. 
Using the calculus of such operators we show that the class of integral kernel operators represented (or representing)
free fields allows the operations of differentiation (similarly as Schwartz distributions)
integration, pointwise Wick product, integration of Wick product integral kernel operators
(including spatial integration), convolution of Wick product integral kernel operators with tempered distributions,
and splitting into advanced and retarded parts of integral kernel operators with causal supports.
Thus, all operations needed for the causal perturbation series have a well-defined mathematical meaning
if understood as operations performed upon integral kernel operators in the sense of Obata.
Therefore, the free fields, understood as integral kernel operators with vector-valued kernels in the sense of Obata, 
can be inserted into the formulas for the higher order contributions to the interacting fields. 
After the insertion we obtain each order term contribution to interacting fields in a form of finite sums of 
well-defined integral kernel operators with vector-valued kernels,  similarly as for the free fields themselves or
for the Wick products of free fields.

But the most essential point is that these formulas for higher-order contributions to 
interacting fields in QED do not lose their rigorous mathematical meaning of finite sums of integral kernel operators 
even if we put in them the intensity-of-interaction function $g$ equal $1$ everywhere over the whole space-time
 if and only if the charged field is massive and the normalization in the splitting of causal distribution is the 
``on mass shell'' normalization. Thus, using the Hida operators, we eliminate the ambiguity in the choice of normalization
and give a theoretical proof that the charged particles are massive.
Thus, with the natural on mass shell normalization and massive charge, and only in this case, 
the contributions still preserve their meaning of integral kernel operators with vector-valued kernels in the adiabatic limit,
which belong to the same general class of integral kernel operators as the Wick products of free fields.
We therefore arrive at the positive solution to Problem (a) in QED.
But at the same time, we obtain the interacting fields in the form of Fock expansions into integral kernel operators
with vector-valued kernels in the sense of \cite{obataJFA}, with a precise estimate of the convergence, which
allows us to give criteria for the convergence of the perturbative series, pertinent to
the white noise calculus and Fock expansions, \emph{i .e} a contribution to the solution of the Problem (b). 

The method is general enough to be capable of application to other QFT with non-Abelian gauge. 

In this manner we obtain causal perturbative QED in which there are no infrared nor ultraviolet divergences
and get insight into problems which were beyond the reach of the approach not usig Hida operators. 
In particular, we hope that we have given a step forward on the way in giving
a rigorous construction of a nontrivial (and realistic) quantum interacting field. This problem 
is one of the the most important unsolved problems in the contemporary analysis and QFT, 
compare \cite{Segal-ProcStone}.

However, we should emphasize, that the causally constructed interacting fields, of course with the intensity 
of interaction function $g=1$ everywhere on the space-time, (in fact all the higher order contributions 
to them) belong to a class of the so-called generalized operators. More precisely they 
are integral kernel operators with vector-valued kernels, transforming continuously 
the nuclear space-time test space $\mathscr{E}$ into the nuclear space 
$\mathscr{L}((\boldsymbol{E}), (\boldsymbol{E})^*)$ of continuous maps from 
the Hida test space $(\boldsymbol{E})$ into its strong dual $(\boldsymbol{E})^*$, \emph{i. e.}
they belong to
\[
\mathscr{L}((\boldsymbol{E}) \otimes \mathscr{E}, \, (\boldsymbol{E})^*) \cong
\mathscr{L}(\mathscr{E}, \, \mathscr{L}((\boldsymbol{E}), (\boldsymbol{E})^*) \, ).
\]
The spaces  $\mathscr{L}((\boldsymbol{E}), (\boldsymbol{E}))$, 
$\mathscr{L}((\boldsymbol{E}) \otimes \mathscr{E}, \, (\boldsymbol{E})^*)$ are endowed with the natural topology of uniform convergence on bounded sets.
In particular each higher order contribution to an interacting field, when evaluated on an element $\phi$ of the space-time test space
$\mathscr{E}$ (\emph{i. e.}``smeared with test function''), is equal to a finite sum of integral kernel operators $\Xi_{l,m}(\kappa_{l,m}(\phi))$ with scalar valued kernels $\kappa_{l,m}(\phi)$. Each such operator $\Xi_{l,m}(\kappa_{l,m}(\phi))$ defines continuous functional
\[
(\boldsymbol{E}) \times (\boldsymbol{E}) \ni \Phi \times \Psi \mapsto
\big \langle \big \langle  \Xi_{l,m}(\kappa_{l,m}(\phi)) \Phi, \Psi  \big \rangle \big \rangle \in \mathbb{C}
\]
with distributional kernel which can be identified with ``matrix elements'' between the many particle 
plane wave states, and which unfortunately are not ordinary numbers, but distributions. Only some 
higher order contributions to interacting fields, when smeared out with test function give generalized operators, which belong to
\[
\mathscr{L}((\boldsymbol{E}) \otimes \mathscr{E}, \, (\boldsymbol{E})) \cong
\mathscr{L}(\mathscr{E}, \, \mathscr{L}((\boldsymbol{E}), (\boldsymbol{E})) \, )
\]
and thus define ordinary operators on the Fock space of free fields, transforming continuously 
the Hida test space $(\boldsymbol{E})$ into itself. This is in particular the case for the first 
order contribution to the interacting electromagnetic potential field (of course with 
the intensity of interaction function $g=1$ everywhere on the space-time).  

Similarly, the causal perturbative series for the scattering operator $S(g)$, for $g$ being 
a space-time test function -- an element of 
a standard nuclear space $\mathscr{E}$,  has the form
\[
\sum \limits_{n,l,m} \Xi_{l,m}\Big(\kappa_{l,m}\big(g^{\otimes \, n} \big)\Big)
\]
into integral kernel operators $\Xi_{l,m}(\kappa_{l,m})$ with scalar-valued distributional kernels 
$\kappa_{l,m}\big(g^{\otimes \, n} \big)$, and with contribution of each fixed order equal to a finite sum 
\[
\sum \limits_{l,m} \Xi_{l,m}\Big(\kappa_{l,m}\big(g^{\otimes \, n}\big)\Big)
\]
of integral kernel operators $\Xi_{l,m}\Big(\kappa_{l,m}\big(g^{\otimes \, n}\big)\Big)$, which in general belong to 
\[
\mathscr{L}((\boldsymbol{E}), \, (\boldsymbol{E})^*).
\]
In case of QFT with massless fields in the interaction Lagrangian density $\mathcal{L}(x)$, 
only part of the integral kernel operators entering higher order contributions are regular enough to be elements of
\[
\mathscr{L}((\boldsymbol{E}), \, (\boldsymbol{E})),
\]
and can be interpreted as ordinary operators on the Fock space transforming continuously the test Hida space $(\boldsymbol{E})$ into itself. All
higer order conributions $S_n(g^{\otimes \, n})$ to $S(g)$ belong to
\[
\mathscr{L}((\boldsymbol{E}), \, (\boldsymbol{E})),
\]
if the free fields underlying the theory are all massive. In QED case no higher order integral kernel operators
$\Xi_{l,m}$ behave so regularly. In general however, for theories including massless fields as the free fields underlying the theory (e.g. QED), 
each $n$-th order contribution $S_{n}(g)= S_n(g^{\otimes \, n})$ to $S(g)$, with $g\in \mathscr{E}$, belongs to $\mathscr{L}((\boldsymbol{E}), \, (\boldsymbol{E})^*)$, and each higher order contribution $S_{n}(g)$ to the scattering operator 
$S(g)$, $g\in \mathscr{E}$,  defines a continuous functional
\[
(\boldsymbol{E}) \times (\boldsymbol{E}) \ni \Phi \times \Psi \mapsto
\big \langle \big \langle  S_{n}(g) \Phi, \Psi  \big \rangle \big \rangle \in \mathbb{C}, \,\,\, \textrm{for each fixed} \, g\in \mathscr{E},
\]
\emph{i. e.} a \emph{distribution}, with distribution kernel which can be canonically identified with the 
distributional ``matrix elements''
\begin{equation}\label{DistributionSmatrixElements}
\big \langle S_{n}(g) \Phi_{{}_{\ldots s,\boldsymbol{\p} \ldots }}, \Psi_{{}_{\ldots s',\boldsymbol{\p}' \ldots }}\big\rangle, \,\,\, 
\textrm{for each fixed} \, g\in \mathscr{E},
\end{equation}
of the (generalized) operator $S_n(g)$, $g\in \mathscr{E}$,  in the non-normalizable many particle plane wave
states
\[
\Phi_{{}_{\ldots s,\boldsymbol{\p} \ldots }} =
\cdots a_{s}(\boldsymbol{\p})^{+} \cdots |0\rangle, 
\,\,\,\,
\Phi_{{}_{\ldots s,\boldsymbol{\p} \ldots }} =
\cdots a_{s'}(\boldsymbol{\p}')^{+} \cdots |0\rangle \in (\boldsymbol{E})^*,
\]
with the creation (Hida) operators $a_{s}(\boldsymbol{\p})^{+}$ in the momentum picture and with 
$|0\rangle = \Psi_{{}_{0}}$ being the vacuum in the Fock space of free fields of the theory.
This is the general situation in causal perturbative QFT we arrive at, when using the Hida white noise operators as the annihilation-creation operators. 
Only part of the integral kernel operators
$\Xi_{l,m}(\kappa_{l,m})$ entering higher order contributions $S_{n}(g)$ to $S(g)$, for each fixed $g\in \mathscr{E}$, belongs to 
$\mathscr{L}((\boldsymbol{E}), \, (\boldsymbol{E}))$, 
and we can compute for them the ordinary matrix elements
\[
\big\langle\Xi_{l,m}(\kappa_{l,m}) \Phi, \Psi \big\rangle,
\,\,\,
\Phi, \Psi \in (\boldsymbol{E}) \subset \mathcal{H}_{\textrm{Fock}}
\]
in the normalizable states $\Phi, \Psi$ belonging to the test Hida space $(\boldsymbol{E})$
densely included into the total Fock space $\mathcal{H}_{\textrm{Fock}}$ of all free fields of the theory, and composing the Gelfand triple
\[
(\boldsymbol{E}) \subset \mathcal{H}_{\textrm{Fock}} \subset (\boldsymbol{E})^*.
\]

Thus, we arrive at the solution of the Adiabatic Limit Problem which, at the first sight, may seem totally unsatisfactory. This may seem so because our scattering matrix $S(g)$, $g\in \mathscr{E}$, and even the separate higher order contributions $S_{n}(g)$, are not ordinary operators, which are in general not well-defined on normalizable states. But in fact we should strongly emphasize here, that we do not need the scattering matrix as an operator acting on normalizable states in the computation of the effective cross-section. Paradoxically $S(g)$ acting on a dense domain $\mathscr{D}$ of normalizable states would be even insufficient for the computation of the effective cross-section,
compare \cite{Bogoliubov_Shirkov}, Chap. IV, \S\S 23-25.  In fact in this computation we need an operator
$S(g)$ for each $g\in \mathscr{E}$, defining the distributional matrix elements (\ref{DistributionSmatrixElements}), \emph{i.e.}
we need $S(g)$ as a continuous operator $(\boldsymbol{E}) \rightarrow (\boldsymbol{E})^*$, depending continuously on $g \in \mathscr{E}$
in order to compute the limit $g \rightarrow g=1$ in the effective cross section. If the dense domain $\mathscr{D}$ would not be endowed 
with the necessary additional nuclear topology structure, which allows us to understand $S(g)$ as a continuous operator defining the distributional kernels
(\ref{DistributionSmatrixElements}), then such $S(g)$ would be useless in computation of the effective
cross section in the limit $g=1$, even if defined on a dense domain $\mathscr{D}$ of normalizable states.  In particular
such operator would be useful in case $\mathscr{D} = (\boldsymbol{E}) \subset \mathcal{H}_{\textrm{Fock}}$, and when it would be a continuous operator $(\boldsymbol{E}) \rightarrow (\boldsymbol{E})$, because such operator is naturally a continuous map $(\boldsymbol{E}) \rightarrow (\boldsymbol{E})^*$, defining the distributional ``matrix elements'' (\ref{DistributionSmatrixElements}), as its kernel, which moreover defines continuous map $\mathscr{E} \ni g \longmapsto S(g) \in 
\mathscr{L}((\boldsymbol{E}), \, (\boldsymbol{E})^*)$. 
This is because in the scattering 
phenomena we are dealing with non-normalizable many particle plane wave states, 
\emph{i. e.} generalized states belonging to $(\boldsymbol{E})^*$. In order to compute the effective cross section we need the ``matrix elements'' (\ref{DistributionSmatrixElements}) which are distributions (and not ordinary numbers), because the many particle plane wave states are non-normalizable generalized states. When evaluating the effective cross-section we do not need to know the amplitude of the absolute probability, but only the amplitude for the registration of a particle with given spin and momentum and mass \emph{per unit volume and unit time} (or per unit time in case of scattering by a static classical field). These circumstances allow us to compute
\[
\big| \big \langle S_{n}(g) \Phi_{{}_{\ldots s,\boldsymbol{\p} \ldots }}, \Psi_{{}_{\ldots s',\boldsymbol{\p}' \ldots }}\big\rangle \big|^2,
\,\,\, g \in \mathscr{E},
\]
with the distribution kernel  
\[
\big \langle S_{n}(g) \Phi_{{}_{\ldots s,\boldsymbol{\p} \ldots }}, \Psi_{{}_{\ldots s',\boldsymbol{\p}' \ldots }}\big\rangle,
\,\,\, g \in \mathscr{E},
\]
behaving regularly enough to be represented by ordinary function, for $g\in \mathscr{E}$, which allows the operation of multiplication
\[
\overline{\big \langle S_{n}(g) \Phi_{{}_{\ldots s,\boldsymbol{\p} \ldots }}, \Psi_{{}_{\ldots s',\boldsymbol{\p}' \ldots }}\big\rangle}
\big \langle S_{n}(g) \Phi_{{}_{\ldots s,\boldsymbol{\p} \ldots }}, \Psi_{{}_{\ldots s',\boldsymbol{\p}' \ldots }}\big\rangle
= \big| \big \langle S_{n}(g) \Phi_{{}_{\ldots s,\boldsymbol{\p} \ldots }}, \Psi_{{}_{\ldots s',\boldsymbol{\p}' \ldots }}\big\rangle \big|^2
\] 
with the continuous dependence on $g$, which in turn allows for the computation of the ``matrix elements''
\[
 \big \langle S_{n}(g=1) \Phi_{{}_{\ldots s,\boldsymbol{\p} \ldots }}, \Psi_{{}_{\ldots s',\boldsymbol{\p}' \ldots }}\big\rangle,
\,\,\, g \in \mathscr{E},
\]
and extraction of the ``residual part''
\[
\big| \big \langle S_{n}(g=1) \Phi_{{}_{\ldots s,\boldsymbol{\p} \ldots }}, \Psi_{{}_{\ldots s',\boldsymbol{\p}' \ldots }}\big\rangle \big|^2,
\]
in the adiabatic limit $g \rightarrow g=1$, and the computation of the effective cross section in the adiabatic limit,
compare e.g. \cite{Bogoliubov_Shirkov}, Chap. IV, \S\S 23-25, even if  
(\ref{DistributionSmatrixElements}) is a distribution. Thus, in order to have a theory in a minimal form needed for the computation of the effective cross section, it is sufficient that the interaction defines, together with the causal rules for the construction of the perturbative series, the scattering matrix
$S(g)\in \mathscr{L}((\boldsymbol{E}), \, (\boldsymbol{E})^*)$, \emph{i.e.} with $S(g)$ as a continuous operator $(\boldsymbol{E}) \rightarrow (\boldsymbol{E})^*$, 
for each fixed $g \in \mathscr{E}$, which moreover defines continuous map $\mathscr{E} \ni g \longmapsto S(g) \in \mathscr{L}((\boldsymbol{E}), \, (\boldsymbol{E})^*)$. The requirement that the scattering operator $S(g=1)$ should be well-defined on normalizable states (intentionally even unitary
in the ordinary sense) is quite unrealistic for realistic interactions of fields on the Minkowski space-time. 

Thus, on using the Hida operators as creation-annihilation operators, we arrive at the causal perturbative 
QFT, in particular QED, which can successfully be applied to the class of generalized states, in particular to the high energy scattering phenomena involving the generalized many particle plane wave states. But there is also another class of generalized states, which can also be experimentally extracted
and related to the infrared problem, and which can successfully be treated with the causal perturbative QED we have just constructed with the help of Hida operators. Namely we consider the generalized (belonging to $(\boldsymbol{E})^*$) homogeneous states of homogeneity degree $-1$ in single particle Fock subspace of the free electromagnetic potential field $A$. Using the homogeneous of degree $-1$ part of the interacting electromagnetic potential field we constructed generalized states which provide a special realisation of the general quantum theory of the electric charge due to \cite{Staruszkiewicz}. There is a unique relationship between the representation structure of the representation of $SL(2, \mathbb{C})$ acting on the specified class of the generalized states, and the value of the fine structure constant. This relationship, in particular, and in general the reconstruction of the relationship of causal perturbation QED with \cite{Staruszkiewicz} is beyond the scope of the approach not using Hida's operators. We have described this relationship in details in Chap \ref{infra}. 

However, it is not the computational aspect, that now we can compute the effective cross sections
for high energy scattering processes involving many particle plane wave states of the elementary fields without any use of renormalization and without infrared infinities, which is of primary importance here. 

The most important thing about the elimination of the infrared and ultraviolet infinities altogether from causal perturbative QFT, in particular from QED, lies in the fact that now we have a mathematical theory, with the basic principles formulated in well-defined mathematical terms. These principles are: 1) the Hamiltonian formulation of classical theory subject to quantization (e.g. classical QED) together with the canonical commutation rules on which the relation between free classical fields and their quantum counterparts is based, 2) the causal rules for the causal perturbative construction of the scattering (generalized) operator $S(g=1)$, 3) causal geometry of space-time (here the Minkowski space-time). The interacting quantum fields are obtained from the scattering operator $S(g=1) = S(\mathcal{L})$, due to the general Bogoliubov rule
\[
A_{{}_{\textrm{int}}}(g=1; x) = \frac{i\delta}{\delta h(x)} 
S(\mathcal{L} +hA)^{-1}S(\mathcal{L})\Big|_{{}_{h=0}}
\]
relating the interacting field $A_{{}_{\textrm{int}}}(g=1)$ to the corresponding free field $A$, compare \cite{Bogoliubov_Shirkov}, \cite{DKS1}, \cite{DutFred}. Here $S(g=1) = S(g\mathcal{L}) =
S(\mathcal{L})$ is the scattering matrix in which the first order term is equal
\[
S_1(g) = \int i\mathcal{L}(x) \, g(x) \, \ud^4 x
\]
understood as an integral kernel operator with the free fields in the interaction Lagrangian density 
$\mathcal{L}(x)$ understood as the integral kernel operators with vector-valued kernels.
Similarly, $S(\mathcal{L} +hA)$ is the scattering (generalized) operator $S(g=1,h)$ in which the first induction step in the causal perturbative construction is equal
\[
S_1(g,h) = \int i\big(g(x) \, \mathcal{L}(x) + h(x) \, A(x)\big) \, \ud^4 x,
\]
also understood as integral kernel operator.
 Here we only must remember that the canonical commutation rules in case of the system of free fields with infinite number of degrees of freedom are rigorously realized through the annihilation-creation operators, which mathematically are understood as white noise Hida operators. Because the Hida operators indeed do fulfil the canonical commutation rules they fit naturally in as the realization of the annihilation-creation operators of quantum mechanics (of the system with infinite number of degrees of freedom). 
 This theory has well-defined range of experimental applicability: it can be applied at least to the two classes of generalized states. The first class involves the many-particle plane wave generalized states in the high energy scattering phenomena and provides the effective cross sections. The second class of generalized states embraces the many particle generalized states constructed by symmetrized/antisymmetrized projective tensor product of homogeneous of degree $-1$ states (of the free field $A$) and the corresponding generalized single particle states of the massive fields coupled to $A$. 
The second class thus embraces the infrared states. 
 We should emphasize here that the first class subsumes the situations in which besides the scattering caused by the fundamental interactions between the quantum fields there also can be included interacting terms coming from unquantized classical fields. In fact this is the success of the theory in the last case which serves as a justification for the causal method of St\"uckelberg-Bogoliubov. But the second class
 embraces essentially only the scattering phenomena. There are very exceptional situations among the ones which include interaction terms with classical unquantized fields, where the classical unquantized fields are static, and one can separate the terms contributing to interacting quantum fields into two sets: a) that giving the contributions to the interacting fields defining the abstract non-interacting quantum fields in the external classical fields (procedure known under the name \emph{external field problem}), and b) contributions coming from the fundamental interactions between the quantum fields (called 
 \emph{radiative corrections}). In the exceptional situations the fields solving external field problem can be given explicitly and can be constructed as free fields, where the single particle states
represents eigenstates of a single particle Hamiltonian with binding potential, which among its spectrum contains the pure point spectrum consisting of bound (\emph{i. e.} normalizable) states. Then the problem can be solved in two steps: first, solution of the external field problem and second, computation of the radiative corrections. In this way the Lamb shift can be computed for bound states of hydrogen-like atoms, as coming from radiative corrections. 
We should emphasize, however, that this method of computing radiative
corrections to bound states does not lie totally within consequent QFT with just interacting quantum fields as the only primary datum, and uses some contributions that are external: namely, we can compute the Green functions, but the many particle wave function tied to the Green function through the Huygens principle cannot be clearly associated to any well-defined state because the  Noether integral with interacting fields corresponding to time translations is not well-defined as an ordinary operator (for QED on the Minkowski space-time).

  Therefore, we should accept that the causal perturbative QED with Hida operators with the corresponding scattering matrix with only the fundamental interactions between quantum fields, and the corresponding interacting fields, as the only primary datum, does not embraces the bound state problem. We need some extra assumptions in order to treat 
some aspects of bound states, compare Subsection \ref{Green}.

The fact that we have a causal perturbative QFT with principles expressed in well-defined mathematical terms cannot be overestimated. Now the rules are under mathematical control and now we can find the sources of difficulties in explaining specific physical problems. This would be impossible within the technique of renormalization which involves handling infinite quantities simply because such a handling is not really a mathematically logical process, and we cannot proceed along any logical-mathematical line from a physical problem to its source identified within the principles. 
We illustrate this by giving two examples. The first is (I) the problem with bound states and the spectrum of stable or meta-stable particles and the second, related to the first, concerns (II) relation of QFT to space-time geometry, compare Subsection \ref{G} and Section \ref{EUandG}.

\begin{center}
{\small (I)}
\end{center}
The Emmy Noether integrals corresponding to the one parameter subgroups of space-time symmetries or to the phase transformations, do exist as ordinary self-adjoint operators in the Fock space for free quantum fields, which we prove using the integral kernel operator analysis. But the same integrals for the interacting fields do not exist as ordinary operators, but only as generalized operators (at least this is so for the separate higher order contributions). They are generalized integral kernel operators transforming continuously the Hida space $(\boldsymbol{E})$ into its strong dual $(\boldsymbol{E})^*$. This is the mathematical consequence of the principles 1) -- 3) stated above. In particular the Noether generator of time translations, the Hamiltonian, is not well-defined as an ordinary self-adjoint operator in the Fock space for the system of interacting fields, but it is only a generalized integral kernel operator belonging to $\mathscr{L}((\boldsymbol{E}), \, (\boldsymbol{E})^*)$. In particular the problem of analysis of normalizable bound eigen-states of the Hamiltonian (say stable particles), or their 
superpositions with small energy uncertainty (say meta-stable particles) cannot be grasped within the principles 1) -- 3), and at least one of these principles will have to be changed in order to account for the existence of stable and metastable particles and generally in order to account for phenomena involved into bound states. We note here that 1) and 2) make sense on general globally causal space-time, at least when there exist four one-parameter groups of space-time symmetries with the corresponding vector fields which span everywhere the tangent space, playing the role analogous to translations (they do not have to commute). 
We have proved that 1) and 2) make sense for more general space-times on the concrete example of the Einstein Universe in Section \ref{EUandG},
compare in particular Subsection \ref{CausalSonEU} and the Emmy Noether integrals of the interacting fields become ordinary operators. 
Because 1) and 2) make sense on general globally causal space-time, and because the high energy scattering experiments  confirm 1) and 2) 
and because these experiments are less sensitive to the global structure of space-time we arrive at the conclusion 
that the geometry of space-time should have to be changed into some other globally causal. 
This throws some light on the problem why we are rather successful in understanding Rutherford-type experiments with deeply inelastic scattering of electron by a nucleon (e.g. the series of the famous experiments of SLAC-MIT cooperation) whenever we confine ourselves to the scattering at the level of many-particle plane wave states of elementary fields for high energies (say inside the nucleon), but at the same time we cannot account for the lower energy scattering involved into bound state production, using the same elementary fields with the same gauge interactions. For more details on this analysis we refer to 
Subsection \ref{G} and Section \ref{EUandG}.

\begin{center}
{\small (II)}
\end{center}
The interacting fields are so much singular on the Minkowski space-time, that even after ``smearing'' with test functions do not give any ordinary operators in the Fock space, but only generalized operators transforming continuously 
$(\boldsymbol{E}) \rightarrow (\boldsymbol{E})^*$. Even the Wick polynomials of free fields (if they contain massless free field factors) behave in the similar singular manner, and even for free fields averages of some physical local quantities, entering e.g. energy-momentum tensor, cannot be sensibly computed even after ``smearing out'' with test function (in case we are using Hida operators as annihilation-creation operators). In particular no sensible quasi-classical limit 
exists for interacting fields, no sensible computation of the averages of the analogues of important classical measurable local quantities, can be performed. Thus, no sensible quasi-classical limit exists for interacting quantum fields. These strange theorems are mathematical conclusions of the above principles 1) -- 3). Put otherwise: from 1) and 2) it follows that no interacting quantum fields can be constructed on the flat Minkowski space-time which have the classical limit, or which have (in Bohr's parlance) correspondence to classical fields. Thus, we arrive at the conclusion, that it follows from the principles 1) and 2) that the quantum fields which have correspondence to their classical counterparts and have classical limits cannot be constructed on the flat Minkowski space-time. 
We are led to the hypothesis that
gravity acts purely as a constraint, restricting only the allowable class of states. \emph{I.e.} (A) by selection of the space-time which
influences the complete set of solutions underlying the free fields of the theory in question, which must live on the space-time, 
without any back-reaction.  (B) By selection of the physical states  with the average value $\big\langle\mathbb{T}(\phi)\big\rangle$
of the Hilbert energy-momentum tensor components $\mathbb{T}(\phi)$ equal to the Einstein geometric tensor components $G(\phi)$ 
with the accuracy equal to the fluctuation
(uncertainty) of the Hilbert energy-momentum tensor $\Delta \mathbb{T}(\phi)$, 
both smeared out with any space-time test function $\phi$.  
In other words: each system of (interacting) quantum fields can 
coexist with the given spacetime geometry if the (interacting) quantum fields constructed according to 1) and 2) with the assumed globally causal geometry of the space-time, are well-behaved interacting fields, being well-defined operator-valued distributions, with the Noether integrals (if symmetry exist)
being self-adjoint operators on the Fock space, and admitting states in which 
$\Delta \mathbb{T}(\phi) \leq \big\langle\mathbb{T}(\phi)\big\rangle - G(\phi) \leq \Delta \mathbb{T}(\phi)$. 
That's all. Of course in order to make this mathematical conclusion to be not empty we need to give at least one non-trivial example of a globally causal space-time, on which the free and the causally and perturbatively constructed interacting fields are regular enough to be elements of $\mathscr{L}\big(\mathscr{E}, \mathscr{L}((\boldsymbol{E}), (\boldsymbol{E}))\big)$ and admit quasi-classical states in which the average values of the smeared-out quantum analogues of the quantities  entering the Hilbert energy-momentum tensor components can sensibly be computed and compared to the smeared out Einstein geometric tensor. Moreover, if we are about to preserve the first two principles 1) and 2), except in changing the globally causal spacetime geometry of the Minkowski space-time into some other globally causal space-time,
we should make this change to preserve agreement with high energy scattering experiments. 
This is strong limitation because theory based on 1) and 2) with the assumption that the many particle plane wave states as \emph{in} and \emph{out} states gives the effective cross sections which are in agreement with experiment although the plane wave states are assumed to live on the flat Minkowski space-time. This is very nontrivial and valuable limitation, which in particular limits the set of allowable ``plane wave'' packets as the sensible single particle states in the Fock spaces of local free quantum fields on the chosen space-time. Unfortunately almost all works concerned with construction of local free quantum fields on globally causal space-times other than the Minkowski space-time ignore this condition. But fortunately there is one exception: in the series of works \cite{SegalZhouQED}, \cite{SegalZhouPhi4}, \cite{PaneitzSegalI}-\cite{PaneitzSegalIII}, in which free quantum fields are constructed on the static Einstein Universe, relation to the scattering phenomena on the Minkowski space-time is seriously accounted for in the form  sufficient for our purposes. These authors do not use 
the Hida operators nor the white noise analysis of integral kernel operators, which is very effective in the investigation of quantum fields. 
Also, the causal perturbative method is not used in \cite{SegalZhouQED}, \cite{SegalZhouPhi4}, \cite{PaneitzSegalI}-\cite{PaneitzSegalIII}. Nonetheless, 
results obtained there already indicate that all local massive free fields on the  Einstein Universe as well as the QED interaction Lagrange 
density behave so regularly as expected. 
In Section \ref{EUandG} we present the analysis (II) in details. In fact, we have developed in detail the causal QFT, including QED,
on the Einstein Universe, based on the white noise analysis. We have proved the analogue theorem as on the Minkowski space-time:
that the Emmy Noether integrals of interacting fields are well-defined operators (at each order separately) if and only if the charged field is massive,
and in this case, the renormalization is unique (the analogue of the splitting of the causal distributions is unique). 
The presented method and proofs are general enough to convince the reader that
the main conclusions remain valid on more general globally causal space-times with non-zero curvature and compact Cauchy surfaces. In particular
this is the case for space-times, which, as smooth manifolds, coincide with the Lie group $\mathbb{R}\times SU(2,\mathbb{C})$, and which, as space-times, are images of
the Einstein Universe under a causal (``conformal'') isomorphism.

\vspace*{1cm}

The following Subsections of Introduction serve as a guideline to the whole work, present the results
of this work more precisely, as well as the whole line of the reasoning, and contain references to the whole remaining part of the work, where the cited results are rigorously formulated and proved.

\subsection{Adiabatic Limit Problem and its solution}\label{AdiabaticLimit}

We keep the causal method of St\"uckelbeg-Bogoliubov-Epstein-Glaser
unchanged, with the only proviso: we insert into the formulas the free fields of the theory
which are constructed with the help of white noise Hida operators -- construction
of free fields which goes back to Berezin and later improved by the Japanese school of
Hida. This allows us to interpret the free fields as integral kernel operators
with vector-valued distribution kernels in the sense of Obata.
The rest part of the work is reduced to application of the white noise calculus of integral kernel
operators, which essentially is reduced to the proof that the operations involved in the causal perturbative
construction of the higher order contributions are well-defined when applied to the integral
kernel operators defined by free fields. The main difficulty lies in the white noise construction of the
free fields, namely the free Dirac and electromagnetic fields $\boldsymbol{\psi}$, $A$, as finite sums
\[
\boldsymbol{\psi} = \Xi_{0,1}(\kappa_{0,1}) + \Xi_{1,0}(\kappa_{1,0}), \,\,\,
A= \Xi_{0,1}(\kappa'_{0,1}) + \Xi_{1,0}(\kappa'_{1,0})
\in \mathscr{L}\big((E) \otimes \mathscr{E}, \, (E)^* \big)
\]
(of two) well-defined integral kernel operators, in the sense of Obata \cite{obataJFA},
with vector valued distributional kernels $\kappa, \kappa'$ which belong respectively to
\[
\mathscr{L}\big(E, \mathscr{E}^* \big),
\]
Here $E$ is the respective nuclear space of restrictions of the Fourier transforms $\widetilde{\varphi}$ of all
space-time test functions $\varphi \in \mathscr{E}$ to the respective orbit $\mathscr{O}$
in the momentum space determining the representation of the $T_4 \circledS SL(2, \mathbb{C})$
acting in the single particle Hilbert space of the respective field, $\boldsymbol{\psi}$ or $A$.
$\mathscr{L}\big(E, \mathscr{E}^* \big)$ denotes the space of all linear continuous operators
$E \rightarrow \mathscr{E}^*$,
\emph{i. e.} $\mathscr{E}^*$-valued distributions over the corresponding orbit $\mathscr{O}$
in the momentum space (recall that $\mathscr{O}$ is equal to the positive energy sheet of the hyperboloid
$p \cdot p = m^2$ in the momentum space in case of field of mass $m$).
We endow $\mathscr{L}\big(E, \mathscr{E}^* \big)$ with the natural topology of uniform convergence
on bounded sets. $(E), (E)^*$ is the nuclear Hida subspace
of the Fock space of the corresponding free field, and its strong dual space.

Moreover, we need a construction of the free fields,
$\boldsymbol{\psi}$, $A$, with as explicit representation of the Poincar\'e group in their Fock spaces as
possible, to which the perturbative expansion can naturally be applied. 
Unfortunately no construction of the two most important fields in the whole of
QFT, namely Dirac $\boldsymbol{\psi}$ field and the e. m. potential $A$ in Gupta-Bleuler Lorentz gauge, 
based on the theory of representations
of $T_4 \circledS SL(2, \mathbb{C})$, has been achieved. This is a well-known fact, compare
\cite{Haag}, p. 48, \cite{lop1}, \cite{lop2}. This is because this problem cannot be solved within the ordinary
unitary representations of the $T_4 \circledS SL(2, \mathbb{C})$ group.
We have been forced to extend the Mackey theory of induced representations over to a more general class
of representations in order to solve this unsolved problem, compare Section \ref{PartIIMackey} for this extension.
But this is not the whole problem, because we additionally need
a white noise constructions of these two free fields $\boldsymbol{\psi}$ and $A$.
This construction was essentially worked out for the simplest massive free scalar field by mathematicians
\cite{HKPS}, and its generalization to other massive fields (if the group theoretical aspect is ignored)
presents no essential difficulties. But concerning the massless fields, such e.g. as $A$,
the white noise construction is far not so obvious and in fact (as to the author's knowledge)
has not been done before. This is because the white noise construction of the massless fields requires the modification
of the single particle nuclear test space $E$. More precisely, the kernels
\[
\kappa'_{0,1},\kappa'_{1,0} \in \mathscr{L}\big(E^*, \mathscr{E}^* \big) \subset \mathscr{L}\big(E, \mathscr{E}^* \big),
\]
and the corresponding field operator $A$ becomes an operator-vaued distribution (in the strong white noise sense), \emph{i.e.} 
\[
A \in  \mathscr{L}\big(\mathscr{E} ,  \mathscr{L}( \, (E), (E)^* ) \big)
\]
if and only if $\mathscr{E} = \mathcal{S}^{00}(\mathbb{R}^{4};\mathbb{C}^4)$, $E = \mathcal{S}^{0}(\mathbb{R}^{3};\mathbb{C}^4)$.
Here $\varphi \in \mathcal{S}^{00}(\mathbb{R}^{n};\mathbb{C}^d)$ if and only if its Fourier transform
$\widetilde{\varphi} \in \mathcal{S}^{0}(\mathbb{R}^{n};\mathbb{C}^d)$, and
$\mathcal{S}^{0}(\mathbb{R}^{n};\mathbb{C}^d)$ is the subspace of $\mathcal{S}(\mathbb{R}^{n};\mathbb{C}^d)$
of all those functions which have all derivatives
vanishing at zero. Correspondingly we have the nuclear algebra $E$
of all restrictions of Fourier transforms to the corresponding orbit $\mathscr{O}$ (positive energy sheet of the cone)
of the elements of the test space
$\mathscr{E} = \mathcal{S}^{00}(\mathbb{R}^{4};\mathbb{C}^4)$, equal to
$E= \mathcal{S}^{0}(\mathbb{R}^3; \mathbb{C}^4)$ (of $\mathbb{C}^4$-valued functions in case of the field $A$,
but for other $r$-component massless fields we will have $\mathbb{C}^r$-valued functions here).
This is related to the singularity of the cone orbit $\mathscr{O}$ at the apex -- the orbit pertinent to the representation associated 
with massless fields, i.e. the positive sheet of the cone in the momentum space
(note that each sheet of the massive hyperboloid $\{p \cdot p = m^2\}$ in the momentum space is everywhere smooth if $m \neq 0$ and everywhere
except $p=0$ if $m=0$).
For massive fields $\boldsymbol{\psi}$,  we can take the ordinary Schwartz spaces for $\mathscr{E},E$, in order to achieve 
\[
\kappa'_{0,1},\kappa'_{1,0} \in \mathscr{L}\big(E^*, \mathscr{E}^* \big) \subset \mathscr{L}\big(E, \mathscr{E} \big),
\,\,\,
\boldsymbol{\psi} \in  \mathscr{L}\big(\mathscr{E} ,  \mathscr{L}( \, (E), (E) ) \big).
\]
The need for the modification of the space-time test space $\mathscr{E}$, when passing to massless fields,
may seem unexpected for those readers which compare it with the construction of massless fields in the sense of Wightman,
which allows the ordinary Schwartz test space also for the massless fields.
We nonetheless choose the white nose construction of free fields as much more adequate mathematical interpretation
of the (free) quantum field. Although, even for the massless fields, we are forced to extend the space-time test space
$\mathscr{E}$ to the whole Schwartz space, keeping $E=\mathcal{S}^{0}$, which results to  
\[
\kappa'_{0,1},\kappa'_{1,0} \in  \mathscr{L}\big(E, \mathscr{E} \big),
\,\,\,
A \in  \mathscr{L}\big(\mathscr{E} ,  \mathscr{L}( \, (E), (E)^* ) \big).
\]
This is because for the causal construction of the $S(g)$, also with interaction containing massless fields, we are using
the intensity of interaction function $g$ coming from the nuclear test space reach enough to contain elements with compact support, and
we need to develope the theory of free fields and their Wick products with space-time test functions with the same flexibility.
Otherwise the causality principle could have not been applied.

Among other things the white noise construction provides a much deeper insight
into the Wick product construction of free fields at the same space-time point, which moreover fits well with the needs 
of the causal perturbative approach. ''Wick product'' construction due to Wightman and G{\aa}rding
(although also rigorous) is not very much useful for the realistic causal perturbative QFT, such as QED.
Again that the Wightman-G{\aa}rding ''Wick product'' is not useful in practical computations
such as the causal perturbative approach, or in construction of conserved currents corresponding to the Noether
theorem (which in fact is the basis for the Canonical Quantization Postulate) has been recognized by Segal
\cite{Segal-NFWP.I}, a prominent analyst
who devoted much part of his research to the mathematical analysis of the Wick product construction.

Thus, we give white noise construction of the free fields $\boldsymbol{\psi}$ and $A$, with the explicit
construction of the representation of $T_4 \circledS SL(2, \mathbb{C})$, compare Sections \ref{e+e-},
\ref{electron}, \ref{positron}, \ref{free-gamma}, \ref{white-noise-proofs}. This construction can be applied
to any other kind of free fields, e.g. all free fields underlying systems of Yang-Mills fields minimally coupled
to charged fields, including the systems arrived at after passing to the broken phase.
As to the author's knowledge it has not been done before.

In fact the white noise construction of the free fields is not a new idea and goes back to Berezin.
It was further developed mainly by Hida and his school.

Having given the free fields, $\boldsymbol{\psi}$ and $A$, constructed as (finite sums of) integral kernel
operators with vector-valued kernels, we show that the operations of differentiation, Wick product
at the same space-time point, integration of the Wick product and its convolution with tempered distribution
are well-defined within the class of integral kernel operators to which the free fields and Wick product
belongs (Subsection \ref{OperationsOnXi}). Next we give the formulation and proof of the Wick theorem
for product in terms of the white noise integral kernel operators in Subsection \ref{WickForProduct}. 
Existence of the product is nontrivial for monomials containing massless factors, as they are more singular
than the massive fields, and it is not obvious that the product operation for such Wick monomials is well-defined 
as a finite sum of integral kernel operators of the white noise calculus. 
We give a proof of the existence of the product for a wide class of integral kernel operators 
of the white noise calculus.
Having the Wick product, and the product opartion for a wide class of 
generalized operators, including normal product polynomials of free fields with tempered translation-invariant
distributions as coefficents, we prove in Subsection \ref{WickForProduct} that the Bogoliubov causality
axioms for the $S$ operator are well-defined, with the higher order contributions regarded as finite
sums of integral kernel operators of the white noise calculus, and define the $S$ operator
up to the ambiguity coming from the ambiguity which is inherent in the splitting of causal distributions
into retarded and advanced parts -- which reflects the ordinary renormalizaton freedom. 
In Section \ref{WickForChronological} we are giving the formulation and proof of the Wick theorem for the 
chronological product in terms of the white noise calculus and the splitting of distributions
into the retarded and advanced parts. Breaking slightly the natural order of the lecture, 
in Subsection \ref{OperationsOnXiIF} we give a proof of the theorem that 
higher-order contributions to interacting fields in the adiabatic limit $g=1$ are well-defined as finite sums 
of generalized operators if and only if the normalization is ``on mass shell'' 
and the mass of the charged field is nonzero.

In particular the formulas for each $n$-th order contributions, with the intensity of the interaction function
$g=1$, are equal to finite sums
\[
\begin{split}
\boldsymbol{\psi}_{{}_{\textrm{int}}}^{(n)}(g=1, x) = \sum \limits_{l,m} \Xi_{l,m}\big(\kappa_{l,m}(x)\big), \\
A_{{}_{\textrm{int}}}^{(n)}(g=1, x) = \sum \limits_{l,m} \Xi_{l,m}\big(\kappa'_{l,m}(x)\big), \\
\end{split}
\]
of integral kernel operators (similarly we have for $\Xi_{l,m}\big(\kappa'_{l,m}(x)\big)$)
\begin{multline*}
\Xi_{l,m}\big(\kappa_{l,m}(x)\big) = \\
\sum \limits_{s_1, \ldots, s_{l+m}} \int \limits_{\mathbb{R}^{3(l+m)}}
\kappa_{l,m}(s_1, \boldsymbol{\p}_{1}, \ldots, s_{l+m}, \boldsymbol{\p}_{l+m}; x) \,
a_{s_{1}}(\boldsymbol{\p}_{1})^{+} \cdots a_{s_{l+m}}(\boldsymbol{\p}_{l+m}) \,
\ud^3 \boldsymbol{\p}_{1} \cdots \ud^3 \boldsymbol{\p}_{l+m},
\end{multline*}
where $a_s(\boldsymbol{\p})^{+}, a_{s}(\boldsymbol{\p})$ are the creation and annihilation
operators, constructed here as Hida operators in the tensor product of the Fock spaces of the
free fields $\boldsymbol{\psi}, A$, in the normal order, with the first $l$ factors
equal to the creation operators and the last $m$ equal to the annihilation operators. Here
\[
\begin{split}
\kappa_{l,m} \in \mathscr{L}\big(E^{\otimes (l+m)}, \mathscr{E}_{1}^{*} \big),
\,\,\,\, \kappa'_{l,m} \in \mathscr{L}\big(E^{\otimes (l+m)}, \mathscr{E}_{1}^{*} \big), \\
\mathscr{E}_{1} = \mathcal{S}(\mathbb{R}^4; \mathbb{C}^4) 
\end{split}
\]
with each factor $E$ in the tensor product $E^{\otimes(l+m)}$ equal
\[
E= \mathcal{S}(\mathbb{R}^3; \mathbb{C}^4)\,\,\, \textrm{or} \,\,\,
E= \mathcal{S}^{0}(\mathbb{R}^3; \mathbb{C}^4).
\]
Each of the operators $\Xi_{l,m}\big(\kappa_{l,m}(x)\big)$, $\Xi_{l,m}\big(\kappa'_{l,m}(x)\big)$ determines
a well-defined integral kernel operator
\[
\begin{split}
\Xi_{l,m}\big(\kappa_{l,m}\big)
\in \mathscr{L}\big((\boldsymbol{E}) \otimes \mathscr{E}_{1} , \, (\boldsymbol{E})^*\big)
\cong \mathscr{L}\big(\mathscr{E}_{1}, \mathscr{L}((\boldsymbol{E}), (\boldsymbol{E})^*) \big), \\
\Xi_{l,m}\big(\kappa'_{l,m}\big)
\in \mathscr{L}\big((\boldsymbol{E}) \otimes \mathscr{E}_{1} , \, (\boldsymbol{E})^*\big)
\cong \mathscr{L}\big(\mathscr{E}_{1}, \mathscr{L}((\boldsymbol{E}), (\boldsymbol{E})^*) \big), 
\end{split}
\]
with vector-valued distribution kernel $\kappa_{l,m}$ , respectively, $\kappa'_{l,m}$, in the sense
of Obata \cite{obataJFA}, where $(\boldsymbol{E})$ is the nuclear Hida subspace in the tensor product of the Fock spaces
of the fields $\boldsymbol{\psi}$ and $A$. The integral kernel operators
$\Xi_{l,m}\big(\kappa_{l,m}(x)\big)$, $\Xi_{l,m}\big(\kappa'_{l,m}(x)\big)$
are uniquely determined by the condition
\[
\begin{split}
\big\langle \big\langle \Xi_{l,m}(\kappa_{l,m})(\Phi \otimes \phi), \Psi \big \rangle \big \rangle
= \langle \kappa_{l,m}(\eta_{\Phi, \Psi}), \phi \rangle,
\,\,\,
\Phi, \Psi \in (\boldsymbol{E}), \phi \in \mathscr{E}_{1}, \\
\big\langle \big\langle \Xi_{l,m}(\kappa'_{l,m})(\Phi \otimes \phi), \Psi \big \rangle \big \rangle
= \langle \kappa'_{l,m}(\eta_{\Phi, \Psi}), \phi \rangle,
\,\,\,
\Phi, \Psi \in (\boldsymbol{E}), \phi \in \mathscr{E}_{1}, 
\end{split}
\]
where
\[
\eta_{\Phi, \Psi}(s_1, \boldsymbol{\p}_{1}, \ldots, s_{l+m}, \boldsymbol{\p}_{l+m}) =
\Big\langle \Big\langle a_{s_{1}}(\boldsymbol{\p}_{1})^{+} \cdots a_{s_{l+m}}(\boldsymbol{\p}_{l+m}) \,
\Phi, \, \Psi \Big\rangle \Big\rangle.
\]
Note that
\[
\eta_{\Phi, \Psi} \in E^{\otimes (l+m)}, \,\,\, \Phi, \Psi \in (\boldsymbol{E}).
\]
with the canonical pairing $\langle\langle \cdot, \cdot \rangle\rangle$ on $(\boldsymbol{E})^* \times (\boldsymbol{E})$.
This results is contained as a particular case of the Theorem \ref{g=1InteractingFieldsQED} of Subsection \ref{OperationsOnXi},
compare also Section \ref{A(1)psi(1)}.

Moreover, the interacting fields, in the adiabatic limit $g=1$, can be understood as Fock expansions
\[
\begin{split}
\boldsymbol{\psi}_{{}_{\textrm{int}}}(g=1) = \sum \limits_{l,m} \Xi_{l,m}\big(\kappa_{l,m}\big), \\
A_{{}_{\textrm{int}}}(g=1) = \sum \limits_{l,m} \Xi_{l,m}\big(\kappa'_{l,m}\big),
\end{split}
\]
into integral kernel operators in the sense of \cite{obataJFA} with all terms $\Xi_{l,m}\big(\kappa_{l,m}(x)\big)$,
$\Xi_{l,m}\big(\kappa'_{l,m}(x)\big)$ equal to integral kernel operators with vector-valued kernels, and all
belonging to the class indicated above. Even more, most of the terms
$\Xi_{l,m}\big(\kappa_{l,m}(x)\big)$, $\Xi_{l,m}\big(\kappa'_{l,m}(x)\big)$ behave even much more ''smoothly''
(although it is not necessary for the theory to work) and belong to
\[
\begin{split}
\Xi_{l,m}\big(\kappa_{l,m}\big)
\in \mathscr{L}\big((\boldsymbol{E}) \otimes \mathscr{E}_{1} , \, (\boldsymbol{E})\big)
\cong \mathscr{L}\big(\mathscr{E}_{1}, \mathscr{L}((\boldsymbol{E}), (\boldsymbol{E})) \big), \\
\Xi_{l,m}\big(\kappa'_{l,m}\big)
\in \mathscr{L}\big((\boldsymbol{E}) \otimes \mathscr{E}_{2} , \, (\boldsymbol{E})\big)
\cong \mathscr{L}\big(\mathscr{E}_{2}, \mathscr{L}((\boldsymbol{E}), (\boldsymbol{E})) \big),
\end{split}
\]
with
\[
\mathscr{E}_{2} = \mathcal{S}^{00}(\mathbb{R}^4; \mathbb{C}^4).
\]
In particular the first order contribution $A_{{}_{\textrm{int}}}^{\mu \,(1)}(g=1)$, given by
\begin{equation}\label{1-ord-A-g=1}
A_{{}_{\textrm{int}}}^{\mu \,(1)}(g=1,x) = -e \, D_{0}^{{}^{\textrm{av}}} \ast {:}\boldsymbol{\psi}^\sharp \gamma_\nu \boldsymbol{\psi}{:} =
-\frac{e}{4 \pi} \int \ud^3 \boldsymbol{x_{1}}
{\textstyle\frac{{:} \boldsymbol{\psi}^+\gamma^0 \gamma^\mu \boldsymbol{\psi} {:} (x_0 - |\boldsymbol{x_1} - \boldsymbol{x}|, \boldsymbol{x_1})}
{|\boldsymbol{x_1} - \boldsymbol{x}|}}
\end{equation}
to the interacting potential field, belongs to
\[
\mathscr{L}\big((\boldsymbol{E}) \otimes \mathscr{E}_{2} , \, (\boldsymbol{E})\big)
\cong \mathscr{L}\big(\mathscr{E}_{2}, \mathscr{L}((\boldsymbol{E}), (\boldsymbol{E})) \big),
\]
if we restrict $\mathscr{E}_{2}$ to $\mathcal{S}^{00}$, and to 
\[
\mathscr{L}\big((\boldsymbol{E}) \otimes \mathscr{E}_{2} , \, (\boldsymbol{E})\big)
\cong \mathscr{L}\big(\mathscr{E}_{2}, \mathscr{L}((\boldsymbol{E}), (\boldsymbol{E})^*) \big),
\]
if we take the whole Schwartz space for $\mathscr{E}_{2}$.
 
Application of the white noise Hida operators, and more generally of the white noise analysis (or infinite dimensional
distribution theory) seems also to be more economical than the conventional method from the purely computational point of view. In particular
having given the kernels $\kappa_{0,1}, \kappa_{1,0}, \kappa'_{0,1}, \kappa'_{1,0}, \ldots$ defining the free
fields of the QFT theory, or the kernels of the lower order contributions $S_k$, $k<n$, 
we can obtain the kernels of the higher order contribution 
of $S_n$ to the scattering operator
by applying very simple operations to the plane wave kernels $\kappa_{0,1}, \kappa_{1,0}, \kappa'_{0,1}, \kappa'_{1,0}, \ldots$
defining the free fields or $S_k$, $k<n$. 
Namely, the operations are the following: pointwise multiplication
of the plane wave kernels, some of them can be taken in the same space-time point, and some spin-momentum variables in the multiplied kernels
can also be equal in the product, integration in the momentum variables and in space-time variables 
and finally symmetrization of the products in the Bose spin-momentum variables and anti-symmetrization
in the Fermi spin-momentum variables and perfoming dispersion integrals giving the retarded/advanced parts. 
Compare the simple recurrence rules for the higher order contributions to the scattering operator
in Theorem \ref{WickThmForChronological}, Subsection \ref{WickForChronological} (on the Minkowski space-time)
and Theorems \ref{WickThmForChronologicalOneMassLessEU} and \ref{WickThmForChronologicalEU}
of Subsection \ref{WickForChronologicalEU} (on the Einstein Universe). These rules seemms to be 
more effective than the more convectional rules obscured with the additional regularization.

A generalization of the interacting fields, with the higher order contributions being the generalized integral kernel operators
with vector-valued kernels, which belong to
\[
\mathscr{L}\big((\boldsymbol{E}) \otimes \mathscr{E} , \, (\boldsymbol{E})^*\big)
\cong \mathscr{L}\big(\mathscr{E}, \mathscr{L}((\boldsymbol{E}), (\boldsymbol{E})^*) \big),
\]
but not to
\[
\mathscr{L}\big((\boldsymbol{E}) \otimes \mathscr{E} , \, (\boldsymbol{E})\big)
\cong \mathscr{L}\big(\mathscr{E}, \mathscr{L}((\boldsymbol{E}), (\boldsymbol{E})) \big),
\]
$\mathscr{E} = \mathcal{S}(\mathbb{R}^4)$, cannot be excluded merely on the ground that the elements of
\[
\mathscr{L}((\boldsymbol{E}), (\boldsymbol{E})^*)
\]
do not operate within the Fock space, and cannot be composed as operators within any
algebra of operators acting within the Fock space. Indeed, 
similar situation we have already for the Wick product ${:}\mathbb{A}_1 \ldots \mathbb{A}_k{:}$ 
of the free fields $\mathbb{A}_1, \ldots$, understood as generalized integral kernel operators with Hida operators
as the creation-annihilation operators,
where $W = {:}\mathbb{A}_1 \ldots \mathbb{A}_k{:}$ is a well-defined (finite sum) of integral kernel operators in
\[
\mathscr{L}\big((\boldsymbol{E}) \otimes \mathscr{E} , \, (\boldsymbol{E})^*\big)
\cong \mathscr{L}\big(\mathscr{E}, \mathscr{L}((\boldsymbol{E}), (\boldsymbol{E})^*) \big)
\]
but not in
\[
\mathscr{L}\big((\boldsymbol{E}) \otimes \mathscr{E} , \, (\boldsymbol{E})\big)
\cong \mathscr{L}\big(\mathscr{E}, \mathscr{L}((\boldsymbol{E}), (\boldsymbol{E})) \big),
\]
if there are massless field factors present in the Wick product. Nonetheless, 
the product 
\[
\Xi(\phi\otimes\varphi) = {:}\mathbb{A}_1 \ldots \mathbb{A}_n{:}(\phi) \,\, {:}\mathbb{A}_1 \ldots \mathbb{A}_k{:}(\varphi), 
\,\,\, \phi \in \mathscr{E} \otimes \mathscr{E}
\]
can still be naturally and uniquely defined as a limit, and gives a well-defined finite sum  
\[
\Xi(\phi\otimes\varphi) = \int 
{:}\mathbb{A}_1(x) \ldots \mathbb{A}_n(x){:} \,\,
{:}\mathbb{A}_1(y) \ldots \mathbb{A}_n(y){:} \,\,
 \phi\otimes \varphi(x, y)
\, \ud^4 x \ud^4 y
\]
of integral kernel operators, compare Subsection \ref{WickForProduct}. Moreover, the class of integral kernel
operators, which admits the operation of product, and which is contained in
\[
\mathscr{L}\big((\boldsymbol{E}) \otimes \mathscr{E}^{\otimes \, k} , \, (\boldsymbol{E})^*\big)
\cong \mathscr{L}\big(\mathscr{E}^{\otimes \, k}, \mathscr{L}((\boldsymbol{E}), (\boldsymbol{E})^*) \big)
\]
but not in
\[
\mathscr{L}\big((\boldsymbol{E}) \otimes \mathscr{E}^{\otimes \, k} , \, (\boldsymbol{E})\big)
\cong \mathscr{L}\big(\mathscr{E}^{\otimes \, k}, \mathscr{L}((\boldsymbol{E}), (\boldsymbol{E})) \big),
\]
includes all operators of the form
\begin{equation}\label{ProductClassGenOpIntro}
t(x_1, \ldots, x_n) \, {:}W_1(x_1)W_2(x_2) \ldots W(x_k){:}, \,\,\, t \in  \mathscr{E}^{* \otimes \, k},
\end{equation}
with translationally invariant $t$, and $W_i$ being Wick products of massless or massive free fields.
For the proof, compare Subsection \ref{WickForProduct}. 
The class, which allows extension of the product operation, can still be enlarged onto convergent Fock expansions
into operators of the class (\ref{ProductClassGenOpIntro}), although (\ref{ProductClassGenOpIntro})
is pretty sufficient for causal QFT with the interaction Lagrangian $\mathcal{L}(x)$ being a Wick polynomial in free
fields, and is sufficient for all known realistic interactions, like in spinor, scalar, $\ldots$, QED's.  
In particular the higher order contributions to interacting fields lie within the class which admints the 
product operation, so that, in principle at least, the Bohr-Rosenfeld gedanken measurement experiment 
can be applied to them, \cite{BohrRosenfeld}, \cite{CompagnoPerisco1}, although in general they belong to
\[
\mathscr{L}\big((\boldsymbol{E}) \otimes \mathscr{E} , \, (\boldsymbol{E})^*\big)
\cong \mathscr{L}\big(\mathscr{E}, \mathscr{L}((\boldsymbol{E}), (\boldsymbol{E})^*) \big)
\]
with $\mathscr{E} = \mathcal{S}(\mathbb{R}^4)$, but not to
\[
\mathscr{L}\big((\boldsymbol{E}) \otimes \mathscr{E} , \, (\boldsymbol{E})\big)
\cong \mathscr{L}\big(\mathscr{E}, \mathscr{L}((\boldsymbol{E}), (\boldsymbol{E})) \big).
\]
Necessity of the extension of QED on the Minkowski space-time, which should include operators transforming 
the Fock space into a wider space, has been pointed out by Dirac in his $15^{\textrm{th}}$ Lindau Lecture, \cite{DiracLindau}. 
It seems remarkable,
that we can construct such extension naturally, just putting the Hida operators for the creation-annihilation
operators for free fields, leaving all the rest of the causal perturbative theory completely unchanged 
and so adding no arbitrary assumptions.

But one point should, however, be emphasized here, that within this generalization
(using Hida operators with the generalized interacting fields existing in the adiabatic limit $g=1$) the bound
state problem cannot be directly treated on the flat Minkowski space-time, as well as the Noether integrals corresponding to interacting fields
cease to be ordinary operators operating within the Fock space (on the Minkowski space-time), compare Subsection \ref{Green}.
We recover also this part of the theory, with well-defined self-adjoint Noether integrals corresponding to interacting fields, 
but only on curved globally hyperbolic space-times with compact Cauchy surfaces.

Already at the level of extending the Noether theorem to the realm of free
quantum fields on the flat Minkowski space-time we encounter expressions like (\ref{1-ord-A-g=1}).
This problem lies at the level of free fields, their Wick products and integrals 
of Wick polynomials of free field operators. We solve the problem of extension of Noether theorem
in this work in Subsection \ref{BSH} for the electromagnetic potential field $A$ and in Subsection
\ref{StandardDiracPsiField} for the Dirac field $\boldsymbol{\psi}$.
In other words this is the problem of rigorous formulation and proof of the
\emph{Bogoliubov-Shirkov Quantization Postulate}.
It is formulated by Bogolubov and Shirkov in \S 9.4 of 1980 Edition of their book \cite{Bogoliubov_Shirkov}
(or in \S 9.2 of the first Russian 1957 Edition)
for free fields including gauge zero-mass fields in the following form: \emph{the operators for the energy-momentum four-vector
(translation generators) $\boldsymbol{P}^\mu$ (in the Fock space of the free fields), the operators of the angular momentum tensor $\boldsymbol{M}$, the charge $\boldsymbol{Q}$, and so on, which are the generators of infinitesimal
transformations of state vectors can be expressed in terms of the quantum free field
(generalized) operators by the same relations as in classical field theory with the (generalized) field operators
arranged in the appropriate order (Wick order)}. One can think of the Quantization Postulate as of a
generalization of the first Noether theorem to the level of free quantum fields. This problem lies among the problems which were unsolved and are concerned with the existence of integrals of local conserved currents corresponding to conserved
symmetries. In the case of zero mass gauge fields, any endeavor of proving the existence of integrals,
expressed in terms of spatial integrals of Wick ordered free fields and their eventual equality to the generator
(ordinary densely and presumably self-adjoint operator on a suitably constructed dense domain)
of the corresponding one-parameter subgroup have permanently been accompanied by infrared divergences,
compare e.g. \cite{Requardt}, \cite{Maison-Reeh-1}, \cite{Maison-Reeh-2},
\cite{Maison}. The particular case of the Postulate concerned with the free electromagnetic
potential free field and
the space-time translation generators $\boldsymbol{P}^\mu$
may thus be formulated in the form of the following equality
\begin{equation}\label{BS-QP}
\boxed{\int \boldsymbol{{:}} T^{0\mu} \boldsymbol{{:}} \, \ud^3 \boldsymbol{\x} = \boldsymbol{P}^\mu = d\Gamma(P^\mu),}
\end{equation}
where under the integral sign there is the (Wick ordered) expression for the $0-\mu$ components
of the energy-momentum tensor formally identical with the classical expression for the energy-momentum
components of the classical electromagnetic field. On the right-hand side we have the generators
$\boldsymbol{P}^\mu = d\Gamma(P^\mu)$ of space-time translations of the Krein-isometric representation coinciding
with the amplification $\Gamma(U^{*-1})$ of the (conjugation) of the {\L}opusza\'nski representation
$U$ to the Krein-Fock space $(\Gamma(\mathcal{H}), \Gamma(\mathfrak{J}))$ of the free
electromagnetic four-potential field\footnote{In our conventions it is the conjugation of the {\L}opusza\'nski representation
and its second quantized amplification which acts in the Fock space of the free electromagnetic four-potential field.}. The crucial difficulty lies in the fact that on the left-hand side we have operator-valued distributions (and not merely unbounded operators),
and their integrals over the spatial coordinates, exactly as in the expression (\ref{1-ord-A-g=1}).
Particularly hard difficulties arise in proving (\ref{BS-QP}) for massless gauge free fields (in fact the problem stayed open in this case, compare e.g. \cite{Requardt}). Segal \cite{Segal-NFWP.I} was not satisfied at all with the analysis of equal time integrals
of Wick products of free fields in each case: massless and massive, and in particular
pointed out that the treatment of similar problems undertaken by Glimm and Jaffe was not
satisfactory.

However, the problem may be solved if the fields are constructed as
with the help of white noise calculus, as particular examples of integral kernel operators.
In this case the integral kernel calculus of Hida-Obata-Sait\^o
for integral kernel operators may be applied to give the result (\ref{BS-QP}): the left-hand side is well-defined continuous operator $(E)\rightarrow (E)^*$, which have an extension to a continuous operator $(E) \rightarrow (E)$, equal on $(E)$ to the right-hand side,
thus to a densely defined operator on the Fock space. Then the standard Riesz-Sz\"okefalvy-Nagy criterion
and the invariance of $(E)$ under translations and unitarity of translations gives the
essential self-adjointness of the operators in (\ref{BS-QP}) on the nuclear space $(E)$ (although the full proof of (\ref{BS-QP}) 
is long and nontrivial, and is provided in Subsection \ref{BSH}).
But similarly assertion that (\ref{1-ord-A-g=1}) is well-defined continuous operator
$(\boldsymbol{E}) \rightarrow (\boldsymbol{E})^*$ requires a considerable amount of technicalities 
which are essentially the same as in the proof of Bogoliubov-Shirkov postulate (\ref{BS-QP})).

\subsection{Ultraviolet and infrared asymptotics of the interacting field}\label{IntAspatialInfty}

Before going on with the problem (b) and even with the problem (a) for fields which contain zero mass gauge fields
(before the interaction is plugged in) there is one nontrivial problem we are confronted with already at the
free field level not encountered when working with non gauge fields. In case of non gauge fields, when the
representation $U$ acting in the single particle subspace $\mathcal{H}$, and thus its amplification
$\Gamma(U)$ in $\Gamma(\mathcal{H})$ is unitary, the free field is essentially uniquely, i.e. up to unitary equivalence, determined by its general properties: i.e. by the transformation rule pertinent to the concrete representation $U$
which already includes the ''generalized charges'' pertinent to the field, for example the allowed spin of the
single particle states, e.t.c.. We should expect of the correctly constructed gauge quantum free fields
that they are likewise essentially uniquely determined by the corresponding ''generalized charged'' structure
pertinent to the field. But in case of gauge zero mass fields, such as the electromagnetic four-vector field,
the representation $U$ (or $U^{*-1}$) and its amplification $\Gamma(U)$ (or $\Gamma(U^{*-1})$) is unbounded
and Krein-isometric. The natural equivalence for such representations is the existence of Krein-isometric mapping
transforming bi-uniquely and continuously the nuclear space $E$ (resp. $(E)$) into itself, and thus by the Banach inverse mapping theorem having the continuous inverse on $E$ (resp. $(E)$), and which intertwines
the representations. Now this equivalence is weaker
in comparison to the case of unitary equivalence of non-gauge fields where the continuous
Hilbert space isometry defining the equivalence, and which is continuous on the respective nuclear space,
can be extended to a bounded operator -- in fact even to a unitary operator. This is the problem we are confronted
with already at the free field level. One consequence of this weaker equivalence is the following.
One can construct two equivalent realizations, say standard and nonstandard, of the local electromagnetic four potential free field based on the common nuclear spaces
$E = \mathcal{S}^{0}(\mathbb{R}^3; \mathbb{C}^4)$ and $(E)$ (regarded as functions on the orbit, i.e. on the positive energy cone without the apex, with the spatial components of the momentum as the natural coordinates on the cone without the apex) in the single particle spaces and in the Fock spaces respectively, which have different
infrared content. Let us formulate this assertion more precisely. The different representatives of free fields of the same equivalence class are constructed by using different inner products and fundamental symmetry operators on $E$ continuous with respect to nuclear topology on $E$, which after completion with respect to the respective inner products give the respective single particle Hilbert spaces of the respective representatives of the field (we give explicit examples in Subsect. \ref{equivalentA-s}). In general different representatives of the same equivalence class of the free field may be constructed in this way. The single particle representations
$U$ in case of the two representatives of the free electromagnetic potential field differ substantially.
In the first case $U$, when restricted to the $SL(2, \mathbb{C})$
subgroup, can be written as a direct integral
(with respect to Hilbert space inner product of the single particle Hilbert space of the corresponding representative of the field)
of representations (in general non-unitary) acting naturally on the functions of
fixed corresponding homogeneity on the cone, and in the second case of the restriction of $U$ to $SL(2, \mathbb{C})$ corresponding to the other representative of the free field no such direct integral decomposition is possible. For a proof compare Subsection \ref{equivalentA-s}. This possibility is no surprise
as the (unbounded) equivalence operator of representations whose representors of Lorentz hyperbolic rotations
are unbounded does not force any bounded equivalence for the action of Lorentz representors.

It is of less importance in
causal perturbative approach to QFT, becaise the pairinig functions in both realizations are the same.

Nonetheless, sensitivity of the infrared asymptotic behavior to the particular choice within one equivalence class 
of the free field cannot be simply ignored. This is because different asymptotic
behavior corresponding to different concrete realizations of the free field within the same equivalence class
survive in the IR asymptotic. The electromagnetic field has nontrivial infrared content corresponding to the Coulomb
interaction, so that its asymptotic behavior may (and in fact does) reflect important physical properties, which are different
for different realizations, and which cannot be ignored. Moreover, it cannot \emph{a priori} be excluded (and even it should be expected)
that this asymptotic behavior is important in fixing the correct choice among different realizations
of the free field within one and the same equivalence class. 
   
In order to solve this problem we recall that there exists a simple and elegant theory
of the quantized homogeneous of degree $-1$ part of the electromagnetic potential field $A$, which
resides at spatial infinity, i.e. at the three-dimensional one-sheet hyperboloid, say the three-dimensional de Sitter space-time, compare \cite{Staruszkiewicz1987} -- \cite{Staruszkiewicz2009}.
At the classical level extraction of the electromagnetic field which resides at spatial infinity is in principle
unique and well-defined, and it is the homogeneous of degree $-1$''part'' of the field $A$
which is free, \cite{GervaisZwanziger}, determined by a scalar $S(x)$ (of ''electric type'')
on de Sitter 3-hyperboloid fulfilling the homogeneous wave equation on de Sitter 3-hyperboloid.
As shown in \cite{Staruszkiewicz1987} or \cite{Staruszkiewicz} its quantization can be performed within a natural
way with the commutation relations based essentially on the two principles: the gauge invariance and the canonical
commutation relations for the conjugated generalized coordinates, \cite{Staruszkiewicz1987} or
\cite{Staruszkiewicz}. As shown in \cite{Staruszkiewicz} the phase of the wave function (of the charge carrying
particle, before the second quantization is performed) is the generalized coordinate conjugated to the total
charge, and at the classical level the phase has been determined in \cite{Staruszkiewicz} as equal to the electric part
$S(x)= -e x^\mu A_\mu(x)$ of the field at infinity, with $A_\mu$ homogeneous of degree $-1$ (in general
distributional solution of d'Alembert equation). The crucial point is that in computing the total
charge we do not need the global solution of the Maxwell equations but need only to know the solution outside the light cone
e.g. knowing the Dirac homogeneous solution of d'Alembert equation (distributional), \cite{Dirac3rdEd} pp.
303-304, which coincides with the ordinary Coulomb potential field outside the light cone is pretty sufficient. In particular the corresponding
field induced on de Sitter 3-hyperboloid by the Dirac homogeneous solution corresponds to the classical Coulomb field
and is determined by the homogeneous of degree $-1$ Coulomb field solution of Maxwell equations at spatial infinity. Therefore,
the standard commutation rules between the phase and the total charge (so identified at spatial infinity
with the respective constants in the general scalar solution of the wave equation on de Sitter 3-hyperboloid)
determine uniquely the commutation rules for the scalar field on de Sitter 3-hyperboloid and include the Coulomb field,
\cite{Staruszkiewicz}. In particular, it contains the total electric charge as an operator acting in the Hilbert space of the quantum phase field $S(x)$, as a scalar field on de Sitter 3-hyperboloid, and explains discrete character of the charge. This theory is remarkable for several reasons. First it is very simple and mathematically transparent. The paper \cite{Staruszkiewicz} does not enter mathematical analysis of the theory, but the harmonic analysis on
$SL(2, \mathbb{C})$ and the theory of continuous functionals on $\mathcal{S}^{00}(\mathbb{R}^4)$ and $\mathcal{S}^{0}(\mathbb{R}^4)$, respectively over space-time and in momentum space, provide the mathematical background for \cite{Staruszkiewicz}, compare Section \ref{infra}.    

In particular the homogeneous of degree $-1$ Dirac's solution of d'Alembert equation is a well-defined distribution over the test space 
$\mathcal{S}^{00}(\mathbb{R}^4)$ whose Fourier transform is a continuous functional 
over $\mathcal{S}^{0}(\mathbb{R}^4)$ with the light cone in the momentum space as the support,
for the proof compare Subsection \ref{DiracHom=-1Sol}. We show in particular that the
support of the Dirac's homogeneous of degree $-1$ solution,
as a distribution on $\mathcal{S}^{00}(\mathbb{R}^4)$, is equal to that part of space-time which lies
outside the light cone. Similar property we have for the transversal homogeneous of degree $-1$
electric type solutions
of d'Alembert equation generated by the Lorentz transforms of the Dirac solution.
These solutions extend over to the 
test function space $\mathcal{S}(\mathbb{R}^4)$. 
These extensions become unique and preserve d'Alembert equation and the support if we add the requirement of preservation
of homogeneity during the extension. This support property has very important physical consequences, compare Sect. \ref{infra}.

We also show (a detailed proof
can be found in Subsection \ref{globalU(1)}) that the standard representation of the commutation relations of Staruszkiewicz theory, proposed in \cite{Staruszkiewicz}, can be characterized (among the infinite family of other possible representations) by the condition that in each reference frame the gauge group $U(1)$ can be reconstructed spectrally in the sense of spectral geometry of Connes, by the phase and the charge operators
$V = e^{iS(u)}, D = (1/e) Q$ of his theory, compare Subsection \ref{globalU(1)}. For other possible non-standard representations of the commutation relations of Staruszkiewicz this would be impossible
with $V= e^{iS(u)}, D = (1/e) \, Q$.
The standard representation of \cite{Staruszkiewicz} is in fact the one which is actually used in
the subsequent papers \cite{Staruszkiewicz1987}--\cite{Staruszkiewicz2009}.
Second, theory presented in \cite{Staruszkiewicz}, \cite{Staruszkiewicz1992ERRATUM},  
involves the fine structure constant and relates it non trivially
to the theory of irreducible unitary representations of $SL(2, \mathbb{C})$, through the unitary representation
of $SL(2, \mathbb{C})$ acting in the Hilbert space of the quantum phase field $S(x)$, a mathematical theory which have attained 
a mature form full of computational devices thanks mainly to Gelfand and his school, Neumark, Harish-Chandra and others. 
This relation is remarkable because it depends on the value of the fine structure constant: only if its square is less than $\pi$
the fine structure enters through the eigenvalue of the discrete series component of the representation of 
$SL(2, \mathbb{C})$ acting in the Hilbert space of the phase $S(x)$. If the square of the fine structure is greater than
$\pi$, the discrete series component is absent, and the fine structure cannot be expressed through the generators of
the representation of $SL(2, \mathbb{C})$. Third, this theory
contains the quantized Coulomb field (at least as it concerns the asymptotic part outside the light cone).
This is perhaps the most remarkable feature of the theory of Staruszkiewicz \cite{Staruszkiewicz},
at least for the perturbative causal approach to QED. Indeed, so far as the gauge electromagnetic 
field was treated with insufficient care the existence of the adiabatic limit was unclear in QED and in particular the status of the Coulomb field
so that the identification of the quantum (interacting) field $A_\textrm{int}(x)$ at spatial infinity was impossible
within the causal perturbative approach due essentially to the troubles with the adiabatic limit.
But with the electromagnetic potential field treated more carefully as a genereralized operator in the sense of \cite{obataJFA},
 with higher order contributions as finite sums of generalized integral kernel operators in the sense of \cite{obataJFA}, 
we restore the adiabatic limit
and at least in principle we can compute $A_{\textrm{int}}(x)$ as a Fock expansion into generalized integral kernel operators
 in which the switching off coupling $g(x) \longrightarrow 1$, so that the interacting field is now a Fock expansion with the ordinary
fine structure constant (and not merely with the ``switching of'' function $g(x)$), with the first order term for $A_{{}_{\textrm{int}}}$ 
equal to an ordinary operator-valued distribution acting in the Fock space of free fields. 
This is of capital importance because now we can compare the homogeneous of degree
zero part of the field $x_\mu A_{\textrm{int}}^\mu(x)$ with the quantum phase field $S(x)$ of Staruszkiewicz theory.

For this plan to be realizable we have to learn how to extract a homogeneous part of a fixed homogeneity $\chi$,
of a quantum (interacting) field. Although this task is still non-trivial there
are several circumstances which both allow the computation to be effective and connect this computation
to important physical phenomena. This is again possible thanks to the fact that the higher order contributions
are equal to finite sums of integral kernel operators $\Xi(\kappa_{\mathpzc{l}\mathpzc{m}})$, which can be decomposed
into direct integrals
\[
\Xi(\kappa_{\mathpzc{l}\mathpzc{m}}) = \int \Xi_{{}_{\chi}}(\kappa_{\chi \, \mathpzc{l}\mathpzc{m}}) \, \ud \chi
\]
of (asymptotically) homogeneous components $\Xi_{{}_{\chi}}(\kappa_{\chi \, \mathpzc{l}\mathpzc{m}})$. This decomposition is canonically
determined by the decomposition of the representation of $SL(2, \mathbb{C})$ acting in the Fock space of free fields.
The kernels $\kappa_{\chi \, \mathpzc{l}\mathpzc{m}}$ of the decomposition are equal to the Fourier transforms $\mathcal{F}$
of the kernels $\kappa_{\mathpzc{l}\mathpzc{m}}$, here with the Fourier transform $\mathcal{F}$ being associated to the decomposition
of $SL(2, \mathbb{C})$ acting in the $(\mathpzc{l}+\mathpzc{m})$-particle Hilbert space.
Let us explain this in more details now, referring to the respective Subsections for the proofs.

Concerning this decomposition and extraction of the (asymptotically) homogeneous part,
of a given contribution to the interacting field, say $x_\mu A_{\textrm{int}}^\mu(x)$,
regarded as finite sum of integral kernel operators $\Xi(\kappa_{\mathpzc{l}\mathpzc{m}})$ we do it gradually.
We start with the free fields $\boldsymbol{\psi}, A, \ldots $, which have the general form of the sum of the two
integral kernel operators (negative and positive frequency components)
\[
\begin{split}
A = \boldsymbol{\psi}^{(-)} + \boldsymbol{\psi}^{(+)} =  \Xi(\kappa_{0,1}) + \Xi(\kappa_{1,0})
\\
\boldsymbol{\psi} = \boldsymbol{\psi}^{(-)} + \boldsymbol{\psi}^{(+)} =  \Xi(\kappa_{0,1}) + \Xi(\kappa_{1,0}),
\\
\ldots.
\end{split}
\]

The general principle is as follows. The action $\alpha\mapsto U(\alpha)$ of $SL(2, \mathbb{C})$ in the single particle Hilbert space
$\mathcal{H}'$ decomposes into the direct integral
\[
\mathcal{H}' = \int \mathcal{H}'_{{}_{\chi}} \ud \chi,
\,\,\,\,
U(\alpha) = \int U(\alpha)_{{}_{\chi}} \ud \chi U(\alpha)
\]
of indecomposable (or irreducible in the unitary case) representations $U(\alpha)_{{}_{\chi}}$ acting in the decomposition Hilbert spaces
$\mathcal{H}'_{{}_{\chi}}$. In case of mass less fields decomposition is determined by the
spectral decomposition of the scaling operator $S_\lambda \widetilde{\phi}(p) = \widetilde{\phi}(\lambda p)$ (in the momentum picture),
for the elements $\widetilde{\phi}$ of $\mathcal{H}'$ representable by the (scalar, four-vector, \emph{e.t.c}) functions
square integrable on the positive energy cone orbit $\mathscr{O}_{{}_{1,0,0,1}}$ in the momentum space.
Here we emphasize the critical difference between the standard and nonstandard realizations
of the free electromagnetic potential free field: for the nonstandard realization this decomposition cannot be constructed,
but it can be constructed for the standard realization, although the action of $SL(2, \mathbb{C})$
is not unitary (with the components indecomposable and even with some components irreducible). For the
proof, compare Subsections \ref{equivalentA-s} and \ref{AS}. The spectrum of this decomposition
is determined by the spectrum of the scaling operator $S_\lambda$, which is a normal operator, in case this decomposition
can be realized (as is the case for the standard free electromagnetic potential field). In case the action
$\alpha\mapsto U(\alpha)$ of $SL(2, \mathbb{C})$ is unitary decomposition is always possible, with the scaling
operator being normal and irreducible unitary components $U(\alpha)_{{}_{\chi}}$. In case of standard free massive fields
for which the representation $\alpha\mapsto U(\alpha)$ of $SL(2, \mathbb{C})$ is unitary, decomposition is always possible
and is determined by the spectral decomposition of the (representors of the) two Casimir operators of the
group $SL(2, \mathbb{C})$. The decomposition Hilbert spaces $\mathcal{H}'_{{}_{\chi}}$ are spanned by the generalized
eigenstates of the scaling operator (in the mass less case) or of the two Casimir operators (in the massive case). These
generalized eigenstates are (non-normalizable scalar, four-vector, bispinor, \emph{e.t.c}) functions on the respective orbit (positive energy
sheet $\mathscr{O}_{{}_{m,0,0,0}}$ in case of a massive field). The decomposition parameter $\chi$ can be interpreted as the
homogeneity degree of the generalized eigenstates of the scaling operator (in the mass less case) or as the
asymptotic homogeneity degree (for the momenta going to infinity) of the generalized eigenstates of the two Casimir operators
(in the massive case). The spectrum of the decomposition (\emph{i.e.} of the scaling or of the Casimir operators)
depends on the specific field, or on the specific representation $\alpha\mapsto U(\alpha)$ of $SL(2, \mathbb{C})$.
In case of the standard free electromagnetic potential field, the spectrum is equal $\chi=-1+i\nu$, $\nu \in \mathbb{R}$,
and coincides with the spectrum for the mass less scalar field. For the massive real scalar field
the spectrum (asymptotic homogeneity) is equal $\chi=-1+i\nu$, $\nu \in \mathbb{R}_+$. For the standard Dirac free
field the spectrum (asymptotic homogeneities) are equal $\chi=-3/2+i\nu$, $\nu \in \mathbb{R}$. For the proofs,
compare Subsections \ref{psichi}, \ref{equivalentA-s}. Decomposition of the action of $SL(2, \mathbb{C})$ in the single
particle riggedd Hilbert space $E\subset \mathcal{H}' \subset E^*$ (with the single particle nuclear space $E$
depending on the field in question) determines naturally the direct integral decomposition of the creation-annihilation
Hida operators $a(\xi)^+, a(\eta)$, $\xi,\eta \in E\subset \mathcal{H}'$
into direct integrals
\[
a(\xi) = \int a_{{}_{\chi}}\big(\xi_{{}_{\chi}}\big) \, \ud\chi,
\,\,\,
a(\xi)^+ = \int a_{{}_{\chi}}\big(\xi_{{}_{\chi}}\big)^+ \, \ud\chi,
\]
of creation annihilation operators $a_{{}_{\chi}}\big(\xi_{{}_{\chi}}\big)^+$, $a_{{}_{\chi}}\big(\xi_{{}_{\chi}}\big)$
in Fock spaces
\[
\Gamma(\mathcal{H}'_{{}_{\chi}})
\]
over the single particle Hilbert spaces $\mathcal{H}'_{{}_{\chi}}$, spanned by the generalized (asymptotically)
homogeneous eigenstates of the scaling or, respectively, Casimir operators. Here
\[
\xi = \int \xi_{{}_{\chi}} \, \ud\chi
\]
is the decomposition of $\xi \in E\subset \mathcal{H}'$. Decomposition of the creation and annihilation Hida operators in turn
determines (for each space-time test function $\phi$) decomposition of the corresponding free field operators
\[
\begin{split}
A(\phi) = \int A(\phi)_{{}_{\chi}} \, \ud\chi,
\\
\boldsymbol{\psi}(\phi) = \int \boldsymbol{\psi}(\phi)_{{}_{\chi}} \, \ud\chi,
\\
\ldots
\end{split}
\]
because $A(\phi)$, $\boldsymbol{\psi}(\phi) $ can be expressed through $a(\widetilde{\phi}|_{\mathscr{O}})$, 
$a(\widetilde{\phi}|_{\mathscr{O}})^+$, with $\widetilde{\phi}|_{\mathscr{O}}$
denoting restriction of the Fourier transform $\widetilde{\phi}$ of the test function $\phi$ 
to the respective orbit, and $\widetilde{\phi}|_{\mathscr{O}} \in E \subset \mathcal{H}'$.
It turns out that
\[
\phi \mapsto A(\phi)_{{}_{\chi}} , \,\, \phi \mapsto \boldsymbol{\psi}(\phi)_{{}_{\chi}} , \,\, \ldots
\]
are well-defined fields, and (discrete) integral kernel operators, over the same space-time test space
as the initial free field $A, \boldsymbol{\psi}, \ldots$, which we denote
by
\[
\phi \mapsto A_{{}_{\chi}}(\phi), \,\, \phi \mapsto \boldsymbol{\psi}_{{}_{\chi}}(\phi), \,\, \ldots
\]
For the proof compare Subsetions \ref{psichi},
\ref{equivalentA-s}, where the component fields and their kernels $A_{{}_{\chi}}(x), \boldsymbol{\psi}_{{}_{\chi}}(x)$
we have determined for the free standard Dirac field and for the free electromagnetic potential field.
Moreover, the component fields $A_{{}_{\chi}}, \boldsymbol{\psi}_{{}_{\chi}}$ are (asymptotically) homogenous
fields with the (asymptotic) homogeneity degree corresponding to the value of the decomposition parameter $\chi$.
In particular the standard free electromagnetic potential field decomposes into homogeneous parts $A_{{}_{\chi}}$
with the homogeneity degree equal $\chi=-1+i\nu$, $\nu \in \mathbb{R}$, and belonging to the spectrum of the scaling operator.
Analogously the free Dirac field decomposes into the asymptotically homogeneous parts $\boldsymbol{\psi}_{{}_{\chi}}$
of asymptotic homogeneity degree $-3/2+i\nu$ determined by the joint spectrum of the two Casimir operators of $SL(2,\mathbb{C})$
acting in the single particle space of the free Dirac field.
Thus, the (asymptotically) homogeneous parts of the free fields are the free fields which live in the Fock spaces
constructed over the single particle Hilbert spaces of the generalized (asymptotically) homogeneous eigenstates,
respectively, of the scaling or the Casimir operators.

Decompositions of the free fields (evaluated at the space-time test
functions $\phi$), determined by the decomposition of the action of $SL(2, \mathbb{C})$ in the single particle space, 
can be better  understood if we regard the free fields as sums of special examples of integral kernel operators
\[
 \Xi(\kappa_{0,1}) + \Xi(\kappa_{1,0})
\]
with vector valued kernels $\kappa_{1,0}$ in the sense of \cite{obataJFA}. In this case the decomposition 
can be written in the form
\[
\Xi(\kappa_{0,1}(\phi)) + \Xi(\kappa_{1,0}(\phi)) =
\int \big[ \Xi_{{}_{\chi}}(\kappa_{\chi \, 0,1}(\phi)) + \Xi_{{}_{\chi}}(\kappa_{\chi \, 1,0}(\phi)) \big] \, \ud\chi
\]
where $\kappa_{\chi \, 0,1}(\phi)$, $\kappa_{\chi \, 1,0}(\phi)$ are equal to the Fourier transforms
$\mathcal{F}[\kappa_{0,1}(\phi)](\chi)$, $\mathcal{F}[\kappa_{1,0}(\phi)](\chi)$
of the distribution kernels $\kappa_{0,1}(\phi)$, $\kappa_{1,0}(\phi)$ $\in E$. Here $\mathcal{F}$
is associated with the decomposition of the action of $SL(2, \mathbb{C})$ in the single particle space $E$. 
But in case of integral kernel operators decomposition is possible for a much wider class of integral
kernel operators $\Xi(\kappa_{\mathpzc{l}\mathpzc{m}})$, 
for a wide class of kernels $\kappa_{\mathpzc{l}\mathpzc{m}}(\phi) \in E^{* \otimes \, (\mathpzc{l}+\mathpzc{m})}$, with the 
kernels of the decomposition component operators equal to the Fourier transforms
$\mathcal{F}\big[\kappa_{\mathpzc{l}\mathpzc{m}}(\phi)\big](\chi, \ldots, \chi)$ restricted to the diagonal,
with $\mathcal{F}$ associated to the decomposition of the action of $SL(2, \mathbb{C})$ in 
$E^{\otimes \, (\mathpzc{l}+\mathpzc{m})}$, \emph{i.e.} with the decomposition
of the representation $U(\alpha)^{\otimes \, (\mathpzc{l}+\mathpzc{m})}$ acting in the $(\mathpzc{l}+\mathpzc{m})$-particle
space. 

Let us give here a more general definition of this decomposition.
Let $\Xi(\kappa_{\mathpzc{l}\mathpzc{m}})$ be a generalized integral kernel operator with vector-valued kernel (here in the Fock space of a
free field, but we can do also for the Fock space of several free fields with decomposable
representation of $SL(2, \mathbb{C})$), compare Subsection \ref{psichi}.
As we have mentioned, each Hida annihilation or creation operator,
evaluated at a fixed element $\xi$ of the nuclear space $E$
\[
a(\xi) = \int \overline{\xi(\boldsymbol{\p})} \, a(\boldsymbol{\p}) \ud \boldsymbol{\p},
\,\,\,\,\,
a^{+}(\xi) = a(\xi)^+ = \int \xi(\boldsymbol{\p}) \, a(\boldsymbol{\p})^+ \ud \boldsymbol{\p},
\]
has direct integral decomposition
\[
a(\xi) = \int a_{{}_{\chi}}\big((\xi)_{{}_{\chi}}\big) \, \ud \chi,
\,\,\,\,\,
a^{+}(\xi) = a(\xi)^+ = \int a_{{}_{\chi}}\big((\xi)_{{}_{\chi}}\big)^{+} \ud \chi.
\]
Correspondingly the product
\[
a(\xi_1)^{+} \ldots a(\xi_n)^{+} = \int a_{{}_{\chi}}(\xi_{1 \, \chi })^{+} \ldots a_{{}_{\chi}}(\xi_{n \, \chi})^{+} \, \ud \chi, \,\,\, \xi_i \in E,
\]
of the creation Hida operators has a well-defined direct integral decomposition. Therefore, each element
\[
\sum\limits_{n} {\textstyle\frac{1}{n!}} a^+(\xi_1) \ldots a^{+}(\xi_n)|0\rangle \in (E)
\]
of the Hida test space has a direct integral decomposition, determined by the decomposition of the representation of
$SL(2, \mathbb{C})$.

Let
\[
\Phi,\Psi \in (E),
\,\,\,\, \Phi = \int \Phi_{{}_{\chi}} \, \ud \chi,
\,\,\,\, \Psi = \int \Psi_{{}_{\chi}} \, \ud \chi
\]
be any two elements of the test Hida space with their direct integral decompositions. Let $\ud \chi$
be a $\sigma$-measure on the spectrum of the decomposition of the representation of $SL(2, \mathbb{C})$ acting
in the single particle Hilbert space $\mathcal{H}'$. We say that the generalized
integral kernel operator $\Xi(\kappa_{\mathpzc{l}\mathpzc{m}})$, eq. (\ref{electron-positron-photon-Xi}), is equal to the direct integral
\[
\Xi(\kappa_{\mathpzc{l}\mathpzc{m}}) = \int \Xi_{{}_{\chi}}(\kappa_{\chi \, \mathpzc{l}\mathpzc{m}}) \, \ud \chi
\]
of (discrete-) integral kernel operators $\Xi_{{}_{\chi}}(\kappa_{\chi \, \mathpzc{l}\mathpzc{m}})$,
acting in the Fock spaces over the single particle Gelfand triples
$E_{{}_{\chi}} \subset \mathcal{H}'_{{}_{\chi}} \subset E_{{}_{\chi}}^{*}$, if
\begin{multline*}
\int \big\langle\big\langle \Xi_{{}_{\chi}}\big(\kappa_{\chi \, \mathpzc{l}\mathpzc{m}}(\phi)\big)
\Phi_{{}_{\chi}}, \Psi_{{}_{\chi}} \big\rangle\big\rangle \, \ud \chi
= \int \Big\langle \kappa_{\chi \, \mathpzc{l}\mathpzc{m}}(\phi), \big(\eta_{{}_{\Phi,\Psi}}\big)_{{}_{\chi}} \Big\rangle \, \ud \chi
\\
=\langle \kappa_{\mathpzc{l}\mathpzc{m}}(\phi), \eta_{{}_{\Phi,\Psi}} \rangle
=\big\langle \big\langle \Xi(\kappa_{\mathpzc{l}\mathpzc{m}}(\phi))\Phi, \Psi \big\rangle \big\rangle
\end{multline*}
for all
\[
\Phi,\Psi \in (E), \phi \in \mathscr{E}.
\]

A wide class of integral kernel operators with vector valued kernels admits decomposition with the decomposition measure depending
solely on the kernels of the decomposed operator. This class includes: 1) the free fields with decomposable action
of $SL(2, \mathbb{C})$ (this is so for the standard free Dirac fields, and the standard electromagnetic field, massive fields
with unitary action of $SL(2, \mathbb{C})$), 2) the Wick polynomials of such free fields, listed in 1), 3) higher order contributions
to interacting fields. For the analysis of decomposable integral kernel operators, we refer to Subsection \ref{psichi}.

As the next step we observe the following behavior of the (asymptotic) homogeneity under the Wick product operation:
The asymptotic homogeneities of the decomposition components 
\[
\big({:} \boldsymbol{\psi} \boldsymbol{\phi}{:}\big)_{{}_{\chi}} 
\]
of the Wick product of two free fields $\boldsymbol{\psi}$, $\boldsymbol{\phi}$, are equal to the sums $\chi_1+\chi_2$ of the 
(asymptotic) homogeneities of the (asymptotically) homogeneous parts (components)  
$\boldsymbol{\psi}_{{}_{\chi_1}}$, $\boldsymbol{\phi}_{{}_{\chi_2}}$ of the fields
$\boldsymbol{\psi}$, $\boldsymbol{\phi}$, for the proof compare Subsection \ref{psichi}.

As the final step we would like to extract the (asymptotically) homogeneous parts of the interacting field, especially
$-x_\mu A_{{}_{\textrm{int}}}^{\mu}(x)$. On the other hand the interacting field itself is beyond our reach,
because, so far we have not\footnote{Although we have given a precise meaning to the limit of the perturbative series,
as the Fock expansion into integral kernel operators in the sense of \cite{obataJFA}.}
yet investigated its convergence. In order to pass over this problem we go back to the
causal perturbative series
for the interacting field $A_{{}_{\textrm{int}}}(x)$ after the adiabatic switching on the interaction
at infinity is performed.
Then we confine attention to each order term separately.
Each higher order therm of the causal perturbation series,
equal to a finite sum of integral kernel operators, we can decompose, using the above stated
decomposition and extract the corresponding (asymptotically) homogeneous parts.

For example in order to compute the first order correction
to the homogeneous part $\big(A_{{}_{\textrm{int}}}^{\mu}(x) \big)_{{}_{\chi = -1}}$
of homogeneity $\chi = -1$ of the interacting field
$A_{{}_{\textrm{int}}}^{\mu}(x)$, in case when the standard free potential field is used,
we compute first the kernels of the first order contribution (Section \ref{A(1)psi(1)})
and represent it in the form of a finite sum of integral kernel operators,
Section \ref{A(1)psi(1)}, Example 1. Next we apply to it the decomposition of an integral kernel
operator using the method given in Subsection \ref{psichi}.
This will give us all (asymptotically) homogeneous components of the first order contribution.
A simple rescaling limit operation will give as the components which are homogeneous
(and not only asymptotically homogeneous), compare Subsection \ref{A(1)chi}.

Here we should note that the decomposition components of the decomposition of representation of $SL(2, \mathbb{C})$
acting in the single particle space of the standard free electromagnetic potential
field, and
determined by the spectral decomposition of scaling operator, are not indecomposable, and can be further decomposed
into the orthogonal direct sum of representations acting, respectively, on the ''electric type'' and ''magnetic type''
homogeneous of degree $\chi=-1+i\nu$ states. Correspondingly each homogeneous of degree
$\chi=-1+i\nu$ component $A_{{}_{\chi}}$ of the free electromagnetic potential field can be further decomposed into the sum
\[
A_{{}_{\chi}} = A_{{}_{\chi}}^{e} + A_{{}_{\chi}}^{\mathfrak{m}}
\]
of, say, the ``electric'' and ``magnetic'' component, acting in the
Fock spaces, respectively, over the ''electric type'' and ''magnetic type'' homogeneous of degree
$\chi=-1+i\nu$ states. For the construction of this decomposition into the ''electric'' and ''magnetic''
parts, compare Subsection \ref{AS}. 

We should emphasize that the above analysis of the homogeneous parts of the interacting e.m. potential
field is possible only if the higher order contributions to the interacting field in the adabatic limit $g=1$
are equal to finite sums of well-defined integral kernel operators with vector-valued kernels in the sense of \cite{obataJFA}.
This is the case in the causal perturbative QED's with the Hida operators as the creation-annihiliation operators
if and only if the charged fields are massive. We have proved it in
Subsection \ref{OperationsOnXi}, Theorems \ref{ExistenceIntFields.g=1.m>0} and \ref{NonExistenceIntFields.g=1.m=0}
on the Mnkowski space-time (compare Thm. \ref{InteractingFieldsAtxOnEU}
and its Corollary, Subection \ref{CausalSonEU} which is its analogue 
on the Einstein Universe).

In Section \ref{infra} we compare the asymptotic (infrared) homogeneous of degree zero and ``electric type'' part
\begin{multline}\label{xA=S}
-e x_\mu\big(A_{{}_{\textrm{int}}}^{e \, \mu}(x)\big)_{{}_{\chi = -1}}
\\
= -e x_\mu \big(A_{{}_{\textrm{int}}}^{\mu \,(1)}(g=1,x)\big)_{{}_{\chi=-1}}
-e x_\mu\big(A^{e \, \mu}(x)\big)_{{}_{\chi = -1}} = S(x), \,\,\,\, x\cdot x <0
\end{multline}
of the field $-ex_\mu A_{{}_{\textrm{int}}}^{\mu}(x)$ with the phase field $S(x)$
of Staruszkiewicz theory.

If the charged fields coupled to the potential are massive, then in extracting the asymptotically (for small momenta, or
at infinity in spacetime, or infrared) homogeneous of degree $-1$ part of the interacting
field $A_{{}_{\textrm{int}}}(x)$, then we expect that only the first and zero order contributions are non-zero:
\[
\big(A_{{}_{\textrm{int}}}^{\mu}(x)\big)_{{}_{\chi = -1}} \,\,\,\,\,\,
= \,\,\,\,\,\, \big(A_{{}_{\textrm{free}}}^{\mu}(x)\big)_{{}_{\chi = -1}} \,\,\,\,\,\, + \,\,\,\,\,\,
\big(A_{{}_{\textrm{int}}}^{(1) \, \mu}(g=1,x)\big)_{{}_{\chi=-1}},
\]
where $\big(A_{{}_{\textrm{int}}}^{\mu \,(1)}(g=1,x)\big)_{{}_{\chi=-1}}$
is the homogeneous of degree $-1$ part of the generalized operator (\ref{1-ord-A-g=1}) defined
as above. This is of capital importance. It means that the asymptotic (infrared) homogeneities 
of the decomposition components $\Xi_{{}_{\chi}}(\kappa_{{}_{\chi} \, \mathpzc{l}\mathpzc{m}})$
of the higher order contributions to the interacting potential, regarded as integral kernel operators $\Xi(\kappa_{\mathpzc{l}\mathpzc{m}})$
with vector valued kernels $\kappa_{\mathpzc{l}\mathpzc{m}}$, have real parts not equal to $-1$. In case of the infrared
asymptotics it is not easily seen, because the (infrared) asymptotic homogeneity degree of the decomposition components
$\Xi_{{}_{\chi}}(\kappa_{{}_{\chi} \, 0,1})+\Xi_{{}_{\chi}}(\kappa_{{}_{\chi} \, 0,1})$
of the massive free fields is not well-defined, \emph{i.e.} it is either zero or infinity
(recall that it is the ultraviolet asymptotic homogeneity degree, \emph{i.e.} for large momenta, which is
invariantly defined for the components $\Xi_{{}_{\chi}}(\kappa_{{}_{\chi} \, 0,1})+\Xi_{{}_{\chi}}(\kappa_{{}_{\chi} \, 0,1})$
of the massive free fields $\Xi(\kappa_{0,1}) + \Xi(\kappa_{0,1})$). Similarly, the infrared asymptotic homogeneity
degree of the massive pairing functions (contrary to the massless paring functions $D_{0}^{(\pm)}$) and
of the vacuum polarization, the self-energy and vacuum distributions $\Pi^\textrm{av}$, $\Pi^\textrm{ret}$, 
$\Sigma^\textrm{ret}$, $\Sigma^\textrm{av}$, $\Upsilon^\textrm{av}$,$\Upsilon^\textrm{ret}$
is not well-defined (for the spinor QED with massive charged field). In the case when the charged field is massive we have not 
any simple automatic rule
telling us if a given Wick product (containing massive free fields) have a nonzero contribution to a part 
of given asymptotic (infrared) homogeneity $\chi$ or not.
Thus, in order to compute the infrared asympotics, we have to compute explicitly the kernels $\kappa_{\mathpzc{l}\mathpzc{m}}$ of the
higher order contributions $\Xi(\kappa_{\mathpzc{l}\mathpzc{m}})$. Next we compute their decomposition components
$\Xi_{{}_{\chi}}(\kappa_{{}_{\chi} \, \mathpzc{l}\mathpzc{m}})$, and finally compute the infrared asymptotic homogenous part
of the distribution kernel $\kappa_{{}_{\chi} \, \mathpzc{l}\mathpzc{m}}$ (using the ordinary Gervais-Zwanziger rescaling).

The case of massless charged fields coupled to the potential seems to be non-realistic, 
as we know from the experimental evidence, and in accordance to the 
Theorems \ref{ExistenceIntFields.g=1.m>0} and \ref{NonExistenceIntFields.g=1.m=0} which we have 
proved in Subsection \ref{OperationsOnXi} on the Minkowski space-time (compare Thm. \ref{InteractingFieldsAtxOnEU}
and its Corollary, Subection \ref{CausalSonEU} which is its analogue 
on the Einstein Universe).

Analysis which higher order contribution gives nonzero part of fixed ultraviolet asymptotic
homogeneity of the interacting potential is simpler, although in this case the higher order contributions
play nontrivial role. In this case decomposition components
$\Xi_{{}_{\chi}}(\kappa_{{}_{\chi} \, 0,1})+\Xi_{{}_{\chi}}(\kappa_{{}_{\chi} \, 0,1})$ of the free (massive or massless) fields
$\Xi(\kappa_{0,1}) + \Xi(\kappa_{0,1})$ have well-defined ultraviolet homogeneity degree $\chi$.
The allowed (asymptotic ultraviolet) homogeneities $\chi$ for the massive fields
coupled to $A_\mu$ are restricted to the spectrum of the decomposition of $SL(2,\mathbb{C})$,
acting in the single particle space of the free field coupled to the potential.
In particular the allowed asymptotic homogeneities
$\chi$ for the scalar massive field are equal: $\chi = -1 +i\nu$, $\nu \in \mathbb{R}$,
or $\chi = -3/2 +i\nu$, $\nu \in \mathbb{R}$, for the standard Dirac field,
for the proof compare Subsection \ref{psichi}.
Similar situation we have for other massive fields.
The homogeneities for the homogeneous parts of the free field $A_\mu$
belong to the spectrum $\chi = -1+i\nu$, $\nu \in \mathbb{R}$,
of the scaling operator (Subsection \ref{equivalentA-s}).
The (asymptotic ultraviolet) homogeneities of the Wick products of several fields,
are equal to the sums of the (asymptotic ultraviolet) homogeneities of the respective fields.
Similarly, the pairing functions of the (massive and massless) fields, as well as the
vacuum polarization distribution $\Pi$, the self-energy distribution $\Sigma$ and the vacuum distribution
$\Upsilon$,
have well-defined ultraviolet asymptotics with well-defined ultraviolet homogeneity degree $-4-\omega$
(with the so-called singularity degree equal $\omega$).
Thus, each factor coming from the retarded (resp. advanced) parts of the pairing of the free spinor field 
with its Dirac conjugation, of the pairing of the free e.m. potential field with itself (commutator) or from the 
distribution functions
$\Sigma^\textrm{ret}$, $\Sigma^\textrm{av}$, $\Pi^\textrm{av}$, $\Pi^\textrm{ret}$, 
$\Upsilon^\textrm{av}$,$\Upsilon^\textrm{ret}$,
 contributes additional (asymptotic ultraviolet) homogeneity $-1,-2$ or $-3,-3$,$-6,-6$,$-8,-8$, 
respectively. This is the case for spinor QED
with the distriutions $\Pi^\textrm{av}$, $\Pi^\textrm{ret}$, 
$\Sigma^\textrm{ret}$, $\Sigma^\textrm{av}$, $\Upsilon^\textrm{av}$, $\Upsilon^\textrm{ret}$ 
given by the formulas (\ref{Piav}), (\ref{Sigmaret}) and (\ref{Upsilon})
of Subsection \ref{OperationsOnXi}. But we have 
analogous situation for any other QED, where we have always finite number of distributions, 
including  retarded (resp. advanced) parts of the pairing functions, 
whose retarded and advanced parts determine (with the corresponding 
Wick products of free fields) the higher order terms, compare Subsection \ref{WickForChronological}. Thus, similar analysis
can be performed also for other QED or, more generally, for several charged fields coupled to the potential.

Namely, the higher order contributions to interacting e.m. potential $A_{{}_{\textrm{int}} \, \mu}(g=1)$
 which have asymptotic ultraviolet homogeneity degree
$-1$ are of odd $n = 2k+1$-order (with $n$ convolutions $\ast$, summation is performed
with respect to the repeated Lorentz indices)
\begin{equation}\label{higherorder.-1}
D_{0}^{{}^{\textrm{av}}} \ast \Pi^{{}^{\textrm{av}} \, \mu_{k-1}}_{\mu} \ast \ldots \ast D_{0}^{{}^{\textrm{av}}} \ast \Pi^{{}^{\textrm{av}} \, \nu}_{\mu_1} \ast
D_{0}^{{}^{\textrm{av}}}  \ast {:}\boldsymbol{\psi}^\sharp \gamma_\nu \boldsymbol{\psi}{:},
\end{equation}
(containing $k$ loop vacuum polarization graph $\Pi$ insertions). In particular for $k=0,1,2, \ldots$
\begin{gather*}
e \, D_{0}^{{}^{\textrm{av}}} \ast {:}\boldsymbol{\psi}^\sharp \gamma_\nu \boldsymbol{\psi}{:},
\\
e^3 \, D_{0}^{{}^{\textrm{av}}} \ast \Pi^{{}^{\textrm{av}} \, \nu}_{\mu} \ast 
D_{0}^{{}^{\textrm{av}}}  \ast {:}\boldsymbol{\psi}^\sharp \gamma_\nu \boldsymbol{\psi}{:},
\\
e^5 \,  D_{0}^{{}^{\textrm{av}}} \ast \Pi^{{}^{\textrm{av}} \, \mu_1}_{\mu} \ast
D_{0}^{{}^{\textrm{av}}}  \ast \Pi^{{}^{\textrm{av}} \, \nu}_{\mu_1} \ast D_{0}^{{}^{\textrm{av}}}  
\ast {:}\boldsymbol{\psi}^\sharp \gamma_\nu \boldsymbol{\psi}{:}
\\ 
\ldots \,\,\,\,.
\end{gather*}

That, in case of massive charged fields coupled to the potential, the higher order corrections do not contribute to the asymptotically (infrared)
homogeneous  of degree $-1$ part of the interacting potential is what one should expect, by comparison with the scattering at the classical level in the infrared regime: the scattered massive charges produce infrared electromagnetic field but the infrared electromagnetic field does not scatter massive charges\footnote{That only first order contribution to
the interacting field at spatial infinity should survive also at the quantum field theory level has been foreseen by Schwinger, as prof. Staruszkiewicz has kindly informed me. Schwinger observes that the only charge carrier fields are massive. 
We have proved this theorem theoretically, compare
Theorems \ref{ExistenceIntFields.g=1.m>0} and \ref{NonExistenceIntFields.g=1.m=0}, Subsection \ref{OperationsOnXi} on the Minkowski space-time 
(and Thm. \ref{InteractingFieldsAtxOnEU} and its Corollary, Subection \ref{CausalSonEU} on the Einstein Universe). 
The infrared photons carry to small an energy to produce pairs sufficient to create massive charge carrying particle. On the other hand we should expect the first order contribution to be nonzero. That there
persist a kind of ''back-reaction'' we should expect by comparison with the ordinary non-relativistic charged quantum particle in the infrared Bremsstrahlung-type infrared field: a nonzero phase shift will persist for each plane wave of the particle which produces nontrivial change of the packet-type wave function of the particle,
compare e.g. \cite{Staruszkiewicz1981}. This may have some reflection in the fact that the homogeneous of degree $-1$ part of the interacting potential
and the contributions to the various homogeneous parts of the interacting charge carrying field are not independent. In particular, the homogeneous of degree
$-1$ part of the free potential contributes to various homogeneous parts of the charge carrying interacting field.}.

Moreover,
$\big(A_{{}_{\textrm{int}}}^{\mu \,(1)}(g=1,x)\big)_{{}_{\chi=-1}}$ and $\big(A_{{}_{\textrm{free}}}^{e \, \mu}(x)\big)_{{}_{\chi = -1}}$ by construction commute. 
Moreover, also the higher order contributions (\ref{higherorder.-1}), whose ultraviolet asymptotic homogeneity
degree is equal $-1$, commute with $A_{{}_{\textrm{free}}}^{e \, \mu}(x)$, so by construction asymptotic of ultraviolet homogenity $-1$
parts of (\ref{higherorder.-1}) commute with $\big(A_{{}_{\textrm{free}}}^{e \, \mu}(x)\big)_{{}_{\chi = -1}}$. 
This again is of capital importance and makes (\ref{xA=S}) still more plausible in case of massive charged fields with 
$-ex_\mu \big(A_{{}_{\textrm{free}}}^{e \, \mu}(x)\big)_{{}_{\chi = -1}}$ corresponding to 
\begin{equation}\label{clmStar}
\sum \limits_{l=1}^{\infty} \sum \limits_{m = -l}^{m= +l}
\{ c_{lm} f_{lm}^{(+)}(x) + \textrm{h.c.} \}
\end{equation}
and $-ex_\mu \big(A_{{}_{\textrm{int}}}^{\mu \,(1)}(g=1,x)\big)_{{}_{\chi=-1}}$ corresponding to 
$-e Q x_0/r$ in the expansion of the quantum phase operator
\[
S(x) = S_0 -e Q x_0/r + \sum \limits_{l=1}^{\infty} \sum \limits_{m = -l}^{m= +l}
\{ c_{lm} f_{lm}^{(+)}(x) + \textrm{h.c.} \}, \,\,\,\, x\cdot x <0 
\]
of the Staruszkiewicz theory (we are using the notation of  \cite{Staruszkiewicz}). The operator $S_0$
in $S(x)$ is that part which cannot be reproduced by 
$-ex_\mu \big(A_{{}_{\textrm{int}}}^{\mu}(x)\big)_{{}_{\chi = -1}}$, which again could have been foreseen
by comparison with the classical theory of infrared fields.

The computation of the (asymptotically ultraviolet) homogeneous of degree $-3$ part of the free current field
(in case of the Dirac field coupled to the potential the free current is equal
${:}\boldsymbol{\psi}^{+}\gamma^0 \gamma^\mu \boldsymbol{\psi}{:}(x)$) is 
given in Subsection \ref{psichi}, Example 4.
In the simpler scalar QED, it can be computed similarly. Only the asymptotically (ultraviolet/infrared) part of the current field 
contributes to the (asymptotically ultraviolet/infrared) homogeneous of degree $-1$ part
of the first order term (\ref{1-ord-A-g=1}).

The Hilbert space of the quantum phase $S(x)$ of
Staruszkiewicz theory has the following structure
\begin{equation}\label{HofS}
\mathcal{H} = \mathscr{H}_0 \otimes \mathcal{H}_0
\end{equation}
(compare Subsection \ref{infra-electric-transversal-generalized-states} and \ref{globalU(1)}).
Here $\mathscr{H}_0$ is the closed subspace of the Hilbert space $\mathcal{H}$ spanned by
\[
e^{im S_0}|0\rangle, m \in \mathbb{Z}.
\]
Note that the direct summand with fixed $m$ spanned by
\[
\Big(e^{imS_0}|0\rangle \Big) \otimes \mathcal{H}_0
\]
in (\ref{HofS}) is the eigenspace of the total charge operator $Q$ corresponding to the eigenvalue $m$.
The direct summand $\mathbb{C} \otimes \mathcal{H}_0 = \mathcal{H}_0$
is the eigenspace corresponding to the eigenvalue zero of $Q$. The Hilbert space $\mathcal{H}_0$
is equal to the Fock space $\mathcal{H}_0 = \Gamma(\mathcal{H}_{0}^{1})$ over
the single particle space $\mathcal{H}_{0}^{1}$ of ''infrared transversal photons'' spanned
by
\[
c_{lm}^{+} |0\rangle.
\]
The representation of $SL(2, \mathbb{C})$ acts on $\mathcal{H}_{0}^{1}$ through the
Gelfand-Minlos-Shapiro irreducible
representation $(l_0=1, l_1 = 0)$ of the principal series and through its amplification
$\Gamma(l_0=1, l_1=0)$ on $\mathcal{H}_0 = \Gamma(\mathcal{H}_{0}^{1})$, and trivially
on the factor $\mathbb{C}$ in (\ref{HofS}), \cite{Staruszkiewicz1995},
\cite{wawrzycki-HilbertOfPhase}.
The factorization property (\ref{HofS}) is preserved (compare Subsection \ref{globalU(1)}) under the representation $U$
of $SL(2, \mathbb{C})$ acting in $\mathcal{H}$:
\begin{multline*}
U \mathcal{H} = \big(U\mathscr{H}_0 U^{-1} \big) \otimes \big(U \mathcal{H}_0 U^{-1}\big) \\
= {\mathscr{H}'}_{0} \otimes \big(\Gamma(l_0=1, l_1=0) \mathcal{H}_0 \Gamma(l_0=1, l_1=0)^{-1}\big)
= {\mathscr{H}'}_{0} \otimes \mathcal{H}_0.
\end{multline*}
But under the action of $U$ only the second factor in (\ref{HofS}) is invariant under $U$
where $U$ acts through $\Gamma(l_0=1, l_1=0)$, as said above.
The first factor in (\ref{HofS}) is transformed under $U$ into another subspace
${\mathscr{H}'}_{0} \subset \mathcal{H}$ spanned by
\[
U e^{im S_0} U^{-1}|0\rangle, m \in \mathbb{Z}.
\] 

Finally, to the tensor product factorization (\ref{HofS}) of the Hilbert space of the phase field $S(x)$
there correspond the tensor product
factorization ${\mathcal{H}'}_1 \otimes {\mathcal{H}'}_0$ of the Hilbert space of the operator
\[
x_\mu\big(A_{{}_{\textrm{int}}}^{\mu}(x)\big)_{{}_{\chi = -1}} \,\,\,\,\,\,
= \,\,\,\,\,\, x_\mu\big(A_{{}_{\textrm{free}}}^{e \, \mu}(x)\big)_{{}_{\chi = -1}} \,\,\,\,\,\, + \,\,\,\,\,\,
x_\mu\big(A_{{}_{\textrm{int}}}^{\mu \,(1)}(g=1,x)\big)_{{}_{\chi=-1}},
\]
where ${\mathcal{H}'}_0$ is the Fock Hilbert space of the field
$x_\mu\big(A_{{}_{\textrm{free}}}^{e \, \mu}(x)\big)_{{}_{\chi = -1}}$
and ${\mathcal{H}'}_1$ is the Hilbert space of the field
$x_\mu\big(A_{{}_{\textrm{int}}}^{\mu \,(1)}(g=1,x)\big)_{{}_{\chi=-1}}$,
by construction equal to a non-invariant (depending on the actually used Lorentz frame) subspace of the Fock space of a homogeneous of degree $-3$ part
of the current field (in the scalar case equal $: \overline{\boldsymbol{\psi}} \overset{\leftrightarrow}{\partial}{}^\mu \boldsymbol{\psi} : (x)$).
It follows (Subsections \ref{infra-electric-transversal-generalized-states}, \ref{globalU(1)}) that the operators (\ref{clmStar}) and $-e Q x_0/r$
on the one hand factorize with respect to the factorization (\ref{HofS}); and on the other hand the operators
$x_\mu\big(A_{{}_{\textrm{free}}}^{e \, \mu}(x)\big)_{{}_{\chi = -1}}$ and
$x_\mu\big(A_{{}_{\textrm{int}}}^{\mu \,(1)}(g=1,x)\big)_{{}_{\chi=-1}}$ factorize with respect to the factorization
${\mathcal{H}'}_1 \otimes {\mathcal{H}'}_0$. (Note here that ${\mathcal{H}'}_1 \otimes {\mathcal{H}'}_0$, as whole, should be Lorentz
invariant, although the factorization itself is expected to be not Lorentz invariant.)
The Fock space ${\mathcal{H}'}_0$ of the field $x_\mu \big(A_{{}_{\textrm{free}}}^{e \, \mu}(x)\big)_{{}_{\chi = -1}}$
can be naturally identified with the Hilbert space $\mathcal{H}_0$
and its action on this space can be naturally identified with the action of the operator
(\ref{clmStar}) on $\mathcal{H}_0$, for the proof compare
Subsection \ref{AS}, \ref{infra-electric-transversal-generalized-states} and \ref{globalU(1)}.
Both $x_\mu \big(A_{{}_{\textrm{free}}}^{e \, \mu}(x)\big)_{{}_{\chi = -1}}$
and (\ref{clmStar}) act as the unit operator on the respective first factors in
${\mathcal{H}'}_1 \otimes {\mathcal{H}'}_0$ and respectively
$\mathscr{H}_0 \otimes \mathcal{H}_0$. Similarly, $-e Q x_0/r$ and
$x_\mu\big(A_{{}_{\textrm{int}}}^{\mu \,(1)}(g=1,x)\big)_{{}_{\chi=-1}}$
act as the unit operator on the respective second factor, for the proof compare
Subsections \ref{infra-electric-transversal-generalized-states} and \ref{globalU(1)}.
Thus, indeed the operators
$-e x_\mu\big(A_{{}_{\textrm{free}}}^{e \, \mu}(x)\big)_{{}_{\chi = -1}}$ and (\ref{clmStar})
understood as operators in the respective Hilbert spaces ${\mathcal{H}'}_1 \otimes {\mathcal{H}'}_0$
and $\mathscr{H}_0 \otimes \mathcal{H}_0$ can be equated, up to a trivial multiplicity: 
\[
-e x_\mu\big(A_{{}_{\textrm{free}}}^{e \, \mu}(x)\big)_{{}_{\chi = -1}} =
\sum \limits_{l=1}^{\infty} \sum \limits_{m = -l}^{m= +l}
\{ c_{lm} f_{lm}^{(+)}(x) + \textrm{h.c.} \}, \,\,\,\,\, x\cdot x <0.
\]
This in particular means that the equality (equivalence)
of the operators $-e Q x_0/r$ and $x_\mu\big(A_{{}_{\textrm{int}}}^{\mu \,(1)}(g=1,x)\big)_{{}_{\chi=-1}}$
in their action on the respective spherically symmetric states ($e^{imS_0}|0\rangle$ on the r.h.s. and corresponding
to them states on the l.h.s.) in the respective first factors would give us the full equality (equivalence)
(\ref{xA=S}) except for the term $S_0$ on the right-hand side of (\ref{xA=S}).
The comparison
\begin{equation}\label{HomogeneousA(1)chi=x0/rQ}
-e x_\mu\big(A_{{}_{\textrm{int}}}^{\mu \,(1)}(g=1,x)\big)_{{}_{\chi=-1}} = -e x_0/r Q, \,\,\,\,\, x\cdot x <0
\end{equation}
we have given in Subsection \ref{A(1)chi}. However, we have not given the complete proof of the last equality (\ref{HomogeneousA(1)chi=x0/rQ})
for general QED, or the full construction of the spherically symmetric states in the Fock space of the left-hand side
operator, corresponding to the states $e^{imS_0}|0\rangle$ of the right-hand side operator, on which the equality
(\ref{HomogeneousA(1)chi=x0/rQ}) is preserved.
Only in the simplest case of the scalar QED we have led the investigation of the last equality (\ref{HomogeneousA(1)chi=x0/rQ})
and the corresponding spherically symmetric states, to an end. 

Although the operator $S_0$ cannot be constructed as a contribution to first order electric homogeneous of degree
zero part of the operator $x_\mu A_{{}_{\textrm{int}}}^{\mu}$, it nonetheless will have to be constructed implicitly
in the Fock space of $-e x_\mu\big(A_{{}_{\textrm{int}}}^{\mu \,(1)}(g=1,x)\big)_{{}_{\chi=-1}}$, 
through the construction of the spherically symmetric states corresponding
to the sates $e^{imS_0}|0\rangle$ in the Hilbert space of the phase field.

Perhaps the most important reason for the comparison of the homogeneous of degree $-1$
part of the interacting potential field with the phase of Staruszkiewicz theory
lies in giving a proof for the universality of the unit of charge.
Namely, in completing the construction of the spherically invariant states in the domain of the operator
$-e x_\mu \big(A_{{}_{\textrm{int}}}^{\mu \,(1)}(g=1,x)\big)_{{}_{\chi=-1}}$ on which indeed it coincides with
$-e Q x_0/r $ on the corresponding spherically symmetric states $e^{imS_0}|0\rangle$ for various massive charged fields (say, scalar, spinor, e.t.c.)
coupled to the potential, with the coupling compatible with gauge invariance, we will identify the coupling constant
and the charge with the respective constant of the Staruszkiewicz theory. More precisely:
if the equality of $-ex_\mu \big(A_{{}_{\textrm{int}}}^{\mu \,(1)}(g=1,x)\big)_{{}_{\chi=-1}}$ to the part
$-e Q x_0/r$ of phase $S(x)$
of Staruszkiewicz theory indeed holds on the spherically symmetric
states of the first operator, which correspond to the states
$e^{imS_0}|0\rangle$ of the second operator, then in the Hilbert space
of the field $x_\mu \big(A_{{}_{\textrm{int}}}^{\mu \,(1)}(g=1,x)\big)_{{}_{\chi=-1}}$
there must exist the operator $e^{iS_0}$
which together with the operator $1/e \, Q$ provides a spectral realization of the global gauge group
$U(1)$. This follows from the fact that this is the case for Strauszkiewicz theory.
Various contributions to $-e Q x_0/r$ coming from various charge carrying fields coupled to the potential
$A$ should give the total charge operator $Q$ which together with the corresponding phase
provides a spectral construction of the global gauge group, as in the case of the Staruszkiewicz theory,
in which $V= e^{iS(u)}, D = 1/e \, Q$ (or $V=e^{iS_0}, 1/e \, Q$) define spectrally the
gauge $U(1)$ group, compare Subsection \ref{globalU(1)}).
This will give us the universality of the unit of charge
because the various contributions to the global charge $Q$ coming from the various charge carrying fields
all should have common spectrum $e \mathbb{Z}$. Otherwise, the total charge operator could not serve as
the Dirac operator for the $U(1)$ manifold, as the contributions coming from various charge carrying fields
would destroy the spectrum $e \mathbb{Z}$ needed for the spectral reconstruction of the global gauge $U(1)$
group, compare Section \ref{globalU(1)}. Thus, the common scale for the electric charge comes from the condition that the infrared fields of each isolated system (involving various charge carrying fields with the couplings to $A$ preserving gauge invariance) provide a spectral description (in their total Hilbert space of infrared states)
of the global gauge group $U(1)$ as in case of
Staruszkiewicz theory, compare Subsection \ref{globalU(1)}. This mechanism forcing universality of the scale of the electric charge still works even for non-standard representation of the commutation rules of the
Staruszkiewicz theory. The only difference would be in changing of the spectrum of the total charge
$Q$ from $e\mathbb{Z}$ into $ce\mathbb{Z}$ for some constant $c>1$, and in changing
$V= e^{iS(u)}, D = 1/e \, Q$ into $V= e^{icS(u)}, D = 1/e \, Q$ in the spectral construction
of $U(1)$, compare Subsection \ref{globalU(1)}
for the definition of the non-standard representation.

Therefore, in order to account for the universality of the electric charge we have to add to the causal perturbative QED axioms
(involving several charged fields coupled to the potential), the following principle: in the Fock space 
over the generalized homogeneous of degree $-1$ states of the potential and
(asymptotically) homogeneous states (of the respective homogeneities in the single particle Hilbert spaces of the fields coupled to the potential),
the operators $e^{iS_0}, 1/e \, Q$ should provide the spectral realization of the gauge group $U(1)$, in the sense explained 
in Subsection \ref{globalU(1)}). Otherwise, we could, in principle, put different electromagnetic couplings for different charged fields.

Our results can also be interpreted as giving an example of a concrete realization of the abstract theory of \cite{Staruszkiewicz}
in the tensor product Fock space over the (asymptotically) homogeneous sates of the (massive)
free charged fields, coupled to the potential, and of the homogeneous of degree $-1$ electric type states of the
free potential itself, in which the homogeneous of degree $-1$ electric type part of the interacting
potential field acts. The proof of universality, given above, should be understood as a proof
within a concrete realization of \cite{Staruszkiewicz} in the causal perturbative QED restricted to the generalized (asymptotically)
homogeneous states, supplemented with the assumption that for the total charge $Q$ in the Fock space over these generalized
(asymptotically) homogeneous states there exists the phase operator $e^{iS_0}$, such that $(e^{iS_0}, Q)$ provide the spectral realization
(up to uniform multiplicity) of the $U(1)$ group. A simpler proof, due to Staruszkiewicz,
of the charge universality can be given, if we use his abstract theory \cite{Staruszkiewicz}, understood as a quantum version of the classical
infrared electric type fields. As such, this theory determines the common spectrum of the total electric charge depending on single
common universal constant. But this universality argument can in fact be put in the close analogy with the universality of spin,
because the theory \cite{Staruszkiewicz} as such, is legitimate for the values of the fine structure constant
smaller than $1$, which is the case in reality, as the quantization of the
infrared classical fields, produced in Bremsstrahlung radiation of realistic particles
becomes legitimate in this regime (Berestecky-Lifshitz-Pitaevsky inequality, \cite{Staruszkiewicz1999}). For
such small (and realistic) value of the fine structure constant there must enter discretely (with Dirac spectral weight concentrated at
a single value of the continuous parameter, numbering the representations of the supplementary series)
the supplementary series component of the $SL(2, \mathbb{C})$, acting in the Hilbert space of the phase field $S(x)$, with the eigenvalue
of the first Casimir operator of $SL(2, \mathbb{C})$ depending on the fine structure constant and on the total charge
of the states pertinent to this supplementary component, compare \cite{Staruszkiewicz1992ERRATUM}
or Subsection \ref{Ustructure}.
Thus, (for sufficiently small values of the fine
structure constant and of the total charge) the total charge and the fine structure constant are related
to the eigenvalue of the first Casimir operator of the supplementary series component of the representation of $SL(2, \mathbb{C})$,
similarly as the value of the total angular momentum is determined by the eigenvalue of the Casimir
operator of $SU(2, \mathbb{C})$ acting in the Hilbert space of any system with spherical symmetry. Because dependence
of the eigenvalue of the first Casimir operator of $SL(2, \mathbb{C})$ of the discrete component of the decomposition,
and belonging to the supplementary series, on the total charge and on the fine structure constant is universal, 
then so is the scale of the electric charge.

\subsection{An algebra of operators in the Fock space closely related to the space-time}\label{CPPS}

Starting with the Fock space of free fields
(with no external classical fields, just for simplicity),
say underlying QED, but the method is general enough to include SM with the Higgs field, we construct
a commutative pre-$C^*$-algebra $\mathcal{A}$ of (bounded) operators in the Fock space.
The algebra
$\mathcal{A}$ (in fact we need a more specific conditions\footnote{It should be an algebra of operators commuting with the Gupta-Bleuler operator, commutative and involutive, with the involution represented by the Krein adjoint equal to the ordinary adjoint as the operators of the algebra commute with the Gupta-Bleuler operator, so let us suppose that it is a C$^*$-algebra or pre-C$^*$-algebra.})
moreover, fulfils the conditions:
\begin{enumerate}

\item[(1)]
$\mathcal{A}$ has the Gelfand spectrum $\Sp \mathcal{A}$ with a smooth finite dimensional manifold structure.

\item[(2)]
The manifold structure and other smooth structures on $\Sp \mathcal{A}$ can be defined as in
\cite{Connes_spectral} by operators acting on a subspace $\mathcal{H}_{{}_{\textrm{inv}}}$ of the Fock space of free fields, invariant for $\mathcal{A}$ and the other operators defining the manifold structure on
$\Sp \mathcal{A}$, which together with $\mathcal{A}, \mathcal{H}_{{}_{\textrm{inv}}}$ respect the conditions
of the spectral ''tuple'' of the manifold $\Sp \mathcal{A}$, as stated in \cite{Connes_spectral}.

\item[(3)]
The algebra $\mathcal{A}$ and the remaining operators of the ''tuple'' are canonically related to representors
of the space-time symmetry group acting in the Fock space, and to the free field operators, so that
each element of $\mathcal{A}$, as well as the remaining operators of the spectral ''tuple'', are canonically expressed in terms of free fields (or Wick polynomials of free fields or their integrals).

\end{enumerate}

The whole point lies in proving the existence of $\mathcal{A}$ and $\mathcal{H}_{{}_{\textrm{inv}}}$ or in constructing the algebra $\mathcal{A}$
together with the remaining operators on $\mathcal{H}_{{}_{\textrm{inv}}}$ defining the manifold structure on $\Sp \mathcal{A}$.
We give in this work explicit construction of $\mathcal{A}$ together with the remaining operators
which respect the conditions (1) -- (3). 

This construction has been proposed earlier \cite{Wawr}.
 
For the algebra $\mathcal{A}$ we should choose an algebra as simply and naturally related to the
Fock space and the free fields as possible. Moreover, the elements of $\Sp \mathcal{A}$ should have
natural physical meaning. In our previous paper \cite{Wawr} we noticed that the space-time
is naturally connected and enters naturally into the construction of free fields, with the spectral reconstruction reducible to the harmonic analysis associated to the decomposition of the representation of the double covering of the Poincar\'e group acting in the Fock space. Thus, space-time manifold
provides a good candidate to be investigated first as $\Sp \mathcal{A}$. 

Two main difficulties in the realization of the plan for the spectral construction of space-time
are the following. We have to find interesting subspace of the Fock space on which the translation
generators act with uniform (in general infinite) multiplicity. This difficulty we have solved by extending
the Mackey theory of inducted representations on Krein-isometric induced representations acting in the single particle states of the realistic fields (e.g. the local electromagnetic potential field).
The needed extension of Mackey theory is presented in Appendix \ref{PartIIMackey}.
We have already encountered necessity of such extension when constructing the zero mass gauge fields, but only here we need this extension with its full power, especially in constructing tensor product representations and their decomposition. The translation generators act indeed with uniform infinite multiplicity on the subspace of the Fock space which is orthogonal to the vacuum and to the single particle subspace.
But there is another difficulty, because we need a subspace of the Fock space on which the joint spectrum
of the translation generators is just the ordinary $\mathbb{R}^4$-manifold. This is however impossible
for the ordinary free fields (e.g. underlying QED), by the very general assumption of positivity of
energy. In particular the joint spectrum of the generators in the subspace orthogonal to the vacuum and to the single particle subspace, is indeed uniform, but is equal to the positive energy sheet of the cone together with its interior,
and cannot be equal to the whole $\mathbb{R}^4$-manifold (of course we exclude additional
unnatural manipulations, such as diffeomorphisms between the interior of the cone and the whole
$\mathbb{R}^4$, not preserving natural metric structure).

In order to solve this problem let us recall the following observation: in construction of the free QFT (underlying e.g. QED)
and its causal perturbation method of introducing interaction one may replace the ''positive energy axiom''
by the ''negative energy axiom'' with the consequently replaced signs in both the Lagrangians
of free fields and in the interaction Lagrangian, in the commutators and with replacement of advanced into retarded
functions and vice versa. That in this case the corresponding Wick product theorem and the perturbation
series may still be constructed on essentially the same grounds as in the ''positive energy theory'' has already been
noted e.g. by Bogoliubov and Shirkov. In particular that the supports of the distributions
in the commutators of free fields still allow in this case
the construction of the Wick product has been noted in \cite{Bogoliubov_Shirkov}, \S II.16.\footnote{Frequently
repeated claim that in QFT positivity of energy is a necessary condition is not strictly true. Taking into account
the causal perturbative method in the standard gauge field theory we may only say that ''definiteness of
energy sign axiom'' is needed in order to build the theory, but instead of ''positive energy'' one may
equally use the opposite sign version. In principle one can construct free fields with Fourier transforms of single particle states 
lying on both (negative and positive) energy sheets of the corresponding orbits (hyperboloids or cones). 
But construction of the scattering operator for such fields (with both energy signs) becomes meaningless, with the troubles e.g. in the Wick theorem. 
Of course from the QFT-point of view the difference between the two energy signs is rather of unimportant and nomenclatural character. Situation is different 
if the relation to gravity is taken into account, but gravity is ignored in QFT, especially in QFT on the flat Minkowski space-time.} We give explicit construction for each field with both energy signs on equal grounds. Let us denote the Fock
space of the underlying free fields with the ''positive energy axiom'' by $\mathcal{H}_{+}$ and respectively
$\mathcal{H}_{-}$ for the free theory with the ''negative energy axiom''. Now we treat the different sign
fields as independent fields which do not interact and
act in the tensor product space $\mathcal{H}_{+} \otimes \mathcal{H}_{-}$. Then we apply the perturbation
to each version (positive and negative) separately in order to obtain the perturbation series
for the composed system. At the end we
may restrict the allowed states to the positive energy states. From the physical point of view this
introduces nothing essentially knew as the positive energy fields do not interact with the negative ones
(even the negative energy fields may have to be chosen free), but for us this gives a considerable profit.
It comes from the fact that $\textrm{Spec} \, (P_1 , \ldots P_4)$ in the subspace orthogonal to the
vacuum and one particle states in $\mathcal{H}_{+} \otimes \mathcal{H}_{-}$ is not only
of uniform (infinite) multiplicity but it is the Minkowski manifold. In fact the last assertion need to
be proved and is in fact reduced to the problem of decomposition of tensor product of ordinary unitary induced representations of the double covering $T_4 \circledS SL(2, \mathbb{C})$ of the Poincar\'e group,
compare Remark \ref{decomposable_L:uniform_mult} of Sect. \ref{decomposable_L}. The
representation of $T_4 \circledS SL(2, \mathbb{C})$ acting in the Fock space
$\mathcal{H}_{+} \otimes \mathcal{H}_{-}$ is a direct sum of representations concentrated on the negative and positive interiors of the cone in the joint spectrum $\textrm{Spec} \, (P_1 , \ldots P_4)$ of the translation generators and the representation concentrated outside the interior of the positive and negative cone in the joint spectrum of the translation generators. We can always modify the second direct summand (concentrated outside the cone
in $\textrm{Spec} \, (P_1 , \ldots P_4)$) which act on the nonphysical subspace of $\mathcal{H}_{+} \otimes \mathcal{H}_{-}$
without altering the physical states acting on the positive (and negative energy states)
without altering the translation generators in the manner which allows to reconstruct the space-time spectrally
on using the harmonic analysis corresponding to the decomposition of the modified representation in the subspace
of $\mathcal{H}_{+} \otimes \mathcal{H}_{-}$ orthogonal to the vacuum and the single particle states.
This is presented in details in Section \ref{constr-of-VF}.
 
Before we explain in more details the spectral construction of space-time (suggested in \cite{Wawr}),
several remarks on the Connes'
spectral format giving a structure of a pseudo-Riemann manifold
to the Gelfand spectrum of a commutative pre-C$^*$-algebra $\mathcal{A}$ of operators acting in a Hilbert space
$\mathcal{H}$ are in order. It has been analysed (in the compact case) by Strohmaier \cite{Stro}, where he recalled a result
of H. Baum \cite{baum} that: 1) the Hilbert space of square integrable sections of the Clifford module naturally associated to the pseudo-Riemann structure on a (not necessary compact) orientable and time orientable pseudo-Riemann manifold admits a fundamental symmetry $\mathfrak{J}$ (induced by a space like reflection) which induces in the space of sections of the module the structure of the Krein space; 2) the natural Dirac operator $D$ associated to the module is not self-adjoint with respect to any natural Hilbert space associated to the pseudo-Riemann manifold, but it is self-adjoint in the Krein sense whenever the ordinary Riemann metric associated to the space like reflection is complete (which is automatic for compact manifolds). The main contribution of \cite{Stro} lies in recognition that for the important class of fundamental symmetries
$\mathfrak{J}$ the operator $(\mathfrak{J} D)^2 + (D\mathfrak{J})^2$ is an ordinary elliptic operator of Laplace-type with respect to a Riemann metric on the pseudo-Riemann manifold, so that a ''Wick-type-rotation-procedure'' using the operator $\mathfrak{J}$ allows us to construct a class of ordinary Riemannian spectral triples naturally associated to the pseudo-Riemann structure with respect to which the manifold is complete. Because the Krein space may be represented as an ordinary Hilbert space $\mathcal{H}$ with an operator $\mathfrak{J}$ which is unitary and selfadjoint, in particular it fulfills $\mathfrak{J}^2 = I$, the results of Strohmaier lie within the general scheme of introducing additional smooth structures on the manifold with the help of
Connes-type-operator format proposed by Fr\"ohlich, Grandjean and Recknagel \cite{frohlich}.
Summing up, we have a tuple\footnote{In our previous paper we have designated the Hilbert space $\mathcal{H}$
accompanying the Krein space $(\mathcal{H}, \mathfrak{J})$ and corresponding to $\mathfrak{J}$
by $\mathfrak{H}_{\mathfrak{J}}$, and the Krein space $(\mathcal{H}, \mathfrak{J})$ just by $\mathfrak{H}$.
We hope this changing of notation, justified in the next section, will not cause any misunderstandings.}
$(\mathcal{A}, D , \mathcal{H})$ acting in the Hilbert space $\mathcal{H}$ which together with a fundamental
symmetry $\mathfrak{J}$ composes a Krein space
$(\mathcal{H}, \mathfrak{J})$, the elements of the involutive algebra $\mathcal{A}$ commute with the (admissible) fundamental symmetry $\mathfrak{J}$, the involution in $\mathcal{A}$ is represented by the Krein-adjoint
$a^\dagger = \mathfrak{J} a^* \mathfrak{J}$ equal to the ordinary adjoint $a^*$, as $\mathcal{A}$ commutes with
$\mathfrak{J}$, $D$ is Krein self-adjoint: $D = \mathfrak{J} D^* \mathfrak{J}$, the operators
$[\mathcal{D}, a]$, $a \in \mathcal{A}$ have bounded extensions; and there exists a self-adjoint operator
$D_\mathfrak{J}$ whose square is equal to the positive self-adjoint operator $1/2 \big( (\mathfrak{J} D)^2
+ (D\mathfrak{J})^2 \big)$ such that $(\mathcal{A}, D_{\mathfrak{J}}, \mathcal{H})$ composes an ordinary spectral triple fulfilling the first five conditions\footnote{With the regularity condition 3 and the orientability condition 4 fulfilled in the slightly stronger form (see \cite{Connes_spectral}, \S 2).} of Connes \cite{Connes_spectral} characterizing the manifold structure spectrally\footnote{Recall that the $\mathfrak{J}$-modulus $[D]_\mathfrak{J}$ of $D$ in the sense of \cite{Stro} is just equal $| D_\mathfrak{J}| = \big( D_\mathfrak{J}^2 \big)^{1/2}$ in our notation;
note also that $(D)_\mathfrak{J}$ in the notation of \cite{Stro} is not in general equal to our $D_\mathfrak{J}$,
but $\big( (D)_\mathfrak{J} \big)^2 = D_\mathfrak{J}^2$.}, which we adopted to the noncompact case
of acyclic manifolds, or more general manifolds with sufficiently simple topology. Moreover we assume the fundamental symmetry $\mathfrak{J}$ to be regular, i.e. lying within the domain of any power of the derivation
$\delta (\cdot) = [D_\mathfrak{J} , \cdot]$, which (by Lemma 13.2 of \cite{Connes_spectral}) is equivalent to the condition that $\mathfrak{J}$ lies within the domain of any power of the derivation
\[
\delta_1 (\cdot) = [D_\mathfrak{J}^2 , \cdot ] (1 + D_\mathfrak{J}^2)^{-1/2},
\]
with $D_\mathfrak{J}^2 = 1/2 \big( (\mathfrak{J} D)^2 + (D\mathfrak{J})^2 \big)$. Assumption\footnote{Several competitive proposals have been proposed for the spectral construction of the pseudo-Riemann
manifold, e. g. Connes and Marcolli \cite{ConMar} proposed to consider operator $D$ which is not self-adjoint
but with self-adjoint $D^2$, but we need more structured situation like that
presented here in order to have the reconstruction
theorem of Connes.} that $(\mathcal{A}, D_{\mathfrak{J}}, \mathcal{H})$ is the spectral triple which respects
the first five conditions of \cite{Connes_spectral}, \S 2 in the slightly
strengthened form (see the assumptions of the Reconstruction Theorem 1.1 of \cite{Connes_spectral}) is crucial in order
to have the reconstruction theorem of Connes applicable (a theorem conjectured in \cite{Connes_manifold} and proved
in \cite{Connes_spectral}). Of course passing to the non compact case will involve new difficulties, like
that concerned with the appropriate choice of the unitization, however we have passed them over for spectral non compact manifolds with sufficiently simple topology, reducing the reconstruction theorem in these cases
to the compact (unital) case proved in \cite{Connes_spectral}, compare Subsections \ref{DirectIntRepVF}-
\ref{VFforFreeFields} and Appendix \ref{AppendixNonCompMani}. Any way
we may assume for our needs that we have the preferred unitization $\mathcal{A}^+$ of $\mathcal{A}$
at hand with the reconstruction problem reduced to the unital case. Indeed we are in the situation
where $\Sp \mathcal{A}$ has the natural $\mathbb{R}^4$-manifold structure with the standard Lorentzian pseudo-metric
tensor and with $\mathcal{A}$ equal to the algebra of complex smooth Schwartz functions, so that
we are at the non compact analogue of the Theorem 11.4 of \cite{Connes_spectral}. It should be stressed that
the non compact version of the reconstruction theorem is important as only under its validity we have the
operator-algebraic characterization of space-time justified, and on the other hand the operator-algebra
format is capable of the deformation/perturbation indicated to above. But we would like to stress here that proving
the reconstruction theorem is a practically independent problem and do not intervene into the operator-algebra construction of the unperturbed $(\mathcal{A}, D, D_{{}_{\mathfrak{J}}}, \mathcal{H}_{{}_{\textrm{inv}}})$
in the invariant subspace $\mathcal{H}_{{}_{\textrm{inv}}}$ of the Fock space, as suggested in
\cite{Wawr}. 
  
Conceptually, in case of $\Sp \mathcal{A}$ equal to the ordinary Minkowski space-time the weak
closure $\mathcal{A}''$ of $\mathcal{A}$ acting on $\mathcal{H}_{{}_{\textrm{inv}}}$
act with finite uniform multiplicity, and the elements of $\mathcal{A}$ are the Fourier transforms of
Schwartz functions
of the translation generators $P^0, \ldots, P^3$ restricted to $\mathcal{H}_{{}_{\textrm{inv}}}$,
which act with uniform finite multiplicity on $\mathcal{H}_{{}_{\textrm{inv}}}$, and with the joint spectrum
being a bona fide standard $\mathbb{R}^4$-manifold. The operator $D_{{}_{\mathfrak{J}}}$ is the Dirac operator corresponding to the ordinary euclidean metric on $\mathbb{R}^4$ and is the equal to the linear combination
$\Gamma^0 P^0 + \ldots + \Gamma^3 P^3$ of the translation generators $P^0, \ldots, P^3$, with the corresponding
representors $\Gamma^0, \ldots, \Gamma^0$ of the generators of the Clifford algebra corresponding to the euclidean metric. Similarly,
for the Dirac operator $D$ corresponding to the Minkowski metric, and the corresponding representors
of the generators of the Clifford algebra corresponding to the Minkowski metric. Thus the corresponding elements of $\mathcal{A}$ are the Schwartz functions of self-adjoint operators $Q^0, \ldots, Q^3$, which together with
$P^0, \ldots, P^3$ compose the von Neumann representation of the canonical pairs $Q^i, P^i$
on $\mathcal{H}_{{}_{\textrm{inv}}}$.

Let us remind the hint of \cite{Wawr} for the construction of
$(\mathcal{A}, D, \mathcal{H})$. It will be convenient
to recall some rudiments of harmonic analysis on smooth manifold $\mathcal{M}$ symmetric for a regular action
under a classical semi simple Lie group $G$ as the construction is in fact an application
of harmonic analysis. Suppose we have a symmetric (uniform) smooth Riemann (or pseudo-Riemann) manifold
$\mathcal{M}$ of dimension $n$, acted on by a Lie group $G$ with a (pseudo-) metric tensor $g$ invariant under $G$.
Then we consider the Hilbert space $\mathcal{H} = L^2 (\mathcal{M}, d\upsilon)$ of square summable functions
with respect to the invariant volume form $d\upsilon$ (in fact we are interested with Hilbert spaces or Krein spaces of
square integrable sections of more general Clifford modules over $T^*\mathcal{M}$, although it is unimportant
in presenting the general idea and its connection to the standard results on harmonic analysis of Gelfand, 
Harish-Chandra and others who actually used $L^2(\mathcal{M}, d\upsilon)$). We consider then the unitary regular right representation
$T$ of $G$ acting in $\mathcal{H}$ and an appropriate algebra $\mathcal{A} = \mathcal{S}(\mathcal{M})$ of functions
of fast decrease with nuclear Fr\'echet topology (just $\mathcal{A} = C^\infty (\mathcal{M})$ for compact $\mathcal{M}$).
We can consider the algebra $\mathcal{S}(\mathcal{M})$ as acting in $\mathcal{H}$ as a multiplication algebra with
pointwise multiplication. The regular representation $T$ induces the transformation
$a \mapsto T_g a T_{g}^{-1}$ which coincides with the ordinary group action $T_g a T_{g}^{-1} (x) = a(xg)$
for functions $a \in \mathcal{S}(\mathcal{M})$. Harmonic analysis (''Fourier transform'' on $\mathcal{M}$)
corresponds to a decomposition of the regular right representation $T$ acting in $\mathcal{H}$ into
direct integral of irreducible subrepresentations. To this decomposition there corresponds a decomposition
of every element $f \in \mathcal{H}$ into direct integral of its components belonging to the irreducible
generalized proper subspaces of the Laplacian -- the ''inverse Fourier integral of $f$''. For example Gelfand, 
Graev and Vilenkin \cite{Gelfand} has done it for the Lobachevsky space $\mathcal{M} = \mathcal{L}^3$
acted on by the $G = SL(2,\mathbb{C})$ and have constructed the appropriate algebra
$\mathcal{S}(\mathcal{M})$. It is important for us that in general the construction of harmonic analysis
together with $\mathcal{S}(\mathcal{M})$ can be given a purely operator-spectral shape. Namely,
we consider a maximal commutative algebra $\widehat{\mathcal{A}}$ generated by representors of one parameter
subgroups (or their appropriate functions). Let $\widehat{\mathcal{A}}$ be generated by $P_1 , P_2 , \ldots P_n$.
Let $\textrm{Spec} \, (P_1 , P_2 , \ldots P_n)$ be their joint spectrum. In particular for the Lobachevsky plane
$\mathcal{M} = \mathcal{L}^2$ acted on by the $G = SL(2, \mathbb{R})$ group we may choose $P_1$ to be the
Casimir operator equal to the Laplacian on the Lobachevsky plane, and for the $P_2$ we may choose the generator
of a one parameter boost subgroup. In this case the inverse Fourier transform and the Fourier transform
relating $f \in \mathcal{H}$ and its Fourier transform $\mathcal{F}f$ may be written as
\[
f(x) = \int \limits_{\textrm{Spec} \, (P_1 , P_2 , \ldots P_n)} \mathcal{F}f(s) \, \Theta(x,s) \, d\nu(s); \,\,\,
\mathcal{F}f(s) = \int \limits_{\mathcal{M}} f(x) \, \Theta(x,s) \, d\upsilon(x),
\]
where $\Theta(\cdot , s)$ is a complete set of common generalized proper functions of the operators
$P_1 , P_2 , \ldots P_n$ corresponding to the point $s$ of their joint spectrum $\Sp (P_1 , P_2 , \ldots P_n)$.
In fact the Fourier transform of \cite{Gelfand} does not have this spectral form because the full
(maximal set of commutative) generators $P_1 , P_2 , \ldots$ (or their functions) have not been explicitly constructed
(besides the Laplacian), which are simultaneously diagonalized by the Fourier transform constructed there.
However, existence of Fourier transforms diagonalizing say the Laplacian on the Lobachevsky plane $\mathcal{L}^2$
and the generator of a one parameter boost subgroup of $SL(2, \mathbb{R})$ follows from the general
theory presented in \cite{GelfandIV}, \cite{Gelfand} (as well as from the papers of Harish-Chandra on harmonic
analysis). Thus, the Fourier transform diagonalizes the algebra of operators $\widehat{\mathcal{A}}$
and the inverse Fourier transform diagonalizes the algebra $\mathcal{A} = \mathcal{S}(\mathcal{M})$.
In this sense the algebras $\mathcal{A}$ and $\widehat{\mathcal{A}}$ are dual to each other.
Note in passing that whenever the commutative algebra $\widehat{\mathcal{A}}$ is not maximal
commutative in the algebra generated
by generators of one parametric subgroups, then the subrepresentations in the direct integral decomposition
of $T$ need not be irreducible, as is the case for example for the double covering
of the Poincar\'e group with $P_1 , \ldots P_4$ equal to the translation generators, where we
have two Casimir generators and one of which is not a function of $P_1 , \ldots P_n$. 

Now suppose that $T$ is the (Krein-)isometric representation $T$ of the double covering of the Poincar\'e
group $G$ acting in the Krein-Fock space $\mathcal{H}_{+} \otimes \mathcal{H}_{-}$ of the
free theory underlying QED or more general gauge fields. In this case we may repeat the above
construction of (space-time) algebra $\mathcal{A} = \mathcal{S}(\mathcal{M})$ of functions now understood
as operators in the appropriate subspace of the Fock
space provided the algebra $\widehat{\mathcal{A}}$ generated by Schwartz functions of generators $P_1 , \ldots P_4$
acts with uniform multiplicity in the subspace. It is the case for the subspace of the Fock space orthogonal
to the vacuum and to the one particle states. Indeed, one can prove that in this case the joint spectral measure
on the joint spectrum $\Sp (P_1 , \ldots P_4) \cong \Sp \widehat{\mathcal{A}}$ is the Lebesgue measure
in case of free fields but moreover the theory of quantum fields,
and especially the theory of free quantum fields, is accompanied by a much stronger
assumption (which is not always explicitly stated), that the joint spectrum of the translation generators
is a subset of the smooth Minkowski space with the pseudo-Riemann (Lorentzian) structure, and in case of the Fock space $\mathcal{H}_{+} \otimes \mathcal{H}_{-}$ of both energy signs it is just the Minkowski manifold, of uniform multiplicity on the subspace orthogonal to the vacuum and single particle states. In case
of the free fields underlying QED it is equal to the closed forward cone and the full Minkowski manifold in case of the Fock space $\mathcal{H}_{+} \otimes \mathcal{H}_{-}$ of free fields of both energy signs, uniform in the subspace orthogonal to the vacuum and the single particle states. In fact the ''positivity of
energy'' assumption of the Wightman axioms \cite{wig} would be meaningless if one was not able to
introduce the Lorentzian manifold structure with the $\Sp (P_1 , \ldots P_4)$
embedded into the manifold. This assumption is of much more profound character than may apparently seem at first
sight. In particular, it enters non trivially into the definition of normal ordering of operator valued distributions, compare e.g. the Bogoliubov analysis of the Wick product theorem as well as the recurrent construction
of the so-called chronological products as operator valued distributions. Thus, in order to construct
the algebra $\mathcal{A}$ we need to know the ''Fourier transform $V_\mathcal{F}$''
connecting it with the algebra
of functions of the generators $P_1 , \ldots P_4$ on the subspace orthogonal to the vacuum and one particle states.

Now it is important that we have fairly explicitly given the ''Fourier transform $V_\mathcal{F}$'' and it is suggested
by the relation between the one particle states in the momentum and position representations.
We extract sub-spaces of the Fock space where $\widehat{\mathcal{A}}$ acts with uniform multiplicity
and invariant for $T$. Then $V_\mathcal{F}$ is achieved in two steps: i) using the uniformity of the algebra
$\widehat{\mathcal{A}}$ we construct the transformation\footnote{Commonly known for the massive states of irreducible representations acting in one particle states which, after Fourier transform, gives local transformation low in the position representation.} $\mathcal{F}_1$ which after the second
step ii) namely the construction of
the von Neumann-Stone representation of the canonical pairs $(P_i , Q_i)$ giving the construction of the
ordinary Fourier transform $\mathcal{F}_2$ (with uniform multiplicity), allows us to apply
$\mathcal{F}_2 \circ \mathcal{F}_1$,
and gives a local transformation rule $T_g a T_g^{-1}(x) = a(xg)$ for $x \in \Sp \mathcal{A}$.
We thus construct $\mathcal{A}$ as $V_\mathcal{F} \widehat{\mathcal{A}}V_\mathcal{F}^{-1}$,
where $V_\mathcal{F}$ is the transform defined by the composition $\mathcal{F}_2 \circ \mathcal{F}_1$.
In case of the irreducible representations $U^{{}_{m}L^{s}}$ (corresponding to the non zero mass $m$ orbits in momentum space and spin $s$) with appropriate multiplicity summed up with an associated Krein-isometric representation
$U^{{}_{m}[L^{s}]_{\Ass}}$ concentrated on the one sheet hyperboloid orbit outside the cone,
the transform $V_\mathcal{F}$ is unbounded and not unitary, which is connected to the pseudo-Riemannian character of the space-time metric,
compare Sect. \ref{constr-of-VF}. Nevertheless $V_\mathcal{F}$ transforms
the direct integral unitary transformation $ U = \oplus_s \{ \int U^{{}_{m}L^{s}} \, \ud m
\oplus \int U^{[{}_{m}L^{s}]_{\Ass}} \, \ud m \}$ into a Krein-unitary
transformation $V_\mathcal{F} U {V_\mathcal{F}}^{-1}$ acting in the invariant subspace
$\mathcal{H}_{ \oplus_s \{ \int U^{{}_{m}L^{s}} \, \ud m \oplus \int U^{[{}_{m}L^{s}]_{\Ass}} \, \ud m \}}$ of the corresponding spectral triple. Analogous transform $V_\mathcal{F}$ can be performed for
the Krein-isometric\footnote{Compare Sect. \ref{def_ind_krein} for definition of Krein-isometric representation.}
representations acting in the subspace orthogonal to the vacuum and to the one particle states of the free photon field,
but this case is more subtle analytically, in particular $V_\mathcal{F}$ is Krein-isomteric but unbounded, nonetheless
it possesses the natural analytic properties, e. g. it preserves the core domain of the original Krein-isometric subrepresentations acting in the Krein space of the free photon field as well as their Krein-isometric character,
compare Sect. \ref{constr-of-VF}.
In fact the first step $\mathcal{F}_1$ is already done when dealing with single particle representations of local fields by the locality assumption of the field. The important point is that we can always construct the associated representation $\int U^{[{}_{m}L^{s}]_{\Ass}} \, \ud m \}$
concentrated outside the cone, which allows us to complete the
construction of $V_\mathcal{F}$ as well as the Dirac operators defining the manifold and Minkowski metric structures of the space-time spectral tuple. For details of the construction compare Subsection
\ref{e1}-\ref{VFforFreeFields}.

Let us remind that the irreducible representations (and their direct sums) acting in the
sub-spaces of one particle states compose the so called Mackey's systems of imprimitivity over the
corresponding orbits in $\Sp (P_1 , \ldots P_4)$ and the representations $T$ are the respective sums of
their symmetrized/antisymmetrized tensor products, which likewise compose systems of imprimitivity
(but no longer corresponding to ergodic orbits). The vector states under the ordinary Fourier transform
do not transform locally but formerly need to be transformed appropriately in order to transform locally after the application of the ordinary Fourier transform. This additional transformation (explicitly known for all
irreducible positive mass and arbitrary spin representations) is constructed at the beginning of
Section \ref{constr-of-VF}. Perhaps we should remind that for zero-mass representations
which act in one particle states of the Fock-Krein space (Gupta-Bleuler space in physics parlance) of the free photon field
are not localized in the sense that they do not allow position measurement operator such as for one particle states in non-relativistic quantum mechanics. But here we are talking about the full four dimensional Fourier transformation
involving integration regions in $\textrm{Spec} \, (P_1 , \ldots P_4)$ intersecting many independent ergodic orbits corresponding
to irreducible representations with the only requirement of locality for the transformed
$V_\mathcal{F}w$ vector states $w \in \mathcal{H}$, i.e. we require the rule
$T_g a T_g^{-1}(x) = a(xg)$ for $x \in \Sp \mathcal{A}$ to be fulfilled, compare the discussion in
the introductory part of Section \ref{constr-of-VF} or \cite{Wawr}.
Therefore, we construct $V_\mathcal{F}$ explicitly first by decomposing the (symmetrized/anti-symmetrized) tensor product
of induced representations acting in the one-particle states into induced representations $\oplus_s U^{{}_{m}L^{s}}$ concentrated on single ergodic orbits $(p^0)^2 - (p^1)^2 - (p^1)^2 - (p^3)^2 = m^2$ in sp($P^0 , \ldots , P^3$). Next we
perform the transformation for each orbit separately which has the property that the ordinary Fourier transform of its image gives a local transformation formula of the ''wave function''. Finally, we integrate over the orbits $m$ (keeping the range of $s$ fixed joining appropriately the spins $s$ with multiplicities depending on $s$) in order to
obtain $V_\mathcal{F}$ in the invariant subspace $\mathcal{H}_{ \oplus_s \{ \int U^{{}_{m}L^{s}} \, \ud m
\oplus \int U^{[{}_{m}L^{s}]_{\Ass}} \, \ud m\}} $ of the Fock-Krein space in which the subrepresentation
$\oplus_s \{ \int U^{{}_{m}L^{s}} \, \ud m
\oplus \int U^{[{}_{m}L^{s}]_{\Ass}} \, \ud m \}$ acts. During this process we explicitly construct the fundamental symmetry $\mathfrak{J}$, the Dirac operator $D$ and the operator $D_\mathfrak{J}$ acting in the invariant subspace
$\mathcal{H}_{ \oplus_s \{ \int U^{{}_{m}L^{s}} \, \ud m
\oplus \int U^{[{}_{m}L^{s}]_{\Ass}} \, \ud m \}} $ and fulfilling the axioms of the spectral triple, i.e. with $(\mathcal{A}, D_\mathfrak{J}, \mathcal{H}_{ \oplus_s \{ \int U^{{}_{m}L^{s}} \, \ud m
\oplus \int U^{[{}_{m}L^{s}]_{\Ass}} \, \ud m\}} )$
fulfilling the axioms for ordinary (non-compact)
spectral triple, and with $D$ and $\mathfrak{J}$ interconnected with the spectral triple in the way indicated to above.
Moreover, the Krein-unitary representation $V_\mathcal{F} \big( \oplus_s \{ \int U^{{}_{m}L^{s}} \, \ud m
\oplus \int U^{[{}_{m}L^{s}]_{\Ass}} \, \ud m \} \big)
{V_\mathcal{F}}^{-1}$ commutes with $D$ on $\mathcal{H}_{ \oplus_s \{ \int U^{{}_{m}L^{s}} \, \ud m
\oplus \int U^{[{}_{m}L^{s}]_{\Ass}} \, \ud m \}} $. Details of the construction of the transform
$V_\mathcal{F}$ and the space-time spectral triple are presented in Section \ref{constr-of-VF}.

Thus, we need to know the action of $T$ in the Fock space as explicitly as possible. In particular, we may look at the
spectral reconstruction of the space-time as if it was a transform
of the Minkowski manifold structure expressed in the algebra-operator
format from the joint spectrum $\Sp (\boldsymbol{P}^0 , \ldots \boldsymbol{P}^3)$ over to the spectrum $
\Sp \mathcal{A} =
\Sp V_\mathcal{F} \widehat{\mathcal{A}}V_\mathcal{F}^{-1}$ with the help of the transformation $V_\mathcal{F}$.

Thus, we have constructed the spectral tuple
$(\mathcal{A}, D, D_{{}_{\mathfrak{J}}}, \mathfrak{J}, \mathcal{H}_{{}_{\textrm{inv}}})$ of space-time
on the invariant sub-spaces $\mathcal{H}_{{}_{\textrm{inv}}}$ of the Fock space
$\mathcal{H}_{+} \otimes \mathcal{H}_{-}$ orthogonal to the vacuum and to the single particle states
(which is invariant likewise for
the translation generators $\boldsymbol{P}^0, \ldots \boldsymbol{P}^3$),
with the elements of $\mathcal{A}$ equal to the Schwartz functions of
\[
(\boldsymbol{Q}^0, \ldots, \boldsymbol{Q}^3) = (V_\mathcal{F} \boldsymbol{P}^0 V_\mathcal{F}^{-1}, \ldots,
V_\mathcal{F} \boldsymbol{P}^3 V_\mathcal{F}^{-1})
\]
composing with $\boldsymbol{P}^0, \ldots \boldsymbol{P}^3$ the von Neumann representation of canonical pairs
$\boldsymbol{P}^i, \boldsymbol{Q}^i$
on $\mathcal{H}_{{}_{\textrm{inv}}}$, and with the Dirac operator $D_{{}_{\mathfrak{J}}}$
equal to $\Gamma^0 P^0 + \ldots + \Gamma^3 P^3$ and similarly
$D = \widehat{\Gamma}^0 P^0 + \ldots + \widehat{\Gamma}^3 P^3$. The matrices $\Gamma^\mu$ and
$\widehat{\Gamma}^\mu$ are the representors of the Cliff ford algebras generators corresponding
to the ordinary euclidean and Minkowski metrics on the manifold $\mathbb{R}^4$ respectively,
determined uniquely by the restriction of the (modified)
representation of $T_4 \circledS SL(2, \mathbb{C})$ to the subspace $\mathcal{H}_{{}_{\textrm{inv}}}$.

In the next step we note that all elements of the spectral tuple
$(\mathcal{A}, D, D_{{}_{\mathfrak{J}}}, \mathfrak{J}, \mathcal{H}_{{}_{\textrm{inv}}})$
are naturally expressible in terms of free fields. This follows by the very construction of the spectral
tuple and from the Bogoliubov-Shirkov Postulate (\ref{BS-QP}), which expresses the translation generators
in terms of the equal time integrals of the Wick product of free fields ${:}T^{0\mu}(x_0, \boldsymbol{\x}){:}$
-- the conserved current.

In principle, we can investigate the perturbation of this spectral tuple by replacing the Wick product
${:}T^{0\mu}(x_0, \boldsymbol{\x}){:}$ in the formula (\ref{BS-QP}) with the causal perturbative series
for the interacting field $\big({:} T^{0\mu}{:}\big)_{{}_{\textrm{int}}}(g=1,x_0, \boldsymbol{\x})$.
But this seems to be of less physical interest, compare Subsection \ref{perturbed(A,H,D)}, where we give
several remarks concerning such a perturbation.

\subsection{On the relation between the space-time geometry and the interacting quantum field}\label{G}

Since the time of Wightman, we know that there is a very deep relation between the free
field construction on a (say highly symmetric, for simplicity, and globally hyperbolic,
for more profound reasons)
space-time, and the geometry of space-time. Roughly speaking the geometry of space-time,
the transformation rule of the field and the existence of the invariant cyclic vacuum state predetermine
essentially the field, including its Hilbert space representation.

Nowadays we know plenty of concrete examples for the flat Minkowski space-time
and for the other symmetric globally hyperbolic space-times, namely the static Einstein Universe
space-time or de Sitter space-time.

In fact we already know that construction of free field(s) on the space-time manifold of the
mentioned class
and construction of harmonic analysis on the space-time treated as a homogeneous manifold, are
essentially equivalent (at least mathematically).

But we maintain that this relation goes deeper and extends over to the interacting fields
as a consequence of the ordinary rules of perturbative causal QED (as well as other theories with non-Abelian gauge).

We come into this conclusion in the following way.

We have shown that if we have a system of free fields, containing a zero mass free field,
e.g. the free fields underlying QED, then space-time and its geometric structure
can be described by operators acting in the Fock space, with the operators
immediately connected to the free field operators. The only thing which may seem non-standard
is that we need to consider the composite Fock space $\mathcal{H}_+ \otimes\mathcal{H}_-$ of the free fields of purely positive (on the Fock space $\mathcal{H}_+$)
and purely negative energy (on the Fock space $\mathcal{H_-}$) sign respectively. We treat the two systems -- of purely positive and purely negative energy signs -- of free
fields as non-interacting and then introduce the interaction perturbatively on each of the two components
(with the two energy signs) separately, according to the ordinary rules of causal QFT.

The choice of the energy sign is a matter of convention and it is not related to any profound physical law
(so far as the gravitational field is ignored).
Even putting the pure negative and pure positive energy sign fields together as a system of independent fields brings nothing essentially new into the conventional
causal QFT, because we assume that the positive energy fields are not coupled to the negative energy fields, so that from the physical point of view this tensoring operation may seem only as an unnecessary complication in description of essentially the same theory, whenever we confine attention to simple tensor states and look only at one of the factors. In this sense we are still within the conventional
causal QFT. 

This technical step allows us to put the relation between the system of fields
on the one hand and the space-time geometry on the other hand, into the
simple form. Indeed, the relation reflects the well known connection between the
space-time and the space of four-momenta, connected by the ordinary Fourier transform.
The algebra $\mathcal{A}$ of smooth rapidly decreasing functions on the space-time becomes identifiable
with the Schwartz functions of the operators $(\boldsymbol{Q}^0, \ldots, \boldsymbol{Q}^3)$,
which together with the (translation generators or) momenta operators
$(\boldsymbol{P}^0, \ldots \boldsymbol{P}^3)$ compose the von Neumann canonical pairs $\boldsymbol{P}^i, \boldsymbol{Q}^i$
on some invariant subspace $\mathcal{H}_{{}_{\textrm{inv}}}$ of the composite Fock space
$\mathcal{H}_{+} \otimes \mathcal{H}_{-}$. The Dirac operators
$D_{{}_{\mathfrak{J}}} = \Gamma^0 P^0 + \ldots + \Gamma^3 P^3$ and
$D = \widehat{\Gamma}^0 P^0 + \ldots + \widehat{\Gamma}^3 P^3$ and the fundamental symmetry operator
$\mathfrak{J}$ compose the spectral tuple, which
gives to the Gelfand spectrum $\textrm{Spec} \, \mathcal{A}$ the structure of the space-time manifold
as in \cite{Connes_spectral}:
\begin{multline}\label{SpacetimeTupleFields}
\Bigg( \,\,\,\,\,\,\,\, \mathcal{A} =
\{f (\boldsymbol{Q}^0, \ldots, \boldsymbol{Q}^3)), f \in \mathcal{S}(\mathbb{R}^4)\} \,\,\,,
\,\,\,\,\,\,\,\,
\mathcal{H}_{{}_{\textrm{inv}}} \,\,\,,
\,\,\,\,\,\,\,\,\,
D_{{}_{\mathfrak{J}}} = \Gamma^0 \boldsymbol{P}^0 + \ldots + \Gamma^3 \boldsymbol{P}^3 \,, \\
D = \widehat{\Gamma}^0 \boldsymbol{P}^0 + \ldots + \widehat{\Gamma}^3 \boldsymbol{P}^3 \,,
\,\,\,\,\,\,\,\,\,
\mathfrak{J} \,\,\,\, \Bigg)
\end{multline}
on the invariant subspace $\mathcal{H}_{{}_{\textrm{inv}}}$
of the Foc space. The constant
matrices $\Gamma_\mu, \widehat{\Gamma}_\mu$ and the fundamental symmetry operator (likewise a constant matrix)
$\mathfrak{J}$
are uniquely determined by the invariant subspace
$\mathcal{H}_{{}_{\textrm{inv}}}$ and the (modified) representation
of $T_4 \circledS SL(2, \mathbb{C})$ acting in $\mathcal{H}_{+}\otimes \mathcal{H}_{-}$
and pertinent to the system of free fields, restricted to the invariant subspace $\mathcal{H}_{{}_{\textrm{inv}}}$.
The translation operators $(\boldsymbol{P}^0, \ldots \boldsymbol{P}^3)$ are uniquely determined
by the (integral of Wick polynomials of) free field operators due to the first Noether
theorem\footnote{Here of course the operator $\boldsymbol{P}^\mu = \boldsymbol{P}^{\mu}_{+} \otimes \boldsymbol{1}
+ \boldsymbol{1} \otimes \boldsymbol{P}^{\mu}_{-}$, where
$\boldsymbol{P}^{\mu}_{+}$ and $ \boldsymbol{P}^{\mu}_{-}$ are the respective generators acting
in the Fock space respectively of the pure positive energy fields and pure negative energy fields.
The same we have for the left-hand side, which is obvious. We have proved the Bogliubov-Shirkov
Postulate, or the first Noether theorem, for the free positive energy electromagnetic potential
field $A_\mu$ in Subsection \ref{BSH}. But from this proof it follows at once its validity for
the composite system of positive and negative energy free fields $A_\mu$. The proof for the (massive)
Dirac field is given in Subsection \ref{StandardDiracPsiField}.} (\ref{BS-QP}):
\begin{equation}\label{BS-QP+-}
\boxed{\int {:} T^{0\mu} {:} \, \ud^3 \boldsymbol{\x} = \boldsymbol{P}^\mu = d\Gamma(P^\mu),}
\end{equation}
which we have proved in Subsection \ref{BSH} (but compare also Subsect. \ref{StandardDiracPsiField}).
Here under the integral sign we have
the expression in which we replace the classical fields by the quantum fields in the Wick ordered form
(with the classical pointwise multiplication replaced with the Wick product of quantum fields).
We call it 'Bogoliubov-Shirkov Quantization Postulate' or Hypothesis, because indeed it can serve as a
principle in quantization of free fields, as proposed in \cite{Bogoliubov_Shirkov}, Chap. 2, \S 9.4.

It should be stressed that this relation between space-time geometry and the system of free fields
is a theorem, which leaves no room for arbitrary manipulations.

If a reader is not convinced to the negative energy free fields with negative energy states on them,
then he can treat them as purely technical tool, which allows the relation between the free fields and
the space-time geometry to be expressed in a pure operator format.

In fact the space-time spectral tuple is reconstructed not so much by the free fields themselves, but by the
representation of the double covering of the Poincar\'e group acting in the Fock space.
But recall, please, that the perturbative series is so constructed that the interacting fields (as the generalised operators)
have the same transformation rule as the corresponding free fields
and with the same representation of the double covering of the Poincar\'e group (understood as acting on the Hida space together with
its linear transposition acting on the strong dual of the Hida space). Therefore, the space-time as well as its spectral-operator
form, determined by the representation of the double covering of the Poincar\'e group, remains unchanged in passing to
interacting fields.

In principle if we switch on the interaction separately in the subsystem
of positive energy free fields and separately into the system of negative energy
free fields according to ordinary rules of causal QFT, keeping the two systems as noninteracting,
we see that we can consider a ``perturbation'' of the spectral tuple (\ref{SpacetimeTupleFields}), \emph{i.e.}
the tuple with replaced coserved Noether densities in the integrals $\boldsymbol{P}^{\mu}$ by the interacting
counterparts giving $\boldsymbol{P}^{\mu}_{\textrm{int}}$ instead of $\boldsymbol{P}^{\mu}$. We expect that we 
obtain in this way the relationship between the interacting fields and the same space-time geometry expressed
spectrally
\begin{multline}\label{SpacetimeTupleIntFields}
\Bigg( \,\,\,\,\,\,\,\, \mathcal{A} = 
\{f (\boldsymbol{Q}^0, \ldots, \boldsymbol{Q}^3)), f \in \mathcal{S}(\mathbb{R}^4)\} \,\,\,,
\,\,\,\,\,\,\,\,
\mathcal{H}_{{}_{\textrm{inv} \, \textrm{int}}} \,\,\,, \\
\,\,\,
D_{{}_{\mathfrak{J} \, \textrm{int}}} = \Gamma^0 \boldsymbol{P}^{0}_{{}_{\, \textrm{int}}} + \ldots + \Gamma^3 \boldsymbol{P}^{3}_{{}_{\, \textrm{int}}} \,, 
D_{{}_{\, \textrm{int}}} = \widehat{\Gamma}^0 \boldsymbol{P}^{0}_{{}_{\, \textrm{int}}} + \ldots 
+ \widehat{\Gamma}^3 \boldsymbol{P}^{3}_{{}_{\, \textrm{int}}} \,,
\,\,\,\,\,\,\,\,\,
\mathfrak{J}  \,\,\,\, \Bigg).
\end{multline}

The conserved Noether integrals have deeper meaning and, in passing to interacting fields,
the conserved densities should be replaced by the interacting counterparts.

In fact: after switching on the interaction, the general rule of the perturbative causal QFT,
which in particular allows us to compute the Lamb shift or the anomalous magnetic moment
of the electron, lies in the replacement of the free fields, $A_{1 \, \mu} = A_{\mu}(x), A_{2} = \ldots,
A_{n}(x) = \psi(x), \ldots$ or their Wick products $A(x) = {:}A_{1} \ldots A_{n} {:} (x)$, with the corresponding
interacting fields $A_{1 \, {}_{\textrm{int}} \, \mu} (g=1, x), \ldots$,
resp. $\big({:}A_{1} \ldots A_{n} {:} \big)_{{}_{\textrm{int}}}(g=1, x)$ defined (after Bogoliubov) by a causal parturbative series:
\[
A_{{}_{\textrm{int}}}(g=1,x) = -\frac{\delta}{i\delta h(x)}S(\mathcal{L}+hA)^{-1}S(\mathcal{L})\big|_{{}_{h=0}},
\]
where $\mathcal{L}$ is the interaction Lagrangian, compare \cite{Bogoliubov_Shirkov}, \cite{DKS1},
\cite{DutFred}.
Indeed, if we apply just the ordinary rules of causal perturbative computation, the same which we apply to
compute the Lamb shift or the magnetic moment of the electron, then we can see that the corresponding operators
$(\boldsymbol{P}^0, \ldots \boldsymbol{P}^3)$, and computed via the Bogolubov-Shirkov Postulate,
$(\boldsymbol{P}^0, \ldots \boldsymbol{P}^3)$ undergo the corresponding perturbation.
Indeed, the Wick polynomials $ {:}T^{0\mu}{:}(x_0, \boldsymbol{\x})$
will have to be replaced by the interacting field
$\big({:}T^{0\mu}{:}\big)_{{}_{\textrm{int}}}(g=1, x_0, \boldsymbol{\x})$.
Thus, if we keep the general rules of perturbative calculation, then the conserved integrals $(\boldsymbol{P}^0, \ldots \boldsymbol{P}^3)$
to be changed by the interaction on exactly the same grounds on which we get the additional Lamb shift in the hydrogen atom and the anomalous magnetic moment of the electron.

Two cases should be distinguished here. Namely:
\begin{enumerate}
\item[I)]
In the first case when computing $\big({:}T^{0\mu}{:}\big)_{{}_{\textrm{int}}}(g=1, x_0, \boldsymbol{\x})$ we insert literally
the free field expression for ${:}T^{0\mu}{:}$, which does not include the interaction Lagrangian term
$\mathcal{L}$ in the formula for $T^{0\mu}$ taken from the Noether theorem for classical free fields.
\item[II)]
The interaction Lagrangian term $\mathcal{L}$ understood as a Wick polynomial in
free fields is included into the full Lagrangian $\mathcal{L}_{{}_{\textrm{Full}}}$
(expressed as a Wick polynomial in free fields) when computing Wick polynomial ${:}T^{0\mu}{:}$. Thus, ${:}T^{0\mu}{:}$ is put equal
to the Wick formal counterpart of the classical expression for $T^{0\mu}$ in the Noether theorem for classical interacting fields,
in which the classical fields are replaced by free fields and their product replaced formally by the Wick product of free fields.
Such Wick polynomial ${:}T^{0\mu}{:}$ including the Wick interaction term is then inserted into
the Bogoliubov-Shirkov perturbative formula
\[
\big({:}T^{0\mu}{:}\big)_{{}_{\textrm{int}}}(g=1,x) = -
\frac{\delta}{i\delta h(x)}S\big(\mathcal{L}+h{:}T^{0\mu}{:}\big)^{-1}S(\mathcal{L})\big|_{{}_{h=0}}.
\]
\end{enumerate}
The second case II) is in agreement with the general definition of the product of interacting fields
within the causal perturbative approach, compare \cite{Bogoliubov_Shirkov}, \cite{DKS1},
\cite{DutFred}. It moreover seems that this is the formula for
\begin{equation}\label{Pint}
\int \big({:}T^{0\mu}{:}\big)_{{}_{\textrm{int}}}(g=1, x_0, \boldsymbol{\x}) \, \ud^3 \boldsymbol{\x}
= \boldsymbol{P}^{\mu}_{\textrm{int}}
\end{equation}
in case II) which gives the correct expression for the conserved integrals,
which corresponds to the classical expression given by the Noether theorem, and is constructed with the same rules which produce the equations of motion for interacting fields, compare \cite{Bogoliubov_Shirkov}, \cite{DKS1}, \cite{DutFred}. The I)-case is not as much justified as the II)-case, but it is helpful to distinguish the I)-case explicitly for some important as well as technical reasons.

In Sections \ref{e+e-}, \ref{white-noise-proofs} and \ref{A(1)psi(1)} (especially Subsection \ref{OperationsOnXi} and \ref{BSH})
we have given a positive solution of the adiabatic limit problem within the causal perturbative QFT (in particular for QED). The solution is based on substitution
of the Hida creation-annihilation operators for the creation-annihilation operators for specified momenta which enter the free fields of the theory. This allows us to interpret the free fields as generalized integral kernel operators transforming continuously the nuclear Hida dense subspace $(E)$ of the Fock space into its strong dual $(E)^*$.
Analysis based on white noise techniques allows us to treat rigorously the causal perturbative series
for interacting fields in the adiabatic limit of the ''intensity of interaction'' function $g=1$. In general
the contributions to interacting fields are generalized integral kernel operators (as the free fields themselves) transforming continuously the Hida dense nuclear subspace
$(\boldsymbol{E})$ of the Fock space into its strong dual
$(\boldsymbol{E})^*$. Using the analysis of Hida, Obata and Sait\^o for generalized integral kernel operators we obtain effective criteria for a generalised integral kernel operator to be
an ordinary operator transforming the Hida subspace into itself.
In particular, we have effective criteria which allows us to check if the contributions
to $\boldsymbol{P}^{\mu}_{\textrm{int}}$ in (\ref{Pint}) are ordinary operators.
In general (depending on free fields of the theory and the interaction Lagrangian) the perturbative contributions to $\boldsymbol{P}^{\mu}_{\textrm{int}}$ given by the causal perturbative
formula (\ref{Pint}) are not ordinary operators acting in the Fock space of free fields, but are generalized operators transforming the Hida space into its strong dual.
$\boldsymbol{P}^{\mu}_{\textrm{int}}$ are not ordinary operators only if the contributions to the interacting fields are generalized operators belonging to
\[
\mathscr{L}\big((\boldsymbol{E}) \otimes \mathscr{E} , \, (\boldsymbol{E})^*\big)
\cong \mathscr{L}\big(\mathscr{E}, \mathscr{L}((\boldsymbol{E}), (\boldsymbol{E})^*) \big).
\]
(compare Subsect. \ref{AdiabaticLimit} of this Introduction) but not to
\[
\mathscr{L}\big((\boldsymbol{E}) \otimes \mathscr{E} , \, (\boldsymbol{E})\big)
\cong \mathscr{L}\big(\mathscr{E}, \mathscr{L}((\boldsymbol{E}), (\boldsymbol{E})) \big)
\]
and thus in such cases when the interacting fields are more singular in comparison
to free field operators which belong to
\[
\mathscr{L}\big((\boldsymbol{E}) \otimes \mathscr{E} , \, (\boldsymbol{E})\big)
\cong \mathscr{L}\big(\mathscr{E}, \mathscr{L}((\boldsymbol{E}), (\boldsymbol{E})) \big).
\]
In particular this means that the ''smeared out'' (by space-time test functions of $\mathscr{E}$) higher order contributions to
interacting fields are not ordinary operators transforming the Hida space
$(\boldsymbol{E})$ into itself (as is the case for free fields) but even after ''smearing''
they stay to be generalized operators transforming continuously the Hida space
$(\boldsymbol{E})$ into its strong dual $(\boldsymbol{E})^*$. This is the case e.g. for causal QED in the Minkowski space-time. In particular this means that in such cases (as causal perturbative QED on the Minkowski space-time) that the Bohr-Rosenfeld measurability analysis
of free fields, cannot be extended over to the interacting fields in causal QED on the Minkowski space-time, nor the ''classical limit'' cannot be extended over interacting fields in causal
QED on Minkowski space-time with its full power analogous to the ''classical limit'' of free fields.

It is not even obvious that the zero order contribution, when we compute the corrections in (\ref{Pint})
in accordance with the I)-st rule distinguished above, and which degenerates to just the equality (\ref{BS-QP+-}), represent ordinary (self-adjoint) operators in the Fock space (for the proof compare Subsections \ref{BSH} and \ref{StandardDiracPsiField}).
Similarly, the question of convergence we have reduced to the convergence of the Fock expansion for
the generalized integral kernel operators (which uses Obata symbol calculus) and its convergence concerning
the question if the limit operator is an ordinary one will need a separate analysis.

We should emphasize the difference between the two above mentioned cases: I) and II), and correspondingly
between the two possible ways of computing the perturbative expansion (\ref{Pint}) for
$\boldsymbol{P}^{\mu}_{\textrm{int}}$. The I)-st one is more regular than the II)-nd one, but I) is physically less interesting.
In the II)-nd case the contributions in (\ref{Pint}) to
$\boldsymbol{P}^{\mu}_{\textrm{int}}$ are in general not ordinary operators in the Fock space but keep the meaning of well-defined generalized operators transforming Hida space into its strong dual (this difficulty is pertinent to the contributions to the zero component $\boldsymbol{P}^{0}_{\textrm{int}}$) and correspond to the well known difficulty in construction of the interacting Hamiltonian for relativistic interacting fields as a well-defined operator in the Fock space of free fields. In particular in case of QED (of course here on Minkowski space-time) the higher order contributions to
$\boldsymbol{P}^{0}_{\textrm{int}}$ are not ordinary operators but generalized operators transforming
Hida space into its strong dual. But for some other QFT, such as scalar field $\varphi$
with $:\varphi^q:$ interaction with sufficiently high and even integer
power $q$ the higher order contributions to $\boldsymbol{P}^{0}_{\textrm{int}}$ are ordinary
(densely defined) operators on the Fock space. Similar conclusions have been arrived
at by I. E. Segal and his coworkers with the help of different methods, compare \cite{SegalZhouQED},
\cite{SegalZhouPhi4}, \cite{PedersenSegalZhou}.

Avoiding usage of the Hamiltonian is the crux of the causal perturbative method. Instead, one starts with the scattering matrix (which again is in general a generalized operator) with the ''intensity of interaction function'' $g$ as a tool for implementing the causality condition. This indeed allows the construction interacting fields, \cite{Bogoliubov_Shirkov}, \cite{DKS1}. The fact that the adiabatic limit problem can be solved
positively for realistic QFT (as we show in Subsect. \ref{OperationsOnXi}, \ref{BSH} and \ref{A(1)psi(1)})
allows us to put the intensity of interaction $g=1$ at the end of computation, and restore the realistic interaction at infinity. In this way we can compute the interacting fields
as generalized integral-kernel operators with vector-valued kernels in the sense of
Obata without any use of the (in fact in general non-existing) operator
$\boldsymbol{P}^{0}_{\textrm{int}}$ (or without any use of the non-existing hamiltonian), or in general without any use of the fundamental Noether first integrals $\boldsymbol{P}^{\mu}_{\textrm{int}}$, $\mu=0,1,2,3$ for interacting fields.
This allows us to treat a class of problems involving generalized states, such as (not normalizable) states of many particles with exact momenta in the scattering processes.
Moreover, we do not have to use the Hamiltonian or the operator $\boldsymbol{P}^{0}_{\textrm{int}}$ in the treatment of the computation of the Lamb shift, as we can compute the equation of motion for the interacting fields using the causal scattering matrix generalized operator, and then construct the so called generalized Dirac operator, compare \cite{Bogoliubov_Shirkov}, \cite{DKS1}. Similarly, we can use the causal perturbative QED on the Minkowski space-time
for the analysis of the infrared states (again using the positive solution of the adiabatic limit problem for QED), compare Section \ref{infra}.
In fact there are no serious obstacles in extending the causal perturbative method on fields living on other space-times, provided they have global causal structure (say globally hyperbolic).

Thus, we have the general causal perturbative QFT governed by the following principles:
\begin{enumerate}
\item[1)]
The Hamiltonian formulation of classical theory subject to quantization (e.g. classical QED) together with the canonical commutation rules on which the relation between free classical fields and their quantum counterparts is based, with the Hida operators as the creation-annihilation operators in the Fock space of free fields of the theory.
\item[2)]
The causal rules for the causal perturbative construction of the scattering
(generalized) operator $S(g=1)$, compare Subsections \ref{MotivationForHida} and \ref{CausalSonEU}.
\item[3)]
Globally causal geometry of space-time.
\end{enumerate}
The interacting quantum fields are obtained from the scattering operator $S(g=1) = S(\mathcal{L})$, due to the general Bogoliubov rule
\[
A_{{}_{\textrm{int}}}(g=1; x) = \frac{i\delta}{\delta h(x)}
S(\mathcal{L} +hA)^{-1}S(\mathcal{L})\Big|_{{}_{h=0}}
\]
relating the interacting field $A_{{}_{\textrm{int}}}(g=1)$ to the corresponding free field $A$, compare \cite{Bogoliubov_Shirkov}, \cite{DKS1}, \cite{DutFred}.

Thus, the free fields are constructed from the single particle Hilbert spaces which compose distributional solutions of the corresponding classical hyperbolic (wave) equations. We need to define the nuclear standard countably Hilbert spaces and their strong duals which together with the single particle Hilbert spaces compose standard Gelfand triples, which can be amplified to the Fock spaces of free fields, and serve in construction of Hida operators as the creation-annihilation operators (in the ''momentum picture'') for the free fields. For simplicity, we assume also that there exist at least four one parameter groups of space-time symmetries,
plying the role of translations, with the corresponding vector fields on
the space-time which span everywhere the tangent space. We obtain in this way a general causal perturbative QFT on a globally causal space-time with the character and analytic properties of the interacting fields strongly depending on the space-time geometry. We give a nontrivial example of the static Einstein Universe
in Section \ref{EUandG}.

Because the causal perturbative QFT, including QED on Minkowski space-time, as summarized in Sect. \ref{AdiabaticLimit}
of Introduction (and presented in more details in Sect. \ref{e+e-}),
with the Hida operators as the creation-annihilation operators for free fields,
can successfully be applied to a class of problems (and allows realistic value of the ''intensity of interaction function'' $g=1$ ) then we propose to give to the principles 1) -- 3) the status
of a full-fledged theory, say causal perturbative QFT on a globally causal space-time.
Although we should remember that
this theory is in general concerned with a more general
class of quantum fields including fields which behave more singularly
than, say, the free fields, which even after ''smearing''
stay to be generalized operators. In general the causal perturbative QFT, with Hida creation-annihilation operators of free fields, gives the interacting
fields as generalized operators which may be singular in that they belong to
\[
\mathscr{L}\big((\boldsymbol{E}) \otimes \mathscr{E} , \, (\boldsymbol{E})^*\big)
\cong \mathscr{L}\big(\mathscr{E}, \mathscr{L}((\boldsymbol{E}), (\boldsymbol{E})^*) \big).
\]
but not to
\[
\mathscr{L}\big((\boldsymbol{E}) \otimes \mathscr{E} , \, (\boldsymbol{E})\big)
\cong \mathscr{L}\big(\mathscr{E}, \mathscr{L}((\boldsymbol{E}), (\boldsymbol{E})) \big),
\]
and do not allow any
sensible correspondence to the classical fields (analogue to that presented by Bohr and Rosenfeld for free electromagnetic or Dirac fields, as we have already indicated above), which is the case on the flat Minkowski space-time. This is even the case for Wick polynomials of massless free fields (with Hida operators) which represent counterparts of the measurable classical quantities entering the energy-momentum density. This is a mathematical consequence of the principles 1) -- 3), with the principle 3) assuming that the space-time is precisely the Minkowski space-time. Put otherwise: from the principles 1) -- 2) it follows that the interacting Dirac and electromagnetic fields, which after smearing-out with a test function gibe ordinary operators on the Fock space, and which have well-defined quasi-classical limit, and well-defined correspondence to the classical fields, cannot exist on the flat Minkowski space-time. Or putting it still otherwise: from the principles 1) -- 3) it follows that the interacting (and even free) Dirac and electromagnetic potential fields, which do have correspondence to classical fields and do posses quasi-classical limit have nontrivial gravitational weight and cannot be constructed on the flat Minkowski space-time.

However, the causal perturbative QFT, even on the Minkowski space-time, has (limited but) well-defined range of experimental applicability, namely the high energy scattering phenomena with the many particle plane-wave generalized states as \emph{in} and \emph{out} states and gives the correct effective cross sections and the second regime of applicability embraces the infrared states. Therefore,
the principles 1) -- 3) should be treated seriously. Further it follows from
1) -- 3) with 3) postulating the Minkowski space-time, that the Hamiltonian or equivalently Noether integral $\boldsymbol{P}^{0}_{\textrm{int}}$, does not exist as an ordinary operator for some realistic QFT on the flat Minkowski space-time, such as QED, or the interacting fields of
the SM with the Higgs field.
This means that in order to account for the existence of stable and meta stable particles as the eigen-states, which are normalizable and thus bound eigen-states (or superpositions of such for meta stable particles) of a well-defined self-adjoint $\boldsymbol{P}^{0}_{\textrm{int}}$, then at least one of the
principles 1), 2) and 3) postulating the Minkowski manifold as the actual space-time, will have to be changed. Because 1) and 2) are of considerable generality, and because the high energy scattering phenomena with the generalized plane wave states of the fundamental free fields are less sensitive to the
global structure of space-time, we arrive at the conclusion that the principles 1) and 2)
should be preserved while the third, 3) postulating the Minkowski manifold as the space-time should be changed. We expect this also because, after Bohr, we do not expect any \emph{ad hoc} fundamental QFT which have no correspondence to classical fields.
Even if for the free fields underlying strong interactions it seems that the ''coherent'' states cannot be physically realized, some thermal
states in which the observable local fields have small fluctuations in comparison to the average values, cannot \emph{a priori} be excluded as unreal.

Thus, first we construct
the causal perturbative QFT with Hida creation-annihilation operators
on a globally causal space-time. Theory obtained at this stage has only limited
applicability to a limited range of problems. In the second stage we apply the criteria
borrowed from the general analysis of integral kernel operators of Hida-Obata-Sait\^o,
and check if the higher order contributions to interacting fields belong to
\[
\mathscr{L}\big((\boldsymbol{E}) \otimes \mathscr{E}_{2} , \, (\boldsymbol{E})\big)
\cong \mathscr{L}\big(\mathscr{E}_{2}, \mathscr{L}((\boldsymbol{E}), (\boldsymbol{E})) \big).
\]
(as do the free fields themselves) and after ''smearing'' with test functions become ordinary operators.

The fact that the Hamiltonian or equivalently Noether integral $\boldsymbol{P}^{0}_{\textrm{int}}$,
does not exist as an ordinary operator for some realistic causal QFT on the flat Minkowski space-time, such as QED,
we do not interpret as a general failure of the general causal QFT expressed by 1) -- 3)
with 3) postulating the Minkowski space-time, but rather as the failure of the postulate which assumes the Minkowski manifold as the actual space-time.
In fact, we propose to keep the general
QFT expressed by 1) -- 3). Rather we conclude from this mathematical fact that some realistic fields such as electromagnetic
potential field and the Dirac field with realistic interaction (given by the minimal coupling) cannot live on the flat Minkowski space-time. Otherwise: although the interacting fields in causal perturbative QED are well-defined generalized operators on the Minkowski space, they are too singular as generalised operators to produce the contributions to $\boldsymbol{P}^{0}_{\textrm{int}}$ in (\ref{Pint}) as ordinary operators, and in particular do not allow us to prolong the relation (\ref{SpacetimeTupleIntFields}) between the space-time geometry and the Noether integrals of the tensor
$\big({:}T^{0\mu}{:}\big)_{{}_{\textrm{int}}}$ to interacting fields.
Similarly, the interacting fields in the causal perturbative QED are too singular to allow the Borhr-Rosenfeld measurability
analysis and classical limit of the interacting fields.
The fact that the causal perturbative QFT,
such as causal perturbative QED, makes sense on the flat Minkowski space-time
(with the realistic ''intensity of interaction function'' $g=1$) we interpret as
indication that the class of problems which can be treated with it do not decisively depend
on the whole of space-time but can be confined to an open and relatively small part of a more realistic space-time manifold, which can successfully be approximated by flat space-time
manifold. This interpretation is nontrivial and can equivalently be expressed by the following statement: some realistic interacting fields do necessary possess gravitational weight and cannot live on flat Minkowski space-time; in particular photons, electrons and positrons interacting in accordance to the minimal coupling do necessary have a nontrivial
gravitational weight and cannot live on the flat Minkowski space-time. Thus with our new interpretation of the failure of $\boldsymbol{P}^{0}_{\textrm{int}}$ as ordinary operator
we open up a new perspective which may throw a new light on the relationship
between QFT and the theory of gravity. It should be stressed that the assertion that interacting electrons, positrons and photons possess a (gravitational) weight has a status of a mathematical conclusion, inferred from the assumption that the causal perturbative QED (with realistic ''intensity of interaction function'' $g=1$) is a fully fledged theory. This suggests that gravity need not be extra postulated but should emerge from the construction of interacting quantum fields on causal (globally hyperbolic) space-times, if we are considering the range of problems which are pertinent to QFT, e.g. scattering phenomena, or phenomena involving
transitions between states differing by small particle numbers. Thus guided solely by the equivalence principle we would arrive at the following 
conclusion:
\begin{enumerate}
\item[]
\emph{We need no extra laws determining the relation of space-time geometry to quantum fields,
if we are considering the range of problems which are pertinent to QFT, e.g. scattering phenomena, 
or phenomena involving
transitions between states differing by small particle numbers. The relation
between them is already contained in the causal perturbative construction 
of the scattering operator and of the quantum interacting fields on a globally causal space-time.}
\end{enumerate}
\emph{A priori} it is possible that an extra law is needed, but it
would be less economoical.  Only in passing to quasi-classical states we need an extra law, 
which is expected to be of constraint-type, and which selects among the quasi-classical states those which have the 
proper average value of the Hilbert energy-momentum tensor components,
equal to the Einstein geometric tensor -- a constraint which plays the role of Einstein equations of the gravitational field.

But moreover: if we are about to preserve the first two principles 1) and 2), except in changing the globally causal space-time
geometry of the Minkowski space-time into some other globally causal space-time,
we should make this change in such a manner which preserves agreement with high energy scattering experiments.
This is a strong limitation because theory based on 1) and 2) with the assumption that the many particle plane wave states as \emph{in} and \emph{out} states gives the effective cross sections which are in agreement with experiment although the plane wave states are assumed to live on the flat Minkowski space-time. In particular this postulates some natural (even unitary) identification of the Minkowski plane wave packets with the corresponding ''plane wave packets''on the actual globally causal space-time. In particular, it can be achieved by assuming the actual space-time to be a natural periodic causal (conformal) covering of the causal compactification (with the corresponding periodic identifications of the boundary) of the Minkowski space-time. We then require the ''plane wave packets'' on the actual space-time as allowable if they are extensions of the plane wave packets on the Minkowski space-time, regarded as causally embedded in the actual space-time (as is the case for the single particle Hilbert spaces of free fields on the static Einstein Universe constructed in \cite{SegalZhouQED}, \cite{PaneitzSegalI}-\cite{PaneitzSegalIII}).

Because we construct the interacting fields using the causal perturbative recurrence rules
which decisively depend on the space-time metric and the causal relations it determines, one could
think of the conclusion, that additional laws joining the quantum fields to gravity would be superfluous, to be merely trivial. But this would be not true.
Two nontrivial problems arise. 

In order to understand these problems correctly we shortly comment the so called ''second
quantization'' procedure. We should emphasize
that whenever we quantize, say the Dirac field or de Broglie's wave field, we \emph{do not understand} the Dirac function or de Brogle's wave function (subject to quantization) as a state which gives us probability distribution of a single particle. We should distinguish between the de Broglie wave function and Dirac wave field understood as a classical matter fields from the wave function
understood as describing a single particle, say a state of an electron (or positron). In the last situation we should call them rather a state or Schr\"odinger function. Quantization of the Dirac wave function
understood as a state of an electron or positron or quantization of the Schr\"odinger
function describing a single particle state would be a nonsense from the physical point of view.
In fact there is no ''second quantization'', but always a quantization of a classical system,
with quantization always performed \emph{once}.
Sometimes the system posses an infinite number of degrees of freedom, which is the case of matter fields,
such as the Dirac matter field understood as de Broglie's wave field with polarization, but the procedure of quantization is always performed once, contrary to what is suggested by the misleading term ''second quantization''. Dirac field understood as a classical de Broglie field with polarization is a totally
different physical object from the Dirac function understood as a single particle state. The first respects
in general nonlinear wave function which formally looks like the Dirac equation with nonlinear
interaction terms. There is no superposition principle to which the Dirac field understood as a classical
de Broglie field can be subjected. In particular Dirac field understood as classical de Brogle field cannot be normalized and cannot be interpreted as a probability. Rather the bilinear sesquilinear expressions build of this field give us density of matter, the classical electric current, energy density, and so on.
To the contrary the Dirac function understood as a single particle state,
is subjected to the quantum superposition law and always normalized fulfills linear Dirac equation,
which represents no matter density but a probability of finding a single particle. The book by Tomonaga \cite{TomonagaI}, \cite{TomonagaII} (one of the best introductions to the physical principles underlying quantum mechanics and qunatum field theory) explains these matters very clearly; we encourage the reader to look especially at the Chapters 6 and 8 of \cite{TomonagaII} for explanation of these matters.
After this explanation, we notice that because Einstein equations of gravitational field can be based naturally on the least action principle (recall the Hilbert's famous derivation of these equations as arising from the least action principle), we have the natural mode of computation of the source, \emph{i. e.} of the Hilbert energy-momentum tensor, on the right (matter source) side of these equations, understood as classical equations. Namely, the Hilbert's energy-momentum tensor should be equal to the variation of
matter part of the total action\footnote{Including the geometric part equal to the integral of the Ricci scalar curvature.} with respect to the space-time metric. This general and natural procedure allows to
include such matter fields as the Dirac field understood as a classical de Broglie field with polarization,
as the source in the Einstein equations of the gravitational field.
We should emphasize that in case of Dirac field (or higher spin fields) we should be careful
and in computing this variation we should further specify the method of variation. For example during this variation we assume that the whole variation of the metric comes from the variation of the initial four
vector fields $X_0, \ldots, X_3$, which represent the \emph{rep\'ere mobile} which is everywhere orthonormal
with respect to the metric. This procedure works if the initial metric (before variation) is non-degenerate in having no isometry Lie groups, because in this case the smooth variation of the orthonormal vector fields $X_0, \ldots, X_3$ translates unambiguously into the variation of the metric.
Of course, we should also emphasize that this procedure would be meaningless for the Dirac
field understood as a state of a single particle, or for the Schr\"odinger wave function. At the same time we would like to emphasize that the de Broglie field or Dirac field understood as a classical de Broglie field is a fully legitimate object as a classical field, compare e.g. \cite{TomonagaII}.
Because the action principle, on which derivation of Einstein equation of the gravitational field is based, is also legitimate we obtain in this way, by inclusion of the Dirac field understood as a classical matter de Broglie field, a system of classical
fields which in principle has non-trivial gravitational weight, and moreover we can in principle compute
this weight using the Einstein equations of the gravitational field with the Hilbert energy-momentum tensor
on the right-hand side. We cannot \emph{a priori} exclude such systems as unphysical.
Nonetheless, we are aware that we enter here a new ground. In particular the fact that the Hilbert energy-momentum tensor
of the Dirac field, or of a more general classical de Brogle field, is not positive-definite in the sense
of \cite{HawkingEllis}, cannot in our opinion be used as a counter-argument for the presented method of computing gravitational weights of such systems, or more precisely that the systems of such classical fields cannot be inserted into the Einstein equations of gravitational fields.
We can only say that the singularity theorems of Penrose-Hawking are not valid for such systems.

After this explanation we can arrive to the mentioned non-trivial problems which have to be faced after we accept that the whole relationship between the matter and space-time geometry should already follow from the causal
perturbative construction of interacting fields on the globally causal space-time. If the construction is possible, then the constructed quantum fields can co-exist with the considered space-time. If not,
the two: the quantum fields and the concrete space-time geometry cannot co-exist. This is all we can know about the relationship of matter and space-time geometry
at the present stage. But the arising problems which need to be solved are the following.

\begin{enumerate}
\item[]
PROBLEM I.
First one has to give at least one nontrivial example of a causal perturbative QED on a nonflat globally causal space-time, for which the contributions to the interacting fields are sufficiently regular and belong to
\[
\mathscr{L}\big((\boldsymbol{E}) \otimes \mathscr{E} , \, (\boldsymbol{E})\big)
\cong \mathscr{L}\big(\mathscr{E}, \mathscr{L}((\boldsymbol{E}), (\boldsymbol{E})) \big),
\]
thus producing, after ''smearing'', ordinary operators in the Fock space, and allowing the analogue of the Bohr-Rosenfeld measurability analysis, with well-defined and self-adjoint Hamiltonian (Noether time translations generator $\boldsymbol{P}^{0}_{\textrm{int}}$)
and the components of the Hilbert energy-momentum tensor.
\end{enumerate}

The second nontrivial problem is concerned with states which are associated to the ''quasi-classical limit'' of the QFT theory in which the fluctuations of the
''smeared-out'' local observable fields are small in comparison to their average values.
\begin{enumerate}
\item[]
PROBLEM II.
Suppose we have interacting fields on a globally causal space-time solving Problem I.
By assumption they are well-behaved quantum fields belonging to
\[
\mathscr{L}\big((\boldsymbol{E}) \otimes \mathscr{E} , \, (\boldsymbol{E})\big)
\cong \mathscr{L}\big(\mathscr{E}, \mathscr{L}((\boldsymbol{E}), (\boldsymbol{E})) \big),
\]
and thus which after ''smearing'' are ordinary operators in the Fock space of free fields.
In some specific states (say ''quasi-classical'') the average values of the ''smeared-out'' quantum
counterparts of the classical Hilbert's energy-momentum tensor expression, and the fluctuation can be computed, and for which
the fluctuation is small in comparison to the average value. The Hilbert's energy-momentum tensor operator behaves in the said quasi-classical
state as the classical counterpart, averaged over the support of the test function.
From Einstein equations for the gravitational field,
this average cannot be accidental and should be equal to the Einstein tensor determined by the space-time metric
and averaged with the test function, provided we can apply Einstein equations to the classical matter fields, including the 
classical Dirac, complex scalar, e.t.c. fields. In this case we have to explain existence of such states, 
and formulate properly the constraint condition which selects, among the \emph{a priori}
possible quasi-classical states, all those for which the Einstein equations of the gravitational field are fulfilled at the level of average
values.
\end{enumerate}

PROBLEM II arises only if we can apply Einstein equations to the classical matter fields, including the 
classical Dirac, complex scalar, e.t.c. fields, so that the additional constraint-type condition will
have to be added to the laws subsumend by the causality axioms for the scattering operator. If the Einstein equations
cannot be so applied, the simpler possibility cannot be excluded, that the full relationship between the quantum fields and
the space-time geometry is subsumend by the causality axioms for the scattering operator.  
In the last case the gravitational constant will have to be associated to the construction
of the quasiclassical states (at least of a specific class of such states).

In Section \ref{EUandG} we give a nontrivial example of a globally causal space-time
on which the higher order contributions to interacting fields of QED behave regularly enough, solving PROBLEM I.
We analyse the PROBLEM II in Subsection \ref{CurvatureQFTandG}. We will see, that there is a deeper
connection between the proper definition of the quasi-classical state and the value of the gravitational
constant. Then we extend the construction of the space-time spectral tuple (\ref{SpacetimeTupleFields})
over the free fields (including those underlying QED) on this space-time.

Our conclusion concerning the linkage between space-time geometry and interacting fields
finds a confirmation in the fact that indeed the construction of convergent
QED on curved globally hyperbolic space-times is possible while QED on the flat Minkowski space-time
is very singular. Indeed, QED on the Einstein Universe is convergent, which we prove in Section \ref{EUandG},
compare also \cite{SegalZhouQED}
(the same can be shown using the harmonic analysis on de Sitter space-time and the method of
\cite{SegalZhouQED} for QED on de Sitter space-time, although it seems that no full proof has been provided
which would be on the same level of rigor). In fact the method of Section \ref{EUandG}, or \cite{SegalZhouQED}
can be used in the proof of convergence of QED on the lower dimensional flat Minkowski
space-time (compare the Schwinger model or \cite{GlimmJaffe}
where a different method based partially
on some intuitive physical ideas is used in showing essentially the same result).
But this again confirms only our conclusion because Einstein
geometric theory of gravity in lower dimensional case is highly degenerate (in particular in three-dimensional
case the Einstein tensor becomes ''proportional'' to the curvature, so that curvature is zero if and only
if the Einstein tensor is zero). Still, we can use the flat toral compactification of the Minkowski
four dimensional space-time. But although the set of allowed modes becomes discrete
on such flat compactification, QED stays as singular as on the ordinary
flat Minkowski space-time.

\subsection{General remark on the notation}

In our formulas the measures in the various Fourier transforms are in general not normalized,
so that in our formulas it frequently happens that a constant factor equal to some power of $2\pi$ is omitted
in order to simplify notation. This is the case for the first half of this work concerned with QFT
on the Minkowski space-time (Sections \ref{constr-of-VF} -- \ref{infra}). The Fourier transform and its inverse
we are using in Sections \ref{constr-of-VF} -- \ref{A(1)psi(1)} read
(up to the ignored factors $(2\pi)^{-n/2}$ in $\mathbb{R}^n$) 
\[
\begin{split}
\widetilde{f}(p) = {\textstyle\frac{1}{(2\pi)^{2}}}
\int\limits_{\mathbb{R}^4} f(x) \, e^{ip \cdot x} \, \ud^4 x,
\\
f(x) = {\textstyle\frac{1}{(2\pi)^{2}}}
\int\limits_{\mathbb{R}^4} \widetilde{f}(p) \, e^{-ip \cdot x} \, \ud^4 p,
\end{split}
\]
for $\mathbb{C}^n$-valued functions
\[
f \in L^2(\mathbb{R}^4; \mathbb{C}^n)\cap L^1(\mathbb{R}^4; \mathbb{C}^n)
\]
or for
\[
f \in \mathcal{S}(\mathbb{R}^4; \mathbb{C}^n).
\]
Here (summation with respect to $\mu,\nu = 0,1,2,3$)
\[
p \cdot x = g^{\mu\nu}p_\mu x_\nu = p_0x_0 - p_1x_1 - p_2x_2 -p_3x_3 = g_{\mu\nu}p^\mu x^\nu = p^0x^0 - p^1x^1-p^2x^2-p^3x^3
\]
with the Minkowski metric components $g^{\mu\nu}$ with signature $1,-1,-1,-1$. We are using various
generalized Dirac operators
\[
D = i\gamma^\mu {\textstyle\frac{\partial}{\partial x^\mu}} = i g^{\mu\nu} \gamma_{\mu}{\textstyle\frac{\partial}{\partial x^\nu}},
\]
with the respective Clifford generators $\gamma^\mu$
\[
\gamma^\mu\gamma^\nu - \gamma^\nu\gamma^\mu = 2g^{\mu\nu} \boldsymbol{1}_{{}_{n}},
\]
acting on multispinor functions
\[
f \in L^2(\mathbb{R}^4; \mathbb{C}^n)
\]
or on
\[
f \in \mathcal{S}(\mathbb{R}^4; \mathbb{C}^n),
\]
or finally on distributions
\[
f \in \mathcal{S}(\mathbb{R}^4; \mathbb{C}^n)^*,
\]
whose Fourier transforms are concentrated on the respective orbits
\[
\begin{split}
\mathscr{O}_{{}_{\pm m, 0,0,0}} = \{p \cdot p = m^2, p_0 >0 \, \textrm{or} \, p_0 <0\},
\\
\mathscr{O}_{{}_{0,m,0,0}} = \{p \cdot p = -m^2 \},
\end{split}
\]
with the convention that its Fourier transform $\widetilde{D}$, equal to the multiplication operator by a matrix function (eventually concentrated on
$\mathscr{O}_{{}_{\pm m, 0,0,0}}$, $\mathscr{O}_{{}_{0,m,0,0}}$) is likewise written through the invariant contraction
\[
\widetilde{Df}(p) = \widetilde{\big[i\gamma^\mu {\textstyle\frac{\partial}{\partial x^\mu}}f\big]}(p) = i\widetilde{\gamma}^\mu p_\mu \widetilde{f}(p).
\]
This requires introduction of the Clifford generators $\widetilde{\gamma}^\mu$ associated to the initial
generators $\gamma^\mu$ in the following manner
\[
\widetilde{\gamma}^0 = \gamma^0, \,\,\,\,\, \widetilde{\gamma}^k = -\gamma^k
\,\,\,\, \textrm{if the signature of} \, g_{\mu\nu} = (1,-1,-1,-1)
\]
or
\[
\widetilde{\gamma}^0 = -\gamma^0, \,\,\,\,\, \widetilde{\gamma}^k = \gamma^k
\,\,\,\, \textrm{if the signature of} \, g_{\mu\nu} = (-1,1,1,1).
\]
Of course if we had started with $\widetilde{\gamma}^\mu$
in the Dirac operator $D=i\widetilde{\gamma}^\mu{\textstyle\frac{\partial}{\partial x^\mu}}$, 
then in the Fourier transformed $\widetilde{D} = \gamma^\mu p_\mu$ Dirac operator we would get
$\widetilde{\widetilde{\gamma^\mu}} = \gamma^\mu$. We sometimes use $\gamma^\mu$ and $\widetilde{\gamma}^\mu$ interchangeably.
This is the case in Subsections \ref{e1}, \ref{e2} and in Section \ref{e+e-}, because we are constructing there
higher spin fields (especially the Dirac field) working mainly with Fourier transforms, and considerably many various conjugations.
In order to simplify notation we have interchanged there $\gamma^\mu$ with $\widetilde{\gamma}^\mu$
in order to eliminate the additional tilde sign in the formulas for operators acting on Fourier 
transformed solutions of the Dirac equation.

In Sections \ref{intro}-\ref{infra}
the Fourier transform of square integrable functions $f$ or test functions in $\mathcal{S}, \mathcal{S}^{00}$
on $\mathbb{R}^n$, associated to the decomposition of the translation group $T_n$, acting naturally and unitarily
on $f$, is denoted by $\widetilde{f}$ or $\mathscr{F}f$. The Fourier transform of square integrable functions
acting on the (scalar, bispinor, four-vector, \emph{e.t.c.}) functions $f$ on the orbits $\mathscr{O}_{\pm m, 0,0,0}$
(positive and negative energy sheets of the massive hyperboloids in momentum space),
$\mathscr{O}_{{}_{\pm 1,0,0,1}}$ (positive and negative energy sheets of the cone in momentum space), 
which is associated to the decomposition of the representation
of $SL(2, \mathbb{C})$ subgroup of the double covering $T_4 \circledS SL(2, \mathbb{C})$
of the Poincar\'e group, is denoted by $\mathcal{F}f$.
In Section \ref{EUandG} the Fourier transform on the Lie groups $\mathbb{R}\times SU(2,\mathbb{C})$
and $\widetilde{\mathbb{S}^{1}}\times SU(2,\mathbb{C})$ ($\widetilde{\mathbb{S}^{1}}$ is the double covering of the unit circle
\emph{i.e.} the circle with radius $2$), and generally on any semisimple Lie group not equal to $T_n$, 
is denoted by $\mathcal{F}f$ or by $\widetilde{f}$.

We assume the Hilbert space inner product to be conjugate linear in the first variable and linear in the second variable,
except Appendix \ref{PartIIMackey}, where it is linear in the first, and conjugate linear in the second variable.

\section{Krein-isometric representations concentrated on single orbits 
and the transform $V_\mathcal{F}$}\label{constr-of-VF}

We intend this and the subsequent Section to play explanatory function giving the motivation for developing
a generalization of Mackey'a theory, presented in Appendix \ref{PartIIMackey}.

In this Section, composed of several Subsections, we assume the results of the mentioned generalization
(Appendix \ref{PartIIMackey}) and use them
in the construction of the transform $V_\mathcal{F}$ (see Introduction) and the associated space-time
spectral triple
acting in the space of the free fields
underlying QED (and more generally of the free fields underlying Standard Model).
In fact the construction is motivated on the well known computational practice connecting
the momentum and position pictures in QFT and is intimately connected to the construction
of single particle wave functions in the position picture which have local transformation rule.
The novelty lies in the application to Krein-isometric representations in Krein spaces and
revealing the spectral geometry lying behind the construction. 

Representations of the double cover $T_{4} \circledS SL(2, \mathbb{C})$ of the
Poincar\'e group\footnote{We denote the representor of $(a,\alpha) \in T_{4} \circledS SL(2, \mathbb{C})$
by $U_{(a,\alpha)}$, and the convention in which the Lorentz transformation $\Lambda(\alpha)$ corresponding
to $\alpha \in SL(2, \mathbb{C})$ is an antihomomorphism, and with the right action of $\Lambda$ on $a \in
\widehat{T_4}$.}
considered here are in general not unitary but Krein-unitary and even
only Krein-isometric (for definitions compare Subsect. \ref{pre} and \ref{def_ind_krein})
with the properties motivated by the properties of representations acting in the Krein-Fock spaces
of the free fields underlying QED (and the Standard Model). The first property is that the
Gupta-Bleuler operator $\mathfrak{J}$ -- plying the role of the fundamental symmetry of the Krein
space (compare Subsect. \ref{pre}) in question, commutes with translations. Consider first
a Krein-isometric representation acting in one particle Krein subspace $(\mathcal{H}, \mathfrak{J})$
(or in its subspace) of the Krein-Fock space in question. Because translations (we mean of course their representors)
commute with $\mathfrak{J}$, they are not only Krein-isometric but unitary with respect to the Hilbert space inner product
of the Krein space $(\mathcal{H}, \mathfrak{J})$ in question.
Let $P^0 , \ldots , P^3$ be the respective generators of the translations (they do exist by the strong continuity assumption posed on the Krein-isometric representation -- physicist's everyday computations involve the generators
and thus our assumption is justified, compare Subsect. \ref{def_ind_krein}). Let $\mathcal{C}$ be the commutative $C^*$-algebra generated
by the functions $f(P^0 , \ldots ,P^3)$ of translation generators $P^0 , \ldots , P^3$,
where $f$ is continuous on $\mathbb{R}^4$ and vanishes at infinity. Let
\begin{equation}\label{dec-sp-P}
\mathcal{H} = \int \limits_{\textrm{Spec} \, (P^0 , \ldots , P^3)} \mathcal{H}_{p} \, \ud \mu (p)
\end{equation}
be the direct integral decomposition of $\mathcal{H}$ corresponding to the algebra $\mathcal{C}$
(in the sense of \cite{von_neumann_dec} or \cite{Segal_dec_I}) with a spectral measure
$\mu$ on the joint spectrum $\textrm{Spec} \, (P^0 , \ldots , P^3)$ of the translation generators.
We may identify $\textrm{Spec} \, (P^0 , \ldots , P^3)$ with a subset of the group $\widehat{T_4}$ dual to the translation group $T_4$. Moreover we may assume that the algebra $\mathcal{C}$ and the spectral measure corresponding to the
above decomposition (\ref{dec-sp-P}) are of uniform multiplicity, compare Theorem \ref{lop_ind:twr.1} of
Subsection \ref{lop_ind}.

Let us denote the translation representor $U_{(a,1)}$ just by $T(a)$ and the
representor $U_{(0,\alpha)}$ of the
$SL(2, \mathbb{C})$ subgroup just by $U(\alpha)$.
The second property of the Krein-isometric representations of the semi-direct products
$T_4 \circledS SL(2, \mathbb{C})$ which are important in QFT is the following.
The restriction $U(\alpha)$, $\alpha \in SL(2, \mathbb{C})$ to the second factor $SL(2, \mathbb{C})$ is locally bounded with respect to the above mentioned direct integral decomposition (\ref{dec-sp-P}) of the Hilbert space $\mathcal{H}$,
determined by the restriction $T(a)$, $a \in G_1 = T_4$, of the representation 
of $T_4 \circledS SL(2, \mathbb{C})$ to the Abelian normal factor $T_4$. More precisely: let $\| \cdot \|$
be the ordinary Hilbert space $\mathcal{H}$ norm, then for every compact subset $\Delta$ of the
dual $\widehat{T_4}$ and every $\alpha \in G_2 = SL(2,\mathbb{C})$ there exists a positive constant
$c_{\Delta, \alpha}$ (possibly depending on $\Delta$ and $\alpha$) such that
\begin{equation}\label{circumstance2}
\|U(\alpha) f \| < c_{\Delta, \alpha} \|f \|,
\end{equation}
for all $f\in \mathcal{H}$ whose spectral support (in the spectral decomposition (\ref{dec-sp-P}))
is contained within the compact set $\Delta$.

It turns out that Mackey's theory of induced representations may be extended on Krein-isometric
representations with the above mentioned properties. In particular the primitive system theorem,
subgroup theorem and Kronecker-producct theorem holds true.  
In particular the uniform multiplicity property holds for the representation acting in the subspace orthogonal to the vacuum and to the one-particle space as it is obtained by direct sum of tensor products of the representation acting in the one-particle subspace (however this is not obvious and requires proof, but compare Remark \ref{decomposable_L:uniform_mult}
of Subsection \ref{decomposable_L} and Subsection \ref{lop_ind}). 

By the multiplication rule in $T_{4} \circledS SL(2, \mathbb{C})$ it follows that
\begin{equation}\label{utu}
T(\Lambda(\alpha^{-1})a) = U(\alpha)^{-1} T(a) U(\alpha),
\end{equation}
such that
\[
U(\alpha)^{-1} P^{\nu} U(\alpha) = \Lambda(\alpha^{-1})_{\mu}^{\nu} P^{\mu} \,\,\,(\textrm{summation \, over} \,\, \mu)
\]
so that $U(\alpha)^{-1} E(S) U(\alpha) = E(\Lambda(\alpha^{-1})S)$ for $S \subset{}$ sp($P^0 , \ldots , P^3$),
i.e. $U(\alpha)$ acts on the joint spectrum of $P^0 , \ldots , P^3$ as the ordinary right action
of the Lorentz transformation $\Lambda(\alpha^{-1})$.
Moreover, we may identify
sp$(P^0 , \ldots , P^3) \subset \widehat{T_4}$ with the orbit $\mathscr{O}_{\bar{p}}$ under the standard action of the Lorentz group of a single point $\bar{p} = \bar{p}(m)$ in the vector space $\mathbb{R}^4$ endowed with the
Minkowski pseudo-metric form $g$ with the signature $(1, -1, -1, -1)$, and with the invariant measure
$\mu_{m}$ on the orbit $\mathscr{O}_{\bar{p}} = \{p: g(p , p) = p\cdot p = m^2\}$ induced by the invariant Lebesgue measure on
$\mathbb{R}^4$ equal to the Haar measure on $\widehat{T_4}$.
Because the fundamental symmetry $\mathfrak{J}$ commutes with $P^0 , \ldots , P^3$ it is decomposable
with respect to the decomposition (\ref{dec-sp-P}), and let $p \mapsto \mathfrak{J}_p$ be its
decomposition with respect to (\ref{dec-sp-P}), i.e.
\[
\mathfrak{J} = \int \limits_{sp(P^0 , \ldots , P^3) \cong \mathscr{O}_{\bar{p}}} \mathfrak{J}_p \,
\ud \mu |_{{}_{\mathscr{O}_{\bar{p}}}} (p)
\]
with $\mathfrak{J}_p$ being a fundamental symmetry in $\mathcal{H}_p$. Because of the uniform multiplicity
$\mathcal{H}_p \cong \mathcal{H}_{\bar{p}}$, $p \in \mathscr{O}_{\bar{p}}$. Moreover, every element
$\widetilde{\psi} \in \mathcal{H}$ may be identified with the function
$\mathscr{O}_{\bar{p}} \ni p \mapsto \widetilde{\psi}(p) \in \mathcal{H}_{\bar{p}}$
equal to the decomposition of $\widetilde{\psi}$ with respect to (\ref{dec-sp-P}).
Therefore, in the notation of von Neumann
\[
\widetilde{\psi} = \int \limits_{sp(P^0 , \ldots , P^3) \cong \mathscr{O}_{\bar{p}}}
\widetilde{\psi}(p) \, \sqrt{\ud \mu |_{{}_{\mathscr{O}_{\bar{p}}}} (p)}.
\]

We may assume that
$\mathfrak{J}_p$ does not depend on $p$ -- which is still sufficient for the representations acting
on one-particle states as well as for the decomposition of their tensor products (the latter assertion
will be proved in the further stages of this work). In this Section we identify
every $\widetilde{\psi} \in \mathcal{H}$ with the corresponding function $p \mapsto \widetilde{\psi}(p)$
-- its decomposition.

Now for each $\alpha \in SL(2, \mathbb{C})$ let us define the following operator $D(\alpha)$ (compare
\cite{Ohnuki, Wigner_Poincare})
\[
D(\alpha) \widetilde{\psi}(p) = \widetilde{\psi}(\Lambda(\alpha)p).
\]
By the Lorentz invariance of the measure $\mu$ on the orbit $\mathscr{O}_{\bar{p}}$ it follows that
$D(\alpha)$ is unitary for every $\alpha \in SL(2, \mathbb{C})$. Moreover, because the components
$\mathfrak{J}_p$ in the decomposition of $\mathfrak{J}$ do not depend on $p \in \mathscr{O}_{\bar{p}}$,
it easily follows that $D(\alpha)$ commutes with $\mathfrak{J}$, so that $D(\alpha)$ is Krein-unitary
for each $\alpha \in SL(2, \mathbb{C})$. Thus, $\alpha \mapsto D(\alpha)$ gives a unitary and Krein-unitary representation
of $SL(2, \mathbb{C})$:
\[
D(\alpha \beta) = D(\alpha) D(\beta).
\]

Let $F$ be any Baire function on sp$(P^0 , \ldots , P^3) = \mathscr{O}_{\bar{p}}$, and let 
$F(P) = F(P^0 , \ldots , P^3)$ be the operator function of $P^0 , \ldots , P^3$, i. e. operator 
\[
F(P)\widetilde{\psi}(p) = F(p) \widetilde{\psi}(p).
\]
An easy computation shows that
\begin{equation}\label{[D,F]}
D(\alpha) \, F(P) = F\big( \Lambda(\alpha)P \big) \, D(\alpha),
\end{equation}
where $F(\Lambda(\alpha)P) = F(\Lambda(\alpha)_\nu^\mu P^\nu)$ (summation with respect to $\nu$).
Joining (\ref{utu}) and (\ref{[D,F]}) it follows that 
\begin{equation}\label{[UD^-1,T]}
\big[ U(\alpha)D(\alpha)^{-1}, T(a) \big] = 0.
\end{equation}
Thus $Q(\alpha) = U(\alpha)D(\alpha)^{-1}$ commutes with the elements of the $C^*$- algebra $\mathcal{C}$ 
and it is decomposable with respect to (\ref{dec-sp-P}) (in other words it is a function of the 
operators $P^0 , \ldots , P^3$). Denote the components $Q(\alpha)_p$ of $Q(\alpha)$ with respect to
this decomposition just by $Q(\alpha , p)$. Recall that they are operators acting in $\mathcal{H}_{\bar{p}}$,
so that 
\[
Q(\alpha) = \int \limits_{\mathscr{O}_{\bar{p}}} Q(\alpha , p) \, \ud \mu |_{{}_{\mathscr{O}_{\bar{p}}}} (p).
\]
Thus in the notation of von Neumann \cite{von_neumann_dec} 
\[
U(\alpha) \widetilde{\psi} = Q(\alpha) D(\alpha) \widetilde{\psi}
= \int \limits \limits_{\mathscr{O}_{\bar{p}}} Q(\alpha , p) \big(D(\alpha)\widetilde{\psi} \big)(p) \, 
\sqrt{\ud \mu |_{{}_{\mathscr{O}_{\bar{p}}}} (p)},
\]  
where $p \mapsto  \big(D(\alpha)\widetilde{\psi} \big)(p)$ is the decomposition of  $D(\alpha)\widetilde{\psi}$,
so that 
\[
p \mapsto \Big( U(\alpha) \widetilde{\psi} \Big)(p) =  Q(\alpha , p) \big(D(\alpha)\widetilde{\psi}\big)(p)
\]
is the decomposition of $U(\alpha) \widetilde{\psi}$.

Because $\alpha \mapsto U(\alpha)$ is a representation it follows that the components $Q(\alpha , p)$ of
$Q(\alpha)$ have the following multiplier property
\[
Q(\delta \alpha), p) = Q(\delta, p) Q(\alpha, \Lambda(\delta)p), \,\,\, p \in \mathscr{O}_{\bar{p}}, 
\alpha,\delta \in SL(2, \mathbb{C}).
\]
In particular

\[
Q(e,p) = 1 , \,\,\, Q(\alpha, p)^{-1} = Q(\alpha^{-1}, \Lambda(\alpha)p).
\]
If we consider any Krein-isometric operator $W$ which preserves the invariant core domain of the Krein-isometric
representation $U$ (i.e. the domain $\mathfrak{D}$ of Subsect. \ref{def_ind_krein}) and which is decomposable
with respect to (\ref{dec-sp-P}) with the decomposition $p \mapsto W(p)$, then 
(with $\widetilde{\Psi} = W \widetilde{\psi}$)
\begin{equation}\label{WUW^-1}
WU(\alpha)W^{-1} \widetilde{\Psi} = \int \limits \limits_{\mathscr{O}_{\bar{p}}} 
W(p) Q(\alpha, p) W(\Lambda(\alpha)p)^{-1} \big(D(\alpha)\widetilde{\Psi}\big)(p) \, 
\sqrt{\ud \mu |_{{}_{\mathscr{O}_{\bar{p}}}} (p)}
\end{equation}
with $WU(\alpha)W^{-1}$ being another Krein-isometric representation, forces 
\begin{equation}\label{Q'}
Q'(\alpha, p)= W(p) Q(\alpha) W(\Lambda(\alpha)p)^{-1}
\end{equation}
to be another multiplier:
\[
Q'(\delta \alpha), p) = Q'(\delta, p) Q'(\alpha, \Lambda(\delta)p), \,\,\, p \in \mathscr{O}_{\bar{p}}, 
\alpha,\beta \in SL(2, \mathbb{C}),
\]
corresponding to the representation $\alpha \mapsto WU(\alpha)W^{-1}$.

Moreover, the core domain $\mathfrak{D}$ have the following \emph{pervasive}\footnote{Term introduced by 
Mackey in \cite{Mackey}.} property that there exist a sequence $\{f_l\}_{l \in \mathbb{N}}$
of elements of $\mathfrak{D}$ such that for all $p \in {}$ sp$(P^0 \ldots , P^3) = \mathscr{O}_{\bar{p}}$
(compare Subsection \ref{dense}, Lemma \ref{lem:dense.6}) $\{f_l (p)\}_{l \in \mathbb{N}}$ is dense in $\mathcal{H}_{\bar{p}} = \mathcal{H}_p$.
This property is preserved in the tensoring process in the sense that the tensor product of the one-particle representations concentrated on the orbits may be decomposed into direct integrals of Krein-isometric representations
concentrated on single orbits $\mathscr{O}_{m}$ (which is proved in the latter part of this paper) in which
the representors of translation generators have uniform multiplicity. The invariant core domains 
of these representations have the pervasive property and the analogue operator $D(\alpha)$ 
connected with each of the representations, and defined analogously as above, have the property that it
preserves the core invariant domain of the corresponding representation.

Now the operator $Q(\alpha,p)$ is Krein-unitary for almost all $p \in \mathscr{O}_{\bar{p}}$.
Indeed, we have
\begin{multline*}
Q(\alpha, p) \mathfrak{J}_{\bar{p}}Q(\alpha, p)^*\mathfrak{J}_{\bar{p}} f_{l} (p) = f_l (p) \,\,\, \textrm{and} \\
\mathfrak{J}_{\bar{p}}Q(\alpha, p)^*\mathfrak{J}_{\bar{p}} f_{l} (p) Q(\alpha, p) = f_l (p)
\,\,\, p \in \mathscr{O}_{\bar{p}}, l \in \mathbb{N}.
\end{multline*}
Because for each $p \in \mathscr{O}_{\bar{p}}$, $\{f_l (p)\}_{l \in \mathbb{N}}$ is dense in $\mathcal{H}_{\bar{p}}$
and because the representation $\alpha \mapsto U(\alpha)$ is locally bounded with respect to the spectral measure
$E$ of $T$ determining the corresponding direct integral decomposition (\ref{dec-sp-P}), i.e. fulfills
(\ref{circumstance2}), then
\[
Q(\alpha, p) \mathfrak{J}_{\bar{p}} Q(\alpha,p)^*\mathfrak{J}_{\bar{p}} = \bold{1} \,\,\, \textrm{and} \,\,\, \\
\mathfrak{J}_{\bar{p}} Q(\alpha,p)^*\mathfrak{J}_{\bar{p}} Q(\alpha,p) = \bold{1},
\]
and $Q(\alpha, p)$ is Krein-unitary for all $p \in \mathscr{O}_{\bar{p}}$.
In case of the single particle representations the restriction $T$
of the representation to translations has finite uniform multiplicity
so that $\mathcal{H}_{\bar{p}}$ has finite dimension and the unitarity of $Q(\alpha, p)$
for almost all $p$ immediately follows independently of the assumption of local
boundedness (\ref{circumstance2}) of $U(\alpha)$ with respect to the decomposition (\ref{dec-sp-P}).

It is well known that each element $p = (p^0 , \ldots , p^3) \in \mathbb{R}^4$ of the dual group
$\widehat{T_4} \supset {}$ sp$(P^0 , \ldots , P^3)$ may be represented by the Hermitian $2 \times 2$ matrix
$\hat{p} = p^0 \bold{1} + p^1 \sigma_1 + p^2 \sigma_2 + p^3 \sigma_3$, where $\sigma_i$
are the Pauli matrices
\[
\sigma_1 =
\left( \begin{array}{cc} 0 & 1 \\

1 & 0 \end{array}\right), \,\,
\sigma_2 =
\left( \begin{array}{cc} 0 & -i \\

i & 0 \end{array}\right), \,\,
\sigma_3 =
\left( \begin{array}{cc} 1 & 0 \\

0 & -1 \end{array}\right),
\]
and with the action of the Lorentz transformation $\Lambda(\alpha)p$ on
$p$ given by $\alpha \hat{p}\alpha^* = \widehat{\Lambda(\alpha^{-1})p}$. Now let $\bar{p}$ be any fixed
point of the orbit $\mathscr{O}_{\bar{p}}$. Now we associate bi-uniquely an element
$\beta(p) \in SL(2, \mathbb{C})$ with every $p \in \mathscr{O}_{\bar{p}}$ such that
$\beta(p)^{-1} \widehat{\bar{p}}{\beta(p)^*}^{-1} = \hat{p}$, i.e. $\Lambda(\beta(p))\bar{p} = p$
and $\Lambda(\beta(p)^{-1})p = \bar{p}$. Of course the function $p \mapsto \beta(p) = \beta_{\bar{p}}(p)$ depends
on the orbit $\mathscr{O}_{\bar{p}}$, but we discard the subscript $\bar{p}$ at $\beta(p)$ in order to simplify
notation, as in the most part of this Section we are concerned with a fixed orbit. In the latter part
of this Section we will integrate the representations over the orbits, but we hope it will not
cause any misunderstandings.

It follows that $\gamma(\alpha,p) = \beta(p) \alpha \beta(\Lambda(\alpha)p)^{-1}$ is an element of the subgroup
$G_{\bar{p}}$ stationary for $\bar{p}$: $\Lambda(\gamma(\alpha,p))\bar{p} = \bar{p}$, or
$\gamma(\alpha,p)\widehat{\bar{p}}\gamma(\alpha,p)^{*} = \widehat{\bar{p}}$. Therefore,
every $\alpha \in SL(2,\mathbb{C})$ has the following factorization:
\[
\alpha = \beta(p)^{-1} \gamma(\alpha,p)\beta(\Lambda(\alpha)p).
\]

Thus, because $Q(\alpha,p)$ is a multiplier we obtain
\begin{multline}\label{WQW^-1}
Q(\alpha, p) = Q\big(\beta(p)^{-1} \gamma(\alpha,p) \beta(\Lambda(\alpha)p, p \big) \\
= Q(\beta(p)^{-1},p) \, Q(\gamma(\alpha, p), \bar{p}) \, Q(\beta(\Lambda(\alpha)p), \bar{p}).
\end{multline}

Now let us introduce the operator $W'$ decomposable with respect to (\ref{dec-sp-P})
whose decomposition function is given by 
\begin{equation}\label{W1}
p \mapsto W'(p) =  Q(\beta(p),\bar{p}).
\end{equation}
Because the components $Q(\alpha,p)$ of $Q(\alpha)$ compose a multiplier, then
$W'(p)^{-1} = Q(\beta(p)^{-1},p)$, so that the operator $W'^{-1}$ has the
decomposition $p \mapsto W'(p)^{-1} = Q(\beta(p)^{-1},p)$. By construction $W'$ is a Krein-isometric
operator which preserves the core domain $\mathfrak{D}$ of the initial representation $U$ and
moreover by (\ref{WQW^-1}) we have:
\[
Q(\alpha, p) = W'(p)^{-1} Q(\gamma(\alpha,p), \bar{p}) W'(\Lambda(\alpha)p).
\] 
Comparison with (\ref{WUW^-1}) and (\ref{Q'}) shows that the original Krein-isometric representation
$U$ is equivalent to the Krein-isometric representation $W'^{-1}UW'$, where\footnote{We have denoted 
$\widetilde{\psi}$ and $W'\widetilde{\psi}$ by the same letter $\widetilde{\psi}$, we hope this will not
cause any misunderstanding.}
\[
p \mapsto W'^{-1}U(\alpha)W' \widetilde{\psi} (p) = Q(\gamma(\alpha,p), \bar{p}) D(\alpha)\widetilde{\psi} (p)
=  Q(\gamma(\alpha,p), \bar{p}) \widetilde{\psi} (\Lambda(\alpha)p)
\]
is the decomposition of $W'^{-1}U(\alpha)W' \widetilde{\psi}$. Note that for $\gamma, \gamma'$
ranging over the subgroup $G_{\bar{p}}$ stationary for $\bar{p}$ we have
\[
Q(\gamma \gamma',\bar{p}) = Q(\gamma, \bar{p})Q(\gamma', \bar{p}),
\]
so that $\gamma \mapsto Q(\gamma, \bar{p})$ is a Krein-unitary representation of the subgroup $G_{\bar{p}}$
of $SL(2, \mathbb{C})$ stationary for $\bar{p}$. Thus the initial representation is equivalent to the representation (we denote it by the same letters $U, T$ as the initial one) whose action on the decomposition functions is given by the following formula
\begin{multline}\label{U,T}
U(\alpha)\widetilde{\psi}(p) = Q(\gamma(\alpha,p),\bar{p}) \widetilde{\psi} (\Lambda(\alpha)p), \\
T(a) \widetilde{\psi}(p) = e^{i a \cdot p} \widetilde{\psi}(p) =  e^{i g(a, p)} \widetilde{\psi}(p), 
\end{multline} 
where $\gamma \mapsto Q(\gamma, \bar{p})$ is a Krein-unitary representation of 
the subgroup $G_{\bar{p}}$ stationary for a fixed point $\bar{p}$ belonging to the orbit 
$\mathscr{O}_{\bar{p}} = {}$ sp$(P^0 , \ldots , P^3)$.

Our next step is to find an explicit formula for the unitary and Krein-unitary transformation
(we denote it by $W''$) which applied to vector states $\widetilde{\psi}$ of the representation
$U_{(a,1)} = T(a)$, $U_{(0,\alpha)} = U(\alpha)$ with $T(a)$ and $U(\alpha)$ given by (\ref{U,T})
gives a transformation formula with a multiplier independent of $p \in \mathscr{O}_{\bar{p}}$, i . e.
$W''$ is such that the Fourier transform 
\begin{equation}\label{F(varphi)}
\varphi (x) = (2\pi)^{-3/2} \int \limits_{\mathscr{O}_{\bar{p}}} \widetilde{\varphi}(p) e^{-ip \cdot x} \, 
\ud \mu |_{{}_{\mathscr{O}_{\bar{p}}}} (p)
\end{equation}
of $\widetilde{\varphi} = W''\widetilde{\psi}$ has a local transformation law, where $\ud \mu (p)$
is the invariant measure induced on the orbit $\mathscr{O}_{\bar{p}}$ by the Lebesgue measure 
on $\mathbb{R}^4$, in particular for the two-sheeted hyperboloid (or for the cone in which case $m = 0$) 
$\mathscr{O}_{\bar{p}}$
\[
\ud \mu |_{{}_{\mathscr{O}_{\bar{p}}}} (\vec{p}) = \frac{\ud^3 \vec{p}}{2  p^0 |_{{}_{\mathscr{O}_{\bar{p}}}}(\vec{p})}
= \frac{\ud^3 \vec{p}}{2 \sqrt{m^2 + \vec{p} \cdot \vec{p}}}.
\]
 
To this end we need a representation $\alpha \mapsto V(\alpha)$ of $SL(2, \mathbb{C})$ 
acting in the Krein space $(\mathcal{H}_{\bar{p}}, \mathfrak{J}_{\bar{p}})$ which extends 
the Krein-unitary representation $\gamma \mapsto Q(\gamma, \bar{p})$ of the subgroup
$G_{\bar{p}} \subset SL(2, \mathbb{C})$ to a representation of the whole $SL(2, \mathbb{C})$
group. $V$ need not be Krein-unitary (resp. unitary in case 
$\gamma \mapsto Q(\gamma, \bar{p})$, $\gamma \in G_{\bar{p}}$, commutes with $\mathfrak{J}_{\bar{p}}$). 
It turns out that such extensions $V$
do exist for the Krein-unitary  
representations  $\gamma \mapsto Q(\gamma, \bar{p})$ concentrated on single orbits associated with decompositions of the 
representations of single particle representations of realistic free fields.  Moreover they are well known for the representations $\gamma \mapsto Q(\gamma, \bar{p})$ associated with the representations concentrated on single orbits which arise in decompositions of tensor
products of representations of single particle spaces of all known realistic free fields 
(in the massive case  $\bar{p} = m, 0,0,0$, $G_{\bar{p}}= SU(2, \mathbb{C})$
and the corresponding representations $\gamma \mapsto Q(\gamma, \bar{p}) = L(\gamma)$ in $\mathcal{H}_{\bar{p}}$ of $G_{\bar{p}}$ are unitary, i.e. commuting with $\mathfrak{J}_{\bar{p}}$
(and of course they are trivially  Krein-unitary).  
In zero mass case $\bar{p} = 1,0,0,1$ the representations $\gamma \mapsto Q(\gamma, \bar{p}) = 
L(\gamma)$ in $\mathcal{H}_{\bar{p}}$ of $G_{\bar{p}} = \widetilde{E}_2$ (double covering of the group of 
symmetries of the euclidean plane) are Krein unitary, but in general we encounter here important examples where $L$ are only Krein-unitary and not unitary, i.e. not commuting with $\mathfrak{J}_{\bar{p}}$.

Having given $V$ let us define the transformation $W''$ whose action on decomposition functions is defined 
in the following manner
\begin{equation}\label{W2}
\widetilde{\varphi}(p) = W'' \widetilde{\psi}(p) = V(\beta(p)^{-1})\widetilde{\psi}(p).
\end{equation}
Then we have
\begin{multline*}
W'' U(\alpha)W''^{-1}\widetilde{\varphi}(p) = V \big( \beta(p)^{-1} \big) V\big( \gamma(\alpha,p) \big) 
V \big(\beta(\Lambda(\alpha)p) \big) \widetilde{\varphi}(\Lambda(\alpha)p) \\
= V \Big( \beta(p)^{-1} \, \beta(p) \, \alpha \, \beta(\Lambda(\alpha)p)^{-1} \, \beta(\Lambda(\alpha)p) \Big) \widetilde{\varphi}(\Lambda(\alpha)p) = V(\alpha) \widetilde{\varphi}(\Lambda(\alpha)p),
\end{multline*}
therefore
\begin{multline}\label{wuw^-1}
W'' U(\alpha)W''^{-1}\widetilde{\varphi}(p) = U'(\alpha) = V(\alpha) \widetilde{\varphi}(\Lambda(\alpha)p),
\\
W''T(a)W''^{-1} \widetilde{\varphi}(p) = T'(a)\widetilde{\varphi}(p) 
= e^{i a \cdot p} \widetilde{\varphi}(p) =  e^{i a \cdot p} \widetilde{\varphi}(p), 
\end{multline}
such that the (inverse) Fourier transform $\varphi$ defined by (\ref{F(varphi)})  of 
$\widetilde{\varphi}$
has a local transformation formula
\begin{equation}\label{loc.tr.x}
U'(\alpha) \varphi(x) = V(\alpha) \varphi (x \Lambda(\alpha^{-1})) 
= V(\alpha) \varphi (x_\nu {\Lambda(\alpha^{-1})_{\mu}}^\nu) , \\
T'(a)\widetilde{\varphi}(p) = \widetilde{\varphi}(p-a)
\end{equation}
(summation with respect to $\nu$), where we have used again the symbol $U'$ for the representation in the space of Fourier transforms $\varphi$ hoping that it will not cause any serious misunderstandings.

Let (in the notation of Segal \cite{Segal_dec_I})
\[
\Big( \mathcal{H} = \int \limits_{sp(P^0 , \ldots , P^3) \cong \mathscr{O}_{\bar{p}}} \mathcal{H}_{\bar{p}} 
\, \ud \mu |_{{}_{\mathscr{O}_{\bar{p}}}} (p), \,\,\,
\mathfrak{J} =  \int \limits_{sp(P^0 , \ldots , P^3) \cong \mathscr{O}_{\bar{p}}} \mathfrak{J}_{\bar{p}} 
\, \ud \mu |_{{}_{\mathscr{O}_{\bar{p}}}} (p) \Big)
\]
be the Krein space of the representation (\ref{U,T}), which we may assume to 
be equal to the Krein space of the initial representation, as the transformation $W'$ given by
(\ref{W1}) preserves te core set $\mathfrak{D}$ of the initial representation and is Krein-isometric.
Let 
\[
\Big( \mathcal{H'} = \int \limits_{sp(P^0 , \ldots , P^3) \cong \mathscr{O}_{\bar{p}}} \mathcal{H'}_{\bar{p}} 
\, \ud \mu |_{{}_{\mathscr{O}_{\bar{p}}}} (p), \,\,\,
\mathfrak{J'} =  \int \limits_{sp(P^0 , \ldots , P^3) \cong \mathscr{O}_{\bar{p}}} \mathfrak{J'}_{p} 
\, \ud \mu |_{{}_{\mathscr{O}_{\bar{p}}}} (p) \Big),
\]
be the Krein space with the Hilbert space inner product in $\mathcal{H'}$ defined by
\begin{equation}\label{innProdWU^LW^-1}
(\widetilde{\varphi}, \widetilde{\varphi}') 
= \int \limits_{sp(P^0 , \ldots , P^3) \cong \mathscr{O}_{\bar{p}}} 
\Big( \widetilde{\varphi}(p), \widetilde{\varphi}'(p) \Big)_{p} 
\, \ud \mu |_{{}_{\mathscr{O}_{\bar{p}}}} (p), 
\end{equation}
where 
\[
\Big( \widetilde{\varphi}(p), \widetilde{\varphi}'(p) \Big)_{p} = 
\Big(  \widetilde{\varphi}(p), 
V(\beta(p))^* V(\beta(p)) \widetilde{\varphi}'(p)  \Big)_{\mathcal{H}_{\bar{p}}}, \,\,\,\,
\widetilde{\varphi}(p), \widetilde{\varphi}'(p) \in \mathcal{H'}_{p},
\]
with the inner product $\big( \cdot , \cdot  \big)_{\mathcal{H}_{\bar{p}}}$ of the
Hilbert space $\mathcal{H}_{\bar{p}}$ (with the convention, assumed everywhere except Appendix \ref{PartIIMackey}, that it is 
conjugate linear in the first variable\footnote{This convention is assumed in most of the physical literature
to which we refer in this work. Because Appendix \ref{PartIIMackey} refers to mathematical literature, we assume 
there the convention mostly assumed by mathematicians: that the inner product is conjugate linear in the second variable});
and let the decomposition components of the fundamental symmetry $\mathfrak{J'}$ be
\[
\mathfrak{J'}_{p} = V(\beta(p)^{-1}) \, \mathfrak{J}_{\bar{p}} \, V(\beta(p)).
\]
We then have the following   

\begin{twr*}
Let the extension $V$  of the representation $L(\cdot) = Q(\cdot, \bar{p})$
be Krein-unitary. Then the second transformation $W''$ defined by (\ref{W2}), 
which transforms $\widetilde{\psi}$ belonging to the 
Krein space $(\mathcal{H}, \mathfrak{J})$
of the initial representation, equal to the Krein space of the representation defined by (\ref{U,T}),
onto  the Krein-Hilbert space $(\mathcal{H'}, \mathfrak{J'})$ of elements $\widetilde{\varphi}$ of the representation (\ref{wuw^-1}), is unitary and Krein-unitary. 
\end{twr*}

Note that we have started our analysis with a Krein-isometric representation
$U^{{}_{\bar{p}} L}$ induced by a Krein-unitary representation $L$ of $G_{\bar{p}}$
and concentrated on the orbit $\mathscr{O}_{\bar{p}}$. The representation space of $L$
we have agreed to denote here by $\mathcal{H}_{\bar{p}}$. For its general definition, compare
Subsection \ref{def_ind_krein}. Then we have applied (although we do not state it explicitly)
to the induced representation $U^{{}_{\bar{p}}L}$
a Krein-unitary operator of Lemma \ref{InducedFormToImprimitivity} of Subsection \ref{lop_ind},
which transforms $U^{{}_{\bar{p}}}$ into the form of \emph{imprimitivity system}, compare
Subsection \ref{lop_ind}, which acts in the Hilbert space of square summable functions
$\mathscr{O}_{\bar{p}} \rightarrow \mathcal{H}_{\bar{p}}$
over the orbit $\mathscr{O}_{\bar{p}} = T_4 \circledS SL(2, \mathbb{C}) \big{/} G_{\bar{p}}$
with respect to the induced invariant measure on the orbit.
This is the form of the induced representation $U^{{}_{\bar{p}}L}$ with which we have
denoted by $U(a,1) = T(a)$, $U(0, \alpha) = U(\alpha)$ and with which we have started this Section.
Then to this representation we apply
the two operators $W'$ and $W''$ given respectively by (\ref{W1}) and (\ref{W2}). Both operators
$W', W''$ are Krein-isometric, but in general only the second one $W''$, namely (\ref{W2}),
is bounded and thus Krein-unitary, the first one $W'$, (\ref{W1}), being in general unbounded.
Indeed, note, please, that the operator $Q(\alpha, \bar{p})$ is Krein-unitary for each fixed
$\alpha \in SL(2, \mathbb{C})$. Being only Krein-unitary allows only finite but in general arbitrary
large value of its operator norm when $\alpha$ is varied over the non compact $SL(2, \mathbb{C})$.
On the other hand the general function $p \mapsto \beta(p)$
associated in the indicated way to the orbit $\mathscr{O}_{\bar{p}}$ is unbounded
in norm and its norm assumes
arbitrarily large values when $p$ is ranging over the non-compact orbit.
Therefore, in general the operator norm of the operator $Q(\beta(p), \bar{p})$ in
the equality (\ref{W1}) defining $W'$
can in principle assume arbitrarily large value when $p$ is ranging over the non-compact orbit
$\mathscr{O}_{\bar{p}}$. This in general can occur even if the representation
$U^{{}_{\bar{p}}L}$ we have started with was unitary, \emph{i. e.} this can happen even if the representation
$U^{{}_{\bar{p}}L}$ is induced by the representation $\gamma \mapsto L(\gamma) = Q(\gamma, \bar{p})$
of the isotropy subgroup $G_{\bar{p}} = SU(2, \mathbb{C})$, $\bar{p} = (m,0,0,0)$.
The fact that the map $\gamma \mapsto Q(\gamma, \bar{p}) = L(\gamma)$ is bounded in norm on the compact
subgroup $G_{\bar{p}} = SU(2, \mathbb{C})$ is obvious but does not imply that the function
$\alpha \mapsto Q(\alpha, \bar{p})$ is bounded in norm on the whole non-compact $SL(2, \mathbb{C})$.
In particular the function $p \mapsto Q(\beta(p))$ will in general be unbounded together
with the unbounded function $p \mapsto \beta(p)$ corresponding to the still non-compact
orbit $\mathscr{O}_{m,0,0,0} \cong SL(2, \mathbb{C})/SU(2, \mathbb{C})=
SL(2, \mathbb{C})/G_{m,0,0,0}$.

In the sequel we need the relation between the
final local representation (\ref{wuw^-1}) acting in Krein-Hilbert space $(\mathcal{H'}, \mathfrak{J'})$
and the initial induced representation $U^{{}_{\bar{p}}L}$ defined as in 
Subsection \ref{def_ind_krein}. For this purpose we
need 

\begin{defin*}
The Krein-isometric (unbounded in general) operator equal to the composition of the Krein-unitary operator $U^{-1}$ of Lemma \ref{InducedFormToImprimitivity} of Subsection \ref{lop_ind}, the first $W'$,(\ref{W1}), and finally the second 
$W''$, (\ref{W2}), will be denoted $W$:
\[
W = W''W'U^{-1}.
\]  
The representation (\ref{wuw^-1}) acting in Krein-Hilbert space $(\mathcal{H'}, \mathfrak{J'})$
is then equal 
\[
W U^{{}_{\bar{p}}L} W^{-1} \,\,\,\,
\textrm{where} \,\,\,
L(\gamma) = Q(\gamma, \bar{p}), \gamma \in G_{\bar{p}}.
\]
\end{defin*}
  
\qed

\begin{rem*}
The components $\mathfrak{J'}_{p}$ of the decomposition of the fundamental symmetry $\mathfrak{J'}$
depend in general on $p \in \mathscr{O}_{\bar{p}}$, because $V(\beta(p))$ -- although being 
Krein-unitary and unitary in the respective Krein space $\mathcal{H'}_{p}, \mathfrak{J'}_{p}$ for all $p$ 
-- are not in general unitary in the Hilbert space $\mathcal{H}_{\bar{p}}$:
\begin{multline*}
\mathfrak{J'}_p = V(\beta(p)^{-1}) \, \mathfrak{J}_{\bar{p}} V(\beta(p)) \\
= \mathfrak{J}_{\bar{p}} V(\beta(p))^* \, \mathfrak{J}_{\bar{p}} \mathfrak{J}_{\bar{p}} V(\beta(p))
= \mathfrak{J}_{\bar{p}} V(\beta(p))^* \, V(\beta(p)), 
\end{multline*}
where $\mathfrak{J}_{\bar{p}}$ does not depend on $p$ and $V(\beta(p))^* V(\beta(p))$ depends on $p$.
\end{rem*}

\begin{rem*}
The following useful general equality holds true
\[
\mathfrak{J}_{\bar{p}} \, \sqrt{V(\beta(p))^* \, V(\beta(p))} \,
= \, \sqrt{V(\beta(p))^* \, V(\beta(p))} \, \mathfrak{J}_{\bar{p}} 
\]
with the foll owning notation 
\[
B(p) =  V(\beta(p))^* \, V(\beta(p)), \,\,\,\,
 \sqrt{B(p}) =  \sqrt{V(\beta(p))^* \, V(\beta(p))}
\]
used in this work for the positive self-adjoint operator
\[
V(\beta(p))^* \, V(\beta(p))
\]
(a finite matrix in the applications which are to follow in this
work) and for its (positive) square root. Of course in general the operator $B(p)$
depends, together with the function $p \mapsto \beta(p)$, on the corresponding orbit 
$\mathscr{O}_{\bar{p}}$. Even for a fixed orbit $\mathscr{O}_{\bar{p}}$ the operator 
$B(p)$ is not unique and depend on the possible various choices of the function $p \mapsto \beta(p)$. 
The exception is the class of orbits $\mathscr{O}_{\bar{p}}$ of $\bar{p}= (m,0,0,0)$,
$m \in \mathbb{R}$, within which the operator $B(p)$ is independent of the possible
choices of $\beta(p)$ and for each orbit $\mathscr{O}_{\bar{p}}$ of this class the operator depends only 
on $m$ and is independent of the choice of the function $p \mapsto \beta(p)$
\end{rem*}
\begin{prop*}
In the Krein-Hilbert space $(\mathcal{H'}, \mathfrak{J'})$ of the Krein-isometric representation
$W U^{{}_{\bar{p}}L} W^{-1}$ given by the local formula (\ref{wuw^-1}), there is a dense nuclear 
subspace $E$ on which the same local formula (\ref{wuw^-1})
defines a representation $\big(W U^{{}_{\bar{p}}L} W^{-1}\big)_{{}_{0}}$ Krein-isometric in $(\mathcal{H'}_{{}_{0}}, \mathfrak{J}_{\bar{p}})$ with 
$\mathcal{H}_{{}_{0}}$ being the closure of $E \subset \mathcal{H'}$ with respect to the 
ordinary inner product given by the formula 
\begin{equation}\label{innProd(WU^LW^-1)_0}
(\widetilde{\varphi}, \widetilde{\varphi}') 
= \int \limits_{sp(P^0 , \ldots , P^3) \cong \mathscr{O}_{\bar{p}}} 
\Big( \widetilde{\varphi}(p), \widetilde{\varphi}'(p) \Big)_{\mathcal{H}_{\bar{p}}}
\, \ud \mu |_{{}_{\mathscr{O}_{\bar{p}}}} (p), 
\end{equation}
i.e. by the formula (\ref{innProdWU^LW^-1}) for the inner product in $\mathcal{H'}$
in which the operator
\[
B(p) =  V(\beta(p))^* \, V(\beta(p))
\]
is put equal to the constant unit operator, and with the fundamental symmetry operator
equal to the operator of multiplication by the constant (matrix) operator
\[
\mathfrak{J}_{\bar{p}}.
\]
The two representations are intertwined by a Krein-isometric operator $M$ which is continuous 
on the nuclear subspace $E$ for the nuclear topology, but is in general unbounded with 
respect to the ordinary Hilbert space norms of the Hilbert spaces $\mathcal{H'}$,
$\mathcal{H'}_{{}_{0}}$. In fact the intertwining operator $M$ giving the Krein-isometric
equivalence coincides with the unit operator on the dense nuclear subspace:
\[
M|_{{}_{E}} = \boldsymbol{1}.
\]

The representation $\big(W U^{{}_{\bar{p}}L} W^{-1}\big)_{{}_{0}}$ is Krein-unitary, whenever $L$
and $V$ are Krein unitary, so that each representor 
$\big(W U^{{}_{\bar{p}}L}_{{}_{\alpha}} W^{-1}\big)_{{}_{0}}$ is a bounded operator in the 
Hilbert space $\mathcal{H'}_{{}_{0}}$ of the representation 
$\big(W U^{{}_{\bar{p}}L} W^{-1}\big)_{{}_{0}}$. In particular 
$\big(W U^{{}_{\bar{p}}L} W^{-1}\big)_{{}_{0}}$ is Krein unitary for finite dimensional Krein-isometric
representations $L,V$.  
\end{prop*}
\qedsymbol
 Besides the Krein-isometric
representation (\ref{wuw^-1}) 
$U' = WU^{{}_{\bar{p}}L}W^{-1}$ acting on the Krein-Hilbert space $(\mathcal{H}', \mathfrak{J}')$,
we have the following natural Krein-isometric representation
\begin{equation}\label{U-ass}
\begin{split}
{}^{{}^{\textrm{acc}}}U'(0,\alpha) \widetilde{\varphi} (p) 
= \sqrt{B(p)}^{-1}V(\alpha)\sqrt{B(\Lambda(\alpha)p)} \widetilde{\varphi} (\Lambda(\alpha)p), \\
{}^{{}^{\textrm{acc}}}T(a) \widetilde{\varphi} (p) =  T(a) \widetilde{\varphi}(p) 
= e^{i a \cdot p}\widetilde{\varphi}(p). 
\end{split}
\end{equation}
accompanying the representation $U' = WU^{{}_{\bar{p}}L}W^{-1}$,
where $\sqrt{B(p)}$ is the (positive) square root of the (positive) matrix 
$B(p), p \in \mathscr{O}_{{}_{\bar{p}}}$. 

This representation is Krein-isometrically equivalent to the 
representation $U' = WU^{{}_{\bar{p}}L}W^{-1}$. 
Indeed, let $\mathcal{S}(\mathbb{R}^4; \mathbb{C}^k)$ be the nuclear Schwarz space 
of $\mathbb{C}^k$-valued functions on the momentum space $\mathbb{R}^4$, 
where $\mathbb{C}^k$ is the Hilbert space in which the representations $V$ and $L$ act. 
Consider the closed nuclear subspace $\mathcal{S}^{0}(\mathbb{R}^4; \mathbb{C}^k)$
of $\mathcal{S}(\mathbb{R}^4; \mathbb{C}^k)$, consisting of all those functions which 
vanish together with all their derivatives at the zero point. Now let for each orbit $\mathscr{O}_{\bar{p}}$
the nuclear space $E$ be equal to the space of restrictions to   $\mathscr{O}_{\bar{p}}$
of the elements of $\mathcal{S}^{0}(\mathbb{R}^4; \mathbb{C}^k)$. We have used here the 
the subspace $\mathcal{S}^{0}(\mathbb{R}^4; \mathbb{C}^k)$ rather than 
$\mathcal{S}(\mathbb{R}^4; \mathbb{C}^k)$, because among the orbits there is the cone in momentum space
with the singularity at the apex. The choice $\mathcal{S}^{0}(\mathbb{R}^4; \mathbb{C}^k)$
allows to include also the cone orbit. However, the analysis of decomposition
of the representation acting in the Fock space of free fields depends essentially only 
on the representations concentrated on the orbits $\mathscr{O}_{m,0,0,0}$, $m \neq 0$,
$m \in \mathbb{R}$ and $\mathscr{O}_{0,m,0,0}$, $m>0$. For these three classes of orbits
we can use the space $E$ of restrictions of elements of the ordinary Schwartz space
$\mathcal{S}(\mathbb{R}^4; \mathbb{C}^k)$ instead of $\mathcal{S}^{0}(\mathbb{R}^4; \mathbb{C}^k)$.
We will explain it more carefully in the later stage of investigation. 

The intertwining operator $C$, understood as an operator
$(\mathcal{H}', \mathfrak{J}') \rightarrow (\mathcal{H}', \mathfrak{J}')$, 
acting in the single particle space is equal
\[
C \widetilde{\varphi}(p) = \sqrt{B(p)}^{-1} \widetilde{\varphi}(p), \,\,\,\,
C^{-1} \widetilde{\varphi}(p) = \sqrt{B(p)} \widetilde{\varphi}(p),
\]
and $C$ transforms bi-uniquely and bi-continuously the nuclear space $E$ onto itself. 
One easily checks that $C$ indeed intertwines 
$U'$ and ${}^{{}^{\textrm{acc}}}U$:
\[
C \, U' \, C^{-1} = {}^{{}^{\textrm{acc}}}U.
\]

Let us introduce another operator $K$:
\[
K \widetilde{\varphi}(p) = \sqrt{B(p)} \widetilde{\varphi}(p), \,\,\,
K^{-1} \widetilde{\varphi}(p) = \sqrt{B(p)}^{-1} \widetilde{\varphi}(p),
\]
understood as a Krein-unitary operator mapping the Krein space $(\mathcal{H}', \mathfrak{J}')$
onto the Krein space $(K\mathcal{H}', K\mathfrak{J}'K^{-1}) 
= (K\mathcal{H}', \mathfrak{J}_{{}_{\bar{p}}}) 
= (\mathcal{H}'_{{}_{0}}, \mathfrak{J}_{{}_{\bar{p}}})$, where the Krein fundamental symmetry
in the Krein space $(K \mathcal{H}', \mathfrak{J}_{{}_{\bar{p}}})$ is equal to the operator
of multiplication by the constant matrix $\mathfrak{J}_{{}_{\bar{p}}}$.
Recall that the Krein fundamental symmetry operator $\mathfrak{J}'$ in the 
Krein-Hilbert space $(\mathcal{H}', \mathfrak{J}')$ is equal to the operator of multiplication
by the matrix 
\[
\mathfrak{J'}_{p} = V(\beta(p))^{-1} \mathfrak{J}_{\bar{p}} V(\beta(p)) = 
\mathfrak{J}_{\bar{p}} B(p),
\]
The operator $K$ gives a Krein-unitary equivalence
between the representation ${}^{{}^{\textrm{acc}}}U$ acting on the Krein space 
$(\mathcal{H}', \mathfrak{J}')$
and defined by the formula (\ref{U-ass}) with the dense nuclear domain
$E$, and the Krein-isometric representation given by formula (\ref{wuw^-1}) and regarded as representation on $E$, but on the Krein space  
$(K \mathcal{H}', \mathfrak{J}_{{}_{\bar{p}}})$ and with the nuclear domain $E$, which differs from 
the Krein space $(\mathcal{H'}, \mathfrak{J'})$
 by the replacement of the matrices
$\sqrt{B(p)}$ and $B(p)$ everywhere with the constant unit matrix $\boldsymbol{1}$. Because on the other hand 
the representation $U'$, defined by (\ref{wuw^-1}),
and the representation ${}^{{}^{\textrm{acc}}}U$, both acting on the Krein space
$(\mathcal{H}', \mathfrak{J}')$ are Krein isometric equivalent (with $C$ defining the equivalence),
then it follows that the representation, defined by (\ref{wuw^-1}),
with the nuclear domain $E$, on the Krein space $(\mathcal{H}', \mathfrak{J}')$ (with
the matrix $B(p) \neq \boldsymbol{1}$) is equivalent to the Krein isometric
representation defined by the same formula (\ref{wuw^-1})) and the same nuclear domain $E$,
but on the Krein space in which the operators $B(p)$ and $\sqrt{B(p)}$ are everywhere replaced
by the constant unit matrices $\boldsymbol{1}$. Thus, $M= CK$ defines the required equivalence.
It immediately follows that $M=\boldsymbol{1}$ on $E$. 

That each representor of the representation $\big(W U^{{}_{\bar{p}}L} W^{-1}\big)_{{}_{0}}$
is a bounded operator, whenever $L$ and $V$ are Krein-unitary, 
easily follows from the transformation formula (\ref{wuw^-1}), 
inner product formula (\ref{innProd(WU^LW^-1)_0}) in $\mathcal{H'}_{{}_{0}}$
and invariance of the measure $\mu |_{{}_{\mathscr{O}_{\bar{p}}}}$ on the orbit
$\mathscr{O}_{\bar{p}}$. Indeed, let $|V(\alpha)|$ be the operator norm 
of the linear operator
$V(\alpha)$ in the Hilbert-Krein space $\mathcal{H}_{\bar{p}}$ of the representations
$V$ and $L$.  Let $\| \cdot \|_{{}_{0}}$ be the Hilbert space norm in $\mathcal{H'}_{{}_{0}}$
uniquely determined by (\ref{innProd(WU^LW^-1)_0}) on the dense nuclear domain $E$. 
Then for any $\widetilde{\varphi} \in E$ we have 
\begin{multline*}
\big\|\big(W U^{{}_{\bar{p}}L}_{\alpha} W^{-1}\big)_{{}_{0}} \widetilde{\varphi} \big\|_{{}_{0}}^{2} \\
= \int \limits_{\mathscr{O}_{\bar{p}}} 
\Big( V(\alpha)\widetilde{\varphi}(\Lambda(\alpha)p), V(\alpha)\widetilde{\varphi}(\Lambda(\alpha)p) \Big)_{\mathcal{H}_{\bar{p}}}
\, \ud \mu |_{{}_{\mathscr{O}_{\bar{p}}}} (p) \\
\leq 
|V(\alpha)|^2
\int \limits_{\mathscr{O}_{\bar{p}}} 
\big\|\widetilde{\varphi}(\Lambda(\alpha)p)\big\|_{\mathcal{H}_{\bar{p}}}^2
\, \ud \mu |_{{}_{\mathscr{O}_{\bar{p}}}} (p) \\
=|V(\alpha)|^2 \int \limits_{\mathscr{O}_{\bar{p}}} 
\big\|\widetilde{\varphi}(p)\big\|_{\mathcal{H}_{\bar{p}}}^2
\, \ud \mu |_{{}_{\mathscr{O}_{\bar{p}}}} (p) =
|V(\alpha)|^2 \| \widetilde{\varphi} \| _{{}_{0}}^2.
\end{multline*}

\qed

The rest of this Section is the application of the above construction of the transformation 
$(\mathcal{H}_{{}_{U^L}}, \mathfrak{J}^L)
\xrightarrow{W}  (\mathcal{H'}, \mathfrak{J'})$ with the properties indicated by the above Theorem, Proposition
and the Remarks stated above. The crucial point being that
the Fourier transform (\ref{F(varphi)}) of the $W$-transformed $\widetilde{\varphi}$ have a local transformation law. 

Before passing to applications, let us introduce several definitions.

\begin{defin*}
Let $U^{{}_{\bar{p}} L}$ and $U^{{}_{\bar{p'}} L'}$ be two Krein-isometric representations
of $T_4 \circledS SL(2, \mathbb{C})$, induced respectively by Krein-unitary representations 
$L$ and $L'$ of the isotropy
subgroups $G_{\bar{p}}$ and $G_{\bar{p'}}$ for the points $\bar{p}, \bar{p'} \in T_4$,
respectively concentrated on their own (in general different) orbits $\mathscr{O}_{\bar{p}}$ and respectively $\mathscr{O}_{\bar{p'}}$.
Representations $U^{{}_{\bar{p}} L}$ and $U^{{}_{\bar{p'}} L'}$
are said to be naturally associated if the extensions $V$ and respectively $V'$ of the 
$SL(2, \mathbb{C})$ subgroup used in the construction of the unitary map $W$, 
presented above, which defines their local versions
$WU^{{}_{\bar{p}} L}W^{-1}$ and $WU^{{}_{\bar{p'}} L'}W^{-1}$ (of course $W$ depends on the representation to which it is applied $U^{{}_{\bar{p}} L}$), are identical.
\end{defin*}

The operator $W$ defined above,
which applied to
the representation $U^{{}_{\bar{p}} L}$, gives a local representation
$WU^{{}_{\bar{p}} L}W^{-1}$, depends on the orbit $\mathscr{O}_{\bar{p}}$, the 
Krein-unitary representation
$L$ of $G_{\bar{p}}$ and on the extension $V$ of $L$, with $V$ having the same 
Hilbert-Krein space as $L$, and coinciding with $L$ on the isotropy subgroup 
$G_{\bar{p}}$. Thus, more adequately we should 
write $W_{{}_{V,\bar{p}}}$ for the operator $W$ constructed here, and we should 
more properly say that
\[
U^{{}_{\bar{p}} L}, \,\,\, U^{{}_{\bar{p'}} L'}
\]
are by definition \emph{naturally associated} iff in their local versions
\[
W_{{}_{V,\bar{p}}} U^{{}_{\bar{p}} L} W_{{}_{V,\bar{p}}}^{-1}, 
\,\,\, W_{{}_{V',\bar{p}}} U^{{}_{\bar{p'}} L'}W_{{}_{V',\bar{p}}}^{-1}
\]
the representations $V$ and $V'$ of $SL(2, \mathbb{C})$ subgroup  are identical. 
In this case we write  
\[
U^{{}_{\bar{p'}} [L]_{\textrm{Ass}}} \,\,\, \textrm{for} \,\,\,
 U^{{}_{\bar{p'}} L'}.
\]

This notation is slightly misleading as the ``association'' relation
concerns the local versions of the induced representations, but correctly reflects the fact 
that the crux lies in existence of a Krein-unitary representation $V$ which extends both 
Krein-unitary representations $L$ and $L'$ inducing them.

\begin{defin*}
The local version 
\[
W_{{}_{V,\bar{p}}} U^{{}_{\bar{p}} L} W_{{}_{V,\bar{p}}}^{-1}
\]
of the induced representation 
\[
U^{{}_{\bar{p}} L}
\]
we call the natural local version of $U^{{}_{\bar{p}} L}$.
\end{defin*}

We extend the definition of associated representations over any pair of local Krein-isometric representations
of $T_4 \circledS SL(2, \mathbb{C})$, concentrated on their own orbits.
Namely, let two local Krein-isometric representations $U'', U'''$
\begin{eqnarray*}
U''(\alpha) \tilde{\phi}(p) = V''(\alpha) \tilde{\phi}(\Lambda(\alpha) p), \,\,\,
\\
T''(a) \tilde{\phi}(p) = e^{ia\cdot p} \tilde{\phi},
\,\,\,\,\,\, \mathscr{O}_{\overline{p''}},
\end{eqnarray*}
and
\begin{eqnarray*}
U'''(\alpha) \tilde{\phi}(p) = V'''(\alpha) \tilde{\phi}(\Lambda(\alpha)p),
\\
T'''(a) \tilde{\phi}(p) = e^{ia\cdot p} \tilde{\phi},
\,\,\,\,\, p \in \mathscr{O}_{\overline{p'''}},
\end{eqnarray*}
of $T_4 \circledS SL(2, \mathbb{C})$ be concentrated on their own orbits
$\mathscr{O}_{\overline{p''}}$ and $\mathscr{O}_{\overline{p'''}}$, respectively. We say that they are \emph{associated}
if the representations $V''$ and $V'''$ of the $SL(2, \mathbb{C})$ subgroup are identical,
irrespectively of how the inner products and Krein-inner products (or corresponding
fundamental symmetries) are defined.

\begin{defin*}
Let $U^{{}_{\bar{p}} L}$ and $U^{{}_{\bar{p'}} L'}$ be two induced Krein-isometric 
representations. Let  $U^{{}_{\bar{p}} L}$ and $U^{{}_{\bar{p'}} L'}$ be Krein-isometrically
equivalent to local representations
\begin{equation}\label{Uloc,U'loc}
\big(U^{{}_{\bar{p}} L}\big)_{loc}, \,\,\, \big(U^{{}_{\bar{p'}} L'}\big)_{loc}
\end{equation}
respectively, which are  not necessary equal to the natural local versions. 
We say that  $U^{{}_{\bar{p}} L}$ and $U^{{}_{\bar{p'}} L'}$ are associated
if  (\ref{Uloc,U'loc}) are associated. In this case we also write
\[
U^{{}_{\bar{p'}} [L]_{\textrm{Ass}}} \,\,\, \textrm{for} \,\,\,
 U^{{}_{\bar{p'}} L'},
\]
although the relation between $L$ and $L'$ may be more complicated in comparison to the
case when the representations are naturally associated (and not merely associated).
\end{defin*}

As we will see it is the \emph{natural association} which is important in the analysis of the representation acting in the Fock space and its relation to space-time structure, 
expressed in operator format.

Before passing to the computation of $V_\mathcal{F}$ let us remind that we are interested in the analysis of the representation
of  $T_{4} \circledS SL(2, \mathbb{C})$ acting in the (Krein-) Hilbert Fock space of free fields. This representation
is obtained by direct sum of (symmetrized/antisymmetrized) tensor products of representations acting in one particle subspace. 
Tensor products of such representations are equal to direct integral of induced representations
concentrated on single orbits. 

Thus, for the positive energy fields the representations in the single particle spaces
are decomposed into induced representations
\[
U^{{}_{\bar{p}} L}
\]
concentrated on the orbit $\mathscr{O}_{\bar{p}} = \mathscr{O}_{(m, 0, 0, 0)}$
with $G_{\bar{p}} = SU(2, \mathbb{C})$ with finite dimensional representation spaces
$\mathcal{H}_{\bar{p}}$ of
the representations $L$ of the isotropy subgroup $G_{\bar{p}} = SU(2, \mathbb{C})$.
In each case the finite-dimensional representations $L$ are naturally unitary
with the Krein fundamental symmetry $\mathfrak{J}_{\bar{p}}$ commuting with $L$, so that $L$
is unitary and Krein unitary.

In order to have a quantum field we need to have the representations, 
into which the single particle Hilbert space decomposes, to be local 
representations of $T_4 \circledS SL(2, \mathbb{C})$.

Thus, in the next step we need a machinery producing local representations out of 
\[
U^{{}_{\bar{p}} L}, \,\,\,\, \bar{p} = (m, 0,0,0), m \in \mathbb{R}, \textrm{or}
\,\,\, \bar{p} = (1,0,0,1)
\]
(we allow both signs of the parameter $m$ because we will use also the construction 
of negative energy free quantum fields).

We have presented above a general natural construction of a local version of Krein-isometric
induced representation
(although we will encounter other more exotic non-unitary and non-Krein-isometric construction of a local representation,
but it seems to be of less importance for physics).
However, the natural local version of the induced representation
is in general non-unitary but Krein-isometric. Surprise lies in the fact that even if we start
with the induced representations of the class
\[
U^{{}_{\bar{p}} L}, \,\,\,\, \bar{p} = (m, 0,0,0), m \in \mathbb{R}
\]
which are unitary ($L$ commutes with $\mathfrak{J}_{\bar{p}}$)
with their natural local versions
\[
W_{{}_{V,\bar{p}}} U^{{}_{\bar{p}} L} W_{{}_{V,\bar{p}}}^{-1}
\]
which are Krein-isometric, but in general are not unitary and even unbounded. This is for example the case
for the induced unitary (and naturally Krein-unitary, compare Subsection \ref{e1})
representation
\begin{equation}\label{InducedVersionOfU^LforDiracPsi}
U^{{}_{m,0,0,0} (L^{{}^{1/2}} \oplus L^{{}^{1/2}})}, \,\,\,\, \bar{p} = (m, 0,0,0), |m| = m_{e}
\end{equation}
whose local and unitary version (unitarily equivalent to (\ref{InducedVersionOfU^LforDiracPsi}), constructed in
Subsection \ref{e1}, which is not equal to the
\emph{natural local version} constructed above)  serves as the
local unitary representation acting in the single particle space of the
ordinary local free quantum Dirac field with the ordinary local transformation formula.

The \emph{natural local version}
\begin{equation}\label{LocalVersionOfU^LforDiracPsi}
W_{{}_{V,m,0,0,0}} U^{{}_{m,0,0,0}\big( L^{{}^{1/2}} \oplus L^{{}^{1/2}}\big) }
W_{{}_{V,m,0,0,0}}^{-1},
\end{equation}
of (\ref{InducedVersionOfU^LforDiracPsi}) is Krein isometric (unbounded) and non-unitary.
Strictly speaking the representation
\begin{equation}\label{LocalVersionOfU^LforDiracPsi0}
\Big( W_{{}_{V,m,0,0,0}} U^{{}_{m,0,0,0}\big( L^{{}^{1/2}} \oplus L^{{}^{1/2}}\big) }
W_{{}_{V,m,0,0,0}}^{-1}\Big)_{{}_{0}}
\end{equation}
associated to the natural local version (\ref{LocalVersionOfU^LforDiracPsi}) as in the last Proposition,
is Krein isometric and non-unitary. In this Section we deal only with decomposition problem of tensor product
of single particle representations and their relation to space-time.
Construction of free fields (based on single particle representations and Hida white noise operators)
and interacting fields (within causal perturbative approach)
is undertaken in the remaining Sections.

The natural local version (\ref{LocalVersionOfU^LforDiracPsi}) of (\ref{InducedVersionOfU^LforDiracPsi}) is no longer unitary
but Krein-isometric and unbounded.

The representation
\[
U^{{}_{m,0,0,0}\big(2L^{{}^{1/2}}\big)} \cong U^{{}_{m,0,0,0}\big(L^{{}^{1/2}}\big)} \oplus U^{{}_{m,0,0,0}\big(L^{{}^{1/2}}\big)}
\]
has the natural Krein structure, induced by its local version (\ref{LocalVersionOfU^LforDiracPsi0}).
In this case the Krein structure, which is induced by the natural local version, pulled back through the Krein-isometric
isomprphism, $W_{{}_{V,m,0,0,0}}$, composed with the Krein-isometric isomorphism
of the last Proposition, can be easily computed explicitly. The Krein fundamental symmetry operator is equal to the operator of multiplication
by a constant matrix (which follows from the general analysis placed above, \emph{i.e.} from the above stated Theorem and Proposition),
which is equal $\boldsymbol{1}$ on the representation space of the first direct summand of
\[
U^{{}_{m,0,0,0}\big(2L^{{}^{1/2}}\big)} \cong U^{{}_{m,0,0,0}\big(L^{{}^{1/2}}\big)} \oplus U^{{}_{m,0,0,0}\big(L^{{}^{1/2}}\big)}.
\]
and equal to $-\boldsymbol{1}$ on the representation space of the second direct summand of
\[
U^{{}_{m,0,0,0}\big(2L^{{}^{1/2}}\big)} \cong U^{{}_{m,0,0,0}\big(L^{{}^{1/2}}\big)} \oplus U^{{}_{m,0,0,0}\big(L^{{}^{1/2}}\big)}.
\]

We also construct in Subsection \ref{e1}, in a less natural manner, a local and non-unitary
version
\begin{equation}\label{LocalVersionOfU^LforDiracPsi00}
\Big(W_{{}_{V,m,0,0,0}} U^{{}_{m,0,0,0}\big(2L^{{}^{1/2}}\big)}
W_{{}_{V,m,0,0,0}}^{-1}\Big)_{{}_{00}},
\end{equation}
such that the unitary and local version, acting in the single particle Dirac field, is equal to a (non-orthogonal) direct summand
of 
\[
\Big(W_{{}_{V,m,0,0,0}} U^{{}_{m,0,0,0}\big(2L^{{}^{1/2}}\big)}
W_{{}_{V,m,0,0,0}}^{-1}\Big)_{{}_{00}} \oplus 
\Big(W_{{}_{V,-m,0,0,0}} U^{{}_{m,0,0,0}\big(2L^{{}^{1/2}}\big)}
W_{{}_{V,m,0,0,0}}^{-1}\Big)_{{}_{00}},
\]
(not equal to the
\emph{natural local version} constructed above) of the representation 
\[
U^{{}_{m,0,0,0}\big(2L^{{}^{1/2}}\big)} \oplus U^{{}_{m,0,0,0}\big(2L^{{}^{1/2}}\big)},
\]
and which is
unitarily equivalent to (\ref{InducedVersionOfU^LforDiracPsi}).

Similarly, starting from the \emph{natural local version} (defined above)
\begin{equation}\label{LocalVersionOfU^LforA}
W_{{}_{V,1,0,0,1}}U^{{}_{(1,0,0,1)}{\L}}W_{{}_{V,1,0,0,1}}^{-1} \\
\,\,\,\,\, \textrm{concentrated on the cone orbit} \,
\mathscr{O}_{(1, 0, 0, 1)},
\end{equation}
of a natural Krein-isometric representation
\[
U^{{}_{(1,0,0,1)}{\L}} \,\, \textrm{concentrated on the cone orbit} \,
\mathscr{O}_{(1, 0, 0, 1)}
\]
induced by a natural Krein-unitary representation, which we call {\L}opusza\'nski
representation, as the single particle space, we construct the ordinary quantum electromagnetic
potential field $A$ in the Gupta-Bleuler local gauge. Strictly speaking we construct two versions of the
free electromegnetic potential field in Sections \ref{free-gamma} and \ref{white-noise-proofs},
one based on the representation (\ref{LocalVersionOfU^LforA}) and the other one based on
the Krein-isometric representation
\begin{equation}\label{LocalVersionOfU^LforA0}
\big(W_{{}_{V,1,0,0,1}}U^{{}_{(1,0,0,1)}{\L}}W_{{}_{V,1,0,0,1}}^{-1}\big)_{{}_{0}} \\
\,\,\,\,\, \textrm{concentrated on the cone orbit} \,
\mathscr{O}_{(1, 0, 0, 1)},
\end{equation}
related to (\ref{LocalVersionOfU^LforA}) as in the last Proposition. The two free local electromagnetic
potential fields have identical Pauli-Jordan (zero-mass) distribution functions, and pairing functions.
Both versions are Krein-isometrically equivalent. They nonetheless have slightly different behavior in the infrared regime (compare Subsect. \ref{equivalentA-s}). The version based on
(\ref{LocalVersionOfU^LforA0}) has identical formula for the field $A$ as the standard one
(in the local Gupta-Bleuler gauge) the only difference is that we are using
the Hida operators as the creation-annihilation operators.

But these are the \emph{natural local versions} (defined above)
\begin{equation}\label{LocU^L-m,0,0,0-and-1,0,0,1}
W_{{}_{V,\bar{p}}} U^{{}_{\bar{p}} L}W_{{}_{V,\bar{p}}}^{-1}, \,\,\,\, \bar{p} = (m, 0,0,0), m \in \mathbb{R}, \textrm{or}
\,\,\, \bar{p} = (1,0,0,1)
\end{equation}
of the induced Krein-isometric (unitary in case $\bar{p} = (m, 0,0,0)$, $m \in \mathbb{R}$) representations
\[
U^{{}_{\bar{p}} L}, \,\,\,\, \bar{p} = (m, 0,0,0), m \in \mathbb{R}, \textrm{or}
\,\,\, \bar{p} = (1,0,0,1),
\]
or rather the Krein-unitary representations
\begin{equation}\label{LocU^L-m,0,0,0-and-1,0,0,1_0}
\Big(W_{{}_{V,\bar{p}}} U^{{}_{\bar{p}} L}W_{{}_{V,\bar{p}}}^{-1}\Big)_{{}_{0}}, \,\,\,\, \bar{p} = (m, 0,0,0), m \in \mathbb{R}, \textrm{or}
\,\,\, \bar{p} = (1,0,0,1)
\end{equation}
related to (\ref{LocU^L-m,0,0,0-and-1,0,0,1}) as in the last Proposition, which enter into direct inegral
decompositions of the natural space-time spectral triples ating on multispinors living on the Minkowski
space-time manifold together with its natural Krein structure.
It turns out that the decompositions of the finite sums of single particle unitary reprsentations 
and their finite direct sums and finite tensor products, which serve as single particle
representations of realistic free massive fields, are naturally Krein isometrically equivalent to the decompositions
of the Krein-isometric representations acting on the natural spectral triples acting on space-time multispinors.

Although we are basically interested in the representations (\ref{LocU^L-m,0,0,0-and-1,0,0,1_0}), their direct sums (as single particle representations) and their tensor products it is important to treat the Krein-unitary
bounded representations (\ref{LocU^L-m,0,0,0-and-1,0,0,1_0}) not as separate in relation to
(\ref{LocU^L-m,0,0,0-and-1,0,0,1}), but instead it is much better to consider instead
(\ref{LocU^L-m,0,0,0-and-1,0,0,1}). This allows both: to treat both cases at once and allows working out an efficient decomposition of tensor products of representations
(\ref{LocU^L-m,0,0,0-and-1,0,0,1_0}). Although the representations (\ref{LocU^L-m,0,0,0-and-1,0,0,1}) may appear at first glance as more singular, being unbounded (and in some respect indeed are), theory of decompositions of their tensor products is much simpler than decomposition of tensor products of the bounded
Krein-unitary representations (\ref{LocU^L-m,0,0,0-and-1,0,0,1_0}), treated by themselves
in their own Krein-Hilbert spaces and their tensor products\footnote{Theory of decompositions of Krein-unitary representations becomes quite involved even in the Pontriagin space, which the degenerate case of Krein space in which the eigenspace of the fundamental symmetry operator $\mathfrak{J}$ corresponding
to $-1$ is finite dimensional, compare \cite{Neumark0}-\cite{Neumark2}}. In our treatment
\[
\Big(W_{{}_{V,\bar{p}}} U^{{}_{\bar{p}} L}W_{{}_{V,\bar{p}}}^{-1}\Big)_{{}_{0}}
\,\,\,\,
\textrm{and} \,\,\,\,
W_{{}_{V,\bar{p}}} U^{{}_{\bar{p}} L}W_{{}_{V,\bar{p}}}^{-1}
\]
always come together, having the common dense nuclear subspace $E$. More adequately we should write
$E_{{}_{\bar{p}, \textrm{dim} \, L}}$ for $E$, as $E$ depends on the orbit $\mathscr{O}_{\bar{p}}$ and
dimension $\textrm{dim} \, L$ of the representation $L$, and is equal to the nuclear space of restrictions
\[
E_{{}_{\bar{p}, \textrm{dim} \, L}} =
\big\{f|_{{}_{\mathscr{O}_{\bar{p}}}}, \,\,\,\, f\in
\mathcal{S}(\mathbb{R}^4; \mathbb{C}^{\textrm{dim} \, L})\big\},
\]
where in practice we have finite dimensional $V,L$ acting in the Krein space $(\mathbb{C}^k, \mathfrak{J}_{\bar{p}})$. This $E_{{}_{\bar{p},\textrm{dim} \, L}}$ works well except the cone orbit
$\mathscr{O}_{1,0,0,1}$
but is sufficient for the decomposition problem we are dealing with in this Section, where
the orbits $\mathscr{O}_{m,0,0,0}$, $m\neq 0$, and $\mathscr{O}_{0,m,0,0}$, $m>0$, are essential.
Representations we encounter in depositions in these Sections, which are concentrated on the two
remaining orbits $\mathscr{O}_{0,0,0,0}$, $\mathscr{O}_{1,0,0,1}$ can be treated separately
without any need for the nuclear subspace in these representations
\[
U^{{}_{\bar{p}} L}, \,\,\,\, \bar{p} = (1,0,0,1), \,\,\,\, \textrm{or}
\,\,\, \bar{p} = (0,0,0,0).
\]
This is because the representations concentrated on
$\mathscr{O}_{0,0,0,0} = (0,0,0,0)$ we encounter here are finite dimensional, and
\[
f|_{{}_{\mathscr{O}_{1,0,0,1}}}, \,\,\,\, f\in
\mathcal{S}(\mathbb{R}^4; \mathbb{C}^{\textrm{dim} \, L})\big\},
\]
belong to the common domain
$\mathcal{H'}_{{}_{0}}\cap\mathcal{H'}$ of the representations
\[
\Big(W_{{}_{V,1,0,0,1}} U^{{}_{1,0,0,1} L}W_{{}_{V,1,0,0,1}}^{-1}\Big)_{{}_{0}}
\,\,\,\,
\textrm{and} \,\,\,\,
W_{{}_{V,1,0,0,1}} U^{{}_{\bar{p}} L}W_{{}_{V,1,0,0,1}}^{-1}
\]
always coming together in our treatment of decompositions.

Having given the pair of representations
\[
\Big(W_{{}_{V,\bar{p}}} U^{{}_{\bar{p}} L}W_{{}_{V,\bar{p}}}^{-1}\Big)_{{}_{0}}
\,\,\,\,
\textrm{and} \,\,\,\, 
W_{{}_{V,\bar{p}}} U^{{}_{\bar{p}} L}W_{{}_{V,\bar{p}}}^{-1}
\]  
we can pass to the Krein-Hilbert space of the first one by taking the closure of the common nuclear space
$E_{{}_{\bar{p},\textrm{dim} \, L}}$ with respect to the inner product (\ref{innProd(WU^LW^-1)_0}) of
$\mathcal{H'}_{{}_{0}}$ (compare the last Proposition). It is important that 
$E_{{}_{\bar{p},\textrm{dim} \, L}}$ lies in the also in the domain of the Krein-isometric (in general unbounded) operator $W_{{}_{V,\bar{p}}}^{-1}$, which  allows us to have a clear and computable linkage to the 
induced representations corresponding to the natural local versions. In particular decomposition of tensor product of natural local versions (\ref{LocU^L-m,0,0,0-and-1,0,0,1}) is reducible to decompositions of the tensor product of the corresponding induced representations
convergent on a dense subdomain of the involved representations. 
 
Now the whole point about the decomposition of the tensor product of representations
(\ref{LocU^L-m,0,0,0-and-1,0,0,1_0}) lies in using the decomposition theory for the corresponding
representations (\ref{LocU^L-m,0,0,0-and-1,0,0,1}), which can be reduced to the decomposition
of tensor products of induced Krein-isometric representations, which can in turn be reduced to the Fubini
theorem for the coset and double coset sub-manifolds in the group $T_4 \circledS SL(2, \mathbb{C})$,
presented in Appendix \ref{PartIIMackey}. We then use the fact that tensor product
of two representations (\ref{LocU^L-m,0,0,0-and-1,0,0,1}) and the tensor product of the
two corresponding representations (\ref{LocU^L-m,0,0,0-and-1,0,0,1_0}) have a common
dense domain, which have simple form
\[
f|_{{}_{\textrm{Sp} \, (P^0, \ldots, P^3)_{{}_{k}}}}, \,\,\,
f \in \mathcal{S}(\mathbb{R}^4; \mathbb{C}^k)\big\},
\]
on the invariant sub-spaces on which the translation generators $P^0, \ldots, P^3$ act with finite
uniform multiplicity $k$ and have joint spectrum (of uniform multiplicity $k$) $\textrm{Sp} \, (P^0, \ldots, P^3)_{{}_{k}}$. Elements of this common space can be decomposed with respect to the decomposition
inherited from the decomposition of tensor product of representations (\ref{LocU^L-m,0,0,0-and-1,0,0,1}).
One should note here that the translation generators are unitary equivalent in the pairs
\[
\Big(W_{{}_{V,\bar{p}}} U^{{}_{\bar{p}} L}W_{{}_{V,\bar{p}}}^{-1}\Big)_{{}_{0}}
\,\,\,\,
\textrm{and} \,\,\,\,
W_{{}_{V,\bar{p}}} U^{{}_{\bar{p}} L}W_{{}_{V,\bar{p}}}^{-1}
\]
and correspondingly in tensor products of the corresponding representations.
Decomposition just constructed cannot be of course prolonged all over the whole Hilbert space of the tensor
product of Krein unitary representations (\ref{LocU^L-m,0,0,0-and-1,0,0,1_0}). But this does not bother us as the constructed decomposition is sufficient for our purposes. Presented decomposition can be understood as an example of construction of generalized spectral measure, presented in \cite{Lanze} and
promoted mainly by Neumark \cite{NeumarkLinOp} and by Gelfand and his school \cite{GelfandIV}.

The fact that the decomposition of tensor product of Krein-unitary local representations 
(\ref{LocU^L-m,0,0,0-and-1,0,0,1_0}) cannot be decomposed in the sense of existence of ordinary 
spectral measure 
in the tensor product of Hilbert spaces $\mathcal{H'}_{{}_{0}}$ of these representations is what we expected, by comparing with ``spectral' analysis of relativistic wave operators on non-compact pseudo-Riemannian manifolds. In general such operators are non-normal and no ordinary spectral measure correspond
to them, although natural Krein-unitary representations are interconnected to them if they have non-trivial invariance under 
a Lie group of isometries of pseudo-Riemannian metric.

Thus, in order to analyze the representation of $T_{4} \circledS SL(2, \mathbb{C})$
in the Fock space of free fields (including the tensor product of Fock spaces of fields with both energy signs) we need to analyze
tensor product of the natural local versions of the induced representations
\[
U^{{}_{\bar{p}} L}, \,\,\,\, \bar{p} = (m, 0,0,0), m \in \mathbb{R}, \,\,\, \textrm{or}
\,\,\, \bar{p} = (1,0,0,1).
\]
Their natural local versions
\[
W U^{{}_{\bar{p}} L} W^{-1}
\]
or more explicitly
\[
W_{{}_{V,\bar{p}}} U^{{}_{\bar{p}} L} W_{{}_{V,\bar{p}}}^{-1},
\]
being Krein-isometrically equivalent to
\[
U^{{}_{\bar{p}} L}, \,\,\,\, \bar{p} = (m, 0,0,0), m \in \mathbb{R}, \,\,\, \textrm{or}
\,\,\, \bar{p} = (1,0,0,1),
\]
allow
reducing the problem of decomposition of tensor products of the natural local versions to the
problem of decomposition of tensor products of the initial induced representations
\[
U^{{}_{\bar{p}} L}, \,\,\,\, \bar{p} = (m, 0,0,0), m \in \mathbb{R}, \,\,\, \textrm{or}
\,\,\, \bar{p} = (1,0,0,1)
\]
(the general theory of decomposition of tensor product of Krein-isometric induced representations
is presented in Appendix \ref{PartIIMackey}). However,
in the class of orbits $\mathscr{O}_{(m, 0,0,0)}$, $m \in \mathbb{R}$
the problem of decomposition of local versions
of the representations can fairly be reduced to the Mackey's original theory
for unitary representations.

In particular tensor product of the unitary representations
concentrated on the orbit $\mathscr{O}_{\bar{p}} = \mathscr{O}_{(m, 0, 0, 0)}$
induced by the spin $s$ representation of the stability group $G_{\bar{p}} = SU(2, \mathbb{C})$
in this case) is equal (compare e.g. \cite{tat}) 
\begin{multline*}
U^{{}_{(m_1,0,0,0)} L^{{}^{s_1}}} \otimes U^{{}_{(m_2,0,0,0)} L^{{}^{s_2}}} = \bigoplus \limits_{s} \,\, [s_1, s_2 ,s] \int 
\limits_{m_1 + m_2 }^{\infty}
U^{{}_{(m,0,0,0)} L^{{}^{s}}} \, \ud m,    \,\,\,\,  m_1 , m_2 > 0 \\
\textrm{or} \\
= \bigoplus \limits_{s} \,\, [s_1, s_2 ,s] \int 
\limits_{- \infty }^{m_1 + m_2}
U^{{}_{(m,0,0,0)} L^{{}^{s}}} \, \ud m,    \,\,\,\,  m_1 , m_2 < 0
\end{multline*}
where $[s_1, s_2 , s]$ is the multiplicity of the representation standing immediately after the sign $[s_1, s_2 , s]$ 
and depending on $s_1 , s_2$ and on the spin $s$ of the integrated representation $U^{{}_{(m,0,0,0)} L^{{}^{s}}}$, 
but independent of the orbit $\mathscr{O}_{(m,0,0,0)}$ of the integrated representation $U^{{}_{(m,0,0,0)} L^{{}^{s}}}$,
and where $\ud m$ is the Lebesgue measure on $\mathbb{R}_+$ resp. $\mathbb{R}_-$.

Passing to the tensor product of the natural local versions
\[
W_{{}_{V_1,m_1,0}}U^{{}_{(m_1,0,0,0)} L^{{}^{s_1}}}W_{{}_{V_2,m_1,0}}^{-1}, 
W_{{}_{V,m_2,0}}U^{{}_{(m_2,0,0,0)} L^{{}^{s_2}}}W_{{}_{V,m_2,0}}^{-1}
\]
of the representations 
\[
U^{{}_{(m_1,0,0,0)} L^{{}^{s_1}}}, U^{{}_{(m_2,0,0,0)} L^{{}^{s_2}}}
\]
we have 
\begin{multline*}
\Big(W_{{}_{V_1,m_1,0}}U^{{}_{(m_1,0,0,0)} L^{{}^{s_1}}}W_{{}_{V_1,m_1,0}}^{-1}\Big) \otimes 
\Big(W_{{}_{V_2,m_2,0}}U^{{}_{(m_2,0,0,0)} L^{{}^{s_2}}}W_{{}_{V_2,m_2,0}}^{-1}\Big)
\\
\big(W_{{}_{V_1,m_1,0}} \otimes 
W_{{}_{V_2,m_2,0}}\big)
\big(U^{{}_{(m_1,0,0,0)} L^{{}^{s_1}}} \otimes U^{{}_{(m_2,0,0,0)} L^{{}^{s_2}}}\big)
\big(W_{{}_{V_1,m_1,0}} \otimes 
W_{{}_{V_2,m_2,0}}\big)^{-1} \\
= \bigoplus \limits_{s} \,\, [s_1, s_2 ,s] \int 
\limits_{m_1 + m_2 }^{\infty}
\big({W_{1\otimes 2}}\big)_{{}_{m}}U^{{}_{(m,0,0,0)} L^{{}^{s}}} \big({W_{1\otimes 2}}\big)_{{}_{m}}^{-1}\, \ud m,    \,\,\,\,  m_1 , m_2 > 0 \\
\textrm{or} \\
= \bigoplus \limits_{s} \,\, [s_1, s_2 ,s] \int 
\limits_{- \infty }^{m_1 + m_2}
\big({W_{1\otimes 2}}\big)_{{}_{m}}U^{{}_{(m,0,0,0)} L^{{}^{s}}}\big({W_{1\otimes 2}}\big)_{{}_{m}}^{-1} \, \ud m,    \,\,\,\,  m_1 , m_2 < 0
\end{multline*}
where 
\begin{multline*}
W_{{}_{V_1,m_1,0}} \otimes 
W_{{}_{V_2,m_2,0}}  
=\int 
\limits_{m_1 + m_2 }^{\infty} \big({W_{1\otimes 2}}\big)_{{}_{m}} \, \ud m, \\
\textrm{or} \\
=\int 
\limits_{m_1 + m_2 }^{\infty} \big({W_{1\otimes 2}}\big)_{{}_{m}} \, \ud m,
\end{multline*}
is the decomposition of the Krein-isometric operator
\[
W_{{}_{V_1,m_1,0}} \otimes W_{{}_{V_2,m_2,0}}.
\]

It is therefore reasonable to consider first the integrated representations (with appropriate multiplicities)
\[
\int U^{{}_{(m,0,0,0)} L^{{}^{s}}} \, \ud m.
\] 

We will construct now invariant sub-spaces $\mathcal{H}_{\textrm{inv}}$ of the Fock space,
 the transformation $V_\mathcal{F}$ on $\mathcal{H}_{\textrm{inv}}$ mentioned in the Subsect.  \ref{G}
 of Introduction, on the invariant sub-spaces $\mathcal{H}_{\textrm{inv}}$, and finally
the spectral tuple (\ref{SpacetimeTupleFields}) on $\mathcal{H}_{\textrm{inv}}$
describing the space-time in operator format.

We proceed by giving concrete examples of invariant sub-spaces $\mathcal{H}_{\textrm{inv}}$
in the tensor product of free fields of both energy signs. We think that going through concrete examples is the best way of showing what is going on. We do not specify the full content of free fields and proceed generally, assuming only that among the free fields there are the free Dirac field and the free
electromagnetic potential field, and we consider the tensor product of the Fock spaces of free fields 
of both energy signs.

We start our investigation with the invariant space 
of the  representation 
\begin{equation}\label{U^m,0,0,0Lstep1}
\int \limits_{\mathbb{R}} 
U^{{}_{(m,0,0,0)} 2L^{{}^{1/2}}} 
\, \ud m \bigoplus  \int \limits_{\mathbb{R}_+} U^{{}_{(0,m,0,0)} \big[2L^{{}^{1/2}}\big]_{\textrm{Ass}}} \, \ud m,
\end{equation}
i.e. a direct integral of positive energy representation and negative energy representation (of Example \ref{e1}) concentrated
resp. on the orbits $\mathscr{O}_{(m,0,0,0)}, p^0 > 0$ and $\mathscr{O}_{(-m,0,0,0)}, p^0 < 0$, both induced by
the spin $1/2$ representation of the stationary subgroup $G_{(m, 0, 0, 0)} = G_{(-m, 0, 0, 0)}
= SU(2, \mathbb{C})$ and both acting with uniform multiplicity two and summed up with the direct integral of naturally associated representations concentrated on $\mathscr{O}_{(0,0,m,0)}$, and described in Example \ref{e2}, induced
by the representation $\big[2L^{1/2}\big]_{\textrm{Ass}}$ of the group $G_{(0,0,m,0)} = SL(2, \mathbb{R})$
associated to the representation $2L^{1/2} = L^{1/2} \oplus L^{1/2}$;
and with the Lebesgue measure $\ud m$ on $\mathbb{R}_+$ (resp. $\mathbb{R}_-$, $\mathbb{R}$).

Strictly speaking we investigate invariant subspace $\mathcal{H}_{\textrm{inv}}'$ of the representation
Krein-isometrically equivalent to (\ref{U^m,0,0,0Lstep1}) with the direct summands replaced with
their natural local versions:
\begin{multline}\label{step1repLoc}
\int \limits_{\mathbb{R}} 
W_{{}_{V,m,0}}U^{{}_{(m,0,0,0)} 2L^{{}^{1/2}}} W_{{}_{V,m,0}}^{-1} 
\, \ud m \bigoplus \\
\bigoplus
  \int \limits_{\mathbb{R}_+}
  W_{{}_{V,0,m}} U^{{}_{(0,m,0,0)} \big[2L^{{}^{1/2}}\big]_{\textrm{Ass}}} W_{{}_{V,0,m}}^{-1} \, \ud m
\end{multline}
with the representations $V$ independent of the obit in this decomposition, and with the representations
$L$ independent of the orbit within the class  $\mathscr{O}_{m,0,0,0}$, $m \in \mathbb{R}$.
 But in order to simplify notation we will write (\ref{U^m,0,0,0Lstep1})
 for the decomposition (\ref{step1repLoc}) of the corresponding natural local versions.
 $\mathcal{H}_{\textrm{inv}}'$ is the invariant subspace in the Fock space of fields 
 (in tensor product of Fock spaces of both energy signs) constructed with the help of the 
 natural local versions (\ref{LocU^L-m,0,0,0-and-1,0,0,1}) as single particle representations. 
 Strictly speaking this is true only for the first direct summand. The second direct summand cannot be 
 reached as a invariant subspace of the Fock space tensor product of free fields of both energy signs,
 but it is concentrated on the nonphysical part of the joint spectrum of translation generators $P^0, \dots,
 P^3$, outside the set-theoretic sum $C_+ \cup C_-$ of positive and negative energy cones. We can therefore 
 modify direct summand concentrated there (outside $C_+ \cup C_-$) without affection the physical content 
of the theory, and regard $\mathcal{H}_{\textrm{inv}}'$ as an invariant subspace of te Fock space (including  
fields with both energy signs) acted on effectively by the representation which coincides with the 
ordinary action of $T_4 \circledS SL(2, \mathbb{C})$  (in the tensor product
 Fock space of both energy signs) on the physical direct summand concentrated on 
$C_+ \cup C_-$. In fact the second (nonphysical) direct summand is uniquely determined 
by the firs direct summand supported on the physical spectrum $C_+ \cup C_-$.

The operator giving the Krein-isometric equivalence between 
(\ref{U^m,0,0,0Lstep1}) and (\ref{step1repLoc}) is equal
\[ 
\int \limits_{\mathbb{R}} W_{{}_{V,m,0}} \, \ud m
\bigoplus \int \limits_{\mathbb{R}_+} W_{{}_{V,0,m}} \, \ud m
\]
with the representations $V$ in this decomposition independent of the orbit. 
$\mathcal{H'}_{\textrm{inv}}$ is an invariant subspace in the Fock space (of both energy signs)
which are constructed from representations (\ref{LocU^L-m,0,0,0-and-1,0,0,1}). 

Then we pass to the investigation of the invariant subspace $\mathcal{H}_{\textrm{inv}}$
of the representation
 \begin{multline}\label{step1repLoc0}
\int \limits_{\mathbb{R}} 
\Big(W_{{}_{V,m,0}}U^{{}_{(m,0,0,0)} 2L^{{}^{1/2}}} W_{{}_{V,m,0}}^{-1}\Big)_{{}_{0}} 
\, \ud m \bigoplus \\
\bigoplus
  \int \limits_{\mathbb{R}_+}
\Big(W_{{}_{V,0,m}} U^{{}_{(0,0,m,0)} \big[2L^{{}^{1/2}}\big]_{\textrm{Ass}}} 
W_{{}_{V,0,m}}^{-1}\Big)_{{}_{0}} \, \ud m
\end{multline}
$\mathcal{H}_{\textrm{inv}}$ is an invariant subspace in the Fock space (of both energy signs)
which are constructed from representations (\ref{LocU^L-m,0,0,0-and-1,0,0,1_0})
or (\ref{LocU^L-m,0,0,0-and-1,0,0,1}), and thus coinciding with the ordinary free fields.

In the following three Subsections \ref{e1}-\ref{1/2VF} we will analyze in details 
the representations (\ref{step1repLoc}) and (\ref{step1repLoc0}),
and construct the transform $V_\mathcal{F}$ on $\mathcal{H}_{\textrm{inv}}$ of Subsect.  \ref{G}
 of Introduction, on the invariant sub-spaces $\mathcal{H}_{\textrm{inv}}$, and finally
the spectral tuple (\ref{SpacetimeTupleFields}) on $\mathcal{H}_{\textrm{inv}}$.

The method is simple and is based on comparison of the representation (\ref{step1repLoc0})
with the Krein-Hilbert space of square summable space-time bispinors on the space-time with respect to the ordinary invariant measure, and the Krein-unitary representation $U^0$ of $T_4 \circledS SL(2, \mathbb{C})$
acting on these bispinors. We show that the representation (\ref{step1repLoc0}) can indeed be identified with
the representation acting on square summable bispinors through a unitary and Krein-unitary equivalence operator 
$V_\mathcal{F}$, and that the translation generators $P^0, \ldots, P^3$ and the operators $Q^1, \ldots, Q^3$ 
in the representation (\ref{step1repLoc0}), which compose with $P^0, \ldots, P^3$ the canonical 
von Neumann pairs acting with uniform multiplicity $4$, are identified through the unitary 
and Krein-unitary equivalence with the operators $i\partial_0, \ldots, i\partial_3$ and operators 
of multiplication by coordinates $x_0, \ldots, x_3$ respectively in the representation acting 
on the square summable bispinors. 

We will show that this result cannot hold if in the representation we replace the 
direct summand representations
\[
\Big(W_{{}_{V,m,0}}U^{{}_{(m,0,0,0)} 2L^{{}^{1/2}}} W_{{}_{V,m,0}}^{-1}\Big)_{{}_{0}} 
\]
in (\ref{step1repLoc0}) by the unnatural local, non-Krein-isometric and non-unitary versions
\[
\Big(W_{{}_{V,m,0,0,0}} U^{{}_{m,0,0,0}\big(2L^{{}^{1/2}}\big)}  
W_{{}_{V,m,0,0,0}}^{-1}\Big)_{{}_{00}}, \,\, m>0
\]
of
\[
U^{{}_{(m,0,0,0)} 2L^{{}^{1/2}}}
\]
constructed in Subsection \ref{e1}.

In general the Pauli-Bogoliubov-Shirkov quantization postulate is preserved by the free massive
fields which are constructed from the unitary representations (not necessarily irreducible)
\[
U^{{}_{m,0,0,0}L},
\]
induced by (not necessarily irreducible) unitary representations ${}_{m,0,0,0}L$ of the small
group $T_{4} \cdot SU(2,\mathbb{C})$ pertinent to the orbit $O_{{}_{m,0,0,0}}$, and Krein isometric with their
natural local versions (\ref{LocU^L-m,0,0,0-and-1,0,0,1_0})
or (\ref{LocU^L-m,0,0,0-and-1,0,0,1}). 
We will show it for the free Dirac field. The Pauli-Bogoliubov-Shirkov quantization postulate is also true for the
free electromagnetic potential field, which we will prove in Subsection \ref{BSH}.

One can compose a more general class of direct integral decomposition sub-representations  
of the representation acting in the Fock space (of fields of both energy signs)
which are unitarily and Krein-unitarily 
identifiable with the square-summable space-time multispinors. This is the case for direct
integral sub-representations if all direct summands in them
are \emph{naturally associated}, as in (\ref{step1repLoc0}).

In the following steps we will construct an infinity  of other more complex examples
of invariant sub-spaces $\mathcal{H}_{\textrm{inv}}$ corresponding to
sub spaces of representations on which translation generators have uniform spectrum equal 
$\mathbb{R}^4$ of arbitrary high (but finite) multiplicity on which the 
spectral space-time tuple can be naturally related to the translation generators. 
The method of construction is similar as for the representation (\ref{step1repLoc0})
and is based on comparison with square summable space-time multispinors based on 
Clifford algebras of arbitrary high dimension. 
The results are similar as for the representations (\ref{step1repLoc}) and 
(\ref{step1repLoc0}).

\subsection{Example 1: Representation $2U^{{}_{(m,0,0,0)} L^{{}^{1/2}}}
=U^{{}_{(m,0,0,0)} 2L^{{}^{1/2}}}$ (spin $1/2$) and the Dirac equation}\label{e1}

Let us consider the special case of the ordinary irreducible unitary representation $U^{{}_{(m,0,0,0)} L^{{}^{1/2}}}$ of
$T_{4} \circledS SL(2, \mathbb{C})$ induced\footnote{In this case the representation
$U^{{}_{(m,0,0,0)} L^{{}^{1/2}}}$ restricted to the subgroup $T_4 \cdot G_{\bar{p}}$ is given by ${}_{(m,0,0,0)} L^{{}^{1/2}} (a \cdot \gamma) =
{}_{\chi_{\bar{p}}} L^{{}^{1/2}}(a \cdot \gamma)
= \chi_{\bar{p}}(a)L^{{}^{1/2}}(\gamma)$, where $\chi_{\bar{p}}$ is the character on $T_4$
equal $\chi_{\bar{p}}(a) = e^{ia \cdot \bar{p}}$. The representation $U^{{}_{(m,0,0,0)} L^{{}^{1/2}}}$
is likewise induced by the representation ${}_{(m,0,0,0)} L^{{}^{1/2}}$ of the subgroup $T_4 \cdot G_{\bar{p}}$
in the sense of Mackey.} by the irreducible unitary spin $1/2$ representation $L^{{}^{1/2}}$
of the small subgroup $G_{\bar{p}} = SU(2, \mathbb{C}) \subset SL(2, \mathbb{C})$
stationary for $\bar{p} = (m, 0, 0, 0)$, and concentrated on the orbit
$\mathscr{O}_{\bar{p}} = \mathscr{O}_{(m,0,0,0)} = \{p: (p^0)^2 - \vec{p} \cdot \vec{p} = m^2 \}$. In this case
the representation $\gamma \mapsto Q(\gamma, \bar{p})$ constructed above is equal to the irreducible
unitary spin $1/2$ representation $L^{{}^{1/2}}$ of $G_{\bar{p}} = SU(2, \mathbb{C})$. In this case it is customary to
choose ($\beta(p)$ is not of course unique)
\[
\beta(p) = m^{-1/2} \big(p^0 \bold{1} - \vec{p} \cdot \vec{\sigma} \big)^{1/2}, \,\,\,\,
\beta(p)^{-1} = m^{-1/2} \big(p^0 \bold{1} + \vec{p} \cdot \vec{\sigma} \big)^{1/2}.
\]
By definition of our representation we choose
\[
SU(2,\mathbb{C}) \ni \gamma \mapsto Q(\gamma, \bar{p}) = L^{{}^{1/2}}(\gamma) = \gamma, \,\,\,\,
SL(2,\mathbb{C}) \ni \alpha \mapsto V(\alpha) = \alpha,
\]
for the construction of the natural local version of the representation $U^{{}_{(m,0,0,0)} L^{{}^{1/2}}}$
and use $\widetilde{\varphi}$ for the elements in its representation space.
In this case $\widetilde{\varphi}$ is given by the formula:
\[
p \mapsto \widetilde{\varphi}(p)
= W'' \widetilde{\psi}(p) = V(\beta(p)^{-1}) \widetilde{\psi}(p) =
W''W'U^{-1}f(p)= Wf(p),
\]
for $\widetilde{\psi}$ in the representation space of the representation $W'U^{-1}U^{{}_{(m,0,0,0)} L^{{}^{1/2}}}UW'^{-1}$ and for $f$ in the representation space of the representation
$U^{{}_{(m,0,0,0)} L^{{}^{1/2}}}$. Recall that here $U^{-1}$ is the Krein-unitary operator of Lemma \ref{InducedFormToImprimitivity} of Subsection \ref{lop_ind}, $W'$ is the Krein-isometric
operator (\ref{W1}) and $W''$ is the Krein-unitary operator (\ref{W2}) and finally $W = W''W'U^{-1}$
is the operator constructed in the introduction to this Section. Here by assumption the Krein fundamental symmetry
$\mathfrak{J}_{\bar{p}} = \boldsymbol{1}_2$ in the representation space $\mathcal{H}_{\bar{p}}
= \mathbb{C}^2$ of $L^{1/2}$, so that all three operators $W'',W',U^{-1}$ are unitary (and thus bounded).

All the operators $ W'',W',U^{-1}$ depend on the current 
orbit and on the representation $V$, and should more adequately be written
\[
W''_{{}_{V,m,0,0,0}}, W'_{{}_{V,m,0,0,0}}, U_{{}_{L^{1/2},m,0,0,0}}. 
\]
Strictly speaking the operator $U$ of Lemma \ref{InducedFormToImprimitivity} of Subsection \ref{lop_ind}
is uniquely determined by the inducing representation (here $L^{1/2}$) and the orbit, but here can be regarded as a function of $V$ extending $L^{1/2}$ and thus uniquely determining its restriction $L^{1/2}$ to
the subgroup $G_{\bar{p}} = SU(2, \mathbb{C})$. 

Similarly, all $\widetilde{\psi}, f$ and $\widetilde{\varphi}$ can be regarded as  functions on the 
orbit $\mathscr{O}_{m,0,0,0}$ and should more properly be written as
\[
\widetilde{\varphi}_{m,0}, \widetilde{\psi}_{m,0}, f_{m,0}.
\]
We discard whenever possible indicating this dependence
on the orbit and $V$ in order to simplify notation. 

$\widetilde{\varphi}$ has the following transformation law
\[
\begin{split}
U(\alpha) \widetilde{\varphi} (p) = \alpha \widetilde{\varphi} (\Lambda(\alpha)p),
T(a) \widetilde{\varphi}(p) = e^{i a \cdot p}\widetilde{\varphi}(p). \\
\end{split}
\]
The Fourier transform (\ref{F(varphi)}) of $\widetilde{\varphi}$ has the local transformation 
law
\[
\begin{split}
U(\alpha)\varphi(x) = \alpha \varphi(\Lambda(\alpha)x)   \\
T(a)\varphi(x) = \varphi(x - a).
\end{split}
\]
The inner product of $\widetilde{\varphi} = W'' \widetilde{\psi}$ and $\widetilde{\varphi'} = 
W'' \widetilde{\psi'}$ is equal
\begin{multline*}
\big(\widetilde{\varphi}, \widetilde{\varphi'} \big) = \int \limits_{\mathscr{O}_{\bar{p}}}
\big(\widetilde{\psi}(p), \widetilde{\psi'}(p) \big) \, \ud \mu |_{{}_{\mathscr{O}_{\bar{p}}}} (p) 
= \int \limits_{\mathscr{O}_{\bar{p}}}
\big( \beta(p) \widetilde{\varphi}(p), \beta(p) \widetilde{\varphi'}(p) \big) \, 
\ud \mu |_{{}_{\mathscr{O}_{\bar{p}}}} (p) \\
= \int \limits_{\mathscr{O}_{\bar{p}}}
\big( \widetilde{\varphi}(p), \frac{1}{m} 
\big( p^0 \bold{1} - \vec{p} \cdot \vec{\sigma} \big) \widetilde{\varphi'}(p) \big)
 \, \ud \mu |_{{}_{\mathscr{O}_{\bar{p}}}} (p)
\end{multline*} 
where the inner product under the integration sign is equal to the inner product of the 
representation space of $ L^{{}^{1/2}}$, equal in this case to the ordinary $\mathbb{C}^2$
Hilbert space.

\vspace*{1cm}

\begin{center}
\small REPRESENTATION $U^{{}_{(m,0,0,0)} 2L^{{}^{1/2}}}$ AND THE CORRESPONDING 
GENERALIZED EIGENSPACE OF DIRAC OPERATOR. A NON-NATURAL AND UNITARY LOCAL VERSION OF 
THE REPRESENTATION 
$U^{{}_{(m,0,0,0)} 2L^{{}^{1/2}}}$
\end{center}

Let us connect the elements $\widetilde{\psi}$ of the space of the irreducible representation 
$W'U^{-1}U^{{}_{(m,0,0,0)} 2L^{{}^{1/2}}}UW'^{-1}$ concentrated on 
the orbit $\mathscr{O}_{(m,0,0,0)}$, $m > 0$, and induced by the spin $1/2$ representation of 
$G_{\bar{p}} = SU(2, \mathbb{C})$ 
with the positive energy solutions of the Dirac equation. 
This connection depends on the fact that besides the representation $ V: SL(2, \mathbb{C}) \ni \alpha \mapsto \alpha$ 
which extends the representation $L^{{}^{1/2}}: SU(2, \mathbb{C}) \ni \gamma \mapsto \gamma$ of the small
subgroup $SU(2, \mathbb{C})$, there exist another natural representation 
$\bar{V}: SL2, \mathbb{C}) \ni \alpha \mapsto {\alpha^*}^{-1}$ extending the representation 
$L^{{}^{1/2}}: SU(2, \mathbb{C}) \ni \gamma \mapsto \gamma$ (indeed ${\gamma^*}^{-1} = \gamma$
for unitary $\gamma \in SL(2, \mathbb{C})$). In this sense $V$ and $\bar{V}$ are conjugate:
$\bar{V}(\alpha) = V({\alpha^*}^{-1})$, $\alpha \in SL(2, \mathbb{C})$. Let us note however that the conjugation
of the representation $V$ depends on the orbit $\mathscr{O}_{(m, 0, 0, 0)}$ in question:
we call the pair of representations $V, \bar{V}$ of $SL(2, \mathbb{C})$ to be conjugate
with respect to the orbit $\mathscr{O}_{\bar{p}}$ if they act in the same space and
are equal on the small subgroup $G_{\bar{p}}$. Then we can introduce two spinors, one $\widetilde{\varphi}$, 
defined by $\widetilde{\varphi}(p) = V (\beta(p)^{-1}) \widetilde{\psi}(p) = \beta(p)^{-1} \widetilde{\psi}(p)$, 
and the contravariant conjugate
spinor $\widetilde{\chi}$, defined analogously by the conjugate representation $\bar{V}$: $\widetilde{\chi}(p) = \bar{V} (\beta(p)^{-1}) \widetilde{\psi}
= \beta(p) \widetilde{\psi}$ with the transformation rule for $\widetilde{\chi}$
\[
\begin{split}
U(\alpha) \widetilde{\chi}(p) = {\alpha^*}^{-1} \widetilde{\chi}(\Lambda(\alpha)p), \\
T(a) \widetilde{\chi}(p) = e^{i a \cdot p}\widetilde{\chi}(p);
\end{split}
\]  
and correspondingly for its Fourier transform $\chi$ (defined as in (\ref{F(varphi)})) 
\[
\begin{split}
U(\alpha) \chi(x) = {\alpha^*}^{-1} \chi(\Lambda(\alpha)x), \\
T(a) \chi(x) = \chi(x - a).
\end{split}
\]    

Consider now the mapping $\widetilde{\psi} \mapsto \widetilde{\phi} = \tvect{\widetilde{\varphi}}{\widetilde{\chi}}$
of the elements $\widetilde{\psi}$ of the representation space of $W'U^{-1}U^{{}_{(m,0,0,0)} L^{{}^{1/2}}}UW'^{-1}$ into the space of bispinors
$\widetilde{\phi} = \tvect{\widetilde{\varphi}}{\widetilde{\chi}}$ given by
\[
\left( \begin{array}{c}   \widetilde{\varphi}(p)   \\
                                           
                                           \widetilde{\chi}(p) \end{array}\right) 
= \left( \begin{array}{cc}   V (\beta(p)^{-1}) & 0  \\
                                           
                                                   0              & \bar{V} (\beta(p)^{-1})  \end{array}\right) 
\left( \begin{array}{c}  \widetilde{\psi}(p)  \\
                                           
                                           \widetilde{\psi}(p) \end{array}\right) 
= \left( \begin{array}{c}   \beta(p)^{-1} \, \widetilde{\psi}(p)   \\
                                           
                                          \beta(p) \, \widetilde{\psi}(p) \end{array}\right), 
\]
with the transformation law
\begin{equation}\label{1/2pUalpha}
U(\alpha) \widetilde{\phi}(p) = 
\left( \begin{array}{cc}   V (\alpha) & 0  \\
                                           
                                                   0              & \bar{V} (\alpha)  \end{array}\right) 
 \widetilde{\phi}(\Lambda(\alpha)p)
= \left( \begin{array}{cc}  \alpha & 0  \\
                                           
                                                   0              & {\alpha^*}^{-1}  \end{array}\right) 
 \widetilde{\phi}(\Lambda(\alpha)p), 
\end{equation}
\begin{equation}\label{1/2pUa}
T(a) \widetilde{\phi}(p) = e^{i a \cdot p}\widetilde{\phi}(p);
\end{equation}
and correspondingly for its Fourier transform $\phi$ (defined as in (\ref{F(varphi)})
with the replacement of $\varphi$ by $\phi$ in (\ref{F(varphi)}))
\begin{equation}\label{1/2xUalpha}
U(\alpha) \phi(x) = 
\left( \begin{array}{cc}  \alpha & 0  \\
                                           
                                                   0              & {\alpha^*}^{-1}  \end{array}\right) 
 \phi(\Lambda(\alpha)x), 
\end{equation}
\begin{equation}\label{1/2xUa}
T(a) \phi(x) = \phi(x - a);
\end{equation}
and with the inner product equal (and independent of the choice of the spacelike surface: $x^0 = constant$ 
of integration):
\begin{multline}\label{sprod-bisp-orb}
(\widetilde{\phi}, \widetilde{\phi'}) = m \int \limits_{x^0 = const.} \Big(\phi(x), \phi'(x) \Big)_{{}_{\mathbb{C}^4}}
\, \ud^3 x = m \int \limits_{\mathscr{O}_{\bar{p}}} \Big(\widetilde{\phi}(p), \widetilde{\phi}'(p) \Big)_{{}_{\mathbb{C}^4}}
\, \frac{\ud^3 \vec{p}}{(2  \varepsilon_{m}(\vec{p}))^{2}} \\
= m \int \limits_{\mathscr{O}_{\bar{p}}} \Big[ \Big(\widetilde{\varphi}(p), \widetilde{\varphi}'(p) \Big)_{{}_{\mathbb{C}^2}}
+ \Big(\widetilde{\chi}(p), \widetilde{\chi}'(p) \Big)_{{}_{\mathbb{C}^2}} \Big] \, 
\frac{\ud^3 \vec{p}}{(2  \varepsilon_{m}(\vec{p}))^{2}} \\
= m \int \limits_{\mathscr{O}_{\bar{p}}} \Big(\widetilde{\psi}(p), 
\Big( V(\beta(p)^{-1})^* \, V(\beta(p)^{-1}) + \bar{V}(\beta(p)^{-1})^* \, \bar{V}(\beta(p)^{-1}) \Big) 
\widetilde{\psi}'(p) \Big)_{{}_{\mathbb{C}^2}}  \, 
\frac{\ud^3 \vec{p}}{(2  \varepsilon_{m}(\vec{p}))^{2}} \\
= m \int \limits_{\mathscr{O}_{\bar{p}}} \Big(\widetilde{\psi}(p), 
\Big(\beta(p)^{-2} + \beta(p)^{2})\Big) 
\widetilde{\psi}'(p) \Big)_{{}_{\mathbb{C}^2}}  \, 
\frac{\ud^3 \vec{p}}{(2  \varepsilon_{m}(\vec{p}))^{2}}
= \int \limits_{\mathscr{O}_{\bar{p}}} \Big(\widetilde{\psi}(p), \widetilde{\psi}'(p) \Big)_{{}_{\mathbb{C}^2}}  \, 
\frac{\ud^3 \vec{p}}{2  \varepsilon_{m}(\vec{p})} \\
= \int \limits_{\mathscr{O}_{\bar{p}}} \Big(\widetilde{\psi}(p), \widetilde{\psi}'(p) \Big)_{\mathcal{H}_{\bar{p}}}  \, 
\ud \mu |_{{}_{\mathscr{O}_{\bar{p}}}} (p) = (\widetilde{\psi}, \widetilde{\psi'}), 
\end{multline}
because 
\[
V(\beta(p)^{-1})^* \, V(\beta(p)^{-1}) + \bar{V}(\beta(p)^{-1})^* \, \bar{V}(\beta(p)^{-1}) = \beta(p)^{-2} + \beta(p)^{2} = \frac{2 \varepsilon_m (\vec{p})}{m} \bold{1}.
\]
Here 
\[
\ud \mu |_{{}_{\mathscr{O}_{\bar{p}}}} (p) 
= \ud \mu |_{{}_{\mathscr{O}_{{}_{m, 0, 0, 0}}}} (p) = \ud \mu_{m,0} (p)  
=  \frac{\ud^3 \vec{p}}{2 \sqrt{m^2 + \vec{p} \cdot \vec{p}}} = \frac{\ud^3 \vec{p}}{2 \varepsilon_m (\vec{p})}
= \frac{\ud^3 \vec{p}}{2 p_0(\vec{p})}, 
\]
\[
\varepsilon_m (\vec{p}) = \sqrt{m^2 + \vec{p} \cdot \vec{p}} = p_0(\vec{p}), \,\,\, \textrm{on} \,\,\, \mathscr{O}_{{}_{m, 0, 0, 0}}
\]
and where in this case $\mathcal{H}_{\bar{p}} = \mathbb{C}^2$.
By construction the above representation acting on bi-spinors is unitary with respect to
the inner product (\ref{sprod-bisp-orb}), and the inner product in the space of Fourier transforms
$\phi$ of bispinors $\widetilde{\phi}$ is given by a standard local integral formula, i.e. as an integration 
over space-like surface of a  density function over the $x$-variables.

Now by the construction of the bispinor $\widetilde{\phi}= \tvect{\widetilde{\varphi}}{\widetilde{\chi}}$:
\[
\begin{split}
\widetilde{\varphi}(p) =  V(\beta(p)^{-1}) \widetilde{\psi}(p) \\
\widetilde{\chi}(p) =  \bar{V} (\beta(p)^{-1}) \widetilde{\psi}(p)
\end{split}
\]   
we have
\[
\begin{split}
\widetilde{\chi}(p) = \bar{V} (\beta(p)^{-1})V(\beta(p)^{-1})^{-1} \widetilde{\varphi}(p) 
= \beta(p)^2 \widetilde{\varphi}(p)\\
\widetilde{\varphi}(p) = V (\beta(p)^{-1}) \bar{V} (\beta(p)^{-1})^{-1}\widetilde{\chi}(p)
 = \beta(p)^{-2} \widetilde{\chi}(p) ,
\end{split}
\]   
or equivalently 
\[
\begin{split}
(\varepsilon_m (\vec{p}) \bold{1} - \vec{p} \cdot \vec{\sigma}) \widetilde{\varphi}(p)= m \widetilde{\chi}(p) \\
(\varepsilon_m (\vec{p}) \bold{1} + \vec{p} \cdot \vec{\sigma}) \widetilde{\chi}(p) = m \widetilde{\varphi}(p);
\end{split}
\]
or in a still more concise notation (summation with respect to $k = 1,2,3$):
\begin{equation}\label{alg-rel-1/2}
\big[ p^0 \gamma^0 - p^k \gamma^k \big] \widetilde{\phi}(p) = m \widetilde{\phi}(p),
\end{equation}
which in the $x$-space of Fourier transformed spinors gives the fulfilment of the Dirac equation
\begin{equation}\label{DiracEqBisp}
(i \widetilde{\gamma}^\mu \partial_\mu ) \phi = m \phi, 
\end{equation} 
\[
\widetilde{\gamma}^0 = \gamma^0, \,\,\,\,\,\,\,\, \widetilde{\gamma}^k = -\gamma^k,
\,\,\,\, \textrm{or} \,\, \tilde{\gamma}^\mu = \gamma^0 \gamma^\mu \gamma^0
\]
with 
\begin{equation}\label{chiralgamma}
\gamma^0 = \left( \begin{array}{cc}   0 &  \bold{1}_2  \\
                                           
                                                   \bold{1}_2              & 0 \end{array}\right), \,\,\,\,
\gamma^k = \left( \begin{array}{cc}   0 &  -\sigma_k  \\
                                           
                                                   \sigma_k             & 0 \end{array}\right),
\end{equation}
being the generators of the representation of the Clifford algebra:
\[
\gamma^\mu \gamma^\nu + \gamma^\nu \gamma^\mu = 2 g^{\mu \nu}
\]
associated to the Minkowski pseudo-metric $g^{\mu \nu}$. Thus the elements $\widetilde{\psi}$ 
of the representation
space of the irreducible representation $W'U^{-1}U^{{}_{(m,0,0,0)} L^{{}^{1/2}}}UW'^{-1}$, $m> 0$ correspond via the indicated 
unitary transformation $\widetilde{\psi} \mapsto \tvect{\varphi}{\chi} = \phi$ to the positive
energy solutions of the Dirac equation.

Now as the elements $\widetilde{\psi}, \widetilde{\phi}$ associated in the indicated way 
to the representation $U^{{}_{(m, 0, 0, 0)} L^{{}^{1/2}}}$ are concentrated on the orbit 
$\mathscr{O}_{\bar{p}} = \mathscr{O}_{(m, 0, 0, 0)}$, and now we consider 
the direct integrals over the orbits of the representations of the type 
\[
W'_{{}_{L^{1/2},m,0}}U_{{}_{L^{1/2},m,0}}^{-1}U^{{}_{m, 0,0,0)} L^{{}^{1/2}}}U_{{}_{L^{1/2},m,0}}{W'}_{{}_{L^{1/2},m,0}}^{-1},
\]
or more generally direct integrals over orbits of 
\[
\oplus_s W'_{{}_{L^s,m,0}}U_{{}_{L^{s},m}}^{-1}U^{{}_{m, 0,0,0)} L^{{}^{s}}}U_{{}_{L^{s},m,0}}{W'}_{{}_{L^s,m,0}}^{-1},
\]
it will be reasonable to reflect the orbit-dependence
of the respective elements $\widetilde{\psi}, \widetilde{\phi}$ by placing the subscript
$(m,0)$ (for $\mathscr{O}_{m,0,0,0}$) respectively at $\widetilde{\psi}, \widetilde{\phi}$: 
i.e. $\widetilde{\psi}_{{}_{m,0}}, \widetilde{\phi}_{{}_{m,0}}$, 
and at the Fourier transform $\phi_{{}_{m,0}}$
\begin{multline*}
\phi_{{}_{m,0}} (x) = (2\pi)^{-3/2} \int \limits_{\mathscr{O}_{\bar{p}}} 
\widetilde{\phi}_{{}_{\mathscr{O}_{\bar{p}}}}(p) e^{-ip \cdot x} \, 
\ud \mu |_{{}_{\mathscr{O}_{\bar{p}}}} (p)
= (2\pi)^{-3/2} \int \limits_{\mathscr{O}_{\bar{p}}} 
\widetilde{\phi}_{{}_{m,0}}(p) e^{-ip \cdot x} \, 
\ud \mu _{{}_{m,0}} (p) \\
= (2\pi)^{-3/2} \int \limits_{\mathscr{O}_{\bar{p}}} \widetilde{\phi}_{{}_{m,0}} (p) e^{-ip \cdot x} \, 
\frac{\ud^3 \vec{p}}{2 \varepsilon_{{}_{m}} (\vec{p})} 
\end{multline*}
of $\widetilde{\phi}_{{}_{0,m}}$.

Let us introduce the notation $V^\oplus$ for the isometric operator mapping the elements $\widetilde{\psi}_{{}_{m,0}}$ of the representation space of the representation 
\[
W'_{{}_{L^{1/2},m,0}}U_{{}_{L^{1/2},m,0}}^{-1}U^{{}_{(m,0,0,0)} L^{{}^{1/2}}}U_{{}_{L^{1/2},m,0}}
{W'_{{}_{L^{1/2},m,0}}}^{-1}
\] 
into the bispinors
$\tvect{\widetilde{\varphi}_{{}_{m,0}}}{\widetilde{\chi}_{{}_{m,0}}} = V^\oplus \widetilde{\psi}_{{}_{m,0}}$
given by the formula 
\[
V^\oplus \widetilde{\psi}_{{}_{m,0}}(p) = \left( \begin{array}{c} \beta(p)^{-1}  \widetilde{\psi}_{{}_{m,0}}(p)  \\

                                                  \beta(p)  \widetilde{\psi}_{{}_{m,0}}(p) \end{array}\right).
\]
We have thus constructed the isometric image  under $V^\oplus$ of the elements $\widetilde{\psi}_{{}_{m,0}}$ of the representation space of the representation 
\[
W'_{{}_{L^{1/2},m,0}}U_{{}_{L^{1/2},m,0}}^{-1}U^{{}_{(m,0,0,0)} L^{{}^{1/2}}}U_{{}_{L^{1/2},m,0}}{W'_{{}_{L^{1/2},m,0}}}^{-1}
\]
onto the closed subspace of the Hilbert space of bispinors $\tvect{\widetilde{\varphi}_{{}_{m,0}}}{\widetilde{\chi}_{{}_{m,0}}}$ with the inner product
\begin{equation}\label{inner-prod-bispin}
(\widetilde{\phi}_{{}_{m,0}}, \widetilde{\phi'}_{{}_{m,0}})  = 
  m \int \limits_{\mathscr{O}_{(m,0,0,0)}} \Big(\widetilde{\phi}_{{}_{m,0}}(p), 
\widetilde{\phi'}_{{}_{m,0}}(p) \Big)_{{}_{\mathbb{C}^4}}
\, \frac{\ud \mu_{{}_{m,0}}}{2 p^0 }. 
\end{equation}
However $V^\oplus$ is not onto the Hilbert space of bispinors concentrated on the orbit $\mathscr{O}_{(m,0,0,0)}$.
The closed image of $V^\oplus$ is characterised by the linear algebraic relation (\ref{alg-rel-1/2}) fulfilled 
at every point $p$ of the orbit by every element of the image, which means that the Fourier transform 
(\ref{F(varphi)}) of every element of its image fulfils
the standard Dirac equation.
This is why we consider the direct sum 
\begin{multline}\label{2UL1/2}
W'_{{}_{L^{1/2},m,0}}U_{{}_{L^{1/2},m,0}}^{-1}U^{{}_{(m,0,0,0)} L^{{}^{1/2}}}U_{{}_{L^{1/2},m,0}}{W'_{{}_{L^{1/2},m,0}}}^{-1} \bigoplus \\ \bigoplus
W'_{{}_{L^{1/2},m,0}}U_{{}_{L^{1/2},m,0}}^{-1}U^{{}_{(m,0,0,0)} L^{{}^{1/2}}}U_{{}_{L^{1/2},m,0}}{W'}_{{}_{L^{1/2},m,0}}^{-1} \\
= 2(W'_{{}_{L^{1/2},m,0}}U_{{}_{L^{1/2},m,0}}^{-1}) 2 U^{{}_{(m,0,0,0)} L^{{}^{1/2}}} 
2(W'_{{}_{L^{1/2},m,0}}U_{{}_{L^{1/2},m,0}}^{-1})^{-1} 
\end{multline}
of two identical copies of the representation 
\[
W'_{{}_{L^{1/2},m,0}}U_{{}_{L^{1/2},m,0}}^{-1}U^{{}_{(m,0,0,0)} L^{{}^{1/2}}}U_{{}_{L^{1/2},m,0}}{W'_{{}_{L^{1/2},m,0}}}^{-1},
\]
 i.e. the representation
\[ 
W'_{{}_{L^{1/2},m,0}}U_{{}_{L^{1/2},m,0}}^{-1}U^{{}_{(m,0,0,0)} L^{{}^{1/2}}}U_{{}_{L^{1/2},m,0}}{W'_{{}_{L^{1/2},m,0}}}^{-1}
\]
acting with uniform multiplicity 2.
Then to the first component, say  ${\widetilde{\psi}_{{}_{m,0}}}^{\oplus}$, of the direct sum Hilbert space of this direct sum representation (\ref{2UL1/2}) we apply the isometric map $V^\oplus$, but to the second component, say 
${\widetilde{\psi}_{{}_{m,0}}}^{\ominus}$
we apply another isometric map $V^\ominus$, which we are going to define now. Namely from what has been said it follows
that the map $V^\ominus$ on the representation space of the representation 
\[
W'_{{}_{L^{1/2},m,0}}U_{{}_{L^{1/2},m,0}}^{-1}U^{{}_{(m,0,0,0)} L^{{}^{1/2}}}U_{{}_{L^{1/2},m,0}}{W'_{{}_{L^{1/2},m,0}}}^{-1},
\]
defined by
\[
V^{\ominus} {\widetilde{\psi}_{{}_{m,0}}}^{\ominus}(p) =  {\widetilde{\phi}_{{}_{m,0}}}^{\ominus}(p)
 = \left( \begin{array}{c} \beta(p)^{-1}  \widetilde{\psi}_{{}_{m,0}}^{\ominus}(p)  \\                                      
                                                  -\beta(p)  {\widetilde{\psi}_{{}_{m,0}}}^{\ominus}(p) \end{array}\right)
\]
(differing from $V^\oplus$ by the minus sign in the second component) is likewise isometric, that every element of its image has the same transformation law, with the only difference that every element in the closed image of $V^\ominus$
fulfils the following linear algebraic relation (with summation with resect to $k=1,2,3$)
\[
\big[ p^0 \gamma^0 - p^k \gamma^k \big] {\widetilde{\phi}_{{}_{m,0}}}^{\ominus}(p) = 
-m {\widetilde{\phi}_{{}_{m,0}}}^{\ominus}(p),
\]
i.e. with the sign of the mass term reversed, which means that the Fourier transform 
$\phi^{\ominus}_{{}_{m,0}}$ of $ {\widetilde{\phi}_{{}_{m,0}}}^{\ominus}$ (given by (\ref{F(varphi)})) 
fulfills the Dirac equation with the sign at the mass term reversed (summation with respect to $\mu=0,1,2,3$):
\[
(i \widetilde{\gamma}^\mu \partial_\mu ) \phi^{\ominus}_{{}_{m,0}} = - m \phi^{\ominus}_{{}_{m,0}},
\] 
\[
\widetilde{\gamma}^0 = \gamma^0, \,\,\,\,\,\,\,\, \widetilde{\gamma}^k = -\gamma^k.
\]
Write  ${\widetilde{\phi}_{{}_{m,0}}}^{\oplus}$ for the image  
$V^{\oplus} \widetilde{\psi}_{{}_{m,0}}^{\oplus}$
of ${\widetilde{\psi}_{{}_{m,0}}}^{\oplus}$ under $V^\oplus$. 
The two images, namely the image of the first direct summand under $V^\oplus$ and the image of the second direct summand
under $V^\ominus$ we do not treat as orthogonal direct summands but literally as their images in one and the same 
Hilbert space of bispinors $\widetilde{\phi}_{{}_{m,0}}$ with the inner product
(\ref{inner-prod-bispin}). Therefore, we define the map $V^{\oplus \ominus}$ which every element
$({\widetilde{\psi}_{{}_{m,0}}}^{\oplus}, {\widetilde{\psi}_{{}_{m,0}}}^{\ominus})$ of the space of the direct sum
representation (\ref{2UL1/2}) maps into the bispinor 
${\widetilde{\phi}_{{}_{m,0}}}^{\oplus} + {\widetilde{\phi}_{{}_{m,0}}}^{\ominus}$ (ordinary sum of bispinors). 
This is well-defined, and as the respective images of $V^\oplus$ and $V^\ominus$ are closed and 
have zero as the only common element, but are not orthogonal. We thus constructed direct sum of two linearly disjoint, 
but not orthogonal, copies
of the representation $W'U^{-1}U^{{}_{(m,0,0,0)} L^{{}^{1/2}}}UW'^{-1}$, by the application of the map $V^{\oplus \ominus}$
to the (not orthogonal, but linearly disjoint) direct sum of two representations (\ref{2UL1/2}). 
Of course because the respective images (of the first summand under $V^\oplus$ and of the second under $V^\ominus$)
are not orthogonal the map $V^{\oplus \ominus}$ is not unitary (but $V^\oplus$ on the first summand and $V^\ominus$
on the second summand are separately unitary). 

Easy verification shows that $V^{\oplus \ominus}$ is onto the Hilbert space of all bispinors $\widetilde{\phi}_{{}_{m,0}}$
concentrated on $\mathscr{O}_{(m,0,0,0)}$ with finite norm 
\[
m \int \limits_{\mathscr{O}_{(m,0,0,0)}} \Big(\widetilde{\phi}_{{}_{m,0}}(p), 
\widetilde{\phi}_{{}_{m,0}}(p) \Big)_{{}_{\mathbb{C}^4}}
\, \frac{\ud \mu_{{}_{m,0}}}{2 p^0 }. 
\]  

Indeed, let $\widetilde{\phi}_{{}_{m,0}} =  
\tvect{\widetilde{\varphi}_{{}_{m,0}}}{\widetilde{\chi}_{{}_{m,0}}}$ be any
such bispinor concentrated on $\mathscr{O}_{(m,0,0,0)}$. Then it is equal $\widetilde{\phi}_{{}_{m,0}}
= {\widetilde{\phi}_{{}_{m,0}}}^{\oplus} + {\widetilde{\phi}_{{}_{m,0}}}^{\ominus} 
= V^\oplus {\widetilde{\psi}_{{}_{m,0}}}^{\oplus} + V^\ominus {\widetilde{\psi}_{{}_{m,0}}}^{\ominus}
= V^{\oplus \ominus} ({\widetilde{\psi}_{{}_{m,0}}}^{\oplus}, {\widetilde{\psi}_{{}_{m,0}}}^{\ominus})$ 
where ${\widetilde{\psi}_{{}_{m,0}}}^{\oplus}(p) = \frac{1}{2} \big\{ \beta(p)  \widetilde{\varphi}_{{}_{m,0}}(p)
+ \beta(p)^{-1}  \widetilde{\chi}_{{}_{m,0}}(p) \big\}$ and 
${\widetilde{\psi}_{{}_{m,0}}}^{\ominus}(p) = \frac{1}{2} \big\{ \beta(p)  \widetilde{\varphi}_{{}_{m,0}}(p)
- \beta(p)^{-1}  \widetilde{\chi}_{{}_{m,0}}(p) \big\}$.
 
From this easily follows the formula for the corresponding orthogonal projections
(strictly speaking idempotents which are not self-adjoint) $P^\oplus$ and $P^\ominus$
on the respective images of $V^\oplus$ and $V^\ominus$. Namely, they are equal to the operators
of multiplications by the respective orthogonal projections $P^\oplus(p)$ and $P^\ominus(p)$,
$p \in \mathscr{O}_{m,0,0,0}$, in $M_{4}(\mathbb{C})$:
\[
P^\oplus(p) = \frac{1}{2}
\left( \begin{array}{cc} 1 & \beta(p)^{-2} \\

\beta(p)^2 & 1 \end{array}\right), \,\,\,\,\,\,
P^\ominus(p) = \frac{1}{2}
\left( \begin{array}{cc} 1 & -\beta(p)^{-2} \\

-\beta(p)^2 & 1 \end{array}\right).
\]
Because of the orthogonality
\[
P^\oplus(p)P^\ominus(p) = P^\ominus(p)P^\oplus(p) = 0, \,\,\,\, P^\oplus(p)+ P^\ominus(p) = \boldsymbol{1}_4
\]
we have orthogonality
\[
P^\oplus P^\ominus = P^\ominus P^\oplus = 0, \,\,\,\, P^\oplus + P^\ominus = \boldsymbol{1}.
\]
Thus any bispinor ${\widetilde{\phi}_{{}_{m,0}}}^{\oplus}$ in the image of $V^\oplus$,
concentrated on $\mathscr{O}_{m,0,0,0}$, is equal to the image
$P^\oplus{\widetilde{\phi}_{{}_{m,0}}}$ of a generic bispinor ${\widetilde{\phi}_{{}_{m,0}}}$
concentrated on the orbit $\mathscr{O}_{m,0,0,0}$
and ${\widetilde{\phi}_{{}_{m,0}}} \in \textrm{Dom} \, P^\oplus$. Similarly, any bispinor
${\widetilde{\phi}_{{}_{m,0}}}^{\ominus}$ in the image of $V^\ominus$
is equal to $P^\ominus{\widetilde{\phi}_{{}_{m,0}}}$ for a generic bispinor ${\widetilde{\phi}_{{}_{m,0}}}$
concentrated on $\mathscr{O}_{m,0,0,0}$ and ${\widetilde{\phi}_{{}_{m,0}}} \in \textrm{Dom} \, P^\ominus$.
Although the closed subspaces $\textrm{Im} \, P^\oplus$ and $\textrm{Im} \, P^\ominus$ are linearly disjoint 
(with zero as the only common element in view of the orthogolality of $P^\oplus, P^\ominus$) they are not othogonal
because  the projections $P^\oplus, P^\ominus$ are not self adjoint.
Thus in general
${\widetilde{\phi}_{{}_{m,0}}}^{\oplus}$ and ${\widetilde{\phi}_{{}_{m,0}}}^{\ominus}$ are not orthogonal,
where ${\widetilde{\phi}_{{}_{m,0}}}^{\oplus}=P^\oplus{\widetilde{\phi}_{{}_{m,0}}}$ 
for some element ${\widetilde{\phi}_{{}_{m,0}}}$ lying
in the domain $\textrm{Dom} \, P^\oplus$ of the (unbounded\footnote{Note that for projectors (idempotents), 
for which by definition $P^2 = P$, the conditions $P^* = P$ and $\| P\| = 1$ are equivalent,
compare e.g. \cite{Lanze}.}) projector (idempotent) $P^\oplus$,
and respectively ${\widetilde{\phi}_{{}_{m,0}}}^{\ominus}=P^\ominus{\widetilde{\phi}_{{}_{m,0}}}$
for some ${\widetilde{\phi}_{{}_{m,0}}} \in \textrm{Dom} \, P^\ominus$.

Note also the simple identities
\begin{equation}\label{SimpleKreinIsometry2(1/2)m,0,0,0}
\begin{split}
\big({\widetilde{\phi}_{{}_{m,0}}}^{\oplus}(p), \mathfrak{J}{\widetilde{\phi}_{{}_{m,0}}}^{\oplus}(p) \big)_{{}_{\mathbb{C}^4}}
= 2 \big({\widetilde{\psi}_{{}_{m,0}}}^{\oplus}(p), {\widetilde{\psi}_{{}_{m,0}}}^{\oplus}(p) \big)_{{}_{\mathbb{C}^2}}
\\
\big({\widetilde{\phi}_{{}_{m,0}}}^{\ominus}(p), \mathfrak{J}{\widetilde{\phi}_{{}_{m,0}}}^{\ominus}(p) \big)_{{}_{\mathbb{C}^4}}
= -2 \big({\widetilde{\psi}_{{}_{m,0}}}^{\ominus}(p), {\widetilde{\psi}_{{}_{m,0}}}^{\ominus}(p) \big)_{{}_{\mathbb{C}^2}}
\end{split}
\end{equation}
\[
p \in \mathscr{O}_{{}_{m,0,0,0}},
\]
where $\mathfrak{J}$ is the operator of multiplication by the constant involutive unitary matrix $\gamma^0$
or $\gamma^1\gamma^2\gamma^3$ (Krein structure).

We have the analogous relation between the elements $({\widetilde{\psi}_{{}_{-m,0}}}^{\oplus}, 
{\widetilde{\psi}_{{}_{-m,0}}}^{\ominus})$ of the representation space of the direct sum 
\begin{multline*}
W'_{{}_{L^{1/2}}}U_{{}_{L^{1/2},-m,0}}^{-1}U^{{}_{(-m,0,0,0)} L^{{}^{1/2}}}U_{{}_{L^{1/2},-m,0}}{W'_{{}_{L^{1/2}}}}^{-1} \bigoplus \\ \bigoplus 
W'_{{}_{L^{1/2}}}U_{{}_{V,-m,0}}^{-1}U^{{}_{(-m,0,0,0)} L^{{}^{1/2}}}U_{{}_{V,-m,0}}{W'_{{}_{L^{1/2}}}}^{-1}
\end{multline*}
of two irreducible representations 
\[
W'_{{}_{L^{1/2},-m,0}}U_{{}_{L^{1/2},-m,0}}^{-1}U^{{}_{(-m,0,0,0)} L^{{}^{1/2}}}U_{{}_{L^{1/2},-m,0}}{W'_{{}_{L^{1/2},-m,0}}}^{-1},
\,\,\, m > 0,
\]
with the Hilbert space of bispinors concentrated on the orbit $\mathscr{O}_{-m,0,0,0}$, equipped with the 
analogous inner product
\begin{equation}\label{InnProdBispOnO_-m,0,0,0}
(\widetilde{\phi}_{{}_{-m,0}}, \widetilde{\phi'}_{{}_{-m,0}})  = 
  m \int \limits_{\mathscr{O}_{(-m,0,0,0)}} \Big(\widetilde{\phi}_{{}_{-m,0}}(p), 
\widetilde{\phi'}_{{}_{-m,0}}(p) \Big)_{{}_{\mathbb{C}^4}}
\, \frac{\ud \mu_{{}_{-m,0}}}{2 |p^0 |}. 
\end{equation}
(they correspond to the negative energy solutions of the Dirac equation being concentrated on the lower branch
of the two-sheeted hyperboloid). Note that in this case we have
\[
\beta(p)^{-2} = \frac{1}{-m} \big(p^0 \bold{1} + \vec{p} \cdot \vec{\sigma} \big)
= \frac{\widehat{p}}{-m}, \,\,\, p \in \mathscr{O}_{-m,0,0,0}, m>0,
\]
or 
\[
\beta(p)^{-2} = \frac{1}{m} \big(-p^0 \bold{1} - \vec{p} \cdot \vec{\sigma} \big), \,\,\,
p^0(\vec{p}) = - \sqrt{\vec{p} \cdot \vec{p} + m^2} = -\varepsilon_m (\vec{p}),
\]
so that
\[
\begin{split}
\beta(p) = m^{-1/2} \big(-p^0 \bold{1} + \vec{p} \cdot \vec{\sigma} \big)^{1/2}
= m^{-1/2} \big(\varepsilon_m (\vec{p}) \bold{1} + \vec{p} \cdot \vec{\sigma} \big)^{1/2}, \,\,\,\,
\\ 
\beta(p)^{-1} =  m^{-1/2} \big(-p^0 \bold{1} - \vec{p} \cdot \vec{\sigma} \big)^{1/2}
=  m^{-1/2} \big(\varepsilon_m (\vec{p}) \bold{1} - \vec{p} \cdot \vec{\sigma} \big)^{1/2}, \\
\textrm{for} \,\,\, p = (p^0, \vec{p}) = (- \sqrt{\vec{p} \cdot \vec{p} + m^2}, \vec{p}) \in \mathscr{O}_{-m,0,0,0},
\end{split} 
\]
with the analogous orthogonal projections $P^\oplus, P^\ominus$ equal to the operators of multiplication
by the mutually orthogonal projections
\[
P^\oplus(p) = \frac{1}{2}
\left( \begin{array}{cc} 1 & \beta(p)^{-2}  \\
                                           
                                                  \beta(p)^2 &  1 \end{array}\right), \,\,\,\,\,\,
P^\ominus(p) = \frac{1}{2}
\left( \begin{array}{cc} 1 & -\beta(p)^{-2}  \\
                                           
                                                  -\beta(p)^2 &  1 \end{array}\right).
\]
Note that by construction the orthogonal projection operators $P^\oplus, P^\ominus$ commute
with the (Fourier transformed Dirac) operator of pointwise  multiplication by the matrix
\[
p^0 \gamma^0 - p^k \gamma^k \,\,\,\,\,\,\, (\textrm{summation with respect to $k=1,2,3$}).
\] 

The analysis being completely analogous may be omitted, although we mention that the role
of ${\widetilde{\phi}_{{}_{-m,0}}}^{\oplus} = P^\oplus\widetilde{\phi}_{{}_{-m,0}}$ and 
${\widetilde{\phi}_{{}_{-m,0}}}^{\ominus} = P^\ominus \widetilde{\phi}_{{}_{-m,0}}$ 
is in a sense reversed.
Namely, the elements ${\widetilde{\phi}_{{}_{-m,0}}}^{\ominus}$ in the image of $V^\ominus$ (with the minus sign in the second component)
are characterised by the following linear algebraic relation ( with summation over $k = 1,2,3$, as usual)
\[
\big[ p^0 \gamma^0 - p^k \gamma^k \big] {\widetilde{\phi}_{{}_{-m,0}}}^{\ominus}(p) 
= m {\widetilde{\phi}_{{}_{-m,0}}}^{\ominus}(p),
\]
i.e. they correspond to the Dirac equation with the ordinary sign at the mass term, and vice versa for
the image ${\widetilde{\phi}_{{}_{-m,0}}}^{\oplus}$ under $V^\oplus$
which are characterised by the algebraic relation
\[
\big[ p^0 \gamma^0 - p^k \gamma^k \big] {\widetilde{\phi}_{{}_{-m,0}}}^{\oplus}(p) 
= - m {\widetilde{\phi}_{{}_{-m,0}}}^{\oplus}(p),
\]
which correspond via the Fourier transform to the solutions of the ordinary Dirac equation with
reversed sign at the mass term.

Note also the analogue identities
\begin{equation}\label{SimpleKreinIsometry2(1/2)-m,0,0,0}
\begin{split}
\big({\widetilde{\phi}_{{}_{-m,0}}}^{\oplus}(p), \mathfrak{J}{\widetilde{\phi}_{{}_{-m,0}}}^{\oplus}(p) \big)_{{}_{\mathbb{C}^4}}
= 2 \big({\widetilde{\psi}_{{}_{-m,0}}}^{\oplus}(p), {\widetilde{\psi}_{{}_{-m,0}}}^{\oplus}(p) \big)_{{}_{\mathbb{C}^2}}
\\
\big({\widetilde{\phi}_{{}_{-m,0}}}^{\ominus}(p), \mathfrak{J}{\widetilde{\phi}_{{}_{-m,0}}}^{\ominus}(p) \big)_{{}_{\mathbb{C}^4}}
= -2 \big({\widetilde{\psi}_{{}_{-m,0}}}^{\ominus}(p), {\widetilde{\psi}_{{}_{-m,0}}}^{\ominus}(p) \big)_{{}_{\mathbb{C}^2}}
\end{split}
\end{equation}
\[
p \in \mathscr{O}_{{}_{-m,0,0,0}},
\]
where $\mathfrak{J}$ is the operator of multiplication by the constant involutive unitary matrix $\gamma^0$
or $\gamma^1\gamma^2\gamma^3$ (Krein structure).

Let us remark at the end of this Subsection that introducing the fundamental symmetry operator $\mathfrak{J}$
into the Hilbert space of bispinors $\widetilde{\phi}_{{}_{m,0}}$ (or $\widetilde{\phi}_{{}_{-m,0}}$)
by the formula 
\[
\big(\mathfrak{J} \widetilde{\phi}_{{}_{\pm m,0}}\big)(p) 
= \gamma^0 \, \widetilde{\phi}_{{}_{\pm m,0}}(p),
\]
or by the formula
\[
\big(\mathfrak{J} \widetilde{\phi}_{{}_{ \pm m,0}}\big)(p) 
= \gamma^1 \gamma^2 \gamma^3 \, \widetilde{\phi}_{{}_{ \pm m,0}}(p)
\]
 we recover that 
${\widetilde{\phi}_{{}_{\pm m,0}}}^{\oplus}$ and ${\widetilde{\phi}_{{}_{ \pm m,0}}}^{\ominus}$
are orthogonal with respect to the Krein-inner product defined by
$\mathfrak{J}$: $\big({\widetilde{\phi}_{{}_{\pm m,0}}}^{\oplus}, 
\mathfrak{J} {\widetilde{\phi}_{{}_{m,0}}}^{\ominus} \big)=0$. 
Accordingly the projections $P^\oplus(p), P^\ominus(p)$, are Krein self adjoint
\[
\mathfrak{J} P^\oplus(p)^{*}\mathfrak{J} = P^\oplus(p),
\,\,\,\,\,\, \mathfrak{J} P^\ominus(p)^{*}\mathfrak{J} = P^\ominus(p), \,\,\,\, p \in \mathscr{O}_{{}_{\pm m,0,0,0}},
\]
for $\mathfrak{J} = \gamma^0$
or Krein skew self adjoint
\[
\mathfrak{J} P^\oplus(p)^{*}\mathfrak{J} = - P^\oplus(p),
\,\,\,\,\,\, \mathfrak{J} P^\ominus(p)^{*}\mathfrak{J} = - P^\ominus(p), \,\,\,\, p \in \mathscr{O}_{{}_{\pm m,0,0,0}},
\]
for $\mathfrak{J} = \gamma^1\gamma^2\gamma^2$.

Therefore the images, respectively, under $V^\oplus$ and $V^\ominus$ are
Krein-orthogonal. 

We shall likewise denote $V^{\oplus \ominus}$ by 
$V^\oplus \oplus V^{\ominus}$.  

The unitary local representation  acting on bispinors according to (\ref{1/2pUalpha}) and (\ref{1/2pUa}),
with the Hilbert space inner product  (\ref{inner-prod-bispin}) and fundamental
symmetry $\mathfrak{J}$, given by the formula 
$\big(\mathfrak{J} \widetilde{\phi}_{{}_{m,0}}\big)(p) 
= \gamma^0 \, \widetilde{\phi}_{{}_{m,0}}(p)$ or by
$\big(\mathfrak{J} \widetilde{\phi}_{{}_{m,0}}\big)(p) 
= \gamma^1 \gamma^2 \gamma^3 \, \widetilde{\phi}_{{}_{m,0}}(p)$, we denote by
\begin{equation}\label{U^2L^1/2_00m>0}
\Big(W_{{}_{V,m,0,0,0}} U^{{}_{m,0,0,0}\big(2L^{{}^{1/2}}\big)}  
W_{{}_{V,m,0,0,0}}^{-1}\Big)_{{}_{00}}, \,\, m>0.
\end{equation}

Analogously the representation, acting similarly on the bispinors concentrated on $\mathscr{O}_{-m,0,0,0}$
$m>0$ with the inner product (\ref{InnProdBispOnO_-m,0,0,0}) and fundamental symmetry
$\mathfrak{J}$, given by the formula 
$\big(\mathfrak{J} \widetilde{\phi}_{{}_{-m,0}}\big)(p) 
= \gamma^0 \, \widetilde{\phi}_{{}_{-m,0}}(p)$ or by the formula
$\big(\mathfrak{J} \widetilde{\phi}_{{}_{m,0}}\big)(p) 
= \gamma^1 \gamma^2 \gamma^3 \, \widetilde{\phi}_{{}_{-m,0}}(p)$, we denote by
\begin{equation}\label{U^2L^1/2_00m<0}
\Big(W_{{}_{V,-m,0,0,0}} U^{{}_{-m,0,0,0}\big(2L^{{}^{1/2}}\big)}  
W_{{}_{V,-m,0,0,0}}^{-1}\Big)_{{}_{00}}, \,\, m>0.
\end{equation}

The representations (\ref{U^2L^1/2_00m>0}) and (\ref{U^2L^1/2_00m<0}) are not Krein-unitary,
because of the non-ivariance of the measures
\[
\frac{\ud \mu_{{}_{m,0}}}{2 p^0 }, \,\,\,\frac{\ud \mu_{{}_{-m,0}}}{2 |p^0 |}
\]
on the respective orbits
\[
\mathscr{O}_{m,0,0,0}, \,\,\, \mathscr{O}_{-m,0,0,0}, \,\,\, m>0.
\]
Of course the direct summands, respectively, of the local representations (\ref{U^2L^1/2_00m>0}) and (\ref{U^2L^1/2_00m<0})
acting separatey on the images $V^\oplus$ and $V^\ominus$,
are unitary and unitary equivalent respectively to
\[
U^{{}_{m,0,0,0}\big(L^{{}^{1/2}}\big)}, \,\,\, U^{{}_{-m,0,0,0}\big(L^{{}^{1/2}}\big)},
\,\,\,m>0.
\]
Because of the non-ivariance of the measures
\[
\frac{\ud \mu_{{}_{m,0}}}{2 p^0 }, \,\,\,\frac{\ud \mu_{{}_{-m,0}}}{2 |p^0 |}.
\]
the maps $V^{\oplus\ominus}$ cannot be used for construction of any invariant Krein structure on
the representation space of the representations
\[
U^{{}_{m,0,0,0}\big(2L^{{}^{1/2}}\big)}, \,\,\, U^{{}_{-m,0,0,0}\big(2L^{{}^{1/2}}\big)},
\,\,\,m>0.
\]

\vspace*{1cm}

\begin{center}
\small 	NATURAL LOCAL VERSION $W_{{}_{V,m,0,0,0}} U^{{}_{m,0,0,0}\big(2L^{{}^{1/2}}\big)}  
W_{{}_{V,m,0,0,0}}^{-1}$ OF THE REPRESENTATION 
$U^{{}_{(m,0,0,0)} 2L^{{}^{1/2}}}$  AND THE CORRESPONDING 
GENERALIZED EIGENSPACE OF DIRAC OPERATOR
\end{center}

Note that $U^{{}_{m,0,0,0}\big(2L^{{}^{1/2}}\big)} = U^{{}_{m,0,0,0}\big(L^{{}^{1/2}} \oplus
L^{{}^{1/2}}\big)} $ is induced by the representation
\[
\gamma \mapsto Q(\gamma,\bar{p}) = \big(L^{1/2} \oplus L^{1/2}\big)(\gamma) = 2L^{1/2}(\gamma) =
\left( \begin{array}{cc}  \gamma & 0  \\
                                           
                                                   0              & \gamma  \end{array}\right) 
= \gamma \oplus \gamma
\]
of the group $SU(2, \mathbb{R})= G_{(m,0,0,0)}$ in the Hilbert space 
$\mathcal{H}_{\bar{p}} = \mathbb{C}^4$ with the standard 
Hilbert space inner product. The Hilbert space $\mathbb{C}^4$ has natural Krein-Hilbert space
structure with the fundamental symmetry  
\[
\mathfrak{J}_{\bar{p}} = \left( \begin{array}{cc}   0 &  \bold{1}_2  \\
                                           
                                                   \bold{1}_2              & 0 \end{array}\right)
                                                   = \gamma^0 
                                                   \,\,\,
                                                   \textrm{or}
                                                   \,\,\,
\mathfrak{J}_{\bar{p}} = \left( \begin{array}{cc}  0 &  i \bold{1}_2  \\
                                           
                                                   -i \bold{1}_2              & 0  \end{array}\right)
= \gamma^1 \gamma^2 \gamma^3 ,
\]
making it a Krein-Hilbert space $(\mathbb{C}^4 = \mathcal{H}_{\bar{p}}, \mathfrak{J}_{\bar{p}})$ such that 
$L^{1/2} \oplus L^{1/2} = 2L^{1/2}$ becomes a Krein-unitary representation in $(\mathbb{C}^4, \mathfrak{J}_{\bar{p}})$. Because $2L^{1/2}$ commutes with the fundamental symmetry $\mathfrak{J}_{\bar{p}}$
so the representation $2L^{1/2}$ is likewise unitary (for the Hilbert space inner product).

In order to construct the natural local version
$W_{{}_{V,m,0,0,0}} U^{{}_{m,0,0,0}\big(2L^{{}^{1/2}}\big)}  
W_{{}_{V,m,0,0,0}}^{-1}$ defined in the introduction to Sect. \ref{constr-of-VF}
we extend $2L^{1/2}$ by the formula
\[
\alpha \mapsto V(\alpha) = 
\left( \begin{array}{cc}  V''(\alpha) & 0  \\
                                           
                                                   0              & \overline{V''}(\alpha) \end{array}\right)
                                                   = 
\left( \begin{array}{cc}  \alpha & 0  \\
                                           
                                                   0              & {\alpha^*}^{-1}  \end{array}\right) 
= \alpha \oplus {\alpha^*}^{-1}
\]
to a representation of $SL(2, \mathbb{C})$. 
Note that here the representations $V''$ and $\overline{V''}$ are conjugated:
$\overline{V''}(\alpha) = V''({\alpha^*}^{-1} )$, with conjugation pertinent to the class of orbits 
$\mathscr{O}_{m,0,0,0}$, \emph{i.e.} $V''$ and $\overline{V''}$ coincide on the
isotropy subgroup $SU(2, \mathbb{R})= G_{(m,0,0,0)}$, common for all $\bar{p} = (m,0,0,0)$, $m \neq 0$.
This conjugation is realised by the automorphism $\alpha \mapsto {\alpha^*}^{-1}$ of $SL(2, \mathbb{C})$.
Both, the initial representation $2L^{1/2}$
of $SU(2,\mathbb{C})$ and its extension $V$ to a representation of $SL(2, \mathbb{C})$
are Krein-unitary in the Krein space $(\mathbb{C}^4, \mathfrak{J}_{\bar{p}})$ with
\[
\mathfrak{J}_{\bar{p}} = \left( \begin{array}{cc}   0 &  \bold{1}_2  \\
                                           
                                                   \bold{1}_2              & 0 \end{array}\right)
                                                   = \gamma^0 
                                                   \,\,\,
                                                   \textrm{or}
                                                   \,\,\,
\mathfrak{J}_{\bar{p}} = \left( \begin{array}{cc}  0 &  i \bold{1}_2  \\
                                           
                                                   -i \bold{1}_2              & 0  \end{array}\right)
= \gamma^1 \gamma^2 \gamma^3 ,
\]
and with the standard inner product in $\mathbb{C}^4$. But note that the extension $V$ is no longer unitary
but only Krein-unitary.

In this case 
\[
B(p) = V(\beta(p))^*V(\beta(p)) =
 \left( \begin{array}{cc}  V''(\beta(p))^*V''(\beta(p)) &  0  \\
                                           
             0              & \overline{V''}(\beta(p))^*\overline{V''}(\beta(p))  \end{array}\right)
\]
\[
         =    \left( \begin{array}{cc} \beta(p)^2 &  0 \\
                                           
                                                   0  & \beta(p)^{-2} \end{array}\right)
\]
\[
= \frac{1}{m}\left( \begin{array}{cc} \boldsymbol{1}_2 &  0 \\
                                           
                                                   0  & \boldsymbol{1}_2 \end{array}\right)                                   
+ \frac{1}{m}\left( \begin{array}{cc} -\vec{p}\cdot \vec{\sigma} &  0 \\
                                                                                              0  & \vec{p}\cdot \vec{\sigma} \end{array}\right).
\]

Now using this extension $V$ of $2L^{1/2}$ we construct the Krein-isometric
operator $W_{{}_{V,m,0,0,0}}$ and the natural local version
\begin{equation}\label{NatLocU^m,0,0,02L^1/2}
W_{{}_{V,m,0,0,0}} U^{{}_{m,0,0,0}\big(2L^{{}^{1/2}}\big)}  
W_{{}_{V,m,0,0,0}}^{-1}
\end{equation}
of the induced Krein-unitary and unitary representation $U^{{}_{m,0,0,0}\big(2L^{{}^{1/2}}\big)}$
as in introduction to Sect. \ref{constr-of-VF}.

To the induced representation $U^{{}_{m,0,0,0}\big(2L^{{}^{1/2}}\big)}$ we apply the Krein-unitary
operator $U^{-1}$ of Lemma \ref{InducedFormToImprimitivity} of Subsection \ref{lop_ind},
which transforms $U^{{}_{m,0,0,0}\big(2L^{{}^{1/2}}\big)}$ into the representation $U^{-1}U^{{}_{m,0,0,0}\big(2L^{{}^{1/2}}\big)}U$
which has the form of \emph{imprimitivity system}, compare
Subsection \ref{lop_ind}, and which acts in the Hilbert space of square summable functions
$\mathscr{O}_{m,0,0,0} \rightarrow \mathcal{H}_{\bar{p}}$
over the orbit $\mathscr{O}_{m,0,0,0} = T_4 \circledS SL(2, \mathbb{C})/G_{m,0,0,0}$
with respect to the induced invariant measure on the orbit. Compare the introductory part of this 
Section \ref{constr-of-VF}. Then we apply to the Krein-isometric representation
$U^{-1}U^{{}_{m,0,0,0}\big(2L^{{}^{1/2}}\big)}U$
 the Krein-isometric operator $W'$ (which depends on the orbit) given by (\ref{W1}) and finally the Krein unitary operator $W''$ depending on the orbit and the extension $V$ given by (\ref{W2}). 
 We have agreed to denote the operator $W''W'U^{-1}$ by $W$, or more precisely 
 by $W_{{}_{V,m,0,0,0}}$ as it depends on $V$ and on the orbit $\mathscr{O}_{m,0,0,0}$. More precisely 
the operator $U$ of Lemma \ref{InducedFormToImprimitivity} of Subsection \ref{lop_ind},
depends on the orbit and on the inducing representation $L$ (here $2L^{1/2}$), similarly as $W'$. 
Therefore we should write
\[
W_{{}_{V,m,0,0,0}} = W''_{{}_{V,m,0,0,0}}W'_{{}_{V,m,0,0,0}}U_{{}_{V,m,0,0,0}}^{-1}
\]
for the operator $W = W''W'U^{-1}$ of introduction to this Section, whenever applied to 
$U^{{}_{m,0,0,0}L}$ with $L$ extended by $V$. This function operators are well-defined because
by construction $V$ extends $L$ and $V$ determines $L$ as the restriction to the
isotropy subgroup  (here $G_{\bar{p}} = G_{m,0,0,0} = SU(2, \mathbb{C})$).

The elements of the representation space of the representation 
\begin{equation}\label{W'U^-1U^m,0,0,02L1/2UW'^-1}
W'_{{}_{V,m,0,0,0}}U_{{}_{V,m,0,0,0}}^{-1}U^{{}_{m,0,0,0}\big(2L^{{}^{1/2}}\big)}
U_{{}_{V,m,0,0,0}}{W'}_{{}_{V,m,0,0,0}}^{-1}
\end{equation}
we have agreed to denote by $\widetilde{\psi}_{m,0}$. Here we have used the superscript $m,0$ 
indicating that they are concentrated on the orbit $\mathscr{O}_{m,0,0,0}$. 
We restrict ourselves to the dense nuclear subspace  
\[
E_{{}_{m,0, \textrm{dim} \, 2L^{1/2}}} = E_{{}_{m,0,4}} = \big\{f|_{{}_{\mathscr{O}_{m,0,0,0}}},
\,\, f \in \mathcal{S}(\mathbb{R}^4; \mathbb{C}^{\textrm{dim}\, 2L^{1/2}})
=  \mathcal{S}(\mathbb{R}^4; \mathbb{C}^4) \big\}
\]
of the representation space of the representation (\ref{W'U^-1U^m,0,0,02L1/2UW'^-1})
transformed into itself (continuously in the nuclear topology) by the operator
$W''_{{}_{V,m,0}}$, equal to the dense nuclear domain of the operator 
$W_{{}_{V,m,0,0,0}}^{-1}$.

The function $p \mapsto \beta(p)$ likewise depends on the orbit (in the separate classes $\mathscr{O}_{m,0,0,0}$, $m \neq 0$ and $\mathscr{O}_{0,m,0,0}$, $m>0$, $\beta$ can be chosen to be given by two different matrix functions depending on the parameter $m$). We nonetheless discard indicating this dependence
and believe that the context will indicate what function $\beta$ is exactly is meant in each context.
Here in Subsection \ref{e1} this the matrix function (depending on $m$) given at the beginning of Sect. \ref{e1}, and $\beta(p)$ depends on the actually analyzed orbit class.

The elements $\widetilde{\phi}_{m,0} = 
W''_{{}_{V,m,0,0,0}}\widetilde{\psi}_{m,0}$, whenever $\widetilde{\psi}_{m,0}$ belong to the dense nuclear domain $E_{{}_{m,0,4}}$, are also in $E_{{}_{m,0,4}}$ are contained in the image of $W_{{}_{V,m,0,0,0}}$
and these are the elements
\[
\widetilde{\phi}_{m,0} = W''_{{}_{V,m,0,0,0}}\widetilde{\psi}_{m,0} = 
\tvect{\widetilde{\varphi}_{{}_{m,0}}}{\widetilde{\chi}_{{}_{m,0}}}
\]
in the representation space of the natural local version (\ref{NatLocU^m,0,0,02L^1/2}) 
of the representation $U^{{}_{(m,0,0,0)} 2L^{{}^{1/2}}}$. They have the local bispinor
transformation law  
\[
U(\alpha) \widetilde{\phi}_{{}_{m,0}}(p) = 
V(\alpha) \widetilde{\phi}_{{}_{m,0}}(\Lambda(\alpha)p)
= \left( \begin{array}{cc}  \alpha & 0  \\
                                           
                                                   0              & {\alpha^*}^{-1}  \end{array}\right) 
 \widetilde{\phi}_{{}_{m,0}}(\Lambda(\alpha)p), 
\]
\[
T(a) \widetilde{\phi}_{{}_{m,0}}(p) = e^{i a \cdot p}\widetilde{\phi}_{{}_{m,0}}(p);
\]
\emph{i.e.} the (local) action of (\ref{NatLocU^m,0,0,02L^1/2}).

Note that according to the introduction to this Section \ref{constr-of-VF}
the inner product in the Hilbert space of representation space of the natural local representation
(\ref{NatLocU^m,0,0,02L^1/2}) is equal
\begin{multline}\label{Inn-Prod-NatLocU^m,0,0,02L^1/2}
(\widetilde{\phi}_{{}_{m,0}}, \widetilde{\phi'}_{{}_{m,0}})  = 
  \int \limits_{\mathscr{O}_{(m,0,0,0)}} 
\Big(  \widetilde{\phi}_{{}_{m,0}}(p), 
V(\beta(p))^* V(\beta(p)) \widetilde{\phi}'_{{}_{m,0}}(p)  \Big)_{{}_{\mathbb{C}^4}}
\, \ud \mu_{{}_{m,0}} \\
= 
\int \limits_{\mathscr{O}_{(m,0,0,0)}} 
\Big(  \widetilde{\phi}_{{}_{m,0}}(p), 
B(p) \widetilde{\phi}'_{{}_{m,0}}(p)  \Big)_{{}_{\mathbb{C}^4}}
\, \ud \mu_{{}_{m,0}}.
\end{multline}

The fundamental symmetry operator in the representation space
of the natural local version (\ref{NatLocU^m,0,0,02L^1/2})  is equal to operator 
of multiplication by the matrix function
\[
\mathfrak{J}_{\bar{p}} V(\beta(p))^* V(\beta(p))= \mathfrak{J}_{\bar{p}} B(p), 
\]
where
\[
\mathfrak{J}_{\bar{p}} = \left( \begin{array}{cc}   0 &  \bold{1}_2  \\
                                           
                                                   \bold{1}_2              & 0 \end{array}\right)
                                                   = \gamma^0 
                                                   \,\,\,
                                                   \textrm{or}
                                                   \,\,\,
\mathfrak{J}_{\bar{p}} = \left( \begin{array}{cc}  0 &  i \bold{1}_2  \\
                                           
                                                   -i \bold{1}_2              & 0  \end{array}\right)
= \gamma^1 \gamma^2 \gamma^3.
\]

The representation (\ref{NatLocU^m,0,0,02L^1/2}) given by the explicit local formula given above
is Krein isometric, but it is no longer unitary, although  $U^{{}_{m,0,0,0}\big(2L^{{}^{1/2}}\big) }$
was.

Let us define the following projectors:
\[
P^\oplus(p) = \frac{1}{2}
\left( \begin{array}{cc} \boldsymbol{1}_2 & V''(\beta(p))^{-1}\overline{V''}(\beta(p))  \\
                                           
                                                  {\overline{V''}(\beta(p)})^{-1}V''(\beta(p)) &  \boldsymbol{1}_2 \end{array}\right)
                                                  =\frac{1}{2}
\left( \begin{array}{cc} \boldsymbol{1}_2 &\beta(p)^{-2}  \\
                                           
                                                  \beta(p)^2  &  \boldsymbol{1}_2 \end{array}\right),  
\]
\[
P^\ominus(p) = \frac{1}{2}
\left( \begin{array}{cc} \boldsymbol{1}_2 & -V''(\beta(p))^{-1}\overline{V''}(\beta(p))  \\
                                           
                                                  -{\overline{V''}(\beta(p)})^{-1}V''(\beta(p)) &  \boldsymbol{1}_2 \end{array}\right)
                                                  =\frac{1}{2}
\left( \begin{array}{cc} \boldsymbol{1}_2 & -\beta(p)^{-2}  \\
                                           
                                                  -\beta(p)^2  &  \boldsymbol{1}_2 \end{array}\right),
\]
and the corresponding projection operators $P^\oplus, P^\ominus$ of pointwise multiplication, respectively, 
by $P^\oplus(p), P^\ominus(p)$. Because of the orthogonality
\[
P^\oplus(p)P^\ominus(p) = P^\ominus(p)P^\oplus(p) = 0, \,\,\,\, P^\oplus(p)+ P^\ominus(p) = \boldsymbol{1}_4
\]
we have orthogonality 
\[
P^\oplus P^\ominus = P^\ominus P^\oplus = 0, \,\,\,\, P^\oplus + P^\ominus = \boldsymbol{1}.
\]
Although note that
\[
P^\oplus(p)^* \neq P^\oplus(p), \,\,\, P^\ominus(p)^* \neq P^\ominus(p) 
\]
in $\mathbb{C}^4$ with respect to  the canonical inner product $(\cdot, \cdot)_{{}_{\mathbb{C}^4}}$ and neither with respect to $(\cdot, B(p)\cdot)_{{}_{\mathbb{C}^4}}$. Similarly  
\[
{P^\oplus}^* \neq P^\oplus, \,\,\, {P^\ominus}^* \neq P^\ominus 
\]
in the Hilbert space of bispinors with respect to the inner product 
(\ref{innLL'}) and neither with respect to the inner product (\ref{innLL'})
with $B(p)$ put equal $\boldsymbol{1}_4$.

These projectors are Krein self adjoint 
\[
\mathfrak{J} P^{\oplus *}\mathfrak{J} = P^\oplus, \,\,\, \mathfrak{J} P^{\ominus *}\mathfrak{J} = P^\ominus,
\]
for $\mathfrak{J}=\gamma^0$ and  Krein skew self adjoint 
\[
\mathfrak{J} P^{\oplus *}\mathfrak{J} = - P^\oplus, \,\,\, \mathfrak{J} P^{\ominus *}\mathfrak{J} = - P^\ominus,
\]
for $\mathfrak{J}=\gamma^1\gamma^2\gamma^3$.

Elements
\[
\tvect{\widetilde{\varphi}}{\widetilde{\chi}}
= 
P^{\oplus} {\widetilde{\phi}_{{}_{m,0}}} 
\]
 of the image of $P^\oplus$ respect the following algebraic relation 
 \[
 \begin{array}{c}  \widetilde{\chi}(p) = \overline{V''}(\beta(p))^{-1}V''(\beta(p)) \widetilde{\varphi} (p)\\
\widetilde{\varphi} (p) = \big(\overline{V''}(\beta(p))^{-1}V''(\beta(p)))^{-1} \widetilde{\chi}(p)                                        
                                \end{array}
 \]
which in more explicit form reads
\[
\big(p^0\gamma_0 - p^1 \gamma^1 - p^2 \gamma^2 - p^3 \gamma^3\big)
\widetilde{\phi} = m \widetilde{\phi}, \,\,\,\, \widetilde{\phi} = 
\tvect{\widetilde{\varphi}}{\widetilde{\chi}} \in \textrm{Im} \, P^\oplus.
\]

Elements
\[
\tvect{\widetilde{\varphi}}{\widetilde{\chi}}
= 
P^{\ominus} {\widetilde{\phi}_{{}_{m,0}}} 
\]
of the image of $P^\ominus$ respect the folowing algebraic relation 
 \[
 \begin{array}{c}  \widetilde{\chi}(p) = -\overline{V''}(\beta(p))^{-1}V''(\beta(p)) \widetilde{\varphi} (p)\\
\widetilde{\varphi} (p) = -\big(\overline{V''}(\beta(p))^{-1}V''(\beta(p)))^{-1} \widetilde{\chi}(p)                                        
                                \end{array}
 \]
which in more explicit form reads
\[
\big(p^0\gamma_0 - p^1 \gamma^1 - p^2 \gamma^2 - p^3 \gamma^3\big)
\widetilde{\phi} = -m \widetilde{\phi}, \,\,\,\, \widetilde{\phi} = 
\tvect{\widetilde{\varphi}}{\widetilde{\chi}} \in \textrm{Im} \, P^\ominus.
\]

\vspace*{1cm}

\begin{center}
\small 	THE REPRESENTATION $\big(W_{{}_{V,m,0,0,0}} U^{{}_{m,0,0,0}(2L^{{}^{1/2}})}  
W_{{}_{V,m,0,0,0}}^{-1}\big)_{{}_{0}}$ 
\end{center}

According to Proposition of introduction to Section \ref{constr-of-VF} we obtain 
the representation
\begin{equation}\label{NatLocU^m,0,0,02L^1/20}
\Big(W_{{}_{V,m,0,0,0}} U^{{}_{m,0,0,0}\big(2L^{{}^{1/2}}\big)}  
W_{{}_{V,m,0,0,0}}^{-1}\Big)_{{}_{0}}
\end{equation}
Krein-isometrically equivalent to  the natural local version (\ref{NatLocU^m,0,0,02L^1/2})
by restriction to the dense nuclear subspace $E_{{}_{m,0,4}}$ in the representation space
of (\ref{NatLocU^m,0,0,02L^1/2}). Then the action of (\ref{NatLocU^m,0,0,02L^1/20}) on
$E_{{}_{m,0,4}}$ is identical as the action of  (\ref{NatLocU^m,0,0,02L^1/2}). The
inner product on $E_{{}_{m,0,4}}$ regarded as a representation subspace of the representation
(\ref{NatLocU^m,0,0,02L^1/20}) is defined by the formula for the inner
product (\ref{Inn-Prod-NatLocU^m,0,0,02L^1/2})
of the representation space of (\ref{NatLocU^m,0,0,02L^1/2}) in which the operator matrix
$B(p)$ is the constant unit matrix. Thus in explicit form the inner product on $E_{{}_{m,0,4}}$
regarded as the representation  subspace of the representation (\ref{NatLocU^m,0,0,02L^1/20})
is equal
\begin{equation}\label{Inn-Prod-NatLocU^m,0,0,02L^1/20}
(\widetilde{\phi}_{{}_{m,0}}, \widetilde{\phi'}_{{}_{m,0}})  = 
  \int \limits_{\mathscr{O}_{(m,0,0,0)}} 
\Big(  \widetilde{\phi}_{{}_{m,0}}(p), 
\widetilde{\phi}'_{{}_{m,0}}(p)  \Big)_{{}_{\mathbb{C}^4}}
\, \ud \mu_{{}_{m,0}}.
\end{equation}
The Hilbert space $\mathcal{H'}_{{}_{0}}$ of the representation  
(\ref{NatLocU^m,0,0,02L^1/20}) is defined as the completion of $E_{{}_{m,0,4}}$ 
with respect to the inner product (\ref{Inn-Prod-NatLocU^m,0,0,02L^1/20}). 

The fundamental symmetry operator on $\mathcal{H'}_{{}_{0}}$ under which 
(\ref{NatLocU^m,0,0,02L^1/20}) is Krein isometric and Krein-isometrically equivalent to 
(\ref{NatLocU^m,0,0,02L^1/2}) is equal to the operator of multiplication by the 
constant matrix operator
\begin{equation}\label{J-NatLocU^m,0,0,02L^1/20}
\mathfrak{J}_{\bar{p}} = \left( \begin{array}{cc}   0 &  \bold{1}_2  \\
                                           
                                                   \bold{1}_2              & 0 \end{array}\right)
                                                   = \gamma^0 
                                                   \,\,\,
                                                   \textrm{or}
                                                   \,\,\,
\mathfrak{J}_{\bar{p}} = \left( \begin{array}{cc}  0 &  i \bold{1}_2  \\
                                           
                                                   -i \bold{1}_2              & 0  \end{array}\right)
= \gamma^1 \gamma^2 \gamma^3.
\end{equation}

The representation space of the representation 
\[
U^{{}_{m,0,0,0}\big(2L^{{}^{1/2}}\big)} = U^{{}_{m,0,0,0}\big(L^{{}^{1/2}}\big)}  \oplus U^{{}_{m,0,0,0}\big(L^{{}^{1/2}}\big)} 
\]
acting on direct sum of spinors $\widetilde{\psi}_{{}_{m,0}}^{\oplus} \oplus \widetilde{\psi}_{{}_{m,0}}^{\ominus}$ concentrated on $\mathscr{O}_{{}_{m,0,0,0}}$ (compare the beginning of Subsection \ref{e1})
with the inner product 
\begin{multline*}
(\widetilde{\psi}_{{}_{m,0}}^{\oplus} \oplus \widetilde{\psi}_{{}_{m,0}}^{\oplus}, \widetilde{\psi}_{{}_{m,0}}^{\oplus} \oplus \widetilde{\psi}_{{}_{m,0}}^{\oplus})
\\
= \int\limits_{\mathscr{O}_{{}_{m,0,0,0}}} \big(\widetilde{\psi}_{{}_{m,0}}^{\oplus}(p),\widetilde{\psi}_{{}_{m,0}}^{\oplus}(p)\big)_{{}_{\mathbb{C}^2}}
\,\, \ud \mu_{{}_{m,0}}(p)
+
\int\limits_{\mathscr{O}_{{}_{m,0,0,0}}} \big(\widetilde{\psi}_{{}_{m,0}}^{\ominus}(p),\widetilde{\psi}_{{}_{m,0}}^{\ominus}(p)\big)_{{}_{\mathbb{C}^2}}
\,\, \ud \mu_{{}_{m,0}}(p),
\end{multline*} 
can be given a natural invariant Krein structure, induced by its natural local version (\ref{NatLocU^m,0,0,02L^1/20}), acting on 
bispinors 
\[
\widetilde{\phi}_{{}_{m,0}} = 
 \widetilde{\phi}_{{}_{m,0}}^{\oplus} \oplus \widetilde{\phi}_{{}_{m,0}}^{\ominus}
= V^{\oplus\ominus}\big(\widetilde{\psi}_{{}_{m,0}}^{\oplus} \oplus \widetilde{\psi}_{{}_{m,0}}^{\ominus}\big)
=
V^\oplus\widetilde{\psi}_{{}_{m,0}}^{\oplus} \oplus V^\ominus \widetilde{\psi}_{{}_{m,0}}^{\ominus}
\]
concentrated on $\mathscr{O}_{{}_{m,0,0,0}}$ 
with the inner product
\[
(\widetilde{\phi}_{{}_{m,0}}, \widetilde{\phi}_{{}_{m,0}})
=
\int\limits_{\mathscr{O}_{{}_{m,0,0,0}}} \big(\widetilde{\phi}_{{}_{m,0}}(p), \widetilde{\phi}_{{}_{m,0}}(p)\big)_{{}_{\mathbb{C}^4}}
\,\, \ud \mu_{{}_{m,0}}(p),
\]
and Krein symmetry operator $\mathfrak{J} = \gamma^0$ or $\gamma^1\gamma^2\gamma^3$.
Indeed, the Krein structure induced on the direct sum $\widetilde{\psi}_{{}_{m,0}}^{\oplus} \oplus \widetilde{\psi}_{{}_{m,0}}^{\ominus}$ in the representation space
of 
\[
U^{{}_{m,0,0,0}\big(2L^{{}^{1/2}}\big)} = U^{{}_{m,0,0,0}\big(L^{{}^{1/2}}\big)}  \oplus U^{{}_{m,0,0,0}\big(L^{{}^{1/2}}\big)} 
\]
is equal to $\boldsymbol{1}$ on the first direct summand and $- \boldsymbol{1}$ on the second direct summand:
\[
\mathfrak{J}\big(\widetilde{\psi}_{{}_{m,0}}^{\oplus} \oplus \widetilde{\psi}_{{}_{m,0}}^{\ominus}\big)
= \big(\widetilde{\psi}_{{}_{m,0}}^{\oplus}\big) \oplus \big(-\widetilde{\psi}_{{}_{m,0}}^{\ominus}\big).
\]
This follows from the general Theorem and Proposition presented in the beginning of this Section, but can also be immediately deduced from the
simple identities (\ref{SimpleKreinIsometry2(1/2)m,0,0,0})
(compare the beginning of Subsection \ref{e1}).  Indeed, using these identities we get 
\begin{multline*}
\big({\widetilde{\phi}_{{}_{m,0}}}, \mathfrak{J}{\widetilde{\phi}_{{}_{m,0}}}\big) =
\big({\widetilde{\phi}_{{}_{m,0}}}^{\oplus}, \mathfrak{J}{\widetilde{\phi}_{{}_{m,0}}}^{\oplus} \big)
+
\big({\widetilde{\phi}_{{}_{m,0}}}^{\ominus}, \mathfrak{J}{\widetilde{\phi}_{{}_{m,0}}}^{\ominus} \big)
\\
=
\int\limits_{\mathscr{O}_{{}_{m,0,0,0}}}
\big({\widetilde{\phi}_{{}_{m,0}}}^{\oplus}(p), \mathfrak{J}{\widetilde{\phi}_{{}_{m,0}}}^{\oplus}(p) \big)_{{}_{\mathbb{C}^4}}
\ud \mu_{{}_{m,0}}(p)
\\
+
\int\limits_{\mathscr{O}_{{}_{m,0,0,0}}}
\big({\widetilde{\phi}_{{}_{m,0}}}^{\ominus}(p), \mathfrak{J}{\widetilde{\phi}_{{}_{m,0}}}^{\ominus}(p) \big)_{{}_{\mathbb{C}^4}}
\ud \mu_{{}_{m,0}}(p)
\end{multline*}
\begin{multline*}
=
2\int\limits_{\mathscr{O}_{{}_{m,0,0,0}}}
\big({\widetilde{\psi}_{{}_{m,0}}}^{\oplus}(p), {\widetilde{\psi}_{{}_{m,0}}}^{\oplus}(p) \big)_{{}_{\mathbb{C}^2}}
\ud \mu_{{}_{m,0}}(p)
\\
-2\int\limits_{\mathscr{O}_{{}_{m,0,0,0}}}
\big({\widetilde{\psi}_{{}_{m,0}}}^{\ominus}(p), {\widetilde{\psi}_{{}_{m,0}}}^{\ominus}(p) \big)_{{}_{\mathbb{C}^2}}
\ud \mu_{{}_{m,0}}(p)
\end{multline*}
\[
=
2\Big(\, \widetilde{\psi}_{{}_{m,0}}^{\oplus} \oplus \widetilde{\psi}_{{}_{m,0}}^{\ominus} \, , \, 
\mathfrak{J}\big(\widetilde{\psi}_{{}_{m,0}}^{\oplus} \oplus \widetilde{\psi}_{{}_{m,0}}^{\ominus}\big) \, \Big).
\]
Thus, it is sufficient to apply the additional multiplication by $\sqrt{2}$ to the map $V^{\oplus\ominus}$ constructed at the beginning of Subsection
\ref{e1} in order to get the required Krein isometry equivalence between
\[
U^{{}_{m,0,0,0}\big(2L^{{}^{1/2}}\big)} = U^{{}_{m,0,0,0}\big(L^{{}^{1/2}}\big)} \oplus U^{{}_{m,0,0,0}\big(L^{{}^{1/2}}\big)}
\]
and (\ref{NatLocU^m,0,0,02L^1/20}) in explicit form expressed immediately through the elements $\psi$ and $\phi$
of the representation spaces of both representations.

\subsection{Example 2: Representation $U^{{}_{(0,m,0,0)} [2L^{{}^{1/2}}]_{\textrm{Ass}}}$
associated to $U^{{}_{(m,0,0,0)} 2L^{{}^{1/2}}}$ (spin $1/2$) and 
concentrated on the orbit $\mathscr{O}_{(0,0,m,0)}$}\label{e2}

As is well known there are no nontrivial finite dimensional unitary representations
of the group $G_{(0,m,0,0)} = SL(2, \mathbb{R})$ stationary for $\bar{p} = (0,m,0,0)$.
The situation is different for Krein-unitary representations. We give an example here.
In this class of orbits we can put
\begingroup\makeatletter\def\f@size{5}\check@mathfonts
\def\maketag@@@#1{\hbox{\m@th\large\normalfont#1}}%
\[
\frac{1}{2} \left( \begin{array}{cc}   
\Big[ \pm (\sin \frac{\theta}{2} - \cos \frac{\theta}{2})(\frac{r}{m} -1)^{\frac{1}{2}} 
+ (\sin \frac{\theta}{2} + \cos \frac{\theta}{2})(\frac{r}{m} + 1)^{\frac{1}{2}}\Big] e^{-i \frac{\vartheta}{2}} & 
i\Big[\pm (\sin \frac{\theta}{2} + \cos \frac{\theta}{2})(\frac{r}{m} -1)^{\frac{1}{2}} 
- (\sin \frac{\theta}{2} - \cos \frac{\theta}{2})(\frac{r}{m} + 1)^{\frac{1}{2}}\Big] e^{i \frac{\vartheta}{2}}  \\
                                           
-i \Big[ (\sin \frac{\theta}{2} - \cos \frac{\theta}{2})(\frac{r}{m} +1)^{\frac{1}{2}} 
\pm (\sin \frac{\theta}{2} + \cos \frac{\theta}{2})(\frac{r}{m} - 1)^{\frac{1}{2}}\Big] e^{-i \frac{\vartheta}{2}} & 
\Big[(\sin \frac{\theta}{2} + \cos \frac{\theta}{2})(\frac{r}{m} +1)^{\frac{1}{2}} 
\mp (\sin \frac{\theta}{2} - \cos \frac{\theta}{2})(\frac{r}{m} - 1)^{\frac{1}{2}}\Big] e^{i \frac{\vartheta}{2}}
  \end{array}\right) 
\]
\endgroup
for $\beta(p)$, where 
\[
p = \left( \begin{array}{c}   p^0  \\
                              p^1  \\
                              p^2   \\          
                              p^3  \end{array}\right) 
= \left( \begin{array}{c}   \pm (r^2 - m^2)^{1/2}  \\
                              r \sin \theta \sin \vartheta  \\
                              r \sin \theta \cos \vartheta   \\          
                              r \cos \vartheta  \end{array}\right) \in \mathscr{O}_{(0,0,m,0)}, \,\,\, 
0 \leq\theta < \pi, 0 \leq \vartheta < 2 \pi, m^2 \leq  r^2, 0 < r. 
\]
Namely the representation
\[
\gamma \mapsto Q(\gamma,\bar{p}) = \big[2L^{{}^{1/2}}\big]_{\textrm{Ass}}(\gamma) =
\left( \begin{array}{cc}  \gamma & 0  \\
                                           
                                                   0              & {\gamma^*}^{-1}  \end{array}\right) 
= \gamma \oplus {\gamma^*}^{-1}
\]
of the group $SL(2, \mathbb{R})= G_{(0,0,m,0)}$, we extend by the formula
\[
\alpha \mapsto V(\alpha) = 
\left( \begin{array}{cc}  \alpha & 0  \\
                                           
                                                   0              & {\alpha^*}^{-1}  \end{array}\right) 
= \alpha \oplus {\alpha^*}^{-1}
\]
to a representation of $SL(2, \mathbb{C})$. Both, the initial representation $\big[2L^{{}^{1/2}}\big]_{\textrm{Ass}}$
of $SL(2,\mathbb{R})$ and its extension $V$ to a representation of $SL(2, \mathbb{C})$
are Krein unitary in the Krein space $(\mathbb{C}^4, \mathfrak{J}_{\bar{p}})$ with
\[
\mathfrak{J}_{\bar{p}} = \left( \begin{array}{cc}   0 &  \bold{1}_2  \\
                                           
                                                   \bold{1}_2              & 0 \end{array}\right)
                                                   = \gamma^0 
                                                   \,\,\,
                                                   \textrm{or}
                                                   \,\,\,
\mathfrak{J}_{\bar{p}} = \left( \begin{array}{cc}  0 &  i \bold{1}_2  \\
                                           
                                                   -i \bold{1}_2              & 0  \end{array}\right)
= \gamma^1 \gamma^2 \gamma^3 ,
\]
and with the standard inner product in $\mathbb{C}^4$.
We will later need to know the operator $V(\beta(p))^*V(\beta(p))$ explicitly -- it is equal
\[
B(p) = V(\beta(p))^*V(\beta(p)) =
\left( \begin{array}{cc}  \beta(p) ^*\beta{p} & 0  \\
                                           
                                               0 &    \big(\beta(p) ^*\beta(p)\big)^{-1}             \end{array}\right)
\]
\[
= \frac{1}{mr}\left( \begin{array}{cccc} r^2 - p^0 p^3 & ip^0 (p^2 + i p^1) & 0 & 0  \\
          -ip^0 (p^2 - i p^1) & r^2 +p^0 p^3 & 0 & 0 \\

           0 & 0 & r^2 + p^0 p^3 & -ip^0 (p^2 + i p^1) \\                                
                                                   0  & 0 & ip^0 (p^2 -ip^1)  & r^2 - p^0 p^3  \end{array}\right), 
\]
($r = (\vec{p}\cdot \vec{p})^{1/2}$, $p\cdot p = (p^0)^2 - \vec{p}\cdot \vec{p}$) with the following set of proper values (counted with multiplicities)
\begin{multline*}
\Big\{ \frac{r}{m} + \Big(\frac{r^2}{m^2} -1\Big)^{1/2}, \frac{r}{m} - \Big(\frac{r^2}{m^2} -1\Big)^{1/2},
\frac{r}{m} - \Big(\frac{r^2}{m^2} -1\Big)^{1/2}, \frac{r}{m} + \Big(\frac{r^2}{m^2} -1\Big)^{1/2}  \Big\} \\
= \Big\{ \frac{(\vec{p}\cdot \vec{p})^{1/2}}{(-p \cdot p)^{1/2}} + \frac{p^0}{(-p \cdot p)^{1/2}}, 
\frac{(\vec{p}\cdot \vec{p})^{1/2}}{(-p \cdot p)^{1/2}} - \frac{p^0}{(-p \cdot p)^{1/2}}, \\
\frac{(\vec{p}\cdot \vec{p})^{1/2}}{(-p \cdot p)^{1/2}} - \frac{p^0}{(-p \cdot p)^{1/2}}, 
\frac{(\vec{p}\cdot \vec{p})^{1/2}}{(-p \cdot p)^{1/2}} + \frac{p^0}{(-p \cdot p)^{1/2}}
  \Big\}
\end{multline*}

Now consider the Krein unitary representation $U^{{}_{0,m,0,0}[ 2L^{{}^{1/2}}]_{\textrm{Ass}} } $ 
of $T_4 \circledS SL(2, \mathbb{C})$ concentrated on the orbit $\mathscr{O}_{(0,0,m,0)}$
of $\bar{p} = (0,0,m,0)$ in $\widehat{T_4}$, induced by the above representation 
$\big[ 2L^{{}^{1/2}}\big]_{\textrm{Ass}}$ of $G_{(0,m,0,0)} = SL(2, \mathbb{R})$.

\vspace*{1cm}

\begin{center}
\small
NATURAL LOCAL VERSION $W_{{}_{V,0,m,0,0}} U^{{}_{0,m,0,0}[ 2L^{{}^{1/2}}]_{\textrm{Ass}} } 
W_{{}_{V,0,m,0,0}}^{-1}$ OF THE REPRESENTATION 
$U^{{}_{(0,m,0,0)} [2L^{{}^{1/2}}]_{\textrm{Ass}}}$  AND THE CORRESPONDING 
GENERALIZED EIGENSPACE OF DIRAC OPERATOR
\end{center}

Using the extension $V$ defined as above of 
$\big[2L^{{}^{1/2}}\big]_{\textrm{Ass}}$ we may construct the natural local version
\begin{equation}\label{NatLoc0,m,0,0,0AssU^2L^1/2}
W_{{}_{V,0,m,0,0}} U^{{}_{0,m,0,0}\big[ 2L^{{}^{1/2}}\big]_{\textrm{Ass}} } 
W_{{}_{V,0,m,0,0}}^{-1}.
\end{equation}

The elements of the representation space of the representation 
\begin{equation}\label{W'U^-1U^0,m,0,0,0AssU^2L^1/2UW'^-1}
W'_{{}_{V,0,m,0,0}}U_{{}_{V,0,m,0,0}}^{-1}U^{{}_{0,m,0,0}\big[ 2L^{{}^{1/2}}\big]_{\textrm{Ass}}}
U_{{}_{V,0,m,0,0}}{W'}_{{}_{V,0,m,0,0}}^{-1}
\end{equation}
we have agreed to denote by $\widetilde{\psi}_{0,m}$. Here we have used the subscript $0,m$ 
indicating that they are concentrated on the orbit $\mathscr{O}_{0,m,0,0}$. 
We restrict ourselves to the dense nuclear subspace  
\[
E_{{}_{0,m, \textrm{dim} \, \big[ 2L^{{}^{1/2}}\big]_{\textrm{Ass}}}} = E_{{}_{0,m,4}} = 
\big\{f|_{{}_{\mathscr{O}_{0,m,0,0}}},
\,\, f \in \mathcal{S}(\mathbb{R}^4; \mathbb{C}^{\textrm{dim}\, \big[ 2L^{{}^{1/2}}\big]_{\textrm{Ass}}})
=  \mathcal{S}(\mathbb{R}^4; \mathbb{C}^4) \big\}
\]
of the representation space of the representation (\ref{W'U^-1U^0,m,0,0,0AssU^2L^1/2UW'^-1})
transformed into itself (continuously in the nuclear topology) by the operator
$W''_{{}_{V,0,m}}$ (defined by (\ref{W2}) with the above $V$ and $\beta(p)$ corresponding to the actual
orbit $\mathscr{O}_{(0,0,m,0)}$), equal to the dense nuclear domain of the operator 
$W_{{}_{V,0,m,0,0}}^{-1}$.

The elements $\widetilde{\phi}_{0,m} = 
W''_{{}_{V,0,m,0,0}}\widetilde{\psi}_{0,m}$, whenever $\widetilde{\psi}_{0,m}$ belong to the dense nuclear domain $E_{{}_{0,m,4}}$, are also in $E_{{}_{0,m,4}}$ are contained in the image of $W_{{}_{V,0,m,0,0}}$
and these are the elements
\[
\widetilde{\phi}_{0,m} = W''_{{}_{V,0,m,0,0}}\widetilde{\psi}_{0,m} = 
\tvect{\widetilde{\varphi}_{{}_{0,m}}}{\widetilde{\chi}_{{}_{0,m}}}
\]
in the representation space of the natural local version (\ref{NatLoc0,m,0,0,0AssU^2L^1/2}) 
of the representation $U^{{}_{0,m,0,0}\big[ 2L^{{}^{1/2}}\big]_{\textrm{Ass}} }$. They have the local bispinor transformation law  
\[
U(\alpha) \widetilde{\phi}_{{}_{0,m}}(p) = 
V(\alpha) \widetilde{\phi}_{{}_{0,m}}(\Lambda(\alpha)p)
= \left( \begin{array}{cc}  \alpha & 0  \\
                                           
                                                   0              & {\alpha^*}^{-1}  \end{array}\right) 
 \widetilde{\phi}_{{}_{0,m}}(\Lambda(\alpha)p), 
\]
\[
T(a) \widetilde{\phi}_{{}_{0,m}}(p) = e^{i a \cdot p}\widetilde{\phi}_{{}_{0,m}}(p);
\]
\emph{i.e.} the (local) action of (\ref{NatLoc0,m,0,0,0AssU^2L^1/2}).

Note that according to the introduction to this Section \ref{constr-of-VF}
the inner product in the Hilbert space of representation space of the natural local representation
(\ref{NatLocU^m,0,0,02L^1/2}) is equal
\begin{multline}\label{Inn-Prod-NatLoc0,m,0,0,0AssU^2L^1/2}
(\widetilde{\phi}_{{}_{0,m}}, \widetilde{\phi'}_{{}_{0,m}})  = 
  \int \limits_{\mathscr{O}_{(0,m,0,0)}} 
\Big(  \widetilde{\phi}_{{}_{0,m}}(p), 
V(\beta(p))^* V(\beta(p)) \widetilde{\phi}'_{{}_{0,m}}(p)  \Big)_{{}_{\mathbb{C}^4}}
\, \ud \mu_{{}_{0,m}} \\
= 
\int \limits_{\mathscr{O}_{(m,0,0,0)}} 
\Big(  \widetilde{\phi}_{{}_{0,m}}(p), 
B(p) \widetilde{\phi}'_{{}_{0,m}}(p)  \Big)_{{}_{\mathbb{C}^4}}
\, \ud \mu_{{}_{0,m}}.
\end{multline}

The fundamental symmetry operator in the representation space
of the natural local version (\ref{NatLoc0,m,0,0,0AssU^2L^1/2})  is equal to operator 
of multiplication by the matrix function
\[
\mathfrak{J}_{\bar{p}} V(\beta(p))^* V(\beta(p))= \mathfrak{J}_{\bar{p}} B(p), 
\]
where
\[
\mathfrak{J}_{\bar{p}} = \left( \begin{array}{cc}   0 &  \bold{1}_2  \\
                                           
                                                   \bold{1}_2              & 0 \end{array}\right)
                                                   = \gamma^0 
                                                   \,\,\,
                                                   \textrm{or}
                                                   \,\,\,
\mathfrak{J}_{\bar{p}} = \left( \begin{array}{cc}  0 &  i \bold{1}_2  \\
                                           
                                                   -i \bold{1}_2              & 0  \end{array}\right)
= \gamma^1 \gamma^2 \gamma^3.
\]

The representation (\ref{NatLoc0,m,0,0,0AssU^2L^1/2}) given by the explicit local formula given above
is Krein isometric, but it is not unitary.

Let us define the following projectors:
\[
P^\oplus(p) = S\frac{1}{2}
\left( \begin{array}{cc} \boldsymbol{1}_2 & \beta(p)^{-1}\overline{\beta(p)}  \\
                                           
                                                  {\overline{\beta(p)}}^{-1}\beta(p) &  \boldsymbol{1}_2 \end{array}\right)S^{-1},  
\]
\[
P^\ominus(p) = S\frac{1}{2}
\left( \begin{array}{cc} \boldsymbol{1}_2 & -\beta(p)^{-1}\overline{\beta(p)}  \\
                                           
                                                  -{\overline{\beta(p)}}^{-1}\beta(p) &  \boldsymbol{1}_2 \end{array}\right)S^{-1},
\]
with the unitary involutive 
\[
S = \left( \begin{array}{cc}   -i\sigma_2  &  0 \\
                                           
                                                   0            & \boldsymbol{1}_2 \end{array}\right).
\]

Because of the orthogonality
\[
P^\oplus(p)P^\ominus(p) = P^\ominus(p)P^\oplus(p) = 0, \,\,\,\, P^\oplus(p)+ P^\ominus(p) = \boldsymbol{1}_4
\]
we have orthogonality 
\[
P^\oplus P^\ominus = P^\ominus P^\oplus = 0, \,\,\,\, P^\oplus + P^\ominus = \boldsymbol{1}.
\]
Although note that
\[
P^\oplus(p)^* \neq P^\oplus(p), \,\,\, P^\ominus(p)^* \neq P^\ominus(p) 
\]
in $\mathbb{C}^4$ with respect to  the canonical inner product $(\cdot, \cdot)_{{}_{\mathbb{C}^4}}$ and neither with respect to $(\cdot, B(p)\cdot)_{{}_{\mathbb{C}^4}}$. Similarly  
\[
{P^\oplus}^* \neq P^\oplus, \,\,\, {P^\ominus}^* \neq P^\ominus 
\]
in the Hilbert space of bispinors with respect to the inner product 
(\ref{Inn-Prod-NatLoc0,m,0,0,0AssU^2L^1/2}) and neither with respect to the inner product 
(\ref{Inn-Prod-NatLoc0,m,0,0,0AssU^2L^1/2})
with $B(p)$ put equal $\boldsymbol{1}_4$.

These projectors are Krein self adjoint (for $\mathfrak{J} = \gamma^0$) 
or Krein skew self adjoint (for $\mathfrak{J} = \gamma^1\gamma^2\gamma^3$).

Elements
\[
\tvect{\widetilde{\varphi}}{\widetilde{\chi}}
= 
P^{\oplus} {\widetilde{\phi}_{{}_{0,m}}} 
\]
 of the image of $P^\oplus$ respect the following algebraic relation 
\[
\big(p^0\gamma_0 - p^1 \gamma^1 - p^2 \gamma^2 - p^3 \gamma^3\big)
\widetilde{\phi} = im \widetilde{\phi}, \,\,\,\, \widetilde{\phi} = 
\tvect{\widetilde{\varphi}}{\widetilde{\chi}} \in \textrm{Im} \, P^\oplus.
\]

Elements
\[
\tvect{\widetilde{\varphi}}{\widetilde{\chi}}
= 
P^{\ominus} {\widetilde{\phi}_{{}_{0,m}}} 
\]
 of the image of $P^\ominus$ respect the following algebraic relation 
\[
\big(p^0\gamma_0 - p^1 \gamma^1 - p^2 \gamma^2 - p^3 \gamma^3\big)
\widetilde{\phi} = -im \widetilde{\phi}, \,\,\,\, \widetilde{\phi} = 
\tvect{\widetilde{\varphi}}{\widetilde{\chi}} \in \textrm{Im} \, P^\ominus.
\]

\begin{center}
\small
THE REPRESENTATION $\Big(W_{{}_{V,0,m,0,0}} U^{{}_{0,m,0,0}[ 2L^{{}^{1/2}}]_{\textrm{Ass}} } 
W_{{}_{V,0,m,0,0}}^{-1}\Big)_{{}_{0}}$ 
\end{center}

According to Proposition of introduction to Section \ref{constr-of-VF} we obtain 
the representation
\begin{equation}\label{NatLoc0,m,0,0,0AssU^2L^1/20}
\Big(W_{{}_{V,0,m,0,0}} U^{{}_{0,m,0,0}\big[ 2L^{{}^{1/2}}\big]_{\textrm{Ass}} } 
W_{{}_{V,0,m,0,0}}^{-1}\Big)_{{}_{0}}
\end{equation}
Krein-isometrically equivalent to  the natural local version (\ref{NatLoc0,m,0,0,0AssU^2L^1/2})
by restriction to the dense nuclear subspace $E_{{}_{0,m,4}}$ in the representation space
of (\ref{NatLoc0,m,0,0,0AssU^2L^1/2}). Then the action of (\ref{NatLoc0,m,0,0,0AssU^2L^1/20}) on
$E_{{}_{0,m,4}}$ is identical as the action of  (\ref{NatLoc0,m,0,0,0AssU^2L^1/2}). The
inner product on $E_{{}_{0,m,4}}$ regarded as a representation subspace of the representation
(\ref{NatLoc0,m,0,0,0AssU^2L^1/20}) is defined by the formula for the inner
product (\ref{Inn-Prod-NatLoc0,m,0,0,0AssU^2L^1/2})
of the representation space of (\ref{NatLoc0,m,0,0,0AssU^2L^1/2}) in which the operator matrix
$B(p)$ is the constant unit matrix. Thus in explicit form the inner product on $E_{{}_{0,m,4}}$
regarded as the representation  subspace of the representation (\ref{NatLoc0,m,0,0,0AssU^2L^1/20})
is equal
\begin{equation}\label{Inn-Prod-NatLoc0,m,0,0,0AssU^2L^1/20}
(\widetilde{\phi}_{{}_{0,m}}, \widetilde{\phi'}_{{}_{0,m}})  = 
  \int \limits_{\mathscr{O}_{(0,m,0,0)}} 
\Big(  \widetilde{\phi}_{{}_{0,m}}(p), 
\widetilde{\phi}'_{{}_{0,m}}(p)  \Big)_{{}_{\mathbb{C}^4}}
\, \ud \mu_{{}_{0,m}}.
\end{equation}
The Hilbert space $\mathcal{H'}_{{}_{0}}$ of the representation  
(\ref{NatLoc0,m,0,0,0AssU^2L^1/20}) is defined as the completion of $E_{{}_{0,m,4}}$ 
with respect to the inner product (\ref{Inn-Prod-NatLoc0,m,0,0,0AssU^2L^1/20}). 

The fundamental symmetry operator on $\mathcal{H'}_{{}_{0}}$ under which 
(\ref{NatLoc0,m,0,0,0AssU^2L^1/20}) is Krein isometric and Krein-isometrically equivalent to 
(\ref{NatLoc0,m,0,0,0AssU^2L^1/2}) is equal to the operator of multiplication by the 
constant matrix operator
\begin{equation}\label{J-NatLoc0,m,0,0,0AssU^2L^1/20}
\mathfrak{J}_{\bar{p}} = \left( \begin{array}{cc}   0 &  \bold{1}_2  \\
                                           
                                                   \bold{1}_2              & 0 \end{array}\right)
                                                   = \gamma^0 
                                                   \,\,\,
                                                   \textrm{or}
                                                   \,\,\,
\mathfrak{J}_{\bar{p}} = \left( \begin{array}{cc}  0 &  i \bold{1}_2  \\
                                           
                                                   -i \bold{1}_2              & 0  \end{array}\right)
= \gamma^1 \gamma^2 \gamma^3.
\end{equation}

In the sequel when constructing  $V_\mathcal{F}$ we will consider the
Hilbert space of ordinary space-time bispinors (and their generalisations with arbitrary 
high dimension of the corresponding Clifford algebra) square integrable 
\[
\int \Big( \phi(x), \phi(x) \Big)_{\mathbb{C}^4} \, \ud^4 x < + \infty,
\]
on the whole 
space-time with respect to the ordinary invariant volume form;
together with the fundamental symmetry $\mathfrak{J}$ defined by the formula 
\[
(\mathfrak{J}\phi)(x) = \gamma^0 \phi(x) 
\,\,\,
\textrm{or}
\,\,\,
(\mathfrak{J}\phi)(x) = \gamma^1 \gamma^2 \gamma^3 \phi (x),
\]
and the representation $U^{0}$ of $T_4 \circledS SL(2, \mathbb{C})$
given by 
\begin{equation}\label{U^0-onSpace-Time-Bispinors}
U^{0}(\alpha) \phi(x) = 
\left( \begin{array}{cc}  \alpha & 0  \\

                                                   0              & {\alpha^*}^{-1}  \end{array}\right) 
 \phi(\Lambda(\alpha)x), 
\end{equation}
\[
T^{0}(a) \phi(x) = \phi(x - a).
\]
This representation is Krein-unitary with respect to the assumed fundamental symmetry  
and Hilbert space structure defined above on space-time bispinors. 
The corresponding (inverse Fourier transformed) Dirac operator 
\[
D = i \widetilde{\gamma}^\mu \partial_\mu \,\,\, (\textrm{summation \, with \, respect \, to} \, \mu)
\]
\[
\widetilde{\gamma}^0= \gamma^0, \,\,\,\,\,\, \widetilde{\gamma}^k = - \gamma^k, \,\, k=1,2,3,
\]
is Krein self-adjoint for the first of the above two fundamental symmetries $\mathfrak{J}= \gamma^0 = \widetilde{\gamma}^0$, 
compare \cite{baum}, and commutes with the representation (\ref{U^0-onSpace-Time-Bispinors}). 
The above Dirac operator $D$ is anti-Krein-selfadjoint with respect to the second fundamental symmetry, given by 
$\gamma^1\gamma^2\gamma^3$, and after multiplication by $i$, \emph{i. e.} $iD$, becomes Krein-self-adjoint for the second fundamental symmetry
$\gamma^1\gamma^2\gamma^3$. Otherwise the second fundamental symmetry is associated to the generators
$i\gamma^\mu$, $\mu = 0, \ldots, 3$, of the Clifford algebra determined by the Minkowski metric
\[
g'^{\mu\nu} = \left( \begin{array}{cccc}   -1 &  0 &0 & 0  \\
                                           0 & 1 & 0 & 0   \\
                                           0 & 0 & 1 & 0   \\
                                                   0 & 0 & 0 & 1   \\ \end{array}\right)                                     
\]
with the opposite signature in comparison to the Minkowski metric
\[
g'^{\mu\nu} = \left( \begin{array}{cccc}   1 &  0 &0 & 0  \\
                                           0 & -1 & 0 & 0   \\
                                           0 & 0 & -1 & 0   \\
                                                   0 & 0 & 0 & -1   \\ \end{array}\right)                                     
\]
we are using.

 Next we consider the 
ordinary Fourier 
transform of this space-time bispinors and the Krein-Hilbert space space of 
restrictions of their Fourier transforms, to the orbits $\mathscr{O}_{m,0,0,0}$, $m \in \mathbb{R}$
and $\mathscr{0}_{0,m,0,0}$, $m >0$, obtained from decomposition through the Fubini theorem, 
on regarding the momentum space $\mathbb{R}^4$
as (disjoint) sum of five manifolds, one zero dimensional single point manifold 
$\mathscr{O}_{0,0,0,0}$, one three dimensional positive energy cone $\mathscr{O}_{1,0,0,1}$,
negative energy cone $\mathscr{O}_{-1,0,0,1}$  and two having  the form of Cartesian product manifolds:
\[
\mathscr{O}_{0,0,0,0} 
\cup \big\{m \times \mathscr{O}_{m,0,0,0}\big\}_{{}_{m \in \mathbb{R}\setminus\{0\}}} \cup 
\big\{m \times \mathscr{O}_{0,m,0,0}\big\}_{{}_{m >0}},
\]
with the invariant measures on the orbits induced by the invariant measure $\ud^4 p$
on $\mathbb{R}^4$ and by the invariant function 
\begin{equation}\label{FUNCTIONm(p)}
m(p) = 
\left\{ \begin{array}{ll}
    \big[\textrm{sign}  (p\cdot p) \, p \cdot p \big]^{1/2}, & p\cdot p <0, \\
    \big[\textrm{sign}  (p\cdot p) \, p \cdot p \big]^{1/2}, & p\cdot p > 0, p^0 > 0, \\
    \textrm{sign} (p^0) \big[\textrm{sign}  (p\cdot p) \, p \cdot p \big]^{1/2}, & p\cdot p <0,
    p^0 <0,
    \end{array} \right.
\end{equation}
on the momentum space $\mathbb{R}^4$. We have
\[
\ud^4p = dm(p) \times \ud \mu_{\mathscr{O}_{m}}(p)
\]
where $\ud \mu_{\mathscr{O}_{m}}(p)$ stands for the induced invariant measure on the orbit
$\mathscr{O}_{0,m,0,0}$ or respectively $\mathscr{O}_{m,0,0,0}$.

Because the measure of the first three manifolds
is zero in the ordinary invariant measure on $\mathbb{R}^4$, then the contributions
of them can be omitted (we will analyze them separately as they are still included into the absolute continuous spectrum of the Dirac operator) in the integration of the Fourier transforms of space-time
bispinors, and thus we are left with integration over two open and disjoint manifolds
both having a natural invariant Cartesian product structure, thus inducing decomposition via 
the Fubini theorem.  

 The Krein-Hilbert space
of Fourier transforms of spacetime bispinors, when restricted to the orbit
$\mathscr{O}_{m,0,0,0}$, $m\neq 0$, or $\mathscr{O}_{0,m,0,0}$, $m>0$,
$\mathscr{O}_{0,,0,0}$, $\mathscr{O}_{1,0,0,1}$  is the  
Krein-Hilbert space of bispinors
 of the local representation 
\begin{equation}\label{FourierDecRepU^0}
\left\{ \begin{array}{ll}
 \Big(W_{{}_{V,m,0,0,0}} U^{{}_{m,0,0,0}(2L^{{}^{1/2}})}  
W_{{}_{V,m,0,0,0}}^{-1}\Big)_{{}_{0}}, 
 & \textrm{for $\mathscr{O}_{m,0,0,0}$, $m\neq 0$}, \\
    \Big(W_{{}_{V,0,m,0,0}} U^{{}_{0,m,0,0}[ 2L^{{}^{1/2}}]_{\textrm{Ass}} } 
W_{{}_{V,0,m,0,0}}^{-1}\Big)_{{}_{0}}, 
& \textrm{for $\mathscr{O}_{0,m,0,0}$, $m> 0$}, \\
   V, 
& \textrm{for $\mathscr{O}_{0,0,0,0}$}, \\
\Big(W_{{}_{V,1,0,0,1}} U^{{}_{1,0,0,1}(2L^{{}^{1/2}})}  
W_{{}_{V,1,0,0,1}}^{-1}\Big)_{{}_{0}}, 
 & \textrm{for $\mathscr{O}_{1,0,0,1}$},
    \end{array} \right.
\end{equation}
with common Krein-unitary representation $V$
\[
\alpha \mapsto V(\alpha) = 
\left( \begin{array}{cc}  \alpha & 0  \\
                                           
                                                   0              & {\alpha^*}^{-1}  \end{array}\right) 
= \alpha \oplus {\alpha^*}^{-1}
\]
of $SL(2, \mathbb{C})$.
And with the inner product on each of the restrictions to the respective orbits 
equal 
\begin{equation}\label{natural-inn}
(\widetilde{\phi}|_{{}_{\mathscr{O}_{\overline{p}}}}, \widetilde{\phi'}|_{{}_{\mathscr{O}_{\overline{p}}}})  = 
  \int \limits_{\mathscr{O}_{(0,m,0,0)}} 
\Big(  \widetilde{\phi}|_{{}_{\mathscr{O}_{\overline{p}}}}(p), 
\widetilde{\phi}'|_{{}_{\mathscr{O}_{\overline{p}}}}(p)  \Big)_{{}_{\mathbb{C}^4}}
\, \ud \mu_{{}_{\mathscr{O}_{\overline{p}}}},
\end{equation}
with 
\[
\ud \mu_{{}_{\mathscr{O}_{\overline{p}}}}
\]
equal to the invariant measure  on the orbit $\mathscr{O}_{\overline{p}}$, and indeed agrees with 
(\ref{Inn-Prod-NatLocU^m,0,0,02L^1/20}) and respectively (\ref{Inn-Prod-NatLoc0,m,0,0,0AssU^2L^1/20}).
The fundamental symmetry operator in each Hilbert space of Fourier transforms restricted to the corresponding
orbit is equal to the operator of multiplication by the constant matrix 
\[
\mathfrak{J}_{\bar{p}} = \left( \begin{array}{cc}   0 &  \bold{1}_2  \\
                                           
                                                   \bold{1}_2              & 0 \end{array}\right)
                                                   = \gamma^0 
                                                   \,\,\,
                                                   \textrm{or}
                                                   \,\,\,
\mathfrak{J}_{\bar{p}} = \left( \begin{array}{cc}  0 &  i \bold{1}_2  \\
                                           
                                                   -i \bold{1}_2              & 0  \end{array}\right)
= \gamma^1 \gamma^2 \gamma^3 ,
\]
and agrees with (\ref{J-NatLocU^m,0,0,02L^1/20}) and respectively with (\ref{J-NatLoc0,m,0,0,0AssU^2L^1/20}),
where the of the two fundamental symmetry operators correspond to the signature $1,-1,-1,-1,-1$
of the Minkowski metric, and the second fundamental symmetry operator corresponds to the signature
$(-1,1,1,1)$ of the Minkowski metric with the Clifford algebra generators $\gamma^\mu$,
$\mu=0,1,2,3$, replaced with $i\gamma^\mu$.

Note that the representation $V$ understood as associated with the orbit 
$\mathscr{O}_{0,0,0,0}= \{(0,0,0,0)\}$ in the third line of the formula (\ref{FourierDecRepU^0}) 
is understood as representation of $T_4 \circledS SL(2, \mathbb{C})$ uniquely determined by the
representation $V$ of $SL(2, \mathbb{C})$ by the canonical extension of $V$ to the whole
$SL(2, \mathbb{C})$ acting trivially on the normal translation subgroup $T_4$:
\[
V(a, \alpha) = V(\alpha), \,\,\, (a, \alpha) \in SL(2, \mathbb{C}).
\]

Note also that here we  restrict ourselves to space-time bispinors 
\[
\phi \in  \mathcal{S}(\mathbb{R}^4; \mathbb{C}^4).
\]

\vspace*{1cm}

\begin{center}
\small REPRESENTATION $U^{{}_{(0,m,0,0)} \big[2L^{{}^{1/2}}\big]_{\textrm{Ass}} }$ ASSOCIATED TO 
$U^{{}_{(m,0,0,0)} 2L^{{}^{1/2}}}$ AND THE CORRESPONDING 
GENERALIZED EIGENSPACE OF DIRAC OPERATOR
\end{center}

Concerning relation of the representation (\ref{Inn-Prod-NatLoc0,m,0,0,0AssU^2L^1/2})
to the eigenspaces of the Dirac operator stated above, it can be likewise recovered by considering a
representation closely related to (\ref{Inn-Prod-NatLoc0,m,0,0,0AssU^2L^1/2}), 
and having identical Hilbert space, in which 
$\big[2L^{{}^{1/2}}\big]_{\textrm{Ass}}$ is replaced with a slightly different representation
$\big[2L^{{}^{1/2}}\big]_{\textrm{Ass}}'$.

Now consider the representation
\[
\gamma \mapsto Q(\gamma,\bar{p}) = \big[2L^{{}^{1/2}}\big]_{\textrm{Ass}}'(\gamma) =
\left( \begin{array}{cc}  \gamma & 0  \\
                                           
                                                   0              & \gamma  \end{array}\right) 
= \gamma \oplus \gamma
\]
of the group $SL(2, \mathbb{R})= G_{(0,m,0,0)}$, we extend by the formula
\[
\alpha \mapsto V'(\alpha) =  
\left( \begin{array}{cc}  V''(\alpha) & 0  \\
                                           
                                                   0              & \overline{V''}(\alpha)  \end{array}\right)       
\]
\[   
=\left( \begin{array}{cc}  \alpha & 0  \\
                                           
                                                   0              & \overline{\alpha}  \end{array}\right)           
= \alpha \oplus \overline{\alpha}
\]
to a representation $V'$ of $SL(2, \mathbb{C})$.

Recall that now for the orbit $\mathscr{O}_{0,m,0,0}$ representations ${V}''$ and $\overline{{V}''}$
of $SL(2, \mathbb{C})$ are conjugated whenever having the same representation space and coinciding on the isotropy subgroup $G_{0,0,m,0} = SL(2, \mathbb{R})$. 
Therefore $\overline{{V}''}(\alpha) = 
{V}''(\overline{\alpha})$, $\alpha \in SL(2, \mathbb{C})$, and $\overline{\alpha}$
is the ordinary complex conjugation of the matrix $\alpha$.

Both, the initial representation $\big[2L^{{}^{1/2}}\big]_{\textrm{Ass}}'$
of $SL(2,\mathbb{R})$ and its extension $V'$ to a representation of $SL(2, \mathbb{C})$
are Krein unitary in the Krein space $(\mathbb{C}^4, \mathfrak{J}_{\bar{p}})$ with
\[
\mathfrak{J}_{\bar{p}} = \left( \begin{array}{cc}   0 &  =-i \sigma_2  \\
                                           
                                                   i\sigma_2              & 0 \end{array}\right)
                                                   = \tilde{\gamma}^0 
                                                   \,\,\,
                                                   \textrm{or}
                                                   \,\,\,
\mathfrak{J}_{\bar{p}} = \left( \begin{array}{cc}  0 &  \sigma_2  \\
                                           
                                                   \sigma_2              & 0  \end{array}\right)
= \tilde{\gamma}^1 \tilde{\gamma}^2 \tilde{\gamma}^3 ,
\]
where 
\[
\sigma_2  =
\left( \begin{array}{cc}  0 &  -i  \\
                                           
                                                   i            & 0  \end{array}\right)
\]
is the second Pauli matrix and with the standard inner product in $\mathbb{C}^4$.
Here $\tilde{\gamma}^\mu \in M_4(\mathbb{C})$ are the matrix generators of the Clifford algebra
corresponding to the Minkowski metric (Dirac matrices) whose explicit form we give below.

It is easily seen that the Krein-Hilbert spaces of the induced representations 
$U^{{}_{0,m,0,0}\big[ 2L^{{}^{1/2}}\big]_{\textrm{Ass}} } 
$ and $U^{{}_{0,m,0,0}\big[ 2L^{{}^{1/2}}\big]_{\textrm{Ass}}' } $
are identical. 
Similarly it is easily seen that the representation Krein-Kilbert spaces 
of their local versions $W_{{}_{V,0,m,0,0}} U^{{}_{0,m,0,0}\big[ 2L^{{}^{1/2}}\big]_{\textrm{Ass}} } 
W_{{}_{V,0,m,0,0}}^{-1}$ and $W_{{}_{V,0,m,0,0}} U^{{}_{0,m,0,0}\big[ 2L^{{}^{1/2}}\big]_{\textrm{Ass}}' } 
W_{{}_{V,0,m,0,0}}^{-1}$ are identical.

Note in particular that by the general construction of introductory part of Sect. \ref{constr-of-VF} the inner product in the Krein-Hilbert space
of the natural local version 
\[
W_{{}_{V,0,m,0,0}} U^{{}_{0,m,0,0}\big[ 2L^{{}^{1/2}}\big]_{\textrm{Ass}}' } 
W_{{}_{V,0,m,0,0}}^{-1}
\]
 is equal
\begin{multline}\label{innLL'}
(\widetilde{\phi}_{{}_{0,m}}, \widetilde{\phi'}_{{}_{0,m}})  = 
  \int \limits_{\mathscr{O}_{(m,0,0,0)}} 
\Big(  \widetilde{\phi}_{{}_{0,m}}(p), 
V'(\beta(p))^* V'(\beta(p)) \widetilde{\phi}'_{{}_{0,m}}(p)  \Big)_{{}_{\mathbb{C}^4}}
\, \ud \mu_{{}_{0,m}} \\
= 
\int \limits_{\mathscr{O}_{(m,0,0,0)}} 
\Big(  \widetilde{\phi}_{{}_{0,m}}(p), 
B'(p) \widetilde{\phi}'_{{}_{0,m}}(p)  \Big)_{{}_{\mathbb{C}^4}}
\, \ud \mu_{{}_{0,m}}.
\end{multline}

Let us define the following projectors:
\[
P^\oplus(p) = \frac{1}{2}
\left( \begin{array}{cc} \boldsymbol{1}_2 & \beta(p)^{-1}\overline{\beta(p)}  \\
                                           
                                                  {\overline{\beta(p)}}^{-1}\beta(p) &  \boldsymbol{1}_2 \end{array}\right),  
\]
\[
P^\ominus(p) = \frac{1}{2}
\left( \begin{array}{cc} \boldsymbol{1}_2 & -\beta(p)^{-1}\overline{\beta(p)}  \\
                                           
                                                  -{\overline{\beta(p)}}^{-1}\beta(p) &  \boldsymbol{1}_2 \end{array}\right).
\]
Because of the orthogonality
\[
P^\oplus(p)P^\ominus(p) = P^\ominus(p)P^\oplus(p) = 0, \,\,\,\, P^\oplus(p)+ P^\ominus(p) = \boldsymbol{1}_4
\]
we have orthogonality 
\[
P^\oplus P^\ominus = P^\ominus P^\oplus = 0, \,\,\,\, P^\oplus + P^\ominus = \boldsymbol{1}.
\]
Although note that
\[
P^\oplus(p)^* \neq P^\oplus(p), \,\,\, P^\ominus(p)^* \neq P^\ominus(p) 
\]
in $\mathbb{C}^4$ with respect to  the canonical inner product $(\cdot, \cdot)_{{}_{\mathbb{C}^4}}$ and neither with respect to $(\cdot, B'(p)\cdot)_{{}_{\mathbb{C}^4}}$. Similarly  
\[
{P^\oplus}^* \neq P^\oplus, \,\,\, {P^\ominus}^* \neq P^\ominus 
\]
in the Hilbert space of bispinors with respect to the inner product 
(\ref{innLL'}) and neither with respect to the inner product (\ref{innLL'})
with $B'(p)$ put equal $\boldsymbol{1}_4$.

These projectors are Krein self adjoint (for $\mathfrak{J} = \gamma^0$) 
or Krein skew self adjoint (for $\mathfrak{J} = \gamma^1\gamma^2\gamma^3$).

Elements
\[
\tvect{\widetilde{\varphi}}{\widetilde{\chi}}
= 
P^{\oplus} {\widetilde{\psi}_{{}_{0,m}}}
\]
 of the image of $P^\oplus$ respect the following algebraic relation 
 \[
 \begin{array}{c}  \widetilde{\chi}(p) = \overline{\beta(p)}^{-1}\beta(p) \widetilde{\varphi} (p)\\
\widetilde{\varphi} (p) = \big(\overline{\beta(p)}^{-1} \beta(p) \big)^{-1} \widetilde{\chi}(p)                                        
                                \end{array}
 \]
which in more explicit form reads
\begin{eqnarray}
-p^0 \sigma_2 -ip^1 \sigma_3 + p^2 \boldsymbol{1}_{2} +i p^3 \sigma_1 \widetilde{\varphi} = 
m \widetilde{\chi}, \\
p^0 \sigma_2 +ip^1 \sigma_3 + p^2 \boldsymbol{1}_{2} -i p^3 \sigma_1 \widetilde{\chi} = 
m \widetilde{\varphi},
\end{eqnarray}
or
\[
\big(p^0\tilde{\gamma}_0 + p^1 \tilde{\gamma}_1 + p^2 \tilde{\gamma}_2 + p^3 \tilde{\gamma}_3\big)
\widetilde{\phi} = im \widetilde{\phi}, \,\,\,\, \widetilde{\phi} = 
\tvect{\widetilde{\varphi}}{\widetilde{\chi}} \in \textrm{Im} \, P^\oplus.
\]
Here
\[
\tilde{\gamma}_0 = \left( \begin{array}{cc}   0 &  -i\sigma_2  \\
                                           
                                                   i\sigma_2             & 0 \end{array}\right), \,\,\,\,
\tilde{\gamma}_1 = \left( \begin{array}{cc}   0 &  \sigma_3  \\
                                           
                                                   -\sigma_3             & 0 \end{array}\right),
\]
\[
\tilde{\gamma}_2 = \left( \begin{array}{cc}   0 &  i  \\
                                           
                                                   i              & 0 \end{array}\right), \,\,\,\,
\tilde{\gamma}_3 = \left( \begin{array}{cc}   0 &  -\sigma_1  \\
                                           
                                                   \sigma_1             & 0 \end{array}\right),
\]
being the generators of the representation of the Clifford algebra:
\[
\tilde{\gamma}_\mu \tilde{\gamma}_\nu + \tilde{\gamma}_\nu \tilde{\gamma}_\mu = 2 g_{\mu \nu}
\]
associated to the Minkowski pseudo-metric $g_{\mu \nu}$ and associated to the Dirac gamma matrices
$\gamma_\mu$ in the chiral representation used in the previous Subsection in the following
manner
\[
S \gamma_\mu S^{-1} = \tilde{\gamma_\mu}, 
\]
with the unitary involutive 
\[
S = \left( \begin{array}{cc}   -i\sigma_2  &  0 \\
                                           
                                                   0            & \boldsymbol{1}_2 \end{array}\right).
\]

Similarly each element in the image of $P^\ominus$ respects the the condition 
\[
\big(p^0\tilde{\gamma}_0 + p^1 \tilde{\gamma}_1 + p^2 \tilde{\gamma}_2 + p^3 \tilde{\gamma}_3\big)
\widetilde{\phi} = -im \widetilde{\phi}, \,\,\,\, \widetilde{\phi} = 
\tvect{\widetilde{\varphi}}{\widetilde{\chi}} \in \textrm{Im} \, P^\ominus.
\]
Of course the operator of multiplication by the constant unitary and Krein-unitary  
matrix $S$ transforms the Hilbert space of bispinors into itself
and the  projectors $P^\oplus, P^\ominus$ into the corresponding projectors
\[
SP^\oplus S^{-1}, \,\,\, SP^\ominus S^{-1},
\]
and each element of the image $\textrm{Im} \, SP^\oplus S^{-1}$ 
(respectively $\textrm{Im} \, SP^\ominus S^{-1}$) ,in the Hilbert space of bispinors 
($W$-images of the representation space of the representation $U^{{}_{(0,m,0,0)} [L^{{}^{1/2}}]_{\textrm{Ass}}'}$ and thus 
likewise of $U^{{}_{(0,m,0,0)} [L^{{}^{1/2}}]_{\textrm{Ass}}}$) fulfils the equation
\[
\big(p^0\gamma_0 + p^1\gamma_1 + p^2\gamma_2 + p^3\gamma_3\big)
\widetilde{\phi} = im \widetilde{\phi}, \,\,\,\, \widetilde{\phi} = 
\tvect{\widetilde{\varphi}}{\widetilde{\chi}} \in \textrm{Im} \, SP^\oplus S^{-1}
\]
or respectively
\[
\big(p^0\gamma_0 + p^1\gamma_1 + p^2\gamma_2 + p^3\gamma_3\big)
\widetilde{\phi} = -im \widetilde{\phi}, \,\,\,\, \widetilde{\phi} = 
\tvect{\widetilde{\varphi}}{\widetilde{\chi}} \in \textrm{Im} \, SP^\ominus S^{-1},
\]
with $\gamma_\mu$ in the chiral representation used in the previous Subsection.

It is easily seen that the images
\[
\textrm{Im} \, SP^\oplus S^{-1}, \,\,\, \textrm{Im} \, SP^\ominus S^{-1},
\]
are invariant for both local representations 
\[
W_{{}_{V', 0,m,0,0}} \,\, U^{{}_{(0,m,0,0)} [L^{{}^{1/2}}]_{\textrm{Ass}}'} \,\, W_{{}_{V', 0,m,0,0}}^{-1},
\,\,\,\,\,\,
W_{{}_{V, 0,m,0,0}} \,\, U^{{}_{(0,m,0,0)} [L^{{}^{1/2}}]_{\textrm{Ass}}} \,\, W_{{}_{V, 0,m,0,0}}^{-1}
\]
acting on the Krein-Hilbert space $W$-image of bispinors.

\subsection{Spin 1/2 and the transform $V_\mathcal{F}$}\label{1/2VF}

Having the class of representations
\[ 
\left\{ \begin{array}{ll}
 \Big(W_{{}_{V,m,0,0,0}} U^{{}_{m,0,0,0}(2L^{{}^{1/2}})}  
W_{{}_{V,m,0,0,0}}^{-1}\Big)_{{}_{0}}, 
 & \textrm{for $\mathscr{O}_{m,0,0,0}$, $m\neq 0$}, \\
    \Big(W_{{}_{V,0,m,0,0}} U^{{}_{0,m,0,0}[ 2L^{{}^{1/2}}]_{\textrm{Ass}} } 
W_{{}_{V,0,m,0,0}}^{-1}\Big)_{{}_{0}}, 
& \textrm{for $\mathscr{O}_{0,m,0,0}$, $m> 0$}, \\
   V, 
& \textrm{for $\mathscr{O}_{0,0,0,0}$}, \\
\Big(W_{{}_{V,1,0,0,1}} U^{{}_{1,0,0,1}(2L^{{}^{1/2}})}  
W_{{}_{V,1,0,0,1}}^{-1}\Big)_{{}_{0}}, 
 & \textrm{for $\mathscr{O}_{1,0,0,1}$},
    \end{array} \right.
\]
we are ready to construct the transform $V_\mathcal{F}$ on the
space of the representation 
\begin{multline}\label{step1LocNaturallyAss}
\int \limits_{-\infty}^{\infty} 
\Big(W_{{}_{V,m,0,0,0}} U^{{}_{m,0,0,0}\big(2L^{{}^{1/2}}\big)}  
W_{{}_{V,m,0,0,0}}^{-1}\Big)_{{}_{0}}
\, \ud m \bigoplus \\ \bigoplus 
\int \limits_{0}^{\infty} 
\Big(W_{{}_{V,0,m,0,0}} U^{{}_{0,m,0,0}\big[ 2L^{{}^{1/2}}\big]_{\textrm{Ass}} } 
W_{{}_{V,0,m,0,0}}^{-1}\Big)_{{}_{0}}
\, \ud m,
\end{multline}
where $\ud m$ is the Lebesgue measure on $\mathbb{R}$.

In fact, we will analyze three more cases at once. In addition to (\ref{step1LocNaturallyAss})
we will analyze
\begin{multline}\label{step1LocAss}
\int \limits_{-\infty}^{\infty} 
W_{{}_{V,m,0,0,0}} U^{{}_{m,0,0,0}(2L^{{}^{1/2}})}  
W_{{}_{V,m,0,0,0}}^{-1}
\, \ud m \bigoplus \\ \bigoplus 
\int \limits_{0}^{\infty} 
W_{{}_{V,0,m,0,0}} U^{{}_{0,m,0,0}[ 2L^{{}^{1/2}}]_{\textrm{Ass}} } 
W_{{}_{V,0,m,0,0}}^{-1}
\, \ud m,
\end{multline}
and:
\begin{multline}\label{step1LocNonNatural}
\int \limits_{-\infty}^{\infty} 
\Big(W_{{}_{V,m,0,0,0}} U^{{}_{m,0,0,0}(2L^{{}^{1/2}})}  
W_{{}_{V,m,0,0,0}}^{-1}\Big)_{{}_{00}}
\, \ud m \bigoplus \\ \bigoplus 
\int \limits_{0}^{\infty} 
\Big(W_{{}_{V,0,m,0,0}} U^{{}_{0,m,0,0}[ 2L^{{}^{1/2}}]_{\textrm{Ass}} } 
W_{{}_{V,0,m,0,0}}^{-1}\Big)_{{}_{0}}
\, \ud m,
\end{multline}
\begin{multline}\label{step1LocNonNatural'}
\int \limits_{-\infty}^{\infty} 
\Big(W_{{}_{V,m,0,0,0}} U^{{}_{m,0,0,0}(2L^{{}^{1/2}})}  
W_{{}_{V,m,0,0,0}}^{-1}\Big)_{{}_{00}}
\, \ud m \bigoplus \\ \bigoplus 
\int \limits_{0}^{\infty} 
W_{{}_{V,0,m,0,0}} U^{{}_{0,m,0,0}[ 2L^{{}^{1/2}}]_{\textrm{Ass}} } 
W_{{}_{V,0,m,0,0}}^{-1}
\, \ud m,
\end{multline}
where 
\[
\Big(W_{{}_{V,m,0,0,0}} U^{{}_{m,0,0,0}\big(2L^{{}^{1/2}}\big)}  
W_{{}_{V,m,0,0,0}}^{-1}\Big)_{{}_{00}}
\]
is the non-natural and unitary local version (\ref{U^2L^1/2_00m>0}) (resp. (\ref{U^2L^1/2_00m<0})) of the representation
\[
U^{{}_{m,0,0,0}\big(2L^{{}^{1/2}}\big)}, \,\,\, m>0, \,\,\,
\textrm{resp.} \,\,\, 
U^{{}_{-m,0,0,0}\big(2L^{{}^{1/2}}\big)}, \,\,\, m>0, \,\,\,
\textrm{resp.} \,\,\, 
\]
constructed in Subsection \ref{e1}.

Let $\widetilde{\psi}$ be any element of the representation space of the representation 
\begin{multline}\label{step1rep}
\int \limits_{-\infty}^{\infty} 
W'_{{}_{V,m,0}}U_{{}_{V,m,0}}^{-1}U^{{}_{m,0,0,0}\big(2L^{{}^{1/2}}\big)}
U_{{}_{V,m,0}}{W'}_{{}_{V,m,0}}^{-1}
\, \ud m \bigoplus \\ \bigoplus 
\int \limits_{0}^{\infty} 
W'_{{}_{V,0,m}}U_{{}_{V,0,m}}^{-1} U^{{}_{0,m,0,0}\big[ 2L^{{}^{1/2}}\big]_{\textrm{Ass}} } 
U_{{}_{V,0,m}}
{W'}_{{}_{V,0,m}}^{-1}
\, \ud m,
\end{multline}
given by the corresponding direct sum of direct integrals of the representations
(\ref{W'U^-1U^m,0,0,02L1/2UW'^-1}) and respectively  (\ref{W'U^-1U^0,m,0,0,0AssU^2L^1/2UW'^-1}). 
Both summands of the direct integral (\ref{step1rep}) may be treated as direct integrals of ordinary Hilbert spaces, equipped additionally with the fundamental symmetry operator equal 
\begin{equation}\label{step1J}
\int \limits_{\mathbb{R}} \mathfrak{J}_{{}_{m,0}} \, \ud m \oplus
\int \limits_{\mathbb{R}_+} \mathfrak{J}_{{}_{0,m}} \, \ud m, \,\,\,\, 
\mathfrak{J}_{{}_{m,0}} = \mathfrak{J}_{{}_{0,m}} = \gamma^0 \,\,\, \textrm{or} \,\,\,
= \gamma^1 \gamma^2 \gamma^3.
\end{equation}
Now into the set theoretical sum of the positive cone $C_+$, $p^0 >0$ and the negative cone $C_-$, $p^0 < 0$ 
we may introduce the coordinates $m$ and $\vec{p}$ in $\mathscr{O}_{(m,0,0,0)}$, and thus we can use $(m, \vec{p})$ instead
of $(p^0 , \vec{p})$, treating the disjoint sum $C_+ \cup C_-$ as a Cartesian product $\mathbb{R} \times \mathbb{R}^3$. 
Similarly, for complementary part $C_{\pm}$ of the joint spectrum of the momentum operators lying outside the disjoint sum
$C_+ \cup C_-$, we may use coordinate $m$ together with the coordinates on $\mathscr{O}_{(0,m,0,0)}$ and treat $C_{\pm}$ as a Cartesian product of $\mathbb{R}_+ \times \mathbb{R}
\times \mathbb{S}^2 \cong C_{\pm}$, as we have described in Subsection \ref{e2}.

By the relation of the direct integral Hilbert space with the generalized Fubini theorem,
and in view of the uniform and constant multiplicity
separately over the classes of orbits $\mathscr{O}_{m,0,0,0,0}$, $m \in \mathbb{R}$ and
$\mathscr{O}_{(0,m,0,0)}$, $m \in \mathbb{R}_+$, it follows that (identifying every
element $\widetilde{\psi}$
with its decomposition function) $\widetilde{\psi}$ may be identified with the pair of measurable functions
(compare e.g. eq. (\ref{dir_int_skew_L^2:decompositions}) or (\ref{dir_int_L^2:decompositions}) of Section
\ref{decomposition})
\begin{equation}\label{psi-psi}
\begin{split}
C_+ \cup C_- \cong \mathbb{R} \times \mathbb{R}^3 \ni (m, p) \mapsto \widetilde{\psi}_{{}_{m,0}}(p) \,\,\, 
p \in \mathscr{O}_{(m,0,0,0)} \cong  \mathbb{R}^3 \\
C_{\pm} \cong \mathbb{R} \times \mathbb{S}^2 \times \mathbb{R} \ni (m, p) \mapsto \widetilde{\psi}_{{}_{0,m}}(p), \,\,\, p
 \in \mathscr{O}_{(0,m,0,0)} \cong  \mathbb{S}^2 \times \mathbb{R},
\end{split}
\end{equation}
such that 
\[
|| \widetilde{\psi} ||^2 = \int \limits_{\mathbb{R}} || \widetilde{\psi} _{{}_{m,0}} ||^2 \, \ud m 
+ \int \limits_{\mathbb{R}_+} || \widetilde{\psi} _{{}_{0,m}} ||^2 \, \ud m < + \infty,
\]
where
\[
\begin{split}
|| \widetilde{\psi} _{{}_{m,0}} ||^2 = \int \limits_{\mathscr{O}_{m,0,0,0}} 
\Big(\widetilde{\psi} _{{}_{m,0}}(p),\widetilde{\psi} _{{}_{m,0}}(p) \Big)_{\mathbb{C}^4} \, \ud \mu_{{}_{m,0}}(p), \\
|| \widetilde{\psi} _{{}_{0,m}} ||^2 = \int \limits_{\mathscr{O}_{0,0,m,0}} 
\Big(\widetilde{\psi} _{{}_{m,0}}(p),\widetilde{\psi} _{{}_{0,m}}(p) \Big)_{\mathbb{C}^4} \, \ud \mu_{{}_{0,m}}(p), \\
\end{split}
\]
and where
\[
\widetilde{\psi} _{{}_{m,0}} = {\widetilde{\psi} _{{}_{m,0}}}^{\oplus} \oplus {\widetilde{\psi} _{{}_{m,0}}}^{\ominus}
\]
we treat as one four-component function (compare Subsection \ref{e1}).
 
By the constructions of Subsection \ref{e1} and \ref{e2} (Examples 1 and 2) we can identify 
elements $\widetilde{\phi}$ of  the representation space of the representations (\ref{step1LocNaturallyAss})
-(\ref{step1LocNonNatural'}) respectively,  with their
decomposition functions given by the pair of measurable functions
\begin{equation}\label{phi-phi}
\begin{split}
C_+ \cup C_- \cong \mathbb{R} \times \mathbb{R}^3 \ni (m, p) \mapsto \widetilde{\phi}_{{}_{m,0}}(p) \,\,\, 
p \in \mathscr{O}_{(m,0,0,0)} \cong  \mathbb{R}^3 \\
C_{\pm} \cong \mathbb{R} \times \mathbb{S}^2 \times \mathbb{R} \ni (m, p) \mapsto \widetilde{\phi}_{{}_{0,m}}(p), \,\,\, p
 \in \mathscr{O}_{(0,0,m,0)} \cong \mathbb{S}^2 \times \mathbb{R},
\end{split}
\end{equation}
such that 
\[
|| \widetilde{\phi} ||^2 = \int \limits_{\mathbb{R}} || \widetilde{\phi} _{{}_{m,0}} ||^2 \, \ud m 
+ \int \limits_{\mathbb{R}_+} || \widetilde{\phi} _{{}_{0,m}} ||^2 \, \ud m < + \infty,
\]
where we have, respectively,
\[
\left. \begin{array}{r}
|| \widetilde{\phi} _{{}_{m,0}} ||^2  =  \int \limits_{\mathscr{O}_{m,0,0,0}} 
\Big(\widetilde{\phi} _{{}_{m,0}}(p),\widetilde{\phi} _{{}_{m,0}}(p) \Big)_{\mathbb{C}^4} 
\, \ud \mu_{{}_{m,0}}(p), \\
|| \widetilde{\phi} _{{}_{0,m}} ||^2 = \int \limits_{\mathscr{O}_{0,m,0,0}} 
\Big(\widetilde{\phi} _{{}_{0,m}}(p), \widetilde{\phi} _{{}_{0,m}}(p) \Big)_{\mathbb{C}^4} \, 
\ud \mu_{{}_{0,m}}(p), 
\end{array} \right\} \,\,\,
\textrm{for rep. (\ref{step1LocNaturallyAss})},
\]
\[
\left. \begin{array}{r}
|| \widetilde{\phi} _{{}_{m,0}} ||^2  =  \int \limits_{\mathscr{O}_{m,0,0,0}} 
\Big(\widetilde{\phi} _{{}_{m,0}}(p),
V(\beta(p))^* V(\beta(p))\widetilde{\phi} _{{}_{m,0}}(p) \Big)_{\mathbb{C}^4} \, 
\ud \mu_{{}_{m,0}}(p), \\
|| \widetilde{\phi} _{{}_{0,m}} ||^2 = \int \limits_{\mathscr{O}_{0,m,0,0}} 
\Big(\widetilde{\phi} _{{}_{0,m}}(p), V(\beta(p))^* V(\beta(p))\widetilde{\phi} _{{}_{0,m}}(p) \Big)_{\mathbb{C}^4} \, 
\ud \mu_{{}_{0,m}}(p), 
\end{array} \right\} \,\,\,
\textrm{for rep. (\ref{step1LocAss})},
\]
\[
\left. \begin{array}{r}
|| \widetilde{\phi} _{{}_{m,0}} ||^2  =  \int \limits_{\mathscr{O}_{m,0,0,0}} 
\Big(\widetilde{\phi} _{{}_{m,0}}(p),\widetilde{\phi} _{{}_{m,0}}(p) \Big)_{\mathbb{C}^4} \, \frac{1}{2 p^0} \, \ud \mu_{{}_{m,0}}(p), \\
|| \widetilde{\phi} _{{}_{0,m}} ||^2 = \int \limits_{\mathscr{O}_{0,m,0,0}} 
\Big(\widetilde{\phi} _{{}_{0,m}}(p), \widetilde{\phi} _{{}_{0,m}}(p) \Big)_{\mathbb{C}^4} \, 
\ud \mu_{{}_{0,m}}(p), 
\end{array} \right\} \,\,\,
\textrm{for rep. (\ref{step1LocNonNatural})},
\]
\[
\left. \begin{array}{r}
|| \widetilde{\phi} _{{}_{m,0}} ||^2  =  \int \limits_{\mathscr{O}_{m,0,0,0}} 
\Big(\widetilde{\phi} _{{}_{m,0}}(p),\widetilde{\phi} _{{}_{m,0}}(p) \Big)_{\mathbb{C}^4} \, \frac{1}{2 p^0} \, \ud \mu_{{}_{m,0}}(p), \\
|| \widetilde{\phi} _{{}_{0,m}} ||^2 = \int \limits_{\mathscr{O}_{0,m,0,0}} 
\Big(\widetilde{\phi} _{{}_{0,m}}(p), V(\beta(p))^* V(\beta(p))\widetilde{\phi} _{{}_{0,m}}(p) \Big)_{\mathbb{C}^4} \, 
\ud \mu_{{}_{0,m}}(p), 
\end{array} \right\} \,\,\,
\textrm{for rep. (\ref{step1LocNonNatural'})},
\]
provided $\widetilde{\phi} _{{}_{m,0}} \in E_{{}_{m,0,4}}$, resp. $\widetilde{\phi} _{{}_{0,m}} \in
E_{{}_{0,m,4}}$. Here $V$ is the fixed representation of $SL(2, \mathbb{C}^4)$ of Examples 1 and 2,
Subsection \ref{e1}, \ref{e2}. Here the function $p\mapsto \beta(p)$ depend on the actual orbit and can likewise be regarded as function of four-momentum, where the parameter $m$ in the formula for $\beta$ (of Subsect. \ref{e1} or \ref{e2}, depending on the orbit) is replaced with the invariant function $p \mapsto m(p)$
given by the formula (\ref{FUNCTIONm(p)}).

Note that the map $\widetilde{\psi} \mapsto \widetilde{\phi}$ is Krein-isometric
in the first two cases of representations (\ref{step1LocNaturallyAss}) and (\ref{step1LocAss})
and even Krein-unitary in the first case (\ref{step1LocNaturallyAss}). In the last two cases situation 
is more complicated. 

In case of representation (\ref{step1LocNonNatural'}) the map 
$\widetilde{\psi} \mapsto \widetilde{\phi}$ is not
unitary on the direct summand over $C_+\cup C_-$ represented by
decomposition functions $m,p \mapsto \widetilde{\psi} _{{}_{m,0}}(p)$ over $C_+\cup C_-$. It is not
Krein-unitary there (for the natural Krein structure defined by (\ref{step1J})).
It is only Krein-isometric in the second direct summand over $C_\pm$
represented by decomposition functions $m,p \mapsto \widetilde{\psi} _{{}_{0,m}}(p)$
over $C_\pm$. 

Similarly, for the representation (\ref{step1LocNonNatural}) the map 
$\widetilde{\psi} \mapsto \widetilde{\phi}$ is not
unitary on the direct summand over $C_+\cup C_-$ represented by
decomposition functions $m,p \mapsto \widetilde{\psi} _{{}_{m,0}}(p)$ over $C_+\cup C_-$. It is not
Krein-unitary there (for the natural Krein structure defined by (\ref{step1J})).
It is Krein-unitary in the second direct summand over $C_\pm$
represented by decomposition functions $m,p \mapsto \widetilde{\psi} _{{}_{0,m}}(p)$
over $C_\pm$.

Thus, any element $\widetilde{\phi}$ of the  representation space of the local representation 
(\ref{step1LocNaturallyAss})-(\ref{step1LocNonNatural'}) respectively, may be identified with a four-component measurable functions on $\mathbb{R}^4$, which
fulfil the following

\begin{center}
\small SUMMABILITY CONDITIONS
\end{center}
depending on the particular representation (\ref{step1LocNaturallyAss})-(\ref{step1LocNonNatural'}).

Elements $\widetilde{\phi}$ of the representation space of the representation
(\ref{step1LocNaturallyAss}) are just all square summable functions with respect to
the ordinary invariant measure $\ud^4 p$. 

Elements $\widetilde{\phi}$ of the representation space of the representation
(\ref{step1LocAss}) are such that $\Big(\widetilde{\phi}(p), V(\beta(p))^* V(\beta(p))\widetilde{\phi}(p) \Big)_{\mathbb{C}^4}$ is summable with respect to the measure $\ud^4 p$. Here
the function $p \mapsto \beta(p)$ is defined by the function $\beta$ depending on the orbit
and given by the formulas of Subsection \ref{e1} resp. \ref{e2} (depending on the class of orbits to which the actual point $p$ belongs) and is regarded as a function of four-momentum $p$ in which the parameter
$m$ is replaced with the invariant function $p \mapsto m(p)$ given by (\ref{FUNCTIONm(p)}). 

Elements $\widetilde{\phi}$ of the representation space of the representation
(\ref{step1LocNonNatural}) are just all functions square summable on $C_+ \cup C_-$ functions with respect to
the ordinary invariant measure $\frac{1}{2|p^{0}|}\ud^4 p$ and square measurable on 
$C_\pm$ with respect to $\ud^4p$.

In case of representation (\ref{step1LocNonNatural'}) elements $\widetilde{\phi}$ of the space of this representation $\widetilde{\phi}$ are square summable on the double cone $C_+ \cup C_-$ with respect to
the measure $\frac{1}{2|p^0|} \, \ud^4 p$ and such that $\Big(\widetilde{\phi}(p), V(\beta(p))^* V(\beta(p))\widetilde{\phi}(p) \Big)_{\mathbb{C}^4}$ is summable with respect to the measure $\ud^4 p$
outside the double cone, where the matrix $V(\beta(p))^* V(\beta(p))$ is given in Subsection \ref{e2}, with
$m$ replaced with $\sqrt{-p \cdot p}$ in the formula for $V(\beta(p))^* V(\beta(p))$.
\qedsymbol

\vspace*{1cm}

On the other hand consider the Hilbert space of bispinors $\phi$ on $\mathbb{R}^4$
equipped with the standard Minkowski metric, which are square integrable:
\begin{equation}\label{inn-x-step1}
\int \Big( \phi(x), \phi(x) \Big)_{\mathbb{C}^4} \, \ud^4 x < + \infty,
\end{equation} 
together with the fundamental symmetry $\mathfrak{J}$ defined by the formula 
\begin{equation}\label{Jstep1}
(\mathfrak{J} \phi)(x) = \gamma^0 \phi(x)
\,\,\,
\textrm{or}
\,\,\,
(\mathfrak{J}\phi)(x) = \gamma^1 \gamma^2 \gamma^3 \phi (x),
\end{equation} 
(the first for the signature $(1,-1,-1,-1)$ of the Minkowski space-time pseudo-metric, the second
for the signature $(-1,1,1,1)$ of the Minkowski space-time pseudo-metric) and the representation 
$U$ of $T_4 \circledS SL(2, \mathbb{C})$
given by 
\begin{equation}\label{Ustep1}
U(\alpha) \phi(x) = 
\left( \begin{array}{cc}  \alpha & 0  \\

                                                   0              & {\alpha^*}^{-1}  \end{array}\right) 
 \phi(\Lambda(\alpha)x), 
\end{equation}
\[
T(a) \phi(x) = \phi(x - a).
\]
This representation is Krein-unitary with respect to the fundamental symmetry (\ref{Jstep1}) and Hilbert space structure
(\ref{inn-x-step1}) defined as above. 
Now the Dirac operator 
\begin{equation}\label{DiracStep1}
D = i \widetilde{\gamma}^\mu \partial_\mu \,\,\, (\textrm{summation \, with \, respect \, to} \, \mu)
\end{equation}
\[
\widetilde{\gamma}^0 = \gamma^0, \,\,\,\,\,\, \widetilde{\gamma}^k = - \gamma^k, \,\, k=1,2,3,
\]
is Krein self adjoint for the first fundamental symmetry given by the operator of multiplication by 
$\gamma^0 \ \widetilde{\gamma}^0$, compare \cite{baum}, and commutes with  the representation (\ref{Ustep1}).
This immediately follows from the following relations (for the firs fundamental symmetry $\mathfrak{J}$ of multiplication by $\widetilde{\gamma}^0
\gamma^0$):
\[
\mathfrak{J}_{\bar{p}} \widetilde{\gamma}^0 = \widetilde{\gamma}^0 \mathfrak{J}_{\bar{p}}, \,\,
\mathfrak{J}_{\bar{p}} \widetilde{\gamma}^k = - \widetilde{\gamma}^k \mathfrak{J}_{\bar{p}},
(\widetilde{\gamma}^0)^* = \widetilde{\gamma}^0, \, (\widetilde{\gamma}^k)^* = - \widetilde{\gamma}^k.
\]
Similarly 
\[
\begin{split}
D^2 = - \widetilde{\gamma}^\mu \widetilde{\gamma}^\nu \partial_\mu \partial_\nu\,\,\, (\textrm{summation \, with \, respect \, to} \, \mu, \nu) \\
= - \bold{1}_4 \, (\partial_0 \partial_0 - \partial_1 \partial_1 - \partial_2 \partial_2 - \partial_3 \partial_3)
\end{split}
\]
commutes with the representation (\ref{Ustep1}) and it is well known that $D^2$ is (essentially) self-adjoint.

One easily checks that  
\[
\frac{1}{2} \Big\{ (D\mathfrak{J})^2 + (\mathfrak{J}D)^2 \Big\}
= - \bold{1}_{4} \, (\partial_0 \partial_0 + \partial_1 \partial_1 + \partial_2 \partial_2 + \partial_3 \partial_3),
\]
so that $\frac{1}{2} \Big\{ (D\mathfrak{J})^2 + (\mathfrak{J}D)^2 \Big\}$ is elliptic and we can
choose
\[
D_\mathfrak{J} = i \widetilde{\Upsilon}^\mu \partial_\mu,
\]
with $\Upsilon^\mu$, defined by
\[
\widetilde{\Upsilon}^0 = \widetilde{\gamma}^0, \,\,\, \widetilde{\Upsilon}^k = i \widetilde{\gamma}^k
\]
and being the generators of the Clifford algebra associated to the ordinary Euclidean metric
$\delta^{\mu\nu}$:
\[
\widetilde{\Upsilon}^\mu \widetilde{\Upsilon}^\nu + \widetilde{\Upsilon}^\nu \widetilde{\Upsilon}^\mu 
= 2 \delta^{\mu\nu} \, \bold{1}_{4}.
\]
For this $D_\mathfrak{J}$, $\mathfrak{J}$, and $D$ we indeed have
\[
\frac{1}{2} \Big\{ (D\mathfrak{J})^2 + (\mathfrak{J}D)^2 \Big\}
= \big( D_\mathfrak{J} \big)^2.
\]

The above Dirac operator $D$ becomes Krein self adjoint for the second fundamental symmetry $\mathfrak{J}$
defined by multiplication by the constant matrix $\gamma^1\gamma^2 \gamma^3$, only after multiplication
by $i$, \emph{i. e.} $iD$ is Krein self adjoint for the second fundamental symmetry $\mathfrak{J}$.
This follows from the relations
\[
\mathfrak{J}_{\bar{p}} \gamma^0 = -\gamma^0 \mathfrak{J}_{\bar{p}}, \,\,
\mathfrak{J}_{\bar{p}} \gamma^k = \gamma^k \mathfrak{J}_{\bar{p}},
\]
for the fundamental symmetry $\mathfrak{J}$ defined by
\[
(\mathfrak{J}\phi)(x) = \gamma^1 \gamma^2 \gamma^3 \phi (x).
\]
This means that the Dirac operator with the Clifford algebra generators $\gamma^\mu$, $i=0,1,2,3$,
replaced with $i\gamma^\mu$, $\mu=0,1,2,3$, is Krein-self adjoint,
or still otherwise speaking, that the second fundamental symmetry 
\[
(\mathfrak{J}\phi)(x) = \gamma^1 \gamma^2 \gamma^3 \phi (x)
\]
corresponds to the opposite signature
$(-1,1,1,1)$ of the space-time Minkowski pseudo-metric.

We may therefore use the ordinary Fourier transform and the rigged Hilbert space technique of Gelfand and his school (\cite{GelfandIV}, Chap. I.4, or \cite{GGG}) to decompose the representation (\ref{Ustep1}) using the generalized eigenvectors (eigenspaces) of the self-adjoint
operator $D^2$, commuting with the representation $U$ defined by (\ref{Ustep1}).
Let us describe shortly the decomposition, basing the whole construction on the Fourier transform
and the generalized Fubini theorem (eq. (\ref{dir_int_skew_L^2:decompositions}) or
(\ref{dir_int_L^2:decompositions}) of Sect.
\ref{decomposition}).
Namely, let $\widetilde{\phi}$ be the ordinary Fourier transform of the square integrable bispinor $\phi$,
so that
\[
\begin{split}
\phi(x) = (2 \pi)^{-1/2} \int \limits_{\mathbb{R}^4} \widetilde{\phi}(p) e^{-i p \cdot x} \,
\ud^4 p \\
= (2 \pi)^{-1/2} \int \limits_{\mathbb{R}} \int \limits_{\mathscr{O}_{(m,0,0,0,)}}
\widetilde{\phi}|_{{}_{\mathscr{O}_{(m,0,0,0)}}}(p) e^{-i p \cdot x} \, \ud m \, \ud \mu_{{}_{m,0}}(p) \\
+ (2 \pi)^{-1/2} \int \limits_{\mathbb{R}_+} \int \limits_{\mathscr{O}_{(0,0,m,0,)}}
\widetilde{\phi}|_{{}_{\mathscr{O}_{(0,0,m,0)}}}(p) e^{-i p \cdot x} \, \ud m \, \ud \mu_{{}_{0,m}}(p)
\end{split}
\]
where $\widetilde{\phi}|_{{}_{\mathscr{O}_{(m,0,0,0)}}}$ (resp. $\widetilde{\phi}|_{{}_{\mathscr{O}_{(0,0,m,0)}}}$)
is the restriction of the Fourier transform $\widetilde{\phi}$ of $\phi$ to the orbit $\mathscr{O}_{(m,0,0,0)}$
(resp. $\mathscr{O}_{(0,0,m,0)}$). We have the Plancherel formula
\[
\begin{split}
||\phi||^2 = ||\widetilde{\phi}||^2 = \int \limits_{\mathbb{R}^4}
\Big( \widetilde{\phi}(p), \widetilde{\phi}(p) \Big)_{\mathbb{C}^4} \, \ud^4 p \\
2 \pi)^{-1/2} \int \limits_{\mathbb{R}} \int \limits_{\mathscr{O}_{(m,0,0,0,)}}
\Big( \widetilde{\phi}|_{{}_{\mathscr{O}_{(m,0,0,0)}}}(p), \widetilde{\phi}|_{{}_{\mathscr{O}_{(m,0,0,0)}}}(p) \Big)_{\mathbb{C}^4} \, \ud m \, \ud \mu_{{}_{m,0}}(p) \\
+ (2 \pi)^{-1/2} \int \limits_{\mathbb{R}_+} \int \limits_{\mathscr{O}_{(m,0,0,0,)}}
\Big( \widetilde{\phi}|_{{}_{\mathscr{O}_{(0,0,m,0)}}}(p), \widetilde{\phi}|_{{}_{\mathscr{O}_{(0,0,m,0)}}}(p) \Big)_{\mathbb{C}^4} \, \ud m \, \ud \mu_{{}_{0,m}}(p).
\end{split}
\]
Thus the Hilbert space of square integrable bispinors $\phi$ is equal to the direct integral
\[
\int \limits_{\mathbb{R}} \mathcal{H}_{{}_{m,0}} \, \ud m
\oplus \int \limits_{\mathbb{R}_+} \mathcal{H}_{{}_{0,m}} \, \ud m
\]
of Hilbert spaces $\mathcal{H}_{{}_{m,0}}$, $\mathcal{H}_{{}_{0,m}}$
of generalized eigenvectors resp.
\[
\phi_{{}_{m,0}} = \int \limits_{\mathscr{O}_{(m,0,0,0,)}} \widetilde{\phi}|_{{}_{\mathscr{O}_{(m,0,0,0)}}}(p)
e^{-i p \cdot x} \, \ud \mu{{}_{m,0}}(p) \in \mathcal{H}_{{}_{m,0}}
\]
and
\[
\phi_{{}_{0,m}} = \int \limits_{\mathscr{O}_{(0,0,m,0,)}} \widetilde{\phi}|_{{}_{\mathscr{O}_{(0,0,m,0)}}}(p)
e^{-i p \cdot x} \, \ud \mu{{}_{0,m}}(p) \in \mathcal{H}_{{}_{0,m}},
\]
of the operator $D^2$, with the norms in $\mathcal{H}_{{}_{m,0}}$, $\mathcal{H}_{{}_{0,m}}$,
which can be read of from the Plancherel formula
\[
\begin{split}
|| \phi_{{}_{m,0}} ||^2 = \int \limits_{\mathscr{O}_{(m,0,0,0,)}}
\Big( \widetilde{\phi}|_{{}_{\mathscr{O}_{(m,0,0,0)}}}(p), \widetilde{\phi}|_{{}_{\mathscr{O}_{(m,0,0,0)}}}(p) \Big)_{\mathbb{C}^4} \, \ud \mu_{{}_{m,0}}(p), \\
|| \phi_{{}_{0,m}} ||^2 = \int \limits_{\mathscr{O}_{(0,0,m,0,)}}
\Big( \widetilde{\phi}|_{{}_{\mathscr{O}_{(0,0,m,0)}}}(p), \widetilde{\phi}|_{{}_{\mathscr{O}_{(0,0,m,0)}}}(p) \Big)_{\mathbb{C}^4} \, \ud \mu_{{}_{0,m}}(p).
\end{split}
\]
Now using the Fubini theorem and the Fourier transform one can easily show that the representation
(\ref{Ustep1}) acting on square summable bispinors $\phi$ may be decomposed into the direct
integral of representations
\[
\int \limits_{\mathbb{R}} T_{{}_{m,0}} \, \ud m
\oplus \int \limits_{\mathbb{R}_+} T_{{}_{m,0}} \, \ud m
\]
by considering the action of the representation (\ref{Ustep1}) on the Fourier transforms
$\phi_{{}_{m,0}}$, $\phi_{{}_{0,m}}$ of the restrictions $\widetilde{\phi}|_{{}_{\mathscr{O}_{(m,0,0,0)}}}$,
$\widetilde{\phi}|_{{}_{\mathscr{O}_{(0,0,m,0)}}}$ of $\widetilde{\phi}$ to the respective orbits, viewed as
decomposition components of $\widetilde{\phi}$ of the direct integral decomposition of the Hilbert space of
square summable Fourier transforms of bispinors given by the Fubini theorem
(eq. (\ref{dir_int_skew_L^2:decompositions}) or (\ref{dir_int_L^2:decompositions}) of Sect.
\ref{decomposition}). The representation $T_{{}_{m,0}}$ (resp. $T_{{}_{0,m}}$)
acts on $\phi_{{}_{0,m}}$ (resp. $\phi_{{}_{0,m}}$) just by the formulas (\ref{Ustep1})
and on $\widetilde{\phi}|_{{}_{\mathscr{O}_{(m,0,0,0)}}}$ (resp.
$\widetilde{\phi}|_{{}_{\mathscr{O}_{(0,0,m,0)}}}$) by the formulas (\ref{1/2pUalpha}) and (\ref{1/2pUa}),
but this time the Hilbert space of allowed functions is slightly different then in Example \ref{e1}.

Note that the Krein-unitary representations $T_{{}_{m,0}}$ (resp. $T_{{}_{0,m}}$) may be further decomposed
$T_{{}_{m,0}} = T^{\oplus}_{{}_{m,0}} \oplus T^{\ominus}_{{}_{m,0}}$
(resp. $T_{{}_{0,m}} = T^{\oplus}_{{}_{0,m}} \oplus T^{\ominus}_{{}_{0,m}}$)
into direct sum of Krein-unitary sub-representations, and acting in
sub-spaces which are not mutually orthogonal (but only Krein-orthogonal). But this decomposition cannot be based on the ordinary spectral
measure corresponding to the self-adjoint invariant operator $D^2$. We need to pass to the
generalized spectral measure (of Neumark-Lanze type, \cite{Lanze}) associated to the
non-normal invariant Dirac operator $D$. This decomposition is valid on the dense nuclear
subdomain of smooth rapidly decreasing bispinors, and the Fourier transform integral
associated to this decomposition is not convergent with respect to the ordinary Hilbert
space norm but with respect to some stronger topology (e.g. the nuclear Schwartz topology of smooth rapidly decreasing
bispinors). This is reflected by the fact that the projection operators $P^\oplus, P^\ominus$ are
unbounded and continuous with respect to the nuclear topology on the corresponding $E_{{}_{m,0,4}}$
or resp. $E_{{}_{0,m,4}}$. 

It is sufficient then to confine attention to the dense nuclear subspace of smooth rapidly decreasing
bispinors $\widetilde{\phi}$. Next we confine the Hilbert spaces of Fourier transforms $\widetilde{\phi}|_{{}_{\mathscr{O}_{(m,0,0,0)}}}$
(resp. $\widetilde{\phi}|_{{}_{\mathscr{O}_{(0,0,m,0)}}}$) to the linear sub-spaces
defining the first subspace $\mathcal{H}^{\oplus}_{{}_{m,0}}$ by the algebraic relation
\[
\big[ p^0 \gamma^0 - p^k \gamma^k \big] \widetilde{\phi}|_{{}_{\mathscr{O}_{(m,0,0,0)}}}(p) =
m \widetilde{\phi}|_{{}_{\mathscr{O}_{(m,0,0,0)}}}(p),
\]
and the second $\mathcal{H}^{\ominus}_{{}_{m,0}}$ by the relation
\[
\big[ p^0 \gamma^0 - p^k \gamma^k \big] \widetilde{\phi}|_{{}_{\mathscr{O}_{(m,0,0,0)}}}(p) =
-m \widetilde{\phi}|_{{}_{\mathscr{O}_{(m,0,0,0)}}}(p)
\]
(resp. $\big[ p^0 \gamma^0 - p^k \gamma^k \big] \widetilde{\phi}|_{{}_{\mathscr{O}_{(0,0,m,0)}}}(p) =
im \widetilde{\phi}|_{{}_{\mathscr{O}_{(0,0,m,0)}}}(p)$ on the first $\mathcal{H}^{\oplus}_{{}_{0,m}}$ and the relation
$\big[ p^0 \gamma^0 - p^k \gamma^k \big] \widetilde{\phi}|_{{}_{\mathscr{O}_{(0,0,m,0)}}}(p) =
-im \widetilde{\phi}|_{{}_{\mathscr{O}_{(0,0,m,0)}}}(p)$ ($m>0$) on the second subspace $\mathcal{H}^{\ominus}_{{}_{0,m}}$).
In this way we have obtained spectral decomposition of the Krein-self-adjoint Dirac operator
$D$ and the corresponding decomposition
\[
\int \limits_{\mathbb{R}} \mathcal{H}^{\oplus}_{{}_{m,0}} \oplus \mathcal{H}^{\ominus}_{{}_{m,0}} \, \ud m
\bigoplus \int \limits_{\mathbb{R}_+} \mathcal{H}^{\oplus}_{{}_{0,m}} \oplus \mathcal{H}^{\ominus}_{{}_{0,m}} \, \ud m
\]
of the Hilbert space acted on by $D$ into the generalized eigenspaces,
corresponding resp. to the eigenvalues $m, -m, im, -im$ ($m>0$). Note however that the above decomposition understood as decomposition in the Hilbert space of square summable space-time bispinors
is justified only if it is restricted to the bispinors belonging to nuclear subspace
$\mathcal{S}(\mathbb{R}^4; \mathbb{C}^4)$ of rapidly decreasing smooth bispinors and converges not in the topology of the Hilbert space but in some stronger topology (e.g. in the Schwartz nuclear topology of smooth rapidly decreasing bispinors).
The operator $D$ acts as the operator of multiplication by $m$ on the
generalized subspace of distributional solutions of the Dirac operator corresponding to the eigenvalue $m$, which span the corresponding
representation in the decomposition
\[
\int \limits_{\mathbb{R}} \mathcal{H}^{\oplus}_{{}_{m,0}} \, \ud m
\]
and separately on the subspace
\[
\int \limits_{\mathbb{R}} \mathcal{H}^{\ominus}_{{}_{m,0}} \, \ud m ,
\]
However Fourier transform corresponding to this decomposition of $D$ cannot be extended on the whole Hilbert space.
Similarly, $D$ acts as the operator of multiplication by $im$ ($m>0$)
on the subspace of distributional solutions of the Dirac equation (corresponding to the eigenvalue $im$)
\[
\int \limits_{\mathbb{R}_+} \mathcal{H}^{\oplus}_{{}_{0,m}} \, \ud m
\]
And separately on the subspace of distributional solutions which span the representations entering
\[
\int \limits_{\mathbb{R}_+} \mathcal{H}^{\ominus}_{{}_{0,m}} \, \ud m,
\]
it acts as the operator of multiplication by $-im$ ($m>0$);
but the Fourier transform corresponding to the decomposition of the operator $D$ cannot be extended all
over the whole Hilbert space.

$D$ still cannot be represented as a multiplication operator on the 
whole closed subspace of the Hilbert space spanned by the last two sub-spaces
-- of course this is not a surprise as $D$ is not normal (does not commute with its adjoint).

\begin{center}
\small DEFINITION OF THE TRANSFORM $V_\mathcal{F}$ ON THE SPACE OF THE REPRESENTATION 
(\ref{step1LocNaturallyAss})-(\ref{step1LocNonNatural'}) 
\end{center}

Thus, any square summable spinor $\phi$ may be identified with its ordinary Fourier transform
$\widetilde{\phi}$, which is likewise square summable. On the other hand any element of the representation
space of the representation (\ref{step1LocNaturallyAss})-(\ref{step1LocNonNatural'}) may, as we have shown above in this Subsection, be identified
with a function $\widetilde{\phi}$, but which is not just square summable over $\mathbb{R}^4$ with respect to the invariant measure $\ud^4 p$, but in general with respect to $\ud^4 p$ multiplied with the additional weight functions stated in the SUMMABILITY CONDITIONS. But both, the Hilbert space of functions
$\widetilde{\phi}$ fulfilling the SUMMABILITY CONDITIONS, and realizing the space of the representation
(\ref{step1LocNaturallyAss})-(\ref{step1LocNonNatural'}) on the one hand and the Hilbert space of square summable Fourier transforms
$\widetilde{\phi}$ of the space-time bispinors $\phi$ have a dense core set in common. For example all continuous functions
$\widetilde{\phi}$ with compact support are in the common domain. We have chosen
the common domain to be equal $\mathcal{S}(\mathbb{R}^4; \mathbb{C}^4)$.
For any element $\widetilde{\phi}$ of the common domain, and thus realizing an element of the space of the representation
(\ref{step1LocNaturallyAss})-(\ref{step1LocNonNatural'}) we define $V_\mathcal{F}\widetilde{\phi}$ to be equal to the square integrable bispinor $\phi$. By definition $V_\mathcal{F}$ likewise has dense image so that $V_\mathcal{F}^{-1}$
is likewise densely defined. By construction $V_\mathcal{F} U V_\mathcal{F}^{-1}$, with $U$ equal to the local representation (\ref{step1LocNaturallyAss})-(\ref{step1LocNonNatural'}), is equal to the
representation (\ref{Ustep1}) restricted to the dense common domain, so that every representor
$V_\mathcal{F} U_{\alpha , a} V_\mathcal{F}^{-1}$ may be extended to a bounded Krein-unitary operator acting on the square integrable bispinors $\phi$.
\qedsymbol

In other words if
\[
T_4 \cong \mathbb{R}^4 \ni p \mapsto \widetilde{\phi}
\]
is the decomposition function representing element
$\widetilde{\phi}$ of the direct integral space of the
representation (\ref{step1LocNaturallyAss})-(\ref{step1LocNonNatural'})
then $V_\mathcal{F}^{-1}$ is well-defined on
this element if the decomposition function $p \mapsto \widetilde{\phi}(p)$ is rapidly
decreasing smooth function, \emph{i. e.} $\widetilde{\phi} \in \mathcal{S}(\mathbb{R}^4; \mathbb{C}^4)$.
In this case $V_\mathcal{F}^{-1}\widetilde{\phi}$
is represented in the direct integral space of the representation
(\ref{step1LocNaturallyAss})-(\ref{step1LocNonNatural'}) by the following decomposition function
\[
p \mapsto V_\mathcal{F}^{-1}\widetilde{\phi}(p)
= \int \limits_{T_4} \widetilde{\phi}(p') \, e^{-ip'\cdot p} \ud^4 p'.
\]
Note that these definitions are correct because the smooth structure we are taking
from the Lie group structure of $T_4 \circledS SL(2, \mathbb{C})$ and the smooth structure
of its coset and double cosset sub-manifolds and the structure of the representation
(\ref{step1LocNaturallyAss})-(\ref{step1LocNonNatural'}) is arrived at by the assumed continuity of the representation,
compare introduction to Section \ref{constr-of-VF}.

In fact there are the operators $Q_0, \ldots, Q_3$ which together with the generators
$P_0, \ldots, P_3$ of translation in the representation (\ref{step1LocNaturallyAss})-(\ref{step1LocNonNatural'}) compose the
canonical Schr\"odinger-von Neumann pairs. We can construct them explicitly having the
representation (\ref{step1LocNaturallyAss})-(\ref{step1LocNonNatural'}) in which the generators
$P_0, \ldots, P_3$ act with uniform multiplicity,
and have the Lebesgue spectrum equal $\mathbb{R}^4 \cong T_4$.
However, in general the summand of the representation (\ref{step1LocNaturallyAss})
-(\ref{step1LocNonNatural'})
concentrated on the unphysical part $C_\pm$, i.e. with the joint spectrum of $P^0, \ldots, P^3$
concentrated outside the light cannot be reached as a sub representation acting in the (tensor product of)
Fock spaces of fields (of both energy signs). But it can be modified there and put equal to the second summand, respectively, of
(\ref{step1LocNaturallyAss})-(\ref{step1LocNonNatural'}).

Indeed,
on the domain of the vectors $\widetilde{\phi}$ of the representation space
of the representation (\ref{step1LocNaturallyAss})-(\ref{step1LocNonNatural'}),
which are represented by decomposition
functions which are smooth and rapidly decreasing, let us define
\begin{equation}\label{SchroedinegVonNeumanP,Q}
\big(P_\mu \widetilde{\phi}\big)(p) = p_\mu \widetilde{\phi}(p), \,\,\,
\big(Q_\mu \widetilde{\phi}\big)(p)
= i \Big(\frac{\partial \widetilde{\phi}}{\partial p^\mu} \Big)(p), \,\,\,
\mu = 0, \ldots 3.
\end{equation}
The Fourier transform $V_\mathcal{F}$, regarded as a map from the representation space of the representation
(\ref{step1LocNaturallyAss})-(\ref{step1LocNonNatural'}) to the representation space of the
representation (\ref{Ustep1}), is not in general unitary being in general even unbounded (depending on the
representation (\ref{step1LocNaturallyAss})-(\ref{step1LocNonNatural'})). Correspondingly
the operators $Q_\mu$, regarded as operators in the Hilbert space of the representation
(\ref{step1rep'}) need not be self-adjoint, as this strongly depends on the (in general complicated)
weight functions in the inner product formula summarized in the SUMMABILITY CONDITIONS. In this case
when $V_\mathcal{F}$ is not unitary
$P^\mu, Q^\mu$ cannot serve as the canonical pairs.

In case of the representation (\ref{step1LocNaturallyAss}) the Fourier transform is by construction
unitary which immediately follows from the inner product formula in the representation space
of the representation (\ref{step1LocNaturallyAss}) and correspondingly from the SUMMABILITY
CONDITIONS. Thus, (\ref{SchroedinegVonNeumanP,Q}) can indeed serve as the canonical Schr\"odinger-Von Neumann
pairs in the representation space of the representation (\ref{step1LocNaturallyAss}), because
this is the case for the operators
\[
\big(\boldsymbol{P}_\mu \phi\big)(x) = i \Big(\frac{\partial \widetilde{\phi}}{\partial x^\mu} \Big)(x),
\,\,\,
\big(\boldsymbol{Q}_\mu \phi\big)(x)
= x_\mu \phi(x), \,\,\,
\mu = 0, \ldots 3,
\]
in the representation space of space-time bispinors of the representation (\ref{Ustep1}),
and $V_\mathcal{F}$ gives the unitary equivalence between them:
\[
\big(V_\mathcal{F} P_\mu V_\mathcal{F}^{-1}\phi\big)(x) = i \Big(\frac{\partial \widetilde{\phi}}{\partial x^\mu} \Big)(x), \,\,\,
\big(V_\mathcal{F} Q_\mu V_\mathcal{F}^{-1} \phi\big)(x)
= x_\mu \phi(x), \,\,\,
\mu = 0, \ldots 3.
\]
In fact in this case
$V_\mathcal{F}$ is even Krein-unitary and gives the unitary and Krein unitary
equivalence between the tuple
\begin{multline}\label{step1tuple}
\Bigg( \,\,\,\,\,\,\,\, \mathcal{A} =
\{f (\boldsymbol{Q}_0, \ldots, \boldsymbol{Q}_3), f \in \mathcal{S}(\mathbb{R}^4)\} \,\,\,,
\,\,\,\,\,\,\,\,
\mathcal{H}_{{}_{\textrm{inv}}} \,\,\,, \\
\,\,\,\,\,\,\,\,\,
D_{{}_{\mathfrak{J}}} = \widetilde{\Upsilon}^0 \boldsymbol{P}_0 + \ldots + \widetilde{\Upsilon}^3 \boldsymbol{P}_3 \,, \\
D = \widetilde{\gamma}^0 \boldsymbol{P}_0 + \ldots + \widetilde{\gamma}^3 \boldsymbol{P}_3 \,,
\,\,\,\,\,\,\,\,\,
\mathfrak{J} \,\,\,\, \Bigg)
\end{multline}
\[
\widetilde{\gamma}^0 = \gamma^0, \,\, \widetilde{\gamma}^k = - \gamma^k,
\]
and with the operators
\[
\boldsymbol{P}_\mu = V_\mathcal{F} P_\mu V_\mathcal{F}^{-1}, \,\,\,\,
\boldsymbol{Q}_\mu = V_\mathcal{F} Q_\mu V_\mathcal{F}^{-1},
\]
on the one hand and the spectral tuple defined by the spacetime bispinors of the representation space
of the (\ref{Ustep1}) and the Dirac operator (\ref{DiracStep1}) on the other hand.

Note here that for the operator $\mathfrak{J}$ defined by
\[
(\mathfrak{J}\phi)(x) = \gamma^0 \phi (x),
\]
compare (\ref{Jstep1}),
and for the Dirac operator $D$ acting on the square summable bispinors $\phi$ we have
\[
\frac{1}{2} \Big\{ (D\mathfrak{J})^2 + (\mathfrak{J}D)^2 \Big\}
= - \bold{1}_4 \, (\partial_0 \partial_0 + \partial_1 \partial_1 + \partial_2 \partial_2 + \partial_3 \partial_3),
\]
so that $\frac{1}{2} \Big\{ (D\mathfrak{J})^2 + (\mathfrak{J}D)^2 \Big\}$ is elliptic, and we can
choose
\[
D_\mathfrak{J} = i \widetilde{\Upsilon}^\mu \partial_\mu,
\]
with $\widetilde{\Upsilon}^\mu$, defined by
\[
\widetilde{\Upsilon}^0 = \widetilde{\gamma}^0, \,\,\, \widetilde{\Upsilon}^k = i \gamma^k
\]
and being the generators of the Clifford algebra associated to the ordinary Euclidean metric
$\delta^{\mu\nu}$:
\[
\widetilde{\Upsilon}^\mu\widetilde{\Upsilon}^\nu + \widetilde{\Upsilon}^\nu\widetilde{\Upsilon}^\mu = 2 \delta^{\mu\nu} \, \bold{1}_4.
\]
In this case we have
\[
\frac{1}{2} \Big\{ (D\mathfrak{J})^2 + (\mathfrak{J}D)^2 \Big\}
= \big( D_\mathfrak{J} \big)^2.
\]

The algebra $\mathcal{A}$ of Schwartz functions acting as multiplication operators
on square integrable bispinors $\phi$ forming a Hilbert space $\mathcal{H}$ of space-time 
bispinors, and the operators 
$D$, $\mathfrak{J}, D_\mathfrak{J}$  fulfil the 
conditions of Introduction (of course here in the subspace associated to the representation 
(\ref{Ustep1})\footnote{Equivalently this is the case for $D$, $\mathfrak{J}, D_\mathfrak{J}$  
on $\mathcal{H}_{{}_{\textrm{inv}}}$ of the spectral tuple associated to the representation
 (\ref{step1LocNaturallyAss})).}, which follows from \cite{Gay} and \cite{Stro}. The strong regularity 
of the spectral triple $\big( \mathcal{A}, \mathcal{H}, D_\mathfrak{J} \big)$ is checked
exactly as in the proof of Theorem 11.4 of \cite{Connes_spectral}.

Recall that
\[
\Gamma^\mu = \widetilde{\Upsilon}^\mu, \,\,\,\, \widehat{\Gamma}^\mu = \widetilde{\gamma}^\mu
\]
in the notation used in Subsection \ref{G} of Introduction, formula (\ref{SpacetimeTupleFields}).

Unbounded character of the transform $V_\mathcal{F}$ in case of the representations
(\ref{step1LocAss}), (\ref{step1LocNonNatural'}) is associated to the fact that it transforms
the representation (\ref{step1LocAss}) or (\ref{step1LocNonNatural'}), which is not unitary nor Krein-unitary,
into the representation
(\ref{Ustep1}) which is Krein-unitary.
Recall that the first summand of
(\ref{step1LocNonNatural'}) is unitary but not Krein $\mathfrak{J}$-unitary and the second summand
is Krein $\mathfrak{J}$-isometric but not Krein-unitary nor unitary. The first summand is not unitarily mapped by $V_\mathcal{F}$, on the second summand $V_\mathcal{F}$ is Krein-isometric, but again not unitary
nor Krein-unitary. We have similar situation in case of representation (\ref{step1LocNonNatural})
where neither of the direct summands, the first one nor the second one, is unitarily mapped
by $V_\mathcal{F}$.

In each case (\ref{step1LocNaturallyAss})-(\ref{step1LocNonNatural'}) there exist $Q^0, \ldots Q^3$
corresponding to the generators $P^0, \ldots, P^3$ which compose with $P^0, \ldots, P^3$
pairs unitarily equivalent with the canonical pairs. But in each case except
(\ref{step1LocNaturallyAss}) this unitary equivalence cannot serve at the same time as a Krein-unitary
equivalence between (\ref{step1LocNaturallyAss}) and the Krein-unitary representation
(\ref{Ustep1}) acting on square summable space-time bispinors, which immediately follows from
the properties of these representations: only the representation
(\ref{step1LocNaturallyAss}) is Krein-unitary among
(\ref{step1LocNaturallyAss})-(\ref{step1LocNonNatural'}).

The nontrivial weight functions of the SUMMABILITY CONDITIONS
and causing the non-unitarity or unboundedness of $V_\mathcal{F}$,
have their source in the non-unitary character of the representation (\ref{Ustep1}) and correspondingly
the non-natural local versions, being unitary, do not match with Krein-unitary, but non-unitary,
character of the representation (\ref{Ustep1}).

\begin{rem}
In the above construction of $V_\mathcal{F}$ we can use symmetry
operators $\mathfrak{J}$ belonging to two complementary classes of fundamental operators. In particular
in the Example 1 and 2 and in this Subsection we can use the operator $\mathfrak{J}$ of multiplication by
$\gamma^0$ or by
$\gamma^1 \gamma^2 \gamma^3$. Both
give rise to the spectral description of the same Minkowski space, and the representation
(\ref{Ustep1}) remains Krein $\mathfrak{J}$-unitary for both $\mathfrak{J}$,
but the two operators $\mathfrak{J}$
correspond to the two essentially equivalent but opposite choices of signatures of the space-time Minkowski pseudo-metric. The fundamental symmetry operator $\mathfrak{J}$ of multiplication by
$\gamma^0$ corresponds to the signature
$(1,-1,-1,-1)$ of the space-time pseudo-metric, and the fundamental symmetry operator $\mathfrak{J}$
of multiplication by $\gamma^1 \gamma^2 \gamma^3$ corresponds to the signature
$(-1,1,1,1)$ of the space-time pseudo-metric.
\label{construction-of-VF}
\end{rem}
\begin{rem}
In constructing the space-time spectral tuple (\ref{step1tuple}) we have borrowed the smooth structure
from the Lie group structure of $T_4\circledS SL(2, \mathbb{C})$ using continuity of the representation
of $T_4\circledS SL(2, \mathbb{C})$ acting in the Fock space. Reconstruction of the space-time smooth structure from the tuple (\ref{step1tuple}) which does not make any immediate use of the smooth structure
of the Lie group $T_4\circledS SL(2, \mathbb{C})$ and continuity conditions of its representation,
is a separate problem, resolved by Connes \cite{Connes_spectral} for compact manifolds.
Here $\mathbb{R}^4$ is non-compact, but we
reduce this problem to the compact case resolved by Connes \cite{Connes_spectral} by introducing
auxiliary operators defined uniquely by the tuple (\ref{step1tuple}), which correspond to the
canonical conformal (stereographic) embedding of $\mathbb{R}^4$ into the standard sphere $\mathbb{S}^4$. We have presented it in more details in Subsection \ref{DirectIntRepVF} and in Appendix \ref{AppendixNonCompMani}.
\end{rem}

\begin{rem}\label{RemarkShortNotation}
In the Examples which are to follow we will generalize the direct integral representation
(\ref{step1LocNaturallyAss})-(\ref{step1LocNonNatural'})
presented in the first two Examples, on which the translation generators act with finite uniform multiplicity and have Lebesgue spectrum (of finite uniform multiplicity) equal $\mathbb{R}^4$.
However, in order to simplify notation, we will use common notation
\[
U^{{}_{\bar{p}} L}
\]
for all representations
\begin{eqnarray*}
W'_{{}_{L,\bar{p}}}U_{{}_{L, \bar{p}}}^{-1} U^{{}_{\bar{p}} L} U_{{}_{L, \bar{p}}}{W'}_{{}_{L,\bar{p}}}^{-1}, \\
W_{{}_{V,\bar{p}}} U^{{}_{\bar{p}} L} W_{{}_{V,\bar{p}}}^{-1}, \\
\big(W_{{}_{V,\bar{p}}} U^{{}_{\bar{p}} L} W_{{}_{V,\bar{p}}}^{-1}\big)_{{}_{0}}, \\
\big(W_{{}_{V,\bar{p}}} U^{{}_{\bar{p}} L} W_{{}_{V,\bar{p}}}^{-1}\big)_{{}_{00}},
\end{eqnarray*}
because keeping the operators
\[
W'_{{}_{L,\bar{p}}}, \,\, U_{{}_{L, \bar{p}}}, \,\, W_{{}_{V,\bar{p}}}
\]
explicitly in notation would give us long expressions with
the representations $L$
\[
L = 2^n \big(L^{1/2}\big)^{\otimes n},
\]
which we encounter, likewise explicitly inserted into the formulas.

We believe that it will be clear from the context in each case which meaning is being used,
because we have presented all details in Examples 1 and 2 for the simplest case (with $n=1$),
and the case with greater $n$ is completely analogous.
\end{rem}

\subsection{Example 3: Representation $U^{{}_{(m,0,0,0)} 4(L^{{}^{1/2}} \otimes L^{{}^{1/2}})}$ (spin 0 and 1) and the generalized Dirac equation}\label{e3}

We start with the construction of 
\[
\big(W_{{}_{V,\bar{p}}} U^{{}_{\bar{p}} L} W_{{}_{V,\bar{p}}}^{-1}\big)_{{}_{00}},
\]
for 
\[
\bar{p}  = (m,0,0,0), \,\,\, L= 4(L^{{}^{1/2}} \otimes L^{{}^{1/2}}),
\]
\[
V(\alpha) = \left( \begin{array}{cccc}  \alpha \otimes \alpha & \bold{0}_4 & \bold{0}_4 & \bold{0}_4  \\
                              \bold{0}_4  & \alpha^{*-1} \otimes \alpha & \bold{0}_4 & \bold{0}_4 \\
                             \bold{0}_4 & \bold{0}_4 & \alpha^{*-1} \otimes \alpha^{*-1} & \bold{0}_4 \\
               \bold{0}_4 & \bold{0}_4 & \bold{0}_4 & \alpha \otimes \alpha^{*-1}  \end{array}\right),
               \,\,\, \alpha \in SL(2, \mathbb{C}).
\]

Consider the unitary representation $U^{{}_{(m,0,0,0)} \big(L^{{}^{1/2}} \otimes L^{{}^{1/2}}\big)}$
unitary equivalent to $U^{{}_{(m,0,0,0)} \big(L^{{}^{0}} \oplus L^{{}^{1}}\big)}
\cong_{U} U^{{}_{(m,0,0,0)} L^{{}^{0}}} \oplus U^{{}_{(m,0,0,0)} L^{{}^{1}}}$, concentrated on the
orbit $\mathscr{O}_{(m,0,0,0)}$, and induced by the unitary representation $L^{{}^{0}} \oplus L^{{}^{1}}
\cong L^{{}^{1/2}} \otimes L^{{}^{1/2}}$: $\gamma \mapsto Q(\gamma, \bar{p}) =
L^{{}^{1/2}} \otimes L^{{}^{1/2}}(\gamma) = \gamma \otimes \gamma$ of the small group
$G_{\bar{p}} = G_{(m,0,0,0)} = SU(2, \mathbb{C})$ stationary for $\bar{p} = (m,0,0,0)$ ($m>0$).
Of course the Hilbert space $\mathcal{H}_{\bar{p}}$ of the representation
$L^{{}^{1/2}} \otimes L^{{}^{1/2}}$ is equal
$\mathbb{C}^2 \otimes \mathbb{C}^2 = \mathbb{C}^4$. In this case
we have four natural extensions $V$ of the representation $L^{{}^{1/2}} \otimes L^{{}^{1/2}}$
acting in $\mathbb{C}^4$, we denote them respectively by $V_1, \ldots V_4$, namely for any
$\alpha \in SL(2, \mathbb{C})$ we have
\[
\begin{split}
V_1 (\alpha) = \alpha \otimes \alpha, \\
V_2 (\alpha) = \alpha^{*-1} \otimes \alpha, \\
V_3 (\alpha) = \alpha^{*-1} \otimes \alpha^{*-1}, \\
V_4 (\alpha) = \alpha \otimes \alpha^{*-1}.
\end{split}
\]
Let $\widetilde{\psi}_{{}_{m,0}}$ be any element of the representation space
of the representation
\[
W'_{{}_{(L^{1/2})^{\otimes 2},m,0}} U_{{}_{(L^{1/2})^{\otimes 2},m,0}}^{-1}
U^{{}_{(m,0,0,0)} (L^{{}^{1/2}})^{\otimes 2}}
U_{{}_{(L^{1/2})^{\otimes 2},m,0}} {W'}_{{}_{(L^{1/2})^{\otimes 2},m,0}}^{-1}
\]
which hereafter will likewise be denoted by
\[
U^{{}_{(m,0,0,0)} (L^{{}^{1/2}} \otimes L^{{}^{1/2}})}
= U^{{}_{(m,0,0,0)} (L^{{}^{1/2}})^{\otimes 2}},
\]
in order to simplify notation (as explained in Remark \ref{RemarkShortNotation}).
We associate with it a positive energy solution of a generalized Dirac equation.

Instead of the bispinor $\widetilde{\phi} = \tvect{\widetilde{\varphi}}{\widetilde{\chi}}$ of Example
\ref{e1}, we will introduce the analogue of it, namely the following (16-component) ''multispinor''
\begin{equation}\label{multispinor}
\widetilde{\phi}_{{}_{m,0}} (p) = \left( \begin{array}{c} V_1 (\beta(p)^{-1})\widetilde{\psi}_{{}_{m,0}}(p)\\
V_2 (\beta(p)^{-1})\widetilde{\psi}_{{}_{m,0}}(p)\\
V_3 (\beta(p)^{-1})\widetilde{\psi}_{{}_{m,0}}(p)\\
V_4 (\beta(p)^{-1})\widetilde{\psi}_{{}_{m,0}}(p) \end{array}\right)
= \left( \begin{array}{c} \widetilde{\varphi}_1 (p)\\
\widetilde{\varphi}_2 (p)\\
\widetilde{\varphi}_3 (p)\\
\widetilde{\varphi}_4 (p) \end{array}\right)
= \Big( V^{\textrm{\ding{192}}} \, \widetilde{\psi}_{{}_{m,0}}\Big)(p),
\end{equation}
we drop the subscript $(m,0)$ at $\widetilde{\varphi}_i $ for simplicity.
Here $V^{\textrm{\ding{192}}}$ and $V^{\textrm{\ding{192}}} \, \widetilde{\psi}_{{}_{m,0}}$
is the immediate analogue of the map $V^\oplus$ and
${\widetilde{\phi}_{{}_{m,0}}}^{\oplus}$ of Example 1. Later on we will introduce the isometric maps $V^{\textrm{\ding{193}}}, V^{\textrm{\ding{194}}}, V^{\textrm{\ding{195}}}$
with the additional minus sign at the respective components in (\ref{multispinor}). This time we will have four (more than just two: $V^\oplus$ and $V^\ominus$) isometric maps of the elements $\widetilde{\psi}_{{}_{m,0}}$ of the representation space
of the representation $U^{{}_{(m,0,0,0)} \big(L^{{}^{1/2}} \otimes L^{{}^{1/2}}\big)}$ into the Hilbert space of
''multispinors'', defined in this Subsection, and the respective multiplicity
of that representation will have to be greater and equal 4 (instead of the previous 2).

Thus, by the general construction of this Section the multispinor $\widetilde{\phi}_{{}_{m,0}}$ has the following 
transformation law 
\begin{equation}\label{0-1pUalpha}
U(\alpha) \widetilde{\phi}_{{}_{m,0}}(p) 
= \left( \begin{array}{cccc}  \alpha \otimes \alpha & \bold{0}_4 & \bold{0}_4 & \bold{0}_4  \\
                              \bold{0}_4  & \alpha^{*-1} \otimes \alpha & \bold{0}_4 & \bold{0}_4 \\
                             \bold{0}_4 & \bold{0}_4 & \alpha^{*-1} \otimes \alpha^{*-1} & \bold{0}_4 \\
               \bold{0}_4 & \bold{0}_4 & \bold{0}_4 & \alpha \otimes \alpha^{*-1}  \end{array}\right) 
 \widetilde{\phi}_{{}_{m,0}}(\Lambda(\alpha)p), 
\end{equation}
\[
T(a) \widetilde{\phi}_{{}_{m,0}}(p) = e^{i a \cdot p}\widetilde{\phi}_{{}_{m,0}}(p);
\]
with the inner product of two multispinors $\widetilde{\phi}_{{}_{m,0}}, \widetilde{\phi}_{{}_{m,0}}'$
given by
\begin{multline}\label{0-1-inn-prod-multspin}
(\widetilde{\phi}_{{}_{m,0}}, \widetilde{\phi'}_{{}_{m,0}}) = 
  \int \limits_{\mathscr{O}_{(m,0,0,0)}} \frac{m^2}{\big(2 p^0(p) \big)^2}
\Big(\widetilde{\phi}_{{}_{m,0}}(p), \widetilde{\phi}_{{}_{m,0}}'(p) \Big)_{{}_{\mathbb{C}^{16}}}
\, \ud \mu_{{}_{m,0}}(p) \\
=  \int \limits_{\mathscr{O}_{(m,0,0,0)}} \frac{m^2}{\big(2 p^0(p) \big)^2}
\Big[ \Big(\widetilde{\varphi}_1 (p), \widetilde{\varphi}'_1(p) \Big)_{{}_{\mathbb{C}^4}}
+ \ldots + \Big(\widetilde{\varphi}_4 (p), \widetilde{\varphi}'_4 (p) \Big)_{{}_{\mathbb{C}^4}} \Big] \, 
\ud \mu_{{}_{m,0}}(p) \\
= \int \limits_{\mathscr{O}_{(m,0,0,0)}} \frac{m^2}{\big(2 p^0(p) \big)^2}
\Big(\widetilde{\psi}(p), 
\Big(\big(\beta(p)^{-2} + \beta(p)^{2}\big) \otimes \big(\beta(p)^{-2} + \beta(p)^{2}\big)\Big) 
\widetilde{\psi}'(p) \Big)_{{}_{\mathbb{C}^4}}  \, 
\ud \mu_{{}_{m,0}}(p)  \\
= \int \limits_{\mathscr{O}_{(m,0,0,0)}} \Big(\widetilde{\psi}_{{}_{m,0}}(p), 
\widetilde{\psi'}_{{}_{m,0}}(p) \Big)_{{}_{\mathbb{C}^4}}  \, 
\ud \mu_{{}_{m,0}}(p)  = (\widetilde{\psi}_{{}_{m,0}}, \widetilde{\psi'}_{{}_{m,0}}), 
\end{multline}
because 
\[
\begin{split}
\beta(p)^{-2} \otimes \beta(p)^{-2} + \beta(p)^{2} \otimes \beta(p)^{-2} + \beta(p)^{2} \otimes \beta(p)^{2} +
\beta(p)^{-2} \otimes \beta(p)^{2} \\
= \big(\beta(p)^{-2} + \beta(p)^{2}\big) \otimes \big(\beta(p)^{-2} + \beta(p)^{2}\big)
 = \frac{\big(2 p^0|_{{}_{\mathscr{O}_{(m,0,0,0)}}} \big)^2}{m^2} \bold{1}_{4};
\end{split}
\]
The Fourier transform $\phi_{{}_{m,0}}$ of $\widetilde{\phi}_{{}_{m,0}}$ (defined by (\ref{F(varphi)})
with $\widetilde{\varphi}$ replaced with $\widetilde{\phi}_{{}_{m,0}}$ in (\ref{F(varphi)}))
has by construction the following local transformation law 
\begin{equation}\label{0-1xUalpha}
U(\alpha) \phi_{{}_{m,0}}(x) 
= \left( \begin{array}{cccc}  \alpha \otimes \alpha & \bold{0}_4 & \bold{0}_4 & \bold{0}_4  \\
                              \bold{0}_4  & \alpha^{*-1} \otimes \alpha & \bold{0}_4 & \bold{0}_4 \\
                             \bold{0}_4 & \bold{0}_4 & \alpha^{*-1} \otimes \alpha^{*-1} & \bold{0}_4 \\
               \bold{0}_4 & \bold{0}_4 & \bold{0}_4 & \alpha \otimes \alpha^{*-1}  \end{array}\right) 
 \phi_{{}_{m,0}} (\Lambda(\alpha)x), 
\end{equation}
\[
T(a) \phi_{{}_{m,0}}(p) = \phi_{{}_{m,0}}(x -a).
\]

Now let us denote by $\beta^2 \otimes \bold{1}_2$ (resp. $\beta^{-2} \otimes \bold{1}_2$,  $\bold{1}_{2} \otimes \beta^2$,
$ \bold{1}_{2} \otimes \beta^{-2} $ ) the (invertible) operator of multiplication by $\beta(p)^2 \otimes \bold{1}_{2}$ 
(regarded as multiplication by the $4 \times 4$ matrix equal to the tensor product of the respective $2 \times 2$ matrices
$\beta(p)^2$ and $\bold{1}_{2}$):
\[
\Big( \big( \beta^2 \otimes \bold{1}_2 \big)\widetilde{\varphi}_i \Big)(p) 
= \big( \beta(p)^2 \otimes \bold{1}_2 \big) \widetilde{\varphi}_i (p),  
\]
(and similarly for $\beta^{-2} \otimes \bold{1}_2, \ldots$).
Then we have
\begin{multline}\label{cyclic-eq}
\widetilde{\varphi}_2 =  \big( \beta^2 \otimes \bold{1}_2 \big) \widetilde{\varphi}_1 ,  \,\,\,\,\,\,\,\,\,\,\,\,
\widetilde{\varphi}_3 = \big( \bold{1}_{2} \otimes \beta^2 \big)  \widetilde{\varphi}_2, \\
\widetilde{\varphi}_4 = \big(\beta^{-2} \otimes \bold{1}_2 \big)  \widetilde{\varphi}_3, \,\,\,\,\,\,\,\,\,\,\,\,
\widetilde{\varphi}_1 = \big( \bold{1}_{2} \otimes \beta^{-2} \big)  \widetilde{\varphi}_4,
\end{multline}
which may be pictured by the following connected and cyclic diagram
\begin{center}
\begin{tikzpicture}

\draw [<-,very thin] (1.932,0.52) arc (15:75:2) ;

\draw [<-,very thin] (-0.52,1.932) arc (105:165:2) ;

\draw [<-,very thin] (-1.932,-0.52) arc (195:255:2) ;

\draw [<-,very thin] (0.52,-1.932) arc (285:345:2) ;

















\node  at (2,0) {$\widetilde{\varphi}_{2}$};

\node  at (0,2) {$\widetilde{\varphi}_{1}$};

\node  at (-2,0) {$\widetilde{\varphi}_{4}$};

\node  at (0,-2) {$\widetilde{\varphi}_{3}$};

\node  at (1.75,1.75) {$\beta^2 \otimes \bold{1}_2$};

\node  at (-1.75,-1.75) {$\beta^{-2} \otimes \bold{1}_2$};

\node  at (1.75,-1.75) {$\bold{1}_2 \otimes \beta^{2}$};

\node  at (-1.75,1.75) {$\bold{1}_2 \otimes \beta^{-2}$};

\node  at (4,0) {$(A)$};










\end{tikzpicture} 
\end{center}
No generalized Dirac equation is connected with the equations (\ref{cyclic-eq}) pictured by the
diagram (A). But the diagonally opposite maps of the diagram (A)
we can joint into pairs of equations which together give two possible generalizations
of the Dirac equation. They can be pictured by the following disconnected diagrams (B)
and (C):
\begin{center}
\begin{tikzpicture}


\draw [<-,very thin] (0.966,0.26) arc (15:165:1) ;
\draw [<-,very thin] (-0.966,-0.26) arc (195:345:1) ;

\draw [<-,very thin] (3.966,0.26) arc (15:165:1) ;
\draw [<-,very thin] (2.034,-0.26) arc (195:345:1) ;

\node  at (-1,0) {$\widetilde{\varphi}_{1}$};
\node  at (1,0) {$\widetilde{\varphi}_{2}$};

\node  at (0,1.25) {$\beta^2 \otimes \bold{1}_2$};
\node  at (0,-1.25) {$\beta^{-2} \otimes \bold{1}_2$};



\node  at (2,0) {$\widetilde{\varphi}_{3}$};
\node  at (4,0) {$\widetilde{\varphi}_{4}$};

\node  at (3,1.25) {$\beta^{-2} \otimes \bold{1}_2$};
\node  at (3,-1.25) {$\beta^2 \otimes \bold{1}_2$};

\node  at (5,0) {$(B)$};



\end{tikzpicture} 
\end{center}

\vspace*{0.5cm}

\begin{center}
\begin{tikzpicture}


\draw [<-,very thin] (0.966,0.26) arc (15:165:1) ;
\draw [<-,very thin] (-0.966,-0.26) arc (195:345:1) ;

\draw [<-,very thin] (3.966,0.26) arc (15:165:1) ;
\draw [<-,very thin] (2.034,-0.26) arc (195:345:1) ;

\node  at (-1,0) {$\widetilde{\varphi}_{4}$};
\node  at (1,0) {$\widetilde{\varphi}_{1}$};

\node  at (0,1.25) {$\bold{1}_2 \otimes \beta^{-2}$};
\node  at (0,-1.25) {$\bold{1}_2 \otimes \beta^{2}$};



\node  at (2,0) {$\widetilde{\varphi}_{3}$};
\node  at (4,0) {$\widetilde{\varphi}_{2}$};

\node  at (3,1.25) {$\bold{1}_2 \otimes \beta^{-2}$};
\node  at (3,-1.25) {$\bold{1}_2 \otimes \beta^{2}$};

\node  at (5,0) {$(C)$};



\end{tikzpicture} 
\end{center}
In particular the diagram (C) corresponds to the following equation fulfilled by the 
multispinor $\widetilde{\phi}_{{}_{m,0}}$:
\[
\left\{ \begin{array}{ll}
\widetilde{\varphi}_4 = & \big( \bold{1}_2 \otimes \beta^{2} \big) \widetilde{\varphi}_1 \\
\widetilde{\varphi}_3 = & \big( \bold{1}_2 \otimes \beta^{2} \big) \widetilde{\varphi}_2 \\
\widetilde{\varphi}_2 = & \big( \bold{1}_2 \otimes \beta^{-2} \big) \widetilde{\varphi}_3 \\
\widetilde{\varphi}_1 = & \big( \bold{1}_2 \otimes \beta^{-2} \big) \widetilde{\varphi}_4 ,
\end{array} \right.
\]
which can be written in the following form (summation with respect to $k = 1,2,3$)
\begin{multline*}
m \widetilde{\phi}_{{}_{m,0}}(p) = \biggl[ 
p^0|_{{}_{\mathscr{O}_{(m,0,0,0)}}}\left( \begin{array}{cccc} \bold{0}_4 & \bold{0}_4 & \bold{0}_4 & \bold{1}_4  \\
                              \bold{0}_4  & \bold{0}_4 & \bold{1}_4 & \bold{0}_4 \\
                             \bold{0}_4 & \bold{1}_4 & \bold{0}_4 & \bold{0}_4 \\
               \bold{1}_4 & \bold{0}_4 & \bold{0}_4 &\bold{0}_4  \end{array}\right)  \\
- p^k \, \left( \begin{array}{cccc} \bold{0}_4 & \bold{0}_4 & \bold{0}_4 & \bold{1}_2 \otimes \sigma_k  \\
                              \bold{0}_4  & \bold{0}_4 & \bold{1}_2 \otimes \sigma_k & \bold{0}_4 \\
                             \bold{0}_4 & -\bold{1}_2 \otimes \sigma_k & \bold{0}_4 & \bold{0}_4 \\
               -\bold{1}_2 \otimes \sigma_k & \bold{0}_4 & \bold{0}_4 &\bold{0}_4  \end{array}\right)   
\biggr]\widetilde{\phi}_{{}_{m,0}}(p) 
\end{multline*}
or in the more concise form 
\[
\big[ \tilde{\gamma}^0 p^0 - p^k \tilde{\gamma}^k \big]\widetilde{\phi}_{{}_{m,0}}(p) 
= m \widetilde{\phi}_{{}_{m,0}}(p), 
\]
where
\[
\tilde{\gamma}^0 = \left( \begin{array}{cccc} \bold{0}_4 & \bold{0}_4 & \bold{0}_4 & \bold{1}_4  \\
                              \bold{0}_4  & \bold{0}_4 & \bold{1}_4 & \bold{0}_4 \\
                             \bold{0}_4 & \bold{1}_4 & \bold{0}_4 & \bold{0}_4 \\
               \bold{1}_4 & \bold{0}_4 & \bold{0}_4 &\bold{0}_4  \end{array}\right) \,\,\,
\tilde{\gamma}^k = \left( \begin{array}{cccc} \bold{0}_4 & \bold{0}_4 & \bold{0}_4 & -\bold{1}_2 \otimes \sigma_k  \\
                              \bold{0}_4  & \bold{0}_4 & -\bold{1}_2 \otimes \sigma_k & \bold{0}_4 \\
                             \bold{0}_4 & \bold{1}_2 \otimes \sigma_k & \bold{0}_4 & \bold{0}_4 \\
               \bold{1}_2 \otimes \sigma_k & \bold{0}_4 & \bold{0}_4 &\bold{0}_4  \end{array}\right),
\]
are generators of a representation of the Clifford algebra associated to
the Minkowski metric $g^{\mu\nu}$:
\[
\tilde{\gamma}^\mu \tilde{\gamma}^\nu + \tilde{\gamma}^\nu \tilde{\gamma}^\mu = 2 g^{\mu\nu} \bold{1}_{16}.
\]
Thus, the Fourier transform $\phi_{{}_{m,0}}$ of $\widetilde{\phi}_{{}_{m,0}}$ (defined by (\ref{F(varphi)})
with $\widetilde{\varphi}$ replaced with $\widetilde{\phi}_{{}_{m,0}}$ in (\ref{F(varphi)}))
fulfills the following generalized Dirac equation
\[
\big[ i\tilde{\gamma}^\mu \partial_\mu \big] \phi_{{}_{m,0}} = m \phi_{{}_{m,0}}.
\]

Similarly as in the Example 1 the isometric map $V^\textrm{\ding{192}}$ from the representation space 
of the representation $U^{{}_{(m,0,0,0)} \big(L^{{}^{1/2}} \otimes L^{{}^{1/2}}\big)}$ into
the Hilbert space of ``multispinors'' with the inner product (\ref{0-1-inn-prod-multspin})
is not onto. Therefore, we consider the representation 
$U^{{}_{(m,0,0,0)} \big(L^{{}^{1/2}} \otimes L^{{}^{1/2}}\big)}$ with uniform multiplicity four:
\begin{multline}\label{4ULm0-1}
U^{{}_{(m,0,0,0)} \big(L^{{}^{1/2}} \otimes L^{{}^{1/2}}\big)} 
\bigoplus U^{{}_{(m,0,0,0)} \big(L^{{}^{1/2}} \otimes L^{{}^{1/2}}\big)} \\
\bigoplus U^{{}_{(m,0,0,0)} \big(L^{{}^{1/2}} \otimes L^{{}^{1/2}}\big)}
\bigoplus U^{{}_{(m,0,0,0)} \big(L^{{}^{1/2}} \otimes L^{{}^{1/2}}\big)} 
= 4 \, U^{{}_{(m,0,0,0)} \big(L^{{}^{1/2}} \otimes L^{{}^{1/2}}\big)}
\end{multline}
and for any element $(\widetilde{\psi}^\textrm{\ding{192}}_{{}_{m,0}}, \widetilde{\psi}^\textrm{\ding{193}}_{{}_{m,0}}, 
\widetilde{\psi}^\textrm{\ding{194}}_{{}_{m,0}}, \widetilde{\psi}^\textrm{\ding{195}}_{{}_{m,0}})$ of the direct sum space
of that representation we define its image $V^\textrm{\ding{192}} \, \widetilde{\psi}^\textrm{\ding{192}}_{{}_{m,0}} + 
V^\textrm{\ding{193}} \, \widetilde{\psi}^\textrm{\ding{193}}_{{}_{m,0}} + 
V^\textrm{\ding{194}} \, \widetilde{\psi}^\textrm{\ding{194}}_{{}_{m,0}} + 
V^\textrm{\ding{195}} \, \widetilde{\psi}^\textrm{\ding{195}}_{{}_{m,0}} = 
\widetilde{\phi}^\textrm{\ding{192}}_{{}_{m,0}} (p) +
\widetilde{\phi}^\textrm{\ding{193}}_{{}_{m,0}} (p) +
\widetilde{\phi}^\textrm{\ding{194}}_{{}_{m,0}} (p)+
\widetilde{\phi}^\textrm{\ding{195}}_{{}_{m,0}} (p)
= V^{\textrm{\ding{192}} \textrm{\ding{193}} \textrm{\ding{194}} \textrm{\ding{195}}}
\big( \widetilde{\psi}^\textrm{\ding{192}}_{{}_{m,0}} \oplus \widetilde{\psi}^\textrm{\ding{193}}_{{}_{m,0}}
\oplus \widetilde{\psi}^\textrm{\ding{194}}_{{}_{m,0}} \oplus \widetilde{\psi}^\textrm{\ding{195}}_{{}_{m,0}} \big)$,
where
\[
\begin{split}
\Big( V^\textrm{\ding{192}} \, \widetilde{\psi}^\textrm{\ding{192}}_{{}_{m,0}} \Big)(p) = \widetilde{\phi}^\textrm{\ding{192}}_{{}_{m,0}} (p) = \left( \begin{array}{ccc} 
 &\beta(p)^{-1} \otimes \beta(p)^{-1} & \widetilde{\psi}^\textrm{\ding{192}}_{{}_{m,0}}(p)\\
 &  \beta(p) \otimes \beta(p)^{-1} & \widetilde{\psi}^\textrm{\ding{192}}_{{}_{m,0}}(p)\\
 & \beta(p) \otimes \beta(p) & \widetilde{\psi}^\textrm{\ding{192}}_{{}_{m,0}}(p)\\
 & \beta(p)^{-1} \otimes \beta(p) & \widetilde{\psi}^\textrm{\ding{192}}_{{}_{m,0}}(p) \end{array}\right), \\
\Big( V^\textrm{\ding{193}} \, \widetilde{\psi}^\textrm{\ding{193}}_{{}_{m,0}} \Big)(p) = \widetilde{\phi}^\textrm{\ding{193}}_{{}_{m,0}} (p) = \left( \begin{array}{ccc} 
 &\beta(p)^{-1} \otimes \beta(p)^{-1} & \widetilde{\psi}^\textrm{\ding{193}}_{{}_{m,0}}(p)\\
- &  \beta(p) \otimes \beta(p)^{-1} & \widetilde{\psi}^\textrm{\ding{193}}_{{}_{m,0}}(p)\\
- & \beta(p) \otimes \beta(p) & \widetilde{\psi}^\textrm{\ding{193}}_{{}_{m,0}}(p)\\
 & \beta(p)^{-1} \otimes \beta(p) & \widetilde{\psi}^\textrm{\ding{193}}_{{}_{m,0}}(p) \end{array}\right), \\
\Big( V^\textrm{\ding{194}} \, \widetilde{\psi}^\textrm{\ding{194}}_{{}_{m,0}} \Big)(p) = \widetilde{\phi}^\textrm{\ding{194}}_{{}_{m,0}} (p) = \left( \begin{array}{ccc} 
- &\beta(p)^{-1} \otimes \beta(p)^{-1} & \widetilde{\psi}^\textrm{\ding{194}}_{{}_{m,0}}(p)\\
 &  \beta(p) \otimes \beta(p)^{-1} & \widetilde{\psi}^\textrm{\ding{194}}_{{}_{m,0}}(p)\\
- & \beta(p) \otimes \beta(p) & \widetilde{\psi}^\textrm{\ding{194}}_{{}_{m,0}}(p)\\
 & \beta(p)^{-1} \otimes \beta(p) & \widetilde{\psi}^\textrm{\ding{194}}_{{}_{m,0}}(p) \end{array}\right), \\
\Big( V^\textrm{\ding{195}} \, \widetilde{\psi}^\textrm{\ding{195}}_{{}_{m,0}} \Big)(p) = \widetilde{\phi}^\textrm{\ding{195}}_{{}_{m,0}} (p) = \left( \begin{array}{ccc} 
 &\beta(p)^{-1} \otimes \beta(p)^{-1} & \widetilde{\psi}^\textrm{\ding{195}}_{{}_{m,0}}(p)\\
 &  \beta(p) \otimes \beta(p)^{-1} & \widetilde{\psi}^\textrm{\ding{195}}_{{}_{m,0}}(p)\\
- & \beta(p) \otimes \beta(p) & \widetilde{\psi}^\textrm{\ding{195}}_{{}_{m,0}}(p)\\
- & \beta(p)^{-1} \otimes \beta(p) & \widetilde{\psi}^\textrm{\ding{195}}_{{}_{m,0}}(p) \end{array}\right), \\
\end{split}
\]
and where, just like in Example 1, we treat the image under $V^\textrm{\ding{192}}$ of the first direct summand,
and similarly the image under $V^\textrm{\ding{193}}$ of the second direct summand, e. t. c., as immersed in one and the same Hilbert space of all multispinors $\widetilde{\phi}_{{}_{m,0}}$ with finite Hilbert space norm defined by
(\ref{0-1-inn-prod-multspin}). The images under $V^\textrm{\ding{192}}, V^\textrm{\ding{193}}, V^\textrm{\ding{194}},
V^\textrm{\ding{195}}$ respectively of the first, second, third and fourth direct summand
are not orthogonal with respect to the inner product (\ref{0-1-inn-prod-multspin}), but they are closed with zero
as the only common element, i.e. zero is the only common element for any two of these images. That
$V^{\textrm{\ding{192}} \textrm{\ding{193}} \textrm{\ding{194}} \textrm{\ding{195}} }$ ((immediate analogue of $V^{\oplus \ominus}$ of Example 4) is onto is easily checked.
Indeed, if
\[
\widetilde{\phi}_{{}_{m,0}} 
= \left( \begin{array}{c}  \widetilde{\varphi}_1 \\
                                                           \widetilde{\varphi}_2 \\
                                                           \widetilde{\varphi}_3 \\
                                                  \widetilde{\varphi}_4  \end{array}\right)
\]     
is any measurable multispinor with finite Hilbert space norm defined by the inner product 
(\ref{0-1-inn-prod-multspin}), then it is the image under $V^{\textrm{\ding{192}} \textrm{\ding{193}} \textrm{\ding{194}} \textrm{\ding{195}}}$ of the element $(\widetilde{\psi}^\textrm{\ding{192}}_{{}_{m,0}}, \widetilde{\psi}^\textrm{\ding{193}}_{{}_{m,0}}, 
\widetilde{\psi}^\textrm{\ding{194}}_{{}_{m,0}}, \widetilde{\psi}^\textrm{\ding{195}}_{{}_{m,0}})$ of the direct sum 
representation space
of the representation $4 \, U^{{}_{(m,0,0,0)} \big(L^{{}^{1/2}} \otimes L^{{}^{1/2}}\big)}$, equal to
\begin{equation}\label{inv-V1234}
\begin{split}
  \begin{array}{cccccc}
\widetilde{\psi}^\textrm{\ding{192}}_{{}_{m,0}} = &
                    \frac{1}{4} \Big\{ \beta \otimes \beta & \widetilde{\varphi}_1 & 
                          + \beta^{-1} \otimes \beta \widetilde{\varphi}_2  &
                              + \beta^{-1} \otimes \beta^{-1} \widetilde{\varphi}_3 &
                                 + \beta \otimes \beta^{-1} \widetilde{\varphi}_4 \Big\} \\ 
\widetilde{\psi}^\textrm{\ding{193}}_{{}_{m,0}} = &
                    \frac{1}{4} \Big\{ \beta \otimes \beta & \widetilde{\varphi}_1 & 
                          - \beta^{-1} \otimes \beta \widetilde{\varphi}_2  &
                              - \beta^{-1} \otimes \beta^{-1} \widetilde{\varphi}_3 &
                                 + \beta \otimes \beta^{-1} \widetilde{\varphi}_4 \Big\} \\
\widetilde{\psi}^\textrm{\ding{194}}_{{}_{m,0}} = &
                      \frac{1}{4} \Big\{ - \beta \otimes \beta & \widetilde{\varphi}_1 & 
                          + \beta^{-1} \otimes \beta \widetilde{\varphi}_2  &
                              - \beta^{-1} \otimes \beta^{-1} \widetilde{\varphi}_3 &
                                 + \beta \otimes \beta^{-1} \widetilde{\varphi}_4 \Big\} \\
\widetilde{\psi}^\textrm{\ding{195}}_{{}_{m,0}} = &
                      \frac{1}{4} \Big\{  \beta \otimes \beta & \widetilde{\varphi}_1 & 
                          + \beta^{-1} \otimes \beta \widetilde{\varphi}_2  &
                              - \beta^{-1} \otimes \beta^{-1} \widetilde{\varphi}_3 &
                                 - \beta \otimes \beta^{-1} \widetilde{\varphi}_4 \Big\}.  
          \end{array}
\end{split}
\end{equation}
That is the map $V^{\textrm{\ding{192}} \textrm{\ding{193}} \textrm{\ding{194}} \textrm{\ding{195}}}$:
\begin{multline*}
V^{\textrm{\ding{192}} \textrm{\ding{193}} \textrm{\ding{194}} \textrm{\ding{195}}} 
\Big( \widetilde{\psi}^\textrm{\ding{192}}_{{}_{m,0}} \oplus \widetilde{\psi}^\textrm{\ding{193}}_{{}_{m,0}} 
\oplus \widetilde{\psi}^\textrm{\ding{194}}_{{}_{m,0}} \oplus \widetilde{\psi}^\textrm{\ding{195}}_{{}_{m,0}} \Big) \\ =
\left( \begin{array}{cccccccc}
   \beta^{-1} \otimes \beta^{-1} &  \widetilde{\psi}^\textrm{\ding{192}}_{{}_{m,0}} & 
                          + \beta^{-1} \otimes \beta^{-1} & \widetilde{\psi}^\textrm{\ding{193}}_{{}_{m,0}}  &
                              - \beta^{-1} \otimes \beta^{-1} & \widetilde{\psi}^\textrm{\ding{194}}_{{}_{m,0}} &
                                 + \beta^{-1} \otimes \beta^{-1} & \widetilde{\psi}^\textrm{\ding{195}}_{{}_{m,0}}  \\
  \beta \otimes \beta^{-1}  & \widetilde{\psi}^\textrm{\ding{192}}_{{}_{m,0}} & 
                          - \beta \otimes \beta^{-1} & \widetilde{\psi}^\textrm{\ding{193}}_{{}_{m,0}}  &
                              + \beta \otimes \beta^{-1} & \widetilde{\psi}^\textrm{\ding{194}}_{{}_{m,0}} &
                                 + \beta \otimes \beta^{-1} & \widetilde{\psi}^\textrm{\ding{195}}_{{}_{m,0}} \\
   + \beta \otimes \beta &  \widetilde{\psi}^\textrm{\ding{192}}_{{}_{m,0}} & 
                          - \beta \otimes \beta & \widetilde{\psi}^\textrm{\ding{193}}_{{}_{m,0}}  &
                              - \beta \otimes \beta & \widetilde{\psi}^\textrm{\ding{194}}_{{}_{m,0}} &
                                 - \beta \otimes \beta & \widetilde{\psi}^\textrm{\ding{195}}_{{}_{m,0}}  \\
   \beta^{-1} \otimes \beta & \widetilde{\psi}^\textrm{\ding{192}}_{{}_{m,0}} & 
                          + \beta^{-1} \otimes \beta & \widetilde{\psi}^\textrm{\ding{193}}_{{}_{m,0}}  &
                              + \beta^{-1} \otimes \beta & \widetilde{\psi}^\textrm{\ding{194}}_{{}_{m,0}} &
                                 - \beta^{-1} \otimes \beta & \widetilde{\psi}^\textrm{\ding{195}}_{{}_{m,0}}. \\ 
          \end{array} \right),
\end{multline*}
has the inverse\footnote{By the Banach inverse mapping theorem the inverse of 
$V^{\textrm{\ding{192}} \textrm{\ding{193}} \textrm{\ding{194}} \textrm{\ding{195}}}$ is likewise bounded, but this can be easily checked directly.} given by (\ref{inv-V1234}). 

Every element of the  image under $V^\textrm{\ding{192}}$ of the 
first direct summand and every element of the image under $V^\textrm{\ding{193}}$ 
of the second direct summand fulfills the algebraic relation at every point $p$ 
of the orbit $\mathscr{O}_{m,0}$, $m>0$ (summation with respect to $k = 1,2,3$):
\[
\Big[ p^0 \widetilde{\gamma}^0 - p^k \widetilde{\gamma}^k \Big] \widetilde{\phi}^\textrm{\ding{192}}_{{}_{m,0}} (p)
= m \widetilde{\phi}^\textrm{\ding{192}}_{{}_{m,0}} (p) \,\,\, \textrm{and} \,\,\, 
\Big[ p^0 \widetilde{\gamma}^0 - p^k \widetilde{\gamma}^k \Big] \widetilde{\phi}^\textrm{\ding{193}}_{{}_{m,0}} (p)
= m \widetilde{\phi}^\textrm{\ding{193}}_{{}_{m,0}} (p).
\]

In turn every element of the image under $V^\textrm{\ding{194}}$ of the the third direct summand and every element in the image under $V^\textrm{\ding{195}}$
of the fourth direct summand fulfills the algebraic relation on $\mathscr{O}_{m,0}$, $m>0$:
\[
\Big[ p^0 \widetilde{\gamma}^0 - p^k \widetilde{\gamma}^k \Big] \widetilde{\phi}^\textrm{\ding{194}}_{{}_{m,0}} (p)
= - m \widetilde{\phi}^\textrm{\ding{194}}_{{}_{m,0}} (p) \,\,\, \textrm{and} \,\,\,
\Big[ p^0 \widetilde{\gamma}^0 - p^k \widetilde{\gamma}^k \Big] \widetilde{\phi}^\textrm{\ding{195}}_{{}_{m,0}} (p)
= - m \widetilde{\phi}^\textrm{\ding{195}}_{{}_{m,0}} (p).
\]
This means that the Fourier transforms (defined by (\ref{F(varphi)})) fulfil the
following generalized Dirac equation with the reversed sign at the mass term on (Fourier transform of) the image of
$V^\textrm{\ding{194}}$ and $V^\textrm{\ding{195}}$:
\[
\begin{split}
\Big[ i \gamma^\mu \partial_\mu \Big] \phi^{\textrm{\ding{192}}}_{{}_{m,0}}
= m \phi^{\textrm{\ding{192}}}_{{}_{m,0}}, \,\,\,\,
\Big[ i \gamma^\mu \partial_\mu \Big] \phi^{\textrm{\ding{193}}}_{{}_{m,0}}
= m \phi^{\textrm{\ding{193}}}_{{}_{m,0}}, \\
\Big[ i \gamma^\mu \partial_\mu \Big] \phi^{\textrm{\ding{194}}}_{{}_{m,0}}
= - m \phi^{\textrm{\ding{194}}}_{{}_{m,0}}, \,\,\,\,
\Big[ i \gamma^\mu \partial_\mu \Big] \phi^{\textrm{\ding{195}}}_{{}_{m,0}}
= - m \phi^{\textrm{\ding{195}}}_{{}_{m,0}}.
\end{split}
\] 
\[
\gamma^0 = \widetilde{\gamma}^0, \,\,\,\,\,  \gamma^k = -\widetilde{\gamma}^k, \,\, k=1,2,3.
\]

We have the analogous relation between the elements  $(\widetilde{\psi}^\textrm{\ding{192}}_{{}_{-m,0}}, 
\widetilde{\psi}^\textrm{\ding{193}}_{{}_{-m,0}}, \widetilde{\psi}^\textrm{\ding{194}}_{{}_{-m,0}}, 
\widetilde{\psi}^\textrm{\ding{195}}_{{}_{-m,0}})$ of the representation space of the direct sum 
\begin{multline}\label{4UL-m0-1}
U^{{}_{(-m,0,0,0)} \big(L^{{}^{1/2}} \otimes L^{{}^{1/2}}\big)} 
\bigoplus U^{{}_{(-m,0,0,0)} \big(L^{{}^{1/2}} \otimes L^{{}^{1/2}}\big)} \\
\bigoplus U^{{}_{(-m,0,0,0)} \big(L^{{}^{1/2}} \otimes L^{{}^{1/2}}\big)}
\bigoplus U^{{}_{(-m,0,0,0)} \big(L^{{}^{1/2}} \otimes L^{{}^{1/2}}\big)} 
= 4 \, U^{{}_{(-m,0,0,0)} \big(L^{{}^{1/2}} \otimes L^{{}^{1/2}}\big)} \\
= U^{{}_{(-m,0,0,0)} 4\big(L^{{}^{1/2}} \otimes L^{{}^{1/2}}\big)}
\end{multline}
of four copies of the irreducible representation $U^{{}_{(-m,0,0,0)} \big(L^{{}^{1/2}} \otimes L^{{}^{1/2}}\big)}$, 
$m > 0$ (concentrated on the orbit $\mathscr{O}_{(-m,0,0,0)}$)  
with the Hilbert space of multispinors concentrated on the orbit $\mathscr{O}_{(-m,0,0,0)}$, equipped with the analogous inner product
\begin{multline}\label{0-1-inn-prod-multspin-m}
(\widetilde{\phi}_{{}_{-m,0}}, \widetilde{\phi'}_{{}_{-m,0}}) = 
  \int \limits_{\mathscr{O}_{(-m,0,0,0)}} \frac{m^2}{\big(2 p^0(p) \big)^2}
\Big(\widetilde{\phi}_{{}_{-m,0}}(p), \widetilde{\phi}_{{}_{-m,0}}'(p) \Big)_{{}_{\mathbb{C}^{16}}}
\, \ud \mu_{{}_{-m,0}}(p); 
\end{multline}
they correspond to the negative energy solutions (of the generalized Dirac equation with the ordinary and with the changed sign at the mass term respectively) being concentrated on the lower branch
of the two-sheeted hyperboloid.

Recall that we are using simplified notation of Remark \ref{RemarkShortNotation} in (\ref{4UL-m0-1}),
discarding the operators 
\[
4W'_{{}_{(L^{1/2})^{\otimes 2},-m,0}} U_{{}_{(L^{1/2})^{\otimes 2},-m,0}}^{-1}
\]
on the left and the operators
\[
4U_{{}_{(L^{1/2})^{\otimes 2},-m,0}} {W'}_{{}_{(L^{1/2})^{\otimes 2},-m,0}}^{-1}
\]
on the right, in order to simplify notation.

The bounded and invertible map $V^{\textrm{\ding{192}} \textrm{\ding{193}} \textrm{\ding{194}} \textrm{\ding{195}}}$
(with bounded inverse) which maps the representation space of the representation  (\ref{4ULm0-1}) 
(resp. of the representation 
(\ref{4UL-m0-1})) onto the space of all multispinors $\widetilde{\phi}_{{}_{m,0}}$ 
(resp. $\widetilde{\phi}_{{}_{-m,0}}$) concentrated on the orbit $\mathscr{O}_{(m,0,0,0)}$
(resp. $\mathscr{O}_{(-m,0,0,0)}$, $m>0$) with finite norm defined by the inner product
(\ref{0-1-inn-prod-multspin}) (resp. (\ref{0-1-inn-prod-multspin-m})) is not unitary, but
restricted to each direct summand is separately isometric. This is because the images under
$V^\textrm{\ding{192}}, V^\textrm{\ding{193}}, \ldots$ of the respective direct summands
are not orthogonal with respect to (\ref{0-1-inn-prod-multspin}) (resp. (\ref{0-1-inn-prod-multspin-m})).
However, if we introduce the fundamental symmetry $\mathfrak{J}$ ($\mathfrak{J}^* = \mathfrak{J}$, 
$\mathfrak{J}^2 = \bold{1}$) into the Hilbert space of multispinors
$\widetilde{\phi}_{{}_{m,0}}$  (resp. $\widetilde{\phi}_{{}_{-m,0}}$) by the formula
\begin{equation}\label{JforU^m0004L^1/2otimes2_00}
\Big( \mathfrak{J} \, \widetilde{\phi}_{{}_{m,0}} \Big)(p) 
= \left( \begin{array}{cccc}  \bold{0}_4 & \bold{0}_4 & \bold{1}_4 & \bold{0}_4  \\
                              \bold{0}_4 & \bold{0}_4 & \bold{0}_4 & \bold{1}_4 \\
                              \bold{1}_4 & \bold{0}_4 & \bold{0}_4 & \bold{0}_4 \\
                              \bold{0}_4 & \bold{1}_4 & \bold{0}_4 & \bold{0}_4 \end{array}\right) 
\big( \widetilde{\phi}_{{}_{m,0}}(p) \big),
\end{equation}
then all the images under $V^\textrm{\ding{192}}, V^\textrm{\ding{193}}, \ldots$
of the respective direct summands are pairwise Krein-$\mathfrak{J}$-orthogonal
and neutral with respect to Krein-inner product. 
We shall denote $V^{\textrm{\ding{192}} \textrm{\ding{193}} \textrm{\ding{194}} \textrm{\ding{195}}}$
by $V^\textrm{\ding{192}} \oplus V^\textrm{\ding{193}} \oplus V^\textrm{\ding{194}}
\oplus V^\textrm{\ding{195}}$. We explain the choice of $\mathfrak{J}$ in the latter part of this
Section.

Construction of the corresponding projectors $P^\oplus(p), P^\ominus(p)$, $P^\oplus, P^\ominus$,
being completely analogous to that presented in Subsections \ref{e1} and \ref{e2}, can be omitted.
These projectors are Krein self adjoint.

Note also the simple identities
\begin{equation}\label{SimpleKreinIsometry4(1/2)x2m,0,0,0}
\begin{split}
\big(\widetilde{\phi}^\textrm{\ding{192}}_{{}_{m,0}}(p), \mathfrak{J}\widetilde{\phi}^\textrm{\ding{192}}_{{}_{m,0}} (p) \big)_{{}_{\mathbb{C}^4}}
= 4 \big({\widetilde{\psi}_{{}_{m,0}}}^\textrm{\ding{192}}(p), {\widetilde{\psi}_{{}_{m,0}}}^\textrm{\ding{192}}(p) \big)_{{}_{\mathbb{C}^2}},
\\
\big(\widetilde{\phi}^\textrm{\ding{193}}_{{}_{m,0}}(p), \mathfrak{J}\widetilde{\phi}^\textrm{\ding{193}}_{{}_{m,0}} (p) \big)_{{}_{\mathbb{C}^4}}
= -4 \big({\widetilde{\psi}_{{}_{m,0}}}^\textrm{\ding{193}}(p), {\widetilde{\psi}_{{}_{m,0}}}^\textrm{\ding{193}}(p) \big)_{{}_{\mathbb{C}^2}},
\\
\big(\widetilde{\phi}^\textrm{\ding{194}}_{{}_{m,0}}(p), \mathfrak{J}\widetilde{\phi}^\textrm{\ding{194}}_{{}_{m,0}} (p) \big)_{{}_{\mathbb{C}^4}}
= 4 \big({\widetilde{\psi}_{{}_{m,0}}}^\textrm{\ding{194}}(p), {\widetilde{\psi}_{{}_{m,0}}}^\textrm{\ding{194}}(p) \big)_{{}_{\mathbb{C}^2}},
\\
\big(\widetilde{\phi}^\textrm{\ding{195}}_{{}_{m,0}}(p), \mathfrak{J}\widetilde{\phi}^\textrm{\ding{195}}_{{}_{m,0}} (p) \big)_{{}_{\mathbb{C}^4}}
= -4 \big({\widetilde{\psi}_{{}_{m,0}}}^\textrm{\ding{195}}(p), {\widetilde{\psi}_{{}_{m,0}}}^\textrm{\ding{195}}(p) \big)_{{}_{\mathbb{C}^2}}
\end{split}
\end{equation}
\[
p \in \mathscr{O}_{{}_{m,0,0,0}}.
\]

Thus, the Hilbert space of the representation
\begin{equation}\label{U^m0004L^1/2otimes2_00}
\big(W_{{}_{V,m,0,0,0}} U^{{}_{m,0,0,0} (4L^{1/2})^{\otimes 2}} W_{{}_{V,m,0,0,0}}^{-1}\big)_{{}_{00}},
\end{equation}
is equal to the linear space of multispinors $\widetilde{\phi}_{{}_{m,0}}$ with finite
norm defined by the inner product (\ref{0-1-inn-prod-multspin}) of this Hilbert space.
The action of the representation (\ref{U^m0004L^1/2otimes2_00}) on the multispinors in this Hilbert space
is given by (\ref{0-1pUalpha}). Into the Hilbert representation space of multispinors of the representation
(\ref{U^m0004L^1/2otimes2_00}) we introduce the fundamental symmetry operator $\mathfrak{J}$,
by the formula (\ref{JforU^m0004L^1/2otimes2_00}). 

Restriction of the the representation (\ref{U^m0004L^1/2otimes2_00}) 
to any one of all four direct summands $\textrm{Im} \, V^{\textrm{\ding{192}}}, \ldots, \textrm{Im} \, V^{\textrm{\ding{195}}}$
is by construction unitarily equivalent
to the respective direct sumands of the induced unitary representation $(\ref{4ULm0-1})$. However,
because $\textrm{Im} \, V^{\textrm{\ding{192}}}, \ldots, \textrm{Im} \, V^{\textrm{\ding{195}}}$
are only Krein orthogonal but are not orthogonal, the map $V^{\textrm{\ding{192}} \textrm{\ding{193}} \textrm{\ding{194}} \textrm{\ding{195}}}$
joining (\ref{4ULm0-1}) and (\ref{U^m0004L^1/2otimes2_00}) is not unitary. Accordingly the represenation (\ref{U^m0004L^1/2otimes2_00}),  as a whole, 
is not unitary. It becomes unitary only after restriction to the said invariant closed subspaces
$\textrm{Im} \, V^{\textrm{\ding{192}}}, \ldots, \textrm{Im} \, V^{\textrm{\ding{195}}}$.

Construction of the natural local version
\[
W_{{}_{V,\bar{p}}} U^{{}_{\bar{p}} L} W_{{}_{V,\bar{p}}}^{-1}
= W_{{}_{V,m,0,0,0}} U^{{}_{m,0,0,0} (4(L^{1/2})^{\otimes 2})} W_{{}_{V,m,0,0,0}}^{-1},
\]
and  of the corresponding representation
\begin{equation}\label{(NatLocU4(1/2)x2m0,0,0)0}
\big(W_{{}_{V,\bar{p}}} U^{{}_{\bar{p}} L} W_{{}_{V,\bar{p}}}^{-1}\big)_{{}_{0}}
= \big(W_{{}_{V,m,0,0,0}} U^{{}_{m,0,0,0} (4(L^{1/2})^{\otimes 2})} W_{{}_{V,m,0,0,0}}^{-1}\big)_{{}_{0}},,
\end{equation}
for 
\[
\bar{p}  = (m,0,0,0), \,\,\, L= 4(L^{{}^{1/2}} \otimes L^{{}^{1/2}}) = 4 \big(L^{{}^{1/2}} \big)^{\otimes 2},
\]
\[
V(\alpha) = \left( \begin{array}{cccc}  \alpha \otimes \alpha & \bold{0}_4 & \bold{0}_4 & \bold{0}_4  \\
                              \bold{0}_4  & \alpha^{*-1} \otimes \alpha & \bold{0}_4 & \bold{0}_4 \\
                             \bold{0}_4 & \bold{0}_4 & \alpha^{*-1} \otimes \alpha^{*-1} & \bold{0}_4 \\
               \bold{0}_4 & \bold{0}_4 & \bold{0}_4 & \alpha \otimes \alpha^{*-1}  \end{array}\right),
               \,\,\, \alpha \in SL(2, \mathbb{C}),
\]
being completely analogous to that presented in Subsection \ref{e1} for 
\[
U^{{}_{m,0,0,0} 2L^{1/2}},
\]
can be omitted. Nonetheless we recall that the representation space
of the representation (\ref{(NatLocU4(1/2)x2m0,0,0)0}) consists of generic multispinors
\[
\widetilde{\phi}_{{}_{m,0}} = V^{\textrm{\ding{192}} \textrm{\ding{193}} \textrm{\ding{194}} \textrm{\ding{195}}}
\big( \widetilde{\psi}^\textrm{\ding{192}}_{{}_{m,0}} \oplus \widetilde{\psi}^\textrm{\ding{193}}_{{}_{m,0}}
\oplus \widetilde{\psi}^\textrm{\ding{194}}_{{}_{m,0}} \oplus \widetilde{\psi}^\textrm{\ding{195}}_{{}_{m,0}} \big),
\]
where $\widetilde{\psi}^\textrm{\ding{192}}_{{}_{m,0}} \oplus \widetilde{\psi}^\textrm{\ding{193}}_{{}_{m,0}}
\oplus \widetilde{\psi}^\textrm{\ding{194}}_{{}_{m,0}} \oplus \widetilde{\psi}^\textrm{\ding{195}}_{{}_{m,0}}$
ranges over the representation space of the representation (\ref{4ULm0-1}), with the inner product, and the invariant Krein inner 
product equal, respectively, to
\[
\begin{split}
(\widetilde{\phi}_{{}_{m,0}}, \widetilde{\phi'}_{{}_{m,0}}) = 
  \int \limits_{\mathscr{O}_{{}_{m,0,0,0}}} 
\Big(\widetilde{\phi}_{{}_{-m,0}}(p), \widetilde{\phi}_{{}_{-m,0}}'(p) \Big)_{{}_{\mathbb{C}^{16}}}
\, \ud \mu_{{}_{m,0}}(p),
\\
(\widetilde{\phi}_{{}_{m,0}}, \mathfrak{J}\widetilde{\phi'}_{{}_{-m,0}}) = 
  \int \limits_{\mathscr{O}_{{}_{m,0,0,0}}} 
\Big(\widetilde{\phi}_{{}_{-m,0}}(p), \mathfrak{J}\widetilde{\phi}_{{}_{m,0}}'(p) \Big)_{{}_{\mathbb{C}^{16}}}
\, \ud \mu_{{}_{m,0}}(p). 
\end{split}
\]

The representation space of the representation (\ref{4ULm0-1})
consisting of direct sums $\widetilde{\psi}^\textrm{\ding{192}}_{{}_{m,0}} \oplus \widetilde{\psi}^\textrm{\ding{193}}_{{}_{m,0}}
\oplus \widetilde{\psi}^\textrm{\ding{194}}_{{}_{m,0}} \oplus \widetilde{\psi}^\textrm{\ding{195}}_{{}_{m,0}}$ of functions  concentrated 
on $\mathscr{O}_{{}_{m,0,0,0}}$ 
with the inner product 
\begin{multline*}
\big(\widetilde{\psi}^\textrm{\ding{192}}_{{}_{m,0}} \oplus \widetilde{\psi}^\textrm{\ding{193}}_{{}_{m,0}}
\oplus \widetilde{\psi}^\textrm{\ding{194}}_{{}_{m,0}} \oplus \widetilde{\psi}^\textrm{\ding{195}}_{{}_{m,0}}, 
\widetilde{\psi}^\textrm{\ding{192}}_{{}_{m,0}} \oplus \widetilde{\psi}^\textrm{\ding{193}}_{{}_{m,0}}
\oplus \widetilde{\psi}^\textrm{\ding{194}}_{{}_{m,0}} \oplus \widetilde{\psi}^\textrm{\ding{195}}_{{}_{m,0}}\big)
\\
= \int\limits_{\mathscr{O}_{{}_{m,0,0,0}}} \big(\widetilde{\psi}^\textrm{\ding{192}}_{{}_{m,0}}(p),
\widetilde{\psi}^\textrm{\ding{192}}_{{}_{m,0}}(p)\big)_{{}_{\mathbb{C}^2}}
\,\, \ud \mu_{{}_{m,0}}(p)
+ \ldots +
\int\limits_{\mathscr{O}_{{}_{m,0,0,0}}} \big( \widetilde{\psi}^\textrm{\ding{195}}_{{}_{m,0}}(p),
 \widetilde{\psi}^\textrm{\ding{195}}_{{}_{m,0}}(p)\big)_{{}_{\mathbb{C}^2}}
\,\, \ud \mu_{{}_{m,0}}(p),
\end{multline*} 
can be given a natural invariant Krein structure, induced by its natural local version (\ref{NatLocU^m,0,0,02L^1/20}), acting on 
generic multispinors 
\[
\widetilde{\phi}_{{}_{m,0}} = V^{\textrm{\ding{192}} \textrm{\ding{193}} \textrm{\ding{194}} \textrm{\ding{195}}}
\big( \widetilde{\psi}^\textrm{\ding{192}}_{{}_{m,0}} \oplus \widetilde{\psi}^\textrm{\ding{193}}_{{}_{m,0}}
\oplus \widetilde{\psi}^\textrm{\ding{194}}_{{}_{m,0}} \oplus \widetilde{\psi}^\textrm{\ding{195}}_{{}_{m,0}} \big),
\]
concentrated on $\mathscr{O}_{{}_{m,0,0,0}}$,
and Krein symmetry operator $\mathfrak{J}$ defined by (\ref{JforU^m0004L^1/2otimes2_00}).
Indeed, the Krein structure induced on the direct sum $\widetilde{\psi}^\textrm{\ding{192}}_{{}_{m,0}} \oplus \widetilde{\psi}^\textrm{\ding{193}}_{{}_{m,0}}
\oplus \widetilde{\psi}^\textrm{\ding{194}}_{{}_{m,0}} \oplus \widetilde{\psi}^\textrm{\ding{195}}_{{}_{m,0}}$ in the representation space
of the representation (\ref{4ULm0-1})
is equal to $\boldsymbol{1}$ on the first direct summand and $- \boldsymbol{1}$ on the second direct summand,
$\boldsymbol{1}$ on the third direct sumand and $-\boldsymbol{1}$ on the fourth:
\[
\mathfrak{J}\big(\widetilde{\psi}^\textrm{\ding{192}}_{{}_{m,0}} \oplus \widetilde{\psi}^\textrm{\ding{193}}_{{}_{m,0}}
\oplus \widetilde{\psi}^\textrm{\ding{194}}_{{}_{m,0}} \oplus \widetilde{\psi}^\textrm{\ding{195}}_{{}_{m,0}}\big)
= \big(\widetilde{\psi}^\textrm{\ding{192}}_{{}_{m,0}}\big) \oplus \big(-\widetilde{\psi}^\textrm{\ding{193}}_{{}_{m,0}}\big)
\oplus \big(\widetilde{\psi}^\textrm{\ding{194}}_{{}_{m,0}}\big) \oplus \big(-\widetilde{\psi}^\textrm{\ding{195}}_{{}_{m,0}}\big).
\]
This follows from the general Theorem and Proposition presented in the beginning of this Section, but we prefer here to deduce it
immediately from the
simple identities (\ref{SimpleKreinIsometry4(1/2)x2m,0,0,0}) analogously as in Subsection \ref{e1}.  
Indeed, using (\ref{SimpleKreinIsometry4(1/2)x2m,0,0,0}) we get 
\begin{multline*}
\big({\widetilde{\phi}_{{}_{m,0}}}, \mathfrak{J}{\widetilde{\phi}_{{}_{m,0}}}\big) =
\big({\widetilde{\phi}_{{}_{m,0}}}^\textrm{\ding{192}}, \mathfrak{J}{\widetilde{\phi}_{{}_{m,0}}}^\textrm{\ding{192}} \big)
+ \ldots +
\big({\widetilde{\phi}_{{}_{m,0}}}^\textrm{\ding{195}}, \mathfrak{J}{\widetilde{\phi}_{{}_{m,0}}}^\textrm{\ding{195}} \big)
\\
=
\int\limits_{\mathscr{O}_{{}_{m,0,0,0}}}
\big({\widetilde{\phi}_{{}_{m,0}}}^\textrm{\ding{192}}(p), \mathfrak{J}{\widetilde{\phi}_{{}_{m,0}}}^\textrm{\ding{192}}(p) \big)_{{}_{\mathbb{C}^4}}
\ud \mu_{{}_{m,0}}(p)
\\
+ \ldots +
\int\limits_{\mathscr{O}_{{}_{m,0,0,0}}}
\big({\widetilde{\phi}_{{}_{m,0}}}^\textrm{\ding{195}}(p), \mathfrak{J}{\widetilde{\phi}_{{}_{m,0}}}^\textrm{\ding{195}}(p) \big)_{{}_{\mathbb{C}^4}}
\ud \mu_{{}_{m,0}}(p)
\end{multline*}
\begin{multline*}
=
4\int\limits_{\mathscr{O}_{{}_{m,0,0,0}}}
\big({\widetilde{\psi}_{{}_{m,0}}}^\textrm{\ding{192}}(p), {\widetilde{\psi}_{{}_{m,0}}}^\textrm{\ding{192}}(p) \big)_{{}_{\mathbb{C}^2}}
\ud \mu_{{}_{m,0}}(p)
\\
-4\int\limits_{\mathscr{O}_{{}_{m,0,0,0}}}
\big({\widetilde{\psi}_{{}_{m,0}}}^\textrm{\ding{193}}(p), {\widetilde{\psi}_{{}_{m,0}}}^\textrm{\ding{193}}(p) \big)_{{}_{\mathbb{C}^2}}
\ud \mu_{{}_{m,0}}(p)
\end{multline*}
\begin{multline*}
+
4\int\limits_{\mathscr{O}_{{}_{m,0,0,0}}}
\big({\widetilde{\psi}_{{}_{m,0}}}^\textrm{\ding{194}}(p), {\widetilde{\psi}_{{}_{m,0}}}^\textrm{\ding{194}}(p) \big)_{{}_{\mathbb{C}^2}}
\ud \mu_{{}_{m,0}}(p)
\\
-4\int\limits_{\mathscr{O}_{{}_{m,0,0,0}}}
\big({\widetilde{\psi}_{{}_{m,0}}}^\textrm{\ding{195}}(p), {\widetilde{\psi}_{{}_{m,0}}}^\textrm{\ding{195}}(p) \big)_{{}_{\mathbb{C}^2}}
\ud \mu_{{}_{m,0}}(p)
\end{multline*}
\[
=
4\Big(\, \widetilde{\psi}_{{}_{m,0}}^\textrm{\ding{192}} \oplus \ldots \oplus \widetilde{\psi}_{{}_{m,0}}^\textrm{\ding{195}} \, , \, 
\mathfrak{J}\big(\widetilde{\psi}_{{}_{m,0}}^\textrm{\ding{192}} \oplus \ldots \oplus \widetilde{\psi}_{{}_{m,0}}^\textrm{\ding{195}}\big) \, \Big).
\]
Thus, it is sufficient to apply the additional multiplication by $2$ to the map 
$V^{\textrm{\ding{192}} \textrm{\ding{193}} \textrm{\ding{194}} \textrm{\ding{195}}}$, constructed above in this Subsection,
in order to get the required Krein isometry equivalence between (\ref{4ULm0-1})
and (\ref{(NatLocU4(1/2)x2m0,0,0)0}) in explicit form expressed immediately through the elements $\psi$ and $\phi$
of the representation spaces of both these representations.

\subsection{Example 4: Representation 
$U^{{}_{(m,0,0,0)} \big[4(L^{{}^{1/2}})^{\otimes 2}\big]_\textrm{Ass}}$ associated to 
$U^{{}_{(m,0,0,0)} 4(L^{{}^{1/2}} \otimes L^{{}^{1/2}})}$
(spin 0 and 1) and concentrated on the orbit $\mathscr{O}_{(0,m,0,0)}$}\label{e4}

Consider the representation
\[
\gamma \mapsto Q(\gamma,\bar{p}) = \big[4(L^{1/2})^{\otimes 2} \big]_\textrm{Ass}(\gamma) =
\left( \begin{array}{cccc}  \gamma \otimes \gamma & \bold{0}_4 & \bold{0}_4 & \bold{0}_4  \\
                             \bold{0}_4   & \gamma^{*-1} \otimes \gamma & \bold{0}_4 & \bold{0}_4 \\
                              \bold{0}_4 & \bold{0}_4 & \gamma^{*-1} \otimes \gamma^{*-1} & \bold{0}_4 \\
                              \bold{0}_4 & \bold{0}_4 & \bold{0}_4   & \gamma \otimes \gamma^{*-1}  \end{array}\right) 
\]
of the group $SL(2, \mathbb{R})= G_{(0,0,m,0)}$, stationary for $\bar{p} = (0,m,0,0)$. We extend it by the formula
\begin{multline*}
\alpha \mapsto V(\alpha) = 
\left( \begin{array}{cccc}  \alpha \otimes \alpha & \bold{0}_4 & \bold{0}_4 & \bold{0}_4  \\
                             \bold{0}_4   & \alpha^{*-1} \otimes \alpha & \bold{0}_4 & \bold{0}_4 \\
                              \bold{0}_4 & \bold{0}_4  & \alpha^{*-1} \otimes \alpha^{*-1} & \bold{0}_4 \\
                              \bold{0}_4 & \bold{0}_4 & \bold{0}_4   & \alpha \otimes \alpha^{*-1}  \end{array}\right) \\
=  (\alpha \otimes \alpha) \oplus (\alpha^{*-1} \otimes \alpha) \oplus (\alpha^{*-1} \otimes \alpha^{*-1})
\oplus (\alpha \otimes \alpha^{*-1})
\end{multline*}
to a representation of $SL(2, \mathbb{C})$. Both, the initial representation $\big[4(L^{1/2})^{\otimes 2} \big]_\textrm{Ass}$
of $SL(2,\mathbb{R})$ and its extension $V$ to a representation of $SL(2, \mathbb{C})$
are Krein unitary in the Krein space $(\mathbb{C}^{16}, \mathfrak{J}_{\bar{p}})$ with
\begin{equation}\label{J_bar(p)for[4L^1/2otimes 2]Ass}
\mathfrak{J}_{\bar{p}} = \left( \begin{array}{cccc}  \bold{0}_4 & \bold{0}_4 & \bold{1}_4 & \bold{0}_4  \\
                              \bold{0}_4 & \bold{0}_4 & \bold{0}_4 & \bold{1}_4 \\
                              \bold{1}_4 & \bold{0}_4 & \bold{0}_4 & \bold{0}_4 \\
                              \bold{0}_4 & \bold{1}_4 & \bold{0}_4 & \bold{0}_4 \end{array}\right),
\end{equation}

and with the standard inner product in $\mathbb{C}^{16}$.

As the orbit $\mathscr{O}_{(0,m,0,0)}$ is exactly the same as in Example 2, we may choose $\beta(p)$
the same as in Example 2. For this $\beta(p)$ we have ($r = \sqrt{\vec{p} \cdot \vec{p}} \geq m >0$):
\[
\begin{split}
\beta(p)^* \beta(p)
= \frac{1}{mr}\left( \begin{array}{cc}  r^2 - p^0 p^3 & p^0 (i p^2 - p^1)\\
                                         p^0 (-ip^2-p^1)  & r^2 + p^0 p^3 \end{array}\right), \\
\big[\beta(p)^* \beta(p)\big]^{-1}
= \frac{1}{mr}\left( \begin{array}{cc}  r^2 + p^0 p^3 &  p^0 (-i p^2 + p^1)\\
                                         p^0 (ip^2+p^1)  & r^2 - p^0 p^3 \end{array}\right)
\end{split}
\]
and
\begingroup\makeatletter\def\f@size{5}\check@mathfonts
\def\maketag@@@#1{\hbox{\m@th\large\normalfont#1}}%
\begin{multline}\label{V(beta)*V(beta)WU^0m00[2L^1/2otimes2]AssW^-1}
V(\beta(p))^* V(\beta(p)) \\
= \left( \begin{array}{cccc}  \beta(p)^* \beta(p) \otimes \beta(p)^* \beta(p) & \bold{0}_4 & \bold{0}_4 & \bold{0}_4  \\
                 \bold{0}_4 & \big[\beta(p)^* \beta(p)\big]^{-1} \otimes  \beta(p)^* \beta(p) & \bold{0}_4 & \bold{0}_4 \\
   \bold{0}_4 & \bold{0}_4 & \big[\beta(p)^* \beta(p)\big]^{-1} \otimes \big[\beta(p)^* \beta(p)\big]^{-1} & \bold{0}_4 \\
  \bold{0}_4 & \bold{0}_4 & \bold{0}_4 & \beta(p)^* \beta(p) \otimes  \big[\beta(p)^* \beta(p)\big]^{-1}\end{array}\right).
\end{multline}\endgroup
Therefore the eigenvalues of $V(\beta(p))^* V(\beta(p))$ are the products $\lambda_1(p) \lambda_1(p)$,
$\lambda_2(p) \lambda_2(p)$, $\lambda_1(p) \lambda_2(p)$, $\lambda_2(p) \lambda _1(p)$ of the eigenvalues
$\lambda_1(p) = \frac{r}{m} + \Big(\big(\frac{r}{m}\big)^2 - 1 \Big)^{1/2}, 
\lambda_2(p) = \frac{r}{m} - \Big(\big(\frac{r}{m}\big)^2 - 1 \Big)^{1/2}$ of $ \beta(p)^* \beta(p)$
($\lambda_1(p), \lambda_2(p)$ are also equal to the eigenvalues of $\big[\beta(p)^* \beta(p)\big]^{-1}$),
each with multiplicity four.

Now consider the Krein unitary representation $U^{{}_{0,m,0,0} \big[4(L^{1/2})^{\otimes 2} \big]_\textrm{Ass}}$ 
of $T_4 \circledS SL(2, \mathbb{C})$ concentrated on the orbit $\mathscr{O}_{(0,m,0,0)}$
of $\bar{p} = (0,m,0,0)$ in $\widehat{T_4}$, induced by the above representation 
$\big[4(L^{1/2})^{\otimes 2} \big]_\textrm{Ass}$ of $G_{(0,m,0,0)} = SL(2, \mathbb{R})$. Using the extension $V$ of 
$\big[4(L^{1/2})^{\otimes 2} \big]_\textrm{Ass}$ we construct wave functions whose Fourier transform has local transformation law,
the same as multispinor $\phi_{{}_{m,0}}$ of Example 3.

The elements\footnote{Note that they have sixteen components.}  of the representation space of 
\[
{W'}_{{}_{V,0,m,0,0}}U_{{}_{[4(L^{1/2})^{\otimes 2}]_\textrm{Ass},0,m,0,0}}^{-1}
U^{{}_{(0,m,0,0)} \big[4(L^{1/2})^{\otimes 2} \big]_\textrm{Ass}}
U_{{}_{[4(L^{1/2})^{\otimes 2}]_\textrm{Ass},0,m,0,0}}{W'}_{{}_{V,0,m,0,0}}^{-1}
\]
which for simplicity we likewise denote
\[
U^{{}_{(0,m,0,0)} \big[4(L^{1/2})^{\otimes 2} \big]_\textrm{Ass}}, 
\]
we have agreed 
to denote  $\widetilde{\psi}_{{}_{0,m}}$ -- they are immediate analogous of the elements 
 $(\widetilde{\psi}^\textrm{\ding{192}}_{{}_{m,0}}, 
\widetilde{\psi}^\textrm{\ding{193}}_{{}_{m,0}}, \widetilde{\psi}^\textrm{\ding{194}}_{{}_{m,0}}, 
\widetilde{\psi}^\textrm{\ding{195}}_{{}_{m,0}})$ (resp.  $(\widetilde{\psi}^\textrm{\ding{192}}_{{}_{-m,0}}, 
\widetilde{\psi}^\textrm{\ding{193}}_{{}_{-m,0}}, \widetilde{\psi}^\textrm{\ding{194}}_{{}_{-m,0}}, 
\widetilde{\psi}^\textrm{\ding{195}}_{{}_{-m,0}})$) of the representation space of the representation 
$U^{{}_{(m,0,0,0)} 4\big(L^{{}^{1/2}} \otimes L^{{}^{1/2}}\big)}$ (resp. $ U^{{}_{(-m,0,0,0)} 4\big(L^{{}^{1/2}} \otimes L^{{}^{1/2}}\big)}$). 
The elements $\widetilde{\varphi}_{{}_{0,m}}$ obtained by the transform $W$, 
using the extension $V$ of $\gamma \mapsto Q(\gamma,\bar{p}) = \big[4(L^{1/2})^{\otimes 2} \big]_\textrm{Ass}(\gamma)$,
are immediate analogues of the multispinor $\widetilde{\phi}_{{}_{0,m}}$ of Example 3
 in having exactly the same transformation law 
\begin{equation}\label{WU^0m00[2L^1/2otimes2]AssW^-1}
U(\alpha) \widetilde{\varphi}_{{}_{0,m}}(p) 
= \left( \begin{array}{cccc}  \alpha \otimes \alpha & \bold{0}_4 & \bold{0}_4 & \bold{0}_4  \\
                              \bold{0}_4  & \alpha^{*-1} \otimes \alpha & \bold{0}_4 & \bold{0}_4 \\
                             \bold{0}_4 & \bold{0}_4 & \alpha^{*-1} \otimes \alpha^{*-1} & \bold{0}_4 \\
               \bold{0}_4 & \bold{0}_4 & \bold{0}_4 & \alpha \otimes \alpha^{*-1}  \end{array}\right) 
 \widetilde{\varphi}_{{}_{0,m}}(\Lambda(\alpha)p),
\end{equation}
\[
T(a) \widetilde{\varphi}_{{}_{0,m}}(p) = e^{i a \cdot p}\widetilde{\varphi}_{{}_{0,m}}(p);
\]
as the multispinor $\widetilde{\phi}_{{}_{m,0}}$ of Example 3 (with the only difference of course that this time they are concentrated on the different orbit $\mathscr{O}_{(0,0,m,0)}$). Analogously we have the transformation law
\[
U(\alpha) \varphi_{{}_{0,m}}(x) 
= \left( \begin{array}{cccc}  \alpha \otimes \alpha & \bold{0}_4 & \bold{0}_4 & \bold{0}_4  \\
                              \bold{0}_4  & \alpha^{*-1} \otimes \alpha & \bold{0}_4 & \bold{0}_4 \\
                             \bold{0}_4 & \bold{0}_4 & \alpha^{*-1} \otimes \alpha^{*-1} & \bold{0}_4 \\
               \bold{0}_4 & \bold{0}_4 & \bold{0}_4 & \alpha \otimes \alpha^{*-1}  \end{array}\right) 
 \varphi_{{}_{0,m}} (\Lambda(\alpha)x), 
\]
\[
T(a) \varphi_{{}_{0,m}}(x) = \varphi_{{}_{0,m}}(x -a).
\]
for the Fourier transform (\ref{F(varphi)}) $\varphi_{{}_{0,m}}$ (with $\mathscr{O}_{\bar{p}} = \mathscr{O}_{(0,m,0,0)}$ in 
(\ref{F(varphi)})) of $\widetilde{\varphi}_{{}_{0,m}}$ exactly the same as the multispinor $\phi_{{}_{m,0}}$ 
(resp. $\phi_{{}_{-m,0}}$) of Example 3. 
We therefore denote $\widetilde{\varphi}_{{}_{0,m}}$ and its Fourier transform in this case immediately by 
$\widetilde{\phi}_{{}_{0,m}}$ and $\phi_{{}_{0,m}}$. They compose, together with the inner product 
\begin{multline}\label{[4L^1/2otimes2]AssiInn-prod-0m00}
(\widetilde{\phi}_{{}_{0,m}}, \widetilde{\phi'}_{{}_{0,m}}) = 
  \int \limits_{\mathscr{O}_{(0,m,0,0)}} 
\Big(\widetilde{\phi}_{{}_{0,m}}(p), 
V(\beta(p))^* V(\beta(p)) \, \widetilde{\phi}_{{}_{0,m}}'(p) \Big)_{{}_{\mathbb{C}^{16}}}
\, \ud \mu_{{}_{0,m}}(p), 
\end{multline}
the Hilbert space of the natural local version
\begin{equation}\label{WU^[4L^1/2otimes2]AssW^-1}
W_{{}_{V,\bar{p}}} U^{{}_{\bar{p}} [L]_\textrm{Ass}} W_{{}_{V,\bar{p}}}^{-1},
\end{equation}
for 
\[
\bar{p}  = (m,0,0,0), \,\,\, L= 4(L^{{}^{1/2}} \otimes L^{{}^{1/2}}) = 4 \big(L^{{}^{1/2}} \big)^{\otimes 2},
\]
\[
V(\alpha) = \left( \begin{array}{cccc}  \alpha \otimes \alpha & \bold{0}_4 & \bold{0}_4 & \bold{0}_4  \\
                              \bold{0}_4  & \alpha^{*-1} \otimes \alpha & \bold{0}_4 & \bold{0}_4 \\
                             \bold{0}_4 & \bold{0}_4 & \alpha^{*-1} \otimes \alpha^{*-1} & \bold{0}_4 \\
               \bold{0}_4 & \bold{0}_4 & \bold{0}_4 & \alpha \otimes \alpha^{*-1}  \end{array}\right),
               \,\,\, \alpha \in SL(2, \mathbb{C}),
\]
The action of (\ref{WU^[4L^1/2otimes2]AssW^-1}) on $\widetilde{\varphi}_{{}_{0,m}}$ viz. 
$\widetilde{\phi}_{{}_{0,m}}$,
is given by (\ref{WU^0m00[2L^1/2otimes2]AssW^-1}), and (\ref{WU^[4L^1/2otimes2]AssW^-1}) is 
Krein-isometric with its natural Krein structure defined by the fundamental symmetry operator
of pointwise multiplication by the matrix
\[
\mathfrak{J}_{\bar{p}} \, V(\beta(p))^* V(\beta(p)),
\]
where $\mathfrak{J}_{\bar{p}}$ is given by (\ref{J_bar(p)for[4L^1/2otimes 2]Ass}).
Here $V(\beta(p))^* V(\beta(p))$ is equal (\ref{V(beta)*V(beta)WU^0m00[2L^1/2otimes2]AssW^-1}).

Construction  of the corresponding representation
\begin{equation}\label{WU^4L^1/2otimes2]AssW^-1_0}
\big(W_{{}_{V,\bar{p}}} U^{{}_{\bar{p}} [L]_\textrm{Ass}} W_{{}_{V,\bar{p}}}^{-1}\big)_{{}_{0}},
\end{equation}
for 
\[
\bar{p}  = (m,0,0,0), \,\,\, L= 4(L^{{}^{1/2}} \otimes L^{{}^{1/2}}) = 4 \big(L^{{}^{1/2}} \big)^{\otimes 2},
\]
\[
V(\alpha) = \left( \begin{array}{cccc}  \alpha \otimes \alpha & \bold{0}_4 & \bold{0}_4 & \bold{0}_4  \\
                              \bold{0}_4  & \alpha^{*-1} \otimes \alpha & \bold{0}_4 & \bold{0}_4 \\
                             \bold{0}_4 & \bold{0}_4 & \alpha^{*-1} \otimes \alpha^{*-1} & \bold{0}_4 \\
               \bold{0}_4 & \bold{0}_4 & \bold{0}_4 & \alpha \otimes \alpha^{*-1}  \end{array}\right),
               \,\,\, \alpha \in SL(2, \mathbb{C}),
\]
now immediately follows (and is completely analogous to that presented in Subsection \ref{e1} for 
\[
U^{{}_{m,0,0,0} 2L^{1/2}}.)
\]
Note that the inner product in the representation space of the
representation (\ref{WU^4L^1/2otimes2]AssW^-1_0}) is equal
\[
(\widetilde{\phi}_{{}_{0,m}}, \widetilde{\phi'}_{{}_{0,m}}) = 
  \int \limits_{\mathscr{O}_{(0,m,0,0)}} 
\Big(\widetilde{\phi}_{{}_{0,m}}(p), 
\, \widetilde{\phi}_{{}_{0,m}}'(p) \Big)_{{}_{\mathbb{C}^{16}}}
\, \ud \mu_{{}_{0,m}}(p), 
\]
and fundamental symmetry $\mathfrak{J}$ is equal to the operator of multiplication by the constant 
matrix
\[
\mathfrak{J}_{\bar{p}}
\]
where $\mathfrak{J}_{\bar{p}}$ is given by (\ref{J_bar(p)for[4L^1/2otimes 2]Ass}).
Recall also  that (\ref{WU^4L^1/2otimes2]AssW^-1_0}) acts identically as 
(\ref{WU^[4L^1/2otimes2]AssW^-1}) on $\widetilde{\phi}_{{}_{0,m}}$ belonging
to the dense nuclear space
\[
E_{{}_{0,m,16}} = \big\{f|_{{}_{\mathscr{O}_{0,m,0,0}}}, \,\,\,
f \in \mathcal{S}(\mathbb{R}^4; \mathbb{C}^{16}) \big\},
\]
common for representation spaces of both representations (\ref{WU^4L^1/2otimes2]AssW^-1_0})
and (\ref{WU^[4L^1/2otimes2]AssW^-1}).

\subsection{Spin 0 and 1 and the transform $V_\mathcal{F}$}\label{0-1VF}

Having the class of induced representations 
\[
U^{{}_{(m,0,0,0)} 4 \big(L^{{}^{1/2}} \big)^{\otimes 2}},
\]
$m \in \mathbb{R}$,
concentrated respectively on $\mathscr{O}_{(m,0,0,0)}$ and the associated representations
$U^{{}_{(0,0,m,0)} \big[ 4 \big(L^{{}^{1/2}} \big)^{\otimes 2} \big]_\textrm{Ass}}$, $m \in \mathbb{R}_+$, concentrated respectively on 
$\mathscr{O}_{(0,m,0,0)}$, or more precisely having the class of representations
\[ 
\left\{ \begin{array}{ll}
 \Big(W_{{}_{V,m,0,0,0}} U^{{}_{m,0,0,0}\big(4(L^{{}^{1/2}})^{\otimes 2}\big)}  
W_{{}_{V,m,0,0,0}}^{-1}\Big)_{{}_{0}}, 
 & \textrm{for $\mathscr{O}_{m,0,0,0}$, $m\neq 0$}, \\
    \Big(W_{{}_{V,0,m,0,0}} U^{{}_{0,m,0,0}\big[4(L^{{}^{1/2}})^{\otimes 2}\big]_{\textrm{Ass}} } 
W_{{}_{V,0,m,0,0}}^{-1}\Big)_{{}_{0}}, 
& \textrm{for $\mathscr{O}_{0,m,0,0}$, $m> 0$}, \\
   V, 
& \textrm{for $\mathscr{O}_{0,0,0,0}$}, \\
\Big(W_{{}_{V,1,0,0,1}} U^{{}_{1,0,0,1}\big(4(L^{{}^{1/2}})^{\otimes 2}\big)}  
W_{{}_{V,1,0,0,1}}^{-1}\Big)_{{}_{0}}, 
 & \textrm{for $\mathscr{O}_{1,0,0,1}$},
    \end{array} \right.
\]
with common representation $V$:
\[
\alpha \mapsto 
V(\alpha) = \left( \begin{array}{cccc}  \alpha \otimes \alpha & \bold{0}_4 & \bold{0}_4 & \bold{0}_4  \\
                              \bold{0}_4  & \alpha^{*-1} \otimes \alpha & \bold{0}_4 & \bold{0}_4 \\
                             \bold{0}_4 & \bold{0}_4 & \alpha^{*-1} \otimes \alpha^{*-1} & \bold{0}_4 \\
               \bold{0}_4 & \bold{0}_4 & \bold{0}_4 & \alpha \otimes \alpha^{*-1}  \end{array}\right),
               \,\,\, \alpha \in SL(2, \mathbb{C}),
\]
of $SL(2, \mathbb{C})$, we are ready to construct the transform $V_\mathcal{F}$ on the
space of the representation 
\begin{equation}\label{step2rep}
 \int \limits_{-\infty}^{\infty} 
U^{{}_{(m,0,0,0)} 4 \big(L^{{}^{1/2}} \big)^{\otimes 2}}
\, \ud m \bigoplus  
\int \limits_{0}^{\infty} 
U^{{}_{(0,0,m,0)} \big[ 4 \big(L^{{}^{1/2}} \big)^{\otimes 2} \big]_\textrm{Ass}}
 \, \ud m.
\end{equation}
Recall however that (\ref{step2rep}) is a shortened notation (of Remark \ref{RemarkShortNotation}) 
for the direct integral representation in which the direct integral summands
\[
U^{{}_{(m,0,0,0)} 4 \big(L^{{}^{1/2}} \big)^{\otimes 2}}, \,\,\,
U^{{}_{(0,0,m,0)} \big[ 4 \big(L^{{}^{1/2}} \big)^{\otimes 2} \big]_\textrm{Ass}}
\]
are understood to be equal 
\[
\big(W_{{}_{V,\bar{p}}} U^{{}_{\bar{p}} L} W_{{}_{V,\bar{p}}}^{-1}\big)_{{}_{0}},
\,\,\,
\big(W_{{}_{V,\bar{p}'}} U^{{}_{\bar{p}'} [L]_\textrm{Ass}} W_{{}_{V,\bar{p}'}}^{-1}\big)_{{}_{0}},
\]
(which we call to be the ``Case 1'') for 
\[
\bar{p}  = (m,0,0,0), \,\,\, L= 4 \big(L^{{}^{1/2}} \big)^{\otimes 2},
\]
or respectively
\[
\bar{p}'  = (0,m,0,0), \,\,\, L = \big[4 \big(L^{{}^{1/2}} \big)^{\otimes 2}\big]_\textrm{Ass},
\]
with common
\[
V(\alpha) = \left( \begin{array}{cccc}  \alpha \otimes \alpha & \bold{0}_4 & \bold{0}_4 & \bold{0}_4  \\
                              \bold{0}_4  & \alpha^{*-1} \otimes \alpha & \bold{0}_4 & \bold{0}_4 \\
                             \bold{0}_4 & \bold{0}_4 & \alpha^{*-1} \otimes \alpha^{*-1} & \bold{0}_4 \\
               \bold{0}_4 & \bold{0}_4 & \bold{0}_4 & \alpha \otimes \alpha^{*-1}  \end{array}\right),
               \,\,\, \alpha \in SL(2, \mathbb{C}).
\]
In fact we may consider three more cases in which the direct integral summads of the first direct summand
and respectively the second direct summand are replaced with
\[
W_{{}_{V,\bar{p}}} U^{{}_{\bar{p}} L} W_{{}_{V,\bar{p}}}^{-1},
\,\,\,
W_{{}_{V,\bar{p}'}} U^{{}_{\bar{p}'} [L]_\textrm{Ass}} W_{{}_{V,\bar{p}'}}^{-1},
\]
which we call to be the ``Case 2'', or with direct integral summands equal
\[
\big(W_{{}_{V,\bar{p}}} U^{{}_{\bar{p}} L} W_{{}_{V,\bar{p}}}^{-1}\big)_{{}_{00}},
\,\,\,
\big(W_{{}_{V,\bar{p}'}} U^{{}_{\bar{p}'} [L]_\textrm{Ass}} W_{{}_{V,\bar{p}'}}^{-1}\big)_{{}_{0}},
\]
which we call to be the ``Case 3'', or finally with the direct integral summands equal
\[
\big(W_{{}_{V,\bar{p}}} U^{{}_{\bar{p}} L} W_{{}_{V,\bar{p}}}^{-1}\big)_{{}_{00}},
\,\,\,
W_{{}_{V,\bar{p}'}} U^{{}_{\bar{p}'} [L]_\textrm{Ass}} W_{{}_{V,\bar{p}'}}^{-1},
\]
which we call to be the ``Case 4''. In each of the cases we may define 
the fundamental symmetry $\mathfrak{J}$ which acts by multiplication by the constant matrix 
\begin{equation}\label{Jstep2}
\left( \begin{array}{cccc}  \bold{0}_4 & \bold{0}_4 & \bold{1}_4 & \bold{0}_4  \\
                              \bold{0}_4 & \bold{0}_4 & \bold{0}_4 & \bold{1}_4 \\
                              \bold{1}_4 & \bold{0}_4 & \bold{0}_4 & \bold{0}_4 \\
                              \bold{0}_4 & \bold{1}_4 & \bold{0}_4 & \bold{0}_4 \end{array}\right).
\end{equation}

Exactly as in the analysis of the representation (\ref{step1LocNaturallyAss})-(\ref{step1LocNonNatural'}) in Subsection \ref{1/2VF}, we show, repeating the proof of Sect. \ref{1/2VF}, that any element 
$\widetilde{\phi}$ of the representation space of the representation (\ref{step2rep}) in each of the Cases 1-4, can be identified with measurable multispinor function $\widetilde{\phi}$
which is square summable with respect
to the measure $\ud^4 p$ appropriately weighted, 
or with $p \mapsto \Big(\widetilde{\phi}(p), V(\beta(p))^* V(\beta(p))\widetilde{\phi}(p) \Big)_{\mathbb{C}^4}$ summable (depending on the actual Case 1, 2, 3 or 4) with the weight simply equal $1$
in Case 1.

On the other hand we consider -- exactly as in Subsection \ref{1/2VF} -- the Hilbert space 
of space-time multispinors $\phi$ square summable with respect to the invariant
measure $\ud^4 x$ on $\mathbb{R}^4$ equipped with the Minkowski pseudo-metric $g^{\mu \nu}$.
Let $L^2 (S, \ud^4 x)$ be the Hilbert space of square summable multispinors $\phi$
with the standard Hilbert space inner product given by (\ref{inn-x-step1}). We equip
$L^2 (S, \ud^4 x)$ with the fundamental symmetry $\mathfrak{J}$ defined by the multiplication
by the matrix 
\[
\mathfrak{J}_{{}_{\overline{p}}}
= \left( \begin{array}{cccc}  \bold{0}_4 & \bold{0}_4 & \bold{1}_4 & \bold{0}_4  \\
                              \bold{0}_4 & \bold{0}_4 & \bold{0}_4 & \bold{1}_4 \\
                              \bold{1}_4 & \bold{0}_4 & \bold{0}_4 & \bold{0}_4 \\
                              \bold{0}_4 & \bold{1}_4 & \bold{0}_4 & \bold{0}_4 \end{array}\right),
\]
giving the Krein space structure $\big(L^2 (S, \ud^4 x), \mathfrak{J}\big)$, together with the following 
Krein unitary representation of $T_4 \circledS SL(2, \mathbb{C})$ in this space:
\begin{equation}\label{Ustep2}
U(\alpha) \phi(x) 
= \left( \begin{array}{cccc}  \alpha \otimes \alpha & \bold{0}_4 & \bold{0}_4 & \bold{0}_4  \\
                              \bold{0}_4  & \alpha^{*-1} \otimes \alpha & \bold{0}_4 & \bold{0}_4 \\
                             \bold{0}_4 & \bold{0}_4 & \alpha^{*-1} \otimes \alpha^{*-1} & \bold{0}_4 \\
               \bold{0}_4 & \bold{0}_4 & \bold{0}_4 & \alpha \otimes \alpha^{*-1}  \end{array}\right) 
 \phi (\Lambda(\alpha)x), 
\end{equation}
\[
T(a) \phi(p) = \phi(x -a).
\]

Now using the ordinary Fourier transform $\widetilde{\phi}$ of $\phi \in L^2 (S, \ud^4 x)$, we may identify 
$L^2 (S, \ud^4 x)$ with the linear space of (equivalence classes -- with functions differing on Lebesgue 
measure zero set being equivalent) of functions $\widetilde{\phi}$ which are square summable with respect
to the invariant measure $\ud^4 p$.
In the Hilbert space of Fourier transforms $\widetilde{\phi}$ of $\phi \in L^2 (S, \ud^4 x)$ the representation
of $T_4 \circledS SL(2, \mathbb{C})$ acts as follows
 \[
U(\alpha) \widetilde{\phi}(p) 
= \left( \begin{array}{cccc}  \alpha \otimes \alpha & \bold{0}_4 & \bold{0}_4 & \bold{0}_4  \\
                              \bold{0}_4  & \alpha^{*-1} \otimes \alpha & \bold{0}_4 & \bold{0}_4 \\
                             \bold{0}_4 & \bold{0}_4 & \alpha^{*-1} \otimes \alpha^{*-1} & \bold{0}_4 \\
               \bold{0}_4 & \bold{0}_4 & \bold{0}_4 & \alpha \otimes \alpha^{*-1}  \end{array}\right) 
 \widetilde{\phi}(\Lambda(\alpha)p), 
\]
\[
T(a) \widetilde{\phi}(p) = e^{i a \cdot p}\widetilde{\phi}(p);
\]
and is of course unitary equivalent to (\ref{Ustep2}), and with the fundamental symmetry
likewise acting as multiplication by the constant matrix (\ref{Jstep2}).

In this case we define the generalized Dirac operator
\[
D = i \gamma^\mu \partial_\mu,
\]
\[
\gamma^0 = \widetilde{\gamma}^0, \,\, \gamma^k = -\widetilde{\gamma}^k, \,\, k=1,2,3,
\]
which is essentially Krein self adjoint in the Krein space $\big(L^2 (S, \ud^4 x), \mathfrak{J}\big)$
(the proof being essentially the same as that in \cite{baum}), and commutes with the representation (\ref{Ustep2}).
Similarly,
\[
\begin{split}
D^2 = - \gamma^\mu \gamma^\nu \partial_\mu \partial_\nu \\
= - \bold{1}_{16} \, (\partial_0 \partial_0 - \partial_1 \partial_1 - \partial_2 \partial_2 - \partial_3 \partial_3)
\end{split}
\]
commutes with (\ref{Ustep2}) and is moreover essentially self adjoint in $L^2 (S, \ud^4 x)$.

Using the following commutation relations, and the following behavior of
$\widetilde{\gamma}^\mu$ under the adjoint operation ($k = 1,2,3$)
\[
\mathfrak{J}_{{}_{\overline{p}}} \widetilde{\gamma}^0 =
\widetilde{\gamma}^0 \mathfrak{J}_{{}_{\overline{p}}} , \,\,\,
\mathfrak{J}_{{}_{\overline{p}}} \widetilde{\gamma}^k =
- \widetilde{\gamma}^k \mathfrak{J}_{{}_{\overline{p}}} , \,\,\,
\big( \widetilde{\gamma}^0 \big)^* = \widetilde{\gamma}^0 , \,\,\,
\big( \widetilde{\gamma}^k \big)^* = - \widetilde{\gamma}^k
\]
one easily checks that
\[
\frac{1}{2} \Big\{ (D\mathfrak{J})^2 + (\mathfrak{J}D)^2 \Big\}
= - \bold{1}_{16} \, (\partial_0 \partial_0 + \partial_1 \partial_1 + \partial_2 \partial_2 + \partial_3 \partial_3),
\]
so that $\frac{1}{2} \Big\{ (D\mathfrak{J})^2 + (\mathfrak{J}D)^2 \Big\}$ is elliptic, and we can
choose
\[
D_\mathfrak{J} = i \Upsilon^\mu \partial_\mu,
\]
with $\Upsilon^\mu$, defined by
\[
\Upsilon^0 = \gamma^0, \,\,\, \Upsilon^k = i \gamma^k
\]
and being the generators of the Clifford algebra associated to the ordinary Euclidean metric
$\delta^{\mu\nu}$:
\[
\Upsilon^\mu \Upsilon^\nu +\Upsilon^\nu \Upsilon^\mu
= 2 \delta^{\mu\nu} \, \bold{1}_{16}.
\]
For this $D_\mathfrak{J}$, $\mathfrak{J}$, and $D$ we indeed have
\[
\frac{1}{2} \Big\{ (D\mathfrak{J})^2 + (\mathfrak{J}D)^2 \Big\}
= \big( D_\mathfrak{J} \big)^2.
\]

The operator $D_\mathfrak{J}$ is the generalized Dirac operator associated to a representation of the Clifford algebra
corresponding to the Riemannian metric, in the sense used in mathematical literature,
compare e.g. \cite{Roe} and the literature cited therein. In particular by general properties of such operators
(compare the results referred in \cite{Roe} and in the literature cited by \cite{Roe}) it follows that
$D_\mathfrak{J}$ is essentially self adjoint. The operator $D$ belongs to the class of operators generalized in the same sense but associated to representation of the Clifford algebra corresponding to pseudo-Riemannian metric.
By the results of \cite{baum} these operators are essentially Krein self adjoint providing some general orientability
conditions and completeness of the pseudo-Riemmannian manifold for the Riemann metric associated to $D_\mathfrak{J}$
are fulfilled -- conditions which are evidently preserved in our case.

The algebra $\mathcal{A}$ of Schwartz functions acting as multiplication operators
on square integrable multispinors $\phi \in L^2(S, \ud^4x)$, and the operators 
$D$, $\mathfrak{J}, D_\mathfrak{J}$  fulfil the 
conditions of Introduction in the subspace associated to the representation 
(\ref{Ustep2})). The strong regularity 
of the spectral triple $\big( \mathcal{A}, \mathcal{H} = L^2(S, \ud^4x), D_\mathfrak{J} \big)$ is checked
exactly as in the proof of Theorem 11.4 of \cite{Connes_spectral}: indeed $D_\mathfrak{J}$ is an elliptic differential
operator of order one on the smooth manifold $\mathbb{R}^4$ with the square
\[
\big(D_\mathfrak{J}\big)^2 = - \bold{1}_{16} \, (\partial_0 \partial_0 + \partial_1 \partial_1 + \partial_2 \partial_2 + \partial_3 \partial_3),
\]
so that the principal symbol of $\big(D_\mathfrak{J}\big)^2$ is a scalar multiple of the identity.

\vspace*{1cm}

\begin{center}
\small DEFINITION OF THE TRANSFORM $V_\mathcal{F}$ ON THE SPACE OF THE REPRESENTATION (\ref{step2rep}) 
\end{center}

Every continuous $\widetilde{\phi}$ multispinor with compact support and thus
realizing an element of the Hilbert space of the representation (\ref{step2rep}) in each Case:
1 through 4, may at the same time be regarded
as an element $\widetilde{\phi}$ equal to the ordinary Fourier transform of some $\phi \in L^2 (S, \ud^4 x)$.
We choose the common linear domain of such $\widetilde{\phi}$ to be just equal
$\mathfrak{D} = \mathcal{S}(\mathbb{R}^4; \mathbb{C}^{16})$.
The elements $\widetilde{\phi}$ of the Hilbert space of the representation (\ref{step2rep})
in each Case: 1, 2, 3, and 4, corresponding to the
elements $\widetilde{\phi}$ of $\mathfrak{D} = \mathcal{S}(\mathbb{R}^4; \mathbb{C}^{16})$
compose a dense domain in the Hilbert space of the representation (\ref{step2rep})
in each of the indicated Cases.
Similarly, the elements $\phi \in L^2 (S, \ud^4 x)$ corresponding to these
$\widetilde{\phi} \in \mathcal{S}(\mathbb{R}^4; \mathbb{C}^{16})$
compose a dense domain in $L^2 (S, \ud^4 x)$.
For any element $\widetilde{\phi}$ corresponding to an element $\widetilde{\phi} \in \mathcal{S}(\mathbb{R}^4; \mathbb{C}^{16})$
we define $V_\mathcal{F} \big( \widetilde{\phi} \big)$ as the square summable multispinor $\phi \in L^2 (S, \ud^4 x)$ -- equal to the ordinary (inverse) Fourier transform of $\widetilde{\phi}$.
\qed

\vspace*{1cm}

Similarly, as for the representation (\ref{step1LocNaturallyAss}) in Subsection \ref{1/2VF}, we show, repeating the proof of Sect. \ref{1/2VF}, that only the representation indicated by the Case 1, is unintarily and Krein
unitarily equivalent to the representation acting on space-time square integrable multispinors.
The unitary and Krein-unitary equivalence being equal $V_\mathcal{F}$ and identifies the translation generators $P^\mu$ with the operators of differentiation
\[
i\frac{\partial}{\partial x_\mu}
\]
in the representation space $L^2 (S, \ud^4 x)$ of square summable space-time multispinors. The operators
$Q^\mu$ acting in the representation space of the Case 1-representation, which compose canonical
pairs with $P^\mu$, are identified with the operators of multiplication by $x^\mu$
in the Hilbert space $L^2 (S, \ud^4 x)$ of square summable space-time multispinors, under the unitary equivalence $V_\mathcal{F}$. The operators $D_\mathfrak{J}, D, \mathfrak{J}$ in $L^2 (S, \ud^4 x)$
are identified under $V_\mathcal{F}$ with the operators
\[
D_\mathfrak{J} = \Upsilon^\mu P_\mu, \,\,\,\,\,\,
D= P_\mu \gamma^\mu, \,\,\,\,\,\,
\gamma^0 = \widetilde{\gamma}^0, \,\,\,\, \gamma^{k} = -\widetilde{\gamma}^k
\]
and respectively with the operator of multiplication by the constant matrix $\mathfrak{J}_{\bar{p}}$
given by (\ref{Jstep2}) or (\ref{J_bar(p)for[4L^1/2otimes 2]Ass}) acting in the representation space
of the Case 1-representation.

No such equivalence of the representation indicated under the Case 2, 3 and 4,
with the representation acting on space-time square integrable multispinors is possible,
similarly as was the case for the representations (\ref{step1LocAss})-(\ref{step1LocNonNatural'}) 
of Subsection  \ref{1/2VF}.

Now we explain the principle connecting the construction of $V_\mathcal{F}$
corresponding to the representations (\ref{step1LocNaturallyAss})-(\ref{step1LocNonNatural'}),
written shortly (notation convention of Remark \ref{RemarkShortNotation})
\begin{equation}\label{step1rep'}
\int \limits_{-\infty}^{\infty}
U^{{}_{m,0,0,0}(2L^{{}^{1/2}})}
\, \ud m \bigoplus
\int \limits_{0}^{\infty}
U^{{}_{0,m,0,0}[ 2L^{{}^{1/2}}]_{\textrm{Ass}} }
\, \ud m,
\end{equation}
with that corresponding to the representation (\ref{step2rep}) (Cases 1-4).
Note that the extension $V$ of the representation $\big[4\big(L^{1/2}\big)^{\otimes 2}\big]_\textrm{Ass}$
of the small group $G_{(0,m,0,0)}$ of Example 4, which induces the representation associated to
the representation
\[
U^{{}_{(m,0,0,0)} \big[4(L^{{}^{1/2}})^{\otimes 2}\big]_\textrm{Ass}},
\]
is equal (where $\otimes$ on the right-hand side is treated here formally as if it was distributive
over $\oplus$ wih ordinarily equality in the distributive law instead of the unitary equivalence)
\begin{multline*}
\alpha \mapsto V(\alpha)
= (\alpha \otimes \alpha) \oplus (\alpha^{*-1} \otimes \alpha) \oplus (\alpha^{*-1} \otimes \alpha^{*-1})
\oplus (\alpha \otimes \alpha^{*-1}) \\
= U \Big( \alpha \oplus \alpha^{*-1} \Big) \otimes \Big( \alpha \oplus \alpha^{*-1} \Big)U^{-1},
\end{multline*}
where $\alpha \mapsto \alpha \oplus \alpha^{*-1}$ is the extension $V$ of the representation
$2\big(L^{1/2}\big)$ which induces the representation associated to
\[
U^{{}_{(m,0,0,0)} 2L^{{}^{1/2}}},
\]
and where $U$ is unitary in $\mathbb{C}^{16}$, which is equal to the compositions of unitary transforms
corresponding to respective inversions of the axes
(base vectors in $\mathbb{C}^{16}$ with the standard inner product). $U$ is easily computable and equal
\begin{equation}\label{clebsch.step2.U}
U =
\left( \begin{array}{cccccccc}
\bold{1}_2 & \bold{0}_2 & \bold{0}_2 & \bold{0}_2 &\bold{0}_2 &\bold{0}_2 & \bold{0}_2 & \bold{0}_2 \\
\bold{0}_2 & \bold{0}_2 & \bold{1}_2 & \bold{0}_2 &\bold{0}_2 &\bold{0}_2 & \bold{0}_2 & \bold{0}_2 \\
\bold{0}_2 & \bold{0}_2 & \bold{0}_2 & \bold{0}_2 &\bold{1}_2 &\bold{0}_2 & \bold{0}_2 & \bold{0}_2 \\
\bold{0}_2 & \bold{0}_2 & \bold{0}_2 & \bold{0}_2 &\bold{0}_2 &\bold{0}_2 & \bold{1}_2 & \bold{0}_2 \\
\bold{0}_2 & \bold{0}_2 & \bold{0}_2 & \bold{0}_2 &\bold{0}_2 &\bold{1}_2 & \bold{0}_2 & \bold{0}_2 \\
\bold{0}_2 & \bold{0}_2 & \bold{0}_2 & \bold{0}_2 &\bold{0}_2 &\bold{0}_2 & \bold{0}_2 & \bold{1}_2 \\
\bold{0}_2 & \bold{1}_2 & \bold{0}_2 & \bold{0}_2 &\bold{0}_2 &\bold{0}_2 & \bold{0}_2 & \bold{0}_2 \\
\bold{0}_2 & \bold{0}_2 & \bold{0}_2 & \bold{1}_2 &\bold{0}_2 &\bold{0}_2 & \bold{0}_2 & \bold{0}_2
\end{array}\right).
\end{equation}
Because the Krein space structure is functorial under tensoring and direct summation, we immediately
obtain all possible fundamental symmetries for which the representation
(\ref{step2rep}), the Case 1, is Krein unitary and the corresponding Dirac operator is essentially Krein-selfadjoint.
Indeed, because the fundamental symmetry for (\ref{step1rep'}), Case (\ref{step1LocNaturallyAss}),
is given by the multiplication by the matrix
$\gamma^1 \gamma^2 \gamma^3$ or by the matrix $\gamma^0$, then we have at least four
possibilities for the fundamental symmetry for
(\ref{step2rep}), Case 1, given by the multiplication by the matrix
$U \big(\gamma^0 \otimes \gamma^0 \big) U^{-1}$
or $U \big(\gamma^1 \gamma^2 \gamma^3 \otimes \gamma^1 \gamma^2 \gamma^3 \big) U^{-1}$ or
$U \big(\gamma^1 \gamma^2 \gamma^3 \otimes \gamma^0 \big) U^{-1}$
or by multiplication by the matrix $U \big(\gamma^0 \otimes \gamma^1 \gamma^2 \gamma^3 \big) U^{-1}$.
In particular $U \big(\gamma^0 \otimes \gamma^0 \big) U^{-1}$ is equal to the matrix (\ref{Jstep2}), and
$U \big(\gamma^1 \gamma^2 \gamma^3 \otimes \gamma^1 \gamma^2 \gamma^3 \big) U^{-1}$ is equal to
\[
\mathfrak{J}_{{}_{\overline{p}}}
= \left( \begin{array}{cccc} \bold{0}_4 & \bold{0}_4 & -\bold{1}_4 & \bold{0}_4 \\
\bold{0}_4 & \bold{0}_4 & \bold{0}_4 & \bold{1}_4 \\
-\bold{1}_4 & \bold{0}_4 & \bold{0}_4 & \bold{0}_4 \\
\bold{0}_4 & \bold{1}_4 & \bold{0}_4 & \bold{0}_4 \end{array}\right).
\]

However, one has to be careful because for some of the above fundamental symmetries (e.g. for the last)
the operator $\widetilde{\gamma}^\mu \partial_\mu$ is Krein-self-adjoint only with the additional
imaginary factor $i$ and correspondingly the commutation relations are changed into:
\[
\mathfrak{J}_{{}_{0}} \widetilde{\gamma}^0 = 
- \widetilde{\gamma}^0 \mathfrak{J}_{{}_{\overline{p}}} , \,\,\,
\mathfrak{J}_{{}_{0}} \widetilde{\gamma}^k =  
\widetilde{\gamma}^k \mathfrak{J}_{{}_{\overline{p}}}. 
\]
Otherwise: $iD$ is Krein self-adjoint and thus $iD$ is the Dirac operator with the 
Clifford algebra generators (generalized Dirac matrices) $\widetilde{\gamma}^\mu$ replaced with
$i\widetilde{\gamma}^\mu$. Therefore, the Krein self-adjoint Dirac operator $iD$  is associated
with the space-time pseudo-metric signature changed into the opposite one $(-1,1,1,1)$.
 
The diagram $B)$ of Subsection \ref{e3} likewise leads to a generalized Dirac operator with the generators 
\begin{equation}\label{gammaB}
\tilde{\gamma}^0 = \left( \begin{array}{cccc} \bold{0}_4 & \bold{1}_4 & \bold{0}_4 & \bold{0}_4  \\

                              \bold{1}_4  & \bold{0}_4 & \bold{0}_4 & \bold{0}_4 \\
                             \bold{0}_4 & \bold{0}_4 & \bold{0}_4 & \bold{1}_4 \\
               \bold{0}_4 & \bold{0}_4 & \bold{1}_4 &\bold{0}_4  \end{array}\right) \,\,\,
\tilde{\gamma}^k = \left( \begin{array}{cccc} \bold{0}_4 & - \sigma_k \otimes \bold{1}_2 & \bold{0}_4 & \bold{0}_4 \\
                             \sigma_k \otimes \bold{1}_2  & \bold{0}_4 & \bold{0}_4 & \bold{0}_4 \\
                             \bold{0}_4 & \bold{0}_4 & \bold{0}_4 & \sigma_k \otimes \bold{1}_2 \\
               \bold{0}_4 & \bold{0}_4 & - \sigma_k \otimes \bold{1}_2 & \bold{0}_4  \end{array}\right),
\end{equation}
of a slightly different representation of the Clifford algebra associated to the same
Minkowski pseudo-metric on $\mathbb{R}^4$; also the signs in the definition of the respective $V^\textrm{\ding{192}},
\ldots$ will have to be changed. This gives rise to another spectral description of one and the same Minkowski 
structure on $\mathbb{R}^4$ on the image of the representation space of (\ref{step2rep}), Case 1, 
under the transform $V_\mathcal{F}$. 

Let us explain the general principle standing behind the construction of $V_\mathcal{F}$ and the associated
generalized Dirac operator $D$ constructed in Subsection \ref{1/2VF} and in this Subsection.

Recall that for the construction of the (generalized) Dirac operator associated
to a representation induced by the representation $\gamma \mapsto Q(\gamma, \bar{p}) = L(\gamma)$ of the small
group $G_{\bar{p}}$ (for $\bar{p} = (m,0,0,0)$) it is crucial to find two conjugate Krein-unitary extensions
$V$ and $\overline{V}$ of $L$ to the whole $SL(2,\mathbb{C})$ group acting in the same space as the
initial representation $L$. Conjugation means here that $V$ and $\overline{V}$
are equal on the small group, and nothing more, thus depends on the class of the orbits $\mathscr{O}_{\bar{p}}$
in question. In particular for the class corresponding to $\bar{p} = (m,0,0,0)$ it can be realized by the
group automorphism $\alpha \mapsto \alpha^{*-1} = \Aut(\alpha)$ of $SL(2, \mathbb{C})$: $\overline{V}(\alpha)
= V(\Aut(\alpha))$. Similarly, for the class
of orbits corresponding to $\bar{p} = (0,0,m,0)$ it can be realized by the group automorphism
$\alpha \mapsto \overline{\alpha}$ (ordinary complex conjugation) of the $SL(2, \mathbb{C})$.
For the orbit of $(1,0,1,0)$ (light cone) it cannot be realized by group automorphism, but this is irrelevant,
because the light cone orbit occupies measure zero set in the space of all orbits (also the second class of orbits is irrelevant for the analysis presented here by the very construction of $V_\mathcal{F}$ as we will soon see).
Next we construct the generalized ''bispinor''
\[
\begin{split}
\widetilde{\varphi} = V(\beta^{-1}) \widetilde{\psi} \\
\widetilde{\chi} = \overline{V}(\beta^{-1}) \widetilde{\psi},
\end{split}
\]
where $\beta: p \mapsto \beta(p)$ is the function corresponding to the orbit of $\bar{p}$ and defined as above
($\beta$ depends on the orbit and is not unique).
From this we obtain the generalized Dirac equation in the momentum space (algebraic relation which after Fourier
transforming passes into a generalized Dirac equation)
\[
\begin{split}
\widetilde{\chi} = \overline{V}(\beta^{-1})V(\beta^{-1})^{-1} \widetilde{\varphi} \\
\widetilde{\varphi} = V(\beta^{-1})\overline{V}(\beta^{-1})^{-1} \widetilde{\chi},
\end{split}
\]
which can be written as
\[
\begin{split}
\widetilde{\chi} = V(\Aut(\beta)^{-1} \beta) \widetilde{\varphi} \\
\widetilde{\varphi} = V\big((\Aut(\beta)^{-1}\beta)^{-1}\big) \widetilde{\chi},
\end{split}
\]
It so happens that the function $p \mapsto \Aut(\beta (p))^{-1} \beta(p)$
(and the function $p \mapsto \big(\Aut(\beta(p))^{-1}\beta(p)\big)^{-1}$)
is a linear function of $p$. Now the function $\beta$ -- even within one and the same
orbit -- is not unique. For example, we have used the standard linear formula for $\beta$ corresponding
to the orbit of $\bar{p} = (m,0,0,0)$ using the known expression for $\beta(p)$ with
the Pauli matrices, which is linear in $p$, and which in addition is self adjoint and positive as a matrix
operator in $\mathbb{C}^2$ with the standard inner product. But we could use instead
the following nonlinear in $p$ expression for $\beta(p)$ in the class of orbits of $\bar{p}
= (m,0,0,0)$, $m>0$:
\[
\frac{1}{m^{1/2}}\left( \begin{array}{cc}
[(r^2 + m^2)^{1/2} -r]^{1/2} \cos \theta/2 \, e^{-i\frac{\vartheta}{2}} &
-i[(r^2 + m^2)^{1/2} -r]^{1/2} \sin \theta/2 \, e^{i\frac{\vartheta}{2}} \\
-i[(r^2 + m^2)^{1/2} +r]^{1/2} \sin \theta/2 \, e^{-i\frac{\vartheta}{2}} &
[(r^2 + m^2)^{1/2} +r]^{1/2} \cos \theta/2 \, e^{i\frac{\vartheta}{2}} \end{array}\right),
\]
where
\[
p = \left( \begin{array}{c}
(r^2 + m^2)^{1/2} \\
r \sin \theta \sin \vartheta \\
r \sin \theta \cos \vartheta \\
r \cos \theta
\end{array}\right), \,\,\, \textrm{and} \,\,\,
r = (\vec{p} \cdot \vec{p})^{1/2}.
\]
It is therefore important that the following simple Lemma holds true:

\begin{lem*}
The function
\[
p \mapsto \Aut(\beta (p))^{-1} \beta(p)
\]
is independent of the choice of the function $\beta: \mathscr{O}_{\bar{p}} \ni
p \mapsto \beta(p)$ fulfilling $\beta(p)^{-1} \widehat{\bar{p}} (\beta(p)^{-1})^{*} = \widehat{p}$
on $\mathscr{O}_{\bar{p}}$, if $\Aut$ is the automorphism of $SL(2, \mathbb{C})$ realizing the conjugation
$\overline{V}$ of the representation $V$. In the class of orbits of $\bar{p} = (m,0,0,0)$
\[
m \Aut(\beta (p))^{-1} \beta(p) = p^0 \bold{1}_2 - \vec{p} \cdot \vec{\sigma};
\]
in the class of orbits of $\bar{p} = (0,0,m,0)$
\[
m \Aut(\beta (p))^{-1} \beta(p) = - p^0 \sigma_2 - i p^1 \sigma_3 + p^2 \bold{1}_2 + i p^3 \sigma_1,
\]
where $\sigma_k$ are the Pauli matrices.
\end{lem*}

We can give a more general form of this Lemma which holds true even in the case when
the conjugation $\overline{V}$ cannot be realized by any automorphism of $SL(2, \mathbb{C})$.
Namely, we have the following simple

\begin{lem*}
If $V$ and $\overline{V}$ is a pair of representations of $SL(2, \mathbb{C})$ acting in the same space and such that
\[
V(\gamma) = \overline{V}(\gamma), \,\,\, \gamma \in G_{\bar{p}}
\]
then the function
\[
p \mapsto \overline{V}(\beta(p)^{-1}) {V(\beta(p)^{-1})}^{-1}
\]
on the orbit $\mathscr{O}_{\bar{p}}$ is independent of the choice of the function $p \mapsto \beta(p)$
fulfilling $\beta(p)^{-1} \widehat{\bar{p}} (\beta(p)^{-1})^{*} = \widehat{p}$
on $\mathscr{O}_{\bar{p}}$.
\end{lem*}

Thus, the above construction of $V_{\mathcal{F}}$ and the associated generalized Dirac operator
$D$ is independent of the choice of the function $\beta$.

\subsection{Direct integrals of higher spin representations and the construction
of $V_\mathcal{F}$}\label{DirectIntRepVF}

Because the Krein space structure is functorial under tensoring and direct summation, we have
utilized the tensor product at the level of the representation of the small group in passing
from the construction of the Dirac operator $D$ and $V_\mathcal{F}$ corresponding to the representation
\[
\int \limits_{-\infty}^{\infty}
U^{{}_{(m,0,0,0)} 2L^{{}^{1/2}}}
\, \ud m \bigoplus \int \limits_{0}^{\infty} U^{{}_{(0,m,0,0)} \big[2L^{{}^{1/2}} \big]_\textrm{Ass} } \, \ud m,
\]
Case (\ref{step1LocNaturallyAss}), to the construction of the generalized Dirac operator $D$ and $V_\mathcal{F}$ corresponding to the representation
\[
\int \limits_{-\infty}^{\infty}
U^{{}_{(m,0,0,0)} 2^2(L^{1/2})^{\otimes 2} }
\, \ud m \bigoplus \int \limits_{0}^{\infty}
U^{{}_{(0,m,0,0)} \big[ 2^2 (L^{1/2})^{\otimes 2}\big]_\textrm{Ass}} \, \ud m,
\]
Case 1 (notation convention of Remark \ref{RemarkShortNotation}).

Of course, we may continue this process of tensoring of the representation $L^{1/2}$ of the small
group $G_{(m,0,0,0)}$ \emph{in infinitum}. We give here general formulas for the generalized Dirac
operator and the fundamental symmetry operator $\mathfrak{J}$ (and implicitly for $V_\mathcal{F}$)
corresponding to
\begin{equation}\label{stepnrep}
\int \limits_{-\infty}^{\infty}
U^{{}_{(m,0,0,0)} 2^n\big(L^{{}^{1/2}}\big)^{\otimes n}}
\, \ud m \bigoplus \int \limits_{0}^{\infty}
U^{{}_{(0,m,0,0)} \big[ 2^n \big(L^{{}^{1/2}}\big)^{\otimes n} \big]_{\Ass}} \, \ud m,
\end{equation}
where we have denoted by $\big(L^{{}^{1/2}}\big)^{\otimes n}$ the n-fold tensor product
$L^{{}^{1/2}} \otimes \ldots \otimes L^{{}^{1/2}}$ of the representation $L^{{}^{1/2}}$, with the convention that
$1$-fold product of $L^{{}^{1/2}}$ is just equal to $L^{{}^{1/2}}$; and by
$\big[2^n \big(L^{{}^{1/2}} \big)^{\otimes n} \big]_{\Ass}$ we have denoted the representation
associated to $2^n \big(L^{{}^{1/2}}\big)^{\otimes n}$ in the way indicated by the two pairs of Examples:
rep. $2L^{{}^{1/2}}$ of $G_{(m,0,0,0)}$ and the associated rep. $\big[2L^{{}^{1/2}}\big]_\textrm{Ass}$
of $G_{(0,m,0,0)}$ (Examples 1 and 2) and the pair:
rep. $2^2 L^{{}^{1/2}} \otimes L^{{}^{1/2}}= 2^2 \big(L^{{}^{1/2}}\big)^{\otimes 2}$ of $G_{(m,0,0,0)}$
and the associated
representation $\big[2^2 \big(L^{{}^{1/2}}\big)^{\otimes 2}\big]_\textrm{Ass}$ of
$G_{(0,m,0,0)}$ (Examples 3 and 4).

Because the number of possible representations of the related Clifford algebra grows with $n$,
we have to choose a fixing rule for the choice of the representation giving a simple formula valid 
for all $n \in \mathbb{N}$.
We start with the recurrence rule fixing the choice of the representation and fixing at the same time 
the formula for the fundamental symmetry.

Let $\gamma \oplus \gamma^{*-1} = \big[\gamma \oplus \gamma^{*-1} \big]^{\otimes 1}$ be the representation $\big[ 2 L^{{}^{1/2}} \big]_\textrm{Ass}(\gamma)
= \big[2 L^{{}^{1/2}} \big]_{\Ass}(\gamma)
= \big[2 \big(L^{{}^{1/2}}\big)^{\otimes 1} \big]_{\Ass}(\gamma)$ associated to $2L^{{}^{1/2}}$
(Example 2) or just $1$-fold product of $\gamma \oplus \gamma^{*-1}$. The extension $V$ of
$\big[2 \big(L^{{}^{1/2}}\big)^{\otimes 1} \big]_{\Ass}$, equal $V(\alpha) = \alpha \oplus \alpha^{*-1}$
we denote here by $V^{(1)}$.
Note that the representation $\big[\big(2^2 L^{{}^{1/2}})^{\otimes 2} \big]_{\Ass}$
associated to $2^2 L^{{}^{1/2}} \otimes L^{{}^{1/2}} = \big(L^{{}^{1/2}}\big)^{\otimes 2}$ is equal to
(where $\otimes$ on the right-hand side is treated in this formula formally as if it was distributive
over $\oplus$ with ordinarily equality in the distributive law instead of the unitary equivalence)
\begin{multline*}
\big[\big(2^2 L^{{}^{1/2}})^{\otimes 2} \big]_{\Ass} (\gamma)
= (\gamma \otimes \gamma) \oplus (\gamma^{*-1} \otimes \gamma) \oplus (\gamma^{*-1} \otimes \gamma^{*-1})
\oplus (\gamma \otimes \gamma^{*-1}) \\
= \Big[ \big[2 \big(L^{{}^{1/2}}\big)^{\otimes 1} \big]_{\Ass}(\gamma) \otimes \gamma \Big] \oplus
\Big[\big[2 \big(L^{{}^{1/2}}\big)^{\otimes 1} \big]_{\Ass, \rev}(\gamma) \otimes \gamma^{*-1} \Big] \,\,\, \textrm{in this order!}
\end{multline*}
where $\big[2 \big(L^{{}^{1/2}}\big)^{\otimes 1} \big]_{\Ass, \rev}(\gamma) $ is equal to $\gamma^{*-1} \oplus \gamma$,
i.e. equal to $\big[2 \big(L^{{}^{1/2}}\big)^{\otimes 1} \big]_{\Ass}(\gamma)$
with the order of direct summands reversed.
Correspondingly we have the following formula for $V$ -- the extension of
$\big[2^2 \big(L^{{}^{1/2}}\big)^{\otimes 2} \big]_{\Ass}$ -- which we denote here by $V^{(2)}$:
\begin{multline}\label{V^(2)}
V^{(2)}(\alpha) =
(\alpha \otimes \alpha) \oplus (\alpha^{*-1} \otimes \alpha) \oplus (\alpha^{*-1} \otimes \alpha^{*-1})
\oplus (\alpha \otimes \alpha^{*-1}) \\
= \Big[\big[2 \big(L^{{}^{1/2}}\big)^{\otimes 1} \big]_{\Ass}(\alpha) \otimes \alpha \Big] \oplus
\Big[\big[2 \big(L^{{}^{1/2}}\big)^{\otimes 1} \big]_{\Ass, \rev}(\alpha) \otimes \alpha^{*-1} \Big] \,\,\, \textrm{in this order!}
\end{multline}
where $\otimes$ on the right-hand side is treated here formally as if it was distributive
over $\oplus$ with ordinarily equality in the distributive law instead of the unitary equivalence.

We have the corresponding four extensions
$V_1(\alpha) = \alpha \otimes \alpha$, $V_2(\alpha) = \alpha^{*-1} \otimes \alpha$,
$V_3(\alpha) = \alpha^{*-1} \otimes \alpha^{*-1}$, $V_4(\alpha) = \alpha \otimes \alpha^{*-1}$,
of the representation $L^{{}^{1/2}} \otimes L^{{}^{1/2}} (\gamma) = \gamma \otimes \gamma$
of $G_{(m,0,0,0)}$, which we use to form the multispinor:
\[
\widetilde{\phi}_{{}_{m,0}} (p) = \left( \begin{array}{c} V_1 (\beta(p)^{-1})\widetilde{\psi}_{{}_{m,0}}(p)\\
V_2 (\beta(p)^{-1})\widetilde{\psi}_{{}_{m,0}}(p)\\
V_3 (\beta(p)^{-1})\widetilde{\psi}_{{}_{m,0}}(p)\\
V_4 (\beta(p)^{-1})\widetilde{\psi}_{{}_{m,0}}(p) \end{array}\right)
\]
preserving the order of (\ref{V^(2)}), i.e. $V_1 (\beta(p)^{-1}) \widetilde{\psi}(p)$ forming the first component,
$V_2 (\beta(p)^{-1}) \widetilde{\psi}(p)$ the second, e.t.c.

The direct summands of (\ref{V^(2)}) have the property that the neighboring summand differ in just one factor,
this holds also for the first and the last summand -- a cyclicity property -- reflected also by the components of the multispinor. We then joint the components of the multispinor into disjoint pairs: $1$-st with the $2$-nd
and $3$-rd with $4$-th -- which is reflected by the diagram $B)$ of Subsection \ref{e3}. Correspondingly to this diagram
we obtain the generators (\ref{gammaB}) of the Clifford algebra as explained in the previous Subsections.
Let the unitary matrix $U$ be such that
\[
V^{(2)}(\alpha) = U (\alpha \oplus \alpha^{*-1})^{\otimes 2} U^{-1} = U V^{(1)}(\alpha)^{\otimes 2}U^{-1}.
\]
Then the fundamental symmetry $\mathfrak{J}$ corresponding to the representation (with respect to which
$\big[ \big(L^{{}^{1/2}}\big)^{\otimes 2} \big]_{\Ass}$ is Krein unitary)
\[
\int \limits_{-\infty}^{\infty}
U^{{}_{(m,0,0,0)} 2^2\big(L^{{}^{1/2}}\big)^{\otimes 2}}
\, \ud m \bigoplus \int \limits_{0}^{\infty}
U^{{}_{(0,0,m,0)} \big[2^2 \big(L^{{}^{1/2}}\big)^{\otimes 2} \big]_{\Ass}} \, \ud m,
\]
(after the operation of tensoring just once) is the operator of multiplication by the matrix
$U(\gamma^0 \otimes \gamma^0)U^{-1}$ or by
$U(\gamma^1 \gamma^2 \gamma^3 \otimes \gamma^1 \gamma^1 \gamma^2 \gamma^3)U^{-1}$, $\ldots$.

In passing to the representation 
\[
\int \limits_{-\infty}^{\infty} 
U^{{}_{(m,0,0,0)} 2^3 \big(L^{{}^{1/2}}\big)^{\otimes 3}}
\, \ud m \bigoplus  \int \limits_{0}^{\infty} 
U^{{}_{(0,m,0,0)} \big[2^3 \big(L^{{}^{1/2}}\big)^{\otimes 3} \big]_{\Ass}} \, \ud m,
\]  
(after the operation of tensoring performed twice) let us note that the representation
 $L^{{}^{1/2}} \otimes L^{{}^{1/2}} \otimes L^{{}^{1/2}} =
\big( L^{{}^{1/2}}\big)^{\otimes 3}$ has $2^3 = 8$ natural extensions (including the representation 
$\big( L^{{}^{1/2}}\big)^{\otimes 3}$ itself): 
$V_1(\gamma) = \gamma \otimes \gamma \otimes \gamma$, $V_2(\gamma) = \gamma^{*-1} \otimes \gamma \otimes \gamma$,
$V_3(\gamma) = \gamma^{*-1} \otimes \gamma^{*-1} \otimes \gamma$, $V_4(\gamma)
 = \gamma \otimes \gamma^{*-1} \otimes \gamma$,
$V_5(\gamma) = \gamma \otimes \gamma^{*-1} \otimes \gamma^{*-1}$, 
$V_6(\gamma) = \gamma^{*-1} \otimes \gamma^{*-1} \otimes \gamma^{*-1}$,
$V_7(\gamma) = \gamma^{*-1} \otimes \gamma \otimes \gamma^{*-1}$, 
$V_8(\gamma) = \gamma \otimes \gamma \otimes \gamma^{*-1}$. We fix their order in the way indicated by the 
subscript, so that the extension $V$ of the representation associated to $\big( L^{{}^{1/2}}\big)^{\otimes 3}$ 
-- we denote it here by $V^{(3)}$ -- is equal
\[
V^{(3)}(\alpha) = \Big[V^{(2)} \otimes \alpha \Big] \oplus \Big[V_{\rev}^{(2)} \otimes \alpha^{*-1} \Big]
\,\,\, \textrm{in this order!}.
\]
where the operation $\otimes$ in this formula is treated as if it was distributive over $\oplus$
(not up to unitary equivalence but with ordinary equality in the distributive law).

Then we joint the components of the multispinor into disjoint pairs: $1$-st with the $2$-nd,
$3$-rd with the $4$-th, and so on, as is reflected by the following diagram\footnote{Note that $\beta$
is the function $p \mapsto \beta(p)$ in the class of orbits $\mathcal{O}_{\bar{p}}$ of $\bar{p}
= (m,0,0,0)$ such that $\beta(p)^{-1} \widehat{\bar{p}} (\beta(p)^{-1})^{*} = \widehat{p}$. Only accidentally
there appears $\beta^2$ and $\beta^{-2}$ in the diagrams below giving the generalized Dirac equation and the generators of a representation of the Clifford algebra of the Minkowski pseudo-metric, because of the special choice of the function
$\beta$ for which $\beta(p)$ is only accidentally self adjoint. For the more general choice of $\beta$ there should appear a more general expression instead of $p \mapsto \beta(p)^2$, namely $p \mapsto \Aut(\beta (p))^{-1} \beta(p)$, with $\Aut$ realizing the conjugation with respect to the class of orbits of $\bar{p}
= (m,0,0,0)$, which -- as we already know -- is independent of the choice of the function $\beta$
corresponding to the mentioned class of orbits.}
\begin{center}
\begin{tikzpicture}


\draw [<-,very thin] (0.966,0.26) arc (15:165:1) ;
\draw [<-,very thin] (-0.966,-0.26) arc (195:345:1) ;

\draw [<-,very thin] (3.966,0.26) arc (15:165:1) ;
\draw [<-,very thin] (2.034,-0.26) arc (195:345:1) ;

\draw [<-,very thin] (6.966,0.26) arc (15:165:1) ;
\draw [<-,very thin] (5.034,-0.26) arc (195:345:1) ;

\draw [<-,very thin] (9.966,0.26) arc (15:165:1) ;
\draw [<-,very thin] (8.034,-0.26) arc (195:345:1) ;

\node  at (-1,0) {$\widetilde{\varphi}_{1}$};
\node  at (1,0) {$\widetilde{\varphi}_{2}$};

\node  at (0,1.25) {$\beta^2 \otimes \bold{1}_2 \otimes \bold{1}_2$};
\node  at (0,-1.25) {$\beta^{-2} \otimes \bold{1}_2 \otimes \bold{1}_2$};



\node  at (2,0) {$\widetilde{\varphi}_{3}$};
\node  at (4,0) {$\widetilde{\varphi}_{4}$};

\node  at (3,1.25) {$\beta^{-2} \otimes \bold{1}_2 \otimes \bold{1}_2$};
\node  at (3,-1.25) {$\beta^{2} \otimes \bold{1}_2 \otimes \bold{1}_2$};

\node  at (6,1.25) {$\beta^{2} \otimes \bold{1}_2 \otimes \bold{1}_2$};
\node  at (6,-1.25) {$\beta^{-2} \otimes \bold{1}_2 \otimes \bold{1}_2$};

\node  at (9,1.25) {$\beta^{-2} \otimes \bold{1}_2 \otimes \bold{1}_2$};
\node  at (9,-1.25) {$\beta^{2} \otimes \bold{1}_2 \otimes \bold{1}_2$};

\node  at (5,0) {$\widetilde{\varphi}_{5}$};
\node  at (7,0) {$\widetilde{\varphi}_{6}$};

\node  at (8,0) {$\widetilde{\varphi}_{7}$};
\node  at (10,0) {$\widetilde{\varphi}_{8}$};



\end{tikzpicture} 
\end{center}

\vspace*{0.5cm}

Correspondingly to this we obtain the generalized Dirac equation and the
corresponding generators of the Clifford algebra:
\begingroup\makeatletter\def\f@size{5}\check@mathfonts
\def\maketag@@@#1{\hbox{\m@th\large\normalfont#1}}%
\[
\widetilde{\gamma}^0 =
\left( \begin{array}{cccccccc}  
\bold{0}_8 & \bold{1}_8 & \bold{0}_8 & \bold{0}_8 &\bold{0}_8 &\bold{0}_8 & \bold{0}_8 & \bold{0}_8  \\
\bold{1}_8 & \bold{0}_8 & \bold{0}_8 & \bold{0}_8 &\bold{0}_8 &\bold{0}_8 & \bold{0}_8 & \bold{0}_8  \\
\bold{0}_8 & \bold{0}_8 & \bold{0}_8 & \bold{1}_8 &\bold{0}_8 &\bold{0}_8 & \bold{0}_8 & \bold{0}_8  \\
\bold{0}_8 & \bold{0}_8 & \bold{1}_8 & \bold{0}_8 &\bold{0}_8 &\bold{0}_8 & \bold{0}_8 & \bold{0}_8  \\
\bold{0}_8 & \bold{0}_8 & \bold{0}_8 & \bold{0}_8 &\bold{0}_8 &\bold{1}_8 & \bold{0}_8 & \bold{0}_8  \\
\bold{0}_8 & \bold{0}_8 & \bold{0}_8 & \bold{0}_8 &\bold{1}_8 &\bold{0}_8 & \bold{0}_8 & \bold{0}_8  \\
\bold{0}_8 & \bold{0}_8 & \bold{0}_8 & \bold{0}_8 &\bold{0}_8 &\bold{0}_8 & \bold{0}_8 & \bold{1}_8  \\
\bold{0}_8 & \bold{0}_8 & \bold{0}_8 & \bold{0}_8 &\bold{0}_8 &\bold{0}_8 & \bold{1}_8 & \bold{0}_8  
                     \end{array}\right)   
\]\endgroup
\begingroup\makeatletter\def\f@size{5}\check@mathfonts
\def\maketag@@@#1{\hbox{\m@th\large\normalfont#1}}%
\[
\widetilde{\gamma}^k =
\left( \begin{array}{cccccccc}  
\bold{0}_8 & -\sigma_k \otimes \bold{1}_2 \otimes \bold{1}_2 & \bold{0}_8 & \bold{0}_8 &\bold{0}_8 &\bold{0}_8 & \bold{0}_8 & \bold{0}_8  \\
\sigma_k \otimes \bold{1}_2 \otimes \bold{1}_2 & \bold{0}_8 & \bold{0}_8 & \bold{0}_8 &\bold{0}_8 &\bold{0}_8 & \bold{0}_8 & \bold{0}_8  \\
\bold{0}_8 & \bold{0}_8 & \bold{0}_8 & \sigma_k \otimes \bold{1}_2 \otimes \bold{1}_2 &\bold{0}_8 &\bold{0}_8 & \bold{0}_8 & \bold{0}_8  \\
\bold{0}_8 & \bold{0}_8 & -\sigma_k \otimes \bold{1}_2 \otimes \bold{1}_2 & \bold{0}_8 &\bold{0}_8 &\bold{0}_8 & \bold{0}_8 & \bold{0}_8  \\
\bold{0}_8 & \bold{0}_8 & \bold{0}_8 & \bold{0}_8 &\bold{0}_8 & -\sigma_k \otimes \bold{1}_2 \otimes \bold{1}_2 & \bold{0}_8 & \bold{0}_8  \\
\bold{0}_8 & \bold{0}_8 & \bold{0}_8 & \bold{0}_8 & \sigma_k \otimes \bold{1}_2 \otimes \bold{1}_2 &\bold{0}_8 & \bold{0}_8 & \bold{0}_8  \\
\bold{0}_8 & \bold{0}_8 & \bold{0}_8 & \bold{0}_8 &\bold{0}_8 &\bold{0}_8 & \bold{0}_8 & \sigma_k \otimes \bold{1}_2 \otimes \bold{1}_2  \\
\bold{0}_8 & \bold{0}_8 & \bold{0}_8 & \bold{0}_8 &\bold{0}_8 &\bold{0}_8 & -\sigma_k \otimes \bold{1}_2 \otimes \bold{1}_2 & \bold{0}_8  
                    \end{array}\right)
\]\endgroup

If the unitary matrix $U$ is such that
\[
V^{(3)}(\alpha) = U (\alpha \oplus \alpha^{*-1})^{\otimes 3}U^{-1} = U V^{(1)}(\alpha)^{\otimes 3}U^{-1}
\]
then the fundamental symmetry corresponding to  the representation 
\[
\int \limits_{-\infty}^{\infty} 
U^{{}_{(m,0,0,0)} 2^3 \big(L^{{}^{1/2}}\big)^{\otimes 3}}
\, \ud m \bigoplus  \int \limits_{0}^{\infty} 
U^{{}_{(0,0,m,0)} \big[2^3 \big(L^{{}^{1/2}}\big)^{\otimes 3} \big]_{\Ass}} \, \ud m,
\]  
is the operator of multiplication by the matrix $U(\gamma^0 \otimes \gamma^0 \otimes \gamma^0)U^{-1}$ or by the matrix
$U(\gamma^0 \otimes \gamma^0 \otimes \gamma^1 \gamma^2 \gamma^3 )U^{-1}$, $\ldots$ or by the matrix 
$U(\gamma^1 \gamma^2 \gamma^3 \otimes \gamma^1 \gamma^2 \gamma^3 \otimes \gamma^1 \gamma^2 \gamma^3 )U^{-1}$.

And generally we fix the order of the direct summands in the possible extensions such that the extension
$V$ of the representation $\big[ \big(L^{{}^{1/2}}\big)^{\otimes n} \big]_{\Ass}$
associated to $\big(L^{{}^{1/2}}\big)^{\otimes n}$, which we denote here by
$V^{(n)}$  -- is defined by induction in the following manner
\[
\begin{split}
1) \,\, V^{(1)}(\alpha) = \alpha \oplus \alpha^{*-1}, \,\,\,\,\,\,\,\,\,\,\,\,\,\,\,\,\,\,\,\,\,\,\,\,\,\,\, \\
2) \,\, \textrm{if} \,\,\, V^{(n-1)} \,\,\textrm{is defined with a fixed order of direct summands then} \\
V^{(n)}(\alpha) = \Big[V^{(n-1)}(\alpha) \otimes \alpha \Big] \oplus \Big[V_{\rev}^{(n-1)}(\alpha) 
\otimes \alpha^{*-1} \Big],
\,\,\, \textrm{(in this order!)}
\end{split}
\]
where the operation $\otimes$ in the formula 2) is treated formally as if it was distributive over
$\oplus$ (not up to unitary equivalences in the distributive law but with ordinary equalities) and
where $V_{\rev}^{(n-1)}(\alpha)$ denotes $V^{(n-1)}(\alpha)$ with the inverse order of direct summands,
e.g. $V_{\rev}^{(1)}(\alpha) =  \alpha^{*-1} \oplus \alpha$. 
We then joint into disjoint pairs the components of the multispinor: the first with the second,
the third with the fourth, and so on, which may be pictured by the following diagram
\begin{center}
\begin{tikzpicture}


\draw [<-,very thin] (0.966,0.26) arc (15:165:1) ;
\draw [<-,very thin] (-0.966,-0.26) arc (195:345:1) ;

\draw [<-,very thin] (3.966,0.26) arc (15:165:1) ;
\draw [<-,very thin] (2.034,-0.26) arc (195:345:1) ;


\draw [<-,very thin] (9.966,0.26) arc (15:165:1) ;
\draw [<-,very thin] (8.034,-0.26) arc (195:345:1) ;

\node  at (-1,0) {$\widetilde{\varphi}_{{}_{1}}$};
\node  at (1,0) {$\widetilde{\varphi}_{{}_{2}}$};

\node  at (0,1.25) {$\beta^2 \otimes \big(\bold{1}_2\big)^{ \otimes (n-1)}$};
\node  at (0,-1.25) {$\beta^{-2} \otimes\big(\bold{1}_2\big)^{ \otimes (n-1)}$};



\node  at (2,0) {$\widetilde{\varphi}_{{}_{3}}$};
\node  at (4,0) {$\widetilde{\varphi}_{{}_{4}}$};

\node  at (3,1.25) {$\beta^{-2} \otimes \big(\bold{1}_2\big)^{ \otimes (n-1)}$};
\node  at (3,-1.25) {$\beta^{2} \otimes \big(\bold{1}_2\big)^{ \otimes (n-1)}$};


\node  at (6,0) {$\bold{\ldots}$};

\node  at (9,1.25) {$\beta^{-2} \otimes \big(\bold{1}_2\big)^{ \otimes (n-1)}$};

\node  at (9,-1.25) {$\beta^{2} \otimes \big(\bold{1}_2\big)^{ \otimes (n-1)}$};


\node  at (8,0) {$\widetilde{\varphi}_{{}_{2^n - 1}}$};
\node  at (10,0) {$\widetilde{\varphi}_{{}_{2^n}}$};



\end{tikzpicture} 
\end{center}
and obtain the corresponding generalized Dirac equation and the
following generators $\widetilde{\gamma}^\mu \in M_{{}_{2^{2n}}}(\mathbb{C})$ 
of a representation of the Clifford algebra associated to the Minkowski pseudo-metric:
\begingroup\makeatletter\def\f@size{5}\check@mathfonts
\def\maketag@@@#1{\hbox{\m@th\large\normalfont#1}}%
\begin{equation}\label{gamma-n-0}
\widetilde{\gamma}^0 =
\left( \begin{array}{ccccccc}  
0 &\bold{1}_{{}_{2^n}}  &  &  &  &  &  0  \\
\bold{1}_{{}_{2^n}} & 0  &  &  &  &  &   \\
 &  & 0 & \bold{1}_{{}_{2^n}} &  &  &   \\
 &  & \bold{1}_{{}_{2^n}} & 0 &  &  &   \\
 &  &  &  & \ddots  &  &   \\
 &  &  &  &  & 0 & \bold{1}_{{}_{2^n}}  \\
 0 &  &  &  &  & \bold{1}_{{}_{2^n}} & 0  
                     \end{array}\right)   
\end{equation}\endgroup
\begingroup\makeatletter\def\f@size{5}\check@mathfonts
\def\maketag@@@#1{\hbox{\m@th\large\normalfont#1}}%
\begin{equation}\label{gamma-n-k}
\widetilde{\gamma}^k =
\left( \begin{array}{ccccccc}  
0 & -\sigma_k \otimes \big(\bold{1}_{{}_{2}}\big)^{{}^{\otimes (n-1)}}  &  &  &  &  & 0  \\
\sigma_k \otimes \big(\bold{1}_{{}_{2}}\big)^{{}^{\otimes (n-1)}} & 0  &  &  &  &  &  \\
 &  & 0  & \sigma_k \otimes \big(\bold{1}_{{}_{2}}\big)^{{}^{\otimes (n-1)}} &  &  &  \\

 &  & -\sigma_k \otimes \big(\bold{1}_{{}_{2}}\big)^{{}^{\otimes (n-1)}} &  &  &  &   \\
 &  &  &  & \ddots  &  &   \\
 &  &  &  &  & 0 & \sigma_k \otimes \big(\bold{1}_{{}_{2}}\big)^{{}^{\otimes (n-1)}}  \\
 0  &  &  &  &  & -\sigma_k \otimes \big(\bold{1}_{{}_{2}}\big)^{{}^{\otimes (n-1)}} & 0  
                     \end{array}\right)   
\end{equation}\endgroup
where the diagonal blocks in $\widetilde{\gamma}^k$
\[
\left( \begin{array}{cc}  
0 & -\sigma_k \otimes \big(\bold{1}_{{}_{2}}\big)^{{}^{\otimes (n-1)}} \\
\sigma_k \otimes \big(\bold{1}_{{}_{2}}\big)^{{}^{\otimes (n-1)}} & 0  
                     \end{array}\right),
\left( \begin{array}{cc}  
0 & \sigma_k \otimes \big(\bold{1}_{{}_{2}}\big)^{{}^{\otimes (n-1)}} \\
-\sigma_k \otimes \big(\bold{1}_{{}_{2}}\big)^{{}^{\otimes (n-1)}} & 0  
                     \end{array}\right), \ldots
\]
change the sign in passing from one to the next, and all the matrix entries which are not explicitly written 
are equal zero.

If the unitary matrix $U$ is such that
\[
V^{(n)}(\alpha) = U (\alpha \oplus \alpha^{*-1})^{\otimes n}U^{-1} = U V^{(1)}(\alpha)^{\otimes n}U^{-1}
\]
then the fundamental symmetry corresponding to  the representation 
\[
\int \limits_{-\infty}^{\infty} 
U^{{}_{(m,0,0,0)} 2^n \big(L^{{}^{1/2}}\big)^{\otimes n}}
\, \ud m \bigoplus  \int \limits_{0}^{\infty} 
U^{{}_{(0,0,m,0)} \big[2^n \big(L^{{}^{1/2}}\big)^{\otimes n} \big]_{\Ass}} \, \ud m,
\]  
is the operator of multiplication by the matrix $U(\gamma^0 \otimes \gamma^0 \otimes \ldots \otimes \gamma^0)U^{-1}$ or by the matrix
(all tensor products here are $n$-fold and embrace all possibilities with the factors equal $\gamma^0$ or 
$\gamma^1 \gamma^2 \gamma^3$)
$U(\gamma^0 \otimes \ldots \otimes \gamma^0 \otimes \gamma^1 \gamma^2 \gamma^3 )U^{-1}$, $\ldots$ or by the matrix 
$U(\gamma^1 \gamma^2 \gamma^3 \otimes \gamma^1 \gamma^2 \gamma^3 \otimes \ldots \otimes \gamma^1 \gamma^2 \gamma^3 )U^{-1}$.

It remains to give the general formula for the unitary matrix $U$ for each $n$, which we denote here by $U^{(n)}$
indicating in this way its dependence on $n$. To this end we introduce some auxiliary definitions. Let 
$(q \leftrightarrows l)$ denote the permutation of the set of $n$ numbers $1, \ldots , n$,
which interchanges $q$-th number with the $l$-th and \emph{vice versa}, and which acts as identity 
on the remaining numbers i.e. an inversion. It will be convenient
to consider the elements of the ring $M_{{}_{2^{2n}}}(\mathbb{C})$ of $2^{2n} \times 2^{2n}$ matrices over the field 
$\mathbb{C}$ as elements of the ring $M_{{}_{2^p}}\big(M_{{}_{2^k}}(\mathbb{C})\big)$ of $2^p \times 2^p$ matrices over the ring of matrices $M_{{}_{2^k}}(\mathbb{C})$ over $\mathbb{C}$, with $p+k = 2n$. For any fixed pair of natural numbers
$p$ and $k$ and any inversion $(q \leftrightarrows l)$ with $q,l \leq 2^p$, we define the $2^p \times 2^p$ matrix 
$U^{{}^{M_{{}_{2^p}}\big(M_{{}_{2^k}}(\mathbb{C})\big)}}\big( q \leftrightarrows l\big)$ over the ring 
$M_{{}_{2^k}}(\mathbb{C})$  
which in the 1-st arrow have the unit matrix $\bold{1}_{{}_{2^k}} \in M_{{}_{2^k}}(\mathbb{C})$
standing in the first column and is equal to the zero matrix  $\bold{0}_{{}_{2^k}}$ for the remaining elements of the first row, if $q \neq 1, l \neq 1$; in the 2-nd column has the unit matrix $\bold{1}_{{}_{2^k}}$ at the second column 
and is equal zero $\bold{0}_{{}_{2^k}}$ for the remaining elements of the second row, if $q \neq 1, l \neq 1$; and so on;
and the only nonzero element of the $q$-th row stands at the $l$-th column ad is equal to unity $\bold{1}_{{}_{2^k}}$,
and the only nonzero element of the $l$-th row stands at the $q$-th column and is equal to unity $\bold{1}_{{}_{2^k}}$. 
Similarly, we can define the matrix $U^{{}^{M_{{}_{2^p}}\big(M_{{}_{2^k}}(\mathbb{C})\big)}}
\big( (q_1 \leftrightarrows l_1) (q_2 \leftrightarrows l_2) \ldots (q_r \leftrightarrows l_r) \big)$ corresponding
to $r$ commutative inversions, defined by disjoint $r$ pairs $(q_1, l_1), (q_2 , l_2) \ldots (q_r , l_r)$ 
of numbers. For example  
\begingroup\makeatletter\def\f@size{5}\check@mathfonts
\def\maketag@@@#1{\hbox{\m@th\large\normalfont#1}}%
\[
U^{{}^{M_{{}_{2^3}}\big(M_{{}_{2^1}}(\mathbb{C})\big)}}
\big( (2 \leftrightarrows 3) (6 \leftrightarrows 7) \big) =
\left( \begin{array}{cccccccc}  
\bold{1}_2 & \bold{0}_2 & \bold{0}_2 & \bold{0}_2 &\bold{0}_2 &\bold{0}_2 & \bold{0}_2 & \bold{0}_2  \\
\bold{0}_2 & \bold{0}_2 & \bold{1}_2 & \bold{0}_2 &\bold{0}_2 &\bold{0}_2 & \bold{0}_2 & \bold{0}_2  \\
\bold{0}_2 & \bold{1}_2 & \bold{0}_2 & \bold{0}_2 &\bold{0}_2 &\bold{0}_2 & \bold{0}_2 & \bold{0}_2  \\
\bold{0}_2 & \bold{0}_2 & \bold{0}_2 & \bold{1}_2 &\bold{0}_2 &\bold{0}_2 & \bold{0}_2 & \bold{0}_2  \\
\bold{0}_2 & \bold{0}_2 & \bold{0}_2 & \bold{0}_2 &\bold{1}_2 &\bold{0}_2 & \bold{0}_2 & \bold{0}_2  \\
\bold{0}_2 & \bold{0}_2 & \bold{0}_2 & \bold{0}_2 &\bold{0}_2 &\bold{0}_2 & \bold{1}_2 & \bold{0}_2  \\
\bold{0}_2 & \bold{0}_2 & \bold{0}_2 & \bold{0}_2 &\bold{0}_2 &\bold{1}_2 & \bold{0}_2 & \bold{0}_2  \\

\bold{0}_2 & \bold{0}_2 & \bold{0}_2 & \bold{0}_2 &\bold{0}_2 &\bold{0}_2 & \bold{0}_2 & \bold{1}_2  
                     \end{array}\right)   
\]\endgroup     
Of course the matrix $U^{{}^{M_{{}_{2^p}}\big(M_{{}_{2^k}}(\mathbb{C})\big)}}
\big( (q_1 \leftrightarrows l_1) (q_2 \leftrightarrows l_2) \ldots (q_r \leftrightarrows l_r) \big)$
can likewise be regarded as unitary $2^{p+k} \times 2^{p+k}$ matrix over $\mathbb{C}$ and below multiplication
of such matrices with different pairs of numbers $p,k$ and $p',k'$ with $p+k = p'+k'$ is understood as 
multiplication of the matrices regarded as $2^{p+k} \times 2^{p+k} = 2^{p'+k'} \times 2^{p'+k'}$ matrices 
over the field $\mathbb{C}$ of complex numbers.

For $n=2$ we have already given
the formula for $U^{(2)}$, namely (\ref{clebsch.step2.U}), and it is equal to the composition
of unitary operators defined by inversions of disjoint subsets of axes, i.e.
as composition of specific involutive unitary operators (matrices):
\[
U^{(2)} = U^{{}^{M_{{}_{2^2}}\big(M_{{}_{2^2}}(\mathbb{C})\big)}}
\big(3 \leftrightarrows 4 \big) U^{{}^{M_{{}_{2^2}}\big(M_{{}_{2^2}}(\mathbb{C})\big)}}
\big(2 \leftrightarrows 3 \big)
U^{{}^{M_{{}_{2^3}}\big(M_{{}_{2^1}}(\mathbb{C})\big)}}
\big( (2 \leftrightarrows 3) (6 \leftrightarrows 7) \big).
\]

Because every factor corresponding to the respective inversion (or to composition of commutative inversions)
is unitary and self-adjoint, then the inverse of every such factor (regarded as matrix over complex numbers)
is equal to its transposition, i.e. to the factor itself, i.e. it is involutive. Therefore, the inverse of
$U^{(2)}$ is just equal to the same composition of factors in the reverse order:
\[
{U^{(2)}}^{-1} = U^{{}^{M_{{}_{2^3}}\big(M_{{}_{2^1}}(\mathbb{C})\big)}}
\big( (2 \leftrightarrows 3) (6 \leftrightarrows 7) \big) U^{{}^{M_{{}_{2^2}}\big(M_{{}_{2^2}}(\mathbb{C})\big)}}
\big(2 \leftrightarrows 3 \big) U^{{}^{M_{{}_{2^2}}\big(M_{{}_{2^2}}(\mathbb{C})\big)}}
\big(3 \leftrightarrows 4 \big).
\]
Now in order to simplify notation let us note the matrix
\[
U^{{}^{M_{{}_{2^p}}\big(M_{{}_{2^k}}(\mathbb{C})\big)}}
\big( (2 \leftrightarrows 3) (2+4 \leftrightarrows 2+4+1) (2+4+4 \leftrightarrows 2+4+4+1)
\ldots ((2^p -2) \leftrightarrows (2^p -1)) \big)
\]
just by $(p,k)$. In order to give a general formula for $U^{(n)}$ we write the matrix
$U^{(n)}$ as a composition $U^{(n)} = U'^{(n)}U''^{(n)}$ of two matrices $U'^{(n)}$
and $U''^{(n)}$, and give separately the formulas for $U'^{(n)}$
and $U''^{(n)}$.
We start with $U''^{(n)}$. In order to define $U''^{(n)}$ we introduce the sequence of representations
$V_{}{}_{(n)}$ such that $V_{}{}_{(n)}(\alpha)$ differs from $V^{(n)}(\alpha)$ only by the order of direct
summands. Namely,
\[
\begin{split}
1) \,\, V_{}{}_{(1)}(\alpha) = \alpha \oplus \alpha^{*-1}, \,\,\,\,\,\,\,\,\,\,\,\,\,\,\,\,\,\,\,\,\,\,\,\,\,\,\, \\
2) \,\, \textrm{if} \,\,\, V_{}{}_{(n-1)} \,\,\textrm{is defined with a fixed order of direct summands then} \\
V_{}{}_{(n)}(\alpha) = \alpha \otimes V_{}{}_{(n-1)}(\alpha) \oplus \alpha^{*-1} \otimes V_{}{}_{(n-1)}(\alpha),
\,\,\, \textrm{(in this order!)}
\end{split}
\]
where the operation $\otimes$ in the formula 2) is treated formally as if it was distributive over
$\oplus$ (with ordinary equality in the distributive law and not merely unitary equivalence).
Then for $U''^{(n)}$ fulfilling
\[
U''^{(n)} \big( \alpha \oplus \alpha^{*-1} \big)^{\otimes n} {U''^{(n)}}^{-1} = V_{}{}_{(n)}(\alpha),
\]
we have the following formula
\begin{multline*}
U''^{(n)} = \overbrace{(n+1,n-1)(n+2,n-2)(n+3,n-3) \ldots (2n-1,1) }^\text{(n-1) factors (i.k)} \\
\overbrace{(n,n)(n+1,n-1)(n+2,n-2) \ldots (2n-3,3) }^\text{(n-2) factors (i,k)} \\
\overbrace{(n-1,n+1)(n,n)(n+1,n-1) \ldots (2n-5,5) }^\text{(n-3) factors (i,k)} \\
\ldots \\
\overbrace{(n-k+2,n+k-2)(n-k+3,n+k-3)(n-k+4,n+k-4) \ldots (2n-2(k+1)+3,2(k+1)-3) }^\text{(n-k) factors (i,k)} \\
\ldots \\
\overbrace{(3,2n-3) }^\text{1 factor}.
\end{multline*}
For example
\begin{multline*}
U''^{(2)} = (3,1), \\
U''^{(3)} = (4,2)(5,1)(3,3), \\
U''^{(4)} = (5,3)(6,2)(7,1)(4,4)(5,3)(3,5),\\
U''^{(5)} = (6,4)(7,3)(8,2)(9,1)(5,5)(6,4)(7,3)(4,6)(5,5)(3,7).
\end{multline*}

Now we define the matrix $U'^{(n)}$ for each $n \in \mathbb{N}$. We need an auxiliary definition. Consider
the permutation of the numbers $1,2,3, \ldots 2^n$ which transforms them into the sequence of numbers with the order 
reversed $2^n, 2^n -1, 2^n -2, \ldots , 1$. Let the permutation be equal to the following composition of inversions
$\inv_1 \circ \inv_2 \circ \ldots \inv_{{}_{2^{n-1}(2^n -1)}}$. We then define the following matrix
\[
\rev^{(n)} = U^{{}^{M_{{}_{2^n}}\big(M_{{}_{2^n}}(\mathbb{C})\big)}}
\big( \inv_1 \big) U^{{}^{M_{{}_{2^n}}\big(M_{{}_{2^n}}(\mathbb{C})\big)}}
\big( \inv_2 \big) \ldots U^{{}^{M_{{}_{2^n}}\big(M_{{}_{2^n}}(\mathbb{C})\big)}}
\big( \inv_{{}_{2^{n-1}(2^n -1)}} \big).
\]
Then 
\[
U'^{(2)} =  U^{{}^{M_{{}_{2^2}}\big(M_{{}_{2^2}}(\mathbb{C})\big)}}
\big( 3 \leftrightarrows 4 \big) \,  U^{{}^{M_{{}_{2^2}}\big(M_{{}_{2^2}}(\mathbb{C})\big)}}
\big( 2 \leftrightarrows 3 \big),
\]
and 
\[
U'^{(n)} = (U'^{(n-1)} \oplus \rev^{(n-1)} U'^{(n-1)}) (2,2n-2)(3,2n-3) \ldots (n,n).
\]

Now in the explicit construction of $V_\mathcal{F}$ on the space of the representation (\ref{stepnrep})
we proceed exactly as in Subsections (\ref{1/2VF}) and (\ref{0-1VF}), with the difference that
instead of two $V^{\oplus}, V^{\ominus}$ (Subsect. \ref{e1} and \ref{1/2VF}) or four $V^\textrm{\ding{192}},
\ldots V^\textrm{\ding{195}}$ (Subsect. \ref{e3} and \ref{0-1VF}), we will have $2^n$
possible embeddings of the representation space of the representation
$U^{{}_{(m,0,0,0)} \big(L^{{}^{1/2}}\big)^{\otimes n}}$ into the corresponding Hilbert space of
multispinors. The argumentation being completely analogous may be omitted without, as we hope, any lost
of information necessary in understanding all relevant things.
Recall however that (\ref{stepnrep}) is a shortened notation (of Remark \ref{RemarkShortNotation})
for the direct integral representation in which the direct integral summands
\[
U^{{}_{(m,0,0,0)} 2^n \big(L^{{}^{1/2}} \big)^{\otimes n}}, \,\,\,
U^{{}_{(0,m,0,0)} \big[ 2^n \big(L^{{}^{1/2}} \big)^{\otimes n} \big]_\textrm{Ass}}
\]
are understood to be, respectively, equal
\[
\big(W_{{}_{V,\bar{p}}} U^{{}_{\bar{p}} L} W_{{}_{V,\bar{p}}}^{-1}\big)_{{}_{0}},
\,\,\,
\big(W_{{}_{V,\bar{p}'}} U^{{}_{\bar{p}'} [L]_\textrm{Ass}} W_{{}_{V,\bar{p}'}}^{-1}\big)_{{}_{0}},
\]
for
\[
\bar{p} = (m,0,0,0), \,\,\, L= 2^n \big(L^{{}^{1/2}} \big)^{\otimes n},
\]
or respectively
\[
\bar{p}' = (0,m,0,0), \,\,\, L = \big[2^n \big(L^{{}^{1/2}} \big)^{\otimes n}\big]_\textrm{Ass},
\]
with common representation
\[
V = V^{(n)}
\]
of $SL(2, \mathbb{C})$.

Instead of tensoring we may likewise apply the operation of direct sum at the level of representations of small groups,
as the Krein structure is functorial under this operation,
in order to obtain a new class of direct integrals of representations for which we can explicitly construct
the transform $V_\mathcal{F}$, the fundamental symmetry $\mathfrak{J}$, and the operators
$D, D_\mathfrak{J}$ and the corresponding spectral triple
$\big(\mathcal{A}, \mathcal{H}, D_\mathfrak{J}\big)$. The construction is even much simpler in comparison to the tensor
operation and is just reduced to the
replacement of the relevant operators by their direct sums. We obtain in this way spectral triples fulfilling the strong regularity condition, because in each case we construct the Clifford module corresponding to the final spectral triple,
and the Hilbert space elements as sections of the module with operators $D, D_\mathfrak{J}$ equal to the generalized Dirac
operators associated to this module,
so that the strong regularity is preserved by the same reason as in Subsection \ref{1/2VF} and \ref{0-1VF}.
This is not entirely trivial as for abstract spectral triples the strong regularity is not in general
preserved under the direct summation process, compare \cite{Connes_spectral}, \S 12.5.

Moreover, the Krein structure is functorial not only with respect to tensoring and direct sum operations
but likewise with respect to symmetrization (antisymmetrization) of the tensor product. Because any 
irreducible unitary representation of the group $G_{(m,0,0,0)} = SU(2,\mathbb{C})$ may be obtained 
as the symmetrized tensor product
$\big(L^{{}^{1/2}}\big)^{\otimes n}$ (compare e.g. \cite{Geland-Minlos-Shapiro}) we obtain 
in this way the construction of $V_\mathcal{F}$
and the corr responding space-time spectral triple for a wide class of representations. In this process
we can go beyond finite sums of tensor products, but we may as well apply infinite direct sums of tensor products,
obtaining in particular spectral triples with the algebra $\mathcal{A}$ acting with uniform infinite
multiplicity. In particular for any unitary representation, equal to the direct integral 
\begin{equation}\label{admissible-U^L}
\int \limits_{-\infty}^{\infty} 
U^{{}_{(m,0,0,0)} L} 
\, \ud m
\end{equation}
of unitary representations $U^{{}_{(m,0,0,0)} L}$ concentrated on the orbits
$\mathscr{O}_{(m,0,0,0)}$ induced by a fixed (not necessary finite) unitary representation $L$ of the
stationary group $G_{(m,0,0,0)} = SU(2, \mathbb{C})$, equal to the direct sum $2 L^{1/2} \oplus 2^2 L^{2/2}
2^{3} L^{3/2} \oplus \ldots$ there exists the associated representation
\[
\int \limits_{0}^{\infty} 
U^{{}_{(0,0,m,0)} \big[ L \big]_{\Ass}} \, \ud m,
\] 
such that one can construct the transform $V_\mathcal{F}$ and the associated spectral tuple
$()$ on the space of the representation
\[
\int \limits_{-\infty}^{\infty} 
U^{{}_{(m,0,0,0)} L} 
\, \ud m
\bigoplus \int \limits_{0}^{\infty} 
U^{{}_{(0,0,m,0)} \big[ L \big]_{\Ass}} \, \ud m.
\]
In particular the construction of the associated representation and $V_\mathcal{F}$ together with the spectral triple 
is likewise possible for the representation
(\ref{admissible-U^L}) in which $L$ decomposes into irreducible components $L^{n/2}$
each with infinite multiplicity.

For example (we discard the symmetrization process for simplicity) the operators $D, D_{{}_{\mathfrak{J}}}$
corresponding to the representation under the transform $V_{\mathcal{F}}$ on the space of the representation
\begin{multline*}
\bigoplus \limits_{n \in \mathbb{N}} \Big\{ \int \limits_{-\infty}^{\infty}
U^{{}_{(m,0,0,0)} 2^n \big(L^{{}^{1/2}}\big)^{\otimes n}}
\, \ud m \bigoplus \int \limits_{0}^{\infty}
U^{{}_{(0,0,m,0)} \big[2^n \big(L^{{}^{1/2}}\big)^{\otimes n} \big]_{\Ass}} \, \ud m \Big\} \\
= \int \limits_{-\infty}^{\infty}
U^{\oplus_{n \in \mathbb{N}} \,\, 2^n \,\, {}_{(m,0,0,0)} \big(L^{{}^{1/2}}\big)^{\otimes n}}
\, \ud m \bigoplus \int \limits_{0}^{\infty}
U^{\oplus_{n \in \mathbb{N}} \,\, {}_{(0,0,m,0)} \big[2^n \big(L^{{}^{1/2}}\big)^{\otimes n} \big]_{\Ass}} \, \ud m
\end{multline*}
are given by the following generators $\widetilde{\gamma}^\mu$ of an (infinite dimensional) representation
of the Clifford algebra associated to the Minkowski metric:
\[
\widetilde{\gamma}^\mu = \bigoplus_{n\in \mathbb{N}} \, {\widetilde{\gamma}_{{}_{(n)}}}^\mu,
\]
where ${\widetilde{\gamma}_{{}_{(n)}}}^\mu$ are the matrices depending on $n$ and defined respectively
by the formulas (\ref{gamma-n-0}) and (\ref{gamma-n-k}). The fundamental symmetry $\mathfrak{J}$ is given by the multiplication by one of the following infinite set of matrices (of infinite order)
\[
U \Big( \gamma^0 \oplus (\gamma^0 \otimes \gamma^0) \oplus (\gamma^0 \otimes \gamma^0 \otimes \gamma^0) \oplus
\ldots \Big) U^{-1},
\]
where each factor $\gamma^0$ in every direct summand $\gamma^0 \otimes \ldots \otimes \gamma^0$ may be replaced with
$\gamma^1 \gamma^2 \gamma^3$, and where
\[
U = \bigoplus_{n\in \mathbb{N}} \, U^{(n)},
\]
with $U^{(n)}$ computed above for each $n \geq 2$ and with $U^{(1)} = \bold{1}_{4}$.

(Even more, not only the operation of direct sum we may apply \emph{in infinitum} but we may perform the infinite tensor product operation, compare \cite{neumann-inf-tensor}.) 

Although one thing should be noted: the algebra $\mathcal{A}$ of Schwartz functions acts with infinite uniform
multiplicity, so that the corresponding ''spectral triple'' $(\mathcal{A}, \mathcal{H}, D_{{}_{\mathfrak{J}}})$
we have just constructed by infinite direct sum operation (at the level of small group representation), is not the ordinary spectral triple. The finiteness axiom (5) of \cite{Connes_spectral}, \S 2 cannot of course be fulfilled in that case, so that the reconstruction theorem,
which tells that the five axioms 1) -- 5) \cite{Connes_spectral} are sufficient for the commutative algebra $\mathcal{A}$ to be the algebra of smooth (Schwartz in non-compact case) functions on a smooth manifold $\Sp \mathcal{A}$ (in our case
$\mathbb{R}^4$), cannot be immediately applied to to our ''spectral triple'' (by application of the method presented in \cite{Connes_spectral}): a substitute for the finiteness axiom is needed in order to
preserve the reconstruction theorem of \cite{Connes_spectral} (the uniform multiplicity of $\mathcal{A}$
seems to be the correct substitute for the regularity axiom).
However, let us remark that the
reconstruction theorem may be proved even if $\mathcal{A}$ acts with infinite uniform multiplicity
whenever the Hilbert space $\mathcal{H}$ is a direct sum of sub-spaces invariant for
$\mathcal{A}, D_{{}_{\mathfrak{J}}}, D$ and the spectral triple (''quadruple'') preserves the five axioms of
\cite{Connes_spectral} together with strong regularity condition on each of the invariant sub-spaces, with identical Riemannian and pseudo-Riemannian metrics corresponding to the invariant sub-spaces. This is the case for our
''spectral triple''. Even more it would be sufficient if there existed just one such invariant subspace, or more generally
a direct sum of more such invariant sub-spaces not necessary summing up to the whole Hilbert space. The last case works even when passing to the problem of deformation of that spectral triple
induced by the perturbation because the causal perturbative series for interacting fields is (likewise in the adiabatic limit), order-by-order, translation ally covariant (we explain this in more details in Remark 2 of the next Subsection).

The additional complication coming from non-compactness
of $\mathbb{R}^4$ brings no additional substantial difficulties in our case where the topology of $\mathbb{R}^4$ is
homologically trivial (acyclic) and the non compactness will not open us to the full complication
of picking out proper unitizations,
or with potentially non-Fredholm character of the sign of the Dirac operator $D_{\mathfrak{J}}$ interconnected to the
non finitely generated character of the cohomology groups, which we must necessarily face in general non-compact manifold.
In the general case the problem of spectral characterization of non-compact manifolds could perhaps be reduced to the simply connected case, but this requires a nontrivial operator-algebraic version
of the universal covering space construction which is (at least in the opinion of the author) still non-trivial
even if we have the spectral characterization of Connes for compact manifolds when the spectral characterization for the
non-compact case is still lacking. The difficulty reflects the fact that uniformization in dimension greater than 2
is still an open problem. Possibly the extension of operator-algebraic axioms
respected by non-compact manifolds, as proposed in \cite{Gay}, together with the condition of the uniform multiplicity
of the representation $(\mathcal{A}'', \mathcal{H})$ of the algebra $\mathcal{A}''$ added in \cite{Connes_spectral} would be sufficient to characterize non-compact manifold, but there are still open questions connected with the correct choice of unitizations, i.e. the problem depends on the appropriate choice
of the ''preferred unitization'' (compare \cite{Gay}). The axioms of \cite{Gay} are not easy to handle
and still the way of proof
that the axioms of \cite{Gay} (together with the uniform multiplicity assumption of \cite{Connes_spectral})
indeed characterize non-compact manifolds in commutative case is not so easy visible (at least for the author).
Although the ''localization idea'' standing behind the axioms seems plausible we propose to replace it by the ''end compactification'' of Freudenthal or eventually a class of compactifications closely related to the end compactification,
in reducing the non-compact case to the compact case proved by Connes \cite{Connes_spectral}, compare the Appendix \ref{AppendixNonCompMani}.

In fact the whole analysis of the (Freudenthal) ends of non-compact manifolds is still not necessary in our special case.
In our homologically trivial, i.e. acyclic, case the minimal unitization
is sufficient in reducing the proof of reconstruction theorem to the unital case worked out in \cite{Connes_spectral}.
Indeed, we require (besides the additional requirement of uniform finite multiplicity of the representation
$(\mathcal{A}, \mathcal{H})$ of $\mathcal{A}$ in $\mathcal{H}$ introduced in \cite{Connes_spectral})
that the operator $D_{\mathfrak{J}}$ constructed above, after multiplication by a self-adjoint ''scaling'' operator $Q$ affiliated with the double commutor $(\mathcal{A}, \mathcal{H})''$ of the representation $(\mathcal{A}, \mathcal{H})$ and addition of a selfadjoint operator $V = \Upsilon^\mu A_\mu$ (''potential'') affiliated with the double commutor $(\mathcal{A}, \mathcal{H})''$ of the representation
$(\mathcal{A}, \mathcal{H})$, i.e. the operator $QD_{\mathfrak{J}} +V$, fulfils all the spectral requirements of the Dirac operator characterizing the compact
case, when restricted to the above mentioned invariant subspace $\mathcal{H}_{inv}$ of $\mathcal{H}$; in other words there exists a (unital) algebra $\mathcal{A}^+$ of operators on $\mathcal{H}_{inv}$ containing
the algebra $A^{+}|_{{}_{\mathcal{H}_{inv}}}$ as an essential ideal such that
\[
\big(A^{+}|_{{}_{\mathcal{H}_{inv}}}, \, \mathcal{H}_{inv}, \, (QD_{\mathfrak{J}} +V)|_{{}_{\mathcal{H}_{inv}}} \big)
\]
respects all Connes conditions \cite{Connes_spectral} necessary and sufficient for
\[
\big(A^{+}|_{{}_{\mathcal{H}_{inv}}}, \, \mathcal{H}_{inv}, \, (QD_{\mathfrak{J}} +V)|_{{}_{\mathcal{H}_{inv}}} \big)
\]
to be identifiable with the spectral triple of a compact Riemannian manifold, compare the Appendix for justification.
This is in fact the requirement saying that the one point compactification of the manifold represented spectrally by
$(\mathcal{A}|_{{}_{\mathcal{H}_{inv}}}, \mathcal{H}_{inv}, D_{\mathfrak{J}}|_{{}_{\mathcal{H}_{inv}}})$ is con formally equivalent to the open Riemannian manifold which possesses
smooth one point compactification being again a Riemannian manifold. Recall the fact that the one point compactification of a simply connected acyclic open manifold homeomorphic to $\mathbb{R}^4$ (in our case just
the standard manifold $\mathbb{R}^4$ with the standard differential structure) gives another manifold which is closed (in our case the
standard sphere $\mathbb{S}^4$)\footnote{Indeed we have the following theorem \cite{Fenille}: \emph{An acyclic and simply connected open n-manifold is homeomorphic to $\mathbb{R}^n$ if and only if its one-point compactification is again a manifold}. This theorem is equivalent to the generalized Poincar\'e conjecture, compare \cite{Fenille}, and as we know the generalized Poincar\'e conjecture holds true in every dimension (for dim = 2 it follows from the classification of 2-manifolds,
for dim = 3 has been proved by Perelman, for $\dim = 4$ by Freedman and for $\dim > 4$ it is a consequence of
the $h$-cobordism
theorem of Smale).}. The binding ''potentials'' $V$ and ''scaling'' operators $Q$ are naturally determined by the geometry (compare the Appendix \ref{AppendixNonCompMani}) and similarly the construction of the corresponding nuclear algebra
$\mathcal{A}$ is well known in distribution theory,
e.g. \cite{Berezansky}, \cite{GelfandIV}, \cite{obata}, \cite{HKPS}, where one uses
the so called Gelfand triple technique associating nuclear algebras (such as $\mathcal{S}(\mathbb{R}^4)$)
with the corresponding self adjoint operators. In fact, we will use these techniques in construction of the free fields as operator-valued distribution in the following Subsections (explicitly in Subsection \ref{free-gamma}).

We should emphasize that the perturbation should in principle preserve the invariance property: at every order of perturbation the existence of the invariant subspace on which the spectral triple preserves the
(strong) version of the five axioms of \cite{Connes_spectral} should be preserved, because the causal perturbation series for interacting fields is covariant, compare 
Subsection \ref{perturbed(A,H,D)}.

Finally, let us turn to the more general case of spectral characterization of
non-compact manifolds (although it is not necessary for
us here). So let $M$ be a space- and time-oriented $n$-dimensional pseudo-Riemannian smooth
(paracompact) manifold. Given a maximal time like sub bundle of $TM$ one can define canonically a Riemannian metric
$g_{{}_{\mathfrak{J}}}$ and a fundamental symmetry $\mathfrak{J}$ in the Hilbert space $\mathcal{H} = L^2 (S)$ of square integrable spinors associated to the Riemannian metric $g_{{}_{\mathfrak{J}}}$ on $M$, \cite{baum}
(the positive Riemannian metric $g_{{}_{\mathfrak{J}}}$ corresponds to the Dirac operator $D_\mathfrak{J}$ introduced earlier).
We have to assume that $M$ with the Riemannian metric $g_{{}_{\mathfrak{J}}}$ induced in such a manner is geodesically complete
(so that $D_\mathfrak{J}$ respects all the conditions of \cite{Gay} put on the Dirac operator).

Note that if the Riemannian manifold $(M, g_{{}_{\mathfrak{J}}})$ is conformally equivalent to a dense open sub-manifold of a compact closed Riemannian manifold $W$, then the compactification
described above may also be applied to $M$, compare the Appendix \ref{AppendixNonCompMani}. Of course the embedding $M \rightarrow W$ cannot preserve
the Riemannian metric\footnote{Complete non compact Riemannian manifold cannot be
isometrically embedded into a compact Riemannian manifold as an open sub-manifold, as isometry preserves
completeness.}
in the sense that the Riemannian metric of the embedded manifold will not
coincide with the Riemannian metric induced from $W$, and this is why we have to introduce the ''scaling'' and ''binding potential'' operators $Q$ and $V$ in order to
re compensate the difference. $\mathcal{A}_W = C^\infty(W)$ and the Dirac operator $D_W$
of the Riemannian manifold respect the ''strong version'' of the five conditions (1)-(5) of \cite{Connes_spectral}.
After the appropriate choice of the potential $V$ and scaling operator $Q$,
$\mathcal{A} \subset \mathcal{A}_W$ is to be identified with an essential ideal of smooth
functions on $W$ vanishing together with all their derivatives on the boundary $\partial M$
of $M$ in $W$ which preserve the regularity condition with respect to the Dirac operator $QD_\mathfrak{J}
+ V$ and the $m$-th characteristic value of the resolvent of $QD_\mathfrak{J} + V$ is $O(m^{-1/n})$.
Thus, the triple $(\mathcal{A}_W \supset \mathcal{A}, \mathcal{H} = L^2 (S), QD_\mathfrak{J} + V)$ respects the necessary and sufficient conditions
of \cite{Connes_spectral} for $(\mathcal{A}_W \supset \mathcal{A}, \mathcal{H} = L^2 (S), QD_\mathfrak{J} + V)$ to be identifiable
with the spectral triple of a closed (compact) manifold. This is the motivation,
compare the Appendix \ref{AppendixNonCompMani}.

The open conformal embedding $(M, g_{{}_{\mathfrak{J}}}) \rightarrow W$ need not be dense.
In particular if the open non-compact manifold $M$ is regular enough to be conformally equivalent to just the interior of a compact manifold $W_1$ with boundary $\partial W_1$, then taking another copy $W_1'$ of $W_1$ an gluing along the common boundary we obtain the compact manifold $W$ into which $M$ and its diffeomorphic copy embeds as two open disjoint sub-manifolds $M, M'$ with $W-(M \sqcup M' ) = \partial W_1$. In this case a unital algebra of operators $A^{+}$ exists such that
$(A^{+} \supset \mathcal{A} \oplus \mathcal{A}', \mathcal{H} \oplus \mathcal{H}' =
L^2 (S) \oplus L^2 (S), (QD_\mathfrak{J} + V) \oplus (Q'D_\mathfrak{J} + V'))$ is a spectral triple which
respects the conditions of Connes \cite{Connes_spectral}, which only doubles the ''multiplicity'' $(QD_\mathfrak{J} + V) \oplus (Q'D_\mathfrak{J} + V')$ of $QD_\mathfrak{J} + V$ but
with the whole motivation unchanged (compare Appendix \ref{AppendixNonCompMani}).
We can still extend over this strategy on
more general oriented and time oriented pseudo-Riemannian manifolds with complete Riemannian metric
$g_{{}_{\mathfrak{J}}}$ by representing $(M, g_{{}_{\mathfrak{J}}})$ as a sequence of compact manifolds which are glued together along the respective common boundaries, compare Appendix \ref{AppendixNonCompMani}.

For a quite general class of manifolds we can realize the nuclear algebra of smooth functions
$\mathcal{A}$ as the nuclear space $K\{M_p\}$
of Gelfand and Shilov \cite{GelfandII}-\cite{GelfandIV} (we use the notation of \cite{GelfandII} and \cite{GelfandIV} here). Construction of $K\{M_p\}$ goes through definition of a countable family of norms
\[
\| \varphi \|_p = \sup_{x \in M, |m| \leq p} M_p (x)|D^m \varphi(x)| \,\,\, (m \in \mathbb{N}^n)
\]
and the elements of $K\{M_p\}$ are smooth functions for which the norms
are finite and where $M_1, M_2, \ldots$
is a sequence of functions such that for each $x \in W$, $1 \leq M_1(x) \leq M_2(x) \leq \ldots$,
which are smooth everywhere on $W$ except the boundary $\partial M$ of $M$ and tend to infinity when approaching
$\partial M$ in $W$, or
when regarded as functions on $M$ they are smooth and tend to infinity when $x$ tends to infinity
(for each number $R>0$ and each natural $p$ there is a compact set $C$ such that $M_p > R$ outside
$C$). Now if the number of the (Freudenthal) ends of the manifold $M$ is finite then we have
a practical method of constructing the functions $M_p$ on $M$ (resp. on $W$), so that the corresponding
space $K\{ M_p \}$ is nuclear and associated canonically with a self adjoint operator (and may serve as well
to construct the core of $QD_\mathfrak{J} + V$ -- in fact we have to compare $D_\mathfrak{J}$ associated to the metric
of $M$ with that $D_W$ induced from the metric of $W$ in order to compute $V$ and $Q$).
Namely, we consider the Nash isometric embedding
$(M, g_{{}_{\mathfrak{J}}}) \rightarrow \mathbb{R}^N$ with appropriate $N$. Because $(M, g_{{}_{\mathfrak{J}}})$ is complete we may assume that this isometric embedding has closed image in $\mathbb{R}^N$ (\cite{Muller}) and in particular for every sequence of points in $M$ which goes to infinity its image
in $\mathbb{R}^N$ goes to infinity. We may choose $N$ large enough to find a point $p_0 \in \mathbb{R}^N$
whose euclidean distance to $M$ in $\mathbb{R}^N$ is greater than $1$. It is known that the function on $M$
which maps $x \in M$ to the euclidean distance of $x$ from $p_0$ is smooth on $M$ (and even
non-degenerate if $p_0$ is not focal). If the number of ends of $M$ is finite then the function just constructed
(with eventual simple rearrangements in some exceptional situations) may serve as the function $M_1$
on $M$, and its $p$-th power may serve as the function $M_p$, $p \in \mathbb{N}$. In quite general situation
of finitely many ends,
for each $p \in \mathbb{N}$ there exists $p' \in \mathbb{N}$ such that the function
\[
x \mapsto \frac{M_p(x)}{M_{p'}(x)}
\]
is square integrable with respect to the volume form associated to the Riemannian metric on $M$,
so that $K\{M_p\}$ is a nuclear algebra of smooth functions on $M$ vanishing at infinity
together with all their derivatives with the rate of the vanishing measured by the functions $M_p$.

But if the number of ends is big enough Nash embedding may behave at infinity in a quite uncontrolled
fashion; in particular one can imagine (in dimension 2 case) a ``surface of a tree trunk and of its brunches'' 
as embedded in $\mathbb{R}^3$ with the number of branches growing fast with the distance from the fixed point $p_0$. There are cases where the above summability condition is difficult to control.

\subsection{Construction of $V_\mathcal{F}$ and of the Dirac operator in the (Krein-) Hilbert space of free fields}\label{VFforFreeFields}

The representation of $T_4 \circledS SL(2, \mathbb{C})$ acting in the Hilbert (or Krein) space of free 
fields (or in the tensor product of such spaces corresponding to free fields with both energy sings as suggested in the Introduction) may be obtained by the direct sum over natural $n$ of (symmetrized or antisymmetrized) $n$-fold tensor products of a fixed (finite) set of induced unitary (or Krein isometric) representations of $T_4 \circledS SL(2, \mathbb{C})$, concentrated on fixed orbits $\mathscr{O}_{(m,0,0,0)} = \mathscr{O}_{\bar{p}}, \bar{p}= (m,0,0,0)$ or 
$\mathscr{O}_{(1,0,0,1)} = \mathscr{O}_{\bar{p}}, \bar{p}= (1,0,0,1)$. 
In particular the induced representation, (atisymmetrized) tensor products of which
give after direct summation the unitary representation of $T_4 \circledS SL(2, \mathbb{C})$ in the Hilbert space of the free positron-electron field, is given in Subsect. \ref{e+e-}. The induced Krein-isometric representation --
we call it  \emph{{\L}opusza\'nski representation}
-- which after symmetrized tensoring and direct summation give the Krein-isometric representation of $T_4 \circledS SL(2, \mathbb{C})$ in the Krein-Fock space of the free photon field will be given in Subsection \ref{free-gamma}. 

We provide first a general analysis of the construction of $V_\mathcal{F}$ and the corresponding
space-time spectral tuple $(\mathcal{A}, \mathcal{H}, D_{{}_{\mathfrak{J}}}, D)$ in the subspace orthogonal to the vacuum and the one particle states in the tensor product
$\mathcal{H}_+ \otimes \mathcal{H}_-$
of state spaces $\mathcal{H}_+, \mathcal{H}_-$ of free fields, with $\mathcal{H}_+$ being the Hilbert-Krein state space
acted on by positive energy free fields and with $\mathcal{H}_-$ being the Hilbert-Krein space
acted on by the negative energy fields. Next we slightly modify the undeformed situation by a (physically irrelevant) modification
of the representation $U$ of $T_4 \circledS SL(2, \mathbb{C})$ acting in the space $\mathcal{H}_+ \otimes \mathcal{H}_-$
on an invariant subspace of nonphysical states in order to simplify the whole situation. 

Note that in dealing with the decomposition of the representation $T_4 \circledS SL(2, \mathbb{C})$ acting in 
$\mathcal{H}_+ \otimes \mathcal{H}_-$ we need to consider decompositions of tensor products of representations concentrated
respectively 1) both on positive energy orbits (positive energy sheet of the two-sheeted hyperboloid), 2) both on negative energy orbits and finally 3) one concentrated on positive and the other on the negative energy orbit.
The first two cases 1) and 2) are from the point of view of their decomposition technique the same. The case 
3) is much more involved, which is mainly connected to the fact that 
in the decomposition of the tensor product there will be present representations concentrated on the 
one-sheet hyperboloid $\mathscr{O}_{(0,0,m,0)}$, and the stationary group $G_{(0,0,m,0)} = SL(2,\mathbb{R})$ 
corresponding to these orbits is not compact.  

Consider first the cases 1) and 2).
The tensor product of (ordinary) unitary
representations concentrated respectively on the orbits $\mathscr{O}_{(m_i,0,0,0)} = \mathscr{O}_{\bar{p}_i},
\bar{p}_i = (m_i,0,0,0)$,$i=1,2$,
and induced by (ordinary) unitary representations of the small group $G_{\bar{p}_i} = G_{(m_i,0,0,0)} = SU(2, \mathbb{C})$,
may be decomposed into a direct integral of representations concentrated on the orbits
$\mathscr{O}_{(m,0,0,0)} = \mathscr{O}_{\bar{p}}, \bar{p}= (m,0,0,0)$, $m \geq m_1 + m_2$ (if both $m_i$ are positive
and $m \leq m_1 + m_2$, if both $m_1$ are negative) induced by unitary representations of the small group
$G_{(m,0,0,0)} \cong SU(2, \mathbb{C})$. This decomposition may effectively be computed by application of the Mackey's
Kronecker product theorem and Fubini theorem together with the Peter-Weyl theory applied to $SU(2, \mathbb{C})$.

Now the analogous formula holds true for the decomposition of the tensor product of Krein-isometric
representations (e.g {\L}opusza\'nski representations) both concentrated on the orbit
$\mathscr{O}_{(1,0,0,1)}$ induced by a Krein-unitary representations of the stationary group
$G_{(1,0,0,1)} = \widetilde{E_2}$ (double covering of the Euclidean group of the Euclidean plane).
The Kronecker product theorem
holds true for the induced Krein-isometric representations (proof of which is the main subject of Sect. \ref{pre}
- \ref{Kronecker_product}). Then by this theorem and by the initial part of this Section \ref{constr-of-VF},
by Sect. \ref{lop_ind} and Sect. \ref{decomposable_L}), it follows that this tensor product may be decomposed into direct integral of Krein-isometric representations concentrated on the orbits $\mathscr{O}_{(m,0,0,0)}$ ,
$m >0$, induced by a fixed Krein-unitary representation $L$ of the stationary group $G_{(m,0,0,0)} \cong SU(2, \mathbb{C})$
in a Krein space.
Because $SU(2, \mathbb{C})$ is compact, then we can define invariant with respect to $L$, non-degenerate,
positive definite Hermitian bilinear form
$(\cdot , \cdot)_1$
\[
(\psi_1,\psi_2)_1 = \int \limits_{SU(2, \mathbb{C})}(L_g \psi_1 , L_g \psi_2) \, \ud g
\]
in the same Krein space of the rep. $L$ (where $(\cdot, \cdot)$ under the integral sign is the ordinary Hilbert space product of the Krein space of the representation $L$), such that $L$ may be treated as unitary representation
of $SU(2, \mathbb{C})$. 

Similarly, we have the analogue decomposition of the tensor product of Krein-isometric
(say {\L}opusza\'nski) representations $U^{{}_{(-1,0,0,1)}{\L}}$ both concentrated on $\mathscr{O}_{(-1,0,0,1)}$.

In case 3) when the signs of $m_i$ are opposite (and the representations are induced by ordinary unitary or Krein-unitary representations), decomposition may be effected in the same way with the use of Mackey (or our Kronecker product theorem
of Sect \ref{Kronecker_product}) and the Fubini theorem for scalar (resp. vector valued) functions (eq. (\ref{dir_int_skew_L^2:decompositions})
of Sect. \ref{decomposition}) and will contain in addition direct integral of representations concentrated on the orbits
$\mathscr{O}_{(0,0,m,0)} = \mathscr{O}_{\bar{p}}, \bar{p}= (0,0,m,0)$, induced by direct integrals or sums of Bargmann's
principal and discrete series of representations (resp. Krein-unitary representations)
of the small group $G_{(0,0,m,0)} \cong SL(2, \mathbb{R})$. Computation is only slightly more laborious in the
ordinary unitary case where it it easily reducible to the decomposition of the regular representation of
$SL(2, \mathbb{R})$ group restricted to the subspace
of generalised spherical functions, and the Plancherel formula
for $SL(2, \mathbb{R})$. The Krein-isometric case is more laborious and is not reducible to the
ordinary unitary harmonic analysis on $SL(2, \mathbb{R})$. First of all the application of our
generalization of Mackey theory of the second Part of our work (particularly the Kronecker product theorem
\ref{twr:Kronecker_product}) gives the decomposition of the tensor product of Krein-isometric
{\L}opusza\'nski representations into direct integral of Krein unitary induced representations
over the orbits of the translation subgroup under the action of the $SL(2, \mathbb{C})$ subgroup. 

Now it follows from our results of Sect. \ref{pre} -- \ref{decomposable_L} that the representation
$U$ of $T_4 \circledS SL(2, \mathbb{C})$ acting in the Hilbert
(or Krein) space $\mathcal{H}_+ \otimes \mathcal{H}_-$ of free fields may be decomposed into
direct integral of unitary (or more generally
Krein-isometric) induced representations concentrated on single orbits, and that after restriction to the invariant subspace
$\mathcal{H}_{1}^{\bot}$ of $\mathcal{H}_+ \otimes \mathcal{H}_-$ orthogonal to the vacuum and single particle states,
the representation restricted to the translation
subgroup is of uniform (infinite) multiplicity (compare Remark \ref{decomposable_L:uniform_mult}).
Consider the restriction $U|_{{}_{\mathcal{H}_{1}^{\bot}}}$ of $T$ to the subspace $\mathcal{H}_{1}^{\bot}$.
It follows that the sub representation $U^+ \oplus U^-$
of $U|_{{}_{\mathcal{H}_{1}^{\bot}}}$, concentrated on the set-theoretical sum $C_+ \cup C_-$ of the forward and
backward cones is (Krein-) unitary equivalent\footnote{With the equivalence defined by a non-singular Krein-isometric map, i.e. with dense domain and image equal to the dense core sets of the equivalent representations.} to the direct integral
\begin{equation}\label{T^+oplusT^-}
U^+ \oplus U^-
\cong \int \limits_{-\infty}^{\infty}
U^{{}_{(m,0,0,0)} L}
\, \ud m
\end{equation}
of Krein-isometric representations $U^{{}_{(m,0,0,0)} L}$ concentrated on the orbits
$\mathscr{O}_{(m,0,0,0)}$ induced by a fixed (Krein-) unitary representation $L$ of the
stationary group $G_{(m,0,0,0)} = SU(2, \mathbb{C})$ such that every direct irreducible summand $L^l$ in the decomposition of $L$ enters with infinite multiplicity. By the preceding paragraph and the
results of the last Subsection it follows that $U^+ \oplus U^-$
possesses the associated representation
\[
[U^+ \oplus U^-]_{\Ass} = \int \limits_{0}^{\infty}
U^{{}_{(0,m,0,0)} [L]_{\Ass}}
\, \ud m
\]
such that the construction of
$V_\mathcal{F}$ and the associated spectral triple may be constructed as in the above Subsection on the space of the representation $U^+ \oplus U^- \oplus [U^+ \oplus U^-]_{\Ass}$. The main open problem which remains to be solved is
to check if the representation $[U^+ \oplus U^-]_{\Ass}$ is Krein-unitary equivalent to the sub representation $U^{+-}$
of $U$ concentrated outside the set-theoretical sum of the forward and backward cones or if the sub representations
$U^+ \oplus U^-$ and $U^{+-}$ are associated.

Now we propose to simplify the whole situation by a modification (below we show that the modification is necessary) of the representation $U$ on the subspace of nonphysical states which does not affect the investigation of the standard theory on physical states.
Namely, we propose to replace the sub representation $U^{+-}$ acting on the invariant subspace $\mathcal{H}^{+-}$
(equal to the image of the spectral projection of the joint spectral decomposition of $P_0, \ldots P_3$
concentrated outside the sum $C_+ \cup C_-$ of positive and negative cones) with the representation
$[U^+ \oplus U^-]_{\Ass}$. By the uniform multiplicity of translation subgroup this modification leaves the representation of the translation subgroup unchanged. Let us justify that we can do it. Let
$\mathcal{H}^+ \subset \mathcal{H}_+ \otimes \mathcal{H}_-$ and $\mathcal{H}^- \subset \mathcal{H}_+ \otimes \mathcal{H}_-$
be the invariant sub spaces of the sub representations $U^+$ and $U^-$. Let $\vac_+ \in \mathcal{H}_+$,
$\vac_- \in \mathcal{H}_-$ be respectively the vacua in the Hilbert-Krein spaces of the positive and negative energy fields. Now $\mathcal{H}_+$ and $\mathcal{H}_-$ may be written in the form of sub spaces invariant for the representation of $T_4 \circledS SL(2, \mathbb{C})$ respectively in $\mathcal{H}_+$ and $\mathcal{H}_-$:
\[
\mathcal{H}_+ = \mathbb{C} \vac_+ \, \oplus \, \mathcal{H}^{\nonvac}_{+}, \,\,\,
\mathcal{H}_- = \mathbb{C} \vac_- \, \oplus \, \mathcal{H}^{\nonvac}_{-}
\]
We define the following sub spaces $\mathcal{H}_+ \otimes \mathbb{C}\vac_- \subset \mathcal{H}_+ \otimes \mathcal{H}_-$,
$\mathbb{C}\vac_+ \otimes \mathcal{H}_- \subset \mathcal{H}_+ \otimes \mathcal{H}_-$ and the subspace
$\mathcal{H}^{\unphys} = \mathcal{H}^{\nonvac}_{+} \otimes \mathcal{H}^{\nonvac}_{-}
\subset \mathcal{H}_+ \otimes \mathcal{H}_-$ invariant for the representation $T$.
By the application of the Mackey and our Kronecker product theorem (Sect \ref{Kronecker_product})
it follows that
\[
\mathcal{H}_+ \otimes \mathbb{C}\vac_- \,\, \subset \,\, \mathcal{H}^+ , \,\,\,
\mathbb{C}\vac_+ \otimes \mathcal{H}_- \,\, \subset \,\, \mathcal{H}^{-}, \,\,\,
\mathcal{H}^{+-} \subset \mathcal{H}^{\unphys};
\]
and moreover we have
\begin{multline*}
\mathcal{H}_+ \otimes \mathcal{H}_- \\
= \big( \mathbb{C} \vac_+ \otimes \mathbb{C} \vac_- \big)
\oplus \big(\mathbb{C} \vac_+ \otimes \mathcal{H}^{\nonvac}_{-} \big)
\oplus \big( \mathcal{H}^{\nonvac}_{+} \otimes \mathbb{C} \vac_- \big)
\oplus \big( \mathcal{H}^{\nonvac}_{+} \otimes \mathcal{H}^{\nonvac}_{-} \big) \\
= \big( \mathcal{H}_{+} \otimes \mathbb{C} \vac_- \big) \oplus
\big(\mathbb{C} \vac_+ \otimes \mathcal{H}^{\nonvac}_{-} \big) \oplus
\big( \mathcal{H}^{\nonvac}_{+} \otimes \mathcal{H}^{\nonvac}_{-} \big) \\
= \big( \mathcal{H}_{+} \otimes \mathbb{C} \vac_- \big) \dot{\oplus}
\big(\mathbb{C} \vac_+ \otimes \mathcal{H}_{-} \big) \oplus
\big( \mathcal{H}^{\nonvac}_{+} \otimes \mathcal{H}^{\nonvac}_{-} \big),
\end{multline*}
where the dot over $\oplus$ means that $\mathcal{H}_{+} \otimes \mathbb{C} \vac_-$ and
$\mathbb{C} \vac_+ \otimes \mathcal{H}_{-}$ have the common nonzero subspace
$\mathbb{C} \vac_+ \otimes \mathbb{C} \vac_-$. Identifying $\mathcal{H}_{+} \otimes \mathbb{C} \vac_-$
with $\mathcal{H}_{+}$ and $\mathbb{C} \vac_+ \otimes \mathcal{H}_{-}$ with $\mathcal{H}_{-}$
we may write the last equality in the following manner (remembering that $\mathcal{H}_+$
and $\mathcal{H}_-$ have the vacuum set in common):
\[
\mathcal{H}_+ \otimes \mathcal{H}_-
= \mathcal{H}_{+} \dot{\oplus} \mathcal{H}_{-} \oplus
\big( \mathcal{H}^{\nonvac}_{+} \otimes \mathcal{H}^{\nonvac}_{-} \big).
\]
We can therefore consistently define the perturbation of the translation generators
(when defining the perturbation of the spectral triple in the next Sect.) on
the subspace $\mathcal{H}_+$ by the ordinary perturbation in the positive energy fields, separately on the
subspace $\mathcal{H}_-$ acted on by negative energy fields, and leave unperturbed on the nonphysical
subspace $\mathcal{H}^{\unphys} = \mathcal{H}^{\nonvac}_{+} \otimes \mathcal{H}^{\nonvac}_{-}$. In this way
the perturbed relevant operators (namely translation generators) will be unchanged in their action in
the invariant subspace $\mathcal{H}^{+-} = \mathcal{H}_{{}_{U^{+-}}}$ so that the sub representation
$U^{+-}$ may be replaced by the sub representation $[U^+ \oplus U^-]_{\Ass}$.

It should be stressed that the proposed modification of the representation $U$ by the indicated replacement of its sub representation $U^{+-}$ is motivated by the simplification of computations, nonetheless it is in a sense forced by the whole situation in which we use the representation $U$ or its restriction to the nuclear Hida's test space (for its definition compare Section \ref{free-gamma}). It is tempting to think of
the original representation $U$, equal to the tensor product of representations of $T_4 \circledS SL(2, \mathbb{C})$
acting respectively in $\mathcal{H}_+$ and $\mathcal{H}_-$, as being more natural.
We add some comments on the additional
technical difficulties encountered in the construction of $V_\mathcal{F}$ and the
associated space-time spectral triple in this non modified case, and show that the construction of the Dirac operator
out of the original non modified representation $U$ along the lines of the previous Subsections would be impossible.

By the application of our (resp. Mackey's) Kronecker product theorem (Sect. \ref{Kronecker_product})
and a generalized Fubini theorem (eq. (\ref{dir_int_skew_L^2:decompositions})) for
vector valued functions we likewise can show that $U^{+-}$ is equivalent to the following direct integral
\begin{equation}\label{T^{+-}}
U^{+-} \cong
\int \limits_{0}^{\infty}
U^{{}_{(0,0,m,0)} \mathbb{L}}
\, \ud m
\end{equation}
of Krein-isometric representations $U^{{}_{(0,0,m,0)} \mathbb{L}}$ concentrated on the orbits
$\mathscr{O}_{(0,0,m,0)}$ induced by a fixed (Krein-) unitary representation $\mathbb{L}$ of the
stationary group $G_{(0,0,1,0)} = SL(2, \mathbb{R})$. The equivalence is defined by Krein isometric map
which is not singular in having dense domain and image both being the core domains of the equivalent representations.

The first important problem is to decompose the Krein-unitary representation $\mathbb{L}$ (present
in the decomposition (\ref{T^{+-}}))
of $SL(2, \mathbb{R})$ into direct integral/sum of indecomposable components.
Possibility of an effective decomposition of $\mathbb{L}$ allows us to resolve at least the Problem (B)
of the following two Problems
(A) and (B):
\begin{enumerate}
\item[(A)]
To check if the Krein-unitary representation $L$ of
$SU(2, \mathbb{C})$ in decomposition (\ref{T^+oplusT^-}) and the representation $\mathbb{L}$ of $SL(2, \mathbb{R})$
in (\ref{T^{+-}}) are associated, i.e. if there exists a Krein-unitary extension $V$ of $L$ to a representation
of $SL(2, \mathbb{C})$ acting in the Krein space of $L$ which at the same time is (Krein-unitary equivalent to)
an extension of $\mathbb{L}$ to a Krein-unitary representation of $SL(2, \mathbb{C})$.
\item[(B)]
To find sub representations of $L'$ and $\mathbb{L}'$ respectively of $L$ and $\mathbb{L}$, which are associated:
there exists a Krein-unitary extension $V$ of $L'$ to a representation
of $SL(2, \mathbb{C})$ acting in the Krein space of $L'$ which at the same time is (Krein-unitary equivalent to)
an extension of $\mathbb{L}'$ to a Krein-unitary representation of $SL(2, \mathbb{C})$.
\end{enumerate}

Indeed, we have quite a huge class of not necessary unitary, but Krein unitary, representations of the
$SL(2, \mathbb{C})$ group, which are effectively decomposable into indecomposable components. Namely, the first class
embraces all finite dimensional representations which are direct sums of irreducible representations in which the conjugate irreducible representations (in the sense of \cite{Geland-Minlos-Shapiro}) appear in pairs, \cite{Geland-Minlos-Shapiro} (recall the very nice property of finite representations of $SL(2, \mathbb{C})$: every finite representation is equal to a direct sum of irreducible representations; for example
this is false for the representations of the group $E_2$ of the euclidean motions of the euclidean plane).
By the preceding Subsections it is clear that in the space of the irreducible ''undotted spinor 1/2'' two-dimensional representation $V_{1,0}$ (for the classification of the irreducible representations of $SL(2, \mathbb{C})$ compare e.g. \cite{Geland-Minlos-Shapiro}) there does not exist fundamental symmetry which makes the representation Krein-unitary. The same holds for the conjugate ''dotted spinor 1/2'' representation $V_{0,1}$. But there exists fundamental symmetry which makes the irreducible tensor product representation $V_{0,1} \otimes V_{1,0}$ Krein unitary. Any irreducible representation
is a symmetrized tensor $n$-fold product $V_{n,0}$ of the spinor 1/2 representation $V_{1,0}$, or symmetrized $m$-fold tensor product $V_{0,m}$ of the conjugate spinor 1/2 representation $V_{0,1}$, or the tensor product
$V_{n,0} \otimes V_{0,m}$, and $V_{n,0} \otimes V_{0,m}$ admits a fundamental symmetry making it Krein unitary iff $n = m$, \cite{Geland-Minlos-Shapiro}. Another class of Krein-unitary representations may be obtained by our generalization of Mackey construction of induced representations applied to the construction of Krein-unitary representations $U^{\chi}$ induced by Krein-unitary representations $\chi$ of the upper triangular subgroup of the $SL(2, \mathbb{C})$ group. There is a natural nuclear space associated with the smooth structure of the upper triangular subgroup coset sub manifolds, giving to the induced representation
the form investigated by Gelfand and Graev \cite{GelfandGraevDec}, together with a nuclear space dense in the Hilbert space of the representation with respect to which the representors are continuous. Application of our subgroup theorem together with the smooth structure of the corresponding double cosets gives a decomposition of the restriction to
the subgroup $SL(2, \mathbb{R})$ of the induced representation $U^{\chi}$ into indecomposable components corresponding to the respective double coset sub manifolds (because the induced representation has the form of the representation investigated in \cite{GelfandGraevDec} with representors transforming a nuclear space into itself with the smooth structure of double coset invariant sub manifolds the Fubini theorem for distributions may be applied for the construction of decomposition, which has already been noticed by Gelfand and Graev \cite{GelfandGraevDec}). Because on the other hand the representation
$\mathbb{L}$ may be decomposed (by what we have mentioned earlier in this Subsection) we can compare and eventually pick up the sub representations of $\mathbb{L}$ which are equal to the restrictions of the induced representations $U^{\chi}$ to the $SL(2, \mathbb{C})$ subgroup. Indeed, $\mathbb{L}$ cannot be decomposed further within the Hilbert space realm, but our Kronecker product theorem for Krein-isometric induced representations allows to continue the decomposition geometrically using the smooth structure of the corresponding double coset sub manifolds and the distributional Fubini theorem, as we have already mentioned. In fact, it is sufficient to notice that the tensor product of the {\L}opusza\'nski representation
with itself as well as the tensor product of the {\L}opusza\'nski representation with a unitary induced representation
of the $T_4 \circledS SL(2, \mathbb{C})$ group can be decomposed in this way. In particular in Sect. \ref{free-gamma}
we will show that the generators of the {\L}opusza\'nski representation are well-defined operators continuously mapping the
corresponding nuclear space into itself as well as the generators of the Krein-isometric representation
of $T_4 \circledS SL(2, \mathbb{C})$ on the Fock space of free fields are well-defined operators transforming continuously the corresponding nuclear
space (Hida's test space) into itself. Similarly, we have the associated nuclear space dense in the Hilbert space of the representation $\mathbb{L}$ with well-defined generators mapping continuously the nuclear space into itself.
In particular the infinitesimal representors of the Casimir operator
\[
Q = (A_{12})^2 - (A_{01})^2 - (A_{02})^2
\]
of the representation $\mathbb{L}$ of $SL(2, \mathbb{R})$ group is well-defined.

In case of the unitary representations, which do not require the fine topology of an invariant dense nuclear space for the
construction of the decomposition, the computation is very effective, because every unitary representation is decomposable into irreducible unitary representations and the irreducible representations of $SL(2, \mathbb{C})$
and $SL(2, \mathbb{R})$ are well known. The only fact which makes a difference in practical computations
for the Krein-unitary case in comparison to the unitary case, is the reachness of the class of indecomposable
(but in general reducible) representations which enter the decomposition of the representation $\mathbb{L}$.
In case of the $SL(2, \mathbb{C})$ group we know all \emph{completely irreducible} representations (finite and infinite dimensional, unitary and non-unitary), where we use complete irreducibility in the sense of Godement \cite{Godement}:
any bounded operator in the Hilbert space of the representation is in the weak closure of the representors of the group ring
corresponding to the representation; in practice: there are no nontrivial invariant proper closed sub spaces
and there are no bounded operators commuting with the representation other than the multiplies of the identity operator.
This classification (due to Neumark for $SL(2, \mathbb{C})$, \cite{NeumarkLorentzIrr}, \cite{Geland-Minlos-Shapiro}
and due to Bargmann for $SL(2, \mathbb{R})$) is however insufficient for us, as we expect to encounter indecomposable although reducible representation (the {\L}opusza\'nski representation itself is indecomposable although reducible and moreover unbounded as a representation in the Hilbert space); where the representation is indecomposable if there are no bounded idempotents (not necessary self adjoint) other than zero and one, which commute with the representation. Thus, indecomposability is weaker than irreducibility, and all the more weaker
than complete irreducibility.\footnote{In case of the additional structure (Krein structure given by a self adjoint and unitary involution) we may consider Krein-orthogonal decompositions with the corresponding Krein-self-adjoint idempotents commuting with the representation.} Thus, in order to solve the Problem (A) we have to compute $\mathbb{L}$ explicitly
as well as its decomposition using our generalization of the Mackey theory along the lines indicated earlier in
this Subsection. To this ''geometric decomposition'' there corresponds the adjoined eigenfunction decomposition
of the non-self-adjoint Casimir operator $Q$ of the representation $\mathbb{L}$.

Suppose that $L' \subset L$ and $\mathbb{L}' \subset \mathbb{L}$ are associated and let $V$ be the corresponding
common extension of the representation $L'$ of $SU(2, \mathbb{C})$ and of the representation $\mathbb{L}'$
of the $SL(2, \mathbb{R})$ to a Krein-unitary representation of the $SL(2, \mathbb{C})$. Now we show that this
does not allow us to perform the construction
of the generators $\widetilde{\gamma}^\mu$ of a representation of the Clifford algebra and the associated generalized Dirac operator as in the previous Subsections. Recall that for the construction of the Dirac operator we need more than just one extension $V$ but there are needed several such extensions $V$ (in fact an infinite number of them is needed in case of infinite dimensional representation $\mathbb{L}'$) which are conjugated to each other.
Let us denote them $V_1, V_2, \ldots$ just as in the previous Subsections.
Recall also that conjugation means here that $V_i$ and $V_k$
are equal on the small group, and nothing more, and thus conjugation depends on the class of the orbits
in question (depend on $p$). In general for neither of the classes of the corresponding orbits (to which $p$ belong)
it can be realized by any group automorphism of $SL(2, \mathbb{C})$.
The extensions $V_1, V_2, \ldots$ define the generalized ''multispinor''
\[
\begin{split}
\widetilde{\varphi}_1 = V_1(\beta^{-1}) \widetilde{\psi}, \\
\ldots \\
\widetilde{\varphi}_k = V_k(\beta^{-1}) \widetilde{\psi}, \\
\widetilde{\varphi}_{k+1} = V_{k+1}(\beta^{-1}) \widetilde{\psi}, \\
\ldots
\end{split}
\]
where $\widetilde{\psi}$ is in fact concentrated on the orbit of some $\bar{p}$ and should be written
$\widetilde{\psi}_{{}_{\bar{p}}}$ in order to make the notation compatible with the previous Subsection, but we omit the subscript for simplicity as in the previous Subsections. Note also that $\beta: p \mapsto \beta(p)$ in the above formula is the function corresponding to the orbit of $\bar{p}$ and defined as in the preceding
Subsections. Recall that $\beta$ depends on the orbit and is not unique.
Let us order the components of the ''multispinor'' and join into disjoint pairs as in the previous Subsections, so that
the successive components belonging to one pair may be mapped into each other:
\[
\begin{split}
\widetilde{\varphi}_{k+1} = V_{k+1}(\beta^{-1})V_k(\beta^{-1})^{-1} \widetilde{\varphi}_k \\
\widetilde{\varphi}_k = V_k(\beta^{-1})V_{k+1}(\beta^{-1})^{-1} \widetilde{\varphi}_{k+1}.
\end{split}
\]
Now although the function $\beta$ -- even within one and the same
orbit -- is not unique,
the functions $p \mapsto V_{k+1}(\beta(p)^{-1})V_k(\beta(p)^{-1})^{-1}$ and
$p \mapsto V_k(\beta(p)^{-1})V_{k+1}(\beta(p)^{-1})^{-1}$ does not depend on the choice of $\beta$
by the last Lemma of the Subsection \ref{0-1VF}.
From this we would obtain a generalized Dirac equation in the momentum space (algebraic relation which after Fourier
transforming passes into a generalized Dirac equation) iff
the function $p \mapsto V_{k+1}(\beta(p)^{-1})V_k(\beta(p)^{-1})^{-1}$
and the function $p \mapsto V_k(\beta(p)^{-1})V_{k+1}(\beta(p)^{-1})^{-1}$
were linear functions of $p$ as in the previous Subsections. Suppose for a while that this is the case and that
we can construct an involutive representation of the Clifford algebra generated by $\widetilde{\gamma}^\mu$
fulfilling
\begin{equation}\label{Clifford-generators-relations}
\widetilde{\gamma}^\mu\widetilde{\gamma}^\nu + \widetilde{\gamma}^\nu\widetilde{\gamma}^\mu = g^{\mu \nu} \bold{1}
\end{equation}
exactly as in the preceding Subsections.
The Clifford algebra corresponding to the Minkowski metric is finite dimensional
and is linearly generated by the following 16 elements: $\bold{1}, \gamma^\mu,
\gamma^\mu \gamma^\nu
(\mu<\nu), \gamma^\mu \gamma^\nu \gamma^\rho (\mu< \nu < \rho),
\gamma^0 \gamma^1 \gamma^2 \gamma^3$.
We can introduce the involution and ordinary operator norm regarding its elements as matrix operators
by the ordinary Hermitian adjoint operation,
which makes it a finite dimensional $C^*$-algebra. It follows that (\ref{Clifford-generators-relations}) defines
its *-representation (by assumption). Therefore, the representation is a direct sum of cyclic representations
(transfinite induction principle).
Because the algebra is finite dimensional, any cyclic representation is a direct and finite sum
of irreducible representations, which are likewise finite (apply just the Gelfand-Neumark-Segal construction
of cyclic representation).
(Even more: any irreducible representation of this algebra must be equivalent to the identity representation generated by
$\gamma^\mu \mapsto \gamma^\mu$, which follows from the Pauli theorem.) Then
because the representation of the Clifford algebra corresponding to Minkowski metric
generated by (\ref{Clifford-generators-relations}) is a direct sum of finite dimensional representations,
we would have therefore obtained
the Dirac operator $D = i \widetilde{\gamma}^0 \partial_0 + \ldots i \widetilde{\gamma}^3 \partial_3$ in full analogy with the preceding Subsections, which meets all relevant conditions. But unfortunately it is impossible. Indeed, already the first application of our generalization of the Mackey theory of induced representation to the decomposition of
tensor product of {\L}opusza\'nski representations with opposite energy signs shows that no finite dimensional representations
can occur in the decomposition\footnote{Stricly speaking our analysis of the representation $\mathbb{L}$
is still on the way, but we have already obtained strong indications that no finite representations can occur
in its decomposition.} of $\mathbb{L}$ of $SL(2, \mathbb{R})$ which contradicts the decomposability of any involutive representation of the Clifford algebra of the Minkowski metric into finite dimensional sub representations.

In case of the modified representation we saw that we can always construct the representation associated to 
(\ref{T^+oplusT^-}), so that the construction of the Dirac operator is possible.  

Below we show another indication that the modification of the representation is necessary when using the representation
$U$ acting in the corresponding nuclear space (and not its adjoint representation in the space adjoint to the nuclear space) appealing to some results of Gelfand, Yaglom, Minlos and Shapiro \cite{Geland-Minlos-Shapiro}. 

Let us consider for a while the following two main possibilities:

\begin{enumerate}

\item[$\alpha)$] All component representations in the decomposition of $\mathbb{L}$ are  completely irreducible.

\item[$\beta)$] Some of the component representations in the decomposition of $\mathbb{L}$ although
being indecomposable\footnote{For definition compare Sect. \ref{decomposable_L}.} are 
nonetheless reducible.

\end{enumerate}
We know from the outset that the first possibility $\alpha)$ has to unfortunately be excluded but nonetheless we may consider a ``maximal'' sub representation of $\mathbb{L}$ which can be written as a direct integral/sum
of completely irreducible components.

In case $\alpha)$ we can relatively easily extend the method of Minlos \cite{Geland-Minlos-Shapiro},
Part II, Section 2.9 to resolve the problem if the Krein-unitary representation $L$ of
$SU(2, \mathbb{C})$ in decomposition (\ref{T^+oplusT^-}) and the representation $\mathbb{L}$ of $SL(2, \mathbb{R})$
in (\ref{T^{+-}}) are associated, i.e. if there exists a Krein-unitary extension $V$ of $L$ to a representation
of $SL(2, \mathbb{C})$ acting in the Krein space of $L$ which at the same time is (Krein-unitary equivalent to)
an extension of $\mathbb{L}$ to a Krein-unitary representation of $SL(2, \mathbb{C})$;
or to find the sub representations $L' \subset L$ and $\mathbb{L}' \subset \mathbb{L}$ which are associated. Recall that
in \cite{Geland-Minlos-Shapiro}, Part II, Section 2.9 there is presented a method of construction
of invariant Hermitian bilinear non-singular forms in the space of not necessary unitary representation $V$ of
$SL(2, \mathbb{C})$ whenever $V$ is a direct sum of irreducible representations. We need to consider the extension
problem for such forms which is similar. The difference is that in case $\alpha)$ we have to deal with
direct integral instead of direct sum of irreducible representations.
It makes no essential change in the method of investigation where the discrete sums will have to be changed
by integrals.

In case $\beta)$ the investigation of the problem cannot be based on the methods of Gelfand, Yaglom
and Minlos, and we have to adhere to the argument presented above.

Observe that the construction of the transform $V_\mathcal{F}$ and the associated Dirac operator are essentially
independent so that we can perform a different but less general method. Namely, having given the common extension $V$ we can perform the transform $V_\mathcal{F}$
from a dense domain of the representation space of the representation\footnote{Restriction of $U$ to the subspace
$\mathcal{H}_{1}^{\bot} \subset \mathcal{H_+ \otimes \mathcal{H}_-}$
orthogonal to the vacuum and single particle states.} $U|_{{}_{\mathcal{H}_{1}^{\bot}}}$ onto a dense subspace of the Hilbert-Krein space of square summable generalized vector valued multispinors $\phi$ (with values in the
Hilbert-Krein space of the representation $V$) with the following local transformation formula
\begin{equation}\label{infinite-multispinor-rep}
\begin{split}
U(\alpha) \phi (x) = V(\alpha) \phi (x\Lambda(\alpha^{-1})), \,\,\, \alpha \in SL(2, \mathbb{C}) \\
T(a) \phi(x) = \phi(x - a), \,\,\, a \in T_4.
\end{split}
\end{equation}
Then we are seeking for the most general (infinite) equation
\begin{equation}\label{relativistic-inv-eq}
i\Gamma_0 \partial_0 \phi + i\Gamma_1 \partial_1 \phi + i\Gamma_2 \partial_2 + i\Gamma_3 \partial_3 \phi
= m \phi, \,\,\, m \in \mathbb{R}
\end{equation}
invariant with respect to the representation (\ref{infinite-multispinor-rep}), where $\Gamma_0 , \ldots \Gamma_3$
are linear operators acting in the space of the representation $V$.
This problem has been exhaustively investigated by Gelfand and Yaglom \cite{GelfandYaglom1}-\cite{GelfandYaglom3},
and also in \cite{Geland-Minlos-Shapiro}, Part II, Chapter II, for the case when $V$ is a direct sum of completely irreducible representations, which are not necessary finite dimensional and not necessary unitary. Again in case $\alpha)$
when the representation $V$ is a direct integral of completely irreducible representations the method of Gelfand and Yaglom
may relatively easily be extended to this case.

Let us suppose first that $V$ is a direct sum
of irreducible representations. We may therefore apply the results cited in
\cite{Geland-Minlos-Shapiro}, Part II, Chapter II. When we have only finite direct summands in the decomposition
of $V$ and among them there are infinite dimensional, then in general case of such $V$ the operators
$\Gamma_0, \ldots \Gamma_3$ (matrices) are unbounded with unbounded sets of eigenvalues, containing
in general the zero eigenvalue with infinite multiplicity, this is the case e.g. for $V$ completely irreducible.
If $V$ is an infinite direct sum of completely irreducible representations, only in very
exceptional cases can the operator matrix
$\Gamma_0$ be bounded with bounded set of eigenvalues, compare \cite{Geland-Minlos-Shapiro}, Part II, Chapter II,
Section 10.8. When passing to the case of $V$ consisting of
direct integral of irreducible summands the flexibility does not raise considerably.
In particular, it is possible to construct
invariant equations with non-singular and bounded $\Gamma_0 , \ldots \Gamma_3$ , and such that
$\Gamma_0, \i \Gamma_1, i\Gamma_2, i\Gamma_3$ are self adjoint (with finite sets of eigenvalues),
compare \cite{Geland-Minlos-Shapiro}, Part II, Chapter II, Sect. 10.8, but the construction is very complicated
and allows practically no flexibility.
In particular when infinite dimensional summands are present there is practically no room for the possibility for including the joint spectrum of $\Gamma_0, i\Gamma_1, i\Gamma_2, i\Gamma_3$
into the two element set $\{1,-1 \}$.
If this was possible then $\Gamma_0 \Gamma_0 = - \Gamma_1 \Gamma_1 = - \Gamma_2 \Gamma_2 = - \Gamma_3 \Gamma_3
= \bold{1}$. Now using the commutation relations
of \cite{Geland-Minlos-Shapiro}, Part II, Chapter II, Sect. 7.2 (page 273) we easily see that
\begin{equation}\label{repGamma}
\Gamma_\mu \Gamma_\nu + \Gamma_\nu \Gamma_\mu = g_{\mu \nu} \bold{1},
\end{equation}
where $[g_{\mu \nu}] = \diag(1,-1,-1,-1)$ are the components of the Minkowski metric.

Now by a general theorem (compare \cite{GelfandYaglom1}-\cite{GelfandYaglom3} or \cite{Geland-Minlos-Shapiro}, Part II,
Chapter II, Sect. 7.2) the ordinary Fourier transform
\begin{equation}\label{F(phi)}
\widetilde{\phi}(p) = (2 \pi)^{-1/2} \int \limits_{\mathbb{R}^4} \phi(x) e^{i p \cdot x} \,
\ud^4 x, \,\,\,\,\,\, p \cdot x = p^0x_0 - p^1 x_1 - p^2 x_2 - p^3 x_3,
\end{equation}
of any square integrable solution (in the distributional sense) $\phi$ of the equation (\ref{relativistic-inv-eq})
is concentrated on the set theoretical sum of orbits $\mathscr{O}_{(\frac{m}{\lambda},0,0,0)}$,
with $\lambda$ ranging over $\Sp \Gamma_0$. And similarly the ordinary Fourier transform
$\widetilde{\phi}$ of the square integrable solution $\phi$ of the equation
\[
i\Gamma_0 \partial_0 \phi + i\Gamma_1 \partial_1 \phi + i\Gamma_2 \partial_2 + i\Gamma_3 \partial_3 \phi
= im \phi, \,\,\, m \in \mathbb{R}
\]
is concentrated on the set theoretical sum of orbits $\mathscr{O}_{(0,0,\frac{m}{\lambda},0)}$,
with $\lambda$ ranging over $\Sp i\Gamma_k$, with $k$ having one of the three possible
values; 1,2,3. As the spectra of $\Gamma_0$ and $i \Gamma_k$ are all equal $\{1,-1 \}$
we obtain the generalized spectral decomposition of the Dirac operator $D$ in full analogy with
the previous Subsections. It follows that (\ref{repGamma}) defines
its *-representation. Because any *-representation of the Clifford algebra corresponding to Minkowski metric
generated by (\ref{repGamma}) is a direct sum of finite dimensional representations, we arrive at the contradiction
because $V$ contains infinite dimensional sub representation. Only in case $V$ can be decomposed into finite dimensional representations (or at least $V$ has a sub representation which can be so decomposed)
the construction can be realised, and we
would have therefore obtained the
Dirac operator $D = i \Gamma_0 \partial_0 + \ldots i \Gamma_3 \partial_3$ in full analogy with the preceding
Subsections, which meets all relevant conditions.

{\bf REMARK 1}.
Concerning our previous paper \cite{Wawr}, we have outlined
the general strategy for the construction of $V_\mathcal{F}$ motivated by harmonic analysis
on homogeneous spaces which are manifolds with ordinary Riemannian metrics -- we have repeated
it in the Introduction to this work. In the case of homogeneous Riemannian manifolds we
considered the algebra $\widehat{\mathcal{A}}$ of (Schwartz) functions of generators $P_0, \ldots P_k$ of commuting one
parameter subgroups of the Lie group acting on the Riemannian manifold, represented on the Hilbert space of square
summable functions on the manifold, and the algebra
$\mathcal{A} = V_\mathcal{F} \widehat{\mathcal{A}} {V_\mathcal{F}}^{-1}$ of ''Fourier transforms''
of the elements of $\widehat{\mathcal{A}}$, where a general Fourier transform on the homogeneous Riemann manifold is used (slightly reformulated in the spirit of Conne-type spectral format). This situation, although preserves the general similarity with our situation
in free QFT, is considerably simpler, regarding the analysis aspect, but concerning the algebraic aspect our situation in QFT is simpler. Namely, in the spectral construction of space-time in the corresponding invariant subspace of the Flock space, the generators of the algebra $\mathcal{A}$ are the
operators $Q^0, \ldots, Q^3$ which together with $P^0, \ldots, P^3$ compose the standard von Neumann
representation of the canonical system of pairs of operators $Q^i, P^i$, acting with finite uniform multiplicity, and thus the construction of $\mathcal{A}$ as the Schwartz functions
of the operators $Q^0, \ldots, Q^3$, is essentially reduced to the Abelian harmonic analysis. The explicit construction of
the corresponding generators on a curved Riemannian manifold is not so easy (in that case $Q^0, \ldots, Q^3$
are the commuting operators simultaneously diagonalized by the general Gelfand-Graev Fourier transform
on the homogeneous Riemannian manifold acting in the Hilbert space of sections of the corresponding Clifford bundle, which we need in order to write the Fourier transform and its inverse purely spectrally in terms of spectra of the operators $P^i, Q^i$, compare \cite{Wawr}).
Concerning analysis our present situation is more complicated.
Namely, we have to check if the sub representation
concentrated on the forward and backward cone (in the spectrum of translation generators) is ''associated''
to the sub representation concentrated outside the set-theoretical sum of back- and forward-cones
in the joint spectrum of translation generators. The second additional complication is that we have
homogeneous pseudo-Riemannian manifold instead of Riemannian, which introduces analytic complications,
namely unbounded and non-unitary character of the transform
$V_\mathcal{F}$.

{\bf REMARK 2}.
The modification of the sub representation $U^{+-}$, concentrated on the one-sheet hyperboloid orbits outside the light cone in the momentum space, of the tensor product representation $U$, ultimately has in our opinion
not merely a technical character.
The modification leaves unchanged the physical states and is essentially uniquely determined by the sub representation
acting on ''physical states''
concentrated on the orbits lying inside the light cone. In fact, it means that
$\mathbb{L}$ should be replaced with a representation which decomposes into finite dimensional representations
(or that $\mathbb{L}$ should be extendable to a representation $V$ of $SL(2, \mathbb{C})$, which decomposes into finite dimensional representations). In fact, it is the simplification of the decomposition problem (avoiding explicit solution of the Problems (A) and (B)) as well as the spectral characterization of Connes of the manifold, which stand behind our choice. In his spectral characterization the module $\cap_{m}\Dom D_\mathfrak{J}^m$ finite and projective over the algebra
$\mathcal{A}$ of coordinate functions, and the representation $(\mathcal{A}, \mathcal{H})$ of $\mathcal{A}$ in the corresponding Hilbert space $\mathcal{H}$ is such that its double commutor $(\mathcal{A}, \mathcal{H})''$ has finite uniform multiplicity.
On the other hand there are spectral triples which likewise characterize smooth manifolds with arbitrary high, and even
infinite, multiplicity with the module $\cap_{m}\Dom D_\mathfrak{J}^m$ projective but infinite (e.g. those constructed in the previous Section). Of course the infinite character of the module $\cap_{m}\Dom D_\mathfrak{J}^m$ and of
the multiplicity of $\mathcal{A}''$ of these examples is somewhat trivial, for there are invariant sub spaces on which they are both finite, but there is no \emph{a priori} reason to exclude the possibility of characterizing smooth manifolds spectrally but with the use of bundles with infinite dimensional fibbers naturally connected with the tangent bundle.
It is rather tempting that Connes' spectral characterization theorem for smooth manifolds is only a (fundamental) example of an infinite family of possible spectral characterizations in which the module $\cap_{m}\Dom D_\mathfrak{J}^m$ is infinite
(although projective) with $\mathcal{H}$ containing infinite dimensional invariant sub spaces $\mathcal{H}_{inv}$ on which
$\cap_{m}\Dom (D_\mathfrak{J}|_{{}_{\mathcal{H}_{inv}}})^m$ is projective but infinite
and with $(\mathcal{A}, \mathcal{H}_{inv})''$ of uniform but infinite multiplicity. Of course the conditions characterizing the manifold spectrally will have to be respectively modified: the crucial part plays the presence of invariant sub spaces on which the Connes conditions are preserved with the finiteness conditions maintained.

For technical reasons we have chosen firstly a simplified situation (in order to make more clear the construction of the spectral construction of the space- time manifold out of the double covering $T_4 \circledS SL(2, \mathbb{C})$ of the
Poincar\'e group representation, especially the translation generators) with the original Connes' spectral characterisation theorem. In this way we are forced to stay within representations $\mathbb{L}$ associated with the representation $L$ in (\ref{T^+oplusT^-}) with the corresponding extension $V$ to a representation
of $SL(2, \mathbb{C})$ decomposing into finite dimensional sub representations. 

On the other hand having the extension $V$ associating $L$ in (\ref{T^+oplusT^-}) with $\mathbb{L}$ in
(\ref{T^{+-}}) we obtain the wave function $\phi$ with values in the space of the representation $V$
with the local transformation formula (\ref{infinite-multispinor-rep}). Now by assumption
$\phi$ and (\ref{infinite-multispinor-rep}) decompose into wave functions with finite-dimensional-valued components
transforming under finite dimensional sub representations of the representation $V$ of $SL(2, \mathbb{C})$.
We obtain all possible states (orthogonal to the vacuum and single particle states) of the free fields under
consideration recognizable as
free particle states with the spin structure inscribed by the sub representations of $V$ of $SL(2, \mathbb{C})$.
It is tempting to assume that only the states with finite-dimensional-valued wave functions $\phi$
are physically relevant, but it is not really the case.

{\bf REMARK 3}.
There are infinite dimensional manifolds connected with abstract Bose-Fermi fields
with the associated K\"ahler-Dirac-type operators connected with them \cite{Asao1}, \cite{Asao2},
\cite{Jaffe-Les-Wei}.
However, our elliptic-type Dirac operator $D_\mathfrak{J}$ (and the associated Dirac operator $D$)
has nothing to do with them. Our construction of the Dirac operator is based on the intimate relation
of the free fields with
the space-time manifold, which is finite dimensional. Our Dirac operator is connected with finite dimensional 
(space-time) manifold as in spectral geometry with $(\mathcal{A}, \mathcal{H}_{inv},\mathfrak{J}, D, D_{\mathfrak{J}})$, regarded as operators 
acting on the respective invariant subspace $\mathcal{H}_{inv}$
of the Fock space.

\subsection{Perturbation of the space-time spectral triple}\label{perturbed(A,H,D)}

The perturbation of the undeformed spectral tuple
$(\mathcal{A}, \mathcal{H}_{inv},\mathfrak{J}, D, D_{\mathfrak{J}})$ of space-time,
should in principle preserve the invariance property. This means that there exists an invariant
subspace $\mathcal{H}_{inv}$ such that at every order of
perturbation of the perturbed spectral triple
\begin{equation}\label{ConnesSpTr}
\big(\mathcal{A}^{+}|_{{}_{\mathcal{H}_{inv}}} \supset
\mathcal{A}|_{{}_{\mathcal{H}_{inv}}}, \, \mathcal{H}_{inv}, \, (QD_{\mathfrak{J}} +V)|_{{}_{\mathcal{H}_{inv}}} \big)
\end{equation}
(corresponding to the perturbed spectral tuple
$(\mathcal{A}, \mathcal{H}_{inv},\mathfrak{J}, D, D_{\mathfrak{J}})$)
preserves the
(strong) version of the five axioms of \cite{Connes_spectral}. This is expected
because the causal perturbation series for interacting fields preserves (in the adiabatic limit) transnational covariance.
Indeed, the Dirac operators $D$ and $D_{\mathfrak{J}}$ of the undeformed space-time ''spectral tuple'',
are build of the translation operators $\boldsymbol{P}^\mu = d\Gamma(P^\mu)$
and the fundamental symmetry operator $\mathfrak{J}$, restricted to the invariant subspace.
For example the Dirac operator $D$ is the restriction to the invariant subspace of the linear combination of the translation operators $\boldsymbol{P}^\mu = d\Gamma(P^\mu)$, with components equal to the constant matrix elements of the generators of the Clifford algebra of the Minkowski metric, as shown above. To the construction of $D_{\mathfrak{J}}$ there enters in addition the corresponding Krein fundamental symmetry $\mathfrak{J}$, commuting with $\boldsymbol{P}^\mu = d\Gamma(P^\mu)$, and constructed as in the previous Subsection. The algebra $\mathcal{A}$ is likewise immediately related to the translation generators, as it is the algebra of Schwartz functions of the operators $Q^0, \ldots Q^3$,
which together with the translation operators $P^0, \ldots P^3$ compose the von Neumann representation of canonical pairs of operators acting with uniform infinite multiplicity, and when restricted to the invariant subspace, the von Neumann representation of canonical pairs $Q^\mu, P^\mu$ acts with finite multiplicity
(one can think of the relation between $P^\mu$-s and $Q$-s in their actions on the invariant subspace as arising from the ordinary Fourier transform of the elements of the invariant subspace, which makes sense because the joint spectrum of the translation generators
on the invariant subspace is the smooth Minkowski space when we consider their action in the subspace of the
composite system of free fields of both signs of the energy, orthogonal to the vacuum and single particle sub spaces). Thus, the elements of $\mathcal{A}$ are Schwartz functions
of the operators $V_\mathcal{F} \boldsymbol{P}^\mu {V_\mathcal{F}}^{-1}$.
The operators $Q, V$ in (\ref{ConnesSpTr}) are the ''scaling'' and the ''potential''
operators of the preceding Subsection, affiliated to the operators $Q^0, \ldots, Q^3$, and thus also are immediately related to the translation generators. On the other
hand in constructing deformation of $(\mathcal{A}, \mathcal{H}_{inv},\mathfrak{J}, D, D_{\mathfrak{J}})$,
we use the relation between the translation generators $\boldsymbol{P}^\mu = d\Gamma(P^\mu)$
expressed by the Wick-polynomial $: T^{0\mu}(x_0, \boldsymbol{\x}) :$ of free fields through the \emph{Bogoliubov-Shirkov Quantization Postulate}:
\[
\int : T^{0\mu}(x_0, \boldsymbol{\x}) : \, \ud^3 \boldsymbol{\x} = \boldsymbol{P}^\mu = d\Gamma(P^\mu).
\]
We give a rigorous formulation of this \emph{Postulate} and its proof in Section \ref{free-gamma}.
The second important observation is that we can, on the same footing as for the expression for
$\boldsymbol{P}^\mu = d\Gamma(P^\mu)$ in the
Bogoliubov-Shirkov Postulate, give a rigorous sense to each order term of the causal perturbative
series for interacting fields in the adiabatic limit $g(x) = 1$. Thus, in principle at least, we can
compute the perturbative series for the translation generators, expressed through the Bogoliubov-Shirkov Postulate, in terms of Wick polynomials of free fields, by replacing the
Wick polynomial field $: T^{0\mu}(x_0, \boldsymbol{\x}) :$ in the expression for $\boldsymbol{P}^\mu = d\Gamma(P^\mu)$ with the corresponding
interacting field $\big( : T^{0\mu}(x_0, \boldsymbol{\x}) :\big)_{{}_{\textrm{int}}}$ expressed in terms of the causal perturbative series. In particular $: T^{0\mu}(x_0, \boldsymbol{\x}) :$, when integrated over Cauchy surface $x_0 = const.$ gives an operator commuting with translation generators.
Because the causal perturbative series for the chronological product
is transitionally covariant, then we expect of
\begin{equation}\label{perturbedP}
\int \big( : T^{0\mu}(x_0, \boldsymbol{\x}) :\big)_{{}_{\textrm{int}}} \, \ud^3 \boldsymbol{\x}
\end{equation}
to become, order-by-order, transitionally invariant in the adiabatic limit $g(x)=1$, i.e. commuting with the translation generators. The most nontrivial part lies in giving the meaning to each finite order term
of approximation in this expression (for the value
of the coupling $g(x)$ equal $1$) of a well-defined self adjoint operator, compare Subsection \ref{BSH}
for the zero order. The general analysis of the higher order contributions to interacting fields
and their spatial integrals
as well-defined integral kernel operators is provided in Subsection \ref{OperationsOnXi}.
Indeed, writing, just for simplicity, the Wick polynomial field (with a fixed $\mu$)
$: T^{0\mu}(x_0, \boldsymbol{\x}) :$ just as $A(x_0, \boldsymbol{\x})$ we have (\cite{DKS1}, or \cite{DutFred} formula (2.8))
\begin{multline*}
\big(A(g=1;x) \big)_{{}_{\textrm{int}}} = A(x) \\
+ \sum \limits_{n=1}^{\infty} \frac{i^n}{n!}
\int d^4 x_1 \ldots d^4 x_n \, R(\mathcal{L}(g=1; x_1) \cdots \mathcal{L}(g=1; x_n); A(x)),
\end{multline*}
with \emph{totally retarded products} $R(\ldots)$ \cite{DKS1} or \cite{DutFred}.
By the translational covariance of the integrand
in this expression in all spacetime variables $x_1, \ldots, x_n, x$, translational invariance of $d^4 x_i$
and translational
invariance of $A(x_0, \boldsymbol{\x})$ when $\ud^3 \boldsymbol{\x}$-integrated over the surface
$x_0 = const.$ (one should remember that the most subtle point lies in giving a strict sense to these expressions for $g=1$ integrated over $x_0 = const.$ as well-defined self adjoint operators, compare Subsection \ref{BSH} where the zero order approximation term is provided in details, i.e. \emph{Bogoliubov-Shirkov-Hypothesis}) we obtain the translational invariance (commutativity with $\boldsymbol{P}^\mu$) of $\ud^3 \boldsymbol{\x}$-integrated field operator
$\big(A(g=1;x_0, \boldsymbol{\x}) \big)_{{}_{\textrm{int}}} = \big( : T^{0\mu}(x_0, \boldsymbol{\x}) :\big)_{{}_{\textrm{int}}}$.

We apply the perturbation to the elements of the algebra, just by substitution
of the perturbed expression for $\boldsymbol{P}^\mu$ in the formula
$V_\mathcal{F} \boldsymbol{P}^\mu {V_\mathcal{F}}^{-1}$, with the ''Fourier transform''
$V_\mathcal{F}$ keeping unchanged. In particular the higher order correction terms in the formula
(\ref{perturbedP}), i.e. higher order corrections to the operators defining the spectral tuple, should, in principle at least, preserve the invariant sub spaces $\mathcal{H}_{inv}$ of the initial undeformed spectral tuple
$(\mathcal{A}, \mathcal{H}_{inv},\mathfrak{J}, D, D_{\mathfrak{J}})$. In particular the perturbation terms to the Dirac operators should be functions of the unperturbed Dirac operators, and similarly the perturbation terms
to the elements of the perturbed algebra $\mathcal{A}$ should be functions of the elements of the
unperturbed algebra $\mathcal{A}$. This means that all five conditions for the commutative spectral triple
of Connes \cite{Connes_spectral}, together with the additional condition of antisymmetricity of the Hochschild cycle and uniform multiplicity of
the action of (the weak closure of) of $\mathcal{A}^{+}$ (containing $\mathcal{\mathcal{A}}$
as an essential ideal, compare Appendix \ref{AppendixNonCompMani})
on $\mathcal{H}_{inv}$, \cite{Connes_spectral},
p. 3, are preserved for the spectral triple (\ref{ConnesSpTr}) at each order.
This would also mean that the perturbation of
$(\mathcal{A}, \mathcal{H}_{inv},\mathfrak{J}, D, D_{\mathfrak{J}})$ is stable
(with the coupling function $g$ in their formal power series
equal to a constant, compare Introduction). But this is true only
under the following prescription I) mentioned in Introduction, Subsection \ref{G}:
\begin{enumerate}
\item[I)]
When computing $\big(:T^{0\mu}:\big)_{{}_{\textrm{int}}}(g=1, x_0, \boldsymbol{\x})$ we insert literally
the free field expression for $:T^{0\mu}:$, which does not include the interaction Lagrangian term
$\mathcal{L}$ in the formula for $T^{0\mu}$ taken from the Noether theorem for classical free fields.
\end{enumerate}

But, as we already indicated in Introduction, Subsection \ref{G}, computation of
$\boldsymbol{P}^{\mu}_{{}_{\textrm{int}}}$ should be rather based on the following
prescription II):
\begin{enumerate}
\item[II)]
The interaction Lagrangian term $\mathcal{L}$ understood as a Wick polynomial in
free fields is included into the full Lagrangian $\mathcal{L}_{{}_{\textrm{Full}}}$ (expressed as a Wick polynomial in free fields) when computing Wick polynomial $:T^{0\mu}:$. Thus, $:T^{0\mu}:$ is put equal
to the Wick formal counterpart of the classical expression for $T^{0\mu}$ in the Noether theorem for classical interacting fields, in which the classical fields are replaced by free fields and their product replaced formally by the Wick product of free fields.
Such Wick polynomial $:T^{0\mu}:$ including the Wick interaction term is then inserted into
the Bogoliubov-Shirkov perturbative formula
\[
\big(:T^{0\mu}:\big)_{{}_{\textrm{int}}}(g=1,x) = -
\frac{\delta}{i\delta h(x)}S\big(\mathcal{L}+h:T^{0\mu}:\big)^{-1}S(\mathcal{L})\big|_{{}_{h=0}}.
\]
\end{enumerate}
The second prescription II) is in agreement with the general definition of the product of interacting fields
within the causal perturbative approach, compare \cite{Bogoliubov_Shirkov}, \cite{DKS1},
\cite{DutFred}. It moreover seems that this is the formula for
\[
\boldsymbol{P}^{\mu}_{{}_{\textrm{int}}} =
\int \big(:T^{0\mu}:\big)_{{}_{\textrm{int}}}(g=1, x_0, \boldsymbol{\x}) \, \ud^3 \boldsymbol{\x}
\]
in case II) which gives the correct expression for the conserved integrals,
which corresponds to the classical expression given by the Noether theorem, and is constructed with the same rules which produce the equations of motion for interacting fields, compare \cite{Bogoliubov_Shirkov}, \cite{DKS1}, \cite{DutFred}.

As we have already indicated in Introduction, Subsection \ref{G},
in the II)-nd case the contributions to
\[
\boldsymbol{P}^{\mu}_{{}_{\textrm{int}}} =
\int \big(:T^{0\mu}:\big)_{{}_{\textrm{int}}}(g=1, x_0, \boldsymbol{\x}) \, \ud^3 \boldsymbol{\x}
\]
$\boldsymbol{P}^{\mu}_{\textrm{int}}$ are in general not ordinary operators in the Fock space but keep the meaning of well-defined generalized operators transforming Hida space into its strong dual (this difficulty is pertinent to the contributions to the zero component $\boldsymbol{P}^{0}_{\textrm{int}}$) and correspond to the well known difficulty in construction of the interacting hamiltonian for relativistic interacting fields as a well-defined operator in the Fock space of free fields. In particular in case of QED (of course here on Minkowski space-time) the higher order contributions to
$\boldsymbol{P}^{0}_{\textrm{int}}$ are not ordinary operators but generalized operators transforming Hida space into its strong dual. But for some other QFT, such as scalar
field $\varphi$ with $:\varphi^q:$ interaction with sufficiently high and even integer
power $q$ the higher order contributions to $\boldsymbol{P}^{0}_{\textrm{int}}$ are ordinary (densely defined) operators on the Fock space. Similar conclusions have been arrived at by I. E. Segal and his coworkers with the help of different methods, compare \cite{SegalZhouQED},
\cite{SegalZhouPhi4}, \cite{PedersenSegalZhou}.

The proof (or even the general analysis of contributions to interacting fields
in the adiabatic limit $g=1$) of these statements
the reader will find in Subsection \ref{OperationsOnXi} and in Section \ref{A(1)psi(1)}.

These negative results (based on the physical prescription II)), which tell that the contributions
to $\boldsymbol{P}^{\mu}_{{}_{\textrm{int}}}$ are not ordinary operators in the Fock space, 
we interpret as a proof that realistic interacting fields,
such as the Dirac and electromagnetic potential fields, with realistic interaction based on the minimal
coupling with $U(1)$-gauge group, cannot live on flat Minkowski space-time, and thus do necessary possess
the gravitational weight, compare Introduction, Subsection \ref{G}.

\section{The representation of $T_4 \circledS SL(2, \mathbb{C})$ in the Hilbert 
space of the free quantum Dirac field. White noise construction. Bogoliubov Postulate}\label{e+e-}

Here we present the construction of quantized free Dirac field, concentrating
mostly on the representation of $T_4 \circledS SL(2, \mathbb{C})$ acting in the Hilbert space
of the field, because this aspect is ignored in the literature.
The second point, so far not presented in the literature, which we undertake here is the
construction of the free quantized Dirac field within the white noise setup of Hida, Obata and
Sait\^o \cite{hida}, \cite{obata-book},
and which is a rigorous realization of the field along the lines suggested (partially heuristically) by
Berezin \cite{Berezin}. White noise construction due to Berezin-Hida, can be regarded as a far-reaching
extension of the definition due to Wightman
\cite{wig} of the (free) field, entering into the analysis of the distributional
(generalised) states.
We should emphasise here that the definition of Wightman is operationally and computationally
much weaker. In general the two definitions are not equivalent.
The main advantage we gain when constructing free fields within the white noise formalism
is that we can give a rigorous meaning to the (free) quantum field
of the so called \emph{integral kernel operator
with vector-valued distributional kernel} (in the sense
\cite{obataJFA} or \cite{obata-book}, Chap. 6.3), which would be impossible within Wightman setup.
This allows to give the meaning of integral kernel operators (with vector-valued kernels)
to the (generalized) operators under the formula (17.1) in \cite{Bogoliubov_Shirkov}, p. 154, or
equivalently to the (generalized) operators (43) of \cite{Epstein-Glaser}, Sect. 4, p. 229.
In particular when constructing free fields according to Berezin-Hida we obtain
Theorem 0 of \cite{Epstein-Glaser} as a corollary to theorems 2.2 and 2.6 of
\cite{hida} and Thm. 3.13 of \cite{obataJFA} with the domain $\mathcal{D}_0$
replaced with the so called Hida test space of white noise functionals.
Moreover, using the Berezin-Hida construction of free fields we gain a rigorous formulation and proof of
the so called ''Wick theorem'',
as stated in \cite{Bogoliubov_Shirkov}, Chap. III. It should be emphasised that Wightman's
definition of the (free) field \cite{wig}, does not provide sufficient computational basis
for any rigorous formulation and proof of the ''Wick theorem'' for free fields as stated in \cite{Bogoliubov_Shirkov}, Chap. III.
Note also that the (free) field constructed within the white noise calculus is well defined at
space-time point as a generalized operator transforming the so called Hida space into its strong dual.

One should note that although the definition of the ``Wick product'' of Wightman and G{\aa}riding
\cite{WightmanGarding} based on the Wightman's definition \cite{wig} of the field, is mathematically rigorous,
it suffers at several crucial points from being computationally ineffective in computations which are important 
from the physical point of view:
\begin{enumerate}
\item[1)] 
The space-time averaging limits
in  Wightman and G{\aa}rding's
\cite{WightmanGarding} definition of the ``Wick product'' are by no means 
canonical and involve a considerable amount
of arbitrariness. 

\item[2)]
Although Wightman and G{\aa}rding
\cite{WightmanGarding} are able to construct their own ''Wick products'' which, after smearing out over space-time domains becomes well-defined densely defined unbounded operators, it would be difficult to investigate the closability questions
for these operators, their eventual self-adjointness,
as well as averaging over space-like (equal-time) surfaces, within the method of Wightman and G{\aa}rding.
But the equal-time averagings are involved through conserved currents when
we consider Noether theorem for free fields -- fundamental from the more conventional, and used by physicists, approach to commutation rules and the more traditional proof of the Pauli theorem for free fields
(compare \cite{Bogoliubov_Shirkov}).

\item[3)]
Wightman and
G{\aa}rding definition of the ''Wick product'' \cite{WightmanGarding} is not a sufficient basis for
the strict formulation
and proof of the ''Wick theorem'' as stated in \cite{Bogoliubov_Shirkov}, Chap. III, so fundamental for the causal approach to QFT which avoids ultraviolet divergences. Note in particular that Theorem 0 of
\cite{Epstein-Glaser} is formulated and proved on the basis of partially heuristic (but solid)
arguments of the more traditional approach presented in \cite{Bogoliubov_Shirkov}, Chap. III, which
uses the free fields at specified space-time points in the intermediate stage, and which are not merely
symbolic in their character (contrary to what we encounter in the Wightman-G{\aa}rding's approach).
White noise construction of free fields on the other hand does provide a sufficient basis for the
rigorous formulation and proof of ''Wick theorem'' for free fields of \cite{Bogoliubov_Shirkov}, Chap. III.

\item[4)]
But most of all when constructing free fields using the white noise formalism, as integral kernel operators with vector-valued kernels, we are able to give a rigorous
meaning to each order term contribution to interacting fields in QED (within the causal perturbative approach), of an integral kernel operator with vector-valued distribution kernel (in the sense \cite{obataJFA}),
which defines a well defined operator valued distribution on the
space-time test space -- a continuous map from the space-time test space
to the linear space of continuous linear operators on the Hida space into its dual
(with the standard topology of uniform convergence on bounded sets).
Each such contribution can be averaged in the states of the Hida subspace and defines a scalar distribution
as a functional of space-time test function.
The crucial point is that these contributions do not loose this rigorous
sense even for the ''coupling space-time function $g$'' put everywhere equal to unity, which allows to avoid both: ultraviolet and infrared infinities in the perturbative (causal) approach to QED.
For a detailed proof of this assertion and analysis of the all higher order contributions to the Dirac and
electromagnetic potential interacting fields, compare Subsection \ref{OperationsOnXi},
Sect. \ref{A(1)psi(1)}. In particular we can
reach in this way a positive solution to the existence problem for the adiabatic limit in QED using a method which is applicable to interactions and fields of more general character, e.g. to the Standard Model.

\end{enumerate}

For these reasons we regard the white noise construction of (free) fields of Berezin-Hida
as integral kernel operators (with vector-valued distributional kernels) as more adequate
mathematical interpretation of the (free) quantum field than the one proposed by Wightman \cite{wig}.

In this Section we present white noise Berezin-Hida construction of the free Dirac field
as an integral kernel operator with  vector-valued distributional kernel in the sense of Obata
\cite{obataJFA}. In the next Section
we give the white noise construction of the free electromagnetic potential field,
which again may be interpreted as integral kernel operator 
with vector-valued distributional kernel in the sense of Obata \cite{obataJFA}.  
   We present the construction of the Dirac field $\boldsymbol{\psi}$
in several steps, keeping the presentation as general as possible, in order to make it to serve as 
an introduction to the construction of (free) local fields within the white noise formalism. 

First, we give definition of the Hilbert space which is subject to second
quantization functor, and then in the remaining four steps quantize it.
The steps are realized in the following Subsections: \ref{FirstStepH},
\ref{electron}, \ref{positron}, \ref{electron+positron}, \ref{psiBerezin-Hida}.
Subsection \ref{psiBerezin-Hida} is the longest, but it contains an introduction to the papers
\cite{hida}, \cite{obataJFA} on integral kernel operators with scalar-valued and respectively
vector-valued distributional kernels in Fermi and Bose Fock spaces (note that \cite{hida}, \cite{obataJFA}
give detailed analysis for the Bose case), which is of use in the remaining part
of the whole work, and which is not so much pertinent to the specific Dirac field
$\boldsymbol{\psi}$, but which is
important for general local fields constructed within the white noise calculus.
In particular, we are using the cited theorems of \cite{hida}, \cite{obataJFA} on integral kernel operators
in the proof of \emph{Bogoliubov-Shirkov Hypothesis} (equivalently the classic Pauli theorem)
for the Dirac field $\boldsymbol{\psi}$ (Subsection \ref{StandardDiracPsiField}) and
for the electromagnetic potential field (Subsection \ref{BSH});
and finally in the analysis of contributions to interacting fields in QED
(Subsection \ref{OperationsOnXi}).

Subsection \ref{OperationsOnXi} is devoted to the proof that the contributions to interacting
fields in causal perturbative spinor QED are well-defined integral kernel operators with vector-valued kernels
in the sense of Obata \cite{obataJFA} whenever we are using in the causal construction of interacting fields
the free fields which themselves are well-defined integral kernel
operators in the sense of Obata. Nonetheless, the Subsection \ref{OperationsOnXi} is of more general
character not pertinent to the special case of spinor QED. It is devoted to
the fundamental operations performed upon the free fields, understood as integral kernel operators with vector-valued kernels,
which serve as fundamental computational rules in construction of the theory, in particular in construction
of the perturbative series for interacting fields such as: Wick product of free fields, derivation and integration
operations. These operations have general character and can be extended over other causal perturbative QFT. 

We add two additional Subsections \ref{psiWightman}
and \ref{StandardDiracPsiField}. Subsection \ref{psiWightman} gives a motivation for using white noise
calculus and for using the construction of fields due to Berezin-Hida, as integral kernel operators with vector-valued kernels. The Subsection \ref{StandardDiracPsiField} contains comparison with the standard realization of the free Dirac field and is devoted to the Bogoliubov-Shirkov Postulate (first Noether theorem for free fields
and the classic Pauli theorem on spin-statistics relation).

In this Section $m > 0$ has the constant value equal to the electron mass.

In Subsections which are to follow we construct the free Dirac field based on the unitary representation:
\begin{multline*}
U^{{}_{m,0,0,0}\big(2L^{{}^{1/2}}\big)} = U^{{}_{m,0,0,0}\big(L^{{}^{1/2}}\big)} \oplus U^{{}_{m,0,0,0}\big(L^{{}^{1/2}}\big)}
\\
=
U^{{}_{m,0,0,0}\big(L^{{}^{1/2}}\big)} \oplus \Big(U^{{}_{-m,0,0,0}\big(L^{{}^{1/2}}\big)}\Big)^\flat, \,\, m>0,
\end{multline*}
or rather its unitarily equivalent local version (for definition of conjugation $(\cdot)^\flat$, compare Subsection \ref{positron})
acting in the single particle Hilbert space.
Construction of the unitarily equivalent local version is based on the analysis of Subsection \ref{e1}.
In Subsection \ref{StandardDiracPsiField} we compare this standard Dirac field with a non-standard Dirac field, based on
a local representation which is not unitary.

In this Section we are using $\tilde{\gamma}^\mu = (\gamma^\mu)^* = \gamma^0 \gamma^\mu\gamma^0$
in the space-time picture, and $\gamma^\mu$ in the momentum picture. This is because we are dealing mainly in the momentum picture
with a considerable number of various conjugations, so in order to simplify notation we wanted to eliminate the
additional tilde $\tilde{(\cdot)}$. Therefore, Dirac equation in spacetime variables reads
\[
i\tilde{\gamma}^\mu \partial_\mu \phi = m \phi, \,\,\,\, \tilde{\gamma}^\mu = (\gamma^\mu)^* = \gamma^0 \gamma^\mu\gamma^0,
\]
with the Clifford gamma generators in momentum picture equal $\gamma^\mu$ and, say in the chiral representation, are given by the formula
(\ref{chiralgamma}).

\subsection{Definition of the Hilbert space $\mathcal{H}$ which is then subject to the second quantization
functor $\Gamma$}\label{FirstStepH}

This is the Hilbert space $\mathcal{H}$ of bispinor solutions $\phi$ (regular function-like distributions
on the Schwartz space $\mathcal{S}(\mathbb{R}^4; \mathbb{C}^4)$ of testing bispinors transforming according to the law (\ref{1/2xUa})) of the Dirac equation
\[
\begin{split}
(i (\gamma^\mu)^* \partial_\mu ) \phi = m \phi, 
\\ 
(\gamma^\mu)^* = \tilde{\gamma}^\mu = \gamma^0\gamma^\mu \gamma^0,
\\
\tilde{\gamma}^0 = (\gamma^0)^* = \gamma^0, \tilde{\gamma}^k = (\gamma^k)^* = -\gamma^k, \,\,\, k=1,2,3,
\end{split}
\]
with the inner product
\begin{equation}\label{Inn-Prod-Single-Dirac}
(\widetilde{\phi}, \widetilde{\phi'}) = 2m \int \limits_{x^0 = const.} \Big(\phi(x), \phi'(x) \Big)_{{}_{\mathbb{C}^4}}
\, \ud^3 x,
\end{equation}
and transformation law (\ref{1/2xUa}), compare e.g. \cite{Scharf} or \cite{Bogoliubov_Shirkov}.
This means that the Fourier transform $\widetilde{\phi}$ of the bispinor $\phi \in \mathcal{H}$ (regular distribution)
is concentrated on the disjoint sum of the positive and negative energy orbits $\mathscr{O}_{m,0,0,0} \sqcup
\mathscr{O}_{-m,0,0,0}$ and $\widetilde{\phi}$ cannot be regarded as ordinary function on the full range of
$p \in \mathbb{R}^4$ of the momentum space. Nonetheless, $\widetilde{\phi}$ is a well-defined
(singular, i.e. non-function-like) distribution in the
Schwartz space
\[
\mathcal{S}(\mathbb{R}^4; \mathbb{C}^4) = \mathcal{S}(\mathbb{R}^4; \mathbb{C})
\oplus \mathcal{S}(\mathbb{R}^4; \mathbb{C}) \oplus \mathcal{S}(\mathbb{R}^4; \mathbb{C}) \oplus
\mathcal{S}(\mathbb{R}^4; \mathbb{C})
\]
of bispinors on $\mathbb{R}^4$ (transforming according to (\ref{1/2pUalpha}) and (\ref{1/2pUa})).
It defines an ordinary bispinor-function $p \mapsto \widetilde{\phi}(p)$ on the disjoint sum $\mathscr{O}_{m,0,0,0} \sqcup \mathscr{O}_{-m,0,0,0}$ of the positive and respectively negative energy orbits, which we denote likewise by the
symbol $\widetilde{\phi}$ (although it makes sense as a function only on the disjoint sum of the respective orbits and not on the whole $\mathbb{R}^4$ space), and which is square integrable with respect to the inner product
(\ref{sprod-bisp-orb}) induced by the above inner product (\ref{Inn-Prod-Single-Dirac}) in $\mathcal{H}$.
Moreover, that part $\widetilde{\phi}\big|_{{}_{\mathscr{O}_{m,0,0,0}}}$ of $\widetilde{\phi} = \widetilde{\phi}\big|_{{}_{\mathscr{O}_{m,0,0,0}}} + \widetilde{\phi}\big|_{{}_{\mathscr{O}_{-m,0,0,0}}}$, $\phi \in \mathcal{H}$, which is concentrated on the positive energy orbit $\mathscr{O}_{m,0,0,0}$ is not
a generic bispinor but a bispinor lying in the image of the projection $P^\oplus$ constructed in Subsection \ref{e1}. Similarly,
that part $\widetilde{\phi}\big|_{{}_{\mathscr{O}_{-m,0,0,0}}}$ of $\widetilde{\phi} = \widetilde{\phi}\big|_{{}_{\mathscr{O}_{m,0,0,0}}} + \widetilde{\phi}\big|_{{}_{\mathscr{O}_{-m,0,0,0}}}$, $\phi \in \mathcal{H}$, which is concentrated on the negative energy orbit $\mathscr{O}_{-m,0,0,0}$ is not
a generic bispinor concentrated on $\mathscr{O}_{-m,0,0,0}$ but a bispinor lying in the image of the projection $P^\ominus$ constructed in Subsection \ref{e1}. 
Therefore the single particle space is precisely equal to the orthogonal direct sum $\mathcal{H}^{\oplus}_{m,0} \oplus \mathcal{H}^{\ominus}_{-m,0}$ 
constructed in Subsection \ref{e1}. 

Namely, for $\phi \in \mathcal{H}$, the action of the Fourier transform $\widetilde{\phi}$ (regarded as a distribution)
on $\widetilde{f} \in \mathcal{S}(\mathbb{R}^4; \mathbb{C}^4)$ is by definition equal to the integration
of the invariant Krein product of the mentioned function $p \mapsto \widetilde{\phi}(p)$ on the disjoint sum of the orbits 
by the restriction of $\widetilde{f}$ to the disjoint sum $\mathscr{O}_{m,0,0,0} \sqcup \mathscr{O}_{-m,0,0,0}$
along $\mathscr{O}_{m,0,0,0} \sqcup \mathscr{O}_{-m,0,0,0}$ with respect to the invariant measure on
$\mathscr{O}_{m,0,0,0} \sqcup \mathscr{O}_{-m,0,0,0} \subset \mathbb{R}^4$ induced by the
invariant measure $\ud^4 p$ on $\mathbb{R}^4$. 
Thus, by definition of the singular distribution
$\delta(F=0)$, where $F$ is a smooth function on $\mathbb{R}^4$ such that $\textrm{grad} \, F \neq 0$
on the surface $F=0$ (compare \cite{GelfandI}, Chap. III), we have
\begin{multline*}
\langle \phi, f \rangle_{{}_{\textrm{invariant pairing}}} = \phi^\sharp(f) 
=\int \big(\phi(x), \mathfrak{J} f(x)\big)_{{}_{\mathbb{C}^4}} \, \ud^4 x 
= \langle \widetilde{\phi}, \widetilde{f} \rangle_{{}_{\textrm{invariant pairing}}}  
\\
= \widetilde{\phi}^\sharp(\widetilde{f}) 
= \int \big(\widetilde{\phi}(p), \mathfrak{J}_{{}_{\bar{p}}} \widetilde{f}(p)\big)_{{}_{\mathbb{C}^4}} \, \ud^4p 
= \int \delta(p\cdot p -m^2) \, \big(\widetilde{\phi}(p), \mathfrak{J}_{{}_{\bar{p}}} \widetilde{f}(p)\big)_{{}_{\mathbb{C}^4}} \, \ud^4 p 
\end{multline*}
\begin{multline*}
=\int \delta(p\cdot p -m^2) \theta(p_0) \, \big(\widetilde{\phi}(p), \mathfrak{J}_{{}_{\bar{p}}} \widetilde{f}(p)\big)_{{}_{\mathbb{C}^4}} \, \ud^4 p
\\
+ 
\int \delta(p\cdot p -m^2) \theta(-p_0) \, \big(\widetilde{\phi}(p), \mathfrak{J}_{{}_{\bar{p}}} \widetilde{f}(p)\big)_{{}_{\mathbb{C}^4}} \, \ud^4 p 
\end{multline*}
\begin{multline*}
=\int \limits_{\mathscr{O}_{m,0,0,0}} \, 
\big(\widetilde{\phi}(p), \mathfrak{J}_{{}_{\bar{p}}} \widetilde{f}|_{{}_{\mathscr{O}_{m,0,0,0}}}(p) \big)_{{}_{\mathbb{C}^4}}
\, \ud \mu_{{}_{m,0}}(p)
\\
+ \int \limits_{\mathscr{O}_{-m,0,0,0}} \, 
\big(\widetilde{\phi}(p), \mathfrak{J}_{{}_{\bar{p}}} \widetilde{f}|_{{}_{\mathscr{O}_{-m,0,0,0}}}(p) \big)_{{}_{\mathbb{C}^4}}
 \, \ud \mu_{{}_{-m,0}}(p),
\end{multline*}
\begin{multline*}
=\int \limits_{\mathscr{O}_{m,0,0,0}} \, 
\big(P^\oplus\widetilde{\phi}(p), \mathfrak{J}_{{}_{\bar{p}}} \widetilde{f}|_{{}_{\mathscr{O}_{m,0,0,0}}}(p) \big)_{{}_{\mathbb{C}^4}}
\, \ud \mu_{{}_{m,0}}(p)
\\
+ \int \limits_{\mathscr{O}_{-m,0,0,0}} \, 
\big(P^\ominus\widetilde{\phi}(p), \mathfrak{J}_{{}_{\bar{p}}} \widetilde{f}|_{{}_{\mathscr{O}_{-m,0,0,0}}}(p) \big)_{{}_{\mathbb{C}^4}}
 \, \ud \mu_{{}_{-m,0}}(p),
\end{multline*}
\begin{multline}\label{InvariantBispinorPairing}
=\int \limits_{\mathscr{O}_{m,0,0,0}} \, 
\big(\widetilde{\phi}(p), \mathfrak{J}_{{}_{\bar{p}}} P^\oplus\widetilde{f}|_{{}_{\mathscr{O}_{m,0,0,0}}}(p) \big)_{{}_{\mathbb{C}^4}}
\, \ud \mu_{{}_{m,0}}(p)
\\
+ \int \limits_{\mathscr{O}_{-m,0,0,0}} \, 
\big(\widetilde{\phi}(p), \mathfrak{J}_{{}_{\bar{p}}} P^\ominus\widetilde{f}|_{{}_{\mathscr{O}_{-m,0,0,0}}}(p) \big)_{{}_{\mathbb{C}^4}}
 \, \ud \mu_{{}_{-m,0}}(p).
\end{multline}
Here $\mathfrak{J}$ is the fundamental symmetry defined by the constant matrix operator $\mathfrak{J}_{{}_{\bar{p}}} = \gamma^0$,
compare Subsection \ref{e1}. $(\cdot,\cdot)_{{}_{\mathbb{C}^4}}$ is the standard inner product in $\mathbb{C}^4$, linear conjugate 
in the first variable. Let us emphasize that we are using invariant pairings. This is important in our context, where we have
the representation of $T_4 \circledS SL(2, \mathbb{C})$ acting on the space of states and distributions. 
Note that only in case of the invariant pairings the action
of the representation on the functional (distribution) representable by ordinary (bispinor) function, defined through the ordinary linear transpose,
coincides with the action of the representation on the ordinary (bispinor) functions. The above invariant pairing is well-defined. 
In our case $\widetilde{\phi} = P^\oplus\widetilde{\phi}$
on the orbit $\mathscr{O}_{m,0,0,0}$ and, respectively, $\widetilde{\phi} = P^\ominus\widetilde{\phi}$
on the orbit $\mathscr{O}_{-m,0,0,0}$. Recall that the multiplication rule by a complex number $\alpha \in \mathbb{C}$
of the solution $\phi$ understood as a distribution in $\mathcal{S}(\mathbb{R}^4; \mathbb{C}^4)$ correspond to the multiplication
by $\overline{\alpha}$ of the function $\phi$, representing distribution $\phi$. Note also that the idempotents
$P^\oplus, P^\ominus$ are equal to operators of multiplication by the idempotent matrix functions
$P^\oplus(p), P^\ominus(p)$ on the respective orbits $\mathscr{O}_{m,0,0,0}$ and $\mathscr{O}_{m,0,0,0}$ determined in Subsection
\ref{e1}, compare also Appendix \ref{fundamental,u,v}. Although
$P^\oplus(p) \neq {P^\oplus(p)}^*, P^\ominus(p) \neq {P^\ominus(p)}^*$ with respect to $(\cdot,\cdot)_{{}_{\mathbb{C}^4}}$, we have 
\[
\gamma^0 {P^\oplus(p)}^* \gamma^0 = P^\oplus(p), \,\, \textrm{on} \,\, \mathscr{O}_{m,0,0,0},
\,\,\,\,\,\,\,\,\,\,
 \gamma^0 {P^\ominus(p)}^* \gamma^0 = P^\ominus(p), \,\, \textrm{on} \,\, \mathscr{O}_{-m,0,0,0}
\]
whence the last equality in (\ref{InvariantBispinorPairing}). Similarly, for the Fourier transform of the Dirac operator we have
\[
\begin{split}
\gamma^0 \big[\gamma^\mu p_\mu - m\boldsymbol{1}_{{}_{4}}\big]^* \gamma^0 = \gamma^\mu p_\mu - m\boldsymbol{1}_{{}_{4}}, 
\,\, \textrm{on} \,\, \mathscr{O}_{m,0,0,0},
\\
 \gamma^0 \big[\gamma^\mu p_\mu - m\boldsymbol{1}_{{}_{4}}\big]^* \gamma^0 = \gamma^\mu p_\mu - m\boldsymbol{1}_{{}_{4}}, 
\,\, \textrm{on} \,\, \mathscr{O}_{-m,0,0,0}.
\end{split}
\]
Fourier transform $\gamma^\mu p_\mu - m\boldsymbol{1}_{{}_{4}}$ of the Dirac operator $i(\gamma^\mu)^*\partial_\mu  - m\boldsymbol{1}_{{}_{4}}$
also commutes with the idempotents $P^\oplus(p), P^\ominus(p)$ on the respective orbits, 
whence $\phi$, understood as a distribution $f \mapsto \phi^\sharp(f)$, fulfills Dirac equation:
\begin{multline*}
\langle \phi, \big[i(\gamma^\mu)^*\partial_\mu  - m\big] f \rangle_{{}_{\textrm{invariant pairing}}} =
\phi^\sharp\big(\big[i(\gamma^\mu)^*\partial_\mu  - m\big]f\big)
\\
=\int \limits_{\mathscr{O}_{m,0,0,0}} \, 
\big(\widetilde{\phi}(p), \mathfrak{J}_{{}_{\bar{p}}} P^\oplus(p) \big[\gamma^\mu p_\mu - m\boldsymbol{1}_{{}_{4}} \big] \widetilde{f}|_{{}_{\mathscr{O}_{m,0,0,0}}}(p) \big)_{{}_{\mathbb{C}^4}}
\, \ud \mu_{{}_{m,0}}(p)
\\
+ \int \limits_{\mathscr{O}_{-m,0,0,0}} \, 
\big(\widetilde{\phi}(p), \mathfrak{J}_{{}_{\bar{p}}} P^\ominus(p) \big[\gamma^\mu p_\mu - m\boldsymbol{1}_{{}_{4}} \big]\widetilde{f}|_{{}_{\mathscr{O}_{-m,0,0,0}}}(p) \big)_{{}_{\mathbb{C}^4}}
 \, \ud \mu_{{}_{-m,0}}(p)
\end{multline*}
\begin{multline*}
=\int \limits_{\mathscr{O}_{m,0,0,0}} \, 
\big(\widetilde{\phi}(p), \mathfrak{J}_{{}_{\bar{p}}} \big[\gamma^\mu p_\mu - m\boldsymbol{1}_{{}_{4}} \big] P^\oplus \widetilde{f}|_{{}_{\mathscr{O}_{m,0,0,0}}}(p) \big)_{{}_{\mathbb{C}^4}}
\, \ud \mu_{{}_{m,0}}(p)
\\
+ \int \limits_{\mathscr{O}_{-m,0,0,0}} \, 
\big(\widetilde{\phi}(p), \mathfrak{J}_{{}_{\bar{p}}}\big[\gamma^\mu p_\mu - m\boldsymbol{1}_{{}_{4}} \big] P^\ominus \widetilde{f}|_{{}_{\mathscr{O}_{-m,0,0,0}}}(p) \big)_{{}_{\mathbb{C}^4}}
 \, \ud \mu_{{}_{-m,0}}(p) = 0, 
\\
\,\,\,\,\, f \in \mathcal{S}(\mathbb{R}^4; \mathbb{C}^4),
\end{multline*}
by the algebraic relation which is preserved by the bispinors lying in the image, respectively, of $P^\oplus$ or $P^\ominus$,
concetrated on $\mathscr{O}_{m,0,0,0}$ or $\mathscr{O}_{-m,0,0,0}$, compare Subsection \ref{e1}. The last equality can also be written
in the form 
\begin{multline*}
=\int \limits_{\mathscr{O}_{m,0,0,0}} \, 
\big(\widetilde{\phi}(p), \mathfrak{J}_{{}_{\bar{p}}} \big[\gamma^\mu p_\mu - m\boldsymbol{1}_{{}_{4}} \big] P^\oplus \widetilde{f}|_{{}_{\mathscr{O}_{m,0,0,0}}}(p) \big)_{{}_{\mathbb{C}^4}}
\, \ud \mu_{{}_{m,0}}(p)
\\
+ \int \limits_{\mathscr{O}_{-m,0,0,0}} \, 
\big(\widetilde{\phi}(p), \mathfrak{J}_{{}_{\bar{p}}}\big[\gamma^\mu p_\mu - m\boldsymbol{1}_{{}_{4}} \big] P^\ominus \widetilde{f}|_{{}_{\mathscr{O}_{-m,0,0,0}}}(p) \big)_{{}_{\mathbb{C}^4}}
 \, \ud \mu_{{}_{-m,0}}(p) 
\end{multline*}
\begin{multline*}
=\int \limits_{\mathscr{O}_{m,0,0,0}} \, 
\big(\big[\gamma^\mu p_\mu - m\boldsymbol{1}_{{}_{4}} \big] \widetilde{\phi}(p), \mathfrak{J}_{{}_{\bar{p}}} P^\oplus \widetilde{f}|_{{}_{\mathscr{O}_{m,0,0,0}}}(p) \big)_{{}_{\mathbb{C}^4}}
\, \ud \mu_{{}_{m,0}}(p)
\\
+ \int \limits_{\mathscr{O}_{-m,0,0,0}} \, 
\big(\big[\gamma^\mu p_\mu - m\boldsymbol{1}_{{}_{4}} \big]\widetilde{\phi}(p), \mathfrak{J}_{{}_{\bar{p}}} P^\ominus \widetilde{f}|_{{}_{\mathscr{O}_{-m,0,0,0}}}(p) \big)_{{}_{\mathbb{C}^4}}
 \, \ud \mu_{{}_{-m,0}}(p) = 0, 
\\
\,\,\,\,\, f \in \mathcal{S}(\mathbb{R}^4; \mathbb{C}^4).
\end{multline*}

Note also, that in principle we can use non-invariant pairings, with the fundamental symmetry
$\mathfrak{J} = \gamma^0$ (Krein structure) removed in the invariat pairings (\ref{InvariantBispinorPairing}) , \emph{i.e}
\begin{multline*}
\langle \phi, f \rangle_{{}_{+}} = \phi^+(f) 
=\int \big(\phi(x), f(x)\big)_{{}_{\mathbb{C}^4}} \, \ud^4 x
= \langle \widetilde{\phi}, \widetilde{f} \rangle = \widetilde{\phi}^+(\widetilde{f})
\\
= \int \big(\widetilde{\phi}(p), \widetilde{f}(p)\big)_{{}_{\mathbb{C}^4}} \, \ud^4p
= \int \delta(p\cdot p -m^2) \, \big(\widetilde{\phi}(p), \widetilde{f}(p)\big)_{{}_{\mathbb{C}^4}} \, \ud^4 p
\end{multline*}
\begin{multline}\label{NonInvariantBispinor+Pairing}
=\int \limits_{\mathscr{O}_{m,0,0,0}} \,
\big(\widetilde{\phi}(p), P^{\oplus *}\widetilde{f}|_{{}_{\mathscr{O}_{m,0,0,0}}}(p) \big)_{{}_{\mathbb{C}^4}}
\, \ud \mu_{{}_{m,0}}(p)
\\
+ \int \limits_{\mathscr{O}_{-m,0,0,0}} \,
\big(\widetilde{\phi}(p),  P^{\ominus *}\widetilde{f}|_{{}_{\mathscr{O}_{-m,0,0,0}}}(p) \big)_{{}_{\mathbb{C}^4}}
\, \ud \mu_{{}_{-m,0}}(p).
\end{multline}
But we have to remember that to the bispinor transformation acting on the solution $\phi$ understood as a distribution with non-invariant pairing,
there corresponds transformation of $\phi$, regarded as a function, which is equal to the conjugated inverse of the ordinary bispinor
transformation. Moreover, $\phi$ understood as a distribution with non-invariant pairing,
$f \mapsto \phi^+(f)$, does not represent any solution of the initial
Dirac equation
\[
i(\gamma^\mu)^* \partial_{\mu} \phi = m \phi,
\]
but a solution of the Dirac equation with conjugated 
\[
\big(\gamma^\mu\big)^{**} = \gamma^0 (\gamma^\mu)^* \gamma^0 = \gamma^\mu
\] 
Clifford
gamma generators, \emph{i.e.}
\[
i\gamma^\mu \partial_{\mu} \phi = m \phi,
\]
whose Fourier transform, concentrated on the respective orbits, is determined by the conjugated idempotents
\[
P^\oplus(p)^* = \gamma^0 {P^\oplus(p)} \gamma^0, \,\, \textrm{on} \,\, \mathscr{O}_{m,0,0,0},
\,\,\,\,\,\,\,\,\,\,
P^\ominus(p)^* = \gamma^0 {P^\ominus(p)} \gamma^0, \,\, \textrm{on} \,\, \mathscr{O}_{-m,0,0,0}
\]
instead of the initial idempotents $P^\oplus(p),P^\ominus(p)$. Thus, the space of Fourier transforms of the solutions of the conjugated
Dirac equation
\[
i\gamma^\mu \partial_{\mu} \phi = m \phi,
\]
is obtained from the Fourier transforms of the solutions of the initial Dirac equation
\[
i(\gamma^\mu)^* \partial_{\mu} \phi = m \phi,
\]
by the application of the fundamental symmetry
operator $\mathfrak{J} = \gamma^0$. Whenever we are using
non-invariant pairing (\ref{NonInvariantBispinorPairing}), the Dirac equation
\[
i(\gamma^\mu)^* \partial_{\mu} \phi = m \phi,
\]
and its conjugation
\[
i\gamma^\mu \partial_{\mu} \phi = m \phi,
\]
always come together. The same situation we have for the free Dirac quantum fields, corresponding to this two conjugated Dirac
equations, which naturally always come together: Dirac quantum field whose kernel respects the initial Dirac equation,
regarded as an operator valued distribution through non-invariant pairing (\ref{NonInvariantBispinor+Pairing}), 
respects the associated equation, and \emph{vice versa}.
But Dirac quantum field whose kernel respects the initial Dirac equation,
regarded as an operator valued distribution through invariant pairing (\ref{InvariantBispinorPairing}), 
resects the same initial equation.
Also, Dirac quantum field whose kernel respects the initial Dirac equation will use the Fock space constructed from the solutions 
(when regarded as functions) of the associated (and not the initial) Dirac equation. 

Finally we can also use another non-invariant pairing
\[
\langle \phi, f \rangle =  \phi(f) = \sum\limits_{a} \int \phi^a(x) f^a(x) \, \ud^4 x 
=\int \big(\overline{\phi(x)}, f(x)\big)_{{}_{\mathbb{C}^4}} \, \ud^4 x
\]
\begin{multline*}
=\int \limits_{\mathscr{O}_{m,0,0,0}} \,
\big(\overline{P^{\oplus}\widetilde{\phi}(p)},\overline{\widetilde{\overline{f}}|_{{}_{\mathscr{O}_{m,0,0,0}}}(p)} \big)_{{}_{\mathbb{C}^4}}
\, \ud \mu_{{}_{m,0}}(p)
\\
+ \int \limits_{\mathscr{O}_{-m,0,0,0}} \,
\big(\overline{P^{\ominus}\widetilde{\phi}(p)},  \overline{\widetilde{\overline{f}}|_{{}_{\mathscr{O}_{-m,0,0,0}}}(p)} \big)_{{}_{\mathbb{C}^4}}
\, \ud \mu_{{}_{-m,0}}(p)
\end{multline*}
\begin{multline}\label{NonInvariantBispinorPairing}
=\int \limits_{\mathscr{O}_{m,0,0,0}} \,
\big(\overline{\widetilde{\phi}(p)}, P^{\oplus}(p)^T\overline{\widetilde{\overline{f}}|_{{}_{\mathscr{O}_{m,0,0,0}}}(p)} \big)_{{}_{\mathbb{C}^4}}
\, \ud \mu_{{}_{m,0}}(p)
\\
+ \int \limits_{\mathscr{O}_{-m,0,0,0}} \,
\big(\overline{\widetilde{\phi}(p)}, P^{\ominus}(p)^T \overline{\widetilde{\overline{f}}|_{{}_{\mathscr{O}_{-m,0,0,0}}}(p)} \big)_{{}_{\mathbb{C}^4}}
\, \ud \mu_{{}_{-m,0}}(p).
\end{multline}
Note that
\[
\langle \phi, \overline{f} \rangle =  \phi(\overline{f}) =
\,\,\,\,\,\,\,\,\,\,\,\,\,\,\,\,\,\,\,\,\,\,\,\,\,\,\,\,\,\,\,\,\,\,\,\,\,\,\,\,\,\,\,\,\,\,\,\,\,\,
\,\,\,\,\,\,\,\,\,\,\,\,\,\,\,\,\,\,\,\,\,\,\,\,\,\,\,\,\,\,\,\,\,\,\,\,\,\,\,\,\,\,\,\,\,\,\,\,\,\,
\,\,\,\,\,\,\,\,\,\,\,\,\,\,\,\,\,\,\,\,\,\,\,\,\,\,\,\,\,\,\,\,\,\,\,\,\,\,\,\,\,\,\,\,\,\,\,\,\,\,
\,\,\,\,\,\,\,\,\,\,\,\,\,\,\,\,\,\,\,\,\,\,\,\,\,\,\,\,
\]
\begin{multline*}
=\int \limits_{\mathscr{O}_{m,0,0,0}} \,
\big(\overline{P^{\oplus}\widetilde{\phi}(p)},\overline{\widetilde{f}|_{{}_{\mathscr{O}_{m,0,0,0}}}(p)} \big)_{{}_{\mathbb{C}^4}}
\, \ud \mu_{{}_{m,0}}(p)
\\
+ \int \limits_{\mathscr{O}_{-m,0,0,0}} \,
\big(\overline{P^{\ominus}\widetilde{\phi}(p)},  \overline{\widetilde{f}|_{{}_{\mathscr{O}_{-m,0,0,0}}}(p)} \big)_{{}_{\mathbb{C}^4}}
\, \ud \mu_{{}_{-m,0}}(p)
\end{multline*}
\begin{multline}\label{NonInvariantBispinorPairingBar}
=\int \limits_{\mathscr{O}_{m,0,0,0}} \,
\big(\overline{\widetilde{\phi}(p)}, P^{\oplus}(p)^T\overline{\widetilde{f}|_{{}_{\mathscr{O}_{m,0,0,0}}}(p)} \big)_{{}_{\mathbb{C}^4}}
\, \ud \mu_{{}_{m,0}}(p)
\\
+ \int \limits_{\mathscr{O}_{-m,0,0,0}} \,
\big(\overline{\widetilde{\phi}(p)}, P^{\ominus}(p)^T \overline{\widetilde{f}|_{{}_{\mathscr{O}_{-m,0,0,0}}}(p)} \big)_{{}_{\mathbb{C}^4}}
\, \ud \mu_{{}_{-m,0}}(p).
\end{multline}
But each $\phi$, regarded as a function
\[
\begin{split}
\phi(x) = \int \limits_{\mathscr{O}_{m,0,0,0}} \, \widetilde{\phi}(p) \, 
e^{-ip\cdot x} \, \ud \mu_{{}_{m,0}}(p) 
+ \int \limits_{\mathscr{O}_{-m,0,0,0}} \, \widetilde{\phi}(p) \, 
e^{-ip\cdot x} \, \ud \mu_{{}_{-m,0}}(p),
\\
\widetilde{\phi} = P^\oplus\widetilde{\phi} \,\, \textrm{on} \,\, \mathscr{O}_{m,0,0,0},
\,\,\,\,\,\,\,\,\,\,\,
\widetilde{\phi} = P^\ominus\widetilde{\phi} \,\, \textrm{on} \,\, \mathscr{O}_{-m,0,0,0},
\end{split}
\] 
which by construction respects the initial Dirac equation
\[
i(\gamma^\mu)^* \partial_{\mu} \phi = m \phi,
\]
determines the distribution, $f \mapsto \phi(f)$, defined by the pairing (\ref{NonInvariantBispinorPairing}), which 
does not respect the initial Dirac equation, but the following associated Dirac equation
\[
\phi\big([-i\overline{\gamma^\mu} \partial_{\mu} - m] f\big) = 0, \,\,\,\, f \in \mathcal{S}(\mathbb{R}^4; \mathbb{C}^4),
\]
with another Clifford generators, equal $-\overline{\gamma^\mu}$. It  follows from 
(\ref{NonInvariantBispinorPairingBar}) that to each solution $\phi$, regarded as a function, of
the initial Dirac equation
\[
i(\gamma^\mu)^* \partial_{\mu} \phi = m \phi,
\]
there corresponds the antilinear distribution $f \mapsto \phi(\overline{f})$, which respects the following Dirac
equation 
\[
\phi\big( \overline{[i\gamma^\mu \partial_\mu -m]f} \big) = 0, \,\,\,\, f \in \mathcal{S}(\mathbb{R}^4; \mathbb{C}^4).
\]
Summing up: for each $\phi$, regarded as a function
\[
\begin{split}
\phi(x) = \int \limits_{\mathscr{O}_{m,0,0,0}} \, \widetilde{\phi}(p) \, 
e^{-ip\cdot x} \, \ud \mu_{{}_{m,0}}(p) 
+ \int \limits_{\mathscr{O}_{-m,0,0,0}} \, \widetilde{\phi}(p) \, 
e^{-ip\cdot x} \, \ud \mu_{{}_{-m,0}}(p),
\\
\widetilde{\phi} = P^\oplus\widetilde{\phi} \,\, \textrm{on} \,\, \mathscr{O}_{m,0,0,0},
\,\,\,\,\,\,\,\,\,\,\,
\widetilde{\phi} = P^\ominus\widetilde{\phi} \,\, \textrm{on} \,\, \mathscr{O}_{-m,0,0,0},
\end{split}
\] 
which by construction respects the initial Dirac equation
\[
\big[i(\gamma^\mu)^* \partial_{\mu} - m\big] \phi = 0,
\]
we have
\begin{equation}\label{PairingsAndDiracEquations}
\begin{split}
\phi^\sharp\big([i(\gamma^\mu)^* \partial_{\mu} - m]\ f \big) =0,  \,\,\,\, f \in \mathcal{S}(\mathbb{R}^4; \mathbb{C}^4),
\\
\phi^+\big([i\gamma^\mu \partial_{\mu} - m]\ f \big) =0,  \,\,\,\, f \in \mathcal{S}(\mathbb{R}^4; \mathbb{C}^4),
\\
\phi\big([-i\overline{\gamma^\mu} \partial_{\mu} - m]\ f \big) =0,  \,\,\,\, f \in \mathcal{S}(\mathbb{R}^4; \mathbb{C}^4),
\\
\phi\big(\overline{[i \gamma^\mu \partial_{\mu} - m]\ f} \big) =0,  \,\,\,\, f \in \mathcal{S}(\mathbb{R}^4; \mathbb{C}^4),
\end{split}
\end{equation}
for the corresponding distributions, defined by $\phi$ and the respective pairing. Note also
that the invariant pairing (\ref{InvariantBispinorPairing}) and the pairing (\ref{NonInvariantBispinor+Pairing}), \emph{i.e.}
$\phi^\sharp(f)$ and  $\phi^+(f)$,
can be expressend through the pairing (\ref{NonInvariantBispinorPairing}), \emph{i.e.} $\phi(f)$, in the following manner
\begin{equation}\label{Pairings}
\phi^\sharp(f) = (\gamma^0\overline{\phi})(f), \,\,\,\, \phi^+(f) = \overline{\phi}(f).
\end{equation}

From now on we agree to denote the ordinary bispinor function $\widetilde{\phi}$ on the disjoint sum
$\mathscr{O}_{m,0,0,0} \sqcup \mathscr{O}_{-m,0,0,0}$ (equal to the distributional Fourier support of the distribution $\widetilde{\phi}$)
by the same symbol $\widetilde{\phi}$ as the distributional Fourier transform $\widetilde{\phi}$
of $\phi \in \mathcal{H}$ (although $\widetilde{\phi}$
makes sense as the ordinary function only on the support of the distribution $\widetilde{\phi}$,
which as a ``function'' is intentionally equal zero outside the support, which makes a precise sense
when $\widetilde{\phi}$ is regarded as distribution defined as above).

In short for $\phi \in \mathcal{H}$ we can write
\[
\begin{split}
\phi(x) = \int \limits_{\mathscr{O}_{m,0,0,0}} \, \widetilde{\phi}(p) \, 
e^{-ip\cdot x} \, \ud \mu_{{}_{m,0}}(p) 
+ \int \limits_{\mathscr{O}_{-m,0,0,0}} \, \widetilde{\phi}(p) \, 
e^{-ip\cdot x} \, \ud \mu_{{}_{-m,0}}(p),
\\
\widetilde{\phi} = P^\oplus\widetilde{\phi} \,\, \textrm{on} \,\, \mathscr{O}_{m,0,0,0},
\,\,\,\,\,\,\,\,\,\,\,
\widetilde{\phi} = P^\ominus\widetilde{\phi} \,\, \textrm{on} \,\, \mathscr{O}_{-m,0,0,0};
\end{split}
\] 
or
\begin{multline}\label{DistributionalSolDiracEq}
\phi(x) = \int \limits_{\mathscr{O}_{m,0,0,0}} \, \widetilde{\phi}(p) \, 
e^{-ip\cdot x} \, \ud \mu_{{}_{m,0}}(p) 
+ \int \limits_{\mathscr{O}_{-m,0,0,0}} \, \widetilde{\phi}(p) \, 
e^{-ip\cdot x} \, \ud \mu_{{}_{-m,0}}(p) \\
= \int \limits_{\mathbb{R}^3} \widetilde{\phi}(\vec{p},|p_0(\vec{p})|) \, e^{-(i|p_0(\vec{p})|t -i\vec{p}\cdot \vec{x}) }
\, \frac{\ud^3 \vec{p}}{2 |p_0(\vec{p})|} + 
\int \limits_{\mathbb{R}^3} \widetilde{\phi}(-\vec{p},-|p_0(\vec{p})|) \,  e^{i|p_0(\vec{p})|t -i\vec{p}\cdot \vec{x}}
\, \frac{\ud^3 \vec{p}}{2 |p_0(\vec{p})|}, \\
 p_0(\vec{p}) = \pm \sqrt{\vec{p} \cdot \vec{p} + m^2},
\end{multline}
\[
\begin{split}
\widetilde{\phi}(\vec{p},|p_0(\vec{p})|) = P^\oplus(\vec{p},|p_0(\vec{p})|\widetilde{\phi}(\vec{p},|p_0(\vec{p})| \,\, \textrm{on} 
\,\, \mathscr{O}_{m,0,0,0},
\\
\widetilde{\phi}(-\vec{p},-|p_0(\vec{p})|) = P^\ominus(-\vec{p},-|p_0(\vec{p})|)\widetilde{\phi}(-\vec{p},-|p_0(\vec{p})|) 
\,\, \textrm{on} \,\, \mathscr{O}_{-m,0,0,0}.
\end{split}
\]
Here of course $p = (p_0(\vec{p}),\vec{p}) = (\sqrt{\vec{p} \cdot \vec{p} + m^2},\vec{p})$ on $\mathscr{O}_{m,0,0,0}$
and $p = (p_0(\vec{p}),\vec{p}) = (-\sqrt{\vec{p} \cdot \vec{p} + m^2},\vec{p})$ on $\mathscr{O}_{-m,0,0,0}$.

In particular for the 
solution $\phi \in \mathcal{H}$ whose Fourier transform $\widetilde{\phi}$ is concentrated on 
the positive energy orbit $\mathscr{O}_{m,0,0,0}$
we have
\begin{multline*}
\phi(x)= \phi(\vec{x},t) = \int \limits_{\mathscr{O}_{m,0,0,0}} \widetilde{\phi}(p) \, 
e^{-ip\cdot x} \, \ud \mu_{{}_{m,0}}(p) \\
= \int \limits_{\mathbb{R}^3} \widetilde{\phi}(\vec{p},p_0(\vec{p})) \, e^{-(ip_0(\vec{p})t -i\vec{p}\cdot \vec{x}) }
\, \frac{\ud^3 \vec{p}}{2 p_0(\vec{p})}, \,\,\,\,\, p_0(\vec{p}) = \sqrt{\vec{p} \cdot \vec{p} + m^2}.
\end{multline*}
Similarly, for the solution $\phi \in \mathcal{H}$ whose Fourier transform is concentrated on the negative energy orbit $\mathscr{O}_{-m,0,0,0}$,
we have:
\begin{multline*}
\phi(x)= \phi(\vec{x},t) = \int \limits_{\mathscr{O}_{-m,0,0,0}} \widetilde{\phi}(p) \, 
e^{-ip\cdot x} \, \ud \mu_{{}_{-m,0}}(p) \\
= \int \limits_{\mathbb{R}^3} \widetilde{\phi}(-\vec{p},-|p_0(\vec{p})|) \,  e^{i|p_0(\vec{p})|t -i\vec{p}\cdot \vec{x}}
\, \frac{\ud^3 \vec{p}}{2 |p_0(\vec{p})|}, \,\,\,\,\, p_0(\vec{p}) = -\sqrt{\vec{p} \cdot \vec{p} + m^2}.
\end{multline*}

We have the following equality for the solutions $\phi,\phi' \in \mathcal{H}$ whose Fourier transforms 
$\widetilde{\phi}, \widetilde{\phi}'$ are 
concentrated on the positive energy orbit $\mathscr{O}_{m,0,0,0}$:
\begin{multline*}
\int \limits_{x^0 =  t = const.} \Big(\phi(\vec{x}, t), \phi'(\vec{x}, t) \Big)_{{}_{\mathbb{C}^4}}
\, \ud^3 x =  \int \limits_{\mathscr{O}_{m,0,0,0}} \Big(\widetilde{\phi}(p), \widetilde{\phi}'(p) \Big)_{{}_{\mathbb{C}^4}}
\, \frac{\ud \mu_{{}_{m,0}}(p)}{2 p_0} = \\
\int \limits_{\mathbb{R}^3} \Big(\widetilde{\phi}(\vec{p}, p_0(\vec{p})), 
\phi'(\vec{p}, p_0(\vec{p})) \Big)_{{}_{\mathbb{C}^4}}
\, \frac{\ud^3 \vec{p}}{(2 p_0(\vec{p}))^2}, \,\,\,\, p_0(\vec{p}) = \sqrt{\vec{p} \cdot \vec{p} + m^2}.
\end{multline*}
Let us emphasize here that the above $T_4\circledS SL(2, \mathbb{C})$-invariant inner product
is compatible with the invariant pairing $\langle \cdot, \cdot \rangle_{{}_{\textrm{invariant pairing}}}$ defined
by the formula (\ref{InvariantBispinorPairing}). Indeed, this follows from the fact that $\widetilde{\phi} = P^\oplus\widetilde{\phi}$
for our positive energy elements (concentrated on $\mathscr{O}_{m,0,0,0}$) 
of the single particle Hilbert space:
\begin{multline*}
\langle \phi, f \rangle_{{}_{\textrm{invariant pairing}}} = \phi^\sharp(f) =
\widetilde{\phi}^\sharp(\widetilde{f}) = \langle \widetilde{\phi}, \widetilde{f} \rangle_{{}_{\textrm{invariant pairing}}} 
\\
= \int \limits_{\mathbb{R}^3} \Big(\widetilde{\phi}(\vec{p}, p_0(\vec{p})), 
\mathfrak{J}_{{}_{\bar{p}}}\widetilde{f}(\vec{p}, p_0(\vec{p})) \Big)_{{}_{\mathbb{C}^4}}
\, \frac{\ud^3 \vec{p}}{2 p_0(\vec{p})}
\end{multline*}
\begin{multline*}
= \int \limits_{\mathbb{R}^3} \Big(P^\oplus\widetilde{\phi}(\vec{p}, p_0(\vec{p})), 
\gamma^0 P^\oplus \widetilde{f}(\vec{p}, p_0(\vec{p})) \Big)_{{}_{\mathbb{C}^4}}
\, \frac{\ud^3 \vec{p}}{2 p_0(\vec{p})}
\\
= \int \limits_{\mathbb{R}^3} \Big(\gamma^0\gamma^0 P^\oplus\widetilde{\phi}(\vec{p}, p_0(\vec{p})), 
\gamma^0 P^\oplus \widetilde{f}(\vec{p}, p_0(\vec{p})) \Big)_{{}_{\mathbb{C}^4}}
\, \frac{\ud^3 \vec{p}}{2 p_0(\vec{p})}
\end{multline*}
\begin{multline*}
= 2m \int \limits_{\mathbb{R}^3} \Big({\textstyle\frac{1}{2 p_0(\vec{p})}} \gamma^0 P^\oplus(\vec{p}, p_0(\vec{p}))^* 
\gamma^0 \gamma^0 P^\oplus \widetilde{\phi}(\vec{p}, p_0(\vec{p})), 
\gamma^0 P^\oplus \widetilde{f}(\vec{p}, p_0(\vec{p})) \Big)_{{}_{\mathbb{C}^4}}
\, \frac{\ud^3 \vec{p}}{2 p_0(\vec{p})}
\\
= 2m \int \limits_{\mathbb{R}^3} \Big({\textstyle\frac{1}{2 p_0(\vec{p})}}\gamma^0 P^\oplus\widetilde{\phi}(\vec{p}, p_0(\vec{p})), 
\gamma^0 P^\oplus(\vec{p}, p_0(\vec{p})) \gamma^0 \gamma^0 P^\oplus \widetilde{f}(\vec{p}, p_0(\vec{p})) \Big)_{{}_{\mathbb{C}^4}}
\, \frac{\ud^3 \vec{p}}{2 p_0(\vec{p})}
\end{multline*}
\begin{multline*}
= 2m\int \limits_{\mathbb{R}^3} \Big({\textstyle\frac{1}{2 p_0(\vec{p})}} \gamma^0 P^\oplus\widetilde{\phi}(\vec{p}, p_0(\vec{p})), 
\gamma^0 P^\oplus \widetilde{f}(\vec{p}, p_0(\vec{p})) \Big)_{{}_{\mathbb{C}^4}}
\, \frac{\ud^3 \vec{p}}{2 p_0(\vec{p})}
\\
= 2m\int \limits_{\mathbb{R}^3} \Big(\widetilde{\phi}(\vec{p}, p_0(\vec{p})), 
P^\oplus \widetilde{f}(\vec{p}, p_0(\vec{p})) \Big)_{{}_{\mathbb{C}^4}}
\, \frac{\ud^3 \vec{p}}{(2 p_0(\vec{p}))^2},
\end{multline*}
Here we have used 1) $\widetilde{\phi} = P^\oplus\widetilde{\phi} = \big(P^\oplus\big)^2\widetilde{\phi}$, 
2) ${P^\oplus(p)}^*\gamma^0 = \gamma^0{P^\oplus(p)}$,
3) the identity $\gamma^0 {P^\oplus(p)}^* = (1/2m)(p_0(p) + \widetilde{H}(p))$,
with $\widetilde{H}(p) = \gamma^0\gamma^1p_1 + \gamma^0\gamma^2p_2 +\gamma^0\gamma^3p_3 + \gamma^0m$
equal to the Fourier transform of the Dirac Hamiltonian operator, equal $p_0(p)$ in action on bispinors 
$\widetilde{\phi} = P^\oplus\widetilde{\phi}, P^\oplus\widetilde{f}$ 
concentrated on the positive energy orbit 
$\mathscr{O}_{m,0,0,0}$ and lying in the image of the  idempotent $P^\oplus$. Next we have used
4) unitarity of $\gamma^0 = {\gamma^{0}}^* = {\gamma^{0}}^{-1}$ in $\mathbb{C}^4$, $\big(\gamma^0\big)^2 = \boldsymbol{1}$.
We have also used 5) the formula for 
the projection $P^\oplus$ acting as multiplication by the
matrix function
\[
P^\oplus(p) = {\textstyle\frac{1}{2m}} \big[\gamma^\mu p_\mu + m\big], \,\,\,\,\,\,
p \in \mathscr{O}_{m,0,0,0},
\]
compare Subsection \ref{e1} and the Appendix \ref{fundamental,u,v}.

Similarly, for the solutions $\phi, \phi' \in \mathcal{H}$ whose Fourier transforms
$\widetilde{\phi}, \widetilde{\phi}'$ are concentrated on the negative energy 
orbit $\mathscr{O}_{-m,0,0,0}$, we have (recall, that we are using here positive measures without sign): 
\begin{multline*}
\int \limits_{x^0 =  t = const.} \Big(\phi(\vec{x}, t), \phi'(\vec{x}, t) \Big)_{{}_{\mathbb{C}^4}}
\, \ud^3 x 
 =  \int \limits_{\mathscr{O}_{-m,0,0,0}} \Big(\widetilde{\phi}(p), \widetilde{\phi}'(p) \Big)_{{}_{\mathbb{C}^4}}
\, \frac{\ud \mu_{{}_{-m,0}}(p)}{2 |p_0|} =\\
= \int \limits_{\mathbb{R}^3} \Big(\widetilde{\phi}(-\vec{p}, -|p_0(\vec{p})|), 
\widetilde{\phi}'(-\vec{p}, -|p_0(\vec{p})|) \Big)_{{}_{\mathbb{C}^4}}
\, \frac{\ud^3 \vec{p}}{(2 p_0)^2} \\=
\int \limits_{\mathbb{R}^3} \Big(\widetilde{\phi}(\vec{p}, p_0(\vec{p})), 
\widetilde{\phi}'(\vec{p}, p_0(\vec{p})) \Big)_{{}_{\mathbb{C}^4}}
\, \frac{\ud^3 \vec{p}}{(2 p_0)^2} \\ =
\int \limits_{\mathscr{O}_{-m,0,0,0}} \Big(\widetilde{\phi}(p), 
\widetilde{\phi}'(p) \Big)_{{}_{\mathbb{C}^4}}
\, \frac{\ud \mu_{{}_{-m,0}}(p)}{2 |p_0|},
 \,\,\,\, p_0(\vec{p}) = -\sqrt{\vec{p} \cdot \vec{p} + m^2},
\end{multline*}
which, analogously as on positive energy solutions, is compatible with the invariant pairing
(\ref{InvariantBispinorPairing}) defined above.

Note that the last expression is equal to the inner product (\ref{InnProdBispOnO_-m,0,0,0})
of the (Fourier transforms of) bispinors $\phi, \phi'$ 
on the Hilbert space of Fourier transforms of bispinors, concentrated on $\mathscr{O}_{-m,0,0,0}$ (up to the irrelevant
constant factor $m>0$), introduced in Subsection \ref{e1}.

Consider now the representation (notation of introduction to Section \ref{constr-of-VF})
\[
W'U^{-1}
U^{{}_{m,0,0,0} L^{{}^{1/2}}}
U{W'}^{-1}
\]
hereafter called (for simplicity of notation)
\begin{equation}\label{m}
U^{{}_{m,0,0,0} L^{{}^{1/2}}}
\end{equation}
of $T_4 \circledS SL(2, \mathbb{C})$, concentrated on the orbit $\mathscr{O}_{m,0,0,0}$.
Now we apply the isometric map $V^\oplus$ to the space of this representation followed by the
Fourier transform (\ref{F(varphi)}) (with the orbit $\mathscr{O}_{\bar{p}}= \mathscr{O}_{m,0,0,0}$),
where $V^\oplus$ is the map defined in Example 1 (Subsection \ref{e1}). Let us denote the composed map
just by $\widetilde{V^\oplus}$. The image of $\widetilde{V^\oplus}$ lies in $\mathcal{H}$. 
Indeed, because of eq. (\ref{sprod-bisp-orb}) 
it is even isometric.

Similarly consider the representation (notation of introduction to Section \ref{constr-of-VF})
\[
W'U^{-1}
U^{{}_{-m,0,0,0} L^{{}^{1/2}}}
U{W'}^{-1}
\]
hereafter called (for simplicity of notation)
\begin{equation}\label{-m}
U^{{}_{-m,0,0,0} L^{{}^{1/2}}}
\end{equation}
of $T_4 \circledS SL(2, \mathbb{C})$, concentrated on the orbit $\mathscr{O}_{-m,0,0,0}$.
To the space of this representation we apply the map $\widetilde{V^\ominus}$ equal to $V^\ominus$ followed by the
Fourier transform (\ref{F(varphi)}) (with the orbit $\mathscr{O}_{\bar{p}}= \mathscr{O}_{-m,0,0,0}$), where
$V^\ominus$ is the map defined in Example 1,
Subsection \ref{e1}. Its image likewise lies in
$\mathcal{H}$ and by the same (\ref{sprod-bisp-orb}) -- which is also valid for $\widetilde{V^\ominus}$ --
it is isometric too. Now the image $\mathcal{H}_{m,0}^{\oplus}$ of the representation
space of the representation (\ref{m}) under
the map $\widetilde{V^\oplus}$ lies in the positive eigenspace subspace
$E_+ \mathcal{H}$ of the essentially self adjoint Dirac hamiltonian operator
$H = -i \gamma^0\gamma^k \partial_k + m \gamma^0 = -i \alpha^k \partial_k + m \gamma^0$ acting on
$\mathcal{H}$, where $E_+$ is the spectral projection corresponding to all positive spectral values
of $H$. Similarly, the image $\mathcal{H}_{-m,0}^{\ominus}$ of the space of the representation (\ref{-m})
under the map $V^\ominus$ lies in the negative eigenspace subspace $E_-\mathcal{H}$ of the
operator $H$. We have $E_+ + E_- = \bold{1}_\mathcal{H}$ and $E_+ E_- = 0$, i.e.
$E_+ \mathcal{H}$ and $E_- \mathcal{H}$ are orthogonal. Therefore, the operator
$\widetilde{V^\oplus} \oplus \widetilde{V^\ominus}$ maps the representation space of the representation
\begin{equation}\label{m+-m}
U^{{}_{m,0,0,0} L^{{}^{1/2}}} \oplus U^{{}_{-m,0,0,0} L^{{}^{1/2}}},
\end{equation}
concentrated on the sum theoretic set $\mathscr{O}_{(m,0,0,0)} \cup \mathscr{O}_{-m,0,0,0}$
of the orbits $\mathscr{O}_{m,0,0,0}$ and $\mathscr{O}_{-m,0,0,0}$, isometrically
into $\mathcal{H}$.

On the other hand the only eigenvalues of the matrix $\gamma^0$ are 1 and -1, so it follows from the
theorem of Section 10.1, Part II,
Chapter II of \cite{Geland-Minlos-Shapiro} (compare also \cite{GelfandYaglom1}-\cite{GelfandYaglom3}),
that the ordinary Fourier transform (\ref{F(phi)}) of any element of $\mathcal{H}$ is
concentrated on the set theoretical sum $\mathscr{O}_{m,0,0,0} \cup \mathscr{O}_{-m,0,0,0}$
of the orbits $\mathscr{O}_{m,0,0,0}$ and $\mathscr{O}_{-m,0,0,0}$. Thus, the operator
$\widetilde{V^\oplus} \oplus \widetilde{V^\ominus}$ regarded as operator on the space of
the representation (\ref{m+-m})
is onto $\mathcal{H}$, and therefore it is unitary, so that
\[
E_+ \mathcal{H} = \mathcal{H}_{m,0}^{\oplus} \,\,\, \textrm{and} \,\,\,
E_- \mathcal{H} = \mathcal{H}_{-m,0}^{\ominus}.
\]
Therefore in the Hilbert space $\mathcal{H} = \mathcal{H}_{m,0}^{\oplus} \oplus \mathcal{H}_{-m,0}^{\ominus}$
there acts the unitary\footnote{Please, note also that the representation
\[
V^\oplus \, U^{{}_{m,0,0,0} L^{{}^{1/2}}} \, (V^\oplus )^{-1} \, \oplus \,
V^\ominus U^{{}_{-m,0,0,0} L^{{}^{1/2}}} \, (V^\ominus)^{-1},
\]
concentrated on $\mathscr{O}_{m,0,0,0} \sqcup \mathscr{O}_{-m,0,0,0}$ is unitary, contrary to the representation
\[
V^{\oplus \ominus} \, \big( U^{{}_{m,0,0,0} L^{{}^{1/2}}} \oplus
U^{{}_{m,0,0,0} L^{{}^{1/2}}} \big) \, (V^{\oplus \ominus})^{-1}
\]
concentrated on $\mathscr{O}_{(m,0,0,0)}$ which is not unitary, but only
similar to a unitary representation through a bounded (non-unitary) invertible similarity operator. This is because $\mathcal{H}^{\oplus}_{m,0}$
and  $\mathcal{H}^{\ominus}_{m,0}$ are not orthogonal, compare Example 1, Subsection \ref{e1}.}
representation
\begin{equation}\label{rep-on-H}
\widetilde{V^\oplus} \, U^{{}_{m,0,0,0} L^{{}^{1/2}}} \, (\widetilde{V^\oplus} )^{-1} \, \oplus \,
\widetilde{V^\ominus} U^{{}_{-m,0,0,0} L^{{}^{1/2}}} \, (\widetilde{V^\ominus})^{-1}
\end{equation}
concentrated on $\mathscr{O}_{m,0,0,0} \cup \mathscr{O}_{-m,0,0,0}$, with
\begin{equation}\label{rep-on-H+}
\widetilde{V^\oplus} \, U^{{}_{m,0,0,0} L^{{}^{1/2}}} \, (\widetilde{V^\oplus} )^{-1}
\end{equation}
acting on $\mathcal{H}_{m,0}^{\oplus}$ and with
\begin{equation}\label{rep-on-H-}
\widetilde{V^\ominus} U^{{}_{-m,0,0,0} L^{{}^{1/2}}} \, (\widetilde{V^\ominus})^{-1}
\end{equation}
acting on $\mathcal{H}_{-m,0}^{\ominus}$.

The representation (\ref{1/2pUalpha}) and (\ref{1/2pUa})  acting on generic bispinors, concentrated on the disjoint
sum  $\mathscr{O}_{(m,0,0,0)} \cup \mathscr{O}_{-m,0,0,0}$ of the orbits, induces precisely the unitary representation
acting in the single particle Hilbert space $\mathcal{H} = \mathcal{H}_{m,0}^{\oplus} \oplus \mathcal{H}_{-m,0}^{\ominus}$,
and commutes, respectively, with the idempotents $P^\oplus$ and $P^\ominus$ on the respective orbits
 $\mathscr{O}_{(m,0,0,0)}$ and $\mathscr{O}_{-m,0,0,0}$. However, the representation 
(\ref{1/2pUalpha}) and (\ref{1/2pUa})  acting on the whole space 
\[
\mathcal{H}_{m,0}^{\oplus} \oplus \mathcal{H}_{m,0}^{\ominus}
\oplus \mathcal{H}_{-m,0}^{\oplus} \oplus \mathcal{H}_{-m,0}^{\ominus}
\]
of generic bispinors, concentrated on the disjoint
sum  $\mathscr{O}_{m,0,0,0} \cup \mathscr{O}_{-m,0,0,0}$ of the orbits,
with the inner product (\ref{sprod-bisp-orb}), is not unitary. This is because the pair of Hilbert spaces 
$\mathcal{H}_{m,0}^{\oplus}$ and  $\mathcal{H}_{m,0}^{\ominus}$ of bispinors concentrated on 
$\mathscr{O}_{m,0,0,0}$ are not orthogonal with respect to  (\ref{sprod-bisp-orb}). Similarly,
the pair of Hilbert spaces 
$\mathcal{H}_{-m,0}^{\oplus}$ and  $\mathcal{H}_{-m,0}^{\ominus}$ of bispinors concentrated on 
$\mathscr{O}_{-m,0,0,0}$ are not orthogonal with respect to  (\ref{sprod-bisp-orb}).
Therefore also the representation   (\ref{1/2pUalpha}) and (\ref{1/2pUa}) acting separately on the Hilbert space 
$\mathcal{H}_{\pm m,0}^{\oplus} \oplus \mathcal{H}_{\pm m,0}^{\ominus}$ of bispinors 
concentrated, respectively, on $\mathscr{O}_{\pm m,0,0,0}$, with the inner product equal (\ref{sprod-bisp-orb}),
likewise is not unitary. The last symbol $\oplus$ between Hilbert spaces with the same sign of $m$ stands for a disjont liner sum, 
but not orthogonal, compare Subsection \ref{e1}, Example 1.

To the Hilbert space $\mathcal{H}$ treated as if it was the single particle space we apply the fermionic
functor of second quantization
$\Gamma$, and obtain the standard absorption and emission operators. Next we split them (i.e. we consider
their restrictions respect to $\mathcal{H}_{m,0}^{\oplus}$ or $\mathcal{H}_{-m,0}^{\ominus}$)
according to the splitting $\mathcal{H} = \mathcal{H}_{m,0}^{\oplus} \oplus \mathcal{H}_{-m,0}^{\ominus}
= E_+ \mathcal{H} \oplus E_- \mathcal{H}$ of the space $\mathcal{H}$,
compare e.g. \cite{Scharf}. We observe then that the absorption and emission operators restricted to
$\mathcal{H}_{m,0}^{\oplus}$ compose a fermionic free field and similarly the restrictions
of the absorption and emission operators restricted to $\mathcal{H}_{-m,0}^{\ominus}$ and that the two sets
of operators commute and are independent in consequence of the orthogonality of the sub spaces
$\mathcal{H}_{m,0}^{\oplus}$ and $\mathcal{H}_{-m,0}^{\ominus}$ (e.g. \cite{Scharf}).
That is, we have two independent fermionic quantizations: the functor $\Gamma$ applied to
$\mathcal{H}_{m,0}^{\oplus}$ and the functor $\Gamma$ applied to $\mathcal{H}_{-m,0}^{\ominus}$
with the tensor product of the two independent sets of annihilation and creation operators acting in
the tensor product of fermionic Fock spaces
$\Gamma\big(\mathcal{H}_{m,0}^{\oplus} \big) \otimes \Gamma\big(\mathcal{H}_{-m,0}^{\ominus} \big)
= \Gamma(\big( \mathcal{H}_{m,0}^{\oplus} \oplus \mathcal{H}_{-m,0}^{\ominus} \big)$.
In order to repair the energy sign
of the free Dirac field on $\Gamma\big(\mathcal{H}_{m,0}^{\oplus} \big) \otimes
\Gamma\big(\mathcal{H}_{-m,0}^{\ominus} \big)$ we interchange the absorption
and emission operators in $\Gamma\big(\mathcal{H}_{-m,0}^{\ominus} \big)$. In this manner we obtain
the following construction which may be described in the following four steps.

\subsection{Application of the Segal second quantization 
functor to the subspace $\mathcal{H}_{m,0}^{\oplus}$}\label{electron}

To the subspace  $\mathcal{H}_{m,0}^{\oplus}$ we apply the Segal's functor $\Gamma$ of fermionic
quantization and obtain the fermionic Fock space
\[ 
\mathcal{H}^{\oplus}_{F} = \Gamma(\mathcal{H}_{m,0}^{\oplus}) = \mathbb{C} \oplus \mathcal{H}_{m,0}^{\oplus} \oplus
\big( \mathcal{H}_{m,0}^{\oplus} \big)^{\widehat{\otimes} 2} \oplus 
\big( \mathcal{H}_{m,0}^{\oplus} \big)^{\widehat{\otimes} 3}
\oplus \ldots;
\] 
with the unitary representation
\begin{multline*} 
\Gamma \Big(\widetilde{V^\oplus} \, U^{{}_{(m,0,0,0)} L^{{}^{1/2}}} \, (\widetilde{V^\oplus} )^{-1} \Big)
= \bigoplus \limits_{n = 0,1,2 \ldots}
\Big(\widetilde{V^\oplus} \, U^{{}_{(m,0,0,0)} L^{{}^{1/2}}} \, (\widetilde{V^\oplus})^{-1} \Big)^{\widehat{\otimes} n},
\end{multline*}
where in the formulas $(\cdot)^{\widehat{\otimes} n}$ stands for $n$-fold antisymmetrized tensor product,
and $(\cdot )^{\widehat{\otimes} n}$ with $n = 0$ applied to the representation gives the trivial representation on 
$\mathbb{C}$ with each representor acting on 
$\mathbb{C}$ as multiplication by 1.

In this and in the following Sections, we will encounter essentially two types of topological vector spaces
and operators acting upon them: 1) \emph{Hilbert spaces} and 2) \emph{nuclear spaces} 
(the Schwartz $\mathcal{S}(\mathbb{R}^n)$ space of test functions on 
$\mathbb{R}^n$ is an example of a nuclear space). Correspondingly we will
use respectively 1) the \emph{Hilbert space tensor product} $\otimes$ 
(if applied to Hilbert spaces, elements of Hilbert spaces 
and operators upon them) and respectively \emph{projective tensor product} $\otimes$ 
(if applied to nuclear spaces, their elements
and operators acting upon them); for definition, and properties of these standard constructions we refer e.g.
to \cite{Murray_von_Neumann}, \cite{treves}, \cite{Schaefer}.

The linear spaces we encounter (Hilbert spaces and nuclear spaces) will be always over $\mathbb{R}$ or over $\mathbb{C}$,
but whenever they are over $\mathbb{C}$ they will be equal to complexifications of real (Hilbert or nuclear)
spaces with naturally defined complex conjugation $\overline{(\cdot)}$ in them.  

Note that by Riesz representation theorem for such Hilbert spaces $\mathcal{H}'$ we have natural identification
of linear continuous functionals on $\mathcal{H}'$ with the elements of the adjoint Hilbert space
$\overline{\mathcal{H}'}$, which in fact becomes an isomorphism of Hilbert spaces if we appropriately introduce the multiplication by a number and the inner product into the space of linear functionals
on $\mathcal{H}'$. Recall that the adjoint space
$\overline{\mathcal{H}'}$ have the same set of elements as $\mathcal{H}'$, but with scalar multiplication
by a number $\alpha \in \mathbb{C}$ and inner
product defined by
\[
\begin{split}
\alpha u \,\,\, \textrm{in} \,\,\,\overline{\mathcal{H}'} \,\,\, = \,\,\,\overline{\alpha}u \,\,\, \textrm{in} \,\,\,
\mathcal{H}', \\
(u,v) \,\,\, \textrm{in} \,\,\,\overline{\mathcal{H}'} \,\,\, = \,\,\, (v,u) \,\,\, \textrm{in} \,\,\,
\mathcal{H}'.
\end{split}
\]
With such a Hilbert space structure on $\overline{\mathcal{H}'}$ the map
$\mathcal{H}' \ni u \mapsto \overline{u} \in \overline{\mathcal{H}'}$
defines a canonical \emph{linear} isomorphism.
In the sequel we will regard the dual space $\mathcal{H}'^*$ as the adjoint space $\overline{\mathcal{H}'}$
with elements the same as elements of $\mathcal{H}'$ (Riesz isomorphism).

For operators on Hilbert spaces we are using the standard notation for
the ordinary adjoint operation with the superscript $*$,
except the annihilation operators, denoting the operators which are adjoint to
them with the superscript $+$ instead $*$ (which is customary in physical literature).
If working with operators $A$ transforming (continuously) one nuclear space into another $E_1 \rightarrow E_2$, we use the superscript $*$ to denote the linear dual (transposed) operator $A^*$:
$E_{2}^* \rightarrow E_{1}^*$, transforming continuously the strong dual space
$E_{2}^*$ into the strong dual space $E_{1}^*$, for definition and general properties
of transposition we again refer to \cite{treves}. For operator $A$ transforming
(continuously) nuclear space into nuclear space we denote by
$A^+$ the operator $\overline{(\cdot)} \circ A^* \circ \overline{(\cdot)}$, i.e. the linear dual
of $A$ composed with complex conjugation (say Hermitian adjoint $=$ linear transposition $+$ complex conjugation).

In the standard way we obtain the map from
$\mathcal{H}_{m,0}^{\oplus} \ni \widetilde{\phi}$ to the families
$a_{\oplus}(\widetilde{\phi}),
a_{\oplus}^+(\widetilde{\phi}) = {a_{\oplus}(\widetilde{\phi})}^+$ of ordinary annihilation and creation operators in the fermionic Fock space
$\Gamma\big(\mathcal{H}_{m,0}^{\oplus}\big)$ fulfilling the canonical anticommutation relations:
\begin{multline*}
\Big\{a_{\oplus}(\widetilde{\phi}), \,\,\,
{a_{\oplus}(\widetilde{\phi}')}^+ \Big\}
= \Big(\widetilde{\phi}, \widetilde{\phi}'\Big)_{{}_{\mathcal{H}_{m,0}^{\oplus}}} \\
= \Big(\widetilde{\phi}, \widetilde{\phi}'\Big) \\
= \int \limits_{x^0 = t = const.} \Big(\phi(\vec{x}, t), \phi'(\vec{x},t) \Big)_{{}_{\mathbb{C}^4}}
\, \ud^3 x \\
= \int \limits_{\mathscr{O}_{m,0,0,0}}
\Big(\widetilde{\phi}(p),
\widetilde{\phi}'(p) \Big)_{{}_{\mathbb{C}^4}}
\, \frac{\ud \mu_{{}_{m,0}}(p)}{2 p_0} \\
= \int \limits_{\mathbb{R}^3} \Big(\widetilde{\phi}(\vec{p}, p_0(\vec{p})), \,\,
\phi'(\vec{p}, p_0(\vec{p})) \Big)_{{}_{\mathbb{C}^4}}
\, \frac{\ud^3 \vec{p}}{(2 p_0(\vec{p}))^2}, \\
p_0(\vec{p}) = \sqrt{\vec{p} \cdot \vec{p} + m^2}.
\end{multline*}
Here and in the rest part of this Section we identify the Hilbert space $\mathcal{H}_{m,0}^{\oplus}
= E_+\mathcal{H}$
of positive energy distributional solutions $\phi$ of the Dirac equation with the ordinary functions
$\widetilde{\phi}$ on the orbit $\mathscr{O}_{m,0,0,0}$
which they induce on the orbit in the manner described above.
Correspondingly we identify the Hilbert space $\mathcal{H}$ of distributional solutions
$\phi$ of Dirac equation with the ordinary functions $\widetilde{\phi}$
on the disjoint sum of orbits $\mathscr{O}_{m,0,0,0} \sqcup \mathscr{O}_{-m,0,0,0}$ ($= \textrm{supp} \, \widetilde{\phi}$
of $\widetilde{\phi}$ regarded as distribution). Similarly, we identify
the Hilbert space $\mathcal{H}_{-m,0}^{\ominus}
= E_-\mathcal{H}$ of negative energy distributional solutions $\phi$ of Dirac equation
with the corresponding ordinary functions on $\mathscr{O}_{m,0,0,0} \sqcup \mathscr{O}_{-m,0,0,0}$
having the support on $\mathscr{O}_{-m,0,0,0}$.

In the later stage of the construction of the free Dirac field we will need a unitary
involutive (and thus self-adjoint) operator
$\In$, which we call \emph{parity number operator}, canonically related to the Fock space construction. 
In order to indicate the relation of the parity number operator $\In$ to the corresponding Fock space
$\Gamma\big(\mathcal{H}_{m,0}^{\oplus}\big)$, we use the subscript $\oplus$: ${\In}_\oplus$.

In order to define ${\In}_\oplus$ recall that 
every element $\Phi \in \Gamma\big(\mathcal{H}_{m,0}^{\oplus}\big)$
may be uniquely represented as the sum 
\begin{equation}\label{GeneralPsiInGamma(H)}
\Phi = \sum \limits_{n \geq 0} \Phi_n  
\end{equation}
over all $n= 0, 1, 2, \ldots $ of the orthogonal components 
$\Phi_n \in \big(\mathcal{H'}\big)^{\widehat{\otimes} n}$
-- the so called $n$-particle states, with 
\begin{equation}\label{NormGeneralPsiInGamma(H)}
\|\Phi\|^2 = \sum \limits_{n \geq 0} \| \Phi_n \|^2 < +\infty. 
\end{equation}
We define on the Fock space a bounded self-adjoint operator ${\In}_\oplus$ -- parity number operator --
which maps a general state $\Phi \in \Gamma\big(\mathcal{H}_{m,0}^{\oplus}\big)$ 
defined by (\ref{GeneralPsiInGamma(H)}) into the following state
\[
{\In}_\oplus \Phi = \sum \limits_{n \geq 0} \, (-1)^n \, \Phi_n. 
\]
It is evident that ${\In}_\oplus$ is unitary and involutive  (thus self-adjoint) 
\[
{\In}_\oplus^2 = \boldsymbol{1}, \,\,\, {\In}_\oplus^* = {\In}_\oplus
\] 
and that ${\In}_\oplus$ anticommutes with the annihilation (and creation) operators:
\[
a_{\oplus}(\widetilde{\phi}) \, {\In}_\oplus =
- {\In}_\oplus \,  a_{\oplus}(\widetilde{\phi}).
\]

Note that the unitary involution $\In$ on general Fock space, and in particular ${\In}_\oplus$,
commutes with any (bounded or even unbounded) operator $B$ which transforms the closed sub spaces of fixed 
particle number into themselves
(in case $B$ is unbounded we assume $\Dom \, B$ to be a linear subspace
or still more generally with $\Dom \, B$ to be closed under operation 
of multiplication by $-1$). In particular $\In$ (or ${\In}_\oplus$) commutes
with any operator of the form 
\[
B = \Gamma(A) = \sum \limits_{n=0}^{\infty} A^{\otimes n},
\]
namely:
\[
\big[\Gamma(A), {\In}_\oplus \big] = 0 \,\,\, \textrm{on} \,\,\,
\Dom \Gamma(A),
\]
irrespectively if $A$ is bounded or not, but with linear $\Dom A$
and $\Dom \Gamma(A)$. This in particular means that the operator 
${\In}_\oplus$ commutes:
\[
\Bigg[\,\,\,
\Gamma \Big(\widetilde{V^\oplus} \, U^{{}_{(m,0,0,0)} L^{{}^{1/2}}} \, (\widetilde{V^\oplus} )^{-1} \Big) \,\,\,,
\,\,\,
{\In}_\oplus \,\,\, \Bigg] = 0
\]
with the representation of $T_4 \circledS SL(2, \mathbb{C})$ acting in the Fock space
$\Gamma\big(\mathcal{H}_{m,0}^{\oplus}\big)$.

\vspace*{1cm}

\begin{rem}\label{TwoRepOfaa^+InFermiFock}
Note that in literature, e.g. \cite{Bratteli-Robinson},  there is frequently used the following
construction of annihilation and creation operators, in a general Fock space (here we concentrate on
the fermionic Fock space) $\Gamma(\mathcal{H}')$. For each $u \in \mathcal{H}'$ of the single particle space
$\mathcal{H}'$  we define the operators $a(u), a^+(u)=a(u)^+$
which by definition act on general element 
\begin{equation}\label{GeneralPsiInGamma(H')}
\Phi = \sum \limits_{n \geq 0} \Phi_n, \,\,\,
\Phi_n \in \mathcal{H}'^{\widehat{\otimes} \, n}
\end{equation}
with
\begin{equation}\label{NormGeneralPsiInGamma(H')}
\|\Phi\|^2 = \sum \limits_{n \geq 0} \| \Phi_n \|^2 < +\infty, 
\end{equation}
of the Fock space
$\Gamma(\mathcal{H}')$, in the following manner
\[
\begin{split}
1) \,\,\,\,\, a(u) \big(\Phi = \Phi_0 \big) = 0, \\
2) \,\,\,\,\, a(u) \Phi =  \sum \limits_{n \geq 0} \, n^{1/2} \, \overline{u} \, \widehat{\otimes}_1 \, \Phi_n, \\
3) \,\,\,\,\, a(u)^+ \Phi =  \sum \limits_{n \geq 0} \,(n+1)^{1/2} \, u \, \widehat{\otimes} \, \Phi_n.  
\end{split}
\]
Here $\widehat{\otimes}$ and
$\widehat{\otimes}_1$ denote respectively the antisymmetrized $n$-fold tensor product and the 
antisymmetrized $1$-contraction, uniquely determined by the formulae
\[
v_{{}_{1}} \, \widehat{\otimes} \, \cdots \widehat{\otimes} \, v_{{}_{n}} =
(n!)^{-1} \sum \limits_{\pi} \textrm{\emph{sign}} \, (\pi) \, v_{{}_{\pi(1)}} \otimes
\cdots \otimes v_{{}_{\pi(n)}} \,\,\,\,
v_{{}_{i}} \in  \in \mathcal{H}',
\]
\[
u \, \widehat{\otimes}_1 v_{{}_{1}} \, \widehat{\otimes} \, \cdots \widehat{\otimes} \, v_{{}_{n}}
= (n!)^{-1} \sum \limits_{\pi} \textrm{\emph{sign}} \, (\pi) \, \langle u,v_{{}_{\pi(1)}} \rangle \, 
 v_{{}_{\pi(2)}} \otimes
\cdots \otimes v_{{}_{\pi(n)}},
 \,\,\,
u \in \mathcal{H}'^{*}, v_{{}_{i}} \in \mathcal{H}',
\]
with the sums ranging over all permutations $\pi$ of the natural numbers $1, \ldots, n$, and with
the evaluation $\langle u,v_{{}_{\pi(1)}} \rangle$ of $u$, understood as a linear functional $\mathcal{H}'^{*}$,
on $v_{{}_{\pi(1)}} \in \mathcal{H}'$ equal
\[
\langle u,v_{{}_{\pi(1)}} \rangle = (\overline{u},v_{{}_{\pi(n)}})
\]
to the inner product of the elements $\overline{u}, v_{{}_{\pi(n)}} \in \mathcal{H}'$. 
Note that in all the relevant physical situations the single particle Hilbert spaces and the corresponding Fock spaces have natural real structure and are equal to complexifications of real Hilbert spaces with naturally defined
complex conjugations $\overline{(\cdot)}$ in them. Recall also that the map 
$\mathcal{H}' \ni u \mapsto \overline{u}$ defines a linear isomorphism of the Hilbert space   
$\mathcal{H}'$ into the adjoint Hilbert space $\overline{\mathcal{H}'}$, which in turn can be identified with the 
Hilbert space of linear functionals on $\mathcal{H}'$, by the Riesz representation theorem.

However, we will interchangeably be using another, unitarily equivalent, realization of the annihilation
and creation operators in the Fock space, which is more frequently used by mathematicians
(and fits well with that used e.g. in \cite{hida}, \cite{obataJFA}, \cite{obata-book}, \cite{HKPS}
for bosons, when adopting their results to the fermion case),
because we will refer to the works \cite{hida}, \cite{obataJFA}, \cite{obata-book},
in the following part of our work. Let us call it the
\emph{modified realization of annihilation-creation operators}
in the Fock space. This realization used by mathematicians is more natural
for the interpretation of the creation and annihilation operators as derivations
(or graded derivations in case of Fermi Fock space) on a nuclear (skew-commutative, or say Grassmann,
in case of Fermi Fock space) algebra of
Hida test functions on an (infinite-dimensional) strong dual space to a nuclear space.

In order to define it we first slightly modify the norm (\ref{NormGeneralPsiInGamma(H')})  of a 
general element  (\ref{GeneralPsiInGamma(H')}) and put for its square instead
\[
\|\Phi\|_{0}^2 = \sum \limits_{n \geq 0} \, n! \, \| \Phi_n \|^2.
\] 
Then we define the annihilation and creation operators through their action on general such element
$\Phi$ given by the following formulae 
\[
\begin{split}
1) \,\,\,\,\, a(u) \big(\Phi = \Phi_0 \big) = 0, \\
2) \,\,\,\,\, a(u) \Phi =  \sum \limits_{n \geq 0} \, n \, \overline{u} \, \widehat{\otimes}_1 \, \Phi_n, \\
3) \,\,\,\,\, a(u)^+ \Phi =  \sum \limits_{n \geq 0} \, u \, \widehat{\otimes} \, \Phi_n.  
\end{split}
\]

The unitary operator:
\[
U \Big( \sum \limits_{n \geq 0} \Phi_n \Big) =
\sum \limits_{n \geq 0} (n!)^{-1/2} \, \Phi_n, \,\,\,
U^{-1} \Big( \sum \limits_{n \geq 0} \Phi_n \Big) =
\sum \limits_{n \geq 0} (n!)^{1/2} \, \Phi_n,
\]
with the convention that $0!=1$,
gives the unitary equivalence between the two realizations of the annihilation and creation operators
in the Fock spaces, as well as of the representations of $T_4 \circledS SL(2, \mathbb{C})$
in the corresponding Fock spaces.
\end{rem}

\vspace*{1cm}

\subsection{Application of the Segal second quantization functor to the space $\mathcal{H}_{-m,0}^{\ominus \, \flat}$ of spinors conjugated to the spinors of the subspace $\mathcal{H}_{-m,0}^{\ominus}$}\label{positron}

In the next step we apply the functor $\Gamma$ of fermionic second quantization to the subspace
$\mathcal{H}_{-m,0}^{\ominus}$ and obtain the fermionic Fock space
\[
\Gamma(\mathcal{H}_{-m,0}^{\ominus}) = \mathbb{C} \oplus \mathcal{H}_{-m,0}^{\ominus} \oplus
\big( \mathcal{H}_{-m,0}^{\ominus} \big)^{\widehat{\otimes} 2} \oplus 
\big( \mathcal{H}_{-m,0}^{\ominus} \big)^{\widehat{\otimes} 3} \oplus \ldots;
\]
but the above mentioned interchange of the emission and absorption operators in 
$\Gamma\big(\mathcal{H}_{-m,0}^{\ominus} \big)$ results in 
replacing the single particle Hilbert space $\mathcal{H}_{-m,0}^{\ominus} = E_-\mathcal{H}$ 
with a conjugated one 
$\mathcal{H}_{-m,0}^{\ominus \, \flat}$ and in replacing of the representation (\ref{rep-on-H-}) acting in $\mathcal{H}_{-m,0}^{\ominus}$ with another conjugated representation acting in the Hilbert space $\mathcal{H}_{-m,0}^{\ominus \, \flat}$. 

This procedure is the well known basis for the solution of the ''negative energy states problem'' in relativistic
quantum field theory, therefore we only sketch briefly the general lines, presenting only the final results
in case of the free quantum Dirac field respecting the Dirac equation.
Namely, the solution is based on the observation that the negative energy solutions
lying in $\mathcal{H}_{-m,0}^{\ominus} = E_-\mathcal{H}$ (classically the negative energy solutions of the
equation which is to be fulfilled by the quantized field, here of the Dirac equation $D \phi = m\phi$ (\ref{DiracEqBisp})), should not be interpreted
as negative energy solutions of the original equation (here Dirac equation), but rather as a
kind of conjugation of \emph{positive} energy solutions of a conjugation of the original (here Dirac) equation,
with the conjugation depending on the actual kind of field. In particular for the scalar (complex) field fulfilling the Klein-Gordon equation the conjugation coincides with the ordinary complex conjugation (but only accidentally).

For (free) Dirac field respecting  Dirac equation the conjugation is slightly more 
complicated and the conjugated equation
does not coincide with the original Dirac equation. In the more general higher spin local fields
the conjugation is similar as for the Dirac equation, and is easy to guess  with its general definition being naturally determined by the general construction of the single particle Hilbert space of the field (with local transformation law). 

Namely, in general case of globally hyperbolic space-time
and a free field, say $\phi$, on it, we can extract the essential points of the construction of the free field
on the flat Minkowski manifold, although the particular computations would be much less easy to handle.
In any case the space-time manifold with its globally hyperbolic causal structure (given by a Lorenzian metric)
is crucial, together with the type of field $\phi$ with its local transformation rule
fixing the associated type of bundle with $\phi$ ranging over its sections, and respecting a hyperbolic differential equation $D\phi = m\phi$.
If a preferable and natural assumptions of analytic type are put
on the pseudo-Riemannian space-time manifold (compare e.g. \cite{Stro}, \cite{HBaum}) then the
Lorenzian metric induces a Krein structure in the space of sections $\phi$ (compare
the formulas (\ref{inn-x-step1}), (\ref{Jstep1}) of Subsect. \ref{1/2VF} in the special case of
flat Minkowski space-time and the Dirac bispinors $\phi$ on it with the transformation law (\ref{Ustep1})).
We expect the corresponding differential operator $D$ to be not merely Krein-self-adjoint, but moreover that it allows orthogonal spectral decomposition similar to that obtained in
Subsection \ref{1/2VF} for the ordinary Dirac operator $D$ (being of ''generalized spectral-type''). This assumption is nontrivial, as in the Krein space
Krein-self-adjoint operator in general does not allow any spectral decomposition of the type obtained
in Subsect. \ref{1/2VF} for $D$ (compare e.g. the classic Dunford-Schwartz analysis of the type of generalized spectral decompositions
of non-normal operators). In particular the method of extension of the construction of a free field on
more general space-times proposed here have a rather restricted domain of validity, and is confined to situations with rather very special kind of corresponding hyperbolic differential operators $D$
allowing ''generalized'' spectral decompositions.
Of course in general the spectral decomposition of $D$ may contain a discrete component,
or even consist of purely discrete part, depending on topology of the space-time manifold,
and in general are not Krein-orthogonal.

Next we consider the generalized eigenspace, which we agreed to denote by $\mathcal{H}$, of the Krein-self-adjoint operator $D$, corresponding to the eigenvalue $m$, and which consists of all
distributional solutions $\phi$ of the equation $D\phi = m\phi$. The closed sub spaces of generalised eigenspaces corresponding to the generalized eigenvalues of $D$ inherit non-degenerate Krein-space structure
from the initial Krein space of sections $\phi$ in which $D$ acts. The restriction of the Krein-self-adjoint operator $D$
to this subspace $\mathcal{H}$ is not only Krein-self-adjoint but likewise self-adjoint with respect to the inherited Krein space and Hilbert space structures on $\mathcal{H}$, with well-defined direct sum structure $\mathcal{H}=
E_+\mathcal{H} \oplus E_\mathcal{H}$ with closed sub spaces $E_{\pm}\mathcal{H}$ which are orthogonal
and with non-degenerate Krein space structure. Moreover, the operator $D$ is of ''generalized spectral-type'' and admits generalized spectral
orthogonal decomposition in the sense of Neumark-Lanze \cite{Lanze}, \cite{NeumarkLinOp}, explicitly
computed in Subsections \ref{e1}-\ref{1/2VF},
with each generalized eigenspace which inherits non-degenerate Hilbert space and Krein space structure.
This is far not the case for general Krein-self-adjoint operator, compare \cite{Bog}.
In particular the space $\mathcal{H}$ of generalized
eigenvectors of $D$ corresponding to the generalized real eigenvalue $m>0$ (say mass) inherits
non-degenerate and natural Krein space structure, in particular Hilbert space structure.
We expect that the space-time manifold, especially its causal structure, allows picking up
the natural discrete operation of time-orientation-reversing in terms of an involutive unitary operator
(say the $\textrm{sign} \, (H)$ = $H|H|^{-1}$ of the Hamiltonian
operator $H$ in $\mathcal{H}$) with the property that the change of time orientation transformation acts
through $\textrm{sign} \, (H)$ as an involutive unitary which exchanges
positive energy subspace $E_+\mathcal{H}$ with the negative
energy subspace $E_-\mathcal{H}$ of $\mathcal{H}$. In case of globally hyperbolic and highly
symmetric space-times with time symmetry (e.g. Einstein Static Universe) this plan
is within our grasp. In particular the harmonic analysis of \cite{PaneitzSegalI}- \cite{PaneitzSegalIII} is sufficiently effective on the Einstein Universe
to allow e.g. construction of QED on it together with the proof of its convergence, compare \cite{SegalZhouQED}.
In general the conjugation corresponding to the division of ''positive'' and ''negative energy''
solution sub spaces $E_+\mathcal{H}$ and $E_+\mathcal{H}$ of the space of distributional solutions
of $D\phi=m\phi$ is easy to guess and is strongly suggested by the geometric context.
Construction of the involutive unitary which corresponds to the division into ''positive'' and ''negative
energy'' solution sub spaces is more tricky when time symmetry is lacking at the space-time geometry level, and reflects the conformal (causal) structure of space-time
in the operator-spectral format. In fact construction of this division involves spectral decomposition of non-normal, Krein-self-adjoint operator $D$, and as we know there are no general theorems which would assure existence of such decompositions nor its sufficiently regular behavior. This is the essential source of difficulty in achieving the honest
division into ''positive'' and ''negative'' frequency modes.
Once a generalized spectral Krein-orthogonal decomposition of $D$, similar to that presented in Subsections
\ref{e1}-\ref{1/2VF} is successful, the involutive unitary and
the corresponding conjugation can be easily guessed. This is the case e.g. for the Einstein Universe, compare
\cite{PaneitzSegalI}- \cite{PaneitzSegalIII}. It can be achieved by explicit expansion
of the general solution of the Dirac equation $D\phi = m\phi$ into ''Einstein spinor modes'' (as called by
Segal and Zhou) and explicit division of the modes into positive and negative frequency parts.
This is a good example to study the relationship of the conformal structure and the corresponding
involutive unitary operator. Still, more interesting case we obtain for de Sitter space-time
lacking time symmetry, but with the sufficiently reach harmonic analysis to study quantum fields on it.
At least one example (of scalar quantum field on the three-dimensional de Sitter space-time), which comes naturally, we will encounter when studying infrared fields in later part
of this work. The generalized regular Krein-isometric decomposition of $D$ (with finite but arbitrary high dimension
of the fiber bundle of sections of the corresponding Clifford module), providing the corresponding
Krein-orthogonal decomposition of the initial Krein space acted on by $D$, serves as the generalization
of the Fourier transform $V_\mathcal{F}$ of Subsections \ref{e1}-\ref{VFforFreeFields} in case
of less symmetric globally hyperbolic space-times.

After this general remark concerning construction of free fields on more general space-time manifolds,
let us back to the construction of the free Dirac field on the flat Minkowski space-time, or more precisely,
to the conjugation, which accompany the division $\mathcal{H} = E_+\mathcal{H} \oplus E_+\mathcal{H}$
into positive and negative energy solutions of the ordinary Dirac equation $D\phi = m\phi$
constructed as above.

Recall once again, that in the current Section we are using $\tilde{\gamma}^\mu = (\gamma^\mu)^* = \gamma^0 \gamma^\mu\gamma^0$
in the space-time picture, and $\gamma^\mu$ in the momentum picture. This is because we are dealing mainly in the momentum picture
with a considerable number of various conjugations, so in order to simplify notation we wanted to eliminate the
additional tilde $\tilde{(\cdot)}$. Therefore, Dirac equation in spacetime variables reads
\[
i\Gamma^\mu \partial_\mu \phi = m \phi, \,\,\,\, \Gamma^\mu = \tilde{\gamma}^\mu = (\gamma^\mu)^* = \gamma^0 \gamma^\mu\gamma^0,
\,\,\, D = i\Gamma^\mu \partial_\mu,
\]
with the Clifford gamma generators in momentum picture equal $\gamma^\mu$ and, say in the chiral representation, are given by the formula
(\ref{chiralgamma}).

As remarked earlier, the negative energy solutions $\phi$ of the Dirac equation $[D-m]\phi=0$ should be interpreted as conjugations of
positive energy solutions $\phi^\flat$ of the conjugated
\begin{equation}\label{DiracConjugatedEq}
-i \partial_\mu \phi^\flat \big(\Gamma^\mu\big)^\flat = m\phi^\flat
\end{equation}
Dirac equation\footnote{In the standard notation used by physicist the conjugated spinor $\phi^\flat$
is written as $\phi^+ = \overline{\phi}^T$, which we have already reserved for the
operator conjugation of operators in the Fock space. The complex conjugation followed by transposition we agree to denote in this section by
using the $+$ superscript interchangeably with the conjugation superscript $\flat$,
which is customary in physical literature concerning Dirac bispinors and Dirac equation.} $[D - m]^\flat \phi^\flat = [D^\flat - m]\phi^\flat = 0$.
The representation space of the conjugated representation is defined as the Hilbert space $\mathcal{H}_{-m,0}^{\ominus \, \flat}$ of conjugated bispinors
\begin{equation}\label{ConjugationForDiracField}
(\widetilde{\phi})^\flat (p) = {\widetilde{\phi}(-p)}^{+}
= \big(\overline{\widetilde{\phi}(-p)} \big)^T
\end{equation}
with
$\widetilde{\phi} = V^{\ominus} {\widetilde{\psi}_{{}_{-m,0}}}$
ranging over the Hilbert space $\mathcal{H}_{-m,0}^{\ominus}$ of bispinors
concentrated on the orbit $\mathscr{O}_{{}_{-m,0,0,0}}$
(i.e. with ${\widetilde{\psi}_{{}_{-m,0}}}$ ranging over the Hilbert space of the representation
\[
U^{{}_{(-m,0,0,0)} L^{{}^{1/2}}}
\]
concentrated on $\mathscr{O}_{{}_{-m,0,0,0}}$, compare Example 1, Subsection \ref{e1}).
Here $(\cdot)^T$ stands for transposition operation and
\[
\big(\Gamma^\mu\big)^\flat = \big(\overline{\Gamma^\mu}\big)^T = \Gamma^{\mu +}.
\]
In the space-time coordinates, i.e. after inverse Fourier transformation, the formula for conjugation
is equivalent to
\[
\phi^\flat(x) = \phi(x)^+ = \big(\overline{\phi(x)} \big)^T.
\]
On the Hilbert space $\mathcal{H}_{-m,0}^{\ominus \, \flat}$ of conjugated bispinors there is defined
the (conjugated) inner product
\begin{multline*}
\big(\phi^\flat, {\phi'}^{\flat}\big)_\flat
= \big((\widetilde{\phi})^\flat, (\widetilde{\phi}')^\flat)_\flat =
(\phi', \phi) \\ =
\int \limits_{x^0 = t = const.} \Big(\phi'(\vec{x}, t), \phi(\vec{x}, t) \Big)_{{}_{\mathbb{C}^4}}
\, \ud^3 x \\
= \int \limits_{\mathbb{R}^3} \Big(\widetilde{\phi}'(-\vec{p}, -|p_0(\vec{p})|),
\widetilde{\phi}(-\vec{p}, -|p_0(\vec{p})|) \Big)_{{}_{\mathbb{C}^4}}
\, \frac{\ud^3 \vec{p}}{(2 p_0)^2} \\=
\int \limits_{\mathbb{R}^3} \Big(\widetilde{\phi}'(\vec{p}, p_0(\vec{p})),
\widetilde{\phi}(\vec{p}, p_0(\vec{p})) \Big)_{{}_{\mathbb{C}^4}}
\, \frac{\ud^3 \vec{p}}{(2 p_0)^2} \\ =
\int \limits_{\mathscr{O}_{-m,0,0,0}} \Big(\widetilde{\phi}'(p),
\phi(p) \Big)_{{}_{\mathbb{C}^4}}
\, \frac{\ud \mu_{{}_{m,0}}(p)}{2 |p_0|} =
\big(\widetilde{\phi}',\widetilde{\phi}\big)_{{}_{\mathcal{H}_{-m,0}^{\ominus}}},
\,\,\,\, p_0(\vec{p}) = -\sqrt{\vec{p} \cdot \vec{p} + m^2}.
\end{multline*}
where $(\cdot, \cdot )$ is the inner product (\ref{Inn-Prod-Single-Dirac}) in the Hilbert space
$\mathcal{H}_{-m,0}^{\ominus} \subset \mathcal{H}$ of distributional solutions (whose Fourier transforms
are concentrated on $\mathscr{O}_{-m,0,0,0}$)
of Dirac equation defined above, which induces, through Fourier transform,
the inner product $\big( \cdot, \cdot \big)_{{}_{\mathcal{H}_{-m,0}^{\ominus}}}$ on their Fourier transforms.
In the Hilbert space $\mathcal{H}_{-m,0}^{\ominus \, \flat}$ there are defined the operations
of multiplication by a number $\alpha \in \mathbb{C}$ and addition by the respective ordinary operations
in $\mathcal{H}_{-m,0}^{\ominus}$, in the following manner
\[
\alpha \cdot (\widetilde{\phi})^\flat = (\overline{\alpha} \widetilde{\phi})^\flat = \alpha (\widetilde{\phi})^\flat, \,\,\,
(\widetilde{\phi})^\flat + (\widetilde{\phi}')^\flat
= (\widetilde{\phi} + \widetilde{\phi})^\flat, \,\,\,
\widetilde{\phi}, \widetilde{\phi}' \in \mathcal{H}_{-m,0}^{\ominus}.
\]

From the formula (\ref{ConjugationForDiracField}) one easily see that the Fourier transforms of 
the conjugated bispinors are concentrated on the positive energy orbit $\mathscr{O}_{m,0,0,0}$
in the momentum space, and thus they are positive energy solutions of the conjugated Dirac
equation (\ref{DiracConjugatedEq}).   

Then on the conjugated Hilbert space $\mathcal{H}_{-m,0}^{\ominus \, \flat}$ (of conjugated bispinors
concentrated on
the positive energy orbit $\mathscr{O}_{{}_{m,0,0,0}}$) there acts naturally the representation
\begin{equation}\label{rep-c-on-Hc}
\big\{\widetilde{V^\ominus} U^{{}_{(-m,0,0,0)} L^{{}^{1/2}}} \,
(\widetilde{V^\ominus})^{-1} \big\}^\flat
\end{equation}
conjugated to
\[
\widetilde{V^\ominus} U^{{}_{(-m,0,0,0)} L^{{}^{1/2}}} \, (\widetilde{V^\ominus})^{-1}
\]
with the general definition of conjugation
\[
U^\flat(\widetilde{\phi})^\flat = (U\widetilde{\phi})^\flat.
\]
Because the spin corresponding to the conjugated representation (\ref{rep-c-on-Hc})
is likewise $1/2$ and the orbit is equal $\mathscr{O}_{m,0,0,0}$, then one can guess
that (\ref{rep-c-on-Hc}) is likewise equivalent to (\ref{m}), by Mackey's classification.
Indeed, one can construct explicit equivalence similarly as $V^\oplus$ in Example 1
(Subsection \ref{e1}) with additional transpositions and complex conjugations in this construction.

Thus, to the space  $\mathcal{H}_{-m,0}^{\ominus \, \flat}$ we apply the Segal's functor $\Gamma$ of fermionic quantization and obtain the fermionic Fock space
\[ 
\mathcal{H}^{\ominus}_{F} = \Gamma\big(\mathcal{H}_{-m,0}^{\ominus \, \flat}\big) 
= \mathbb{C} \oplus \mathcal{H}_{-m,0}^{\ominus \flat} \oplus
\big( \mathcal{H}_{-m,0}^{\ominus \, \flat} \big)^{\widehat{\otimes} 2} \oplus 
\big( \mathcal{H}_{-m,0}^{\ominus \, \flat} \big)^{\widehat{\otimes} 3}
\oplus \ldots;
\] 
with the unitary representation
\begin{multline*} 
\Gamma \Bigg(\big\{\widetilde{V^\ominus} U^{{}_{(-m,0,0,0)} L^{{}^{1/2}}} \, (\widetilde{V^\ominus})^{-1} \big\}^{\flat} \Bigg)
= \bigoplus \limits_{n = 0,1,2 \ldots}
\Bigg(\big\{\widetilde{V^\ominus} U^{{}_{(-m,0,0,0)} L^{{}^{1/2}}} \, (\widetilde{V^\ominus})^{-1} \big\}^{\flat} \Bigg)^{\widehat{\otimes} n}.
\end{multline*}

The conjugation $\big({\widetilde{\phi}} \big)^\flat$ of
the bispinor function concentrated on $\mathscr{O}_{-m,0,0,0}$
will sometimes be denoted by ${\widetilde{\phi}}^{\flat}$
in order to simplify notation.
We construct in the standard manner the map
\[
\mathcal{H}_{-m,0}^{\ominus \, \flat} \ni
{\widetilde{\phi}}^{\flat} \,\, \longrightarrow \,\,\,
a_{\ominus}\big({\widetilde{\phi}}^{\flat}\big), \,\,\,
a_{\ominus}^+\big({\widetilde{\phi}}^{\flat}\big)
= {a_{\ominus}\big({\widetilde{\phi}}^{\flat}\big)}^+
\]
from $\mathcal{H}_{-m,0}^{\ominus \, \flat}$ to the families of (ordinary operators, not distributions) of
annihilation and creation operators acting in the fermionic Fock space
$\Gamma\big(\mathcal{H}_{-m,0}^{\ominus \, \flat}\big)$, fulfilling the canonical anticommutation relations:
\begin{multline*}
\Big\{a_{\ominus}\big({\widetilde{\phi}}^{\flat}\big), \,\,\,
{a_{\ominus}\big(\widetilde{\phi}'^{\flat}\big)}^+ \Big\}
= \big({\widetilde{\phi}}^\flat,
\widetilde{\phi}'^{\flat}\big)_{{}_{\mathcal{H}_{-m,0}^{\ominus \, \flat}}} \\
= \big({\widetilde{\phi}}^\flat,
\widetilde{\phi}'^{\flat}\big)_\flat \\
= \Big(\widetilde{\phi}',
\widetilde{\phi}\Big)_{{}_{\mathcal{H}_{-m,0}^{\ominus}}} \\
= \int \limits_{\mathscr{O}_{-m,0,0,0}} \Big(\widetilde{\phi}'(p),
\widetilde{\phi}(p) \Big)_{{}_{\mathbb{C}^4}}
\, \frac{\ud \mu_{{}_{-m,0}}(p)}{2 |p_0|} \\
= \int \limits_{\mathbb{R}^3}
\Big(\widetilde{\phi}'(-\vec{p}, -|p_0(\vec{p})|), \,\,
\widetilde{\phi}(-\vec{p}, -|p_0(\vec{p})|) \Big)_{{}_{\mathbb{C}^4}}
\, \frac{\ud^3 \vec{p}}{(2 |p_0(\vec{p})|)^2}, \\
p_0(\vec{p}) = -\sqrt{\vec{p} \cdot \vec{p} + m^2}.
\end{multline*}

In particular the representation of the group $T_4 \circledS SL(2, \mathbb{C})$ which acts in the Fock 
space $\mathcal{H}^{\ominus}_{F}$ is equal 
\[
\Gamma \Bigg(\big\{\widetilde{V^\ominus} U^{{}_{(-m,0,0,0)} L^{{}^{1/2}}} \, (\widetilde{V^\ominus})^{-1} \big\}^\flat \Bigg). 
\]

Of course on the Fock space $\mathcal{H}^{\ominus}_{F}=\Gamma\big(\mathcal{H}_{-m,0}^{\ominus \, \flat}\big)$
we have the corresponding parity number (unitary and involutive) operator  ${\In}_\ominus$  fulfilling 
\[
{\In}_\ominus^2 = \boldsymbol{1}, \,\,\,  {\In}_\ominus^* = {\In}_\ominus,
\] 
and such that ${\In}_\ominus$ anticommutes with the annihilation (and creation) operators:
\[
\Big\{ a_{\ominus}\big((\widetilde{\phi}|_{{}_{\mathscr{O}_{-m,0,0,0}}})^\flat\big), \, {\In}_\ominus \Big\} = 0.
\]

Of course the operator ${\In}_\ominus$ commutes:
\[
\Bigg[\,\,\,
\Gamma \Big(\big\{\widetilde{V^\ominus} U^{{}_{(-m,0,0,0)} L^{{}^{1/2}}} \, (\widetilde{V^\ominus})^{-1} \big\}^{\flat} \Big) 
\,\,\,,
\,\,\,
{\In}_\ominus \,\,\, \Bigg] =0
\]
with the representation of $T_4 \circledS SL(2, \mathbb{C})$ acting in the Fock space
$\Gamma\big(\mathcal{H}_{-m,0}^{\oplus \, \flat}\big)$ and with any operator of the form
$\Gamma(A)$ (bounded or unbounded with linear $\Dom \, \Gamma(A)$
in $\Gamma\big(\mathcal{H}_{-m,0}^{\oplus \, \flat}\big)$).

\subsection{The Fock-Hilbert space $\mathcal{H}_F$ of the free Dirac 
field $\boldsymbol{\psi}$}\label{electron+positron}

The Hilbert space $\mathcal{H}_F$ of the free Dirac field is defined as the application
of the Fermi second quantization functor $\Gamma$ to the ``single particle'' Hilbert space 
$\mathcal{H}' = \mathcal{H}_{m,0}^{\oplus} \oplus \mathcal{H}_{-m,0}^{\ominus \, \flat}$--orthogonal sum of the 
Hilbert spaces $\mathcal{H}_{m,0}^{\oplus}$ and $\mathcal{H}_{-m,0}^{\ominus \, \flat}$. Therefore,
by the known property of the functor $\Gamma$, 
it is equal to the tensor product
\[
\mathcal{H}_F = \mathcal{H}^{\oplus}_{F} \otimes \mathcal{H}^{\ominus}_{F}
= \Gamma\big(\mathcal{H}_{m,0}^{\oplus}\big) \otimes \Gamma\big(\mathcal{H}_{-m,0}^{\ominus \, \flat}\big) 
= \Gamma\big(\mathcal{H}_{m,0}^{\oplus} \oplus \mathcal{H}_{-m,0}^{\ominus \, \flat}\big)
\]
of the fermion Fock
spaces $\mathcal{H}^{\oplus}_{F}= \Gamma\big(\mathcal{H}_{m,0}^{\oplus}\big)$ and 
$\mathcal{H}^{\ominus}_{F}= \Gamma\big(\mathcal{H}_{-m,0}^{\ominus \, \flat}\big)$ with the representation
\[
\bigg[ \bigoplus \limits_{n=0,1,2, \ldots}
\Big(\widetilde{V^\oplus} \, U^{{}_{(m,0,0,0)} L^{{}^{1/2}}} \, 
(\widetilde{V^\oplus})^{-1} \Big)^{\widehat{\otimes} n}\bigg] \otimes 
\bigg[  
\bigoplus \limits_{n=0,1,2, \ldots} \Big(\big\{\widetilde{V^\ominus} U^{{}_{(-m,0,0,0)} L^{{}^{1/2}}} \, 
(\widetilde{V^\ominus})^{-1} \big\}^\flat  \Big)^{\widehat{\otimes} n} \bigg]
\] 
of the group $T_4 \circledS SL(2, \mathbb{C})$ acting in the Hilbert space $\mathcal{H}_F$. Here we should
emphasize that $\Gamma\big(\mathcal{H}_{m,0}^{\oplus} \oplus \mathcal{H}_{-m,0}^{\ominus \, \flat}\big)$
is understood in the sense that each simple tensor in $\Gamma\big(\mathcal{H}_{m,0}^{\oplus} \oplus \mathcal{H}_{-m,0}^{\ominus \, \flat}\big)$
is \emph{separately} antisymmetrized in all factors coming from
$\mathcal{H}_{m,0}^{\oplus}$ and \emph{separately} in all factors coming from $\mathcal{H}_{-m,0}^{\ominus \, \flat}$.

Now observe that 
\[
\big\{\widetilde{V^\ominus} \, U^{{}_{(-m,0,0,0)} L^{{}^{1/2}}} \, (\widetilde{V^\ominus})^{-1} \big\}^\flat
= (\widetilde{V^\ominus})^{+ \, -1} \, \{ U^{{}_{(-m,0,0,0)} L^{{}^{1/2}}} \}^\flat \, (\widetilde{V^\ominus})^+. 
\]

Because by Mackey's construction of induced representation it follows that
\[
\big\{ U^{{}_{(-m,0,0,0)} L^{{}^{1/2}}} \big\}^\flat
= {S}^{-1} \, U^{{}_{(m,0,0,0)} L^{{}^{1/2}}} \, S
\]
with some (involutive) unitary operator $S$, we have
\[
\big\{\widetilde{V^\ominus} U^{{}_{(-m,0,0,0)} L^{{}^{1/2}}} \, (\widetilde{V^\ominus})^{-1} \big\}^\flat
= {U_0}^{-1} \,  U^{{}_{(m,0,0,0)} L^{{}^{1/2}}} \, U_0 , \,\,\, U_0 = S \, (\widetilde{V^\ominus})^+.
\]
Thus the joint spectrum of the translation generators of the representation acting
in the Hilbert space $\mathcal{H}_F$ of the free Dirac field thus constructed is concentrated on the
positive energy cone $C_+$, i.e. it is a positive energy field.

Into the Fock-Hilbert space $\mathcal{H}_F$ of the free Dirac field we again introduce in the standard manner
the families 
\[
\mathcal{H}_{m,0}^{\oplus} \oplus \mathcal{H}_{-m,0}^{\ominus \, \flat} \ni 
\widetilde{\phi}_1 \oplus \widetilde{\phi}_2 \,\,\, \longrightarrow 
a'\big(\widetilde{\phi}_1 \oplus \widetilde{\phi}_2\big), \,\,\, 
a'^+\big(\widetilde{\phi}_1 \oplus \widetilde{\phi}_2\big) 
= {a'\big(\widetilde{\phi}_1 \oplus \widetilde{\phi}_2\big)}^+,
\] 
fulfilling canonical anticummutation relations
\begin{multline}\label{AntiComRelFor-a-InH_F}
\Big\{a'\big(\widetilde{\phi}_1 \oplus \widetilde{\phi}_2\big), 
{a'\big(\widetilde{\phi}'_{1} \oplus \widetilde{\phi}'_{2}\big)}^+ \Big\} = \\ 
\Big(\, \widetilde{\phi}_1 \oplus \widetilde{\phi}_2, \,\,\, 
\widetilde{\phi}'_{1} \oplus \widetilde{\phi}'_{2} \,\Big)_{{}_{\mathcal{H}_{m,0}^{\oplus} \oplus \mathcal{H}_{-m,0}^{\ominus \, \flat}}} 
=\Big( \widetilde{\phi}_1,  
\widetilde{\phi}'_{1}\Big)_{{}_{\mathcal{H}_{m,0}^{\oplus}}} 
+
\Big( \widetilde{\phi}_2,  
\widetilde{\phi}'_{2}\Big)_{{}_{\mathcal{H}_{-m,0}^{\ominus \, \flat}}},
\end{multline}
where $(\cdot, \cdot)_{{}_{\mathcal{H}}}$ stands for the inner product on the Hilbert 
space $\mathcal{H}$. Here $\widetilde{\phi}_{1}, \widetilde{\phi}'_{1} \in \mathcal{H}_{m,0}^{\oplus}$
and $\widetilde{\phi}_{2}, \widetilde{\phi}'_{2} \in \mathcal{H}_{-m,0}^{\ominus \, \flat}$.

It follows that\footnote{Note that the equality $\Gamma(\mathcal{H}_1 \oplus \mathcal{H}_2)
= \Gamma(\mathcal{H}_1) \otimes \Gamma(\mathcal{H}_2)$ expresses in fact existence of a
\emph{canonical} unitary isomorphism respecting the relevant Fock structure
with particular importance of the canonical nature of the identification (a mere existence of a
unitary map, here in the context of separable Hilbert spaces, is trivial and would tell us nothing as there is plenty of such maps devoid of any relevance).
The point is that the identification makes the following equality to hold
\[
a(u\oplus v) = a_{{}_{1}}(u) \otimes {\In}_{{}_{2}} + \boldsymbol{1} \otimes a_{{}_{2}}(v),
\]
for the corresponding annihilation and creation operators:
$a(u\oplus v), a(u\oplus v)^+$ acting in
$\Gamma(\mathcal{H}_1 \oplus \mathcal{H}_2)$, $a_{{}_{1}}(u), a_{{}_{1}}(u)^+$ acting in
$\Gamma(\mathcal{H}_1)$
and $a_{{}_{2}}(v), a_{{}_{2}}(v)^+$ in $\Gamma(\mathcal{H}_2)$. Recall that ${\In}_2$ is the
involutive unitary (and self-adjoint) parity number operator in Fock space $\Gamma(\mathcal{H}_2)$.
In fact in case of the fermionic Fock spaces we have two canonical choices for the identification
of the spaces $\Gamma(\mathcal{H}_1 \oplus \mathcal{H}_2)$ and
$\Gamma(\mathcal{H}_1) \otimes \Gamma(\mathcal{H}_2)$. The second identification makes the following equality to hold
\[
a(u\oplus v) = a_{{}_{1}}(u) \otimes \boldsymbol{1} + {\In}_{{}_{1}} \otimes a_{{}_{2}}(v)
\]
with the parity number involution ${\In}_{{}_{1}}$ of the Fock space $\Gamma(\mathcal{H}_1)$.
Thus, in particular we can use the other canonical identification, where instead of
(\ref{a_plusInH_F}), (\ref{a_minusInH_F}), (\ref{aInH_F}) we had
\[
\begin{split}
a'\big(\widetilde{\phi}_1 \oplus 0 \big) =
a_\oplus\big(\widetilde{\phi}_1\big) \otimes \boldsymbol{1},
\,\,\
\widetilde{\phi}_{1} \in \mathcal{H}_{m,0}^{\oplus}, \\
a'\big(0 \oplus \widetilde{\phi}_2\big) =
{\In}_\oplus \otimes a_\ominus\big(\widetilde{\phi}_2\big),
\,\,\
\widetilde{\phi}_{2} \in \mathcal{H}_{-m,0}^{\ominus \, \flat}, \\
a'\big(\widetilde{\phi}_1 \oplus \widetilde{\phi}_2) = a_\oplus\big(\widetilde{\phi}_1\big) \otimes
\boldsymbol{1}
+ {\In}_\oplus \otimes a_\ominus\big(\widetilde{\phi}_2\big).
\end{split}
\]

In case of the boson Fock spaces we have essentially one canonical identification
of the Fock spaces $\Gamma(\mathcal{H}_1 \oplus \mathcal{H}_2)$ and
$\Gamma(\mathcal{H}_1) \otimes \Gamma(\mathcal{H}_2)$ which makes the following equality
to hold
\[
a(u\oplus v) = a_{{}_{1}}(u) \otimes \boldsymbol{1} + \boldsymbol{1} \otimes a_{{}_{2}}(v).
\]
Therefore during the construction of a field with integer spin,
which is not essentially neutral (with antiparticles), when the fermionic functor $\Gamma$
is replaced with bosonic
and the anticommutation relations are replaced with commutation relations, the
involutive unitary and self adjoint operators ${\In}_\oplus$ and ${\In}_\ominus$ are replaced here
with the unital operator $\boldsymbol{1}$.}
\begin{equation}\label{a_plusInH_F}
a'\big(\widetilde{\phi}_1 \oplus 0 \big) =
a_\oplus\big(\widetilde{\phi}_1\big) \otimes {\In}_\ominus,
\,\,\
\widetilde{\phi}_{1} \in \mathcal{H}_{m,0}^{\oplus},
\end{equation}
\begin{equation}\label{a_minusInH_F}
a'\big(0 \oplus \widetilde{\phi}_2\big) =
\boldsymbol{1} \otimes a_\ominus\big(\widetilde{\phi}_2\big),
\,\,\
\widetilde{\phi}_{2} \in \mathcal{H}_{-m,0}^{\ominus \, \flat}
\end{equation}
and
\begin{equation}\label{aInH_F}
a'\big(\widetilde{\phi}_1 \oplus \widetilde{\phi}_2) = a_\oplus\big(\widetilde{\phi}_1\big) \otimes {\In}_\ominus
+ \boldsymbol{1} \otimes a_\ominus\big(\widetilde{\phi}_2\big).
\end{equation}
Here ${\In}_\ominus$ is the parity number (involutive and self-adjoint unitary)
operator in the Fock space
$\Gamma\big(\mathcal{H}_{-m,0}^{\ominus \, \flat}\big)$
anticommuting with $a_\ominus\big(\widetilde{\phi}_2\big)$.
The operators $a_\oplus(\widetilde{\phi}_1)$ act on $\Gamma\big(\mathcal{H}_{m,0}^{\oplus}\big)$
and $a_\ominus(\widetilde{\phi}_2)$, ${\In}_\ominus$ act on $\Gamma\big(\mathcal{H}_{-m,0}^{\ominus \, \flat}\big)$.

In order to simplify notation the operators (\ref{a_plusInH_F}) and (\ref{a_minusInH_F})
understood as operators in the total Fock space
\[
\mathcal{H}_F = \mathcal{H}^{\oplus}_{F} \otimes \mathcal{H}^{\ominus}_{F}
= \Gamma\big(\mathcal{H}_{m,0}^{\oplus}\big) \otimes \Gamma\big(\mathcal{H}_{-m,0}^{\ominus \, \flat}\big) 
= \Gamma\big(\mathcal{H}_{m,0}^{\oplus} \oplus \mathcal{H}_{-m,0}^{\ominus \, \flat}\big)
\] 
of the free Dirac field will likewise be denoted by $a_\oplus(\widetilde{\phi}_1)$
and $a_\ominus(\widetilde{\phi}_2)$, where $\widetilde{\phi}_1$ and $\widetilde{\phi}_2$ 
are understood as elements $\widetilde{\phi}_1\oplus0$ and $0\oplus \widetilde{\phi}_2$
of the Hilbert space $\mathcal{H}_{m,0}^{\oplus} \oplus \mathcal{H}_{-m,0}^{\ominus \, \flat}$
respectively, especially when the context suggest with what Fock
space we are working.

Note in particular that 
for the operators (\ref{a_plusInH_F}) and (\ref{a_minusInH_F}),
understood as operators on $\mathcal{H}_F$ and denoted simply by
$a_\oplus(\widetilde{\phi}_1)$ and $a_\ominus(\widetilde{\phi}_2)$, we have the following
canonical aticommutation relations (which follow from (\ref{AntiComRelFor-a-InH_F}))
\begin{equation}\label{AnticommutationRelationsInDiracFockSpace}
\begin{split}
\big\{a_\oplus(\widetilde{\phi}_{1}), a_\oplus(\widetilde{\phi}'_{1})^+\big\} =
\big(\widetilde{\phi}_{1},\widetilde{\phi}'_{1}\big)_{{}_{\mathcal{H}_{m,0}^{\oplus}}}, \\
\big\{a_\ominus(\widetilde{\phi}_{2}), a_\ominus(\widetilde{\phi}'_{2})^+\big\} =
\big(\widetilde{\phi}_{2},\widetilde{\phi}'_{2}\big)_{{}_{\mathcal{H}_{-m,0}^{\ominus \, \flat}}}, \\
\big\{a_\oplus(\widetilde{\phi}_{1}), a_\oplus(\widetilde{\phi}'_{1})\big\} =
\big\{a_\ominus(\widetilde{\phi}_{2}), a_\ominus(\widetilde{\phi}'_{2})\big\} = 0, \\
\big\{a_\oplus(\widetilde{\phi}_{1}), a_\ominus(\widetilde{\phi}'_{2})^+\big\} = 
\big\{a_\oplus(\widetilde{\phi}_{1}), a_\ominus(\widetilde{\phi}'_{2}) \big\} = 0,
\end{split}
\end{equation}
where again $\widetilde{\phi}_1, \widetilde{\phi}_1$ and $\widetilde{\phi}_2,\widetilde{\phi}'_2$ 
are understood respectively as elements $\widetilde{\phi}_1\oplus0, \widetilde{\phi}'_1\oplus0$ and 
$0\oplus \widetilde{\phi}_2, 0\oplus \widetilde{\phi}'_2$
of the Hilbert space $\mathcal{H}_{m,0}^{\oplus} \oplus \mathcal{H}_{-m,0}^{\ominus \, \flat}$.

Similarly, we may construct the Fock Hilbert space of the negative energy Dirac field
exchanging the absorption and emission operators
in the fermionic Fock space $\Gamma\big(\mathcal{H}_{m,0}^{\oplus} \big)$. The resulting representation
will differ by the interchanged role of the representations
$\widetilde{V^\oplus} \, U^{{}_{(m,0,0,0)} L^{{}^{1/2}}} \, (\widetilde{V^\oplus} )^{-1}$ and
$\widetilde{V^\ominus} U^{{}_{(-m,0,0,0)} L^{{}^{1/2}}} \, (\widetilde{V^\ominus})^{-1}$, i.e.
with the following representation
\[
\bigg[ \bigoplus \limits_{n=0,1,2, \ldots}
\Big(\big\{\widetilde{V^\oplus} \, U^{{}_{(m,0,0,0)} L^{{}^{1/2}}} \,
(\widetilde{V^\oplus} )^{-1} \big\}^\flat \Big)^{\widehat{\otimes} n}\bigg] \otimes
\bigg[
\bigoplus \limits_{n=0,1,2, \ldots} \Big(\widetilde{V^\ominus} U^{{}_{(-m,0,0,0)} L^{{}^{1/2}}} \,
(\widetilde{V^\ominus})^{-1} \Big)^{\widehat{\otimes} n} \bigg]
\]
in the Hilbert space of the free negative energy Dirac field with the joint spectrum of the translation generators concentrated on the negative energy cone $C_-$.
Now the conjugation of the representation acts on the conjugations of positive energy
bispinor solutions, i.e. concentrated on the negative energy orbit.

The functor $\Gamma$ allows us to
have a clear insight into the structure of the representation of $T_4 \circledS SL(2, \mathbb{C})$ 
acting in $\mathcal{H}_F$, as by construction it behaves functorially under the application
of $\Gamma$, applied separately to $\mathcal{H}_{m,0}^{\oplus}$ and $\mathcal{H}_{-m,0}^{\ominus \, \flat}$, and preserves
the structure $\mathcal{H}_F  
= \Gamma\big(\mathcal{H}_{m,0}^{\oplus} \big) \otimes \Gamma\big(\mathcal{H}_{-m,0}^{\ominus \, \flat} \big)$
because both $\mathcal{H}_{m,0}^{\oplus}$ and $\mathcal{H}_{-m,0}^{\ominus}$ are invariant for the representation 
of $T_4 \circledS SL(2, \mathbb{C})$ in the single particle Hilbert space 
$\mathcal{H}' = \mathcal{H}_{m,0}^{\oplus} \oplus \mathcal{H}_{-m,0}^{\ominus \, \flat}$.
In particular by the general properties of $\Gamma$ the representation of $T_4 \circledS SL(2, \mathbb{C})$
acting in $\mathcal{H}_F$ is naturally equivalent to the representation (in the positive energy case) 
\begin{multline*}
\Gamma\Big( U^{{}_{(m,0,0,0)} L^{{}^{1/2}}} \Big) \otimes \Gamma\Big( U^{{}_{(m,0,0,0)} L^{{}^{1/2}}} \Big) \\
= \bigg[ \bigoplus \limits_{n=0,1,2, \ldots}
\Big( U^{{}_{(m,0,0,0)} L^{{}^{1/2}}} \,  \Big)^{\widehat{\otimes} n}\bigg] \otimes 
\bigg[ 
 \bigoplus \limits_{n=0,1,2, \ldots} \Big( U^{{}_{(m,0,0,0)} L^{{}^{1/2}}} \Big)^{\widehat{\otimes} n} \bigg]
\end{multline*} 
and to the representation (in the negative energy case)
\begin{multline*}
\Gamma\Big( U^{{}_{(-m,0,0,0)} L^{{}^{1/2}}} \Big) \otimes \Gamma\Big( U^{{}_{(-m,0,0,0)} L^{{}^{1/2}}} \Big) \\
= \bigg[ \bigoplus \limits_{n=0,1,2, \ldots}
\Big( U^{{}_{(-m,0,0,0)} L^{{}^{1/2}}} \,  \Big)^{\widehat{\otimes} n}\bigg] \otimes 
\bigg[ 
 \bigoplus \limits_{n=0,1,2, \ldots} \Big( U^{{}_{(-m,0,0,0)} L^{{}^{1/2}}} \Big)^{\widehat{\otimes} n} \bigg],
\end{multline*} 
with the equivalence given by the unitary operator $\Gamma(V^\oplus) \otimes 
\Gamma\big(S \, (\widetilde{V^\ominus})^+\big)$  
in the positive energy case or by $\Gamma\big(S \, (\widetilde{V^\oplus})^+ \big) 
\otimes \Gamma(\widetilde{V^\ominus)}$ in the negative energy case.

Recall also the simple functorial property of $\Gamma$: for any group representations $U_1$ and $U_2$,
$\Gamma(U_1 \oplus U_2)$ is naturally equivalent to $\Gamma(U_1) \otimes \Gamma(U_2)$. Thus, the Hilbert space
$\mathcal{H}_{F}$ is naturally equivalent to the ordinary (in the mathematical sense) Fock space
with the representation of $T_4 \circledS SL(2, \mathbb{C})$ in the single particle Hilbert space
$\mathcal{H}' = \mathcal{H}_{m,0}^{\oplus} \oplus \mathcal{H}_{-m,0}^{\ominus \, \flat}$ equivalent to $U^{{}_{(m,0,0,0)} L^{{}^{1/2}}} \oplus U^{{}_{(m,0,0,0)} L^{{}^{1/2}}}$.

As the final remark note, please, that the presence of the ${\In}_\oplus$ (respectively ${\In}_\ominus$) parity number operator 
in the annihilation operators of the positron-electron field is pertinent to the fact that the electron states and positron states are the
states of a particle and the corresponding antiparticle, and cannot serve as the single particle states of 
two independent fields. In the situation in which $\mathcal{H}_{m,0}^{\oplus}$ and $\mathcal{H}_{-m,0}^{\ominus \, \flat}$
could serve as the single particle of two independent fields, the creation operators would be similar, but
without the parity number operator ${\In}_\oplus$ (respectively ${\In}_\ominus$), so that in this case
the creation/annihilation operators of the states $\mathcal{H}_{m,0}^{\oplus}$ would simply
commute with creation/annihilation operators of the states $\mathcal{H}_{-m,0}^{\ominus \, \flat}$.
This is not the case for the electron-positron Dirac spinor field and for any other charged field, with
$\mathcal{H}_{m,0}^{\oplus}$ and $\mathcal{H}_{-m,0}^{\ominus \, \flat}$ serving as the single particle and, respectively, 
antiparticle states. In case, $\mathcal{H}_1$ and $\mathcal{H}_2$
served as the single particle states of two independent fields, without ${\In}_2$ (respectively ${\In}_1$)
in the creation-annihilation operators, we would have unique natural isomorphism between
$\Gamma(\mathcal{H}_1 \oplus \mathcal{H}_2)$
and $\Gamma(\mathcal{H}_1) \otimes \Gamma(\mathcal{H}_2)$, given in the last
footnote, with the parity number operator ${\In}_2$ (respectively ${\In}_1$) put equal $\boldsymbol{1}$.

\subsection{Quantum Dirac free field $\boldsymbol{\psi}$ as a Wightman 
operator-valued distribution}\label{psiWightman}

In order to construct quantum Dirac field, $\boldsymbol{\psi}$, we need a more subtle structure than just the Fock space,
as the quantum field is something which could be called suggestively ``operator-valued distribution'',
and which in turn is motivated by the classic analysis of measurement of quantum fields due to Bohr and Rosenfeld. In fact the precise mathematical interpretation is in fact still on the way.  
Intentionally (direction initiated by Wightman) quantum field, say $\boldsymbol{\psi}$, 
is regarded as a map  
$f \mapsto \boldsymbol{\psi}(f)$ with $\boldsymbol{\psi}(f)$, intentionally equal 
\begin{equation}\label{psi(f)-symbollical}
\boldsymbol{\psi}(f) =
\int \boldsymbol{\psi}(x) f(x) \, \ud^4 x = \sum_{a} \int \boldsymbol{\psi}^a (x) f^a(x) \, \ud^4 x,
\end{equation}
which maps continuously
a specified test space (here the Schwartz's space $\mathcal{S}(\mathbb{R}^4; \mathbb{C}^4)$ 
of bispinors $f$ on the space-time) 
into a specified class of (in general unbounded) operators 
$L(\mathcal{D})$ on a dense domain $\mathcal{D}$ of the Hilbert space, i.e. of the Fock space 
$\mathcal{H}_{F}= \Gamma\big(\mathcal{H}_{m,0}^{\oplus}\big) \otimes \Gamma\big(\mathcal{H}_{-m,0}^{\ominus \, \flat}\big) = \Gamma\big(\mathcal{H}_{m,0}^{\oplus} \oplus \mathcal{H}_{-m,0}^{\ominus \,\flat}\big)$
in case of the field $\boldsymbol{\psi}$ in question,
with a specified sequentially complete topology on $L(\mathcal{D})$ respecting the nuclear theorem
and a nuclear topology on the test space, compare \cite{wig} and \cite{Woronowicz}
for a more detailed treatment. This should be regarded as the first step toward the precise mathematical interpretation of the notion of quantum field
introduced by the founders of QED, and in fact this is one possible approach, most popular among mathematical 
physicists working within the  ``axiomatic approach to QFT''. There is also another possible approach, 
initiated by Berezin \cite{Berezin} and developed by mathematicians \cite{hida}, \cite{obataJFA}, 
\cite{obata-book}. Although Wightman's definition of the quantum
(free) field does not fit well with the causal approach to QFT, we give a general remark on it before passing
to the Berezin-Hida white noise construction -- more adequate here.

Before continuing our short excurse into Wightman approach, let us note that the formula 
(\ref{psi(f)-symbollical}) is the immediate quantum field analogue of the non invariant pairing $\phi(f)$, \emph{i.e.} 
(\ref{NonInvariantBispinorPairing}), for the classical (non quantized) solutions $\phi$ of the Dirac equation.
We have of course, besides (\ref{psi(f)-symbollical}), the natural analogues of the invariant pairing (\ref{InvariantBispinorPairing})
and, respectively,  of the non invariant pairing (\ref{NonInvariantBispinorPairing}): 
\begin{equation}\label{psi++(f)psi+(f)}
\begin{split}
\boldsymbol{\psi}^\sharp(f) =
\int \boldsymbol{\psi}^\sharp (x) f(x) \, \ud^4 x = \sum_{ab} \int \big[\gamma^{0}\big]^{ab} \boldsymbol{\psi}^b (x)^{+} f^a(x) \, \ud^4 x,
\\
\boldsymbol{\psi}^+(f) =
\int \boldsymbol{\psi}(x)^{+ \, T} f(x) \, \ud^4 x = \sum_{a} \int \boldsymbol{\psi}^a (x)^+ f^a(x) \, \ud^4 x.
\end{split}
\end{equation}  

In the Wightman's construction of a (free) quantum field the integral expression
(\ref{psi(f)-symbollical}), and especially the quantum
field $\boldsymbol{\psi}(x)$ at a specified space-time point, has only symbolic character, lacking any immediate meaning even when
considering free field(s), such as $\boldsymbol{\psi}$. This is just like the symbol $\psi(x)$ for a symbolic evaluation at $x$ of a ''function'' which symbolizes (when -- again symbolically -- integrated with a test function $f$) the value at $f$ of a proper distribution -- singular generalized function.
In particular when considering a free field
$\boldsymbol{\psi}$, the value $\boldsymbol{\psi}(f)$ for a space-time test (say bispinor
function $f \in \mathcal{S}(\mathbb{R}^4; \mathbb{C}^4)$) is obtained through the creation and annihilation operators evaluated at
the Fourier transform $\widetilde{f}$ restricted to the orbit
$\mathscr{O}$ pertinent to the representation defining the
field(s) $\boldsymbol{\psi}$
(in case of presence of antiparticles the representation is not irreducible and
evaluation of the creation operator, acting over the Fock space over the single particle Hilbert space of conjugated solutions is involved, and even in general one has to consider many orbits in presence of more complicated fields or several fields\footnote{One can consider even spectral measure of translation generators concentrated on the set of orbits with a finite range of possible mass parameters and the corresponding field which is called in this case a \emph{generalized free field}. We describe the case of the quantum Dirac field in details below.}).
The expression (\ref{psi(f)-symbollical}) is given a meaning
whenever applied to the vectors of the allowed domain $\mathcal{D}$,
only very indirectly, utilizing the quantity $\boldsymbol{\psi}(f)$, $f \in \mathcal{S}(\mathbb{R}^4; \mathbb{C}^4)$, which must be defined as the primary datum,
together with the appropriate domain $\mathcal{D}$, compare \cite{wig}, \S 3-3. For the free Dirac field
$\boldsymbol{\psi}$, the expression
$\boldsymbol{\psi}(f)$, $f \in \mathcal{S}(\mathbb{R}^4; \mathbb{C}^4)$, is defined through the creation
$a_{\ominus}\big((P^\ominus\widetilde{f}|_{{}_{\mathscr{O}}})^\flat\big)^+$ and annihilation
$a_{\oplus}\big(P^\oplus\widetilde{f}|_{{}_{\mathscr{O}}}\big)$ operators:
\begin{equation}\label{psi(f)=a_+(f)+a_-(f^c)^+}
\boldsymbol{\psi}(\overline{f}) = a_{\oplus}\big(P^\oplus\widetilde{f}|_{{}_{\mathscr{O}_{m,0,0,0}}}\big) +
a_{\ominus}\Big(\big(P^\ominus\widetilde{f}|_{{}_{\mathscr{O}_{-m,0,0,0}}}\big)^\flat\Big)^+,
\end{equation}
evaluated respectively at $P^\oplus\widetilde{f}|_{{}_{\mathscr{O}}}$ and
$\big(P^\ominus\widetilde{f}|_{{}_{\mathscr{O}_{-m,0,0,0}}}\big)^\flat$. Here $\widetilde{f}$ is the ordinary
Fourier transform of space-time bispinor $f$, and $\widetilde{f}|_{{}_{\mathscr{O}_{m,0,0,0}}}$,
$\widetilde{f}|_{{}_{\mathscr{O}_{-m,0,0,0}}}$ the respective restrictions of $\widetilde{f}$ to
the orbits $\mathscr{O}_{m,0,0,0}$, $\mathscr{O}_{-m,0,0,0}$:
\[
\widetilde{f}|_{{}_{\mathscr{O}_{m,0,0,0}}}(p_0, \boldsymbol{\p}) = \widetilde{f}(\sqrt{|\boldsymbol{\p}|^2 +m^2},
\boldsymbol{\p}), \,\,\,
\widetilde{f}|_{{}_{\mathscr{O}_{-m,0,0,0}}}(p_0, \boldsymbol{\p}) = \widetilde{f}(-\sqrt{|\boldsymbol{\p}|^2 +m^2},
\boldsymbol{\p}).
\]
Here $P^\oplus$ is the projection operator acting on bispinors
$\widetilde{f}|_{{}_{\mathscr{O}_{m,0,0,0}}}$ concentrated on $\mathscr{O}_{m,0,0,0}$
and projecting on the Hilbert space $\mathcal{H}_{m,0}^{\oplus}$, defined in Subsection \ref{e1}.
$P^\ominus$ is the projection operator which projects bispinors
$\widetilde{f}|_{{}_{\mathscr{O}_{-m,0,0,0}}}$ concentrated on $\mathscr{O}_{-m,0,0,0}$ on
the Hilbert space $\mathcal{H}_{-m,0}^{\ominus}$, and defined in Subsection
\ref{e1}, so that
\[
\begin{split}
P^\oplus\widetilde{f}|_{{}_{\mathscr{O}_{m,0,0,0}}}(p) \overset{\textrm{df}}{=}
P^\oplus(p)\widetilde{f}(p), \,\,\,\, p = (\sqrt{|\boldsymbol{\p}|^2 +m^2},
\boldsymbol{\p}) \in \mathscr{O}_{m,0,0,0}, \\
P^\ominus\widetilde{f}|_{{}_{\mathscr{O}_{-m,0,0,0}}}(p) \overset{\textrm{df}}{=}
P^\ominus(p)\widetilde{f}(p), \,\,\,\, p = (-\sqrt{|\boldsymbol{\p}|^2 +m^2},
\boldsymbol{\p}) \in \mathscr{O}_{-m,0,0,0}.
\end{split}
\]
Finally $(\cdot)^\flat$ stands for the conjugation defined in Subsection \ref{positron}.
By construction $P^\oplus\widetilde{f}|_{{}_{\mathscr{O}}}$ and
$\big(P^\ominus\widetilde{f}|_{{}_{\mathscr{O}_{-m,0,0,0}}}\big)^\flat$
belong respectively to $\mathcal{H}_{m,0}^{\oplus}$ and $\mathcal{H}_{-m,0}^{\ominus \, \flat}$ whenever $f \in \mathcal{S}(\mathbb{R}^4; \mathbb{C}^4)$,
and thus belong to the single particle Hilbert space
$\mathcal{H}_{m,0}^{\oplus} \oplus \mathcal{H}_{-m,0}^{\ominus \, \flat}$,
so that the expressions $a_\ominus\big((P^\ominus\widetilde{f}|_{{}_{\mathscr{O}_{-m,0,0,0}}})^\flat\big)^+$ and
$a_\oplus(P^\oplus\widetilde{f}|_{{}_{\mathscr{O}}})$ make sense.
Moreover, both operators $P^\oplus, P^\ominus$ of multiplication by the projectors
$P^\oplus(p)$, $p \in \mathscr{O}_{m,0,0,0}$ and respectively $P^\ominus(p)$, $p \in \mathscr{O}_{-m,0,0,0}$,
commute by construction with the Fourier transformed
Dirac operator of pointwise multiplication by the matrix $p_0\gamma^0 - p_k\gamma^k$ (summation with respect to
$k=1,2,3$) on the Hilbert spaces $\mathcal{H}_{m,0}^{\oplus}$ and $\mathcal{H}_{-m,0}^{\ominus}$ of bispinors $\widetilde{f}|_{{}_{\mathscr{O}_{m,0,0,0}}}$ and respectively $\widetilde{f}|_{{}_{\mathscr{O}_{-m,0,0,0}}}$
concentrated respectively on $\mathscr{O}_{m,0,0,0}$ and $\mathscr{O}_{-m,0,0,0}$, so that
\begin{equation}\label{psi(bar(D*f))=0}
\boldsymbol{\psi}\big(\overline{(i(\gamma^\mu)^* \partial_\mu -m \boldsymbol{1})f}\big) = 0, \,\,\,\,\,
f \in \mathcal{S}(\mathbb{R}^4; \mathbb{C}^4),
\end{equation}
and the field $\boldsymbol{\psi}$ fulfils the free Dirac equation as expected,
because the algebraic relation
\begin{equation}\label{AlgRelDiracEq}
\begin{split}
\big[p_0\gamma^0 - p_k\gamma^k - m\boldsymbol{1}\big]P^\oplus\widetilde{f}|_{{}_{\mathscr{O}_{m,0,0,0}}}(p) =0,
\,\,\,\,\,\,\,\,\,\,
p = (p_0, \boldsymbol{\p}) \in \mathscr{O}_{m,0,0,0}
\\
\big[p_0\gamma^0 - p_k\gamma^k - m\boldsymbol{1}\big]P^\ominus\widetilde{f}|_{{}_{\mathscr{O}_{-m,0,0,0}}}(p) =0,
\,\,\,\,\,\,\,\,\,\,
p = (p_0, \boldsymbol{\p}) \in \mathscr{O}_{-m,0,0,0},
\end{split}
\end{equation}
holds on the Hilbert spaces $\mathcal{H}_{m,0}^{\oplus}$ and $\mathcal{H}_{-m,0}^{\ominus}$ of bispinors
$P^\oplus\widetilde{f}|_{{}_{\mathscr{O}_{m,0,0,0}}}$ and respectively $P^\ominus\widetilde{f}|_{{}_{\mathscr{O}_{-m,0,0,0}}}$,
concentrated on $\mathscr{O}_{m,0,0,0}$ and respectively on $\mathscr{O}_{-m,0,0,0}$, compare Subsection \ref{e1}.
Indeed, that $\boldsymbol{\psi}$ fulfills the homogeneous Dirac equation, can also be immediately seen by noting
that the Fourier transformed
operator defining homogeneous Dirac equation is equal to pointwise multiplication by the matrix (summation
with respect to $k=1,2,3$)
\[
\big[p_0\gamma^0 - p_k\gamma^k - m\boldsymbol{1}_{{}_{4}}\big] = \big[ \slashed{p} - m \big]
\]
and that the projection operators $P^{\oplus}, P^\ominus$, commuting with it, are equal to operators
of multiplication by the projection matrices
\[
\begin{split}
P^{\oplus}(p) = \frac{1}{2m} \big[ \slashed{p} + m \big], \,\,\, p \in \mathscr{O}_{m,0,0,0}, \\
P^{\ominus}(p) = \frac{1}{2m} \big[ \slashed{p} + m \big], \,\,\, p \in \mathscr{O}_{-m,0,0,0},
\end{split}
\]
compare Appendix \ref{fundamental,u,v}, formula (\ref{covariantPplusPminus}). From this
and from the fact that
\[
\begin{split}
\big[ \slashed{p} + m \big]\big[ \slashed{p} - m \big] =
\big[ \slashed{p} - m \big]\big[ \slashed{p} + m \big] = [p \cdot p - m^2] \, \boldsymbol{1}_{{}_{4}} = 0, \,\,\,
p \in \mathscr{O}_{m,0,0,0}, \\
\big[ \slashed{p} + m \big]\big[ \slashed{p} - m \big] =
\big[ \slashed{p} - m \big]\big[ \slashed{p} + m \big] = [p \cdot p - m^2] \, \boldsymbol{1}_{{}_{4}} = 0, \,\,\,
p \in \mathscr{O}_{-m,0,0,0},
\end{split}
\]
the commutativity of $\big[p_0\gamma^0 - p_k\gamma^k - m\boldsymbol{1}_{{}_{4}}\big]$
with $P^\oplus(p)$ on $\mathscr{O}_{m,0,0,0}$ and with $P^\ominus(p)$ on $\mathscr{O}_{-m,0,0,0}$,
as well as the relations (\ref{AlgRelDiracEq}) are easily seen to hold, so that
our assertion follows.

Note that in the formula (\ref{psi(f)=a_+(f)+a_-(f^c)^+}) we have used the simplified notation
for the operator (\ref{a_plusInH_F}) and for the operator adjoint to (\ref{a_minusInH_F}).
For the operator $a_{\oplus}\big(P^\oplus\widetilde{f}|_{{}_{\mathscr{O}_{m,0,0,0}}}\big)$
in the formula (\ref{psi(f)=a_+(f)+a_-(f^c)^+}) the reader should read
\begin{equation}\label{a'Inpsi}
a'\big(P^\oplus\widetilde{f}|_{{}_{\mathscr{O}_{m,0,0,0}}} \oplus 0\big) =
a_{\oplus}\big(P^\oplus\widetilde{f}|_{{}_{\mathscr{O}_{m,0,0,0}}}\big) \otimes {\In}_\ominus
\end{equation}
and for the operator $a_{\ominus}\Big(\big(P^\ominus\widetilde{f}|_{{}_{\mathscr{O}_{-m,0,0,0}}}\big)^\flat \Big)^+$
in (\ref{psi(f)=a_+(f)+a_-(f^c)^+}) the reader should read
\begin{equation}\label{a'^+Inpsi}
a'\Big( 0 \oplus \big(P^\ominus\widetilde{f}|_{{}_{\mathscr{O}_{-m,0,0,0}}}\big)^\flat \Big)^+ =
\boldsymbol{1} \otimes a_{\ominus}\Big(\big(P^\ominus\widetilde{f}|_{{}_{\mathscr{O}_{-m,0,0,0}}}\big)^\flat \Big)^+.
\end{equation}
On the left-hand sides of the last two formulas we have the standard annihilation and creation
operators $a'(u\oplus v), a'(u\oplus v)^+$ acting on the Fock space
\[
\mathcal{H}_F = \Gamma\big( \mathcal{H}_{m,0}^{\oplus} \oplus \mathcal{H}_{-m,0}^{\ominus \, \flat} \big)
= \Gamma\big( \mathcal{H}_{m,0}^{\oplus} \big) \otimes \Gamma \big(\mathcal{H}_{-m,0}^{\ominus \, \flat} \big)
\]
of the free Dirac field introduced in Subsection \ref{electron+positron}.
On the right-hand sides of the last two formulas we have the annihilation and creation operators
$a_{\oplus}\big(P^\oplus\widetilde{f}|_{{}_{\mathscr{O}_{m,0,0,0}}}\big)$ and
$a_{\ominus}\Big(\big(P^\ominus\widetilde{f}|_{{}_{\mathscr{O}_{-m,0,0,0}}}\big)^\flat\Big)^+$
acting respectively in the Fock spaces $\Gamma\big(\mathcal{H}_{m,0}^{\oplus}\big)$
and $\Gamma\big(\mathcal{H}_{-m,0}^{\ominus \, \flat}\big)$, and defined respectively in Subsections \ref{electron}
and \ref{positron}. For definition of the unitary involutive (and thus self-adjoint)
operator\footnote{The operator ${\In}_\ominus$ is replaced with the unital operator in case of integer spin (non-neutral) field.} ${\In}_\ominus$ we refer to Subsections \ref{electron} and \ref{positron}. 

Thus, the formula (\ref{psi(f)=a_+(f)+a_-(f^c)^+}) should properly be written as 
\begin{equation}\label{psi(f)=a_+(f)+a_-(f^c)^+proper}
\boldsymbol{\psi}(\overline{f}) = a'\big(P^\oplus\widetilde{f}|_{{}_{\mathscr{O}_{m,0,0,0}}} \oplus 0\big) + 
a'\Big( 0 \oplus \big(P^\ominus\widetilde{f}|_{{}_{\mathscr{O}_{-m,0,0,0}}}\big)^\flat \Big)^+.
\end{equation}

It should be stressed however that the structure
$\mathcal{H}_F = \mathcal{H}^{\oplus}_{F} \otimes \mathcal{H}^{\ominus}_{F}
= \Gamma\big(\mathcal{H}_{m,0}^{\oplus} \big) \otimes \Gamma\big(\mathcal{H}_{-m,0}^{\ominus \, \flat} \big)$ of the
Hilbert space of the free
quantum Dirac field $\boldsymbol{\psi}$, as well as the tensor product form of the operators
(\ref{a'Inpsi}) and (\ref{a'^+Inpsi}) in (\ref{psi(f)=a_+(f)+a_-(f^c)^+proper})
does not mean that the quantum Dirac field may be treated as system of two independent fields of electrons and positrons.
Indeed, the quantized Dirac field, equal to the linear combination (\ref{psi(f)=a_+(f)+a_-(f^c)^+proper}) of operators
\footnote{Both treated as tensor product operators on $\Gamma\big(\mathcal{H}_{m,0}^{\oplus} \big) \otimes \Gamma\big(\mathcal{H}_{-m,0}^{\ominus \, \flat} \big)$, the first having the second factor trivial and
equal to the fundamental unitary involution ${\In}_\ominus$ and vice versa for the second, with the first factor
trivial and equal to the unit operator.},
cannot be treated as a sum of field operators respectively in $\Gamma\big(\mathcal{H}_{m,0}^{\oplus} \big)$
and $\Gamma\big(\mathcal{H}_{-m,0}^{\ominus \flat} \big)$ simply because the arguments
\[
P^\oplus\widetilde{f}|_{{}_{\mathscr{O}_{m,0,0,0}}}
\,\,\, \textrm{and} \,\,\
\big(P^\ominus\widetilde{f}|_{{}_{\mathscr{O}_{-m,0,0,0}}}\big)^\flat
\]
in the operators (\ref{a'Inpsi}) and (\ref{a'^+Inpsi}) entering the formula
(\ref{psi(f)=a_+(f)+a_-(f^c)^+proper}) for $\boldsymbol{\psi}(f)$ are not independent.
Indeed, by choosing a function $f$ from the test space $\mathcal{S}(\mathbb{R}^4; \mathbb{C}^4)$
we predetermine the restrictions
\[
\widetilde{f}|_{{}_{\mathscr{O}_{m,0,0,0}}}
\,\,\, \textrm{and} \,\,\
\widetilde{f}|_{{}_{\mathscr{O}_{-m,0,0,0}}}
\]
of its Fourier transform to the orbits $\mathscr{O}_{m,0,0,0}$ and $\mathscr{O}_{-m,0,0,0}$, which cannot be
varied independently one from another. This dependence, imposed on
\[
\widetilde{f}_1 = \widetilde{f}|_{{}_{\mathscr{O}_{m,0,0,0}}}
\,\,\, \textrm{and} \,\,\
\widetilde{f}_2 = \widetilde{f}|_{{}_{\mathscr{O}_{-m,0,0,0}}}
\]
by the fact that they come from restrictions to the orbits of the Fourier transform of one and the same
$f$, cannot be realized by any natural relation put on the two \emph{a priori} independent fields of
electrons and positrons, and realized through (\ref{a'Inpsi}) and (\ref{a'^+Inpsi}) with two independent
arguments $f$, respectively, in (\ref{a'Inpsi}) and (\ref{a'^+Inpsi}). Besides, in the creation-annihilation 
operator of the electron-positron Dirac field there is used the parity operator, 
which causes the electron creation operator to anticommutate 
with the proton creation operator, which of course would be impossible if electrons 
and protons composed independent fields.

Comparing (\ref{psi(bar(D*f))=0}) with the last formula in (\ref{PairingsAndDiracEquations}) we should expect that the field
(\ref{psi(f)=a_+(f)+a_-(f^c)^+proper}) represents Dirac field $\boldsymbol{\psi}$ whose kernel $\boldsymbol{\psi}(x)$
in (\ref{psi(f)-symbollical})
respects the Dirac equation
\[
\big[i\gamma^\mu \partial_\mu -m\big]\boldsymbol{\psi}(x) = 0
\]
and not the equation
\[
\big[i(\gamma^\mu)^* \partial_\mu -m\big]\boldsymbol{\psi}(x) = 0.
\]
But, of course, we can perform analogue construction of the annihilation-creation operators $a'(\cdot),a'(\cdot)^+$, with the single
particle Hilbert spaces $\mathcal{H}_{m,0}^{\oplus}, \mathcal{H}_{-m,0}^{\ominus}$ of positive and negative energy solutions
$\phi$ of the Dirac equation
\[
\big[i\gamma^\mu \partial_\mu -m\big]\phi(x) = 0,
\]
by replacing simply the Clifford generators $\gamma^\mu$ with $(\gamma^\mu)^*$, used in the momentum picture
(and \emph{vice versa} in the space-time picture). This in particular will result in replacing the projectors 
$P^\oplus(p), P^\ominus(p)$ with their conjugations
\[
P^\oplus(p)^* = \gamma^0 P^\oplus(p) \gamma^0, \,\,\,\,\, P^\ominus(p)^* = \gamma^0 P^\ominus(p)\gamma^0.
\]
Then the corresponding Dirac field operator
\begin{equation}\label{psi(f)=a_+(f)+a_-(f^c)^+proper*}
\underset{*}{\boldsymbol{\psi}}(\overline{f}) = a'\big(P^{\oplus*}\widetilde{f}|_{{}_{\mathscr{O}_{m,0,0,0}}} \oplus 0\big) +
a'\Big( 0 \oplus \big(P^{\ominus*}\widetilde{f}|_{{}_{\mathscr{O}_{-m,0,0,0}}}\big)^\flat \Big)^+
\end{equation}
will have the kernel $\underset{*}{\boldsymbol{\psi}}(x)$
in (\ref{psi(f)-symbollical}) which
respects the Dirac equation
\[
\big[i(\gamma^\mu)^* \partial_\mu -m\big]\underset{*}{\boldsymbol{\psi}}(x) = 0.
\]
We have to remember that the Dirac field operators (\ref{psi(f)=a_+(f)+a_-(f^c)^+proper}) and (\ref{psi(f)=a_+(f)+a_-(f^c)^+proper*})
are different (with different Clifford gamma generators) and with different annihilation-creation operators $a'(\cdot),a'(\cdot)^+$. 
By construction, both (\ref{psi(f)=a_+(f)+a_-(f^c)^+proper}) and (\ref{psi(f)=a_+(f)+a_-(f^c)^+proper*}), have identical local and unitary
transformation law. We will construct the kernels $\boldsymbol{\psi}(x), \underset{*}{\boldsymbol{\psi}}(x)$ in 
(\ref{psi(f)-symbollical}) of both fields (\ref{psi(f)=a_+(f)+a_-(f^c)^+proper}) and (\ref{psi(f)=a_+(f)+a_-(f^c)^+proper*})
in Subsection \ref{psiBerezin-Hida} using white noise analysis -- a task which is outside the scope of Wightman approach.

But let us return to Wightman's approach.
The domain $\mathcal{D}$ of the field $\boldsymbol{\psi}$, due to the interpretation initiated by Wightman, is not determined uniquely
but in any case contains at least the
domain $\mathcal{D}_0$ which arises by the action of
polynomial expressions in
\[
\boldsymbol{\psi}(f_1), \boldsymbol{\psi}(f_2), \ldots, \,\,\,\,\,
f_i \in \mathcal{S}(\mathbb{R}^4; \mathbb{C}^4)
\]
on the vacuum $|0\rangle = \Psi_0$.
However, we know that the domain must be considerably larger if $L(\mathcal{D})$
is supposed to satisfy kernel theorem
in accordance to the result of \cite{Woronowicz}.
In particular, it must contain the domain called $\mathcal{D}_1$ in \cite{wig}, p. 107,
but it is even not clear for the free field determined by an irreducible representation corresponding to a single orbit that $L(\mathcal{D}_1)$
satisfies the theorem on kernel as stated in \cite{Woronowicz}. We only know,
by the result of \cite{Woronowicz}, that such domain
$\mathcal{D}$ exists on which $L(\mathcal{D})$ satisfies the theorem on kernel (with the ''strong topology''
on $L(\mathcal{D})$), and contains the domain called $\mathcal{D}_1$ in \cite{wig}, p. 107.

More generally for any $f \in \mathcal{S}(\mathbb{R}^{4k}) =
\mathcal{S}(\mathbb{R}^{4})^{\otimes k}$ and for any system of free fields
$\boldsymbol{\psi}_1, \ldots, \boldsymbol{\psi}_k$ one can give a meaning of a well-defined vector in the
dense domain $\mathcal{D}$ of the Fock space of the total system
to the expression of the form
\begin{equation}\label{PsiInD}
\Psi = \int \ud^4 x_1 \ldots \ud^4x_k \, f(x_1, \ldots, x_k ) \, \boldsymbol{\psi}_1(x_1) \ldots
\boldsymbol{\psi}_k(x_k) \, \Psi_0,
\end{equation}
and then for any field $\boldsymbol{\psi}$ of the considered system of free fields and for any
$\Psi$ of the form (\ref{PsiInD}) one can give a meaning by a limit process to the expression
\begin{equation}\label{psi(f)Psi}
\boldsymbol{\psi}(f)\Psi
\end{equation}
thus giving a meaning to $\boldsymbol{\psi}_1(x_1) \ldots
\boldsymbol{\psi}_k(x_k)$ of an operator-valued distribution over the test space
$\mathcal{S}(\mathbb{R}^{4})^{\otimes k}$ on the domain containing all vectors
of the form (\ref{PsiInD}), compare \cite{wig}, \S 3-3. This is achieved by noting first
that
\[
(\Psi_0, \boldsymbol{\psi}_1(f_1) \ldots
\boldsymbol{\psi}_k(f_k) \, \Psi_0)
\]
is a well-defined and separately continuous multilinear functional of the arguments
$f_i$ in the nuclear topology on the Schwartz space $\mathcal{S}(\mathbb{R}^4)$.
Thus, by the ordinary Schwartz kernel theorem it follows that there exists a unique distribution
$\mathscr{W}(x_1, \ldots, x_k)$ such that
\[
\int \mathscr{W}(x_1, \ldots, x_k) f_1(x_1)f_2(x_2) \ldots f_k(x_k) \, \ud^4x_1 \ldots \ud^4x_k
= (\Psi_0, \boldsymbol{\psi}_1(f_1) \ldots
\boldsymbol{\psi}_k(f_k) \, \Psi_0)
\]
for any $f_i \in \mathcal{S}(\mathbb{R}^4)$. Using this fact (as in \cite{wig}, p. 107)
we next show that the states
\[
\Psi_J = \sum_{j=1}^{J} \boldsymbol{\psi}_1(f_{1j}) \ldots
\boldsymbol{\psi}_k(f_{kj}) \, \Psi_0
\]
converge in norm of the Fock space whenever the functions
\[
f_J(x_1, \ldots, x_k) = \sum_{j=1}^{J} f_{1j}(x_1)f_2(x_2) \ldots f_{kj}(x_k)
\]
converge to $f$ in $\mathcal{S}(\mathbb{R}^4)^k = \mathcal{S}(\mathbb{R}^{4k})$. The limit
of $\Psi_J$ is defined as the vector $\Psi$ giving the meaning to the expression
(\ref{PsiInD}). The value (\ref{psi(f)Psi}) is defined as the limit of $\boldsymbol{\psi}(f)\Psi_J$,
and gives a well-defined ''operator-valued'' distribution by the pre-closed character
of the operators $\boldsymbol{\psi}(f)$ on the domains $\mathcal{D}_0 \subset \mathcal{D}_1$, compare
\cite{Woronowicz}.

In Wightman approach it is the formula (\ref{psi(f)=a_+(f)+a_-(f^c)^+}) which gives
the meaning to the symbolic expression (\ref{psi(f)-symbollical}) when applied to the
elements of the domain $\mathcal{D}$.

For a given free field (or a system of free fields
$\boldsymbol{\psi}_1, \boldsymbol{\psi}_2, \ldots, \boldsymbol{\psi}_k$) one can give, within the mentioned Wightman approach, a meaning to the expression
\begin{equation}\label{:psi(f):-symbollical}
\boldsymbol{{:}}\partial^{\alpha_1}\boldsymbol{\psi}_1 \ldots \partial^{\alpha_k}\boldsymbol{\psi}_k\boldsymbol{{:}}(f)
= \int \boldsymbol{{:}}\partial^{\alpha_1}\boldsymbol{\psi}_1(x) \ldots \partial^{\alpha_k}\boldsymbol{\psi}_k(x)\boldsymbol{{:}} \, f(x) \, \ud^4 x
\end{equation}
as a limit, giving an operator-valued distribution \cite{WightmanGarding}. However, here for definition of the ''Wick product'' due to \cite{WightmanGarding} and using Wightman's definition of the field the limit process involved here is devoid of any natural choice, as the ''Wick product field'' of Wightman and G{\aa}rding is obtained from an operator-valued distribution
in several space-time variables, and then as a limit we obtain operator valued distribution in just one space-time
variable. Such definition involves a considerable amount of unnatural and rather arbitrary choices
in selecting a (class of) limit(s) of passing from test function spaces in just one space-time variable
to the test space in several space-time variables, compare \cite{WightmanGarding} for one possible
choice\footnote{For the opposite direction, i.e. for passing from distribution of one variable to distribution of several variables, we would have the natural choice given by the map defined by the restriction to the diagonal, which is continuous between the test spaces. Reverse direction is by no means natural nor unique. The reader should also note that the ''definition'' of the
Wick product in \cite{wig}, \S 3-2, p. 104, which merely says:
\[
\boldsymbol{{:}}\partial^{\alpha_1}\boldsymbol{\psi}(x) \partial^{\alpha_k}\boldsymbol{\psi}(x)\boldsymbol{{:}}
=\lim \limits_{x_1, x_2 \rightarrow x} \Big[\partial^{\alpha}\boldsymbol{\psi}(x_1) \partial^{\beta}\boldsymbol{\psi}(x_2) - \big(\Psi_0, \partial^{\alpha}\boldsymbol{\psi}(x_1) \partial^{\beta}\boldsymbol{\psi}(x_2) \Psi_0\big)\Big],
\]
and
\begin{multline*}
\boldsymbol{{:}}\partial^{\alpha}\boldsymbol{\psi}(x) \partial^{\beta}\boldsymbol{\psi}(x)
\partial^{\gamma}\boldsymbol{\psi}(x)\boldsymbol{{:}}
=\lim \limits_{x_1, x_2 \rightarrow x} \Big[\partial^{\alpha}\boldsymbol{\psi}(x_1) \partial^{\beta}\boldsymbol{\psi}(x_2) \\
-\big(\Psi_0, \partial^{\alpha}\boldsymbol{\psi}(x_1) \partial^{\beta}\boldsymbol{\psi}(x_2)
\partial^{\gamma}\boldsymbol{\psi}(x_3) \Psi_0\big) \partial^{\gamma}\boldsymbol{\psi}(x_3) \\
-\big(\Psi_0, \partial^{\alpha}\boldsymbol{\psi}(x_1) \partial^{\gamma}\boldsymbol{\psi}(x_3)
\partial^{\gamma}\boldsymbol{\psi}(x_3) \Psi_0\big) \partial^{\beta}\boldsymbol{\psi}(x_2) \\
-\big(\Psi_0, \partial^{\beta}\boldsymbol{\psi}(x_2) \partial^{\gamma}\boldsymbol{\psi}(x_3)
\partial^{\gamma}\boldsymbol{\psi}(x_3) \Psi_0\big) \partial^{\alpha}\boldsymbol{\psi}(x_1)
\Big], \\
\,\,\,\, \textrm{and so on} \, \ldots
\end{multline*}
is again only heuristic, and strictly speaking is meaningless
as a definition of operator-valued distribution,
as it involves limit process of passing from test space of one space-time variable to test space of several space-time variables, which is not specified there. The reader which would like to know the concrete choice
of the possible limit process involved there which is meant by the authors will have to consult
the paper \cite{WightmanGarding}.} of the limit process.

Unfortunately the method of \cite{WightmanGarding} is not efficient (for boson, and particularly
for massless fields) in the investigation of the closability
of the operator (\ref{:psi(f):-symbollical}) or its eventual self-adjointness nor for the proof of the
``Wick theorem'' \cite{Bogoliubov_Shirkov}, Chap. III, useful in the causal perturbative approach to QED.
Similarly, the space-time averaging as presented in \cite{WightmanGarding} is not applicable to the averaging
over space-like Cauchy hypersurfaces of their ``Wick product fields'', necessary in construction of the conserved currents appearing in the Noether theorem for free fields.
In particular the Quantization Postulate for free fields
as formulated in \cite{Bogoliubov_Shirkov}, Chap. 2, \S 9.4, cannot be simply treated with Wightman-G{\aa}rding method, and for zero mass fields this Postulate seems to be intractable with
Wightman-G{\aa}rding method\footnote{The mentioned weaknesses of Wightman-G{\aa}rding definition of the
''Wick product'' have also been noted by I. E. Segal, compare e.g. \cite{Segal-NFWP.I}, \cite{Segal-ProcStone}.}.

\subsection{Motivation for introduction of Hida operators into causal perturbative QFT}\label{MotivationForHida}

This is somewhat unsatisfactory because the causal method, which is successful in
avoiding ultraviolet infinities (also avoiding infrared infinities for the adiabatically
switched off interaction at infinity), expresses the scattering operator $S(g)$ in terms
of time ordered products of Wick polynomials of free fields, and is substantially based
on the ''Wick theorem'' for free fields as stated in \cite{Bogoliubov_Shirkov}, Chap. III.
These Wick polynomial fields are constructed with the help of the ''Wick theorem'' applied to
the expressions including products of normally ordered factors in free fields, representing interaction terms, and some of the annihilation-creation operators as additional factors, with the annihilation-creation operators not necessary ''normally ordered''. Wick theorem allows us to equate such expressions with the corresponding expressions containing solely ''normally ordered products'' of creation-annihilation operators multiplied by the respective pairings. The point is that we explicitly utilize the commutation generalized functions and the Wick theorem and only at the end prove e.g. that the product of several factors with the space-time variable in each factor treated as independent of the variables of other factors and with each factor being normally ordered product of free fields representing interacting terms, is a well-defined operator valued distribution on the domain $\mathcal{D}_0$ in the Wightman sense.
Essentially this ''theorem'' allows to treat the (generalized) operators
of the type (compare Theorem 0 in \cite{Epstein-Glaser})
\begin{equation}\label{IntKerOpFreeWickField}
\kappa(x_1, \ldots, x_k) \, \boldsymbol{{:}}\partial^{\alpha_1}\boldsymbol{\psi}_1(x_1) \ldots \partial^{\alpha_k}\boldsymbol{\psi}_k(x_k)\boldsymbol{{:}}
\end{equation}
with numerical,''translationally invariant'' ($\kappa(x_1+a, \ldots, x_k+a)=\kappa(x_1, \ldots, x_k)$), distributions\footnote{In fact we are interested here in distributions $\kappa$ which arise
as tensor products of the pairings of the corresponding free fields $\partial^{\alpha_i}\boldsymbol{\psi}_i$
and when the interaction does not contain derivatives we may confine attention in
(\ref{IntKerOpFreeWickField}) to the case where derivatives are absent, i.e. with all the multiindices
$\alpha_i = 0$. In particular all such distributions have the mentioned invariance property.}
$\kappa \in \mathcal{S}(\mathbb{R}^{4k})^* =
\big(\mathcal{S}(\mathbb{R}^4)^*\big)^{\otimes k}$
which, when integrated with test functions $f \in \mathcal{S}(\mathbb{R}^{4k}) =
\mathcal{S}(\mathbb{R}^4)^{\otimes k}$, define an operator valued distribution
\begin{equation}\label{IntKerOpFreeWickField(f)}
f \rightarrow \int \, f(x_1, \ldots, x_k) \, \kappa(x_1, \ldots, x_k) \, :\partial^{\alpha_1}\boldsymbol{\psi}_1(x_1) \ldots \partial^{\alpha_k}\boldsymbol{\psi}_k(x_k): \, \ud^4x_1 \ldots \ud^4x_k.
\end{equation}
Thus in practical computations we proceed from the ``kernel'' of the ``operator distribution'' to the distribution, and not in the reverse direction which is pertinent to the Wightman approach in which the ``kernel''is only symbolic and difficult to handle.
It is therefore not satisfactory
that already at the free field level the ``Wick theorem'' in the form needed for the causal perturbative approach is not clearly related to the free field defined according to Wightman \cite{wig}.

In spite of this inconvenience, ``Wick theorem'' of \cite{Bogoliubov_Shirkov}, Chap III, provides
partially heuristic (but honest) basis for construction of ``operator-valued distributions'' of the type
(\ref{IntKerOpFreeWickField(f)}), compare Theorem 0 of \cite{Epstein-Glaser}.
This turned up to be effective in the realization of the causal
approach program of St\"uckelberg-Bogoliubov. As realized later by Epstein and Glaser \cite{Epstein-Glaser}
the causal approach of St\"uckelberg-Bogoliubov provides a perturbative method which avoids ultraviolet infinities (and also infrared but with the nonphysical adiabatically switched off interaction at infinity which, especially in case of QED, needs a further analysis of the behavior of the theory when the physical interaction is restored, say by adiabatic switching on the interaction at infinity). The essential improvement of the causal method of St\"uckelberg-Bogoliubov added by Epstein and Glaser
is the careful splitting of the operator-valued distributions of the type (\ref{IntKerOpFreeWickField(f)})
with causally supported distribution kernels $\kappa$ into the retarded and advanced parts -- a task which we
encounter in the causal construction of the perturbative series. Epstein and Glaser
\cite{Epstein-Glaser} reduce this task
to the splitting of the numerical causally supported distribution kernels $\kappa$
into the retarded and advanced part.

Let us shortly summarize the causal perturbative approach due to
St\"uckelberg-Bogoliubov-Epstein-Glaser on the example of QED. This approach, contrary to that based on the Hamiltonian, is not based on the Schr\"odinger-Tomonaga equation in the interaction picture, with the main motivation lying in avoiding the problem with the singular character of the interaction Hamiltonian. In causal approach this is the scattering operator which plays the fundamental role, and the remaining quantities, \emph{i. e.} effective cross sections, and the local interacting fields as well as the definition of the product of interacting fields are obtained from the scattering operator. The time evolution is encoded in the general principles put on the scattering matrix $S$, with the causality condition implemented by the ``switching off space-time function $g$'' multiplying the Lagrange interaction density $\mathcal{L}$, expressed as Wick polynomial of free fields. The scattering operator is treated as a (say operator-valued) functional of $g \mapsto S(g)$. The functional $S(g)$ is constructed perturbatively, and written as a formal functional power series in $g$
\[
S(g) = \boldsymbol{1} + \sum \limits_{n=1}^{\infty} {\textstyle\frac{1}{n!}} \int \ud^4 x_1 \cdots \ud^4 x_n
S_n(x_1, \ldots, x_n) \, g(x_1) \ldots g(x_n)
\]
\[
S^{-1}(g) = \boldsymbol{1} + \sum \limits_{n=1}^{\infty} {\textstyle\frac{1}{n!}} \int \ud^4 x_1 \cdots \ud^4 x_n
\overline{S_n}(x_1, \ldots, x_n) \, g(x_1) \ldots g(x_n)
\]
The conditions put on $S(g)$ are the following.
\emph{Causality}:
\[
(\textrm{I}) \,\,\,\,\,\,\,\,\,\,\,\,
S(g_1 + g_2) = S(g_2)S(g_1), \,\,\,\textrm{whenever}
\,\,\,
\textrm{supp} \, g_1 \preceq \textrm{supp} \, g_2
\]
where the space-time region $G_1$ is said to causally precede $G_2$, in short $G_1 \preceq G_2$,
iff $\big(G_1 + \overline{V_{-}}\big) \cap G_2 =\emptyset$ or equivalently iff $G_1 \cap \big(G_2 + \overline{V_{+}}\big) = \emptyset$. Here $\overline{V_{-}}, \overline{V_{+}}$ are the closures of the interiors of the backward or forward light cones.
\emph{Covariance}
\[
(\textrm{II}) \,\,\,\,\,\,\,\,\,\,\,\,
U_{a,\Lambda} S(g) U_{a,\Lambda}^{-1} = S(T_{{}_{a,\Lambda}}g), \,\,\,\,
T_{{}_{a,\Lambda}}g(x) = g(\Lambda(x + a))
\]
where $U$ is the representation of $T_4 \circledS SL(2, \mathbb{C})$ acting in the Fock space of all free fields of the theory. 
Note here that $T$ is a representation of the opposite $T_4 \circledS SL(2, \mathbb{C})^{\textrm{Op}}$ group, giving the 
antihomomorphic map of $T_4 \circledS SL(2, \mathbb{C})$ into the linear operators. If we interchange $U$ and $U^{-1}$
in (II), then we get $T$ on the right-hand-side, which is a represenation of $T_4 \circledS SL(2, \mathbb{C})$.
Unitarity (Krein-isometricity in case when gauge fields are present, as is the case e.g. for QED)
\[
(\textrm{III}) \,\,\,\,\,\,\,\,\,\,\,\,
\eta S(g)^{+} \eta = S(g)^{-1}
\]
where $\eta= \eta^* = \eta^{-1}$ is the Krein-fundamental symmetry (or the Gupta-Bleuler operator in the particular case of QED). Finally, we are using the
\emph{Bohr's correspondence principle} (quasi-classical limit, compare \cite{Bogoliubov_Shirkov}),
which allows us to identify the first order contribution $S_1(x)$ to the scattering operator
functional $S$:
\[
(\textrm{IV}) \,\,\,\,\,\,\,\,\,\,\,\,
S_1(x) = i \mathcal{L}(x)
\]
where $\mathcal{L}(x)$ is the Lagrange interaction density of the theory in question,
expressed as a Wick polynomial of free fields underlying the theory. In case of spinor QED\footnote{Denoted frequently $\mathcal{L}(x) = \, \boldsymbol{{:}}\overline{\boldsymbol{\psi}}(x) \gamma^\mu \boldsymbol{\psi}(x) A_\mu(x)\boldsymbol{{:}}$, with the Dirac conjugated bispinor field
$\boldsymbol{\psi}^\sharp = \boldsymbol{\psi}^+ \gamma_0$ written
$\overline{\boldsymbol{\psi}}$ in physical literature.},
$\mathcal{L}(x) = \, \boldsymbol{{:}}\boldsymbol{\psi}^\sharp(x) \gamma^\mu \boldsymbol{\psi}(x) A_\mu(x)\boldsymbol{{:}}$. 
Here $\boldsymbol{\psi}^\sharp$ stands for the Dirac conjugation
$\boldsymbol{\psi}^\sharp = \boldsymbol{\psi}^+\gamma_0$,
with the Hermitian conjugation $\boldsymbol{\psi}^+ = \big((\boldsymbol{\psi}^1)^+, \ldots, (\boldsymbol{\psi}^4)^+\big)$. Here $(\cdot)^+$ is the ordinary Hermitian conjugation, if applied to ordinary operators in the Fock space, and the Hermitian conjugation can be extended naturally on the other classes of generalized operators we will encounter (\emph{e.g.} operator-valued distributions, or Hida generalized operators we will encounter in the sequel, and of course this extension need to be precisely defined, and definition will be provided in the next Subsection for the generalized Hida operators). The conditions (I) -- (IV), expressed in terms of the ''generalized operator kernels''
$S_n(x_1, \ldots, x_n)$ reads
\begin{align*}
(\textrm{I}) & \,\,\,\,\,\,\,\,\,\,\,\, &
S_{n}(x_1, \ldots, x_n) = S_{k}(x_1, \ldots, x_k)S_{n-k}(x_{k+1}, \ldots, x_n), \\
& & \,\,\,\,\,\,\, \textrm{if} \, \{x_{k+1}, \ldots, x_m\} \preceq \{x_1, \ldots, x_k\}
\\
(\textrm{II} )& \,\,\,\,\,\,\,\,\,\,\,\, &
U_{a,\Lambda} S_n(x_1, \ldots, x_n)U_{a,\Lambda}^{-1} = S_n(\Lambda^{-1}x_1 - a, \ldots, \Lambda^{-1}x_n - a),
\\
(\textrm{III} )& \,\,\,\,\,\,\,\,\,\,\,\, &
\eta S_n(x_1, \ldots, x_n)^{+} \eta = \overline{S_n}(x_1, \ldots, x_n), \\
(\textrm{IV}) & \,\,\,\,\,\,\,\,\,\,\,\, &
S_1(x) = i \mathcal{L}(x).
\end{align*}

Please, note here, that in (II) we have used $\Lambda = \Lambda(\alpha)$, 
which is a homomorphism $\alpha \rightarrow \Lambda(\alpha)$
of $SL(2,\mathbb{C})$ onto the proper orthochronous Lorentz group. In Section \ref{constr-of-VF} we have used antihomomorphism 
$\alpha \rightarrow \Lambda(\alpha)$, so that in the right-hand side of (II) we would have 
\[
S_n(\Lambda x_1 - a, \ldots, \Lambda x_n - a)
\] 
if we were using  $\Lambda = \Lambda(\alpha)$ with the antihomomorphism 
$\alpha \rightarrow \Lambda(\alpha)$, as in Section \ref{constr-of-VF}.

These conditions (I) -- (IV) should be understood as candidates for axioms of a theory which is to be formulated in well-defined mathematical terms. Also, the mathematical character of the operator
$S(g)$, as well as these axioms should be understood properly. In fact the particular mathematical choices still remain to lie in very front of us, but have not been done yet. So we need to be very careful now. 
At the present stage of the theory
we should proceed at the most general level, just taking care only not to fall into evident conflict with the computational practice we perform whenever we compute the effective cross section. For this purpose the operator $S(g)$ \emph{need not be an ordinary operator} acting on normalizable states, and even the separate contributions
\[
\int \ud^4 x_1 \cdots \ud^4 x_n
S_n(x_1, \ldots, x_n) \, g(x_1) \ldots g(x_n)
\]
of fixed order need not be ordinary operators. Before making further mathematical specifications and particular choices, let us look at the concrete example of QED. In that case the
``operator kernels'' $S_n(x_1, \ldots, x_n)$ have the general form
\begin{multline}\label{QEDgeneralS_n(X)andA'_n(X)}
S_n(x_1, \ldots, x_n) = \\ = \sum \limits_{k, a_j, b_i, \mu_m} t^{k, a_j b_i \mu_m}_{n}(x_1, \ldots, x_n) \,\,\, \boldsymbol{{:}} \sqcap_{j}
\boldsymbol{\psi}^{\sharp \, a_j}(x_j) \sqcap_{i} \boldsymbol{\psi}^{b_i}(x_i) \boldsymbol{{:}} \,\, \boldsymbol{{:}} \sqcap_{m} A_{\mu_m}(x_m)\boldsymbol{{:}}
\end{multline}
where the sum over $k$ is finite with the number of terms depending on the order $n$, and with
total number of factors under Wick products less than or equal $n$ and depending on the particular $k$.
The distribution kernels $t^{k}_{n}$ are transitionally invariant and have causal supports and are determined by the axioms (I)--(IV).
Thus, the $n$-th order contribution $S_n(g)$ to $S(g)$ written in the momentum picture has the general form
\begin{multline}\label{generalS_n(g)}
{\textstyle\frac{1}{n!}} \int \ud^4 x_1 \cdots \ud^4 x_n
S_n(x_1, \ldots, x_n) \, g(x_1) \ldots g(x_n) = \\ =
\sum \limits_{0\leq l+m \leq n} \int
\kappa_{l,m}^{(n)}(\boldsymbol{\p}_1, \ldots, \boldsymbol{\p}_{l+m}; g^{\otimes n}) \,
a_{s_1}(\boldsymbol{\p}_1)^{+} \cdots a_{s_{l+m}}(\boldsymbol{\p}_{l+m}),
\end{multline}
where $a_{s_i}(\boldsymbol{\p}_i)^{+}, \ldots a_{s_j}(\boldsymbol{\p}_j)$ are the creation-annihilation operators of the free fields underlying the theory in the momentum representation, which respect canonical commutation/anticommutation relations.
Here the distributional kernels $\kappa_{l,m}^{(n)}(\boldsymbol{\p}_1, \ldots, \boldsymbol{\p}_{l+m}; g^{\otimes n}) $ are equal
\begin{multline}\label{kappa_l,m(p1,...p_l+m;g)}
\kappa_{l,m}^{(n)}(\boldsymbol{\p}_1, \ldots, \boldsymbol{\p}_{l+m}; g^{\otimes n}) = \\
= \int \ud^4x_1 \cdots \ud^4 x_n \,
\kappa_{l,m}^{(n)}(\boldsymbol{\p}_1, \ldots, \boldsymbol{\p}_{l+m}; x_1, \ldots, x_n)
g^{\otimes n}(x_1, \ldots, x_n) \\ =
\int \ud^4x_1 \cdots \ud^4 x_n \,
\kappa_{l,m}^{(n)}(\boldsymbol{\p}_1, \ldots, \boldsymbol{\p}_{l+m}; x_1, \ldots, x_n)
g(x_1) \cdots g(x_n),
\end{multline}
with $\kappa_{l,m}^{(n)}(\boldsymbol{\p}_1, \ldots, \boldsymbol{\p}_{l+m}; x_1, \ldots, x_n)$
representing kernels of distributions $\kappa_{l,m}$ belonging to
\[
E^{* \otimes(l+m)} \otimes \mathscr{E}^{* \otimes n} \cong
\mathscr{L}(E^{\otimes(l+m)}; \mathscr{E}^{*\otimes n}) \cong
\mathscr{L}(\mathscr{E}^{\otimes n}; E^{* \otimes (l+m)})
\]
which may be regarded as vector-valued distributions $\kappa_{l,m}$ over the $n$-fold tensor product
\[
\mathscr{E}^{\otimes n}
\]
of space-time test function space with values in distribution space
\[
\big(E^{\otimes(l+m)} \big)^* = E^{* \otimes(l+m)}
\]
over the $l+m$-fold tensor product of the restrictions of Fourier transforms of space-time test functions to the respective orbits in momentum space defining the single particle Hilbert spaces of the respective free fields of the theory. Of course $E$ will thus depend on the particular orbit and the corresponding species of the free field, but we disregard this dependence in notation here in order to simplify the notation. Also, there will appear nontrivial analysis in case of light cone orbits corresponding to massless free fields, but we do not enter this point here, because it is not essential here in explaining the general line of our strategy. The importance of these subtleties will however appear at the further stage.

Using the Wick theorem as in\footnote{Compare also Theorem 0 in \cite{Epstein-Glaser}.} \cite{Bogoliubov_Shirkov}, Chap. III \S 18, and the explicit form of the commutation generalized functions as kernels we can show that (\ref{QEDgeneralS_n(X)andA'_n(X)}) is a kernel of an operator valued distribution in the Wightman sense on the domain $\mathcal{D}_0$, or that the map $g \mapsto S_n(g)$ with $S_n(g)$ equal (\ref{generalS_n(g)}) is an operator valued distribution on the domain $\mathcal{D}_0$
in the Wightman sense. Although
we should emphasize once again that the Wick product is not understood in this ''proof'' in the
G{\aa}rding-Wightman sense, but rather that the Wick products are reconstructed from their kernels using the explicit form of the commutation generalized functions and the Wick theorem as stated in \cite{Bogoliubov_Shirkov}, Chap. III. Note, please, that the Wick theorem and its proof in
\cite{Bogoliubov_Shirkov}, Chap. III utilizes the kernels of generalized operators, \emph{i. e.}
free field operators at particular space-time points and the commutation rules for the annihilation-creation operators at particular points in the momentum representation. This kind of computation is rather intractable within the G{\aa}rding-Wightman approach, in which the ''kernels'' or fields at specified space-time point, so hardly used in \cite{Bogoliubov_Shirkov}, are only symbolic quantities.
Nonetheless, ''proof'' that (\ref{QEDgeneralS_n(X)andA'_n(X)})
is a kernel of a well-defined operator-valued distribution on $\mathcal{D}_0$ in the Wightman sense and based on the Wick theorem of \cite{Bogoliubov_Shirkov} seems
to be honest, and concerning only this respect our remarks are rather only pedantic. We only like to emphasize that the computational practice in \cite{Bogoliubov_Shirkov} involves the free fields at particular space-time points explicitly
and the creation-annihilation operators at specific momenta, and only afterwards with the explicit formulas for operator kernels, the statements concerning the analytic properties of Wick products are rather rigorously proved.

But we should emphasize that we cannot \emph{a priori} insist that the free fields we should understand mathematically precisely in the sense of Wightman and similarly we cannot \emph{a priori} insist that the operator valued functional $g \mapsto S(g)$ we should understand as the operator valued distribution in the sense of Wightman. In fact looking at the effective cross section computation within the causal approach for the physical interaction with $g=1$, as presented e.g. in \cite{Bogoliubov_Shirkov}, Chap IV, \S \S 24, 25, we can even see that $g \mapsto S(g)$ cannot be understood as Wightman operator-valued distribution with $g$ in the Schwartz test space, if we are about to give precise mathematical basis for the computation of the effective cross section. Indeed, because $g=1$ does not belong to the Schwartz space, we expect rather that $S(g=1)$ is in general meaningless if $g \mapsto S(g)$ is understood as Wightmann operator valued distribution. In spite of some partial results, which try to find some sense of the limit $g=1$ of the expression $S(g)$
still understood as distribution in the Wightman sense, we know that for realistic interactions of fields including massless fields (as QED) the limit $g=1$ cannot be given any sensible mathematical meaning, although compare some partial results in \cite{epstein-glaser-al}, \cite{BlaSen}, \cite{duch}.

Before giving the mathematical interpretation of free quantum fields and of the functionals
$g \mapsto S_n(g)$, $g \mapsto S(g)$, which in our opinion is more adequate, and allows to convert the axioms (I) -- (IV) into
precise mathematical statements as well the formal Wick theorem of \cite{Bogoliubov_Shirkov}
into a mathematical theorem, which give rigorous mathematical basis for the computation of effective
cross sections, we stay for a while with the mathematical interpretation of $g \mapsto S(g)$
as the Wightman operator distribution, after Epstein and Glaser \cite{Epstein-Glaser}, 
with $g$ which cannot be put equal $g=1$ in the formulas for $S_n(g^{\otimes \, n})$ and $S(g)$. 
We do this in order to remind shortly the ingenious contribution of Epstein and Glaser \cite{Epstein-Glaser}, 
who were the first which proved that indeed the conditions (I) -- (IV) are sufficient to construct $S_n(x_1, \ldots, x_n)$, 
provided we do have all $S_k(x_1, \ldots, x_k)$ for $k=1, \ldots, n-1$. 
In fact, they have used (I) -- (IV) in a weaker form with the covariance condition (II) restricted 
only to the covariance under space-time translations.
We will simplify notation after Epstein and Glaser \cite{Epstein-Glaser}, and for sets of
space-time variables $\{x_1, \ldots, x_n\}$, with each $x_k \in \mathcal{M}$ representing a space-time point, 
we will use capital characters $X$, and respectively
$S(X)$ and $\overline{S}(X)$ for $S_n(x_1, \ldots, x_n)$ and $\overline{S_n}(x_1, \ldots, x_n)$,
with the indices $n$ omitted in $S(X)$ and $\overline{S}(X)$, 
as they are understood to be always equal to the number of elements in $X$. 
Order in $X$ understood as argument of $S_n$ can be ignored in view of the symmetricity 
of each $S_n$. So let us assume we have all $S_k(x_1, \ldots, x_k)$ for $k=1, \ldots, n-1$. 
Epstein and Glaser introduce the following operator-valued distributions definig continuous maps
\[
\begin{split}
A'_{(n)}(x_1, \ldots, x_{n-1}, x_n) = \sum \limits_{P_2} \overline{S}(X)S(Y,x_n), \\
R'_{(n)}(x_1, \ldots, x_{n-1}, x_n) = \sum \limits_{P_2} S(Y,x_n)\overline{S}(X),
\end{split}
\]
where the sums run over all divisions $P_2$ of the set $\{x_1, \ldots, x_{n-1} \}$
into two disjoint subsets $X$ and $Y$:
\[
\{x_1, \ldots, x_{n-1} \} = X \sqcup Y,
\,\,\, \textrm{with} \,\,\,
X \neq \emptyset.
\]
Thus by assumption $A'_{(n)}$ and $R'_{(n)}$ are known. Next Epstein and Glaser
introduce the following operator-valued distributions
\[
\begin{split}
A_{(n)}(x_1, \ldots, x_{n-1}, x_n) = \sum \limits_{P_{2}^{0}} \overline{S}(X)S(Y,x_n)
= \sum \limits_{P_2} \overline{S}(X)S(Y,x_n) + S(x_1, \ldots, x_n), \\
R_{(n)}(x_1, \ldots, x_{n-1}, x_n) = \sum \limits_{P_{2}^{0}} S(Y,x_n)\overline{S}(X)
= \sum \limits_{P_2} S(Y,x_n)\overline{S}(X) + S(x_1, \ldots, x_n),
\end{split}
\]
where now, summation is extended over all divisions $P_{2}^{0}$ of the set
$\{x_1, \ldots, x_{n-1} \}$
into two disjoint subsets $X$ and $Y$, which include the empty set $X= \emptyset$. Note that
\[
D_{(n)} = R'_{(n)} - A'_{(n)} = R_{(n)} - A_{(n)}.
\]
In order to finish presentation of the essential point of Epstein-Glaser contribution, let us introduce after \cite{Epstein-Glaser} higher dimensional generalization of the backward and forward cones:
\[
\Gamma_{\pm}^{(n)}(y) = \big\{X \in \mathcal{M}^n: x_j - y \in \overline{V_{\pm}} \big\}, \,\,\,\,\,
X = \{x_1, \ldots, x_n \}.
\]
Then it is shown in \cite{Epstein-Glaser} that
\[
\begin{split}
\textrm{supp} \, R_{(n)}(x_1, \ldots, x_{n-1}, x_n) \subseteq \Gamma_{+}^{(n-1)}(x_n),
\\
\textrm{supp} \, A_{(n)}(x_1, \ldots, x_{n-1}, x_n) \subseteq \Gamma_{-}^{(n-1)}(x_n),
\\
\textrm{supp} \, D_{(n)}(x_1, \ldots, x_{n-1}, x_n) \subseteq \Gamma_{+}^{(n-1)}(x_n)
\sqcup \Gamma_{-}^{(n-1)}(x_n),
\\
\end{split}
\]
and moreover that each $D_{(n)}$ can be (almost) uniquely
splitted into sum of operator distributions each having the support, respectively, in
$\Gamma_{+}^{(n-1)}(x_n)$ or in $\Gamma_{-}^{(n-1)}(x_n)$ and that this splitting can be made explicitly
and independently of the conditions (I) -- (IV). The essential point is that $R_{(n)}$ and
$A_{(n)}$ can be separately computed as the spitting of $D_{(n)}$ into the advanced $A_{(n)}$ and retarded
$R_{(n)}$ parts, so that
\[
\begin{split}
S_n(x_1, \ldots, x_{n-1}, x_n) = A_{(n)}(x_1, \ldots, x_{n-1}, x_n) - A'_{(n)}(x_1, \ldots, x_{n-1}, x_n)
\\
\,\,\,
\textrm{or equivalently}
\,\,\,
\\
S_n(x_1, \ldots, x_{n-1}, x_n) = R_{(n)}(x_1, \ldots, x_{n-1}, x_n) - R'_{(n)}(x_1, \ldots, x_{n-1}, x_n)
\end{split}
\]
and the inductive step from $n-1$ to $n$ can be computed without encountering any infinities and without any need for renormalization. In practical computations, in case of QED, the distributions $S_n(x_1, \ldots, x_n)$, $R_{(n)}(x_1, \ldots, x_n)$, $A_{(n)}(x_1, \ldots, x_n)$, $R'_{(n)}(x_1, \ldots, x_n)$ and $A'_{(n)}(x_1, \ldots, x_n)$, all have the general form (\ref{QEDgeneralS_n(X)andA'_n(X)}) with the respective translationally invariant and causally supported scalar distribution kernels $t^{k}_{n}$. These distributions, when evaluated at the test function $g^{\otimes n}$, all have the general form
(\ref{generalS_n(g)}) with the corresponding distributions $\kappa_{l,m}^{(n)}$.
Therefore, the splitting problem is reduced to the splitting of scalar distributions $t^{k}_{n}$, or respectively $\kappa_{l,m}^{(n)}$, closely related to the tensor product of pairing functions of the respective free fields underlying the theory in question.
That $S_n(x_1, \ldots, x_n)$, $R_{(n)}(x_1, \ldots, x_n)$, $A_{(n)}(x_1, \ldots, x_n)$, $R'_{(n)}(x_1, \ldots, x_n)$ and $A'_{(n)}(x_1, \ldots, x_n)$ are e.g. well-defined operator distributions in Wightman sense on $\mathcal{D}_0$, we convince ourselves by utilizing the formal Wick theorem of \cite{Bogoliubov_Shirkov} at the level of operator kernels, e.g. free fields at specified space-time points and the commutation functions (''pairings''), as we have already said.
We should remark here, that this causal method, due to Bogoliubov-Epstein-Glaser uses in fact additional axiom (V), which allows
to compute $S_n$ up to a finite set of constants, depending on the singularity degree of the products of pairings, which we encounter
in the Wick decomposition of the product $\mathcal{L}(x)\mathcal{L}(y)$. Namely, we assume that the splitting
is ``maximally natural'', \emph{i.e.} coinciding with the natural formula given by the ordinary multiplication by the step theta function,
whenever the natural formula makes sense, \emph{i.e.} whenever can be applied to a test function. 
This means that we assume that the singularity degree is preserved during the splitting, \emph{i.e.} retarded part of a distribution
should have the same singularity degree as the distribution itself. Sometimes this assumption is called ``preservation of the scaling degree''.  

Now we are ready to give the basis which allows us to solve the Adiabatic Limit Problem. This is in fact the choice of a concrete mathematical interpretation for the axioms (I) -- (IV) and (V). Namely, we propose
to use the Hida white noise operators, \cite{hida}, \cite{HKPS}, \cite{obata-book}, as the creation-annihilation operators $a_{s_i}(\boldsymbol{\p}_i)^{+}, \ldots a_{s_j}(\boldsymbol{\p}_j)$ of free fields
in the momentum picture. All the rest of the causal principles (I) -- (IV) and (V), we keep totally unchanged.
This in fact means that we have to construct the free fields of the theory using the creation-annihilation operators which are the Hida operators. Because the Hida operators indeed respect the canonical commutation relations, this mathematical realization fits in naturally into the laws of QFT, and in particular into the axioms (I) -- (IV) and (V). Dipper discussion of these axioms we give in Subsections \ref{WickForProduct} and \ref{WickForChronological}. 
This allows us to treat the operators
$a_{s}(\boldsymbol{\p})^{+}, a_{s}(\boldsymbol{\p})$ at each fixed point $\boldsymbol{\p}$ as a well-defined operator transforming continuously the test Hida space into its strong dual. Similarly, this will allow us to treat the free fields at fixed space-time point as well-defined operators transforming continuously the test Hida space into its strong dual. Moreover, using Hida operators as creation-annihilation operators will allow us to interpret the free fields of the theory, in case of QED
the Dirac $\boldsymbol{\psi}$ and electromagnetic potential free field $A$,
\[
\boldsymbol{\psi} = \Xi_{0,1}({}^{1}\kappa_{0,1}) + \Xi_{1,0}({}^{1}\kappa_{1,0}), \,\,\,
A = \Xi_{0,1}({}^{2}\kappa_{0,1}) + \Xi_{1,0}({}^{2}\kappa_{1,0}),
\]
as integral kernel generalized operators with vector valued kernels (in the sense of Obata \cite{obataJFA}, \cite{obata-book})
\[
\begin{split}
{}^{1}\kappa_{0,1}, {}^{1}\kappa_{1,0} \in \mathscr{L}(E, \mathscr{E}^{*})
\cong E^{*} \otimes \mathscr{E}^{*}, \\
{}^{2}\kappa_{0,1}, {}^{2}\kappa_{1,0} \in \mathscr{L}(E, \mathscr{E}^{*})
\cong E^{*} \otimes \mathscr{E}^{*},
\end{split}
\]
with the integral kernels of these distributions, which are equal to ordinary functions:
\[
\begin{split}
{}^{2}\kappa_{0,1}(\nu, \boldsymbol{\p}; \mu, x) =
{\textstyle\frac{g_{\nu \mu}}{(2\pi)^{3/2}\sqrt{2p^0(\boldsymbol{\p})}}}
e^{-ip\cdot x}, \,\,\,\,\,\,
p = (|p_0(\boldsymbol{\p})|, \boldsymbol{\p}), \, p\cdot p=0, \\
{}^{2}\kappa_{1,0}(\nu, \boldsymbol{\p}; \mu, x) =
{\textstyle\frac{g_{\nu \mu}}{(2\pi)^{3/2}\sqrt{2p^0(\boldsymbol{\p})}}}
e^{ip\cdot x},
\,\,\,\,\,\,
p \cdot p = 0,
\end{split}
\]
\[
{}^{1}\kappa_{0,1}(s, \boldsymbol{\p}; a,x) = \left\{ \begin{array}{ll}
(2\pi)^{-3/2}u_{s}^{a}(\boldsymbol{\p})e^{-ip\cdot x}, \,\,\, \textrm{$p = (|p_0(\boldsymbol{\p})|, \boldsymbol{\p}), \, p \cdot p = m^2$} & \textrm{if $s=1,2$}
\\
0 & \textrm{if $s=3,4$}
\end{array} \right.,
\]
\[
{}^{1}\kappa_{1,0}(s, \boldsymbol{\p}; a,x) = \left\{ \begin{array}{ll}
0 & \textrm{if $s=1,2$}
\\
(2\pi)^{-3/2} v_{s-2}^{a}(\boldsymbol{\p})e^{ip\cdot x}, \,\,\, \textrm{$p \cdot p = m^2$} & \textrm{if $s=3,4$}
\end{array} \right.
\]
which are in fact the respective plane wave solutions of d'Alembert and of Dirac equation, which span the corresponding generalized eigen-solution sub spaces. 
Here $g_{\nu\mu}$ are the components of the space-time Minkowski metric tensor. All the important operations, which can be performed upon the free field operators, now understood as integral kernel operators, include: 1) differentiation of the free field operator, 2) Wick product
(with space-time variables in each factor treated as independent), 3) Wick product of free fields and their differentials at fixed space-time point, 4) splitting of integral kernel operators with causal support into retarded and advanced parts. These operations give another integral kernels operations, which can be realized by the corresponding operations performed upon the corresponding kernels $\kappa_{l,m}$: 1) differentiation of the kernel, well-defined for $\kappa_{l,m}$ being a distribution, 2) projective tensor product of the corresponding kernels, realized by ordinary product of functions representing the kernels with independent variables in each factor, 3) operation of pointwise multiplication (in space-time variables) performed on the functions corresponding to the kernels. 4) Splitting of the corresponding
kernel distribution with causal support. In particular the formal Wick theorem
of \cite{Bogoliubov_Shirkov}, Chap. III, becomes a mathematical theorem (when using Hida operators), which becomes a particular case of the Theorem 4.5.1 \cite{obata-book}, which in case of Wick theorem simplifies to the case of finite decomposition of a generalised operator into integral kernel operators (with normally ordered creation-annihilation Hida operators). The higher order contributions $S_n(g^{\otimes \, n})$ (\ref{generalS_n(g)}) evaluated at Schwartz space-time test function $g$
become finite sums of integral kernel operators, and $S(g)$ becomes to be equal to a Fock expansion into integral kernel operators transforming continuously the Hida test space into its strong dual in the sense \cite{obata-book} or \cite{obataJFA}. Moreover, the particular higher order contributions to the expansions
of more general scattering operators with additional interaction therms and the higher order contributions to the Bogoliubov-Shirkov functional derivatives of these scattering operators defining interacting fields gain the meaning of a (finite sum of) integral kernel operators with vector valued kernels, and the interacting fields gain the meaning of generalized operators given by Fock expansions into integral kernel operators.
The point is that now, when using Hida operators as creation-annihilation operators, each higher order contribution $S_n(g^{\otimes \, n})\in \mathscr{L}((\boldsymbol{E}), (\boldsymbol{E})^*)$, \emph{i.e.} becomes a continuous map $(\boldsymbol{E}) \longmapsto (\boldsymbol{E})$ or 
$(\boldsymbol{E}) \longmapsto (\boldsymbol{E})^*$, for each fixed element $g$ of the
Schwartz test space $\mathcal{S}(\mathbb{R}^4)$ of functions on the space-time, which moreover defines a continuous map
$\mathcal{S}(\mathbb{R}^4)^{\otimes \, n} \longmapsto \mathscr{L}((\boldsymbol{E}), (\boldsymbol{E})^*)$, which is sufficient for the computation of the effective cross-section
in the adiabatic limit $g=1$, as explained in Introduction.
 
In this approach the function $g$ in the distributional kernel (\ref{kappa_l,m(p1,...p_l+m;g)}) in the causal perturbative formula for the interacting fields
and in the scattering operator $S(g)$,
still belongs to the Schwartz space, but in particular
the interacting field operators gain interpretation of Fock expansions into integral kernel operators with vector-valued kernels
in the sense of \cite{obataJFA} even with $g=1$. 

The Wick product and product operations can be extended
on a wide class of generalized operators in the space
\[
\mathscr{L}\big(\mathscr{E}, \mathscr{L}((\boldsymbol{E}), (\boldsymbol{E})^*)\big),
\]
which are needed for the construction of the scattering operator with Hida operators and massless fields in $\mathcal{L}$.

We obtain in this manner theory with the continuous scattering operator function with the higher order contributions
\[
\mathcal{S}(\mathbb{R}^4)^{\otimes \, n} \ni g^{\otimes \, n} \longmapsto S_n(g^{\otimes \, n}) \in
\mathscr{L}((\boldsymbol{E}), (\boldsymbol{E})^*),
\]
or 
\[
\mathcal{S}(\mathbb{R}^4)^{\otimes \, n} \ni g^{\otimes \, n} \longmapsto S(g^{\otimes \, n}) \in
\mathscr{L}((\boldsymbol{E}), (\boldsymbol{E})),
\]
depending if there are or there are no massless fields in $\mathcal{L}$.
The particular contributions
$S_n(g^{\otimes \, n})$, $g \in \mathcal{S}(\mathbb{R}^4)$, 
are generalized operators transforming continuously the Hida space into 
itself or into its strong dual. The higher order contributions to the interacting quantum field operators in the adiabatic limit $g=1$ are generalized operators transforming the test Hida space into its strong dual. This allows us to apply the theory only to the generalized states, e.g. the many particle plane wave states, and to the computation of the effective cross sections, provided we are interested in the high energy processes involving the plane wave states of the elementary free fields of the theory in question. This is sufficient for the computation of the effective cross sections in high energy scattering processes. Problems involving bound, and thus normalizable, states are beyond the scope of this theory, as we have already said in Introduction.

In the following Sections we construct the free quantum Dirac field and the free quantum electromagnetic potential field with the help of Hida operators as the creation-annihilation operators. Next we prove the statements mentioned above, together with the proof that the interacting Dirac and electromagnetic potential fields are well-defined generalized operators transforming continuously the Hida space into its strong dual, with the higher order contributions equal to well-defined integral kernel operators with vector valued kernels in the sense of Obata \cite{obata-book}.

Here we only emphasize a novelty in comparison to the free fields in Wightman sense, that
when constructing free fields with the help of Hida operators the standard nuclear spaces
$E$ which compose the standard Gelfand triples
\[
E \subset \mathcal{H}' \subset E^*
\]
in the single particle Hilbert spaces $\mathcal{H}'$ of the corresponding free fields (Dirac field $\boldsymbol{\psi}$ and $A$ in case of spinor QED) depend on the type of field and for massless fields are different in comparison to massive fields. $E$ is unitarily equivalent (with the equivalence preserving the nuclear topology) to the space of restrictions $\widetilde{\phi}\big|_{{}_{\mathscr{O}}}$ to the corresponding orbits
\[
\mathscr{O} = \{p: p_0 \geq 0, p \cdot p = 0 \,\, \textrm{or} \,\, = m^2 \},
\]
of the Fourier transforms $\widetilde{\phi}$ of space-time test functions $\phi \in \mathscr{E}$, composing likewise a standard countably Hilbert nuclear space $\mathscr{E}$. It turns out that when constructing massless free fields with the help of Hida operators, the space-time test space cannot be equal to the Schwartz space of scalar, four-vector, spinor,
\emph{e. t. c.} functions but need to be chosen differently if we want the free fields are to be 
regular operators lying in
\[
\mathscr{L}\big(\mathscr{E}, \mathscr{L}((\boldsymbol{E}), (\boldsymbol{E}))\big).
\]
When the free field is understood in the sense of Wightman, $\mathscr{E}$ can be put equal to the Schwartz space irrespectively if the free 
field is massive or massless. For the higher order contributions to interacting fields in the adiabatic limit $g=1$, which are more singular and are in
\[
\mathscr{L}\big(\mathscr{E}, \mathscr{L}((\boldsymbol{E}), (\boldsymbol{E})^*)\big),
\]
the space-time test space $\mathscr{E}$ can be chosen to be the ordinary Schwartz space.

\subsection{Quantum Dirac free field $\boldsymbol{\psi}$ 
as an integral kernel operator with vector-valued distributional kernel 
within the white noise 
construction of Berezin-Hida-Obata}\label{psiBerezin-Hida}

In constructing the quantum free Dirac field $\boldsymbol{\psi}$ according to Berezin-Hida,
we proceed in a sense in a totally opposite direction in comparison to Wightman.
Namely, Wightman restricts the arguments 
$u\oplus v \in \mathcal{H}'= \mathcal{H}_{m,0}^{\oplus} \oplus \mathcal{H}_{-m,0}^{\ominus \, \flat}$ 
of the operators $a'(u\oplus v), a'(u\oplus v)^+$
in (\ref{psi(f)=a_+(f)+a_-(f^c)^+proper}) 
to the nuclear subspace $E \cong \mathcal{S}(\mathbb{R}^3; \mathbb{C}^4)$ 
of all those $u\oplus v$ for which $u$ are equal to 
\[
u = P^\oplus\widetilde{f}|_{{}_{\mathscr{O}_{+m,0,0,0}}}, \,\,\
f \in \mathcal{S}(\mathbb{R}^4; \mathbb{C}^4)
\]
and 
\[
v = \big(P^\ominus\widetilde{f}|_{{}_{\mathscr{O}_{-m,0,0,0}}}\big)^\flat, \,\,\
f \in \mathcal{S}(\mathbb{R}^4; \mathbb{C}^4).
\]
In the following steps he keeps the arguments $u\oplus v$ of the annihilation
and creation operators $a'(u\oplus v), a'(u\oplus v)^+$ within the nuclear space $E$, and with the domain
$\mathcal{D}$ of the operators $a'(u\oplus v), a'(u\oplus v)^+$ which is not uniquely nor naturally determined.

According to Berezin-Hida we choose quite an opposite direction: we extend the domain of the arguments
$u\oplus v$ of the creation
and annihilation operators $a'(u\oplus v), a'(u\oplus v)^+$ to include also generalized states
(elements of the strong dual $E^* \cong \mathcal{S}(\mathbb{R}^3; \mathbb{C}^4)^*$ -- tempered distributions)
$u\oplus v$, like the plane wave solutions. This is exactly what is needed (and used but at the formal level)
in the (formal) proof of the so called
''Wick theorem'' for free fields, presented in
\cite{Bogoliubov_Shirkov}, Chap. III. By utilizing the rigorous construction of the Hida operators
$a'(u\oplus v), a'(u\oplus v)^+$ we convert this formal proof into a rigorous one.

This is achieved in the following manner. First we introduce the nuclear space $E$ as above, which composes
with the single particle Hilbert space 
$\mathcal{H}'= \mathcal{H}_{m,0}^{\oplus} \oplus \mathcal{H}_{-m,0}^{\ominus \, \flat}$,
a Gelfand triple
\[
\left. \begin{array}{ccccc}              E         & \subset &  \mathcal{H'} & \subset & E^*        \\
                                                   &         & \parallel      &         &  \\
                        &   & \mathcal{H}_{m,0}^{\oplus} \oplus \mathcal{H}_{-m,0}^{\ominus\, \flat} &  &  \end{array}\right..
\]  
We should do it in such a manner which allows lifting of this construction to the second quantized level
with the corresponding Gelfand triple
\[
\left. \begin{array}{ccccc}              (E)         & \subset &  \Gamma(\mathcal{H'}) & \subset & (E)^*        \\
                                                   &         & \parallel      &         &  \\
                        &   & \Gamma\big(\mathcal{H}_{m,0}^{\oplus} \oplus \mathcal{H}_{-m,0}^{\ominus \, \flat}\big) &  &  \end{array}\right.,
\]  
with a nuclear (Hida) dense subspace $(E)$ in the Fock space 
$\Gamma(\mathcal{H}')= \Gamma\big(\mathcal{H}_{m,0}^{\oplus} \oplus \mathcal{H}_{-m,0}^{\ominus \, \flat}\big)$.
For each $u\oplus v \in E^*$ the annihilation operators $a'(u\oplus v)$ become operators 
continuously transforming the nuclear dense space $(E)$ into itself. Because the inclusion
of $(E)$ into the strong dual $(E)^*$ is continuous, the operators $a'(u\oplus v)$ 
can be naturally regarded as continuous operators $(E) \rightarrow (E)^*$.
By construction the creation operators $a'(u\oplus v)^+$, $u\oplus v \in E^*$, are equal $\overline{(\cdot)}
\circ a'(u\oplus v)^*  \circ \overline{(\cdot)}$, \emph{i.e.} to the linear duals $a'(u\oplus v)^*$ of the
annihilation operators  $a'(u\oplus v)$ composed with complex conjugation, 
and thus transform continuously the strong dual space $(E)^*$ into itself, and can be naturally regarded as continuous operators $(E) = (E)^{**} \rightarrow (E)^*$ (because $(E)$ is reflexive).
For  $u\oplus v \in E$ the operators $a'(u\oplus v), a'(u\oplus v)^+$ become operators 
transforming continuously the nuclear dense space $(E)$ into itself and thus belong to
$\mathscr{L}\big((E), (E)\big)$. Moreover,
the maps
\[
\begin{split}
E \ni u\oplus v \, \longmapsto a'(u\oplus v) \in \mathscr{L}\big((E),(E)\big), \\
E \ni u\oplus v \, \longmapsto a'(u\oplus v)^+ \in \mathscr{L}\big((E),(E)\big),
\end{split}
\]
are continuous when $\mathscr{L}\big((E),(E)\big)$ -- the linear space of linear continuous operators
from $(E)$ into $(E)$ -- is given the natural nuclear topology of 
uniform convergence on bounded sets. 

Therefore, it is important to have the Gelfand triple $E \subset \mathcal{H}' \subset E^*$ in the form which
allows its lifting to the Fock space and the construction of
the Hida test space $(E)$ composing the Gelfand triple $(E) \subset \Gamma(\mathcal{H}') \subset (E)^*$.
This is in particular the case when we have the nuclear space $E \subset \mathcal{H}'$ in the standard form,
\cite{obata-book}.
Namely, let $(\mathscr{O}, \ud\mu_{{}_{\mathscr{O}}})$ be a topological space $\mathscr{O}$ with a
Baire (or Borel) measure $\ud\mu_{{}_{\mathscr{O}}}$. Then we assume that $\mathcal{H'}$ is naturally unitarily $U$
equivalent to the Hilbert
space of $\mathbb{C}$-valued measurable (equivalence classes
modulo equality almost everywhere) and square summable functions $L^2(\mathscr{O}, \ud\mu_{{}_{\mathscr{O}}})$.
Next we assume that $E \subset \mathcal{H}'$ is naturally unitarily equivalent, with the same unitary equivalence $U$ which also defines an isomorphism of $E$ with the standard countably Hilbert nuclear space
$\mathcal{S}_{A}(\mathscr{O}; \mathbb{C}) \subset L^2(\mathscr{O}, \ud\mu_{{}_{\mathscr{O}}};\mathbb{C})$,
composing a Gelfand triple
\[
\left. \begin{array}{ccccc} \mathcal{S}_{A}(\mathscr{O}; \mathbb{C}) & \subset & L^2(\mathscr{O}, \ud\mu_{{}_{\mathscr{O}}}; \mathbb{C}) & \subset & \mathcal{S}_{A}(\mathscr{O}; \mathbb{C})^*
\end{array}\right.,
\]
and fulfilling the Kubo-Takenaka conditions. For standard construction of a nuclear space
$\mathcal{S}_{A}(\mathscr{O}; \mathbb{C}) \subset L^2(\mathscr{O}, \ud\mu_{{}_{\mathscr{O}}};\mathbb{C})$
as arising from a standard (self-adjoint with nuclear or Hilbert Schmidt $A^{-r}$
for some positive $r$) operator $A$ on
$L^2(\mathscr{O}, \ud\mu_{{}_{\mathscr{O}}};\mathbb{C})$, fulfilling Kubo-Takenaka conditions, compare
\cite{obata-book}, or Subsection \ref{white-setup}.

In this situation we have the natural lifting of the Gelfand triple over to the Fock space:
\[
\left. \begin{array}{ccccc} \big( \mathcal{S}_{A}(\mathscr{O}; \mathbb{C})\big) & \subset & \Gamma\big(L^2(\mathscr{O}, \ud\mu_{{}_{\mathscr{O}}}; \mathbb{C})\big) &\subset & \big( \mathcal{S}_{A}(\mathscr{O}; \mathbb{C}) \big)^*
\end{array}\right.,
\]
constructed from the standard operator $\Gamma(A)$ in
$\Gamma\big(L^2(\mathscr{O}, \ud\mu_{{}_{\mathscr{O}}}; \mathbb{C})\big)$. That the operator
$\Gamma(A)$ will be standard whenever $A$ is, also for the fermionic functor $\Gamma$ and under the same
assumptions for $A$ as in the boson case, can be proved in exactly the same way as in \cite{obata-book}, Lemma 3.1.2,
for the bosonic case (the proof is even simpler in Fermi case because the occupation numbers assume only the values $0$
or $1$ in this case).

Eventually we have the initial standard Gelfand triple in the single particle
Hilbert space $\mathcal{H}'$ given in the standard form only up to a unitary isomorphism:
\[
\left. \begin{array}{ccccc} \mathcal{S}_{A}(\mathscr{O};\mathbb{C}) & \subset & L^2(\mathscr{O};\mathbb{C}) & \subset & \mathcal{S}_{A}(\mathscr{O}; \mathbb{C})^* \\
\downarrow \uparrow & & \downarrow \uparrow & & \downarrow \uparrow \\
E & \subset & \mathcal{H'} & \subset & E^* \\
& & \parallel & & \\
& & \mathcal{H}_{m,0}^{\oplus} \oplus \mathcal{H}_{-m,0}^{\ominus\, \flat} & &
\end{array}\right.,
\]
with the vertical arrows indicating the unitary operator (and its inverse) $U: \mathcal{H}' \rightarrow L^2(\mathscr{O};\mathbb{C})$ whose restriction to $E$ defines an isomorphism
$U: E \rightarrow \mathcal{S}_{A}(\mathscr{O}; \mathbb{C})$ of nuclear spaces
and whose linear transposition $U^*$ defines isomorphism
$\mathcal{S}_{A}(\mathscr{O}; \mathbb{C})^* \rightarrow E^*$.
The nuclear space $E\subset \mathcal{H}'$ then corresponds to the standard operator
$U^{-1}AU$ on $\mathcal{H}'$, and can be constructed from it (compare \cite{obata-book},
Subsection \ref{white-setup}).

The last Gelfand triples can be lifted to the corresponding Fock spaces together with the corresponding
isomorphisms determined by the unitary operator $\Gamma(U)$: its restriction to $(E) \subset
\Gamma\big(L^2(\mathscr{O}; \mathbb{C})\big)$ transforming continuously
$(E) \rightarrow \big(\mathcal{S}_{A}(\mathscr{O}; \mathbb{C})\big)$,
or linear transposition of this restriction, defining the isomorphism
$(E)^* \rightarrow \big(\mathcal{S}_{A}(\mathscr{O}; \mathbb{C}))^*$:
\[
\left. \begin{array}{ccccc} \big(\mathcal{S}_{A}(\mathscr{O}; \mathbb{C})\big) & \subset & \Gamma\big(L^2(\mathscr{O};\mathbb{C})\big) & \subset & \big(\mathcal{S}_{A}(\mathscr{O}; \mathbb{C}))^* \\
\downarrow \uparrow & & \downarrow \uparrow & & \downarrow \uparrow \\
(E) & \subset & \Gamma(\mathcal{H'}) & \subset & (E)^* \\
& & \parallel & & \\
& & \Gamma\big(\mathcal{H}_{m,0}^{\oplus} \oplus \mathcal{H}_{-m,0}^{\ominus\,\flat}\big) & &
\end{array}\right..
\]
In this case we have the following relations for the annihilation (and correspondingly creation)
operators
\begin{multline}\label{G(U)^+a(U(u+v))G(U)=a'(u+v)}
\Gamma(U)^{+} \, a\big(U^{+-1}(u\oplus v)\big) \, \Gamma(U)= a'(u\oplus v), \\
\Gamma(U)^{+} \, a\big(U^{+-1}(u\oplus v)\big)^+ \, \Gamma(U)= a'(u\oplus v)^+, \\
u\oplus v \in E^*.
\end{multline}
Here the Hida operators $a'(u\oplus v), a'(u\oplus v)^+$ coincide with the ordinary annihilation and
creation operators
$a'(u\oplus v), a'(u\oplus v)^+$ (defined in Subsection \ref{electron+positron}) on the Hida
subspace $(E) \subset \Gamma(\mathcal{H'})
\subset (E)^*$ of the Fock space
$\Gamma(\mathcal{H'}) = \Gamma\big(\mathcal{H}_{m,0}^{\oplus} \oplus \mathcal{H}_{-m,0}^{\ominus\, \flat}\big)$,
whenever $u\oplus v \in E \subset \mathcal{H'}= \mathcal{H}_{m,0}^{\oplus} \oplus \mathcal{H}_{-m,0}^{\ominus c}
\subset E^*$. Similarly, $a(w),a(w)^+$ coincide with the standard annihilation and creation operators
on the Hida subspace $\big(\mathcal{S}_{A}(\mathscr{O}; \mathbb{C})\big)$
of the Fock space $\Gamma\big(L^2(\mathscr{O};\mathbb{C})\big)$, whenever
$w \in \mathcal{S}_{A}(\mathscr{O}; \mathbb{C}) \subset L^2(\mathscr{O};\mathbb{C}) \subset
\mathcal{S}_{A}(\mathscr{O}; \mathbb{C})^*$.
In this case we can restrict the creation and annihilation operators
$a'(u\oplus v), a'(u\oplus v)^+$ to the Hida subspace $(E)$
and regard them as elements of $\mathscr{L}\big((E),(E)\big)$
(and respectively
$a(w),a(w)^+ \in \mathscr{L}\big((\mathcal{S}_{A}(\mathscr{O}; \mathbb{C})), (\mathcal{S}_{A}(\mathscr{O}; \mathbb{C}))\big)$) and similarly restrict the linear
dual composed with complex conjugation $\Gamma(U)^{+} = \overline{(\cdot)} \circ \Gamma(U)^{*} \circ \overline{(\cdot)}
: \big(\mathcal{S}_{A}(\mathscr{O}; \mathbb{C})\big)^* \rightarrow (E)^*$
to the subspace $(E)$, where it coincides with the ordinary inverse
$\Gamma(U)^{-1}$ of the unitary operator $\Gamma(U)$, and with the inverse $U^{+-1} = \overline{(\cdot)} \circ
U^{*-1} \circ\overline{(\cdot)}$ of the linear dual
$U^*: \mathcal{S}_{A}(\mathscr{O}; \mathbb{C})^* \rightarrow E^*$ to $U$ composed with conjugations
degenerating to $U^{+-1} = U$ on the subspace $E \subset E^*$.
In this particular case the general formula (\ref{G(U)^+a(U(u+v))G(U)=a'(u+v)}) degenerates to
\begin{multline}\label{G(U)^+a(U(u+v))G(U)=a'(u+v)degenerated}
\Gamma(U)^{-1} \, a\big(U(u\oplus v)\big) \, \Gamma(U)= a'(u\oplus v), \\
\Gamma(U)^{-1} \, a\big(U(u\oplus v)\big)^+ \, \Gamma(U)= a'(u\oplus v)^+, \\
u\oplus v \in E \subset E^*.
\end{multline}
But the formula (\ref{G(U)^+a(U(u+v))G(U)=a'(u+v)}) is valid generally for the
operators $a'(u\oplus v), a'(u\oplus v)^+ \in \mathscr{L}\big((E), (E)^*\big)$,
\[
a(w), a(w)^+ \in \mathscr{L}\Big(\big(\mathcal{S}_{A}(\mathscr{O}; \mathbb{C})\big), \big(\mathcal{S}_{A}(\mathscr{O}; \mathbb{C})\big)^*\Big),
\]
understood in the sense of Hida with $u\oplus v \in E^*$, or
respectively $w \in \mathcal{S}_{A}(\mathscr{O}; \mathbb{C})^*$, and with $\Gamma(U)$
understood as a continuous isomorphism
\[
(E) \, \longrightarrow \, \big(\mathcal{S}_{A}(\mathscr{O}; \mathbb{C})\big)
\]
of nuclear spaces in the first formula of (\ref{G(U)^+a(U(u+v))G(U)=a'(u+v)})
and with $\Gamma(U)^+ = \overline{(\cdot)} \circ \Gamma(U)^* \circ \overline{(\cdot)}$
as its continuous dual isomorphism
\[
\big(\mathcal{S}_{A}(\mathscr{O}; \mathbb{C})\big)^* \, \longrightarrow \, (E)^*
\]
composed with complex conjugation in
(\ref{G(U)^+a(U(u+v))G(U)=a'(u+v)}). Below we give generalized operators
$a'(u\oplus v), a'(u\oplus v)^+$ (and respectively $a(w),a(w)^+$), due to Hida, which make sense
also for $u\oplus v$ (respectively $w$), lying in the space dual to $E$, respectively
dual to $\mathcal{S}_{A}(\mathscr{O}; \mathbb{C})$.

In order to simplify notation we agree to write the last isomorphisms
(\ref{G(U)^+a(U(u+v))G(U)=a'(u+v)}) (and their particular case
(\ref{G(U)^+a(U(u+v))G(U)=a'(u+v)degenerated})) induced by $U$
simply identifying the corresponding operators, namely
\begin{multline}\label{a(U(u+v))=a'(u+v)}
\begin{split}
a\big(U^{+-1}(u\oplus v)\big) = a'(u\oplus v), \,\,\,
a\big(U^{+-1}(u\oplus v)\big)^+ = a'(u\oplus v)^+, \,\,\,
u\oplus v \in E^*, \\
a\big(U(u\oplus v)\big) = a'(u\oplus v), \,\,\,
a\big(U(u\oplus v)\big)^+ = a'(u\oplus v)^+, \,\,\,
u\oplus v \in E \subset E^*,
\end{split}
\end{multline}
as operators transforming continuously Hida spaces into their strong duals
(in the first case) or as operators transforming continuously Hida spaces
into Hida spaces (in the second case).

Note that in our case the initial Gelfand triple
$E  \subset \mathcal{H}_{m,0}^{\oplus} \oplus \mathcal{H}_{-m,0}^{\ominus \, \flat}
\subset E^*$ over the single particle Hilbert space  
$\mathcal{H'} = \mathcal{H}_{m,0}^{\oplus} \oplus \mathcal{H}_{-m,0}^{\ominus \, \flat}$ 
does not have the standard form, because the single particle Hilbert space $\mathcal{H}'$
does not have the form $L^2(\mathscr{O}, \ud\mu_{{}_{\mathscr{O}}};\mathbb{C})$.
Indeed, note that the Hilbert space 
\begin{multline*}
L^2(\mathbb{R}^3, \ud^3 \boldsymbol{\p}/(2p_0(\boldsymbol{\p}))^2; \mathbb{C}^4)
= \oplus_{1}^{4} L^2(\mathbb{R}^3, \ud^3 \boldsymbol{\p}/(2p_0(\boldsymbol{\p}))^2; \mathbb{C}) \\
= L^2(\mathbb{R}^3 \sqcup \mathbb{R}^3 \sqcup \mathbb{R}^3 \sqcup \mathbb{R}^3, \ud^3 \boldsymbol{\p}/(2p_0(\boldsymbol{\p}))^2; \mathbb{C})
\end{multline*}
does have the required form $L^2(\mathscr{O}, \ud\mu_{{}_{\mathscr{O}}};\mathbb{C})$,
with 
\[
\mathscr{O} = \mathbb{R}^3 \sqcup \mathbb{R}^3 \sqcup \mathbb{R}^3 \sqcup \mathbb{R}^3
\]
equal to the disjoint sum of four copies of $\mathbb{R}^3$ and the direct sum measure 
$\ud\mu_{{}_{\mathscr{O}}}$ coinciding with $\frac{\ud^3 \boldsymbol{\p}}{(2p_0(\boldsymbol{\p}))^2}$
on each copy $\mathbb{R}^3$. But recall that although in our case the values $\widetilde{\phi}(p)$ of
the bispinors $\widetilde{\phi} \in \mathcal{H}_{m,0}^{\oplus}$ concentrated on the positive energy orbit
$\mathscr{O}_{m,0,0,0}$ range over $\mathbb{C}^4$, nonetheless $\mathcal{H}_{m,0}^{\oplus}$ does not have the standard form
\[
L^2(\mathbb{R}^3, \ud^3 \boldsymbol{\p}/(2p_0(\boldsymbol{\p}))^2; \mathbb{C}^4),
\]
because for each fixed $\boldsymbol{p}$ the vectors $\widetilde{\phi}(\boldsymbol{\p}, p_0(\boldsymbol{\p}))$,
with $\widetilde{\phi}$ ranging over $\mathcal{H}_{m,0}^{\oplus}$, 
do not span $\mathbb{C}^4$, but are equal to the image 
$\textrm{Im} \, P^\oplus(\boldsymbol{\p}, p_0(\boldsymbol{\p})) \neq \mathbb{C}^4$, 
for $p = (\boldsymbol{\p}, p_0(\boldsymbol{\p})) \in \mathscr{O}_{m,0,0,0}$,
because $\textrm{rank} \, P^\oplus(\boldsymbol{\p}, p_0(\boldsymbol{\p})) = 2 \neq 4$ 
(compare Subsection \ref{e1}, where the projection operator $P^\oplus$ of pointwise multiplication by
$P^\oplus(p)$, $p \in \mathscr{O}_{m,0,0,0}$, acting on bispinors concentrated on the orbit 
$\mathscr{O}_{m,0,0,0}$ is defined). 

Similarly, $\mathcal{H}_{-m,0}^{\ominus \, \flat}$ does not have the standard form
\[
L^2(\mathbb{R}^3, \ud^3 \boldsymbol{\p}/(2p_0(\boldsymbol{\p}))^2; \mathbb{C}^4)
\]
in spite of the fact that the conjugations $\widetilde{\phi}^\flat \in \mathcal{H}_{-m,0}^{\ominus \, \flat}$ of the 
bispinors $\widetilde{\phi} \in \mathcal{H}_{-m,0}^{\ominus}$ concentrated on the negative energy orbit 
$\mathscr{O}_{-m,0,0,0}$ take their values in $\mathbb{C}^4$, 
because $\{\widetilde{\phi}(\boldsymbol{\p}, p_0(\boldsymbol{\p})), \widetilde{\phi} \in \mathcal{H}_{-m,0}^{\ominus}\} = \textrm{Im} \, P^\ominus(\boldsymbol{\p}, p_0(\boldsymbol{\p})) \neq \mathbb{C}^4$ with 
$\textrm{rank} \, P^\ominus(\boldsymbol{\p}, p_0(\boldsymbol{\p})) = 2 \neq 4$, for 
$p = (\boldsymbol{\p}, p_0(\boldsymbol{\p})) \in \mathscr{O}_{-m,0,0,0}$. 

But there exists a natural unitary isomorphism $U$ (in fact a class of such natural $U$)
\[
U: \,\, \mathcal{H'} = \mathcal{H}_{m,0}^{\oplus} \oplus \mathcal{H}_{-m,0}^{\ominus \, \flat} \,\,\,
\longrightarrow \,\,\,
L^2(\mathbb{R}^3, \ud^3 \boldsymbol{\p}; \mathbb{C}^4) 
\]
between the single particle Hilbert space $\mathcal{H}'$ and the Hilbert space 
\[
L^2(\mathbb{R}^3, \ud^3 \boldsymbol{\p}; \mathbb{C}^4)  = \oplus L^2(\mathbb{R}^3, \ud^3 \boldsymbol{\p}; \mathbb{C})
=L^2\big(\mathbb{R}^3 \sqcup \mathbb{R}^3 \sqcup \mathbb{R}^3 \sqcup \mathbb{R}^3, \ud^3 \boldsymbol{\p};
\mathbb{C}\big),
\]
which moreover restricts to an isomorphism between the nuclear spaces 
of Schwartz bispinors in $E  \subset \mathcal{H}'$
and Schwartz functions in $\mathcal{S}(\mathbb{R}^3; \mathbb{C}^4)
=\mathcal{S}_{A}(\mathbb{R}^3; \mathbb{C}^4) \subset L^2(\mathbb{R}^3, \ud^3 \boldsymbol{\p}; \mathbb{C}^4)$. 

Indeed, for $\widetilde{\phi} \in \mathcal{H}_{m,0}^{\oplus}$,
$\widetilde{\phi}' \in \mathcal{H}_{-m,0}^{\ominus}$ we put
\begin{multline}\label{isomorphismU}
U\Big( \widetilde{\phi} \oplus (\widetilde{\phi}')^\flat\Big) \overset{\textrm{df}}{=}
(\widetilde{\phi})_{1+} \oplus (\widetilde{\phi})_{2+} \oplus (\widetilde{\phi}')_{1-}
\oplus (\widetilde{\phi}')_{2-} \\ =
(\widetilde{\phi})_{1} \oplus (\widetilde{\phi})_{2} \oplus (\widetilde{\phi}')_{3}
\oplus (\widetilde{\phi}')_{4} \,\,\, \in \oplus_{1}^{4} L^2(\mathbb{R}^3; \mathbb{C}) =
L^2(\mathbb{R}^3; \mathbb{C}^4),
\end{multline}
where
\[
\begin{split}
(\widetilde{\phi})_{1}(\boldsymbol{\p}) = (\widetilde{\phi})_{1+}(\boldsymbol{\p}) \overset{\textrm{df}}{=}
\frac{1}{2p_0(\boldsymbol{\p})} u_1(\boldsymbol{\p})^+ \widetilde{\phi}(p_0(\boldsymbol{\p}), \boldsymbol{\p}),
\,\,\,\,\, p_0(\boldsymbol{\p}) = \sqrt{|\boldsymbol{\p}|^2 + m^2}, \\
(\widetilde{\phi})_{2}(\boldsymbol{\p}) = (\widetilde{\phi})_{2+}(\boldsymbol{\p}) \overset{\textrm{df}}{=}
\frac{1}{2p_0(\boldsymbol{\p})} u_2(\boldsymbol{\p})^+ \widetilde{\phi}(p_0(\boldsymbol{\p}), \boldsymbol{\p}),
\,\,\,\,\, p_0(\boldsymbol{\p}) = \sqrt{|\boldsymbol{\p}|^2 + m^2},
\end{split}
\]
and
\begin{multline*}
(\widetilde{\phi}')_{3}(\boldsymbol{\p}) = (\widetilde{\phi}')_{1-}(\boldsymbol{\p}) \overset{\textrm{df}}{=}
\overline{\frac{1}{2|p_0(\boldsymbol{\p})|} v_1(\boldsymbol{\p})^+
\widetilde{\phi}'(-|p_0(\boldsymbol{\p})|, -\boldsymbol{\p})} \\ =
\overline{\frac{1}{2|p_0(\boldsymbol{\p})|} v_1(\boldsymbol{\p})^+
\big(\overline{(\widetilde{\phi}')^\flat(|p_0(\boldsymbol{\p})|, \boldsymbol{\p})}\big)^T}, \\
\,\,\,\,\, p_0(\boldsymbol{\p}) = - \sqrt{|\boldsymbol{\p}|^2 + m^2},
\end{multline*}
\begin{multline*}
(\widetilde{\phi}')_{4}(\boldsymbol{\p}) =
(\widetilde{\phi}')_{2-}(\boldsymbol{\p}) \overset{\textrm{df}}{=}
\overline{\frac{1}{2|p_0(\boldsymbol{\p})|} v_2(\boldsymbol{\p})^+
\widetilde{\phi}'(-|p_0(\boldsymbol{\p})|, -\boldsymbol{\p})} \\ =
\overline{\frac{1}{2|p_0(\boldsymbol{\p})|} v_2(\boldsymbol{\p})^+
\big(\overline{(\widetilde{\phi}')^\flat(|p_0(\boldsymbol{\p})|, \boldsymbol{\p})}\big)^T}, \\
\,\,\,\,\, p_0(\boldsymbol{\p}) = - \sqrt{|\boldsymbol{\p}|^2 + m^2}
\end{multline*}
Here $u_s(\boldsymbol{\p}), v_s(-\boldsymbol{\p})$, $s=1,2$, are the Fourier transforms
of the complete system of solutions of the Dirac equation, given by the formula (\ref{chiral,u,v})
of Appendix \ref{fundamental,u,v} in the so-called chiral representation of Dirac gamma
matrices (which we have used in Subsection \ref{e1}); or by the formula (\ref{standard,u,v})
of Appendix \ref{fundamental,u,v} in the so-called standard representation of the Dirac gamma
matrices. It follows that for any $(\widetilde{\phi})_{1}=(\widetilde{\phi})_{1+},
(\widetilde{\phi})_{2} = (\widetilde{\phi})_{2+}, (\widetilde{\phi}')_{3} = (\widetilde{\phi}')_{1-},
(\widetilde{\phi}')_{4} = (\widetilde{\phi}')_{2-} \in L^2(\mathbb{R}^3; \mathbb{C})$ we have
\begin{equation}\label{isomorphismU^-1}
U^{-1} \Big( (\widetilde{\phi})_{1+} \oplus (\widetilde{\phi})_{2+} \oplus (\widetilde{\phi}')_{1-}
\oplus (\widetilde{\phi}')_{2-}\Big) \overset{\textrm{df}}{=}
\widetilde{\phi} \oplus (\widetilde{\phi}')^\flat \,\,\, \in \mathcal{H}_{m,0}^{\oplus} \oplus \mathcal{H}_{-m,0}^{\ominus \, \flat},
\end{equation}
where
\[
\widetilde{\phi}(p_0(\boldsymbol{\p}), \boldsymbol{\p}) \overset{\textrm{df}}{=}
\sum_{s=1,2} 2p_0(\boldsymbol{\p}) \, (\widetilde{\phi})_{s+} (\boldsymbol{\p}) \, u_s(\boldsymbol{\p}),
\,\,\,\,\, p_0(\boldsymbol{\p}) = \sqrt{|\boldsymbol{\p}|^2 + m^2} \\
\]
and
\begin{multline*}
\big((\widetilde{\phi}')^\flat(|p_0(\boldsymbol{\p})|, \boldsymbol{\p})\big)^T =
\overline{\widetilde{\phi}'(-|p_0(\boldsymbol{\p})|, -\boldsymbol{\p})}
\overset{\textrm{df}}{=}
\sum_{s=1,2} 2|p_0(\boldsymbol{\p})| \, (\widetilde{\phi}')_{s-} (\boldsymbol{\p}) \,
\overline{v_s(\boldsymbol{\p})}, \\
p_0(\boldsymbol{\p}) = - \sqrt{|\boldsymbol{\p}|^2 + m^2}.
\end{multline*}
That $U^{-1}$ is indeed equal to the inverse of the operator $U$ follows immediately from the
relations (\ref{E_+Phi=Phi}) for $\widetilde{\phi} \in \mathcal{H}_{m,0}^{\oplus}$
and from the relations (\ref{E_-Phi=Phi}) for $\widetilde{\phi}' \in \mathcal{H}_{-m,0}^{\ominus}$
of Appendix \ref{fundamental,u,v}. That $U^{-1}$ is isometric follows immediately from the orthonormality
relations (\ref{u^+u=delta}) for $u_s(\boldsymbol{\p}), v_s(\boldsymbol{\p})$, $s=1,2$.
That $U$ is isometric follows immediately from the
relations (\ref{E_+Phi=Phi}) for $\widetilde{\phi} \in \mathcal{H}_{m,0}^{\oplus}$
and from the relations (\ref{E_-Phi=Phi}) for $\widetilde{\phi}' \in \mathcal{H}_{-m,0}^{\ominus}$
of Appendix \ref{fundamental,u,v}.
That $U$ transforms isomorphically the indicated nuclear spaces follows from the fact
that the components of $u_s(\boldsymbol{\p}), v_s(\boldsymbol{\p})$, $s=1,2$, are all multipliers
of the Schwartz algebra $\mathcal{S}(\mathbb{R}^3; \mathbb{C})$.

Note here that there are more than just one canonical choice of the solutions
$u_s(\boldsymbol{\p}), v_s(\boldsymbol{-\p})$, $s=1,2$, with smooth components
belonging to the algebra of multipliers or even convolutors of $\mathcal{S}(\mathbb{R}^3; \mathbb{C})$.
Indeed, having given one choice $u_s(\boldsymbol{\p}), v_s(\boldsymbol{-\p})$, $s=1,2$,
we can apply the unitary operator to $u_s(\boldsymbol{\p}), v_s(\boldsymbol{-\p})$, $s=1,2$,
of multiplication by a unitary matrix with components smoothly
depending on $\boldsymbol{\p}$ and belonging to the algebra of multipliers of $\mathcal{S}(\mathbb{R}^3; \mathbb{C})$,
and which rotates the initial $u_s(\boldsymbol{\p}), v_s(\boldsymbol{-\p})$, $s=1,2$,
within the $2$-dimensional images respectively of $P^{\oplus}(p_0(\boldsymbol{\p}), \boldsymbol{\p})$
or $P^{\ominus}(-|p_0(\boldsymbol{\p})|, \boldsymbol{\p})$. We obtain in this way
various isomorphisms $U$ and the corresponding unitary equivalent realizations of the Dirac field. 

Recall, please, that the nuclear Schwartz space $\mathcal{S}(\mathbb{R}^3; \mathbb{C}^4)$
can be obtained as a standard countably Hilbert nuclear space
\[
\mathcal{S}(\mathbb{R}^3; \mathbb{C}^4)
=\mathcal{S}_{A}(\mathbb{R}^3; \mathbb{C}^4) \subset L^2(\mathbb{R}^3, \ud^3 \boldsymbol{\p}; \mathbb{C}^4)
= \oplus_{1}^{4} L^2(\mathbb{R}^3, \ud^3 \boldsymbol{\p}; \mathbb{C})
\]
with the standard operator $A$ on
\[
L^2(\mathbb{R}^3, \ud^3 \boldsymbol{\p}; \mathbb{C}^4)
= \oplus_{1}^{4} L^2(\mathbb{R}^3, \ud^3 \boldsymbol{\p}; \mathbb{C})
\]
equal to the direct sum
\begin{equation}\label{AinL^2(R^3;C^4)}
A = \oplus H_{(3)}
\end{equation}
of four copies of the three-dimensional oscillator Hamiltonian operator
\[
H_{(3)} = - \Delta_{{}_{\boldsymbol{\p}}} + \boldsymbol{\p} \cdot \boldsymbol{\p} + 1
\]
on
\[
L^2(\mathbb{R}^3, \ud^3 \boldsymbol{\p}; \mathbb{C}),
\]
compare e.g. Appendix A.3 of \cite{Hida1}, or \cite{Simon}.

Summing up we will construct the Gelfand triples
\begin{equation}\label{SinglePartGelfandTriplesForPsi}
\left. \begin{array}{ccccc} & & L^2(\mathbb{R}^3 \sqcup \mathbb{R}^3 \sqcup \mathbb{R}^3 \sqcup \mathbb{R}^3, \ud^3 \boldsymbol{\p}; \mathbb{C}) & & \\
& & \parallel & & \\
\mathcal{S}_{A}(\mathbb{R}^3; \mathbb{C}^4) & \subset & \oplus L^2(\mathbb{R}^3; \mathbb{C}) & \subset & \mathcal{S}_{A}(\mathbb{R}^3; \mathbb{C}^4)^* \\
\downarrow \uparrow & & \downarrow \uparrow & & \downarrow \uparrow \\
E & \subset & \mathcal{H'} & \subset & E^* \\
& & \parallel & & \\
& & \mathcal{H}_{m,0}^{\oplus} \oplus \mathcal{H}_{-m,0}^{\ominus \, \flat} & &
\end{array}\right.,
\end{equation}
related by vertical isomorphisms induced by the unitary operator (\ref{isomorphismU})
\[
U: \mathcal{H'} = \mathcal{H}_{m,0}^{\oplus} \oplus \mathcal{H}_{-m,0}^{\ominus \, \flat}
\longrightarrow \oplus L^2(\mathbb{R}^3; \mathbb{C})
\]
with restriction to the nuclear space $E$ mapping isomorphically
\[
E \longrightarrow \mathcal{S}_{A}(\mathbb{R}^3; \mathbb{C}^4) = \mathcal{S}(\mathbb{R}^3; \mathbb{C}^4)
\]
with $A$ defined by (\ref{AinL^2(R^3;C^4)}). The first triple has the standard form,
and can be lifted with the help of $\Gamma(A)$. Thus, we may define in the standard form the
Hida operators $a(w), a(w)^+$ in the Fock space $\Gamma\big(\oplus L^2(\mathbb{R}^3; \mathbb{C})\big)$.
The corresponding Hida operators $a'(u\oplus v),a'(u\oplus v)^+$ in the Fock space
$\Gamma(\mathcal{H}')$ of the free Dirac field need not be separately constructed, and
can be expressed with the help of the standard Hida operators $a(w), a(w)^+$ in the Fock space
$\Gamma\big(\oplus L^2(\mathbb{R}^3; \mathbb{C})\big)$, by utilizing the isomorphism induced by $U$.
Namely, Hida operators $a'(u\oplus v),a'(u\oplus v)^+$ can be expressed by the Hida operators
$a(w), a(w)^+$ as in the formula (\ref{a(U(u+v))=a'(u+v)}), namely:
\[
\begin{split}
a\big(U^{+-1}(u\oplus v)\big) = a'(u\oplus v), \,\,\,
a\big(U^{+-1}(u\oplus v)\big)^+ = a'(u\oplus v)^+, \,\,\,
u\oplus v \in E^*, \\
a\big(U(u\oplus v)\big) = a'(u\oplus v), \,\,\,
a\big(U(u\oplus v)\big)^+ = a'(u\oplus v)^+, \,\,\,
u\oplus v \in E \subset E^*.
\end{split}
\]

The plan of the rest part of this Subsection is the following.
First, we give the white noise construction of the
Hida operators $a(w), a(w)^+$ obtained by lifting to the Fock space of the first (standard)
Gelfand triple in (\ref{SinglePartGelfandTriplesForPsi}). In the next step we utilize
the natural unitary isomorphism $U$ given by (\ref{isomorphismU}),
which induces the isomorphism of the Gelfand triples in (\ref{SinglePartGelfandTriplesForPsi}).
Namely, using the unitary isomorphism $U$ and the Hida operators
$a(w), a(w)^+$ corresponding to the lifting of the first triple in
(\ref{SinglePartGelfandTriplesForPsi}) we compute the
Hida operators $a'(u\oplus v),a'(u\oplus v)^+$ in the Fock space
$\Gamma(\mathcal{H}')$ (which enter into the Dirac field (\ref{psi(f)=a_+(f)+a_-(f^c)^+proper})),
using the formula (\ref{a(U(u+v))=a'(u+v)}).

Let us concentrate now on the first (standard) of the Gelfand triples 
in (\ref{SinglePartGelfandTriplesForPsi}) and its lifting to the Fock space
$\Gamma\big(\oplus L^2(\mathbb{R}^3; \mathbb{C})\big)$, together with the Hida
definition of the Hida operators 
$a(w), a(w)^+$, $w \in \mathcal{S}_{A}(\mathbb{R}^3)^* = \mathcal{S}(\mathbb{R}^3)^*$.
We only recall definition and some basic facts, referring e.g. to \cite{obata-book}, \cite{HKPS},
\cite{obataJFA}, \cite{Shimada}, for more information. 

We are using here the \emph{modified realization of annihilation-creation operators} in the Fock space,
defined in the Remark \ref{TwoRepOfaa^+InFermiFock} of Subsection \ref{electron}.
It fits well with that used by Hida, Obata, Sait\^o, \cite{hida}, \cite{obata-book}, \cite{obataJFA},
for boson case, when adopting the results of \cite{hida}, \cite{obata-book}, \cite{obataJFA}, concerning integral kernel
operators, to fermion case.

\begin{rem}\label{TretmentFermiIntegKerOp=TretmentBoseIntegKerOp}
It should be emphasized here that the results of
\cite{hida}, \cite{obata-book}, \cite{obataJFA}, concerning the so-called integral kernel
operators and their Fock expansions, can be proved without any essential changes
also for the fermi case
after \cite{hida}, \cite{obata-book}, \cite{obataJFA}. Note that these
theorems (e.g. Lemma 2.2, Thm. 2.2, Thm. 2.6. of \cite{hida}, or Thm. 3.13 of \cite{obataJFA})
could have been formulated and proved as well for the so-called general Fock space
\[
\Gamma_\textrm{general}(\mathcal{H}) = \bigoplus_{n=0}^{\infty} \mathcal{H}^{\otimes n}
\]
without summarizing or antisymmetrizing the tensor products. In particular symmertization
(antisymmetrization) plays no fundamental role in the proof of these theorems, which are based on
the norm estimations of the $m$-contractions $\otimes_m, \otimes^m$. Their eventual symmetrizations
$\widehat{\otimes}_m, \widehat{\otimes}^m$
(or antisymmetrizations), which arise in the latter stage when restricting attention to the boson (or fermion)
case, has nothing to do with these estimations and allows to state the analogous results for
boson as well as for the fermion case.

Although differences between the Fermi and Bose case which arise have nothing to do with the analysis of
integral kernel operators (in which we are mostly interested),
we should mention here some of them. The fundamental difference is that the algebra structure of the
nuclear Hida test space, determined by the tensor product, is not commutative but skew commutative, due to the atisymmetricity of the tensors in the Fermi Fock space, and cannot be naturally realized as a nuclear function
space on the strong dual $E^*$ with multiplication defined by pointwise multiplication
(because such multiplication is always commutative). In connection with this we have no natural isomorphism
of the Fermi Fock space to the space of square integrable functions on $E^*$ with the Gaussian measure
on $E^*$ (no Wiener-It\^o-Segal decomposition based on commutative infinite-dimensional measure space is possible).
Of course a mere existence of a unitary map between the Fermi Fock space and an $L^2$ space over a Gaussian
measure space is trivial, but there are plenty of such maps devoid of any relevance. \emph{Naturality} of the
Wiener-It\^o-Segal decomposition for the Bose case is crucial. In order to keep a \emph{natural nature}, e.g. preserving
the algebra structure of the Hida test space (now skew commutative), in extending Wiener-It\^o-Segal decomposition
to the Fermi case, a non-commutative extension
of abstract integration is needed, and has been provided by Segal (note however that Segal
\cite{Segal-TnsorAlg-II} is not using
a non-commutative extension of ordinary measure -- but of a weak distribution on a Hilbert space).
Because these questions concerning non-commutative character of the multiplicative structure of the
Hida test space in case of fermi
case are not immediately related to the calculus of Fock expansions of integral kernel operators,
developed in \cite{hida}, \cite{obata-book}, \cite{obataJFA}, we do not enter these questions
in our work. In particular we do not exploit in any substantial manner the fact that Hida annihilation operators can be interpreted as graded derivations on the $\mathbb{Z}_2$ graded skew commutative nuclear algebra of Hida test functionals.
The only practical consequence of this fact we feel in computations concerning integral kernel operators is that we
confine ourselves to skew-symmetric kernels (in variables corresponding to Fermi
Hida creation-annihilation operators) in order to keep one-to-one correspondence between the
kernels and corresponding operators.

But there is a relevant tool for computations which must be treated in slightly different manner in the two cases
-- Bose and Fermi case. Namely, the \emph{symbol calculus}, initiated by Berezin \cite{Berezin}, \cite{BerezinMS},
\cite{BerezinCMP} and developed mainly
by Obata \cite{obata}, \cite{obataJFA}, must be realized in a slightly different manner for Fermi case in
comparison with the Bose case. In order to adopt the symbol calculus of Obata
to the Fermi case it is convenient first to divide the Fermi Fock space $\Gamma(\mathcal{H}')$
into the sub spaces $\Gamma_+(\mathcal{H}')$ of even elements
\[
\Phi = \sum \limits_{n=0}^{\infty} \Phi_n,
\]
(with even $n$ in this decomposition), and $\Gamma_-(\mathcal{H}')$ of odd
elements $\Phi$ (with $n$ odd in this decomposition).
Similarly, we do for the nuclear spaces
$(E) = (E)_+ \oplus (E)_-, (E)^* = (E)_{+}^{*} \oplus (E)_{-}^{*}$. Next we note that
for $\zeta \in E^{\widehat{\otimes} \, 2}$ (and generally $\zeta \in E^{\widehat{\otimes} m}$
with even $m$) the exponential maps of the type
\begin{equation}\label{exp+}
\zeta \mapsto \Phi_{\zeta}^{+} = \sum \limits_{n=0}^\infty \frac{1}{(2n)!} \zeta^{\widehat{\otimes} \, n}
\in (E)_+,
\end{equation}
are well-defined and continuous. Using exponential map we construct the so called $S$-transform,
which allows to realize the even-grade subspace of the Fermi Fock space as a reproducing kernel Hilbert space, and then we utilize the Obata symbol
for even operators, i.e. transforming $(E)_+ \rightarrow (E)_{+}^{*}$
and $(E)_- \rightarrow (E)_{-}^{*}$. The odd operators, i.e. transforming
$(E)_+ \rightarrow (E)_{-}^{*}$ and $(E)_- \rightarrow (E)_{+}^{*}$ are reduced to even by
multiplication by one Hida (creation, respectively annihilation) operator.
Finally, we note that any continuous operator $(E) \rightarrow (E)^*$ is naturally a
direct sum of an even and an odd operator; compare \cite{Berezin}, \cite{BerezinMS},
\cite{BerezinCMP}, \cite{obata-book}, \cite{Shimada}. However, in general not all elements of the images of the exponential maps of the type (\ref{exp+})
in the even-grade part of the Fermi Fock space deserve the name
of \emph{coherent} states.
In Fermi case the Berezin symbols of the simplest products of two Fermi-Hida operators, $\partial_{{}_{\boldsymbol{\p}}} \partial_{{}_{\boldsymbol{\q}}}$,
$\partial_{{}_{\boldsymbol{\q}}}^{*} \partial_{{}_{\boldsymbol{\p}}}^{*}$, $\partial_{{}_{\boldsymbol{\p}}}^{*}\partial_{{}_{\boldsymbol{\q}}}$,
are in general equal to the values $\xi(\boldsymbol{\p},\boldsymbol{\q}), \overline{\xi(\boldsymbol{\p},\boldsymbol{\q})},
\eta(\boldsymbol{\p},\boldsymbol{\q})$ at $(\boldsymbol{\p},\boldsymbol{\q})$ of functions
$\xi \in E^{\widehat{\otimes} \, 2}$, $\eta= \overline{\eta} \in E^{\otimes \, 2}$, associated by non-local
integral-type formulas to the functions $\zeta, \overline{\zeta} \in E^{\widehat{\otimes} \, 2}$
in the exponential states $\Phi_{\overline{\zeta}}^{+}, \Phi_{\zeta}^{+}$.
Moreover, the Berezin symbols of the Wick products of the generating even operators $\partial_{{}_{\boldsymbol{\p}}} \partial_{{}_{\boldsymbol{\q}}}$,
$\partial_{{}_{\boldsymbol{\q}}}^{*}\partial_{{}_{\boldsymbol{\p}}}^{*}$, $\partial_{{}_{\boldsymbol{\p}}}^{*}\partial_{{}_{\boldsymbol{\q}}}$
are not in general equal to the products of Berezin symbols of the generating even operators,
$\partial_{{}_{\boldsymbol{\p}}} \partial_{{}_{\boldsymbol{\q}}}$,
$\partial_{{}_{\boldsymbol{\q}}}^{*}\partial_{{}_{\boldsymbol{\p}}}^{*}$, $\partial_{{}_{\boldsymbol{\p}}}^{*}\partial_{{}_{\boldsymbol{\q}}}$.
In particular the multiplicative law is not preserved by symbols determined by $\zeta, \overline{\zeta}$ equal to finite sums
\[
\zeta = \bigoplus\limits_{i=1}^{N} \xi_i \widehat{\otimes} \eta_i, \,\,\,\,\,\,\,\, \xi_i, \eta_i \in E,
\]
of simple antisymmetric tensors $\xi_i \widehat{\otimes} \eta_i$
in the exponential states $\Phi_{\overline{\zeta}}^{+}, \Phi_{\zeta}^{+}$.
In particular, the symbols of even integral kernel
operators in Fermi case are not equal to the evaluations of the kernels at the tensor products $\xi^{\otimes \, \ell} \otimes \overline{\xi}^{\otimes \, m}
\otimes \eta^{\otimes \, n}$ of $\xi, \overline{\xi}, \eta$ associated
to the symbols of the generating even operators.
Such symbol calculus of even generalized operators, associated to an even exponential, requires additional investigation of the symbols
of simple integral kernel operators.

The need for such symbols would arise in situations in which we had a Fock expansion of a generalized operator into integral kernel operators. It would be hard to apply convergence criteria of the symbol calculus without the symbols of the particular contributions
easily manageable in computations at our disposal.

Suppose the additional condition is fulfilled: the subspace of those $\zeta, \eta \in E^{\widehat{\otimes} \, 2}$ for which the Berezin symbol
\[
\widehat{\Xi_+}(\overline{\eta}; \zeta)
= \frac{\langle\langle \Phi_{\overline{\eta}}^{+}, \Xi_+ \Phi_{\zeta}^{+} \rangle\rangle}{\langle\langle \Phi_{\overline{\eta}}^{+}, \Phi_{\zeta}^{+}\rangle\rangle},
\]
of even operators $\Xi_+$, determined by the exponential map, behaves multiplicatively under the Wick product
of even integral kernel operators, is sufficiently reach, say with dense image $\Phi_\zeta \in (E)_+$.
In this case the Fermi symbol calculus would be very effective, and the symbol of Wick products
of even integral kernel operators would be easily computable and equal to evaluations
of the corresponding kernels at the tensor products of basis functions $\xi, \overline{\xi}, \eta$
determined by the symbols of the generating even operators $\partial_{{}_{\boldsymbol{\p}}} \partial_{{}_{\boldsymbol{\q}}}$,
$\partial_{{}_{\boldsymbol{\q}}}^{*} \partial_{{}_{\boldsymbol{\p}}}^{*}$, $\partial_{{}_{\boldsymbol{\p}}}^{*}\partial_{{}_{\boldsymbol{\q}}}$.
Additional work is required to find such $\zeta$ which respects this ''natural'' multiplicativity requirement whose images,
say \emph{coherent} states, under the even exponential map $\zeta \mapsto \Phi_\zeta$,
compose a dense set in $(E)_+$. The very existence of such $\zeta$ is an open problem
(we only know that such $\zeta$ do not exist in finite dimensional Fock spaces,
and it is hardly possible their existence for the infinite dimensional case). In case there are no such $\zeta$ with multiplicative symbols
in general, a preparatory investigation of symbols of integral kernel operators is needed before passing to application
of the symbol calculus to the convergence of Fock expansions. Because the operators
\[
a(\xi)
= \int \xi(\boldsymbol{\p}) \, \partial_{{}_{\boldsymbol{\p}}} \, \ud^3 \boldsymbol{\p},
\,\,\,\,\,\,
a^{+}(\xi)
= \int \xi(\boldsymbol{\p}) \, \partial_{{}_{\boldsymbol{\p}}}^{*} \, \ud^3 \boldsymbol{\p},
\,\,\,\, \xi \in E
\]
are bounded in Fermi case, with the bound equal to the $L^2$ norm of $\xi$, the symbols of the integral kernel operators still can be effectively investigated
even without any \emph{coherent} states at hand.
Because we do not use the symbol calculus for Fermi fields in this work,
we do not enter into the investigation of the determination of the symbols of integral kernel operators. In the investigation of Fock expansions we
are using the said boundedness of $a(\xi), a^{+}(\xi)$, and consider the investigated integral kernel operators
on the Fock states being finite direct sums of simple tensors (which are of course dense in the Fock space). No symbol
calculus is needed in Fermi case. Existence or nonexistence of \emph{coherent} states in Fermi case (in the above sense) is more interesting
rather from the physical point of view, without any relevance for the investigation of convergence of Fock expansions.
\end{rem}

Let $|\cdot |_0$, $(\cdot, \cdot)_0$ denote the standard $L^2$ norm and inner product on
\[
L^2(\mathbb{R}^3, \ud^3 \boldsymbol{\p}; \mathbb{C}^4)
= \oplus_{1}^{4} L^2(\mathbb{R}^3, \ud^3 \boldsymbol{\p}; \mathbb{C})
\]
and by the same symbol $|\cdot|_0$, after \cite{hida} and \cite{obata-book},
we denote the Hilbert space norm on the Hilbert space tensor
product
\[
L^2(\mathbb{R}^3, \ud^3 \boldsymbol{\p}; \mathbb{C}^4)^{\otimes n},
\]
as well as its restriction to the antisymmetrized tensor product
\[
L^2(\mathbb{R}^3, \ud^3 \boldsymbol{\p}; \mathbb{C}^4)^{\widehat{\otimes} \, n}.
\]
Recall that
\[
|f|_k = |(A^{\otimes n})^k f|_0 \,\,\,\,\,
f \in \Dom \, (A^{\otimes n})^k \subset L^2(\mathbb{R}^3, \ud^3 \boldsymbol{\p}; \mathbb{C}^4)^{\otimes n}
\]
(in particular well-defined for $f \in \mathcal{S}_{A}(\mathbb{R}^3; \mathbb{C}^4)^{\widehat{\otimes} \, n}$).

Let $\|\cdot\|_0$, $((\cdot, \cdot))_0$ denote the Hilbert space norm and the corresponding inner product
on Fock space defined by the formula (convention used by \cite{hida}, \cite{obataJFA}, compare Remark \ref{TwoRepOfaa^+InFermiFock} of Subsection \ref{electron})
\[
\| \Phi\|_{0}^2 = \sum \limits_{n=0}^{\infty} \, n! \, |\Phi_n |_{0}^2
\]
for $\Phi$ with decomposition
\[
\Phi = \sum \limits_{n=0}^{\infty} \Phi_n, \,\,\, \textrm{with} \,\,\,
\Phi_n \in L^2(\mathbb{R}^3, \ud^3 \boldsymbol{\p}; \mathbb{C}^4)^{\widehat{\otimes} \, n}.
\]
Recall that by definition
\[
\|\Phi\|_k = \|\Gamma(A)^k \Phi \|_0 \,\,\, \textrm{and} \,\,\,
|\Phi_n|_k = |(A^{\otimes n})^k \Phi_n|_0
\]
for $\Phi \in \Gamma\big(L^2(\mathbb{R}^3, \ud^3 \boldsymbol{\p}; \mathbb{C}^4) \big)$
and $\Phi_n \in L^2(\mathbb{R}^3, \ud^3 \boldsymbol{\p}; \mathbb{C}^4)^{\widehat{\otimes} \, n}$.

It follows in particular that the general element
\begin{equation}\label{HidaPhi}
\Phi = \sum \limits_{n=0}^{\infty} \Phi_n, \,\,\, \textrm{with} \,\,\,
\| \Phi\|_{0}^2 = \sum \limits_{n=0}^{\infty} \, n! \, |\Phi_n |_{0}^2  < \infty,
\end{equation}
of the Fock space 
$\Gamma\big(L^2(\mathbb{R}^3, \ud^3 \boldsymbol{\p}; \mathbb{C}^4) \big)$
belongs to the Hida test space $\big(\mathcal{S}_{A}(\mathbb{R}^3; \mathbb{C}^4)\big)
\subset \Gamma\big(L^2(\mathbb{R}^3, \ud^3 \boldsymbol{\p}; \mathbb{C}^4) \big)$
iff $\Phi_n \in \mathcal{S}_{A}(\mathbb{R}^3; \mathbb{C}^4)^{\widehat{\otimes} \, n}$ 
for all $n= 0, 1,2, \ldots$ and
\[
\sum \limits_{n=0}^{\infty} \, n! \, |\Phi_n|_k < \infty \,\,\,
\textrm{for all} \,\,\, k\geq 0.
\] 
In this case 
\begin{equation}\label{||Phi||_kPhiInHida}
\|\Phi\|_{k}^2 = \sum \limits_{n=0}^{\infty} \, n! \, |\Phi_n|_k < \infty \,\,\,
\textrm{for all} \,\,\, k\geq 0.
\end{equation}

Note that the norms
\[
\|\Phi\|_{k} = \|\Gamma(A)^k \Phi \|_0 \,\,\, \textrm{with} \,\,\,
\Phi \in \big(\mathcal{S}_{A}(\mathbb{R}^3; \mathbb{C}^4)\big)
\]
are well-defined on the Hida space $\big(\mathcal{S}_{A}(\mathbb{R}^3; \mathbb{C}^4)\big)
\subset \Gamma\big(L^2(\mathbb{R}^3, \ud^3 \boldsymbol{\p}; \mathbb{C}^4) \big)$
also for $k$ equal to any negative integer. Completion of $\big(\mathcal{S}_{A}(\mathbb{R}^3; \mathbb{C}^4)\big)$
with respect to the Hilbertian norm
\[
\|\cdot \|_{-k} = \|\Gamma(A)^{-k} \cdot \|_0 \,\,\, \textrm{with fixed} \,\,\,
k \in \mathbb{N}
\]
is equal to a Hilbert space, which we denote
\begin{equation}\label{Hida-k}
\Big(\mathcal{S}_{A}(\mathbb{R}^3; \mathbb{C}^4)\Big)_{-k},
\end{equation}
and which is also equal to the completion of $\Dom \, \Gamma(A)^{-k}$ (equal to the whole Fock space
$\Dom \, \Gamma(A)^{-k} = \Gamma\big(L^2(\mathbb{R}^3, \ud^3 \boldsymbol{\p}; \mathbb{C}^4) \big)$ for
$k=0,1,2, \ldots $) with respect to the norm $\|\cdot \|_{-k}$.
The Hilbert space (\ref{Hida-k}) is for each $k\geq0$ canonically isomorphic, including the case $k=0$,
(Riesz isomorphism) to the Hilbert space dual
of the Hilbert space
\begin{equation}\label{Hida+k}
\Big(\mathcal{S}_{A}(\mathbb{R}^3; \mathbb{C}^4)\Big)_{k},
\end{equation}
compare \cite{obata-book}. Recall that the Hilbert space (\ref{Hida+k}) is equal to the completion of the domain
$\Dom \, \Gamma(A)^{k}$ with respect to the
norm $\| \cdot\|_k$. The Hilbert spaces
\[
\Big(\mathcal{S}_{A}(\mathbb{R}^3; \mathbb{C}^4)\Big)_{-k}, \,\,\, k = 0, 1, 2, \ldots
\]
compose an inductive system, \cite{GelfandIV}, \cite{obata-book}, with natural continuous inclusions
\begingroup\makeatletter\def\f@size{5}\check@mathfonts
\def\maketag@@@#1{\hbox{\m@th\large\normalfont#1}}%
\begin{equation}\label{IndSystemHida}
\left. \begin{array}{cccccccc} \Big(\mathcal{S}_{A}(\mathbb{R}^3; \mathbb{C}^4)\Big)_{-0}&
\subset & \Big(\mathcal{S}_{A}(\mathbb{R}^3; \mathbb{C}^4)\Big)_{-1} & \subset &
\Big(\mathcal{S}_{A}(\mathbb{R}^3; \mathbb{C}^4)\Big)_{-2} & \subset \ldots \subset &
\Big(\mathcal{S}_{A}(\mathbb{R}^3; \mathbb{C}^4)\Big)^*& \\
\parallel &&&&&&& \\
\overline{\Gamma\big(L^2(\mathbb{R}^3, \ud^3 \boldsymbol{\p}; \mathbb{C}^4) \big)}&&&&&&& \\
\parallel &&&&&&& \\
\Gamma\big(L^2(\mathbb{R}^3, \ud^3 \boldsymbol{\p}; \mathbb{C}^4) \big)^*&&&&&&& \end{array}\right..
\end{equation}
\endgroup
which is dual to the projective system
\begingroup\makeatletter\def\f@size{5}\check@mathfonts
\def\maketag@@@#1{\hbox{\m@th\large\normalfont#1}}%
\begin{equation}\label{ProjSystemHida}
\left. \begin{array}{cccccccc} \big(\mathcal{S}_{A}(\mathbb{R}^3; \mathbb{C}^4)\big)&
\subset \ldots & \ldots \subset \Big(\mathcal{S}_{A}(\mathbb{R}^3; \mathbb{C}^4)\Big)_{2} & \subset &
\Big(\mathcal{S}_{A}(\mathbb{R}^3; \mathbb{C}^4)\Big)_{1}& \subset &
& \Big(\mathcal{S}_{A}(\mathbb{R}^3; \mathbb{C}^4)\Big)_0 \\
&&&&&&& \parallel \\
&&&&&&& \Gamma\big(L^2(\mathbb{R}^3, \ud^3 \boldsymbol{\p}; \mathbb{C}^4) \big) \end{array}\right..
\end{equation}
\endgroup
defining the Hida space $\big(\mathcal{S}_{A}(\mathbb{R}^3; \mathbb{C}^4)\big)$.
The two systems (\ref{ProjSystemHida}) and (\ref{IndSystemHida}) can be joined into single system of Hilbert spaces
with comparable and compatible norms, by using the natural isomorphism of the dual to the adjoint space
\[
\Gamma\big(L^2(\mathbb{R}^3, \ud^3 \boldsymbol{\p}; \mathbb{C}^4) \big)^* \cong
\overline{\Gamma\big(L^2(\mathbb{R}^3, \ud^3 \boldsymbol{\p}; \mathbb{C}^4) \big)} =
\Big(\mathcal{S}_{A}(\mathbb{R}^3; \mathbb{C}^4)\Big)_{-0}
\]
to the Hilbert space
\[
\Gamma\big(L^2(\mathbb{R}^3, \ud^3 \boldsymbol{\p}; \mathbb{C}^4) \big)
= \Big(\mathcal{S}_{A}(\mathbb{R}^3; \mathbb{C}^4)\Big)_{0}
\]
(Riesz isomorphism, compare \cite{GelfandIV}, \cite{obata-book}),
and noting that the elements of the Hilbert space $H$ and its adjoint space $\overline{H}$ are the same:
\begingroup\makeatletter\def\f@size{5}\check@mathfonts
\def\maketag@@@#1{\hbox{\m@th\large\normalfont#1}}%
\begin{multline*}
\left. \begin{array}{ccccccccc} \big(\mathcal{S}_{A}(\mathbb{R}^3; \mathbb{C}^4)\big)&
\subset \ldots & \ldots \subset \Big(\mathcal{S}_{A}(\mathbb{R}^3; \mathbb{C}^4)\Big)_{2} & \subset &
\Big(\mathcal{S}_{A}(\mathbb{R}^3; \mathbb{C}^4)\Big)_{1}& \subset &
& \Big(\mathcal{S}_{A}(\mathbb{R}^3; \mathbb{C}^4)\Big)_0 & = \\
&&&&&&& \parallel& \\
&&&&&&& \Gamma\big(L^2(\mathbb{R}^3, \ud^3 \boldsymbol{\p}; \mathbb{C}^4) \big)& \end{array}\right.
\,\,\,
\\
\,\,\,
\left. \begin{array}{ccccccccc} = &\Big(\mathcal{S}_{A}(\mathbb{R}^3; \mathbb{C}^4)\Big)_{-0}&
\subset & \Big(\mathcal{S}_{A}(\mathbb{R}^3; \mathbb{C}^4)\Big)_{-1} & \subset &
\Big(\mathcal{S}_{A}(\mathbb{R}^3; \mathbb{C}^4)\Big)_{-k} & \subset \ldots \subset &
\Big(\mathcal{S}_{A}(\mathbb{R}^3; \mathbb{C}^4)\Big)^*& \\
& \parallel &&&&&&& \\
&\overline{\Gamma\big(L^2(\mathbb{R}^3, \ud^3 \boldsymbol{\p}; \mathbb{C}^4) \big)}&&&&&&& \end{array}\right.
\end{multline*}
\endgroup

The strong dual $\big(\mathcal{S}_{A}(\mathbb{R}^3; \mathbb{C}^4)\big)^*$ 
of the Hida space $\big(\mathcal{S}_{A}(\mathbb{R}^3; \mathbb{C}^4)\big)$ is equal
to the inductive limit of the system (\ref{IndSystemHida}). 
Recall that the Hida space $\big(\mathcal{S}_{A}(\mathbb{R}^3; \mathbb{C}^4)\big)$
itself is equal to the projective limit 
of the system (\ref{ProjSystemHida}), compare \cite{obata-book}.

Similarly, as for the elements of Hida (or Fock) space, likewise 
each element $\Phi \in \big(\mathcal{S}_{A}(\mathbb{R}^3; \mathbb{C}^4)\big)^*$ 
of the strong dual to the Hida space has a unique decomposition
\begin{equation}\label{Hida*Phi}
\Phi = \sum \limits_{n=0}^{\infty} \Phi_n, \,\,\, \textrm{with} \,\,\,
\Phi_n \in \big(\mathcal{S}_{A}(\mathbb{R}^3; \mathbb{C}^4)^{\widehat{\otimes} \, n} \big)^*.
\end{equation}
In this case there exists a natural $k$ such that 
\[
\|\Phi\|_{-k}^2 = \sum \limits_{n=0}^{\infty} \, n! \, |\Phi_n|_{-k}^{2} < \infty.
\]

Note that we have natural real and complex structure on the spaces we encounter here with well-defined complex conjugation
$\overline{(\cdot)}$. In particular, if we denote the dual pairings 
on $\mathcal{S}_{A}(\mathbb{R}^3; \mathbb{C}^4)^* \times \mathcal{S}_{A}(\mathbb{R}^3; \mathbb{C}^4)$
and on 
$\big(\mathcal{S}_{A}(\mathbb{R}^3; \mathbb{C}^4)\big)^* \times \big(\mathcal{S}_{A}(\mathbb{R}^3; \mathbb{C}^4)\big)$
by $\langle \cdot, \cdot \rangle$ and respectively by $\langle \langle \cdot, \cdot \rangle \rangle$
then we have
\[
\begin{split}
\langle \xi, \eta \rangle = (\overline{\xi}, \eta)_0, \,\,\,
\textrm{for} \,\,\,
\xi \in \mathcal{S}_{A}(\mathbb{R}^3; \mathbb{C}^4) \subset \mathcal{S}_{A}(\mathbb{R}^3; \mathbb{C}^4)^*,
\eta \in \mathcal{S}_{A}(\mathbb{R}^3; \mathbb{C}^4), \\
\langle\langle \Psi, \Phi \rangle \rangle = (( \, \overline{\Psi} \, , \Phi \, ))_0, \,\,\,
\textrm{for} \,\,\,
\Psi \in \big(\mathcal{S}_{A}(\mathbb{R}^3; \mathbb{C}^4)\big) 
\subset \big(\mathcal{S}_{A}(\mathbb{R}^3; \mathbb{C}^4)\big)^*,
\Phi \in \big(\mathcal{S}_{A}(\mathbb{R}^3; \mathbb{C}^4)\big).
\end{split}
\]

Now we are ready to define the Hida operators $a(w), a(w)^+$, $w \in \mathcal{S}_{A}(\mathbb{R}^3; \mathbb{C}^4)^*$
in the Fock space $\Gamma\big(L^2(\mathbb{R}^3, \ud^3 \boldsymbol{\p}; \mathbb{C}^4) \big)$
corresponding to the first (standard) Gelfand triple in (\ref{SinglePartGelfandTriplesForPsi}).

Namely, for each $w \in \mathcal{S}_{A}(\mathbb{R}^3; \mathbb{C}^4)^*$,
and each general element (\ref{HidaPhi}) of the Hida space 
we define Hida annihilation operator $a(w)$ which by definition acts on the element
$\Phi$ given by (\ref{HidaPhi}) according to the following formula
\[
\begin{split}
1) \,\,\,\,\, a(w) \big(\Phi = \Phi_0 \big) = 0, \\
2) \,\,\,\,\, a(w) \Phi =  \sum \limits_{n \geq 0} \, n \, \overline{w} \, \widehat{\otimes}_1 \, \Phi_n.   
\end{split}
\]

Now we define the Hida creation operator $a(w)^+$, $w \in \mathcal{S}_{A}(\mathbb{R}^3; \mathbb{C}^4)^*$,
transforming the strong dual $\big(\mathcal{S}_{A}(\mathbb{R}^3; \mathbb{C}^4)\big)^*$ of the Hida space
into itself. Namely, let $w \in \mathcal{S}_{A}(\mathbb{R}^3; \mathbb{C}^4)^*$ and let $\Phi$
be any general element (\ref{Hida*Phi}) of the strong dual 
$\big(\mathcal{S}_{A}(\mathbb{R}^3; \mathbb{C}^4)\big)^*$.
The action of the Hida creation
operator $a(w)^+$, $w \in \mathcal{S}_{A}(\mathbb{R}^3; \mathbb{C}^4)^*$, on such $\Phi$
is by definition equal 
\[
\begin{split}
a(w)^+ \Phi =  \sum \limits_{n \geq 0}  \, w \, \widehat{\otimes} \, \Phi_n.  
\end{split}
\]
Here as well as in the definition of the Hida annihilation operator
the tensor product $\otimes$ and its $1$-contraction $\otimes_1$
(antisymmetrized  $\widehat{\otimes}$, $\widehat{\otimes}_1$) is equal to the 
projective tensor product over the respective nuclear spaces:
\begin{multline*}
\mathcal{S}_{A}(\mathbb{R}^3; \mathbb{C}^4)^*, \mathcal{S}_{A}(\mathbb{R}^3; \mathbb{C}^4)^{\otimes n},
\mathcal{S}_{A}(\mathbb{R}^3; \mathbb{C}^4)^{\widehat{\otimes} \, n}, \\
\big(\mathcal{S}_{A}(\mathbb{R}^3; \mathbb{C}^4)^{\otimes n}\big)^*,
\big(\mathcal{S}_{A}(\mathbb{R}^3; \mathbb{C}^4)^{\widehat{\otimes} \, n} \big)^*, 
\end{multline*}
In this case (of nuclear spaces) tensor product is essentially unique with the projective tensor product coinciding
with the equicontinuous tensor product. 
Recall that 
\[
v_{{}_{1}} \, \widehat{\otimes} \, \cdots \widehat{\otimes} \, v_{{}_{n}} =
(n!)^{-1} \sum \limits_{\pi} \textrm{sign} \, (\pi) \, v_{{}_{\pi(1)}} \otimes
\cdots \otimes v_{{}_{\pi(n)}},
\]
with $v_{{}_{i}}$ in the respective space, and that the antisymmetrized $1$-contraction $\widehat{\otimes}_1$
is uniquely determined by the formula
\begin{multline*}
u \, \widehat{\otimes}_1 v_{{}_{1}} \, \widehat{\otimes} \, \cdots \widehat{\otimes} \, v_{{}_{n}}
= (n!)^{-1} \sum \limits_{\pi} \textrm{sign} \, (\pi) \, \langle u,v_{{}_{\pi(1)}} \rangle \, 
 v_{{}_{\pi(2)}} \otimes
\cdots \otimes v_{{}_{\pi(n)}}, \\
u \in \mathcal{S}_{A}(\mathbb{R}^3; \mathbb{C}^4)^{*}, v_{{}_{i}} \in \mathcal{S}_{A}(\mathbb{R}^3; \mathbb{C}^4),
\end{multline*}
with the sums ranging over all permutations $\pi$ of the natural numbers $1, \ldots, n$, and with
the evaluation $\langle u,v_{{}_{\pi(1)}} \rangle$ of $u$ 
on $v_{{}_{\pi(1)}}$, which restricts to
\[
\langle u,v_{{}_{\pi(1)}} \rangle = (\overline{u},v_{{}_{\pi(n)}})_0 \,\,\,
\textrm{whenever} \,\,\,
u \in \mathcal{S}_{A}(\mathbb{R}^3; \mathbb{C}^4) \subset \mathcal{S}_{A}(\mathbb{R}^3; \mathbb{C}^4)^{*}.
\] 

It follows that $a(w)$, $w \in \mathcal{S}_{A}(\mathbb{R}^3; \mathbb{C}^4)^*$,
transforms continuously the Hida space into the Hida space
\[
a(w): \, \big(\mathcal{S}_{A}(\mathbb{R}^3; \mathbb{C}^4)\big) \, \longrightarrow \,
\big(\mathcal{S}_{A}(\mathbb{R}^3; \mathbb{C}^4)\big),
\]
for a proof compare e.g. \cite{obata-book}, \cite{Shimada}. By composing it with the natural continuous
inclusion $\big(\mathcal{S}_{A}(\mathbb{R}^3; \mathbb{C}^4)\big) 
\subset \big(\mathcal{S}_{A}(\mathbb{R}^3; \mathbb{C}^4)\big)^*$, we can also regard the Hida
annihilation operator $a(w)$, $w \in \mathcal{S}_{A}(\mathbb{R}^3; \mathbb{C}^4)^*$, as  a continuous operator
\[
a(w): \, \big(\mathcal{S}_{A}(\mathbb{R}^3; \mathbb{C}^4)\big) \, \longrightarrow \,
\big(\mathcal{S}_{A}(\mathbb{R}^3; \mathbb{C}^4)\big)^*.
\]

It follows by general property of transposition, \cite{treves}, that 
$a(w)^*$, $w \in \mathcal{S}_{A}(\mathbb{R}^3; \mathbb{C}^4)^*$, 
maps continuously the strong dual of the Hida space into itself
\[
a(w)^*: \, \big(\mathcal{S}_{A}(\mathbb{R}^3; \mathbb{C}^4)\big)^* \, \longrightarrow \,
\big(\mathcal{S}_{A}(\mathbb{R}^3; \mathbb{C}^4)\big)^*.
\] 
By composing it with the dual 
\[
\big(\mathcal{S}_{A}(\mathbb{R}^3; \mathbb{C}^4)\big) \cong \big(\mathcal{S}_{A}(\mathbb{R}^3; \mathbb{C}^4)\big)^{**} 
\subset \big(\mathcal{S}_{A}(\mathbb{R}^3; \mathbb{C}^4)\big)^*
\]
of the natural inclusion $\big(\mathcal{S}_{A}(\mathbb{R}^3; \mathbb{C}^4)\big) 
\subset \big(\mathcal{S}_{A}(\mathbb{R}^3; \mathbb{C}^4)\big)^*$, we can regard the Hida 
creation operator $a(w)^*$, $w \in \mathcal{S}_{A}(\mathbb{R}^3; \mathbb{C}^4)^*$,
as a continuous operator 
\[
a(w)^*: \, \big(\mathcal{S}_{A}(\mathbb{R}^3; \mathbb{C}^4)\big) \, \longrightarrow \,
\big(\mathcal{S}_{A}(\mathbb{R}^3; \mathbb{C}^4)\big)^*.
\]
It turns out that 
\[
a(w)^+  = \overline{(\cdot)} \circ a(w)^* \circ \overline{(\cdot)}, \,\,\,
w \in \mathcal{S}_{A}(\mathbb{R}^3; \mathbb{C}^4)^*,
\] 
for $a(w)^*, a(w)^+$
understood as maps of the strong dual of the Hida space into itself 
(or respectively as maps transforming the 
Hida space into its strong dual); compare \cite{obata-book}, \cite{Shimada}.  

\begin{rem*}
Note that in fact the definition of the Hida operator used by mathematicians
is slightly different in comparison to ours with the additional complex conjugation
\[
\textrm{mathematicians's} \,\,\, a(w) = \,\,\, \textrm{ours} \,\,\,a(\overline{w}).
\]
In particular ours $a(w)$ is anti-linear in $w$, which is the convention accepted in physical
literature. This is the conjugation $A^+ = \overline{(\cdot)} \circ A^{^*} \circ \overline{(\cdot)}$
equal to the linear transpose composed with complex conjugations, which connects the Hida generalized
annihilation $a(w)$ and creation operators $a(w)^+$, due to the convention which we have accepted,
and which is used by physicists. In the convention accepted by mathematicians it is the ordinary linear
transpose which connects the generalized Hida annihilation
$a(w)$ and creation operators $a(w)^*$.
\end{rem*}

In the mathematical literature the fact that the Hida annihilation operator $a(w)$ is a ($\mathbb{Z}_2$-graded
in Fermi case) derivation on the Hida nuclear algebra (with the multiplication defined by the antisymmetrized tensor
product $\widehat{\otimes}$) is reflected by the following notation introduced by Hida:
\[
D_{w} \overset{\textrm{df}}{=} a(w), w \in \mathcal{S}_{A}(\mathbb{R}^3; \mathbb{C}^4)^* = \mathcal{S}(\mathbb{R}^3; \mathbb{C}^4)^*.
\]
(here the convention used by mathematicians is better because their
\[
D_{w} \overset{\textrm{df}}{=} a(w)
\]
is linear in $w$, and in bose case when the Hida space is realized as commutative algebra of functions on
$\mathcal{S}_{A}(\mathbb{R}^3; \mathbb{C}^4)^*$, the Hida annihilation operator $a(w)$ is indeed
equal to the G{\aa}teaux derivation in the direction of $w$ and not in direction
$\overline{w}$).

Recall that $\mathcal{S}_{A}(\mathbb{R}^3; \mathbb{C}^4) = \mathcal{S}(\mathbb{R}^3; \mathbb{C}^4)
= \oplus_{1}^{4} \mathcal{S}(\mathbb{R}^3; \mathbb{C})$
we regard as the nuclear space of complex valued functions $f$ on four disjoint copies
of $\mathbb{R}^3$ whose restrictions $f_s$ to each $s$-th copy coincide with
the Schwartz functions in
$\mathcal{S}_{H_{(3)}}(\mathbb{R}^3; \mathbb{C}) = \mathcal{S}(\mathbb{R}^3; \mathbb{C})$. In particular for each
value of the discrete index $s \in \{1,2,3,4\}$, corresponding to each copy, and for each point
$\boldsymbol{\p} \in \mathbb{R}^3$, we have well-defined Dirac delta-functional
$\delta_{s,\boldsymbol{\p}} \in \mathcal{S}_{A}(\mathbb{R}^3; \mathbb{C}^4)^*
= \mathcal{S}(\mathbb{R}^3; \mathbb{C}^4)^*$
defined by
\[
\delta_{s,\boldsymbol{\p}} (f) = f_s(\boldsymbol{\p}),
\]
i.e. the evaluation of the restriction of $f$ to the $s$-th copy of $\mathbb{R}^3$ at the point
$\boldsymbol{\p}$ of that copy. Simply speaking $\delta_{s,\boldsymbol{\p}}$ is the evaluation functional at
fixed point $(s,\boldsymbol{\p})$ of the disjoint sum
$\mathbb{R}^3 \sqcup \mathbb{R}^3 \sqcup \mathbb{R}^3 \sqcup \mathbb{R}^3$.

The generalized Hida annihilation and creation operators $a(w), a(w)^+$ evaluated
at $w=\delta_{s,\boldsymbol{\p}}$ equal to the Dirac delta functionals $\delta_{s,\boldsymbol{\p}}$ have special importance,
and have special notation in mathematical literature
\[
\partial_{s, \boldsymbol{\p}} \overset{\textrm{df}}{=} D_{{}_{\delta_{s,\boldsymbol{\p}}}}
\overset{\textrm{df}}{=} a(\delta_{s,\boldsymbol{\p}}), \,\,\,
\partial_{s, \boldsymbol{\p}}^+ = D_{{}_{\delta_{s,\boldsymbol{\p}}}}^+
= a(\delta_{s,\boldsymbol{\p}})^+
\]
reflecting the derivation-like character of these generalized Hida operators, and are called
\emph{Hida's differential operators}.
But we have also widely used notation for operators in physical literature, with whom the Hida differential
operators should be identified. Namely, generalized Hida
operators should be identified with the operators frequently written by physicists in the following manner
\[
\begin{split}
a_s(\boldsymbol{\p}) \overset{\textrm{df}}{=} D_{{}_{\delta_{s,\boldsymbol{\p}}}} \overset{\textrm{df}}{=}
\partial_{s, \boldsymbol{\p}}
\overset{\textrm{df}}{=} a(\delta_{s,\boldsymbol{\p}}), \\
a_s(\boldsymbol{\p})^+ \overset{\textrm{df}}{=} D_{{}_{\delta_{s,\boldsymbol{\p}}}}^+ \overset{\textrm{df}}{=}
\partial_{s, \boldsymbol{\p}}^+ \overset{\textrm{df}}{=} a(\delta_{s,\boldsymbol{\p}})^+.
\end{split}
\]
More precisely the operators
$a_s(\boldsymbol{\p}), a_s(\boldsymbol{\p})^+$ for $s=1,2$
should be identified with the operators
$b_s(\boldsymbol{\p}), b_s(\boldsymbol{\p})^+$ for $s=1,-1$
of the book \cite{Scharf}, p. 82 (or with the operators
$\overset{*}{a}_{s}^{-}(\boldsymbol{\p}), {a}_{s}^{+}(\boldsymbol{\p})$, $s=1,2$,
of the book \cite{Bogoliubov_Shirkov}, p. 123)).
The operators $a_s(\boldsymbol{\p}), a_s(\boldsymbol{\p})^+$ for $s=3,4$
should respectively be identified with the operators $d_{s}(\boldsymbol{\p}), d_{s}(\boldsymbol{\p})^+$
for $s=1,-1$, of the book \cite{Scharf}, p. 82 (or respectively with the operators
${a}_{s}^{-}(\boldsymbol{\p}), \overset{*}{a}_{s}^{+}(\boldsymbol{\p})$, $s= 1,2$, of the book
\cite{Bogoliubov_Shirkov}, p. 123).

Note that because the Dirac delta functional $\delta_{s,\boldsymbol{\p}}$ is real
$\overline{\delta_{s,\boldsymbol{\p}}} = \delta_{s,\boldsymbol{\p}}$ (i.e. commutes with complex conjugation),
then
\[
a(\delta_{s,\boldsymbol{\p}})^+ = \partial_{s, \boldsymbol{\p}}^+ =
a(\delta_{s,\boldsymbol{\p}})^* = \partial_{s, \boldsymbol{\p}}^*,
\]
so that for Hida's differential operators the linear adjunction $\partial_{s, \boldsymbol{\p}}^*$ coincides 
with the Hermitean adjunction $\partial_{s, \boldsymbol{\p}}^+$.

We may thus summarize the notation used here with that used by other authors
in the following table
\begin{center}
\begin{tabular}{c|c|c|c}
\hline
& {\small Hida-Obata \cite{obata-book}} & Scharf \cite{Scharf} & 
Bogoliubov-Shirkov \cite{Bogoliubov_Shirkov} \\
\hline
$a_{s=1}(\boldsymbol{\p}) \overset{\textrm{df}}{=}
a(\delta_{s=1,\boldsymbol{\p}})$ & $\partial_{s=1, \boldsymbol{\p}}$ & $b_{s=1}(\boldsymbol{\p})$ &
$\overset{*}{a}_{s=1}^{-}(\boldsymbol{\p})$ \\
\hline
$a_{s=2}(\boldsymbol{\p}) \overset{\textrm{df}}{=}
a(\delta_{s=2,\boldsymbol{\p}})$ & $\partial_{s=2, \boldsymbol{\p}}$ & $b_{s=-1}(\boldsymbol{\p})$ &
$\overset{*}{a}_{s=2}^{-}(\boldsymbol{\p})$ \\
\hline
$a_{s=3}(\boldsymbol{\p}) \overset{\textrm{df}}{=}
a(\delta_{s=3,\boldsymbol{\p}})$ & $\partial_{s=3, \boldsymbol{\p}}$ & $d_{s=1}(\boldsymbol{\p})$ &
${a}_{s=1}^{-}(\boldsymbol{\p})$ \\
\hline
$a_{s=4}(\boldsymbol{\p}) \overset{\textrm{df}}{=}
a(\delta_{s=4,\boldsymbol{\p}})$ & $\partial_{s=4, \boldsymbol{\p}}$ & $d_{s=-1}(\boldsymbol{\p})$ &
${a}_{s=2}^{-}(\boldsymbol{\p})$ \\
\hline
$a_{s=1}(\boldsymbol{\p})^+ \overset{\textrm{df}}{=}
a(\delta_{s=1,\boldsymbol{\p}})^+$ & $\partial_{s=1, \boldsymbol{\p}}^*$ & $b_{s=1}(\boldsymbol{\p})^+$ &
${a}_{s=1}^{+}(\boldsymbol{\p})$ \\
\hline
$a_{s=2}(\boldsymbol{\p})^+ \overset{\textrm{df}}{=}
a(\delta_{s=2,\boldsymbol{\p}})^+$ & $\partial_{s=2, \boldsymbol{\p}}^*$ & $b_{s=-1}(\boldsymbol{\p})^+$ &
${a}_{s=2}^{+}(\boldsymbol{\p})$ \\
\hline
$a_{s=3}(\boldsymbol{\p})^+ \overset{\textrm{df}}{=}
a(\delta_{s=3,\boldsymbol{\p}})^+$ & $\partial_{s=3, \boldsymbol{\p}}^*$ & $d_{s=1}(\boldsymbol{\p})^+$ &
$\overset{*}{a}_{s=1}^{+}(\boldsymbol{\p})$ \\
\hline
$a_{s=4}(\boldsymbol{\p})^+ \overset{\textrm{df}}{=}
a(\delta_{s=4,\boldsymbol{\p}})^+$ & $\partial_{s=4, \boldsymbol{\p}}^*$ & $d_{s=-1}(\boldsymbol{\p})^+$ &
$\overset{*}{a}_{s=2}^{+}(\boldsymbol{\p})$ \\
\hline
\end{tabular}
\end{center}

Now we remind some basic results of the calculus of integral kernel operators
constructed mainly by Hida, Obata, and Sait\^o, which we will use here and in the following
Sections (especially in Section \ref{A(1)psi(1)}). 

Before doing it we make a general remark concerning norm estimations of the left $\widehat{\otimes_l}$ and right
$\widehat{\otimes^l}$ antisymmetrized (or symmetrized) $l$-contractions
(compare \cite{obata-book})
\[
|\widehat{f} \widehat{\otimes^l} \widehat{g}|_k, |\widehat{F} \widehat{\otimes}^l \widehat{g}|_{-k},
|\widehat{F} \widehat{\otimes_l} \widehat{g}|_{-k},
\,\,\,
\widehat{F} \in
\Big(\mathcal{S}_{A}(\mathbb{R}^3; \mathbb{C}^4)^{\widehat{\otimes} \, (l+m)} \Big)^{*},
\widehat{f}, \widehat{g} \in \mathcal{S}_{A}
(\mathbb{R}^3; \mathbb{C}^4)^{\widehat{\otimes} \, (l+n)}.
\]
Namely passing from estimations for the norms
\[
|f\otimes^l g|_k, |F \otimes^l g|_{-k}, |F \otimes_l g|_{-k}, \,\,\, \textrm{for} \,\,\,
F \in
\Big(\mathcal{S}_{A}(\mathbb{R}^3; \mathbb{C}^4)^{\otimes (l+m)} \Big)^{*},
f, g \in \mathcal{S}_{A}
\mathbb{R}^3; \mathbb{C}^4)^{\otimes (l+n)},
\]
with non antisymmetrized (or non symmetrized $F$, $f$ and $g$),
summarized in Prop. 3.4.3, Lemma 3.4.4, 3.4.5 of \cite{obata-book}, to estimations with
symmetrized or antisymmetrized $\widehat{F}$, $\widehat{f}$ and $\widehat{g}$ we note that
we have
\begin{multline*}
F \widehat{\otimes}^l g = F \otimes^l g = \pm F\otimes_l g = \pm F \widehat{\otimes}_l g, \\
\textrm{for} \,\,\,
F \in
\Big(\mathcal{S}_{A}(\mathbb{R}^3; \mathbb{C}^4)^{\widehat{\otimes} (l+m)} \Big)^{*},
g \in \mathcal{S}_{A}(\mathbb{R}^3; \mathbb{C}^4)^{\widehat{\otimes} (l+n)},
\end{multline*}
and
\[
|\widehat{f}|_{k} \leq |f|_{k}, \,\,\, f \in
\mathcal{S}_{A}(\mathbb{R}^3; \mathbb{C}^4)^{\otimes n}, k \in \mathbb{Z},
\]
in each case: for symmetrization as well as for antisymmetrization $\widehat{(\cdot)}$.
This allows to restate the estimations for non symmetrized/antisymmetrized
$F$, $f$ and $g$ (summarized in Prop. 3.4.3, Lemma 3.4.4, 3.4.5 of \cite{obata-book})
in the form of propositions
analogous to Prop. 3.4.7, 3.4.8, 3.4.9 in \cite{obata-book} for the contractions
of antisymmetrized $\widehat{F},\widehat{G},\widehat{g}, \widehat{f}$ on exactly the same footing
as for symmetrized $\widehat{F},\widehat{G},\widehat{g}, \widehat{f}$ (as we have already mentioned
in Remark \ref{TretmentFermiIntegKerOp=TretmentBoseIntegKerOp}). In particular
theorems concerning integral kernel operators and Fock expansions, in both cases
1) of scalar-valued kernels \cite{hida}, \cite{obata}, and 2) of vector-valued kernels \cite{obataJFA},
can be stated and proved exactly as in \cite{hida}, \cite{obata}, \cite{obataJFA}
also for the fermi case. The only difference which arises in fermi case (compared to the bose case)
comes from additional factor $(-1)$ depending on the degree of the involved tensors.
In particular, we should note that for nonsymmetrized
$F \in \Big( \mathcal{S}_{A}(\mathbb{R}^3; \mathbb{C}^4)^{\otimes k} \Big)^*$,
$G \in \Big( \mathcal{S}_{A}(\mathbb{R}^3; \mathbb{C}^4)^{\otimes l} \Big)^*$,
and $h \in \mathcal{S}_{A}(\mathbb{R}^3; \mathbb{C}^4)^{\otimes (k+l+m)}$,
we have
\[
F \otimes_k \big(G \otimes_l h \big) =
\big(G \otimes F \big) \otimes_{k+l} h \,\,\, \textrm{in this order!}
\]
and thus by antisymmetrization $\widehat{(\cdot)}$ we get
\begin{multline*}
\widehat{F} \widehat{\otimes_k} \big(\widehat{G} \widehat{\otimes_l} \widehat{h} \big) =
\big(\widehat{G} \widehat{\otimes} \widehat{F} \big) \widehat{\otimes_{k+l}} \widehat{h} =
(-1)^{(\textrm{deg} \, \widehat{F})(\textrm{deg} \, \widehat{G})} \,
\big(\widehat{F} \widehat{\otimes} \widehat{G} \big) \widehat{\otimes_{k+l}} \widehat{h}, \\
\,\,\,
\textrm{deg} \, \widehat{F} \overset{\textrm{df}}{=} k,
\textrm{deg} \, \widehat{G} \overset{\textrm{df}}{=} l;
\end{multline*}
(instead of Proposition 3.4.8 of \cite{obata-book} with symmetrization $\widehat{(\cdot)}$ in bose case, where the factor
$(-1)^{(\textrm{deg} \, \widehat{F})(\textrm{deg} \, G)}$ degenerates to $1$).

Similarly, we have for $F \in \Big( \mathcal{S}_{A}(\mathbb{R}^3; \mathbb{C}^4)^{\otimes  l} \Big)^*$,
$G \in \Big( \mathcal{S}_{A}(\mathbb{R}^3; \mathbb{C}^4)^{\otimes  m} \Big)^*$,
and $f \in \mathcal{S}_{A}(\mathbb{R}^3; \mathbb{C}^4)^{\otimes  (l+n)}$
\[
\langle F \otimes_l f, G \otimes_m g \rangle = \langle F \otimes G, f \otimes^n g \rangle.
\]
Again passing to the sub spaces of antisymmetrized tensors we obtain
\[
\langle \widehat{F} \widehat{\otimes_l} \widehat{f}, \widehat{G} 
\widehat{\otimes_m} \widehat{g} \rangle = \langle \, \widehat{F} \, \widehat{\otimes} \, \widehat{G}, 
\widehat{f} \widehat{\otimes^n} \widehat{g} \rangle = 
(-1)^{m (\textrm{deg} \, \widehat{f})} \, \langle \, \widehat{F} \, \widehat{\otimes} \, \widehat{G}, 
\, \widehat{f} \, \widehat{\otimes_n} \, \widehat{g} \rangle,
\]
(instead of Prop. 3.4.9 in \cite{obata-book} with symmetrization $\widehat{(\cdot)}$ for Bose case).

The replacements of symmetrization $\widehat{(\cdot)}$ with antisymmetrization
$\widehat{(\cdot)}$ (with the appropriate factors $-1$) in the analysis of 
integral kernel operators in \cite{obata-book}, are rather obvious, thus we 
leave the detailed inspection to the reader as an exercise. 
We mention only some particular cases in explicit form.

In particular, we have the following analogue of Thm 4.1.7 of \cite{obata-book}. 
\begin{twr}\label{Dy1...Dym}
Let $\Phi \in \big(\mathcal{S}_{A}(\mathbb{R}^3; \mathbb{C}^4)\big)$ be any element of the Hida space,
and let
\[
\Phi = \sum \limits_{n=0}^{\infty} \Phi_n, \,\,\,
\Phi_n \in \mathcal{S}_{A}(\mathbb{R}^3; \mathbb{C}^4)^{\widehat{\otimes} \, n}
\]
be its decomposition (thus fulfilling (\ref{||Phi||_kPhiInHida})). Then for 
\[
y_1, \dots y_m \in \mathcal{S}_{A}(\mathbb{R}^3; \mathbb{C}^4)^*
\]
we have
\[
D_{y_1} \cdots D_{y_m} \Phi =
\sum \limits_{n=0}^{\infty} \, (-1)^{m-1} \frac{(n+m)!}{n!} \, (\overline{y_1} \widehat{\otimes}
\cdots \widehat{\otimes} \overline{y_m}) \widehat{\otimes}_m \Phi_{m+n}. 
\]
Moreover,  for any $k\geq 0$, $q>0$ and 
$\Phi \in \big(\mathcal{S}_{A}(\mathbb{R}^3; \mathbb{C}^4)\big)$ we have
\[
\| D_{y_1} \cdots D_{y_m} \Phi \|_{k} \leq
\rho^{-q/2} m^{m/2} \Bigg( \frac{\rho^{-q}}{-2qe\textrm{ln} \rho}\Bigg)^{m/2}
|y_1|_{{}_{-(k+q)}} \cdots |y_m|_{{}_{-(k+q)}} \, \| \Phi \|_{k+q}.
\]
\end{twr}

Here 
\[
\rho \overset{\textrm{df}}{=} \|A^{-1}\|_{\textrm{op}} = \lambda_{0}^{-1}, \,\,\,
\lambda_{0} = \textrm{inf Spec} \, A >1,
\]
which we achieve by eventually  adding the unit operator to the ordinary $3$-dimensional oscillator 
Hamiltonian operator and taking the sum as the direct summand $H_{(3)}$  in $A$ defined by (\ref{AinL^2(R^3;C^4)}). 

Using this theorem (analogue of Thm. 4.1.7 of \cite{obata-book}) as well as the mentioned above
analogue of Prop. 3.4.9 of \cite{obata-book} as does Obata in \cite{obata-book})
we prove in particular the following (analogue of Lemma 4.3.1 in \cite{obata-book} or 
Lemma 2.1 in \cite{hida}):

\begin{lem}\label{etaPhiPsi}
For any elements $\Phi, \Psi \in \big(\mathcal{S}_{A}(\mathbb{R}^3; \mathbb{C}^4)\big)$
of the Hida space we put ($s_i, t_i \in \{1, \ldots, 4\}$, $\boldsymbol{\q}_i, \boldsymbol{\p}_i \in \mathbb{R}^3$)
\[
\eta_{{}_{\Phi, \Psi}}(s_1,\boldsymbol{\q}_1, \ldots, s_l, \boldsymbol{\q}_l, t_1, \boldsymbol{\p}_1, 
\ldots, t_m, \boldsymbol{\p}_m) 
= \big\langle \big\langle \partial_{s_1, \boldsymbol{\q}_1}^* \cdots \partial_{s_l, \boldsymbol{\q}_l}^* 
\partial_{t_1, \boldsymbol{\p}_1} \cdots \partial_{t_m, \boldsymbol{\p}_m} \, \Phi, \, \Psi \big\rangle \big\rangle ,
\]
then for any $k>0$ we have
\[
|\eta_{{}_{\Phi, \Psi}}|_{{}_{k}} \leq \rho^{-k} \big(l^lm^m\big)^{1/2}
\Bigg( \frac{\rho^{-k}}{-2ke \textrm{ln} \rho} \Bigg)^{(l+m)/2} \, \| \Phi \|_{k} \| \Psi \|_k.
\]
In particular, $\eta_{{}_{\Phi, \Psi}} \in \mathcal{S}_{A}(\mathbb{R}^3; \mathbb{C}^4)^{\otimes (l+m)}$.
\end{lem}

This allows analysis of an important class of \emph{integral kernel operators} $\Xi_{l,m}(\kappa_{l,m}) \in 
\mathscr{L}\big( \, \big(\mathcal{S}_{A}(\mathbb{R}^3; \mathbb{C}^4)\big)  \, ,
 \, \big(\mathcal{S}_{A}(\mathbb{R}^3; \mathbb{C}^4)\big)^* \, \big)$,
corresponding to $\kappa_{l,m} \in \big(\mathcal{S}_{A}(\mathbb{R}^3; \mathbb{C}^4)^{\otimes (l+m)}\big)^*
= \big(\mathcal{S}(\mathbb{R}^3; \mathbb{C}^4)^{\otimes (l+m)}\big)^*
= \mathcal{S}(\mathbb{R}^3; \mathbb{C}^4)^{* \, \otimes (l+m)}$, and written 
\begin{multline}\label{Xilm(kappalm)Fermi}
\Xi_{l,m}(\kappa_{l,m})  \\
= \sum \limits_{s_1, \ldots s_l, t_1, \ldots t_m =1}^{4} \int \limits_{(\mathbb{R}^3)^{l+m}} \,
\kappa_{l,m}(s_1,\boldsymbol{\q}_1, \ldots, s_l, \boldsymbol{\q}_l, t_1, \boldsymbol{\p}_1, 
\ldots, t_m, \boldsymbol{\p}_m) \,\times \\
\times
\partial_{s_1, \boldsymbol{\q}_1}^* \cdots \partial_{s_l, \boldsymbol{\q}_l}^* 
\partial_{t_1, \boldsymbol{\p}_1} \cdots \partial_{t_m, \boldsymbol{\p}_m} \,
\ud^3 \boldsymbol{\q}_1 \ldots \ud^3 \boldsymbol{\q}_l \ud^3 \boldsymbol{\p}_1 \ldots \ud^3 \boldsymbol{\p}_m.
\end{multline}

\begin{twr}\label{Xi_l,m}
Namely (compare Thm.4.3.2 in \cite{obata-book} or Thm. 2.2. of \cite{hida}) 
for any $\kappa_{l,m} \in \big(\mathcal{S}_{A}(\mathbb{R}^3; \mathbb{C}^4)^{\otimes (l+m)}\big)^*
= \big(\mathcal{S}(\mathbb{R}^3; \mathbb{C}^4)^{\otimes (l+m)}\big)^*$ there exists (uniquely corresponding to
$\kappa_{l,m}$ if $\kappa_{l,m}$ is antisymmetric: 
$\kappa_{l,m} \in \big(\mathcal{S}(\mathbb{R}^3; \mathbb{C}^4)^{\widehat{\otimes} \, l}
\otimes \mathcal{S}(\mathbb{R}^3; \mathbb{C}^4)^{\widehat{\otimes} \, m}\big)^*$  in Fermi case, 
or symmetric in bose case) continuous operator $\Xi_{l,m}(\kappa_{l,m}) \in 
\mathscr{L}\big( \, \big(\mathcal{S}_{A}(\mathbb{R}^3; \mathbb{C}^4)\big)  \, ,
 \, \big(\mathcal{S}_{A}(\mathbb{R}^3; \mathbb{C}^4)\big)^* \, \big)$, written as in (\ref{Xilm(kappalm)Fermi}),
such that
\[
\big\langle \big\langle \Xi_{l,m}(\kappa_{l,m}) \Phi, \, \Psi \big\rangle\big\rangle
= \langle \kappa_{l,m}, \eta_{{}_{\Phi, \Psi}} \rangle, \,\,\,\,\,\,
\Phi, \Psi \in \big(\mathcal{S}_{A}(\mathbb{R}^3; \mathbb{C}^4)\big),
\] 
where
\[
\eta_{{}_{\Phi, \Psi}}(s_1,\boldsymbol{\q}_1, \ldots, s_l, \boldsymbol{\q}_l, t_1, \boldsymbol{\p}_1, 
\ldots, t_m, \boldsymbol{\p}_m) 
= \big\langle \big\langle \partial_{s_1, \boldsymbol{\q}_1}^* \cdots \partial_{s_l, \boldsymbol{\q}_l}^* 
\partial_{t_1, \boldsymbol{\p}_1} \cdots \partial_{t_m, \boldsymbol{\p}_m} \, \Phi, \, \Psi \big\rangle \big\rangle. 
\]
Moreover, for any $k>0$ with $|\kappa_{l,m}|_{-k} <\infty$ it holds
\[
\| \Xi_{l,m}(\kappa_{l,m}) \Phi\|_{-k} \leq 
\rho^{-k} \big(l^lm^m\big)^{1/2}
\Bigg( \frac{\rho^{-k}}{-2ke \textrm{\emph{ln}} \rho} \Bigg)^{(l+m)/2} \, |\kappa_{l,m}|_{{}_{-k}} \, \| \Phi \|_{k} .
\]
\end{twr}

We have the following important theorem (Thm. 4.3.9 of  \cite{obata-book}, Thm. 2.6 of \cite{hida})
which provides necessary and sufficient condition for the integral kernel operator
(\ref{Xilm(kappalm)Fermi}) to be continuous not merely as an operator on the Hida space into its strong dual,
but likewise as operator transforming continuously the Hida space into itself (thus becoming ordinary densely defined
operator in the Fock space):
\begin{twr}\label{Xi_l,m:Hida->Hida}
Let $\kappa_{l,m} \in \big(\mathcal{S}_{A}(\mathbb{R}^3; \mathbb{C}^4)^{\otimes (l+m)}\big)^*$.
Then 
\[
\Xi_{l,m}(\kappa_{l,m}) \in 
\mathscr{L}\big( \, \big(\mathcal{S}_{A}(\mathbb{R}^3; \mathbb{C}^4)\big)  \, ,
 \, \big(\mathcal{S}_{A}(\mathbb{R}^3; \mathbb{C}^4)\big) \, \big)
\]
 if and only if 
 $\kappa_{l,m} \in \mathcal{S}_{A}(\mathbb{R}^3; \mathbb{C}^4)^{\otimes l} \otimes \big(\mathcal{S}_{A}(\mathbb{R}^3; \mathbb{C}^4)^{\otimes m}\big)^*$. In that case, for any $k \in \mathbb{Z}$, $q>0$ with 
$\alpha+\beta \leq 2q$, it holds
\begin{multline*}
\| \Xi_{l,m}(\kappa_{l,m}) \Phi\|_{k}  \\ \leq 
\rho^{-q/2} \big(l^lm^m\big)^{1/2}
\Bigg( \frac{\rho^{-\alpha/2}}{-\alpha e \textrm{\emph{ln}} \rho} \Bigg)^{l/2} \, 
\Bigg( \frac{\rho^{-\beta/2}}{-\beta e \textrm{\emph{ln}} \rho} \Bigg)^{m/2} \, 
|\kappa_{l,m}|_{{}_{l,m;k,-(k+q)}}
\| \Phi \|_{k+q}, 
\end{multline*}
for all $\Phi \in \big(\mathcal{S}_{A}(\mathbb{R}^3; \mathbb{C}^4)\big)$. 
\end{twr}

Here for $f \in \big(\mathcal{S}_{A}(\mathbb{R}^3; \mathbb{C}^4)^{\otimes (l+m)}\big)^*$
we have defined after \cite{obata-book}, Chap. 3.4
\[
|f|_{{}_{l,m;k,q}} \overset{\textrm{df}}{=}
\Bigg( \sum \limits_{\boldsymbol{\textrm{i}}, \boldsymbol{\textrm{j}}} |\langle f,
e(\boldsymbol{\textrm{i}}) \otimes e(\boldsymbol{\textrm{j}}) \rangle |^{2} |e(\boldsymbol{\textrm{i}})|_{{}_{k}}^{2}
|e(\boldsymbol{\textrm{j}})|_{{}_{q}}^{2} \Bigg)^{1/2}, \,\,\, k,q \in \mathbb{R}.
\]
Recall that here we have used (after \cite{obata-book}) the multiindex notation
\[
\begin{split}
e(\boldsymbol{\textrm{i}}) = e_{{}_{i_1}} \otimes \cdots \otimes e_{{}_{i_l}}, \,\,\,\,
\boldsymbol{\textrm{i}} = (i_1, \ldots, i_l), \\
e(\boldsymbol{\textrm{j}}) = e_{{}_{j_1}} \otimes \cdots \otimes e_{{}_{j_m}}, \,\,\,\,
\boldsymbol{\textrm{j}} = (j_1, \ldots, j_m), \\
\end{split}
\]
with $\{e_{{}_{j}}\}_{j=0}^{\infty}$ being the complete orthonormal system in
\[
L^2(\mathbb{R}^3; \mathbb{C}^4)
= L^2(\mathbb{R}^3 \sqcup \mathbb{R}^3 \sqcup \mathbb{R}^3 \sqcup \mathbb{R}^3; \mathbb{C})
\]
of eigenvectors of the operator $A$ defined by (\ref{AinL^2(R^3;C^4)}):
$Ae_{{}_{j}} = \lambda_{{}_{j}} e_{{}_{j}}$, which belong to the nuclear Schwartz space
\[
e_{{}_{j}} \in \mathcal{S}_{A}(\mathbb{R}^3; \mathbb{C}^4) =
\mathcal{S}_{A}(\mathbb{R}^3 \sqcup \mathbb{R}^3 \sqcup \mathbb{R}^3 \sqcup \mathbb{R}^3; \mathbb{C}).
\]
In our case
\[
\mathbb{R}^3 \sqcup \mathbb{R}^3 \sqcup \mathbb{R}^3 \sqcup \mathbb{R}^3 \ni
(s, \boldsymbol{\p}) \longmapsto e_{{}_{j}}(s, \boldsymbol{\p}) = \varepsilon_{{}_{j}}(\boldsymbol{\p}), \,\,\, s \in \{1,2,3,4\},
\]
where $\{\varepsilon_{{}_{j}}\}_{j=0}^{\infty}$ is the system of products $\varepsilon_{{}_{j}}
= h_{{}_{n_j}}h_{{}_{m_j}}h_{{}_{l_j}}$, $\lambda_j = \mu_{{}_{n_j}} + \mu_{{}_{m_j}} + \mu_{{}_{l_j}} +1$
of Hermite functions --
composing the complete orthonormal system of eigenfunctions of the Hamiltonian operator
$H_{(3)}$ in $L^2(\mathbb{R}^3; \mathbb{C})$ of the three-dimensional oscillator (here
$\mu_{{}_{i}}$ is the eigenvalue corresponding to the Hermite function $h_{{}_{i}}$
of the one dimensional oscillator Hamiltonian $H_{(1)}$). When considering the white noise
construction of zero mass fields we will likewise encounter another family of nuclear spaces
$\mathcal{S}_{A}(\mathbb{R}^3, \mathbb{C}^4) = \mathcal{S}^{0}(\mathbb{R}^3, \mathbb{C}^4)$, or
$\mathcal{S}_{A}(\mathbb{R}^3, \mathbb{C}^n) = \mathcal{S}^{0}(\mathbb{R}^3, \mathbb{C}^n)$
with another standard operator $A = \oplus A^{(3)}$ on
$L^2(\mathbb{R}^3; \mathbb{C}^4) = \oplus L^2(\mathbb{R}^3; \mathbb{C})$, or on
$L^2(\mathbb{R}^3; \mathbb{C}^n) = \oplus L^2(\mathbb{R}^3; \mathbb{C})$, with $A^{(3)} \neq H_{(3)}$.

In particular, we have the following Corollary (the fermi analogue of Prop. 4.3.10 of \cite{obata-book}) 
\begin{cor}\label{D_xi=int(xiPartial)}
For $y \in \mathcal{S}_{A}(\mathbb{R}^3, \mathbb{C}^4)^*$ it holds that
\[
D_{\overline{y}} = \Xi_{0,1}(y) = \sum \limits_{s = 1}^{4} \int \limits_{\mathbb{R}^3} y(s, \boldsymbol{\p}) 
\partial_{s, \boldsymbol{\p}} \ud^3 \boldsymbol{\p}, \,\,\,\,
D_{y}^+ = \Xi_{1,0}(y) = \sum \limits_{s = 1}^{4} \int \limits_{\mathbb{R}^3} y(s, \boldsymbol{\p}) 
\partial_{s, \boldsymbol{\p}}^{*} \ud^3 \boldsymbol{\p}.
\]
In particular,
\[
\partial_{s, \boldsymbol{\p}} = \Xi_{0,1}(\delta_{s, \boldsymbol{\p}}), \,\,\,\,\,
\partial_{s, \boldsymbol{\p}}^{*} = \Xi_{1,0}(\delta_{s, \boldsymbol{\p}}).
\] 
For $y \in \mathcal{S}_{A}(\mathbb{R}^3, \mathbb{C}^4) \subset \mathcal{S}_{A}(\mathbb{R}^3, \mathbb{C}^4)^*$
\[
\Xi_{0,1}(y), \Xi_{1,0}(y) \in \mathscr{L}\Big( \, \big( \mathcal{S}_{A}(\mathbb{R}^3, \mathbb{C}^4)\big) \, , \, \big( \mathcal{S}_{A}(\mathbb{R}^3, \mathbb{C}^4)\big)  \, \Big)
\]
and the linear maps
\[
\begin{split}
\mathcal{S}_{A}(\mathbb{R}^3, \mathbb{C}^4) \ni y \longmapsto 
\Xi_{0,1}(y) = D_{\overline{y}} \in 
\mathscr{L}\Big( \, \big( \mathcal{S}_{A}(\mathbb{R}^3, \mathbb{C}^4)\big),  \, 
\big( \mathcal{S}_{A}(\mathbb{R}^3, \mathbb{C}^4)\big)  \, \Big) \\
\mathcal{S}_{A}(\mathbb{R}^3, \mathbb{C}^4) \ni y \longmapsto 
\Xi_{1,0}(y) = D_{y}^+ \in \mathscr{L}\Big( \, \big( \mathcal{S}_{A}(\mathbb{R}^3, \mathbb{C}^4)\big),
\, \big( \mathcal{S}_{A}(\mathbb{R}^3, \mathbb{C}^4)\big)  \, \Big)
\end{split}
\]
are continuous. 

Moreover, for $y_1, \ldots, y_m \in \mathcal{S}_{A}(\mathbb{R}^3, \mathbb{C}^4)^*$ it holds
\begin{multline*}
D_{\overline{y_1}} \cdots D_{\overline{y_m}} = \Xi_{0,m}(y_1 \otimes \cdots \otimes y_m) \\ =
\Xi_{0,m}(y_1 \, \widehat{\otimes} \,  \cdots \,  \widehat{\otimes} \, y_m)  \\ =
\sum \limits_{s_1, \ldots, s_m = 1}^{4} \int \limits_{(\mathbb{R}^3)^m} \, 
y_1(s_1, \boldsymbol{\p}_1) \cdots y_1(s_m, \boldsymbol{\p}_m) \, 
\partial_{s_1, \boldsymbol{\p}_1} \cdots \partial_{s_m, \boldsymbol{\p}_m} \,
\ud^3 \boldsymbol{\p}_1 \cdots \ud^3 \boldsymbol{\p}_m \\ =
(m!)^{-1} \sum \limits_{\pi \in \mathfrak{S}_m} \, \textrm{\emph{sign}} \, \pi \,\,
\sum \limits_{s_{1}, \ldots, s_{m} = 1}^{4} \int \limits_{(\mathbb{R}^3)^m} \, 
y_1(s_{{}_{\pi(1)}}, \boldsymbol{\p}_{{}_{\pi(1)}}) \cdots y_m
(s_{{}_{\pi(m)}}, \boldsymbol{\p}_{{}_{\pi(m)}}) \,
\times \\
\times \,
\partial_{s_1, \boldsymbol{\p}_1} \cdots 
\partial_{s_m, \boldsymbol{\p}_m} \,
\ud^3 \boldsymbol{\p}_1 \cdots \ud^3 \boldsymbol{\p}_m,
\end{multline*}
where $\pi$ runs over the set $\mathfrak{S}_m$ of all permutations of the numbers $1,2,\ldots, m$.
\end{cor}

Note that because for $y,y' \in \mathcal{S}_{A}(\mathbb{R}^3, \mathbb{C}^4)^*$,
$\xi,\xi' \in \mathcal{S}_{A}(\mathbb{R}^3, \mathbb{C}^4)$ all the operators
\[
\begin{split}
D_{\overline{y}} = \Xi_{0,1}(y) = \sum \limits_{s = 1}^{4} \int \limits_{\mathbb{R}^3} y(s, \boldsymbol{\p}) 
\partial_{s, \boldsymbol{\p}} \ud^3 \boldsymbol{\p}, \,\, \textrm{and} \,\,
D_{\xi}^+ = \Xi_{1,0}(\xi) = \sum \limits_{s = 1}^{4} \int \limits_{\mathbb{R}^3} \xi(s, \boldsymbol{\p}) 
\partial_{s, \boldsymbol{\p}}^{*} \ud^3 \boldsymbol{\p}, \\
D_{\overline{y'}} = \Xi_{0,1}(y') = \sum \limits_{s = 1}^{4} \int \limits_{\mathbb{R}^3} y'(s, \boldsymbol{\p}) 
\partial_{s, \boldsymbol{\p}} \ud^3 \boldsymbol{\p}, \,\, \textrm{and} \,\,
D_{\xi'}^+ = \Xi_{1,0}(\xi') = \sum \limits_{s = 1}^{4} \int \limits_{\mathbb{R}^3} \xi'(s, \boldsymbol{\p}) 
\partial_{s, \boldsymbol{\p}}^{*} \ud^3 \boldsymbol{\p}, 
\end{split}
\]
belong to $\mathscr{L}\Big( \, \big( \mathcal{S}_{A}(\mathbb{R}^3, \mathbb{C}^4)\big) \, , \, \big( \mathcal{S}_{A}(\mathbb{R}^3, \mathbb{C}^4)\big)  \, \Big)$ then their products as operators transforming Hida space into Hida space
are meaningful. We have in this case the canonical anticommutation rules
\begin{equation}\label{[Xi_01, Xi_1,0]}
\big\{ \Xi_{0,1}(y), \Xi_{1,0}(\xi) \big\} = \langle y, \xi \rangle \, \boldsymbol{1},
\,\,\,
\big\{ \Xi_{0,1}(y), \Xi_{0,1}(y') \big\} 
= \big\{ \Xi_{1,0}(\xi), \Xi_{1,0}(\xi') \big\} = 0,
\end{equation}
or
\[
\big\{ D_{\overline{y}}, D_{\xi}^+ \big\} = \langle y, \xi \rangle \, \boldsymbol{1},
\,\,\,
\big\{ D_{\overline{y}}, D_{\overline{y'}} \big\} 
= \big\{ D_{\xi}^+, D_{\xi'}^+ \big\} = 0.
\]
They are frequently written in the form (which should be understood properly
in a rigorous sense explained below)
\begin{equation}\label{[partial, partial^*]}
\big\{\partial_{s, \boldsymbol{\p}}, \partial_{s', \boldsymbol{\p}'}^* \big\} = 
\delta_{s, \boldsymbol{\p}}(s', \boldsymbol{\p}'), \,\,\, 
\big\{\partial_{s, \boldsymbol{\p}}, \partial_{s', \boldsymbol{\p}'} \big\} = 
\big\{\partial_{s, \boldsymbol{\p}}^*, \partial_{s', \boldsymbol{\p}'}^* \big\} = 0, 
\end{equation}
or using the notation of physicists
\begin{multline*}
\big\{ a_{s}(\boldsymbol{\p}), a_{s'}(\boldsymbol{\p}')^+ \big\} = 
\delta_{ss'} \delta(\boldsymbol{\p} -\boldsymbol{\p}'), \,\,\,
\big\{ a_{s}(\boldsymbol{\p}), a_{s'}(\boldsymbol{\p}') \big\} = 
\big\{ a_{s}(\boldsymbol{\p})^+, a_{s'}(\boldsymbol{\p}')^+ \big\} = 0, \\
s,s' \in \{1,2,3,4\}
\end{multline*}
or (like in \cite{Scharf}, p. 82)
\begin{equation}\label{[b,b^+],[d,d^+]}
\boxed{
\begin{split}
\big\{ b_{s}(\boldsymbol{\p}), b_{s'}(\boldsymbol{\p}')^+ \big\} = 
\delta_{ss'} \delta(\boldsymbol{\p} -\boldsymbol{\p}'), \,\,\,
\big\{ b_{s}(\boldsymbol{\p}), b_{s'}(\boldsymbol{\p}') \big\} = 
\big\{ b_{s}(\boldsymbol{\p})^+, b_{s'}(\boldsymbol{\p}')^+ \big\} =0, \\
\big\{ d_{s}(\boldsymbol{\p}), d_{s'}(\boldsymbol{\p}')^+ \big\} = 
\delta_{ss'} \delta(\boldsymbol{\p} -\boldsymbol{\p}'), \,\,\,
\big\{ d_{s}(\boldsymbol{\p}), d_{s'}(\boldsymbol{\p}') \big\} = 
\big\{ d_{s}(\boldsymbol{\p})^+, d_{s'}(\boldsymbol{\p}')^+ \big\} =0,\\
\big\{ b_{s}(\boldsymbol{\p}), d_{s'}(\boldsymbol{\p}')^+ \big\} = 0, \,\,\, s,s'=1,-1,
\end{split}
}
\end{equation}
with the obvious identifications 
\[
\begin{split}
D_{y}= a(y) = a(y|_{{}_{s=1}} \oplus y|_{{}_{s=2}} \oplus y|_{{}_{s=3}} \oplus y|_{{}_{s=4}}) \\ =
b(y|_{{}_{s=1}} \oplus y|_{{}_{s=2}} \oplus 0 \oplus 0) + 
d(0 \oplus 0 \oplus y|_{{}_{s=3}} \oplus y|_{{}_{s=4}}) \\
a(y) = \sum \limits_{s = 1}^{4} \int \limits_{\mathbb{R}^3} \overline{y(s, \boldsymbol{\p})} 
a_{s} (\boldsymbol{\p}) \ud^3 \boldsymbol{\p}, \\
b(y|_{{}_{s=1}} \oplus y|_{{}_{s=2}} \oplus 0 \oplus 0) = 
\sum \limits_{s = 1}^{2} \int \limits_{\mathbb{R}^3} \overline{y(s, \boldsymbol{\p})} 
a_{s} (\boldsymbol{\p}) \ud^3 \boldsymbol{\p} =
\sum \limits_{s = 1}^{2} \int \limits_{\mathbb{R}^3} \overline{y(s, \boldsymbol{\p})} 
b_{-2s+3} (\boldsymbol{\p}) \ud^3 \boldsymbol{\p}, \\
d(0 \oplus 0 \oplus y|_{{}_{s=3}} \oplus y|_{{}_{s=4}}) = 
\sum \limits_{s = 3}^{4} \int \limits_{\mathbb{R}^3} \overline{y(s, \boldsymbol{\p})} 
a_{s} (\boldsymbol{\p}) \ud^3 \boldsymbol{\p} =
\sum \limits_{s = 3}^{4} \int \limits_{\mathbb{R}^3} \overline{y(s, \boldsymbol{\p})} 
d_{-2s+7} (\boldsymbol{\p}) \ud^3 \boldsymbol{\p}
\end{split}
\]
for 
\[
y \in \mathcal{S}_{A}(\mathbb{R}^3, \mathbb{C}^4)^*.
\]

The relations (\ref{[partial, partial^*]}) or equivalently (\ref{[b,b^+],[d,d^+]})
should be interpreted properly. Namely the first set of relations (\ref{[Xi_01, Xi_1,0]})
in the particular case $y,\xi \in \mathcal{S}_{A}(\mathbb{R}^3, \mathbb{C}^4)$
reduces to 
\[
\big\{ \Xi_{0,1}(y), \Xi_{1,0}(\xi) \big\} = ( \overline{y}, \xi)_0 \, \boldsymbol{1}
\]
with the inner product $(\cdot, \cdot)_0$ on $L^2(\mathbb{R}^3; \mathbb{C}^4)$.
Using the continuity of the inner product $(\cdot, \cdot)_0$
in the nuclear topology of $\mathcal{S}_{A}(\mathbb{R}^3, \mathbb{C}^4)
\subset L^2(\mathbb{R}^3; \mathbb{C}^4)$ (compare \cite{GelfandIV}, Ch. I.4.2) and the fact that 
$\mathcal{S}_{A}(\mathbb{R}^3, \mathbb{C}^4)$ is a Frech\'et space,
by Proposition 1.3.11 of \cite{obata-book}, it follows that the bilinear map $y\times \xi \mapsto 
(\overline{y}, \xi)_{0} \bold{1}$
defines an operator-valued distribution:
\begin{multline*}
\mathcal{S}_{A}(\mathbb{R}^3, \mathbb{C}^4) \otimes \mathcal{S}_{A}(\mathbb{R}^3, \mathbb{C}^4) \ni \zeta \mapsto \Xi_{0,0}(\zeta)  \\
= \int \limits_{\mathbb{R}^{3} \times \mathbb{R}^{3}}  
\zeta (s,\boldsymbol{\p}, s' ,\boldsymbol{\p}')
\tau(s,\boldsymbol{\p}, s', \boldsymbol{\p}')\, \bold{1} \, \ud^3 p \ud^3 p'  = 
\tau(\zeta) \bold{1}
\end{multline*}
where $\tau \in (\mathcal{S}_{A}(\mathbb{R}^3, \mathbb{C}^4) \otimes \mathcal{S}_{A}(\mathbb{R}^3, \mathbb{C}^4))^*$ 
is defined by
\[
\langle \tau, y \otimes \xi \rangle = (\overline{y}, \xi)_0
= \langle y, \xi \rangle, \,\,\, y, \xi \in \mathcal{S}_{A}(\mathbb{R}^3, \mathbb{C}^4), 
\]
therefore we have
\begin{multline*}
\Xi_{0,0}(y \otimes \xi) 
= \big\{ \Xi_{0,1}(y), \Xi_{1,0}(\xi) \big\} \\
= \sum \limits_{s,s'} \, \int \limits_{\mathbb{R}^{3} \times \mathbb{R}^{3}}  
y \otimes \xi (s',\boldsymbol{\p}', s,\boldsymbol{\p}) \,\,
\delta_{s s'} \, \delta(\boldsymbol{\p}- \boldsymbol{\p}') \, \bold{1} \, \ud^3 p \ud^3 p' \\
= \sum \limits_{s,s'} \, \int \limits_{\mathbb{R}^{3} \times \mathbb{R}^{3}}  
y(s',\boldsymbol{\p}') \, \xi(s,\boldsymbol{\p}) \,\,
\delta_{s s'} \, \delta(\boldsymbol{\p}- \boldsymbol{\p}') \, \bold{1} \, \ud^3 p \ud^3 p',
\end{multline*}
and
\[
\big\{\partial_{s, \boldsymbol{\p}}, \partial_{s', \boldsymbol{\p}'}^* \big\} 
= \delta_{s s'} \, \delta(\boldsymbol{\p} - \boldsymbol{\p}') \bold{1}.
\]

Note here that within the white noise construction of Hida
the operators $\partial_{s,\boldsymbol{\p}}, \partial_{s, \boldsymbol{\p}}^*$
are well-defined at each point $(s,\boldsymbol{\p}) \in \sqcup \, \mathbb{R}^3
= \mathbb{R}^3 \sqcup \mathbb{R}^3 \sqcup \mathbb{R}^3 \sqcup \mathbb{R}^3$, and 
there is no need for treating them as operator-valued distributions
when using the calculus for integral kernel operators.

The exceptional situations, which involve more factors
$\partial_{s,\boldsymbol{\p}}, \partial_{s, \boldsymbol{\p}}^*$ in non ''normal'' order,
in which we are forced to treat them as distributions
are however easily and naturally grasped within the white noise calculus.
The first such situation where we need to use distributional interpretation we encounter when
trying to give proper meaning to (\ref{[partial, partial^*]}) or equivalently
(\ref{[b,b^+],[d,d^+]}) which formally involve both
\begin{equation}\label{HiaPartial*Partial,PartialPartial*}
\partial_{s', \boldsymbol{\p}'}^* \partial_{s, \boldsymbol{\p}} \,\,\,
\textrm{and} \,\,\,
\partial_{s, \boldsymbol{\p}} \partial_{s', \boldsymbol{\p}'}^*,
\end{equation}
with more than just one factor of the type $\partial_{s,\boldsymbol{\p}}, \partial_{s, \boldsymbol{\p}}^*$
containing both $\partial_{s,\boldsymbol{\p}}$ and the adjoint operator
$\partial_{s, \boldsymbol{\p}}^*$. Note that the first of the expressions (that in the ''normal'' order)
in (\ref{HiaPartial*Partial,PartialPartial*}) is meaningful as a continuous operator transforming the Hida space into its dual. But the second expression in (\ref{HiaPartial*Partial,PartialPartial*}) is meaningless
as a generalised operator on the Hida space (or its dual). Nonetheless both expressions
in (\ref{HiaPartial*Partial,PartialPartial*}) are well-defined as operator-valued distributions.
Indeed, the corresponding maps
\[
\chi \times \xi \longmapsto \Xi_{1,0}(\xi) \circ \Xi_{0,1}(\chi), \,\,\,\,
\chi \times \xi \longmapsto \Xi_{0,1}(\chi) \circ \Xi_{1,0}(\xi)
\]
are bilinear and separately continuous as maps
\[
\mathcal{S}_{A}(\mathbb{R}^3, \mathbb{C}^4) \times
\mathcal{S}_{A}(\mathbb{R}^3, \mathbb{C}^4) \longrightarrow
\mathscr{L}\Big( \, \big( \mathcal{S}_{A}(\mathbb{R}^3, \mathbb{C}^4)\big),
\, \big( \mathcal{S}_{A}(\mathbb{R}^3, \mathbb{C}^4)\big) \, \Big).
\]
Therefore because $\mathcal{S}_{A}(\mathbb{R}^3, \mathbb{C}^4)$ is Frech\'et then by Proposition 1.3.11
of \cite{obata-book} there exist the corresponding operator-valued distributions, written
\begin{multline}\label{Distribution:Partial^*Partial}
\chi \otimes \xi \longmapsto \\
\sum \limits_{s,s' = 1}^{4} \int \limits_{\mathbb{R}^3} \chi \otimes \xi(s', \boldsymbol{\p}', s, \boldsymbol{\p}) \,
\partial_{s', \boldsymbol{\p}'}^* \partial_{s, \boldsymbol{\p}} \, \ud^3 \boldsymbol{\p}' \ud^3 \boldsymbol{\p}
= \Xi_{1,1}(\chi \otimes \xi) = \Xi_{1,0}(\xi) \circ \Xi_{0,1}(\chi),
\end{multline}
and
\begin{equation}\label{Distribution:PartialPartial^*}
\chi \otimes \xi \longmapsto
\sum \limits_{s,s' = 1}^{4} \int \limits_{\mathbb{R}^3} \chi \otimes \xi(s', \boldsymbol{\p}', s, \boldsymbol{\p}) \,
\partial_{s, \boldsymbol{\p}} \partial_{s', \boldsymbol{\p}'}^* \, \ud^3 \boldsymbol{\p}' \ud^3 \boldsymbol{\p}
= \Xi_{0,1}(\chi) \circ \Xi_{1,0}(\xi),
\end{equation}
continuous as maps
\[
\mathcal{S}_{A}(\mathbb{R}^3, \mathbb{C}^4)^{\otimes 2} \longrightarrow
\mathscr{L}\Big( \, \big( \mathcal{S}_{A}(\mathbb{R}^3, \mathbb{C}^4)\big),
\, \big( \mathcal{S}_{A}(\mathbb{R}^3, \mathbb{C}^4)\big) \, \Big).
\]
Here in the formula (\ref{Distribution:PartialPartial^*}) the
''distributional integral kernel'', say operator-valued distribution
$\partial_{s, \boldsymbol{\p}} \partial_{s', \boldsymbol{\p}'}^*$, has only formal meaning,
and cannot be interpreted as any actual generalized operator on the Hida space.
But the integral in the formula (\ref{Distribution:Partial^*Partial})
represents an integral kernel operator so that the equalities in the formula
(\ref{Distribution:Partial^*Partial}) is actually a theorem which can immediately be
checked by application of definition of Hida operators. But likewise the operator
$\Xi_{0,1}(\chi) \circ \Xi_{1,0}(\xi)$ in the formula
(\ref{Distribution:PartialPartial^*}), transforming continuously the Hida space into itself,
can be expressed as a (here finite) sum of integral kernel operators. This follows from the general theorem,
\cite{obata} Thm. 6.1 or \cite{obata-book}, Thm 4.5.1 (which can as well be proved for Fermi case without any essential
changes in the proof of \cite{obata}, \cite{obata-book}). However, our case is so simple that the corresponding
decomposition of the operator $\Xi_{0,1}(\chi) \circ \Xi_{1,0}(\xi)$ into the sum of integral kernel
operators can be proven to be equal
\begin{multline}\label{:PartialPartial*:}
\chi \otimes \xi \longmapsto \Xi_{0,1}(\chi) \circ \Xi_{1,0}(\xi) \\ =
- \Xi_{1,1}(\chi \otimes \xi)
\,\,\,\,
+
\,\,\,\,
\Xi_{0,0}(\chi \otimes \xi) \\
- \sum \limits_{s,s' = 1}^{4} \int \limits_{\mathbb{R}^3} \chi \otimes \xi(s', \boldsymbol{\p}', s, \boldsymbol{\p}) \,
\partial_{s', \boldsymbol{\p}'}^* \partial_{s, \boldsymbol{\p}} \, \ud^3 \boldsymbol{\p}' \ud^3 \boldsymbol{\p}
\,\,\,\,
+
\,\,\,\,
( \overline{\chi}, \xi)_0 \, \boldsymbol{1} \\
= - \sum \limits_{s,s' = 1}^{4} \int \limits_{\mathbb{R}^3} \chi \otimes \xi(s', \boldsymbol{\p}', s, \boldsymbol{\p}) \,
\partial_{s', \boldsymbol{\p}'}^* \partial_{s, \boldsymbol{\p}} \, \ud^3 \boldsymbol{\p}' \ud^3 \boldsymbol{\p}
\\
+
\sum \limits_{s,s' = 1}^{4} \, \int \chi \otimes \xi(s', \boldsymbol{\p}', s, \boldsymbol{\p}) \,
\big\{\partial_{s, \boldsymbol{\p}}, \partial_{s', \boldsymbol{\p}'}^* \big\} \,
\ud^3 \boldsymbol{\p}' \ud^3 \boldsymbol{\p},
\end{multline}
using the definition of Hida operators and the relations (\ref{[Xi_01, Xi_1,0]}).

The operator-valued distribution (\ref{:PartialPartial*:}) is called the normal order form distribution
$\boldsymbol{:} \partial_{s, \boldsymbol{\p}} \partial_{s', \boldsymbol{\p}'}^* \boldsymbol{:} + pairing$
of the operator-valued distribution (\ref{Distribution:PartialPartial^*}) symbolized by
$\partial_{s, \boldsymbol{\p}} \partial_{s', \boldsymbol{\p}'}^*$, which is written symbolically
\[
\partial_{s, \boldsymbol{\p}} \partial_{s', \boldsymbol{\p}'}^* = \,\,\,
\boldsymbol{:} \partial_{s, \boldsymbol{\p}} \partial_{s', \boldsymbol{\p}'}^* \boldsymbol{:}
\,\,\, + pairing = - \partial_{s', \boldsymbol{\p}'}^* \partial_{s, \boldsymbol{\p}}
+
\big\{\partial_{s, \boldsymbol{\p}}, \partial_{s', \boldsymbol{\p}'}^* \big\}
\]

Similar situation we have for decomposition of the operator-valued distributions
involving more factors
\begin{equation}\label{...Pariali...Partial*j...}
\cdots \partial_{s_i,\boldsymbol{\p}_i} \cdots \cdots \partial_{s_j, \boldsymbol{\p}_j}^* \cdots
\end{equation}
of the type $\partial_{s,\boldsymbol{\p}}, \partial_{s, \boldsymbol{\p}}^*$, not necessary normally ordered,
into a sum of components with ''normally'' ordered Hida's differential operators, and similarly
as in the ''Wick theorem'' in \cite{Bogoliubov_Shirkov}, Chap. III.
Note that although reduction of such distributions into ''normal form'' follows from the general theorem
for decompositions of the corresponding operators
\begin{equation}\label{...Xi01(chii)...Xi10(xij)...}
\cdots \circ \Xi_{0,1}(\chi_i) \circ \cdots \cdots \circ \Xi_{1,0}(\xi_j) \circ \cdots
\end{equation}
transforming continuously the Hida space into itself into sums of integral kernel operators
(\cite{obata} Thm. 6.1 or \cite{obata-book}, Thm 4.5.1 ),
the simple operator (\ref{...Xi01(chii)...Xi10(xij)...}) can be decomposed
by induction, using the definition of Hida operators and the relations (\ref{[Xi_01, Xi_1,0]}).
We may also compute decompositions of more involved distributions than (\ref{...Pariali...Partial*j...})
which contain ''normally ordered'' factors $\partial_{s, \boldsymbol{\p}}^* \partial_{s, \boldsymbol{\p}}$
with both $\partial_{s, \boldsymbol{\p}}^*$ and $ \partial_{s, \boldsymbol{\p}}$ evaluated
at the same point $(s,\boldsymbol{\p})$, as well-defined distributions:
\begin{equation}\label{...Pariali...Partial*pjPartialpj...}
\cdots \partial_{s_i,\boldsymbol{\p}_i} \cdots \cdots
\partial_{s_j, \boldsymbol{\p}_j}^*\partial_{s_j, \boldsymbol{\p}_j} \cdots
\end{equation}
with the corresponding operators
\begin{equation}\label{...Xi01(chii)...Xi11(xij(tau))...}
\cdots \circ \Xi_{0,1}(\chi_i) \circ \cdots \cdots \circ \Xi_{1,1}\big((\xi_j\otimes 1)\tau\big) \circ \cdots
\end{equation}
transforming continuously the Hida space into itself. Here $\tau \in \mathcal{S}_{A}(\mathbb{R}^3, \mathbb{C}^4) \otimes \mathcal{S}_{A}(\mathbb{R}^3, \mathbb{C}^4)^*$
is uniquely determined by the formula
\[
\langle \tau, y \otimes \xi \rangle
= \langle y, \xi \rangle = (\overline{y}, \xi )_0, \,\,\, y, \xi \in \mathcal{S}_{A}(\mathbb{R}^3, \mathbb{C}^4).
\]
By Theorem \ref{Xi_l,m:Hida->Hida} the operator $\Xi_{1,1}\big((\xi_j \otimes 1)\tau \big)$, with
$\xi_j\in \mathcal{S}_{A}(\mathbb{R}^3, \mathbb{C}^4)$, belongs to
\[
\mathscr{L}\Big( \, \big( \mathcal{S}_{A}(\mathbb{R}^3, \mathbb{C}^4)\big),
\, \big( \mathcal{S}_{A}(\mathbb{R}^3, \mathbb{C}^4)\big) \, \Big),
\]
and the map
\[
\mathcal{S}_{A}(\mathbb{R}^3, \mathbb{C}^4) \ni \xi_j \longmapsto
\Xi_{1,1}\big((\xi_j \otimes 1)\tau\big) \in \mathscr{L}\Big( \, \big( \mathcal{S}_{A}(\mathbb{R}^3, \mathbb{C}^4)\big),
\, \big( \mathcal{S}_{A}(\mathbb{R}^3, \mathbb{C}^4)\big) \, \Big)
\]
is continuous, similarly as for the remaining integral kernel operators $\Xi_{0,1}(\chi_i), \ldots$
in (\ref{...Xi01(chii)...Xi11(xij(tau))...}), so that indeed
(\ref{...Xi01(chii)...Xi11(xij(tau))...})
determines a well-defined distribution transforming continuously
\[
\mathcal{S}_{A}(\mathbb{R}^3, \mathbb{C}^4)^{\otimes n} \longrightarrow
\mathscr{L}\Big( \, \big( \mathcal{S}_{A}(\mathbb{R}^3, \mathbb{C}^4)\big),
\, \big( \mathcal{S}_{A}(\mathbb{R}^3, \mathbb{C}^4)\big) \, \Big).
\]
By the general theorem (\cite{obata} Thm. 6.1 or \cite{obata-book}, Thm 4.5.1 ) the operator
(\ref{...Xi01(chii)...Xi11(xij(tau))...}) can be uniquely decomposed into (here finite)
sum of integral kernel operators, thus providing the decomposition of the distribution
(\ref{...Pariali...Partial*pjPartialpj...}) into sum of components, each in the ''normal order''.
We do not enter here into the investigation of the ''Wick theorem'' for distributions expressed as
simple monomials in the Hida differential operators, as the presented remarks are pretty sufficient
for the simple case of monomials (\ref{...Pariali...Partial*j...}) or (\ref{...Pariali...Partial*pjPartialpj...})
in Hida operators. In this case the Wick theorem is an immediate consequence of the
definition of Hida operators and commutation rules. Note also that the Fock expansion of such a monomial (\ref{...Pariali...Partial*j...})
or respectively (\ref{...Pariali...Partial*pjPartialpj...}), into integral kernel operators $\Xi_{l,m}(\kappa_{l,m})$ becomes finite
if among the integral kernel operators $\Xi_{l,m}(\kappa_{l,m})$ into which we expand the monomial we count for the scalar operators
\[
\Xi_{0,0}(\kappa_{0,0}), \,\,\,
\Xi_{0,0}(\kappa_{0,0}(\zeta))= \langle\kappa_{0,0}, \zeta \rangle \, \boldsymbol{1}
\]
with
\[
\kappa_{0,0} \in \mathcal{S}_{A}(\mathbb{R}^3, \mathbb{C}^4)^{* \,\, \otimes (l+m)}.
\]
and
\[
\zeta \in \mathcal{S}_{A}(\mathbb{R}^3, \mathbb{C}^4)^{\otimes (l+m)}.
\]
Here $l+m$ is equal to the number of factors in (\ref{...Pariali...Partial*j...})
or respectively (\ref{...Pariali...Partial*pjPartialpj...}).

Moreover, the operator distribution, \emph{i.e} the continuous map 
\[
\mathcal{S}_{A}(\mathbb{R}^3, \mathbb{C}^4)^{\otimes n} \longrightarrow 
\mathscr{L}\Big( \, \big( \mathcal{S}_{A}(\mathbb{R}^3, \mathbb{C}^4)\big),
\, \big( \mathcal{S}_{A}(\mathbb{R}^3, \mathbb{C}^4)\big)  \, \Big)
\]
determined by the ``kernel operator'' (\ref{...Pariali...Partial*j...}):
\[
 \cdots \partial_{s_i,\boldsymbol{\p}_i}  \cdots  \partial_{s_j, \boldsymbol{\p}_j}^* \cdots
\]
can be extended to a continuous map
\begin{equation}\label{...Pariali...Partial*j...:continuous...E*...E...->(Hida->Hida)}
 \cdots \otimes \mathcal{S}_{A}(\mathbb{R}^3, \mathbb{C}^4)^{*}
\cdots \otimes \mathcal{S}_{A}(\mathbb{R}^3, \mathbb{C}^4) \otimes \cdots 
\longrightarrow 
\mathscr{L}\Big( \, \big( \mathcal{S}_{A}(\mathbb{R}^3, \mathbb{C}^4)\big),
\, \big( \mathcal{S}_{A}(\mathbb{R}^3, \mathbb{C}^4)\big)  \, \Big)
\end{equation}
only in those variables $s_i,\boldsymbol{\p}_i$ which correspond to the annihilation Hida operators
$\partial_{s_i,\boldsymbol{\p}_i}$ in (\ref{...Pariali...Partial*j...}). Indeed in case of just two Hida factors
\[
\partial_{s, \boldsymbol{\p}} \partial_{s', \boldsymbol{\p}'}^*
\]
in (\ref{...Pariali...Partial*j...}) we have already seen it, because in this case the operator $\Xi$ determined by the 
``operator distribution'' (\ref{...Pariali...Partial*j...}) and evaluated at $\zeta = \chi \otimes \xi$ is equal 
\[
\Xi(\chi \otimes \xi) = -\Xi_{1,1}(\chi\otimes \xi) + \Xi_{0,0}(\chi \otimes \xi).
\]
The said extendibility of the operator $\Xi$ follows in this case because the operator
\[
\chi \otimes \xi \longrightarrow  \Xi_{0,0}(\chi \otimes \xi)
\]
can be extended to a continuous map 
\[
\mathcal{S}_{A}(\mathbb{R}^3, \mathbb{C}^4)^{*}
\otimes \mathcal{S}_{A}(\mathbb{R}^3, \mathbb{C}^4)  
\longrightarrow 
\mathscr{L}\Big( \, \big( \mathcal{S}_{A}(\mathbb{R}^3, \mathbb{C}^4)\big),
\, \big( \mathcal{S}_{A}(\mathbb{R}^3, \mathbb{C}^4)\big)  \, \Big)
\]
because $\tau$, uniquely determined by the formula
\[
\langle \tau, y \otimes \xi \rangle  
= \langle y, \xi \rangle = (\overline{y}, \xi )_0, \,\,\, y, \xi \in \mathcal{S}_{A}(\mathbb{R}^3, \mathbb{C}^4),
\]
can be canonically extended to an element $\tau \in \mathcal{S}_{A}(\mathbb{R}^3, \mathbb{C}^4)^* \otimes \mathcal{S}_{A}(\mathbb{R}^3, \mathbb{C}^4)$.
But also the operator 
\[
\chi \otimes \xi \longrightarrow  \Xi_{1,1}(\chi \otimes \xi)
\]
determined by the normal ordered operator kernel $\partial_{s', \boldsymbol{\p}'}^*\partial_{s, \boldsymbol{\p}}$ can be extended to a continuous map 
\[
\mathcal{S}_{A}(\mathbb{R}^3, \mathbb{C}^4)^{*}
\otimes \mathcal{S}_{A}(\mathbb{R}^3, \mathbb{C}^4)  
\longrightarrow 
\mathscr{L}\Big( \, \big( \mathcal{S}_{A}(\mathbb{R}^3, \mathbb{C}^4)\big),
\, \big( \mathcal{S}_{A}(\mathbb{R}^3, \mathbb{C}^4)\big)  \, \Big)
\]
by Theorem \ref{Xi_l,m:Hida->Hida}. Therefore the extedibility property (\ref{...Pariali...Partial*j...:continuous...E*...E...->(Hida->Hida)})
follows in case there are two Hida factors in (\ref{...Pariali...Partial*j...}). But because the variables $(s_i, \boldsymbol{\p}_i)$
corresponding to different Hida operators  in (\ref{...Pariali...Partial*j...}) are assumed independent, then the extendibility
property (\ref{...Pariali...Partial*j...:continuous...E*...E...->(Hida->Hida)}) in case there are more Hida operators present
easily follows by induction on repeated application of the canonical anti-commutation relations.

Similar extedibility property (\ref{...Pariali...Partial*j...:continuous...E*...E...->(Hida->Hida)}) also is true for the 
distribution operator determined by the more involved ``operator kernel'' (\ref{...Pariali...Partial*pjPartialpj...})
with the additional condition that the extension to the dual factor  $\mathcal{S}_{A}(\mathbb{R}^3, \mathbb{C}^4)^*$
is possible if it corresponds to variable $(s_i, \boldsymbol{\p}_i)$ with the Hida annihilation operator 
$\partial_{s_i,\boldsymbol{\p}_i}$ in (\ref{...Pariali...Partial*pjPartialpj...}), but this annihilation operator
cannot enter (\ref{...Pariali...Partial*pjPartialpj...}) together with creation operator $\partial_{s_i,\boldsymbol{\p}_i}^*$
with the same $(s_i, \boldsymbol{\p}_i)$.

Of course the said extedibility property and the analogous Wick theorem of the distribution kernel (\ref{...Pariali...Partial*j...})
holds true also for bose Hida operators by the analogue of Theorem \ref{Xi_l,m:Hida->Hida} for bose integral kernel operators,
compare Thm. 2.6 of \cite{hida}. 

In fact the ''Wick theorem'' of \cite{Bogoliubov_Shirkov}, Chap III, involves the free field operators
and not merely the (simpler) operators
$a(\delta_{s,\boldsymbol{\p}}) = \partial_{s, \boldsymbol{\p}} = a_s(\boldsymbol{\p}), a(\delta_{s,\boldsymbol{\p}})^+ = \partial_{s', \boldsymbol{\p}'}^*
= a_{s'}(\boldsymbol{\p}')^+$. It is true that Wick theorem for free field operators may be immediately reduced to the
Wick theorem for the corresponding $\partial_{s, \boldsymbol{\p}} = a_s(\boldsymbol{\p}), \partial_{s', \boldsymbol{\p}'}^*
= a_{s'}(\boldsymbol{\p}')^+$ by utilizing the corresponding unitary isomorphisms $U$ (relating the standard Gelfand triples over the corresponding $L^2(\mathbb{R}^3; \mathbb{C}^n)$ with that over the single particle Hilbert spaces), in our case of Dirac field the isomorphism $U$ relating the Gelfand triples
(\ref{SinglePartGelfandTriplesForPsi}), which serves to construct the field out of the standard Hida operators through the formula (\ref{a(U(u+v))=a'(u+v)}). However, starting with ''Wick theorem'' for the standard Hida differential operators
would not be the correct succession for doing things,
because we are interested in very special kind
of distributions to be decomposed, which arise as polynomials of free fields containing concrete form of (Wick ordered)
interacting term (or terms).
Therefore, we should first construct explicitly the free fields in therms of Hida differential operators
(as special kinds of integral kernel operators, with vector-valued kernels), and then prove ''Wick theorem''
for polynomials of free fields containing the Wick ordered polynomials as interaction terms.
Indeed, in the further stage of our analysis (Subsection \ref{WickForProduct}) we will see that the ''Wick theorem'' for the monomials
(\ref{...Pariali...Partial*j...}) in Hida operators together with the extendibility property (\ref{...Pariali...Partial*j...:continuous...E*...E...->(Hida->Hida)})
is sufficient for the proof of the Wick theorem for ''product''
of Wick monomials in free fields and their derivations, as stated formally in \cite{Bogoliubov_Shirkov}, Chap. III, \S 17.2.
In particular no excursion into the ''Wick theorem'' for more involved monomials, e.g. of the form (\ref{...Pariali...Partial*pjPartialpj...}),
is needed for the Wick theorem for ''products'' of \cite{Bogoliubov_Shirkov}.

Here we have only taken the opportunity to emphasize the proper mathematical basis for the
''Wick theorem for product of Wick monomials'' as stated in \cite{Bogoliubov_Shirkov}, Chap. III, \S 17.2, which
becomes a particular case of the general theorem, \cite{obata} Thm. 6.1 or \cite{obata-book}, Thm 4.5.1 (extended on generalized operators in the tensor product of several Fock -- Bose and Fermi -- spaces) on decomposition
of operators transforming continuously the Hida space into itself into a series of integral kernel
operators. 

Summing up the discussion of the relations (\ref{[partial, partial^*]}) or equivalently (\ref{[b,b^+],[d,d^+]})
and of the ''Wick theorem for Hida differential operators'',
we should emphasize that (\ref{[partial, partial^*]}) or (\ref{[b,b^+],[d,d^+]})
should be understood as equalities of operator valued distributions, transforming continuously
\[
\mathcal{S}_{A}(\mathbb{R}^3, \mathbb{C}^4)^{\otimes 2} \longrightarrow
\mathscr{L}\Big( \, \big( \mathcal{S}_{A}(\mathbb{R}^3, \mathbb{C}^4)\big),
\, \big( \mathcal{S}_{A}(\mathbb{R}^3, \mathbb{C}^4)\big) \, \Big),
\]
which even can be extended to continuous maps
\[
\mathcal{S}_{A}(\mathbb{R}^3, \mathbb{C}^4)^* \otimes \mathcal{S}_{A}(\mathbb{R}^3, \mathbb{C}^4) \longrightarrow
\mathscr{L}\Big( \, \big( \mathcal{S}_{A}(\mathbb{R}^3, \mathbb{C}^4)\big),
\, \big( \mathcal{S}_{A}(\mathbb{R}^3, \mathbb{C}^4)\big) \, \Big).
\]

\begin{rem}\label{TretmentFermiIntegKerOp=TretmentBoseIntegKerOpII}
Continuing our Remark \ref{TretmentFermiIntegKerOp=TretmentBoseIntegKerOp}, with the Fermi-Hida
operators $\partial_{{}_{s,\boldsymbol{\p}}} = a_s(\boldsymbol{p})$, $\partial_{{}_{s,\boldsymbol{\p}}}^{*} = a_{s}(\boldsymbol{p})^{+}$
now at our disposal, we recall here that Berezin proposed in \cite{BerezinCMP} the following even exponential state
\[
\Phi_{\zeta}^{+} = \sum \limits_{n=0}^\infty \frac{(-1)^n}{2^nn!} \zeta^{\widehat{\otimes} \, n}
\in (E)_+, \,\,\,\,\,\,\,\,\,\, \zeta \in E^{\widehat{\otimes} \, 2},
\]
instead of (\ref{exp+}). Using the Fermi-Hida operators, this exponential state can be written in the following manner
\begin{equation}\label{BerezinExponentialState}
\Phi_{\zeta}^{+}= e^{-{\textstyle\frac{1}{2}} \Sigma_{s,r} \, \int \zeta(s,\boldsymbol{\p},r,\boldsymbol{\q}) a_{s}(\boldsymbol{p})^{+}
a_{r}(\boldsymbol{\q})^{+} \, \ud^3 \boldsymbol{\p} \, \ud^3 \boldsymbol{\q}} \Phi_0.
\end{equation}
Let us associate with any $\zeta\in E\widehat{\otimes}E$ an integral kernel operator $\zeta$:
\[
\zeta f(s,\boldsymbol{\p}) =
\sum\limits_{r} \int \zeta(s,\boldsymbol{\p}, r,\boldsymbol{\q}) \, f(r, \boldsymbol{\q}) \, \ud^3 \boldsymbol{\q},
\,\,\,\,\,
f \in L^2(\mathbb{R}^4),
\]
with the function $\zeta \in E\widehat{\otimes}E$
as the kernel of the operator $\zeta$ on $L^2(\mathbb{R}^4)$, denoted likewise by $\zeta$. Because the function
\[
\zeta \in E\widehat{\otimes}E \subset L^2(\mathbb{R}^4) \otimes L^2(\mathbb{R}^4) = L^2(\mathbb{R}^4 \times \mathbb{R}^4),
\]
then the corresponding integral kernel operator $\zeta$ corresponding to $\zeta \in E\widehat{\otimes}E$ is of Hilbert-Schmidt class.
By the results of \cite{BerezinCMP} the Berezin symbols
\begin{multline*}
\textrm{\emph{Symbol}} \, \Big[a_{s}(\boldsymbol{\p})^{+}
a_{r}(\boldsymbol{\q})^{+}\Big](\overline{\zeta},\zeta)
=
{\textstyle\frac{\langle\langle \Phi_{\overline{\zeta}}, a_{s}(\boldsymbol{p})^{+}
a_{r}(\boldsymbol{\q})^{+}\Phi_\zeta \rangle\rangle}{\langle\langle \Phi_{\overline{\zeta}}, \Phi_\zeta\rangle\rangle}},
\\
\textrm{\emph{Symbol}} \, \Big[a_{r}(\boldsymbol{\q})
a_{s}(\boldsymbol{\p})\Big](\overline{\zeta},\zeta),
\,\,\,
\textrm{\emph{Symbol}} \, \Big[a_{r}(\boldsymbol{\q})^{+}
a_{s}(\boldsymbol{\p})\Big](\overline{\zeta},\zeta),
\end{multline*}
respectively, of the operators
\[
a_{s}(\boldsymbol{\p})^{+}
a_{r}(\boldsymbol{\q})^{+}, \,\,\,
a_{r}(\boldsymbol{\q})
a_{s}(\boldsymbol{\p}), \,\,\,
a_{r}(\boldsymbol{\q})^{+}
a_{s}(\boldsymbol{\p}), \,\,\,
\]
are, respectively, equal to the kernels of the integral kernel operators
\[
-\zeta(\boldsymbol{1} - \zeta\overline{\zeta})^{-1}, \,\,\,\,\,\, -\overline{\zeta(\boldsymbol{1} - \zeta\overline{\zeta})^{-1}},
\,\,\,\,\,\,
\overline{\zeta}(\boldsymbol{1} - \zeta\overline{\zeta})^{-1}\zeta
\]
evaluated at $(r,\boldsymbol{\q}; s,\boldsymbol{\p})$. Here the bar over the integral kernel operator denotes
the integral kernel operator with the ordinary complex-conjugated kernel.
Let us denote the kernels of these integral kernel operators evaluated at
$(r,\boldsymbol{\q}; s,\boldsymbol{\p})$, by
\begin{multline*}
-\big(\zeta(\boldsymbol{1} - \zeta\overline{\zeta})^{-1}\big)(r,\boldsymbol{\q}; s,\boldsymbol{\p}) \,\,\,
-\big(\overline{\zeta(\boldsymbol{1} - \zeta\overline{\zeta})^{-1}\big)(r,\boldsymbol{\q}; s,\boldsymbol{\p})},
\\
\big(\overline{\zeta}(\boldsymbol{1} - \zeta\overline{\zeta})^{-1}\zeta\big)(r,\boldsymbol{\q}; s,\boldsymbol{\p}).
\end{multline*}

It is easily seen that the Berezin symbol, defined by the
exponential state $\Phi_{\zeta}^{+}$ is multiplicative under the Wick product of even integral kernel operators if and only if
\begin{multline*}
\textrm{\emph{Symbol}} \, \Big[a_{s}(\boldsymbol{\p})^{+}
a_{r}(\boldsymbol{\q})^{+}\Big](\overline{\zeta},\zeta) \,\, \times \,\,
\textrm{\emph{Symbol}} \, \Big[a_{r'}(\boldsymbol{\q}')
a_{s'}(\boldsymbol{\p}')\Big](\overline{\zeta},\zeta)
\\
=
\textrm{\emph{Symbol}} \, \Big[a_{s}(\boldsymbol{\p})^{+}
a_{r'}(\boldsymbol{\q}')\Big](\overline{\zeta},\zeta) \,\, \times \,\,
\textrm{\emph{Symbol}} \, \Big[a_{r}(\boldsymbol{\q})^{+}
a_{s'}(\boldsymbol{\p}')\Big](\overline{\zeta},\zeta)
\end{multline*}
For the exponential map (\ref{BerezinExponentialState}) this cannot be the case for all $\zeta$ because
the pointwise product
\[
-\zeta(\boldsymbol{1} - \zeta\overline{\zeta})^{-1} \,\,\times \,\, -\overline{\zeta(\boldsymbol{1} - \zeta\overline{\zeta})^{-1}}
\]
of the kernels of the operators
\[
-\zeta(\boldsymbol{1} - \zeta\overline{\zeta})^{-1}, \,\,\,\,\,\, -\overline{\zeta(\boldsymbol{1} - \zeta\overline{\zeta})^{-1}},
\]
evaluated, respectively, at $(r,\boldsymbol{\q}; s,\boldsymbol{\p})$ and $(r',\boldsymbol{\q}'; s',\boldsymbol{\p}')$ 
is not equal to pointwise product of the kernels of the operator
\[
\overline{\zeta}(\boldsymbol{1} - \zeta\overline{\zeta})^{-1}\zeta;
\]
evaluated at $(s,\boldsymbol{\p}; r',\boldsymbol{\q}')$ and at $(r,\boldsymbol{\q}; s',\boldsymbol{\p}')$.
Equivalently, at least for some $\zeta$
\begin{multline*}
(\zeta(\boldsymbol{1} - \zeta\overline{\zeta})^{-1})(r,\boldsymbol{\q}; s,\boldsymbol{\p}) \,\, \times \,\,
\big(\overline{\zeta(\boldsymbol{1} - \zeta\overline{\zeta})^{-1}\big)(r',\boldsymbol{\q}'; s',\boldsymbol{\p}')}
\\
\neq
\big(\overline{\zeta}(\boldsymbol{1} - \zeta\overline{\zeta})^{-1}\zeta\big)(s,\boldsymbol{\p}; r',\boldsymbol{\q}')
\,\, \times \,\,
\big(\overline{\zeta}(\boldsymbol{1} - \zeta\overline{\zeta})^{-1}\zeta\big)(r',\boldsymbol{\q}'; s',\boldsymbol{\p}').
\end{multline*}

In particular, it is immediately seen
that the multiplicative law of the symbol is not preserved for each $\zeta$ equal to a finite sum
\[
\zeta = \xi_1 \widehat{\otimes} \eta_1 \oplus \ldots \oplus \xi_k \widehat{\otimes} \eta_k, \,\,\,\, \xi_i, \eta_i \in E,
\]
of simple antisymmetric tensors $\xi_i \widehat{\otimes} \eta_i$. By the anticommutation relations
the exponential state $\Phi_{\zeta}$ for such $\zeta$ is a state with a finite particle number, with the particle number less
than $N<\infty$. Application of the operator
\[
a_{r_1}(\boldsymbol{\q}_1)
a_{s_1}(\boldsymbol{\p}_1) \ldots a_{r_{M}}(\boldsymbol{\q}_{M})
a_{s_{M}}(\boldsymbol{\p}_{M}), \,\,\,\, \textrm{with} \, M>N/2,
\]
to the exponential state $\Phi_\zeta$ gives zero, whence the symbol
of the operator
\[
a_{r_1}(\boldsymbol{\q}_1)
a_{s_1}(\boldsymbol{\p}_1) \ldots a_{r_{M}}(\boldsymbol{\q}_{M})
a_{s_{M}}(\boldsymbol{\p}_{M}), \,\,\, M>N/2,
\]
determined by the exponential state $\Phi_{\zeta}^{+}$, is equal zero.
Because, as is easily seen, the Berezin symbol of each of the operators
$a_{r_i}(\boldsymbol{\q}_i)a_{s_i}(\boldsymbol{\p}_i)$, determined by the exponential state $\Phi_{\zeta}^{+}$,
is, in general for such $\zeta$, not equal zero, then the multiplicativity law cannot be preserved for the
symbols determined by such $\zeta$. Indeed, in particular
\[
\textrm{\emph{Symbol}} \, \Big[a_{r}(\boldsymbol{\q})a_{s}(\boldsymbol{\p})\Big](\overline{\zeta}, \zeta)
=
{\textstyle\frac{\xi \widehat{\otimes} \eta(r,\boldsymbol{\q}; s,\boldsymbol{\p})}{1 + 1/4(\xi \widehat{\otimes} \eta, \xi \widehat{\otimes} \eta)_0}},
\,\,\,\,\,
\zeta = \xi \widehat{\otimes} \eta.
\]
Thus for each fixed $\zeta$ equal to a finite sum of simple tensors the equality
\begin{equation}\label{a(q)a(p)Phizeta=xi(q,p)Phizeta}
a_{r}(\boldsymbol{\q})a_{s}(\boldsymbol{\p}) \Phi_{\zeta}^{+} = \xi(r,q; s,p) \, \Phi_{\zeta}^{+}, \,\,\,\,\,\,\,\,\,
\boldsymbol{\p}, \boldsymbol{\q}
\in \mathbb{R}, \, s,r =1, \ldots, 4,
\end{equation}
cannot be fulfilled for any fixed $\xi \in E^{\widehat{\otimes} \, 2}$ (depending on $\zeta$). It is an open problem
if there exist such $\zeta$ (equal to infinite series of simple tensors) for which (\ref{a(q)a(p)Phizeta=xi(q,p)Phizeta})
is fulfilled. We suspect there are no such $\zeta$ even in the Fock space of infinite dimension, although we have no proof.

We have, of course, similar situation for the exponential map (\ref{exp+}). The particular choice of the exponential map
is not essential, and should be chosen in such a manner which simplifies the formula for the symbols
of the products of Hida operators.

But, as we have already said, no symbol calculus is needed for the investigation of convergence of Fock expansions
in the Fermi Fock space. Therefore, we do not
enter into the investigation of the existence of such $\zeta \in E^{\widehat{\otimes} \, 2}$ which determine
$\Phi_{\zeta}^{+}$ with symbols of even operators
which are multiplicative under the Wick product of even operators. It is evident that such
$\zeta$, if they exist at all, are necessary given by infinite series of simple tensors,
and thus in Fermi Fock spaces $\Gamma(\mathcal{H})$ over finite dimensional
Hilbert spaces $\mathcal{H}$ there are no coherent states $\Phi_{\zeta}^{+}$ with multiplicative symbols.

Fortunately existence of the \emph{coherent} states with multiplicative symbols (if they exist at all in Fermi Fock space)
is completely irrelevant for the investigation of convergence of Fock expansions into integral kernel operators.
This is because the operator norms
\[
\| a(\xi) \|_{{}_{\textrm{\emph{Op}}}} = \| a^+(\xi)\|_{{}_{\textrm{\emph{Op}}}} = \|\xi\|_{{}_{0}},
\,\,\,\,\, \xi \in E,
\]
are finite and the operators
\begin{align*}
a(\xi) = \sum \limits_{s=1}^{4} \int \xi(s, \boldsymbol{\p}) \, a_{s}(\boldsymbol{\p}) \, \ud^3 \boldsymbol{\p}
= \sum \limits_{s=1}^{4} \int \xi(s, \boldsymbol{\p}) \, \partial_{{}_{s,\boldsymbol{\p}}} \, \ud^3 \boldsymbol{\p},
\\
a^{+}(\xi) =
\sum \limits_{s=1}^{4} \int \xi(s, \boldsymbol{\p}) \, a_{s}^{+}(\boldsymbol{\p}) \, \ud^3 \boldsymbol{\p}
= \sum \limits_{s=1}^{4} \int \xi(s, \boldsymbol{\p}) \, \partial_{{}_{s,\boldsymbol{\p}}}^{*} \, \ud^3 \boldsymbol{\p}
\end{align*}
are bounded in Fermi case. Thus, we investigate the generalized operator, even or odd, and their expansions
into integral kernel operators in their action
on the states $\Phi_N$ which are equal to finite direct sums of simple tensors (instead of $\Phi_{\overline{\zeta}}^{+}, \Phi_{\zeta}^{+}$).
The above formula for the operator norm of $a(\xi), a^{+}(\xi)$, together with the Swartz inequality
gives estimations of
\[
|\langle\langle \Phi'_{N'} , \Xi_{\ell,m}(\kappa_{\ell,m}) \Phi_N \rangle \rangle|
\]
for the integral kernel operators $\Xi_{\ell,m}(\kappa_{\ell,m})$.
\end{rem}

Now having given the Hida operators $a(\delta_{s,\boldsymbol{\p}}) = \partial_{s, \boldsymbol{\p}} = a_s(\boldsymbol{\p}), \partial_{s', \boldsymbol{\p}'}^* = a_{s'}(\boldsymbol{\p}')^+$, $a(w), a(w)^*$, $w \in
\mathcal{S}_{A}(\mathbb{R}^3, \mathbb{C}^4)\big)^*$ corresponding to the Fock lifting $\Gamma$
of the first standard Gelfand triple in (\ref{SinglePartGelfandTriplesForPsi}), we can now utilize the unitary isomorphism $U$, given by (\ref{isomorphismU}),
relating the triples in (\ref{SinglePartGelfandTriplesForPsi}), and then construct the free Dirac field
as Hida generalized operator, using $a(\delta_{s,\boldsymbol{\p}}) = \partial_{s, \boldsymbol{\p}} = a_s(\boldsymbol{\p}),
a(\delta_{s,\boldsymbol{\p}})^+ = \partial_{s', \boldsymbol{\p}'}^* = a_{s'}(\boldsymbol{\p}')^+$,
$a(w), a(w)^*$, $w \in \mathcal{S}_{A}(\mathbb{R}^3, \mathbb{C}^4)\big)^*$ and the formula
(\ref{a(U(u+v))=a'(u+v)}):
\begin{multline*}
\boldsymbol{\psi}(\overline{\phi}) = a'\big(P^\oplus\widetilde{\phi}|_{{}_{\mathscr{O}_{m,0,0,0}}} \oplus 0\big) +
a'\Big( 0 \oplus \big(P^\ominus\widetilde{\phi}|_{{}_{\mathscr{O}_{-m,0,0,0}}}\big)^\flat \Big)^+ \\
= a\Big(U\big(P^\oplus\widetilde{\phi}|_{{}_{\mathscr{O}_{m,0,0,0}}} \oplus 0\big)\Big) +
a\Bigg(U\Big( 0 \oplus \big(P^\ominus\widetilde{\phi}|_{{}_{\mathscr{O}_{-m,0,0,0}}}\big)^\flat \Big)\Bigg)^+,
\end{multline*}
for
\begin{multline*}
0 \oplus \big(P^\ominus\widetilde{\phi}|_{{}_{\mathscr{O}_{-m,0,0,0}}}\big)^\flat, \\
\,\,\, \textrm{and} \,\,\,
P^\oplus\widetilde{\phi}|_{{}_{\mathscr{O}_{m,0,0,0}}} \oplus 0 \in
E, \phi \in \mathscr{E} = \mathcal{S}(\mathbb{R}^4; \mathbb{C}^4) =
\mathcal{S}_{\oplus H_{(4)}}(\mathbb{R}^4, \mathbb{C}^4)\big).
\end{multline*}

This would be the case for the Dirac field (\ref{psi(f)=a_+(f)+a_-(f^c)^+proper}), if the annihilation-creation
operators $a'(\cdot), a'(\cdot)^+$, are exatly the operators used in (\ref{psi(f)=a_+(f)+a_-(f^c)^+proper}),
and the fundamental system of solutions $u,v$ used in the construction of the $U$ isomorphism in (\ref{isomorphismU})
is that corresponding to the gamma generators $\gamma^\mu$ in the momentum picture (given in Subsection \ref{fundamental,u,v}), 
corresponding to (\ref{psi(f)=a_+(f)+a_-(f^c)^+proper}). That is $\boldsymbol{\p} \mapsto u(\boldsymbol{\p}), \boldsymbol{\p} \mapsto v(-\boldsymbol{\p})$, 
span in the momentum picture the single particle positive and negative energy solutions of the Fock space
$\Gamma(\mathcal{H}^{\oplus}_{m,0}\oplus\mathcal{H}^{\ominus}_{-m,0})$, associated to  $a'(\cdot), a'(\cdot)^+$ and 
(\ref{psi(f)=a_+(f)+a_-(f^c)^+proper}). Of course this construction can be repeated for the Dirac
field $\underset{*}{\boldsymbol{\psi}}$, given by (\ref{psi(f)=a_+(f)+a_-(f^c)^+proper*}):
\begin{multline*}
\underset{*}{\boldsymbol{\psi}}(\overline{\phi}) = a'\big(P^{\oplus *}\widetilde{\phi}|_{{}_{\mathscr{O}_{m,0,0,0}}} \oplus 0\big) +
a'\Big( 0 \oplus \big(P^{\ominus *}\widetilde{\phi}|_{{}_{\mathscr{O}_{-m,0,0,0}}}\big)^\flat \Big)^+ \\
= a\Big(U\big(P^{\oplus *}\widetilde{\phi}|_{{}_{\mathscr{O}_{m,0,0,0}}} \oplus 0\big)\Big) +
a\Bigg(U\Big( 0 \oplus \big(P^{\ominus *}\widetilde{\phi}|_{{}_{\mathscr{O}_{-m,0,0,0}}}\big)^\flat \Big)\Bigg)^+,
\end{multline*}
for
\begin{multline*}
0 \oplus \big(P^{\ominus *}\widetilde{\phi}|_{{}_{\mathscr{O}_{-m,0,0,0}}}\big)^\flat, \\
\,\,\, \textrm{and} \,\,\,
P^{\oplus *}\widetilde{\phi}|_{{}_{\mathscr{O}_{m,0,0,0}}} \oplus 0 \in
E, \phi \in \mathscr{E} = \mathcal{S}(\mathbb{R}^4; \mathbb{C}^4) =
\mathcal{S}_{\oplus H_{(4)}}(\mathbb{R}^4, \mathbb{C}^4)\big),
\end{multline*}
but in the formulas for $\underset{*}{\boldsymbol{\psi}}$ the annihilation-creation
operators $a'(\cdot), a'(\cdot)^+$ are exatly the operators $a'(\cdot), a'(\cdot)^+$ used in (\ref{psi(f)=a_+(f)+a_-(f^c)^+proper*})
and the fundamental system of solutions $u,v$ used in the construction of the $U$ isomorphism in (\ref{isomorphismU})
is that 
\[
\underset{*}{u} = \gamma^0 u, \,\,\,\, \underset{*}{v} = \gamma^0 v
\]
corresponding to the gamma generators $(\gamma^\mu)^* = \gamma^0 \gamma^\mu \gamma^0$ in the momentum picture, 
and corresponding to (\ref{psi(f)=a_+(f)+a_-(f^c)^+proper*}).

But the (free) Dirac field $\boldsymbol{\psi}$ or $\underset{*}{\boldsymbol{\psi}}$  (and in general quantum free field) is naturally an integral
kernel operator with well-defined kernel equal to integral kernel operator
\begin{multline*}
\boldsymbol{\psi}^a(x) = \sum_{s=1}^{4} \, \int \limits_{\mathbb{R}^3}
\kappa_{0,1}(s, \boldsymbol{p}; a, x) \,\, \partial_{s, \boldsymbol{\p}} \, \ud^3 \boldsymbol{\p}
+
\sum_{s=1}^{4} \, \int \limits_{\mathbb{R}^3}
\kappa_{1,0}(s, \boldsymbol{p}; a, x) \,\, \partial_{s, \boldsymbol{\p}}^* \, \ud^3 \boldsymbol{\p} \\
= \Xi_{0,1}\big(\kappa_{0,1}(a,x)\big) + \Xi_{1,0}\big(\kappa_{1,0}(a,x)\big),
\end{multline*}
\begin{multline*}
\underset{*}{\boldsymbol{\psi}}^a(x) = \sum_{s=1}^{4} \, \int \limits_{\mathbb{R}^3}
\kappa^{*}_{0,1}(s, \boldsymbol{p}; a, x) \,\, \partial_{s, \boldsymbol{\p}} \, \ud^3 \boldsymbol{\p}
+
\sum_{s=1}^{4} \, \int \limits_{\mathbb{R}^3}
\kappa^{*}_{1,0}(s, \boldsymbol{p}; a, x) \,\, \partial_{s, \boldsymbol{\p}}^* \, \ud^3 \boldsymbol{\p} \\
= \Xi_{0,1}\big(\kappa^{*}_{0,1}(a,x)\big) + \Xi_{1,0}\big(\kappa^{*}_{1,0}(a,x)\big),
\end{multline*}
with vector-valued distributional kernels $\kappa_{lm}(a,x), \kappa^{*}_{lm}(a,x)$ representing distributions
\begin{multline*}
\kappa_{lm}, \kappa^{*}_{lm} \in \mathscr{L}\big( \mathcal{S}_{A}(\mathbb{R}^3, \mathbb{C}^4)^{\otimes(l+m)}, \,\,
\mathscr{L}(\mathscr{E}, \mathbb{C}) \big) \cong
\mathscr{L}\big( \mathcal{S}_{A}(\mathbb{R}^3, \mathbb{C}^4)^{\otimes(l+m)}, \,\,
\mathscr{E}^* \big) \\
\cong \big( \mathcal{S}_{A}(\mathbb{R}^3, \mathbb{C}^4)^{\otimes(l+m)} \big)^* \otimes
\mathscr{E}^* \cong
\mathscr{L}\Big( \mathscr{E}, \,\, \big(\mathcal{S}_{A}(\mathbb{R}^3, \mathbb{C}^4)^{\otimes(l+m)}\big)^*
\Big),
\end{multline*}
in the sense of Obata \cite{obataJFA}. In fact, we have used the standard
nuclear space $\mathcal{S}_{A}(\mathbb{R}^3, \mathbb{C}^4)$ instead of the isomorphic
nuclear space $E$, because we have discarded the isomorphism $\Gamma(U)$ in
(\ref{G(U)^+a(U(u+v))G(U)=a'(u+v)}) or in
(\ref{G(U)^+a(U(u+v))G(U)=a'(u+v)degenerated})), and realize the Hida operators $a'$
in the Fock lifting of the standard Gelfand triple in
(\ref{SinglePartGelfandTriplesForPsi}). We will find such
$\mathscr{L}\big( \mathscr{E}, \,\, \mathbb{C} \big) \cong
\mathscr{E}^*$-valued
distribution kernels $\kappa_{0,1}, \kappa_{1,0}, \kappa^{*}_{0,1}, \kappa^{*}_{1,0} \in
\mathscr{L}\big( \mathscr{E}, \,\, \mathcal{S}_{A}(\mathbb{R}^3, \mathbb{C}^4)^*
\big) \cong \mathscr{L}\big( \mathcal{S}_{A}(\mathbb{R}^3, \mathbb{C}^4), \,\,
\mathscr{L}(\mathscr{E}, \mathbb{C}) \big)$ that
\begin{multline}\label{psi=IntKerOpVectValKer'}
\boldsymbol{\psi}(\overline{\phi}) = a'\big(P^\oplus\widetilde{\phi}|_{{}_{\mathscr{O}_{m,0,0,0}}} \oplus 0\big) +
a'\Big( 0 \oplus \big(P^\ominus\widetilde{\phi}|_{{}_{\mathscr{O}_{-m,0,0,0}}}\big)^\flat \Big)^+ \\
= a\Big(U\big(P^\oplus\widetilde{\phi}|_{{}_{\mathscr{O}_{m,0,0,0}}} \oplus 0\big)\Big) +
a\Bigg(U\Big( 0 \oplus \big(P^\ominus\widetilde{\phi}|_{{}_{\mathscr{O}_{-m,0,0,0}}}\big)^\flat \Big)\Bigg)^+ \\
= \sum_{s=1}^{4} \, \int \limits_{\mathbb{R}^3}
\kappa_{0,1}(\overline{\phi})(s, \boldsymbol{p}) \,\, \partial_{s, \boldsymbol{\p}} \, \ud^3 \boldsymbol{\p}
+
\sum_{s=1}^{4} \, \int \limits_{\mathbb{R}^3}
\kappa_{1,0}(\overline{\phi})(s, \boldsymbol{p}) \,\, \partial_{s, \boldsymbol{\p}}^* \, \ud^3 \boldsymbol{\p} \\
= \Xi_{0,1}\big(\kappa_{0,1}(\overline{\phi})\big) + \Xi_{1,0}\big(\kappa_{1,0}(\overline{\phi})\big)
= \sum\limits_{a} \int \boldsymbol{\psi}^{a}(x) \, \overline{\phi^a(x)} \ud^4x, \,\,\,\,\,\,
\phi \in \mathscr{E} = \mathcal{S}(\mathbb{R}^4; \mathbb{C}^4).
\end{multline}
and
\begin{multline}\label{psi=IntKerOpVectValKer}
\underset{*}{\boldsymbol{\psi}}(\overline{\phi}) = a'\big(P^{\oplus *}\widetilde{\phi}|_{{}_{\mathscr{O}_{m,0,0,0}}} \oplus 0\big) +
a'\Big( 0 \oplus \big(P^{\ominus *}\widetilde{\phi}|_{{}_{\mathscr{O}_{-m,0,0,0}}}\big)^\flat \Big)^+ \\
= a\Big(U\big(P^{\oplus *}\widetilde{\phi}|_{{}_{\mathscr{O}_{m,0,0,0}}} \oplus 0\big)\Big) +
a\Bigg(U\Big( 0 \oplus \big(P^{\ominus *}\widetilde{\phi}|_{{}_{\mathscr{O}_{-m,0,0,0}}}\big)^\flat \Big)\Bigg)^+ \\
= \sum_{s=1}^{4} \, \int \limits_{\mathbb{R}^3}
\kappa^{*}_{0,1}(\overline{\phi})(s, \boldsymbol{p}) \,\, \partial_{s, \boldsymbol{\p}} \, \ud^3 \boldsymbol{\p}
+
\sum_{s=1}^{4} \, \int \limits_{\mathbb{R}^3}
\kappa^{*}_{1,0}(\overline{\phi})(s, \boldsymbol{p}) \,\, \partial_{s, \boldsymbol{\p}}^* \, \ud^3 \boldsymbol{\p} \\
= \Xi_{0,1}\big(\kappa^{*}_{0,1}(\overline{\phi})\big) + \Xi_{1,0}\big(\kappa^{*}_{1,0}(\overline{\phi})\big)
= \sum\limits_{a} \int \underset{*}{\boldsymbol{\psi}}^{a}(x) \, \overline{\phi^a(x)} \ud^4x, \,\,\,\,\,\,
\phi \in \mathscr{E} = \mathcal{S}(\mathbb{R}^4; \mathbb{C}^4).
\end{multline}
Here $\kappa_{0,1}, \kappa_{1,0}, \kappa^{*}_{0,1}, \kappa^{*}_{1,0} \in
\mathscr{L}\big( \mathscr{E}, \,\, \mathcal{S}_{A}(\mathbb{R}^3, \mathbb{C}^4)^*
\big) \cong \mathscr{L}\big( \mathcal{S}_{A}(\mathbb{R}^3, \mathbb{C}^4), \,\,
\mathscr{L}(\mathscr{E}, \mathbb{C}) \big)$ are vector valued distributions represented with the
following distribution kernels
\[
\boxed{
\kappa_{0,1}(s, \boldsymbol{\p}; a,x) = \left\{ \begin{array}{ll}
{\textstyle\frac{1}{2}}\underset{*}{u}{}_{s}^{a}(\boldsymbol{\p})e^{-ip\cdot x} \,\,\, \textrm{with $p = (|p_0(\boldsymbol{\p})|, \boldsymbol{\p}) \in \mathscr{O}_{m,0,0,0}$} & \textrm{if $s=1,2$}
\\
0 & \textrm{if $s=3,4$}
\end{array} \right.,
}
\]
\[
\boxed{
\kappa_{1,0}(s, \boldsymbol{\p}; a,x) = \left\{ \begin{array}{ll}
0 & \textrm{if $s=1,2$}
\\
-{\textstyle\frac{1}{2}}\underset{*}{v}{}_{s-2}^{a}(\boldsymbol{\p})e^{ip\cdot x} \,\,\, \textrm{with $p = (|p_0(\boldsymbol{\p})|, \boldsymbol{\p}) \in \mathscr{O}_{m,0,0,0}$} & \textrm{if $s=3,4$}
\end{array} \right.
}
\]
\begin{equation}\label{kappa_0,1}
\boxed{
\kappa^{*}_{0,1}(s, \boldsymbol{\p}; a,x) = \left\{ \begin{array}{ll}
{\textstyle\frac{1}{2}}u_{s}^{a}(\boldsymbol{\p})e^{-ip\cdot x} \,\,\, \textrm{with $p = (|p_0(\boldsymbol{\p})|, \boldsymbol{\p}) \in \mathscr{O}_{m,0,0,0}$} & \textrm{if $s=1,2$}
\\
0 & \textrm{if $s=3,4$}
\end{array} \right.,
}
\end{equation}
\begin{equation}\label{kappa_1,0}
\boxed{
\kappa^{*}_{1,0}(s, \boldsymbol{\p}; a,x) = \left\{ \begin{array}{ll}
0 & \textrm{if $s=1,2$}
\\
-{\textstyle\frac{1}{2}}
v_{s-2}^{a}(\boldsymbol{\p})e^{ip\cdot x} \,\,\, \textrm{with $p = (|p_0(\boldsymbol{\p})|, \boldsymbol{\p}) \in \mathscr{O}_{m,0,0,0}$} & \textrm{if $s=3,4$}
\end{array} \right.
}
\end{equation}
\[
\underset{*}{u} = \gamma^0 u, \,\,\,\,\, \underset{*}{v} = \gamma^0 v.
\]
Here $\kappa_{0,1}(\phi), \kappa_{1,0}(\phi), \kappa^{*}_{0,1}(\phi), \kappa^{*}_{1,0}(\phi)$ denote the kernels representing distributions
in $\mathcal{S}_{A}(\mathbb{R}^3, \,\, \mathbb{C}^4)^*$ which are defined in the standard manner
\[
\kappa_{0,1}(\phi)(s, \boldsymbol{\p})
= \sum_{a=1}^{4} \int \limits_{\mathbb{R}^3}
\kappa_{0,1}(s, \boldsymbol{\p}; a,x) \phi^{a}(x) \, \ud^4 x
\]
and analogously for $\kappa_{1,0}(\phi), \kappa^{*}_{0,1}(\phi), \kappa^{*}_{1,0}(\phi)$, and such that
\[
\begin{split}
\kappa_{0,1}, \kappa^{*}_{0,1}: \mathscr{E} \ni \phi \longmapsto \kappa_{0,1}(\phi)
\in \mathcal{S}_{A}(\mathbb{R}^3, \,\, \mathbb{C}^4)^*, \\
\kappa_{1,0}, \kappa^{*}_{1,0}: \mathscr{E} \ni \phi \longmapsto \kappa_{1,0}(\phi)
\in \mathcal{S}_{A}(\mathbb{R}^3, \,\, \mathbb{C}^4)^*
\end{split}
\]
belong to $\mathscr{L}\big( \mathscr{E}, \,\, \big(\mathcal{S}_{A}(\mathbb{R}^3, \mathbb{C}^4)^*
\big) \cong \mathscr{L}\big( \mathcal{S}_{A}(\mathbb{R}^3, \mathbb{C}^4), \,\,
\mathscr{L}(\mathscr{E}, \mathbb{C}) \big)$. We should emphasise here that in case of free fields the vector-valued distributions $\kappa_{0,1}, \kappa_{1,0}$ are regular function like distributions with
distribution kernels $\kappa_{0,1}(s, \boldsymbol{\p}; a,x), \kappa_{0,1}(s, \boldsymbol{\p}; a,x)$
equal to ordinary functions, determining functions (and similarly for $\kappa^{*}_{l,m}$)
\begin{equation}\label{multiplicative-kappa_{0,1},kappa_{1,0}}
\begin{split}
\Bigg( \, (a,x) \mapsto \kappa_{0,1; s, \boldsymbol{\p}}(a,x)
\overset{\textrm{df}}{=}
\kappa_{0,1}(s, \boldsymbol{\p}; a,x) \, \Bigg) \in \mathcal{O}_{M} \subset \mathscr{E}^*,
\,\,\,(s,\boldsymbol{\p}) \in \sqcup \, \mathbb{R}^3, \\
\Bigg( \, (a,x) \mapsto \kappa_{1,0; s, \boldsymbol{\p}}(a,x)
\overset{\textrm{df}}{=}
\kappa_{1,0}(s, \boldsymbol{\p}; a,x) \, \Bigg) \in \mathcal{O}_{M} \subset \mathscr{E}^*,
\,\,\,(s,\boldsymbol{\p}) \in \sqcup \, \mathbb{R}^3, \\
\Bigg( \, (s,\boldsymbol{\p}) \mapsto \kappa_{0,1; a,x }(s, \boldsymbol{\p})
\overset{\textrm{df}}{=}
\kappa_{0,1}(s, \boldsymbol{\p}; a,x) \, \Bigg) \in \mathcal{O}_{M, A}
\subset \mathcal{S}_{A}(\mathbb{R}^3, \mathbb{C}^4)^*, \\
\Bigg( \, (s,\boldsymbol{\p}) \mapsto \kappa_{1,0; a,x }(s, \boldsymbol{\p})
\overset{\textrm{df}}{=}
\kappa_{1,0}(s, \boldsymbol{\p}; a,x) \, \Bigg) \in \mathcal{O}_{M, A}
\subset \mathcal{S}_{A}(\mathbb{R}^3, \mathbb{C}^4)^*,
\end{split}
\end{equation}
which belong respectively to the function algebra of multipliers $\mathcal{O}_{M}$ of the nuclear algebra
$\mathscr{E} = \mathcal{S}(\mathbb{R}^4; \mathbb{C}^4) =
\mathcal{S}_{\oplus H_{(4)}}(\mathbb{R}^3, \mathbb{C}^4)$ (in the first two cases),
and respectively to the algebra of multipliers $\mathcal{O}_{M, A}$ of the nuclear algebra
$\mathcal{S}_{A}(\mathbb{R}^3, \mathbb{C}^4) = \mathcal{S}(\mathbb{R}^3, \mathbb{C}^4)$
(in the last two cases). These statements can be understood in the sense that for each fixed value of the respective
discrete index, $a$ or $s$, the functions $x \mapsto \kappa_{l,m}(s, \boldsymbol{\p}; a,x)$
or $\boldsymbol{\p} \mapsto \kappa_{0,1}(s, \boldsymbol{\p}; a,x)$, belong respectively to the algebra of multipliers of
$\mathcal{S}(\mathbb{R}^4; \mathbb{C}) =
\mathcal{S}_{H_{(4)}}(\mathbb{R}^3, \mathbb{C})$ or convolutors of $\mathcal{S}_{H_{(3)}}(\mathbb{R}^3, \mathbb{C}) = \mathcal{S}(\mathbb{R}^3, \mathbb{C})$. But according to our general prescription, we should also note that
$\mathscr{E} = \mathcal{S}(\mathbb{R}^4; \mathbb{C}^4) =
\mathcal{S}_{\oplus H_{(4)}}(\mathbb{R}^3, \mathbb{C}^4) = \mathcal{S}_{\oplus H_{(4)}}(\sqcup \mathbb{R}^4; \mathbb{C})$
can be treated as nuclear algebra of $\mathbb{C}$-valued functions on the disjoint sum $\sqcup \mathbb{R}^4$
of four disjoint copies of $\mathbb{R}^4$, with the natural pointwise multiplication rule of any two such functions.
So that the algebra $\mathcal{O}_{M}$ of multipliers is well-defined and coincides with all those functions whose restrictions to each copy $\mathbb{R}^4$ belongs to the algebra of multipliers of
$\mathcal{S}(\mathbb{R}^4; \mathbb{C}) =
\mathcal{S}_{H_{(4)}}(\mathbb{R}^3, \mathbb{C})$.
The algebra of convolutors $\mathcal{O}_{C}$ of $\mathscr{E}$, is also well-defined with the ordinary Fourier transform exchanging the convolution and pointwise multiplication if we define action of translation $T_{b}$, $b \in \mathbb{R}^4$
on $(a, x) \in \sqcup \mathbb{R}^4$ as equal $T_{b}(a,x) = (a, x + b)$. Similarly, the algebras
$\mathcal{O}_{M, A}(\mathbb{R}^3; \mathbb{C}^4)$,
$\mathcal{O}_{M, A}(\mathbb{R}^3; \mathbb{C}^4)$, of multipliers and convolutors of
$\mathcal{S}_{A}(\mathbb{R}^3, \mathbb{C}^4) = \mathcal{S}(\mathbb{R}^3, \mathbb{C}^4)
= \mathcal{S}(\sqcup \mathbb{R}^3, \mathbb{C})$ are well-defined, where the last is the algebra of all such functions
on $\sqcup \mathbb{R}^4$ with restrictions to each copy $\mathbb{R}^3$ belonging to $\mathcal{S}(\mathbb{R}^3; \mathbb{C})
= \mathcal{S}_{H_{(3)}}(\mathbb{R}^3; \mathbb{C})$. 

Note in particular that the integrals in the pairings
\begin{multline*}
\langle \kappa_{0,1}(\phi), \xi \rangle
= \sum_{s=1}^{4} \, \int \limits_{ \mathbb{R}^4 \times \mathbb{R}^3}
\kappa_{0,1}(\phi)(s, \boldsymbol{p}) \,\, \xi(s, \boldsymbol{\p}) \, \ud^3 \boldsymbol{\p} \\
= \sum_{s=1}^{4} \, \sum_{a=1}^{4} \, \int \limits_{\mathbb{R}^3}
\kappa_{0,1}(s, \boldsymbol{p}; a, x) \, \phi^{a}(x) \,\, \xi(s, \boldsymbol{\p}) \, \ud^4 x \, \ud^3 \boldsymbol{\p},
\,\,\, \xi \in \mathcal{S}_{A}(\mathbb{R}^3, \mathbb{C}^4), \phi \in \mathscr{E},
\end{multline*}
are not merely symbolic but actual well-defined Lebesgue integrals.\footnote{Here for the case of the Dirac field.
But we have analogous situation for other fields with the standard Hilbert space
$L^2(\mathbb{R}^3; \mathbb{C}^4)$ and the standard operator $A$
in (\ref{SinglePartGelfandTriplesForPsi}) possibly replaced with corresponding standard
$L^2(\mathbb{R}^3; \mathbb{C}^n)$ and $A= \oplus H_{(3)}$ or $=\oplus A^{(3)}$. In this case
$\mathcal{S}_{A= \oplus H_{(3)}}(\mathbb{R}^3; \mathbb{C}^n) = \mathcal{S}(\mathbb{R}^3; \mathbb{C}^n)$
or $\mathcal{S}_{A = \oplus A^{(3)}}(\mathbb{R}^3; \mathbb{C}^n) = \mathcal{S}^{0}(\mathbb{R}^3; \mathbb{C}^n)$,
$\mathscr{E} = \mathcal{S}_{\oplus H_{(4)}}(\mathbb{R}^4;\mathbb{C}^n) = \mathcal{S}(\mathbb{R}^4;\mathbb{C}^n)$ or
$\mathscr{E} = \widetilde{\mathcal{S}_{\oplus A_{(4)}}(\mathbb{R}^4;\mathbb{C}^n)}
= \widetilde{\mathcal{S}^{0}(\mathbb{R}^4;\mathbb{C}^n)} = \mathcal{S}^{00}(\mathbb{R}^4;\mathbb{C}^n)$
(compare the next Section)
and with the corresponding unitary isomorphism $U$ joining the corresponding Gelfand triples analogous to
(\ref{SinglePartGelfandTriplesForPsi}). In this case the summation with respect to the indices $s,a$
runs over $\{1, 2, \ldots, n\}$.} 

We have the following
\begin{lem}\label{kappa_0,1(barphi),kappa_1,0(barphi)}
Let $\phi \in \mathscr{E} = \mathcal{S}(\mathbb{R}^4; \mathbb{C}^4)$
and $\kappa^{*}_{0,1}, \kappa^{*}_{1,0}$ be the vector-valued distributions (\ref{kappa_0,1})
and respectively (\ref{kappa_1,0}). Then 
\[
\begin{split}
\kappa^{*}_{0,1}(\overline{\phi})(s, \boldsymbol{\p}) = 
\overline{\big(P^{\oplus *}\widetilde{\phi}|_{{}_{\mathscr{O}_{m,0,0,0}}}\big)_{s+}(\boldsymbol{\p})}=
\overline{\big(P^{\oplus *}\widetilde{\phi}|_{{}_{\mathscr{O}_{m,0,0,0}}}\big)_{s}(\boldsymbol{\p})}, \,\,\,
s=1,2, \\
\kappa^{*}_{0,1}(\overline{\phi})(s, \boldsymbol{\p}) = 0, \,\,\, s=3,4, \\
\kappa^{*}_{1,0}(\overline{\phi})(s, \boldsymbol{\p}) = 0, \,\,\, s=1,2, \\
\kappa^{*}_{1,0}(\overline{\phi})(s, \boldsymbol{\p}) = 
\big(P^{\ominus *}\widetilde{\phi}|_{{}_{\mathscr{O}_{-m,0,0,0}}}\big)_{s}(\boldsymbol{\p}), \,\,\,
s=3,4,
\end{split}
\]  
where $\big(P^{\oplus *}\widetilde{\phi}|_{{}_{\mathscr{O}_{m,0,0,0}}}\big)_{s}$
stands for the $s$-th component of 
\[
U\Big( P^{\oplus *}\widetilde{\phi}|_{{}_{\mathscr{O}_{m,0,0,0}}} \oplus 0 \Big), \,\,\, \textrm{for} \,\, s= 1,2
\]
or respectively $\big(P^{\ominus *}\widetilde{\phi}|_{{}_{\mathscr{O}_{-m,0,0,0}}}\big)_{s}$ stands for the $s$-th component of 
\[
U\Big( 0 \oplus \big(P^{\ominus *}\widetilde{\phi}|_{{}_{\mathscr{O}_{-m,0,0,0}}}\big)^\flat \Big), \,\,\, \textrm{for} \,\, s= 3,4
\]
in the image of the unitary isomorphism (\ref{isomorphismU}) computed with the hepl of the fundamental solutions $\underset{*}{u} = \gamma^0u, \underset{*}{v}
= \gamma^0 v$ corresponding to the single particle Hilbert space of the Fock space corresponding to the operators
$a'(\cdot), a'(\cdot)^+$ of the field $\underset{*}{\boldsymbol{\psi}}$ given by (\ref{psi(f)=a_+(f)+a_-(f^c)^+proper*}).
\end{lem}
\qedsymbol \,
We have by definition for $s=1,2$ (up to the irrelvant constant $m$, which would be cancelled if it was counted for in the definition of the
$U$ isomorphism (\ref{isomorphismU}) according to our definition of the inner product in the single particle Hilbert space)
\begin{multline*}
\kappa^{*}_{0,1}(\overline{\phi})(s, \boldsymbol{\p})
= \sum_{a=1}^{4} 
{\textstyle\frac{1}{2}} u_{s}^{a}(\boldsymbol{\p})
\overline{\int \limits_{\mathbb{R}^4} \phi^a(x)e^{ip\cdot x} \, \ud^4 x}
=  \sum_{a=1}^{4} {\textstyle\frac{1}{2}}
u_{s}^{a}(\boldsymbol{\p})
\overline{\widetilde{\phi}^a(p_0(\boldsymbol{\p}), \boldsymbol{\p})} \\ 
= \sum_{a=1}^{4} {\textstyle\frac{1}{2}}
\overline{\overline{u_{s}^{a}(\boldsymbol{\p})}
\widetilde{\phi}^a(p_0(\boldsymbol{\p}), \boldsymbol{\p}) } = 
{\textstyle\frac{1}{2}}
\overline{u_{s}(\boldsymbol{\p})^+ 
\widetilde{\phi}(p_0(\boldsymbol{\p}), \boldsymbol{\p})} 
\end{multline*}
\begin{multline*}
={\textstyle\frac{1}{2}}
\overline{
\big(u_{s}(\boldsymbol{\p},  \widetilde{\phi}(p_0(\boldsymbol{\p}), \boldsymbol{\p}) \big)_{{}_{\mathbb{C}^4}}}
\\
=
{\textstyle\frac{1}{2p_0(\boldsymbol{\p})}}
\overline{\underset{*}{u}{}_{s}(\boldsymbol{\p})^+
\big(P^{\oplus *}\widetilde{\phi}\big)(p_0(\boldsymbol{\p}), \boldsymbol{\p}) } =
\overline{\big(P^{\oplus *}\widetilde{\phi}|_{{}_{\mathscr{O}_{m,0,0,0}}}\big)_{s}(\boldsymbol{\p})}, \,\,\,
\textrm{for $s=1,2$}.
\end{multline*}
Here the first five equalities follow by definition, the sixth equality follows from the property
(\ref{u^+P^plusPhi=u^+Phi''}) or from (\ref{u^+P^plusPhi=u^+Phi'}) and (\ref{u^+P^plusPhi=u^+Phi'})  jointly (compare Appendix \ref{fundamental,u,v}) of
$u_s(\boldsymbol{\p}), \, \underset{*}{u}{}_s(\boldsymbol{\p}) = \gamma^0 u_s(\boldsymbol{\p})$. 
The last term $\overline{\big(P^{\oplus *}\widetilde{\phi}|_{{}_{\mathscr{O}_{m,0,0,0}}}\big)_{s}}$ 
is equal to the complex conjugation of 
the $s$-th direct summand in 
\[
U\Big( P^{\oplus *}\widetilde{\phi}|_{{}_{\mathscr{O}_{m,0,0,0}}} \oplus 0 \Big), \,\,\, \textrm{for} \,\, s= 1,2
\]
by definition (\ref{isomorphismU}) of the unitary isomorphism $U$ determined by the fundamental solutions 
$\underset{*}{u} = \gamma^0 u, \, \underset{*}{v} = \gamma^0 v$, and computed for the Fock space 
of the field $\underset{*}{\boldsymbol{\psi}}$. 

Similarly, we have  by definition for $s=3,4$ (up to the constant factor $m$ by the same reason as for $s=1,2$)
\begin{multline*}
\kappa^{*}_{1,0}(\overline{\phi})(s, \boldsymbol{\p})
\\
=
- \sum_{a=1}^{4} 
{\textstyle\frac{1}{2}}
v_{s-2}^{a}(\boldsymbol{\p})
\overline{\int \limits_{\mathbb{R}^4} \phi^a(x)e^{-ip\cdot x} \, \ud^4 x}
=  -\sum_{a=1}^{4} 
{\textstyle\frac{1}{2}}
v_{s-2}^{a}(\boldsymbol{\p})
\overline{\widetilde{\phi}^a(-|p_0(\boldsymbol{\p})|, -\boldsymbol{\p})} \\ 
= -\sum_{a=1}^{4}
{\textstyle\frac{1}{2}}
\overline{
\overline{v_{s-2}^{a}(\boldsymbol{\p})}
\widetilde{\phi}^a(-|p_0(\boldsymbol{\p})|, -\boldsymbol{\p}) } = 
- {\textstyle\frac{1}{2}}
\overline{
v_{s-2}(\boldsymbol{\p})^+ 
\widetilde{\phi}(-|p_0(\boldsymbol{\p})|, -\boldsymbol{\p})} 
\end{multline*}
\begin{multline*}
= -{\textstyle\frac{1}{2}}
\overline{
\big(v_{s-2}(\boldsymbol{\p}), \widetilde{\phi}(-|p_0(\boldsymbol{\p})|, -\boldsymbol{\p}) \big)_{{}_{\mathbb{C}^4}}
}
\\
=
\overline{
{\textstyle\frac{1}{2|p_0(\boldsymbol{\p})|}}
 \underset{*}{v}{}_{s-2}(\boldsymbol{\p})^+
\big(P^{\ominus *}\widetilde{\phi}\big)(-|p_0(\boldsymbol{\p})|, -\boldsymbol{\p}) } =
\big(P^{\ominus *}\widetilde{\phi}|_{{}_{\mathscr{O}_{-m,0,0,0}}}\big)_{s}(\boldsymbol{\p}), \,\,\,
\textrm{for $s=3,4$}.
\end{multline*}
Here the equalities follow by definition, except the sixth equality, which follows from the property
(\ref{v^+P^minusPhi=v^+Phi''}) or from the properties (\ref{v^+P^minusPhi=v^+Phi}) and (\ref{v^+P^minusPhi=v^+Phi'}) jointly 
(compare Appendix \ref{fundamental,u,v}) of the fundamental solutions $v_s(\boldsymbol{\p}), \, \underset{*}{v}{}_s(\boldsymbol{\p}),
= \gamma^0 v_s(\boldsymbol{\p})$. 
The last term $\big(P^{\ominus *}\widetilde{\phi}|_{{}_{\mathscr{O}_{-m,0,0,0}}}\big)_{s}$ is equal to the  
$s$-th direct summand in 
\[
U\Big( 0 \oplus \big(P^{\ominus *}\widetilde{\phi}|_{{}_{\mathscr{O}_{-m,0,0,0}}}\big)^\flat \Big), \,\,\, 
\textrm{for} \,\, s= 3,4,
\]
by definition (\ref{isomorphismU}) of the unitary isomorphism $U$ 
determined by the fundamental solutions $\underset{*}{u} = \gamma^0 u, \, \underset{*}{v} = \gamma^0 v$ 
for the Fock space of the field $\underset{*}{\boldsymbol{\psi}}$.

The rest part:
\[
\begin{split}
\kappa^{*}_{0,1}(\overline{\phi})(s, \boldsymbol{\p}) = 0, \,\,\, s=3,4, \\
\kappa^{*}_{1,0}(\overline{\phi})(s, \boldsymbol{\p}) = 0, \,\,\, s=1,2, \\
\end{split}
\] 
of our Lemma follows immediately from definition (\ref{kappa_0,1})
and respectively (\ref{kappa_1,0}) of the distributions
$\kappa^{*}_{0,1}, \kappa^{*}_{1,0}$.
\qed

From Lemma \ref{kappa_0,1(barphi),kappa_1,0(barphi)} and from (\ref{a(U(u+v))=a'(u+v)})
it follows
\begin{lem}\label{psi=integKerOpVecValProof} 
Let $\kappa^{*}_{0,1}$ and $\kappa^{*}_{1,0}$ be the vector-valued distributions
(\ref{kappa_0,1}) and respectively (\ref{kappa_1,0}). Then
the equality (\ref{psi=IntKerOpVectValKer}) holds true:
\begin{multline*}
\underset{*}{\boldsymbol{\psi}}(\overline{\phi}) = a'\big(P^{\oplus *}\widetilde{\phi}|_{{}_{\mathscr{O}_{m,0,0,0}}} \oplus 0\big) + 
a'\Big( 0 \oplus \big(P^{\ominus *}\widetilde{\phi}|_{{}_{\mathscr{O}_{-m,0,0,0}}}\big)^\flat \Big)^+ \\
=  a\Big(U\big(P^{\oplus *}\widetilde{\phi}|_{{}_{\mathscr{O}_{m,0,0,0}}} \oplus 0\big)\Big) + 
a\Bigg(U\Big( 0 \oplus \big(P^{\ominus *}\widetilde{\phi}|_{{}_{\mathscr{O}_{-m,0,0,0}}}\big)^\flat \Big)\Bigg)^+ \\
 = \sum_{s=1}^{4} \, \int \limits_{\mathbb{R}^3} 
\kappa^{*}_{0,1}(\overline{\phi})(s, \boldsymbol{p}) \,\, \partial_{s, \boldsymbol{\p}} \, \ud^3 \boldsymbol{\p}
+
\sum_{s=1}^{4} \, \int \limits_{\mathbb{R}^3} 
\kappa^{*}_{1,0}(\overline{\phi})(s, \boldsymbol{p}) \,\, \partial_{s, \boldsymbol{\p}}^* \, \ud^3 \boldsymbol{\p} \\
= \Xi_{0,1}\big(\kappa^{*}_{0,1}(\overline{\phi})\big) + \Xi_{1,0}\big(\kappa^{*}_{1,0}(\overline{\phi})\big), \,\,\,\,\,\,
\phi \in \mathscr{E} = \mathcal{S}(\mathbb{R}^4; \mathbb{C}^4).
\end{multline*}
\end{lem}
\qedsymbol \, 
Indeed, we have
\begin{multline*}
\sum_{s=1}^{4} \int \limits_{\mathbb{R}^4} \kappa^{*}_{0,1}(\overline{\phi})(s, \boldsymbol{\p}) \, 
\partial_{s, \boldsymbol{\p}} \, \ud^3 \boldsymbol{p} =
\sum_{s=1}^{2} \int \limits_{\mathbb{R}^3} 
\overline{\big( P^{\oplus *} \widetilde{\phi}|_{{}_{\mathscr{O}_{m,0,0,0}}} \big)_{s}(\boldsymbol{\p})} \, \partial_{s, \boldsymbol{\p}}
\, \ud^3 \boldsymbol{\p} \\
= a\Big( \,  \big( P^{\oplus *} \widetilde{\phi}|_{{}_{\mathscr{O}_{m,0,0,0}}} \big)_{1} \oplus
\big( P^{\oplus *} \widetilde{\phi}|_{{}_{\mathscr{O}_{m,0,0,0}}} \big)_{2} \oplus 0 \oplus 0 \Big)
= a\Big( \,  U \big( P^{\oplus *} \widetilde{\phi}|_{{}_{\mathscr{O}_{m,0,0,0}}} \oplus 0 \big) \, \Big) \\ =
a'\Big( P^{\oplus *} \widetilde{\phi}|_{{}_{\mathscr{O}_{m,0,0,0}}} \oplus 0 \Big),
\end{multline*}
with the operators $a'(\cdot),a'(\cdot)^+$ and the isomorphism $U$ corresponding to the field
$\underset{*}{\boldsymbol{\psi}}$. Here the first three equalities  follow from Lemma \ref{kappa_0,1(barphi),kappa_1,0(barphi)},
and Corollary \ref{D_xi=int(xiPartial)}, the last equality follows from 
(\ref{a(U(u+v))=a'(u+v)}).

Similarly, we have
\begin{multline*}
\sum_{s=1}^{4} \int \limits_{\mathbb{R}^4} \kappa^{*}_{1,0}(\overline{\phi})(s, \boldsymbol{\p}) \, 
\partial_{s, \boldsymbol{\p}}^{*} \, \ud^3 \boldsymbol{p} =
\sum_{s=3}^{4} \int \limits_{\mathbb{R}^3} 
\big( P^{\ominus *} \widetilde{\phi}|_{{}_{\mathscr{O}_{-m,0,0,0}}} \big)_{s}(\boldsymbol{\p}) \, 
\partial_{s, \boldsymbol{\p}}^{*}
\, \ud^3 \boldsymbol{\p} \\
= a\Big( \,  0 \oplus 0 \oplus  \big( P^{\ominus *} \widetilde{\phi}|_{{}_{\mathscr{O}_{m,0,0,0}}} \big)_{3} \oplus
\big( P^{\ominus *} \widetilde{\phi}|_{{}_{\mathscr{O}_{m,0,0,0}}} \big)_{4} \, \Big)
= a\Big( \,  U \big( 0 \oplus (P^{\ominus *} \widetilde{\phi}|_{{}_{\mathscr{O}_{-m,0,0,0}}})^\flat  \big) \, \Big) \\ =
a'\Big( 0 \oplus (P^{\ominus *} \widetilde{\phi}|_{{}_{\mathscr{O}_{-m,0,0,0}}})^\flat \Big),
\end{multline*}
with the operators $a'(\cdot),a'(\cdot)^+$ and the isomorphism $U$ corresponding to the field
$\underset{*}{\boldsymbol{\psi}}$. 
Here the first three equalities  follow from Lemma \ref{kappa_0,1(barphi),kappa_1,0(barphi)},
and Corollary \ref{D_xi=int(xiPartial)}, the last equality follows from 
(\ref{a(U(u+v))=a'(u+v)}).
\qed

Proof of the equalities (\ref{psi=IntKerOpVectValKer'}) being completely analogous to the presented proof
of (\ref{psi=IntKerOpVectValKer}) may be omitted. They represent the quantum field counterpart of the
non-invariant pairing (\ref{NonInvariantBispinorPairing}), and represent the rigorous form
of the non-invariant pairing (\ref{psi(f)-symbollical}) for the Dirac fields $\underset{*}{\boldsymbol{\psi}}$
and $\boldsymbol{\psi}$. We have also the quantum field analogues for the remaining pairings, e.g. the invariant pairing
(\ref{InvariantBispinorPairing}).
In particular, we have the following rigorous quantum field invariant pairing (\ref{psi++(f)psi+(f)}):
\begin{multline*}
\underset{*}{\boldsymbol{\psi}}^\sharp(\phi) = a'\big(\gamma^0 P^{\oplus}\widetilde{\phi}|_{{}_{\mathscr{O}_{m,0,0,0}}} \oplus 0\big)^+ + 
a'\Big( 0 \oplus \big(\gamma^0 P^{\ominus}\widetilde{\phi}|_{{}_{\mathscr{O}_{-m,0,0,0}}}\big)^\flat \Big) \\
=  a\Big(U\big(\gamma^0 P^{\oplus}\widetilde{\phi}|_{{}_{\mathscr{O}_{m,0,0,0}}} \oplus 0\big)\Big)^+ + 
a\Bigg(U\Big( 0 \oplus \big(\gamma^0 P^{\ominus}\widetilde{\phi}|_{{}_{\mathscr{O}_{-m,0,0,0}}}\big)^\flat \Big)\Bigg) \\
 = \sum_{s=1}^{4} \, \int \limits_{\mathbb{R}^3} 
\kappa^{* \sharp }_{0,1}(\phi)(s, \boldsymbol{p}) \,\, \partial_{s, \boldsymbol{\p}} \, \ud^3 \boldsymbol{\p}
+
\sum_{s=1}^{4} \, \int \limits_{\mathbb{R}^3} 
\kappa^{* \sharp }_{1,0}(\phi)(s, \boldsymbol{p}) \,\, \partial_{s, \boldsymbol{\p}}^* \, \ud^3 \boldsymbol{\p} \\
= \Xi_{0,1}\big(\kappa^{* \sharp}_{0,1}(\phi)\big) + \Xi_{1,0}\big(\kappa^{* \sharp}_{1,0}(\phi)\big)
= \int \underset{*}{\boldsymbol{\psi}}^\sharp(x) \, \phi(x) \ud^4x, \,\,\,\,\,\,
\phi \in \mathscr{E} = \mathcal{S}(\mathbb{R}^4; \mathbb{C}^4),
\end{multline*}
with the operators $a'(\cdot),a'(\cdot)^+$ and the isomorphism $U$ corresponding to the field
$\underset{*}{\boldsymbol{\psi}}$;

\begin{multline*}
\boldsymbol{\psi}^\sharp(\phi) = a'\big(\gamma^0 P^{\oplus *}\widetilde{\phi}|_{{}_{\mathscr{O}_{m,0,0,0}}} \oplus 0\big)^+ +
a'\Big( 0 \oplus \big(\gamma^0 P^{\ominus *}\widetilde{\phi}|_{{}_{\mathscr{O}_{-m,0,0,0}}}\big)^\flat \Big) \\
= a\Big(U\big(\gamma^0 P^{\oplus *}\widetilde{\phi}|_{{}_{\mathscr{O}_{m,0,0,0}}} \oplus 0\big)\Big)^+ +
a\Bigg(U\Big( 0 \oplus \big(\gamma^0 P^{\ominus *}\widetilde{\phi}|_{{}_{\mathscr{O}_{-m,0,0,0}}}\big)^\flat \Big)\Bigg) \\
= \sum_{s=1}^{4} \, \int \limits_{\mathbb{R}^3}
\kappa^{\sharp}_{0,1}(\phi)(s, \boldsymbol{p}) \,\, \partial_{s, \boldsymbol{\p}} \, \ud^3 \boldsymbol{\p}
+
\sum_{s=1}^{4} \, \int \limits_{\mathbb{R}^3}
\kappa^{\sharp}_{1,0}(\phi)(s, \boldsymbol{p}) \,\, \partial_{s, \boldsymbol{\p}}^* \, \ud^3 \boldsymbol{\p} \\
= \Xi_{0,1}\big(\kappa^{\sharp}_{0,1}(\phi)\big) + \Xi_{1,0}\big(\kappa^{\sharp}_{1,0}(\phi)\big)
= \int \boldsymbol{\psi}^\sharp(x) \, \phi(x) \ud^4x, \,\,\,\,\,\,
\phi \in \mathscr{E} = \mathcal{S}(\mathbb{R}^4; \mathbb{C}^4),
\end{multline*}
with the operators $a'(\cdot),a'(\cdot)^+$ and the isomorphism $U$ corresponding to the field
$\boldsymbol{\psi}$. Here
\[
\begin{split}
\kappa^{\sharp}_{l,m}(s, \boldsymbol{p}; a,x) 
= \sum\limits_{b}\big[\gamma^0\big]^{ab}
\overline{
\kappa_{m,l}(s, \boldsymbol{p}; b,x) 
},
\\
\kappa^{* \sharp}_{l,m}(s, \boldsymbol{p}; a,x) 
= \sum\limits_{b}\big[\gamma^0\big]^{ab}
\overline{
\kappa^{*}_{m,l}(s, \boldsymbol{p}; b,x) 
},
\end{split}
\]
\[
l,m = 0,1 \,\, \textrm{or} \,\, 1,0.
\]

Although in the sequel we formulate and prove several lemmas for the kernels $\kappa_{l,m}$, $l,m=0,1$ or $l,m=1,0$, of the Dirac field
$\boldsymbol{\psi}$ they are obviously true for the kernels  $\kappa^{*}_{l,m}$, $l,m=0,1$ or $1,0$,  
of the field $\underset{*}{\boldsymbol{\psi}}$, and for Dirac field with any other gamma Clifford generators.

Let $\mathcal{O}_C = \mathcal{O}_C(\mathbb{R}^4; \mathbb{C}^4)$ be the predual of 
the Schwartz algebra of convolutors $\mathcal{O}'_{C} = \mathcal{O}'_{C}(\mathbb{R}^4; \mathbb{C}^4)$,
which means that each component of each element of $\mathcal{O}_C$ belongs to the Horv\'ath
predual $\mathcal{O}_{C}(\mathbb{R}^4; \mathbb{C})$ of the ordinary Schwartz convolution
algebra $\mathcal{O}'_{C}(\mathbb{R}^4; \mathbb{C})$. For detailed construction and definition of
$\mathcal{O}'_{C}(\mathbb{R}^4; \mathbb{C})$ and $\mathcal{O}_{C}(\mathbb{R}^4; \mathbb{C})$,
compare \cite{Schwartz}, \cite{Horvath} or \cite{Kisynski}, or finally compare the summary of their
properties presented in Appendix \ref{convolutorsO'_C}.

The following Lemma holds true (and we have in general analogous Lemma for a local field
understood as a sum of integral kernel operators with vector-valued kernels) 
\begin{lem}\label{kappa0,1,kappa1,0psi}
For the $\mathscr{L}(\mathscr{E},\mathbb{C})$-valued (or $\mathscr{E}^*$ -valued) distributions 
$\kappa_{0,1}, \kappa_{1,0}$, given by (\ref{kappa_0,1}) and (\ref{kappa_1,0}),
in the equality (\ref{psi=IntKerOpVectValKer}) defining the Dirac $\psi$ field, for the 
the pairings $\kappa_{0,1}(\xi)$, $\kappa_{1,0}(\xi)$ of
$\kappa_{0,1}, \kappa_{1,0}$ with single particle test functions $\xi$,
we have
\begin{multline*}
\Bigg( \, (a,x) \mapsto \sum_{s} \, \int \limits_{\mathbb{R}^3}
\kappa_{0,1}(s, \boldsymbol{\p}; a,x)\, \xi(s,\boldsymbol{\p}) \,\ud^3\boldsymbol{\p} = \kappa_{0,1}(\xi)(a,x) \,\, \Bigg) 
\in \mathcal{O}_C \subset \mathcal{O}_M \subset \mathscr{E}^*, 
\\
\xi \in \mathcal{S}_{A}(\mathbb{R}^3, \mathbb{C}^4), 
\\
\Bigg( \, (a,x) \mapsto \sum_{s} \, \int \limits_{\mathbb{R}^3}
\kappa_{1,0}(s, \boldsymbol{\p}; a,x) \, \xi(s, \boldsymbol{\p}) \,\ud^3\boldsymbol{\p} = \kappa_{1,0}(\xi)(a,x) \,\, \Bigg)  
\in \mathcal{O}_C \subset \mathcal{O}_M \subset \mathscr{E}^*, 
 \\ 
\xi \in \mathcal{S}_{A}(\mathbb{R}^3, \mathbb{C}^4), 
\end{multline*}
are smooth, with their Fourier transforms $\widetilde{\kappa_{0,1}(\xi)}$, $\widetilde{\kappa_{1,0}(\xi)}$  
concentrated, respectively, on the negative or positive energy orbit 
$\mathscr{O}_{{}_{\pm m, 0,0,0}}$, and equal there 
\[
\begin{split}
\widetilde{\kappa_{0,1}(\xi)}(p_0(\boldsymbol{\p}), \boldsymbol{\p})
= \sum_{s =1}^{2} \, |p_0(\boldsymbol{\p})|{\textstyle\frac{\underset{*}{u}{}_{s}^{a}(\boldsymbol{\p})}{2}}
 \, \xi(s,\boldsymbol{\p}),
\\
\widetilde{\kappa_{1,0}(\xi)}(-p_0(\boldsymbol{\p}), -\boldsymbol{\p})
=-\sum_{s =3}^{4} \, |p_0(\boldsymbol{\p})| {\textstyle\frac{\underset{*}{v}{}_{s-2}^{a}(\boldsymbol{\p})}{2}}
 \, \xi(s,\boldsymbol{\p}),
\end{split}
\]
\[
\,\,\,\, p_0(\boldsymbol{\p}) = \sqrt{|\boldsymbol{\p}|^2 + m^2}, 
\] 
to elements of $\mathcal{S}_{A}(\mathbb{R}^3; \mathbb{C}^4)$.
In particular $\kappa_{0,1}(\xi)$, $\kappa_{1,0}(\xi)$ have all space-time derivatives bounded. For the pairings of
$\kappa_{0,1}, \kappa_{1,0}$ with space-time test functions $\phi$ we have
\begin{multline*}
\Bigg( \, (s,\boldsymbol{\p}) \mapsto \sum_{a} \, \int \limits_{\mathbb{R}^4}
\kappa_{0,1}(s, \boldsymbol{\p}; a,x) \, \phi^a(x) \,\ud^4x = \kappa_{0,1}(\phi)(s,\boldsymbol{\p}) \,\, \Bigg) \in 
\mathcal{S}_{A}(\mathbb{R}^3, \mathbb{C}^4), \,\, \phi \in \mathscr{E}, \\
\Bigg( \, (s,\boldsymbol{\p}) \mapsto \sum_{a} \, \int \limits_{\mathbb{R}^4}
\kappa_{1,0}(s, \boldsymbol{\p}; a,x) \, \phi^a(x) \,\ud^4x = \kappa_{1,0}(\phi)(s,\boldsymbol{\p}) \,\, \Bigg) \in 
\mathcal{S}_{A}(\mathbb{R}^3, \mathbb{C}^4), 
 \,\, \phi \in \mathscr{E}. 
\end{multline*}

The maps
\[
\begin{split}
\kappa_{0,1}: \mathcal{S}_{A}(\mathbb{R}^3, \mathbb{C}^4) \ni \xi \longmapsto \kappa_{0,1}(\xi) 
\in \mathcal{O}_M(\mathbb{R}^4), \\
\kappa_{1,0}: \mathcal{S}_{A}(\mathbb{R}^3, \mathbb{C}^4) \ni \xi \longmapsto \kappa_{1,0}(\xi) 
\in \mathcal{O}_M(\mathbb{R}^4),
\end{split}
\]
(for $\kappa_{0,1}, \kappa_{1,0}$ understood as maps 
$\mathscr{L}\big( \mathcal{S}_{A}(\mathbb{R}^3, \mathbb{C}^4), \,\,
\mathscr{L}(\mathscr{E}, \mathbb{C})  \big) \cong 
\mathscr{L}\big( \mathcal{S}_{A}(\mathbb{R}^3, \mathbb{C}^4), \,\,
\mathscr{E}^*  \big)$) with the Schwartz' operator topology on $\mathcal{O}_M$.

Moreover the maps 
\[
\begin{split}
\kappa_{0,1}: \mathscr{E} \ni \phi \longmapsto \kappa_{0,1}(\phi) 
\in \mathcal{S}_{A}(\mathbb{R}^3, \,\, \mathbb{C}^4), \\
\kappa_{1,0}: \mathscr{E} \ni \phi \longmapsto \kappa_{1,0}(\phi) 
\in \mathcal{S}_{A}(\mathbb{R}^3, \,\, \mathbb{C}^4)
\end{split}
\]
are continuous (for $\kappa_{0,1}, \kappa_{1,0}$ understood as maps in 
\[
\mathscr{L}\big( \mathscr{E}, \,\, \big(\mathcal{S}_{A}(\mathbb{R}^3, \mathbb{C}^4)^* 
\big) \cong \mathscr{L}\big( \mathcal{S}_{A}(\mathbb{R}^3, \mathbb{C}^4), \,\,
\mathscr{L}(\mathscr{E}, \mathbb{C})  \big)) 
\]
and, equivalently,
the maps $\xi \longmapsto \kappa_{0,1}(\xi)$, $\xi \longmapsto \kappa_{1,0}(\xi)$ can be extended to
continuous maps
\[
\begin{split}
\kappa_{0,1}: \mathcal{S}_{A}(\mathbb{R}^3, \mathbb{C}^4)^* \ni \xi \longmapsto \kappa_{0,1}(\xi) 
\in \mathscr{E}^*, \\
\kappa_{1,0}: \mathcal{S}_{A}(\mathbb{R}^3, \mathbb{C}^4)^* \ni \xi \longmapsto \kappa_{1,0}(\xi) 
\in \mathscr{E}^*,
\end{split}
\]
(for $\kappa_{0,1}, \kappa_{1,0}$ understood as maps 
$\mathscr{L}\big( \mathcal{S}_{A}(\mathbb{R}^3, \mathbb{C}^4), \,\,
\mathscr{L}(\mathscr{E}, \mathbb{C})  \big) \cong 
\mathscr{L}\big( \mathcal{S}_{A}(\mathbb{R}^3, \mathbb{C}^4), \,\,
\mathscr{E}^*  \big)$). Therefore 
not only $\kappa_{0,1}, \kappa_{1,0}
\in \mathscr{L}\big( \mathcal{S}_{A}(\mathbb{R}^3, \mathbb{C}^4), \,\,
\mathscr{L}(\mathscr{E}, \mathbb{C})  \big)$, but both $\kappa_{0,1}, \kappa_{1,0}$
can be (uniquely) extended to elements of 
\[
\mathscr{L}\big( \mathcal{S}_{A}(\mathbb{R}^3, \mathbb{C}^4)^*, \,\,
\mathscr{L}(\mathscr{E}, \mathbb{C})  \big) \cong 
\mathscr{L}\big( \mathcal{S}_{A}(\mathbb{R}^3, \mathbb{C}^4)^*, \,\,
\mathscr{E}^*  \big)  \cong
\mathscr{L}\big( \mathscr{E}, \,\, 
\mathcal{S}_{A}(\mathbb{R}^3, \mathbb{C}^4)  \big).
\]
\end{lem}
\qedsymbol \,
That for each $\xi \in \mathcal{S}_{A}(\mathbb{R}^3, \mathbb{C}^4)$ the functions
$\kappa_{0,1}(\xi), \kappa_{1,0}(\xi)$ given by (here $x = (x_0, \boldsymbol{\x})$)
\begin{multline}\label{Dirackappa01(xi)}
(a,x) \mapsto \kappa_{0,1}(\xi)(a,x) = \sum_{s=1}^{4} \, \int \limits_{\mathbb{R}^3}
\kappa_{0,1}(s, \boldsymbol{\p}; a,x)\, \xi(s,\boldsymbol{\p}) \,\ud^3\boldsymbol{\p} 
\\
=
\sum_{s =1}^{2} \, \int \limits_{\mathbb{R}^3}
{\textstyle\frac{\underset{*}{u}{}_{s}^{a}(\boldsymbol{\p})}{2}}
 \, \xi(s,\boldsymbol{\p}) 
e^{-ip_0(\boldsymbol{\p})x_0 + i\boldsymbol{\p} \cdot \boldsymbol{\x}} \, \ud^3 \boldsymbol{\p}
\\
= \int \delta(p^2-m^2)\theta(p_0) \zeta^a(p) e^{-ip\cdot x} \, \ud^4p, \,\,\,\, \zeta^a(p_0,\boldsymbol{\p}) =  
\sum_{s =1}^{2} \, |p_0|{\textstyle\frac{\underset{*}{u}{}_{s}^{a}(\boldsymbol{\p})}{2}}
 \, \xi(s,\boldsymbol{\p}) 
\end{multline}
\begin{multline}\label{Dirackappa10(xi)}
(a,x) \mapsto \kappa_{1,0}(\xi)(a,x) =  \sum_{s=1}^{4} \, \int \limits_{\mathbb{R}^3}
\kappa_{1,0}(s, \boldsymbol{\p}; a,x)\, \xi(s,\boldsymbol{\p}) \,\ud^3\boldsymbol{\p}
\\
 =
- \sum_{s=3}^{4} \, \int \limits_{\mathbb{R}^3}
{\textstyle\frac{\underset{*}{v}{}_{s-2}^{a}(\boldsymbol{\p})}{2}}
 \, \xi(s,\boldsymbol{\p}) 
e^{i|p_0(\boldsymbol{\p})|x_0 - i\boldsymbol{\p} \cdot \boldsymbol{\x}} \, \ud^3 \boldsymbol{\p}
\\
= \int \delta(p^2-m^2)\theta(-p_0) \zeta^a(-p)e^{-ip\cdot x} \, \ud^4p, \,\,\,\,\, \zeta^a(p_0,\boldsymbol{\p}) =  
-\sum_{s =3}^{4} \, |p_0|{\textstyle\frac{\underset{*}{v}{}_{s-2}^{a}(\boldsymbol{\p})}{2}}
 \, \xi(s,\boldsymbol{\p}),
\end{multline}
have Fourier transforms $\widetilde{\kappa_{0,1}(\xi)}$, $\widetilde{\kappa_{0,1}(\xi)}$ concentrated respectively
on $\mathscr{O}_{{}_{\pm m, 0,0,0}}$ and equal there to elements $\zeta\big|_{\mathscr{O}_{{}_{\pm m, 0,0,0}}}$
of $\mathcal{S}_{A}(\mathbb{R}^3; \mathbb{C}^4)
= \mathcal{S}(\mathbb{R}^3; \mathbb{C}^4)$ is obvious from the formulas (\ref{Dirackappa01(xi)}) and (\ref{Dirackappa10(xi)}).
In particular it is obvius that $\kappa_{0,1}(\xi)$ and $\kappa_{1,0}(\xi)$ are smooth with all derivatives being bounded.
In particular that $\kappa_{0,1}(\xi)$ and $\kappa_{1,0}(\xi)$ 
belong to $\mathcal{O}_C \subset \mathcal{O}_M \subset \mathscr{E}^*$ is immediate. 
Indeed, that they are smooth is obvious, similarly as it is obvious the existence of such a natural $N$ (it is sufficient to take here $N=0$)
that for each multi-index $\alpha \in \mathbb{N}^4$ the functions
\[
(a,x) \mapsto (1 + |x|^2)^{-N} |\partial^{\alpha}\kappa_{0,1}(\xi)(a,x)|, \,\,\,
(a,x) \mapsto (1 + |x|^2)^{-N} |\partial^{\alpha}\kappa_{1,0}(\xi)(a,x)|
\]
are bounded. Here $\partial^{\alpha}\kappa_{l,m}(\xi)$ denotes the ordinary derivative of 
the function $\kappa_{l,m}(\xi)$
of $|\alpha| = \alpha_0 + \alpha_1+\alpha_2+ \alpha_3$ order with respect to space-time 
variables $x= (x_0, x_1, x_2, x_3)$; and here 
$|x|^2= (x_{0})^2 + (x_{1})^2 + (x_{2})^2+ (x_{3})^2$.
The  statement $\kappa_{0,1}(\xi)$, $\kappa_{0,1}(\xi)$  $\in \mathcal{O}_C(\mathbb{R}^4; \mathbb{C})$ 
means that if we fix the value of the discrete index $a$ in the above functions 
\[
(a,x) \mapsto \kappa_{0,1}(\xi)(a,x), \,\,\,\,\, (a,x) \mapsto \kappa_{1,0}(\xi)(a,x), 
\]
then we obtain functions which belong to the Horv{\'a}t's predual of the algebra $\mathcal{O}'_C(\mathbb{R}^4; \mathbb{C})$ of convolutors
of the algebra 
\[\mathcal{S}(\mathbb{R}^4; \mathbb{C}) = \mathcal{S}_{H_{(4)}}(\mathbb{R}^4; \mathbb{C})
\]
of $\mathbb{C}$-valued functions, compare Appendix \ref{convolutorsO'_C}.

That $\xi \mapsto \kappa_{0,1}(\xi), \kappa_{1,0}(\xi) \in \mathcal{O}_M(\mathbb{R}^4)$ are continuous
follows easily if we use the system of norms $p_{{}_{m, \omega}}$, $\omega \in \mathcal{S}(\mathbb{R}^4)$, $m= 0,1,2, \ldots$ 
\[
p_{{}_{m, \omega}}(f) = \underset{|\alpha| \leq m}{\textrm{sup}} \big|\omega. \partial^\alpha f \big|_{{}_{L^\infty}}, \,\,\, 
f \in \mathcal{O}_M(\mathbb{R}^4),
\]
defining the topology in $\mathcal{O}_M(\mathbb{R}^4)$, compare \cite{Larcher}. The proof
that for each $\omega \in \mathcal{S}(\mathbb{R}^4)$ and natural $m$ there exist a constant $C_m$
and a norm $|\cdot |_{q(m)} = | A^{q(m)} \cdot |_{{}_{L^2}}$, 
from the system of norms defining the nuclear space $\mathcal{S}_{A}(\mathbb{R}^3; \mathbb{C}^4)
= \mathcal{S}(\mathbb{R}^3; \mathbb{C}^4)$, with
$A = \oplus H_{(3)}$, such that
\[
p_{{}_{m, \omega}}\big(\kappa_{\ell,m}(\xi)\big) \leq C_m \big|\omega\big|_{{}_{\infty}} |\xi|_{q(m)}
\]
is simple in this massive case. We  have presented a full proof of this statement for the less
trivial massless case in Lemma \ref{kappa0,1,kappa1,0ForA}, Subsection \ref{A=Xi0,1+Xi1,0}.
We refer the reader there for a comparison. Because the components of the functions of momentum
$\boldsymbol{\p} \mapsto u(\boldsymbol{\p})/p_0(\boldsymbol{\p}), \, v(\boldsymbol{\p})/p_0(\boldsymbol{\p})$ 
defining the kernels $\kappa_{0,1}, \kappa_{1,0}$ are multipliers 
of $\mathcal{S}_{A^{(3)}}(\mathbb{R}^3; \mathbb{C})
= \mathcal{S}(\mathbb{R}^3; \mathbb{C})$, then the proof given there can be repeated without any difficulty
in this simpler massive case.

Consider now the functions (in both formulas below the variable $p = (|p_0(\boldsymbol{\p})|, \boldsymbol{\p})$ 
is restricted to the \emph{positive} energy orbit $\mathscr{O}_{m,0,0,0}$)
\[
\begin{split}
(s, \boldsymbol{\p}) \mapsto \kappa_{0,1}(\phi)(s, \boldsymbol{\p})
= \sum_{a=1}^{4} 
{\textstyle\frac{\underset{*}{u}{}_{s}^{a}(\boldsymbol{\p})}{2}}
\int \limits_{\mathbb{R}^3}
\phi^{a}(x) e^{-ip \cdot x} \, \ud^4 x \\
=  \sum_{a=1}^{4} 
{\textstyle\frac{\underset{*}{u}{}_{s}^{a}(\boldsymbol{\p})}{2}}
\widetilde{\phi}^{a}|_{{}_{\mathscr{O}_{-m,0,0,0}}}(-p), \\
(s, \boldsymbol{\p}) \mapsto \kappa_{1,0}(\phi)(s, \boldsymbol{\p})
= -\sum_{a=1}^{4} 
{\textstyle\frac{\underset{*}{v}{}_{s}^{a}(\boldsymbol{\p})}{2}}
\int \limits_{\mathbb{R}^3}
\phi^{a}(x) e^{ip \cdot x} \, \ud^4 x \\
=  -\sum_{a=1}^{4} 
{\textstyle\frac{\underset{*}{v}{}_{s}^{a}(\boldsymbol{\p})}{2}}
\widetilde{\phi}^{a}|_{{}_{\mathscr{O}_{m,0,0,0}}}(p),
\end{split}
\] 
with $\phi \in \mathcal{S}(\mathbb{R}^4; \mathbb{C}^4)$. That both functions
$\kappa_{0,1}(\phi), \kappa_{1,0}(\phi)$ depend continuously on $\phi$ as maps 
\[
\mathscr{E} = \mathcal{S}(\mathbb{R}^4; \mathbb{C}^4)
\longrightarrow \mathcal{S}_{A}(\mathbb{R}^3, \,\, \mathbb{C}^4) = \mathcal{S}(\mathbb{R}^3, \,\, \mathbb{C}^4)
\]
follows from: 1) continuity of the Fourier transform as a map on the Schwartz space, as well as 2)
from the continuity of the restriction to the orbits $\mathscr{O}_{m,0,0,0}$ and $\mathscr{O}_{-m,0,0,0}$
(with $m \neq 0$) regarded as a map from $\mathcal{S}(\mathbb{R}^4;\mathbb{C})$ into 
$\mathcal{S}(\mathbb{R}^3;\mathbb{C})$, and finally 3) from the fact that the 
functions $\boldsymbol{\p} \mapsto \frac{u_{s}^{a}(\boldsymbol{\p})}{2|p_0(\boldsymbol{\p})|}$ and 
$\boldsymbol{\p} \mapsto \frac{v_{s}^{a}(\boldsymbol{\p})}{2|p_0(\boldsymbol{\p})|}$
are multipliers of the Schwartz algebra $\mathcal{S}(\mathbb{R}^3;\mathbb{C})$,
compare Appendix \ref{fundamental,u,v} and Appendix \ref{convolutorsO'_C}. 
\qed

\begin{rem*}
Note here that the continuity of the maps
\[
\begin{split}
\kappa_{0,1}: \mathscr{E} \ni \phi \longmapsto \kappa_{0,1}(\phi) 
\in \mathcal{S}_{A}(\mathbb{R}^3, \,\, \mathbb{C}^4), \\
\kappa_{1,0}: \mathscr{E} \ni \phi \longmapsto \kappa_{1,0}(\phi) 
\in \mathcal{S}_{A}(\mathbb{R}^3, \,\, \mathbb{C}^4)
\end{split}
\]
is based on the continuity of the restriction to the orbits 
$\mathscr{O}_{m,0,0,0}$ and $\mathscr{O}_{-m,0,0,0}$, regarded as a map
$\widetilde{\mathscr{E}} = \mathcal{S}(\mathbb{R}^4;\mathbb{C}) \rightarrow \mathcal{S}(\mathbb{R}^3;\mathbb{C})$
between the ordinary Schwartz spaces. This continuity breaks down for the orbit 
equal to the light cone $\mathscr{O}_{1,0,0,1}$, because of the singularity at the apex.
Therefore the space-time test space 
\[
\mathscr{E} = \widetilde{\mathcal{S}_{\oplus A^{(4)}}(\mathbb{R}^4); \mathbb{C}^n)} =
\mathcal{S}^{00}(\mathbb{R}^4; \mathbb{C}^n) \neq \mathcal{S}(\mathbb{R}^4;\mathbb{C}^n)
\]
cannot be equal $\mathcal{S}(\mathbb{R}^4;\mathbb{C}^n)$
and the standard operator $A\neq \oplus H_{(3)}$ with
\[
\mathcal{S}_{A}(\mathbb{R}^3, \,\, \mathbb{C}^n) = 
\mathcal{S}_{\oplus A^{(3)}}(\mathbb{R}^3); \mathbb{C}^n) 
= \mathcal{S}^{0}(\mathbb{R}^3, \,\, \mathbb{C}^n)
\neq \mathcal{S}(\mathbb{R}^3;\mathbb{C}^n),
\]
for fields based on representations pertinent to the light cone orbit $\mathscr{O}_{1,0,0,1}$,
if the continuity of the said maps
$\phi \rightarrow \kappa_{0,1}(\phi)$, $\phi \rightarrow \kappa_{1,0}(\phi)$ is to be preserved. 
But the said continuity of the map $\phi \rightarrow \kappa_{1,0}(\phi)$ is necessary and sufficient
(as we will soon see, compare Corollary \ref{continuity-kappa_0,1,kappa_1,0=psiOpValDistr}) 
for the field 
$\boldsymbol{\psi} = \Xi_{0,1}(\kappa_{1,0}) + \Xi_{1,0}(\kappa_{1,0})$  to be continuous
\[
\phi \longmapsto \Xi_{0,1}(\kappa_{1,0}(\phi) + \Xi_{1,0}(\kappa_{1,0}(\phi))
\] 
as a map in
\[
\mathscr{L}\Big( \mathscr{E}, \, \mathscr{L}\big((E), \, (E) \big) \Big),
\]
i.e. necessary and sufficient condition for $\boldsymbol{\psi} = \Xi_{0,1}(\kappa_{1,0}) + \Xi_{1,0}(\kappa_{1,0})$
to be a well-defined operator valued
distribution. Therefore, the space-time test function space $\mathscr{E}$
for zero mass fields must be modified and cannot coincide with the ordinary Schwartz space.
This is at least the case for zero mass fields constructed as above as
integral kernel operators with vector-valued kernels in the sense of Obata
\cite{obataJFA}, within the white noise formalism, compare Thm. \ref{ZeromassTestspace}
of Subsection \ref{A=Xi0,1+Xi1,0}.
When using the Wightman definition of quantum field, no such modification of the test function space
is necessary in passing to zero mass fields. But Wightman's definition is not very much useful
for the traditional perturbative approach to QED and other realistic perturbative QFT.
For definition of the standard operators $A^{(m)}$ and the nuclear spaces
$\mathcal{S}_{\oplus_{1}^{n} A^{(m)}}(\mathbb{R}^m; \mathbb{C}^n) = \mathcal{S}^{0}(\mathbb{R}^m; \mathbb{C}^n)$
and their Fourier transform images $\mathcal{S}^{00}(\mathbb{R}^m; \mathbb{C}^n)$
we refer to Section \ref{white-noise-proofs}.

\end{rem*}

Therefore, before
giving the construction of the Dirac field $\boldsymbol{\psi}$ as an integral kernel operator with vector-valued kernel we should give here general theorems on integral kernel operators (\ref{Xilm(kappalm)Fermi})
\begin{multline*}
\Xi_{l,m}(\kappa_{l,m}(a,x)) \\
= \sum \limits_{s_1, \ldots s_l, t_1, \ldots t_m =1}^{4} \int \limits_{(\mathbb{R}^3)^{l+m}} \,
\kappa_{l,m}(s_1,\boldsymbol{\q}_1, \ldots, s_l, \boldsymbol{\q}_l, t_1, \boldsymbol{\p}_1,
\ldots, t_m, \boldsymbol{\p}_m; a,x) \,\times \\
\times
\partial_{s_1, \boldsymbol{\q}_1}^* \cdots \partial_{s_l, \boldsymbol{\q}_l}^*
\partial_{t_1, \boldsymbol{\p}_1} \cdots \partial_{t_m, \boldsymbol{\p}_m} \,
\ud^3 \boldsymbol{\q}_1 \ldots \ud^3 \boldsymbol{\q}_l \ud^3 \boldsymbol{\p}_1 \ldots \ud^3 \boldsymbol{\p}_m,
\end{multline*}
for which
\begin{multline*}
\Xi_{l,m}(\kappa_{l,m}(\phi)) \\
= \sum \limits_{s_1, \ldots s_l, t_1, \ldots t_m =1}^{4} \int \limits_{(\mathbb{R}^3)^{l+m}} \,
\kappa_{l,m}(\phi)(s_1,\boldsymbol{\q}_1, \ldots, s_l, \boldsymbol{\q}_l, t_1, \boldsymbol{\p}_1,
\ldots, t_m, \boldsymbol{\p}_m) \,\times \\
\times
\partial_{s_1, \boldsymbol{\q}_1}^* \cdots \partial_{s_l, \boldsymbol{\q}_l}^*
\partial_{t_1, \boldsymbol{\p}_1} \cdots \partial_{t_m, \boldsymbol{\p}_m} \,
\ud^3 \boldsymbol{\q}_1 \ldots \ud^3 \boldsymbol{\q}_l \ud^3 \boldsymbol{\p}_1 \ldots \ud^3 \boldsymbol{\p}_m,
\end{multline*}
are equal to integral kernel operators (\ref{Xilm(kappalm)Fermi}) with scalar valued kernels
$\kappa_{l,m}(\phi) \in \big(\mathcal{S}_{A}(\mathbb{R}^3, \,\, \mathbb{C}^4)^{\otimes(l+m)}\big)^*$,
and with
\begin{multline*}
\kappa_{l,m} \in
\mathscr{L}\big( \mathscr{E}, \,\, \big(\mathcal{S}_{A}(\mathbb{R}^3, \mathbb{C}^4)^{\otimes(l+m)}\big)^*
\big) \cong \mathscr{L}\big( \mathcal{S}_{A}(\mathbb{R}^3, \mathbb{C}^4)^{\otimes(l+m)}, \,\,
\mathscr{L}(\mathscr{E}, \mathbb{C}) \big) \\
= \mathscr{L}\big( \mathcal{S}_{A}(\mathbb{R}^3, \mathbb{C}^4)^{\otimes(l+m)}, \,\,
\mathscr{E}^* \big),
\end{multline*}
worked out by Obata \cite{obataJFA}, \cite{obata-book}, Chap. 6.3. Obata provided detailed analysis of the Bose case,
but in a manner easily adopted to the Fermi case, and moreover he analyzed slightly more general case
of integral kernel operators with
$\mathscr{L}\big( \mathscr{E}, \,\, \mathscr{E}^* \big)$-valued distributions
\[
\kappa_{l,m} \in
\mathscr{L}\big( \mathcal{S}_{A}(\mathbb{R}^3, \mathbb{C}^4)^{\otimes(l+m)}, \,\,
\mathscr{L}(\mathscr{E}, \mathscr{E}^*) \big).
\]
We only need to analyze the
special case
of $\mathscr{L}\big( \mathscr{E}, \,\, \mathbb{C} \big) \cong \mathscr{E}^*$-valued distribution kernels
\[
\kappa_{l,m} \in
\mathscr{L}\big( \mathcal{S}_{A}(\mathbb{R}^3, \mathbb{C}^4)^{\otimes(l+m)}, \,\,
\mathscr{L}(\mathscr{E}, \mathbb{C}) \big) \cong
\mathscr{L}\big( \mathcal{S}_{A}(\mathbb{R}^3, \mathbb{C}^4)^{\otimes(l+m)}, \,\,
\mathscr{E}^* \big).
\]

In fact in realistic QFT, such as QED, we have several free fields, coupled with Lagrangian equal to a
Wick polynomial of free fields (we have in view the causal perturbative approach). Therefore,
we need to consider a generalization of \cite{obataJFA} to the case of integral kernel operators in tensor product
of, say $N$, (Fermi and/or Bose) Fock spaces $\Gamma(\mathcal{H}'_{i})$ over the corresponding single particle Hilbert spaces $\mathcal{H}'_{i}$, the corresponding standard Gelfand triples
\[
\left. \begin{array}{ccccc}   & & L^2( \sqcup \mathbb{R}^3, \ud^3 \boldsymbol{\p}; \mathbb{C}) & & \\
 & & \parallel & & \\
           \mathcal{S}_{A_i}(\mathbb{R}^3; \mathbb{C}^{{}^{r_i}})         & \subset & \oplus_{{}_{1}}^{{}^{r_i}} L^2(\mathbb{R}^3; \mathbb{C}) & \subset & \mathcal{S}_{A_{i}}(\mathbb{R}^3; \mathbb{C}^{{}^{r_i}})^*        \\
                               \downarrow \uparrow &         & \downarrow \uparrow      &         & \downarrow \uparrow  \\
                                         E_i         & \subset &  \mathcal{H}'_{i} & \subset & E_{i}^*        \\
                                                   
\end{array}\right., \,\,\, i=1,2, \dots, N,
\]
(the analogues of (\ref{SinglePartGelfandTriplesForPsi})) with the corresponding unitary isomorphisms $U_i$
(analogues of the isomorphism $U$ joining the Gelfand triples (\ref{SinglePartGelfandTriplesForPsi})). 
We only need to analyze the 
special case
of $\mathscr{L}\big( \mathscr{E}, \,\, \mathbb{C} \big) \cong \mathscr{E}^*$-valued distribution kernels
\begin{equation}\label{general-vect-valued--kappa_(lm)}
\kappa_{l,m} \in 
\mathscr{L}\big(\mathcal{S}_{A_{{}_{n_1}}}(\mathbb{R}^3, \mathbb{C}^{r_{{}_{1}}}) \otimes \cdots \otimes 
\mathcal{S}_{A_{{}_{n_i}}}(\mathbb{R}^3, \mathbb{C}^{r_{{}^{i}}}) \otimes \cdots 
\otimes \mathcal{S}_{A_{{}_{n_{l+m}}}}(\mathbb{R}^3, \mathbb{C}^{r_{{}^{l+m}}}), \,\,
\mathscr{L}(\mathscr{E}, \mathbb{C})  \big).
\end{equation}
Here 
\begin{multline}\label{general-mathscr(E)}
\mathscr{E} = \mathcal{S}_{B}\big(\sqcup \mathbb{R}^W; \mathbb{C}\big) 
= \mathcal{S}_{B_{{}_{p_1}}}(\mathbb{R}^4; \mathbb{C}^{q_{{}_{1}}}) \otimes \cdots \otimes
\mathcal{S}_{B_{{}_{p_M}}}(\mathbb{R}^4; \mathbb{C}^{q_{{}_{M}}}) \\
\subset L^2\big(\sqcup \mathbb{R}^W; \mathbb{C}\big)
= L^2(\mathbb{R}^4; \mathbb{C}^{q_{{}_{1}}})\otimes \dots \otimes 
L^2(\mathbb{R}^4; \mathbb{C}^{q_{{}_{M}}}),
\end{multline}
with
\begin{multline*}
B = B_{{}_{p_1}} \otimes \cdots \otimes B_{{}_{p_M}}, \,\,\,
p_k \in \{1,2\},
\\
\textrm{on} \,\,\  
L^2\big(\sqcup \mathbb{R}^W; \mathbb{C}\big) = L^2\big(\mathbb{R}^4; \mathbb{C}^{q_1}) \otimes \cdots 
\otimes L^2\big(\mathbb{R}^4; \mathbb{C}^{q_M}\big), \\ 
\,\,\, W= 4M,  \,\,\,\,\,\,\,\, q_k, M =1, 2, \ldots, \\
\,\,\,\,\ \sqcup \mathbb{R}^W = \textrm{$q_1 q_2 \cdots q_M$ disjoint copies of $\mathbb{R}^W$}
\end{multline*}
Moreover, we have only two possibilities for $A_i,B_i$, $i=1,2$, on each respective 
$L^2(\mathbb{R}^3, \mathbb{C}^{r_i}), L^2(\mathbb{R}^4, \mathbb{C}^{q_i})$:
\[
\begin{split}
\mathcal{S}_{A_{n_{{}_{i}}}}(\mathbb{R}^3; \mathbb{C}^{r_{{}_{i}}}) = \mathcal{S}_{\oplus H_{(3)}}(\mathbb{R}^3; \mathbb{C}^{r_{{}_{i}}})
= \mathcal{S}(\mathbb{R}^3; \mathbb{C}^{r_{{}_{i}}}), \,\,\, \textrm{or} \\
\mathcal{S}_{A_{n_{{}_{i}}}}(\mathbb{R}^3; \mathbb{C}^{r_{{}_{i}}}) = 
\mathcal{S}_{\oplus A^{(3)}}(\mathbb{R}^3; \mathbb{C}^{r_{{}_{i}}})
= \mathcal{S}^{0}(\mathbb{R}^3; \mathbb{C}^{r_{{}_{i}}}), \,\,\, \\
 \mathcal{S}_{B_{p_{{}_{i}}}}(\mathbb{R}^4; \mathbb{C}^{q_{{}_{i}}}) = \mathcal{S}_{\oplus H_{(4)}}(\mathbb{R}^4; \mathbb{C}^{q_{{}_{i}}})
= \mathcal{S}(\mathbb{R}^4; \mathbb{C}^{q_{{}_{i}}}), \,\,\, \textrm{or} \\
\mathcal{S}_{B_{p_{{}_{i}}}}(\mathbb{R}^4; \mathbb{C}^{q_{{}_{i}}}) = \widetilde{\mathcal{S}_{\oplus A^{(4)}}(\mathbb{R}^4; \mathbb{C}^{q_{{}_{i}}})}
= \mathcal{S}^{00}(\mathbb{R}^4; \mathbb{C}^{q_{{}_{i}}}).
\end{split} 
\]
Here we have the nuclear spaces $\mathcal{S}^{00}(\mathbb{R}^4; \mathbb{C}^n),
\mathcal{S}^{0}(\mathbb{R}^3; \mathbb{C}^n)$, and
the standard operators $A^{(n)}$ in $L^2(\mathbb{R}^n, \mathbb{C})$, constructed in
Subsections \ref{dim=1}-\ref{SA=S0} and \ref{WhiteNoiseA}).
$H_{(4)}$ is the Hamiltonian operator on $L^2(\mathbb{R}^4; \mathbb{C})$ of the
$4$-dimensional oscillator, compare Appendix \ref{asymptotics}.
Here $\widetilde{(\cdot)} = \mathscr{F}(\cdot)$
stands for the Fourier transform image. Note that
\[
\mathcal{S}_{\oplus A^{(4)}}(\mathbb{R}^4; \mathbb{C}^{q}) = \mathcal{S}^{0}(\mathbb{R}^4; \mathbb{C}^{q})
\]
is the nuclear subspace of all those functions in $\mathcal{S}(\mathbb{R}^4; \mathbb{C}^{q})$
which together with all their derivatives vanish at zero, so that
$\mathcal{S}^{00}(\mathbb{R}^4; \mathbb{C}^{q})$ is the nuclear space
of Fourier transforms of all such functions, compare Subsections \ref{dim=1}-\ref{SA=S0}.

For QED, it is sufficient to confine attention to just one case of all $r_i =4$ in
(\ref{general-vect-valued--kappa_(lm)}) and the case of integral kernel operators in the tensor product of two Fock liftings of the standard Gelfand triples
$\mathcal{S}_{A_i}(\mathbb{R}^3; \mathbb{C}^4) \subset L^2(\mathbb{R}^3;\mathbb{C}^4)
\subset \mathcal{S}_{A_i}(\mathbb{R}^3; \mathbb{C}^4)^*$, $i=1,2$, both over $L^2(\mathbb{R}^3;\mathbb{C}^4)$.
Namely: one Fermi Fock lifting of the standard triple in (\ref{SinglePartGelfandTriplesForPsi}), corresponding to the Dirac field, with the standard operators $A_1 = \oplus H_{(3)},
B_1 = \oplus H_{(4)}$ defined above,
and one boson Fock lifting of the standard triple in (\ref{3-Gelfand-triples}) (Subsection \ref{WhiteNoiseA})
corresponding to the electromagnetic potential field with the standard operators
$A_2 =\oplus A^{(3)}, B_2 = \mathscr{F}^{-1}\oplus A^{(4)}\mathscr{F}$ constructed in Subsection \ref{WhiteNoiseA}.
Then we consider the standard Hida space $(\boldsymbol{E})= (E_1) \otimes (E_2)$ as arising from the standard
(with nuclear inverse) operator $\Gamma_{\textrm{Fermi}}(A_1) \otimes \Gamma_{\textrm{Bose}}(A_2)$ in the
tensor product Fock space
$\Gamma_{\textrm{Fermi}}\big( L^2(\mathbb{R}^3;\mathbb{C}^4)\big)
\otimes \Gamma_{\textrm{Bose}}\big( L^2(\mathbb{R}^3;\mathbb{C}^4)\big)$ and equal to the tensor product of the
Hida spaces
\[
(E_i) = \big( \mathcal{S}_{A_i}(\mathbb{R}^3; \mathbb{C}^4) \big).
\]
The corresponding Bose Hida differential operators acting on
$(E_2) \subset \Gamma_{\textrm{Bose}}\big( L^2(\mathbb{R}^3;\mathbb{C}^4)\big)$ (constructed in the next Section)
we denote here by $\partial_{\mu, \boldsymbol{\p}}$, $\mu \in \{0,1,2,3\}$, $\boldsymbol{\p} \in \mathbb{R}^3$.
We use the Greek indices notation for the discrete parameter $\mu$ in order to distinguish them
from the Fermi Hida differential operators $\partial_{s, \boldsymbol{\p}}$ acting on
$(E_1) \subset \Gamma_{\textrm{Fermi}}\big( L^2(\mathbb{R}^3;\mathbb{C}^4)\big)$. In fact the Hida differential
operators as acting on
$(\boldsymbol{E}) = (E_1) \otimes (E_2) \subset \Gamma_{\textrm{Fermi}}\big( L^2(\mathbb{R}^3;\mathbb{C}^4)\big)
\otimes \Gamma_{\textrm{Bose}}\big( L^2(\mathbb{R}^3;\mathbb{C}^4)\big)$ should be understood respectively as
equal $\partial_{s, \boldsymbol{\p}} \otimes \boldsymbol{1}$ and
$\boldsymbol{1} \otimes \partial_{\mu, \boldsymbol{\p}}$. However, in order to simplify notation we will
likewise write for them simply $\partial_{s, \boldsymbol{\p}}$ and $\partial_{\mu, \boldsymbol{\p}}$.
Of course in this notation $E_1, \mathcal{H}'_{1}$ is the standard nuclear space
$E_1= \mathcal{S}_{\oplus H_{(3)}}(\mathbb{R}^3; \mathbb{C}^4)$
and the single particle Hilbert space $\mathcal{H}'$ in (\ref{SinglePartGelfandTriplesForPsi}); and
$E_2, \mathcal{H}'_{2}$ is the nuclear space $E_2 = E = \mathcal{S}_{\oplus A^{(3)}}(\mathbb{R}^3; \mathbb{C}^4)$
and the single particle Hilbert space $\mathcal{H}'$ in
(\ref{3-Gelfand-triples}) of Subsection \ref{WhiteNoiseA}.

Of course one can consider the generalization of \cite{obataJFA} for vector-valued kernels for integral kernel operators
on tensor product of any finite number of standard Fermi and/or Bose Fock spaces with the respective tensor product
of the corresponding standard Gelfand triples. Having in view only the QED case we confine attention to the tensor product
of just two mentioned above Fock spaces and the tensor product of the corresponding standard Gelfand triples
(\ref{SinglePartGelfandTriplesForPsi})) and (\ref{3-Gelfand-triples}).
We consider integral kernel operators $\Xi_{l,m}(\kappa_{l,m})$ for general $\mathscr{L}\big( \mathscr{E}, \,\, \mathbb{C} \big) \cong \mathscr{E}^*$-valued kernel
\[
\kappa_{l,m} \in
\mathscr{L}\big(\underbrace{\mathcal{S}_{A_{i_{{}_{1}}}}(\mathbb{R}^3, \mathbb{C}^{4}) \otimes \cdots \otimes
\mathcal{S}_{A_{i_{{}_{l+m}}}}(\mathbb{R}^3, \mathbb{C}^{4})}_{\textrm{$(l+m)$-fold tesor product}}, \,\,
\mathscr{L}(\mathscr{E}, \mathbb{C}) \big),
\]
with
\[
A_{i_{{}_{k}}} = A_1 = \oplus_{1}^{4} H_{(3)} \,\,\, \textrm{or} \,\,\, A_{i_{{}_{k}}} = A_2 = \oplus_{0}^{3}
A^{(3)} \,\,\,
\textrm{on} \,\,\, L^2(\mathbb{R}^3; \mathbb{C}^4) = \oplus L^2(\mathbb{R}^3; \mathbb{C}).
\]
In this case $\Xi_{l,m}(\kappa_{l,m})$, if expressed as integral kernel operator
\begin{multline*}
\Xi_{l,m}(\kappa_{l,m}) \\
= \sum \limits_{ s_{i_{{}_{k}}}, \mu_{i_{{}_{k}}} }
\int \limits_{(\mathbb{R}^3)^{l+m}}
\kappa_{l,m}(\overbrace{s_{i_{{}_{1}}},\boldsymbol{\p}_{i_{{}_{1}}}, \ldots,
\mu_{l}, \boldsymbol{\p}_l}^{\textrm{jointly $l$ terms
$s_{i_{{}_{k}}},\boldsymbol{\p}_{i_{{}_{k}}}$ or $\mu_{i_{{}_{k}}}, \boldsymbol{\p}_{i_{{}_{k}}}$}},
\underbrace{s_{i_{{}_{l+1}}}, \boldsymbol{\p}_{i_{{}_{l+1}}},
\ldots, \mu_{i_{{}_{l+m}}}, \boldsymbol{\p}_{i_{{}_{l+m}}}}_{\textrm{jointly $m$ terms
$s_{i_{{}_{k}}},\boldsymbol{\p}_{i_{{}_{k}}}$ or $\mu_{i_{{}_{k}}}, \boldsymbol{\p}_{i_{{}_{1}}}$}})
\,\times \\
\times
\overbrace{\partial_{s_{i_{{}_{1}}}, \boldsymbol{\p}_{i_{{}_{1}}}}^* \cdots
\partial_{\mu_{i_{{}_{l}}}, \boldsymbol{\p}_{i_{{}_{l}}}}^*}^{\textrm{jointly $l$ terms
$\partial_{s_{i_{{}_{k}}},\boldsymbol{\p}_{i_{{}_{k}}}}^*$ or
$\partial_{\mu_{i_{{}_{k}}}, \boldsymbol{\p}_{i_{{}_{k}}}}^*$}}
\underbrace{\partial_{s_{i_{{}_{l+1}}}, \boldsymbol{\p}_{i_{{}_{l+1}}}} \cdots
\partial_{\mu_{i_{{}_{l+m}}}, \boldsymbol{\p}_{i_{{}_{l+m}}}}}_{\textrm{jointly $m$ terms
$\partial_{s_{i_{{}_{k}}},\boldsymbol{\p}_{i_{{}_{k}}}}$ or
$\partial_{\mu_{i_{{}_{k}}}, \boldsymbol{\p}_{i_{{}_{k}}}}$}} \,
\ud^3 \boldsymbol{\p}_{i_{{}_{1}}} \ldots \ud^3 \boldsymbol{\p}_{i_{{}_{l}}}
\ud^3 \boldsymbol{\p}_{i_{{}_{l+1}}} \ldots \ud^3 \boldsymbol{\p}_{i_{{}_{l+m}}}
\end{multline*}
\begin{multline*}
= \sum \limits_{ s_{i_{{}_{k}}}, \mu_{i_{{}_{k}}}, t_{j_{{}_{k}}}, \nu_{j_{{}_{k}}}}
\int \limits_{(\mathbb{R}^3)^{l+m}}
\kappa_{l,m}(s_{i_{{}_{1}}},\boldsymbol{\q}_{i_{{}_{1}}}, \ldots,
\mu_{i_{{}_{l}}}, \boldsymbol{\q}_{i_{{}_{l}}},
t_{j_{{}_{1}}}, \boldsymbol{\p}_{j_{{}_{1}}},
\ldots, \nu_{j_{{}_{m}}}, \boldsymbol{\p}_{j_{{}_{m}}})
\,\times \\
\times
\partial_{s_{i_{{}_{1}}}, \boldsymbol{\q}_{i_{{}_{1}}}}^* \cdots
\partial_{\mu_{i_{{}_{l}}}, \boldsymbol{\q}_{i_{{}_{l}}}}^*
\partial_{t_{i_{{}_{1}}}, \boldsymbol{\p}_{i_{{}_{1}}}} \cdots
\partial_{\nu_{j_{{}_{m}}}, \boldsymbol{\p}_{j_{{}_{m}}}} \,
\ud^3 \boldsymbol{\q}_{i_{{}_{1}}} \ldots \ud^3 \boldsymbol{\q}_{i_{{}_{l}}} \ud^3
\boldsymbol{\p}_{j_{{}_{1}}} \ldots \ud^3 \boldsymbol{\p}_{j_{{}_{m}}},
\end{multline*}
transforming $(\boldsymbol{E}) \otimes \mathscr{E}$ into $(\boldsymbol{E})$, is understood as follows
(compare \cite{obataJFA}): the operators
$\partial_{s,\boldsymbol{\p}}^*, \partial_{\mu, \boldsymbol{\p}}^*$
and $\partial_{s,\boldsymbol{\p}}, \partial_{\mu, \boldsymbol{\p}}$ as operators
on $(\boldsymbol{E}) \otimes \mathscr{E} = (E_1) \otimes(E_2) \otimes \mathscr{E}$ are, respectively,
shortened notation for $\big((\partial_{s,\boldsymbol{\p}} \otimes \boldsymbol{1}) \otimes
\boldsymbol{1}_{{}_{\mathscr{E}}}\big)^*, \big((\boldsymbol{1} \otimes \partial_{\mu, \boldsymbol{\p}}) \otimes
\boldsymbol{1}_{{}_{\mathscr{E}}}\big)^*$
and $(\partial_{s,\boldsymbol{\p}} \otimes \boldsymbol{1}) \otimes
\boldsymbol{1}_{{}_{\mathscr{E}}}, (\boldsymbol{1} \otimes \partial_{\mu, \boldsymbol{\p}}) \otimes
\boldsymbol{1}_{{}_{\mathscr{E}}}$, and $\kappa_{l,m}$ is an
$\mathscr{L}\big( \mathscr{E}, \,\, \mathbb{C} \big) \cong \mathscr{E}^*$-valued distribution
on $(\mathbb{R}^3)^{(l+m)}$, i.e. on the test space $E_{i_1} \otimes \cdots \otimes E_{i_{l+m}}$ ($(l+m)$-fold tensor product) and this distribution $\kappa_{l,m}$ in the above formula for the integral kernel operator should be
identified with $\boldsymbol{1}_{{}_{(\boldsymbol{E})}} \otimes \kappa_{l,m}$.

Now any element $\Phi \in (\boldsymbol{E}) = (E_1) \otimes (E_2)$ has the unique absolutely convergent
decomposition (compare \cite{obataJFA}, Prop. 2.3)
\begin{equation}\label{PhiIn(E_1)otimes(E_2)}
\Phi = \sum_{n=0}^{\infty} \Phi_n, \,\,\,
\Phi_n \in \bigoplus_{n_1+n_2 =n} E_{1}^{\widehat{\otimes} \, n_1} \otimes E_{2}^{\widehat{\otimes} \, n_2},
\end{equation}
(here the tensor product $E_{1}^{\widehat{\otimes} \, n_1}$ is antisymmetrized $\widehat{\otimes}$
and symmetrized $\widehat{\otimes}$ in $E_{2}^{\widehat{\otimes} \, n_2}$).
For any element
\[
\Phi \otimes \phi \in (\boldsymbol{E}) \otimes \mathscr{E} = (E_1) \otimes (E_2) \otimes \mathscr{E}
\]
and any $\mathscr{L}(\mathscr{E}, \mathbb{C})$-valued distribution
\[
\kappa_{l,m} \in \mathscr{L} \big(\overbrace{E_{i_1} \otimes \cdots
\otimes E_{i_{l+m}}}^{\textrm{$(l+m)$ terms $E_{i_j}$, $i_j\in \{1,2\}$}},
\,\, \mathscr{L}(\mathscr{E}, \mathbb{C}) \big)
\cong \mathscr{L} \big(E_{i_1} \otimes \cdots \otimes E_{j_{l+m}}, \,\, \mathscr{E}^* \big).
\]
we put after \cite{obataJFA}
\[
\Xi_{l,m}(\kappa_{l,m}) (\Phi \otimes \phi)
= \sum_{n=0}^{\infty} \kappa_{l,m} \otimes_m (\Phi_{n+m} \otimes \phi).
\]
Note that here $\otimes_m$ denotes the $m$-contraction of $\Phi_{n+m} \otimes \phi$
with the $\mathscr{L}(\mathscr{E}, \mathbb{C})$-valued distribution uniquely determined
(after \cite{obataJFA}) by the formula
\[
\begin{split}
\langle \kappa_{l,m} \otimes_m (f_0 \otimes \phi), g_0 \rangle =
\langle \kappa_{l,m}(g_0 \otimes_n f_0), \phi \rangle, \\
f_0 \in E_{{}_{j_{{}_{1}}}} \otimes \cdots \otimes E_{{}_{j_{{}_{m}}}}
\otimes E_{{}_{i_{{}_{1}}}}\otimes E_{i_{n}}, \\
g_0 \in E_{{}_{j_{{}_{1}}}} \otimes \cdots \otimes E_{{}_{j_{{}_{m}}}}
\otimes E_{{}_{i_{{}_{1}}}} \otimes \ldots \otimes E_{{}_{i_{{}_{n}}}}, \,\,\,\,
\phi \in \mathscr{E}.
\end{split}
\]
It follows that for any
\begin{multline*}
\kappa_{l,m} \in \mathscr{L} \big(\overbrace{E_{i_1} \otimes \cdots
\otimes E_{i_{l+m}}}^{\textrm{$(l+m)$ terms $E_{i_k}$, $i_k \in \{1,2\}$}},
\,\, \mathscr{L}(\mathscr{E}, \mathbb{C}) \big) \\
\cong \mathscr{L} \big(E_{i_1} \otimes \cdots \otimes E_{i_{l+m}}, \,\, \mathscr{E}^* \big) \\ \cong
\mathscr{L} \big(\mathscr{E}, \,\, \big( E_{i_1} \otimes \cdots \otimes E_{i_{l+m}} \big)^* \big),
\end{multline*}
the operator $\Xi_{l,m}(\kappa_{l,m})$, defined by contraction $\otimes_m$ with $\kappa_{l,m}$,
belongs to
\[
\mathscr{L}\big((\boldsymbol{E}) \otimes \mathscr{E}, (\boldsymbol{E})^* \big) \cong
\mathscr{L}\big( \mathscr{E}, \, \mathscr{L}((\boldsymbol{E}), \, (\boldsymbol{E})^*) \big)
\]
with a precise norm estimation (compare Theorems 3.6 and 3.9 of \cite{obataJFA}).
Moreover, $\Xi_{l,m}(\kappa_{l,m})$ is uniquely determined by the formula
\begin{equation}\label{VectValotimesXi=intKerOp}
\big\langle \big\langle \Xi_{l,m}(\kappa_{l,m})(\Phi \otimes \phi), \Psi \big \rangle \big \rangle
= \langle \kappa_{l,m}(\eta_{\Phi, \Psi}), \phi \rangle,
\,\,\,
\Phi, \Psi \in (\boldsymbol{E}), \phi \in \mathscr{E},
\end{equation}
or equivalently
\begin{equation}\label{VectValotimesXi=intKerOp'}
\big\langle \big\langle \Xi_{l,m}(\kappa_{l,m})(\Phi \otimes \phi), \Psi \big \rangle \big \rangle
= \langle \kappa_{l,m}(\phi), \eta_{\Phi, \Psi} \rangle
= \langle \kappa_{l,m}(\eta_{\Phi, \Psi}), \phi \rangle,
\,\,\,
\Phi, \Psi \in (\boldsymbol{E}), \phi \in \mathscr{E},
\end{equation}
for $\kappa_{l,m}$ understood as an element of
\[
\mathscr{L} \big(E_{i_1} \otimes \cdots \otimes E_{i_{l+m}}, \,\, \mathscr{E}^* \big)
\,\,\, \textrm{or} \,\,\,
\mathscr{L} \big(\mathscr{E} , \,\, \big(E_{i_1} \otimes \cdots \otimes E_{i_{l+m}} \big)^* \, \big)
\cong \mathscr{L} \big(E_{i_1} \otimes \cdots \otimes E_{i_{l+m}}, \,\, \mathscr{E}^* \big)
\]
respectively in the first case (\ref{VectValotimesXi=intKerOp}) and in the second case
(\ref{VectValotimesXi=intKerOp'}).
Here
\[
\eta_{\Phi, \Psi}(w_{i_1}, \ldots w_{i_l}, w_{i_{l+1}}, \ldots w_{i_{l+m}})
= \big\langle \big\langle \partial_{w_{i_1}}^* \cdots \partial_{w_{i_l}}^* \partial_{w_{i_{l+1}}} \cdots
\partial_{w_{i_{l+m}}} \Phi, \Psi
\big \rangle \big \rangle,
\]
and $w_{i_k} = (s_{i_k},\boldsymbol{\q}_{i_k})$ if $E_{i_k} = E_1$
or $w_{i_k} = (\mu_{i_k},\boldsymbol{\q}_{i_k})$ if $E_{i_k} = E_2$.

Note that 
\[
\eta_{\Phi, \Psi} \in E_{i_1} \otimes \cdots \otimes E_{i_{l+m}}.
\]
The formula (\ref{VectValotimesXi=intKerOp}), or equivalently (\ref{VectValotimesXi=intKerOp'}), 
justifies the identification of $\Xi_{l,m}(\kappa_{l,m})$, 
defined through the $m$-contraction $\otimes_m$ with vector valued distribution $\kappa_{l,m}$, 
with the integral kernel operator 
\begin{multline}\label{electron-positron-photon-Xi}
\Xi_{l,m}(\kappa_{l,m}) = 
\int \limits_{(\sqcup \mathbb{R}^3)^{(l+m)}}
\kappa_{l,m}(w_{i_1}, \ldots w_{i_l}, w_{i_{l+1}}, \ldots w_{i_{l+m}}) 
\, \\ \times
\partial_{w_{i_1}}^* \cdots \partial_{w_{i_l}}^* \partial_{w_{i_{l+1}}} \cdots \partial_{w_{i_{l+m}}}
\ud w_{i_1} \cdots \ud w_{i_l} \ud w_{i_{l+1}} \cdots \ud w_{i_{l+m}} = \\
\int \limits_{(\sqcup \mathbb{R}^3)^{(l+m)}}
\kappa_{l,m}(w_{i_1}, \ldots w_{i_l}, u_{j_{1}}, \ldots u_{j_{m}}) 
\,
\partial_{w_{i_1}}^* \cdots \partial_{w_{i_l}}^* \partial_{u_{j_{1}}} \cdots \partial_{u_{j_{m}}}
\ud w_{i_1} \cdots \ud w_{i_l} \ud u_{j_{1}} \cdots \ud u_{j_{m}} 
\end{multline} 
defined by $\mathscr{L}(\mathscr{E}, \mathbb{C})$-valued distribution
kernel $\kappa_{l,m}$.
Here of course
\[
\begin{split}
\int \limits_{\sqcup \mathbb{R}^3} f(w) \ud w \overset{\textrm{df}}{=}  \sum_{s=1}^{4} \, \int \limits_{\mathbb{R}^3} 
f(s, \boldsymbol{\p}) \ud^3 \boldsymbol{\p} \,\,\, \textrm{for} \,\,\, w = (s, \boldsymbol{\p}), \\
\int \limits_{\sqcup \mathbb{R}^3} f(w) \ud w \overset{\textrm{df}}{=}  \sum_{\mu=0}^{3} \, \int \limits_{\mathbb{R}^3} 
f(\mu, \boldsymbol{\p}) \ud^3 \boldsymbol{\p} \,\,\, \textrm{for} \,\,\, w = (\mu, \boldsymbol{\p}),
\end{split}
\]
and we have put $u_{j_k} = w_{i_{l+k}}$, $k=1,2, \ldots, m$.

In our work we are especially interested in (the generalization of) Thm. 3.13 of \cite{obataJFA},
which gives necessary and sufficient condition for the $\mathscr{L}\big( \mathscr{E}, \,\, \mathbb{C} \big)
\cong \mathscr{E}^*$-valued
distribution $\kappa_{l,m}$ in order that the corresponding $\Xi_{l,m}(\kappa_{l,m})$ be a continuous
operator from $(\boldsymbol{E}) \otimes \mathscr{E}$ into $(\boldsymbol{E})$, thus belonging
to
\[
\mathscr{L}\big((\boldsymbol{E}) \otimes \mathscr{E}, (\boldsymbol{E})\big) \cong
\mathscr{L}\Big(\mathscr{E}, \,\, \mathscr{L}\big((\boldsymbol{E}), \, (\boldsymbol{E})\big) \Big)
\]
and thus determining a well-defined operator-valued distribution on the test space $\mathscr{E}$.

We formulate the generalization of Thm. 3.13 over to our tensor product of Fock spaces
and the corresponding tensor product of Gelfand triples (\ref{SinglePartGelfandTriplesForPsi})
and (\ref{3-Gelfand-triples}) of Subsection \ref{WhiteNoiseA}.
We will use the (generalization of) Theorem 3.13 and Proposition 3.12 of \cite{obataJFA}
for the construction of free fields and in Subsection \ref{OperationsOnXi} and Section \ref{A(1)psi(1)}
when analysing the perturbative corrections (within the
causal method of St\"uckelberg-Bogoliubov) to interacting fields, as integral kernel operators with
$\mathscr{E}^*$-valued kernels, in QED.

Exactly as for the analysis of
integral kernel operators with scalar valued kernels, also the results and proofs of \cite{obataJFA} for
integral kernel operators with vector-valued kernels can be easily adopted to the Fermi case, as well as for
the more general case of several Bose and Fermi fields on the tensor product of the corresponding Fock spaces.

We have the following generalization of Thm. 3.13 of \cite{obataJFA}:
\begin{twr}\label{obataJFA.Thm.3.13}
Let 
\[
\kappa_{l,m} \in \mathscr{L} \big(\overbrace{E_{i_1} \otimes \cdots 
\otimes E_{i_{l+m}}}^{\textrm{$(l+m)$ terms $E_{i_j}$, $i_j\in \{1,2\}$}}, 
\,\, \mathscr{L}(\mathscr{E}, \mathbb{C}) \big)
\cong \mathscr{L} \big(E_{i_1} \otimes \cdots \otimes E_{i_{l+m}}, \,\, \mathscr{E}^* \big).
\]
Then 
\[
\Xi_{l,m}(\kappa_{l,m}) \in 
\mathscr{L}\big((\boldsymbol{E}) \otimes \mathscr{E}, (\boldsymbol{E})\big) \cong 
\mathscr{L}\Big(\mathscr{E}, \,\, \mathscr{L}\big((\boldsymbol{E}), \, (\boldsymbol{E})\big) \Big)
\]
if and only if the bilinear map
\begin{multline*}
\xi \times \eta \mapsto \kappa_{l,m}(\xi \otimes \eta), 
\\
\xi \in \overbrace{E_{i_1} \otimes \cdots 
\otimes E_{i_l}}^{\textrm{first $l$ terms $E_{i_j}$, $i_j \in \{1,2\}$}}, \\
\eta \in \overbrace{E_{i_{l+1}} \otimes \cdots 
\otimes E_{i_{l+m}}}^{\textrm{last $m$ terms $E_{i_j}$, $i_j\in \{1,2\}$}},
\end{multline*}
can be extended to a separately continuous bilinear map from
\[
\Big( \overbrace{E_{i_1} \otimes \cdots 
\otimes E_{i_l}}^{\textrm{first $l$ terms $E_{i_j}$}} \Big)^*
\times
\Big( \overbrace{E_{i_{l+1}} \otimes \cdots 
\otimes E_{i_{l+m}}}^{\textrm{last $m$ terms $E_{i_j}$}} \Big)
\,\,\, \textrm{into} \,\,\,\mathscr{L}(\mathscr{E}, \mathbb{C}) = \mathscr{E}^*.
\]
This is the case if and only if for any $k\geq 0$ there exist 
$r \in \mathbb{R}$ such that $|\kappa_{l,m}|_{{}_{l,m;k,r;k}} < \infty$;
and moreover in this case for any $k \in \mathbb{R}$ and $q_0 < q_1 < q$ we have
\begin{multline*}
\|\Xi_{l,m}(\kappa_{l,m}) (\Phi \otimes \phi)\|_{{}_{k}} \leq \rho^{-q/2} \delta^{-1} \sigma^{2} \sqrt{l^l m^m} \Delta_{q_1}^{(l+m)/2} \\
\times 
|\kappa_{l,m}|_{{}_{l,m;k+1, -(k+q+1);k+1}} \|\Phi\|_{{}_{k+q+2}}, \,\,\,
\Phi \in (\boldsymbol{E}), \phi \in \mathscr{E}.
\end{multline*}
\end{twr}

Here for any linear map
\[
\kappa_{l,m}: \overbrace{E_{i_1} \otimes \cdots
\otimes E_{i_{l+m}}}^{\textrm{$(l+m)$ terms $E_{i_j}$, $i_j\in \{1,2\}$}}
\longrightarrow \mathscr{L}(\mathscr{E}, \mathbb{C}) = \mathscr{E}^*
\]
and $k,q,r \in \mathbb{R}$ we put (after \cite{obataJFA}):
\begin{multline*}
|\kappa_{l,m}|_{{}_{l,m;kq;r}} = \textrm{sup} \Bigg\{\sum_{\textrm{i}, \textrm{j}}
|\langle \kappa_{l,m}(e(\textrm{i}) \otimes e(\textrm{j})), \phi\rangle|^2 |e(\textrm{i})|_{{}_{k}}^2
|e(\textrm{j})|_{{}_{q}}^2, \\
\phi \in \mathscr{E}, |\phi|_{{}_{-r}} \leq 1 \Bigg\}^{1/2}.
\end{multline*}
Note that we are using the multi-index notation
\[
e(\textrm{i}) = e_{{}_{i_{{}_{1}}}} \otimes \cdots \otimes e_{{}_{i_{{}_{l}}}} \in
E_{{}_{i_{{}_{1}}}} \otimes \cdots \otimes E_{{}_{i_{{}_{l}}}}, \,\,\,
\textrm{i} = (i_{{}_{1}}, \ldots, i_{{}_{l}})
\]
\begin{multline*}
e(\textrm{j}) = e_{{}_{j_{{}_{1}}}} \otimes \cdots \otimes e_{{}_{j_{{}_{m}}}} =
e_{{}_{i_{{}_{l+1}}}} \otimes \cdots \otimes e_{{}_{i_{{}_{l+m}}}} \in
E_{{}_{i_{{}_{l+1}}}} \otimes \cdots \otimes E_{{}_{i_{{}_{l+m}}}}, \\
\,\,\,\,\,\, \textrm{j} = (j_{{}_{1}}, \ldots, j_{{}_{m}}) = (i_{{}_{l+1}}, \ldots, i_{{}_{l+m}}),
\end{multline*}
but now $e_{{}_{i_{{}_{k}}}}$ is the element of the complete orthonormal system of eigenvectors
of the standard operator $A_1$ whenever $e_{{}_{i_{{}_{k}}}} \in E_{{}_{i_{{}_{k}}}} = E_1$ or of the standard
operator $A_2$ whenever $e_{{}_{i_{{}_{k}}}} \in E_{{}_{i_{{}_{k}}}} = E_2$.
Note also that with the system of eigenvalues (counted with multiplicity)
\[
\lambda_{i0}, \lambda_{i1}, \lambda_{i2}, \ldots \,\,\,\, \textrm{of $A_i$},
\]
we have put here
\[
\delta_i = \Bigg(\sum_{j=0}^{\infty} \lambda_{ij} \Bigg)^{1/2}
= \|A_{i}^{-1} \|_{\textrm{HS}} < \infty,
\,\,\,\,
\delta^{-1} = \underset{i=1,2}{\textrm{max}} \,\, \delta_{i}^{-1},
\]
for the maximum of the Hilbert-Schmidt norms of the nuclear operators $A_{i}^{-1}$, $i=1,2$.
Similarly, here
\[
\rho = \underset{i=1,2}{\textrm{max}} \,\, \|A_{i}^{-1} \|_{\textrm{op}}
\]
for the operator norm $\| \cdot \|_{\textrm{op}}$. Here
\[
\Delta_q = \underset{i=1,2}{\textrm{max}} \,\, \Delta_{q_1,i}, \,\,\, q> \underset{i=1,2}{\textrm{max}} \,\, q_{0i} = q_0
\]
where for $i=1,2$
\[
\Delta_{q,i} = \frac{\delta_i}{-e \rho_{i}^{q/2} \textrm{ln}(\delta_{i}^2 \rho_{i}^q)}, \,\,\,
q> q_{0i} = \textrm{inf} \, \{q> 0, \delta_{i}^2 \rho_{i}^{q} \leq 1 \}
\]
is a finite constant uniquely determined by the standard
operator $A_i$, $i=1,2$, if $q>q_{0,i}$ for the positive constant $q_{0i}$ again depending on $A_i$,
compare \cite{obataJFA}, p. 210.
Recall that
\[
\delta_i = \Bigg(\sum_{j=0}^{\infty} \lambda_{ij} \Bigg)^{1/2}
= \|A_{i}^{-1} \|_{\textrm{HS}} \,\,\,\,
\rho_i = \|A_{i}^{-1} \|_{\textrm{op}}.
\]
Finally
\[
\sigma = (\textrm{inf \, Spec} B)^{-1} = \|B^{-1}\|_{\textrm{op}}
\]
for the standard operator $B = B_{{}_{p_1}} \otimes \cdots \otimes B_{{}_{p_M}}$,
$p_k \in \{1,2\}$ on $\otimes_{k=1}^{M}L^2(\mathbb{R}^4; \mathbb{C}^{q_k})$, defining the nuclear test space
\begin{multline*}
\mathscr{E} = \mathcal{S}_{B}(\sqcup \mathbb{R}^{4M}; \mathbb{C}) \\
= \mathcal{S}_{B_{{}_{p_1}}}(\mathbb{R}^4; \mathbb{C}^{q_1}) \otimes \cdots \otimes
\mathcal{S}_{B_{{}_{p_M}}}(\mathbb{R}^4; \mathbb{C}^{q_M}) \subset L^2(\sqcup \mathbb{R}^{4M}; \mathbb{C})
= \otimes_{k=1}^{M}L^2(\mathbb{R}^4; \mathbb{C}^{q_k})
\end{multline*}
(we need the general case with $M>1$ for the analysis of Wick products of $M$ free fields
or of their space-time derivatives or of their separate components).
Recall once more that here
\[
\begin{split}
B_{p_k} = \oplus H_{(4)} \,\,\,\, \textrm{on} \,\,\,\,
\oplus_{k=1}^{q_k} L^2(\mathbb{R}^4; \mathbb{C}) = L^2(\mathbb{R}^4; \mathbb{C}^{q_k}), \,\,\, \textrm{for} \,\,\,
p_k = 1 \\
B_{p_k} = \mathscr{F}^{-1} \oplus A^{(4)} \mathscr{F} \,\,\,\, \textrm{on} \,\,\,\,
\oplus_{k=1}^{q_k} L^2(\mathbb{R}^4; \mathbb{C}) = L^2(\mathbb{R}^4; \mathbb{C}^{q_k}),\,\,\, \textrm{for} \,\,\,
p_k = 2
\end{split}
\]
with the Hamiltonian operator $H_{(4)}$ on $L^2(\mathbb{R}^4; \mathbb{C})$ of the
$4$-dimensional oscillator, compare Appendix \ref{asymptotics}. The standard operator
$A^{(4)}$ on $L^2(\mathbb{R}^4; \mathbb{C})$ is defined in Subsection \ref{dim=n}.
\begin{equation}\label{mathscrE_1,mathscrE_2}
\begin{split}
\mathscr{E}_{p_k} = \mathcal{S}_{B_{p_k}}(\mathbb{R}^4; \mathbb{C}^{q_k}) =
\mathcal{S}_{\oplus H_{(4)}}(\mathbb{R}^4; \mathbb{C}^{q_k}) = \mathcal{S}(\mathbb{R}^4; \mathbb{C}^{q_k}),
\,\,\, p_k = 1 \\
\mathscr{E}_{p_k} = \mathcal{S}_{B_{p_k}}(\mathbb{R}^4; \mathbb{C}^{q_k}) =
\mathcal{S}_{\mathscr{F}\oplus A^{(4)}\mathscr{F}^{-1}}(\mathbb{R}^4; \mathbb{C}^{q_k})
= \mathcal{S}^{00}(\mathbb{R}^4; \mathbb{C}^{q_k}), \,\,\, p_k = 2.
\end{split}
\end{equation}
Recall that
\begin{multline*}
|\phi|_{{}_{-r}} \overset{\textrm{df}}{=} \big|B^{-r} \phi \big|_0 =
\big| (B_{{}_{p_1}} \otimes \cdots \otimes B_{{}_{p_M}})^{-r} \phi\big|_0
\\
= \big|(B_{{}_{p_1}} \otimes \cdots \otimes B_{{}_{p_M}})^{-r} \phi \big|_{{}{\otimes_{k=1}^{M}
L^2(\mathbb{R}^4; \mathbb{C}^{q_k})}},
\,\,\,\,\,\,
\phi \in \mathscr{E}, r \in \mathbb{R}.
\end{multline*}
Recall that in computation of the operator or Hilbert-Schmidt norm the unitary Fourier transform $\mathscr{F}$
in definition of $B_2$ can be ignored and the respective norms can be simply computed for
$\oplus A^{(4)}$.

From Theorem \ref{obataJFA.Thm.3.13} we obtain the following
\begin{cor}\label{continuity-kappa_0,1,kappa_1,0=psiOpValDistr}
The Dirac free field
\[
\boldsymbol{\psi} = \Xi_{0,1}(\kappa_{0,1}) + \Xi_{1,0}(\kappa_{1,0}) \in
\mathscr{L}\big( (E) \otimes \mathscr{E}, \, (E)^* \big) \cong
\mathscr{L}\big( \mathscr{E}, \,\, \mathscr{L}( (E), (E)^*) \big)
\]
understood as integral kernel operator
with vector-valued distributions
\[
\kappa_{0,1}, \kappa_{1,0} \in \mathscr{L}\big( \mathcal{S}_{A}(\mathbb{R}^3, \mathbb{C}^4), \,\,
\mathscr{E}^* \big) \cong \mathcal{S}_{A}(\mathbb{R}^3, \mathbb{C}^4)^* \otimes \mathscr{E}^*
\]
belongs to $\mathscr{L}\big( (E) \otimes \mathscr{E}, \, (E) \big) \cong
\mathscr{L}\big( \mathscr{E}, \,\, \mathscr{L}( (E), (E)) \big)$, i.e.
\[
\boldsymbol{\psi} = \Xi_{0,1}(\kappa_{0,1}) + \Xi_{1,0}(\kappa_{1,0}) \in
\mathscr{L}\big( (E) \otimes \mathscr{E}, \, (E) \big) \cong
\mathscr{L}\big( \mathscr{E}, \,\, \mathscr{L}( (E), (E)) \big),
\]
if and only if the map $\phi \mapsto \kappa_{1,0}(\phi)$
belongs to
\[
\mathscr{L}\big( \mathscr{E}, \,\, \mathcal{S}_{A}(\mathbb{R}^3, \mathbb{C}^4) \, \big),
\]
i.e. if and only if $\kappa_{1,0}$ can be extended to a map belonging to
\begin{multline*}
\mathscr{L}\big( \mathcal{S}_{A}(\mathbb{R}^3, \mathbb{C}^4)^*, \,\,
\mathscr{E}^* \big) \cong \mathcal{S}_{A}(\mathbb{R}^3, \mathbb{C}^4) \otimes \mathscr{E}^* \\
\cong \mathscr{E}^* \otimes \mathcal{S}_{A}(\mathbb{R}^3, \mathbb{C}^4)
\cong \mathscr{L}\big( \mathscr{E}, \,\, \mathcal{S}_{A}(\mathbb{R}^3, \mathbb{C}^4) \, \big).
\end{multline*}
\end{cor}
Here of course we have the special case of Theorem \ref{obataJFA.Thm.3.13} with the tensor product
of the two Fock spaces (corresponding to the Dirac field and the electromagnetic potential field)
degenerated to just one Fock space -- that corresponding to the Dirac field, and with
the Hida space $(\boldsymbol{E}) = (E_1) \otimes (E_2)$ degenerated to just the Hida space
$(E_1) \overset{df}{=} (E) \overset{df}{=} \big(\mathcal{S}_{A}(\mathbb{R}^3;\mathbb{C}^4)\big) =
\big(\mathcal{S}_{\oplus H_{(3)}}(\mathbb{R}^3;\mathbb{C}^4)\big)$ corresponding to the Dirac field, with
the standard operator $A =A_1= \oplus H_{(3)}$ given by (\ref{AinL^2(R^3;C^4)}); and finally with $M=1$
and $B$ degenerated to $B_1$ with the nuclear test space $\mathscr{E}$ degenerated to
\[
\mathscr{E} = \mathcal{S}_{B}(\sqcup \mathbb{R}^4; \mathbb{C})
= \mathcal{S}_{B_1}(\mathbb{R}^4; \mathbb{C}^4) = \mathcal{S}_{\oplus H_{(4)}}(\mathbb{R}^4; \mathbb{C}^4)
= \mathcal{S}(\mathbb{R}^4; \mathbb{C}^4) = \mathscr{E}_1
\]
of (\ref{mathscrE_1,mathscrE_2}).

Equivalently we may consider here the integral kernel operator 
$\boldsymbol{\psi} = \Xi_{0,1}(\kappa_{0,1}) + \Xi_{1,0}(\kappa_{1,0})$ as acting 
in the said tensor product of two Fock spaces, having the form of sum of tensor
product operators on $(\boldsymbol{E}) = (E_1) \otimes (E_2)$ with the second factor operators acting on the
second factor $(E_2)$ trivially as the unit operator, in accordance with the identification
of the operator
\[
\partial_w = \left\{ \begin{array}{ll}
\partial_{s, \boldsymbol{\p}} \otimes \boldsymbol{1}, & \textrm{if $w = (s, \boldsymbol{\p})$ refers to Fermi variables},
\\
\boldsymbol{1} \otimes \partial_{\mu, \boldsymbol{\p}}, &
\textrm{if $w = (\mu, \boldsymbol{\p})$ refers to Bose variables},
\end{array} \right.
\]
in the general formula (\ref{electron-positron-photon-Xi}). But now we
have to replace the general formula (\ref{electron-positron-photon-Xi}) defining
the operators $\Xi_{0,1}(\kappa_{0,1}),
\Xi_{1,0}(\kappa_{1,0})$ giving the Dirac field, with another one
in which the integration variables are
restricted only to the Fermi variables. This is not the special case
of (\ref{electron-positron-photon-Xi}) for $l=0, m=1$ (or $l=1, m=0$)
of an integral operator in the tensor product of Fock spaces, because this is not true
that the kernels $\kappa_{0,1}, \kappa_{1,0}$ inserted into the general formula
(\ref{electron-positron-photon-Xi}) cancel out the unwanted boson
variables. Thus, $\boldsymbol{\psi} = \Xi_{0,1}(\kappa_{0,1}) + \Xi_{1,0}(\kappa_{1,0})$ considered
as acting in the said tensor product of two Fock spaces is a special integral kernel operator with
integration variables restricted to fermion variables. Similarly, we have for the electromagnetic
potential field, if considered as integral kernel operator in the said tensor product
of Fock spaces: it is an exceptional integral kernel operator with the integration variables
in the general formula (\ref{electron-positron-photon-Xi}) restricted only to boson variables.

From Corollary \ref{continuity-kappa_0,1,kappa_1,0=psiOpValDistr}
and Lemma \ref{kappa0,1,kappa1,0psi} it follows
\begin{cor}\label{psi=intKerOpVectVal=OpValDistr}
Let
\[
\boldsymbol{\psi} = \Xi_{0,1}(\kappa_{0,1}) + \Xi_{1,0}(\kappa_{1,0}) \in
\mathscr{L}\big( (E) \otimes \mathscr{E}, \, (E)^* \big) \cong
\mathscr{L}\big( \mathscr{E}, \,\, \mathscr{L}( (E), (E)^*) \big)
\]
be the Dirac field understood as an integral kernel operator with vector-valued kernels
\[
\kappa_{0,1}, \kappa_{1,0} \in \mathscr{L}\big( \mathcal{S}_{A}(\mathbb{R}^3, \mathbb{C}^4), \,\,
\mathscr{E}^*  \big) \cong \mathcal{S}_{A}(\mathbb{R}^3, \mathbb{C}^4)^* \otimes \mathscr{E}^*,
\]
defined by (\ref{kappa_0,1}) and (\ref{kappa_1,0}). Then the Dirac field operator
\[
\boldsymbol{\psi}  = \boldsymbol{\psi}^{(-)} + \boldsymbol{\psi}^{(+)} = \Xi_{0,1}(\kappa_{0,1}) + \Xi_{1,0}(\kappa_{1,0}),
\]
belongs to $\mathscr{L}\big( (E) \otimes \mathscr{E}, \, (E) \big) \cong
\mathscr{L}\Big( \mathscr{E}, \,\, \mathscr{L}\big( (E), (E)\big) \, \Big)$, i.e.
\[
\boldsymbol{\psi} = \Xi_{0,1}(\kappa_{0,1}) + \Xi_{1,0}(\kappa_{1,0}) \in
\mathscr{L}\big( (E) \otimes \mathscr{E}, \, (E) \big) \cong
\mathscr{L}\Big( \mathscr{E}, \,\, \mathscr{L}\big( (E), (E)\big) \, \Big),
\]
which means in particular that the Dirac field $\boldsymbol{\psi}$, understood as a sum 
$\boldsymbol{\psi} = \Xi_{0,1}(\kappa_{0,1}) + \Xi_{1,0}(\kappa_{1,0})$ of 
two integral kernel operators with vector-valued kernels, 
defines an operator valued distribution through the continuous map
\[
\mathscr{E} \ni \varphi \longmapsto
\Xi_{0,1}\big(\kappa_{0,1}(\varphi)\big) + \Xi_{1,0}\big(\kappa_{1,0}(\varphi)\big)
\in \mathscr{L}\big( (E), (E)\big).
\]
\end{cor}
Note here that the last Corollary \ref{psi=intKerOpVectVal=OpValDistr} follows immediately 
from the proved equality (\ref{psi=IntKerOpVectValKer}), i.e. Lemma
\ref{psi=integKerOpVecValProof}, Corollary \ref{D_xi=int(xiPartial)}, and continuity of the restriction
to the orbit $\mathscr{O}_{m,0,0,0}$ regarded as a map $\mathcal{S}(\mathbb{R}^4; \mathbb{C})
\rightarrow \mathcal{S}(\mathbb{R}^4; \mathbb{C})$.

We have introduced the decomposition of the Dirac field operator $\boldsymbol{\psi}$ into the positive and negative
frequency parts after the classic physical tradition
\[
\boldsymbol{\psi}^{(-)} \overset{df}{=} \Xi_{0,1}(\kappa_{0,1}), \,\,\,\,
\boldsymbol{\psi}^{(+)} \overset{df}{=} \Xi_{1,0}(\kappa_{1,0}).
\]  

Thus, as a Corollary to Thm. \ref{obataJFA.Thm.3.13} we have obtained the Dirac field
$\boldsymbol{\psi}$ as a sum of two integral kernel operators with vector valued
kernels $\kappa_{0,1}, \kappa_{1,0}$ (\ref{kappa_0,1}) and (\ref{kappa_1,0}).
But as we have seen the (free) Dirac field $\boldsymbol{\psi}$ (and in general a quantum free field
understood as sum of integral kernel operators with vector-valued kernels) is naturally an integral
kernel operator with well-defined kernel equal to (scalar) integral kernel operator
\begin{equation}\label{psi(x)}
\boxed{
\begin{split}
\boldsymbol{\psi}^a(x) = \sum_{s=1}^{4} \, \int \limits_{\mathbb{R}^3} 
\kappa_{0,1}(s, \boldsymbol{p}; a, x) \,\, \partial_{s, \boldsymbol{\p}} \, \ud^3 \boldsymbol{\p}
+
\sum_{s=1}^{4} \, \int \limits_{\mathbb{R}^3} 
\kappa_{1,0}(s, \boldsymbol{p}; a, x) \,\, \partial_{s, \boldsymbol{\p}}^* \, \ud^3 \boldsymbol{\p} \\
= \boldsymbol{\psi}^{(-) \, a}(x) + \boldsymbol{\psi}^{(+) \, a}(x)
= \Xi_{0,1}\big(\kappa_{0,1}(a,x)\big) + \Xi_{1,0}\big(\kappa_{1,0}(a,x)\big)  \\ =
\sum_{s=1}^{2} \, \int \limits_{\mathbb{R}^3} 
{\textstyle\frac{1}{2}} \underset{*}{u}{}_{s}^{a}(\boldsymbol{\p})e^{-ip\cdot x} \,\, \partial_{s, \boldsymbol{\p}} \, \ud^3 \boldsymbol{\p}
+
\sum_{s=1}^{2} \, \int \limits_{\mathbb{R}^3} 
{\textstyle\frac{-1}{2}}
\underset{*}{v}{}_{s}^{a}(\boldsymbol{\p})e^{ip\cdot x} \,\, \partial_{s+2, \boldsymbol{\p}}^* \, \ud^3 \boldsymbol{\p} \\ =
\sum_{s=1}^{2} \, \int \limits_{\mathbb{R}^3} 
{\textstyle\frac{1}{2}} \underset{*}{u}{}_{s}^{a}(\boldsymbol{\p})e^{-ip\cdot x} \,\, a_{s}(\boldsymbol{\p}) \, \ud^3 \boldsymbol{\p}
+
\sum_{s=1}^{2} \, \int \limits_{\mathbb{R}^3} 
{\textstyle\frac{-1}{2}}\underset{*}{v}{}_{s}^{a}(\boldsymbol{\p})e^{ip\cdot x} \,\, a_{s+2}(\boldsymbol{\p})^+ \, \ud^3 \boldsymbol{\p}
\\ =
\sum_{s=1}^{2} \, \int \limits_{\mathbb{R}^3} 
{\textstyle\frac{1}{2}}\underset{*}{u}{}_{s}^{a}(\boldsymbol{\p})e^{-ip\cdot x} \,\, b_{s}(\boldsymbol{\p}) \, \ud^3 \boldsymbol{\p}
+
\sum_{s=1}^{2} \, \int \limits_{\mathbb{R}^3} 
{\textstyle\frac{-1}{2}}\underset{*}{v}{}_{s}^{a}(\boldsymbol{\p})e^{ip\cdot x} \,\, d_{s}(\boldsymbol{\p})^+ \, \ud^3 \boldsymbol{\p}.
\end{split}
}
\end{equation}
\[
\,\,\,\,\,\,\,\,\,\,\,\, \textrm{with $p = (|p_0(\boldsymbol{\p})|, \boldsymbol{\p}) \in \mathscr{O}_{m,0,0,0}$},
\]
and where we have put $b_{s=1} (\boldsymbol{\p}), b_{s=2} (\boldsymbol{\p}),
d_{s=1} (\boldsymbol{\p}), d_{s=2} (\boldsymbol{\p})$, respectively, for the operators
$b_{s=1} (\boldsymbol{\p}), b_{s=-1} (\boldsymbol{\p}), d_{s=1} (\boldsymbol{\p}), d_{s=-1} (\boldsymbol{\p})$
used in \cite{Scharf}, p. 82, just changing the names of the
summation index from $\{1,-1 \}$ into $\{1, 2\}$.
Similarly we have for the Dirac field
\begin{equation}\label{*psi(x)}
\boxed{
\begin{split}
\underset{*}{\boldsymbol{\psi}}^a(x) = \sum_{s=1}^{4} \, \int \limits_{\mathbb{R}^3} 
\kappa^{*}_{0,1}(s, \boldsymbol{p}; a, x) \,\, \partial_{s, \boldsymbol{\p}} \, \ud^3 \boldsymbol{\p}
+
\sum_{s=1}^{4} \, \int \limits_{\mathbb{R}^3} 
\kappa^{*}_{1,0}(s, \boldsymbol{p}; a, x) \,\, \partial_{s, \boldsymbol{\p}}^* \, \ud^3 \boldsymbol{\p} \\
= \boldsymbol{\psi}^{(-) \, a}(x) + \boldsymbol{\psi}^{(+) \, a}(x)
= \Xi_{0,1}\big(\kappa^{*}_{0,1}(a,x)\big) + \Xi_{1,0}\big(\kappa^{*}_{1,0}(a,x)\big)  \\ =
\sum_{s=1}^{2} \, \int \limits_{\mathbb{R}^3} 
{\textstyle\frac{1}{2}} u_{s}^{a}(\boldsymbol{\p})e^{-ip\cdot x} \,\, \partial_{s, \boldsymbol{\p}} \, \ud^3 \boldsymbol{\p}
+
\sum_{s=1}^{2} \, \int \limits_{\mathbb{R}^3} 
{\textstyle\frac{-1}{2}}
v_{s}^{a}(\boldsymbol{\p})e^{ip\cdot x} \,\, \partial_{s+2, \boldsymbol{\p}}^* \, \ud^3 \boldsymbol{\p} \\ =
\sum_{s=1}^{2} \, \int \limits_{\mathbb{R}^3} 
{\textstyle\frac{1}{2}} u_{s}^{a}(\boldsymbol{\p})e^{-ip\cdot x} \,\, a_{s}(\boldsymbol{\p}) \, \ud^3 \boldsymbol{\p}
+
\sum_{s=1}^{2} \, \int \limits_{\mathbb{R}^3} 
{\textstyle\frac{-1}{2}} v_{s}^{a}(\boldsymbol{\p})e^{ip\cdot x} \,\, a_{s+2}(\boldsymbol{\p})^+ \, \ud^3 \boldsymbol{\p}
\\ =
\sum_{s=1}^{2} \, \int \limits_{\mathbb{R}^3} 
{\textstyle\frac{1}{2}}u_{s}^{a}(\boldsymbol{\p})e^{-ip\cdot x} \,\, b_{s}(\boldsymbol{\p}) \, \ud^3 \boldsymbol{\p}
+
\sum_{s=1}^{2} \, \int \limits_{\mathbb{R}^3} 
{\textstyle\frac{-1}{2}}v_{s}^{a}(\boldsymbol{\p})e^{ip\cdot x} \,\, d_{s}(\boldsymbol{\p})^+ \, \ud^3 \boldsymbol{\p},
\end{split}
}
\end{equation}
\[
\,\,\,\,\,\,\,\,\,\,\,\, \textrm{with $p = (|p_0(\boldsymbol{\p})|, \boldsymbol{\p}) \in \mathscr{O}_{m,0,0,0}$},
\]
with the Fourier transforms of the fundamental solutions
\[
\underset{*}{u} = \gamma^0 u, \,\,\, \underset{*}{v} = \gamma^0 v
\]
and with the explicit formulas for $u,v$ given in the Appendix \ref{fundamental,u,v}. Note that 
\[
\big[i\gamma^\mu \partial_\mu - m \big] \boldsymbol{\psi}(x) = 0
\]
and 
\[
\big[i(\gamma^\mu)^* \partial_\mu - m \big] \underset{*}{\boldsymbol{\psi}}(x) = 0
\]
with 
\[
\widetilde{\gamma^\mu} = (\gamma^\mu)^* = \gamma^0 \gamma^\mu\gamma^0.
\]

Here the expressions in (\ref{psi(x)}) or in (\ref{*psi(x)}), for each fixed space-time point $x$, are not merely symbolic,
but they are meaningful integral kernel
operators transforming continuously the Hida space $(E)$ into its strong dual $(E)^*$, and moreover even the integral
signs in these expressions are not merely symbolic, but are meaningful (pointwise) Pettis integrals
(compare \cite{HKPS}, or Subsection \ref{WhiteNoiseA}).

Note here that it is immediately seen that the kernel distributions $\kappa_{0,1}(a,x)$, $\kappa_{1,0}(a,x)$, with $(a,x)$ fixed and treated as a fixed parameter,
and defined by the kernel functions (similarly for $\kappa^{*}_{l,m}$)
\[
\kappa_{0,1}(s, \boldsymbol{p}; a, x), \,\,\,\,
\kappa_{1,0}(s, \boldsymbol{p}; a, x) \,\,\, \textrm{with fixed value of} \, (a,x),
\]
define continuous functionals on $E \overset{df}{=} \mathcal{S}_{A}(\mathbb{R}^3;\mathbb{C}^4) = 
\mathcal{S}_{\oplus H_{(3)}}(\mathbb{R}^3;\mathbb{C}^4) = \mathcal{S}(\mathbb{R}^3;\mathbb{C}^4)$, \emph{i.e.}
\[
\kappa_{0,1}(a,x), \kappa_{1,0}(a,x) \in E^*, \,\,\, \textrm{for any fixed value of} \, (a,x).
\]
Therefore, it follows from Thm. \ref{Xi_l,m} (\emph{i.e.} from the fermionic analogue of Thm. 2.2 of \cite{hida}), that each component of the Dirac field evaluated at fixed space-time point $x$ is a well-defined integral
kernel operator transforming continuously the Hida space $(E)$ into its strong dual $(E)^*$, and each component of the negative frequency part,
evaluated at $x$, transforms continuously the Hida space into itself, \emph{i.e.}
\begin{align*}
\boldsymbol{\psi}^a(x) \in \mathscr{L}\big( (E), (E)^* \big),
\,\,\, \textrm{for any fixed value of} \, (a,x), \\
\boldsymbol{\psi}^{(-) \, a}(x) \in \mathscr{L}\big( (E), (E) \big),
\,\,\, \textrm{for any fixed value of} \, (a,x).
\end{align*}

We see that there is an additional irrelevant constant factor $\pm 1/2$ in the integrand 
for the local free Dirac field $\boldsymbol{\psi}(x), \underset{*}{\boldsymbol{\psi}}(x)$ 
in our formula (\ref{psi(x)}) and (\ref{*psi(x)}) in comparison to the standard formula for the free quantum
Dirac field used in other books, compare e.g. \cite{Scharf} formula (2.2.33)\footnote{In the formula (2.2.33)
of \cite{Scharf} the summation sign over $s$ has been lost (of course by a trivial misprint).}
or the formula (7.32) of \cite{Bogoliubov_Shirkov} (with the respective amplitudes $a_{\nu}^{\pm}$ replaced with
the creation-annihilation operators). In fact an additional mass term should be present in 
 our formula (\ref{isomorphismU}) for the isomorphism $U$ due to the formula for the single particle inner product, so that the mass
$m$ cancells out in the formula 
for the field (\ref{psi(x)}) or (\ref{*psi(x)}). However, we have been
not scrupulous enough to account for in (\ref{isomorphismU}) the additional term $2m$ in the inner product of the single particle
space inner product of Subsection \ref{FirstStepH} (although we have cancelled $m$ as if it was accounted for in (\ref{isomorphismU})).
Also, the other additional constant factors, 
and equal to the respective powers of $2\pi$ appear in the literature, 
which are lost in our formula because we have not normalized the measures when using Fourier transformations.
The additional minus sign in the negative energy contribution term is irrelevant, and has no influence on the 
pairing functions in the scattering operator in spinor causal perturbative QED, and can simply be ignored. 

In Subsections \ref{OperationsOnXi} -- \ref{WickForChronological} and Section \ref{A(1)psi(1)}
we will use the free Dirac field  $\underset{*}{\boldsymbol{\psi}}(x)$, given by
(\ref{*psi(x)}), and will simply write for it  $\boldsymbol{\psi}(x)$, without the asterisk subscript, with the kenels
$\kappa_{0,1}^{*}$, $\kappa_{1,0}^{*}$ defined by the fundamental solutions $u,v$, and will sometimes simply write $\kappa_{0,1}$,
$\kappa_{1,0}$ or (in Subsection \ref{OperationsOnXi}) ${}^{1}\kappa_{0,1}$,
${}^{1}\kappa_{1,0}$ or (in Subsection \ref{WickForChronological}) ${}^{(1)}\kappa_{0,1}$,
${}^{(1)}\kappa_{1,0}$, for them. 
In Subsections \ref{StandardDiracPsiField}, \ref{psichi} and \ref{A(1)chi} we will use the Dirac free field (\ref{psi(x)})
with the kernels $\kappa_{0,1}$, $\kappa_{1,0}$  defined by the fundamental solutions $\underset{*}{u} = \gamma^0 u, \underset{*}{v}= \gamma^0 v$,
and construct its decomposition and of the Wick product fields associated to it. We will use simplified normalization, 
given by (\ref{kappa_0,1}) and (\ref{kappa_1,0}) without the constant $1/2$, so that the kernels should be multiplied by the additional factor $(2\pi)^{-3/2}$
in order to restore the correct normalization.

In Subsection \ref{StandardDiracPsiField} we will construct a nonstandard local Dirac field,
given by the formula (\ref{standardpsi(x)}) of Subsection \ref{StandardDiracPsiField}, with non-unitary Lorentz
transformations (keeping unitarity of translations and rotations) and compare it with the standard Dirac
field $\boldsymbol{\psi}(x)$ or $\underset{*}{\boldsymbol{\psi}}(x)$, given by (\ref{psi(x)}) or (\ref{*psi(x)}).
Although they are mutually unitary isomorphic in a sense
explained in Subsection \ref{StandardDiracPsiField}, nonetheless, there are
important differences between these two realizations of the field $\boldsymbol{\psi}$.
We explain them in more details in Subsection
\ref{StandardDiracPsiField}.

\subsection{Comparison of the standard free Dirac field $\boldsymbol{\psi}$
with a nonstandard Dirac field.
Bogoliubov-Shirkov quantization postulate}\label{StandardDiracPsiField}

In this and the following Sections we will use the standard Dirac field $\boldsymbol{\psi}$,
given by the formula (\ref{psi(x)}) of Subsection \ref{psiBerezin-Hida}.

In accordance to the formula  (\ref{psi(x)}) for the free Dirac field $\boldsymbol{\psi}(x)$, we have:
\begin{multline}\label{psi(x)'}
\boldsymbol{\psi}(x)  =
\sum_{s=1}^{2} \, \int \limits_{\mathbb{R}^3} 
{\textstyle\frac{1}{2}}
\underset{*}{u}{}_{s}(\boldsymbol{\p})e^{-ip\cdot x} \,\, b_{s}(\boldsymbol{\p}) \, 
\ud^3 \boldsymbol{\p} \\
+
\sum_{s=1}^{2} \, \int \limits_{\mathbb{R}^3} 
{\textstyle\frac{-1}{2}}
\underset{*}{v}{}_{s}(\boldsymbol{\p})e^{ip\cdot x} \,\, d_{s}(\boldsymbol{\p})^+ \, 
\ud^3 \boldsymbol{\p},
\end{multline}
\[
\underset{*}{u} =  \gamma^0u \,\,\, \underset{*}{v} = \gamma^0 v.
\]
Here we construst an example of a non-standard local Dirac field, with the ordibary local bispinor transformation law, 
but which is nonunitary for the hyperbolic rotations (Lorentz transformations), and unitary for translations and rotations.

This nonstandard field will have the additional weight $|p_0(\boldsymbol{\p})|$ in comparison to the standard formula 
(\ref{psi(x)'}) or (\ref{*psi(x)}):
\begin{multline}\label{standardpsi(x)}
\boldsymbol{\psi}(x)  =
\sum_{s=1}^{2} \, \int \limits_{\mathbb{R}^3} 
|p_0(\boldsymbol{\p})| \,\,
\underset{*}{u}{}_{s}(\boldsymbol{\p})e^{-ip\cdot x} \,\, b_{s}(\boldsymbol{\p}) \, 
\ud^3 \boldsymbol{\p}
\\
+
\sum_{s=1}^{2} \, \int \limits_{\mathbb{R}^3} 
(-1)|p_0(\boldsymbol{\p})| \,\,
\underset{*}{v}{}_{s}(\boldsymbol{\p})e^{ip\cdot x} \,\, d_{s}(\boldsymbol{\p})^+ \, 
\ud^3 \boldsymbol{\p}.
\end{multline}

Nonetheless, the standard quantum Dirac field $\boldsymbol{\psi}$ given by (\ref{psi(x)'}) ,
is unitarily isomorphic to the Dirac field $\boldsymbol{\psi}$ given by (\ref{standardpsi(x)}).
Indeed, the unitary equivalence between
standard $\boldsymbol{\psi}$ and nonstandard (\ref{standardpsi(x)}) is realized by the lifting to the Fock space of the unitary
operator $\mathbb{U}$, and its inverse $\mathbb{U}^{-1}$, of pointwise multiplication by the function
$\boldsymbol{\p} \mapsto |2p_0(\boldsymbol{\p})|^{-1}$ and respectively
$\boldsymbol{\p} \mapsto |2p_0(\boldsymbol{\p})|$ regarded as unitary operators on the respective single particle Hilbert spaces of the realisations of the field 
$\boldsymbol{\psi}$: first is the space
$\mathcal{H}'= \mathcal{H}_{m,0}^{\oplus} \oplus \mathcal{H}_{-m,0}^{\ominus \, \flat}$ used by us for the standard field and the second
$\mathbb{U}\mathcal{H}'$ is almost identical with ours, the only change is that we are using the ordinary measure
$\ud^{3} \boldsymbol{\p}$ on the orbits
$\mathscr{O}_{m,0,0,0}$,
$\mathscr{O}_{-m,0,0,0}$ instead of $\frac{\ud^{3} \boldsymbol{\p}}{|2p_0(\boldsymbol{\p})|^{2}}$,
in constructing Hilbert spaces of bispinors whose Fourier transforms
are concentrated respectively on $\mathscr{O}_{m,0,0,0}$, $\mathscr{O}_{-m,0,0,0}$
and are component-wise square summable with respect to $\ud^{3}\boldsymbol{\p}$.
Therefore, the corresponding function
$|2p_0(\boldsymbol{\p})|^{-1}$ is just equal to the square root of the Radon-Nikodym derivation of the measure
$\frac{\ud^{3} \boldsymbol{\p}}{|2p_0(\boldsymbol{\p})|^{2}}$ on the orbits
$\mathscr{O}_{m,0,0,0}$, $\mathscr{O}_{-m,0,0,0}$ used by us (compare Subsection \ref{e1})
with respect to the new one $\ud^{3} \boldsymbol{\p}$.
Under this redefinition of measure on the orbits the formulas for
$u_{s}(\boldsymbol{\p}), v_{s}(-\boldsymbol{\p})$ remain unchanged, similarly as the formulas for the projectors 
$P^\oplus, P^\oplus(p), P^\ominus, P^\ominus(p), E_\pm, E_\pm(p)$ (compare Appendix \ref{fundamental,u,v})
remain unchanged. The nuclear space $E$ in the corresponding Gelfand triples (\ref{SinglePartGelfandTriplesForPsi})
will remain unchanged with the single particle Hilbert space $\mathcal{H}'$ replaced of course
by $\mathbb{U}\mathcal{H}'$. The formula (\ref{isomorphismU}) for the unitary isomorphism $U$ joining the
Gelfand triple $E \subset \mathbb{U}\mathcal{H}' \subset E^*$ with the standard Gelfand triple
$\mathcal{S}_{A}(\mathbb{R}^3; \mathbb{C}^4) \subset L^2(\mathbb{R}^3; \mathbb{C}^4) \subset
\mathcal{S}_{A}(\mathbb{R}^3; \mathbb{C}^4)^*$ will remain almost the same with the only difference that the additional
factor $1/|2p_0(\boldsymbol{\p})|$ will be absent in it, and accordingly the factor $2|p_0(\boldsymbol{\p})|$
will be absent in the formula for $U^{-1}$. It is readily seen now that the construction of Subsection
\ref{psiBerezin-Hida},
with the mentioned modification of the measure, will indeed produce the nonstandard formula
(\ref{standardpsi(x)}) for the Dirac field.

Note that the unitary operators $\mathbb{U}$, and $\Gamma(\mathbb{U})$, are well-defined as unitary isomorphisms for fields
understood as integral kernel operators with vector-valued kernels, because the operator
$\mathbb{U}$ of multiplication by the function
$\boldsymbol{\p} \mapsto |2p_0(\boldsymbol{\p})|^{-1}$ transforms
$\mathcal{S}(\mathbb{R}^3; \mathbb{C})$ continuously, and even isomorphically, into itself
and induces the isomorphism of the Gelfand triples
\[
\left. \begin{array}{ccccc}   & & \mathcal{H}_{m,0}^{\oplus} \oplus \mathcal{H}_{-m,0}^{\ominus \, \flat} & & \\
 & & \parallel & & \\
           E        & \subset & \mathcal{H}' & \subset & E^*        \\
                               \downarrow \uparrow &         & \mathbb{U}\downarrow \uparrow \mathbb{U}^{-1}      &         & \downarrow \uparrow  \\
                                         E   & \subset &  \mathcal{H}'' = \mathbb{U}\mathcal{H}' & \subset & E^*        \\ 
\end{array}\right..
\]
Let us denote the standard annihilation and creation operators 
over the Fock space $\Gamma(\mathbb{U}\mathcal{H}')$ by $a''(u \oplus v), a''(u \oplus v)^+$. They are constructed 
exactly as the operators $a'(u\oplus v), a'(u\oplus v)^+$
in Subsections \ref{electron}-\ref{electron+positron} with the only change that the weight 
$1/|2p_0(\boldsymbol{\p})|^2$ in the inner products will be absent, and analogously we extend them
over to $u\oplus v \in E^*$ using the corresponding isomorphism
\[
\left. \begin{array}{ccccc}   & & L^2(\sqcup \mathbb{R}^3; \mathbb{C}) & & \\
 & & \parallel & & \\
           \mathcal{S}_{A}(\mathbb{R}^3; \mathbb{C}^4)        & \subset & L^2(\mathbb{R}^3; \mathbb{C}^4) & \subset & \mathcal{S}_{A}(\mathbb{R}^3; \mathbb{C}^4)^*        \\
                               \downarrow \uparrow &         & U \downarrow \uparrow U^{-1}      &         & \downarrow \uparrow  \\
                                         E   & \subset &  \mathcal{H}'' = \mathbb{U}\mathcal{H}' & \subset & E^*        \\ 
\end{array}\right., 
\]
of the triple $E \subset \mathbb{U}\mathcal{H}'  \subset  E^*$ with the standard Gelfand triple,
and with $U,U^{-1}$ given by the formula (\ref{isomorphismU}) with the factors 
$1/|2p_0(\boldsymbol{\p})|$ (resp. $2|p_0(\boldsymbol{\p})$) removed. 
Then if $\boldsymbol{\psi}$ is the nonstandard Dirac field (\ref{standardpsi(x)}) we have
\begin{equation}\label{standardpsi(x)''}
\boldsymbol{\psi}(f) = a''\big(P^\oplus\widetilde{f}|_{{}_{\mathscr{O}_{m,0,0,0}}} \oplus 0\big) + 
a''\Big( 0 \oplus \big(P^\ominus\widetilde{f}|_{{}_{\mathscr{O}_{-m,0,0,0}}}\big)^\flat \Big)^+,
\,\,\, f \in \mathcal{S}(\mathbb{R}^4; \mathbb{C}^4)
\end{equation}
correspondingly to the formula
\begin{equation}\label{psi(x)''}
\boldsymbol{\psi}(f) = a'\big(P^\oplus\widetilde{f}|_{{}_{\mathscr{O}_{m,0,0,0}}} \oplus 0\big) + 
a'\Big( 0 \oplus \big(P^\ominus\widetilde{f}|_{{}_{\mathscr{O}_{-m,0,0,0}}}\big)^\flat \Big)^+,
\,\,\, f \in \mathcal{S}(\mathbb{R}^4; \mathbb{C}^4)
\end{equation}
for the standard free Dirac field (\ref{psi(x)'}) constructed in Subsection \ref{psiBerezin-Hida}, and with the 
following isomorphism
\begin{multline}\label{G(bbU)^+a(bbU(u+v))G(bbU)=a'(u+v)}
 a'\big(\mathbb{U}^{+}(u\oplus v)\big) =  a''(u\oplus v), \\
 a'\big(\mathbb{U}^{+}(u\oplus v)\big)^+ =  a''(u\oplus v)^+, \\
u\oplus v \in E^*,
\end{multline} 
\begin{multline}\label{G(bbU)^+a(bbU(u+v))G(bbU)=a'(u+v)'}
 a'\big(\mathbb{U}^{-1}(u\oplus v)\big) =  a''(u\oplus v), \\
a'\big(\mathbb{U}^{-1}(u\oplus v)\big)^+ =  a''(u\oplus v)^+, \\
u\oplus v \in E \subset E^*.
\end{multline} 
joining the Hida operators $a'(u\oplus v)$ and $a''(u\oplus v)$.

Of course the plane waves defining the vector-valued distributional kernels
$\kappa_{0,1}, \kappa_{1,0}$ defining the nonstandard Dirac field (\ref{standardpsi(x)})
as integral kernel operator
\[
\boldsymbol{\psi} = \Xi_{0,1}(\kappa_{0,1}) + \Xi_{1,0}(\kappa_{1,0})
\]
are equal
\begin{equation}\label{skappa_0,1}
\boxed{
\kappa_{0,1}(s, \boldsymbol{\p}; a,x) = \left\{ \begin{array}{ll}
|p_0(\boldsymbol{\p})|
u_{s}^{a}(\boldsymbol{\p})e^{-ip\cdot x} \,\,\,  & \textrm{if $s=1,2$}
\\
0 & \textrm{if $s=3,4$}
\end{array} \right.,
}
\end{equation}
\begin{equation}\label{skappa_1,0}
\boxed{
\kappa_{1,0}(s, \boldsymbol{\p}; a,x) = \left\{ \begin{array}{ll}
0 & \textrm{if $s=1,2$}
\\
-|p_0(\boldsymbol{\p})|
v_{s-2}^{a}(\boldsymbol{\p})e^{ip\cdot x} \,\,\,  & \textrm{if $s=3,4$}
\end{array} \right.
}
\end{equation}
\[
\textrm{with $p = (|p_0(\boldsymbol{\p})|, \boldsymbol{\p}) \in \mathscr{O}_{m,0,0,0}$}
\]

We claim that if the orthonormality conditions (\ref{u^+u=delta}) for 
$u_{s}(\boldsymbol{\p}), v_{s}(-\boldsymbol{\p})$,
$s=1,2$ (compare Appendix \ref{fundamental,u,v}) are to be preserved, then it is the formula
(\ref{psi(x)'}) for the standard free Dirac field $\boldsymbol{\psi}(x)$ which defines the 
Dirac field with the local and unitary transformation formula, as an immediate consequence of the locality of the 
transformation law (\ref{1/2xUalpha}) and (\ref{1/2xUa}). The locality of 
(\ref{1/2xUalpha}) and (\ref{1/2xUa}) is in turn an immediate consequence of the fact that there are no momentum 
dependent multipliers in the 
transformation law (\ref{1/2pUalpha}) and (\ref{1/2pUa}) acting on the Fourier transforms of 
bispinors concentrated respectively on $\mathscr{O}_{m,0,0,0}$ (elements of $\mathcal{H}_{m,0}^{\oplus}$)
or on $\mathscr{O}_{-m,0,0,0}$ (elements of $\mathcal{H}_{-m,0}^{\ominus}$). 

Namely, recall that the representation $U(a,\alpha)$ of $(a,\alpha) \in T_4 \circledS SL(2, \mathbb{C})$
acts on the Fourier transform $\widetilde{\phi} \in \mathcal{H}_{m,0}^{\oplus}$ 
(concentrated on $\mathscr{O}_{m,0,0,0}$) of bispinor $\phi$ through the formulas (\ref{1/2pUalpha}) and (\ref{1/2pUa})
and on $\phi$ through (\ref{1/2xUalpha}) and (\ref{1/2xUa}). Similarly,
$U'(a,\alpha)^c$ act on $(\widetilde{\phi}')^c \in \mathcal{H}_{-m,0}^{\ominus \, \flat}$ by the conjugation
of the representation $U'(a,\alpha)$ acting on the bispinor $\widetilde{\phi}' \in \mathcal{H}_{-m,0}^{\ominus}$
by the same formula (\ref{1/2pUalpha}) and (\ref{1/2pUa}) and on $\phi'$
through the formula (\ref{1/2xUalpha}) and (\ref{1/2xUa}). On writing
$\boldsymbol{U}(a,\alpha) = U(a,\alpha) \oplus U'(a,\alpha)^c$ for the representation of
$(a,\alpha) \in T_4 \circledS SL(2, \mathbb{C})$ acting in the single particle Hilbert space
$\mathcal{H}_{m,0}^{\oplus} \oplus \mathcal{H}_{-m,0}^{\ominus \, \flat}$ of
the field (\ref{psi(x)'}), we have
\begin{equation}\label{transformationGeneralLocalpsi}
\Gamma(\boldsymbol{U}(a,\alpha)) \boldsymbol{\psi}(f) \Gamma(\boldsymbol{U}(a,\alpha))^{-1}
= \boldsymbol{\psi}\big(U(a,\alpha)f\big)
\end{equation}
where $U(a, \alpha)$ acts on $f \in \mathcal{S}(\mathbb{R}^4; \mathbb{C}^4)$ and gives $U(a, \alpha)f$
in the same fashion as in  (\ref{1/2xUalpha}) and (\ref{1/2xUa}). In particular\footnote{Recall that
here $\Lambda: \alpha \rightarrow \Lambda(\alpha)$ is an antihomomorphism.}
\begin{equation}\label{Ualphaxf}
U(\alpha) f(x) = 
\left( \begin{array}{cc}  \alpha & 0  \\
                                           
                                                   0              & {\alpha^*}^{-1}  \end{array}\right) 
 f(x\Lambda(\alpha^{-1}))
= \left( \begin{array}{cc}  \alpha & 0  \\
                                           
                                                   0              & {\alpha^*}^{-1}  \end{array}\right) 
 f(\Lambda(\alpha)x), 
\end{equation}
\begin{equation}\label{Uaxf}
T(a) f(x) = f(x - a).
\end{equation}
In particular the field (\ref{psi(x)'}) transforms locally, and in particular translations act on (\ref{psi(x)'})
in the standard fashion
\begin{equation}\label{transformationTranslationGeneralLocalpsi}
\Gamma(\boldsymbol{U}(a,0)) \boldsymbol{\psi}(f) \Gamma(\boldsymbol{U}(a,0))^{-1}
= \boldsymbol{\psi}\big(U(a,0)f\big) = \boldsymbol{\psi}\big(T(a) f\big) 
\end{equation}

It is easily seen that the operator of multiplication by the function
$\boldsymbol{\p} \mapsto |p_0(\boldsymbol{\p})|^{-1}$ in action on
$\mathcal{H}_{m,0}^{\oplus}$ and on $\mathcal{H}_{-m,0}^{\ominus}$ (compare Subsection \ref{e1})
commutes with the translation operator (\ref{1/2pUa}) and with the operators
(\ref{1/2pUalpha}) representing spatial rotations (because $|p_0(\boldsymbol{\p})|
= \sqrt{|\boldsymbol{\p}|^2 + m^2}$ is invariant under rotations). Therefore, both free Dirac fields:
the standard (\ref{psi(x)}) and the nonstandard one (\ref{standardpsi(x)}), transform locally and identically under
translations and spatial rotations. Namely, for $(a,\alpha) = (a,0) \in T_4 \circledS SL(2, \mathbb{C})$
or for $(a,\alpha) = (0, \alpha) \in T_4 \circledS SU(2, \mathbb{C}) \subset T_4 \circledS SL(2, \mathbb{C})$
i.e. for translations or spatial rotations, we have
\[
\Gamma\big(\mathbb{U}\boldsymbol{U}(a,\alpha)\mathbb{U}^{-1}\big) \boldsymbol{\psi}(f)
\Gamma\big(\mathbb{U}\boldsymbol{U}(a,0) \mathbb{U}^{-1}\big)^{-1}
= \boldsymbol{\psi}\big(U(a,\alpha)f\big)
\]
with the standard local formula for the transformation formula (\ref{Ualphaxf}), (\ref{Uaxf})
for space-time transformed bispinor $U(a,\alpha)f$, and for the nonstandard Dirac quantum field (\ref{standardpsi(x)}) with the representation
\[
\Gamma\big(\mathbb{U}\boldsymbol{U}(a,\alpha)\mathbb{U}^{-1}\big)
\]
acting in its Fock space
\[
\Gamma\big(\mathbb{U}\big) \big( \mathcal{H}_{m,0}^{\oplus} \oplus \mathcal{H}_{-m,0}^{\ominus \, \flat}\big)
= \Gamma\big(\mathbb{U}(\mathcal{H}_{m,0}^{\oplus} \oplus \mathcal{H}_{-m,0}^{\ominus \, \flat}) \big),
\]
and with the representation
\[
\mathbb{U}\boldsymbol{U}(a,\alpha)\mathbb{U}^{-1}
\]
acting in its single particle Hilbert space
\[
\mathcal{H}'' = \mathbb{U}\big(\mathcal{H}_{m,0}^{\oplus} \oplus \mathcal{H}_{-m,0}^{\ominus \, \flat}\big)
= \mathbb{U} \mathcal{H}'.
\]

Note that for the bispinor $\underset{\circ}{\widetilde{\phi}} = \mathbb{U} \widetilde{\phi}$,
$\widetilde{\phi} \in \mathcal{H}_{m,0}^{\oplus}$, such that  
$\underset{\circ}{\widetilde{\phi}} \oplus 0 \in \mathcal{H}''$, concentrated on
$\mathscr{O}_{m,,0,0,0}$, or $0 \oplus {\underset{\circ}{\widetilde{\phi}}}^c  \in \mathcal{H}''$, 
$\underset{\circ}{\widetilde{\phi}} = \mathbb{U} \widetilde{\phi}$,
$\widetilde{\phi} \in \mathcal{H}_{-m,0}^{\ominus}$,
concentrated on $\mathscr{O}_{-m,0,0,0}$, 
we have
\[
\mathbb{U} U(\alpha)\mathbb{U}^{-1} \underset{\circ}{\widetilde{\phi}}(p) = \Bigg|\frac{p_0(\Lambda(\alpha)p)}{p_0(p)}\Bigg|
\left( \begin{array}{cc}  \alpha & 0  \\
                                           
                                                   0              & {\alpha^*}^{-1}  \end{array}\right) 
 \underset{\circ}{\widetilde{\phi}}(\Lambda(\alpha)p), 
\]
\[
\mathbb{U} T(a) \mathbb{U}^{-1} \underset{\circ}{\widetilde{\phi}}(p) 
= e^{i a \cdot p}\underset{\circ}{\widetilde{\phi}}(p).
\]

Therefore, for the Lorentz transformations (\ref{1/2pUalpha}) situation is
different for the two mentioned
realizations of the Dirac free field. Namely, the standard field (\ref{psi(x)'}) by construction transforms locally as
a bispinor field also under Lorentz transformations. But the operator $\mathbb{U}$
of pointwise multiplication by the function $\boldsymbol{\p} \mapsto |p_0(\boldsymbol{\p})|^{-1}$ does not commute with
the operator $U(\alpha)$ for $\alpha \notin SU(2, \mathbb{C})$ given by (\ref{1/2pUalpha}),
and moreover it is immediately seen
that transformation formula $\mathbb{U} U(\alpha) \mathbb{U}^{-1}$ gains nontrivial momentum dependent
multiplier
\[
|p_0(\Lambda(\alpha)p)/p_0(p)| \neq 1
\]
for $\alpha \notin SU(2, \mathbb{C})$.
This additional multiplier means that $\mathbb{U}\boldsymbol{U}(a,\alpha)\mathbb{U}^{-1}$ in action on
the elements of $\mathcal{H}''$, viewed as distributional Fourier
transforms of positive (respectively conjugations of negative) energy solutions
$\mathscr{F}^{-1}\widetilde{\underset{\circ}{\phi}}$ of Dirac equation,
concentrated respectively on $\mathscr{O}_{m,0,0,0,0}$ or $\mathscr{O}_{-m,0,0,0}$,
induce non-local transformation law on $\mathscr{F}^{-1}\widetilde{\underset{\circ}{\phi}}$. Alternatively
this additional multiplier, however, can be viewed as coming from the non-invariance of the ordinary euclidean measure
$\ud^3 \boldsymbol{\p}$ under Lorentz transformation on the respective orbits $\mathscr{O}_{m,0,0,0}$ and
$\mathscr{O}_{-m,0,0,0}$, which assures locality of Lorentz transformations not for the ordinary inverse Fourier transformed elements of
$\mathcal{H}''$ but for the inverse Fourier transform of the elements
$\mathbb{U}^{-1}\underset{\circ}{\widetilde{\phi}}$, $\underset{\circ}{\widetilde{\phi}} \in \mathcal{H}''$.
Namely, consider the following formula
\begin{multline*}
\phi(x) = \int \limits_{\mathscr{O}_{m,0,0,0}} \widetilde{\phi}(p) e^{-ip \cdot x} \, \ud \mu_{{}_{\mathscr{O}_{m,0,0,0}}}(p)
= \int \limits_{\mathbb{R}^3} \frac{\widetilde{\phi}(\boldsymbol{\p}, p_0(\boldsymbol{\p}))}{p_0(\boldsymbol{\p})}
e^{-ip \cdot x} \, \ud^3 \boldsymbol{\p} \\
= \int \limits_{\mathbb{R}^3} \mathbb{U}\widetilde{\phi}(\boldsymbol{\p}) e^{-ip \cdot x}
\, \ud^3 \boldsymbol{\p}
= \int \limits_{\mathbb{R}^3} \underset{\circ}{\widetilde{\phi}}(\boldsymbol{\p}) e^{-ip \cdot x}
\, \ud^3 \boldsymbol{\p},
\end{multline*}
for the positive energy solutions. We have analogue formula for negative energy solutions.
Consider now the local transformation formula for $U(\alpha)\phi$ with $\phi$ expressed by the above formula.
We will get
\begin{multline*}
U(\alpha)\phi(x) = 
\left( \begin{array}{cc}  \alpha & 0  \\  
                                                   0              & {\alpha^*}^{-1}  \end{array}\right) 
\phi(\Lambda(\alpha)x) \\
=\left( \begin{array}{cc}  \alpha & 0  \\                                          
                                                   0              & {\alpha^*}^{-1}  \end{array}\right) 
\int \limits_{\mathbb{R}^3} \underset{\circ}{\widetilde{\phi}}(\boldsymbol{\p}) e^{-ip \cdot \Lambda x}
 \, \ud^3 \boldsymbol{\p} \\ =
\left( \begin{array}{cc}  \alpha & 0  \\                                           
                                                   0              & {\alpha^*}^{-1}  \end{array}\right) 
\int \limits_{\mathbb{R}^3} \underset{\circ}{\widetilde{\phi}}(\Lambda \boldsymbol{\p}) e^{-ip \cdot x}
 \, \ud^3 \Lambda \boldsymbol{\p} \\ =
\left( \begin{array}{cc}  \alpha & 0  \\                                           
                                                   0              & {\alpha^*}^{-1}  \end{array}\right) 
\int \limits_{\mathbb{R}^3} \underset{\circ}{\widetilde{\phi}}(\Lambda \boldsymbol{\p}) e^{-ip \cdot x}
 \, \Bigg|\frac{\ud^3 \Lambda \boldsymbol{\p}}{\ud^3 \boldsymbol{\p}} \Bigg| \, \ud^3 \boldsymbol{\p}.
\end{multline*}
Taking into account the invariance property
\[
\frac{\ud^3 \Lambda \boldsymbol{\p}}{|p_0(\Lambda \boldsymbol{\p})|} =
\frac{\ud^3 \boldsymbol{\p}}{|p_0(\boldsymbol{\p})|} \,\,\, \Longleftrightarrow \,\,\,
\Bigg| \frac{\ud^3 \Lambda \boldsymbol{\p}}{\ud^3 \boldsymbol{\p}} \Bigg| = 
\frac{|p_0(\Lambda \boldsymbol{\p})|}{|p_0(\boldsymbol{\p})|},
\]
we obtain 
\[
U(\alpha)\phi(x)  =
\left( \begin{array}{cc}  \alpha & 0  \\                                           
                                                   0              & {\alpha^*}^{-1}  \end{array}\right) 
\int \limits_{\mathbb{R}^3} \underset{\circ}{\widetilde{\phi}}(\Lambda \boldsymbol{\p}) e^{-ip \cdot x}
 \, \frac{|p_0(\Lambda \boldsymbol{\p})|}{|p_0(\boldsymbol{\p})|} \, \ud^3 \boldsymbol{\p},
\,\,\,\,\, p \in \mathcal{O}_{m,0,0,0},
\]
\emph{i.e.} again the assertion that the transformation 
$\mathbb{U} U(\alpha)\mathbb{U}^{-1} \underset{\circ}{\widetilde{\phi}}$ 
of $\underset{\circ}{\widetilde{\phi}} = \mathbb{U}\widetilde{\phi}$ is accompanied by the ordinary local 
bispinor transformation $U(\alpha) \phi$ of $\phi$, but not of $\mathscr{F}^{-1}\widetilde{\underset{\circ}{\phi}}$. 
Here, of course, we have used the spatial momentum coordinates $\boldsymbol{\p} = (p_1,p_2,p_3)$ 
on the respective orbits $\mathscr{O}_{{}_{\pm m, 0,0,0}}$ and put shortly
$\widetilde{\phi}(\boldsymbol{\p})$ for $\widetilde{\phi}(\boldsymbol{\p}, p_0(\boldsymbol{\p}))$
and, respectively, $\underset{\circ}{\widetilde{\phi}}(\boldsymbol{\p})$ for $\underset{\circ}{\widetilde{\phi}}(\boldsymbol{\p}, p_0(\boldsymbol{\p}))$.
In this coordinates  $\boldsymbol{\p} = (p_1,p_2,p_3)$  on the orbits the action of the $SL(2, \mathbb{C})$ group reads
\begin{multline}\label{SL(2,C)actionInSpatialMomentumCoordinates}
\Lambda \boldsymbol{\p} = \Lambda(\alpha)\boldsymbol{\p}
\\ 
\coloneqq 
\Bigg( 
\, \sum\limits_{k=1}^{3} \Lambda^{k}_{1} \, p_k + \Lambda^{0}_{1} \, p_0(\boldsymbol{\p}) \, ,
\, \sum\limits_{k=1}^{3} \Lambda^{k}_{2} \, p_k + \Lambda^{0}_{2} \, p_0(\boldsymbol{\p}) \, ,
\, \sum\limits_{k=1}^{3} \Lambda^{k}_{3} \, p_k + \Lambda^{0}_{3} \, p_0(\boldsymbol{\p}) \,
\Bigg)
\end{multline}
\[
\Lambda = \Lambda(\alpha), \,\,\,\, p_0(\boldsymbol{\p}) = \pm \sqrt{|\boldsymbol{\p}|^2 + m^2} 
\,\,\,\, \textrm{on} \,\,\,\, \mathscr{O}_{{}_{\pm m, 0,0,0}}.
\]

Similar relation we obtain for the conjugations of the negative energy solutions whose Fourier transforms are 
concentrated on $\mathscr{O}_{-m,0,0,0}$. Therefore, if 
$f \in \mathcal{S}(\mathbb{R}^4; \mathbb{C}^4)$
is a space-time test bispinor, then the transformation $\mathbb{U} U(\alpha)\mathbb{U}^{-1}$ (or its conjugation)
in action on 
\[
P^\oplus \mathbb{U}\widetilde{f}|_{{}_{\mathscr{O}_{m,0,0,0}}} \,\,\, \textrm{or resp.} \,\,\,
\big(P^\ominus \mathbb{U}\widetilde{f}|_{{}_{\mathscr{O}_{-m,0,0,0}}}\big)^\flat
\]
induces local bispinor transformation on $f$. This would be false for the action 
of $\mathbb{U} U(\alpha)\mathbb{U}^{-1}$ (or its conjugation) on 
\[
P^\oplus \widetilde{f}|_{{}_{\mathscr{O}_{m,0,0,0}}} \,\,\, \textrm{or resp.} \,\,\,
\big(P^\ominus \widetilde{f}|_{{}_{\mathscr{O}_{-m,0,0,0}}}\big)^\flat.
\]
Thus we see again that it is the standard field (\ref{psi(x)'}), or equivalently the field 
(\ref{psi(x)''}), which transforms locally as ordinary bispinor under the Fock lifting of
$U(\alpha)$ (summed up with its conjugation). The nonstandard field (\ref{standardpsi(x)}), or equivalently the field
(\ref{standardpsi(x)''}), transforms non-locally under the Fock lifting
of the unitary representation $\mathbb{U} U(\alpha)\mathbb{U}^{-1}$ (summed up with its conjugation). 
 Correspondingly the nonstandard Dirac quantum field
(\ref{standardpsi(x)}) transforms non-locally under Lorentz transformations, if the unitarity is to be preserved. 
Locality under proper Lorentz transformations of the nonstandard field (\ref{standardpsi(x)})
can be restored, but then the unitarity of the Lorentz transformations will have to be abandoned.
Now we explain this in more details and, in the later part of this Subsection, its connection to the so called
Noether theorem for free fields.

Let use the Cartesian spatial momentum components $\boldsymbol{\p}$ on the respective orbits $\mathscr{O}_{{}_{\pm m, 0,0,0}}$.
Let us put shortly
$\widetilde{\phi}(\boldsymbol{\p})$ for $\widetilde{\phi}(\boldsymbol{\p}, p_0(\boldsymbol{\p}))$, with
$p_0(\boldsymbol{\p}) = \pm \sqrt{|\boldsymbol{\p}|^2 + m^2}$ on $\mathscr{O}_{{}_{\pm m, 0,0,0}}$, for the bispinors
$\widetilde{\phi}$ concentrated, repectively, on $\mathscr{O}_{{}_{\pm m, 0,0,0}}$, and lying,
respectively, in the images of $P^\oplus$ or $P^\ominus$.  
Indeed, locality for the transformation of the nonstandard field (\ref{standardpsi(x)}) can be restored, 
if instead of the transformation $\mathbb{U} U(\alpha)\mathbb{U}^{-1}$,
which is unitary in the single particle Hilbert space of positive and negative energy solutions
of the nonstandard field (\ref{standardpsi(x)}) with respect to the inner product
\begin{equation}\label{StandardScalarProduct}
(\widetilde{\phi},\widetilde{\phi}') = \int\limits_{\mathscr{O}_{{}_{m,0,0,0}} \sqcup \mathscr{O}_{{}_{-m,0,0,0}}}
\big(\widetilde{\phi}(\boldsymbol{\p}),\widetilde{\phi}'(\boldsymbol{\p})\big)_{{}_{\mathbb{C}^4}} \, \ud^3 \boldsymbol{\p},
\end{equation}
we restore the ordinary local transformation $U(\alpha)$ given by (\ref{1/2pUalpha}) and (\ref{1/2pUa}), 
which in the $x$-variables has the ordinary local form of bispinor transformation (\ref{Ualphaxf}) and (\ref{Uaxf}). But the scalar product
(\ref{StandardScalarProduct}) cannot be invariant under
$U(\alpha)$, acting on the bispinor functions $\widetilde{\phi}$, concentrated 
on $\mathscr{O}_{{}_{m,0,0,0}} \sqcup \mathscr{O}_{{}_{-m,0,0,0}}$, and lying in the images, respectively, 
of $P^\oplus$ on $\mathscr{O}_{{}_{m,0,0,0}}$ and $P^\ominus$ on $\mathscr{O}_{{}_{-m,0,0,0}}$.
Indeed, as we already know, the inner product
\begin{equation}\label{InvariantScalarProduct}
(\widetilde{\phi},\widetilde{\phi}') = \int\limits_{\mathscr{O}_{{}_{m,0,0,0}} \sqcup \mathscr{O}_{{}_{-m,0,0,0}}}
\big(\widetilde{\phi}(\boldsymbol{\p}),\widetilde{\phi}'(\boldsymbol{\p})\big)_{{}_{\mathbb{C}^4}} \, 
{\textstyle\frac{\ud^3 \boldsymbol{\p}}{(2p_0(\boldsymbol{\p}))^2}},
\end{equation}
in the single particle space $\mathcal{H}^{\oplus}_{m,0} \oplus \mathcal{H}^{\ominus}_{-m,0}$ of positive and negative energy solutions
is the one which is invariant under $U(\alpha)$, and serves as the inner product in the single particle space of the
standard field (\ref{psi(x)'}), or equivalently of the field (\ref{psi(x)''}). But the same $U(\alpha)$
cannot preserve both inner products (\ref{StandardScalarProduct}) and (\ref{InvariantScalarProduct}).  
 Indeed, let us suppose that both (\ref{StandardScalarProduct}) and (\ref{InvariantScalarProduct}),
are invariant under $U(\alpha)$. Let 
\[
\widetilde{\phi} \in P^\oplus\big[\mathcal{S}(\mathbb{R}^3; \mathbb{C}^4) \big] \oplus 
P^\ominus\big[\mathcal{S}(\mathbb{R}^3; \mathbb{C}^4) \big],
\]
so that $\widetilde{\phi}$ belongs to a dense subspace of the single particle Hilbert spaces
of both fields, and both inner products (\ref{StandardScalarProduct}) and (\ref{InvariantScalarProduct})
are well-defined for such $\widetilde{\phi}$.
We will get
\begin{multline*}
\int\limits_{\mathscr{O}_{{}_{m,0,0,0}} \sqcup \mathscr{O}_{{}_{-m,0,0,0}}}
\big(U(\alpha)\widetilde{\phi}(\boldsymbol{\p}),U(\alpha)\widetilde{\phi}(\boldsymbol{\p})\big)_{{}_{\mathbb{C}^4}} 
\, {\textstyle\frac{\ud^3 \boldsymbol{\p}}{(2p_0(\boldsymbol{\p}))^2}}
\\
=\int\limits_{\mathscr{O}_{{}_{m,0,0,0}} \sqcup \mathscr{O}_{{}_{-m,0,0,0}}}
\big(\widetilde{\phi}(\boldsymbol{\p}),\widetilde{\phi}(\boldsymbol{\p})\big)_{{}_{\mathbb{C}^4}} \, 
{\textstyle\frac{\ud^3 \boldsymbol{\p}}{(2p_0(\boldsymbol{\p}))^2}}
= \int\limits_{\mathscr{O}_{{}_{m,0,0,0}} \sqcup \mathscr{O}_{{}_{-m,0,0,0}}}
\Big({\textstyle\frac{1}{2p_0}}\widetilde{\phi}(\boldsymbol{\p}),{\textstyle\frac{1}{2p_0}}\widetilde{\phi}(\boldsymbol{\p})\Big)_{{}_{\mathbb{C}^4}} 
\, \ud^3 \boldsymbol{\p}
\end{multline*}
\begin{multline*}
= \int\limits_{\mathscr{O}_{{}_{m,0,0,0}} \sqcup \mathscr{O}_{{}_{-m,0,0,0}}}
\Big(U(\alpha)\big({\textstyle\frac{1}{2p_0}}\widetilde{\phi}\big)(\boldsymbol{\p}),
U(\alpha)\big({\textstyle\frac{1}{2p_0}}\widetilde{\phi}\big)(\boldsymbol{\p})\Big)_{{}_{\mathbb{C}^4}} 
\, \ud^3 \boldsymbol{\p}
\\
= \int\limits_{\mathscr{O}_{{}_{m,0,0,0}} \sqcup \mathscr{O}_{{}_{-m,0,0,0}}}
\Big({\textstyle\frac{p_0(\boldsymbol{\p})}{p_0(\Lambda\boldsymbol{\p})}}\Big)^2
\Big(U(\alpha)\widetilde{\phi}(\boldsymbol{\p}),
U(\alpha)\widetilde{\phi}(\boldsymbol{\p})\Big)_{{}_{\mathbb{C}^4}} 
\, {\textstyle\frac{\ud^3 \boldsymbol{\p}}{(2p_0(\boldsymbol{\p}))^2}},
\end{multline*}
for all $\alpha \in SL(2, \mathbb{C})$ and all
\[
\widetilde{\phi} \in P^\oplus\big[\mathcal{S}(\mathbb{R}^3; \mathbb{C}^4) \big] \oplus 
P^\ominus\big[\mathcal{S}(\mathbb{R}^3; \mathbb{C}^4) \big].
\]
Thus, for all such $\widetilde{\phi}$ and all $\alpha \in SL(2, \mathbb{C})$ and all corresponding hyperbolic rotations $\Lambda(\alpha)$
\begin{equation}\label{contradiction}
\int\limits_{\mathscr{O}_{{}_{m,0,0,0}} \sqcup \mathscr{O}_{{}_{-m,0,0,0}}}
\Big[1 -  \Big({\textstyle\frac{p_0(\boldsymbol{\p})}{p_0(\Lambda\boldsymbol{\p})}} \Big)^2 \, \Big]
\big(U(\alpha)\widetilde{\phi}(\boldsymbol{\p}),U(\alpha)\widetilde{\phi}(\boldsymbol{\p})\big)_{{}_{\mathbb{C}^4}} 
\, {\textstyle\frac{\ud^3 \boldsymbol{\p}}{(2p_0(\boldsymbol{\p}))^2}} = 0.
\end{equation}
Now let $\alpha \notin SU(2,\mathbb{C})$ be any fixed element of $SL(2, \mathbb{C})$, and let $\big(\boldsymbol{\p}, p_0(\boldsymbol{\p})\big)$
be any fixed point of the orbit $\mathscr{O}_{{}_{m,0,0,0}}$ or $\mathscr{O}_{{}_{-m,0,0,0}}$ with coordinates $\boldsymbol{\p}$. 
Suppose that 
\[
p_0(\boldsymbol{\p}) \neq p_0(\Lambda(\alpha)\boldsymbol{\p}),
\]
for these $\alpha, \boldsymbol{\p}$.  But this would be in contradiction to (\ref{contradiction}), because
choosing $\widetilde{\phi} = P^\oplus f$ or $\widetilde{\phi} = P^\ominus f$, with $f\in \mathcal{S}(\mathbb{R}^3; \mathbb{C}^4)$
concentrated arbitrarily closely around the point $\Lambda(\alpha)\boldsymbol{\p}$ of $\mathscr{O}_{{}_{m,0,0,0}}$ or $\mathscr{O}_{{}_{-m,0,0,0}}$,
i.e. with 
\[
\big(U(\alpha)\widetilde{\phi}(\boldsymbol{\p}),U(\alpha)\widetilde{\phi}(\boldsymbol{\p})\big)_{{}_{\mathbb{C}^4}} > 0,
\]
we would get a nonzero value of the left-hand side integral in (\ref{contradiction}). Therefore assumption that both
(\ref{StandardScalarProduct}) and (\ref{InvariantScalarProduct}) are invariant under $U(\alpha)$ leads us to a contradiction
that
\[
p_0(\boldsymbol{\p}) = p_0(\Lambda(\alpha)\boldsymbol{\p}),  \,\,\,\,
\alpha \notin SU(2,\mathbb{C}),  \alpha \in SL(2,\mathbb{C}) \,\,\,\,\, \textrm{(contradiction)},
\]
because the last equality is impossible for the the hyperbolic rotations, \emph{i.e.} Lorentz transformations. 
Thus, both (\ref{StandardScalarProduct}) and (\ref{InvariantScalarProduct}) cannot be invariant under $U(\alpha)$
given by (\ref{1/2pUalpha}) and (\ref{1/2pUa}). Because
(\ref{InvariantScalarProduct}) is invariant under $U(\alpha)$, we have just proved that 
(\ref{StandardScalarProduct}) is not invariant under the local transformation $U(\alpha)$
given by (\ref{1/2pUalpha}) and (\ref{1/2pUa}).
Therefore, if we restore the local transformation $U(\alpha)$, given by (\ref{1/2pUalpha}) and (\ref{1/2pUa}),
the nonstandard field (\ref{standardpsi(x)}) will keep the local ordinary transformation of a bispinor field. But in this case
the transformation $U(\alpha)$ is no longer unitary with respect to the inner product 
(\ref{StandardScalarProduct}) in the single particle Hilbert space of the nonstandard field 
 (\ref{standardpsi(x)}). Strictly speaking we have proved it for the single particle positive and negative 
energy states of the field  (\ref{standardpsi(x)}). But from the proved lack of unitarity it immedately follows
that $U(\alpha)$ cannot be unitary in the single particle Hilbert space of positive energy solutions
and that its conjugation $U^\flat(\alpha)$ cannot be unitary on the Hilbert space of conjugations of the
negative enery states of the nonstandard field (\ref{standardpsi(x)}). In the next Subsection
we will analyze in more details the non-unitary representation $\alpha \mapsto U(\alpha)$ of $SL(2, \mathbb{C})$ regarded
as acting in the single particle Hilbert space of the nonstanard field
(\ref{standardpsi(x)}) with the inner product (\ref{StandardScalarProduct}) and as the unitary representation in the single
particle space of the standard field (\ref{psi(x)'}) with the invariant inner product (\ref{InvariantScalarProduct}).

Although the Dirac free fields (\ref{psi(x)'}) and (\ref{standardpsi(x)}) are unitarily isomorphic,
in the sense of the isomorphism (\ref{G(bbU)^+a(bbU(u+v))G(bbU)=a'(u+v)}) or (\ref{G(bbU)^+a(bbU(u+v))G(bbU)=a'(u+v)'}),
joining the corresponding Hida operators $a', a''$, 
there are some further and important differences between them. 

The first concerns locality
under the proper Lorentz transformations, already explained. The field (\ref{psi(x)'}) is constructed from the
direct sum of two (equivalent) irreducible unitary representations, giving the local transformation law for the elements
of the single particle Hilbert
space regarded as the space of (regular distributional) solutions of the Dirac equation, whose Fourier transforms
compose $\mathcal{H}'$ and are concentrated on the orbit $\mathscr{O}_{m,0,0,0}$ or eventually are equal to conjugations
of bispinors concentrated on the orbit $\mathscr{O}_{-m,0,0,0}$. The nonstandard field (\ref{standardpsi(x)})
is constructed from different non-unitary representation, not equivalent with it, which
assures the local and unitary  transformation law of the elements of the single particle space, 
understood as solutions of the Dirac equation, but only under the translation subgroup or spatial rotations.
It is a general paradigm that the locality of the transformation under the full $T_4 \circledS SL(2, \mathbb{C})$ is
the fundamental assumption, and whenever we are able to construct a free field out of a representation
of $T_4 \circledS SL(2, \mathbb{C})$ it is customary to put the additional requirement of locality of the transformation
law induced by the representation. This is the case within causal perturbative QFT.  It is true that, in the realm of causal perturbatve approach to QFT,
it is the covariance under translations only (with the standard local transformation formula) which plays the crucial 
role in the existence proof of the causal perturbative series of the scattering operator $S$. 
Concerning the scattering operator, its contruction is not unique, if we take into account the Bogliubov
axioms (I)-(IV) plus the axiom (V) of the order preservation, compare Subsections \ref{MotivationForHida}, \ref{WickForProduct}. 
Here Lorentz covarince
and locality under Lorentz trnsformations intervene. It is by the Lorentz covariance condition that a considarable
part of the freedom is eliminated in the construction of $S$. 
Lorentz covariance would make no sense in the case of a non-local Lorentz transformation, 
by the very local switching off/on mechanism of the interaction, which could not be reconciled with the non-locality
of the transformation. Thus here Loretz covariance intervenes rather dramatically.  
As to the interactig fields
their existence in the adiabatic limit 
is possible in the theory with the Hida operators as the creation-annihilation operators, if and only if the splitting
of causal distributions is ``natural'', and again only Lorentz covariance condition elimites possible freedom here.  

The Lorentz covariance postulate for $S$ could in principle be abandoned, even though the free fields were Lorentz covariant,
although it seems it would be an artificial construction. 
We would get a theory with undefined parameters in the splitting of causal distributions into the retarded and advanced part.
In this sense non-local transformation under Lorentz transformations can in principle be allowed, because we can in principle abandon
the Lorentz covariance condition for $S$, using instead covariance under translations only. 

 The local Lorentz covariance for the free fields, and Lorentz covariance for $S$ turns out to be optional
in the sense explainded above, with the important caveat that the theory without Lorentz covariance for $S$ 
would contain indefinite parameters. Thus, axioms (I) - (V) without Lorentz covariance 
but only with translational covariance, would not give a fully defined theory.
 
Moreover, it is known that also for determination of the commutation rules for free fields according to the classic
procedure due to Pauli-Bogoliubov-Shirkov, it is the so called Noether theorem for translations which is
sufficient in derivation of these rules (compare \cite{Bogoliubov_Shirkov}, where it is understood as an example of the
Bohr's \emph{correspondence principle}). Below we explain the relation of the (\ref{psi(x)'}) and (\ref{standardpsi(x)}) 
fields to the translation generator, which shows a dramatic difference between the fields (\ref{psi(x)'}) and (\ref{standardpsi(x)}).
At least from the causal perturbative approach, both
(\ref{psi(x)'}) and (\ref{standardpsi(x)}) could be accepted at the axiomatic level, although lead to completely different
theories. In principle at least 
(\ref{standardpsi(x)}) could be accepted, but without Lorentz covariance in the axioms for (I)-(V)
for $S$, leading to a theory which is incomplete.

Although (\ref{psi(x)}) and (\ref{standardpsi(x)}) are unitarily isomorphic, they
have \emph{different} ''commutation generalized functions'' as well as \emph{different} ''pairing functions'',
which enter the causal perturbative series
accordingly to different anti-commutation rules
\[
\begin{split}
\big\{a'(u \oplus v), a'(u'\oplus v')^+ \big\} = \big(u \oplus v, \, u' \oplus v' \big)_{{}_{\mathcal{H}'}}, \,\,\,\,
\,\,\,
u \oplus v \in E, \\
\big\{a''(u \oplus v), a''(u'\oplus v')^+ \big\} = \big(u \oplus v, \, u' \oplus v' \big)_{{}_{\mathbb{U}\mathcal{H}'}},
\,\,\,
u \oplus v \in E
\end{split}
\]
with different inner products: with the additional weight $|2p_0(\boldsymbol{\p})|^{-2}$
in the formula for $\big(\cdot, \cdot \big)_{{}_{\mathcal{H}'}}$ in comparison to
$\big( \cdot, \cdot \big)_{{}_{\mathbb{U}\mathcal{H}'}}$, where the weight $|2p_0(\boldsymbol{\p})|^{-2}$
is absent.

That locality and unitarity under Lorentz transformations cannot be reconciled for nonstandard Dirac 
fields, such as e.g. (\ref{standardpsi(x)}), is non-trivial.
Also, the consequent construction of the Dirac field based on the 
the theory of representations of $T_4 \circledS SL(2, \mathbb{C})$ an white-noise analysis, which we have given
in Subsections \ref{FirstStepH} --\ref{psiBerezin-Hida}, is non-trivial.
The lack of the adequate group theoretical construction of the Dirac field has been noted e.g. by Haag
\cite{Haag}, p. 48.

But there is also another difference between (\ref{psi(x)'}) and (\ref{standardpsi(x)}), which can invariantly be
expressed by recalling to the first Noether theorem applied to the free quantum fields. We devote the rest
part of this Subsection to the Noether theorem restricted to translations and Lorentz transformations and its 
relation to the fields (\ref{psi(x)'}) and (\ref{standardpsi(x)}).

Let us recall the Noether theorem for free fields after \cite{Bogoliubov_Shirkov}, 
Chap. 2, \S 9.4 (in 1980 Ed.), where it is called 
the \emph{Quantization Postulate}: 

\emph{The operators for the energy-momentum four-vector
$\boldsymbol{P}$, and the angular momentum tensor $\boldsymbol{M}$, the charge
$\boldsymbol{Q}$, and so on, which are the generators of the corresponding 
symmetry transformations of state vectors, can be expressed in terms of the operator functions 
of the fields by the same relations as in classical field theory with the operators
arranged in the normal order}.
 
Let us start our analysis with translations.

Here we confine our attention to the Dirac field $\boldsymbol{\psi}$ given by 
(\ref{standardpsi(x)}) and, respectively, by  (\ref{psi(x)'}).
Let $T^{0 \mu}$ be the $0-\mu$-components of the energy-momentum tensor for the free ``classic'' Dirac field
$\psi$ corresponding to translations via Emmy Noether theorem (compare \cite{Bogoliubov_Shirkov})
expressed in terms of $\psi(x)$ and of its derivatives $\partial_\nu \psi(x)$. According to this theorem the spatial 
integral
\[
\int T^{0\mu} \, \ud^3 \boldsymbol{\x}= \frac{i}{2}  \int  \Bigg(
\psi^\sharp(x)\gamma^0 \frac{\partial \psi}{\partial x_\mu}(x) 
- \frac{\partial \psi^\sharp}{\partial x_\mu}(x) \gamma^0 \psi(x)  \Bigg) \, \ud^3 \boldsymbol{\x}, 
\] 
is equal to the conserved integral corresponding to the translational symmetry, i.e. energy-momentum components
of the field $\psi$. Here $\psi^\sharp(x) = \psi(x)^\sharp$ (frequently written $\overline{\psi}(x)$) stands for the Dirac adjoint $\psi(x)^+\gamma^0$. 
We replace the classical field $\psi$ in the above integral formally by the quantum field 
$\boldsymbol{\psi}$ with the counterpart of Dirac adjoint appropriately defined (see below) and with the product 
under the integral sign defined as the Wick product of the fields at the 
same space-time point (compare preceding Subsection \ref{OperationsOnXi}).  

Recall that in both cases, (\ref{psi(x)'}) and (\ref{standardpsi(x)}), we realize the field operators as the integral 
kernel operators with the corresponding vector-valued distributions $\kappa_{0,1}, \kappa_{1,0}$,
over the standard Gelfand triple
$E_1 = \mathcal{S}_{A}(\mathbb{R}^3; \mathbb{C}^4) \subset L^{2}(\mathbb{R}^3; \mathbb{C}^4) \subset E_{1}^{*}$ in
both cases (\ref{psi(x)'}) and (\ref{standardpsi(x)}).

Thus, we are going to check if 
\[
\int \boldsymbol{:} T^{0\mu} \boldsymbol{:} \, \ud^3 \boldsymbol{\x} = \boldsymbol{P}^\mu = d\Gamma(P^\mu),
\] 
where $P^\mu$, $\mu = 0,1,2,3$, are the translation generators of the representation  
$U \boldsymbol{U}(a,\alpha)U^{-1}$, acting in $U\mathcal{H}' = L^2(\mathbb{R}^3; \mathbb{C}^4)$ 
(in the first case (\ref{psi(x)'}))
or $U \mathbb{U}\boldsymbol{U}(a,\alpha)\mathbb{U}^{-1}U^{-1}$ in the same $U\mathbb{U}\mathcal{H}'
= L^2(\mathbb{R}^3; \mathbb{C}^4)$ standard Hilbert space (in the second case 
(\ref{standardpsi(x)})),
and with $\boldsymbol{P}^\mu = d\Gamma(P^\mu)$, $\mu = 0,1,2,3$, equal to the generators of translations of the representation 
\[
\Gamma\Big(U\boldsymbol{U}(a, \alpha)U^{-1}\Big) \,\,\, \textrm{or resp.}, \,\,\,
\Gamma\Big(U \mathbb{U}\boldsymbol{U}(a,\alpha)\mathbb{U}^{-1}U^{-1}\Big)
\]
of $T_4 \circledS SL(2, \mathbb{C})$, both acting in the Fock space $\Gamma(U\mathcal{H}') = 
\Gamma(L^2(\mathbb{R}^3; \mathbb{C}^4))$
(in the second case corresponding to (\ref{standardpsi(x)}) we also have 
$\Gamma(U\mathbb{U}\mathcal{H}') = \Gamma(L^2(\mathbb{R}^3; \mathbb{C}^4))$ with the isomorphism 
$U$ given by the modification of (\ref{isomorphismU})
in which we remove the factor $1/p_0(\boldsymbol{\p})$, with the removal being compensated by the presence 
of $\mathbb{U}$). Note that in the first case (\ref{psi(x)'})
the unitary operator is given by the formula (\ref{isomorphismU}), and in the second case $U$
is given by the similar formula with the weight factor $1/p_0(\boldsymbol{\p})$ omitted.

Equivalently Bogoliubov-Shirkov Quantization Postulate for $\boldsymbol{\psi}$ demands the equality 

\begin{equation}\label{BSPpsi}
\frac{i}{2}  \int  \boldsymbol{:} \Bigg(
\boldsymbol{\psi}^\sharp(x)\gamma^0 \frac{\partial \boldsymbol{\psi}}{\partial x_\mu}(x)
- \frac{\partial \boldsymbol{\psi}^\sharp}{\partial x_\mu}(x) \gamma^0 \boldsymbol{\psi}(x)  \Bigg)
\boldsymbol{:}  \, \ud^3 \boldsymbol{\x}
=
d\Gamma(P^\mu), \,\,\, \textrm{in this order!}
\end{equation}
to hold.

The whole point about the Quantization Postulate (or Emmy Noether theorem for free fields) 
is that the operators $\boldsymbol{P}^\mu = d\Gamma(P^\mu)$ may be computed in 
therms of Wick polynomials in free fields -- integral kernel operators -- to which we know how to apply the 
perturbative series in the sense of Bogoliubov-Epstein-Glaser. 
In  checking its validity for the Dirac field we proceed in two steps. 
In the first step we show that for each $\mu = 0,1,2,3$, 
there exist a distribution $\kappa^\mu \in E_1 \otimes E_{1}^*$ 
such that the corresponding integral kernel operator  $\Xi_{1,1}(\kappa^\mu)$
  is equal to $\boldsymbol{P}^\mu = d\Gamma(P^\mu)$. Then according to the rule giving the Wick product 
of free fields at the same point as integral kernel operator with vector valued kernel as well as 
the rule giving its spatial integral as an integral kernel operator with scalar kernel, given in the preceding Subsection, 
we show that the left hand side integral kernel operator is equal to the right hand side integral kernel operator $\Xi_{1,1}(\kappa^\mu)$ 
in (\ref{BSPpsi}) for the standard field (\ref{psi(x)'}). It turns out that 
(\ref{BSPpsi}) does not hold for the nonstandard local field (\ref{standardpsi(x)}).

It is easily seen that the representors $U\boldsymbol{U}(a, \alpha)U^{-1}$ and respectively 
\[
U\mathbb{U}\boldsymbol{U}(a, \alpha)\mathbb{U}^{-1}U^{-1}
\]
are continuous as operators $E_1 \to E_1$, 
in case of both the representations of
$T_4\circledS SL(2, \mathbb{S})$: 
\begin{enumerate}
\item[1)]
for the representation $U\boldsymbol{U}(a, \alpha)U^{-1}$ acting in $U\mathcal{H}' = L^2(\mathbb{R}^3; \mathbb{C}^4)$,
with $U$ given by (\ref{isomorphismU}), corresponding to the field
(\ref{psi(x)'}), 
\item[2)]
for the representation $U\mathbb{U}\boldsymbol{U}(a, \alpha)\mathbb{U}^{-1}U^{-1}$, acting
in $U\mathbb{U}\mathcal{H}' = L^2(\mathbb{R}^3; \mathbb{C}^4)$, with $U$ 
given by (\ref{isomorphismU}) without the factor
$1/p_0(\boldsymbol{\p})$, which is compensated here by the operator $\mathbb{U}$, and
corresponding to the field (\ref{standardpsi(x)}).
\end{enumerate}
 
In particular this holds
for the translation subgroup representors.
And the translation representors in both of the representations are unitary and act identically 
on the common nuclear space $E_1 = \mathcal{S}_{A}(\mathbb{R}^3; \mathbb{C}^4)$.
Therefore the translation subgroup in both cases of representations 
compose the subgroup of the Yoshizawa group $U\big(E_1; L^2(\mathbb{R}^3; \mathbb{C}^4)\big)$.  
The Yoshizawa group $U\big(E_1; L^2(\mathbb{R}^3; \mathbb{C}^4)\big)$ is the group of unitary operators on 
$L^2(\mathbb{R}^3; \mathbb{C}^4)$ which induce homeomorphisms
of the test function space $E_1= \mathcal{S}_{A}(\mathbb{R}^3; \mathbb{C}^4)$ with respect to the nuclear topology of 
$E_1$. In other words the translation representors in both representations compose automorphisms of the Gelfand triple 
$E_1 \subset L^2(\mathbb{R}^3; \mathbb{C}^4) \subset E_{1}^*$. Moreover,
any one parameter subgroup $\{T_\theta\}_{\theta \in \mathbb{R}}$ of translations in both considered representations 
is differentiable, i.e. $\lim_{\theta \to 0} (T_{\theta}\xi - \xi)/\theta = X\xi$ converges in $E_1$.  
Let us consider the one parameter 
subgroup of translations along the $\mu$-th axis and write in this case $X^\mu$ for $X$,
where in our case $X^\mu$ is the operator $M_{ip^\mu}$ of multiplication by the function 
$\boldsymbol{\p} \to ip^\mu(\boldsymbol{\p})$, and where 
$(p^0(\boldsymbol{\p}), \ldots p^3(\boldsymbol{\p}))
 = (\sqrt{\boldsymbol{\p} \cdot \boldsymbol{\p} + m^2}, \boldsymbol{\p}) \in \mathscr{O}_{(1,0,0,1)}$. 
Existence of the limit is equivalent to 
\begin{multline}\label{T-theta-differentiability-psi}
\lim \limits_{\theta \to 0} \bigg| \frac{T_{\theta}\xi - \xi}{\theta} - X^\mu \xi  \bigg|_k^2 \\ =
\lim \limits_{\theta \to 0}
\int \bigg( \frac{A^k \Big(e^{i\theta p^\mu} -1 
- i \theta p^\mu \Big)\xi(\boldsymbol{\p})}{\theta} \, , \,\, 
\frac{A^k \Big(e^{i\theta p^\mu} -1 
- i \theta p^\mu \Big)\xi(\boldsymbol{\p})}{\theta}
\bigg)_{{}_{\mathbb{C}^4}} \,\, 
\ud^3 \boldsymbol{\p} \,\, = 0, \\
 k = 0, 1, 2, \ldots, \,\,\, \xi \in E_1,
\end{multline}
where $p^\mu$, $\mu = 0,1,2,3$, in the exponent are the functions 
$\boldsymbol{\p} \mapsto (p^\mu(\boldsymbol{\p})) = (\sqrt{\boldsymbol{\p} \cdot \boldsymbol{\p}+m^2}, \boldsymbol{\p})$
and where $A$ is the standard operator (\ref{AinL^2(R^3;C^4)}) used in the construction of the standard
Gelfand triple $E_1 = \mathcal{S}_{A}(\mathbb{R}^3; \mathbb{C}^4) \subset L^2(\mathbb{R}^3; \mathbb{C}^4)
\subset E_{1}^*$. Explicit calculation shows that (\ref{T-theta-differentiability-psi}) is fulfilled. Therefore  
$\{T_\theta\}_{\theta \in \mathbb{R}}$ is differentiable subgroup and by the Banach-Steinhaus
theorem the linear operators $X^\mu$, $\mu = 0,1,2,3$, 
are continuous as operators $E_1 \to E_1$ and finally by Proposition 3.1 of \cite{hida}
every such subgroup is regular in the sense of \cite{hida}, \S  3.

For every operator $X$ which is continuous as the operator $E_1 \to E_1$ we define $\Gamma(X)$
and $d \Gamma(X)$ on $(E_1)$. Let $\Phi \in (E_1)$ be any element of the Hida space with decomposition (\ref{HidaPhi}) corresponding to the Gelfand triple
$E_1 = \mathcal{S}_{A}(\mathbb{R}^3; \mathbb{C}^4) \subset L^2(\mathbb{R}^3; \mathbb{C}^4)
\subset E_{1}^*$, i.e. with the pairing
$\langle \cdot, \cdot \rangle$ induced by the inner product
$(\cdot, \cdot)_{{}_{L^2(\mathbb{R}^3; \mathbb{C}^4)}}$ in $L^2(\mathbb{R}^3; \mathbb{C}^4)$.
Then we define
\[
\Gamma(X)\Phi
= \sum \limits_{n=0}^{\infty} \, X^{\otimes n} \Phi_n;
\]
\[
d\Gamma(X)\Phi
= \sum \limits_{n=0}^{\infty} n \, (X \otimes I^{\otimes (n-1)}) \, \Phi_n.
\]
In this case it is easily seen that the Theorem 4.1 of \cite{hida} is easily adopted to our Fermi case
and that $\{\Gamma(T_\theta ) \}_{\theta \in \mathbb{R}}$, with the generator $X^\mu$, is a regular
one parameter subgroup with the generator $d\Gamma(X^\mu)$ which continuously maps $(E)$
into itself.

In this situation it is not difficult to see that for each $\mu =0,1,2,3$, the proof of Proposition 4.2
and Theorem 4.3 of \cite{hida} is applicable in the Fermi case to any of the one parameter translation subgroups
of the mentioned representations, in particular
for any of the translation subgroup along the direction of the
$\mu$-th axis, $\mu =0,1,2,3$, there exists
a symmetric distribution $\kappa^\mu \in E_1 \otimes E_{1}^*$ such that
\begin{equation}\label{dGammaX=Berexin-int-psi}
d \Gamma(X^\mu) = \Xi_{1,1}(\kappa^\mu) = \sum_{s,s'} \, \int \limits_{\mathbb{R}^3 \times \mathbb{R}^3}
\kappa^\mu(\boldsymbol{\p}', s', \boldsymbol{\p},s) \,\,
\partial_{\boldsymbol{\p}',s'}^*
\partial_{\boldsymbol{\p},s}
\,\, \ud^3 \boldsymbol{\p}' \ud^3 \boldsymbol{\p},
\end{equation}
and $\kappa^\mu \in E_1 \otimes E_{1}^*$ fulfils
\begin{equation}\label{kappa-distribution-P-psi}
\langle \kappa^\mu, \zeta \otimes \xi \rangle = \langle \zeta, X^\mu \xi \rangle,
\,\,\, \zeta, \xi \in E_1.
\end{equation}
Because the pairings $\langle \cdot, \cdot \rangle$ in the formula are induced by
the inner product $(\cdot, \cdot)_{{}_{L^2(\mathbb{R}^3; \mathbb{C}^4)}}$ in $L^2(\mathbb{R}^3; \mathbb{C}^4)$,
and because $X^\mu$ is the operator of multiplication by $ip^\mu(\boldsymbol{\p})$, we have
\[
(\overline{\zeta}, \, X^\mu \xi)_{{}_{\oplus L^2(\mathbb{R}^3)}} = \langle \zeta, X^\mu \xi \rangle
= \langle X^\mu \xi , \zeta \rangle = \langle \xi , X^\mu \zeta \rangle,
\,\,\, \zeta, \xi \in E,
\]
so that
\[
\langle \kappa^\mu, \zeta \otimes \xi \rangle = \langle \kappa^\mu, \xi \otimes \zeta \rangle,
\,\,\, \zeta, \xi \in E,
\]
and $\kappa^\mu$ is indeed symmetric.

On the other hand the pairing $\langle \cdot, \cdot \rangle$
on left-hand side of (\ref{kappa-distribution-P-psi}) expressed in terms
of the kernel $\kappa^\mu (\boldsymbol{\p}', \boldsymbol{\p})$ is likewise induced by the inner
product $(\cdot, \cdot)_{{}_{\oplus L^2(\mathbb{R}^3)}}$ in $L^2(\mathbb{R}^3; \mathbb{C}^4)$. Therefore,
we have
\[
\langle \kappa^\mu, \zeta \otimes \xi \rangle = \sum \limits_{s,s'}
\int \limits_{\mathbb{R}^3 \times \mathbb{R}^3}
\kappa^\mu(\boldsymbol{\p}', s', \boldsymbol{\p},s) \,\,
\zeta(\boldsymbol{\p}',s') \xi(\boldsymbol{\p},s)
\,\, \ud^3 \boldsymbol{\p}' \ud^3 \boldsymbol{\p}.
\]
Joining this with (\ref{kappa-distribution-P-psi}) we obtain
\[
\kappa^\mu(\boldsymbol{\p}', s' \boldsymbol{\p}, s)
= i p^\mu(\boldsymbol{\p}) \delta_{s\,s'} \delta(\boldsymbol{\p}'- \boldsymbol{\p}).
\]

Therefore, we get 
\begin{equation}\label{d(Gamma(P)-psi}
\boldsymbol{P}^\mu = d \Gamma(P^\mu) = \sum_{s,s'} \, \int \limits_{\mathbb{R}^3 \times \mathbb{R}^3}
p^\mu(\boldsymbol{\p}) \,\, \delta_{s  \, s'} \delta(\boldsymbol{\p}'- \boldsymbol{\p}) \,\,\, 
\partial_{\boldsymbol{\p}', s'}^{*} \partial_{\boldsymbol{\p}, s}
\,\,\, \ud^3 \boldsymbol{\p}' \ud^3 \boldsymbol{\p},
\end{equation}
which is customary to be written as
\begin{multline}\label{d(Gamma(P^0)-psi}
\boldsymbol{P}^0 = d \Gamma(P^0) = \sum \limits_s \int \limits_{\mathbb{R}^3}
|p^0(\boldsymbol{\p})| \,\, 
\partial_{\boldsymbol{\p}, s}^{*} \partial_{\boldsymbol{\p}, s}
\,\,\, \ud^3 \boldsymbol{\p} \\
= \sum \limits_{s=1,2} \int \limits_{\mathbb{R}^3}
|p^0(\boldsymbol{\p})| \,\, 
b_s(\boldsymbol{\p})^{+} b_s(\boldsymbol{\p})
\,\,\, \ud^3 \boldsymbol{\p}
+
\sum \limits_{s=1,2} \int \limits_{\mathbb{R}^3}
|p^0(\boldsymbol{\p})| \,\, 
d_s(\boldsymbol{\p})^{+} d_s(\boldsymbol{\p})
\,\,\, \ud^3 \boldsymbol{\p},
\end{multline}
\begin{multline}\label{d(Gamma(P^i)-psi}
\boldsymbol{P}^i = d \Gamma(P^i) = \sum \limits_s \int \limits_{\mathbb{R}^3}
p^i(\boldsymbol{\p}) \,\, 
\partial_{\boldsymbol{\p}, s}^{*} \partial_{\boldsymbol{\p}, s}
\,\,\, \ud^3 \boldsymbol{\p} \\
= \sum \limits_{s=1,2} \int \limits_{\mathbb{R}^3}
p^i(\boldsymbol{\p}) \,\, 
b_s(\boldsymbol{\p})^{+} b_s(\boldsymbol{\p})
\,\,\, \ud^3 \boldsymbol{\p}
+
\sum \limits_{s=1,2} \int \limits_{\mathbb{R}^3}
p^i(\boldsymbol{\p}) \,\, 
d_s(\boldsymbol{\p})^{+} d_s(\boldsymbol{\p})
\,\,\, \ud^3 \boldsymbol{\p}.
\end{multline}
Both operators $d \Gamma(P^\mu)$ and $\Xi_{1,1}(-i\kappa^\mu)$ transform (continuously)
the nuclear, and thus perfect, space $(E_1)$ into itself and both being equal and symmetric 
on $(E_1)$ have self-adjoint extension to self-adjoint operator
in the Fock space $\Gamma(L^2(\mathbb{R}^3; \mathbb{C}^4))$, again by the classical criterion of 
\cite{Riesz-Szokefalvy}
(p. 120 in Russian Ed. 1954). In general the criterion of Riesz-Sz\"okefalvy-Nagy
does not exclude existence of more than just one self-adjoint extension, but
in our case it is unique. Indeed, because for each $\mu=0,1,2,3$, the one-parameter unitary
group generated by $d \Gamma(P^\mu)$ leaves invariant the dense nuclear space $(E_1)$, then by general
theory, e.g. Chap. 10.3., it follows that  $d \Gamma(P^\mu)$ with domain $(E_1)$ is essentially self adjoint
(admits unique self adjoint extension).

Now applying the Rules II and V' of Subsection \ref{OperationsOnXi} to the left-hand side of  
(\ref{BSPpsi}) with $\boldsymbol{\psi}$ equal to the standard Dirac free field (\ref{psi(x)'}),
understood as an integral kernel operator
\[
\boldsymbol{\psi} = \Xi_{0,1}(\kappa_{0,1}) + \Xi_{1,0}(\kappa_{1,0})
\]
with the kernels $\kappa_{0,1}, \kappa_{1,0}$, (\ref{kappa_0,1}) and (\ref{kappa_1,0}), 
Subsection \ref{psiBerezin-Hida}, we immediately get 
the result equal to (\ref{d(Gamma(P)-psi}) or equivalently (\ref{d(Gamma(P^0)-psi}),
(\ref{d(Gamma(P^i)-psi}). Thus we arrive at the following
\begin{prop*}
The standard free Dirac field $\boldsymbol{\psi}$, equal (\ref{psi(x)'}), satisfies
the Bogoliubov-Shirkov Quantization Postulate (\ref{BSPpsi}) for translations: 
\[
\frac{i}{2}  \int  \boldsymbol{:} \Bigg(
\boldsymbol{\psi}^\sharp(x)\gamma^0 \frac{\partial \boldsymbol{\psi}}{\partial x_\mu}(x)
- \frac{\partial \boldsymbol{\psi}^\sharp}{\partial x_\mu}(x) \gamma^0 \boldsymbol{\psi}(x)  \Bigg)
\boldsymbol{:}  \, \ud^3 \boldsymbol{\x}
=
d\Gamma(P^\mu).
\]
\end{prop*}

On the other hand if we apply the Rules II and V' of Subsection \ref{OperationsOnXi} to the left-hand side of  
(\ref{BSPpsi}) with $\boldsymbol{\psi}$ equal to the nonstandard local Dirac free field (\ref{standardpsi(x)}),
understood as an integral kernel operator
\[
\boldsymbol{\psi} = \Xi_{0,1}(\kappa_{0,1}) + \Xi_{1,0}(\kappa_{1,0})
\]
with the kernels $\kappa_{0,1}, \kappa_{1,0}$, (\ref{skappa_0,1}) and (\ref{skappa_1,0}), 
we obtain an integral kernel operator not equal
to (\ref{d(Gamma(P)-psi}) or, equivalently, not equal to (\ref{d(Gamma(P^0)-psi}),
(\ref{d(Gamma(P^i)-psi}). Thus, we arrive at the following
\begin{prop*}
The Bogoliubov-Shirkov Quantization Postulate (\ref{BSPpsi}) for translations is not satisfied
by the nonstandard local Dirac field (\ref{standardpsi(x)}). 
\end{prop*}

Now let us consider Lorentz transformations. The Noether integral
generator corresponding to Lorentz transformations is equal
\begin{equation}\label{d(Gamma(M)-psi}
\frac{i}{2} \int \boldsymbol{:} \Bigg(
\boldsymbol{\psi}(x)^{+}x^\mu\frac{\partial \boldsymbol{\psi}}{\partial x_\nu}(x)
- \boldsymbol{\psi}(x)^{+} x^\nu\frac{\partial \boldsymbol{\psi}}{\partial x_\mu}(x)
+ \frac{1}{2} \boldsymbol{\psi}(x)^{+} \gamma^\mu \gamma^\nu \boldsymbol{\psi}(x) \Bigg)
\boldsymbol{:} \, \ud^3 \boldsymbol{\x} = \boldsymbol{M}^{\mu\nu}
\end{equation}
Again applying the Rules II and V' of Subsection \ref{OperationsOnXi} we arrive at the following
(infinitesimal form of) local transformation formula
\[
i[\boldsymbol{M}^{\mu\nu}, \boldsymbol{\psi}^a] = \Sigma_{b}^{a \mu \nu}\boldsymbol{\psi}^b +
(x^\mu \partial^\nu - x^\nu \partial^\mu)\boldsymbol{\psi}^a
\]
for the standard Dirac free field $\boldsymbol{\psi}$ equal (\ref{psi(x)'}). It generates the ordinary
local bispinor transformation formula $\boldsymbol{U}(a, \alpha)$ in the single particle Hilbert space
$\mathcal{H}''$ of the standard Dirac field (\ref{psi(x)'}), which coincides with the
unitary representation $\boldsymbol{U}(a, \alpha)$, and which is unitary
if regarded as representation in the single particle Hilbert space
$\mathcal{H}'$. In particular
$\boldsymbol{M}^{\mu \nu} = d \Gamma(M^{\mu \nu})$, regarded as operator in the Fock space
$\Gamma(\mathcal{H}'')$ of the standard Dirac free field (\ref{standardpsi(x)}), generates a unitary
transformation. Therefore, the generator $\boldsymbol{M}^{\mu \nu} = d \Gamma(M^{\mu \nu})$ given by te Noether integral
(\ref{d(Gamma(M)-psi}) corresponding to the Lorentz transformations, and computed for the standard Dirac
field (\ref{psi(x)'}) is self-adjoint. It is not the case for the nonstandard Dirac field 
(\ref{standardpsi(x)}).

We therefore have the following alternative: we can save locality of the transformation of the nonstandard Dirac
fields, e.g. (\ref{standardpsi(x)}), but unitarity of the Lorentz transformations have to be abandoned for them.
Alternatively we have the unitary representation
$\Gamma(\mathbb{U}\boldsymbol{U}(a, \alpha) \mathbb{U}^{-1})$ in the Fock space $\Gamma(\mathcal{H}'')$
of the nonstandard Dirac field (\ref{standardpsi(x)}), but locality of the Lorentz transformations is lost.
In each case the generators of the representation do not coincide with the Noether
integrals (with Wick ordered products) for the nonstandard Dirac fields, e.g. (\ref{standardpsi(x)}).

This alternative has not been discovered before. One reason lies in the fact that there are the white noise
techniques which allow us to construct equal time integrals of Wick products of free fields,
and to investigate their self-adjointness. As far as we know nobody has applied them before
to the realistic fields, and in particular to the analysis of Wick product fields and their Cauchy
integrals. On the other hand the approach more popular among mathematical physicists, \emph{i. e.}
due to Wightman-G{\aa}rding, is not effective here, which was recognized by Segal \cite{Segal-NFWP.I},
p. 455. In particular self-adjointness or, respectively, non-self-adjointnes of the Lorentz transformations generator $\boldsymbol{M}^{\mu\nu}$
for the standard (\ref{psi(x)'}), or respectively, for nonstandard Dirac field (\ref{standardpsi(x)}), 
given by the Noether integral formula (\ref{d(Gamma(M)-psi}),
could have not been proved by such founders of Quantum Field Theory like Pauli or Schwinger.
Proved alternative also confirms, among other things, the role of Noether conserved integrals in relation to commutation rules, 
which was used by those founders in selection of the correct quantization procedure and construction of free quantum fields. 
Also, rigorous construction of the standard Dirac field (\ref{psi(x)'}) (as well as the nonstandard versions, e.g. (\ref{standardpsi(x)})) 
has not been performed before. In particular, it escaped the classification of free fields based on local unitary
irreducible representations of the double covering of the Poincar\'e group given in \cite{lop1}
or \cite{lop2}. This lack was also recognized by
Haag \cite{Haag}, p. 48. The local bispinor field (\ref{psi(x)'}) has the standard local
and unitary bispinor transformation formula, and coincides with the standard Dirac field. 
Note that the standard Dirac field (\ref{psi(x)'}) is a field which is
obtained through the canonical quantization, \emph{i.e.} it is uniquely determined by the condition
that it satisfies the Bogoliubov-Shirkov Quantization Postulate (\ref{BSPpsi}) for translations.

Note that the Wick product of the Dirac field components is skew-commutative, therefore the order is important 
in (\ref{BSPpsi}).

We end this Subsection with a remark on the Pauli theorem on spin-statistics relation. It
is based on the properties of the ''classical'', \emph{i.e.} before ''quantization'', fields.
Essentially it says that the energy component of the Noether energy-momentum tensor
is not positive definite for half-odd-integer free ''classical'' fields, compare e.g. \cite{Geland-Minlos-Shapiro}
and Pauli's book cited there. Technically speaking,
generic half-odd-integer spin field (solution of equations of motion), when Fourier decomposed and inserted
into the Noether energy integral, gives formally the expression (\ref{d(Gamma(P^0)-psi}),
but with operators $b_s(\boldsymbol{\p}), d_s(\boldsymbol{\p})$ replaced with
the Fourier coefficients and with the opposite sign at the second term in
(\ref{d(Gamma(P^0)-psi}). Pauli then joined this result with the canonical quantization procedure,
equivalent to the Pauli-Bogoliubov-Shirkov Quantization Postulate (\ref{BSPpsi})
for translations. Because the Wick product of Fermi fields in (\ref{d(Gamma(P)-psi})
repairs the sign of the second term in the ``classical'' counterpart of (\ref{d(Gamma(P^0)-psi}),
Pauli arrived at the spin-statistics relation:
half-odd-integer spin ``classical'' (free) fields should be quantized with the
canonical anticommutation relations.

The so-called ``spin-statistis theorem'' due to Wightman is different and in fact gives the relation
between the commutation relation of smeared out fields, within his axiomatic definition of 
a quantum field, and  the representation defining a local transformation rule of the field. 
In Wightman's proof no relation with ``classical'' fields and with positivity of the energy-momentum 
of ``classical'' fields  intervenes. In this sense Pauli's spin-statistics theorem is different pointing 
out that such relation exists, and in this sense reveals what is untouched in the Wightman's version 
of spin-statistics theorem.

\subsection{Comparison of the two free Dirac fields continued. Decomposition of the representation of $SL(2, \mathbb{C})$
pertinent to these fields}\label{psichi}

In this Subsection
we will analyze the representation $\alpha \mapsto U(\alpha)$ of $SL(2, \mathbb{C})$, 
given by (\ref{1/2pUalpha}) and (\ref{1/2pUa}), regarded as 
\begin{enumerate}
\item[1)]
the unitary representation in the single
particle Hilbert space $\mathcal{H}^{\oplus}_{m,0} \oplus \mathcal{H}^{\ominus}_{-m,0}$ of the standard field (\ref{psi(x)'}) with the invariant inner product (\ref{InvariantScalarProduct}), 
\item[]
and as 
\item[2)]
the non-unitary representation acting in the single particle Hilbert space $\mathcal{H}^{\oplus}_{m,0} \oplus \mathcal{H}^{\ominus}_{-m,0}$ of the nonstandard field
(\ref{standardpsi(x)}) with the inner product (\ref{StandardScalarProduct}).
\end{enumerate}

In case 1) we will decompose the representation $U(\alpha)$ of $SL(2,\mathbb{C})$ into a direct integral of indecomposable components. In case 1), when $U(\alpha)$
is unitary, decomposition components are irreducible, without any closed invariant subspaces. 
In case 2), when
$U(\alpha)$ is not unitary, the very existence of the corresponding decomposition into components $U_{{}_{\chi}}(\alpha)$, say indecomposable (no nontrivial bounded idempotent commuting with $U_{{}_{\chi}}$ exists for each component $U_{{}_{\chi}}$), remains to be an open problem. It seems that in case 2)
no such decomposition exists. 

This will allow us (in case 1)) to construct decomposition of the standard Dirac field
\[
\boldsymbol{\psi} = \Xi(\kappa_{0,1}) + \Xi(\kappa_{1,0})
\]
(\ref{psi(x)'}) into direct integral of spinor fields
\[
\boldsymbol{\psi}_{{}_{\chi}} =  \Xi_{{}_{\chi}}(\kappa_{\chi \, 0,1}) + \Xi(\kappa_{\chi \, 1,0})
\]
of asymptotic homogeneity $\chi$, with the asymptotic homogeneities $\chi$
determined by the direct integral decomposition of $U(\alpha)$. Moreover, decomposition 
of the representation of $SL(2,\mathbb{C})$ acting in the single particle space determines canonically
a direct integral decomposition
of a wide class of more general integral kernel operators $\Xi(\kappa_{lm})$ acting in the Fock space of the Dirac field. This decomposition
is determined by the Fourier transform $\mathcal{F}$ of the distribution kernel $\kappa_{l,m}$, where
the Fourier transform $\mathcal{F}$ is that associated with the decomposition of the tensor product $U(\alpha)^{\otimes(l+m)}$
of the representation  $U(\alpha)$ of $SL(2,\mathbb{C})$. We present this decomposition in detail. It has the expected
(asymptotic) homogeneity property: asymptotic homogeneity degrees of the decomposition components of a Wick product of, say, the conjugated 
Dirac field with the Dirac field itself, are equal to the sums of the (asymptotic) homogeneity degrees of the decomposition
components of the conjugated Dirac field and the Dirac field itself. The same decomposition construction is valid for integral
kernel operators  $\Xi(\kappa_{lm})$ in more general Fock spaces of several free fields, provided the representation of 
$SL(2, \mathbb{C})$ acting in the single particle space is decomposable. This is the case for the Fock spaces
of free fields underlying realistic causal perturbative QFT, e.g. spinor or scalar QED, and for the finite sums of integral kernel
operators $\Xi(\kappa_{lm})$ representing higher order contributions to interacting fields.

We will use the Cartesian spatial momentum $\boldsymbol{\p}$ coordinates on the respective
orbits $\mathscr{O}_{{}_{\pm m, 0,0,0}}$, and the bispinors $\widetilde{\phi}$
concentrated on the orbits will regard as functions of $\boldsymbol{\p}$
writing simply $\widetilde{\phi}(\boldsymbol{\p})$ for $\widetilde{\phi}(\boldsymbol{\p}, p_0(\boldsymbol{\p}))$
with $p_0(\boldsymbol{\p}) = \pm \sqrt{|\boldsymbol{\p}|^2 + m^2}$ on $\mathscr{O}_{{}_{\pm m,0,0,0}}$.
In both cases, 1) and 2), the bispinors $\widetilde{\phi} \in \mathcal{H}^{\oplus}_{m,0} \oplus \mathcal{H}^{\ominus}_{-m,0}$ are concentrated on $\mathscr{O}_{{}_{m,0,0,0}} \sqcup \mathscr{O}_{{}_{-m,0,0,0}}$
and lie, respectively, in the images of the idempotents $P^\oplus$ (on $\mathscr{O}_{{}_{m,0,0,0}}$)
and $P^\ominus$ (on $\mathscr{O}_{{}_{-m,0,0,0}}$), with the invariant inner product (\ref{InvariantScalarProduct})
\[
(\widetilde{\phi},\widetilde{\phi}') = \int\limits_{\mathscr{O}_{{}_{m,0,0,0}} \sqcup \mathscr{O}_{{}_{-m,0,0,0}}}
\big(\widetilde{\phi}(\boldsymbol{\p}),\widetilde{\phi}'(\boldsymbol{\p})\big)_{{}_{\mathbb{C}^4}} \,
{\textstyle\frac{\ud^3 \boldsymbol{\p}}{(2p_0(\boldsymbol{\p}))^2}},
\,\,\,\,\, \textrm{case 1)}
\]
and with the non-invariant inner product
(\ref{StandardScalarProduct})
\[
(\widetilde{\phi},\widetilde{\phi}') = \int\limits_{\mathscr{O}_{{}_{m,0,0,0}} \sqcup \mathscr{O}_{{}_{-m,0,0,0}}}
\big(\widetilde{\phi}(\boldsymbol{\p}),\widetilde{\phi}'(\boldsymbol{\p})\big)_{{}_{\mathbb{C}^4}} \,
\ud^3 \boldsymbol{\p}.
\,\,\,\,\, \textrm{case 2)}
\]
In both cases the transformation rule
\begin{equation}\label{U(alpha)inCartesian-p}
U(\alpha) \widetilde{\phi}(\boldsymbol{\p}) =
\left( \begin{array}{cc} \alpha & 0 \\
0 & {\alpha^*}^{-1} \end{array}\right)
\widetilde{\phi}(\Lambda(\alpha)\boldsymbol{\p}),
\end{equation}
is, by assumption, the same, with the action $\boldsymbol{\p} \mapsto \Lambda(\alpha)\boldsymbol{\p}$ of
$SL(2, \mathbb{C})$ in the respective orbit $\mathscr{O}_{{}_{\pm m, 0,0,0}}$
given by the formula (\ref{SL(2,C)actionInSpatialMomentumCoordinates}) in the coordinates $\boldsymbol{\p}$.

Before we pass to the construction of generators of $U(\alpha)$ in both cases, 1) and 2),
we should fix appropriately their dense nuclear domains, determined by the single particle
Gelfand triples of the field (\ref{psi(x)'}) -- in case 1), and of the field (\ref{standardpsi(x)}) -- in case 2).
Recall that we have used the unitary isomorphism
\[
U: \mathcal{H}^{\oplus}_{m,0} \oplus \mathcal{H}^{\ominus \, \flat}_{-m,0} \longrightarrow L^2(\mathbb{R}^3; \mathbb{C}^4)
\]
of Subsection \ref{psiBerezin-Hida}, given by (\ref{isomorphismU}),
which maps unitarily the Hilbert space $\mathcal{H}^{\oplus}_{m,0} \oplus \mathcal{H}^{\ominus \, \flat}_{-m,0}$
onto the standard Hilbert space $L^2(\mathbb{R}^3; \mathbb{C}^4)$. Recall, please, that in case 2) the additional
factors $\big(2p_0(\boldsymbol{\p})\big)^{-1}, 2p_0(\boldsymbol{\p})$ in the formula for $U$ and $U^{-1}$ will be absent,
accordingly to the difference in the formulas for the single particle inner products (\ref{InvariantScalarProduct})
-- in case 1) and (\ref{StandardScalarProduct}) -- in case 2). The nuclear space
\[
E = E^\oplus \oplus E^{\ominus \flat} \subset \mathcal{H}^{\oplus}_{m,0} \oplus \mathcal{H}^{\ominus \, \flat}_{-m,0}
\]
is equal to the inverse image $U^{-1}\mathcal{S}_{{}_{A}}$ of the standard nuclear space $\mathcal{S}_{{}_{A}}$
\[
\mathcal{S}(\mathbb{R}^3; \mathbb{C}^4) = \mathcal{S}_{{}_{A}}(\mathbb{R}^3; \mathbb{C}^4) = E^+ \oplus (E^{-})^{\flat},
\,\,\,\,\,\, A = \oplus_{1}^{4} H_{{}_{(3)}}.
\]
Here we regard the nuclear Schwartz space of $\mathbb{C}^4$-valued functions as the following direct sum of nuclear spaces of
$\mathbb{C}$-valued Schwartz functions
\begin{multline*}
\mathcal{S}(\mathbb{R}^3;\mathbb{C}) \oplus \mathcal{S}(\mathbb{R}^3;\mathbb{C}) \oplus \mathcal{S}(\mathbb{R}^3;\mathbb{C})
\oplus \mathcal{S}(\mathbb{R}^3;\mathbb{C})
\\
=
\mathcal{S}_{{}_{A}}(\mathbb{R}^3;\mathbb{C}) \oplus \mathcal{S}_{{}_{A}}(\mathbb{R}^3;\mathbb{C}) \oplus \mathcal{S}_{{}_{A}}(\mathbb{R}^3;\mathbb{C})
\oplus \mathcal{S}_{{}_{A}}(\mathbb{R}^3;\mathbb{C}),
\\
A = H_{{}_{(3)}},
\end{multline*}
and note that the operation of conjugation ${\widetilde{\phi}}^{{}^{\,\, \flat}}$ of negative energy bispinors (compare Subsection
\ref{positron}) correspond to the operation of complex conjugation together with sign inversion in the argument of the last two
components of the image $U\big(\widetilde{\phi} \oplus {\widetilde{\phi}}^{{}^{ \,\, \flat}}\big)$. Namely, if
\begin{multline*}
U\big(\widetilde{\phi}' \oplus {\widetilde{\phi}}^{{}^{\,\, \flat}}\big)(\boldsymbol{\p}) =
(\widetilde{\phi}')_{{}_{1}}(\boldsymbol{\p}) \oplus (\widetilde{\phi}')_{{}_{2}}(\boldsymbol{\p})
\oplus (\widetilde{\phi})_{{}_{3}}(\boldsymbol{\p}) \oplus (\widetilde{\phi})_{{}_{4}}(\boldsymbol{\p}),
\end{multline*}
then
\begin{multline*}
U\big(\widetilde{\phi}' \oplus \widetilde{\phi}\big)(\boldsymbol{\p}) =
(\widetilde{\phi}')_{{}_{1}}(\boldsymbol{\p}) \oplus (\widetilde{\phi}')_{{}_{2}}(\boldsymbol{\p})
\oplus \overline{(\widetilde{\phi})_{{}_{3}}(-\boldsymbol{\p})} \oplus \overline{(\widetilde{\phi})_{{}_{4}}(-\boldsymbol{\p})}.
\end{multline*}
Therefore
\[
(f \oplus g) = (f_1,f_2, g_1, g_2) \in E^+ \oplus E^{-} \subset L^2(\mathbb{R}^3; \mathbb{C}^4) =
L^2(\mathbb{R}^3; \mathbb{C}^2)
\oplus
L^2(\mathbb{R}^3; \mathbb{C}^2)
\]
if and only if the function
\begin{multline*}
\boldsymbol{\p} \longmapsto \big(f \oplus g^\flat \big)(\boldsymbol{\p})\big) =
\big(f_1(\boldsymbol{\p}), f_2(\boldsymbol{\p}), g_{3}^{\flat}(\boldsymbol{\p}), g_{4}^{\flat}(\boldsymbol{\p})\big)
\\
=
\big(f_1(\boldsymbol{\p}), f_2(\boldsymbol{\p}), \overline{g_{3}(-\boldsymbol{\p})}, \overline{g_{4}(-\boldsymbol{\p})} \, \big)
\end{multline*}
belongs to
\begin{multline*}
\Big(\mathcal{S}(\mathbb{R}^3;\mathbb{C}) \oplus \mathcal{S}(\mathbb{R}^3;\mathbb{C})\Big)
\oplus
\Big( \mathcal{S}(\mathbb{R}^3;\mathbb{C}) \oplus \mathcal{S}(\mathbb{R}^3;\mathbb{C}) \Big)
\\
=
\mathcal{S}(\mathbb{R}^3;\mathbb{C}) \oplus \mathcal{S}(\mathbb{R}^3;\mathbb{C}) \oplus \mathcal{S}(\mathbb{R}^3;\mathbb{C})
\oplus \mathcal{S}(\mathbb{R}^3;\mathbb{C}).
\end{multline*}
Thus the nuclear space
\[
E^\oplus \oplus E^\ominus = U^{-1}\big(E^+ \oplus E^- \big)
\]
and also composes a Gelfand triple (rigged Hilbert space)
\[
E^\oplus \oplus E^\ominus \,\,\,\,\,\,\,\,\,\, \subset \,\,\,\,\,\,\,\,\,\, \mathcal{H}^{\oplus}_{m,0} \oplus \mathcal{H}^{\ominus}_{-m,0}
\,\,\,\,\,\,\,\,\,\,
\subset \,\,\,\,\,\,\,\,\,\, \big(E^\oplus \oplus E^\ominus\big)^* = E^{\oplus *} \oplus E^{\ominus *}.
\]
Using the formula for decomposition $\widetilde{\phi}_{{}_{\boldsymbol{m},0}}
= \widetilde{\phi}_{{}_{\boldsymbol{m},0}}^\oplus + \widetilde{\phi}_{{}_{\boldsymbol{m},0}}^\ominus$
of a generic bispinor $\widetilde{\phi}_{{}_{\boldsymbol{m},0}}$ concentrated on $\mathscr{O}_{{}_{\boldsymbol{m}, 0,0,0}}$ and the analogue
decomposition $\widetilde{\phi}_{{}_{-\boldsymbol{m},0}}
= \widetilde{\phi}_{{}_{-\boldsymbol{m},0}}^\oplus + \widetilde{\phi}_{{}_{-\boldsymbol{m},0}}^\ominus$
of a generic bispinor $\widetilde{\phi}_{{}_{-\boldsymbol{m},0}}$ concentrated on $\mathscr{O}_{{}_{-\boldsymbol{m}, 0,0,0}}$,
given in Subsection \ref{e1} and the formula (\ref{isomorphismU}) for the isomorphism $U$, we easily see that
\[
\begin{split}
E^\oplus = P^\oplus \mathcal{S}(\mathbb{R}^3; \mathbb{C}^4)
= P^\oplus \Big[\mathcal{S}(\mathbb{R}^3; \mathbb{C}^4)\big|_{{}_{\mathscr{O}_{{}_{\boldsymbol{m}, 0,0,0}}}}\Big]
\\
E^\ominus = P^\ominus \mathcal{S}(\mathbb{R}^3; \mathbb{C}^4)
= P^\ominus \Big[\mathcal{S}(\mathbb{R}^3; \mathbb{C}^4)\big|_{{}_{\mathscr{O}_{{}_{-\boldsymbol{m}, 0,0,0}}}}\Big]
\end{split}
\]
with
\[
\mathcal{S}(\mathbb{R}^3; \mathbb{C}^4)\big|_{{}_{\mathscr{O}_{{}_{\pm \boldsymbol{m}, 0,0,0}}}}
\]
equal to the space of Schwartz $\mathbb{C}^4$-valued functions on $\mathbb{R}^4$, restricted to the respective orbit, which is
isomorphic to the Schwartz $\mathbb{C}^4$-valued functions on $\mathbb{R}^3$.
Equivalently
\[
\begin{split}
\widetilde{\phi} \in E^\oplus \,\,\,
\textrm{if and only if} \,\,
\widetilde{\phi} = P^\oplus h\big|_{{}_{\mathscr{O}_{{}_{\boldsymbol{m}, 0,0,0}}}}, \,\,\, h \in \mathcal{S}(\mathbb{R}^4; \mathbb{C}^4),
\\
\widetilde{\phi} \in E^\ominus \,\,\,
\textrm{if and only if} \,\,
\widetilde{\phi} = P^\ominus h\big|_{{}_{\mathscr{O}_{{}_{-\boldsymbol{m}, 0,0,0}}}}, \,\,\, h \in \mathcal{S}(\mathbb{R}^4; \mathbb{C}^4).
\end{split}
\]

We have thus established the appropriate nuclear invariant domain $E^\oplus \oplus E^\ominus$ of the generators, determined
by the single particle Gelfand triple, respectively, in each case: 1) and 2).

Let us consider the generators $A_{\mu\nu}$ of one-parameter subgroups of rotations or Lorentz hyperbolic rotations,
in the respective planes $23,31, 12, 01, 02, 03$ defined by
\[
A_{\mu\nu}\widetilde{\phi}(\boldsymbol{\p}) = {\textstyle\frac{d}{dt}}U(\Lambda_{\mu\nu}(\alpha_{{}_{\mu\nu}}(t))\widetilde{\phi}(\boldsymbol{\p})\Big|_{{}_{t=0}},
\,\,\,\,\,\,\,  \widetilde{\phi} \in E^\oplus \oplus E^\ominus.
\]
In both cases, 1) and 2), they have identical form, and are given by the following explicit formulas
\[
A_{0k} = -{\textstyle\frac{1}{2}}\gamma^0\gamma^k +p_0(\boldsymbol{\p})\partial_{{}_{p_k}}, \,\,\,\, k=1,2,3
\]

\[
A_{ik} = -{\textstyle\frac{1}{2}}\gamma^i\gamma^k + p_i\partial_{{}_{p_k}} -p_k\partial_{{}_{p_i}}, \,\,\,\, i,k = 1,2,3
\]
on the dense common domain $E^\oplus \oplus E^\ominus$,
which is isomorphic to a standard nuclear (and perfect) space, and which is dense in the Hilbert space $\mathcal{H}^{\oplus}_{m,0} \oplus \mathcal{H}^{\ominus}_{-m,0}$
in both cases: 1) and 2) and composes a Gelfand triple (rigged Hilbert space in the terminology of  \cite{GelfandIV})
\[
E^\oplus \oplus E^\ominus  \subset \mathcal{H}^{\oplus}_{m,0} \oplus \mathcal{H}^{\ominus}_{-m,0}
\subset {E^\oplus}^* \oplus {E^\ominus}^*  
\]
in both cases: 1) and 2).

The two Casimir operators, also given by the same formula  on $E^\oplus \oplus E^\ominus$ are equal
\begin{multline*}
Q = \big(A_{23}\big)^2 + \big(A_{31}\big)^2 + \big(A_{12}\big)^2
-\big(A_{01}\big)^2 - \big(A_{02}\big)^2 - \big(A_{03}\big)^2
\\
=
-{\textstyle\frac{3}{2}} \boldsymbol{1}_{{}_{4}} - \sum\limits_{ik} \gamma^i\gamma^k [p_i\partial_{{}_{p_k}} -p_k\partial_{{}_{p_i}}]
+ \sum\limits_{ik}  [p_i\partial_{{}_{p_k}} -p_k\partial_{{}_{p_i}}]^2
\\
+\sum\limits_{k} p_0(\boldsymbol{\p})\gamma^0\gamma^k \partial_{{}_{p_k}}
+ \sum\limits_{k} p_k\partial_{{}_{p_k}} 
+ p_0(\boldsymbol{\p})^2 \sum\limits_{k} \partial_{{}_{p_k}}^{2} 
\end{multline*}
where in the sums $k=1,2,3$ and $ik = 23,31,12$. 

\begin{multline*}
R = A_{23}A_{01}+A_{31}A_{02}+A_{12}A_{03} 
\\
= 
{\textstyle\frac{3}{4}}\gamma^0\gamma^1\gamma^2\gamma^3 
- \sum\limits_{ikj} p_0(\boldsymbol{\p})\gamma^i\gamma^k \partial_{{}_{p_j}}
- \sum\limits_{ikj} \gamma^0\gamma^j [p_i\partial_{{}_{p_k}} -p_k\partial_{{}_{p_i}}]
\end{multline*}
where in the sum $ikj = 231,312,123$.

Note that
\[
\begin{split}
[A_{\mu\nu}, P^\oplus] = 0 \,\, \textrm{on} \,\, E^\oplus \,\, \textrm{(concentrated on $\mathscr{O}_{{}_{m, 0,0,0}}$)}
\\
[A_{\mu\nu}, P^\ominus] = 0 \,\, \textrm{on} \,\, E^\ominus \,\, \textrm{(concentrated on $\mathscr{O}_{{}_{-m, 0,0,0}}$)},
\end{split}
\]
as expected, because the representation $U(\alpha)$ transforms bispinor solutions of the Dirac 
equation again into solutions of the Dirac equation, \emph{i.e.} commutes with $P^\oplus$
in action on bispinors concentrated on $\mathscr{O}_{{}_{m, 0,0,0}}$ and, respectively,
commutes with $P^\ominus$ in action on bispinors concentrated on $\mathscr{O}_{{}_{-m, 0,0,0}}$.

In the unitary case 1) decomposition of $U(\alpha)$ will be based on the commutative decomposition $C^*$-algebra
$C$ generated by the two commuting Casimir operators $Q$ and $R$, compare Theorems 1 and 2 in \cite{Segal_dec_I}.
In fact, we will construct this decomposition after \cite{GelfandIV}, Chap. I.4, using the generalized eigenstates of the two
Casimir operators, $Q,R$, self adjoint in the unitary case 1).

That the Casimir operators $Q,R$ are self adjoint in case 1) immediately follows from the fact that the representation
$U(\alpha)$, given by (\ref{1/2pUalpha}) and (\ref{1/2pUa}), or by (\ref{U(alpha)inCartesian-p}) in the coordinates
$\boldsymbol{\p}$ on $\mathscr{O}_{{}_{\pm m, 0,0,0}}$, is unitary with respect to the
invariant inner product (\ref{InvariantScalarProduct}). But this can also be proved independently by showing that
$Q$ and $R$, or the generators $M_{\mu\nu} = iA_{\mu\nu}$, are symmetric on the dense perfect space $E^\oplus \oplus E^\ominus$.
Unfortunately the inner product (\ref{InvariantScalarProduct}) is quite useless for this task, because the Hilbet spaces
$\mathcal{H}^{\oplus}_{m,0}$ and $\mathcal{H}^{\ominus}_{-m,0}$ consist of bispinors $\widetilde{\phi}$
concentrated, respectively, on $\mathscr{O}_{{}_{\pm m, 0,0,0}}$, which in addition lie in the domain
of the idempotents $P^\oplus$ or $P^\ominus$ with $\textrm{rank} \, P^\oplus = \textrm{rank} \, P^\ominus = 2 \neq 4$,
and these idempotents are not self adjoint. Thus, we need to write the operator $iA_{\mu\nu}$, say on $E^\oplus$, in the form
of the operator $P^+ iA_{\mu\nu}P^+$ which coincides with $iA_{\mu\nu}$ on $E^\oplus$ and which is zero on each bispinor
$\widetilde{\phi}$ concentrated on $\mathscr{O}_{{}_{m, 0,0,0}}$, which is orthogonal to $E^\oplus$
with respect to (\ref{InvariantScalarProduct}).
Then symmetricity of $P^+ iA_{\mu\nu}P^+$ on the generic bispinors in $\mathcal{S}(\mathbb{R}^3, \mathbb{C}^4)$
concentrated on $\mathscr{O}_{{}_{m, 0,0,0}}$, with respect to (\ref{InvariantScalarProduct}), will prove
symmetricity of $iA_{\mu\nu}$ on $E^\oplus \subset \mathcal{H}^{\oplus}_{m,0}$,
and thus self adjointness of $iA_{\mu\nu}$ in $\mathcal{H}^{\oplus}_{m,0}$.
Similarly, we would have to proceed with $iA_{\mu\nu}$ on $E^\ominus$.
The fact that
\[
P^\oplus iA_{\mu\nu}P^\oplus = iA_{\mu\nu}P^\oplus = P^\oplus iA_{\mu\nu} = iA_{\mu\nu},
\,\,\, \textrm{on} \,\, E^\oplus
\]
(and similarly for $P^\ominus iA_{\mu\nu}P^\ominus$ on $E^\ominus$) does not help because in general $P^\oplus iA_{\mu\nu}P^\oplus$
is nonzero in action on generic smooth bispinors concentrated on $\mathscr{O}_{{}_{m, 0,0,0}}$ and orthogonal to $E^\oplus$.
Analogous situation we have for $P^\ominus iA_{\mu\nu}P^\ominus$ acting on generic smooth bispinors concentrated on $\mathscr{O}_{{}_{-m, 0,0,0}}$.
Difficulty lies in the fact explained in Subsection \ref{e1}: the representation $U(\alpha)$,
regarded as acting on generic bispinors concentrated on $\mathscr{O}_{{}_{\pm m, 0,0,0}}$
with inner product (\ref{InvariantScalarProduct}), is equal to the
\emph{non-orthogonal} direct sum of unitary representations acting on the non-orthogonal direct sum
$\mathcal{H}^{\oplus}_{\pm m,0} \oplus \mathcal{H}^{\ominus}_{\pm m,0}$. Therefore, $U(\alpha)$, regarded
as acting on generic bispinors (not necessarily lying in the images of $P^\oplus, P^\ominus$)
is not unitary for (\ref{InvariantScalarProduct}). We need the self adjoint projections $P^+,P^-$ on $\mathcal{H}^{\oplus}_{m,0}, \mathcal{H}^{\ominus}_{-m,0}$.
Then symmetricity of $P^\pm iA_{\mu\nu}P^\pm$ on the generic bispinors in $\mathcal{S}(\mathbb{R}^3, \mathbb{C}^4)$
concentrated on $\mathscr{O}_{{}_{\pm m, 0,0,0}}$, with respect to (\ref{InvariantScalarProduct}), will prove
symmetricity of $iA_{\mu\nu}$ on $E^\oplus \oplus E^\ominus \subset \mathcal{H}^{\oplus}_{m,0} \oplus \mathcal{H}^{\ominus}_{-m,0}$,
and thus self adjointness of $iA_{\mu\nu}$ in $\mathcal{H}^{\oplus}_{m,0} \oplus \mathcal{H}^{\ominus}_{-m,0}$.

We bypass this difficulty using the unitary isomorphism $U$ of Subsection \ref{psiBerezin-Hida}, given by (\ref{isomorphismU}),
which maps unitarily the Hilbert space $\mathcal{H}^{\oplus}_{m,0} \oplus \mathcal{H}^{\ominus \, \flat}_{-m,0}$
onto the standard Hilbert space $L^2(\mathbb{R}^3; \mathbb{C}^4)$. We will apply it
to $\mathcal{H}^{\oplus}_{m,0} \oplus \mathcal{H}^{\ominus}_{-m,0}$, with the second argument
not conjugated, \emph{i.e.} to the direct sum of Hilbert spaces of positive and negative energy
solutions of the Dirac equation. We do it in both cases, 1) and 2),
and introduce the following notation
\[
\left( \begin{array}{c}   f_{{}_{1}}(\boldsymbol{\p}) \\
                                           
                                              f_{{}_{2}}(\boldsymbol{\p})      \end{array}\right)
=
\left( \begin{array}{c}   (\widetilde{\phi})_{{}_{1}}(\boldsymbol{\p}) \\
                                           
                                           (\widetilde{\phi})_{{}_{2}}(\boldsymbol{\p})         \end{array}\right)
\,\,\,\,
\left( \begin{array}{c}   g_{{}_{1}}(\boldsymbol{\p}) \\
                                           
                                              g_{{}_{2}}(\boldsymbol{\p})      \end{array}\right)
=
\left( \begin{array}{c}   \overline{(\widetilde{\phi}')_{{}_{3}}(-\boldsymbol{\p})} \\
                                           
                                           \overline{(\widetilde{\phi}')_{{}_{4}}(-\boldsymbol{\p})}         \end{array}\right)
\]
for the four-component elements 
\[
\Big((\widetilde{\phi})_{{}_{1}}, \ldots,  (\widetilde{\phi})_{{}_{4}} \big) \in L^2(\mathbb{R}^3; \mathbb{C}^4)
\]
of the standard Hilbert space $L^2(\mathbb{R}^3; \mathbb{C}^4)$.  Recall once again, please, that in case 2) the additional
factors $\big(2p_0(\boldsymbol{\p})\big)^{-1}, 2p_0(\boldsymbol{\p})$ in the formula for $U$ will be absent
due to the absence of the factor $\big(2p_0(\boldsymbol{\p})\big)^{-2}$ in (\ref{StandardScalarProduct}). 

We then have
\[
\begin{split}
\big(UA_{\mu\nu}U^{-1}f\big)_{{}_{r}}(\boldsymbol{\p}) = \sum\limits_{s=1,2}
{\textstyle\frac{1}{2p_0(\boldsymbol{\p})}}
\Big(
u_{{}_{r}}(\boldsymbol{\p}),
A_{\mu\nu}(2p_0f_{}{}_{s}u_{{}_{s}})(\boldsymbol{\p})
\Big)_{{}_{\mathbb{C}^4}},
\\
\textrm{in case 1)},
\end{split}
\]

\[
\begin{split}
\big(UA_{\mu\nu}U^{-1}g\big)_{{}_{r}}(\boldsymbol{\p}) = \sum\limits_{s=1,2}
{\textstyle\frac{1}{2p_0(\boldsymbol{\p})}}
\Big(
\check{v}_{{}_{r}}(\boldsymbol{\p}),
A_{\mu\nu}(2p_0g_{}{}_{s}\check{v}_{{}_{s}})(\boldsymbol{\p})
\Big)_{{}_{\mathbb{C}^4}},
\,\,\,\, \check{v}(\boldsymbol{\p}) = v(-\boldsymbol{\p})
\\
\textrm{in case 1)},
\end{split}
\]
for $A_{\mu\nu}$ understood as acting on $E^\oplus$ in the first formula, and as acting on $E^\ominus$
in the second. Similarly,

\[
\big(UA_{\mu\nu}U^{-1}f\big)_{{}_{r}}(\boldsymbol{\p}) = \sum\limits_{s=1,2}
\Big(
u_{{}_{r}}(\boldsymbol{\p}),
A_{\mu\nu}(f_{}{}_{s}u_{{}_{s}})(\boldsymbol{\p})
\Big)_{{}_{\mathbb{C}^4}}
\,\,\,\,\,\,\, 
\textrm{in case 2)},
\]

\[
\big(UA_{\mu\nu}U^{-1}g\big)_{{}_{r}}(\boldsymbol{\p}) = \sum\limits_{s=1,2}
\Big(
\check{v}_{{}_{r}}(\boldsymbol{\p}),
A_{\mu\nu}(g_{}{}_{s}\check{v}_{{}_{s}})(\boldsymbol{\p})
\Big)_{{}_{\mathbb{C}^4}}
\,\,\,\,\,\,\, 
\textrm{in case 2)},
\]
where $A_{\mu\nu}$ is understood as acting on $E^\oplus$ in the first, and on $E^\ominus$, in the second formula.
Here $u,v$ are the Fourier transforms of the fundamental solutions, compare Subsections \ref{fundamental,u,v}
and \ref{e1}, and $f \oplus g \in E^+ \oplus E^-$ with $f \in E^+$ and $g \in E^-$.

In more explicit form:
\begin{multline*}
U A_{0k} U^{-1}f (\boldsymbol{\p}) = 
p_0(\boldsymbol{\p})
\left( \begin{array}{cc}  \big(u_{{}_{1}}(\boldsymbol{\p}), \partial_{{}_{p_k}}u_{{}_{1}}(\boldsymbol{\p})\big)_{{}_{\mathbb{C}^4}} & 
\big(u_{{}_{1}}(\boldsymbol{\p}), \partial_{{}_{p_k}}u_{{}_{2}}(\boldsymbol{\p})\big)_{{}_{\mathbb{C}^4}}  \\
 \big(u_{{}_{2}}(\boldsymbol{\p}), \partial_{{}_{p_k}}u_{{}_{1}}(\boldsymbol{\p})\big)_{{}_{\mathbb{C}^4}}              
& \big(u_{{}_{2}}(\boldsymbol{\p}), \partial_{{}_{p_k}}u_{{}_{2}}(\boldsymbol{\p})\big)_{{}_{\mathbb{C}^4}}  \end{array}\right) 
\left( \begin{array}{c} 
 f_{{}_{1}}  \\
f_{{}_{2}}  \end{array}\right) (\boldsymbol{\p})
\\
+ {\textstyle\frac{p_k}{2p_0(\boldsymbol{\p})}}f(\boldsymbol{\p})
+  p_0(\boldsymbol{\p}) \partial_{{}_{p_k}}f(\boldsymbol{\p}), 
\end{multline*}
\[
\begin{split}
p_0(\boldsymbol{\p}) = \sqrt{|\boldsymbol{\p}|^2 + m^2}
\\
\textrm{in case 1)}
\end{split}
\]

\begin{multline*}
U A_{0k} U^{-1}g(\boldsymbol{\p}) 
\\
= 
p_0(\boldsymbol{\p})
\left( \begin{array}{cc}  \big(v_{{}_{1}}(\boldsymbol{-\p}), \partial_{{}_{p_k}}v_{{}_{1}}(-\boldsymbol{\p})\big)_{{}_{\mathbb{C}^4}} & 
\big(v_{{}_{1}}(-\boldsymbol{\p}), \partial_{{}_{p_k}}v_{{}_{2}}(-\boldsymbol{\p})\big)_{{}_{\mathbb{C}^4}}  \\
 \big(v_{{}_{2}}(-\boldsymbol{\p}), \partial_{{}_{p_k}}v_{{}_{1}}(-\boldsymbol{\p})\big)_{{}_{\mathbb{C}^4}}              
& \big(v_{{}_{2}}(-\boldsymbol{\p}), \partial_{{}_{p_k}}v_{{}_{2}}(-\boldsymbol{\p})\big)_{{}_{\mathbb{C}^4}}  \end{array}\right) 
\left( \begin{array}{c} 
 g_{{}_{1}}  \\
g_{{}_{2}}  \end{array}\right) (\boldsymbol{\p})
\\
+ {\textstyle\frac{p_k}{2p_0(\boldsymbol{\p})}}g(\boldsymbol{\p})
+  p_0(\boldsymbol{\p}) \partial_{{}_{p_k}}g(\boldsymbol{\p}), 
\end{multline*}
\[
\begin{split}
p_0(\boldsymbol{\p}) = -\sqrt{|\boldsymbol{\p}|^2 + m^2}
\\
\textrm{in case 1)}
\end{split}
\]

\begin{multline*}
U A_{0k} U^{-1}f (\boldsymbol{\p}) = 
p_0(\boldsymbol{\p})
\left( \begin{array}{cc}  \big(u_{{}_{1}}(\boldsymbol{\p}), \partial_{{}_{p_k}}u_{{}_{1}}(\boldsymbol{\p})\big)_{{}_{\mathbb{C}^4}} & 
\big(u_{{}_{1}}(\boldsymbol{\p}), \partial_{{}_{p_k}}u_{{}_{2}}(\boldsymbol{\p})\big)_{{}_{\mathbb{C}^4}}  \\
 \big(u_{{}_{2}}(\boldsymbol{\p}), \partial_{{}_{p_k}}u_{{}_{1}}(\boldsymbol{\p})\big)_{{}_{\mathbb{C}^4}}              
& \big(u_{{}_{2}}(\boldsymbol{\p}), \partial_{{}_{p_k}}u_{{}_{2}}(\boldsymbol{\p})\big)_{{}_{\mathbb{C}^4}}  \end{array}\right) 
\left( \begin{array}{c} 
 f_{{}_{1}}  \\
f_{{}_{2}}  \end{array}\right) (\boldsymbol{\p})
\\
- {\textstyle\frac{p_k}{2p_0(\boldsymbol{\p})}}f(\boldsymbol{\p})
+  p_0(\boldsymbol{\p}) \partial_{{}_{p_k}}f(\boldsymbol{\p}), 
\end{multline*}
\[
\begin{split}
p_0(\boldsymbol{\p}) = \sqrt{|\boldsymbol{\p}|^2 + m^2}
\\
\textrm{in case 2)}
\end{split}
\]

\begin{multline*}
U A_{0k} U^{-1}g(\boldsymbol{\p}) 
\\
= 
p_0(\boldsymbol{\p})
\left( \begin{array}{cc}  \big(v_{{}_{1}}(\boldsymbol{-\p}), \partial_{{}_{p_k}}v_{{}_{1}}(-\boldsymbol{\p})\big)_{{}_{\mathbb{C}^4}} & 
\big(v_{{}_{1}}(-\boldsymbol{\p}), \partial_{{}_{p_k}}v_{{}_{2}}(-\boldsymbol{\p})\big)_{{}_{\mathbb{C}^4}}  \\
 \big(v_{{}_{2}}(-\boldsymbol{\p}), \partial_{{}_{p_k}}v_{{}_{1}}(-\boldsymbol{\p})\big)_{{}_{\mathbb{C}^4}}              
& \big(v_{{}_{2}}(-\boldsymbol{\p}), \partial_{{}_{p_k}}v_{{}_{2}}(-\boldsymbol{\p})\big)_{{}_{\mathbb{C}^4}}  \end{array}\right) 
\left( \begin{array}{c} 
 g_{{}_{1}}  \\
g_{{}_{2}}  \end{array}\right) (\boldsymbol{\p})
\\
- {\textstyle\frac{p_k}{2p_0(\boldsymbol{\p})}}g(\boldsymbol{\p})
+  p_0(\boldsymbol{\p}) \partial_{{}_{p_k}}g(\boldsymbol{\p}), 
\end{multline*}
\[
\begin{split}
p_0(\boldsymbol{\p}) = -\sqrt{|\boldsymbol{\p}|^2 + m^2}
\\
\textrm{in case 2)}
\end{split}
\]

\begin{multline*}
U A_{ik} U^{-1}f (\boldsymbol{\p}) = 
-{\textstyle\frac{1}{2}}
\left( \begin{array}{cc}  \big(u_{{}_{1}}(\boldsymbol{\p}), \gamma^i \gamma^ku_{{}_{1}}(\boldsymbol{\p})\big)_{{}_{\mathbb{C}^4}} & 
\big(u_{{}_{1}}(\boldsymbol{\p}), \gamma^i \gamma^k u_{{}_{2}}(\boldsymbol{\p})\big)_{{}_{\mathbb{C}^4}}  \\
 \big(u_{{}_{2}}(\boldsymbol{\p}), \gamma^i \gamma^k u_{{}_{1}}(\boldsymbol{\p})\big)_{{}_{\mathbb{C}^4}}              
& \big(u_{{}_{2}}(\boldsymbol{\p}), \gamma^i \gamma^k u_{{}_{2}}(\boldsymbol{\p})\big)_{{}_{\mathbb{C}^4}}  \end{array}\right) 
\left( \begin{array}{c} 
 f_{{}_{1}}  \\
f_{{}_{2}}  \end{array}\right) (\boldsymbol{\p})
\\
+\left( \begin{array}{cc}  \big(u_{{}_{1}}(\boldsymbol{\p}), [p_i\partial_{{}_{p_k}} -p_k\partial_{{}_{p_i}}] u_{{}_{1}}(\boldsymbol{\p})\big)_{{}_{\mathbb{C}^4}} & 
\big(u_{{}_{1}}(\boldsymbol{\p}), [p_i\partial_{{}_{p_k}} -p_k\partial_{{}_{p_i}}] u_{{}_{2}}(\boldsymbol{\p})\big)_{{}_{\mathbb{C}^4}}  \\
 \big(u_{{}_{2}}(\boldsymbol{\p}), [p_i\partial_{{}_{p_k}} -p_k\partial_{{}_{p_i}}] u_{{}_{1}}(\boldsymbol{\p})\big)_{{}_{\mathbb{C}^4}}              
& \big(u_{{}_{2}}(\boldsymbol{\p}), [p_i\partial_{{}_{p_k}} -p_k\partial_{{}_{p_i}}] u_{{}_{2}}(\boldsymbol{\p})\big)_{{}_{\mathbb{C}^4}}  \end{array}\right) 
\left( \begin{array}{c} 
 f_{{}_{1}}  \\
f_{{}_{2}}  \end{array}\right) (\boldsymbol{\p})
\\
+ [p_i\partial_{{}_{p_k}} -p_k\partial_{{}_{p_i}}]f(\boldsymbol{\p}), 
\end{multline*}
\[
\begin{split}
p_0(\boldsymbol{\p}) = \sqrt{|\boldsymbol{\p}|^2 + m^2}
\\
\textrm{in case 1) and 2)}
\end{split}
\]

\begin{multline*}
U A_{ik} U^{-1}g (\boldsymbol{\p}) = 
-{\textstyle\frac{1}{2}}
\left( \begin{array}{cc}  \big(v_{{}_{1}}(-\boldsymbol{\p}), \gamma^i \gamma^kv_{{}_{1}}(-\boldsymbol{\p})\big)_{{}_{\mathbb{C}^4}} & 
\big(v_{{}_{1}}(-\boldsymbol{\p}), \gamma^i \gamma^k v_{{}_{2}}(-\boldsymbol{\p})\big)_{{}_{\mathbb{C}^4}}  \\
 \big(v_{{}_{2}}(-\boldsymbol{\p}), \gamma^i \gamma^k v_{{}_{1}}(-\boldsymbol{\p})\big)_{{}_{\mathbb{C}^4}}              
& \big(v_{{}_{2}}(-\boldsymbol{\p}), \gamma^i \gamma^k v_{{}_{2}}(-\boldsymbol{\p})\big)_{{}_{\mathbb{C}^4}}  \end{array}\right) 
\left( \begin{array}{c} 
 g_{{}_{1}}  \\
g_{{}_{2}}  \end{array}\right) (\boldsymbol{\p})
\\
+\left( \begin{array}{cc}  \big(v_{{}_{1}}(-\boldsymbol{\p}), [p_i\partial_{{}_{p_k}} -p_k\partial_{{}_{p_i}}] v_{{}_{1}}(-\boldsymbol{\p})\big)_{{}_{\mathbb{C}^4}} & 
\big(v_{{}_{1}}(-\boldsymbol{\p}), [p_i\partial_{{}_{p_k}} -p_k\partial_{{}_{p_i}}] v_{{}_{2}}(-\boldsymbol{\p})\big)_{{}_{\mathbb{C}^4}}  \\
 \big(v_{{}_{2}}(-\boldsymbol{\p}), [p_i\partial_{{}_{p_k}} -p_k\partial_{{}_{p_i}}] v_{{}_{1}}(-\boldsymbol{\p})\big)_{{}_{\mathbb{C}^4}}              
& \big(v_{{}_{2}}(-\boldsymbol{\p}), [p_i\partial_{{}_{p_k}} -p_k\partial_{{}_{p_i}}] v_{{}_{2}}(-\boldsymbol{\p})\big)_{{}_{\mathbb{C}^4}}  \end{array}\right) 
\left( \begin{array}{c} 
 g_{{}_{1}}  \\
g_{{}_{2}}  \end{array}\right) (\boldsymbol{\p})
\\
+ [p_i\partial_{{}_{p_k}} -p_k\partial_{{}_{p_i}}]g(\boldsymbol{\p}), 
\end{multline*}
\[
\begin{split}
p_0(\boldsymbol{\p}) = -\sqrt{|\boldsymbol{\p}|^2 + m^2}
\\
\textrm{in case 1) and 2)}
\end{split}
\]

In deriving the formulas for $UA_{0k}U^{-1}$ we have used the identities
\[
\begin{split}
\big(u_{{}_{s}}(\boldsymbol{\p}), \gamma^0\gamma^ku_{{}_{s}}(\boldsymbol{\p})\big)_{{}_{\mathbb{C}^4}}
= {\textstyle\frac{p_k}{p_0(\boldsymbol{\p})}}, \,\,\,\, s=1,2
\\
\big(u_{{}_{s}}(\boldsymbol{\p}), \gamma^0\gamma^ku_{{}_{s'}}(\boldsymbol{\p})\big)_{{}_{\mathbb{C}^4}} =0,
\,\,\, s\neq s',
\end{split}
\]
as well as the analogue identities for $v_{{}_{s}}(-\boldsymbol{\p})$.

Because the fundamental solutions $u,v$ respect at each point $\boldsymbol{\p}$ the orthonormality relations (\ref{u^+u=delta}),
compare Section \ref{fundamental,u,v}, then by differentiating both sides of (\ref{u^+u=delta})
we get
\[
\begin{split}
\big(u_{{}_{r}}(\boldsymbol{\p}), \partial_{{}_{p_k}}u_{{}_{s}}(\boldsymbol{\p})\big)_{{}_{\mathbb{C}^4}}
=-
\big(\partial_{{}_{p_k}}u_{{}_{r}}(\boldsymbol{\p}), u_{{}_{s}}(\boldsymbol{\p})\big)_{{}_{\mathbb{C}^4}},
\\
\big(v_{{}_{r}}(-\boldsymbol{\p}), \partial_{{}_{p_k}}v_{{}_{s}}(-\boldsymbol{\p})\big)_{{}_{\mathbb{C}^4}}, 
= -
\big(\partial_{{}_{p_k}}v_{{}_{r}}(-\boldsymbol{\p}), v_{{}_{s}}(-\boldsymbol{\p})\big)_{{}_{\mathbb{C}^4}}. 
\end{split}
\]
We have also ${\gamma^0}^* = \gamma^0, {\gamma^k}^* = - \gamma^k$, $\gamma^\mu\gamma^\nu +\gamma^\nu\gamma^\mu = 2g^{\mu\nu}$.
Using this we see that in case 1) $UiA_{\mu\nu}U^{-1}$ in action on 
\[
f,g \in E^+, E^- \subset L^{2}(\mathbb{R}^3; \mathbb{C}^2)
\]
are symmetric on the perfect, even nuclear, spaces, respectively, 
\[
E^+, E^- \subset  L^{2}(\mathbb{R}^3; \mathbb{C}^2).
\]
Similarly, in case 2)  the operators $UiA_{ik}U^{-1}$, $i,k = 1,2,3$, in action on 
\[
f,g \in E^+, E^- \subset L^{2}(\mathbb{R}^3; \mathbb{C}^2) 
\]
are symmetric. But it is evident that in case 2) $UiA_{0k}U^{-1}$, $k = 1,2,3$, in action on 
\[
f,g \in E^+, E^- \subset L^{2}(\mathbb{R}^3; \mathbb{C}^2), 
\]
are not symmetric. Accordingly, in case 2), $UiA_{0k}U^{-1}$ is not self adjoint, and the representation $UU(\alpha)U^{-1}$
of the hyperbolic rotations $\alpha \in SL(2, \mathbb{C})$, is not unitary in case 2). Only
the representors $UU(\alpha)U^{-1}$ of spatial rotations $\alpha \in SU(2, \mathbb{C}) \subset SL(2, \mathbb{C})$
are unitary in case 2), in accordance to the result proved in the preceding Subsection. In case 1) the operators
$UiA_{\mu\nu}U^{-1}$
have self adjoint extensions, and of course, this is true separately on each direct summand
\[
E^+, E^- \subset L^{2}(\mathbb{R}^3; \mathbb{C}^2)
\]
because $E^+, E^-$ are perfect, invariant for $UiA_{0k}U^{-1}$ and densely included into
$L^{2}(\mathbb{R}^3; \mathbb{C}^2)$. These operators are even essentially self adjoint in case 1). Thus,
we have proved unitarity of $UU(\alpha)U^{-1}$ in case 1), independently of the proof following from the
general representation theory, applied as in Subsection \ref{e1}. From this it follows that
the two Casimir operators $Q$ and $R$ are self adjoint in case 1), but are not self adjoint in case 2).

Moreover, using the standard Hilbert space representation we can easily compute the operators
$\big[UiA_{\mu\nu}U^{-1}\big]^*$ adjoint to $UiA_{0k}U^{-1}$, in case 2), and then transform them
back, using the unitary isomorphism $U$, and compute their action on bispinors in $E^\oplus \oplus E^\ominus$.
The result is:
\[
\begin{split}
A_{0k}^{*} = - A_{0k} - {\textstyle\frac{2p_k}{p_0(\boldsymbol{\p})}} \boldsymbol{1},
\\
A_{ik}^{*} = - A_{ik}, \,\,\,\, \textrm{on} \,\, E^\oplus \oplus E^\ominus
\\
\textrm{in case 2)}.
\end{split}
\]
From this it easily follows that on $E^\oplus \oplus E^\ominus$
\[
[Q^*,Q] \neq 0, \,\,\, [R^*, R] \neq 0,
\,\,\,\,\,
\textrm{in case 2)}.
\]

But (denoting the self adjoint extension of $A_{\mu\nu},Q,R$, by the same symbol $A_{\mu\nu},Q,R$)
\[
\begin{split}
A_{0k}^{*} = - A_{0k}, \,\,\,\, A_{ik}^{*} = - A_{ik},
\\
Q^* = Q, \,\,\,\, R^* = R
\\
\textrm{in case 1)}.
\end{split}
\]

From this it immediately follows that the Casimir operators $Q,R$ not only are not self adjoint in case 2),
but they are not normal operators in this case. In particular their generalized states cannot be used in the 
construction of the direct integral decomposition of the Hilbert space $\mathcal{H}^{\oplus}_{m,0} \oplus \mathcal{H}^{\ominus}_{-m,0}$,
acted on by the representation $U(\alpha)$, into invariant (generalized) subspaces in case 2). Indeed, if it
was possible, then the Casimir operators, $Q,R$, in the non-unitary case 2), would be unitarily equivalent to the
operators of multiplication by functions, which contradicts the proved result, that in case 2), $Q,R$
are not normal.

\begin{center}
{\small DECOMPOSITION OF $\alpha\mapsto U(\alpha)$ IN CASE 1)}
\end{center}

Let us concentrate now on the case 1) and construct the direct integral decomposition:
\begin{equation}\label{DecompositionH++H-}
\mathcal{H}^{\oplus}_{m,0} \oplus \mathcal{H}^{\ominus}_{-m,0} =
\int
\big(\mathcal{H}^{\oplus}_{m,0}\big)_{{}_{\chi}} \oplus \big(\mathcal{H}^{\ominus}_{-m,0}\big)_{{}_{\chi}}
\, \ud \chi
= \int \big(\mathcal{H}^{\oplus}_{m,0}\big)_{{}_{\chi}} \, \ud \chi
\oplus \int \big(\mathcal{H}^{\ominus}_{-m,0}\big)_{{}_{\chi}} \, \ud \chi
\end{equation}
\[
U(\alpha)
=
\int U_{{}_{\oplus \, \chi}}(\alpha) \, \ud \chi
\oplus \int U_{{}_{\ominus \, \chi}}(\alpha) \, \ud \chi
\]
into irreducible invariant subspaces $\big(\mathcal{H}^{\oplus}_{m,0}\big)_{{}_{\chi}} \oplus \big(\mathcal{H}^{\ominus}_{-m,0}\big)_{{}_{\chi}}$
with the irreducible components $U_{{}_{\oplus \, \chi}}(\alpha)$ and
$U_{{}_{\ominus \, \chi}}(\alpha)$ acting, respectively, upon $\big(\mathcal{H}^{\oplus}_{m,0}\big)_{{}_{\chi}}$ and
$\big(\mathcal{H}^{\ominus}_{-m,0}\big)_{{}_{\chi}}$
Decomposition (\ref{DecompositionH++H-}), in case 1), is determined by the spectral decomposition of the Casimir operators, $Q,R$. Accordingly, the range of the
decomposition parameter $\chi$ is determined by the joint spectrum $\textrm{Spec} \, (Q,R)$ of the two Casimir operators
$Q$ and $R$. We construct this decomposition using the rigged Hilbert space structure, determined by the single particle
Gelfand triple, defined above, and the method of \cite{GelfandIV}, Chap. I.4. In accordance to the construction
of decomposition presented there, each direct integral component Hilbert space $\big(\mathcal{H}^{\oplus}_{m,0}\big)_{{}_{\chi}}$
and $\big(\mathcal{H}^{\ominus}_{-m,0}\big)_{{}_{\chi}}$ consists of generalized states, \emph{i.e.} elements
$F^{{}^{\oplus}}_{{}_{\chi}} \in E^{\oplus *}$ and $F^{{}^{\oplus}} \in E^{\oplus *}_{{}_{\chi}}$ which are equal to generalized
eigenstates of the two Casimir operators:
\[
\begin{split}
F^{{}^{\oplus}}_{{}_{\chi}}\big(Q \widetilde{\phi} \big) = \lambda_{{}_{Q}}(\chi) F^{{}^{\oplus}}_{{}_{\chi}}\big(\widetilde{\phi} \big),
\,\,\,\,\,\,
F^{{}^{\oplus}}_{{}_{\chi}}\big(R \widetilde{\phi} \big) = \lambda_{{}_{R}}(\chi) F^{{}^{\oplus}}_{{}_{\chi}}\big(\widetilde{\phi} \big),
\\
\textrm{for all $\widetilde{\phi} \in E^\oplus$},
\\
F^{{}^{\ominus}}_{{}_{\chi}}\big(Q \widetilde{\phi} \big) = \lambda_{{}_{Q}}(\chi) F^{{}^{\ominus}}_{{}_{\chi}}\big(\widetilde{\phi} \big),
\,\,\,\,\,\,
F^{{}^{\ominus}}_{{}_{\chi}}\big(R \widetilde{\phi} \big) = \lambda_{{}_{R}}(\chi) F^{{}^{\ominus}}_{{}_{\chi}}\big(\widetilde{\phi} \big),
\\
\textrm{for all $\widetilde{\phi} \in E^\ominus$},
\end{split}
\]
Each element $\widetilde{\phi} \in E^\oplus$ or, respectively, $\widetilde{\phi} \in E^\ominus$, determines
a functional $\widetilde{\phi}_{{}_{\chi}}$, respectively, on $\big(\mathcal{H}^{\oplus}_{m,0}\big)_{{}_{\chi}}$
or, respectively, on $\big(\mathcal{H}^{\ominus}_{-m,0}\big)_{{}_{\chi}}$, by the rule
\[
\begin{split}
\big(\mathcal{H}^{\oplus}_{m,0}\big)_{{}_{\chi}} \ni F^{{}^{\oplus}}_{{}_{\chi}} \longmapsto
\widetilde{\phi}_{{}_{\chi}}\big(F^{{}^{\oplus}}_{{}_{\chi}}\big) \coloneqq
F^{{}^{\oplus}}_{{}_{\chi}}\big(\widetilde{\phi}\big),
\\
\big(\mathcal{H}^{\ominus}_{-m,0}\big)_{{}_{\chi}} \ni F^{{}^{\ominus}}_{{}_{\chi}} \longmapsto
\widetilde{\phi}_{{}_{\chi}}\big(F^{{}^{\ominus}}_{{}_{\chi}}\big) \coloneqq
F^{{}^{\oplus}}_{{}_{\chi}}\big(\widetilde{\phi}\big).
\end{split}
\]
These functionals, $\widetilde{\phi}_{{}_{\chi}}$, respectively, for $\widetilde{\phi} \in E^\oplus$ or $\widetilde{\phi} \in E^\ominus$
define the decomposition components
\[
\widetilde{\phi} = \int \widetilde{\phi}_{{}_{\chi}} \ud \chi
\]
of $\widetilde{\phi} \in E^\oplus \subset \mathcal{H}^{\oplus}_{m,0}$ or
of $\widetilde{\phi} \in E^\ominus \subset \mathcal{H}^{\ominus}_{-m,0}$ in the direct integral decomposition (\ref{DecompositionH++H-}).

The eigenstates or distributions $F^{{}^{\oplus}}_{{}_{\chi}} \in E^{\oplus *}$ and $F^{{}^{\oplus}} \in E^{\oplus *}_{{}_{\chi}}$,
are representable by ordinary functions $\boldsymbol{\p} \mapsto F^{{}^{\oplus}}_{{}_{\chi}}(\boldsymbol{\p})$
and $\boldsymbol{\p} \mapsto F^{{}^{\oplus}}(\boldsymbol{\p})$ on the respective hyperboloids
$\mathscr{O}_{{}_{\pm \boldsymbol{m}, 0,0,0}}$, which moreover fulfill
\[
F^{{}^{\oplus}}_{{}_{\chi,lm}} = P^\oplus F^{{}^{\oplus}}_{{}_{\chi,lm}},
\,\,\,\
F^{{}^{\ominus}}_{{}_{\chi,lm}} = P^\ominus F^{{}^{\ominus}}_{{}_{\chi,lm}}.
\]
That is, their value, if regarded as functionals, on the test function $\widetilde{\phi} \in E^\oplus$
or $\phi \in E^\ominus$, is given by ordinary integral along $\mathscr{O}_{{}_{\pm \boldsymbol{m}, 0,0,0}}$ of
$\widetilde{\phi}$ multiplied by the respective function $F^{{}^{\oplus}}_{{}_{\chi}}$ or $F^{{}^{\ominus}}_{{}_{\chi}}$.
Therefore, the condition of being the generalized eigenstate $F^{{}^{\oplus}}_{{}_{\chi}}$ or $F^{{}^{\ominus}}_{{}_{\chi}}$,
expressed in terms of the functions $F^{{}^{\oplus}}_{{}_{\chi}}$,
$F^{{}^{\ominus}}_{{}_{\chi}}$, is equivalent to the system of partial differential equations
for the functions $F^{{}^{\oplus}}_{{}_{\chi}}$ or $F^{{}^{\ominus}}_{{}_{\chi}}$, corresponding to the
differential operators representing $Q$ and $R$.
To each fixed pair of eigenvalues $\big(\lambda_{{}_{Q}}(\chi),\lambda_{{}_{r}}(\chi)\big)$ lying in the joint
spectrum of $Q,R$ corresponds a complete orthonormal system $\big\{F^{{}^{\oplus}}_{{}_{\chi \, lm}} \big\}$ and, respectively,
$\big\{F^{{}^{\ominus}}_{{}_{\chi \, lm}} \big\}$, which spans the invariant Hilbert space
$\big(\mathcal{H}^{\oplus}_{m,0}\big)_{{}_{\chi}}$ and $\big(\mathcal{H}^{\ominus}_{-m,0}\big)_{{}_{\chi}}$,
of the irrrdubile direct integral component $U_{{}_{\oplus \, \chi}}(\alpha)$ and $U_{{}_{\ominus \, \chi}}(\alpha)$ of the representation,
$U(\alpha)$, corresponding to the decomposition parameter $\chi$.
From the general theory of irreducible unitary representations of $SL(2, \mathbb{C})$, \cite{Geland-Minlos-Shapiro}
or \cite{NeumarkLorentzBook}, we know that each such irreducible representation, when restricted to the compact
subgroup $SU(2, \mathbb{C})$ decomposes into direct sum
\[
L^{l_0} \oplus L^{l_0 +1} \oplus L^{l_0 +2} \oplus \ldots
\]
of irreducible representations, into which each irreducible representation with the weight $l = l_0, l_0+1,l_0+2, \ldots$
enters once, beginning with some minimal value of the weight equal $l_0$.
Thus, we are looking for the systems of smooth functions $\big\{F^{{}^{\oplus}}_{{}_{\chi \, lm}} \big\}$ and, respectively,
$\big\{F^{{}^{\ominus}}_{{}_{\chi \, lm}} \big\}$, which fulfill the following differential equations
\begin{equation}\label{DiffEigenEqn1}
\begin{split}
Q F^{{}^{\oplus}}_{{}_{\chi,lm}} = \lambda_{{}_{Q}} \, F^{{}^{\oplus}}_{{}_{\chi,lm}},
\,\,
R F^{{}^{\oplus}}_{{}_{\chi,lm}} = \lambda_{{}_{R}} \, F^{{}^{\oplus}}_{{}_{\chi,lm}},
\,\, \textrm{on $\mathscr{O}_{{}_{\boldsymbol{m}, 0,0,0}}$},
\\
Q F^{{}^{\ominus}}_{{}_{\chi,lm}} = \lambda_{{}_{Q}} \, F^{{}^{\ominus}}_{{}_{\chi,lm}},
\,\,
R F^{{}^{\ominus}}_{{}_{\chi,lm}} = \lambda_{{}_{R}} \, F^{{}^{\ominus}}_{{}_{\chi,lm}},
\,\, \textrm{on $\mathscr{O}_{{}_{-\boldsymbol{m}, 0,0,0}}$},
\end{split}
\end{equation}
\[
l = l_0, l_0+1, \ldots, \,\,\,\,\, -l \leq m \leq l
\]
\begin{equation}\label{DiffEigenEqn2}
\begin{split}
\Big[-\big(A_{23}\big)^2 - \big(A_{31}\big)^2 - \big(A_{12}\big)^2 \Big]F^{{}^{\oplus}}_{{}_{\chi,lm}} =
l(l+1) F^{{}^{\oplus}}_{{}_{\chi,lm}},
\,\,\,\, iA_{12} F^{{}^{\oplus}}_{{}_{\chi,lm}} = m F^{{}^{\oplus}}_{{}_{\chi,l,m}},
\\
\Big[-\big(A_{23}\big)^2 - \big(A_{31}\big)^2 - \big(A_{12}\big)^2 \Big]F^{{}^{\ominus}}_{{}_{\chi,lm}} =
l(l+1) F^{{}^{\ominus}}_{{}_{\chi,l,m}},
\,\,\,\, iA_{12} F^{{}^{\ominus}}_{{}_{\chi,lm}} = m F^{{}^{\ominus}}_{{}_{\chi,lm}}.
\end{split}
\end{equation}
The eigenvalues $\lambda_{{}_{Q}},\lambda_{{}_{R}}$ of the operators $Q,R$,
are associated to the pair of numbers $(l_0, l_1)$, uniquely characterizing the irreducible
representation, in the following manner (\cite{Geland-Minlos-Shapiro}, \cite{NeumarkLorentzBook})
\[
\lambda_{{}_{Q}} = -2(l_{0}^{2}+l_{1}^{2} - 1), \,\,\,\,\, \lambda_{{}_{R}} = -4il_0l_1.
\]
with $\chi$ depending on $(\lambda_{{}_{Q}}, \lambda_{{}_{R}}) \in \textrm{Spec}(Q,R)$.

The whole problem lies of course in the correct choice of the set of eigenvalues $\lambda_{{}_{Q}}, \lambda_{{}_{R}}$,
or in determination of the joint spectrum  of $Q,R$. Only in this case the systems of generalized 
states $\big\{F^{{}^{\oplus}}_{{}_{\chi \, lm}} \big\}$ and, respectively,
$\big\{F^{{}^{\ominus}}_{{}_{\chi \, lm}} \big\}$ are complete, and the Plancherel formula holds. Existence of such 
a complete system follows in the unitary case 1) from the general theory given in \cite{GelfandIV}, Chap. I.4.

Recall, that the generalized states $\big\{F^{{}^{\oplus}}_{{}_{\chi \, lm}} \big\}$ and, respectively,
$\big\{F^{{}^{\ominus}}_{{}_{\chi \, lm}} \big\}$, similarly as the ordinary normalizable single particle states, 
are given by ordinary functions on the respective orbits
(positive and negative energy hyperboloids). Recall also that the $U$ isomorphism (\ref{isomorphismU}) is naturally
defined on the generalized states, and in terms of the corresponding representing functions, is given by the same
formula (\ref{isomorphismU}). We will denote the linear dual (transposition) of the isomorphism $U$, acting on the generalized
states, by the same symbol $U$.

The invariant valuations, expressed in explicit form, read
\begin{equation}\label{InvPairingPsi}
\begin{split}
F^{{}^{\oplus}}_{{}_{\chi,lm}}\big(\widetilde{\phi}\big)
= \Big\langle F^{{}^{\oplus}}_{{}_{\chi,lm}}, \widetilde{\phi} \Big\rangle_{{}_{\textrm{invariant pairing}}}
= \int \limits_{\mathscr{O}_{\boldsymbol{m},0,0,0}} \big(F^{{}^{\oplus}}_{{}_{\chi,lm}}(\boldsymbol{\p}), \gamma^0 \widetilde{\phi}(\boldsymbol{\p}) \big)_{{}_{\mathbb{C}^4}}
\, \ud \mu_{{}_{\mathscr{O}_{\boldsymbol{m},0,0,0}}}(p)
\\
=\int \limits_{\mathscr{O}_{\boldsymbol{m},0,0,0}} \big(F^{{}^{\oplus}}_{{}_{\chi,lm}}(p), P^\oplus(\boldsymbol{\p})\widetilde{\phi}(\boldsymbol{\p}) \big)_{{}_{\mathbb{C}^4}}
\, {\textstyle\frac{\ud^3 \boldsymbol{\p}}{(2p_0(\boldsymbol{\p}))^2}}
\\
F^{{}^{\ominus}}_{{}_{\chi,lm}}\big(\widetilde{\phi}\big)
= \Big\langle F^{{}^{\ominus}}_{{}_{\chi,lm}}, \widetilde{\phi} \Big\rangle_{{}_{\textrm{invariant pairing}}}
= \int \limits_{\mathscr{O}_{-\boldsymbol{m},0,0,0}} \big(F^{{}^{\ominus}}_{{}_{\chi,lm}}(\boldsymbol{\p}), \gamma^0 \widetilde{\phi}(\boldsymbol{\p}) \big)_{{}_{\mathbb{C}^4}}
\, \ud \mu_{{}_{\mathscr{O}_{-\boldsymbol{m},0,0,0}}}(p)
\\
=\int \limits_{\mathscr{O}_{-\boldsymbol{m},0,0,0}} \big(F^{{}^{\ominus}}_{{}_{\chi,lm}}(\boldsymbol{\p}), P^\ominus(\boldsymbol{\p}) \widetilde{\phi}(\boldsymbol{\p}) \big)_{{}_{\mathbb{C}^4}}
\, {\textstyle\frac{\ud^3 \boldsymbol{\p}}{(2p_0(\boldsymbol{\p}))^2}}.
\end{split}
\end{equation}
Here we have used the same identities as in Subsection \ref{FirstStepH} when proving the compatibility of the invariant pairing
with the invariant inner product in the single particle Hilbert space. Note, please, that for any $\widetilde{\phi}$
smooth on the respective orbit, not necessarily lying in the image of the idempotent $P^\oplus$ or $P^\ominus$, we have
\[
F^{{}^{\oplus}}_{{}_{\chi,lm}}\big(\widetilde{\phi}\big) = F^{{}^{\oplus}}_{{}_{\chi,lm}}\big(P^\oplus\widetilde{\phi}\big),
\,\,\,\, 
F^{{}^{\ominus}}_{{}_{\chi,lm}}\big(\widetilde{\phi}\big) = F^{{}^{\ominus}}_{{}_{\chi,lm}}\big(P^\ominus\widetilde{\phi}\big).
\]

We will be seeking the solutions in the general form
\[
\begin{split}
F^{{}^{\oplus}}_{{}_{\chi,lm}}(\boldsymbol{\p}) 
= \sum\limits_{s=1}^{2} |p_0(\boldsymbol{\p})| F^{{}^{\oplus}}_{{}_{\chi,lms}}(\boldsymbol{\p}) \, u_{{}_{s}}(\boldsymbol{\p})
\\
F^{{}^{\ominus}}_{{}_{\chi,lm}}(\boldsymbol{\p}) 
= \sum\limits_{s=3}^{4} |p_0(\boldsymbol{\p})| F^{{}^{\ominus}}_{{}_{\chi,lms}}(\boldsymbol{\p}) \, v_{{}_{s-2}}(-\boldsymbol{\p})
\end{split}
\]
so that 
\[
\begin{split}
F^{{}^{\oplus}}_{{}_{\chi,lms}}(\boldsymbol{\p}) 
= {\textstyle\frac{1}{2|p_0(\boldsymbol{\p})|}} \big(u_{{}_{s}}(\boldsymbol{\p}), F^{{}^{\oplus}}_{{}_{\chi,lm}}(\boldsymbol{\p}) \big)_{{}_{\mathbb{C}^4}}
=
\big(UF^{{}^{\oplus}}_{{}_{\chi,lm}}\big)_{{}_{s}}, \,\,\, s=1,2
\\
F^{{}^{\ominus}}_{{}_{\chi,lms}}(\boldsymbol{\p}) 
= 
\overline{
{\textstyle\frac{1}{2|p_0(\boldsymbol{\p})|}} \big(v_{{}_{s-2}}(\boldsymbol{\p}), F^{{}^{\ominus}}_{{}_{\chi,lm}}(-\boldsymbol{\p}) \big)_{{}_{\mathbb{C}^4}}
}
\,\,\,\,\,\,\,\,\,\,\,\,\,\,\,\,\,\,\,\,\,\,\,\,\,\,\,\,\,\,\,\,\,\,\,\,\,\,\,\,\,\,\,\,\,\,\,\,\,\,\,\,\,\,\,\,\,
\\
=
\overline{
{\textstyle\frac{1}{2|p_0(\boldsymbol{\p})|}} \big(v_{{}_{s-2}}(\boldsymbol{\p}), 
\overline{F^{{}^{\ominus \, \flat \, T}}_{{}_{\chi,lm}}(\boldsymbol{\p})} \big)_{{}_{\mathbb{C}^4}}
}
=
\big(U\overline{F^{{}^{\ominus \, \flat \, T}}_{{}_{\chi,lm}}}\big)_{{}_{s}}, \,\,\, s=3,4.
\end{split}
\]
Here $U$ is the natural extension of the $U$ isomorphism (\ref{isomorphismU}) over the generalized states,
also representable by ordinary functions on the respective orbits $\mathscr{O}_{{}_{\pm \boldsymbol{m},0,0,0}}$.

Let $\widetilde{\phi} \in E^\oplus \oplus E^\ominus \subset \mathcal{H}^{\oplus}_{m,0} \oplus \mathcal{H}^{\ominus}_{-m,0}$.
By the general theory the following Plancherel formula holds
\begin{multline*}
\|\widetilde{\phi} \|^2 =  
\int\limits_{\mathscr{O}_{{}_{\boldsymbol{m},0,0,0}} \sqcup \mathscr{O}_{{}_{-\boldsymbol{m},0,0,0}}}
\big|\widetilde{\phi}(\boldsymbol{\p})\big|_{{}_{\mathbb{C}^4}}^{2} \, 
{\textstyle\frac{\ud^3 \boldsymbol{\p}}{(2p_0(\boldsymbol{\p}))^2}}
\\
=
\int\limits_{\mathscr{O}_{{}_{\boldsymbol{m},0,0,0}}}
\big|\widetilde{\phi}(\boldsymbol{\p})\big|_{{}_{\mathbb{C}^4}}^{2} \, 
{\textstyle\frac{\ud^3 \boldsymbol{\p}}{(2p_0(\boldsymbol{\p}))^2}}
+
\int\limits_{\mathscr{O}_{{}_{-\boldsymbol{m},0,0,0}}}
\big|\widetilde{\phi}(\boldsymbol{\p})\big|_{{}_{\mathbb{C}^4}}^{2} \, 
{\textstyle\frac{\ud^3 \boldsymbol{\p}}{(2p_0(\boldsymbol{\p}))^2}}
\end{multline*}
\begin{multline}\label{PlancherelForSingleHforpsi}
= \sum \limits_{l,m} \,\, \int \big| F^{{}^{\oplus}}_{{}_{\chi,lm}}(\widetilde{\phi}) \big|^2 \, d \chi  
+
\sum \limits_{l,m} \,\, \int \big| F^{{}^{\ominus}}_{{}_{\chi,lm}}(\widetilde{\phi}) \big|^2 \, d \chi  
\end{multline}
\[
l = l_0, l_0+1, \ldots, \,\,\, -l \leq m \leq l, \,\,\,\, s = 1,2 \,\,  \textrm{or} \,\, 3,4.
\]
The indicated generalized states 
$F^{{}^{\oplus}}_{{}_{\chi,lm}} \in E^{\oplus *}$, $F^{{}^{\ominus}}_{{}_{\chi,lm}} \in E^{\ominus *}$
compose, for each fixed $\chi$, a complete orthonormal systems, respectively, in $\big(\mathcal{H}^{\oplus}_{m,0}\big)_{{}_{\chi}}$
and $\big(\mathcal{H}^{\ominus}_{-m,0}\big)_{{}_{\chi}}$.

In order to determine the joint spectrum of $Q,R$, and the corresponding $F^{{}^{\oplus}}_{{}_{\chi,lm}}$, $F^{{}^{\ominus}}_{{}_{\chi,lm}}$
explicitly, so that the above Plancherel formula holds, we determine
the asymptotic behavior (for large momenta $\boldsymbol{\p}$) of the complete systems of solutions
$F^{{}^{\oplus}}_{{}_{\chi,lm}}$, $F^{{}^{\ominus}}_{{}_{\chi,lm}}$.
In order to establish the asymptotic behavior, \emph{i.e} for $|\boldsymbol{\p}|$ going to infinity, of these complete systems
we observe that the massive orbits (both sheets of massive hyperboloid) $\mathscr{O}_{{}_{\pm \boldsymbol{m},0,0,0}}$,
become indistinguishable from the corresponding sheets $\mathscr{O}_{{}_{\pm 1,0,0,1}}$ of the cone orbits.
Similarly, the projectors $P^\oplus, P^\ominus$, and the fundamental solutions $u,v$, become,
for momenta $\boldsymbol{\p}$ with $|\boldsymbol{\p}|$ going to infinity, equal to to the projectors
$P^\oplus, P^\ominus$, and the fundamental solutions $u,v$ which correspond to the massless Dirac equation. Accordingly,
the formulas for the generators $M_{\mu\nu} = iA_{\mu\nu}$, and the Casimir operators $Q,R$, regarded as local differential
operators, become for large $|\boldsymbol{\p}|$, equal to the respective operators of massless Dirac equation, with
$p_0(\boldsymbol{\p}) \rightarrow |\boldsymbol{\p}|$.
On the other hand the Plancherel formula for positive and, respective, negative energy bispinor solutions,
concentrated on the cone $\mathscr{O}_{{}_{\pm 1,0,0,1}}$, of the massless Dirac equation, can be easily established.
Indeed, we are using the generalized Pontrjagin duality for the multiplicative group of positive reals and the additive group
of all real numbers, and the Plancherel theorem for the unit sphere, exactly as in Subsection \ref{equivalentA-s}.
The only difference in construction of the complete system of generalized states on bispinors concentrated on the cone
in comparison to the construction of Subsection \ref{equivalentA-s}, is that we are using the fundamental solutions
$u$ on the positive energy cone, which in case of massless bispinors
are homogeneous of degree zero functions on the positive energy cone, and use them as the homogeneous vectors $w$ in \ref{equivalentA-s}.
Similarly, we are using the fundamental solutions $v$ on the negative energy cone (which in case of massless bispinors are homogeneous
of degree zero functions on the cone) instead of the homogeneous vector functions $w$.
It turns out that the complete systems of generalized states (in the momentum picture, of course), regarded as functions on the cone
in the momentum space, for the massless
bispinors, are homogeneous of degree $-1/2 -i\nu$, $\nu \in \mathbb{R}$ but regarded as distributions in the full momentum space $\mathbb{R}^4$,
are homogeneous of degree $3/2 - i\nu -4$. Therefore, their inverse Fourier transforms, regarded as distributions
over Minkowski space-time, are homogeneous of degree $-3/2 + i \nu$, $\nu \in \mathbb{R}$.
We can therefore infer that the generalized states
$F^{{}^{\oplus}}_{{}_{\chi,lm}}$, $F^{{}^{\ominus}}_{{}_{\chi,lm}}$, composing complete
systems of generalized state spaces of bispinors concentrated on
$\mathscr{O}_{{}_{\pm \boldsymbol{m},0,0,0}}$, are asymptotically (for large momenta) homogeneous of degree
$-1/2 - i\nu$, $\nu \in \mathbb{R}$, when regarded as functions of the Cartesian coordinates $\boldsymbol{\p}$
on the respective orbit $\mathscr{O}_{{}_{\pm \boldsymbol{m},0,0,0}}$. This means that the limits
\[
\underset{\lambda\rightarrow +\infty}{\textrm{lim}} \lambda^{{}^{-\chi}}F^{{}^{\oplus}}_{{}_{\chi,lm}}(\lambda \boldsymbol{\p}),
\,\,\,\,\,\,\,\,
\underset{\lambda\rightarrow +\infty}{\textrm{lim}} \lambda^{{}^{-\chi}}F^{{}^{\ominus}}_{{}_{\chi,lm}}(\lambda \boldsymbol{\p})
\]
exist for each $\chi = -1/2 -i \nu$, $\nu \in \mathbb{R}$, and each $\boldsymbol{\p} \in \mathbb{R}^3$.
Note here that although homogeneity is not invariant under the representation $U(\alpha)$, or not Lorentz invariant
on the massive orbits $\mathscr{O}_{{}_{\pm \boldsymbol{m},0,0,0}}$, the asymptotic homogeneity is invariant, and can serve as an important invariant
in the decomposition of the representation of $SL(2, \mathbb{C})$ for the massive fields (transforming under unitary representation)
and plays the same role on $\mathscr{O}_{{}_{\pm \boldsymbol{m},0,0,0}}$ as the homogeneity does
on the cone $\mathscr{O}_{{}_{\pm 1,0,0,1}}$ for the massless fields.

Having established the asymptotic behavior of $F^{{}^{\oplus}}_{{}_{\chi,lm}}$, $F^{{}^{\ominus}}_{{}_{\chi,lm}}$,
which as functions of $\boldsymbol{\p}$ are asymptotically (for large momenta) homogeneous of degree $-1/2 - i \nu$, $\nu \in \mathbb{R}$
we can go back to the differential equations (\ref{DiffEigenEqn1}) and (\ref{DiffEigenEqn2}) for the functions
$F^{{}^{\oplus}}_{{}_{\chi,lm}}$, $F^{{}^{\ominus}}_{{}_{\chi,lm}}$. We then rewrite
the system of differential equations (\ref{DiffEigenEqn1}) and (\ref{DiffEigenEqn2}) in the spherical coordinates
and use the method of separation of variables and insert the asymptotic homogeneity
$\chi = -1/2 - i\nu$, with a fixed value of $\nu$. For each fixed $\chi = -1/2-i\nu$ of this form there correspond the eigenvalues
$\lambda_{{}_{Q}}(\chi),\lambda_{{}_{R}}(\chi)$ of $Q$ and $R$, and are determined by the
allowed asymptotics $\chi$ of the functions $F^{{}^{\oplus}}_{{}_{\chi,lm}}$, $F^{{}^{\ominus}}_{{}_{\chi,lm}}$.
The computation being a repetition of the known procedures of the theory of special functions (or the associated representation theory),
compare e.g. \cite{Geland-Minlos-Shapiro}, can be omitted. The recurrence rules for
$F^{{}^{\oplus}}_{{}_{\chi,lm}}$, $F^{{}^{\ominus}}_{{}_{\chi,lm}}$ can be established exactly as in \cite{Geland-Minlos-Shapiro}.
The function $F^{{}^{\oplus}}_{{}_{\chi,lm}}$ or $F^{{}^{\ominus}}_{{}_{\chi,lm}}$, should be identified with the vector state
which is denoted by $\xi_{lm}$ in \cite{Geland-Minlos-Shapiro} in the representation space of the irreducible representation denoted
with the pair $(l_0=1/2,l_1=i\nu)$, $\nu \in \mathbb{R}$, in \cite{Geland-Minlos-Shapiro} corresponding
to $\chi$, and with relation between $(l_0=1/2,l_1 = i\nu)$ and $\chi=-1/2 -i\nu$ determined by the system of equations (\ref{DiffEigenEqn1}) and (\ref{DiffEigenEqn2})
and the asymptotics $\chi=-1/2-i\nu$ of $F^{{}^{\oplus}}_{{}_{\chi,lm}}$, $F^{{}^{\ominus}}_{{}_{\chi,lm}}$.

Let $\widetilde{\phi} \in E^\oplus$ or, respectively,  $\widetilde{\phi} \in E^\ominus$.
By the Riesz representation theorem there exists the unique element $\widetilde{\phi}_{{}_{\chi}} \in \big(\mathcal{H}^{\oplus}_{m,0}\big)_{{}_{\chi}}$
and, respectively, unique element $\widetilde{\phi}_{{}_{\chi}} \in \big(\mathcal{H}^{\ominus}_{-m,0}\big)_{{}_{\chi}}$, such that
\begin{equation}\label{Riesz-GelfandComponents}
\begin{split}
F^{{}^{\oplus}}_{{}_{\chi}}(\widetilde{\phi}) = \Big(F^{{}^{\oplus}}_{{}_{\chi}}, \widetilde{\phi}_{{}_{\chi}} \Big)_{{}_{\chi}},
\,\,\,\,\,\, \textrm{for all} \,\,\, F^{{}^{\oplus}}_{{}_{\chi}} \in \big(\mathcal{H}^{\oplus}_{m,0}\big)_{{}_{\chi}}
\\
F^{{}^{\ominus}}_{{}_{\chi}}(\widetilde{\phi}) = \Big(F^{{}^{\ominus}}_{{}_{\chi}}, \widetilde{\phi}_{{}_{\chi}} \Big)_{{}_{\chi}},
\,\,\,\,\,\, \textrm{for all} \,\,\,  F^{{}^{\ominus}}_{{}_{\chi}} \in \big(\mathcal{H}^{\ominus}_{-m,0}\big)_{{}_{\chi}}.
\end{split}
\end{equation}
These 
\[
\widetilde{\phi}_{{}_{\chi}} \in \big(\mathcal{H}^{\oplus}_{m,0}\big)_{{}_{\chi}},
\,\,\,\,\,\,\,
\widetilde{\phi}_{{}_{\chi}} \in \big(\mathcal{H}^{\ominus}_{-m,0}\big)_{{}_{\chi}}
\]
provide decomposition components of the elements
\[
\widetilde{\phi} \in E^\oplus \subset \mathcal{H}^{\oplus}_{m,0},
\,\,\,\,\,\,\,
\widetilde{\phi} \in E^\ominus \subset \mathcal{H}^{\ominus}_{-m,0},
\]
in the direct integral decomposition (\ref{DecompositionH++H-}).

Note that, using $\phi \in \mathcal{S}(\mathbb{R}^4; \mathbb{C}^4)$, the Plancherel formula can be rewritten
in the following form:
\begin{multline}\label{PlancherelForSingleHforpsi'} 
\int\limits_{\mathscr{O}_{{}_{m,0,0,0}}}
\Big| \, \widetilde{\phi}\big|_{{}_{\mathscr{O}_{{}_{m,0,0,0}}}}(\boldsymbol{\p}) \, \Big|_{{}_{\mathbb{C}^4}}^{2} \, 
{\textstyle\frac{\ud^3 \boldsymbol{\p}}{(2p_0(\boldsymbol{\p}))^2}}
+
\int\limits_{\mathscr{O}_{{}_{-m,0,0,0}}}
\Big| \, \widetilde{\phi}\big|_{{}_{\mathscr{O}_{{}_{-m,0,0,0}}}}(\boldsymbol{\p}) \, \Big|_{{}_{\mathbb{C}^4}}^{2} \, 
{\textstyle\frac{\ud^3 \boldsymbol{\p}}{(2p_0(\boldsymbol{\p}))^2}}
\\
= \sum \limits_{l,m} \,\, \int
 \Big| F^{{}^{\oplus}}_{{}_{\chi,lm}}\Big(\widetilde{\phi}\big|_{{}_{\mathscr{O}_{{}_{\boldsymbol{m},0,0,0}}}}\Big) \Big|^2 \, d \chi  
+
\sum \limits_{l,m} \,\, \int
\Big| F^{{}^{\ominus}}_{{}_{\chi,lm}}\Big(\widetilde{\phi}\big|_{{}_{\mathscr{O}_{{}_{-\boldsymbol{m},0,0,0}}}}\Big) \Big|^2 \, d \chi  
\end{multline}

\vspace*{1cm}

\begin{center}
{\small DECOMPOSITION OF $\boldsymbol{\psi} = \Xi(\kappa_{0,1}) + \Xi(\kappa_{1,0})$ IN CASE 1)}
\end{center}

For each fixed $\chi$ we introduce the annihilation-creation operators ${a_{\oplus}}_{{}_{\chi}}(\cdot), {a_{\oplus}}_{{}_{\chi}}(\cdot)^+$
over the decomposition Hilbert space $\big(\mathcal{H}^{\oplus}_{m,0}\big)_{{}_{\chi}}$
and annihilation-creation operators ${a_{\ominus}}_{{}_{\chi}}(\cdot), {a_{\ominus}}_{{}_{\chi}}(\cdot)^+$
over $\big(\mathcal{H}^{\ominus}_{-m,0}\big)_{{}_{\chi}}^{\flat}$, exactly as we did for the Hilbert spaces
$\mathcal{H}^{\oplus}_{m,0}$ and $\mathcal{H}^{\ominus \, \flat}_{-m,0}$ in Subsections \ref{electron} and \ref{positron}.
By definition, they respect the following canonical anticommutation relations
\[
\begin{split}
\Big[
{a_{\oplus}}_{{}_{\chi}}\big(F^{{}^{\oplus}}_{{}_{\chi,lm}}\big)
{a_{\oplus}}_{{}_{\chi}}\big(F^{{}^{\oplus}}_{{}_{\chi,l'm'}}\big)^+
\Big]_+
=
\Big(F^{{}^{\oplus}}_{{}_{\chi,lm}}, F^{{}^{\oplus}}_{{}_{\chi,l'm'}} \Big)_{{}_{\chi}}
 =  \delta_{{}_{l \, l'}} \delta_{{}_{m \, m'}}
\\
\Big[
{a_{\oplus}}_{{}_{\chi}}\big(F^{{}^{\oplus}}_{{}_{\chi,lm}}\big)
{a_{\oplus}}_{{}_{\chi}}\big(F^{{}^{\oplus}}_{{}_{\chi,l'm'}}\big)
\Big]_+=
\Big[
{a_{\oplus}}_{{}_{\chi}}\big(F^{{}^{\oplus}}_{{}_{\chi,lm}}\big)^+
{a_{\oplus}}_{{}_{\chi}}\big(F^{{}^{\oplus}}_{{}_{\chi,l'm'}}\big)^+
\Big]_+=0
\\
\Big[
{a_{\ominus}}_{{}_{\chi}}\Big( F^{{}^{\ominus \, \flat}}_{{}_{\chi,lm}} \Big),
{a_{\ominus}}_{{}_{\chi}}\Big( F^{{}^{\ominus \, \flat }}_{{}_{\chi,l'm'}}\Big)^+
\Big]_+
= 
\Big(F^{{}^{\ominus \, \flat}}_{{}_{\chi,lm}}, F^{{}^{\ominus \, \flat}}_{{}_{\chi,l'm'}} \Big)_{{}_{\chi}}^{{}^{\flat}}
 =  \delta_{{}_{l \, l'}} \delta_{{}_{m \, m'}} 
\\
\Big[
{a_{\ominus}}_{{}_{\chi}}\Big(\big(F^{{}^{\ominus}}_{{}_{\chi,lm}}\big)^{\flat}\Big),
{a_{\ominus}}_{{}_{\chi}}\Big(\big(F^{{}^{\ominus}}_{{}_{\chi,l'm'}}\big)^{\flat}\Big)
\Big]_+ = 0
\\
\Big[
{a_{\ominus}}_{{}_{\chi}}\Big(\big(F^{{}^{\ominus}}_{{}_{\chi,lm}}\big)^{\flat}\Big)^+,
{a_{\ominus}}_{{}_{\chi}}\Big(\big(F^{{}^{\ominus}}_{{}_{\chi,l'm'}}\big)^{\flat}\Big)^+
\Big]_+ =0
\end{split}
\]
For each fixed $\chi$ we introduce the annihilation-creation operators $a'_{{}_{\chi}}\big(\xi\oplus \eta^\flat \big), a'_{{}_{\chi}}\big(\xi \oplus \eta^\flat \big)^+$
over the decomposition Hilbert space
$\big(\mathcal{H}^{\oplus}_{m,0}\big)_{{}_{\chi}} \oplus \big(\mathcal{H}^{\ominus}_{-m,0}\big)_{{}_{\chi}}^{\flat}$, exactly
as we did in Subsection \ref{electron+positron} for $\mathcal{H}^{\oplus}_{m,0} \oplus \mathcal{H}^{\ominus \, \flat}_{-m,0}$:
\[
a'_{{}_{\chi}}(\xi \oplus 0) \coloneqq {a_{\oplus}}_{{}_{\chi}}(\xi) \otimes  \textrm{In}_{\ominus}, 
\,\,\, a'_{{}_{\chi}}(0 \oplus \eta^\flat) \coloneqq \boldsymbol{1}  \otimes {a_{\oplus}}_{{}_{\chi}}(\eta^\flat),
\]
\[
\xi \oplus \eta \in \big(\mathcal{H}^{\oplus}_{m,0}\big)_{{}_{\chi}} \oplus \big(\mathcal{H}^{\ominus}_{-m,0}\big)_{{}_{\chi}}.
\]

By construction, if
\[
\begin{split}
\xi = \int \xi_{{}_{\chi}} \, \ud \chi \,\,\, \xi \in \mathcal{H}^{\oplus}_{m,0},
\\
\eta = \int \eta_{{}_{\chi}} \, \ud \chi \,\,\, \eta \in \mathcal{H}^{\ominus}_{-m,0},
\end{split}
\]
are the respective direct integral decompositions of $\xi, \eta$, then 
\[
a'\big(\xi\oplus \eta^\flat \big) = 
\int a'_{{}_{\chi}}\big(\xi_{{}_{\chi}} \oplus \eta^{\flat}_{{}_{\chi}}\big) \,  \ud \chi. 
\]
In particular, for each test function $\phi \in \mathscr{E} = \mathcal{S}(\mathbb{R}^4; \mathbb{C}^4)$, the Dirac field operator
\begin{multline*}
\boldsymbol{\psi}(\overline{\phi}) = a_{\oplus}\big(P^\oplus\widetilde{\phi}|_{{}_{\mathscr{O}_{m,0,0,0}}}\big)\otimes \textrm{In}_{\ominus} +
\boldsymbol{1} \otimes a_{\ominus}\Big(\big(P^\ominus\widetilde{\phi}|_{{}_{\mathscr{O}_{-m,0,0,0}}}\big)^\flat\Big)^+
\\
=
a'\big(P^\oplus\widetilde{\phi}|_{{}_{\mathscr{O}_{m,0,0,0}}} \oplus 0\big) +
a'\Big(0 \oplus \big(P^\ominus\widetilde{\phi}|_{{}_{\mathscr{O}_{-m,0,0,0}}}\big)^\flat\Big)^+
\end{multline*}
undergoes the following direct integral decomposition
\begin{multline*}
\boldsymbol{\psi}(\overline{\phi})
= \int \boldsymbol{\psi}(\overline{\phi})_{{}_{\chi}} \, \ud\chi
\\
= 
\int \Bigg[{a_{\oplus}}_{{}_{\chi}}\big(\big(P^\oplus\widetilde{\phi}|_{{}_{\mathscr{O}_{m,0,0,0}}}\big)_{{}_{\chi}}\big)\otimes \textrm{In}_{\ominus}  +
\boldsymbol{1} \otimes {a_{\ominus}}_{{}_{\chi}}\Big(\big(P^\ominus\widetilde{\phi}|_{{}_{\mathscr{O}_{-m,0,0,0}}}\big)^{\flat}_{{}_{\chi}}\Big)^{+} \Bigg] \, \ud\chi
\end{multline*}
\[
= 
\int \Bigg[a'_{{}_{\chi}}\big(\big(P^\oplus\widetilde{\phi}|_{{}_{\mathscr{O}_{m,0,0,0}}}\big)_{{}_{\chi}} \oplus 0\big)  +
a'_{{}_{\chi}}\Big(0 \oplus \big(P^\ominus\widetilde{\phi}|_{{}_{\mathscr{O}_{-m,0,0,0}}}\big)^{\flat}_{{}_{\chi}}\Big)^{+} \Bigg] \, \ud\chi.
\]
This decomposition can be rewritten as a direct integral 
\begin{multline*}
\boldsymbol{\psi}(\overline{\phi}) 
=
\int \boldsymbol{\psi}_{{}_{\chi}}(\overline{\phi}) \, \ud\chi
\\
= 
\int \Bigg[{a_{\oplus}}_{{}_{\chi}}\big(\big(P^\oplus\widetilde{\phi}|_{{}_{\mathscr{O}_{m,0,0,0}}}\big)_{{}_{\chi}}\big)  \otimes \textrm{In}_\ominus +
\boldsymbol{1} \otimes {a_{\ominus}}_{{}_{\chi}}\Big(\big(P^\ominus\widetilde{\phi}|_{{}_{\mathscr{O}_{-m,0,0,0}}}\big)^{\flat}_{{}_{\chi}}\Big)^{+} \Bigg] \, \ud\chi
\end{multline*}
of actual fields $\overline{\phi} \mapsto \boldsymbol{\psi}_{{}_{\chi}}(\overline{\phi})$, over the same space-time test space
$\mathscr{E} = \mathcal{S}(\mathbb{R}^4; \mathbb{C}^4)$ as the original Dirac field $\overline{\phi} \mapsto \boldsymbol{\psi}(\overline{\phi})$.
Let us compute its kernel $\boldsymbol{\psi}_{{}_{\chi}}(x)$.

In order to simplify notation, let us introduce the following abbreviated notation for the above stated operators:
\[
\begin{split}
b'_{{}_{\chi \, lm}} \coloneqq {a_{\oplus}}_{{}_{\chi}}\big(F^{{}^{\oplus}}_{{}_{\chi,lms}}\big) \otimes \textrm{In}_{\ominus}
= a'_{{}_{\chi}}\big(F^{{}^{\oplus}}_{{}_{\chi,lm}} \oplus 0 \big),
\\
d'_{{}_{\chi \, lm}} \coloneqq \boldsymbol{1} \otimes {a_{\ominus}}_{{}_{\chi}}\Big(\big(F^{{}^{\ominus \, \flat}}_{{}_{\chi,lm}}\big)_{{}_{\chi}}\Big)
=  a'_{{}_{\chi}}\big(0 \oplus F^{{}^{\ominus \flat}}_{{}_{\chi,lm}} \big),
\end{split}
\]
They have the canonical anticommutation relations stated above:
\[
\begin{split}
\big[b'_{{}_{\chi \, lm}}, {b'}_{{}_{\chi \, l'm'}}^+ \big]_+ =    \delta_{{}_{l \, l'}} \delta_{{}_{m \, m'}},
\\
\big[b'_{{}_{\chi \, lms}}, b'_{{}_{\chi \, l'm'}} \big]_+ = \big[{b'}_{{}_{\chi \, lms}}^+, {b'}_{{}_{\chi \, l'm'}}^+ \big]_+ = 0
\\
\big[d'_{{}_{\chi \, lm}}, {d'}_{{}_{\chi \, l'm'}}^+ \big]_+ =   \delta_{{}_{l \, l'}} \delta_{{}_{m \, m'}},  
\\
\big[d'_{{}_{\chi \, lm}}, d'_{{}_{\chi \, l'm'}} \big]_+ = \big[{d'}_{{}_{\chi \, lm}}^+, {d'}_{{}_{\chi \, l'm'}}^+ \big]_+ = 0
\end{split}
\]
with all the other pairs of these operators anticommuting.

Let $\phi \in \mathscr{E} = \mathcal{S}(\mathbb{R}^4; \mathbb{C}^4)$. Using (\ref{Riesz-GelfandComponents})
we construct the following  developments
\begin{equation}\label{DefinitionE+chi,E-chiInH+chi,H-chi}
\begin{split}
\big(P^\oplus\widetilde{\phi}|_{{}_{\mathscr{O}_{\boldsymbol{m},0,0,0}}}\big)_{{}_{\chi}}
= \sum\limits_{lm} c^{\oplus}_{{}_{lm}} F^{{}^{\oplus}}_{{}_{\chi,lm}}, 
\\
\big(P^\ominus\widetilde{\phi}|_{{}_{\mathscr{O}_{-\boldsymbol{m},0,0,0}}}\big)_{{}_{\chi}}
= 
\sum\limits_{lm} c^{\ominus}_{{}_{lm}} F^{{}^{\ominus}}_{{}_{\chi,lm}}
\\
\big(P^\ominus\widetilde{\phi}|_{{}_{\mathscr{O}_{-\boldsymbol{m},0,0,0}}}\big)_{{}_{\chi}}^{\flat}
= 
\sum\limits_{lm} \overline{c^{\ominus}_{{}_{lm}}} \, F^{{}^{\ominus \,\, \flat}}_{{}_{\chi,lm}}
\end{split}
\end{equation}
in the orthonormal basis $\{F^{{}^{\oplus}}_{{}_{\chi,lm}} \}$ or respectively $\{F^{{}^{\ominus}}_{{}_{\chi,lm}}\}$
with
\[
\begin{split}
c^{\oplus}_{{}_{lm}} = \Big(
F^{{}^{\oplus}}_{{}_{\chi, \, lm}},
\big(P^\oplus\widetilde{\phi}|_{{}_{\mathscr{O}_{\boldsymbol{m},0,0,0}}}\big)_{{}_{\chi}}
\Big)_{{}_{\chi}}
= F^{{}^{\oplus}}_{{}_{\chi, \, lm}}(P^\oplus\widetilde{\phi}|_{{}_{\mathscr{O}_{\boldsymbol{m},0,0,0}}})
= F^{{}^{\oplus}}_{{}_{\chi, \, lm}}(\widetilde{\phi}|_{{}_{\mathscr{O}_{\boldsymbol{m},0,0,0}}})
\\
c^{\ominus}_{{}_{lm}}
= \Big(
F^{{}^{\ominus}}_{{}_{\chi, \, lm}},
\big(P^\ominus\widetilde{\phi}|_{{}_{\mathscr{O}_{-\boldsymbol{m},0,0,0}}}\big)_{{}_{\chi}} 
\Big)_{{}_{\chi}}
= F^{{}^{\ominus}}_{{}_{\chi, \, lm}}(P^\ominus\widetilde{\phi}|_{{}_{\mathscr{O}_{-\boldsymbol{m},0,0,0}}})
= F^{{}^{\ominus}}_{{}_{\chi, \, lm}}(\widetilde{\phi}|_{{}_{\mathscr{O}_{-\boldsymbol{m},0,0,0}}}) 
\end{split}
\]
which are not only convergent in the respective Hilbert space component  $\big(\mathcal{H}^{\oplus}_{\boldsymbol{m},0}\big)_{{}_{\chi}}$ or
$\big(\mathcal{H}^{\ominus}_{-\boldsymbol{m},0}\big)_{{}_{\chi}}$, with respect to the Hilbert space norm
$\|  \cdot \|^{2}_{{}_{\chi}} = ( \cdot, \cdot)_{{}_{\chi}}$:
\[
\begin{split}
\sum\limits_{lm} \big|c^{\oplus}_{{}_{lm}} \big|^2 
= \Big\| \big(P^\oplus\widetilde{\phi}|_{{}_{\mathscr{O}_{\boldsymbol{m},0,0,0}}}\big)_{{}_{\chi}} \Big\|_{{}_{\chi}}^{2}
< + \infty,
\\
\sum\limits_{lm} \big|c^{\ominus}_{{}_{lm}} \big|^2 
= \Big\| \big(P^\ominus\widetilde{\phi}|_{{}_{\mathscr{O}_{-\boldsymbol{m},0,0,0}}}\big)_{{}_{\chi}} \Big\|_{{}_{\chi}}^{2}
< + \infty,
\end{split}
\]
but moreover the sequences $\{ c^{\oplus}_{{}_{lm}} \}, \{ c^{\oplus}_{{}_{lm}} \}$ in (\ref{DefinitionE+chi,E-chiInH+chi,H-chi})
are rapidly decreasing. This is the case for $\phi \in \mathscr{E} = \mathcal{S}(\mathbb{R}^4; \mathbb{C}^4)$ ranging
over the nuclear space $\mathscr{E} = \mathcal{S}(\mathbb{R}^4; \mathbb{C}^4)$, with $\phi \in \cap_{n\in \mathbb{N}} \textrm{Dom} \, A^n$
for the standard operator $A$ defining the standard nuclear space $\mathscr{E}$. In this case 
$P^\oplus\widetilde{\phi}|_{{}_{\mathscr{O}_{\boldsymbol{m},0,0,0}}} \in E^\oplus$, and 
$P^\ominus\widetilde{\phi}|_{{}_{\mathscr{O}_{-\boldsymbol{m},0,0,0}}} \in E^\ominus$ and lie respectively in the domain
of each positive power of the operators $A_{{}_{\oplus}} = U^{-1} \oplus_{1}^{2} H_{(3)}U$, 
$A_{{}_{\ominus}} = U^{-1} \oplus_{3}^{4} H_{(3)}U$ defining the nuclear spaces $E^\oplus, E^\ominus$. For $\phi$ ranging over 
the nuclear space $\mathscr{E}$ the first two decompositions in (\ref{DefinitionE+chi,E-chiInH+chi,H-chi}), range, respectively,
over the nuclear spaces $E^{\oplus}_{{}_{\chi}}, E^{\oplus}_{{}_{\chi}}$, which together with 
$\big(\mathcal{H}^{\oplus}_{\boldsymbol{m},0}\big)_{{}_{\chi}}$, $\big(\mathcal{H}^{\ominus}_{-\boldsymbol{m},0}\big)_{{}_{\chi}}$
compose Gelfand triples
\[
E^{\oplus}_{{}_{\chi}} \oplus E^{\ominus}_{{}_{\chi}} \,\,\, \subset \,\,\,
\big(\mathcal{H}^{\oplus}_{\boldsymbol{m},0}\big)_{{}_{\chi}} \oplus \big(\mathcal{H}^{\ominus}_{-\boldsymbol{m},0}\big)_{{}_{\chi}}
\,\,\, \subset \,\,\,
E^{\oplus \, *}_{{}_{\chi}} \oplus E^{\ominus \, *}_{{}_{\chi}}.
\]
Because the Hida creation, annihilation operators, ${a_{\oplus}}_{{}_{\chi}}^+,{a_{\oplus}}_{{}_{\chi}}, {a_{\ominus}}_{{}_{\chi}}^+,
{a_{\ominus}}_{{}_{\chi}}$, are continuous, regarded as the maps
\[
\begin{split}
E^{\oplus}_{{}_{\chi}} \ni \xi \longmapsto 
{a_{\oplus}}_{{}_{\chi}}(\xi) \in \mathscr{L}\big( \, \big(E^{\oplus}_{{}_{\chi}}\big) \, , \,  \big(E^{\oplus}_{{}_{\chi}}\big) \,  \big),
\\
E^{\oplus}_{{}_{\chi}} \ni \xi \longmapsto 
{a_{\oplus}}_{{}_{\chi}}(\xi)^+ \in \mathscr{L}\big( \, \big(E^{\oplus}_{{}_{\chi}}\big) \, , \,  \big(E^{\oplus}_{{}_{\chi}}\big)^* \,  \big),
\end{split}
\]
\[
\begin{split}
E^{\ominus \flat}_{{}_{\chi}} \ni \xi \longmapsto 
{a_{\ominus}}_{{}_{\chi}}(\xi) \in \mathscr{L}\big( \, \big(E^{\ominus \flat}_{{}_{\chi}}\big) \, ,
 \,  \big(E^{\ominus \, \flat }_{{}_{\chi}}\big) \,  \big),
\\
E^{\ominus \, \flat}_{{}_{\chi}} \ni \xi \longmapsto 
{a_{\ominus}}_{{}_{\chi}}(\xi)^+ \in \mathscr{L}\big( \, \big(E^{\ominus \flat}_{{}_{\chi}}\big) \, , \,  \big(E^{\ominus \flat}_{{}_{\chi}}\big)^* \,  \big),
\end{split}
\]
then we can pull the summation out of the argument of  
${a_{\oplus}}_{{}_{\chi}}(\xi )$, 
${a_{\ominus}}_{{}_{\chi}}(\eta)^+$, in the developments
\[
\xi = \sum\limits_{lm} c^{\oplus}_{{}_{lm}} F^{{}^{\oplus}}_{{}_{\chi,lm}},
\,\,\, \eta = \sum\limits_{lm} \overline{c^{\ominus}_{{}_{lm}}} \,\, F^{{}^{\ominus \, \flat}}_{{}_{\chi,lm}}
\]
representing elements $\xi = \big(P^\oplus\widetilde{\phi}|_{{}_{\mathscr{O}_{\boldsymbol{m},0,0,0}}}\big)_{{}_{\chi}} \in E^{\oplus}_{{}_{\chi}}$
and $\eta = \big(P^\ominus\widetilde{\phi}|_{{}_{\mathscr{O}_{-\boldsymbol{m},0,0,0}}}\big)_{{}_{\chi}}^{\flat} \in E^{\ominus \flat}_{{}_{\chi}}$ 
of the respective nuclear spaces, $E^{\oplus}_{{}_{\chi}}, E^{\ominus \flat}_{{}_{\chi}}$:
\[
\begin{split}
{a_{\oplus}}_{{}_{\chi}}(\xi) = \sum\limits_{lm} \overline{c^{\oplus}_{{}_{lm}}} \,\,
{a_{\oplus}}_{{}_{\chi}}\big( F^{{}^{\oplus}}_{{}_{\chi,lm}}\big), 
\\
{a_{\ominus}}_{{}_{\chi}}(\eta )^+ = \sum\limits_{lm} \overline{c^{\ominus}_{{}_{lm}}} \,\,
{a_{\ominus}}_{{}_{\chi}}\big( F^{{}^{\oplus \, \flat}}_{{}_{\chi,lm}}\big)^+. 
\end{split}
\]

Therefore, for $\phi \in \mathscr{E} = \mathcal{S}(\mathbb{R}^4; \mathbb{C}^4)$ we have
\begin{multline*}
\boldsymbol{\psi}_{{}_{\chi}}(\overline{\phi}) \coloneqq 
{a_{\oplus}}_{{}_{\chi}}\big(\big(P^\oplus\widetilde{\phi}|_{{}_{\mathscr{O}_{\boldsymbol{m},0,0,0}}}\big)_{{}_{\chi}}\big)\otimes \textrm{In}_\ominus +
\boldsymbol{1} \otimes {a_{\ominus}}_{{}_{\chi}}\Big(\big(P^\ominus\widetilde{\phi}|_{{}_{\mathscr{O}_{-\boldsymbol{m},0,0,0}}}\big)^{\flat}_{{}_{\chi}}\Big)^{+} 
\\
= 
\sum\limits_{lm} \overline{c^{\oplus}_{{}_{lm}}} \,\, b'_{{}_{\chi, \, lm}}
+
\sum\limits_{lm} \overline{c^{\ominus}_{{}_{lm}}} \,\, {d'}_{{}_{\chi, \, lm}}^+
\\
=
\sum\limits_{lm} \overline{F^{{}^{\oplus}}_{{}_{\chi, \, lm}}(\widetilde{\phi}|_{{}_{\mathscr{O}_{\boldsymbol{m},0,0,0}}})}  \,\,\, b'_{{}_{\chi, \, lm}}
+
\sum\limits_{lm} \overline{F^{{}^{\ominus}}_{{}_{\chi, \, lm}}(\widetilde{\phi}|_{{}_{\mathscr{O}_{-\boldsymbol{m},0,0,0}}})}  \,\,\, {d'}_{{}_{\chi, \, lm}}^+
\end{multline*}
\begin{equation}\label{psichi(barphi)}
\,\,\,\,\,\,\,\,\,\,\,\,\,\,\,\,\,\,\,\,\,\,\,\,\,\,\,\,\,\,\,\,\,\,\,\,\,\,\,\,\,\,\,\,\,\,\,\,\,\,\,\,\,\,\,\,\,\,\,\,
=
\sum\limits_{lm} \overline{f^{{}^{\oplus}}_{{}_{\chi, \, lm}}(\phi)}  \,\,\, b'_{{}_{\chi, \, lm}}
+
\sum\limits_{lm} \overline{f^{{}^{\ominus}}_{{}_{\chi, \, lm}}(\phi)}  \,\,\, {d'}_{{}_{\chi, \, lm}}^+.
\end{equation}
Here $f^{{}^{\oplus}}_{{}_{\chi, \, lm}}$ and $f^{{}^{\ominus}}_{{}_{\chi, \, lm}}$ are the distributions
in $\mathcal{S}(\mathbb{R}^4; \mathbb{C}^4)^*$, equal to the inverse Fourier transforms of the distributions
\[
\begin{split}
\mathcal{S}(\mathbb{R}^4; \mathbb{C}^4) \ni \widetilde{\phi} \longmapsto 
F^{{}^{\oplus}}_{{}_{\chi, \, lm}}(\widetilde{\phi}|_{{}_{\mathscr{O}_{\boldsymbol{m},0,0,0}}}) 
\\
\mathcal{S}(\mathbb{R}^4; \mathbb{C}^4) \ni  \widetilde{\phi} \longmapsto 
F^{{}^{\ominus}}_{{}_{\chi, \, lm}}(\widetilde{\phi}|_{{}_{\mathscr{O}_{-\boldsymbol{m},0,0,0}}}).
\end{split}
\]
$f^{{}^{\oplus}}_{{}_{\chi, \, lm}}$ and $f^{{}^{\ominus}}_{{}_{\chi, \, lm}}$ are regular for massive fields,
with the pairings, $f^{{}^{\oplus}}_{{}_{\chi, \, lm}}(\phi)$, $f^{{}^{\ominus}}_{{}_{\chi, \, lm}}(\phi)$,
$\phi \in \mathcal{S}(\mathbb{R}^4; \mathbb{C}^4)$, representable by integration of the test function $\phi$
multiplied by ordinary functions, $x \mapsto f^{{}^{\oplus}}_{{}_{\chi, \, lm}}(x)$
and $x \mapsto f^{{}^{\ominus}}_{{}_{\chi, \, lm}}(x)$,
However in general $f^{{}^{\oplus}}_{{}_{\chi, \, lm}}$ and $f^{{}^{\ominus}}_{{}_{\chi, \, lm}}$, e.g.
for mass less complex field, the spherically symmetric
$f^{{}^{\oplus}}_{{}_{\chi, \, 00}}$ and $f^{{}^{\ominus}}_{{}_{\chi, \, 00}}$ have strong singularity
at the light cone. In each case $f^{{}^{\oplus}}_{{}_{\chi, \, lm}}$ and $f^{{}^{\ominus}}_{{}_{\chi, \, lm}}$
cannot be multiplied, as in general the product leads to functions which are not locally integrable,
so in particular the Wick products of the decomposition component
fields $\boldsymbol{\psi}^{\sharp}_{{}_{\chi}}(x)$, $\boldsymbol{\psi}_{{}_{\chi}}(x)$
cannot be computed by the ordinary product of the corresponding
kernels. For the component fields of the space-time derivatives of the free fields situation is even worse,
and their Wick product cannot be computed through the multiplication of their kernels,
as in general the product of distributions is not well-defined.

Using the pairings
\[
f^{{}^{\oplus}}_{{}_{\chi, \, lm}}(\phi) = \int f^{{}^{\oplus}}_{{}_{\chi, \, lm}}(x)\phi(x) \, \ud^4x 
= \sum\limits_{a} \int f^{{}^{\oplus \, a}}_{{}_{\chi, \, lm}}(x)\phi^a(x) \, \ud^4x,
\]
\[
\boldsymbol{\psi}_{{}_{\chi}}(\phi) = \int \boldsymbol{\psi}_{{}_{\chi}}(x) \phi(x) \, \ud^4x 
= \sum\limits_{a} \int \boldsymbol{\psi}^{a}_{{}_{\chi}}(x) \phi^{a}(x) \, \ud^4x 
\]
and (\ref{psichi(barphi)}) we obtain the following formula 
\begin{equation}\label{psichi(x)}
\boldsymbol{\psi}_{{}_{\chi}}(x) = \sum\limits_{lm} \overline{ f^{{}^{\oplus}}_{{}_{\chi, \, lm}}(x) } \,\,\, b'_{{}_{\chi, \, lm}}
+
\sum\limits_{lm} \overline{f^{{}^{\ominus}}_{{}_{\chi, \, lm}}(x) } \,\,\, {d'}_{{}_{\chi, \, lm}}^+
\end{equation}
with distributional positive $f^{{}^{\oplus}}_{{}_{\chi, \, lm}}$ and negative energy solutions 
$f^{{}^{\ominus}}_{{}_{\chi, \, lm}}$ of the Dirac equation. Their Fourier transforms, regarded as distributions,
are equal
\[
\begin{split}
\mathcal{S}(\mathbb{R}^4; \mathbb{C}^4) \ni  \widetilde{\phi} \longmapsto 
F^{{}^{\oplus}}_{{}_{\chi, \, lm}}(\widetilde{\phi}|_{{}_{\mathscr{O}_{\boldsymbol{m},0,0,0}}}) 
\\
\mathcal{S}(\mathbb{R}^4; \mathbb{C}^4) \ni  \widetilde{\phi} \longmapsto 
F^{{}^{\ominus}}_{{}_{\chi, \, lm}}(\widetilde{\phi}|_{{}_{\mathscr{O}_{-\boldsymbol{m},0,0,0}}}).
\end{split}
\]
Note that the bispinor functions $\overline{f^{{}^{\oplus}}_{{}_{\chi, \, lm}}(x)}$ and
$\overline{f^{{}^{\ominus}}_{{}_{\chi, \, lm}}(x)}$ are smooth everywhere outside the light cone,
and respect there the same Dirac equation, as the kernel $\boldsymbol{\psi}(x)$ of the original
Dirac field $\boldsymbol{\psi}$. $\overline{f^{{}^{\oplus}}_{{}_{\chi, \, lm}}}$ and
$\overline{f^{{}^{\ominus}}_{{}_{\chi, \, lm}}}$, regarded as distributions, fulfill this Dirac equation everywhere.

Therefore, $\boldsymbol{\psi}_{{}_{\chi}}(x) $, which respects the same Dirac equation
as $\overline{f^{{}^{\oplus}}_{{}_{\chi, \, lm}}(x)}$ and
$\overline{f^{{}^{\ominus}}_{{}_{\chi, \, lm}}(x)}$, respects the same Dirac equation as 
the kernel $\boldsymbol{\psi}(x)$. By the invariance of the pairings, the representation components
$U_{{}_{\chi}}(\alpha)$ act on the bispinor functions $f^{{}^{\oplus}}_{{}_{\chi, \, lm}}(x)$ and
$f^{{}^{\ominus}}_{{}_{\chi, \, lm}}(x)$ as the ordinary local bispinor transformation.

Moreover, the functions $x \mapsto f^{{}^{\oplus}}_{{}_{\chi, \, lm}}(x)$, $x \mapsto f^{{}^{\ominus}}_{{}_{\chi, \, lm}}(x)$ are asymptotically homogeneous
of asymptotic homogeneity degree $\overline{\chi}$, where $\chi$ is the decomposition parameter which ranges over the set 
\[
\{\chi = -3/2-i\nu, \nu \in \mathbb{R}\}. 
\]
This means that the limits
\[
\underset{\lambda\rightarrow +\infty}{\textrm{lim}} \,
{\textstyle\frac{1}{\lambda^{3/2+i\nu}}}
f^{{}^{\oplus}}_{{}_{\chi=-3/2+i\nu, \, lm}}\big({\textstyle\frac{1}{\lambda}}x\big),
\,\,\,\,\,
\underset{\lambda\rightarrow +\infty}{\textrm{lim}} \,
{\textstyle\frac{1}{\lambda^{3/2+i\nu}}}
f^{{}^{\ominus}}_{{}_{\chi=-3/2+i\nu, \, lm}}\big({\textstyle\frac{1}{\lambda}}x\big)
\]
exist are nonzero distributions in the $x$-variable and represent homogeneous of degree $-3/2-i\nu$ functions of $x$, and each of the limits
(in the distribution sense)
\[
\underset{\lambda\rightarrow +\infty}{\textrm{lim}} \,
{\textstyle\frac{1}{\lambda^{s}}}
f^{{}^{\oplus}}_{{}_{\chi=-3/2+i\nu, \, lm}}\big({\textstyle\frac{1}{\lambda}}x\big),
\,\,\,\,
\underset{\lambda\rightarrow +\infty}{\textrm{lim}} \,
{\textstyle\frac{1}{\lambda^{s}}}
f^{{}^{\ominus}}_{{}_{\chi=-3/2+i\nu, \, lm}}\big({\textstyle\frac{1}{\lambda}}x\big)
\]
is zero, $\infty$ or does not exist for each value of $s \neq 3/2+i\nu$, $\nu \in \mathbb{R}$.

These asymptotic properties of  $f^{{}^{\oplus}}_{{}_{\chi, \, lm}}$,
$f^{{}^{\ominus}}_{{}_{\chi, \, lm}}$, understood as distributions
on the space-time test functions 
\[
\phi \in \mathscr{E} = \mathcal{S}(\mathbb{R}^4; \mathbb{C}) = \mathcal{S}_{{}_{\oplus H_{(4)}}}(\mathbb{R}^4; \mathbb{C}^4)
\]
can be expressed in the following form.
The limits
\[
\underset{\lambda\rightarrow +\infty}{\textrm{lim}} \,
\lambda^{5/2-i\nu} f^{{}^{\oplus}}_{{}_{\chi=-3/2+i\nu, \, lm}}(S_\lambda \phi),
\,\,\,\,
\underset{\lambda\rightarrow +\infty}{\textrm{lim}} \,
\lambda^{5/2-i\nu} f^{{}^{\ominus}}_{{}_{\chi=-3/2+i\nu, \, lm}}(S_\lambda \phi)
\]
do exist for each $\phi \in \mathscr{E}$, 
and are nonzero homogeneous distribution of homogeneity degree $\chi = -3/2 -i\nu$, and 
each of the limits
\[
\underset{\lambda\rightarrow +\infty}{\textrm{lim}} \,
\lambda^{s} f^{{}^{\oplus}}_{{}_{\chi=-3/2+i\nu, \, lm}}(S_\lambda \phi),
\,\,\,\,
\underset{\lambda\rightarrow +\infty}{\textrm{lim}} \,
\lambda^{s} f^{{}^{\ominus}}_{{}_{\chi=-3/2+i\nu, \, lm}}(S_\lambda \phi)
\]
is zero, $\infty$ or does not exist for each $\phi \in \mathscr{E}$ and each $s \neq 5/2-i\nu$.
Here $S_\lambda$ is the scaling operator which acts on the test function 
$\phi$ by the following rule $S_\lambda \phi(x) \coloneqq \phi(\lambda x)$.

These asymptotic properties of the fundamental solutions $f^{{}^{\oplus}}_{{}_{\chi, \, lm}}$,
$f^{{}^{\ominus}}_{{}_{\chi, \, lm}}$
follow from the asymptotic homogeneity at infinity of homogeneity degree $-1/2 +i\nu$,
of the corresponding functions $F^{{}^{\oplus}}_{{}_{\chi, \, lm}}$,
$F^{{}^{\ominus}}_{{}_{\chi, \, lm}}$ on $\mathscr{O}_{{}_{\boldsymbol{m},0,0,0}}$
and, respectively, on $\mathscr{O}_{{}_{-\boldsymbol{m},0,0,0}}$. Namely,
the limits
\[
\underset{\lambda\rightarrow +\infty}{\textrm{lim}} \,
\lambda^{1/2-i\nu}
F^{{}^{\oplus}}_{{}_{\chi=-3/2+i\nu, \, lm}}\big(\lambda \boldsymbol{\p}\big),
\,\,\,\,\,
\underset{\lambda\rightarrow +\infty}{\textrm{lim}} \,
\lambda^{1/2-i\nu}
F^{{}^{\ominus}}_{{}_{\chi=-3/2+i\nu, \, lm}}\big(\lambda \boldsymbol{\p}\big)
\]
exist in the sense of distributions and represent nonzero homogeneous of degree $-1/2+i\nu$ functions of $\boldsymbol{\p}$, and each of the limits
\[
\underset{\lambda\rightarrow +\infty}{\textrm{lim}} \,
{\textstyle\frac{1}{\lambda^{s}}}
F^{{}^{\oplus}}_{{}_{\chi=-3/2-i\nu, \, lm}}\big({\textstyle\frac{1}{\lambda}}\boldsymbol{\p}\big),
\,\,\,\,
\underset{\lambda\rightarrow +\infty}{\textrm{lim}} \,
{\textstyle\frac{1}{\lambda^{s}}}
f^{{}^{\ominus}}_{{}_{\chi=-3/2-i\nu, \, lm}}\big({\textstyle\frac{1}{\lambda}}\boldsymbol{\p}\big)
\]
regarded as a distribution, is zero, $\infty$ or does not exist for each value of $s \neq 3/2+i\nu$, $\nu \in \mathbb{R}$.

These asymptotic properties of the functions $\boldsymbol{\p} \mapsto F^{{}^{\oplus}}_{{}_{\chi, \, lm}}(\boldsymbol{\p})$,
$\boldsymbol{\p} \mapsto F^{{}^{\ominus}}_{{}_{\chi, \, lm}}(\boldsymbol{\p})$, understood as distributions
on the space-time test functions 
\[
\widetilde{\phi} \in \widetilde{\mathscr{E}} = \mathcal{S}(\mathbb{R}^4; \mathbb{C}) = \mathcal{S}_{{}_{\oplus H_{(4)}}}(\mathbb{R}^4; \mathbb{C}^4)
\]
can be expressed in the following form.
The limits
\[
\underset{\lambda\rightarrow +\infty}{\textrm{lim}} \,
\lambda^{-3/2-i\nu} F^{{}^{\oplus}}_{{}_{\chi=-3/2+i\nu, \, lm}}(S_{1/\lambda}\widetilde{\phi}),
\,\,\,\,
\underset{\lambda\rightarrow +\infty}{\textrm{lim}} \,
\lambda^{-3/2-i\nu} F^{{}^{\ominus}}_{{}_{\chi=-3/2+i\nu, \, lm}}(\phi)
\]
do exist for each $\widetilde{\phi} \in \widetilde{\mathscr{E}}=\mathscr{E}$, 
and are nonzero homogeneous distributions of homogeneity degree $\chi = -3/2 -i\nu-4$, and 
each of the limits
\[
\underset{\lambda\rightarrow +\infty}{\textrm{lim}} \,
\lambda^{s} F^{{}^{\oplus}}_{{}_{\chi=-3/2-i\nu, \, lm}}(\phi),
\,\,\,\,
\underset{\lambda\rightarrow +\infty}{\textrm{lim}} \,
\lambda^{s} F^{{}^{\ominus}}_{{}_{\chi=-3/2-i\nu, \, lm}}(\phi)
\]
is zero, $\infty$ or does not exist for each $\widetilde{\phi} \in \mathscr{E}$ and each $s \neq -3/2-i\nu$.
Indeed, using the fact that the Fourier transform $\widetilde{\lambda^4 S_\lambda}$ of the operator
$\lambda^4 S_\lambda$ is equal $S_{1/\lambda}$, regarded as an operator on the test space $\mathscr{E} = \mathcal{S}(\mathbb{R}^4; \mathbb{C}^4)$,
\emph{i.e.}:
\[
S_{1/\lambda} = \widetilde{\lambda^4 S_\lambda},
\] 
we obtain from the asymptotic homogeneity properties of $F^{{}^{\oplus}}_{{}_{\chi=-3/2-i\nu, \, lm}}$,
$F^{{}^{\ominus}}_{{}_{\chi=-3/2-i\nu, \, lm}}$, the asymptotic homogeneity properties of $f^{{}^{\oplus}}_{{}_{\chi=-3/2-i\nu, \, lm}}$, 
$f^{{}^{\ominus}}_{{}_{\chi=-3/2-i\nu, \, lm}}$.

The Dirac field $\boldsymbol{\psi}$, which we have decomposed above, is also given by the formula (\ref{*psi(x)}), Subsection \ref{psiBerezin-Hida}, 
and is a well-defined integral kernel generalized operator
\[
\begin{split}
\boldsymbol{\psi}^a(x) = \sum_{s=1}^{2} \, \int \limits_{\mathbb{R}^3} 
\kappa_{0,1}(s, \boldsymbol{p}; a, x) \,\, b_{s}(\boldsymbol{\p})\, \ud^3 \boldsymbol{\p}
+
\sum_{s=1}^{2} \, \int \limits_{\mathbb{R}^3} 
\kappa_{1,0}(s, \boldsymbol{p}; a, x) \,\, d_{s}(\boldsymbol{\p})^+ \, \ud^3 \boldsymbol{\p} \\
= \boldsymbol{\psi}^{(-) \, a}(x) + \boldsymbol{\psi}^{(+) \, a}(x)
= \Xi_{0,1}\big(\kappa_{0,1}(a,x)\big) + \Xi_{1,0}\big(\kappa_{1,0}(a,x)\big)  \\ 
=
\sum_{s=1}^{2} \, \int \limits_{\mathbb{R}^3} 
\underset{*}{u}_{s}^{a}(\boldsymbol{\p})e^{-ip\cdot x} \,\, b_{s}(\boldsymbol{\p}) \, \ud^3 \boldsymbol{\p}
+
\sum_{s=1}^{2} \, \int \limits_{\mathbb{R}^3} 
\underset{*}{v}_{s}^{a}(\boldsymbol{\p})e^{ip\cdot x} \,\, d_{s}(\boldsymbol{\p})^+ \, \ud^3 \boldsymbol{\p},
\end{split}
\]
with vector valued distributional kernels 
\[
\kappa_{0,1}(s, \boldsymbol{\p}; a, x) = \underset{*}{u}_{s}^{a}(\boldsymbol{\p})e^{-ip\cdot x},
\,\,\,
\kappa_{1,0}(s, \boldsymbol{\p}; a, x)  = \underset{*}{v}_{s}^{a}(\boldsymbol{\p})e^{ip\cdot x},
\]
\[
p = (p_0(\boldsymbol{\p}), \boldsymbol{\p}) \in \mathscr{O}_{{}_{\boldsymbol{m},0,0,0}},
\]
in the sense of \cite{obataJFA}, compare Subsection \ref{psiBerezin-Hida}. 
Recall, please, that $\underset{*}{u}= \gamma^0 u$ and $\underset{*}{v}= \gamma^0 v$. 
Here we are using the direct sum decomposition
\[
E^\oplus \oplus E^{\ominus \, \flat} = E, \,\,\, E^{+} \oplus E^{- \, \flat} = \mathcal{S}_{H_{(3)}}(\mathbb{R}^3; \mathbb{C}^4)
\]
and regard the kernels $\kappa_{0,1}$ and $\kappa_{1,0}$
as the continuous maps
\[
\kappa_{0,1}: \mathscr{E} \ni \phi \longmapsto \kappa_{0,1}(\phi) \in E^{+},
\,\,\,\,\,\,
\kappa_{0,1}: \mathscr{E} \ni \phi \longmapsto \kappa_{0,1}(\phi) \in E^{- \, \flat}
\]
with the index $s$ in the argument of the functions
\[
\begin{split}
(s,\boldsymbol{\p}) \longmapsto \kappa_{0,1}(\phi)(s,\boldsymbol{\p}) = \sum\limits_{a}\underset{*}{u}_{s}^{a}\widetilde{\phi^a}(-p_0(\boldsymbol{\p}), -\boldsymbol{\p}),
\\
\,\,\,\,
(s,\boldsymbol{\p}) \longmapsto \kappa_{0,1}(\phi)(s,\boldsymbol{\p}) = 
\sum\limits_{a}\underset{*}{v}_{s}^{a}\widetilde{\phi^a}(p_0(\boldsymbol{\p}), \boldsymbol{\p}),
\end{split}
\]
assuming only two values $s=1$ or $2$. Using the isomorphism\footnote{Note that with the normalization of the kernels
changed, with the factor $1/2$ omitted, the factor $1/2$ and $2$ will have to be omitted also in the formula (\ref{isomorphismU}) for the isomorphism 
$U$ and its inverse. In fact the mass factor $\boldsymbol{m}$ should be restored in it together with the mass term in the single particle inner product, which cancels out in the kernels, and a factor 
$(2\pi)^{-3/2}$ which should be present in the kernels of the Dirac field, and accordingly, in the formula for the isomorphism $U$.} (\ref{isomorphismU}) of Subsection \ref{psiBerezin-Hida}
(denoting its restriction to each summand, separately, $E^\oplus$, $E^{\ominus \, \flat}$, or $E^{\oplus \, *}$, $E^{\ominus \, \flat *}$, 
by the same symbol $U$,
and the restriction of its inverse  to each summand, separately, $E^+$, $E^{- \, \flat}$, $E^{+ \, *}$, $E^{- \, \flat \, *}$ 
by the same symbol $U^{-1}$) we see that (by the analogue of Lemma \ref{kappa_0,1(barphi),kappa_1,0(barphi)}, Subsection \ref{psiBerezin-Hida},
computed for the kernels with $\underset{*}{u} = \gamma^0 u$, $\underset{*}{v}= \gamma^0 v$ replaced with $u,v$)
\[
P^\oplus\widetilde{\phi}\big|_{{}_{\mathscr{O}_{{}_{\boldsymbol{m},0,0,0}}}} = U^{-1}\overline{\kappa_{0,1}(\overline{\phi})},
\,\,\,\,\,
P^\ominus\widetilde{\phi}\big|_{{}_{\mathscr{O}_{{}_{\boldsymbol{-m},0,0,0}}}} = \big[U^{-1}\kappa_{0,1}(\overline{\phi})\big]^\flat,
\]
for $\phi \in \mathscr{E}$. Thus
\[
\begin{split}
f^{\oplus}_{{}_{\chi\, lm}}(\phi) = \overline{\big\langle
\overline{\kappa_{0,1}(\overline{\phi})}, \,
UF^{\oplus}_{{}_{\chi \, lm}} \big\rangle}
=
\big\langle UF^{\oplus}_{{}_{\chi \, lm}}, \,
\overline{\kappa_{0,1}(\overline{\phi})} \big\rangle
= \big\langle F^{\oplus}_{{}_{\chi \, lm}}, \,
U^{-1}\overline{\kappa_{0,1}(\overline{\phi})} \big\rangle
\\
f^{\ominus}_{{}_{\chi\, lm}}(\phi)=
\big\langle
\kappa_{1,0}(\overline{\phi}), \, 
UF^{\ominus \, \flat}_{{}_{\chi \, lm}} 
\big\rangle
=
\big\langle
U^{-1}\kappa_{1,0}(\overline{\phi}), \, 
F^{\ominus \, \flat}_{{}_{\chi \, lm}} 
\big\rangle
=
\overline{
\big\langle
F^{\ominus \, \flat}_{{}_{\chi \, lm}}, \,
U^{-1}\kappa_{1,0}(\overline{\phi}) 
\big\rangle
}
\end{split}
\]
if $\langle \, \cdot \,, \, \cdot \, \rangle$ stands for the natural extension of the invariant inner product
(conjugate linear in the first variable), or, for the \emph{invariant
pairing} (conjugate linear in the first variable), defined above:
\[
\big\langle \, \cdot \,, \, \cdot \, \big\rangle \overset{\textrm{df}}{=}
\big\langle \, \cdot, \, \cdot \,
\big\rangle_{{}_{\textrm{inv. pairing}}}.
\]
The canonical \emph{bilinear} pairing, which we have denoted by the same sign $\langle \, \cdot \,, \, \cdot \, \rangle$
in Sections \ref{psiBerezin-Hida}-\ref{WickForChronological}, is equal to 
\[
\big\langle \, \overline{\,\, \cdot \,\,}, \, \cdot \,
\big\rangle_{{}_{\textrm{inv. pairing}}}.
\]
Thus, if we want to express all formulas in terms of the canonical and \emph{bilinear} pairing
we need to apply the additional conjugation in the first variable of the \emph{invariant
pairing} (conjugate linear in the first variable).
The direct integral decomposition of the Dirac field $\boldsymbol{\psi}$ constructed above, we can regard as a direct integral decomposition
of the integral kernel operator 
\[
\overline{\phi} \mapsto \boldsymbol{\psi}(\overline{\phi})
= \Xi_{0,1}\big(\kappa_{0,1}(\overline{\phi})\big) + \Xi_{1,0}\big(\kappa_{1,0}(\overline{\phi})\big) 
\]
into direct integral of integral kernel operators $\overline{\phi} \mapsto \boldsymbol{\psi}_{{}_{\chi}}(\overline{\phi})$,
with the integral kernel operators $\boldsymbol{\psi}_{{}_{\chi}}(\overline{\phi})$ equal 
\begin{multline*}
\boldsymbol{\psi}_{{}_{\chi}}(\overline{\phi}) =
\sum\limits_{lm} 
\kappa_{\chi \, 0,1}^{+}(\overline{\phi})(lm)
\,\,\, b'_{{}_{\chi, \, lm}}
+
\sum\limits_{lm} \kappa_{\chi \, 1,0}^{-}(\overline{\phi})(lm)
\,\,\, {d'}_{{}_{\chi, \, lm}}^+
\\
=
\sum\limits_{lm} 
\big\langle
\overline{\kappa_{0,1}(\overline{\phi})}, \,
UF^{\oplus}_{{}_{\chi \, lm}} \big\rangle  \,\,\, b'_{{}_{\chi, \, lm}}
+
\sum\limits_{lm} \big\langle
UF^{\ominus \, \flat}_{{}_{\chi \, lm}}, \,
\kappa_{1,0}(\overline{\phi}) 
\big\rangle  \,\,\, {d'}_{{}_{\chi, \, lm}}^+,
\end{multline*}
with the distributional kernels of the distributions (here with the invariant pairing $\langle \cdot,\cdot\rangle$)
\[
\begin{split}
\overline{\phi} \longmapsto \kappa_{\chi \, 0,1}^{+}(\overline{\phi})(lm) \overset{\textrm{df}}{=} 
\big\langle
\overline{\kappa_{0,1}(\overline{\phi})}, \,
UF^{\oplus}_{{}_{\chi \, lm}} \big\rangle = 
\big\langle
U^{-1}\overline{\kappa_{0,1}(\overline{\phi})}, \,
F^{\oplus}_{{}_{\chi \, lm}} \big\rangle
=
\overline{
\big \langle
F^{\oplus}_{{}_{\chi \, lm}}, \,
U^{-1}\overline{\kappa_{0,1}(\overline{\phi})}
\big\rangle
}
\\
\overline{\phi} \longmapsto \kappa_{\chi \, 1,0}^{-}(\overline{\phi})(lm) \overset{\textrm{df}}{=} 
\big\langle
UF^{\ominus \, \flat}_{{}_{\chi \, lm}}, \,
\kappa_{1,0}(\overline{\phi}) 
\big\rangle
=
\big\langle
F^{\ominus \, \flat}_{{}_{\chi \, lm}}, \,
U^{-1}\kappa_{1,0}(\overline{\phi}) 
\big\rangle
=\overline{
\big\langle
U^{-1}\kappa_{1,0}(\overline{\phi}), \, 
F^{\ominus \, \flat}_{{}_{\chi \, lm}}
\big\rangle,
}
\end{split}
\]
respectively, equal
\[
\overline{f^{\oplus}_{{}_{\chi \, lm}}(x)}
\,\,\,\,\,\,\, \textrm{and} \,\,\,\,\,\,\,
\overline{f^{\ominus}_{{}_{\chi \, lm}}(x)}.
\]
Recall, please, that the decomposition component 
\[
\Big(P^\oplus\widetilde{\phi}\big|_{{}_{\mathscr{O}_{{}_{\boldsymbol{m},0,0,0}}}}\Big)_{{}_{\chi}},
\,\,\,\,\, \textrm{or} \,\,\,\,\,
\Big(P^\ominus\widetilde{\phi}\big|_{{}_{\mathscr{O}_{{}_{\boldsymbol{-m},0,0,0}}}}\Big)_{{}_{\chi}},
\]
for $\phi \in \mathscr{E}$, and for a fixed value $\chi$ of the decomposition parameter, can be understood as the value of the Fourier transform
\[
\begin{split}
\mathcal{F}\Big( P^\oplus\widetilde{\phi}\big|_{{}_{\mathscr{O}_{{}_{\boldsymbol{m},0,0,0}}}}\Big)(\chi) = \Big(P^\oplus\widetilde{\phi}\big|_{{}_{\mathscr{O}_{{}_{\boldsymbol{m},0,0,0}}}}\Big)_{{}_{\chi}},
\\
\textrm{or} \,\,
\mathcal{F}\Big( P^\ominus\widetilde{\phi}\big|_{{}_{\mathscr{O}_{{}_{-\boldsymbol{m},0,0,0}}}}\Big)(\chi) =
\Big(P^\ominus\widetilde{\phi}\big|_{{}_{\mathscr{O}_{{}_{-\boldsymbol{m},0,0,0}}}}\Big)_{{}_{\chi}},
\end{split}
\]
at $\chi$, determined by the decomposition of the representation of $SL(2, \mathbb{C})$ acting respectively
on the positive or negative energy solutions. Indeed, the direct integral decomposition of the Hilbert space
determined by the decomposition of the representation of the translation group determines the ordinary Fourier transform. Decomposition
of the representation of $SL(2, \mathbb{C})$ acting on the single particle positive or the (conjugated) negative energy solutions
determines its own Fourier transform, $\mathcal{F}$, which is no longer scalar-valued even when determined by the decomposition into irreducible
sub representations and even in the single particle scalar field, because $SL(2, \mathbb{C})$ is not Abelian, with infinite dimensional irreducible
unitary representations (compare e.g. \cite{GelfandV}), which in case of the Abelian translation group, are all one dimensional. 

For each fixed $\chi$ the value of the Fourier transform
\[
\begin{split}
\mathcal{F}\Big( P^\oplus\widetilde{\phi}\big|_{{}_{\mathscr{O}_{{}_{\boldsymbol{m},0,0,0}}}}\Big)(\chi) = \Big(P^\oplus\widetilde{\phi}\big|_{{}_{\mathscr{O}_{{}_{\boldsymbol{m},0,0,0}}}}\Big)_{{}_{\chi}},
\\
\textrm{or} \,\,
\mathcal{F}\Big( P^\ominus\widetilde{\phi}\big|_{{}_{\mathscr{O}_{{}_{-\boldsymbol{m},0,0,0}}}}\Big)(\chi) =
\Big(P^\ominus\widetilde{\phi}\big|_{{}_{\mathscr{O}_{{}_{-\boldsymbol{m},0,0,0}}}}\Big)_{{}_{\chi}},
\\
\textrm{or} \,\,
\mathcal{F}\Big( P^\ominus\widetilde{\phi}\big|_{{}_{\mathscr{O}_{{}_{-\boldsymbol{m},0,0,0}}}}^{\flat}\Big)(\chi) =
\Big(P^\ominus\widetilde{\phi}\big|_{{}_{\mathscr{O}_{{}_{-\boldsymbol{m},0,0,0}}}}^{\flat}\Big)_{{}_{\chi}}
= \Big(P^\ominus\widetilde{\phi}\big|_{{}_{\mathscr{O}_{{}_{-\boldsymbol{m},0,0,0}}}}\Big)_{{}_{\chi}}^{\flat}
,
\end{split}
\]
is, as we have already seen, equal to an antilinear (conjugate linear) functional over
$E^{\oplus}_{{}_{\chi}}$ or, respectively, over $E^{\ominus}_{{}_{\chi}}$ (or $E^{\ominus \, \flat}_{{}_{\chi}}$), and thus,
cannot be identified with any ordinary number.

\vspace*{1cm}

\begin{center}
{\small DECOMPOSITION OF $\Xi(\kappa_{\mathpzc{l},\mathpzc{m}})$ IN CASE 1)}
\end{center}

Now we extend this decomposition over more general integral kernel operators
$\Xi(\kappa_{lm})$, eq. (\ref{electron-positron-photon-Xi}), Subsection \ref{psiBerezin-Hida},
with vector-valued distributional kernels $\kappa_{lm}$ in the sense of definition (\ref{VectValotimesXi=intKerOp'}) of
Subsection \ref{psiBerezin-Hida} or \cite{obataJFA}, which evaluated at test function $\phi \in \mathscr{E}$
can be written
\begin{multline*}
\Xi(\kappa_{\mathpzc{l},\mathpzc{m}}(\phi)) \\
= \sum \limits_{s_1, \ldots s_l, t_1, \ldots t_m =1}^{4} \int \limits_{(\mathbb{R}^3)^{l+m}} \,
\kappa_{\mathpzc{l},\mathpzc{m}}(\phi)(s_1,\boldsymbol{\q}_1, \ldots, s_\mathpzc{l}, \boldsymbol{\q}_\mathpzc{l}, t_1, \boldsymbol{\p}_1,
\ldots, t_\mathpzc{m}, \boldsymbol{\p}_\mathpzc{m}) \,\times \\
\times
a_{s_1}(\boldsymbol{\q}_1)^+ \cdots a_{s_\mathpzc{l}}(\boldsymbol{\q}_\mathpzc{l})^+
a_{t_1}(\boldsymbol{\p}_1) \cdots a_{t_\mathpzc{m}}(\boldsymbol{\p}_\mathpzc{m}) \,
\ud^3 \boldsymbol{\q}_1 \ldots \ud^3 \boldsymbol{\q}_\mathpzc{l} \ud^3 \boldsymbol{\p}_1 \ldots \ud^3 \boldsymbol{\p}_\mathpzc{m},
\end{multline*}
In doing this
we use the integral kernel operators in the Fock space of the free Dirac field,
but this construction is general and can be realized as well for integral kernel operators
acting in the Fock space of a mass less field, even in case the $SL(2,\mathbb{C})$
subgroup acts through a non-unitary representation in the single particle Hilbert space,
provided it allows a well-defined direct integral decomposition (which is the case for the free electromagnetic potential
field, compare Subsection \ref{equivalentA-s}, and for all mass less fields with unitary action of $SL(2,\mathbb{C})$ in the single particle
Hilbert space).

We proceed generally, with the total nuclear space
\[
E = E^\oplus \oplus E^{\ominus \, \flat},
\]
equal to the image of the standard nuclear space
\[
\mathcal{S}_{H_{(3)}}(\mathbb{R}^3; \mathbb{C}^4) = E^{+} \oplus E^{- \, \flat}
\]
under the unitary isomorphism $U$ defined by eq. (\ref{isomorphismU}) of Subsection \ref{psiBerezin-Hida} (in case of the Dirac field). 
We have analogous construction for other charged fields and the analogous isomorphism $U$.
In case of the neutral field construction simplifies, because in this case the single particle Hilbert space consists solely of the positive
energy solutions, without the direct summand consisting of conjugated negative energy solutions. Analogous decomposition
of a wide class of integral kernel operators $\Xi(\kappa_{lm})$ can be constructed in the Fock spaces of several free
fields underlying a causal perturbative QFT, e.g. QED, provided the action of $SL(2,\mathbb{C})$ in the full single space admits direct integral
decomposition. In case of the full Fock space of spinor, scalar, \emph{e.t.c.} QED the action of $SL(2,\mathbb{C})$
admits canonical direct integral decomposition into indecomposable components. This follows from the following two facts.
1) the action of $SL(2, \mathbb{C})$ in the single particle Hilbert space of massive (and massless, although massless case seems non-physical, compare 
Subsection \ref{OperationsOnXi}) charged fields is unitary for all realistic charged free fields underlying realistic QED-s. 2) The action of $SL(2,\mathbb{C})$
acting in the single particle space of the standard realization of the free electromagnetic potential field is decomposable.
Decomposition is determined by the spectral decomposition of the scaling operator, for the proof compare Subsection \ref{equivalentA-s}.
Each component of this decomposition can be further decomposed into indecomposable ''electric'' and ''magnetic'' type components,
compare Subsection \ref{AS}. But the general principle on which the canonical decomposition of integral kernel operators
$\Xi(\kappa_{lm})$ is based, and which is canonically determined by the decomposition of the action of $SL(2, \mathbb{C})$ in the single
particle space, can be shown by using the particular example of the integral kernel operators $\Xi(\kappa_{lm})$
acting in the Fock space of the free Dirac field. No essentially new ingredients are needed for decomposition
of $\Xi(\kappa_{lm})$, which act in the Fock space of several free fields, say underlying QED, or more involved causal
perturbative QFT.

Thus, let us return to $\Xi(\kappa_{\mathpzc{l}\mathpzc{m}})$ acting in the Fock space of the standard free Dirac field (case 1)).
After Subsection \ref{psiBerezin-Hida}, we will denote the nuclear space $E$ and its standard form, $UE = \mathcal{S}_{H_{(3)}}(\mathbb{R}^3; \mathbb{C}^4)$,
by the same symbol $E$, whenever it does not lead to misunderstandings, at the general level, in order to simplify notation.
However, in the final formulas the isomorphism operator $U$ and its inverse $U^{-1}$, as well as their tensor powers $U^{\otimes \, n}$,
$(U^{-1})^{\otimes \, n}$ will have to appear in the explicit form, whenever we are using the ordinary realizations of the single particle Hilbet spaces
$\mathcal{H}' = \mathcal{H}^\oplus \oplus\mathcal{H}^{\ominus \flat}$,
of the ordinary free fields together with the basic functionals $F^{{}^{\oplus}}_{{}_{\chi,lm}}$, $F^{{}^{\ominus}}_{{}_{\chi,lm}}$
in $E^{\oplus \, *} \supset \mathcal{H}^\oplus$,$E^{\ominus \, *}\supset\mathcal{H}^\ominus$, because they are not in the standard form.

Let $\Xi(\kappa_{\mathpzc{l}\mathpzc{m}})$ be a generalized integral kernel operator with vector-valued kernel (here in the Fock space of the
free Dirac field), compare Subsection \ref{psiBerezin-Hida}. As we have seen, each Hida annihilation or creation operator,
evaluated at a fixed element $\xi$ of the nuclear space $E$
\[
a(\xi) = \sum\limits_{s=1}^{4} \overline{\xi(s,\boldsymbol{\p})} \, a_{s}(\boldsymbol{\p}) \ud \boldsymbol{\p},
\,\,\,\,\,
a^{+}(\xi) = a(\xi)^+ = \sum\limits_{s=1}^{4} \xi(s,\boldsymbol{\p}) \, a_{s}(\boldsymbol{\p})^+ \ud \boldsymbol{\p},
\]
has direct integral decomposition
\[
a(\xi) = \int a_{{}_{\chi}}\big((\xi)_{{}_{\chi}}\big) \, \ud \chi,
\,\,\,\,\,
a^{+}(\xi) = a(\xi)^+ = \int a_{{}_{\chi}}\big((\xi)_{{}_{\chi}}\big)^{+} \ud \chi.
\]
Correspondingly the product
\[
a(\xi_1)^{+} \ldots a(\xi_n)^{+} = \int a_{{}_{\chi}}(\xi_{\chi \, 1})^{+} \ldots a_{{}_{\chi}}(\xi_{\chi \, n})^{+} \, \ud \chi, \,\,\, \xi_i \in E,
\]
of the creation Hida operators has a well-defined direct integral decomposition. Therefore, each element
\[
\sum\limits_{n} a^+(\xi_1) \ldots a^{+}(\xi_n)|0\rangle \in (E)
\]
of the Hida test space has a direct integral decomposition, determined by the decomposition of the representation of
$SL(2, \mathbb{C})$.
\begin{defin}\label{DecompositionOfXi}
Let
\[
\Phi,\Psi \in (E),
\,\,\,\, \Phi = \int \Phi_{{}_{\chi}} \, \ud \chi,
\,\,\,\, \Psi = \int \Psi_{{}_{\chi}} \, \ud \chi
\]
be any two elements of the test Hida space with their direct integral decompositions. Let $\ud \chi$
be a $\sigma$-measure on the spectrum of the decomposition of the representation of $SL(2, \mathbb{C})$ acting
in the single particle Hilbert space $\mathcal{H}'$. We say that the generalized
integral kernel operator $\Xi(\kappa_{\mathpzc{l}\mathpzc{m}})$, eq. (\ref{electron-positron-photon-Xi}), is equal to the direct integral
\[
\Xi(\kappa_{\mathpzc{l}\mathpzc{m}}) = \int \Xi_{{}_{\chi}}(\kappa_{\chi \, \mathpzc{l}\mathpzc{m}}) \, \ud \chi
\]
of (discrete-) integral kernel operators $\Xi_{{}_{\chi}}(\kappa_{\chi \, \mathpzc{l}\mathpzc{m}})$,
acting in the Fock spaces over the single particle Gelfand triples
$E_{{}_{\chi}} \subset \mathcal{H}'_{{}_{\chi}} \subset E_{{}_{\chi}}^{*}$, if
\begin{multline*}
\int \big\langle\big\langle \Xi_{{}_{\chi}}\big(\kappa_{\chi \, \mathpzc{l}\mathpzc{m}}(\phi)\big)
\Phi_{{}_{\chi}}, \Psi_{{}_{\chi}} \big\rangle\big\rangle \, \ud \chi
= \int \Big\langle \kappa_{\chi \, \mathpzc{l}\mathpzc{m}}(\phi), \big(\eta_{{}_{\Phi,\Psi}}\big)_{{}_{\chi}} \Big\rangle \, \ud \chi
\\
=\langle \kappa_{\mathpzc{l}\mathpzc{m}}(\phi), \eta_{{}_{\Phi,\Psi}} \rangle
=\big\langle \big\langle \Xi(\kappa_{\mathpzc{l}\mathpzc{m}}(\phi))\Phi, \Psi \big\rangle \big\rangle
\end{multline*}
for all
\[
\Phi,\Psi \in (E), \phi \in \mathscr{E}.
\]
\qed
\end{defin}
Compare Subsection \ref{psiBerezin-Hida}. Here we are using the canonical \emph{bilinear}
pairings $\langle \cdot, \cdot \rangle$ and $\langle\langle \cdot, \cdot \rangle\rangle$.
Recall that (compare definition (\ref{VectValotimesXi=intKerOp}) and (\ref{VectValotimesXi=intKerOp'}) of Subsection
\ref{psiBerezin-Hida}) $\eta_{{}_{\Phi, \Psi}}$ is equal to the function
\begin{multline*}
\eta_{{}_{\Phi, \Psi}}(s_1,\boldsymbol{\q}_1, \ldots, s_\mathpzc{l}, \boldsymbol{\q}_\mathpzc{l}, t_1, \boldsymbol{\p}_1,
\ldots, t_\mathpzc{m}, \boldsymbol{\p}_\mathpzc{m})
\\
=
\big\langle\big\langle a_{s_1}(\boldsymbol{\q}_1)^+ \cdots a_{s_\mathpzc{l}}(\boldsymbol{\q}_\mathpzc{l})^+
a_{t_1}(\boldsymbol{\p}_1) \cdots a_{t_\mathpzc{m}}(\boldsymbol{\p}_\mathpzc{m}) \, \Phi, \,\, \Psi \big\rangle\big\rangle
\end{multline*}
and that
\[
\eta_{{}_{\Phi, \Psi}} \in E^{\widehat{\otimes} \, (\mathpzc{l}+\mathpzc{m})},
\]
so that $\eta_{{}_{\Phi, \Psi}}$ can be canonically identified with an $(\mathpzc{l}+\mathpzc{m})$-particle state belonging to
the test Hida space $(E)$, and has the canonical direct integral decomposition
\[
\eta_{{}_{\Phi, \Psi}} = \int \big(\eta_{{}_{\Phi, \Psi}} \big)_{{}_{\chi}} \, \ud \chi,
\]
but here with the spectral measure $\ud \chi$ on the spectrum $\textrm{Spec}$ of the decomposition of the representation
of $SL(2, \mathbb{C})$ acting in the single particle space, equal to a measure $\ud\chi = w(\nu) \ud \nu$
absolutely continuous with respect to the ordinary Lebesgue measure
$\ud \nu$ on the spectrum $\textrm{Spec} = \{\chi = -1/2+i\nu, \nu \in \mathbb{R}_+\}$ in case of the Dirac field, with a simple weight function
$w(\nu)$ equal to a power function of $\nu$. Note that in general case for any free field (whenever the representation of $SL(2, \mathbb{C})$
in the single particle space is decomposable) $\textrm{Spec} = \{\chi = -r +i\nu, \nu \in \mathbb{R}\}$ or
$\textrm{Spec} = \{\chi = -r +i\nu, \nu \in \mathbb{R}_+\}$
with a fixed real $r>0$ and $\ud\chi = w(\nu) \ud \nu$ with a power function $w(\nu)$.
For the massive scalar field $\textrm{Spec} = \{\chi = -1 +i\nu, \nu >0\}$, and $w(\nu) = \nu^2$.
For the mass less scalar field $\textrm{Spec} = \{\chi = -1 +i\nu, \nu \in \mathbb{R}\}$, and $w(\nu) = 1$.
This is the case e.g. if the representation of $SL(2,\mathbb{C})$ in the single particle space is unitary,
or even for the free electromagnetic potential field (and still for more general mass less gauge free fields) with Krein isometric,
but non-unitary representation of $SL(2,\mathbb{C})$, compare Subsection \ref{equivalentA-s}.

This means that
\[
\kappa_{\chi \, \mathpzc{l}\mathpzc{m}}(\phi) = \mathcal{F}[\kappa_{\mathpzc{l}\mathpzc{m}}(\phi)](\chi, \chi, \ldots, \chi)
\]
is equal to the Fourier transform of the distribution
\[
\kappa_{\mathpzc{l}\mathpzc{m}}(\phi) \in E^{* \widehat{\otimes} (\mathpzc{l}+\mathpzc{m})},
\]
restricted to the diagonal. Here the Fourier transform $\mathcal{F}$ is that determined by the decomposition
of the $(\mathpzc{l}+\mathpzc{m})$-fold tensor product (eventually antisymmetrized in Fermi case or symmetrized in Bose case)
of the representation of $SL(2,\mathbb{C})$ acting in $E$, which acts on the tensor
product space $E^{\widehat{\otimes} \, (\mathpzc{l}+\mathpzc{m})}$ regarded as the nuclear space of functions
on the $(\mathpzc{l}+\mathpzc{m})$-fold Cartesian product
\[
\mathscr{O}_{{}_{\boldsymbol{m},0,0,0}} \times \ldots \times \mathscr{O}_{{}_{\boldsymbol{m},0,0,0}},
\]
and then restricted to the diagonal, \emph{i.e.} with all values of the decomposition variables equal $\chi$.

For each fixed $\chi$ belonging to the spectrum of decomposition
the Fourier transform
\[
\kappa_{\chi \, \mathpzc{l}\mathpzc{m}}(\phi) = \mathcal{F}[\kappa_{\mathpzc{l}\mathpzc{m}}(\phi)](\chi, \chi, \ldots, \chi)
\]
is equal to the $(\mathpzc{l}+\mathpzc{m})$-conjugate-linear functional
over $E_{{}_{\chi}}^{\widehat{\otimes} (\mathpzc{l}+\mathpzc{m})}$.
Using the canonical orthonormal basis
\[
\{F_{{}_{\chi \, \mathpzc{l}\mathpzc{m}}}\} = \{ UF^{\oplus}_{{}_{\chi \, \mathpzc{l}'\mathpzc{m}'}}\}
\cup \{ UF^{\ominus \, \flat}_{{}_{\chi \, \mathpzc{l}''\mathpzc{m}''}}\}
\]
in each factor space $E_{{}_{\chi}}$ we can decompose $\kappa_{\chi \, \mathpzc{l}\mathpzc{m}}(\phi)$
into the basis simple tensors:
\begin{multline*}
\kappa_{\chi \, \mathpzc{l}\mathpzc{m}}(\phi) = \mathcal{F}[\kappa_{\mathpzc{l}\mathpzc{m}}(\phi)](\chi, \chi, \ldots, \chi)
\\
= \sum\limits_{l_1m_1 \ldots l_{\mathpzc{l}+\mathpzc{m}} m_{\mathpzc{l}+\mathpzc{m}}}
c_{\chi \, l_1m_1 \ldots l_{\mathpzc{l}+\mathpzc{m}} m_{\mathpzc{l}+\mathpzc{m}}}(\phi) \,\,
F_{{}_{\chi, \, l_1 m_1}} \widehat{\otimes} \ldots \widehat{\otimes}
F_{{}_{\chi, \, l_{\mathpzc{l}+\mathpzc{m}} m_{\mathpzc{l}+\mathpzc{m}}}},
\\
c_{\chi \, l_1m_1 \ldots l_{\mathpzc{l}+\mathpzc{m}} m_{\mathpzc{l}+\mathpzc{m}}}(\phi) =
\big\langle \mathcal{F}[\kappa_{\mathpzc{l}\mathpzc{m}}(\phi)](\chi, \chi, \ldots, \chi), \, F_{{}_{\chi, \, l_1 m_1}} \widehat{\otimes} \ldots \widehat{\otimes}
F_{{}_{\chi, \, l_{\mathpzc{l}+\mathpzc{m}} m_{\mathpzc{l}+\mathpzc{m}}}} \big\rangle.
\end{multline*}
Here each $F_{{}_{\chi, \, l_i m_i}}$ is naturally understood as the antilinear functional
\begin{equation}\label{AntilinFchilm}
E_{{}_{\chi}} \ni \eta_{{}_{\chi}} = F_{{}_{\chi, \, l_i m_i}}(\eta_{{}_{\chi}}) = \big(\eta_{{}_{\chi}}, \, F_{{}_{\chi, \, l_i m_i}} \big)_{{}_{\chi}}.
\end{equation}
In particular
\begin{multline}\label{Xichi(kappachilm(phi))}
\Xi_{{}_{\chi}}\big(\kappa_{\chi \, \mathpzc{l}\mathpzc{m}}(\phi)\big)
\\
=
\sum\limits_{l_1m_1 \ldots l_{\mathpzc{l}+\mathpzc{m}} m_{\mathpzc{l}+\mathpzc{m}}}
\big\langle \mathcal{F}[\kappa_{lm}(\phi)](\chi, \ldots, \chi), \, F_{{}_{\chi, \, l_1 m_1}} \widehat{\otimes} \ldots \widehat{\otimes}
F_{{}_{\chi, \, l_{\mathpzc{l}+\mathpzc{m}} m_{\mathpzc{l}+\mathpzc{m}}}} \big\rangle \,\, \times \,\,
\\
\times
\,\,
a_{{}_{\chi \, l_1m_1}}^{+} \cdots a_{{}_{\chi \, l_\mathpzc{l}m_\mathpzc{l}}}^{+}
a_{{}_{\chi \, l_{\mathpzc{l}+1}m_{\mathpzc{l}+1}}} \cdots a_{{}_{\chi \, l_{\mathpzc{l}+\mathpzc{m}}m_{\mathpzc{l}+\mathpzc{m}}}},
\end{multline}
where
\[
a_{{}_{\chi \, l_im_i}}^{+} = a_{{}_{\chi}}(F_{{}_{\chi, l_im_i}})^{+},
\,\,\,
a_{{}_{\chi \, l_im_i}} = a_{{}_{\chi}}(F_{{}_{\chi, l_im_i}}),
\]
are the Hida creation-annihilation operators
acting in the Fock spaces over the single particle Gelfand triples
$E_{{}_{\chi}} \subset \mathcal{H}'_{{}_{\chi}} \subset E_{{}_{\chi}}^{*}$,
which respect the canonical anticommutation relations
(here in case of the generalized integral kernel operators $\Xi(\kappa_{\mathpzc{l}\mathpzc{m}})$ acting in the Fock space of the free Dirac field,
or commutation relations in case of the integral kernel operators $\Xi(\kappa_{\mathpzc{l}\mathpzc{m}})$ acting in the Fock space of a Bose field).

That we allow general $\sigma$-measures $\ud \chi$ on the spectrum of the decomposition of the representation
of $SL(2, \mathbb{C})$ acting in $E$ in Definition \ref{DecompositionOfXi} allows us to extend substantially the class
of decomposable integral kernel operators $\Xi(\kappa_{\mathpzc{l}\mathpzc{m}})$. 
If the measure $\ud \chi$ in Definition \ref{DecompositionOfXi} was restricted and just equal to the spectral
measure of the decomposition of the representation
of $SL(2, \mathbb{C})$ acting in $E$, then the class of decomposable $\Xi(\kappa_{\mathpzc{l}\mathpzc{m}})$ would be considerably smaller.
It would contain in particular the positive and negative frequency parts of the free fields (in our particular case the 
negative and positive frequency parts of the free Dirac field) and generally the integral kernel operators $\Xi(\kappa_{\mathpzc{l}\mathpzc{m}})$
with the Fourier transforms $\mathcal{F}[\kappa_{\mathpzc{l}\mathpzc{m}}(\phi)](\chi, \chi, \ldots, \chi)$ of 
the distributions 
\[
\kappa_{\mathpzc{l}\mathpzc{m}}(\phi) \in E^{* \otimes (\mathpzc{l}+\mathpzc{m})}, \phi \in \mathscr{E},
\]
which can be supported by measures absolutely continuous with respect to the spectral measure, and thus which can be represented
by the integral
\[
\langle  \kappa_{\mathpzc{l}\mathpzc{m}}(\phi), \eta_{{}_{\Phi,\Psi}} \rangle =
\int \Big\langle \mathcal{F}[\kappa_{\mathpzc{l}\mathpzc{m}}(\phi)](\chi, \chi, \ldots, \chi), \big(\eta_{{}_{\Phi,\Psi}}\big)_{{}_{\chi}} \Big\rangle \, \ud \chi,
\]
with  $\ud \chi$ equal to the spectral measure of the decomposition of $SL(2, \mathbb{C})$ acting in $E$. 
In order to extend the class of decomposable $\Xi(\kappa_{\mathpzc{l}\mathpzc{m}})$ we allow any $\sigma$-measure on 
$\textrm{Spec}$ in Definition \ref{DecompositionOfXi}. Let us explain this in more details. Consider, for simplicity of notation,
the integral kernel operator of the form $\Xi(\kappa_{01})$ or $\Xi(\kappa_{10})$. In this case
\[
\eta_{{}_{\Phi, \Psi}} \in E
\]
and the equality 
\[
\langle  \kappa_{10}(\phi), \eta_{{}_{\Phi,\Psi}} \rangle =
\int \Big\langle \kappa_{\chi \, 10}(\phi), \big(\eta_{{}_{\Phi,\Psi}}\big)_{{}_{\chi}} \Big\rangle \, \ud \chi
\]
with the spectral measure $\ud \chi$  of the decomposition of $SL(2, \mathbb{C})$ acting in $E$ means that 
\[
\kappa_{\chi \, 10}(\phi) = \mathcal{F}[\kappa_{10}(\phi)](\chi),
\]
is the Fourier transform $\mathcal{F}[\kappa_{10}(\phi)](\chi)$ of 
$\kappa_{10}(\phi) \in E^{*}$ supported by the measure absolutely continuous with respect to the spectral 
measure $\ud \chi$. Of course here $\mathcal{F}$ is the Fourier transform associated with decomposition of $SL(2, \mathbb{C})$
actng in $E$.  But among the distributions $\kappa_{10}$ with $\kappa_{10}(\phi) \in E^{*}$ there are such  $\kappa_{10}(\phi)$ whose Fourier
transforms, besides the absolutely continuous measure support, possess components with discrete measures, supported on the measure
zero subsets of the spectrum of decomposition  of $SL(2, \mathbb{C})$ acting in $E$ (with respect to the spectral measure), for which the equality 
\[
\langle  \kappa_{10}(\phi), \eta_{{}_{\Phi,\Psi}} \rangle =
\int\limits_{{}_{\textrm{Spec}}} \Big\langle \mathcal{F}[\kappa_{10}(\phi)](\chi), \big(\eta_{{}_{\Phi,\Psi}}\big)_{{}_{\chi}} \Big\rangle \, \ud \chi,
\,\,\,\, \Phi,\Psi \in (E), \, \phi \in E,
\]
with the spectral measure $\ud \chi$  of the decomposition of $SL(2, \mathbb{C})$ acting in $E$,  cannot hold.

EXAMPLE 1.
In order to illustrate this fact we use the simpler situation of the decomposition of the
translation group $T_1$ (analogue of $SL(2, \mathbb{C})$) acting on the reals $\mathbb{R}$, and on the $\mathbb{C}$-valued functions
$f$ in $\mathcal{S}(\mathbb{R}; \mathbb{C}) \subset L^2(\mathbb{R}; \mathbb{C})$ with
$\mathcal{S}(\mathbb{R}; \mathbb{C})$ being the analogue of $E$ and with $L^2(\mathbb{R}; \mathbb{C})$
being the analogue of the single particle Hilbert space. In this case the Fourier transform $\mathscr{F}$
associated to the decomposition of the translation group $T_1$ is determined by the corresponding direct integral
decomposition of $\mathcal{S}(\mathbb{R}; \mathbb{C}) \subset L^2(\mathbb{R}; \mathbb{C})$ and $L^2(\mathbb{R}; \mathbb{C})$ into one dimensional
nuclear and Hilbert spaces $\mathbb{C}$ (being the analogue
of $E_{{}_{\chi}}$ and $\mathcal{H}'_{{}_{\chi}}$).
Each $f \in L^2(\mathbb{R}; \mathbb{C})$, and in particular each $f \in \mathcal{S}(\mathbb{R}; \mathbb{C})$,
has the direct integral decomposition determined by the decomposition of $T_1$ acting in $\mathcal{S}(\mathbb{R}; \mathbb{C})$
with decomposition components
\[
(f)_{{}_{\chi}} = \widetilde{f}(\chi) = \mathscr{F}f(\chi), \,\,\, \chi \in \mathbb{R}
\]
given by the values of the ordinary Fourier transform $\widetilde{f} = \mathscr{F}f$, compare \cite{GelfandIV}.
Decomposition of the elements $f$ of the nuclear space $\mathcal{S}(\mathbb{R}; \mathbb{C})$ also determines canonically
decompositions of the functionals $F$ in $\mathcal{S}(\mathbb{R}; \mathbb{C})^{*}$. We say that
$\chi \mapsto F_{{}_{\chi}} \in \mathbb{C}$ is the decomposition of the functional $F\in \mathcal{S}(\mathbb{R}; \mathbb{C})^{*}$
if
\begin{multline*}
\langle F, f\rangle = \int\limits_{\textrm{Spec} = \mathbb{R}} F_{{}_{\chi}} \mathscr{F}{f}(\chi) \, \ud \chi =
\int\limits_{{}_{\textrm{Spec} = \mathbb{R}}} \big\langle F_{{}_{\chi}}, \mathscr{F}{f}(\chi)\big\rangle \, \ud \chi
\\
=
\int\limits_{{}_{\textrm{Spec} = \mathbb{R}}} \big\langle \mathscr{F}F(\chi), (f)_{{}_{\chi}}\big\rangle \, \ud \chi
, \,\,\,\,\,\,
f \in \mathcal{S}(\mathbb{R}; \mathbb{C}).
\end{multline*}
If the measure $\ud \chi$ here coincides with the ordinary Lebesgue measure on the spectrum $\textrm{Spec}
=\mathbb{R}$ of decomposition of $T_1$, \emph{i.e.} with the spectral measure of the decomposition of $T_1$ acting in $\mathcal{S}(\mathbb{R}; \mathbb{C})$,
then the class of decomposable $F$ becomes quite restricted, and includes precisely those distributions
$F$ whose Fourier transforms $\mathscr{F}F$ are representable by ordinary functions. But this restriction is unnatural because
for a much larger class of functionals $F$ the last equality still holds with the measure $\ud \chi$ on $\textrm{Spec}
=\mathbb{R}$ determined solely by the functional $F$. Consider for example the functional $F$ represented by the Heaviside step
function $\theta \notin L^2(\mathbb{R}; \mathbb{C})$. In this case we have
\begin{multline*}
\langle \theta, f\rangle = \int\limits_{{}_{\textrm{Spec} = \mathbb{R}}} {\textstyle\frac{i}{\sqrt{2\pi} \, \chi}} \, \mathscr{F}{f}(\chi) \,\, \ud\chi +
\sqrt{{\textstyle\frac{\pi}{2}}} \, \mathscr{F}{f}(\chi=0)
\\
=
\int\limits_{{}_{\textrm{Spec} = \mathbb{R}}} \Big\langle {\textstyle\frac{i}{\sqrt{2\pi} \, \chi}}, \, (f)_{{}_{\chi}}\Big\rangle \,\, \ud\chi +
\Big\langle \sqrt{{\textstyle\frac{\pi}{2}}}, (f)_{{}_{\chi=0}}\Big\rangle
, \,\,\,\,\,\,\, f \in \mathcal{S}(\mathbb{R}; \mathbb{C}).
\end{multline*}
Here $\ud\chi$ is the ordinary Lebesgue measure on $\mathbb{R}$ equal to spectral measure of the decomposition
of $T_1$ acting in $\mathcal{S}(\mathbb{R}; \mathbb{C})$. $\big\langle \sqrt{{\textstyle\frac{\pi}{2}}}, (f)_{{}_{\chi=0}}\big\rangle$
is the value of the functional
defined by multiplication by $\sqrt{\tfrac{\pi}{2}}$
on $(f)_{{}_{\chi=0}} \in \mathbb{C}$.
We see that if we include the discrete Dirac
measure concentrated on single point set $\{0\} \subset \mathbb{R} = \textrm{Spec}$ on the spectrum $\textrm{Spec}
=\mathbb{R}$ of decomposition of $T_1$, then decomposition can be extended over the functional $F=\theta$
with the spectral measure replaced by the measure which can be decomposed into the
absolutely continuous spectral measure $\ud\chi$ and the Dirac measure concentrated on the single point set $\{0\}$. \qed

We have similar situation for the Fourier transform $\mathcal{F}[\kappa_{10}(\phi)]$ of distributions $\kappa_{10}(\phi) \in E^*$
determined by the decomposition of $SL(2, \mathbb{C})$ acting in $E\subset \mathcal{H}'$, and for the Fourier transform
$\mathcal{F}[\kappa_{\mathpzc{l}\mathpzc{m}}(\phi)]$ of distributions 
$\kappa_{\mathpzc{l}\mathpzc{m}}(\phi) \in E^{* \widehat{\otimes} \, (\mathpzc{l}+\mathpzc{m})}$
determined by the decomposition of the tensor product (eventually antisymmetrized or symmetrized) of the representation 
of $SL(2, \mathbb{C})$ acting in 
$E^{\otimes \,(\mathpzc{l}+\mathpzc{m})} \subset \mathcal{H}'^{ \widehat{\otimes} \, (\mathpzc{l}+\mathpzc{m})}$. 
A wide class of generalized integral kernel operators $\Xi(\kappa_{\mathpzc{l}\mathpzc{m}})$ 
with vector-valued kernels allows decomposition in the sense of Definition \ref{DecompositionOfXi} with the measure 
$\ud\chi$ on the spectrum of decomposition of $SL(2, \mathbb{C})$ acting in $E$, determined
solely by the kernel $\kappa_{\mathpzc{l}\mathpzc{m}}$. Such decomposition in general includes discrete components.

Thus, at a first sight at least, it seems that in order to have an effective method which allows checking if a given
$\Xi(\kappa_{\mathpzc{l}\mathpzc{m}})$ is decomposable and allows effective computation of the kernels
\begin{multline}\label{cchilimi}
c_{\chi \, l_1m_1 \ldots l_{\mathpzc{l}+\mathpzc{m}} m_{\mathpzc{l}+\mathpzc{m}}}(\phi) =
\big\langle \mathcal{F}[\kappa_{\mathpzc{l}\mathpzc{m}}(\phi)](\chi, \chi, \ldots, \chi), \, F_{{}_{\chi, \, l_1 m_1}} \widehat{\otimes} \ldots \widehat{\otimes}
F_{{}_{\chi, \, l_{\mathpzc{l}+\mathpzc{m}} m_{\mathpzc{l}+\mathpzc{m}}}} \big\rangle
\\
= \kappa_{{}_{\chi \, \mathpzc{l}\mathpzc{m}}}(\phi)(l_1m_1, \ldots, l_{\mathpzc{l}+\mathpzc{m}}m_{\mathpzc{l}+\mathpzc{m}})
\end{multline}
in (\ref{Xichi(kappachilm(phi))}) of the decomposition discrete-integral operator components $\Xi_{{}_{\chi}}(\kappa_{\chi \, \mathpzc{l}\mathpzc{m}})$,
we need to have the theory of Fourier transforms $\mathcal{F}[\kappa_{\mathpzc{l}\mathpzc{m}}(\phi)]$
of distributions
\[
\kappa_{\mathpzc{l}\mathpzc{m}}(\phi) \in E^{*\widehat{\otimes} (\mathpzc{l}+\mathpzc{m})}
\]
with the Fourier transform $\mathcal{F}$ associated to the decomposition of the representation of $SL(2, \mathbb{C})$
acting in $E^{\widehat{\otimes} (\mathpzc{l}+\mathpzc{m})}$ -- the (antisymmetrized or symmetrized) tensor product of the representation
of $SL(2,\mathbb{C})$ acting in $E$. We know that such theory exists (\cite{GGG}, \cite{GelfandV}, \cite{GelfandIV},
\cite{Geland-Minlos-Shapiro}, \cite{NeumarkLorentzBook}),
but has so far been not developed
in all computational details which would be needed for effective computation of the Fourier transform $\mathcal{F}$
of a wide class of distributions in $E^{*\widehat{\otimes} (\mathpzc{l}+\mathpzc{m})}$,
and in particular we do not have at our disposal the corresponding ''Fourier transform tables''
although the complete theoretical basis has been already developed in the cited works\footnote{For the scalar field case,
and $\mathcal{F}$ associated to the decomposition of the $SL(2, \mathbb{C})$ acting on $E^\oplus \oplus E^{\ominus \flat} =
U^{-1}[\mathcal{S}(\mathbb{R}^3) \oplus \mathcal{S}(\mathbb{R}^3)]$,
where $\mathcal{S}(\mathbb{R}^3)$ is the Schwartz space of scalar functions
in the Cartesian coordinates $\boldsymbol{\p}$ on the orbit $\mathscr{O}_{{}_{\boldsymbol{m}, 0,0,0}}$ -- the Lobachevsky space
of curvature $-\boldsymbol{m}$, we have some further theoretical results (we restrict $\mathcal{F}$ separately to $E^\oplus$ or $E^{\ominus \, \flat}$
and put $\boldsymbol{m}=1$), e.g.
the Paley-Wiener theorem and some Paley-Wiener-type theorems, compare e.g. \cite{Andersen}, \cite{Helgason},
and references therein.}.
Fortunately the computationally fully-fledged form of this harmonic analysis is not absolutely necessary for investigation of decomposability
of $\Xi(\kappa_{\mathpzc{l}\mathpzc{m}})$ and for the computation
of (\ref{cchilimi}) for the integral kernel operators $\Xi(\kappa_{\mathpzc{l}\mathpzc{m}})$ which are decomposable
in the sense of Definition \ref{DecompositionOfXi}. The computation of the coefficients
(\ref{cchilimi}) can essentially be reduced in this case to the valuations of the kernels $\kappa_{\mathpzc{l}\mathpzc{m}}$ at the tensor
products of the fundamental basic functionals $F_{{}_{\chi \, lm}} \in E^*$. Indeed, for the free field integral kernel operators
\[
\Xi(\kappa_{0,1}) + \Xi(\kappa_{1,0})
\]
or more generally for the kernels $\kappa_{\mathpzc{l}\mathpzc{m}}$ in the finite sums
\[
\sum\limits_{\mathpzc{l}\mathpzc{m}} \Xi(\kappa_{\mathpzc{l}\mathpzc{m}})
\]
representing Wick products of free fields or higher order contributions to interacting fields
the kernels
\[
\kappa_{\mathpzc{l}\mathpzc{m}}(\phi) \in E^{*\widehat{\otimes} (\mathpzc{l}+\mathpzc{m})}, \,\,\, \phi \in \mathscr{E}
\]
are quite regular distributions which can be represented by ordinary (bounded and smooth, except submanifolds of lower dimension) functions
on the $(\mathpzc{l}+\mathpzc{m})$-fold Cartesian product
\[
\mathscr{O}_{{}_{\boldsymbol{m},0,0,0}} \times \ldots \times \mathscr{O}_{{}_{\boldsymbol{m},0,0,0}}
\]
(some factors being replaced with the cone
$\mathscr{O}_{{}_{1,0,0,1}}$ in case of the Fock space including mass less fields and some mass
parameters $\boldsymbol{m}$ may differ accordingly to the masses of the free fields), decrease
in each of the momentum variables separately (\emph{i.e.} with all remaining variables being arbitrarily fixed).
In particular for the Wick product of massive free fields the kernels
\[
\kappa_{\mathpzc{l},\mathpzc{m}=0}(\phi) \in E^{\widehat{\otimes} l}
\]
so that $\kappa_{\mathpzc{l},\mathpzc{m}=0}$, regarded as a functional over $E^{\widehat{\otimes}\mathpzc{l}}$, can be extended over
$E^{*\widehat{\otimes} \mathpzc{l}}$. In particular such kernel can be evaluated at the (antisymmetrized or respectively, symmetrized) tensor product
of the basic functionals $F_{{}_{\chi \, lm}} \in E^*$:
\[
\langle \kappa_{\mathpzc{l},\mathpzc{m}=0}(\phi), F_{{}_{\chi \, l_1m_1}} \widehat{\otimes} \ldots
\widehat{\otimes} F_{{}_{\chi \, l_\mathpzc{l}m_\mathpzc{l}}} \rangle < \infty.
\]
Even the kernels $\kappa_{\mathpzc{l},\mathpzc{m}=0}(\phi)$ of Wick products of mass less fields, although do not belong to $E^{\widehat{\otimes} l}$,
can be evaluated at (antisymmetrized or, respectively, symmetrized) tensor product
of the basic functionals $F_{{}_{\chi \, l_im_i}} \in E^*$. More generally if the kernel
$\kappa_{\mathpzc{l}\mathpzc{m}}(\phi)$ can be evaluated at the (antisymmetrized or respectively, symmetrized) $(\mathpzc{l}+\mathpzc{m})$-fold tensor product
of the basic functionals $F_{{}_{\chi \, l_im_i}} \in E^*$, then, as is easily seen, the corresponding
coefficient (\ref{cchilimi}) can be computed through this evaluation:
\begin{multline}\label{cchilimi}
c_{\chi \, l_1m_1 \ldots l_{\mathpzc{l}+\mathpzc{m}} m_{\mathpzc{l}+\mathpzc{m}}}(\phi) =
\big\langle \mathcal{F}[\kappa_{\mathpzc{l}\mathpzc{m}}(\phi)](\chi, \chi, \ldots, \chi), \, F_{{}_{\chi, \, l_1 m_1}} \widehat{\otimes} \ldots \widehat{\otimes}
F_{{}_{\chi, \, l_{\mathpzc{l}+\mathpzc{m}} m_{\mathpzc{l}+\mathpzc{m}}}} \big\rangle
\\
= \kappa_{{}_{\chi \, \mathpzc{l}\mathpzc{m}}}(\phi)(l_1m_1, \ldots, l_{\mathpzc{l}+\mathpzc{m}}m_{\mathpzc{l}+\mathpzc{m}})
=
\big\langle \kappa_{\mathpzc{l}\mathpzc{m}}(\phi), \overline{F_{{}_{\chi, \, l_1 m_1}}} \widehat{\otimes} \ldots \widehat{\otimes}
F_{{}_{\chi, \, l_{\mathpzc{l}+\mathpzc{m}} m_{\mathpzc{l}+\mathpzc{m}}}} \big\rangle.
\end{multline}
Here, in the first line, $F_{{}_{\chi, \, l_i m_i}}$ are understood as the antilinear functionals (\ref{AntilinFchilm}) on $E_{{}_{\chi}}$.
In the second line
\[
\overline{F_{{}_{\chi, \, l_i m_i}}} , \ldots, F_{{}_{\chi, \, l_j m_j}}, \ldots
\]
are understood as distributions in $E^{*}$,
defined by the complex conjugation of the functions $F_{{}_{\chi, \, l_i m_i}}$ or, respectively, by the functions $F_{{}_{\chi, \, l_j m_j}}$
themselves.
In the second line $F_{{}_{\chi, \, l_i m_i}}$ are complex conjugated for the first $i=1, \ldots, \mathpzc{l}$ variables of the kernel
$\kappa_{\mathpzc{l}\mathpzc{m}}$, corresponding to the creation operators, and the remaining
\[
F_{{}_{\chi, \, l_{\mathpzc{l}+1} m_{\mathpzc{l}+1}}}, \ldots, F_{{}_{\chi, \, l_{\mathpzc{l}+\mathpzc{m}} m_{\mathpzc{l}+\mathpzc{m}}}},
\]
which correspond to the last $\mathpzc{m}$ variables $i=\mathpzc{l}+1, \ldots, \mathpzc{l}+\mathpzc{m}$, of the kernel
$\kappa_{\mathpzc{l}\mathpzc{m}}$, and corresponding to the annihilation operators, enter without complex conjugation.

In general, each distribution $\kappa_{\mathpzc{l}\mathpzc{m}}(\phi)$
is a limit of more regular distributions $\kappa_{\mathpzc{l}\mathpzc{m}}(\phi)_{{}_{\epsilon}}$ (we can
even choose $\kappa_{\mathpzc{l}\mathpzc{m}}(\phi)_{{}_{\epsilon}} \in E^{\widehat{\otimes} (\mathpzc{l}+\mathpzc{m})}$, but sometimes it is more convenient to choose
them, say, in $\mathcal{H}'^{\widehat{\otimes} (\mathpzc{l}+\mathpzc{m})}$) which converge to $\kappa_{\mathpzc{l}\mathpzc{m}}(\phi)$ in
$E^{*\widehat{\otimes} (\mathpzc{l}+\mathpzc{m})}$
if $\epsilon \rightarrow 0$. The point
is that $\kappa_{\mathpzc{l}\mathpzc{m}}(\phi)_{{}_{\epsilon}}$ can be evaluated at the tensor product
of the basic functionals $F_{{}_{\chi \, l_im_i}} \in E^*$, and the Fourier transform $\mathcal{F}$ of
$\kappa_{\mathpzc{l}\mathpzc{m}}(\phi)_{{}_{\epsilon}}$ can be easily computed through the evaluation of $\kappa_{\mathpzc{l}\mathpzc{m}}(\phi)_{{}_{\epsilon}}$
at the basic functionals $F_{{}_{\chi \, l_im_i}} \in E^*$. Assuming that the Fourier transform
$\mathcal{F}[\kappa_{\mathpzc{l}\mathpzc{m}}(\phi)](\chi, \ldots, \chi)$
of $\kappa_{\mathpzc{l}\mathpzc{m}}(\phi)$ is supported by the measure absolutely continuous with respect to the spectral measure
and
\begin{equation}\label{chi->Fkappa(phi)(chi,...,chi)OrdinaryFuncion}
\textrm{Spec} \ni \chi \longmapsto
\big\langle \mathcal{F}[\kappa_{\mathpzc{l}\mathpzc{m}}(\phi)](\chi, \chi, \ldots, \chi), \, F_{{}_{\chi, \, l_1 m_1}} \widehat{\otimes} \ldots \widehat{\otimes}
F_{{}_{\chi, \, l_{\mathpzc{l}+\mathpzc{m}} m_{\mathpzc{l}+\mathpzc{m}}}} \big\rangle
\end{equation}
is an ordinary function on $\textrm{Spec}$ for each sequence of indices $l_1m_i, \ldots, l_{\mathpzc{l}+\mathpzc{m}} m_{\mathpzc{l}+\mathpzc{m}}$
ranging over the indices numbering the basic functionals the coefficients (\ref{cchilimi}) can be easily computed.
In this case the limit
\begin{multline}\label{SimpleLimitkappaepsilonlm(phi)->kappalm(phi)}
\big\langle \mathcal{F}[\kappa_{\mathpzc{l}\mathpzc{m}}(\phi)](\chi, \chi, \ldots, \chi), \, F_{{}_{\chi, \, l_1 m_1}} \widehat{\otimes} \ldots \widehat{\otimes}
F_{{}_{\chi, \, l_{\mathpzc{l}+\mathpzc{m}} m_{\mathpzc{l}+\mathpzc{m}}}} \big\rangle
\\
= \underset{\epsilon \rightarrow 0}{\textrm{lim}}
\big\langle \kappa_{\mathpzc{l}\mathpzc{m}}(\phi)_{{}_{\epsilon}}, \overline{F_{{}_{\chi, \, l_1 m_1}}} \widehat{\otimes} \ldots \widehat{\otimes}
F_{{}_{\chi, \, l_{\mathpzc{l}+\mathpzc{m}} m_{\mathpzc{l}+\mathpzc{m}}}} \big\rangle
\end{multline}
converges for each $\chi$ and represents the function (\ref{chi->Fkappa(phi)(chi,...,chi)OrdinaryFuncion}).
But we should be careful, as in general the limit
$\kappa_{\mathpzc{l}\mathpzc{m}}(\phi)_{{}_{\epsilon}} \overset{\epsilon \rightarrow 0}{\rightarrow}\kappa_{\mathpzc{l}\mathpzc{m}}(\phi)$
assures the existence of the limit of the Fourier transforms as the limit of Fourier transforms
of distributions and means that for any
\[
\eta \in E^{\widehat{\otimes} (\mathpzc{l}+\mathpzc{m})} \,\,\, \textrm{with} \,\,\, \eta = \int\limits_{{}_{\textrm{Spec}}} (\eta)_{{}_{\chi}} \, \ud \chi
\]
with its decomposition (here with the spectral measure $\ud \chi$) and any $\phi \in \mathscr{E}$ we have
(for decomposable $\Xi(\kappa_{\mathpzc{l}\mathpzc{m}})$)
\begin{multline}\label{GeneralFTLimit}
\underset{\epsilon \rightarrow 0}{\textrm{lim}}
\int \big\langle \mathcal{F}[\kappa_{\mathpzc{l}\mathpzc{m}}(\phi)_{{}_{\epsilon}}](\chi, \chi, \ldots, \chi), \, (\eta)_{{}_{\chi}} \big\rangle \,
\ud \chi
\\
= \int \big\langle \mathcal{F}[\kappa_{\mathpzc{l}\mathpzc{m}}(\phi)](\chi, \chi, \ldots, \chi), \, (\eta)_{{}_{\chi}} \big\rangle \,
\ud \chi'
\end{multline}
and on the right-hand side of the last formula $\ud \chi'$ is a $\sigma$-measure which \emph{a priori} may differ from the spectral measure
$\ud \chi$ of the decomposition of $SL(2, \mathbb{C})$ acting in $E$, on the left-hand side. If these measures indeed are different,
then the simple formula (\ref{SimpleLimitkappaepsilonlm(phi)->kappalm(phi)}) cannot be used. Thus, the whole point lies
in determination if the measure $\ud \chi'$ (which should be identified with the measure $\ud \chi$ in Definition \ref{DecompositionOfXi})
associated to the kernel $\kappa_{\mathpzc{l}\mathpzc{m}}(\phi)$
possess components which are not absolutely continuous with respect to the spectral measure. In order to establish it
we need to analyze the last limit (\ref{GeneralFTLimit}), and in case there are such additional discrete components, the simple formula
(\ref{SimpleLimitkappaepsilonlm(phi)->kappalm(phi)}) will have to be
accordingly modified. If $\Xi(\kappa_{\mathpzc{l}\mathpzc{m}})$ is not decomposable the limit on left-hand side of
(\ref{GeneralFTLimit}) cannot be written in the form which is present on the right-hand side of (\ref{GeneralFTLimit})
with any $\sigma$-measure $\ud \chi'$ on $\textrm{Spec}$.

Thus, in general, if the integral representing the pairing
\begin{equation}\label{Pairing<kappa,Fx...xF>}
\big\langle \kappa_{\mathpzc{l}\mathpzc{m}}(\phi), \overline{F_{{}_{\chi, \, l_1 m_1}}} \widehat{\otimes} \ldots \widehat{\otimes}
F_{{}_{\chi, \, l_{\mathpzc{l}+\mathpzc{l}} m_{\mathpzc{l}+\mathpzc{m}}}} \big\rangle
\end{equation}
is divergent, we will have to analyze the limit (\ref{GeneralFTLimit}).
If it has the form (\ref{GeneralFTLimit}) then $\Xi(\kappa_{\mathpzc{l}\mathpzc{m}})$ is decomposable.
If no discrete components, in addition to the
spectral measure appear, then we can use the simple limit formula (\ref{SimpleLimitkappaepsilonlm(phi)->kappalm(phi)}).
If the pairing (\ref{Pairing<kappa,Fx...xF>}) is convergent, then we can use still simpler formula (\ref{cchilimi}).
In particular, the formula (\ref{cchilimi}) can be used for the integral kernel operator
\[
\Xi(\kappa_{0,1}) +\Xi(\kappa_{1,0})
\]
representing free fields, as in this case $\kappa_{0,1}(\phi)$, $\kappa_{1,0}(\phi)$ belong
to $E$, if $\phi \in \mathscr{E}$, and, $\kappa_{0,1}(\phi)$, $\kappa_{1,0}(\phi)$, as functionals
over $E$, can be extended over $E^{*}$ (compare Subsection \ref{psiBerezin-Hida}). We have seen this
for the free Dirac field above in this Subsection, and the proof for the free electromagnetic potential field is given
in Subsection \ref{equivalentA-s}. In case when the $\sigma$-measure associated to $\kappa_{\mathpzc{l}\mathpzc{m}}(\phi)$
coincides with the spectral measure, \emph{e.g.} in case $\kappa_{\mathpzc{l}\mathpzc{m}}(\phi)$ can be extended to functionals
over $E^{*\otimes (\mathpzc{l}+\mathpzc{m})}$, the decomposition of $\Xi(\kappa_{\mathpzc{l}\mathpzc{m}}(\phi))$, $\phi \in \mathscr{E}$,
which, in this case, becomes an ordinary operator, coincides with the ordinary direct integral decomposition. In more general
case, there may appear additional discrete components and the decomposition cannot be understood as the ordinary
decomposition of an operator. This is what could have been expected, as generally $\Xi(\kappa_{\mathpzc{l}\mathpzc{m}}(\phi))$
is not an ordinary operator, even if $\phi \in \mathscr{E}$, but a generalization of a distribution
and of an operator, and its decomposition is similar to the decomposition of a distribution,
as we have explained above.

It would therefore be desirable to reduce the formula (\ref{GeneralFTLimit}) to a form suitable for practical calculations.
To this end we note that it is sufficient to put the simple tensor
\[
\eta = \eta_{{}_{1}} \widehat{\otimes} \ldots \widehat{\otimes}  \eta_{{}_{\mathpzc{l}+\mathpzc{m}}}
\]
into (\ref{GeneralFTLimit}), which gives 
\begin{multline}\label{GeneralFTLimitExpanded}
\underset{\epsilon \rightarrow 0}{\textrm{lim}} 
\sum\limits_{l_1m_1\ldots l_{\mathpzc{l}+\mathpzc{m}}m_{\mathpzc{l}+\mathpzc{m}}}
\int \Big\{
F_{{}_{\chi, \, l_1 m_1}}(\eta_{{}_{1}}) \ldots 
\overline{F_{{}_{\chi, \, l_{\mathpzc{l}+\mathpzc{m}} m_{\mathpzc{l}+\mathpzc{m}}}}}( \eta_{{}_{\mathpzc{l}+\mathpzc{m}}}) \, \times
\\
\,\,\,\,\,\,\,\,\,\,\,\,\,\,\,\,\,\,\,\,\,\,\,\,\,\,\,\,  \times \,
\big\langle \kappa_{\mathpzc{l}\mathpzc{m}}(\phi)_{{}_{\epsilon}},   \overline{F_{{}_{\chi, \, l_1 m_1}}} \widehat{\otimes} \ldots \widehat{\otimes} 
F_{{}_{\chi, \, l_{\mathpzc{l}+\mathpzc{m}} m_{\mathpzc{l}+\mathpzc{m}}}} \big\rangle  \Big\} \,
\ud \chi
\\
= 
\sum\limits_{l_1m_1\ldots l_{\mathpzc{l}+\mathpzc{m}}m_{\mathpzc{l}+\mathpzc{m}}}
\int \Big\{
F_{{}_{\chi, \, l_1 m_1}}(\eta_{{}_{1}}) \ldots 
\overline{F_{{}_{\chi, \, l_{\mathpzc{l}+\mathpzc{m}} m_{\mathpzc{l}+\mathpzc{m}}}}}( \eta_{{}_{\mathpzc{l}+\mathpzc{m}}}) \, \times
\\
\,\,\,\,\,\,\,\,\,\,\,\,\,\,\,\,\,\,\,\,\,\,\,\,\,\,\,\,  \times \,
\big\langle \mathcal{F}[\kappa_{\mathpzc{l}\mathpzc{m}}(\phi)](\chi, \ldots, \chi), \,  
F_{{}_{\chi, \, l_1 m_1}} \widehat{\otimes} \ldots \widehat{\otimes} 
F_{{}_{\chi, \, l_{\mathpzc{l}+\mathpzc{m}} m_{\mathpzc{l}+\mathpzc{m}}}}
 \big\rangle \Big\} \,
\ud \chi'
\end{multline}
in expanded form. 
Here $\overline{F_{{}_{\chi, \, l_i m_i}}}(\eta_{{}_{1}})$ we have canonical \emph{bilinear} pairings with the functionals defined by complex 
conjugations of the basic functions $F_{{}_{\chi, \, l_i m_i}}$, whenever 
the corresponding argument $F_{{}_{\chi, \, l_i m_i}}$ of $\kappa_{\mathpzc{l}\mathpzc{m}}(\phi)_{{}_{\epsilon}}$ enters without complex
conjugation, and \emph{vice versa}.

In order to simplify the expression (\ref{GeneralFTLimitExpanded}) we observe that there exists a system of functions
$\eta_{{}_{lm}} \in E$ such that
\[
F_{{}_{\chi, \, l' m'}}(\eta_{{}_{lm}}) = 0, \,\, \textrm{if} \,\, l\neq l' \, \textrm{or} \, m\neq m'
\]
and
\[
F_{{}_{\chi, \, l' m'}}(\eta_{{}_{lm}}) \neq 0, \,\, \textrm{if} \,\, l =l' \, \textrm{and} \, m= m'.
\]
Indeed recall please that the generalized eigen-functions $F^{\oplus}_{{}_{\chi, lm}}$, $F^{\ominus}_{{}_{\chi, \,lm}}$
of the two Casimir operators $Q,R$ are determined through (\ref{DiffEigenEqn1}). We solve them by using the standard separation of variables,
the same which is used for the solution of the Dirac equation with the spherically symmetric
Coulomb potential in the spherical coordinates $(r,\theta, \phi)$ of
\[
\boldsymbol{\p} = (r\sin \theta \cos \phi, r \sin \theta \sin \phi, r\cos \theta),
\]
and by inserting $F^{\oplus}_{{}_{\chi, lm}}$, $F^{\ominus}_{{}_{\chi, \,lm}}$ in the form of a product
bispinor $F^{\oplus}_{{}_{\chi, lm}}(r,\theta,\phi) =
F_{{}_{lm}}^{\oplus}(\theta,\phi)G^{\oplus}_{{}_{\chi, lm}}(r)$,
$F^{\ominus}_{{}_{\chi, \,lm}}(r,\theta,\phi)=F_{{}_{lm}}^{\ominus}(\theta,\phi)G^{\ominus}_{{}_{\chi, lm}}(r)$ meaning that each component
of these bispinors are equal to product
of component functions, the first one being the components depending solely on the angle coordinates and the second one on the radius
$r = |\boldsymbol{\p}|$. From (\ref{DiffEigenEqn2}) it follows that $F_{{}_{lm}}^{\oplus}(\theta,\phi)$
and $F_{{}_{lm}}^{\ominus}(\theta,\phi)$ are the bispinor
spherical harmonics with $l_0=1/2$. By the formula for the invariant pairing (\ref{InvPairingPsi}) it is sufficient to put
(with the additional complex conjugation in case we are using the canonical \emph{bilinear} pairing $\langle \cdot, \cdot \rangle$)
\[
\eta_{{}_{lm}}^{\oplus}(r,\theta,\phi) = F_{{}_{lm}}^{\oplus}(\theta,\phi)g_{{}_{lm}}^{\oplus}(r),
\,\,\, \eta_{{}_{lm}}^{\ominus}(r,\theta,\phi) = F_{{}_{lm}}^{\ominus}(\theta,\phi)g_{{}_{lm}}^{\ominus}(r),
\]
with any smooth rapidly decreasing $g_{{}_{lm}}^{\oplus}$, $g_{{}_{lm}}^{\ominus}$, when regarded
as functions of $r$. Note that in each case (of the positive and negative energy solutions) we have only
two independent components of the radial factor $G^{\oplus}_{{}_{\chi, lm}}$, $g^{\oplus}_{{}_{\chi, lm}}$ or, respectively, $G^{\oplus}_{{}_{\chi, lm}}$,
$g^{\oplus}_{{}_{\chi, lm}}$. The same is true for the angular factor, which also has only two independent components.
In order to map these functions back to the standard nuclear space we apply the isomorphism $U$:
\[
\eta_{{}_{lm}} = U\eta_{{}_{lm}}^{\oplus}, \,\,\, \textrm{and, respectively,} \,\,\,
\eta_{{}_{lm}} = U\eta_{{}_{lm}}^{\ominus}.
\]
For such $\eta_{{}_{lm}}$ the Fourier transforms $\mathcal{F}\eta_{{}_{lm}}$ of $\eta_{{}_{\chi}}$, regarded as functions
\[
\chi \mapsto F_{{}_{\chi, lm}}(\eta_{{}_{lm}})
\]
of $\chi$, are smooth and rapidly decreasing in $\chi$ and the Fourier transform $\mathcal{F}\eta_{{}_{lm}}(\chi)$
of $\eta_{{}_{lm}}$ is equal
\begin{equation}\label{F(etalm)}
\mathcal{F}\eta_{{}_{lm}}(\chi) = F_{{}_{\chi, lm}}(\eta_{{}_{lm}}) F_{{}_{\chi, lm}}.
\end{equation}
Recall once again that the value $\mathcal{F}\eta(\chi)$ of the Fourier transform at fixed $\chi$
is not a scalar (as for $\mathscr{F}$ associated to the decomposition of the Abelian group $T_n$)
but it is a conjugate linear functional over $E_{{}_{\chi}}$ (which for the Abelian group $T_n$ degenerates to $\mathbb{C}$
with $\mathbb{C}$-valued $\mathscr{F}\eta(\chi)$). Accordingly, in the formula (\ref{F(etalm)})
the second $F_{{}_{\chi, lm}}$ is understood as the antilinear functional (\ref{AntilinFchilm}). But for the valuation
$F_{{}_{\chi, lm}}(\eta_{{}_{lm}}) =
\langle F_{{}_{\chi, lm}}, \eta_{{}_{lm}} \rangle$ in (\ref{F(etalm)}) we have two \emph{a priori} possible choices:
(I) with the canonical \emph{bilinear} pairing $\langle \cdot, \cdot \rangle$ or (II) with
the invariant valuation $\langle \cdot, \cdot \rangle$, with the additional complex conjugation in the first argument. Accordingly, we will have the
corresponding inverse Fourier transform $\mathcal{F}^{-1}$ formulas. The choice is not arbitrary, and in fact both
$\mathcal{F}$ have to be used: depending on which variable (orbit $\mathscr{O}_{{}_{\boldsymbol{m}, 0,0,0}}$) of the kernel $\kappa_{\mathpzc{l}\mathpzc{m}}$
the test function $\eta_{{}_{lm}}$ depends (lives): (I) the variable (orbit) corresponding
to creation operators, or, (II) annihilation operators, as we explain it below. In the first case (I) we chose the first possibility (I)
for $\mathcal{F}$, in case (II) the second (II) $\mathcal{F}$.

Similar systems of $\eta_{{}_{lm}} \in E$ exist in the single particle
nuclear spaces $E$ of other free fields. In particular for the single particle nuclear space $E$
of the free electromagnetic potential field we can put
\[
\eta_{{}_{slm}}(r,\theta,\phi) = F^{{}^{s}}_{{}_{\chi, \, lm}}(\theta,\phi)f_{{}_{slm}}(r),
\]
where
\[
F^{{}^{s}}_{{}_{\chi, lm}}(r,\theta,\phi) = F^{{}^{s}}_{{}_{\chi, \, lm}}(\theta,\phi)r^{\chi}
\]
are the basic functionals of Subsection \ref{equivalentA-s} and $f_{{}_{slm}}$ are smooth rapidly decreasing
$\mathbb{C}$-valued functions of $r$ belonging to $S^{0}(\mathbb{R})$.

Any system of such $\eta_{{}_{lm}} \in E$ span $E$, so it is sufficient to use
\begin{equation}\label{basicetalm}
\eta = \eta_{{}_{l_1m_1}} \widehat{\otimes} \ldots \widehat{\otimes} \, 
\eta_{{}_{l_{\mathpzc{l}+\mathpzc{m}}m_{\mathpzc{l}+\mathpzc{m}}}} \in E^{\widehat{\otimes} \, (\mathpzc{l}+\mathpzc{m})}
\end{equation}
in (\ref{GeneralFTLimit}).  

Inserting $\eta$ of the form (\ref{basicetalm}) into (\ref{GeneralFTLimitExpanded}) 
we get the following simplified condition (\ref{GeneralFTLimit}): 
for each $\phi \in \mathscr{E}$
and for each $l_im_i$ ranging separately over the indices of the system $\{F_{{}_{\chi, \, l_{i} m_{i}}} \}$ of the basic functionals
we have
\begin{multline}\label{GeneralFTLimitPractical}
\underset{\epsilon \rightarrow 0}{\textrm{lim}} 
\int \Big\{
\big\langle \kappa_{\mathpzc{l}\mathpzc{m}}(\phi)_{{}_{\epsilon}},   \overline{F_{{}_{\chi, \, l_1 m_1}}} \widehat{\otimes} \ldots \widehat{\otimes} 
F_{{}_{\chi, \, l_{\mathpzc{l}+\mathpzc{m}} m_{\mathpzc{l}+\mathpzc{m}}}} \big\rangle \, \times 
\\
\times \,
F_{{}_{\chi, \, l_1 m_1}}(\eta_{{}_{l_1m_1}}) \ldots 
\overline{F_{{}_{\chi, \, l_{\mathpzc{l}+\mathpzc{m}} m_{\mathpzc{l}+\mathpzc{m}}}}}(\eta_{{}_{l_{\mathpzc{l}+\mathpzc{m}}m_{\mathpzc{l}+\mathpzc{m}}}}) \Big\}
\ud \chi
\\
= 
\int \Big\{
\big\langle \mathcal{F}[\kappa_{lm}(\phi)](\chi, \ldots, \chi), \,  
F_{{}_{\chi, \, l_1 m_1}} \widehat{\otimes} \ldots \widehat{\otimes} 
F_{{}_{\chi, \, l_{\mathpzc{l}+\mathpzc{m}} m_{\mathpzc{l}+\mathpzc{m}}}}
 \big\rangle \, \times
\\
\times \,
F_{{}_{\chi, \, l_1 m_1}}(\eta_{{}_{l_1m_1}}) \ldots 
\overline{F_{{}_{\chi, \, l_{\mathpzc{l}+\mathpzc{m}} m_{\mathpzc{l}+\mathpzc{m}}}}}(\eta_{{}_{l_{\mathpzc{l}+\mathpzc{m}}m_{\mathpzc{l}+\mathpzc{m}}}}) \Big\} \,
\ud \chi'.
\end{multline}
Introducing explicitly the Fourier transform
\begin{multline*}
\mathcal{F}[\eta](\chi_1, \ldots, \chi_{\mathpzc{l}+\mathpzc{m}}) = 
\mathcal{F}\big[\eta_{{}_{l_1m_1}} \widehat{\otimes} \ldots \widehat{\otimes} \, 
\eta_{{}_{l_{\mathpzc{l}+\mathpzc{m}}m_{\mathpzc{l}+\mathpzc{m}}}} \big](\chi_1, \ldots, \chi_{\mathpzc{l}+\mathpzc{m}})
\\
= F_{{}_{\chi_1, \, l_1 m_1}}(\eta_{{}_{l_1m_1}})
\ldots 
\overline{F_{{}_{\chi_{\mathpzc{l}+\mathpzc{m}}, \, l_{\mathpzc{l}+\mathpzc{m}} m_i}}}(\eta_{{}_{l_{\mathpzc{l}+\mathpzc{m}}m_{\mathpzc{l}+\mathpzc{m}}}}) 
\,
F_{{}_{\chi_1, \, l_1 m_1}} \widehat{\otimes} \ldots \widehat{\otimes} \,
F_{{}_{\chi_{\mathpzc{l}+\mathpzc{m}}, \, l_{\mathpzc{l}+\mathpzc{m}} m_{\mathpzc{l}+\mathpzc{m}}}} 
\end{multline*}
of 
\[
\eta =\eta_{{}_{l_1m_1}} \widehat{\otimes} \ldots \widehat{\otimes} \, 
\eta_{{}_{l_{\mathpzc{l}+\mathpzc{m}}m_{\mathpzc{l}+\mathpzc{m}}}} \in E^{\widehat{\otimes} \, (\mathpzc{l}+\mathpzc{m})},
\]
here with the Fourier transform $\mathcal{F}$ which is associated to the decomposition of $SL(2, \mathbb{C})$
acting in $E^{\widehat{\otimes} \, (\mathpzc{l}+\mathpzc{m})}$, and restricted to the diagonal $\chi_1= \ldots = \chi_{\mathpzc{l}+\mathpzc{m}}
=\chi$,
we can write the last condition in the original short form (\ref{GeneralFTLimit}): for each $\phi \in \mathscr{E}$
and for each $\eta$ of the form (\ref{basicetalm}),  
\begin{multline}\label{GeneralFTLimitFTform}
\underset{\epsilon \rightarrow 0}{\textrm{lim}} 
\int 
\big\langle \mathcal{F}\big[\kappa_{\mathpzc{l}\mathpzc{m}}(\phi)_{{}_{\epsilon}}\big],  \mathcal{F}\eta(\chi,\ldots,\chi) \big\rangle \,
\ud \chi
\\
= 
\int \Big\{
\big\langle \mathcal{F}[\kappa_{\mathpzc{l}\mathpzc{m}}(\phi)](\chi, \ldots, \chi), \,  
\mathcal{F}\eta(\chi, \ldots, \chi)
 \big\rangle \,
\ud \chi'.
\end{multline}

Condition (\ref{GeneralFTLimitPractical}) can be easily applied in practical computations, and in particular the limit
(\ref{GeneralFTLimitPractical})  can be used for  computation of
the Fourier transform $ \mathcal{F}[\kappa_{lm}(\phi)](\chi, \ldots, \chi)$ of a wide class of distributions
$\kappa_{\mathpzc{l}\mathpzc{m}}(\phi)$. The functions
\[
\textrm{Spec} \ni \chi \mapsto F_{{}_{\chi} \, l_1m_1}(\eta_{{}_{l_1m_1}}) \ldots 
\overline{F_{{}_{\chi \, l_{\mathpzc{l}+\mathpzc{m}}m_{\mathpzc{l}+\mathpzc{m}}}}}(\eta_{{}_{l_{\mathpzc{l}+\mathpzc{m}}m_{\mathpzc{l}+\mathpzc{m}}}})
\]
play the role of test functions, so that the computation of the kernel 
\[
\big\langle \mathcal{F}[\kappa_{\mathpzc{l}\mathpzc{m}}(\phi)](\chi, \ldots, \chi), \,  
F_{{}_{\chi, \, l_1 m_1}} \widehat{\otimes} \ldots \widehat{\otimes} 
F_{{}_{\chi, \, l_{\mathpzc{l}+\mathpzc{m}} m_{\mathpzc{l}+\mathpzc{m}}}}
 \big\rangle
\]
reduces to the computation essentially the same which we encounter in the computation of the distributional kernel 
in the theory of ordinary distributions of single variable $\chi \in \textrm{Spec}$. 

In general the limit on the left-hand side in (\ref{GeneralFTLimitPractical}) cannot be written
in the form standing on the right-hand side of (\ref{GeneralFTLimitPractical}) with a function like kernel, 
and any measure $\ud \chi'$ eventually different from the spectral measure
$\ud \chi$ of the decomposition of $SL(2, \mathbb{C})$ acting in $E$. 
In particular a discrete component may appear on the right-hand side
equal to evaluation of the derivatives of the test function
\[
\chi \mapsto F_{{}_{\chi} \, l_1m_1}(\eta_{{}_{l_1m_1}}) \ldots 
\overline{F_{{}_{\chi \, l_{\mathpzc{l}+\mathpzc{m}}m_{\mathpzc{l}+\mathpzc{m}}}}}(\eta_{{}_{l_{\mathpzc{l}+\mathpzc{m}}m_{\mathpzc{l}+\mathpzc{m}}}})
\]
at some points $\chi =\chi_1, \ldots, \chi_k$.
In this case the generalized operator $\Xi(\kappa_{\mathpzc{l}\mathpzc{m}})$
is not decomposable in the sense of the above Definition \ref{DecompositionOfXi}. In case in which
this limit has the form standing on the right-hand side of (\ref{GeneralFTLimitPractical})
the generalized operator $\Xi(\kappa_{\mathpzc{l}\mathpzc{m}})$ is decomposable, with the kernel on the right-hand side 
which is an ordinary function
\[
\textrm{Spec} \ni \chi \longmapsto \big\langle \mathcal{F}[\kappa_{\mathpzc{l}\mathpzc{m}}(\phi)](\chi, \ldots, \chi), \,  
F_{{}_{\chi, \, l_1 m_1}} \widehat{\otimes} \ldots \widehat{\otimes} 
F_{{}_{\chi, \, l_{\mathpzc{l}+\mathpzc{m}} m_{l+m}}}
 \big\rangle,
\] 
but in general with the measure $\ud \chi'$ different from the spectral measure $\ud \chi$. 

EXAMPLE 2. In order to illustrate this, we recall once again the simpler one dimensional case of the ordinary
Fourier transform $\mathscr{F}$ associated to the decomposition of the translation group $T_1$
acting in $\mathcal{S}(\mathbb{R}; \mathbb{C}) \subset L^2(\mathbb{R}; \mathbb{C}) \subset \mathcal{S}(\mathbb{R}; \mathbb{C})^*$.
Consider the case of the step function $\theta_{\epsilon}$ equal zero for negative real numbers and for the real numbers greater 
than $1/\epsilon$, and equal $1$ otherwise. This $\theta_{\epsilon} \in L^2(\mathbb{R}; \mathbb{C})$
converges to $\theta$ in $\mathcal{S}(\mathbb{R}; \mathbb{C})^*$ if $\epsilon \rightarrow 0$. We have
\[
\langle \theta_{\epsilon}, F_{{}_{\chi}} \rangle =  \mathscr{F}[\theta_{\epsilon}](\chi) 
= {\textstyle\frac{i}{\sqrt{2\pi} \, \chi}}-{\textstyle\frac{ie^{i\chi/\epsilon}}{\sqrt{2\pi} \, \chi}},
\,\,\,\, F_{{}_{\chi}}(x) = e^{i\chi x}, \,\, \chi \in \textrm{Spec} = \mathbb{R},
\]  
with $\theta_{\epsilon}$ regarded as a functional which can be evaluated 
\[
\langle \theta_{\epsilon}, F_{{}_{\chi}} \rangle = \int \theta_{\epsilon}(x)  e^{i\chi x}
\, \ud x
\]
at the basic plane wave functional 
$F_{{}_{\chi}}(x) = e^{i\chi x}$, because $\theta_{\epsilon} \in L^1(\mathbb{R}; \mathbb{C})$. 
We see that for each fixed
$\chi \in \textrm{Spec} = \mathbb{R}$ the numerical limit
\[
\underset{\epsilon \rightarrow 0}{\textrm{lim}}\langle \theta_{\epsilon}, F_{{}_{\chi}} \rangle
\]
does not exist. But for each $f \in \mathcal{S}(\mathbb{R}; \mathbb{C})$ the limit 
\begin{multline*}
\underset{\epsilon \rightarrow 0}{\textrm{lim}} \int\limits_{{}_{\textrm{Spec} = \mathbb{R}}}  \langle \theta_{\epsilon}, F_{{}_{\chi}} \rangle
\, \mathscr{F}f(\chi) \, \ud \chi
=
\int\limits_{{}_{\textrm{Spec} = \mathbb{R}}} \Big\langle {\textstyle\frac{i}{\sqrt{2\pi} \, \chi}}, \, \mathscr{F}f(\chi) \Big\rangle \,\, \ud\chi +
\Big\langle \sqrt{{\textstyle\frac{\pi}{2}}}, \mathscr{F}f(\chi=0)\Big\rangle
\\
= \int\limits_{{}_{\textrm{Spec} = \mathbb{R}}} \Big\langle \mathscr{F}\theta(\chi), \, \mathscr{F}f(\chi) \Big\rangle \,\, \ud\chi' 
= \langle \theta, f \rangle
\end{multline*}
does exist, and it is the analogue of the limit (\ref{GeneralFTLimitPractical}) or (\ref{GeneralFTLimitFTform}), 
but with the measure $\ud \chi' = \ud \chi \oplus \ud \theta$
which decomposes into the absolutely continuous spectral measure $\ud \chi$ and the Dirac measure $\ud \theta$
concentrated at the single point set $\{0\}$. But if instead of the functional $\theta$ $\in \mathcal{S}(\mathbb{R}; \mathbb{C})^*$
we consider, say, the functional $x$ represented by the function $x$, 
and regarded as a limit of $x_{{}_{\epsilon}} \in L^{1}(\mathbb{R}; \mathbb{C})$, 
we get on the right-hand side of the last formula
the purely discrete component concentrated at $\chi=0$ which, up to a constant factor, 
has the form of evaluation $\tfrac{d\mathscr{F}f}{d\chi}(\chi=0)$ of the derivation $\tfrac{d\mathscr{F}f}{d\chi}$ at $\chi=0$.
Thus, the analogue of the left-hand side limit  (\ref{GeneralFTLimitFTform}) for  the functional $x$
$\in \mathcal{S}(\mathbb{R}; \mathbb{C})^*$ cannot be written in the form
analogous to that on the right-hand side of (\ref{GeneralFTLimitFTform}) and the functional $x$ cannot be decomposed
in the sense indicated above. For the functional $|x|^{-1/2}$ $\in \mathcal{S}(\mathbb{R}; \mathbb{C})^*$,
regarded as a limit of $|x|^{-1/2}_{{}_{\epsilon}} \in L^{1}(\mathbb{R}; \mathbb{C})$, we obtain the form
analogous to (\ref{GeneralFTLimitFTform}) with $\ud \chi' = \ud \chi$ and with $\mathscr{F}\big[|x|^{-1/2}\big](\chi)
= |\chi|^{-1/2}$, so that $|x|^{-1/2}$ $\in \mathcal{S}(\mathbb{R}; \mathbb{C})^*$ 
is decomposable with the corresponding measure coinciding with the spectral measure $\ud \chi$,
although $|x|^{-1/2}$ $\notin L^{1}(\mathbb{R}; \mathbb{C})$, and $|x|^{-1/2}$ $\notin L^{2}(\mathbb{R}; \mathbb{C})$. 
For the functional  $1$ $\in \mathcal{S}(\mathbb{R}; \mathbb{C})^*$, represented by the constant function $1$ and
regarded as a limit of $1_{{}_{\epsilon}} \in L^{1}(\mathbb{R}; \mathbb{C})$, we obtain the form
analogous to (\ref{GeneralFTLimitFTform}) with the Dirac delta measure $\ud \chi' = \ud \theta$
concentrated at $\chi=0$, and with $\mathscr{F}\big[1\big](\chi=0) = \sqrt{2\pi}$. Thus, the functional
$1$ is decomposable with the corresponding measure equal to the Dirac delta measure concentrated on 
$\{0\} \subset \textrm{Spec} = \mathbb{R}$. \qed

The formula for the Fourier transform $\mathcal{F}\big[\kappa_{\mathpzc{l}\mathpzc{m}}(\phi)\big](\chi_1, \ldots, \chi_{\mathpzc{l}+\mathpzc{m}})$
of $\kappa_{\mathpzc{l}\mathpzc{m}}(\phi)$ $\in E^{\widehat{\otimes} \, (\mathpzc{l}+\mathpzc{m})}$
or the formula (\ref{GeneralFTLimitPractical}) follow immediately from the Fourier transform formula and 
from the formula for its inverse, provided we use the proper choice of the two possibilities for $\mathcal{F}$
on each factor $E$ in $E^{\widehat{\otimes} \, (\mathpzc{l}+\mathpzc{m})}$.
The proper choice is not arbitrary, and the correct $\mathcal{F}$ can be established by application of 
the decomposition method, which we have used for the free Dirac field, to the cass of operators
$\Xi(\kappa_{\mathpzc{l}\mathpzc{m}}(\phi))$ with kernels 
\[
\kappa_{\mathpzc{l}\mathpzc{m}}(\phi) \in E^{\widehat{\otimes} \, (\mathpzc{l}+\mathpzc{m})}
\,\,\, \textrm{for} \,\,\, \phi \in \mathscr{E}.
\]
For such operators $\Xi(\kappa_{\mathpzc{l}\mathpzc{m}}(\phi))$ we can use exactly the same decomposition method,
which we have used to the Dirac field, as in this case $\Xi(\kappa_{\mathpzc{l}\mathpzc{m}}(\phi))$ is an ordinary
operator for each $\phi \in \mathscr{E}$ and the kernel $\kappa_{\mathpzc{l}\mathpzc{m}}(\phi)$
can be evaluated at the basic functionals  for each $\phi \in \mathscr{E}$. Below we give the explicit formula for the correct $\mathcal{F}$
without the explicit computation, because it is completely analogous to the computation we have already given
in decomposition of the free Dirac field.

We are still using the canonical \emph{bilinear} pairing $F(\eta) = \langle F, \eta \rangle$.
Let $\eta \in E$. As we have already mentioned  we have two \emph{a priori} possible canonical choices of the 
Fourier transform on $E$, associated to the decomposition of $SL(2, \mathbb{C})$ acting in $E$. Then for the 
first choice (I) of the Fourier transform of  $\eta \in E$
\[
\mathcal{F}\eta(\chi) = 
\sum\limits_{lm} F_{{}_{\chi \, lm}}(\eta) F_{{}_{\chi \, lm}}
\]
we have the inverse formula
\begin{multline*}
\eta(s,\boldsymbol{\p})=
\sum\limits_{lm} \int
\overline{F_{{}_{\chi \, lm}}(s,\boldsymbol{\p})} F_{{}_{\chi \, lm}}(\eta)
 \, \ud\chi
\\
=
\sum\limits_{lm} \int
\overline{F_{{}_{\chi \, lm}}(s,\boldsymbol{\p})} \, \big\langle \mathcal{F}\eta(\chi), \, F_{{}_{\chi \, lm}}\big\rangle
 \, \ud\chi.
\end{multline*}
More generally, let $\eta  \in E^{\widehat{\otimes} \, k}$. For the Fourier transform
\begin{multline*}
\mathcal{F}\eta(\chi_1, \ldots, \chi_k) = 
\\
= 
\sum\limits_{l_1m_1, \ldots l_km_k}
\big(
F_{{}_{\chi_1, \, l_1 m_1}} \widehat{\otimes} 
\ldots \widehat{\otimes} \,
F_{{}_{\chi_{k}, \, l_k m_k}}
\big)(\eta) 
\,
F_{{}_{\chi_1, \, l_1 m_1}} \widehat{\otimes} 
\ldots \widehat{\otimes} \,
F_{{}_{\chi_{k}, \, l_k m_k}}
\end{multline*}
we have the inverse formula
\begin{multline*}
\eta(s_1, \boldsymbol{\p}_1, \ldots, s_k, \boldsymbol{\p}_k)
=
\sum\limits_{l_1m_1, \ldots l_km_k}
\int
\overline{
F_{{}_{\chi_1, \, l_1 m_1}}(s_1, \boldsymbol{\p}_1)} 
\ldots \,
\overline{
F_{{}_{\chi_{k}, \, l_k m_k}}(s_k, \boldsymbol{\p}_k)
}
\, \times \,
\\
\times \,
\big(
F_{{}_{\chi_1, \, l_1 m_1}} \widehat{\otimes} 
\ldots \widehat{\otimes} \,
F_{{}_{\chi_{k}, \, l_k m_k}}
\big)(\eta) \,
\ud\chi_1 \ldots \ud\chi_k
\end{multline*}
\begin{multline*}
=
\sum\limits_{l_1m_1, \ldots l_km_k}
\int
\overline{
F_{{}_{\chi_1, \, l_1 m_1}}(s_1, \boldsymbol{\p}_1)} 
\ldots \,
\overline{
F_{{}_{\chi_{k}, \, l_k m_k}}(s_k, \boldsymbol{\p}_k)
}
\, \times \,
\\
\times \,
\Big\langle
\mathcal{F}\eta(\chi_1, \ldots, \chi_k),
\,\,
F_{{}_{\chi_1, \, l_1 m_1}} \widehat{\otimes} 
\ldots \widehat{\otimes} \,
F_{{}_{\chi_{k}, \, l_k m_k}}
\Big\rangle \,
\ud\chi_1 \ldots \ud\chi_k,
\end{multline*}
here with $\mathcal{F}$ associated with the decomposition of $SL(2,\mathbb{C})$ acting in $E^{\widehat{\otimes} \, k}$.
In these formulas for the Fourier transform and its inverse, $\ud\chi$ and $\ud\chi_i$ denote the spectral
measure of the decomposition of $SL(2,\mathbb{C})$ acting in $E$. 

But for $\eta \in E$ we have the second canonical choice (II) for the Fourier transform 
of  $\eta \in E$:
\[
\mathcal{F}\eta(\chi) = 
\sum\limits_{lm} \overline{F_{{}_{\chi \, lm}}}(\eta) F_{{}_{\chi \, lm}}
\]
with the inverse formula
\begin{multline*}
\eta(s,\boldsymbol{\p})=
\sum\limits_{lm} \int
F_{{}_{\chi \, lm}}(s,\boldsymbol{\p}) \overline{F_{{}_{\chi \, lm}}}(\eta)
 \, \ud\chi
\\
=
\sum\limits_{lm} \int
F_{{}_{\chi \, lm}}(s,\boldsymbol{\p}) \, \big\langle \mathcal{F}\eta(\chi), \, F_{{}_{\chi \, lm}}\big\rangle
 \, \ud\chi.
\end{multline*}
Accordingly, for the Fourier transform of  $\eta  \in E^{\widehat{\otimes} \, (\mathpzc{l}+\mathpzc{m})}$
we have $2^{(\mathpzc{l}+\mathpzc{m})}$ \emph{a priori} possible choices, depending on  
each choice, (I) or (II), on each tensor product factor $E$. 
Let 
\[
\eta = \eta_{{}_{1}} \widehat{\otimes}  \ldots \widehat{\otimes} \, \eta_{{}_{\mathpzc{l}}} \otimes  \eta_{{}_{\mathpzc{l}+1}}
\widehat{\otimes} \ldots \widehat{\otimes} \, \eta_{{}_{\mathpzc{l}+\mathpzc{m}}}.
\]
Whenever computing $\mathcal{F}\big[\kappa_{\mathpzc{l}\mathpzc{m}}(\phi)\big]$ we need to pick up the correct 
choice for $\mathcal{F}\eta$. The correct choice for $\mathcal{F}\eta$ is the following
\begin{multline}\label{CorrectFT}
\mathcal{F}\eta(\chi_1, \ldots, \chi_{\mathpzc{l}+\mathpzc{m}}) = 
\\
= 
\sum\limits_{l_1m_1, \ldots l_{\mathpzc{l}+\mathpzc{m}}m_{\mathpzc{l}+\mathpzc{m}}}
\Big(
F_{{}_{\chi_1, \, l_1 m_1}} \widehat{\otimes} 
\ldots \widehat{\otimes} \,
F_{{}_{\chi_{\mathpzc{l}}, \, l_{\mathpzc{l}} m_{\mathpzc{l}}}}
\otimes 
\overline{F_{{}_{\chi_{\mathpzc{l}+1}, \, l_{\mathpzc{l}+1} m_{\mathpzc{l}+1}}}}
\widehat{\otimes} \ldots
\widehat{\otimes} \,
\overline{F_{{}_{\chi_{\mathpzc{l}+\mathpzc{m}}, \, l_{\mathpzc{l}+\mathpzc{m}} m_{\mathpzc{l}+\mathpzc{m}}}}}
\Big)(\eta) 
\, \times
\\
\times \,
F_{{}_{\chi_1, \, l_1 m_1}} \widehat{\otimes} 
\ldots \widehat{\otimes} \,
F_{{}_{\chi_{\mathpzc{l}+\mathpzc{m}}, \, l_{\mathpzc{l}+\mathpzc{m}} m_{\mathpzc{l}+\mathpzc{m}}}}
\end{multline}
with the inverse formula
\begin{multline}\label{InverseCorrectFT}
\eta(s_1, \boldsymbol{\p}_1, \ldots, s_{\mathpzc{l}+\mathpzc{m}}, \boldsymbol{\p}_{\mathpzc{l}+\mathpzc{m}})
=
\sum\limits_{l_1m_1, \ldots l_{\mathpzc{l}+\mathpzc{m}}m_{\mathpzc{l}+\mathpzc{m}}}
\int
\overline{
F_{{}_{\chi_1, \, l_1 m_1}}(s_1, \boldsymbol{\p}_1)} 
\ldots \,
F_{{}_{\chi_{k}, \, l_k m_k}}(s_k, \boldsymbol{\p}_k)
\, \times \,
\\
\times \,
\Big(
F_{{}_{\chi_1, \, l_1 m_1}} \widehat{\otimes} 
\ldots \widehat{\otimes} \,
\overline{F_{{}_{\chi_{\mathpzc{l}+\mathpzc{m}}, \, l_{\mathpzc{l}+\mathpzc{m}} m_{\mathpzc{l}+\mathpzc{m}}}}} \,
\Big)(\eta) \,
\ud\chi_1 \ldots \ud\chi_{\mathpzc{l}+\mathpzc{m}}
\end{multline}
\begin{multline*}
=
\sum\limits_{l_1m_1, \ldots l_km_k}
\int
\overline{
F_{{}_{\chi_1, \, l_1 m_1}}(s_1, \boldsymbol{\p}_1)} 
\ldots \,
F_{{}_{\chi_{\mathpzc{l}+\mathpzc{m}}, \, l_{\mathpzc{l}+\mathpzc{m}} m_{\mathpzc{l}+\mathpzc{m}}}}
(s_{\mathpzc{l}+\mathpzc{m}}, \boldsymbol{\p}_{\mathpzc{l}+\mathpzc{m}})
\, \times \,
\\
\times \,
\Big\langle
\mathcal{F}\eta(\chi_1, \ldots, \chi_{\mathpzc{l}+\mathpzc{m}}),
\,\,
F_{{}_{\chi_1, \, l_1 m_1}} \widehat{\otimes} 
\ldots \widehat{\otimes} \,
F_{{}_{\chi_{\mathpzc{l}+\mathpzc{m}}, \, l_{\mathpzc{l}+\mathpzc{m}} m_{\mathpzc{l}+\mathpzc{m}}}}
\Big\rangle \,
\ud\chi_1 \ldots \ud\chi_{\mathpzc{l}+\mathpzc{m}},
\end{multline*}

By definition of the Fourier transform $\mathcal{F}\big[\kappa_{\mathpzc{l}\mathpzc{m}}(\phi)_{{}_{\epsilon}}\big]$  
of $\kappa_{\mathpzc{l}\mathpzc{m}}(\phi)_{{}_{\epsilon}} \in E^{\widehat{\otimes} \, (\mathpzc{l}+\mathpzc{m})}$,
regarded as a distribution, we have
\begin{multline}\label{FTkappaepsilon}
\Big\langle \kappa_{\mathpzc{l}\mathpzc{m}}(\phi)_{{}_{\epsilon}}, \eta \Big\rangle
\\
=
\int 
\Big\langle \mathcal{F}\big[
\kappa_{\mathpzc{l}\mathpzc{m}}(\phi)_{{}_{\epsilon}}
\big](\chi_1,\ldots,\chi_{\mathpzc{l}+\mathpzc{m}}), \, \mathcal{F}\eta(\chi_1,\ldots,\chi_{\mathpzc{l}+\mathpzc{m}}) \Big\rangle \,
\ud \chi_i \ldots \ud\chi_{\mathpzc{l}+\mathpzc{m}},
\end{multline}
for 
\begin{equation}\label{CanonicalSimpleTensoreta}
\eta = \eta_{{}_{l_1m_1}} \otimes \ldots 
\otimes \eta_{{}_{l_{\mathpzc{l}+\mathpzc{m}}m_{\mathpzc{l}+\mathpzc{m}}}}.
\end{equation}
On the other hand 
\begin{multline}\label{PairingInt.kappalmepsilon}
\Big\langle \kappa_{\mathpzc{l}\mathpzc{m}}(\phi)_{{}_{\epsilon}}, \eta_{{}_{l_1m_1}} \widehat{\otimes} \ldots 
\widehat{\otimes} \eta_{{}_{l_{\mathpzc{l}+\mathpzc{m}}m_{\mathpzc{l}+\mathpzc{m}}}} \Big\rangle
\\
=
\sum\limits_{s_1 \ldots s_{\mathpzc{l}+\mathpzc{m}}}
\int \Big\{
\kappa_{\mathpzc{l}\mathpzc{m}}(\phi)_{{}_{\epsilon}}(s_1, \boldsymbol{\p}_1, \ldots, s_{\mathpzc{l}+\mathpzc{m}}, \boldsymbol{\p}_{\mathpzc{l}+\mathpzc{m}})
\, \times \,
\\
\times \,
\eta_{{}_{l_1m_1}}(s_1, \boldsymbol{\p}_1) \ldots 
\eta_{{}_{l_{\mathpzc{l}+\mathpzc{m}}m_{\mathpzc{l}+\mathpzc{m}}}}(s_{\mathpzc{l}+\mathpzc{m}}, \boldsymbol{\p}_{\mathpzc{l}+\mathpzc{m}})
\Big\}
\, \ud \boldsymbol{\p}_{1} \ldots 
\ud \boldsymbol{\p}_{\mathpzc{l}+\mathpzc{m}}.
\end{multline}
Inserting here $\eta$ of the form (\ref{CanonicalSimpleTensoreta}), eventually antisymmetrized or symmetrized, 
expressed through the inverse formula (\ref{InverseCorrectFT}),
\begin{multline*}
\eta(s_1, \boldsymbol{\p}_1, \ldots, s_{\mathpzc{l}+\mathpzc{m}}, \boldsymbol{\p}_{\mathpzc{l}+\mathpzc{m}}) =
\big(\eta_{{}_{l_1m_1}} \otimes \ldots 
\otimes \eta_{{}_{l_{\mathpzc{l}+\mathpzc{m}}m_{\mathpzc{l}+\mathpzc{m}}}}\big)
(s_1, \boldsymbol{\p}_1, \ldots, s_{\mathpzc{l}+\mathpzc{m}}, \boldsymbol{\p}_{\mathpzc{l}+\mathpzc{m}})=
\\
=
\eta_{{}_{l_1m_1}}(s_1, \boldsymbol{\p}_1) \ldots 
\eta_{{}_{l_{\mathpzc{l}+\mathpzc{m}}m_{\mathpzc{l}+\mathpzc{m}}}}(s_{\mathpzc{l}+\mathpzc{m}}, \boldsymbol{\p}_{\mathpzc{l}+\mathpzc{m}})
=
\int
\overline{F_{{}_{\chi_1 \, l_im_i}}(s_1,\boldsymbol{\p}_1)} F_{{}_{\chi_1 \, l_1m_1}}(\eta_{{}_{l_1m_1}})
\ldots \, \times
\\
\times \,
F_{{}_{\chi_{\mathpzc{l}+\mathpzc{m}} \, l_{\mathpzc{l}+\mathpzc{m}}m_{\mathpzc{l}+\mathpzc{m}}}}
(s_{\mathpzc{l}+\mathpzc{m}},\boldsymbol{\p}_{\mathpzc{l}+\mathpzc{m}}) 
\overline{F_{{}_{\chi_{\mathpzc{l}+\mathpzc{m}} \, l_{\mathpzc{l}+\mathpzc{m}}m_{\mathpzc{l}+\mathpzc{m}}}}}
(\eta_{{}_{l_{\mathpzc{l}+\mathpzc{m}}m_{\mathpzc{l}+\mathpzc{m}}}})
 \, \ud\chi_1 \ldots \ud\chi_{\mathpzc{l}+\mathpzc{m}},
\end{multline*}
(eventually antisymmetrized or symmetrized) we get
\begin{multline*}
\mathcal{F}\big[
\kappa_{\mathpzc{l}\mathpzc{m}}(\phi)_{{}_{\epsilon}}
\big](\chi_1,\ldots,\chi_{\mathpzc{l}+\mathpzc{m}}) 
\\
= 
\big\langle \kappa_{\mathpzc{l}\mathpzc{m}}(\phi)_{{}_{\epsilon}},   \overline{F_{{}_{\chi_1, \, l_1 m_1}}} \widehat{\otimes} \ldots \widehat{\otimes} 
F_{{}_{\chi_{\mathpzc{l}+\mathpzc{m}}, \, l_{\mathpzc{l}+\mathpzc{m}} m_{\mathpzc{l}+\mathpzc{m}}}} \big\rangle
\,
F_{{}_{\chi_1, \, l_1 m_1}} \widehat{\otimes} \ldots \widehat{\otimes} 
F_{{}_{\chi_{\mathpzc{l}+\mathpzc{m}}, \, l_{\mathpzc{l}+\mathpzc{m}} m_{\mathpzc{l}+\mathpzc{m}}}} ,
\end{multline*}
or 
\begin{multline*}
\mathcal{F}\big[
\kappa_{\mathpzc{l}\mathpzc{m}}(\phi)_{{}_{\epsilon}}
\big](\chi,\ldots,\chi) 
\\
=
\big\langle \kappa_{\mathpzc{l}\mathpzc{m}}(\phi)_{{}_{\epsilon}},   \overline{F_{{}_{\chi, \, l_1 m_1}}} \widehat{\otimes} \ldots \widehat{\otimes} 
F_{{}_{\chi, \, l_{\mathpzc{l}+\mathpzc{m}} m_{\mathpzc{l}+\mathpzc{m}}}} \big\rangle
\,
F_{{}_{\chi, \, l_1 m_1}} \widehat{\otimes} \ldots \widehat{\otimes} 
F_{{}_{\chi, \, l_{\mathpzc{l}+\mathpzc{m}} m_{\mathpzc{l}+\mathpzc{m}}}},
\end{multline*}
and (\ref{GeneralFTLimitPractical}).

Frequently, 
(e.g. for the invariant kernels $\kappa_{\mathpzc{l}\mathpzc{m}} \in E^{* \widehat{\otimes} \, (\mathpzc{l}+\mathpzc{m})}$),
it is sufficient to insert $\eta$ of the form (\ref{CanonicalSimpleTensoreta}) expressed through the inverse
formula into the pairing integral
\begin{multline*}
\Big\langle \kappa_{\mathpzc{l}\mathpzc{m}}, \eta_{{}_{l_1m_1}} \otimes \ldots 
\otimes \eta_{{}_{l_{\mathpzc{l}+\mathpzc{m}}m_{\mathpzc{l}+\mathpzc{m}}}} \Big\rangle
\\
=
\sum\limits_{s_1 \ldots s_{\mathpzc{l}+\mathpzc{m}}}
\int \Big\{
\kappa_{\mathpzc{l}\mathpzc{m}}(s_1, \boldsymbol{\p}_1, \ldots, s_{\mathpzc{l}+\mathpzc{m}}, \boldsymbol{\p}_{\mathpzc{l}+\mathpzc{m}})
\, \times \,
\\
\times \,
\eta_{{}_{l_1m_1}}(s_1, \boldsymbol{\p}_1) \ldots 
\eta_{{}_{l_{\mathpzc{l}+\mathpzc{m}}m_{\mathpzc{l}+\mathpzc{m}}}}(s_{\mathpzc{l}+\mathpzc{m}}, \boldsymbol{\p}_{\mathpzc{l}+\mathpzc{m}})
\Big\}
\, \ud \boldsymbol{\p}_{1} \ldots 
\ud \boldsymbol{\p}_{\mathpzc{l}+\mathpzc{m}}
\end{multline*}
for the computation of $\mathcal{F}[\kappa_{\mathpzc{l}\mathpzc{m}}]$. The same situation we have for a wide class
of vector-valued kernels $\kappa_{\mathpzc{l}\mathpzc{m}}$, where it is sufficient to insert
$\eta$, expressed through the inverse formula, into the pairing integral
\begin{multline*}
\Big\langle \kappa_{\mathpzc{l}\mathpzc{m}}(\phi), \eta_{{}_{l_1m_1}} \otimes \ldots 
\otimes \eta_{{}_{l_{\mathpzc{l}+\mathpzc{m}}m_{\mathpzc{l}+\mathpzc{m}}}} \Big\rangle
\\
=
\sum\limits_{s_1 \ldots s_{\mathpzc{l}+\mathpzc{m}}}
\int \Big\{
\kappa_{\mathpzc{l}\mathpzc{m}}(\phi)(s_1, \boldsymbol{\p}_1, \ldots, s_{\mathpzc{l}+\mathpzc{m}}, \boldsymbol{\p}_{\mathpzc{l}+\mathpzc{m}})
\, \times \,
\\
\times \,
\eta_{{}_{l_1m_1}}(s_1, \boldsymbol{\p}_1) \ldots 
\eta_{{}_{l_{\mathpzc{l}+\mathpzc{m}}m_{\mathpzc{l}+\mathpzc{m}}}}(s_{\mathpzc{l}+\mathpzc{m}}, \boldsymbol{\p}_{\mathpzc{l}+\mathpzc{m}})
\Big\}
\, \ud \boldsymbol{\p}_{1} \ldots 
\ud \boldsymbol{\p}_{\mathpzc{l}+\mathpzc{m}},
\end{multline*}
for the computation of $\mathcal{F}[\kappa_{\mathpzc{l}\mathpzc{m}}(\phi)]$.

EXAMPLE 3. As an example we consider an integral kernel operator $\Xi(\kappa_{1,1})$, with a scalar-valued kernel
$\kappa_{1,1}$ in the Fock space of the scalar complex (charged) massive free field
\begin{multline}\label{complexscalarpsi}
\boldsymbol{\psi}(x) =
{\textstyle\frac{1 }{(2\pi)^{3/2}}}
\int {\textstyle\frac{e^{-ip\cdot x} }{\sqrt{2p_0(\boldsymbol{\p})}}}
\, b(\boldsymbol{\p}) \ud^3\boldsymbol{\p}
+
{\textstyle\frac{1 }{(2\pi)^{3/2}}}
\int {\textstyle\frac{e^{ip\cdot x} }{\sqrt{2p_0(\boldsymbol{\p})}}}
\, d(\boldsymbol{\p})^{+} \ud^3\boldsymbol{\p}
\\
=
\Xi(\kappa_{0,1}(x)) +\Xi(\kappa_{1,0}(x)),
\end{multline}
\[
\kappa_{0,1}(\boldsymbol{\p}; x) = {\textstyle\frac{e^{-ip\cdot x} }{(2\pi)^{3/2}\sqrt{2p_0(\boldsymbol{\p})}}},
\,\,\,
\kappa_{1,0}(\boldsymbol{\p}; x) = {\textstyle\frac{e^{ip\cdot x} }{(2\pi)^{3/2}\sqrt{2p_0(\boldsymbol{\p})}}},
\]
and with the Hida annihilation-creation operators $b(\boldsymbol{\p})$, $b(\boldsymbol{\p})^+$,
$d(\boldsymbol{\p})$, $d(\boldsymbol{\p})^+$
which respect the canonical commutation rules
\[
[b(\boldsymbol{\p}), b(\boldsymbol{\p}')^+]_{-} = [d(\boldsymbol{\p}), d(\boldsymbol{\p}')^+]_{-} = \delta(\boldsymbol{\p}-\boldsymbol{\p}')
\]
with all remaining pairs of the creation-annihilation operators commuting.
For simplicity, we put
the mass of the field $\boldsymbol{m}=1$. The single particle nuclear space
\[
E^\oplus \oplus E^{\ominus \, \flat}
\]
is equal to the direct sum of positive energy solutions and conjugated negative energy solutions, which are Schwartz
$\mathbb{C}$-valued functions in the Cartesian coordinates $\boldsymbol{\p}$ of the spatial momentum $\boldsymbol{\p}$ on the positive
energy sheet $\mathscr{O}_{{}_{1,0,0,0}}$ of mass
$\boldsymbol{m}=1$ hyperboloid, with the invariant measure
\[
\ud \mu_{{}_{\mathscr{O}_{{}_{1,0,0,0}}}}(\boldsymbol{\p}) = {\textstyle\frac{\ud^3\boldsymbol{\p}}{p_0(\boldsymbol{\p})}},
\,\,\,\, p_0(\boldsymbol{\p}) = \sqrt{|\boldsymbol{\p}|^2+1^2}
\]
on $\mathscr{O}_{{}_{1,0,0,0}}$. The single particle Hilbert space $\mathcal{H}' = \mathcal{H}^\oplus \oplus \mathcal{H}^{\ominus \, \flat}$
is the direct sum of square integrable
functions $f$ on $\mathscr{O}_{{}_{1,0,0,0}}$ with respect to the invariant measure
and conjugations $g^\flat$ of square integrable functions $g$ on $\mathscr{O}_{{}_{-1,0,0,0}}$. Recall
that here in the scalar case the conjugation degenerates to $g\flat(p) = \overline{g(-p)}$, compare Subsection \ref{psiBerezin-Hida}.
$\mathcal{H}' = \mathcal{H}^\oplus \oplus \mathcal{H}^{\ominus \, \flat}$ is almost in the standard form
(in this scalar field case), so that the unitary isomorphism $U$, given by (\ref{isomorphismU}) of Subsection \ref{psiBerezin-Hida}
for the Dirac field, degenerates just to the operator of multiplication by $\tfrac{1}{\sqrt{p_0(\boldsymbol{\p})}}$.
Due to the direct sum decomposition of the single particle nuclear space and its standard form
\[
UE^\oplus \oplus UE^{\ominus \, \flat} = E^+ \oplus E^-,
\]
each kernel $\kappa_{\mathpzc{l},\mathpzc{m}} \in E^{*\widehat{\otimes} \, (\mathpzc{l}+\mathpzc{m})}$
can be decomposed into the finite direct sum of kernels, respectively, in $E^{\pm *\widehat{\otimes} \, (\mathpzc{l}+\mathpzc{m})}$.
Now we consider an invariant scalar-valued kernel $\kappa_{1,1}$ of the general form depending
on real positive parameter $t$:
\[
\begin{split}
\kappa_{1,1}(+,\boldsymbol{\p}_1,+,\boldsymbol{\p}_2 ) =
{\textstyle\frac{1}{\sqrt{p_0(\boldsymbol{\p}_1)}}}{\textstyle\frac{1}{\sqrt{p_0(\boldsymbol{\p}_2)}}}\kappa_t(\boldsymbol{\p}_1,\boldsymbol{\p}_2),
\\
\kappa_{1,1}(-,\boldsymbol{\p}_1,-,\boldsymbol{\p}_2 ) = \kappa_{1,1}(+,\boldsymbol{\p}_1,-,\boldsymbol{\p}_2 )
= \kappa_{1,1}(-,\boldsymbol{\p}_1,+,\boldsymbol{\p}_2 ) =0,
\end{split}
\]
which means that all components of the kernel $\kappa_{1,1}$ are zero
except the component in $E^{+*}\otimes E^{+*}$ so that the corresponding operator
$\Xi(\kappa_{1,1})$ contains only the product $b(\boldsymbol{\p}_1)^+b(\boldsymbol{\p}_2)$
of annihilation-creation operators of the particle, not involving the creation-annihilation operators
$d(\boldsymbol{\p}_1)^+,d(\boldsymbol{\p}_2)$
of the anti-particle. Let
\[
\kappa_t(\boldsymbol{\p}_1,\boldsymbol{\p}_2) = e^{-t\{\lambda(\boldsymbol{\p}_1,\boldsymbol{\p}_2)
\textrm{coth} [\lambda(\boldsymbol{\p}_1,\boldsymbol{\p}_2)] -1\}}
\]
be an invariant kernel on $\mathscr{O}_{{}_{1,0,0,0}}$, which can be identified with the Lobachevsky space
with constant curvature equal $-\boldsymbol{m}=-1$. Here $\lambda(\boldsymbol{\p}_1,\boldsymbol{\p}_2)$ is the
hyperbolic angle between the corresponding points $p_1, p_2 \in \mathscr{O}_{{}_{1,0,0,0}}$:
\[
\textrm{cosh}[\lambda(\boldsymbol{\p}_1,\boldsymbol{\p}_2)] = p_1\cdot p_2
= p_0(\boldsymbol{\p}_1)p_0(\boldsymbol{\p}_2) - \boldsymbol{\p}_1 \cdot \boldsymbol{\p}_2.
\]

To the basic functionals $F^{\oplus}_{{}_{\chi \, lm}}, F^{\ominus \, \flat}_{{}_{\chi \, lm}}$,
associated to the decomposition of $SL(2, \mathbb{C})$ acting respectively in $E^{\oplus}$ or $E^{\ominus \, \flat}$,
we associate the system of their standard form counterparts (as explained above)
\[
\{F_{{}_{\chi \, lm}} \} = \{UF^{\oplus}_{{}_{\chi \, lm}} \} \cup \{UF^{\ominus \, \flat}_{{}_{\chi \, lm}} \}
=\{F^{+}_{{}_{\chi \, lm}} \} \cup \{F^{-}_{{}_{\chi \, lm}} \}.
\]
Similarly for the elements $f\oplus g^{\flat} \in \mathcal{H}'$ we have their standard form counterparts
\[
Uf\oplus Ug^{\flat} \in L^2(\mathbb{R}^3; \mathbb{C}) \oplus L^2(\mathbb{R}^3; \mathbb{C}),
\]
compare Subsection \ref{psiBerezin-Hida}.
Note that for the canonical \emph{bilinear} pairing, defined by the invariant integration or, respectively, the
standard integration $\ud^3 \boldsymbol{\p}$, we have
\[
F^{\oplus}_{{}_{\chi \, lm}} (f) = UF^{\oplus}_{{}_{\chi \, lm}} (Uf) = F^{+}_{{}_{\chi \, lm}}(\eta),
\,\,\,
F^{\ominus \flat}_{{}_{\chi \, lm}} (g^\flat) = UF^{\ominus \flat}_{{}_{\chi \, lm}}(Ug^\flat) = F^{-}_{{}_{\chi \, lm}}(\xi),
\]
where
\[
\eta = Uf, \,\,\, \xi = Ug.
\]

For $t>1$, the Fourier transform $\mathcal{F}[\kappa_{1,1}]$ can be found just by insertion of
$\eta_{{}_{1}} \otimes \eta_{{}_{2}}$ expressed through the  inverse Fourier transform
into the pairing 
\begin{equation}\label{<kappa1,1,eta1xeta2>}
\big\langle \kappa_{1,1},  \eta_{{}_{1}} \otimes \eta_{{}_{2}} \big\rangle,
\end{equation}
and for $t>1$, by the analytic continuation of the distribution $\mathcal{F}[\kappa_{1,1}]$
computed for $t>1$, and which has been done in \cite{Staruszkiewicz1992ERRATUM}.  

Using the basic functionals $F^{\oplus}_{{}_{\chi \, lm}}$ we can rewrite the Fourier transform
of $f \in E^\oplus$ on the Lobachevsky space $\mathscr{O}_{{}_{1,0,0,0}}$
(found in \cite{GelfandV}, where we have used $\nu$ instead of $\nu/2$ in \cite{GelfandV},
with the variables $k,k'$ ranging over the unit sphere $\mathbb{S}^2$ in the cone $k\cdot k =0$,
with the ordinary rotationally invariant measure $d^2k$ on $\mathbb{S}^2$
and with $p \in \mathscr{O}_{{}_{1,0,0,0}}$)
in the following form:
\begin{multline*}
\mathcal{F}f(k;\nu) = \int\limits_{\mathscr{O}_{{}_{1,0,0,0}}} \ud \mu_{{}_{\mathscr{O}_{{}_{1,0,0,0}}}}(p) f(p) (k\cdot p)^{i\nu-1}
\\
=
\sum\limits_{lm}
\int\limits_{\mathscr{O}_{{}_{1,0,0,0}}} \ud \mu_{{}_{\mathscr{O}_{{}_{1,0,0,0}}}}(p) \int\limits_{\mathbb{S}^2} d^2k' f(p) (k'\cdot p)^{i\nu-1}
\overline{Y_{{}_{lm}}(k')} \,\,\, Y_{{}_{lm}}(k)
\\
=
\sum\limits_{lm} F^{\oplus}_{{}_{\chi=-1+i\nu \, lm}}(f) \,\, Y_{{}_{lm}}(k),
\end{multline*}
where (in case of the scalar field) the basic functionals are defined through the following (asymptotically)
homogeneous functions
\[
F^{\oplus}_{{}_{\chi=-1+i\nu \, lm}}(p) = \int\limits_{\mathbb{S}^2} d^2k f(p) (k\cdot p)^{i\nu-1}
\overline{Y_{{}_{lm}}(k)}, \,\,\,\, p \in \mathscr{O}_{{}_{1,0,0,0}}.
\]
Using $F^{\oplus}_{{}_{\chi \, lm}}$ we can rewrite the inverse formula in the following form
\[
f(p) = \sum\limits_{lm} \int\limits_{\textrm{Spec}} \ud\chi \overline{F^{\oplus}_{{}_{\chi=-1+i\nu, \, lm}}(p)} F^{\oplus}_{{}_{\chi=-1+i\nu, \, lm}}(f), \,\,\,
\ud\chi = {\textstyle\frac{1}{(2\pi)^3}} \nu^2 \ud \nu,
\]
with $\textrm{Spec} = \{\chi=-1+i\nu, \nu \in \mathbb{R}_+ \}$.
Here
$p = \big(p_0(\boldsymbol{\p}), \boldsymbol{\p}\big) \in \mathscr{O}_{{}_{1,0,0,0}}$.

Let $f_{{}_{1}},f_{{}_{2}} \in E^\oplus$ and $\eta_{{}_{1}} = Uf_{{}_{1}}$, $\eta_{{}_{2}} = Uf_{{}_{2}}$.

In case $t>1$ we insert the inverse formula into the pairing integral (\ref{<kappa1,1,eta1xeta2>}). The result is that 
obtained in \cite{Staruszkiewicz1992ERRATUM}, which can be rephrased in the following form
\begin{multline*}
\int\limits_{\mathscr{O}_{{}_{1,0,0,0}}} \kappa_t(\boldsymbol{\p}_1,\boldsymbol{\p}_2) 
f_1(\boldsymbol{\p}_1)f_2(\boldsymbol{\p}_2) \ud \mu_{{}_{\mathscr{O}_{{}_{1,0,0,0}}}}(\boldsymbol{\p}_1)
\ud \mu_{{}_{\mathscr{O}_{{}_{1,0,0,0}}}}(\boldsymbol{\p}_2) 
\\
= \int \kappa_{1,1}(+,\boldsymbol{\p}_1,+,\boldsymbol{\p}_2) \eta_{{}_{1}}(\boldsymbol{\p}_1)\eta_{{}_{2}}(\boldsymbol{\p}_2)
\ud^3 \boldsymbol{\p}_1 \ud^3\boldsymbol{\p}_2
\end{multline*}
\begin{multline}\label{GeneralFTLimitExpandedAS>1}
=
\sum\limits_{l_1m_1l_2m_2} \, \int\limits_{\textrm{Spec}} K(\chi;t) \delta_{{}_{l_1l_2}}\delta_{{}_{m_1m_2}} F^{\oplus}_{{}_{\chi l_1m_1}}(f_{{}_{1}})
\overline{F^{\oplus}_{{}_{\chi l_2m_2}}}(f_{{}_{2}}) \, \ud \chi
\\
=
\sum\limits_{l_1m_1l_2m_2} \, \int\limits_{\textrm{Spec}} K(\chi;t) \delta_{{}_{l_1l_2}}\delta_{{}_{m_1m_2}} F_{{}_{\chi l_1m_1}}(\eta_{{}_{1}})
\overline{F_{{}_{\chi l_2m_2}}}(\eta_{{}_{2}}) \, \ud\chi
\end{multline}
for $t>1$, and (by analytic continuation with $t = t+i0$ treated as a variable in the compex plane)
\begin{multline}\label{GeneralFTLimitExpandedAS<1}
\int \kappa_{1,1}(+,\boldsymbol{\p}_1,+,\boldsymbol{\p}_2) \eta_{{}_{1}}(\boldsymbol{\p}_1)\eta_{{}_{2}}(\boldsymbol{\p}_2)
\ud^3 \boldsymbol{\p}_1 \ud^3\boldsymbol{\p}_2
\\
=
\sum\limits_{l_1m_1l_2m_2} \, \int\limits_{\textrm{Spec}} K(\chi;t) \delta_{{}_{l_1l_2}}\delta_{{}_{m_1m_2}} F_{{}_{\chi l_1m_1}}(\eta_{{}_{1}})
\overline{F_{{}_{\chi l_2m_2}}}(\eta_{{}_{2}}) \, \ud\chi
\\
+
{\textstyle\frac{(1-t^2)(2e)^t}{16\pi^2}} \sum\limits_{l_1m_1l_2m_2} 
\int {\textstyle\frac{d^2kd^2q}{(k\cdot q)^t}}\overline{Y_{{}_{l_1m_1}}(k)}Y_{{}_{l_2m_2}}(q)
\,\,
 F_{{}_{\chi=-2+t, \, l_1m_1}}(\eta_{{}_{1}})
\overline{F_{{}_{\chi=-2+t, \, l_2m_2}}}(\eta_{{}_{2}}), 
\end{multline}
for $0<t<1$, and note that sums here are not zero only for the positive energy basic functionals
$F^{+}_{{}_{\chi=-2+t, \, l_2m_2}}$. Here 
\[
K(\chi=-1+i\nu;t) = -{\textstyle\frac{4\pi}{\nu}}t^2e^t \sum\limits_{n=-\infty}^{\infty}
{\textstyle\frac{[\nu+i(2n+1-t)]^{n-1}}{[\nu+i(2n+1+t)]^{n+2}}}, \,\,\, \textrm{for} \,\, 0<t<1
\]
and
\[
K(\chi=-1+i\nu;t) = {\textstyle\frac{4\pi}{\nu}} \int\limits_{\lambda=0}^{+\infty} \textrm{sinh} \lambda \sin(\nu \lambda)
e^{-t(\lambda \textrm{coth} \lambda -1)} \,\,\, \textrm{for} \,\, 1<t.
\]

Comparing (\ref{GeneralFTLimitExpandedAS>1}) and (\ref{GeneralFTLimitExpandedAS<1}) with (\ref{GeneralFTLimitExpanded}) we see that $\kappa_{1,1}$ is decomposable for $t>1$ and not decomposable for $0<t<1$. Moreover, when $t>1$
\[
\big\langle \mathcal{F}[\kappa_{1,1}(\phi)](\chi, \chi), \,  
F_{{}_{\chi, \, l_1 m_1}} \otimes
F_{{}_{\chi, \, l_{2} m_{2}}} \big\rangle = K(\chi;t) \delta_{{}_{l_1l_2}}\delta_{{}_{m_1m_2}}. 
\]
When $0<t<1$ we get an additional discrete contribution, concentrated at $\chi = -2+t \notin \textrm{Spec}$,
so that (\ref{GeneralFTLimitExpandedAS<1}) cannot be written in the form standing on the right-hand side of (\ref{GeneralFTLimitExpanded}).

This means that the generalized integral kernel operator
\begin{equation}\label{Xi(kappa1,1AS)}
\Xi(\kappa_{1,1}) = \int {\textstyle\frac{e^{-t\{\lambda(\boldsymbol{\p}_1,\boldsymbol{\p}_2) 
\textrm{coth} [\lambda(\boldsymbol{\p}_1,\boldsymbol{\p}_2)] -1\}} }{\sqrt{p_0(\boldsymbol{\p}_1)}\sqrt{p_0(\boldsymbol{\p}_2)}}}
\, b(\boldsymbol{\p}_1)^+ b(\boldsymbol{\p}_2) \,  \ud^3\boldsymbol{\p}_1 \ud^3\boldsymbol{\p}_2
\end{equation}
in the Fock space of the scalar complex field (\ref{complexscalarpsi}) (of mass $\boldsymbol{m}=1$) is decomposable for $1<t$, and decomposes
into the (discrete-) integral kernel operators
\[
\sum\limits_{l_1m_1l_2m_2} K(\chi;t) \delta_{{}_{l_1l_2}}\delta_{{}_{m_1m_2}} b_{{}_{\chi, \, l_1m_1}}^+ b_{{}_{\chi, \, l_2m_2}}
=  \sum\limits_{lm} K(\chi;t) \, b_{{}_{\chi, \, lm}}^+ b_{{}_{\chi, \, lm}}, \,\,\, \chi \in \textrm{Spec},
\]
with the decomposition measure coinciding with the spectral measure $\ud\chi$;  and (\ref{Xi(kappa1,1AS)}) is not 
decomposable for $0<t<1$. Recall, that 
\[
b_{{}_{\chi, \, lm}} = b_{{}_{\chi}}(F^{\oplus}_{{}_{\chi, \, lm}}),
\,\, b_{{}_{\chi, \, lm}}^{+} = b_{{}_{\chi}}(F^{\oplus}_{{}_{\chi, \, lm}})^{+} 
\]
respect the canonical commutation rules in the Fock space $\Gamma(\mathcal{H}^{\oplus}_{{}_{\chi}})$ over the decomposition Hilbert space
component $\mathcal{H}^{\oplus}_{{}_{\chi}}$, defined through the creation-annihilation Hida operators
\[
b_{{}_{\chi}}(F^{\oplus}_{{}_{\chi, \, lm}})^{+}, \,\,\, b_{{}_{\chi}}(F^{\oplus}_{{}_{\chi, \, lm}})
\,\,\, \textrm{in} \,\,\, \Gamma(\mathcal{H}^{\oplus}_{{}_{\chi}}).
\]
Here $\mathcal{H}^{\oplus}_{{}_{\chi}}$ are the direct integral decomposition components of $\mathcal{H}^\oplus$ determined by the 
decomposition of $SL(2, \mathbb{C})$ acting in the single particle nuclear and single particle Hilbert space $E^\oplus \subset \mathcal{H}^\oplus$ 
of positive energy solutions of the complex scalar field (\ref{complexscalarpsi}). 

Note also that for one of the momentum variables, say $\boldsymbol{\p}_1$, fixed, the kernel $\kappa_{1,1}$ of (\ref{Xi(kappa1,1AS)}) decreases relatively slowly
with the variable $\boldsymbol{\p}_2$ going to infinity
\[
{\textstyle\frac{e^{-t\{\lambda(\boldsymbol{\p}_1,\boldsymbol{\p}_2) 
\textrm{coth} [\lambda(\boldsymbol{\p}_1,\boldsymbol{\p}_2)] -1\}} }{\sqrt{p_0(\boldsymbol{\p}_1)}\sqrt{p_0(\boldsymbol{\p}_2)}}}
\sim 
{\textstyle\frac{e^t}{\sqrt{p_0(\boldsymbol{\p}_1)}\sqrt{p_0(\boldsymbol{\p}_2)} (p_1\cdot p_2)^t}}, 
\,\,\, \boldsymbol{\p}_2 \rightarrow \infty,
\]
but still (\ref{Xi(kappa1,1AS)}) is decomposable (if $t>1$). 
Therefore by comparison to the kernel $\kappa_{1,1}$ of (\ref{Xi(kappa1,1AS)}) (or, respectively $\kappa_{1,1} \otimes \kappa_{1,1},
\ldots$)  we can see that higher order contributions
$\Xi(\kappa_{\mathpzc{l},\mathpzc{m}})$ to interacting fields, with $\kappa_{\mathpzc{l},\mathpzc{m}}(\phi)$
decreasing not slower than $\kappa_{1,1}^{\otimes{\textstyle\frac{\mathpzc{l}+\mathpzc{m}}{2}}}$ 
(with, say, $\mathpzc{l}+\mathpzc{m}$ even) should be decomposable. 
The operator (\ref{Xi(kappa1,1AS)}) is quite singular with the kernel $\kappa_{1,1}$ not belonging to $E \otimes E^*$, even for $t>1$, 
and it is not an ordinary operator transforming the Hida space into itself 
but it transforms continuously the Hida space into its dual 
(by the results of \cite{hida} or Thm. \ref{Xi_l,m} and \ref{Xi_l,m:Hida->Hida} of Subsection \ref{psiBerezin-Hida} for the Bose case). 

The same sigular character we encounter for some of the higher order contributions
$\Xi(\kappa_{\mathpzc{l},\mathpzc{m}})$ to interacting fields, with $\kappa_{\mathpzc{l},\mathpzc{m}}(\phi)$ 
$\notin E^{\otimes \, \mathpzc{l}} \otimes E^{* \otimes \mathpzc{m}}$, even for $\phi \in \mathscr{E}$. 

Note, please, that the pairing integral
\[
\big\langle \kappa_{1,1}, F_{{}_{\chi, \, l_1m_1}} \otimes F_{{}_{\chi, \, l_2m_2}}\big\rangle
\]
for the kernel $\kappa_{1,1}$ of (\ref{Xi(kappa1,1AS)}) is divergent, so the simple formula (\ref{cchilimi}) for
$\mathcal{F}[\kappa_{1,1}]$ cannot be used. Because the decomposition measure of (\ref{Xi(kappa1,1AS)}) coincides with the
spectral measure, the Fourier transform $\mathcal{F}[\kappa_{1,1}](\chi,\chi)$ could have also been computed through the limit formula
(\ref{SimpleLimitkappaepsilonlm(phi)->kappalm(phi)}). Frequently it is more effective just to insert
the inverse Fourier formula for $\eta_{{}_{1}} \otimes \eta_{{}_{2}} \in E\otimes E$ into the pairing integral
\[
\big\langle \kappa_{1,1}(\phi), \eta_{{}_{1}} \otimes \eta_{{}_{2}} \big\rangle
\]
as we did in case of the kernel $\kappa_{1,1}$ of the operator (\ref{Xi(kappa1,1AS)}).
\qed

The limit approximation $\kappa_{\mathpzc{l},\mathpzc{m}}(\phi)_{{}_{\epsilon}} \in 
E^{\widehat{\otimes} \, \mathpzc{l}} \otimes E^{\widehat{\otimes} \mathpzc{m}}$, 
or $\kappa_{\mathpzc{l},\mathpzc{m}}(\phi)_{{}_{\epsilon}} \in L^2(\mathbb{R}^{3}; \mathbb{C}^d)^{\otimes \, (\mathpzc{l}+\mathpzc{m})}$, 
in (\ref{SimpleLimitkappaepsilonlm(phi)->kappalm(phi)}),
depends on the specific
$\kappa_{\mathpzc{l},\mathpzc{m}}(\phi)$, and its choice is determined by computational convenience. In principle,
in case the decomposition measure associated to the kernel $\kappa_{\mathpzc{l},\mathpzc{m}}(\phi)$
coincides with the spectral measure $\ud\chi$, we can always use
\begin{multline*}
\kappa_{\mathpzc{l},\mathpzc{m}}(\phi)_{{}_{\epsilon}}(s_{1},\boldsymbol{\p}_{1}, \ldots,
s_{\mathpzc{l}+\mathpzc{m}},\boldsymbol{\p}_{\mathpzc{l}+\mathpzc{m}}) 
\\
= 
e^{-\epsilon\{|\boldsymbol{\p}_{1}|^2 + \ldots + |\boldsymbol{\p}_{\mathpzc{l}+\mathpzc{m}}|^2\}}
\kappa_{\mathpzc{l},\mathpzc{m}}(\phi)(s_{1},\boldsymbol{\p}_{1}, \ldots,
s_{\mathpzc{l}+\mathpzc{m}},\boldsymbol{\p}_{\mathpzc{l}+\mathpzc{m}}),
\,\,\, \epsilon >0
\end{multline*}
but it is not always computationally convenient, so we are using the regularizing
factor in the form which makes the computation of the limit as simple as possible.

EXAMPLE 4.
Finally, as an example, we analyze decomposition of the Wick product operator ${:}\boldsymbol{\psi}^{\sharp} \gamma^\mu \boldsymbol{\psi}{:}$
with the Wick product kernel ${:}\boldsymbol{\psi}^{\sharp}(x) \gamma^\mu \boldsymbol{\psi}(x){:}$,
where $\boldsymbol{\psi}$ is the free Dirac field. This is the electric current field associated to the free
Dirac field. 

We are using the direct sum decomposition $E = E^+\oplus E^{-} = UE^{\oplus} \oplus UE^{\ominus \, \flat}$ of the single particle
nuclear space and of the single particle Hilbert space of the Dirac field. Correspondingly
each kernel $\kappa_{\mathpzc{l},\mathpzc{m}}(\phi) \in E^{*\widehat{\otimes} \, \mathpzc{l}}\otimes E^{*\widehat{\otimes} \, \mathpzc{m}}$
can be decomposed into the components $\kappa_{\mathpzc{l},\mathpzc{m}}^{\pm \pm}(\phi) \in E^{\pm *\widehat{\otimes} \, \mathpzc{l}}
\otimes E^{\pm *\widehat{\otimes} \, \mathpzc{m}}$. We are still using the canonical \emph{bilinear} pairing $\langle \cdot, \cdot\rangle$ on $E^* \times E$ and
its tensor product extension over $E^{*\otimes \mathpzc{l}} \times E^{\otimes \mathpzc{l}}$.  

Regarded as a finite sum of integral kernel operators,  $\Xi(\kappa_{\mathpzc{l},\mathpzc{m}}(\varphi))$, evaluated at the space-time
four-vector test function $\varphi \in \mathcal{S}(\mathbb{R}^4; \mathbb{C}^4)$, it can be written as
(summation with respect to the  Lorentz index $\mu$)
\begin{multline*}
\sum\Xi(\kappa_{\mathpzc{l},\mathpzc{m}}(\varphi))= 
{:}\boldsymbol{\psi}^{\sharp} \gamma^\mu \boldsymbol{\psi}{:}(\varphi_\mu) 
= \sum\limits_{\mu} \int {:}\boldsymbol{\psi}^{\sharp} \gamma^\mu \boldsymbol{\psi}{:}(x) \, \varphi_\mu(x) \, \ud^4 x
\\
=
\sum\limits_{s',s}\int \kappa_{1,1}^{++}(\varphi)(s',\boldsymbol{\p}',s,\boldsymbol{\p}) \,\, {b'}_{{}_{s'}}(\boldsymbol{\p}')^{+} b'_{{}_{s}}(\boldsymbol{\p})
\,\,
\ud^3 \boldsymbol{\p}' \, \ud^3 \boldsymbol{\p} 
\end{multline*}
\begin{multline*}
+
\sum\limits_{s',s} \int \kappa_{1,1}^{--}(\varphi)(s',\boldsymbol{\p}',s,\boldsymbol{\p}) \,\, {d'}_{{}_{s'}}(\boldsymbol{\p}')^{+} d'_{{}_{s}}(\boldsymbol{\p})
\,\,
\ud^3 \boldsymbol{\p}' \, \ud^3 \boldsymbol{\p}
\\
+
\sum\limits_{s',s}\int \kappa_{2,0}^{+-}(\varphi)(s'\boldsymbol{\p}',s,\boldsymbol{\p})  \,\, {b'}_{{}_{s'}}(\boldsymbol{\p}')^{+} {d'}_{{}_{s}}(\boldsymbol{\p})^{+} 
\,\,
\ud^3 \boldsymbol{\p}' \, \ud^3 \boldsymbol{\p}
\end{multline*}
\[
+
\sum\limits_{s',s} \int \kappa_{0,2}^{-+}(\varphi)(s',\boldsymbol{\p}',s,\boldsymbol{\p}) \,\, {d'}_{{}_{s'}}(\boldsymbol{\p}') b'_{{}_{s}}(\boldsymbol{\p})
\,\,
\ud^3 \boldsymbol{\p}' \, \ud^3 \boldsymbol{\p}. 
\]
Here (compare Subsection \ref{psiBerezin-Hida}) 
\[
\begin{split}
\kappa_{2,0}^{+-}(\phi)(s',\boldsymbol{\p}',s,\boldsymbol{\p}) 
= \Big({\kappa}^{\sharp}_{1,0}\dot{\otimes} \gamma^\mu\kappa_{1,0}\Big)(s',\boldsymbol{\p}',s,\boldsymbol{\p}, x)
\\
= \underset{*}{u}{}_{{}_{s'}}(\boldsymbol{\p}')^+\gamma^0\gamma^\mu \underset{*}{v}{}_{{}_{s}}(\boldsymbol{\p})
e^{i(p_0(\boldsymbol{\p}') + p_0(\boldsymbol{\p}))x_0 - i(\boldsymbol{\p}'+\boldsymbol{\p})\cdot \boldsymbol{\x}},  
\\
\kappa_{0,2}^{-+}(\phi)(s',\boldsymbol{\p}',s,\boldsymbol{\p}) 
= \Big({\kappa}^{\sharp}_{0,1}\dot{\otimes} \gamma^\mu\kappa_{0,1}\Big)(s',\boldsymbol{\p}',s,\boldsymbol{\p}, x) 
\\
= \underset{*}{v}{}_{{}_{s'}}(\boldsymbol{\p}')^+\gamma^0\gamma^\mu \underset{*}{u}{}_{{}_{s}}(\boldsymbol{\p})
e^{i(-p_0(\boldsymbol{\p}') - p_0(\boldsymbol{\p}))x_0 - i(-\boldsymbol{\p}'-\boldsymbol{\p})\cdot \boldsymbol{\x}}, 
\end{split}
\]
and
\[
\begin{split}
\kappa_{1,1}^{++}(s',\boldsymbol{\p}',s,\boldsymbol{\p}; \mu, x) 
= 
\Big({\kappa}^{\sharp}_{1,0}\overline{\dot{\otimes}} \gamma^\mu\kappa_{0,1}\big)(s',\boldsymbol{\p}',s,\boldsymbol{\p}, x)
\\
= \underset{*}{u}{}_{{}_{s'}}(\boldsymbol{\p}')^+\gamma^0\gamma^\mu \underset{*}{u}{}_{{}_{s}}(\boldsymbol{\p})
e^{i(p_0(\boldsymbol{\p}') - p_0(\boldsymbol{\p}))x_0 - i(\boldsymbol{\p}'-\boldsymbol{\p})\cdot \boldsymbol{\x}},
\\
\kappa_{1,1}^{--}(s',\boldsymbol{\p}',s,\boldsymbol{\p}; \mu, x) 
= 
\Big({\kappa}^{\sharp}_{0,1}\overline{\dot{\otimes}} \gamma^\mu\kappa_{1,0}\Big)(s',\boldsymbol{\p}',s,\boldsymbol{\p}, x)
\\
= \underset{*}{v}{}_{{}_{s'}}(\boldsymbol{\p}')^+\gamma^0\gamma^\mu \underset{*}{v}{}_{{}_{s}}(\boldsymbol{\p})
e^{i(p_0(\boldsymbol{\p}) - p_0(\boldsymbol{\p}'))x_0 - i(\boldsymbol{\p}-\boldsymbol{\p}')\cdot \boldsymbol{\x}}.
\end{split}
\]
so that (summation with respect to the  Lorentz index $\mu$)
\[
\begin{split}
\kappa_{2,0}^{+-}(\phi)(s',\boldsymbol{\p}',s,\boldsymbol{\p}) 
= \Big({\kappa}^{\sharp}_{1,0}\dot{\otimes} \gamma^\mu\kappa_{1,0}\Big)(\phi_\mu)(s',\boldsymbol{\p}',s,\boldsymbol{\p})
\\
= \underset{*}{u}{}_{{}_{s'}}(\boldsymbol{\p}')^+\gamma^0\gamma^\mu \underset{*}{v}{}_{{}_{s}}(\boldsymbol{\p})
\widetilde{\phi_\mu}(p_0(\boldsymbol{\p}') + p_0(\boldsymbol{\p})), \boldsymbol{\p}'+\boldsymbol{\p}),  
\\
\kappa_{0,2}^{-+}(\phi)(s',\boldsymbol{\p}',s,\boldsymbol{\p}) 
= \Big({\kappa}^{\sharp}_{0,1}\dot{\otimes} \gamma^\mu\kappa_{0,1}\Big)(\phi_\mu)(s',\boldsymbol{\p}',s,\boldsymbol{\p}) 
\\
= \underset{*}{v}{}_{{}_{s'}}(\boldsymbol{\p}')^+\gamma^0\gamma^\mu \underset{*}{u}{}_{{}_{s}}(\boldsymbol{\p})
\widetilde{\phi_\mu}(-p_0(\boldsymbol{\p}') - p_0(\boldsymbol{\p}), -\boldsymbol{\p}'-\boldsymbol{\p}), 
\end{split}
\]
and
\[
\begin{split}
\kappa_{1,1}^{++}(\phi)(s',\boldsymbol{\p}',s,\boldsymbol{\p}) 
= 
\Big({\kappa}^{\sharp}_{1,0}\overline{\dot{\otimes}} \gamma^\mu\kappa_{0,1}\big)(\phi_\mu)(s',\boldsymbol{\p}',s,\boldsymbol{\p})
\\
= \underset{*}{u}{}_{{}_{s'}}(\boldsymbol{\p}')^+\gamma^0\gamma^\mu \underset{*}{u}{}_{{}_{s}}(\boldsymbol{\p})
\widetilde{\phi_\mu}(p_0(\boldsymbol{\p}') - p_0(\boldsymbol{\p}), \boldsymbol{\p}'-\boldsymbol{\p}),
\\
\kappa_{1,1}^{--}(\phi)(s',\boldsymbol{\p}',s,\boldsymbol{\p}) 
= 
\Big({\kappa}^{\sharp}_{0,1}\overline{\dot{\otimes}} \gamma^\mu\kappa_{1,0}\Big)(\phi_\mu)(s',\boldsymbol{\p}',s,\boldsymbol{\p})
\\
= \underset{*}{v}{}_{{}_{s'}}(\boldsymbol{\p}')^+\gamma^0\gamma^\mu \underset{*}{v}{}_{{}_{s}}(\boldsymbol{\p})
\widetilde{\phi_\mu}(p_0(\boldsymbol{\p}) - p_0(\boldsymbol{\p}'), \boldsymbol{\p}-\boldsymbol{\p}').
\end{split}
\]

It is easily seen that
\[
\kappa_{2,0}^{+-}(\phi) \in E^+\otimes E^-, \,\,\,
\kappa_{0,2}^{-+}(\phi) \in E^-\otimes E^+,
\]
so that the kernels $\kappa_{2,0}^{+-}(\phi)$, $\kappa_{2,0}^{+-}(\phi)$
can be evaluated at the basic functionals.
Therefore the Fourier transforms
\[
\begin{split}
\Bigg\langle
\mathcal{F}\big[\kappa_{2,0}^{+-}(\phi)\big](\chi,\chi), \,\,
F^{+}_{{}_{\chi, \, l'm'}} \, \otimes \, 
F^{-}_{{}_{\chi, \, lm}}
\Bigg\rangle
\\
\Bigg\langle
\mathcal{F}\big[\kappa_{0,2}^{-+}(\phi)\big](\chi,\chi), \,\,
F^{-}_{{}_{\chi, \, l'm'}} \, \otimes \, 
F^{+}_{{}_{\chi, \, lm}}
\Bigg\rangle,
\end{split}
\]
expressed immediately in terms of $F^{{}^{\oplus}}_{{}_{\chi, \, lm}}$, $F^{{}^{\ominus \, \flat}}_{{}_{\chi, \, lm}}$
and the isomorphism $U$, eq. (\ref{isomorphismU}) of Subsection \ref{psiBerezin-Hida}, which extends naturally on the functionals 
given by ordinary functions, are respectively equal
\begin{equation}\label{F(k+-2,0)F(k-+0,2)}
\begin{split}
\kappa_{\chi\, 2,0}^{+-}(\varphi)(l'm',lm) 
= \Big\langle
\kappa_{2,0}^{+-}(\phi), \,\,
 \overline{U F^{{}^{\oplus}}_{{}_{\chi, \, l'm'}}}
\, \widehat{\otimes} \,  \overline{UF^{{}^{\ominus \, \flat}}_{{}_{\chi, \, lm}}}
 \Big\rangle,
\\
\kappa_{\chi \, 0,2}^{-+}(\varphi)(l'm',lm) 
= 
\Big\langle 
\kappa_{0,2}^{-+}(\phi), \,\,
U F^{{}^{\ominus \, \flat}}_{{}_{\chi, \, l'm'}}
\, \widehat{\otimes} \, UF^{{}^{\oplus}}_{{}_{\chi, \, lm}} \,\,
\Big\rangle,
\end{split}
\end{equation}

It is easily seen that
\[
\kappa_{1,1}^{++}(\phi) \in E^+\otimes E^{+*}, \,\,\,
\kappa_{1,1}^{--}(\phi) \in E^-\otimes E^{-*},
\]
so that the Wick product operator is regular and defines
a continuous map
\[
\mathcal{S}(\mathbb{R}^4;\mathbb{C}^4) \ni \phi 
\longmapsto {:}\boldsymbol{\psi}^{\sharp} \gamma^\mu \boldsymbol{\psi}{:}
\in \mathscr{L}((E), (E))
\]
from the space-time test space into the  nuclear space of continuous operators, transforming
continuously the Hida test space $(E)$ into itself. In particular this Wick product is a well-defined
operator valued distribution. But the kernels $\kappa_{1,1}^{++}(\phi)$, $\kappa_{1,1}^{--}(\phi)$ cannot be evaluated at the 
tensor products of the basic functionals
and the Fourier transforms
\[
\begin{split}
\Bigg\langle
\mathcal{F}\big[\kappa_{1,1}^{++}(\phi)\big](\chi,\chi), \,\,
F^{+}_{{}_{\chi, \, l'm'}} \, \otimes \, 
F^{+}_{{}_{\chi, \, lm}}
\Bigg\rangle
\\
\Bigg\langle
\mathcal{F}\big[\kappa_{1,1}^{--}(\phi)\big](\chi,\chi), \,\,
F^{-}_{{}_{\chi, \, l'm'}} \, \otimes \, 
F^{-}_{{}_{\chi, \, lm}}
\Bigg\rangle,
\end{split}
\]
are, respectively, equal
\begin{equation}\label{F(k++1,1)}
\kappa_{\chi \, 1,1}^{++}(\varphi)(l'm',lm) 
= \underset{\epsilon \rightarrow 0}{\textrm{lim}} 
\Big\langle 
\kappa_{1,1}^{++}(\phi)_{{}_{\epsilon}}, \,\,
\overline{U F^{{}^{\oplus}}_{{}_{\chi, \, l'm'}}} \, \otimes \, 
UF^{{}^{\oplus}}_{{}_{\chi, \, lm}}
\Big\rangle,
\end{equation}
\begin{equation}\label{F(k--1,1)}
\kappa_{\chi \, 1,1}^{--}(\varphi)(l'm',lm) 
= 
\underset{\epsilon \rightarrow 0}{\textrm{lim}} 
\Big\langle 
\kappa_{1,1}^{--}(\phi)_{{}_{\epsilon}}, \,\,
U F^{{}^{\ominus \, \flat}}_{{}_{\chi, \, l'm'}} \, \,  \, \otimes \,
\overline{UF^{{}^{\ominus \, \flat}}_{{}_{\chi, \, lm}}}
\Big\rangle.
\end{equation}

Thus the Wick product operator ${:}\boldsymbol{\psi}^{\sharp} \gamma^\mu \boldsymbol{\psi}{:}$
has the direct integral decomposition
\[
{:}\boldsymbol{\psi}^{\sharp} \gamma^\mu \boldsymbol{\psi}{:}(\varphi_\mu) = 
\int\limits_{\textrm{Spec}} {:}\boldsymbol{\psi}^{\sharp} \gamma^\mu \boldsymbol{\psi}{:}(\varphi_\mu)_{{}_{\chi}} \,
\ud \chi
\]
with the spectral measure $\ud \chi$, and each decomposition component
\[
 {:}\boldsymbol{\psi}^{\sharp} \gamma^\mu \boldsymbol{\psi}{:}(\varphi_\mu)_{{}_{\chi}}
\]
of the operator 
\[
{:}\boldsymbol{\psi}^{\sharp} \gamma^\mu \boldsymbol{\psi}{:}(\varphi_\mu)
\]
being here equal to a well-defined field operator
\[
\varphi \longmapsto {:}\boldsymbol{\psi}^{\sharp} \gamma^\mu \boldsymbol{\psi}{:}(\varphi_\mu)_{{}_{\chi}}
\]
which we write in the form
\[
\varphi \longmapsto {:}\boldsymbol{\psi}^{\sharp} \gamma^\mu \boldsymbol{\psi}{:}_{{}_{\chi}}(\varphi_\mu).
\]
Here
\begin{multline*}
{:}\boldsymbol{\psi}^{\sharp} \gamma^\mu \boldsymbol{\psi}{:}_{{}_{\chi}}(\varphi_\mu) 
\\
=
\sum\limits_{l',m',l,m} \kappa_{\chi \, 1,1}^{++}(\varphi)(l'm',lm) \,\, {b'}_{{}_{\chi, \, l'm'}}^{+} b'_{{}_{\chi, \, lm}} +
\sum\limits_{l',m',l,m} \kappa_{\chi \, 1,1}^{--}(\varphi)(l'm',lm) \,\, {d'}_{{}_{\chi, \, l'm'}}^{+} d'_{{}_{\chi, \, lm}} 
\\
+
\sum\limits_{l',m',l,m} \kappa_{\chi \, 2,0}^{+-}(\varphi)(l'm',lm) \,\, {b'}_{{}_{\chi, \, l'm'}}^{+} {d'}_{{}_{\chi, \, lm}}^{+} +
\sum\limits_{l',m',l,m} \kappa_{\chi \, 0,2}^{-+}(\varphi)(l'm',lm) \,\, {d'}_{{}_{\chi, \, l'm'}} b'_{{}_{\chi, \, lm}}. 
\end{multline*}

It is natural to expect this operator to be closely related to the ''Wick product''
${:}\boldsymbol{\psi}_{{}_{\chi}}^{\sharp}(x) \gamma^\mu \boldsymbol{\psi}_{{}_{\chi}}(x){:}$. However, in general this ''Wick product''
is not well-defined as the functions representing the distributions $\overline{f^{{}^{\oplus}}_{{}_{\chi, \, lm}}(x)}$ and
$\overline{f^{{}^{\ominus}}_{{}_{\chi, \, lm}}(x)}$ in (\ref{psichi(x)})
cannot be multiplied. But if the product of their components was well-defined and represented locally integrable
functions, and the order of integration in (\ref{F(k+-2,0)F(k-+0,2)}) could have been changed,
then (\ref{F(k+-2,0)F(k-+0,2)}) would be equal 
\[
\begin{split}
\Big\langle
\kappa_{2,0}^{+-}(\phi), \,\,
 \overline{U F^{{}^{\oplus}}_{{}_{\chi, \, l'm'}}}
\, \widehat{\otimes} \,  \overline{UF^{{}^{\ominus \, \flat}}_{{}_{\chi, \, lm}}}
 \Big\rangle 
=
\int {f^{{}^{\ominus}}_{{}_{\chi, \, l'm'}}(x)}^{T} \gamma^0\gamma^\mu \overline{f^{{}^{\oplus}}_{{}_{\chi, \, lm}}(x)}
\, \phi_\mu(-x) \, \ud^4 x
\\
\Big\langle 
\kappa_{0,2}^{-+}(\phi), \,\,
U F^{{}^{\ominus \, \flat}}_{{}_{\chi, \, l'm'}}
\, \widehat{\otimes} \, UF^{{}^{\oplus}}_{{}_{\chi, \, lm}} \,\,
\Big\rangle
=
\int {f^{{}^{\oplus}}_{{}_{\chi, \, l'm'}}(x)}^{T} \gamma^0\gamma^\mu \overline{f^{{}^{\ominus}}_{{}_{\chi, \, lm}}(x)}
\, \phi_\mu(-x) \, \ud^4 x,
\end{split}
\]
as can be easily checked, by simple computation. Similarly, if the pairing integrals (\ref{F(k++1,1)})
and (\ref{F(k--1,1)}) would be convergent, just with $\kappa_{1,1}^{++}(\phi)$ and $\kappa_{1,1}^{--}(\phi)$,
and the product of the functions $\overline{f^{{}^{\oplus}}_{{}_{\chi, \, lm}}(x)}$ and
$\overline{f^{{}^{\oplus}}_{{}_{\chi, \, lm}}(x)}$, as well as the product 
of the functions $\overline{f^{{}^{\ominus}}_{{}_{\chi, \, lm}}(x)}$ and
$\overline{f^{{}^{\ominus}}_{{}_{\chi, \, lm}}(x)}$,
representing the corresponding distributions, was well-defined then the order of integration
could have been changed and one would obtain
\[
\begin{split}
\Big\langle 
\kappa_{1,1}^{++}(\phi), \,\,
\overline{U F^{{}^{\oplus}}_{{}_{\chi, \, l'm'}}} \, \otimes \, 
UF^{{}^{\oplus}}_{{}_{\chi, \, lm}}
\Big\rangle
=
\int {f^{{}^{\oplus}}_{{}_{\chi, \, l'm'}}(x)}^{T} \gamma^0\gamma^\mu \overline{f^{{}^{\oplus}}_{{}_{\chi, \, lm}}(x)}
\, \phi_\mu(-x) \, \ud^4 x
\\
\Big\langle 
\kappa_{1,1}^{--}(\phi), \,\,
U F^{{}^{\ominus \, \flat}}_{{}_{\chi, \, l'm'}} \, \,  \, \otimes \,
\overline{UF^{{}^{\ominus \, \flat}}_{{}_{\chi, \, lm}}}
\Big\rangle
=
\int {f^{{}^{\ominus}}_{{}_{\chi, \, l'm'}}(x)}^{T} \gamma^0\gamma^\mu \overline{f^{{}^{\ominus}}_{{}_{\chi, \, lm}}(x)}
\, \phi_\mu(-x) \, \ud^4 x.
\end{split}
\]
Unfortunately for the most interesting Wick product field operators, e.g.
${:}\boldsymbol{\psi}^{\sharp} \gamma^\mu \boldsymbol{\psi}{:}$,
the last integrals are not well-defined, and the kernels
${:}\boldsymbol{\psi}^{\sharp} \gamma^\mu \boldsymbol{\psi}{:}_{{}_{\chi}}(x)$
of the decomposition component fields
${:}\boldsymbol{\psi}^{\sharp} \gamma^\mu \boldsymbol{\psi}{:}_{{}_{\chi}}(\varphi_\mu)$
cannot be computed through the ''Wick product''
${:}\boldsymbol{\psi}_{{}_{\chi}}^{\sharp}(x) \gamma^\mu \boldsymbol{\psi}_{{}_{\chi}}(x){:}$.
However, for the field
operators or integral kernel operators with more regular behavior of
the kernels of their decomposition components, decomposition components
of the Wick products are equal to the Wick products of the components (up to space-time reflection).
In particular this is the case for the Wick products of the field which we obtain from the free Dirac field
by application of an additional weight factor in its kernels $\kappa_{1,0}$
$\kappa_{0,1}$, which sufficiently rapidly decreases in momentum space.

It should also be noted that the (asymptotic) homogeneity of the distribution kernels
${:}\boldsymbol{\psi}^{\sharp} \gamma^\mu \boldsymbol{\psi}{:}_{{}_{\chi}}(x)$ of
${:}\boldsymbol{\psi}^{\sharp} \gamma^\mu \boldsymbol{\psi}{:}_{{}_{\chi}}$, given by 
(\ref{F(k+-2,0)F(k-+0,2)}), (\ref{F(k++1,1)})
and (\ref{F(k--1,1)}) is equal $\overline{\chi}+\chi = -3$, as expected, as the (asymptotic)
homogeneity of $\boldsymbol{\psi}^{\sharp}$ is equal $\overline{\chi} = -3/2+ i\nu$ and the
(asymptotic) homogeneity of  $\boldsymbol{\psi}$ is equal $\chi = -3/2-i\nu$.   
\qed

\section{Fundamental rules for computations involving free fields understood 
as integral kernel operators with vector-valued kernels}\label{OperationsOnXi}

In this Subsection we give several useful computational rules, performed upon integral kernel operators
$\Xi_{l,m}(\kappa_{l,m})$ determined by $\mathscr{L}(\mathscr{E}, \mathbb{C})$-valued distributions,
$\kappa_{l,m}$, respecting the extendibility condition of Theorem \ref{obataJFA.Thm.3.13} of the preceding
Subsection \ref{psiBerezin-Hida} (or respectively of Theorem 3.13 of \cite{obataJFA}). This property allows to treat
such $\Xi_{l,m}(\kappa_{l,m})$ as well-defined operator-valued distributions on the standard nuclear test space
$\mathscr{E}$, which in our case will always be equal to the tensor product
\[
\mathscr{E} = \mathscr{E}_{{}_{n_1}} \otimes \cdots \otimes \mathscr{E}_{{}_{n_M}}, \,\,\, n_k \in \{1,2\},
\]
of $M$ space-time test spaces
$\mathscr{E}_1, \mathscr{E}_2$ given by (\ref{mathscrE_1,mathscrE_2}), Subsection \ref{psiBerezin-Hida},
with $M =1$ and $p_k$ put equal $n_k$.
We encounter the cases with $M =1$ and (operator-valued distributions with one space-time variable)
or with $M>1$ space-time variables.
In fact the integral kernel operators which are of importance for us are of still more special character,
being obtainable from the integral kernel operators defined by the free fields underlying the considered
Quantum Field Theory, as a result of special operations: composition of Wick product, differentiation,
integration and convolution with pairing functions.

Having in view the causal perturbative QED we confine attention to integral kernel operators
$\Xi_{l,m}(\kappa_{l,m})$ in the tensor product of just two Fock spaces -- the first one fermionic 
and corresponding to the Dirac field and the second one bosonic and corresponding to the electromagnetic potential field, compare 
Subsection \ref{psiBerezin-Hida}. 
Thus considered here integral kernel operators $\Xi_{l,m}(\kappa_{l,m})$ act on the Hida space
$(\boldsymbol{E}) = (E_1) \otimes (E_2) \subset \Gamma_{\textrm{Fermi}}\big( L^2(\mathbb{R}^3;\mathbb{C}^4)\big)
\otimes \Gamma_{\textrm{Bose}}\big( L^2(\mathbb{R}^3;\mathbb{C}^4)\big)$, constructed as in the previous 
Subsection \ref{psiBerezin-Hida}. We have also formulated the Thm. \ref{obataJFA.Thm.3.13}, 
Subsection \ref{psiBerezin-Hida}, for the said tensor product of the two mentioned above Fock spaces. 
Of course analogous Theorem and corresponding rules of calculation
with integral kernel operators $\Xi_{l,m}(\kappa_{l,m})$ are valid on tensor product of 
more than just two indicated Fock spaces. 

The space $E_1 = \mathcal{S}_{A_1}(\mathbb{R}^3; \mathbb{C}^4)
= \mathcal{S}(\mathbb{R}^3; \mathbb{C}^4)$ 
with index $1$ and the standard operator $A_1 = A$ (\ref{AinL^2(R^3;C^4)}) refers to the standard 
nuclear space in (\ref{SinglePartGelfandTriplesForPsi})), corresponding to the Dirac field, with the space-time 
test space $\mathscr{E}_1 = \mathcal{S}_{\oplus H_{(4)}}(\mathbb{R}^4; \mathbb{C}^4) 
= \mathcal{S}(\mathbb{R}^4; \mathbb{C}^4)$.
The space  $E_2 = \mathcal{S}_{A_2}(\mathbb{R}^3; \mathbb{C}^4)
= \mathcal{S}^{0}(\mathbb{R}^3; \mathbb{C}^4)$ 
with index $2$ is the nuclear space $E$ determined by the standard operator 
$A_2 = \oplus_{0}^{3} A^{(3)} = A$, which enters the triple in (\ref{3-Gelfand-triples}), and which serves to define 
the free quantum electromagnetic potential field, Subsection \ref{WhiteNoiseA}, with the space-time 
test space $\mathscr{E}_2 = \mathcal{S}_{\mathscr{F}^{-1}\oplus A^{(4)}\mathscr{F}}(\mathbb{R}^4; \mathbb{C}^4) 
= \mathcal{S}^{00}(\mathbb{R}^4; \mathbb{C}^4)$.

The vector-valued distributions $\kappa_{0,1}, \kappa_{1,0} \in \mathscr{L}(E_1, \mathscr{E}_{1}^{*})$
determined by the plane wave kernels
(\ref{kappa_0,1}) and (\ref{kappa_1,0}), defining the free Dirac field as the integral kernel operator
\[
\boldsymbol{\psi} = \Xi_{0,1}(\kappa_{0,1}) + \Xi_{1,0}(\kappa_{1,0})
= \boldsymbol{\psi}^{(-)} + \boldsymbol{\psi}^{(+)},
\]
and in general the vector-valued plane-wave distributions
$\kappa_{0,1}, \kappa_{1,0}, \ldots$ defining all free quantum fields of
the theory play a fundamental role in the theory.
In QED we enc outer, besides the plane waves (\ref{kappa_0,1}) and (\ref{kappa_1,0}), also
the plane waves $\kappa_{0,1}, \kappa_{1,0} \in \mathscr{L}(E_2, \mathscr{E}_{2}^{*})$
(\ref{kappa_0,1kappa_1,0A'}), Subsection \ref{equivalentA-s}, defining the free quantum electromagnetic
potential field:
\[
A = \Xi_{0,1}(\kappa_{0,1}) + \Xi_{1,0}(\kappa_{1,0}) = A^{(-)} + A^{(+)},
\]
if we change slightly the convention (used by mathematicians) of Subsection
\ref{psiBerezin-Hida} and use for $\partial_{w}^*$ in the general integral kernel operator
(\ref{electron-positron-photon-Xi}),
on the tensor product of Fock spaces of the Dirac field $\boldsymbol{\psi}$ and the electromagnetic potential field $A$,
the operators $\eta\partial_{\mu, \boldsymbol{\p}}^{*} \eta$ whenever
$w = (\mu, \boldsymbol{\p})$ corresponds to the photon variables $\mu, \boldsymbol{\p}$
in (\ref{electron-positron-photon-Xi}),
instead of the ordinary transposed operators $\partial_{\mu, \boldsymbol{\p}}^{*}$.
Here $\eta$ is the Gupta-Bleuler operator. This convention fits well with notation used by physicists,
as they are using the Krein-adjoined annihilation operators of the photon variables in Fock
normal expansions.

Indeed, in terms of these kernels $\kappa_{0,1}, \kappa_{0,1}, \ldots$
all important quantities of the theory are expressed:
\begin{enumerate}
\item[1)]
The Wick polynomials of free fields are expressed through
(symmetrized in Bose variables or respectively antisymmetrized in Fermi variables) tensor product operation
performed upon the plane wave kernels $\kappa_{0,1}, \kappa_{1,0}, \ldots$
defining the free fields of the theory,
\item[2)]
Wick polynomial of free fields at the same space-time point are expressed through the symmetrized or antisymmetrized
in $\xi_{1}, \ldots , \xi_M$
operation of pointwise product $\kappa_{l_1,m_1}(\xi_1) \cdot \kappa'_{l_1,m_1}(\xi_1) \cdot \ldots \cdot \kappa^{(M)}_{l_M,m_M}(\xi_M)$ utilizing the fact that
$\kappa_{0,1}(\xi), \kappa_{1,0}(\xi), \kappa'_{0,1}(\xi), \kappa'_{1,0}(\xi), \ldots$,
with $\xi_i \in \mathcal{S}_{A_i}(\mathbb{R}^3, \mathbb{C}^4)$ belong to the algebra of multipliers
of the respective nuclear algebra $\mathscr{E}_i= \mathcal{S}_{B_i}(\mathbb{R}^4; \mathbb{C}^4)$
(equal $\mathcal{S}(\mathbb{R}^4; \mathbb{C}^4)$ or respectively $\mathcal{S}^{00}(\mathbb{R}^4; \mathbb{C}^4)$)
of spaces of space-time test functions, and the fact that the maps
\begin{multline*}
E_{i} \times E_{j} \ni \xi \times \zeta
\mapsto \kappa_{1,0}(\xi)\cdot \kappa'_{1,0}(\zeta) \in \mathscr{E}_{k}^*, \\
i,j,k \in \{1, 2\},
\end{multline*}
are jointly continuous in the ordinary nuclear topology on $E_{i}$ and strong dual topology on
$\mathscr{E}_{k}^*$ which secures the Wick product to be a well-defined integral kernel operator belonging to
\[
\mathscr{L}((\boldsymbol{E}) \otimes \mathscr{E}, \, (\boldsymbol{E})^*)
\]
for $\mathscr{E}$ equal to the test function space $\mathscr{E}_{1}= \mathcal{S}(\mathbb{R}^4)$ as well as for
$\mathscr{E}_{2}= \mathcal{S}^{00}(\mathbb{R}^4)$.
Moreover, if among the integral kernel operators defined by the plane waves defining free fields there are no
factors corresponding to zero mass free fields, then
\begin{multline*}
E_{i}^{*} \times E_{j}^* \subset E_{i} \times E_{j} \ni \xi \times \zeta
\mapsto \kappa_{1,0}(\xi)\cdot \kappa'_{1,0}(\zeta) \in \mathscr{E}_{k}^*, \\
i,j,k \in \{1, 2\},
\end{multline*}
defined through ordinary pointwise product $\cdot$, are hypocontinuous in the topology inherited from
the strong dual topology on $E_{i}^{*}$, and strong dual topology on $\mathscr{E}_{j}^{*}$,
which secures in this case the Wick product to be an integral kernel operator which belongs even to
\[
\mathscr{L}((\boldsymbol{E}) \otimes \mathscr{E}, \, (\boldsymbol{E})) \cong
\mathscr{L}(\mathscr{E}, \, \mathscr{L}((\boldsymbol{E}), (\boldsymbol{E}))
\]
for $\mathscr{E}$ equal to the test function space $\mathscr{E}_{1}= \mathcal{S}(\mathbb{R}^4)$ as well as for
$\mathscr{E}_{2}= \mathcal{S}^{00}(\mathbb{R}^4)$.
\item[3)]
The perturbative contributions to interacting fields are expressed through convolutions of the kernels corresponding to
Wick polynomials of free fields with the respective ''retarded and advanced parts of causally supported generalized functions'', 
and utilizing the fact that
$\kappa_{0,1}(\xi_{n_1}), \kappa_{0,1}(\xi_{n_2}), \kappa'_{0,1}(\xi_{n_3}), \ldots$, and their pointwise products
with $\xi_{n_{k}} \in \mathcal{S}_{A_{n_k}}(\mathbb{R}^3, \mathbb{C}^4)$ belong to the algebra of convolutors
of the respective nuclear algebra $\mathscr{E}_{n_k}$ ($n_k \in \{1,2\}$).
\end{enumerate}

In all these constructions we apply the Theorem \ref{obataJFA.Thm.3.13}, and check validity of the condition stated in this Theorem, asserting that the constructed integral kernel operator belongs to
\[
\mathscr{L}((\boldsymbol{E}) \otimes \mathscr{E}, \, (\boldsymbol{E})) \cong
\mathscr{L}(\mathscr{E}, \, \mathscr{L}((\boldsymbol{E}), (\boldsymbol{E}))
\]
and defines an operator-valued distribution on the corresponding test space $\mathscr{E}$.
Alternatively we check that the constructed operator $\Xi(\kappa)$ has the kernel which respect
weaker condition (\ref{general-vect-valued--kappa_(lm)})
\[
\kappa \in \mathscr{L}(E_{n_1}, \ldots, E_{n_{l+m}}, \, \mathscr{E}^*),
\]
which means by the generalization to tensor product of Fock spaces of Theorem 3.9 (compare Subsection
\ref{psiBerezin-Hida}) that the integral kernel operator belongs to
\[
\mathscr{L}((\boldsymbol{E}) \otimes \mathscr{E}, \, (\boldsymbol{E})^*) \cong
\mathscr{L}(\mathscr{E}, \, \mathscr{L}((\boldsymbol{E}), (\boldsymbol{E})^*).
\]

In general, it cannot be asserted\footnote{Below we give a proof that the Wick product of integral kernel operators corresponding to zero mass fields or their derivatives does not belong to
\[
\mathscr{L}((\boldsymbol{E}) \otimes \mathscr{E}, \, (\boldsymbol{E}))
\,\,\,
\textrm{but does belong to} \,\, \mathscr{L}((\boldsymbol{E}) \otimes \mathscr{E}, \, (\boldsymbol{E})^*).
\]
} that the integral kernel operator $\Xi$ represented by the Wick product $\Xi$ of integral kernel operators defined
by free fields belonging to
\[
\mathscr{L}((\boldsymbol{E}) \otimes \mathscr{E}, \, (\boldsymbol{E})),
\]
belongs to
\[
\mathscr{L}((\boldsymbol{E}) \otimes \mathscr{E}, \, (\boldsymbol{E})).
\]
This would be true only for the Wick product (at the fixed space-time point) $\Xi$ of integral kernel operators
corresponding to massive free fields
(such as Dirac field) or their derivatives. But if among the factors in the Wick product there are present
integral kernel operators corresponding to zero mass fields (or their derivatives), then their Wick product (at the fixed space-time point) $\Xi$
represents a general integral kernel operator (with vector valued kernel) $\Xi(\kappa)$ which belongs to
\[
\mathscr{L}((\boldsymbol{E}) \otimes \mathscr{E}, \, (\boldsymbol{E})^*).
\]
Therefore for any test function $\phi \in \mathscr{E}$ this Wick product operator $\Xi(\kappa)$ can be
evaluated $\langle \langle \Xi(\kappa)(\Phi \otimes \phi), \, \Psi \rangle \rangle
= \langle \langle \Xi(\kappa(\phi))\Phi, \, \Psi \rangle \rangle$ at $\Phi \otimes \phi$ and
$\Phi, \Psi \in (\boldsymbol{E})$, and for fixed $\Phi, \Psi \in (\boldsymbol{E})$ represents a
scalar distribution (as a function of $\phi \in \mathscr{E}$ compare (\ref{VectValotimesXi=intKerOp}) or
(\ref{VectValotimesXi=intKerOp})). Otherwise: for any test function $\phi \in \mathscr{E}$ the Wick
product operator $\Xi(\kappa(\phi))$ can be evaluated at $\Phi, \Psi \in (\boldsymbol{E})$, and gives the value
$\langle \langle \Xi(\kappa(\phi))\Phi, \, \Psi \rangle \rangle$, which is equal to a distribution (as a functional of the space-time test function $\phi$).
This is what might have been expected since the very work of Wick himself or from the
analysis of Bogoliubov and Shirkov \cite{Bogoliubov_Shirkov}, which already suggested that the general Wick product
of free fields determines, at each fixed space-time point, is a well-defined sesquilinear form for states ranging over a suitable dense domain.

But what is most important each order contribution to interacting Dirac and electromagnetic potential field, has
the form of a finite sum of integral kernel operators
\[
\Xi_{l,m}(\kappa_{l,m}) \in \mathscr{L}((\boldsymbol{E}) \otimes \mathscr{E}, \, (\boldsymbol{E})^*),
\]
respectively with
$\mathscr{E}_{i}^{*}$ -valued kernels $\kappa_{l,m}$, $i=1,2$, exactly as for the Wick polynomials of free fields
(at fixed space-time point), ant thus represent objects of the same class as the Wick polynomials of free fields,
i.e. finite sums of well-defined integral kernel operators with vector-valued kernels.
Moreover, the full interacting Dirac field and the interacting electromagnetic field (in all orders) have the form of Fock expansions (in the sense of \cite{obataJFA})
\[
\sum \limits_{l,m =0}^{\infty} \Xi_{l,m}(\kappa_{l,m}),
\]
which can be subject to computationally effective convergence criteria of \cite{obataJFA}, utilizing
symbol calculus of Obata.

Thus, all operators considered by the theory: free fields, Wick products of their derivatives, and contributions
to interacting fields are all finite sums of integral kernel operators in the sense of Obata \cite{obataJFA}
introduced in Subsection \ref{psiBerezin-Hida}. Among them the free fields operators, their derivatives
and Wick polynomials of derivatives of massive fields behave most ''smoothly'' and belong to
\[
\mathscr{L}((\boldsymbol{E}) \otimes \mathscr{E}, \, (\boldsymbol{E})) \cong
\mathscr{L}(\mathscr{E}, \, \mathscr{L}((\boldsymbol{E}), (\boldsymbol{E})).
\]
General Wick polynomials of derivatives of free fields (including zero mass fields) and
contributions to interacting fields, of which we can say that belong to the general class
of integral kernel operators, belong to
\[
\mathscr{L}((\boldsymbol{E}) \otimes \mathscr{E}, \, (\boldsymbol{E})^*) \cong
\mathscr{L}(\mathscr{E}, \, \mathscr{L}((\boldsymbol{E}), (\boldsymbol{E})^*),
\]
and are in this sense slightly more singular integral kernel operators
than the free fields themselves. In particular, we cannot say that they are operator-valued
distributions in the white noise sense but nonetheless, when evaluated at fixed elements of Hida
subspace of the Fock space, they represent scalar-valued distributions on the space-time
test function space $\mathscr{E}_2$ or $\mathscr{E}_1$.

Thus, we start with the fundamental integral kernel operators $\Xi_{0,1}(\kappa_{0,1})$, $\Xi_{1,0}(\kappa_{1,0})$
defined by the free fields of the theory.
But we should distinguish the free field integral kernel operators
\[
\boldsymbol{\psi} = \Xi_{0,1}(\kappa_{0,1}) + \Xi_{1,0}(\kappa_{1,0}), \,\,\,
A = \Xi_{0,1}(\kappa_{0,1}) + \Xi_{1,0}(\kappa_{1,0}),
\]
acting in their own (resp. fermionic or bosonic) Fock spaces from the corresponding free field integral kernel
operators
\[
\boldsymbol{\psi} = \Xi_{0,1}({}^{1}\kappa_{0,1}) + \Xi_{1,0}({}^{1}\kappa_{1,0}), \,\,\,
A = \Xi_{0,1}({}^{2}\kappa_{0,1}) + \Xi_{1,0}({}^{2}\kappa_{1,0}),
\]
both acting in the tensor product Fock space. In the last case the integral kernel operators
$\Xi_{0,1}({}^{1}\kappa_{0,1}), \Xi_{1,0}({}^{1}\kappa_{1,0})$ are defined by the integral
formula (\ref{electron-positron-photon-Xi}) in which the integration is restricted to Fermi variables $w$
only, and the operators $\Xi_{0,1}({}^{1}\kappa_{0,1}), \Xi_{1,0}({}^{1}\kappa_{1,0})$ act trivially as unit
operators on the second factor. Here ${}^{1}\kappa_{0,1}, {}^{1}\kappa_{0,1}$ are exactly the kernels
(\ref{kappa_0,1}) and (\ref{kappa_1,0}) corresponding to the Dirac field, and denoted with the additional
left-handed-superstript $1$, in order to distinguish them from the kernels ${}^{2}\kappa_{0,1}, {}^{2}\kappa_{0,1}$
(\ref{kappa_0,1kappa_1,0A'}), Subsection \ref{equivalentA-s}, in
$A = \Xi_{0,1}({}^{2}\kappa_{0,1}) + \Xi_{1,0}({}^{2}\kappa_{1,0})$ acting trivially on the first factor
in the tensor product of Fock spaces, and defined by the formula (\ref{electron-positron-photon-Xi})
in which the integration is restricted to Bose variables $w$ only.

And generally kernels $\kappa_{0,1}, \kappa_{1,0}$ respecting the condition of Lemma \ref{kappa0,1,kappa1,0psi},
Subsection \ref{psiBerezin-Hida}, corresponding to integral kernel operators
which act trivially as unit operators on the second bosonic Fock space factor with integration in their definition restricted to Fermi variables, will be denoted by
${}^{1}\kappa_{0,1}, {}^{1}\kappa_{0,1}$ with the additional superscript $1$; and vice versa for kernels corresponding to integral kernel operators acting trivially on the first fermionic Fock space factor with integration in their definition restricted to boson variables,
denoted by ${}^{2}\kappa_{0,1}, {}^{2}\kappa_{0,1}$ with the additional left-handed- superscript $2$.

Thus, we start with the following fundamental integral kernel operators 
\[
\Xi_{0,1}({}^{1}\kappa_{0,1}), \Xi_{1,0}({}^{1}\kappa_{1,0}), 
\Xi_{0,1}({}^{2}\kappa_{0,1}), \Xi_{1,0}({}^{2}\kappa_{1,0}), 
\]
determined by the free fields of the theory and their derivatives, corresponding to 
vector-valued distributions
\[
\begin{split}
{}^{1}\kappa_{0,1}, {}^{1}\kappa_{1,0} \in \mathscr{L}(E_{1}, \mathscr{E}_{1}^{*}) 
\cong  E_{1}^{*} \otimes \mathscr{E}_{1}^{*}, \\
{}^{2}\kappa_{0,1}, {}^{2}\kappa_{1,0} \in \mathscr{L}(E_{2}, \mathscr{E}_{2}^{*}) 
\cong  E_{2}^{*} \otimes  \mathscr{E}_{2}^{*}, 
\end{split}
\]
which have the property that they can be (uniquely) extended to elements (denoted by the same symbols) 
\[
\begin{split}
{}^{1}\kappa_{0,1}, {}^{1}\kappa_{1,0} \in \mathscr{L}(E_{1}^{*}, \mathscr{E}_{1}^{*}) 
\cong  E_{1} \otimes \mathscr{E}_{1}^{*}, \\
{}^{2}\kappa_{0,1}, {}^{2}\kappa_{1,0} \in \mathscr{L}(E_{2}^{*}, \mathscr{E}_{2}^{*}) 
\cong  E_{2}  \otimes \mathscr{E}_{2}^{*}, \\
{}^{1}\kappa_{0,1}(\xi), {}^{1}\kappa_{1,0}(\xi) \in \mathcal{O}_C = \mathcal{O}_{CB_1} 
 \,\,\, \textrm{if $\xi \in E_1$}, \\
{}^{2}\kappa_{0,1}(\xi), {}^{2}\kappa_{1,0}(\xi) \in \mathcal{O}_C
 \,\,\, 
\textrm{if $\xi \in E_2$},
\end{split}
\] 
compare Lemma \ref{kappa0,1,kappa1,0psi}, Subsection \ref{psiBerezin-Hida}
(for the kernels defining Dirac field), and respectively Lemma \ref{kappa0,1,kappa1,0ForA},
Subsection \ref{A=Xi0,1+Xi1,0} for the kernels defining the electromagnetic potential field.
Here $\mathcal{O}'_{C}(\mathbb{R}^4), \mathcal{O}'_{CB_2}(\mathbb{R}^4)$ denote the algebras of convolutors, respectively, of $\mathcal{S}_{B_1}(\mathbb{R}^4) = \mathcal{S}(\mathbb{R}^4), \mathcal{S}_{B_2}(\mathbb{R}^4)
= \mathcal{S}^{00}(\mathbb{R}^4)$, and $\mathcal{O}_{C}(\mathbb{R}^4), \mathcal{O}_{CB_2}(\mathbb{R}^4)$
are their preduals, compare Appendix \ref{convolutorsO'_C}. 
Because all the spaces $E_{i}, E_{i}^{*}, \mathscr{E}_{i}, \mathscr{E}_{i}^{*}$, $i=1,2$,
are nuclear then we have natural topological inclusions
\[
\mathscr{L}(E_{i}^{*}, \mathscr{E}_{i}^{*}) 
\cong  E_{i} \otimes \mathscr{E}_{i}^{*} \subset 
E_{i}^{*} \otimes \mathscr{E}_{i}^{*} \cong \mathscr{L}(E_{i}, \mathscr{E}_{i}^{*}), \,\,\,i=1,2 
\]
induced by the natural topological inclusions $E_{i} \subset E_{i}^{*}$ in both cases: if we endow $E_{i}$
with the topologies on $E_i$ inherited from $E_{i}^{*}$ and with their ordinary nuclear 
topologies, compare Prop. 43.7 and its Corollary  
in \cite{treves}. In the first case we obtain isomorphic inclusions by the cited Proposition,
as in case of nuclear spaces the projective tensor product coincides with the equicontinuous
and thus with the essentially unique tensor product in this category of linear topological 
spaces, compare \cite{treves}. Therefore, we simply have
\[
\begin{split}
{}^{1}\kappa_{0,1}, {}^{1}\kappa_{1,0} \in \mathscr{L}(E^{*}_{1}, \mathscr{E}_{1}^{*}) 
\cong  E_{1} \otimes \mathscr{E}_{1}^{*}, \\
{}^{2}\kappa_{0,1}, {}^{2}\kappa_{1,0} \in \mathscr{L}(E^{*}_{2}, \mathscr{E}_{2}^{*}) 
\cong  E_{2}  \otimes \mathscr{E}_{2}^{*}, \\
{}^{1}\kappa_{0,1}(\xi), {}^{1}\kappa_{1,0}(\xi) \in \mathcal{O}_C  = \mathcal{O}_{CB_1}
\,\,\, \textrm{if $\xi \in E_1$}, \\
{}^{2}\kappa_{0,1}(\xi), {}^{2}\kappa_{1,0}(\xi) \in  \mathcal{O}_C  \,\,\, 
\textrm{if $\xi \in E_2$}.
\end{split}
\] 
Recall that in case of kernels ${}^{1}\kappa_{0,1}, {}^{1}\kappa_{1,0}$,
respectively, ${}^{2}\kappa_{0,1}, {}^{2}\kappa_{1,0}$, defining the free fields 
$\boldsymbol{\psi}, A$ we have the space-time
test spaces $\mathscr{E}_{1}$, respectively, $\mathscr{E}_{2}$, 
given by the formula (\ref{mathscrE_1,mathscrE_2}) with $p_k = n_k=1$, and respectively, 
$p_k= n_k = 2$ and with $q_k = 4$ and $M=1$ in (\ref{mathscrE_1,mathscrE_2}).

In fact, we have two possible realizations of the free Dirac field $\boldsymbol{\psi}$,
having different commutation functions and pairings, which nonetheless are \emph{a priori}
equally good form the point of view of causal perturbative approach. This will be explained in
Subsection \ref{StandardDiracPsiField}. Thus besides the plane wave distributions
${}^{1}\kappa_{0,1}, {}^{1}\kappa_{0,1}$ defined by (\ref{kappa_0,1}) and (\ref{kappa_1,0}),
Subsection \ref{psiBerezin-Hida}, we can use (\ref{skappa_0,1}) and (\ref{skappa_1,0}) of
Subsection \ref{StandardDiracPsiField}. Similarly, we have two possibilities
for the realization of the free electromagnetic potential field $A$, both having the same commutation and pairing
functions, but with slightly different behavior in the infrared regime. This will be explained in
Subsection \ref{equivalentA-s}. Namely besides the formulas
(\ref{kappa_0,1kappa_1,0A'}) for ${}^{2}\kappa_{0,1}, {}^{2}\kappa_{0,1}$ we can use
(\ref{kappa_0,1kappa_1,0A}), Subsection \ref{A=Xi0,1+Xi1,0}. Correspondingly we have \emph{a priori}
four versions of perturbative QED, and although it seems that they all should be essentially equivalent,
they all should be subject to a systematic investigation. The formulas
(\ref{kappa_0,1kappa_1,0A}) and (\ref{skappa_0,1}) and (\ref{skappa_1,0}) are the standard
(in the Gupta-Bleuler gauge of QED) but the remaining three possibilities should also be
seriously considered.

Here we give definition and general rules in forming Wick product of integral kernel operators
\begin{equation}\label{FreeFieldOp}
\Xi_{l_1,m_1}\Big({}^{{}^{n_1}}_{{}_{1}}\kappa_{l_1,m_1}\Big), \ldots
\Xi_{l_M,m_M}\Big({}^{{}^{n_M}}_{{}_{M}}\kappa_{l_M,m_M}\Big)
\end{equation}
with general (not necessary equal to plane wave distributions defining the free fields, as we have in view
e.g. also their spatio-temporal-derivative fields)
\[
{}^{{}^{n_k}}_{{}_{k}}\kappa_{l_k,m_k}
\in \mathscr{L}(E_{{}_{n_k}}, \mathscr{E}^{*}_{{}_{n_k}}) \cong
E^{*}_{{}_{n_k}} \otimes \mathscr{E}^{*}_{{}_{n_k}}, \,\,\,\, k=1,2, \ldots M
\]
extendable to
\[
{}^{{}^{n_k}}_{{}_{k}}\kappa_{l_k,m_k}
\in \mathscr{L}(E^{*}_{{}_{n_k}}, \mathscr{E}^{*}_{{}_{n_k}}) \cong
E_{{}_{p_k}} \otimes \mathscr{E}^{*}_{{}_{n_k}}
\]
and with the property that
\[
{}^{{}^{n_k}}_{{}_{k}}\kappa_{l_k,m_k}(\xi) \in \mathcal{O}_C, \,\,\, \xi \in E_{{}_{n_k}}.
\]
Here
\[
n_k = \left\{ \begin{array}{l}
1 \\
\textrm{or} \\
2
\end{array} \right., \,\,\,\, \textrm{and} \,\,\,
(l_k, m_k) = \left\{ \begin{array}{l}
(0,1) \\
\textrm{or} \\
(1,0)
\end{array} \right.
\]
and the integral kernel operator
\[
\Xi_{l_k,m_k}\Big({}^{{}^{n_k}}_{{}_{k}}\kappa_{l_k,m_k}\Big),
\]
regarded as the operator on the said tensor product of Fock spaces, has the exceptional form
(similarly as for the operators defined by the free fields $A$ and $\boldsymbol{\psi}$)
that the integration in the general formula (\ref{electron-positron-photon-Xi})
for this operator is restricted to fermion variables, if $n_k = 1$, or to bose variables, if
$n_k = 2$.

We then define the Wick product
\[
\boldsymbol{}: \Xi_{l_1,m_1}\Big({}^{{}^{n_1}}_{{}_{1}}\kappa_{l_1,m_1}\Big) \cdots
\Xi_{l_M,m_M}\Big({}^{{}^{n_M}}_{{}_{M}}\kappa_{l_M,m_M}\big) \boldsymbol{:}
\]
of $M$ such operators as the ordinary product of these operators, but rearranged
in such a manner that all operators
\[
\Xi_{l_k,m_k}\Big({}^{{}^{n_k}}_{{}_{k}}\kappa_{l_k,m_k}\Big)
\]
with $(l_k, m_k) = (1,0)$ stand to the left of all operators
\[
\Xi_{l_k,m_k}\Big({}^{{}^{n_k}}_{{}_{k}}\kappa_{l_k,m_k}\Big)
\]
with $(l_k, m_k) = (0,1)$, multiplied in addition by the factor $(-1)^p$
with $p$ equal to the parity of the
permutation performed upon Fermi operators, having $n_k = 1$ and corresponding to the Fermi variables,
required to bring the operators into the required ''normal'' order. 

\begin{center}
{\small RULE I}
\end{center}
\emph{We have the following computational rule}
\begin{multline*}
\boldsymbol{}: \Xi_{l_1,m_1}\Big({}^{{}^{n_1}}_{{}_{1}}\kappa_{l_1,m_1}\Big) \cdots
\Xi_{l_M,m_M}\Big({}^{{}^{n_1}}_{{}_{M}}\kappa_{l_M,m_M}\big) \boldsymbol{:} \\
= \Xi_{l,m}(\kappa_{lm}),
\\
l = l_1 + \cdots l_M, \,\,\, m = m_1 + \cdots m_M
\end{multline*}
\emph{where}
\[
\kappa_{l,m} = \Big({}^{{}^{n_1}}_{{}_{1}}\kappa_{l_1,m_1} \Big) \overline{\otimes} \cdots \overline{\otimes} \,\,
\Big( {}^{{}^{n_1}}_{{}_{M}}\kappa_{l_M,m_M} \Big)
\]
\emph{stands for the ordinary tensor product
\begin{multline*}
\Big({}^{{}^{n_1}}_{{}_{1}}\kappa_{l_1,m_1}\Big) \otimes \cdots \otimes \,\,
\Big({}^{{}^{n_1}}_{{}_{M}}\kappa_{l_M,m_M}\Big) \in
E_{{}_{n_1}} \otimes \mathscr{E}^{*}_{{}_{n_1}} \otimes \cdots \otimes
E_{{}_{n_M}} \otimes \mathscr{E}^{*}_{{}_{n_M}} \\
\cong E_{{}_{n_1}} \otimes \cdots \otimes E_{{}_{n_M}} \otimes
\mathscr{E}^{*}_{{}_{n_1}} \otimes \cdots \otimes \mathscr{E}^{*}_{{}_{n_M}} \\
\cong \mathscr{L}(E^{*}_{{}_{n_1}} \otimes \cdots \otimes E^{*}_{{}_{n_M}}, \,\,
\mathscr{E}^{*}_{{}_{n_1}} \otimes \cdots \otimes \mathscr{E}^{*}_{{}_{n_M}})
\end{multline*}
1) separately symmetrized with respect to all Bose variables, lying among the first $l$ variables,
2) separately symmetrized with respect to all Bose variables, lying among the last $m$
variables, 3) separately antisymmetrized with respect to all Fermi variables which
lie among the first $l$ variables, 4) separately antisymmetrized with respect
to all Fermi variables lying among the
last $m$ variables, finally 5) the result multiplied by the factor
$(-1)^p$, where $p$ is the parity of the permutation performed upon the Fermi operators necessary to rearrange them into
the order in which they stand in the general formula (\ref{electron-positron-photon-Xi})
for $\Xi_{l,m}(\kappa_{l,m})$. Here by definition $n_k$ is counted among the first $l$ variables
iff the corresponding $(l_k, m_k) = (1,0)$, and $n_k$ is counted among last $m$
variables iff the corresponding $(l_k, m_k) = (0,1)$.}

This is effective computational rule because in practical situations, e.g. for the Wick product of
integral kernel operators defined by free fields of the theory, the tensor product of the corresponding kernels
may be represented by ordinary products of the functions representing kernels:
\begin{multline*}
\Big({}^{{}^{n_1}}_{{}_{1}}\kappa_{l_1,m_1}\Big) \otimes \cdots \otimes \,\,
\Big({}^{{}^{n_1}}_{{}_{M}}\kappa_{l_M,m_M}\Big)(w_1, \ldots, w_M; X_1, \ldots, X_M) \\ =
\Big({}^{{}^{n_1}}_{{}_{1}}\kappa_{l_1,m_1}\Big)(w_1, X_1) \cdots
\Big({}^{{}^{n_1}}_{{}_{M}}\kappa_{l_M,m_M}\Big)(w_M, X_M), \\
X_k = \left\{ \begin{array}{ll}
(a_k,x_k), & \textrm{for $X_k$ corresponding to Fermi variables $w_k= (s_k, \boldsymbol{\p}_k)$} \\
\textrm{or} & \\
(\mu_k,x_k), & \textrm{for $X_k$ corresponding to Bose variables $w_k= (\nu_k, \boldsymbol{\p}_k)$}
\end{array} \right., \\
w_k = \left\{ \begin{array}{ll}
(s_k,\boldsymbol{\p}_k), & \textrm{for Fermi variables $w_k$} \\
\textrm{or} & \\
(\nu_k,\boldsymbol{\p}_k), & \textrm{for Bose variables $w_k$}
\end{array} \right.,
\\
\textrm{$x_k$ denotes for each $k$ space-time coordinates variable}, \\
s_k \in \{1,2,3,4\}, \,\,\, \mu_k, \nu_k \in \{0,1,2,3\}, a_k \in \{1,2,3,4\}.
\end{multline*}
In case of Wick product integral kernel operators corresponding to fixed components of the fields,
the respective values of $\mu_k$ and $a_k$ will be correspondingly fixed, and the test spaces $\mathscr{E}_{n_k}$
will be equal (\ref{mathscrE_1,mathscrE_2}) with $q_k = 1$, \emph{i.e.} scalar test spaces.
Thus, the symmetrized/antisymmetrized
tensor product $\overline{\otimes}$ of the kernels corresponding to free fields can be easily and explicitly
computed, by the indicated symmetrizations and antisymmetrizations applied to the kernel functions:
\[
\Big({}^{{}^{n_1}}_{{}_{1}}\kappa_{l_1,m_1}\Big) \otimes \cdots \otimes \,\,
\Big({}^{{}^{n_1}}_{{}_{M}}\kappa_{l_M,m_M}\Big)(w_1, \ldots, w_M; X_1, \ldots, X_M),
\]
remembering that the variable $(w_k, X_k)$ is counted among the first $l$ variables
iff $(l_k, m_k) = (1,0)$, and the variable $(w_k, X_k)$ is counted among the last $m$ variables
iff $(l_k, m_k) = (0,1)$.

The Rule I can be justified by utilizing the fact that
\[
\Xi_{l_k,m_k}\Big({}^{{}^{n_k}}_{{}_{k}}\kappa_{l_k,m_k}(X_k)\Big),
\]
exist pointwisely as Pettis integral for each fixed point $X_k$, with the scalar distribution
\[
{}^{{}^{n_k}}_{{}_{k}}\kappa_{l_k,m_k}(X_k)
\]
(with fixed $X_k$) represented by the scalar function
\[
w_k \longmapsto {}^{{}^{n_k}}_{{}_{k}}\kappa_{l_k,m_k}(w_k,X_k)
\]
kernel, as in the proof of Bogoliubov-Shirkov Hypothesis in Subsection \ref{BSH}.

From the Rule I it easily follows that the Wick product
of the class of integral kernel operators (\ref{FreeFieldOp}), subsuming free field operators, is a well-defined
(sum of) integral kernel operator(s) $\Xi(\kappa_{l,m})$ with the kernel(s)
\begin{equation}\label{StrongkappaProperty}
\kappa_{l,m} \in \mathscr{L}(E^{*}_{{}_{n_1}} \otimes \cdots \otimes E^{*}_{{}_{n_M}}, \,\,
\mathscr{E}^{*}_{{}_{n_1}} \otimes \cdots \otimes \mathscr{E}^{*}_{{}_{n_M}}), \,\,\, M= l+m
\end{equation}
and thus with
\[
\Xi_{l,m}(\kappa_{l,m}) \in
\mathscr{L}\big((\boldsymbol{E}) \otimes \mathscr{E}, (\boldsymbol{E})\big) \cong
\mathscr{L}\Big(\mathscr{E}, \,\, \mathscr{L}\big((\boldsymbol{E}), \, (\boldsymbol{E})\big) \Big)
\]
by Thm. \ref{obataJFA.Thm.3.13}, Subsection \ref{psiBerezin-Hida}, for
\begin{equation}\label{tensormathscrE}
\mathscr{E} = \mathscr{E}^{*}_{{}_{n_1}} \otimes \cdots \otimes \mathscr{E}^{*}_{{}_{n_M}}.
\end{equation}
In particular it defines an operator-valued distribution on the tensor product
(\ref{tensormathscrE}) of space-time test function spaces $\mathscr{E}_1, \mathscr{E}_2$ with
$\mathscr{E}_{{}_{n_k}} = \mathscr{E}_1$ iff $n_k = 1$ and $\mathscr{E}_{{}_{n_k}} = \mathscr{E}_2$ iff
$n_k = 2$ (respectively for the Fermi operator or Bose operator in the Wick product).

It is easily seen that we get in this way a graded Wick algebra which subsumes in particular all
finite sums of integral kernel operators $\Xi_{l,m}(\kappa_{l,m})$
with kernels $\kappa_{l,m}$ having the property (\ref{StrongkappaProperty}).
Let
\[
\Xi(\kappa'_{l',m'}) \,\,\, \textrm{and} \,\,\,
\Xi_{l'',m''}(\kappa''_{l'',m''})
\]
be two such operators with
\[
\begin{split}
\kappa'_{l',m'} \in \mathscr{L}(E^{*}_{{}_{n'_1}} \otimes \cdots \otimes E^{*}_{{}_{n'_{M'}}}, \,\,
\mathscr{E}^{*}_{{}_{n'_1}} \otimes \cdots \otimes \mathscr{E}^{*}_{{}_{n'_{M'}}}), \\
\kappa''_{l'',m''} \in \mathscr{L}(E^{*}_{{}_{n''_1}} \otimes \cdots \otimes E^{*}_{{}_{n''_{M''}}}, \,\,
\mathscr{E}^{*}_{{}_{n''_1}} \otimes \cdots \otimes \mathscr{E}^{*}_{{}_{n''_{M''}}})
\end{split}
\]
It is easily seen that
we have the following rule for Wick product of such operators
\[
\boldsymbol{:} \Xi(\kappa'_{l',m'}) \, \Xi_{l'',m''}(\kappa''_{l'',m''}) \boldsymbol{:} =
\Xi_{l,m}(\kappa_{l,m}), \,\,\, l = l' + l'', \, m = m' + m'',
\]
where
\[
\kappa_{l,m} = \kappa'_{l',m'} \, \overline{\otimes} \,\, \kappa''_{l'',m''}
\]
is equal to the ordinary tensor product
\begin{multline*}
\kappa'_{l',m'} \, \otimes \,\, \kappa''_{l'',m''} \\
\in E_{{}_{n'_1}} \otimes \cdots \otimes E_{{}_{n'_{M'}}} \otimes
E_{{}_{n''_1}} \otimes \cdots \otimes E_{{}_{n''_{M''}}} \otimes
\mathscr{E}^{*}_{{}_{n'_1}} \otimes \cdots \otimes \mathscr{E}^{*}_{{}_{n'_{M'}}} \otimes
\mathscr{E}^{*}_{{}_{n''_1}} \otimes \cdots \otimes \mathscr{E}^{*}_{{}_{n''_{M''}}} \\ \cong
\mathscr{L}(E^{*}_{{}_{n'_1}} \otimes \cdots \otimes E^{*}_{{}_{n'_{M'}}} \otimes
E^{*}_{{}_{n''_1}} \otimes \cdots \otimes E^{*}_{{}_{n''_{M''}}}, \,\,
\mathscr{E}^{*}_{{}_{n'_1}} \otimes \cdots \otimes \mathscr{E}^{*}_{{}_{n'_{M'}}} \otimes
\mathscr{E}^{*}_{{}_{n''_1}} \otimes \cdots \otimes \mathscr{E}^{*}_{{}_{n''_{M''}}}),
\end{multline*}
\begin{enumerate}
\item[1)]
multiplied by $(-1)^p$ where $p$ is the parity of the permutation which has to be applied to the
Fermi operators lying among the Hida operators put in the order
\[
\partial_{w'_1}^* \cdots \partial_{w'_{M'}} \partial_{w''_{1}}^* \cdots \partial_{w''_{M''}}
\]
in which the Hida operators are put formally together in the order in which they stand in the general formula
(\ref{electron-positron-photon-Xi}) for $\Xi_{l',m'}(\kappa'_{l',m'})$ (first) and in the general formula
(\ref{electron-positron-photon-Xi}) for $\Xi_{l'',m''}(\kappa''_{l'',m''})$ (second), in order to rearrange them into
the order in which they stand in the general formula (\ref{electron-positron-photon-Xi})
for $\Xi_{l,m}(\kappa''_{l,m})$
\item[2)]
separately symmetrized with respect to all Bose variables which lie within the first $l$ variables,
\item[3)]
separately symmetrized with respect to all Bose variables which lie within the last $m$
variables,
\item[4)]
separately antisymmetrized with respect to all Fermi variables which lie among the first $l$
variables,
\item[5)]
separately antisymmetrized with respect to all Fermi variables which lie among the last
$m$ variables,
\item[6)]
the $n'_k$-th or respectively $n''_k$-th variable is counted as lying among the first $l$ variables if it lies
among the first $l'$ variables in $\kappa'_{l',m'}$ or among the first $l''$ variables
of the kernel $\kappa''_{l'',m''}$. The remaining variables are counted as the last $m$ variables.
\end{enumerate}

In fact Wick product is well-defined on a much larger class of integral kernel operators
$\Xi_{l,m}(\kappa_{l,m})$, because for its validity it is sufficient that the kernels $\kappa_{l,m}$
respect the condition of Theorem \ref{obataJFA.Thm.3.13}, considerably weaker than the condition
(\ref{StrongkappaProperty}). In this wider class of operators the last rule for computation
of the Wick product remains true.

A much more interesting case we encounter when among the integral kernel operators
(\ref{FreeFieldOp}) there are present such, which are equal to Wick polynomials
of free fields at one and the same space-time point. Now we give general definition of
such a Wick product of (fixed components of) free fields at one and the same space-time point, and show that the
corresponding integral kernel operator lies among the class which can be placed into the
above Wick product. The resulting integral kernel operator $\Xi$ will be a finite sum of well-defined
integral kernel operators $\Xi(\kappa_{l,m})$ with the kernel(s)
\begin{equation}\label{StrongkappaProperty}
\kappa_{l,m} \in \mathscr{L}(E_{{}_{n_1}} \otimes \cdots \otimes E_{{}_{n_M}}, \,\,
\mathscr{E}^{*}_{{}_{n_1}} \otimes \cdots \otimes \mathscr{E}^{*}_{{}_{n_M}}), \,\,\, M= l+m
\end{equation}
and thus with
\[
\Xi_{l,m}(\kappa_{l,m}) \in
\mathscr{L}\big((\boldsymbol{E}) \otimes \mathscr{E}, (\boldsymbol{E})^*\big) \cong
\mathscr{L}\Big(\mathscr{E}, \,\, \mathscr{L}\big((\boldsymbol{E}), \, (\boldsymbol{E})^*\big) \Big)
\]
by the generalization of Thm. 3.9 of \cite{obataJFA} to the tensor product of Fock spaces,
compare Subsection \ref{psiBerezin-Hida}. Therefore, Wick product of free fields (or their derivatives)
$\Xi$ at the fixed space-time point belongs to the general class of finite sums of integral kernel operators with vector-valued kernels,
which in general does not belong to
\[
\mathscr{L}\big((\boldsymbol{E}) \otimes \mathscr{E}, (\boldsymbol{E})\big) \cong
\mathscr{L}\Big(\mathscr{E}, \,\, \mathscr{L}\big((\boldsymbol{E}), \, (\boldsymbol{E})\big) \Big)
\]
if among the factors in the Wick product (at fixed point) there are zero mass fields or their derivatives.
But if among the factors there are no factors corresponding to zero mass fields (or their derivatives)
then the resulting integral kernel operator $\Xi$ -- Wick product at fixed point -- will be a finite sum of
well-defined integral kernel operators $\Xi(\kappa_{l,m})$ with the kernels respecting the condition of Thm.
\ref{obataJFA.Thm.3.13}, \emph{i. e.} with
\[
\Xi_{l,m}(\kappa_{l,m}) \in
\mathscr{L}\big((\boldsymbol{E}) \otimes \mathscr{E}, (\boldsymbol{E})\big) \cong
\mathscr{L}\Big(\mathscr{E}, \,\, \mathscr{L}\big((\boldsymbol{E}), \, (\boldsymbol{E})\big) \Big)
\]
by the generalization of Thm. 3.13 of \cite{obataJFA} to the tensor product of Fock spaces,
compare Thm. \ref{obataJFA.Thm.3.13} of Subsection \ref{psiBerezin-Hida}, and
with $\mathscr{E}_{1}^{*}$-valued or respectively
$\mathscr{E}_{2}^*$-valued distribution kernels, for both nuclear space-time test function spaces:
$\mathscr{E}_{1}$ and for $\mathscr{E}_{2}$ given by the special case of (\ref{mathscrE_1,mathscrE_2})
with $M=1$ and $q_k=1$ in it, \emph{i.e.}
\[
\begin{split}
\mathscr{E}_{1} = \mathcal{S}_{H_{(4)}}(\mathbb{R}^4;\mathbb{C})=
\mathcal{S}(\mathbb{R}^4; \mathbb{C}) \,\,\, \textrm{or} \\
\mathscr{E}_{2} = \mathcal{S}_{\mathscr{F}A^{(4)}\mathscr{F}^{-1}}(\mathbb{R}^4;\mathbb{C})=
\mathcal{S}^{00}(\mathbb{R}^4; \mathbb{C}).
\end{split}
\]

For the need of causal perturbative construction of interacting fields it is sufficient to confine attention to integral kernel operators representing the respective components of free fields, of their spatio-temporal derivatives, their Wick products, their integrals with pairing functions (e.g. convolutions of Wick products of spatio-temporal
derivatives of fixed components of free fields with pairing distributions, \emph{i. e.}'' pairing functions'').
Therefore, we confine ourselves to fixed components of the free fields and of their spatio-temporal derivatives
and thus to scalar-valued space-time test
function spaces $\mathscr{E}_{1}= \mathcal{S}(\mathbb{R}^4; \mathbb{C})$ or respectively
$\mathscr{E}_{2} = \mathcal{S}^{00}(\mathbb{R}^4; \mathbb{C})$.
Correspondingly to this we consider integral kernel operators with the vector-valued kernels corresponding
to fixed components of free fields which can be represented
by the functions
\begin{equation}\label{KernelsOfFreeFieldComponents}
\begin{split}
{}^{1}\kappa_{0,1}(w;X) = {}^{1}\kappa_{0,1}(s, \boldsymbol{\p}; a, x), \,\,\,
{}^{1}\kappa_{1,0}(w;X) = {}^{1}\kappa_{0,1}(s, \boldsymbol{\p}; a, x) \,\,\, \textrm{or} \\
{}^{2}\kappa_{0,1}(w;X) = {}^{2}\kappa_{0,1}(\nu, \boldsymbol{\p}; \mu, x), \,\,\,
{}^{2}\kappa_{1,0}(w;X) = {}^{2}\kappa_{0,1}(\nu, \boldsymbol{\p}; \mu, x),
\end{split}
\end{equation}
with fixed values of the discrete indices $a, \mu$. To this class (\ref{KernelsOfFreeFieldComponents}) of kernels
we add their spatio-temporal derivatives
\begin{equation}\label{KernelsOfDerivativesOfFreeFieldComponents}
\begin{split}
\partial^{\alpha} \,\, {}^{1}\kappa_{0,1}(w;X) = \partial^{\alpha} \,\, {}^{1}\kappa_{0,1}(s, \boldsymbol{\p}; a, x), \,\,\,
\partial^{\alpha} \,\, {}^{1}\kappa_{1,0}(w;X) = \partial^{\alpha} \,\, {}^{1}\kappa_{0,1}(s, \boldsymbol{\p}; a, x) \,\,\, \textrm{or} \\
\partial^{\alpha} \,\, {}^{2}\kappa_{0,1}(w;X) = \partial^{\alpha} \,\, {}^{2}\kappa_{0,1}(\nu, \boldsymbol{\p}; \mu, x), \,\,\,
\partial^{\alpha} \,\, {}^{2}\kappa_{1,0}(w;X) = \partial^{\alpha} \,\, {}^{2}\kappa_{0,1}(\nu, \boldsymbol{\p}; \mu, x), \\
\textrm{where} \\
\alpha = (\alpha_0, \alpha_1, \alpha_2, \alpha_3) \in \mathbb{N}_{0}^{4} \,\,\, \textrm{and} \,\,\,
\partial^{\alpha} = \frac{\partial^{|\alpha_0|}}{(\partial x_0)^{\alpha_0}}
\frac{\partial^{|\alpha_1|}}{(\partial x_1)^{\alpha_1}}\frac{\partial^{|\alpha_2|}}{(\partial x_2)^{\alpha_2}}
\frac{\partial^{|\alpha_3|}}{(\partial x_3)^{\alpha_3}}
\end{split}
\end{equation}

\begin{defin}\label{K_0}
The class $\mathfrak{K}_0$ of kernels we are considering in the sequel consists of the plane wave kernels
(\ref{KernelsOfFreeFieldComponents}) defining the free fields of the theory and of 
their spatio-temporal derivatives (\ref{KernelsOfDerivativesOfFreeFieldComponents}), with fixed values
of the indices $a, \mu, \alpha$. 
\end{defin}

It is immediately seen that if for any $\kappa_{0,1}, \kappa_{1,0} \in \mathfrak{K}_0$ we fix the free field index $a$ and the multi-index
$\alpha$ of space-time derivatives as well as the space-time point $x$, then the obtained functions
\begin{align*}
(s, \boldsymbol{\p}) \longrightarrow \kappa_{0,1}(s, \boldsymbol{\p}; \alpha, a, x), \\
(s, \boldsymbol{\p}) \longrightarrow \kappa_{1,0}(s, \boldsymbol{\p}; \alpha, a, x), \\
\kappa_{0,1}, \kappa_{1,0} \in \mathfrak{K}_0 \,\,\, \textrm{with fixed} \, (\alpha, a, x)
\end{align*}
represent well-defined continuous functionals on the corresponding nuclear spaces
$E_{{}_{n_i}}$. Denoting so obtained functionals by $\kappa_{0,1}(\alpha, a, x) , \kappa_{1,0}(\alpha, a, x) $ we have
\[
\kappa_{0,1}(\alpha, a, x) , \kappa_{1,0}(\alpha, a, x) \in E_{{}_{n_i}}^{*}.
\]
Therefore by Theorem 2.2 of \cite{hida} or its generalization embracing fermionic case -- Thm. \ref{Xi_l,m} of Subsection \ref{psiBerezin-Hida} --
we obtain the following
\begin{prop*}
Each component of a free field, its negative or positive frequency part or their spatio-temporal derivative
evaluated at fixed space-time point $x$ represent well-defined
integral kernel operator transforming the Hida space into its strong dual. Moreover, the negative frequency parts of free
fields and their spatio-temporal derivatives evaluated at fixed space-time point, represent integral
kernel operators transforming continuously the Hida space into itself, and thus are ordinary (densely defined)
operators on the Fock space.
\end{prop*}

Upon the integral kernel operators determined by the vector valued kernels
$\mathfrak{K}_0$ we perform the operations of Wick product (Rule I), Wick products at the same space-time point
(Rule II), spatio-temporal derivations (Rule III), integrations (IV and V) and finally convolutions with pairing functions
(Rule VI). Correspondingly to each of the said operations there exists the corresponding Rule performed upon the kernels, corresponding to the operators. Of course the operations performed upon the kernels
in $\mathfrak{K}_0$ and determined by the Rules will extend the initial class $\mathfrak{K}_0$.
We use a general notation
\[
{}^{{}^{n}}_{{}_{k}}\kappa_{l,m}(s, \boldsymbol{\p};x), \,\, n = 1
\]
for a kernel
\[
\partial^{\alpha} \,\, {}^{1}\kappa_{l,m}(s, \boldsymbol{\p}; a, x), \,\,\, (l,m) = (0,1) \,\, \textrm{or} \,\, =(1,0)
\]
with fixed indices $a, \alpha$ and
with ${}^{1}\kappa_{0,1}(s, \boldsymbol{\p}; a, x)$ equal to the plane wave kernel defining the free Dirac field.
Similarly, we will denote simply by
\[
{}^{{}^{n}}_{{}_{k}}\kappa_{l,m}(\nu, \boldsymbol{\p}; x), \,\, n = 2
\]
the kernel
\[
\partial^{\alpha} \,\, {}^{2}\kappa_{l,m}(\nu, \boldsymbol{\p}; \mu, x), \,\,\, (l,m) = (0,1) \,\, \textrm{or} \,\, =(1,0)
\]
with fixed indices $\mu, \alpha$ and
with ${}^{2}\kappa_{l,m}(\nu, \boldsymbol{\p}; \mu, x)$ equal to the plane wave kernel defining the free
electromagnetic potential field.

Assuming 
\[
{}^{{}^{n_k}}_{{}_{k}}\kappa_{l_k,m_k} \in \mathfrak{K}_0, \,\,\, k = 1, \ldots, M,
\]
we consider the following Wick monomials,
i.e. Wick products at the same space-time point, of the following operators 
\begin{equation}\label{FreeFieldOpAtx}
\Xi_{l_1,m_1}\Big({}^{{}^{n_1}}_{{}_{1}}\kappa_{l_1,m_1}\Big), \ldots 
\Xi_{l_M,m_M}\Big({}^{{}^{n_M}}_{{}_{M}}\kappa_{l_M,m_M}\Big)
\end{equation}
with general (not necessary equal to plane wave distributions defining the free fields, as we have in view 
also their spatio-temporal-derivative fields) kernels
\[
{}^{{}^{n_k}}_{{}_{k}}\kappa_{l_k,m_k}
\in \mathscr{L}(E_{{}_{n_k}}, \mathscr{E}^{*}_{{}_{n_k}}) \cong 
E^{*}_{{}_{n_k}} \otimes \mathscr{E}^{*}_{{}_{n_k}}, , \,\,\,\, k=1,2, \ldots M
\]
representable by ordinary functions, respecting the conditions expressed in Lemma \ref{kappa0,1,kappa1,0psi}, Subsection \ref{psiBerezin-Hida} or respectively Lemma \ref{kappa0,1,kappa1,0ForA},
Subsection \ref{A=Xi0,1+Xi1,0}, \emph{i.e.} extendable to elements
\begin{equation}\label{ExtedibilityCondition}
{}^{{}^{n_k}}_{{}_{k}}\kappa_{l_k,m_k}
\in \mathscr{L}(E^{*}_{{}_{n_k}}, \mathscr{E}^{*}_{{}_{n_k}}) \cong 
E_{{}_{n_k}} \otimes \mathscr{E}^{*}_{{}_{n_k}}
\end{equation}
with the property that
\begin{equation}\label{ConvolutabilityCondition}
{}^{{}^{n_k}}_{{}_{k}}\kappa_{l_k,m_k}(\xi) \in \mathcal{O}_C(\mathbb{R}^4; \mathbb{C}), 
\,\,\, \xi \in E_{{}_{n_k}}.
\end{equation}
Here 
\[
n_k = \left\{ \begin{array}{l}
1 \\
\textrm{or} \\
2
\end{array} \right., \,\,\,\, \textrm{and} \,\,\,
(l_k, m_k) = \left\{ \begin{array}{l}
(0,1) \\
\textrm{or} \\
(1,0)
\end{array} \right.
\]
and the integral kernel operator 
\[
\Xi_{l_k,m_k}\Big({}^{{}^{n_k}}_{{}_{k}}\kappa_{l_k,m_k}\Big),
\]
regarded as the operator on the said tensor product of Fock spaces, has the exceptional form
(similarly as for the operators defined by the free fields $A$ and $\boldsymbol{\psi}$)
that the integration in the general formula (\ref{electron-positron-photon-Xi})
for this operator is restricted to fermion variables, if $n_k = 1$, or to bose variables, if
$n_k = 2$. 

Validity of (\ref{ExtedibilityCondition}) and (\ref{ConvolutabilityCondition}) for spatio-temporal
derivatives of the plane wave kernels (\ref{KernelsOfFreeFieldComponents}) can be proved exactly as for
kernels (\ref{KernelsOfFreeFieldComponents}) themselves by repeating the argument of the proof of
Lemma \ref{kappa0,1,kappa1,0psi}, Subsection \ref{psiBerezin-Hida} or respectively 
Lemma \ref{kappa0,1,kappa1,0ForA},
Subsection \ref{A=Xi0,1+Xi1,0}. 
 
In fact in construction of interacting fields in the standard spinor QED it would be sufficient to consider only 
the kernels  (\ref{KernelsOfFreeFieldComponents}) and the kernels which arise by performing upon them the respective operations determined by the Rules I - VI, except the III-rd, given below. This is because no spatio-temporal derivatives of free fields enter the interaction Lagrangian in spinor QED, but only free fields themselves. But in case of scalar QED the interaction Lagrangian contains derivatives of 
free fields, so in that case spatio-temporal derivatives of the kernels determining the scalar free field has 
to be taken into consideration.  

So let
\[
{}^{{}^{n_k}}_{{}_{k}}\kappa_{l_k,m_k} \in \mathfrak{K}_0, \,\,\, k = 1, \ldots, M.
\]
Then for each fixed space-time point $x$ the scalar integral kernel operators
\begin{equation}\label{FreeFieldOpAtx}
\Xi_{l_1,m_1}\Big({}^{{}^{n_1}}_{{}_{1}}\kappa_{l_1,m_1}(x)\Big), \ldots
\Xi_{l_M,m_M}\Big({}^{{}^{n_M}}_{{}_{M}}\kappa_{l_M,m_M}(x)\big)
\end{equation}
determined by scalar kernel functions
\[
{}^{{}^{n_k}}_{{}_{k}}\kappa_{l_k,m_k}(x): w_{n_k} \longmapsto
{}^{{}^{n_k}}_{{}_{k}}\kappa_{l_k,m_k}(w_{n_k}; x),
\]
are well defined generalized operators transforming continuously the Hida space $(\boldsymbol{E})$
into its strong dual $(\boldsymbol{E})^*$, and exist pointwisely as Pettis integrals
(\ref{electron-positron-photon-Xi}) with integration in (\ref{electron-positron-photon-Xi})
restricted to fermi variables, iff $n_k = 1$, or to bose variables, iff $n_k = 2$, compare
Subsection \ref{BSH}.
Moreover for each fixed $x$ there exist a well defined Wick product of the operators
(\ref{FreeFieldOpAtx})
\begin{equation}\label{WickFreeFieldOpAtx}
\boldsymbol{:}\Xi_{l_1,m_1}\Big({}^{{}^{n_1}}_{{}_{1}}\kappa_{l_1,m_1}(x)\Big), \ldots
\Xi_{l_M,m_M}\Big({}^{{}^{n_1}}_{{}_{M}}\kappa_{l_M,m_M}(x)\big) \boldsymbol{:}
\end{equation}
defined as the ordinary product of these operators, but rearranged in the so called ''normal'' order,
in which all operators
\begin{equation}\label{Xi0,1}
\Xi_{l_k,m_k}\Big({}^{{}^{n_k}}_{{}_{k}}\kappa_{l_k,m_k}(x)\Big)
\end{equation}
with $(l_k, m_k) = (1,0)$ stand to the left of all operators
\begin{equation}\label{Xi1,0}
\Xi_{l_k,m_k}\Big({}^{{}^{n_k}}_{{}_{k}}\kappa_{l_k,m_k}(x)\Big)
\end{equation}
with $(l_k, m_k) = (0,1)$, multiplied in addition by the factor $(-1)^p$
with $p$ equal to the parity of the
permutation performed upon Fermi operators, having $n_k = 1$ and corresponding to the Fermi variables,
required to bring the operators into the required ''normal'' order.

\begin{center}
{\small RULE II}
\end{center}
\emph{We have the following computational rule}
\begin{multline*}
\boldsymbol{}: \Xi_{l_1,m_1}\Big({}^{{}^{n_1}}_{{}_{1}}\kappa_{l_1,m_1}(x)\Big) \cdots
\Xi_{l_M,m_M}\Big({}^{{}^{n_1}}_{{}_{M}}\kappa_{l_M,m_M}(x)\big) \boldsymbol{:} \\
= \Xi_{l,m}(\kappa_{lm}(x)),
\\
l = l_1 + \cdots l_M, \,\,\, m = m_1 + \cdots m_M
\end{multline*}
\emph{where the ordinary function representing the kernel $\kappa_{l,m}$}
\[
\kappa_{l,m}(w_1, \ldots, w_M; x) =
\Big({}^{{}^{n_1}}_{{}_{1}}\kappa_{l_1,m_1} \Big) \overline{\dot{\otimes}} \cdots \overline{\dot{\otimes}} \,\,
\Big( {}^{{}^{n_M}}_{{}_{M}}\kappa_{l_M,m_M} \Big)(w_1, \ldots, w_M; x)
\]
\emph{is equal to the ordinary product
\begin{multline*}
\Big({}^{{}^{n_1}}_{{}_{1}}\kappa_{l_1,m_1}\Big) \dot{\otimes} \cdots \dot{\otimes} \,\,
\Big({}^{{}^{n_M}}_{{}_{M}}\kappa_{l_M,m_M}\Big)(w_1, \ldots, w_M; x) \\ =
\Big({}^{{}^{n_1}}_{{}_{1}}\kappa_{l_1,m_1}\Big)(w_1;x)
\cdots \Big({}^{{}^{n_M}}_{{}_{M}}\kappa_{l_M,m_M}\Big)(w_M;x),
\end{multline*}
1) separately symmetrized with respect to all Bose variables, lying among the first $l$ variables,
2) separately symmetrized with respect to all Bose variables, lying among the last $m$
variables, 3) separately antisymmetrized with respect to all Fermi variables which
lie among the first $l$ variables, 4) separately antisymmetrized with respect
to all fermi variables lying among the
last $m$ variables, finally 5) the result multiplied by the factor
$(-1)^p$, where $p$ is the parity of the permutation performed upon the Fermi operators necessary to rearrange them into
the order in which they stand in the general formula (\ref{electron-positron-photon-Xi})
for $\Xi_{l,m}(\kappa_{l,m})$. Here by definition $n_k$ is counted among the first $l$ variables
iff the corresponding $(l_k, m_k) = (1,0)$, and $n_k$ is counted among last $m$
variables iff the corresponding $(l_k, m_k) = (0,1)$.}

Again the Rule II can be justified by using the fact that the operators (\ref{Xi0,1})
exist pointwisely as Pettis integrals, and represent operators mapping continuously 
the strong dual $(\boldsymbol{E})^*$ of the Hida space into its strong dual $(\boldsymbol{E})^*$
(continuous as well as operators $(\boldsymbol{E}) \rightarrow (\boldsymbol{E})^*$), and similarly
we have for the operators (\ref{Xi1,0}), representing continuous operators 
$(\boldsymbol{E}) \rightarrow (\boldsymbol{E})$ 
(as well  continuous as operators $(\boldsymbol{E}) \rightarrow (\boldsymbol{E})^*$). The proof, using essentially the same arguments as that used in the proof of Bogoliubov-Shirkov Hypothesis in Subsection \ref{BSH}, can be omitted,
compare Subsection \ref{BSH}.

From the Rule II it easily follows that the Wick product (\ref{WickFreeFieldOpAtx}) determines
integral kernel operator
\[
\Xi_{l,m}(\kappa_{l,m}) =
\Xi_{l,m} \Bigg( \Big({}^{{}^{n_1}}_{{}_{1}}\kappa_{l_1,m_1} \Big) \overline{\dot{\otimes}} \cdots \overline{\dot{\otimes}} \,\, \Big( {}^{{}^{n_M}}_{{}_{M}}\kappa_{l_M,m_M} \Big) \Bigg)
\]
with vector valued kernel
\begin{multline}\label{kappaInE^*xE^*x...xE^*xmathscrE^*}
\kappa_{l,m} = \Big({}^{{}^{n_1}}_{{}_{1}}\kappa_{l_1,m_1} \Big) \overline{\dot{\otimes}} \cdots \overline{\dot{\otimes}} \,\, \Big( {}^{{}^{n_M}}_{{}_{M}}\kappa_{l_M,m_M} \Big) \\
\in
E^{*}_{{}_{n_1}} \otimes \cdots \otimes
E^{*}_{{}_{n_M}} \otimes \mathscr{E}^{*}_{{}_{i}}
\cong \mathscr{L}(E_{{}_{n_1}} \otimes \cdots \otimes E_{{}_{n_M}}, \,\,
\mathscr{E}^{*}_{{}_{i}}), \,\,\, i=1,2,
\end{multline}
and, when all $n_k =1$ (\emph{i.e.} all ${}^{{}^{n_k}}_{{}_{k}}\kappa_{l_k,m_k}$ are the plane wave kernels corresponding to derivatives of the Dirac field), defines the bilinear map
\begin{multline}\label{Bilin.kappa.kappaInExExE^*}
\xi \times \eta \mapsto \kappa_{l,m}(\xi \otimes \eta),
\\
\xi \in \overbrace{E_{i_1} \otimes \cdots
\otimes E_{i_l}}^{\textrm{first $l$ terms $E_{i_j}$, $i_j \in \{1,2\}$}}, \\
\eta \in \overbrace{E_{i_{l+1}} \otimes \cdots
\otimes E_{i_{l+m}}}^{\textrm{last $m$ terms $E_{i_j}$, $i_j\in \{1,2\}$}},
\end{multline}
which can be extended to a separately continuous bilinear map from
\begin{equation}\label{kappa.kappaInExExE^*}
\Big( \overbrace{E_{i_1} \otimes \cdots
\otimes E_{i_l}}^{\textrm{first $l$ terms $E_{i_j}$}} \Big)^*
\times
\Big( \overbrace{E_{i_{l+1}} \otimes \cdots
\otimes E_{i_{l+m}}}^{\textrm{last $m$ terms $E_{i_j}$}} \Big)
\,\,\, \textrm{into} \,\,\,\mathscr{L}(\mathscr{E}, \mathbb{C}) = \mathscr{E}^*.
\end{equation}
Thus in each case
\begin{multline*}
\Xi_{l,m}(\kappa_{l,m}) =
\Xi_{l,m} \Bigg( \Big({}^{{}^{n_1}}_{{}_{1}}\kappa_{l_1,m_1} \Big) \overline{\dot{\otimes}} \cdots \overline{\dot{\otimes}} \,\, \Big( {}^{{}^{n_M}}_{{}_{M}}\kappa_{l_M,m_M} \Big) \Bigg) \\
\in \mathscr{L}\big((\boldsymbol{E}) \otimes \mathscr{E}_i, (\boldsymbol{E})^{*} \big) \cong
\mathscr{L}\Big(\mathscr{E}_i, \,\, \mathscr{L}\big((\boldsymbol{E}), \, (\boldsymbol{E})^{*}\big) \Big),
\,\,\, i = 1,2,
\end{multline*}
by Theorem 3.9 of \cite{obataJFA} (or its generalization to the case of tensor product of Fock spaces,
compare Subsection \ref{psiBerezin-Hida}).

In case in which there are no factors 
\[
\Xi_{l_k, m_k}\Big({}^{{}^{n_k}}_{{}_{k}}\kappa_{l_1,m_1}\Big) \,\,\, \textrm{with} \,\,\, n_k = 2
\]
\emph{i.e.} no factors corresponding to the (derivatives) of the zero mass free fields of the theory, e.g. of the electromagnetic potential field in case of QED, we have
\begin{multline*}
\Xi_{l,m}(\kappa_{l,m})  = 
\Xi_{l,m} \Bigg( \Big({}^{{}^{n_1}}_{{}_{1}}\kappa_{l_1,m_1} \Big) \overline{\dot{\otimes}} \cdots \overline{\dot{\otimes}} \,\, \Big( {}^{{}^{n_M}}_{{}_{M}}\kappa_{l_M,m_M} \Big)  \Bigg) \\
\in \mathscr{L}\big((\boldsymbol{E}) \otimes \mathscr{E}_i, (\boldsymbol{E})\big) \cong 
\mathscr{L}\Big(\mathscr{E}_i, \,\, \mathscr{L}\big((\boldsymbol{E}), \, (\boldsymbol{E})\big) \Big),
\,\,\, i = 1,2,
\end{multline*}
by Theorem  \ref{obataJFA.Thm.3.13}, Subsection \ref{psiBerezin-Hida} (generalisation of
Thm. 3.13 in \cite{obataJFA}).

Indeed, we use several technical Lemmas which allow us to show (\ref{kappaInE^*xE^*x...xE^*xmathscrE^*})
as well as the extedibility
(\ref{kappa.kappaInExExE^*}) property of the bilinear map (\ref{Bilin.kappa.kappaInExExE^*})
in case in which the zero mass terms are absent. 
We need the following technical definition

\begin{defin}\label{mathfrakS_i}
Let $\mathfrak{S}_i$, $i=1,2$, denote the family of subsets of $E_{i} \subset E_{i}^{*}$
which are bounded in the topology on  $E_{i}$ 
induced by the strong dual topology on $E_{i}^{*}$. Otherwise: 
$\mathfrak{S}_i$ is the family of intersections of all sets bounded in
the strong dual space $E_{i}^{*}$ with the subset $E_{i}$ of $E_{i}^{*}$.
\end{defin}

\begin{lem}\label{Hypocont.Ofkappa.kappa}
Let
\[
{}^{{}^{1}}_{{}_{1}}\kappa_{1,0}, {}^{{}^{1}}_{{}_{2}}\kappa_{1,0} \in \mathfrak{K}_0,
\]
i.e. let the above two kernels be equal to fixed components of plane wave kernels
defining the massive free fields of the theory (\emph{i. e.} the Dirac field in case of QED),
or to their spatio-temporal derivatives $\partial^{\alpha}$
with fixed value of the multi-index $\alpha \in \mathbb{N}_{0}^{4}$.
Then the map
\[
E_{1}^{*} \times E_{1}^{*} \supset E_{1} \times E_{1} \ni \xi_1 \times \xi_2
\longmapsto {}^{{}^{1}}_{{}_{1}}\kappa_{1,0}(\xi_1)
\cdot {}^{{}^{1}}_{{}_{2}}\kappa_{1,0}(\xi_2) \in \mathscr{E}_{k}^{*},
\]
is $\big(\mathfrak{S}_{1}, \mathfrak{S}_{1}\big)$-hypocontinuous as a map
\[
E_{1} \times E_{1} \longrightarrow \mathscr{E}_{k}^{*}, \,\,\, k =1,2
\]
with the topology on $E_{1} \subset E_{1}^{*}$, induced by the strong dual topology
on $E_{1}^{*}$, and with the strong dual topology on $\mathscr{E}_{k}^{*}$, $k=1,2$.
\end{lem}

\qedsymbol \, (An outline of the proof)
$\mathscr{E}_2 = \mathcal{S}^{00}(\mathbb{R}^4; \mathbb{C})$ is continuously
inserted into $\mathcal{S}(\mathbb{R}^4; \mathbb{C})$, and thus 
the strong dual $\mathscr{E}_{1}^{*} = \mathcal{S}(\mathbb{R}^4; \mathbb{C})^*$ is continuously inserted 
into the strong dual $\mathscr{E}_{2}^{*} = \mathcal{S}^{00}(\mathbb{R}^4; \mathbb{C})^*$,
for the proof compare Subsection \ref{SA=S0}. It is therefore sufficient to prove the Lemma for the case
$\mathscr{E}_{1}^{*} = \mathcal{S}(\mathbb{R}^4; \mathbb{C})^{*}$ with  $k=1$.

Consider for example the case of the plane wave kernel $\kappa_{1,0}$ given by the formula
(\ref{kappa_1,0}),
Subsection \ref{psiBerezin-Hida} or (\ref{skappa_1,0}) of
Subsection \ref{StandardDiracPsiField} which defines (one of the two \emph{a priori} possible)
Dirac free fields (the analysis of their fixed spatio-temporal derivation components is identical).

Recall that for $\phi \in \mathscr{E}_1 = \mathcal{S}(\mathbb{R}^4; \mathbb{C})$,
$\xi_1, \xi_2 \in E_1 = \mathcal{S}(\mathbb{R}^3; \mathbb{C}^4)$
(here we fix once for all the spinor indices $a_1, a_2$ and in case of spatio-temporal derivatives
$\partial^{\alpha_1}\kappa_{1,0}$ and $\partial^{\alpha_2}\kappa_{1,0}$ the additional multi-indices
$\alpha_1, \alpha_2 \in \mathbb{N}_{0}^{4}$ would also be fixed) we have
\begin{multline*}
\langle \kappa_{1,0}(\xi_1) \cdot\kappa_{1,0}(\xi_2), \phi \rangle =\\
\sum \limits_{s_1, s_2} \int \limits_{\mathbb{R}^3 \times \mathbb{R}^3 \times \mathbb{R}^4}
\kappa_{1,0}(s_1, \boldsymbol{\p}_1; a_1, x) \cdot \kappa_{1,0}(s_2, \boldsymbol{\p}_2; a_2, x)
\, \xi_{1}(s_1, \boldsymbol{\p}_1) \xi_{2}(s_2, \boldsymbol{\p}_1) \phi(x) \,
\ud^3 \boldsymbol{\p}_1 \, \ud^3 \boldsymbol{\p}_2 \, \ud^4x.
\end{multline*}
\[
\begin{split}
\kappa_{1,0}(\xi_1)(a_1,x) =
\sum \limits_{s_1}\int \limits_{\mathbb{R}^3}
\kappa_{1,0}(s_1, \boldsymbol{\p}_1; a_1, x) \, \xi_{1}(s_1, \boldsymbol{\p}_1)
\, \ud^3 \boldsymbol{\p}_1, \\
\kappa_{1,0}(\xi_2)(a_2,x) =
\sum \limits_{s_2}\int \limits_{\mathbb{R}^3}
\kappa_{1,0}(s_2, \boldsymbol{\p}_2; a_2, x) \, \xi_{2}(s_2, \boldsymbol{\p}_2)
\, \ud^3 \boldsymbol{\p}_1.
\end{split}
\]

Next we show that if $\xi_1 \in E_1 = \mathcal{S}(\mathbb{R}^3; \mathbb{C})$ ranges over a set 
$S \in \mathfrak{S}_1$, i.e. over $S \subset E_1 \subset E_{1}^{*}$ bounded in the strong 
dual topology on $E_{1}^{*}$, and if $\phi \in \mathscr{E}_{1} = \mathcal{S}(\mathbb{R}^4; \mathbb{C})$
ranges over a set $B \subset \mathscr{E}_{1} = \mathcal{S}(\mathbb{R}^4; \mathbb{C})$ 
bounded in $\mathscr{E}_{1} = \mathcal{S}(\mathbb{R}^4; \mathbb{C})$ (with respect to the ordinary 
nuclear Schwartz topology on $\mathcal{S}(\mathbb{R}^4;\mathbb{C})$, then the set
$B^{+}(S, B)$ of functions (spinor indices $a_1, a_2$ are fixed)
\[
(s_2, \boldsymbol{\p}_2) \longmapsto
\sum \limits_{s_1}
\int \limits_{\mathbb{R}^3 \times \mathbb{R}^4} 
\kappa_{1,0}(s_1, \boldsymbol{\p}_1; a_1, x) \cdot \kappa_{1,0}(s_2, \boldsymbol{\p}_2; a_2, x)
\, \xi_1(s_1, \boldsymbol{\p}_1) \, \phi(x) \, \ud^3 \boldsymbol{\p}_1 \, \ud^4x
\]
and the set $B^{+}(B, S)$ of functions 
\[
(s_1, \boldsymbol{\p}_1) \longmapsto
\sum \limits_{s_1}
\int \limits_{\mathbb{R}^3 \times \mathbb{R}^4} 
\kappa_{1,0}(s_1, \boldsymbol{\p}_1; a_1, x) \cdot \kappa_{1,0}(s_2, \boldsymbol{\p}_2; a_2, x)
\, \xi_2(s_2, \boldsymbol{\p}_2) \, \phi(x) \, \ud^3 \boldsymbol{\p}_2 \, \ud^4x
\]
with $\xi_2$ ranging over $S \in \mathfrak{S}_1$
and $\phi \in B$
are bounded in $E_1= \mathcal{S}(\mathbb{R}^3; \mathbb{C}^4)$. The proof, being a simple
verification of definition of boundedness, can be omitted, but we encourage 
the reader to perform the computations explicitly.

Next we observe that for any $S \in \mathfrak{S}_1$ and any strong zero-neighborhood $W(B, \epsilon)$ in
$\mathscr{E}_{1}^{*} = \mathcal{S}(\mathbb{R}^4; \mathbb{C})^{*}$, determined by a bounded
set $B$ in $\mathscr{E}_1 = \mathcal{S}(\mathbb{R}^4; \mathbb{C})$ and $\epsilon >0$, for
the strong zero-neighborhoods $W\big(B^{+}(S,B), \, \epsilon\big)$ and $W\big(B^{+}(B,S), \, \epsilon\big)$
we have
\[
|\langle \kappa_{1,0}(\xi_1) \cdot\kappa_{1,0}(\xi_2), \phi \rangle| < \epsilon
\]
whenever
\[
\xi_1 \in S, \,\,\,\, \xi_2 \in W\big(B^{+}(S,B), \, \epsilon\big)
\]
or whenever
\[
\xi_1 \in W\big(B^{+}(B,S), \, \epsilon\big), \,\,\,\, \xi_2 \in S.
\]
Put otherwise
\[
\begin{split}
\kappa_{1,0}(S) \cdot \kappa_{1,0}\Big(V\big(B^{+}(S,B), \, \epsilon\big)\Big) \subset W(B, \epsilon), \\
\kappa_{1,0}\Big( V\big(B^{+}(B,S), \, \epsilon\big) \Big) \cdot \kappa_{1,0}(S) \subset W(B, \epsilon).
\end{split}
\]
\qed

\begin{lem}\label{Cont.Ofkappa.kappa}
\begin{enumerate}
\item[1)]
Let $\phi \in \mathscr{E}_1 = \mathcal{S}(\mathbb{R}^4; \mathbb{C})$ and let
$\widetilde{\phi}$ be equal to its Fourier transform
\[
\widetilde{\phi}(p) = \int \limits_{\mathbb{R}^4} \phi(x) \, e^{i p \cdot x} \, \ud^4 x.
\]
Then if $\phi \in \mathcal{S}(\mathbb{R}^4; \mathbb{C})$ ranges over a bounded set $B$
in the Schwartz space $\mathcal{S}$, equivalently, if $\widetilde{\phi}$ ranges over a
bounded set $\widetilde{B}$ in $\mathcal{S}(\mathbb{R}^4; \mathbb{C})$, then there exists
a constant $C_{B}$ depending on $B$ such that
\[
|\widetilde{\phi}(\boldsymbol{\p} \pm \boldsymbol{\p}', p_0(\boldsymbol{\p}) \pm p'_{0}(\boldsymbol{\p}'))|
\leq C_{B}, \,\,\,\,\,\,
\boldsymbol{\p}, \boldsymbol{\p}' \in \mathbb{R}^3, \phi \in B
\]
in each case
\[
\begin{split}
p_0(\boldsymbol{\p}) = \sqrt{|\boldsymbol{\p}|^2 + m}, \,\,\, \textrm{or} \,\,\,
p_0(\boldsymbol{\p}) = \sqrt{|\boldsymbol{\p}|^2} = |\boldsymbol{\p}| \\
p'_{0}(\boldsymbol{\p}') = \sqrt{|\boldsymbol{\p}'|^2 + m}, \,\,\, \textrm{or} \,\,\,
p'_{0}(\boldsymbol{\p}') = \sqrt{|\boldsymbol{\p}'|^2} = |\boldsymbol{\p}'|.
\end{split}
\]
\item[2)]
Let
\[
{}^{{}^{n_1}}_{{}_{1}}\kappa_{l_1,m_1}, {}^{{}^{n_2}}_{{}_{2}}\kappa_{l_2,m_2} \in \mathfrak{K}_0,
\,\,\, (l_k, m_k) \in \{(0,1), (1,0) \}, n_k \in \{1, 2\}, k =1,2,
\]
i.e. let the above two kernels be equal to fixed components of plane wave kernels
defining free fields of the theory, or to their spatio-temporal derivatives $\partial^{\alpha}$
with fixed value of the multi-index $\alpha \in \mathbb{N}_{0}^{4}$.
Then the map
\[
E_{n_1} \times E_{n_2} \ni \xi_1 \times \xi_2
\longmapsto {}^{{}^{n_1}}_{{}_{1}}\kappa_{l_1,m_1}(\xi_1)
\cdot {}^{{}^{n_2}}_{{}_{2}}\kappa_{l_2,m_2}(\xi_2) \in \mathscr{E}_{k}^{*},
\]
is continuous as a map
\[
E_{n_1} \times E_{n_2} \longrightarrow \mathscr{E}_{k}^{*}, \,\,\, k =1,2
\]
with the ordinary nuclear topology on $E_{n_k}$, $k=1,2$, and with the strong
dual topology on $\mathscr{E}_{k}^{*}$, $k=1,2$.
\end{enumerate}
\end{lem}
\qedsymbol \,
The first part 1) is obvious.

Concerning 2) we will use the following two facts.
\begin{enumerate}
\item[I)]
The functions
\[
\boldsymbol{\p} \rightarrow \frac{P(\boldsymbol{\p})}{p_0(\boldsymbol{\p})}
= \frac{P(\boldsymbol{\p})}{\sqrt{|\boldsymbol{\p}|^2 + m}}, \,\,\, m \neq 0
\]
with $P(\boldsymbol{\p})$ being equal to polynomials in four real variables
$(\boldsymbol{\p}, p_0(\boldsymbol{\p}))= (p_1, p_2, p_3, \sqrt{|\boldsymbol{\p}|^2 + m})$
are multipliers of the Schwartz algebra
$\mathcal{S}(\mathbb{R}^3; \mathbb{C})$, compare \cite{Schwartz} or Appendix \ref{convolutorsO'_C}.
\item[II)]
The functions
\[
\boldsymbol{\p} \rightarrow \frac{P(\boldsymbol{\p})}{p_0(\boldsymbol{\p})}
= \frac{P(\boldsymbol{\p})}{|\boldsymbol{\p}|},
\]
with $P(\boldsymbol{\p})$ being equal to polynomials in four real variables
$(\boldsymbol{\p}, p_0(\boldsymbol{\p}))= (p_1, p_2, p_3, |\boldsymbol{\p}|)$
are multipliers of the nuclear algebra $\mathcal{S}^{0}(\mathbb{R}^3; \mathbb{C})$,
for a proof compare Subsections \ref{dim=1} - \ref{SA=S0}.
\end{enumerate}

Recall that in case of QED we have
\[
\begin{split}
E_1 = \mathcal{S}_{A_1}(\mathbb{R}^3; \mathbb{C}^4) 
= \mathcal{S}(\mathbb{R}^3; \mathbb{C}^4) = \oplus \mathcal{S}(\mathbb{R}^3; \mathbb{C}) 
\,\,\, \textrm{and} \\
E_2 = \mathcal{S}_{A_2}(\mathbb{R}^3; \mathbb{C}^4) 
= \mathcal{S}^{0}(\mathbb{R}^3; \mathbb{C}^4) = \oplus \mathcal{S}^{0}(\mathbb{R}^3; \mathbb{C}).
\end{split}
\]
with $A_2 = \oplus_{0}^{3} A^{(3)}$ and $A^{(3)}$ on $L^2(\mathbb{R}^3; \mathbb{C})$ constructed in Subsection
\ref{dim=n}, and with $A_1 = \oplus_{1}^{4} H_{(3)}$ equal to the direct sum of four copies of the three-dimensional
oscillator Hamiltonian, \emph{i. e.} $A_1$ is equal to the operator $A$ given by (\ref{AinL^2(R^3;C^4)}). 

In particular let us consider the distribution defined by the kernel
\begin{equation}\label{kappa.kappa(x)}
\kappa_{1,0} \overset{\cdot}{\otimes} \kappa_{1,0}(\nu_1, \boldsymbol{p}_1, \nu_2, \boldsymbol{p}_2; x) = 
\kappa_{1,0}(\nu_1, \boldsymbol{p}_1; \mu, x) \cdot \kappa_{1,0}(\nu_2, \boldsymbol{\p}_2; \lambda, x), \,\,\, 
\textrm{with fixed $\mu, \lambda$}
\end{equation}
and with $\kappa_{1,0}$ equal to the plane wave kernel defining the free electromagnetic potential field,
and given by the formula (\ref{kappa_0,1kappa_1,0A'}), Subsection \ref{equivalentA-s}.
For each $\xi_1, \xi_2 \in E_2 = \mathcal{S}^{0}(\mathbb{R}^4; \mathbb{C})$ the value
 of the distribution 
\begin{multline*}
\kappa_{0,1} \overset{\cdot}{\otimes} \kappa_{1,0}(\xi_1 \otimes \xi_2)(x) = 
\kappa_{1,0}(\xi_1)(\mu,x) \cdot \kappa_{1,0}(\xi_2)(\lambda, x) \\
= \int \limits_{\mathbb{R}^3 \times \mathbb{R}^3} 
\frac{\ud^3 \boldsymbol{\p}_1 \, \ud^3 \boldsymbol{\p}_2}{|\boldsymbol{\p}_1| |\boldsymbol{\p}_2|}
\xi_{1}^{\mu}(\boldsymbol{\p}_1) \xi_{2}^{\lambda}(\boldsymbol{\p}_2) \, e^{i(p_1 + p_2) \cdot x}, \\
 \,\,\, 
\xi_1 \otimes \xi_2 (\boldsymbol{p}_1 \times \boldsymbol{\p}_2) = \xi_1(\boldsymbol{\p}_1)\xi_2(\boldsymbol{\p}_2)
\end{multline*}
 on $\phi \in \mathcal{S}(\mathbb{R}^4; \mathbb{C})$ is equal
\[
\langle \kappa_{1,0}(\xi_1) \cdot \kappa_{1,0}(\xi_{2}), \phi \rangle
=  \int \limits_{\mathbb{R}^3 \times \mathbb{R}^3} 
\frac{\ud^3 \boldsymbol{\p}_1 \, \ud^3 \boldsymbol{\p}_2}{|\boldsymbol{\p}_1| |\boldsymbol{\p}_2|}
\xi_{1}^{\mu}(\boldsymbol{\p}_1) \xi_{2}^{\lambda}(\boldsymbol{\p}_2) \, 
\widetilde{\phi}(\boldsymbol{\p}_1 + \boldsymbol{\p}_2, |\boldsymbol{\p}_1| + |\boldsymbol{\p}_2|).
\]
Now let $\xi_1$, $\xi_2$ range respectively over the bounded sets $B_1$ and $B_2$ in $E_2 = 
\mathcal{S}^{0}(\mathbb{R}^3; \mathbb{C}^4)$. Let $\phi$ range over a bounded set $B$ in 
$\mathcal{S}(\mathbb{R}^4; \mathbb{C})$, equivalently, $\widetilde{\phi}$ range over a bounded
set $\widetilde{B}$ in $\mathcal{S}(\mathbb{R}^4; \mathbb{C})$. 
Because the function 
\[
\boldsymbol{\p} \mapsto \frac{1}{|\boldsymbol{\p}|}
\]
is a multiplier of the nuclear algebra $\mathcal{S}^{0}(\mathbb{R}^3; \mathbb{C})$
(Subsections \ref{diffSA} and \ref{SA=S0}) then the sets of functions
\[
\begin{split}
B'_1 = \big\{\xi'_{1}, \xi_1 \in B_1 \big\}  \,\,\,\,\,\,\,  \textrm{where} \,\,\,
\xi'_1 (\boldsymbol{\p}_1) = \frac{\xi_1(\boldsymbol{\p}_1)}{|\boldsymbol{\p}_1|}, \\
B'_2 = \big\{\xi'_{2}, \xi_2 \in B_2 \big\}  \,\,\,\,\,\,\,  \textrm{where} \,\,\,
\xi'_2 (\boldsymbol{\p}_2) = \frac{\xi_2(\boldsymbol{\p}_2)}{|\boldsymbol{\p}_2|},
\end{split}
\]
are bounded in $E_2 = \mathcal{S}(\mathbb{R}^3; \mathbb{C}^4)$, and the set $B'_{1} \otimes B'_{2}$
is bounded in $E_2 \otimes E_2$. 
This in particular means that each of the norms (values  of the indices $\mu, \nu \in \{0,1,2,3\}$
are fixed and $\zeta^{(q)}$ denotes derivative of $q$-th order $q \in \mathbb{N}_{0}^{6}$ of a function 
$\zeta$ on $\mathbb{R}^{6}$)
\[
\rceil \rceil \xi_{1}^{\mu} \otimes \xi_{2}^{\lambda} \lceil \lceil_{{}_{m}} \overset{\textrm{df}}{=}
\underset{|q| \leq m}{\textrm{sup}}(1 + |\boldsymbol{\p}_1 \times \boldsymbol{2}|^2)^m 
\Big| \big(\xi_{1}^{\mu} \otimes \xi_{2}^{\lambda} \big)^{(q)}\Big|
\]
is separately bounded on $B'_{1} \otimes B'_{2}$, \emph{i. e.} for each
$m= 0,1,2, \ldots$ there exists a finite constant $C'_{{}_{m}}$ such that
\[
\rceil \rceil \xi_{1}^{\mu} \otimes \xi_{2}^{\lambda} \lceil \lceil_{{}_{m}}
\leq C'_{{}_{m}}, \,\,\, \xi_1 \in B'_{1}, \xi_2 \in B'_{2},
\]
and moreover for each $m = 0,1,2, \ldots$ there exists $m'(m) \in \mathbb{N}_{0}$ and $C(m) < \infty $ such that
\begin{equation}\label{||.||<C|.|.|.|}
\Bigg\rceil \Bigg\rceil
\frac{(1+|\boldsymbol{\p}_1 \times \boldsymbol{\p}_2|^2)^4}{|\boldsymbol{\p}_1| \, |\boldsymbol{\p}_2|}
\xi_1 \otimes \xi_2
\Bigg\lceil \Bigg\lceil_{{}_{m}} \leq \,\, C(m) \,
\rceil \xi_1 \lceil_{{}_{m'}}  \rceil \xi_2 \lceil_{{}_{m'}}
\end{equation}
where $\{ \rceil \cdot \lceil_{{}_{m}} \}_{m \in \mathbb{N}_{0}}$ is one of the equivalent systems of norms defining 
$\mathcal{S}^{0}(\mathbb{R}^3; \mathbb{C})$ and given in Subsection \ref{SA=S0}, 
compare Subsection \ref{SA=S0}.

Now using the part 1) of the Lemma and the inequality (\ref{||.||<C|.|.|.|}) we obtain the following inequalities 
(with fixed values of the indices $\mu$ and $\lambda$ in each factor 
$\kappa_{1,0}(\xi_1)$ and $\kappa_{1,0}(\xi_1)$)
\begin{multline*}
|\langle \kappa_{1,0}(\xi_1) \cdot \kappa_{1,0}(\xi_2), \phi \rangle|
= \Bigg|\int \limits_{\mathbb{R}^3 \times \mathbb{R}^3} 
\frac{\ud^3 \boldsymbol{\p}_1 \, \ud^3 \boldsymbol{\p}_2}{|\boldsymbol{\p}_1| |\boldsymbol{\p}_2|}
\xi_{1}^{\mu}(\boldsymbol{\p}_1) \xi_{2}^{\lambda}(\boldsymbol{\p}_2) \, 
\widetilde{\phi}(\boldsymbol{\p}_1 + \boldsymbol{\p}_2, |\boldsymbol{\p}_1| + |\boldsymbol{\p}_2|) \Bigg| \\
\leq 
\int \limits_{\mathbb{R}^3 \times \mathbb{R}^3} 
\frac{\ud^3 \boldsymbol{\p}_1 \, \ud^3 \boldsymbol{\p}_2}{|\boldsymbol{\p}_1| |\boldsymbol{\p}_2|}
|\xi_{1}^{\mu}(\boldsymbol{\p}_1) \xi_{2}^{\lambda}(\boldsymbol{\p}_2)| \, 
|\widetilde{\phi}(\boldsymbol{\p}_1 + \boldsymbol{\p}_2, |\boldsymbol{\p}_1| + |\boldsymbol{\p}_2|)| \\
\leq 
C_B \, \int \limits_{\mathbb{R}^3 \times \mathbb{R}^3} 
\frac{\ud^3 \boldsymbol{\p}_1 \, \ud^3 \boldsymbol{\p}_2}{|\boldsymbol{\p}_1| |\boldsymbol{\p}_2|}
|\xi_{1}^{\mu}(\boldsymbol{\p}_1) \xi_{2}^{\lambda}(\boldsymbol{\p}_2)| \\
\end{multline*}
\begin{multline}\label{|<kappa(xi).kappa(xi),phi>|}
\leq
C_B \, \int \limits_{\mathbb{R}^3 \times \mathbb{R}^3} 
\ud^3 \boldsymbol{\p}_1 \, \ud^3 \boldsymbol{\p}_2 \frac{1}{(1+|\boldsymbol{\p}_1 \times \boldsymbol{\p}_2|^2)^{4}}
\frac{(1+|\boldsymbol{\p}_1 \times \boldsymbol{\p}_2|^2)^{4}|\xi_{1}^{\mu}(\boldsymbol{\p}_1) \xi_{2}^{\lambda}(\boldsymbol{\p}_2)|}{|\boldsymbol{\p}_1| |\boldsymbol{\p}_2|} \\
\leq 
C_B \, \Bigg| \frac{1}{(1+|\boldsymbol{\p}_1 \times \boldsymbol{\p}_2|^2)^{4}} \Bigg|_{{}_{L^2(\mathbb{R}^6)}} \,
\Bigg| \frac{(1+|\boldsymbol{\p}_1 \times \boldsymbol{\p}_2|^2)^{4}}{|\boldsymbol{\p}_1| |\boldsymbol{\p}_2|}
\xi_{1}^{\mu} \otimes \xi_{2}^{\lambda}
\Bigg|_{{}_{\infty}} \\
 \leq C' \Bigg\rceil \Bigg\rceil
\frac{(1+|\boldsymbol{\p}_1 \times \boldsymbol{\p}_2|^2)^4}{|\boldsymbol{\p}_1||\boldsymbol{\p}_2|}
\xi_{1}^{\mu} \otimes \xi_{2}^{\lambda}
\Bigg\lceil \Bigg\lceil_{{}_{4}} \\
\, \leq \,\, C' C(4) \,
\rceil \xi_{1}^{\mu} \lceil_{{}_{m'}}  \rceil \xi_{2}^{\lambda} \lceil_{{}_{m'}}
\end{multline}
for some finite $m' \in \mathbb{N}_0$. 

Therefore, for any strong zero-neighborhood $V(B, \epsilon)$ in $\mathcal{S}(\mathbb{R}^4; \mathbb{C})^*$
determined by a bounded subset $B$ in $\mathcal{S}(\mathbb{R}^4; \mathbb{C})$ and $\epsilon >0$
there exist zero-neighborhoods $V_1$ and $V_2$ in $E_2 = \mathcal{S}^{0}(\mathbb{R}^3; \mathbb{C}^4)$
such that
\[
|\langle \kappa_{1,0}(\xi_1) \cdot \kappa_{1,0}(\xi_2), \phi \rangle| \leq \epsilon,
\,\,\,\,
\xi_1 \in V_1, \xi_2 \in V_2, \phi \in B,
\]
or equivalently
\[
\kappa_{1,0}(\xi_1) \cdot \kappa_{1,0}(\xi_2) \in V(B, \epsilon), \,\,\,
\xi_1 \in V_1, \xi_2 \in V_2,
\]
if we define
\[
V_1 = \Bigg\{\xi, \rceil \xi^{\mu} \lceil_{{}_{m'}} < \sqrt{\frac{\epsilon}{C' C(4)}} \Bigg\}, \,\,\,
V_2 = \Bigg\{\xi, \rceil \xi^{\lambda} \lceil_{{}_{m'}} < \sqrt{\frac{\epsilon}{C' C(4)}} \Bigg\},
\]
which follows from the inequalities (\ref{|<kappa(xi).kappa(xi),phi>|}).

The same proof holds if we replace one or both the kernels $\kappa_{1,0}$
by the kernel $\kappa_{0,1}$ defined by (\ref{kappa_0,1kappa_1,0A'}),
Subsection \ref{equivalentA-s}, or by their derivatives because for any polynomial
$P(\boldsymbol{\p}_1, \boldsymbol{\p}_2)$ in eight real variables
\[
(\boldsymbol{\p}_1, p_{10}(\boldsymbol{\p}_1), \boldsymbol{\p}_2, p_{20}(\boldsymbol{\p}_2))
= (\boldsymbol{\p}_1, |\boldsymbol{\p}_1|, \boldsymbol{\p}_2, |\boldsymbol{\p}_2|)
\]
and for each $m = 0,1,2, \ldots$ there exists $m'(m) \in \mathbb{N}_{0}$ and $C(m) < \infty $ such that
\begin{equation}\label{||P(p1p2).||<C|.|.|.|}
\Bigg\rceil \Bigg\rceil
\frac{(1+|\boldsymbol{\p}_1 \times \boldsymbol{\p}_2|^2)^4 P(\boldsymbol{\p}_1, \boldsymbol{\p}_2)}{|\boldsymbol{\p}_1|
\, |\boldsymbol{\p}_2|}
\xi_{1}^{\mu} \otimes \xi_{2}^{\lambda}
\Bigg\lceil \Bigg\lceil_{{}_{m}} \leq \,\, C(m) \,
\rceil \xi_1 \lceil_{{}_{m'}} \rceil \xi_2 \lceil_{{}_{m'}}.
\end{equation}
Analogous proof can be repeated for all $\kappa_{1,0}, \kappa_{0,1}$ defined by
(\ref{kappa_0,1kappa_1,0A}), Subsection \ref{A=Xi0,1+Xi1,0} (for plane wave kernels defining the free electromagnetic
potential field) and their derivatives; or for plane wave kernels
(\ref{kappa_0,1}) and (\ref{kappa_1,0}),
Subsection \ref{psiBerezin-Hida} or (\ref{skappa_0,1}) and (\ref{skappa_1,0}) of
Subsection \ref{StandardDiracPsiField} (for kernels defining the Dirac field)
and their derivatives. We have to remember that if the kernel corresponds to the electromagnetic potential field
then the nuclear space on which it is defined is equal $E_2 = \mathcal{S}^{0}(\mathbb{R}^3; \mathbb{C}^4)$
and if the kernel corresponds to the Dirac field then it is defined on the nuclear space $E_1 = \mathcal{S}(\mathbb{R}^3; \mathbb{C}^4)$. In the last case we can use the standard system of norms defining the Schwartz topology
on $\mathcal{S}(\mathbb{R}^3; \mathbb{C})$. In particular if both factors\footnote{$(l_k,m_k) = (1,0)$ or $(l_k,m_k) = (0,1)$ for $k = 1, 2$.}
$\kappa_{l_1,m_1}(\xi_1)$
and $\kappa_{l_2,m_2}(\xi_2)$ in the pointwise product $\kappa_{l_1,m_1}(\xi_1) \cdot \kappa_{l_2,m_2}(\xi_2)$
correspond to kernels defining a fixed component of the Dirac field (or its fixed component derivative) then
we are using the inequality (\ref{||P(p1p2).||<C|.|.|.|}) with the same system of norms
$\{ \rceil \rceil \cdot \lceil \lceil_{{}_{m}} \}_{m \in \mathbb{N}_{0}}$ on the left-hand side
but with the system of norms
$\{ \rceil \cdot \lceil_{{}_{m}} \}_{m \in \mathbb{N}_{0}}$ replaced by the standard system of norms defining
the Schwartz topology on $\mathcal{S}(\mathbb{R}^3; \mathbb{C})$
and with
\[
\frac{(1+|\boldsymbol{\p}_1 \times \boldsymbol{\p}_2|^2)^4 P(\boldsymbol{\p}_1, \boldsymbol{\p}_2)}{|\boldsymbol{\p}_1|
\, |\boldsymbol{\p}_2|}
\]
in (\ref{||P(p1p2).||<C|.|.|.|}) replaced by
\[
\frac{(1+|\boldsymbol{\p}_1 \times \boldsymbol{\p}_2|^2)^4 P(\boldsymbol{\p}_1, \boldsymbol{\p}_2)}
{\sqrt{|\boldsymbol{\p}_1|^2 +m} \, \sqrt{|\boldsymbol{\p}_2|^2+m}} \,\,\,
\textrm{or} \,\,\,
(1+|\boldsymbol{\p}_1 \times \boldsymbol{\p}_2|^2)^4 P(\boldsymbol{\p}_1, \boldsymbol{\p}_2)
\]
with
$P(\boldsymbol{\p}_1, \boldsymbol{\p}_2)$ equal to any polynomial
in eight real variables
\[
(\boldsymbol{\p}_1, p_{10}(\boldsymbol{\p}_1), \boldsymbol{\p}_2, p_{20}(\boldsymbol{\p}_2))
= (\boldsymbol{\p}_1, \sqrt{|\boldsymbol{\p}_1|^2 +m}, \boldsymbol{\p}_2, \sqrt{|\boldsymbol{\p}_2|^2+m}).
\]
If the first factor $\kappa_{l_1,m_1}(\xi_1)$ corresponds to a fixed component of the free Dirac field
(or its fixed component derivative)
and the second factor $\kappa_{l_2,m_2}(\xi_2)$ corresponds to a fixed component of the free electromagnetic
potential field (or its fixed component derivative) then we are using the inequality
(\ref{||P(p1p2).||<C|.|.|.|}) with the same system of norms
$\{ \rceil \rceil \cdot \lceil \lceil_{{}_{m}} \}_{m \in \mathbb{N}_{0}}$ on the left-hand side,
the same system of norms
$\{ \rceil \xi_2 \lceil_{{}_{m}} \}_{m \in \mathbb{N}_{0}}$ defining the nuclear topology
$S^{0}(\mathbb{R}^3; \mathbb{C})$ (inherited from $\mathcal{S}(\mathbb{R}^3; \mathbb{C})$, compare Subsections
\ref{dim=1}-\ref{SA=S0}) but with the system of norms $\{ \rceil \xi_1 \lceil_{{}_{m}} \}_{m \in \mathbb{N}_{0}}$
replaced by any standard which defines the Schwartz topology on $\mathcal{S}(\mathbb{R}^3; \mathbb{C})$,
and with
\[
\frac{(1+|\boldsymbol{\p}_1 \times \boldsymbol{\p}_2|^2)^4 P(\boldsymbol{\p}_1, \boldsymbol{\p}_2)}{|\boldsymbol{\p}_1|
\, |\boldsymbol{\p}_2|}
\]
in (\ref{||P(p1p2).||<C|.|.|.|}) replaced by
\[
\frac{(1+|\boldsymbol{\p}_1 \times \boldsymbol{\p}_2|^2)^4 P(\boldsymbol{\p}_1, \boldsymbol{\p}_2)}
{\sqrt{|\boldsymbol{\p}_1|^2 +m} \, |\boldsymbol{\p}_2|} \,\,\,
\textrm{or} \,\,\,
\frac{(1+|\boldsymbol{\p}_1 \times \boldsymbol{\p}_2|^2)^4 P(\boldsymbol{\p}_1, \boldsymbol{\p}_2)}
{|\boldsymbol{\p}_2|}
\]
with $P(\boldsymbol{\p}_1, \boldsymbol{\p}_2)$ equal to any polynomial
in eight real variables
\[
(\boldsymbol{\p}_1, p_{10}(\boldsymbol{\p}_1), \boldsymbol{\p}_2, p_{20}(\boldsymbol{\p}_2))
= (\boldsymbol{\p}_1, \sqrt{|\boldsymbol{\p}_1|^2 +m}, \boldsymbol{\p}_2, |\boldsymbol{\p}_2|).
\]
\qed

\begin{lem}\label{kappaBarDotOtimeskappa}
Let
\[
{}^{{}^{n_k}}_{{}_{k}}\kappa_{l_k,m_k} \in \mathfrak{K}_0, \,\,\, k = 1, \ldots, M.
\]
i.e. we have the kernels belonging to the class\footnote{Recall that each element of $\mathfrak{K}_0$ 
is equal to a component of a plane wave kernel defining free field of the theory or to its spatio-temporal derivative
$\partial^{\alpha}$ with fixed $\alpha$, compare Definition \ref{K_0}.} 
$\mathfrak{K}_0$. 
\begin{enumerate}
\item[1)]
Then it follows in particular that
\[
{}^{{}^{n_k}}_{{}_{k}}\kappa_{l_k,m_k}
\in \mathscr{L}(E_{{}_{n_k}}, \mathscr{E}^{*}_{{}_{n_k}}) \cong 
E^{*}_{{}_{n_k}} \otimes \mathscr{E}^{*}_{{}_{n_k}}, \,\,\, k=1, \ldots, M,
\]
are regular vector-valued distributions defined by ordinary functions, which 
fulfil the condition (\ref{ExtedibilityCondition}), \emph{i.e.} are extendable to elements 
\[
{}^{{}^{n_k}}_{{}_{k}}\kappa_{l_k,m_k}
\in \mathscr{L}(E^{*}_{{}_{n_k}}, \mathscr{E}^{*}_{{}_{n_k}}) \cong 
E_{{}_{n_k}} \otimes \mathscr{E}^{*}_{{}_{n_k}}, \,\,\, k=1, \ldots, M,
\]
and the evaluations
\[
{}^{{}^{n_k}}_{{}_{k}}\kappa_{l_k,m_k}(\xi), \,\,\, \xi \in E_{{}_{n_k}},
\]
have Fourier transforms concentrated on the respective orbit $\mathscr{O}_{{\pm m,0,0,0}{}}$ or $\mathscr{O}_{{\pm 1,0,0,1}{}}$, 
corresponding to the respective field, they are smooth with each space-time derivative being bounded, so in particular they have
the property (\ref{ConvolutabilityCondition}) that
\[
{}^{{}^{n_k}}_{{}_{k}}\kappa_{l_k,m_k}(\xi) \in \mathcal{O}_C(\mathbb{R}^4; \mathbb{C}), 
\,\,\, \xi \in E_{{}_{n_k}}.
\]
\item[2)]
The  ``pointwise'' multiplicative tensor product $\dot{\otimes}$ of these distributions, defining kernels of the Wick products
of the free fields of the class $\mathfrak{K}_0$,
defined as in Rule II, gives a vector valued kernel 
\[
\kappa_{l,m} = \Big({}^{{}^{n_1}}_{{}_{1}}\kappa_{l_1,m_1} \Big) \overline{\dot{\otimes}} \cdots \overline{\dot{\otimes}} \,\, \Big( {}^{{}^{n_M}}_{{}_{M}}\kappa_{l_M,m_M} \Big) 
\]
such that its evaluation
\[
\kappa_{l,m}(\xi) =  \Big({}^{{}^{n_1}}_{{}_{1}}\kappa_{l_1,m_1} \Big)(\xi_1) \,\, \cdots \,\, \Big( {}^{{}^{n_M}}_{{}_{M}}\kappa_{l_M,m_M} \Big)(\xi_{{}_{M}}) 
\]
is equal to the product of the functions 
\[
{}^{{}^{n_k}}_{{}_{k}}\kappa_{l_k,m_k}(\xi_k)
\]
of 1), in particular
\[
\kappa_{l,m}(\xi) \in \mathcal{O}_{C}(\mathbb{R}^4; \mathbb{C}) \subset \mathcal{O}_{M}(\mathbb{R}^4; \mathbb{C}),
\]
for any
\[
\xi = \xi_1 \otimes \ldots \otimes \xi_{{}_{M}} \in \overbrace{E_{i_1} \otimes \cdots 
\otimes E_{i_l}}^{\textrm{first $l$ terms $E_{i_j}$, $i_j \in \{1,2\}$}} \otimes \underbrace{E_{i_{l+1}} \otimes \cdots 
\otimes E_{i_{l+m}}}^{\textrm{last $m$ terms $E_{i_j}$, $i_j\in \{1,2\}$}} \,\, l+m=M.
\]
\item[3)]
The  ``pointwise'' multiplicative tensor product $\dot{\otimes}$ of these distributions,
defined as in Rule II, gives a vector valued kernel 
\begin{multline*}
\kappa_{l,m} = \Big({}^{{}^{n_1}}_{{}_{1}}\kappa_{l_1,m_1} \Big) \overline{\dot{\otimes}} 
\cdots \overline{\dot{\otimes}} \,\, \Big( {}^{{}^{n_M}}_{{}_{M}}\kappa_{l_M,m_M} \Big) \\
\in
E^{*}_{{}_{n_1}} \otimes \cdots \otimes 
E^{*}_{{}_{n_M}} \otimes \mathscr{E}^{*}_{{}_{i}} 
\cong \mathscr{L}(E_{{}_{n_1}} \otimes \cdots \otimes  E_{{}_{n_M}}, \,\,
\mathscr{E}^{*}_{{}_{i}}), \,\,\, i=1,2.
\end{multline*}
\item[4)] 
If all $n_1, \ldots, n_M$ are equal $1$, \emph{i. e.} if all factors
\[
\Xi_{l_k, m_k}\Big({}^{{}^{n_k}}_{{}_{k}}\kappa_{l_1,m_1}\Big) \,\,\, \textrm{with} \,\,\, n_k = 1
\]
correspond to (derivatives) of the free massive fields of the theory (\emph{i. e.} 
derivatives of the Dirac free field in case of spinor QED), then the bilinear map
\begin{multline*}
\xi \times \eta \mapsto \kappa_{l,m}(\xi \otimes \eta), 
\\
\xi \in \overbrace{E_{i_1} \otimes \cdots 
\otimes E_{i_l}}^{\textrm{first $l$ terms $E_{i_j}$, $i_j \in \{1,2\}$}}, \\
\eta \in \overbrace{E_{i_{l+1}} \otimes \cdots 
\otimes E_{i_{l+m}}}^{\textrm{last $m$ terms $E_{i_j}$, $i_j\in \{1,2\}$}},
\end{multline*}
can be extended to a separately continuous bilinear map from
\[
\Big( \overbrace{E_{i_1} \otimes \cdots 
\otimes E_{i_l}}^{\textrm{first $l$ terms $E_{i_j}$}} \Big)^*
\times
\Big( \overbrace{E_{i_{l+1}} \otimes \cdots 
\otimes E_{i_{l+m}}}^{\textrm{last $m$ terms $E_{i_j}$}} \Big)
\,\,\, \textrm{into} \,\,\,\mathscr{L}(\mathscr{E}, \mathbb{C}) = \mathscr{E}^*.
\]
\end{enumerate}
\end{lem}

\qedsymbol \, 
The first two parts 1) and 2) follow from Lemma \ref{kappa0,1,kappa1,0psi}, Subsection \ref{psiBerezin-Hida} and respectively Lemma \ref{kappa0,1,kappa1,0ForA},
Subsection \ref{A=Xi0,1+Xi1,0}.

Concerning 3) it is sufficient to consider the case $M=2$. But the case $M=2$ follows immediately
from the part 2) of Lemma \ref{Cont.Ofkappa.kappa}.

Concerning 4) it is sufficient to consider the case $M=2$. 
Let us consider first the case in which the first factor has $(l_1,m_1) =(1,0)$ and the second $(l_2, m_2) = (0,1)$.
That the map
\[
\xi_1 \times \xi_2 \longmapsto 
{}^{{}^{1}}_{{}_{1}}\kappa_{1,0} \overset{\cdot}{\otimes} {}^{{}^{1}}_{{}_{2}}\kappa_{0,1} (\xi_1 \otimes \xi_2)
=
{}^{{}^{1}}_{{}_{1}}\kappa_{1,0}(\xi_1) \cdot {}^{{}^{1}}_{{}_{2}}\kappa_{0,1}(\xi_2) 
\]
can be extended to a map which is separately continuous as a map
\[
E_{1}^{*} \times E_{1} \mapsto \mathscr{E}_{k}^{*}, \,\,\, k= 1,2
\]
follows immediately from the extendable property (\ref{ExtedibilityCondition})
asserted in the first part of our Lemma and from the property (\ref{ConvolutabilityCondition}) 
which assures that
\[
{}^{{}^{n_k}}_{{}_{k}}\kappa_{l_k,m_k}(\xi) \in \mathcal{O}_C(\mathbb{R}^4; \mathbb{C}), 
\,\,\, \xi \in E_{{}_{n_k}}.
\]
and in particular assures that
\[
{}^{{}^{n_k}}_{{}_{k}}\kappa_{l_k,m_k}(\xi), \,\,\, \xi \in E_{{}_{n_k}}
\]
is contained within the algebra of multipliers of $\mathscr{E}_{k}$, $k=1,2$ and 
of $\mathscr{E}_{k}^{*}$. This is because $\mathcal{O}_C(\mathbb{R}^4; \mathbb{C})$ is contained
in both the algebras of multipliers $\mathcal{O}_{MB_1} = \mathcal{O}_{M}, \mathcal{O}_{MB_2}$, 
respectively, of $\mathscr{E}_1, \mathscr{E}_2$,
compare Subsections \ref{diffSA}, \ref{SA=S0} and Appendix \ref{convolutorsO'_C}.  
In particular the operator of pointwise multiplication by a fixed 
\[
{}^{{}^{n_k}}_{{}_{k}}\kappa_{l_k,m_k}(\xi), \,\,\, \xi \in E_{{}_{n_k}}
\]
transforms continuously $\mathscr{E}_{k}$, $k=1,2$ and 
$\mathscr{E}_{k}^{*}$, $k=1,2$ into themselves.

Let us consider now the case $M=2$ in which both factors have $(l_1,m_1) = (l_2,m_2) = (1,0)$:
\begin{equation}\label{kappaDirac.kappaDirac}
\xi_1 \times \xi_2 \longmapsto 
{}^{{}^{1}}_{{}_{1}}\kappa_{1,0} \overset{\cdot}{\otimes} {}^{{}^{1}}_{{}_{2}}\kappa_{1,0} (\xi_1 \otimes \xi_2)
=
{}^{{}^{1}}_{{}_{1}}\kappa_{1,0}(\xi_1) \cdot {}^{{}^{1}}_{{}_{2}}\kappa_{1,0}(\xi_2) 
\end{equation}
and the plane wave kernels 
\[
{}^{{}^{1}}_{{}_{1}}\kappa_{1,0}, {}^{{}^{1}}_{{}_{2}}\kappa_{1,0}
\]
correspond to some fixed components of the Dirac field or its fixed component derivative.
In this case the above map (\ref{kappaDirac.kappaDirac}) coincides with a particular case of the map
of Lemma \ref{Hypocont.Ofkappa.kappa}. 
From Lemma \ref{Hypocont.Ofkappa.kappa}
and the Proposition of Chap III \S 5.4, p. 90 of \cite{Schaefer}, it follows
that the $\big(\mathfrak{S}_{n_1}, \mathfrak{S}_{n_2}\big)$-hypocontinuous map
\[
E_{n_1} \times E_{n_2} \longrightarrow \mathscr{E}_{k}^{*}, \,\,\, k =1,2
\]
of Lemma \ref{Hypocont.Ofkappa.kappa}, can be uniquely extended to 
$\big(\mathfrak{S}_{n_1}^{*}, \mathfrak{S}_{n_2}^{*}\big)$-hypocontinuous map
\[
E_{n_1}^{*} \times E_{n_2}^{*} \longrightarrow \mathscr{E}_{k}^{*}, \,\,\, k =1,2
\]
with respect to the strong dual topology on each indicated space, where 
$\mathfrak{S}_{n_k}^{*}$, $k=1,2$, is the family of all bounded sets on 
strong dual space $E_{n_k}^{*}$, which simply means that the map of Lemma
\ref{Hypocont.Ofkappa.kappa} can be uniquely extended to a hypocontinuous map
\[
E_{n_1}^{*} \times E_{n_2}^{*} \longrightarrow \mathscr{E}_{k}^{*}, \,\,\, k =1,2
\]
or in particular to separately continuous map
\[
E_{n_1}^{*} \times E_{n_2}^{*} \longrightarrow \mathscr{E}_{k}^{*}, \,\,\, k =1,2
\]
with respect to the strong dual topology. Because $E_{n_k}^{*}$, $\mathscr{E}_{k}^{*}$, $k=1,2$
are all equal to strong dual spaces of reflexive Fr\'echet spaces 
$E_{n_k}$, $\mathscr{E}_{k}$, then by Thm. 41.1 the map of Lemma \ref{Hypocont.Ofkappa.kappa}
can be uniquely extended to (jointly) continuous map
\[
E_{n_1}^{*} \times E_{n_2}^{*} \longrightarrow \mathscr{E}_{k}^{*}, \,\,\, k =1,2
\]
with respect to the strong dual topology.
\qed

Before continuing we give a commentary concerning the proof of 4), case $M=2$ of the last Lemma.
Namely, in this proof we can proceed as in the proof of the second part of
Lemma \ref{kappa0,1,kappa1,0psi}, Subsection \ref{psiBerezin-Hida} or respectively of
Lemma \ref{kappa0,1,kappa1,0ForA},
Subsection \ref{A=Xi0,1+Xi1,0}. Namely,
\[
{}^{{}^{1}}_{{}_{1}}\kappa_{1,0} \overset{\cdot}{\otimes} {}^{{}^{1}}_{{}_{2}}\kappa_{1,0}
\]
we can treat as an element of
\[
\mathscr{L}(\mathscr{E}_{{}_{i}}, \, E_{{}_{n_1}}^{*}
\otimes E_{{}_{n_2}}^{*}) \cong
\mathscr{L}(E_{{}_{n_1}} \otimes E_{{}_{n_2}}, \,\,
\mathscr{E}^{*}_{{}_{i}}).
\]
Assertion 4), case $M=2$, will be proved if we show that
\[
{}^{{}^{1}}_{{}_{1}}\kappa_{1,0} \overset{\cdot}{\otimes} {}^{{}^{1}}_{{}_{2}}\kappa_{1,0}
\in
\mathscr{L}(\mathscr{E}_{{}_{i}}, \, E_{{}_{n_1}}^{*}
\otimes E_{{}_{n_2}}^{*})
\]
actually belongs to
\[
\mathscr{L}(\mathscr{E}_{{}_{i}}, \, E_{{}_{n_1}}
\otimes E_{{}_{n_2}}).
\]
Similarly
\[
{}^{{}^{2}}_{{}_{1}}\kappa_{1,0} \overset{\cdot}{\otimes} {}^{{}^{2}}_{{}_{2}}\kappa_{1,0}
\in
\mathscr{L}(\mathscr{E}_{{}_{i}}, \, E_{{}_{n_1}}^{*}
\otimes E_{{}_{n_2}}^{*}) \cong
\mathscr{L}(E_{{}_{n_1}} \otimes E_{{}_{n_2}}, \,\,
\mathscr{E}^{*}_{{}_{i}}).
\]
would be extendible to an element of
\[
\mathscr{L}(E_{{}_{n_1}}^{*} \otimes E_{{}_{n_2}}^{*}, \,\,
\mathscr{E}^{*}_{{}_{i}}) \cong
\mathscr{L}(\mathscr{E}_{{}_{i}}, \, E_{{}_{n_1}}
\otimes E_{{}_{n_2}})
\]
if
\[
{}^{{}^{2}}_{{}_{1}}\kappa_{1,0} \overset{\cdot}{\otimes} {}^{{}^{2}}_{{}_{2}}\kappa_{0,1}
\in \mathscr{L}(\mathscr{E}_{{}_{i}}, \, E_{{}_{n_1}}^{*}
\otimes E_{{}_{n_2}}^{*})
\]
actually belongs to
\[
\mathscr{L}(\mathscr{E}_{{}_{i}}, \, E_{{}_{n_1}}
\otimes E_{{}_{n_2}}).
\]
This however is impossible because if both kernels ${}^{{}^{2}}_{{}_{1}}\kappa_{1,0},
{}^{{}^{2}}_{{}_{2}}\kappa_{1,0}$ are associated to a fixed component of the free zero mass electromagnetic
potential field (or its derivative ), then easy computation shows that
${}^{{}^{2}}_{{}_{1}}\kappa_{1,0} \overset{\cdot}{\otimes} {}^{{}^{2}}_{{}_{2}}\kappa_{1,0}(\phi)$,
$\phi \in \mathscr{E}_2$, has the following general form
\[
{}^{{}^{2}}_{{}_{1}}\kappa_{1,0} \overset{\cdot}{\otimes} {}^{{}^{2}}_{{}_{2}}\kappa_{1,0}(\phi)
(\boldsymbol{\p}_1, \boldsymbol{\p}_2) =
M_{1}^{\nu_1}(\boldsymbol{\p}_1) M_{2}^{\nu_2}(\boldsymbol{\p}_2) \widetilde{\phi}(\boldsymbol{\p}_1 + \boldsymbol{\p}_2,
p_{10}(\boldsymbol{\p}_1) + p_{20}(\boldsymbol{\p}_2)),
\]
where $M_{i}^{\nu_i}$ is a multiplier of $E_{{}_{n_i}} = \mathcal{S}_{{}_{A_{i}}}(\mathbb{R}^3; \mathbb{C}^4)$, $i=1,2$, and
\[
p_{10}(\boldsymbol{\p}_1) = |\boldsymbol{\p}_1|, \,\,\, p_{20}(\boldsymbol{\p}_2) = |\boldsymbol{\p}_2|.
\]
We can now easily see that
\[
{}^{{}^{2}}_{{}_{1}}\kappa_{1,0} \overset{\cdot}{\otimes} {}^{{}^{2}}_{{}_{2}}\kappa_{1,0}(\phi)
\]
cannot even belong to $\mathscr{C}^{\infty}(\mathbb{R}^3 \times \mathbb{R}^3; \mathbb{C}^{8})$,
so all the more it cannot belong to
$\mathcal{S}(\mathbb{R}^3; \mathbb{C}^4) \otimes \mathcal{S}(\mathbb{R}^3; \mathbb{C}^4) = E_1 \otimes E_1$
or to $\mathcal{S}^{0}(\mathbb{R}^3; \mathbb{C}^4) \otimes \mathcal{S}^{0}(\mathbb{R}^3; \mathbb{C}^4)
= E_2 \otimes E_2$ or to $E_1 \otimes E_2$ or finally to $E_2 \otimes E_1$.
In particular
\begin{equation}\label{discon.2kappa.2kappa.map}
\phi \longmapsto
{}^{{}^{2}}_{{}_{1}}\kappa_{1,0} \overset{\cdot}{\otimes} {}^{{}^{2}}_{{}_{2}}\kappa_{1,0}(\phi)
\end{equation}
cannot be continuous as a map
\[
\mathscr{E}_i \longmapsto E_{{}_{n_1}}
\otimes E_{{}_{n_2}}.
\]
From this it follows that
\[
{}^{{}^{2}}_{{}_{1}}\kappa_{1,0} \overset{\cdot}{\otimes} {}^{{}^{2}}_{{}_{2}}\kappa_{1,0}
\]
cannot be extended to an element of
\[
\mathscr{L}(E_{{}_{n_1}}^{*} \otimes E_{{}_{n_2}}^{*}, \,\,
\mathscr{E}^{*}_{{}_{i}}).
\]

Of course, we have the same situation if any one of the factors
\[
{}^{{}^{n_k}}_{{}_{k}}\kappa_{l_k,m_k}, \,\,\, k =1,2, \,\,\, (l_k,m_k) = (1,0) \, \textrm{or} \, (0,1)
\]
has $n_k =2$ and thus corresponds to massless field, say $n_2 = 2$. In this case  we have similar expression
\[
{}^{{}^{1}}_{{}_{1}}\kappa_{l_1,m_1} \overset{\cdot}{\otimes} {}^{{}^{2}}_{{}_{2}}\kappa_{l_2,m_2}(\phi)
(\boldsymbol{\p}_1, \boldsymbol{\p}_2)  =
M_{1}^{\nu_1}(\boldsymbol{\p}_1) M_{2}^{\nu_2}(\boldsymbol{\p}_2) \widetilde{\phi}(\pm \boldsymbol{\p}_1 \pm \boldsymbol{\p}_2,
\pm p_{10}(\boldsymbol{\p}_1) \pm p_{20}(\boldsymbol{\p}_2)),
\]
where $M_{i}^{\nu_i}$ is a multiplier of $E_{{}_{n_i}}$, $i=1,2$, the sign $\pm$ depends on the particular
choice of $(l_k,m_k)$ equal $(0,1)$ or $(1,0)$, and 
\[
p_{10}(\boldsymbol{\p}_1) = \sqrt{|\boldsymbol{\p}_1|^2 + m^2},  \,\,\, p_{20}(\boldsymbol{\p}_2) = |\boldsymbol{\p}_2|.
\]
It is immediately seen that ${}^{{}^{1}}_{{}_{1}}\kappa_{l_1,m_1} \overset{\cdot}{\otimes} {}^{{}^{2}}_{{}_{2}}\kappa_{l_2,m_2}(\phi)$
cannot be even smooth on $\mathbb{R}^3 \times \mathbb{R}^3$ in view of the form of the zero-mass particle energy $p_{02}$ as the function
of momentum $\boldsymbol{\p}_2$. We therefore get the following 
\begin{prop*}
Wick product of free fields, or their spatio-temporal derivatives, does not belong to
\[
\mathscr{L}\big(\mathscr{E}, \, \mathscr{L}((\boldsymbol{E}), (\boldsymbol{E}))\big)
\cong \mathscr{L}\big((\boldsymbol{E}) \otimes \mathscr{E}, \, (\boldsymbol{E})\big),
\]
if among the factors there are massless fields or their spatio-temporal derivatives.
\end{prop*} 

But we should note here that the Wick products of free fields or their derivatives, behaves equally well,
irrespectively if there are mas less factors present or not, if evaluated at fixed space-time point. In each case
the Wick product evaluated at fixed space-time point represent well-defined integral kernel operator transforming
continuously the Hida space into its strong dual. Indeed, this follows by the simple observation that if we fix the space-time
point $x$ and the field index $a$ in the kernel function
\begin{align*}
{}^{{}^{n_1}}_{{}_{1}}\kappa_{l_1,m_1}
\overset{\cdot}{\otimes}
{}^{{}^{n_2}}_{{}_{k}}\kappa_{l_2,m_2} \big(\nu_1 \boldsymbol{\p}_1, \nu_2 \boldsymbol{\p}_2; a,x \big)
\\
\,\,\, k =1,2, \,\,\, (l_k,m_k) = (1,0) \, \textrm{or} \, (0,1), \,\,\, \textrm{and} \, (a,x) \, \textrm{fixed},
\end{align*}
then we get
\begin{multline*}
{}^{{}^{n_1}}_{{}_{1}}\kappa_{l_1,m_1}
\overset{\cdot}{\otimes}
{}^{{}^{n_2}}_{{}_{k}}\kappa_{l_2,m_2} \big(\nu_1 \boldsymbol{\p}_1, \nu_2 \boldsymbol{\p}_2; a,x \big) = \\ =
M_{1}^{\nu_1}(\boldsymbol{\p}_1) e^{\pm i\boldsymbol{\p}_1 \mp i p_{10}(\boldsymbol{\p}_1)} M_{2}^{\nu_2}(\boldsymbol{\p}_2) 
e^{\pm i \boldsymbol{\p}_2  \mp i p_{20}(\boldsymbol{\p}_1)},
\end{multline*}
and in any case the functions
\begin{align*}
\boldsymbol{\p}_1 \longrightarrow M_{1}^{\nu_1}(\boldsymbol{\p}_1) e^{\pm i\boldsymbol{\p}_1  \mp i p_{10}(\boldsymbol{\p}_1)}, \\
\boldsymbol{\p}_2 \longrightarrow M_{1}^{\nu_1}(\boldsymbol{\p}_2) e^{\pm i\boldsymbol{\p}_2 \mp i p_{10}(\boldsymbol{\p}_2)},
\end{align*}
are multipliers of the corresponding nuclear algebras $E_{{}_{n_k}}$, $k=1,2$, with the corresponding
\[
p_{10}(\boldsymbol{\p}_1) = |\boldsymbol{\p}_1|, \,\,\, p_{20}(\boldsymbol{\p}_2) = |\boldsymbol{\p}_2|,
\]
or
\[
p_{10}(\boldsymbol{\p}_1) = \sqrt{|\boldsymbol{\p}_1|^2 + m^2}, \,\,\, p_{20}(\boldsymbol{\p}_2) = \sqrt{\boldsymbol{\p}_2|^2 + m^2}.
\]
It is easily seen that the same holds for the kernels corresponding to space-time derivatives of free fields.
Therefore, we arrive at the
\begin{prop*}
In general the Wick product of space-time derivatives of massive or massless fields evaluated at fixed space-time point represents
well-defined integral kernel operator transforming continuously the Hida space into its strong dual, \emph{i.e.} it belongs to
\[
\mathscr{L}((\boldsymbol{E}), (\boldsymbol{E})^*).
\]
\end{prop*}

Of course, from the last Lemma, part 3), it follows that the Wick product at the same point of
any number of zero mass or massive fields is a well-defined integral kernel operator belonging to
\[
\mathscr{L}\big(\mathscr{E}, \, \mathscr{L}((\boldsymbol{E}), (\boldsymbol{E})^*)\big)
\cong \mathscr{L}\big((\boldsymbol{E}) \otimes \mathscr{E}, \, (\boldsymbol{E})^*\big)
\]
in the sense of Obata \cite{obataJFA} with vector-valued kernel.
We therefore have the following
\begin{prop*}
\begin{enumerate}
\item[1)]
For the Wick product at te same space-time point $x$
\begin{multline*}
\boldsymbol{}: \Xi_{l_1,m_1}\Big({}^{{}^{n_1}}_{{}_{1}}\kappa_{l_1,m_1}(x)\Big) \cdots
\Xi_{l_M,m_M}\Big({}^{{}^{n_M}}_{{}_{M}}\kappa_{l_M,m_M}(x)\big) \boldsymbol{:} \\
= \Xi_{l,m}(\kappa_{lm}(x)), \,\,\,
{}^{{}^{n_k}}_{{}_{k}}\kappa_{l_k,m_k} \in \mathfrak{K}_0
\end{multline*}
of the integral kernel operators corresponding to the free fields of the theory
or their derivatives we have
\begin{multline*}
\kappa_{l,m} = \Big({}^{{}^{n_1}}_{{}_{1}}\kappa_{l_1,m_1} \Big) \overline{\dot{\otimes}}
\cdots \overline{\dot{\otimes}} \,\, \Big( {}^{{}^{n_M}}_{{}_{M}}\kappa_{l_M,m_M} \Big) \\
\in
E^{*}_{{}_{n_1}} \otimes \cdots \otimes
E^{*}_{{}_{n_M}} \otimes \mathscr{E}^{*}_{{}_{i}}
\cong \mathscr{L}(E_{{}_{n_1}} \otimes \cdots \otimes E_{{}_{n_M}}, \,\,
\mathscr{E}^{*}_{{}_{i}}), \,\,\, i=1,2.
\end{multline*}
Thus by (the generalization to tensor product of Fock spaces of) Thm. 3.9 of \cite{obataJFA}
\begin{multline*}
\boldsymbol{}: \Xi_{l_1,m_1}\Big({}^{{}^{n_1}}_{{}_{1}}\kappa_{l_1,m_1}\Big) \cdots
\Xi_{l_M,m_M}\Big({}^{{}^{n_M}}_{{}_{M}}\kappa_{l_M,m_M}\Big) \boldsymbol{:} \\
= \Xi_{l,m}(\kappa_{lm}) \in
\mathscr{L}\big((\boldsymbol{E}) \otimes \mathscr{E}_{{}_{i}}, \, (\boldsymbol{E})^* \big)
\cong \mathscr{L}\big(\mathscr{E}_{{}_{i}}, \, \mathscr{L}((\boldsymbol{E}), (\boldsymbol{E})^*) \big)
\end{multline*}
\item[2)]
If all $n_k = 1$, \emph{i.e.} among the factors
\[
\Xi_{l_1,m_1}\Big({}^{{}^{n_k}}_{{}_{k}}\kappa_{l_k,m_k}(x)\Big)
\]
there are no integral kernel operators corresponding to massless free fields
(electromagnetic potential field in case of QED) or their derivatives,
then (by 4) of the preceding Lemma) the bilinear map
\begin{multline*}
\xi \times \eta \mapsto \kappa_{l,m}(\xi \otimes \eta),
\\
\xi \in \overbrace{E_{i_1} \otimes \cdots
\otimes E_{i_l}}^{\textrm{first $l$ terms $E_{i_j}$, $i_j \in \{1,2\}$}}, \\
\eta \in \overbrace{E_{i_{l+1}} \otimes \cdots
\otimes E_{i_{l+m}}}^{\textrm{last $m$ terms $E_{i_j}$, $i_j\in \{1,2\}$}},
\end{multline*}
can be extended to a separately continuous bilinear map from
\[
\Big( \overbrace{E_{i_1} \otimes \cdots
\otimes E_{i_l}}^{\textrm{first $l$ terms $E_{i_j}$}} \Big)^*
\times
\Big( \overbrace{E_{i_{l+1}} \otimes \cdots
\otimes E_{i_{l+m}}}^{\textrm{last $m$ terms $E_{i_j}$}} \Big)
\,\,\, \textrm{into} \,\,\,\mathscr{L}(\mathscr{E}, \mathbb{C}) = \mathscr{E}^*.
\]
Thus by Thm. \ref{obataJFA.Thm.3.13}, Subsection \ref{psiBerezin-Hida}
\begin{multline*}
\boldsymbol{}: \Xi_{l_1,m_1}\Big({}^{{}^{n_1}}_{{}_{1}}\kappa_{l_1,m_1}\Big) \cdots
\Xi_{l_M,m_M}\Big({}^{{}^{n_M}}_{{}_{M}}\kappa_{l_M,m_M}\Big) \boldsymbol{:} \\
= \Xi_{l,m}(\kappa_{lm}) \in
\mathscr{L}\big((\boldsymbol{E}) \otimes \mathscr{E}_{{}_{i}}, \, (\boldsymbol{E}) \big)
\cong \mathscr{L}\big(\mathscr{E}_{{}_{i}}, \, \mathscr{L}((\boldsymbol{E}), (\boldsymbol{E})) \big)
\end{multline*}
\end{enumerate}
\end{prop*}

Now we pass to the operation of differentiation with respect to space-time coordinates. Suppose we have an
integral kernel operator $\Xi_{l,m}(\kappa_{l,m})$ with vector-valued
kernel
\[
\kappa_{l,m} \in \mathscr{L} \big(\mathscr{E} , \,\, \big(E_{i_1} \otimes \cdots \otimes E_{i_{l+m}} \big)^* \, \big)
\cong \mathscr{L} \big(E_{i_1} \otimes \cdots \otimes E_{i_{l+m}}, \,\, \mathscr{E}^* \big)
\]
with the operator
\[
\Xi_{l,m}(\kappa_{l,m}) \in \mathscr{L}\big(\mathscr{E}, \, \mathscr{L}((\boldsymbol{E}), (\boldsymbol{E})^*)\big)
\cong \mathscr{L}\big((\boldsymbol{E}) \otimes \mathscr{E}, \, (\boldsymbol{E})^*\big)
\]
uniquely determined by
\begin{multline*}
\big\langle \big\langle \Xi_{l,m}(\kappa_{l,m})(\Phi \otimes \phi), \, \Psi \big \rangle \big \rangle
=\big\langle \big\langle \Xi_{l,m}\big(\kappa_{l,m}(\phi)\big)\Phi, \, \Psi \big \rangle \big \rangle \\
= \langle \kappa_{l,m}(\phi), \eta_{\Phi, \Psi} \rangle
= \langle \kappa_{l,m}(\eta_{\Phi, \Psi}), \phi \rangle,
\,\,\,
\Phi, \Psi \in (\boldsymbol{E}), \phi \in \mathscr{E},
\end{multline*}
compare (\ref{VectValotimesXi=intKerOp'}) Subsection \ref{psiBerezin-Hida}. Suppose moreover
that
\[
\begin{split}
\mathscr{E} = \mathscr{E}_{1} = \mathcal{S}_{H_{(4)}}(\mathbb{R}^4;\mathbb{C})=
\mathcal{S}(\mathbb{R}^4; \mathbb{C}) \,\,\, \textrm{or} \\
\mathscr{E} = \mathscr{E}_{2} = \mathcal{S}_{\mathscr{F}A^{(4)}\mathscr{F}^{-1}}(\mathbb{R}^4;\mathbb{C})=
\mathcal{S}^{00}(\mathbb{R}^4; \mathbb{C}).
\end{split}
\]
Let for $\kappa_{l,m}$ understood as an element of
\[
\mathscr{L} \big(E_{i_1} \otimes \cdots \otimes E_{i_{l+m}}, \,\, \mathscr{E}^* \big) \cong
\mathscr{L} \big(\mathscr{E} , \,\, \big(E_{i_1} \otimes \cdots \otimes E_{i_{l+m}} \big)^* \, \big)
\]
we have
\[
\kappa_{l,m}(\xi_1 \otimes \cdots \otimes \xi_{l+m}) \in \mathcal{O}_C(\mathbb{R}^4; \mathbb{C}),
\,\,\,\
\xi_{k} \in E_{i_k}, i_k \in \{1, 2\}.
\]
We moreover include into consideration the special cases of integral kernel operators
\begin{equation}\label{FreeFieldXi-s}
\Xi_{0,1}({}^{1}\kappa_{0,1}), \Xi_{1,0}({}^{1}\kappa_{1,0}),
\Xi_{0,1}({}^{2}\kappa_{0,1}), \Xi_{1,0}({}^{2}\kappa_{1,0}),
\end{equation}
determined by the free fields of the theory with the integration in the general formula
(\ref{electron-positron-photon-Xi}) is restricted, respectively, only to Fermi or only to Bose
variables, and the Wick products of (\ref{FreeFieldXi-s}) at the same space-time point
(representing ordinary integral kernel operators (\ref{electron-positron-photon-Xi})
with vector-valued kernels and integration with integration in general ranging over both, Bose and Fermi,
variables if the Wick product involves both, Bose and Fermi, field components).

Then we can define the space-time derivative 
\[
\big(\frac{\partial}{\partial x^\mu} \Xi_{l,m}\big)(\kappa_{l,m})
\]
as the integral kernel operator uniquely determined by the condition
\begin{multline*}
\Big\langle \Big\langle \Big(\frac{\partial}{\partial x^\mu}\Xi_{l,m}\Big)(\kappa_{l,m})(\Phi \otimes \phi), 
\, \Psi  \Big \rangle \Big \rangle
=\Big\langle \Big\langle \Xi_{l,m}\Big(\Big(\frac{\partial}{\partial x^\mu}\kappa_{l,m}\big)(\phi)\Big)\Phi, \, \Psi  \Big \rangle \Big \rangle \\
= -\Big\langle \Big\langle \Xi_{l,m}\Big(\kappa_{l,m}\Big(\Big(\frac{\partial}{\partial x^\mu}\phi\big)\big)\Phi, \, \Psi  \Big \rangle \Big \rangle 
=  \Big\langle \Big(\frac{\partial}{\partial x^\mu}\kappa_{l,m}\Big)(\phi),  \eta_{\Phi, \Psi} \Big \rangle \\
= - \Big\langle \kappa_{l,m}\Big(\frac{\partial}{\partial x^\mu}\phi\Big),  \eta_{\Phi, \Psi} \Big \rangle
= - \langle \kappa_{l,m}(\eta_{\Phi, \Psi}), \frac{\partial}{\partial x^\mu}\phi \rangle,
\,\,\,
\Phi, \Psi \in (\boldsymbol{E}), \phi \in \mathscr{E}.
\end{multline*}

\begin{center}
{\small RULE III'}
\end{center}
\emph{We have the following computational rule}
\[
\Big(\frac{\partial}{\partial x^\mu} \Xi_{l,m}\Big)(\kappa_{l,m})
=
\Xi_{l,m}\Big(\frac{\partial}{\partial x^\mu}\kappa_{l,m}\Big)
\]
\emph{for $\kappa_{l,m}$ understood as an element of}
\[
\mathscr{L} \big(\mathscr{E} , \,\, \big(E_{i_1} \otimes \cdots \otimes E_{i_{l+m}} \big)^* \, \big)
\cong \mathscr{L} \big(E_{i_1} \otimes \cdots \otimes E_{i_{l+m}}, \,\, \mathscr{E}^* \big).
\]

Thus, the operation of space-time differentiation performed on $\Xi(\kappa_{l,m})$
corresponds, via the Rule III', to the operation of differentiation performed upon the
vector-valued distributional kernel $\kappa_{l,m}$, understood as an
$\big(E_{i_1} \otimes \cdots \otimes E_{i_{l+m}} \big)^*$-valued distribution on the test function
space $\mathscr{E}$. Again the Rule III' can be justified by utilizing the fact that
\begin{multline}\label{electron-positron-photon-Xi(x)}
\Xi_{l,m}\big(\kappa_{l,m}(x)\big) =
\int \limits_{(\sqcup \mathbb{R}^3)^{(l+m)}}
\kappa_{l,m}(w_{i_1}, \ldots w_{i_l}, w_{i_{l+1}}, \ldots w_{i_{l+m}}; x)
\, \\ \times
\partial_{w_{i_1}}^* \cdots \partial_{w_{i_l}}^* \partial_{w_{i_{l+1}}} \cdots \partial_{w_{i_{l+m}}}
\ud w_{i_1} \cdots \ud w_{i_l} \ud w_{i_{l+1}} \cdots \ud w_{i_{l+m}} = \\
= \int \limits_{(\sqcup \mathbb{R}^3)^{(l+m)}}
\kappa_{l,m}(w_{i_1}, \ldots w_{i_l}, u_{j_{1}}, \ldots u_{j_{m}}; x) \,\, \times \\
\times \,\,
\partial_{w_{i_1}}^* \cdots \partial_{w_{i_l}}^* \partial_{u_{j_{1}}} \cdots \partial_{u_{j_{m}}}
\ud w_{i_1} \cdots \ud w_{i_l} \ud u_{j_{1}} \cdots \ud u_{j_{m}}
\end{multline}
exists pointwisely as a Pettis integral, just repeating the arguments
in construction of space-time derivatives of the free electromagnetic potential field
during the proof of Bogoliubov-Shirkov Quantization Postulate, compare Subsection \ref{BSH}.
Moreover, during this proof we have given justification of the following Rules IV, V
and VI.

For the integral kernel operator (\ref{electron-positron-photon-Xi(x)}) we have 

\begin{center}
{\small RULE IV'}
\end{center}
\[
\int \limits_{\mathbb{R}^4} \Xi_{l,m}\big(\kappa_{l,m}(x)\big) \, \ud^4 x
= \Xi_{l,m}\Bigg(\int \limits_{\mathbb{R}^4}\kappa_{l,m}(x) \ud^4 x \Bigg).
\]

\begin{center}
{\small RULE V'}
\end{center}
\[
\int \limits_{\mathbb{R}^4} \Xi_{l,m}\big(\kappa_{l,m}(\boldsymbol{\x}, x_0)\big) \, \ud^3 \boldsymbol{\x}
= \Xi_{l,m}\Bigg(\int \limits_{\mathbb{R}^4}\kappa_{l,m}(\boldsymbol{\x}, x_0) \, \ud^3 \boldsymbol{\x} \Bigg).
\]

Let $S \in \mathcal{S}(\mathbb{R}^4; \mathbb{C})^*$ then
\begin{center}
{\small RULE VI}
\end{center}
\begin{multline*}
S \ast \Xi_{l,m}(\kappa_{l,m})(x) = 
\int \limits_{\mathbb{R}^4} S(x-y) \Xi_{l,m}\big(\kappa_{l,m}(y)\big) \, \ud^4 y \\
= \Xi_{l,m}\Bigg(\int \limits_{\mathbb{R}^4} S(x-y)\kappa_{l,m}(y) \, \ud^4 y \,  \Bigg)
= \Xi_{l,m}\big( S \ast \kappa_{lm}(x) \big).
\end{multline*}
\emph{Here}
\begin{multline*}
S \ast \kappa_{lm}(\xi_1, \ldots, \xi_{l+m})(x)  \\
= \int \limits_{\mathbb{R}^4} S(x-y)\kappa_{l,m}(w_{i_1}, 
\ldots, w_{i_{l+m}}; y) \, \xi_{i_1}(w_{i_1}), 
\ldots, \xi_{i_{l+m}}(w_{i_{l+m}}) \ud^4 y,
\,\,\,\, \xi_{i_k} \in E_{i_k}
\end{multline*}
\emph{is well-defined for $S$ equal to the product of pairings or to the retarded or advanced part of the causal combinantions of products
of pairings and $\kappa_{l,m}$ are the kernels of the Wick product of the free fields or their derivatives. 
This can be checked by explicit computation, because by Lemma} \ref{kappaBarDotOtimeskappa}
\[
\kappa_{l,m}(\xi_1 \otimes \cdots \otimes \xi_{l+m}) \in \mathcal{O}_C(\mathbb{R}^{4}; \mathbb{C}),
\]
\emph{is equal to the product of smooth functions whose Furier transforms are concetrated on the orbits of the free fields
and are equal there to the elements of the respective spaces $E_i = \mathcal{S}_{A_i}(\mathbb{R}^3)$  
and by definition is equal to the (kernel of the) distribution 
$S \ast (\kappa_{lm}(\xi_1, \ldots, \xi_{l+m}))$. 
The only exception is when $\kappa_{lm} = \kappa_{0,1},\kappa_{1,0}$ is the kernel of a free field
with the mass equal to the mass of the free field whose commutation function (or its retarded or 
advanced part) is equal $S$, in which case the convolution $S \ast (\kappa_{lm}(\xi))$ is not well-defined.}

The Rules III', IV', V', VI are also valid in case of more than just one space-time variable $x$. 
In order to see it we can repeat the proof replacing $\mathscr{E}$ (previously equal to $\mathscr{E}_1 = 
\mathcal{S}(\mathbb{R}^4; \mathbb{C})$ or $\mathscr{E}_2 = \mathcal{S}^{00}(\mathbb{R}^4; \mathbb{C})$)
by $\mathscr{E}$ equal to tensor product of several $\mathscr{E}_1$ or $\mathscr{E}_2$.
In this case we would obtain  more generally with
\[
\kappa_{l,m}(\xi_1 \otimes \cdots \otimes \xi_{l+m};x_1, \ldots, x_n) \in \mathcal{O}_C(\mathbb{R}^{4n}; \mathbb{C})
\]
the integral kernel operator
\begin{multline}\label{electron-positron-photon-Xi(x1...xn)}
\Xi_{l,m}\big(\kappa_{l,m}(x_1, \ldots, x_n)\big) = 
\int \limits_{(\sqcup \mathbb{R}^3)^{(l+m)}}
\kappa_{l,m}(w_{i_1}, \ldots w_{i_l}, w_{i_{l+1}}, \ldots w_{i_{l+m}}; x_1, \ldots, x_n) 
\, \times  \\ \times \,
\partial_{w_{i_1}}^* \cdots \partial_{w_{i_l}}^* \partial_{w_{i_{l+1}}} \cdots \partial_{w_{i_{l+m}}}
\ud w_{i_1} \cdots \ud w_{i_l} \ud w_{i_{l+1}} \cdots \ud w_{i_{l+m}} = \\
= \int \limits_{(\sqcup \mathbb{R}^3)^{(l+m)}}
\kappa_{l,m}(w_{i_1}, \ldots w_{i_l}, u_{j_{1}}, \ldots u_{j_{m}}; x_1, \ldots, x_n) \,\, \times \\
\times \, \partial_{w_{i_1}}^* \cdots \partial_{w_{i_l}}^* \partial_{u_{j_{1}}} \cdots \partial_{u_{j_{m}}}
\ud w_{i_1} \cdots \ud w_{i_l} \ud u_{j_{1}} \cdots \ud u_{j_{m}} 
\end{multline} 
existing pointwisely as a Pettis integral and with the following Rules:

\begin{center}
{\small RULE III}
\end{center}
\[
\Big(\frac{\partial^{n}}{\partial x_{1}^{\mu_1}\cdots \partial x_{n}^{\mu_n}} \Xi_{l,m}\Big)(\kappa_{l,m})
=
\Xi_{l,m}\Big(\frac{\partial^{n}}{\partial x_{1}^{\mu_1}\cdots \partial x_{n}^{\mu_n}}\kappa_{l,m}\Big)
\]
\emph{for $\kappa_{l,m}$ understood as an element of}
\[
\mathscr{L} \big(\mathscr{E} , \,\, \big(E_{i_1} \otimes \cdots \otimes E_{i_{l+m}} \big)^* \, \big)
\cong \mathscr{L} \big(E_{i_1} \otimes \cdots \otimes E_{i_{l+m}}, \,\, \mathscr{E}^* \big).
\]
\emph{with}
\[
\mathscr{E} = \mathscr{E}_{n_1} \otimes \cdots \otimes \mathscr{E}_{n_n}, \,\,\, n_k \in \{1,2\}.
\]

\begin{center}
{\small RULE IV}
\end{center}
\[
\int \limits_{\mathbb{R}^{4n}} \Xi_{l,m}\big(\kappa_{l,m}(x_1, \ldots, x_n)\big) \, \ud^4 x_1 \ldots \ud^4 x_n
= \Xi_{l,m}\Bigg(\int \limits_{\mathbb{R}^{4n}}\kappa_{l,m}(x_1, \ldots, x_n) \ud^4 x_1 \ldots \ud^4 x_n \Bigg).
\]

\begin{center}
{\small RULE V}
\end{center}
\begin{multline*}
\int \limits_{\mathbb{R}^{3n}} \Xi_{l,m}\big(\kappa_{l,m}(\boldsymbol{\x}_1, x_{10}, \ldots \boldsymbol{\x}_n, x_{n0})\big) \, \ud^3 \boldsymbol{\x}_1 \cdots \ud^3 \boldsymbol{\x}_n \\
= \Xi_{l,m}\Bigg(\int \limits_{\mathbb{R}^4}\kappa_{l,m}(\boldsymbol{\x}_1, x_{10}, \ldots \boldsymbol{\x}_n, x_{n0}) \, \ud^3 \boldsymbol{\x}_1 \ldots \ud^3 \boldsymbol{\x}_n \Bigg).
\end{multline*}

Now concerning the Rule VI for more space-time variables we can repeatedly combine the convolutions
of several distributions $S \in \mathcal{S}(\mathbb{R}^4; \mathbb{C})^*$ each in one space-time variable,
with the Wick product operation provided the corresponding kernels $\kappa_{l,m}$ obtained in the intermediate steps
are well-defined elements of $\mathscr{L}(E_{i_1}\otimes \cdots \otimes, \,\, \mathscr{E}_{n_1}^{*} \otimes \cdots)$
with
\[
\kappa_{l,m}(\xi_{i_1} \otimes \cdots )( x_{n_1}, \ldots) \in \mathcal{O}_C.
\]

Namely, we have the following useful Lemma which allows us to operate with convolutions
of integral kernel operators with tempered distributions $S \in \mathcal{S}(\mathbb{R}^4; \mathbb{C})^*$:

\begin{lem}\label{S*Xi} 
Let $S \in \mathcal{S}(\mathbb{R}^4; \mathbb{C})^*$, and let 
\[
\kappa_{l,m} \in \mathscr{L} \big(\mathscr{E} , \,\, \big(E_{i_1} \otimes \cdots \otimes E_{i_{l+m}} \big)^* \, \big)
\cong \mathscr{L} \big(E_{i_1} \otimes \cdots \otimes E_{i_{l+m}}, \,\, \mathscr{E}^* \big)
\]
with the kernel
\[
\kappa_{l,m} = \Big({}^{{}^{n_1}}_{{}_{1}}\kappa_{l_1,m_1} \Big) \overline{\dot{\otimes}} \cdots \overline{\dot{\otimes}} \,\, \Big( {}^{{}^{n_M}}_{{}_{M}}\kappa_{l_M,m_M} \Big)
\]
corresponding to the Wick product (at the same space-time point $x$) 
\[
\Xi_{l,m}(\kappa_{lm}(x)) =
\boldsymbol{}: \Xi_{l_1,m_1}\Big({}^{{}^{n_1}}_{{}_{1}}\kappa_{l_1,m_1}(x)\Big) \cdots 
\Xi_{l_M,m_M}\Big({}^{{}^{n_1}}_{{}_{M}}\kappa_{l_M,m_M}(x)\big) \boldsymbol{:} 
\]
of the integral kernel operators 
\[
\Xi_{l_k,m_k}\Big({}^{{}^{n_k}}_{{}_{k}}\kappa_{l_k,m_k}(x)\Big), \,\,\,\,
{}^{{}^{n_k}}_{{}_{k}}\kappa_{l_k,m_k} \in \mathfrak{K}_0.
\]
Let the integral kernel $S \ast \kappa_{l,m}$  be equal
\begin{multline*}
\langle S \ast \kappa_{l,m}(\xi_{i_1} \otimes \cdots \otimes \xi_{i_{l+m}}), \phi \rangle
=  \int \limits_{\mathbb{R}^4} S \ast \kappa_{lm}(\xi_1, \ldots, \xi_{l+m})(x) \, \phi(x) \, \ud^4 x \\
\int \limits_{\mathbb{R}^4 \times \mathbb{R}^4} S(x-y)\kappa_{l,m}(w_{i_1}, 
\ldots, w_{i_{l+m}}; y) \, \xi_{i_1}(w_{i_1}), 
\ldots, \xi_{i_{l+m}}(w_{i_{l+m}}) \, \phi(x) \,
\ud w_{i_1} \cdots \ud w_{i_{l+m}}
\ud^4 y \ud^4 x, \\
\,\,\,\, \xi_{i_k} \in E_{i_k}, \, \phi \in \mathscr{E} = \mathcal{S}(\mathbb{R}^4; \mathbb{}C) 
\,\, \textrm{or} \,\, \mathscr{E} = \mathcal{S}^{00}(\mathbb{R}^4; \mathbb{C}). 
\end{multline*}
Then
\begin{enumerate}
\item[1)]
 the kernel
\[
S_n \ast S_{n-1} \ast \ldots \ast S_1 \ast \kappa_{l,m} \in \mathscr{L} \big(E_{i_1} \otimes \cdots \otimes E_{i_{l+m}}, \,\, \mathscr{E}^* \big)
\]
if 
\[
\kappa_{l,m} = \Big({}^{{}^{n_1}}_{{}_{1}}\kappa_{l_1,m_1} \Big) \overline{\dot{\otimes}} \cdots \overline{\dot{\otimes}} \,\, \Big( {}^{{}^{n_M}}_{{}_{M}}\kappa_{l_M,m_M} \Big), \,\, l+m =M,
\,\,\,\,\,
{}^{{}^{n_k}}_{{}_{k}}\kappa_{l_k,m_k} \in \mathfrak{K}_0
\]
and each $S_i$ is equal to the product of pairings or to the retarded or advanced part of the causal combinations of products
of pairings and $M>1$, which we encounter as the higher order contributions to interacting fields in spinor QED with the intensity of interaction
function $g$ put equal $g=1$ and with the ``natural'' splitting of the causal distributions in the computation of the scattering operator. 
The contributions to interacting fields
whose kernels do not fulfil this condition cancel out and are identically zero for each intensity of interaction function
$g\in \mathcal{S}(\mathbb{R}^4)$, before passing to the limit $g \rightarrow 1$.

\item[2)]
Let moreover, in case $M=1$, ${}^{{}^{n_1}}_{{}_{1}}\kappa_{l_1,m_1} = \kappa_{0,1},\kappa_{1,0}$ be equal to the kernel of a free field
with a mass $m_{i_1}$. If further among the distributions $S_n, S_{n-1}, \ldots, S_1$ there are no
(retarded or advanced parts of the) commutation functions of a free field of mass $m_2 = m_{i_1}$, then the convolutions 
\[
S_{n} \ast \ldots \ast S_1 \ast \kappa_{0,1}(\xi), S_{n} \ast \ldots \ast S_1 \ast \kappa_{1,0}(\xi) , \,\,\,\, \xi \in E,
\]
are well-defined and
\[
S_{n} \ast \ldots \ast S_1 \ast \kappa_{0,1}, \, S_{n} \ast \ldots \ast S_1 \ast \kappa_{1,0} \in \mathscr{L} \big(E_{i_1}, \,\, \mathscr{E}^* \big),
\]
for the kernels of the generalized integral kernel operators which we encounter as the higher order contributions to interacting fields 
in spinor QED with the intensity of interaction function $g$ put equal $g=1$. The contributions whose kernels do not respect this condition
cancel out for the ``natural'' splitting used in the computation of the scattering operator, before passing to the limit $g\rightarrow 1$.
\item[3)]
If $\, {}^{{}^{n_1}}_{{}_{1}}\kappa_{l_1,m_1} = \kappa_{0,1},\kappa_{1,0}$ is the kernel of a free field
with the mass not equal to the mass of the free field whose commutation function (or its retarded or 
advanced part) is equal $S$ then the convolutions
\[
S \ast \kappa_{0,1}, S \ast \kappa_{1,0} ,  \in \mathscr{L} \big(E_{i_1}^*, \,\, \mathscr{E}^* \big) 
 = \mathscr{L} \big( \mathscr{E}, E_{i_1} \big)
\subset \mathscr{L} \big(E_{i_1}, \mathscr{E}^* \big).
\]
are well-defined.
\item[4)]
If $\, {}^{{}^{n_1}}_{{}_{1}}\kappa_{l_1,m_1} = \kappa_{0,1},\kappa_{1,0}$ is the kernel of a free field
with the mass equal to the mass of the free field whose commutation function (or its retarded or 
advanced part) is equal $S$ then the convolutions
\[
S \ast \kappa_{0,1}, S \ast \kappa_{1,0},  
\]
are not well-defined.

\end{enumerate}
\end{lem}

\qedsymbol \,
The map 
\begin{multline*}
E_1 \otimes \ldots \otimes E_{{}_{M}} \ni
\xi_1 \otimes \ldots \xi_{{}{M}} \longmapsto \kappa_{l,m}(\xi_1 \otimes \ldots \xi_{{}{M}}) 
\\
= \Big({}^{{}^{n_1}}_{{}_{1}}\kappa_{l_1,m_1} \Big) \dot{\otimes} \cdots \dot{\otimes} \,\, \Big( {}^{{}^{n_M}}_{{}_{M}}\kappa_{l_M,m_M} \Big)
(\xi_1 \otimes \ldots \xi_{{}{M}})
\\
= \Big({}^{{}^{n_1}}_{{}_{1}}\kappa_{l_1,m_1} \Big)(\xi_1) \,\, \cdots  \,\, \Big( {}^{{}^{n_M}}_{{}_{M}}\kappa_{l_M,m_M} \Big)(\xi_{{}{M}})
\in \mathcal{S}(\mathbb{R}^4)^*
\end{multline*}
for
\[
{}^{{}^{n_k}}_{{}_{k}}\kappa_{l_k,m_k} \in \mathfrak{K}_0
\]
is continuous, which follows from this product form and from Lemma \ref{kappa0,1,kappa1,0psi}, Subsection \ref{psiBerezin-Hida} and respectively Lemma \ref{kappa0,1,kappa1,0ForA}, Subsection \ref{A=Xi0,1+Xi1,0}. Because, for fixed tempered distributions
$S_1, \ldots, S_n$, the map
\[
E_{i_1} \otimes \cdots \otimes E_{i_{l+m}} \ni \xi \longmapsto S_n \ast \ldots \ast S_1 \ast \kappa_{l,m}(\xi) \in \mathcal{S}(\mathbb{R}^4)^*
\]
is continuous with respect to the ordinary strong dual topology on $\mathcal{S}(\mathbb{R}^4)^*$
whenever the convolution
\[
S_n \ast \ldots \ast S_1 \ast \kappa_{l,m}(\xi)
\] 
is well-defined (Banach-Steinhaus theorem), 
then we need only prove the very existence of the said convolutions. But the asserted existence (eventually non-existence) 
follows immediatelly from the explicit formulas
for these convolutions, for the plane wave kernels
\[
{}^{{}^{n_k}}_{{}_{k}}\kappa_{l_k,m_k} \in \mathfrak{K}_0
\]
of the free fields (or their derivatives) and for the advanced and retarded parts of the causal combinations $S_1, \ldots, S_n$
of the products of pairings of the free fields and with the assumed``natural'' splitting  into retarded and advanced parts
(below in this Subsection we are giving explicit examples together with the explicit
formulas for these convolutions). 

Indeed, let us show that continuity follows from the very existence of the convolution.

It is sufficient to consider the case $\mathscr{E} = \mathscr{E}_{1} = \mathcal{S}(\mathbb{R}^4; \mathbb{C})$,
because $\mathscr{E}_{1}^{*}$ is continuously embedded into 
$\mathscr{E}_{2}^{*} = \mathcal{S}^{00}(\mathbb{R}^4; \mathbb{C})^{*}$, compare Subsection \ref{SA=S0}.

Because the Schwartz' algebra $\mathcal{O}'_{C}(\mathbb{R}^4; \mathbb{C})$ of convolutors
of $\mathcal{S}(\mathbb{R}^4; \mathbb{C})^*$
(for definition of $\mathcal{O}'_{C}$ compare e.g. \cite{Schwartz} or Appendix \ref{convolutorsO'_C})
is dense in $\mathcal{S}(\mathbb{R}^4; \mathbb{C})^*$ in the strong dual topology, then for $\epsilon >0$
we can find $S_{i \, \epsilon} \in \mathcal{O}'_{C}$ such that 
\[
\underset{\epsilon \rightarrow 0}{\textrm{lim}}S_{i \, \epsilon} = S_i, \,\,\, i= 1, \ldots, n,
\]
in the strong topology of the dual space $\mathcal{S}(\mathbb{R}^4; \mathbb{C})^*$ of tempered distributions.
Let 
\[
S_\epsilon = S_{n \, \epsilon} \ast \ldots \ast S_{1 \, \epsilon}.
\]
Let $\xi$ be any element of
\[
E_{i_1} \otimes \cdots \otimes E_{i_{l+m}}.
\]
For $\epsilon >0$ we define the following linear operator $\Lambda_\epsilon$
\[
\Lambda_\epsilon(\xi) \overset{\textrm{df}}{=}
S_\epsilon \ast \kappa_{l,m}(\xi), \,\,\,
\xi \in E_{i_1} \otimes \cdots \otimes E_{i_{l+m}},
\]
on 
\[
E_{i_1} \otimes \cdots \otimes E_{i_{l+m}}.
\]
Because $S_\epsilon \in \mathcal{O}'_{C}$, $\epsilon>0$, and because 
\[
\kappa_{l,m} \in \mathscr{L} \big(E_{i_1} \otimes \cdots \otimes E_{i_{l+m}}, \,\, \mathscr{E}^* \big),
\]
then for each $\epsilon >0$ the operator 
\[
\Lambda_\epsilon: E_{i_1} \otimes \cdots \otimes E_{i_{l+m}} \longrightarrow
\mathscr{E}^*
\]
is continuous, i.e. 
\[
\Lambda_\epsilon \in \mathscr{L} \big(E_{i_1} \otimes \cdots \otimes E_{i_{l+m}}, \,\, \mathscr{E}^* \big).
\]

Let $\xi \in E_{i_1} \otimes \cdots \otimes E_{i_{l+m}}$.
By the existence assumption, for each $\xi \in E_{i_1} \otimes \cdots \otimes E_{i_{l+m}}$
\[
\underset{\epsilon \rightarrow 0}{\textrm{lim}} \Lambda_\epsilon(\xi) = 
\underset{\epsilon \rightarrow 0}{\textrm{lim}} S_\epsilon \ast \kappa_{l,m}(\xi) \,\,\,
\]
in strong dual topology of $\mathscr{E}^*$
exists 
\[
\underset{\epsilon \rightarrow 0}{\textrm{lim}} \Lambda_\epsilon(\xi) = S \ast \kappa_{l,m}(\xi),
\]
where the notation $S \ast \kappa_{l,m}(\xi)$,  used for this limit, is rather symbolic
(here we do not pretend to give any deeper justification for this notation).

Because  $E_{i_1} \otimes \cdots \otimes E_{i_{l+m}}$ is a complete Fr\'echet space then by the Banach-Steinhaus
theorem (e.g. Thm. 2.8 of \cite{Rudin}) it follows that
$S \ast \kappa_{l,m}$ is a continuous linear operator 
$E_{i_1} \otimes \cdots \otimes E_{i_{l+m}} \rightarrow \mathscr{E}^{*}$, i.e.
\[
S \ast \kappa_{l,m} \in \mathscr{L} \big(E_{i_1} \otimes \cdots \otimes E_{i_{l+m}}, \,\, \mathscr{E}^* \big).
\]
If $\mathscr{E} = \mathcal{S}^{00}(\mathbb{R}^4; \mathbb{C})$ then $S$ can be extended over to an element of
$\mathcal{S}^{00}(\mathbb{R}^4; \mathbb{C})^*$ (Hahn-Banach theorem), and the above proof can be repeated, 
because the algebra of convolutors of $\mathcal{S}^{00}(\mathbb{R}^4; \mathbb{C})^*$ is dense in 
$\mathcal{S}^{00}(\mathbb{R}^4; \mathbb{C})^*$ and contains $\mathcal{O}_{C}(\mathbb{R}^4; \mathbb{C})$
(compare Subsection \ref{diffSA}, \ref{SA=S0} and Appendix \ref{convolutorsO'_C}). 
This completes the proof of continuity. 

\qed

\begin{rem*}
We should emphasize here that the fact that the space $E_2$ is equal
\[
\mathcal{S}^{0}(\mathbb{R}^3; \mathbb{C}^4) \neq \mathcal{S}(\mathbb{R}^3; \mathbb{C}^4)
\]
intervenes here non trivially. For the wrong space $\mathcal{S}(\mathbb{R}^3; \mathbb{C}^4)$
used for $E_2$ this Lemma would be false. But this Lemma
is important for the construction of higher order contributions to interacting
fields understood as well-defined integral kernel operators with vector-valued kernels.
Analogue situation we encounter for any other zero mass field for which the corresponding space
$E_2$ must be equal $\mathcal{S}^{0}(\mathbb{R}^3; \mathbb{C}^r)$.
\end{rem*}

From the Rule VI, Lemma \ref{S*Xi} and the generalization of Theorem 3.6, 3.9 of \cite{obataJFA} to the general Fock space, including the Fermi case,
compare Subsection \ref{psiBerezin-Hida} (in \cite{obataJFA} it is treated the Bose case),  it follows the following

\begin{lem}\label{corS*Xi}
\item[1)]
Let
\[
\kappa_{l,m} = \Big({}^{{}^{n_1}}_{{}_{1}}\kappa_{l_1,m_1} \Big) \overline{\dot{\otimes}} \cdots \overline{\dot{\otimes}} \,\, \Big( {}^{{}^{n_M}}_{{}_{M}}\kappa_{l_M,m_M} \Big), \,\, l+m =M,
\,\,\,\,\,
{}^{{}^{n_k}}_{{}_{k}}\kappa_{l_k,m_k} \in \mathfrak{K}_0,
\]
and each $S_i$ is equal to the product of pairings or to the retarded or advanced part of the causal combinations of products
of pairings and $M>1$, 
which we encounter as the kernels of the higher order contributions to interacting fields in spinor QED with the intensity of interaction
function $g$ put equal $g=1$ and with the ``natural'' splitting of the causal distributions in the computation of the scattering operator.
Then the operator
\begin{multline*}
S_n \ast \ldots \ast S_1 \ast \Xi_{l,m}(\kappa_{l,m})(x) 
\\
= 
\int \limits_{[\mathbb{R}^4]^{\times n}} S_n(x-y_n) S_{n-1}(y_{n}-y_{n-1}) \ldots S_1(y_2-y_1) \Xi_{l,m}\big(\kappa_{l,m}(y_1)\big) \, \ud^4 y_1 \ldots \ud^4y_n \\
= \Xi_{l,m}\Bigg(\int \limits_{[\mathbb{R}^4]^{\times n}} S_n(x-y_n) S_{n}(y_{n-1}-y_{n-2}) \ldots S_1(y_2-y_1) \kappa_{l,m}(y_1) \, \ud^4 y_1 \ldots \ud^4y_n \,  \Bigg)
\\
= \Xi_{l,m}\big( S_n \ast \ldots \ast S_1 \ast \kappa_{lm}(x) \big) 
\end{multline*}
defines integral kernel operator 
\[
\Xi_{l,m}\big( S_n \ast \ldots \ast S_1 \ast \kappa_{lm} \big) 
\in \mathscr{L}\big((\boldsymbol{E}) \otimes \mathscr{E}, \, (\boldsymbol{E})^*\big) \cong
\mathscr{L}\big(\mathscr{E}, \, \mathscr{L}((\boldsymbol{E}), (\boldsymbol{E})^*)\big)
\]
with the vector-valued kernel
\[
S_n \ast \ldots \ast S_1 \ast \kappa_{lm} \in \mathscr{L}\big(\mathscr{E}, \, \big(E_{i_1} \otimes \cdots \otimes E_{i_{l+m}} \big)^* \, \big)
\cong \mathscr{L} \big(E_{i_1} \otimes \cdots \otimes E_{i_{l+m}}, \,\, \mathscr{E}^* \big).
\]
The contributions to interacting fields
whose kernels do not fulfil this condition cancel out and are identically zero for each intensity of interaction function
$g\in \mathcal{S}(\mathbb{R}^4)$, before passing to the limit $g\rightarrow 1$.

\item[2)]
Let moreover, for the higher order contributions, in case $M=1$, ${}^{{}^{n_1}}_{{}_{1}}\kappa_{l_1,m_1} = \kappa_{0,1},\kappa_{1,0}$ 
be equal to the kernel of a free field
with a mass $m_{i_1}$. If further among the distributions $S_n, S_{n-1}, \ldots, S_1$ there are no
(retarded or advanced parts of the) commutation functions of a free field of mass $m_2 = m_{i_1}$, then
\begin{multline*}
S_n \ast \ldots \ast S_1 \ast \Xi_{0,1}(\kappa_{0,1})(x) 
\\
= 
\int \limits_{[\mathbb{R}^4]^{\times n}} S_n(x-y_n) S_{n-1}(y_{n}-y_{n-1}) \ldots S_1(y_2-y_1) \Xi_{0,1}\big(\kappa_{0,1}(y)\big) \, \ud^4 y_1 \ldots \ud^4 y_n 
\\
= \Xi_{0,1}\Bigg(\int \limits_{[\mathbb{R}^4]^{\times n}} S_n(x-y_n) S_{n-1}(y_{n}-y_{n-1}) \ldots S_1(y_2-y_1)\kappa_{0,1}(y) \, \ud^4 y_1 \ldots \ud^4 y_n \,  \Bigg)
\\
= \Xi_{0,1}\big( S_n \ast \ldots \ast S_1 \ast \kappa_{0,1}(x) \big) 
\end{multline*}
defines integral kernel operator 
\[
\Xi_{0,1}\big( S_n \ast \ldots \ast S_1 \ast \kappa_{lm} \big) 
\in  \mathscr{L}\big((\boldsymbol{E}) \otimes \mathscr{E}, \, (\boldsymbol{E})^*\big) \cong
\mathscr{L}\big(\mathscr{E}, \, \mathscr{L}((\boldsymbol{E}), (\boldsymbol{E})^*)\big)
\]
with the vector-valued kernel
\[
S_n \ast \ldots \ast S_1 \ast \kappa_{0,1} \in \mathscr{L} \big(E_{i_1}, \,\, \mathscr{E}^* \big);
\]
and similarly for the kernel $\kappa_{1,0}$.
The contributions to interacting fields
whose kernels do not fulfil this condition are identically zero,  if we use the ``natural'' splitting of the causal distributions
in the construction of the scattering operator.

\item[3)] 
If  moreover, in case $M=1$, $\, {}^{{}^{n_1}}_{{}_{1}}\kappa_{l_1,m_1} = \kappa_{0,1},\kappa_{1,0}$ is the kernel of a free field
with the mass not equal to the mass of the free field whose commutation function (or its retarded or 
advanced part) is equal $S$ then the integral kernel operators
\[
S \ast \Xi_{0,1}(\kappa_{0,1}) = \Xi_{0,1}\big( S \ast \kappa_{0,1} \big), \, 
S \ast \Xi_{1,0}(\kappa_{1,0}) = \Xi_{0,1}\big( S \ast \kappa_{1,0} \big),
\]
are well-defined and
\begin{multline*}
S \ast \Xi_{0,1}(\kappa_{0,1}) = \Xi_{0,1}\big( S \ast \kappa_{0,1} \big), \, 
S \ast \Xi_{1,0}(\kappa_{1,0}) = \Xi_{0,1}\big( S \ast \kappa_{1,0} \big) 
\\
\in \mathscr{L}\big(\mathscr{E}, E_{i_1}\big) = \mathscr{L}\big(E_{i_1}^{*}, \mathscr{E}^* \, \big)
\subset \mathscr{L} \big(E_{i_1}, \,\, \mathscr{E}^* \big).
\end{multline*}

\item[4)]
If $\, {}^{{}^{n_1}}_{{}_{1}}\kappa_{l_1,m_1} = \kappa_{0,1},\kappa_{1,0}$ is the kernel of a free field
with the mass equal to the mass of the free field whose commutation function (or its retarded or 
advanced part) is equal $S$ then the integral kernel operators
\[
S \ast \Xi_{0,1}(\kappa_{0,1}) = \Xi_{0,1}\big( S \ast \kappa_{0,1} \big), \, 
S \ast \Xi_{1,0}(\kappa_{1,0}) = \Xi_{0,1}\big( S \ast \kappa_{1,0} \big), 
\]
are not well-defined.
\end{lem}

\subsection{Interactig fields in QED. Adiabatic limit}\label{OperationsOnXiIF}

\begin{twr}\label{g=1InteractingFieldsQED}
Let
\[
\boldsymbol{\psi}(x) = \Xi_{0,1}\big({}^{1}\kappa_{0,1}(x)\big)
+ \Xi_{1,0}\big({}^{1}\kappa_{1,0}(x)\big), \,\,\,
A = \Xi_{0,1}\big({}^{2}\kappa_{0,1}(x)\big) + \Xi_{1,0}\big({}^{2}\kappa_{1,0}(x)\big),
\]
be the integral kernel operators defining the free fields of the spinor QED with massive charged spinor field $\boldsymbol{\psi}$
and with ``natural'' splitting applied in the computation of the scattering operator.
Let $g \in \mathcal{S}(\mathbb{R}^4)$ and let
\[
\boldsymbol{\psi}_{{}_{\textrm{int}}}^{a}(g, x) =
\boldsymbol{\psi}^{a}(x) + \sum \limits_{n=1}^{\infty} {\textstyle\frac{1}{n!}}
\int \limits_{\mathbb{R}^{4n}} \ud^4x_1 \cdots \ud^4 x_n \boldsymbol{\psi}^{a \, (n)}(x_1, \ldots, x_n; x)
\, g(x_1) \ldots g(x_n),
\]
with
\[
\boldsymbol{\psi}^{a \, (1)}(x_1; x) =
-e S_{{}_{\textrm{ret}}}^{aa_1}(x-x_1) \gamma^{\nu_1 \, a_1a_2} \boldsymbol{\psi}^{a_2}(x_1)A_{\nu_1}(x_1),
\]
\begin{multline*}
\boldsymbol{\psi}^{a \, (2)}(x_1, x_2; x) = \\
e^2 \Bigg\{ S_{{}_{\textrm{ret}}}^{aa_1}(x-x_1) \gamma^{\nu_1 \, a_1a_2}S_{{}_{\textrm{ret}}}^{a_2a_3}(x_1-x_2)
\gamma^{\nu_2 \, a_3a_4} \, {:}\boldsymbol{\psi}^{a_4}(x_2) A_{\nu_1}(x_1)A_{\nu_2}(x_2){:} \\
- S_{{}_{\textrm{ret}}}^{aa_1}(x-x_1) \gamma^{\nu_1 \, a_1a_2} \,
{:}\boldsymbol{\psi}^{a_2}(x_1)
\boldsymbol{\psi}^{\sharp \, a_3}(x_2) \gamma_{\nu_1}^{a_3a_4} \boldsymbol{\psi}^{a_4}(x_2){:} \,
D^{{}^{\textrm{ret}}}_{0}(x_1-x_2) \\
+S_{{}_{\textrm{ret}}}^{aa_1}(x-x_1) \Sigma_{{}_{\textrm{ret}}}^{a_1a_2}(x_1-x_2)\boldsymbol{\psi}^{a_2}(x_2)
\Bigg\} \,\,\, + \,\,\, \Bigg\{ x_1 \longleftrightarrow x_2 \Bigg\},
\end{multline*}
\[
\textrm{e. t. c.}
\]
and let
\[
{A_{{}_{\textrm{int}}}}_{\mu}(g, x) =
A_{\mu}(x) + \sum \limits_{n=1}^{\infty} {\textstyle\frac{1}{n!}}
\int \limits_{\mathbb{R}^{4n}} \ud^4x_1 \cdots \ud^4 x_n A_{\mu}^{\, (n)}(x_1, \ldots, x_n; x)\, g(x_1) \ldots g(x_n),
\]
with
\[
A_{\mu}^{\, (1)}(x_1;x) = -e D^{{}^{\textrm{av}}}_{0}(x_1-x) \,
{:}\boldsymbol{\psi}^{\sharp \, a_1}(x_1) \gamma_{\mu}^{a_1a_2} \boldsymbol{\psi}^{a_2}(x_1){:},
\]
\begin{multline*}
A_{\mu}^{\, (2)}(x_1, x_2; x) =
e^2 \Bigg\{
{:}\boldsymbol{\psi}^{\sharp \, a_1}(x_1)
\Big(
\gamma_{\mu}^{a_1a_2} S_{{}_{\textrm{ret}}}^{a_2a_3}(x_1-x_2) \gamma^{\nu_1 \, a_3a_4}
D^{{}^{\textrm{av}}}_{0}(x_1-x) A_{\nu_1}(x_2) \\
+ \gamma^{\nu_1 \, a_1a_2}S_{{}_{\textrm{av}}}^{a_2a_3}(x_1-x_2) \gamma_{\mu}^{a_3a_4}
D^{{}^{\textrm{av}}}_{0}(x_2-x)A_{\nu_1}(x_1)
\Big) \boldsymbol{\psi}^{a_4}(x_2){:} \\
+ D^{{}^{\textrm{av}}}_{0}(x_1-x) {\Pi^{{}^{\textrm{av}}}}_{\mu}^{\nu_1}(x_2-x_1)A_{\nu_1}(x_2)
\Bigg\} \,\,\, + \,\,\, \Bigg\{ x_1 \longleftrightarrow x_2 \Bigg\}
\end{multline*}
\[
\textrm{e. t. c.}
\]
be equal to the formulas for (fixed components $a$ and $\mu$) of interacting Dirac and electromagnetic
fields $\boldsymbol{\psi}_{{}_{\textrm{int}}}$ and $A_{{}_{\textrm{int}}}$ in the causal
St\"uckelberg-Bogoliubov spinor QED, \cite{DutFred}, \cite{DKS1} or \cite{Scharf}. 
Here the summation convention is used:
summation is to be understood with respect to all repeated twice spinor or Lorentz
index, and $\boldsymbol{\psi}^{\sharp} = \boldsymbol{\psi}^+ \gamma_0$ is the Dirac conjugation
of the bispinor generalized free field operator\footnote{$\boldsymbol{\psi}^{\sharp}$ is denoted with
bar $\overline{\boldsymbol{\psi}}$ in \cite{DutFred}, \cite{DKS1} or \cite{Scharf}.}.

If the free fields $\boldsymbol{\psi}(x)$, $A(x)$ in these formulas for
$\boldsymbol{\psi}_{{}_{\textrm{int}}}$ and $A_{{}_{\textrm{int}}}$
are understood as integral kernel operators
\[
\boldsymbol{\psi}(x) = \Xi_{0,1}\big({}^{1}\kappa_{0,1}(x)\big)
+ \Xi_{1,0}\big({}^{1}\kappa_{1,0}(x)\big), \,\,\,
A = \Xi_{0,1}\big({}^{2}\kappa_{0,1}(x)\big) + \Xi_{1,0}\big({}^{2}\kappa_{1,0}(x)\big),
\]
and correspondingly the operations of Wick product $: \cdot :$ and integrations $\ud^4x_1, \ldots \ud^4x_n$
involved in the formulas for $\boldsymbol{\psi}_{{}_{\textrm{int}}}$ and $A_{{}_{\textrm{int}}}$
are understood as Wick products and integrations of integral kernel operators with
vector valued distributional kernels (which as we know have the properties expressed by the Rules
I-VI), then the limits ${A_{{}_{\textrm{int}}}}_{\mu}^{(n)}(g=1, x)$ and $\boldsymbol{\psi}_{{}_{\textrm{int}}}^{a \,(n)}(g=1)$  
(when $g\rightarrow 1$ in $\mathcal{S}(\mathbb{R}^4)^*$) of each $n$-th order term contribution
\[
\begin{split}
\boldsymbol{\psi}_{{}_{\textrm{int}}}^{a \,(n)}(g, x) =
\frac{1}{n!}
\int \limits_{\mathbb{R}^{4n}} \ud^4x_1 \cdots \ud^4 x_n \boldsymbol{\psi}^{a \, (n)}(x_1, \ldots, x_n; x) \, g(x_1) \ldots g(x_n), \\
{A_{{}_{\textrm{int}}}}_{\mu}^{\, (n)}(g, x) =
\frac{1}{n!}
\int \limits_{\mathbb{R}^{4n}} \ud^4x_1 \cdots \ud^4 x_n A_{\mu}^{\, (n)}(x_1, \ldots, x_n; x) \, g(x_1) \ldots g(x_n),
\end{split}
\]
respectively, to the interacting field $\boldsymbol{\psi}_{{}_{\textrm{int}}}^{a}(g=1, x)$
and ${A_{{}_{\textrm{int}}}}_{\mu}(g=1, x)$ is equal to a finite sum
\[
\sum \limits_{l,m} \Xi(\kappa_{l,m}(x)) \,\,\, \textrm{respectively} \,\,\,
\sum \limits_{l,m} \Xi(\kappa'_{l,m}(x))
\]
of integral kernel operators
\[
\Xi_{l,m}(\kappa_{lm}(x)), \,\,\, \textrm{respectively} \,\,\,
\Xi(\kappa'_{l,m}(x))
\]
which define integral kernel operators
\[
\begin{split}
\Xi_{l,m}(\kappa_{lm}) \in \mathscr{L}\big((\boldsymbol{E}) \otimes \mathscr{E}_1, \,(\boldsymbol{E})^* \big)
\cong \mathscr{L}\big(\mathscr{E}_1, \, \mathscr{L}((\boldsymbol{E}), (\boldsymbol{E})^*) \big), \\
\textrm{respectively} \\
\Xi_{l,m}(\kappa'_{lm}) \in \mathscr{L}\big((\boldsymbol{E}) \otimes \mathscr{E}_2, \,(\boldsymbol{E})^* \big)
\cong \mathscr{L}\big(\mathscr{E}_2, \, \mathscr{L}((\boldsymbol{E}), (\boldsymbol{E})^*) \big)
\end{split}
\]
with vector-valued distributional kernels
\[
\begin{split}
\kappa_{l,m} \in \mathscr{L} \big(E_{i_1} \otimes \cdots \otimes E_{i_{l+m}}, \,\, \mathscr{E}_{1}^* \big) \\
\kappa'_{l,m} \in \mathscr{L} \big(E_{i_1} \otimes \cdots \otimes E_{i_{l+m}}, \,\, \mathscr{E}_{2}^* \big).
\end{split}
\]
Thus each $n$-th order term contribution $\boldsymbol{\psi}_{{}_{\textrm{int}}}^{a, \,(n)}(g=1)$
and ${A_{{}_{\textrm{int}}}}_{\mu}^{\, (n)}(g=1)$, respectively, to interacting fields
$\boldsymbol{\psi}_{{}_{\textrm{int}}}^{a}(g=1)$
and ${A_{{}_{\textrm{int}}}}_{\mu}(g=1)$ is equal
\[
\begin{split}
\boldsymbol{\psi}_{{}_{\textrm{int}}}^{a, \,(n)}(g=1) \,\, = \,\,
\sum \limits_{l,m} \Xi(\kappa_{l,m}), \\
{A_{{}_{\textrm{int}}}}_{\mu}^{\, (n)}(g=1) \,\, = \,\,
\sum \limits_{l,m} \Xi(\kappa'_{l,m}),
\end{split}
\]
to a finite sum of well-defined integral kernel operators $\Xi(\kappa_{l,m}), \Xi(\kappa'_{l,m})$
with vector-valued distributional kernels $\kappa_{l,m}, \kappa'_{l,m}$ in the sense of Obata
\cite{obataJFA} (compare Subsection \ref{psiBerezin-Hida}).
\end{twr}

\qedsymbol \, 
The proof follows by induction and the repeated application of the Rules I-VI and the 
fundamental Lemma \ref{S*Xi} and Lemma \ref{corS*Xi}. 
\qed

In spinor QED we do not encounter, among the higher order contributions to interacting fields in the adiabatic limit $g=1$, the integral
kernel operators of the form counted as case 4) of Lemmas \ref{S*Xi} and \ref{corS*Xi}. There are however higher order
terms equal to repeated convolutions of the commutation and/or retarded/advanced parts of the causal combinations of the products of pairings
with just one free field operator (case $M=1$ of Lemmas \ref{S*Xi} and \ref{corS*Xi}) which however do not respect the condition
of case 2) of these Lemmas. All contributions of this form to $\boldsymbol{\psi}_{{}_{\textrm{int}}}$ and $A_{{}_{\textrm{int}}}$,
have even  $2k$-th order and have the general form
\begin{equation}\label{LoopTermsINpsi(2k)andA(2k)}
\ldots \underbrace{\big(S_{{}_{\textrm{ret}, \textrm{av}}} \ast \Sigma_{{}_{\textrm{ret}, \textrm{av}}} \ast\big)}_\textrm{$k$ terms} \ldots  \ast \boldsymbol{\psi}, \,\,\, \textrm{and respectively} \,\,\,
\ldots \underbrace{\big(D_{0}^{{}^{\textrm{av}, \textrm{ret}}} \ast \Pi^{{}^{\textrm{av}, \textrm{ret}} \, \nu_k}_{\mu_k} \ast \big)}_\textrm{$k$ terms} \ldots \ast A.
\end{equation}
Here 
\begin{gather*}
\Pi^{\textrm{ret} \, \mu \nu}(x) = C_{2}^{\textrm{ret} \, \mu \nu}(x), \,\,\, \Pi^{\textrm{av} \, \mu \nu}(x) = - C_{2}^{\textrm{ret} \mu \nu}(-x),
\\
C_2(x-y) = \big({}^{1}\kappa_{0,1}^{\sharp}  \dot{\otimes}  \gamma^\mu {}^{1}\kappa_{0,1}\big)
\otimes_2 
\big({}^{1}\kappa_{1,0}^{\sharp} \dot{\otimes} \gamma_{\nu}\,\, {}^{1}\kappa_{1,0}\big)(x,y),
\\
\Sigma^{\textrm{ret}}(x) = K_{2}^{\textrm{ret}}(x), \,\,\, \Sigma^{\textrm{av}}(x) = K_{2}^{\textrm{ret}}(-x),
\\
K_{2}^{ab}(x-y) = \big(\gamma^\mu \,\, {}^{1}\kappa_{0,1}^{\sharp}  \dot{\otimes} {}^{2}\kappa_{0,1 \,\, \mu}\big)
\otimes_2 
\big(\gamma^\nu \,\, {}^{1}\kappa_{1,0}^{\sharp} \dot{\otimes} {}^{2}\kappa_{1,0 \,\, \nu}\big)(a,x,b,y),
\\
\Upsilon^{\textrm{ret}}(x) = C_{3}^{\textrm{ret}}(x), \,\,\, \Upsilon^{\textrm{av}}(x) = - C_{3}^{\textrm{ret}}(-x),
\\
C_3(x-y) = \big({}^{1}\kappa_{0,1}^{\sharp}  \dot{\otimes}  \gamma^\mu {}^{1}\kappa_{0,1} \dot{\otimes} {}^{2}\kappa_{0,1 \,\, \mu}\big)
\otimes_3 
\big({}^{1}\kappa_{1,0}^{\sharp} \dot{\otimes} \gamma_{\nu} \,\, {}^{1}\kappa_{1,0} \dot{\otimes} {}^{2}\kappa_{1,0 \,\, \nu}\big).
\end{gather*}
Here ${}^{2}\kappa_{1,0 \,\, \mu}(x) = {}^{2}\kappa_{1,0}(\mu, x)$,
${}^{2}\kappa_{0,1 \,\, \mu}(x) = {}^{2}\kappa_{0,1}(\mu, x)$ denote the positive and negative frequency kernels of the free
e.m. potential field $A_\mu(x)$, ${}^{1}\kappa_{1,0}(a,x), {}^{1}\kappa_{1,0}(a,x)$ the positive and negative frequency kernels of 
the free spinor free field $\boldsymbol{\psi}^{a}(x)$, ${}^{1}\kappa^{\sharp}_{1,0}(a,x), {}^{1}\kappa^{\sharp}_{1,0}(a,x)$ 
the positive and negative frequency kernels of 
the Dirac-conjugated spinor free field $\boldsymbol{\psi}^{\sharp \, a}(x)$ and finally 
\[
\gamma^{\mu} \,\, {}^{1}\kappa_{1,0}(a,x) = \sum\limits_{b} \big[\gamma^{\mu} \big]^{ab} \,\, {}^{1}\kappa_{1,0}(b,x).
\]
Summations are performed with respect to the repeated Lorentz indices.

For example the third term in the second order contributions to interacting fields $\boldsymbol{\psi}_{{}_{\textrm{int}}}$ and $A_{{}_{\textrm{int}}}$ have, in each case, the form
\begin{equation}\label{LoopTermsINpsi(2)andA(2)}
S_{{}_{\textrm{ret}}} \ast \Sigma_{{}_{\textrm{ret}}} \ast \boldsymbol{\psi}, \,\,\, \textrm{and respectively} \,\,\,
D_{0}^{{}^{\textrm{av}}} \ast \Pi^{{}^{\textrm{av}} \, \nu}_{\mu} \ast A_\nu.
\end{equation}
But here in both these convolution operators we have the retarded/advanced part of the commutation function, $D_0$ or, respectively, $S$,
of the same field with which it is convoluted. However, for the natural choice of the normalization in the causal Epstein-Glaser splitting
of the causal combinations of products of pairing distributions\footnote{Recall that in general the splitting is not unique and contains freedom with the number of the
arbitrary constants depending on the singularity degree at zero of the causal distribution which is to be splitted.} all these
terms, in particular (\ref{LoopTermsINpsi(2)andA(2)}), are well-defined. In fact both terms (\ref{LoopTermsINpsi(2)andA(2)}),
and \emph{a fortiori} all terms (\ref{LoopTermsINpsi(2k)andA(2k)}),
become equal to zero for the standard normalization. This normalization in the choice of the splitting is not arbitrary
and in particular follows if we put the condition that the natural equations of motion for the interacting fields $\boldsymbol{\psi}_{{}_{\textrm{int}}}(g;x)$,
$A_{{}_{\textrm{int}}}(g;x)$ are fulfilled, when regarded perturbatively term by term in the expansion into integral kernel operators
regarded as a series of integral kernel operators in functional powers of the intensity of interaction function $g$,
compare \cite{DKS1}. Here we are giving another justification for the choice of the natural normalization in the splitting procedure:
for any other normalization in the splitting the terms of the type under case 2) of Lemmas \ref{S*Xi} and \ref{corS*Xi} in which the condition put on the mass
in 2) is violated, e.g. (\ref{LoopTermsINpsi(2)andA(2)}) and (\ref{LoopTermsINpsi(2k)andA(2k)}), are no longer well-defined
and the convolutions in such terms become meaningless as generalized integral kernel operators with $g=1$
for the other non-natural normalizations in the splitting. Strictly speaking the existence of (\ref{LoopTermsINpsi(2)andA(2)}),
forcing existence of (\ref{LoopTermsINpsi(2k)andA(2k)}) determines the splitting for the vacuum polarization distribution $\Pi, \Pi^{\textrm{av}, \textrm{av}}$
and self-energy distribution $\Sigma, \Sigma^{\textrm{av}, \textrm{av}}$. In order to fix the splitting in the computation of the  
vacuum graph distribution $\Upsilon$, and its retarded and advanced part, we need to impose the condition of the existence of the adiabatic limit
\[
\underset{g\rightarrow 1}{\textrm{lim}} \big\langle \Phi_0, S_n(g) \Phi_0 \big\rangle
\]
for the  vacuum $\Phi_0$ in the full Fock space of all free fields underlying QED (the free spinor and e.m. potential fields)
and for each $n$-th order contribution $S_n$ to the scattering operator,
compare \cite{Scharf}.

In fact the higher order sub-contributions are of two kinds. To the first kind belong all those sub-contributions, which make sense as
integral kernel operators simply for $g=1$. They correspond to the connected Feynman (amputed) graphs.
To the second kind of $n$-th order sub-contributions
\[
F(x_1, \ldots, x_n, x) \,\,\,
\textrm{in}
\,\,\,
A_{{}_{\textrm{int}}}^{(n)}(x_1, \ldots, x_n, x),
\,\,\,
\boldsymbol{\psi}_{{}_{\textrm{int}}}^{(n)}(x_1, \ldots, x_n, x),
\]
belong the sub-contributions for which we cannot
simply put $g=1$ in
\[
\int F(x_1, \ldots, x_n, x) g(x_1) \ldots g(x_n) \, d^4 x_1 \ldots d^4 x_n,
\]
with the space-time variables $X = \{x_1, \ldots, x_n, x \}$ divided into several disjoint subsets, $X_1, \ldots, X_k$
each separate subset of space-time variables entering as variables of the kernel $\kappa_i(X_i)$ of the corresponding separate factor $F_i(X_i)$ of
the kernels
\[
\kappa(X) = \kappa_1(X_1) \ldots \kappa_k(X_k)
\]
of the contribution
\[
F(X) = F_1(X_1) \ldots F_k(X_k).
\]
The kernels of $F(X)$ are equal to the sums of product of the kernels of the factors $F_1(X_1), \ldots, F_k(X_k)$.
The kernels $\kappa_i(X_i)$ of the factor $F_i(X_i)$, with $X_i = \{x_{i1}, \ldots, x_{ij_n}, x \}$,
containing the space-time variable $x$, when
$d^4x_{i1} \ldots d^4x_{ij_i} = dX_i$-integrated with respect to the corresponding set of space-time variables $\{x_{i1}, \ldots, x_{ij_n} \}$, except $x$,
becomes equal to a convolution
\[
\Bigg(\int \kappa_i(X_i) \, dX_i\Bigg)(x) = S_{i1} \ast \ldots \ast S_{ij_{i}}(x)
\]
with all $S_{i1, \ldots}$ being equal to some tempered distributions (products of advanced or retarded parts of the basic distributions
in differences of the corresponding space-time variables), possibly with the last one $S_{ij_i}$
being a product of the kernels of the free fields in some space-time variables $x_{j_i}$.
Such contributions,
\[
F_1(X_1) \ldots F_{k}(X_{k})
\]
when  $d^4 x_1 \ldots d^4x_n$-integrated with $g=1$,
have kernels equal to the sum of the integrals
\[
\int \kappa_1(X_1) \ldots \kappa_{k}(X_{k}) dX_1 \ldots dX_k
\]
which are not convergent. They correspond to disconnected Feynman (amputed)
graphs. An example of the $3$-rd order sub-contribution
to $A_{{}_{\textrm{int}} \, \mu}$ of the second kind (corresponding to a disconnected graph) would be obtained for $g=1$ by integrating
\begin{multline}\label{TypicalDivergentContribution}
\Upsilon^{\textrm{av}}(x_1-x_3) \,\,  {:}\boldsymbol{\psi}^\sharp \gamma_\mu \boldsymbol{\psi} {:} (x_2) \,\, D_{0}^{\textrm{av}}(x_2-x)
\\
-
\,\,
\Upsilon^{\textrm{av}}(x_1-x_3) \,\, {:}\boldsymbol{\psi}^\sharp \gamma_\mu \boldsymbol{\psi} {:} (x_2) \,\, D_{0}^{\textrm{av}}(x_2-x)
\\
+ \,\,\,\, \ldots
\end{multline}
(where dots represent the sum of similar paired differences for all permutations of the variables $x_2$, $x_2$, $x_3$)
with respect to $d^4x_1 d^4x_2 d^4 x_3$, which, with each term in (\ref{TypicalDivergentContribution})
taken separately,  would of course be divergent (for $g=1$). But for each $g\in \mathcal{S}(\mathbb{R}^4)$
the contribution (\ref{TypicalDivergentContribution}) is identically zero, being equal to the $d^4x_1 d^4x_2 d^4 x_3$-integral
of (\ref{TypicalDivergentContribution}) multiplied by the (symmetric) function
\[
g^{\otimes \, 3}(x_1,x_2, x_3) = g(x_1)g(x_2)g(x_3)
\]
in the space-time variables $x_1,x_2,x_3$, with the pairs of terms in the differences in (\ref{TypicalDivergentContribution})
cancelling out each other.
Because all contributions, with the variables factoring in the similar way into disjoint sets of variables
entering into respective various factors, always enter in pairs cancelling each other, similarly as in (\ref{TypicalDivergentContribution}),
when integrated with  the (symmetric) $g^{\otimes \, n}$, then the contributions of second kind are zero
for each $g \in \mathcal{S}(\mathbb{R}^4)$ and have zero contribution in the limit $g \rightarrow 1$.
Thus all contributions corresponding to disconnected graphs drop out before passing to the limit $g \rightarrow 0$.
However, because of the presence of the sub-contributions of the second kind, we cannot simply state that the interacting fields make
sense (for the ``natural'' splitting) in spinor massive QED simply for $g$ put equal $1$, but instead it is the case for the
limit $g \rightarrow 1$ with $g$ tending to $1$ in the strong dual space $\mathcal{S}(\mathbb{R}^4)^*$.

Below, in this Subsection, we are giving
detailed analysis of all contributions to interacting fields in the adiabatic limit up to order 2,
including (\ref{LoopTermsINpsi(2)andA(2)}), proving that they make sense only
for the ``natural'' splitting, and in this case (\ref{LoopTermsINpsi(2)andA(2)}) are equal to zero, and thus also all
(\ref{LoopTermsINpsi(2k)andA(2k)}) are identically zero. In addition, we will analyze kernels of sub-contributions to interacting fields
in the adiabatic limit with arbitrary
many vacuum polarization loop insertions to interacting fields of odd $n=2k+1$-order (analysis for arbitrary many self-energy loop insertions
is analogous).

Note that each $n$-th order contribution $\boldsymbol{\psi}_{{}_{\textrm{int}}}^{a, \,(n)}(g=1)$
and ${A_{{}_{\textrm{int}}}}_{\mu}^{\, (n)}(g=1)$ to interacting fields
$\boldsymbol{\psi}_{{}_{\textrm{int}}}^{a}(g=1)$
and ${A_{{}_{\textrm{int}}}}_{\mu}(g=1)$ belongs to the same general class of (finite sums of) integral
kernel operators (with vector-valued kernels) as the Wick products (at fixed space-time point)
of massless fields. In fact some contributions to interacting fields are finite
sums of integral kernel operators which even belong to a much better
behaved class of integral kernel operators, which belong to
\[
\begin{split}
\mathscr{L}\big((\boldsymbol{E}) \otimes \mathscr{E}_1, \,(\boldsymbol{E}) \big)
\cong \mathscr{L}\big(\mathscr{E}_1, \, \mathscr{L}((\boldsymbol{E}), (\boldsymbol{E})) \big), \\
\textrm{respectively} \\
\mathscr{L}\big((\boldsymbol{E}) \otimes \mathscr{E}_2, \,(\boldsymbol{E}) \big)
\cong \mathscr{L}\big(\mathscr{E}_2, \, \mathscr{L}((\boldsymbol{E}), (\boldsymbol{E})) \big).
\end{split}
\]
In particular one can show that the first order contribution
${A_{{}_{\textrm{int}}}}_{\mu}^{\, (1)}(g=1)$ to the interacting electromagnetic
potential field ${A_{{}_{\textrm{int}}}}_{\mu}(g=1)$ belongs to
\[
\mathscr{L}\big((\boldsymbol{E}) \otimes \mathscr{E}_2, \,(\boldsymbol{E}) \big)
\cong \mathscr{L}\big(\mathscr{E}_2, \, \mathscr{L}((\boldsymbol{E}), (\boldsymbol{E})) \big).
\]
Let us emphasize here that the Wick product (at the same space-time point)
of massless free fields (or containing such among the factors) does not belong to
\[
\begin{split}
\mathscr{L}\big((\boldsymbol{E}) \otimes \mathscr{E}_1, \,(\boldsymbol{E}) \big)
\cong \mathscr{L}\big(\mathscr{E}_1, \, \mathscr{L}((\boldsymbol{E}), (\boldsymbol{E})) \big), \\
\textrm{respectively} \\
\mathscr{L}\big((\boldsymbol{E}) \otimes \mathscr{E}_2, \,(\boldsymbol{E}) \big)
\cong \mathscr{L}\big(\mathscr{E}_2, \, \mathscr{L}((\boldsymbol{E}), (\boldsymbol{E})) \big).
\end{split}
\]
But we know that such product, as an integral kernel operator with vector-valued kernel,
belongs to
\[
\begin{split}
\mathscr{L}\big((\boldsymbol{E}) \otimes \mathscr{E}_1, \,(\boldsymbol{E})^* \big)
\cong \mathscr{L}\big(\mathscr{E}_1, \, \mathscr{L}((\boldsymbol{E}), (\boldsymbol{E})^*) \big), \\
\textrm{respectively} \\
\mathscr{L}\big((\boldsymbol{E}) \otimes \mathscr{E}_2, \,(\boldsymbol{E})^* \big)
\cong \mathscr{L}\big(\mathscr{E}_2, \, \mathscr{L}((\boldsymbol{E}), (\boldsymbol{E})^*) \big).
\end{split}
\]
Similarly, we know that each order term contribution to interacting fields is a finite sum of
integral kernel operators, which belongs to
\[
\begin{split}
\mathscr{L}\big((\boldsymbol{E}) \otimes \mathscr{E}_1, \,(\boldsymbol{E})^* \big)
\cong \mathscr{L}\big(\mathscr{E}_1, \, \mathscr{L}((\boldsymbol{E}), (\boldsymbol{E})^*) \big), \\
\textrm{respectively} \\
\mathscr{L}\big((\boldsymbol{E}) \otimes \mathscr{E}_2, \,(\boldsymbol{E})^* \big)
\cong \mathscr{L}\big(\mathscr{E}_2, \, \mathscr{L}((\boldsymbol{E}), (\boldsymbol{E})^*) \big).
\end{split}
\]
But at least some of them, e.g. the first order contribution
$\boldsymbol{\psi}_{{}_{\textrm{int}}}^{a, \,(1)}(g=1)$ to the interacting Dirac field
$\boldsymbol{\psi}_{{}_{\textrm{int}}}^{a}(g=1)$, do not belong to
\[
\mathscr{L}\big((\boldsymbol{E}) \otimes \mathscr{E}_1, \,(\boldsymbol{E}) \big)
\cong \mathscr{L}\big(\mathscr{E}_1, \, \mathscr{L}((\boldsymbol{E}), (\boldsymbol{E})) \big).
\]
Nonetheless, the contributions to interacting fields are finite sums
of integral kernel operators which belong to the same general class as the integral kernel operators
which are equal to Wick products (at the same space-time point) of massless free fields.

One can even show that if the Wick products (at the same space-time point) of
free fields (including massless fields) were equal to finite sums of integral kernel
operators belonging to
\[
\begin{split}
\mathscr{L}\big((\boldsymbol{E}) \otimes \mathscr{E}_1, \,(\boldsymbol{E}) \big)
\cong \mathscr{L}\big(\mathscr{E}_1, \, \mathscr{L}((\boldsymbol{E}), (\boldsymbol{E})) \big), \\
\textrm{respectively} \\
\mathscr{L}\big((\boldsymbol{E}) \otimes \mathscr{E}_2, \,(\boldsymbol{E}) \big)
\cong \mathscr{L}\big(\mathscr{E}_2, \, \mathscr{L}((\boldsymbol{E}), (\boldsymbol{E})) \big),
\end{split}
\]
then the same would be true of the contributions to interacting fields.
But the assumption about the Wick product necessary to infer this conclusion
is however false (compare the corresponding Proposition of this Subsection).

Here we give the explicit form of the distributions $S$:
$S_{{}_{\textrm{ret}}}^{a_1a_2}(x_1-x_2)$, $S_{{}_{\textrm{av}}}^{a_1a_2}(x_1-x_2)$,
$g^{\mu \nu}D^{{}^{\textrm{av}}}_{0}(x_1-x_2)$, $g^{\mu \nu}D^{{}^{\textrm{ret}}}_{0}(x_1-x_2)$, $\Sigma_{{}_{\textrm{av}}}^{ab}(x_1-x_2)$, 
$\Sigma_{{}_{\textrm{ret}}}^{ab}(x_1-x_2)$, ${\Pi^{{}^{\textrm{av}}}}_{\mu \nu}(x_1-x_2)$,
${\Pi^{{}^{\textrm{ret}}}}_{\mu \nu}(x_1-x_2)$, $\ldots$
which are present in spinor QED in the last Theorem. It is important only that they are well-defined
tempered distributions $S$, for which the convolutions of Lemma \ref{S*Xi} exist, and it is important that higher order contributions are
of the form of repeated convolution operation with these tempered distributions (in fact a finite number of such distributions
for each QFT case, here we have spinor QED). 
The said distributions, their retarded and advanced parts are obtained by the Epstein-Glaser splitting into the 
retarded and advanced parts of the causally supported scalar distributions which appear
in the higher order contributions to the scattering
generalized operator. Explicit computation and explicit form of them
the reader will find in the monograph \cite{Scharf}. In Subsection \ref{WickForChronological} we give a simple method for the 
computation of this set of distributions for general causal QFT. For the sake of completeness we write here the Fourier transforms of the complete finite 
set of these distributions for the spinor QED:
\begin{multline}\label{Pi}
\widetilde{{\Pi}_{\mu \nu}}(p) = 
(2\pi)^{-4} \big({\textstyle\frac{p_\mu p_\nu}{p^2}} - g_{\mu\nu}\big) \widetilde{\Pi}(p),
\\
\widetilde{\Pi}(p) =
{\textstyle\frac{1}{3}} p^4
\int\limits_{4m^2}^{\infty} {\textstyle\frac{s+2m^2}{s^2(p^2-s+i0)}}\sqrt{1-{\textstyle\frac{4m^2}{s}}} ds,
\end{multline}
\begin{multline}\label{Piav}
\widetilde{{\Pi^{{}^{\textrm{av}}}}_{\mu \nu}}(p) = 
(2\pi)^{-4} \big({\textstyle\frac{p_\mu p_\nu}{p^2}} - g_{\mu\nu}\big) \widetilde{\Pi{{}^{\textrm{av}}}}(p),
\\
\widetilde{\Pi{{}^{\textrm{av}}}}(p) =
{\textstyle\frac{1}{3}} p^4
\int\limits_{4m^2}^{\infty} {\textstyle\frac{s+2m^2}{s^2(p^2-s- \, i \,  p_0 \, 0)}}\sqrt{1-{\textstyle\frac{4m^2}{s}}} ds,
\end{multline}
\begin{multline}\label{Sigma}
\widetilde{\Sigma}(p) = 
(2\pi)^{-4}
\Big\{
\big(1-{\textstyle\frac{m^2}{p^2}}\big)
\big[\textrm{ln} \big|1- {\textstyle\frac{p^2}{m^2}} \big| -i\pi \, \theta(p^2 - m^2)\big]
\, \times 
\\
\times \, 
\big[ 
m-{\textstyle\frac{\slashed{p}}{4}}\big(1+{\textstyle\frac{m^2}{p^2}}\big)\big]
+ {\textstyle\frac{m^2}{p^2}}{\textstyle\frac{\slashed{p}}{4}}
-{\textstyle\frac{m}{4}}
+{\textstyle\frac{1}{8}}(\slashed{p}-m)
\Big\},
\end{multline}
\begin{multline}\label{Sigmaret}
\widetilde{\Sigma_{{}_{\textrm{ret}}}}(p) = 
(2\pi)^{-4}
\Big\{
\big(1-{\textstyle\frac{m^2}{p^2}}\big) \big[\textrm{ln} \big|1- {\textstyle\frac{p^2}{m^2}} \big| -i\pi \, \textrm{sgn} \, p_0 \,\, \theta(p^2 - m^2)\big]
\, \times 
\\
\times \, 
\big[ 
m - {\textstyle\frac{\slashed{p}}{4}}\big(1+{\textstyle\frac{m^2}{p^2}}\big)
\big]
+ {\textstyle\frac{m^2}{p^2}}{\textstyle\frac{\slashed{p}}{4}}
-{\textstyle\frac{m}{4}}
+{\textstyle\frac{1}{8}}(\slashed{p}-m)
\Big\},
\end{multline}
\begin{multline}\label{Upsilon}
\widetilde{\Upsilon}(p)= \widetilde{\Upsilon''}(p) -\widetilde{\Upsilon'}(p),
\\
\widetilde{\Upsilon''}(p) =
i (2\pi)^{-6} m^4 
\Bigg\{
{\textstyle\frac{5p^4}{48m^4}} + {\textstyle\frac{2p^2}{3m^2}} + 1 
+\big(3- {\textstyle\frac{4m^2}{p^2}}\big) \, \textrm{ln}^2\Big(\sqrt{{\textstyle\frac{-p^2}{4m^2}}} + \sqrt{1-{\textstyle\frac{p^2}{4m^2}}}\Big)
\\
+
\big({\textstyle\frac{p^4}{24m^4}} + {\textstyle\frac{p^2}{12m^2}} + 1\big) \sqrt{1-{\textstyle\frac{4m^2}{p^2}}}
\textrm{ln}{\textstyle\frac{\sqrt{1-{\textstyle\frac{4m^2}{p^2}}}-1}{\sqrt{1-{\textstyle\frac{4m^2}{p^2}}}+1}}
\Bigg\}, 
\end{multline}
\begin{multline*}
\widetilde{\Upsilon'}(p) =
(2\pi)^{-5} \, \theta(p^2 - 4m^2) \, \theta(-p_0) \, \times
\\
\times \,
\Big\{
\big({\textstyle\frac{p^4}{24}} + {\textstyle\frac{m^2}{12}} p^2 + m^4\big) \sqrt{1-{\textstyle\frac{4m^2}{p^2}}}
\\
+ {\textstyle\frac{m^4}{p^2}}(4m^2-3p^2)
\textrm{ln}\Big(\sqrt{{\textstyle\frac{p^2}{4m^2}}} + \sqrt{{\textstyle\frac{p^2}{4m^2}}-1}\Big)
\Big\}. 
\end{multline*}
Here $p^4 = (p \cdot p)^2$, and $p^2 = p\cdot p$ is the Lorentz invariant square with the Feynman slash
notation $\slashed{p} = p_\mu \gamma^\mu$. The formula for $\widetilde{\Upsilon''}$
is valid for time-like $p$ with $p^2>0$, and the general formula follows by analytic continuation. The remaining 
distributions are the standard pairings and commutation functions of the free fields as well as their retarded and advanced parts, 
so there is no necessity to write them explicitly here.

Let us give several explicit formulas for the convolution kernels
\[
S \ast {}^{i}\kappa_{0,1}(\xi), \,\,\,\, S \ast \big[ {}^{i}\kappa_{0,1} \dot{\otimes} {}^{j}\kappa_{0,1}(\xi_1 \otimes \xi_2) \big]= 
S \ast \big[ {}^{i}\kappa_{0,1}(\xi_1) \cdot {}^{j}\kappa_{0,1}(\xi_2) \big], \,\,\,\, \ldots
\]
where $S$ range over the basic distributions and ${}^{i}\kappa_{0,1}, {}^{i}\kappa_{1,0}$ range over the positive and negative energy
plane wave kernels of the free fields in spinor (or various) QED's. Let for example $S$ be equal to the retarded part
\[
S_{\textrm{ret}}(x) = \theta(x)S(x)
=
{\textstyle\frac{1}{(2\pi)^4}}
\int
{\textstyle\frac{m+\slashed{k}}{m^2 - k^2-i\epsilon k_{0}}} e^{-ik\cdot x} \, \ud^4 k
\]
\emph{i.e.} the retarded part of the
(anti)commutation function $S(x)$ for the Dirac conjugated and the Dirac field itself, and let ${}^{2}\kappa_{0,1}$, ${}^{2}\kappa_{1,0}$ be the
positive and negative energy plane wave distribution kernels of the free electromagnetic potential field. Then 
\begin{multline}\label{Sret*2kappa01}
S_{\textrm{ret}} \ast \big[{}^{2}\kappa_{0,1}(\xi)\big](y) = \int S_{\textrm{ret}}(y-x) \,\, \cdot \,\,\, {}^{2}\kappa_{0,1}(\xi)(\mu, x)  \, \ud^4 x 
\\
=
\int \ud^3 \boldsymbol{\p} {\textstyle\frac{\xi^\mu(\boldsymbol{\p})}{|\boldsymbol{\p}|}}
{\textstyle\frac{m+\slashed{p}}{m^2}} e^{-i|\boldsymbol{\p}|y_0 + i \boldsymbol{\p} \cdot \boldsymbol{\y}},
\end{multline}
\begin{multline}\label{Sret*2kappa10}
S_{\textrm{ret}} \ast \big[{}^{2}\kappa_{1,0}(\xi)\big](y) = \int S_{\textrm{ret}}(y-x) \,\, \cdot \,\,\, {}^{2}\kappa_{1,0}(\xi)(\mu, x)  \, \ud^4 x 
\\
=
\int \ud^3 \boldsymbol{\p} {\textstyle\frac{\xi^\mu(\boldsymbol{\p})}{|\boldsymbol{\p}|}}
{\textstyle\frac{m+\slashed{p}}{m^2}} e^{i|\boldsymbol{\p}|y_0 - i \boldsymbol{\p} \cdot \boldsymbol{\y}},
\end{multline}
(up to a constant factor equal to a power of $2\pi$). Similarly, for the negative and positive energy plane wave kernels ${}^{1}\kappa_{0,1}, {}^{1}\kappa_{1,0}$ 
defining the free Dirac field and for the negative and positive energy plane wave kernels ${}^{2}\kappa_{0,1}, {}^{2}\kappa_{1,0}$
defining the free electromagnetic potential field
and for $\xi_1=\chi, \xi_2=\zeta$, we have
\begin{multline}\label{Sret*[2kappa.1kappa]}
S_{\textrm{ret}} \ast \big[ {}^{2}\kappa_{0,1}(\zeta) \cdot {}^{1}\kappa_{0,1}(\chi) \big] (y)
= \int S_{\textrm{ret}}(y-x) \,\, \big[ {}^{2}\kappa_{0,1}(\zeta) \cdot {}^{1}\kappa_{0,1}(\chi) \big] (x)  \, \ud^4 x
\\
=
\sum\limits_{s}
\int \ud^3 \boldsymbol{\p} \ud^3\boldsymbol{\p} 
{\textstyle\frac{\zeta^\mu(\boldsymbol{\p}')\chi_s(\boldsymbol{\p}) (m+\slashed{p} +\slashed{p'})u_s(\boldsymbol{\p})}{|\boldsymbol{\p}'|(\langle \boldsymbol{\p}' | \boldsymbol{\p} \rangle -|\boldsymbol{\p}'|p_0(\boldsymbol{\p}) )}} 
e^{-i(|\boldsymbol{\p}'|+p_0(\boldsymbol{\p}))y_0 + i (\boldsymbol{\p}'+ \boldsymbol{\p}) \cdot \boldsymbol{\y}}
\end{multline}
\[
p = (p_0(\boldsymbol{\p}), \boldsymbol{\p}), \,\,\, p' = (|\boldsymbol{\p}'|, \boldsymbol{\p}')
\]
with the analogous expressions for the various combinations of the positive and negative energy kernels,
with the corresponding signs in front of the momenta in the numerator of the integrand and eventually with the
Fourier transforms of the positive energy solutions $u_s$ replaced with the Fourier transforms of the negative energy solutions
$v_s$ of the free Dirac equation. Similarly, for the retarded part 
\[
D_{0}^{\textrm{ret}}(x) = \theta(x)D_{0}(x)
=
-{\textstyle\frac{1}{(2\pi)^4}}
\int
{\textstyle\frac{1}{k^2+i\epsilon k_{0}}} e^{-ik\cdot x} \, \ud^4 k
\]
of the massless scalar commutation Pauli-Jordan function, we have the followong convolution kernels (up to a power of $2\pi$)
\begin{multline}\label{D0ret*1kappa01}
D_{0}^{\textrm{ret}} \ast \big[{}^{1}\kappa_{0,1}(\xi)\big](y) = \int D_{0}^{\textrm{ret}}(y-x) \,\, \cdot\,\,\, {}^{1}\kappa_{0,1}(\xi)(x)  \, \ud^4 x
\\
 =
-
\sum\limits_{s}\int \ud^3 \boldsymbol{\p} \xi_s(\boldsymbol{\p})
{\textstyle\frac{u_s(\boldsymbol{\p})}{m^2}} e^{-ip_0(\boldsymbol{\p})y_0 + i \boldsymbol{\p} \cdot \boldsymbol{\y}},
\end{multline}
\begin{multline}\label{D0ret*1kappa10}
D_{0}^{\textrm{ret}} \ast \big[{}^{1}\kappa_{1,0}(\xi)\big](y) = \int D_{0}^{\textrm{ret}}(y-x) \,\, \cdot\,\,\, {}^{1}\kappa_{1,0}(\xi)(x)  \, \ud^4 x
\\
 = 
-
\sum\limits_{s}\int \ud^3 \boldsymbol{\p} \xi_s(\boldsymbol{\p})
{\textstyle\frac{u_s(\boldsymbol{\p})}{m^2}} e^{ip_0(\boldsymbol{\p})y_0 - i \boldsymbol{\p} \cdot \boldsymbol{\y}}.
\end{multline}
Similarly
\begin{multline}\label{Sigmaret*1kappa01}
\Sigma_{\textrm{ret}} \ast \big[{}^{1}\kappa_{0,1}(\xi)\big](y) = \int\Sigma_{\textrm{ret}}(y-x) \,\, \cdot\,\,\, {}^{1}\kappa_{0,1}(\xi)(x)  \, \ud^4 x
\\
 =
\sum\limits_{s}\int \ud^3 \boldsymbol{\p} \xi_s(\boldsymbol{\p}) 
\widetilde{\Sigma_{\textrm{ret}}}(p_0(\boldsymbol{\p}),\boldsymbol{\p}) u_s(\boldsymbol{\p})
e^{-ip_0(\boldsymbol{\p})y_0 + i \boldsymbol{\p} \cdot \boldsymbol{\y}},
\end{multline}
\begin{multline}\label{Sigmaret*1kappa10}
\Sigma_{\textrm{ret}} \ast \big[{}^{1}\kappa_{1,0}(\xi)\big](y) = \int\Sigma_{\textrm{ret}}(y-x) \,\, \cdot\,\,\, {}^{1}\kappa_{1,0}(\xi)(x)  \, \ud^4 x
\\
 =
\sum\limits_{s}\int \ud^3 \boldsymbol{\p} \xi_s(\boldsymbol{\p}) 
\widetilde{\Sigma_{\textrm{ret}}}(-p_0(\boldsymbol{\p}),-\boldsymbol{\p}) v_s(\boldsymbol{\p}) 
e^{ip_0(\boldsymbol{\p})y_0 - i \boldsymbol{\p} \cdot \boldsymbol{\y}}.
\end{multline}
\begin{multline}\label{Piav*2kappa01}
\Pi^{\textrm{av}}_{\mu\nu} \ast \big[{}^{2}\kappa_{0,1}(\xi^\nu)\big](y) 
= \int \Pi^{\textrm{av}}_{\mu\nu}(y-x) \,\, \cdot \,\,\, {}^{2}\kappa_{0,1}(\xi^\nu)(\mu, x)  \, \ud^4 x 
\\
=
\int \ud^3 \boldsymbol{\p} {\textstyle\frac{\xi^\nu(\boldsymbol{\p})}{|\boldsymbol{\p}|}}
\widetilde{\Pi^{\textrm{av}}_{\mu\nu}}(|\boldsymbol{\p}|, \boldsymbol{\p}) e^{-i|\boldsymbol{\p}|y_0 + i \boldsymbol{\p} \cdot \boldsymbol{\y}},
\end{multline}
\begin{multline}\label{Piav*2kappa10}
\Pi^{\textrm{av}}_{\mu\nu} \ast \big[{}^{2}\kappa_{1,0}(\xi^\nu)\big](y) = \int \Pi^{\textrm{av}}_{\mu\nu}(y-x) 
\,\, \cdot \,\,\, {}^{2}\kappa_{1,0}(\xi^\nu)(\mu, x)  \, \ud^4 x 
\\
=
\int \ud^3 \boldsymbol{\p} {\textstyle\frac{\xi^\nu(\boldsymbol{\p})}{|\boldsymbol{\p}|}}
\widetilde{\Pi^{\textrm{av}}_{\mu\nu}}(-|\boldsymbol{\p}|, -\boldsymbol{\p}) e^{i|\boldsymbol{\p}|y_0 - i \boldsymbol{\p} \cdot \boldsymbol{\y}}.
\end{multline}
In all these formulas $p_0(\boldsymbol{\p}) = \sqrt{|\boldsymbol{\p}|^2 + m^2}$ with $m$ equal to the mass of the charged field (here the spinor Dirac
field).

Now we note that all the convolution kernels (\ref{Sret*2kappa01})-(\ref{Piav*2kappa10}) are smooth \emph{very slowly increasing} functions
of the space-time variable
$y$, whose derivatives are all bounded, in particular they all belong to the Horv{\'a}th's predual
$\mathcal{O}_C(\mathbb{R}^4)$ of the Schwartz convolutor algebra
$\mathcal{O}'_{C}(\mathbb{R}^4)$ of \emph{rapidly decreasing distributions}.
For definition of $\mathcal{O}_C$ and $\mathcal{O}'_C$,
compare Appendix \ref{convolutorsO'_C}. Indeed this immediately follows from the explicit expressions
(\ref{Sret*2kappa01})-(\ref{D0ret*1kappa10}) and the same proof which has been applied in
Lemma \ref{kappa0,1,kappa1,0psi}, Subsection \ref{psiBerezin-Hida} or respectively Lemma \ref{kappa0,1,kappa1,0ForA},
Subsection \ref{A=Xi0,1+Xi1,0}. Indeed, this follows for (\ref{Sret*2kappa01}) and (\ref{Sret*2kappa10}),
because $\xi \in \mathcal{S}^{0}(\mathbb{R}^3)$ there and the functions $u_s$ and $v_s$ are bounded.
The case (\ref{D0ret*1kappa01}) and (\ref{D0ret*1kappa10}) follows because $\xi \in \mathcal{S}(\mathbb{R}^3)$ there.
The case (\ref{Sigmaret*1kappa01}) and (\ref{Sigmaret*1kappa10}) follows because $\xi \in \mathcal{S}(\mathbb{R}^3)$ there
and $\widetilde{\Sigma_{\textrm{ret}}}(\pm p_0(\boldsymbol{\p}),\pm \boldsymbol{\p})$ is locally integrable function of
$\boldsymbol{\p}$ which grows at most polynomially at infinity and $u_s$
and $v_s$ are bounded functions of $\boldsymbol{\p}$.
The case (\ref{Piav*2kappa01}) and (\ref{Piav*2kappa10}) follows because $\xi \in \mathcal{S}^{0}(\mathbb{R}^3)$ there
and $\widetilde{\Pi^{\textrm{av}}_{\mu\nu}}(\pm|\boldsymbol{\p}|, \pm\boldsymbol{\p})$ is locally integrable function of
$\boldsymbol{\p}$ which grows at most polynomially at infinity.
Finally the case (\ref{Sret*[2kappa.1kappa]}) follows because $\chi \in \mathcal{S}(\mathbb{R}^3)$ and
$\zeta \in \mathcal{S}^{0}(\mathbb{R}^3)$ there, the matrix elements of $(m\pm\slashed{p} \pm\slashed{p'})u_s(\boldsymbol{\p})$,
$(m\pm\slashed{p} \pm\slashed{p'})v_s(\boldsymbol{\p})$ grows not faster than polynomially with $\boldsymbol{\p}$, $\boldsymbol{\p}'$
going to infinity and by using the following elementary estimation
\begin{equation}\label{EstimationForm>0}
\Big|{\textstyle\frac{1}{|\boldsymbol{\p}'|(\langle \boldsymbol{\p}' | \boldsymbol{\p} \rangle -|\boldsymbol{\p}'|p_0(\boldsymbol{\p}) )}}  \Big| <
{\textstyle\frac{4(m+|\boldsymbol{\p}|)}{m^2|\boldsymbol{\p}'|^2}},
\end{equation}
of course applicable only if the charged field is massive with the mass $m \neq 0$.
Therefore, the integrand in (\ref{Sret*[2kappa.1kappa]}) is absolutely integrable, and remains to be absolutely integrable for the
expression analogous to (\ref{Sret*[2kappa.1kappa]}) representing any derivative in $y$ of (\ref{Sret*[2kappa.1kappa]}), similarly as
for (\ref{Sret*2kappa01})-(\ref{D0ret*1kappa10}) and (\ref{Sigmaret*1kappa01})-(\ref{Piav*2kappa10});
compare Lemma \ref{kappa0,1,kappa1,0psi}, Subsection \ref{psiBerezin-Hida} or respectively Lemma \ref{kappa0,1,kappa1,0ForA},
Subsection \ref{A=Xi0,1+Xi1,0}.

For example, using this convolution kernels (\ref{Sret*2kappa01})-(\ref{Piav*2kappa10}) we prove that the higher order contributions
(here we restrict ourselves to the first and second order) contributions
to interacting fields are well-defined finite sums of generalized integral kernel operators with vector-vaued kernels in the sense of Obata
\cite{obataJFA}, (particular case of Lemmas \ref{S*Xi} and \ref{corS*Xi}). Indeed, it follows from the existence of the convolutions
with their continuity following immediately from the fact that
\[
\big[ {}^{i}\kappa_{0,1} \dot{\otimes} \,\,  {}^{j}\kappa_{0,1}(\xi_i \otimes \xi_j) \big]=
{}^{i}\kappa_{0,1}(\xi_i) \cdot {}^{j}\kappa_{0,1}(\xi_j), \ldots
\in \mathcal{O}_C(\mathbb{R}^4) \subset \mathcal{O}_M(\mathbb{R}^4)
\]
(dots stand for this expression with all possible combinations of the positive
${}^{i}\kappa_{1,0}$ and negative ${}^{i}\kappa_{0,1}$ energy plane wave kernels and arbitrary number of factors), which depend continuously
on $\xi_i \otimes \xi_j$, $\xi_i \otimes \xi_j \otimes \ldots$, $\ldots$,
which in turn immediately follows from Lemma \ref{kappa0,1,kappa1,0psi}, Subsection \ref{psiBerezin-Hida} and Lemma \ref{kappa0,1,kappa1,0ForA},
Subsection \ref{A=Xi0,1+Xi1,0}, and from the following product formula
(for various combinations of the positive ${}^{i}\kappa_{1,0}$ and negative
${}^{i}\kappa_{0,1}$ energy plane wave kernels)
\[
\big[ {}^{i}\kappa_{0,1} \dot{\otimes} \,\, {}^{j}\kappa_{0,1} \dot{\otimes} \ldots  \dot{\otimes} \,\, {}^{n}\kappa_{0,1}\big] (\xi_i \otimes \xi_j \otimes
\ldots \otimes \xi_n) =
{}^{i}\kappa_{0,1}(\xi_i) \cdot {}^{j}\kappa_{0,1}(\xi_2) \cdot \ldots \cdot {}^{n}\kappa_{0,1}(\xi_n),
\]
and the explicit formula for
\[
{}^{i}\kappa_{0,1}(\xi_i), {}^{i}\kappa_{1,0}(\xi_i)
\]
given in  Lemma \ref{kappa0,1,kappa1,0psi}, Subsection \ref{psiBerezin-Hida} and Lemma \ref{kappa0,1,kappa1,0ForA},
Subsection \ref{A=Xi0,1+Xi1,0}. Recall that $\xi_i$ in ${}^{i}\kappa_{0,1}(\xi_i)$ belongs to $\mathcal{S}(\mathbb{R}^3)$ or
$\mathcal{S}^{0}(\mathbb{R}^3)$, respectively, for the massive (Dirac) field ($i=1$) or massless photon field ($i=2$).
The second, and even more important fact, is that the corresponding convolutions, e.g. (\ref{Sret*[2kappa.1kappa]}) exist,
which, using for example existence of  (\ref{Sret*[2kappa.1kappa]}), we  prove that the first
order contribution to the interacting Dirac field in the adiabatic limit $g=1$ is a finite sum
of well-defined integral kernel operators with vector-valued kernels in the sense
of \cite{obataJFA} explained in this and the previous Subsection.
Also, for $\xi_1=\zeta, \xi_2=\chi$ in $E_1 = \mathcal{S}(\mathbb{R}^3)$,
\begin{multline}\label{D0av*[1kappa.1kappa]}
D_{0}^{\textrm{av}} \ast \big[ {}^{1}\kappa_{0,1}^{\sharp}(\zeta) \cdot {}^{1}\kappa_{0,1}(\chi) \big] (y)
= \int D_{0}^{\textrm{av}}(y-x) \,\, \big[ {}^{1}\kappa_{0,1}^{\sharp}(\zeta) \cdot {}^{1}\kappa_{0,1}(\chi) \big] (\mu, x)  \, \ud^4 x
\\
=
\sum\limits_{s,s'}
\int \ud^3 \boldsymbol{\p} \ud^3\boldsymbol{\p}'
{\textstyle\frac{\zeta_{s}(\boldsymbol{\p})v_{s}(\boldsymbol{\p})^{+}\chi_s(\boldsymbol{\p}') u_{s'}(\boldsymbol{\p}')}{((p_0(\boldsymbol{\p})+p_0(\boldsymbol{\p}'))^2
-|\boldsymbol{\p}+\boldsymbol{\p}'|^2 -i\epsilon (p_0(\boldsymbol{\p})+p_0(\boldsymbol{\p}'))}}
e^{-i(p_0(\boldsymbol{\p})+p_0(\boldsymbol{\p}'))y_0 + i (\boldsymbol{\p}+ \boldsymbol{\p}') \cdot \boldsymbol{\y}}
\end{multline}
with the analogous expressions for the various combinations of the positive and negative energy kernels,
with the corresponding signs in front of the momenta  $\pm (p_0(\boldsymbol{\p}),\boldsymbol{\p})$, and  
$\pm (p_0(\boldsymbol{\p}'), \boldsymbol{\p}')$ of the integrand and eventually with the
Fourier transforms of the positive energy solutions $u_s$ replaced with the Fourier transforms of the negative energy solutions
$v_s$ of the free Dirac equation. Therefore, the convolution (\ref{D0av*[1kappa.1kappa]}), in the limit $\epsilon \rightarrow 0^+$,
even with various combinations of the signs in front of the momenta, is a well-defined distribution,
because the function of $(\boldsymbol{\p}, \boldsymbol{\p}')$ in the numerator
belongs to $E_{1}^{\otimes \, 2} = \mathcal{S}(\mathbb{R}^3)^{\otimes \, 2}$. Therefore the first order contribution
to the interacting potential field in the adiabatic limit $g=1$ is equal to a finite sum of integral kernel operators.

As we have already proved (Lemma \ref{S*Xi}), continuity of the convolution kernels follows automatically from the  
Lemma \ref{kappa0,1,kappa1,0psi}, Subsection \ref{psiBerezin-Hida} and Lemma \ref{kappa0,1,kappa1,0ForA},
Subsection \ref{A=Xi0,1+Xi1,0}. But, as an example, let us prove it independently for the kernels
(\ref{Sret*[2kappa.1kappa]}) defining the first order contribution to the interacting Dirac field.
In fact, we should insert additional gamma matrix $\gamma^{\mu}$ into (\ref{Sret*[2kappa.1kappa]}) in order to obtain
actually the kernels of the operators which give this first order contribution, which of course does not matter for the continuity
question.   
To this end let $W(B, \varepsilon)$ be a strong zero-neighborhood in $\mathscr{E}^* = \mathcal{S}(\mathbb{R}^4)^*$,
determined by a bounded set $B$ in $\mathscr{E} = \mathcal{S}(\mathbb{R}^4)$ and $\varepsilon >0$.
We will construct zero-neighborhoods $V^0$ and $V$, respectively, in $E_2 = \mathcal{S}^{0}(\mathbb{R}^3)$
and in $E_1 = \mathcal{S}(\mathbb{R}^3)$ such that
\begin{equation}\label{kappapsi(1)(V0,V)inW(B,e)}
S_{\textrm{ret}} \ast \big[ \gamma^{\nu'} \, {}^{1}\kappa_{0,1}(\chi) \cdot {}^{2}\kappa_{0,1}(\zeta_{\nu'}) \big] \subset W(B,\varepsilon)
\end{equation}
for all $\zeta \in V^0$ and $\chi \in V$. We have the following estimation for $\phi$ ranging over $B$:
\begin{multline*}
\Big| \big\langle S_{\textrm{ret}} \ast \big[ \gamma^{\nu'} \, {}^{1}\kappa_{0,1}(\chi) \cdot {}^{2}\kappa_{0,1}(\zeta_{\nu'}) \big], \phi \big\rangle \Big|
\\
\leq
\sum\limits_{s, \nu'}
\int \ud^3 \boldsymbol{\p} \ud^3\boldsymbol{\p}
{\textstyle\frac{|(m+\slashed{p} +\slashed{p'})\gamma^{\nu'}u_s(\boldsymbol{\p})\widetilde{\phi}(-|\boldsymbol{\p}'|-p_0(\boldsymbol{\p}), -\boldsymbol{\p}'- \boldsymbol{\p})|}{|\boldsymbol{\p}'|(\langle \boldsymbol{\p}' | \boldsymbol{\p} \rangle -|\boldsymbol{\p}'|p_0(\boldsymbol{\p}) )}}
|\zeta_{\nu'}(\boldsymbol{\p}')| \, |\chi_s(\boldsymbol{\p})|
\end{multline*}

\begin{multline*}
\leq
{\textstyle\frac{4}{m^2}}
\sum\limits_{s, \nu'}
\int \ud^3 \boldsymbol{\p} \ud^3\boldsymbol{\p}
|(m+\slashed{p} +\slashed{p'})\gamma^{\nu'}u_s(\boldsymbol{\p})\widetilde{\phi}(-|\boldsymbol{\p}'|-p_0(\boldsymbol{\p}), -\boldsymbol{\p}'- \boldsymbol{\p})|
\,\, \times
\\
\times \,\,
\, (m+|\boldsymbol{\p}|) |\chi_s(\boldsymbol{\p})|
{\textstyle\frac{|\zeta_{\nu'}(\boldsymbol{\p}')|}{|\boldsymbol{\p}'|}}
\,\,
\leq
\,\,
\sum\limits_{s, \nu'} c^{s\nu'}
\int \ud^3 \boldsymbol{\p} (m+|\boldsymbol{\p}|) \, |\chi_s(\boldsymbol{\p})| \,
\int \ud^3\boldsymbol{\p}' {\textstyle\frac{|\zeta_{\nu'}(\boldsymbol{\p}')|}{|\boldsymbol{\p}'|}}
\end{multline*}
\begin{multline*}
=
\sum\limits_{s, \nu'} c^{s\nu'}
\int \ud^3 \boldsymbol{\p}
(1+|\boldsymbol{\p}|^2)^3
\, |\chi_s(\boldsymbol{\p})| \,
{\textstyle\frac{(m+|\boldsymbol{\p}|)}{(1+|\boldsymbol{\p}|^2)^3}}
\int \ud^3\boldsymbol{\p}' {\textstyle\frac{|\zeta_{\nu'}(\boldsymbol{\p}')|}{|\boldsymbol{\p}'|}} (1+|\boldsymbol{\p}'|^2)^2
{\textstyle\frac{1}{(1+|\boldsymbol{\p}'|^2)^2}}
\\
\leq
\sum\limits_{s, \nu'} c^{s\nu'} \Big\| {\textstyle\frac{1}{(1+|\boldsymbol{\p}'|^2)^2}} \Big\|_{1}
\Big\| {\textstyle\frac{(m+|\boldsymbol{\p}|)}{(1+|\boldsymbol{\p}|^2)^3}} \Big\|_{1}
p_0(\zeta_{\nu'}) \, p_1(\chi_s) = \sum\limits_{s, \nu'} c^{s\nu'} c_1 c_2 \,\,  p_0(\zeta_{\nu'})p_1(\chi_s)
\end{multline*}
where in the second inequality we have used the elementary estimation (\ref{EstimationForm>0}).
Here the functions $u_s$ and $\widetilde{\phi}$ are contracted with respect to the spinor index:
$u_s\widetilde{\phi} = \Sigma_a u_{s}^{a}\widetilde{\phi}_a$ and
\[
c^{s\nu'} = \underset{\boldsymbol{\p}',\boldsymbol{\p}' \in \mathbb{R}^3, \phi \in B}{\textrm{sup}}
{\textstyle\frac{4}{m^2}}
|(m+\slashed{p} +\slashed{p'})\gamma^{\nu'}u_s(\boldsymbol{\p})\widetilde{\phi}(-|\boldsymbol{\p}'|-p_0(\boldsymbol{\p}), -\boldsymbol{\p}'- \boldsymbol{\p})|
\]
\[
p_0(\zeta_{\nu'}) = \underset{\boldsymbol{\p}' \in \mathbb{R}^3}{\textrm{sup}}
{\textstyle\frac{|\zeta_{\nu'}(\boldsymbol{\p}')|}{|\boldsymbol{\p}'|}} (1+|\boldsymbol{\p}'|^2)^2
\]
and
\[
p_1(\chi_s) = \underset{\boldsymbol{\p} \in \mathbb{R}^3}{\textrm{sup}}
(1+|\boldsymbol{\p}|^2)^3
\, |\chi_s(\boldsymbol{\p})|
\]
\[
c_1 = \Big\| {\textstyle\frac{1}{(1+|\boldsymbol{\p}'|^2)^2}} \Big\|_{1}, \,\,\,
c_2 = \Big\| {\textstyle\frac{(m+|\boldsymbol{\p}|)}{(1+|\boldsymbol{\p}|^2)^3}} \Big\|_{1},
\]
Therefore, we can put
\[
V^0 = \{\zeta: p_0(\zeta_{\nu'}) < {\textstyle\frac{\sqrt{\varepsilon}}{C}}  \},
\,\,\, V = \{\chi: p_0(\chi_{s}) < {\textstyle\frac{\sqrt{\varepsilon}}{C}}  \},
\,\,\, C = \sum\limits_{s, \nu'} c^{s\nu'} c_1c_2
\]
because, by the results of Subsections \ref{dim=1}-\ref{SA=S0}, $p_0$ is a continuous seminorm on $\mathcal{S}^{0}(\mathbb{R}^3)$,
and it is a well-known fact that
$p_1$  is a continuous seminorm on $\mathcal{S}(\mathbb{R}^3)$.
Because
\[
\Big| \big\langle S_{\textrm{ret}} \ast \big[ \gamma^{\nu'} \, {}^{1}\kappa_{0,1}(\chi) \cdot {}^{2}\kappa_{0,1}(\zeta_{\nu'}) \big], \phi \big\rangle \Big|
< \varepsilon
\]
for all $\zeta \in V^0$, $\chi \in V$ and $\phi \in B$, then (\ref{kappapsi(1)(V0,V)inW(B,e)}) follows and the kernel(s)
of the first order contribution to the interacting Dirac field are continuous and belong to
\[
\mathscr{L}(E_1\otimes E_2; \mathscr{E}^*) = \mathscr{L}\big(\mathcal{S}^{0}(\mathbb{R}^3) \otimes \mathcal{S}(\mathbb{R}^3); \mathcal{S}(\mathbb{R}^4)^* \big).
\]

Because we have mass $m>0$ of the charged field (here Dirac field) in the denominator in the above estimations,
coming of course from the elementary estimation (\ref{EstimationForm>0}), then one could perhaps be inclined to think
that the first order contribution to the interacting (Dirac) field is a finite sum of well-defined integral kernel
operators only if the charged field is massive. This conclusion would be false, however, and the mass $m$ in the denominator
is only an artifact of the estimation method, based on (\ref{EstimationForm>0}), which is applicable only if
$m\neq 0$. In fact the first order contributions to interacting fields make sense as sums of generalized integral kernel operators
for $g=1$ also for spinor QED with the massless Dirac field. 

However, for the second and higher order contributions to interacting fields for spinor QED with massless
Dirac field, situation is different. In this massless case the higher order contributions to interacting fields
are meaningless even as the sums of generalized integral kernel operators with vector-valued kernels
in the sense of \cite{obataJFA}.

On the curved space-times with compact Cauchy surfaces, where the interacting
fields behave more regularly, the situation is similar, where the massive charged fields are even more 
strongly preferable from the mathematical point of view, and where it can proved within causal QED with Hida 
operators as the creation-annihilation operators, see Chapter \ref{CausalSonEU}, Thm. \ref{InteractingFieldsAtxOnEU}
and its Corollary. 
 
Let us look at the second order contributions. In particular, having given
the convolution kernels (\ref{Sret*[2kappa.1kappa]}), with all possible combinations of positive and negative energy plane wave kernels
\[
\begin{split}
S_{\textrm{ret}} \ast \big[ {}^{2}\kappa_{0,1}(\zeta) \cdot {}^{1}\kappa_{0,1}(\chi) \big], \,\,
S_{\textrm{ret}} \ast \big[ {}^{2}\kappa_{0,1}(\zeta) \cdot {}^{1}\kappa_{1,0}(\chi) \big], \,\,
\\
S_{\textrm{ret}} \ast \big[ {}^{2}\kappa_{1,0}(\zeta) \cdot {}^{1}\kappa_{0,1}(\chi) \big], \,\,
S_{\textrm{ret}} \ast \big[ {}^{2}\kappa_{1,0}(\zeta) \cdot {}^{1}\kappa_{1,0}(\chi) \big],
\end{split}
\]
we can easily see then, that all convolutions 
\[
\begin{split}
S_{\textrm{ret}} \ast \big[ {}^{2}\kappa_{0,1}(\zeta) \cdot {}^{1}\kappa_{0,1}(\chi) \big] \cdot {}^{2}\kappa_{0,1}(\xi), \,\,
S_{\textrm{ret}} \ast \big[ {}^{2}\kappa_{0,1}(\zeta) \cdot {}^{1}\kappa_{1,0}(\chi) \big] \cdot {}^{2}\kappa_{0,1}(\xi), \,\,
\\
S_{\textrm{ret}} \ast \big[ {}^{2}\kappa_{1,0}(\zeta) \cdot {}^{1}\kappa_{0,1}(\chi) \big]\cdot {}^{2}\kappa_{0,1}(\xi), \,\,
S_{\textrm{ret}} \ast \big[ {}^{2}\kappa_{1,0}(\zeta) \cdot {}^{1}\kappa_{1,0}(\chi) \big]\cdot {}^{2}\kappa_{0,1}(\xi), \,\,
\\
S_{\textrm{ret}} \ast \big[ {}^{2}\kappa_{0,1}(\zeta) \cdot {}^{1}\kappa_{0,1}(\chi) \big] \cdot {}^{2}\kappa_{1,0}(\xi), \,\,
S_{\textrm{ret}} \ast \big[ {}^{2}\kappa_{0,1}(\zeta) \cdot {}^{1}\kappa_{1,0}(\chi) \big] \cdot {}^{2}\kappa_{1,0}(\xi), \,\,
\\
S_{\textrm{ret}} \ast \big[ {}^{2}\kappa_{1,0}(\zeta) \cdot {}^{1}\kappa_{0,1}(\chi) \big]\cdot {}^{2}\kappa_{1,0}(\xi), \,\,
S_{\textrm{ret}} \ast \big[ {}^{2}\kappa_{1,0}(\zeta) \cdot {}^{1}\kappa_{1,0}(\chi) \big]\cdot {}^{2}\kappa_{1,0}(\xi), \,\,
\end{split}
\]
are well-defined. In particular
\begin{multline*}
\Big(S_{\textrm{ret}} \ast \big[ {}^{2}\kappa_{0,1}(\zeta) \cdot {}^{1}\kappa_{0,1}(\chi) \big] \cdot {}^{2}\kappa_{0,1}(\xi)\Big)(y)
\\
=
\sum\limits_{s,s'}
\int \ud^3 \boldsymbol{\p} \ud^3\boldsymbol{\p}' \ud^3\boldsymbol{\p}''
{\textstyle\frac{[m+\slashed{p}+\slashed{p'}+\slashed{p''}]\zeta^\mu(\boldsymbol{\p}')\chi_s(\boldsymbol{\p}) (m+\slashed{p} +\slashed{p'})u_s(\boldsymbol{\p})\xi^\mu(\boldsymbol{\p}'')}{[m^2 - (p+p'+p'')\cdot (p+p'+p'') -i\epsilon(p_0+p'_{0}+p'''_{0})]|\boldsymbol{\p}'|(\langle \boldsymbol{\p}' | \boldsymbol{\p} \rangle -|\boldsymbol{\p}'|p_0(\boldsymbol{\p}) )|\boldsymbol{\p}''|}} \,\, \times
\\
\times \,\,
e^{-i(p+p'+p'')\cdot y}
\end{multline*}
where
\[
p=(p_0,\boldsymbol{\p}) = (p_0(\boldsymbol{\p}), \boldsymbol{\p}), \,\,\, p' = (p'_{0},\boldsymbol{\p}') = (p_0(\boldsymbol{\p}'), \boldsymbol{\p}'), 
\,\,\, p'' = (p''_{0},\boldsymbol{\p}'') = (|\boldsymbol{\p}''|, \boldsymbol{\p}''),
\]
is a well-defined distribution in $y$-variable, because 
the numerator of the integrand, regarded as a function of 
$\boldsymbol{\p} \times \boldsymbol{\p}' \times \boldsymbol{\p}''$,
belongs to 
\[
E_1 \otimes E_2 \otimes E_2 = \mathcal{S}(\mathbb{R}^3) \otimes \mathcal{S}^0(\mathbb{R}^3)  \otimes \mathcal{S}^{0}(\mathbb{R}^3).
\] 
Recall that here $\zeta \in E_2 = \mathcal{S}^{0}(\mathbb{R}^3)$, $\chi \in E_1 = \mathcal{S}(\mathbb{R}^3)$ and $\xi \in E_2 = \mathcal{S}^{0}(\mathbb{R}^3)$. 
Therefore, by the very existence of these convolutions and the continuity proof of Lemma \ref{S*Xi} and Lemma \ref{corS*Xi}
it follows that the first term 
\begin{multline}\label{firstInpsi(2)}
{\textstyle\frac{e^2}{2}} \int\limits_{[\mathbb{R}^4]^{\times 2}} \ud^4 x_1 \ud^4 x_2  \Bigg[ 
\\
 \Big\{S_{{}_{\textrm{ret}}}^{aa_1}(x-x_1) \gamma^{\nu_1 \, a_1a_2}S_{{}_{\textrm{ret}}}^{a_2a_3}(x_1-x_2)
\gamma^{\nu_2 \, a_3a_4} \, {:}\boldsymbol{\psi}^{a_4}(x_2) A_{\nu_1}(x_1)A_{\nu_2}(x_2){:}
\Big\} 
\\
+ \Big\{x_1 \longleftrightarrow x_2 \Big\} 
\Bigg] 
\end{multline}
in the second order contribution $\boldsymbol{\psi}^{(2)}(g=1,x)$ in Theorem \ref{g=1InteractingFieldsQED}
defines a finite sum of integral kernel operators with vector-valued kernels in the sense of \cite{obataJFA}
(explained in Subsetion \ref{psiBerezin-Hida}). 

Similarly, because 
the convolution kernels (\ref{D0ret*1kappa01}) and (\ref{D0ret*1kappa10})
\[
D_{0}^{\textrm{ret}} \ast \big[{}^{1}\kappa_{0,1}(\xi)\big], \,\,
D_{0}^{\textrm{ret}} \ast \big[{}^{1}\kappa_{1,0}(\xi)\big]
\]
both are smooth functions with each derivative being bounded, then, as is easily seen
\begin{equation}\label{kappaInsecondInpsi(2)}
\begin{split}
D_{0}^{\textrm{ret}} \ast \big[{}^{1}\kappa_{0,1}(\xi)\big] \cdot {}^{1}\kappa_{0,1}^{\sharp}(\chi) \cdot {}^{1}\kappa_{0,1}(\zeta), 
D_{0}^{\textrm{ret}} \ast \big[{}^{1}\kappa_{0,1}(\xi)\big] \cdot {}^{1}\kappa_{1,0}^{\sharp}(\chi) \cdot {}^{1}\kappa_{0,1}(\zeta), \,\,
\\
D_{0}^{\textrm{ret}} \ast \big[{}^{1}\kappa_{1,0}(\xi)\big] \cdot {}^{1}\kappa_{0,1}^{\sharp}(\chi) \cdot {}^{1}\kappa_{0,1}(\zeta), 
D_{0}^{\textrm{ret}} \ast \big[{}^{1}\kappa_{1,0}(\xi)\big] \cdot {}^{1}\kappa_{1,0}^{\sharp}(\chi) \cdot {}^{1}\kappa_{0,1}(\zeta), \,\,
\\
D_{0}^{\textrm{ret}} \ast \big[{}^{1}\kappa_{0,1}(\xi)\big] \cdot {}^{1}\kappa_{0,1}^{\sharp}(\chi) \cdot {}^{1}\kappa_{1,0}(\zeta), 
D_{0}^{\textrm{ret}} \ast \big[{}^{1}\kappa_{0,1}(\xi)\big] \cdot {}^{1}\kappa_{1,0}^{\sharp}(\chi) \cdot {}^{1}\kappa_{1,0}(\zeta), \,\,
\\
D_{0}^{\textrm{ret}} \ast \big[{}^{1}\kappa_{1,0}(\xi)\big] \cdot {}^{1}\kappa_{0,1}^{\sharp}(\chi) \cdot {}^{1}\kappa_{1,0}(\zeta), \,\,
D_{0}^{\textrm{ret}} \ast \big[{}^{1}\kappa_{1,0}(\xi)\big] \cdot {}^{1}\kappa_{1,0}^{\sharp}(\chi) \cdot {}^{1}\kappa_{1,0}(\zeta), \,\,
\end{split}
\end{equation}
all are smooth with each derivative bounded, so allthemore define well-defined distributions in space-time variable.
Therefore by the very existence of these convolutions and the continuity proof of Lemma \ref{S*Xi} and Lemma \ref{corS*Xi}
\begin{multline}\label{secondInpsi(2)}
-{\textstyle\frac{e^2}{2}} \int\limits_{[\mathbb{R}^4]^{\times 2}} \ud^4 x_1 \ud^4 x_2  \Bigg[ 
\\
\Big\{
S_{{}_{\textrm{ret}}}^{aa_1}(x-x_1) \gamma^{\nu_1 \, a_1a_2} \,
{:}\boldsymbol{\psi}^{a_2}(x_1)
\boldsymbol{\psi}^{\sharp \, a_3}(x_2) \gamma_{\nu_1}^{a_3a_4} \boldsymbol{\psi}^{a_4}(x_2){:} \,
D^{{}^{\textrm{ret}}}_{0}(x_1-x_2)
\Big\} 
\\
+ \Big\{x_1 \longleftrightarrow x_2 \Big\} 
\Bigg] 
\end{multline}
in the second order contribution $\boldsymbol{\psi}^{(2)}(g=1,x)$ in Theorem \ref{g=1InteractingFieldsQED}
defines a finite sum of integral kernel operators with vector-valued kernels in the sense of \cite{obataJFA}. 

In the standard normalization in 
the computation of the splitting of the causal distributions, \emph{i.e.} for the standard normalization
of the vacuum polarization distribution $\Pi$ and the self-energy distribution $\Sigma$
it is immediately seen that all the convolution kernels
(\ref{Sigmaret*1kappa01}), (\ref{Sigmaret*1kappa10}) and  
(\ref{Piav*2kappa01}), (\ref{Piav*2kappa10}) are identically zero. Therefore, 
the third term 
\[
{\textstyle\frac{e^2}{2}} \int\limits_{[\mathbb{R}^4]^{\times 2}} \ud^4 x_1 \ud^4 x_2  \Bigg[ 
\Big\{
S_{{}_{\textrm{ret}}}^{aa_1}(x-x_1) \Sigma_{{}_{\textrm{ret}}}^{a_1a_2}(x_1-x_2)\boldsymbol{\psi}^{a_2}(x_2)
\Big\} 
+ \Big\{x_1 \longleftrightarrow x_2 \Big\} 
\Bigg] 
\]
drops out and becomes identically zero, similarly as the third term in the second order contribution to the 
interacting electromagnetic potential field.

We should emphasize here that with any other normalization the convolution kernels (\ref{Sigmaret*1kappa01})-(\ref{Piav*2kappa10})
will be non-zero and in this case the third term (both in the second order contribution to the interacting Dirac field and the electromagnetic
potential field) would be not well-defined as a generalized integral kernel operator. 
In particular the kernels (evaluated at $\xi\in E_1$) of the integral kernel operators
associated to the third term in the second order contribution to the interacting Dirac field 
are equal to the following convolutions
\[
S_{\textrm{ret}} \ast \Sigma_{\textrm{ret}} \ast \big[{}^{1}\kappa_{0,1}(\xi)\big],
\,\,\, S_{\textrm{ret}} \ast \Sigma_{\textrm{ret}} \ast \big[{}^{1}\kappa_{1,0}(\xi)\big].
\]
In particular evaluation of the first of these kernels at the space-time test function $\phi \in \mathcal{S}(\mathbb{R}^4)$ would be given by the
limit $\epsilon \rightarrow 0^+$ of the valuation integral (here $p = (p_0(\boldsymbol{\p}), \boldsymbol{\p})$)
\begin{multline*}
\Big\langle S_{\textrm{ret}} \ast \Sigma_{\textrm{ret}} \ast \big[{}^{1}\kappa_{0,1}(\xi)\big], \phi \Big\rangle
\\
=
\underset{\epsilon\rightarrow0^+}{\textrm{lim}}
\sum\limits_{s}\int \ud^3 \boldsymbol{\p} \ud^4 y \xi_s(\boldsymbol{\p}) 
{\textstyle\frac{
(m+\slashed{p})\widetilde{\Sigma_{\textrm{ret}}}(p_0(\boldsymbol{\p}),\boldsymbol{\p}) u_s(\boldsymbol{\p})}{-i\epsilon p_0(\boldsymbol{\p})}}
e^{-ip_0(\boldsymbol{\p})y_0 + i \boldsymbol{\p} \cdot \boldsymbol{\y}} \phi(y) 
\\
=
\underset{\epsilon\rightarrow0^+}{\textrm{lim}}
\sum\limits_{s}\int \ud^3 \boldsymbol{\p} \xi_s(\boldsymbol{\p}) 
{\textstyle\frac{
(m+\slashed{p})\widetilde{\Sigma_{\textrm{ret}}}(p_0(\boldsymbol{\p}),\boldsymbol{\p}) u_s(\boldsymbol{\p})}{-i\epsilon p_0(\boldsymbol{\p})}}
 \widetilde{\phi}(-p_0(\boldsymbol{\p}), -\boldsymbol{\p})
= +\infty \,\, \textrm{or} \,\, -\infty, 
\end{multline*}
which is meaningless as a value of the vector-valued distribution on the test functions. Difficulty comes of course from the fact that both the
Fourier transforms $\widetilde{S_{\textrm{ret}}}$  and $\widetilde{{}^{1}\kappa_{0,1}(\xi)}$ are concentrated on the
same orbit $\mathscr{O}_{{}_{\pm m,0,0,0}}$ -- the mass shell -- in the momentum space. In the standard normalization difficulty disappears
because in this normalization 
\[
\widetilde{\Sigma_{\textrm{ret}}}(p_0(\boldsymbol{\p}), \boldsymbol{\p}) u_s(\boldsymbol{\p}) = 0 \,\,\, 
\textrm{for each} \,\, \boldsymbol{\p} \in \mathbb{R}^3,
\]
and similarly on the negative energy orbit with the Fourier transforms of the fundamental positive energy solutions $u_s$ replaced
with the negative energy solutions $v_s$,
so that the above valuation limit is well-defined in the standard normalization and is equal zero.
 
We have analogous situation with the third term 
\[
D_{0}^{{}^{\textrm{av}}} \ast \Pi^{{}^{\textrm{av}} \, \mu\nu} \ast A_\nu
\]
in the second order contribution to the interacting electromagnetic potential field,
where both $\widetilde{D_0}$ and $\widetilde{{}^{2}\kappa_{0,1}(\xi)}$ are concentrated on the light cone
in the momentum space. Only with the standard normalization in which
\[
\widetilde{\Pi^{\textrm{av}}_{\mu\nu}}(\pm|\boldsymbol{\p}|, \pm \boldsymbol{\p})
= 0 \,\,\, 
\textrm{for each} \,\, \boldsymbol{\p} \in \mathbb{R}^3,
\]
this term is well-defined and in fact becomes identically zero. For other normalizations in the splitting determining
the vacuum polarization distribution $\Pi$ the third term 
\[
D_{0}^{{}^{\textrm{av}}} \ast \Pi^{{}^{\textrm{av}} \, \mu\nu} \ast A_\nu
\]
in the second order contribution to the interacting electromagnetic potential
field would be meaningless as a generalized integral kernel operator in the adiabatic limit $g=1$. 

Thus we have alredy shown that the full second order contribution $\boldsymbol{\psi}^{(2)}(g=1,x)$
to $\boldsymbol{\psi}_{{}_{\textrm{int}}}(g=1,x)$
is equal to a finite sum of well-defined integral kernel operators with vector-valued kernels in the sense of \cite{obataJFA}
(explained in Subsetion \ref{psiBerezin-Hida}) in spinor QED with massive charged (here Dirac) field and ``natural''
splitting in the construction of the scattering operator.

Proof that the full second order contribution $A^{(2)}(g=1)$ to the interacting potential field  
$A_{{}_{\textrm{int}}}(g=1)$ is equal to a finite sum of well-defined integral kernel operators with vector-valued kernels in the sense of \cite{obataJFA}
is identical.

Note that it does not follow from the general properties of the Schwartz algebra of multipliers and convolutors (\cite{Schwartz} or Appendix \ref{convolutorsO'_C})
that the convolution kernels (\ref{Sret*2kappa01})-(\ref{D0ret*1kappa10}) are well-defined tempered distributions, because
\[
\kappa(\xi) = {}^{i}\kappa_{0,1}(\xi_i), {}^{i}\kappa_{1,0}(\xi_i) \notin \mathcal{O}'_C(\mathbb{R}^4)
\]
although these $\kappa(\xi)$ belong to $\mathcal{O}_C(\mathbb{R}^4)$. It follows immediately from the fact that
the Fourier transforms $\widetilde{\kappa(\xi)}$ are concentrated on the orbit of the field defined by the kernel
$\kappa$, compare Lemma \ref{kappa0,1,kappa1,0psi}, Subsection \ref{psiBerezin-Hida} and Lemma \ref{kappa0,1,kappa1,0ForA},
Subsection \ref{A=Xi0,1+Xi1,0}. Therefore by the Schwartz's Fourier exchange theorem (compare Appendix \ref{convolutorsO'_C})
it follows that $\kappa(\xi)$ cannot belong to $\mathcal{O}'_C(\mathbb{R}^4)$ because
$\widetilde{\kappa(\xi)} \notin \mathcal{O}_M(\mathbb{R}^4)$. Thus, it is not obvious that
the convolutions (\ref{Sret*2kappa01})-(\ref{D0ret*1kappa10}) exist.

However, the situation for the second and higher order contributions to interacting fields for spinor QED with massless
Dirac field, situation is different,
because the standard normalization in the choice of the splitting,
in the computation of $\Pi^{\mu\nu}$, $\Pi^{\textrm{av} \, \mu\nu}$, $\Pi^{\textrm{ret} \, \mu\nu}$,
$\Sigma^{\mu\nu}$, $\Sigma^{\textrm{av} \, \mu\nu}$, $\Sigma^{\textrm{ret} \, \mu\nu}$,
$\Upsilon^{\mu\nu}$, $\Upsilon^{\textrm{av} \, \mu\nu}$, $\Upsilon^{\textrm{ret} \, \mu\nu}$ cannot be retained.
In particular
\[
\widetilde{\Pi^{\textrm{av} \, \mu\nu}}\Big|_{{}_{p\cdot p =0}}(p) \neq 0,
\]
and for no choice of the splitting we can achieve
\[
\widetilde{\Pi^{\textrm{av} \, \mu\nu}}\Big|_{{}_{p\cdot p =0}}(p) = 0.
\]
Recall that for the massless charged  (here spinor) field the products of parings with positive singularity degree
which we are about to split, have their Fourier transforms which have the singularity of the jump-type discontinuity
on the cone surface $p\cdot p =0$ in the momentum space. Therefore, the central splitting with the normalization point
$p'=0$ cannot be used (compare \cite{Scharf} or Subsection \ref{WickForChronological}).
By the analysis shown above, the convolution kernels
(\ref{Sigmaret*1kappa01}), (\ref{Sigmaret*1kappa10}) and  
(\ref{Piav*2kappa01}), (\ref{Piav*2kappa10}) cease to be identically zero if $m=0$, and, by the above analysis, the convolutions
\[
D_{0}^{{}^{\textrm{av}}} \ast \Pi^{{}^{\textrm{av}} \, \mu\nu} \ast A_\nu,
\]
equal to the third term in the second order contribution to the interacting electromagnetic potential field,
ceases to be well-defined even as the finite sum of generalized integral kernel operators in the sense
of \cite{obataJFA}. Indeed, let us remind the computation of the splitting in case the charged spinor field is massless,
with $m=0$.
In this case the Fourier transform
of the causal combinations of the product of pairings whose retarded part enters into the
vacuum polarization distribution has the jump singularity $\theta(p^2)$ on the cone in momentum space, and the normalization point
in the computation of this retarded part (compare \cite{Scharf}, pp. 178-181) cannot be chosen at zero.
In particular, we cannot naively put $m=0$ in the formula for the vacuum
polarization in the massive spinor QED in order to compute the vacuum polarization for the massless spinor QED.
Therefore, in QED with massless spinor
field we compute the vacuum polarization using the normalization point $p'$ (denoted by $q$ in \cite{Scharf}, pp. 178-181)
shifted from zero. In particular
for the formula valid in the positive energy cone and in the open domain including it, we use the normalization point
$p'$ somewhere within the positive energy cone in the momentum space. The result is the following
\begin{multline}\label{FTPim=0}
\widetilde{\Pi_{\mu \nu}}(p) =
(2\pi)^{-4} \big({\textstyle\frac{p_\mu p_\nu}{p^2}} - g_{\mu\nu}\big) \widetilde{\Pi}(p),
\\
\widetilde{\Pi}(p) =
{\textstyle\frac{2\pi}{3}} p^2 \theta(p^2) \Bigg[
\textrm{sgn}(p_0)\big[(p-p')\cdot p\big]^3
{\textstyle\frac{i}{2\pi}}
\int\limits_{-\infty}^{+\infty} {\textstyle\frac{t^2 \, dt}{(1-t+i0)(tp^2+p\cdot p' +i0)^3}}
+ \theta(-p_0)
\Bigg],
\end{multline}
and $p'^2>0$ so $p'$ cannot be put equal to zero. The absolute value of this function is everywhere
bounded by a polynomial,
but has the jump singularity $\theta(p^2)$ at the cone $p^2=0$. These properties are preserved
by the splitting, as is easily seen by the dispersion formula
\begin{equation}\label{dispesionFm=0}
\widetilde{d^{\textrm{ret}}}(p) =
\big[(p-p')\cdot p\big]^3
{\textstyle\frac{i}{2\pi}}
\int\limits_{-\infty}^{+\infty} {\textstyle\frac{\widetilde{d}(-tp) \, dt}{(1-t+i0)(tp^2+p\cdot p' +i0)^3}}
\end{equation}
in case $m=0$ (and analogously for the advanced part),
for the splitting of $d$ with singularity degree $\omega=2$ (as is the case for $\Pi_{\mu \nu}$).

Now we give existence proof the kernels of typical contributions to interacting fields which contain arbitrary many
vacuum polarization loop insertions in the adiabatic limit (simply for $g=1$) in both cases with $m\neq 0$
and in case $m=0$ for spinor QED with the charged spinor free field of mass $m$. Analysis of the kernels
of the contributions containing arbitrary many self-energy loop insertions is completely analogous.

Namely, let us consider the following examples of even $n = 2k+1$-order sub-contributions (with $n$ convolutions $\ast$, summation is performed
with respect to the repeated Lorentz indices)
\[
D_{0}^{{}^{\textrm{av}}} \ast \Pi^{{}^{\textrm{av}} \, \mu_k}_{\mu} \ast \ldots \ast D_{0}^{{}^{\textrm{av}}} \ast \Pi^{{}^{\textrm{av}} \, \nu}_{\mu_1} \ast
D_{0}^{{}^{\textrm{av}}}  \ast {:}\boldsymbol{\psi}^\sharp \gamma_\nu \boldsymbol{\psi}{:},
\]
(containing $k$ loop vacuum polarization graph $\Pi$ insertions) to interacting e.m. potential $A_{{}_{\textrm{int}} \, \mu}(g=1)$
and their kernels equal to the $\epsilon \rightarrow 0$ limits of the kernels of the form (with various combinations of the positive and negative
frequency kernels $\kappa_{1,0}^{\sharp}$, $\kappa_{0,1}^{\sharp}$ of the Dirac conjugated field and $\kappa_{1,0}$,$\kappa_{0,1}$ of the Dirac field
and the corresponding signs $\pm$ in front of the four-momenta $p_1,p_2$)
\begin{multline*}
\Big(D_{0 \, \epsilon}^{{}^{\textrm{av}}} \ast \Pi^{{}^{\textrm{av}} \, \mu_k}_{\mu} \ast \ldots \ast D_{0 \, \epsilon}^{{}^{\textrm{av}}} \ast \Pi^{{}^{\textrm{av}} \, \mu_1}_{\nu} \ast D_{0 \, \epsilon}^{{}^{\textrm{av}}}  \ast \big[\kappa_{\mathpzc{l}_1,\mathpzc{m}_1}^{\sharp}(\xi_1) \gamma^\nu \dot{\otimes} \kappa_{\mathpzc{l}_2,\mathpzc{m}_2}(\xi_2)\big]\Big)(x)
\\
=
\sum\limits_{s_1,s_2} \int \ud^3 \boldsymbol{\p}_1\ud^3 \boldsymbol{\p}_2
{\textstyle\frac{\xi_1(s_1, \boldsymbol{\p}_1)\xi_2(s_2, \boldsymbol{\p}_2) u_{s_1}^{\pm}(\boldsymbol{\p}_1)^\sharp \gamma^\nu u_{s_2}^{\mp}(\boldsymbol{\p}_2)}{[(\pm p_1\pm p_2)^2 +i \epsilon \, (\pm p_{10} \pm p_{20})]^{k+1}}} \,\, \times
\\
\times \,\,
\widetilde{\Pi^{{}^{\textrm{av}} \, \mu_k}_{\mu}}(\pm p_1 \pm p_2) \widetilde{\Pi^{{}^{\textrm{av}} \, \mu_{k-1}}_{\mu_k}}(\pm p_1 \pm p_2)
\ldots \widetilde{\Pi^{{}^{\textrm{av}} \, \mu_{1}}_{\nu}}(\pm p_1+ \pm p_2)
e^{i(\pm p_1\pm p_2)\cdot x}
\end{multline*}
\[
p_1, p_2 \in \mathscr{O}_{{}_{m,0,0,0}} = \{p: \,\,\, p\cdot p = m^2, \, p_0>0  \}.
\]
Here $\xi_1, \xi_2 \in E$ of the single particle Hilbert space of the free Dirac field, i.e. $\xi_1, \xi_2 \in \mathcal{S}(\mathbb{R}^3)$,
$u_{s}^{+}(\boldsymbol{\p}) = u_s(\boldsymbol{\p}$, $u_{s}^{-}(\boldsymbol{\p}) = v_s(\boldsymbol{\p}$ are the Fourier transforms of the basic solutions of the free Dirac equation given in Appendix \ref{fundamental,u,v}, $\gamma^\mu$ are the Dirac gamma matrices (\ref{chiralgamma}), and finally
$u_{s}^{\pm}(\boldsymbol{\p})^\sharp$ is the Dirac conjugation of the spinor $u_{s}^{\pm}(\boldsymbol{\p})$. The plus sign stands everywhere
in $\pm p_1$ and in $u_{s_1}^{\pm}(\boldsymbol{\p}_1)^\sharp$ whenewer $(\mathpzc{l}_1,\mathpzc{m}_1) = (1,0)$. The minus
sign stands everywhere
in $\pm p_1$ and in $u_{s_1}^{\pm}(\boldsymbol{\p}_1)^\sharp$ whenewer $(\mathpzc{l}_1,\mathpzc{m}_1) = (0,1)$. Analogously,
the plus sign stands everywhere
in $\pm p_2$ and minus sign in $u_{s_2}^{\pm}(\boldsymbol{\p}_2)$ whenewer $(\mathpzc{l}_2,\mathpzc{m}_2) = (1,0)$. The minus
sign stands everywhere
in $\pm p_2$ and plus sign in $u_{s_2}^{\pm}(\boldsymbol{\p}_2)$ whenewer $(\mathpzc{l}_2,\mathpzc{m}_2) = (0,1)$.

For the ``natural'' normalization in the Epstein-Glaser splitting, and in case $m\neq 0$, the singularity appearing in the limit
\[
{\textstyle\frac{1}{[p^2+\epsilon p_0]^{k+1}}} \overset{\epsilon\rightarrow 0}{\longrightarrow} {\textstyle\frac{1}{(p^{2})^{k+1}}}
-  \textrm{sgn} \, (p_0) \, {\textstyle\frac{i\pi(-1)^{k}}{k!}}\delta^{(k)}(p^2),
\]
is cancelled by the Fourier transform $\widetilde{\Pi^{\textrm{av} \, \mu\nu}}$ of $\Pi^{\textrm{av} \, \mu\nu}$, as $\widetilde{\Pi^{\textrm{av} \, \mu\nu}}
= (\tfrac{p^\mu p^\nu}{p^2} -g^{\mu\nu}) \widetilde{\Pi}(p)$ with a regular $\widetilde{\Pi}$ in the vicinity of the cone $p^2=0$, and equal there to 
$\widetilde{\Pi}(p) = [p^2]^2g_0(p)$ with still regular $g_0$ there.

This is the case for spinor QED with massive charged field. With the ``natural'' splitting this convolution kernels exist as tempered distributions 
in the space-time variable $x$. Namely we have the following
\begin{lem}
The $\epsilon\rightarrow 0$ limit of the kernels
\begin{multline}\label{++}
\Big(D_{0 \, \epsilon }^{{}^{\textrm{av}}} \ast \Pi^{{}^{\textrm{av}} \, \mu_k}_{\mu} \ast \ldots \ast D_{0 \, \epsilon}^{{}^{\textrm{av}}} \ast \Pi^{{}^{\textrm{av}} \, \mu_1}_{\nu} \ast D_{0 \, \epsilon}^{{}^{\textrm{av}}}  \ast \big[\kappa_{1,0}^{\sharp}(\xi_1) \gamma^\nu \dot{\otimes} \kappa_{1,0}(\xi_2)\big]\Big)(x)
\\
=
\sum\limits_{s_1,s_2} \int \ud^3 \boldsymbol{\p}_1\ud^3 \boldsymbol{\p}_2
{\textstyle\frac{\xi_1(s_1, \boldsymbol{\p}_1)\xi_2(s_2, \boldsymbol{\p}_2) u_{s_1}^{+}(\boldsymbol{\p}_1)^\sharp \gamma^\nu u_{s_2}^{-}(\boldsymbol{\p}_2)}{[(p_1+ p_2)^2 +i\epsilon (p_{10} + p_{20})]^{k+1}}} \,\, \times
\\
\times \,\,
\widetilde{\Pi^{{}^{\textrm{av}} \, \mu_k}_{\mu}}(p_1+p_2) \widetilde{\Pi^{{}^{\textrm{av}} \, \mu_{k-1}}_{\mu_k}}(p_1+ p_2)
\ldots \widetilde{\Pi^{{}^{\textrm{av}} \, \mu_{1}}_{\nu}}(p_1+ p_2)
e^{i(p_1+p_2)\cdot x}
\end{multline}
\begin{multline}\label{--}
\Big(D_{0\, \epsilon}^{{}^{\textrm{av}}} \ast \Pi^{{}^{\textrm{av}} \, \mu_k}_{\mu} \ast \ldots \ast D_{0\, \epsilon}^{{}^{\textrm{av}}} \ast \Pi^{{}^{\textrm{av}} \, \mu_1}_{\nu} \ast D_{0 \, \epsilon}^{{}^{\textrm{av}}}  \ast \big[\kappa_{0,1}^{\sharp}(\xi_1) \gamma^\nu \dot{\otimes} \kappa_{0,1}(\xi_2)\big]\Big)(x)
\\
=
\sum\limits_{s_1,s_2} \int \ud^3 \boldsymbol{\p}_1\ud^3 \boldsymbol{\p}_2
{\textstyle\frac{\xi_1(s_1, \boldsymbol{\p}_1)\xi_2(s_2, \boldsymbol{\p}_2) u_{s_1}^{-}(\boldsymbol{\p}_1)^\sharp \gamma_\nu u_{s_2}^{+}(\boldsymbol{\p}_2)}{[(-p_1- p_2)^2 +i\epsilon (-p_{10} - p_{20})]^{k+1}}} \,\, \times
\\
\times \,\,
\widetilde{\Pi^{{}^{\textrm{av}} \, \mu_k}_{\mu}}(-p_1-p_2) \widetilde{\Pi^{{}^{\textrm{av}} \, \mu_{k-1}}_{\mu_k}}(-p_1-p_2)
\ldots \widetilde{\Pi^{{}^{\textrm{av}} \, \mu_{1}}_{\nu}}(-p_1- p_2)
e^{i(-p_1-p_2)\cdot x}
\end{multline}
\begin{multline}\label{+-}
\Big(D_{0\, \epsilon}^{{}^{\textrm{av}}} \ast \Pi^{{}^{\textrm{av}} \, \mu_k}_{\mu} \ast \ldots \ast D_{0\, \epsilon}^{{}^{\textrm{av}}} \ast \Pi^{{}^{\textrm{av}} \, \mu_1}_{\nu} \ast D_{0 \, \epsilon}^{{}^{\textrm{av}}}  \ast \big[\kappa_{1,0}^{\sharp}(\xi_1) \gamma^\nu \dot{\otimes} \kappa_{0,1}(\xi_2)\big]\Big)(x)
\\
=
\sum\limits_{s_1,s_2} \int \ud^3 \boldsymbol{\p}_1\ud^3 \boldsymbol{\p}_2
{\textstyle\frac{\xi_1(s_1, \boldsymbol{\p}_1)\xi_2(s_2, \boldsymbol{\p}_2) u_{s_1}^{+}(\boldsymbol{\p}_1)^\sharp \gamma_\nu u_{s_2}^{+}(\boldsymbol{\p}_2)}{[(p_1- p_2)^2 +i\epsilon (p_{10} - p_{20})]^{k+1}}} \,\, \times
\\
\times \,\,
\widetilde{\Pi^{{}^{\textrm{av}} \, \mu_k}_{\mu}}(p_1-p_2) \widetilde{\Pi^{{}^{\textrm{av}} \, \mu_{k-1}}_{\mu_k}}(p_1- p_2)
\ldots \widetilde{\Pi^{{}^{\textrm{av}} \, \mu_{1}}_{\nu}}(p_1- p_2)
e^{i(p_1-p_2)\cdot x}
\end{multline}
\begin{multline}\label{-+}
\Big(D_{0\, \epsilon}^{{}^{\textrm{av}}} \ast \Pi^{{}^{\textrm{av}} \, \mu_k}_{\mu} \ast \ldots \ast D_{0\, \epsilon}^{{}^{\textrm{av}}} \ast \Pi^{{}^{\textrm{av}} \, \mu_1}_{\nu} \ast D_{0 \, \epsilon}^{{}^{\textrm{av}}}  \ast \big[\kappa_{0,1}^{\sharp}(\xi_1) \gamma^\nu \dot{\otimes} \kappa_{1,0}(\xi_2)\big]\Big)(x)
\\
=
\sum\limits_{s_1,s_2} \int \ud^3 \boldsymbol{\p}_1\ud^3 \boldsymbol{\p}_2
{\textstyle\frac{\xi_1(s_1, \boldsymbol{\p}_1)\xi_2(s_2, \boldsymbol{\p}_2) u_{s_1}^{-}(\boldsymbol{\p}_1)^\sharp \gamma_\nu u_{s_2}^{-}(\boldsymbol{\p}_2)}{[(-p_1+ p_2)^2 +i\epsilon (-p_{10} + p_{20})]^{k+1}}} \,\, \times
\\
\times \,\,
\widetilde{\Pi^{{}^{\textrm{av}} \, \mu_k}_{\mu}}(-p_1+p_2) \widetilde{\Pi^{{}^{\textrm{av}} \, \mu_{k-1}}_{\mu_k}}(-p_1+ p_2)
\ldots \widetilde{\Pi^{{}^{\textrm{av}} \, \mu_{1}}_{\nu}}(-p_1+ p_2)
e^{i(-p_1+p_2)\cdot x}
\end{multline}
\[
p_1, p_2 \in \mathscr{O}_{{}_{m,0,0,0}}
\]
exist as tempered distributions 
in the space-time variable $x$ 
with the ``natural'' splitting applied in the computation of the vacuum polarization distribution $\Pi^{\textrm{av} \, \mu\nu}$ and with massive
spinor field ($m \neq 0$) in causal perturbative spinor QED with the Hida operators as the creation-annihilation operators. 
\label{kernelsg=1withkVPloopsm>0}
\end{lem}

\qedsymbol \,\,\,
We need to show that for any Schwartz test function $\phi$ of $x$ the kernel, with $\epsilon >0$, integrated with $\phi$ 
has a limit $\epsilon \rightarrow 0$. After  this integration we are left with the integrals which in fact not only exist
in the limit  $\epsilon \rightarrow 0$, but in fact are convergent for  $\epsilon$ simply put equal zero. They are, respectively, of the following form
\begin{multline}\label{++int}
\sum\limits_{s_1,s_2} \int \ud^3 \boldsymbol{\p}_1\ud^3 \boldsymbol{\p}_2
{\textstyle\frac{\xi_1(s_1, \boldsymbol{\p}_1)\xi_2(s_2, \boldsymbol{\p}_2) u_{s_1}^{+}(\boldsymbol{\p}_1)^\sharp \gamma^\nu u_{s_2}^{-}(\boldsymbol{\p}_2)}{[(p_1+ p_2)^2]^{k+1}}} \,\, \times
\\
\times \,\,
\widetilde{\Pi^{{}^{\textrm{av}} \, \mu_k}_{\mu}}(p_1+p_2) \widetilde{\Pi^{{}^{\textrm{av}} \, \mu_{k-1}}_{\mu_k}}(p_1+ p_2)
\ldots \widetilde{\Pi^{{}^{\textrm{av}} \, \mu_{1}}_{\nu}}(p_1+ p_2)
\widetilde{\phi}(p_1+p_2)
\end{multline}
\begin{multline}\label{--int}
\sum\limits_{s_1,s_2} \int \ud^3 \boldsymbol{\p}_1\ud^3 \boldsymbol{\p}_2
{\textstyle\frac{\xi_1(s_1, \boldsymbol{\p}_1)\xi_2(s_2, \boldsymbol{\p}_2) u_{s_1}^{-}(\boldsymbol{\p}_1)^\sharp \gamma^\nu u_{s_2}^{+}(\boldsymbol{\p}_2)}{[(-p_1- p_2)^2]^{k+1}}} \,\, \times
\\
\times \,\,
\widetilde{\Pi^{{}^{\textrm{av}} \, \mu_k}_{\mu}}(-p_1-p_2) \widetilde{\Pi^{{}^{\textrm{av}} \, \mu_{k-1}}_{\mu_k}}(-p_1-p_2)
\ldots \widetilde{\Pi^{{}^{\textrm{av}} \, \mu_{1}}_{\nu}}(-p_1- p_2)
\widetilde{\phi}(-p_1-p_2)
\end{multline}
\begin{multline}\label{+-int}
\sum\limits_{s_1,s_2} \int \ud^3 \boldsymbol{\p}_1\ud^3 \boldsymbol{\p}_2
{\textstyle\frac{\xi_1(s_1, \boldsymbol{\p}_1)\xi_2(s_2, \boldsymbol{\p}_2) u_{s_1}^{+}(\boldsymbol{\p}_1)^\sharp \gamma^\nu u_{s_2}^{+}(\boldsymbol{\p}_2)}{[(p_1- p_2)^2]^{k+1}}} \,\, \times
\\
\times \,\,
\widetilde{\Pi^{{}^{\textrm{av}} \, \mu_k}_{\mu}}(p_1-p_2) \widetilde{\Pi^{{}^{\textrm{av}} \, \mu_{k-1}}_{\mu_k}}(p_1- p_2)
\ldots \widetilde{\Pi^{{}^{\textrm{av}} \, \mu_{1}}_{\nu}}(p_1- p_2)
\widetilde{\phi}(p_1-p_2)
\end{multline}
\begin{multline}\label{-+int}
\sum\limits_{s_1,s_2} \int \ud^3 \boldsymbol{\p}_1\ud^3 \boldsymbol{\p}_2
{\textstyle\frac{\xi_1(s_1, \boldsymbol{\p}_1)\xi_2(s_2, \boldsymbol{\p}_2) u_{s_1}^{-}(\boldsymbol{\p}_1)^\sharp \gamma^\nu u_{s_2}^{-}(\boldsymbol{\p}_2)}{[(-p_1+ p_2)^2]^{k+1}}} \,\, \times
\\
\times \,\,
\widetilde{\Pi^{{}^{\textrm{av}} \, \mu_k}_{\mu}}(-p_1+p_2) \widetilde{\Pi^{{}^{\textrm{av}} \, \mu_{k-1}}_{\mu_k}}(-p_1+ p_2)
\ldots \widetilde{\Pi^{{}^{\textrm{av}} \, \mu_{1}}_{\nu}}(-p_1+ p_2)
\widetilde{\phi}(-p_1+p_2)
\end{multline}
with 
\[
p_i(\boldsymbol{\p}_i) = (p_0(\boldsymbol{\p}_i), \boldsymbol{\p}_i) = (\sqrt{|\boldsymbol{\p}_i|^2 + m^2}, \boldsymbol{\p}_i) \in \mathscr{O}_{{}_{m,0,0,0}} = \{p: p^2=m^2, p_0 >0\},
\]
corresponding, respectively, to the valuation of the kernels (\ref{++}), (\ref{--}), (\ref{+-}), (\ref{-+}) at the space-time
test function $\phi \in \mathcal{S}(\mathbb{R}^4)$. 

We show that in each case the integrand
\begin{multline*}
{\textstyle\frac{\xi_1(s_1, \boldsymbol{\p}_1)\xi_2(s_2, \boldsymbol{\p}_2) u_{s_1}^{\pm}(\boldsymbol{\p}_1)^+\gamma^\nu u_{s_2}^{\mp}(\boldsymbol{\p}_2)}{[(\pm p_1\pm p_2)^2 ]^{k+1}}} \,\, \times
\\
\times \,\,
\widetilde{\Pi^{{}^{\textrm{av}} \, \mu_k}_{\mu}}(\pm p_1\pm p_2) \widetilde{\Pi^{{}^{\textrm{av}} \, \mu_{k-1}}_{\mu_k}}(\pm p_1\pm p_2)
\ldots \widetilde{\Pi^{{}^{\textrm{av}} \, \mu_{1}}_{\nu}}(\pm p_1\pm p_2)
\widetilde{\phi}(\pm p_1\pm p_2)
\end{multline*}
with 
\[
p_1, p_2 \in \mathscr{O}_{{}_{m,0,0,0}},
\]
regarded as a function of $(\boldsymbol{\p}_1, \boldsymbol{\p}_2)$, is absolutely integrable.

In this integrand the function
\begin{equation}\label{f}
f(s_1,s_2, \nu, \boldsymbol{\p}_1, \boldsymbol{\p}_2) = 
\xi_1(s_1, \boldsymbol{\p}_1)\xi_2(s_2, \boldsymbol{\p}_2) u_{s_1}^{\pm}(\boldsymbol{\p}_1)^\sharp \gamma^\nu u_{s_2}^{\mp}(\boldsymbol{\p}_2) \,
\widetilde{\phi}\big(\pm p_1(\boldsymbol{\p}_1)\pm p_2(\boldsymbol{\p}_2)\big)
\end{equation}
is a Schwartz function of $(\boldsymbol{\p}_1, \boldsymbol{\p}_2) \in \mathbb{R}^6$, because $u_{s}^{+} = u_{s}$, $u_{s}^{-} = v_s$ 
are smooth and bounded functions of $\boldsymbol{\p}_i$
and $\widetilde{\phi}\big(p_1(\boldsymbol{\p}_1)+p_2(\boldsymbol{\p}_2)\big)$ is a smooth bounded function of 
$(\boldsymbol{\p}_1, \boldsymbol{\p}_2) \in \mathbb{R}^6$ and both $\xi_1, \xi_2$ are Schwartz functions.

Recall that in the ``natural'' splitting we have (\ref{Piav}),
and in the domain $p^2 < 4m^2$ covering the cone $p^2=0$ in the momentum, we have
\begin{multline*}
\widetilde{{\Pi^{{}^{\textrm{av}}}}_{\mu \nu}}(p) = 
(2\pi)^{-4} \big({\textstyle\frac{p_\mu p_\nu}{p^2}} - g_{\mu\nu}\big) \widetilde{\Pi{{}^{\textrm{av}}}}(p),
\\
\widetilde{\Pi{{}^{\textrm{av}}}}(p) =
{\textstyle\frac{1}{3}} p^4
\int\limits_{4m^2}^{\infty} {\textstyle\frac{s+2m^2}{s^2(p^2-s)}}\sqrt{1-{\textstyle\frac{4m^2}{s}}} ds,
\end{multline*}
so that in the domain $p^2 < 4m^2$, $\widetilde{\Pi^{\textrm{av} \ \mu \nu}}$ is smooth, equal to the product of the homogeneous of degree zero tensor of second rank 
\begin{equation}\label{ScaleInv}
(2\pi)^{-4} \big({\textstyle\frac{p_\mu p_\nu}{p^2}} - g_{\mu\nu}\big)
\end{equation}
and a smooth function 
\[
\widetilde{\Pi{{}^{\textrm{av}}}}(p) =
{\textstyle\frac{1}{3}} p^4
\int\limits_{4m^2}^{\infty} {\textstyle\frac{s+2m^2}{s^2(p^2-s)}}\sqrt{1-{\textstyle\frac{4m^2}{s}}} ds,
\,\,\,\, p^2 < 4m^2.
\]
We immediately see that $\widetilde{\Pi{{}^{\textrm{av}}}}(p) = [p^2]^2 g_0(p)$ with $g_0$ which is smooth for $p^2<4m^4$,
and it is easily seen that it is also smooth for $p^2>4m^4$. 

We need to show that $g_0$ stays bounded whenever $p^2 \rightarrow {4m^2}^{\pm}$ in the upper and lower case limit. To this end we need 
to compute the integral for $\widetilde{\Pi{{}^{\textrm{av}}}}$ in (\ref{Piav}) explicitly, using the Euler substitution 
\[
{\textstyle\frac{s}{m^2}} = - {\textstyle\frac{(1-\eta)^2}{\eta}}
\]
with the new integration variable $\eta$. This task has beed done in ref. \cite{Scharf}, p. 205, where in particular it was shown that
$g_0$ is a function of $p^2$, stays locally integrable with at most polynomial growth and limits of $g_0(p)$, whenever $p^2 \rightarrow {4m^2}^{\pm}$, stay bounded. 
In particular the product of any
number of factors $\widetilde{\Pi{{}^{\textrm{av}}}}$ stays locally integrable with at most polynomial growth and the same 
is true for the factor $g_0$.

The convergence of (\ref{++int}) and (\ref{--int}) is now trivial because for $p_1,p_2 \in \mathscr{O}_{}{}_{m,0,0,0}$
\[
(p_1+p_2)^2 = (-p_1-p_2)^2 > 2m^2.
\]
Also the convergence of all (\ref{+-int}) and (\ref{-+int}) becomes now trivial for $k>1$ in (\ref{+-int}) and (\ref{-+int}).
Indeed, because
\[
\Big({\textstyle\frac{p_\mu p^\nu}{p^2}} - \delta_{\mu}^{\nu} \Big) 
\Big({\textstyle\frac{p_\nu p^\sigma}{p^2}} - \delta_{\nu}^{\sigma} \Big)
=
(-1)
\Big({\textstyle\frac{p_\mu p^\sigma}{p^2}} - \delta_{\mu}^{\sigma} \Big)
\]
then
\[
\widetilde{\Pi^{{}^{\textrm{av}} \, \mu_k}_{\mu}}(p) \widetilde{\Pi^{{}^{\textrm{av}} \, \mu_{k-1}}_{\mu_k}}(p)
\ldots \widetilde{\Pi^{{}^{\textrm{av}} \, \mu_{1}}_{\nu}}(p)
=
(-1)^{k+1}
\Big({\textstyle\frac{p_\mu p_\nu}{p^2}} - g_{\mu \nu} \Big) 
\big[\widetilde{\Pi^{{}^{\textrm{av}}}}(p)\big]^k.
\]
The integrand of (\ref{+-int}) takes the form
\begin{multline*}
{\textstyle\frac{\xi_1(s_1, \boldsymbol{\p}_1)\xi_2(s_2, \boldsymbol{\p}_2) u_{s_1}^{+}(\boldsymbol{\p}_1)^+\gamma_\nu u_{s_2}^{+}(\boldsymbol{\p}_2)}{[(p_1- p_2)^2 ]^{k+1}}} \,\, \times
\\
\times \,\,
(-1)^{k+1}
\Big({\textstyle\frac{(p_1-p_2)_\mu (p_1-p_2)_\nu}{(p_1-p_2)^2}} - g_{\mu \nu} \Big) 
\big[\widetilde{\Pi^{{}^{\textrm{av}}}}(p_1 - p_2)\big]^k
\widetilde{\phi}(p_1 - p_2)
\end{multline*}
and in this case with $k>1$ we can cancel just one $(p_1 -p_2)^2$ in each factor $\widetilde{\Pi^{{}^{\textrm{av}}}}(p_1 -p_2)$ with one $(p_1 -p_2)^2$ in the denominator,
and in two of the factors $\widetilde{\Pi^{{}^{\textrm{av}}}}(p_1 -p_2)$ another one $(p_1 -p_2)^2$ can be cancelled respectively with the last one coming from 
the photon propagator in the denominator and in the denominator in the factor
\begin{equation}\label{ScaleInv(p1-p2)}
\Big({\textstyle\frac{(p_1-p_2)_\mu (p_1-p_2)_\nu}{(p_1-p_2)^2}} - g_{\mu \nu} \Big),
\end{equation}
and still we are left with a locally integrable function and absolutely integrable function (being equal to the product of a Schwartz function and a locally integrable function going to infinity not faster than polynomially). Analysis of the case (\ref{-+int}) with $k>1$ is of course identical.

For the proof of convergence of (\ref{+-int}) and (\ref{-+int}) with $k=1$
 we need to  analyze more carefully the function 
\[
{\textstyle\frac{(p_1 -p_2)_\mu (p_1 -p_2)_\nu}{(p_1 -p_2)^2}}, \,\,\, p_1, p_2 \in \mathscr{O}_{{}_{m,0,0,0}},
\]
as a function of $(\boldsymbol{\p}_1, \boldsymbol{\p}_1) \in \mathbb{R}^6$. In general, for any four-momenta $p$,
the homogeneous of degree zero function (\ref{ScaleInv}) is not locally bounded, streaming, in general, to infinity when
$p^2 \rightarrow 0$. Similarly, the function (\ref{ScaleInv(p1-p2)}), regarded as a function of two unrestricted
four-momenta $(p_1,p_2)$ is not locally bounded when $(p_1-p_2)^2$ approaches zero. But here comes the essential point:
if the four-momenta
$(p_1,p_2)$ are restricted to the positive energy hyperboloid $\{p:p^2=m^2, p_0>0\} = \mathscr{O}_{{}_{m,0,0,0}}$,
\emph{i.e.} $p_i$ are functions, respectively, of the corresponding spatial momenta $\boldsymbol{\p}_i$,
such that
\[
p_i(\boldsymbol{\p}_i) = \big(\sqrt{|\boldsymbol{\p}_i|^2 + m^2}, \boldsymbol{\p}_i \big) \in \mathscr{O}_{{}_{m,0,0,0}},
\]
then (\ref{ScaleInv(p1-p2)}), regarded as a function of the spatial momenta $(\boldsymbol{\p}_1, \boldsymbol{\p}_2) \in \mathbb{R}^6$,
becomes locally integrable of at most polynomial growth in the variables $(\boldsymbol{\p}_1, \boldsymbol{\p}_2) \in \mathbb{R}^6$.
And this is the case only if $m \neq 0$.

In order to show it, we first need to prove the estimation
\begin{equation}\label{estimation1}
0 \leq {\textstyle\frac{|\boldsymbol{\p}_1-\boldsymbol{\p}_2|^2}{-(p_1-p_2)^2}}
\leq {\textstyle\frac{|\boldsymbol{\p}_1|^2+|\boldsymbol{\p}_2|^2 + 2m^2}{2m^2}}
\end{equation}
for $p_1,p_2 \in \mathscr{O}_{{}_{m,0,0,0}}$ and for
\[
{\textstyle\frac{|\boldsymbol{\p}_1-\boldsymbol{\p}_2|^2}{-(p_1-p_2)^2}}
\]
regarded as the function of the spatial momenta
$(\boldsymbol{\p}_1, \boldsymbol{\p}_2)$.
As is easily checked, $(p_1-p_2)^2 \leq 0$ for all $(\boldsymbol{\p}_1, \boldsymbol{\p}_2)$
and $(p_1-p_2)^2 = 0$ if and only if
$\boldsymbol{\p}_1-\boldsymbol{\p}_2 =0$. Thus zeros of $(p_1-p_2)^2$ coincide with the diagonal linear subspace
$\{(\boldsymbol{\p}, \boldsymbol{\p}), \boldsymbol{\p} \in \mathbb{R}^3\}$ in $\mathbb{R}^3 \times \mathbb{R}^3$.
Next we investigate the left and the right hand side functions
\[
{\textstyle\frac{|\boldsymbol{\p}_1-\boldsymbol{\p}_2|^2}{-(p_1-p_2)^2}}
\,\,\,\,
\textrm{and}
\,\,\,\,
{\textstyle\frac{|\boldsymbol{\p}_1|^2+|\boldsymbol{\p}_2|^2 + 2m^2}{2m^2}}
\]
along the straight lines perpendicular to the diagonal (in
 $\mathbb{R}^3 \times \mathbb{R}^3 = \mathbb{R}^6$ regarded as the Euclidean space) and use the fact that the right
hand side function is rotationally invariant in $\mathbb{R}^3 \times \mathbb{R}^3 = \mathbb{R}^6$. Elementary application
of the de l'Hospital rule shows that the limit value of the left hand side function along the straight line perpendicular
to the diagonal at $(\boldsymbol{\p}, \boldsymbol{\p})$, when it reaches the diagonal point $(\boldsymbol{\p}, \boldsymbol{\p})$,
is finite. The limit value when the perpendicular straight line reaches the point $(\boldsymbol{\p}, \boldsymbol{\p})$
depends on the direction $(\boldsymbol{\n}, -\boldsymbol{\n})$ of the perpendicular straight line, and its maximum value
\[
{\textstyle\frac{|\boldsymbol{\p}|^2 +m^2}{m^2}}
\]
is reached when the direction of $\boldsymbol{\n}$ coincides with the direction of $\boldsymbol{\p}$, its minimal value $1$
is reached when the direction $\boldsymbol{\n}$ is perpendicular to $\boldsymbol{\p}$. Comparison of the left and right-hand side functions
along the lines perpendicular to the diagonal, and rotational symmetry of the right-hand side function, shows
the validity of the estimation (\ref{estimation1}). Alternatively we can use the following elementary inequality
\[
-m^2 + \sqrt{ \mathpzc{x}^2 + m^2} \sqrt{ \mathpzc{y}^2 + m^2} -  \lambda  \mathpzc{x} \mathpzc{y}  
\geq  {\textstyle\frac{m^2 \big( \mathpzc{x}^2 + \mathpzc{y}^2 - 2 \lambda \mathpzc{x} \mathpzc{y}  \big)}{\mathpzc{x}^2+\mathpzc{y}^2 + 2m^2}},
\]
valid for real $\mathpzc{x},\mathpzc{y}$, $m, \lambda$ in the range  $-\infty < \mathpzc{x},\mathpzc{y},m < +\infty$,
and $-1 \leq \lambda \leq 1$. In order to prove this inequality, we multiply both sides by the denominator.
Then, after transferring the terms without the root to the right-hand side and squaring,
we reduce the above inequality to the following inequality for the polynomial  
\begin{multline*}
Q(\mathpzc{x},\mathpzc{y}) =
( \mathpzc{x}^2+\mathpzc{y}^2 + 2m^2)^2 ( \mathpzc{x}^2 + m^2 ) ( \mathpzc{y}^2 + m^2 )  
\\
- \Big[m^2 (\mathpzc{x}^2 + \mathpzc{y}^2 - 2 \lambda \mathpzc{x} \mathpzc{y} ) + m^2 ( \mathpzc{x}^2+\mathpzc{y}^2 + 2m^2) + \lambda \mathpzc{x} \mathpzc{y}
( \mathpzc{x}^2+\mathpzc{y}^2 + 2m^2) \Big]^2 \geq 0
\end{multline*}
in the real variables $\mathpzc{x},\mathpzc{y}$ with the parameters $m, \lambda$. We prove the positivity of $Q$ along the sraight lines
$\mathpzc{y} = a \mathpzc{x}$, by considering the polynomial $P(\mathpzc{x}) = Q(\mathpzc{x}, a\mathpzc{x})$ of the single
real variable $\mathpzc{x}$ with the real parameters $a,m, \lambda$. Writing explicitly $P(\mathpzc{x})$ we see that for
all $\mathpzc{x}$,$a$ and in the assumed domain $-1 \leq \lambda \leq 1$ of $\lambda$:
\begin{multline*}
P(\mathpzc{x}) = (1-\lambda^2)a^2(a^2+1)^2 \mathpzc{x}^8 \,\,\,\,
 + \,\,\,\, m^2(a^6 -4a^5\lambda + 7a^4 - 8a^3\lambda +7a^2-4a\lambda+1)\mathpzc{x}^6
\\
+ m^4(1-4a^3\lambda+6a^2-4a\lambda +1) \mathpzc{x}^4
\\
\geq
(1-\lambda^2)a^2(a^2+1)^2 \mathpzc{x}^8 \,\,\,\,
 + \,\,\,\, m^2(a^6 -4a^5 + 7a^4 - 8a^3 +7a^2-4a+1)\mathpzc{x}^6
\\
+ m^4(1-4a^3+6a^2-4a +1) \mathpzc{x}^4
\\
=
(1-\lambda^2)a^2(a^2+1)^2 \mathpzc{x}^8 \,\, + \,\,  m^2(a-1)^4(a^2+1)\mathpzc{x}^6 \,\,  + \,\, m^4(a-1)^4\mathpzc{x}^4 \geq 0,
\end{multline*}
so that $P(\mathpzc{x})$ is of degree $8$ in which there are only even degrees present, and such that all
its coefficients are positive in the full range  $-\infty < a,m < + \infty$, $-1 \leq \lambda \leq 1$, of the parameters.
Next we use the last elementary inequality in the subrange $0 < \mathpzc{x},\mathpzc{y}$ with the interpretation
\[
\mathpzc{x} = |\boldsymbol{\p}_1|, \,\,\,\, \mathpzc{y}=|\boldsymbol{\p}_2|,
\,\,\,\, \lambda  \mathpzc{x} \mathpzc{y} = \langle \boldsymbol{\p}_1, \boldsymbol{\p}_2 \rangle
\]
with $\lambda$ equal to the cosine of the angle between $\boldsymbol{\p}_1$ and $\boldsymbol{\p}_2$, immediately obtaining the inequality
(\ref{estimation1}).

From the estimation (\ref{estimation1}) it follows that the function 
\[
{\textstyle\frac{|\boldsymbol{\p}_1-\boldsymbol{\p}_2|^2}{-(p_1-p_2)^2}}, \,\,\, p_1, p_2 \in \mathscr{O}_{{}_{m,0,0,0}},
\]
is locally integrable (even with its absolute value majorized everywhere by a single polynomial) of polynomial growth in the variables  $(\boldsymbol{\p}_1, \boldsymbol{\p}_2)$ and also
\[
\sqrt{{\textstyle\frac{|\boldsymbol{\p}_1-\boldsymbol{\p}_2|^2}{-(p_1-p_2)^2}}}
=
{\textstyle\frac{|\boldsymbol{\p}_1-\boldsymbol{\p}_2|}{\sqrt{-(p_1-p_2)^2}}},
 \,\,\, p_1, p_2 \in \mathscr{O}_{{}_{m,0,0,0}},
\] 
is locally integrable (even with its absolute value majorized everywhere by a single polynomial) of polynomial growth in the variables  $(\boldsymbol{\p}_1, \boldsymbol{\p}_2)$.

Because, for $p_1, p_2 \in \mathscr{O}_{{}_{m,0,0,0}}$, 
\begin{multline*}
{\textstyle\frac{(p_1 -p_2)_0 (p_1 -p_2)_0}{(p_1 -p_2)^2}}
= -\Big({\textstyle\frac{\sqrt{|\boldsymbol{\p}_1|^2+m^2}-\sqrt{|\boldsymbol{\p}_2|^2+m^2}}{\sqrt{-(p_1-p_2)^2}}} \Big)^2
= {\textstyle\frac{(p_1-p_2)^2+|\boldsymbol{\p}_1-\boldsymbol{\p}_2|^2}{(p_1-p_2)^2}}
\\
=
1 + {\textstyle\frac{|\boldsymbol{\p}_1-\boldsymbol{\p}_2|^2}{(p_1-p_2)^2}},
\end{multline*}
then by (\ref{estimation1}) it follows that the $\mu,\nu=0,0$ component of the function 
(\ref{ScaleInv(p1-p2)}), with $p_1, p_2 \in \mathscr{O}_{{}_{m,0,0,0}}$, and regarded as a function of 
the variables  $(\boldsymbol{\p}_1, \boldsymbol{\p}_2)$, is locally integrable 
(even with its absolute value majorized everywhere by a single polynomial) of polynomial growth.
We have, thus, also shown that the function
\[
{\textstyle\frac{\sqrt{|\boldsymbol{\p}_1|^2+m^2}-\sqrt{|\boldsymbol{\p}_2|^2+m^2}}{\sqrt{-(p_1-p_2)^2}}} 
\]
of the variables  $(\boldsymbol{\p}_1, \boldsymbol{\p}_2)$ is also locally integrable 
(even with its absolute value majorized everywhere by a single polynomial) of polynomial growth.
Therefore, also the other components $\mu,\nu$ of the function 
(\ref{ScaleInv(p1-p2)}), with $p_1, p_2 \in \mathscr{O}_{{}_{m,0,0,0}}$, and regarded as a function of 
the variables  $(\boldsymbol{\p}_1, \boldsymbol{\p}_2)$, are locally integrable 
(even with its absolute value majorized everywhere by a single polynomial) of polynomial growth, because
\[
\begin{split}
{\textstyle\frac{(p_1 -p_2)_0 (p_1 -p_2)_k}{(p_1 -p_2)^2}} =
-{\textstyle\frac{\sqrt{|\boldsymbol{\p}_1|^2+m^2}-\sqrt{|\boldsymbol{\p}_2|^2+m^2}}{\sqrt{-(p_1-p_2)^2}}} 
{\textstyle\frac{|\boldsymbol{\p}_1-\boldsymbol{\p}_2|}{\sqrt{-(p_1-p_2)^2}}}
{\textstyle\frac{(\boldsymbol{\p}_1-\boldsymbol{\p}_2)_k}{|\boldsymbol{\p}_1-\boldsymbol{\p}_2|}}
\\
{\textstyle\frac{(p_1 -p_2)_i (p_1 -p_2)_k}{(p_1 -p_2)^2}} =
{\textstyle\frac{|\boldsymbol{\p}_1-\boldsymbol{\p}_2|^2}{(p_1-p_2)^2}}
{\textstyle\frac{(\boldsymbol{\p}_1-\boldsymbol{\p}_2)_i}{|\boldsymbol{\p}_1-\boldsymbol{\p}_2|}}
{\textstyle\frac{(\boldsymbol{\p}_1-\boldsymbol{\p}_2)_k}{|\boldsymbol{\p}_1-\boldsymbol{\p}_2|}},
\end{split}
\]
for $\mu=0$, and $\mu=k=1,2,3$ or  $\mu=i=1,2,3$ and $\mu=k=1,2,3$.  

Therefore, in the integrand 
\begin{multline*}
{\textstyle\frac{\xi_1(s_1, \boldsymbol{\p}_1)\xi_2(s_2, \boldsymbol{\p}_2) u_{s_1}^{+}(\boldsymbol{\p}_1)^+\gamma^\nu u_{s_2}^{+}(\boldsymbol{\p}_2)}{[(p_1- p_2)^2 ]^{k+1}}} \,\, \times
\\
\times \,\,
(-1)^{k+1}
\Big({\textstyle\frac{(p_1-p_2)_\mu (p_1-p_2)_\nu}{(p_1-p_2)^2}} - g_{\mu \nu} \Big) 
\big[\widetilde{\Pi^{{}^{\textrm{av}}}}(p_1 - p_2)\big]^k
\widetilde{\phi}(p_1 - p_2)
\end{multline*}
 of (\ref{+-int}), which in case $k=1$ becomes equal 
\begin{multline*}
{\textstyle\frac{\xi_1(s_1, \boldsymbol{\p}_1)\xi_2(s_2, \boldsymbol{\p}_2) u_{s_1}^{+}(\boldsymbol{\p}_1)^+\gamma^\nu u_{s_2}^{+}(\boldsymbol{\p}_2)}
{[(p_1- p_2)^2 ]^{2}}} \,\, \times
\\
\times \,\,
\Big({\textstyle\frac{(p_1-p_2)_\mu (p_1-p_2)_\nu}{(p_1-p_2)^2}} - g_{\mu \nu} \Big) 
\widetilde{\Pi^{{}^{\textrm{av}}}}(p_1 - p_2)
\widetilde{\phi}(p_1 - p_2)
\end{multline*}
the factor 
\[
\Big({\textstyle\frac{(p_1-p_2)_\mu (p_1-p_2)_\nu}{(p_1-p_2)^2}} - g_{\mu \nu} \Big) 
\]
is locally integrable 
(even with its absolute value majorized everywhere by a single polynomial) of polynomial growth in $(\boldsymbol{\p}_1, \boldsymbol{\p}_2)$, and in the factor
\[
\widetilde{\Pi^{{}^{\textrm{av}}}}(p_1 - p_2) = [(p_1 - p_2)^2]^2 g_0(p_1 - p_2)
\]
also $g_0(p_1 - p_2)$, as the function of  $(\boldsymbol{\p}_1, \boldsymbol{\p}_2)$, is majorized by a polynomial. We can therefore cancel 
$[(p_1- p_2)^2 ]^{2}$ in the denominator coming from the photon propagator with that $[(p_1 - p_2)^2]^2$ in 
$\widetilde{\Pi^{{}^{\textrm{av}}}}(p_1 - p_2)$, and after this we are left with the following integrand
\begin{multline*}
\xi_1(s_1, \boldsymbol{\p}_1)\xi_2(s_2, \boldsymbol{\p}_2) u_{s_1}^{+}(\boldsymbol{\p}_1)^+\gamma^\nu u_{s_2}^{+}(\boldsymbol{\p}_2)
\widetilde{\phi}(p_1 - p_2)
 \,\, \times
\\
\times \,\,
\Big({\textstyle\frac{(p_1-p_2)_\mu (p_1-p_2)_\nu}{(p_1-p_2)^2}} - g_{\mu \nu} \Big). 
g_0(p_1 - p_2)
\end{multline*}

Because
\[
\xi_1(s_1, \boldsymbol{\p}_1)\xi_2(s_2, \boldsymbol{\p}_2) u_{s_1}^{+}(\boldsymbol{\p}_1)^+\gamma^\nu u_{s_2}^{+}(\boldsymbol{\p}_2)
\widetilde{\phi}(p_1 - p_2)
\]
is a Schwartz function of  $(\boldsymbol{\p}_1, \boldsymbol{\p}_2)$, then we are left with an absolutely integrable
(and locally integrable) integrand and the integral  (\ref{+-int}), also in case $k=1$,
is absolutely convergent. 

Because $(-p_1+p_2)^2 = (p_1-p_2)^2$, the proof of convergence of the integral  (\ref{-+int}), also in case $k=1$,
is identical.

Thus, the Lemma is now fully proved. 
\qed

Proof of the absolute convergence of the evaluation integrals of the kernels of the
higher order contributions with arbitrary many ``self-energy'' loop insertions is analogous,
provided the self-energy distribution and its advanced and retarded parts $\Sigma, \Sigma_{\textrm{ret}}, \Sigma_{\textrm{av}}$,
are computed with the ``natural'' normalization in the splitting and $m\neq 0$.

\vspace*{2cm}

\begin{center}
{\scriptsize RELATION OF THIS PROOF TO A ONE-DIMENSIONAL DISTRIBUTION. A GENERALIZATION}
\end{center}

\vspace*{0.5cm}

Let us now give a more elegant interpretation of this proof giving its relation to the distribution theory.
More precisely, we relate this proof to the limit distribution
\begin{equation}\label{[p2+i0]k}
{\textstyle\frac{1}{[p^2+\epsilon p_0]^{k+1}}} \overset{\epsilon\rightarrow 0}{\longrightarrow} {\textstyle\frac{1}{(p^{2})^{k+1}}}
-  \textrm{sgn} \, (p_0) \, {\textstyle\frac{i\pi(-1)^{k}}{k!}}\delta^{(k)}(p^2),
\end{equation}
in the single real variable $p^2=p\cdot p$.
For the properties of the limit distribution,
compare Vol. I of the famous course \cite{GelfandI}, p. 60 on generalized functions.
Recall, that $1/(p^2)^k$ in (\ref{[p2+i0]k}) is the standard distribution $x^{-k}$ in the single real variable  $x=p^2$,
and defined in  \cite{GelfandI}, p. 51.

In order to do it  we will introduce new integration variables $\upsilon_1, \ldots, \upsilon_6$
instead of $(\boldsymbol{\p}_1, \boldsymbol{\p}_2)$ in the integrals
\begin{multline}\label{<epsilonkernels,phi>}
\Big \langle D_{0 \epsilon}^{{}^{\textrm{av}}} \ast \Pi^{{}^{\textrm{av}} \, \mu_k}_{\mu} \ast \ldots \ast D_{0 \epsilon}^{{}^{\textrm{av}}} \ast \Pi^{{}^{\textrm{av}} \, \mu_1}_{\nu} \ast D_{0 \, \epsilon}^{{}^{\textrm{av}}}  \ast \big[\kappa_{\mathpzc{l}_1,\mathpzc{m}_1}^{\sharp}(\xi_1) \gamma^\nu \dot{\otimes} \kappa_{\mathpzc{l}_2,\mathpzc{m}_2}(\xi_2)\big], \,\, \phi \Big \rangle
\\
=
\sum\limits_{s_1,s_2} \int \ud^3 \boldsymbol{\p}_1\ud^3 \boldsymbol{\p}_2
{\textstyle\frac{\xi_1(s_1, \boldsymbol{\p}_1)\xi_2(s_2, \boldsymbol{\p}_2) u_{s_1}^{\pm}(\boldsymbol{\p}_1)^\sharp \gamma^\nu u_{s_2}^{\mp}(\boldsymbol{\p}_2)}{[(\pm p_1\pm p_2)^2 +i\epsilon (\pm p_{10} \pm p_{20})]^{k+1}}} \,\, \times
\\
\times \,\,
\widetilde{\Pi^{{}^{\textrm{av}} \, \mu_k}_{\mu}}(\pm p_1 \pm p_2) \widetilde{\Pi^{{}^{\textrm{av}} \, \mu_{k-1}}_{\mu_k}}(\pm p_1 \pm p_2)
\ldots \widetilde{\Pi^{{}^{\textrm{av}} \, \mu_{1}}_{\nu}}(\pm p_1 \pm p_2)
\widetilde{\phi}(\pm p_1\pm p_2)
\end{multline}
\[
p_1, p_2 \in \mathscr{O}_{{}_{m,0,0,0}} = \{p: \,\,\, p\cdot p = m^2, \, p_0>0  \},
\]
with $\epsilon >0$, and let us focus on the less trivial case
\begin{multline*}
\Big \langle D_{0 \epsilon}^{{}^{\textrm{av}}} \ast \Pi^{{}^{\textrm{av}} \, \mu_k}_{\mu} \ast \ldots \ast D_{0 \epsilon}^{{}^{\textrm{av}}} \ast \Pi^{{}^{\textrm{av}} \, \mu_1}_{\nu} \ast D_{0 \, \epsilon}^{{}^{\textrm{av}}}  \ast \big[\kappa_{1,0}^{\sharp}(\xi_1) \gamma^\nu \dot{\otimes} \kappa_{0,1}(\xi_2)\big], \,\, \phi \Big \rangle
\\
=
\sum\limits_{s_1,s_2} \int \ud^3 \boldsymbol{\p}_1\ud^3 \boldsymbol{\p}_2
{\textstyle\frac{\xi_1(s_1, \boldsymbol{\p}_1)\xi_2(s_2, \boldsymbol{\p}_2) u_{s_1}^{+}(\boldsymbol{\p}_1)^\sharp \gamma^\nu u_{s_2}^{+}(\boldsymbol{\p}_2)}{[(p_1- p_2)^2 +i\epsilon (p_{10} - p_{20})]^{k+1}}} \,\, \times
\\
\times \,\,
\widetilde{\Pi^{{}^{\textrm{av}} \, \mu_k}_{\mu}}(p_1 -p_2) \widetilde{\Pi^{{}^{\textrm{av}} \, \mu_{k-1}}_{\mu_k}}(p_1 - p_2)
\ldots \widetilde{\Pi^{{}^{\textrm{av}} \, \mu_{1}}_{\nu}}(p_1- p_2)
\widetilde{\phi}(p_1-p_2)
\end{multline*}

Because $(p_1-p_2)^2$, regarded as the function of $(\boldsymbol{\p}_1, \boldsymbol{\p}_2)$ is zero only on the
measure zero diagonal hypersurface $\boldsymbol{\p}_1 = \boldsymbol{\p}_2$ in $\mathbb{R}^6$ and its gradient
\[
\Big({\textstyle\frac{\partial}{\partial \boldsymbol{\p}_1}}(p_1-p_2)^2, {\textstyle\frac{\partial}{\partial \boldsymbol{\p}_2}}(p_1-p_2)^2 \Big)
\]
is zero only on the hypersurface $(p_1-p_2)^2 =0$, then we can use a new set of (improper) coordinates
$(\upsilon_1 = (p_1-p_2)^2, \ldots, \upsilon_6)$ as the new integration variables. The absolute value of the determinant
of the Jacobian matrix
\[
\Big| \textrm{det} {\textstyle\frac{\partial (\boldsymbol{\p}_1, \boldsymbol{\p}_2)}{\partial (\upsilon_1, \ldots, \upsilon_6)}} \Big|
\]
is regular (finite) everywhere, except the singularity hypersurface  $(p_1-p_2)^2 =0$, where it has locally integrable
singularity, namely around $\upsilon_1=(p_1-p_2)^2 =0$ it behaves like
\begin{equation}\label{Jac|dp/dv|}
\Big| \textrm{det} {\textstyle\frac{\partial (\boldsymbol{\p}_1, \boldsymbol{\p}_2)}{\partial (\upsilon_1, \ldots, \upsilon_6)}} \Big|(\upsilon_1, \ldots, \upsilon_6)
= {\textstyle\frac{f(\upsilon_1, \ldots, \upsilon_6)}{\sqrt{\upsilon_1}}}
\end{equation}
with a smooth function $f$, $f \neq 0$ around $\upsilon_1=0$,
and absolutely integrable whenever multiplied by any absolutely integrable function (multiplier of
the $L_1\cap L^2$ algebra).
That the singularity must be locally integrable follows immediately from the fact
that the constant function equal $1$ everywhere is locally integrable, and the change of variables
cannot, of course, change this property, whence the local integrability of the absolute value 
of the determinant of the Jacobian matrix follows. Indeed, $\tfrac{1}{\sqrt{\upsilon_1}}$ is locally integrable
around $\upsilon_1=0$ as a function of the real variable $\upsilon_1$.

Next, from the proved estimation it follows that each Lorentz component $\mu,\nu$ of the function (\ref{ScaleInv(p1-p2)}),
when expressed by the new variables $(\upsilon_1, \ldots, \upsilon_6)$ is everywhere smooth, except $\upsilon_1=0$, 
everywhere locally integrable, with absolute value bounded everywhere
by a second degree positive polynomial in the variables $(\boldsymbol{\p}_1, \boldsymbol{\p}_2)$,
and has the singularity at $\upsilon_1 =0$ of the finite-jump-type in each compact domain in
the variables $(\upsilon_1, \ldots, \upsilon_6)$. The function
\begin{equation}\label{sgn(p10-p20)}
\textrm{sgn}(p_{10}-p_{20})
\end{equation}
is bounded and also has a finite-jump-type singularity at $\upsilon_1 =0$ in
the variables $(\upsilon_1, \ldots, \upsilon_6)$. Thus, the product of the functions
(\ref{ScaleInv(p1-p2)}), (\ref{Jac|dp/dv|}) and (\ref{sgn(p10-p20)}) by the $k$-th
power of $\widetilde{\Pi^{\textrm{av}}}(p_1-p_2) = [(p_1-p_2)^2]^2 g_0(p_1-p_2) = \upsilon_{1}^{2}h(\upsilon_{1})$
with a function $h$ of $\upsilon_1$ smooth around $\upsilon_1=0$, gives always a function
wich is of $\mathscr{C}^{k+1}(\mathbb{R}^6)$-class around the hypersurface $\upsilon_1=0$, in case $k>1$, and
of $\mathscr{C}^{k}(\mathbb{R}^6)$-class around $\upsilon_1=0$, in case $k=1$.
This is of course thanks to the fact that in the ``natural'' normalization $\widetilde{\Pi^{\textrm{av}}}(p_1-p_2)$
has zero of second order in the variable $(p_1-p_2)^2 = \upsilon_1$ at  $(p_1-p_2)^2 = \upsilon_1=0$, and its $k$-th
power has zero of sufficiently high order canceling the jump-type singularities of the other two factors
(\ref{ScaleInv(p1-p2)}), (\ref{sgn(p10-p20)}) and the singularity $\sim 1/\sqrt{\upsilon_1}$ of the factor (\ref{Jac|dp/dv|})
at  $(p_1-p_2)^2 = \upsilon_1=0$.
We arrive with the total integrand
\[
{\textstyle\frac{1}{[\upsilon_1 + i \epsilon (p_{10} - p_{20})]^{k+1}}}
\varphi_1(s_1,s_2,\upsilon_1, \ldots, \upsilon_6)
\]
in which $\varphi_1$ as well as
\[
\textrm{sgn}(p_{10}-p_{20})\varphi_1 = \varphi_2
\]
are absolutely integrable and of $\mathscr{C}^{k+1}(\mathbb{R}^6)$-class if $k>1$ (or $\mathscr{C}^{k}(\mathbb{R}^6)$-class if $k=1$)
around the hypersurface $\upsilon_1=0$.
Moreover, derivatives of $\varphi_1$ and $\varphi_2$, up to $k+1$-th order if $k>1$ (and up to $k$-th order if $k=1$),
in the first new variable $\upsilon_1$ at $(p_1-p_2)^2 = \upsilon_1 =0$
are equal zero.
Therefore, introducing the families of functions
\[
\begin{split}
\mathpzc{f}_{{}_{(s_1,s_2,\upsilon_2, \ldots, \upsilon_6)}}(\upsilon_1) =  \varphi_1(s_1,s_2,\upsilon_1, \upsilon_2, \ldots, \upsilon_6),
\\
\mathpzc{g}_{{}_{(s_1,s_2,\upsilon_2, \ldots, \upsilon_6)}}(\upsilon_1) =  \varphi_2(s_1,s_2,\upsilon_1, \upsilon_2, \ldots, \upsilon_6),
\end{split}
\]
we obtain in the limit $\epsilon \rightarrow 0$:
\begin{multline*}
\Big \langle D_{0}^{{}^{\textrm{av}}} \ast \Pi^{{}^{\textrm{av}} \, \mu_k}_{\mu} \ast \ldots \ast D_{0}^{{}^{\textrm{av}}} \ast \Pi^{{}^{\textrm{av}} \, \mu_1}_{\nu} \ast D_{0}^{{}^{\textrm{av}}}  \ast \big[\kappa_{1,0}^{\sharp}(\xi_1) \gamma^\nu \dot{\otimes} \kappa_{0,1}(\xi_2)\big], \, \phi \Big \rangle
\\
=
\sum\limits_{s_1,s_2} \int \ud^3 \boldsymbol{\p}_1\ud^3 \boldsymbol{\p}_2
{\textstyle\frac{\xi_1(s_1, \boldsymbol{\p}_1)\xi_2(s_2, \boldsymbol{\p}_2) u_{s_1}^{+}(\boldsymbol{\p}_1)^\sharp \gamma^\nu u_{+}^{\pm}(\boldsymbol{\p}_2)}{[(p_1- p_2)^2 + i 0 (p_{10} - p_{20})]^{k+1}}} \,\, \times
\\
\times \,\,
\widetilde{\Pi^{{}^{\textrm{av}} \, \mu_k}_{\mu}}(p_1 -p_2) \widetilde{\Pi^{{}^{\textrm{av}} \, \mu_{k-1}}_{\mu_k}}(p_1 - p_2)
\ldots \widetilde{\Pi^{{}^{\textrm{av}} \, \mu_{1}}_{\nu}}(\pm p_1- \pm p_2)
\widetilde{\phi}(p_1-p_2)
\\
=
\sum\limits_{s_1,s_2} \int d\upsilon_1 \ldots d\upsilon_6 {\textstyle\frac{1}{[\upsilon_1 + i 0 \, (p_{10} - p_{20})]^{k+1}}}
\varphi(s_1,s_2,\upsilon_1, \ldots, \upsilon_6)
\\
= \sum\limits_{s_1,s_2} \int d\upsilon_2 \ldots d\upsilon_6 \Big \langle {\textstyle\frac{1}{\upsilon_{1}^{k+1}}}, \,\,
 \mathpzc{f}_{{}_{(s_1,s_2,\upsilon_2, \ldots, \upsilon_6)}}(\upsilon_1)\Big \rangle
\\
-  \sum\limits_{s_1,s_2} \int d\upsilon_2 \ldots d\upsilon_6
 \Big \langle {\textstyle\frac{i\pi(-1)^{k}}{k!}}\delta^{(k)}(\upsilon_1), \, \mathpzc{g}_{{}_{(s_1,s_2,\upsilon_2, \ldots, \upsilon_6)}}(\upsilon_1) \Big \rangle,
\end{multline*}
where in the last integrals evaluation $\langle \cdot ,  \cdot \rangle$ of the distributions
$1/\upsilon_{1}^{k+1}, \delta^{(k)}(\upsilon_1)$ on the functions
\[
\upsilon_1 \longmapsto \mathpzc{f}_{{}_{(s_1,s_2,\upsilon_2, \ldots, \upsilon_6)}}(\upsilon_1),
\,\,\,
\upsilon_1 \longmapsto \mathpzc{g}_{{}_{(s_1,s_2,\upsilon_2, \ldots, \upsilon_6)}}(\upsilon_1)
\]
are understood, and are well-defined, because for each $s_1,s_2,\upsilon_2, \ldots, \upsilon_6$,
both $\mathpzc{f}_{{}_{(s_1,s_2,\upsilon_2, \ldots, \upsilon_6)}}$ and $\mathpzc{g}_{{}_{(s_1,s_2,\upsilon_2, \ldots, \upsilon_6)}}$
are absolutely integrable and of $\mathscr{C}^{k+1}(\mathbb{R})$-class around $\upsilon_1=0$ if $k>1$ (and of $\mathscr{C}^k(\mathbb{R})$-class
if $k=1$).
Recall that the distributions $1/\upsilon_{1}^{k+1}$, $\delta^{(k)}(\upsilon_1)$ are precisely the distributions
which arise in the limit
\[
{\textstyle\frac{1}{[\upsilon_1 + i \epsilon]^{k+1}}}
\overset{\epsilon\rightarrow 0}{\longrightarrow}
{\textstyle\frac{1}{\upsilon_{1}^{k+1}}}
-
{\textstyle\frac{i\pi(-1)^{k}}{k!}}\delta^{(k)}(\upsilon_1).
\]
Let us remind that for even $k+1 = 2\mathpzc{m}$
\begin{multline}\label{even-}
 \Big \langle {\textstyle\frac{1}{\upsilon^{k+1}}}, \,\, \mathpzc{f}(\upsilon) \Big\rangle
= \int\limits_{-\infty}^{0} \upsilon^{-(k+1)} \Big\{ \mathpzc{f}(\upsilon) + \mathpzc{f}(-\upsilon)
\\
-2\Big[  
\mathpzc{f}(0) + {\textstyle\frac{\upsilon^2}{2!}}\mathpzc{f}''(0) + \ldots +
{\textstyle\frac{\upsilon^{k-1}}{(k-1)!}}\mathpzc{f}^{(k-1)}(0)
\Big]
\Big\} d\upsilon,
\end{multline}
with only the even degree derivatives of $\mathpzc{f}$ at zero subtracted, and
for odd $k+1 = 2\mathpzc{m} +1$
\begin{multline}\label{odd-}
 \Big \langle {\textstyle\frac{1}{\upsilon^{k+1}}}, \,\, \mathpzc{f}(\upsilon) \Big\rangle
= \int\limits_{-\infty}^{0} \upsilon^{-(k+1)}  \Big\{ \mathpzc{f}(\upsilon) - \mathpzc{f}(-\upsilon)
\\
-2\Big[  
\upsilon \mathpzc{f}'(0) + {\textstyle\frac{\upsilon^3}{3!}}\mathpzc{f}'''(0) + \ldots +
{\textstyle\frac{\upsilon^{k-1}}{(k+1)!}}\mathpzc{f}^{(k+1)}(0)
\Big]
\Big\}  d\upsilon,
\end{multline}
with only the odd degree derivatives of $\mathpzc{f}$ at zero subtracted. Let us note that here we have two equivalent conventions,
and we can equivalently replace the integrations from $-\infty$ to $0$ by the integrations from $0$ to $+\infty$ in (\ref{even-})
and (\ref{odd-}), as the integrands in (\ref{even-}) and (\ref{odd-}) are in each case even functions of the real variable
$\upsilon$, compare the formulas (5) and (6) in \cite{GelfandI}, p. 51. In our case the variable
$\upsilon = \upsilon_1$ is always negative or positive (depending on the particular kernel), so the convention actually used
will depend on the actual range of the variable $\upsilon = \upsilon_1$.

We see now that indeed, if the ``splitting'' is natural, the valuations of the distributions  
 $1/\upsilon_{1}^{k+1}$, $\delta^{(k)}(\upsilon_1)$ in (\ref{+-int-distribution})
are well-defined and in particular for odd $k$ and even $k+1$, the limit $\epsilon \rightarrow 0$ is in general equal
\begin{multline}\label{+-int-distribution}
\Big \langle D_{0}^{{}^{\textrm{av}}} \ast \Pi^{{}^{\textrm{av}} \, \mu_k}_{\mu} \ast \ldots \ast D_{0}^{{}^{\textrm{av}}} \ast \Pi^{{}^{\textrm{av}} \, \mu_1}_{\nu} \ast D_{0}^{{}^{\textrm{av}}}  \ast \big[\kappa_{1,0}^{\sharp}(\xi_1) \gamma^\nu \dot{\otimes} \kappa_{0,1}(\xi_2)\big], \,\, \phi \Big \rangle
\\
=
\sum\limits_{s_1,s_2}
\int\limits_{-\infty}^{0}d\upsilon_1 \int d\upsilon_2 \ldots d\upsilon_6 \, {\textstyle\frac{1}{\upsilon_{1}^{k+1}}}
 \Big\{ \varphi_1(s_1,s_2, \upsilon_1, \upsilon_2, \ldots, \upsilon_6)
+ \varphi_1(s_1,s_2,-\upsilon_1, \upsilon_2, \ldots, \upsilon_6)
\\
-2\Big[  
\varphi_1(s_1,s_2, 0, \upsilon_2, \ldots, \upsilon_{6})
+ {\textstyle\frac{\upsilon^2}{2!}}\partial_{\upsilon_1}^{2}\varphi_1(s_1,s_2,0, \upsilon_2, \ldots, \upsilon_{6}) + \ldots
\\
\ldots +
{\textstyle\frac{\upsilon^{k-1}}{(k-1)!}}\partial_{\upsilon_1}^{k-1}\varphi_1(s_1,s_2,0, \upsilon_2, \ldots, \upsilon_{6})
\Big]
\Big\}
\\
- {\textstyle\frac{i\pi}{k!}} \sum\limits_{s_1,s_2} \int d\upsilon_2 \ldots d\upsilon_6 \,
\partial_{\upsilon_1}^{k}\varphi_2(s_1,s_2,0, \upsilon_2, \ldots, \upsilon_{6}).
\end{multline}
But in our case all derivatives
\[
\partial_{\upsilon_1}\varphi_i(s_1,s_2,0, \upsilon_2, \ldots, \upsilon_{6}), \ldots,
\partial_{\upsilon_1}^{k}\varphi_i(s_1,s_2,0, \upsilon_2, \ldots, \upsilon_{6})
\]
vanish for $k=1$, or even up to $k+1$ order, if $k>1$,
so that in our case the above limit $\epsilon \rightarrow 0$ value of the valuation integral
will not contain any terms involving these derivatives:
\begin{multline*}
\Big \langle D_{0}^{{}^{\textrm{av}}} \ast \Pi^{{}^{\textrm{av}} \, \mu_k}_{\mu} \ast \ldots \ast D_{0}^{{}^{\textrm{av}}} \ast \Pi^{{}^{\textrm{av}} \, \mu_1}_{\nu} \ast D_{0}^{{}^{\textrm{av}}}  \ast \big[\kappa_{1,0}^{\sharp}(\xi_1) \gamma^\nu \dot{\otimes} \kappa_{0,1}(\xi_2)\big], \, \phi \Big \rangle
\\
=
\sum\limits_{s_1,s_2}
\int\limits_{-\infty}^{0}d\upsilon_1 \int \upsilon_2 \ldots d\upsilon_6 \, {\textstyle\frac{1}{\upsilon_{1}^{k+1}}}
\big\{ \varphi_1(s_1,s_2, \upsilon_1, \upsilon_2, \ldots, \upsilon_6)
+ \varphi_1(s_1,s_2,-\upsilon_1, \upsilon_2, \ldots, \upsilon_6)
\big\}
\\
=\sum\limits_{s_1,s_2}
\int d\upsilon_1 \upsilon_2 \ldots d\upsilon_6 \, {\textstyle\frac{1}{\upsilon_{1}^{k+1}}} \varphi_1(s_1,s_2, \upsilon_1, \upsilon_2, \ldots, \upsilon_6),
\end{multline*}
with the integral absolutely convergent just with $\epsilon$ simply put equal to zero, as we have already seen. In our case
however, also the last integral is taken over the domain $0 \geq \upsilon_1$, as $(p_1-p_2)^2 = (-p_1+p_2)^2 \leq 0$ for the kernels
(\ref{+-}) and (\ref{-+}), in which we have used the versions of the formulas (\ref{even-}), (\ref{odd-}) with integrals restricted to the negative half-line.
But for the kernels (\ref{++}) and (\ref{--}) the variable $\upsilon_1$ in the valuation integral will be restricted to $2m^2 \leq \upsilon_1$, because
$(p_1+p_2)^2 =  (-p_1-p_2)^2 \geq 2m^2$ in spinor QED with massive charged spinor field,
so the functions $\varphi_1$, $\varphi_2$ in the integrand identically vanish in the open neighborhood of the hypersurface
$\upsilon_1=0$ and the regularizing terms disappear in  (\ref{+-int-distribution}) (with the $d\upsilon_1$-integration limits 
$(-\infty,0)$ replaced with $(0,+\infty)$)
and again the integral is absolutely convergent also for
 the kernels (\ref{++}) and (\ref{--}) evaluated with test function
with $\epsilon$ simply put equal zero. Note that for the kernels (\ref{++}) and (\ref{--})
we use the versions of the formulas (\ref{even-}) and  (\ref{odd-})
with the $d\upsilon_1$-integrals restricted to the positive half-line, with exactly the same convention
as that used in \cite{GelfandI}, p. 51. Of course the formulas (\ref{even-}) and  (\ref{odd-})
 become relevant for the analysis of the kernels (\ref{++}) and (\ref{--})
only in case of QED with massless charged field with $m=0$ in (\ref{even-}) and  (\ref{odd-})  and
(\ref{+-int-distribution}), where the $d\upsilon_1$-integration limits $(-\infty,0)$  in (\ref{even-}), (\ref{odd-})
and in (\ref{+-int-distribution}) are replaced with $(0,+\infty)$.

Let us emphasize that the Fourier transform of the retarded part of the causal combinations of the product of pairings which enters into the
vacuum polarization distribution is determined up to the polynomial in $p$ which is of second degree.
In particular in principle we can add to it the constant term of the form
$g^{\mu\nu}$. But this will destroy the vanishing of the differentials and even differentiability of the functions $\varphi_1$
and $\varphi_2$ at $\upsilon_1= (p_1-p_2) = 0$, so that the above limit will no longer exist. Of course, after this modification,
the integrals (\ref{++int})-(\ref{-+int}), will no longer be convergent, but from what we have shown above, and what is more important,
the limits $\epsilon \rightarrow 0$ will no longer exist after this modification. Still, in principle (freedom in the splitting),
we can add to $\widetilde{\Pi^{\mu\nu}}(p)$, or to $\widetilde{\Pi^{\textrm{av} \, \mu\nu}}(p)$ the term
of the form $f(p^2) g^{\mu\nu}$ with  $f$ which has zero of at least second order at zero. But the
kernels of some even order sub contributions are well-defined only after the stronger condition
\[
{\textstyle\frac{\widetilde{\Pi}(p)}{p^2}}\Big|_{{}_{p^2=0}} = 0,  \,\,\, \widetilde{\Pi}(0) = 0,
\]
compare discussion of the $2$-nd and $2k$-th order sub-contributions (\ref{LoopTermsINpsi(2)andA(2)})
and (\ref{LoopTermsINpsi(2k)andA(2k)}) given above.

This method of the proof, based on the limit distribution (\ref{[p2+i0]k}) in the single real variable, is more general  
and is suitable for the
investigation of the other QED's with massless ($m=0$) charged field in which the integrals (\ref{++int})-(\ref{-+int})
with $\epsilon$ put just equal zero are divergent. In these theories we have to investigate the limits $\epsilon \rightarrow 0$ of the
integrals (\ref{<epsilonkernels,phi>}), using the the formula (\ref{+-int-distribution})
and it's analogue for the odd $k+1$ and, eventually, with the integrations
\[
\int\limits_{-\infty}^{0} d \upsilon_1
\,\,\,\,
\textrm{replaced with}
\,\,\,\,
\int\limits_{0}^{+\infty} d \upsilon_1.
\qed
\]

\vspace*{1cm}

\begin{center}
{\scriptsize MASSLESS CHARGED FIELD}
\end{center}

\vspace*{0.5cm}

Let us consider for example the kernels (\ref{++})-(\ref{-+}) of the analogue odd $n=2k+1$-order contributions with vacuum polarization loop
insertions for spinor QED with massless Dirac field, \emph{i.e.} with $m=0$. In this case the Fourier transform
of the causal combinations of the product of pairings whose retarded part enters into the
vacuum polarization distribution has the jump singularity $\theta(p^2)$ on the cone in momentum space, and the normalization point
in the computation of this retarded part cannot be chosen at zero. In particular we cannot naively put $m=0$ in the formula for the vacuum
polarization in the massive spinor QED in order to compute the vacuum polarization for the massless spinor QED.
Therefore, in QED with massless spinor
field we compute the vacuum polarization using the normalization point $p'$ shifted from zero. In particular
for the formula valid in the positive energy cone and in the open domain including it, we use the normalization point
$p'$ somewhere within the positive energy cone in the momentum space. The result is given by (\ref{FTPim=0}) in which
$p'^2>0$ so $p'$ cannot be put equal to zero. The absolute value of this function (\ref{FTPim=0}) is everywhere
bounded by a polynomial,
but it has the jump singularity $\theta(p^2)$ at the cone $p^2=0$. These properties are preserved
by the splitting, as is easily seen by the dispersion formula (\ref{dispesionFm=0}) for the splitting in case $m=0$
and singularity degree of $d$ equal $\omega=2$ (which is the case for $\Pi^{\mu\nu}$). Thus, also
$\widetilde{\Pi^{\textrm{av}}_{\mu \nu}}$ is everywhere majorized by a polynomial, and is given by the analogous formula
with the analogous scalar factor $\widetilde{\Pi^{\textrm{av}}}$ with the jump-type
singularity $\theta(p^2)$ at the cone $p^2=0$ in the momentum space. This jump cannot be removed by using the
freedom in the splitting, which in the momentum space is determined up to the polynomial
of degree equal to the singularity degree of the splitted distribution (in the considered case equal $\omega =2$).

Let us regularize
the singular  $\theta$ everywhere in $\widetilde{\Pi^{\textrm{av}}_{\mu \nu}}$ and in $\widetilde{\Pi^{\textrm{av}}}$,
including
\[
\textrm{sgn}(p_0) = \theta(p_0) - \theta(-p_0),
\]
by a smooth
function $\theta_{\varepsilon}$ parametrized by $\varepsilon>0$, and such that
\[
\theta_{\varepsilon} \overset{\varepsilon \rightarrow 0}{\longrightarrow} \theta
\]
in the tempered distributions of the single real variable. Let us denote by
$\widetilde{\Pi^{\textrm{av}}_{\varepsilon \, \mu \nu}}$ and $\widetilde{\Pi^{\textrm{av}}_{\varepsilon}}$
the functions $\widetilde{\Pi^{\textrm{av}}_{\mu \nu}}$
and, respectively, $\widetilde{\Pi^{\textrm{av}}}$
with $\theta$ replaced by the smooth $\theta_{\varepsilon}$.
We insert $\widetilde{\Pi^{\textrm{av}}_{\varepsilon \, \mu \nu}}$ instead of $\widetilde{\Pi^{\textrm{av}}_{\mu \nu}}$
into the valuation integrals (\ref{<epsilonkernels,phi>}) for the spinor QED with massless Dirac field
($m=0$).
Now the four-momenta $p_1,p_1$  in the valuation integrals (\ref{<epsilonkernels,phi>}) range over the
positive energy sheet of the cone $p^2=0$ in the momentum space, and, as before, becomes a function
of the spatial momenta $(\boldsymbol{\p}_1, \boldsymbol{\p}_2)$.

It is immediately seen that
\[
(p_1\pm p_2)^2 = \pm 2\big(|\boldsymbol{\p}_1| |\boldsymbol{\p}_2|-\langle \boldsymbol{\p}_1|\boldsymbol{\p}_2\rangle \big)
\]
is greater than or equal to zero for the upper sign and less than or equal to zero for the lower sign,
and its gradient (in the variables $(\boldsymbol{\p}_1, \boldsymbol{\p}_2)$)
is nonzero whenever $(p_1\pm p_2)^2 \neq 0$. We can therefore use the analogue change of variables
$\upsilon_1= (p_1\pm p_2)^2, \upsilon_2, \ldots, \upsilon_6$ in the valuation integrals (\ref{<epsilonkernels,phi>})
with the locally integrable singularity around $\upsilon_1=0$ of the absolute value of the determinant of the Jacobian matrix
\[
\Big| \textrm{det}  {\textstyle\frac{\partial (\boldsymbol{\p}_1, \boldsymbol{\p}_2)}{\partial (\upsilon_1, \ldots, \upsilon_6)}} \Big|(\upsilon_1, \ldots, \upsilon_6)
= {\textstyle\frac{f(\upsilon_1, \ldots, \upsilon_6)}{\sqrt{\upsilon_1}}}
\]
of the type $\sim \upsilon_{1}^{-1/2}$, with a smooth multiplier $f$ of the $L^1 \cap L^2$ algebra, $f \neq 0$
around $\upsilon_1=0$, of the same type as before. This time, with $m=0$, the scale invariant
function (\ref{ScaleInv(p1-p2)}) is not locally integrable, with the singularity
\[
\sim {\textstyle\frac{1}{\upsilon_1}}
\]
around $\upsilon_1=0$, and thus the factor
$(p_1\pm p_2)^2 = \upsilon_1$ in $\widetilde{\Pi^{\textrm{av}}_{\varepsilon}}$
has to be cancelled with the denominator in the factor (\ref{ScaleInv(p1-p2)}).
After this cancellation we are still left
with a locally integrable integrand
\[
{\textstyle\frac{1}{[\upsilon_1 + i \epsilon (p_{10} - p_{20})]^{k+1}}}
\varphi_{1 \, \varepsilon}(s_1,s_2,\upsilon_1, \ldots, \upsilon_6)
\]
in which $\varphi_{1 \, \varepsilon}$ as well as
\[
\textrm{sgn}(p_{10}-p_{20})\varphi_{1 \, \varepsilon} = \varphi_{2 \, \varepsilon}
\]
are absolutely integrable of $\mathscr{C}^1$-class in $\upsilon_1$ around $\upsilon_1=0$ ony if $k = 2$, and of
$\mathscr{C}^{k-1}$-class in $\upsilon_1$ around $\upsilon_1=0$ for $k>1$.
But in case of $k=1$  the functions $\varphi_{i \, \varepsilon}$ still are not smooth
around $\upsilon_1=0$, having there the singularity $\sim \upsilon_{1}^{-1/2}$ coming from the Jacobian. Therefore, we still have
to ``regularize'' the determinant of the Jacobian
\[
\Big| \textrm{det}_{\varepsilon}  {\textstyle\frac{\partial (\boldsymbol{\p}_1, \boldsymbol{\p}_2)}{\partial (\upsilon_1, \ldots, \upsilon_6)}} \Big|(\upsilon_1, \ldots, \upsilon_6)
= {\textstyle\frac{f(\upsilon_1, \ldots, \upsilon_6)}{\sqrt{\upsilon_1} + \varepsilon^2}}
\]
in order to get $\varphi_{i \, \varepsilon}$ of $\mathscr{C}^1$-class in $\upsilon_1$ around $\upsilon_1=0$ for $k=1$,
and of $\mathscr{C}^{k+1}$-class in $\upsilon_1$ around $\upsilon_1=0$ for $k>1$.
After this regularization
 the limit $\epsilon \rightarrow 0$ of the integrals (\ref{<epsilonkernels,phi>}) with
$\widetilde{\Pi^{\textrm{av}}_{\mu \nu}}$ replaced by
$\widetilde{\Pi^{\textrm{av}}_{\varepsilon \, \mu \nu}}$ and with regularized Jacobian and with $\varepsilon>0$ kept fixed,
exists
and is equal (\ref{+-int-distribution}) with $\varphi_{i}$ replaced with $\varphi_{i \, \varepsilon}$ and
eventually with the integrals in $\upsilon_1$ from $-\infty$ to $0$
replaced with the integrals in $\upsilon_1$ from $0$ to $+\infty$. There are also further replacements, namely
the FT of the basic solutions $u_{s_i}^{\pm}$ of the massive Dirac equation are replaced by the scale invariant
FT of the basic solutions of the massless Dirac field, and a less obvious replacement: the functions
$\xi_1$, $\xi_2$ belong now, for the massless spinor field, to the nuclear space $\mathcal{S}^{0}(\mathbb{R}^3)$
of all those Schwartz functions with all derivatives vanishing at zero. Only in this case massless field composes
a well-defined free field as an operator-valued distribution in the white noise sense (for the proof
compare Subsection \ref{A=Xi0,1+Xi1,0}).

But finally we have to pass to the limit $\varepsilon \rightarrow 0$ in the expression
(\ref{+-int-distribution}) with $\varphi_{i}$ replaced with $\varphi_{i \, \varepsilon}$ and 
eventually with the integrals in $\upsilon_1$ from $-\infty$ to $0$
replaced with the integrals in $\upsilon_1$ from $0$ to $+\infty$. But the limit
of the derivatives of $\varphi_{i \, \varepsilon}$ at $\upsilon_1=0$ tend to plus or minus infinity,
and it is easily seen that, for the general test function $\phi$, the limit $\varepsilon \rightarrow 0$ does not exist, as the 
$\partial_{\upsilon_1}$ derivatives of $\varphi_{i \, \varepsilon}$ at $\upsilon_1=0$ of various degrees tend to infinity with various 
asymptotic degrees. 

Therefore, the $\epsilon \rightarrow 0$ limit of the last Lemma does not exist in spinor QED with the massless charged spinor field,
and this is the case for all possible choices in the splitting. 
The same holds for other QED's with massless charged fields.

Alternatively one can give a very short argument proving this result. Because the FT of the vacuum polarization in QED
with massless charged field is not smooth at the cone $p^2=0$, having the jump $\theta(p^2)$ there, and this singularity
cannot be repaired by any choice of the
splitting (addition of any polynomial in momenta of second degree), then we are confronted with the valuation of the
distribution
\[
{\textstyle\frac{1}{[\upsilon + i \epsilon]^{k+1}}}
\overset{\epsilon\rightarrow 0}{\longrightarrow}
{\textstyle\frac{1}{\upsilon^{k+1}}}
-
{\textstyle\frac{i\pi(-1)^{k}}{k!}}\delta^{(k)}(\upsilon)
\]
in single real variable $\upsilon =(p_1\pm p_2)^2$ at the ``test function'' which has the jump-type and $\sim \tfrac{1}{\sqrt{\upsilon}}$-type
singularity at $\upsilon=(p_1\pm p_2)^2=0$, which, as we know from the distribution theory,
is not well-defined, or alternatively: there is no sensible way of definition of the product of the theta function $\theta(\upsilon)$-distribution
(or the $\tfrac{1}{\sqrt{\upsilon}}$-function-type-distribution)
and the derivatives of the Dirac delta distribution $\delta^{(k)}(\upsilon)$.

We have thus proved the following

\begin{lem*}
For any choice in the splitting applied in the computation of the vacuum polarization distribution $\Pi^{\textrm{av} \, \mu\nu}$
the $\epsilon\rightarrow 0$ limits of the kernels
\begin{multline*}
\Big(D_{0 \, \epsilon}^{{}^{\textrm{av}}} \ast \Pi^{{}^{\textrm{av}} \, \mu_k}_{\mu} \ast \ldots \ast D_{0 \, \epsilon}^{{}^{\textrm{av}}} \ast \Pi^{{}^{\textrm{av}} \, \mu_1}_{\nu} \ast D_{0 \, \epsilon}^{{}^{\textrm{av}}}  \ast \big[\kappa_{1,0}^{\sharp}(\xi_1) \gamma^\nu \dot{\otimes} \kappa_{1,0}(\xi_2)\big]\Big)(x)
\\
=
\sum\limits_{s_1,s_2} \int \ud^3 \boldsymbol{\p}_1\ud^3 \boldsymbol{\p}_2
{\textstyle\frac{\xi_1(s_1, \boldsymbol{\p}_1)\xi_2(s_2, \boldsymbol{\p}_2) u_{s_1}^{+}(\boldsymbol{\p}_1)^\sharp \gamma^\nu u_{s_2}^{-}(\boldsymbol{\p}_2)}{[(p_1+ p_2)^2 +i\epsilon (p_{10} + p_{20})]^{k+1}}} \,\, \times
\\
\times \,\,
\widetilde{\Pi^{{}^{\textrm{av}} \, \mu_k}_{\mu}}(p_1+p_2) \widetilde{\Pi^{{}^{\textrm{av}} \, \mu_{k-1}}_{\mu_k}}(p_1+ p_2)
\ldots \widetilde{\Pi^{{}^{\textrm{av}} \, \mu_{1}}_{\nu}}(p_1+ p_2)
e^{i(p_1+p_2)\cdot x}
\end{multline*}
\begin{multline*}
\Big(D_{0 \, \epsilon}^{{}^{\textrm{av}}} \ast \Pi^{{}^{\textrm{av}} \, \mu_k}_{\mu} \ast \ldots \ast D_{0 \, \epsilon}^{{}^{\textrm{av}}} \ast \Pi^{{}^{\textrm{av}} \, \mu_1}_{\nu} \ast D_{0 \, \epsilon}^{{}^{\textrm{av}}}  \ast \big[\kappa_{0,1}^{\sharp}(\xi_1) \gamma^\nu \dot{\otimes} \kappa_{0,1}(\xi_2)\big]\Big)(x)
\\
=
\sum\limits_{s_1,s_2} \int \ud^3 \boldsymbol{\p}_1\ud^3 \boldsymbol{\p}_2
{\textstyle\frac{\xi_1(s_1, \boldsymbol{\p}_1)\xi_2(s_2, \boldsymbol{\p}_2) u_{s_1}^{-}(\boldsymbol{\p}_1)^\sharp \gamma^\nu u_{s_2}^{+}(\boldsymbol{\p}_2)}{[(-p_1- p_2)^2 +i\epsilon (-p_{10} - p_{20})]^{k+1}}} \,\, \times
\\
\times \,\,
\widetilde{\Pi^{{}^{\textrm{av}} \, \mu_k}_{\mu}}(-p_1-p_2) \widetilde{\Pi^{{}^{\textrm{av}} \, \mu_{k-1}}_{\mu_k}}(-p_1-p_2)
\ldots \widetilde{\Pi^{{}^{\textrm{av}} \, \mu_{1}}_{\nu}}(-p_1- p_2)
e^{i(-p_1-p_2)\cdot x}
\end{multline*}
\begin{multline*}
\Big(D_{0 \, \epsilon}^{{}^{\textrm{av}}} \ast \Pi^{{}^{\textrm{av}} \, \mu_k}_{\mu} \ast \ldots \ast D_{0 \, \epsilon}^{{}^{\textrm{av}}} \ast \Pi^{{}^{\textrm{av}} \, \mu_1}_{\nu} \ast D_{0 \, \epsilon}^{{}^{\textrm{av}}}  \ast \big[\kappa_{1,0}^{\sharp}(\xi_1) \gamma^\nu \dot{\otimes} \kappa_{0,1}(\xi_2)\big]\Big)(x)
\\
=
\sum\limits_{s_1,s_2} \int \ud^3 \boldsymbol{\p}_1\ud^3 \boldsymbol{\p}_2
{\textstyle\frac{\xi_1(s_1, \boldsymbol{\p}_1)\xi_2(s_2, \boldsymbol{\p}_2) u_{s_1}^{+}(\boldsymbol{\p}_1)^\sharp \gamma^\nu u_{s_2}^{+}(\boldsymbol{\p}_2)}{[(p_1- p_2)^2 +i\epsilon (p_{10} - p_{20})]^{k+1}}} \,\, \times
\\
\times \,\,
\widetilde{\Pi^{{}^{\textrm{av}} \, \mu_k}_{\mu}}(p_1-p_2) \widetilde{\Pi^{{}^{\textrm{av}} \, \mu_{k-1}}_{\mu_k}}(p_1- p_2)
\ldots \widetilde{\Pi^{{}^{\textrm{av}} \, \mu_{1}}_{\nu}}(p_1- p_2)
e^{i(p_1-p_2)\cdot x}
\end{multline*}
\begin{multline*}
\Big(D_{0 \, \epsilon}^{{}^{\textrm{av}}} \ast \Pi^{{}^{\textrm{av}} \, \mu_k}_{\mu} \ast \ldots \ast D_{0 \, \epsilon}^{{}^{\textrm{av}}} \ast \Pi^{{}^{\textrm{av}} \, \mu_1}_{\nu} \ast D_{0 \, \epsilon}^{{}^{\textrm{av}}}  \ast \big[\kappa_{0,1}^{\sharp}(\xi_1) \gamma^\nu \dot{\otimes} \kappa_{1,0}(\xi_2)\big]\Big)(x)
\\
=
\sum\limits_{s_1,s_2} \int \ud^3 \boldsymbol{\p}_1\ud^3 \boldsymbol{\p}_2
{\textstyle\frac{\xi_1(s_1, \boldsymbol{\p}_1)\xi_2(s_2, \boldsymbol{\p}_2) u_{s_1}^{-}(\boldsymbol{\p}_1)^\sharp \gamma^\nu u_{s_2}^{-}(\boldsymbol{\p}_2)}{[(-p_1+ p_2)^2 +i\epsilon (-p_{10} + p_{20})]^{k+1}}} \,\, \times
\\
\times \,\,
\widetilde{\Pi^{{}^{\textrm{av}} \, \mu_k}_{\mu}}(-p_1+p_2) \widetilde{\Pi^{{}^{\textrm{av}} \, \mu_{k-1}}_{\mu_k}}(-p_1+ p_2)
\ldots \widetilde{\Pi^{{}^{\textrm{av}} \, \mu_{1}}_{\nu}}(-p_1+ p_2)
e^{i(-p_1+p_2)\cdot x}
\end{multline*}
\[
p_1, p_2 \in \mathscr{O}_{{}_{m,0,0,0}}
\]
do not exist as tempered distributions 
in the space-time variable $x$ for causal perturbative spinor QED with the Hida operators as the creation-annihilation operators and
with massless spinor field ($m = 0$). 
\qed
\end{lem*}

Analogous Lemma holds for other QED's with massless charged fields. This is because the commutation functions of the
various type of charged free fields are equal to the action of invariant linear differential operators on one and the same 
commutation function of the scalar free field. 

Thus, by the last two existence/nonexistence Lemmas, the Rule VI, 
Lemma \ref{S*Xi} of this Subsection and the generalization of Theorems 3.6, 3.9 of \cite{obataJFA} to the general Fock space, 
including the Fermi case, compare Subsection \ref{psiBerezin-Hida} (in \cite{obataJFA} it is treated the Bose case),
we obtain
the last Theorem for the existence of the adiabatic limit of interacting fields in QED with massive charged field
and nonexistence of this limit if the charged field is massless. 

\vspace*{2cm}

Summing up:

\vspace{1cm}

\begin{twr}
For causal perturbative spinor QED on the Minkowski space-time with the Hida operators as the creation-annihilation 
operators and with massive charged field the higher order contributions to
interacting fields in the adiabatic limit $g\rightarrow 1$ are well-defined as sums of generalized integral kernel operators
with vector valued kernels in the sense of Obata, and this is the case only for the ``natural''
choice in the Epstein-Glaser splitting in the construction of the scattering operator.
\label{ExistenceIntFields.g=1.m>0}
\end{twr}

\vspace{0.5cm}

But:

\vspace{0.5cm}

\begin{twr}
For causal perturbative QED on the Minkowski space-time with the Hida operators as 
the creation-annihilation operators and with massless charged field the higher order contributions to
interacting fields  in the adiabatic limit $g\rightarrow 1$ are not well-defined, 
even as sums of generalized integral kernel operators in the sense of Obata, and for
no choice in the Epstein-Glaser splitting in the construction of the scattering operator.
\label{NonExistenceIntFields.g=1.m=0}
\end{twr}

\vspace{0.5cm}

Essentially, the same argument allows us to prove the following:

\vspace{1cm}

\begin{twr}
For causal perturbative QFT on the Minkowski space-time with each of the Wick 
monomials in the interaction Lagrangian $\mathcal{L}$ containing at most one massless field,
with the Hida operators as the creation-annihilation 
operators, the higher order contributions to
interacting fields in the adiabatic limit $g\rightarrow 1$ are well-defined as sums of generalized integral kernel operators
with vector valued kernels in the sense of \cite{obataJFA}, and this is the case only for the ``natural'', \emph{i.e.}
``on mass-shell'' normalization, in the Epstein-Glaser splitting in the construction of the scattering operator.
\label{ExistenceIntFields.g=1.m>0QFT}
\end{twr}

\vspace{0.5cm}

But:

\vspace{0.5cm}

\begin{twr}
For causal perturbative QFT on the Minkowski space-time with some of the Wick 
monomials in the interaction Lagrangian $\mathcal{L}$ containing more than one massless field,
with the Hida operators as the creation-annihilation 
operators, the higher order contributions to
interacting fields in the adiabatic limit $g\rightarrow 1$ are not well-defined as sums of generalized integral kernel operators
with vector valued kernels, and for
no choice in the Epstein-Glaser splitting in the construction of the scattering operator.
\label{NonExistenceIntFields.g=1.m=0QFT}
\end{twr}

\vspace{0.5cm}

We obtain as a corollary that Yang-Mills gauge field coupled to a charged matter fields
should be trated only in the broken phase, with the spontaneously broken symmetry
in order to avoid the singularities of Theorem \ref{NonExistenceIntFields.g=1.m=0QFT}. 
But moreover, we obtain a further corollary conerning the neutrino mass, when applying   
Theorems \ref{ExistenceIntFields.g=1.m>0QFT} and \ref{NonExistenceIntFields.g=1.m=0QFT}
to the Glashow-Weinberg-Salam system of inteacting fields with broken symmetry. In this system
we still have the Wick monomials in the interaction Lagrangian $\mathcal{L}$, which are of second
order in the neutrino fields. In order to avoid singularities of Theorem \ref{NonExistenceIntFields.g=1.m=0QFT},
we should assume nonzero mass of the neutrino.

Whether the gluons have zero mass or not is experimentally 
open question for now. 
But also, rejecting the QFT's of the class of Theorem \ref{NonExistenceIntFields.g=1.m=0QFT} as nonphysical,
we expect that also in QCD we should have the spontaneous symmetry breaking with
nonzero gauge fields. 

Based on this assumption, we also arrive at the conclusion that, e.g., the scalar QED is nonphysical. 
This is in agreement with the experiment, as the only observed spin-zero, electrically charged particles,
are not elementary and unstable.

In this way we get a theoretical proof, within causal perturbative QED on the Minkowski space-time
with Hida operators as the creation-annihilation operators, 
that charged particles 
should be massive.

The behavior of each higher order term $S_n$ to the scattering operator, evaluated at $g\in \mathscr{E}$
\[
S_n\big(g \mathcal{L}\big) = 
\int \ud^4 x_1 \cdots \ud^4 x_n S_n(x_1, \ldots, x_n) \,
g(x_1) \cdots g(x_n), \,\,\, g\in \mathscr{E}
\]
need not be analyzed separately, and its behavior can be inferred from the behavior of the higher order contributions
to the interacting fields
\[
\mathbb{A}_{{}_{\textrm{int}}}(x) = \mathbb{A}_{{}_{\textrm{int}}}(g,x),
\]
where
\begin{multline*}
\mathbb{A}_{{}_{\textrm{int}}}(g,x) = S^{-1}(g\mathcal{L})
\frac{\delta S(g\mathcal{L}+h\mathbb{A})}{\delta h(x)}\Bigg{|}_{{}_{h=0}}
\\
= \mathbb{A}(x) + \sum \limits_{n=1}^{\infty} {\textstyle\frac{1}{n!}}
\int \ud^4 x_1 \cdots \ud^4 x_n \mathbb{A}^{(n)}(x_1, \ldots, x_n,x) \,
g(x_1) \cdots g(x_n),
\end{multline*}
and where $\mathbb{A}$ is the free electromagnetic field or the free Dirac field. Indeed, the behavior of the higher order contributions
to the scattering operator $S_n\big(g \mathcal{L}\big)$ can be obtained, e. g. by putting for the
free electromagnetic potential operator $\mathbb{A}^\mu(x) = A^\mu(x) = \boldsymbol{1}$ into the formula for
$\mathbb{A}_{{}_{\textrm{int}}}(g,x) = \mathbb{A}_{{}_{\textrm{int}}}(g,x)$, with arbitrary $g \in \mathscr{E}$.
In particular from the last
Theorem (or repeated application of Lemma \ref{S*Xi}) it follows the following
\begin{cor*}
For each fixed $g \in \mathscr{E}$, and $n \in \mathbb{N}$
\[
S_n\big(g \mathcal{L}\big) \in \mathscr{L}((\boldsymbol{E}), \, (\boldsymbol{E})^*)
\]
and the map
\[
\mathscr{E} \ni g \longmapsto S_n\big(g \mathcal{L}\big) \in \mathscr{L}((\boldsymbol{E}), \, (\boldsymbol{E})^*)
\]
is continuous.
\end{cor*}

The last Corollary is sufficient for the computation of the effective cross-sections for the \emph{in} and \emph{out}
states which are of the form of many particle plane-wave states, in the adiabatic limit $g=1$, as we have explained in Introduction,
without any need for handling infrared of ultraviolet infinities.
The reader is encouraged to consult the computation of the effective 
cross-section presented in \cite{Bogoliubov_Shirkov}, \S\S 24.5 and 25 
or in Subsection \ref{EffCrossSection} of the present work.

Note that $S_n(x_1, \ldots, x_n)$ in the last Corollary can be look upon as a definition of the
''chronological products'' $S_n(x_1, \ldots, x_n) = T(i\mathcal{L}(x_1) \cdots i\mathcal{L}(x_1))$
of the Lagrange interaction density $\mathcal{L}(x)$, evaluated at $x_1, \ldots, x_n$. Even more,
each higher order therm $S_n(x_1, \ldots, x_n)$ in the causal perturbation series has the form of sums of normally
ordered products (finite sums of integral kernel operators), which can heuristically be looked at as if the
''Wick theorem for chronological product'' had automatically been done in the causal construction of $S_n(x_1, \ldots, x_n)$.
Note that in the last Corollary the ''chronological products'' $S_n(x_1, \ldots, x_n) = T(i\mathcal{L}(x_1) \cdots i\mathcal{L}(x_1))$
depend in fact on the particular choices in the Epstein-Glaser splitting of the
causally supported tempered distributions performed in each inductive step of the causal construction of the chronological product $S_n(x_1, \ldots, x_n)$,
and the last Theorem and its Corollary hold true for each of the particular choices in the splittings.
This also shows that the axioms (I)-(IV) of Subsection \ref{psiBerezin-Hida} do not determine $S_n(x_1, \ldots, x_n)$
uniquely but only within the flexibility in decomposition of the causal tempered distributions into retarded and advanced parts,
defined by the pairing functions of the theory in question. This arbitrariness can be further eliminated by the requirement posed on the
interacting fields, which should respect the corresponding equations of motion, compare in particular \cite{DKS1}, \cite{DutFred}.
Also, the computation of the splitting is involved into the analysis of the quasi-asymptotics, which unfortunately need to be
performed separately at each order.

For this reason we give in Subsection \ref{WickForChronological} a construction of a particular and ''natural'' example
of chronological product (which is equivalent to making particular choices in the splitting) and which respects (I)-(IV),
and which is closely motivated by the heuristic definition of the ''chronological product'' used in \cite{Bogoliubov_Shirkov}.
This will allow us to avoid the analysis of quasi asymtotics, and provide a new effective method for the computation
of the perturbative series for $S_n(x_1, \ldots, x_n)$.
In Subsection \ref{WickForProduct} we also give again a proof of a strengthened version of the last Corollary for the chronological
product constructed from the axioms (I)-(IV) and using Epstein-Glaser splitting.

{\bf REFERENCES}. \,
The essential part of this Section, especially the application of the Hida operators and white noise calculus for the elimination 
of the freedom in the splitting of causal distributions $D_{{}_{(n)}}$ into retarded and advanced parts (elimination in the choice of renormalization)
and the assiociated mass relations subsumed in Theorems \ref{ExistenceIntFields.g=1.m>0} and \ref{NonExistenceIntFields.g=1.m=0}
for QED, has been published in \cite{wawrzyckiIF}. In \cite{wawrzyckiYM} we give further perspective for the extension of this
results on the Yang-Mills fields coupled minimally to charged fields with asymptotic freedom (including the case
with spontaneous symmetry breaking).  

That Theorems \ref{NonExistenceIntFields.g=1.m=0}, \ref{NonExistenceIntFields.g=1.m=0QFT} 
should hold on the Minkowski space-time has been suggested to me by prof. D. Kazakov, 
when he saw the analogue result (Thm. \ref{InteractingFieldsAtxOnEU}
and its Corollary, Subection \ref{CausalSonEU}), that I have proved for the  Einstein Universe space-time. 
Moreover, he turned my attention to the fact that the same argument should provide a theoretical 
proof that neutrino has nonzero mass.

\section{Wick's theorem for ``products'' and the Scattering Operator}\label{WickForProduct}

Finally, let us return to the Wick product theorem for free fields. In the intermediate stage of the computations
of the scattering operator and interacting fields a so-called Wick theorem (\cite{Bogoliubov_Shirkov}, \S 17.2) is used for
decomposition of the ``product''
\begin{equation}\label{Wick(x)Wick(y)}
\boldsymbol{{:}} \mathbb{A}_{{}_{1}}^{a}(x) \ldots \mathbb{A}_{{}_{N}}^{a}(x) \boldsymbol{{:}} \,
\boldsymbol{{:}} \mathbb{A}_{{}_{N+1}}^{b}(y) \ldots \mathbb{A}_{{}_{M}}^{b}(y) \boldsymbol{{:}}
\end{equation}
of Wick product monomials
\begin{equation}\label{WickMonomialsInGeneralFreeFields}
\boldsymbol{{:}} \mathbb{A}_{{}_{1}}^{a}(x) \ldots \mathbb{A}_{{}_{N}}^{a}(x) \boldsymbol{{:}}
\,\,\,
\textrm{and}
\,\,\,
\boldsymbol{{:}} \mathbb{A}_{{}_{N+1}}^{b}(y) \ldots \mathbb{A}_{{}_{M}}^{b}(y) \boldsymbol{{:}}
\end{equation}
in fixed components $\mathbb{A}_{{}_{k}}$ of free fields, each separately evaluated at the same space-time point $x$ or respectively, $y$,
into the sum of Wick monomials (each in the so called ''normal order'').

The point lies in the correct definition of such ''product'', because each factor evaluated respectively at $x$ or $y$,
represents a generalized integral kernel operator transforming continuously the Hida space $(\boldsymbol{E})$ into its strong dual
$(\boldsymbol{E})^*$, so that the product cannot be understood as ordinary operator composition, and therefore a correct definition is here required.
Recall that $(\boldsymbol{E}) = (E_1) \otimes \ldots \otimes (E_M)$ is the Hida test space in the total Fock space of the free fields involved in the product (in fact
in the total Fock space of all free fields underlying the QFT in question), and
\[
E_k \subset \mathcal{H}_k \subset E_{k}^{*}
\]
are the single particle Gelfand triples of the corresponding fields $\mathbb{A}_{{}_{k}}$ , $k=1, \ldots, M$, with the corresponding
second quantized versions of the Gelfand triples
\[
(E_k) \subset \Gamma(\mathcal{H}_k) \subset (E_{k})^{*}
\]
and Hida test spaces $(E_k)$.

\subsection{Product of monomials of massive free fields}

The crucial point is that the free fields and their Wick products define (finite sums of) integral kernel operators with vector-valued
kernels in the sense of \cite{obataJFA}, as we have explained above, and the ''product'' can be given as a distributional
kernel operator. Indeed, from what we have already shown, it follows that each factor (\ref{WickMonomialsInGeneralFreeFields}) separately
represents an integral kernel operator which belongs to
\[
\mathscr{L}\big(\mathscr{E}_{n_1}, \, \mathscr{L}((\boldsymbol{E}), \, (\boldsymbol{E})) \big), \,\,\, i =1 \, \textrm{or} \, 2,
\]
if among the factors $\mathbb{A}_{{}_{k}}$, $1 \leq k \leq M$, there are no massless fields (or their derivatives).
This means that the first factor in (\ref{WickMonomialsInGeneralFreeFields}) defines the corresponding continuous map
\[
\mathscr{E}_{n_1} \ni \phi \longmapsto \Xi'(\phi) =
\sum\limits_{\substack{\ell',m' \\ \ell'+m'=N}}
\Xi'_{\ell',m'}\big(\kappa'_{\ell',m'}(\phi)\big)
\in \mathscr{L}\big( (\boldsymbol{E}), (\boldsymbol{E})\big), \,\,\, i =1 \, \textrm{or} \, 2,
\]
and similarly the second factor in (\ref{WickMonomialsInGeneralFreeFields}) defines continuous map
\[
\mathscr{E}_{n_2} \ni \varphi \longmapsto \Xi''(\varphi) =
\sum\limits_{\substack{\ell'',m'' \\ \ell''+m''=M-N}}
\Xi''_{\ell'',m'''}\big({\kappa''}_{\ell'',m''}(\varphi)\big)
\in \mathscr{L}\big( (\boldsymbol{E}), (\boldsymbol{E})\big),
\]
where $\mathscr{E}_{n_i} = \mathcal{S}_{{}_{B_{p_i}}}(\mathbb{R}^4; \mathbb{C}^4)$
$i= 1,2$, $p_i, p_j= 1,2$ (compare Subsection \ref{psiBerezin-Hida} for the definition of the standard operators
$B_{p_i}$ on the corresponding standard Hilbert spaces). Both factors $\Xi'$ and $ \Xi'''$
are equal to finite sums of integral kernel operators with $\mathscr{E}_{n_1}^{*}$- or $\mathscr{E}_{n_2}^{*}$-valued distributional kernels
$\kappa'_{\ell',m'}, \kappa''_{\ell'',m''}$. In this case both factors $\Xi'(\phi)$ and $ \Xi''(\varphi)$, when evaluated at the test functions $\phi, \varphi$,
are ordinary operators on the Fock space transforming continuously the Hida space $(\boldsymbol{E})$ into itself,
and thus can be composed $\Xi'(\phi) \circ \Xi''(\varphi)$ as operators, giving the composition operator
\[
\Xi'(\phi) \circ \Xi''(\varphi) \in \mathscr{L}\big( (\boldsymbol{E}), (\boldsymbol{E})\big),
\]
defining the map
\[
\mathscr{E}_{n_1} \otimes \mathscr{E}_{n_2} \ni \phi \otimes \varphi \longmapsto \Xi'(\phi) \circ \Xi''(\varphi)
\in \mathscr{L}\big( (\boldsymbol{E}), (\boldsymbol{E})\big),
\]
which by construction is separately continuous in the arguments $\phi \in \mathscr{E}_{n_1}$ and $\varphi \in \mathscr{E}_{n_2}$. Because
$\mathscr{E}_{n_1}, \mathscr{E}_{n_2}$ are complete Fr\'echet spaces, then by Proposition 1.3.11 of \cite{obataJFA} there exist
the corresponding operator-valued continuous map (say operator-valued distribution)
\begin{multline}\label{Xi(phi)circXi(varphi)distribution}
\phi \otimes \varphi \longmapsto
\Xi(\phi \otimes \varphi) \overset{\textrm{df}}{=} \\ \overset{\textrm{df}}{=}
\sum\limits_{a,b}
\int\limits_{\big[\mathbb{R}^4\big]^{\times \, 2}}
\boldsymbol{{:}} \mathbb{A}_{{}_{1}}^{a}(x) \ldots \mathbb{A}_{{}_{N}}^{a}(x) \boldsymbol{{:}} \,
\boldsymbol{{:}} \mathbb{A}_{{}_{N+1}}^{b}(y) \ldots \mathbb{A}_{{}_{M}}^{b}(y) \boldsymbol{{:}} \,
\phi^a \otimes \varphi^b(x,y) \, \ud^4x \, \ud^4 y
\\
= \Xi'(\phi) \circ \Xi''(\varphi).
\end{multline}
In particular the operator map $\phi \otimes \varphi \mapsto \Xi(\phi \otimes \varphi )$ defines a
generalized operator
\[
\Xi \in \mathscr{L}\big(\mathscr{E}_{n_1} \otimes \mathscr{E}_{n_2}, \, \mathscr{L}((\boldsymbol{E}), \, (\boldsymbol{E}) \big)
\cong \mathscr{L}((\boldsymbol{E}) \otimes \mathscr{E}_i \otimes \mathscr{E}_j, \, (\boldsymbol{E}))
\]
which by Theorem 4.8 of \cite{obataJFA} possesses unique Fock expansion
\begin{equation}\label{FockExpansionOfWick(x)Wick(y)}
\Xi = \sum\limits_{\ell, m} \Xi_{\ell, m}({\kappa}_{\ell, m}),
\end{equation}
into integral kernel operators $\Xi_{\ell, m}({\kappa}_{\ell, m})$
with $\mathscr{E}_{n_1}^{*} \otimes \mathscr{E}_{n_2}^{*} = \mathscr{L}(\mathscr{E}_{n_1} \otimes \mathscr{E}_{n_2}, \mathbb{C})$-valued
kernels ${\kappa}_{\ell, m}$. This Fock expansion (\ref{FockExpansionOfWick(x)Wick(y)}) becomes finite, with $\ell+ m \leq M$ with
one pair of indices $(\ell, m)=(0,0)$
in it, and its computation can be easily reduced to the commutation operation for Hida monomials (\ref{...Pariali...Partial*j...})
if among the integral kernel operators $\Xi_{\ell, m}({\kappa}_{\ell, m})$, we count for also the scalar
integral kernel operator
\[
\Xi_{0, 0}(\kappa_{0,0}(\phi \otimes \varphi)) = \kappa_{0,0}(\phi \otimes \varphi) \, \boldsymbol{1}
= \langle \kappa_{0,0}, \phi \otimes \varphi \rangle \, \boldsymbol{1},
\]
with the kernel $\kappa_{0, 0} \in \mathscr{E}_{n_1}^{*} \otimes \mathscr{E}_{n_{2}}^{*}$, $i = 1,2$, as well as the remaining kernels $\kappa_{\ell,m}$
of the Fock expansion, determined by the kernels $\kappa_{0,1}, \kappa_{1,0}$ defining the free fields $\mathbb{A}_{{}_{k}}$ involved into the Wick product. 

This gives us the Wick theorem for the ``product'' (\ref{Wick(x)Wick(y)}) in case in which all factors
$\mathbb{A}_{{}_{k}}$ are massive free fields or their derivatives.

Indeed, the finite Fock expansion (\ref{FockExpansionOfWick(x)Wick(y)}) , \emph{i.e.} ''Wick theorem for tensor product (\ref{Wick(x)Wick(y)})'',
immediately follows from the ''Wick theorem for monomials (\ref{...Pariali...Partial*j...}) of Hida operators, in which there are present Bose- 
as well as the Fermi-Hida operators, correspondingly to the free field factors $\mathbb{A}_{{}_{k}}$, $1 \leq k \leq M$ in the product (\ref{Wick(x)Wick(y)}).
As we have shown in Subsection \ref{psiBerezin-Hida}, the monomials (\ref{...Pariali...Partial*j...})
\begin{equation}\label{...ai...a^+j...}
\ldots a_{s_i}(\boldsymbol{p}_i) \ldots a_{s_j}(\boldsymbol{p}_j)^{+} \ldots
\end{equation}
in Hida annihilation-creation operators $\partial_{{}_{s_{i}, \, \boldsymbol{p}_{i}}} = a_{s_{i}}(\boldsymbol{p}_{i})$,
$\partial_{{}_{s_{i}, \, \boldsymbol{p}_{i}}}^{*} = a_{s_{i}}(\boldsymbol{p}_{i})^{+}$, are well-defined
operator distributions, or kernels of continuous operators
\begin{equation}\label{...Ei*...Ej...}
E_1 \otimes \ldots \otimes E_M \longrightarrow \mathscr{L}\big((\boldsymbol{E}), (\boldsymbol{E}) \big),
\end{equation}
which can be extended to continuous maps
\begin{equation}\label{...ai...aj+...:cont:...Ei*...Ej...->(Hida->Hida)}
\ldots \otimes E_{i}^{*} \otimes \ldots \otimes E_j \ldots \longrightarrow \mathscr{L}\big((\boldsymbol{E}), (\boldsymbol{E}) \big)
\end{equation}
on duals $E_{i}^{*}$ in (\ref{...Ei*...Ej...}) for all those factors $E_{i}$ and the corresponding variables $(s_{i}, \, \boldsymbol{p}_{i})$, which correspond
to the annihilation Hida operators $\partial_{{}_{s_{i}, \, \boldsymbol{p}_{i}}} = a_{s_{i}}(\boldsymbol{p}_{i})$ in (\ref{...ai...a^+j...}),
compare the extendibility property (\ref{...Pariali...Partial*j...:continuous...E*...E...->(Hida->Hida)}) of Subsection \ref{psiBerezin-Hida}.
The canonical commutation rules imply that each such monomial (\ref{...ai...a^+j...}) in Hida operators can be written
as finite sum of normally ordered Hida operators, or decomposed into integral kernel operators with normally ordered Hida operators,
compare Subsection \ref{psiBerezin-Hida}. In particular
\begin{multline}\label{aia^+j = a^+jai+pairing}
a_{s'_{i}}(\boldsymbol{p}'_{i}) a_{s''_{j}}(\boldsymbol{p}''_{j})^{+} = \mp a_{s''_{j}}(\boldsymbol{p}''_{j})^{+} a_{s'_{i}}(\boldsymbol{p}'_{i})
+ \big[ a_{s'_{i}}(\boldsymbol{p}'_{i}), a_{s''_{j}}(\boldsymbol{p}''_{j})^{+}\big]_{\pm}
\\
=
\mp a_{s''_{j}}(\boldsymbol{p}''_{j})^{+} a_{s'_{i}}(\boldsymbol{p}'_{i})
+ \delta_{s'_{i} \,\, s''_{j}} \, \delta(\boldsymbol{p}'_{i} - \boldsymbol{p}''_{j}).
\end{multline}
This formula can be easily extended over to the case of more than just two Hida factors, compare Subsection \ref{psiBerezin-Hida}.
We joint this fact with the observation that
the composition $\Xi'(\phi) \circ \Xi''(\varphi)$ (compare (\ref{Xi(phi)circXi(varphi)distribution})) of the integral kernel operators
\[
\Xi'(\phi) =
\sum\limits_{\substack{\ell',m' \\ \ell'+m'=N}}
\Xi'_{\ell',m'}\big(\kappa'_{\ell',m'}(\phi)\big)
\]
and
\[
\Xi''(\varphi) =
\sum\limits_{\substack{\ell'',m'' \\ \ell''+m''=M-N}}
\Xi''_{\ell'',m''}\big({\kappa''}_{\ell'',m''}(\varphi)\big)
\]
also defines a well-defined distribution. In particular consider a concrete additive term
$\Xi'_{\ell',m'}\big(\kappa'_{\ell',m'}(\phi)\big) \circ \Xi''_{\ell'',m''}\big({\kappa'}_{\ell'',m''}(\varphi)\big)$
of the total sum $\Xi'(\phi) \circ \Xi''(\varphi)$. It follows that
\[
\mathscr{E}_{n_1} \otimes \mathscr{E}_{n_2} \ni
\phi \otimes \varphi \longmapsto \Xi'_{\ell',m'}\big(\kappa'_{\ell',m'}(\phi)\big) \circ \Xi''_{\ell'',m''}\big({\kappa''}_{\ell'',m''}(\varphi)\big)
\in \mathscr{L} \big((\boldsymbol{E}), (\boldsymbol{E}) \big)
\]
is continuous and $\Xi'(\phi) \circ \Xi''(\varphi)$ is equal to the finite sum of the following operators
\begin{multline}\label{X(kappa')circXi(kappa'')Full}
\Xi'_{\ell',m'}\big(\kappa'_{\ell',m'}(\phi)\big) \circ \Xi''_{\ell'',m''}\big({\kappa''}_{\ell'',m''}(\varphi)\big)
\\
=
\sum \limits_{s'_{i}, s'_{j}}
\int
\kappa'_{\ell',m'}(\phi)(s'_{1}, \boldsymbol{\p}'_{1}, \ldots, s'_{l'+m'}, \boldsymbol{\p}'_{l'+m'})
\kappa''_{\ell'',m''}(\phi)(s''_{1}, \boldsymbol{\p}''_{1}, \ldots, s''_{l''+m''}, \boldsymbol{\p}''_{l''+m''}) \,\,\,
\times
\\
\times \,\,\,
a_{s'_{1}}(\boldsymbol{p}'_{1})^{+} \cdots a_{s'_{l'}}(\boldsymbol{p}'_{l'})^{+}
a_{s'_{l'}}(\boldsymbol{p}'_{l'+1}) \cdots a_{s'_{l'+m'}}(\boldsymbol{p}'_{l'+m'}) \,\,\, \times
\\
\times \,\,\,
a_{s''_{1}}(\boldsymbol{p}''_{1})^{+} \cdots a_{s''_{l''}}(\boldsymbol{p}''_{l''})^{+}
a_{s''_{l''}}(\boldsymbol{p}''_{l''+1})
\cdots a_{s'_{l'+m'}}(\boldsymbol{p}''_{l''+m''})
\,\,\, \times
\\
\times \,\,\,
\ud^3 \boldsymbol{\p}'_{1} \cdots \ud^3 \boldsymbol{\p}'_{l'+m'} \ud^3 \boldsymbol{\p}''_{1} \dots \ud^3 \boldsymbol{\p}''_{l''+m''}
\end{multline}
with $\kappa'_{\ell',m'}$ ranging over the finite set of kernels of the Fock expansion of the
operator $\Xi'$, and with $\kappa''_{\ell'',m''}$ ranging over the finite set of kernels of the Fock expansion of the
operator $\Xi''$. 

Note that according to Subsection \ref{OperationsOnXi}
\begin{multline*}
\kappa'_{\ell',m'} \in \mathscr{L}(\mathscr{E}_{n_1}, \, E_1 \otimes \ldots E_{l'} \otimes E_{l'+1}^{*} \otimes \ldots E_{l'+m'}^{*})
\\
\cong \mathscr{L}( E_{1}^{*} \otimes \ldots E_{l'}^{*} \otimes E_{l'+1} \otimes \ldots E_{l'+m'}, \, \mathscr{E}_{n_1}^{*}),
\,\,\,\, l'+m'=N
\end{multline*}
and 
\begin{multline*}
\kappa''_{\ell'',m''} \in \mathscr{L}(\mathscr{E}_{n_2}, \, E_{N+1} \otimes \ldots E_{N+l''} \otimes E_{N+l''+1}^{*} \otimes \ldots E_{N+l''+m''}^{*})
\\
\cong \mathscr{L}( E_{N+1}^{*} \otimes \ldots  E_{N+l''}^{*} \otimes E_{N+l''+1} \otimes \ldots E_{N+l''+m''}, \, \mathscr{E}_{n_{2}}^{*}),
\,\,\,\, N+l''+m'' = M.
\end{multline*}
Therefore
\begin{equation}\label{kappa'InExE*}
\kappa'_{\ell',m'}(\phi) \in  E_1 \otimes \ldots E_{l'} \otimes E_{l'+1}^{*} \otimes \ldots E_{l'+m'}^{*},
\,\,\,\,\,\,\, l'+m'=N
\end{equation}
and 
\begin{equation}\label{kappa''InExE*}
\kappa''_{\ell'',m''}(\varphi) \in E_{N+1} \otimes \ldots E_{N+l''} \otimes E_{N+l''+1}^{*} \otimes \ldots E_{N+l''+m''}^{*},
\,\,\,\,\,\,\, N+l''+m'' = M.
\end{equation}

The operator expression (\ref{X(kappa')circXi(kappa'')Full}) is well-defined in accordance to the extendibility property
(\ref{...ai...aj+...:cont:...Ei*...Ej...->(Hida->Hida)}). Indeed, note that by (\ref{kappa'InExE*}) and (\ref{kappa''InExE*})
the variables of the kernel in (\ref{X(kappa')circXi(kappa'')Full}) corresponding to Hida creation operators
correspond to smooth factors $ E_1 \otimes \ldots E_{l'}$ and $ E_{N+1} \otimes \ldots E_{N+l''}$, and the variables corresponding
to the annihilation Hida operators correspond to the distributional factors $E_{l'+1}^{*} \otimes \ldots E_{l'+m'}^{*}$
and $E_{N+l''+1}^{*} \otimes \ldots E_{N+l''+m''}^{*}$, in accordance with (\ref{...ai...aj+...:cont:...Ei*...Ej...->(Hida->Hida)}),
so (\ref{X(kappa')circXi(kappa'')Full}) is well-defined. Because $\Xi'(\phi)$ and $\Xi''(\varphi)$ are ordinary operators
they can be composed $\Xi'(\phi) \circ \Xi''(\varphi)$, and it can be proved by a method applied in the proof of the
Bogoliubov-Shirkov Quantization Postulate in Subsection \ref{BSH}, that (\ref{X(kappa')circXi(kappa'')Full})
is indeed equal $\Xi'(\phi) \circ \Xi''(\varphi)$.

Note that in the expression (\ref{X(kappa')circXi(kappa'')Full}) we need apply the commutation relation only to those pairs of
Hida operators $a_{s'_{i}}(\boldsymbol{p}'_{i})$ and $a_{s''_{j}}(\boldsymbol{p}''_{j})^{+}$ with
$(s'_{i},\boldsymbol{p}'_{i})$ lying among the last $m'$ variables of the kernel $\kappa'_{\ell',m'}(\phi)$
and with $(s''_{j},\boldsymbol{p}''_{j})$ lying among the first $l''$ variables of the kernel $\kappa''_{\ell'',m''}(\phi)$,
and with both $a_{s'_{i}}(\boldsymbol{p}'_{i})$ and $a_{s''_{j}}(\boldsymbol{p}''_{j})^{+}$ corresponding to the same
free field factor $\mathbb{A}_{{}_{k}}$, \emph{i.e.} not commuting (or anti-commuting in case of the Fermi field):
\begin{multline}\label{X(kappa')circXi(kappa'')Short}
\Xi'_{\ell',m'}\big(\kappa'_{\ell',m'}(\phi)\big) \circ \Xi''_{\ell'',m''}\big({\kappa''}_{\ell'',m''}(\varphi)\big)
\\
=
\sum \limits_{s'_{i}, s'_{j}}
\int
\kappa'_{\ell',m'}(\phi)(\ldots, s'_{i}, \boldsymbol{\p}'_{i}, \ldots )
\kappa''_{\ell'',m''}(\varphi)(\ldots, s''_{j}, \boldsymbol{\p}''_{j}, \ldots) \,\,\,
\times
\\
\times \,\,\,
\cdots a_{s'_{i}}(\boldsymbol{p}'_{1}) \cdots
\cdots a_{s''_{j}}(\boldsymbol{p}''_{j})^{+} \cdots
\,\,\, \times
\\
\times \,\,\,
\cdots \ud^3 \boldsymbol{\p}'_{i} \cdots \ud^3 \boldsymbol{\p}''_{j} \cdots.
\end{multline}
We see from (\ref{aia^+j = a^+jai+pairing}) that each such pair of Hida operators $a_{s'_{i}}(\boldsymbol{p}'_{i})$ and $a_{s''_{j}}(\boldsymbol{p}''_{j})^{+}$
in (\ref{X(kappa')circXi(kappa'')Short}) gives two terms, first which puts them into the normal order and the second commutation/anticommutation
term, which actually acts as the ordinary contraction of the variables $(s'_{i},\boldsymbol{p}'_{i})$ and $(s''_{j},\boldsymbol{p}''_{j})$
corresponding to the considered pair of Hida operators. This contraction is well-defined because by (\ref{kappa'InExE*})
and (\ref{kappa''InExE*}) the kernel $\kappa'_{\ell',m'}(\phi)$ belongs to $E_{i}^{*}$ in the variable $(s'_{i},\boldsymbol{p}'_{i})$
and the kernel $\kappa''_{\ell'',m''}(\varphi)$ belongs to $E_{j}$ in the variable $(s''_{j},\boldsymbol{p}''_{j})$
so the contraction can indeed be performed in (\ref{X(kappa')circXi(kappa'')Short}).

In particular the contribution to the scalar term $\Xi_{0,0}(\kappa_{0,0}(\phi \otimes \varphi))$ coming from 
\[
\Xi'_{\ell',m'}\big(\kappa'_{\ell',m'}(\phi)\big) \circ \Xi''_{\ell'',m''}\big({\kappa''}_{\ell'',m''}(\varphi)\big)
\]
is non-zero if and only if $\ell'=0$, $m''=0$, $m'=l''$ and to each of the Hida operators $a_{s'_{i}}(\boldsymbol{p}'_{i})$
there correspond bi-uniquely a Hida operator $a_{s''_{j}}(\boldsymbol{p}''_{j})^{+}$ corresponding to the same free field factor.
Then the contribution to the kernel $\kappa_{0,0}(\phi\otimes\varphi)$ coming from this composition is equal to the contraction
(\emph{i.e.} dual pairing)
\[
\big\langle\kappa'_{0,m'}(\phi), {\kappa''}_{\ell''=m',0}(\varphi)\big\rangle,
\] 
and $\kappa_{0,0}(\phi\otimes\varphi) = \langle \kappa_{0,0}, \phi\otimes\varphi \rangle$ is equal to the sum of all
contractions
\[
\big\langle\kappa'_{0,m'}(\phi), {\kappa''}_{\ell''=m',0}(\varphi)\big\rangle,
\]
where $\kappa'_{0,m'},  {\kappa''}_{\ell''=m',0}$ range over the pairs of  kernels $\kappa'_{0,m'}, {\kappa''}_{\ell''=m',0}$, 
such that to each Hida operator $a_{s'_{i}}(\boldsymbol{p}'_{1})$ there exists exactly one Hida
operator  $a_{s''_{j}}(\boldsymbol{p}''_{j})^{+}$ corresponding to the same free field.
Of course in this sum the kernels $\kappa'_{0,m'}$ range over the kernels of the operators of the Fock expansion of $\Xi'$ 
and respectively the kernels $\kappa''_{l''=m',0}$ range over the kernels of the operators of the Fock expansion of $\Xi''$.
In particular this shows that indeed  $\kappa_{0,0} \in \mathscr{E}_{n_1}^{*} \otimes \mathscr{E}_{n_{2}}^{*}$, $n_i = 1,2$.

Note that before we apply the commutation formula (\ref{aia^+j = a^+jai+pairing}) to the indicated Hida operators
$a_{s'_{i}}(\boldsymbol{p}'_{i})$ and $a_{s''_{j}}(\boldsymbol{p}''_{j})^{+}$
in (\ref{X(kappa')circXi(kappa'')Short}), we need to put these operators next to each other. To achieve this we need to perform
several permutations of the annihilation Hida operator
$a_{s'_{i}}(\boldsymbol{p}'_{i})$ with the other annihilation operators, which correspond to the last $m'$ variables
of the kernel $\kappa'_{\ell',m'}(\phi)$, and similarly we need to perform several commutation operations
of the creation operator $a_{s''_{j}}(\boldsymbol{p}''_{j})^{+}$ with the other creation operators which correspond to the
first $l''$ variables of the kernel $\kappa''_{\ell'',m''}(\varphi)$. The creation Hida operators commute or anti-commute
between themselves,
and similarly the annihilation Hida operators commute or anti-commute between themselves. Then the operation of putting
$a_{s'_{i}}(\boldsymbol{p}'_{i})$ and $a_{s''_{j}}(\boldsymbol{p}''_{j})^{+}$ next to each other brings in only an
over all factor $(-1)^c$ with the power $c$ equal to the sum of all Fermi Hida commutations which need to be performed during the
operation of putting $a_{s'_{i}}(\boldsymbol{p}'_{i})$ and $a_{s''_{j}}(\boldsymbol{p}''_{j})^{+}$ next to each other.
In particular to each contraction corresponding to the second term in (\ref{aia^+j = a^+jai+pairing}) and to the indicated
above operators $a_{s'_{i}}(\boldsymbol{p}'_{i})$ and $a_{s''_{j}}(\boldsymbol{p}''_{j})^{+}$ in (\ref{X(kappa')circXi(kappa'')Short})
there correspond unique factor $(-1)^c$, by which we multiply the contraction. In particular
each term
\[
\big\langle\kappa'_{0,m'}(\phi), {\kappa''}_{\ell''=m',0}(\varphi)\big\rangle,
\]
of the sum of the contractions
representing the contribution to $\kappa_{0,0}(\phi\otimes\varphi)$ also need to be multiplied by an overall
factor $(-1)^c$ with $c$ equal to the total number of Fermi commutations involved in the computation of this contraction, so more
properly we should write
\[
\kappa_{0,0}(\phi\otimes\varphi) =
(-1)^c \,
\sum_{\substack{\kappa'_{0,m'} \\ \kappa''_{\ell''=m',0}}} \big\langle\kappa'_{0,m'}(\phi), {\kappa''}_{\ell''=m',0}(\varphi)\big\rangle,
\]
where $\kappa'_{0,m'}$ and $\kappa''_{\ell''=m',0}$ range over the kernels of the Fock expansions of $\Xi'$ and $\Xi''$.

Similarly, the contribution 
\[
\Xi_{\ell,m}\big(\kappa_{\ell,m}(x,y)\big) = \Xi_{\ell'+\ell'',m'+m''}\big(\kappa_{\ell'+\ell'',m'+m''}(x,y)\big),
\]
to the normal order term 
\[
\boldsymbol{{:}} \mathbb{A}_{{}_{1}}^{a}(x) \ldots \mathbb{A}_{{}_{N}}^{a}(x)
\mathbb{A}_{{}_{N+1}}^{b}(y) \ldots \mathbb{A}_{{}_{M}}^{b}(y)  \boldsymbol{{:}}
\]
coming from 
\[
\Xi'_{\ell',m'}\big(\kappa'_{\ell',m'}(\phi)\big) \circ \Xi''_{\ell'',m''}\big({\kappa''}_{\ell'',m''}(\varphi)\big)
\]
is computed if for all indicated pairs of Hida operators $a_{s'_{i}}(\boldsymbol{p}'_{i})$ and $a_{s''_{j}}(\boldsymbol{p}''_{j})^{+}$
in (\ref{X(kappa')circXi(kappa'')Short}) we choose the first normal order term contribution 
$\mp a_{s''_{j}}(\boldsymbol{p}''_{j})^{+}a_{s'_{i}}(\boldsymbol{p}'_{i})$ of (\ref{aia^+j = a^+jai+pairing}). 
In this way we obtain the contribution equal to the Wick product 
\[
\Xi_{\ell,m}\big(\kappa_{\ell,m}(\phi \otimes \varphi)\big) =
\Xi_{\ell'+\ell'',m'+m''}\big(\kappa'_{\ell',m'}\overline{\otimes}{\kappa''}_{\ell'',m''}(\phi \otimes \varphi)\big),
\]
or
\begin{equation}\label{kappa'timeskappa''}
\Xi_{\ell'+\ell'',m'+m''}\big(\kappa'_{\ell',m'}\overline{\otimes}{\kappa''}_{\ell'',m''}\big)
\in \mathscr{L}\big(\mathscr{E}_{n_1} \otimes \mathscr{E}_{n_{2}}, \,\, \mathscr{L}((\boldsymbol{E}), (\boldsymbol{E}))\big),
\end{equation}
\[
\kappa_{\ell,m} = \kappa_{\ell'+\ell'',m'+m''} = \kappa'_{\ell',m'}\overline{\otimes}{\kappa''}_{\ell'',m''},
\]
\[
\ell = \ell'+\ell'', \, m = m'+m'',
\]
of the integral kernel operators
\[
\Xi'_{\ell',m'}\big(\kappa'_{\ell',m'}\big) 
\,\,\,\,
\textrm{and}
\,\,\,\,
\Xi''_{\ell'',m''}\big({\kappa''}_{\ell'',m''}\big).
\] 
Similarly, we can explicitly compute all remaining terms $\Xi_{\ell,m}(\kappa_{\ell,m})$ of the finite Fock expansion
(\ref{FockExpansionOfWick(x)Wick(y)}) of the ``product'' (\ref{Wick(x)Wick(y)}), which are given by similar formula
as given in (\ref{kappa'timeskappa''}) with the eventual additional contractions of, say $q$ pairs,
of variables in $\kappa'_{\ell',m'}\overline{\otimes}{\kappa''}_{\ell'',m''}$:
\[
\kappa_{\ell,m} = \kappa_{\ell'+\ell''-q,m'+m''-q} = (-1)^{c(q)} \, \kappa'_{\ell',m'}\overline{\otimes_{q}} \, {\kappa''}_{\ell'',m''},
\]
\[
\ell = \ell'+\ell''-q, \, m = m'+m''-q.
\]

In this way we easily see e.g. that $\kappa_{0, 0} \in \mathscr{E}_{n_1}^{*} \otimes \mathscr{E}_{n_{2}}^{*}$, 
and more generally that for the kernels $\kappa_{\ell,m}$, $(\ell,m) \neq (0,0)$, of the finite Fock expansion (\ref{FockExpansionOfWick(x)Wick(y)})
\begin{multline*}
\kappa_{\ell,m} \in \mathscr{L}(\mathscr{E}_{n_1} \otimes \mathscr{E}_{n_2}, \, E_{j_1} \otimes \ldots E_{j_l} \otimes E_{j_{l+1}}^{*} \otimes \ldots 
E_{j_{l+m}}^{*})
\\
\cong \mathscr{L}( E_{j_1}^{*} \otimes \ldots E_{j_l}^{*} \otimes E_{j_{l+1}} \otimes \ldots E_{j_{l+m}}, \, \mathscr{E}_{n_1}^{*} \otimes  \mathscr{E}_{n_2}^{*}).
\end{multline*}

Thus, we can summarize the results in the Wick theorem for massive fields. Namely, if the fields $\mathbb{A}_{{}_{k}}$, $k=1, \ldots, M$
in (\ref{Wick(x)Wick(y)}) are massive, then the operators
\begin{eqnarray*}
\Xi'(\phi) = \sum \limits_{a} \int\limits_{\mathbb{R}^4}
\boldsymbol{{:}} \mathbb{A}_{{}_{1}}^{a}(x) \ldots \mathbb{A}_{{}_{N}}^{a}(x) \boldsymbol{{:}} \, \phi^a(x) \,\,\,\, \ud^4x
\in \mathscr{L}((\boldsymbol{E}),(\boldsymbol{E})),
\\
\Xi''(\varphi) = \sum \limits_{b} \int\limits_{\mathbb{R}^4}
\boldsymbol{{:}} \mathbb{A}_{{}_{N+1}}^{b}(y) \ldots \mathbb{A}_{{}_{M}}^{b}(y) \boldsymbol{{:}} \,\,\,\, \varphi^b(y) \, \ud^4y
\in \mathscr{L}((\boldsymbol{E}),(\boldsymbol{E})),
\\
\Xi(\phi \otimes \varphi) =
\sum \limits_{a,b}
\int\limits_{\big[\mathbb{R}^4\big]^{\times \, 2}}
\boldsymbol{{:}} \mathbb{A}_{{}_{1}}^{a}(x) \ldots \mathbb{A}_{{}_{N}}^{a}(x) \boldsymbol{{:}} \,\,
\boldsymbol{{:}} \mathbb{A}_{{}_{N+1}}^{b}(y) \ldots \mathbb{A}_{{}_{M}}^{b}(y) \boldsymbol{{:}} \,\, \times
\\ \times \,\,\,\,\,\,\,
\phi^a \otimes \varphi^b(x,y) \, \ud^4x \ud^4y \,\,\, \in \mathscr{L}((\boldsymbol{E}),(\boldsymbol{E})),
\end{eqnarray*}
and moreover
\begin{eqnarray*}
\Xi' \in \mathscr{L}(\mathscr{E}_{n_1}, \mathscr{L}((\boldsymbol{E}),(\boldsymbol{E}))\big),
\\
\Xi''\in \mathscr{L}(\mathscr{E}_{n_2}, \mathscr{L}((\boldsymbol{E}),(\boldsymbol{E}))\big),
\\
\Xi\in \mathscr{L}(\mathscr{E}_{n_1} \otimes \mathscr{E}_{n_2}, \mathscr{L}((\boldsymbol{E}),(\boldsymbol{E}))\big),
\end{eqnarray*}
and the following Wick theorem for massive fields holds
\begin{twr}
\[
\Xi =
\sum_{\substack{\kappa'_{\ell',m'} \\ \kappa''_{\ell'',m''}}} \sum \limits_{0\leq q \leq \textrm{min} \{m',\ell''\}} (-1)^{c(q)} \,
\Xi_{\ell'+\ell''-q,m'+m''-q}\big( \kappa'_{\ell',m'}\overline{\otimes_{q}} \, {\kappa''}_{\ell'',m''}\big),
\]
where the kernels $\kappa'_{\ell',m'}, \kappa''_{\ell'',m''}$ range, respectively, over the kernels of finite Fock expansions
of the operators $\Xi', \Xi''$. The (symmetrized/antisymmetrized) $q$-contractions $\overline{\otimes_{q}}$ are performed
upon the pairs of variables in which the first element of the contracted pair lies among the last $m''$ variables of the kernel
$\kappa'_{\ell',m'}$ and the second variable
of the contracted pair lies among the first $l''$ variables of the kernel $\kappa''_{\ell'',m''}$, and to both variables
of the contracted pair correspond respectively annihilation and creation operator of one and the same free field. The number
$c(q)$ is equal to the number of Fermi commutations performed in the contraction $\overline{\otimes_{q}}$.
\label{WickThmForMassiveFields}
\end{twr}

That e.g. $\kappa_{0,0}$ is indeed a well-defined distribution is not visible in the conventional formulation of the (formal/symbolic version of the) Wick theorem.
This is because the kernel $\kappa_{0,0}$ is given there in the form of the pointwise products (at one and the same point with respect to space-time variables)
of the kernels of the pairing functions, which are singular, \emph{i.e.} proper tempered distributions not equal to the Schwartz multipliers. Therefore,
it is not obvious how the combination of products of the kernels of the pairing positive and negative frequency functions
giving $\kappa_{0,0}$, indeed define well-defined distribution $\kappa_{0,0}$. The same difficulty we encounter in all terms of the
formal Wick expansion of (\ref{Wick(x)Wick(y)}) of the conventional approach, except the first term
\[
\boldsymbol{{:}} \mathbb{A}_{{}_{1}}^{a}(x) \ldots \mathbb{A}_{{}_{N}}^{a}(x)
\mathbb{A}_{{}_{N+1}}^{b}(y) \ldots \mathbb{A}_{{}_{M}}^{b}(y) \boldsymbol{{:}}
\]
which does not contain any ``pairings'' and except all the terms with at most one ``pairing''.
Therefore, in order to give a distributional sense to $\kappa_{0,0}$ given in the conventional approach
formally by the ``formal pointwise product''
of pairing distribution functions it is customary in physical literature to apply a procedure called ``regularization''. It is in fact the
``regularization'' process which gives the meaning to the particular terms of the formal Wick expansion of (\ref{Wick(x)Wick(y)})
and not the formal expansion itself summarized in the so called ``Wick theorem for product'' (\ref{Wick(x)Wick(y)})
in terms of ``pairings'', as given e.g. in \cite{Bogoliubov_Shirkov}, \S 17.2. There are several ``regularization'' prescriptions:
e.g. Feynman's regularization, t'Hooft-Veltman, dimensional, or Pauli-Villars regularization with the auxiliary massive fields, compare
\cite{Bogoliubov_Shirkov}.

Let us recall the 
conventional formulation of the (formal) Wick theorem with ``pairings''(compare \cite{Bogoliubov_Shirkov}, \S 17.2):
\begingroup\makeatletter\def\f@size{5}\check@mathfonts
\def\maketag@@@#1{\hbox{\m@th\large\normalfont#1}}%
\begin{multline*}
\boldsymbol{{:}} \mathbb{A}_{{}_{1}}^{a}(x) \ldots \mathbb{A}_{{}_{N}}^{a}(x) \boldsymbol{{:}} \,\,
\boldsymbol{{:}} \mathbb{A}_{{}_{N+1}}^{b}(y) \ldots \mathbb{A}_{{}_{M}}^{b}(y)  \boldsymbol{{:}}
\\
=
\boldsymbol{{:}} \mathbb{A}_{{}_{1}}^{a}(x)  \ldots \mathbb{A}_{{}_{N}}^{a}(x) 
 \mathbb{A}_{{}_{N+1}}^{b}(y) \ldots \mathbb{A}_{{}_{M}}^{b}(y)  \boldsymbol{{:}}
\\
+
\sum_{\substack{1\leq i \leq N \\ 1 \leq j \leq M-N}}
\boldsymbol{{:}} \mathbb{A}_{{}_{1}}^{a}(x) \ldots \quad \underbracket{\mathbb{A}_{{}_{i}}^{a}(x) \ldots \mathbb{A}_{{}_{N}}^{a}(x) 
 \mathbb{A}_{{}_{N+1}}^{b}(y) \ldots \mathbb{A}_{{}_{N+j}}^{b}}(y) \ldots \mathbb{A}_{{}_{M}}^{b}(y)  \boldsymbol{{:}}
\\
+
\sum_{\substack{1\leq i \leq N \\ 1 \leq j \leq M-N \\ 1 \leq k \leq N \\ 1 \leq n \leq M-N}} 
\boldsymbol{{:}} \mathbb{A}_{{}_{1}}^{a}(x) \ldots \quad \underbracket{\mathbb{A}_{{}_{i}}^{a}(x) \ldots 
\quad \underbracket{\mathbb{A}_{{}_{k}}^{a}(x) \ldots  \mathbb{A}_{{}_{N}}^{a}(x) 
 \mathbb{A}_{{}_{N+1}}^{b}(y) \ldots \mathbb{A}_{{}_{N+j}}^{b}}(y) \ldots \mathbb{A}_{{}_{N+n}}^{b}}(y)  \ldots \mathbb{A}_{{}_{M}}^{b}(y)  \boldsymbol{{:}}
\\
+ \ldots
\end{multline*}
\endgroup
Here we have the expansion of the tensor product operator distribution (\ref{Wick(x)Wick(y)}) into the Wick ordered terms,
first being the Wick product of all factors $\mathbb{A}_{{}_{k}}^{a}(x)$, second equal to the sum of all terms
equal to the Wick product af all factors with one pairing, with the pairing ranging over all possible pairings of one element
coming from the first Wick factor of (\ref{Wick(x)Wick(y)}) and the second element of the second Wick factor of (\ref{Wick(x)Wick(y)}).
Next we have the sum of such terms with two such pairings, then sum of terms with three such parings, and so on. Thus, we have
here the expansion of (\ref{Wick(x)Wick(y)}) into the Wick product of all $\mathbb{A}_{{}_{k}}^{a}(x)$, including all
possible pairings, with the proviso that pairings of fields $\mathbb{A}_{{}_{k}}^{a}(x)$ coming from one and the same normal order factor
of (\ref{Wick(x)Wick(y)}) are excluded.

Recall that having a free field given as a sum of integral kernel operators with vector valued kernels $\kappa_{0,1}, \kappa_{1,0}$
in the sense \cite{obataJFA}, explained above:
\[
\mathbb{A} = \Xi_{0,1}(\kappa_{0,1}) +\Xi_{1,0}(\kappa_{1,0}) = \mathbb{A}^{(-)} + \mathbb{A}^{(+)},
\]
we have
\[
-i D^{(-) \, ab}(x,y) \,\, \boldsymbol{1} \, = [\mathbb{A}^{(-) \, a}(x), \mathbb{A}^{(+) \, b}(y)]_{\mp} =
\quad \underbracket{\mathbb{A}^{a}(x) \mathbb{A}^{b}}(y)
\]
and it is easily seen that for space-time test functions $\phi, \varphi \in \mathscr{E}$ of the field $\mathbb{A}$
\[
-i\big\langle D^{(-)}, \phi \otimes \varphi \big\rangle = \big\langle \kappa_{0,1}(\phi), \kappa_{1,0}(\varphi) \big\rangle
= \kappa_{0,1}(\phi) \otimes_1 \, \kappa_{1,0}(\varphi)
\]
is a proper distribution in $\mathscr{E}^* \otimes \mathscr{E}^*$, which does not belong to multipliers of $\mathscr{E}^* \otimes \mathscr{E}^*$.
Because of the translational invariance of the unit operator,
\[
D^{(-) \, ab}(x,y) = D^{(-) \, ab}(x-y)
\]
with $D^{(-)} \in \mathscr{E}^*$, which is a proper distribution which in particular does not belong to multipliers of $\mathscr{E}^*$.

Recall also that in the conventional approach the expression
\[
\boldsymbol{{:}} \mathbb{A}_{{}_{1}}^{a}(x) \ldots \quad \underbracket{\mathbb{A}_{{}_{i}}^{a}(x) \ldots \mathbb{A}_{{}_{N}}^{a}(x) 
 \mathbb{A}_{{}_{N+1}}^{b}(y) \ldots \mathbb{A}_{{}_{N+j}}^{b}}(y) \ldots \mathbb{A}_{{}_{M}}^{b}(y)  \boldsymbol{{:}}
\]
is defined formally as
\[
(-1)^{p_{ij}} \, \quad \underbracket{\mathbb{A}_{{}_{i}}^{a}(x) \mathbb{A}_{{}_{N+j}}^{b}}(y) \,
\boldsymbol{{:}} \mathbb{A}_{{}_{1}}^{a}(x)  \ldots \overbrace{\mathbb{A}_{{}_{i}}^{a}(x)}^{\textrm{deleted}} \ldots  
 \ldots \overbrace{\mathbb{A}_{{}_{N+j}}^{b}(y)}^{\textrm{deleted}} \ldots  \mathbb{A}_{{}_{M}}^{b}(y)  \boldsymbol{{:}}
\]
where $p_{ij}$ is the number of Fermi fields inversions in passing from the order of fields
\[
 \mathbb{A}_{{}_{1}}^{b}(y) \ldots \mathbb{A}_{{}_{N}}^{a}(x)\mathbb{A}_{{}_{N+1}}^{b}(y) \ldots \mathbb{A}_{{}_{M}}^{b}(y)  
\]
to the order
\[
\mathbb{A}_{{}_{i}}^{a}(x) \mathbb{A}_{{}_{N+j}}^{b}(y) 
\mathbb{A}_{{}_{1}}^{a}(x)  \ldots \overbrace{\mathbb{A}_{{}_{i}}^{a}(x)}^{\textrm{deleted}} \ldots  \mathbb{A}_{{}_{N}}^{a}(x) 
 \mathbb{A}_{{}_{N+1}}^{b}(y) \ldots \overbrace{\mathbb{A}_{{}_{N+j}}^{b}(y)}^{\textrm{deleted}} \ldots  \mathbb{A}_{{}_{M}}^{b}(y). 
\]
In particular if the scalar contribution in the Wick expansion is nonzero (all fields $\mathbb{A}_{{}_{k}}$ in (\ref{Wick(x)Wick(y)})
can be paired), then for the scalar distribution $\kappa_{0,0}$ we obtain the formal pointwise product
of all pairings taken at the same space-time point $x-y$, which is not well-defined as a distribution. This would be well-defined if all,
or all but one, of the multiplied pairings $D^{(-)}$ where multipliers of $\mathscr{E}^*$, which for free fields on the Minkowski
space-time is impossible (recall that the positive and negative frequency parts of the Pauli-Jordan function of the scalar massive field
is a proper tempered distribution, compare e.g. \cite{Bogoliubov_Shirkov} or Subsection \ref{splitting}).
Although, in particular, the contraction
\[
\Big\langle \kappa_{0,1} \dot{\otimes} \kappa_{0,1},  \kappa_{1,0} \dot{\otimes} \kappa_{1,0}  \Big\rangle 
=
\big(\kappa_{0,1} \dot{\otimes} \kappa_{0,1} \big) \otimes_2 \, \big(\kappa_{1,0} \dot{\otimes} \kappa_{1,0}\big)
\]
is a well defined element of $\mathscr{E}^{* \,\, \otimes 2}$, it is not true that its kernel 
\[
\big( \kappa_{0,1} \dot{\otimes} \kappa_{0,1}\big) \otimes_2 \big(\kappa_{1,0} \dot{\otimes} \kappa_{1,0} \big)(a,b,c,d,x,y)
\]
can be defined through the pointwise product of the kernels 
\[
D^{(-) \, ab}(x,y) = D^{(-) \, ab}(x-y) = \big(\kappa_{0,1} \otimes_1 \, \kappa_{1,0}\big)(a,b,x,y),
\]
because the product
\[
D^{(-) \, ab}(x-y) \, D^{(-) \, cd}(x-y) 
\]
is, in general, meaningless as a kernel of a distribution. This is why in the conventional approach ''regularization'' was used.
The point is that kernels $\kappa_{\ell,m}$ of the Fock expansion (\ref{FockExpansionOfWick(x)Wick(y)}) can be obtained in well-defined
mathematical terms by the operations of pointed tensor product $\dot{\otimes}$, ordinary tensor product $\otimes$, $q$-contraction
$\otimes_q$, symmetrization and antisymmetrization,
from the kernels $\kappa_{0,1}, \kappa_{1,0}$ defining the free fields $\mathbb{A}_{{}_{k}}$, but the kernels
$\kappa_{\ell,m}$ of the Fock expansion (\ref{FockExpansionOfWick(x)Wick(y)}) cannot be obtained from the
pairing distributions of the fields $\mathbb{A}_{{}_{k}}$ without the intermediate step, in which we are 
about to multiply products of pairings. Using the white noise construction of the free fields and the kernels
$\kappa_{0,1}, \kappa_{1,0}$ defining the free fields $\mathbb{A}_{{}_{k}}$, we can
forget about any ''regularization'', at least for massive fields $\mathbb{A}_{{}_{k}}$. Indeed, because
contraction of the kernels is well-defined, then so is the product of pairings, which can be defined through it. So we have rigorously
shown, using white noise, that the pairings are exceptional, and their products are well-defined
distributions. The fact that ``regularization'' can be avoided in the computation of the product of pairings in the Wick theorem
was also noticed in \cite{Scharf}, but in a rather informal manner. Thus here we give a mathematical justification for the 
practical computations of the products of pairings as given in \cite{Scharf}.

\subsection{Product of monomials of massless free fields}

This form of Wick theorem, \emph{i.e.} Theorem \ref{WickThmForMassiveFields}, is however
insufficient in realistic QFT, such as QED, because in the causal construction of the scattering operator or causal construction of interacting
fields from the scattering operator, the  Wick factors (\ref{WickMonomialsInGeneralFreeFields}) necessary include the Lagrange interaction density
\[
\mathcal{L}(x) = \boldsymbol{{:}} \boldsymbol{\psi}(x)^{+}\gamma_{0} \gamma^\mu \boldsymbol{\psi}(x) A_\mu(x) \boldsymbol{{:}} 
\]
and necessary include the massless electromagnetic potential field $A$ as one of the factors $\mathbb{A}_{{}_{k}}$
in the Wick products which have to be considered. In particular, in the second inductive computational step (second order contribution) 
in the causal perturbative construction of the 
scattering operator we need to consider the``product'' (compare \cite{Scharf}, \cite{Bogoliubov_Shirkov}, \cite{Epstein-Glaser})
\begin{equation}\label{L(x)L(y)}
\mathcal{L}(x) \mathcal{L}(y)
\end{equation}
and apply Wick theorem of \cite{Bogoliubov_Shirkov} in order to write it in the form of ``normally ordered''
operators. 

But Wick Theorem \ref{WickThmForMassiveFields} for massive fields can be generalized over to the case
of product (\ref{Wick(x)Wick(y)}) in which each of the normal factors (\ref{WickMonomialsInGeneralFreeFields})
contains massless fields $\mathbb{A}_{{}_{k}}$.

In this situation, when among the factors $\mathbb{A}_{{}_{k}}$ in (\ref{Wick(x)Wick(y)}) there are present mass-less 
fields (or their derivatives) in (\ref{Wick(x)Wick(y)}), as e.g. in (\ref{L(x)L(y)}) of QED -- a particular case of (\ref{Wick(x)Wick(y)}) -- 
then we replace the $\mathscr{E}_{2}^{*}$-valued kernels $\kappa_{0,1}, \kappa_{1,0}$ defining the massless factors by their massive counterparts.
In practice, we just replace the zero mass energy functions
\[
p_{0}(\boldsymbol{\p}) = |\boldsymbol{\p}|
\]
in the massless kernels $\kappa_{0,1}, \kappa_{1,0}$ by the massive energy functions
\[
p_{0}(\boldsymbol{\p}) = \sqrt{|\boldsymbol{\p}|^2 + \epsilon^2}
\]
and obtain in this manner the kernels $\kappa_{\epsilon \,\, 0,1}, \kappa_{\epsilon \,\, 1,0}$ with exponents $e^{\mp ip\cdot x}$
with $p$ ranging over the positive energy massive hyperboloid $\mathscr{O}_{{}_{\epsilon,0,0,0}}$, the nuclear massless single particle spaces
$E_i$ corresponding to the massless fields we keep unchanged, but the space-time test spaces
\[
\mathscr{E}_i =  \mathcal{S}^{00}(\mathbb{R}^4)\subset  \mathcal{S}(\mathbb{R}^4),
\]
corresponding to these fields, we can, and we do, enlarge to  $\mathscr{E}_i =  \mathcal{S}(\mathbb{R}^4)$, although the regularity of the corresponding
field operator will have to be weakened. In the limit $\epsilon \rightarrow 0$, the replacement of the space-time test space
does not make any difference, as the regularity of pointwise Wick product operators, in both cases,  is weakened and in both cases
the pointwise Wick products transform continuously the space-time test space into operators transforming $(E)$ into $(E)^*$.  
Taking into account application to causal QFT, the space-time test space should be enlarged to the Schwartz test space $\mathcal{S}(\mathbb{R}^4)$ 
as it contains all smooth functions of compact support, which is needed for the implementation of causality. 

Then we construct the analogue  $\Xi_{\epsilon}$ of the
product (\ref{Wick(x)Wick(y)}) in which all the massless free field kernels are replaced with massive ``counterparts'' 
$\kappa_{\epsilon \,\, 0,1}, \kappa_{\epsilon \,\, 1,0}$ (more precisely:
we replace $|\boldsymbol{\p}|$ with $(|\boldsymbol{\p}|^2+\epsilon^2)^{1/2}$ in the massless kernles $\kappa_{0,1}, \kappa_{1,0}$). Next, we compute
the Fock expansion of $\Xi_{\epsilon}$:
\begin{equation}\label{FockExpansionXiepsilon}
\Xi_{\epsilon} = 
\sum\limits_{\ell, m} \Xi_{\epsilon \,\, \ell, m}({\kappa}_{\epsilon \,\,  \ell, m})
\,\,\,\,\,\,\,
\in \,\,\,\,\,\,\, \mathscr{L}((E) \otimes \mathscr{E}_{1} \otimes \mathscr{E}_{2}, \, (E)), \,\,\, \ell+ m \leq n+q,
\end{equation}
into integral kernel operators
with $\mathscr{E}_{1}^{*} \otimes \mathscr{E}_{2}^{*} = \mathscr{L}(\mathscr{E}_{1} \otimes \mathscr{E}_{2}, \mathbb{C})$-valued
kernels ${\kappa}_{\epsilon \,\, \ell, m}$, exactly as above for the massive fields, with $\mathscr{E}_{1}$ and  $\mathscr{E}_{2}$ 
being the ordinary Schwartz spaces, as was for the purely massive case.

Next we observe that
${\kappa}_{\epsilon \,\, \ell, m} \rightarrow {\kappa}_{\ell, m}, {\kappa}_{\epsilon \,\, 0,0}
\rightarrow {\kappa}_{0,0}$ converge in
\[
\mathscr{L}(E^{\otimes (\ell+m)}; \mathscr{E}_{1}^{*} \otimes \mathscr{E}_{2}^{*}), \,\,\, \ell+ m \leq n+q,
\]
and, respectively (for the scalar contractions), in $\mathscr{E}_{1}^{*} \otimes \mathscr{E}_{2}^{*}$,
when $\epsilon \rightarrow 0$, with $\mathscr{E}_{1}$, $\mathscr{E}_{2}$, equal to the ordinary Schwartz
space-time test spaces $\mathcal{S}(\mathbb{R}^4;\mathbb{C}^{d_1})$, $\mathcal{S}(\mathbb{R}^4;\mathbb{C}^{d_2}), \ldots$, 
depending on the fields in (\ref{Wick(x)Wick(y)}), in which the spaces
$\mathcal{S}^{00}(\mathbb{R}^4; \mathbb{C}^{d_i})$ corresponding to the massless fields 
are replaced with $\mathcal{S}(\mathbb{R}^4)$.  
By Thm. 3.9 and Theorem 4.8 of \cite{obataJFA} (or, respectively, by their Fermi-Fock analogues) the operator ''product'' $\Xi_{\epsilon}$
converges to an operator
\begin{equation}\label{XimasslessProduct}
\Xi \in \mathscr{L}(\,\, (E) \otimes \mathscr{E}_{1} \otimes \mathscr{E}_{2} \,\, , \,\,\, (E)^* \,\,)
\end{equation}
when $\epsilon \rightarrow 0$, which in general does not belong to
\[
\mathscr{L}(\,\,(E) \otimes \mathscr{E}_{1} \otimes \mathscr{E}_{2} \,\, , \,\,\, (E) \,\,).
\]
This operator, when evaluated at a fixed element $\phi \otimes \varphi \in \mathscr{E}_{1} \otimes \mathscr{E}_{2}$,
gives an operator in $\mathscr{L}((E), \, (E)^*)$, 
but this time, its value cannot be written as operator composition.
Again, by Thms. 3.9 and 4.8 of \cite{obataJFA} or their fermionic analogues,
which are applicable to general operators  $\Xi$ of the class (\ref{XimasslessProduct}),
possesses unique (here finite, because we include the scalar term $\Xi_{0,0}$)
Fock expansion
\[
\Xi =
\sum\limits_{\ell, m} \Xi_{\ell, m}({\kappa}_{\ell, m})
\,\,\,\,
\in \,\,\,\, \mathscr{L}((E) \otimes \mathscr{E}_{1} \otimes \mathscr{E}_{2}, \, (E)^*), \,\,\, \ell+ m \leq n+q,
\]
which in fact gives the rigorous version of the Wick theorem stated in
\cite{Bogoliubov_Shirkov}, \S 17, as ''The Wick's Theorem for Ordinary Products''\footnote{In the English Edition we read there: ''Wick Theorem for
Normal Products'', but ''The Wick's Theorem for Ordinary Products'' would be a better translation of the Russian original.}.
Generalization of this theorem to the ''products''
(\ref{Wick(x)Wick(y)}) containing a greater number of normally ordered Wick product factors (\ref{WickMonomialsInGeneralFreeFields}),
is obvious. This gives the mathematical justification for the Wick theorem stated
in \cite{Bogoliubov_Shirkov}, and shows that indeed the decomposition, or Fock expansion, can be
computed through the kernels $\kappa_{0,1}^{{}^{k}}, \kappa_{1,0}^{{}^{k}}$ defining the free field factors $\mathbb{A}_{{}_{k}}$,
because indeed they can be effectively computed through the operator
products (with massless factors replaced with the massive counterparts) and by the observation that in the zero-mass limit
of the massive pairings we indeed get the kernels of the massless fields. We thus have a generalization
of the Wick Theorem \ref{WickThmForMassiveFields} to the case of (\ref{Wick(x)Wick(y)}) in which some free fields
$\mathbb{A}_{{}_{k}}$, or even all, are massless. Although in this case, convergence of the contractions $\otimes_q$
is not automatic and does not follow from the very construction. Nonetheless, 
we can generalize the notion of contraction by the indicated above limit process, replacing the massless
kernels by the massive counterparts and passing to the limit. 
Namely, for the kernels $\kappa'_{\ell',m'}, \kappa''_{\ell'',m''}$ of the Fock
expansions, respectively, of the operators $\Xi', \Xi''$, we define the \emph{limit contraction} $\otimes|_{{}_{q}}$
\begin{equation}\label{kappaepsilonconvergence}
\kappa'_{\ell',m'} \otimes|_{{}_{q}} \, \kappa''_{\ell'',m''} \overset{\textrm{df}}{=}
\textrm{lim}_{\epsilon \rightarrow 0} \,\,\,
\kappa'_{\epsilon \,\, \ell',m''} \otimes_q \, \kappa''_{\epsilon \,\, \ell'',m''}
\end{equation}
which exists in
\[
\mathscr{L}\big(\mathscr{E}_{1}\otimes \mathscr{E}_{2} , E^{* \otimes (\ell'+\ell''-q + m'+m''-q)} \big)
\cong
\mathscr{L}\big(E^{\otimes (\ell'+\ell''-q + m'+m''-q)}, \, \mathscr{E}_{1}^{*}\otimes \mathscr{E}_{2}^{*} \big).
\]

Before we give a general existence proof of (\ref{kappaepsilonconvergence}) in the said space and Wick theorem
for massless fields, let us note the following lemmas
\begin{lem}
The free massless field (e.g. e.m. potential) and massive fields (e.g. spinor field) 
operators $\mathbb{A}=A,\boldsymbol{\psi}$ are operator-valued distributions, \emph{i.e.} belong 
to $\mathscr{L}\big(\mathscr{E}, \mathscr{L}((E),(E))\big)$, \emph{i.e.}
\[
\kappa_{0,1}, \kappa_{1,0} \in \mathscr{L}(E^*, \mathscr{E}^{*}) = \mathscr{L}(\mathscr{E}, E) \subset \mathscr{L}(E, \mathscr{E}^{*})
\cong E^{*} \otimes \mathscr{E}^{*},
\]
if and only if
\[
\begin{split}
E= \mathcal{S}^{0}(\mathbb{R}^3; \mathbb{C}^{d}) = \mathcal{S}_{A_{(3)}}(\mathbb{R}^3;
\mathbb{C}^d), \,\, \mathscr{E} = \mathcal{S}^{00}(\mathbb{R}^4; \mathbb{C}^d) = \mathcal{S}_{\widetilde{A}_{(4)}}(\mathbb{R}^4;
\mathbb{C}^d)
\,\,\,\, \textrm{for} \,\,\, A
\\
E= \mathcal{S}(\mathbb{R}^3; \mathbb{C}^d) = \mathcal{S}_{\widetilde{H}_{(3)}}(\mathbb{R}^3;
\mathbb{C}^d), \,\, \mathscr{E} = \mathcal{S}(\mathbb{R}^4; \mathbb{C}^d) = \mathcal{S}_{H_{(4)}}(\mathbb{R}^4; \mathbb{C}^d)
\,\,\,\, \textrm{for} \,\,\, \boldsymbol{\psi}.
\end{split}
\]
Moreover
\begin{equation}\label{kappa(0,1)(xi)inOM}
\kappa_{0,1}, \kappa_{1,0} \in \mathscr{L}(E, \mathcal{O}_M)
\end{equation}
and for each $\xi \in E$, $\kappa_{0,1}(\xi), \kappa_{1,0}(\xi)$ are smooth having all derivatives bounded.
\qed
\label{Cont.free.field.kernels}
\end{lem}

\begin{lem}
The functions $u,$ in the kernels
\begin{multline*}
\kappa_{{}_{0,1}}(s, \boldsymbol{\p}; a,x) = u_{{}_{s \,\,\,a}}(\boldsymbol{\p})e^{-ip_i\cdot x},
\,\,\,
\kappa_{{}_{1,0}}(s, \boldsymbol{\p}; a,x) = v_{{}_{s \,\,\,a}}(\boldsymbol{\p})e^{ip_i\cdot x},
\\
p = (\boldsymbol{\p}, p_{0}(\boldsymbol{\p})) = (\boldsymbol{\p}, \sqrt{|\boldsymbol{\p}|^2+ \mathfrak{m}^2})
\end{multline*}
of the free fields, are multipliers of the single particle nuclear spaces $E$
of the corresponding free fields $\mathbb{A}$.
\qed
\label{FreeFieldMultipliers}
\end{lem}
\begin{lem}
The kernels $\kappa_{{}_{l,m}}$ of the Wick product ${:}\mathbb{A}^{{}^{1}}(x) \ldots \mathbb{A}^{{}^{n}}(x) {:}$ of free fields
map continuously
\[
E^{\otimes n } \ni \xi = \xi_1 \otimes \ldots \otimes \xi_n \longmapsto
\kappa_{{}_{l,m}}(\xi) =
\kappa_{{}_{l_1,m_1}}^{{}^{1}}(\xi_1)
\ldots
 \kappa_{{}_{l_n,m_n}}^{{}^{n}}(\xi_n)  \in \mathcal{O}_{M}(\mathbb{R}^4)
\]
and each space-time derivative of $\kappa_{{}_{l,m}}(\xi)$ is bounded.
\label{DistrProdWickProd}
\end{lem}
\qedsymbol \,
Lemma \ref{DistrProdWickProd} immediately follows from (\ref{kappa(0,1)(xi)inOM}) of lemma \ref{Cont.free.field.kernels}
and from the product formula
\[
\kappa^{{}^{1}}_{l_1, m_1} \dot{\otimes} \ldots \dot{\otimes} \kappa^{{}^{n}}_{l_n, m_n}(\xi_1 \otimes \ldots \xi_m)
= \kappa^{{}^{1}}_{l_1, m_1}(\xi_1) \cdots \kappa^{{}^{n}}_{l_n, m_n}(\xi_n), 
\,\,\,\,
(l_i,m_i) = (0,1) \,\, \textrm{or} \,\, (l_i,m_i) = (1,0), \xi_i \in E_i. 
\qed
\] 
For the proof of these lemmas compare our white noise construction of the free Dirac and e.m. potential fields, 
Subsections \ref{psiBerezin-Hida}, \ref{WhiteNoiseA}.

We can thus generalize  the Wick Theorem \ref{WickThmForMassiveFields} over to massless fields $\mathbb{A}_{{}_{k}}$, by replacing the
ordinary contractions $\otimes_q$ with the \emph{limit contractions} $\otimes|_{{}_{q}}$ in it.
\begin{twr}
In case there are massless fields $\mathbb{A}_{{}_{k}}$ in each of the normal factors (\ref{WickMonomialsInGeneralFreeFields})
of the ``product'' (\ref{Wick(x)Wick(y)}) we obtain the following Fock expansion, or Wick theorem, 
for the generalized ``product'' operator $\Xi = \textrm{lim}_{{}_{\epsilon \rightarrow 0}} \, \Xi_\epsilon$.

The operators
\begin{eqnarray*}
\Xi'(\phi) = \sum \limits_{a} \int\limits_{\mathbb{R}^4}  
\boldsymbol{{:}} \mathbb{A}_{{}_{1}}^{a_1}(x) \ldots \mathbb{A}_{{}_{N}}^{a_N}(x) \boldsymbol{{:}} \, \phi^a(x) \,\,\,\, \ud^4x
\in \mathscr{L}((E),(E)^*),
\\
\Xi''(\varphi) = \sum \limits_{b} \int\limits_{\mathbb{R}^4}  
\boldsymbol{{:}} \mathbb{A}_{{}_{N+1}}^{b_{N+1}}(y) \ldots \mathbb{A}_{{}_{M}}^{b_M}(y)  \boldsymbol{{:}}  \,\,\,\, \varphi^b(y) \, \ud^4y
\in \mathscr{L}((E),(E)^*),
\\
\Xi(\phi \otimes \varphi) = 
\sum \limits_{a,b}
\int\limits_{\big[\mathbb{R}^4\big]^{\times \, 2}} 
\boldsymbol{{:}} \mathbb{A}_{{}_{1}}^{a_1}(x) \ldots \mathbb{A}_{{}_{N}}^{a_N}(x) \boldsymbol{{:}} \,\,
\boldsymbol{{:}} \mathbb{A}_{{}_{N+1}}^{b_{N+1}}(y) \ldots \mathbb{A}_{{}_{M}}^{b_M}(y)  \boldsymbol{{:}} \,\, \times
\\ \times \,\,\,\,\,\,\,
\phi^a \otimes \varphi^b(x,y) \, \ud^4x \ud^4y \,\,\, \in \mathscr{L}((E),(E)^*),
\end{eqnarray*} 
and moreover
\begin{eqnarray*}
\Xi' \in \mathscr{L}(\mathscr{E}_{1}, \mathscr{L}((E),(E)^*)\big),
\\
\Xi''\in \mathscr{L}(\mathscr{E}_{2}, \mathscr{L}((E),(E)^*)\big),
\\
\Xi\in \mathscr{L}(\mathscr{E}_{1} \otimes \mathscr{E}_{2}, \mathscr{L}((E),(E)^*)\big),
\end{eqnarray*}
and the following Wick theorem holds
\[
\Xi =
\sum_{\substack{\kappa'_{\ell',m'} \\ \kappa''_{\ell'',m''}}} \sum \limits_{0\leq q \leq \textrm{min} \, \{m',\ell''\}}  \,
\Xi_{\ell'+\ell''-q,m'+m''-q}\big( \kappa'_{\ell',m'} \otimes|_{{}_{q}} \, {\kappa''}_{\ell'',m''}\big),
\]
where the kernels $\kappa'_{\ell',m'}, \kappa''_{\ell'',m''}$ range, respectively, over the  kernels of finite Fock expansions 
of the operators $\Xi', \Xi''$. The limit $q$-contractions $\otimes|_{{}_{q}}$ are performed
upon the pairs of variables in which the first element of the contracted pair lies among the last $m'$ variables of the kernel 
$\kappa'_{\ell',m'}$ and the second variable
of the contracted pair lies among the first $l''$ variables of the kernel $\kappa''_{\ell'',m''}$, and to both variables
of the contracted pair correspond respectively annihilation and creation operator of one and the same free field. The limit contraction 
$\kappa_{l,m}= \kappa'_{\ell',m'} \otimes|_{{}_{q}} \, {\kappa''}_{\ell'',m''}$ should be symmetrized in Bose variables and antisymmetrized in Fermi variables
in order to achieve one-to-ne correspondence between kernels and corresponding integral kernel operators and this should be done separately
for the first variables $l$ and separately for the last $m$ variables. 
Here $\mathscr{E}_{1}$ and $\mathscr{E}_{2}$, are the  ordinary Schwartz spaces 
$\mathcal{S}(\mathbb{R}^4; \mathbb{C}^{d_1})$, $\mathcal{S}(\mathbb{R}^4; \mathbb{C}^{d_2})$,  depending on
the free fields in the Wick product. 
\label{WickThmForMasslessFields}
\end{twr}

\begin{rem*}
We should also emphasize here that although in principle the \emph{limit contraction} $\otimes|_{{}_{q}}$ 
is computed as the limit $\textrm{lim}_{\epsilon \rightarrow 0}$
of the contraction with the massless kernels $\kappa_{0,1}^{{}^{k}},\kappa_{1,0}^{{}^{k}}$ replaced by the massive counterparts
$\kappa_{\epsilon \,\, 0,1}^{{}^{k}},\kappa_{\epsilon \,\, 1,0}^{{}^{k}}$, the integral representing ordinary contraction with
$\kappa_{\epsilon \,\, 0,1}^{{}^{k}},\kappa_{\epsilon \,\, 1,0}^{{}^{k}}$ in it remains convergent even if
we put $\epsilon$ simply equal zero in it, and represents the \emph{limit contraction}.
This we have proved below
and, as we will see, the proof 
remains valid if the number of basic plane wave kernels is arbitrary large.
Thus, the \emph{limit contraction} can be expressed by ordinary formulas for contraction integrals
expressed through the original kernels $\kappa_{0,1}^{{}^{k}},\kappa_{1,0}^{{}^{k}}$ of the free fields $\mathbb{A}^{{}^{k}}$.
Therefore, in the sequel we will denote the limit contraction  $\otimes|_{{}_{q}}$ simply by $\otimes{{}_{q}}$.
\qed
\end{rem*}

{\bf Proof of theorem \ref{WickThmForMasslessFields}}.
For the proof it is sufficient to show the convergence (\ref{kappaepsilonconvergence}). Without loosing generality, we assume that all 
considered free field kernels are massless. We start with a proof of the remark, \emph{i.e.}
we show absolute convergence of the integral representing  $\otimes_{q}$-contractions, staring with scalar contractions 
$\kappa_{\epsilon \,\, 0,0}$, in which all spin-momentum variables are contracted, 
with $\epsilon$ simply put equal zero in it, giving a norm estimation for the absolute value of this integral:
\begin{multline}\label{masslesskappa01.masslesskappa10...q'-contraction...masslesskappa10.masslesskappa10}
\kappa^{{}^{q_1}}_{0,1} \dot{\otimes} \kappa^{{}^{q_2}}_{0,1} \ldots \dot{\otimes} \kappa^{{}^{q_q}}_{0,1} (\phi)
\, \otimes|_{{}_{q}}  \,\, \kappa^{{}^{q_1}}_{1,0} \dot{\otimes} \kappa^{{}^{q_2}}_{1,0} \ldots \dot{\otimes} \kappa^{{}^{q_q}}_{1,0} (\varphi) 
= \overset{q}{\underset{i=1}{\dot{\otimes}}} \kappa^{{}^{q_i}}_{0,1}(\phi)
\, \otimes|_{{}_{q}} \,\,
\overset{q}{\underset{i=1}{\dot{\otimes}}} \kappa^{{}^{q_i}}_{1,0}(\varphi)
\\
=\sum \limits_{s_1, \ldots, s_{q'}}
\int
u^{{}^{q_1}}_{{}_{s_{q_1}}}(\boldsymbol{\p}_{{}_{q_1}}) \ldots u^{{}^{q_q}}_{{}_{s_{q_q}}}(\boldsymbol{\p}_{q_q}) 
\widetilde{\phi}\big(-\boldsymbol{\p}_{{}_{q_1}}- \ldots - \boldsymbol{\p}_{{}_{q_q}}, \, -|\boldsymbol{\p}_{{}_{q_1}}|
\ldots - |\boldsymbol{\p}_{{}_{q_q}}| \big)
\,\, \times
\\
\times \,\,\,
v^{{}^{q_1}}_{{}_{s_{q_1}}}(\boldsymbol{\p}_{{}_{q_1}}) \ldots v^{{}^{q_q}}_{{}_{s_{q_q}}}(\boldsymbol{\p}_{{}_{q_q}}) 
\widetilde{\varphi}\big(\boldsymbol{\p}_{{}_{q_1}}+ \ldots + \boldsymbol{\p}_{{}_{q_q}}, |\boldsymbol{\p}_{{}_{q_1}}|
\ldots +|\boldsymbol{\p}_{{}_{q_q}}| \big)
\,\, \ud^3 \boldsymbol{\p}_{{}_{q_1}} \ldots  \ud^3 \boldsymbol{\p}_{{}_{q_q}}.
\end{multline}
Indeed, because for the Schwartz function $\phi \in \mathscr{E}_1$
its Fourier transform $\widetilde{\phi}$ also is a Schwatrz function, ranging over a bounded set, whenever so does $\phi$, 
we can use this fact to estimate 
(\ref{masslesskappa01.masslesskappa10...q'-contraction...masslesskappa10.masslesskappa10}). Namely the function
\[
V_{(4)}(\boldsymbol{\p}, p_0) = 1+ p_{0}^2 + |\boldsymbol{\p}|^2  
\]
is a multiplier of  the Schwartz space  $\widetilde{\mathscr{E}_1}$. 
For each natural $n$, and $\phi$ ranging over a bounded set $B$ in the Schwartz space, or $\widetilde{\phi}$
ranging over  a bounded set $\widetilde{B}$ in the Schwartz space $\widetilde{\mathscr{E}_1}$,
the set of functions $V_{(4)}^{n} \cdot \widetilde{\phi}$
ranges over a bounded set in the Schwartz space $\widetilde{\mathscr{E}_1}$, depending only on $B$ and $n$, 
on which the countable set of norms
$| \cdot |_1$, $|\cdot |_2$, $\ldots$, 
defining  $\mathscr{E}_1$ are finite. For any fixed number $q$ of functions $u^{{}^{q_i}},v^{{}^{q_i}}$ determining kernels of free (say massless) 
fields there exists a natural $n$ such that
\begin{equation}\label{V(4)}
V_{(4)}^{-n}(\boldsymbol{\p}_{{}_{q_1}} + \ldots + \boldsymbol{\p}_{{}_{q_q}}, \, |\boldsymbol{\p}_{{}_{q_1}}|+ \ldots 
+ |\boldsymbol{\p}_{{}_{q_q}}|) u^{{}^{q_1}}(\boldsymbol{\p}_{{}_{q_1}})
v^{{}^{q_1}}(\boldsymbol{\p}_{{}_{q_1}}) \ldots u^{{}^{q_q}}(\boldsymbol{\p}_{{}_{q_q}})v^{{}^{q_q}}(\boldsymbol{\p}_{{}_{q_q}}) 
\end{equation}
is absolutely integrable regarded as the function of $\boldsymbol{\p}_{{}_{q_1}}, \ldots, \boldsymbol{\p}_{{}_{q_q}}$. Denoting the $L_1$-norm
of the function (\ref{V(4)}) by $c$, we have the following inequality for the absolute value of (\ref{masslesskappa01.masslesskappa10...q'-contraction...masslesskappa10.masslesskappa10})
\begin{equation}\label{AbsoluteConvergenceKappa00}
\Big|\overset{q}{\underset{i=1}{\dot{\otimes}}} \kappa^{{}^{q_i}}_{0,1}(\phi)
\, \otimes|_{{}_{q}} \,\,
\overset{q}{\underset{i=1}{\dot{\otimes}}} \kappa^{{}^{q_i}}_{1,0}(\varphi) \Big| \,\, \leq 
\,\,\, c \, \underset{p\in \mathbb{R}^4}{\textrm{sup}} \big|\widetilde{\phi}(p) \big| \,\,\,
\,\,\ \underset{p\in \mathbb{R}^4}{\textrm{sup}} \big|V_{(4)}^{n}. \widetilde{\varphi}(p) \big|
\leq
c \, \big|\widetilde{\phi}\big|_{{}_{k}} \,\,  \big|\widetilde{\varphi}\big|_{{}_{n}}
=c \, \big|\phi\big|_{{}_{k}} \, \big|\varphi\big|_{{}_{n}},
\end{equation}
for all sufficiently large natural $k,n$, with the last norms $|\cdot|_{{}_{k}} = |\widetilde{H_{{}_{(4)}}}^k \cdot|_{{}_{L^2}}$, 
$|\cdot|_{{}_{k}} = |H_{{}_{(4)}}^k \cdot|_{{}_{L^2}}$, on the Fourier image $\widetilde{\mathscr{E}_i}$, $i=1,2$, of the Schwartz space $\mathscr{E}_i$ and on 
the Schwartz space $\mathscr{E}_i$, coming from the system of Hilbertian norms on $\mathscr{E}_i$.
Note that the estimation (\ref{AbsoluteConvergenceKappa00}) is valid for the integral 
(\ref{masslesskappa01.masslesskappa10...q'-contraction...masslesskappa10.masslesskappa10}) in which the integrand is replaced
by its absolute value, so that (\ref{masslesskappa01.masslesskappa10...q'-contraction...masslesskappa10.masslesskappa10})
is absolutely convergent.

Let us pass now to the general $\otimes_{q}$-contraction. Let, for simplicity of notation,
the last $q$ spin-momentum variables $s_{{}_{q_1}},\boldsymbol{\p}_{{}_{q_1}}, \ldots, s_{{}_{q_q}},\boldsymbol{\p}_{{}_{q_q}}$ 
in the first kernel $\kappa'_{l, m+q}(\phi) = \overset{l}{\underset{i=1}{\dot{\otimes}}} \kappa^{{}^{l_i}}_{1,0} 
\overset{m}{\underset{i=1}{\dot{\otimes}}} \kappa^{{}^{m_i}}_{0,1}
\overset{q}{\underset{i=1}{\dot{\otimes}}} \kappa^{{}^{q_i}}_{0,1}(\phi)$  be contracted with the
first $q$ spin-momentum variables $s_{{}_{q_1}},\boldsymbol{\p}_{{}_{q_1}}, \ldots, s_{{}_{q_q}},\boldsymbol{\p}_{{}_{q_q}}$  
in the second kernel $\kappa''_{q+\ell, \mathpzc{m}}(\varphi) =\overset{q}{\underset{i=1}{\dot{\otimes}}} \kappa^{{}^{q_i}}_{1,0}
\overset{\ell}{\underset{i=1}{\dot{\otimes}}} \kappa^{{}^{\ell_i}}_{1,0}
\overset{\mathpzc{m}}{\underset{i=1}{\dot{\otimes}}} \kappa^{{}^{\mathpzc{m}_i}}_{0,1}(\varphi)$:
\begin{multline}\label{(l+m+q)q(q+ell+em)Contraction}
\overset{l}{\underset{i=1}{\dot{\otimes}}} \kappa^{{}^{l_i}}_{1,0} 
\overset{m}{\underset{i=1}{\dot{\otimes}}} \kappa^{{}^{m_i}}_{0,1}
\overset{q}{\underset{i=1}{\dot{\otimes}}} \kappa^{{}^{q_i}}_{0,1}(\phi)
\, \otimes|_{{}_{q}} \,\,
\overset{q}{\underset{i=1}{\dot{\otimes}}} \kappa^{{}^{q_i}}_{1,0}
\overset{\ell}{\underset{i=1}{\dot{\otimes}}} \kappa^{{}^{\ell_i}}_{1,0}
\overset{\mathpzc{m}}{\underset{i=1}{\dot{\otimes}}} \kappa^{{}^{\mathpzc{m}_i}}_{0,1}(\varphi)
\\
=\sum\limits_{s_{{}_{q_1}}, \ldots, s_{{}_{q_q}}} \bigintsss \Bigg[
\\
\prod\limits_{i=1}^{l} v^{{}^{l_i}}_{{}_{s_{{}_{l_i}}}}(\boldsymbol{\p}_{{}_{l_i}})
\prod\limits_{i=1}^{m} u^{{}^{m_i}}_{{}_{s_{{}_{m_i}}}}(\boldsymbol{\p}_{{}_{m_i}})
\prod\limits_{i=1}^{q} u^{{}^{q_i}}_{{}_{s_{{}_{q_i}}}}(\boldsymbol{\p}_{{}_{q_i}})
\widetilde{\phi}\left(\sum\limits_{i=1}^{l}\boldsymbol{\p}_{{}_{l_i}}- \sum\limits_{i=1}^{m}\boldsymbol{\p}_{{}_{m_i}}
- \sum\limits_{i=1}^{q} \boldsymbol{\p}_{{}_{q_i}}, \sum\limits_{i=1}^{l}|\boldsymbol{\p}_{{}_{l_i}}|- \sum\limits_{i=1}^{m}|\boldsymbol{\p}_{{}_{m_i}}|
- \sum\limits_{i=1}^{q} |\boldsymbol{\p}_{{}_{q_i}}|\right) \,\, \times
\\
\times \,\, 
\prod\limits_{i=1}^{q} v^{{}^{q_i}}_{{}_{s_{{}_{q_i}}}}(\boldsymbol{\p}_{{}_{q_i}})
\prod\limits_{i=1}^{\ell} v^{{}^{\ell_i}}_{{}_{s_{{}_{m_i}}}}(\boldsymbol{\p}_{{}_{\ell_i}})
\prod\limits_{q=1}^{\mathpzc{m}} u^{{}^{\mathpzc{m}_i}}_{{}_{s_{{}_{\mathpzc{m}_i}}}}(\boldsymbol{\p}_{{}_{\mathpzc{m}_i}})
\widetilde{\varphi}\left(\sum\limits_{i=1}^{q}\boldsymbol{\p}_{{}_{q_i}} + \sum\limits_{i=1}^{\ell}\boldsymbol{\p}_{{}_{\ell_i}}
-\sum\limits_{i=1}^{\mathpzc{m}} \boldsymbol{\p}_{{}_{\mathpzc{m}_i}}, \sum\limits_{i}^{q} |\boldsymbol{\p}_{{}_{q_i}}|+ 
\sum\limits_{i}^{\ell}|\boldsymbol{\p}_{{}_{\ell_i}}|
-\sum\limits_{i}^{\mathpzc{m}} |\boldsymbol{\p}_{{}_{\mathpzc{m}_i}}|\right) 
\\
\Bigg] \prod\limits_{i=1}^{q} \ud^3 \boldsymbol{\p}_{{}_{q_i}} 
\end{multline}
being the generalized function in $\overset{l}{\underset{i=1}{\otimes}} E_{{}_{l_i}}^*
\overset{m}{\underset{i=1}{\otimes}} E_{{}_{m_i}}^*
\overset{\ell}{\underset{i=1}{\otimes}} E_{{}_{\ell_i}}^*
\overset{\mathpzc{m}}{\underset{i=1}{\otimes}} E_{{}_{\mathpzc{m}_i}}^*$ of the non contracted spin-momenta variables
\begin{equation}\label{NonContractedVariables}
(\ldots, s_{{}_{l_i}},\boldsymbol{\p}_{{}_{l_i}}, \ldots, s_{{}_{m_i}},\boldsymbol{\p}_{{}_{m_i}}, \ldots, 
s_{{}_{\ell_i}},\boldsymbol{\p}_{{}_{\ell_i}}, \ldots, 
s_{{}_{\mathpzc{m}_i}},\boldsymbol{\p}_{{}_{\mathpzc{m}_i}}, \ldots),
\end{equation}
and with all free field kernels being massless. Replacement of $\widetilde{\phi}, \widetilde{\varphi}$ with their translations
$T_{{}_{w}}\widetilde{\phi} = \widetilde{e_w\phi}, T_{{}_{\upsilon}}\widetilde{\varphi} = \widetilde{e_{\upsilon}\varphi}$, 
with $w,\upsilon, e_{w}, e_{\upsilon}$ given below, in the proof of (\ref{AbsoluteConvergenceKappa00}),
gives a proof of the absolute convergence of the integral (\ref{(l+m+q)q(q+ell+em)Contraction}), majorized at infinity by a fixed 
polynomial in the non-contracted variables and depending only on the norms of $\phi,\varphi$, compare our analysis 
of the limit (\ref{kappaepsilonconvergence}) given below. We pass now to the analysis of the limit (\ref{kappaepsilonconvergence}).

Replacing $|\boldsymbol{\p}_i|$ by $|\boldsymbol{\p}_i|_{{}_{\epsilon}} = \sqrt{|\boldsymbol{\p}_i|^2+\epsilon^2}$ in
(\ref{(l+m+q)q(q+ell+em)Contraction}) and  subtracting (\ref{(l+m+q)q(q+ell+em)Contraction})  we obtain the following estimation
for the absolute value of such obtained difference 
\begin{multline}\label{q-q'-contractionEstimation}
\Big|
\overset{l}{\underset{i=1}{\dot{\otimes}}} \kappa^{{}^{l_i}}_{\epsilon \, 1,0} 
\overset{m}{\underset{i=1}{\dot{\otimes}}} \kappa^{{}^{m_i}}_{\epsilon \, 0,1}
\overset{q}{\underset{i=1}{\dot{\otimes}}} \kappa^{{}^{q_i}}_{\epsilon \, 0,1}(\phi)
\, \otimes|_{{}_{q}} \,\,
\overset{q}{\underset{i=1}{\dot{\otimes}}} \kappa^{{}^{q_i}}_{\epsilon \, 1,0}
\overset{\ell}{\underset{i=1}{\dot{\otimes}}} \kappa^{{}^{\ell_i}}_{\epsilon \, 1,0}
\overset{\mathpzc{m}}{\underset{i=1}{\dot{\otimes}}} \kappa^{{}^{\mathpzc{m}_i}}_{\epsilon \, 0,1}(\varphi)
\\
-
\overset{l}{\underset{i=1}{\dot{\otimes}}} \kappa^{{}^{l_i}}_{1,0} 
\overset{m}{\underset{i=1}{\dot{\otimes}}} \kappa^{{}^{m_i}}_{0,1}
\overset{q}{\underset{i=1}{\dot{\otimes}}} \kappa^{{}^{q_i}}_{0,1}(\phi)
\, \otimes|_{{}_{q}} 
\overset{q}{\underset{i=1}{\dot{\otimes}}} \kappa^{{}^{q_i}}_{1,0}
\overset{\ell}{\underset{i=1}{\dot{\otimes}}} \kappa^{{}^{\ell_i}}_{1,0}
\overset{\mathpzc{m}}{\underset{i=1}{\dot{\otimes}}} \kappa^{{}^{\mathpzc{m}_i}}_{0,1}(\varphi)
\Big|
\\
\leq
\epsilon \, c \,
\left(\sum\limits_{0=|\gamma|\leq |\alpha|\leq k} |w^\gamma|{\alpha \choose \gamma}P_{{}_{l,m}} \cdot \, \big|\phi\big|_{{}_{k}}\right)
\left(\sum\limits_{0=|\gamma|\leq |\alpha| \leq n}|\upsilon^\gamma|{\alpha \choose \gamma}P_{{}_{\ell,\mathpzc{m}}} \cdot \, \big|\varphi\big|_{{}_{n}}\right),
\end{multline}
for all $\phi \in \mathscr{E}_1$, $\varphi \in \mathscr{E}_2$ and with $k,n \in \mathbb{N}$ and finite $c>0$, independent of $\phi,\varphi$. 
Here
\begin{gather*}
w = 
\Bigg( \overbrace{\sum\limits_{i=1}^{l}\boldsymbol{\p}_{{}_{l_i}}- \sum\limits_{i=1}^{m}\boldsymbol{\p}_{{}_{m_i}}}^{\boldsymbol{w}} \,\,\,\, , 
\,\,\,\,\,\,\,\,\,\,\,
\overbrace{\sum\limits_{i=1}^{l}|\boldsymbol{\p}_{{}_{l_i}}|- \sum\limits_{i=1}^{m}|\boldsymbol{\p}_{{}_{m_i}}|}^{w_0}  \Bigg) 
= (\boldsymbol{w},w_0) = (w_1,w_2,w_3,w_0),
\\
\upsilon = 
\Big( \underbrace{\sum\limits_{i=1}^{\ell}\boldsymbol{\p}_{{}_{\ell_i}}
-\sum\limits_{i=1}^{\mathpzc{m}} \boldsymbol{\p}_{{}_{\mathpzc{m}_i}}}_{\boldsymbol{\upsilon}} \,\,\,\, , 
\,\,\,\,\,\,\,\,\,\,\,
\underbrace{\sum\limits_{i}^{\ell}|\boldsymbol{\p}_{{}_{\ell_i}}|
-\sum\limits_{i=1}^{\mathpzc{m}} |\boldsymbol{\p}_{{}_{\mathpzc{m}_i}}|}_{\upsilon_0}  \Big)
= (\boldsymbol{\upsilon},\upsilon_0) = (\upsilon_1,\upsilon_2,\upsilon_3,\upsilon_0),
\end{gather*}
and $\alpha,\beta, \gamma$, are $4$-component multiindices, for example $\gamma= (\gamma_0,\gamma_1, \gamma_2, \gamma_3)$ and
\[
w^{\gamma} = {w}_{0}^{\gamma_0}{w}_{1}^{\gamma_1} {w}_{2}^{\gamma_2} {w}_{3}^{\gamma_3}, 
\,\,\,\,\,
\upsilon^{\gamma} = {\upsilon}_{0}^{\gamma_0}{\upsilon}_{1}^{\gamma_1} {\upsilon}_{2}^{\gamma_2} {\upsilon}_{3}^{\gamma_3},
\,\,\,\,\,
{\alpha \choose \gamma} = {\alpha_0 \choose \gamma_0} {\alpha_1 \choose \gamma_1}{\alpha_2 \choose \gamma_2}{\alpha_3 \choose \gamma_3}.   
\]
\[
P_{{}_{l,m}} = \prod\limits_{i=1}^{l} \left| v^{{}^{l_i}}_{{}_{s_{{}_{l_i}}}}(\boldsymbol{\p}_{{}_{l_i}})\right| \,\,
\prod\limits_{i=1}^{m} \left| u^{{}^{m_i}}_{{}_{s_{{}_{m_i}}}}(\boldsymbol{\p}_{{}_{m_i}}) \right|,
\,\,\,
P_{{}_{\ell,\mathpzc{m}}} = \prod\limits_{i=1}^{\ell} \left| v^{{}^{\ell_i}}_{{}_{s_{{}_{\ell_i}}}}(\boldsymbol{\p}_{{}_{\ell_i}}) \right| \,\,
\prod\limits_{i=1}^{\mathpzc{m}} \left| u^{{}^{\mathpzc{m}_i}}_{{}_{s_{{}_{\mathpzc{m}_i}}}}(\boldsymbol{\p}_{{}_{\mathpzc{m}_i}}) \right|.
\]
In case of scalar $\otimes_q$-contraction, \emph{i.e.} when $|l|= |m| = |\ell| =|\mathpzc{m}|=0$ with all spin-momentum variables contracted,
$P_{{}_{l,m}},P_{{}_{\ell,\mathpzc{m}}}, w^\gamma, \upsilon^\gamma$ for $\gamma =0$ degenerate to the constant equal $1$,
and  $w^\gamma, \upsilon^\gamma =0$ for $\gamma >0$  in (\ref{q-q'-contractionEstimation}).

Indeed, let $e_{w}(x) = e^{iw\cdot x}$. 
Then $\widetilde{e_{w}\phi}(\boldsymbol{\p}, p_0 \big) = \widetilde{\phi}\big(\boldsymbol{\p}+ \boldsymbol{w}, p_0 + w_0 \big)$.
We then use the functions $e_{w}\phi$, $e_{\upsilon}\varphi$, expressing the left-hand-side of 
(\ref{q-q'-contractionEstimation}) as follows 
\begin{multline*}
\Bigg| \sum\limits_{\ldots} \int \cdots u^{{}^{q_i}}_{{}_{\epsilon}} \cdots \widetilde{e_{w}\phi}(-\ldots, -\ldots + \Delta) 
\cdots v^{{}^{q_i}}_{{}_{\epsilon}} \cdots \widetilde{e_{\upsilon}\varphi}(\ldots,\ldots + \Delta)
\\
-  \sum\limits_{\ldots} \int \cdots u^{{}^{q_i}} \cdots \widetilde{e_{w}\phi}(\ldots,\ldots) 
\cdots v^{{}^{q_i}} \cdots \widetilde{e_{\upsilon}\varphi}(\ldots,\ldots)\Bigg|
\end{multline*}
\begin{multline}\label{MeanValue1}
\leq
\Bigg| 
\sum\limits_{\ldots} \int \cdots u^{{}^{q_i}}_{{}_{\epsilon}} \cdots \widetilde{e_{w}\phi}(-\ldots,-\ldots + \Delta) 
\cdots v^{{}^{q_i}}_{{}_{\epsilon}} \cdots \widetilde{e_{\upsilon}\varphi}(\ldots,\ldots + \Delta)
\\
-  \sum\limits_{\ldots} \int \cdots u^{{}^{i}} \cdots \widetilde{e_{w}\phi}(-\ldots,-\ldots+\Delta) 
\cdots v^{{}^{q_i}} \cdots \widetilde{e_{\upsilon}\varphi}(\ldots,\ldots+\Delta)
\Bigg|
\end{multline}
\begin{multline}\label{MeanValue2}
\,\,\,\, +
\Bigg| \sum\limits_{\ldots} \int \cdots u^{{}^{q_i}} \cdots \widetilde{e_{w}\phi}(-\ldots,-\ldots + \Delta) 
\cdots v^{{}^{q_i}} \cdots \widetilde{e_{\upsilon}\varphi}(\ldots,\ldots + \Delta)
\\
-  \sum\limits_{\ldots} \int \cdots u^{{}^{q_i}} \cdots \widetilde{e_{w}\phi}(\ldots,\ldots) 
\cdots v^{{}^{q_i}} \cdots \widetilde{e_{\upsilon}\varphi}(\ldots,\ldots)\Bigg|,
\end{multline}  
where the first dots $(\ldots,\ldots)$ in the arguments of the functions $\widetilde{e_{w}\phi},\widetilde{e_{\upsilon}\varphi}$ 
denote the sum of all the spatial components of the contracted momenta, \emph{i.e.} integrated, and, the second dots
in $(\ldots,\ldots)$ representing the sum of zero components of contracted momenta;  
$u^{{}^{i}}_{{}_{\epsilon}}, v^{{}^{i}}_{{}_{\epsilon}}$ denote the multipliers 
$u^{{}^{i}}, v^{{}^{i}}$  defining the kernels of the massless fields
 in which $|\boldsymbol{\p}_i|$ are replaced with $|\boldsymbol{\p}_i|_{{}_{\epsilon}}$. Here
\[
\Delta = \sum\limits_{i=1}^{l}\Delta_{{}_{l_i}}-\sum\limits_{i=1}^{m}\Delta_{{}_{m_i}}-\sum\limits_{i=1}^{q}\Delta_{{}_{q_i}},
\Delta = \sum\limits_{i=1}^{q}\Delta_{{}_{q_i}}+\sum\limits_{i=1}^{\ell}\Delta_{{}_{\ell_i}}-\sum\limits_{i=1}^{\mathpzc{m}}\Delta_{{}_{\mathpzc{m}_i}},
\,\, \textrm{resp. in} \,\, \widetilde{\phi}(-\ldots,-\ldots+\Delta), \widetilde{\varphi}(\ldots,\ldots+\Delta),
\]
with
\[
0 \leq |\boldsymbol{\p}_i|_{{}_{\epsilon}} - |\boldsymbol{\p}_i| 
= \Delta_i = {\textstyle\frac{\epsilon^2}{\sqrt{|\boldsymbol{\p}_i|^2 +\epsilon^2} + |\boldsymbol{\p}_i|}} \leq \epsilon,
\]
denotes the total increment of the zero component. Next we use the mean value theorem 
\begin{gather*}
\widetilde{e_{w}\phi}(-\ldots,-\ldots + \Delta) = \widetilde{e_{w}\phi}(-\ldots,-\ldots) 
+ \partial_{{}_{p_0}}\widetilde{e_{w}\phi}(-\ldots,-\ldots + \lambda_1 \Delta)\Delta,
\\
\widetilde{e_{\upsilon}\varphi}(\ldots,\ldots + \Delta) = \widetilde{e_{w}\phi}(\ldots,\ldots) 
+ \partial_{{}_{p_0}}\widetilde{e_{w}\phi}(\ldots,\ldots + \lambda_2 \Delta)\Delta, 
\end{gather*}
with smooth functions $\lambda_k$ of $\boldsymbol{\p}_i$, such that $0 \leq \lambda_k \leq 1$
(and divide the range $\mathbb{R}^3$ of integration in each variable $\boldsymbol{\p}_{{}_{q_i}}$, into two subsets 
$|\boldsymbol{\p}_{{}_{q_i}}| \leq 1$, $|\boldsymbol{\p}_{{}_{q_i}}| > 1$) in (\ref{MeanValue1}), (\ref{MeanValue2}).
Then, we insert under the integration the multiplier $V_{(4)}^{n}(\ldots,\ldots)V_{(4)}^{-n}(\ldots,\ldots)$ given by (\ref{V(4)}) 
and proceeding like in (\ref{AbsoluteConvergenceKappa00}) we obtain
\begin{multline*}
\big| \kappa'_{\epsilon \, l, m+q}(\phi)  \, \otimes|_{{}_{q}}  \,\, \kappa''_{\epsilon \, q+\ell, \mathpzc{m}}(\varphi)
- \kappa'_{l, m+q}(\phi)  \, \otimes|_{{}_{q}}  \,\, \kappa''_{q+\ell, \mathpzc{m}}(\varphi) \big|
\\
\leq c \, P_{{}_{l,m}} \,\underset{p\in \mathbb{R}^4}{\textrm{sup}}\big|\widetilde{e_{w}\phi}(p) \big| 
\,\,P_{{}_{\ell,\mathpzc{m}}} \, \underset{p\in \mathbb{R}^4}{\textrm{sup}} \big|V_{(4)}^{n}\cdot \widetilde{e_{\upsilon}\varphi}(p) \big|
\leq
c \, P_{{}_{l,m}} \cdot \big|\widetilde{e_{w}\phi}\big|_{{}_{k'}} \,P_{{}_{\ell,\mathpzc{m}}} \cdot  \big|\widetilde{e_{\upsilon}\varphi}\big|_{{}_{n'}}
\\
=c \, P_{{}_{l,m}} \cdot \big|e_{w}\phi\big|_{{}_{k'}} \, P_{{}_{\ell,\mathpzc{m}}} \cdot \big|e_{\upsilon}\varphi\big|_{{}_{n'}}.
\end{multline*}
Using the equivalent system of norms 
\[
| \phi |_{{}_{k}} = \underset{|\alpha|,|\beta| \leq k, x\in\mathbb{R}^4}{\textrm{sup}} \, |x^\beta \partial^\alpha \phi(x)|,
\,\,\,\,\,\,
| \varphi |_{{}_{n}} = \underset{|\alpha|,|\beta| \leq n, x\in\mathbb{R}^4}{\textrm{sup}} \, |x^\beta \partial^\alpha \varphi(x)|,
\]
in this estimation, we obtain (\ref{q-q'-contractionEstimation}). 

From (\ref{q-q'-contractionEstimation}) it follows that the absolute value of
\begin{equation}\label{F(p^q',p^(q'+1),...)/|phi||varphi|epsilon}
{\textstyle\frac{1}{\big|\phi \big|_{{}_{k}} \big|\varphi \big|_{{}_{n}} \epsilon }}
\Big[
\kappa'_{\epsilon \, l, m+q}(\phi)  \, \otimes|_{{}_{q}}  \,\, \kappa''_{\epsilon \, q+\ell, \mathpzc{m}}(\varphi)
- \kappa'_{l, m+q}(\phi)  \, \otimes|_{{}_{q}}  \,\, \kappa''_{q+\ell, \mathpzc{m}}(\varphi)
\Big],
\end{equation}
regarded as the function of the non-contracted variables (\ref{NonContractedVariables}),
is bounded by the absolute value of a fixed multiplier (lemma \ref{FreeFieldMultipliers}) of the nuclear space (\ref{E^(q'+1)x ...xE^q}), 
independent of non-zero $\phi,\varphi$, $\epsilon$, 
bounded at infinity by a fixed polynomial in these variables. Therefore,  (\ref{F(p^q',p^(q'+1),...)/|phi||varphi|epsilon})
is a multiplier of the nuclear space
\begin{equation}\label{E^(q'+1)x ...xE^q}
\overset{l}{\underset{i=1}{\otimes}} E_{{}_{l_i}}
\overset{m}{\underset{i=1}{\otimes}} E_{{}_{m_i}}
\overset{\ell}{\underset{i=1}{\otimes}} E_{{}_{\ell_i}}
\overset{\mathpzc{m}}{\underset{i=1}{\otimes}} E_{{}_{\mathpzc{m}_i}},
\end{equation}
so that there exists a norm $|\cdot|_i$, among the norms $|\cdot|_1, |\cdot|_2, \ldots$, 
defining (\ref{E^(q'+1)x ...xE^q}), and a finite constant $c$, both independent of $\phi, \varphi, \xi$,  
such that 
\begin{equation}\label{BasicInequalityForWickEpsilon->0}
\Big|\Big\langle
\Big[
\kappa'_{\epsilon \, l, m+q}(\phi)  \, \otimes|_{{}_{q}}  \,\, \kappa''_{\epsilon \, q+\ell, \mathpzc{m}}(\varphi)
- \kappa'_{l, m+q}(\phi)  \, \otimes|_{{}_{q}}  \,\, \kappa''_{q+\ell, \mathpzc{m}}(\varphi)
\Big],
\, \xi
\Big\rangle
\Big|
\leq \epsilon \, c \, \big|\phi \big|_{{}_{k}} \big|\varphi \big|_{{}_{n}}  |\xi|_i,
\end{equation}
for all $\phi \in \mathscr{E}_1, \varphi \in \mathscr{E}_2$ and $\xi$ in (\ref{E^(q'+1)x ...xE^q}).
The inequality (\ref{BasicInequalityForWickEpsilon->0}) means that
\[
\kappa'_{\epsilon \,\, l, m+q}  \, \otimes|_{{}_{q}}  \,\, \kappa''_{\epsilon \,\, q+\ell, \mathpzc{m}}
\overset{\epsilon \rightarrow 0}{\longrightarrow}
\kappa'_{l, m+q} \, \otimes|_{{}_{q}}  \,\, \kappa''_{q+\ell, \mathpzc{m}}
\]
in
\[
\mathscr{L}\left(\overset{l}{\underset{i=1}{\otimes}} E_{{}_{l_i}}
\overset{m}{\underset{i=1}{\otimes}} E_{{}_{m_i}}
\overset{\ell}{\underset{i=1}{\otimes}} E_{{}_{\ell_i}}
\overset{\mathpzc{m}}{\underset{i=1}{\otimes}} E_{{}_{\mathpzc{m}_i}}, \mathscr{E}_{1}^{*}\otimes \mathscr{E}_{2}^{*}\right),
\]
which proves convergence (\ref{kappaepsilonconvergence}).
\qed

Of course the analogue Wick theorem decomposition holds for the (tensor) product (\ref{Wick(x)Wick(y)}) of more than just two normally ordered
Wick product factors $W_1(x)$ and $W_2(x)$ of the form (\ref{WickMonomialsInGeneralFreeFields}). Repeating the analogous decomposition for more
than just two factors, we easily see that the analogous decomposition holds for the product  
\begin{equation}\label{Wicki(xi)...Wick(xn)NormalKernelDecomposition}
t(x_1, \ldots, x_n) \, {:}W_1(x_1)W_2(x_2) \ldots W_n(x_n){:} \,\,\,\, d(y_1, \ldots, y_k) \, {:}P_1(y_1)W_2(y_2) \ldots P_k(y_k){:}
\end{equation}
where each $W_i(x_i), P_j(y_j)$ are Wick products of free fields (or their derivatives)  at $x_i$ or $y_j$,
and $t(x_1, \ldots, x_n)$, $d(y_1, \ldots, y_k) $, are the translationally invariant scalar $\otimes_q$-contractions (products of pairings) 
of massive or massless free fields, 
and with the kernels $\kappa_{l,m}$ of the Wick decomposition of the operator (\ref{Wicki(xi)...Wick(xn)NormalKernelDecomposition}) equal 
to the contractions
\[
\kappa_{{}_{l,m}}(\phi\otimes \varphi) = \sum\limits_{\kappa'_{{}_{l',m'}},\kappa''_{{}_{l'',m''}}, k}
\kappa'_{{}_{l',m'}}(\phi) \otimes_{k} \kappa''_{{}_{l'',m''}}(\varphi)
\]
\[
\phi \in \mathscr{E}^{\otimes \, n}, \,\, \varphi \in \mathscr{E}^{\otimes \, k}
\]
where in this sum $\kappa'_{{}_{l',m'}},\kappa''_{{}_{l'',m''}}$ range over the kernels respectively
of the operators
\begin{gather*}
t(x_1, \ldots, x_n) \, {:}W_1(x_1)W_2(x_2) \ldots W_n(x_n){:}  
\\
 \textrm{and} 
\\
d(y_1, \ldots, y_k) \, {:}P_1(y_1)W_2(y_2) \ldots P_k(y_k){:}
\end{gather*}
and
\[
l'+l''-k=l, \,\,\, m'+m''-k=m, 
\]
and where the contractios $\kappa'_{{}_{l',m'}}(\phi) \otimes_{k} \kappa''_{{}_{l'',m''}}(\varphi)$ 
are performed upon all $k$ pairs of spin-momenta variables which can be contracted, in which the first variable in the pair
lies among the last $m'$ variables in $\kappa'_{{}_{l',m'}}(\phi)$ and corresponds to an annihilation operator variable 
and the second one lies upon the first $l''$ variables in  $\kappa''_{{}_{l'',m''}}(\varphi)$ and corresponds to the 
creation operator variable of \emph{the same free field}. 
All these contractions are given by absolutely convergent sums/integrals with respect 
to the contracted variables.  After the contraction, 
the kernels should be symmetrized in Boson spin-momentum
variables and antisymmetrized in the Fermion spin-momentum variables
in order to keep one-to-one correspondence between the kernels and operators.

\begin{twr}
If all free fields in the Wick products $W_i(x_i)$ are massive, then the  Wick decomposition of the operator 
product (\ref{Wicki(xi)...Wick(xn)NormalKernelDecomposition}) is equal to finite sum of integral kernel operators 
which belong to
\[
\mathscr{L}\big(\mathscr{E}^{\otimes \,(n+k)}, \,\mathscr{L}((E),(E))\big)
\]
also in case the products of pairings $t(x_1, \ldots, x_n)$, $d(y_1, \ldots, y_k) $,
include the pairings of the massless fields (or their derivatives). 

Namely, in case $t(x_1, \ldots, x_n)$, $d(y_1, \ldots, y_k) $, are products of pairings of possibly massless free fields,
\emph{i.e.} scalar $\otimes_q$-contractions of $\dot{\otimes}$-products of free field kernels, 
and the free fields in $W_i(x)$, $P_j(y)$ are massive, the factors
\begin{gather*}
t(x_1, \ldots, x_n) \, {:}W_1(x_1)W_2(x_2) \ldots W_n(x_n){:} 
\\
\textrm{and} 
\\ 
d(y_1, \ldots, y_k) \, {:}P_1(y_1)P_2(y_2) \ldots P_k(y_k){:}
\end{gather*}
in (\ref{Wicki(xi)...Wick(xn)NormalKernelDecomposition}) 
represent generalized operators which belong, respectively, to 
\[
\mathscr{L}\big(\mathscr{E}^{\otimes \,n}, \,\mathscr{L}((E),(E))\big)
\,\,\, \textrm{and}
\,\,\,
\mathscr{L}\big(\mathscr{E}^{\otimes \,k}, \,\mathscr{L}((E),(E))\big),
\]
so that the product 
\[
t(x_1, \ldots, x_n) \, {:}W_1(x_1)W_2(x_2) \ldots W_n(x_n){:}  \,\,\,
d(y_1, \ldots, y_k) \, {:}P_1(y_1)P_2(y_2) \ldots P_k(y_n){:}
\]  
belongs to
\[
\mathscr{L}\big(\mathscr{E}^{\otimes \,(n+k)}, \,\mathscr{L}((E),(E))\big), 
\]
and for which the above Wick decomposition theorem likewise holds true. 
Here $\mathscr{E} = \mathcal{S}(\mathbb{R}^4; \mathbb{C}^d)$. 
The same holds true if we replace $t,d$ with any translationally invariant tempered distributions, in particular
if we replace $t,d$ with $\textrm{ret} \, t$, $\textrm{ret} \, d$ for any causal translationally invariant tempered distributions
of finite singularity order. 
$t,d$.
\label{ProductTheorem}
\end{twr}

For simplicity of notation
we assumed $W_1, \ldots. P_1, \ldots$, to have the same number $d$ of components -- inessential assumption. 
$ret$ denote the retarded part of a translationally invariant causal
distribution having finite singularity order.

{\bf Proof of theorem \ref{ProductTheorem}}.
Theorem \ref{ProductTheorem} follows from the repeated application of Lemma \ref{rett(x,y):W'(x)W''(y):InSxS->L((E),(E))Contq1>2}
given below.
\qed

In order to simplify notation we introduce the following abbreviation for scalar contractions (of possibly massless kernels) 
\begin{equation}\label{(x)q}
\kappa_q =
\overset{q}{\underset{i=1}{\dot{\otimes}}} \kappa^{{}^{q_i}}_{0,1}
\, \otimes_{{}_{q}} \,\,
\overset{q}{\underset{i=1}{\dot{\otimes}}} \kappa^{{}^{q_i}}_{1,0},
\,\,\,
\kappa_k =
\overset{k}{\underset{i=1}{\dot{\otimes}}} \kappa^{{}^{k_i}}_{0,1}
\, \otimes_{{}_{k}} \,\,
\overset{k}{\underset{i=1}{\dot{\otimes}}} \kappa^{{}^{k_i}}_{1,0},
\,\,\,
\kappa_j =
\overset{j}{\underset{i=1}{\dot{\otimes}}} \kappa^{{}^{j_i}}_{0,1}
\, \otimes_{{}_{j}} \,\,
\overset{k}{\underset{i=1}{\dot{\otimes}}} \kappa^{{}^{j_i}}_{1,0},
\,\,\,
\ldots
\end{equation}
Using thm. 3.13 of \cite{obataJFA} it follows that for the proof of theorem \ref{ProductTheorem} it is sufficient to show that 
(we have indicated the space-time variables $x_r$ with respect to which the corresponding pointwise products $\dot{\otimes}$ are computed)
\begin{gather}
\underbrace{\kappa_q}_{\textrm{funct. of $x_1,x_2$}} \cdot
\underbrace{\overset{\ell}{\underset{i=1}{\dot{\otimes}}} \kappa^{{}^{\ell_i}}_{1,0}
\overset{\mathpzc{m}}{\underset{i=1}{\dot{\otimes}}} \kappa^{{}^{\mathpzc{m}_i}}_{0,1}}_{\textrm{funct. of $x_2$}},
\label{x1(x2)}
\\
\underbrace{\kappa_q}_{\textrm{funct. of $x_1,x_2$}} \cdot
\underbrace{\overset{l}{\underset{i=1}{\dot{\otimes}}} \kappa^{{}^{l_i}}_{1,0} 
\overset{m}{\underset{i=1}{\dot{\otimes}}} \kappa^{{}^{m_i}}_{0,1}}_{\textrm{funct. of $x_1$}}
\otimes
\underbrace{\overset{\ell}{\underset{i=1}{\dot{\otimes}}} \kappa^{{}^{\ell_i}}_{1,0}
\overset{\mathpzc{m}}{\underset{i=1}{\dot{\otimes}}} \kappa^{{}^{\mathpzc{m}_i}}_{0,1}}_{\textrm{funct. of $x_2$}},
\label{x1x2}
\\
\Big[\underbrace{\kappa_q}_{\textrm{$x_1,x_3$}}
\underbrace{\kappa_k}_{\textrm{$x_2,x_3$}}
\Big] \cdot
\underbrace{\overset{l}{\underset{i=1}{\dot{\otimes}}} \kappa^{{}^{l_i}}_{1,0} 
\overset{m}{\underset{i=1}{\dot{\otimes}}} \kappa^{{}^{m_i}}_{0,1}}_{\textrm{$x_1$}}
\,
\otimes
\,
\underbrace{\overset{\ell}{\underset{i=1}{\dot{\otimes}}} \kappa^{{}^{\ell_i}}_{1,0}
\overset{\mathpzc{m}}{\underset{i=1}{\dot{\otimes}}} \kappa^{{}^{\mathpzc{m}_i}}_{0,1}}_{\textrm{$x_2$}}
\,
\otimes
\,
\underbrace{\overset{\mathfrak{l}}{\underset{i=1}{\dot{\otimes}}} \kappa^{{}^{\mathfrak{l}_i}}_{1,0}
\overset{\mathfrak{m}}{\underset{i=1}{\dot{\otimes}}} \kappa^{{}^{\mathfrak{m}_i}}_{0,1}}_{\textrm{$x_3$}},
\label{x1x2x3}
\\
\Big[\underbrace{\kappa_q}_{\textrm{$x_1,x_n$}}
\underbrace{\kappa_k}_{\textrm{$x_2,x_n$}}
\ldots
\underbrace{\kappa_j}_{\textrm{$x_{n-1},x_n$}}
\Big] \cdot
\underbrace{\overset{l}{\underset{i=1}{\dot{\otimes}}} \kappa^{{}^{l_i}}_{1,0} 
\overset{m}{\underset{i=1}{\dot{\otimes}}} \kappa^{{}^{m_i}}_{0,1}}_{\textrm{$x_1$}}
\,
\otimes
\,
\underbrace{\overset{\ell}{\underset{i=1}{\dot{\otimes}}} \kappa^{{}^{\ell_i}}_{1,0}
\overset{\mathpzc{m}}{\underset{i=1}{\dot{\otimes}}} \kappa^{{}^{\mathpzc{m}_i}}_{0,1}}_{\textrm{$x_2$}}
\, 
\otimes
\,
\ldots 
\, \otimes
\,
\underbrace{\overset{\mathfrak{l}}{\underset{i=1}{\dot{\otimes}}} \kappa^{{}^{\mathfrak{l}_i}}_{1,0}
\overset{\mathfrak{m}}{\underset{i=1}{\dot{\otimes}}} \kappa^{{}^{\mathfrak{m}_i}}_{0,1}}_{\textrm{$x_n$}},
\label{x1...xn}
\end{gather}
equal, respectively, to the kernels
\begin{gather*}
\overset{q}{\underset{i=1}{\dot{\otimes}}} \kappa^{{}^{q_i}}_{0,1}
\, \otimes_{{}_{q}} \,\,
\overset{q}{\underset{i=1}{\dot{\otimes}}} \kappa^{{}^{q_i}}_{1,0}
\overset{\ell}{\underset{i=1}{\dot{\otimes}}} \kappa^{{}^{\ell_i}}_{1,0}
\overset{\mathpzc{m}}{\underset{i=1}{\dot{\otimes}}} \kappa^{{}^{\mathpzc{m}_i}}_{0,1}
\\
\overset{l}{\underset{i=1}{\dot{\otimes}}} \kappa^{{}^{l_i}}_{1,0} 
\overset{m}{\underset{i=1}{\dot{\otimes}}} \kappa^{{}^{m_i}}_{0,1}
\overset{q}{\underset{i=1}{\dot{\otimes}}} \kappa^{{}^{q_i}}_{0,1}
\, \otimes_{{}_{q}} \,\,
\overset{q}{\underset{i=1}{\dot{\otimes}}} \kappa^{{}^{q_i}}_{1,0}
\overset{\ell}{\underset{i=1}{\dot{\otimes}}} \kappa^{{}^{\ell_i}}_{1,0}
\overset{\mathpzc{m}}{\underset{i=1}{\dot{\otimes}}} \kappa^{{}^{\mathpzc{m}_i}}_{0,1},
\\
\overset{l}{\underset{i=1}{\dot{\otimes}}} \kappa^{{}^{l_i}}_{1,0} 
\overset{m}{\underset{i=1}{\dot{\otimes}}} \kappa^{{}^{m_i}}_{0,1}
\overset{q}{\underset{i=1}{\dot{\otimes}}} \kappa^{{}^{q_i}}_{0,1}
\, \otimes_{{}_{q}} \,\,
\overset{q}{\underset{i=1}{\dot{\otimes}}} \kappa^{{}^{q_i}}_{1,0}
\overset{\ell}{\underset{i=1}{\dot{\otimes}}} \kappa^{{}^{\ell_i}}_{1,0}
\overset{\mathpzc{m}}{\underset{i=1}{\dot{\otimes}}} \kappa^{{}^{\mathpzc{m}_i}}_{0,1}
\overset{k}{\underset{i=1}{\dot{\otimes}}} \kappa^{{}^{k_i}}_{0,1}
\, \otimes_{{}_{k}} \,\,
\overset{k}{\underset{i=1}{\dot{\otimes}}} \kappa^{{}^{k_i}}_{1,0}
\overset{\mathfrak{l}}{\underset{i=1}{\dot{\otimes}}} \kappa^{{}^{\mathfrak{l}_i}}_{1,0}
\overset{\mathfrak{m}}{\underset{i=1}{\dot{\otimes}}} \kappa^{{}^{\mathfrak{m}_i}}_{0,1},
\\
\overset{l}{\underset{i=1}{\dot{\otimes}}} \kappa^{{}^{l_i}}_{1,0} 
\overset{m}{\underset{i=1}{\dot{\otimes}}} \kappa^{{}^{m_i}}_{0,1}
\overset{q}{\underset{i=1}{\dot{\otimes}}} \kappa^{{}^{q_i}}_{0,1}
\, \otimes_{{}_{q}} \,\,
\overset{q}{\underset{i=1}{\dot{\otimes}}} \kappa^{{}^{q_i}}_{1,0}
\overset{\ell}{\underset{i=1}{\dot{\otimes}}} \kappa^{{}^{\ell_i}}_{1,0}
\overset{\mathpzc{m}}{\underset{i=1}{\dot{\otimes}}} \kappa^{{}^{\mathpzc{m}_i}}_{0,1}
\overset{k}{\underset{i=1}{\dot{\otimes}}} \kappa^{{}^{k_i}}_{0,1}
\, \otimes_{{}_{k}} \,
\ldots
\,
\otimes_{{}_{j}} \,
\overset{j}{\underset{i=1}{\dot{\otimes}}} \kappa^{{}^{j_i}}_{1,0}
\overset{\mathfrak{l}}{\underset{i=1}{\dot{\otimes}}} \kappa^{{}^{\mathfrak{l}_i}}_{1,0}
\overset{\mathfrak{m}}{\underset{i=1}{\dot{\otimes}}} \kappa^{{}^{\mathfrak{m}_i}}_{0,1},
\end{gather*}
with one $\otimes_{{}_{q}}$, two $\otimes_{{}_{q}}, \otimes_{{}_{k}}$, and $n-1$ contracted tensor products 
$\otimes_{{}_{q}}, \otimes_{{}_{k}}, \ldots, \otimes_{{}_{j}}$, belong, respectively, to
\begin{gather}
\mathscr{L}\left(
\left(\overset{\ell}{\underset{i=1}{\otimes}} E_{{}_{\ell_i}}\right)^*
\overset{\mathpzc{m}}{\underset{i=1}{\otimes}} E_{{}_{\mathpzc{m}_i}}, 
\mathscr{E}_{1}^{*}\otimes \mathscr{E}_{2}^{*}\right), \,\,\,
\label{E*xE,S*xS*}
\\
\mathscr{L}\left(
\left(\overset{l}{\underset{i=1}{\otimes}} E_{{}_{l_i}}
\overset{\ell}{\underset{i=1}{\otimes}} E_{{}_{\ell_i}}\right)^*
\overset{m}{\underset{i=1}{\otimes}} E_{{}_{m_i}}
\overset{\mathpzc{m}}{\underset{i=1}{\otimes}} E_{{}_{\mathpzc{m}_i}}, 
\mathscr{E}_{1}^{*}\otimes \mathscr{E}_{2}^{*}\right), \,\,\,
\label{E*..xE..,S*xS*}
\\
\mathscr{L}\left(
\left(\overset{l}{\underset{i=1}{\otimes}} E_{{}_{l_i}}
\overset{\ell}{\underset{i=1}{\otimes}} E_{{}_{\ell_i}}
\overset{\mathfrak{l}}{\underset{i=1}{\otimes}} E_{{}_{\mathfrak{l}_i}}\right)^*
\overset{m}{\underset{i=1}{\otimes}} E_{{}_{m_i}}
\overset{\mathpzc{m}}{\underset{i=1}{\otimes}} E_{{}_{\mathpzc{m}_i}}
\overset{\mathfrak{m}}{\underset{i=1}{\otimes}} E_{{}_{\mathfrak{m}_i}}, \,\,\,
\mathscr{E}_{1}^{*}\otimes \mathscr{E}_{2}^{*}\otimes \mathscr{E}_{3}^{*}\right),
\label{E*..xE..,S*xS*xS*}
\\ 
\mathscr{L}\left(
\left(\overset{l}{\underset{i=1}{\otimes}} E_{{}_{l_i}}
\overset{\ell}{\underset{i=1}{\otimes}} E_{{}_{\ell_i}}
\otimes \ldots
\otimes
\overset{\mathfrak{l}}{\underset{i=1}{\otimes}} E_{{}_{\mathfrak{l}_i}}\right)^*
\overset{m}{\underset{i=1}{\otimes}} E_{{}_{m_i}}
\overset{\mathpzc{m}}{\underset{i=1}{\otimes}} E_{{}_{\mathpzc{m}_i}}
\otimes
\ldots
\otimes
\overset{\mathfrak{m}}{\underset{i=1}{\otimes}} E_{{}_{\mathfrak{m}_i}}, \,\,\,
\mathscr{E}_{1}^{*}\otimes \dots \otimes \mathscr{E}_{n}^{*}\right),
\label{E*..xE..,S*x...xS*}
\end{gather}
provided all non paired free field kernels are massive.

And, moreover, we have to show that it remains true if we replace the scalar contractions
\begin{gather}
t(x_1,x_2) = \kappa_q(x_1,x_2) = \kappa(x_1-x_2)
\label{kappax1x2}
\\
 t(x_1,x_2,x_3) = \kappa_q(x_1,x_3)\kappa_k(x_2,x_3) = \kappa(x_1-x_3,x_2-x_3)
\label{kappax1x2x3}
\\
 t(x_1,\ldots, x_n) = \kappa_q(x_1,x_n)\kappa_k(x_2,x_n) \ldots \kappa_p(x_{n-1}-x_n) 
= \kappa(x_1-x_n, x_2-x_{n}, \ldots, x_{n-1}-x_n),
\label{kappax1...xn}
\end{gather}
respectively, in (\ref{x1(x2)}) -- (\ref{x1...xn}),  with any traslationally invariat 
tempered distributions $t$, respectively, in 
\[
\mathscr{E}_{1}^{*}\otimes \mathscr{E}_{2}^{*}, \,\,
\mathscr{E}_{1}^{*}\otimes \mathscr{E}_{2}^{*} \otimes \mathscr{E}_{3}^{*}, \mathscr{E}_{1}^{*}\otimes \ldots \otimes \mathscr{E}_{n}^{*}.
\]
In particular, it is true for $t = \textrm{ret} \, d$ with 
any causal translationally invariant tempered distribution $d$ of finite singularity order.

\begin{lem}
Under the assumption that all non-paired kernels of free fields are massive, respectively, in (\ref{x1(x2)}) -- (\ref{x1...xn}), 
the kernels (\ref{x1(x2)})--(\ref{x1...xn})
belong, respectively, to (\ref{E*xE,S*xS*}) -- (\ref{E*..xE..,S*x...xS*}), 
and the integral kernel operators corresponding, respectively, to the kernels (\ref{x1(x2)}) -- (\ref{x1...xn})
belong, respectively, to
\[
\mathscr{L}\left(\mathscr{E}_{1} \otimes \mathscr{E}_{2}, \mathscr{L}((E), (E))\right), \,\, \ldots, \,\,
\mathscr{L}\left(\mathscr{E}_{1} \otimes \ldots \otimes \mathscr{E}_{n}, \mathscr{L}((E), (E))\right),
\,\,\,\,\,\, \mathscr{E}_i = \mathcal{S}(\mathbb{R}^4) \otimes \mathbb{C}^{d_i}.
\]
This remains true if we replace the products $t$ (\ref{kappax1x2}) -- (\ref{kappax1...xn})  of scalar $\otimes_q$-contractions 
in (\ref{x1(x2)}) -- (\ref{x1...xn}) with any translationally invariant tempered distributions $t$.
\label{rett(x,y):W'(x)W''(y):InSxS->L((E),(E))Contq1>2}
\end{lem}

\qedsymbol \,
We give a proof at once with $t$ equal to any translationally invariant tempered distribution $t$.
Analysis for $n>1$ contracted tensor products is reduced to only minor modification in the analysis of the case $n=1$, 
thus, we start with the case $n=1$, \emph{i.e.} with the kernel (\ref{x1x2}).  Then we extend the proof on 
(\ref{x1...xn}).

Let $\kappa_q(x,y)$ in (\ref{x1x2}) be equal $t(x,y) = \kappa(x-y)$, $\kappa \in \mathscr{E}_{1}^* = \mathcal{S}(\mathbb{R}^4)\otimes \mathbb{C}^{d_1}$.

Without losing generality, we can use the space-time test functions $\chi \in \mathscr{E}_1 \otimes \mathscr{E}_2$ of the form
\[
\chi(x,y) = \phi(x-y)\varphi(y), \,\,\,\, \phi \in \mathscr{E}_1 = \mathcal{S}(\mathbb{R}^4) \otimes \mathbb{C}^{d_1}, \,\,
\varphi \in \mathscr{E}_2= \mathcal{S}(\mathbb{R}^4) \otimes \mathbb{C}^{d_2}.
\]
For such $\chi$, $\langle t, \chi\rangle = \langle \kappa, \phi \rangle \widetilde{\varphi}(0) 
= \langle \widetilde{\kappa}, \widetilde{\phi} \rangle \widetilde{\varphi}(0)$.
 
In particular, there exists $k\in \mathbb{N}$ and finite positive $C_k$, 
such that
\begin{equation}\label{retkappa2Estimation}
|\langle \kappa, \phi \rangle| = \Big|\int\kappa(x)\phi(x) \, \ud^4 x\Big| = \Big|\int\widetilde{\kappa}(p)\widetilde{\phi}(p) \, \ud^4 p\Big| \leq C_{{}_{k}} \big|\phi \big|_{{}_{k}},
\end{equation}
where
\[
\big|\cdot \big|_{{}_{1}}, \big|\cdot \big|_{{}_{2}}, \ldots
\]
is the countable system of Hilbertian norms defining the Schwartz nuclear topology on $\mathscr{E}_1 = \mathcal{S}(\mathbb{R}^4) \otimes \mathbb{C}^{d_1}$.

Let $T_{{}_{w}}\widetilde{\phi}(p) = \widetilde{\phi}(p+w)$.
From  (\ref{retkappa2Estimation}), it follows existence of $k\in \mathbb{N}, 0<c< +\infty$, independent of $\phi$, such that 
\begin{equation}\label{|inttheta.Omega'phi(p',p'',...)| dpk1dpk2<infty}
\left| \left\langle \widetilde{\kappa}, T_{{}_{w}}\widetilde{\phi}  \right\rangle \right| \leq 
c \, \sum\limits_{0=|\gamma| \leq |\alpha| \leq k}|w^\gamma|{\alpha \choose \gamma} \,\, \big|\phi\big|_{{}_{k}},
\end{equation}
for any $w$. In the sequel we will use 
\[
w = 
\Bigg(\,\,\,\, \overbrace{\sum\limits_{i=1}^{l}\boldsymbol{p}_{{}_{l_i}}- \sum\limits_{i=1}^{m}\boldsymbol{p}_{{}_{m_i}}}^{\boldsymbol{w}} \,\,\,\, , 
\,\,\,\,\,\,\,\,\,\
\overbrace{ \sum\limits_{i=1}^{l}p_0(\boldsymbol{p}_{{}_{l_i}}) - \sum\limits_{i=1}^{m}p_0(\boldsymbol{p}_{{}_{m_i}})}^{w_0} \,\,\,\,  \Bigg)
= (w_1,w_2,w_3,w_0),
\]
and $\alpha,\beta, \gamma$, are $4$-component multiindices with
\[
w^{\gamma} = {w}_{0}^{\gamma_0}{w}_{1}^{\gamma_1} {w}_{2}^{\gamma_2} {w}_{3}^{\gamma_3}, \,\,\,
{\alpha \choose \gamma} = {\alpha_0 \choose \gamma_0} {\alpha_1 \choose \gamma_1}{\alpha_2 \choose \gamma_2}{\alpha_3 \choose \gamma_3}.   
\]

Indeed, let
\[
e_{w}(x) = e^{iw\cdot x}.
\]
Using the system of norms 
\[
| \phi |_{{}_{k}} = \underset{|\alpha|,|\beta| \leq k, x\in\mathbb{R}^4}{\textrm{sup}} \, |x^\beta \partial^\alpha \phi(x)|
\]
in (\ref{retkappa2Estimation}), and applying (\ref{retkappa2Estimation})
to $e_{w}\phi$, instead of $\phi$, we get (\ref{|inttheta.Omega'phi(p',p'',...)| dpk1dpk2<infty}). 
From (\ref{|inttheta.Omega'phi(p',p'',...)| dpk1dpk2<infty}) it follows that  
\begin{equation}\label{|inttheta.Omega'phi(p',p'',...)| dpk1dpk2Bounded}
\left| \left\langle \widetilde{\kappa}, T_{{}_{w}}\widetilde{\phi}  \right\rangle \right|
\end{equation}
regarded as the function of the 
variables
\[
\left(\ldots, s_{{}_{l_i}}, \boldsymbol{p}_{{}_{l_i}}, \ldots,s_{{}_{m_i}}, \boldsymbol{p}_{{}_{m_i}}, \ldots  \right) =
\overset{l}{\underset{i=1}{\times}} (s_{{}_{l_i}}, \boldsymbol{p}_{{}_{l_i}}) 
\overset{m}{\underset{i=1}{\times}} (s_{{}_{m_i}}, \boldsymbol{p}_{{}_{m_i}}),
\]
contained in $w$, is bounded by the absolute value of a fixed multiplier of $\overset{l}{\underset{i=1}{\otimes}} E_{{}_{l_i}}
\overset{m}{\underset{i=1}{\otimes}} E_{{}_{m_i}}$ , \emph{i.e.}  by a fixed polynomial in these variables, 
whenever $\phi$ ranges over a bounded set in $\mathscr{E}_1$ (recall that for massive fields $E_i$ are equal to Schwartz test spaces).

We have
\[
\Big \langle \,\, t \, \cdot \, \Big[
\overset{l}{\underset{i=1}{\dot{\otimes}}} \kappa^{{}^{l_i}}_{1,0} 
\overset{m}{\underset{i=1}{\dot{\otimes}}} \kappa^{{}^{m_i}}_{0,1}
\overset{\ell}{\underset{i=1}{\dot{\otimes}}} \kappa^{{}^{\ell_i}}_{1,0}
\overset{\mathpzc{m}}{\underset{i=1}{\dot{\otimes}}} \kappa^{{}^{\mathpzc{m}_i}}_{0,1}
\Big](\chi), \, \xi_1 \otimes \xi_2 \, \Big\rangle
\]
\begin{multline*}
=\sum\limits_{s_{{}_{l_i}}, s_{{}_{\ell_i}}, s_{{}_{\mathpzc{m}_i}}} \bigintsss \Bigg[ 
\prod\limits_{i=1}^{l} v^{{}^{l_i}}_{{}_{s_{{}_{l_i}}}}(\boldsymbol{\p}_{{}_{l_i}})
\prod\limits_{i=1}^{m} u^{{}^{m_i}}_{{}_{s_{{}_{m_i}}}}(\boldsymbol{\p}_{{}_{m_i}})
\prod\limits_{i=1}^{\ell} v^{{}^{\ell_i}}_{{}_{s_{{}_{m_i}}}}(\boldsymbol{\p}_{{}_{\ell_i}})
\prod\limits_{q=1}^{\mathpzc{m}} u^{{}^{\mathpzc{m}_i}}_{{}_{s_{{}_{\mathpzc{m}_i}}}}(\boldsymbol{\p}_{{}_{\mathpzc{m}_i}})
\,
\left\langle \widetilde{\kappa}, 
T_{{}_{w}} \widetilde{\phi} \right\rangle
 \,\, \times
\\
\times \,\, 
\widetilde{\varphi}\left(\sum\limits_{i=1}^{l}\boldsymbol{\p}_{{}_{l_i}} + \sum\limits_{i=1}^{\ell}\boldsymbol{\p}_{{}_{\ell_i}}
-\sum\limits_{i=1}^{m}\boldsymbol{\p}_{{}_{m_i}} 
-\sum\limits_{i=1}^{\mathpzc{m}} \boldsymbol{\p}_{{}_{\mathpzc{m}_i}}, \sum\limits_{i}^{l} p_0(\boldsymbol{\p}_{{}_{l_i}})
+ \sum\limits_{i}^{\ell}p_0(\boldsymbol{\p}_{{}_{\ell_i}})
-\sum\limits_{i=1}^{m}p_0(\boldsymbol{\p}_{{}_{m_i}})
-\sum\limits_{i}^{\mathpzc{m}} p_0(\boldsymbol{\p}_{{}_{\mathpzc{m}_i}})\right) 
\\
\xi_1\left(
\overset{l}{\underset{i=1}{\times}} (s_{{}_{l_i}}, \boldsymbol{p}_{{}_{l_i}} )
\overset{\ell}{\underset{i=1}{\times}} (s_{{}_{\ell_i}}, \boldsymbol{p}_{{}_{\ell_i}}) 
  \right)
\xi_2\left(
\overset{m}{\underset{i=1}{\times}} (s_{{}_{m_i}}, \boldsymbol{p}_{{}_{m_i}})
\overset{\mathpzc{m}}{\underset{i=1}{\times}} (s_{{}_{\mathpzc{m}_i}}, \boldsymbol{p}_{{}_{\mathpzc{m}_i}}) 
\right)
\Bigg] 
\prod\limits_{i}
\ud^3 \boldsymbol{\p}_{{}_{l_i}} 
\ud^3 \boldsymbol{\p}_{{}_{m_i}} 
\ud^3 \boldsymbol{\p}_{{}_{\ell_i}}  
\ud^3 \boldsymbol{\p}_{{}_{\mathpzc{m}_i}}  
\end{multline*}

Because for any fixed tempered distribution, say 
$t$,
the map
\[
\mathcal{O}_M(\mathbb{R}^4\times \mathbb{R}^4) 
\ni \zeta  \longmapsto 
\zeta \cdot t
\in
\mathcal{S}(\mathbb{R}^4\times \mathbb{R}^4)^*
\]
is continuous, then by lemma \ref{DistrProdWickProd}, the kernel 
\begin{equation}\label{tx1x2}
\kappa_{l+\ell,m+\mathpzc{m}} =
t \, \cdot \, \Big[
\overset{l}{\underset{i=1}{\dot{\otimes}}} \kappa^{{}^{l_i}}_{1,0} 
\overset{m}{\underset{i=1}{\dot{\otimes}}} \kappa^{{}^{m_i}}_{0,1}
\overset{\ell}{\underset{i=1}{\dot{\otimes}}} \kappa^{{}^{\ell_i}}_{1,0}
\overset{\mathpzc{m}}{\underset{i=1}{\dot{\otimes}}} \kappa^{{}^{\mathpzc{m}_i}}_{0,1}
\Big]
\end{equation}
 defines a continuous map
\[
\overset{l}{\underset{i=1}{\otimes}} E_{{}_{l_i}}
\overset{\ell}{\underset{i=1}{\otimes}} E_{{}_{\ell_i}}
\overset{m}{\underset{i=1}{\otimes}} E_{{}_{m_i}}
\overset{\mathpzc{m}}{\underset{i=1}{\otimes}} E_{{}_{\mathpzc{m}_i}} \ni
\xi \longmapsto  \kappa_{l+\ell,m+\mathpzc{m}}(\xi) \in \mathscr{E}_{1}^* \otimes \mathscr{E}_{2}^*.
\]

Let us give here the proof that the kernel 
(\ref{tx1x2})
can be extended to a separately continuous map
\[
\left(\overset{l}{\underset{i=1}{\otimes}} E_{{}_{l_i}}^*
\overset{\ell}{\underset{i=1}{\otimes}} E_{{}_{\ell_i}}^*\right)
\times
\left(\overset{m}{\underset{i=1}{\otimes}} E_{{}_{m_i}}
\overset{\mathpzc{m}}{\underset{i=1}{\otimes}} E_{{}_{\mathpzc{m}_i}}\right)
 \longrightarrow \mathscr{E}_{1}^* \otimes \mathscr{E}_{2}^*
\]
over the distributions in $\overset{l}{\underset{i=1}{\otimes}} E_{{}_{l_i}}^*\overset{\ell}{\underset{i=1}{\otimes}} E_{{}_{\ell_i}}^*$
in the variables $\overset{l}{\underset{i=1}{\times}} (s_{{}_{l_i}},\boldsymbol{\p}_{{}_{l_i}}) 
\overset{\ell}{\underset{i=1}{\times}} (s_{{}_{\ell_i}},\boldsymbol{\p}_{{}_{\ell_i}})$.

We note that both $\phi,\varphi$ range over bounded sets in $\mathscr{E}_1,\mathscr{E}_2$,
whenever $\chi$ ranges over a bounded set $B$ in  $\mathscr{E}_1 \otimes \mathscr{E}_2 = \mathcal{S}(\mathbb{R}^4\times \mathbb{R}^4)
\otimes \mathbb{C}^{d_1+d_2}$. Next, we use (\ref{|inttheta.Omega'phi(p',p'',...)| dpk1dpk2<infty}) or, equivalently, 
(\ref{|inttheta.Omega'phi(p',p'',...)| dpk1dpk2Bounded}), in the proof of boundedness of the following sets of functions.
First, we define the following set $B'(S',B)$
of functions
\begin{multline*}
\overset{m}{\underset{i=1}{\times}} (s_{{}_{m_i}}, \boldsymbol{p}_{{}_{m_i}}) 
\overset{\mathpzc{m}}{\underset{i=1}{\times}} (s_{{}_{\mathpzc{m}_i}}, \boldsymbol{p}_{{}_{\mathpzc{m}_i}}) 
\longmapsto
\\
\sum\limits_{s_{{}_{\ell_i}, s_{{}_{\mathpzc{m}_i}}}} \bigintsss \Bigg[ 
\prod\limits_{i=1}^{l} v^{{}^{l_i}}_{{}_{s_{{}_{l_i}}}}(\boldsymbol{\p}_{{}_{l_i}})
\prod\limits_{i=1}^{m} u^{{}^{m_i}}_{{}_{s_{{}_{m_i}}}}(\boldsymbol{\p}_{{}_{m_i}})
\prod\limits_{i=1}^{\ell} v^{{}^{\ell_i}}_{{}_{s_{{}_{m_i}}}}(\boldsymbol{\p}_{{}_{\ell_i}})
\prod\limits_{q=1}^{\mathpzc{m}} u^{{}^{\mathpzc{m}_i}}_{{}_{s_{{}_{\mathpzc{m}_i}}}}(\boldsymbol{\p}_{{}_{\mathpzc{m}_i}})
\,
\left\langle \widetilde{\kappa}, 
T_{{}_{w}} \widetilde{\phi} \right\rangle
 \,\, \times
\\
\times \,\, 
\widetilde{\varphi}\left(\sum\limits_{i=1}^{l}\boldsymbol{\p}_{{}_{l_i}} + \sum\limits_{i=1}^{\ell}\boldsymbol{\p}_{{}_{\ell_i}}
-\sum\limits_{i=1}^{m}\boldsymbol{\p}_{{}_{m_i}} 
-\sum\limits_{i=1}^{\mathpzc{m}} \boldsymbol{\p}_{{}_{\mathpzc{m}_i}}, \sum\limits_{i}^{l} p_0(\boldsymbol{\p}_{{}_{l_i}})
+ \sum\limits_{i}^{\ell}p_0(\boldsymbol{\p}_{{}_{\ell_i}})
-\sum\limits_{i=1}^{m}p_0(\boldsymbol{\p}_{{}_{m_i}})
-\sum\limits_{i}^{\mathpzc{m}} p_0(\boldsymbol{\p}_{{}_{\mathpzc{m}_i}})\right) 
\\
\xi_1\left(
\overset{l}{\underset{i=1}{\times}} (s_{{}_{l_i}}, \boldsymbol{p}_{{}_{l_i}})
\overset{\ell}{\underset{i=1}{\times}} (s_{{}_{\ell_i}}, \boldsymbol{p}_{{}_{\ell_i}}) 
  \right)
\Bigg] 
\prod\limits_{i=1}^{l} \ud^3 \boldsymbol{\p}_{{}_{l_i}} 
\prod\limits_{i=1}^{\ell} \ud^3 \boldsymbol{\p}_{{}_{\ell_i}} 
\end{multline*}
and note that it is bounded in the topology induced from the strong dual topology on 
$\overset{m}{\underset{i=1}{\otimes}} E_{{}_{m_i}}^*
\overset{\mathpzc{m}}{\underset{i=1}{\otimes}} E_{{}_{\mathpzc{m}_i}}^*$
$=\mathcal{S}(\mathbb{R}^3)^{* \otimes \, (m+\mathpzc{m})}\otimes \mathbb{C}^{n_1}$, whenever $\xi_1$ ranges over 
a set $S'$ in $\overset{l}{\underset{i=1}{\otimes}} E_{{}_{l_i}}
\overset{\ell}{\underset{i=1}{\otimes}} E_{{}_{\ell_i}}$
$=\mathcal{S}(\mathbb{R}^3)^{\otimes \, (l+\ell)}\otimes \mathbb{C}^{n_2}$
which is bounded with respect to the strong
dual topology on $\overset{l}{\underset{i=1}{\otimes}} E_{{}_{l_i}}^*
\overset{\ell}{\underset{i=1}{\otimes}} E_{{}_{\ell_i}}^*$
$=\mathcal{S}(\mathbb{R}^3)^{* \otimes \, (l+\ell)}\otimes \mathbb{C}^{n_2}$
and whenever $\chi$ ranges over a bounded set $B$ in $\mathscr{E}_1 \otimes \mathscr{E}_1 = \mathcal{S}(\mathbb{R}^4 \times \mathbb{R}^4)
\otimes \mathbb{C}^{d_1+d_2}$. 
Let us denote the family of subsets in $\overset{l}{\underset{i=1}{\otimes}} E_{{}_{l_i}}
\overset{\ell}{\underset{i=1}{\otimes}} E_{{}_{\ell_i}}$
which are bounded with respect to  the strong dual topology inherited from 
$\overset{l}{\underset{i=1}{\otimes}} E_{{}_{l_i}}^*
\overset{\ell}{\underset{i=1}{\otimes}} E_{{}_{\ell_i}}^*$
by $\mathfrak{S}$.

Because $\overset{m}{\underset{i=1}{\otimes}} E_{{}_{m_i}}
\overset{\mathpzc{m}}{\underset{i=1}{\otimes}} E_{{}_{\mathpzc{m}_i}}$
satisfies the first axiom of countability, then
there exists a zero-neighborhood $V\big(B'(B,S'),\epsilon\big)$  in 
$\overset{m}{\underset{i=1}{\otimes}} E_{{}_{m_i}}
\overset{\mathpzc{m}}{\underset{i=1}{\otimes}} E_{{}_{\mathpzc{m}_i}}$
such that (\cite{GelfandII}, p. 45) 
\[
|\langle \xi_2, f \rangle| <\epsilon, \,\,\, \xi_2 \in V\big(B'(B,S'),\epsilon\big), \,\,\,\,f  \in B'(B,S'). 
\]

Next, using (\ref{|inttheta.Omega'phi(p',p'',...)| dpk1dpk2<infty}), 
or, equivalently, (\ref{|inttheta.Omega'phi(p',p'',...)| dpk1dpk2Bounded}), we note that the following set $B^+(B,S)$ of functions
\begin{multline*}
\overset{l}{\underset{i=1}{\times}} (s_{{}_{l_i}}, \boldsymbol{p}_{{}_{l_i}})
\overset{\ell}{\underset{i=1}{\times}} (s_{{}_{\ell_i}}, \boldsymbol{p}_{{}_{\ell_i}}) 
\longmapsto
\\
\sum\limits_{s_{{}_{m_i}, s_{{}_{\mathpzc{m}_i}}}} \bigintsss \Bigg[ 
\prod\limits_{i=1}^{l} v^{{}^{l_i}}_{{}_{s_{{}_{l_i}}}}(\boldsymbol{\p}_{{}_{l_i}})
\prod\limits_{i=1}^{m} u^{{}^{m_i}}_{{}_{s_{{}_{m_i}}}}(\boldsymbol{\p}_{{}_{m_i}})
\prod\limits_{i=1}^{\ell} v^{{}^{\ell_i}}_{{}_{s_{{}_{m_i}}}}(\boldsymbol{\p}_{{}_{\ell_i}})
\prod\limits_{q=1}^{\mathpzc{m}} u^{{}^{\mathpzc{m}_i}}_{{}_{s_{{}_{\mathpzc{m}_i}}}}(\boldsymbol{\p}_{{}_{\mathpzc{m}_i}})
\,
\left\langle \widetilde{\kappa}, 
T_{{}_{w}} \widetilde{\phi} \right\rangle
 \,\, \times
\\
\times \,\, 
\widetilde{\varphi}\left(\sum\limits_{i=1}^{l}\boldsymbol{\p}_{{}_{l_i}} + \sum\limits_{i=1}^{\ell}\boldsymbol{\p}_{{}_{\ell_i}}
-\sum\limits_{i=1}^{m}\boldsymbol{\p}_{{}_{m_i}} 
-\sum\limits_{i=1}^{\mathpzc{m}} \boldsymbol{\p}_{{}_{\mathpzc{m}_i}}, \sum\limits_{i}^{l} p_0(\boldsymbol{\p}_{{}_{l_i}})
+ \sum\limits_{i}^{\ell}p_0(\boldsymbol{\p}_{{}_{\ell_i}})
-\sum\limits_{i=1}^{m}p_0(\boldsymbol{\p}_{{}_{m_i}})
-\sum\limits_{i}^{\mathpzc{m}} p_0(\boldsymbol{\p}_{{}_{\mathpzc{m}_i}})\right) 
\\
\xi_2\left(
\overset{m}{\underset{i=1}{\times}} (s_{{}_{m_i}}, \boldsymbol{p}_{{}_{m_i}}) 
\overset{\mathpzc{m}}{\underset{i=1}{\times}} (s_{{}_{\mathpzc{m}_i}}, \boldsymbol{p}_{{}_{\mathpzc{m}_i}}) 
  \right)
\Bigg] 
\prod\limits_{i=1}^{m} \ud^3 \boldsymbol{\p}_{{}_{m_i}} 
\prod\limits_{i=1}^{\mathpzc{m}} \ud^3 \boldsymbol{\p}_{{}_{\mathpzc{m}_i}} 
\end{multline*}
is bounded in 
$\overset{l}{\underset{i=1}{\otimes}} E_{{}_{l_i}}
\overset{\ell}{\underset{i=1}{\otimes}} E_{{}_{\ell_i}}$, whenever $\xi_2$ ranges over 
a set $S$ bounded in 
$\overset{m}{\underset{i=1}{\otimes}} E_{{}_{m_i}}
\overset{\mathpzc{m}}{\underset{i=1}{\otimes}} E_{{}_{\mathpzc{m}_i}}$
and whenever $\chi$ ranges over a bounded set $B$ in $\mathscr{E}_1 \otimes \mathscr{E}_2$. 
Let us denote the family of subsets in 
$\overset{m}{\underset{i=1}{\otimes}} E_{{}_{m_i}}
\overset{\mathpzc{m}}{\underset{i=1}{\otimes}} E_{{}_{\mathpzc{m}_i}}$
which are bounded with respect to  the initial topology in 
$\overset{m}{\underset{i=1}{\otimes}} E_{{}_{m_i}}
\overset{\mathpzc{m}}{\underset{i=1}{\otimes}} E_{{}_{\mathpzc{m}_i}}$,
by $\mathfrak{I}$.

Next, for any $S'\in \mathfrak{S}$, $S\in \mathfrak{I}$ and any strong zero-neighborhood $W(B,\epsilon)$
determined by a bounded set $B$ in 
$\mathscr{E}_1 \otimes \mathscr{E}_2$, 
for the zero-neighborhood $V\big(B'(B,S'),\epsilon\big)$ in 
$\overset{m}{\underset{i=1}{\otimes}} E_{{}_{m_i}}
\overset{\mathpzc{m}}{\underset{i=1}{\otimes}} E_{{}_{\mathpzc{m}_i}}$, 
and the strong zero-neighborhood $W\big(B^+(S,B),\epsilon\big)$ in 
\[
\left(\overset{l}{\underset{i=1}{\otimes}} E_{{}_{l_i}}^*
\overset{\ell}{\underset{i=1}{\otimes}} E_{{}_{\ell_i}}^*\right)
\cap
\left(\overset{l}{\underset{i=1}{\otimes}} E_{{}_{l_i}}
\overset{\ell}{\underset{i=1}{\otimes}} E_{{}_{\ell_i}}\right)
=
\left(\mathcal{S}(\mathbb{R}^3)^{* \otimes \, (l+\ell)} \cap \mathcal{S}(\mathbb{R}^3)^{\otimes \, (l+\ell)}\right) \otimes \mathbb{C}^{n_1+n_2}
\]
we have
\[
\Bigg|
\Big \langle \, t \cdot \, \Big[
\overset{l}{\underset{i=1}{\dot{\otimes}}} \kappa^{{}^{l_i}}_{1,0} 
\overset{m}{\underset{i=1}{\dot{\otimes}}} \kappa^{{}^{m_i}}_{0,1}
\overset{\ell}{\underset{i=1}{\dot{\otimes}}} \kappa^{{}^{\ell_i}}_{1,0}
\overset{\mathpzc{m}}{\underset{i=1}{\dot{\otimes}}} \kappa^{{}^{\mathpzc{m}_i}}_{0,1}
\Big](\xi_1 \otimes \xi_2), 
\, \chi \, \Big\rangle
\Bigg| < \epsilon
\]
whenever 
\[
\xi_2 \in S, 
\,\,\,\,\,
\xi_1 \in W\big(B^+(S,B),\epsilon\big)
\]
or whenever
\[
\xi_2 \in V\big(B'(B,S'),\epsilon\big),
\,\,\,\,\,
\xi_1 \in S'.
\]
Put otherwise, 
\[
t \, \cdot \,  \Big[
\overset{l}{\underset{i=1}{\dot{\otimes}}} \kappa^{{}^{l_i}}_{1,0} 
\overset{m}{\underset{i=1}{\dot{\otimes}}} \kappa^{{}^{m_i}}_{0,1}
\overset{\ell}{\underset{i=1}{\dot{\otimes}}} \kappa^{{}^{\ell_i}}_{1,0}
\overset{\mathpzc{m}}{\underset{i=1}{\dot{\otimes}}} \kappa^{{}^{\mathpzc{m}_i}}_{0,1}
\Big](\xi_1 \otimes \xi_2)
\in W(B,\epsilon)
\]
whenever 
\[
\xi_2 \in S, 
\,\,\,\,\,
\xi_1 \in W\big(B^+(S,B),\epsilon\big)
\]
or whenever
\[
\xi_2 \in V(B'(B,S'),\epsilon),
\,\,\,\,\,
\xi_1 \in S'.
\]
Thus, the map
\begin{multline}\label{ret(kappa...)x2(kappa...)hypocont}
\left(\overset{l}{\underset{i=1}{\otimes}} E_{{}_{l_i}}
\overset{\ell}{\underset{i=1}{\otimes}} E_{{}_{\ell_i}}\right)
\times
\left(\overset{m}{\underset{i=1}{\otimes}} E_{{}_{m_i}}
\overset{\mathpzc{m}}{\underset{i=1}{\otimes}} E_{{}_{\mathpzc{m}_i}}\right) \ni
\xi_1 \times \xi_2
\longmapsto
\\
t \, \cdot \, \Big[
\overset{l}{\underset{i=1}{\dot{\otimes}}} \kappa^{{}^{l_i}}_{1,0} 
\overset{m}{\underset{i=1}{\dot{\otimes}}} \kappa^{{}^{m_i}}_{0,1}
\overset{\ell}{\underset{i=1}{\dot{\otimes}}} \kappa^{{}^{\ell_i}}_{1,0}
\overset{\mathpzc{m}}{\underset{i=1}{\dot{\otimes}}} \kappa^{{}^{\mathpzc{m}_i}}_{0,1}
\Big](\xi_1 \otimes \xi_2)
\\
\in \mathscr{E}_{1}^{*} \otimes \mathscr{E}_{2}^{*} = \mathcal{S}(\mathbb{R}^4\times \mathbb{R}^4)^* \otimes \mathbb{C}^{d_1+d_2}
\end{multline}
is $(\mathfrak{S},\mathfrak{I})$-hypocontinuous as a map
\[
\left(\overset{l}{\underset{i=1}{\otimes}} E_{{}_{l_i}}
\overset{\ell}{\underset{i=1}{\otimes}} E_{{}_{\ell_i}}\right)
\times
\left(\overset{m}{\underset{i=1}{\otimes}} E_{{}_{m_i}}
\overset{\mathpzc{m}}{\underset{i=1}{\otimes}} E_{{}_{\mathpzc{m}_i}}\right) \longrightarrow
\mathscr{E}_{1}^{*} \otimes \mathscr{E}_{2}^{*} 
\]
with the topology on 
$\overset{l}{\underset{i=1}{\otimes}} E_{{}_{l_i}}
\overset{\ell}{\underset{i=1}{\otimes}} E_{{}_{\ell_i}}$
induced by the strong dual topology of 
$\overset{l}{\underset{i=1}{\otimes}} E_{{}_{l_i}}^*
\overset{\ell}{\underset{i=1}{\otimes}}^* E_{{}_{\ell_i}}^* \supset \overset{l}{\underset{i=1}{\otimes}} E_{{}_{l_i}}
\overset{\ell}{\underset{i=1}{\otimes}} E_{{}_{\ell_i}}$, and with the initial topology
on $\overset{m}{\underset{i=1}{\otimes}} E_{{}_{m_i}}
\overset{\mathpzc{m}}{\underset{i=1}{\otimes}} E_{{}_{\mathpzc{m}_i}}$
 and the ordinary strong dual topology on $\mathscr{E}_{1}^{*} \otimes \mathscr{E}_{2}^{*}$.

By the Proposition of Chap. III, \S 5.4, p. 90 of \cite{Schaefer}, we see that the map (\ref{ret(kappa...)x2(kappa...)hypocont})
can be uniquely extended to a $(\mathfrak{S}^*,\mathfrak{I})$-hypocontinuous map 
\[
\left(\overset{l}{\underset{i=1}{\otimes}} E_{{}_{l_i}}^*
\overset{\ell}{\underset{i=1}{\otimes}} E_{{}_{\ell_i}}^*\right)
\times
\left(\overset{m}{\underset{i=1}{\otimes}} E_{{}_{m_i}}
\overset{\mathpzc{m}}{\underset{i=1}{\otimes}} E_{{}_{\mathpzc{m}_i}}\right) \longrightarrow
\mathscr{E}_{1}^{*} \otimes \mathscr{E}_{2}^{*},
\]
where $\mathfrak{S}^*$ is the family of all bounded sets in 
$\overset{l}{\underset{i=1}{\otimes}} E_{{}_{l_i}}^*
\overset{\ell}{\underset{i=1}{\otimes}} E_{{}_{\ell_i}}^*$
with respect to the strong dual topology. In particular the map (\ref{ret(kappa...)x2(kappa...)hypocont})
can be uniquely extended to a separately continuous map 
\[
\left(\overset{l}{\underset{i=1}{\otimes}} E_{{}_{l_i}}^*
\overset{\ell}{\underset{i=1}{\otimes}} E_{{}_{\ell_i}}^*\right)
\times
\left(\overset{m}{\underset{i=1}{\otimes}} E_{{}_{m_i}}
\overset{\mathpzc{m}}{\underset{i=1}{\otimes}} E_{{}_{\mathpzc{m}_i}}\right) \longrightarrow
\mathscr{E}_{1}^{*} \otimes \mathscr{E}_{2}^{*},
\]
which was to be proved, because by Theorem 3.13 of \cite{obataJFA} (in the Bose case, or by its Fermi generalization), assertion of our Lemma follows. 
This eds the proof for (\ref{x1x2}).

In passing to (\ref{x1...xn}) with translationally invariant $t(x_1,\ldots,x_n) = \kappa(x_1-x_n,\ldots,x_{n-1}-x_n)$ 
we note that we can analogously, and without losing generality, 
use the space-time test functions $\chi \in \mathscr{E}_1 \otimes \ldots \otimes \mathscr{E}_n$ of the form
\begin{multline*}
\chi(x_1, \ldots, x_n) = \phi_{{}_{1}}(x_1-x_n) \ldots \phi_{{}_{n-1}}(x_{n-1}-x_n)\varphi(x_n), \,\,\,\, \phi_{{}_{i}} \in \mathscr{E}_i 
= \mathcal{S}(\mathbb{R}^4) \otimes \mathbb{C}^{d_i}, 
\\
\varphi \in \mathscr{E}_2= \mathcal{S}(\mathbb{R}^4) \otimes \mathbb{C}^{d_n}.
\end{multline*}
The above proof remains unchanged, with only minor and obvious replacement 
of $\phi$ with $\phi = \phi_{{}_{1}} \otimes \ldots \otimes \phi_{{}_{n-1}}$ and $y$ with $x_n$. The formula for the pairing of the kernel
(\ref{x1...xn}) with the respective $\chi \otimes \xi_1\otimes \xi_2$ is identical with that given above for the pairing of (\ref{x1x2})
with the obvious replacement of the sum of momenta in the argument of $\widetilde{\varphi}$ enlarged respectively to 
\[
\overset{l}{\underset{i}{\sum}}... + \overset{\ell}{\underset{i}{\sum}}... + \ldots + \overset{\mathfrak{l}}{\underset{i}{\sum}}...
-\overset{m}{\underset{i}{\sum}}... - \overset{\mathpzc{m}}{\underset{i}{\sum}}... - \ldots - \overset{\mathfrak{m}}{\underset{i}{\sum}}...
\]
and with the obvious modification of $w$ with sum of momenta in it respectively enlarged to the sum of momenta corresponding to all
the contracted tensor product factors in (\ref{x1...xn}) except the last one taken with $+$ sign for $1,0$ and $-$ sign for $0,1$
free field kernel factors, and with additional multipliers $u,v$ corresponding to the additional free field kernels. 

Concerning  (\ref{x1(x2)}), the proof remains the same as above provided $\ell,\mathpzc{m} \neq 0$ and in case $\mathpzc{m}=0$ extendibility of (\ref{x1(x2)})
to an element of (\ref{E*xE,S*xS*}) is obvious.   
\qed

We have also the following 
\begin{lem}
Let translationally invariant $t_\epsilon \in \mathscr{E}_{1}^* \otimes \ldots \otimes \mathscr{E}_{n}^*$, $t_\epsilon(x_1,\ldots,x_n) = \kappa_\epsilon(x_1-x_n, \ldots, x_{n-1}-x_n)$, $\kappa_\epsilon \in \mathscr{E}_{1}^* \otimes \ldots \otimes \mathscr{E}_{n-1}^*$, converge to a translationally invariant $t$ in $\mathscr{E}_{1}^* \otimes \ldots \otimes \mathscr{E}_{n}^*$ when $\epsilon \rightarrow 0$. The following  kernel
\[
\kappa_{{}_{\epsilon \,\,\, l+\ell+ \ldots +\mathfrak{l}, \,\, m+ \mathpzc{m} + \ldots + \mathfrak{m}}} =
\big[ t_\epsilon \big] \cdot
\underbrace{\overset{l}{\underset{i=1}{\dot{\otimes}}} \kappa^{{}^{l_i}}_{\epsilon \, 1,0} 
\overset{m}{\underset{i=1}{\dot{\otimes}}} \kappa^{{}^{m_i}}_{\epsilon \, 0,1}}_{\textrm{$x_1$}}
\,
\otimes
\,
\underbrace{\overset{\ell}{\underset{i=1}{\dot{\otimes}}} \kappa^{{}^{\ell_i}}_{\epsilon \, 1,0}
\overset{\mathpzc{m}}{\underset{i=1}{\dot{\otimes}}} \kappa^{{}^{\mathpzc{m}_i}}_{\epsilon \, 0,1}}_{\textrm{$x_2$}}
\, 
\otimes
\,
\ldots 
\, \otimes
\,
\underbrace{\overset{\mathfrak{l}}{\underset{i=1}{\dot{\otimes}}} \kappa^{{}^{\mathfrak{l}_i}}_{\epsilon \, 1,0}
\overset{\mathfrak{m}}{\underset{i=1}{\dot{\otimes}}} \kappa^{{}^{\mathfrak{m}_i}}_{\epsilon \, 0,1}}_{\textrm{$x_n$}},
\]
in which only the non contracted plane wave massless kernels are replaced with their massive counterparts,
converges to
\[
\kappa_{{}_{l+\ell+ \ldots +\mathfrak{l}, \,\, m+ \mathpzc{m} + \ldots + \mathfrak{m}}} =
\big[ t \big] \cdot
\underbrace{\overset{l}{\underset{i=1}{\dot{\otimes}}} \kappa^{{}^{l_i}}_{1,0} 
\overset{m}{\underset{i=1}{\dot{\otimes}}} \kappa^{{}^{m_i}}_{0,1}}_{\textrm{$x_1$}}
\,
\otimes
\,
\underbrace{\overset{\ell}{\underset{i=1}{\dot{\otimes}}} \kappa^{{}^{\ell_i}}_{1,0}
\overset{\mathpzc{m}}{\underset{i=1}{\dot{\otimes}}} \kappa^{{}^{\mathpzc{m}_i}}_{0,1}}_{\textrm{$x_2$}}
\, 
\otimes
\,
\ldots 
\, \otimes
\,
\underbrace{\overset{\mathfrak{l}}{\underset{i=1}{\dot{\otimes}}} \kappa^{{}^{\mathfrak{l}_i}}_{1,0}
\overset{\mathfrak{m}}{\underset{i=1}{\dot{\otimes}}} \kappa^{{}^{\mathfrak{m}_i}}_{0,1}}_{\textrm{$x_n$}},
\]
in
\[
\mathscr{L}\left(
\overset{l}{\underset{i=1}{\otimes}} E_{{}_{l_i}}
\overset{\ell}{\underset{i=1}{\otimes}} E_{{}_{\ell_i}}
\otimes \ldots
\otimes
\overset{\mathfrak{l}}{\underset{i=1}{\otimes}} E_{{}_{\mathfrak{l}_i}}
\overset{m}{\underset{i=1}{\otimes}} E_{{}_{m_i}}
\overset{\mathpzc{m}}{\underset{i=1}{\otimes}} E_{{}_{\mathpzc{m}_i}}
\otimes
\ldots
\otimes
\overset{\mathfrak{m}}{\underset{i=1}{\otimes}} E_{{}_{\mathfrak{m}_i}}, \,\,\,
\mathscr{E}_{1}^{*}\otimes \dots \otimes \mathscr{E}_{n}^{*}\right),
\]
if $\epsilon \rightarrow 0$.
\label{auxiliaryProductLemma}
\end{lem}

\qedsymbol \,
The proof follows from the inequality, which assures existence of finite $c$ and natural $k,n,j$,
such that
\[
\big|
\langle  
\kappa_{{}_{\epsilon \,\,\, l+\ell+ \ldots +\mathfrak{l}, \,\, m+ \mathpzc{m} + \ldots + \mathfrak{m}}}(\phi)
-
\kappa_{{}_{l+\ell+ \ldots +\mathfrak{l}, \,\, m+ \mathpzc{m} + \ldots + \mathfrak{m}}}(\phi), \, \xi
\rangle
\big|
\leq
\epsilon \, c \, \big|\phi  \big|_{{}_{k}}  |\xi|_{{}_{j}}
\]
for all $\phi \in \mathscr{E}_1 \otimes \ldots \otimes \mathscr{E}_n$, $\mathscr{E}_i = \mathcal{S}(\mathbb{R}^4; \mathbb{C}^{d_i})$ and for all 
\[
\xi \in \overset{l}{\underset{i=1}{\otimes}} E_{{}_{l_i}}
\overset{\ell}{\underset{i=1}{\otimes}} E_{{}_{\ell_i}}
\otimes \ldots
\otimes
\overset{\mathfrak{l}}{\underset{i=1}{\otimes}} E_{{}_{\mathfrak{l}_i}}
\overset{m}{\underset{i=1}{\otimes}} E_{{}_{m_i}}
\overset{\mathpzc{m}}{\underset{i=1}{\otimes}} E_{{}_{\mathpzc{m}_i}}
\otimes
\ldots
\otimes
\overset{\mathfrak{m}}{\underset{i=1}{\otimes}} E_{{}_{\mathfrak{m}_i}}.
\]
We do not present derivation of this inequality as it is the same as derivation of (\ref{BasicInequalityForWickEpsilon->0}).
\qed

The above series of Wick decomposition Theorems \ref{WickThmForMassiveFields}-\ref{ProductTheorem} and 
lemma \ref{auxiliaryProductLemma},
gives us a class of generalized operators $\Xi'$, equal to finite sums of integral kernel operators
\[
\Xi' = \sum\limits_{l',m'} \Xi_{l',m'}(\kappa'_{l',m'}) \in \mathscr{L}\big(\mathscr{E}, \mathscr{L}((E),(E)^*) \big), 
\,\,\, \mathscr{E}_{{}_{'}} = \mathcal{S}(\mathbb{R}^{4k'})
\]
for which the product operation is well-defined and the Wick decomposition through the normal ordering is applicable to their products, 
although in general this class includes
operators in 
\[
\mathscr{L}\big(\mathscr{E}_{{}_{'}}, \mathscr{L}((E),(E)^*) \big)
\]
which do not belong to
\[
\mathscr{L}\big(\mathscr{E}_{{}_{'}}, \mathscr{L}((E),(E)) \big).
\]
Namely, we have the following
\begin{twr}
The class of generalized integral kernel operators with $\mathcal{S}(\mathbb{R}^{4k})^*$-valued kernels $\kappa_{l,m}$, which admits the 
operation of (tensor) product, includes the operators  
\begin{equation}\label{ProductClassGenOp}
t(x_1, \ldots, x_n) \, {:}W_1(x_1)W_2(x_2) \ldots W(x_k){:}, \,\,\, t \in  \mathcal{S}(\mathbb{R}^{4k})^* = \mathcal{S}(\mathbb{R}^{4})^{* \, \otimes \, k},
\end{equation}
with translationally invariant $t$, and $W_i$ being Wick products of massless or massive free fields. 
\label{ClassWithProductTheorem}
\end{twr}
\qedsymbol \,
By Theorems \ref{WickThmForMassiveFields}-\ref{ProductTheorem} and 
lemma \ref{auxiliaryProductLemma}, for operators $\Xi', \Xi''$ in this class 
there exist $\epsilon$-approximations
\begin{gather*}
\Xi'_{{}_{\epsilon}} = \sum\limits_{l',m'} \Xi_{l,m}(\kappa'_{\epsilon \, l,m}) \in \mathscr{L}\big(\mathscr{E}_{{}_{'}}, \mathscr{L}((E),(E)) \big), 
\\
\Xi''_{{}_{\epsilon}} = \sum\limits_{l'',m''} \Xi_{l,m}(\kappa''_{\epsilon \, l'',m''}) 
\in \mathscr{L}\big(\mathscr{E}_{{}_{''}}, \mathscr{L}((E),(E)) \big), 
\end{gather*}
(here with $\mathscr{E}_{{}_{'}}$ and $\mathscr{E}_{{}_{''}}$ being some Schwartz spaces of $\mathbb{C}^{d_i}$-valued functions) for which 
\[
\Xi'_{{}_{\epsilon}} \longrightarrow \Xi',
\,\,\,
\Xi''_{{}_{\epsilon}} \longrightarrow \Xi''
\]
in 
\[
\mathscr{L}\big(\mathscr{E}_{{}_{(i)}}, \mathscr{L}((E),(E)^*) \big)
\]
and moreover, the Fock decomposition is naturally applicable
to their operator product $\Xi_{{}_{\epsilon}}$
\[
\Xi_{{}_{\epsilon}} (\phi\otimes\varphi) \overset{\textrm{df}}{=} \Xi'_{{}_{\epsilon}} (\phi) \Xi''_{{}_{\epsilon}}(\varphi) 
\]
and such that the kernels
\[
\kappa'_{\epsilon, l',m'} \otimes_q \kappa''_{\epsilon \, l'',m''}
\]
of the Wick decomposition of the operator product $\Xi_{{}_{\epsilon}}$ converge to some kernels
\[
\kappa'_{l',m'} \otimes_q \kappa''_{l'',m''} \in \mathscr{L}(E^{\otimes(l'+l''+m'+m''-2q)}, \mathscr{E}_{{}_{'}}^{*}\otimes \mathscr{E}_{{}_{''}}^{*})
\]
 representing a well-defined finite sum $\Xi$ of generalized operators. 
Thus, by thm. 3.9 of \cite{obataJFA}, in this class the product operator $\Xi_{{}_{\epsilon}}$ converges
\[
\Xi_{{}_{\epsilon}} \longmapsto \Xi \in \mathscr{L}\big(\mathscr{E}_{{}_{'}}\otimes \mathscr{E}_{{}_{''}}, \mathscr{L}((E),(E)^*) \big)
\]
to an operator $\Xi$ in 
\[
\mathscr{L}\big(\mathscr{E}_{{}_{'}}\otimes \mathscr{E}_{{}_{''}}, \mathscr{L}((E),(E)^*) \big).
\]
In practice, we construct the $\epsilon$-approximation within the class indicated above, just by replacing $|\boldsymbol{\p}|$ by 
$(|\boldsymbol{\p}|^2+\epsilon^2)^{1/2}$ in the
free massless field plane wave kernels.
\qed

Definition of product, given in the above proof, 
can be generalized over a still more general finite sums of integral kernel operators
$\Xi', \Xi''$, respectively in $\mathscr{L}\big(\mathscr{E}_{{}_{(i)}}, \mathscr{L}((E),(E)^*) \big)$, 
$\mathscr{E}_{{}_{(i)}}= \mathcal{S}(\mathbb{R}^{4k_i})$,
provided only they possess the $\epsilon$-approximations 
with the kernels $\kappa'_{\epsilon, l',m'} \otimes_q \kappa''_{\epsilon \, l'',m''}$
of the Wick decompositions of their products
converging in the sense defined above. In fact the class of  generalized operators, on which product operation is well-defined, 
can still be extended over infinite sums 
of integral kernel operators of the class (\ref{ProductClassGenOp}), provided the infinite sums represent Fock expansions convergent in the sense
of \cite{obataJFA}.  In application to causal QFT, where the Lagrangian interaction density operator $\mathcal{L}(x)$
is equal to a Wick polynomial in free fields of finite degree, the class  (\ref{ProductClassGenOp}), allowing
the product operation, is sufficient.

\subsection{Axioms for $S(g)$ with Hida operators}\label{axiomsS}

Repeated application of the Wick Theorem \ref{WickThmForMassiveFields}, \ref{WickThmForMasslessFields}, \ref{WickThmForMassiveFields}
or \ref{ProductTheorem}  and Lemma \ref{auxiliaryProductLemma}, 
allow us to compute the higher order contributions $S_n(x_1, \ldots, x_n)$ to the scattering operator.
If there are massless fields in the Lagrangian of interaction $\mathcal{L}(x)$, then 
Theorem \ref{ProductTheorem} allows us to define 
\[
S_n(g^{\otimes \, n})S_k(h^{\otimes \, k}), \,\,\, g,h \in \mathcal{S}(\mathbb{R}^4) \,\, \textrm{with mutually causal supp},
\]
and the corresponding (tensor) product kernel 
\[
S_n(x_1, \ldots, x_n)S_n(y_1, \ldots, y_k)
\]
at each inductive step
of computation due to Bogoliubov-Epstein-Glaser
whenever the supports of $g,h$ are mutually causal, compare \cite{Epstein-Glaser} or Subsection \ref{MotivationForHida}. 
Indeed, we replace all the exponents of non contracted massless kernels of free fields
in the Wick decompositions of $S_n(x_1, \ldots, x_n)$ and of $S_k(y_1, \ldots, y_k)$,  with the corresponding massive exponents
with mass $\epsilon$.  After this operation we obtain
the operators for which the (tensor) product can naturally be defined by the operator product, which follows
from Theorem \ref{ProductTheorem}. Next we pass to the limit $\epsilon \rightarrow 0$. 
This limit exists, as a consequence of repeated application of the Lemma \ref{auxiliaryProductLemma}. 
The Epstein-Glaser generalized operators $D_{(n)}$ of Subsection \ref{MotivationForHida} have causal support 
with the advanced and retarded distributions 
$A_{(n)}$, $R_{(n)}$, but the product 
\[
S_n(g^{\otimes \, n})S_k(h^{\otimes \, k}), \,\,\, g,h \in \mathcal{S}(\mathbb{R}^4) \,\, \textrm{with mutually causal supp},
\]
which underlines the implementation of causality, has to be understood in the above stated limit sense,
and correspondingly the (tensor) product 
\[
S_n(x_1, \ldots, x_n)S_k(y_1, \ldots, y_k)
\]
associated to the limit definition. 
 
In computation of the second order contribution $S_2(x,y)$ it is sufficent to use Theorem \ref{WickThmForMassiveFields}
or \ref{WickThmForMasslessFields}.
In particular, using the Wick Theorem \ref{WickThmForMasslessFields} or \ref{WickThmForMassiveFields} 
for the (tensor) product (\ref{Wick(x)Wick(y)}) of generalized operators,
we can write the kernel (\ref{Wick(x)Wick(y)}) in the following normal decomposition:
\begingroup\makeatletter\def\f@size{5}\check@mathfonts
\def\maketag@@@#1{\hbox{\m@th\large\normalfont#1}}%
\begin{multline}\label{Wick(x)Wick(y)NormalKernelDecomposition}
\boldsymbol{{:}} \mathbb{A}_{{}_{1}}^{a}(x) \ldots \mathbb{A}_{{}_{N}}^{a}(x) \boldsymbol{{:}} \,\,
\boldsymbol{{:}} \mathbb{A}_{{}_{N+1}}^{b}(y) \ldots \mathbb{A}_{{}_{M}}^{b}(y)  \boldsymbol{{:}}
\\
= \,\,\,\,\,\,\,\,
\boldsymbol{{:}} \mathbb{A}_{{}_{1}}^{a}(x)  \ldots \mathbb{A}_{{}_{N}}^{a}(x) 
 \mathbb{A}_{{}_{N+1}}^{b}(y) \ldots \mathbb{A}_{{}_{M}}^{b}(y)  \boldsymbol{{:}}
\\
+
\sum_{\substack{1\leq i \leq N \\ 1 \leq j \leq M-N}}
t^{ij}(x-y) \,
\boldsymbol{{:}} \mathbb{A}_{{}_{1}}^{a}(x)  \ldots \overbrace{\mathbb{A}_{{}_{i}}^{a}(x)}^{\textrm{deleted}} \ldots  
 \ldots \overbrace{\mathbb{A}_{{}_{N+j}}^{b}(y)}^{\textrm{deleted}} \ldots  \mathbb{A}_{{}_{M}}^{b}(y)  \boldsymbol{{:}}
\\
+
\sum_{\substack{1\leq i \leq N \\ 1 \leq j \leq M-N \\ 1 \leq k \leq N \\ 1 \leq n \leq M-N}} 
t^{inkj}(x-y) \,\,\, \times
\\
\times \,\,\,
\boldsymbol{{:}} \mathbb{A}_{{}_{1}}^{a}(x)  \ldots \overbrace{\mathbb{A}_{{}_{i}}^{a}(x)}^{\textrm{deleted}} \ldots  
\overbrace{\mathbb{A}_{{}_{k}}^{a}(x)}^{\textrm{deleted}} \ldots
 \ldots \overbrace{\mathbb{A}_{{}_{N+j}}^{b}(y)}^{\textrm{deleted}} \ldots
\ldots \overbrace{\mathbb{A}_{{}_{N+n}}^{b}(y)}^{\textrm{deleted}}
\ldots  \mathbb{A}_{{}_{M}}^{b}(y)  \boldsymbol{{:}}
\\
+ \ldots + \overbrace{t^{in\ldots kj}(x-y)}^{\kappa_{0,0}(x-y)} \,\, \boldsymbol{1},
\end{multline}
\endgroup
with the well-defined and causally supported distributions
\[
t^{ij}, t^{inkj}, \ldots, t^{in\ldots kj} \in \mathscr{E}^*
\]
defining the kernels 
\[
t^{ij}(x-y), t^{inkj}(x-y), \ldots, t^{in\ldots kj}(x-y)
\]
of translationally invariant distributions in $\mathscr{E}^*\otimes \mathscr{E}^*$. Note that the distributions
\[
t^{ij}, t^{inkj}, \ldots, t^{in\ldots kj} \in \mathscr{E}^*
\]
can be easily read-off from the rigorous version of the Wick Theorem \ref{WickThmForMasslessFields} or \ref{WickThmForMassiveFields},
and are computed from the kernels $\kappa_{0,1}, \kappa_{1,0}$ defining the free fields $\mathbb{A}_{{}_{k}}$ in (\ref{Wick(x)Wick(y)})
by the repeated application of the pointed product $\dot{\otimes}$, contractions $\otimes_q$ or limit contraction $\otimes|_{{}_{q}}$,
and finally by symmetrization and antisymmetrization. 

In particular putting $M=2N$ and $\mathbb{A}_{{}_{k}} = \mathbb{A}_{{}_{N+k}}$, $k=1, \ldots, N$, with
\[
i\mathcal{L}(x) = \boldsymbol{{:}} \mathbb{A}_{{}_{1}}^{a}(x) \ldots \mathbb{A}_{{}_{N}}^{a}(x) \boldsymbol{{:}}
\] 
and 
\[
i\mathcal{L}(y) = \boldsymbol{{:}} \mathbb{A}_{{}_{N+1}}^{b}(y) \ldots \mathbb{A}_{{}_{N+N=M}}^{b}(y)  \boldsymbol{{:}}
\]
in (\ref{Wick(x)Wick(y)NormalKernelDecomposition}) with $\mathcal{L}$ equal to the Lagrange interaction density of a QFT theory, 
the formula (\ref{Wick(x)Wick(y)NormalKernelDecomposition}) gives us the normal decomposition of the operator valued distribution
\[
A'_{(2)}(x,y) = -i^2 \mathcal{L}(x) \mathcal{L}(y) = \overline{S_1}(x)S_1(y) = - S_1(x)S_1(y),
\]
used in the construction of the kernel $S_2(x,y)$ of the second order contribution to the scattering matrix, compare
\ref{MotivationForHida}. Replacing $x$ and $y$ in the formula (\ref{Wick(x)Wick(y)NormalKernelDecomposition})
computed in this special case, we obtain from (\ref{Wick(x)Wick(y)NormalKernelDecomposition}) the normal decomposition
of the operator valued distribution 
\[
R'_{(2)}(x,y) = -i^2 \mathcal{L}(y) \mathcal{L}(x) = S_1(y) \overline{S_1}(x) = - S_1(y)S_1(x),
\]
needed in the computation of the second order term $S_2(x,y)$, compare
\ref{MotivationForHida}. Thus for the distribution kernel (compare
\ref{MotivationForHida})
\[
D_{(2)} = R'_{(2)} - A'_{(2)},
\]
we again obtain the formula similar to (\ref{Wick(x)Wick(y)NormalKernelDecomposition}), in fact the formula
(\ref{Wick(x)Wick(y)NormalKernelDecomposition}) plus the same expression (\ref{Wick(x)Wick(y)NormalKernelDecomposition}) with $x$ and $y$ interchanged. 
Thus, for the retarded part $R_{(2)}$ of $D_{(2)}$ we obtain again the formula (\ref{Wick(x)Wick(y)NormalKernelDecomposition})
minus the same expression with $x$ and $y$ interchanged, and with the scalar distributions 
\[
t^{ij}, t^{inkj}, \ldots, t^{in\ldots kj} 
\]
replaced by their retarded parts, computed according to the Epstein-Glaser splitting prescription, working
for causally supported distributions. Similar expression (\ref{Wick(x)Wick(y)NormalKernelDecomposition})
minus the expression (\ref{Wick(x)Wick(y)NormalKernelDecomposition}) with interchanged $x$ and $y$
and with the scalar distributions 
\[
t^{ij}, t^{inkj}, \ldots, t^{in\ldots kj} 
\]
replaced with their advanced parts, we obtain for the advanced part $A_{(2)}$ of the distribution $D_{(2)}$.

Therefore, the second order contribution
\[
S_2(x,y) = A_{(2)}(x,y) - A'_{(2)}(x,y)
\]
or equivalently
\[
S_2(x,y) = R_{(2)}(x,y) - R'_{(2)}(x,y),
\]
again has the general form (\ref{Wick(x)Wick(y)NormalKernelDecomposition}) with the scalar distributions
\[
t^{ij}, t^{inkj}, \ldots, t^{in\ldots kj}
\]
replaced with the respective advanced and retarded parts, in any case with well-defined translationally
invariant and causally supported distributions. Let for example
\begin{equation}\label{ContributionToS2}
\textrm{ret} \, t^{ij}(x-y) \,
\boldsymbol{{:}} \mathbb{A}_{{}_{1}}^{a}(x) \ldots \overbrace{\mathbb{A}_{{}_{i}}^{a}(x)}^{\textrm{deleted}} \ldots
\ldots \overbrace{\mathbb{A}_{{}_{N+j}}^{b}(y)}^{\textrm{deleted}} \ldots \mathbb{A}_{{}_{M}}^{b}(y) \boldsymbol{{:}}
\end{equation}
be one of the finite contributions to $S_2(x,y)$.
Let
\[
\sum \limits_{\ell',m'} \Xi'_{\ell',m'}(\kappa'_{\ell',m'}(x)) =
\boldsymbol{{:}} \mathbb{A}_{{}_{1}}^{a}(x) \ldots \overbrace{\mathbb{A}_{{}_{i}}^{a}(x)}^{\textrm{deleted}} \ldots \mathbb{A}_{{}_{N}}^{a}(x) \boldsymbol{{:}}
\]
and
\[
\sum \limits_{\ell'',m''} \Xi''_{\ell'',m''}(\kappa''_{\ell'',m''}(y)) =
\boldsymbol{{:}} \mathbb{A}_{{}_{N+1}}^{b}(y) \ldots \overbrace{\mathbb{A}_{{}_{N+j}}^{b}(y)}^{\textrm{deleted}} \ldots \mathbb{A}_{{}_{2N}}^{b}(y) \boldsymbol{{:}}.
\]
Therefore the value of the contribution
to
\[
S_2(g)=S_2(1 \otimes g_2) = \int S_2(x,y) \big(1_{{}_{\mathbb{R}^4}} \otimes g_2\big) (x, y) \, \ud^4 x \, \ud^4 y
= \int S_2(x,y) g_2(y) \, \ud^4 x \, \ud^4 y,
\]
\[
g(x,y) = \big(1_{{}_{\mathbb{R}^4}} \otimes g_2\big) (x, y) = 1_{{}_{\mathbb{R}^4}}(x) g_2(y) = g_2(y), \,\, g_2 \in \mathscr{E},
\]
and coming from (\ref{ContributionToS2}) is equal to the finite sum
of integral kernel operators
\begin{multline*}
\sum\limits_{\ell,m}\Xi_{\ell,m}(\kappa_{\ell,m}(g_n))
\\
=
\int
\textrm{ret} \, t^{ij}(x-y) \,
\boldsymbol{{:}} \mathbb{A}_{{}_{1}}^{a}(x) \ldots \overbrace{\mathbb{A}_{{}_{i}}^{a}(x)}^{\textrm{deleted}} \ldots
\ldots \overbrace{\mathbb{A}_{{}_{N+j}}^{b}(y)}^{\textrm{deleted}} \ldots \mathbb{A}_{{}_{M}}^{b}(y) \boldsymbol{{:}}
\, g_2(y)
\,
\ud^4 x \, \ud^4 y
\end{multline*}
with
\[
\kappa_{\ell,m} = \big(\textrm{ret} t^{ij} \ast \kappa'_{\ell',m'}\big) \dot{\otimes} \kappa''_{\ell'',m''},
\,\,\, \ell = \ell'+\ell'', m = m' + m''.
\]

Thus, we see from Lemma \ref{S*Xi} of Subsection \ref{OperationsOnXi} that
\[
\sum\limits_{\ell,m}\Xi_{\ell,m}(\kappa_{\ell,m}(g_n)) \in  \mathscr{L}(\boldsymbol{E}), (\boldsymbol{E})^*)
\]
and 
\[
\mathscr{E} \ni g_2 \longrightarrow \sum\limits_{\ell,m}\Xi_{\ell,m}(\kappa_{\ell,m}(g_2)) \in \mathscr{L}(\boldsymbol{E}), (\boldsymbol{E})^* ),
\]
is continuous. Note also that 
\[
\kappa'_{\ell',m'}(\xi), \kappa''_{\ell'',m''}(\eta) \in \mathcal{O}_C(\mathbb{R}^4) \subset \mathcal{O}_M(\mathbb{R}^4)
\]
for 
\[
\xi \in E_{n_1} \otimes \ldots \otimes E_{n_{\ell'+m'}},
\,\,\, \textrm{respectively} \,\,\,
\eta \in E_{n'_{1}} \otimes \ldots \otimes E_{n'_{\ell''+m''}},
\]
\emph{i.e.} for elements of tensor products of the nuclear spaces $E_{n_k} \subset \mathcal{H}_{n_k} \subset E_{n_k}^{*}$ composing the single particle Gelfand triples
of the fields
\[
\mathbb{A}_{{}_{1}}^{a}(x)  \ldots \overbrace{\mathbb{A}_{{}_{i}}^{a}(x)}^{\textrm{deleted}} \ldots \mathbb{A}_{{}_{N}}^{b}(x)
\]
or, respectively, composing the Gelfand triples  $E_{n'_{k}} \subset \mathcal{H}_{n'_{k}} \subset E_{n'_{k}}^{*}$ in the single particle Hilbert spaces  
of the fields
\[
\mathbb{A}_{{}_{N+1}}^{b}(y) \ldots \overbrace{\mathbb{A}_{{}_{N+j}}^{b}(y)}^{\textrm{deleted}} \ldots  \mathbb{A}_{{}_{2N}}^{b}(y),
\]
compare Lemma \ref{kappaBarDotOtimeskappa} of Subsection \ref{OperationsOnXi}. Moreover, $\kappa'_{\ell',m'}(\xi), \kappa''_{\ell'',m''}(\eta)$ 
depend continuously, respectively, on $\xi$ and $\eta$, in the strong dual topology of $\mathscr{E}^*$ in the image. 
Therefore, the kernel $\kappa_{\ell, m}(\xi \otimes \eta)$ of the operator 
$\Xi_{\ell,m}(\kappa_{\ell, m}(\xi \otimes \eta))$, being equal
\[
\kappa_{\ell, m}(\xi \otimes \eta)(x,y) = \textrm{ret} t^{ij}(x-y) \kappa'_{\ell',m'}(\xi)(x) \kappa''_{\ell'',m''}(\eta)(y)
\]
represents a continuous map
\[
\xi \otimes \eta \longmapsto \kappa_{\ell, m}(\xi \otimes \eta) \in \mathscr{E}^{* \otimes \, 2}
\]
and thus by Thm. 3.6 and 3.9 of \cite{obataJFA} (compare also Subsection \ref{psiBerezin-Hida} for the case including Fermi fields)
\[
\sum\limits_{\ell,m}\Xi_{\ell,m}(\kappa_{\ell,m}(g_2)) \in  \mathscr{L} \big( \mathscr{E}^{\otimes \, 2}, \, 
\mathscr{L}(\boldsymbol{E}), (\boldsymbol{E})^*) \big).
\]

Presented analysis is the same for all remaining contributions to $S_2(1\otimes g_2)$
and by the same Lemma  \ref{S*Xi} of Subsection \ref{OperationsOnXi}
\[
S_2(1 \otimes g_2) = \int S_2(x,y) g_2(y) \, \ud^4 x \, \ud^4 y \in \mathscr{L}(\boldsymbol{E}), (\boldsymbol{E})^*)
\]
and 
\[
 \mathscr{E} \ni g_2 \longrightarrow S(1 \otimes g_2) \in \mathscr{L}(\boldsymbol{E}), (\boldsymbol{E})^*)
\]
is continuous,
including the case with massless fields $\mathbb{A}_{{}_{k}}$ in the interaction Lagrange density
\[
\mathcal{L}(x) = \boldsymbol{{:}} \mathbb{A}_{{}_{1}}^{a}(x) \ldots \mathbb{A}_{{}_{N}}^{a}(x) \boldsymbol{{:}},
\] 
as e.g. in QED, and moreover
\[
S_2 \in  \mathscr{L} \big( \mathscr{E}^{\otimes \, 2}, \, 
\mathscr{L}(\boldsymbol{E}), (\boldsymbol{E})^*) \big).
\]

Of course the Wick Theorem \ref{WickThmForMasslessFields} or \ref{WickThmForMassiveFields} can be applied in the computation of the higher order
terms $S_n$ of the scattering operator (compare Subsection \ref{MotivationForHida})
and now it should be clear that the repeated application of Lemmas \ref{S*Xi} and \ref{kappaBarDotOtimeskappa} of Subsection \ref{OperationsOnXi}
gives
\[
S_n(1 \otimes 1 \ldots \otimes 1 \otimes g_n) \in \mathscr{L}(\boldsymbol{E}), (\boldsymbol{E})^*), \,\,\,\, g_n \in \mathscr{E},
\]
including the case with massless factors $\mathbb{A}_{{}_{k}}$ in the interaction density $\mathcal{L}$.

Moreover,
\[
\mathscr{E} \ni g_n \longrightarrow S_n(1 \otimes 1 \ldots \otimes 1 \otimes g_n) \in \mathscr{L}(\boldsymbol{E}), (\boldsymbol{E})^*),
\]
is continuous, also in case there are massless factors $\mathbb{A}_{{}_{k}}$ in the interaction density $\mathcal{L}$, and moreover
\[
S_n \in \mathscr{L} \big(\mathscr{E}^{\otimes \, n}, \, \mathscr{L}(\boldsymbol{E}), (\boldsymbol{E})^*) \big).
\]

Therefore, we can summarize this Subsection with a strengthened form of the Corollary
of Thm. \ref{g=1InteractingFieldsQED} of Subsection \ref{OperationsOnXi}:
\begin{twr}
Let $S_n$ be the $n$-th order contribution to the scattering integral kernel operator. Then
\begin{enumerate}
\item[1)]
\[
\mathscr{E} \ni g_n \longrightarrow S_n(1 \otimes 1 \ldots \otimes 1 \otimes g_n) \in \mathscr{L}(\boldsymbol{E}), (\boldsymbol{E})^*),
\]
is continuous,
including the case with massless factors $\mathbb{A}_{{}_{k}}$ in the interaction density $\mathcal{L}$.
\item[2)]
\[
S_n \in \mathscr{L} \big(\mathscr{E}^{\otimes \, n}, \, \mathscr{L}(\boldsymbol{E}), (\boldsymbol{E})^*) \big),
\]
in the case with massless factors $\mathbb{A}_{{}_{k}}$ in the interaction density $\mathcal{L}$.
\item[3)]
\[
S_n \in \mathscr{L} \big(\mathscr{E}^{\otimes \, n}, \, \mathscr{L}(\boldsymbol{E}), (\boldsymbol{E})) \big),
\]
in the case with only massive factors $\mathbb{A}_{{}_{k}}$ in the interaction density $\mathcal{L}$.
\end{enumerate}
\label{Sin(SE)xn->((E)->(E)*)}
\end{twr}
\qedsymbol \, 3) follows from Theorem \ref{ProductTheorem}. 
\qed

Note here that Theorem \ref{Sin(SE)xn->((E)->(E)*)} gives at the same time the so-called Wick theorem for the chronological product
\[
S_n(x_1, \ldots, x_n) = i^n \, T\big(\mathcal{L}(x_1) \ldots \mathcal{L}(x_n) \big)
\]
because each $S_n$ is given here in the form of finite sum of integral kernel operators with vector valued kernels,
and thus we represent $S_n$ as finite decomposition into ''normally ordered operators''. Here the causal perturbative
construction of $S_n$ and the Wick theorem for the chronological product come together.

In the more conventional approach, which uses ''renormalization technique'', we proceed in two formal stages
both not quite clear mathematically. We apply the ''formal/symbolic Wick theorem''\footnote{Which already is not quite logical mathematically,
because the ``regularization technique'' works only for the ``product'' but not for the ``formal chronological product''.}
to the formal expression
\begin{multline*}
S_n(x_1, \ldots, x_n) = i^n \, T\big(\mathcal{L}(x_1) \ldots \mathcal{L}(x_n) \big) \\ = i^{n}
\sum\limits_{\pi} \theta(t_{\pi(1)} -t_{\pi(2)}) \theta(t_{\pi(2)} -t_{\pi(3)}) \ldots \theta(t_{\pi(n-1)} -t_{\pi(n)}) \,
\mathcal{L}(x_{\pi(1)}) \ldots \mathcal{L}(x_{\pi(n)}), \\
\,\,\,\,\,\,\,\,\,\,\,\,\,\,\,\,\,\,\,\,\,\,\,\,\,\,\,\,\,\,\,\,
x_{\pi(k)} = (t_{\pi(k)}, \boldsymbol{\x}_{\pi(k)}), \,\,\, \pi \in \textrm{Permutations of} \,\{1, \ldots, n\}.
\end{multline*}
and arrive at a symbolic expression for the ``chronological product'', and which is called ``Wick theorem for chronological product'',
with the normally ordered operators, but multiplied by expressions which are
not well-defined as distributions. Even if the products of pairing functions we repair with limits of products of regularized pairings,
these limits (well-defined distributions) are again disturbed there by multiplication by the step theta function, which is
a problematic operation as the step theta function is not a multiplier of the considered distribution spaces.
Therefore, in the conventional approach, we need to apply the ``renormalization
technique'' to the amplitudes computed formally from the formal expression for the ``chronological product'', which
is not really a mathematically logical process. We will analyze the construction of chronological product with the help
of multiplication by the step theta function in more details in the next Subsection. We explain there more precisely that indeed
the fact that Fourier transform of the step theta function is not any convolutor of the Schwartz tempered distribution space
is indeed the reason of the difficulty in the
conventional construction of the chronological product. Equivalently (Fourier exchange theorem) the difficulty comes from the fact
that the theta function is not any multiplier of the Schwartz space.

In our mathematically rigorous approach the chronological product in Theorem \ref{Sin(SE)xn->((E)->(E)*)}
is a well-defined integral kernel operator,
so that we can forget about ''renormalization technique''.
We should emphasize however that the scattering operator in Theorem \ref{Sin(SE)xn->((E)->(E)*)}
is not uniquely determined by the causal axioms (I)-(IV). The non-uniqueness comes from the non-uniqueness
of the Epstein-Glaser splitting. But this arbitrariness can be eliminated by imposing the requirement
of the existence of the adiabatic limit for higher order contributions to interacting fields, which is possible
only if we interpret the creation-annihilation operators as the Hida operators, 
compare the end of Subsection \ref{OperationsOnXi}.

In the next Subsection we eliminate the arbitrariness coming from the splitting by using a natural definition
of the chronological product, and make the computation of the scattering operator more systematical which moreover
can be subject to an algorithmic calculus.

Before passing to the next Subsection let us emphasize that 
the axioms  for the scattering operator-valued distribution
$g \rightarrow S(g)$ are the same as in the Bogliubov-Epstein-Glaser formulation,
i.e. Bogliubov's axioms (I)-(IV)  plus the inductive step from $S_1$, ..., $S_{n-1}$ to $S_{n}$,
emerging from (I)-(IV),  and are given in Subsection \ref{MotivationForHida}.

This is because the only replacement we are doing is the use  of the free fields as the  (white noise)
integral kernel generalized operators with vector valued (in the distribution space over space-time as the values) kernels.
Put otherwise: we make one refinement interpreting the creation-annihilation operators at specific momenta as the Hida operators.
We introduce no additions. These white noise free fields are also distributions  with values in $\mathscr{L}(\boldsymbol{E}), (\boldsymbol{E}))$,
but, we consider also the more general class of distributions, includng all the operators (\ref{ProductClassGenOp}),
with values in $\mathscr{L}(\boldsymbol{E}), (\boldsymbol{E})^*)$
and the respective integral kernel operators, which allow the product operation, although they in general belong to
$\mathscr{L}(\boldsymbol{E}), (\boldsymbol{E})^*)$-valued distributions but in general not to
$\mathscr{L}(\boldsymbol{E}), (\boldsymbol{E}))$-valued disributions. 
This class includes the interaction lagrangian $\mathcal{L}$ operator as well as the
the higher order contributions $S_n$ to the scattering operator for all known and realistic QFT.
Because the Wick product operation of the integral
kernel operators within this class is a well-defined integral kernel operator, again within this class,
as well as the ``product'' operation for any two integral kernel operators within this class is a well-defined integral kernel
operator within this class, or a generalized distribution with values in 
$\mathscr{L}(\boldsymbol{E}), (\boldsymbol{E}))$ or  in $\mathscr{L}(\boldsymbol{E}), (\boldsymbol{E})^*)$,
then axioms admit the said mathematical refinement, and are identical (except the mathematical refinement in the definition of the allowed class
of generalized operators) as for the Bogoliubov-Epstein-Glaser theory. As we have seen in this Subsection we
have well-defined product and Wick product operations within the generalized integral kernel operators of the class
includng all the operators (\ref{ProductClassGenOp}) --
all we need to formulate the axioms and for computations. This theory uses only these two operations
within the class including all the operators of the form (\ref{ProductClassGenOp}).

In expanded form:

\begin{center}
{\small AXIOMS FOR $S(g)$ WITH THE HIDA OPERATORS}
\end{center}

The scattering operator and its formal inverse is constructed as the series ($\otimes$ stands for the projective tensor product)
\[
\begin{split}
S(g) = \boldsymbol{1} + {\textstyle\frac{1}{1!}} S_1(g) + {\textstyle\frac{1}{2!}} S_2(g\otimes g) + {\textstyle\frac{1}{3!}} S_3(g\otimes g \otimes g) + \ldots
\\
S(g)^{-1} = \boldsymbol{1} + {\textstyle\frac{1}{1!}} \overline{S}_1(g) + {\textstyle\frac{1}{2!}} \overline{S}_2(g \otimes g)
+ {\textstyle\frac{1}{3!}} \overline{S}_3(g\otimes g \otimes g) + \ldots,
\end{split}
\]
where each $S_n(g^{\otimes \, n})$ is a finite sum of integral kernel operators with vector valued kernels
of the class includng all the operators (\ref{ProductClassGenOp}) 
(with values in the symmetric distributions in $n$ space-time variables $x_1$, ..,$x_n$ for the situation with the interaction Lagrangian
density operator equal to a Wick polynomial with each monomial  containing even number of Fermi fields) mapping continuously
\begin{multline*}
g^{\otimes \, n} \longmapsto \mathscr{L}(\boldsymbol{E}), (\boldsymbol{E})^*)
\,\,
\textrm{or}
\,\,
\mathscr{L}(\boldsymbol{E}), (\boldsymbol{E})),
\\
\,\,\,\,\,\,
\textrm{\tiny depnding if there are present or not massless fields in the int. Lagrangian}
\end{multline*}
 which is a generalized distribution
\[
S_n(x_1, \ldots, x_n)
\,\,\,\,\,\,
\textrm{\tiny symmetric if the Wick int. Lagrangian is even in Fermi fields},
\]
in $n$ space-time variables $x_1$, $\ldots$, $x_n$, which maps each test element
$g^{\otimes \, n}$ continuously into $\mathscr{L}(\boldsymbol{E}), (\boldsymbol{E})^*)$ or $\mathscr{L}(\boldsymbol{E}), (\boldsymbol{E}))$.
In particular the product operation
\[
S_{n}(x_1, \ldots, x_{n}) S_k(y_1, \ldots ,y_k)
\]
is well-defined as a (finite sum of) generalized integral kernel operators with vector-valued kernels
(with values in the distribution spaces in $n+k$ space-time variables $x_1$, $\ldots$, $x_n$, $y_1$, $\ldots$, $y_k$,
eventually symmetric
in the variables $x_1$, $\ldots$, $x_{n}$ and separately in the variables $y_1$, $\ldots$, $y_k$).
Above in this Subsection we have proved that this product operation
is well-defined within the class includng all the operators (\ref{ProductClassGenOp}), and is sufficient
if the Wick interaction Lagrangian $\mathcal{L}(x)$ is a Wick polynomial in free fields, regarded as the integral kernel operators.
In fact the class of allowed operators can still be extended onto convergent Fock decompositions into 
integral kernel operators of the said class, and accordingly the interaction Lagrangian $\mathcal{L}(x)$
can have an infinite Fock expansion form. But we do not enter this situation having in mind the practical
application to realistic interactions with  $\mathcal{L}(x)$ equal to a Wick polynomial in free fields.
Indeed, if all free fields in the Lagrangian interaction Wick polynomial $\mathcal{L}(x)$
are massive, then from Theorem \ref{ProductTheorem} it follows that 
\[
S_n(x_1, \ldots, x_n)
\]
is a finite sum of generalized integral kernel operators with the value space
\[
\mathscr{L}(\boldsymbol{E}), (\boldsymbol{E})).
\]
But also in case there are present massless field factors in $\mathcal{L}(x)$ the product
in (I) is well-defined by the limit operation. Namely, we replace all nancontracted (or non paired) massless free fields
in the Wick products in 
\[
S_{n}(x_1, \ldots, x_{n}), \,\,\, S_k(y_1, \ldots ,y_k)
\]
with the massive counterparts, and then pass to the zero mass limt, exactly as in the proof
of the Wick decomposition theorem for the Wick products of massless free fields.  

Bogoliubov's axioms (I)-(IV) (compare Subsection \ref{MotivationForHida}) 
make perfect sense after this
refinement in the definition of free fields and $S_n$ as generalized distributions within the said class 
of integral kernel operators
or as (finite sums of) integral kernel  generalized operators with vector-valued kernels including the operators
of the form (\ref{ProductClassGenOp}).
Bogoliubov's axioms (I)-(IV), formulated in terms of generalized integral kernel operators
(or generalized distributions) of the said class $S_n(x_1, ..,x_n)$ read
\[
\textrm{(I)} \,\,\,\,\,\,\,\,\,\,\,\,\,\,\,\,\,\,\,\,   
              S_n(x_1, \ldots, x_n) = S_k(x_1, \ldots, x_k)S_{n-k}(x_{k+1}, \ldots, x_n) 
\]
whenever $\{x_{k+1}, \ldots, x_n \} \preceq \{x_1, \ldots, x_k\}$. 
Here for two subsets $X,Y \subset \mathbb{R}^4$ we say, after Bogoliubiv, Epstein and Glaser,
that $Y \preceq X$ if and only if 
\[
X \cap (Y +\overline{V_{-}}) = \emptyset \,\,\, \Longleftrightarrow \,\,\,
Y \cap (X + \overline{V_{+}}) = \emptyset,
\]
or $X$ does not intersect the past causal shadow of $Y$, or equivalently,
$Y$ does not intersect the future causal shadow of $X$, or, equivalently,
$X$ and $Y$ are separated by a space-like hyperplane, and in some Lorentz frame
all points of $Y$ lie in the past of all points of $X$,
compare \cite{Bogoliubov_Shirkov}, \cite{Epstein-Glaser} or Subsection \ref{MotivationForHida}. 

\[
\textrm{(II)}  \,\,\,\,\,\,\,\,\,\,\,\,\,\,\,\,\,\,\,\,      
 U_{a,\Lambda} S_n(x_1, ..,x_n) U_{a, \Lambda}^{+} = S_n(\Lambda^{-1}x_1 - a, .., \Lambda^{-1}x_n -a)
\]
where $a, \Lambda \mapsto U_{a,\Lambda}$ is the Krein isometric or unitary representation of the  $T_4 \circledS SL(2, \mathbb{C})$ group.
Here $U_{a,\Lambda}^{+}$ is  the (Krein or Hilbert space adjoint). The left-operator $U_{a,\Lambda}$ is the equal to the representor
of the representation acting in the test Hida space or the representation acting in the strong dual $(\boldsymbol{E})^*$
to the Hida space $(\boldsymbol{E})$, wich is dual to $U_{a,\Lambda}$ acting in the Hida test space or the
representation, depending on the value space  
$\mathscr{L}(\boldsymbol{E}), (\boldsymbol{E}))$ or $\mathscr{L}(\boldsymbol{E}), (\boldsymbol{E})^*)$
of the generalized distribution $S_n$. The adjoint operation is equal to the ordinary Hermitian adjoint or the
the linear dual accompanied by the complex conjugation and the Gupta Bleuler operation $\eta$ (or its dual), depending on the value
space  $\mathscr{L}(\boldsymbol{E}), (\boldsymbol{E}))$ or $\mathscr{L}(\boldsymbol{E}), (\boldsymbol{E})^*)$
of the generalized distribution $S_n$.  If no gauge fields are present, then $\eta =1$ and the formal Krein isometricity
is replaced with the formal unitarity of $U_{a,\Lambda}$ and $S(g)$. Please, note here, that in (II) we have used 
$\Lambda = \Lambda(\alpha)$, which is a homomorphism $\alpha \rightarrow \Lambda(\alpha)$
of $SL(2,\mathbb{C})$ onto the proper orthochronous Lorentz group. In Section \ref{constr-of-VF} we have used antihomomorphism 
$\alpha \rightarrow \Lambda(\alpha)$, so that in the right-hand side of (II) we would have 
\[
S_n(\Lambda x_1 - a, \ldots, \Lambda x_n - a)
\] 
if we were using  $\Lambda = \Lambda(\alpha)$ with the antihomomorphism $\alpha \rightarrow \Lambda(\alpha)$, as in Section \ref{constr-of-VF}.

\[
\textrm{(III)}  \,\,\,\,\,\,\,\,\,\,\,\,\,\,\,\,\,\,\,\,        
\overline{S}_n(x_1, \ldots, x_n) = \eta S_n(x_1, \ldots, x_n)^{+} \eta
\]
where $S_n(x_1, \ldots, x_n)^{+}$ is the linear dual of $S_n(x_1, \ldots, x_n)$ accompanied by the complex conjugation,
and the left-$\eta$ is equal do the linear dual of the Gupta-Bleuler operator $\eta$ in case
$\mathscr{L}(\boldsymbol{E}), (\boldsymbol{E})^*)$ is the value space of $S_n$.

\[
\textrm{(IV)}  \,\,\,\,\,\,\,\,\,\,\,\,\,\,\,\,\,\,\,\,                    
S_1(x_1) = i \mathcal{L}(x_1)
\]
where $\mathcal{L}(x_1)$ is the interaction Lagrangian density operator equal to a Wick polynomial in free fields
(understood as the sums of two integral kernel
generalized operators), also understood as the finite sum of integral kernel operators
with vector-valued kernels (with the values in the distribution space in one space-time variable $x_1$).

Finally, we have the following axiom
\begin{enumerate}
\item[(V)] \,\,\,\,\,\,\,\,\,\,\,\,
The value of the retarded part of a vector valued kernel should coincide with the natural formula given by the multiplication by the step theta 
function on a space-time test function, whenever the natural formula is meaningful for this test function.
\,\,\,\,\,\,\,\,\,\,\,\,\,\,\,\,\,\,\,\,\,\,\,\,\,\,\,\,\,\,\,\,\,\,\,\,\,\,\,\,\,\,\,\,\,\,\,\,\,\,\,\,\,\,\,\,\,\,\,\,
\end{enumerate}

\vspace*{0.5cm}

It means that the singularity degree of the retarded part of a kernel
should coincide with the singularity degree of this kernel, for the kernels of the above defined class of generalized integral kernel operators.
Some authors call axiom (V) ``preservation of the Steinmann scaling degree''.
We look more carefully at axiom (V) in Subsection \ref{WickForChronological}, and look more carefully at the contruction
of $S_2$ with Hida operators from (I)-(V), in case in which there are massless fields present in the interaction Lagrangian $\mathcal{L}$
including spinor QED.

The Epstein-Glaser inductive step construction, using (I)-(V), makes perfect sense for the generalized
distributions $S_n$, understood as integral kernel operators, using only the product and Wick product operations, compare
please Subsection \ref{MotivationForHida}.

In fact the same causal pertubative construction is possible in the more general situation, and in this more
general formulation it was proposed in \cite{Bogoliubov_Shirkov}. Namely, inspired by Schwinger's theory of local
observables \cite{Schwinger}, there are introduced in \cite{Bogoliubov_Shirkov} the higher order contributions to any
interacting field, which in the absence of interaction
is given by a Wick polynomial $\mathbb{A}$ in free fields by the variational formula of the
functional $S(g,h)$ equal to the scattering operator in which
\[
g(x) \mathcal{L}(x)
\]
is replaced with
\[
g(x) \mathcal{L}(x) + h(x) \mathbb{A}(x),
\]
with the additional term $h(x) \mathbb{A}(x)$ in the interaction Lagrangian $g(x)\mathcal{L}(x)$, compare
\cite{Bogoliubov_Shirkov}, \cite{Epstein-Glaser}, or \cite{Scharf}, \cite{DKS1}. This introduces additional
component $h$ in the switching-off function $(g,h)$, which, similarly as $g$, is a scalar if the Wick product field
$\mathbb{A}(x)$ is a scalar (Wick polynomial of necessary even degree in Fermi fields). However, in the situation
when $\mathbb{A}(x)$ is of odd degree in Fermi fields,
e.g. when computing higher order contributions to an interacting Fermi free field, e.g. to the spinor field
$\boldsymbol{\psi}$ or to its Dirac conjugation $\boldsymbol{\psi}^\sharp$ in spinor QED, we consider the following additional
terms in the interaction Lagrangian (with the switching-off function)
\[
h(x)^\sharp \boldsymbol{\psi}(x)
\,\,\,
\textrm{or}
\,\,\,
\boldsymbol{\psi}^\sharp(x) h(x)
\]
with
\[
S_{1,0}(g) = \int \mathcal{L}(x) g(x) \, \ud^4x
\]
and with
\[
\begin{split}
S_{0,1}(\overline{h}) = \int h(x)^\sharp \boldsymbol{\psi}(x) \, d^4x = \int \underset{a}{\Sigma} \, \overline{h^{a}(x)} \big[\gamma^0 \boldsymbol{\psi}\big]^{a}(x)
\, d^4x
\\
\textrm{or}
\\
S_{0,1}(h) = \int \boldsymbol{\psi}^\sharp(x) h(x) \, d^4x  = \int  \underset{a}{\Sigma} \, \boldsymbol{\psi}^{\sharp \, a}(x)  h^a(x) \, d^4x
\end{split}
\]
with the four components of
\[
h(x) = \big(h^1(x), \ldots, h^4(x)\big), 
\]
compose \emph{generators of the Grassmann algebra with the inner product} in the sense of \cite{Berezin}, Chap. III.3.3, III.3.4, 
eventually multiplied pointwisely, respectively,
by the functions $\phi^1, \ldots, \phi^4 \in \mathcal{S}(\mathbb{R}^4)$, which can have compact support. 

In short, we start with a nuclear countably Hilbert space $\mathcal{E}^1$
which composes together with its strong dual $\mathcal{E}^{1 \,*}$ a Gelfand triple $\mathcal{E}^{1} \subset \mathcal{H}^{1} \subset \mathcal{E}^{1 \, *}$.
For example, we can use the abstract realization 
of the Gelfand triple 
\begin{equation}\label{GelfandTripleGAIP}
\mathcal{E}^1 \subset \mathcal{H}^{1} \subset \mathcal{E}^{1 \,*}, 
\end{equation}
associated 
with a positive selfadjoint operator $A$ in a separable (in our work we are using only separable Hilbert spaces) infinite dimensional complex Hilbert space 
$\mathcal{H}^1$ with $\textrm{Inf Spec} \, A >1$, whose some negative power $A^{-r}$, $r>0$, is of Hilbert-Schmidt class,
\cite{GelfandIV}, \cite{HKPS}, \cite{obata-book}, \cite{Hida}, Subsection \ref{white-setup}.
In this abstract realization the spaces $\mathcal{E}^1$ and $\mathcal{E}^{1 \, *}$ are constructed as the projective and inductive limits
\[
\mathscr{E}^1 = \underset{n\in \mathbb{Z}}{\bigcap} \overline{\textrm{Dom} \, A^n}, \,\,\,\,\,
\mathscr{E}^{1 \, *} = \underset{n\in \mathbb{Z}}{\bigcup} \overline{\textrm{Dom} \, A^n}
\]
of the Hilbert spaces $\overline{\textrm{Dom} \, A^n}$ with inner products
\[
\big(\cdot , \cdot \big)_{{}_{n}} = \big(A^n\cdot , A^n \cdot \big)_{{}_{\mathcal{H}^1}} \,\,\,
\textrm{on}
\,\,\,
\overline{\textrm{Dom} \, A^n} = \mathcal{E}^{1}_{n},
\,\,\,
n \in \mathbb{Z}.
\]
Then the abstract \emph{Grassmann algebra with the inner product},
can be realized through the external algebra 
\[
{\bigwedge}\mathcal{E}^{1 \, *} = \bigoplus_{n=1}^{\infty} \mathcal{E}^{1 \, * \widehat{\otimes} \, n} = \bigoplus_{n=1}^{\infty} \mathcal{E}^{n \, *} 
\]
over $\mathcal{E}^{1 \, *}$ with the Grassmann product given by the wedge product, and regarded as the direct sum of topological vector spaces, 
equal to the atisymmetrized $n$-fold tensor products $\mathcal{E}^{n \, *} = \mathcal{E}^{1 \, *} \widehat{\otimes} \ldots \widehat{\otimes} \mathcal{E}^{1 \, *}$ 
 of $\mathcal{E}^{1 \, *}$ (unique, as for nuclear spaces
the equicontinuous and projective tensor products coincide). Recall that the construction of
\[
(\mathcal{E}^{1}) =\bigwedge \mathcal{E}^{1 \, } \,\,\, \subset  \,\,\, \Gamma(\mathcal{H}^1) \,\,\, \subset \,\,\, (\mathcal{E}^{1})^* =\bigwedge \mathcal{E}^{1 \, *}
\]
can also be understood as application of the Fermi second quantization functor $\Gamma$ applied to
the Gelfand triple $\mathcal{E}^{1} \subset \mathcal{H}^{1} \subset \mathcal{E}^{1 \, *}$, with 
$\bigwedge\mathcal{E}^{1 \, }$ equal to the Hida test space in $\Gamma(\mathcal{H}^1)$ and with
$\bigwedge \mathcal{E}^{1 \, *}$ equal to the space of Hida generalized functionals,
compare Subsection \ref{psiBerezin-Hida}. 
We then consider the disjoint sum
\[
M = \mathbb{R}^4 \sqcup \mathbb{R}^4 \sqcup \mathbb{R}^4 \sqcup \mathbb{R}^4 
\]  
of four copies of $\mathbb{R}^4$ with the measure $dx$ on $M$ coinciding with the ordinary invariant measure $\ud^4x$ on each
copy $\mathbb{R}^4$ and a unitary isomorphism $\iota$ of  $L^2(M, dx)$ onto $\mathcal{H}^1$, which can be written in the form
\begin{gather*}
L^2(M, dx) \ni \varphi \longmapsto  \iota(\varphi) = \sum\limits_{a=1}^{4} \int \iota^a(x) \varphi^a(x) \, \ud^4x = 
 \int \iota(x) \varphi(x) \,  dx \in \mathcal{H}^1,
\\
\big(\iota(\varphi),\iota(\varphi)\big)_{{}_{\mathcal{H}^1}} = \sum\limits_{a=1}^{4} \int |\varphi^a(x)|^2 \, \ud^4x = \int |\varphi(x)|^2 \, dx, 
\end{gather*}
where, for each $(a,x) \in M$, $h^a(x)$ (with the index $a$ numbering the copies of $\mathbb{R}^4$ and here with $x \in \mathbb{R}^4$)
is a generalized state in $\mathcal{H}^{1}$ in the sense of \cite{GelfandIV}, not belonging to  $\mathcal{H}^{1}$ but to $\mathcal{E}^{1 \, *}$.
This unitary isomorphism $\iota$ plays the same role as the unitary isomorphism $U$, eq. (\ref{isomorphismU}), of Subsection \ref{psiBerezin-Hida} and converts
the abstract Gelfand triple $\mathcal{E}^{1} \subset \mathcal{H}^{1} \subset \mathcal{E}^{1 \, *}$ associated to the operator
$A$ on $\mathcal{H}^1$ into the standard Gelfand triple on the measure space $(M, dx)$ assosiated to the operator
$\iota^{-1}A\iota$ on $L^2(M)$. 
Let $\mathcal{E}^{1}(M) \subset L^2(M)$ be the set of all functions corresponding to the elements of $\mathcal{E}^1$ under the mapping $h$. 
It follows that each element $\varphi_0 \in \mathcal{E}^{1}(M)$ defines on $\mathcal{E}^{1}(M)$ a continuous functional $\Phi_0$  by virtue of the formula
\[
\Phi_0(\varphi) = \big(\varphi_0,\varphi\big) = \int \overline{\varphi_0(x)} \varphi(x) \, dx.
\]
This follows by putting $\varphi_0 = \iota(f_0)$, $\varphi = \iota(f)$, $f_0,f \in \mathcal{E}^1$, 
because $\mathcal{E}^{1} \subset \mathcal{H}^{1} \subset \mathcal{E}^{1 \, *}$ compose a Gelfand triple, and 
\[
F_0(f) = \big(f_0, f\big)_{{}_{\mathcal{H}^1}}  = \big(f_0, f\big)_{{}_{0}} 
\]
defines a continuous functional on $\mathcal{E}^1$ for each fixed $f_0 \in \mathcal{H}^1 = \mathcal{E}^{1}_{0}$.  
This allows us to identify each element of  $\mathcal{E}^{1}(M) = \iota^{-1}\mathcal{E}^{1}$ with a continuous functional 
on $\mathcal{E}^{1}(M) = \iota^{-1}\mathcal{E}^{1}$, and thus allows us to
introduce the strong dual topology of $\mathcal{E}^{1 \, *}$ on $\mathcal{E}^{1}(M)$. 
The completion $\mathcal{E}^{1 \, *}(M)$ 
of $\mathcal{E}^{1}(M)$ with respect to this topology coincides with $\iota^{-1}\mathcal{E}^{1 \, *}$.
The isomorphism $\iota$ can be extended by continuity
(and is defined by duality through $\iota$ mapping $\mathcal{E}^1(M)$ onto $\mathcal{E}^1$) to the isomorphic 
mapping $\iota: \mathcal{E}^{1 \, *}(M) \rightarrow \mathcal{E}^{1 \, *}$ (compare Subsection \ref{psiBerezin-Hida}).  
By construction each element $\varphi_0$ of $ \mathcal{E}^{1 \, *}(M)$ belongs to $h^{-1}\mathcal{E}^{1 \, *}$. Also,
by construction
$\mathcal{E}^1(M)$ and $\mathcal{E}^{1 \, *}(M)$ is equal, respectively, to the projective and inductive limit
\[
\mathscr{E}^1(M) = \underset{n\in \mathbb{Z}}{\bigcap} \overline{\textrm{Dom} \, \big[\iota^{-1}A\iota\big]^n}, \,\,\,\,\,
\mathscr{E}^{1 \, *}(M) = \underset{n\in \mathbb{Z}}{\bigcup} \overline{\textrm{Dom} \, \big[\iota^{-1}A\iota\big]^n}
\]
of the Hilbert spaces $\overline{\textrm{Dom} \, \big[\iota^{-1}A\iota\big]^n}$ with inner products
\[
\big(\cdot , \cdot \big)_{{}_{n}} = \Big( \big[\iota^{-1}A\iota\big]^n \,\,\,\, \cdot \,\, , \,\, \big[\iota^{-1}A\iota\big]^n \,\,\, \cdot \, \Big)_{{}_{L^2}} \,\,\,
\textrm{on}
\,\,\,
\overline{\textrm{Dom} \, \big[\iota^{-1}A\iota\big]^n} = \mathcal{E}^{1}_{n}(M),
\,\,\,
n \in \mathbb{Z}.
\]
We moreover assume that the image $\iota^{-1}(\mathcal{E}^1) = \mathcal{E}^1(M)$ of $\mathcal{E}^1$ respects
the Kubo-Takenaka conditions (H1)-(H3) of Subsection \ref{white-setup} and that for each Poincar\'e
transformation $(b,\Lambda)$ the ordinary unitary transformation
$\mathcal{U}_{{}_{b,\Lambda}}$ on $L^2(M) = L^2(\mathbb{R}^4;\mathbb{C}^4)$: 
\begin{equation}\label{mathcalU}
\big(\mathcal{U}_{{}_{b,\Lambda}}\varphi\big)^a(x) = \varphi^a(\Lambda(x +b)) , \,\,\,x \in\mathbb{R}^4, \,\,\,a \in \{1,\ldots,4\},
\end{equation}
is a continuous map $\mathcal{E}^1(M) \rightarrow \mathcal{E}^1(M)$. Let us consider in $L^2(M)$ the
commutative ring of multiplications by smooth bounded real functions. To this ring there corresponds, through $\iota$,
the commutative ring $\mathfrak{A}$  of self adjoint opertors in $\mathcal{H}^1$. Then $\iota(x) \in \mathcal{E}^{1*}$, 
$x\in M$, are the generalized  eigenfunctions common to all operators of the ring $\mathfrak{A}$, all lying in 
$\mathcal{E}^{1*}$ which composes with $\mathcal{H}^1$ the rigged Hilbert space (or Gelfand triple) (\ref{GelfandTripleGAIP})
in the sense explained in \cite{GelfandIV}.  

Under this circumstances, each element
$f$ of the space $\mathcal{E}^{1 \, *}$ is represented in the form
\begin{equation}\label{h-isomorphism}
f= \iota(\varphi) = \int \iota(x) \varphi(x) \,  dx, \,\,\, \varphi \in \mathcal{E}^{1 \, *}(M).
\end{equation}
The condition that $\iota$ is isometric can be expressed in terms of $\iota(x)$, $x \in M$, in the following form
\[
\big(\iota(x), \iota(y)\big)_{{}_{\mathcal{H}^1}} = \delta(x,y).
\]
The set of the generalized elemets $\iota(x)$, $x\in M$, is called \emph{generalized orthonormal basis in} $\mathcal{H}^1$. 
In the exceptional case in which the measure space $(M,dx)$ is purely discrete (so that (H1)-(H3) are fulfilled automatically, condition (\ref{mathcalU})
is not applicable in this case) and consists of the countable set of points $x_n$, $n \in \mathbb{N}$,
each having measure $1$, the values $\iota(x_n) = \iota_n$ belong to $\mathcal{H}^1$ and compose an orthonormal complete system in $\mathcal{H}^1$
and the integral (\ref{h-isomorphism}) is converted into discrete sum
\[
f= \iota(\varphi) = \sum\limits_{n} \iota_n \varphi_n, \,\,\,\,\,\,\,\, \varphi_n = \varphi(x_n).
\]
Then using the isomorphism (\ref{h-isomorphism}) we consider two elements
\[
f_1 = \int \iota(x) \varphi_1(x) \,  dx \in \mathcal{E}^{1 \, *}
\,\,\,\,
\textrm{and}
\,\,\,\,
f_2 = \int \iota(y) \varphi_2(y) \,  dy \in \mathcal{E}^{1 \, *},
\]
and form their Grassmann product
\[
f_1f_2 = \int \iota(x) \varphi_1(x) \, \int \iota(y) \varphi_2(y) \,  dy \in \mathcal{E}^{2 \, *}.
\]
We define the Grassmann product $\iota(x)\iota(y)$ of the generalized elements $\iota(x)$ and $\iota(y)$ putting
\[
f_1f_2  = \int \iota(x) \varphi_1(x) dx \, \int \iota(y) \varphi_2(y) \,  dy
\\
= \int \iota(x)\iota(y) \varphi_1(x) \varphi_2(y) \,  dxdy,
\]
\emph{i.e.} by the lifting $\Gamma(\iota)$ (application of the Fermi second quantization functor $\Gamma$) 
of the isomorphism $\iota$ from $\mathcal{E}^{1 \, *}(M) \rightarrow \mathcal{E}^{1 \, *}$ up to
the external algebra $\bigwedge \mathcal{E}^{1 \, *}(M) \rightarrow \bigwedge \mathcal{E}^{1 \, *}$, similarly 
as we did with the isomorphism $U$ and its lifting $\Gamma(U)$ in Subsection \ref{psiBerezin-Hida}.  
Because for the Grassmann algebra we have
\[
f_1f_2 = -f_2f_1, \,\,\,\, f_1(f_2f_3) = (f_1f_2)f_3,
\]
then
\[
\iota(x)\iota(y) = - \iota(y)\iota(x), \,\,\,\, \iota(x)(\iota(y)\iota(z)) = (\iota(x)\iota(y))\iota(z).
\]
More generally, let 
\[
\iota^n(\varphi) = \int \iota^n(x_1, \ldots, x_n) \, \varphi(x_1, \ldots, x_n) \, dx_1 \ldots dx_n, \,\,\, \varphi \in \mathcal{E}^{n \, *}(M),
\]
be the restriction of the $n$-fold tensor product $\iota^{\otimes \, n}$ of the isomorphism $\iota$, eq. (\ref{h-isomorphism}),
to the subspace $\mathcal{E}^{n \, *}(M)$ of skew tensors, equal to the restriction of $\Gamma(\iota)$ to $\mathcal{E}^{n \, *}(M)$.
Here $\iota^n(x_1, \ldots, x_n)$ is antisymmetric in $x_1, \ldots, x_n$. 
Then we put for the Grassmann product
\[
\iota(x_1) \ldots \iota(x_n) \overset{\textrm{df}}{=} \iota^n(x_1, \ldots, x_n)
\] 
of the generalized states $\iota(x_1), \ldots, \iota(x_n)$. Therefore, each element
\[
f = f^1 + f^2 + \ldots \in \underset{n}{\oplus} \, \mathcal{E}^{n \, * }
\]
of the abstract Grassmann algebra with inner product can be represented by
\[
f = \int \iota(x) \varphi^1(x) \, dx + \int \iota(x_1)\iota(x_2) \varphi^2(x_1,x_2) \, dx_1 dx_2 + \ldots, \,\,\,\, \varphi^n \in \mathcal{E}^{n \, *}(M).
\]
In particular 
\begin{gather}
\mathcal{E}^{n \, *}(M) \ni \varphi^n \longmapsto \int \iota(x_1)\ldots \iota(x_n) \varphi^n(x_1, \ldots, x_n) \, dx_1 \ldots dx_n \in \mathcal{E}^{n \, *},
\label{iota-n-isomorphism}
\\
x_i \in M,
\nonumber
\end{gather}
represents a continuous isomorphism, the restriction of $\iota$ to $\mathcal{E}^{n \, *}(M)$.

We consider the space $\mathscr{E}^1$ of Grassmann-valued functions 
\begin{gather*}
h(x) = \phi(x)\iota(x)
= \phi \cdot \iota(x), 
\\
\phi \in  \mathcal{S}(M; \mathbb{C}) = \mathcal{S}(\mathbb{R}^4; \mathbb{C}^4), x\in M,
\end{gather*}
as the \emph{Grassmann-valued test functions}, \emph{i.e.}
we consider the space $\mathscr{E}^1$ of Grassmann-valued functions $h$ on $M$:
\begin{gather*}
\big(h(1,x), \ldots, h(4,x) \big)
\\
= \big(h^1(x), \ldots, h^4(x)\big) =  \big(\phi^1(x)\iota^1(x), \ldots, \phi^4(x)\iota^4(x)\big)
= \phi \cdot \iota(x), 
\\
\phi \in  \mathcal{S}(\mathbb{R}^4; \mathbb{C}^4), x\in \mathbb{R}^4, a \in \{1, \ldots, 4\},
\end{gather*}
as the Grassmann-valued test functions. The linear space $\mathscr{E}^1$ of the test Grassmann-valued functions we can
endow with the topology of the Schwartz test space $\mathcal{S}(\mathbb{R}^4; \mathbb{C}^4)$, identifying each element 
$h = \phi \cdot \iota$ with the corresponding Schwartz function $\phi \in \mathcal{S}(\mathbb{R}^4; \mathbb{C}^4)$.

More generally we consider the space $\mathscr{E}^n$ of Grassmann-valued functions 
\begin{gather}
h^n(a_1,x_1, \ldots, a_n,x_n)=  \iota^{a_1}(x_1) \ldots \iota^{a_n}(x_n)
\phi(a_1, x_1, \ldots, a_n, x_n)
\nonumber
\\ = \iota^n \cdot \phi(a_1,x_1, \ldots, a_n,x_n),
\label{h^n}
\\
\phi \in \mathcal{S}(\mathbb{R}^4; \mathbb{C}^4)^{\otimes \, n}_{{}_{\textrm{sym}}}, x_i \in \mathbb{R}^4, a_i \in \{1, \ldots, 4\},
\nonumber
\end{gather}
and endow it with the topology of the symmetric tensor product space
$\mathcal{S}(\mathbb{R}^4, \mathbb{C}^4)^{\otimes \, n}_{{}_{\textrm{sym}}}$ using the identification
of $h^n = \iota^n \cdot \phi$ with $\phi \in \mathcal{S}(\mathbb{R}^4, \mathbb{C}^4)^{\otimes \, n}_{{}_{\textrm{sym}}}$. 

We define the pairing of a distribution $\kappa \in \mathcal{E}^{n \, *}(M)$ with $h^n$, equal
\begin{gather}
h^n(a_1,x_1, \ldots, a_n,x_n)= h^{a_1}(x_1) \ldots  h^{a_n}(x_n) =  \iota^{a_n}(x_n) \ldots \iota^{a_n}(x_n)
\phi^{a_1}(x_1) \ldots \phi^{a_n}(x_n), 
\label{h^nPower}
\\
\phi \in  \mathcal{S}(\mathbb{R}^4; \mathbb{C}^4),
\nonumber
\end{gather}
by putting
\begin{multline*}
\big\langle\kappa, h^n \big\rangle
\overset{\textrm{df}}{=} 
\sum\limits_{a_1, \ldots, a_n}\int h^{a_1}(x_1)\ldots h^{a_n}(x_n) \kappa(a_1,x_1, \ldots, a_n,x_n) \, \ud^4x_1 \ldots \ud^4dx_n =
\\
\sum\limits_{a_1, \ldots, a_n}\int \iota^{a_1}(x_1)\ldots \iota^{a_n}(x_n) 
\phi^{a_1}(x_1) \ldots \phi^{a_n}(x_n) \kappa(a_1,x_1, \ldots, a_n,x_n) \, \ud^4x_1 \ldots \ud^4x_n
\in \mathcal{E}^{n \, *},
\\
x_i \in \mathbb{R}^4,
\end{multline*}
for $h^n$ of the form (\ref{h^nPower}) or
\begin{multline*}
\big\langle\kappa, h^n \big\rangle
\overset{\textrm{df}}{=} 
\sum\limits_{a_1, \ldots, a_n}\int h^n(a_1, x_1, \ldots, a_n, x_n) \kappa(a_1,x_1, \ldots, a_n,x_n) \, \ud^4x_1 \ldots \ud^4dx_n =
\\
\sum\limits_{a_1, \ldots, a_n}\int \iota^{a_1}(x_1)\ldots \iota^{a_n}(x_n) 
\phi(a_1, x_1 \ldots, a_n,x_n) \kappa(a_1,x_1, \ldots, a_n,x_n) \, \ud^4x_1 \ldots \ud^4x_n
\in \mathcal{E}^{n \, *},
\\
x_i \in \mathbb{R}^4,
\end{multline*}
for $h^n$ of the more general form (\ref{h^n}).
In slightly shorter notation: 
\begin{gather*}
\big\langle\kappa, h^n \big\rangle
\overset{\textrm{df}}{=} 
\int h(x_1)\ldots h(x_n) \kappa(x_1, \ldots, x_n) \, dx_1 \ldots dx_n 
\\
=
\int \iota(x_1)\ldots \iota(x_n) \phi(x_1)\ldots \phi(x_n) \kappa(x_1, \ldots, x_n) \, dx_1 \ldots dx_n \in \mathcal{E}^{n \, *},
\\
x_i \in M,
\end{gather*}
for $h^n$ of the form (\ref{h^nPower}) or
\begin{gather*}
\big\langle\kappa, h^n \big\rangle
\overset{\textrm{df}}{=} 
\int h^n(x_1, \ldots, x_n) \kappa(x_1, \ldots, x_n) \, dx_1 \ldots dx_n 
\\
=
\int \iota(x_1)\ldots \iota(x_n) \phi(x_1, \ldots, x_n) \kappa(x_1, \ldots, x_n) \, dx_1 \ldots dx_n \in \mathcal{E}^{n \, *},
\\
x_i \in M,
\end{gather*}
for $h^n$ of the more general form (\ref{h^n}). Therefore the value $\kappa(h^n) = \big\langle\kappa, h^n \big\rangle$ of a distribution 
$\kappa \in \mathcal{E}^{n \, *}(M)$ on the Grassmann-valued test function $h^n \in \mathscr{E}^n$ is not a complex number
but an element of the abstract Grassmann algebra $\mathcal{E}^{n \, *} \subset \underset{n} \oplus \, \mathcal{E}^{n \, *}$, and, by 
continuity of the map (\ref{iota-n-isomorphism}), gives a continuous map
\[
\mathscr{E}^n \ni h^n \longmapsto \kappa(h^n) = \big\langle\kappa, h^n \big\rangle \in \mathcal{E}^{n \, *}.
\]
Therefore, each scalar antisymmetric distribution $\kappa \in \mathcal{E}^{n \, *}(M)$ can also be regarded as a Grassmann-valued 
distribution, when it is regaded as a functional of the Grassmann-valued test functions $h^n \in \mathscr{E}^n$,
\emph{i.e.} $ \kappa \in \mathscr{L}(\mathscr{E}^n, \mathcal{E}^{n \, *})$.

More generally, each distribution $\kappa$ with the kernel
\[
\kappa(x_1, \ldots, x_n, y_1, \ldots, y_p), \,\,\, x_1 \in \mathbb{R}^4, y_j \in M
\]
antisymmetric in $y_j$, belonging to $\mathscr{E}^{*\otimes \, n} \otimes \mathcal{E}^{p \, *}(M)$, $\mathscr{E} = \mathcal{S}(\mathbb{R}^4;\mathbb{C})$,
defines also a continuous map
\[
\mathscr{E}^{\otimes \, n} \otimes \mathscr{E}^p \ni \phi \otimes h^p \longmapsto \kappa(\phi \otimes h^p) = 
\big\langle\kappa, \phi \otimes h^p \big\rangle \in \mathcal{E}^{p \, *},
\]
\emph{i.e.} $ \kappa \in \mathscr{L}(\mathscr{E}^{\otimes \, n} \otimes \mathscr{E}^p, \mathcal{E}^{p \, *})$. Therefore,
$\kappa$, when evaluated at $\phi  \otimes h^p$, where $h^p$ is a Grassmann-valued test function, is not scalar valued but
has the values in $\mathcal{E}^{p \, *}$ -- the subspace $\mathcal{E}^{p \, *}$ of $p$-degree of the abstract Grassmann algebra  
$\underset{n} \oplus \, \mathcal{E}^{n \, *}$ with inner product. 

In particular each $\mathscr{E}^{*\otimes \, n} \otimes \mathcal{E}^{p \, *}(M)$-valued, or equivalently,
each $\mathscr{L}(\mathscr{E}^{\otimes \, n} \otimes \mathcal{E}^{p}(M), \mathbb{C})$-valued distribution
\begin{multline}\label{kappa-C-distribution-valued}
\kappa_{\mathpzc{l}, \mathpzc{m}} \in 
\mathscr{L} \big(E_1 \otimes \ldots \otimes E_\mathpzc{l} \otimes E_{\mathpzc{l}+1} \ldots \otimes E_{\mathpzc{l}+\mathpzc{m}} , \,  
\mathscr{L}(\mathscr{E}^{\otimes \, n} \otimes \mathcal{E}^{p}(M), \mathbb{C})\big)
\\
\cong
\mathscr{L} \big(E_1 \otimes \ldots \otimes E_\mathpzc{l} \otimes E_{\mathpzc{l}+1} \ldots \otimes E_{\mathpzc{l}+\mathpzc{m}} , \,  
\mathscr{E}^{*\otimes \, n} \otimes \mathcal{E}^{p \, *}(M)\big)
\end{multline}
defines also a vector valued kernel 
\begin{multline}\label{kappa-Grassmann-distribution-valued}
\kappa_{\mathpzc{l}, \mathpzc{m}} \in 
\mathscr{L} \big(E_1 \otimes \ldots \otimes E_\mathpzc{l} \otimes E_{\mathpzc{l}+1} \ldots \otimes E_{\mathpzc{l}+\mathpzc{m}} , \, 
\mathscr{L}(\mathscr{E}^{\otimes \, n} \otimes \mathscr{E}^p, \mathcal{E}^{p \, *}) \big)
\\
\cong 
E_{1}^{*} \otimes \ldots \otimes E_{\mathpzc{l}}^{*} \otimes E_{\mathpzc{l}+1}^{*} \ldots \otimes E_{\mathpzc{l}+\mathpzc{m}}^{*} 
\otimes \mathscr{E}^{* \otimes \, n} \otimes \big[\mathscr{E}^p \big]^* \otimes  \mathcal{E}^{p \, *}
\\
\cong
\mathscr{L} \big(\mathscr{E}^{\otimes \, n} \otimes \mathscr{E}^p, 
\, (E_1 \otimes \ldots \otimes E_\mathpzc{l} \otimes E_{\mathpzc{l}+1} \ldots \otimes E_{\mathpzc{l}+\mathpzc{m}})^* \otimes \mathcal{E}^{p \, *} \big)
\end{multline}
of an integral kernel operator 
\begin{multline}\label{classOfXi}
\Xi(\kappa_{\mathpzc{l}, \mathpzc{m}}) \in \mathscr{L} \big((\boldsymbol{E})\otimes \mathscr{E}^{\otimes \, n} \otimes \mathscr{E}^p, \,  (\boldsymbol{E})^* \otimes \mathcal{E}^{p \, *}\big)
\\
\cong 
(\boldsymbol{E})^* \otimes \mathscr{E}^{* \otimes \, n} \otimes  [\mathscr{E}^p]^* \otimes  (\boldsymbol{E})^* \otimes \mathcal{E}^{p \, *}
\cong 
\mathscr{L}\big((\boldsymbol{E}), (\boldsymbol{E})^*\big) \otimes \mathscr{L}(\mathscr{E}^{\otimes \, n} \otimes \mathscr{E}^p, \mathcal{E}^{p \, *})
\end{multline}
for which we have the following $\mathcal{E}^{p \, *}$-valued pairing 
\begin{multline*}
\big\langle\kappa_{\mathpzc{l}, \mathpzc{m}}(\eta_{{}_{\Phi,\Psi}}), g^{\otimes \, n} \otimes h^p \big\rangle
\overset{\textrm{df}}{=} 
\sum\limits_{a_1, \ldots, a_p} \int  \Big\{
\\
g(x_1)\ldots g(x_n) h(a_1,y_1)\ldots h(a_p,y_p) \kappa_{\mathpzc{l}, \mathpzc{m}}(\eta_{{}_{\Phi,\Psi}})(x_1, \ldots, x_n, a_1,y_1, \ldots, a_p, y_p)
\\ 
\Big\}
\ud^4x_1 \ldots \ud^4x_n\ud^4y_1 \ldots \ud^4y_n 
\end{multline*}
\begin{multline*}
=
\sum\limits_{a_1, \ldots, a_p} \int  \Big\{ g(x_1)\ldots g(x_n) h(a_1,y_1)\ldots h(a_p,y_p) \, \times
\\
\times \,
 \kappa_{\mathpzc{l}, \mathpzc{m}}(p_1 \ldots, p_\mathpzc{l}, q_1, \ldots, q_\mathpzc{m}; x_1, \ldots, x_n, a_1,y_1, \ldots, a_p, y_p) \eta_{{}_{\Phi,\Psi}}(p_1 \ldots, p_\mathpzc{l}, q_1, \ldots, q_\mathpzc{m})
\\ 
\Big\}
\ud^4x_1 \ldots \ud^4x_n\ud^4y_1 \ldots \ud^4y_n dp_1 \ldots dp_\mathpzc{l}dq_1 \ldots dq_\mathpzc{m} 
\end{multline*}
\begin{multline}\label{pairingOfkappa}
=
\sum\limits_{a_1, \ldots, a_p} \int  \Big\{ 
\\
 \kappa_{\mathpzc{l}, \mathpzc{m}}(g^{\otimes \, n} \otimes h^p)(p_1 \ldots, p_\mathpzc{l}, q_1, \ldots, q_\mathpzc{m}) 
\eta_{{}_{\Phi,\Psi}}(p_1 \ldots, p_\mathpzc{l}, q_1, \ldots, q_\mathpzc{m})
\\ 
\Big\}
dp_1 \ldots dp_\mathpzc{l}dq_1 \ldots dq_\mathpzc{m} 
\\
=
\big\langle\kappa_{\mathpzc{l}, \mathpzc{m}}(g^{\otimes \, n} \otimes h^p ), \eta_{{}_{\Phi,\Psi}} \big\rangle
\in \mathcal{E}^{p \, *}
\end{multline}
for $\kappa_{\mathpzc{l}, \mathpzc{m}}$ understood, respectively, as an element of
\[
\mathscr{L} \big(E_1 \otimes \ldots \otimes E_\mathpzc{l} \otimes E_{\mathpzc{l}+1} \ldots \otimes E_{\mathpzc{l}+\mathpzc{m}} , \, 
\mathscr{L}(\mathscr{E}^{\otimes \, n} \otimes \mathscr{E}^p, \mathcal{E}^{p \, *}) \big)
\]
or
\[
\mathscr{L} \big(\mathscr{E}^{\otimes \, n} \otimes \mathscr{E}^p, 
\, (E_1 \otimes \ldots \otimes E_\mathpzc{l} \otimes E_{\mathpzc{l}+1} \ldots \otimes E_{\mathpzc{l}+\mathpzc{m}})^* \otimes \mathcal{E}^{p \, *} \big),
\]
and for $p_i = (s_i, \boldsymbol{p}_i)$ and $q_j = (\nu_j, \boldsymbol{q}_j)$ denoting the spin-momentum variables, and
with $\int \ldots dp_i \ldots dq_j \ldots$ denoting integrations and summations 
$\Sigma_{s_i\ldots \nu_j}\int \ldots \ud^3\boldsymbol{p}_i \ldots \ud^3\boldsymbol{q}_j \ldots$
with respect to spin-momentum variables.
Here
\[
\eta_{{}_{\Phi,\Psi}} \in E_1 \otimes \ldots \otimes E_\mathpzc{l} \otimes E_{\mathpzc{l}+1} \ldots \otimes E_{\mathpzc{l}+\mathpzc{m}},
\]
for ech pair $\Phi, \Psi$  of elements of the Hida test space $(\boldsymbol{E})$ and is defined by
\[
\eta_{{}_{\Phi,\Psi}}(p_1 \ldots, p_\mathpzc{l}, q_1, \ldots, q_\mathpzc{m}) 
= \big\langle\big\langle a(p_1)^+ \ldots a(p_\mathpzc{l})^+ a(q_1) \ldots  a(p_\mathpzc{m}) \Phi, \Psi \big\rangle\big\rangle,
\]
$E_\mathpzc{i} \subset \mathcal{H}_\mathpzc{i} \subset E_{\mathpzc{i}}^{*}$ are the single particle Gelfand triples of the free
fields, and $(\boldsymbol{E})$ is the test Hida space in the total Fock space of the free fields underlying the actual QFT, equa
to the tensor product of all free fields underlying the theory (tensor product of the Dirac spinor field and the e.m potential field
in case of spinor QED), compare Subsection \ref{psiBerezin-Hida}.

Recall, that the integral kernel operator
\begin{multline}\label{Xi(kappa_l,m)}
\Xi(\kappa_{\mathpzc{l}, \mathpzc{m}}) =
\int  \kappa_{\mathpzc{l}, \mathpzc{m}}(p_1 \ldots, p_\mathpzc{l}, q_1, \ldots, q_\mathpzc{m})  \,\, \times
\\
\times \,\, 
a(p_1)^+ \ldots a(p_\mathpzc{l})^+ a(q_1) \ldots  a(p_\mathpzc{m}) \,\,\,\,
dp_1 \ldots dp_\mathpzc{l}dq_1 \ldots dq_\mathpzc{m} 
\end{multline}
corresponding to the above given vector-valued kernel $\kappa_{\mathpzc{l}, \mathpzc{m}}$ given by (\ref{kappa-C-distribution-valued}),
regarded as an element of (\ref{kappa-Grassmann-distribution-valued}), is uniquely determined by the condition
(for $\mathcal{E}^{p \, *}$-valued pairings)
\begin{equation}\label{X(kappa_l,m)-through-E^*p-valuedPairing}
\big\langle\big\langle \Xi(\kappa_{\mathpzc{l}, \mathpzc{m}}) (\Phi \otimes \phi \otimes h^p), \Psi \big\rangle\big\rangle
= \big\langle\kappa_{\mathpzc{l}, \mathpzc{m}}(\phi \otimes h^p ), \eta_{{}_{\Phi,\Psi}} \big\rangle
= \big\langle\kappa_{\mathpzc{l}, \mathpzc{m}}(\eta_{{}_{\Phi,\Psi}}), \phi \otimes h^p \big\rangle,
\end{equation}
for all
\[
\phi \otimes h^p \in \mathscr{E}^{\otimes \, n} \otimes \mathscr{E}^p, \,\,\,\, \textrm{and} \,\,\,\, \Phi, \Psi \in (\boldsymbol{E}),
\]
or, using scalar pairigs,
\[
\big\langle\big\langle \Xi(\kappa_{\mathpzc{l}, \mathpzc{m}}) (\Phi \otimes \phi \otimes h^p), \Psi \otimes \varphi \big\rangle\big\rangle
= \big\langle\kappa_{\mathpzc{l}, \mathpzc{m}}(\phi \otimes h^p ), \eta_{{}_{\Phi,\Psi}} \otimes \varphi \big\rangle
= \big\langle\kappa_{\mathpzc{l}, \mathpzc{m}}(\eta_{{}_{\Phi,\Psi}}), \phi \otimes h^p \otimes \varphi \big\rangle,
\]
for all
\[ 
\phi \otimes h^p \in \mathscr{E}^{\otimes \, n} \otimes \mathscr{E}^p, \varphi \in \mathcal{E}^p \,\,\,\, \textrm{and} \,\,\,\, \Phi, \Psi \in (\boldsymbol{E}),
\]
compare \cite{obataJFA} or eqs. (\ref{VectValotimesXi=intKerOp})-(\ref{electron-positron-photon-Xi}), Subsection \ref{psiBerezin-Hida}.

The same $\kappa_{\mathpzc{l}, \mathpzc{m}}$ given by (\ref{kappa-C-distribution-valued}), regarded as an element
of (\ref{kappa-C-distribution-valued}), but not as any element of (\ref{kappa-Grassmann-distribution-valued}),
admits the analogue pairing (\ref{pairingOfkappa}) which is $\mathbb{C}$-valued, provided $h^p$ we replace with ordinary test function
in $\mathcal{E}^p(M)$, and determines the corresponding integral kernel operator (\ref{Xi(kappa_l,m)}) by the condition
(\ref{X(kappa_l,m)-through-E^*p-valuedPairing}) for scalar valued pairings (because now the Grassmann-valued test function $h^p$
is replced with a $\mathbb{C}$-valued test function in  $\mathcal{E}^p(M)$).

The operator valued distribution
\[
S(g,h) = S(g\mathcal{L} + h^\sharp \boldsymbol{\psi})
\,\,\,
\textrm{or}
\,\,\
S(g,h) = S(g\mathcal{L} + \boldsymbol{\psi}^\sharp h),
\]
is now constructed perturbatively exactly as before in the form of the series
\begin{multline*}
S(g,h) = \boldsymbol{1} + {\textstyle\frac{1}{1!}} \big[ S_{1,0}(g)+S_{0,1}(h) \big]
+ {\textstyle\frac{1}{2!}} \big[ S_{2,0}(g^{\otimes\, 2}) + S_{1,1}(g \otimes h) + S_{0,2}(h ^2) \big]
\\
+ {\textstyle\frac{1}{3!}} \big[ S_{3,0}(g^{\otimes \, 3}) + S_{2,1}(g^{\otimes \, 2} \otimes h)
+  S_{1,2}(g\otimes h^2) + S_{0,3}(h^3) \big]
+ \ldots
\\
= \sum\limits_{n=0}^{+\infty} {\textstyle\frac{1}{n!}} S_n\big(g,h\big)
\end{multline*}
\begin{multline*}
S(g,h)^{-1} = \boldsymbol{1} + {\textstyle\frac{1}{1!}} \big[ \overline{S}_{1,0}(g)+\overline{S}_{0,1}(h) \big]
+ {\textstyle\frac{1}{2!}} \big[ \overline{S}_{2,0}(g\otimes g) + \overline{S}_{1,1}(g \otimes h) + \overline{S}_{0,2}(h^2) \big]
\\
+ {\textstyle\frac{1}{3!}} \big[ \overline{S}_{3,0}(g\otimes g \otimes g) + \overline{S}_{2,1}(g\otimes g \otimes h)
+  \overline{S}_{1,2}(g\otimes h^2) + \overline{S}_{0,3}(h^3) \big]
+ \ldots
\\
= \sum\limits_{n=0}^{+\infty} {\textstyle\frac{1}{n!}} \overline{S}_n\big(g,h\big)
\end{multline*}
with the only difference that now we replace the interaction Lagrangian $g\mathcal{L}$
by $g\mathcal{L} + h^\sharp \boldsymbol{\psi}$ (or $g\mathcal{L} + \boldsymbol{\psi}^\sharp h$) and with
\[
S_1(g,h) = S_{1,0}(g)+S_{0,1}(h),
\]
\[
S_n\big(g,h\big) = \,\,\, \sum\limits_{p=0}^{n}
S_{n-p,p}\big(g^{\otimes \, (n-p)} \otimes h^p\big),
\]
\[
\overline{S}_n\big(g,h\big) = \,\,\, \sum\limits_{p=0}^{n}
\overline{S}_{n-p,p}\big(g^{\otimes \, (n-p)} \otimes h^p\big),
\]
where $S_{n,p}$
and $\overline{S}_{n,p}$
are mathematically interpretad as finite sums of integral kernel operators (\ref{Xi(kappa_l,m)})
with the kernels (\ref{kappa-C-distribution-valued}), regarded as elements of
(\ref{kappa-Grassmann-distribution-valued}).
The operators $S_{n,p}\big(g^{\otimes \, n} \otimes h^p\big)$
and $\overline{S}_{n,p}\big(g^{\otimes \, n} \otimes h^p\big)$
are mathematically interpretad as the finite sum of integral kernel operators (\ref{Xi(kappa_l,m)})
with the kernels $\kappa_{\mathpzc{l},\mathpzc{m}}$ evaluated at $g^{\otimes \, n} \otimes h^p$:
\[
S_{n,p}\big(g^{\otimes \, n} \otimes h^p\big)
= \sum\limits_{\mathpzc{l},\mathpzc{m}}\Xi\big(\kappa_{\mathpzc{l},\mathpzc{m}}(g^{\otimes \, n} \otimes h^p) \big),
\]
and analogously for $\overline{S}_{n,p}\big(g^{\otimes \, n} \otimes h^p\big)$.
Here
\[
g\otimes \ldots \otimes g \otimes h^p = g^{\otimes \, n} \otimes h^p \in \mathscr{E}^{\otimes \, n} \otimes \mathscr{E}^p
\]
with $h^p$ being the Grassmann-valued test function (\ref{h^nPower}).

The operator-valued generalized
$n$-th order distributions
\begin{multline*}
S_{n,0}(x_1, \ldots, x_n), S_{n-1,1}(x_1, \ldots, x_{n-1}, a_1, y_1), \ldots,
\\
\ldots, S_{n-p,p}(x_1, \ldots, x_{n-p}, a_{1}, y_{1}, \ldots, a_p, y_p), \ldots, S_{0,n}(a_1,y_1, \ldots, a_n,y_n)
\end{multline*}
contain different numbers $p$ of space-time variables corresponding to the variables
of the Grassmann-valued switching-off function $h$. Each such component
changes sign whenever the variables  $a_q,y_q$ and  $a_j,y_j$ of $h^p$ are exchanged and both correspond to the variables of the Grassmenn-valued switching-off
function $h$, and remains the same under exchange of the space-time variables corresponding to the $g$ switching-off function. Thus
now  
\[
S_{n,p}(x_1, \ldots, x_{n}, a_1,y_1, \ldots, a_p, y_p)
\]
has mixed symmetry
\begin{gather*}
S_{n,p}(x_1, \ldots, x_i, \ldots, x_j, \ldots, x_{n}, a_1,y_1, \ldots, a_p, y_p)
\\
= S_{n,p}(x_1, \ldots, x_i, \ldots, x_j, \ldots, x_{n}, a_1,y_1, \ldots, a_p, y_p),
\\
S_{n,p}(x_1, \ldots, x_i, \ldots, x_{n}, a_1,y_1, \ldots, a_q, y_q, \ldots, a_p, y_p)
\\
= S_{n,p}(x_1, \ldots, a_q, y_q, \ldots, x_{n}, a_1,y_1, \ldots, x_i, \ldots, a_p, y_p),
\\
S_{n,p}(x_1, \ldots, x_{n}, a_1,y_1, \ldots, a_q, y_q, \ldots,  a_j, y_j, \ldots, a_p, y_p)
\\
= - S_{n,p}(x_1, \ldots, x_{n}, a_1,y_1, \ldots, a_j, y_j, \ldots,  a_q, y_q, \ldots, a_p, y_p),
\end{gather*}
but going always intact with the symmetry of the corresponding
switching-off test-function $g^{\otimes \, n} \otimes h^p$.

The causality axiom for
\[
S(g,h) = S(g\mathcal{L} + h^\sharp \boldsymbol{\psi})
\,\,\,
\textrm{or}
\,\,\
S(g,h) = S(g\mathcal{L} + \boldsymbol{\psi}^\sharp h),
\]
for the scattering operator-valued distribution $S(g,h)$ is implemented with the help of the switching-off function
\[
(g,h) \in \mathscr{E}\oplus\mathscr{E}^1 = \mathcal{S}(\mathbb{R}^4; \mathbb{C})\oplus\mathscr{E}^1
\]
and reads
\[
S(g_1+g_2,h_1+h_2) = S(g_2,h_2)S(g_1,h_1)
\,\,\,\,
\textrm{\tiny whenever $\textrm{supp} \,(g_1,h_1) \preceq  \textrm{supp} \,(g_2,h_2)$},
\]
and is equivalent to the above-stated axiom (I) in terms of $S_{n,p}$. In the shortened notation
for the space-time variables
$X = \{x_1, \ldots, x_{n}\}$, $Y = \{a_1, y_1, \ldots, a_p, y_p\}$
\begin{multline*}
\textrm{(I)} \,\,\,\,\,\,\,\,\,\,\,\,\,\,\,\,\,\,\,\,   
S_{n,p}(X,Y) = S_{n,p}(X_1,X_2,Y_1,Y_2)
\\
= (-1)^{\textrm{sgn}(Y_2,Y_1)}S_{k,q}(X_2,Y_2) S_{n-k,p-q}(X_1,Y_1),
\end{multline*}
whenever $X_1 \cup Y_1 \preceq X_2 \cup Y_2$, or in expanded form,
\begin{multline*}
              S_{n,p}(x_1, \ldots, x_{n}, a_1, y_1, \ldots, a_p, y_p)
=
(-1)^{\textrm{sgn}(p-q+1, p-q+2, \ldots, p, 1,2, \ldots, p-q)} \,\, \times
\\
\times
S_{k,q}(x_{n-k+1}, \ldots, x_{n}, a_{p-q+1}, y_{p-q+1}, \ldots, a_p, y_p) S_{n-k,p-q}(x_{1}, \ldots, x_{n-k},
a_{1}, y_{1}, \ldots a_{p-q}, y_{p-q})
\end{multline*}
whenever $\{x_{1}, \ldots, x_{n-k}, y_{1}, \ldots, y_{p-q} \} \preceq \{x_{n-k+1}, \ldots, x_{n}, y_{p-q+1}, \ldots, y_p\}$
(compare \cite{Bogoliubov_Shirkov}, \cite{Epstein-Glaser}, \cite{DKS1}, \cite{Scharf}). Here $\textrm{sgn}(Y_2,Y_2)$
is the sign of the permutation of the variables $Y$, which transforms $Y = (Y_1,Y_2))$ into $(Y_2,Y_1)$.

The Epstein-Glaser formula for the computation of
\[
S_{m,p}(x_1, \ldots, x_{k},y_1, \ldots, y_p), \,\,\, x_i \in \mathbb{R}^4, y_j \in M,
\]
(with $m+p=n$) from
\[
S_{k,q}(x_1, \ldots, x_{k}, a_1,y_1, \ldots, a_q, y_q),
\,\,\,
k+ q \leq n -1,
\]
is exactly the same as for $S(g)$ in Subsection \ref{MotivationForHida},
compare \cite{Epstein-Glaser} or \cite{DKS1}. Let the number $p\leq n$ of variables corresponding to the 
variables of the Grassmann-valued test function $h$ be fixed, and let us compute inductively $S_{m,p}$ for $m+p = n$. We consider the set
of variables $\{x_1, \ldots, x_n\} = \{ Z,x_n\}$ containing precisely $p\leq n$ variables corresponding to the Grassmann-valued test function $h$.
Next, we construct after \cite{Epstein-Glaser} and \cite{Scharf} 
\begin{gather*}
A'_{(n)}(Z,x_n) = \sum\limits_{{}_{X\sqcup Y = Z, X\neq \emptyset}}
(-1)^{\textrm{sgn}(X,Y,x_n)}\overline{S}(X)S(Y,x_n),
\\
R'_{(n)}(Z,x_n) = \sum\limits_{{}_{X\sqcup Y = Z, X\neq \emptyset}}
(-1)^{\textrm{sgn}(Y,x_n, X)}
S(Y,x_n)\overline{S}(X),
\\
Z = \{x_1,\ldots, x_{n-1}\}, \,\,\,\, x_{i}\in\mathbb{R}^4 \,\, \textrm{or} \,\, x_i \in M
\end{gather*}
as in Subsection \ref{MotivationForHida}, replacing everywhere $S_k$, $k\leq n-1$, by $S_{k,q}$, with $k+q \leq n-1$,
just as Epstein and Glaser did in \cite{Epstein-Glaser}. The only modification is that in each product
\[
\overline{S}(X)S(Y,x_n) \,\,\,
\textrm{or}
\,\,\,
S(Y,x_n)\overline{S}(X)
\]
in $A'_{(n)}(Z,x_n)$ or $R'_{(n)}(Z,x_n)$ is multilplied by $(-1)^{\textrm{sgn}(X,Y,x_n)}$
or $(-1)^{\textrm{sgn}(Y,x_n, X)}$.
Here $\textrm{sgn}(X,Y,x_n)$ or $\textrm{sgn}(Y,x_n, X)$ is the number of inversions of the variables 
corresponding to the variables of the Grassmann-valued test function $h$ in the permutation
\[
Z,x_n \longrightarrow X,Y,x_n 
\,\,\,
\textrm{or}
\,\,\,
Z,x_n \longrightarrow Y,x_n,X, 
\] 
of the variables
\[
\{Z,x_n\} = \{x_1, \ldots, z_{n-1}, x_n\} \,\,\,\textrm{in $\mathbb{R}^4$ or in $M$}
\]
and we have put 
\begin{gather*}
S(X) = S_{k,q}(x_1, \ldots, x_{k}, x_{k+1}, \ldots, x_{k+q}),
\\
X = \{x_1, \ldots, x_{k}, x_{k+1}, \ldots, x_{k+q}\}, 
\\
 \textrm{$x_i\in \mathbb{R}^4$ for $i\leq k$ and $x_i\in M$ for $k \leq i\leq k+q$}
\,\,\,
k+ q \leq n -1,
\end{gather*}
for brevity.

By inductive assumption all $A'_{(n)}$ and $R'_{(n)}$ are known, and 
\begin{gather*}
A_{(n)}(Z,x_n) = A'_{(n)}(Z,x_n) + S(Z,x_n) =  \sum\limits_{{}_{X\sqcup Y = Z}}
(-1)^{\textrm{sgn}(X,Y,x_n)}\overline{S}(X)S(Y,x_n),
\\
R_{(n)}(Z,x_n) = R'_{(n)}(Z,x_n) + S(Z,x_n) =  \sum\limits_{{}_{X\sqcup Y = Z}}
(-1)^{\textrm{sgn}(Y,x_n, X)}
S(Y,x_n)\overline{S}(X),
\end{gather*}
have causal support in the sense of Subsection \ref{MotivationForHida} and give, respectively the retarded and advanced part
of
\[
D_{(n)} = R'_{(n)} - A'_{(n)} = A_{(n)}-R_{(n)},
\]
and can be computed indepenetly on using the splitting of causal distributions, with the ambiguity in the splitting 
depending on the singularity degree, which can be eliminated on imposing natural existence condition for the interacting 
fields in the adiabatic limit, compare
Sybsections \ref{OperationsOnXi} and \ref{WickForChronological}.
Therefore 
\begin{gather*}
S(Z, x_n) =  A_{(n)}(Z,x_n) - A'_{(n)}(Z, x_n)
\\
\textrm{or}
\\
S(Z, x_n) =  R_{(n)}(Z,x_n) - R'_{(n)}(Z, x_n),
\end{gather*}
and $S_{m,p}$, for $m+p=n$ can be computed out of the $S_{k,q}$, $k+p<n$. In particular kernels $S_{n,0}(X)$ of the higher order 
contributions containing no variables corresponding to the variables of the Grassmann-valued test function $h$
coincide with the kernels $S_n(X)$ of the higher order contributions $S_n(g^{\otimes \, n})$ for the scattering operator 
$S(g)$ computed wihout any additional terms in the Lagrangian. All kernels $S_{n,1}$ of the higher order contributions
$S_{n,1}$ containing precisely one variable of $h$ (with the rest of variables corresponding to the space-time variables of $g^{\otimes \, n}$)
can be computed by the analogous formulas in which the additional factors $(-1)^{\textrm{sgn}(X,Y,x_n)}$
and $(-1)^{\textrm{sgn}(Y,x_n, X)}$ disapppear. We can choose $x_n$ to be the only variable equal to the variable of $h$, with
\begin{gather*}
A'_{(n)}(Z,x_n) = \sum\limits_{{}_{X\sqcup Y = Z, X\neq \emptyset}}
\overline{S}(X)S(Y,x_n),
\\
R'_{(n)}(Z,x_n) = \sum\limits_{{}_{X\sqcup Y = Z, X\neq \emptyset}}
S(Y,x_n)\overline{S}(X),
\end{gather*}
\begin{gather*}
A_{(n)}(Z,x_n) = A'_{(n)}(Z,x_n) + S(Z,x_n) =  \sum\limits_{{}_{X\sqcup Y = Z}}
\overline{S}(X)S(Y,x_n),
\\
R_{(n)}(Z,x_n) = R'_{(n)}(Z,x_n) + S(Z,x_n) =  \sum\limits_{{}_{X\sqcup Y = Z}}
S(Y,x_n)\overline{S}(X),
\end{gather*}
without $(-1)^{\textrm{sgn}(X,Y,x_n)}$
and $(-1)^{\textrm{sgn}(Y,x_n, X)}$, because here the sets $X,Y$ of variables contain in this case only the space-time variables
corresponding to the variables of $g$ and $x_n$ is the only variable of $h$. The kernels $S_{k,0}(Z)$ and
$S_{k-1,1}(Z,x_{k})$, $k\leq n$, respectively, without and with only one variable equal to the variable of $h$ and equal $x_{k}$, are used in the computation
of the  $n$-th order contributions to the interacting fields $\boldsymbol{\psi}^{\sharp}_{{}_{\textrm{int}}}$ 
and $\boldsymbol{\psi}_{{}_{\textrm{int}}}$, compare Section \ref{A(1)psi(1)}.

Besides (I) we are using the analogous axioms (II)-(V), except the formal unitarity (or formal Krein-isometricity) (III),
as now the added term $S_{0,1}$ is not self-adjoint or Krein-selfadjoint, and which suffice for the computation
of all $S_{n,p}$. 

Here we should emphasize that the axioms (I)-(V), for $S(g,h)$, are consistent (we are not using (III) for $S(g,h)$). It is not obvious as
the axiom (I), as well as the Epstein-Glaser inductive step, involve the product operation. 
It is however well-defined, and in fact follows from the proved Lemmas used in the construction
of the product $S(g_1)S(g_2)$. Let us explain it. The operators
\begin{equation}\label{S_n,p=Xi(kappa_l,m)}
S_{n,p} = \sum\limits_{\mathpzc{l},\mathpzc{m}}\Xi(\kappa_{\mathpzc{l},\mathpzc{m}})
\in \mathscr{L}\big((\boldsymbol{E}), (\boldsymbol{E})^*\big) \otimes \mathscr{L}(\mathscr{E}^{\otimes \, n} \otimes \mathscr{E}^p, \mathcal{E}^{p \, *})
\end{equation}
are equal to finite sums of integral kernel operators $\Xi(\kappa_{\mathpzc{l},\mathpzc{m}})$, eq. (\ref{Xi(kappa_l,m)}),
of the class (\ref{classOfXi}), with
\[
S_{n,p}\big(g^{\otimes \, n} \otimes h^p\big)
= \sum\limits_{\mathpzc{l},\mathpzc{m}}\Xi\big(\kappa_{\mathpzc{l},\mathpzc{m}}(g^{\otimes \, n} \otimes h^p) \big)
\in
\mathscr{L}\big((\boldsymbol{E}), (\boldsymbol{E})^*\big) \otimes  \mathcal{E}^{p \, *}
\]
and are given by the kernels $\kappa_{\mathpzc{l},\mathpzc{m}}$ of the class 
(\ref{kappa-C-distribution-valued}), which are regarded as kernels of the class (\ref{kappa-Grassmann-distribution-valued}).
But we observe that the class of the kernels  $\kappa_{\mathpzc{l},\mathpzc{m}}$ we encounter in computation of $S(g,h)$
is similar as that we had in $S(g)$, except that now the kernels are antisymmetric in the
variables of $h$. In fact, the products of pairings are identical, as we are operating with the same
free fields underlying the theory in question, and which we have in the total interaction Lagrangian, exept
that now in the subset of bispinor-space-time variables these products are antisymmetrized, and symmetrized in remaining space-time variables.  
In particular, using the result proved obove, we observe that
if there are no massless fields in the total interaction Lagrangian, or if we raplace the exponents of the kernels of the massless fields
in the total interaction Lagrangian with the exponents of massive kernels, then the kernels $\kappa_{\mathpzc{l},\mathpzc{m}}$, regarded as the elements
of  (\ref{kappa-C-distribution-valued}), can be extended to separately continuous maps
\[
\big(E_1 \otimes \ldots \otimes E_\mathpzc{l}\big)^* \times \big(E_{\mathpzc{l}+1} \ldots \otimes E_{\mathpzc{l}+\mathpzc{m}}\big) 
\longrightarrow  
\mathscr{E}^{*\otimes \, n} \otimes \mathcal{E}^{p*}(M).
\]
This follows by the repeated application of the Lemma \ref{rett(x,y):W'(x)W''(y):InSxS->L((E),(E))Contq1>2}.
Therefore these kernels $\kappa_{\mathpzc{l},\mathpzc{m}}$, regarded as kernels of the class  
(\ref{kappa-Grassmann-distribution-valued}), can be extended to separately continuous maps
\[
\big(E_1 \otimes \ldots \otimes E_\mathpzc{l}\big)^* \times \big(E_{\mathpzc{l}+1} \ldots \otimes E_{\mathpzc{l}+\mathpzc{m}}\big) 
\longrightarrow  
\mathscr{L}(\mathscr{E}^{\otimes \, n} \otimes \mathscr{E}^p, \mathcal{E}^{p \, *}).
\]
Using this fact as in the proof of Theorem \ref{obataJFA.Thm.3.13}, Subsection \ref{psiBerezin-Hida} (compare
Thm. 3.13  in \cite{obataJFA} or Thm. 2.6 in \cite{hida} for the purely Bose case) we see that the operators
(\ref{S_n,p=Xi(kappa_l,m)}) belong to
\[
\mathscr{L}\big((\boldsymbol{E}), (\boldsymbol{E})\big) \otimes \mathscr{L}(\mathscr{E}^{\otimes \, n} \otimes \mathscr{E}^p, \mathcal{E}^{p \, *}),
\]
so that
\[
S_{n,p}\big(g^{\otimes \, n} \otimes h^p\big)
= \sum\limits_{\mathpzc{l},\mathpzc{m}}\Xi\big(\kappa_{\mathpzc{l},\mathpzc{m}}(g^{\otimes \, n} \otimes h^p) \big)
\in
\mathscr{L}\big((\boldsymbol{E}), (\boldsymbol{E})\big) \otimes  \mathcal{E}^{p \, *},
\]
if all free fields in the interaction Lagrangian are massive, or, if we replace the exponents of the kernels of massless
fields with the exponents of massive kernels, as above in definition of the product $S(g_1)S(g_2)$.  
Therefore the product 
\[
S_{n_1,p_1}\big(g^{\otimes \, n_1} \otimes h^{p_1}\big)S_{n_2,p_2}\big(g^{\otimes \, n_2} \otimes h^{p_2}\big)
\in \mathscr{L}\big((\boldsymbol{E}), (\boldsymbol{E})\big) \otimes  \mathcal{E}^{(p_1+p_2) \, *}
\]
of the operators 
\[
S_{n_1,p_1}\big(g^{\otimes \, n_1} \otimes h^{p_1}\big) \in \mathscr{L}\big((\boldsymbol{E}), (\boldsymbol{E})\big) \otimes  \mathcal{E}^{p_1 \, *}
\,\,\,\,\,\,
\textrm{and}
\,\,\,\,\,\,
S_{n_2,p_2}\big(g^{\otimes \, n_2} \otimes h^{p_2}\big)
\in
\mathscr{L}\big((\boldsymbol{E}), (\boldsymbol{E})\big) \otimes  \mathcal{E}^{p_2 \, *}
\]
is well-defined in this case, and is given by the operator composition on the first tensor product factor, and by the  
Grassmann product on the second tensor product factor. This is so if there are no massless fields
in the interaction Lagrangian, or if we replace the exponents of the massless kernels with the exponents of massive kernels,
as we did above in the construction of $S(g)$ for the interaction Lagrangian containing massless fields.
Next, using Lemma \ref{auxiliaryProductLemma},
we pass to the zero mass limit in the exponents of the kernels of the massless fields, in which we have replaced the exponents
by the exponents of the massive kernels, as we did above in the construction of the product $S_{n_1}(g^{\otimes \, n_1})S_{n_2}(g^{\otimes \, n_2})$.
By the Lemmas \ref{rett(x,y):W'(x)W''(y):InSxS->L((E),(E))Contq1>2} and \ref{auxiliaryProductLemma} the limit 
\[
S_{n_1,p_1}\big(g^{\otimes \, n_1} \otimes h^{p_1}\big)S_{n_2,p_2}\big(g^{\otimes \, n_2} \otimes h^{p_2}\big)
\in \mathscr{L}\big((\boldsymbol{E}), (\boldsymbol{E})^*\big) \otimes  \mathcal{E}^{(p_1+p_2) \, *}
\]
of the product of the operators 
\begin{gather*}
S_{n_1,p_1}\big(g^{\otimes \, n_1} \otimes h^{p_1}\big) \in \mathscr{L}\big((\boldsymbol{E}), (\boldsymbol{E})\big) \otimes  \mathcal{E}^{p_1 \, *}
\\
\textrm{and}
\\
S_{n_2,p_2}\big(g^{\otimes \, n_2} \otimes h^{p_2}\big)
\in
\mathscr{L}\big((\boldsymbol{E}), (\boldsymbol{E})\big) \otimes  \mathcal{E}^{p_2 \, *},
\end{gather*}
in which we have applied the said replacement of the exponents of the kernels of the massless
fields with exponents of massive fields,
exists in 
\[
\mathscr{L}\big((\boldsymbol{E}), (\boldsymbol{E})^*\big) \otimes  \mathcal{E}^{(p_1+p_2) \, *},
\]
because the corresponding limits of the kernels of the products, regarded, respectively, as 
$\mathscr{E}^{*\otimes \, n_1} \otimes \mathcal{E}^{p_1*}(M)$-valued, 
$\mathscr{E}^{*\otimes \, n_2} \otimes \mathcal{E}^{p_2*}(M)$-valued and
$\mathscr{E}^{*\otimes \, (n_1+n_2)} \otimes \mathcal{E}^{(p_1+p_2)*}(M)$-valued
kernels exist, by the above proved  Lemmas \ref{rett(x,y):W'(x)W''(y):InSxS->L((E),(E))Contq1>2} and \ref{auxiliaryProductLemma}.

In fact, when considering charged fields (quantized complex fields), such as the Dirac field, we are using not just 
the abstract Grassmann algebra $\oplus_n \mathcal{E}^{n*}$ with inner product, 
but instead the abstract \emph{Grassman algebra} $\oplus_n \mathcal{E}^{n*}$ \emph{with inner product and with involution} 
$\overline{\,\,\cdot \,\,}$ in the sense of Berezin
\cite{Berezin}, which respects the following conditions
\begin{enumerate}
\item[(1)]
$\overline{\overline{f}} = f$,
\item[(2)]
$\overline{f_1f_2} = \overline{f_2}\overline{f_1}$,
\item[(3)]
$\overline{\alpha f} = \overline{\alpha}\overline{f}$, $\alpha \in \mathbb{C}$,
\item[(4)]
If the inner product $(f,g)$ is well-defined for $f,g$, then it is well-defined for $\overline{f},\overline{g}$
and $(f,g) = (\overline{g}, \overline{f})$,
\item[(5)]
The space $\mathcal{E}^{1*}$ has direct sum decomposition
$\mathcal{E}^{1*} = F \oplus \overline{F}$, such that $F \cap \mathcal{H}^1$ and $\overline{F}\cap \mathcal{H}^1$
are unitarily isomorphic and orthogonal.
\end{enumerate}

It can likewise be realized as the Gelfand triple
\[
\begin{array}{ccccc}
(\mathcal{E}^{1 \, }) & \,\,\,\,\,\,\,\,\, \subset   & \Gamma(\mathcal{H}^1) =\Gamma(\mathcal{H}^{01} \oplus \mathcal{H}^{10})  & \,\,\,\,\,\,\,\,\, \subset & (\mathcal{E}^{1})^*
\\
\parallel & & \parallel & & \parallel
\\
\bigwedge \mathcal{E}^{01} \,\, \widehat{\otimes} \,\, \bigwedge \mathcal{E}^{10} & & \Gamma(\mathcal{H}^{01}) \widehat{\otimes} \Gamma(\mathcal{H}^{10}) &&
\bigwedge \mathcal{E}^{01*} \,\, \widehat{\otimes} \,\, \bigwedge \mathcal{E}^{10*}
\end{array}
\]
arising from the application of the Fermi second quantized functor $\Gamma$ to the Gelfand triple
\begin{equation}\label{GelfandTripleGAIPI}
\mathcal{E}^{1} =\mathcal{E}^{01} \oplus \mathcal{E}^{10} \,\,\,  \subset \,\,\,  \mathcal{H}^{1} = \mathcal{H}^{01} \oplus \mathcal{H}^{10}
\,\,\, \subset \mathcal{E}^{1*} =\mathcal{E}^{01*} \oplus \mathcal{E}^{10*} 
\end{equation}
equal to the direct sum of two copies of the Gelfand triples (\ref{GelfandTripleGAIP}) used above, so that
$\mathcal{H}^{10}=\overline{\mathcal{H}^{01}}$ are both equal to $\mathcal{H}^1$ in (\ref{GelfandTripleGAIP}),
$\mathcal{E}^{10} = \overline{\mathcal{E}^{01}}$ are both equal to $\mathcal{E}^{1}$ in (\ref{GelfandTripleGAIP}),
and  $\mathcal{E}^{10*} = \overline{\mathcal{E}^{01*}}$ are both equal to $\mathcal{E}^{1*}$ in (\ref{GelfandTripleGAIP}),
and with the standard operator $A\oplus A$ of (\ref{GelfandTripleGAIPI}), where $A$ is the standard operator 
of (\ref{GelfandTripleGAIP}). Here $\overline{\mathcal{H}^{01}}$ is the Hilbert space
conjugate to $\mathcal{H}^{01}$, consisting of the same elements as  $\mathcal{H}^{01}$ with the same addition as in 
$\mathcal{H}^{01}$, but with multiplication by a scalar and with inner product defined by
\begin{enumerate}
\item[1)]
\textrm{$\alpha u$ in $\overline{\mathcal{H}^{01}}$ is equal to $\overline{\alpha} u$ in $\mathcal{H}^{01}$}, 
\item[2)]
\textrm{$(u,v)$ in $\overline{\mathcal{H}^{01}}$ is equal to $(v,u)$ in $\mathcal{H}^{01}$},
\end{enumerate}
compare e.g. \cite{Mackey2}, p. 243.
Note that any unitary map $U:\mathcal{H}^{01} \longmapsto \mathcal{H}^{01}$, \emph{e.g.} identity map $\textrm{Id}$,
becomes conjugate linear and anti-unitary when regarded as a map  $\overline{U}:\mathcal{H}^{01} \longmapsto \overline{\mathcal{H}^{01}}$
or as a map $\overline{U}:\overline{\mathcal{H}^{01}} \longmapsto \mathcal{H}^{01}$,
and $\overline{\textrm{Id}}$ defines involution in  
\[
\mathcal{H}^{1} = \mathcal{H}^{01} \oplus \mathcal{H}^{10} = \mathcal{H}^{1} = \mathcal{H}^{01} \oplus \overline{\mathcal{H}^{01}}.
\]
Note however that here $\Gamma(\mathcal{H}^{01} \oplus \mathcal{H}^{10})$ is understood in a slightly different manner than
in Subsection \ref{electron+positron} or \ref{psiBerezin-Hida}, where the simple tensors in $\Gamma(\mathcal{H}^{01} \oplus \mathcal{H}^{10})$
of the elements of $\mathcal{H}^{01}$ and $\mathcal{H}^{10}$, are \emph{separately} antisymmetrized in the factors lying in  
$\mathcal{H}^{01}$ and \emph{separately} antisymmetrized in the factors lying in  $\mathcal{H}^{10}$. This meaning
of $\Gamma(\mathcal{H}^{01} \oplus \mathcal{H}^{10})$ is used in physics whenever the superpositions of two states, one lying in
$\mathcal{H}^{01}$ and the other one in $\mathcal{H}^{10}$, is meaningless, as in  Subsection \ref{electron+positron} or \ref{psiBerezin-Hida}
or more generally whenever $\mathcal{H}^{01}$ and  $\mathcal{H}^{10}$ are single particle states of different kind of fields,
with $\Gamma(\mathcal{H}^{01} \oplus \mathcal{H}^{10}) =\Gamma(\mathcal{H}^{01}) \otimes \Gamma(\mathcal{H}^{10})$, 
understood as their common Fock space. Here we antisymmetrize all simple tensors in
$\Gamma(\mathcal{H}^{01} \oplus \mathcal{H}^{10})$ in all single particle factors lying in $\mathcal{H}^{01}$ and $\mathcal{H}^{10}$ 
on equal grounds. Thereofore, it perhaps would be better to write for this functor $\bigwedge$ instead of $\Gamma$. 

The standard operator determining (\ref{GelfandTripleGAIPI}) is equal $\Gamma(A\oplus A)$ and the Grassmann product in the 
abstract Grassmann algebra 
\[
\bigwedge \mathcal{E}^{1 \, *} = \bigwedge \Big(\mathcal{E}^{01*} \oplus \mathcal{E}^{10*}\Big) = 
\bigwedge \mathcal{E}^{01*} \,\,\, \widehat{\otimes} \,\,\, \bigwedge \mathcal{E}^{10*},
\]
is given by the wedge product, as before. By construction
\[
\bigwedge \mathcal{E}^{1 \, *} = \bigwedge \mathcal{E}^{01*} \widehat{\otimes} \bigwedge \mathcal{E}^{10*}
= \underset{n}{\oplus} \mathcal{E}^{n*} = \underset{n}{\oplus}\underset{p+q=n}{\oplus}\mathcal{E}^{pq*},
\,\,\,\,
\mathcal{E}^{pq}= \big[\mathcal{E}^{10}\big]^{\wedge \, p} \wedge \big[\mathcal{E}^{01}\big]^{\wedge \, q}.
\]

By construction, the restriction of the involution $\overline{\textrm{Id}}$
to $\mathcal{E}^{1}$ is continuous on $\mathcal{E}^1$ and can be lifted to a continuous map
$\bigwedge \mathcal{E}^{1} \rightarrow \bigwedge \mathcal{E}^{1}$ by application of the Fermi functor, with the additional modification that in action
on a homogeneous simple tensor 
\[
f_1 \wedge \ldots \wedge f_p \wedge \overline{f_{p+1}} \wedge \overline{f_{p+q}} \in \mathcal{E}^{pq}, \,\,\, f_i,f_j \in \mathcal{E}^{10},
\,\,\, \overline{f_j} \in \mathcal{E}^{01}, 
\]  
it is not merely equal to $\overline{\textrm{Id}}^{\otimes \, (p+q)}$ but in addition we apply the inverse order of the
tensor factors
\begin{multline*}
\mathcal{E}^{pq} \ni
\overline{f_1 \wedge \ldots \wedge f_p \wedge \overline{f_{p+1}} \wedge \ldots \wedge \overline{f_{p+q}}} 
\\
\overset{\textrm{df}}{=}
f_{p+1} \wedge \ldots \wedge f_{p+q} \wedge\overline{f_{1}} \wedge \ldots \wedge \overline{f_p} \in \mathcal{E}^{qp}.
\end{multline*}
This involution is a continuous map
\[
\bigwedge \mathcal{E}^{1} \longrightarrow \bigwedge \mathcal{E}^{1}
\]
which defines, by duality, the continuous involution
\[
\bigwedge \mathcal{E}^{1*} \longrightarrow \bigwedge \mathcal{E}^{1*}
\]
in the Grassmann algebra $\underset{n}{\oplus} \mathcal{E}^{n*}$. 

In the subspace $\mathcal{H}^{01}$ we introduce the generalized basis $\iota(x)$, $x\in M$, as before.
In the subspace $\mathcal{H}^{10} = \overline{\mathcal{H}^{01}}$ we introduce the generalized basis $\overline{\iota(x)}$, $x\in M$,
adjoint to $\iota(x)$, for which
\[
\overline{f} = \int \overline{\iota(x)} \,\,\, \overline{\varphi(x)} \, dx, \,\,\, \overline{f} \in \mathcal{E}^{10*} = \overline{\mathcal{E}^{01*}},
\]
whenever
\[
f = \int \iota(x) \varphi(x) \, dx, \,\,\, f \in \mathcal{E}^{01*}.
\]
The union $\{\overline{\iota(x)},\iota(x), x,y \in M\}$ is the generalized basis of $\mathcal{E}^{1*}$, and provides a system
of \emph{generators} of the Grassmann algebra $\bigwedge \mathcal{E}^{1*} = \underset{n}{\oplus} \mathcal{E}^{n*}$
with involution $\overline{\,\,\cdot \,\,}$. Each element 
$f$ and its adjoint $\overline{f}$ of the Grassmann algebra have the following representation
\begin{gather*}
f = \sum\limits_{p,q} \int \overline{\iota(x_1)} \ldots \overline{\iota(x_p)} \iota(y_1) \ldots \iota(y_q)  \,\, \times
\\
\times \,\, \varphi^{pq}(x_1, \ldots, x_p, y_1, \ldots, y_q) \, dx_1 \ldots dx_p dy_1 \ldots dy_q \in \mathcal{E}^{pq*},
\\
\overline{f} = \sum\limits_{p,q} \int \overline{\iota(y_q)}\ldots \overline{\iota(y_1)} \iota(x_p) \ldots \iota(x_p)  \,\, \times
\\
\times \,\, \overline{\varphi^{pq}(x_1, \ldots, x_p, y_1, \ldots, y_q)} \, dx_1 \ldots dx_p dy_1 \ldots dy_q \in \mathcal{E}^{qp*},
\end{gather*}
with the generalized functions $\varphi^{pq}\in \mathcal{E}^{pq \, *}(M)$ antisymmetric in all variables $x_1, \ldots, y_q \in M$.

We define the space $\mathscr{E}^{pq}$ of Grassmann-valued test functions
\begin{gather}
h^{pq}(a_1,x_1, \ldots, a_p,x_p, b_1,y_1, \ldots, b_q,y_q)
\\
=  
\overline{\iota^{a_1}(x_1)} \ldots \overline{\iota^{a_p}(x_p)} \iota^{b_1}(y_1) \ldots \iota^{b_q}(y_q)
\phi(a_1, x_1, \ldots, a_p, x_p, b_1, y_1, \ldots, b_q, y_q)
\nonumber
\\ = \overline{\iota}^p\iota^q \cdot \phi(a_1,x_1, \ldots, b_q,y_q),
\label{h^pq}
\\
\phi \in \mathcal{S}(\mathbb{R}^4; \mathbb{C}^4)^{\otimes \, (p+q)}_{{}_{\textrm{sym}}}, x_i \in \mathbb{R}^4, a_i,b_j \in \{1, \ldots, 4\},
\nonumber
\end{gather}
and endow it with the topology of the symmetric tensor product space
$\mathcal{S}(\mathbb{R}^4, \mathbb{C}^4)^{\otimes \, (p+q)}_{{}_{\textrm{sym}}}$ using the identification
of $h^{pq} = \overline{\iota}^p\iota^q \cdot \phi$ with $\phi \in \mathcal{S}(\mathbb{R}^4, \mathbb{C}^4)^{\otimes \, (p+q)}_{{}_{\textrm{sym}}}$. 
We define the pairing of a distribution $\kappa \in \mathcal{E}^{pq \, *}(M)$ with $h^{pq}$ in the same way as before. 

Using integral kernel operators with 
$\mathcal{E}^{pq \, *}(M)$-valued kernels $\kappa_{\mathpzc{l}, \mathpzc{m}}$, regarded naturally as 
$\mathscr{L}(\mathscr{E}^{\otimes \, pq} \otimes \mathscr{E}^{pq}, \mathcal{E}^{pq \, *}) \big)$-valued
kernels, we can restore the axiom (III) of formal Krein-unitarity by considering
\[
S(g,\overline{h},h) = S(g\mathcal{L} + h^\sharp \boldsymbol{\psi}+ \boldsymbol{\psi}^\sharp h),
\]
exactly as before in the form of the series
\begin{multline*}
S(g,\overline{h},h) = \boldsymbol{1} + {\textstyle\frac{1}{1!}} \big[ S_{1,0,0}(g)+S_{0,1,0}(\overline{h}) + S_{0,0,1}(h) \big]
\\
+ {\textstyle\frac{1}{2!}} \big[ S_{2,0,0}(g^{\otimes\, 2}) + S_{1,1,0}(g \otimes h^{10}) + S_{1,0,1}(g \otimes h^{01}) + S_{0,1,1}(h ^{11}) \big]
+ \ldots
\\
= \sum\limits_{n=0}^{+\infty} {\textstyle\frac{1}{n!}} S_n\big(g,\overline{h},h\big)
\end{multline*}
\begin{multline*}
S(g,\overline{h},h)^{-1} = \boldsymbol{1} + {\textstyle\frac{1}{1!}} \big[ \overline{S}_{1,0,0}(g)+\overline{S}_{0,1,0}(\overline{h}) 
+ \overline{S}_{0,0,1}(h) \big]
\\
+ {\textstyle\frac{1}{2!}} \big[ \overline{S}_{2,0,0}(g^{\otimes\, 2}) + \overline{S}_{1,1,0}(g \otimes h^{10}) 
+ \overline{S}_{1,0,1}(g \otimes h^{01}) + \overline{S}_{0,1,1}(h ^{11}) \big]
+ \ldots
\\
= \sum\limits_{n=0}^{+\infty} {\textstyle\frac{1}{n!}} \overline{S}_n\big(g,\overline{h},h\big)
\end{multline*}
with the only difference that now we replace the interaction Lagrangian $g\mathcal{L}$
by the total sum $g\mathcal{L} + h^\sharp \boldsymbol{\psi} + \boldsymbol{\psi}^\sharp h$ and with
\[
S_1(g,\overline{h},h) = S_{1,0,0}(g)+S_{0,1,0}(\overline{h})+S_{0,0,1}(h),
\]
\[
S_n\big(g,\overline{h},h\big) = \,\,\, \sum\limits_{m+p+q=n}
S_{m,p,q}\big(g^{\otimes \, m} \otimes h^{pq}\big),
\]
\[
\overline{S}_n\big(g,\overline{h},h\big) = \,\,\, \sum\limits_{m+p+q=n}
\overline{S}_{m,p,q}\big(g^{\otimes \, m} \otimes h^{pq}\big),
\]
where $S_{n,p,q}$
and $\overline{S}_{n,p,q}$
are mathematically interpretad as finite sums of integral kernel operators (\ref{Xi(kappa_l,m)})
with the $\mathcal{E}^{pq \, *}(M)$-valued kernels , regarded naturally as 
$\mathscr{L}(\mathscr{E}^{\otimes \, pq} \otimes \mathscr{E}^{pq}, \mathcal{E}^{pq \, *}) \big)$-valued
kernels.
The operators $S_{n,p,q}\big(g^{\otimes \, n} \otimes h^{pq}\big)$
and $\overline{S}_{n,p,q}\big(g^{\otimes \, n} \otimes h^{pq}\big)$
are mathematically interpretad as the finite sum of integral kernel operators (\ref{Xi(kappa_l,m)})
with the kernels $\kappa_{\mathpzc{l},\mathpzc{m}}$ evaluated at $g^{\otimes \, n} \otimes h^{pq}$:
\[
S_{n,p,q}\big(g^{\otimes \, n} \otimes h^{pq}\big)
= \sum\limits_{\mathpzc{l},\mathpzc{m}}\Xi\big(\kappa_{\mathpzc{l},\mathpzc{m}}(g^{\otimes \, n} \otimes h^{pq}) \big),
\]
and analogously for $\overline{S}_{n,p,q}\big(g^{\otimes \, n} \otimes h^{pq}\big)$.
Here
\[
g\otimes \ldots \otimes g \otimes h^{pq} = g^{\otimes \, n} \otimes h^{pq} \in \mathscr{E}^{\otimes \, n} \otimes \mathscr{E}^{pq}
\]
with $h^{pq}$ being the Grassmann-valued test function (\ref{h^pq}) of the special product form
\begin{multline*}
h^{pq}(a_1,x_1, \ldots, a_p,x_p, b_1,y_1, \ldots, b_q,y_q)
= \overline{h^{a_1}(x_1)} \ldots \overline{ h^{a_p}(x_p)}h^{b_1}(y_1) \ldots  h^{b_q}(y_q) 
\\
=  \overline{\iota^{a_1}(x_1)} \ldots \overline{\iota^{a_p}(x_p)} \iota^{b_1}(y_1) \ldots \iota^{b_q}(y_q)
\overline{\phi^{a_1}(x_1)} \ldots \overline{\phi^{a_p}(x_p)}\phi^{b_1}(y_1)\phi^{b_q}(y_q)
\end{multline*}
with 
\[
\phi(a_1,x_1, \ldots, b_q,y_q) =  \overline{\phi^{a_1}(x_1)} \ldots \overline{\phi^{a_p}(x_p)}\phi^{b_1}(y_1) \ldots \phi^{b_q}(y_q)
\]
in (\ref{h^pq}) and with $\overline{h} = h^{10}$, $h=h^{01}$.

In order to formulate the covariance axiom (II) for $S(g,\overline{h},h)$, we note first that
the unitary transformation $\mathcal{U}$, (\ref{mathcalU}), acting in $\iota^{-1}\mathcal{H}^{01} = L^2(M)$ and acting by the same formula, in $\overline{\iota}^{-1}\mathcal{H}^{10}
= \overline{\iota}^{-1}\overline{\mathcal{H}^{01}} = \overline{L^2(M)}$ in the second orthogonal copy of $L^2(M)$, 
are both unitary and both give mappings $\mathcal{E}^{01}(M) \rightarrow \mathcal{E}^{01}(M)$
and $\mathcal{E}^{10}(M) \rightarrow \mathcal{E}^{10}(M)$ which are continuous, and their direct sum $\mathcal{U}\oplus \mathcal{U}$
is a unitary and continuous map
\[
\mathcal{E}^{1}(M) = \mathcal{E}^{01}(M) \oplus \mathcal{E}^{10}(M) \longrightarrow
\mathcal{E}^{1}(M) = \mathcal{E}^{01}(M) \oplus \mathcal{E}^{10}(M).
\] 
It can be lifted to a continuous map $\Gamma(\mathcal{U}\oplus \mathcal{U}) =  \Gamma(\mathcal{U})\widehat{\otimes}\Gamma(\mathcal{U})$ 
of the test Hida space $\oplus_n \mathcal{E}^{n}(M)$ onto itself, and by
duality, to the continuous map of $\oplus_n \mathcal{E}^{n*}(M)$ onto itself.  It induces a continuous representation $\boldsymbol{\mathcal{U}}$,
through the isomorphism $\iota$ and its lifting $\Gamma(\iota)$, on the Grassman algebra $\oplus_n \mathcal{E}^{n*}$ with involution.
It can be expressed in terms of the action on the generalized basis  
 $\{\overline{\iota(x)},\iota(x), x,y \in M\}$ in the form
\begin{equation}\label{mathcalUiota}
\boldsymbol{\mathcal{U}}_{{}_{b,\Lambda}} \big(\overline{\iota^a(x)} \big)= \overline{\iota^a(\Lambda^{-1}x - b)},
\,\,\,\,
\boldsymbol{\mathcal{U}}_{{}_{b,\Lambda}} \big(\iota^a(x)\big) = \iota^a(\Lambda^{-1}x - b).
\end{equation}
Finally, the Poincar\'e group acts naturally on the Grassmann-valued test functions $h^{pq} = \overline{\iota}^p\iota^q \cdot \phi$ 
through the bispinor transformation on the corresponding Schwartz test functions $\phi$. 
If 
\begin{equation}\label{FieldTransformationLaw}
U_{{}_{b,\Lambda}} \boldsymbol{\psi}^a(x) U_{{}_{b,\Lambda}}^{-1} = \sum\limits_{c} V^{ac} \boldsymbol{\psi}^c(\Lambda^{-1}x -b)
\end{equation}
is the transformation of the spinor field, then we put
\begin{equation}\label{TestTransformationLaw}
T_{{}_{b,\Lambda}} \phi^a(x) = \sum\limits_{c} \big[V^{-1}\big]^{ac} \phi^{c}(\Lambda(x+b))
\end{equation}
for the transformation law of the Schwartz test functions $\phi$ in $h^{10}, h^{01}$
and for the transformation law of $h= h^{01}$ and $\overline{h} = h^{10}$ we put
\[
T_{{}_{b,\Lambda}} h = \iota \cdot T_{{}_{b,\Lambda}} \phi, \,\,\,
T_{{}_{b,\Lambda}} \overline{h} = \overline{\iota} \cdot T_{{}_{b,\Lambda}} \phi,
\] 
and similarly for the general $h^{pq} = \overline{\iota}^p\iota^q \cdot \phi$ 
by application of the tensor product of this $T_{{}_{b,\Lambda}}$ to $\phi \in \mathcal{S}(\mathbb{R}^4;\mathbb{C}^4)^{\otimes \, (p+q)}_{{}_{\textrm{sym}}}$
in  $h^{pq} = \overline{\iota}^p\iota^q \cdot \phi$. Note here that in defining the transformation 
\[
T_{{}_{b,\Lambda}} h^{pq} = \overline{\iota}^p\iota^q \cdot T_{{}_{b,\Lambda}}^{\otimes \, (p+q)}\phi
\]
we apply the transformation only to the corresponding Schwartz test function $\phi$ with the generalized basis elements
$\overline{\iota(x)}, \iota(x)$ being fixed.

Note that for $g^{\otimes \, n} \otimes h^{pq} \in \mathscr{E}^{\otimes \, n}\otimes \mathscr{E}^{pq}$
\begin{equation}\label{DomainS_n,p,q(g^nh^pq)}
S_{n,p,q}(g^{\otimes \, n} \otimes h^{pq}) \in \mathscr{L}((\boldsymbol{E}),(\boldsymbol{E})^* ) \otimes \mathcal{E}^{pq*}
\cong \mathscr{L}((\boldsymbol{E}),(\boldsymbol{E})^* \otimes \mathcal{E}^{pq*})
\end{equation}
and the Poincar\'e covariance axiom (II) (compare Subsection \ref{MotivationForHida}) now reads
\[
\textrm{(II)}  \,\,\,\,\,\,\,\,\,\,\,\,\,\,\,\,\,\,\,\,   
S\big(T_{{}_{b,\Lambda}}g,\, T_{{}_{b,\Lambda}}\overline{h}, \, T_{{}_{b,\Lambda}}h \big)
= \big(U_{{}_{b,\Lambda}} \otimes \boldsymbol{\mathcal{U}}_{{}_{b,\Lambda}}\big) \, S(g,\overline{h},h) \,  U_{{}_{b,\Lambda}}^{-1}.
\]
Here, as before, the left $U$ is the continuous representation in $(\boldsymbol{E})^*$, dual to the continuous
Krein-isometric representation acting in the test Hida space $(\boldsymbol{E})$, and $\boldsymbol{\mathcal{U}}$
is the representation in $\oplus_{p,q} \mathcal{E}^{pq*}$, dual to the unitary representation $\boldsymbol{\mathcal{U}}$
acting in $\oplus_{p,q} \mathcal{E}^{pq}$, and defined as above. This axiom (II) can be expressed in terms
of 
\begin{gather*}
S_{n,p,q}(x_1, \ldots, x_n, a_1,y_1, \ldots, a_p,y_p, b_1,z_1, \ldots, b_q, z_q), 
\\
\,\,\,\,\, a_i,b_j, \in \{1, \dots, 4 \}, 
\,\,\, x_i,y_j,z_k \in \mathbb{R}^4,
\end{gather*}
analogously as in (II) above:
\begin{multline*}
\textrm{(II)}  \,\,\,\,\,\,\,\,\,\,\,\,\,\,\,\,\,\,\,\,      
 U_{a,\Lambda} S_{n,p,q}(x_1, \ldots, x_n, a_1,y_1, \ldots, a_p,y_p, b_1,z_1, \ldots, b_q, z_q) U_{a, \Lambda}^{+} 
\\
= \sum\limits_{a'_1, \ldots, b'_q}
\overline{V^{a_1 a'_1}} \ldots \overline{V^{a_p a'_p}} V^{b_1 b'_1} \ldots V^{b_q b'_q} \,\,\, \times 
\\
\times \,\,\,
S_{n,p,q}\big(\Lambda^{-1}x_1 + a, .., \Lambda^{-1}x_n +a, a'_1, \Lambda^{-1}y_1 + a, \ldots , a'_p, \Lambda^{-1}y_p +a,
b'_1, \Lambda^{-1}z_1 + a, \ldots , b'_q, \Lambda^{-1}z_q +a\big).
\end{multline*}
In this last form of (II) the operators $S_{n,p,q}(x_1,\ldots)$ are understood
as finite sums of integral kernel operators with 
$\mathcal{E}^{pq \, *}(M)$-valued kernels $\kappa_{\mathpzc{l}, \mathpzc{m}}$. Please, note here, that in (II) 
we have used $\Lambda = \Lambda(\alpha)$, 
which is a homomorphism $\alpha \rightarrow \Lambda(\alpha)$
of $SL(2,\mathbb{C})$ onto the proper orthochronous Lorentz group. In Section \ref{constr-of-VF} we have used antihomomorphism 
$\alpha \rightarrow \Lambda(\alpha)$, so that in the right-hand side of (II) we would have 
$\Lambda x_1 - a, \ldots$, if we were using  $\Lambda = \Lambda(\alpha)$ with the antihomomorphism 
$\alpha \rightarrow \Lambda(\alpha)$, as in Section \ref{constr-of-VF}.

In order to formulate the Krein isometricity axiom (III) for $S(g,\overline{h},h)$ with Grassmann-valued test functions $\overline{h},h$, 
the adjoint operation for $S_{n,p}(g^{\otimes \, n}\otimes h^{pq})$ is now accompanied by the Gupta-Bleuler operator $\eta$, with the left-hand side 
$\eta$ tensored now with the involution operator $\overline{(\cdot)}$ in the Grassmann algebra $\oplus_{p,q}\mathcal{E}^{pq*}$ with involution,
so that 
\[
\textrm{(III)}  \,\,\,\,\,\,\,\,\,\,\,\,\,\,\,\,\,\,\,\,     
\big(\eta \otimes \overline{\,\, \cdot \,\,} \big) S_{n,p,q}(g^{\otimes \, n}\otimes h^{pq})^+ \eta =
\eta S_{n,p,q}(g^{\otimes \, n}\otimes \overline{h^{pq}})^+ \eta = \overline{S}_{n,q,p}(g^{\otimes \, n}\otimes h^{qp})
\]
or
\[
\textrm{(III)}  \,\,\,\,\,\,\,\,\,\,\,\,\,\,\,\,\,\,\,\,     
\big(\eta \otimes \overline{\,\, \cdot \,\,} \big) S(g,\overline{h},h)^+ \eta = S^{-1}(g,\overline{h},h)
\]
Here the adjoint operation $(\cdot)^+$ in action on (\ref{DomainS_n,p,q(g^nh^pq)}) 
is understood as the adjoint operation acting only in the first tensor product factor
\[
\mathscr{L}((\boldsymbol{E}),(\boldsymbol{E})^* ),
\]
which results in changing in each summand $\Xi(\kappa_{\mathpzc{l},\mathpzc{m}})$ of 
\[
S_{n,p,q}
= \sum\limits_{\mathpzc{l},\mathpzc{m}}\Xi\big(\kappa_{\mathpzc{l},\mathpzc{m}} \big),
\]
the kernel  $\kappa_{\mathpzc{l},\mathpzc{m}}$, regarded as $\mathcal{E}^{pq \, *}(M)$-valued kernel,
with the kernel
\[
\kappa^{+}_{\mathpzc{l},\mathpzc{m}}(\ldots) = \overline{\kappa_{\mathpzc{m},\mathpzc{l}}(\ldots)},
\]
with the bar understood here as the complex conjugation, so that it coincides with ordinary adjoint
of an integral kernel operator $\Xi(\kappa_{\mathpzc{l},\mathpzc{m}})$  in the Fock space,
with $\kappa_{\mathpzc{l},\mathpzc{m}}$ regarded as $\mathcal{E}^{pq \, *}(M)$-valued kernel, and which results
in complex conjugation of the kernel and exchange of the creation and annihilation operators. 
Dots in the last formula denote $\mathpzc{l} + \mathpzc{m}$ spin-momentum and $p+q$ space-time and discrete spinor component variables
of the kernel $\kappa_{\mathpzc{l},\mathpzc{m}}$ regarded as a distribution.

The Krein-isometricity axiom (III), expressed in terms of 
\[
S_{n,p,q}(x_1, \ldots, x_n, a_1,y_1, \ldots, a_p,y_p, b_1,z_1, \ldots, b_q, z_q), \,\,\, a_i,b_j, \in \{1, \dots, 4 \}, 
\,\,\, x_i,y_j,z_k \in \mathbb{R}^4,
\]
reads
\begin{multline*}
\textrm{(III)}  \,\,\,\,\,\,\,\,\,\,\,\,\,\,\,\,\,\,\,\,   
\eta S_{n,p,q}(x_1, \ldots, x_n, a_1,y_1, \ldots, a_p,y_p, b_1,z_1, \ldots, b_q, z_q)^+\eta
\\
= \overline{S}_{n,p,q}(x_1, \ldots, x_n, a_1,y_1, \ldots, a_p,y_p, b_1,z_1, \ldots, b_q, z_q).
\end{multline*}
In this last form of (III) the operators $S_{n,p,q}(x_1,\ldots)$ are understood
as finite sums of integral kernel operators with 
$\mathcal{E}^{pq \, *}(M)$-valued kernels $\kappa_{\mathpzc{l}, \mathpzc{m}}$.

In case the Wick polynomial field $\mathbb{A}$ is not a scalar, but is of even degree in Fermi fields (thus also in this case
the additional switching-off function $h$
has more than one component) the additional test  switching-off function $h$ is regarded as ordinary Schwartz
$\mathbb{C}^d$-valued function, where $d$ is the number of components of the field $\mathbb{A}$, transforming analogously
as in (\ref{TestTransformationLaw}) with the matrix $V$ coming from the trasformation law of the field $\mathbb{A}$ -- the analogue
of (\ref{FieldTransformationLaw}),
and with the product
\[
h^{\mu_1}(y_{1}) h^{\mu_2}(y_{2}) \ldots h^{\mu_p}(y_{p})
\]
now understood as the ordinary pointwise product, symmetric in the variavles $(\mu_1,y_{1}, \ldots, \mu_p, y_{p})$
and with
\begin{multline*}
\big(\underbrace{g\otimes \ldots \otimes g}_{\textrm{$k$ factors}} \otimes  
h^p\big)(x_1, \ldots, x_k, \mu_1, y_1, \ldots, \mu_p, y_p)
\\
= g(x_1) \ldots g(x_k) h^{\mu_1}(y_{1}) h^{\mu_2}(y_{2}) \ldots h^{\mu_p}(y_{p}),
\end{multline*}
as is the case e.g. for the computation of the higher order contributions to the interactng e.m.
potential field, where in this case $\mathbb{A}_\mu(x) = A_\mu(x)$
and (summation with respect to the Lorentz index $\mu$ is understood)
\[
S_{0,1}(h) = \int h^\mu(x) A_\mu(x) \, d^4 x,
\]
with $h$ which can be interpreted physically as the classical electric current density field, compare \cite{Bogoliubov-Shirkov}
and \cite{Schwinger}. In this case the product  
\begin{multline*}
\big(g\otimes \ldots \otimes g \otimes  
h^p\big)(x_1, \ldots, x_k, \mu_1, y_1, \ldots, \mu_p, y_p)
\\
=
\big(g\otimes \ldots \otimes g \otimes  
h \otimes \ldots \otimes h\big)(x_1, \ldots, x_k, \mu_1, y_1, \ldots, \mu_p, y_p)
\\
= g(x_1) \ldots g(x_k) h^{\mu_1}(y_{1}) h^{\mu_2}(y_{2}) \ldots h^{\mu_p}(y_{p}),
\end{multline*}
is symmetric
\begin{gather*}
g(x_1) \ldots g(x_k) h^{\mu_1}(y_{1}) \ldots h^{\mu_i}(y_{i}) \ldots h^{\mu_j}(y_{j}) \ldots h^{\mu_p}(y_{p})
\\
=
g(x_1) \ldots g(x_k) h^{\mu_1}(y_{1}) \ldots h^{\mu_j}(y_{j}) \ldots h^{\mu_i}(y_{i})   \ldots h^{\mu_p}(y_{p}),
\\
g(x_1) \ldots g(x_i) \ldots g(x_j) \ldots g(x_k) h^{\mu_1}(y_{1}) \ldots h^{\mu_p}(y_{p})
\\
 = g(x_1) \ldots g(x_j) \ldots g(x_i) \ldots g(x_k) h^{\mu_1}(y_{1}) \ldots h^{\mu_p}(y_{p}),
\end{gather*}
and $S_{k,p}$ have the corresponding symmetry property. In this case all axioms (I)-(V) are fulfilled
by $S(g,h)$, but now the factors $(-1)^{\textrm{sgn}(Y_2,Y_1)}$, $(-1)^{\textrm{sgn}(X,Y,x_n)}$,
$(-1)^{\textrm{sgn}(Y,x_n,X)}$ in (I) and in $A'_{(n)}$,$A_{(n)}$, $R'_{(n)}$, $R_{(n)}$ are, of course, absent
and
\[
S_{n,p}\big(g^{\otimes \, n} \otimes  
h^p\big) \in \mathscr{L}\big((\boldsymbol{E}), (\boldsymbol{E})^*\big)
\] 
(here in case when we have massless fields in the interaction Lagrangian), 
with the proof that the product operation
\[
S_{n_1,p_1}\big(g^{\otimes \, n_1} \otimes  
h^{p_1}\big)S_{n_2,p_2}\big(g^{\otimes \, n_2} \otimes  
h^{p_2}\big) \in \mathscr{L}\big((\boldsymbol{E}), (\boldsymbol{E})^*\big)
\]
is well-defined as the limit operation, first by replacing the exponents of the kernels of the massless
fields with exponents of massive kernels, and then by passing to the zero mass limit in the exponent, 
exactly the same as for the product 
\[
S_{n_1}(g^{\otimes \, n_1})S_{n_2}(g^{\otimes \, n_2})  \in \mathscr{L}\big((\boldsymbol{E}), (\boldsymbol{E})^*\big),
\]
in case we have massless fields in the interaction Lagrangian. 

We may summarize the results in the following

\begin{twr}
The Bogoliubov causal axioms (I)-(V) for the scattering generalized operator $S$ are consistent and, moreover, 
the higher order constributions $S_n$ to $S$ can be computed by induction with respect to the order $n$.
\label{ConsistencyComputability}
\end{twr}
\qedsymbol \,
Suppose we have given $S_k$ for $k=1, \ldots, n-1$, which respect (I)-(V). 
After \cite{Epstein-Glaser} we apply the following inductive step construction of $S_n$. 
By assumption the following distributions are known
\begin{equation}\label{A'_(n),R'_(n)}
\begin{split}
A'_{(n)}(Z, j_n,x_n) = \sum\limits_{X\sqcup Y=Z, X\neq \emptyset} (-1)^{{}^{s(X,Y,j_n,x_n)}} \overline{S}(X)S(Y,j_nx_n), \\
R'_{(n)}(Z, j_n,x_n) = \sum\limits_{X\sqcup Y=Z, X\neq \emptyset} (-1)^{{}^{s(Y,j_n,x_n,X)}} S(Y,j_nx_n)\overline{S}(X),
\end{split}
\end{equation}
where the sums run over all divisions $X\sqcup Y=Z$ of the set $Z$ of variables $\{j_1,x_1, \ldots, j_{n-1},x_{n-1} \}$
into two disjoint subsets $X$ and $Y$:
\[
\{j_1,x_1, \ldots, j_{n-1},x_{n-1} \} = X \sqcup Y,
\,\,\, \textrm{with} \,\,\,
X \neq \emptyset,
\]
with each pair $j_i,x_i$ treated as a single element in $Z$.
Here $s(X,Y,j_n,x_n)$ is equal to the sign of permutation of the Grassmann variables in the permutation from the order 
$Z,j_n,x_n$ of variables to the order $X,Y,j_n,x_n$. Then we construct
\begin{multline*}
A_{(n)}(Z, j_n,x_n) = \sum \limits_{X\sqcup Y=Z} (-1)^{{}^{s(X,Y,j_n,x_n)}}  \overline{S}(X)S(Y,j_n,x_n)
\\
= \sum \limits_{X\sqcup Y=Z, X\neq \emptyset} (-1)^{{}^{s(X,Y,j_n,x_n)}} \overline{S}(X)S(Y,j_n,x_n) + S(j_1,x_1, \ldots, j_n,x_n), 
\end{multline*}
\begin{multline*}
R_{(n)}(Z, j_n,x_n) = \sum \limits_{X\sqcup Y=Z} (-1)^{{}^{s(Y,j_n,x_n,X)}} S(Y,j_n,x_n)\overline{S}(X)
\\
= \sum \limits_{X\sqcup Y=Z, X\neq \emptyset} (-1)^{{}^{s(Y,j_n,x_n,X)}} S(Y,j_n,x_n)\overline{S}(X) + S(j_1,x_1, \ldots, j_n,x_n),
\end{multline*}
where now, the first summation is extended over all divisions of the set $Z$
into two disjoint subsets $X$ and $Y$, which include the empty set $X= \emptyset$. Note that
\[
D_{(n)} = R'_{(n)} - A'_{(n)} = R_{(n)} - A_{(n)}.
\]
In order to finish presentation of the essential point of the inductive step, 
let us introduce after \cite{Epstein-Glaser} higher dimensional generalization of the backward and forward cones:
\[
\Gamma_{\pm}^{(n)}(y) = \big\{X \in \mathcal{M}^n: x_j - y \in \overline{V_{\pm}} \big\}, \,\,\,\,\,
X = \{x_1, \ldots, x_n \}.
\]
Then, from the axioms (I)-(V), which are assumed to be fulfilled by all $S_k$, $k\leq n-1$, 
it is shown in \cite{Epstein-Glaser} (compare also \cite{Scharf}) that for $n>2$
\[
\begin{split}
\textrm{supp} \, R_{(n)}(j_1,x_1, \ldots, j_{n-1},x_{n-1}, j_nx_n) \subseteq \Gamma_{+}^{(n-1)}(j_n,x_n),
\\
\textrm{supp} \, A_{(n)}(j_1,x_1, \ldots, j_{n-1},x_{n-1}, j_{n},x_n) \subseteq \Gamma_{-}^{(n-1)}(j_n,x_n),
\\
\textrm{supp} \, D_{(n)}(j_1,x_1, \ldots, j_{n-1},x_{n-1}, j_{n},x_n) \subseteq \Gamma_{+}^{(n-1)}(j_n,x_n)
\sqcup \Gamma_{-}^{(n-1)}(j_n,x_n),
\\
\end{split}
\]
and moreover that each $D_{(n)}$ can be 
splitted into sum of operator distributions each having the support, respectively, in
$\Gamma_{+}^{(n-1)}(x_n)$ or in $\Gamma_{-}^{(n-1)}(x_n)$ and that this splitting can be made explicitly
and independently of the axioms (I) -- (V). The splitting is determined up a number of free constants depending on the 
singularity order of the kernels of $D_{(n)}$.  
The essential point is that $R_{(n)}$ and
$A_{(n)}$ can be separately computed as the spitting of $D_{(n)}$ into the advanced $A_{(n)}$ and retarded
$R_{(n)}$ parts, so that
\[
\begin{split}
S_n(j_1,x_1, \ldots, j_{n}, x_n) = A_{(n)}(j_1,x_1, \ldots, j_{n}, x_n) - A'_{(n)}(j_1,x_1, \ldots, j_{n}, x_n)
\\
\,\,\,
\textrm{or equivalently}
\,\,\,
\\
S_n(j_1,x_1, \ldots, j_{n}, x_n) = R_{(n)}(j_1,x_1, \ldots, j_{n}, x_n) - R'_{(n)}(j_1,x_1, \ldots, j_{n}, x_n)
\end{split}
\]
and the inductive step from $n-1$ to $n$ can be computed without encountering any infinities and without any need for renormalization. 
Here we emphasize that the proof \cite{Epstein-Glaser} of the support properties remain completely 
unchanged for $S_n$ understood as finite sums of integral kernel
operators with vector valued kernels. By assumption $S_1$ is equal to a Wick polynomial in free fields, and thus belongs
to the class of Theorem \ref{ClassWithProductTheorem}. By these theorems 
$D_{(2)}(j_1,x_1,j_2,x_2) = S(j_1,x_1)\overline{S}(j_2,x_2) - (-1)^{s(j_2,x_2,j_1,x_1)}\overline{S}(j_2,x_2)S(j_1,x_1)$
makes sense, with the products $S(j_1,x_1)\overline{S}(j_2,x_2)$, $\overline{S}(j_2,x_2)S(j_1,x_1)$ computed 
by application of the Wick theorem \ref{WickThmForMasslessFields}, 
understood in the limit sense when massless fields are in $S_1(j,x) = S(j,x)$. Because the commutation functions
are translationally invariant and have causal support, 
it then follows from the Wick theorem \ref{WickThmForMasslessFields} that $D_{(2)}$ has causal support (compare \cite{Bogoliubov_Shirkov}, \S 17.3) 
and $S_2 = R_{(2)} - R'_{(2)}$ can be computed using the splitting of $D_{(2)}$ into the retarded and advanced part. 
Because (causal combinations of) the contractions
(\ref{(x)q}) (products of pairings) are translationally invariant and have finite singularity order, 
they can be splitted into retarded and advanced part (up to a finite linear combination
of derivatives of the Dirac delta functional). Tensor products of distributions of finite order singularity, 
is again finite order singularity distribution, and the singularity is preserved in the splitting -- axiom (V). From this
it follows that  $S_2 = R_{(2)} - R'_{(2)}$ can be computed (up to a finite set of constants), and moreover, in each step
we apply Wick theorem to the operators of the class of Theorem \ref{ClassWithProductTheorem},
and eventually compute $ret$ or $av$ parts of their causal combinations,  which again leads us into the class 
of Theorem \ref{ClassWithProductTheorem}. Therefore the products in the axiom
(I), as well as in the operators $A'_{(n)},R'_{(n)}$ are well-defined 
by Theorem \ref{ClassWithProductTheorem}. 
\qed

From theorem \ref{ConsistencyComputability} it follows
\begin{cor}
The  $n$-order contributions $W_{{}_{j \,\, \textrm{int}}}^{(n)}$ to interacting 
fields $W_{{}_{j \,\, \textrm{int}}}$ can be computed as finite sums
of integral kernel operators with vector valued kernels. 
\end{cor}
\qedsymbol \,
Indeed (using the notation of the last proof) it is easily seen that, with the set of variables $Z=\{0,x_1,0,x_2, \ldots, 0,x_n\}$,
the distributional kernels  of  $W_{{}_{j \,\, \textrm{int}}}^{(n)}$ are equal
\[
W_{{}_{j \,\, \textrm{int}}}^{(n)}(x_1, \ldots, x_n,j,x) = \textstyle{\frac{1}{i}} A_{(n+1)}(Z,j,x) 
= \textstyle{\frac{1}{i}} \textrm{av} \, D_{(n+1)}(Z,j,x).
\qed 
\]

We end with several remarks. First, note, please, that axiom (V) is used only to fix the splitting, which makes it
unique up to the number of constants depending on the singularity order. Without this axiom, the splitting would not be fixed even 
up to this freedom and even for distributions of finite order, which we have in QFT.
Next, we note that the singularity order of the contractions (\ref{(x)q}), 
can immediately and easily be computed from the formula 
(\ref{masslesskappa01.masslesskappa10...q'-contraction...masslesskappa10.masslesskappa10}) with 
$|\boldsymbol{p}_i|$ replaced, eventually, by the respective $p_0(\boldsymbol{p}_i)$. 
Further, also the explicit formula for the Fourier transform of the causal combinations of (\ref{(x)q}) can be easily 
computed, by noting the fact that the causal combinations of 
(\ref{masslesskappa01.masslesskappa10...q'-contraction...masslesskappa10.masslesskappa10})
(regarded as acting on space-time test space) is the limit
of the integrals in spin-momenta of the respective ordinary product of the free field kernels, with the range
of integration going to infinity (compare \cite{Scharf}), with 
simple dispersion formula for the retarded part of causal combinations of (\ref{(x)q}) in terms of its Fourier transform. 
Thus, the $\dot{\otimes}$ products of free field kernels, their contractions, and $ret$, $av$ parts of their
causal combinations can be practically computed with a simple dispersion formula, and inductively for
causal combinations of scalar contractions of higher order contributions.

Theory is renormalizable, e.g. QED, if and only if the singularity order $\omega$ of $D_{n}$, equal to the singularity order of $S_n$, 
is bounded, with a bound independent of the order $n$, which, in case of contributions to $D_{n}$ of the form proportional to 
${:}\partial^{\alpha_1}\mathbb{A}^{{}^{1}} \ldots \partial^{\alpha_k}\mathbb{A}^{{}^{k}}{:}$ (\emph{i.e.} with ``external lines''
$\partial^{\alpha_1}\mathbb{A}^{{}^{1}}, \ldots, \partial^{\alpha_k}\mathbb{A}^{{}^{k}}$), the bound is equal to $4$ minus the number
of external lines, counted with the weight $w(i)$ depending on the spin of the field $\mathbb{A}^{{}^{i}}$
in the ``external line'' $\partial^{\alpha_i}\mathbb{A}^{{}^{i}}$, and minus the number of derivatives in external lines.

The presented theory can, in principle, be generalized to include the case of interaction Lagrangian 
\[
\mathcal{L}(x) = \sum\limits_{j=0}^{k} g_{{}_{j}}(x) \mathcal{L}_{{}_{j}}(x) = g_{{}_{0}}(x) \mathcal{L}_{{}_{0}}(x) +
\sum\limits_{j=1}^{k} g_{{}_{j}}(x) W_{{}_{j}}(x), \,\,\,\, g_{{}_{j}} \in \mathcal{S}(\mathbb{R}^4; \mathbb{R})
\]
with the Wick products $W_{{}_{j}}$ replaced by infinite Fock expansions \cite{obataJFA} into integral kernel
operators of the class (\ref{ProductClassGenOp}), involving finite number of different kinds of free fields. Analysis of renormalizability
would be much more difficult in case $\mathcal{L}_{{}_{0}}$ was equal to the said infinite Fock expansion. 
So far, we have rather had no serious physical indications for developing theory in this direction.

Thus, our axioms for $S(g)$ are the same as in the original Bogoliubov's formulation. The only modification
lies in the usage of precise mathematical realization of the creation-annihilation operators as the Hida operators.
The fact that in case we are using Grassmann valued swiching off test function $h$, the scattering operator $S(g,\overline{h},h)$
necessary takes values no simply in the Hida space or its dual, but in the Grassmann-valued Hida test functions or in generalized 
Grassmann-valued Hida test functions, is not the modification pertinent to the usage of Hida operators. This fact follows by the very usage
of the Grassmann-valued test functions, and goes back to Schwinger \cite{Schwinger}, and is undertaken in \cite{Bogoliubov_Shirkov}.
Thus usage of the Grassmann-valued Hilbert space vectors or generalized vectors comes from the very usage of the Grassmann-valued test functions. 
 
Also,  the axiom (III) is the same as in Bogoliubov's
perturbative causal theory.

The usage of the Hida operators (specification of the distribution spaces actually used)
has nothing to do with the formal unitarity or with the formal  Krein-isometricity of the
scattering distribution-operator.

Bogoliubov-Shirkov book \cite{Bogoliubov_Shirkov} also uses formal Krein-isometric (and not formal unitary)
scattering operator distribution $S(g)$ in case of QED (and this is the case in general
for gauge theory in which unitary gauge is non-local, as in BRST more general case).
In \cite{Bogoliubov_Shirkov} the Gupta-Bleuler gauge is used in the causal construction of $S(g)$,
which is non-unitary but Krein-isometric.
This choice of the gauge is not a matter of taste, and follows from
the fact that in this gauge we are left with local interaction and local transformation formula.
The unitary gauges (like the Coulomb gauge, or the axial gauge)
are non-local and are in clash with the local implementation of the switching
on/off multiplication by $g$ mechanism, fundamental for the implementation of the causality axiom.
If one insists on the unitarity of the gauge, then Bogoliubov's causality axiom, which allow avoiding
UV divergence, cannot be applied and one is left with UV-infinities within the renormalization
approach.

In the Bogoliubov-Shirkov book the Gupta-Bleuler operator
$\eta$ does not appear explicitly in the axioms, but instead $\eta$ is incorporated into the modified
Krein-conjugation (which is no longer Hermitian) of the annhilation-creation operators
of the gauge field (e.m. potential) and which is modified and uses $\eta$.  In the above and further computations
we are also using this convention, which allows simplification of notation, and we discard the repeated
$\eta$ in the integral kernel operators.  However, at the level of axioms we have decided to
write $\eta$ explicitly.

{\bf REFERENCES}. \,
The content of Section \ref{WickForProduct} has been published in \cite{wawrzyckiInfinite} and \cite{wawrzyckiModPhys}.

\section{Wick's theorem for the natural chronological product}\label{WickForChronological}

The first motivation of this Subsection is to work out a method for the computation of the $n$-th order contribution $S_n$ 
\[
S_n(x_1, \ldots, x_n) = i^n T\big[ \mathcal{L}(x_1) \ldots  \mathcal{L}(x_n) \big], 
\]
to the
scattering operator $S$, or to the chronological product
which would be subjectable to algorithmic calculations and which would be effective.

The second motivation of this Subsection is to investigate more precisely the source of difficulties in the conventional
construction of the chronological product which uses the operation of pointwise multiplication by the step theta function
\[
\theta: \mathbb{R}^4 \ni x \mapsto \theta(x) = \theta(x_0),
\]
where the last $\theta$ is the ordinary theta function on the reals, for the construction of the retarded and advanced parts
of the causal scalar distributions.
We will show in this Subsection that the conventional construction of the retarded and advanced parts of the causal distributions
in the chronological product constructed in the previous Subsection,
could be defined through the multiplication by the step theta function if and only if the orbits $\mathscr{O}_k$
of all, or all but one, of the free fields $\mathbb{A}_k$ in each of the Wick monomials
\[
\boldsymbol{{:}} \mathbb{A}_1(x) \ldots \mathbb{A}_N(x) \boldsymbol{{:}}
\]
in the Lagrange interaction density
\[
\mathcal{L}(x) = \ldots + \boldsymbol{{:}} \mathbb{A}_1(x) \ldots \mathbb{A}_N(x) \boldsymbol{{:}} + \ldots,
\]
where compact, \emph{i.e.}
\begin{equation}\label{CompactOrbits}
\mathscr{O}_k \,\, \textrm{compact}
\end{equation}
for all $k \in \{1, \ldots, N \}$ except at most one $k$.

Recall that the orbit $\mathscr{O}_k$
of a free field $\mathbb{A}_k$ is equal to the support of the Fourier transform of the fundamental solutions
composing the single particle Hilbert space of the field $\mathbb{A}_k$.
Of course in case of free fields on the Minkowski space-time which respect linear Poincar\'e invariant hyperbolic
equations of motion, the orbit $\mathscr{O}_k$ is equal to an actual orbit of a single point (equal $(m,0,0,0)$ or $(1,0,0,1)$
in case od massive or massless positive energy field) in the momentum space
under the action of the Lorentz group. Thus, in this particular Poincar\'e invariant hyperbolic and linear fields on the Minkowski space-time
the orbits must be non-compact (one sheet of the two-sheet hyperbola, or one sheet of the cone), and the condition
(\ref{CompactOrbits}) is not interesting for such fields on the flat Minkowski space-time, and presents nothing unexpected
for physicists (ultraviolet divergences disappear if we cut out all integrations in momentum space at some finite
value of the momentum).

As we will see the condition (\ref{CompactOrbits}) arises because the Fourier transform
$\widetilde{\theta}$ of the theta function $\theta$ is not any element of the Schwartz convolutor algebra $\mathcal{O}'_C(\mathbb{R}^4)$.
Put otherwise,  the condition (\ref{CompactOrbits}) arises because  the operation of convolution 
\begin{equation}\label{tildephi->tildetheta*tildephi}
\widetilde{\phi} \mapsto \widetilde{\theta} \ast \widetilde{\phi}
\end{equation}
is not a map\footnote{Continuity of (\ref{tildephi->tildetheta*tildephi}) is not essential here.} which transforms the Schwartz 
nuclear test space of rapidly decreasing functions on the additive Lie group $\mathbb{R}^4$ into itself
(but unfortunately, as we know, it leads us out of the test space, compare Appendix \ref{convolutorsO'_C}). 
Equivalently, using the Fourier exchange theorem on $\mathbb{R}^4$,
the condition (\ref{CompactOrbits}) arises because
 the operation of pointwise multiplication
\[
\phi \mapsto \theta . \phi
\]
by the theta function is not a map which transforms the Schwartz test space of rapidly decreasing functions into itself
(but, as we know, this map leads us out of the Schwartz nuclear test space, again continuity is not essential here,
but also the continuity is of course lost, compare Appendix \ref{convolutorsO'_C}). 

Third motivation of this Subsection is the following: having given the sufficiency condition
(\ref{CompactOrbits})
we can see that for space-times with compact Cauchy surfaces the ''natural'' construction of the chronological product
using the retarded and advanced parts constructed through multiplication by the (analogue) of the step theta function
has mathematically well-defined sense without any need for normalization, in case there is at most one massless free field
in the interaction Lagrange density $\mathcal{L}$ (as e.g. in QED).
In particular the Einstein Universe is not only static
(no difficulties with frequency division) and not only has the Cauchy surfaces compact, but all fundamental solutions composing the single particle
Hilbert spaces of free fields on this space-time have common time period, so effectively the free fields on the Einstein Universe (which is a Lie group)
live on its compactification (which is a compact Lie group), and all massive fields on this space-time have compact (in fact finite,
because the momentum space becomes discrete in this case)
orbits, compare \cite{SegalZhouQED}, \cite{PaneitzSegalI} -\cite{PaneitzSegalIII}, or Subsection \ref{WhiteNoiseFreeFieldsonEU}.
On the compact Lie group the convolution
map (\ref{tildephi->tildetheta*tildephi}), with the analogue (periodic) step theta function $\theta$,
also leads us out of the test space (Fourier exchange thm.). But because the massive fields on the Einstein Universe have
finite orbits (supports of the Fourier transforms of the fundamental solutions composing single particle Hilbert spaces of massive
free fields), then the natural construction of the chronological product can indeed be
constructed as a well-defined mathematical object on the Einstein Universe, provided each Wick monomial in $\mathcal{L}$
contains at most one massless free field factor.
This we will prove in details in Section \ref{EUandG}.
This opportunity has important implications on such space-times, because it
gives a procedure for the computation of the higher order contribution to the scattering
operator, which avoids using the Epstein-Glaser splitting procedure,
and which allows to simplify computation of higher order terms and make them subjectable to algorithmic calculations.

Because the orbits of the free fields on the Minkowski space-time, which respect Poincar\'e invariant equations,
are necessary non-compact, the splitting of Epstein-Glaser seems to be unavoidable on the Minkowski space-time.
Therefore, on the flat Minkowski space-time any effective and mathematically well-defined
method of calculation has to be based on the Epstein-Glaser splitting. The natural chronological product based on the step theta function
is not sufficient here, although also on the Minkowski space-time it will serve for us as the starting point.

The method which we apply in this Subsection is based on the intuitive
definition of the chronological product, used e.g. in \cite{Bogoliubov_Shirkov}, which we convert into a mathematically
rigorous construction. The idea is very simple. We replace the step theta function $\theta$ in the heuristic definition
of the chronological product by a one-parameter family $\theta_\varepsilon$ of smooth functions. Then we apply the
Wick Theorem of Subsection \ref{WickForProduct} with the kernels depending on $\theta_\varepsilon$,
and with the kernels in the Wick decomposition containing various contractions $\otimes_q$. By the Epstein-Glaser
splitting we know that all terms with contractions $\otimes_q$ with $q>1$ can be convergent, when $\theta_\varepsilon
\rightarrow \theta$, at most only on a closed
subspace of the space-time test space. Indeed, we show that families
$\theta_\varepsilon$ exist for which such convergence hold and with the limit independent of the choice of the family.
This allows to determine the limit kernels of the chronological product up to a finite number of constants
in each order contribution to the chronological product. Further extension of such constructed kernels 
all over the whole space-time test space
 cannot be determined uniquely by the causality axioms (I)-(V) for the scattering operator, Subsection \ref{WickForProduct}.
The arbitrary constants, which arise in such extension, can then be determined by further and natural
condition, independent of the causality axioms (I)-(V) for the scattering operator itself, namely, the condition
of existence of the adiabatic limit for the higher order contributions to interacting fields, well-defined 
as the integral kernel operators (compare Subsecion \ref{OperationsOnXi}).

As we did for the construction of the ''product'' of Wick ordered factors including massless fields in Subsection \ref{WickForProduct},
we give here a similar construction of a '' natural'' chronological product, which is essentially based on the step theta
function $\theta(x) \overset{\textrm{df}}{=} \theta(x_0)=\theta(t)$, where the last $\theta$ is the ordinary step $\theta$-function on $\mathbb{R}$:
\[
\theta(t)=
\left\{ \begin{array}{ll}
1, & 0 \leq t, \\
0, & t <0
\end{array} \right.,
\]
and is immediately motivated by the heuristic definition used in \cite{Bogoliubov_Shirkov}:
\begin{multline}\label{thetaS_n}
S_n(x_1, \ldots, x_n) = i^n \, T\big(\mathcal{L}(x_1) \ldots \mathcal{L}(x_n) \big) \\ = i^{n}
\sum\limits_{\pi} \theta(t_{\pi(1)} -t_{\pi(2)}) \theta(t_{\pi(2)} -t_{\pi(3)}) \ldots \theta(t_{\pi(n-1)} -t_{\pi(n)}) \,
\mathcal{L}(x_{\pi(1)}) \ldots \mathcal{L}(x_{\pi(n)}), \\
\,\,\,\,\,\,\,\,\,\,\,\,\,\,\,\,\,\,\,\,\,\,\,\,\,\,\,\,\,\,\,\,
x_{\pi(k)} = (t_{\pi(k)}, \boldsymbol{\x}_{\pi(k)}), \,\,\, \pi \in \textrm{Permutations of} \,\{1, \ldots, n\}.
\end{multline}
We give to the expression (\ref{thetaS_n}) a rigorous meaning in case the Wick monomial, or each of the Wick monomials
in the interaction Lagrange density Wick polynomial $\mathcal{L}$ contains massless free field factors or not. Let for example
\[
\mathcal{L}(x) = \boldsymbol{{:}} \boldsymbol{\psi}(x)^{+}\gamma_{0} \gamma^\mu \boldsymbol{\psi}(x) A_\mu(x) \boldsymbol{{:}},
\]
as in spinor QED, but our analysis is general, and we can replace $\mathcal{L}$ with any Wick polynomial of free fields
with each of the Wick monomials possibly containing massless free field factors.

We apply the causal method of Bogoliubov-Epstein-Glaser of
Subsection \ref{MotivationForHida}. In particular for the second order contribution the formula
reduces to
\[
S_2(x,y) = R_{(2)}(x,y) - R'_{(2)}(x,y),
\] 
where $R_{(2)}(x,y)$ is the retarded part of the causally supported
distribution $D_{(2)}(x,y)$. In particular all scalar factors in terms of the Wick decomposition of 
\[
D_{(2)}(x,y)= S(y)\overline{S}(x) - \overline{S}(x)S(y) = -i^2\big[\mathcal{L}(y)\mathcal{L}(x) - \mathcal{L}(x)\mathcal{L}(y)\big]
\]
(those with pairings in the Wick decomposition of
$D_{(2)}(x,y)$) are indeed causal distributions, \emph{i.e.} distributions of one space-time
variable $x-y$, which is supported within the light cone. Therefore the method of splitting of causal distributions
into the retarded and advanced part (due to Epstein-Glaser \cite{Epstein-Glaser} or \cite{Scharf})
can indeed be used for the computation of $S_2$. 

To the Bogoliubov-Epstein-Glaser method we add a very small improvement.
We observe that the general splitting method, to which Epstein and Glaser or Scharf refer in \cite{Epstein-Glaser} or \cite{Scharf},
works not only for the strictly causally supported distributions. In fact it is the local property around $x-y=0$
of a distribution $C(x-y)$ which is to be splitted, which is important for the splitting, and is governed
by a single number $\omega$ pertinent to the distribution $C$, and called \emph{singular order $\omega$ of} $C$ at zero
(or at infinity for the Fourier transformed $\widetilde{C}$). 
Namely, for the plane wave kernels 
\begin{equation}\label{FFkernels}
\kappa^{(1)}_{0,1}, \kappa^{(1)}_{1,0}, \ldots, \kappa^{(q)}_{0,1}, \kappa^{(q)}_{1,0},
\end{equation}
of the free fields we show in this Subsection that all contraction distributions (products of pairings, compare Subsection \ref{WickForProduct}) 
\begin{equation}\label{ProductsOfPairings}
\Big(\kappa^{(1)}_{0,1} \dot{\otimes} \ldots \dot{\otimes} \kappa^{(q)}_{0,1}\Big)
\otimes_q
\Big(\kappa^{(1)}_{0,1} \dot{\otimes} \ldots \dot{\otimes} \kappa^{(q)}_{0,1}\Big)(x,y) = C(x-y)
\end{equation}
have finite order $\omega$ of singularity at zero and can be splitted as in the cited works. In general 
 (\ref{ProductsOfPairings}) is not causally suported. 
This splitting of (\ref{ProductsOfPairings}) is possible because the following differences or sums (depending on the parity of the number $q$) 
of such $\otimes_q$-contractions (\ref{ProductsOfPairings}) (being equal to products of pairings, compare Subsection \ref{WickForProduct})
\begin{multline}\label{CausalDifferencesOfProductsOfPairings}
\Big(\kappa^{(1)}_{0,1} \dot{\otimes} \ldots \dot{\otimes} \kappa^{(q)}_{0,1}\Big)
\otimes_q
\Big(\kappa^{(1)}_{1,0} \dot{\otimes} \ldots \dot{\otimes} \kappa^{(q)}_{1,0}\Big)(x,y)
\\
-(-1)^q
\Big(\kappa^{(1)}_{1,0} \dot{\otimes} \ldots \dot{\otimes} \kappa^{(q)}_{1,0}\Big)
\otimes_q
\Big(\kappa^{(1)}_{0,1} \dot{\otimes} \ldots \dot{\otimes} \kappa^{(q)}_{0,1}\Big)(x,y)
\\
= C(x-y) -(-1)^q C(y-x)
\end{multline}
have always causal support, to which the Epstein-Glaser splitting can be applied. The condition
for the convergence of the retarded part of the difference or sum (\ref{CausalDifferencesOfProductsOfPairings}) 
of the contractions (\ref{ProductsOfPairings}) implies convergence for the retarded part of  (\ref{ProductsOfPairings}).
The only diffrence is that now the retarded part of each $C(x-y)$ and $C(y-x)$ in (\ref{CausalDifferencesOfProductsOfPairings})
taken separately, is not Lorentz invariant with the time-like unit versor $v$ of the reference frame in the theta function
$\theta(v\cdot x)$ giving the support $\textrm{supp} \theta$ of the retarded part. Only in the sum/diffrene of the retarded 
parts of $C(x-y)$ and $C(y-x)$ in (\ref{CausalDifferencesOfProductsOfPairings}) the dependence on $v$ drops out, with the retarded part
of the whole distribution (\ref{CausalDifferencesOfProductsOfPairings}) being Lorentz invariant.

Using this fact, to the second order contribution formula
\begin{equation}\label{S2(x1,x2)}
S_2(x_1,x_2) = i^2\theta(x_1-x_2)\mathcal{L}(x_1)\mathcal{L}(x_2)+i^2\theta(x_2-x_1)\mathcal{L}(x_2)\mathcal{L}(x_1)
\end{equation}
can be given a rigorous sense, as the application of the Wick formula (of Subsection \ref{WickForProduct}) to the operators
\[
\mathcal{L}(x_1)\mathcal{L}(x_2) \,\,\,\, \textrm{and} \,\,\,\,
\mathcal{L}(x_2)\mathcal{L}(x_1)
\]
gives the normal order product of free fields plus the terms with pairings (contractions) which can be splitted
into the retarded and advanced parts by the Epstein-Glaser method. In particular we can compute the retarded part of 
all the pairing distributions involved into the Wick decomposition of 
\[
\mathcal{L}(x_1)\mathcal{L}(x_2) \,\,\,\, \textrm{and} \,\,\,\,
\mathcal{L}(x_2)\mathcal{L}(x_1).
\]
Thus, in the Wick formula of Subsection \ref{WickForProduct} for
\[
\mathcal{L}(x_1)\mathcal{L}(x_2) \,\,\,\, \textrm{and} \,\,\,\,
\mathcal{L}(x_2)\mathcal{L}(x_1)
\]
each term, containing the scalar causal distribution factor
(coming from the pairing, \emph{i.e.} $\otimes_q$-contraction) multiplied with $\theta(x_1-x_2)$ or $\theta(x_2-x_1)$
we interpret as the replacement of the pairing (contraction) with the retarded part of the scalar distribution
or as the retarded part of the $\otimes_q$-contraction.
The term without pairings is trivial
as $\theta(x_1-x_2)$ is a well-defined translationally invariant distribution and the normal product
of free field operators in the variables $x_1$ and $x_2$ multiplied by $\theta(x_1-x_2)$ is well-defined.
The only difference in comparison to the standard formal approach is that we use the Epstein-Glaser splitting
into the retarded and advanced part of contractions (pairings) instead of the naive multiplication by $\theta$. In fact the splitting is non-trivial
only in case of $\otimes_q$ contractions with $q>1$. For $\otimes_0 = \otimes$-contraction or $\otimes_1$-contraction
the splitting is trivial and can be realized through the ordinary multiplication by $\theta$. The contributions
with $\otimes_q$ contractions with $q>1$ correspond to the contributions represented by the loop-graphs insertions,
which in the standard renormalization approach are divergent. There are essentially only three types of such divergent insertions
in case of QED.

Moreover, using this method of computation of $S_2$ we have the same singularity orders $\omega$ in the splitiing as in 
the Epstein-Glaser formula for $S_2$, which uses $R_{(2)}$ and $D_{(2)}$ , with the same range of freedom in the splitting.
We obtain the same $S_2$ as in the Epstein-Glaser method, using the same standard normalization conditions
determining the splitting, e.g. by imposing the standard normalization conditions. Having the contractions (which we encounter
in the computation of $S_2$) and their splitting into retarded and advanced parts fixed, we determine the higher order
contributions $S_n$.

For the computation of the higher order contribution $S_n(x_1, \ldots, x_n)$ we apply the Epstein-Glaser causal method,
summarized in Subsection \ref{MotivationForHida} and \ref{WickForProduct}, supplemented by the above-mentioned very small improvement.
This computation method of $S_n(x_1, \ldots, x_n)$ is reduced to the application
of the Wick theorem of Subsection \ref{WickForProduct} 
to the product of at most two normally ordered operators (which we treat as the generalized integral kernel operators),
as the operator $D_{(n)}(x_1, \ldots, x_n)$ is equal to the finite sum of products of at most two normally ordered
operators, compare the formula for
\[
D_{(n)} = R'_{(n)}-A'_{(n)}
\]
in Subsection \ref{MotivationForHida}. Moreover, the operator $D_{(n)}(x_1, \ldots, x_n)$
has causal support, in the sense explained in Subsection \ref{MotivationForHida},
so that the computation of the retarded part
\[
R_{(n)}(x_1, \ldots, x_n) = \textrm{ret} \, D_{(n)}(x_1, \ldots, x_n)
\]
of $D_{(n)}$ allows us to compute
\[
S_{n}(x_1, \ldots, x_n) = R_{(n)}(x_1, \ldots, x_n) - R'_{(n)}(x_1, \ldots, x_n)
\]
by repeating the computation of  
\[
S_2(x,y) = R_{(2)}(x,y) - R'_{(2)}(x,y),
\] 
but instead of computing the retadred part of causal distributions of single space-time variable
we are confronted with the computation of causal distributions of $n$ space-time variables, which
is given by the dispersion formula analogous to the one valid for one space-time variable.

So let us concentrate on the general mathematical analysis and computation of $S_2$, making only general remarks concerning the
higher order generalized operators $D_{(n)}$, $R'_{(n)}$, $A'_{(n)}$ and $S_n$ which all can be reduced to the application of the Wick theorem
for (finite sum of) products of at most two normally ordered (generalized integral kernel) operators and in addition the 
application of the splitting of the causal combinations of the kernels of the Wick product $D_{(n)}$. These kernels
are equal to the contractions of the kernels of $S_k$, $k < n$.

The basic distributions in one space-time variable $x_1-x_2$ are the products of
non-zero pairings (contractions $\otimes_q$) of the free fields entering into the Wick formula (Subsection \ref{WickForProduct})
for the (tensor) product generalized operator $\mathcal{L}(x_1)\mathcal{L}(x_2)$, where $\mathcal{L}(x)$
is the interaction Lagrange density of the QFT in question, together with the particular choice of their splitting
into the retarded and advanced parts. The particular choice of the splitting will have to be used already in the computation
of the second order contribution $S_2$ outlined above, so we will especially concentrate on the analysis of the computation
of $S_2$. The computation of the higher order terms brings no new ingredients from the point of view of analysis of distributions
(we apply the same dispersion formula for the splitting of causal distributions into retarded and advanced part).
Choice of the splitting can be naturally fixed, by the requirement that the higher order contributions of interacting fields in the adiabatic limit
$g=1$ are finite sums of well-defined integral kernel operators with vector-valued kernels in the sense \cite{obataJFA},
together with the natural requirements of gauge invariance and natural normalization
of the $S$-operator, as we have shown in Subsection \ref{OperationsOnXi}.

Therefore we consider the product $\mathcal{L}(x_1)\mathcal{L}(x_2)$ of two normally ordered generalized integral kernel
operators $\mathcal{L}(x_1)$, $\mathcal{L}(x_2)$, with vector valued distributional kernels 
\begin{equation}\label{kernelsOfL(x)L(y)}
\kappa_{l,m} = \kappa'_{l',m'} \otimes_{q'} \kappa''_{l'',m''}
\end{equation}
of the product $\mathcal{L}(x_1)\mathcal{L}(x_2)$,
and the kernels $\kappa'_{l',m'}$, $\kappa''_{l'',m''}$ ranging over the kernels of the operator
$\mathcal{L}(x)$, compare Subsection \ref{WickForProduct}. The possible values of $q'$ range from zero up to the degree
of the Wick polynomial $\mathcal{L}$. Denoting the kernels
of the free fields which are involved in the Wick polynomial $\mathcal{L}$, respectively, by
\[
\kappa'_{0,1}, \kappa'_{1,0},\kappa''_{0,1}, \kappa''_{1,0}, \ldots 
\] 
we easily see that the kernels  (\ref{kernelsOfL(x)L(y)}) have the general form
\[
\Big[
\big(\kappa'_{0,1} \dot{\otimes}  \kappa''_{0,1} \dot{\otimes} \ldots \kappa^{(q)}_{0,1} \big)
\, \otimes_{q'} \,\, \big(\kappa'_{1,0} \dot{\otimes} \kappa''_{1,0}
\dot{\otimes} \ldots \kappa^{(q)}_{1,0} \big)\Big]
\]
\begin{multline*}
=
\Big[
\big(\kappa^{(k_1)}_{0,1} \dot{\otimes} \kappa^{(k_2)}_{0,1} \ldots \dot{\otimes} \kappa^{(k_{q'})}_{0,1} \big)
\, \otimes_{q'} \,\, \big(\kappa^{(k_1)}_{1,0} \dot{\otimes} \kappa^{(k_2)}_{1,0} \ldots \dot{\otimes} \kappa^{(k_{q'})}_{1,0}  \big)\Big] \, \times
\\
\times
 \,
\Big[
\big(\kappa'_{0,1} \dot{\otimes} \kappa''_{0,1} \dot{\otimes} \widehat{\ldots} \kappa^{(q)}_{0,1} \big)
\, \otimes \,\, \big(\kappa'_{1,0} \dot{\otimes} \kappa''_{1,0} \dot{\otimes} \widehat{\ldots} \kappa^{(q)}_{1,0}  \big)
\Big]
\end{multline*}
Hat means that the non contracted kernels are deleted. The product $\mathcal{L}(x_1)\mathcal{L}(x_2)$ still
makes sense and is a well-defined element of
\[
\mathcal{L} \in \mathscr{L} \big(\mathscr{E}\otimes \mathscr{E}, \mathscr{L}((\boldsymbol{E}), (\boldsymbol{E})^*) \big)
\]
whenever massless fields are present in $\mathcal{L}$, even though  in this case
\[
\mathcal{L} \in \mathscr{L} \big(\mathscr{E}, \mathscr{L}((\boldsymbol{E}), (\boldsymbol{E})^*) \big),
\]
for a proof, compare Subsection \ref{WickForProduct}. The product operator $\mathcal{L}(x_1)\mathcal{L}(x_2)$ is defined by a limit 
process on replacing the massless fields by their massive counterparts and then by passing with the auxiliary mass to the zero limit.
If all free fields are massive in $\mathcal{L}$, then 
\[
\mathcal{L} \in \mathscr{L} \big(\mathscr{E}, \mathscr{L}((\boldsymbol{E}), (\boldsymbol{E})) \big),
\]
and the product $\mathcal{L}(x_1)\mathcal{L}(x_2)$ represents a finite sum of generalized integral kernel operators with vector
valued kernels, all belonging to
\[
\mathcal{L} \in \mathscr{L} \big(\mathscr{E}\otimes \mathscr{E}, \mathscr{L}((\boldsymbol{E}), (\boldsymbol{E})) \big),
\]
for a proof, compare Section \ref{WickForProduct}.

Next we consider the retarded part
\begin{equation}\label{retL(x)L(y)}
\textrm{ret} \, \mathcal{L}(x_1)\mathcal{L}(x_2)
\end{equation}
which we symbolically write as
\[
\theta(x_1-x_2)\mathcal{L}(x_1)\mathcal{L}(x_2),
\]
of the product $\mathcal{L}(x_1)\mathcal{L}(x_2)$, expressed through the Wick formula of Subsection \ref{WickForProduct}.
This will allow us to compute $S_2$ given by (\ref{S2(x1,x2)}).  The same Wick formula can be applied 
to the sum $D_{(n)}$ of products of at most two normally ordered operators. Becuse $D_{(n)}$ is causally
supported (compare Subsections \ref{MotivationForHida} and \ref{WickForProduct}, references therein or \cite{Scharf}) then we can repeat
analogue computation of the retarded part\footnote{Here with the multidimensional theta
$\theta(x_1, \ldots, x_n) \overset{\textrm{df}}{=}\theta(x_1-x_n)\ldots\theta(x_{n-1}-x_n)$.} $R_{(n)} = \theta D_{(n)}$ of $D_{(n)}$,
with the analogue analysis also valid for $S_n = R_{(n)}-R'_{(n)}$, $n>2$. 

The kernels of (\ref{retL(x)L(y)}) have the general form
\[
\textrm{ret} \,\,
\Big[
\big(\kappa'_{0,1} \dot{\otimes}  \kappa''_{0,1} \dot{\otimes} \ldots \kappa^{(q)}_{0,1} \big)
\, \otimes_{q'} \,\, \big(\kappa'_{1,0} \dot{\otimes} \kappa''_{1,0}
\dot{\otimes} \ldots \kappa^{(q)}_{1,0} \big)\Big]
\]
\begin{multline}\label{specialRetkappa'timesq'kappa''}
=
\textrm{ret} \,\,
\Big[
\big(\kappa^{(k_1)}_{0,1} \dot{\otimes} \kappa^{(k_2)}_{0,1} \ldots \dot{\otimes} \kappa^{(k_{q'})}_{0,1} \big)
\, \otimes_{q'} \,\, \big(\kappa^{(k_1)}_{1,0} \dot{\otimes} \kappa^{(k_2)}_{1,0} \ldots \dot{\otimes} \kappa^{(k_{q'})}_{1,0}  \big)\Big] \, \times
\\
\times
 \,
\Big[
\big(\kappa'_{0,1} \dot{\otimes} \kappa''_{0,1} \dot{\otimes} \widehat{\ldots} \kappa^{(q)}_{0,1} \big)
\, \otimes \,\, \big(\kappa'_{1,0} \dot{\otimes} \kappa''_{1,0} \dot{\otimes} \widehat{\ldots} \kappa^{(q)}_{1,0}  \big),
\Big]
\end{multline}
with the integral kernel operators $\Xi(\kappa_{l,m})$, corresponding to the kernels $\kappa_{l,m}$ equal (\ref{specialRetkappa'timesq'kappa''}),
which belong to
\[
\mathcal{L} \in \mathscr{L} \big(\mathscr{E}\otimes \mathscr{E}, \mathscr{L}((\boldsymbol{E}), (\boldsymbol{E})) \big),
\]
or, respecively, to
\[
\mathcal{L} \in \mathscr{L} \big(\mathscr{E}\otimes \mathscr{E}, \mathscr{L}((\boldsymbol{E}), (\boldsymbol{E})^*) \big),
\]
depending on whether all non contracted kernels are massive in (\ref{specialRetkappa'timesq'kappa''}) or not.

The scalar distribution (scalar $\otimes_{q'}$ contraction) in (\ref{specialRetkappa'timesq'kappa''}) 
\begin{equation}\label{specialscalarq'contraction}
\textrm{ret} \,\,
\Big[
\big(\kappa^{(k_1)}_{0,1} \dot{\otimes} \kappa^{(k_2)}_{0,1} \ldots \dot{\otimes} \kappa^{(k_{q'})}_{0,1} \big)
\, \otimes_{q'} \,\, \big(\kappa^{(k_1)}_{1,0} \dot{\otimes} \kappa^{(k_2)}_{1,0} \ldots \dot{\otimes} \kappa^{(k_{q'})}_{1,0}  \big)\Big](x_1,x_2)
\end{equation}
we compute by the ordinary multiplication by the step theta function
\begin{multline*}
\textrm{ret} \,\,
\Big[
\big(\kappa^{(k_1)}_{0,1} \dot{\otimes} \kappa^{(k_2)}_{0,1} \ldots \dot{\otimes} \kappa^{(k_{q'})}_{0,1} \big)
\, \otimes_{q'} \,\, \big(\kappa^{(k_1)}_{1,0} \dot{\otimes} \kappa^{(k_2)}_{1,0} \ldots \dot{\otimes} \kappa^{(k_{q'})}_{1,0}  \big)\Big]
\\
=
\theta(x_1-x_2) \,\,
\Big[
\big(\kappa^{(k_1)}_{0,1} \dot{\otimes} \kappa^{(k_2)}_{0,1} \ldots \dot{\otimes} \kappa^{(k_{q'})}_{0,1} \big)
\, \otimes_{q'} \,\, \big(\kappa^{(k_1)}_{1,0} \dot{\otimes} \kappa^{(k_2)}_{1,0} \ldots \dot{\otimes} \kappa^{(k_{q'})}_{1,0}  \big)\Big](x_1,x_2)
\\
\textrm{if} \,\, q'=0, \,\, \textrm{or} \,\, q'=1,
\end{multline*}
\emph{i.e.} if the singularity degree $\omega$ at zero of the distribution (scalar $\otimes_{q'=0}$, $\otimes_{q'=1}$-contraction)
\[
\big(\kappa^{(k_1)}_{0,1} \dot{\otimes} \kappa^{(k_2)}_{0,1} \ldots \dot{\otimes} \kappa^{(k_{q'})}_{0,1} \big)
\, \otimes_{q'} \,\, \big(\kappa^{(k_1)}_{1,0} \dot{\otimes} \kappa^{(k_2)}_{1,0} \ldots \dot{\otimes} \kappa^{(k_{q'})}_{1,0}  \big)
\]
is negative, and can be mulitplied by the step theta function.
If the singularity degree $\omega$ of 
\[
\kappa_{q'}(x_1-x_2)
=
\big(\kappa^{(k_1)}_{0,1} \dot{\otimes} \kappa^{(k_2)}_{0,1} \ldots \dot{\otimes} \kappa^{(k_{q'})}_{0,1} \big)
\, \otimes_{q'} \,\, \big(\kappa^{(k_1)}_{1,0} \dot{\otimes} \kappa^{(k_2)}_{1,0} \ldots \dot{\otimes} \kappa^{(k_{q'})}_{1,0}  \big)(x_1,x_2)
\]
is zero, or positive, which is the case for $q'>1$, then we define (\ref{specialscalarq'contraction})
on the subspace of finite codimension of all space-time test functions $\chi\in \mathscr{E}^{\otimes \, 2}$, $\mathscr{E} = \mathcal{S}(\mathbb{R}^4)$,
$\chi(x_1,x_2) = \phi(x_1-x_2)\varphi(x_2)$, $\phi, \varphi \in \mathscr{E}$,  for which all derivatives of 
$\phi$, up to order $\omega$, vanish at zero. Denoting the continuous idemponent operator projecting
$\mathscr{E}$ on the subspace of functions with vanishing derivatives at zero up to order $\omega$, by $\Omega'$,
we define (\ref{specialscalarq'contraction}), first by projecting  $\Omega'$ and then by multiplication by $\theta$
function 
\[
\langle \textrm{ret} \, \kappa_{q'}, \phi\rangle = \langle \kappa_{q'}, \theta.\Omega'\phi \rangle,
\]
and putting $\textrm{ret} \, \kappa_{q'}(x_1-x_2)$ for (\ref{specialscalarq'contraction}). This definition is
not unique and we can add to  $\textrm{ret} \, \kappa_{q'}$ a distribution 
\[
\sum\limits_{|\alpha|=0}^{\omega} C_\alpha \delta^{(\alpha)},
\]
with arbitrary (in principle) $C_\alpha$, which is zero on the $\textrm{Im} \, \Omega'$
and is most general on the finite dimensional subspace $\textrm{Ker} \, \Omega'$, giving the most general 
retarded part of $\kappa_{q'}$ on the  whole test space
\[
\mathscr{E} = \textrm{Im} \, \Omega' \, \oplus \, \textrm{Ker} \, \Omega',
\]
and correspondingly, giving the most general (\ref{specialscalarq'contraction}) which together with
\begin{multline*}
\textrm{av} \,\,
\Big[
\big(\kappa^{(k_1)}_{0,1} \dot{\otimes} \kappa^{(k_2)}_{0,1} \ldots \dot{\otimes} \kappa^{(k_{q'})}_{0,1} \big)
\, \otimes_{q'} \,\, \big(\kappa^{(k_1)}_{1,0} \dot{\otimes} \kappa^{(k_2)}_{1,0} \ldots \dot{\otimes} \kappa^{(k_{q'})}_{1,0}  \big)
\Big]
\\
=
\big(\kappa^{(k_1)}_{0,1} \dot{\otimes} \kappa^{(k_2)}_{0,1} \ldots \dot{\otimes} \kappa^{(k_{q'})}_{0,1} \big)
\, \otimes_{q'} \,\, \big(\kappa^{(k_1)}_{1,0} \dot{\otimes} \kappa^{(k_2)}_{1,0} \ldots \dot{\otimes} \kappa^{(k_{q'})}_{1,0}  \big)
\\
-
\textrm{ret} \,\,
\Big[
\big(\kappa^{(k_1)}_{0,1} \dot{\otimes} \kappa^{(k_2)}_{0,1} \ldots \dot{\otimes} \kappa^{(k_{q'})}_{0,1} \big)
\, \otimes_{q'} \,\, \big(\kappa^{(k_1)}_{1,0} \dot{\otimes} \kappa^{(k_2)}_{1,0} \ldots \dot{\otimes} \kappa^{(k_{q'})}_{1,0}  \big)
\Big]
\end{multline*}
gives the most general splitting of 
\[
\big(\kappa^{(k_1)}_{0,1} \dot{\otimes} \kappa^{(k_2)}_{0,1} \ldots \dot{\otimes} \kappa^{(k_{q'})}_{0,1} \big)
\, \otimes_{q'} \,\, \big(\kappa^{(k_1)}_{1,0} \dot{\otimes} \kappa^{(k_2)}_{1,0} \ldots \dot{\otimes} \kappa^{(k_{q'})}_{1,0}  \big)
\]
into the difference of the retarded and advanced parts. 

Introducing the continuous idempotent operator $\Omega$: 
\[
\Omega\chi(x_1,x_2) = \phi(x_1-x_2)\varphi(x_2),
\]
which projects the test functions $\chi(x_1,x_2) = \phi(x_1-x_2)\varphi(x_2)$ onto the subspace
of $\chi(x_1,x_2) = \phi'(x_1-x_2)\varphi(x_2)$ in which all derivatives of $\phi'$ vanish a zero
up to order $\omega$, we can write 
\begin{equation}\label{DoubleLimitContraction}
\textrm{ret} \, \kappa'_{l',m'} \otimes_{q'} \kappa''_{l'',m''} = \theta \kappa'_{l',m'} \otimes_{q'} \kappa''_{l'',m''} \circ \Omega
= \theta \kappa'_{l',m'} \otimes||_{q'} \kappa''_{l'',m''},  
\end{equation}
where we have introduced the ``limit contraction''
\[
\kappa'_{l',m'} \otimes||_{q'} \kappa''_{l'',m''}(\chi) =
\kappa'_{l',m'} \otimes_{q'} \kappa''_{l'',m''}(\Omega \chi),
\]
on the test functions $\chi(x_1,x_2) = \phi(x_1-x_2)\varphi(x_2)$, $\phi,\varphi \in \mathscr{E}$ and where 
the multiplication by $\theta$ in (\ref{DoubleLimitContraction}) is inderstood as the multiplication by the 
following functon $(x_1,x_2) \mapsto\theta(x_1-x_2)$ 
of two space-time variables $x_2,x_2$.

Similarly, we have for the kernels of the operator $\theta D_{(n)}$, $n>2$
which are given by the analogue expression. In the last case $\kappa'_{l',m'}$
and $\kappa''_{l'',m''}$ run independently over the kernels, 
respectively, of the operators $\overline{S_{l}}$ and $S_{k}$, $l+k=n$, $l>0$. But in this case $\theta D_{(n) \, \epsilon}$ 
the function $\theta$
is understand as equal to the product of the one-dimensional $\theta$-functions
evaluated at $x_1-x_n, \ldots, x_{n-1}-x_n$:
\[
\theta(x_1-x_n, x_2-x_n, \ldots, x_{n-1}-x_n) =  \theta(x_1-x_n) \ldots \theta(x_{n-1}-x_n).
\]
In this case we consider elements $\chi \in \mathscr{E}^{\otimes \, n}$  of the form
\begin{multline}\label{FormOfchi}
\chi(x_1, \ldots, x_n) = \phi(x_1 -x_n, x_2-x_n, \ldots, x_{n-1} -x_{n})\varphi(x_n), 
\\
= (\phi \otimes \varphi) \circ L^{-1}(x_1, \ldots, x_n), \,\,\,\,\,\,
\phi \in \mathscr{E}^{n-1}, \varphi\in \mathscr{E}
\end{multline}
which respect the condition
\begin{equation}\label{Dalphaphi|0=0,dim=k-1}
D^\alpha\phi(0) = 0, \,\,\, 0 \leq |\alpha| \leq \omega, 
\end{equation}
or, equivalently, the condition
\begin{equation}\label{Der|0chi=0,dim=n}
D^\alpha_{{}_{x}}\chi(x_1=x_n, \ldots, x_{n-1} = x_n, y) = 0, \,\,\, 0 \leq |\alpha| \leq \omega, 
\end{equation}
\[
x= (x_1, \ldots, x_{n-1}) \in \big[\mathbb{R}^4\big]^{n-1}, \,\,\, y = x_n \in \mathbb{R}^4.
\]
Here $L^{-1}$ is the invertible linear map on $\big[\mathbb{R}^{4}\big]^{\times \, n}$ given by:
\[
L^{-1}: (x_1, \ldots, x_n) \longmapsto (x_1-x_n, x_2 - x_n, \ldots, x_{n-1}-x_{n}, x_{n}).
\]
Here $\omega$ depends only on the \emph{singularity degree} at zero of the vector-valued distribution
\begin{equation}\label{chronologicalkappal,m}
\kappa'_{l',m'} \otimes_{{}_{q}} \, \kappa''_{l'',m''} \,\,
\end{equation}
$\phi \in \mathscr{E}^{\otimes \, n}$ in the first $n-1$  space-time variables
$(x_1-x_n, x_2-x_n, \ldots, x_{n-1}-x_n)$. 
This singularity degree is in fact equal to the singularity degree (in the sense of \cite{Epstein-Glaser})
of a scalar-valued translationally invariant (with
causally supported singular part) distribution canonically related to (\ref{chronologicalkappal,m}), which follows from the canonical
factorization of (\ref{chronologicalkappal,m}) into a scalar-valued factor, which captures all contractions,
and a vector-valued factor, without any contractions. Below we also explain how this factorization appears.

Let, for each multi-index $\alpha$, such that $0 \leq |\alpha| \leq \omega$, 
$\omega_{{}_{o \,\, \alpha}} \in \mathscr{E}^{\otimes \, (n-1)}$ on $\mathbb{R}^{4(n-1)}$ be such 
functions that\footnote{Such functions $\omega_{{}_{o \,\, \alpha}} \in \mathscr{E}^{\otimes \, (n-1)}$, $0 \leq |\alpha| \leq \omega$, do exist. 
Indeed let
\[
f_{{}_{\alpha}}(x) = 
{\textstyle\frac{x^\alpha}{\alpha!}}, \,\,\, x=(x_1, \ldots, x_{n-1}) \in \mathbb{R}^{4(n-1)}, 0 \leq |\alpha| \leq \omega,
\]
\[
\alpha ! = \prod \limits_{i=1}^{n-1} \prod \limits_{\mu=0}^{3} \alpha_{i\mu} !, 
\,\,\, x^\alpha = \prod \limits_{i=1}^{n-1} \prod \limits_{\mu=0}^{3} (x_{1\mu})^{\alpha_{1\mu}}, \,\,\, \mu = 0,1,2,3.
\]
It is not difficult to see that there exists $w \in \mathscr{E}^{n-1} = \mathcal{S}(\mathbb{R}^4)^{\otimes(n-1)}$,
which is equal $1$ on some neighborhood of zero.
Indeed,  let $w \in \mathscr{C}^\infty(\mathbb{R}^{4(n-1)})$, which is equal $1$ on some neighborhood of zero and zero outside some larger
neighborhood of zero (such a function does exist, compare  e.g. \cite{Rudin}).
Then we can put
\[
\omega_{{}_{o \,\, \alpha}} \overset{\textrm{df}}{=} f_{{}_{\alpha}}.w.
\]}
\[
D^\beta \omega_{{}_{o \,\, \alpha}} (0) = \delta^{\beta}_{\alpha}, \,\,\,\, 0 \leq |\alpha|, |\beta| \leq \omega.
\]
Let for any $\phi \in \mathscr{E}^{\otimes \, (n-1)}$ 
\begin{equation}\label{OmegaphiR(k-1)}
\Omega' \phi = \phi - \sum \limits_{0\leq |\alpha| \leq \omega} D^\alpha \phi(0) \, \omega_{{}_{o \,\, \alpha}} 
\end{equation}

Of course $\Omega'$ is a continuous operator $\mathscr{E}^{\otimes \, (n-1)} \rightarrow \mathscr{E}^{\otimes \, (n-1)}$,
whose image has finite codimension in $\mathscr{E}^{\otimes \, (n-1)}$ and consists of all functions $\Omega' \phi$
such that
\[
\big(D^\alpha \Omega' \phi\big)(0) = 0, \,\,\,  0 \leq |\alpha| \leq \omega.
\]
Next, let $\phi \in \mathscr{E}^{\otimes \, (n-1)}$ and $\varphi \in \mathscr{E}$, and 
let 
\begin{multline*}
\chi(x_1, \ldots, x_n) \overset{\textrm{df}}{=} \phi(x_1-x_n, x_2 - x_n, \ldots, x_{n-1} - x_n) \varphi(x_n)
\\
= (\phi \otimes \varphi) \circ L^{-1}(x_1, \ldots, x_n).
\end{multline*}

Let, for any such $\chi$, the following function be defined
\begin{multline}\label{SimpleFormulaForOmega}
\Omega\chi(x_1, \ldots, x_n) \overset{\textrm{df}}{=} \Omega' \phi(x_1-x_n, x_2 - x_n, \ldots, x_{n-1} - x_n) \varphi(x_n)
\\
= (\Omega' \phi \otimes \varphi) \circ L^{-1}(x_1, \ldots, x_n),
\end{multline}
which by construction respects the condition (\ref{Der|0chi=0,dim=n}). 
More generally for any 
\[
\chi \in \mathscr{E}^{\otimes \, n}
\]
we define the function
\begin{multline*}
\chi^\natural(x_1, \ldots, x_n) = \chi(x_1+ x_n, x_2+ x_n, \ldots , x_{n-1} + x_n, x_n)
\\
=
\chi \circ L(x_1, \ldots, x_n),
\end{multline*}
where $L$ is the linear map
\[
(x_1, \ldots, x_n) \longmapsto (x_1+ x_n, \ldots, x_{n-1} +x_n, x_n).
\]

It is immediately seen that
\[
\chi^\natural \in \mathscr{E}^{\otimes \, n}
\]
and there exists a series 
\[
\chi^\natural = \sum \limits_{j} \phi_j \otimes \varphi_j, \,\,\,\,\,\,\, \chi = \sum \limits_{j} (\phi_j \otimes \varphi_j ) \circ L^{-1},
\]
of simple tensors $\phi_j \otimes \varphi_j$, 
$\phi_j \in \mathscr{E}^{\otimes \, (n-1)}$ and $\varphi_j \in \mathscr{E}$,
converging in $\mathscr{E}^{\otimes \, n}$. We define 
\begin{multline*}
\Omega \chi(x_1, \ldots, x_n) \overset{\textrm{df}}{=} \sum \limits_{j} \big(\Omega' \phi_j \big)(x_1-x_n, x_2-x_n, \ldots, x_{n-1}-x_n) 
\varphi_j(x_n)
\\
\overset{\textrm{df}}{=} \sum \limits_{j} \big(\Omega' \phi_j \otimes \varphi_j \big) \circ L^{-1}(x_1, \ldots, x_n),
\end{multline*}
so that
\begin{multline*}
\Omega(\chi) \circ L(x_1, \ldots, x_n) = \sum \limits_{j} \big(\Omega' \phi_j \big)(x_1, \ldots, x_{n-1}) 
\varphi_j(x_n)
\\
=
\sum \limits_{j} \big[(\Omega'\phi_j) \otimes \varphi_j \big](x_1, \ldots, x_n),
\end{multline*}
with $\Omega' \phi_j$
given by (\ref{OmegaphiR(k-1)}). Writing the same otherwise
\begin{multline*}
\Omega(\chi) \circ L (x_1, \ldots, x_n) = \chi \circ L(x_1, \ldots, x_n)
- \sum\limits_{|\beta|=0}^{\omega}  \omega_{{}_{0 \,\, \beta}}(x_1, \ldots, x_{n-1}) \,\, \times
\\
\times \,\,
D^{\beta}_{{}_{x_1, \ldots, x_{n-1}}} (\chi \circ L)(x_1=0, \ldots, x_{n-1}=0,x_n).
\end{multline*}

Equivalently
\begin{multline}\label{Omega-n-variables}
\Omega \chi(x_1, \ldots, x_n) 
= \chi(x_1, \ldots, x_n) -  \sum \limits_{|\beta|=0}^{\omega}
 \omega_{{}_{0 \,\, \beta}}(x_1-x_n, x_2-x_n, \ldots, x_{n-1}-x_n) \,\, \times
\\
\times \,\,
D^{\beta}_{{}_{x_1, \ldots, x_{n-1}}} \chi(x_1=x_n, x_2=x_n, \ldots, x_{n-1}=x_{n}, x_n).
\end{multline}

Existence of the kernels (\ref{specialRetkappa'timesq'kappa''}) of $S_2$, and more generally of $S_n$,
 and their corresponding continuity, which assures
\begin{gather*}
S_n \in  \mathscr{L} \big(\mathscr{E}^{\otimes \, n}, \mathscr{L}((\boldsymbol{E}), (\boldsymbol{E})) \big),
\,\,\, 
\textrm{if no massless fields are in $\mathcal{L}$}
\\
S_n \in \mathscr{L} \big(\mathscr{E}^{\otimes \, n}, \mathscr{L}((\boldsymbol{E}), (\boldsymbol{E})^*) \big)
\,\,\, 
\textrm{if massless fields are in $\mathcal{L}$},
\end{gather*}
 we have already proved in Subsection \ref{WickForProduct}, where we have used the fact that there exists well-defined
splitting of the scalar $\otimes_{q'}$-contractions (\ref{specialscalarq'contraction}) into retarded and advanced parts, 
and which was shown earlier for the causal symmetric or antisymmetric combinations of the products (\ref{CausalDifferencesOfProductsOfPairings}) 
of pairings in \cite{Epstein-Glaser}, and which moreover is unique if $q\leq 1$, and non unique if $g>1$, with the non uniqueness
depending on a finite number $N(\omega)$ of constants, with $N(\omega)$ depending on the singularity degree $\omega$ at zero of (\ref{specialscalarq'contraction}). 
Although this is already a well-established result, we present the splitting of the scalar contractions (products of pairings)
(\ref{specialscalarq'contraction}) or their causal symmetric or antisymmetric parts (\ref{CausalDifferencesOfProductsOfPairings}),
and then, using this fact, we show once again existence and respective continuity of the kernels (\ref{specialRetkappa'timesq'kappa''}).

Having in view applications to realistic QFT, like QED, we are especially concentrated here on the case in which the
interaction Lagrangian $\mathcal{L}$ contains massless free fields.

Before we return to the analysis of the kernels (\ref{specialRetkappa'timesq'kappa''})
of the second order contribution $S_2$ to the scattering operator
let us consider kernels of a more general product 
\begin{equation}\label{L(x1)...L(xk)}
\mathcal{L}(x_1) \ldots \mathcal{L}(x_k) 
\end{equation}
of more than just two factors and apply to it the Wick theorem of Subsection \ref{WickForProduct}.
The factor operators
$\mathcal{L}(x_i)$ can be replaced by any other Wick monomials of free fields in our analysis.
We explain now how the factorization, metioned above, appears.
We are also going to give a simple proof that the product (\ref{L(x1)...L(xk)}) in 
\[
\mathcal{L} \in \mathscr{L} \big(\mathscr{E}^{\otimes \, k}, \mathscr{L}((\boldsymbol{E}), (\boldsymbol{E})^*) \big),
\]
of Wick product operators $\mathcal{L}(x_i)$ of free, possibly massless fields, can be multiplied by a tempered distribution, 
giving again a finite sum of generalized integral kernel operators in
\[
\mathcal{L} \in \mathscr{L} \big(\mathscr{E}^{\otimes \, k}, \mathscr{L}((\boldsymbol{E}), (\boldsymbol{E})^*) \big).
\]
If all the free fields are massive in (\ref{L(x1)...L(xk)}) then the operator is more regular, and still kan be multiplied by 
any tempered translationally invariant distribution, which after this operation wiill give an operator 
in 
 \[
\mathcal{L} \in \mathscr{L} \big(\mathscr{E}^{\otimes \, k}, \mathscr{L}((\boldsymbol{E}), (\boldsymbol{E})) \big),
\]
which we have already proved in Subsection \ref{WickForProduct}, where the proof of this stronger result
is slightly more involved.

Namely, each of the 
kernels $\kappa^{i}_{\ell_{i},m_{i}}$ (defining the finite Fock decomposition into integral kernel operators 
$\Xi(\kappa^{i}_{l,m}(x_i))$ of the 
interaction operator $\mathcal{L}(x_i)$) in (\ref{chronologicalkappal,m}) is equal to the simple pointwise product (dot product
$\dot{\otimes}$, the momentum variables of the kernels are not written explicitly) or to the sum of the following simple products
\begin{equation}\label{kappa'30}
\kappa^{i}_{\ell_{i},m_{i}}(x_i) = \kappa^{(1_i)}_{1,0} \dot{\otimes} \kappa^{(2_i)}_{1,0} \dot{\otimes} \kappa^{(3_i)}_{1,0}(x_i)
= \kappa^{(1_i)}_{1,0}(x_i) \kappa^{(2_i)}_{1,0}(x_i) \kappa^{(3_i)}_{1,0}(x_i),
\end{equation}
in case $ l_i =3, m_i=0$, or 
\begin{align}
\kappa^{i}_{\ell_{i},m_{i}}(x_i) & = \kappa^{(1_i)}_{1,0} \dot{\otimes} \kappa^{(2_i)}_{1,0} \dot{\otimes} \kappa^{(3)}_{0,1}(x_i)
= \kappa^{(1)}_{1,0}(x_i)  \kappa^{(2)}_{1,0}(x_i)  \kappa^{(3)}_{0,1}(x_i), \label{kappa'21.1}
\\
\kappa^{i}_{\ell_{i},m_{i}}(x_i) & = \kappa^{(1_i)}_{1,0} \dot{\otimes} \kappa^{(2_i)}_{0,1} \dot{\otimes} \kappa^{(3_i)}_{1,0}(x_i)
= \kappa^{(1_i)}_{1,0}(x_i)  \kappa^{(2_i)}_{0,1}(x_i)  \kappa^{(3_i)}_{1,0}(x_i), \label{kappa'21.2}
\\
\kappa^{i}_{\ell_{i},m_{i}}(x_i) & = \kappa^{(1_i)}_{0,1} \dot{\otimes} \kappa^{(2_i)}_{1,0} \dot{\otimes} \kappa^{(3_i)}_{1,0}(x_i)
= \kappa^{(1_i)}_{0,1}(x_i) \kappa^{(2_i)}_{1,0}(x_i) \kappa^{(3_i)}_{1,0}(x_i), \label{kappa'21.3}
\end{align}
in case $l_i =2, m_i=1$,
or 
\begin{align}
\kappa^{i}_{\ell_{i},m_{i}}(x_i) & = \kappa^{(1_i)}_{1,0} \dot{\otimes} \kappa^{(2_i)}_{0,1} \dot{\otimes} \kappa^{(3_i)}_{0,1}(x_i)
= \kappa^{(1_i)}_{1,0}(x_i)  \kappa^{(2_i)}_{0,1}(x_i)  \kappa^{(3_i)}_{0,1}(x_i), \label{kappa'12.1}
\\
\kappa^{i}_{\ell_{i},m_{i}}(x_i) & = \kappa^{(1_i)}_{0,1} \dot{\otimes} \kappa^{(2_i)}_{1,0} \dot{\otimes} \kappa^{(3_i)}_{0,1}(x_i)
= \kappa^{(1_i)}_{0,1}(x_i)  \kappa^{(2_i)}_{1,0}(x_i)  \kappa^{(3_i)}_{0,1}(x_i), \label{kappa'12.2}
\\
\kappa^{i}_{\ell_{i},m_{i}}(x_i) & = \kappa^{(1_i)}_{0,1} \dot{\otimes} \kappa^{(2_i)}_{0,1} \dot{\otimes} \kappa^{(3_i)}_{1,0}(x_i)
=\kappa^{(1_i)}_{0,1}(x_i)  \kappa^{(2_i)}_{0,1}(x_i)  \kappa^{(3_i)}_{1,0}(x_i), \label{kappa'12.3}
\end{align}
in case $l_i =1, m_i=2$,
or, respectively
\begin{equation}\label{kappa'03}
\kappa^{i}_{\ell_{i},m_{i}}(x_i) = \kappa^{(1_i)}_{0,1} \dot{\otimes} \kappa^{(2_i)}_{0,1} \dot{\otimes} \kappa^{(3_i)}_{0,1}(x_i)
=  \kappa^{(1_i)}_{0,1}(x_i)  \kappa^{(2_i)}_{0,1}(x_i)  \kappa^{(3_i)}_{0,1}(x_i),
\end{equation}
in case $l_i =0, m_i=3$.
Each of the kernels 
\[
\kappa^{(1_i)}_{0,1}, \kappa^{(1_i)}_{1,0}, \kappa^{(2_i)}_{0,1}, \kappa^{(2_i)}_{1,0},
\kappa^{(3_i)}_{0,1}, \kappa^{(3_i)}_{1,0}
\]
is equal to the plane wave kernel which defines one of the free fields in the interaction Lagrange density operator
$\mathcal{L}$. In case of spinor QED interaction $\mathcal{L}$ each contribution $\kappa^{i}_{\ell_{i},m_{i}}$ is thus given in the form of dot product
$\dot{\otimes}$ of three  plane wave kernels, and the first factor among them is equal to the kernel 
\[
\kappa^{(1_i)}_{0,1} = \kappa^{\sharp}_{0,1}=\gamma^0\overline{\kappa_{1,0}} \,\,\, \textrm{or} \,\,\, 
\kappa^{(1_i)}_{1,0} =\kappa^{\sharp}_{1,0}=\gamma^0\overline{\kappa_{0,1}}
\]
which defines the free Dirac-conjugated spinor field $\boldsymbol{\psi}^\sharp$ (where
$\kappa_{0,1},\kappa_{1,0}$ are the kernels of the Dirac field $\boldsymbol{\psi}$, compare Subsection \ref{psiBerezin-Hida}),
the second factor
\[
\kappa^{(2_i)}_{0,1} \,\,\, \textrm{or} \,\,\, \kappa^{(2_i)}_{1,0}
\]
is the one defining the free electromagnetic potential field, and the third
\[
\kappa^{(3_i)}_{0,1} \,\,\, \textrm{or} \,\,\, \kappa^{(3_i)}_{1,0}
\]
is equal to that which defines the free Dirac spinor field. 

In case of more general Wick monomial $\mathcal{L}$ of $N$ free fields 
each contribution to $\kappa^{i}_{\ell_{i},m_{i}}$ will be given in the form of the dot product of $N$ plane wave kernels, each defining the respective free
field entering the Lagrange density interaction operator $\mathcal{L}$.

Recall that bar $\overline{\dot{\otimes}}$ over the pointwise product $\dot{\otimes}$ (pointwise with respect to the space-time variable)
in
\[
\kappa^{(1_i)}_{0,1} \overline{\dot{\otimes}} \kappa^{(2_i)}_{0,1} \overline{\dot{\otimes}} \kappa^{(3_i)}_{0,1}(x_i),
\]
means that all momentum variables in the product
\[
\kappa^{(1_i)}_{0,1} \dot{\otimes} \kappa^{(2_i)}_{0,1} \dot{\otimes} \kappa^{(3_i)}_{0,1}(x_i)
= 
\kappa^{(1_i)}_{0,1}(x_i)  \kappa^{(2_i)}_{0,1}(x_i)  \kappa^{(3_i)}_{0,1}(x_i),
\] 
are respectively symmetrized in Bose momentum and spin variables and antisymmetrized in Fermi spin and momentum variables, according to the ordinary prescription
of Subsection \ref{OperationsOnXi}, in order to keep one-to-one relation between integral kernel operator and its kernel.

However, in practical computations it is convenient to use non-symmetrized products in the intermediate steps, and perform
the symmetrization and antisymmetrization at the very end, in order to keep the one-to-one relation between the kernel and the operator.
We will do so from now on.

Note also that the full contribution $\Xi_{\ell,m}(\kappa^{i}_{\ell,m})$ to the Fock expansion of $\mathcal{L}$ with fixed $\ell,m$, $\ell+m=3$,
is equal to the sum of three terms (\ref{kappa'21.1}) - (\ref{kappa'21.3}) in case $\ell =2, m=1$. Similarly, full contribution
$\Xi_{\ell,m}(\kappa^{i}_{\ell,m})$ to $\mathcal{L}$ with $\ell =1, m=2$ is equal to the sum of three terms
(\ref{kappa'12.1}) - (\ref{kappa'12.3}). But it is sufficient to investigate
and (\ref{chronologicalkappal,m}) with the kernels $\kappa^{i}_{\ell_i, m_i}$ equal to the simple space-time
pointwise products $\dot{\otimes}$ (\ref{kappa'30}) - (\ref{kappa'03}) of the plane wave kernels defining the free fields of the theory.
Note also that the limit contractions $\otimes|_{{}_{q_i}}$, when expressed through the momentum integrals of the kernels are given by the ordinary
contraction integrals which are absolutely convergent, even if some plane wave kernels are mass less, as explained in Subsection \ref{WickForProduct}.
Therefore, the limit process, which in principle is involved into the definition of the \emph{limit contraction} $\otimes|_{{}_{q_i}}$,
is trivial and the contraction integral remains meaningful even with the massless kernels
put literally into the integral. We therefore will sometimes be simply writing $\otimes_{q_i}$ for $\otimes|_{{}_{q_i}}$ even if there are
dot products of massless plane wave kernels involved into the contraction,
as the ordinary contraction formula still works, when expressed through the integrals of the kernels, and in particular we will simply write 
\[
(-1)^{c(q_1)+\ldots +c(q_{k-1})}
\kappa^{1}_{\ell_{1},m_{1}}\overline{\otimes_{q_1}} \, \kappa^{2}_{\ell_{2},m_{2}} \,\,
\overline{\otimes_{q_2}}
\,\, \ldots  \,\,
\overline{\otimes_{q_{k-2}}} \, 
 \kappa^{k-1}_{\ell_{k-1},m_{k-1}}\overline{\otimes_{q_{k-1}}} \,  \kappa^{k}_{\ell_{k},m_{k}}
\]
for the kernels of (\ref{L(x1)...L(xk)}). 

In case of spinor QED the only pairs of plane wave kernels which can be contracted are the following
\begin{eqnarray*}
\kappa^{(1_i)}_{0,1} \,\,\, \textrm{and} \,\,\, \kappa^{(3_j)}_{1,0},
\\
\kappa^{(3_i)}_{0,1} \,\,\, \textrm{and} \,\,\, \kappa^{(1_j)}_{1,0},
\\
\kappa^{(2_i)}_{0,1} \,\,\, \textrm{and} \,\,\, \kappa^{(2_j)}_{1,0},
\end{eqnarray*}
\[
i \neq j.
\]
In case of QED and the particular expression (\ref{chronologicalkappal,m}) the only contracted pairs are
\begin{eqnarray*}
\kappa^{(1_i)}_{0,1} \,\,\, \textrm{and} \,\,\, \kappa^{(3_{i+1})}_{1,0},
\\
\kappa^{(3_i)}_{0,1} \,\,\, \textrm{and} \,\,\, \kappa^{(1_{i+1})}_{1,0},
\\
\kappa^{(2_i)}_{0,1} \,\,\, \textrm{and} \,\,\, \kappa^{(2_{i+1})}_{1,0},
\end{eqnarray*}
Of course $q_i \leq \textrm{min}\{ m_i, \ell_{i+1} \}$, as by assumption in
\[
\kappa^{i}_{\ell_{i},m_{i}} \otimes|_{{}_{q_{i}}} \,  \kappa^{i+1}_{\ell_{i+1},m_{i+1}}
\]
we contract $q_i$ among some of the ``last $m_i$'' momentum variables of $\kappa^{i}_{\ell_{i},m_{i}}$
with $q_i$ among some of the ``first $\ell_{i+1}$'' momentum variables of $\kappa^{i+1}_{\ell_{i+1},m_{i+1}}$.
In spinor QED case the contraction numbers $q_i$ range between $0$ and $3$. 

The only difference, respectively, between
\begin{eqnarray*}
\kappa^{(1_i)}_{0,1} \,\,\, \textrm{and} \,\,\, \kappa^{(1_j)}_{1,0},
\\
\kappa^{(2_i)}_{0,1} \,\,\, \textrm{and} \,\,\, \kappa^{(2_j)}_{1,0},
\\
\kappa^{(3_i)}_{0,1} \,\,\, \textrm{and} \,\,\, \kappa^{(3_j)}_{1,0},
\end{eqnarray*}
is that the space-time variable and spin-momentum variables of the first kernel
are $x_i, s_i,\boldsymbol{\p}_i$ and the space-time and spin-momentum variables of the second
are $x_j, s_j, \boldsymbol{\p}_j$ and are  regarded as independent. 
In case the kernels, respectively,
\begin{eqnarray*}
\kappa^{(1_i)}_{0,1} \,\,\, \textrm{and} \,\,\, \kappa^{(3_j)}_{1,0},
\\
\kappa^{(3_i)}_{0,1} \,\,\, \textrm{and} \,\,\, \kappa^{(1_j)}_{1,0},
\\
\kappa^{(2_i)}_{0,1} \,\,\, \textrm{and} \,\,\, \kappa^{(2_j)}_{1,0},
\end{eqnarray*}
are contracted, the spin-momentum variables are put equal in both contracted kernels $s_i=s_j$,
$\boldsymbol{\p}_i = \boldsymbol{\p}_j$ and summed up over $s_i=s_j$ and $\ud^3 \boldsymbol{\p}_i$-integrated. 

In particular the kernel contributions $\kappa^{i}_{\ell_i, m_i}$ (which are of the general simple product form 
(\ref{kappa'30}) - (\ref{kappa'03}) stated above) and which are present in the expression 
(\ref{chronologicalkappal,m}) need to have the special form, which is consistent with the contractions
$q_1, \ldots, q_{n-1}$. The pairs of the contracted kernels are not necessary subsequent. But, in particular, if the contracted
kernels are subsequent then each pair of the subsequent kernels
$\kappa^{i}_{\ell_i, m_i}$ and $\kappa^{i+1}_{\ell_{i+1}, m_{i+1}}$ have common $q_i$ variables which can be contracted. 
Namely the kernel $\kappa^{i}_{\ell_i, m_i}$
has some $q_i$ among the last $m_i$ variables which can be contracted with some $q_i$ of the first $\ell_{i+1}$ variables of the next kernel
$\kappa^{i+1}_{\ell_{i+1}, m_{i+1}}$.  Therefore, in the analysis of (\ref{chronologicalkappal,m}) we should restrict our attention to the kernels
$\kappa^{i}_{\ell_i, m_i}$  which have the following special simple product form
\begin{multline*}
\kappa^{i}_{\ell_i,m_i} = \overbrace{\kappa^{(I_{i}^{1})}_{1,0} \dot{\otimes} 
\ldots \dot{\otimes}  \kappa^{(I_{i}^{\ell_i})}_{1,0}}^{\textrm{$\ell_i$ terms}}
\dot{\otimes}  \overbrace{\kappa^{(O_{i}^{1})}_{0,1} \dot{\otimes} 
\ldots \dot{\otimes}  \kappa^{(O_{i}^{m_i})}_{0,1}}^{\textrm{$m_i$ terms}}
\\
=
\big(\underbrace{\kappa^{(I_{i}^{1})}_{1,0} \dot{\otimes}
\ldots \dot{\otimes}  \kappa^{(I_{i}^{q_{i-1}})}_{1,0}}_{\textrm{$q_{i-1}$ terms}}\big)
\dot{\otimes} \big(\underbrace{\kappa^{(O_{i}^{1})}_{0,1} 
\dot{\otimes}  \ldots \overline{\dot{\otimes}}  \kappa^{(O_{i}^{q_i})}_{0,1}}_{\textrm{$q_i$ terms}}\big) \dot{\otimes}
\\
\dot{\otimes}
\big(\underbrace{\kappa^{(I_{i}^{q_{i-1}+1})}_{1,0} \dot{\otimes}
\ldots \dot{\otimes}  \kappa^{(I_{i}^{\ell_i})}_{1,0}}_{\textrm{$\ell_i-q_{i-1}$ terms}}\big)
\dot{\otimes}  \big(\underbrace{\kappa^{(O_{i}^{q_i +1})}_{0,1} 
\dot{\otimes}  \ldots \overline{\dot{\otimes}}  \kappa^{(O_{i}^{m_i})}_{0,1}}_{\textrm{$m_i-q_i$ terms}}\big)
\end{multline*}
\begin{multline*}
=
\big(\overbrace{\kappa^{(I_{i}^{1})}_{1,0} \dot{\otimes} 
\ldots \dot{\otimes} \kappa^{(I_{i}^{q_{i-1}})}_{1,0}}^{\kappa^{(\overline{O_{i-1}^{1}})}_{1,0} \dot{\otimes}
\ldots \overline{\dot{\otimes}}  \kappa^{(\overline{O_{i-1}^{q_{i-1}}})}_{1,0}}\big)
\dot{\otimes}  \big(\overbrace{\kappa^{(O_{i}^{1})}_{0,1} 
\dot{\otimes}  \ldots \dot{\otimes} \kappa^{(O_{i}^{q_i})}_{0,1}}^{\kappa^{(\overline{I_{i+1}^{1}})}_{0,1} 
\dot{\otimes}  \ldots \dot{\otimes}  \kappa^{(\overline{I_{i+1}^{q_i}})}_{0,1}}\big) \dot{\otimes}
\\
\dot{\otimes}
\big(\underbrace{\kappa^{(I_{i}^{q_{i-1}+1})}_{1,0} \dot{\otimes}
\ldots \dot{\otimes}  \kappa^{(I_{i}^{\ell_i})}_{1,0}}_{\textrm{$\ell_i-q_{i-1}$ terms}}\big)
\dot{\otimes}  \big(\underbrace{\kappa^{(O_{i}^{q_i +1})}_{0,1} 
\dot{\otimes}  \ldots \dot{\otimes}  \kappa^{(O_{i}^{m_i})}_{0,1}}_{\textrm{$m_i-q_i$ terms}}\big)
\end{multline*}

In particular the vector-valued distribution (\ref{chronologicalkappal,m}) can be written in the form
\begin{multline*}
\kappa^{1}_{\ell_{1},m_{1}}\overline{\otimes|_{{}_{q_1}}} \, \kappa^{2}_{\ell_{2},m_{2}} \,\,
\overline{\otimes|_{{}_{q_2}}}
\,\, \ldots  \,\,
\overline{\otimes|_{{}_{q_{n-2}}}} \, 
 \kappa^{k-1}_{\ell_{k-1},m_{k-1}}\overline{\otimes|_{{}_{q_{k-1}}}} \,  \kappa^{k}_{\ell_{k},m_{k}}
\\
=
\prod \limits_{i=1}^{k-1} 
\big(\underbrace{\kappa^{(O_{i}^{1})}_{0,1} 
\dot{\otimes}  \ldots \dot{\otimes}  \kappa^{(O_{i}^{q_i})}}_{\textrm{$q_i$ terms}}\big)
\otimes_{q_i}
\big(\underbrace{\kappa^{(I_{i+1}^{1})}_{1,0} \dot{\otimes}
\ldots \dot{\otimes}  \kappa^{(I_{i+1}^{q_i})}_{1,0}}_{\textrm{$q_i$ terms}}\big)
\\
\bigotimes \limits_{i=1}^{k}
\big(\underbrace{\kappa^{(I_{i}^{q_{i-1}+1})}_{1,0} \overline{\dot{\otimes}} 
\ldots \overline{\dot{\otimes}}  \kappa^{(I_{i}^{\ell_i})}_{1,0}}_{\textrm{$\ell_i-q_{i-1}$ terms}}\big)
\overline{\dot{\otimes}}  \big(\underbrace{\kappa^{(O_{i}^{q_i +1})}_{0,1} 
\overline{\dot{\otimes}}  \ldots \overline{\dot{\otimes}}  \kappa^{(O_{i}^{m_i})}_{0,1}}_{\textrm{$m_i-q_i$ terms}}\big)
\end{multline*}
\begin{multline*}
=
\prod \limits_{i=1}^{k-1} 
\big(\underbrace{\kappa^{(O_{i}^{1})}_{0,1} 
\dot{\otimes}  \ldots \dot{\otimes}  \kappa^{(O_{i}^{q_i})}}_{\textrm{$q_i$ terms}}\big)
\otimes_{q_i}
\big(\underbrace{\kappa^{(\overline{O_{i}^{1}})}_{1,0} \dot{\otimes}  
\ldots \dot{\otimes}  \kappa^{(\overline{O_{i}^{q_i}})}_{1,0}}_{\textrm{$q_i$ terms}}\big)
\\
\bigotimes \limits_{i=1}^{k}
\big(\underbrace{\kappa^{(I_{i}^{q_{i-1}+1})}_{1,0} \overline{\dot{\otimes}} 
\ldots \overline{\dot{\otimes}}  \kappa^{(I_{i}^{\ell_i})}_{1,0}}_{\textrm{$\ell_i-q_{i-1}$ terms}}\big)
\overline{\dot{\otimes}}  \big(\underbrace{\kappa^{(O_{i}^{q_i +1})}_{0,1} 
\overline{\dot{\otimes}}  \ldots \overline{\dot{\otimes}}  \kappa^{(O_{i}^{m_i})}_{0,1}}_{\textrm{$m_i-q_i$ terms}}\big),
\end{multline*}
where
\[
q_0 \overset{\textrm{df}}{=} 0.
\]
Here for each $i$
\[
(O_{i}^{1}, O_{i}^{2}, \ldots, O_{i}^{m_i}), \,\,\,\,
(I_{i}^{1}, I_{i}^{2}, \ldots, I_{i}^{\ell_i})
\]
are sequences with the values 
\[
O_{i}^{j}, I_{i}^{j} \in \{1_i, 2_i, 3_i \}
\]
in case of spinor QED. In case of more general Wick monomial $\mathcal{L}$ of $N$ free fields 
\[
O_{i}^{j}, I_{i}^{j} \in \{1_i, 2_i, 3_i, \ldots, N_i \}.
\]
The bar over the index $O_{i}^{j} \in \{1_i, 2_i, 3_i \}$ or $I_{i}^{j} \in \{1_i, 2_i, 3_i \}$ denotes the index
with which it can be contracted, \emph{i.e.} in case of spinor QED
\[
\overline{1_i} = 3_i, \,\, \overline{3_i} = 1_i, \,\, \overline{2_i} = 2_i.
\]
By definition the space-time variable of the kernel, respectively,
\begin{eqnarray*}
\kappa^{(O_{i}^{1})}_{0,1} 
\dot{\otimes}  \ldots \dot{\otimes} \kappa^{(O_{i}^{q_i})}_{0,1} \,\,\,
\textrm{is} \,\,\, x_i,
\\
\kappa^{(I_{i+1}^{1})}_{1,0} \dot{\otimes} 
\ldots \dot{\otimes}  \kappa^{(I_{i+1}^{q_i})}_{1,0} \,\,\,
\textrm{is} \,\,\, x_{i+1},
\\
\kappa^{(\overline{O_{i}^{1}})}_{1,0} \dot{\otimes} 
\ldots \dot{\otimes}  \kappa^{(\overline{O_{i}^{q_i}})}_{1,0}
= \kappa^{(I_{i+1}^{1})}_{1,0} \dot{\otimes} 
\ldots \dot{\otimes}  \kappa^{(I_{i+1}^{q_i})}_{1,0}
\,\,\,
\textrm{is} \,\,\, x_{i+1},
\\
\big(\kappa^{(I_{i}^{q_{i-1}+1})}_{1,0} \dot{\otimes} 
\ldots \dot{\otimes}  \kappa^{(I_{i}^{\ell_i})}_{1,0}\big)
\dot{\otimes}  \big(\kappa^{(O_{i}^{q_i +1})}_{0,1} 
\dot{\otimes}  \ldots \dot{\otimes}  \kappa^{(O_{i}^{m_i})}_{0,1}\big)
\,\,\,
\textrm{is} \,\,\, x_{i}.
\end{eqnarray*}
Summing up the vector-valued distribution (\ref{chronologicalkappal,m}) can be canonically written as the product
\begin{multline}\label{FactorizationOfchronologicalkappal,m}
\kappa^{1}_{\ell_{1},m_{1}}\overline{\otimes|_{{}_{q_1}}} \, \kappa^{2}_{\ell_{2},m_{2}} \,\,
\overline{\otimes|_{{}_{q_2}}}
\,\, \ldots  \,\,
\overline{\otimes|_{{}_{q_{k-2}}}} \, 
 \kappa^{k-1}_{\ell_{k-1},m_{n-1}}\overline{\otimes|_{{}_{q_{k-1}}}} \,  \kappa^{k}_{\ell_{k},m_{k}}
\\
=
\prod \limits_{i=1}^{k-1} 
\big(\underbrace{\kappa^{(O_{i}^{1})}_{0,1} 
\dot{\otimes}  \ldots\dot{\otimes}  \kappa^{(O_{i}^{q_i})}_{0,1}}_{\textrm{$q_i$ terms}}\big)
\otimes_{q_i}
\big(\underbrace{\kappa^{(\overline{O_{i}^{1}})}_{1,0} \dot{\otimes} 
\ldots \dot{\otimes}  \kappa^{(\overline{O_{i}^{q_i}})}_{1,0}}_{\textrm{$q_i$ terms}}\big)
\\
\bigotimes \limits_{i=1}^{k}
\big(\underbrace{\kappa^{(I_{i}^{q_{i-1}+1})}_{1,0} \overline{\dot{\otimes}} 
\ldots \overline{\dot{\otimes}}  \kappa^{(I_{i}^{\ell_i})}_{1,0}}_{\textrm{$\ell_i-q_{i-1}$ terms}}\big)
\overline{\dot{\otimes}}  \big(\underbrace{\kappa^{(O_{i}^{q_i +1})}_{0,1} 
\overline{\dot{\otimes}}  \ldots \overline{\dot{\otimes}}  \kappa^{(O_{i}^{m_i})}_{0,1}}_{\textrm{$m_i-q_i$ terms}}\big),
\end{multline}
where
\[
q_0 \overset{\textrm{df}}{=} 0,
\]
for the scalar valued translationally invariant 
\begin{equation}\label{ScalarPartOfchronologicalkappal,m}
\prod \limits_{i=1}^{k-1} 
\big(\underbrace{\kappa^{(O_{i}^{1})}_{0,1} 
\dot{\otimes}  \ldots \dot{\otimes}  \kappa^{(O_{i}^{q_i})}_{0,1}}_{\textrm{$q_i$ terms}}\big)
\otimes_{q_i} 
\big(\underbrace{\kappa^{(\overline{O_{i}^{1}})}_{1,0} \dot{\otimes} 
\ldots \dot{\otimes}  \kappa^{(\overline{O_{i}^{q_i}})}_{1,0}}_{\textrm{$q_i$ terms}}\big)
\end{equation}
which captures all contractions and a vector-valued tensor product distribution 
of dot products of plane wave kernels without any contractions
\begin{equation}\label{VectorPartOfchronologicalkappal,m}
\bigotimes \limits_{i=1}^{k}
\big(\underbrace{\kappa^{(I_{i}^{q_{i-1}+1})}_{1,0} \overline{\dot{\otimes}} 
\ldots \overline{\dot{\otimes}}  \kappa^{(I_{i}^{\ell_i})}_{1,0}}_{\textrm{$\ell_i-q_{i-1}$ terms}}\big)
\overline{\dot{\otimes}}  \big(\underbrace{\kappa^{(O_{i}^{q_i +1})}_{0,1} 
\overline{\dot{\otimes}}  \ldots \overline{\dot{\otimes}}  \kappa^{(O_{i}^{m_i})}_{0,1}}_{\textrm{$m_i-q_i$ terms}}\big).
\end{equation}

In case  (\ref{ScalarPartOfchronologicalkappal,m}) is causally supported
the singularity degree of the vector-valued distribution (\ref{chronologicalkappal,m}) is defined
as the singularity degree of the scalar-valued distribution (\ref{ScalarPartOfchronologicalkappal,m}) canonically related to
(\ref{chronologicalkappal,m}), and defined in the sense of  \cite{Epstein-Glaser}. 

We should emphasize here that the Wick Theorem \ref{WickThmForMasslessFields}, Subsection \ref{WickForProduct}, assures the contraction kernel
(\ref{FactorizationOfchronologicalkappal,m}) to be a well-defined element of
\[
\mathscr{L}\big(E_{1}^{\otimes \, M(q)} \otimes E_{2}^{\otimes \, M(q)}, \,  \mathscr{E}^{* \, \otimes \, k}\big)
\cong \mathscr{L}\big(\mathscr{E}^{\otimes \, k}, \, E_{1}^{* \, \otimes \, M(q)} \otimes E_{2}^{* \, \otimes \, M(q)}\big)
\]
\[
M(q) +N(q) = N_{{}_{\mathcal{L}}}k - q_1 - \ldots - q_{n-1}.
\]
It can be uniquely extended (Theorem \ref{WickThmForMassiveFields}, Subsection \ref{WickForProduct}) to an element of 
\[
\mathscr{L}\big(E_{1}^{* \, \otimes \, M(q)} \otimes E_{2}^{\otimes \, M(q)}, \,  \mathscr{E}^{* \, \otimes \, k}\big)
\cong \mathscr{L}\big(\mathscr{E}^{\otimes \, k}, \, E_{1}^{\otimes \, M(q)} \otimes E_{2}^{* \, \otimes \, M(q)}\big)
\]
in case only massive plane wave kernels $\kappa^{(1)}_{0,1}, \kappa^{(1)}_{1,0}, \ldots $ were in
(\ref{FactorizationOfchronologicalkappal,m}) (which is not the case in particular for QED $\mathcal{L}$).
The Wick Theorem \ref{WickThmForMasslessFields}, Subsection \ref{WickForProduct}, working for general $\mathcal{L}$
with massless fields and including QED interaction, also assures the
factorization (\ref{FactorizationOfchronologicalkappal,m}) to be well-defined as an element of
\[
\mathscr{L}\big(E_{1}^{\otimes \, M(q)} \otimes E_{2}^{\otimes \, M(q)}, \,  \mathscr{E}^{* \, \otimes \, k}\big)
\cong \mathscr{L}\big(\mathscr{E}^{\otimes \, k}, \, E_{1}^{* \, \otimes \, M(q)} \otimes E_{2}^{* \, \otimes \, M(q)}\big).
\]
But we shall prove it also more directly now. Indeed, for the dot product 
\[
\kappa^{(1)}_{0,1} \dot{\otimes} \ldots  \dot{\otimes} \kappa^{(N)}_{0,1} \dot{\otimes}
\kappa^{(N+1)}_{1,0} \dot{\otimes} \ldots \dot{\otimes} \ldots \dot{\otimes} \kappa^{(M)}_{1,0} 
\]
of massive or massless plane wave kernels
\[
\kappa^{(1)}_{0,1}, \ldots, \kappa^{(N)}_{0,1},
\kappa^{(N+1)}_{1,0}, \ldots, \kappa^{(M)}_{1,0}, 
\]
corresponding to free fields with single particle Gelfand triples
\[
E_{{}_{(1)}} \subset \mathcal{H}_{{}_{(1)}} \subset E_{{}_{(1)}}^{*}, \,\,\,\, \ldots
E_{{}_{(M)}} \subset \mathcal{H}_{{}_{(M)}} \subset E_{{}_{(M)}}^{*},
\]
the map (for each fixed component of the dot product)
\begin{multline*}
E_{{}_{(1)}} \otimes \ldots \otimes E_{{}_{(M)}} \ni
\xi_{{}_{(1)}} \otimes \ldots \otimes \xi_{{}_{(M)}} \longmapsto 
\\
\longmapsto 
\big(\kappa^{(1)}_{0,1} \dot{\otimes} \ldots  \dot{\otimes} \kappa^{(N)}_{0,1} \dot{\otimes}
\kappa^{(N+1)}_{1,0} \dot{\otimes} \ldots \dot{\otimes} \ldots \dot{\otimes} \kappa^{(M)}_{1,0}\big)
(\xi_{{}_{(1)}} \otimes \ldots \otimes \xi_{{}_{(M)}}) \in \mathcal{O}_{M}(\mathbb{R}^4; \mathbb{C})
\end{multline*}
is continuous for the Schwartz' operator topology in $\mathcal{O}_{M}$.
This means that the map 
\begin{multline*}
E_{{}_{(1)}} \otimes \ldots \otimes E_{{}_{(M)}} \ni
\xi_{{}_{(1)}} \otimes \ldots \otimes \xi_{{}_{(M)}} \longmapsto 
\\
\longmapsto 
\big(\kappa^{(1)}_{0,1} \dot{\otimes} \ldots  \dot{\otimes} \kappa^{(N)}_{0,1} \dot{\otimes}
\kappa^{(N+1)}_{1,0} \dot{\otimes} \ldots \dot{\otimes} \ldots \dot{\otimes} \kappa^{(M)}_{1,0}\big)
(\xi_{{}_{(1)}} \otimes \ldots \otimes \xi_{{}_{(M)}}) 
\\
\in \textrm{Algebra of multipliers of }\mathscr{E}^{*}
\end{multline*}
is continuous in both cases $\mathscr{E} = \mathcal{S}(\mathbb{R}^4)$, $\mathscr{E}= \mathcal{S}^{00}(\mathbb{R}^4) \subset  \mathcal{S}(\mathbb{R}^4)$
(by Subsection \ref{SA=S0}) with the Schwartz' operator topology in the algebra of multipliers of $\mathscr{E}^{*}$.

Indeed the continuity 
\[
E_{{}_{(1)}} \ni \xi_{{}_{(i)}} \longmapsto \kappa^{(i)}_{\ell,m}(\xi_{{}_{(i)}}) \in \mathcal{O}_{M}(\mathbb{R}^4; \mathbb{C}), \,\,\, (\ell,m) = (0,1), (1,0),
\]
has been proved in Lemma \ref{kappa0,1,kappa1,0psi}, Subsection \ref{psiBerezin-Hida} (massive case)
and in Lemma \ref{kappa0,1,kappa1,0ForA}, Subsection \ref{A=Xi0,1+Xi1,0} (massless case, including the free electromagnetic
potential field). The general case for dot products  of more plane wave kernels follows from the obvious product formula
\begin{multline*}
\big(\kappa^{(1)}_{0,1} \dot{\otimes} \ldots  \dot{\otimes} \kappa^{(N)}_{0,1} \dot{\otimes}
\kappa^{(N+1)}_{1,0} \dot{\otimes} \ldots \dot{\otimes} \ldots \dot{\otimes} \kappa^{(M)}_{1,0}\big)
(\xi_{{}_{(1)}} \otimes \ldots \otimes \xi_{{}_{(M)}}) 
\\
= 
\kappa^{(1)}_{0,1}(\xi_{{}_{(1)}})\cdot \ldots \cdot \kappa^{(N)}_{0,1}(\xi_{{}_{(N)}}) \cdot
\kappa^{(N+1)}_{1,0}(\xi_{{}_{(N+1)}}) \cdot \ldots \cdot \kappa^{(M)}_{1,0}(\xi_{{}_{(M)}})
\end{multline*}
and the fact that each factor
\[
\kappa^{(i)}_{\ell,m}(\xi_{{}_{(i)}}),  \,\,\,\,\,\, (\ell,m) = (0,1), (1,0),
\]
is a multiplier of $\mathscr{E}^*$ continuously depending on $\xi_{{}_{(i)}}$.  

In particular, writing
\[
\kappa_{\ell,m}, \,\,\,\,\,\, \ell= \ell_1 +\ldots +\ell_k - q_1 - \ldots - q_{k-1}, \,\,\, m = m_1 + \ldots + m_k - q_{k-1} - \ldots - q_{k-1}
\]
for the kernel of the vector valued factor (\ref{VectorPartOfchronologicalkappal,m}) without contractions,
we see that it defines a continuous map
\[
E_{1}^{\otimes \, M(q)} \otimes E_{2}^{\otimes \, M(q)} \ni
\xi \longmapsto \kappa_{\ell,m}(\xi) \in \mathcal{O}_M\big([\mathbb{R}^4]^{\times \, k}\big)
\]
into the Schwartz' algebra $\mathcal{O}_M\big([\mathbb{R}^4]^{\times \, k}\big)$ of multipliers of
$\mathcal{S}\big([\mathbb{R}^4]^{\times \, k}\big)^*$, with the Schwartz operator topology on
$\mathcal{O}_M\big([\mathbb{R}^4]^{\times \, k}\big)$. Because the scalar factor
(\ref{ScalarPartOfchronologicalkappal,m}) is a well-defined element of $\mathscr{E}^{* \otimes \, k}$
then it follows that the kernel of the whole product (\ref{FactorizationOfchronologicalkappal,m})
defines a continuous map
\begin{multline*}
E_{1}^{\otimes \, M(q)} \otimes E_{2}^{\otimes \, M(q)} \ni
\xi \longmapsto
\\
\longmapsto
\big(\underbrace{\kappa^{(O_{i}^{1})}_{0,1}
\dot{\otimes} \ldots \dot{\otimes} \kappa^{(O_{i}^{q_i})}_{0,1}}_{\textrm{$q_i$ terms}}\big)
\otimes_{q_i}
\big(\underbrace{\kappa^{(\overline{O_{i}^{1}})}_{1,0} \dot{\otimes}
\ldots \dot{\otimes} \kappa^{(\overline{O_{i}^{q_i}})}_{1,0}}_{\textrm{$q_i$ terms}}\big) \cdot
\kappa_{\ell,m}(\xi) \in \mathscr{E}^{* \otimes \, k}
\end{multline*}
and the product is well-defined because $\kappa_{\ell,m}(\xi)$ is a multiplier of $\mathscr{E}^{* \otimes \, k}$.
In particular the kernel function (\ref{FactorizationOfchronologicalkappal,m}) can be written as the
ordinary product of the scalar kernel (depending only on space-time variables) and of a vector-valued
kernel depending on space-time and momentum variables. 

From this it also immediately follows that the multiplication of a Wick product of free fields, say respectively of
$x_1, \ldots, x_n$ space-time variables, by the product $\theta(x_1-x_n)\theta(x_2-x_n) \ldots \theta(x_{n-1}-x_n)$
of $\theta$-functions is a well-defined integral kernel operator with vector-valued kernel. 

The factorization (\ref{FactorizationOfchronologicalkappal,m}) of the contraction kernel
understood as an element of
\[
\mathscr{L}\big(\mathscr{E}^{\otimes \, k}, \, E_{1}^{*\, \otimes \, M(q)} \otimes E_{2}^{* \, \otimes \, M(q)}\big)
\cong
\mathscr{L}\big(E_{1}^{\otimes \, M(q)} \otimes E_{2}^{\otimes \, M(q)}, \,\, \mathscr{E}^{* \otimes \, k} \big),
\]
has practical computational
consequences. In particular its support, understood as $E_{1}^{*\, \otimes \, M(q)} \otimes E_{2}^{* \, \otimes \, M(q)}$-valued
distribution on the space-time test space $\mathscr{E}^{\otimes \, n}$, has the support which is not greater than
the support of the scalar factor (\ref{ScalarPartOfchronologicalkappal,m}). It is immediately seen
that the scalar factor (\ref{ScalarPartOfchronologicalkappal,m}) is translationally invariant. Moreover, the pairing functions
\[
-i\big[\mathbb{A}^{(i)(-)}, \mathbb{A}^{(j)(+)} \big]_\mp = \kappa^{(i)}_{0,1} \otimes_1 \kappa^{(j)}_{1,0} = D^{(+)}
\]
of the free fields $\mathbb{A}^{ (i)}, \mathbb{A}^{ (j)}$
are equal to the $1$-contractions
\[
\kappa^{(i)}_{0,1} \otimes_1 \kappa^{(j)}_{1,0}
\]
of the corresponding kernels, have the following general form (for the massive scalar field of mass $m$, compare \cite{Bogoliubov_Shirkov})
\begin{multline}\label{D(+)}
D^{(+)}(x) = {\textstyle\frac{1}{4\pi}}\varepsilon(x_0)\delta(x\cdot x)
- {\textstyle\frac{m i}{8\pi \sqrt{x\cdot x}}} \theta(x\cdot x)
\big[
N_1(m \sqrt{x\cdot x}) - i\varepsilon(x_0) J_1(m\sqrt{x\cdot x}) 
\big]
\\
-\theta(-x\cdot x) {\textstyle\frac{m i}{4\pi^2 \sqrt{-x\cdot x}}} K_1(m\sqrt{- x\cdot x}), 
\end{multline}
\begin{multline}\label{D(-)}
D^{(-)}(x) = {\textstyle\frac{1}{4\pi}}\varepsilon(x_0)\delta(x\cdot x)
+ {\textstyle\frac{m i}{8\pi \sqrt{x\cdot x}}} \theta(x\cdot x)
\big[
N_1(m \sqrt{x\cdot x}) + i\varepsilon(x_0) J_1(m\sqrt{x\cdot x}) 
\big]
\\
+\theta(-x\cdot x) {\textstyle\frac{m i}{4\pi^2 \sqrt{-x\cdot x}}} K_1(m\sqrt{- x\cdot x}).
\end{multline}
Here
\[
\varepsilon(x_0) = \textrm{sgn}(x_0) =  \theta(x_0) - \theta(-x_0),
\]
$J_\nu$ are the Bessel functions of the first kind, $N_\nu$ -- the Neumann functions 
(or the Bessel functions of the second kind) and $K_\nu$ -- the Hankel functions (or the Bessel functions
of the third kind). In order to obtain the pairings for other massive fields we only need to apply a corresponding invariant differential
operator to (\ref{D(+)}) and (\ref{D(-)}), e.g. the operator $i\gamma^\mu \partial_\mu +m$ in case of the Dirac field,
or pass to the limit $m\rightarrow 0$ in case of a massless field. Therefore in each case the singular part of  $D^{(+)}$ and of $D^{(-)}$
is located at the light cone. 

From (\ref{D(+)}) and (\ref{D(-)}) we see that the commutation functions, \emph{i.e.} the sums $D^{(+)}+D^{(-)}$ of the pairing functions, 
have causal support lying within the closure of the interiors of the past
and forward light cones. Using the support property of (\ref{D(+)}) and (\ref{D(-)}), and thus of the pairings, or the $1$-contractions, 
one can rather easily prove that the scalar factors in the Wick monomials of the Wick decomposition of
\[
D_{(2)}(x_1,x_2) = \mathcal{L}(x_2)\mathcal{L}(x_1) - \mathcal{L}(x_1)\mathcal{L}(x_2),
\]
are causally supported. These causal scalar factors
are linear combinations (in this case of two factors or products of pairings) of the form
(\ref{ScalarPartOfchronologicalkappal,m}). More generally, using (\ref{D(+)}) and (\ref{D(-)}) one can show that the
the difference of the products of even number of pairings (or sum of products of odd number of pairings), has causal support.

As we have proved in Subsection \ref{WickForProduct}, the scalar factor distribution (\ref{ScalarPartOfchronologicalkappal,m}) 
with contraction $q=q_1>1$ (here we have $k=2$
and only one contraction) can be written as a $\epsilon\rightarrow 0$ limit of the product of
of a natural $\epsilon$-approximations of $1$-contractions (the orbits of the free fields on the Minkowski space-time are non-compact,
as we have already mentioned in Subsection \ref{WickForProduct}),
we can nonetheless use the support properties of $1$-contractions (\ref{D(+)}) and (\ref{D(-)}) 
to investigate the support of (\ref{ScalarPartOfchronologicalkappal,m})
applying a limit procedure to the cut out at finite momemta of the $1$-contractions. Namely, we
cut out each of the orbits at a finite radius of the momentum. After this cut of the momentum the scalar
factor (\ref{ScalarPartOfchronologicalkappal,m}), equal to the product of $q_i$-contractions
(\ref{ScalarPartOfchronologicalkappal,m}), can be written as the product of $1$-contractions
with the cut out momenta in their integral expressions (the reader may compare this method
with the regularization prescription of the more conventional approach, \cite{Bogoliubov_Shirkov}, \S 19.1). 
Next we evaluate the product on a test function
and finally we pass to the infinite radius limit. 
It is regular enough at zero to have well defined splitting into retarded and advanced part (as we will prove in this Section)
with the freedom depending on finite number of constants depending solely on the singularity degree $\omega$
of (\ref{ScalarPartOfchronologicalkappal,m}). We do not enter here into the detailed analysis of the support
of the products of pairings or their causal symmetric or antisymmetric parts, as these properties are known.

Because the support of the operator $D_{(n)}$ is causal, then, both, in case of the computation of the retarded part
\[
\theta(x_1-x_2)\mathcal{L}(x_1)\mathcal{L}(x_2) 
\]
of $\mathcal{L}(x_1)\mathcal{L}(x_2)$ in
\[
S_2(x_1,x_2) = \theta(x_1-x_2)\mathcal{L}(x_1)\mathcal{L}(x_2) + \theta(x_2-x_1)\mathcal{L}(x_2)\mathcal{L}(x_1), 
\]
as well as in the computation
of the retarded part $R_{(n)}=\theta D_{(n)}$ of $D_{(n)}$ in $S_n = R_{(n)} - R'_{(n)}$ we can use
the splitting method of \cite{Scharf} (inspired by \cite{Epstein-Glaser}) applied to
the scalar factors (\ref{ScalarPartOfchronologicalkappal,m}) of the kernels of 
$\mathcal{L}(x_1)\mathcal{L}(x_2)$ or, respectively, of 
$D_{(n)}$, into the retarded and advanced parts, although each of these factors taken separately is not causal.
Recall, please, that only full scalar factors of the contributions collecting all terms in  $D_{(n)}$ which multiply 
a fixed Wick monomial, are causal, by the causality of $D_{(n)}$.
The terms without contractions in the Wick product expansion of $\mathcal{L}(x_1)\mathcal{L}(x_2)$
or, respectively, of $D_{(n)}(x_1, \ldots, x_n)$, can be multiplied by $\theta(x_1-x_2)$, or respectively, by
$\theta(x_1-x_n) \ldots \theta(x_{n-1}-x_n)$, as we have already seen. 

The singularity degree $\omega$ of (\ref{ScalarPartOfchronologicalkappal,m}) with $k=2$ (for the kernels of $\mathcal{L}(x_1)\mathcal{L}(x_2)$), 
in the sense of \cite{Epstein-Glaser}, 
determines canonically the subspace of $\mathscr{E}^k$ (here with $k=2$ for the kernels of
$\mathcal{L}(x_1)\mathcal{L}(x_2)$, or $k>2$ for the kernels of $D_{(k)}$) 
on which the scalar distribution 
(\ref{ScalarPartOfchronologicalkappal,m}), multiplied by the step theta function $\theta(x_1-x_2)$ (or, respectively, by 
$\theta(x_1-x_k) \ldots \theta(x_{k-1}-x_k)$ for  $R_{(k)}=\theta D_{(k)}$),
is given  by an ordinary contraction integral which is absolutely convergent, and defines a retarded part of 
(\ref{ScalarPartOfchronologicalkappal,m}) on this subspace. Moreover, by translational
invariance it determines
the retarded part of  (\ref{ScalarPartOfchronologicalkappal,m}) up to 
\begin{equation}\label{SimpleEpsteinGlaserReminder}
\delta^{\omega}_{{}_{1, \ldots, k-1; \,\, k}}(x_1, \ldots, x_k)
\overset{\textrm{df}}{=}
\sum \limits_{|\alpha|=0}^{\omega} C_\alpha D^\alpha \big[ \delta(x_1 - x_k) \delta(x_2-x_k) \ldots \delta(x_{k-1} -x_k)\big]
\end{equation}
with a finite sequence of arbitrary constants
$C_0, \ldots, C_\omega$, compare \cite{Epstein-Glaser}.

Let us analyze in more details the particular case of the \emph{limit contraction} $\otimes||_{{}_{q}}$
(\ref{DoubleLimitContraction})  in which we have only two kernels $\theta \kappa'_{\ell',m'}$
and $\kappa''_{\ell'',m''}$. Let $\kappa'_{\ell',m'}$
and $\kappa''_{\ell'',m''}$ be kernels of the Fock expansion, respectively, of $\mathcal{L}(x)$ and $\mathcal{L}(y)$.  
Recall that we replace them with $\theta_\varepsilon \kappa'_{\epsilon \,\, \ell',m'}$,
$\kappa''_{\epsilon'' \,\, \ell'',m''}$,
then make the ordinarily $q$-contraction $\otimes_q$, $\theta_\varepsilon \kappa'_{\epsilon \,\, \ell',m'} \otimes_q \,
\kappa''_{\epsilon'' \,\, \ell'',m''}$, next project on the closed subspace of the space-time test space $\mathscr{E}^{\otimes \, 2}$ with the projection $\Omega$, 
depending on the singularity degree of the kernel\footnote{Recall that we use the limit contraction sign $\otimes|_{{}_{q}}$ exchangeably with
the ordinary contraction sign $\otimes_q$ because the contraction integral is convergent even for kernels being equal to pointed products of massless kernels,
compare Subsection \ref{WickForProduct}.}
\[
\kappa'_{\ell',m'} \otimes|_q  \,  \kappa''_{\ell'',m''} = \kappa'_{\ell',m'} \otimes_q  \,  \kappa''_{\ell'',m''},
\]
and finally integrate.

Namely, for the kernels $\kappa'_{\ell',m'}, \kappa''_{\ell'',m''}$  of the Fock 
expansion of the operator $\Xi' = \mathcal{L}$,  we have
\begin{equation}\label{kappa'otimes||kappa'}
\theta \kappa'_{\ell',m'} \otimes||_{{}_{q}}  \,  \kappa''_{\ell'',m''} \overset{\textrm{df}}{=}
\theta \kappa'_{\epsilon \,\, \ell',m'} \otimes_q \, \kappa''_{\epsilon \,\, \ell'',m''} \circ \Omega
\end{equation}
in
\[
\mathscr{L}\big(\mathscr{E} \otimes \mathscr{E} , E_{j_1}^{*} \ldots  \otimes \ldots E_{j_{\ell'+\ell''-q}}^{*} \otimes E_{j_{\ell'+\ell''-q+1}}^{*} 
\ldots \otimes \ldots E_{j_{\ell'+\ell''+m'+m''-q}}^{*}  \big)
\]
\[
\cong \mathscr{L}\big(E_{j_1} \ldots  \otimes \ldots E_{j_{\ell'+\ell''-q}} \otimes E_{j_{\ell'+\ell''-q+1}} 
\ldots \otimes \ldots E_{j_{\ell'+\ell''+m'+m''-q}}, \, \mathscr{E}^* \otimes \mathscr{E}^*   \big),
\]
which, can also be undertand as a limit, in which all exponents of the non contracted massless free field kernels
are replaced with the exponents of massive kernels, next integration and finally by passing to the limit zero with
the auxiliary mass.  This limit is convergent, as we have shown in the previous Subsection, where existence of the retarded part 
of the scalar contractions (\emph{i.e.} products of pairings, in which no non contracted varaibles are present) was assumed. This 
assumption was not arbitrary, as it was proved to hold earlier by Epstein and Glaser, compare also \cite{Scharf}.

The limit (\ref{kappa'otimes||kappa'}) does exist because of the presence of the operator $\Omega$ which projects on the closed subspace
on which the convergence holds. If the operator $\Omega$ was removed from (\ref{kappa'otimes||kappa'}), then the limit
(\ref{kappa'otimes||kappa'}) would not be existing in general for the kernel 
\[
\kappa'_{\ell',m'} \otimes_q  \,  \kappa''_{\ell'',m''}
\]
with $q>1$, whose singularity degree $\omega>0$.

Let us explain this fact in more details in case of particular but generic examples. In these examples we consider 
$\kappa'_{\ell',m'}, \kappa''_{\ell'',m''}$ being equal not only to simple dot products of just three massless or massive plane wave kernels 
(as in QED), but a more general situation of simple products of arbitrary large number of plane wave kernels, massless or massive.
We start with the analysis of the limit (\ref{kappa'otimes||kappa'}) in case in which the projection operator $\Omega$ can be removed.

Let in particular $\mathbb{A}', \mathbb{A}''$ be two massless free fields
\[
\mathbb{A}' = \Xi_{0,1}(\kappa'_{0,1}) + \Xi_{1,0}(\kappa'_{1,0}) = \mathbb{A}'^{(-)} + \mathbb{A}'^{(+)},
\]
\[
\mathbb{A}'' = \Xi_{0,1}(\kappa''_{0,1}) + \Xi_{1,0}(\kappa''_{1,0}) = \mathbb{A}''^{(-)} + \mathbb{A}''^{(+)},
\]
with the corresponding single particle Gelfand triples: $E_{{}_{'}} \subset \mathcal{H}' \subset E_{{}_{'}}^{*}$
and $E_{{}_{''}} \subset \mathcal{H}'' \subset E_{{}_{''}}^{*}$ and space-time test spaces $\mathscr{E}_{{}_{'}}, \mathscr{E}_{{}_{'}}$,
and the respective orbits $\mathscr{O}'=\mathscr{O}''= \mathscr{O}_{1,0,0,1}$ in momentum space.
Recall that the standard nuclear spaces
\[
E_{{}_{''}} = \mathcal{S}^{0}(\mathbb{R}^3; \mathbb{C}^{d''})
=\mathcal{S}_{\oplus A_{(3)}}(\mathbb{R}^3; \mathbb{C}^{d''}) , E_{{}_{'}} = \mathcal{S}^{0}(\mathbb{R}^3; \mathbb{C}^{d'})
=\mathcal{S}_{\oplus A_{(3)}}(\mathbb{R}^3; \mathbb{C}^{d'}),
\]
\[
\mathscr{E}_{{}_{''}} = \mathcal{S}^{00}(\mathbb{R}^4; \mathbb{C}^{d''})
=\mathcal{S}_{\mathscr{F}^{-1}\oplus A_{(4)}\mathscr{F}}(\mathbb{R}^4; \mathbb{C}^{d''})
, \mathscr{E}_{{}_{'}} = \mathcal{S}^{00}(\mathbb{R}^4; \mathbb{C}^{d'})
=\mathcal{S}_{\mathscr{F}^{-1}\oplus A_{(4)}\mathscr{F}}(\mathbb{R}^4; \mathbb{C}^{d'}),
\]
has been constructed in Section \ref{white-noise-proofs}, and that
\[
\kappa'_{0,1}, \kappa'_{1,0} \in \mathscr{L} \big(\mathscr{E}_{{}_{'}}, E_{{}_{'}} \big) \cong E_{{}_{'}} \otimes \mathscr{E}_{{}_{'}}^{*},
\]
\[
\kappa''_{0,1}, \kappa''_{1,0} \in \mathscr{L} \big(\mathscr{E}_{{}_{''}}, E_{{}_{''}} \big) \cong E_{{}_{''}} \otimes \mathscr{E}_{{}_{''}}^{*}.
\]
\emph{i.e.} the maps
\begin{eqnarray*}
\mathscr{E}_{{}_{''}} \ni \phi \mapsto \kappa''_{0,1}(\phi) = u'' \check{\widetilde{\phi}}\big|_{{}_{\mathscr{O}_{1,0,0,1}}}
\in E_{{}_{''}},
\\
\mathscr{E}_{{}_{''}} \ni \phi \mapsto \kappa''_{1,0}(\phi) = v'' \widetilde{\phi}\big|_{{}_{\mathscr{O}_{1,0,0,1}}}
\in E_{{}_{''}},
\\
\mathscr{E}_{{}_{'}} \ni \phi \mapsto \kappa'_{0,1}(\phi) = u' \check{\widetilde{\phi}}\big|_{{}_{\mathscr{O}_{1,0,0,1}}}
\in E_{{}_{'}},
\\
\mathscr{E}_{{}_{'}} \ni \phi \mapsto \kappa'_{1,0}(\phi) = v' \widetilde{\phi}\big|_{{}_{\mathscr{O}_{1,0,0,1}}}
\in E_{{}_{'}}
\end{eqnarray*}
are continuous. Because $\boldsymbol{\p} \mapsto u''(\boldsymbol{\p})$, $\boldsymbol{\p} \mapsto v''(\boldsymbol{\p})$,
$\boldsymbol{\p} \mapsto u'(\boldsymbol{\p})$, $\boldsymbol{\p} \mapsto v'(\boldsymbol{\p})$, are multipliers resp. of $E_{{}_{''}}$, $E_{{}_{'}}$, and because restriction to the respective orbit $\mathscr{O}'=\mathscr{O}''= \mathscr{O}_{1,0,0,1}$ is continuous as a map between the respective test nuclear spaces (Subsection \ref{Lop-on-E}), then
\begin{eqnarray*}
\theta \kappa''_{0,1}(\phi) = u'' \check{\widetilde{\theta\phi}}\big|_{{}_{\mathscr{O}_{1,0,0,1}}}
= u'' \check{\widetilde{\theta} \ast \widetilde{\varphi}}\big|_{{}_{\mathscr{O}_{1,0,0,1}}}
\in E_{{}_{''}}^{*},
\\
\theta \kappa''_{1,0}(\phi) = v'' \widetilde{\theta \phi}\big|_{{}_{\mathscr{O}_{1,0,0,1}}}
= v'' \widetilde{\theta} \ast \widetilde{\phi}\big|_{{}_{\mathscr{O}_{1,0,0,1}}}
\in E_{{}_{''}}^{*},
\\
\theta \kappa'_{0,1}(\phi) = u' \check{\widetilde{\theta \phi}}\big|_{{}_{\mathscr{O}_{1,0,0,1}}}
= u' \check{\widetilde{\theta} \ast \widetilde{\phi}}\big|_{{}_{\mathscr{O}_{m,0,0,0}}}
\in E_{{}_{'}}^{*},
\\
\theta \kappa'_{1,0}(\phi) = v' \widetilde{\theta \phi}\big|_{{}_{\mathscr{O}_{1,0,0,1}}}
= v' \widetilde{\theta} \ast \widetilde{\phi}\big|_{{}_{\mathscr{O}_{1,0,0,1}}}.
\in E_{{}_{'}}^{*}
\end{eqnarray*}
Here $\theta(x) = \theta(x_0)$ is the standard step theta function.
It follows thus in particular that
\[
\theta \kappa''_{0,1}, \theta \kappa''_{1,0} \in \mathscr{L} \big(\mathscr{E}_{{}_{''}}, E_{{}_{''}}^{*} \big) \cong E_{{}_{''}}^{*} \otimes \mathscr{E}_{{}_{''}}^{*}.
\]
Here $\theta \kappa'_{0,1}$ represents distribution with the kernel
\[
\theta \kappa''_{0,1}(s'', \boldsymbol{\p}''; a,x) \overset{\textrm{df}}{=} \theta(x_0) \, \kappa''_{0,1}(s'', \boldsymbol{\p}''; a,x).
\]
Similarly if we replace $\theta$ function $x \mapsto \theta(x) = \theta(x_0)$ by its translation $\theta_y: x \mapsto \theta(x-y)$, we have
\begin{eqnarray*}
\theta_y \kappa'_{0,1}(\phi) = \kappa'_{0,1}(\theta_y \phi)
= e^{ip_0y_0} u' \check{\widetilde{\theta} \ast \widetilde{\phi}}\big|_{{}_{\mathscr{O}_{1,0,0,1}}}
\textrm{in multipliers of} \,  E_{{}_{''}}^{*} \otimes \mathscr{E}^*,
\\
\theta_y \kappa'_{1,0}(\phi) = e^{-ip_0y_0} v' \widetilde{\theta} \ast \widetilde{\phi}\big|_{{}_{\mathscr{O}_{1,0,0,1}}}
\textrm{in multipliers of} \, E_{{}_{''}}^{*}\otimes \mathscr{E}^*,
\\
\theta_y \kappa''_{0,1}(\phi)
= e^{ip_0y_0} u'' \check{\widetilde{\theta} \ast \widetilde{\phi}}\big|_{{}_{\mathscr{O}_{1,0,0,1}}}
\textrm{in multipliers of} \, E_{{}_{'}}^{*}\otimes \mathscr{E}^*,
\\
\theta \kappa''_{1,0}(\phi)
= e^{-ip_0y_0} v'' \widetilde{\theta} \ast \widetilde{\phi}\big|_{{}_{\mathscr{O}_{1,0,0,1}}}
\textrm{in multipliers of} \, E_{{}_{'}}^{*}\otimes \mathscr{E}^*,
\end{eqnarray*}
depending continuously on $\phi$.
Note here that, because of the restriction to the orbit, the zero momentum component $p_0$ is the function of the spatial
momentum $\boldsymbol{\p}$, so that $(p_0, \boldsymbol{\p})$ lies on the respective orbit, an thus $p_0 = p_0(\boldsymbol{\p}) = | \boldsymbol{p}|$
such that $(p_0(\boldsymbol{\p}), \boldsymbol{\p})$ belongs to $\mathscr{O}_{1,0,0,1}$.
This follows because the convolution of an element of $\mathscr{E}_{{}_{'}}$ with an element of $\mathscr{E}_{{}_{'}}^*$ is a well-defined
element of $\mathscr{E}_{{}_{'}}^*$, and restriction to the orbit $\mathscr{O}_{1,0,0,1}$ is a continuous
map $\mathscr{E}_{{}_{'}} \rightarrow {E}_{{}_{'}}$, compare Section \ref{white-noise-proofs}. Similarly,
the convolution of an element of $\mathscr{E}_{{}_{''}}$ with an element of $\mathscr{E}_{{}_{''}}^*$ is a well-defined
element of $\mathscr{E}_{{}_{''}}^*$, and restriction to the cone orbit $\mathscr{O}_{1,0,0,1}$ is a continuous
map $\mathscr{E}_{{}_{''}} \rightarrow {E}_{{}_{'}}$, compare Section \ref{white-noise-proofs}, and especially Subsection \ref{Lop-on-E}.
Therefore, the following contractions make sense
\begin{multline*}
\theta_{\varepsilon \, y} \kappa'_{0,1} \otimes_1 \kappa'_{1,0}(\phi \otimes \varphi)
\overset{\textrm{df}}{=}
\\
\int \limits_{\mathscr{O}_{1,0,0,1}} \theta_\varepsilon(x_0-y_0) \kappa'_{0,1}(s', \boldsymbol{\p}'; a,x) \kappa'_{1,0}(s', \boldsymbol{\p}'; b,y)
\, \phi^a \otimes \varphi^b(x,y) \, \ud^3 \boldsymbol{\p}' \, \ud^4 x \ud^4 y
\\
\overset{\varepsilon \rightarrow 0}{\longrightarrow}
\int \limits_{\mathscr{O}_{1,0,0,1}} \theta(x_0-y_0) \kappa'_{0,1}(s'', \boldsymbol{\p}'; a,x) \kappa'_{1,0}(s', \boldsymbol{\p}'; b,y)
\, \phi^a \otimes \varphi^b(x,y) \, \ud^3 \boldsymbol{\p}' \, \ud^4 x \ud^4 y
\\
\int \limits_{\mathscr{O}_{1,0,0,1}} \theta(x_0-y_0) \kappa'_{0,1}(s', \boldsymbol{\p}'; a,x) \phi^a(x) \, \kappa'_{1,0}(s', \boldsymbol{\p}'; b,y)
\, \varphi^b(y) \, \ud^3 \boldsymbol{\p}' \, \ud^4 x \ud^4 y
\\
=
\big( \theta_{y} \kappa'_{0,1} \otimes_1 \kappa'_{1,0}\big)(\phi \otimes \varphi)
=
\int \textrm{ret} \, [\mathbb{A}'^{(-) \, a}(x), \mathbb{A}'^{(+) \, b}(y)] \, \phi^a \otimes \varphi^b(x,y) \, \ud^4 x \ud^4 y,
\end{multline*}
and give well-defined distributions, or
\[
\theta_{\varepsilon, y} \kappa'_{0,1} \otimes_1 \kappa'_{1,0}
\overset{\varepsilon \rightarrow 0}{\longrightarrow}
\theta_y \kappa'_{0,1} \otimes_1 \kappa'_{1,0} =
\textrm{ret} \quad \underbracket{\mathbb{A}' \mathbb{A}'}.
\]
Similarly
\[
\theta_{\varepsilon, y} \kappa''_{0,1} \otimes_1 \kappa''_{1,0}
\overset{\varepsilon \rightarrow 0}{\longrightarrow}
\theta_y \kappa''_{0,1} \otimes_1 \kappa''_{1,0} = \,\,
\textrm{ret} \quad \underbracket{\mathbb{A}'' \mathbb{A}''}.
\]
Below we give another proof.

This particular case of $1$-contraction of massless kernels is however trivial. Let us pass to the case of $2$-contraction
of two kernels each being equal to the dot product $\dot{\otimes}$ of two massless kernels, one being multiplied
by the theta function.
In order to analyze this case we have to look more carefully at the concrete form of the
distribution $\widetilde{\theta} \ast \widetilde{\phi}$, for any element
\[
\widetilde{\phi} \in \widetilde{\mathscr{E}} = \mathcal{S}(\mathbb{R}^4; \mathbb{C}^d).
\]
It turns out that each coefficient of the distribution $\widetilde{\theta} \ast \widetilde{\phi}$ can be represented by
a function which is bounded everywhere on $\mathbb{R}^4$, and smooth.
In order to show it we need to analyze more carefully the Fourier transform $\widetilde{\theta}$
of the step $\theta$ function. The Fourier transform of $\theta$, understood as a distribution, and thus all the more being a
tempered distribution, has been computed in \cite{GelfandI}, p. 360. Namely, recall that our theta function on $\mathbb{R}^4$
can be understood as
\[
\theta(x) = \theta(\boldsymbol{\x},t) = 1_{{}_{\mathbb{R}^3}} \otimes \theta(\boldsymbol{\x},t) = 1_{{}_{\mathbb{R}^3}}(\boldsymbol{\x}) \theta(t),
\]
where on the right $\theta$ is the ordinary step theta function on $\mathbb{R}$ and $1_{{}_{\mathbb{R}^3}}$ is the constant function
on $\mathbb{R}^3$ everywhere equal $1$.
We see that
\[
\widetilde{\theta}(p) = \widetilde{\theta}(\boldsymbol{\p}, p_0) = \delta(\boldsymbol{\p}) \widetilde{\theta}(p_0)
\]
where $\widetilde{\theta}$ is understood as the Fourier transform of the theta function $\theta \in \mathcal{S}(\mathbb{R}; \mathbb{C})$.
Thus using the Table of Fourier Transforms of \cite{GelfandI}, p. 360, we see that
\[
\widetilde{\theta}(p) = \widetilde{\theta}(\boldsymbol{\p}, p_0) = \delta(\boldsymbol{\p}) \frac{i}{p_0}
+ \pi \delta(\boldsymbol{\p})\delta(p_0).
\]
Here the value of the homogeneous distribution $\tfrac{i}{p_0}$ on the test function $\xi \in \mathcal{S}(\mathbb{R}; \mathbb{C})$
is defined through the \emph{regularised} integral with any $\Delta >0$
\[
\textrm{reg} \, \int {\textstyle\frac{i}{p_0}} \xi(p_0) \, dp_0 \overset{\textrm{df}}{=}
\int \limits_{-\infty}^{-\Delta} {\textstyle\frac{i}{p_0}} \xi(p_0) \, dp_0
+\int \limits_{-\Delta}^{\Delta} {\textstyle\frac{i}{p_0}} \big\{\xi(p_0) - \xi(0)\big\} \, dp_0
+\int \limits_{\Delta}^{+\infty} {\textstyle\frac{i}{p_0}} \xi(p_0) \, dp_0,
\]
compare \cite{GelfandI}. Therefore, for any
\begin{equation}\label{phiInS00}
\widetilde{\phi} \in \widetilde{\mathscr{E}}
= \mathcal{S}(\mathbb{R}^4; \mathbb{C}^d),
\end{equation}
we have
\begin{multline}\label{F(theta.phi)}
\widetilde{\theta.\phi}(\boldsymbol{\p}', p'_0) = \widetilde{\theta} \ast \widetilde{\phi}(\boldsymbol{\p}', p'_0)
\\
=
\textrm{reg} \, \int {\textstyle\frac{i}{p_0}} \widetilde{\phi}(\boldsymbol{\p}', p'_{0} - p_0) \, \ud p_0
+ \pi \widetilde{\phi}(\boldsymbol{\p}', p'_0)
\\
=
\int \limits_{-\infty}^{-\Delta} {\textstyle\frac{i}{p_0}} \widetilde{\phi}(\boldsymbol{\p}', p'_{0} - p_0) \, dp_0
+\int \limits_{-\Delta}^{\Delta} {\textstyle\frac{i}{p_0}} \big\{\widetilde{\phi}(\boldsymbol{\p}', p'_{0} - p_0)
- \widetilde{\phi}(\boldsymbol{\p}', p'_{0})\big\} \, dp_0
\\
+\int \limits_{\Delta}^{+\infty} {\textstyle\frac{i}{p_0}} \widetilde{\phi}(\boldsymbol{\p}', p'_{0} - p_0) \, dp_0
+ \pi \widetilde{\phi}(\boldsymbol{\p}', p'_0).
\end{multline}
Now we can see that $\widetilde{\theta} \ast \widetilde{\phi}$ is bounded and smooth
on $\mathbb{R}^4$ because of (\ref{phiInS00}), and $\widetilde{\theta} \ast \widetilde{\phi} \in L^2$ because
$\theta. \phi \in L^2$ and the Fourier transform is unitary. Similarly,
\begin{multline*}
\widetilde{\theta_y} \ast \widetilde{\phi}(\boldsymbol{\p}', p'_0)
\\
=
\textrm{reg} \,
\int {\textstyle\frac{ie^{-ip_0y_0}}{p_0}} \widetilde{\phi}(\boldsymbol{\p}', p'_{0} - p_0) \, dp_0
+ \pi \widetilde{\phi}(\boldsymbol{\p}', p'_0)
\\
=
\textrm{reg} \,
\int {\textstyle\frac{ie^{ip_0y_0}}{p_0}} \widetilde{\phi}(\boldsymbol{\p}', p'_{0} + p_0) \, dp_0
+ \pi \widetilde{\phi}(\boldsymbol{\p}', p'_0)
\end{multline*}
\begin{multline*}
=
\int \limits_{-\infty}^{-\Delta} {\textstyle\frac{ie^{ip_0y_0}}{p_0}} \widetilde{\phi}(\boldsymbol{\p}', p'_{0} + p_0) \, dp_0
+\int \limits_{-\Delta}^{\Delta} {\textstyle\frac{ie^{ip_0y_0}}{p_0}} \big\{\widetilde{\phi}(\boldsymbol{\p}', p'_{0} + p_0) -
\widetilde{\phi}(\boldsymbol{\p}', p'_{0})\big\} \, dp_0
\\
+\int \limits_{\Delta}^{+\infty} {\textstyle\frac{ie^{ip_0y_0}}{p_0}} \widetilde{\phi}(\boldsymbol{\p}', p'_{0} + p_0) \, dp_0
+ \pi \widetilde{\phi}(\boldsymbol{\p}', p'_0),
\end{multline*}
smooth and uniformly bounded on $\mathbb{R}^4$, with respect to $y\in \mathbb{R}$, \emph{i.e.} with the bound (for each fixed $\phi$)
independent of $y$. In particular for each function $\phi$ which respects (\ref{phiInS00}) the function
\begin{multline}\label{BoundFtheta*Fphi(p+p'',|p'|+|p''|)}
\widetilde{\theta_y} \,\, \ast \,\, \widetilde{\phi}\big(-\boldsymbol{\p}' - \boldsymbol{\p}'', \, -|\boldsymbol{\p}'| - |\boldsymbol{\p}''|\big)
\\
=
\int \limits_{-\infty}^{-\Delta} {\textstyle\frac{ie^{ip_0y_0}}{p_0}} \widetilde{\phi}\big(-\boldsymbol{\p}' - \boldsymbol{\p}', \,
-|\boldsymbol{\p}'| - |\boldsymbol{\p}''| + p_0\big) \, dp_0
\\
+\int \limits_{-\Delta}^{\Delta} {\textstyle\frac{ie^{ip_0y_0}}{p_0}} \Big\{\widetilde{\phi}\big(-\boldsymbol{\p}' - \boldsymbol{\p}'', \, -|\boldsymbol{\p}'| - |\boldsymbol{\p}''| + p_0 \big) - \widetilde{\phi}\big (-\boldsymbol{\p}'- \boldsymbol{\p}'', \, -|\boldsymbol{\p}'| - |\boldsymbol{\p}''|\big) \Big\} \, dp_0
\\
+\int \limits_{\Delta}^{+\infty} {\textstyle\frac{ie^{ip_0y_0}}{p_0}}
\widetilde{\phi}\big(-\boldsymbol{\p}'- \boldsymbol{\p}'', \, -|\boldsymbol{\p}'| - |\boldsymbol{\p}''| + p_0\big) \, dp_0
+ \pi \widetilde{\phi}\big(-\boldsymbol{\p}' - \boldsymbol{\p}'', \, -|\boldsymbol{\p}'| - |\boldsymbol{\p}''|\big)
\\
\overset{\textrm{df}}{=} g_\phi(\boldsymbol{\p}', \boldsymbol{\p}'',y) \overset{\textrm{df}}{=} f[\phi]_{\boldsymbol{\p}', \boldsymbol{\p}''}(y),
\end{multline}
is a bounded function (with respect to $\boldsymbol{\p}' \times \boldsymbol{\p}'' \times y$) and even smooth with respect to
$\boldsymbol{\p}' \times \boldsymbol{\p}''$
except the points $(\boldsymbol{\p}'=0, \boldsymbol{\p}'')$ or $(\boldsymbol{\p}', \boldsymbol{\p}''=0)$ and smooth in $y$.
In particular each component of $f[\phi]_{\boldsymbol{\p}', \boldsymbol{\p}''}$, regarded as a function of $y$, is a multiplier of the Schwartz
algebra $\mathcal{S}(\mathbb{C})$ (with operation of pointwise multiplication, compare Appendix \ref{convolutorsO'_C})
and belongs to $\mathcal{O}_M(\mathbb{R}^4; \mathbb{C})$.
It follows that the Fourier transform $\widetilde{f[\phi]_{\boldsymbol{\p}', \boldsymbol{\p}''}}$
of each component of $f[\phi]_{\boldsymbol{\p}', \boldsymbol{\p}''}$, regarded as a function of $y$ (with $\phi \in \mathcal{S}^{00}, \boldsymbol{\p}', \boldsymbol{\p}''$ fixed but arbitrary)
\begin{multline}\label{F(f[phi])}
\widetilde{f[\phi]_{\boldsymbol{\p}', \boldsymbol{\p}''}}(\boldsymbol{\p}, p_0)
= {\textstyle\frac{i}{p_0}} \widetilde{\phi}\big(-\boldsymbol{\p}' - \boldsymbol{\p}', \,
-|\boldsymbol{\p}'| - |\boldsymbol{\p}''| + p_0\big) \, \delta(\boldsymbol{\p})
\\
+ \pi \, \widetilde{\phi}\big(-\boldsymbol{\p}' - \boldsymbol{\p}'', \, -|\boldsymbol{\p}'| - |\boldsymbol{\p}''|\big) \delta(\boldsymbol{\p})\delta(p_0)
\end{multline}
and where each component of $\widetilde{f[\phi]_{\boldsymbol{\p}', \boldsymbol{\p}''}}$, regarded as a functional (distribution),
is equal to the functional whose evaluation on $\xi \in \mathcal{S}(\mathbb{R}^4; \mathbb{C})$ is given by the integral
\[
\big\langle \widetilde{f[\phi]_{\boldsymbol{\p}', \boldsymbol{\p}''}}, \xi \big\rangle
= \textrm{reg}_{p_0} \,
\int \widetilde{f[\phi]_{\boldsymbol{\p}', \boldsymbol{\p}''}}(\boldsymbol{\p}, p_0) \, \xi(\boldsymbol{\p}, p_0) \, \ud^3 \boldsymbol{p} \, \ud p_0
\]
\emph{regularised} in the variable $p_0$, compare \cite{GelfandI}, \S 1.7. Thus, for
\[
\phi,\varphi \in \mathscr{E}= \mathcal{S}(\mathbb{R}^4; \mathbb{C}^d)
\]
or equivalently, for
\[
\widetilde{\phi},\widetilde{\varphi} \in \widetilde{\mathscr{E}} = \mathcal{S}(\mathbb{R}^4; \mathbb{C}^d),
\]
we can compute the convolution $\widetilde{f[\phi]_{\boldsymbol{\p}', \boldsymbol{\p}''}} \,\, \ast \,\, \widetilde{\varphi}$
of any fixed component of $\widetilde{f[\phi]_{\boldsymbol{\p}', \boldsymbol{\p}''}}$ with any fixed component
of $\widetilde{\varphi}$ and this convolution belongs to $\mathcal{S}(\mathbb{R}^4; \mathbb{C}^4)$
(because each component of $\widetilde{f[\phi]_{\boldsymbol{\p}', \boldsymbol{\p}''}} \in \mathcal{O}'_{C}(\mathbb{R}^4; \mathbb{C})$),
and moreover we have
\begin{multline}\label{F(f[phi])*varphi}
\Big(\widetilde{f[\phi]_{\boldsymbol{\p}', \boldsymbol{\p}''}} \,\, \ast \,\, \widetilde{\varphi}\Big) (\bar{\boldsymbol{\p}}, \bar{p}_{0})
= \textrm{reg}_{{}_{p_0}} \, \int \widetilde{f[\phi]_{\boldsymbol{\p}', \boldsymbol{\p}''}}(\boldsymbol{\p}, p_{0})
\widetilde{\varphi}(\bar{\boldsymbol{\p}}-\boldsymbol{\p}, \bar{p}_{0}-p_0)
\, \ud^3 \boldsymbol{p} \, \ud p_0
\\
= \textrm{reg} \, \int
{\textstyle\frac{i}{p_0}} \widetilde{\phi}\big(-\boldsymbol{\p}' - \boldsymbol{\p}', \,
-|\boldsymbol{\p}'| - |\boldsymbol{\p}''| + p_0\big) \widetilde{\varphi}(\bar{\boldsymbol{\p}}, \bar{p}_{0}- p_0) \, \ud p_0
\\
+ \pi \, \widetilde{\phi}\big(-\boldsymbol{\p}' - \boldsymbol{\p}'', \, -|\boldsymbol{\p}'| - |\boldsymbol{\p}''|\big)
\widetilde{\varphi}\big(\bar{\boldsymbol{\p}}, \, \bar{p}_{0}\big)
\end{multline}
In this formula the components $a,b$ of the convoluted and multiplied functions are fixed but arbitrary, \emph{i.e.} the components $a,b$ of the functions
$\widetilde{f^{a}[\phi]_{\boldsymbol{\p}', \boldsymbol{\p}''}}(\ldots)$ and $\widetilde{\varphi^b}(\ldots)$ are fixed, but arbitrary,
similarly the products $\widetilde{\phi}(\ldots)\widetilde{\varphi}(\ldots)$ are understood as
$\widetilde{\phi^a}(\ldots)\widetilde{\varphi^b}(\ldots)$ with fixed components $a,b$.
The indices $a,b$ are not written explicitly in order to simplify
notation.

Note also that 
\begin{multline*}
\theta_y \kappa'_{0,1} \dot{\otimes} \kappa''_{0,1}(\phi)(\boldsymbol{\p}', \boldsymbol{\p}'')
=\kappa'_{0,1} \dot{\otimes} \kappa''_{0,1}(\theta_y\phi)(\boldsymbol{\p}', \boldsymbol{\p}'')
\\
= \kappa'_{0,1} \dot{\otimes} \kappa''_{0,1}(\theta_y \phi)(\boldsymbol{\p}', \boldsymbol{\p}'')
= u'(\boldsymbol{\p}') u'(\boldsymbol{\p}') \widetilde{\theta_y \phi}\big(-\boldsymbol{\p}'- \boldsymbol{\p}'', -p'_{0}(\boldsymbol{\p}')
- p''_{0}(\boldsymbol{\p}'')\big)
\\
= u'(\boldsymbol{\p}') u''(\boldsymbol{\p}'') \widetilde{\theta_y} \ast \widetilde{\phi}\big(-\boldsymbol{\p}'- \boldsymbol{\p}'', -|\boldsymbol{\p}'|
- |\boldsymbol{\p}''|\big) = u'(\boldsymbol{\p}') u''(\boldsymbol{\p}'') f[\phi]_{\boldsymbol{\p}', \boldsymbol{\p}''}(y),
\end{multline*}
and 
\[
\kappa'_{1,0} \dot{\otimes} \kappa''_{1,0}(\varphi)(\boldsymbol{\p}', \boldsymbol{\p}'')
= v'(\boldsymbol{\p}') v''(\boldsymbol{\p}'') \widetilde{\varphi}\big(\boldsymbol{\p}'+ \boldsymbol{\p}'', |\boldsymbol{\p}'|
+|\boldsymbol{\p}''|\big),
\]
understood as equalities for matrix-valued functions. 

Let
\begin{equation}\label{phi,varphiInS00}
\widetilde{\phi}, \widetilde{\varphi} \in \widetilde{\mathscr{E}} 
= \mathcal{S}(\mathbb{R}^4; \mathbb{C}^{d'd''}),
\end{equation}
or equivalently 
\[
\phi, \varphi \in \mathscr{E}= \mathcal{S}(\mathbb{R}^4; \mathbb{C}^{d'd''}).
\]
Let $x$ be the space-time variable in the kernel
\[
\kappa'_{0,1} \dot{\otimes} \kappa''_{0,1}
\]
and let $y$ be the space-time variable in the kernel
\[
\kappa'_{1,0} \dot{\otimes} \kappa''_{1,0}.
\]
Then for general $\phi,\varphi$ which respect (\ref{phi,varphiInS00}) and by (\ref{F(f[phi])}), (\ref{F(f[phi])*varphi}) the contraction
\begin{multline*}
\theta_y\big(\kappa'_{0,1} \dot{\otimes} \kappa''_{0,1}\big)
\, \otimes_2  \,\, \big(\kappa'_{1,0} \dot{\otimes} \kappa''_{1,0}\big)(\phi \otimes \varphi) 
\\
=
\sum \limits_{s',s'',a,b,c,d}
\int 
\theta_y(x)\kappa'_{0,1} \dot{\otimes} \kappa''_{0,1}(s',\boldsymbol{\p}',s'', \boldsymbol{\p}''; a,b,x) 
\,\, \times 
\\
\times \,\,
\kappa'_{1,0} \dot{\otimes} \kappa''_{1,0}(s',\boldsymbol{\p}', s'',\boldsymbol{\p}''; c,d,y) \phi^{ab}(x)\varphi^{cd}(y)
\, \ud^3 \boldsymbol{\p}' \, \ud^3 \boldsymbol{\p}' \, \ud^4 x \, \ud^4 y
\end{multline*}
\begin{multline*}
=
\sum \limits_{s',s'',a,b,c,d}
\int 
\theta(x-y)\kappa'_{0,1} \dot{\otimes} \kappa''_{0,1}(s',\boldsymbol{\p}',s'', \boldsymbol{\p}''; a,b,x)
\,\, \times 
\\
\times \,\,
\kappa'_{1,0} \dot{\otimes} \kappa''_{1,0}(s',\boldsymbol{\p}', s'',\boldsymbol{\p}''; c,d,y) \phi^{ab}(x)\varphi^{cd}(y)
\, \ud^3 \boldsymbol{\p}' \, \ud^3 \boldsymbol{\p}' \, \ud^4 x \, \ud^4 y
\end{multline*}
\begin{multline*}
= 
\sum \limits_{s',s'',c,d}
\int 
\theta_y\kappa'_{0,1} \dot{\otimes} \kappa''_{0,1}(\phi)(\boldsymbol{\p}', \boldsymbol{\p}'')
\,\, \times 
\\
\times \,\,
\kappa'_{1,0} \dot{\otimes} \kappa''_{1,0}(s',\boldsymbol{\p}', s'',\boldsymbol{\p}''; c,d, y) \varphi^{cd}(y)
\, \ud^3 \boldsymbol{\p}' \, \ud^3 \boldsymbol{\p}' \, \ud^4 y
\end{multline*}
\begin{multline*}
= 
\sum \limits_{s',s'',c,d}
\int 
u'_{s'}(\boldsymbol{\p}') u''_{s''}(\boldsymbol{\p}'')f[\phi]_{\boldsymbol{\p}', \boldsymbol{\p}''}(y)
\,\, \times 
\\
\times \,\,
\kappa'_{1,0} \dot{\otimes} \kappa''_{1,0}(s',\boldsymbol{\p}', s'',\boldsymbol{\p}''; c,d, y) \varphi^{cd}(y)
\, \ud^3 \boldsymbol{\p}' \, \ud^3 \boldsymbol{\p}' \, \ud^4 y
\end{multline*}
\begin{multline*}
= 
\sum \limits_{s',s''}
\int 
u'_{s'}(\boldsymbol{\p}') u''_{s''}(\boldsymbol{\p}'') 
v'_{s'}(\boldsymbol{\p}') v''_{s''}(\boldsymbol{\p}'') 
\, \times 
\\
\times \,\,
\big( \widetilde{f[\phi]_{\boldsymbol{\p}', \boldsymbol{\p}''}} \,\,
\ast \,\, \widetilde{\varphi}\big) \big(\boldsymbol{\p}'+ \boldsymbol{\p}'', \, |\boldsymbol{\p}'| + | \boldsymbol{\p}''| \big)
\, \ud^3 \boldsymbol{\p}' \, \ud^3 \boldsymbol{\p}'
\end{multline*}
\begin{multline}\label{theta(x-y)masslesskappa01.masslesskappa10contractionmasslesskappa10.masslesskappa10}
= 
\sum \limits_{s',s''}
\textrm{reg}_{{}_{p_0}}
\int 
u'_{s'}(\boldsymbol{\p}') u''_{s''}(\boldsymbol{\p}'') 
v'_{s'}(\boldsymbol{\p}') v''_{s''}(\boldsymbol{\p}'') 
\, \times 
\\
\times \,\, \Bigg\{
{\textstyle\frac{i}{p_0}}  \widetilde{\phi}\big(-\boldsymbol{\p}' - \boldsymbol{\p}'', \,
-|\boldsymbol{\p}'| - |\boldsymbol{\p}''| + p_0\big) 
\widetilde{\varphi}\big(\boldsymbol{\p}' +\boldsymbol{\p}'', \, |\boldsymbol{\p}'| + | \boldsymbol{\p}''| -p_0\big)
\Bigg\}
\, \ud^3 \boldsymbol{\p}' \, \ud^3 \boldsymbol{\p}'' \, \ud p_0 
\\
+
\sum \limits_{s',s''}
\int 
u'_{s'}(\boldsymbol{\p}') u''_{s''}(\boldsymbol{\p}'') 
v'_{s'}(\boldsymbol{\p}') v''_{s''}(\boldsymbol{\p}'') 
\, \times 
\\
\times \,\, \Bigg\{ \pi \,
\widetilde{\phi}\big(-\boldsymbol{\p}' - \boldsymbol{\p}'', \,
-|\boldsymbol{\p}'| - |\boldsymbol{\p}''| \big) 
\widetilde{\varphi}\big(\boldsymbol{\p}' +\boldsymbol{\p}'', \, |\boldsymbol{\p}'| + | \boldsymbol{\p}''| \big)
\Bigg\}
\, \ud^3 \boldsymbol{\p}' \, \ud^3 \boldsymbol{\p}'' 
\end{multline}
\begin{multline*}
=
\sum \limits_{s',s''}
\int 
u'_{s'}(\boldsymbol{\p}') u''_{s''}(\boldsymbol{\p}'') 
v'_{s'}(\boldsymbol{\p}') v''_{s''}(\boldsymbol{\p}'') 
\, \times 
\\
\times \,\, 
\Big[\widetilde{\theta} \, \ast \, \big(\widetilde{\phi}.\check{\widetilde{\varphi}}\big) \Big]\big(-\boldsymbol{\p}' - \boldsymbol{\p}'', \,
-|\boldsymbol{\p}'| - |\boldsymbol{\p}''| \big) 
\, \ud^3 \boldsymbol{\p}' \, \ud^3 \boldsymbol{\p}' 
\end{multline*}
\begin{multline*}
=
\sum \limits_{s',s''}
\int 
u'_{s'}(\boldsymbol{\p}') u''_{s''}(\boldsymbol{\p}'') 
v'_{s'}(\boldsymbol{\p}') v''_{s''}(\boldsymbol{\p}'') 
\, \times 
\\
\times \,\, 
\Big[\mathscr{F}\big(\theta. (\phi\ast\check{\varphi})\big) \Big]\big(-\boldsymbol{\p}' - \boldsymbol{\p}'', \,
-|\boldsymbol{\p}'| - |\boldsymbol{\p}''| \big) 
\, \ud^3 \boldsymbol{\p}' \, \ud^3 \boldsymbol{\p}' 
\end{multline*}
integral is divergent. 

Indeed the divergence of the integral 
\begin{multline*}
\sum \limits_{s',s''}
\textrm{reg}_{{}_{p_0}}
\int 
u'_{s'}(\boldsymbol{\p}') u''_{s''}(\boldsymbol{\p}'') 
v'_{s'}(\boldsymbol{\p}') v''_{s''}(\boldsymbol{\p}'') 
\, \times 
\\
\times \,\, \Bigg\{
{\textstyle\frac{i}{p_0}}  \widetilde{\phi}\big(-\boldsymbol{\p}' - \boldsymbol{\p}'', \,
-|\boldsymbol{\p}'| - |\boldsymbol{\p}''| + p_0\big) 
\widetilde{\varphi}\big(\boldsymbol{\p}' +\boldsymbol{\p}'', \, |\boldsymbol{\p}'| + | \boldsymbol{\p}''| -p_0\big)
\Bigg\}
\, \ud^3 \boldsymbol{\p}' \, \ud^3 \boldsymbol{\p}' \, \ud p_0 
\end{multline*}
\begin{multline*}
=
\sum \limits_{s',s''}
\textrm{reg}_{{}_{p_0}}
\int 
u'_{s'}(\boldsymbol{\p}') u''_{s''}(\boldsymbol{\p}'') 
v'_{s'}(\boldsymbol{\p}') v''_{s''}(\boldsymbol{\p}'') 
\, \times 
\\
\times \,\, \Bigg\{
{\textstyle\frac{i}{p_0}}  \widetilde{\phi}\big(-\boldsymbol{\p}' - \boldsymbol{\p}'', \,
-|\boldsymbol{\p}'| - |\boldsymbol{\p}''| + p_0\big) 
\check{\widetilde{\varphi}}\big(\boldsymbol{\p}' +\boldsymbol{\p}'', \, |\boldsymbol{\p}'| + | \boldsymbol{\p}''| + p_0\big)
\Bigg\}
\, \ud^3 \boldsymbol{\p}' \, \ud^3 \boldsymbol{\p}' \, \ud p_0 
\end{multline*}
\begin{multline}\label{Int.tildephi.checktildevarphi}
=
\sum \limits_{s',s''}
\textrm{reg}_{{}_{p_0}}
\int 
u'_{s'}(\boldsymbol{\p}') u''_{s''}(\boldsymbol{\p}'') 
v'_{s'}(\boldsymbol{\p}') v''_{s''}(\boldsymbol{\p}'') 
\, \times 
\\
\times \,\, 
{\textstyle\frac{i}{p_0}}  \check{\widetilde{\phi}}.\widetilde{\varphi}\big(\boldsymbol{\p}' + \boldsymbol{\p}'', \,
|\boldsymbol{\p}'| + |\boldsymbol{\p}''| + p_0\big) 
\, \ud^3 \boldsymbol{\p}' \, \ud^3 \boldsymbol{\p}' \, \ud p_0 
\end{multline}
follows, e.g. for $\phi,\varphi$ whose coefficients are all of the Gaussian form 
\[
\widetilde{\varphi}(\boldsymbol{\p}, p_0) = e^{-|\boldsymbol{\p}|^2 - p_{0}^{2}}, 
\,\,\,\,\,\, \widetilde{\phi}(\boldsymbol{\p}, p_0)  = \widetilde{\phi}(-\boldsymbol{\p}, -p_0) = \check{\widetilde{\phi}}(\boldsymbol{\p}, p_0),
\]
and thus for $\widetilde{\phi},\widetilde{\varphi}$ belonging to the Schwartz space of rapidly decreasing functions.
Indeed, one can easily see the divergence on using the
ordinary formula for the Gaussian integrals. Not only the integral is not absolutely convergent, but also is divergent if we perform the integration
in the specific order: first space-time integrations (that is how the Fourier transforms $\widetilde{\phi},\widetilde{\varphi}$ appear instead of the initial
$\phi, \varphi$), then integration with respect to $\ud p_0$, and finally $\ud^3 \boldsymbol{\p}', \ud^3 \boldsymbol{\p}''$.
This succession of integrations is necessary in order to have the final distribution with correct supports (in case this integration would
be convergent at all).
Now the signs of the variables in the product 
\[
\widetilde{\phi}\big(-\boldsymbol{\p}' - \boldsymbol{\p}'', \,
-|\boldsymbol{\p}'| - |\boldsymbol{\p}''| + p_0\big) 
\widetilde{\varphi}\big(\boldsymbol{\p}' +\boldsymbol{\p}'', \, |\boldsymbol{\p}'| + | \boldsymbol{\p}''| -p_0 \big)
\]
do not ``cooperate properly'', and it is no longer true that this product decreases rapidly in the variables
$\boldsymbol{\p}' \times \boldsymbol{\p}'' \times p_0$, whenever they tend to infinity (in any direction).

Recall that in each integration $\ud^3 \boldsymbol{\p}', \ud^3 \boldsymbol{\p}''$
the singularities at zero
\begin{eqnarray*}
u'(\boldsymbol{\p}')v'(\boldsymbol{\p}') \sim \frac{1}{|\boldsymbol{\p}'|}
\\
u''(\boldsymbol{\p}'')v''(\boldsymbol{\p}'') \sim \frac{1}{|\boldsymbol{\p}''|},
\end{eqnarray*}
are locally integrable without any need for regularization and do not cause any difficulty.
Thus, indeed the integrals in (\ref{theta(x-y)masslesskappa01.masslesskappa10contractionmasslesskappa10.masslesskappa10})
are divergent and cannot represent any continuous functional in $\mathscr{E}^* \otimes \mathscr{E}^*$, where
\[
\mathscr{E} = \mathcal{S}^{00}(\mathbb{R}^4; \mathbb{C}^{d'd''}).
\]
We have analogous divergence of the integral
\begin{multline}\label{theta(x-y)masslesskappa01.masslesskappa10contractionmasslesskappa10.masslesskappa10q}
\sum \limits_{s',s'', \ldots}
\textrm{reg}_{{}_{p_0}}
\int
u'_{s'}(\boldsymbol{\p}') u''_{s''}(\boldsymbol{\p}'') u''_{s''}(\boldsymbol{\p}'') \ldots
v'_{s'}(\boldsymbol{\p}') v''_{s''}(\boldsymbol{\p}'') v''''_{s''}(\boldsymbol{\p}''') \ldots
\, \times
\\
\times \,\, \Bigg\{
{\textstyle\frac{i}{p_0}} \widetilde{\phi}\big(-\boldsymbol{\p}' - \boldsymbol{\p}'' - \boldsymbol{\p}''' - \ldots, \,
-|\boldsymbol{\p}'| - |\boldsymbol{\p}''| - |\boldsymbol{\p}'''| - \ldots + p_0\big) \, \times
\\
\times \,
\widetilde{\varphi}\big(\boldsymbol{\p}' +\boldsymbol{\p}'' +\boldsymbol{\p}''' + \ldots, \, |\boldsymbol{\p}'|
+ | \boldsymbol{\p}''| + | \boldsymbol{\p}'''| +\ldots -p_0\big)
\Bigg\}
\, \ud^3 \boldsymbol{\p}' \, \ud^3 \boldsymbol{\p}'' \, \ud^3 \boldsymbol{\p}''' \ldots \, \ud p_0
\\
+
\sum \limits_{s',s'', \ldots}
\int
u'_{s'}(\boldsymbol{\p}') u''_{s''}(\boldsymbol{\p}'') u''_{s''}(\boldsymbol{\p}'') \ldots
v'_{s'}(\boldsymbol{\p}') v''_{s''}(\boldsymbol{\p}'') v''''_{s''}(\boldsymbol{\p}''') \ldots
\, \times
\\
\times \,\, \Bigg\{ \pi \,
\widetilde{\phi}\big(-\boldsymbol{\p}' - \boldsymbol{\p}''- \boldsymbol{\p}''' - \ldots, \,
-|\boldsymbol{\p}'| - |\boldsymbol{\p}''| - |\boldsymbol{\p}'''| - \ldots \big) \, \times
\\
\times \,
\widetilde{\varphi}\big(\boldsymbol{\p}' +\boldsymbol{\p}''+\boldsymbol{\p}''' + \ldots, \, |\boldsymbol{\p}'|
+ | \boldsymbol{\p}''| +|\boldsymbol{\p}'''| + \ldots \big)
\Bigg\}
\, \ud^3 \boldsymbol{\p}' \, \ud^3 \boldsymbol{\p}'' \, \ud^3 \boldsymbol{\p}''' \ldots
\end{multline}
giving the $q$-contraction $\otimes_q$ of two kernels, the first being the dot product $\dot{\otimes}$ of $q$ massless kernels
$\kappa'_{0,1}, \kappa''_{0,1}, \kappa'''_{0,1}, \ldots$ with space-time variable $x$ and multiplied by $\theta(x-y)$, and the second being the dot
product $\dot{\otimes}$ of $q$ massless kernels $\kappa'_{1,0}, \kappa''_{1,0}, \kappa'''_{1,0}, \ldots$ with common space-time variable $y$.
We get the analogous integral for this $q$-contraction $\otimes||_{{}{q}}$ and with the analogous integrand products
\begingroup\makeatletter\def\f@size{5}\check@mathfonts
\def\maketag@@@#1{\hbox{\m@th\large\normalfont#1}}%
\[
\widetilde{\phi}\big(-\boldsymbol{\p}' - \boldsymbol{\p}'' - \boldsymbol{\p}''' - \ldots, \,\,
-|\boldsymbol{\p}'| - |\boldsymbol{\p}''| - |\boldsymbol{\p}'''| - \ldots \big)
\widetilde{\varphi}\big(\boldsymbol{\p}' +\boldsymbol{\p}'' + \boldsymbol{\p}''', \,\,\, +|\boldsymbol{\p}'| + | \boldsymbol{\p}''|
+ | \boldsymbol{\p}'''| + \ldots \big)
\]
\endgroup
and
\begingroup\makeatletter\def\f@size{5}\check@mathfonts
\def\maketag@@@#1{\hbox{\m@th\large\normalfont#1}}%
\[
\widetilde{\phi}\big(-\boldsymbol{\p}' - \boldsymbol{\p}' - \boldsymbol{\p}''' - \ldots, \,\,
-|\boldsymbol{\p}'| - |\boldsymbol{\p}''| - |\boldsymbol{\p}'''| - \ldots + p_0\big)
\widetilde{\varphi}\big(\boldsymbol{\p}' +\boldsymbol{\p}'' +\boldsymbol{\p}''' + \ldots, \,\,\, +|\boldsymbol{\p}'| + | \boldsymbol{\p}''|
+| \boldsymbol{\p}'''| + \ldots -p_0 \big)
\]
\endgroup
and signs of variables in the second product do not ''cooperate properly', and the second product does not decrease sufficiently rapidly
in the variables
$\boldsymbol{\p}' \times \boldsymbol{\p}''\times \boldsymbol{\p}''' \times \ldots \times p_0$,
whenever they tend to infinity (in any direction), and the corresponding integral becomes divergent.
Also in each integration $\ud^3 \boldsymbol{\p}', \ud^3 \boldsymbol{\p}'', \ud^3 \boldsymbol{\p}''', \ldots$
the singularities at zero
\begin{eqnarray*}
u'(\boldsymbol{\p}')v'(\boldsymbol{\p}') \sim \frac{1}{|\boldsymbol{\p}'|}
\\
u''(\boldsymbol{\p}'')v''(\boldsymbol{\p}'') \sim \frac{1}{|\boldsymbol{\p}''|},
\\
u'''(\boldsymbol{\p}''')v'''(\boldsymbol{\p}''') \sim \frac{1}{|\boldsymbol{\p}'''|},
\\
\vdots
\end{eqnarray*}
are all locally integrable without any need for regularization, and do not cause any difficulty.

Note here that the $2$-contraction integral (\ref{theta(x-y)masslesskappa01.masslesskappa10contractionmasslesskappa10.masslesskappa10})
or the $q$-contraction integral (\ref{theta(x-y)masslesskappa01.masslesskappa10contractionmasslesskappa10.masslesskappa10q}) for
$q>2$, becomes absolutely convergent for abstract fields if all (or all except one) of the kernels
$\kappa'_{0,1}, \kappa''_{0,1}, \ldots$ correspond to abstract fields
which have compact orbits, because in this case all the integrations $\ud^3 \boldsymbol{\p}', \ud^3 \boldsymbol{\p}'', \ldots$
in (\ref{theta(x-y)masslesskappa01.masslesskappa10contractionmasslesskappa10.masslesskappa10q})
become integrations over compact sets, in accordance with what we have asserted at the beginning of this Subsection.
Because the case of compact orbits is not interesting on the Minkowski space-time we do not enter now into further details of this
abstract situation, and go back to the ordinary free fields with ordinary non-compact orbits. Note however that although the map
\[
\phi \longrightarrow \theta.\phi
\]
does not transform continuously the space-time test space $\mathscr{E}$ into the test space $\mathscr{E}$, this map behaves much better
on the subspace of all those $\phi \in \mathscr{E}$ which vanish at zero together with all their derivatives up
to some fixed order. Indeed, any such $\phi$ can be written as a finite sum
\[
\phi = \sum\limits_{|\alpha|=\omega+1} x^\alpha\phi_\alpha, \,\,\, \phi_\alpha \in \mathscr{E},
\]
and the evaluation of a distribution $\kappa$ of positive singularity order $\omega$  at such $\phi$
can thus be written as a finite sum
\[
\langle \kappa, \phi\rangle = \sum\limits_{|\alpha|=\omega+1} \langle x^\alpha \kappa, \phi_\alpha \rangle 
\]
of valuations $\langle x^\alpha \kappa, \phi_\alpha \rangle$ of distributions $x^\alpha\kappa$
which have the order $\omega-|\alpha|$, negative if $|\alpha|>\omega$, so that $x^\alpha\kappa$
becomes regular enough to allow multiplication by the step theta function. This means that 
$\kappa$ can be multiplied by theta function, providing we restrict the test space to the subspace of
functions whose derivatives vanish up to singularity order $\omega$ of $\kappa$. 
This is the basis for the Epstein-Glaser splitting method to which we pass now, and which
must be applied in case of the $q$-contraction integral
(\ref{theta(x-y)masslesskappa01.masslesskappa10contractionmasslesskappa10.masslesskappa10q}) with $q\geq 2$.

The divergence of the evaluation integral (\ref{theta(x-y)masslesskappa01.masslesskappa10contractionmasslesskappa10.masslesskappa10})
of $2$ contraction $\otimes_2$
and of its analogue (\ref{theta(x-y)masslesskappa01.masslesskappa10contractionmasslesskappa10.masslesskappa10q})
for the $q$-contraction $\otimes_q$ with $q$ plane-wave factors $\kappa'_{0,1}, \kappa''_{0,1} \kappa'''_{0,1}, \ldots$
and $\kappa'_{1,0}, \kappa''_{1,0} \kappa'''_{1,0}, \ldots$ simply means that the distributions
\begin{multline}\label{[A'(-),A'(+)]^q}
\,\,\,\,\,\,\,\,\,\,\,\,\,\,\,\,\,\,\,\,\,\,\,\,\,\,\,\,\,\,\,\,\,\,\,\,\,\,\,\,\,\,\,\,\,\,\,\,\,\,\,\,\,\,\,\,\,\,\,\,\,\,\,\,\,\,\,\,
\big(\kappa'_{0,1} \dot{\otimes} \kappa''_{0,1} \big)
\, \otimes_2 \,\, \big(\kappa'_{1,0} \dot{\otimes} \kappa''_{1,0} \big)
\\
\big(\kappa'_{0,1} \dot{\otimes} \kappa''_{0,1} \dot{\otimes} \kappa'''_{0,1}\big)
\, \otimes_3 \,\, \big(\kappa'_{1,0} \dot{\otimes} \kappa''_{1,0} \dot{\otimes} \kappa'''_{1,0} \big),
\\
\vdots
\\
\big(\kappa'_{0,1} \dot{\otimes} \kappa''_{0,1} \dot{\otimes} \kappa'''_{0,1} \dot{\otimes} \ldots \kappa^{(q)}_{0,1} \big)
\, \otimes_q \,\, \big(\kappa'_{1,0} \dot{\otimes} \kappa''_{1,0} \dot{\otimes} \kappa'''_{1,0} \dot{\otimes} \ldots \kappa^{(q)}_{1,0} \big),
\\
\vdots \,\,\,\,\,\,\,\,\,\,\,\,\,\,\,\,\,\,\,\,\,\,\,\,\,\,\,\,\,\,\,\,\,\,\,\,\,\,\,\,\,\,\,\,\,\,\,\,\,\,\,\,\,\,\,\,\,\,\,\,\,\,\,\,\,\,\,\,
\,\,\,\,\,\,\,\,\,\,\,\,\,\,\,
\end{multline}
with the kernels
\begin{multline*}
\big(\kappa'_{0,1} \dot{\otimes} \kappa''_{0,1} \big)
\, \otimes_2 \,\, \big(\kappa'_{1,0} \dot{\otimes} \kappa''_{1,0}\big)(x,y)
\\
= 
\,\, \mathclap{\displaystyle\int}\mathclap{\textstyle\sum} \,\,\,\,\,
u'(\boldsymbol{\p}') v'(\boldsymbol{\p}') u''(\boldsymbol{\p}'') v''(\boldsymbol{\p}'')
e^{-i(|\boldsymbol{p}'| +|\boldsymbol{p}''|)(x_0-y_0) +i(\boldsymbol{p}' + \boldsymbol{p}'')\cdot (\boldsymbol{\x} - \boldsymbol{\y})}
\ud^3 \boldsymbol{p}'\ud^3 \boldsymbol{p}''
\\
= \kappa_2(x-y)
\end{multline*}
\begin{multline*}
\big(\kappa'_{0,1} \dot{\otimes} \kappa''_{0,1} \dot{\otimes} \kappa'''_{0,1}\big)
\, \otimes_3 \,\, \big(\kappa'_{1,0} \dot{\otimes} \kappa''_{1,0} \dot{\otimes} \kappa'''_{1,0} \big)(x,y)
\\
=
\,\, \mathclap{\displaystyle\int}\mathclap{\textstyle\sum} \,\,\,\,\,
u'(\boldsymbol{\p}') v'(\boldsymbol{\p}') u''(\boldsymbol{\p}'') v''(\boldsymbol{\p}'') u'''(\boldsymbol{\p}''') v'''(\boldsymbol{\p}''')
\,\, \times
\\
\times \,\,
e^{-i(|\boldsymbol{p}'| +|\boldsymbol{p}''|+ |\boldsymbol{p}'''|)(x_0-y_0) +i(\boldsymbol{p}' + \boldsymbol{p}''+ \boldsymbol{p}''')
\cdot (\boldsymbol{\x} - \boldsymbol{\y})} \, \ud^3 \boldsymbol{p}'\ud^3 \boldsymbol{p}''\ud^3 \boldsymbol{p}'''
= \kappa_3(x-y),
\end{multline*}
\[
\vdots
\]
\begin{multline}\label{kernel_q(x-y)}
\big(\kappa'_{0,1} \dot{\otimes} \kappa''_{0,1} \dot{\otimes} \kappa'''_{0,1} \dot{\otimes} \ldots \dot{\otimes} \kappa^{(q)}_{0,1} \big)
\, \otimes_q \,\, \big(\kappa'_{1,0} \dot{\otimes} \kappa''_{1,0} \dot{\otimes} \kappa'''_{1,0} \dot{\otimes} \ldots \dot{\otimes} \kappa^{(q)}_{1,0} \big)(x,y)
\\
=
\,\, \mathclap{\displaystyle\int}\mathclap{\textstyle\sum} \,\,\,\,\,
u'(\boldsymbol{\p}') v'(\boldsymbol{\p}') u''(\boldsymbol{\p}'') v''(\boldsymbol{\p}'') u'''(\boldsymbol{\p}''') v'''(\boldsymbol{\p}''')
\ldots u^{(q)}(\boldsymbol{\p}^{(q)}) v^{(q)}(\boldsymbol{\p}^{(q)}) \times
\\
\times \,\,
e^{-i(|\boldsymbol{p}'| +|\boldsymbol{p}''|+ |\boldsymbol{p}'''| + \ldots + |\boldsymbol{\p}^{(q)}|)(x_0-y_0)
+i(\boldsymbol{p}' + \boldsymbol{p}''+ \boldsymbol{p}''' + \ldots + \boldsymbol{\p}^{(q)})
\cdot (\boldsymbol{\x} - \boldsymbol{\y})} \, \ud^3 \boldsymbol{p}' \ldots \ud^3 \boldsymbol{p}^{(q)}
\\
= \kappa_q(x-y),
\end{multline}
\[
\vdots
\]
and with $q\geq 2$ are all singular of strictly positive order $\omega$ (depending on $q$) at $x=y$ (in the terminology of Epstein and Glaser, \cite{Epstein-Glaser}),
or have quasi-asymptotics of strictly positive order $\omega$ (in terminology used in \cite{Scharf}). Here the summation
and itegration is with respect to all spin momentum variables $s', \boldsymbol{p}', \ldots, s^{(q)}, \boldsymbol{p}^{(q)}$, athough the spin variables
$s', s'', \ldots, s^{(q)}$ are not written explicitly.
But in case of $q=1$-contraction $\otimes_1$ the contraction integral
(\ref{theta(x-y)masslesskappa01.masslesskappa10contractionmasslesskappa10.masslesskappa10q}) becomes convergent because in this case only one
integration is performed among the possible $q$ integrations $\ud^3 \boldsymbol{\p}' \ud^3 \boldsymbol{\p}'' \ldots $ in
(\ref{theta(x-y)masslesskappa01.masslesskappa10contractionmasslesskappa10.masslesskappa10q})
with only one regularized integration $\ud p_0$ being applied in it.
Similarly, the contraction integral
(\ref{theta(x-y)masslesskappa01.masslesskappa10contractionmasslesskappa10.masslesskappa10q})
becomes convergent in case of $0$-contraction $\otimes_0 = \otimes$, as in this case no integration of all possible
integrations $\ud^3 \boldsymbol{\p}' \ud^3 \boldsymbol{\p}'' \ldots $ is performed
with only one regularized integration $\ud p_0$ being applied in it.
This means that the distributions
\[
\big(\kappa'_{0,1} \dot{\otimes} \kappa''_{0,1} \dot{\otimes} \kappa'''_{0,1} \dot{\otimes} \ldots \kappa^{(q)}_{0,1} \big)
\, \otimes_{q'} \,\, \big(\kappa'_{1,0} \dot{\otimes} \kappa''_{1,0} \dot{\otimes} \kappa'''_{1,0} \dot{\otimes} \ldots \kappa^{(q)}_{1,0} \big),
\]
are of strictly negative order if $q'=0$ or $q'=1$ (here $q \geq q'$),
with the evaluations of their retarded parts on test functions $\phi \otimes \varphi$
given by the corresponding absolutely converging contraction integrals
(\ref{theta(x-y)masslesskappa01.masslesskappa10contractionmasslesskappa10.masslesskappa10q})
with at most one integration performed among the possible $q$ integrations $\ud^3 \boldsymbol{\p}' \ud^3 \boldsymbol{\p}'' \ldots$
in (\ref{theta(x-y)masslesskappa01.masslesskappa10contractionmasslesskappa10.masslesskappa10q}).
In particular for each
\[
\chi \otimes \varphi \in  \mathcal{S}(\mathbb{R}^4; \mathbb{C}^{d'd''}) \otimes \mathcal{S}(\mathbb{R}^4; \mathbb{C}^{d'd''}) =
\mathcal{S}(\mathbb{R}^4\times \mathbb{R}^4; \mathbb{C}^{d'd''d'd''}),
\]
of the form
\[
\chi(x,y) = \phi(x-y)\varphi(y), \,\,\,\, \phi, \varphi \in \mathcal{S}(\mathbb{R}^4; \mathbb{C}^{d'd''})
\]
and
\[
\xi \in E_{{}_{'}} \otimes E_{{}_{'''}} \otimes\ldots E_{{}_{(q)}} \otimes
E_{{}_{'}} \otimes E_{{}_{'''}} \otimes\ldots E_{{}_{(q)}},
\]
the zero contraction integral is convergent and equal
\begin{multline}\label{<kappatimes0kappa(phitimesvarphi),xi>}
\Big \langle
\theta_y \big(\kappa'_{0,1} \dot{\otimes} \kappa''_{0,1} \dot{\otimes} \kappa'''_{0,1} \dot{\otimes} \ldots \kappa^{(q)}_{0,1} \big)
\, \otimes_{q'=0} \,\, \big(\kappa'_{1,0} \dot{\otimes} \kappa''_{1,0}
\dot{\otimes} \kappa'''_{1,0} \dot{\otimes} \ldots \kappa^{(q)}_{1,0} \big)(\chi), \xi \Big\rangle
\\
=
\Big \langle
\theta_y\big(\kappa'_{0,1} \dot{\otimes} \kappa''_{0,1} \dot{\otimes} \kappa'''_{0,1} \dot{\otimes} \ldots \kappa^{(q)}_{0,1} \big)
\, \otimes \,\, \big(\kappa'_{1,0} \dot{\otimes} \kappa''_{1,0}
\dot{\otimes} \kappa'''_{1,0} \dot{\otimes} \ldots \kappa^{(q)}_{1,0} \big)(\chi), \xi \Big\rangle
\end{multline}
\begin{multline*}
=
\sum\limits_{s',s'', \ldots, r', r'', \ldots}
\int
u'_{s'}(\boldsymbol{\p}') u''_{s''}(\boldsymbol{\p}'') u'''_{s'''}(\boldsymbol{\p}''') \ldots
v'_{r'}(\boldsymbol{\q}') v''_{s''}(\boldsymbol{\q}'') v'''_{r'''}(\boldsymbol{\q}''') \ldots
\, \times
\\
\times \,\,
\widetilde{\theta.\phi}\big(-\boldsymbol{\p}' - \boldsymbol{\p}''- \boldsymbol{\p}''' - \ldots, \,
-|\boldsymbol{\p}'| - |\boldsymbol{\p}''| - |\boldsymbol{\p}'''| - \ldots \big) \, \times
\end{multline*}
\[
\times \,
\widetilde{\varphi}\big(-\boldsymbol{\p}' - \boldsymbol{\p}''- \ldots
+\boldsymbol{\q}' +\boldsymbol{\q}''+ \ldots, \,\,
\,-|\boldsymbol{\p}'| - |\boldsymbol{\p}''| - \ldots
+|\boldsymbol{\q}'| + | \boldsymbol{\q}''| + \ldots \big)
\]
\[
\times \,
\xi(s',\boldsymbol{\p}', s'',\boldsymbol{\p}'', \ldots, r',\boldsymbol{\q}', r'',\boldsymbol{\q}'', \ldots)
\, \ud^3 \boldsymbol{\p}' \, \ud^3 \boldsymbol{\p}'' \, \ud^3 \boldsymbol{\p}''' \ldots
\, \ud^3 \boldsymbol{\q}' \, \ud^3 \boldsymbol{\q}'' \, \ud^3 \boldsymbol{\q}''' \ldots
\]
But $\widetilde{\theta.\phi} = \mathscr{F}(\theta.\phi) \in L^2$ is smooth and bounded as the function of $(\boldsymbol{\p}, p_0) \in \mathbb{R}^4$,
as we have already shown in (\ref{F(theta.phi)}), and of course
$\widetilde{\varphi} \in \mathcal{S}(\mathbb{R}^4)$, therefore
\begin{multline*}
\Big|\Big \langle
\theta_y \big(\kappa'_{0,1} \dot{\otimes} \kappa''_{0,1} \dot{\otimes} \kappa'''_{0,1} \dot{\otimes} \ldots \kappa^{(q)}_{0,1} \big)
\, \otimes_{q'=0} \,\, \big(\kappa'_{1,0} \dot{\otimes} \kappa''_{1,0}
\dot{\otimes} \kappa'''_{1,0} \dot{\otimes} \ldots \kappa^{(q)}_{1,0} \big)(\chi), \xi \Big\rangle\Big|
\\
\,\,\,\,\,\,\,\,\,\,\,\,\,\,\,\,\,\,\,\,\,\,\,\,\,\,\,\,\,\,\,\,\,\,\,\,\,\,\,\,\,\, \leq \,\,\,
\underset{p \in \mathbb{R}^4}{\textrm{sup}} \, |\widetilde{\theta.\phi}(p)| \,\,\,
\underset{p \in \mathbb{R}^4}{\textrm{sup}} \, |\widetilde{\varphi}(p)| \,\,\, \times
\\
\times \,\,
\sum\limits_{s',s'', \ldots, r', r'', \ldots}
\int
\Big|
u'_{s'}(\boldsymbol{\p}') u''_{s''}(\boldsymbol{\p}'') u'''_{s'''}(\boldsymbol{\p}''') \ldots
v'_{r'}(\boldsymbol{\q}') v''_{s''}(\boldsymbol{\q}'') v'''_{r'''}(\boldsymbol{\q}''') \ldots \,\, \times
\\
\times \,
\xi(s',\boldsymbol{\p}', s'', \boldsymbol{\p}'', \ldots, r', \boldsymbol{\q}', r'', \boldsymbol{\q}'', \ldots)
\Big|
\, \ud^3 \boldsymbol{\p}' \, \ud^3 \boldsymbol{\p}'' \, \ud^3 \boldsymbol{\p}''' \ldots
\, \ud^3 \boldsymbol{\q}' \, \ud^3 \boldsymbol{\q}'' \, \ud^3 \boldsymbol{\q}''' \ldots
\end{multline*}
Note that for $\phi\in\mathcal{S}(\mathbb{R}^4)$
\begin{multline*}
\big| \widetilde{\theta.\phi}\big|_{{}_{L^2}}
= \big| \theta.\phi \big|_{{}_{L^2}} < \infty,
\\
\big| D^\alpha\mathscr{F}(\theta.\phi)\big|_{{}_{L^2}}
= \Big| \mathscr{F}^{-1}\big[ D^\alpha\mathscr{F}(\theta.\phi)\big] \Big|_{{}_{L^2}}=
\\
= \Big| \mathscr{F}^{-1}\big[\mathscr{F}(x^\alpha.\theta.\phi)\big] \Big|_{{}_{L^2}}
= \big| (x^\alpha.\theta.\phi) \big|_{{}_{L^2}} < \infty,
\end{multline*}
because $\mathscr{F}$ is the unitary operator which converts $-i\partial_{{}_{p_{\mu}}}$ into the multiplication by $x_\mu$ and
\[
x^\alpha.\theta.\phi \in L^2(\mathbb{R}^4) \cap L^1(\mathbb{R}^4), \,\,\, \textrm{for} \,\,\, \phi\in\mathcal{S}(\mathbb{R}^4), \,\,
\alpha \in \mathbb{N}^4.
\]
Here $\alpha$ is the multiindex with the Schwartz' multiindex notation for derivation operator
\[
D^\alpha = \partial_{{}_{p_{0}}}^{\alpha_0}\partial_{{}_{p_{1}}}^{\alpha_1}\partial_{{}_{p_{2}}}^{\alpha_2}\partial_{{}_{p_{3}}}^{\alpha_3},
\,\,\,\,\, x^\alpha = x_{0}^{\alpha_0}x_{1}^{\alpha_1}x_{2}^{\alpha_2}x_{3}^{\alpha_3},
\]
and
\[
x^\alpha.\theta.\phi(x) = x^\alpha\theta(x)\phi(x), \,\,\, x\in\mathbb{R}^4.
\]
From this it follows that also (momentum) derivatives of $\widetilde{\theta.\phi}$ are square integrable, and absolutely integrable,
of course regarded as the functions of the four real independent momentum variables
$p= (p_0,p_1,p_2,p_3) = (p_0, \boldsymbol{\p})$, in which we do not insert the function $p_0= p_0(\boldsymbol{\p})$.
Therefore, it follows from the proof of the third Lemma of Subsection \ref{SA=S0} that
\begin{multline*}
\underset{p \in \mathbb{R}^4}{\textrm{sup}} \, |(\widetilde{\theta.\phi})(p)|^2 \leq \int\limits_{\mathbb{R}^4} \big| \widetilde{\theta.\phi}\big|^2 \, \ud^4 p
+ \sum\limits_{\mu} \int\limits_{\mathbb{R}^4} \big| \partial_{{}_{p_\mu}}\widetilde{\theta.\phi}\big|^2 \, \ud^4 p
\\
+ \sum\limits_{\mu_1 \neq \mu_2} \int\limits_{\mathbb{R}^4} \big| \partial_{{}_{p_{\mu_1}}}\partial_{{}_{p_{\mu_2}}}\widetilde{\theta.\phi}\big|^2 \, \ud^4 p
+ \ldots \int\limits_{\mathbb{R}^4} \big| \partial_{{}_{p_{0}}} \partial_{{}_{p_{1}}}\partial_{{}_{p_{2}}} \partial_{{}_{p_{3}}}\widetilde{\theta.\phi}\big|^2 \, \ud^4 p
\end{multline*}
\[
\mu,\mu_i = 0,1,2,3.
\]
Because the Fourier transform is unitary, then
\begin{multline*}
\underset{p \in \mathbb{R}^4}{\textrm{sup}} \, |(\widetilde{\theta.\phi})(p)|^2 \leq \int\limits_{\mathbb{R}^4} \big|\theta.\phi \big|^2 \, \ud^4 x
+ \sum\limits_{\mu} \int\limits_{\mathbb{R}^4} \big| x_\mu \theta.\phi\big|^2 \, \ud^4 x
\\
+ \sum\limits_{\mu_1 \neq \mu_2} \int\limits_{\mathbb{R}^4} \big| x_{\mu_1} x_{\mu_2} \theta.\phi\big|^2 \, \ud^4 x
+ \ldots \int\limits_{\mathbb{R}^4} \big| x_0 x_1 x_2 x_3 \theta.\phi\big|^2 \, \ud^4 x.
\end{multline*}
Writing
\[
W(x) = 1 + \sum\limits_{\mu} \big(x_\mu \big)^2 + \sum\limits_{\mu_1\neq\mu_2} \big( x_{\mu_1} x_{\mu_2} \big)^2
+ \ldots + \big(x_0x_1x_2x_3 \big)^2
\]
we thus see that
\[
\underset{p \in \mathbb{R}^4}{\textrm{sup}} \, |(\widetilde{\theta.\phi})(p)|^2 \leq \big|W. \theta.\phi \big|_{{}_{L^2}}^2.
\]
Similarly, because $\widetilde{\varphi} \in \mathcal{S}^{0}(\mathbb{R}^4) \subset \mathcal{S}(\mathbb{R})$, then by the
results of Subsection \ref{SA=S0}
\[
\underset{p \in \mathbb{R}^4}{\textrm{sup}} \, |\widetilde{\varphi}(p)|^2
\leq \big|\big(A^{(4)}\big)^{m} \widetilde{\varphi} \big|_{{}_{L^2}}^{2} = |\varphi|_{m}^{2},
\]
for some norm $|\cdot|_m$ defining the topology of the nuclear space $\mathcal{S}^{00}(\mathbb{R}^4) = \mathscr{F}\mathcal{S}^{0}(\mathbb{R}^4)$.
Finally, by the results of Subsectons \ref{diffSA} and \ref{SA=S0}, the function
\begin{multline*}
(s',\boldsymbol{\p}', s'',\boldsymbol{\p}'', \ldots, \ldots, r',\boldsymbol{\q}', r'',\boldsymbol{\q}'', r''',\boldsymbol{\q}''', \ldots)
\longmapsto
\\
\longmapsto
u'_{s'}(\boldsymbol{\p}') u''_{s''}(\boldsymbol{\p}'') u'''_{s'''}(\boldsymbol{\p}''') \ldots
v'_{r'}(\boldsymbol{\q}') v''_{r''}(\boldsymbol{\q}'') v'''_{r'''}(\boldsymbol{\q}''') \ldots
\end{multline*}
is a multiplier of the standard nuclear space
\[
E_{{}_{'}} \otimes E_{{}_{'''}} \otimes\ldots E_{{}_{(q)}} \otimes
E_{{}_{'}} \otimes E_{{}_{'''}} \otimes\ldots E_{{}_{(q)}}
\]
so that, in particular, the function
\begin{multline*}
(s',\boldsymbol{\p}', s'',\boldsymbol{\p}'', \ldots, \ldots, r', \boldsymbol{\q}', r'',\boldsymbol{\q}'', r''',\boldsymbol{\q}''', \ldots)
\longmapsto
\\
\longmapsto
u'_{s'}(\boldsymbol{\p}') u''_{s''}(\boldsymbol{\p}'') u'''_{s'''}(\boldsymbol{\p}''') \ldots
v'_{r'}(\boldsymbol{\q}') v''_{r''}(\boldsymbol{\q}'') v'''_{r'''}(\boldsymbol{\q}''') \ldots
\, \times
\\
\times \,
\xi(s', \boldsymbol{\p}', s'', \boldsymbol{\p}'', \ldots, r', \boldsymbol{\q}', r'', \boldsymbol{\q}'', r''', \boldsymbol{\q}''', \ldots)
\end{multline*}
is again an element of the same nuclear space
\[
E_{{}_{'}} \otimes E_{{}_{'''}} \otimes\ldots E_{{}_{(q)}} \otimes
E_{{}_{'}} \otimes E_{{}_{'''}} \otimes\ldots E_{{}_{(q)}},
\]
of functions which rapidly decrease when the variables tend to infinity, and moreover
\begin{multline*}
\sum\limits_{s',s'', \ldots, r',r'', \ldots}
\int
\Big|
u'_{s'}(\boldsymbol{\p}') u''_{s''}(\boldsymbol{\p}'') u'''_{s'''}(\boldsymbol{\p}''') \ldots
v'_{r'}(\boldsymbol{\q}') v''_{s''}(\boldsymbol{\q}'') v'''_{r'''}(\boldsymbol{\q}''') \ldots \,\, \times
\\
\times \,
\xi(s', \boldsymbol{\p}', s'', \boldsymbol{\p}'', \ldots, r', \boldsymbol{\q}', r'', \boldsymbol{\q}'', \ldots)
\Big|
\, \ud^3 \boldsymbol{\p}' \, \ud^3 \boldsymbol{\p}'' \, \ud^3 \boldsymbol{\p}''' \ldots
\, \ud^3 \boldsymbol{\q}' \, \ud^3 \boldsymbol{\q}'' \, \ud^3 \boldsymbol{\q}''' \ldots
\leq
c \, |\xi|_n
\end{multline*}
for some norm $| \cdot |_n$ defining the standard nuclear space
\[
E_{{}_{'}} \otimes E_{{}_{'''}} \otimes\ldots E_{{}_{(q)}} \otimes
E_{{}_{'}} \otimes E_{{}_{'''}} \otimes\ldots E_{{}_{(q)}}.
\]
Indeed one can choose a function equal to the inverse $M= 1/Q$ of a polynomial $Q$,
which multiplied by the product function $u'u''\ldots v'v''\ldots$ gives a function
\[
{\textstyle\frac{u'u''\dots v'v''\ldots}{Q}} \in L^2\big(\big[\mathbb{R}^3\big]^{2q})
\]
and which is a multiplier of
\[
E_{{}_{'}} \otimes E_{{}_{'''}} \otimes\ldots E_{{}_{(q)}} \otimes
E_{{}_{'}} \otimes E_{{}_{'''}} \otimes\ldots E_{{}_{(q)}},
\]
compare Subsections \ref{dim=1}-\ref{SA=S0}.
In this case
\[
|M\xi|_{{}{L^2}} \leq c' |\xi|_{n}
\]
for some natural $n$ and a constant $c'$, because multiplication by $M$
is a continuous map in
\[
E_{{}_{'}} \otimes E_{{}_{'''}} \otimes\ldots E_{{}_{(q)}} \otimes
E_{{}_{'}} \otimes E_{{}_{'''}} \otimes\ldots E_{{}_{(q)}}.
\]
Next using the Schwartz inequality we obtain
\begin{multline*}
\sum\limits_{s',s'', \ldots, r',r'', \ldots}
\int
\Big|
u'_{s'}(\boldsymbol{\p}') u''_{s''}(\boldsymbol{\p}'') u'''_{s'''}(\boldsymbol{\p}''') \ldots
v'_{r'}(\boldsymbol{\q}') v''_{s''}(\boldsymbol{\q}'') v'''_{r'''}(\boldsymbol{\q}''') \ldots \,\, \times
\\
\times \,
\xi(s', \boldsymbol{\p}', s'', \boldsymbol{\p}'', \ldots, r', \boldsymbol{\q}', r'', \boldsymbol{\q}'', \ldots)
\Big|
\, \ud^3 \boldsymbol{\p}' \, \ud^3 \boldsymbol{\p}'' \, \ud^3 \boldsymbol{\p}''' \ldots
\, \ud^3 \boldsymbol{\q}' \, \ud^3 \boldsymbol{\q}'' \, \ud^3 \boldsymbol{\q}''' \ldots
\leq
c \, |\xi|_n
\end{multline*}
with
$c = c'c''$ where $c'' = \big|(u'u''\ldots v'v''\ldots)/Q \big|_{{}{L^2}}$.

Therefore, we arrive at the inequality
\begin{multline}\label{|<(thetakappax0kappa)(chi),xi>|<|phi||chi||xi|}
\Big|\Big \langle  
\theta_y \big(\kappa'_{0,1} \dot{\otimes} \kappa''_{0,1} \dot{\otimes} \kappa'''_{0,1} \dot{\otimes} \ldots \kappa^{(q)}_{0,1} \big)
\, \otimes_{q'=0} \,\, \big(\kappa'_{1,0} \dot{\otimes} \kappa''_{1,0}  
\dot{\otimes} \kappa'''_{1,0} \dot{\otimes} \ldots \kappa^{(q)}_{1,0} \big)(\chi), \xi \Big\rangle\Big|
\\
\leq c' \big|W. \theta.\phi \big|_{{}_{L^2}}  |\varphi|_{m} |\xi|_n
\leq c \big|\phi \big|_{{}_{k}}  |\varphi|_{{}_{m}} |\xi|_n.
\end{multline}

This means that
\[
\theta_{y} \big(\kappa'_{0,1} \dot{\otimes} \kappa''_{0,1} \dot{\otimes} \kappa'''_{0,1} \dot{\otimes} \ldots \kappa^{(q)}_{0,1} \big)
\, \otimes_{q'=0} \,\, \big(\kappa'_{1,0} \dot{\otimes} \kappa''_{1,0}  
\dot{\otimes} \kappa'''_{1,0} \dot{\otimes} \ldots \kappa^{(q)}_{1,0} \big)
\]
is a well-defined element of 
\[
\mathscr{L}\big( \mathscr{E}^{\otimes \, 2}, 
 \, E_{{}_{'}}^{*} \otimes E_{{}_{'''}}^{*} \otimes\ldots E_{{}_{(q)}}^{*} \otimes 
E_{{}_{'}}^{*} \otimes E_{{}_{'''}}^{*} \otimes\ldots E_{{}_{(q)}}^{*} \big).
\]

Similarly, we have uniform convergence for the retarded part of the $1$-contraction
defined through ordinary multiplication by the theta function (as we have already indicated above).
Indeed, using (\ref{F(theta.phi)}), we easily show that for any $u,v$,
defining a free field plane wave kernels (we do not write explicitly the discrete indices)
\[
\kappa_{0,1}(\boldsymbol{\p},x) = u(\boldsymbol{\p})e^{-ip\cdot x}, \,\,
\kappa_{1,0}(\boldsymbol{\p},x) = v(\boldsymbol{\p})e^{ip\cdot x}, \,\,\, 
p = (\boldsymbol{\p}',p_0(\boldsymbol{\p}')) \in \mathscr{O}, 
\]
there exist Hilbertian norms  $|\cdot|_n$, $|\cdot|_k$,
among the countable system of norms $|\cdot|_1, |\cdot|_2, \ldots$, independent of $\phi$, definining, respectively, $\mathscr{E}$, $\widetilde{\mathscr{E}}$, 
and  finite constants $c_k$ $c_n$, , independent of $\phi$, such that 
\begin{equation}\label{IntCone|(theta.phi)|d3p}
\int \big| u(\boldsymbol{\p})v(\boldsymbol{\p}) \widetilde{\theta.\phi}(-\boldsymbol{\p}, -| \boldsymbol{\p}|) \big| \, \ud^3 \boldsymbol{\p} 
\leq c_k \big|\widetilde{\phi} \big|_k  \leq c_n \big|\phi\big|_{{}_{n}},
\end{equation}
for all $\phi \in \mathscr{E}$.
Here we have put $p_0(\boldsymbol{\p}') = | \boldsymbol{\p}'|$, as for the massless case, 
but the same statement holds in the massive case, and summation over the discrete contracted indices is subsumed under the integral sign.

From (\ref{IntCone|(theta.phi)|d3p}) we see
that the retaded part of any pairing (scalar $\otimes_1$-contraction with no free non contracted variables) 
can be defined through the pointwise multiplication by the $\theta$ function. This immediately follows from  (\ref{IntCone|(theta.phi)|d3p})
because (integral symbol stands for integration and summation over the contracted variable, 
notation we will frequently use below)
\begin{equation}\label{thetakappa0,1times1kappa1,0(chi)}
\big[\theta_y \kappa'_{0,1} \, \otimes_1 \,\, \kappa'_{1,0} \big](\chi)
=
\widetilde{\varphi}(0)\,
\int  u'(\boldsymbol{\p}')v'(\boldsymbol{\p}')\widetilde{\theta.\phi}(-\boldsymbol{\p}', -| \boldsymbol{\p}'|)
 \, \ud^3 \boldsymbol{\p}',
\end{equation}
for
\[
\chi(x-y)=\phi(x-y)\varphi(y), \,\,\, \phi,\varphi \in \mathscr{E},
\]
so that there exist natural $n,k$ and a finite $c,c'$, independent of $\chi$, such that
\begin{equation}\label{|thetakappatimes1kappa(chi)|<|chi|m}
\Big|
\big[\theta_y \kappa'_{0,1} \, \otimes_1 \,\, \kappa'_{1,0} \big](\chi)
\Big| 
\leq c \big|\phi\big|_{{}_{n}} | \widetilde{\varphi}(0)|
\leq
c' \big|\phi\big|_{{}_{n}} \big| \varphi \big|_{{}_{i}}
\end{equation}
for all $\phi,\varphi$, ranging over $\mathscr{E}$ equal to
\[
\mathscr{E} = \mathcal{S}(\mathbb{R}^4).
\]

The inequality, analogue to (\ref{|<(thetakappax0kappa)(chi),xi>|<|phi||chi||xi|})
with $q'=1$, \emph{i.e.} valid for general $\otimes_1$-contraction, with $q'\leq q$, now immediately follows
from the case (\ref{|thetakappatimes1kappa(chi)|<|chi|m}), with $q'=q =1$, or, equivalently,
from (\ref{IntCone|(theta.phi)|d3p}). Indeed, from the continuity
of the map (hat means that the space, kernel or variable, which is contracted, is deleted)
\begin{multline}\label{productkernelsmultipliers}
E_{{}_{'}} \otimes E_{{}_{''}} \otimes  \ldots  \otimes E_{{}_{(q)}}
\otimes
E_{{}_{'}} \otimes E_{{}_{''}}\otimes \ldots  \otimes E_{{}_{(q)}} \ni
\xi \longmapsto
\\
\longmapsto
\Big[\big(\kappa'_{0,1} \dot{\otimes} \kappa''_{0,1} \dot{\otimes} \ldots \dot{\otimes}  \kappa^{(q)}_{0,1} \big)
\, \otimes \,\, \big(\kappa'_{1,0} \dot{\otimes} \kappa''_{1,0}
\dot{\otimes} \ldots  \kappa^{(q)}_{1,0} \big)\Big](\widehat{\xi}) \in \mathcal{O}_{M}(\mathbb{R}^4\times \mathbb{R}^4)
\end{multline}
or
\begin{multline*}
E_{{}_{'}} \otimes E_{{}_{''}} \otimes  \widehat{\ldots} \otimes  E_{{}_{(q)}}
\otimes
E_{{}_{'}} \otimes E_{{}_{''}}\otimes  \widehat{\ldots} \otimes  E_{{}_{(q)}}
\ni \widehat{\xi} \longmapsto
\\
\longmapsto
\Big[\big(\kappa'_{0,1} \dot{\otimes} \kappa''_{0,1} \dot{\otimes} \ldots \widehat{\kappa^{(k)}_{0,1}} \dot{\otimes} \ldots \kappa^{(q)}_{0,1} \big)
\, \otimes \,\, \big(\kappa'_{1,0} \dot{\otimes} \kappa''_{1,0}
\dot{\otimes} \ldots \widehat{\kappa^{(k=j)}_{1,0}} \ldots \kappa^{(q)}_{1,0} \big)\Big](\widehat{\xi}) 
\\
\in \mathcal{O}_{M}(\mathbb{R}^4\times \mathbb{R}^4),
\end{multline*}
which, in turn, is an immediate consequence of the Lemmas \ref{kappa0,1,kappa1,0psi}, Subsection \ref{psiBerezin-Hida} 
and \ref{kappa0,1,kappa1,0ForA} of Subsection \ref{A=Xi0,1+Xi1,0}, and from the fact that 
\[
\theta_y\kappa^{(k)}_{0,1}
\, \otimes_{1} \,\, \kappa^{(k)}_{1,0}
\]
is a well-defined distribution in $\mathscr{E}^{* \otimes \, 2}$, we obtain the analogue of 
(\ref{|<(thetakappax0kappa)(chi),xi>|<|phi||chi||xi|})
for $q'=1$, $q\geq 1$. 

For the same reason it is sufficient to
investigate the scalar distributions (\ref{[A'(-),A'(+)]^q}) and their retarded parts, because for $q'\leq q$,
we have the following product formula
\begin{multline}\label{ProductFormula}
\big(\kappa'_{0,1} \dot{\otimes} \kappa''_{0,1} \dot{\otimes} \ldots \dot{\otimes} \kappa^{(q)}_{0,1} \big)
\, \otimes_{q'} \,\, \big(\kappa'_{1,0} \dot{\otimes} \kappa''_{1,0} \dot{\otimes} \ldots \dot{\otimes} \kappa^{(q)}_{1,0} \big)
\\
=
\big(\kappa'_{0,1} \dot{\otimes} \kappa''_{0,1} \dot{\otimes} \ldots \dot{\otimes} \kappa^{(q')}_{0,1} \big)
\, \otimes_{q'} \,\, \big(\kappa'_{1,0} \dot{\otimes} \kappa''_{1,0} \dot{\otimes} \ldots \dot{\otimes} \kappa^{(q')}_{1,0} \big)
\, \times
\\
\times  \, 
\big(\kappa^{(q'+1)}_{0,1} \dot{\otimes} \ldots \dot{\otimes} \kappa^{(q)}_{0,1} \big)
\, \otimes \,\, \big(\kappa^{q'+1}_{1,0} \dot{\otimes} \ldots \dot{\otimes}  \kappa^{(q)}_{1,0} \big),
\end{multline}
where the first pairs of the kernels $\kappa'_{0,1}$ and $\kappa'_{1,0}, \ldots$, up to  $\kappa^{(q')}_{0,1}$
and $\kappa^{(q')}_{1,0}$ are assumed to be contracted. 

Similarly, the inequalty (\ref{|<(thetakappax0kappa)(chi),xi>|<|phi||chi||xi|})
 could have been deduced from the continuity of the map
(\ref{productkernelsmultipliers}) and from the fact that
$\theta(y-x)$ is a well-defined distribution in $\mathscr{E}^{*\otimes \, 2}$.

Therefore, we primarily are concentrated
on the scalar contractions  (\ref{[A'(-),A'(+)]^q}) in which all spin-momemtum variables are contracted.

However, we give now an independent proof of the analogue of 
(\ref{|<(thetakappax0kappa)(chi),xi>|<|phi||chi||xi|})
for $q'=1$, $q\geq 1$, using the case $q'=q =1$, \emph{i.e.} (\ref{|thetakappatimes1kappa(chi)|<|chi|m}) or,
equivalently using (\ref{IntCone|(theta.phi)|d3p}).

From (\ref{IntCone|(theta.phi)|d3p}) it easily follows the inequality
\begin{multline}\label{int|TranslationFthetaphi|<|W|}
\Bigg|
\sum\limits_{s^{(k)}} \int \Big\{ u^{(k)}_{s^{(k)}}(\boldsymbol{\p}^{(k)}) v^{(k)}_{s^{(k)}}(\boldsymbol{\p}^{(k)}) \,\, \times
\\ \times \,\,
\widetilde{\theta.\phi}\big(-\boldsymbol{\p}' - \boldsymbol{\p}''- \boldsymbol{\p}''' - \ldots - \boldsymbol{\p}^{(k)} - \ldots, \,
-|\boldsymbol{\p}'| - |\boldsymbol{\p}''| - |\boldsymbol{\p}'''| - \ldots - |\boldsymbol{\p}^{(k)}| - \ldots \big) \, 
\ud^3 \boldsymbol{\p}^{(k)}
\Bigg|
\\
\leq
\sum\limits_{|\gamma|,|\alpha|=0}^{n}|w^\gamma|{\alpha \choose \gamma} \,\, \big|\phi\big|_{{}_{n}}.
\end{multline}
Here
\[
w = \big(-\boldsymbol{\p}' - \boldsymbol{\p}''- \ldots -\cancel{\boldsymbol{\p}^{(k)}} -\boldsymbol{\p}^{(k)} - \ldots , \,
-|\boldsymbol{\p}'| - |\boldsymbol{\p}''| -\ldots  - \cancel{|\boldsymbol{\p}^{(k)}|} - |\boldsymbol{\p}^{(k+1)}| - \ldots  \big),
\]
and $\alpha,\beta, \gamma$, are $4$-component multiindices, for example $\gamma= (\gamma_0,\gamma_1, \gamma_2, \gamma_3)$ and
\[
w^{\gamma} = {w}_{0}^{\gamma_0}{w}_{1}^{\gamma_1} {w}_{2}^{\gamma_2} {w}_{3}^{\gamma_3},
\]
\[
{\alpha \choose \gamma} = {\alpha_0 \choose \gamma_0} {\alpha_1 \choose \gamma_1}{\alpha_2 \choose \gamma_2}{\alpha_3 \choose \gamma_3}.   
\]

Indeed, let
\[
e_{w}(x) = e^{iw\cdot x}.
\]
Then
\[
\widetilde{\theta.e_{w}.\phi}(-\boldsymbol{\p}, -| \boldsymbol{\p}|) =  
\widetilde{e_{w}.\theta.\phi}(-\boldsymbol{\p}, -| \boldsymbol{\p}|) 
= 
\widetilde{\theta.\phi}(-\boldsymbol{\p}+ \boldsymbol{w}, -| \boldsymbol{\p}| + w_0).
\]
Using the system of norms 
\[
| \phi |_{{}_{n}} = \underset{|\alpha|,|\beta| \leq n, x\in\mathbb{R}^4}{\textrm{sup}} \, |x^\beta \partial^\alpha \phi(x)|
\]
in (\ref{IntCone|(theta.phi)|d3p}), and applying (\ref{IntCone|(theta.phi)|d3p})
to $e_{p'}\phi$, instead of $\phi$, with
\[
w = \big(-\boldsymbol{\p}' - \boldsymbol{\p}''- \ldots -\cancel{\boldsymbol{\p}^{(k)}} -\boldsymbol{\p}^{(k)} - \ldots , \,
-|\boldsymbol{\p}'| - |\boldsymbol{\p}''| -\ldots  - \cancel{|\boldsymbol{\p}^{(k)}|} - |\boldsymbol{\p}^{(k+1)}| - \ldots  \big),
\]
we easily get (\ref{int|TranslationFthetaphi|<|W|}). Thus the function
\begin{multline}\label{smoothFofPolynomialGrowth}
F\big(\boldsymbol{\p}', \boldsymbol{\p}'', \ldots, \widehat{\boldsymbol{\p}^{(k)}}, \boldsymbol{\p}^{(k)}, \ldots \big)
\\
=
\sum\limits_{s^{(k)}} \int \Big\{ u^{(k)}_{s^{(k)}}(\boldsymbol{\p}^{(k)}) v^{(k)}_{s^{(k)}}(\boldsymbol{\p}^{(k)}) \,\, \times
\\ \times \,\,
\widetilde{\theta.\phi}\big(-\boldsymbol{\p}' - \boldsymbol{\p}''- \boldsymbol{\p}''' - \ldots - \boldsymbol{\p}^{(k)} - \ldots, \,
-|\boldsymbol{\p}'| - |\boldsymbol{\p}''| - |\boldsymbol{\p}'''| - \ldots - |\boldsymbol{\p}^{(k)}| - \ldots \big) \, 
\ud^3 \boldsymbol{\p}^{(k)}
\end{multline}
is of polynomial growth. Moreover, absolute value of the function
\[
{\textstyle\frac{1}{|\phi|_n}} F
\]
is majorized by a polynomial independent of $\phi$.

In the massive case, with 
\[
p_0(\boldsymbol{\p}')= |\boldsymbol{\p}'|, \,\,\,\,\, p_0(\boldsymbol{\p}'') = |\boldsymbol{\p}''|, \ldots
\]
replaced by
\[
p_0(\boldsymbol{\p}')= \sqrt{|\boldsymbol{\p}'|^2+m^2}, \,\,  p_0(\boldsymbol{\p}'') = \sqrt{|\boldsymbol{\p}''|^2+m^2}, \ldots
\]
the function (\ref{smoothFofPolynomialGrowth}) is smooth everywhere, and it is a multiplier of
\[
E_{{}_{'}} \otimes E_{{}_{''}} \otimes   \widehat{\ldots} \otimes  E_{{}_{(q)}}
=
\mathcal{S}(\mathbb{R}^3) \otimes \widehat{\ldots} \otimes \mathcal{S}(\mathbb{R}^3),
\]
with the hat symbol meaning that the spaces corresponding to the contracted variables are withdrawn.
More generally, hat over a variable, integration $\ud^3 \boldsymbol{\p}$, space, function, $\ldots$ will mean that this
variable, space, function, $\ldots$ is withdrawn.
Because any inverse power of $|\boldsymbol{\p}|$ is a multiplier of $\mathcal{S}^{0}(\mathbb{R}^3)$, compare 
Subsections \ref{diffSA} and \ref{SA=S0}, then also in the massless case the function (\ref{smoothFofPolynomialGrowth}), 
is a multiplier of the corresponding nuclear space
\[
E_{{}_{'}} \otimes E_{{}_{''}} \otimes   \widehat{\ldots} \otimes  E_{{}_{(q)}}
\\
=
\mathcal{S}^{0}(\mathbb{R}^3) \otimes \widehat{\ldots} \otimes \mathcal{S}^{0}(\mathbb{R}^3),
\]
and by the same reason also the functions
\[
u'u'' \widehat{\ldots} u^{(q)} F
\,\,\,\, \textrm{and}
\,\,\,\,
v'v''\widehat{\ldots} v^{(q)},
\]
are mulipliers of 
\begin{equation}\label{E'...^...E(q)}
E_{{}_{'}} \otimes E_{{}_{''}} \otimes   \widehat{\ldots} \otimes  E_{{}_{(q)}}.
\end{equation}
Again the hat symbol in $u'u'' \widehat{\ldots}$, $v'v''\widehat{\ldots} v^{(q)}$ means that $u^{(k)}$, respectively, $v^{(k)}$,  
corresponding to the contracted kernels,
are withdrawn from the products  $u'u''\ldots$, $v'v''\ldots$.

Thus in each case, massive and massless, for each natural $n$ there exists a natural $m$ and a finite constant $c_m$ such that
\[
|u'u'' \widehat{\ldots}u^{(q)}{\textstyle\frac{1}{|\phi|_n}} F\widehat{\xi}|_n
\leq c_m |\widehat{\xi}|_m,
\,\,\,
|v'v'' \widehat{\ldots}v^{(q)}\widehat{\xi}|_n
\leq c_m |\widehat{\xi}|_m,
\]
where $|\cdot|_1, |\cdot|_2, \ldots$ is the countable system of norms defining (\ref{E'...^...E(q)}) and with
$c_m$ independent of $\phi$.
Note, please, that 
\[
u', v', \,\,\, u'',v'', \,\,\, \ldots
\]
are multipliers of the respective nuclear spaces
\[
E_{{}_{'}}, \,\,\,\,\,\,\,  E_{{}_{''}}, \,\,\,\,\,\,\,  \ldots.
\]
They are smooth (except zero in the massless case) and bounded (outside a zero
neighborhood in the massless case). But in case of the electromagnetic potential field or, more generally,
massless free field, the product $u^{(k)}_{s^{(k)}}. v^{(k)}_{s^{(k)}}$ summed out with respect to $s^{(k)}$ 
has locally integrable singularity at zero (remember that for the e.m. potential field the zero component summand 
gets minus sign in accordance to the non-Hermitian conjugation of the creation-annihilation operators
in the Gupta-Bleuler gauge).  In each case
the singularity is irrevant as by construction they are always multipliers of $E_{{}_{(k)}}$.

Similarly, the product  function
\[
u'u'' \widehat{\ldots}u^{(q)}Fv'v'' \widehat{\ldots}v^{(q)}
\] 
of the variables (discrete indices are not written explicitly)
\begin{equation}\label{p',p'',...^...,k',k'',...^...}
\big(\boldsymbol{\p}', \boldsymbol{\p}'', \ldots , \ldots, \widehat{\boldsymbol{\p}^{(k)}}, \ldots, \boldsymbol{\p}^{(q)},
\boldsymbol{\q}', \boldsymbol{\q}'', \ldots, \widehat{\boldsymbol{\p}^{(j)}}, \boldsymbol{\q}^{(j)}, \ldots, \boldsymbol{\q}^{(q)} \big)
\end{equation}
is a multiplier of
\begin{equation}\label{E'x...xEqxE'x...xEq}
E_{{}_{'}} \otimes E_{{}_{''}} \otimes   \widehat{\ldots} \otimes  E_{{}_{(q)}}
\otimes
E_{{}_{'}} \otimes E_{{}_{''}}\otimes \widehat{\ldots} \otimes  E_{{}_{(q)}}.
\end{equation}
Similarly for any smooth function $f$ of polynomial growth in the variables (\ref{p',p'',...^...,k',k'',...^...})  
\[
f u'u'' \widehat{\ldots}u^{(q)}Fv'v'' \widehat{\ldots}v^{(q)}
\]
is a multiplier of (\ref{E'x...xEqxE'x...xEq}). Choosing $f$ such that $f^{-1} \in L^1$, and denoting $c = \big| f^{-1}\big|_{{}_{L^1}}$,
we see that for each $\widehat{\xi}$ in (\ref{E'x...xEqxE'x...xEq})
\begin{multline}\label{|u'.v'...1/|phi|F.xi|n<|xi|m}
\big| u'u'' \widehat{\ldots}u^{(q)}{\textstyle\frac{1}{|\phi|_n}} Fv'v'' \widehat{\ldots}v^{(q)} \widehat{\xi} \big|_{{}_{L^1}}
=
\big| f^{-1} f u'u'' \widehat{\ldots}u^{(q)}{\textstyle\frac{1}{|\phi|_n}} Fv'v'' \widehat{\ldots}v^{(q)} \widehat{\xi} \big|_{{}_{L^1}}
\\
\leq c \underset{(\boldsymbol{\p}', \ldots, \boldsymbol{\q}^{(q)})\in \mathbb{R}^{3(q-1)}\times \mathbb{R}^{3(q-1)}}{\textrm{sup}} \, 
\big| \big(fu'u'' \widehat{\ldots}u^{(q)}{\textstyle\frac{1}{|\phi|_n}} Fv'v'' \widehat{\ldots}v^{(q)} \widehat{\xi}\big)(\boldsymbol{\p}', \widehat{\ldots}, \boldsymbol{\q}^{(q)}) |
\\
\leq
c'c \big|\widehat{\xi}\big|_{{}_{m}}
\end{multline}
with $c$ independent of $\phi$ and $\widehat{\xi}$. The last inequality is the conseqence of the fact
that the sup norm indicated here is continuous on (\ref{E'x...xEqxE'x...xEq}), compare Subsection \ref{SA=S0}.
Note, please, that here in the measure defining the $L^1$ norm summation over the discrete indices is included
by definition.

Thus, for $\phi, \varphi \in \mathscr{E} = \mathcal{S}(\mathbb{R}^4)$ the integral
\begin{multline*}
\sum\limits_{s^{(k)}} \int \Big\{ u^{(k)}_{s^{(k)}}(\boldsymbol{\p}^{(k)}) v^{(k)}_{s^{(k)}}(\boldsymbol{\p}^{(k)}) \,\, \times
\\ \times \,\,
\widetilde{\theta.\phi}\big(-\boldsymbol{\p}' - \boldsymbol{\p}''- \boldsymbol{\p}''' - \ldots - \boldsymbol{\p}^{(k)} - \ldots, \,
-|\boldsymbol{\p}'| - |\boldsymbol{\p}''| - |\boldsymbol{\p}'''| - \ldots - |\boldsymbol{\p}^{(k)}| - \ldots \big) \, \times
\\
\times \,\,
\widetilde{\varphi}\big(-\boldsymbol{\p}' - \boldsymbol{\p}''-\ldots - \cancel{\boldsymbol{\p}^{(k)}} - \ldots
+\boldsymbol{\q}' +\boldsymbol{\q}''+ \ldots + \cancel{\boldsymbol{\p}^{(k)}} + \boldsymbol{\q}^{(j+1)} +\ldots, \,\,
\,-|\boldsymbol{\p}'| - |\boldsymbol{\p}''| - \ldots - \cancel{|\boldsymbol{\p}^{(k)}|} - \ldots
\\
\ldots +|\boldsymbol{\q}'| + | \boldsymbol{\q}''| + \ldots + \cancel{|\boldsymbol{\p}^{(k)}|} + |\boldsymbol{\q}^{(j+1)}| +\ldots\big)
\Big\}
\,
\ud^3 \boldsymbol{\p}^{(k)}
\end{multline*}
is convergent and equal to the product of the multiplier $F$ by a smooth
bounded function $G$ of all non contracted variables $\boldsymbol{\p}'$, $\ldots$, $\boldsymbol{\q}^{(q)}$. 
The absolute value of the function $F$ is bounded by a polynomial of $\boldsymbol{\p}'$, $\boldsymbol{\p}''$, $\ldots$, $\boldsymbol{\p}^{(q)}$, which is fixed 
and the function 
\begin{multline}\label{defG}
G(\boldsymbol{\p}', \widehat{\ldots}, \boldsymbol{\q}^{(q)}) 
\\
= \widetilde{\varphi}(-\boldsymbol{\p}' - \widehat{\ldots}- \boldsymbol{\p}^{(q)} + \boldsymbol{\q}' + \widehat{\ldots} +  \boldsymbol{\q}^{(q)},  
-|\boldsymbol{\p}'| - \widehat{\ldots} - |\boldsymbol{\p}^{(q)}| + |\boldsymbol{\q}'| + \widehat{\ldots} +  |\boldsymbol{\q}^{(q)}|) 
\end{multline}
of all the non contracted  variables $\boldsymbol{\p}'$, $\ldots$, $\boldsymbol{\q}^{(q)}$ is uniformly 
bounded whenever $\phi$, $\varphi$ range over any fixed bounded sets in $\mathscr{E}$. 
Indeed 
\begin{multline}\label{sup|Fvaphi|<|varphi|}
\underset{(\boldsymbol{\p}', \widehat{\ldots}, \boldsymbol{\q}^{(q)})\in \mathbb{R}^{3(q-1)}\times \mathbb{R}^{3(q-1)}}{\textrm{sup}} \, 
|G(\boldsymbol{\p}', \widehat{\ldots}, \boldsymbol{\q}^{(q)})|
\\
\leq \underset{p\in \mathbb{R}^4}{\textrm{sup}} \, |\widetilde{\varphi}(p)| 
\leq c''' \big|\widetilde{\varphi} \big|_{{}_{j}} \leq c'' \big|\varphi \big|_{{}_{i}},
\end{multline}
compare Subsection \ref{SA=S0}.

Thus, from (\ref{|u'.v'...1/|phi|F.xi|n<|xi|m})
\begin{multline}\label{|u'.v'...1/|phi|FG.xi|n<|xi|m}
\big| u'u'' \widehat{\ldots}u^{(q)}{\textstyle\frac{1}{|\phi|_n}} Fv'v'' \widehat{\ldots}v^{(q)} G \widehat{\xi} \big|_{{}_{L^1}}
\\
\leq 
\,\,\,\,\,\,
\underset{(\boldsymbol{\p}', \widehat{\ldots}, \boldsymbol{\q}^{(q)})}{\textrm{sup}} \,\,
|G(\boldsymbol{\p}', \widehat{\ldots}, \boldsymbol{\q}^{(q)})| \,\,\,\,\,\,\,\,\,\,\,\,\,
\big| u'u'' \widehat{\ldots}u^{(q)}{\textstyle\frac{1}{|\phi|_n}} Fv'v'' \widehat{\ldots}v^{(q)} \widehat{\xi} \big|_{{}_{L^1}}
\\
\leq
cc'c'' \big|\widehat{\xi}\big|_{{}_{m}}
 \big|\varphi \big|_{{}_{i}}
\end{multline}
with $c.c',c''$ independent of $\phi,\varphi, \widehat{\xi}$.

From this 
absolute and uniform (on bounded sets in the space-time test space) 
convergence of the retarded part of evaluation of the $1$-contraction integral easily follows, which is
defined through the pointwise product by the theta function.

Indeed, for each
\[
\chi(x,y) = \phi(x-y)\varphi(y), \,\,\,\,\, \phi, \varphi \in \mathscr{E},
\]

the absolute value of the evaluation of the retared part of the $1$-contraction 
\[
\Big \langle
\theta_y \big(\kappa'_{0,1} \dot{\otimes} \kappa''_{0,1} \dot{\otimes} \kappa'''_{0,1} \dot{\otimes} \ldots \kappa^{(q)}_{0,1} \big)
\, \otimes_{1} \,\, \big(\kappa'_{1,0} \dot{\otimes} \kappa''_{1,0}
\dot{\otimes} \kappa'''_{1,0} \dot{\otimes} \ldots \kappa^{(q)}_{1,0} \big)(\chi), \, \widehat{\xi} \, \Big\rangle
\]
\begin{multline*}
=
\sum \limits_{s', s'', \ldots, r',r'', \ldots \widehat{r^{(j)}}, \ldots }
\int
u'_{s'}(\boldsymbol{\p}') u''_{s''}(\boldsymbol{\p}'') u'''_{s'''}(\boldsymbol{\p}''') \ldots u^{(k)}_{s^{(k)}}(\boldsymbol{\p}^{(k)}) \ldots \, \times
\\
\times \,
v'_{r'}(\boldsymbol{\q}') v''_{r''}(\boldsymbol{\q}'') v'''_{r'''}(\boldsymbol{\q}''') \ldots v^{(k)}_{s^{(k)}}(\boldsymbol{\p}^{(k)}))
v^{(j+1)}_{r^{(j+1)}}(\boldsymbol{\q}^{(j+1)})) \ldots
\, \times
\end{multline*}
\begin{multline*}
\times \,\,
\widetilde{\theta.\phi}\big(-\boldsymbol{\p}' - \boldsymbol{\p}''- \boldsymbol{\p}''' - \ldots - \boldsymbol{\p}^{(k)} - \ldots, \,
-|\boldsymbol{\p}'| - |\boldsymbol{\p}''| - |\boldsymbol{\p}'''| - \ldots - |\boldsymbol{\p}^{(k)}| - \ldots \big) \, \times
\end{multline*}
\begin{multline*}
\times \,
\widetilde{\varphi}\big(-\boldsymbol{\p}' - \boldsymbol{\p}''-\ldots - \cancel{\boldsymbol{\p}^{(k)}} - \ldots
+\boldsymbol{\q}' +\boldsymbol{\q}''+ \ldots + \cancel{\boldsymbol{\p}^{(k)}} + \boldsymbol{\q}^{(j+1)} +\ldots, \,\,
\,-|\boldsymbol{\p}'| - |\boldsymbol{\p}''| - \ldots - \cancel{|\boldsymbol{\p}^{(k)}|} - \ldots
\\
\ldots +|\boldsymbol{\q}'| + | \boldsymbol{\q}''| + \ldots + \cancel{|\boldsymbol{\p}^{(k)}|} + |\boldsymbol{\q}^{(j+1)}| +\ldots\big)
\end{multline*}
\begin{multline*}
\times \,
\xi(s',\boldsymbol{\p}', s'', \boldsymbol{\p}'', \ldots, \widehat{s^{(k)}, \boldsymbol{\p}^{(k)}}, \ldots, \boldsymbol{\q}', \boldsymbol{\q}'', \boldsymbol{\q}''', \ldots,
\widehat{r^{(j)},\boldsymbol{\q}^{(j)}}, \ldots ) \, \times
\\
\, \times
\, \ud^3 \boldsymbol{\p}' \, \ud^3 \boldsymbol{\p}'' \, \ud^3 \boldsymbol{\p}''' \ldots
\, \ud^3 \boldsymbol{\q}' \, \ud^3 \boldsymbol{\q}'' \, \ud^3 \boldsymbol{\q}''' \ldots \widehat{\ud^3 \boldsymbol{\q}^{(j)}} \ldots
\end{multline*}
can be estimated:
\[
\Big|\Big \langle
\theta_y \big(\kappa'_{0,1} \dot{\otimes} \kappa''_{0,1} \dot{\otimes} \kappa'''_{0,1} \dot{\otimes} \ldots \kappa^{(q)}_{0,1} \big)
\, \otimes_{1} \,\, \big(\kappa'_{1,0} \dot{\otimes} \kappa''_{1,0}
\dot{\otimes} \kappa'''_{1,0} \dot{\otimes} \ldots \kappa^{(q)}_{1,0} \big)(\chi), \, \widehat{\xi} \, \Big\rangle\Big|
\]
\begin{multline*}
\leq
\sum \limits_{s',s'', \ldots, r',r'',\ldots, \widehat{r^{(j)}}, \ldots} \,\,
\int
\Big|
u'_{s'}(\boldsymbol{\p}') u''_{s''}(\boldsymbol{\p}'') u'''_{s'''}(\boldsymbol{\p}''') \ldots u^{(k)}_{s^{(k)}}(\boldsymbol{\p}^{(k)}) \ldots \, \times
\\
\times \,
v'_{r'}(\boldsymbol{\q}') v''_{r''}(\boldsymbol{\q}'') v'''_{r'''}(\boldsymbol{\q}''') \ldots v^{(k)}_{s^{(k)}}(\boldsymbol{\p}^{(k)})
v^{(j+1)}_{r^{(j+1)}}(\boldsymbol{\q}^{(j+1)}) \ldots
\, \times
\end{multline*}
\[
\times \,\,
\widetilde{\theta.\phi}\big(-\boldsymbol{\p}' - \boldsymbol{\p}''- \boldsymbol{\p}''' - \ldots - \boldsymbol{\p}^{(k)} - \ldots, \,
-|\boldsymbol{\p}'| - |\boldsymbol{\p}''| - |\boldsymbol{\p}'''| - \ldots - |\boldsymbol{\p}^{(k)}| - \ldots \big) \, \times
\]
\begin{multline*}
\times \,
\widetilde{\varphi}\big(-\boldsymbol{\p}' - \boldsymbol{\p}''-\ldots - \cancel{\boldsymbol{\p}^{(k)}} - \ldots
+\boldsymbol{\q}' +\boldsymbol{\q}''+ \ldots + \cancel{\boldsymbol{\p}^{(k)}} + \boldsymbol{\q}^{(j+1)} +\ldots, \,\,
\,-|\boldsymbol{\p}'| - |\boldsymbol{\p}''| - \ldots - \cancel{|\boldsymbol{\p}^{(k)}|} - \ldots
\\
\ldots +|\boldsymbol{\q}'| + | \boldsymbol{\q}''| + \ldots + \cancel{|\boldsymbol{\p}^{(k)}|} + |\boldsymbol{\q}^{(j+1)}| +\ldots\big)
\end{multline*}
\begin{multline*}
\times \,
\xi(s',\boldsymbol{\p}', s'', \boldsymbol{\p}'', \ldots, \widehat{s^{(k)}, \boldsymbol{\p}^{(k)}}, \ldots, \boldsymbol{\q}', \boldsymbol{\q}'', \boldsymbol{\q}''', \ldots,
\widehat{r^{(j)},\boldsymbol{\q}^{(j)}}, \ldots ) \,
\Big|
\, \times
\\
\, \times
\, \ud^3 \boldsymbol{\p}' \, \ud^3 \boldsymbol{\p}'' \, \ud^3 \boldsymbol{\p}''' \ldots
\, \ud^3 \boldsymbol{\q}' \, \ud^3 \boldsymbol{\q}'' \, \ud^3 \boldsymbol{\q}''' \ldots \widehat{\ud^3 \boldsymbol{\q}^{(j)}} \ldots
\end{multline*}
\begin{equation}\label{<kappatimes1kappa(phitimesvarphi),xi>}
= \big|u'u''\widehat{...}Fv'v''\widehat{...}G\widehat{\xi}\big|_{L^1}
\leq cc'c''  \big|\phi \big|_{{}_{n}} \big|\varphi\big|_{{}_{i}} |\widehat{\xi}|_{{}_{m}},
\end{equation}
on using (\ref{|u'.v'...1/|phi|FG.xi|n<|xi|m}) and (\ref{sup|Fvaphi|<|varphi|}), with $F$ given by (\ref{smoothFofPolynomialGrowth})
and  $G$ given by (\ref{defG}).
The hat-character $\widehat{\cdot}$ means that the contracted variable, or integration, or function $u^{(k)}$, $v^{(k)}$,
corresponding to the contracted variable, is deleted.
We assumed here that $(s^{(k)}, \boldsymbol{\p}^{(k)})$ and $(r^{(j)},\boldsymbol{\q}^{(j)})$ is the pair of the contracted variables.
Here $\widehat{\xi}$ is any element of the nuclear space
\[
E_{{}_{'}} \otimes E_{{}_{''}} \otimes   \widehat{\ldots} \otimes  E_{{}_{(q)}}
\otimes
E_{{}_{'}} \otimes E_{{}_{''}}\otimes \widehat{\ldots} \otimes  E_{{}_{(q)}}
\]
which we have defined through
\[
\xi \in
E_{{}_{'}} \otimes E_{{}_{''}} \otimes  \ldots  \otimes E_{{}_{(q)}}
\otimes
E_{{}_{'}} \otimes E_{{}_{''}}\otimes \ldots  \otimes E_{{}_{(q)}}
\]
by deleting the paired $(s^{(k)}, \boldsymbol{\p}^{(k)})$ and $(r^{(j)},\boldsymbol{\q}^{(j)})$ momentum variables and
where the hat over the space means that it is deleted from the tensor product.

The inequality (\ref{<kappatimes1kappa(phitimesvarphi),xi>}) means that
\begin{equation}\label{uniform(theta-thetaepsi)kappa...x1kappa...->0}
\theta_{y} \big(\kappa'_{0,1} \dot{\otimes} \kappa''_{0,1} \dot{\otimes} \kappa'''_{0,1} \dot{\otimes} \ldots \kappa^{(q)}_{0,1} \big)
\, \otimes_{q'=1} \,\, \big(\kappa'_{1,0} \dot{\otimes} \kappa''_{1,0}  
\dot{\otimes} \kappa'''_{1,0} \dot{\otimes} \ldots \kappa^{(q)}_{1,0} \big)
\end{equation}
is well-defined element of
\[
\mathscr{L}\big( \mathscr{E}^{\otimes \, 2}, 
 \, E_{{}_{'}}^{*} \otimes E_{{}_{''}}^{*}  \otimes \ldots \otimes \widehat{E_{{}_{(k)}}^{*}}  \otimes \ldots  \otimes \ldots E_{{}_{(q)}}^{*}  
E_{{}_{'}}^{*} \otimes E_{{}_{''}}^{*}\otimes \ldots \otimes \widehat{E_{{}_{(j)}}^{*}}  \otimes \ldots  \otimes \ldots E_{{}_{(q)}}^{*} \big).
\]

Recall that the convergence of the $0$-contraction and $1$-contraction
we have proved above in case all the plane-wave kernels $\kappa'_{0,1}, \kappa'_{1,0}, \kappa'_{0,1}, \kappa'_{1,0}, \ldots $ in
(\ref{<kappatimes0kappa(phitimesvarphi),xi>}) are massless and determining free massless fields
$\mathbb{A}', \mathbb{A}'', \ldots$.
Of course, and even all the more, the same convergence of the $0$-contraction
and $1$-contraction holds true if some massless kernels
$\kappa'_{0,1}, \kappa'_{1,0}, \kappa'_{0,1}, \kappa'_{1,0}, \ldots $ in (\ref{<kappatimes0kappa(phitimesvarphi),xi>}),
or all, we replace with massive kernels. The massive case is easier with $\mathscr{E} = \mathcal{S}(\mathbb{R}^4)$
and with smooth and bounded functions $u',v',u'',v'', \ldots$ of the momentum, without any singularity at zero
and with the orbits $\mathscr{O}_{m',0,0,0}, \mathscr{O}_{m'',0,0,0}, \ldots$ being equal the one sheet of the two-sheet
hyperboloid in the momentum space $\mathbb{R}^4$, and thus being smooth sub-manifolds of $\mathbb{R}^4$.

The fact that the distributions (\ref{[A'(-),A'(+)]^q}) have strictly positive singularity degree $\omega$
means that if we consider the value of the contraction integral
(\ref{theta(x-y)masslesskappa01.masslesskappa10contractionmasslesskappa10.masslesskappa10})
for a more general element
\begin{equation}\label{xiInS00timesS00}
\chi  \in \mathcal{S}(\mathbb{R}^4; \mathbb{C}^{d'd''}) \otimes \mathcal{S}(\mathbb{R}^4; \mathbb{C}^{d'd''}) =
\mathcal{S}(\mathbb{R}^4\times \mathbb{R}^4; \mathbb{C}^{d'd''d'd''}),
\end{equation}
instead of a simple tensor $\phi \otimes \varphi$, which moreover fulfils
\begin{equation}\label{Dalphaphi(x,x)=0,alpha=<omega}
\big(\chi \circ L\big) (x=0,y) = 0, \,\,\,
D_{{}_{x}}^{\alpha} \big(\chi \circ L\big)(x=0,y) = 0, \,\, 1 \leq |\alpha| \leq \omega,
\end{equation}
in the Schwartz multi-index notation for partial derivatives in the variables $x = (x_0, \ldots x_3)$,
where $L$ is the following linear invertible transformation
\[
\mathbb{R}^4 \times \mathbb{R}^4 \ni (x,y) \longmapsto L(x,y) = (x+y, y)
\in \mathbb{R}^4 \times \mathbb{R}^4,
\]
and where $\omega$ is the singularity degree of the distribution
\[
\big(\kappa'_{0,1} \dot{\otimes} \kappa''_{0,1} \big)
\, \otimes_2 \,\, \big(\kappa'_{1,0} \dot{\otimes} \kappa''_{1,0} \big) \subset
\mathcal{S}(\mathbb{R}^4\times \mathbb{R}^4; \mathbb{C}^{d'd''d'd''})^*,
\]
(in the sense of \cite{Epstein-Glaser}) then the integral
(\ref{theta(x-y)masslesskappa01.masslesskappa10contractionmasslesskappa10.masslesskappa10})
(with $\phi \otimes \varphi(x,y)$ replaced with $\chi(x,y)$)
\begin{multline}\label{[A'(-),A'(+)]^2}
\int
u'(\boldsymbol{\p}') v'(\boldsymbol{\p}') u''(\boldsymbol{\p}'') v''(\boldsymbol{\p}'')
e^{-i(|\boldsymbol{p}'| +|\boldsymbol{p}''|)(x_0-y_0) +i(\boldsymbol{p}' + \boldsymbol{p}'')\cdot (\boldsymbol{\x} - \boldsymbol{\y})}
\,\, \times
\\
\times \,\,
\theta(x-y) \, \chi(x,y)
\, \ud^3 \boldsymbol{p}' \, \ud^3 \boldsymbol{p}'' \, \ud^4 x \ud^4 y
\end{multline}
becomes absolutely convergent. Analogous statement holds for the $q$ contraction integrals, with the respective
order $\omega$ depending on $q$ and on the plane wave kernels $\kappa'_{0,1}, \kappa''_{0,1}, \ldots$
in the two contracted kernels, one being equal to dot product $\dot{\otimes}$ of $q$
plane waves $\kappa'_{0,1}, \kappa''_{0,1}, \ldots$ and multiplied by $\theta(x-y)$ and the second kernel being equal to the
dot product $\dot{\otimes}$ of $q$
plane waves $\kappa'_{1,0}, \kappa''_{1,0}, \ldots$.
This can be seen relatively simply for their causal symmetric/antisymmetric parts if we use the translation invariance of the respective distributions
(\ref{[A'(-),A'(+)]^q}) and causal support, on using their ultraviolet quasiasymptotis. 
By translational invariance the integral (\ref{[A'(-),A'(+)]^2})
for $\chi$ which respect (\ref{Dalphaphi(x,x)=0,alpha=<omega}), is equal to the integral
\begin{multline}\label{[A'(-),A'(+)]^2(x)}
\int
u'(\boldsymbol{\p}') v'(\boldsymbol{\p}') u''(\boldsymbol{\p}'') v''(\boldsymbol{\p}'')
e^{-i(|\boldsymbol{p}'| +|\boldsymbol{p}''|)x_0 +i(\boldsymbol{p}' + \boldsymbol{p}'')\cdot \boldsymbol{\x}}
\,\, \times
\\
\times \,\,
\theta(x) \, \phi(x)
\, \ud^3 \boldsymbol{p}' \, \ud^3 \boldsymbol{p}'' \, \ud^4 x \, \widetilde{\varphi}(0)
\end{multline}
\begin{multline*}
=
\int
u'(\boldsymbol{\p}') v'(\boldsymbol{\p}') u''(\boldsymbol{\p}'') v''(\boldsymbol{\p}'')
e^{-i(|\boldsymbol{p}'| +|\boldsymbol{p}''|)x_0 +i(\boldsymbol{p}' + \boldsymbol{p}'')\cdot \boldsymbol{\x}}
\,\, \times
\\
\times \,\,
\widetilde{\theta \phi}(-\boldsymbol{\p}'-\boldsymbol{\p}'', -|\boldsymbol{\p}'| - |\boldsymbol{\p}''|)
\, \ud^3 \boldsymbol{\p}' \, \ud^3 \boldsymbol{\p}'' \,\, \widetilde{\varphi}(0)
\end{multline*}
for any
\[
\phi \in 
\mathcal{S}(\mathbb{R}^4; \mathbb{C}^{d'd''}),
\]
which fulfills
\begin{equation}\label{up-to-omegaorder-derivatives-of-phi=0}
\phi(0) = 0, \,\,\,
D^{\alpha} \phi(0) = 0, \,\, 1 \leq |\alpha| \leq \omega,
\end{equation}
in the Schwartz multi-index notation for partial derivatives, and where
$\omega$ is the singularity degree of the distribution $\kappa_2$, such that
\begin{multline}\label{kappa2(x-y)}
\kappa_2(x-y) = \big(\kappa'_{0,1} \dot{\otimes} \kappa''_{0,1} \big)
\, \otimes_2 \,\, \big(\kappa'_{1,0} \dot{\otimes} \kappa''_{1,0} \big)(x,y)
\\
= \big(\kappa'_{0,1} \dot{\otimes} \kappa''_{0,1} \big)
\, \otimes_2 \,\, \big(\kappa'_{1,0} \dot{\otimes} \kappa''_{1,0} \big)(x-y),
\end{multline}
and which, by definition, coincides with the singularity degree of the distribution
\[
\big(\kappa'_{0,1} \dot{\otimes} \kappa''_{0,1} \big)
\, \otimes_2 \,\, \big(\kappa'_{1,0} \dot{\otimes} \kappa''_{1,0} \big)(x,0)
\]
at $x=0$.

Instead of the $\otimes_q$-contractions (\ref{[A'(-),A'(+)]^q}) themselves, we concentrate our attention first
on their symmetric (or antisymmetric) parts, which have causal support. Using (\ref{D(+)}) and (\ref{D(-)}), we can prove that
\begin{multline*}
 \big(\kappa'_{0,1} \dot{\otimes} \kappa''_{0,1} \dot{\otimes} \kappa'''_{0,1} \dot{\otimes} \ldots \dot{\otimes} \kappa^{(q)}_{0,1} \big)
\, \otimes_{q} \,\, \big(\kappa'_{1,0} \dot{\otimes} \kappa''_{1,0}  \dot{\otimes} \kappa'''_{1,0} \dot{\otimes} 
\ldots \dot{\otimes} \kappa^{(q)}_{1,0} \big)(x,y)
\\
- (-1)^q
\big(\kappa'_{1,0} \dot{\otimes} \kappa''_{1,0} \dot{\otimes} \kappa'''_{1,0} \dot{\otimes} \ldots \dot{\otimes} \kappa^{(q)}_{1,0} \big)
\, \otimes_{q} \,\, \big(\kappa'_{0,1} \dot{\otimes} \kappa''_{0,1}  \dot{\otimes} \kappa'''_{0,1} \dot{\otimes} 
\ldots \dot{\otimes} \kappa^{(q)}_{0,1} \big)(x,y)
\\
= \kappa_q(x-y) - (-1)^q \kappa_q(y-x) = \big[\kappa_q - (-1)^q \check{\kappa_q}\big](x-y)
\end{multline*}
is always causally supported.

Then, for 
\[
\chi(x,y) = \phi(x-y)\varphi(y), \,\,\,\, \phi,\varphi \in \mathscr{E} = \mathcal{S}(\mathbb{R}^4),
\]
and with 
\[
\phi \in \mathscr{E}= \mathcal{S}(\mathbb{R}^4) 
\]
which fulfil (\ref{up-to-omegaorder-derivatives-of-phi=0}), the valuation  integral
\begin{multline}\label{thetaASkappa2(chi)}
\Big[
\theta_y\big(\kappa'_{0,1} \dot{\otimes} \kappa''_{0,1} \dot{\otimes} \dots \dot{\otimes} \kappa^{(q)}_{0,1}\big)
 \otimes_q \big(\kappa'_{1,0} \dot{\otimes} \kappa''_{1,0} \dot{\otimes} \ldots 
\dot{\otimes} \kappa^{(q)}_{1,0}\big)
\\
- 
\theta_y\big(\kappa'_{1,0} \dot{\otimes} \kappa''_{1,0} \dot{\otimes} \dots \dot{\otimes} \kappa^{(q)}_{1,0} \big) \otimes_2\big(\kappa'_{0,1} \dot{\otimes} \kappa''_{0,1} \dot{\otimes} \dots \dot{\otimes} \kappa^{(q)}_{0,1} \big)
\Big] (\chi)
\end{multline}
becomes equal
\begin{multline*}
\int
[\kappa_{q}(x-y) -\kappa_q(y-x)]
\theta(x-y) \, \phi(x-y)\varphi(y) \, \ud^4 x \, \ud^4 y
\\
=
\widetilde{\varphi}(0) \,\, \Big\langle \theta \big[\kappa_q - (-1)^q \check{\kappa_q}\big], \phi \Big\rangle 
= \widetilde{\varphi}(0) \,\, \Big\langle \big[\kappa_q - (-1)^q \check{\kappa_q}\big], \theta \phi \Big\rangle
\end{multline*}
\begin{multline}\label{Convergent[A',A''][A'', A'']theta}
=
\widetilde{\varphi}(0) \,\,
\int
u'(\boldsymbol{\p}') v'(\boldsymbol{\p}') u''(\boldsymbol{\p}'') v''(\boldsymbol{\p}'') \ldots 
\,
\widetilde{\theta\phi}(-\boldsymbol{p}' - \boldsymbol{p}'' - \ldots, \,-|\boldsymbol{p}'| -|\boldsymbol{p}''|-\ldots) 
\, \ud^3 \boldsymbol{p}' \, \ud^3 \boldsymbol{p}'' \ldots  \ud^3 \boldsymbol{p}^{(q)}
\\
-
\widetilde{\varphi}(0) \,\,
\int
u'(\boldsymbol{\p}') v'(\boldsymbol{\p}') u''(\boldsymbol{\p}'') v''(\boldsymbol{\p}'') \ldots 
\,
\widetilde{\theta\phi}(\boldsymbol{p}' + \boldsymbol{p}''+ \ldots, \,|\boldsymbol{p}'| +|\boldsymbol{p}''| + \ldots) 
\, \ud^3 \boldsymbol{p}' \, \ud^3 \boldsymbol{p}''\ldots \ud^3 \boldsymbol{p}^{(q)}. 
\end{multline}
which can also be understood (and which is non-trivial) as a limit $\lambda \rightarrow +\infty$ of
\begin{equation}\label{Convergent[A',A''][A'', A'']Slambda(chi0)}
\widetilde{\varphi}(0) \,\, \Big\langle (S_{\lambda}\chi_0) \big[\kappa_q - (-1)^q \check{\kappa_q}\big], \phi \Big\rangle 
\end{equation}
where for any function $h$ on $\mathbb{R}^4$ and real $\lambda$, $(S_\lambda h)(x) = h(\lambda x)$, 
and where $\chi_0(x) = f(x_0)$
with an everywhere smooth non decreasing function $f$, equal zero for $x_0 \leq 0$, $0\leq f \leq 1$ for $0\leq x_0 \leq 1$,
and $f=1$ for $x_0\geq 1$. Thus $S_\lambda \chi_0$ converges to $\theta(x)$ in $\mathcal{S}(\mathbb{R}^4)^*$ if $\lambda \rightarrow +\infty$,
with the fixed choice of the timelike unit versor $v$ of the reference frame equal $v =(1,0,0,0)$.  
More generally, for any other unit timelike $v\cdot v=1$ versor $v$ of the reference frame, we can consider
$\chi_0(x) = f(v\cdot x)$, with $S_\lambda \chi_0$ converging to $\theta(v\cdot x)$ in $\mathcal{S}(\mathbb{R}^4)^*$ if $\lambda \rightarrow +\infty$.

First we show convergence of the limit $\lambda \rightarrow \infty$
of (\ref{Convergent[A',A''][A'', A'']Slambda(chi0)}),
for each $\phi\in \mathscr{E}$ fulfilling (\ref{up-to-omegaorder-derivatives-of-phi=0}), with $\omega$ not less than the singularity degree
of $\kappa_q - (-1)^q \check{\kappa_q}$. For this purpose, it is sufficient to determine
the quasi-asymptotics together with its homogeneity of the distribution $\kappa_q$ 
at zero (in $x$ or at infinity in $p$ for its Fourier transform).
Let for any test function $\phi \in \mathcal{S}(\mathbb{R}^4)$ and $\lambda>0$, $S_\lambda\phi(x) = \phi(\lambda x)$.
Let us remind that the distribution $\kappa_{q \,0}$ is equal to the quasi-asymptotic part $\kappa_{q \,0}$
of $\kappa_q \in \mathcal{S}(\mathbb{R}^4)^*$ at zero if there exists a number $\omega$, 
called singularity degree at zero (in $x$ and at $\infty$
in $p$), such that the limit
\[
\underset{\lambda \rightarrow +\infty}{\textrm{lim}} \,\,
\big({\textstyle\frac{1}{\lambda}}\big)^\omega \big\langle \kappa_q, S_{\lambda}\phi \big\rangle
= 
\underset{\lambda \rightarrow +\infty}{\textrm{lim}} \,\,
\big({\textstyle\frac{1}{\lambda}}\big)^\omega \big({\textstyle\frac{1}{\lambda}}\big)^4
 \big\langle \widetilde{\kappa_q}, S_{1/\lambda} \widetilde{\phi} \big\rangle
=
\big\langle \kappa_{q \,0},\phi\big\rangle
\]
exists for each $\phi\in \mathcal{S}(\mathbb{R}^4)$ and is not equal zero for all $\phi\in \mathcal{S}(\mathbb{R}^4)$;
$\kappa_{q \,0}$ represents in fact the asymtoticaly homogeneous
part of $\kappa_{q} $, which is of ultraviolet homogeneity degree $\omega$. 
It is easily seen that $\kappa_{q \,0}$ is homogeneous of homogeneity degree $\omega$.

The singularity degrees $s', s'', \ldots$,  of the functions $u',v'$, $u'',v'',\ldots$ of $\boldsymbol{\p}'$,
$\boldsymbol{\p}'', \ldots$, defining the plane wave kernels
\begin{gather*}
\kappa'_{0,1}(\boldsymbol{\p}'; x) = u'(\boldsymbol{\p}') e^{-ip'\cdot x},
\, \kappa'_{1,0}(\boldsymbol{\p}'; x) = v'(\boldsymbol{\p}') e^{ip'\cdot x}, 
\\
\kappa''_{0,1}(\boldsymbol{\p}''; x) = u''(\boldsymbol{\p}'') e^{-ip''\cdot x},
\, \kappa''_{1,0}(\boldsymbol{\p}'; x) = v'(\boldsymbol{\p}'') e^{ip'''\cdot x}, 
\\
\ldots
\end{gather*}
 of the free fields $\mathbb{A}', \mathbb{A}'', \ldots$ 
and regarded as Fourier transform 
of distributions in $\mathbb{R}^3$, can be easily computed.  By definition $s',s'', \ldots$, is the number for which,
respectively,
\begin{gather*}
\underset{\lambda \rightarrow +\infty}{\textrm{lim}} \,\,
\big({\textstyle\frac{1}{\lambda}}\big)^{s'}u'(\boldsymbol{\lambda \p}'),  
\,\,\,\,\,\,\,\,\,\,\,\,
\underset{\lambda \rightarrow +\infty}{\textrm{lim}} \,\,
\big({\textstyle\frac{1}{\lambda}}\big)^{s'}v'(\boldsymbol{\lambda\p}'),
\\
\underset{\lambda \rightarrow +\infty}{\textrm{lim}} \,\,
\big({\textstyle\frac{1}{\lambda}}\big)^{s''}u''(\lambda\boldsymbol{\p}''), 
\,\,\,\,\,\,\,\,\,\,\,\,
\underset{\lambda \rightarrow +\infty}{\textrm{lim}} \,\,
\big({\textstyle\frac{1}{\lambda}}\big)^{s''}v''(\lambda\boldsymbol{\p}''),
\\
\ldots
\end{gather*}
exists in distributional sense and is non zero.
Becuse the functions $u',v'$, $u'',v'',\ldots$, respectively, of $\boldsymbol{\p}'$, $\boldsymbol{\p}', \ldots$, are ordinary functions, analytic,
except eventually at zero in the massless case, then in practice it is sufficient to check for what choice of the number
$s',s'', \ldots$, the above limit exists pointwisely and represents a
nonzero function, respectively, of $\boldsymbol{\p}'$, $\boldsymbol{\p}'', \ldots$. It is easily seen
that the numbers $s',s'', \ldots$, respectively, are the same, for $u'$ and $v'$, then for $u''$ and $v''$, 
$\dots$, and the same for the massive field and for its massless counterpart. In particular for the
free spinor Dirac field $s = 0$. For the free electromagnetic potential field $s=-1/2$. For the scalar field $s = -1/2$.

Because
\begin{multline*}
\big\langle \kappa_q, \phi \big\rangle = \big\langle \widetilde{\kappa_q}, \widetilde{\phi} \big\rangle  =
\int 
u'(\boldsymbol{\p}') v'(\boldsymbol{\p}') u''(\boldsymbol{\p}'') v''(\boldsymbol{\p}'') \ldots 
u^{(q)}(\boldsymbol{\p}^{(q)}) v^{(q)}(\boldsymbol{\p}^{(q)}) \, \times 
\\
\times
\,
\widetilde{\phi}(-\boldsymbol{p}' - \boldsymbol{p}'' - \ldots, \,-p_0(\boldsymbol{p}') -p_0(\boldsymbol{p}'') - \ldots)
\, \ud^3 \boldsymbol{p}' \, \ud^3 \boldsymbol{p}''  \ldots \boldsymbol{p}^{(q)}
\end{multline*}
then the singularity degree $\omega$ at zero of the scalar $\otimes_q$-contraction, or product 
\[
\kappa_q(x-y) =
\big(\kappa'_{0,1} \dot{\otimes} \kappa''_{0,1} \dot{\otimes} \kappa'''_{0,1} \dot{\otimes} \ldots \dot{\otimes} \kappa^{(q)}_{0,1} \big)
\, \otimes_{q} \,\, \big(\kappa'_{1,0} \dot{\otimes} \kappa''_{1,0}  \dot{\otimes} \kappa'''_{1,0} \dot{\otimes} 
\ldots \dot{\otimes} \kappa^{(q)}_{1,0} \big)(x,y)
\]
of $q$ pairings, respectively, of $q$ free
fields $\mathbb{A}', \mathbb{A}'', \ldots$, is equal
\[
\omega = 2(s'+s'' + \ldots +s^{(q)}) +3q - 4. 
\]

In each QFT we have the finite set of basic $\otimes_1$, $\otimes_2, \ldots$, $\otimes_q$-contractions 
$\kappa_1$, $\kappa_2, \ldots$, $\kappa^{(q)}$,
which enter into the Wick product decomposition of the product $\mathcal{L}(x)\mathcal{L}(y)$  of the
interaction Lagrangian of the theory. The whole theory can be constructed with the help of the advanced and retarded 
parts of the basic $\kappa_1$, $\kappa_2, \ldots$, $\kappa^{(q)}$. The maximal number $q$ depends on the degree of the Wick
polynomial $\mathcal{L}(x)$, which in case of QED is equal $3$.   
In spinor QED we have the following basic $\otimes_1$,$\otimes_2$ and $\otimes_3$-contractions: 
\begin{enumerate}
\item[1)]
the one-contraction $\kappa_1$ of the Dirac field
with the conjugated Dirac field, with $s' = 0$, $q=1$ and with $\omega = 2(0) +3-4 = -1$, 
\item[2)]
the one-contraction $\kappa_1$ of the electromagnetic potential with itself, with $s'=-1/2$, $s' = +1/2$ , 
$q=1$, and with  $\omega= 2(-1/2) +3 -4= -2$, 

\item[4)]
the two-contraction $\kappa_2$ containing two contractions of the Dirac field
with the conjugated Dirac field with $s'=s''=0$, $q=2$ and $\omega = 2(0+0) + 3\cdot 2 - 4 = 2$,

\item[5)]
the two-contraction $\kappa_2$ containing one contraction of the Dirac field
with the conjugated Dirac field and one contraction of electromagnetic potential with itself,
with $s'=0$, $s''=-1/2$, $q=2$,
and with $\omega = 2(0-1/2) +3\cdot 2 - 4 = 1$,

\item[6)]
the three-contraction $\kappa_3$ containing two contractions of the Dirac field
with the conjugated Dirac field and one contracion 
of the electromagnetic potential with itself, with $s'=s''=0$, $s'''=-1/2$, 
$q=3$, and $\omega = 2(0+0-1/2) +3\cdot 3 -4 = 4$. 
\end{enumerate}

The singularity degree $\omega$ of the symmetric or antisymmetric part 
\[
\kappa_q - (-1)^q \check{\kappa_q} \in \mathscr{E}^*,
\]
of the product $\kappa_q$ of pairings,  is the same as the singularity degree of the respective $\kappa_q$. 

Now we observe that in order to show the convergence of (\ref{Convergent[A',A''][A'', A'']Slambda(chi0)})
of the limit $\lambda\rightarrow +\infty$ for each $\phi\in\mathscr{E}$ fulfilling 
(\ref{up-to-omegaorder-derivatives-of-phi=0}), it is sufficient to show convergence of the limit
$n \rightarrow +\infty$ of
\begin{equation}\label{Convergent[A',A''][A'', A'']Slambda^n(chi0)}
\Big\langle (S_{\lambda^n}\chi_0) \big[\kappa_q - (-1)^q \check{\kappa_q}\big], \phi \Big\rangle 
\end{equation}
 for each fixed $\lambda >1$ and $\phi\in\mathscr{E}$ fulfilling 
(\ref{up-to-omegaorder-derivatives-of-phi=0}), or, equivalently, that (\ref{Convergent[A',A''][A'', A'']Slambda^n(chi0)})
is a Cauchy sequence in $n$, for each fixed $\lambda >1$ and $\phi\in\mathscr{E}$ fulfilling 
(\ref{up-to-omegaorder-derivatives-of-phi=0}). We present here a proof due to \cite{Scharf}, pp. 175-177, and going back to \cite{Epstein-Glaser},
that (\ref{Convergent[A',A''][A'', A'']Slambda^n(chi0)}) is a Cauchy sequence. 

Because 
\[
\underset{\lambda \rightarrow +\infty}{\textrm{lim}} \,\,
\big({\textstyle\frac{1}{\lambda}}\big)^\omega \Big\langle [\kappa_q - (-1)^q \check{\kappa_q}], S_{\lambda}\phi \Big\rangle
=
\Big\langle [\kappa_{q \,0}-(-1)^q \check{\kappa_{q \, 0}}],\phi\big\rangle
\]
exists and is finite for each $\phi\in\mathscr{E}$, then for each $h \in \mathscr{E}$
\[
\underset{\lambda \rightarrow +\infty}{\textrm{lim}} \,\,
\big({\textstyle\frac{1}{\lambda}}\big)^\omega \Big\langle h[\kappa_q - (-1)^q \check{\kappa_q}], S_{\lambda}\phi \Big\rangle
=
\Big\langle [\kappa_{q \,0}-(-1)^q \check{\kappa_{q \, 0}}],\phi\big\rangle \, h(0)
\]
exists and is finite for each $\phi\in\mathscr{E}$, and thus 
\begin{multline*}
\underset{\lambda \rightarrow +\infty}{\textrm{lim}} \,\,
\big({\textstyle\frac{1}{\lambda}}\big)^{\omega-|\alpha|} \Big\langle x^\alpha h [\kappa_q - (-1)^q \check{\kappa_q}], S_{\lambda} \phi \Big\rangle
\\
=
\underset{\lambda \rightarrow +\infty}{\textrm{lim}} \,\,
\big({\textstyle\frac{1}{\lambda}}\big)^{\omega-|\alpha|} \Big\langle h [\kappa_q - (-1)^q \check{\kappa_q}], 
S_{\lambda} \big (\textstyle{\frac{x^\alpha}{\lambda^{|\alpha|}}} \phi\big) \Big\rangle
\\
=
\underset{\lambda \rightarrow +\infty}{\textrm{lim}} \,\,
\big({\textstyle\frac{1}{\lambda}}\big)^{\omega} \Big\langle h [\kappa_q - (-1)^q \check{\kappa_q}], 
S_{\lambda} \big (x^\alpha \phi\big) \Big\rangle
=
\Big\langle [\kappa_{q \,0}-(-1)^q \check{\kappa_{q \, 0}}], x^\alpha\phi \Big\rangle \,\, h(0)
\end{multline*}
exists and is finite for each $h,\phi\in\mathscr{E}$.
Next, for each $\phi\in\mathscr{E}$ fulfilling 
(\ref{up-to-omegaorder-derivatives-of-phi=0}) there exists multiindices $\alpha$, $|\alpha| = \omega+1$, and $\phi_\alpha \in \mathscr{E}$
such that (summation is understood over $\alpha$ with $|\alpha| = \omega+1$)
\[
\phi = x^\alpha \phi_\alpha.
\]
Let $\psi$ be an auxiliary smooth bounded function equal $1$ on the closure of the future and past light cones, 
\emph{i.e.} on the support of $\kappa_q - (-1)^q \check{\kappa_q}$,
equal zero outside an open $\epsilon$-neighborhood of the light cone 
(sum of open $\epsilon$-balls centered at the points of the closed future and past light cones). Note also
that for each  $\lambda>0$, also $S_\lambda\psi$ is smooth, bounded, equal $1$ on the cone
(the support of $\kappa_q - (-1)^q \check{\kappa_q}$), and equal zero outside the  $\epsilon/\lambda$-neighborhood of the cone. 
Then, for each $\lambda \geq 1$ the intersection
\[
\textrm{supp} \, \psi \cap \textrm{supp} \, [S_{\lambda}\chi_0 - \chi_0]
\]
has compact closure and 
\[
[S_{\lambda}\chi_0 - \chi_0] \psi \in \mathcal{D}(\mathbb{R}^4) \subset \mathcal{S}(\mathbb{R}^4).
\]   
Therefore, using the causal support of $\kappa_q - (-1)^q \check{\kappa_q}$,  for each fixed $\lambda >1$ and $\phi\in\mathscr{E}$ fulfilling 
(\ref{up-to-omegaorder-derivatives-of-phi=0}) we get
\begin{multline*}
\Big\langle (S_{\lambda^{m+n}}\chi_0) \big[\kappa_q - (-1)^q \check{\kappa_q}\big], \phi \Big\rangle 
-
\Big\langle (S_{\lambda^m}\chi_0) \big[\kappa_q - (-1)^q \check{\kappa_q}\big], \phi \Big\rangle 
\\
=
\Big\langle (S_{\lambda^{m+n}}\chi_0 -S_{\lambda^m}\chi_0) x^\alpha\big[\kappa_q - (-1)^q \check{\kappa_q}\big], \phi_\alpha \Big\rangle 
\\
=
\Big\langle (S_{\lambda^{m+n}}\chi_0 -S_{\lambda^m}\chi_0) x^\alpha\big[\kappa_q - (-1)^q \check{\kappa_q}\big], \psi \phi_\alpha \Big\rangle 
\end{multline*}
\begin{multline*}
= \Big\langle 
\phi_\alpha x^\alpha \big[\kappa_q - (-1)^q \check{\kappa_q}\big],
\, 
(S_{\lambda^{m+n}}\chi_0 -S_{\lambda^m}\chi_0)
\psi \Big\rangle 
\\
= \Big\langle 
\phi_\alpha x^\alpha \big[\kappa_q - (-1)^q \check{\kappa_q}\big],
\, 
S_{\lambda^m}(S_{\lambda^{n}}\chi_0 -\chi_0).\psi\big]
\Big\rangle 
\end{multline*}
\begin{multline*}
=
\sum\limits_{j=0}^{n-1}
\Big\langle 
\phi_\alpha x^\alpha \big[\kappa_q - (-1)^q \check{\kappa_q}\big],
\, 
S_{\lambda^j}S_{\lambda^m}\big[(S_{\lambda}\chi_0 -\chi_0)\big] \, \psi
\Big\rangle 
\\
=
\sum\limits_{j=0}^{n-1}
\Big\langle 
\phi_\alpha x^\alpha \big[\kappa_q - (-1)^q \check{\kappa_q}\big],
\, 
S_{\lambda^j}S_{\lambda^m}\big[(S_{\lambda}\chi_0 -\chi_0)\big] \, S_{\lambda^j}S_{\lambda^m} \psi
\Big\rangle 
\end{multline*}
\begin{multline*}
=
\sum\limits_{j=0}^{n-1}
\Big\langle 
\phi_\alpha x^\alpha \big[\kappa_q - (-1)^q \check{\kappa_q}\big],
\, 
S_{\lambda^j}S_{\lambda^m}\big[(S_{\lambda}\chi_0 -\chi_0)\psi\big]
\Big\rangle 
\\
=
\sum\limits_{j=0}^{n-1}
\big(\lambda^{j+m}\big)^{\omega-|\alpha|}
\big({\textstyle\frac{1}{\lambda^{j+m}}}\big)^{\omega-|\alpha|}
\Big\langle 
\phi_\alpha x^\alpha \big[\kappa_q - (-1)^q \check{\kappa_q}\big],
\, 
S_{\lambda^j}S_{\lambda^m}\big[(S_{\lambda}\chi_0 -\chi_0)\psi\big]
\Big\rangle 
\end{multline*}
Now, because
\begin{multline*}
\big({\textstyle\frac{1}{\lambda^{j+m}}}\big)^{\omega-|\alpha|}
\Big\langle 
\phi_\alpha x^\alpha \big[\kappa_q - (-1)^q \check{\kappa_q}\big],
\, 
S_{\lambda^j}S_{\lambda^m}\big[(S_{\lambda}\chi_0 -\chi_0)\psi\big]
\Big\rangle 
\\
\overset{m,j\rightarrow\infty}{\longrightarrow}
\Big\langle 
 \big[\kappa_q - (-1)^q \check{\kappa_q})\big]_0,
\, 
x^\alpha (S_{\lambda}\chi_0 -\chi_0)\psi
\Big\rangle \,\, \phi_\alpha(0)
\end{multline*}
are convergent, then for each fixed $\lambda>1$, $\phi\in\mathscr{E}$, the set of the numbers
\[
\big({\textstyle\frac{1}{\lambda^{j+m}}}\big)^{\omega-|\alpha|}
\Big\langle 
\big[\phi_\alpha x^\alpha (\kappa_q - (-1)^q \check{\kappa_q})\big],
\, 
S_{\lambda^j}S_{\lambda^m}\big[(S_{\lambda}\chi_0 -\chi_0)\psi\big]
\Big\rangle, \,\,\, j,m \in \mathbb{N}
\]
is bounded. Let $B$ denote an upper bound of their absolute values. Thus, putting $\varepsilon = |\alpha|-\omega$,
\begin{multline*}
\Big|
\Big\langle (S_{\lambda^{m+n}}\chi_0) \big[\kappa_q - (-1)^q \check{\kappa_q}\big], \phi \Big\rangle 
-
\Big\langle (S_{\lambda^m}\chi_0) \big[\kappa_q - (-1)^q \check{\kappa_q}\big], \phi \Big\rangle 
\Big|
\\
\leq 
B
\sum\limits_{j=0}^{n-1}
\big(\lambda^{j+m}\big)^{\omega-|\alpha|}
= B
\sum\limits_{j=0}^{n-1}
\big(\lambda^{j+m}\big)^{-\varepsilon}
= B \lambda^{-\varepsilon m}
\,
{\textstyle\frac{1-\lambda^{-\varepsilon n}}{1-\lambda^{-\varepsilon}}} \,\,\,
\overset{m,n\rightarrow\infty}{\longrightarrow} 0,
\end{multline*}
and (\ref{Convergent[A',A''][A'', A'']Slambda^n(chi0)}) is a Cauchy sequence in $n$
for each $\phi\in\mathscr{E}$ fulfilling 
(\ref{up-to-omegaorder-derivatives-of-phi=0}). This also proves
convergence of (\ref{Convergent[A',A''][A'', A'']Slambda(chi0)})
if  $\lambda\rightarrow +\infty$ for each $\phi\in\mathscr{E}$ fulfilling 
(\ref{up-to-omegaorder-derivatives-of-phi=0}).

If the integral (\ref{Convergent[A',A''][A'', A'']theta}) is convergent, then
it has to be equal to the limit $\lambda\rightarrow +\infty$ of (\ref{Convergent[A',A''][A'', A'']Slambda(chi0)}),
on the subspace of test functions $\phi$ which respect (\ref{up-to-omegaorder-derivatives-of-phi=0}).
Indeed, by construction the limit has singularity degree $\omega$, the same as $\kappa_q-(-1)^q\check{\kappa_q}$
and the same as $\kappa_q$. It is easily seen that also the singularity degree
of (\ref{Convergent[A',A''][A'', A'']theta}), provided it is convergent, also must be equal to the singularity 
degree $\omega$ of $\kappa_q$. Because (\ref{Convergent[A',A''][A'', A'']theta}) and the 
 limit $\lambda\rightarrow +\infty$ of (\ref{Convergent[A',A''][A'', A'']Slambda(chi0)}) are both retarded parts
of the same causal distribution $\kappa_q-(-1)^q\check{\kappa_q}$, they differ at most by a distribution supported at zero:
\[
\sum\limits_{\alpha} C_\alpha \delta^{(\alpha)}
\]
and because, moreover, they both have the same singularity degree $\omega$, this sum cannot
involve terms with $|\alpha|>\omega$, thus their difference can at most be equal 
\[
\sum\limits_{|\alpha|=0}^{\omega} C_\alpha \delta^{(\alpha)}.
\] 
But the last distribution is zero on the subspace of test functions $\phi$ respecting (\ref{up-to-omegaorder-derivatives-of-phi=0}).

Convergence of the limit $\lambda\rightarrow +\infty$ of (\ref{Convergent[A',A''][A'', A'']Slambda(chi0)}) only shows existence
of the retarded part of $\kappa_q-(-1)^q\check{\kappa_q}$ on the subspace of $\phi$ which respect (\ref{up-to-omegaorder-derivatives-of-phi=0}).
For the practical computations this result is not very much interesting yet, as we need a practical method for computation
of the retarded part, or its Fourier transform, explicitly. For this task the formula (\ref{Convergent[A',A''][A'', A'']theta})
is useful, because it can be converted into a practical formula.

Therefore we concentrate now on the proof of convergence of
(\ref{Convergent[A',A''][A'', A'']theta}) or, equivalently, (\ref{thetaASkappa2(chi)}), for
\[
\chi(x,y) = \phi(x-y)\varphi(y), \,\,\,\,\,\,\,\, \phi, \varphi \in \mathscr{E},
\]
with $\phi$ which respects (\ref{up-to-omegaorder-derivatives-of-phi=0}), with $\omega$ in (\ref{up-to-omegaorder-derivatives-of-phi=0})
equal at least to the singularity order of the distribution
\[
\kappa_q(x)-(-1)^q\kappa_q(-x) = \big[\kappa_q-(-1)^q \check{\kappa_q} \big](x).
\]

Presented proof of convergence of the limit $\lambda\rightarrow +\infty$ of (\ref{Convergent[A',A''][A'', A'']Slambda(chi0)}), 
based on the singularity degree, is not very much interesting also because this proof strongly uses causality of 
$\kappa_q-(-1)^q -\check{\kappa_q}$, property which no longer holds for each
term $\kappa_q$ or $\check{\kappa_q}(x) = \kappa_q(-x)$, taken separately. Practical computations, which we need to perform, give
more effective construction of the retarded part, and which is inspired by the integral formula, and allows to prove
that the following integral is convergent in distributional sense and represents well-defined functional of $\chi$ 
\begin{equation}\label{thetakappaq(chi)}
\Big[
\theta_y\big(\kappa'_{0,1} \dot{\otimes} \kappa''_{0,1} \dot{\otimes} \dots \dot{\otimes} \kappa^{(q)}_{0,1}\big)
 \otimes_q \big(\kappa'_{1,0} \dot{\otimes} \kappa''_{1,0} \dot{\otimes} \ldots 
\dot{\otimes} \kappa^{(q)}_{1,0}\big)
\Big] (\chi)
\end{equation}
equal
\[
\int
\kappa_{q}(x-y)
\theta(x-y) \, \phi(x-y)\varphi(y) \, \ud^4 x \, \ud^4 y
=
\widetilde{\varphi}(0) \,\, \big\langle \kappa_q, \phi \big\rangle 
= \widetilde{\varphi}(0) \,\, \big\langle\kappa_q, \theta \phi \big\rangle
\]
\begin{multline}\label{ConvergentProdPairings[A',A''][A'', A'']theta}
=
\widetilde{\varphi}(0) \,\,
\int \big\{
u'(\boldsymbol{\p}') v'(\boldsymbol{\p}') u''(\boldsymbol{\p}'') v''(\boldsymbol{\p}'') \ldots 
\,
\widetilde{\theta\phi}(-\boldsymbol{p}' - \boldsymbol{p}'' - \ldots, \,-|\boldsymbol{p}'| -|\boldsymbol{p}''|-\ldots) 
\\
\big\}
\, \ud^3 \boldsymbol{p}' \, \ud^3 \boldsymbol{p}'' \ldots  \ud^3 \boldsymbol{p}^{(q)},
\end{multline}
for 
\[
\chi(x,y) = \phi(x-y)\varphi(y), \,\,\,\, \phi,\varphi \in \mathscr{E} = \mathcal{S}(\mathbb{R}^4),
\]
and with 
\[
\phi \in \mathscr{E}= \mathcal{S}(\mathbb{R}^4) 
\]
which fulfil (\ref{up-to-omegaorder-derivatives-of-phi=0}). After \cite{Scharf} we briefly report a practical method for calculation which, 
as far as we know, goes back to \cite{Scharf}. Namely, the Fourier transform of the scalar $\otimes_q$-contraction
$\kappa_q$ can be computed explicitly quite easily, on using the completeness relations of the plane wave kernels. 
Below in this Subsection, as an example, we give Fourier transforms $\widetilde{\kappa_q}$ of all basic scalar $\kappa_q$, $q=1,2,3$, $\otimes_q$-contractions 
for QED. It turns out that for $q>1$ (the most interesting case)  $\widetilde{\kappa_q}$ are regular, \emph{i.e.} function-like,
distributions, analytic in $p$ everywhere, except the the finite set of characteristic submanifolds of the type $p\cdot p = const.$, or $p_0=0$,
where they have $\theta(p\cdot p - const.)$, $\theta(p_0)$-type finite jump. Next  we observe
that each $\phi \in \mathscr{E}$, which fulfils (\ref{up-to-omegaorder-derivatives-of-phi=0}), can be written
as an element of the image of a continuous idempotent operator $\Omega'$ acting on $\mathscr{E} = \mathcal{S}(\mathbb{R}^4)$,
and  (\ref{thetakappaq(chi)}) can be understood as defined on $\chi(x,y)=(\Omega'\phi)(x-y)\varphi(y)$,
now for $\phi,\varphi$ ranging over general elements in $\mathscr{E}$. 
Indeed, let, for each multi-index $\alpha$, such that $0 \leq |\alpha| \leq \omega$,
$\omega_{{}_{o \,\, \alpha}} \in \mathscr{E}$ on $\mathbb{R}^{4}$ be such
functions that\footnote{As we have already mentioned such functions $\omega_{{}_{o \,\, \alpha}} \in \mathscr{E}$, $0 \leq |\alpha| \leq \omega$ do exist.
Indeed, let
\[
f_{{}_{\alpha}}(x) =
{\textstyle\frac{x^\alpha}{\alpha!}}, \,\,\, x \in \mathbb{R}^{4}, 0 \leq |\alpha| \leq \omega,
\]
\[
\alpha ! = \prod \limits_{\mu=0}^{3} \alpha_{\mu} !,
\,\,\, x^\alpha = \prod \limits_{\mu=0}^{3} (x_{\mu})^{\alpha_{\mu}}, \,\,\, \mu = 0,1,2,3.
\]
It is well-known fact that there exists $w \in \mathscr{E} = \mathcal{S}(\mathbb{R}^4)$,
which is equal $1$ on some neighborhood of zero.
Then we can put
\[
\omega_{{}_{o \,\, \alpha}} \overset{\textrm{df}}{=} f_{{}_{\alpha}}.w.
\]}
\[
D^\beta \omega_{{}_{o \,\, \alpha}} (0) = \delta^{\beta}_{\alpha}, \,\,\,\, 0 \leq |\alpha|, |\beta| \leq \omega.
\]
Then, for any $\phi \in \mathscr{E}$ we put
\[
\Omega' \phi = \phi - \sum \limits_{0 \leq |\alpha| \leq \omega} D^\alpha \phi(0) \,\, \omega_{{}_{o \,\, \alpha}}
\,\,\,
=
 \phi - \sum \limits_{0 \leq |\alpha| \leq \omega} D^\alpha \phi(0) \,\, {\textstyle\frac{x^\alpha}{\alpha!}}.w
\]
Now, the integral (\ref{thetakappaq(chi)}), with $\chi(x,y)=(\Omega'\phi)(x-y)\varphi(y)$ and $\phi, \varphi \in \mathscr{E}$, becomes equal 
\[
\widetilde{\varphi}(0) \,\, \big\langle\kappa_q, \theta \Omega'\phi \big\rangle = 
\big\langle \widetilde{\kappa_q}, \widetilde{\theta} \ast \widetilde{\Omega'\phi} \big\rangle, 
\] 
and is expected to be a well-defined continuous functional of $\widetilde{\phi},\widetilde{\varphi} \in \mathcal{S}(\mathbb{R}^4)$,
or equivalently
\[
\big\langle \widetilde{\kappa_q}, \widetilde{\theta} \ast \widetilde{\Omega'\phi} \big\rangle, 
\]
should give a well-defined functional of $\phi \in \mathcal{S}(\mathbb{R}^4)$. It defines
(\ref{thetakappaq(chi)}) for all $\chi(x,y)=\phi(x-y)\varphi(y)$ with
$\phi \in \textrm{Im} \, \Omega'$, and because $\textrm{Ker} \, \Omega' \neq\{0\}$ for $q>1$, its definition
is not unique and can be extended by addition of any functional which is zero 
on $\chi(x-y)\varphi(y) = \phi(x-y)\varphi(y)$ with
$\phi \in \textrm{Im} \, \Omega'$. Equivalently the retarded part
\begin{gather}
\textrm{ret} \, \kappa_q \overset{\textrm{df}}{=} \kappa_q \circ (\theta \Omega'),
\label{Def(retkappaq)}
\\
\big\langle \textrm{ret} \, \kappa_q, \phi \big\rangle
= \big\langle \widetilde{\textrm{ret} \, \kappa_q}, \widetilde{\phi} \big\rangle
=\big\langle\kappa_q, \theta \Omega'\phi \big\rangle  
=\big\langle \widetilde{\kappa_q}, \widetilde{\theta} \ast \widetilde{\Omega'\phi} \big\rangle
\label{FT(retkappaq)}
\end{gather}
of $\kappa_q$ is not unique, and we can add to it
\begin{equation}\label{freedom}
\sum\limits_{|\alpha|=0}^{\omega} C_\alpha \delta^{\alpha},
\end{equation} 
which is most general on $\textrm{Ker} \, \Omega'$ and which is zero on $\textrm{Im} \, \Omega'$. 
In order to show (\ref{Def(retkappaq)}) is a well-defined tempered distribution, we proceed after \cite{Scharf},
using the explicit formula for the function $\widetilde{\kappa_q}$ in (\ref{FT(retkappaq)}). 

Since
\[
\widetilde{x^\alpha w}(p) = (iD_{p})^\alpha \widetilde{w}(p),
\]
and
\[
D^\alpha\phi(0) = (-1)^\alpha \big\langle D^\alpha \delta, \phi \big\rangle 
= (-1)^\alpha \big\langle \widetilde{D^\alpha\delta}, \widetilde{\phi} \big\rangle
=
 (2\pi)^{-4/2} \big\langle (ip)^\alpha, \widetilde{\phi}  \big\rangle,
\]
immediately from (\ref{FT(retkappaq)}) we obtain
\begin{multline*}
\big\langle \widetilde{\textrm{ret} \, \kappa_q}, \widetilde{\phi} \big\rangle
= \big\langle \widetilde{\kappa_q}, \widetilde{\theta \Omega'\phi}\big\rangle
= 
(2\pi)^{-4/2} \Bigg\langle \widetilde{\kappa_q},  \widetilde{\theta}
\\
\ast \,\, \Big[ \widetilde{\phi} - \sum\limits_{|\alpha|=0}^{\omega} {\textstyle\frac{1}{\alpha!}} (iD_{p})^\alpha \widetilde{w}
(2\pi)^{-4/2} \big\langle (ip')^\alpha, \widetilde{\phi} \big\rangle \Big] \Bigg\rangle
\\
=
(2\pi)^{-4/2} \Big\langle \widetilde{\theta} \ast \widetilde{\kappa_q}, \,\,\,
\widetilde{\phi} - \sum\limits_{|\alpha|=0}^{\omega} {\textstyle\frac{1}{\alpha!}} (iD_{p})^\alpha \widetilde{w}
(2\pi)^{-4/2} \big\langle (ip')^\alpha, \widetilde{\phi} \,\, \Big\rangle
\end{multline*}
were the convolution $\widetilde{\theta} \ast \widetilde{\kappa_q}$ is defined only on the subtracted $\widetilde{\phi}$,
\emph{i.e.} on Fourier transforms of test functions with all derivatives vanishing  at zero up to order $\omega$. Interchanging the integration
variables $p'$ and $p$ in the subtracted terms we get
\begin{multline*}
\big\langle \widetilde{\textrm{ret} \, \kappa_q}, \widetilde{\phi} \big\rangle 
= (2\pi)^{-4/2} \int d^4 k \, \widetilde{\theta}(k) \,\, \Big\langle \widetilde{\kappa_q}(p-k)
\\
- (2\pi)^{-4/2} \sum\limits_{|\alpha|=0}^{\omega} {\textstyle\frac{(-1)^\alpha}{\alpha!}}p^\alpha 
\int d^4 p' \,  \widetilde{\kappa_q}(p'-k) D_{p'}^{\alpha}\widetilde{w}, \, \widetilde{\phi} \Big\rangle.
\end{multline*}
By partial integration in the $p'$ variable we arrive at the formula
\begin{equation}\label{FT(retkappaq)1}
\widetilde{\textrm{ret} \, \kappa_q}(p) = {\textstyle\frac{1}{(2\pi)^2}} \int d^4k \, \widetilde{\theta}(k)
\Bigg[
\widetilde{\kappa_q}(p-k) - {\textstyle\frac{1}{(2\pi)^2}} \sum\limits_{|\alpha|=0}^{\omega}
{\textstyle\frac{p^\alpha}{\alpha!}}\int d^4 p' \, D^\alpha \widetilde{\kappa_q}(p'-k)\widetilde{w}(p')  
\Bigg],
\end{equation}
which, in general should be understood in the distributional sense (converging when
integrated in $p$ variable with a test function of $p$). But in practical computations we are interesting
in the regularity domain, outside the characteristic submanifold $p\cdot p = const.$, where $\widetilde{\textrm{ret} \, \kappa_q}$
is represented by ordinary function of $p$ and where 
this integral should be convergent in the ordinary sense. 
Then we assume that there exists a point $p''$ around which (\ref{thetakappaq(chi)}) is regular, and moreover
possess all derivatives in the usual sense (should not be mixed with distributional) at $p''$, 
up to order $\omega$ equal to the singularity degree. This is not entirely arbitrary assumption, because
if the formula (\ref{thetakappaq(chi)}) makes any sense at all as a functional of $\phi$, it must have support
in the half space (or forward cone in case we replace $\kappa_q$ with the causal $\kappa_q - (-1)^q\check{\kappa_q}$),
so its Fourier transform should be a boundary value of an analytic function, regular all over $\mathbb{R}^4$,
but compare the comment we give below on this topic.
Next, we subtract  the first 
Taylor terms of $\widetilde{\textrm{ret} \, \kappa_q}$ up to order $\omega$ around the regularity point $p''$ 
\begin{equation}\label{FT(retkappa)p''}
\big[\widetilde{\textrm{ret} \, \kappa_q}\big]_{{}_{{}_{p''}}}(p) = \widetilde{\textrm{ret} \, \kappa_q}(p)  
- \sum\limits_{|\beta|=0}^{\omega}
{\textstyle\frac{(p-p'')^\beta}{\beta!}} D^\beta \widetilde{\textrm{ret} \, \kappa_q}(p'') 
\end{equation}
and get another possible retarded part $\big[\widetilde{\textrm{ret} \, \kappa_q}\big]_{{}_{{}_{p''}}}$, 
Fourier transformed, of $\kappa_q$, as the added terms have the form (\ref{freedom}). 
By construction (\ref{FT(retkappa)p''}) is so normalized, that all its derivatives vanish up to order $\omega$
at $p''$. This point is called \emph{normalization point}.
Next we compute $D^\beta \widetilde{\textrm{ret} \, \kappa_q}$ from the formula (\ref{FT(retkappaq)1}), and substitute
into the formula (\ref{FT(retkappa)p''}). Using
\[
\sum\limits_{\beta\leq \alpha} {\textstyle\frac{(p-p'')^\beta}{\beta!}}
 {\textstyle\frac{{p''}^{\alpha-\beta}}{(\alpha-\beta)!}} 
= 
{\textstyle\frac{1}{\alpha!}}
\sum\limits_{\beta\leq \alpha} {\alpha \choose \beta} (p-p'')^\beta {p''}^{\alpha-\beta}
= {\textstyle\frac{p^\alpha}{\alpha!}},
\]
we see that all terms with the auxiliary function $\widetilde{w}$
drop out and we obtain for $\big[\widetilde{\textrm{ret} \, \kappa_q}\big]_{{}_{{}_{p''}}}$ the formula
\begin{equation}\label{FT(retkappaq)2}
\big[\widetilde{\textrm{ret} \, \kappa_q}\big]_{{}_{{}_{p''}}}(p) = {\textstyle\frac{1}{(2\pi)^2}} \int d^4k \widetilde{\theta}(k)
\Bigg[
\widetilde{\kappa_q}(p-k) - \sum\limits_{|\beta|=0}^{\omega}
{\textstyle\frac{(p-p'')^\beta}{\beta!}} D^\beta \widetilde{\kappa_q}(p''-k) 
\Bigg],
\end{equation}
by construction so normalized that all derivatives of $\big[\widetilde{\textrm{ret} \, \kappa_q}\big]_{{}_{{}_{p''}}}$ 
vanish at $p''$ up to order $\omega$, compare \cite{Scharf}, p. 179. 
We need to rewrite the last integral (\ref{FT(retkappaq)2}) with $\widetilde{\kappa_q}$ taken at the same point in the two terms
in (\ref{FT(retkappaq)2}), in order to combine the two terms into a single term, as these terms taken separately
are divergent. At this point the case in which we have causal $\kappa_q-(-1)^q\check{\kappa_q}$ and Lorentz convariant, instead of
$\kappa_q$, is simpler, because $\kappa_q-(-1)^q\check{\kappa_q}$ has causal support and is Lorentz convariant. 
In this causal case we can use the special $\theta(x) = \theta(x_0) = \theta(v\cdot x)$ with $v=(1,0,0,0)$, and insert
\begin{equation}\label{FTtheta}
\widetilde{\theta}(k_0,\boldsymbol{k}) = 2\pi {\textstyle\frac{i}{k_0+i\epsilon}} \,
\delta(\boldsymbol{k}),
\end{equation}
and use Lorentz frame in which $p=(p_0,0,0,0)$ and then using integration by parts we remove the differentiainion
operation in momentum variables  off $\widetilde{\kappa_q}$ in the second term in (\ref{FT(retkappaq)2}).  
Next we use invariance and analytic continuation to extend 
$\widetilde{\textrm{ret} \, \kappa_q}$ all over $p$.
Because  $\kappa_q$ is not causal, nor Lorentz invariant, we need to consider separately the case
$p=(p_0,0,0,0)$ inside the positive and negative cone and $p=(0,0,0,p_3)$ outside the cone in the momentum space
as well as $\theta(x) = \theta(v\cdot x)$ with a more general timelike unit $v$
in
\[
\widetilde{\theta}(k_0,\boldsymbol{k}) = 2\pi {\textstyle\frac{i}{k_0 +i\epsilon v_0}} \,
\delta(\boldsymbol{k}- k_0{\textstyle\frac{\boldsymbol{v}}{v_0}} ),
\]
in order to reconstruct $\widetilde{\textrm{ret} \, \kappa_q}$, now depending on $v$, as $\widetilde{\kappa_q}$
is not causal. 

Still proceeding after \cite{Scharf}, let us concentrate now on the computation of the retarded part the causal $d= \kappa_q-(-1)^q\check{\kappa_q}$,which
is simpler and moreover, on such QFT in which the normalization point $p''$ can be put equal zero. 
This is e.g. the case for QED's with massive charged fields. 
We reconstruct $\widetilde{\textrm{ret} \, d}(p)$ first for $p$ in the cone $V^+$.  
In this case computations simplify, as we can put $p''=0$, and moreover
we use Lorentz covariance of the causal $d$, and choose a Lorentz frame in which $p=(p_0,0,0,0)$.
Moreover, because $d$ is causal its retarded part is independent of the choice of the time like
unit versor $v$ in the theta function $\theta(v\cdot x)$, so that we can choose $v=(1,0,0,0)$, thus assuming that it is
always parallel  to $p\in V^+$, and thus varies with $p$. After these choices, the last formula
(\ref{FT(retkappaq)2}) for the retarded part, applied to $d$ instead of $\kappa_q$, and with substituted
Fourier transform (\ref{FTtheta}), reads
\begin{multline*}
\widetilde{\textrm{ret} \, d}(p_0,0,0,0) 
\\
= {\textstyle\frac{i}{2\pi}} \int dk_0 
{\textstyle\frac{1}{k_0+i\epsilon}}
\Bigg[
\widetilde{d}(p_0-k_0,0,0,0) - \sum\limits_{a=0}^{\omega}
{\textstyle\frac{(p_0)^a}{a!}}(-1)^a D_{k_0}^{a} \widetilde{d}(q_0-k_0,0,0,0)\big|_{{}_{q_0=0}} 
\Bigg].
\end{multline*}
Integrating the subtracted terms by parts and then using the intergration variable $k'_{0}= k_0-p_0$ in the first term, we get
\[
\widetilde{\textrm{ret} \, d}(p_0) = {\textstyle\frac{i}{2\pi}} \int dk'_{0} 
\,
\Bigg[
{\textstyle\frac{1}{p_0 + k'_{0}+i\epsilon}}
 - \sum\limits_{a=0}^{\omega}
{\textstyle\frac{(p_0)^a}{a!}} 
{\textstyle\frac{\partial^a}{\partial {k'}_{0}^{a}}}
{\textstyle\frac{1}{k'_{0}+i\epsilon}}
\Bigg] \,
\widetilde{d}(-k'_{0}),
\]
with
\begin{equation}
{\textstyle\frac{1}{p_0 + k'_{0}+i\epsilon}}
 - \sum\limits_{a=0}^{\omega}
{\textstyle\frac{(p_0)^a}{a!}} 
{\textstyle\frac{\partial^a}{\partial {k'}_{0}^{a}}}
{\textstyle\frac{1}{k'_{0}+i\epsilon}} 
=
\Big({\textstyle\frac{-p_0}{k'_{0}+i\epsilon}}\Big)^{\omega+1}
{\textstyle\frac{1}{p_0 +k'_{0}+i\epsilon}}.
\end{equation}
To recover the formula for arbitrary $p\in V^+$ we introduce the variable $t=k_0/p_0$ and obtain
\[
\widetilde{\textrm{ret} \, d}(p) = {\textstyle\frac{i}{2\pi}} \int\limits_{-\infty}^{+\infty} d t
\,
{\textstyle\frac{\widetilde{d}(tp),}{(t-i\epsilon)^{\omega+1}(1-t+i\epsilon)}},
\] 
or
\begin{equation}\label{FT(retCausalkappaq)}
\mathscr{F} \Big(\textrm{ret} \, \big[\kappa_q-(-1)^q\check{\kappa_q}\big]\Big)(p) = {\textstyle\frac{i}{2\pi}} \int\limits_{-\infty}^{+\infty} d t
\,
{\textstyle\frac{\mathscr{F}\big(\kappa_q-(-1)^q\check{\kappa_q}\big)(tp),}{(t-i\epsilon)^{\omega+1}(1-t+i\epsilon)}},
\,\,\,\, p \in V^+,
\end{equation}
and with  normalization point $p''=0$.
To compute the retarded part outside the cone $V^+$ we have several possibilities, e.g. we can use the analytic 
continuation method, or choose the normalization point $p''$ outside the cone and repeat the computation.

Below in this Subsection we give explicit formulas for $\widetilde{\textrm{ret} \, \kappa_q}$ and
\[
\widetilde{\textrm{ret} \, d} = \mathscr{F}\big\{\textrm{ret} \, [\kappa_q-(-1)^q\check{\kappa_q}]\big\},
\]
for all basic $\kappa_q$, $q=1,2,3$ in case of spinor QED, as an example.

Anyway, explicit formula, obtained by the method outlined above,
shows that $\big[\widetilde{\textrm{ret} \, \kappa_q}\big]_{{}_{{}_{p''}}}$ 
as well as that
\[
\widetilde{\textrm{ret} \, \kappa_q}  = \big[\widetilde{\textrm{ret} \, \kappa_q}\big]_{{}_{{}_{p''}}} - \sum\limits_{|\alpha|=0}^{\omega}
C_\alpha p^\alpha,
\]
for some constants $C_\alpha$, is a well-defined tempered distribution. 
Analogously
\[
\mathscr{F}\big(\textrm{ret} \, [\kappa_q - (-1)\check{\kappa_q}]\big) 
= \Big[\mathscr{F}\big(\textrm{ret} \, [\kappa_q - (-1)\check{\kappa_q}]\big)\Big]_{{}_{{}_{p''}}} - \sum\limits_{|\alpha|=0}^{\omega}
C_\alpha p^\alpha
\]
is a well-defined tempered distribution. Moreover, replacing $\theta(x)$ with $-\theta(x)$ we compute
in the same manner the Fourier transformed advanced part $\widetilde{\textrm{av} \, \kappa_q}$, which ideed
composes together with $\widetilde{\textrm{av} \, \kappa_q}$ the splitting
\[
\widetilde{\kappa_q}= \widetilde{\textrm{ret} \, \kappa_q} - \widetilde{\textrm{av} \, \kappa_q}
\]
of $\kappa_q$ into a retadred and advanced part. 
 
If there exists a regularity point $p''$ for (\ref{FT(retkappaq)1}), which we have used above as an assumption, 
then this would give a proof that  (\ref{thetakappaq(chi)}) is a well-defined
distribution, because its Fourier transform would be equal to a well-defined distribution $\widetilde{\textrm{ret} \, \kappa_q}$. 
Similarly, for the causal combination $\kappa_q - (-1)\check{\kappa_q}$, if there exist a regularity point
$p''$ for (\ref{FT(retkappaq)1}) with $\kappa_q$ replaced with $\kappa_q - (-1)\check{\kappa_q}$, then
its Fourier transform, being given by a formula 
\[
\mathscr{F}\big(\textrm{ret} \, [\kappa_q - (-1)\check{\kappa_q}]\big) 
\]
obtained by the method outlined above and defining a function-like distribution, would be a well-defined distribution.

We know that this is not entirely arbitrary assumption, as for any distribution with cone-shaped support, as e.g. for the retarded
parts, their Fourier transform has to be a boundary value of an analytic function, regular in the tube $\mathbb{R}^4+iV$,
with $V$ being equal to the corresponding support of  $\textrm{ret} \, \kappa_q$, or, $\textrm{ret} \, [\kappa_q - (-1)\check{\kappa_q}]$,
by a well-known theorem on distributions with cone-shaped supports, compare \cite{Reed_SimonII}, Theorem IX.16.
The expression, given by the formula (\ref{thetaASkappa2(chi)}) or (\ref{thetakappaq(chi)}), 
if it makes any sense at all, evidently must be
supported within the future light cone or half-space, respectively, in case of $\kappa_q - (-1)\check{\kappa_q}$
or $\kappa_q$. Therefore, if  (\ref{thetakappaq(chi)}) or (\ref{thetaASkappa2(chi)}) make any sense at all,
their Fourier transform must be equal  (\ref{FT(retkappaq)2}) or (\ref{FT(retkappaq)2}) with $\kappa_q$
replaced by $\kappa_q - (-1)\check{\kappa_q}$, up to a distribution (\ref{freedom}).
Below we give explicit expressions for the Fourier transform  $\widetilde{\kappa_q}$ for particular
examples of products of pairings. Becuse $\widetilde{\kappa_q}$ is analytic except the characteristic submanifold, where it has finite jump,
so in general it is expected that (\ref{FT(retkappaq)1}) indeed possess regularity points $p''$ whenever the integral  (\ref{FT(retkappaq)1})
is convergent. This is indeed the case for the final formula $\textrm{ret} \, \kappa_q$ obtained by the outlined method.  
  
Finally, we replace $\theta(x\cdot v)$, in the above computation of $\widetilde{\textrm{ret} \, \kappa_q}$, with
\begin{gather*}
\theta_{{}_{\lambda}}(x \cdot v) = {\textstyle\frac{1}{2}} +{\textstyle\frac{1}{\pi}} \textrm{arctan} \, (\lambda(x \cdot v) ) 
\\
\widetilde{\theta_{{}_{\lambda}}}(k) = 
2\pi \big[\sqrt{{\textstyle\frac{\pi}{2}}} \delta(k_0) + {\textstyle\frac{1}{\sqrt{2\pi}}}
{\textstyle\frac{ik_0}{k_{0}^{2} +v_{0}^2}} e^{-|k_0|/(\lambda v_0)} \big] \, \delta\big(\boldsymbol{k} - k_0 {\textstyle\frac{\boldsymbol{v}}{v_0}}\big)
\\
=
2\pi \Big[ {\textstyle\frac{i}{k_0+i\epsilon v_0}} - {\textstyle\frac{ik_0}{k_{0}^{2}+\epsilon^2 v_{0}^{2}}}
\big(e^{-|k_0|/(\lambda v_0)}\big) \Big] \, \delta\big(\boldsymbol{k} - k_0 {\textstyle\frac{\boldsymbol{v}}{v_0}}\big),
\end{gather*}
obtaining $\widetilde{\kappa_q}_{{}_{\lambda}}$ and show convergence 
\[
\big\langle \widetilde{\kappa_q}_{{}_{\lambda}} ,\widetilde{\phi} \big\rangle \overset{\lambda \rightarrow \infty}{\longrightarrow}
\big\langle \widetilde{\kappa_q} ,\widetilde{\phi} \big\rangle,
\]
for each test function $\widetilde{\phi}$. The integral (\ref{thetakappaq(chi)}) with 
 $\theta(x\cdot v)$ replaced by $\theta_{{}_{\lambda}}(x \cdot v)$, coincides with
$\widetilde{\varphi}(0) \, \big\langle \widetilde{\kappa_q}_{{}_{\lambda}} ,\widetilde{\phi}\big\rangle$. In this distributional
sense the integral (\ref{thetakappaq(chi)}) is convergent.

In particular, we arrive at the conclusion that there exist natural $k$ and finite $c$, such that 
\begin{multline}\label{existencethetakappaq}
\Bigg|
\Big[
\theta_y\big(\kappa'_{0,1} \dot{\otimes} \kappa''_{0,1} \dot{\otimes} \kappa'''_{0,1} \dot{\otimes} \ldots \dot{\otimes} \kappa^{(q)}_{0,1} \big)
\, \otimes_{q} \,\, \big(\kappa'_{1,0} \dot{\otimes} \kappa''_{1,0}  \dot{\otimes} \kappa'''_{1,0} \dot{\otimes} 
\ldots \dot{\otimes} \kappa^{(q)}_{1,0} \big)
\Big] (\chi)
\Bigg|
\\
\leq c \big|\phi \big|_{{}_{k}}\big|\varphi \big|_{{}_{k}}
\end{multline}
for all test functions $\chi(x,y) = \phi(x-y)\varphi(y)$, $\phi,\varphi\in \mathscr{E}$, 
with $\phi$ which respects (\ref{up-to-omegaorder-derivatives-of-phi=0}) 
and with $\omega$ in (\ref{up-to-omegaorder-derivatives-of-phi=0})
equal at least to the singularity order of $\kappa_q$, in fact equal to the singularity order of
\[
\kappa_q(x)-(-1)^q\kappa_q(-x).
\]

Therefore, the valuation integral
\[
\Big( \theta_y \big(\kappa'_{0,1} \dot{\otimes} \kappa''_{0,1} \dot{\otimes} \kappa'''_{0,1} \dot{\otimes} \ldots \dot{\otimes} \kappa^{(q)}_{0,1} \big)
\, \otimes_q \,\, \big(\kappa'_{1,0} \dot{\otimes} \kappa''_{1,0}  \dot{\otimes} \kappa'''_{1,0} \dot{\otimes} 
\ldots \dot{\otimes} \kappa^{(q)}_{1,0} \big) \big)(\chi)
\]
is convergent for
\[
\chi(x,y) = \phi(x-y)\varphi(y), \,\,\,\,\,\,\,\, \phi, \varphi \in \mathscr{E},
\]
and with $\phi$ which respect  (\ref{up-to-omegaorder-derivatives-of-phi=0})
with $\omega$ in (\ref{up-to-omegaorder-derivatives-of-phi=0})
equal at least to the singularity order of $\kappa_q$.

Still more generally, with $q'\leq q$, using the result for $q'= q$ as in passing from $q'=q$ to $q'<q$
with $q'=1$, or better, using the continuity (\ref{productkernelsmultipliers}), the product formula
(\ref{ProductFormula}) and the fact that the retarded part of the scalar contraction (with $q'=q$)
is a well-defined distribution
we arrive at the inequality
\[
\Big|\Big\langle \Big( \theta_y \big(\kappa'_{0,1} \dot{\otimes} \kappa''_{0,1} \dot{\otimes} \kappa'''_{0,1} \dot{\otimes} \ldots \dot{\otimes} \kappa^{(q)}_{0,1} \big)
\, \otimes_{q'} \,\, \big(\kappa'_{1,0} \dot{\otimes} \kappa''_{1,0}  \dot{\otimes} \kappa'''_{1,0} \dot{\otimes} 
\ldots \dot{\otimes} \kappa^{(q)}_{1,0} \big) \Big)(\chi), \,\, \widehat{\xi} \, \Big\rangle\Big|
\]
\begin{equation}\label{InequalityForq'<qContraction}
\leq c \big|\phi \big|_{{}_{k}}\big|\varphi \big|_{{}_{k}} |\widehat{\xi}|_m, 
\end{equation}
for
\[
\chi(x,y) = \phi(x-y)\varphi(y), \,\,\, \phi,\varphi \in \mathscr{E}
\]
with $\phi$ which respects
 (\ref{up-to-omegaorder-derivatives-of-phi=0}), with $\omega$ equal at least to the singularity order
of $\kappa_{q'}$.
Here $|\cdot|_{m}$ being one of the norms defining the nuclear topology of 
\[
E_{{}_{'}} \otimes E_{{}_{''}}  \otimes  \widehat{\ldots}   \otimes  E_{{}_{(q)}}  
E_{{}_{'}} \otimes E_{{}_{''}}\otimes  \widehat{\ldots}    \otimes  E_{{}_{(q)}}. 
\]
Hat-character $\widehat{\ldots}$ means that $q'$ contracted pairs of nuclear spaces 
$E_{{}_{'}}^{*}, E_{{}_{''}}^{*}, \ldots$ are removed from the tensor product.

Recall that here we have considered the contraction distributions (\ref{[A'(-),A'(+)]^q}) with the massless
plane wave distribution kernels $\kappa'_{0,1}, \kappa'_{1,0}, \kappa''_{0,1}, \kappa''_{1,0}, \ldots$
defining massless free fields $\mathbb{A}', \mathbb{A}'', \ldots$. Of course exactly the same analysis is valid
if some free fields, $\mathbb{A}'$, or all, are replaced with massive fields $\mathbb{A}'$. In this case the functions
$p_0(\boldsymbol{\p}') = |\boldsymbol{\p}'|$ in the exponents of the contraction integrals are of course replaced
by the functions $p_0(\boldsymbol{\p}')$ such that $(p_0(\boldsymbol{\p}'), \boldsymbol{\p}') \in \mathscr{O}' = \mathscr{O}_{m',0,0,0}$
for the orbit $\mathscr{O}' = \mathscr{O}_{m',0,0,0}$ corresponding to the massive field $\mathbb{A}'$. Similarly,
the multipliers $u'(\boldsymbol{\p}'),v'(\boldsymbol{\p}')$ are replaced with the multipliers which are present in the massive
kernels
\[
\kappa'_{0,1}(\boldsymbol{\p}';x)= u'(\boldsymbol{\p}') e^{-ip'x}, \,\,\,\,
\kappa'_{1,0}(\boldsymbol{\p}'; x) = v'(\boldsymbol{\p}') e^{ip'x},
\,\,\,\,\, p'= (p_0(\boldsymbol{\p}'), \boldsymbol{\p}') \in \mathscr{O}'.
\]

Inspired by \cite{Epstein-Glaser} we can extend the contraction kernels
\[
\theta_{y} \big(\kappa'_{0,1} \dot{\otimes} \kappa''_{0,1} \dot{\otimes} \kappa'''_{0,1} \dot{\otimes} \ldots \dot{\otimes} \kappa^{(q')}_{0,1} \big)
\, \otimes_{q'} \,\, \big(\kappa'_{1,0} \dot{\otimes} \kappa''_{1,0} \dot{\otimes} \kappa'''_{1,0} \dot{\otimes} \ldots \dot{\otimes} \kappa^{(q)}_{1,0} \big)
\]
with $q'\leq q$, over all elements $\chi$, which respect (\ref{xiInS00timesS00}). In order to do it, we use the previously defined linear continuous
operator on $\mathscr{E}$
\[
\Omega' \phi = \phi - \sum \limits_{0 \leq |\alpha| \leq \omega} D^\alpha \phi(0) \,\, \omega_{{}_{o \,\, \alpha}}.
\]
We extend this operator on functions in $\mathscr{E}^{\otimes \, 2}$ according to the general formula sated above in this Subsection.
Namely, let for any $\chi \in \mathscr{E}^{\otimes \, 2}$ the function $\chi^\natural \in \mathscr{E}^{\otimes \, 2}$ be defined by the rule
\[
\chi^\natural(x,y) = \chi(x+y,y) = (\chi \circ L) (x,y),
\]
and let
\[
\chi^\natural = \sum \limits_{j} \phi_j \otimes \varphi_j
\]
be its expansion into simple tensors $\phi_j \otimes \varphi_j$ in $\mathscr{E}^{\otimes \, 2}$, so that
\[
\chi^\natural(x,y) = \sum \limits_{j} \phi_j(x)\varphi_j(y).
\]
Then for any $\chi$ which respects (\ref{xiInS00timesS00}), $\Omega \chi$ be defined by the formula
\[
\Omega \chi(x,y) = \sum \limits_{j} \big(\Omega' \phi_j\big)(x-y)\varphi_j(y),
\]
\[
\Omega (\chi) \circ L(x,y) = \sum \limits_{j} \big(\Omega' \phi_j\big)(x)\varphi_j(y).
\]
Equivalently
\[
\Omega (\chi) \circ L (x,y) = (\chi \circ L)(x,y) - \, \sum \limits_{|\beta|=0}^{\omega} \omega_{{}_{o \,\, \beta}} (x)
\,\, D_{{}_{x}}^{\beta} \big(\chi \circ L\big) (x=0,y)
\]
\[
x,y \in \mathbb{R}^4.
\]
From this the explicit formula for $\Omega \chi (x,y)$ immediately follows
for any $\chi$ which respects (\ref{xiInS00timesS00}):
\begin{equation}\label{EpsteinGlaserOmega}
\Omega \chi (x,y) = \chi(x,y)
- \, \sum \limits_{|\beta|=0}^{\omega} \omega_{{}_{o \,\, \beta}} (x-y)
\,\,\, D_{{}_{x}}^{\beta}\chi (x=y,y).
\end{equation}

By the results obtained above for each contraction distribution 
(\ref{[A'(-),A'(+)]^q})
\[
\big(\kappa'_{0,1} \dot{\otimes} \kappa''_{0,1} \dot{\otimes} \kappa'''_{0,1} \dot{\otimes} \ldots \dot{\otimes} \kappa^{(q)}_{0,1} \big)
\, \otimes_q \,\, \big(\kappa'_{1,0} \dot{\otimes} \kappa''_{1,0}  \dot{\otimes} \kappa'''_{1,0} \dot{\otimes} \ldots \dot{\otimes} \kappa^{(q)}_{1,0} \big)
\]
the contraction integral 
\[
\Big\langle \theta_y \big(\kappa'_{0,1} \dot{\otimes} \kappa''_{0,1} \dot{\otimes} \kappa'''_{0,1} \dot{\otimes} \ldots \dot{\otimes} \kappa^{(q)}_{0,1} \big)
\, \otimes||_{{}_{q}} \,\, \big(\kappa'_{1,0} \dot{\otimes} \kappa''_{1,0}  \dot{\otimes} \kappa'''_{1,0} \dot{\otimes} \ldots \dot{\otimes} \kappa^{(q)}_{1,0} \big), \,\,
\chi \Big\rangle
\]
\[
\overset{\textrm{df}}{=}
\Big\langle \theta_y \big(\kappa'_{0,1} \dot{\otimes} \kappa''_{0,1} \dot{\otimes} \kappa'''_{0,1} \dot{\otimes} \ldots \dot{\otimes} \kappa^{(q)}_{0,1} \big)
\, \otimes_q \,\, \big(\kappa'_{1,0} \dot{\otimes} \kappa''_{1,0}  \dot{\otimes} \kappa'''_{1,0} \dot{\otimes} \ldots \dot{\otimes} \kappa^{(q)}_{1,0} \big), \,\,
\Omega \chi \Big\rangle
\]
\begin{multline}\label{Int[A'(-),A'(+)]^qOmegaxi}
=
\int 
u'(\boldsymbol{\p}') v'(\boldsymbol{\p}') u''(\boldsymbol{\p}'') v''(\boldsymbol{\p}'') \ldots u^{(q)}(\boldsymbol{\p}^{(q)}) v^{(q)}(\boldsymbol{\p}^{(q)}) 
\, \times
\\
\times \,
e^{-i(|\boldsymbol{p}'| + |\boldsymbol{p}''| + \ldots +|\boldsymbol{p}^{(q)}|)(x_0-y_0) +i(\boldsymbol{p}' 
+ \boldsymbol{p}'' + \ldots + \boldsymbol{p}^{(q)})\cdot (\boldsymbol{\x} - \boldsymbol{\y})}
\,\, \times 
\\
\times \,\,
\theta(x-y) \, \Omega \chi(x,y)
\, \ud^3 \boldsymbol{p}' \, \ud^3 \boldsymbol{p}'' \ldots \ud^3 \boldsymbol{p}^{(q)} \, \ud^4 x \ud^4 y 
\end{multline}
\[
=
\int \kappa_q(x-y) \theta(x-y) \, \Omega \chi(x,y) \,\, \ud^4 x \ud^4 y
\]
\[
=
\int \kappa_q(x) \theta(x) \, \Omega (\chi \circ L)(x,y) \,\, \ud^4 x \ud^4 y
\]
\begin{multline*}
=
\int 
u'(\boldsymbol{\p}') v'(\boldsymbol{\p}') u''(\boldsymbol{\p}'') v''(\boldsymbol{\p}'') \ldots u^{(q)}(\boldsymbol{\p}^{(q)}) v^{(q)}(\boldsymbol{\p}^{(q)}) 
\, \times
\\
\times \,
e^{-i(|\boldsymbol{p}'| + \ldots +|\boldsymbol{p}^{(q)}|)x_0 +i(\boldsymbol{p}' 
+ \ldots + \boldsymbol{p}^{(q)})\cdot \boldsymbol{\x}}
\,\, \times 
\\
\times \,\,
\theta(x) \, \Omega (\chi \circ L) (x,y)
\, \ud^3 \boldsymbol{p}' \, \ud^3 \boldsymbol{p}'' \ldots \ud^3 \boldsymbol{p}^{(q)} \, \ud^4 x \ud^4 y 
\end{multline*}
converges absolutely and uniformly  for $\chi$ ranging over any bounded 
set in  $\mathcal{S}(\mathbb{R}^4; \mathbb{C}^{d'd''}) \otimes \mathcal{S}(\mathbb{R}^4; \mathbb{C}^{d'd''})$.

Therefore, 
\[
 \big(\theta_{y} \kappa'_{0,1} \dot{\otimes} \kappa''_{0,1} \dot{\otimes} \kappa'''_{0,1} \dot{\otimes} \ldots \dot{\otimes} \kappa^{(q)}_{0,1} \big)
\, \otimes_q \,\, \big(\kappa'_{1,0} \dot{\otimes} \kappa''_{1,0}  \dot{\otimes} \kappa'''_{1,0} \dot{\otimes} \ldots \dot{\otimes} \kappa^{(q)}_{1,0} \big)
\circ \Omega
\]
exists in 
\[
\mathscr{L}\big(\mathscr{E}^{\otimes \, 2}, \mathbb{C} \big) = \mathscr{E}^{* \, \otimes \, 2}.
\]
and analogously for $q' \leq q$
\[
\theta_{y} \big(\kappa'_{0,1} \dot{\otimes} \kappa''_{0,1} \dot{\otimes} \kappa'''_{0,1} \dot{\otimes} \ldots \dot{\otimes} \kappa^{(q')}_{0,1} \big)
\, \otimes_{q'} \,\, \big(\kappa'_{1,0} \dot{\otimes} \kappa''_{1,0}  \dot{\otimes} \kappa'''_{1,0} \dot{\otimes} \ldots \dot{\otimes} \kappa^{(q)}_{1,0} \big)
\circ \Omega
\]
in 
\[
\mathscr{L}\big(\mathscr{E}^{\otimes \, 2}, \,\, E_{{{}_{(q'+1)}}}^{*} \otimes \ldots \otimes E_{{{}_{(q)}}}^{*}\big);
\]
provided 
$\omega$ in (\ref{EpsteinGlaserOmega}) is
equal at least to the singularity degree at zero of the distribution
\[
 \big( \kappa'_{0,1} \dot{\otimes} \ldots \kappa^{(q')}_{0,1} \big)
\, \otimes_{q'}  \,\, \big( \kappa'_{1,0} \dot{\otimes} 
 \ldots \dot{\otimes}  \kappa^{(q')}_{1,0} \big)
\] 
in the sense of \cite{Epstein-Glaser}, explained above. 

However, note that the extension 
\[
\big(\theta_{y} \kappa'_{0,1} \dot{\otimes} \kappa''_{0,1} \dot{\otimes} \kappa'''_{0,1} \dot{\otimes} \ldots \dot{\otimes} \kappa^{(q)}_{0,1} \big)
\, \otimes_q \,\, \big(\kappa'_{1,0} \dot{\otimes} \kappa''_{1,0}  \dot{\otimes} \kappa'''_{1,0} \dot{\otimes} \ldots \dot{\otimes} \kappa^{(q)}_{1,0} \big)
\circ \Omega
\]
of the contraction distribution  
\[
\big(\theta_{y} \kappa'_{0,1} \dot{\otimes} \kappa''_{0,1} \dot{\otimes} \kappa'''_{0,1} \dot{\otimes} \ldots \dot{\otimes} \kappa^{(q)}_{0,1} \big)
\, \otimes_q \,\, \big(\kappa'_{1,0} \dot{\otimes} \kappa''_{1,0}  \dot{\otimes} \kappa'''_{1,0} \dot{\otimes} \ldots \dot{\otimes} \kappa^{(q)}_{1,0} \big)
\]
defined on the image of $\Omega$,
consisting of all those test functions $\chi$ which respect (\ref{Dalphaphi(x,x)=0,alpha=<omega}),
to the space of all test functions, is not unique. The singularity 
degree $\omega$ at zero of the retarded (and advanced) part of the 
product of pairings distribution ($\otimes_q$-contraction)
\[
\big(\theta_{y} \kappa'_{0,1} \dot{\otimes} \kappa''_{0,1} \dot{\otimes} \kappa'''_{0,1} \dot{\otimes} \ldots \dot{\otimes} \kappa^{(q)}_{0,1} \big)
\, \otimes_q \,\, \big(\kappa'_{1,0} \dot{\otimes} \kappa''_{1,0}  \dot{\otimes} \kappa'''_{1,0} \dot{\otimes} \ldots \dot{\otimes} \kappa^{(q)}_{1,0} \big)
\]
is the same as the singularity degree $\omega$ of that distribution at zero. It is determined up to a distribution 
supported at zero, and we can add a distribution
\[
\sum \limits_{|\alpha| =0}^{\omega} C_\alpha D^{\alpha}_{{}_{x}}\delta(x-y)
\]
to 
\begin{multline*}
\big( \theta_{y}\kappa'_{\epsilon \,\, 0,1} \dot{\otimes} \kappa''_{0,1} 
\dot{\otimes} \kappa'''_{0,1} \dot{\otimes} \ldots \big)
\, \otimes||_{{}_{q}}  \,\, \big( \kappa'_{\epsilon \,\, 1,0} \dot{\otimes} \kappa''_{1,0} 
\dot{\otimes} \kappa'''_{1,0} \dot{\otimes} \ldots \big)
\\
= \big(\theta_{y} \kappa'_{0,1} \dot{\otimes} \kappa''_{0,1} \dot{\otimes} \kappa'''_{0,1} \dot{\otimes} \ldots \dot{\otimes} \kappa^{(q)}_{0,1} \big)
\, \otimes_q \,\, \big(\kappa'_{1,0} \dot{\otimes} \kappa''_{1,0}  \dot{\otimes} \kappa'''_{1,0} \dot{\otimes} \ldots \dot{\otimes} \kappa^{(q)}_{1,0} \big)
\circ \Omega
\end{multline*}
or, equivalently, we can add the expression
\[
\sum \limits_{|\alpha| =0}^{\omega} (-1)^{|\alpha|} C_\alpha \int D^{\alpha}_{{}_{x}}\Big|_{{}_{x=0}}\chi(x+y,y) \, \ud^4 y
\]
to the contraction given by the absolutely
convergent integral (\ref{Int[A'(-),A'(+)]^qOmegaxi}), with arbitrary constants $C_\alpha$ in it. This is because
\[
\sum \limits_{|\alpha| =0}^{\omega} C_\alpha D^{\alpha}\delta(x)
\]
is the most general continuous functional on the kernel of $\Omega'$, equal zero on the image of $\Omega'$, with $\omega$ less
than or equal to the singularity order.

Note, please, that $\phi \in \mathscr{E}$ lies in the image $\textrm{Im} \, \Omega'$ 
if and only if  $\phi \in \textrm{ker} \, [\boldsymbol{1} - \Omega']$.
Therefore, the image $\textrm{Im} \, \Omega'$ of the continuous idempotent operator $\Omega' = \Omega'^{2}$, being equal
\[
\textrm{Im} \, \Omega' = \textrm{ker} \, [\boldsymbol{1} - \Omega']
\]
is closed in $\mathscr{E}$.
Thus,
\[
\mathscr{E} = \textrm{ker} \, \Omega' \oplus \textrm{Im} \, \Omega'
= \textrm{ker} \, \Omega' \oplus \textrm{ker} \, [\boldsymbol{1} - \Omega']
\]
is equal to the direct sum of closed sub-spaces $\textrm{Im} \, \Omega'$ and $\textrm{ker} \, \Omega'$. 
It is immediately seen that for each fixed $\omega$ 
the functional (distribution) 
\[
\delta^{\omega}_{{}_{1;2}}(x) = \sum \limits_{|\alpha| =0}^{\omega} C_\alpha D^{\alpha}\delta(x)
\]
is the most general functional on
$\textrm{Ker} \, \Omega' \subset \mathscr{E}$ which vanishes on $\textrm{Im} \, \Omega' \subset \mathscr{E}$.  
Because in addition $\delta^{\omega}_{{}_{1;2}}$ is equal identically zero on $\textrm{Im} \, \Omega'$ then the most general extension of the
retarded part 
\[
\kappa_q \circ \theta.\Omega'
\]
of $\kappa_q$ from $\textrm{Im} \Omega' \subset \mathscr{E}$  all over the whole $\mathscr{E}$
is equal to the sum
\begin{equation}\label{freedomInsplitting}
\kappa_q \circ \theta.\Omega' + \delta^{\omega}_{{}_{1;2}}.
\end{equation}

Equivalently the most general translationally invariant extension of the retarded part
\[
\big(\theta_{y} \kappa'_{0,1} \dot{\otimes} \kappa''_{0,1} \dot{\otimes} \kappa'''_{0,1} \dot{\otimes} \ldots \dot{\otimes} \kappa^{(q)}_{0,1} \big)
\, \otimes_q \,\, \big(\kappa'_{1,0} \dot{\otimes} \kappa''_{1,0}  \dot{\otimes} \kappa'''_{1,0} \dot{\otimes} \ldots \dot{\otimes} \kappa^{(q)}_{1,0} \big)
\circ \Omega
\]
of 
\[
\big(\kappa'_{0,1} \dot{\otimes} \kappa''_{0,1} \dot{\otimes} \kappa'''_{0,1} \dot{\otimes} \ldots \dot{\otimes} \kappa^{(q)}_{0,1} \big)
\, \otimes_q \,\, \big(\kappa'_{1,0} \dot{\otimes} \kappa''_{1,0}  \dot{\otimes} \kappa'''_{1,0} \dot{\otimes} \ldots \dot{\otimes} \kappa^{(q)}_{1,0} \big)
\] 
from the subspace $\textrm{Im} \, \Omega \subset \mathscr{E}^{\otimes \, 2}$ all over the whole 
\[
\mathscr{E}^{\otimes \, 2} = \textrm{ker} \, \Omega \oplus \textrm{Im} \, \Omega
= \textrm{ker} \, \Omega \oplus \textrm{ker} \, [\boldsymbol{1} - \Omega]
\]
is equal to the sum
\[
\big(\theta_{y} \kappa'_{0,1} \dot{\otimes} \kappa''_{0,1} \dot{\otimes} \kappa'''_{0,1} \dot{\otimes} \ldots \dot{\otimes} \kappa^{(q)}_{0,1} \big)
\, \otimes_q \,\, \big(\kappa'_{1,0} \dot{\otimes} \kappa''_{1,0}  \dot{\otimes} \kappa'''_{1,0} \dot{\otimes} \ldots \dot{\otimes} \kappa^{(q)}_{1,0} \big)
\circ \Omega + \delta^{\omega}_{{}_{1;2}},
\]
where $\delta^{\omega}_{{}_{1;2}}$ regarded as distribution, or functional, of two variables $x,y$ is equal
\[
\delta^{\omega}_{{}_{1;2}}(x,y) = \sum \limits_{|\alpha| =0}^{\omega} C_\alpha D^{\alpha}\delta(x-y).
\]

The same additional translationally invariant functional $\delta^{\omega}_{{}_{1;2}}$ will have to be added to the advanced part
\begin{multline*}
-\check{\theta}_{y}\kappa'_{\epsilon \,\, 0,1} \dot{\otimes} \dots  \dot{\otimes} \kappa^{(q)}_{\epsilon \,\, 0,1} \big)
\, \otimes_q  \,\, \big(\kappa'_{\epsilon \,\, 1,0} \dot{\otimes} \ldots \dot{\otimes} \kappa^{(q)}_{\epsilon \,\, 1,0} \big)  \circ \Omega
\\
= 
-\check{\theta}_{y}\kappa'_{0,1} \dot{\otimes} \ldots  \dot{\otimes} \kappa^{(q)}_{0,1} \big)
\, \otimes||_{{}_{q}}  \,\, \big(\kappa'_{\epsilon \,\, 1,0} \dot{\otimes} \ldots \dot{\otimes} \kappa^{(q)}_{1,0} \big)
\end{multline*}
where $\check{\theta}(x) = \theta(-x)$.

Recall that here $\theta_y(x) = \theta(x-y)$, $\theta_{\varepsilon \,\, y}(x) = \theta_\varepsilon(x-y)$, 
where $x$ is the space-time variable in the kernels 
\[
\kappa'_{\epsilon \,\, 0,1} \dot{\otimes} \kappa''_{\epsilon \,\, 0,1} \dot{\otimes} \kappa'''_{\epsilon \,\, 0,1} \dot{\otimes} \ldots
\,\,\,\, \textrm{and}
\,\,\,\,\,\,\,\,
\kappa'_{0,1} \dot{\otimes} \kappa''_{0,1} \dot{\otimes} \kappa'''_{0,1} \dot{\otimes} \ldots 
\]
and $y$ is the space-time variable in the kernels
\[
\kappa'_{\epsilon \,\, 1,0} \dot{\otimes} \kappa''_{\epsilon \,\, 1,0} \dot{\otimes} \kappa'''_{\epsilon \,\, 1,0} \dot{\otimes} \ldots
\,\,\,\,\, \textrm{and}
\,\,\,\,\,\,\,\,
\kappa'_{1,0} \dot{\otimes} \kappa''_{1,0} \dot{\otimes} \kappa'''_{1,0} \dot{\otimes} \ldots.
\]

Similarly, we show that for the kernels $\kappa'_{\ell',m'}, \kappa''_{\ell'',m''}$  of the Fock 
expansion of the operator $\Xi' = \mathcal{L}$, the \emph{double limit contraction} $\otimes||_{{}_{q}}$
\begin{multline*}
\theta_{{}_{'}} \,\, \kappa'_{\ell',m'} \otimes||_{{}_{q}}  \,   \kappa''_{\ell'',m''} \overset{\textrm{df}}{=}
\underset{\varepsilon, \epsilon \rightarrow 0}{\textrm{lim}} \,\,\,
\theta_{{}_{\varepsilon \,\, '}} \,\, \kappa'_{\epsilon \,\, \ell',m''} \otimes_q \,  \kappa''_{\epsilon \,\, \ell'',m''} \circ \Omega
= \theta_{{}_{ '}} \,\, \kappa'_{\ell',m''} \otimes_q \,  \kappa'';_{\ell'',m''} \circ \Omega
\end{multline*}
exists in the ordinary topology of uniform convergence on bounded sets in
\[
\mathscr{L}\big(\mathscr{E} \otimes \mathscr{E} , E_{j_1}^{*} \ldots  \otimes \ldots E_{j_{\ell'+\ell''-q}}^{*} \otimes E_{j_{\ell'+\ell''-q+1}}^{*} 
\ldots \otimes \ldots E_{j_{\ell'+\ell''+m'+m''-q}}^{*}  \big)
\]
\[
\cong \mathscr{L}\big(E_{j_1} \ldots  \otimes \ldots E_{j_{\ell'+\ell''-q}} \otimes E_{j_{\ell'+\ell''-q+1}} 
\ldots \otimes \ldots E_{j_{\ell'+\ell''+m'+m''-q}}, \, \mathscr{E}^* \otimes \mathscr{E}^*   \big),
\]
so that we can define 
\[
\theta_{{}_{'}} \,\, \kappa'_{\ell',m'} \otimes||_{{}_{q}}  \,   \kappa'_{\ell'',m''} \overset{\textrm{df}}{=}
\underset{\varepsilon, \epsilon \rightarrow 0}{\textrm{lim}} \,\,\,
\theta_{{}_{\varepsilon \,\, '}} \,\, \kappa'_{\epsilon \,\, \ell',m'} \otimes_q \,  \kappa''_{\epsilon \,\, \ell'',m''} \circ \Omega
\]
as existing in
\[
\mathscr{L}\big(\mathscr{E} \otimes \mathscr{E} , E_{j_1}^{*} \ldots  \otimes \ldots E_{j_{\ell'+\ell''-q}}^{*} \otimes E_{j_{\ell'+\ell''-q+1}}^{*} 
\ldots \otimes \ldots E_{j_{\ell'+\ell''+m'+m''-q}}^{*}  \big)
\]
\[
\cong \mathscr{L}\big(E_{j_1} \ldots  \otimes \ldots E_{j_{\ell'+\ell''-q}} \otimes E_{j_{\ell'+\ell''-q+1}} 
\ldots \otimes \ldots E_{j_{\ell'+\ell''+m'+m''-q}}, \, \mathscr{E}^* \otimes \mathscr{E}^*   \big),
\]
and determined up to the Epstein-Glaser reminder kernel of the general form given above.

Here we have put $\theta_{{}_{'}}$ for the function $\theta_{{}_{'}}(x',x'') = \theta(x'-x'')$
with the space-time variable $x'$ in the kernel $\kappa'_{\ell',m'}$ and with the space-time variable $x''$ in the kernel
$\kappa''_{\ell'',m''}$. 
 Similarly, for  $\theta_{{}_{\varepsilon \,\, '}}$ denoting the function 
$\theta_{{}_{\varepsilon \,\, '}}(x',x'') = \theta_\varepsilon(x'-x'')$, 
with the space-time variable $x'$ in the kernel $\kappa'_{\epsilon \,\, \ell',m'}$ and with the space-time variable $x''$ in the kernel
$\kappa''_{\epsilon \,\, \ell'',m''}$.

We should emphasize here that the above formula for the splitting 
\[
\kappa_q-(-1)^q\check{\kappa_q} = \textrm{ret} \, \big[ \kappa_q-(-1)^q\check{\kappa_q} \big]
- \textrm{av} \, \big[\kappa_q-(-1)^q\check{\kappa_q} \big]
\]
of the causal distribution (or the causal combination of the 
scalar $\otimes_q$-contractions) $\kappa_q-(-1)^q\check{\kappa_q}$, into the part supported in the future and, respectively,
past light cones, uses in fact one more assumption: that the retarded and advanced parts of
$\kappa_q-(-1)^q\check{\kappa_q}$
should have the same singularity order as $\kappa_q-(-1)^q\check{\kappa_q}$ does. 
Similarly in our computation the retarded and advanced parts of $\kappa_q$ have the same singularity order as 
$\kappa_q$. Some authors call this assumtion
\emph{preservation of the Steinmann scaling degree}. In fact, in our, or rather Epstein-Glaser-Scharf computation of the retarded part,
we proceed in two steps. First we find the maximal subspace $\Omega'\mathscr{E}$ of the test space $\mathscr{E}$, 
on which the retarded (and advanced) part is defined through the natural formula, given by the ordinary multiplication 
by the step theta function and put zero on the complementary subspace $[\boldsymbol{1}-\Omega']\mathscr{E}$,
on which the natural formula does not work. This subspace $\Omega'\mathscr{E}$
turns out to be of finite codimension, meaning that its complementary 
subspace $[\boldsymbol{1}-\Omega']\mathscr{E}$, on which this formula does not work, has finite dimension. 
Then we extend this formula, by addition of the most general functional on $[\boldsymbol{1}-\Omega']\mathscr{E}$, 
which is zero on the subspace $\Omega'\mathscr{E}$.
Any such addition is supported at zero and can be added to the retarded and advanced part. Abandoning preservation of the 
singularity degree would allow addition of a functional (also supported at zero), but which would not be zero on $\Omega'\mathscr{E}$
and thus, which would modify the natural formula there, so the resulting retarded part no loner would be equal to the extension of the natural formula 
given by the multiplication by theta function on $\Omega'\mathscr{E}$. Thus our ``additional assumption'' is rather this: \emph{the retarded
part should coincide with the natural formula given by multiplication by the step theta function on a test function, whenever
the natural formula is meaningfull for this test function}. Thus, speaking intuitively: the splitiing should be ``maximally natural''.
As we have shown in  Subsection \ref{OperationsOnXi}, the remaining freedom in this  ``maximally natural'' splitting
(which preserves singularity order) can be completely eleminated by the requirement that the higher order contributions to interacting fields,
in the adiabatic limit $g=1$,
should be equal to finite sums of well-defined integral kernel operators with vector valued kernels in the sense of \cite{obataJFA},
and which holds for the so called ``natural normalization'' of the splitting.

The formula
\[
\textrm{ret} \, \kappa_q =  \kappa_{q} \circ \theta.\Omega'
\]
replaces the naive multiplication $\theta \kappa_q$ by theta function. The point is that this formula
can also serve for the practical computation of $\textrm{ret}\kappa_q$, to which the method shown in \cite{Scharf}
works pretty well, although  $\kappa_q$ in general is not causal.
This can be done effectively for the Fourier transform $\widetilde{\textrm{ret}\kappa_q}$. Namely, denoting the 
space-time test function by $\phi$ we explicitly compute the evaluation integral
\[
\big\langle \widetilde{\textrm{ret} \, \kappa_q}, \widetilde{\phi}  \big\rangle =
\big\langle \textrm{ret} \, \kappa_q, \phi  \big\rangle = \big\langle \theta \kappa_q  \circ \Omega', \phi  \big\rangle
= \big\langle \kappa_q, \theta \Omega'\phi  \big\rangle =
\big\langle \widetilde{\kappa_q}, \widetilde{\theta} \ast \widetilde{\Omega'\phi}  \big\rangle,  
\]
which initially contains the auxiliary function(s) $w$ in
\[
\omega_{{}_{o \,\, \alpha}} = {\textstyle\frac{x^\alpha}{\alpha!}} w,
\]
used in the definition of the idempotent $\Omega'$. But using the fact that $\widetilde{\textrm{ret} \, \kappa_q}$
must have regularity regions, where it is given by ordinary function, 
with points at which it posses all derivatives, we can eliminate the auxiliary function(s) $w$  
in $\omega_{{}_{o \,\, \alpha}}$ by subtracting all terms up to order $\omega$ in the Taylor expansion
of $\widetilde{\textrm{ret} \, \kappa_q}$  around such a point (here chosen to be
zero, which is posiible e.g. for spinor QED with massive charged field), 
with the very small cost that the obtained solution 
\begin{equation}\label{DispersionFormulaForRetarded}
\widetilde{\textrm{ret} \, \kappa_q}(p) = {\textstyle\frac{i}{2\pi}}  \int\limits_{-\infty}^{+\infty}
{\textstyle\frac{dt}{(t+i0)}}\big[\widetilde{\kappa_q}(p-tv) -
\sum\limits_{|\alpha|=0}^{\omega} {\textstyle\frac{p^\alpha}{\alpha!}} D^\alpha \widetilde{\kappa_q}(-tv)
\big],
\end{equation}
of the splitting problem is the one with specific normalization which is smooth around zero in
momentum space, with all derivatives (in momentum space) vanishing at zero up to order equal to the singularity
degree $\omega$ of $\kappa_q$. 
This cost is very small, as adding
\[
\sum \limits_{|\alpha| =0}^{\omega} C_\alpha p^{\alpha}
\]
to (\ref{DispersionFormulaForRetarded}) we obtain the most general solution of the splitting problem.
This choice of normalization is possible in spinor QED with the massive charged
field. For QED with massless charged field the normalization point in momenta will have to be shifted
from the zero point $p'=0$ to a point $p'\neq 0$ around which $\widetilde{\kappa_q}$ and $\widetilde{\textrm{ret} \, \kappa_q}$ 
are regular function-like distributions
and have all derivatives up to $\omega$ at $p' \neq 0$ 
and the respective formula $\widetilde{\textrm{ret} \, \kappa_q}$  
becomes slightly different
\[
\widetilde{\textrm{ret} \, \kappa_q}(p) = {\textstyle\frac{i}{2\pi}} \int\limits_{-\infty}^{+\infty}
{\textstyle\frac{dt}{(t+i0)}}\big[\widetilde{\kappa_q}(p-tv) -
\sum\limits_{|\alpha|=0}^{\omega} {\textstyle\frac{(p-p')^\alpha}{\alpha!}} D^\alpha \widetilde{\kappa_q}(p'-tv)
\big]
\]
giving solution with all derivatives up to order $\omega$ of $\widetilde{\textrm{ret} \, \kappa_q}$
equal zero at $p'\neq 0$.
Because in general the $\otimes_q$-contraction $\kappa_q$ is not causally supported, then its splitting into the retarded 
part (\ref{DispersionFormulaForRetarded}) and the advanced part is not invariant and depends on the unit time-like versor
$v$ of the coordinate system, which determines the $\theta$ function
\[
\theta(x) = \theta(v\cdot x), \,\,\, v=(1,0,0,0),
\]
with the ordinary $\theta$ function on the reals on the right hand side, so that (\ref{DispersionFormulaForRetarded})
in general depends on $v$.  

Dispersion integral (\ref{DispersionFormulaForRetarded}) has also its higher dimensional
analogue, \cite{DKS1'}, \cite{Scharf}.

Alternatively, one can consider all the time only the scalar causal distributions, equal to the causal
combinantions of $\otimes_q$-contractions, which are the scalar factors in the Wick monomials
entering the Wick decomposition of the causally supported distribution $D_{(n)}(x_1, \ldots, x_n)$ in the Bogoliubov-Epstein-Glaser
method, as reported in Subsection \ref{MotivationForHida}. Anyway, we are
confronted with the computation of the retarded and advanced parts of a set of multidimensional causal
distributions, which is enlarged together with the order, and cannot confine the splitting problem to a finite
set of causal distributions.

In particular for the causal distributions $d$ which are present in the Wick monomials
of the Wick decomposition of
\[
D_{(2)}(x_1,x_2) = \mathcal{L}(x_2)\mathcal{L}(x_1) - \mathcal{L}(x_1)\mathcal{L}(x_2)
\] 
these distributions $d$ are equal to the causal combinations of contractions $\kappa_q$,
because each $D_{(n)}$ is causally supported. Because these $d$ are causally supported
(within the closed future and past light cone), then the above computation of the retarded
part $\textrm{ret}d = \theta d \circ \Omega'$ simplifies and we arrive
at the formula
\begin{equation}\label{InvariantDispersionFormulaForRetarded}
\widetilde{\textrm{ret} \, d}(p) ={\textstyle\frac{i}{2\pi}} \int\limits_{-\infty}^{+\infty}
dt {\textstyle\frac{\widetilde{d}(tp)}{(t-i0)^{\omega +1}(1-t +i0)}}
\end{equation}
which is invariant, in particular independed of the time-like versor $v$ of the used Lorentz
coordinate system.  This (\ref{InvariantDispersionFormulaForRetarded}) is a solution
of the splitting problem with specific normalization point at zero, 
which is smooth around zero in momentum space, with all derivatives (in momentum space) of $\widetilde{d}$ vanishing at zero up to order 
equal to the singularity degree $\omega$ of $d$. Adding
\[
\sum \limits_{|\alpha| =0}^{\omega} C_\alpha p^{\alpha}
\]
to (\ref{InvariantDispersionFormulaForRetarded}) we obtain the most general solution of the splitting problem.
Again this choice of normalization is possible in spinor QED with the massive charged
field. Exactly the same formula (\ref{InvariantDispersionFormulaForRetarded}), with the normalization point at zero, holds 
for the retarded part of a causal distribution $d$ of more than one space-time variable, 
provided the Fourier transform $\widetilde{d}$ of $d$ is smooth around zero, \cite{DKS1'}, \cite{Scharf}. 
We have an analogue dispersion formula with the normalization point $p'\neq 0$ for causal $d$,
which have smooth $\widetilde{d}$ around $p'$, \cite{DKS3}.    
For QED with massless charged field the normalization point in momenta will have to be shifted
from the zero point $p'=0$ to a point $p'\neq 0$ around which $\widetilde{d}$ is a regular function-like distribution
and has all derivatives up to $\omega$  equal zero at $p' \neq 0$.
 Dispersion integral formula  (\ref{InvariantDispersionFormulaForRetarded})
is valid for time-like momenta with $p\cdot p >0$, and in order to compute
$\widetilde{\textrm{ret} \, d}(p)$ for $p\cdot p <0$ we use the analytic continuation.

Therefore, if we introduce the \emph{double limit contraction} $\otimes||_{{}_{q}}$, 
which defines the retarded part of the contraction kernel,
up to the reminder additive term 
with the arbitrary constants $C_\alpha$ in 
the reminder,
then we can summarize our analysis of the ``natural''
chronological product in the following Wick theorem for the ``natural'' chronological product. 
\begin{twr}
Let each of the Wick monomials
of the Wick polynomial $\mathcal{L}$ contains (or not) massless free field factors and be of degree at most $N$.
In particular $\mathcal{L}$ for spinor QED interaction has $N=3$. 
Then for $\phi, \phi_k \in \mathscr{E} = \mathcal{S}(\mathbb{R}^4)$ 
the operators
\[
\Xi'(\phi) =  \int\limits_{\mathbb{R}^4}  \mathcal{L}(x) \, \phi(x) \,\,\,\, \ud^4x
\in \mathscr{L}((\boldsymbol{E}),(\boldsymbol{E})^*),
\]
\begin{multline*}
\Xi_{\varepsilon,\epsilon}(\phi_1 \otimes \phi_2)
\\
=
\sum\limits_{\pi}
\int\limits_{\big[\mathbb{R}^4\big]^{\times \, 2}} 
\theta_{{}_{\varepsilon \,\, \pi(2)}} \mathcal{L}_\epsilon(x_{\pi(1)}) 
\mathcal{L}_\epsilon(x_{\pi(2)}) \,\,
\phi_{1} \otimes \phi_{2}(x_1, x_2) \, \ud^4x_1 \ud^4x_2 
\\
=
\sum \limits_{\ell +m \leq Nn} \Xi_{\ell,m}\big( \kappa_{\varepsilon, \epsilon \,\, \ell,m}(\phi_1 \otimes \phi_2)\big)
\in \mathscr{L}((\boldsymbol{E}),(\boldsymbol{E}))
\end{multline*}
or, respectively, for $n>2$,
\begin{multline*}
\Xi_{\varepsilon,\epsilon}(\phi_1 \otimes \ldots \phi_n)
\\
=
\int\limits_{\big[\mathbb{R}^4\big]^{\times \, n}} 
\big[
\theta_{{}_{\varepsilon}}(x_1-x_n) \ldots \theta_{{}_{\varepsilon}}(x_{n-1}-x_n) \, D_{(n)}(x_1, \ldots, x_n)
-R'_{(n)}(x_1, \ldots, x_n)
\big]
\,\, \times
\\
\times \,\,\,
\phi_{1} \otimes \ldots \otimes \phi_{n}(x_1, \ldots, x_n) \, \ud^4x_1 \ldots \ud^4x_n 
\\
=
\sum \limits_{\ell +m \leq Nn} \Xi_{\ell,m}\big( \kappa_{\varepsilon, \epsilon \,\, \ell,m}(\phi_1 \otimes \ldots \phi_n)\big)
\in \mathscr{L}((\boldsymbol{E}),(\boldsymbol{E}))
\end{multline*}

\begin{multline*}
\Xi^\Omega(\phi_1 \otimes \ldots \otimes \phi_n) =
\\
= 
\int\limits_{\big[\mathbb{R}^4\big]^{\times \, n}} 
i^n \, T\big(\mathcal{L}(x_1) \ldots \mathcal{L}(x_n) \big) \,
\Omega \phi_{1} \otimes \ldots \otimes \phi_{n}(x_1, \ldots, x_n) \, \ud^4x_1 \ldots \ud^4x_n 
\\
\overset{\textrm{df}}{=} 
\underset{\varepsilon,\epsilon \rightarrow 0}{\textrm{lim}}
\Xi_{\varepsilon,\epsilon}^{\Omega}(\phi_1 \otimes \ldots \phi_n)
\,\,\, \in \mathscr{L}((\boldsymbol{E}),(\boldsymbol{E})^*),
\end{multline*} 
and moreover
\begin{eqnarray*}
\Xi' \in \mathscr{L}(\mathscr{E}, \mathscr{L}((\boldsymbol{E}),(\boldsymbol{E})^*)\big),
\\
\Xi_{\varepsilon,\epsilon}, \Xi_{\varepsilon,\epsilon}^{\Omega} \in \mathscr{L}(\mathscr{E}^{\otimes \, n}, \mathscr{L}((\boldsymbol{E}),(\boldsymbol{E}))\big)
\\
\Xi^\Omega \in \mathscr{L}(\mathscr{E}^{\otimes \, n}, \mathscr{L}((\boldsymbol{E}),(\boldsymbol{E})^*)\big).
\end{eqnarray*}
\[
 \Xi_{\varepsilon,\epsilon}^{\Omega} = \sum \limits_{\ell +m \leq Nn} 
\Xi_{\ell,m}\big( \kappa_{\varepsilon, \epsilon \,\, \ell,m} \circ \Omega_{{}_{\ell,m}}\big);
\]
(in QED $N=3$). The following Wick theorem holds 
\[
\textrm{in case} \, n=2
\]
\[
\Xi^\Omega =
i^2 \sum \limits_{\pi}\sum_{\substack{\kappa^{\pi(i)}_{\ell_{\pi(i)},m_{\pi(i)}}}} \,\,\,
\sum \limits_{q} \,\,\, (-1)^{c(q)} 
\,\, 
\Xi_{{}_{\ell_{\pi(1)}+\ell_{\pi(2)} -q, \,\, m_{\pi(1)}+m_{\pi(2)} -q}},
\]
where
\[
\Xi_{{}_{\ell_{\pi(1)}+ \ell_{\pi(2)} -q, \,\, m_{\pi(1)}+m_{\pi(2)} -q}}
=
\Xi\big( \theta_{{}_{\pi(2)}} \kappa^{\pi(1)}_{\ell_{\pi(1)},m_{\pi(1)}}\otimes||_{{}_{q}} \, 
\kappa^{\pi(2)}_{\ell_{\pi(2)},m_{\pi(2)}} \big).
\]
Here the kernels $\kappa^{\pi(i)}_{\ell_{\pi(i)},m_{\pi(i)}}$ range independently over the  kernels of the finite Fock expansion
of the operator $\Xi' = \mathcal{L}(x_{\pi(i)})$. The (symmetrized/antisymmetrized) double limit $q_i$-contractions $\otimes||_{{}_{q_{i}}}$ are performed
upon the pairs of variables in which the first element of the contracted pair lies among the last $m_{\pi(i)}$ variables of the kernel 
$\theta_{{}_{\pi(2)}}\kappa^{\pi(1)}_{\ell_{\pi(1)},m_{\pi(1)}}$ and the second variable
of the contracted pair lies among the first $l_{\pi(2)}$ variables of the kernel 
$\kappa^{\pi(2)}_{\ell_{\pi(2)},m_{\pi(2)}}$, and to both variables
of the contracted pair correspond, respectively, to pairs of the annihilation and creation operator of the 
free fields with non-zero pairing. The number 
$c(q)$ is equal to the number of Fermi commutations performed 
in the double limit contraction $\otimes||_{{}_{q}}$. 
\[
\textrm{in case} \, n\geq 2
\]
\begin{multline*}
\Xi^\Omega =
\\
\sum_{\substack{\kappa^{i}_{\ell_{i},m_{i}} \\ i\in\{1,2\}, \, q}} \,\,\,
(-1)^{c(q)}
\Xi\Big(\textrm{ret} \, \big(\kappa^{1}_{\ell_{1},m_{1}}\otimes_{{}_{q}} \, 
\kappa^{2}_{\ell_{2},m_{2}} \big) \Big)
-\sum_{\substack{\kappa^{i}_{\ell_{i},m_{i}} \\ i\in\{1,2\}, \, q}} \,\,\,
(-1)^{c(q)}
\Xi\Big(\textrm{ret} \, \big(\kappa^{2}_{\ell_{2},m_{2}}\otimes_{{}_{q}} \, 
\kappa^{1}_{\ell_{1},m_{1}} \big) \Big)
\\
- 
\sum_{\substack{\kappa^{i}_{\ell_{i},m_{i}} \\ i\in\{1,2\}, \, q}} \,\,\,
(-1)^{c(q)}
\Xi\Big(\kappa^{1}_{\ell_{1},m_{1}}\otimes_{{}_{q}} \, 
\kappa^{2}_{\ell_{2},m_{2}} \Big).
\end{multline*}
The kernels
$\kappa^{1}_{\ell_{1},m_{1}}$  range over the kernels of $S(Y,x_n)$ and
$\kappa^{2}_{\ell_{2},m_{2}}$ range over the kernels of $\overline{S}(X)$, with $X \sqcup Y= \{x_1, \ldots, x_{n-1}\}$,
$X \neq \emptyset$. 
\label{WickThmForChronological}
\end{twr}

Several remarks are in order. 

{\bf REMARKS}. 1. The kernels 
\[
\kappa^{1}_{\ell_{1},m_{1}}\overline{\otimes_{{}_{q}}} \, 
\kappa^{2}_{\ell_{2},m_{2}},
\,\,\,\,
\kappa^{2}_{\ell_{2},m_{2}}\overline{\otimes_{{}_{q}}} \, 
\kappa^{1}_{\ell_{1},m_{1}}
\]
are not in general causal but can be grouped into sums representing the kernels of the terms proportional
to a fixed normal product term in $D_{{}_{(n)}}$, which are causal. But in general, the dispersion formula, 
giving the retarded part of such a causal sum, can be applied to each term separately, and 
\[
\textrm{ret} \, \big(\kappa^{1}_{\ell_{1},m_{1}}\overline{\otimes_{{}_{q}}} \, 
\kappa^{2}_{\ell_{2},m_{2}} \big),
\,\,\,
\textrm{ret} \, \big(\kappa^{2}_{\ell_{2},m_{2}}\overline{\otimes_{{}_{q}}} \, 
\kappa^{1}_{\ell_{1},m_{1}} \big)
\]
represent such contributions coming from each term separately, as explained above in detail in case
of order $n=2$. This however, does not give any computational simplification, except for giving a formula 
of kernels of $S_n$ using kernels of $S_k$, $k<n$. This is because of causality and Lorentz invariance
substantially simplify the computation of the dispersion integrals, and in practice, we always use the 
causal sums of kernels in the dispersion integrals. 

2. Theory is renormalizable 
if the singularity order $\omega$ of each contribution to $D_n$
is bounded by a constant independent of $n$ and equal $4$ minus the number of external lines, counted with a weight depending on the spin
of the external line, and minus the number of derivatives in external lines. 
E.g. for the spinor QED interaction Lagrangian $\mathcal{L}$, $\omega \leq 4 - (3/2\mathfrak{f}+\mathfrak{k})$, 
with $\mathfrak{f}$ equal to the total number of fermion external lines and $\mathfrak{k}$ equal to the total number of external photon lines,
so that QED is renormalizable.

3. The kernels and the corresponding integral kernel operators in Theorem \ref{WickThmForChronological}
are well-defined mathematically without any need for infinite renormalization.

4. We should note that the contractions
\begin{eqnarray*}
\big(\theta_y\kappa'_{0,1} \dot{\otimes} \kappa''_{0,1}\big)
\, \otimes||_{{}_{2}} \,\, \big(\kappa'_{1,0} \dot{\otimes} \kappa''_{1,0} \big)
\\
\big(\theta_y\kappa'_{0,1} \dot{\otimes} \kappa''_{0,1} \dot{\otimes} \kappa'''_{0,1} \big)
\, \otimes||_{{}_{3}} \,\, \big(\kappa'_{1,0} \dot{\otimes} \kappa''_{1,0} \dot{\otimes} \kappa'''_{1,0} \big),
\\
\vdots
\end{eqnarray*}
well-defined distributions, which we have analyzed in details in
and (\ref{Int[A'(-),A'(+)]^qOmegaxi}), correspond, in the conventional approach based on renormalization, to the formal expressions which in
\cite{Bogoliubov_Shirkov} are denoted by
\begingroup\makeatletter\def\f@size{5}\check@mathfonts
\def\maketag@@@#1{\hbox{\m@th\large\normalfont#1}}%
\begin{eqnarray*}
\theta(x-y) \, \textrm{lim}_{M\rightarrow \infty}
\Big[\textrm{reg}_M \big\{\quad\underbracket{\mathbb{A}' \mathbb{A}'}(x-y)\big\} \,\,
\textrm{reg}_M \big\{\quad\underbracket{\mathbb{A}'' \mathbb{A}''}(x-y)\big\} \Big],
\\
\theta(x-y) \, \textrm{lim}_{M\rightarrow \infty}
\Big[
\textrm{reg}_M \big\{\quad\underbracket{\mathbb{A}' \mathbb{A}'}(x-y)\big\} \,\,
\textrm{reg}_M \big\{\quad\underbracket{\mathbb{A}'' \mathbb{A}''}(x-y)\big\}\,\,
\textrm{reg}_M \big\{\quad\underbracket{\mathbb{A}''' \mathbb{A}'''}(x-y)\big\}
\Big],
\\
\vdots
\end{eqnarray*}
\endgroup
\emph{i.e.} to the retarded parts of the limits of the products of regularized pairing functions
of the corresponding free fields $\mathbb{A}', \mathbb{A}'',\mathbb{A}''', \ldots$. If we compute the limits of products of
regularized pairings as well-defined distributions, then the computation of the retarded pars through the ordinary pointwise
multiplication by the theta function is of course problematic because the condition (\ref{CompactOrbits}) is not fulfilled
on the Minkowski space-time for Poincar\'e invariant free fields of realistic QFT,
and the theta function is not a multiplier
of the considered distribution spaces. Therefore, the amplitudes computed formally from the formal chronological product,
in the ordinary case with the ordinary free fields with non-compact orbits,
in which we multiply distributions by the step theta function (as is done in the conventional approach using renormalization),
are mathematically meaningless expressions with infinities, and the renormalization technique is then used in order to extract finite
expressions. The careful construction of the retarded and advanced parts
of the limits of products of regularized pairing functions, and based on the Epstein-Glaser splitting, allows to avoid divergences
in the ordinary case with non-compact orbits,
and this is the Epstein-Glaser contribution, based on the Bogoliubov causal perturbative approach.

5. We shall emphasize here that the white noise analysis, in fact the theory of Fock expansions of generalized operators due to Hida,
Obata and Sait\^o, not only gives unique kernels $S_n(x_1, \ldots, x_n)$ of the $n$-th order contributions $S_n$ to the scattering
operator but also gives the interpretation to the $n$-th order contributions $S_n$
as particular cases of generalized operators
\[
S_n = \Xi \in \mathscr{L}( (\boldsymbol{E}) \otimes \mathscr{E}^{\otimes \, n} , (\boldsymbol{E})^* )
\]
which, when evaluated at $\phi = \phi_1 \otimes \ldots \phi_n \in \mathscr{E}^{\otimes \, n}$, give
integral kernel operators
\[
\Xi(\phi) = \sum\limits_{\substack{\ell,m \\ \ell+m \leq 3n}}
\Xi_{\ell,m}\big({\kappa}_{\ell,m}(\phi)\big)
\in \mathscr{L}( (\boldsymbol{E}) , (\boldsymbol{E})^* )
\]
with scalar-valued kernels
\[
{\kappa}_{\ell,m}(\phi) \in E_{{}_{n_1}}^{*} \otimes \ldots \otimes E_{{}_{n_{3n}}}^{*}
\cong \mathscr{L}(E_{{}_{n_1}} \otimes \ldots \otimes E_{{}_{n_{3n}}}, \, \mathbb{C}).
\]
Because this $\Xi(\phi)$ gives the $n$-th order contribution $S_n\big(g\mathcal{L}\big)$ to the scattering operator, evaluated at $\phi
= g^{\otimes \, n}$, then in particular we obtain the Corollary to Theorem \ref{g=1InteractingFieldsQED} of Subsection \ref{OperationsOnXi}.
Thus we have just proven that each higher order contribution $S_n: \mathscr{E}^{\otimes \, n} \ni g^{\otimes \, n} \longmapsto S_n(g)$,
restricted to the diagonal $\phi =\phi_1 \otimes \ldots \otimes \phi_n$, with $\phi_1 = \ldots = \phi_n =g$, defines an operator
\[
g^{\otimes \, n} \longmapsto S_n\big(g\mathcal{L}\big) \,\,\, \textrm{which belongs to} \,\,
\mathscr{L}( (\boldsymbol{E}) \otimes \mathscr{E}^{\otimes \, n} , (\boldsymbol{E})^* );
\]
and thus we have proved again Corollary of Thm. \ref{g=1InteractingFieldsQED} of Subsection \ref{OperationsOnXi}
or the first part of Thm. \ref{Sin(SE)xn->((E)->(E)*)} of Subsection \ref{WickForProduct}, for the ''natural''
chronological product defined here.

6. Analogously we give the rigorous meaning to the ``natural'' anti-chronological ``product''
\begin{equation}\label{barthetaS_n}
\overline{S}_n(x_1, \ldots, x_n) = (-i)^n \, T^{+}\big(\mathcal{L}(x_1), \ldots, \mathcal{L}(x_n) \big)  = 
\eta \big[S_n(x_1, \ldots, x_n)^{+}\big] \eta.
\end{equation}

7. The contribution $\Xi_{\ell,m} = \Xi(\kappa_{\ell,m})$ to the $n$-th order scattering operator $S_n$, and coming from the kernel
$
\kappa_{\ell,m}$, correspond to the $n$-vertex Feynman diagram
and thus to the $n$-th order diagram with $\ell+m$ external lines, namely $m$ outgoing external lines and $\ell$ ingoing external lines.
In particular the total $n$-th order contribution to the vacuum expectation
\[
S_0 = \langle\langle \Phi_0 S_n(g), \Phi_0 \rangle\rangle
\]
is given by the value
\[
\kappa_{00}(g), \,\,\,\ g \in \mathscr{E}^{\otimes \, n}
\]
of the kernel $\kappa_{00}$ giving the scalar contribution $\Xi_{0,0} = \Xi(\kappa_{0,0}) = \kappa_{0,0} \boldsymbol{1}$
to the chronological product in the Wick expansion of $S_n$ in Theorem \ref{WickThmForChronological}.

Note that the contribution to $S_n$ corresponding to the kernel $\kappa_{\ell,m}$
is represented by a ``loop'' diagram if and only if
some contraction numbers $q_1, \ldots, q_{n-1}$  in $\kappa_{\ell,m}$ are greater than $1$. Of course in the conventional approach based on renormalization
these are the last contributions with ``loop'' diagrams which bring in the ultraviolet infinities
and ultraviolet infinities of the approach based on renormalization have their origin in the careless splitting of the causal distributions
into advanced and retarded parts, which have positive singularity degree, with the help of ordinary multiplication
by the step theta function. We have avoided ultraviolet infinities by using mathematically well-defined integral kernel operators,
which among other things indicate the necessity of the presence of the projection operator $\Omega$ in front of a kernel
of positive singularity degree $\omega$, when we change
the order of the following operations: 1) passing to the limit $\varepsilon,\epsilon \rightarrow 0$ and 2) application of the Wick theorem.

8. Having given the ``natural'' chronological product $T[\ldots]$ of any Wick monomials in free fields we can, in particular,
construct perturbatively the many particle higher order Schwinger Green functions
\begin{multline}\label{SymbolocGreenFunction}
G(g; x_1, \ldots, x_k) = {\textstyle\frac{1}{S_0}}
\Big\langle \Big\langle \Phi_{{}_{0}}, T\big[u_1(x_1)u_2(x_2) \ldots u_k(x_k) S(g) \big] \Phi_{{}_{0}} \Big\rangle \Big\rangle,
\\
S_0 = \langle \langle \Phi_{{}_{0}}, S(g) \Phi_{{}_{0}} \rangle \rangle,
\end{multline}
corresponding to the quanta of the free fields $u_1(x_1),u_2(x_2), \ldots, u_k(x_k)$, compare \cite{Bogoliubov_Shirkov}, \S 37,
without encountering any ultraviolet or infrared infinities. Using the Wick Theorem \ref{WickThmForChronological}
for the ``natural'' chronological product we can recover the Schwinger equation for the higher order Green function
using the method presented in \cite{Bogoliubov_Shirkov}, \S 37. Compare also Subsection \ref{Green} for more details on
Green functions and computation of the Lamb shift. Recall here that (\ref{SymbolocGreenFunction}) is only an
abbreviated notation for a mathematically well-defined expression, written explicitly in Subsection \ref{Green} of Section \ref{A(1)psi(1)}.

9. Wick Theorem \ref{WickThmForChronological} for the ``natural'' chronological product is sufficient for the
computation of the effective cross section between the generalized (unnormalizable) plane wave states in the adiabatic limit. Indeed,
we can construct a limit
\[
g_\epsilon \overset{\epsilon \rightarrow 0}{\longrightarrow} 1_{{}_{[\mathbb{R}^4]^{\times \, n}}}
\,\,\,\, \textrm{in} \,\,\,\,
\mathscr{E}^{* \, \otimes \, n},
\]
such that the set $g_\epsilon$, $0 \leq \epsilon \leq \epsilon_0$, remains bounded in $\mathscr{E}^{* \, \otimes \, n}$,
and
\[
S_n(g_{1/k}), \,\,\, k \rightarrow \infty
\]
possesses a subsequence convergent in
\[
\mathscr{L}( (\boldsymbol{E}), (\boldsymbol{E})^* )
\]
and defining $S_n(1) \in \mathscr{L}( (\boldsymbol{E}), (\boldsymbol{E})^* )$.
The choice of the converging subsequence is irrelevant for the value of the effective cross section
of generalized states. For more details on the effective cross section
compare Subsection \ref{EffCrossSection}.
\qed

\subsection{Second order contribution $S_2$ for spinor QED}

We compute the basic distributions for spinor QED with the standard realizations of the e.m. potential and Dirac free fields
(constructed with the Hida operators as the creation-annihilation operators).
These are the retarded and advanced parts of the $\otimes_q$-contractions, $q=1,2,3$, which appear in the Wick decomposition
of the product $\mathcal{L}(x_1)\mathcal{L}(x_2)$,
where $\mathcal{L}(x) = e {:} \boldsymbol{\psi}^{\sharp}\gamma^\mu \boldsymbol{\psi}A_\mu{:}(x)$ 
is the interaction Lagrangian density generalized operator for spinor QED.

Let $\kappa_{l,m}$, $\kappa^{\sharp}_{l,m}$, $\kappa'_{l,m}$, $l,m = 0,1$, 
be the plane wave kernels defining, respecively, the free Dirac field, its Dirac conjugation and the e.m. potential field.
The $\otimes_1$-contractions (pairings) are the following:
\begin{multline*}
\sum\limits_{a,b}\int \big[\boldsymbol{\psi}^{(-) \, a}(x), \boldsymbol{\psi}^{\sharp (+) \, b}(x)]_{{}_{+}} \, \phi_a(x)\varphi_b(y) \ud^4x \ud^4y
\\
=
\sum\limits_{a,b}\int \quad\underbracket{\boldsymbol{\psi}^{a}(x) \boldsymbol{\psi}^{\sharp \, b}}(x) \, \phi_a(x)\varphi_b(y) \ud^4x \ud^4y
\\
= \sum\limits_{a,b,s}\int \kappa_{0,1}(s, \boldsymbol{\p};a,x)  \kappa_{1,0}^{\sharp}(s, \boldsymbol{\p};b,y)
\, \phi_a(x)\varphi_b(y) \, \ud^3\boldsymbol{\p} \, \ud^4x \ud^4y 
\\
= \kappa_{0,1}(\phi) \otimes_1 \kappa_{1,0}^{\sharp}(\varphi) = \big\langle \kappa_{0,1}(\phi), \kappa_{1,0}^{\sharp}(\varphi) \big\rangle
= -i S^{(-)}(\phi\otimes \varphi),
\end{multline*}
\[
\kappa_{1,0}^{\sharp}(\varphi) \otimes_1 \kappa_{0,1}(\phi) =
\kappa_{0,1}(\phi) \otimes_1 \kappa_{1,0}^{\sharp}(\varphi) = \big\langle \kappa_{0,1}(\phi), \kappa_{1,0}^{\sharp}(\varphi) \big\rangle
= -i S^{(-)}(\phi\otimes \varphi),
\]
\[
\kappa_{1,0}(\phi) \otimes_1 \kappa_{0,1}^{\sharp}(\varphi)=
\kappa_{0,1}^{\sharp}(\varphi) \otimes_1 \kappa_{1,0}(\phi) = \big\langle \kappa_{0,1}^{\sharp}(\varphi), \kappa_{1,0}(\phi) \big\rangle
= -i S^{(+)}(\phi\otimes \varphi),
\]
Similarly we have the contraction formula for the pairing
\begin{multline*}
\quad\underbracket{
A_{\mu}(x)
A_{\nu}}(y) = i g_{\mu\nu} D_{0}^{(-)}(x-y) = \left[A_{\mu}^{(-)}(x),A_{\nu}^{(+)}(y) \right]_{-}
\\
= \kappa'_{0,1}(\mu,x) \otimes_1 \kappa'_{1,0}(\nu,y)=\kappa'_{0,1}(\nu,x) \otimes_1 \kappa'_{1,0}(\mu,y),
\\
\left[A_{\mu}^{(+)}(x),A_{\nu}^{(-)}(y) \right]_{-} = i g_{\mu\nu} D_{0}^{(+)}(x-y) 
= \kappa'_{1,0}(\mu,x) \otimes_1 \kappa'_{0,1}(\nu,y)=\kappa'_{1,0}(\nu,x) \otimes_1 \kappa'_{0,1}(\mu,y),
\\\quad\underbracket{
A_{\nu}(y)
A_{\mu}}(x) 
=\kappa'_{0,1}(\nu,y) \otimes_1 \kappa'_{1,0}(\mu,x) = \kappa'_{1,0}(\mu,x) \otimes_1 \kappa'_{0,1}(\nu,y).
\end{multline*}

We have the following $\otimes_2$-contractions:
\begin{multline*}
-\int \textrm{tr}\big[\gamma^\mu S^{(-)}(x-y) \gamma^\nu S^{(+)}(y-x)] \, \phi_\mu(x)\varphi_\nu(y) \ud^4x \ud^4y
\\
= 
\sum_{\substack{a,b,c,d \\ 
s,s'}} 
\int
\kappa^{\sharp}_{0,1}(s', \boldsymbol{\p}';a,x)\gamma^{\mu}_{ab}\kappa_{0,1}(s, \boldsymbol{\p}; b,x)
\kappa^{\sharp}_{1,0}(s, \boldsymbol{\p};c,y)\gamma^{\nu}_{cd}\kappa_{1,0}(s', \boldsymbol{\p}'; d,y) 
\, \times
\\
\times
\, \phi_\mu(x)\varphi_\nu(y) \ud^3\boldsymbol{\p} \ud^3\boldsymbol{\p}' \ud^4x \ud^4y
\end{multline*}
\begin{multline*}
=
\big(\kappa^{\sharp}_{0,1} \gamma^\mu \dot{\otimes} \kappa_{0,1}  \big)(\phi_\mu) \otimes_2
(\kappa^{\sharp}_{1,0} \dot{\otimes} \gamma^\nu \kappa_{1,0}\big)(\varphi_\nu)
\\
=
\sum_{\substack{a,b,c,d 
}} \int
\gamma^{\mu}_{ab} \gamma^{\nu}_{cd}
\quad\underbracket{
\boldsymbol{\psi}^{\sharp}_{a}(x)
\boldsymbol{\psi}_{d}}(y)
\quad\underbracket{
\boldsymbol{\psi}_{b}(x)
\boldsymbol{\psi}^{\sharp}_{c}}(y)
\, \phi_\mu(x)\varphi_\nu(y) \ud^4x \ud^4y
\end{multline*}
or 
\begin{multline*}
\sum\limits_{a,b,c,d}
\gamma^{\mu}_{ab} \gamma^{\nu}_{cd}
\quad\underbracket{
\boldsymbol{\psi}^{\sharp}_{a}(x)
\boldsymbol{\psi}_{d}}(y)
\quad\underbracket{
\boldsymbol{\psi}_{b}(x)
\boldsymbol{\psi}^{\sharp}_{c}}(y)
\\
=
\big(\kappa^{\sharp}_{0,1} \gamma^\mu \dot{\otimes} \kappa_{0,1}\big) \otimes_2
(\kappa^{\sharp}_{1,0} \dot{\otimes} \gamma^\nu \kappa_{1,0}\big)(x, y) = \kappa^{\mu\nu}_{2}(x-y)
\end{multline*}
Similarly
\begin{multline*}
\sum\limits_{a,d,\mu}\int \big[\gamma^\mu S^{(-)}\gamma_\mu\big]^{ad}(x-y)D_{0}^{(-)}(x-y) \phi_a(x)\varphi_d(y) \ud^4x \ud^4y
\\
= 
\sum_{\substack{a,b,c,d \\ \mu, \nu, s,s'}} 
\int
g_{\mu\nu}
\gamma^{\mu}_{ab}\kappa_{0,1}(s, \boldsymbol{\p}; b,x) \kappa'_{0,1}(s', \boldsymbol{\p}';\mu,x) \kappa^{\sharp}_{1,0}(s, \boldsymbol{\p};c,y)\gamma^{\nu}_{cd}
\kappa'_{1,0}(s', \boldsymbol{\p}';\nu,y)
\, \times
\\
\times
\, \phi_a(x)\varphi_d(y) \ud^3\boldsymbol{\p} \ud^3\boldsymbol{\p}' \ud^4x \ud^4y
\end{multline*}
\begin{multline*}
=
\sum\limits_{\mu, \nu}
\big(\gamma^\mu\kappa_{0,1} \dot{\otimes} \kappa'_{0,1 \,\, \mu}\big)(\phi) \otimes_2
(\kappa^{\sharp}_{1,0} \gamma^\nu \dot{\otimes} \kappa'_{1,0 \,\, \nu}\big)(\varphi)
\\
=
\sum_{\substack{a,b,c,d \\ \mu, \nu}} \int
\gamma^{\mu}_{ab} \gamma^{\nu}_{cd}
\quad\underbracket{
\boldsymbol{\psi}_{b}(x)
\boldsymbol{\psi}^{\sharp}_{c}}(y)
\quad\underbracket{
A_{\mu}(x)
A_{\nu}}(y)
\, \phi_a(x)\varphi_d(y) \ud^4x \ud^4y
\end{multline*}
or
\begin{multline*}
\sum_{\substack{b,c \\ \mu, \nu}}
\gamma^{\mu}_{ab} \gamma^{\nu}_{cd}
\quad\underbracket{
\boldsymbol{\psi}_{b}(x)
\boldsymbol{\psi}^{\sharp}_{c}}(y)
\quad\underbracket{
A_{\mu}(x)
A_{\nu}}(y)
\\
=
\sum\limits_{\mu\nu}
\big(\gamma^\mu\kappa_{0,1} \dot{\otimes} \kappa'_{0,1 \,\, \mu}\big) \otimes_2
(\kappa^{\sharp}_{1,0} \gamma^\nu  \dot{\otimes}\kappa'_{1,0 \,\, \nu}\big)(a,x,d,y) = \kappa^{(-) \, ad}_{2}(x-y)
\end{multline*}
\begin{multline*}
\sum_{\substack{b,c \\ \mu, \nu}}
\gamma^{\nu}_{ab} \gamma^{\mu}_{cd}
\quad\underbracket{
\boldsymbol{\psi}^{\sharp}_{c}(y)
\boldsymbol{\psi}_{b}}(x)
\quad\underbracket{
A_{\nu}(y)
A_{\mu}}(x)
\\
=
\sum\limits_{\mu\nu}
\big(\gamma^\nu\kappa_{1,0} \dot{\otimes} \kappa'_{1,0 \,\, \mu}\big) \otimes_2
(\kappa^{\sharp}_{0,1} \gamma^\mu  \dot{\otimes} \kappa'_{0,1 \,\, \nu}\big)(a,x,d,y) = \kappa^{(+) \, ad}_{2}(x-y)
\end{multline*}

Finally we have one $\otimes_3$-contraction:
\begin{multline*}
\sum_{\substack{a,b,c,d \\ \mu, \nu}} 
\gamma^{\mu}_{ab} \gamma^{\nu}_{cd}
\quad\underbracket{
\boldsymbol{\psi}^{\sharp}_{a}(x)
\boldsymbol{\psi}_{d}}(y)
\quad\underbracket{
\boldsymbol{\psi}_{b}(x)
\boldsymbol{\psi}^{\sharp}_{c}}(y)
\quad\underbracket{
A_{\mu}(y)
A_{\nu}}(x)
\\
= -i \textrm{tr}\big[\gamma^\mu S^{(-)}(x-y) \gamma_\mu S^{(+)}(y-x)]D_{0}^{(-)}(y-x)
\\
=
\sum\limits_{\mu\nu}
\big(\kappa^{\sharp}_{0,1} \dot{\otimes}\gamma^\mu \kappa_{0,1} \dot{\otimes} \kappa'_{0,1 \,\,\mu}\big) \otimes_3
(\kappa^{\sharp}_{1,0} \dot{\otimes} \gamma_\mu \kappa_{1,0} \dot{\otimes} \kappa'_{1,0 \,\,\nu}\big)(x, y) = \kappa_3(x-y)
\end{multline*}

Let us introduce $C^{\mu\nu}_{2}(x-y)= e^2[\kappa^{\mu\nu}_{2}(x-y)-\kappa^{\nu\mu}_{2}(y-x)] , 
K_{2}^{ad}(x-y)= e^2[\kappa^{(-) \, ad}_{2}(x-y) + \kappa^{(+) \, ad}_{2}(x-y)],
C_3(x-y) = e^2[\kappa_3(x-y) - \kappa_3(y-x)]$. It follows that $\kappa^{\mu\nu}_{2}(x-y)$ is a negative frequency $\otimes_2$-contraction,
and it is easily seen that $\kappa^{\nu\mu}_{2}(y-x)$ is the correspondinding positive frequency contraction. Similarly $\kappa_3(y-x)$ is 
the positive frequency $\otimes_3$-contraction correspnding to the negative frequency contraction $\kappa_{3}(x-y)$. Thus,
$C^{\mu\nu}_{2},K_{2}^{ad}(x-y), C_3$ are causally supported.  Recall that
\[
\kappa_{1,0} \otimes_1 \kappa_{0,1}
= \kappa_{0,1} \otimes_1 \kappa_{1,0}
= \kappa^{\sharp}_{1,0} \otimes_1 \kappa_{0,1}^{\sharp}
= \kappa^{\sharp}_{0,1} \otimes_1 \kappa_{1,0}^{\sharp}
=0,
\] 
so that in the contractions the kernels $\kappa_{0,1}, \kappa_{1,0}$ are always contracted, respectively, with the opposite frequency 
Dirac conjugated kernels $\kappa^{\sharp}_{1,0}, \kappa^{\sharp}_{0,1}$ and the kernels $\kappa'_{0,1}, \kappa'_{1,0}$, respectively, 
with the corresponding opposite frequency kernels  $\kappa'_{1,0}, \kappa'_{0,1}$.

By the application of the Wick theorem, we have 
(summation with repeated indices is understood)
\begin{multline*}
D_{(2)}(x_1,x_2) = S_1(x_2)S_1(x_1)-S_1(x_1)S_1(x_2) = -\mathcal{L}(x_2)\mathcal{L}(x_1)+\mathcal{L}(x_1)\mathcal{L}(x_2) 
\\
= e^2\Big[ 
-i \gamma^{\mu}_{ab}\gamma_{\mu \, cd} \left( D^{(-)}_{0}(x_2-x_1) - D^{(-)}_{0}(x_1-x_2) \right) \, 
{:} \boldsymbol{\psi}^{\sharp a}(x_1) \boldsymbol{\psi}^{b}(x_1)\boldsymbol{\psi}^{\sharp c}(x_2) \boldsymbol{\psi}^{d}(x_2){:}
\\
-i\big[\gamma^\mu S^{(-)}(x_1-x_2)\gamma^\nu + \gamma^\mu S^{(+)}(x_1-x_2)\gamma^\nu\big]_{ab} \, 
{:}\boldsymbol{\psi}^{\sharp a}(x_1) \boldsymbol{\psi}^{b}(x_2) 
A_\mu(x_1)A_\nu(x_2){:} 
\\
+i\big[\gamma^\mu S^{(-)}(x_2-x_1)\gamma^\nu+\gamma^\mu S^{(+)}(x_2-x_1)\gamma^\nu\big]_{ab} \, 
{:}\boldsymbol{\psi}^{\sharp a}(x_2) \boldsymbol{\psi}^{b}(x_1) 
A_\mu(x_2)A_\nu(x_1){:} \Big]
\\
+ K_{2}^{ab}(x_1-x_2) \,  {:}\boldsymbol{\psi}^{\sharp a}(x_1) \boldsymbol{\psi}^{b}(x_2){:}
- K_{2}^{ba}(x_2-x_1) \, {:}\boldsymbol{\psi}^{\sharp b}(x_2) \boldsymbol{\psi}^{a}(x_1){:}
\\
+C^{\mu\nu}_{2}(x_1-x_2) \, {:}A_\mu(x_1)A_\nu(x_2){:}
+ C_3(x_1-x_2) \, \boldsymbol{1} ,
\end{multline*}

Using the contraction formula in explicit form and the completeness relations for the plane wave kernels $\kappa_{0,1}$, $\kappa_{1,0}$ 
of the free fields we can compute the Fourier transforms 
(compare \cite{Scharf}) of these contraction distributions in explicit form:
\begin{multline*}
\widetilde{C_{2}^{\mu\nu}}(p) =- {\textstyle\frac{e^2(2\pi)^{-3}}{3}} \big({\textstyle\frac{p^\mu p^\nu}{p^2}} - g^{\mu\nu}\big)
(p^2+2m^2)\sqrt{1-{\textstyle\frac{4m^2}{p^2}}}\theta(p^2-4m^2) \, \textrm{sgn}(p_0), 
\\
\widetilde{K_{2}}(p) = e^2 (2\pi)^{-3} \theta(p^2-m^2)  \, \textrm{sgn}(p_0) \, \big(1- {\textstyle\frac{m^2}{p^2}}\big)
\big[ m - {\textstyle\frac{\slashed{p}}{4}}\big(1+{\textstyle\frac{m^2}{p^2}}\big) \big], 
\\
\widetilde{C_{3}}(p) = - e^2 (2\pi)^{-5} \theta(p^2-4m^2) \, \textrm{sgn}(p_0) \, \Big[ 
\big({\textstyle\frac{p^4}{24}} + {\textstyle\frac{m^2}{12}}p^2 + m^4\big)\sqrt{1-{\textstyle\frac{4m^2}{p^2}}}
\\
+ {\textstyle\frac{m^4}{p^2}}(4m^2-3p^2) \textrm{ln} \big(\sqrt{{\textstyle\frac{p^2}{4m^2}}} + \sqrt{{\textstyle\frac{p^2}{4m^2}}-1}\big)
\Big], 
\end{multline*}
with the singularity degree $\omega$, respectively, equal $2,1,4$.

As we have seen computation of the retarded and advanced parts of the $\otimes_1$-contractions, or pairings, can be achieved by the 
ordinary multiplication by the $\theta$ function and this gives a non-invariant formula because the pairings are not causally supported. 
In fact their singularity degree $\omega$ at zero is negative (equal $-1$ for the Dirac field or $-2$ for the e.m. potential field), 
which together with the formulas (\ref{D(+)}) and (\ref{D(-)}) justifies this fact, although they are not causal. We have given a proof 
of this fact which is not based on the singularity degree. 
In order to compute the retarded parts $\textrm{ret} \, C_{2}, \textrm{ret} \, K_{2}$ and 
$\textrm{ret} \, C_3$ of the contractions $C_{2}, K_{2}$ and $C_3$
we just insert the above formulas for the Fourier transforms $\widetilde{C_2}, \widetilde{K_2}$ and $\widetilde{C_3}$  
into the dispersion formula (\ref{DispersionFormulaForRetarded}).

In particular we can compute the kernel of the second order contribution to the scattering operator and obtain the result
\begin{multline*}
S_2(x_1,x_2)= -e^2\Big[ {:} \boldsymbol{\psi}^{\sharp}(x_1)\gamma^\mu \boldsymbol{\psi}(x_1)A_\mu(x_1)
\boldsymbol{\psi}^{\sharp}(x_2)\gamma^\nu \boldsymbol{\psi}(x_2)A_\nu(x_2){:}
\\
-i \gamma^{\mu}_{ab}\gamma_{\mu \, cd} D^{c}_{0}(x_1-x_2) \, 
{:} \boldsymbol{\psi}^{\sharp a}(x_1) \boldsymbol{\psi}^{b}(x_1)\boldsymbol{\psi}^{\sharp c}(x_2) \boldsymbol{\psi}^{d}(x_2){:}
\\
+i\big[\gamma^\mu S^c(x_1-x_2)\gamma^\nu\big]_{ab} \, 
{:}\boldsymbol{\psi}^{\sharp a}(x_1) \boldsymbol{\psi}^{b}(x_2) 
A_\mu(x_1)A_\nu(x_2){:} 
\\
+i\big[\gamma^\mu S^c(x_2-x_1)\gamma^\nu\big]_{ab} \, 
{:}\boldsymbol{\psi}^{\sharp a}(x_2) \boldsymbol{\psi}^{b}(x_1) 
A_\mu(x_2)A_\nu(x_1){:} 
\Big]
\\
- i\Sigma_{ab}(x_2-x_1) \, {:}\boldsymbol{\psi}^{\sharp a}(x_2) \boldsymbol{\psi}^{b}(x_1){:}
- i\Sigma_{ab}(x_1-x_2) \,  {:}\boldsymbol{\psi}^{\sharp a}(x_1) \boldsymbol{\psi}^{b}(x_2){:}
\\
-i\Pi^{\mu\nu}(x_1-x_2) \, {:}A_\mu(x_1)A_\nu(x_2){:}
+ \Upsilon(x_1-x_2) \, \boldsymbol{1} ,
\end{multline*}
where 
\[
S^c(x)= S_{\textrm{ret}}(x) - S^{(+)}(x) = S_{\textrm{av}}(x) + S^{(-)}(x), 
\,\,\,\,
D^{c}_{0}(x) = D^{\textrm{ret}}_{0}(x)-D^{(+)}_{0}(x),
\]
 and the explicit expressions 
for the Fourier transforms of the distributions $\Pi, \Sigma, \Upsilon$,
are given by the formulas (\ref{Pi})-(\ref{Upsilon}). Recall, that the (always causal) Pauli-Jordan commutation functions
are equal $D(x) = D^{(-)}(x) + D^{(+)}(x)$, \emph{e.g.} $S_{ab}(x) = S^{(-)}_{ab}(x) + S^{(+)}_{ab}(x)$ and that
$D^{(-)}_{0}(x_2-x_1) - D^{(-)}_{0}(x_1-x_2) = -D_0(x_1-x_2)$,
because $D^{(-)}_{0}(x_2-x_1) = - D^{(-)}_{0}(x_1-x_2)$. In particular
\begin{multline*}
\Pi^{\mu\nu}(x) = \textrm{ret} \, C_{2}^{\mu\nu}(x) + \textrm{ret} \, C_{2}^{\mu\nu}(-x),
\,\,\, \Upsilon(x) = \textrm{ret} \, C_{3}^{\mu\nu}(x) + \textrm{ret} \, C_{3}^{\mu\nu}(-x),
\\
\Sigma(x) = \textrm{ret} \, K_{2}(x) - \textrm{ret} \, K_{2}(-x). 
\end{multline*}

$S_2(x_1,x_2)$ can also be obtained as equal to
\begin{multline*}
S_2(x_1,x_2) = \textrm{ret} \, D_{(2)}(x_1,x_2) -R'_{(2)}(x_1,x_2) 
\\
= \textrm{ret} \, \big[\mathcal{L}(x_2)\mathcal{L}(x_1) -\mathcal{L}(x_1)\mathcal{L}(x_2) \big] 
- \mathcal{L}(x_2)\mathcal{L}(x_1)
\end{multline*}
in which case we consider the causal scalar distributions $d$ in front of the Wick monomials in the Wick decomposition of $D_{(2)}$
and apply the invariant formula (\ref{InvariantDispersionFormulaForRetarded}) for the computation of $\widetilde{\textrm{ret} \, d}$,
compare \cite{Scharf}.

For example in order to get the vacuum polarization term 
\[
-i\Pi^{\mu\nu}(x_1-x_2) \, {:}A_\mu(x_1)A_\nu(x_2){:},
\]
we apply the Wick theorem 
to the operators
\[
\mathcal{L}(x_2)\mathcal{L}(x_1) - \mathcal{L}(x_1)\mathcal{L}(x_2) 
\,\,\,\,\,
\textrm{and}
\,\,\,\,\,
\mathcal{L}(x_2)\mathcal{L}(x_1)
\]
and collect all terms proportional to
${:}A_\mu(x_1)A_\nu(x_2){:}$. Next, using the formula (\ref{InvariantDispersionFormulaForRetarded}), we compute the retarded part of the scalar factor 
$d^{\mu\nu}(x_1-x_2) = C_{2}^{\mu\nu}(x_1-x_2)$ which multiplies ${:}A_\mu(x_1)A_\nu(x_2){:}$ (which is causal) in
\[
\mathcal{L}(x_2)\mathcal{L}(x_1) - \mathcal{L}(x_1)\mathcal{L}(x_2),
\]
and subtract the scalar factor $e^2 \kappa^{\nu\mu}_{2}(x_2-x_1)$ proportional to
${:}A_\mu(x_1)A_\nu(x_2){:}$ in the Wick decomposition of $\mathcal{L}(x_2)\mathcal{L}(x_1)$, finally getting
\[
-i\Pi^{\mu\nu}(x_1-x_2) = \textrm{ret} \, d^{\mu\nu}(x_1-x_2) -e^2 \kappa^{\nu\mu}_{2}(x_2-x_1).
\]

We perform computations in the momentum space for the Fourier transform 
$\widetilde{\textrm{ret} \, d^{\mu\nu}}(p)- e^2\widetilde{c_{2}^{\nu\mu}}(-p)= \widetilde{\textrm{ret} \, C_{2}^{\mu\nu}}(p)- e^2\widetilde{\kappa_{2}^{\nu\mu}}(-p)$ of
$ret \, d^{\mu\nu}(x) - e^2 \kappa^{\nu\mu}_{2}(-x)$, 
\emph{i.e.} apply the formula (\ref{InvariantDispersionFormulaForRetarded}) with $\widetilde{d^{\mu\nu}}(p)= \widetilde{C_{2}^{\mu\nu}}(p)$
substituted for $\widetilde{d}$ with singularity order $\omega=2$.
Note here that although in general the product of tempered distributions is not a well-defined distribution,
the products $d$ of pairings in the Wick decomposition of $D_{(n)}$, $R'_{(n)}$, $A'_{(n)}$,
are always well-defined and their valuations on the test functions is given by the contraction
of the tensor products of values of the products of kernels of free fields at the test functions, 
and which are always given by absolutely converget integrals,
compare our previous paper. 

Analogously, in order to compute the self-energy term
\[
-i {:}\psi^\sharp(x_1) \Sigma(x_1-x_2) \psi(x_2){:} 
\]
in $S_2(x_1,x_2)$ we collect all terms in the Wick decomposition proportional to 
${:}\psi^{\sharp \, a}(x_1) \psi^{b}(x_2){:}$
and repeat the above stated operations. We do similarly with all terms proportional to each fixed Wick monomial
in the Wick decomposition of 
\[
\mathcal{L}(x_2)\mathcal{L}(x_1) - \mathcal{L}(x_1)\mathcal{L}(x_2)
\,\,\,\,\,
\textrm{and}
\,\,\,\,\,
\mathcal{L}(x_2)\mathcal{L}(x_1),
\]
computing in this way the full second order contribution $S_2(x_1,x_2)$.

1) For the vacuum polarization term 
\[
-i \Pi^{\mu\nu}(x_1-x_2) \, {:}A_\mu(x_1)A_\nu(x_2){:}
\]
in $S_2(x_1,x_2)$ we get with the ``natural'' splitting
\begin{equation}\label{Pi}
\widetilde{{\Pi}_{\mu \nu}}(p) = 
(2\pi)^{-4} \big({\textstyle\frac{p_\mu p_\nu}{p^2}} - g_{\mu\nu}\big) \widetilde{\Pi}(p),
\,\,\,\,\,\,\,\,\,\,\,\,\,\,\,\,\,\,\,\,\,\,
\widetilde{\Pi}(p) =
{\textstyle\frac{e^2}{3}} p^2 p^2
\int\limits_{4m^2}^{\infty} {\textstyle\frac{s+2m^2}{s^2(p^2-s+i0)}}\sqrt{1-{\textstyle\frac{4m^2}{s}}} ds.
\end{equation}

2) For the self-energy term 
\[
-i {:}\psi^\sharp(x_1) \Sigma(x_1-x_2) \psi(x_2){:} 
\]
in $S_2(x_1,x_2)$ we get with the ``natural'' splitting
\begin{equation}\label{Sigma}
\widetilde{\Sigma}(p) = 
e^2 (2\pi)^{-4}
\left\{
\left[\textrm{ln} \big|1- {\textstyle\frac{p^2}{m^2}} \big| -i\pi \, \theta(p^2 - m^2)\right]
\, 
\left[ 
m\left(1-\textstyle{\frac{m^2}{p^2}} \right)-{\textstyle\frac{\slashed{p}}{4}}\left(1-{\textstyle\frac{m^4}{p^4}}\right)\right]
+ {\textstyle\frac{m^2}{p^2}}{\textstyle\frac{\slashed{p}}{4}}
-m
+{\textstyle\frac{\slashed{p}}{8}}
\right\},
\end{equation}

3) For the vacuum term 
\[
\Upsilon(x_1-x_2) \, \boldsymbol{1}
\]
in $S_2(x_1,x_2)$ we get with the ``natural'' splitting
\begin{multline}\label{Upsilon}
\widetilde{\Upsilon}(p)= \widetilde{\Upsilon''}(p) -\widetilde{\Upsilon'}(p),
\\
\widetilde{\Upsilon''}(p) =
i e^2 (2\pi)^{-6} m^4 
\Bigg\{
{\textstyle\frac{5p^4}{48m^4}} + {\textstyle\frac{2p^2}{3m^2}} + 1 
+\big(3- {\textstyle\frac{4m^2}{p^2}}\big) \, \textrm{ln}^2\Big(\sqrt{{\textstyle\frac{-p^2}{4m^2}}} + \sqrt{1-{\textstyle\frac{p^2}{4m^2}}}\Big)
\\
+
\big({\textstyle\frac{p^4}{24m^4}} + {\textstyle\frac{p^2}{12m^2}} + 1\big) \sqrt{1-{\textstyle\frac{4m^2}{p^2}}}
\textrm{ln}{\textstyle\frac{\sqrt{1-{\textstyle\frac{4m^2}{p^2}}}-1}{\sqrt{1-{\textstyle\frac{4m^2}{p^2}}}+1}}
\Bigg\}, 
\end{multline}
\begin{multline*}
\widetilde{\Upsilon'}(p) =
e^2 (2\pi)^{-5} \, \theta(p^2 - 4m^2) \, \theta(-p_0) \, 
\Big\{
\big({\textstyle\frac{p^4}{24}} + {\textstyle\frac{m^2}{12}} p^2 + m^4\big) \sqrt{1-{\textstyle\frac{4m^2}{p^2}}}
\\
+ {\textstyle\frac{m^4}{p^2}}(4m^2-3p^2)
\textrm{ln}\Big(\sqrt{{\textstyle\frac{p^2}{4m^2}}} + \sqrt{{\textstyle\frac{p^2}{4m^2}}-1}\Big)
\Big\}. 
\end{multline*}

On the globally causal space-times with compact Cauchy surfaces and non-zero curvature
situation is much better and, as we will see in Section \ref{EUandG}, not only regularization is not necessary for the construction
of the ``product'' of Wick-ordered factors, even containing massless fields, but even the ``natural'' chronological product can be more immediately
constructed without any need of regularization by the usage of the (analogue of the) step theta function $\theta$, at least when
we have at most one massless free field in the interaction Lagrange density $\mathcal{L}$. This follows from the fact that the ``momentum
space'' becomes discrete, and the basic plane wave kernels defining free fields, their pointed products and contractions,
become, naturally, equal to convergent series of smooth functions to which the operation of multiplication by (the analogue of) the step theta function
can be applied with much flexibility. The abstract splitting of Epstein-Glaser can also be applied there, but it becomes unique in case we have at most one massless
field factor in the interaction Lagrangian $\mathcal{L}$, and in this case we can avoid the whole quasi-asymptotic analysis
in the splitting and express the chronological product in terms of largely algebraic and combinatorial-type operations performed upon the basic plane wave kernels,
similarly as in Theorem \ref{WickThmForChronological}.
The most stringent fact about the causally constructed interacting fields on such space-times is that all higher order contributions
to them become ordinary operators on the Fock space transforming continuously the Hida space into itself, when evaluated at
test functions and even when evaluated at fixed space-time points, provided we have at most one massless field in $\mathcal{L}$
(as in QED). This, as we have already seen, is far not the case on the flat Minkowski space-time.

\section{The representation of $T_4 \circledS SL(2, \mathbb{C})$ in the Krein-Hilbert 
space of the free electromagnetic potential field. Bogoliubov Postulate}\label{free-gamma}

We give here a mathematically rigorous quantization of the vector potential of free
electromagnetic field based on the
Krein-isometric, but non-unitary, {\L}opusza\'nski representation in a Krein
space, i.e. in the ordinary Hilbert space equipped with involutive unitary operator $\mathfrak{J}$,
called fundamental symmetry. We construct the field using the white noise setup
of Berezin-Hida, with the field which makes rigorous sense of the white-noise generalized operator
of Hida, when evaluated at specified space-time point. This setup allows us to treat rigorously
the Wick theorem in the form needed for causal perturbative approach, heuristically (but honestly)
formulated by Bogoliubov and Shirkov \cite{Bogoliubov_Shirkov}, Chap. III.
The plan of this Section is the following. First we
define the {\L}opusza\'nski representation. Next we define the Hilbert space with a fundamental symmetry
$\mathfrak{J}'$ to which we then apply the Segal's functor $\Gamma$ of second bosonic quantization.
Next using the creation and annihilation densely defined and pre-closed operators (not distributions)
in the Fock space we make a short excursion toward the Wightman operator valued distributions fulfilling the ordinary commutation
rules (with standard Wigtman functions and Green functions) and the Gupta-Bleuler operator.
Next
using the white noise calculus of Hida, Obata and Sait\^o, \cite{hida} in the Fock space
of the field $A_\mu$ we give the white noise construction of the field $A_\mu(x)$ at specified space-time point as generalized Hida operator,
and compare this construction with the field $A_\mu$ in Wightman sense.
Finally, we give a rigorous mathematical formulation of the
Bogoliubov-Shirkov \emph{Quantization Postulate for Free Fields} together with the proof using the white noise
techniques of \cite{hida}.

In the standard treatments (including the mathematically oriented papers devoted to quantized
free electromagnetic potential field
and generally gauge field) not only 1) the white noise construction of massless fields is not presented
but likewise 2) the group
theoretical aspect is almost totally ignored.

These circumstances, 1) and 2), have at least one unpleasant consequence that our manipulations with Wick polynomial 
of free fields (among them $A_\mu$) which we encounter in the casual perturbative series, 
are not under full control and are partially based
on heuristic arguments (compare the ``Wick theorem'' for free fields in \cite{Bogoliubov_Shirkov}
and Theorem 0 in \cite{Epstein-Glaser}).  

The first omission, namely 1), comes from the fact that adaptation of the white noise construction
of Hida to massless field (such as $A_\mu$) requires a test space which differs from the ordinary
Schwartz space being equal to its closed subspace, which is connected to the singularity
of the cone in momentum space--the orbit connected to the representation pertinent to zero mass field.
This is accompanied by a necessary
additional analysis, which to the new test space (in momentum space) must give the so-called standard form
$\mathcal{S}_{A}(\mathbb{R}^3) \subset L^2(\mathbb{R}^3)$ of Gelfand \cite{GelfandIV},
\cite{obata-book}, as arising from a standard nuclear operator $A$ on $L^2(\mathbb{R}^3)$.
It is well known that in case of the Schwartz space
$\mathcal{S}(\mathbb{R}^3)$ the operator $A$ can be taken to be the ordinary quantum mechanical
Hamiltonian in $L^2(\mathbb{R}^3)$ of the three-dimensional oscillator. Because the adaptation
of the white noise calculus to zero mass field requires a considerable long additional
analysis (in particular for choosing the operator $A$ in sufficiently clever and easily manageable
form), then the white noise formulation of zero mass field is ignored by mathematicians (so far as the author is aware),
and only the massive case is taken into account as an example of application of white noise techniques to quantum fields,
compare e.g. \cite{HKPS}. Therefore, we present in details the white noise construction of the field $A_\mu(x)$. 

Construction of the field $A_\mu$ within Wightman approach does not require abandoning of the ordinary
Schwartz space of vector-valued functions on the space-time as the test space, but we are not interested here with Wightman
fields, because they are not satisfactory for the needs of the causal perturbative approach. 

Concerning the second omission, 2), it is interrelated to the fact that the general construction of the transform
from the momentum to the position wave function of photon with local transformation law cannot be consequently
pursued within a rigorous group representation theoretical fashion without the generalization of the Mackey's theory of induced representations to the case of Krein-isometric representations. Such mathematical theory has
so far been lacking. Therefore, the standard mathematical presentations for massive fields
(or massless but non gauge fields, which do not require the Krein space or Gupta-Bleuler or BRST formalism)
cover firmly the group theoretical aspect,
but for zero mass gauge fields no presentation has so far appeared which covers these important
group theoretical aspects.
The group theoretical aspect is well presented for arbitrary spin and massive free fields, but not for the field $A_\mu$ and the other free gauge fields underlying the standard model. In order to cover this omission we have constructed the required generalization of Mackey's theory of induced representations some time ago, and insert into this work as Section \ref{PartIIMackey}. 
 
Even the non gauge (no redundant degrees of freedom) but zero mass or massive fields are not treated with sufficient care if concerning the operator distributional aspect, which allows to treat clearly the
``Wick theorem'' of \cite{Bogoliubov_Shirkov}, Chap. III or Theorem 0 of \cite{Epstein-Glaser}
(Wightman approach is not satisfactory here).  

In the zero mass case, when the field is constructed as generalized white noise operator of Hida (which provides satisfactory base for the Wick theorem),
the Schwartz space of rapidly decreasing functions as test function space is not the correct space.
The situation for the photon field is still more delicate as the representation of the double covering of the Poincar\'e group cannot be unitary and even it is not bounded. There are various realizations
of the Fock space for the photon field in the Fock-Hilbert space equipped with the Gupta-Bleuler
operator $\eta$, however in all cases
(at least all known to the author) the proposed realization of the Fock space obscures the concrete shape of the
(non-unitary) representation of the group $T_4 \circledS SL(2, \mathbb{C})$ in the Krein-Fock space of the
photon field.

We hope this Section to cover these omissions.
Additional weight functions in passing from operators to generalized white noise operators
(operator valued distributions), will have to appear in order to preserve clear insight into the action
of $T_4 \circledS SL(2, \mathbb{C})$. The additional weight functions are related to the unitary and Krein-unitary operator $W$ of the introductory part of Section
\ref{constr-of-VF}, which relates the representation space of the initially defined Krein-isometric induced representation to the space of the equivalent representation, having the properties that Fourier transform of every element of the representation space of the equivalent representation has a local transformation formula. The extension of the Mackey theory, presented in Section \ref{PartIIMackey}, allows us to compute them explicitly as well as to analyze the representation of $T_4 \circledS SL(2, \mathbb{C})$
in the resulting Krein-Fock space of the photon field.

We give here a Gupta-Bleuler realization of the free quantum electromagnetic potential field $A_\mu(x)$
as a generalized white noise generalized operator of Hida
and its Fock space with a clear structure of the representation of 
$T_4 \circledS SL(2, \mathbb{C})$ and the correct test function nuclear space.

\subsection{Definition of the {\L}opusza\'nski representation}\label{DefLopRep}

The construction o of this representation may be treated as one more example of application of the construction
and the theorem placed at the introductory part of Section \ref{constr-of-VF}. 

Consider the orbit $\mathscr{O}_{(1,0,0,1)}$ of $\bar{p} = (1,0,0,1)$, i.e. positive energy surface of the cone (without
the apex $(0,0,0,0)$). The subgroup $G_{(1,0,0,1)} \subset SL(2, \mathbb{C})$ of matrices\footnote{We hope
this notation will not cause them mixed with Dirac's $\gamma$'s.}
\[
\gamma = (z, \phi) = \left( \begin{array}{cc} e^{i\phi/2} & e^{i \phi/2}z \\

0 & e^{-i\phi/2} \end{array}\right), \,\,\,
0 \leq \phi < 4\pi , \,\,\, z \in \mathbb{C}
\]
is stationary for $(1,0,0,1)$ and is isomorphic to the double covering group\footnote{Equal to the semidirect
product $T_2 \circledS \widetilde{\mathbb{S}^1}$ of the two-dimensional translation group $T_2$ and the double covering of
the circle group $\mathbb{S}^1$.} $\widetilde{E_2}$ of the Euclidean group $E_2$ of the Euclidean plane. 

As is well known there are no irreducible unitary representations of $G_{(1,0,0,1)}$ besides the infinite dimensional,
induced by the characters of the Abelian normal subgroup $T_2$ of $G_{(1,0,0,1)}$ (numbered by a positive real number),
and the one dimensional induced by the characters of the Abelian subgroup $\widetilde{\mathbb{S}^1}$ and obtained by
lifting to $G_{(1,0,0,1)}$ te one dimensional character representations of $G_{(1,0,0,1)}/T_2 \cong
\widetilde{\mathbb{S}^1}$. And no standard combinations performed on them (direct summation, tensoring, conjugation)
can produce after a natural extension $V$ of the resulting representation to the whole $SL(2, \mathbb{C})$ the representation giving the ordinary transformation of a real four- vector in Minkowski space (after the natural homomorphic map connecting $SL(2, \mathbb{C})$ to the homogeneous Lorenz group).\footnote{This in particular means that no local zero mass four vector free quantum field can exist with unitary representation $T_4 \circledS SL(2, \mathbb{C})$
in the Hilbert space of this field. For scalar field this of course would be possible.
Systematic work with conclusions going into this
direction was initiated by {\L}opusza\'nski, \cite{lop1}, \cite{lop2}.
This is quite unexpected at first sight when compared
to the construction of other local fields. Note that finite dimensional and unitary representation of the Lorentz group must be equal to (finite) direct sum of the trivial representation. In particular the four-vector transformation (at fixed point) must necessarily be non-unitary.
This also holds for the transformation of Dirac bispinor at fixed point under Lorentz group, which is
likewise non-unitary. Nonetheless, unitarity of representation of
$T_4 \circledS SL(2, \mathbb{C})$ in the Hilbert space of the standard free quantum Dirac field
is preserved.}. The situation is different when passing to Krein-unitary representations of
$G_{(1,0,0,1)}$.

Namely consider the following representation ${\L}$ of $G_{(1,0,0,1)}$ 
\[
{\L}_\gamma = S ( \gamma \otimes \overline{\gamma}) S^{-1}, \gamma \in G_{(1,0,0,1)}, 
\]
in $\mathbb{C}^4$, where 
\[
S =
\left( \begin{array}{cccc} \sqrt{2} & 0 & 0 & \sqrt{2} \\
                                0 & \sqrt{2} & \sqrt{2} & 0     \\
                                0 & i\sqrt{2} & -i \sqrt{2} & 0  \\
                                     \sqrt{2} & 0 & 0 & -\sqrt{2}  \end{array}\right)
\] 
is unitary in $\mathbb{C}^4$, and where $\overline{\gamma}$ means the ordinary complex conjugation:
if 
\[
\gamma
= \left( \begin{array}{cc} a & b \\ 
                                     c & d  \end{array}\right), \,\,\, \textrm{then} \,\,\,
\overline{\gamma}
= \left( \begin{array}{cc} \overline{a} & \overline{b} \\ 
                                     \overline{c} & \overline{d}  \end{array}\right).
\] 

If we introduce to $\mathbb{C}^4$ the ordinary inner product and the following fundamental symmetry operator
\begin{equation}\label{J-barp}
\mathfrak{J}_{\bar{p}} = \mathfrak{J}_{(1,0,0,1)} =
\left( \begin{array}{cccc} -1 & 0 & 0 & 0 \\
                                0 & 1 & 0 & 0     \\
                                0 & 0 & 1 & 0  \\
                                     0 & 0 & 0 & 1  \end{array}\right)
\end{equation}  
then the representation ${\L}$ of $ G_{(1,0,0,1)} = G_{\bar{p}}$ becomes Krein-unitary in the Krein space
$(\mathbb{C}^4, \mathfrak{J}_{\bar{p}})$:
\[
{\L} \, \mathfrak{J}_{\bar{p}} \, {\L}^* \, \mathfrak{J}_{\bar{p}} = \bold{1}_4, \,\,\, \textrm{and} \,\,\, 
\mathfrak{J}_{\bar{p}} \, {\L}^* \, \mathfrak{J}_{\bar{p}} \, {\L}  = \bold{1}_4, 
\]
where ${{\L}_\gamma}^*$ denotes the ordinary adjoint operator of ${\L}_\gamma$ with respect to the ordinary inner product
in $\mathbb{C}^4$.

The function $p \mapsto \beta(p)$, fulfilling $\beta(p)^{-1} \, \widehat{\bar{p}} \, (\beta(p)^{-1})^{*} = \widehat{p}$
on the orbit $\mathscr{O}_{(1,0,0,1)}$, may be chosen to be equal
\[
\beta(p) 
= \left( \begin{array}{cc} r^{{}^{-1/2}}\cos \frac{\theta}{2} e^{-i\frac{\vartheta}{2}} & 
-ir^{{}^{-1/2}}\sin \frac{\theta}{2} e^{i\frac{\vartheta}{2}} \\ 
 -ir^{{}^{1/2}}\sin \frac{\theta}{2} e^{-i\frac{\vartheta}{2}} & 
r^{{}^{1/2}}\cos \frac{\theta}{2} e^{i\frac{\vartheta}{2}}  \end{array}\right),
\] 
where
\begin{multline}\label{cone-r,theta,phi-coordinates}
p = 
\left( \begin{array}{c} p^0 \\
                               p^1     \\
                            p^2    \\
                               p^3       \end{array}\right) 
= \left( \begin{array}{c} r \\
                               r \sin \theta \sin \vartheta     \\
                            r \sin \theta \cos \vartheta     \\
                                r \cos \theta       \end{array}\right) \in \mathscr{O}_{(1,0,0,1)}, 
\,\,\, 0 \leq \theta < \pi, 
0 \leq \vartheta < 2 \pi, r > 0. 
\end{multline}
Now we construct, like in the introductory part of Section \ref{constr-of-VF}, the Krein-isometric representation of
$T_4 \circledS SL(2, \mathbb{C})$ induced
by the Krein-unitary representation ${\L}$, putting there ${\L}_\gamma$ for $Q(\gamma, \bar{p})$
with $\bar{p} = (1,0,0,1)$. Let us denote the representation by $U^{{}_{(1,0,0,1)}{\L}}$
and call the \emph{{\L}opusza\'nski representation}. By Section \ref{lop_ind},
it is Krein-unitary equivalent to the Krein-isometric representation of $T^4 \circledS SL(2,\mathbb{C})$
induced\footnote{In the sense of definition placed in
Section \ref{def_ind_krein}, which is a generalization of the Mackey's induced representation.}
by the representation ${}_{(1,0,0,1)}{\L} = {}_{\chi_{\bar{p}}}{\L}$:
\[
a \cdot \gamma \mapsto \chi_{\bar{p}}(a) \, {\L}_\gamma,
\]
of the subgroup $T_4 \cdot G_{\bar{p}} \subset T_4 \circledS SL(2,\mathbb{C})$.

Now we define the following extension
\[
V(\alpha) = S( \alpha \otimes \overline{\alpha}) S^{-1}, \alpha \in SL(2, \mathbb{C}),
\]
of the representation ${\L}$, to the whole $SL(2, \mathbb{C})$ group, which is likewise
Krein-unitary in $(\mathbb{C}^4, \mathfrak{J}_{\bar{p}})$:
\[
V(\alpha) \, \mathfrak{J}_{\bar{p}} \, {V(\alpha)}^* \, \mathfrak{J}_{\bar{p}} = \bold{1}_4,
\,\,\, \textrm{and} \,\,\,
\mathfrak{J}_{\bar{p}} \, {V(\alpha)}^* \, \mathfrak{J}_{\bar{p}} \, V(\alpha) = \bold{1}_4,
\,\,\, \alpha \in SL(2, \mathbb{C}).
\]
Moreover $\alpha \mapsto V(\alpha)$
gives a natural homomorphism of the $SL(2, \mathbb{C})$ onto the proper orthochronous Lorentz group
in the Minkowski vector space, i.e. each $V(\alpha)$, $\alpha \in SL(2, \mathbb{C})$, is a real Lorentz
transformation.
It is customary to write $V(\alpha)$ as the corresponding Lorentz transformation $\Lambda(\alpha)$.
Because we have already occupied the notation $\Lambda(\alpha)$ for a natural antihomomorphism
$\Lambda$, we have $V(\alpha) = \Lambda(\alpha^{-1})$ in our notation.

With the extension $V$ at our disposal, we apply
to the elements $\widetilde{\psi}$  of the 
space  of the {\L}opusza\'nski representation 
\[
{W'}_{{}_{V,1,0,0,1}}U_{{}_{{\L},1,0,0,1}}^{-1}
U^{{}_{(1,0,0,1)}{\L}} 
U_{{}_{{\L},1,0,0,1}} {W'}_{{}_{V,1,0,0,1}}^{-1}
\]
the Krein unitary and unitary
transformation ${W''}_{{}_{V,1,0,0,1}}: \widetilde{\psi} \mapsto \widetilde{\varphi}$, as in the introductory part of Section
\ref{constr-of-VF},
having the property that the Fourier transform (\ref{F(varphi)}) $\varphi$ have the local transformation
law. In order to simplify notation we write just $W'U^{-1}$ for the Krein-isometric operator
\[
{W'}_{{}_{V,1,0,0,1}}U_{{}_{{\L},1,0,0,1}}^{-1},
\]
and just $W$ for the Krein-isometric operator
\[
W_{{}_{V,1,0,0,1}} = {W''}_{{}_{V,1,0,0,1}}{W'}_{{}_{V,1,0,0,1}}U_{{}_{{\L},1,0,0,1}}^{-1}
\]
defined in the introductory part of Section \ref{constr-of-VF}.

Namely, the representation $WU^{{}_{(1,0,0,1)}{\L}}W^{-1}$, which we have called
the \emph{natural local version} of $U^{{}_{(1,0,0,1)}{\L}}$ in introduction to Section \ref{constr-of-VF}, acts as follows
\begin{equation}\label{Lop-rep--on-tildevarphi}
\begin{split}
WU_{0,\alpha}^{{}_{(1,0,0,1)}{\L}}W^{-1} \widetilde{\varphi} (p) =
U(\alpha) \widetilde{\varphi} (p) = V(\alpha) \widetilde{\varphi} (\Lambda(\alpha)p), \\
WU_{a,1}^{{}_{(1,0,0,1)}{\L}}W^{-1} \widetilde{\varphi} (p) = T(a) \widetilde{\varphi}(p)
= e^{i a \cdot p}\widetilde{\varphi}(p).
\end{split}
\end{equation}
Therefore the Fourier transform (\ref{F(varphi)}) $\varphi$ of $\widetilde{\varphi} = W \widetilde{\psi}$ has the following local transformation law
\[
\begin{split}
U(\alpha)\varphi(x) = V(\alpha) \varphi(\Lambda(\alpha)x)
= \Lambda(\alpha^{-1}) \varphi(\Lambda(\alpha)x), \,\,\,
T(a)\varphi(x) = \varphi(x - a).
\end{split}
\]
of a four-vector field on the Minkowski manifold. Because by construction $\widetilde{\varphi}$ are concentrated
on the orbit $\mathscr{O}_{(1,0,0,1)}$, it follows that the elements $\varphi \in \mathcal{H}''$
are the positive energy (distributional) solutions of the ordinary wave equation with zero mass
\[
\partial^\mu \partial_\mu \varphi = 0.
\]

In this Section and the following Section we construct the free quantum electromagnetic potential 
field based on the natural local version
\[
WU^{{}_{(1,0,0,1)}{\L}}W^{-1}.
\] 
In Subsection \ref{equivalentA-s} we compare this field with that based on the representation
denoted by
\[
\Big(WU^{{}_{(1,0,0,1)}{\L}}W^{-1} \Big)_{{}_{0}}
\]
in introduction to Section \ref{constr-of-VF}.

Because the light cone (in the momentum space) is not an ordinary sub-manifold in $\mathbb{R}^4$ (for the standard
manifold structure on $\mathbb{R}^4$) the last sentences need an explanation. Namely, consider a manifold $\mathscr{O}$ of dimension less
that $4$ (or less that $n$) in $\mathbb{R}^4$ (or in $\mathbb{R}^n$) with the measure ${\ud}\mu|_{{}_{\mathscr{O}}}(p)$
on $\mathscr{O}$ induced from the ordinary invariant measure on $\mathbb{R}^4$ (or from $\mathbb{R}^n$).
Let $f$ be a function on $\mathscr{O}$ which is locally integrable w.r.t. ${\ud} \mu|_{{}_{\mathscr{O}}}(p)$,
or is a multiplier of $\mathcal{D}(\mathscr{O})$ or of
$\mathcal{S}(\mathscr{O})$.
For the mostly used nuclear topological
test function spaces, e.g. the space of smooth functions of compact support $\mathcal{D}(\mathbb{R}^4)$
(or $\mathcal{D}(\mathbb{R}^n)$) or the Schwartz test function space $\mathcal{S}(\mathbb{R}^4)$ (or
$\mathcal{S}(\mathbb{R}^n)$) the simplest distribution $f$ concentrated on the manifold $\mathscr{O}$ of dimension less
that $4$ (or less that $n$) in $\mathbb{R}^4$ (or in $\mathbb{R}^n$) defined by
\begin{equation}\label{(f,phi)-sol}
(f,\phi) = \int \limits_{\mathscr{O}} \, f(p) \phi|_{{}_{\mathscr{O}}}(p) \, \ud \mu|_{{}_{\mathscr{O}}}(p),
\,\,\, \phi \in \mathcal{D}(\mathbb{R}^4) \, \textrm{or} \, \in \mathcal{S}(\mathbb{R}^4)
\end{equation}
is a well-defined continuous functional on $\mathcal{D}(\mathbb{R}^4)$ or $\mathcal{S}(\mathbb{R}^4)$
(or $\mathcal{D}(\mathbb{R}^n)$ or $\mathcal{S}(\mathbb{R}^n)$),
as in this case the map $\phi \mapsto \phi|_{{}_{\mathscr{O}}}$ is a continuous map from
$\mathcal{D}(\mathbb{R}^4)$ (or $\mathcal{S}(\mathbb{R}^4)$) into
$\mathcal{D}(\mathscr{O})$ (or $\mathcal{S}(\mathscr{O})$), where
$\phi|_{{}_{\mathscr{O}}}$ is the restriction of the function $\phi$ to the sub-manifold $\mathscr{O}$,
compare e.g. \cite{GelfandI}, Chapter III (although continuity of the map
$\phi \mapsto \phi|_{{}_{\mathscr{O}}}$ is not absolutely necessary for the continuity of
the said functional).

In our case $\mathscr{O}= \mathscr{O}_{\bar{p}}$ is the ``positive'' (or ``negative energy'') 
light cone without the apex in the momentum space, for which 
the manifold structure fails at the tip of the light cone. In particular 
$\phi \mapsto \phi|_{{}_{\mathscr{O}}}$ is not continuous as the map of 
$\mathcal{D}(\mathbb{R}^4)$  (or $\mathcal{S}(\mathbb{R}^4)$) into 
$\mathcal{D}(\mathbb{R}^3)$  (or $\mathcal{S}(\mathbb{R}^3)$) with the spatial 
momentum components as the natural coordinate map on the cone, which is easily checked. 

Although continuity of $\phi \mapsto \phi|_{{}_{\mathscr{O}}}$  is not necessary for the said functional 
(\ref{(f,phi)-sol}) to stay continuous
(and in fact it will be likewise continuous for the ordinary Schwartz space
although $\phi \mapsto \phi|_{{}_{\mathscr{O}}}$ is not continuous as a map 
$\mathcal{S}(\mathbb{R}^4) \rightarrow \mathcal{S}(\mathbb{R}^3)$) we 
a proiri allow two possibilities: 1) one which uses test space different from that of Schwartz, but which 
saves continuity of $\phi \mapsto \phi|_{{}_{\mathscr{O}}}$ and the other one 
2) which uses the ordinary Schwartz space
but the continuity of $\phi \mapsto \phi|_{{}_{\mathscr{O}}}$ is lost.

In case 1) we are using the test space (correct for the white noise construction of the field $A_\mu$,
as we will see in the later part of our presentation)
in the momentum space as equal to the closed subspace
$\mathcal{S}^{0}(\mathbb{R}^4)$ of $\mathcal{S}(\mathbb{R}^4)$ consisting of all those elements
of $\mathcal{S}(\mathbb{R}^4)$ for which their values and all their derivatives vanish
at the zero point, and its inverse Fourier transform $\mathscr{F}^{-1}$ image $\mathcal{S}^{00}(\mathbb{R}^4)$ as the test function space over space-time. In this case $\phi \mapsto \phi|_{{}_{\mathscr{O}}}$,
as a map $\mathcal{S}^{0}(\mathbb{R}^4) \rightarrow \mathcal{S}^{0}(\mathbb{R}^3)$,
will be continuous, as well as the functional (\ref{(f,phi)-sol}).
Namely, for $f$ locally integrable on $\mathscr{O}$
or for $f$ being a multiplier of the nuclear algebra $\mathcal{S}^{0}(\mathbb{R}^3) \cong \mathcal{S}^{0}(\mathscr{O})$
the functional defined by
\begin{equation}\label{Distr.Wave.Eq.Sol}
(f,\mathscr{F}\phi) = \int \limits_{\mathscr{O}} \, f(p) (\mathscr{F}\phi)|_{{}_{\mathscr{O}}}(p) \,
\ud \mu|_{{}_{\mathscr{O}}}(p),
\,\,\, \mathscr{F}\phi \in \mathcal{S}^{0}(\mathbb{R}^4) \, \textrm{and} \,
\phi \in \mathcal{S}^{00}(\mathbb{R}^4)
\end{equation}
is a continuous functional on $\mathcal{S}^{00}(\mathbb{R}^4)$, if understood as a map
$\phi \mapsto (f,\mathscr{F}\phi) = (\mathscr{F} f,\phi)$, and
$\mathscr{F} \phi \mapsto (f,\mathscr{F}\phi)$ is a continuous functional on $\mathcal{S}^{0}(\mathbb{R}^4)$,
because in this case
$\phi \mapsto \phi|_{{}_{\mathscr{O}}}$ maps continuously
$\mathcal{S}^{0}(\mathbb{R}^4)$ into $\mathcal{S}^{0}(\mathbb{R}^3)$. For the proof, compare Subsection \ref{Lop-on-E}.
In particular for $\widetilde{\varphi} \in \mathcal{H}'$ the function $f = \widetilde{\varphi}$
on the light cone $\mathscr{O} = \mathscr{O}_{\bar{p}}$ in the momentum representation belonging to the representation space of the {\L}opusza\'nski
representation defines a distribution on $\mathcal{S}^{00}(\mathbb{R}^4)$ whose Fourier transform
$\mathscr{F}$ is concentrated on the positive energy light cone and is given by
the distribution (\ref{Distr.Wave.Eq.Sol}) with $f = \widetilde{\varphi}$.

Therefore, distributional four-vector solutions $\varphi \in \mathcal{H}''$ of d'Alembert equation 
whose Fourier transforms $f$ defined by  (\ref{Distr.Wave.Eq.Sol}) which correspond to ordinary functions 
$\widetilde{\varphi} \in \mathcal{H}'$, are rather of special character. 
In case of more general distributions (or distributions on smooth functions of compact supports) 
which are solutions of the wave equation we may only say that their Fourier transforms are 
concentrated on the light cone
in the momentum picture, but nothing more. In particular in general such distributional solution
defines after Fourier transformation a distribution which is not regular--function like-- distribution on 
the orbit $\mathscr{O}$, i.e. on the test space
$\mathcal{S}(\mathscr{O}) \cong \mathcal{S}(\mathbb{R}^3)$
of functions on the ''positive'' or ''negative'' energy light cone in the momentum space. 

But when using the test function
spaces (correct for the white noise construction of $A_\mu(x)$)
$\mathcal{S}^{0}(\mathbb{R}^4)$, $\mathcal{S}^{0}(\mathbb{R}^3)$, $\mathcal{S}^{00}(\mathbb{R}^4)$,
we gain a natural relationship between distributions $S \in
\mathcal{S}(\mathscr{O})^* \cong \mathcal{S}(\mathbb{R}^3)^*$
(i.e. generalised states in $\mathcal{H}'$) and distributional solutions
$F\in \mathcal{S}^{00}(\mathbb{R}^4)$ of wave equation, given by the formula:
\[
F(\phi) = S(\widetilde{\phi}|_{{}_{\mathscr{O}}}), \,\,\, \phi \in \mathcal{S}^{00}(\mathbb{R}^4),
\]
well-defined because the restriction to the cone orbit $\mathscr{O} = \mathscr{O}_{\bar{p}}$
\[
\phi \,\,\, \longmapsto \,\,\, \phi|_{{}_{\mathscr{O}}}
\]
maps continuously $\mathcal{S}^{0}(\mathbb{R}^4)$ into $\mathcal{S}^{0}(\mathbb{R}^3)$.

Similarly, we have a well-defined restriction map
\begin{equation}\label{tilde(varphi)}
\mathscr{F}F \mapsto \mathscr{F} F|_{{}_{\mathscr{O}}}
\end{equation}
for $F$, $\mathscr{F}F$ understood as elements of $\mathcal{S}^{00}(\mathbb{R}^4)^*$,
$\mathcal{S}^{0}(\mathbb{R}^4)^*$ respectively, defined by
\[
\mathscr{F} F|_{{}_{\mathscr{O}}}(\phi) = F\big((\mathscr{F} \phi)|_{{}_{\mathscr{O}}}\big).
\]

There is no such correspondence between the generalized states $S$ on the light cone orbit (of the representation
concentrated on the light cone) in the momentum space and the distributional solutions of 
the wave equation when using the Schwartz test function space. It seems that this important fact has escaped due attention 
of mathematical physicists, and was one of the stumbling blocks in the correct understanding 
of representation theory aspect of the zero mass fields, and in particular of the 
electromagnetic four-potential field $A^\mu$ and the infrared states. We will show in the latter part of this work connections of this fact to the infrared problems within the causal perturbative approach of Bogoliubov.

\vspace*{0.5cm}

\noindent \fbox{%
       \parbox{\textwidth}{%

{\bf NOTATION}. In what follows we will use the sign $\mathscr{F}(\cdot)$ for the ordinary
 Fourier transform in $\mathbb{R}^n$  (with the sign at $ip_0x_0$ opposite with respect
to the sign at $i\boldsymbol{\p} \cdot \boldsymbol{\x}$ in the exponent in case of $\mathbb{R}^4$
understood as the Minkowski space) interchangeably with the sign $\widetilde{\cdot}$ in order
to shorten expressions which otherwise would contain too many $\mathscr{F}$-signs to be of
reasonable size, which arise in our proofs. We shall trust to the context or explanatory remarks
which will make clear what is meant in each instance. In particular, it is clear that under the
integral sign for the integration over $\mathscr{O}_{\bar{p}}$, as in the formula (\ref{F(varphi)}), 
understood as  four dimensional inverse Fourier formula of a distributional solution $\varphi \in \mathcal{H}$, 
the function $\widetilde{\varphi}$ is understood as the function 
$p \mapsto f(p) = \widetilde{\varphi}(p)$ on the orbit
$\mathscr{O}_{\bar{p}}$ which determines the four dimensional distributional Fourier transform
$f=\widetilde{\varphi}$ of $\varphi$, given by the formula (\ref{Distr.Wave.Eq.Sol}). For $\varphi$
which is an ordinary square integrable function on $\mathbb{R}^4$ the function $\widetilde{\varphi}$ in the formula (\ref{F(varphi)}) is understood as the restriction 
 $\widetilde{\varphi}|_{{}_{\mathscr{O}_{\bar{p}}}}$
of the ordinary $4$-dimensional Fourier transform $\widetilde{\varphi}$ of $\varphi$ to the orbit
$\mathscr{O}_{\bar{p}}$, and the formula (\ref{F(varphi)}) itself is not understood as the full inverse Fourier integral but merely as the restriction of the full inverse integral to the orbit $\mathscr{O}_{\bar{p}}$. Otherwise, when the context does not fix the meaning of $\widetilde{\cdot}$
the restriction sign has to be written explicitly.
}%
}

\vspace*{0.5cm}

However, we should emphasize that we have the second possibility here, 2). Namely,
even when $\mathscr{F}\phi, \phi$
in (\ref{Distr.Wave.Eq.Sol}) belong to $\mathcal{S}(\mathbb{R}^4)$
and the function $p\mapsto f(p)$ in (\ref{Distr.Wave.Eq.Sol}) defined on the cone $\mathscr{O}
= \mathscr{O}_{\bar{p}}$ is a multiplier of
$\mathcal{S}(\mathbb{R}^3) \cong \mathcal{S}(\mathscr{O})$ or if $p\mapsto f(p)$
is measurable and fulfills (for some natural $M>1$ and $N$)
\[
\int \limits_{\mathscr{O}} \, \big|(1 + p_0(p)^2)^{-N}f(p)\big|^{M} \,
\ud \mu|_{{}_{\mathscr{O}}}(p) \, < \infty
\]
(which is the case for example for $(p\mapsto f(p)) \, \in \mathcal{H}'$), then the formula (\ref{Distr.Wave.Eq.Sol})
still represents a continuous functional of $\mathscr{F}\phi$ and of $\phi$, when regarded as a functional
on $\mathcal{S}(\mathbb{R}^4)$. This is in particular the case for the zero mass
Pauli-Jordan function $\boldsymbol{D}_0$ and its Fourier transform $\widetilde{\boldsymbol{D}_0}$,
compare Subsection \ref{splitting}.
Although continuity of the map $\phi \mapsto \phi|_{{}_{\mathscr{O}}}$
is lost now, when regarded on the Schwartz spaces, the functional (\ref{Distr.Wave.Eq.Sol})
stays continuous on the Schwartz space.
However, if we are using the ordinary Schwartz space
for the distributions concentrated on the light cone $\mathscr{O}
= \mathscr{O}_{\bar{p}}$ or $\mathscr{O}
= \mathscr{O}_{\bar{p}} \sqcup \mathscr{O}_{-\bar{p}}$
of the form (\ref{Distr.Wave.Eq.Sol}), say $f|_{{}_{p \cdot p=0}}(p)\delta(p \cdot p)$, $\delta(p\cdot p)$,
there will arise additional complications when trying to incorporate the formal rules of differentiation,
with the need of regularization, compare \cite{GelfandI} and Subsections \ref{splitting}, \ref{Lop-on-E}.
Treatment of these distributions becomes much more transparent and simpler when
using the test spaces $\mathcal{S}^{0}(\mathbb{R}^4)$, $\mathcal{S}^{00}(\mathbb{R}^4)$.

To any wave function $\widetilde{\varphi}$ on the light cone from the Hilbert space of the
{\L}opusza\'nski representation there correspond a well-defined regular
(function-like) distributional solution $\varphi$
of the wave equation which can be regarded either as element of $\mathcal{S}^{00}(\mathbb{R}^4)^*$
or $\mathcal{S}(\mathbb{R}^4)^*$.

At the group theoretical level the two possible choices of space-time test spaces:
$\mathcal{S}(\mathbb{R}^4)$ or $\mathcal{S}^{00}(\mathbb{R}^4)$, are equally well.
Even in the construction of the free field $A_\mu$ within Wightman approach we can equally
use $\mathcal{S}(\mathbb{R}^4)$ as well as $\mathcal{S}^{00}(\mathbb{R}^4)$. But when using
the white noise construction of the field $A_\mu(x)$ we have only one possible choice
of the space-time test space and, as we will see, it must be equal
$\mathcal{S}^{00}(\mathbb{R}^4)$.

After giving the two \emph{a priori} possible distributional interpretations of the elements
$\widetilde{\varphi}$ of the
Hilbert space $\mathcal{H}'$ (or $\varphi \in \mathcal{H}''$) of the {\L}opusza\'nski representation, 
we go back to $\mathcal{H}'$ itself and give further details of its structure. 

The explicit form of the Krein space structure $(\mathcal{H}', \mathfrak{J}')$ of the representation space of the
representation $WU^{{}_{(1,0,0,1)}{\L}}W^{-1}$ can be obtained by substitution of the explicit formulas for 
the function $p \mapsto \beta(p)$ and the extension $V$ into the formulas written at the introductory part
of Section \ref{constr-of-VF}.

In particular the inner product of $\widetilde{\varphi} = W \widetilde{\psi}$ and $\widetilde{\varphi'} =
W \widetilde{\psi'}$ is equal
\begin{multline}\label{inn-Lop-1-space}
(\widetilde{\varphi}, \widetilde{\varphi}')
= \int \limits_{sp(P^0 , \ldots , P^3) \cong \mathscr{O}_{\bar{p}}}
\Big( \widetilde{\varphi}(p), \widetilde{\varphi}'(p) \Big)_{p}
\, \ud \mu |_{{}_{\mathscr{O}_{\bar{p}}}} (p) \\
= \int \limits_{sp(P^0 , \ldots , P^3) \cong \mathscr{O}_{\bar{p}}}
\Big( \widetilde{\varphi}(p),
V(\beta(p))^* V(\beta(p)) \widetilde{\varphi}'(p) \Big)_{\mathcal{H}_{\bar{p}}}
\, \ud \mu |_{{}_{\mathscr{O}_{\bar{p}}}} (p) \\
= \int \limits_{\mathscr{O}_{\bar{p}}}
\Big( \widetilde{\varphi}(p),
B(p) \widetilde{\varphi}'(p) \Big)_{\mathbb{C}^4}
\, \ud \mu |_{{}_{\mathscr{O}_{\bar{p}}}} (p), \\
= \int \limits_{\mathbb{R}^3}
\Big( \widetilde{\varphi}(\vec{p}, p^0(\vec{p})),
(B\widetilde{\varphi}')(\vec{p}, p^0(\vec{p})) \Big)_{\mathbb{C}^4}
\, \ud^3 p = (\widetilde{\varphi}, B \widetilde{\varphi}')_{{}_{\oplus L^2(\mathbb{R}^3)}} , \\
\ud \mu |_{{}_{\mathscr{O}_{\bar{p}}}} (\vec{p}) = \frac{\ud^3 p}{2p^0(\vec{p})}, \, p^0(\vec{p}) = (\vec{p}\cdot \vec{p})^{{}^{1/2}},
\end{multline}
where we have introduced the matrix
\[
B(p) = V(\beta(p))^* V(\beta(p))
\]
depending on $p \in \mathscr{O}_{\bar{p}}$, strictly positive (invertible) on $\mathscr{O}_{\bar{p}}$
and the operator $B$ of pointwise multiplication by the matrix
\begin{equation}\label{operatorB}
\frac{1}{2p^0(\vec{p})} B\big(\vec{p}, p^0(\vec{p})\big),
\end{equation}
on the Hilbert space $\oplus L^2(\mathbb{R}^3)$ with respect to the ordinary invariant Lebesgue measue
$\ud^3 \boldsymbol{\p}$ on $\mathbb{R}^3$ (the direct sum $\oplus$ is over the four components of the
function $\widetilde{\varphi}$),
in order to simplify notation of the formulas which are to follow in the remaining part of this Subsection.
 
The fundamental symmetry operator $\mathfrak{J'}$ is given by the pointwise multiplication by the following 
operator
\begin{equation}\label{operatorJ'}
\mathfrak{J'}_{p} = V(\beta(p))^{-1} \mathfrak{J}_{\bar{p}} V(\beta(p)).
\end{equation} 
Because  for each $p \in \mathscr{O}_{\bar{p}}$ the matrix operator  $V(\beta(p))$ (and the same of course 
holds for $V(\beta(p))^*$)   is by construction Krein-unitary 
in the Krein space $(\mathbb{C}^4, \mathfrak{J}_{\bar{p}})
 = (\mathcal{H}_{\bar{p}}, \mathfrak{J}_{\bar{p}})$ of the representation ${\L}$, 
then the Krein product in $(\mathcal{H}', \mathfrak{J'})$ is given by the following formula
\begin{multline}\label{Kr-inn-Lop-1-space}
(\widetilde{\varphi}, \mathfrak{J'} \widetilde{\varphi}') 
= \int \limits_{sp(P^0 , \ldots , P^3) \cong \mathscr{O}_{\bar{p}}} 
\Big( \widetilde{\varphi}(p), V(\beta(p))^* V(\beta(p)) \mathfrak{J'}_p \widetilde{\varphi}'(p)  
 \Big)_{\mathcal{H}_{\bar{p}}} 
\, \ud \mu |_{{}_{\mathscr{O}_{\bar{p}}}} (p) \\
= \int \limits_{\mathscr{O}_{\bar{p}}} 
\Big( \widetilde{\varphi}(p), V(\beta(p))^* V(\beta(p)) V(\beta(p))^{-1} \mathfrak{J}_{\bar{p}} V(\beta(p)) \widetilde{\varphi}'(p) \Big)_{\mathbb{C}^4} 
\, \ud \mu |_{{}_{\mathscr{O}_{\bar{p}}}} (p) \\
= \int \limits_{\mathscr{O}_{\bar{p}}} 
\Big( \widetilde{\varphi}(p), \mathfrak{J}_{\bar{p}} \widetilde{\varphi}'(p) \Big)_{\mathbb{C}^4} 
\, \ud \mu |_{{}_{\mathscr{O}_{\bar{p}}}} (p), \\
\end{multline}
because $V(\beta(p))^* \mathfrak{J}_{\bar{p}} V(\beta(p)) = \mathfrak{J}_{\bar{p}}$.

Introducing the coordinates $\vec{p}$ on $\mathscr{O}_{\bar{p}}$ and regarding any function
$p \mapsto \widetilde{\varphi}(p)$ on $\mathscr{O}_{\bar{p}}$ as a function 
$\vec{p} \mapsto \widetilde{\varphi}(\vec{p}) = \widetilde{\varphi}\big(\vec{p}, p^0(\vec{p})\big)$ with $p^0(\vec{p})$
as in (\ref{inn-Lop-1-space}), the last formula (\ref{Kr-inn-Lop-1-space}) may be written as
\begin{multline}\label{Kr-inn-Lop-1-space'}
(\widetilde{\varphi}, \mathfrak{J'} \widetilde{\varphi}') 
= (\widetilde{\varphi}, B \mathfrak{J}' \widetilde{\varphi}')_{{}_{\oplus L^2(\mathbb{R}^3)}}
= (\sqrt{B}\widetilde{\varphi}, \sqrt{B} \mathfrak{J}' \widetilde{\varphi}')_{{}_{\oplus L^2(\mathbb{R}^3)}} 
= (\widetilde{\varphi}, \mathfrak{J}_{\bar{p}} \widetilde{\varphi}')_{{}_{\oplus L^2(\mathbb{R}^3, 
\ud \mu |_{{}_{\mathscr{O}_{\bar{p}}}})}} ,
\end{multline}  
where the last inner product 
\[
(\cdot, \cdot)_{{}_{\oplus L^2(\mathbb{R}^3, \ud \mu)}}
\]
is with respect to the measure
\[
\ud \mu = \frac{\ud^3 p}{2p^0(\vec{p})}, \,\, p^0(\vec{p}) = (\vec{p}\cdot \vec{p})^{{}^{1/2}},
\]
on $\mathbb{R}^3$, and where $B$ is the positive self-adjoint operator on $\oplus L^2(\mathbb{R}^3)$ introduced above and 
$\sqrt{B}$ is its square root equal to the operator of pointwise multiplication by the matrix
\[
\frac{1}{\sqrt{2p^0(\vec{p})}} \sqrt{B\big(\vec{p}, p^0(\vec{p})\big)},
\]
with $\sqrt{B\big(\vec{p}, p^0(\vec{p})\big)}$  being the square root of the positive matrix
$B\big(\vec{p}, p^0(\vec{p})\big)$.

Krein-isometric and Krein-unitary representations in a Krein space $(\mathcal{H}, \mathfrak{J})$ allows the specific
kind of conjugation, which is trivial for ordinary unitary representations when $\mathfrak{J} = \bold{1}_\mathcal{H}$. Namely, for every representation $U$ of this kind in the Krein space $(\mathcal{H}, \mathfrak{J})$,
the ordinary Hilbert space adjoint
operation $^*$ and passing to the inverse, i.e. $U^{*-1} = \mathfrak{J}U\mathfrak{J}$, is well-defined, which is nontrivial
for Krein-isometric representation, compare Sect. \ref{def_ind_krein}. Moreover, $U^{*-1} = \mathfrak{J}U\mathfrak{J}$ defines another Krein-isometric (resp. Krein unitary) representation with respect to the same Krein structure, compare Subsection \ref{def_ind_krein}, which is unitary and Krein-unitary equivalent to the initial representation $U$, with the equivalence given by the fundamental symmetry $\mathfrak{J}$
itself, and $\mathfrak{J}$ is by construction unitary and Krein-unitary. 

In particular together with the Krein-isometric representation $WU^{{}_{(1,0,0,1)}{\L}}W^{-1}$ in
the Krein space $(\mathcal{H}', \mathfrak{J}')$ just constructed, there acts in the same Krein space
$(\mathcal{H}', \mathfrak{J}')$ the naturally conjugate Krein isometric representation
\begin{equation}\label{conjugated-Lop-rep}
\big[WU^{{}_{(1,0,0,1)}{\L}}W^{-1}\big]^{*-1} = \mathfrak{J'} WU^{{}_{(1,0,0,1)}{\L}}W^{-1} \mathfrak{J'}
\end{equation}
unitary and Krein-unitary equivalent to $WU^{{}_{(1,0,0,1)}{\L}}W^{-1}$, with the equivalence
given by the fundamental symmetry $ \mathfrak{J'}$ itself. Because we have explicitly computed
$\mathfrak{J'}$ and $WU^{{}_{(1,0,0,1)}{\L}}W^{-1}$ we also know the explicit formula for the action of 
$\big[WU^{{}_{(1,0,0,1)}{\L}}W^{-1}\big]^{*-1}$. Namely, we have 
\begin{multline*}
\big[WU^{{}_{(1,0,0,1)}{\L}}W^{-1}\big]^{*-1} \widetilde{\varphi}(p) =
\big( \mathfrak{J'} WU^{{}_{(1,0,0,1)}{\L}}W^{-1} \mathfrak{J'} \big) \widetilde{\varphi}(p) \\ 
= V(\beta(p))^{-1} V(\beta(p))^{*-1} V(\alpha)^{*-1} V(\beta(\Lambda(\alpha)p))^{*} 
V(\beta(\Lambda(\alpha)p)) \widetilde{\varphi}(\Lambda(\alpha)p).
\end{multline*}

Before passing to quantization, we give here several formulas which will be useful in further computations.

First let us note the simple formula for the Krein inner product in the Krein space 
$(\mathcal{H''}, \mathfrak{J''})$ of all  Fourier transforms $\varphi$, given by (\ref{F(varphi)}), of the
elements $\widetilde{\varphi}$ of the Krein space $(\mathcal{H'}, \mathfrak{J'})$ of the representation
$WU^{{}_{(1,0,0,1)}{\L}}W^{-1}$. Namely, easy computation gives
\begin{multline}\label{krein-prod-1-photon}
(\varphi, \mathfrak{J''}\varphi) = 
i \int \limits_{t = const.} \Big\{
 \overline{\varphi(x)}\partial_t\big(\mathfrak{J}_{\bar{p}} \varphi'\big)(x) 
- \overline{\partial_t \varphi(x)} \mathfrak{J}_{\bar{p}} \varphi'(x) \Big\} \, \ud^3 x \\
= - i g^{\mu \nu} \int \limits_{t = const.} \Big\{
 \overline{\varphi_\mu (x)}\partial_t \varphi'_\nu\big(x) 
- \overline{\partial_t \varphi_\mu(x)} \varphi_\nu (x) \Big\} \, \ud^3 x
\end{multline}
Next we give explicit formulas for $V(\beta(p))^{-1}$, $B(p) = V(\beta(p))^* V(\beta(p))$ and 
$\sqrt{B(p)}$, $p \in \mathscr{O}_{(1,0,0,1)}$, and give their useful properties.
\begin{multline*}
V(\beta(p))^{-1} \\
= \left( \begin{array}{cccc} 
\frac{r^{-1} + r}{2} & 0 & 0 & -\frac{r^{-1} - r}{2} \\
-\frac{r^{-1} - r}{2} \frac{p^1}{r} & \frac{p^2}{\sqrt{(p^1)^2 + (p^2)^2}} &  \frac{p^1}{\sqrt{(p^1)^2 + (p^2)^2}} \frac{p^3}{r} & \frac{r^{-1} + r}{2} \frac{p^1}{r}  \\
-\frac{r^{-1} - r}{2} \frac{p^2}{r} & - \frac{p^1}{\sqrt{(p^1)^2 + (p^2)^2}} & \frac{p^2}{\sqrt{(p^1)^2 + (p^2)^2}}  \frac{p^3}{r} &  \frac{r^{-1} + r}{2} \frac{p^2}{r}  \\
-\frac{r^{-1} - r}{2} \frac{p^3}{r} & 0 & -\frac{\sqrt{(p^1)^2 + (p^2)^2}}{r} & \frac{r^{-1} + r}{2} \frac{p^3}{r} \end{array}\right) \\
= \left( \begin{array}{cccc} 
\frac{r^{-1} + r}{2} & 0 & 0 & -\frac{r^{-1} - r}{2} \\
-\frac{r^{-1} - r}{2} \sin \theta \sin \vartheta & \cos \vartheta & \cos \theta \sin \vartheta & \frac{r^{-1} + r}{2} \sin \theta \sin \vartheta \\
-\frac{r^{-1} - r}{2} \sin \theta \cos \vartheta & -\sin \vartheta & \cos \theta \cos \vartheta & \frac{r^{-1} + r}{2} \sin \theta \cos \vartheta  \\
-\frac{r^{-1} - r}{2} \cos \theta & 0 & - \sin \theta & \frac{r^{-1} + r}{2} \cos \theta \end{array}\right). 
\end{multline*}
\begin{multline}\label{Bmatrix}
B(p) = V(\beta(p))^* V(\beta(p)) = \\
\left( \begin{array}{cccc} 
\frac{r^{-2} + r^2}{2} & \frac{r^{-2} - r^2}{2r}p^1 & \frac{r^{-2} - r^2}{2r}p^2 & \frac{r^{-2} - r^2}{2r}p^3 \\
\frac{r^{-2} - r^2}{2r} p^1&\frac{r^{-2} + r^2 -2}{2r^2}p^1 p^1 +1 & \frac{r^{-2} + r^2 -2}{2r^2}p^1 p^2 & \frac{r^{-2} + r^2 -2}{2r^2} p^1 p^3  \\
\frac{r^{-2} - r^2}{2r}p^2 & \frac{r^{-2} + r^2 -2}{2r^2}p^2 p^1 &\frac{r^{-2} + r^2 -2}{2r^2}p^2 p^2 +1 & \frac{r^{-2} + r^2 -2}{2r^2} p^2 p^3  \\
\frac{r^{-2} - r^2}{2r}p^3 & \frac{r^{-2} + r^2 -2}{2r^2}p^3 p^1 & \frac{r^{-2} + r^2 -2}{2r^2}p^3 p^2 &\frac{r^{-2} + r^2 -2}{2r^2}p^3 p^3 +1  \end{array}\right) 
\end{multline}
\begingroup\makeatletter\def\f@size{5}\check@mathfonts
\def\maketag@@@#1{\hbox{\m@th\large\normalfont#1}}%
\begin{multline*}
= 
\left( 
\begin{array}{ccc} 
\frac{r^{-2} + r^2}{2} & \frac{r^{-2} - r^2}{2}\sin \theta \sin \vartheta &  \\
\frac{r^{-2} - r^2}{2}\sin \theta \sin \vartheta & \frac{r^{-2} + r^2}{2}\sin^2 \theta \sin^2 \vartheta +\cos^2 \theta \sin^2 \vartheta +\cos^2 \vartheta &  \ldots \\
\frac{r^{-2} - r^2}{2}\sin \theta \cos \vartheta & \frac{r^{-2} + r^2}{2}\sin^2 \theta \cos \vartheta \sin \vartheta +\cos^2 \theta \sin \vartheta \cos \vartheta - \sin \vartheta \cos \vartheta &   \\
\frac{r^{-2} - r^2}{2} \cos \theta & \frac{r^{-2} + r^2}{2} \sin \theta \cos \theta \sin \vartheta
- \sin \theta \cos \theta \sin \vartheta &   \end{array} \right. \\
\end{multline*}
\begin{multline*}
\,\,\,\,\,\,\,\,\,\,\,\,\,\,\,\,\,\,\,\,\,\,\,\,\,\,\,\,\,\,\,\,\,\,\,\,\,\,\,\,\,\,\,\,\,\,\,\,\,\,\,\,\,\,\,\,\,
\left.
\begin{array}{ccc} 
 & \frac{r^{-2} - r^2}{2}\sin \theta \cos \vartheta & \frac{r^{-2} - r^2}{2} \cos \theta \\
 \ldots & \frac{r^{-2} + r^2}{2}\sin^2 \theta \cos \vartheta \sin \vartheta +\cos^2 \theta \sin \vartheta \cos \vartheta - \sin \vartheta \cos \vartheta & \frac{r^{-2} + r^2}{2} \sin \theta \cos \theta \sin \vartheta
- \sin \theta \cos \theta \sin \vartheta \\
 & \frac{r^{-2} + r^2}{2}\sin^2 \theta \cos^2 \vartheta +\cos^2 \theta \cos^2 \vartheta +\sin^2 \vartheta & \frac{r^{-2} + r^2}{2} \sin \theta \cos \theta \cos \vartheta - \sin \theta \cos \theta \cos \vartheta  \\
 & \frac{r^{-2} + r^2}{2} \sin \theta \cos \theta \cos \vartheta - \sin \theta \cos \theta \cos \vartheta & \frac{r^{-2} + r^2}{2} \cos^2 \theta + \sin^2 \theta  \end{array} \right). 
\end{multline*}\endgroup   

The orthonormal (with respect to the ordinary inner product in $\mathbb{C}^4$) system $\{w_\lambda(p) \}$ of eigenvectors   of the operator matrix $B(p) =  V(\beta(p))^* V(\beta(p))$ in $\mathbb{C}^4$, corresponding to the eigenvalues
$\lambda(p) \in \{1,1, r^{-2}, r^2 \}$  has the form
\begin{multline}\label{eigen-vectors-matrixB}
{w_{{}_1}}^+(p) = \left( \begin{array}{c} 0 \\
                               \frac{p^2}{\sqrt{(p^1)^2 + (p^2)^2}}     \\
                             \frac{-p^1}{\sqrt{(p^1)^2 + (p^2)^2}}   \\
                               0       \end{array}\right), 
{w_{{}_1}}^- (p)= \left( \begin{array}{c} 0 \\
                               \frac{p^1 p^3}{\sqrt{(p^1)^2 + (p^2)^2}r}     \\
                             \frac{p^2 p^3}{\sqrt{(p^1)^2 + (p^2)^2}r}   \\
                              - \frac{\sqrt{(p^1)^2 + (p^2)^2}}{r}       \end{array}\right), \\ 
w_{{}_{r^{-2}}}(p) = \left( \begin{array}{c} \frac{1}{\sqrt{2}} \\
                              \frac{1}{\sqrt{2}}\frac{p^1}{r}     \\
                           \frac{1}{\sqrt{2}}\frac{p^2}{r}    \\
                             \frac{1}{\sqrt{2}}\frac{p^3}{r}       \end{array}\right),
w_{{}_{r^2}}(p) = \left( \begin{array}{c} \frac{1}{\sqrt{2}} \\
                              -\frac{1}{\sqrt{2}}\frac{p^1}{r}     \\
                           -\frac{1}{\sqrt{2}}\frac{p^2}{r}    \\
                             -\frac{1}{\sqrt{2}}\frac{p^3}{r}       \end{array}\right) 
\end{multline} 
There are two transversal eigenvectors ${w_{{}_1}}^+(p), {w_{{}_1}}^-(p)$ to the constant eigenvalue 1,
both of pure space direction and both orthogonal to the space part $(0,\vec{p})$ of the momentum direction of the
corresponding momentum $p = (p^0, \vec{p}) \in \mathscr{O}_{(1,0,0,1)}$. The eigenvector
$w_{{}_{r^{-2}}}(p)$ corresponding to the eigenvalue $r^{-2} = (p^0)^{-2} = (\vec{p} \cdot \vec{p})^{-1}$,
has the same direction as the
corresponding momentum $p = (p^0, \vec{p}) \in \mathscr{O}_{(1,0,0,1)}$, and
$w_{{}_{r^2}}(p)$ has the same direction as $(p^0, -\vec{p})$, where
$p =(p^0, \vec{p}) \in \mathscr{O}_{(1,0,0,1)}$
is the corresponding momentum. Note that the linear combinations $w_{{}_{r^{-2}}}(p) + w_{{}_{r^2}}(p)$
and $w_{{}_{r^{-2}}}(p) - w_{{}_{r^2}}(p)$ give respectively the purely time-like vector
of direction the same as $(p^0, 0)$ and a purely longitudinal
vector of direction the same as $(0, \vec{p})$, where $p = (p^0, \vec{p}) \in \mathscr{O}_{(1,0,0,1)}$ is the corresponding
momentum vector.

The square root of $B(p) = V(\beta(p))^* V(\beta(p))$ is equal
\begin{multline}\label{sqrtB}
\sqrt{V(\beta(p))^* V(\beta(p))} = \sqrt{B(p)} \\
= \left( \begin{array}{cccc} 
\frac{r^{-1} + r}{2} & \frac{r^{-1} - r}{2} \frac{p^1}{r} & \frac{r^{-1} - r}{2} \frac{p^2}{r} & \frac{r^{-1} - r}{2} \frac{p^3}{r} \\
\frac{r^{-1} - r}{2} \frac{p^1}{r} & \frac{r^{-1} + r -2}{2} \frac{p^1}{r} \frac{p^1}{r} +1 &  \frac{r^{-1} + r -2}{2} \frac{p^1}{r} \frac{p^2}{r} & \frac{r^{-1} + r - 2}{2} \frac{p^1}{r} \frac{p^3}{r} \\
\frac{r^{-1} - r}{2} \frac{p^2}{r} & \frac{r^{-1} + r -2}{2} \frac{p^2}{r} \frac{p^1}{r} & \frac{r^{-1} + r -2}{2}  \frac{p^2}{r} \frac{p^2}{r} +1 &  \frac{r^{-1} + r - 2}{2} \frac{p^2}{r} \frac{p^3}{r} \\
\frac{r^{-1} - r}{2} \frac{p^3}{r} & \frac{r^{-1} + r -2}{2} \frac{p^3}{r} \frac{p^1}{r} & \frac{r^{-1} + r -2}{2} \frac{p^3}{r}\frac{p^2}{r} & \frac{r^{-1} + r -2}{2} \frac{p^3}{r} \frac{p^3}{r} + 1\end{array}\right) 
\end{multline}
\begingroup\makeatletter\def\f@size{5}\check@mathfonts
\def\maketag@@@#1{\hbox{\m@th\large\normalfont#1}}%
\begin{multline*}
= \left( \begin{array}{cccc} 
\frac{r^{-1} + r}{2} & \frac{r^{-1} - r}{2} \sin \theta \sin \vartheta & \frac{r^{-1} - r}{2} \sin \theta \cos \vartheta & \frac{r^{-1} - r}{2} \cos \theta \\
\frac{r^{-1} - r}{2} \sin \theta \sin \vartheta & \frac{r^{-1} + r -2}{2} \sin^2 \theta \sin^2 \vartheta  +1 &  \frac{r^{-1} + r -2}{2} \sin^2 \theta \sin \vartheta \cos \vartheta & \frac{r^{-1} + r -2}{2} \sin \theta \cos \theta \sin \vartheta \\
\frac{r^{-1} - r}{2} \sin \theta \cos \vartheta & \frac{r^{-1} + r -2}{2} \sin^2 \theta \cos \vartheta \sin \vartheta & \frac{r^{-1} + r -2}{2}  \sin^2 \theta \cos^2 \vartheta +1 &  \frac{r^{-1} + r -2}{2} \sin \theta \cos \theta \cos \vartheta \\
\frac{r^{-1} - r}{2} \cos \theta & \frac{r^{-1} + r -2}{2} \sin \theta \cos \theta \sin \vartheta & \frac{r^{-1} + r -2}{2} \sin \theta \cos \vartheta \cos \vartheta & \frac{r^{-1} + r -2}{2} \cos^2 \theta +1 \end{array}\right). 
\end{multline*}\endgroup 

By construction $V(\beta(p))$, $V(\beta(p))^* = V(\beta(p))^{T}$ and their inverses are at every
$p \in \mathscr{O}_{(1,0,0,1)}$ Krein unitary, as matrix operators in $(\mathcal{H}_{\bar{p}}, \mathfrak{J}_{\bar{p}})
= (\mathbb{C}^4, \mathfrak{J}_{\bar{p}})$, i.e. they are real Lorentz transformations. It is less trivially, but may be checked directly that for every $p \in \mathscr{O}_{(1,0,0,1)}$ the operator $\sqrt{B(p)}$ is also Krein unitary
in $(\mathbb{C}^4, \mathfrak{J}_{\bar{p}})$. Thus, we have the formulas
\begin{multline}\label{BJBJ=1}
V(\beta(p)) \, \mathfrak{J}_{\bar{p}} \, {V(\beta(p))}^* \, \mathfrak{J}_{\bar{p}} = \bold{1}_4,
\,\,\, \textrm{and} \,\,\,
\mathfrak{J}_{\bar{p}} \, {V(\beta(p))}^* \, \mathfrak{J}_{\bar{p}} \, V(\beta(p)) = \bold{1}_4, \,\,\,\textrm{and} \\
\sqrt{B(p)} \, \mathfrak{J}_{\bar{p}} \, \sqrt{B(p)} \, \mathfrak{J}_{\bar{p}} = \bold{1}_4,
\,\,\, p \in \mathscr{O}_{(1,0,0,1)}.
\end{multline}

Although the properties are simple consequences of definitions (possibly except the last one),
they will be of use in further computations.

\subsection{Definition of the Krein-Hilbert space $(\mathcal{H'}, \mathfrak{J'})$ which is then 
subject to the second quantization functor $\Gamma$}\label{SingleKreinLopRep}

Now to the Hilbert space $\mathcal{H'}$, or more precisely to the Krein space $(\mathcal{H'}, \mathfrak{J'})$
of the representation $WU^{{}_{(1,0,0,1)}{\L}}W^{-1}$ and \emph{eo ipso} of the representation
\[
\big[WU^{{}_{(1,0,0,1)}{\L}}W^{-1}\big]^{*-1},
\]
we apply the Segal's bosonic second quantization
functor $\Gamma$. The Krein space $(\mathcal{H'}, \mathfrak{J'}) = \big( W \mathcal{H}, W\mathfrak{J}W^{-1} \big)$
of the elements $\widetilde{\varphi} = W\widetilde{\psi}$ of the representation
\[
WU^{{}_{(1,0,0,1)}{\L}}W^{-1}
\]
may be identified, via the Fourier transform (\ref{F(varphi)}) with the Hilbert space $\mathcal{H''}$, or more precisely
with the Krein space $(\mathcal{H''}, \mathfrak{J''})$ of positive energy solutions $\phi$
of the wave equation
\begin{equation}\label{nonloc-evol-wave}
g^{\mu \nu}\partial_\mu \partial_\nu \phi = 0,
\end{equation}
as a consequence of the fact that $\widetilde{\varphi} \in \mathcal{H'}$ are concentrated on the cone $\mathscr{O}_{(1,0,0,1)} = \mathscr{O}_{(1,0,0,1)}$. Although it is well known that the equation (\ref{nonloc-evol-wave}) only apparently gives a local law for dynamics in terms of a local equation. Indeed, because only the positive energy solutions\footnote{In the construction of the positive energy field via the second quantization functor applied to the space $(\mathcal{H'}, \mathfrak{J'})$. In the construction of the negative energy field the roles of positive and negative energy is interchanged.} are admitted the quantities $\varphi$ and $\partial_t\varphi$ are not independent
on a fixed time surface.
The differentiation $\partial_t$ in momentum space is equal to the operator of multiplication by $-i \sqrt{\vec{p}\cdot \vec{p}}$, which in position picture at fixed time corresponds to a convolution with the nonlocal integral kernel\footnote{Already the definition of the kernel necessities a special care, and may be defined in the distributional sense }
\[
K(\vec{x} - \vec{x'}) = -i (2 \pi)^{-3/2} \int \limits_{\mathbb{R}^3} \,
\sqrt{\vec{p}\cdot \vec{p}} \,\, e^{i \vec{p} \cdot (\vec{x} - \vec{x'})} \ud^3 p,
\]
exactly as for the spineless massive particles (compare e.g. \cite{Haag}, I. 3.3.).

Unfortunately the inner product $(\varphi, \varphi') = (\widetilde{\varphi}, \widetilde{\varphi'})$,
when expressed in terms of $\varphi$ and $\varphi'$, involves unpleasant kernel. This is however not so important
as the inner product plays the (important but) only technical role of controlling all the analytical subtleties.
It is the Krein inner product $(\varphi, \mathfrak{J''}\varphi') = (\widetilde{\varphi}, \mathfrak{J'}
\widetilde{\varphi'})$ which serves to compute probabilities on the subspace of physical states
on which it is positive definite, and it is nice to have the relatively simple and explicit formula
(\ref{krein-prod-1-photon}) for the Krein-inner product in the Krein space $(\mathcal{H''}, \mathfrak{J''})$
expressed in terms of position wave functions $\varphi, \varphi'$.

It should be stressed that already the elements $\widetilde{\varphi}$ of the single particle space of the
{\L}opusza\'nski representation (and its conjugation) in the momentum picture do not in general fulfil
the condition $p^\mu \widetilde{\varphi}_\mu = 0$, so that in general their Fourier transforms
$\varphi$ do not preserve the Lorentz condition $\partial^\mu \varphi_\mu = 0$. This corresponds to the well known
fact that the Lorentz condition cannot be preserved as an operator equation. It can be preserved in the sense of
the Krein-product average on a subspace of Lorentz states which arise from the closed subspace
$\mathcal{H}_{\textrm{tr}}$ of the so-called transversal states together with all their images under the action
of the {\L}opusza\'nski representation and its conjugation. We are now going to define the closed subspace
$\mathcal{H}_{\textrm{tr}}$.

Note that the operator $B$ of multiplication by the positive self adjoint matrix (\ref{operatorB})
is self adjoint in the Hilbert space $\oplus L^2(\mathbb{R}^3) = L^2(\mathbb{R}^3, \mathbb{C}^4)$
with respect to the ordinary invariant Lebesgue measure
$\ud^3 \boldsymbol{\p}$ on $\mathbb{R}^3$ (the direct sum $\oplus$ is over the four components of the
function $\widetilde{\varphi}$), and that the Hilbert space inner product in the single-photon
state space $\mathcal{H}'$
is equal $(\cdot , \cdot) = (\cdot, B \cdot)_{{}_{L^2(\mathbb{R}^3, \mathbb{C}^4)}}$. The unitary operator
which has the direct integral decomposition
\begin{equation}\label{diagonalizingU}
\int \limits_{\mathscr{O}_{\bar{p}}} U_{\boldsymbol{\p}} \, \ud^3 \boldsymbol{\p}
\end{equation}
(in the integral we use the spatial momentum coordinates $\boldsymbol{\p}$ on
the cone $\mathscr{O}_{\bar{p}}$, and the integral may be treated as an integral on $\mathbb{R}^3$)
with each component $U_{\boldsymbol{\p}}$ being a unitary matrix operator in $\mathbb{C}^4$ transforming the standard basis
in $\mathbb{C}^4$ into the basis\footnote{Here $p \in \mathscr{O}_{\bar{p}}$ is regarded as the standard function of
spatial momentum coordinates $\boldsymbol{\p}$.} ${w_{{}_1}}^+(p), {w_{{}_1}}^-(p), w_{{}_{r^{-2}}}(p), w_{{}_{r^2}}(p)$
of eigenvectors of the Hermitian matrix $B(p)$. It is easily seen that (\ref{diagonalizingU})
transforms the operator $B$, regarded as an operator in $L^2(\mathbb{R}^3, \mathbb{C}^4)$, into the orthogonal direct
sum of four multiplication operators on the measure space. Two first components of this direct sum are the multiplication
operators by the constant function equal to unity everywhere, the next direct summand is the operator of multiplication
by $\frac{1}{2}r^{-3}$ and the third orthogonal direct summand is the multiplication operator by $\frac{1}{2}r$
(recall that $r$ is the following function: $r(\boldsymbol{\p}) = | \boldsymbol{\p} | =
\sqrt{\boldsymbol{\p} \cdot \boldsymbol{\p}}$). Therefore,
the operator $B$ treated as an operator in $\mathcal{H}'$ is likewise unitarily equivalent to a
direct sum of multiplication operators and thus self-adjoint. And similarly $B$ as the operator in
$\mathcal{H}'$ has a pure point
spectrum $\{1\}$ consisting of just one element 1, and a continuous spectrum equal $\mathbb{R}_+$.
Indeed, any element $\widetilde{\varphi} \in \mathcal{H}'$ may be uniquely written as the
following linear combination
\begin{equation}\label{eigenspaceB}
\widetilde{\varphi}(p) = {w_{{}_1}}^+(p) \, f_+(p) \, + \,
{w_{{}_1}}^-(p) \, f_-(p) \,+ \, w_{{}_{r^{-2}}}(p)\, f_{0+}(p)
\, + \, w_{{}_{r^2}}(p) \, f_{0-}(p)
\end{equation}
where $f_+, f_-, f_{0+}, f_{0-}$
are scalar functions on the light cone $\mathscr{O}_{\bar{p}}$.
The first two functions $f_+, f_-$
run over the set of all square integrable functions on the light cone
$\mathscr{O}_{\bar{p}}$ with respect to the invariant measure $\ud \mu |_{{}_{\mathscr{O}_{\bar{p}}}} =
\frac{\ud^3 \boldsymbol{\p}}{| \boldsymbol{\p} |}$.
The functions $f_{0+}$ range over all functions on $\mathscr{O}_{\bar{p}}$ square integrable with respect to the measure
$\frac{\ud^3 \boldsymbol{\p}}{| \boldsymbol{\p} |^3}$, and finally
$f_{0-}$ range over all square integrable functions with respect to the measure\footnote{The measures
$\frac{\ud^3 \boldsymbol{\p}}{| \boldsymbol{\p} |^3}$ and
$| \boldsymbol{\p} | \ud^3 \boldsymbol{\p}$ are of course not invariant on the cone, but note that the ordinary
Hilbert space inner product which they define on $\mathcal{H}'$ \emph{is not} the inner product preserved by
the {\L}opusz\'nski representation. The representation preserves the Krein-inner product.}
$| \boldsymbol{\p} | \ud^3 \boldsymbol{\p}$.

Note that the four elements
\[
{w_{{}_1}}^+ \, f_+, \,\,
{w_{{}_1}}^- \, f_-, \,\, w_{{}_{r^{-2}}} \, f_{0+},
\,\, w_{{}_{r^2}} \, f_{0-}
\]
of $\mathcal{H}'$ on the right-hand side of (\ref{eigenspaceB}) define orthogonal decomposition
$\mathcal{H}'$ into closed invariant subspaces of the self=adjoint operator $B$, treated as an operator in
$\mathcal{H}'$. Moreover, by the formula for the Krein inner-product in $\mathcal{H}', \mathfrak{J}'$
(compare (\ref{Kr-inn-Lop-1-space}) and (\ref{Kr-inn-Lop-1-space'})) the closed subspace spanned by the
elements
\[
w_{{}_{r^{-2}}} \, f_{0+},
\,\, w_{{}_{r^2}} \, f_{0-},
\]
the closed subspace spanned by ${w_{{}_1}}^+ \, f_+, $ and te closed subspace spanned by
${w_{{}_1}}^- \, f_-$ are also mutually Krein-orthogonal.

Let $\mathcal{H}_{\textrm{tr}}$ be the closed subspace of the Hilbert space $\mathcal{H}'$ corresponding
to the pure point spectrum $\{1\}$ of the operator $B$ in $\mathcal{H}'$. Then
$\mathcal{H}_{\textrm{tr}}$ is spanned by the elements
\[
{w_{{}_1}}^+ \, f_+ \, + \,
{w_{{}_1}}^- \,f_-
\]
and the inner product of any two members of $\mathcal{H}_{\textrm{tr}}$ is equal to the Krein-inner product
which easily follows from the construction.
Thus, by construction for every element of $\mathcal{H}'$ existence and uniqueness of the projection on
$\mathcal{H}_{\textrm{tr}}$ with respect to the Krein-inner product $(\cdot , \mathfrak{J}' \cdot)$
follows\footnote{Recall that for a general subspace
in a Krein space neither the existence, nor the uniqueness of the projection of a vector on the subspace
with respect to the Krein-inner-product is guaranteed. Thus, its existence and uniqueness
as well as the existence of the corresponding Krein-self-adjoint idempotent need to be proved.}.

It is important to understand that the properties of $\mathcal{H}_{\textrm{tr}}$ are of fundamental importance
for the construction of the physical space of transversal states and contrary to ordinary 
Hilbert space the stated above properties of the subspace $\mathcal{H}_{\textrm{tr}}$ are by far
not shared by a general (even closed) subspaces of a Krein space.

Because the inner product $(\cdot, \cdot)$ of $\mathcal{H}'$ is just equal to the positive inner product which
corresponds through the fundamental symmetry $\mathfrak{J}'$ to the Krein-inner product
$(\cdot, \mathfrak{J}' \cdot)$ (in the notation
of \cite{Bog} $(\cdot, \cdot)_{\mathfrak{J}'} = (\cdot, \mathfrak{J}' \mathfrak{J}' \cdot)
= (\cdot, \cdot)$) it follows that the subspace $\mathcal{H}_{\textrm{tr}}$ is uniformly
positive in the sense of \cite{Bog}, V.5. Being a closed subspace $\mathcal{H}_{\textrm{tr}}$
is regular in the sense of \cite{Bog}, therefore by \cite{Bog}, Ch. V. the subspace
$\mathcal{H}_{\textrm{tr}}$ is orthocomplemented with respect to the Krein-inner-product
$(\cdot, \mathfrak{J}' \cdot)$ and admits unique projection on $\mathcal{H}_{\textrm{tr}}$
with respect to the Krein-inner-product, which is bounded (boundedness, closedness, continuity always refer to the
ordinary Hilbert space inner product of $\mathcal{H}'$ or in general to the corresponding Hilbert space).
Thus, by \cite{Bog} there exist bounded Krein-self-adjoint idempotent $P$ (i.e. $P^2 = P$,
$P^\dagger = P$, where $P^\dagger = \mathfrak{J}' P^* \mathfrak{J}'$ with the ordinary adjoint $P^*$
in the Hilbert space $\mathcal{H}'$) with range $P\mathcal{H}' = \mathcal{H}_{\textrm{tr}}$. 

Now we define the elements of $\mathcal{H}_{\textrm{tr}}$ as the physical transversal states.
But it turns out that in order to account for the Lorentz covariance and the gauge freedom
we cannot stay within $\mathcal{H}_{\textrm{tr}}$. The {\L}opusza\'nski
representation and the representation conjugate to it, whenever applied to a vector $\widetilde{\varphi}$ of 
$\mathcal{H}_{\textrm{tr}}$, in general transform it into a vector $\widetilde{\varphi}''$
which does not lie in $\mathcal{H}_{\textrm{tr}}$. But the amazing property of these
representations is that always 
\begin{equation}\label{propertyLop}
\widetilde{\varphi}'' = \widetilde{\varphi}' + \widetilde{\varphi}_0
\end{equation}
for a unique vector $\widetilde{\varphi}' \in \mathcal{H}_{\textrm{tr}}$ and a unique 
$\widetilde{\varphi}_0$ whose Krein-inner-product norm vanishes:
\[
(\widetilde{\varphi}_0, \mathfrak{J}'\widetilde{\varphi}_0) = 0, \,\,\, 
\] 
where $(\cdot, \cdot)$ is the inner product in $\mathcal{H}'$,
and which is Krein-orthogonal to $\mathcal{H}_{\textrm{tr}}$: 
\[
(\widetilde{\varphi}_0, \mathfrak{J}' \widetilde{\varphi}''') = 0, \,\,\,
\widetilde{\varphi}''' \in \mathcal{H}_{\textrm{tr}}
\] 
(both $\widetilde{\varphi}'$ and $\widetilde{\varphi}_0$ in general depend on $\widetilde{\varphi}$
and on the applied transformation). Because the Krein-norm of 
$\widetilde{\varphi}'' = \widetilde{\varphi}' + \widetilde{\varphi}_0$
is equal to the Krein norm of $\widetilde{\varphi}'$, and the Krein inner product on 
$\mathcal{H}_{\textrm{tr}}$ coincides with the ordinary inner product on $\mathcal{H}'$,
and the representations are Krein-isometric, then it follows that the transformation
$\widetilde{\varphi} \mapsto \widetilde{\varphi}'$ which they generate on $\mathcal{H}_{\textrm{tr}}$
is isometric with respect to the ordinary Hilbert space-inner product induced on 
$\mathcal{H}_{\textrm{tr}}$ by the Krein inner product.

Moreover, by construction of the dense core domain $\mathfrak{D}$ of the induced representation
(compare Subsection \ref{def_ind_krein} of Section \ref{PartIIMackey} or \cite{wawrzycki-mackey}, Sect. 2), 
to which the {\L}opusza\'nski representation is equivalent,
it follows that $\mathfrak{D}$ is likewise dense in the subspace $\mathcal{H}_{\textrm{tr}}$.
It is easily seen because in our case $\mathfrak{D}$ consists of all those $\widetilde{\varphi} \in \mathcal{H}'$
which are continuous functions on the cone with compact support, and all of them when projected on
$\mathcal{H}_{\textrm{tr}}$ include all the functions of the form
\[
{w_{{}_1}}^+ \, f_+ \, + \,
{w_{{}_1}}^- \, f_-
\]
with $f_+, f_-$ continuous of compact support, which are obviously dense in $\mathcal{H}_{\textrm{tr}}$.
Therefore, the representations generated by the action modulo nonphysical states by the
Krein representation (and its conjugation) on the transversal subspace $\mathcal{H}_{\textrm{tr}}$
is not only Hilbert-space isometric but can be uniquely extended to an ordinary unitary
representation on $\mathcal{H}_{\textrm{tr}}$. This is really amazing in view of the quite singular character
of the {\L}opusza\'nski representation (and its conjugation) for which representor of any boost
is unbounded (with respect to the Hilbert space norm of $\mathcal{H}'$).
We have shown in \cite{wawrzycki-photon} that the {\L}opusza\'nski representation
$WU^{{}_{(1,0,0,1)}{\L}}W^{-1}$ and its conjugation
$\mathfrak{J}'WU^{{}_{(1,0,0,1)}{\L}}W^{-1}\mathfrak{J}'$ does have the property (\ref{propertyLop}).
In fact during the proof in \cite{wawrzycki-photon} we have given explicit construction
of the unitary representation
\[
\mathbb{U}(\alpha)\left( \begin{array}{c} f_{+} \\ 
                 f_{-} \end{array}\right)(p)
= \left( \begin{array}{cc} \cos \Theta(\alpha, p) & \sin \Theta(\alpha, p) \\ 
                          -\sin \Theta(\alpha, p) & \cos \Theta(\alpha, p)  \end{array}\right)
\left( \begin{array}{c} f_{+}(\Lambda(\alpha) p) \\ 
                 f_{-}(\Lambda(\alpha) p) \end{array}\right),
\]
\[
\mathbb{T}(a)\left( \begin{array}{c} f_{+} \\ 
                 f_{-} \end{array}\right)(p)
= e^{ia\cdot p}
\left( \begin{array}{c} f_{+}(p) \\ 
                 f_{-}(p) \end{array}\right).
\]
generated on the physical subspace $\mathcal{H}_{\textrm{tr}}$. 
Recall that $(f_+, f_-)$ compose the Hilbert space $\mathcal{H}_{\textrm{tr}}$ of all pairs of 
functions on the cone
which are square integrable with respect to the invariant measure on the cone. Applying to this Hilbert space
and to the unitary representation $\mathbb{U},\mathbb{T}$ in $\mathcal{H}_{\textrm{tr}}$ the unitary transformation $\mathcal{U}: \mathcal{H}_{\textrm{tr}} \rightarrow \mathcal{H}_{\textrm{tr}}$ defined by 
\[
\mathcal{U}\left( \begin{array}{c} f_{1} \\ 
                 f_{-1} \end{array}\right)(p)
= \left( \begin{array}{cc} \frac{-i}{\sqrt{2}} & \frac{-i}{\sqrt{2}} \\ 
                          \frac{1}{\sqrt{2}} & \frac{-1}{\sqrt{2}}  \end{array}\right)
\left( \begin{array}{c} f_{+}(p) \\ 
                 f_{-}(p) \end{array}\right),
\]
we obtain
\begin{equation}\label{alpha-helicity+1-1}
\mathcal{U}^{-1}\mathbb{U}(\alpha) \mathcal{U} \left( \begin{array}{c} f_{1} \\ 
                 f_{-1} \end{array}\right)(p)
= \left( \begin{array}{cc}  e^{i \Theta(\alpha, p)} & 0 \\ 
                          0 &  e^{-i \Theta(\alpha, p)} \end{array}\right)
\left( \begin{array}{c} f_{1}(\Lambda(\alpha) p) \\ 
                 f_{-1}(\Lambda(\alpha) p) \end{array}\right),
\end{equation}
\begin{equation}\label{a-helicity+1-1}
\mathcal{U}^{-1} \mathbb{T}(a)\mathcal{U}\left( \begin{array}{c} f_{1} \\ 
                 f_{-1} \end{array}\right)(p)
= e^{ia\cdot p}
\left( \begin{array}{c} f_{1}(p) \\ 
                 f_{-1}(p) \end{array}\right).
\end{equation} 

Concerning the explicit formula for the phase
$\Theta(\alpha, p)$ we give its form for the
several particular rotations and Lorentz transformations, which are sufficient for the reconstruction
of the general formula for $\Theta$. To this end let $\alpha_{\mu\nu}$ denote the element of $SL(2, \mathbb{C})$
which corresponds through the natural homomorphism $\Lambda$ to the rotation
$\Lambda(\alpha_{\mu\nu})$ in the $\mu-\nu$ plane. In particular $\Lambda(\alpha_{03})$ denotes
the Lorentz rotation in the $0-3$ plane and $\Lambda(\alpha_{23})$ stands for the ordinary
spatial rotation in the $2-3$ plane, i.e. around the first axis. We agree to use
$\lambda$ to denote the hyperbolic angle of the Lorentz rotations $\Lambda(\alpha_{0k})$,
and let $\theta$ be the angle of the spatial rotations $\Lambda(\alpha_{ik})$.  
Then in particular we have    
\begin{multline*}
\Lambda({\alpha_{03}}^{-1}) w_{1}^{+}(\Lambda(\alpha_{03})p) = w_{1}^{+}(p), \\
\Lambda({\alpha_{03}}^{-1}) w_{1}^{-}(\Lambda(\alpha_{03})p) =  w_{1}^{-}(p) +  
 \frac{\sinh \lambda \sqrt{(p^1)^2 + (p^2)^2}}{p^0 \cosh \lambda  + p^3 \sinh \lambda} \, w_{{}_{r^{-2}}}(p);
\end{multline*}
\[
\begin{split}
\Lambda({\alpha_{12}}^{-1}) w_{1}^{+}(\Lambda(\alpha_{12})p) = w_{1}^{+}(p), \\
\Lambda({\alpha_{12}}^{-1}) w_{1}^{-}(\Lambda(\alpha_{12})p) =  w_{1}^{-}(p);
\end{split}
\]
\begin{multline*}
\Lambda({\alpha_{23}}^{-1}) w_{1}^{+}(\Lambda(\alpha_{23})p) = \\
\frac{1}{\sqrt{(p^1)^2 + (p^2 \cos \theta + p^3 \sin \theta)^2}} 
\Big( \frac{p^2 p^3 \sin \theta}{\sqrt{(p^1)^2 + (p^2)^2}} + \sqrt{(p^1)^2 + (p^2)^2} \cos \theta \Big) w_{1}^{+}(p) \\
+ \frac{r p^1 \sin \theta}{\sqrt{(p^1)^2 + (p^2 \cos \theta + p^3 \sin \theta)^2} \, \sqrt{(p^1)^2 + (p^2)^2}}
 w_{1}^{-}(p), 
\end{multline*}

\begin{multline*}
\Lambda({\alpha_{23}}^{-1}) w_{1}^{-}(\Lambda(\alpha_{23})p) = \\ 
\frac{-r p^1 \sin \theta}{\sqrt{(p^1)^2 + (p^2 \cos \theta + p^3 \sin \theta)^2} \, \sqrt{(p^1)^2 + (p^2)^2}}
 w_{1}^{+}(p) \\
+ \frac{1}{\sqrt{(p^1)^2 + (p^2 \cos \theta + p^3 \sin \theta)^2}} 
\Big( \frac{p^2 p^3 \sin \theta}{\sqrt{(p^1)^2 + (p^2)^2}} + \sqrt{(p^1)^2 + (p^2)^2} \cos \theta \Big) w_{1}^{-}(p).
\end{multline*}
The formulas for $\Lambda({\alpha_{13}}^{-1}) w_{1}^{+}(\Lambda(\alpha_{13})p)$ and 
$\Lambda({\alpha_{13}}^{-1}) w_{1}^{-}(\Lambda(\alpha_{13})p)$ are obtained 
by interchanig $p^1$ and $p^2$ with each other in the last two formulas respectively.

From the above formulas it follows for example that 
\begin{multline*}
\,\,\,\,\,\,\,\,\,\,\,\,\,\,\,\,\,\,\,\,\,\,\,\,\,\,\,\,\,\,\,\,\,\,\,\,\,\,\,\,\,\,\,\,\,\,\,\,\,\,\,
\sin \Theta(\alpha_{03},p) = 0, \,\,\, \cos \Theta(\alpha_{03},p) = 1, \\
\sin \Theta(\alpha_{12},p) = 0, \,\,\, \cos \Theta(\alpha_{12},p) = 1, \\
\sin \Theta(\alpha_{23},p) = 
\frac{r p^1 \sin \theta}{\sqrt{(p^1)^2 + (p^2 \cos \theta + p^3 \sin \theta)^2} \, \sqrt{(p^1)^2 + (p^2)^2}}; \\
 \cos \Theta(\alpha_{23},p) 
\,\,\,\,\,\,\,\,\,\,\,\,\,\,\,\,\,\,\,\,\,\,\,\,\,\,\,\,\,\,\,\,\,\,\,\,\,\,\,\,\,\,\,\,\,\,\,\,\,\,\,
\,\,\,\,\,\,\,\,\,\,\,\,\,\,\,\,\,\,\,\,\,\,\,\,\,\,\,\,\,\,\,\,\,\,\,\,\,\,\,\,\,\,\,\,\,\,\,\,\,\,\,\,\,\,
\,\,\,\,\,\,\,\,\,\,\,\,\,\,\,\,\,\,\,\,\,\,\,\,\,\,\,\,\,\,\,\,\,\,\,\,\,\,\,\,\,\,\,\,\,\,\,\,\,\,\,\,\,\,
\,\,\,\,\,\,\,\,\,\,\,\,\,\,\,\,\,\,\,\,\,\,\,\,\,\,\,\,\,\,
\\
=
\frac{1}{\sqrt{(p^1)^2 + (p^2 \cos \theta + p^3 \sin \theta)^2}} 
\Big( \frac{p^2 p^3 \sin \theta}{\sqrt{(p^1)^2 + (p^2)^2}} + \sqrt{(p^1)^2 + (p^2)^2} \cos \theta \Big).
\end{multline*} 
The formulas for $\sin \Theta(\alpha_{13},p)$ and 
$\cos \Theta(\alpha_{13},p)$ are obtained 
by interchanig $p^1$ and $p^2$ with each other in the last two formulas respectively.

For derivation of the above formulas for the phase
$\Theta(\alpha, p)$, see \cite{wawrzycki-photon}. In the papers \cite{wawrzycki-photon},
\cite{wawrzycki2016bialynicki} we have compared the physical single particle space $\mathcal{H}_{\textrm{tr}}$
with the above unitary representation acting upon it 
with the single particle photon space used by other authors, for example
with the single particle photon space used in the works of Bialynicki-Birula \cite{bialynicki-2},
\cite{Bialynicki} and have shown there that they are identical.

\subsection{Application of the second quantization functor $\Gamma$ to the Krein-Hilbert space
$(\mathcal{H'}, \mathfrak{J'})$}\label{Gamma-applied-to-(H',J')}

Now we apply the second quantization functor $\Gamma$ of Segal to the one particle Krein space
$(\mathcal{H'}, \mathfrak{J'})$,  and \emph{eo ipso} to the Krein space $(\mathcal{H''}, \mathfrak{J''})$,
or which amounts to the same thing,  to the ordinary Hilbert space $\mathcal{H'}$ equipped with the fundamental 
symmetry $\mathfrak{J'}$. We adopt here the convention (which is customary in the physical literature) that
the ordinary Hilbert space adjoint of the operator $a$ in the resulting Fock space is written as $a^+$.

Thus, we obtain the Fock space
\[ 
\Gamma(\mathcal{H'}) = \mathbb{C} \oplus \mathcal{H'} \oplus \big[\mathcal{H'}\big]^{\otimes 2}_{S}
\oplus \big[\mathcal{H'}\big]^{\otimes 3}_{S} \oplus \ldots 
\]
as the direct sum of symmetrized n-fold tensor products $\big[\mathcal{H'}\big]^{\otimes n}_{S}$ 
of $\mathcal{H'}$ and the Hilbert space $\mathbb{C}$
generated by the vacuum $\Omega$. We will use interchangeably the notation 
$\big[\mathcal{H'}\big]^{\widehat{\otimes} n}$ for the symmetrized $n$-fold tensor product
$\big[\mathcal{H'}\big]^{\otimes n}_{S}$ 
of $\mathcal{H'}$.

In particular introducing the projection operator $P_+$ onto the symmetric tensors
in the $n$-fold tensor product $\big[\mathcal{H'}\big]^{\otimes 3}$ we have 
\begin{multline*}
\underset{1}{\widetilde{\varphi}} \, \widehat{\otimes} \, \underset{2}{\widetilde{\varphi}} \, \widehat{\otimes} \, \ldots 
\, \widehat{\otimes} \, \underset{n}{\widetilde{\varphi}} =
\big(\underset{1}{\widetilde{\varphi}} \otimes \underset{2}{\widetilde{\varphi}} \otimes  \ldots \otimes \underset{n}{\widetilde{\varphi}} \big)_S \\
= P_+ \big(\underset{1}{\widetilde{\varphi}} \otimes \underset{2}{\widetilde{\varphi}} \otimes  \ldots \otimes \underset{n}{\widetilde{\varphi}} \big) = (n!)^{-1} \sum \limits_{\pi}
\underset{\pi(1)}{\widetilde{\varphi}} \otimes \underset{\pi(2)}{\widetilde{\varphi}} \otimes  \ldots \otimes 
\underset{\pi(n)}{\widetilde{\varphi}},
\end{multline*}
where the sum is over all permutations $\pi$ of the numbers $1, 2, \ldots n$. Every element $\Phi \in \Gamma(\mathcal{H'})$
may be represented as the sum 
\begin{equation}\label{GeneralPhi}
\Phi = \sum \limits_{n \geq 0} \Phi_n  
\end{equation}
over all $n= 0, 1, 2, \ldots $ of the orthogonal components $\Phi_n \in \big[\mathcal{H'}\big]^{\otimes n}_{S}$
-- $n$-particle states, with 
\begin{equation}\label{NormGeneralPhi}
\| \Phi\|^2 = \sum \limits_{n \geq 0} \| \Phi_n \|^2 < +\infty. 
\end{equation}
The domain $\Dom N$ of the number operator $N$ is defined as the linear set of all those
$\Phi  = \sum \Phi_n$ for which $\sum_{n\geq 0} n^2 \|\Phi_n \|^2 < + \infty$, 
and on the domain $\Dom N$, $N$ is defined as
\[
N \Phi = \sum \limits_{n \geq 0} n \Phi_n.
\]
Thus we see that $N$ is an operator of multiplication by a measurable function on a direct
sum measure space, i.e. it is self adjoint operator.

For each $\widetilde{\varphi} \in \mathcal{H'}$ we define operators $a'(\widetilde{\varphi})$ and 
$a'^+ (\widetilde{\varphi})$ by setting
\[
1) \,\,\,
a' (\widetilde{\varphi})\Phi_0 = 0, \,\,\, 
a'^+ (\widetilde{\varphi})\Phi_0 = \widetilde{\varphi}, 
\]
\[
2) \,\,\,
a' (\widetilde{\varphi}) \big(\underset{1}{\widetilde{\varphi}} \otimes \underset{2}{\widetilde{\varphi}} \otimes  \ldots \otimes \underset{n}{\widetilde{\varphi}} \big)_S = n^{1/2} \, (n!)^{-1} \sum \limits_{\pi}
(\widetilde{\varphi},\underset{\pi(1)}{\widetilde{\varphi}}) \underset{\pi(2)}{\widetilde{\varphi}} \otimes  
\underset{\pi(3)}{\widetilde{\varphi}} \otimes \ldots \otimes 
\underset{\pi(n)}{\widetilde{\varphi}}, 
\]
\[
3) \,\,\,
a'^+ (\widetilde{\varphi}) \big(\underset{1}{\widetilde{\varphi}} \otimes \underset{2}{\widetilde{\varphi}} \otimes  \ldots \otimes \underset{n}{\widetilde{\varphi}} \big)_S = (n+1)^{1/2}
\big( \widetilde{\varphi} \otimes \underset{1}{\widetilde{\varphi}} \otimes \underset{2}{\widetilde{\varphi}} \otimes \ldots \otimes \underset{n}{\widetilde{\varphi}}\big)_S.
\]
Or put otherwise, using the symmetrized $1$-contraction $\widehat{\otimes}_{1}$ (one can just put the right-hand side
of the formula 2) as the definition of $1$-contraction $\widehat{\otimes}_{1}$)
we have  
\[
1) \,\,\,
a' (\widetilde{\varphi})\Phi^{(0)} = 0, \,\,\, 
a'^+ (\widetilde{\varphi})\Phi^{(0)} = \widetilde{\varphi}, 
\]
\[
2) \,\,\,
a' (\widetilde{\varphi}) \,\, \underset{1}{\widetilde{\varphi}} \, \widehat{\otimes} \, \underset{2}{\widetilde{\varphi}} \, \widehat{\otimes}  \, \ldots \, \widehat{\otimes} \, \underset{n}{\widetilde{\varphi}} 
= n^{1/2} \,\,  \overline{\widetilde{\varphi}} \, \widehat{\otimes}_{1} \,
\underset{1}{\widetilde{\varphi}} \, \widehat{\otimes} \, \underset{2}{\widetilde{\varphi}} \, \widehat{\otimes} \,
 \ldots \, \widehat{\otimes} \, \underset{n}{\widetilde{\varphi}}, 
\]
\[
3) \,\,\,
a'^+ (\widetilde{\varphi}) \,\, \underset{1}{\widetilde{\varphi}} \, \widehat{\otimes} \, \underset{2}{\widetilde{\varphi}} \, \widehat{\otimes} \, \ldots \, \widehat{\otimes} \, \underset{n}{\widetilde{\varphi}} 
 = (n+1)^{1/2}
\widetilde{\varphi} \, \widehat{\otimes} \, \underset{1}{\widetilde{\varphi}} \, \widehat{\otimes} \,
\underset{2}{\widetilde{\varphi}} \, \widehat{\otimes}  \, \ldots \, \widehat{\otimes} \, \underset{n}{\widetilde{\varphi}}.
\]

It follows that 
\[
\| a'(\widetilde{\varphi}) \Phi^{(n)} \| \leq n^{1/2} \| \widetilde{\varphi} \| \|\Phi^{(n)} \|, \,\,\,
\| a'^+ (\widetilde{\varphi}) \Phi^{(n)} \| \leq (n+1)^{1/2}( \| \widetilde{\varphi} \| \| \Phi^{(n)} \|,
\]
so that $a(\widetilde{\varphi})$ and 
$a'^+ (\widetilde{\varphi})$ have extensions to the common domain $\Dom \, (N)^{1/2}$ of the selfadjoint 
operator $N^{1/2}$ and for all $\Phi, \Psi \in \Dom \, (N)^{1/2}$
\[
(a'^+ (\widetilde{\varphi}) \Phi, \Psi) = (\Phi, a'(\widetilde{\varphi}) \Psi),
\]
so that $a'(\widetilde{\varphi})$ possesses a densely defined adjoint operator $a'(\widetilde{\varphi})^+$
which is equal to an extension of $a'^+ (\widetilde{\varphi})$, and the operators 
$a'(\widetilde{\varphi})$ and $a'^+ (\widetilde{\varphi})$ on $\Dom \, (N)^{1/2}$ are preclosed. 
We thus obtain
the two canonical linear maps $\mathcal{H'} \ni \widetilde{\varphi} \mapsto a(\overline{\widetilde{\varphi}})$
and $\mathcal{H'} \ni \widetilde{\varphi} \mapsto a^+(\widetilde{\varphi})$ -- the annihilation and creation operator
valued maps, such that for each $\widetilde{\varphi} \in \mathcal{H'}$, $a'(\widetilde{\varphi})$ is densely defined closable operator, 
i.e. with densely defined adjoint $a'(\widetilde{\varphi})^+$ and with the closure of the operator 
\[
a'(\widetilde{\varphi}) + a'(\widetilde{\varphi})^+
\]  
being self adjoint. Denoting the commutator by $[\cdot, \cdot]$ we have for any $\Phi$ in the
dense domain of the self adjoint\footnote{The square root of the particle number operator $N$.} operator $N^{1/2}$, common for the domain of all $a'(\widetilde{\varphi}), a'^+(\widetilde{\varphi})$, and all $\widetilde{\varphi}, \widetilde{\varphi'}  
\in \mathcal{H'}$:
\begin{multline*}
[a'(\widetilde{\varphi}), a'(\widetilde{\varphi'})^+] \Phi = (\widetilde{\varphi}, \widetilde{\varphi'}) \Phi \\
= \Big[ \int \limits_{\mathscr{O}_{\bar{p}}} 
\Big(  \widetilde{\varphi}(p), 
B(p) \widetilde{\varphi}'(p)  \Big)_{\mathbb{C}^4}
\, \ud \mu |_{{}_{\mathscr{O}_{\bar{p}}}} (p)\Big] \Phi \\
= \Big[ \int \limits_{\mathbb{R}^3} 
\Big(  \widetilde{\varphi}(p), 
B(p) \widetilde{\varphi}'(p)  \Big)_{\mathbb{C}^4}
\, \frac{\ud^3 p}{2(\vec{p}\cdot \vec{p})^{{}^{1/2}}} \Big] \Phi, 
\end{multline*}
which we write simply as\footnote{In the remaining part of this section everywhere in the integral 
$\int_{\mathscr{O}_{\bar{p}}} \ldots \frac{dp}{2p^0(p)} = \int_{\mathbb{R}^3} \ldots \frac{\ud^3 p}{2(\vec{p}\cdot \vec{p})^{{}^{1/2}}}$ 
we will write the last integral simply as 
$\int_{\mathbb{R}^3} \ldots \frac{\ud^3 p}{2p^0}$ and understand $p^0$ as the function 
$p^0(\vec{p}) = (\vec{p}\cdot \vec{p})^{{}^{1/2}}$; similarly for any function of $p^0$ under the integral
sign over the orbit $\mathscr{O}_{\bar{p}}$.} 
\[
[a'(\widetilde{\varphi}), a'(\widetilde{\varphi'})^+] = (\widetilde{\varphi}, \widetilde{\varphi'})
= \int \limits_{\mathbb{R}^3}
\Big(  \widetilde{\varphi}(p), 
B(p) \widetilde{\varphi}'(p)  \Big)_{\mathbb{C}^4}
\, \frac{\ud^3 p}{2(\vec{p}\cdot \vec{p})^{{}^{1/2}}}.
\]

Into the Fock space $\Gamma(\mathcal{H'})$ we introduce the fundamental symmetry operator
\[
\eta = \Gamma(\mathfrak{J'}) = 1_\mathbb{C} \oplus \mathfrak{J'} \oplus \big[\mathfrak{J'} \otimes \mathfrak{J'}\big]_S
\oplus \big[\mathfrak{J'} \otimes \mathfrak{J'} \otimes \mathfrak{J'} \big]_S \oplus \ldots 
\]
and the representation
\begin{multline*}
\Gamma\Big(\big[WU^{{}_{(1,0,0,1)}{\L}}W^{-1}\big]^{*-1}\Big) \\ =
1_\mathbb{C} \oplus \big[WU^{{}_{(1,0,0,1)}{\L}}W^{-1}\big]^{*-1} \oplus  
\bigg[\big[WU^{{}_{(1,0,0,1)}{\L}}W^{-1}\big]^{*-1} \bigg]^{\otimes 2}_S \oplus 
\bigg[\big[WU^{{}_{(1,0,0,1)}{\L}}W^{-1}\big]^{*-1} \bigg]^{\otimes 3}_S \oplus \,\, \bold{\ldots} \\
= \Gamma(\mathfrak{J'}) \, \Gamma\Big( WU^{{}_{(1,0,0,1)}{\L}}W^{-1} \Big) \, \Gamma (\mathfrak{J'}),
\end{multline*}
of $T_4 \circledS SL(2, \mathbb{C})$, which is Krein-isometric in the Krein-Fock space
$\big( \Gamma(\mathcal{H'}), \, \Gamma(\mathfrak{J'}) \big)$.

\vspace*{1cm}

\begin{rem}\label{TwoRepOfaa^+InBoseFock}
In the sequel we will likewise be using a unitary equivalent construction of annihilation
and creation operators in the Fock space, which is frequently used in mathematical literature
(in particular by Hida, Obata and Sait\^o in their works, \cite{hida}, \cite{obata-book}), and which is better
whenever we are using the Wiener-It\^o-Segal chaos decomposition, where the annihilation operators
at fixed points gain the geometric interpretation of derivations on a nuclear algebra of test
functions on a strong dual of a nuclear space.

For this purpose we redefine slightly the norm of (\ref{GeneralPhi})
by puting its square equal
\[
\|\Phi\|_{0}^2 = \sum \limits_{n \geq 0} n! \| \Phi_n \|^2
\]
instead of of (\ref{NormGeneralPhi}), and replace the norm
of the $n$-particle component $\Phi_n \in \mathcal{H}'^{\widehat{\otimes}\, n}$ by
\[
(n!)^{1/2} \| \Phi_n \|_{{}_{\otimes n}} = (n!)^{1/2} \| \Phi_n \|.
\]
Next we define the annihilation and creation operators by the formulas
\[
1) \,\,\,
a' (\widetilde{\varphi})\Phi_0 = 0, \,\,\,
a'^+ (\widetilde{\varphi})\Phi_0 = \widetilde{\varphi},
\]
\[
2) \,\,\,
a' (\widetilde{\varphi}) \, \, \underset{1}{\widetilde{\varphi}} \, \widehat{\otimes} \,
\underset{2}{\widetilde{\varphi}} \, \widehat{\otimes} \, \ldots \widehat{\otimes} \, \underset{n}{\widetilde{\varphi}}
= n \,\, \overline{\widetilde{\varphi}} \, \widehat{\otimes}_{1} \,
\underset{1}{\widetilde{\varphi}} \, \widehat{\otimes} \, \underset{2}{\widetilde{\varphi}} \, \widehat{\otimes}
\, \ldots \, \widehat{\otimes} \, \underset{n}{\widetilde{\varphi}},
\]
\[
3) \,\,\,
a'^+ (\widetilde{\varphi}) \,\, \underset{1}{\widetilde{\varphi}} \, \widehat{\otimes} \, \underset{2}{\widetilde{\varphi}} \, \widehat{\otimes} \, \ldots \, \widehat{\otimes} \, \underset{n}{\widetilde{\varphi}}
= \, \widetilde{\varphi} \, \widehat{\otimes} \, \underset{1}{\widetilde{\varphi}} \, \widehat{\otimes} \, \underset{2}{\widetilde{\varphi}} \, \widehat{\otimes} \, \ldots \, \widehat{\otimes} \underset{n}{\widetilde{\varphi}}.
\]

Here $\widehat{\otimes}_1$ is the symmetrized $1$-contraction defined uniquely by
\[
\widetilde{\varphi} \, \widehat{\otimes}_{1} \,
\underset{1}{\widetilde{\varphi}} \, \widehat{\otimes} \, \underset{2}{\widetilde{\varphi}} \, \widehat{\otimes}
\, \ldots \, \widehat{\otimes} \, \underset{n}{\widetilde{\varphi}} =
(n!)^{-1} \sum \limits_{\pi}
\langle \widetilde{\varphi},\underset{\pi(1)}{\widetilde{\varphi}} \rangle \underset{\pi(2)}{\widetilde{\varphi}} \otimes
\underset{\pi(3)}{\widetilde{\varphi}} \otimes \ldots \otimes
\underset{\pi(n)}{\widetilde{\varphi}}, \,\,\,\, \widetilde{\varphi},\underset{i}{\widetilde{\varphi}} \in \mathcal{H}',
\subset \mathcal{H}'
\]
with the elements $\overline{\widetilde{\varphi}}$ of the adjoint space $\overline{\mathcal{H}'}$ identified with the elements of the dual space $\mathcal{H}'^{*}$ through the Riesz isomorphism
$\widetilde{\varphi} \mapsto \overline{\widetilde{\varphi}}$ as before,
and with the pairing $\langle \cdot , \cdot \rangle$
\[
\langle \widetilde{\varphi},\underset{\pi(1)}{\widetilde{\varphi}} \rangle
= (\overline{\widetilde{\varphi}},\underset{\pi(1)}{\widetilde{\varphi}})
\]
with $(\cdot,\cdot)$ equal to the Hilbert space inner product of
$\widetilde{\varphi} \in \mathcal{H}'$ and $\underset{\pi(1)}{\widetilde{\varphi}} \in \mathcal{H}'$
in the single particle Hilbert space $\mathcal{H}'$.

Note that the unitary operator:
\[
U \Big( \sum \limits_{n \geq 0} \Phi_n \Big) =
\sum \limits_{n \geq 0} (n!)^{-1/2} \, \Phi_n, \,\,\,
U^{-1} \Big( \sum \limits_{n \geq 0} \Phi_n \Big) =
\sum \limits_{n \geq 0} (n!)^{1/2} \, \Phi_n,
\]
with the convention that $0!=1$,
gives the unitary equivalence between the two realizations of the annihilation and creation operators
in the Fock spaces, as well as of the representations of $T_4 \circledS SL(2, \mathbb{C})$
in the corresponding Fock spaces.
\end{rem}

\vspace*{1cm}

\subsection{Wightman operator valued distributions compared to the white noise
generalized operators in case of the electromagnetic potential field}\label{WightmanField}

Following Streater's and Wightman's suggestion \cite{wig}, 2.2 page 104, adopted to our situation
of the Krein-isometric, non-unitary representation
with the fundamental symmetry operator $\eta' = \Gamma(\mathfrak{J'})$, we define the
operator valued distribution
\begin{equation}\label{op-distr-A'}
\varphi \mapsto A(\varphi) = a'(\overline{\check{\widetilde{\varphi}}|_{{}_{\mathscr{O}}}}) +
\eta a'(\widetilde{\varphi}|_{{}_{\mathscr{O}}})^+ \eta,
\end{equation}
where $a(\widetilde{\varphi}|_{{}_{\mathscr{O}}})$ and $a(\widetilde{\varphi}|_{{}_{\mathscr{O}}})^+$ are the annihilation and creation operators
of the Fock space constructed as above, and where $\overline{\widetilde{\varphi}}$
is the ordinary complex conjugation of the function $\widetilde{\varphi}$,
$\widetilde{\varphi}(-p)
= \check{\widetilde{\varphi}}(p)$ (so that $\check{\widetilde{\varphi}}(p) = \overline{\widetilde{\varphi}(p)}$
whenever $\varphi$ is real valued)
and where $\widetilde{\varphi}|_{{}_{\mathscr{O}}}$ are ranging
over the appropriate
nuclear topological space $E \subset \mathcal{H'}$ of functions on the cone and $\varphi$
ranging over the appropriate test space of functions over space-time, which are to be defined below.
$\widetilde{\varphi}$ denotes ordinary Fourier transform
\[
\widetilde{\varphi}(p) = \int \limits_{\mathbb{R}^4} \varphi(x) \, e^{ip \cdot x} \, \ud^4 x
\]
of a test function $\varphi$
on the space-time, and $\widetilde{\varphi}|_{{}_{\mathscr{O}}}$ denotes restriction of the ordinary (four dimensional) Fourier transform to the cone $\mathscr{O}$. In the sequel we will sometimes write shortly
$\widetilde{\varphi}$ instead $\widetilde{\varphi}|_{{}_{\mathscr{O}}}$ for the argument of
the annihilation or creation operator in order to simplify notation, but we should remember that the
restriction to the cone of the ordinary four dimensional Fourier transform in necessary for the argument of creation/annihilation operator in the momentum picture.
Also, the appropriate domain $\mathscr{D}$ of the involutive algebra of the (unbounded)
operators $A(\varphi)$, $\widetilde{\varphi}|_{{}_{\mathscr{O}}} \in E$, and the appropriate topology in the
linear space $\mathcal{L}(\mathscr{D})$ of the operators $A(\varphi)$,
$\widetilde{\varphi}|_{{}_{\mathscr{O}}} \in E$ (and with $\varphi$ ranging over the space-time test space)
should be properly defined (which we do below).

Two points should be noted before passing to the details of the construction.
First, let us remind that the original Streater's and Wightmans'
suggestion was concerned with non-gauge field with ordinary unitary ($\mathfrak{J} = \mathfrak{J'} =\bold{1}$,
$\eta = \bold{1}$) representation
$U^{{}_{(m,0,0,0)}L^s}$ instead of $U^{{}_{(1,0,0,1)}{\L}}$, and that already in \cite{wig} it was noticed that it is necessary to pass to the
Hilbert space $\mathcal{H'} = W\mathcal{H}$ of the representation $WU^{{}_{(m,0,0,0)}L^s}W^{-1}$
with the property that the Fourier transforms (\ref{F(varphi)}) $\varphi$ of the elements
$\widetilde{\varphi} = W\widetilde{\psi} \in \mathcal{H'}$ have local transformation law in order to obtain the quantum
operator valued distributional field with the local transformation formula.
The second point worth to be noted here is that Streater and Wightman leaved as an exercise
all details of the proof that such a field (\ref{op-distr-A'})
is indeed an operator valued distribution and preserves the axioms of \cite{wig}, 3.2\footnote{Of course in case of the problem originally stated
by Streater and Wightman for non gauge field with $\mathfrak{J} = \bold{1}$
and $\eta = \bold{1}$.}
with $\varphi$ ranging over the space of all functions for which $\widetilde{\varphi}$ belong
to $\mathcal{S}(\mathbb{R}^4)$. Thus, $\varphi$ in their definition compose the ordinary
Schwartz space $\mathcal{S}(\mathbb{R}^4)$ (of scalar, spinor, vector, e.t.c. depending on the field) irrespectively if the field in question is massless
(with the orbit $\mathscr{O}$ equal to the positive energy cone) or massive (with the orbit
$\mathscr{O}$ equal to the positive energy sheet of the two-sheeted hyperboloid).
That this choice of the test space in the Wightmann approach works well in both cases,
is related to the fact that (in the momentum picture) the map which a test function $\widetilde{\varphi}$
sends into the integral along the orbit $\mathscr{O}$ (with the invariant measure on $\mathscr{O}$ induced from the ordinary invariant measure in the ambient space $\mathbb{R}^4$) of its restriction $\widetilde{\varphi}|_{{}_{\mathscr{O}}}$ to the orbit $\mathscr{O}$, is a well-defined continuous functional on the ordinary
Schwartz space $\mathcal{S}(\mathbb{R}^4)$ in both cases: for the positive sheet $\mathscr{O}$ of the cone and
for the positive energy sheet $\mathscr{O}$ of the hyperboloid (compare the positive energy part of the Fourier transforms of the zero mass and for the massive Pauli-Jordan function, or Subsections \ref{Lop-on-E} and
\ref{splitting}).
Although the standard kernel theorem of Laurent Schwartz for $\mathcal{S}(\mathbb{R}^n)$ is sufficient for the
proof that the so-called Wightman functions
do exist, Wightman realization \cite{wig} of the (free) field is nonetheless insufficient for of the standard Wick theorem for free fields \cite{Bogoliubov_Shirkov}, Chap III and for the construction Wick polynomials of operator valued distributions in the form which is of fundamental use for example in the St\"uckelberg-Bogoliubov causal method of constructing the perturbation series. This was noticed by Irving
E. Segal \cite{Segal-ProcStone}. A substantially more elaborate theory of nuclear spaces embracing a whole set
of properties which run under the name of ''kernel-type theorems'' is needed here together with their functorial behavior under the second quantization functor for the construction of free fields which is more adequate here
-- namely the white noise construction due to Berezin-Hida.

Interesting contribution
toward the appropriate generalization
of the Schwartz's kernel theorem had been found and proved by Woronowicz in \cite{Woronowicz} seven years after the publication of \cite{wig}. However, we proceed along two different ways, one initiated by Berezin, and formalized by Hida and his school and construct s nuclear space $E \subset \mathcal{H'}$
and construct Wick products of fields understood as generalized operators within the white noise setup -- which is much more than just operator valued distribution in Wightman sense.
White noise method is based on the construction of the nuclear space $E \subset \mathcal{H'}$
with the help of an essentially self adjoint differential
operator\footnote{Of course this operator $A$ should not be mixed with the quantum four-potential field (\ref{op-distr-realA'}), and we hope that the objects are so much different that it will be clear from the context what me mean in each case by using the symbol $A$.} $A$ in $\mathcal{H'}$ such that $A^{-1}$ is compact of Hilbert-Schmidt class and $A^{-2}$ being a trace class, i.e. nuclear, and with $E$ being countably Hilbert and nuclear -- a general recipe worked out by Gelfand and his school \cite{GelfandIV}), compare also \cite{HKPS}. The whole point is that the construction
of $E$ may be lifted to the Fock space with the help of the second quantization functor $\Gamma$
and the white noise calculus may be applied.

In fact the classical four-potential field is real, and we confine ourselves in the formula (\ref{op-distr-A'})
to real functions: $\varphi = \overline{\varphi}$, so that $\widetilde{\varphi}(-p)
= \check{\widetilde{\varphi}}(p) = \overline{\widetilde{\varphi}(p)}$, with the operator valued distribution
\begin{equation}\label{op-distr-realA'}
\varphi \mapsto A(\varphi) = a'(\widetilde{\varphi})
+ \eta a'(\widetilde{\varphi})^+ \eta,
\,\,\, \textrm{if} \,\,\,
\varphi = \overline{\varphi},
\end{equation}
where the restriction to the cone sign was omitted in the arguments $\overline{\widetilde{\varphi}}$
of creation and annihilation operators $a', a'^{+}$.

Now let us define the self adjoint operator $\sqrt{B}$ of pointwise multiplication by the matrix
$\frac{1}{\sqrt{2 p^0(p)}}\sqrt{B(p)}$, where $\sqrt{B(p)}$ is the square root (\ref{sqrtB}) of the positive matrix
$B(p) = V(\beta(p))^* V(\beta(p))$, compare eq. (\ref{Bmatrix}), in the Hilbert space $\mathcal{H'}$ of the representations
$WU^{{}_{(1,0,0,1)}{\L}}W^{-1}$ and $\big[WU^{{}_{(1,0,0,1)}{\L}}W^{-1}\big]^{*-1}$.
Similarly, we define the operator $\sqrt{B}^{-1}$ on $\mathcal{H'}$ as the operator of pointwise multiplication
by the matrix $\sqrt{2 p^0 (p)} \sqrt{B(p)}^{-1}$.

By reasons explained above (and in Subsections \ref{psiWightman}, \ref{psiBerezin-Hida}) we therefore use 
the appropriate Gelfand triple $E \subset \mathcal{H'} \subset E^*$ and its lifting to the second quantized level $(E) \subset \Gamma(\mathcal{H'}) \subset (E)^*$ and the white noise 
construction of the Hida annihilation and creation generalized operators and the operators
$\widetilde{\varphi} \mapsto a'(\widetilde{\varphi})$ and 
$\widetilde{\varphi} \mapsto a'^+ (\widetilde{\varphi})$ 
in the formula (\ref{op-distr-realA'}) through the integral-kernel operators. 
Namely, we construct the Hida operators (details are given in the following Subsections) -- the creation-annihilation generalized operators
$a'(\vec{p})$, $a'(\vec{p})^+$ at points $\vec{p}$, which transform continuously the Hida space $(E)$ into its dual $(E)^*$ and respect the canonical commutation relations
\[
[a'^{\mu} (\vec{p}), a'^{\nu} (\vec{p'})^+] = \frac{1}{2 p^0(\vec{p})}B(\vec{p})^{\mu \nu} 
\, \delta(\vec{p} - \vec{p'}).
\]
Next we define  $\widetilde{\varphi} \mapsto a'(\widetilde{\varphi})$ and 
$\widetilde{\varphi} \mapsto a'^+ (\widetilde{\varphi})$ through the 
(special type of) the so called \emph{integral kernel operators}
\begin{equation}\label{WhiteNoiseA'}
\begin{split}
\widetilde{\varphi} \mapsto a' (\overline{\widetilde{\varphi}}) 
= \int \limits_{\mathbb{R}^3} \widetilde{\varphi}^\mu (\vec{p}) a'^{\mu}(\vec{p}) \, \ud^3 p, \,\,\, (\textrm{summation with respect to} \,\,\, \mu) \\
\widetilde{\varphi} \mapsto a' (\widetilde{\varphi})^+ 
= \int \limits_{\mathbb{R}^3} \widetilde{\varphi}^\mu (\vec{p}) a'^{+ \, \mu}(\vec{p}) \, \ud^3 p, \,\,\, 
(\textrm{summation with respect to} \,\,\, \mu),
\end{split} 
\end{equation}
which define continuous maps from the nuclear space $E$ to the nuclear space $\mathscr{L}((E), (E))$
of continuous linear operators $(E)\rightarrow (E)$, endowed with the nuclear
topology of uniform convergence on bounded sets ($(E)$ is a nuclear space),
compare \cite{hida}, \cite{obata-book}, \cite{luo} or \cite{huang}. 
The nuclear space $E \subset \mathcal{H'}$ and the whole Gelfand triple
$E \subset \mathcal{H'} \subset E^*$ does not have the standard form (compare \cite{hida}, \cite{obata-book}), 
but we will use the fact that it is canonically isomorphic (in the sense defined in Subsection 
\ref{psiBerezin-Hida}) to the standard Gelfand triple
\begin{equation}\label{StandardGelfandTripleForSinglePhoton}
\left. \begin{array}{ccccc}             \mathcal{S}^{0}(\mathbb{R}^3)           & \subset &  L^2(\mathbb{R}^3;\mathbb{C}^4) & \subset & E^*        \\
                                        \parallel           &         & \parallel      &         &  \\

            \mathcal{S}_{\oplus A^{(3)}}(\mathbb{R}^3)            &   & L^2(\mathbb{R}^3 \sqcup \mathbb{R}^3 \sqcup\mathbb{R}^3 \sqcup\mathbb{R}^3 ;\mathbb{C}) &  &  \end{array}\right..
\end{equation}  
with the standard operator
\[
\oplus A^{(3)} \,\,\, \textrm{on} \,\,\,
L^2(\mathbb{R}^3;\mathbb{C}^4) = \oplus L^2(\mathbb{R}^3;\mathbb{C})
\]
equal to the direct sum of four copies of a standard operator $A^{(3)}$
on
\[
L^2(\mathbb{R}^3;\mathbb{C}),
\]
constructed in the following Subsections (compare Subsection \ref{dim=n}).
Then we construct the Hida generalised annihilation and creation operators $a^\mu(\vec{p}), a^\mu(\vec{p})^+$
for the Fock space $\Gamma\big(L^2(\mathbb{R}^3;\mathbb{C}^4)\big)$ using the lifting to the Fock space of
the standard triple (\ref{StandardGelfandTripleForSinglePhoton}). Next we construct the Hida generalized annihilation
and creation operators $a'^{\mu}(\vec{p}), a'^{\mu}(\vec{p})$ in the Fock space lifting
of the Gelfand triple $E \subset \mathcal{H'} \subset E^*$, using the Hida operators
$a'^{\mu}(\vec{p}), a'^{\mu}(\vec{p})$ and the unitary isomorphism between the triple
$E \subset \mathcal{H'} \subset E^*$ and the standard triple (\ref{StandardGelfandTripleForSinglePhoton})
in the way already explained in Subsection \ref{psiBerezin-Hida}.

But if we define the operator valued distributions
$\widetilde{\varphi} \mapsto a'(\widetilde{\varphi})$ and $\widetilde{\varphi} \mapsto a'^+ (\widetilde{\varphi})$ 
through the above maps, with $\widetilde{\varphi}$ being the Fourier transform of space-time
test function $\varphi$ we should recall
that in fact we have to insert into the formula (\ref{op-distr-realA'})
the restriction to the orbit $\mathscr{O}_{1,0,0,1}$ -- here the positive energy sheet of the light
cone in the momentum space, which is more explicitly written in the formula
(\ref{op-distr-A'}). But this makes sense if the restriction to the orbit
\[
\widetilde{\varphi} \,\, \longrightarrow \,\,\,\widetilde{\varphi}|_{{}_{\mathscr{O}}}
\]
defines a continuous map from the nuclear space of Fourier transformed test functions,
to the nuclear space $E$ in the single particle Hilbert space. Here we see
in particular that $E$ cannot be equal to the Schwartz space of functions in $\mathcal{S}(\mathbb{R}^4)$
restricted to the cone $\mathscr{O}_{1,0,0,1}$ (say with the spatial momentum coordinates as the coordinates on the cone). This is because the map defined by the restriction to the cone
$\mathscr{O}_{1,0,0,1}$ is not continuous as a map
$\mathcal{S}(\mathbb{R}^4) \rightarrow \mathcal{S}(\mathbb{R}^3)$ for the ordinary nuclear topology
of Schwartz.

We observe also that the ordinary creation and annihilation generalized operators (in the sense
of white noise calculis, compare \cite{hida}, \cite{obata-book} or \cite{luo}) 
$a^\mu(\vec{p})$, $a^\mu(\vec{p})^+$ at specified points $\vec{p}$ (much more than just 
operator valued distributions in Wightman sense)  fulfilling (as generalized operators, \cite{hida},
\cite{obata-book}, \cite{luo})
\begin{equation}\label{[a,a+]}
[a^\mu (\vec{p}), a^\nu (\vec{p'})^+] = \delta^{\mu \nu} \delta(\vec{p} - \vec{p'}),
\end{equation}
may only be defined as the following operator valued distributions (generalized operators) 
\begin{equation}\label{creation-annihilation-distributions}
\widetilde{\varphi} \mapsto a'(\sqrt{B}^{-1}\overline{\widetilde{\varphi}}) =  a(\overline{\widetilde{\varphi}}) \,\,\, \textrm{and} \,\,\,
\widetilde{\varphi} \mapsto a'(\sqrt{B}^{-1}\widetilde{\varphi})^+ = a(\widetilde{\varphi})^+.
\end{equation}
(We have used the prime sign at the  operator valued distributions $a', a'^+$ in order to distinguish them from the ordinary operator valued distributions $a, a^+$ fulfilling (\ref{[a,a+]}), as they are indeed different.) 
The operators $a'(\sqrt{B}^{-1}\overline{\widetilde{\varphi}})$  and $\widetilde{\varphi} \mapsto a'(\sqrt{B}^{-1}\widetilde{\varphi})^+$  may be strictly defined with the help of white noise calculus as the 
special type of integral kernel operators 
\cite{luo}, \cite{hida} or \cite{huang} (motivated by the construction of the Fock expansion into normal
operators initiated by Berezin \cite{Berezin})
\begin{equation}\label{WhiteNoise-a'a}
\begin{split}
\widetilde{\varphi} \mapsto a' (\sqrt{B}^{-1}\overline{\widetilde{\varphi}}) = a (\overline{\widetilde{\varphi}})
= \int \limits_{\mathbb{R}^3} \widetilde{\varphi}^\mu (\vec{p}) a^\mu(\vec{p}) \, \ud^3 p, \,\,\, (\textrm{summation with respect to} \,\,\, \mu) \\
\widetilde{\varphi} \mapsto a' (\sqrt{B}^{-1}\widetilde{\varphi})^+ =  a (\widetilde{\varphi})^+ 
= \int \limits_{\mathbb{R}^3} \widetilde{\varphi}^\mu (\vec{p}) a^{+ \, \mu}(\vec{p}) \, \ud^3 p, \,\,\, 
(\textrm{summation with respect to} \,\,\, \mu)
\end{split} 
\end{equation}    
This is of course possible only if the operators $\sqrt{B}$ and $\sqrt{B}^{-1}$ transform the nuclear space in question 
$E \subset \mathcal{H'}$ into the nuclear space $E$ and do this in a continuous manner with respect
to the nuclear topology on $E$. 

Therefore, we see again that the Schwartz space $\mathcal{S}(\mathbb{R}^3)$ of rapidly decreasing
smooth functions on $\mathbb{R}^3$ which is sufficient for massive non gauge free fields
is inappropriate in the case of the free photon field and cannot serve as the test function nuclear
space $E \subset \mathcal{H'}$ in the white noise construction of the field,
because the operators $\sqrt{B}$ and $\sqrt{B}^{-1}$ are the pointwise multiplication by matrices which have
the singularities of the type $r^{-1/2} = \frac{1}{(\vec{p} \cdot \vec{p})^{1/4}}$. As is easily seen in general
a function of $\mathcal{S}(\mathbb{R}^3)$ non-vanishing at zero, after multiplication by $r^{-1/2} = \frac{1}{(\vec{p} \cdot \vec{p})^{1/4}}$ will not stay in $\mathcal{S}(\mathbb{R}^3)$ and all the more the multiplication by
$r^{-1/2} = \frac{1}{(\vec{p} \cdot \vec{p})^{1/4}}$ cannot be continuous from $\mathcal{S}(\mathbb{R}^3)$
into $\mathcal{S}(\mathbb{R}^3)$. The same of course holds for the operators $\sqrt{B}$ and $\sqrt{B}^{-1}$
which cannot be continuous $\mathcal{S}(\mathbb{R}^3) \to \mathcal{S}(\mathbb{R}^3)$.

We should emphasize a substantial point in the white noise construction of the field (especially the
potential field $A$) which distinguishes it from the Wighman approach. Namely, in the Wightman approach it is
the quantity (\ref{op-distr-A'}) which is fundamental, with the expressions like
(\ref{WhiteNoiseA'}) or (\ref{WhiteNoise-a'a}) having only symbolic character, in fact definable only through
(\ref{WhiteNoiseA'}) and the appropriate choice of the domain $\mathscr{D}$
in the Fock space, consisting at least of the images of polynomial expressions in operators
(\ref{op-distr-A'}) (with $\varphi \in \mathcal{S}(\mathbb{R}^4)$) acting on the vacuum.
In the white noise construction adopted here we proceed in a sense in the opposite direction:
these are the expressions (\ref{WhiteNoiseA'}) or (\ref{WhiteNoise-a'a}) which are more fundamental, and we utilize the fact that (\ref{WhiteNoiseA'}) or (\ref{WhiteNoise-a'a}) define well-defined
continuous maps $E \rightarrow \mathscr{L}\big((E), (E)\big)$ -- in particular defining operator valued
distributions (and much more than just distributions). But in this approach it is of fundamental importance
that the map of Fourier transforms $\widetilde{\varphi}$ into their restrictions to the
respective orbit $\mathscr{O}$ is continuous as a map $\widetilde{\varphi} \mapsto
\widetilde{\varphi}|_{{}_{\mathscr{O}}} \in E$ from the correct test space of functions
$\varphi$ over space-time to the nuclear space $E$, in order to have well-defined distribution
(\ref{WhiteNoiseA'}). Thus, in white noise approach (adopted here) this has a dramatic consequence
for the choice of the correct space-time test space: for massive fields it can be chosen to be equal
to the ordinary Schwartz space $\mathcal{S}(\mathbb{R}^4)$, but for zero mass fields it has to be changed,
(because of the singularity of the cone at the apex). In the Wightman approach this singularity plays
no essential role (at least at the level of construction of a free zero mass field) and in fact
the space-time test space for massive as well as for zero mass fields
can be chosen to be equal to the ordinary Schwartz space $\mathcal{S}(\mathbb{R}^4)$.
This insensitivity of the Wightmann approach has considerably high prise: his approach is
practically useless for the rigorous formulation and proof of the ''Wick theorem'' of
\cite{Bogoliubov_Shirkov}, Chap. III, in the form needed
in the perturbative causal approach to QFT (e.g. QED). Therefore, we have chosen to construct free fields
(including $A_\mu$) within the white noise approach of Berezin-Hida, which provides a sufficient
basis for the said ''Wick theorem''. 

Our task is to construct the correct nuclear test function space $E \subset \mathcal{H'}$ such that
the operators $\sqrt{B}$ and $\sqrt{B}^{-1}$ will preserve $E$ invariant and will be continuous
as operators $E \to E$ with respect to the nuclear topology of $E$.      

Namely, we define $E$ to be equal to the subspace $\mathcal{S}^0(\mathbb{R}^3) \subset \mathcal{S}(\mathbb{R}^3)
\subset \mathcal{H'}$ of all smooth rapidly decreasing
functions $\widetilde{\varphi}$ such that all their partial derivatives of any order
vanish at the zero point: $D^\alpha \widetilde{\varphi}(0) = 0$.
$\mathcal{S}^0(\mathbb{R}^3)$ as the intersection of kernels of continuous maps
$\mathcal{S}(\mathbb{R}^3) \to \mathcal{S}(\mathbb{R}^3)$ is a closed linear subspace
of the nuclear space $\mathcal{S}(\mathbb{R}^3)$.
By \cite{GelfandIV}, I.3.4, $\mathcal{S}^{0}(\mathbb{R}^3)$ as a closed subspace of a nuclear space
is nuclear. It is easily seen that $\mathcal{S}^{0}(\mathbb{R}^3)$ is dense in $L^2(\mathbb{R}^3, \ud^3x)$
and in $\mathcal{H'}$, as it contains all smooth functions of compact support with the support not containing
the zero point.

The linear operator $\sqrt{B}$ (the same holds for the operator $\sqrt{B}^{-1}$) is symmetric on the subspace 
$\mathcal{S}^0(\mathbb{R}^3) \subset \mathcal{H'}$ and transforms $\mathcal{S}^0(\mathbb{R}^3)$ into
$\mathcal{S}^0(\mathbb{R}^3)$. Such an operator is automatically continuous as an operator
$\mathcal{S}^0(\mathbb{R}^3) \to \mathcal{S}^0(\mathbb{R}^3)$, compare \cite{GelfandIII}, page 190. On the same footing 
$\sqrt{B}^{-1}$ is continuous $\mathcal{S}^0(\mathbb{R}^3) \to \mathcal{S}^0(\mathbb{R}^3)$.
Exactly on the same footing the operators of pointwise multiplication by the following functions 
$r^{-1/2}(\vec{p}) = \frac{1}{(\vec{p} \cdot \vec{p})^{1/4}}$, $r^{1/2}(\vec{p}) = (\vec{p} \cdot \vec{p})^{1/4}$,
$r^{-1}(\vec{p}) = \frac{1}{(\vec{p} \cdot \vec{p})^{1/2}}$, $r(\vec{p}) = (\vec{p} \cdot \vec{p})^{1/2}$ on $\mathcal{H'}$
preserve $\mathcal{S}^{0}(\mathbb{R}^3) \subset \mathcal{H'}$ and are all 
continuous as operators $\mathcal{S}^{0}(\mathbb{R}^3) \to \mathcal{S}^{0}(\mathbb{R}^3)$.

Now consider another subspace $\mathcal{S}^{00}(\mathbb{R}^3) \subset \mathcal{S}(\mathbb{R}^3)
\subset \mathcal{H'}$ of all those functions $\varphi$ whose ordinary Fourier transforms $\mathscr{F}$:
\[
\mathscr{F}\varphi(\vec{p}) = \int \varphi(\vec{x}) e^{i\vec{p} \cdot \vec{x}} \, \ud^3 x
\]
are in $\mathcal{S}^0(\mathbb{R}^3)$.
It is of course linear and again as the inverse image under a continuous map
$\mathcal{S}(\mathbb{R}^3) \mapsto \mathcal{S}(\mathbb{R}^3)$ of a closed set $\mathcal{S}^0(\mathbb{R}^3)$
is likewise closed and again, by \cite{GelfandIV}, I.3.4, nuclear. Joining the continuity of
$\mathscr{F}^{-1}: \mathcal{S}(\mathbb{R}^3) \to \mathcal{S}(\mathbb{R}^3)$ with the continuity of
the operator of multiplication by the function $r^{-1} = \frac{1}{(\vec{p} \cdot \vec{p})^{1/2}}$:
$\mathcal{S}^{0}(\mathbb{R}^3) \to \mathcal{S}^{0}(\mathbb{R}^3)$ we easily see the continuity of the Fourier transform
$\widetilde{\varphi} \mapsto \varphi$ defined by (\ref{F(varphi)}) and regarded as operator $\mathcal{S}^{0}(\mathbb{R}^3) \to \mathcal{S}^{00}(\mathbb{R}^3)$ (with the coordinates on the orbit $\mathscr{O}_{(1,0,0,1)}$
equal to the three spatial components $\vec{p}$ of momentum), as well as its onto character.
By the Banach inverse mapping theorem the inverse
map $\varphi \mapsto \widetilde{\varphi}$ is likewise continuous when regarded as the map
$\mathcal{S}^{00}(\mathbb{R}^3) \to \mathcal{S}^{0}(\mathbb{R}^3)$ of nuclear spaces.
It is easily seen that $\mathcal{S}^{00}(\mathbb{R}^3)$ is likewise equal to the inverse image under the inverse
map $\varphi \mapsto \widetilde{\varphi}$ of Fourier
transform defined by (\ref{F(varphi)}) of the closed subspace $\mathcal{S}^{0}(\mathbb{R}^3)$. 

It likewise is easily seen that the operators $\mathfrak{J}'$, $WU_{{}_{a,\alpha}}^{{}_{(1,0,0,1)}{\L}}W^{-1}$ and
$\big[WU_{{}_{a,\alpha}}^{{}_{(1,0,0,1)}{\L}}W^{-1}\big]^{*-1}$, $(a,\alpha) \in T_4 \circledS SL(2, \mathbb{C})$
preserve $\mathcal{S}^0(\mathbb{R}^3)$ (with the coordinates on the orbit $\mathscr{O}_{(1,0,0,1)}$
equal to the three spatial components $\vec{p}$ of momentum) and are continuous as operators
$\mathcal{S}^0(\mathbb{R}^3) \to \mathcal{S}^0(\mathbb{R}^3)$, although they are discontinuous with respect to the Hilbert space norm of $\mathcal{H'}$.

The definition of the nuclear space $\mathcal{S}^{0}(\mathbb{R}^3)$ has natural extension to higher dimensions
$\mathcal{S}^{0}(\mathbb{R}^{n})$. Namely, we consider the linear subspace of functions
$\widetilde{\varphi} \in \mathcal{S}(\mathbb{R}^{n})$ whose all derivatives vanish at zero $D^\alpha \widetilde{\varphi}(0) = 0$ and the nuclear subspace $\mathcal{S}^{00}(\mathbb{R}^{n}) \subset \mathcal{S}(\mathbb{R}^{n})$
equal to the inverse image of $\mathcal{S}^{0}(\mathbb{R}^{n})$ under the ordinary Fourier transform in
$\mathbb{R}^{n}$.

In particular for any two elements 
$\underset{1}{\widetilde{\varphi}}$ and $\underset{2}{\widetilde{\varphi}}$ of $\mathcal{S}^0(\mathbb{R}^4)$ 
define 
\[
\underset{1}{\varphi} \otimes \underset{2}{\varphi} \,\big(\underset{1}{x},  \underset{2}{x} \big) 
=  \underset{1}{\varphi} (\underset{1}{x}) \, \underset{2}{\varphi} (\underset{2}{x});
\] 
and similarly
\[
\underset{1}{\widetilde{\varphi}} \otimes \underset{2}{\widetilde{\varphi}} \,\big(\underset{1}{p},  \underset{2}{p} \big) 
=  \underset{1}{\widetilde{\varphi}} (\underset{1}{p}) \, \underset{2}{\widetilde{\varphi}} (\underset{2}{p}).
\]
Because
\begin{multline*}
\underset{1}{\varphi} \otimes \underset{2}{\varphi} \,\big(\underset{1}{x},  \underset{2}{x} \big)
= \int \limits_{\mathbb{R}^4 \times \mathbb{R}^4} 
\underset{1}{\widetilde{\varphi}} \otimes \underset{2}{\widetilde{\varphi}} \,
\big(\underset{1}{p},  \underset{2}{p} \big) \,\,\,
e^{i \underset{1}{p} \cdot \underset{1}{x}} e^{i \underset{2}{p} \cdot \underset{2}{x}}  \,\,\,
\ud^4 \underset{1}{p}  \, \times \ud^4 \underset{2}{p}  \\
= \int \limits_{\mathbb{R}^4 \times \mathbb{R}^4} 
\underset{1}{\widetilde{\varphi}} (\underset{1}{p}) \, \underset{2}{\widetilde{\varphi}} (\underset{2}{p}) \,\,\,
e^{i \underset{1}{p} \cdot \underset{1}{x}} e^{i \underset{2}{p} \cdot \underset{2}{x}}  \,\,\,
\ud^4 \underset{1}{p}  \, \times \ud^4 \underset{2}{p},
\end{multline*}
then from
\[
\underset{1}{\widetilde{\varphi}}, \underset{2}{\widetilde{\varphi}} \in \mathcal{S}^0(\mathbb{R}^4), 
\]
it follows that 
\begin{multline*}
\underset{1}{\varphi} \otimes \underset{2}{\varphi} \in \mathcal{S}^{00}(\mathbb{R}^4) \otimes
\mathcal{S}^{00}(\mathbb{R}^4) \subset  \mathcal{S}^{00}(\mathbb{R}^4 \times \mathbb{R}^4), \\
\underset{1}{\widetilde{\varphi}} \otimes \underset{2}{\widetilde{\varphi}} 
\in \mathcal{S}^0(\mathbb{R}^4) \otimes \mathcal{S}^0(\mathbb{R}^4) \subset \mathcal{S}^0(\mathbb{R}^4 \times \mathbb{R}^4).
\end{multline*}
Because the topology of the closed nuclear sub-spaces 
$\mathcal{S}^{00}(\mathbb{R}^{n}) \subset \mathcal{S}(\mathbb{R}^{n})$
and $\mathcal{S}^{0}(\mathbb{R}^{n}) \subset \mathcal{S}(\mathbb{R}^{n})$ is that inherited from 
the nuclear space $\mathcal{S}(\mathbb{R}^{n})$, then it follows that the bilinear maps
\begin{multline}\label{contOtimesS^0xS^0toS^0}
\otimes: \mathcal{S}^{0}(\mathbb{R}^{n}) \times \mathcal{S}^{0}(\mathbb{R}^{m}) \to 
\mathcal{S}^0(\mathbb{R}^n) \otimes \mathcal{S}^0(\mathbb{R}^n) \subset \mathcal{S}^{0}(\mathbb{R}^{(n+m}) \,\,\, \textrm{and} \\
\otimes: \mathcal{S}^{00}(\mathbb{R}^{n}) \times \mathcal{S}^{00}(\mathbb{R}^{m}) \to 
\mathcal{S}^{00}(\mathbb{R}^n) \otimes \mathcal{S}^{00}(\mathbb{R}^n) \subset
\mathcal{S}^{00}(\mathbb{R}^{(n+m})
\end{multline}
are (jointly) continuous (compare also the Grothendieck's characterisation of nuclear  
topological linear spaces \cite{Grothendieck}). Indeed, the topology of 
$\mathcal{S}^0(\mathbb{R}^n) \otimes \mathcal{S}^0(\mathbb{R}^n) \subset \mathcal{S}^{0}(\mathbb{R}^{(n+m})$
is stronger than the topology of $\mathcal{S}^{0}(\mathbb{R}^{(n+m})$, so the inclusion
$\mathcal{S}^0(\mathbb{R}^n) \otimes \mathcal{S}^0(\mathbb{R}^n) \subset \mathcal{S}^{0}(\mathbb{R}^{(n+m})$
is continuous. From the (joint) continuity of the mapping 
$\otimes: \mathcal{S}^{0}(\mathbb{R}^{n}) \times \mathcal{S}^{0}(\mathbb{R}^{m}) \to 
\mathcal{S}^0(\mathbb{R}^n) \otimes \mathcal{S}^0(\mathbb{R}^n)$ it follows the (joint) 
continuity of the composite mapping (\ref{contOtimesS^0xS^0toS^0}). Similarly for the continuity of the second mapping
in (\ref{contOtimesS^0xS^0toS^0}).  

We define the domain $\mathscr{D}$, of all  $a'(\widetilde{\varphi}|_{{}_{\mathscr{O}}}), 
a'(\widetilde{\varphi}|_{{}_{\mathscr{O}}})^+$,
$\widetilde{\varphi}|_{{}_{\mathscr{O}}} \in \mathcal{S}^0(\mathbb{R}^3)$ to consist of all those 
$\Phi = \sum \limits_{n=0}^{\infty}
\Phi^{(n)}$ which belong to the \emph{Hida test functional} space $(E)$ (for definition and construction of
the nuclear space $(E)$ and its strong dual $(E)^*$, compare Subsections 
\ref{white-noise-proofs} and \ref{WhiteNoiseA}
and \cite{obata-book} for a more detailed study) . 

On the linear spaces $\mathscr{L}\big((E), (E)\big)$
$\mathscr{L}\big((E), (E)^*\big)$ of all linear and continuous operators $(E) \rightarrow (E)$ and resp.
$(E) \rightarrow (E)^*$, we define the topology of uniform convergence on bounded sets. This topology 
on $\mathscr{L}\big((E), (E)\big)$ and $\mathscr{L}\big((E), (E)^*\big)$ is nuclear (recall that $(E)$
and $(E)^*$ are nuclear spaces). 

In this situation the generalized kernel theorem (compare \cite{obata-book}, \cite{obataJFA}, \cite{Schaefer,
\cite{treves}} is applicable to the bilinear separately continuous maps
\[
\mathcal{S}^{00}(\mathbb{R}^{4}) \times \mathcal{S}^{00}(\mathbb{R}^{4})
\ni \underset{1}{\varphi} \times \underset{2}{\varphi} \mapsto
A\big( \underset{1}{\varphi} \big) A\big( \underset{2}{\varphi} \big) \in \mathscr{L}\big((E),(E)\big)
\]
and
\[
\mathcal{S}^{00}(\mathbb{R}^{4}) \times \mathcal{S}^{00}(\mathbb{R}^{4})
\ni \underset{1}{\varphi} \times \underset{2}{\varphi} \mapsto
a'\big(\sqrt{B}^{-1} \underset{1}{\widetilde{\varphi}}|_{{}_{\mathscr{O}}} \big) a'\big( \sqrt{B}^{-1}
\underset{2}{\widetilde{\varphi}}|_{{}_{\mathscr{O}}} \big)
\in \mathscr{L}\big((E),(E)\big), \,\,\, \textrm{e.t.c.}.
\]
Here the operators $A\big( \underset{i}{\varphi} \big),
a'\big(\sqrt{B}^{-1} \underset{i}{\widetilde{\varphi}}|_{{}_{\mathscr{O}}} \big)$, e.t.c. with
$\underset{i}{\varphi} \in \mathcal{S}^{00}(\mathbb{R}^{4})$, are defined through the integral kernel operators of the type (\ref{WhiteNoiseA'})
or (\ref{WhiteNoise-a'a}) and belong to $\mathscr{L}\big((E),(E)\big)$ (compare \cite{hida}, \cite{obata-book}).

We can apply here the results of \cite{hida} and obtain bilinear maps between nuclear spaces
in the indicated manner because the restriction
\[
\mathcal{S}^{0}(\mathbb{R}^{4}) \ni \widetilde{\varphi} \to \widetilde{\varphi}|_{{}_{\mathscr{O}}}
\in \mathcal{S}^{0}(\mathbb{R}^{3})
\]
is continuous as a map of the nuclear space $\mathcal{S}^{0}(\mathbb{R}^{4})$ onto 
$\mathcal{S}^{0}(\mathbb{R}^{3})$. 

With these definitions the maps $\varphi \mapsto A(\varphi)$, 
$\mathcal{S}^{00}(\mathbb{R}^{4}) \ni \varphi \mapsto a'(\overline{\widetilde{\varphi}}|_{{}_{\mathscr{O}}})$, 
$\mathcal{S}^{00}(\mathbb{R}^{4}) \ni \varphi \mapsto a'(\widetilde{\varphi}|_{{}_{\mathscr{O}}})^+$, 
$\mathcal{S}^{00}(\mathbb{R}^{4}) \ni \varphi \mapsto a(\overline{\widetilde{\varphi}}|_{{}_{\mathscr{O}}})$, 
$\mathcal{S}^{00}(\mathbb{R}^{4}) \ni \varphi \mapsto a(\widetilde{\varphi}|_{{}_{\mathscr{O}}})^+$, are continuous maps on 
the nuclear space $\mathcal{S}^{00}(\mathbb{R}^{4})$ into the nuclear space $\mathscr{L}\big((E),(E)\big)$.

From nuclearity of $\mathcal{S}^{00}(\mathbb{R}^{4})$ and $\mathscr{L}\big((E),(E)\big)$ it follows by generalized kernel theorem (compare \cite{obata-book}, \cite{treves}, \cite{GelfandIV}), that the bilinear separately continuous functional
\[
\mathcal{S}^{00}(\mathbb{R}^{4}) \times \mathcal{S}^{00}(\mathbb{R}^{4})
\ni \underset{1}{\varphi} \times \underset{2}{\varphi} \mapsto
\big(\Psi_0, \big[ A\big( \underset{1}{\varphi} \big), \, A\big( \underset{2}{\varphi} \big) \big] \Psi_0\big)
\]
defines a numerical distribution on $\mathcal{S}^{00}(\mathbb{R}^{4}) \otimes \mathcal{S}^{00}(\mathbb{R}^{4})$ and that the bilinear continuous map
\begin{equation}\label{Pauli-Jordan}
\mathcal{S}^{00}(\mathbb{R}^{4}) \times \mathcal{S}^{00}(\mathbb{R}^{4})
\ni \underset{1}{\varphi} \times \underset{2}{\varphi} \mapsto
\big[A\big( \underset{1}{\varphi} \big), \, A\big( \underset{2}{\varphi} \big) \big] \in
\mathscr{L}\big((E),(E)\big)
\end{equation}
defines in the canonical manner a continuous linear map
\[
\mathcal{S}^{00}(\mathbb{R}^{4}) \otimes \mathcal{S}^{00}(\mathbb{R}^{4})
\,\,\, \ni \underset{1}{\varphi} \otimes \underset{2}{\varphi} \longrightarrow \,\,\,
[A,A](\underset{1}{\varphi} \otimes \underset{2}{\varphi}) \in \mathscr{L}\big((E),(E)\big),
\]
(thus operator valued distribution), such that
\[
\big[A\big( \underset{1}{\varphi} \big), \, A\big( \underset{2}{\varphi} \big) \big] =
[A,A](\underset{1}{\varphi} \otimes \underset{2}{\varphi}), \,\,\,
\underset{i}{\varphi} \in \mathcal{S}^{00}(\mathbb{R}^{4}).
\]
It follows that the said numerical distribution
\[
\underset{1}{\varphi} \otimes \underset{2}{\varphi} \, \longmapsto \,
(\Psi_0, [A,A](\underset{1}{\varphi} \otimes \underset{2}{\varphi}) \Psi_0)
\]
on $\mathcal{S}^{00}(\mathbb{R}^{4}) \otimes \mathcal{S}^{00}(\mathbb{R}^{4}) \subset
\mathcal{S}(\mathbb{R}^{4}) \otimes \mathcal{S}(\mathbb{R}^{4}) =
\mathcal{S}(\mathbb{R}^{8})$ is equal (on $\mathcal{S}^{00}(\mathbb{R}^{4}) \otimes \mathcal{S}^{00}(\mathbb{R}^{4})$)
to the standard distribution which can be represented by the integral with the kernel equal to
$ig^{\mu \nu} D_0 (x- y)$, with $D_0$ equal to the Pauli-Jordan distribution function.
Moreover, if we introduce the
Gupta-Bleuler operator $\eta$ as above, then for the standard annihilation-creation operator
valued distributions $a^\mu(\vec{p}), a^\mu(\vec{p})^+$ we obtain the correct commutation rules:
\begin{equation}\label{[eta,a]}
a^0(\varphi)\eta = - \eta a^0(\varphi), \,\,\, a^k(\varphi) \eta = \eta a^k(\varphi),
\end{equation}
the field (\ref{op-distr-realA'}) have the local transformation law and transforms as a four-potential field
under the representation $\Gamma\Big(\big[WU^{{}_{(1,0,0,1)}{\L}}W^{-1}\big]^{*-1}\Big)$ and fulfills the ordinary
massless wave equation.

\section{Proof of the statements of the last Section. Hida's white noise approach}\label{white-noise-proofs}

Now we give the proof of these statements. But in the proof we proceed in the ''reverse direction'':
we start with the standard realization of the Fock space based on the ordinary application
of the second quantization functor $\Gamma$ to the four component functions $\widetilde{\varphi}$
square integrable with respect to the ordinary Lebesgue measure on $\mathbb{R}^3$, i.e. we apply $\Gamma$ to the Hilbert space
\[
\oplus L^2(\mathbb{R}^3; \mathbb{C}) = L^2(\mathbb{R}^3; \mathbb{C}^4)
\]
regarded as one particle Hilbert space (summation is over four copies of $L^2(\mathbb{R}^3 ; \mathbb{C})$
corresponding to the four components of the functions $\widetilde{\varphi}$). Thus, we start with the generalized
Hida operators, the creation-annihilation operators $a^\mu(\vec{p}), a^\mu(\vec{p})^+$ in the momentum picture,
respecting the ordinary canonical commutation relations (\ref{[a,a+]}), with the given
Gupta-Bleuler operator $\eta = \Gamma(\mathfrak{J}_{\bar{p}})$, where $\mathfrak{J}_{\bar{p}}$ is the operator acting in
$L^2(\mathbb{R}^3; \mathbb{C}^4)$ as the operator of multiplication by the constant matrix (\ref{J-barp})
having the ordinary commutation rules (\ref{[a,a+]}). The reason for doing so is the standard form
of the Gelfand triple $E = \mathcal{S}_{A}(\mathbb{R}^3; \mathbb{R})
\subset L^2(\mathbb{R}^3;\mathbb{R}^4) \subset E^*$,
and was already justified in Subsection \ref{psiBerezin-Hida}.
We construct them in a mathematically rigorous manner \cite{obata-book}, \cite{Hida1}, \cite{HKPS}, \cite{luo} as generalized operators with the help of white noise calculus.
In particular, we need to construct the appropriate Gelfand triple
\[
E \subset \oplus L^2(\mathbb{R}^3) \subset E^*
\]
with the countably Hilbert nuclear space $E$ using an essentially self adjoint differential operator (should not be mixed with (\ref{op-distr-realA'})) $A$ in $\oplus L^2(\mathbb{R}^3; \mathbb{C}) = L^2(\mathbb{R}^3; \mathbb{C}^4)$
with $A^{-1}$ compact of Hilbert-Schmidt
class (as in \cite{GelfandIV} or \cite{hida}, compare also \cite{HKPS})
such that the operators $\sqrt{B}$ and $\sqrt{B}^{-1}$, the operators of multiplication by the following functions
$r^{-1/2}(\vec{p}) = \frac{1}{(\vec{p} \cdot \vec{p})^{1/4}}$, $r^{1/2}(\vec{p}) = (\vec{p} \cdot \vec{p})^{1/4}$,
the operator of differentiation, and the representors of the {\L}opusza\'nski representation and its conjugation are all continuous as operators $E \to E$ and with $E$ containing $\mathcal{S}^0 (\mathbb{R}^3)$ as a subset. In fact, we will show that $\mathcal{S}^0 (\mathbb{R}^3) = E$. Then in a canonical manner using the ordinary Fourier transform $\mathscr{F}$ we construct the corresponding Gelfand triple $\mathbb{E} \subset \oplus L^2(\mathbb{R}^3) \subset \mathbb{E}^*$
by replacing $A$ with $\mathscr{F} \, A \, \mathscr{F}^{-1}$ in the position picture
(with $\mathbb{E}$ plying the role of $\mathcal{S}^{00}(\mathbb{R}^3)$, in fact we will show that
$\mathcal{S}^{00}(\mathbb{R}^3) = \mathbb{E}$) so that the ordinary Fourier transform
$\mathscr{F}$ is continuous and onto as an operator $E \to \mathbb{E}$ 
and so that the triples are connected in the following manner 
\begin{equation}\label{2-Gelfand-triples}
\left. \begin{array}{ccccc}              E         & \subset & \oplus L^2(\mathbb{R}^3) & \subset & E^*        \\
                               \downarrow \uparrow &         & \downarrow \uparrow      &         & \downarrow \uparrow  \\
                             \mathbb{E} & \subset & \oplus L^2(\mathbb{R}^3) & \subset & \mathbb{E}^* \end{array}\right.,
\end{equation}  
with the vertical arrows representing the ordinary Fourier transform $\mathscr{F}$ and its inverse
which are continuous and invertible between the indicated spaces; and with the transform $\widetilde{\varphi} \mapsto \varphi$ defined by (\ref{F(varphi)}) continuous and onto when regarded as a map $E \to \mathbb{E}$. We then apply the second quantization functor $\Gamma$ to the diagram (\ref{2-Gelfand-triples}) and the white noise calculus to the construction of generalised field operators exactly as in \cite{hida}, compare also \cite{huang} or \cite{luo} 
(for a friendly handbook presentation the reader may consult \cite{HKPS}, where the construction 
with the operator $A$ equal to the one dimensional oscillator Hamiltonian is presented in detail). 

Having obtained this we then prove that\footnote{The sign indicating restriction of $\widetilde{\varphi}$
$\widetilde{\varphi}'$, $\underset{n}{\widetilde{\varphi}}$, e.t.c. to the cone $\mathscr{O}$ 
is omitted for simplicity.} (compare Subsection \ref{psiBerezin-Hida})
\begin{equation}\label{proof-first}
[a(\sqrt{B}\widetilde{\varphi}), a(\sqrt{B}\widetilde{\varphi}')^+] = (\widetilde{\varphi}, \widetilde{\varphi}')
\end{equation}
where $(\cdot , \cdot)$ in the last expression is the inner product in $\mathcal{H'}$.
Then we prove that
\begin{multline}\label{proof-second}
\eta a(\sqrt{B}\underset{1}{\widetilde{\varphi}})^+ \, 
a(\sqrt{B}\underset{2}{\widetilde{\varphi}})^+
\, \dots \, 
a(\sqrt{B}\underset{n}{\widetilde{\varphi}})^+ \Omega \\ =
a(\sqrt{B}\mathfrak{J'}\underset{1}{\widetilde{\varphi}})^+ \, 
a(\sqrt{B}\mathfrak{J'}\underset{2}{\widetilde{\varphi}})^+
\, \dots \, 
a(\sqrt{B}\mathfrak{J'}\underset{n}{\widetilde{\varphi}})^+ \Omega;
\end{multline}
i.e. that Gupta-Bleuler operator $\eta$ is indeed implemented by $\Gamma(\mathfrak{J'})$ in the Fock space
$\Gamma(\mathcal{H'})$ constructed above, as by the relation between $a$ and $a'$ the last equality
(\ref{proof-second}) may be written as
\[
\eta a'(\underset{1}{\widetilde{\varphi}})^+ \, 
a'(\underset{2}{\widetilde{\varphi}}))^+
\, \dots \, 
a'(\underset{n}{\widetilde{\varphi}}))^+ \Omega =
a'(\mathfrak{J'}\underset{1}{\widetilde{\varphi}})^+ \, 
a'(\mathfrak{J'}\underset{2}{\widetilde{\varphi}})^+
\, \dots \, 
a'(\mathfrak{J'}\underset{n}{\widetilde{\varphi}})^+ \Omega.
\]

Then we define the representation
$\bold U$ of the group $T_4 \circledS SL(2,\mathbb{C})$ in the following manner
\begin{multline}\label{proof-third}
\bold{U}_{a,\alpha} a(\sqrt{B}\underset{1}{\widetilde{\varphi}})^+ \, 
a(\sqrt{B}\underset{2}{\widetilde{\varphi}}))^+
\, \dots \, 
a(\sqrt{B}\underset{n}{\widetilde{\varphi}}))^+ \Omega \\ =
a(\sqrt{B} U'_{a,\alpha} \underset{1}{\widetilde{\varphi}})^+ \, 
a(\sqrt{B} U'_{a,\alpha} \underset{2}{\widetilde{\varphi}})^+
\, \dots \, 
a(\sqrt{B} U'_{a,\alpha} \underset{n}{\widetilde{\varphi}})^+ \Omega; \\
U'_{a,\alpha} = \big[WU_{a,\alpha}^{{}_{(1,0,0,1)}{\L}}W^{-1}\big]^{*-1}, 
\underset{i}{\widetilde{\varphi}} \in E;
\end{multline}  
that is we define $\bold U$ so that by the correspondence between $a$ and $a'$ the representation may indeed be
identified with the representation $\Gamma\Big(\big[WU^{{}_{(1,0,0,1)}{\L}}W^{-1}\big]^{*-1}\Big)$
in the Fock space $\Gamma(\mathcal{H'})$ defined as above, on the indicated domain. 
Indeed, by the relation between the fields $a$ and $a'$ the last formula (\ref{proof-third}) is equivalent to 
\begin{multline*}
\bold{U}_{a,\alpha} a'(\underset{1}{\widetilde{\varphi}})^+ \, 
a'(\underset{2}{\widetilde{\varphi}})^+
\, \dots \, 
a'(\underset{n}{\widetilde{\varphi}})^+ \Omega \\ =
a'( U'_{a,\alpha} \underset{1}{\widetilde{\varphi}})^+ \, 
a'( U'_{a,\alpha} \underset{2}{\widetilde{\varphi}})^+
\, \dots \, 
a'( U'_{a,\alpha} \underset{n}{\widetilde{\varphi}})^+ \Omega; \\
U'_{a,\alpha} = \big[WU^{{}_{(1,0,0,1)}{\L}}W^{-1}\big]^{*-1}, 
\underset{i}{\widetilde{\varphi}} \in E.
\end{multline*}

In the next step we prove that the field ($\varphi = \overline{\varphi}$)
\[
\mathcal{S}^{00}(\mathbb{R}^{4}) \ni \varphi \mapsto A(\varphi)
= a(\sqrt{B}\, \overline{\widetilde{\varphi}}) + \eta a(\sqrt{B} \, \overline{\widetilde{\varphi}})^+ \eta
\]
(which by construction may be identified with the field (\ref{op-distr-realA'}), as by construction
$a'(\widetilde{\varphi}) = a(\sqrt{B} \, \widetilde{\varphi})$,
$a'(\widetilde{\varphi})^+ = a(\sqrt{B} \, \widetilde{\varphi})^+$) has the local transformation law
\begin{equation}\label{proof-fourth}
\bold{U}_{a,\alpha} A(\varphi) \bold{U}_{a,\alpha}^{-1} = A(\varphi'),
\,\,\, \varphi'(x) = \Lambda(\alpha)^T \, \varphi(x\Lambda(\alpha^{-1}) - a),
\end{equation}
and fulfills the massless wave equation.
Finally, we prove that (\ref{Pauli-Jordan}) defines the distribution
$ig^{\mu \nu} D_0 (x- y)$, with $D_0$ equal to the Pauli-Jordan function.

This approach has several advantages. First we are dealing with the ordinary annihilation and creation operator valued
distributions (generalized operators) $a^\mu(\vec{p}), a^\mu(\vec{p})^+$ in the momentum picture together
with the Gupta-Bleuler operator $\eta = \Gamma(\mathfrak{J}_{\bar{p}})$, where $\mathfrak{J}_{\bar{p}}$ equal to the operator acting in
$L^2(\mathbb{R}^3; \mathbb{C}^4)$ as the operator of multiplication by the constant matrix (\ref{J-barp}),
which is customary in the physical literature. It is therefore better to construct the quantum local
electromagnetic four-potential field $A$ using these more standard tools, then construct the field from
the outset without indicating any interrelation with the existing formalism. The second advantage is
that using the white noise calculus we will be able to formulate and prove the Bogoliubov
\emph{Quantization Postulate for Free Quantum Fields}
(\cite{Bogoliubov_Shirkov}, \S 9.4, page 89 of the second ed.), as a
mathematical theorem. Bogoliubov and Shirkov used the \emph{Postulate} as a guiding rule in constructing free quantum fields
(including gauge fields). We then made a heavy use of this \emph{Postulate} in the latter part when constructing
the perturbation (deformation) of the undeformed spectral space-time tuple
$(\mathcal{A}, \mathcal{H}, D_\mathfrak{J}, \mathfrak{J}, D)$, constructed in the previous Subsections
\ref{e1} - \ref{VFforFreeFields}.
Third advantage is that the Wick product of generalized operators and the Berezin-type integrals of
generalized operators may be precisely constructed with the quantum white noise calculus (compare \cite{hida}, \cite{obata}, \cite{luo}),
giving the mathematical justification
to the formal manipulations with such integrals as are presented e.g. in the cited Bogoliubov and Shirkov book.
Thus, the existent white noise techniques reduce the whole problem to the appropriate construction of the
Gelfand triple $E \subset \oplus L^2(\mathbb{R}^3) \subset E^*$.

\subsection{Standard setup of white noise calculus}\label{white-setup}

For a real vector space $E$ we write $E_\mathbb{C}$ for its complexification. If $E$ is a topological vector space,
we always assume its dual $E^*$ to carry the strong dual topology, and the linear space $\mathscr{L}(E,F)$ of linear continuous maps $E \rightarrow F$ to carry the topology of uniform convergence on bounded sets. For topological
vector spaces $E$ and $F$ which are nuclear we always write $E \otimes F$ for the projective tensor product
$E \otimes_{\pi}F$,
i.e. the completion of the algebraic tensor product $E \otimes_{\textrm{alg}}F$ with respect to the
$\pi$-topology -- the strongest locally convex linear topology on $E \otimes_{\textrm{alg}}F$ such that
the canonical bilinear map $E \times F \rightarrow E \otimes_{\textrm{alg}}F$ is continuous; recall that for nuclear spaces the projective tensor product is an essentially unique construction and in particular the projective tensor
product coincides with the equicontinuous tensor product, which is false for linear spaces which are not nuclear.
Whenever $\mathcal{H}$, $\mathcal{H}'$ are Hilbert spaces we always write $\mathcal{H} \otimes \mathcal{H}'$
for their Hilbert space tensor product. Recall that for Hilbert spaces $\mathcal{H}$, $\mathcal{H}'$
their Hilbert space tensor product $\mathcal{H} \otimes \mathcal{H}'$,
their projective tensor product $\mathcal{H} \otimes_{\pi} \mathcal{H}'$ and their equicontinuous tensor product
$\mathcal{H} \otimes_{\varepsilon} \mathcal{H}'$ are all different
whenever both factors $\mathcal{H}$, $\mathcal{H}'$ have infinite dimension (in which case $\mathcal{H}$ and
$\mathcal{H}'$ are not nuclear vector spaces).

In what follows extensive use is made of the Gaussian measures
on real $\mathcal{H}_\mathbb{R}$ Hilbert spaces and the Minlos theorem for such measures. 
Reality in this construction is important. 

On the other hand the complex Hilbert spaces encountered in our proof have always naturally inscribed
complex structures being equal to the complexification of real Hilbert spaces of real valued (or direct sums
of real valued) square integrable functions on measure spaces of locally compact topological (or even differentiable)
manifolds. Therefore, when dealing with such complex Hilbert spaces $\mathcal{H}$ we always assume that
\[
\mathcal{H} = (\mathcal{H}_\mathbb{R})_\mathbb{C} = \mathcal{H}_\mathbb{R} \oplus i\mathcal{H}_\mathbb{R},
\]
where $\mathcal{H}_\mathbb{R}$ is a real Hilbert space $L^2(\mathscr{O}; \mathbb{R})$ (of $\mathbb{R}$-valued square integrable functions on a topological measure space $\mathscr{O}$), with the real canonical $\mathbb{R}$-bilinear
form $\langle \cdot, \cdot \rangle$ on $\mathcal{H}_{\mathbb{R}}^{*} \times \mathcal{H}_\mathbb{R}$, which by the Riesz' representation theorem can be identified with the inner product $(\cdot, \cdot)_0$ on
$\mathcal{H}_{\mathbb{R}} \times \mathcal{H}_\mathbb{R} \cong \mathcal{H}_{\mathbb{R}}^{*} \times \mathcal{H}_\mathbb{R}$. We thus assume that
$\mathcal{H} \times \mathcal{H}$ is equipped
with the natural $\mathbb{C}$-bilinear form $\langle \cdot, \cdot \rangle$ -- equal to the unique extension
of the natural $\mathbb{R}$-bilinear form $\langle \cdot , \cdot \rangle$ on
$\mathcal{H}_{\mathbb{R}}^{*} \times \mathcal{H}_\mathbb{R}$ to the complexification $\mathcal{H} \times \mathcal{H}
= (\mathcal{H}_\mathbb{R})_\mathbb{C} \times (\mathcal{H}_\mathbb{R})_\mathbb{C}$. Thus,
if $\overline{\xi}$ denotes the complex conjugation of $\xi \in \mathcal{H} = (\mathcal{H}_\mathbb{R})_\mathbb{C}$
induced by the natural complex structure of $\mathcal{H}$ as a complexification of $\mathcal{H}_\mathbb{R}$,
then for the inner product norm $| \cdot |$ associated with the strictly positive Hermitian sesquilinear inner product
$(\cdot, \cdot)$ on $\mathcal{H}$, we have
\[
| \xi |^2 = (\xi , \xi) = \langle \overline{\xi}, \xi \rangle.
\]

Let $L^2(\mathscr{O}; \mathbb{R})$ be a real separable Hilbert space of square integrable (classes) of functions
on a locally compact topological space $\mathscr{O}$ with a countably additive Radon regular measure
$\ud \mu_{{}_{\mathscr{O}}}$,
with the standard Hilbert space $L^2$-norm $| \cdot |_0$ and the canonical associated $\mathbb{R}$-bilinear
form $\langle \cdot, \cdot \rangle $ on $L^2(\mathscr{O}; \mathbb{R})^* \times L^2(\mathscr{O}; \mathbb{R})$.
We shall be mostly concerned with a Gelfand triple $E \subset L^2(\mathscr{O}; \mathbb{R}) \subset E^*$
constructed from a standard operator $A$ on $L^2(\mathscr{O}; \mathbb{R})$, in short
$(A, L^2(\mathscr{O}; \mathbb{R}))$. Here we call after \cite{obata} an operator $A$ on $L^2(\mathscr{O}; \mathbb{R})$
to be standard if the domain $\Dom A \subset L^2(\mathscr{O}; \mathbb{R})$ of $A$ contains a complete orthonormal
basis $\{e_j\}_{j =0, 1, \ldots}$ for $L^2(\mathscr{O}; \mathbb{R})$ such that
\begin{enumerate}
\item[(A1)]
$Ae_j = \lambda_j e_j$ for $\lambda_j \in \mathbb{R}$;

\item[(A2)]
$1 < \lambda_0 \leq \lambda_1 \leq \lambda_2 \leq \ldots \rightarrow + \infty$;
\item[(A3)]
\[
\| A^{-1} \|_{H.S.} = \Big(\sum \limits_{j = 0}^{+\infty} {\lambda_{j}}^{-2}\Big)^{1/2} < + \infty,
\]
\end{enumerate} 
where $\| \cdot \|_{H.S.}$ stands for the Hilbert-Schmidt operator norm. In particular
\[
0 < \rho \overset{\textrm{df}}{=} {\lambda_0}^{-1} = \| A^{-1} \| < 1.
\]
For any $m \in \mathbb{Z}$ we define $E_m$ to be completion of $\Dom A^m$ with respect to the norm
\[
| \xi |_m = |A^m \xi |_0, \,\,\, \xi \in \Dom A^m,
\]
where for $m <0$, $\Dom A^m = H = L^2(\mathscr{O}; \mathbb{R})$. In this way we obtain a chain of Hilbert spaces
$\{ E_m \}_{m \in \mathbb{Z}}$ with inner products
$(\cdot , \cdot )_m \overset{\textrm{df}}{=} (A^m \cdot, A^m \cdot)_0$ and corresponding Hilbertian norms $| \cdot |_m
= \sqrt{(\cdot ,\cdot)_m}$, joined by natural topological inclusions
\[
\ldots E_m \subset \ldots \subset E_q \subset \ldots \subset H = E_0 = L^2(\mathscr{O}; \mathbb{R})
\subset \ldots \subset E_{-q} \subset E_{-m} \subset \ldots
\]
for $0 \leq q \leq m$.
We have the following theorem
\begin{twr*}
If $A$ is a standard operator on a Hilbert space $H = L^2(\mathscr{O}; \mathbb{R})$, then the Hilbertian
norms are compatible in the sense Gelfand-Shilov, $E = \cap_{m\geq 0} E_m$ with the countable Hilbert space topology
defined by the countable system of norms $\{ | \cdot |_m\}_{m \in \mathbb{N}}$, equal to the projective limit topology
of the system of Hilbert spaces $E_m$, is a countably Hilbert nuclear Fr\'echet space. The dual $E^*$ equal as a linear set $E^* = \cup_{m \in \mathbb{N}} E_{-m} = E^* = \cup_{m \in \mathbb{N}} {E_{m}}^*$ and equipped with the strong dual topology
is equal to the inductive limit topology $\ind \lim \limits_{m \rightarrow +\infty} {E_m}^* =
\ind \lim \limits_{m \rightarrow +\infty} E_{-m}$, and with the strong topology $b(E^*, E)$ on $E^*$ coinciding
with the Mackey topology $\tau(E^*, E)$ on $E^*$.
\end{twr*}
For the proof compare e.g. \cite{GelfandIV}, \cite{Berezansky}, \cite{obataJFA}. For the construction of the countably Hilbert space, conditions much weaker than (A1)-(A3) would be sufficient, even for maintaining nuclearity the condition
(A2) may be weakened, namely instead of (A2) it would be sufficient that $\inf \Sp A >0$, which among other things
assures existence of such domain for $A$ that $A$ will have dense range and bounded inverse. We have strengthened
the condition after Hida and Obata in order to make possible the lifting $(E) \subset \Gamma(H) \subset (E)^*$
to the boson Fock space $\Gamma(H)$ of the initial Gelfand triple $E \subset H \subset E^*$
after Hida, with the standard $(A, H)$ replaced by the likewise standard $(\Gamma(A), \Gamma(H))$, compare e.g.
\cite{obataJFA}. Similarly, the condition (A3) may be weakened while keeping the whole assertion of the last theorem,
namely it would be sufficient to assume that for some natural number $k$
\[
\| A^{-k} \|_{H.S.} = \Big(\sum \limits_{j = 0}^{+\infty} {\lambda_{j}}^{-2k}\Big)^{1/2} < + \infty,
\]
i.e that $A^{-k}$ is of Hilbert-Schmidt class.
We accept after Obata \cite{obata} the following notation
$\mathcal{S}_{A}(\mathscr{O}; \mathbb{R})$ for the nuclear space $E$ of the Gelfand triple $E \subset H = L^2(\mathscr{O}; \mathbb{R}) \subset E^*$, constructed as above from the standard operator $(A, H)$
on a separable Hilbert space $H = L^2(\mathscr{O}; \mathbb{R})$, and $\mathcal{S}_{A}(\mathscr{O}; \mathbb{R})^*$ for its
strong dual. Indeed, $\mathcal{S}_{A}(\mathscr{O}; \mathbb{R})$ plays in analysis the role of the nuclear
Schwartz space $\mathcal{S}$ of rapidly decreasing functions and the dual $\mathcal{S}_{A}(\mathscr{O}; \mathbb{R})^*$
plays the role of tempered distributions. Note in particular that each
$\xi \in \mathcal{S}_{A}(\mathscr{O}; \mathbb{R})$ determines a function on $\mathscr{O}$ up to
$\mu_{{}_{\mathscr{O}}}$-null set.

For the construction of the generalized operators, which realize the creation 
and the annihilation operator valued distributions, 
the Dirac evaluation functional plays a crucial role. We therefore restrict ourselves 
to situations in which the following Kubo and Takenaka conditions (H1)-(H3) are preserved.
\begin{enumerate}
\item[(H1)]
For each 
$\xi \in \mathcal{S}_{A}(\mathscr{O}; \mathbb{R}) \subset L^2(\mathscr{O}; \mu_{{}_{\mathscr{O}}} ; \mathbb{R})$ there exists a unique continuous function $\tilde{\xi}$
on $\mathscr{O}$ such that $\xi(p) = \tilde{\xi}(p)$, for $\mu_{{}_{\mathscr{O}}}$-a.e. $p \in \mathscr{O}$.
In this case we identify each $\xi \in \mathcal{S}_{A}(\mathscr{O}, \mathbb{R})$ with its unique continuous 
representative without using the tilde $\sim$ sign. 
\item[(H2)]
For each $p \in \mathscr{O}$ the evaluation map $\delta_{p}$: $\mathcal{S}_{A}(\mathscr{O}; \mathbb{R}) \ni 
\xi \mapsto \xi(p)$ is a continuous functional on $\mathcal{S}_{A}(\mathscr{O}; \mathbb{R})$, i.e.
$\delta_p \in \mathcal{S}_{A}(\mathscr{O}; \mathbb{R})^*$.
\item[(H3)]
The map $ \mathscr{O} \ni p \mapsto \delta_p \in \mathcal{S}_{A}(\mathscr{O}; \mathbb{R})^*$
is continuous.
\end{enumerate}  

Let $A$ be any essentially self adjoint operator on a Hilbert space $H$, with the domain $\Dom A$. We introduce
the operator
\begin{multline*}
d \Gamma_n (A) = A \otimes \boldsymbol{1} \otimes \ldots \otimes \boldsymbol{1}
+ \boldsymbol{1} \otimes A \otimes \boldsymbol{1} \otimes \ldots \boldsymbol{1} \ldots
+ \boldsymbol{1} \otimes \ldots \otimes \boldsymbol{1} \otimes A \\
= \sum \limits_{k = 0}^{n-1} \boldsymbol{1}^{\otimes k} \otimes A \otimes \boldsymbol{1}^{\otimes(n-k-1)},
\end{multline*}
on the domain
\[
\Dom d \Gamma_n (A) = \Dom A \otimes_{\textrm{alg}} \ldots \otimes_{\textrm{alg}} \Dom A
= (\Dom A)^{\otimes_{\textrm{alg}} n},
\]
in the $n$-fold Hilbert space tensor product $H \otimes \ldots \otimes H = H^{\otimes n}$
which is likewise essentially self adjoint in $H^{\otimes n}$, which remains essentially self adjoint if we replace the $n$-fold
tensor products by symmetrized or anti-symmetrized $n$-fold tensor products, compare e.g. Thm. VIII.3 and its Corollary
in \cite{Reed_Simon}. In particular
\[
d \Gamma_2 (A) = A \otimes \boldsymbol{1} + \boldsymbol{1} \otimes A, \,\,\,
\Dom ( \Gamma_2 (A)) = \Dom A \otimes_{\textrm{alg}} \Dom A
\]
is ess. self adjoint on $H \otimes H$.

Recall that we have the following propositions
\begin{prop*}
Let for $i = 1,2$, $\mathscr{O}_i$ be locally compact topological spaces with Borel measures
$\mu_{{}_{\mathscr{O}_i}}$. Let $A_i$ be standard operators on
$H_i = L^2(\mathscr{O}_i; \mu_{{}_{\mathscr{O}_i}} ; \mathbb{R})$ with domains $\Dom A_i$ respectively.
Then $A_1 \otimes A_2$ is a standard operator on
the Hilbert space tensor product $H_1 \otimes H_2$ with domain $\Dom A_1 \otimes_{\textrm{alg}} \Dom A_2$
and
\[
\mathcal{S}_{A_1 \otimes A_2}(\mathscr{O}_1 \times \mathscr{O}_2; \mathbb{R}) =
\mathcal{S}_{A_1}(\mathscr{O}_1; \mathbb{R}) \otimes \mathcal{S}_{A_2}(\mathscr{O}_2; \mathbb{R}),
\]
under the identification $L^2(\mathscr{O}_1, \mu_{{}_{\mathscr{O}_1}}; \mathbb{R}) \otimes
L^2(\mathscr{O}_2, \mu_{{}_{\mathscr{O}_2}}; \mathbb{R}) = L^2(\mathscr{O}_1 \times \mathscr{O}_2,
\mu_{{}_{\mathscr{O}_1}} \times \mu_{{}_{\mathscr{O}_1}}; \mathbb{R})$; and where the tensor product of the nuclear spaces on the right is the projective tensor product (equal to the equicontinuous tensor product).
If moreover the nuclear spaces $\mathcal{S}_{A_i}(\mathscr{O}_i; \mathbb{R})$ preserve the Kubo-Takenaka
conditions (H1)-(H3) then the projective tensor product
$\mathcal{S}_{A_1}(\mathscr{O}_1; \mathbb{R}) \otimes \mathcal{S}_{A_2}(\mathscr{O}_2; \mathbb{R})
= \mathcal{S}_{A_1 \otimes A_2}(\mathscr{O}_1 \times \mathscr{O}_2; \mathbb{R})$ also preserves the conditions
(H1)-(H3) of Kubo-Takenaka.
\end{prop*}
\begin{prop*}
Let $(A, H = L^2(\mathscr{O}, \mu_{{}_{\mathscr{O}}}; \mathbb{R})$ be standard, so that the construction of the corresponding countably Hilbert nuclear space $\mathcal{S}_{A}(\mathscr{O}; \mathbb{R})$ and the Gelfand triple
$\mathcal{S}_{A_1}(\mathscr{O}_1, \mathbb{R}) \subset H \subset \mathcal{S}_{A_1}(\mathscr{O}_1; \mathbb{R})^*$
is possible.
Then $(d \Gamma_2 (A), H \otimes H)$ is standard and fulfils (A1)-(A3), so that the countably Hilbert nuclear Fr\'echet space $\mathcal{S}_{d \Gamma_2 (A)}(\mathscr{O} \times \mathscr{O}; \mathbb{R})$ and the corresponding Gelfand triple
can be constructed, and moreover
\[
\mathcal{S}_{d \Gamma_2 (A)}(\mathscr{O} \times \mathscr{O}; \mathbb{R}) =
\mathcal{S}_{A \otimes A}(\mathscr{O} \times \mathscr{O}; \mathbb{R}) =
\mathcal{S}_{A}(\mathscr{O}, \mathbb{R}) \otimes \mathcal{S}_{A}(\mathscr{O}; \mathbb{R}),
\]
where $\mathcal{S}_{A}(\mathscr{O}; \mathbb{R}) \otimes \mathcal{S}_{A}(\mathscr{O}; \mathbb{R})$ on the right-hand side stands for the projective tensor product, equal to the equicontinuous tensor product, as
the space $\mathcal{S}_{A}(\mathscr{O}; \mathbb{R})$ is nuclear. If moreover the nuclear space
$\mathcal{S}_{A}(\mathscr{O}; \mathbb{R})$ preserves the conditions (H1)-(H2) of Kubo-Takenaka,
then $\mathcal{S}_{d \Gamma_2 (A)}(\mathscr{O} \times \mathscr{O}; \mathbb{R})$ preserves the conditions
(H1)-(H3) of Kubo-Takenaka.
\end{prop*}
Similarly we have
\[
\mathcal{S}_{A_1 \otimes \boldsymbol{1} + \boldsymbol{1} \otimes A_2}(\mathscr{O}_1 \times \mathscr{O}_2; \mathbb{R}) =
\mathcal{S}_{A_1}(\mathscr{O}_1; \mathbb{R}_2) \otimes \mathcal{S}_{A_2}(\mathscr{O}; \mathbb{R}),
\]
for standard $A_1, A_2$, with the projective (and thus also equicontinuous) tensor product of the
nuclear spaces on the right.
Of course the same holds not only for $\mathbb{R}$-valued nuclear function spaces
$\mathcal{S}_{A}(\mathscr{O}; \mathbb{R})$ constructed from standard
$(A, H = L^2(\mathscr{O}, \mu_{{}_{\mathscr{O}}} ; \mathbb{R}))$ but likewise for $\mathbb{R}^n$-valued
nuclear function spaces
$\mathcal{S}_{A}(\mathscr{O}; \mathbb{R}^n)$ constructed from standard, i.e. fulfilling (A1)-(A2),
$(A, H = L^2(\mathscr{O}, \mu_{{}_{\mathscr{O}}} ; \mathbb{R}^n)) = \big( \,\oplus_{i = 1}^{n} A_i, \,
\oplus_{i = 1}^{n} L^2(\mathscr{O}, \mu_{{}_{\mathscr{O}}} ; \mathbb{R}) \, \big)$,
as they can be regarded as direct sums of nuclear $\mathbb{R}$-valued function spaces
\begin{equation}\label{splitS-space}
\mathcal{S}_{\oplus_{i = 1}^{n} A_i}(\sqcup \mathscr{O}; \mathbb{R})=
\mathcal{S}_{\oplus_{i = 1}^{n} A_i}(\mathscr{O}; \mathbb{R}^n)
= \bigoplus_{i=1}^{n} \mathcal{S}_{A_i}(\mathscr{O}; \mathbb{R}),
\end{equation}
constructed from the standard
\begin{multline*}
\big( \,\oplus_{i = 1}^{n} A_i,
\oplus_{i = 1}^{n} L^2(\mathscr{O}, \mu_{{}_{\mathscr{O}}} ; \mathbb{R}) \, \big)
= \big( \oplus A_i, L^2(\sqcup \mathscr{O}, \mu_{{}_{\sqcup \mathscr{O}}} ; \mathbb{R}) \, \big)
\\
=
\big( \oplus A_i, \oplus L^2(\mathscr{O}, \mu_{{}_{\mathscr{O}}} ; \mathbb{R}) \, \big)
= \big( \oplus A_i, L^2(\mathscr{O}, \mu_{{}_{\mathscr{O}}} ; \mathbb{R}^n) \, \big).
\end{multline*}
Here $\sqcup \mathscr{O}$ is the disjoint sum of $n$ copies of $\mathscr{O}$
with the natural measure $\mu_{{}_{\sqcup \mathscr{O}}}$ on $\sqcup \mathscr{O}$
equal to the measure $\mu_{{}_{\mathscr{O}}}$ on each copy $\mathscr{O}$. For the proof compare e.g. \cite{obataJFA}, \cite{obata.Cont.Version.Thm}.

Now let $E = \mathcal{S}_{A}(\mathscr{O}, \mathbb{R})$ be the nuclear space and $E \subset H 
\subset E^*$ be the corresponding Gelfand triple constructed from a 
standard $(A, H = L^2(\mathscr{O}, \mu_{{}_{\mathscr{O}}}; \mathbb{R}))$ on a real Hilbert space
$H = L^2(\mathscr{O}, \mu_{{}_{\mathscr{O}}}; \mathbb{R})$. By Minlos theorem, \cite{GelfandIV}, there exists a unique Gaussian measure $\mu$ on the space $ E^* = \mathcal{S}_{A}(\mathscr{O}, \mathbb{R})^*$ dual to 
$E = \mathcal{S}_{A}(\mathscr{O}, \mathbb{R})$ such that 
\[
\int \limits_{E^*} e^{i \langle F, \xi \rangle} \ud \mu (F)
= e^{-\frac{1}{2} |\xi|_{0}^{2}},
\,\,\, \xi \in E, \, F \in E^*,
\]
where $\langle \cdot, \cdot \rangle$ is the dual pairing between $E^* = \mathcal{S}_{A}(\mathscr{O}, \mathbb{R})^*$
and $E = \mathcal{S}_{A}(\mathscr{O}, \mathbb{R})$. 
Let $(L^2) = L^2(E^*, \mu; \mathbb{R})$ be the space of square integrable 
functions (white noise functionals in general nonlinear) on the Radon measure space 
$(E^* = \mathcal{S}_{A}(\mathscr{O}, \mathbb{R})^*, \mu)$
with the $L^2$-norm $\| \cdot \|_0$, $L^2$-inner product $((\cdot, \cdot))_0$, and the canonical $\mathbb{R}$-bilibear form $\langle \langle \cdot, \cdot \rangle \rangle $ on $(L^2)^* \times (L^2)$ naturally induced by the inner product 
$((\cdot, \cdot))_0$
via the Riesz representation theorem. Let $\Gamma(H)$ be the real boson Fock space over the real Hilbert space $H$;
and let $\Gamma(A)$ be the second quantized operator 
\[
\Gamma(A) = \bigoplus_{n=0}^{\infty} A^{\otimes n}
\]
on $\Gamma(H)$. Then by the Wiener-It\^o-Segal chaos decomposition theorem the Hilbert space 
$(L^2) = L^2(E^*, \mu; \mathbb{R})$ is naturally isomorphic (unitary equivalent)
to the bosonic Fock space $\Gamma(H)$ with the natural action of the second quantized operator
$\Gamma(A)$ on $(L^2) = L^2(E^*, \mu; \mathbb{R})$ given
by the natural isomorphism. In particular, we have the following
\begin{prop*}
If the operator $A$ is standard on $H= L^2(\mathscr{O}, \mu_{{}_{\mathscr{O}}}; \mathbb{R})$ then the operator
$\Gamma(A)$ is standard on $(L^2) = L^2(E^*, \mu; \mathbb{R})$,
so that the Gelfand triple 
\[
\mathcal{S}_{\Gamma(A)}\big(E^*\big) \subset 
L^2(E^*, \mu; \mathbb{R}) \subset
\mathcal{S}_{\Gamma(A)}\big(E^*\big)^*
\]
can be constructed. Let us denote the last (``second quantized'') Gelfand triple by
\[
(E) \subset (L^2) \subset (E)^*.
\] 
If moreover
the nuclear space $E = \mathcal{S}_{A}(\mathscr{O}, \mathbb{R})$ corresponding to the standard 
$(A, H= L^2(\mathscr{O}, \mu_{{}_{\mathscr{O}}}; \mathbb{R}))$ preserves the Kubo-Takenaka conditions (H1)-(H3),
then $\mathcal{S}_{\Gamma(A)}\big(E^*\big) = (E)$ also preserves the conditions (H1)-(H3) of Kubo-Takenaka. 
\end{prop*}
For the proof compare e.g.
\cite{obataJFA}, \cite{obata.Cont.Version.Thm}, \cite{obata}.

Note that the complexification $H_\mathbb{C}= L^2(\mathscr{O}, \mu_{{}_{\mathscr{O}}}; \mathbb{R})_\mathbb{C}$ of 
$H= L^2(\mathscr{O}, \mu_{{}_{\mathscr{O}}}; \mathbb{R})$ is equal 
$H= L^2(\mathscr{O}, \mu_{{}_{\mathscr{O}}}; \mathbb{C})$ and the complexification of the real Gelfand triple
$E \subset H \subset E^*$ gives a Gelfand triple $E_\mathbb{C} \subset H_\mathbb{C} \subset {E_{\mathbb{C}}}^*$
for the complex Hilbert space $H_\mathbb{C}$. Similarly, the complexification $(L^2)_\mathbb{C} 
= L^2(E^*, \mu; \mathbb{R})_\mathbb{C}$ of $(L^2) = L^2(E^*, \mu; \mathbb{R})$ is equal 
$(L^2)_\mathbb{C} = L^2(E^*, \mu; \mathbb{C})$ and because for the Fock spaces $\Gamma(H)$ and 
$\Gamma(H)_\mathbb{C}$ we have $\Gamma(H_\mathbb{C}) = \Gamma(H)_\mathbb{C}$, then the Wiener-It\^o-Segal
decomposition can be lifted over to the complex Fock space and by the complexification of the 
Gelfand triple $(E) \subset (L^2) \subset (E)^*$ we obtain a Gelfand triple
$(E)_\mathbb{C} \subset (L^2)_\mathbb{C} \subset (E)^*$ for the complex Fock space isomorphic to
$L^2)\mathbb{C} = L^2(E^*, \mu; \mathbb{C})$; compare e.g. \cite{obata-book}, \cite{hida}, \cite{HKPS}.

In what follows the natural bilinear form on $E^*\times E$ as well as its natural amplification 
to $(E^{\otimes n})^* \times (E^{\otimes n})$, and its natural extension to the 
$\mathbb{C}$-bilinear form on $(E_{{}_{\mathbb{C}}}^{\otimes n})^* \times (E_{{}_{\mathbb{C}}}^{\otimes n})$,
will be denoted by one and the same symbol $\langle \cdot, \cdot \rangle$. Similarly, for the natural
bilinear form on $((E)^{\otimes n})^* \times ((E)^{\otimes n})$ and its unique extension
to the $\mathbb{C}$-bilinear pairing on 
$((E)_{{}_{\mathbb{C}}}^{\otimes n})^* \times ((E)_{{}_{\mathbb{C}}}^{\otimes n})$ we will always write
$\langle \langle \cdot, \cdot \rangle \rangle$, so that 
\[
\begin{split}
|\phi|_{0}^{2} = (\phi, \phi)_0 = \langle \overline{\phi}, \phi \rangle, \,\,\, \phi \in E_\mathbb{C},
\\
\|\Phi\|_{0}^{2} = ((\phi, \phi))_0 = \langle \langle \overline{\Phi}, \Phi \rangle \rangle, 
\,\,\, \Phi \in (E)_\mathbb{C} \subset (L^2)_\mathbb{C},
\end{split}
\]
where $\overline{\phi}$ and $\overline{\Phi}$ is the complex conjugation given by the natural complex structure 
respectively in $H_\mathbb{C}$ and $(L^2)_\mathbb{C}$.

Now the key point is the use of generalized continuous operators $(E)_\mathbb{C} \rightarrow {(E)_\mathbb{C}}^*$
instead of staying within the Hilbert Fock space, and use the symbol theory for such operators, in particular Fock
expansions, worked out by Hida, Obata Saito and others. In particular $(E)_\mathbb{C}$
is a nuclear Fr\'echet algebra under pointwise multiplication (note that the elements
of $(E)$ and $(E)_\mathbb{C}$ are in a canonical way realized as functions)
and if we define after Hida the following operators $\partial_p$
\[
\partial_p \Phi \overset{\textrm{df}}{=}
\lim \limits_{\theta \rightarrow 0} \frac{\Phi(\vartheta + \theta \delta_p) - \Phi(\vartheta)}{\theta},
\,\,\,
\vartheta \in E^*, \, \Phi \in (E)_\mathbb{C},
\]
then it turns out that $\partial_{p}$ for each $p \in \mathscr{O}$ is a continuous derivation
$(E)_\mathbb{C} \rightarrow (E)_\mathbb{C}$, and all the more, a continuous operator
$((E)_\mathbb{C} \rightarrow (E)_\mathbb{C})$. By the canonical continuous inclusion
$(E)_\mathbb{C} \rightarrow {(E)_{\mathbb{C}}}^*$, $\partial_p$ can be naturally regarded as a continuous
operator $(E)_\mathbb{C} \rightarrow {(E)_{\mathbb{C}}}^*$. Its linear adjoint ${\partial_p}^*$ is likewise
a continuous operator $(E)_\mathbb{C} \rightarrow {(E)_{\mathbb{C}}}^*$, by the reflexivity of $(E)_\mathbb{C}$.
It turns out that the operators $\partial_p$ and ${\partial_p}^*$ realize the annihilation and creation
operators at a point $p \in \mathscr{O}$ and satisfy the canonical commutation relations, \cite{Hida1}, \cite{HKPS},
\cite{hida}, \cite{obata-book}.

Below we use this framework to construct the free quantum electromagnetic four-potential field. As we have already indicated
the correct one particle test space necessary for the construction of the field is not the Schwartz
space $\mathcal{S}(\mathbb{R}^3; \mathbb{C}^4)$, but the closed subspace $\mathcal{S}^0(\mathbb{R}^3; \mathbb{C}^4)$.
We construct the appropriate standard operator $A$ in $L^2(\mathbb{R}^3, \ud^3 \p; \mathbb{C}^4)$
in $L^2(\mathbb{R}^3,\ud^3 p; \mathbb{C}^4)$ so that
\[
\mathcal{S}_{A} (\mathbb{R}^3; \mathbb{C}^4) = \mathcal{S}^0(\mathbb{R}^3; \mathbb{C}^4).
\]
Because
\[
\mathcal{S}^0(\mathbb{R}^3; \mathbb{C}^4) = \bigoplus_{n=1}^{4} \mathcal{S}^0(\mathbb{R}^3; \mathbb{R})_\mathbb{C}
\]
and on the other hand we have the property (\ref{splitS-space}) it is sufficient that we construct
just a standard scalar operator $A''' = A^{(3)}$ on $H = L^2(\mathbb{R}^3, \ud^3 \p; \mathbb{R})$
such that
\[
\mathcal{S}_{A^{(3)}} (\mathbb{R}^3; \mathbb{R}) = \mathcal{S}^0(\mathbb{R}^3; \mathbb{R})
\]
and put
\[
A = \oplus A^{(3)} \,\,\, \textrm{on} \,\, L^2(\mathbb{R}^3, \ud^3 \p; \mathbb{R}^4) = \oplus L^2(\mathbb{R}^3, \ud^3 \p; \mathbb{R})
\]
so that
\[
\mathcal{S}^0(\mathbb{R}^3 ; \mathbb{C}^4) = \mathcal{S}_{A}(\mathbb{R}^3; \mathbb{R}^4)_{\mathbb{C}}
\]
(summation is over the four components of the functions in $\mathcal{S}_{A}(\mathbb{R}^3; \mathbb{R}^4)$
or respectively in $L^2(\mathbb{R}^3, \ud^3 \p; \mathbb{R}^4)$).
It is important to construct $\mathcal{S}^0(\mathbb{R}^3 ; \mathbb{C}^4)$ as a standard countably Hilbert nuclear space from a standard $\big(A = \oplus A^{(3)}, \oplus H)$ because in this situation construction of the ''second quantized Gelfand triple'' in Fock space is possible as well as the
application of the white noise technics. Moreover, it is important that
$\mathcal{S}^0(\mathbb{R}^3) = \mathcal{S}_{A'''}(\mathbb{R}^3) \subset \mathcal{H}'$ is invariant under the action of the {\L}opusza\'nski representation and its conjugation, so that each representor of these representations is a continuous map of $E = \mathcal{S}^0(\mathbb{R}^3) = \mathcal{S}_{A'''}(\mathbb{R}^3) \subset \mathcal{H}'$ into itself in the nuclear topology. Joining this fact with the results of Hida, Obata and Saito \cite{hida} we give a proof of the existence of the generators for these representations as well as a proof of the Bogoliubov-Shirkov quantization postulate.

In fact, we will have to construct the whole family of standard operators
$A^{(n)}$ in $L^2(\mathbb{R}^3, \ud^3 \p, \mathbb{R})$ such that 
\[
\mathcal{S}_{A^{(n)}} (\mathbb{R}^n; \mathbb{R}) = \mathcal{S}^0(\mathbb{R}^n; \mathbb{R}).
\]

It should be noted however that 
\[
\mathcal{S}^0(\mathbb{R}^n; \mathbb{R}) \otimes \mathcal{S}^0(\mathbb{R}^m; \mathbb{R})
\subset \mathcal{S}^0(\mathbb{R}^{n+m}; \mathbb{R})
\]
but  
\[
\mathcal{S}^0(\mathbb{R}^n; \mathbb{R}) \otimes \mathcal{S}^0(\mathbb{R}^m; \mathbb{R})
\neq \mathcal{S}^0(\mathbb{R}^{n+m}; \mathbb{R}).
\]

In particular
\[
\mathcal{S}^0(\mathbb{R}; \mathbb{R}) \otimes \mathcal{S}^0(\mathbb{R}; \mathbb{R})
\otimes \mathcal{S}^0(\mathbb{R}; \mathbb{R}) \neq
\mathcal{S}^0(\mathbb{R}^3; \mathbb{R})
\]
so that the single particle test space $E_\mathbb{C} = \mathcal{S}^0(\mathbb{R}^3; \mathbb{R}^4)_\mathbb{C}$ needed
for the construction of the electromagnetic four-potential field cannot be constructed by simple direct summation,
tensoring and complexification of the real scalar nuclear space $\mathcal{S}^0(\mathbb{R}; \mathbb{R})$
on $\mathbb{R}$. In particular $\mathcal{S}^0(\mathbb{R}; \mathbb{C}^4) \otimes \mathcal{S}^0(\mathbb{R}; \mathbb{C}^4)
\otimes \mathcal{S}^0(\mathbb{R}; \mathbb{C}^4)$ is much too small and is not invariant for the
{\L}opusza\'nski representation and its conjugation; in particular the representations are only densely defined and unbounded on $\mathcal{S}^0(\mathbb{R}; \mathbb{C}^4) \otimes \mathcal{S}^0(\mathbb{R}; \mathbb{C}^4)
\otimes \mathcal{S}^0(\mathbb{R}; \mathbb{C}^4)$.
This means that the appropriate standard operator $A^{(n)}$
on $L^2(\mathbb{R}^n, \ud^n \p ; \mathbb{R})$ need to be constructed separately for each dimension $n$,
in particular construction of $A^{(3)}$ is not reducible to simple tensoring of the operator $A^{(1)}$
in dimension $1$.

However, there is a common way of construction and investigation of the whole family of
nuclear spaces $\mathcal{S}_{A^{(n)}} (\mathbb{R}^n; \mathbb{R}) = \mathcal{S}^0(\mathbb{R}^n; \mathbb{R})$.
Namely, we reduce the investigation and construction of the multipliers, convolutors, continuous functionals and differential operators
on $\mathcal{S}_{A^{(n)}} (\mathbb{R}^n; \mathbb{R})$ to the properties of the multipliers, convolutors, functionals and differentiation operation on the nuclear space
\[
\mathcal{S}(\mathbb{R}; \mathbb{R}) \otimes \mathcal{S}_{\Delta_{\mathbb{S}^{n-1}}}(\mathbb{S}^{n-1};\mathbb{R})
= \mathcal{S}_{H_{(1)}
\otimes \boldsymbol{1} + \boldsymbol{1} \otimes \Delta_{\mathbb{S}^{n-1}}}(\mathbb{R} \times \mathbb{S}^{n-1};\mathbb{R}).
\]
We do it for $\mathcal{S}_{A^{(n)}} (\mathbb{R}^n; \mathbb{R})$
before the proof of the equality $\mathcal{S}_{A^{(n)}} (\mathbb{R}^n; \mathbb{R}) =
\mathcal{S}^0(\mathbb{R}^n; \mathbb{R})$ and in the reduction process just mentioned we use the following facts
1) first:
\[
\mathcal{S}_{H_{(1)}}(\mathbb{R}) = \mathcal{S}(\mathbb{R})
\]
and 2) second:
\[
A^{(n)} = U\big( H_{(1)}
\otimes \boldsymbol{1} + \boldsymbol{1} \otimes \Delta_{\mathbb{S}^{n-1}}\big)U^{-1}
\]
for a unitary operator
$U: L^2(\mathbb{R}^n) \rightarrow L^2(\mathbb{R} \times \mathbb{S}^{n-1})$, $n>1$, so that
\begin{multline*}
\mathcal{S}_{A^{(n)}}(\mathbb{R}^n) = \mathcal{S}_{U\big( H_{(1)}
\otimes \boldsymbol{1} + \boldsymbol{1} \otimes \Delta_{\mathbb{S}^{n-1}}\big)U^{-1}}(\mathbb{R}^n)
= U \Big( \mathcal{S}_{H_{(1)}
\otimes \boldsymbol{1} + \boldsymbol{1} \otimes \Delta_{\mathbb{S}^{n-1}}}(\mathbb{R} \times \mathbb{S}^{n-1}) \Big)U^{-1}
\end{multline*}
is isomorphic to
\[
\mathcal{S}_{H_{(1)}
\otimes \boldsymbol{1} + \boldsymbol{1} \otimes \Delta_{\mathbb{S}^{n-1}}}(\mathbb{R} \times \mathbb{S}^{n-1};\mathbb{R})
\]
where $H_{(n)}$ is the self adjoint extension of
\[
H_{(n)} = \Gamma_{n}(H_{(1)}) = - \Delta_{\mathbb{R}^n} + r^n + n, \,\,\,
H_{(1)} = - \frac{d^2}{dp^2} + p^2 +1,
\]
i.e. (the double) of the $n$-dimensional quantum harmonic oscillator Hamiltonian. In particular the spectra (counting with multiplicity) of the operators
$A^{(n)}$ and $H_{(1)}
\otimes \boldsymbol{1} + \boldsymbol{1} \otimes \Delta_{\mathbb{S}^{n-1}}$ are identical for each dimension $n>1$
(definition of $A^{(1)}$ is different and is not unitarily equivalent to $H_{(1)}$, but the asymptotics of the spectra of the operators $A^{(1)}$ and $H_{(1)}$ are sufficiently similar). In fact the whole point is to construct
$A^{(n)}$ through a construction of the corresponding complete orthonormal systems using the Szeg\"o-von Neumann method
in such a manner that the asymptotics of the spectrum of $A^{(n)}$ is close enough to the asymptotics of the
spectrum of the operator $H_{(1)}
\otimes \boldsymbol{1} + \boldsymbol{1} \otimes \Delta_{\mathbb{S}^{n-1}}$ (and close enough to the asymptotics of the quantum harmonic oscillator Hamiltonian operator $H_{(1)}$ for $n=1$) for each dimension $n >1$.
In case $n = 1$ the nuclear space $\mathcal{S}_{A^{(1)}}(\mathbb{R})$ is isomorphic to the direct sum of two copies
of $\mathcal{S}(\mathbb{R})$ (corresponding to the fact that the $0$-sphere $\mathbb{S}^{0} \subset \mathbb{R}$ consists of just two points), but in this case when $A^{(1)}$ is not unitary equivalent to $H_{(1)}$ each eigenvalue of $H_{(1)}$ is
at the same time an eigenvalue of $A^{(1)}$ but appears with multiplicity two in $\Sp A^{(1)}$. The asymptotics of the spectra of $A^{(1)}$
and $H_{(1)}$ are still close enough to each other for allowing the reduction of the problem of investigation of multipliers, convolutors, or more general continuous operators and continuous functionals on $\mathcal{S}_{A^{(1)}}(\mathbb{R})$
to the problem of determination multipliers, convolutors, $\ldots$ of the ordinary
Schwartz space $\mathcal{S}_{H_{(1)}}(\mathbb{R}) = \mathcal{S}(\mathbb{R})$.

Then we show that for $\xi \in \mathcal{S}_{A^{(n)}} (\mathbb{R}^n; \mathbb{R}) = \mathcal{S}^0(\mathbb{R}^n; \mathbb{R})$ the restriction to the cone $(p_1)^2 - (p_1)^2 - \ldots (p_{n})^2 = 0$, $p_1>0$ (or $p_1<0$) in $\mathbb{R}^n$
gives a map $\mathcal{S}^0(\mathbb{R}^n; \mathbb{R}) \rightarrow \mathcal{S}^0(\mathbb{R}^{n-1}; \mathbb{R})$
which is a continuous map for the nuclear topologies. Note in particular that for the ordinary Schwartz test spaces 
$\mathcal{S}(\mathbb{R}^n)$ and $\mathcal{S}(\mathbb{R}^{n-1})$ this is false.

Thus in our presentation we give the full analysis of the one dimensional case 
$\mathcal{S}^0(\mathbb{R}) = \mathcal{S}_{A^{(1)}}(\mathbb{R})$, then we give the full construction
of the respective standard $A^{(n)}$. Finally the identity of the spectra of $A^{(n)}$ and $H_{(1)} \oplus 
\boldsymbol{1} + \boldsymbol{1} \oplus \Delta_{\mathbb{S}^{n-1}}$ (resp. sufficient similarity of the asymptotic 
behavior of the spectra of $A^{(1)}$ and $H_{(1)}$) allows us to reduce the investigation 
of $\mathcal{S}_{A^{(n)}}(\mathbb{R}^n)$ to the case $\mathcal{S}(\mathbb{R})$ 
and $\mathcal{S}_{\Delta_{\mathbb{S}^{n-1}}}(\mathbb{S}^{n-1})
= \mathscr{C}^{\infty}(\mathbb{S}^{n-1})$. 

All the nuclear spaces $\mathcal{S}_{A^{(n)}} (\mathbb{R}^n, \mathbb{R}) = \mathcal{S}^0(\mathbb{R}^n, \mathbb{R})$
have the additional symmetry, resembling that of inversion in the complex plane, interchanging the point at infinity with 
the distinguished zero point at which all derivatives af all elements of these spaces vanish. This additional symmetry is absent in the ordinary Schwartz test space. 
The paradox is that the one dimensional case $A^{(1)} = A'$ is the most subtle case as the distinguished zero point
in $\mathbb{R}$ at which all derivatives of all elements of $\mathcal{S}^0(\mathbb{R})$ vanish dissect
the whole space $\mathbb{R}$ into disjoint peaces, which is not the case in higher dimensions, where
$\mathbb{R}^n \diagdown \{0\}$, $n>1$, is connected.

\subsection{The space $\mathcal{S}_{A^{(1)}}(\mathbb{R})$. Construction of $A^{(1)} = A'$}\label{dim=1}

We construct now the essentially self adjoint differential operator $A'$ on $L^2(\mathbb{R}; \mathbb{R})$
with the indicated properties which serves to construct the Gelfand triple. 
We construct in fact a scalar operator $A'$ on $L^2(\mathbb{R})$, 
with $A'^{-1}$ being compact of Hilbert-Schmidt class
and the Gelfand triple $E \subset L^2(\mathbb{R}) \subset E^*$ corresponding to it with $E = \mathcal{S}_{A'}(\mathbb{R})$ being countably Hilbert nuclear as in \cite{GelfandIV} or \cite{Hida1}, \cite{HKPS}, \cite{obata-book},
with the properties such that the 
operators of multiplication by the following functions $p \mapsto f(p)$: $f(p) = p - p^{-1}$,
$f(p) = |p|$, $f(p) = |p|^{-1/2}$, $f(p) = |p|^{1/2}$, and the operator of differentiation are all continuous
as operators $E \to E$ with the nuclear topology on $E$. In this Subsection we prove Lemmas used in all higher 
dimensions in the Subsection which are to follow.
The proof that $\mathcal{S}^0(\mathbb{R}) = \mathcal{S}_{A^{(1)}}(\mathbb{R})$ we postpone to the following Subsections
where a general proof of the equality $\mathcal{S}^0(\mathbb{R}^n) = \mathcal{S}_{A^{(n)}}(\mathbb{R}^n)$ for all dimensions $n$ will be given.

In constructing $A'$ we start with the construction of an orthonormal system $\{u_n , u'_n\}_{n\in \mathbb{N}}$
of functions $u_n ,u'_{n}\in \mathcal{S}^0 (\mathbb{R})$ which is complete in $L^2(\mathbb{R})$.
In order to construct them consider the following smooth double covering map $p \mapsto t(p): \mathbb{R} \backslash \{0\} \to \mathbb{R}$,
given by the formula $t(p) = p - p^{-1}$.
\begin{center}
\begin{tikzpicture}[yscale=1]

\draw[thin, ->] (-2,0) -- (2,0);
\draw[thin, ->] (0,-2) -- (0,2);
\draw[thin, domain=-2:2] plot(\x, {\x});
\draw[ultra thick, domain=-2.5:-0.3] plot (\x, {\x - pow(\x,-1)});
\draw[ultra thick, domain= 0.3: 2.5] plot (\x, {\x - pow(\x,-1)});
\node [right] at (2,-0.2) {$p$};
\node [left] at (-0.3,2.3) {$t(p)$};
\end{tikzpicture}
\end{center}
Because the map $p \mapsto t(p)$ does not preserve the Lebesgue measure on $\mathbb{R}$ then the transformation
$f \mapsto g$, which the function $t \mapsto f(t)$ belonging to $L^2(\mathbb{R})$ transforms into the
function $p \mapsto g(p) = f(t(p))$ is not isometric from $L^2(\mathbb{R})$ into $L^2(\mathbb{R})$.
But the non invariance of the Lebesgue measure under $p \mapsto t(p)$ may be compensated for by the additional factor
equal to the square root of the Radon-Nikodym derivative of the transformed Lebesgue measure with respect to the original
Lebesgue measure and the $2$-valuedness may be compensated by the factor $2^{-1/2}$, such that the transform
\begin{equation}\label{von-neumann-trick}
f \mapsto Uf, \,\,\,
\textrm{with} \,\,\,
Uf(p) = \sqrt{2}^{-1}(1 + p^{-2})^{1/2}f(t(p)),
\end{equation}
becomes an isometric operator $L^2(\mathbb{R}) \to L^2(\mathbb{R})$:
\begin{multline*}
\int \limits_{-\infty}^{+\infty} \overline{Uf(p)}Uf(p) \, \ud p = 
\frac{1}{2}\int \limits_{-\infty}^{+\infty} \overline{f(t(p))}f(t(p)) \,
(1+p^{-2}) \, \ud p \\ = 
\frac{1}{2}\int \limits_{-\infty}^{0} \overline{f(t(p))}f(t(p)) \, \overbrace{(1+p^{-2}) \, \ud p}^\text{d$t$}
\, + \, \frac{1}{2}\int \limits_{0}^{+\infty} \overline{f(t(p))}f(t(p)) \, \overbrace{(1+p^{-2}) \, \ud p}^\text{d$t$} \\ =
\frac{1}{2}\int \limits_{-\infty}^{+\infty} \overline{f(t)}f(t) \, \ud t
\, + \, \frac{1}{2}\int \limits_{-\infty}^{+\infty} \overline{f(t)}f(t) \, \ud t \\
= \int \limits_{-\infty}^{+\infty} \overline{f(t)}f(t) \, \ud t.
\end{multline*} 
Of course in consequence of the double-covering character of the map $\mathbb{R} \backslash \{0\}
\ni p \mapsto t(p) \in \mathbb{R}$, the operator $U$ cannot be unitary (onto) operator.

In particular applying the isometric operator (\ref{von-neumann-trick}) to the system of Hermite
functions
\begin{multline*}
h_n(t) = \frac{1}{\sqrt{\pi^{1/2}n! 2^n}}e^{t^2/2}\bigg(\frac{d}{dt}\bigg)^n e^{-t^2} \\
= \frac{1}{\sqrt{\pi^{1/2}n! 2^n}}H_n(t) e^{-t^2/2}, \,\,\, \\
H_n \textrm{--Hermite polynomials}, n= 0,1,2, \ldots
\end{multline*}
we obtain an orthonormal (incomplete) system in $L^2(\mathbb{R})$:
\begin{multline}\label{u-n}
u_n(p) = Uh_n(p) = \sqrt{2}^{-1}(1 + p^{-2})^{1/2}h_n(t(p)) \\
= \sqrt{2}^{-1}(1 + p^{-2})^{1/2} U_n(p) e^{-(p^2 + p^{-2}) +1}, n= 0,1,2, \ldots,
\end{multline}
with the system of rational functions $U_n$ which can be obtained by application of the Gram-Schmidt
orthonormalization process to the set of linearly independent functions
$1, p-p^{-1}, (p-p^{-1})^2, (p-p^{-1})^3, \ldots$ with respect to measure $w(p) \ud p$ on $\mathbb{R}$
with the weight function
\[
w(p) = \frac{1}{2}(1+p^{-2})e^{-p^2 - p^{-2} +1}.
\]
It is easily checked that $u_n \in \mathcal{S}^0(\mathbb{R})$. Because we know the simple essentially self adjoint
differential operator for which the Hermite functions provide a complete system of eigenvectors
-- the one dimensional quantum oscillator Hamiltonian operator (in fact we add the unit operator in order
to reach $\inf \Sp H_{(1)} > 1$ and $\inf \Sp A' > 1$ but this trivial modification is unimportant here in
the construction of the complete system corresponding to $A'$)
\begin{equation}\label{oscillatorH}
H_{(1)} = - \bigg( \frac{d}{dt} \bigg)^2 + t^2,
\end{equation}
we can easily
construct the corresponding operator $A'$ for which the system (\ref{u-n}) is the system of eigenvectors, as the Radon-Nikodym derivative $(1+p^{-2})$ is relatively simple and smooth function on $\mathbb{R} \backslash \{0\}$. Namely, the operator is equal
\begin{multline}\label{Sturm-LiouvilleA'}
A' = \, - (1+p^{-2})^{-2}\bigg( \frac{d}{dp} \, \bigg)^{2} \,\, - \,\, 4 p^{-3}(1+p^{-2})^{-3} \, \frac{d}{dp} \\
\, + \big(p^2 - 2 + p^{-2} - 2 (1+ p^{-2})^{-4}p^{-6} - 3(1+ p^{-2})^{-3}p^{-4} \big).
\end{multline}
$A'$ is constructed in such a manner that we have
\[
A' u_n = \lambda_n u_n, \,\,\, \lambda_n = 2n +1
\]
with the eigenvalues $\lambda_n$ exactly the same as for the one dimensional quantum oscillator for the corresponding
Hermite functions $h_n$.

Now we find the missing eigenfunctions $u'_n$ of $A'$, not contained in the system $\{u_n\}$ computed above.
To this end consider the map $p \mapsto t(p) = p - p^{-1}$ now treated as an one-to-one map of the
disjoint sum $\mathbb{R}_+ \sqcup \mathbb{R}_-$ onto the disjoint sum $\mathbb{R} \sqcup \mathbb{R}$.
Again compensating for the measure non invariance by the square root of the Radon-Nikodym derivative
(now the factor $2^{-1/2}$ is absent as the map is one-to-one and onto) we obtain a unitary map
\[
\mathbb{U}: L^2(\mathbb{R}) \oplus L^2(\mathbb{R}) \ni
f_1 \oplus f_2 \mapsto g_+ \oplus g_-
\in L^2(\mathbb{R}_+) \oplus L^2(\mathbb{R}_+),
\]
which the pair of functions $(f_1 , f_2) \in L^2(\mathbb{R})$ sends into the following pair
of functions $(g_+, g_-) = \mathbb{U}(f_1,f_2)$ respectively in $L^2(\mathbb{R}_+), L^2(\mathbb{R}_-)$:
\[
\mathbb{U}\big(f_1 ,f_2\big)(p)
= \Big( 1_{{}_{\mathbb{R}_+}}(p) \, (1 + p^{-2})^{1/2} f_1 (t(p)), \,
1_{{}_{\mathbb{R}_-}}(p) \, (1 + p^{-2})^{1/2} f_2 (t(p)) \Big),
\]
where $1_{{}_{\mathbb{R}_+}}, 1_{{}_{\mathbb{R}_-}}$ are the characteristic functions on $\mathbb{R}$
of the subsets $\mathbb{R}_+ \subset \mathbb{R}$ and $\mathbb{R}_- \subset \mathbb{R}$ respectively.
In particular if $f_1 \in L^2(\mathbb{R})$ runs over a complete orthonormal system in $L^2(\mathbb{R})$
then the first component of $\mathbb{U}\big(f_1 ,f_2\big)$ runs over a complete orthonormal system
in $L^2(\mathbb{R}_+)$, and similarly if $f_2 \in L^2(\mathbb{R})$ runs over a complete orthonormal system in
$L^2(\mathbb{R})$ then the second component of $\mathbb{U}\big(f_1 ,f_2\big)$ runs over a complete orthonormal system
in $L^2(\mathbb{R}_-)$. In particular if $\{ t \mapsto h_n(t) \}_{n=0,1, \ldots}$ are Hermite functions, then
\begin{equation}\label{un+}
p \mapsto 1_{{}_{\mathbb{R}_+}}(p) \, (1 + p^{-2})^{1/2} h_n (t(p)) = u^{+}_{n} (p)
\end{equation}
is a complete orthonormal system of functions in $L^2(\mathbb{R}_+)$, and
\begin{equation}\label{un+}
p \mapsto 1_{{}_{\mathbb{R}_-}}(p) \, (1 + p^{-2})^{1/2} h_n (t(p)) = u^{-}_{n} (p)
\end{equation}
is a complete orthonormal system in $L^2(\mathbb{R}_-)$. It is easily seen that the following
extensions $u^{0+}_n, u^{-0}_n$ of the functions $u^{+}_{n} (p), u^{-}_{n} (p)$ to the whole real axis
\[
u^{0+}_n = \left\{ \begin{array}{ll}
0, & p \leq 0, \\
u^{+}_{n} (p), & p > 0
\end{array} \right. \,\,\,\, \text{and} \,\,\,\,
u^{-0}_n = \left\{ \begin{array}{ll}
u^{-}_{n} (p), & p < 0, \\
0, & p \geq 0
\end{array} \right.
\]
belong to $\mathcal{S}^{0}(\mathbb{R})$, and that $\{u^{0+}_n, u^{-0}_n \}_{n=0,1,2,\ldots}$
is a complete orthonormal system in $L^2(\mathbb{R})$. It follows by construction that
\[
A' u^{0+}_n = \lambda_n u^{0+}_n, \,\,\, A' u^{-0}_n = \lambda_n u^{-0}_n, \,\,\,
u_n = 2^{-1/2}u^{-0}_n + 2^{-1/2}u^{0+}_n.
\]
Therefore each eigenvalue $\lambda_n$ is of multiplicity two, with both
$u^{0+}_n, u^{-0}_n$ being independent orthogonal eigenfunctions of $A'$ to the same eigenvalue
$\lambda_{n} = 2n +1$. In the sequel we will be using the following orthonormal system
$\{u_n , u'_n \}$ of eigenfunctions of $A'$:
\[
u_n = 2^{-1/2} u^{-0}_n + 2^{-1/2}u^{0+}_n, \,\,\,
u'_n = -2^{-1/2}u^{-0}_n + 2^{-1/2}u^{0+}_n,
\]
of course complete in $L^2(\mathbb{R})$.

It is indeed easily seen that the operator $A'$ maps the nuclear (perfect) space
$\mathcal{S}^0(\mathbb{R})$ into itself and remains symmetric when restricted
to $\mathcal{S}^0(\mathbb{R})$. Because $\mathcal{S}^0(\mathbb{R})$ is densely included into
the Hilbert space $L^2(\mathbb{R})$, then by the known theorem of Riesz and Sz\"okefalvy-Nagy,
\cite{Riesz-Szokefalvy} (p. 120 in Russian Ed. 1954), or \cite{GelfandIII}, p. 192, the operator
$A'$ can be extended to a selfadjoint operator on $L^2(\mathbb{R})$ -- as expected by its very construction.
Indeed, $A'$ by construction has the complete orthonormal system of eigenvectors all
belonging to the nuclear space $\mathcal{S}^0(\mathbb{R})$. Therefore, it is unitarily equivalent to
an operator which acts as multiplication by a locally measurable function $a'$ operator $M_{a'}$
on $L^2(M, \ud \mu)$ with the discrete measure space $M, \ud\mu$ (corresponding to the discrete spectrum of $A'$) on a dense domain $\textrm{Dom} \, A'$ corresponding to the elements of $\mathcal{S}^0(\mathbb{R})$.
As such it is essentially selfadjoint on $\textrm{Dom} \, A' = \mathcal{S}^0(\mathbb{R})$, i.e. possess
only one self adjoint extension.
Let us note the extension by the same symbol $A'$.

Presented method of constructing the complete orthonormal system $\{u_n , u'_n \}$ in $L^2(\mathbb{R})$
is well known and is attributed by Szeg\"o to von Neumann, \cite{Szego}, p. 108. 
Our slight modification of the von Neumann method by introducing
the intermediate double covering map and the corresponding isometric operator (\ref{von-neumann-trick}) in constructing the corresponding Sturm-Liouville operator with singular point at zero, may easily be extended to obtain solutions 
of the Sturm-Liouville problem with any number $n$ of singular points lying between $-\infty$ and $\infty$, 
with the use of the intermediate $(n+1)$-fold-covering maps and the corresponding isometry operators in obtaining eigenfunctions and spectra with generally 
uniform $n+1$ multiplicity.  

Because 
\[
\sum \limits_{n=0}^{\infty} \lambda_{n}^{-2} < + \infty
\] 
and each eigenvalue $\lambda_n$ of $A'$ has the same finite multiplicity (equal 2), then the operator $A'^{-1}$ 
is of Hilbert-Schmidt class, as desired. 

Now using the positive self adjoint operator $A'$ on $L^2(\mathbb{R}; \mathbb{R})$ we construct the Gelfand triple
$E \subset L^2(\mathbb{R}) \subset E^*$. Namely, for $k \in \mathbb{N}$ we put $E_k$ for the completion of the domain
$\Dom A'^k$ of $A'^k$ with respect to the norm $| \cdot |_k = | A'^k \cdot |_0$, where $| \cdot |_0$ is the ordinary Hilbert space $L^2$-norm in $L^2(\mathbb{R})$ (it is convenient to put $(\cdot,\cdot )_0$ for the inner product in $L^2(\mathbb{R})$).
It follows that $E_k$ is a Hilbert space with the norm
$| \cdot |_k = | A'^k \cdot |_0$, equal to the completion of the space $\mathcal{S}^0(\mathbb{R})$ with respect to
the norm $| \cdot |_k$. Let the dual space $E_{k}^{*}$ with the dual norm $| \cdot |_{-k}$ be
denoted by $E_{-k}$. The norms $| \cdot |_k$ are compatible in the sense of Gelfand-Shilov \cite{GelfandII}, which easily follows e.g. from the closeness of the graph of the self adjoint operator $A'$. Then the projective limit
$E = \cap_{k} E_k$ of $E_k$ is countably Hilbert nuclear Fr\'echet space and its dual $E^*$ with the strong topology is the inductive limit $E^* = \cup_{k} E_{-k}$ of $E_{-k}$,
compare \cite{GelfandIV}, \cite{Hida1} or \cite{obataJFA}, \cite{obata-book}, \cite{HKPS},
with the natural inclusion maps
\[
E \subset \ldots \ldots \subset E_k \subset \ldots \subset E_0 = L^2(\mathbb{R}) \subset \ldots
\subset E_{-k} \subset \ldots \ldots \subset E^*.
\]
In particular the completeness of $E$ follows from the equality $E = \cap_{k} E_k$ and the simple necessary and sufficient condition for completeness of a countably normed space given in \cite{GelfandII}. ChI \S 3.2.

Now let $\widetilde{\varphi} \in E \subset L^2(\mathbb{R})$. From the completeness 
of the orthonormal system $\{u_n, u'_n\}$
it follows that the series
\begin{equation}\label{CnC'n}
\widetilde{\varphi} = \sum \limits_n C_n(\widetilde{\varphi})u_n 
+ \sum \limits_n C'_n(\widetilde{\varphi})u'_n,
\end{equation}
where 
\begin{multline}
C_n(\widetilde{\varphi}) = (u_n, \widetilde{\varphi}) = \int \limits_{\mathbb{R}}
u_n(p) \widetilde{\varphi}(p) \, \ud p, \\
C'_n(\widetilde{\varphi}) = (u'_n, \widetilde{\varphi}) = \int \limits_{\mathbb{R}}
u'_n(p) \widetilde{\varphi}(p) \, \ud p,
\end{multline}
converges in $L^2(\mathbb{R})$. 

\begin{lem}\label{Cn-E-converge}
In this case, i.e. when $\widetilde{\varphi} \in E = \mathcal{S}_{A'}(\mathbb{R})$, 
it follows that the series (\ref{CnC'n})
is convergent in the nuclear topology of $E$.
\end{lem}

\qedsymbol \,

Compare \cite{Reed_Simon}, Appendix to Ch. V.3,
where the proof in the particular case of the nuclear space
$E = \mathcal{S}(\mathbb{R}) = \mathcal{S}_{H_{(1)}}(\mathbb{R})$ is outlined, however the same holds for the
general construction with oscillator Hamiltonian operator $H_{(1)}$ replaced with any standard operator $A'$,
in particular it holds true for $\mathcal{S}_{A'}(\mathbb{R})$ with our operator $A'$.

Indeed, having given the complete orthonormal system $\{u_n, u'_n\}$), we may define a self adjoint operator $A'^N$,
$N\in \mathbb{N}$ (in fact for $N \in \mathbb{Z}$ with $\Dom A'^N = L^2(\mathbb{R})$ for $N < 0$).
By construction (compare also the spectral theorem) the operator $A'^N$ is unitarily equivalent
to the operator $M_{{}_{a'^N}}$ of multiplication by a function $a'^N$ on the $L^2(\mathcal{M})$ space
of square measurable functions on the discrete measure space $\mathcal{M} = \{1,1',2,2', \ldots$, with $a' (n) = a'(n') = \lambda_n$, with the $L^2$ norm
\[
|b|^2 = \sum \limits_{n \in \mathbb{N}} |b(n)|^2 + \sum \limits_{n' \in \mathbb{N}} |b(n')|^2, \,\,\,
b \in L^2(\mathcal{M}).
\]
Because the domain $\Dom M_{{}_{a'^N}} \subset L^2(\mathcal{M})$ consists of all those sequences
$\{b(n), b(n')\}_{{}_{n,n' \in \mathbb{N} \sqcup \mathbb{N}}}$ on $\mathbb{N} \sqcup \mathbb{N}$ for which
\[
\sum \limits_{n \in \mathbb{N}} |a'^N(n)\, b(n)|^2 + \sum \limits_{n' \in \mathbb{N}} |a'^N(n')\, b(n')|^2 < + \infty,
\]
and because $A'^N: E \rightarrow E \subset L^2(\mathbb{R})$, then $A'^N \widetilde{\varphi} \in L^2(\mathbb{R})$
and $\widetilde{\varphi} \in \Dom A'^N$. Therefore,
\begin{equation}\label{ineq(*)}
\begin{split}
\sum \limits_n \lambda_{n}^{2N} C_n(\widetilde{\varphi})
+ \sum \limits_n \lambda_{n'}^{2N} C'_n(\widetilde{\varphi}) < +\infty \,\,\, \textrm{and}
\\
A'^N \widetilde{\varphi} = \sum \limits_n \lambda_{n}^{N} C_n(\widetilde{\varphi}) u_n
+ \sum \limits_n \lambda_{n'}^{N} C'_n(\widetilde{\varphi})u'_n.
\end{split}
\end{equation}
In particular
\[
\sup \limits_{n, n' \in \mathbb{N}} \{ \lambda_{n}^{N} |C_n(\widetilde{\varphi})|,
\lambda_{n'}^{N} |C'_n(\widetilde{\varphi})| \} < +\infty, \,\,\,
\textrm{for all} \,\,\, N \in \mathbb{N}.
\]
Let now
\[
\widetilde{\varphi}_{M} =
\sum \limits_{n=1}^{M} C_n(\widetilde{\varphi})u_n
+ \sum \limits_{n=1}^{M} C'_n(\widetilde{\varphi})u'_n.
\]
Then from (\ref{ineq(*)}) we get
\[
| A'^m (\widetilde{\varphi}_{M} - \widetilde{\varphi}_{L}) |_{{}_{L^2(\mathbb{R})}}^{2} =
\sum \limits_{n= M+1}^{L} \lambda_{n}^{2m} |C_n(\widetilde{\varphi})|^2
+ \sum \limits_{n=M+1}^{L} \lambda_{n'}^{2m} |C'_n(\widetilde{\varphi})|^2 \longrightarrow 0
\]
when $M, L \rightarrow 0$. Thus, $\{\widetilde{\varphi}_{M}\}_{{}_{M \in \mathbb{N}}} \subset E$
is a Cauchy sequence with respect to each of the norms $| \cdot |_m = | A'^m \cdot |_0
= | A'^m \cdot |_{{}_{L^2(\mathbb{R})}}^{2}$, $m \in \mathbb{N}$. Therefore,
$\{\widetilde{\varphi}_{M}\}_{{}_{M \in \mathbb{N}}}$ is a Cauchy sequence in $E$, \cite{GelfandII}.
Because $E= \cap_{k} E_m$ as a countably Hilbert space with compatible norms $| \cdot |_m$ is complete,
then there exists the limit point $\widetilde{\varphi}_0 \in E$ for the sequence $\{\widetilde{\varphi}_{M}\}_{{}_{M \in \mathbb{N}}}$, and all the more $\{\widetilde{\varphi}_{M}\}_{{}_{M \in \mathbb{N}}}$ converges
to $\widetilde{\varphi}_0$ in $E_0 = L^2(\mathbb{R})$ with respect to the $L^2$-norm $| \cdot |_0$,
\cite{GelfandII}. Because $\{\widetilde{\varphi}_{M}\}_{{}_{M \in \mathbb{N}}}$ converges in the $L^2$-norm
$| \cdot |_0$ in $E_0 = L^2(\mathbb{R})$ to $\widetilde{\varphi}$, then $\widetilde{\varphi}_0 = \widetilde{\varphi}$
and (\ref{CnC'n}) converges to $\widetilde{\varphi}$ in the nuclear topology of $E$. 

\qed

\begin{cor*}
The space $\mathcal{S}^0(\mathbb{R})$ is dense in $\mathcal{S}_{A^{(1)}}(\mathbb{R}) = \mathcal{S}_{A'}(\mathbb{R})$
in the nuclear topology of $\mathcal{S}_{A'}(\mathbb{R})$.
\end{cor*}
\qedsymbol \,
Indeed, the elements $u_n$ and $u'_n$ of the complete system of eigenfunctions of the operator
$A^{(1)} = A'$ belong to $\mathcal{S}^0(\mathbb{R})$ by the very construction. Our corollary
now follows from Lemma \ref{Cn-E-converge}.
\qed

Because for $\widetilde{\varphi} \in E$ the series (\ref{CnC'n}) is convergent in the
nuclear topology of $E$, and by construction the operator $A'$ is continuous as an operator
$E \to E$ in the nuclear topology of $E$, compare \cite{GelfandIV}, p. 109, it follows that for any 
$N \in \mathbb{N}$ 
\[
A'^N \widetilde{\varphi} = \sum \limits_n \lambda_{n}^{N} C_n(\widetilde{\varphi})u_n 
+ \sum \limits_n \lambda_{n}^{N} C'_n(\widetilde{\varphi})u'_n.
\] 
Therefore, the norm $| \widetilde{\varphi}|_{N}$ squared is equal
\[
| \widetilde{\varphi}|_{N}^{2} = \sum \limits_n \lambda_{n}^{2N} |C_n(\widetilde{\varphi})|^2
+ \sum \limits_n \lambda_{n}^{2N} |C'_n(\widetilde{\varphi})|^2.
\]

Now before we prove the continuity of the formerly indicated maps as operators $E \to E$
in the nuclear topology, let us recall a property of the Gelfand's construction of 
$E \subset L^2(\mathbb{R}) \subset E^*$ connected with a unitary change of the 
positive symmetric differential 
operator $A'$ and with its invariant sub-spaces.

Namely, if we replace the operator $A'$ with an operator $U_0 A' U_{0}^{-1}$ on $\mathcal{H} = U_0 (L^2(\mathbb{R}))$,
unitrarily equivalent to $A'$, and will similarly construct the Gelfand triple $E' \subset \mathcal{H}
\subset E'^{*}$ corresponding to $U_0 A' U_{0}^{-1}$ then by the very construction, the operators $U_0$
and $U_{0}^{-1}$ are continuous as operators respectively $E \to E'$ and $E' \to E$
with the nuclear topology.
We will use the property in this and in the following Subsections in passing from momentum to position picture
with the unitary operator $U_0$ equal to the ordinary Fourier transform $\mathscr{F}$.
But having this fact in mind we illustrate this property using another operator $U_0$,
which allows us to make a slightly closer insight into the structure of the nuclear space $E$.
Namely, the Gelfand triple
\[
\mathcal{S}(\mathbb{R}) \subset L^2(\mathbb{R}) \subset \mathcal{S}(\mathbb{R})^{*}
\]
with the nuclear Schwartz space $\mathcal{S}(\mathbb{R})$ and its dual $\mathcal{S}(\mathbb{R})^{*}$,
equal to the space of tempered distributions may be constructed in exactly the same manner if we use the
operator $H_{(1)} = - d^2/dt^2 + t^2 + 1$ equal to the one dimensional oscillator Hamiltonian operator
(\ref{oscillatorH}) instead of
the operator\footnote{Recall that we add the unit operator to the operator (\ref{Sturm-LiouvilleA'}) in order to
achieve $\inf \Sp A' > 1$.} $A'$ given by (\ref{Sturm-LiouvilleA'}), and with the Hermite functions $\{h_n\}$
instead of $\{u_n, u'_n\}$, compare \cite{GelfandIV}, \cite{HKPS}, p. 484, \cite{Reed_Simon},
\cite{Simon}, \cite{Hida1}, \cite{Hida2}, \cite{Hida3}. Now although the spectra of the operators
$H_{(1)}$ and $A'$ are identical they are nevertheless not unitarily equivalent, because each $\lambda_n$
in the spectrum of $H_{(1)}$ has multiplicity one and the same eigenvalue $\lambda_n$ has multiplicity two
in the spectrum of $A'$. However, by construction the operator $A'$ has two orthogonal invariant sub-spaces
$E_{0I}$ and $E_{0II}$, the first spanned by $\{ u_n\}$ and the second by $\{u'_n\}$,
which together span the whole Hilbert space $L^2(\mathbb{R}) = E_{0I} \oplus E_{0II}$
and such that both restrictions $A'_I$ and $A'_{II}$ separately of the operator $A'$ to the invariant sub-spaces
$E_{0I}$, $E_{0II}$ are unitarily equivalent to the operator $H_{(1)}= - d^2/dt^2 + t^2 + 1$ 
given by\footnote{After addition of the unit operator.} (\ref{oscillatorH})
-- which is evident because each $A'_I , A'_{II}$ has the same discrete spectrum as the operator
$H_{(1)}$ with the multiplicity of each eigenvalue equal one.
Therefore, if we consider the Hilbert space $L^2(\mathbb{R}) \oplus L^2(\mathbb{R})$ with the self-adjoint operator
$H_{(1)} \oplus H_{(1)}$ with $H_{(1)}$ equal to (the self-adjoint extension of) (\ref{oscillatorH}), then the operator
\begin{equation}\label{UplusU'}
U_0 = U \oplus U'
\end{equation}
gives the unitary equivalence $U_0A'U_{0}^{-1} = H_{(1)} \oplus H_{(1)}$, where $U$ is given by (\ref{von-neumann-trick})
and is unitary if treated as operator $L^2(\mathbb{R}) \to U\big( L^2(\mathbb{R}) \big) = E_{0I}$;
and the unitary operator $U': L^2(\mathbb{R}) \to E_{0II}$ is defined as follows
\begin{equation}\label{U'f}
U'f(p) = ( 1_{{}_{\mathbb{R}_+}}(p) - 1_{{}_{\mathbb{R}_-}}(p)) \, \sqrt{2}^{-1}(1 + p^{-2})^{1/2}f(t(p)).
\end{equation}
In particular
\[
u'_n = U'h_n.
\]
Therefore the nuclear space $E$ is isomorphic to the direct sum $E_I \oplus E_{II}$ of nuclear closed sub-spaces
$E_I, E_{II}$ each isomorphic to the Schwartz space $\mathcal{S}(\mathbb{R})$ of rapidly decreasing functions
with the isomerism given by $U \oplus U'$.
This fact is frequently useful in checking if a concrete linear functional on $E$ is continuous
in the nuclear topology of $E$, i.e. if it actually belongs to the dual space $E^*$, because it reduces the
problem to checking if a given functional is in $\mathcal{S}(\mathbb{R})^*$. In particular, we use the fact
in a simple proof that the Dirac delta function is an element of $E^*$.

Now after this digression, let us back to the proof of the continuity of the operators of multiplication 
by the following functions $p \mapsto f(p)$: $f(p) = p - p^{-1}$,
$f(p) = |p|$, $f(p) = |p|^{-1/2}$, $f(p) = |p|^{1/2}$, and of the differentiation operator,
regarded as operators $E \to E$ with the nuclear topology on $E$.

In doing so we reduce the problem to the properties of Hermite functions, their connection to the
Schwartz space $\mathcal{S}(\mathbb{R})$, and several properties of the multipliers and convolutors
of the algebra $\mathcal{S}(\mathbb{R})$.   For this purpose we need a technical lemma. 

In order to simplify notation let us define for any function $f$, such that for 
any function $u_n, u'_n \in L^2(\mathbb{R})$ of the orthonormal system of eigenfunctions of $A'$, 
the function $fu_n$ and the function $fu'_n$ is in $L^2(\mathbb{R})$, the following
matrix elements of the operator $M_f$ of multiplication by the function $f$ in the basis
$\{ u_n, u'_n\}$:
\[
\begin{split}
\langle n |f(p) | m \rangle = \int \limits_{-\infty}^{+\infty} u_n(p) f(p) u_m(p) \, \ud p, \\
\langle n' |f(p) | m' \rangle = \int \limits_{-\infty}^{+\infty} u'_n(p) f(p) u'_m(p) \, \ud p, \\
\langle n |f(p) | m' \rangle = \int \limits_{-\infty}^{+\infty} u_n(p) f(p) u'_m(p) \, \ud p \\
\langle n' |f(p) | m \rangle = \int \limits_{-\infty}^{+\infty} u'_n(p) f(p) u_m(p) \, \ud p.
\end{split}
\]
And similarly the matrix elements of the operator of multiplication by the function $f$,
but in the orthonormal basis $\{h_n \}$ of Hermite functions (provided the Hermite functions
are in the domain of the operator) we denote by
\[
( n |f(t) | m ) = \int \limits_{-\infty}^{+\infty} h_n(t) f(t) h_n(t) \, \ud t.
\]   
Note in passing that 
\[
C'_n(f\widetilde{\varphi}) = \sum_m \langle n' |f(p) | m \rangle C_m(\widetilde{\varphi})
+ \sum_m \langle n' |f(p) | m' \rangle C'_m(\widetilde{\varphi}),
\]
\[
C_n(f\widetilde{\varphi}) = \sum_m \langle n |f(p) | m \rangle C_m(\widetilde{\varphi})
+ \sum_m \langle n |f(p) | m' \rangle C'_m(\widetilde{\varphi}).
\]
And similarly for Hermite functions
\[
C_{n}^{0}(f\widetilde{\varphi}) = \sum_m \langle n |f(t) | m \rangle C_{m}^{0}(\widetilde{\varphi}),
\]
where 
\[
C_{n}^{0}(\widetilde{\varphi}) = (h_n, \widetilde{\varphi}) = \int \limits_{\mathbb{R}}
h_n(t) \widetilde{\varphi}(t) \, \ud t.
\]
And generally for two operators $Op_1$, $Op_2$ in $L^2(\mathbb{R})$, the matrix representing their composition 
$Op_1 \circ Op_2$ is equal to the matrix multiplication of the matrices corresponding respectively to $Op_1$ and $Op_2$, 
providing the basis elements $\{u_n , u'_n\}$ (resp. $\{h_n\}$) are in the domain of the
operators $Op_1$, $Op_2$ and $Op_1 \circ Op_2$. 

\begin{lem*}
\[
\langle n |p-p^{-1} | m \rangle = \langle n' |p-p^{-1} | m' \rangle
= \bigg(\frac{n+1}{2}\bigg)^{1/2} \delta_{m \,\, n+1} + \bigg(\frac{n}{2}\bigg)^{1/2} \delta_{m \,\,n-1},
\]
\[
\langle n | p-p^{-1} | m' \rangle = \langle n' |f(p) | m \rangle = 0.
\]
For each $N \in \mathbb{N}$ there exist $N^0$ and $c_N >0$ independent of $\widetilde{\varphi} \in E$, and
depending on the operator in question, i.e. on the
function $f(p)$ equal respectively $p$, $|p|, |p|^{-1/2}, |p|^{1/2}$, such that
\begin{multline*}
\sum_n \lambda_n^{2N} \bigg| \sum_{m_1,m_2} \langle n |f(p) | m_1 \rangle \langle n |f(p) | m_2 \rangle
C_{m_1}(\widetilde{\varphi}) C_{m_2}(\widetilde{\varphi}) \bigg| \\
< \, c_N \sum_n \lambda_n^{2(N+N^0)} | C_n(\widetilde{\varphi})|^2,
\end{multline*}
\begin{multline*}
\sum_n \lambda_n^{2N} \bigg| \, \sum_{m_1,m_2} \langle n |f(p) | m_1' \rangle \langle n |f(p) | m_2' \rangle
C'_{m_1}(\widetilde{\varphi}) C'_{m_2}(\widetilde{\varphi}) \bigg| \\
< \, c_N \, \sum_n \lambda_n^{2(N+N^0)} | C'_n(\widetilde{\varphi})|^2,
\end{multline*}
\begin{multline*}
\sum_n \lambda_n^{2N} \bigg| \sum_{m_1,m_2} \langle n' |f(p) | m_1 \rangle \langle n' |f(p) | m_2 \rangle
C_{m_1}(\widetilde{\varphi}) C_{m_2}(\widetilde{\varphi}) \bigg| \\
< \, c_N \, \sum_n \lambda_n^{2(N+N^0)} | C_n(\widetilde{\varphi})|^2,
\end{multline*}
\begin{multline*}
\sum_n \lambda_n^{2N} \bigg| \, \sum_{m_1,m_2} \langle n' |f(p) | m_1' \rangle \langle n' |f(p) | m_2' \rangle
C'_{m_1}(\widetilde{\varphi}) C'_{m_2}(\widetilde{\varphi}) \bigg| \\
< \, c_N \, \sum_n \lambda_n^{2(N+N^0)} | C'_n(\widetilde{\varphi})|^2;
\end{multline*}
similarly for the differentiation operator
\begin{multline*}
\sum_n \lambda_n^{2N} \Bigg| \sum_{m_1,m_2} \bigg\langle n \bigg| \frac{d}{dp} \bigg| m_1 \bigg\rangle \bigg\langle n
\bigg| \frac{d}{dp} \bigg| m_2 \bigg\rangle
C_{m_1}(\widetilde{\varphi}) C_{m_2}(\widetilde{\varphi}) \Bigg| \\
< c_N \, \sum_n \lambda_n^{2(N+N^0)} | C_n(\widetilde{\varphi})|^2,
\end{multline*}
\begin{multline*}
\sum_n \lambda_n^{2N} \Bigg| \, \sum_{m_1,m_2} \bigg\langle n \bigg|\frac{d}{dp} \bigg| m_1' \bigg\rangle \bigg\langle n
\bigg|\frac{d}{dp} \bigg| m_2' \bigg\rangle
C'_{m_1}(\widetilde{\varphi}) C'_{m_2}(\widetilde{\varphi}) \Bigg| \\
< c_N \, \sum_n \lambda_n^{2(N+N^0)} | C'_n(\widetilde{\varphi})|^2,
\end{multline*}
\begin{multline*}
\sum_n \lambda_n^{2N} \Bigg| \sum_{m_1,m_2} \bigg\langle n' \bigg| \frac{d}{dp} \bigg| m_1 \bigg\rangle \bigg\langle n'
\bigg| \frac{d}{dp} \bigg| m_2 \bigg\rangle
C_{m_1}(\widetilde{\varphi}) C_{m_2}(\widetilde{\varphi}) \Bigg| \\
< c_N \, \sum_n \lambda_n^{2(N+N^0)} | C_n(\widetilde{\varphi})|^2,
\end{multline*}
\begin{multline*}
\sum_n \lambda_n^{2N} \Bigg| \, \sum_{m_1,m_2} \bigg\langle n' \bigg|\frac{d}{dp} \bigg| m_1' \bigg\rangle \bigg\langle n'
\bigg|\frac{d}{dp} \bigg| m_2' \bigg\rangle
C'_{m_1}(\widetilde{\varphi}) C'_{m_2}(\widetilde{\varphi}) \Bigg| \\
< c_N \, \sum_n \lambda_n^{2(N+N^0)} | C'_n(\widetilde{\varphi})|^2,
\end{multline*}
and similarly for the ordinary Fourier transform operator $\mathscr{F}$:
$\mathscr{F}f(p) = \int f(x)e^{ixp} \ud x$ and $\varphi \in \mathcal{S}^{00}(\mathbb{R})$:
\begin{multline*}
\sum_n \lambda_n^{2N} \bigg| \sum_{m_1,m_2} \big\langle n \big| \mathscr{F}| m_1 \big\rangle \big\langle n
\big| \mathscr{F} \big| m_2 \big\rangle
C_{m_1}(\varphi) C_{m_2}(\varphi) \bigg| \\
< c_N \, \sum_n \lambda_n^{2(N+N^0)} | C^{0}_{n}(\varphi)|^2.
\end{multline*}
\begin{multline*}
\sum_n \lambda_n^{2N} \bigg| \, \sum_{m_1,m_2} \big\langle n \big|\mathscr{F} \big| m_1' \big\rangle \big\langle n
\big|\mathscr{F} \big| m_2' \big\rangle
C'_{m_1}(\varphi) C'_{m_2}(\varphi) \bigg| \\
< c_N \, \sum_n \lambda_n^{2(N+N^0)} | C^{0}_{n}(\varphi)|^2.
\end{multline*}
\begin{multline*}
\sum_n \lambda_n^{2N} \bigg| \sum_{m_1,m_2} \big\langle n' \big| \mathscr{F}| m_1 \big\rangle \big\langle n'
\big| \mathscr{F} \big| m_2 \big\rangle
C_{m_1}(\varphi) C_{m_2}(\varphi) \bigg| \\
< c_N \, \sum_n \lambda_n^{2(N+N^0)} | C^{0}_{n}(\varphi)|^2.
\end{multline*}
\begin{multline*}
\sum_n \lambda_n^{2N} \bigg| \, \sum_{m_1,m_2} \big\langle n' \big|\mathscr{F} \big| m_1' \big\rangle \big\langle n'
\big|\mathscr{F} \big| m_2' \big\rangle
C'_{m_1}(\varphi) C'_{m_2}(\varphi) \bigg| \\
< c_N \, \sum_n \lambda_n^{2(N+N^0)} | C^{0}_{n}(\varphi)|^2.
\end{multline*}
\end{lem*}

\qedsymbol \, 
The idea of the proof is simple. Namely we use the isometric maps $U$ and $U'$ defined respectively by (\ref{von-neumann-trick}) and (\ref{U'f}) to express the matrix elements of the lemma\footnote{Which are computed in 
the basis $\{u_n, u'_n \}$.} of the indicated 
multiplication (and eventually differentiation or Fourier transform) operators in terms of 
matrix elements in the basis $\{h_n\}$
of Hermite functions of another multiplication operators (eventually composed with the differentiation or Fourier transform operator) 
by another functions which turns out to be multipliers of the nuclear algebra $\mathcal{S}(\mathbb{R})$
and thus are continuous as operators $\mathcal{S}(\mathbb{R}) \to \mathcal{S}(\mathbb{R})$. 
The inequality of the lemma then follows from the mentioned continuity of the operators expressed in therms of the norms 
$| \cdot |_N = | H_{(1)}^{N} \cdot |_{{}_{L^2(\mathbb{R})}}$, where $H_{(1)}$ is given by the one dimensional oscillator 
Hamiltonian (\ref{oscillatorH}) (recall that the nuclear topology of $\mathcal{S}(\mathbb{R})$ is equivalent to a countably Hilbert nuclear Frechet space such that $\mathcal{S}(\mathbb{R})$ and the dual space $\mathcal{S}(\mathbb{R})^*$ of tempered distributions  can be  constructed as a Gelfand triple $\mathcal{S}(\mathbb{R}) \subset L^2(\mathbb{R}) \subset \mathcal{S}(\mathbb{R})^*$
with the help of the  Hamiltonian operator $H_{(1)}$ of the one dimensional oscillator, compare \cite{GelfandIV}, 
\cite{HKPS}, p. 484, \cite{Reed_Simon}, \cite{Simon}, \cite{Hida1}, \cite{Hida2}, \cite{Hida3}.

For the first part of the Lemma note, please, that ($t(p) = p - p^{-1}$)
\begin{multline*}
\langle n |p-p^{-1} | m \rangle 
= \int \limits_{-\infty}^{+\infty} u_n(p) \, (p - p^{-1}) \, u_n(p) \, \ud p \\
= \int \limits_{-\infty}^{0} u_n(p) \, (p - p^{-1}) \, u_n(p) \, \ud p 
+  \int \limits_{0}^{+\infty} u_n(p) \, (p - p^{-1}) \, u_n(p) \, \ud p \\
= \frac{1}{2} \int \limits_{-\infty}^{0} h_n(t(p)) \, t(p) \, h_n(t(p)) \, (1 + p^{-2}) \, \ud p
+ \frac{1}{2} \int \limits_{0}^{+\infty} h_n(t(p)) \, t(p) \, h_n(t(p)) \, (1 + p^{-2}) \, \ud p \\
= \frac{1}{2} \int \limits_{-\infty}^{-\infty} h_n(t) \, t \, h_n(t) \, \ud t
+ \frac{1}{2} \int \limits_{-\infty}^{+\infty} h_n(t) \, t \, h_n(t) \, \ud t \\
=  \int \limits_{-\infty}^{+\infty} h_n(t) \, t \, h_n(t) \, \ud t 
= (n |t| m) = \bigg(\frac{n+1}{2}\bigg)^{1/2} \delta_{m \,\, n+1} + \bigg(\frac{n}{2}\bigg)^{1/2} \delta_{m \,\,n-1}, 
\end{multline*} 
where the last equality follows from the well known property of Hermite functions. Now because $u'_n$ can be 
constructed from $u_n$ by changing the sign of the value of $u_n$ for all $p<0$, then the rest of the first part of the lemma easily follows from the above equality.

Now we express the remaining matrix elements $\langle n |f(p) | m \rangle$
in terms of  matrix elements $( n |g(t) | m )$ in the basis $\{ h_n\}$  of the corresponding  multiplication 
operators by another functions $t \mapsto g(t)$.   
\begin{multline*}
\langle n |p | m \rangle = \langle n' |p | m' \rangle  
= \int \limits_{-\infty}^{+\infty} u_n(p) \, p \, u_n(p) \, \ud p 
= \int \limits_{-\infty}^{0} u_n(p) \, p \, u_n(p) \, \ud p 
+  \int \limits_{0}^{+\infty} u_n(p) \, p \, u_n(p) \, \ud p \\
= \frac{1}{2} \int \limits_{-\infty}^{0} h_n(t(p)) \, \frac{t(p) - \sqrt{t(p)^2 +4}}{2} 
\, h_n(t(p)) \, (1 + p^{-2}) \, \ud p \\
+ \frac{1}{2} \int \limits_{0}^{+\infty} h_n(t(p)) \,  \frac{t(p) + \sqrt{t(p)^2 +4}}{2} \, h_n(t(p)) 
\, (1 + p^{-2}) \, \ud p \\
= \frac{1}{2} \int \limits_{-\infty}^{-\infty} h_n(t) \,  \frac{t - \sqrt{t^2 +4}}{2} \, h_n(t) \, \ud t
+ \frac{1}{2} \int \limits_{-\infty}^{+\infty} h_n(t) \, \frac{t + \sqrt{t^2 +4}}{2} \, h_n(t) \, \ud t \\
=  \frac{1}{2} \int \limits_{-\infty}^{+\infty} h_n(t) \, t \, h_n(t) \, \ud t 
= \frac{1}{2} (n |t| m) = \frac{1}{2} \bigg(\frac{n+1}{2}\bigg)^{1/2} \delta_{m \,\, n+1} 
+ \frac{1}{2} \bigg(\frac{n}{2}\bigg)^{1/2} \delta_{m \,\,n-1}. 
\end{multline*}   
Similarly we get 
\[ 
\langle n |p | m' \rangle  = \langle n' |p | m \rangle 
= \frac{1}{2} (n |\sqrt{t^2 + 4}| m), 
\]
\begin{multline*}
\langle n |p^{-1} | m \rangle  = \langle n' |p^{-1} | m' \rangle 
= \frac{1}{2} (n |-t| m) \\ = -\frac{1}{2} \bigg(\frac{n+1}{2}\bigg)^{1/2} \delta_{m \,\, n+1} 
- \frac{1}{2} \bigg(\frac{n}{2}\bigg)^{1/2} \delta_{m \,\,n-1}, 
\end{multline*}
\[
\langle n |p^{-1} | m' \rangle  = \langle n' |p^{-1} | m \rangle 
= \frac{1}{2} \big(n \big|\sqrt{t^2 + 4} \big| m \big), 
\]
\[
\langle n \big| \, |p| \, \big| m \rangle  = \langle n' \big| \, |p| \, \big| m' \rangle 
= \frac{1}{2} \big(n \big|\sqrt{t^2 + 4} \big| m \big), 
\]
\[
\langle n \big| \, |p| \, \big| m' \rangle  = \langle n' | \, |p| \, | m \rangle 
= \frac{1}{2} (n | t | m), 
\]
\[
\big\langle n \big| \, |p|^{-1/2} \, \big| m \big\rangle  = \big\langle n' \big|\, |p|^{-1/2} \,  \big| m' \big\rangle 
= \frac{1}{2} \bigg(n \bigg|\sqrt{\sqrt{t^2 + 4} +t} + \sqrt{\sqrt{t^2 + 4} -t} \bigg| m \bigg), 
\]
\[
\big\langle n \big| \, |p|^{-1/2} \, \big| m' \big\rangle  = \big\langle n' \big| \, |p|^{-1/2} \, \big| m \big\rangle 
= \frac{1}{2} \bigg(n \bigg|\sqrt{\sqrt{t^2 + 4} +t} - \sqrt{\sqrt{t^2 + 4} -t} \bigg| m \bigg), 
\]
\[
\bigg\langle n \bigg| \, \frac{d}{dp} \, \bigg| m \rangle  
= \bigg\langle n' \bigg| \, \frac{d}{dp} \, \bigg| m' \bigg\rangle 
= \frac{1}{2} \bigg(n \bigg|\frac{t}{\sqrt{t^2 + 4}} + (t^2 + 4)\frac{d}{dt} \bigg| m \bigg),
\] 
\[
\bigg\langle n \bigg| \, \frac{d}{dp} \, \bigg| m' \rangle  
= \bigg\langle n' \bigg| \, \frac{d}{dp} \, \bigg| m \bigg\rangle 
= \frac{1}{2} \bigg(n \bigg|-\frac{t^2 +1}{t^2 + 4} - t\sqrt{t^2 + 4}\frac{d}{dt} \bigg| m \bigg),
\] 
\[
\big\langle n \big| \, \mathscr{F} \, \big| m \rangle  = \big\langle n' \big| \, \mathscr{F} \, \big| m' \rangle 
= \big( n \big| \, Op_1 + Op_2 \, \big| m \big),
\]
\[
\big\langle n \big| \, \mathscr{F} \, \big| m' \rangle  = \big\langle n' \big| \, \mathscr{F} \, \big| m \rangle 
= \big( n \big| \, Op_1 - Op_2 \, \big| m \big),
\] 
where $\Op_1$ and $\Op_2$ are the following operators $\mathcal{S}(\mathbb{R}) \to \mathcal{S}(\mathbb{R})$:
\[
Op_1 \, h(p) = \int \limits_{\-\infty}^{+\infty} h(t) \frac{1}{\sqrt{t^2 +4 - t\sqrt{t^2 +4}}} 
e^{-ip\frac{t + \sqrt{t^2+4}}{2}} \, \ud t,
\]
\[
Op_2 \, h(p) = \int \limits_{\-\infty}^{+\infty} h(t) \frac{1}{\sqrt{t^2 +4 + t\sqrt{t^2 +4}}} 
e^{-ip\frac{t - \sqrt{t^2+4}}{2}} \, \ud t.
\]

It is easily seen that all the functions $t \mapsto g(t)$ in $(m |g(t)|n)$ obtained above are multipliers
of the algebra $\mathcal{S}(\mathbb{R})$, i.e. they are smooth and all their derivatives grows not faster than
polynomially at infinity, so that all of them are members of the algebra
$\mathcal{O}_M$ of all smooth functions $g$ such that for each $n \in \mathbb{N}$
there exists $k \in \mathbb{N}$ such that $(1 + |t|^2)^{-k} \frac{d^n g }{dt^n} \in C_0(\mathbb{R})$,
where $C_0(\mathbb{R})$ is the algebra of continuous functions on $\mathbb{R}$ tending to $0$ at infinity.
It is widely known that $\mathcal{O}_M$ is the space of multipliers of $\mathcal{S}(\mathbb{R})$
(it is even algebra under pointwise product). Therefore, the operators $M_g$ of multiplication by those
functions are continuous as operators $\mathcal{S}(\mathbb{R}) \to \mathcal{S}(\mathbb{R})$. 

Similarly, it is not difficult to check that the operators $Op_1$, $Op_2$, 
\[
\frac{t}{\sqrt{t^2 + 4}} + (t^2 + 4)\frac{d}{dt} \,\,\, \textrm{and}
\,\,\,
-\frac{t^2 +1}{t^2 + 4} - t\sqrt{t^2 + 4}\frac{d}{dt} 
\] 
are continuous as operators $\mathcal{S}(\mathbb{R}) \to \mathcal{S}(\mathbb{R})$. 

The Gelfand triple $\mathcal{S}(\mathbb{R}) \subset L^2(\mathbb{R}) \subset \mathcal{S}(\mathbb{R})^*$
is constructed with the help of the Hamiltonian operator of the one dimensional oscillator
with the nuclear topology  $\mathcal{S}(\mathbb{R})$ given by the norms 
\[
| h |_{N}^{2}  = | {H_{(1)}}^N h |_{0}^{2} = \sum_n \lambda_{n}^{2N} |C^{0}_{n}(h)|^2, \,\,\,
h = \sum_n C^{0}_{n}(h) h_n 
\]
where $h_n$ are the Hermite functions.
Because $\mathcal{S}(\mathbb{R}) = \cap_N \mathcal{S}_N(\mathbb{R})$, where
$\mathcal{S}_N(\mathbb{R})$ is the completion of $\mathcal{S}_N(\mathbb{R})$ with respect to
the norm $| \cdot |_{N}$, thus the necessary and sufficient condition for the function
$h$ to be an element of $\mathcal{S}(\mathbb{R})$ is that 
\[
\sum_n \lambda_{n}^{N} |C^{0}_{n}(h)|^2 < +\infty  \,\,\, \textrm{for all} \,\,\,N \in \mathbb{N}. 
\]

Because each of the indicated operators $Op:$
$\mathcal{S}(\mathbb{R}) \to \mathcal{S}(\mathbb{R})$ (multiplication
operator $M_g$ by the  functions $g$ obtained above and differentiation operator)
is continuous, then for each of them and for any $N \in \mathbb{N}$ there exist
(independent of $h \in \mathcal{S}(\mathbb{R})$) $N^0 \in \mathbb{N}$ and $c_N >0$
such that
\begin{multline*}
| Op \, h |_{N}^{2} = \sum_n \lambda_n |C^{0}_{n}(Op \, h)|^2 
= \sum_n \lambda_{n}^{2N} |C^{0}_{n}(Op \,h)^2| \\
= \sum_n \lambda_{n}^{2N} \bigg| \sum_{m_1,m_2}  (n |Op | m_1 )  (n |Op | m_2 )
C^{0}_{m_1}(h) C^{0}_{m_2}(h) \bigg| \\
\leq \, c_N \, \sum_n \lambda_n^{2(N+N^0)} | C^{0}_{n}(h)|^2 
\, = \, c_N \, | h |_{N + N^0}^{2}.
\end{multline*}
Note that the last inequality holds for any sequence $\{C^{0}_n\}_{n = 0,1,\ldots}$
for which
\[
\sum_n \lambda_{n}^{N} |C^{0}_{n}|^2 < +\infty  \,\,\, \textrm{for all} \,\,\,N \in \mathbb{N}. 
\]
Now because $E = \cap_N E_N$, it follows that for any $\widetilde{\varphi} \in E$ the sequence
$\{C_n(\widetilde{\varphi})\}_{n = 0,1,\ldots}$ as well as the sequence 
$\{C'_n(\widetilde{\varphi})\}_{n = 0,1,\ldots}$ fulfils the last condition. But this is equivalent
to the assertion of the Lemma, as the matrix elements $(n |Op | m )$ are equal to the matrix
elements $\langle n |\bold{Op} | m \rangle$ (resp. $\langle n |\bold{Op} | m' \rangle$,
$\langle n' |\bold{Op} | m \rangle$) for the operators $\bold{Op}$ in the assertion of the Lemma.

\qed

\vspace*{0.5cm}

From the Lemma it immediately follows that the operators $M_f$ of multiplication
by the functions $p \mapsto f(p)$ of the Lemma are continuous as operators
$E \to E$. Indeed, for any $N \in \mathbb{N}$, there exist (independent of $\widetilde{\varphi} \in E$)
$N^0 \in \mathbb{N}$ and $c_N >0$ such that  
\begin{multline*}
| f\widetilde{\varphi}|_{N}^{2} = \sum \limits_n \lambda_{n}^{2N} |C_n(f\widetilde{\varphi})|^2
+ \sum \limits_n \lambda_{n}^{2N} |C'_n(f\widetilde{\varphi})|^2 \\
= \sum \limits_n \lambda_{n}^{2N} |C_n(f\widetilde{\varphi})^2|
+ \sum \limits_n \lambda_{n}^{2N} |C'_n(f\widetilde{\varphi})^2| 
\end{multline*}
\begin{multline*}
= \, \sum_n \lambda_n^{2N} \bigg| \sum_{m_1,m_2} \langle n |f(p) | m_1 \rangle \langle n |f(p) | m_2 \rangle
C_{m_1}(\widetilde{\varphi}) C_{m_2}(\widetilde{\varphi})  \\
+ \, \sum_{m_1,m_2} \langle n |f(p) | m_1' \rangle \langle n |f(p) | m_2' \rangle
C'_{m_1}(\widetilde{\varphi}) C'_{m_2}(\widetilde{\varphi}) \\
+ \, \sum_{m_1,m_2} \langle n |f(p) | m_1 \rangle \langle n |f(p) | m_2' \rangle
C_{m_1}(\widetilde{\varphi}) C'_{m_2}(\widetilde{\varphi}) \\
+ \, \sum_{m_1,m_2} \langle n |f(p) | m_1' \rangle \langle n |f(p) | m_2 \rangle
C'_{m_1}(\widetilde{\varphi}) C_{m_2}(\widetilde{\varphi})\bigg| 
\end{multline*}
\begin{multline*}
+ \, \sum_n \lambda_n^{2N} \bigg| \sum_{m_1,m_2} \langle n' |f(p) | m_1 \rangle \langle n' |f(p) | m_2 \rangle
C_{m_1}(\widetilde{\varphi}) C_{m_2}(\widetilde{\varphi})  \\
+ \, \sum_{m_1,m_2} \langle n' |f(p) | m_1' \rangle \langle n' |f(p) | m_2' \rangle
C'_{m_1}(\widetilde{\varphi}) C'_{m_2}(\widetilde{\varphi}) \\
+ \, \sum_{m_1,m_2} \langle n' |f(p) | m_1 \rangle \langle n' |f(p) | m_2' \rangle
C_{m_1}(\widetilde{\varphi}) C'_{m_2}(\widetilde{\varphi}) \\
+ \, \sum_{m_1,m_2} \langle n' |f(p) | m_1' \rangle \langle n' |f(p) | m_2 \rangle
C'_{m_1}(\widetilde{\varphi}) C_{m_2}(\widetilde{\varphi})\bigg|
\end{multline*}
\begin{multline*}
\leq \,\, 2 \, \sum_n \lambda_n^{2N} \bigg| \sum_{m_1,m_2} \langle n |f(p) | m_1 \rangle \langle n |f(p) | m_2 \rangle
 C_{m_1}(\widetilde{\varphi}) C_{m_2}(\widetilde{\varphi}) \bigg| \\
+ \,\, 
2 \, \sum_n \lambda_n^{2N} \bigg| \sum_{m_1,m_2} \langle n |f(p) | m_1' \rangle \langle n |f(p) | m_2' \rangle
C'_{m_1}(\widetilde{\varphi}) C'_{m_2}(\widetilde{\varphi}) \bigg| \\
\, + \,\,
2 \, \sum_n \lambda_n^{2N} \bigg| \sum_{m_1,m_2} \langle n' |f(p) | m_1 \rangle \langle n' |f(p) | m_2 \rangle
C_{m_1}(\widetilde{\varphi}) C_{m_2}(\widetilde{\varphi}) \bigg| \\
+ \,\,
2 \, \sum_n \lambda_n^{2N} \bigg| \, \sum_{m_1,m_2} \langle n' |f(p) | m_1' \rangle \langle n' |f(p) | m_2' \rangle
C'_{m_1}(\widetilde{\varphi}) C'_{m_2}(\widetilde{\varphi}) \bigg| \\
< \,  c_N \Big[ \sum_n \lambda_n^{2(N+N^0)} | C_n(\widetilde{\varphi})|^2 
+ \,  \sum_n \lambda_n^{2(N+N^0)} | C'_n(\widetilde{\varphi})|^2 \Big]  \\
= \, c_N \, | \widetilde{\varphi} |_{N+N^0}^{2}, 
\end{multline*} 
where the first inequality follows from the inequality\footnote{Special case of the Cauchy-Schwartz inequality in the Hilbert space $\mathbb{C}^2$.} $|ab + ba| \leq |a^2| + |b^2|$, $a,b \in \mathbb{C}$ for
\[
a = \sum_{m} \langle n |f(p) | m\rangle C_{m}(\widetilde{\varphi}) \,\,\,\,\,\, 
b = \sum_{m} \langle n |f(p) | m' \rangle C'_{m}(\widetilde{\varphi}) 
\]
or respectively
\[
a = \sum_{m} \langle n' |f(p) | m \rangle C_{m}(\widetilde{\varphi})  \,\,\,\,\,\, 
b = \sum_{m} \langle n' |f(p) | m' \rangle C'_{m}(\widetilde{\varphi}); 
\]
and the second inequality follows from the Lemma. Continuity of the differentiation operator
$E \to E$ follows from the Lemma in exactly the same manner.

Note that in the proof of the last Lemma a close similarity of the spectra of the operators
$H_{(1)}$ and $A^{(1)} = A'$ play a crucial role. In this one dimensional case their spectra are equal
but each eigenvalue (common for $H_{(1)}$ and $A'$) appears twice in $\Sp A'$. From this it follows
the following fact used in the proof of the last Lemma: If $\{\lambda_{n}^{0}\}_{n \in \mathbb{N}}
= \Sp H_{(1)}$ and $\{\lambda_{n}\}_{n \in \mathbb{N}}
= \Sp A^{(1)} = \Sp A'$, then a sequence $\{C_n\}_{n \in \mathbb{N}}$ of numbers
fulfils
\[
\sum \limits_{m \in \mathbb{N}} {(\lambda_{m}^{0})}^{N} |C_m|^2 < +\infty, \,\,\,
N \in \mathbb{N}
\]
if and only if
\[
\sum \limits_{m \in \mathbb{N}} {(\lambda_{m})}^{N} |C_m|^2 < +\infty, \,\,\,
N \in \mathbb{N}.
\]
We will construct the standard operators $A^{(n)} = U\big(H_{(1)} \otimes \boldsymbol{1} + \boldsymbol{1} \otimes \Delta_{\mathbb{S}^{n-1}}\big)U^{-1}$ in $L^2(\mathbb{R}^n)$ for higher dimensions $n$ which have spectra identical with the spectra of the corresponding
operators $H_{(1)} \otimes \boldsymbol{1} + \boldsymbol{1} \otimes \Delta_{\mathbb{S}^{n-1}}$ on $L^2(\mathbb{R} \times \mathbb{S}^{n-1})$. We therefore reduce the problem of investigation of continuous operators on $\mathcal{S}_{A^{(n)}}(\mathbb{R}^n)$ to the investigation of the continuous operators on $\mathcal{S}_{H_{(1)} \otimes \boldsymbol{1} + \boldsymbol{1} \otimes \Delta_{\mathbb{S}^{n-1}}} (\mathbb{R} \times \mathbb{S}^{n-1})$ exactly as we have reduced the investigation of continuous operators on
$\mathcal{S}_{A^{(1)}}(\mathbb{R})$ to the investigation of the continuous operators on $\mathcal{S}_{H_{(1)}}(\mathbb{R})
= \mathcal{S}(\mathbb{R})$ using the similarity of the spectra of $A^{(1)}$ and $H_{(1)}$. Moreover, because
\[
\mathcal{S}_{H_{(1)} \otimes \boldsymbol{1} + \boldsymbol{1} \otimes \Delta_{\mathbb{S}^{n-1}}} (\mathbb{R} \times \mathbb{S}^{n-1}) = \mathcal{S}_{H_{(1)}}(\mathbb{R}) \otimes \mathcal{S}_{\Delta_{\mathbb{S}^{n-1}}}(\mathbb{S}^{n-1})
= \mathcal{S}(\mathbb{R}) \otimes \mathscr{C}^{\infty}(\mathbb{S}^{n-1}),
\]
we reduce the whole problem to the determination of continuous operators, functionals (and convolutors)
on $\mathcal{S}(\mathbb{R})$ and $\mathcal{S}_{\Delta_{(\mathbb{S}^{n-1})}}(\mathbb{S}^{n-1}) = \mathscr{C}^{\infty}(\mathbb{S}^{n-1})$.



By construction of the nuclear space $\mathbb{E}$ in the position picture it follows that the ordinary 
Fourier transform $\mathscr{F}$ and its inverse $\mathscr{F}^{-1}$ are continuous when regarded
as operators $\mathbb{E} \to E$ and $E \to \mathbb{E}$.  

We have shown that the spaces $E$ and $\mathcal{S}(\mathbb{R}) \oplus \mathcal{S}(\mathbb{R})$
are isomorphic as nuclear spaces with the isomorphism given by the transform $U\oplus U'$
with $U$ and $U'$ given respectively by (\ref{von-neumann-trick}) and (\ref{U'f}). Exactly the same
proof with the additional use of the ordinary Fourier transform $\mathscr{F}$ gives the isomorphism
of $\mathbb{E}$ and $\mathcal{S}(\mathbb{R}) \oplus \mathcal{S}(\mathbb{R})$. These isomorphisms are useful
in checking if a concrete functional on $E$ or $\mathbb{E}$ is continuous in reducing the problem to checking if a concrete functional is continuous on $\mathcal{S}(\mathbb{R})$.
In particular, we give a proof that the Dirac delta functional $\delta_{p_{{}_0}}: E \ni g \mapsto
g(p_{{}_0}) \in \mathbb{C}$
is continuous on $E$. Indeed, assume first that $p_{{}_0} \neq 0$. Using the explicit formulas for the unitary
operators\footnote{Of course
$U$ and $U'$ are unitary as operators $L^2(\mathbb{R})
\to E_{0I}$ and $L^2(\mathbb{R}) \to E_{0II}$; treated as operators $L^2(\mathbb{R}) \to L^2(\mathbb{R})$
they are only isometric.}
$U: L^2(\mathbb{R})
\to E_{0I} = U(L^2(\mathbb{R}))$ (eq. (\ref{von-neumann-trick})) and $U': L^2(\mathbb{R})
\to E_{0II} = U'(L^2(\mathbb{R}))$ (eq. (\ref{U'f})), we easily see that
\[
\mathcal{S}(\mathbb{R}) \oplus \mathcal{S}(\mathbb{R}) \xrightarrow{U \oplus U'}
E = E_I \oplus E_{II} \xrightarrow{\delta_{p_{{}_0}}} \mathbb{C}
\]
is continuous. Because $U \oplus U'$ is an isomorphism of the nuclear spaces it follows that
$\delta_{p_0}$ is continuous on $E$, i.e. $\delta_{p_0} \in E^*$, $p_{{}_0} \neq 0$. That $\delta_{p_{{}_0}}$
is continuous also for $p_{{}_0} = 0$ is trivial as it is easily seen that in this case $\delta_{p_{{}_0}}$ is equal
to the zero functional.

Using the same isomorphism one can likewise easily show that the maps $p \mapsto \delta_p \in E^*$, $p \in \mathbb{R}$
and $x \mapsto \delta_x \in \mathbb{E}^*$, $x \in \mathbb{R}$, are continuous. Because the Lebesgue measure on
$\mathbb{R}$ is perfect then for every $\widetilde{\varphi} \in E$ there exists a unique continuous function on
$\mathbb{R}$ which coincides with
$\widetilde{\varphi}$ up to a Lebesgue null function, and the same holds for any element $\varphi$ of $\mathbb{E}$.
Therefore, the spaces $E$ and $\mathbb{E}$ fulfil the conditions (H1)-(H3) of \S1 of \cite{hida} and \cite{obata}.
In particular any element of the topological projective $n$-fold tensor product $E^{\otimes n}$ is a continuous
function on $\mathbb{R}^{n}$ and the same holds for elements of the projective tensor product $\mathbb{E}^{\otimes n}$.
However, that the Kubo-Takenaka conditions (H1)-(H3) of Subection \ref{white-setup} (or
\S1 of \cite{hida} and \cite{obata}) are fulfilled for $E = \mathcal{S}_{A'}(\mathbb{R}) =
\mathcal{S}_{A^{(1)}}(\mathbb{R})$ immediately follows from the simple criterion given in the Proposition of the
Appendix in \cite{obata.Cont.Version.Thm}. We will apply this criterion in higher dimensions. 

It is known that the pointwise multiplication defines a (jointly) continuous bilinear map 
$E \times E \to E$ (compare e.g. \cite{obata}). 

That $\delta_{p_{{}_0}} \in \mathcal{S}^0(\mathbb{R})^*$ 
is obvious as $\delta_{p_{{}_0}} \in \mathcal{S}(\mathbb{R})^*$ and $\mathcal{S}^0(\mathbb{R})$
is a closed subspace of $\mathcal{S}(\mathbb{R})$ with the topology inherited from $\mathcal{S}(\mathbb{R})$.
And similarly it is obvious that $\delta_{x_{{}_0}} \in \mathcal{S}^{00}(\mathbb{R})^*$.

Later on we will show a stronger result than just the preservation of (H1)-(H3).
Namely, we will show in the subsequent Subsections that $\mathcal{S}^{0}(\mathbb{R})
= E = \mathcal{S}_{A^{(1)}}(\mathbb{R})$
(resp. $\mathcal{S}^{00}(\mathbb{R}) = \mathbb{E}$), and still more generally, that
$\mathcal{S}^{0}(\mathbb{R}^n) = \mathcal{S}_{A^{(n)}}(\mathbb{R}^n)$,
in store of elements and in topology. Note the case
$\mathcal{S}^{0}(\mathbb{R}^3) = \mathcal{S}_{A^{(3)}}(\mathbb{R}^3)$
is crucial for the applicability of the white noise calculus in the construction
of massless fields.

At the end of this Subsection let us note that the operator $A'$ and correspondingly
$E = \mathcal{S}_{A'}(\mathbb{R})$ has an extra unitary involutive symmetry $\Inv: L^2(\mathbb{R}) \rightarrow L^2(\mathbb{R})$:
\[
\Inv g(p) = |p|^{-1} g(p^{-1}), \,\,\, g \in L^2(\mathbb{R}),  
\]
which is closely related with the geometric inversion with respect to the unit sphere. An analogous $\Inv$
exists in higher dimensions. It is easily checked that $\Inv$ is unitary  
and 
\[
\Inv = {\Inv}^{-1}, \,\,\, \textrm{or} \,\,\, \Inv \circ \Inv = \boldsymbol{1}, \,\,\,
\textrm{and} \,\,\, \Inv A' \Inv = A'.
\] 
By the last Lemma or by the last equality it easily follows that $\Inv E \subset E$. 

Let $\{u_n, u'_m\}_{n,m \in\mathbb{N}}$ be the complete orthonormal system in $L^2(\mathbb{R})$ corresponding
to the operator $A'$, constructed in this Subsection. If $u_n$ or $u'_n$ corresponds to even Hermite
function then we write $u_{n}^{\oplus}$ or ${u'}_{n}^{\oplus}$; if they correspond to odd Hermite function then we write
$u_{n}^{\ominus}$ or ${u'}_{n}^{\ominus}$, respectively. 

One immediately checks that
\[
\begin{split}
\Inv u_{n}^{\oplus} = u_{n}^{\oplus}, \,\,\,\,\,\, \Inv u_{n}^{\ominus} = - u_{n}^{\ominus}, \\
\Inv {u'}_{n}^{\oplus} = {u'}_{n}^{\oplus}, \,\,\,\,\,\, \Inv {u'}_{n}^{\ominus} = -{u'}_{n}^{\ominus}.
\end{split}
\]

\subsection{Construction of $A^{(n)}$, $n >1$}\label{dim=n}

In order to construct the self adjoint operator $A^{(n)}$ in $L^2(\mathbb{R}^n, \ud^n p; \mathbb{R})$,
we give a construction of the complete orthonormal system in $L^2(\mathbb{R}^n, \ud^n p; \mathbb{R})$
defining $A^{(n)}$. Note that we are using the ordinary invariant Lebesgue measure $\ud^n p$ in the euclidean space $\mathbb{R}^n$. We use the original von Neumann's method, \cite{Szego}, p. 108 for construction of the
complete system, without any additional modification using double (or multiple) covering maps,
needed in the one dimensional case. Thus, the construction is even simpler than for dimension $1$,
with the only irrelevant difference in comparison to \cite{Szego} that the corresponding unitary map 
is constructed in two steps as a composition of two unitary maps.

Namely, we consider the euclidean space $\mathbb{R}^n$ as naturally embedded hyperplane in the $n+1$ dimensional
euclidean space $\mathbb{R}^{n+1} = \mathbb{R} \times \mathbb{R}^n$, with the ordinary (standard) sub-manifold, 
metric and measure structures inherited from the ordinary (standard) manifold, 
metric and measure structures of the euclidean space $\mathbb{R}^{n+1}$ and the coordinates $(t;p) = (t; p_1, \ldots , p_n) =(t; r, \phi_1, \ldots , \phi_{n-1})$ in $\mathbb{R}^{n+1} = \mathbb{R} \times \mathbb{R}^n$, where
$(p_1, \ldots , p_n)$ are the ordinary Cartesian coordinates in $\mathbb{R}^n$ and where $(r, \phi_1, \ldots , \phi_{n-1})$
are the standard generalized spherical coordinates in $\mathbb{R}^n$, with $r>0, 0 \leq \phi_1 < \pi, \ldots , 0 \leq \phi_{n-2} < \pi, 0 \leq \phi_{n-1} < 2 \pi$.

In the euclidean space we consider another sub-manifold,
namely the ``cylinder'' $\mathbb{R} \times \mathbb{S}^{n-1}$, where $\mathbb{S}^{n-1}$ 
is the unit $n-1$-sphere in the euclidean space $\mathbb{R}^n$ regarded as a sub-manifold naturally embedded in $\mathbb{R}^{n+1} 
= \mathbb{R} \times \mathbb{R}^{n}$. On the cylinder manifold $\mathbb{R} \times \mathbb{S}^{n-1}$ we are 
using the natural ''spherical'' coordinates
$(t, \phi_1, \ldots \phi_{n-1})$ with the spherical coordinates $(\phi_1, \ldots \phi_{n-1})$ on the
unit sphere $\mathbb{S}^{n-1}$. The manifold, metric, and measure structures on $\mathbb{R} \times \mathbb{S}^{n-1}$ 
are the ordinary ones, which we regard as inherited from $\mathbb{R}^{n+1}$ by the embedding of  
$\mathbb{R} \times \mathbb{S}^{n-1}$ into 
$\mathbb{R}^{n+1} = \mathbb{R} \times \mathbb{R}^n$. Besides the two sub-manifolds $\mathbb{R} \times \mathbb{S}^{n-1}$
and $\mathbb{R}^n$ with the indicated structures inherited from $\mathbb{R}^{n+1} = \mathbb{R} \times \mathbb{R}^n$,
we consider a third funnel-shape sub-manifold $\mathbb{F}$ in $\mathbb{R}^{n+1} = \mathbb{R} \times \mathbb{R}^n$,
defined by the equation $t = r - r^{-1}$, where $r$ is the radial coordinate in $\mathbb{R}^n$ regarded as embedded in 
$\mathbb{R}^{n+1} = \mathbb{R} \times \mathbb{R}^n$ as hyperplane.

\begin{center}
\begin{tikzpicture}[yscale=1]
    \draw[thin, ->] (0,0) -- (3.5,0);
    \draw[thin, ->] (0,0) -- (0,2);
    \draw[thin, <-, domain=-2:0] plot(\x, {1.2*\x});
    \draw[ultra thick, rotate around = {40:(3.5,0)}, domain=1.65:3.4] plot(\x, {1.2*\x -4.2});
\draw[ultra thick, rotate around = {40:(3.5,0)}, domain=3.6:4.73] plot(\x, {1.2*\x -4.2});
\draw[thin, rotate around = {-10:(-2,0)}, domain=-4.15:-1.8] plot(\x, {1.2*\x +2.4});
    \draw[ultra thick] (0.99,0.1) -- (0.99,-2.9);
    \draw[ultra thick] (-0.99,-0.1) -- (-0.99,-3.1);

\draw[thin, rotate around = {-5:(-2,-2.4)}] (-1.9,-2.4) -- (3.5,-2.4);
\draw[thin, rotate around = {-5:(-2,-2.4)}] (-4.6,-2.4) -- (-2.1,-2.4);

\draw[thin, dashed] (0.99,0.1) -- (0.99,2.1);
 \draw[thin, dashed] (-0.99,-0.1) -- (-0.99,1.9);

\draw[ultra thick] (0.99,1) -- (0.99,2.1);
 \draw[ultra thick] (-0.99,0.7) -- (-0.99,1.9);

\draw[thin, rotate around = {3:(1.5,1.8)}] (2.5,1.8) -- (3.5,1.8);

    \draw[thin, dashed, domain=-2.4:-0.8] plot (\x, {pow(\x,-1)-\x});
\draw[ultra thick, domain=-2.4:-1.45] plot (\x, {pow(\x,-1)-\x});
\draw[ultra thick, domain=-0.8:-0.3] plot (\x, {pow(\x,-1)-\x});

\draw[thin, dashed, domain=-2.5:-0.195] plot (\x, {-pow(-\x,-0.7)-\x-0.2});
\draw[ultra thick, domain=-1:-0.195] plot (\x, {-pow(-\x,-0.7)-\x-0.2});
\draw[ultra thick, domain=-2.5:-1.75] plot (\x, {-pow(-\x,-0.7)-\x-0.2});

\draw[ultra thick, domain=-2.4:-0.8] plot (\x, {-pow(-\x,-0.2)-\x-0.2});
\draw[very thick, domain=-0.8:-0.125] plot (\x, {-pow(-\x,-0.55)-\x-0.2});

 \draw[ultra thick, domain= 0.3: 2.4] plot (\x, {\x - pow(\x,-1)});
\draw[ultra thick, rotate=10] (0,0) ellipse (1cm and 0.5cm);
\draw[ultra thick, rotate around = {10:(0,2)}] (0,2) ellipse (1cm and 0.5cm);
\draw[ultra thick, rotate around = {10:(0,2)}] (0,2) ellipse (2.5cm and 1.25cm);
\draw[ultra thick, rotate around = {10:(0,-3)}] (0,-3) ellipse (1cm and 0.5cm);
\draw[ultra thick, rotate around = {10:(0,-3)}] (0,-3) ellipse (0.3cm and 0.15cm);

\draw[fill] (0.8,2.4) circle (0.1);
\draw[fill] (2,3) circle (0.1);
\draw[fill] (2,1) circle (0.1);

\draw[thin, <-|] (1,2.5) -- (1.8,2.9);
\draw[thin, dashed, <-] (2,2.8) -- (2,1.5);
\draw[thin] (2,1.5) -- (2,1.15);

\draw [<-,very thin] (3.6,-2.8) to [out=-70,in=130] (4.5, -2.8);
\draw [->,very thin] (-2,-2.9) to [out=-70,in=130] (-1.1,-2.9);
\draw [->,very thin] (-3.4,2.2) to [out=-70,in=130] (-2.4,2.2);

\node [right] at (4.5, -2.8) {$\mathbb{R}^n$};
\node [left] at (0,2) {$t$};
\node [left] at (-2,-2.9) {$\mathbb{R} \times \mathbb{S}^{n-1}$};
\node [left] at (-3.4,2.2) {$\mathbb{F}$};
\end{tikzpicture}
\end{center}

Consider the natural projections $g: \mathbb{R}^n \ni (r,\phi_1, \ldots \phi_{n-1}) \mapsto (t(r), \phi_1, \ldots \phi_{n-1}) \in \mathbb{F}$
and $\mathbb{F} \ni  (t, \phi_1, \ldots \phi_{n-1}) \mapsto (t, \phi_1, \ldots \phi_{n-1}) 
\in \mathbb{R} \times \mathbb{S}^{n-1}$, where
\[
t(r) = r - r^{-1},
\]
which are in fact diffeomorphisms respectively $g:  \mathbb{R}^n \rightarrow \mathbb{F}$ and 
$\mathbb{F}\rightarrow \mathbb{R} \times \mathbb{S}^{n-1}$ between the indicated manifolds. 
But although we have already fixed the metrics and measures 
\[
\ud^n p = r^{n-1} \ud r \ud \mu_{\mathbb{S}^{n-1}} =  r^{n-1} \sin^{n-2} \phi_1 \sin^{n-3} \phi_{2} \ldots 
\sin \phi_{n-2} \ud r \ud \phi_1 \ldots \ud \phi_{n-1}
\]
and
\[
\ud t \ud \mu_{\mathbb{S}_{n-1}}
\] 
respectively  on $\mathbb{R}^n$ and $\mathbb{R} \times \mathbb{S}^{n-1}$ as inherited from 
$\mathbb{R}^{n+1}$, we do not define the metric and measure on the funnel $\mathbb{F} \subset \mathbb{R}^{n+1}$
as inherited from $\mathbb{R}^{n+1}$. Instead, we define the metric and measure on $\mathbb{F}$
as the one pulled back from the euclidean hyperplane $\mathbb{R}^n$ by the projection (diffeomorphism)
$g: \mathbb{R}^n \rightarrow \mathbb{F}$. In particular the measure so defined on $\mathbb{F}$
has the form  $\ud \mu_{\mathbb{F}} =  \nu_n (t) \ud t \ud \mu_{\mathbb{S}_{n-1}}$. 
Below we give the formula for the density function $\nu_n$
for each dimension $n>1$ explicitly. Using the mentioned projections 
(diffeomorphisms) $\mathbb{R}^n  \rightarrow \mathbb{F}$ and $\mathbb{F} \rightarrow \mathbb{R} \times \mathbb{S}^{n-1}$
we define the corresponding two unitary maps $U_2: L^2(\mathbb{F}, \ud \mu_{\mathbb{F}}) \rightarrow 
L^2(\mathbb{R}^n, {\ud}^n p)$ and $U_1: L^2(\mathbb{R} \times \mathbb{S}^{n-1}, \ud t \times \ud \mu_{\mathbb{S}^{n-1}})
\rightarrow L^2(\mathbb{F}, \ud \mu_{\mathbb{F}})$, given by the following formulas
\[
\begin{split}
U_1 f(t, \phi_1, \ldots \phi_{n-1}) = \frac{1}{\sqrt{\nu_n (t)}} f(t, \phi_1, \ldots \phi_{n-1}), 
\,\,\, f \in L^2(\mathbb{R} \times \mathbb{S}^{n-1}, \ud t \times \ud \mu_{\mathbb{S}^{n-1}}), \\
U_2 f(r, \phi_1, \ldots \phi_{n-1}) = f (t(r), \phi_1, \ldots \phi_{n-1}), \,\,\,
f \in L^2(\mathbb{F}, \ud \mu_{\mathbb{F}});
\end{split}
\]  
where in the first formula there is present the additional factor 
\[
\frac{1}{\sqrt{\nu_n (t)}}
\]
equal to the square of the Radon-Nikodym derivative of the original measure with respect to that transformed under the 
diffeomorphic projection $\mathbb{F}\rightarrow \mathbb{R} \times \mathbb{S}^{n-1}$, absent in the second formula because by the very construction the projection $g:  \mathbb{R}^n \rightarrow \mathbb{F}$ preserves the metric and the measure, so that the corresponding Radon-Nikodym derivative is equal $1$.

In the the Hilbert space 
\[
L^2(\mathbb{R} \times \mathbb{S}^{n-1}, \ud t \times \ud \mu_{\mathbb{S}^{n-1}})
= L^2(\mathbb{R} , \ud t) \otimes L^2(\mathbb{S}^{n-1}, \ud \mu_{\mathbb{S}^{n-1}})
\]
we consider the 
self adjoint operator $A = H_{(1)} \otimes \boldsymbol{1} + \boldsymbol{1} \otimes \Delta_{\mathbb{S}^{n-1}}$,
where $H_{(1)}$ is the hamiltonian of the one dimensional harmonic oscillator 
\[
H_{(1)} = - \bigg( \frac{d}{dt} \bigg)^2 + t^2 + 1,
\]
and $\Delta_{\mathbb{S}^{n-1}}$ is the Laplace operator on the unit $(n-1)$-sphere $\mathbb{S}^{n-1}$ 
(after addition of the unit operator). It is not difficult to see that an appropriate negative integer power $-k$ of $\Delta_{\mathbb{S}^{n-1}}$ (after addition of a constant) is Hilbert-Smidt operator, or that $k$-th power of $\Delta_{\mathbb{S}^{n-1}}$ (after addition of constant $c\boldsymbol{1}$ with $c$ lying in the resolvent set of $\Delta_{\mathbb{S}^{n-1}}$) is a standard operator on $L^2(\mathbb{S}^{n-1}, \ud \mu_{\mathbb{S}^{n-1}})$ . Indeed it follows from the general properties of the resolvents of Laplace operators on compact manifolds, but one can check it by an explicit calculation using the following
\begin{enumerate}
\item[\bf{Fact}]
\[
\{ \lambda = l(l+n_{{}_{0}}-2), l= 0,1,2, \ldots\} = \Sp \Delta_{\mathbb{S}^{n_{{}_{0}}-1}}
\]
with the multiplicity of each $\lambda = l(l+n_{{}_{0}}-2)$ equal to
\[
{l+n_{{}_{0}}-1 \choose n_{{}_{0}}-1} 
- {l+n_{{}_{0}}-3 \choose n_{{}_{0}}-1},
\]
compare e.g. \cite{Shubin}, Ch. III. \S 22.
\end{enumerate}
 It is likewise easy to verify that 
$\mathcal{S}_{\Delta_{\mathbb{S}^{n-1}}}(\mathbb{S}^{n-1}) = \mathscr{C}^{\infty}(\mathbb{S}^{n-1})$,
where the system of norms given by $|{\Delta_{\mathbb{S}^{n-1}}}^k \cdot |_{{}_{L^2(\mathbb{S}^{n-1})}}$ is equivalent to the system of norms given by the suprema $\sup \limits_{s \in \mathbb{S}^{n-1}}$ of the absolute value 
$|\partial_{k_{1}} \ldots \partial_{k_{m}} f(s)|$ of derivatives $\partial_{k}$ with respect to one parameter groups of diffeomorphisms generated by one parameter subgroups of $SO(n)$ naturally acting on $\mathbb{S}^{n-1}$.
It is not difficult to see that an equivalent system of norms on the nuclear space 
$\mathscr{C}^{\infty}(\mathbb{S}^{n-1})$ is given by the suprema of the absolute values of derivatives of any order with respect to the coordinates of the two maps of compact domains obtained by the steregraphic projections form the ``north'' and ``south'' poles\footnote{Another proof that 
$\mathscr{C}^{\infty}(\mathbb{S}^1)$ with the system of norms indicated here is a nuclear countably Hilbert space may be found e.g. in \cite{GelfandIV}, Ch. 3.6. The proof presented there may likewise be easily adopted to the more general case
$\mathscr{C}^{\infty}(\mathbb{S}^{n-1})$.}.  Moreover
because $H_{(1)}$ and $\Delta_{\mathbb{S}^{n-1}}$ are standard then by the Propositions of Subsect. \ref{white-setup} it follows that 
$A = H_{(1)} \otimes \boldsymbol{1} + \boldsymbol{1} \otimes \Delta_{\mathbb{S}^{n-1}}$
is standard and 
\[
\mathcal{S}_{A = H_{(1)} \otimes \boldsymbol{1} + \boldsymbol{1} \otimes \Delta_{\mathbb{S}^{n-1}}}
(\mathbb{R} \times \mathbb{S}^{n-1}) =
\mathcal{S}_{H_{(1)}}(\mathbb{R}) \otimes \mathcal{S}_{\Delta_{\mathbb{S}^{n-1}}}(\mathbb{S}^{n-1})
= \mathcal{S}(\mathbb{R}) \otimes \mathscr{C}^{\infty}(\mathbb{S}^{n-1}),
\]
with the projetive tensor product of the nuclear spaces on the right.

In the next step we apply the unitary operator 
$U = U_2 \circ U_1: L^2(\mathbb{R} \times \mathbb{S}^{n-1}, \ud t \times \ud \mu_{\mathbb{S}^{n-1}})
\rightarrow L^2(\mathbb{R}^n, \ud^n p)$ to the standard operator 
$A = H_{(1)} \otimes \boldsymbol{1} + \boldsymbol{1} \otimes \Delta_{\mathbb{S}^{n-1}}$  
on $L^2(\mathbb{R} \times \mathbb{S}^{n-1}, \ud t \times \ud \mu_{\mathbb{S}^{n-1}})$ 
in order to construct the desired standard operator
\[
A^{(n)} = U\big( H_{(1)} \otimes \boldsymbol{1} + \boldsymbol{1} \otimes \Delta_{\mathbb{S}^{n-1}} \big)U^{-1}
\] 
on $L^2(\mathbb{R}^n, \ud^n p)$. Let 
\[
e_{n,m}(t, \phi_1, \ldots \phi_{n-1} ) =  h_n \otimes Y_m (t, \phi_1, \ldots \phi_{n-1}) =
h_n(t) Y_m (\phi_1, \ldots \phi_{n-1})
\] 
be the complete orthonormal system of eigenfunctions of the operator $A = H_{(1)} \otimes \boldsymbol{1} + \boldsymbol{1} \otimes \Delta_{\mathbb{S}^{n-1}}$ in 
\[
L^2(\mathbb{R} \times \mathbb{S}^{n-1}, \ud t \times  \ud \mu_{\mathbb{S}^{n-1}})
= L^2(\mathbb{R} , \ud t) \otimes L^2(\mathbb{S}^{n-1}, \ud \mu_{\mathbb{S}^{n-1}});
\]
note that
$h_n$ are the Hermite functions -- the eigenfunctions of $H_{(1)}$ and $Y_{m}$ are the eigenfunctions of the
Laplace operator $\Delta_{\mathbb{S}^{n-1}}$ on $L^2(\mathbb{S}^{n-1}, \ud \mu_{\mathbb{S}^{n-1}})$. 
The unitary operator  $U = U_2 U_1$ applied to the complete orthonormal system $\{e_{n,m}\}$
of eigenfunctions of the self adjoint operator 
$A= H_{(1)} \otimes \boldsymbol{1} + \boldsymbol{1} \otimes \Delta_{\mathbb{S}^{n-1}}$
gives the complete othonormal system
\[
Ue_{n,m} = u_{n,m}
\] 
in $L^2(\mathbb{R}^n, \ud^n p)$ of the self adjoint standard operator $A^{(n)}$, as the unitary equivalence
$UAU^{-1}$ preserves the requirements (A1)-(A3) fulfilled by $A$. 

It is obvious by the very construction that the rotation transformations naturally acting in $L^2(\mathbb{R}^n, \ud^n p)$
as unitary operators compose unitary symmetries of the operator $A^{(n)}$, i.e. $A^{(n)}$ is rotationally symmetric.
Thus, the corresponding nuclear space $\mathcal{S}_{A^{(n)}}(\mathbb{R}^n)$ is invariant under rotations which, as unitary operators on $L^2(\mathbb{R}^n, \ud^n p)$, transform continuously $E = \mathcal{S}_{A^{(n)}}(\mathbb{R}^n)$ 
into itself. Note that this is not the case for example for the nuclear space 
\[
\mathcal{S}_{A^{(1)}}(\mathbb{R}) \otimes \mathcal{S}_{A^{(1)}}(\mathbb{R}) \otimes \mathcal{S}_{A^{(1)}}(\mathbb{R})
\subset L^2(\mathbb{R}^3, \ud^3 \p)
\]  
which is not invariant under rotations.

In the later part of this work we show that $E = \mathcal{S}_{A^{(3)}}(\mathbb{R}^3) \subset \mathcal{H}'$ 
is not only invariant 
under rotations but under the full {\L}opusza\'nski representation and its conjugation.

Easy computation shows that 
\[
\begin{split}
\nu_2(t) = \frac{t + \sqrt{t^2 + 4}}{t^2 + 4 -t \sqrt{t^2 + 4}}, \\
 \nu_3(t) = \frac{1}{2}\frac{(t + \sqrt{t^2 + 4})^2}{t^2 + 4 -t \sqrt{t^2 + 4}},\\
\nu_n(t) = \frac{1}{2^{n-2}}\frac{(t + \sqrt{t^2 + 4})^{n-1}}{t^2 + 4 -t \sqrt{t^2 + 4}}.
\end{split}
\]
For each dimension $n$ there exists the unitary involutive symmetry ${\Inv}_{(n)}$ of $A^{(n)}$
 in $L^2(\mathbb{R}^n, \ud^n p)$ and of the corresponding nuclear space $\mathcal{S}_{A^{(n)}}(\mathbb{R}^n)$
transforming continuously $\mathcal{S}_{A^{(n)}}(\mathbb{R}^n)$ into itself. 
Namely we have
\[
\begin{split}
{\Inv}_{(n)} f (r, \phi_1, \ldots \phi_{n-1}) = r^{-n} f(r^{-1}, \phi_1, \ldots \phi_{n-1}),
\,\,\, f \in L^2(\mathbb{R}^n, \ud^n p).
\end{split}
\]

The general formula for the differential operator $A^{(n)}$ in the spherical coordinates 
can be explicitly written at once without any computation for arbitrary $n$, so writing the explicit
formula in the spherical coordinates would be aimless. The formula for 
$A^{(n)}$ in the Cartesian coordinates can likewise be explicitly written, but it is more complicated.
In particular, we have 
\begin{multline*}
A^{(3)} = A''' = \Big\{- \frac{r^2}{r^2 +1} + r^2 \Big\} \Big( \sum \limits_{i,j=1}^{3} \frac{x_i}{r}\frac{x_j}{r}
\frac{\partial}{\partial x_i} \frac{\partial}{\partial x_j}\Big) \\
- r^2 \Delta_{\mathbb{R}^3} + \Big\{- \frac{r^3(r^2 + 4)}{(r^2 +1)^3} + 2r \Big\} \Big( \sum \limits_{i = 1}^{3} 
\frac{x_i}{r}\frac{\partial}{\partial x_i} \Big) \\
+ \Big\{ \frac{r^2(r^2 + 4)(r^2 - 2)}{4(r^2 + 1)^4} + r^2 + r^{-2} \Big\}.
\end{multline*}

Note that the operator $A^{(n)}$ is well-defined and symmetric on the nuclear (and thus perfect)
space $\mathcal{S}^0(\mathbb{R}^n)$ and transforms $\mathcal{S}^0(\mathbb{R}^n)$ into itself. By the already cited criterion of Riesz and Sz\"okefalvy-Nagy $A^{(n)}$ possesses an extension to a self adjoint operator
in $L^2(\mathbb{R}^n, \ud^n p)$, as expected by the very construction of the operator $A^{(n)}$.
Because moreover $A^{(n)}$ with domain $\mathcal{S}^0(\mathbb{R}^n)$ possess by construction the complete orthonormal system belonging to 
$\mathcal{S}^0(\mathbb{R}^n)$, ten it is diagonalizable, and thus essentially self adjoint. This means 
that $A^{(n)}$ with domain $\mathcal{S}^0(\mathbb{R}^n)$ has exactly one self adjoint extension,
let us denote it by the same sign $A^{(n)}$.
Because $A^{(n)} \big( \mathcal{S}^0(\mathbb{R}^n) \big) \subset \mathcal{S}^0(\mathbb{R}^n)$,  
then it follows that 
\[
\mathcal{S}^0(\mathbb{R}^n) \subset \mathcal{S}_{A^{(n)}}(\mathbb{R}^n).
\] 
The opposite inclusion will be shown latter.

Note that the unitary operator $U = U_2U_1$ constructed above defines in a canonical manner
a natural isomorphism of the corresponding nuclear spaces
\[
\mathcal{S}_{A}(\mathbb{R}^n)=
\mathcal{S}_{H_{(1)}\otimes \boldsymbol{1} + \boldsymbol{1} \otimes
\Delta_{\mathbb{S}^{n-1}}}(\mathbb{R} \times \mathbb{S}^{n-1})
= \mathcal{S}_{H_{(1)}}(\mathbb{R}) \otimes \mathcal{S}_{\Delta_{\mathbb{S}^{n-1}}}(\mathbb{S}^{n-1})
\]
and
\[
\mathcal{S}_{UAU^{-1}}(\mathbb{R}^n) = \mathcal{S}_{A^{(n)}}(\mathbb{R}^n).
\]
Note further that the restriction to the cone $(p_1)^2 - (p_2)^2 - \ldots - (p_n)^2 = 0$ and $p_1 >0$ (or $p_1 <0$)
defines a map on $\mathcal{S}_{A^{(n)}}(\mathbb{R}^n)$, which through the above canonical isomorphism
correspond to the map on $\mathcal{S}_{H_{(1)}\otimes \boldsymbol{1} + \boldsymbol{1} \otimes \Delta_{\mathbb{S}^{n-1}}}(\mathbb{R} \times \mathbb{S}^{n-1})$, which a function in $\mathcal{S}_{H_{(1)}\otimes \boldsymbol{1} + \boldsymbol{1} \otimes \Delta_{\mathbb{S}^{n-1}}}(\mathbb{R} \times \mathbb{S}^{n-1})$
sends into its restriction to the sub-manifold of $\mathbb{R}\times \mathbb{S}^{n-1}$ given by
$\phi_1 = \pi /4$ (or $\phi_1 = 3/4 \pi$ respectively). In particular the map on $\mathcal{S}_{A^{(n)}}(\mathbb{R}^n)$, given by the restriction to the cone $(p_1)^2 - (p_2)^2 - \ldots - (p_n)^2 = 0$ and $p_1 >0$ (or $p_1 <0$), sends an element of $\mathcal{S}_{A^{(n)}}(\mathbb{R}^n)$ into an element
of $\mathcal{S}_{A^{(n-1)}}(\mathbb{R}^{n-1})$, if and only if the corresponding map on
$\mathcal{S}_{H_{(1)}\otimes \boldsymbol{1} + \boldsymbol{1} \otimes \Delta_{\mathbb{S}^{n-1}}}(\mathbb{R} \times \mathbb{S}^{n-1})$ defined by the restriction to the latitude $\phi_1 = \pi /4$ (or $\phi_1 = 3/4 \pi$ respectively) sends the corresponding element of $\mathcal{S}_{H_{(1)}\otimes \boldsymbol{1} + \boldsymbol{1} \otimes \Delta_{\mathbb{S}^{n-1}}}(\mathbb{R} \times \mathbb{S}^{n-1})$
into an element of $\mathcal{S}_{H_{(1)}\otimes \boldsymbol{1} + \boldsymbol{1} \otimes \Delta_{\mathbb{S}^{n-2}}}(\mathbb{R} \times \mathbb{S}^{n-2})$. Because on the other hand
\[
\mathcal{S}_{H_{(1)}\otimes \boldsymbol{1} + \boldsymbol{1} \otimes \Delta_{\mathbb{S}^{n-1}}}(\mathbb{R} \times \mathbb{S}^{n-1})
= \mathcal{S}_{H_{(1)}}(\mathbb{R}) \otimes \mathcal{S}_{\Delta_{\mathbb{S}^{n-1}}}(\mathbb{S}^{n-1})
\]
has the natural tensor product structure it is easily seen that the map defined by the restriction to the latitude $\phi_1 = \pi /4$ is a (continuous) map
\[
\mathcal{S}_{H_{(1)}\otimes \boldsymbol{1} + \boldsymbol{1} \otimes \Delta_{\mathbb{S}^{n-1}}}(\mathbb{R} \times \mathbb{S}^{n-1}) \rightarrow \mathcal{S}_{H_{(1)}\otimes \boldsymbol{1} + \boldsymbol{1} \otimes \Delta_{\mathbb{S}^{n-2}}}(\mathbb{R} \times \mathbb{S}^{n-2})
\]
if and only if the map of $\mathcal{S}_{\Delta_{\mathbb{S}^{n-1}}}(\mathbb{S}^{n-1})$, given by the restriction to the latitude $\phi_1 = \pi/4$, is a map transforming (continuously)
$\mathcal{S}_{\Delta_{\mathbb{S}^{n-1}}}(\mathbb{S}^{n-1})$ into
$\mathcal{S}_{\Delta_{\mathbb{S}^{n-2}}}(\mathbb{S}^{n-2})$. That the last map is indeed a continuous map
easily follows from the fact that any one parameter group of diffeomorphisms corresponding to a one parameter subgroup
of $SO(n-1)$ acting naturally on the sub-manifold of $\mathbb{S}^{n-1}$ given by the equation $\phi_1 = \pi/4$
(or $\phi_1 = 3 \pi/4$), i.e. on the $(n-2)$-sphere, is a restriction of a one parameter subgroup of
$SO(n) \supset SO(n-1)$ to the sub-manifold $\phi_1 = \pi /4$ (resp. $\phi_1 = 3\pi /4$). The statement likewise easily follows from the fact that
the system of norms on $\mathcal{S}_{\Delta_{\mathbb{S}^{n-1}}}(\mathbb{S}^{n-1})
= \mathscr{C}^{\infty}(\mathbb{S}^{n-1})$, defined by the suprema of the absolute values of the derivatives with respect to the coordinates of the two maps given by the steregraphic projections, gives a system of norms equivalent to the original
system given by $|\cdot |_k = |(\Delta_{\mathbb{S}^{n-1}})^k \cdot |_{{}_{L^2(\mathbb{S}^{n-1})}}$. Therefore,
the map defined by the restriction to the cone defines a continuous map $\mathcal{S}_{A^{(n)}}(\mathbb{R}^n)
\rightarrow \mathcal{S}_{A^{(n-1)}}(\mathbb{R}^{n-1})$ in the nuclear topology.
Thus, we have proven the following

\begin{lem*}
The map defined by the restriction of a function on $\mathbb{R}^n$ to the cone 
$(p_1)^2 - (p_2)^2 - \ldots - (p_n)^2 = 0$ and $p_1 >0$ (or $p_1 <0$) is a map which continuously
transforms $\mathcal{S}_{A^{(n)}}(\mathbb{R}^n)$ into $\mathcal{S}_{A^{(n-1)}}(\mathbb{R}^{n-1})$.
\end{lem*}

Note that the restriction to the cone is not continuous as a map
$\mathcal{S}(\mathbb{R}^n) \rightarrow \mathcal{S}(\mathbb{R}^{n-1})$. Indeed, the restriction to the cone
leads to the elimination of one coordinate, $p_1$, which has to be expressed as non-trivial square root
of the sum of squares of the remaining Cartesian coordinates. Such a function leads us out of $\mathcal{S}(\mathbb{R}^{n-1})$, and in particular the differentiation operation with respect to the remaining coordinates leads to a singularity at the zero point, which is of course impossible for any element of $\mathcal{S}(\mathbb{R}^{n-1})$. This is connected to the fact that the ordinary Schwartz space
$\mathcal{S}(\mathbb{R}^n)$ does not have any natural tensor product structure of the form $\mathcal{S}_{C}(\mathbb{R})
\otimes \mathcal{S}_{\Delta_{\mathbb{S}^{n-1}}}(\mathbb{S}^{n-1})$ and is not naturally isomorphic to such a tensor product
of nuclear spaces $\mathcal{S}_{C}(\mathbb{R})$ and $\mathcal{S}_{\Delta_{\mathbb{S}^{n-1}}}(\mathbb{S}^{n-1})$
for any standard operator $C$ on $L^2(\mathbb{R})$.

\subsection{Multipliers, convolutors and differentiation
operation on $\mathcal{S}_{A^{(n)}}(\mathbb{R}^n)$}\label{diffSA}

In this Subsection we reduce the problem of investigation of multipliers, convolutors, differentiation
operation, continuous functionals, $\ldots$ on $\mathcal{S}_{A^{(n)}}(\mathbb{R}^n)$ to the investigation
of multipliers, convolutors, differentiation
operation, continuous functionals, $\ldots$ on
\[
\mathcal{S}(\mathbb{R}) \otimes \mathscr{C}^{\infty}(\mathbb{S}^{n-1}) =
\mathcal{S}_{H_{(1)}}(\mathbb{R}) \otimes \mathcal{S}_{\Delta_{\mathbb{S}^{n-1}}}(\mathbb{S}^{n-1}) =
\mathcal{S}_{H_{(n)}\otimes \boldsymbol{1} + \boldsymbol{1} \otimes \Delta_{\mathbb{S}^{n-1}}}(\mathbb{R} \times \mathbb{S}^{n-1}).
\]
We do it exactly as we did in Subsection \ref{dim=1} by application of the Lemma completely analogous to the Lemma
of Subsection \ref{dim=1}, using the identity of the spectra of the operators $A^{(n)}
= U\Big( H_{(n)} \otimes \boldsymbol{1}
+ \boldsymbol{1} \otimes \Delta_{\mathbb{S}^{n-1}} \Big)U^{-1}$ and $H_{(n)} \otimes \boldsymbol{1}
+ \boldsymbol{1} \otimes \Delta_{\mathbb{S}^{n-1}}$.

This method uses the natural tensor product structure of $\mathcal{S}_{A^{(n)}}(\mathbb{R}^n)$ inherited form the
product structure of the manifold $\mathbb{R} \times \mathbb{S}^{n-1}$ by the unitary operator
$U =U_2U_1$ of Subsection \ref{dim=n} transforming $\mathcal{S}_{H_{(n)}\otimes \boldsymbol{1} + \boldsymbol{1} \otimes \Delta_{\mathbb{S}^{n-1}}}(\mathbb{R} \times \mathbb{S}^{n-1})$ onto $\mathcal{S}_{A^{(n)}}(\mathbb{R}^n)$. This method prefers the generalized spherical coordinates, although theorems referring to the Cartesian coordinates can likewise be reached in this way.
We thus reduce the problem to the investigation of the simpler nuclear spaces $\mathcal{S}(\mathbb{R})$ and
$\mathscr{C}^{\infty}(\mathbb{S}^{n-1})$.

Before we proceed to the details, let us make a general remark that the presented method admits generalizations.
For example, we can consider the two complete orthonormal systems corresponding respectively to the operators
$H_{(n)}$ and $A^{(n)}$ in $L^2(\mathbb{R}^n)$. Then we can define a unitary operator (the analogue of the operator
$U = U_2U_1$ of Subsection \ref{dim=n} by associating each element of the first complete orthonormal system to a corresponding element
in the second orthonormal system. Although $U H_{(n)}U^{-1} \neq A^{(n)}$ (exactly as in Subsection \ref{dim=1},
where $H_{(1)}$ and $A^{(1)}$ are not unitarily equivalent) the asymptotics
of the spectra of the operators $A^{(n)}$ and $H_{(n)}$ are close enough for the applicability of the reduction method
of Subsection \ref{dim=1}, as we have shown in the Appendix \ref{asymptotics}. In this manner we reduce the problem of determination continuous operators on $\mathcal{S}_{A^{(n)}}(\mathbb{R}^n)$ to the determination
of continuous operators on the Schwartz space
\[
\mathcal{S}(\mathbb{R}^n) = \mathcal{S}_{H_{(n)}}(\mathbb{R}^n)
= \mathcal{S}_{\Gamma_{n}(H_{(1)})}(\mathbb{R}^n) = \Big( \mathcal{S}_{H_{(1)}}(\mathbb{R}) \Big)^{\otimes n}
= \Big( \mathcal{S}(\mathbb{R}) \Big)^{\otimes n}.
\]
This is likewise quite effective method for investigation of the family of nuclear spaces
$\mathcal{S}_{A^{(n)}}(\mathbb{R}^n)$.

For our purposes the first method preferring the spherical coordinates is sufficient and seems to be simpler, 
as the transformation between the complete systems $Ue_{n,m}$ and $e_{n,m}$ is simpler than the transformation expressing 
$Ue_{n,m}$ in terms of the complete system of eigenfunctions of $H_{(n)}$.
For example, we have

\begin{lem*}
The functions
\[
\begin{split}
 r^{-1}:\, (r, \phi_1, \ldots \phi_{n-1}) \mapsto r^{-1} \,\, \textrm{or} \,\,
(p_1, \ldots p_n) \mapsto \big((p_1)^2 + \ldots + (p_n)^2 \big)^{-1/2}, \\
r: \,(r, \phi_1, \ldots \phi_{n-1}) \mapsto r \,\, \textrm{or} \,\,
(p_1, \ldots p_n) \mapsto \big((p_1)^2 + \ldots + (p_n)^2 \big)^{1/2}, \\
p_i \, :(p_1, \ldots p_n) \mapsto p_i,
\end{split}
\]
and more generally, the functions $r^{-\frac{1}{k}}$, $k \in \mathbb{N}$, are all multipliers of the nuclear algebra 
$\mathcal{S}_{A^{(n)}}(\mathbb{R}^n)$.
\end{lem*}

\qedsymbol \, 
 Let $e_{n,m} (t,\phi_1, \ldots \phi_{n-1}) = h_n \otimes Y_{m}(t, \phi_1, \ldots \phi_{n-1})= h_n(t) Y_{m}(\phi_1,
\ldots \phi_{n-1})$
be the complete orthonormal system of the eigenfunctions of the operator 
$H_{(1)} \otimes \boldsymbol{1} + \boldsymbol{1} \otimes \Delta_{\mathbb{S}^{n-1}}$ on
$L^2(\mathbb{R}\times \mathbb{S}^{n-1}, \ud t \times \ud \mu_{\mathbb{S}^{n-1}})$, note also that we are using spherical coordinates. Let $Ue_{n,m}$ be the complete system of the operator $A^{(n)}$ in $L^2(\mathbb{R}^n, \ud^n p)$, where $U = U_2 U_1$ is the unitary transformation of Subsection \ref{dim=n}. Recall that $t(r) = r - r^{-1}$, compare Subsection \ref{dim=n}. To simplify notation let us note the density function on the $(n-1)$-sphere $\mathbb{S}^{n-1}$ by $\omega$, so that 
\[
\omega(\phi_1, \ldots \phi_{n-2}) = \sin^{n-2} \phi_1 \sin^{n-3} \phi_{2} \ldots 
\sin \phi_{n-2}, 
\]
\[
\ud \mu_{\mathbb{S}^{n-1}} = \omega(\phi_1, \ldots \phi_{n-2}) \ud \phi_1 \ldots \ud \phi_{n-1},
\]
\[
\ud^n p = r^{n-1} \ud r \ud \mu_{\mathbb{S}^{n-1}} = r^{n-1} \omega \, \ud r \ud \phi_1 \ldots \ud \phi_{n-1},
\]
and 
\[
\ud t \times \ud \mu_{\mathbb{S}^{n-1}} =  \omega \, \ud t \ud \phi_1 \ldots \ud \phi_{n-1}.
\]
Then for the matrix elements of the operator of multiplication 
by the function $r^{-1}$ we obtain 
\begin{multline*}
\Big\langle \, {}_{nm} \, \Big| r^{-1} \Big| \, {}_{n'm'} \, \Big\rangle \\
= \int \limits_{\mathbb{R}^3} \overline{U(h_n \otimes Y_{m})(r,\phi_1, \ldots \phi_{n-1})} \, r^{-1} \, U(h_{n'} \otimes Y_{m'})(r,\phi_1, \ldots \phi_{n-1}) \,\, 
\overbrace{r^{n-1} \omega \, \ud r \ud \phi_1 \ldots \ud \phi_{n-1}}^{\ud^n p} \\
= \int \limits_{\mathbb{R}^3} \overline{\frac{1}{\sqrt{\nu_{{}_{n}}(t(r))}}h_{n'} (t(r)) Y_{m'}(\phi_1, \ldots \phi_{n-1})} 
\underbrace{r^{-1}}_{\frac{-t(r)+\sqrt{t(r)^2 + 4}}{2}} \,\, 
\times  \\ \times \,\,
\frac{1}{\sqrt{\nu_{{}_{n}}(t(r))}}h_{n'} (t(r)) Y_{m'}(\phi_1, \dots ,\phi_{n-1}) 
\underbrace{r^{n-1} \omega}_{\nu_{{}_{n}}(t(r)) \, \omega \,  \Big|\det \frac{\partial(t,\phi_1, \ldots, \phi_{n-1})}{\partial(r,\phi_1, \ldots ,\phi_{n-1})} \Big|} 
dr d \phi_1 \ldots  d\phi_{n-1} 
\end{multline*}
\begin{multline*}
= \int \limits_{\mathbb{R}^3} \overline{\frac{1}{\sqrt{\nu_{{}_{n}}(t(r))}}h_n (t(r)) Y_{m}(\phi_1, \ldots \phi_{n-1})} 
\frac{-t(r)+\sqrt{t(r)^2 + 4}}{2}  \,\,
\times  \\ \times \,\,
\frac{1}{\sqrt{\nu_{{}_{n}}(t(r))}}h_{n'} (t(r)) Y_{m'}(\phi_1, \ldots, \phi_{n-1}) 
\nu_{{}_{n}}(t(r)) \omega \Big|\det \frac{\partial(t,\phi_1, \ldots \phi_{n-1})}{\partial(r,\phi_1, \ldots \phi_{n-1})} \Big| 
dr d \phi_1 \ldots d\phi_{n-1} 
\end{multline*}
\begin{multline*}
= \int \limits_{\mathbb{R}\times \mathbb{S}^2} \overline{h_n (t) Y_{m}(\phi_1, \ldots, \phi_{n-1})} 
\frac{-t+\sqrt{t^2 + 4}}{2}  
h_{n'} (t) Y_{m'}(\phi_1, \ldots, \phi_{n-1}) 
 \omega \, dt d \phi_1 \ldots d\phi_{n-1} \\ 
= \Big( \, {}_{nm} \, \Big| \frac{-t+\sqrt{t^2 + 4}}{2} \Big| \, {}_{n'm'} \, \Big) 
\end{multline*}
\begin{multline*}
= \int \limits_{\mathbb{R}} \overline{h_n (t)} \,
\frac{-t+\sqrt{t^2 + 4}}{2}  h_{n'} (t) dt \, \cdot
\int \limits_{\mathbb{S}^2} \overline{Y_m(\phi_1, \ldots, \phi_{n-1})} Y_{m'}(\phi_1, \ldots, \phi_{n-1}) 
 \omega \, d \phi_1 \ldots d\phi_{n-1} \\
= \delta_{mm'} \,\, \int \limits_{\mathbb{R}} \overline{h_n (t)} \,
\frac{-t+\sqrt{t^2 + 4}}{2}  h_{n'} (t) dt 
= \delta_{mm'} \,\, \Big( n \Big| \frac{-t+\sqrt{t^2 + 4}}{2}  \Big|n' \Big).
\end{multline*}
Because the function
\[
g_1: \, t \mapsto \frac{-t+\sqrt{t^2 + 4}}{2} 
\]
is a multiplier of the algebra $\mathcal{S}(\mathbb{R}) = \mathcal{S}_{H_{(1)}}(\mathbb{R})$, then the function 
\[
g_2: \, (t,\theta, \phi_1, \ldots \phi_{n-1}) \mapsto \frac{-t+\sqrt{t^2 + 4}}{2} 
\]
is a multiplier of the algebra 
\[
\mathcal{S}_{H_{(1)}}(\mathbb{R}) \otimes \mathcal{S}_{\Delta_{\mathbb{S}^{n-1}}}(\mathbb{S}^{n-1})
= \mathcal{S}_{H_{(1)} \otimes \boldsymbol{1} + \boldsymbol{1}\otimes \Delta_{\mathbb{S}^{n-1}}}
(\mathbb{R} \times \mathbb{S}^{n-1}).
\]
Indeed: note that the operator $M_{{}_{g_2}}$ of multiplication by the function $g_2$, acting on $\mathcal{S}_{H_{(1)}}(\mathbb{R}) \otimes \mathcal{S}_{\Delta_{\mathbb{S}^{n-1}}}(\mathbb{S}^{n-1})$, is equal to the tensor product

$M_{{}_{g_2}} = M_{{}_{g_1}} \otimes \boldsymbol{1}$ of  the operator $M_{{}_{g_1}}$
of multiplication by the function $g_1$ and of the operator $\boldsymbol{1}$, acting respectively on 
$\mathcal{S}_{H_{(1)}}(\mathbb{R})$ and 
$\mathcal{S}_{\Delta_{\mathbb{S}^{n-1}}}(\mathbb{S}^{n-1})$, which by Proposition 43.6 of \cite{treves}
is a continuous operator 
\[
\mathcal{S}_{H_{(1)}}(\mathbb{R}) \otimes \mathcal{S}_{\Delta_{\mathbb{S}^{n-1}}}(\mathbb{S}^{n-1}) \rightarrow
\mathcal{S}_{H_{(1)}}(\mathbb{R}) \otimes \mathcal{S}_{\Delta_{\mathbb{S}^{n-1}}}(\mathbb{S}^{n-1}). 
\]
Because the spectra of the operators
$H_{(1)} \otimes \boldsymbol{1} + \boldsymbol{1}\otimes \Delta_{\mathbb{S}^{n-1}}$ and $A^{(n)}
= U \big( H_{(1)} \otimes \boldsymbol{1} + \boldsymbol{1}\otimes \Delta_{\mathbb{S}^{n-1}} \big)U^{-1}$ are identical,
as the operators are unitarily equivalent, then we may proceed as in the proof of the second Lemma of Subsection \ref{dim=1}
and show that the function
\[
r^{-1}: (r,\phi_1, \ldots \phi_{n-1}) \mapsto r^{-1}
\]
is a multiplier of the algebra 
\[
\mathcal{S}_{A^{(n)}}(\mathbb{R}^n) = \mathcal{S}_{U\big(H_{(1)} \otimes \boldsymbol{1} + \boldsymbol{1}\otimes 
\Delta_{\mathbb{S}^{n-1}}\big)U^{-1}}(\mathbb{R}^n).
\]

And similarly because for each $k \in \mathbb{N}$ the function
\[
t \mapsto \Big(\frac{-t+\sqrt{t^2 + 4}}{2}\Big)^{\frac{1}{k}} 
\]
is a multiplier of the algebra $\mathcal{S}(\mathbb{R}) = \mathcal{S}_{H_{(1)}}(\mathbb{R})$, then the function 
\[
r^{-\frac{1}{k}}: (r,\phi_1, \ldots \phi_{n-1}) \mapsto r^{-\frac{1}{k}}
\]
is a multiplier of the algebra $\mathcal{S}_{A^{(n)}}(\mathbb{R}^n)$.

Similarly, because
\[
r = \frac{t(r)+\sqrt{t(r)^2 + 4}}{2}
\]
and the function 
\[
t \mapsto \frac{t+\sqrt{t^2 + 4}}{2} 
\]
is a multiplier of the algebra $\mathcal{S}(\mathbb{R}) = \mathcal{S}_{H_{(1)}}(\mathbb{R})$, then the function 
$r$ is a multiplier of the algebra $\mathcal{S}_{A^{(n)}}(\mathbb{R}^n)$. 

Further, the functions 
\[
\left\{ \begin{array}{l}
s_1: (\phi_1, \ldots \phi_{n-1}) \mapsto \cos \phi_1, \\

s_2: (\phi_1, \ldots \phi_{n-1}) \mapsto \sin \phi_1 \cos \phi_2, \\
s_3: (\phi_1, \ldots \phi_{n-1}) \mapsto \sin \phi_1 \sin \phi_2 \cos \phi_3, \\
\ldots, \\
s_{n-1}(\phi_1, \ldots \phi_{n-1}) \mapsto \sin \phi_1 \ldots \sin \phi_{n-2} \cos \phi_{n-1}, \\
s_n: (\phi_1, \ldots \phi_{n-1}) \mapsto \sin \phi_1 \ldots \sin \phi_{n-2} \sin \phi_{n-1},
\end{array} \right.
\]
 are easily checked  to be 
multipliers of the nuclear algebra 
$\mathcal{S}_{\Delta_{\mathbb{S}^{n-1}}}(\mathbb{S}^{n-1}) = \mathscr{C}^{\infty}(\mathbb{S}^{n-1})$
when using the stereographic projection maps and supremum norms mentioned in Subsection \ref{dim=n}.
Because of the tensor product structure of the algebra 
\[
\mathcal{S}_{H_{(1)}}(\mathbb{R}) \otimes \mathcal{S}_{\Delta_{\mathbb{S}^{n-1}}}(\mathbb{S}^{n-1})
= \mathcal{S}_{H_{(1)} \otimes \boldsymbol{1} + \boldsymbol{1}\otimes \Delta_{\mathbb{S}^{n-1}}}
(\mathbb{R} \times \mathbb{S}^{n-1}).
\]
it follows, again by Proposition 43.6 of \cite{treves}, that the functions 

\[
\left\{ \begin{array}{l}
g_1: (t,\phi_1, \ldots \phi_{n-1}) \mapsto \cos \phi_1, \\
g_2: (t,\phi_1, \ldots \phi_{n-1}) \mapsto \sin \phi_1 \cos \phi_2, \\
g_3: (t,\phi_1, \ldots \phi_{n-1}) \mapsto \sin \phi_1 \sin \phi_2 \cos \phi_3, \\
\ldots, \\
g_{n-1}: (t,\phi_1, \ldots \phi_{n-1}) \mapsto \sin \phi_1 \ldots \sin \phi_{n-2} \cos \phi_{n-1}, \\
g_n: (t,\phi_1, \ldots \phi_{n-1}) \mapsto \sin \phi_1 \ldots \sin \phi_{n-2} \sin \phi_{n-1},
\end{array} \right.
\]
are all multipliers of the algebra 
\[
\mathcal{S}_{H_{(1)}}(\mathbb{R}) \otimes \mathcal{S}_{\Delta_{\mathbb{S}^{n-1}}}(\mathbb{S}^{n-1})
= \mathcal{S}_{H_{(1)} \otimes \boldsymbol{1} + \boldsymbol{1}\otimes \Delta_{\mathbb{S}^{n-1}}}
(\mathbb{R} \times \mathbb{S}^{n-1}).
\]
Thus the functions 
\[
\left\{ \begin{array}{l}
f_1 :(r,\phi_1, \ldots \phi_{n-1}) \mapsto \cos \phi_1, \\
f_2: (r,\phi_1, \ldots \phi_{n-1}) \mapsto \sin \phi_1 \cos \phi_2, \\
f_3: (r,\phi_1, \ldots \phi_{n-1}) \mapsto \sin \phi_1 \sin \phi_2 \cos \phi_3, \\
\ldots, \\
f_{n-1}: (r,\phi_1, \ldots \phi_{n-1}) \mapsto \sin \phi_1 \ldots \sin \phi_{n-2} \cos \phi_{n-1}, \\
f_n: (r,\phi_1, \ldots \phi_{n-1}) \mapsto \sin \phi_1 \ldots \sin \phi_{n-2} \sin \phi_{n-1},
\end{array} \right.
\]
are multipliers of the algebra $\mathcal{S}_{A^{(n)}}(\mathbb{R}^n)$, which again may be easily checked by the 
computation of the matrix elements $\langle nm | f_i | n'm' \rangle = (nm|g_i|n'm')$ 
of the operators of multiplication by the functions $(r, \phi_1, \ldots, \phi_{n-1}) \mapsto f_i(r, \phi_1, \ldots, \phi_{n-1})$ and
$(t, \phi_1, \ldots, \phi_{n-1}) \mapsto g_i(t, \phi_1, \ldots, \phi_{n-1})$ and using the identity of the spectra of the operators
$A^{(n)}$ and $H_{(1)} \otimes \boldsymbol{1} + \boldsymbol{1} \otimes \Delta_{\mathbb{S}^{n-1}}$.

On the other hand in the spherical coordinates we have
\[
\left\{ \begin{array}{l}
p_1 = r \cos \phi_1, \\
p_2 = r \sin \phi_1 \cos \phi_2, \\
p_3 = r \sin \phi_1 \sin \phi_2 \cos \phi_3, \\
\ldots, \\
p_{n-1} = r \sin \phi_1 \ldots \sin \phi_{n-2} \cos \phi_{n-1}, \\
p_{n} = r \sin \phi_1 \ldots \sin \phi_{n-2} \sin \phi_{n-1},
\end{array} \right.
\]
and because composition of continuous operators is continuous, then the operators of multiplication
by the Cartesian coordinates are all continuous maps of $\mathcal{S}_{A^{(n)}}(\mathbb{R}^n)$
into itself.

\qed

Note that using the atlas on $\mathbb{S}^{n-1}$ consisting of the two stereographic maps
$(+)$ and $(-)$,
corresponding respectively to the projection from the ''North Pole'' and from the ''South Pole'',
we can easily prove that the functions $s_i$, $i = 1, \ldots n$ in the proof of the last Lemma
are smooth, i.e. that they are smooth
functions in the stereographic maps $(+)$ and $(-)$. Of course the domains of the maps $(+)$
and $(-)$ are compact (in fact they can be chosen to be compact $(n-1)$-balls around the origin in the euclidean
space $\mathbb{R}^{n-1}$. Using the norms in $\mathcal{S}_{\Delta_{\mathbb{S}^{n-1}}}(\mathbb{S}^{n-1})
= \mathscr{C}^{\infty}(\mathbb{S}^{n-1})$ given by the suprema of the absolute values of derivatives of any order in the coordinates of the two maps $(+)$ and $(-)$ of compact domains, we can easily show not only that the mentioned functions are multipliers
(or that they belong to $\mathcal{S}_{\Delta_{\mathbb{S}^{n-1}}}(\mathbb{S}^{n-1})$), but likewise
that the differential operator $\sin \phi_1\partial_{\phi_{1}}$ of differentiation with respect to the first ''latitude'' spherical
coordinate $\phi_1$, followed by the operator of multiplication by $\sin \phi_1$, is an operator mapping continuously
$\mathcal{S}_{\Delta_{\mathbb{S}^{n-1}}}(\mathbb{S}^{n-1})$ into itself.

For the simplicity of notation, consider the case of the $2$-sphere $\mathbb{S}^{2}$ with $\phi_1 = \theta$
and $\phi_2 = \phi$. The higher dimensional case is completely analogous.
Formulae connecting the coordinates $z = x + iy$ and $\zeta = u + iv$ of the two stereographic maps $(+)$ and $(-)$ with the spherical and Cartesian coordinates are very simple.
In particular for the map $(+)$
\[
\left\{ \begin{array}{l}
x = \frac{\sin \phi_1}{1 - \cos \phi_1} \cos \phi_2 , \\
y = \frac{\sin \phi_1}{1 - \cos \phi_1} \sin \phi_2, \\
\end{array} \right.
\]
and
\[
\left\{ \begin{array}{lll}
2 \frac{x^2 + y^2}{x^2 + y^2 +1} -1 = & \cos \phi_1 = & s_1(\phi_1, \phi_2)=p_1, \\
\frac{2x}{x^2 + y^2 +1} = & \sin \phi_1 \cos \phi_2 = & s_2(\phi_1, \phi_2)=p_2, \\
\frac{2y}{x^2 + y^2 +1} = & \sin \phi_1 \sin \phi_2 = & s_3(\phi_1, \phi_2)=p_3. \\
\end{array} \right.
\,\,\, (p_1)^2 + (p_2)^2 + (p_3)^2 = 1.
\]
Therefore in the map $(+)$ all the functions $s_i$, and in particular the function $\cos \phi_1$, are smooth. The same holds in the map $(-)$,
and the representations of the functions $s_i$, and in particular of the function $\cos \phi_1$, in the maps $(+)$ and $(-)$ glue together and compose smooth functions on the manifold $\mathbb{S}^2$. In particular $\cos \phi_1$
is a multiplier of the algebra $\mathcal{S}_{\Delta_{\mathbb{S}^2}}(\mathbb{S}^2)$.
Easy computation shows that
\[
\sin \phi_1 \, \partial_{\phi_1} = - x \partial_x - y \partial_y \,\,\, \textrm{in the map} \, (+)
\]
and similarly we get in the second stereographic map $(-)$, so that the representations of the operator
$\sin \phi_1 \, \partial_{\phi_1}$ in the maps $(+)$ and $(-)$ glue smoothly to an operator on
$\mathbb{S}^2$ which maps continuously
the nuclear space $\mathcal{S}_{\Delta_{\mathbb{S}^2}}(\mathbb{S}^2)$ into itself.
Exactly the same reasoning repeated for the $n$-dimensional stereogrpahic projections shows the validity
of the following

\begin{lem*}
1) \, The function $\cos \phi_1$, and in general $s_i$,$i= 1, \ldots n$, represent a smooth function on 
the standard manifold $\mathbb{S}^{n-1}$, and in particular
it is a multiplier of the algebra $\mathcal{S}_{\Delta_{\mathbb{S}^{n-1}}}(\mathbb{S}^{n-1})$. 2) \, The differential
operator $\sin \phi_1 \partial_{\phi_1}$ maps continuously the nuclear space $\mathcal{S}_{\Delta_{\mathbb{S}^{n-1}}}(\mathbb{S}^{n-1})$ into itself. 
\end{lem*}

Note that because the spherical coordinates fail at $\phi_1 = 0$ or $\pi$, and at $\phi_i = 0$ or $2\pi$,
for $i> 1$ as a manifold map on the standard differential manifold $\mathbb{S}^{n-1}$,
functions that are smooth in terms
of these coordinates need not be smooth as functions on the manifold $\mathbb{S}^{n-1}$ for $n>2$.
In particular $\cos \phi_1$ is smooth on the manifold $\mathbb{S}^{n-1}$, but for example
$\sin \phi_1$, $\sin \phi_i$, $\cos \phi_i$, $i>1$ are not smooth on the manifold $\mathbb{S}^{n-1}$, $n>2$
(with the standard $(n-1)$-sphere manifold structure for each $n$). Similarly, we have for differential
operators, for example the operator $\partial_{\phi_1}$ is not continuous as an operator on
the nuclear space $\mathcal{S}_{\Delta_{\mathbb{S}^{n-1}}}(\mathbb{S}^{n-1}) = \mathscr{C}^{\infty}(\mathbb{S}^{n-1})$,
$n>2$. The spherical functions $Y_{l}^{m}$ on $\mathbb{S}^2$ (and the generalized spherical functions $Y_m$ on
$\mathbb{S}^{n-1}$ -- the eigenfunctions of the Laplace operator $\Delta_{\mathbb{S}^{n-1}}$ on $\mathbb{S}^{n-1}$)
are smooth as functions on the manifold $\mathbb{S}^2$(resp. on the standard $\mathbb{S}^{n-1}$), but this
is a very nontrivial fact, and cannot be inferred from the smoothness of trigonometric functions, but
follows for example by the general properties of Laplace operators on smooth compact manifolds,
or more generally, by the \emph{regularity property} of elliptic operators on smooth manifolds.
It is rather amazing that the singularities of trigonometric functions as functions on the manifold
$\mathbb{S}^2$ expressed in spherical coordinates cancel out in $Y_{l}^{m}$ (and generally in $Y_m$
as functions on the standard manifold $\mathbb{S}^{n-1}$).

\begin{lem*}
The operator $\partial_r$ continuously maps the nuclear space $\mathcal{S}_{A^{(n)}}(\mathbb{R}^n)$ into itself.
\end{lem*}

\qedsymbol \, 

Again we proceed like in the proof of the first Lemma of this Subsection and as in the proof of the second  Lemma
of Subsection \ref{dim=1}, by computing the matrix elements $\big\langle nm \big| \partial_r \big| n'm' \big\rangle$
of the operator $\partial_r$ in the basis $Ue_{n,m}$ of eigenfunctions of the operator $A^{(n)}$
in $L^2(\mathbb{R}^n, \ud^n p)$, and express them in terms of the matrix elements  
$\big( nm \big| \partial_r \big| n'm' \big)$ of another operator $Op$ in the basis $e_{n,m} = h_n \otimes Y_m$
of eigenfunctions of the operator $H_{(1)} \otimes \boldsymbol{1} + \boldsymbol{1} \otimes \Delta_{\mathbb{S}^{n-1}}$.
The clue is that $Op$ turns out to be an operator mapping $\mathcal{S}_{H_{(1)} \otimes \boldsymbol{1} + \boldsymbol{1}\otimes \Delta_{\mathbb{S}^{n-1}}}
(\mathbb{R} \times \mathbb{S}^{n-1})$ continuously into itself. Namely computing 
$\big\langle nm \big| \partial_r \big| n'm' \big\rangle$ as in the proof of the first Lemma of this Subsection 
we get
\begin{multline*}
\Big\langle \, {}_{nm} \, \Big| \frac{\partial}{\partial_r} \Big| \, {}_{n'm'} \, \Big\rangle 
= \big({}_{nm} \big| Op \big| {}_{n'm'}\big)\\
= \Big( \, {}_{nm} \, \Big| -\frac{1}{2} \frac{1}{\nu_{{}{n}}(t)}\frac{d\nu_{{}{n}}(t)}{dt}
\frac{t^2 + 4 - t\sqrt{t^2 + 4}}{2} + \frac{t^2 + 4 - t\sqrt{t^2 + 4}}{2} \frac{\partial}{\partial t} \Big| \, {}_{n'm'} \, \Big) \\
= \delta_{mm'} \,\, \int \limits_{\mathbb{R}} \overline{h_n (t)} \,
-\frac{1}{2} \frac{1}{\nu_{{}{n}}(t)}\frac{d\nu_{{}{n}}(t)}{dt}
\frac{t^2 + 4 - t\sqrt{t^2 + 4}}{2} + \frac{t^2 + 4 - t\sqrt{t^2 + 4}}{2} \frac{d}{d t}
  h_{n'} (t) dt \\
= \delta_{mm'} \,\, \Big( n \Big| 
-\frac{1}{2} \frac{1}{\nu_{{}{n}}(t)}\frac{d\nu_{{}{n}}(t)}{dt}
\frac{t^2 + 4 - t\sqrt{t^2 + 4}}{2} + \frac{t^2 + 4 - t\sqrt{t^2 + 4}}{2} \frac{d}{d t}
  \Big|n' \Big) \\
= \delta_{mm'} \big(n \big| Op_{{}_{t}} \big| n'\big).
\end{multline*}
$Op_{{}_{t}}$ is the following operator
\[
-\frac{1}{2} \frac{1}{\nu_{{}{n}}(t)}\frac{d\nu_{{}{n}}(t)}{dt}
\frac{t^2 + 4 - t\sqrt{t^2 + 4}}{2} + \frac{t^2 + 4 - t\sqrt{t^2 + 4}}{2} \frac{d}{d t}
\]
acting on the functions of one real variable. Because the functions
\[
t \mapsto \frac{t^2 + 4 - t\sqrt{t^2 + 4}}{2}
\]
and 
\[
t \mapsto \frac{1}{\nu_{{}{n}}(t)}\frac{d\nu_{{}{n}}(t)}{dt}
\]
are   multipliers of the algebra of Schwartz functions $\mathcal{S}(\mathbb{R})
= \mathcal{S}_{H_{(1)}}(\mathbb{R})$ for each natural $n>2$, and because the operator of differentiation $d/dt$ continuously
maps $\mathcal{S}(\mathbb{R})$ into itself, then the operator $Op_{{}_{t}}$ maps continuously
$\mathcal{S}(\mathbb{R})
= \mathcal{S}_{H_{(1)}}(\mathbb{R})$ into itself. Because the operator $Op$ in the above formula, defined on the
tensor product algebra
\[
\mathcal{S}_{H_{(1)}}(\mathbb{R}) \otimes \mathcal{S}_{\Delta_{\mathbb{S}^{n-1}}}(\mathbb{S}^{n-1})
= \mathcal{S}_{H_{(1)} \otimes \boldsymbol{1} + \boldsymbol{1}\otimes \Delta_{\mathbb{S}^{n-1}}}
(\mathbb{R} \times \mathbb{S}^{n-1})
\]
is equal to 
\[
Op = Op_{{}_{t}} \otimes \boldsymbol{1},
\]
then again by Proposition 43.6 of \cite{treves} $Op$ maps continuously 
$\mathcal{S}_{H_{(1)}}(\mathbb{R}) \otimes \mathcal{S}_{\Delta_{\mathbb{S}^{n-1}}}(\mathbb{S}^{n-1})$
into itself. Because spectra of the operators $A^{(n)}$ and 
$H_{(1)} \otimes \boldsymbol{1} + \boldsymbol{1}\otimes \Delta_{\mathbb{S}^{n-1}}$ are identical, then
we may proceed like in Subsection \ref{dim=1} and show that the operator $\partial_r$
maps continuously $\mathcal{S}_{A^{(n)}}(\mathbb{R}^n) = 
\mathcal{S}_{U(H_{(1)} \otimes \boldsymbol{1} + \boldsymbol{1}\otimes \Delta_{\mathbb{S}^{n-1}})U^{-1}}
(\mathbb{R} \times \mathbb{S}^{n-1})$
into itself.

\qed

\begin{lem*}
The differential operator $\sin \phi_1 \partial_{\phi_1}$ as an operator on $\mathcal{S}_{A^{(n)}}(\mathbb{R}^n)$
maps continuously the nuclear space $\mathcal{S}_{A^{(n)}}(\mathbb{R}^n)$ into itself. 
\end{lem*}

\qedsymbol \,
The operator $\sin \phi_1 \partial_{\phi_1}$ as an operator on the nuclear space
\[
\mathcal{S}_{H_{(1)}}(\mathbb{R}) \otimes \mathcal{S}_{\Delta_{\mathbb{S}^{n-1}}}(\mathbb{S}^{n-1})
= \mathcal{S}_{H_{(1)} \otimes \boldsymbol{1} + \boldsymbol{1}\otimes \Delta_{\mathbb{S}^{n-1}}}
(\mathbb{R} \times \mathbb{S}^{n-1}).
\]
is equal to
\[
\boldsymbol{1} \otimes \sin \phi_1 \partial_{\phi_1}
\]
where the operator in the second factor is understood as the operator $\sin \phi_1 \partial_{\phi_1}$
on $\mathcal{S}_{\Delta_{\mathbb{S}^{n-1}}}(\mathbb{S}^{n-1})$, which is continuous, as we have already shown
in one of the preceding Lemmas of this Subsection. Thus, again by Proposition 43.6 of \cite{treves}
the operator $\sin \phi_1 \partial_{\phi_1}$ as an operator on the nuclear space
$\mathcal{S}_{H_{(1)}}(\mathbb{R}) \otimes \mathcal{S}_{\Delta_{\mathbb{S}^{n-1}}}(\mathbb{S}^{n-1})$
is continuous. Again by the identity of the spectra of the operators
$A^{(n)}$ and
$H_{(1)} \otimes \boldsymbol{1} + \boldsymbol{1}\otimes \Delta_{\mathbb{S}^{n-1}}$, we infer the continuity
of the operator $\sin \phi_1 \partial_{\phi_1}$ as an operator on $\mathcal{S}_{A^{(n)}}(\mathbb{R}^n)$, because
the matrix elements $\langle nm | \sin \phi_1 \partial_{\phi_1} | n'm' \rangle $, in the basis $Ue_{n,m}$,
of the operator $\sin \phi_1 \partial_{\phi_1}$
understood as a mapping on $\mathcal{S}_{A^{(n)}}(\mathbb{R}^n) =
\mathcal{S}_{U\big(H_{(1)} \otimes \boldsymbol{1} + \boldsymbol{1}\otimes \Delta_{\mathbb{S}^{n-1}}\big)U^{-1}}
(\mathbb{R} \times \mathbb{S}^{n-1})$
are equal to the matrix elements $( nm | \sin \phi_1 \partial_{\phi_1} | n'm' )$, in the basis
$e_{n,m}$, of the operator $\sin \phi_1 \partial_{\phi_1}$ understood as an operator on
$\mathcal{S}_{H_{(1)} \otimes \boldsymbol{1} + \boldsymbol{1}\otimes \Delta_{\mathbb{S}^{n-1}}}
(\mathbb{R} \times \mathbb{S}^{n-1})$.

\qed

\begin{lem*}
The operators $\partial_i = \frac{\partial}{\partial p_i}$, $i = 1, \ldots , n$, of differentiation 
with respect to Cartesian coordinates map continuously the nuclear space 
$\mathcal{S}_{A^{(n)}}(\mathbb{R}^n)$ into itself.
\end{lem*}

\qedsymbol \,
As we have already noted, the rotations $R$ act naturally as unitary operators $U_{{}_{R}}$ in $L^2(\mathbb{R}^n, \ud^n p)$,
and by the very construction the operator $A^{(n)}$ is symmetric with respect to rotations
$A^{(n)} = U_{{}_{R}} \big( A^{(n)}\big) {U_{{}_{R}}}^{-1}$. Thus, each $U_{{}_{R}}$ transforms
continuously $\mathcal{S}_{A^{(n)}}(\mathbb{R}^n)$ into itself, i.e. continuously with respect
to the nuclear topology. Therefore, it is sufficient to prove our Lemma for the differential operator
\[
\frac{\partial}{\partial p_1} = \cos \phi_1 \frac{\partial}{\partial r}
- \frac{\sin \phi_1}{r} \frac{\partial}{\partial \phi_1}.
\]
Now by the preceding Lemmas the operators of multiplication by the functions
$r^{-1}$ and $\cos \phi_1$ and the differential operators $\partial_r$ and $\sin \phi_1 \partial_{\phi_1}$
all map continuously the nuclear space $\mathcal{S}_{A^{(n)}}(\mathbb{R}^n)$ into itself.
Because composition of continuous maps is continuous, our Lemma is proved.

\qed

\begin{lem*}
The nuclear spaces $\mathcal{S}_{A^{(n)}}(\mathbb{R}^n)$ respect Kubo-Takenaka conditions
(H1)-(H3) of Subsection \ref{white-setup}.
\end{lem*}

\qedsymbol \, 
By construction the eigenfunctions $Ue_{n,m}= u_{n,m}$ of the operator $A^{(n)}$, corresponding to the eigenvalues $\lambda_{nm}$,
are continuous (even smooth). Likewise by construction there exists an open covering
$\mathbb{R}^n = \cup_\gamma \Omega_\gamma$ with the property that for each $\gamma$ there exists $\alpha(\gamma) >0$ such that
for each $\gamma$
\[
\sup\{{(\lambda_{nm}})^{-\alpha(\gamma)} \, |u_{n,m}(p_1, \ldots p_n)|, (p_1, \ldots p_n)\in \Omega_\gamma, n,m = 1,2, \ldots   \} < \infty.
\]
By the Proposition of the Appendix of \cite{obata.Cont.Version.Thm} the conditions (H1)-(H3)
are fulfilled.

\qed

In particular by (H1) each element of $\mathcal{S}_{A^{(n)}}(\mathbb{R}^n) \subset L^2(\mathbb{R}^n)$ as a class of functions differing on null sets
may be represented by a unique continuous function on $\mathbb{R}^n$. However by the very construction it follows that
every element of $\mathcal{S}_{A^{(n)}}(\mathbb{R}^n) \subset L^2(\mathbb{R}^n)$, which is a class of equivalent functions, has a unique
representative being a smooth function. Indeed: note that this is true for
$\mathcal{S}_{H_{(1)}}(\mathbb{R}) = \mathcal{S}(\mathbb{R}^{n})$ and for
$\mathcal{S}_{A^{(1)}}(\mathbb{S}^{n-1})= \mathscr{C}^{\infty}(\mathbb{S}^{n-1})$, and on the other hand the unitary operator $U = U_2 U_1$ of Subsection \ref{dim=n} is constructed from measure space transformations -- the
maps from the cylinder $\mathbb{R} \times \mathbb{S}^{n-1}$ onto the funnel $\mathbb{F}$ and from the funnel
onto the hyperplane $\mathbb{R}^n$ -- which are at the same time diffeomorphisms for the standard manifold
structure of the said manifolds: $\mathbb{R} \times \mathbb{S}^{n-1}$, $\mathbb{F}$, $\mathbb{R}^n$.
Analogously we have for $\textrm{dim} = n=1$ and the operator $U_0 = U \oplus U'$.
Thus, the part of the last Lemma concerning the condition (H1) tells us nothing new. But the remaining conditions
(H2) and (H2) are less trivial. In particular from the last Lemma it follows that for each $p_0 \in \mathbb{R}^n$ the Dirac
delta function $\delta_{p_0}$ is an element of the space $\mathcal{S}_{A^{(n)}}(\mathbb{R}^n)^*$ dual to the nuclear space
$\mathcal{S}_{A^{(n)}}(\mathbb{R}^n)$. In fact, we will show much more in the next Subsection, namely that
$\mathcal{S}_{A^{(n)}}(\mathbb{R}^n) = \mathcal{S}^0(\mathbb{R}^n)$ in store of elements and
in their nuclear topologies.

From now on we identify the elements of the nuclear space  $\mathcal{S}_{A^{(n)}}(\mathbb{R}^n)$, i.e. classes of equivalent functions, with the smooth functions representing them uniquely, and thus regard $\mathcal{S}_{A^{(n)}}(\mathbb{R}^n)$ as ordinary smooth function spaces. In fact we have already done it implicitly in the proof of the preceding Lemmas of this Subsection concerned with differential operators.

\subsection{The equality $\mathcal{S}_{A^{(n)}}(\mathbb{R}^n) = \mathcal{S}^0(\mathbb{R}^n)$}\label{SA=S0}

In this Subsection we will use the multi-index notation of Schwartz. Namely, $q$ will stand for
$q = (q_1, q_2, \ldots , q_n) \in \mathbb{N}^n$ and in this case $|q| = q_1 + \ldots + q_n$, and the symbol
$D^q$ will stand for the differentiation operation
$D^q = \frac{\partial^{|q|}}{\partial p_{q_1} \ldots \partial p_{q_1}}$
with respect to Cartesian coordinates,
as well as the symbol $\varphi^{(q)}$ for $\varphi^{(q)} = D^q \varphi$. In general the symbol $(q)$
or $(n)$ with parenthesis in the superscript will always be understood in this manner, the exception
being the symbol for the operator $A^{(n)}$.

Because $A^{(n)}$ transforms $\mathcal{S}^0(\mathbb{R}^n)$ into itself, then it easily follows that 
$\mathcal{S}^0(\mathbb{R}^n) \subset \mathcal{S}_{A^{(n)}}(\mathbb{R}^n)$ regarding the store of elements
(topology is for a while ignored in this inclusion relation). 

Now let $\varphi \in \mathcal{S}_{A^{(n)}}(\mathbb{R}^n) \subset L^2(\mathbb{R}^n, \ud^n \p)$. From the completeness 
of the orthonormal system $\{u_{n,m} = Ue_{n,m}\}$ of eigenfunctions of the operator $A^{(n)}$
it follows that the series
\begin{equation}\label{Cnm}
\varphi = \sum \limits_{n,m} C_{n,m}(\varphi)u_{n,m},
\end{equation}
where 
\[
C_{n,m}(\varphi) = \big\langle u_{n,m} \big| \varphi \big\rangle = \int \limits_{\mathbb{R}^n}
u_{n,m}(\p) \varphi(p) \, \ud^n \p, 
\]
converges in $L^2(\mathbb{R}^n, \ud^n \p)$. 

\begin{lem*}
In this case, i.e. when $\varphi \in \mathcal{S}_{A^{(n)}}(\mathbb{R}^n)$, 
it follows that the series (\ref{Cnm})
converges in the nuclear topology of $\mathcal{S}_{A^{(n)}}(\mathbb{R}^n)$.
\end{lem*}

\qedsymbol \,

Proof is exactly the same as the proof of the first Lemma of Subsection \ref{dim=1}.
\qed

Because by construction the eigenfunctions $u_{n,m} = Ue_{n,m}$ of $A^{(n)}$ belong to 
$\mathcal{S}^0(\mathbb{R}^n)$ then from the last Lemma it follows 
\begin{lem*}
The space $\mathcal{S}^0(\mathbb{R}^n)$ is dense in $\mathcal{S}_{A^{(n)}}(\mathbb{R}^n)$
with respect to the nuclear topology of $\mathcal{S}_{A^{(n)}}(\mathbb{R}^n)$.
\end{lem*}

In what follows we show that $\mathcal{S}^0(\mathbb{R}^n)$ with the topology inherited from
$\mathcal{S}_{A^{(n)}}(\mathbb{R}^n)$ is complete, which by the last Lemma gives the
equality $\mathcal{S}_{A^{(n)}}(\mathbb{R}^n) = \mathcal{S}^0(\mathbb{R}^n)$ in store of elements.
For the proof of the completeness we compare the system of norms 
$| \cdot |_m = \big|\, \big(A^{(n)}\big)^m \, \cdot \, \big|_{{}_{L^2(\mathbb{R}^n)}}$ on $\mathcal{S}^0(\mathbb{R}^n)$ 
inherited from $\mathcal{S}_{A^{(n)}}(\mathbb{R}^n)$, with the system of norms of a class
of countably normed spaces $K\{M_{{}_{m}}\}$ of smooth functions described by Gelfand and Shilov
in their classic book \cite{GelfandII}. We choose the system $\{M_{{}_{m}}\}$ such that
$K\{M_{{}_{m}}\} = \mathcal{S}^0(\mathbb{R}^n)$ in store of elements and show that the system of norms on 
$K\{M_{{}_{m}}\} = \mathcal{S}^0(\mathbb{R}^n)$ is equivalent to the system of norms 
$| \cdot |_m = \big|\, \big(A^{(n)}\big)^m \, \cdot \, \big|_{{}_{L^2(\mathbb{R}^n)}}$
inherited from $\mathcal{S}_{A^{(n)}}(\mathbb{R}^n)$. In particular by the completeness of the space
$K\{M_{{}_{m}}\}$ (proven in \cite{GelfandII}) the completeness of the topology on $\mathcal{S}^0(\mathbb{R}^n)$
inherited from $\mathcal{S}_{A^{(n)}}(\mathbb{R}^n)$ will thus follow. 

For the proof of the equality 
of the topology on $\mathcal{S}^0(\mathbb{R}^n)$ inherited from $\mathcal{S}_{A^{(n)}}(\mathbb{R}^n)$ 
and the topology inherited from $\mathcal{S}(\mathbb{R}^n)$ we have to compare the system of norms 
$| \cdot |_m = \big|\, \big(A^{(n)}\big)^m \, \cdot \, \big|_{{}_{L^2(\mathbb{R}^n)}}$ on 
$\mathcal{S}^0(\mathbb{R}^n)$ inherited from $\mathcal{S}_{A^{(n)}}(\mathbb{R}^n)$, with the system of norms 
inherited from $\mathcal{S}(\mathbb{R}^n)$ and use the closed graph theorem for maps of Frech\'et spaces.

Let us start by introducing the following two systems of norms on $\mathcal{S}^0(\mathbb{R}^n)$
\begin{equation}\label{|||.|||}
||| \varphi |||_{m}^{2} = \sum \limits_{k \in \mathbb{Z}, 0 \leq |k|, |q| \leq m } \,\, \int \limits_{\mathbb{R}^n} \big|r^k \varphi^{(q)} \big|^2
\, \ud^n \p
\end{equation}
and
\begin{equation}\label{||.||}
|| \varphi ||_m = \sup \limits_{k \in \mathbb{Z}, 0 \leq |k|, |q| \leq m, \p \in \mathbb{R}^n }
\big|r^k \varphi^{(q)}(\p) \big|.
\end{equation}
Note that in the formulas (\ref{|||.|||}), (\ref{||.||}) for the norms the index $k$ is an integer,
which may be positive as well as negative. They are well-defined on $\mathcal{S}^0(\mathbb{R}^n)$,
in particular for $\varphi \in \mathcal{S}^0(\mathbb{R}^n)$ the function $r^k \varphi^{(q)}$ is not only finite and smooth but even $r^k \varphi^{(q)} \in \mathcal{S}^0(\mathbb{R}^n)$, and in particular
$r^k \varphi^{(q)} \in L^2(\mathbb{R}^n, \ud^n \p)$.

In our first step we show that the two systems of norms (\ref{|||.|||}) and (\ref{||.||}) are equivalent
on  $\mathcal{S}^0(\mathbb{R}^n)$. 

\begin{lem*}
If $\varphi \in \mathcal{S}(\mathbb{R}^3)$ and ${\p}_0\in \mathbb{R}^3$ then
\begin{multline*}
|\varphi({\p}_0)|^2 \leq \int \limits_{\mathbb{R}^3} | \varphi |^2 \, \ud^3 \p \\
+ \int \limits_{\mathbb{R}^3} | \partial_1\varphi |^2 \, \ud^3 \p 
+ \int \limits_{\mathbb{R}^3} | \partial_2 \varphi |^2 \, \ud^3 \p  
+ \int \limits_{\mathbb{R}^3} | \partial_3 \varphi |^2 \, \ud^3 \p \\
+ \int \limits_{\mathbb{R}^3} | \partial_1 \partial_2 \varphi |^2 \, \ud^3 \p
+  \int \limits_{\mathbb{R}^3} | \partial_1 \partial_3 \varphi |^2 \, \ud^3 \p
+  \int \limits_{\mathbb{R}^3} | \partial_2 \partial_3 \varphi |^2 \, \ud^3 \p \\
+  \int \limits_{\mathbb{R}^3} | \partial_1 \partial_2 \partial_3 \varphi |^2 \, \ud^3 \p.
\end{multline*}
And more generally if  $\varphi \in \mathcal{S}(\mathbb{R}^n)$ and ${\p}_0\in \mathbb{R}^n$ then
\begin{multline*}
|\varphi({\p}_0)|^2 \leq \int \limits_{\mathbb{R}^n} | \varphi |^2 \, \ud^n \p \\
+ \int \limits_{\mathbb{R}^n} | \partial_1\varphi |^2 \, \ud^n \p 
+ \ldots 
+ \int \limits_{\mathbb{R}^n} | \partial_n \varphi |^2 \, \ud^n \p \\
+  \int \limits_{\mathbb{R}^n} | \partial_1 \partial_2 \varphi |^2 \, \ud^n \p
+  \int \limits_{\mathbb{R}^n} | \partial_1 \partial_3 \varphi |^2 \, \ud^n \p + \ldots \\
\ldots \\
+  \int \limits_{\mathbb{R}^n} | \partial_1 \partial_2 \ldots \partial_n \varphi |^2 \, \ud^n \p.
\end{multline*}
\end{lem*}

\qedsymbol \,
For $\varphi \ \in \mathcal{S}(\mathbb{R})$
\[
\varphi(p_0) = \int \limits_{-\infty}^{p_0} \frac{d}{d p}\varphi(p) \, \ud p.
\]
Thus
\[
|\varphi(p_0)| = \Bigg| \int \limits_{-\infty}^{p_0} \frac{d}{d p}\varphi(p) \, \ud p \Bigg|
\leq \int \limits_{-\infty}^{+ \infty} \Big| \frac{d}{d p}\varphi(p) \Big| \, \ud p.
\]
Similarly for $\varphi \in \mathcal{S}(\mathbb{R}^3)$
\begin{multline*}
\varphi(p_{10}, p_{20}, p_{30}) = \int \limits_{-\infty}^{p_{10}}
\frac{\partial}{\partial p_{1}}\varphi(p_{1}, p_{20}, p_{30}) \, \ud p_{1}
= \int \limits_{-\infty}^{p_{10}} \int \limits_{-\infty}^{p_{20}}
\frac{\partial}{\partial p_{2}} \frac{\partial}{\partial p_{1}}\varphi(p_{1}, p_{2}, p_{30}) \, \ud p_{1} \ud p_{2} \\
= \int \limits_{-\infty}^{p_{10}} \int \limits_{-\infty}^{p_{20}} \int \limits_{-\infty}^{p_{30}}
\frac{\partial}{\partial p_{3}} \frac{\partial}{\partial p_{2}} \frac{\partial}{\partial p_{1}}\varphi(p_{1}, p_{2},
p_{3}) \, \ud p_{1} \ud p_{2} \ud p_{3},
\end{multline*}
from which it follows
\[
| \varphi({\p}_0) | \leq \int \limits_{\mathbb{R}^3} \Big| \partial_1 \partial_2 \partial_3 \varphi(\p) \Big|
\, \ud^3 \p.
\]
Because for $\varphi \in \mathcal{S}(\mathbb{R}^3)$ also
$\varphi^2 = \varphi \cdot \varphi \in \mathcal{S}(\mathbb{R}^3)$, then
\begin{multline*}
|\varphi({\p}_0)|^2 = |\varphi^2(({\p}_0)| \leq
\int \limits_{\mathbb{R}^3} \Big| \partial_1 \partial_2 \partial_3 \varphi^2 \Big|
\, \ud^3 \p \\
= \int \limits_{\mathbb{R}^3} \Big| \partial_1 \partial_2 \big\{ 2 \varphi \partial_3 \varphi \big\} \Big|
\, \ud^3 \p \\
= \int \limits_{\mathbb{R}^3} \Big| \partial_1 \big\{ 2 \partial_2 \varphi \partial_3 \varphi + 2 \varphi \partial_2 \partial_3 \varphi \big\} \Big|
\, \ud^3 \p \\
= \int \limits_{\mathbb{R}^3} \Big| 2 \partial_1 \partial_2 \varphi \partial_3 \varphi
+ 2 \partial_2 \varphi \partial_1 \partial_3 \varphi
+ 2 \partial_1 \varphi \partial_2 \partial_3 \varphi
+ 2 \varphi \partial_1 \partial_2 \partial_3 \varphi \Big|
\, \ud^3 \p \\
\leq
\int \limits_{\mathbb{R}^3} 2 |\partial_1 \partial_2 \varphi| \, |\partial_3 \varphi| \, \ud^3 \p
+ \int \limits_{\mathbb{R}^3} 2 |\partial_2 \varphi| \, |\partial_1 \partial_3 \varphi| \, \ud^3 \p \\
+ \int \limits_{\mathbb{R}^3} 2 |\partial_1 \varphi| \, |\partial_2 \partial_3 \varphi| \, \ud^3 \p
+ \int \limits_{\mathbb{R}^3} 2 |\varphi| \, | \partial_1 \partial_2 \partial_3 \varphi|
\, \ud^3 \p,
\end{multline*}
so that by the application of the elementary inequality $2|a|\, |b| \leq |a|^2 + |b|^2$ valid for any pair of real or complex numbers $a,b$ to each integrand separately we obtain the three-dimensional assertion of our Lemma.

The proof of the general $n$-dimensional case is completely analogous.

\qed

\begin{lem*}
The systems $\{|||\cdot |||_m\}_{{}_{m \in \mathbb{N}}}$ and 
$\{||\cdot ||_m\}_{{}_{m \in \mathbb{N}}}$ of norms on $\mathcal{S}^0(\mathbb{R}^n)$, 
given by the formulas (\ref{|||.|||}) and (\ref{||.||}) respectively, 
are equivalent in the sense of \cite{GelfandII}.
\end{lem*}

\qedsymbol \, 
That for any $m \in \mathbb{N}$ there exists such a positive and finite constant $c_m$
that
\[
||| \varphi |||_m \leq c_m || \varphi ||_m, \,\,\, \varphi \in  \mathcal{S}^0(\mathbb{R}^n)
\]
is obvious, so that the system of norms $||\cdot ||_m$ is stronger that the system of norms 
$||| \cdot |||_m$. Indeed, each of the integrand functions $\big|r^k \varphi^{(q)}\big|^2$ in 
$||| \varphi |||_{m}^{2}$  can be represented as the product 
\[
\big[r^k(1+r^{m'(m)})^{-1}\big] \big[\big(1+r(p)^{m'(m)}\big) \big|\varphi^{(q)}\big|^2 \big] 
\]
with $m'(m)$ depending on $m$, and such that
\[
p \longmapsto \big[r(p)^k(1+r(p)^{m'(m)})^{-1}\big] 
\]
is absolutely integrable, \emph{i.e.} in $L^1$. The above assertion then holds for 
$c_m$ equal to the square root of  the $L^1$ norm of this absolutely integrable function 
(compare also \cite{Reed_Simon}). 

The proof of the converse statement is less trivial. But applying the last Lemma to the function
$r^k \varphi^{(m)}$, which for $\varphi \in \mathcal{S}^0(\mathbb{R}^n)$ and $k \in \mathbb{Z}$
likewise belongs to $\mathcal{S}^0(\mathbb{R}^n) \subset \mathcal{S}(\mathbb{R}^n)$, we easily show existence
of a positive and finite constant $c_{k,q}$ such that for each $\varphi \in \mathcal{S}^0(\mathbb{R}^n)$
\[
\big| r^k \varphi^{(q)}(\p) \big|^2 \leq c_{k,q} \,
\sum \limits_{\nu \in \mathbb{Z}, |\nu|, |(\alpha)| \leq |q| + |k| +n} \,\,
\int \limits_{\mathbb{R}^n} \big| r^\nu \varphi^{\alpha} \big|^2 \, \ud^n \p
= c_{k, q} ||| \varphi |||_{|q| + |k| +n}^2.
\]
Thus it follows that for each $m \in \mathbb{N}$ there exists natural
$m'(m) = 2m + n > m$, and a positive and finite number $c'_m$ such that
\[
|| \varphi ||_m \leq c'_m ||| \varphi |||_{2m + n} \,\,\, \textrm{for all} \,\,
\varphi \in \mathcal{S}^0(\mathbb{R}^n),
\]
so that the two systems of norms $||\cdot ||_m$ and $||| \cdot |||_m$
on $\mathcal{S}^0(\mathbb{R}^n)$ are equivalent in the sense of \cite{GelfandII},
Ch. I.3.6, pp. 28-30: each norm of the first system is weaker than some norm of the second system
and \emph{vice versa}.

\qed

\begin{lem*}
The system of norms:
\[
| \cdot |_m = \big|\, \big(A^{(n)}\big)^m \, \cdot \, \big|_{{}_{L^2(\mathbb{R}^n)}}
\]
on $\mathcal{S}^0(\mathbb{R}^n)$ induced from $\mathcal{S}_{A^{(n)}}(\mathbb{R}^n)$ is equivalent
in the sense of \cite{GelfandII} with the system of norms (\ref{|||.|||}):
\[
||| \varphi |||_{m}^{2} = \sum \limits_{k \in \mathbb{Z}, 0  \leq |k|,  |q| \leq m } \,\, \int \limits_{\mathbb{R}^n} \big|r^k \varphi^{(q)} \big|^2
\, \ud^n \p
\]
on $\mathcal{S}^0(\mathbb{R}^n)$.
\end{lem*}

\qedsymbol \,
Existence for each $m \in \mathbb{N}$ of a positive number $c_m$ such that for all
$\varphi \in \mathcal{S}^0(\mathbb{R}^n)$ the inequalities
\[
| \varphi |_{m}^{2} \leq c_m \, ||| \varphi |||_{m+2}^{2}
\]
are fulfilled follows from the explicit form of the operator $A^{(n)}$. It likewise follows from the continuity of $A^{(n)}$ as an operator transforming
$\mathcal{S}^0(\mathbb{R}^n) \subset \mathcal{S}_{A^{(n)}}(\mathbb{R}^n)$
into $\mathcal{S}^0(\mathbb{R}^n) \subset \mathcal{S}_{A^{(n)}}(\mathbb{R}^n)$.
Thus, the system of norms $|\cdot|_m$ is weaker than the system of norms $||| \cdot |||_m$.

The proof of the converse statement is less trivial and uses the results of the previous Subsection.
Namely, by the results of the Subsection \ref{diffSA} the operators
\[
\varphi \mapsto r^k \varphi^{(q)}, \,\,\, k \in \mathbb{Z}, q \in \mathbb{N}^n,
\]
map continuously the nuclear space $\mathcal{S}_{A^{(n)}}(\mathbb{R}^n)$ into itself and
transform the subspace $\mathcal{S}^0(\mathbb{R}^n)$ into itself. Thus, for each $k \in \mathbb{Z}$
and $q \in \mathbb{N}^n$ there exists such an $m' = m'(k,q) \in \mathbb{N}$ that
\[
\int \limits_{\mathbb{R}^n} \big| r^k \varphi^{(q)}(\p) \big|^2 \, \ud^n \p
= \big| r^k \varphi^{(q)} \big|_{{}_{L^2(\mathbb{R}^n)}}^{2} = \big| r^k \varphi^{(q)} \big|_{0}^{2}
\leq c_{k,q} \big| \varphi \big|_{m'(p,q)}^{2},
\]
for all $\varphi \in \mathcal{S}^0(\mathbb{R}^n)$. In particular
\[
|||\varphi |||_m \leq \max \limits_{|k|, |q| \leq m} \{c_{k,q}\} \, \big| \varphi \big|_{{}{\max\{m'(k,q)\}}}
\]
where $\max$ in the subscript $\max\{m'(k,q)\}$ is taken over all $k,q$ such that $|k|, |q| \leq m$;
so that our Lemma is proved.
\qed

Therefore, joining the last two Lemmas we see that system of norms $|\cdot|_m$ inherited from
$\mathcal{S}_{A^{(n)}}(\mathbb{R}^n)$
and the norms $|| \cdot ||_m$ defined by (\ref{||.||}) on $\mathcal{S}^0(\mathbb{R}^n)$ are equivalent
and can be used on $\mathcal{S}^0(\mathbb{R}^n)$ interchangeably.
At this point we turn to the class of countably normed spaces $K\{M_{{}_{m}}\}$ of smooth
functions of Gelfand-Shilov \cite{GelfandII}.

Recall that for the construction of the space $K\{M_{{}_{m}}\}$ one first assigns a sequence 
of functions $\{M_{{}_{m}}\}_{{}{m = 0, 1, \ldots}}$ on the fundamental space 
which is a manifold, in our case $\mathbb{R}^n$, which for each $\p \in \mathbb{R}^n$
satisfy the inequalities $1 \leq M_{{}_{0}}(\p) \leq M_{{}_{1}}(\p) \leq \ldots$,
taking on finite or simultaneously infinite values,  
and continuous everywhere where they are finite. By definition, the space $K\{M_{{}_{m}}\}$ consists
of all infinitely differentiable functions $\varphi$ on the fundamental space, in our case
on $\mathbb{R}^n$, for which the product functions
\[
\p \mapsto M_{{}_{m}}(\p) \varphi^{(q)}(\p), \,\,\, |q| \leq m, m= 0,1, \ldots
\]
are everywhere continuous and bounded in the whole fundamental space, in our case in the whole
$\mathbb{R}^n$. The norms in $K\{M_{{}_{m}}\}$ are defined by the formulas
\begin{equation}\label{normMm}
\rceil \varphi \lceil_{{}_{m}} = \sup \limits_{|q| \leq m, \p \in \mathbb{R}^n} \, 
M_{{}_{m}}(\p)\big| \varphi^{(q)}(\p) \big|,
\,\,\,\,
m = 0, 1, 2, \ldots.
\end{equation}

In particular, we have simple
\begin{lem*}
For
\begin{equation}\label{M}
M_{{}_{m}}(\p) = (r + r^{-1})^m, 
\end{equation}
(recall that $\p = (p_1, \ldots , p_n)$ and 
$r = \big((p_1)^2 + \ldots + (p_1)^2  \big)^{\frac{1}{2}}$) we have
\[
K\{M_{{}_{m}}\} = \mathcal{S}^0(\mathbb{R}^n) \,\,
\textrm{in store of elements}
\]
(topology is ignored here).
\end{lem*}

\qedsymbol \,
Indeed, for
\begin{equation}\label{M''M'}
M''_{{}_{m}}(\p) = \sup \limits_{k \in \mathbb{N}, 0 \leq k \leq m} r^k
\,\, \textrm{and} \,\,
M'_{{}_{m}}(\p) = (1 + r)^m
\end{equation}
we have
\[
K\{M''_{{}_{m}}\} = K\{M'_{{}_{m}}\} = \mathcal{S}(\mathbb{R}^n)
\]
in store of elements and in topology (for spaces of type
$K\{M_{{}_{m}}\}$ equality in store of elements implies equality of topologies), 
for the proof compare the method of \cite{GelfandII}, Ch. II \S 2.4, easily adopted to our case. 
From this it easily follows that for any 
$\varphi \in \mathcal{S}^0(\mathbb{R}^n)$ the function
$M_{{}_{m}} \varphi^{(q)} \in \mathscr{C}(\mathbb{R}^n)$ and each norm 
$\rceil \varphi \lceil_{{}_{m}}$, $m= 0, 1, \ldots$ is finite; which means that 
$\varphi \in K\{M_{{}_{m}}\}$. 

Conversely: every $\varphi \in K\{M_{{}_{m}}\}$ is by construction smooth and for each $\varphi \in K\{M_{{}_{m}}\}$
\[
\varphi^{(q)}(0) = 0, \,\, q \in \mathbb{N}^n,
\] 
 
\[
\sup \limits_{k \in \mathbb{N}, 0 \leq k, |q| \leq m, \p \in \mathbb{R}^n} r^k | \varphi^{(q)}(\p) | < + \infty,
\,\,\,
m = 0,1, \ldots
\]
as well as the functions $M''_{{}_{m}} \varphi^{(q)}$ are continuous, so that
$\varphi \in \mathcal{S}(\mathbb{R}^n)$. Therefore, 
$\varphi \in \mathcal{S}^0(\mathbb{R}^n)$.
\qed

\begin{lem*}
The topology of the countably normed space $K\{M_{{}_{m}}\}= \mathcal{S}^0(\mathbb{R}^n)$
defined by the sequence of functions (\ref{M}) and the system of norms (\ref{normMm})
coincides with the topology on $\mathcal{S}^0(\mathbb{R}^n)$ defined by the system of norms
(\ref{||.||}), and thus with the topology on $\mathcal{S}^0(\mathbb{R}^n)$ inherited from
$\mathcal{S}_{A^{(n)}}(\mathbb{R}^n)$.
\end{lem*}

\qedsymbol \,
For the proof it will be sufficient to show that the system of norms (\ref{||.||})
is equivalent to the system of norms (\ref{normMm}) with $M_{{}_{m}}$ defined by 
(\ref{M}). But this equivalence easily follows from the formulas (\ref{||.||}), (\ref{M})
and (\ref{normMm}). Indeed:
\begin{multline*}
\rceil \varphi \lceil_{{}_{m}} = \sup \limits_{|q| \leq m, \p \in \mathbb{R}^n} \, 
(r + r^{-1})^m \, \big| \varphi^{(q)}(\p) \big| \\
\leq \sup \limits_{|q| \leq m, \p \in \mathbb{R}^n} \, 
\Bigg( r^m +  {m \choose 1}r^{m-1}r^{-1} + \ldots + {m\choose m}r^{-m}\Bigg) \, 
\big| \varphi^{(q)}(\p) \big| \\
\leq 
\sup \limits_{|q| \leq m, \p \in \mathbb{R}^n} \, r^m \, \big| \varphi^{(q)}(\p) \big| 
+ \sup \limits_{|q| \leq m, \p \in \mathbb{R}^n} \, {m \choose 1}r^{m-1}r^{-1} \, \big| \varphi^{(q)}(\p) \big| \\
\ldots + \sup \limits_{|q| \leq m, \p \in \mathbb{R}^n} \, {m\choose m}r^{-m} \, \big| \varphi^{(q)}(\p) \big| \\
\leq
(m+1) \max \limits_{0 \leq j \leq m} \Bigg\{ {m \choose j} \Bigg\} \,\, 
\sup \limits_{|k|,|q| \leq m, \p \in \mathbb{R}^n} \, r^k \, \big| \varphi^{(q)}(\p) \big| \\
= (m+1) \max \limits_{0 \leq j \leq m} \Bigg\{ {m \choose j} \Bigg\} \,\, 
|| \varphi ||_m
\end{multline*}
Conversely:
\begin{multline*}
|| \varphi ||_m = \sup \limits_{|k|,|q| \leq m, \p \in \mathbb{R}^n} \, r^k \, \big| \varphi^{(q)}(\p) \big| \\
\leq 
\sup \limits_{|q| \leq m, \p \in \mathbb{R}^n} \, 
(r + r^{-1})^m \, \big| \varphi^{(q)}(\p) \big| 
= \rceil \varphi \lceil_{{}_{m}}, 
\end{multline*}
because 
\[
0 < r^k < (r + r^{-1})^{|k|} \leq (r + r^{-1})^m
\]
for $|k| \leq m$, $k \in \mathbb{Z}$.
\qed

From the last Lemma we get the following

\begin{lem*}
The linear set $\mathcal{S}^0(\mathbb{R}^n)$ with the topology inherited from 
$\mathcal{S}_{A^{(n)}}(\mathbb{R}^n)$ is a complete linear topological space. In particular, it follows
that $\mathcal{S}^0(\mathbb{R}^n)$ with the topology inherited from 
$\mathcal{S}_{A^{(n)}}(\mathbb{R}^n)$ is a Fr\'echet space, and 
$\mathcal{S}^0(\mathbb{R}^n) = \mathcal{S}_{A^{(n)}}(\mathbb{R}^n)$ in store of elements.
\end{lem*}

\qedsymbol \,
By the results of \cite{GelfandII}, Chap. II, the countably normed space $K\{M_{{}_{m}}\}= \mathcal{S}^0(\mathbb{R}^n)$
defined by the sequence of functions (\ref{M}) and topology defined by the corresponding system of 
norms (\ref{normMm}) is complete. By the last Lemma this topology on  $K\{M_{{}_{m}}\}= \mathcal{S}^0(\mathbb{R}^n)$
coincides with the topology inherited from $\mathcal{S}_{A^{(n)}}(\mathbb{R}^n)$. Thus, the last topology
is complete. From the second Lemma of this Subsection and completeness of the topology on $\mathcal{S}^0(\mathbb{R}^n)$ 
inherited from $\mathcal{S}_{A^{(n)}}(\mathbb{R}^n)$ it follows the equality
$\mathcal{S}^0(\mathbb{R}^n) = \mathcal{S}_{A^{(n)}}(\mathbb{R}^n)$ in store of elements.
 Because the topology on $\mathcal{S}^0(\mathbb{R}^n)$ 
inherited from $\mathcal{S}_{A^{(n)}}(\mathbb{R}^n)$ 
is by construction countably normed and locally convex, then it is a linear 
Fr\'echet topology.
\qed

The operation of differentiation $\partial_i$
with respect to Cartesian coordinates transforms  $\mathcal{S}(\mathbb{R}^n) = \mathcal{S}_{H_{(n)}}(\mathbb{R}^n)$ continuously into itself, and the Dirac delta functional maps  $\mathcal{S}(\mathbb{R}^n)$
continuously into $\mathbb{C}$. Thus, the subspace 
$\mathcal{S}^0(\mathbb{R}^n) \subset \mathcal{S}(\mathbb{R}^n)$
as the intersection of the kernels of continuous maps $\delta_0 \circ D^q$ of
$\mathcal{S}(\mathbb{R}^n)$ into complex numbers is a closed subspace
of $\mathcal{S}(\mathbb{R}^n)$. 
The subspace $\mathcal{S}^0(\mathbb{R}^n)$ with the topology inherited from $\mathcal{S}(\mathbb{R}^n)$
is a nuclear space, \cite{GelfandIV}, \cite{treves}. 
Let $\mathcal{S}^{00}(\mathbb{R}^n)$ be the Fourier image of $\mathcal{S}^0(\mathbb{R}^n)$
in $\mathcal{S}(\mathbb{R}^n)$. Because the Fourier transform and its inverse are continuous maps
of $\mathcal{S}(\mathbb{R}^n)$ onto $\mathcal{S}(\mathbb{R}^n)$, then $\mathcal{S}^{00}(\mathbb{R}^n)$ 
is likewise a closed subspace of $\mathcal{S}(\mathbb{R}^n)$. 

Therefore, the space $\mathcal{S}^0(\mathbb{R}^n)$ is a nuclear space with the topology inherited form 
$\mathcal{S}(\mathbb{R}^n) = \mathcal{S}_{H_{(n)}}(\mathbb{R}^n) = \mathcal{S}_{H_{(1)}^{\otimes n}}(\mathbb{R})$
and $\mathcal{S}^{00}(\mathbb{R}^n)$ as well is a nuclear space with the topology inherited from 
$\mathcal{S}(\mathbb{R}^n)$ and moreover we have the following simple
\begin{lem*}
The space $\mathcal{S}^0(\mathbb{R}^n)$ with the topology inherited from $\mathcal{S}(\mathbb{R}^n)$
is a Fr\'echet space. The space $\mathcal{S}^{00}(\mathbb{R}^n)$ with the topology inherited from 
$\mathcal{S}(\mathbb{R}^n)$ is a Fr\'echet space. 
\end{lem*}

\qedsymbol \,
The nuclear space $\mathcal{S}(\mathbb{R}^n) = \mathcal{S}_{H_{(n)}}(\mathbb{R}^n)$ as a countably Hilbert and complete space is a Fr\'echet space. Because any closed subspace of a Fr\'echet space
$F$ with the topology induced from $F$ is a Fr\'echet space, compare e.g. \cite{treves}, Part I, \S 10, our Lemma is proved.    
\qed

\begin{lem*}
The system of norms (\ref{||.||}):
\[
|| \varphi ||_m = \sup \limits_{k \in \mathbb{Z}, 0 \leq |k|,  |q| \leq m, \p \in \mathbb{R}^n } 
r^k \big|\varphi^{(q)}(\p) \big|, \,\,\,\
m = 0,1, \ldots
\]
on $\mathcal{S}^0(\mathbb{R}^n)$ is stronger\footnote{The term ``stronger'' used by us does not exclude the possibility of equivalence.} that than the system of norms
\begin{equation}\label{normM'm}
\rceil \varphi \lceil'_{{}_{m}} = \sup \limits_{|q| \leq m, \p \in \mathbb{R}^n} \, 
(1 + r)^{m} \, \big| \varphi^{(q)}(\p) \big|,
\,\,\,\,
m = 0, 1, 2, \ldots 
\end{equation}
on $\mathcal{S}^0(\mathbb{R}^n)$ inherited from $K\{M'_{{}_{m}}\} = \mathcal{S}(\mathbb{R}^n)$. 
\end{lem*}

\qedsymbol \,
The system of norms (\ref{normM'm}), i.e. $\rceil \cdot \lceil'_{{}_{m}}$, on $\mathcal{S}^0(\mathbb{R}^n) \subset
\mathcal{S}(\mathbb{R}^n) = K\{M'_{{}_{m}}\}$ is determined by the corresponding system
of functions (\ref{M''M'})
\[
M'_{{}_{m}}(\p) = (1 + r(\p))^m.
\]
On the other hand we have already shown that the system of norms (\ref{||.||}), i.e. 
\[
|| \varphi ||_m = \sup \limits_{k \in \mathbb{Z}, 0 \leq |k|,  |q| \leq m, \p \in \mathbb{R}^n } 
r^k \big|\varphi^{(q)}(\p) \big|, \,\,\,\
m = 0,1, \ldots
\]
on
$\mathcal{S}^0(\mathbb{R}^n) = K\{M_{{}_{m}}\}$, is equivalent to the system of
norms (\ref{normMm}), i.e. $\rceil \cdot \lceil_{{}_{m}}$, associated with the system $\{M_{{}_{m}}\}$ of functions
(\ref{M}):
\[
M_{{}_{m}}(\p) = (r(\p) + r(\p)^{-1})^m.
\] 
But for each $m \in \mathbb{N}$ there exists $c_m > 0$ such that 
\[
0 < c_m < \frac{M_{{}_{m}}(\p)}{M'_{{}_{m}}(\p)} = \frac{(r + r^{-1})^m}{(1+r)^m}, \,\,\,
\p \in \mathbb{R}^n;
\]
for example one can put
\[
c_m = \Bigg(\frac{2 + 2\sqrt{2}}{3 + 2 \sqrt{2}}\Bigg)^m.
\]
Therefore 
\begin{multline*}
\rceil \varphi \lceil'_{{}_{m}} = \sup \limits_{|q| \leq m, \p \in \mathbb{R}^n} \, 
M'_{{}_{m}}(\p) \, \big| \varphi^{(q)}(\p) \big|
\leq \frac{1}{c_m} \, \sup \limits_{|q| \leq m, \p \in \mathbb{R}^n} \, 
M_{{}_{m}}(\p) \, \big| \varphi^{(q)}(\p) \big| \\
= \frac{1}{c_m} \, \rceil \varphi \lceil_{{}_{m}}
= \frac{1}{c_m} \, (m+1) \max \limits_{0 \leq j \leq m} \Bigg\{ {m \choose j} \Bigg\} \,\, 
|| \varphi ||_m.
\end{multline*}
\qed

Joining the last three Lemmas with the continuity of the Fourier transform $\mathscr{F}$ and its inverse $\mathscr{F}^{-1}$
as maps $\mathcal{S}(\mathbb{R}^n) \rightarrow \mathcal{S}(\mathbb{R}^n)$ and with the closed graph theorem
we obtain 
\begin{lem*}
The Fourier transform $\mathscr{F}: \mathcal{S}^{00}(\mathbb{R}^n) \rightarrow \mathcal{S}^{0}(\mathbb{R}^n)$
is continuous, if $\mathcal{S}^{00}(\mathbb{R}^n)$ is equipped with the topology inherited from 
$\mathcal{S}(\mathbb{R}^n)$ and the linear space $\mathcal{S}^{0}(\mathbb{R}^n)$ is equipped
with the topology inherited from $\mathcal{S}_{A^{(n)}}(\mathbb{R}^n)$.
\end{lem*}

\qedsymbol \,
Let $\phi_j \xrightarrow{j \rightarrow +\infty} \phi$ in the topology on $\mathcal{S}^{00}(\mathbb{R}^n)$ inherited from
$\mathcal{S}(\mathbb{R}^n)$ and let $\mathscr{F} \phi_j \xrightarrow{j \rightarrow +\infty} \varphi$ 
in the topology on $\mathcal{S}^{0}(\mathbb{R}^n)$ inherited from $\mathcal{S}_{A^{(n)}}(\mathbb{R}^n)$. 
Because the norms $| \cdot |_m$ on 
$\mathcal{S}^{0}(\mathbb{R}^n)$ inherited from $\mathcal{S}_{A^{(n)}}(\mathbb{R}^n)$ are equivalent to the norms
$|| \cdot ||_m$ given by (\ref{||.||}), then by the last Lemma it follows that 
$\mathscr{F} \phi_j \xrightarrow{j \rightarrow +\infty} \varphi$ in the topology on
$\mathcal{S}^{0}(\mathbb{R}^n)$ inherited from $\mathcal{S}(\mathbb{R}^n)$. Because 
$\mathscr{F}: \mathcal{S}^{00}(\mathbb{R}^n) \rightarrow \mathcal{S}^{0}(\mathbb{R}^n)$ is continuous
in the topologies on $\mathcal{S}^{00}(\mathbb{R}^n)$  and $\mathcal{S}^{0}(\mathbb{R}^n)$ inherited
from $\mathcal{S}(\mathbb{R}^n)$ then the graph of the map $\mathscr{F}$ is closed
in $\mathcal{S}^{00}(\mathbb{R}^n) \times \mathcal{S}^{0}(\mathbb{R}^n)$ in the product topology
of the topologies on $\mathcal{S}^{00}(\mathbb{R}^n)$  and $\mathcal{S}^{0}(\mathbb{R}^n)$ inherited
from $\mathcal{S}(\mathbb{R}^n)$. Therefore,
\[
\varphi = \mathscr{F} \phi.
\]
If follows from this that the graph of $\mathscr{F}$, on the product
$\mathcal{S}^{00}(\mathbb{R}^n) \times \mathcal{S}^{0}(\mathbb{R}^n)$ 
of the topology  on $\mathcal{S}^{00}(\mathbb{R}^n)$
inherited from $\mathcal{S}(\mathbb{R}^n)$ and the topology on
$\mathcal{S}^{0}(\mathbb{R}^n)$ inherited from $\mathcal{S}_{A^{(n)}}(\mathbb{R}^n)$,
is closed. Because by the preceding Lemmas the said topologies, i.e. the topology 
on $\mathcal{S}^{00}(\mathbb{R}^n)$
inherited from $\mathcal{S}(\mathbb{R}^n)$ and the topology on
$\mathcal{S}^{0}(\mathbb{R}^n)$ inherited from $\mathcal{S}_{A^{(n)}}(\mathbb{R}^n)$
are Fr\'echet topologies, then we can apply the closed graph theorem, \cite{Rudin}, Thm. 2.15, 
which says in this case 
that $\mathscr{F}$ is continuous in these topologies.
\qed

We obtain from the last Lemma and from the inverse mapping theorem (or the open mapping theorem, 
\cite{Rudin}, Thm. 2.11, Corollary 2.12) the following
 
\begin{prop*}
The topology on $\mathcal{S}^{0}(\mathbb{R}^n)$ inherited from $\mathcal{S}_{A^{(n)}}(\mathbb{R}^n)$
coincides with the topology on $\mathcal{S}^{0}(\mathbb{R}^n)$ inherited from $\mathcal{S}(\mathbb{R}^n)$;
thus 
\[
\boxed{\mathcal{S}_{A^{(n)}}(\mathbb{R}^n) = \mathcal{S}^{0}(\mathbb{R}^n) 
\,\, \textrm{in store of elements and in topology}.}
\]
\end{prop*}

We thus can apply the theory of Gelfand and Shilov for the class of spaces which they
denote by $K\{M_{{}_{m}}\}$ in \cite{GelfandII}. In particular the functions of compact support are dense
in $\mathcal{S}^{0}(\mathbb{R}^n) = \mathcal{S}_{{}_{A^{(n)}}}(\mathbb{R}^n)$.
In particular using the system (\ref{M}) of functions $M_m(p)$ for
$\mathcal{S}^{0}(\mathbb{R}^n) = K\{M_k\}$, compare \cite{GelfandII}, Theorem of Chap. II.4.2, we obtain the following corollary
\begin{prop*}
Each continuous functional $\widetilde{F}$ in $\mathcal{S}^{0}(\mathbb{R}^n)^*$ is a finite sum
of (distributional) derivatives all of a fixed order $k$ of continuous functions $\widetilde{F}_q$
with the speed of growth not faster than the power $r^{nk}$ when $r \rightarrow \infty$
and not faster than $r^{-nk}$ when $r \rightarrow 0$ with $k$ depending on $\widetilde{F}$:

\begin{equation}\label{tilFqtilF-representation(tilF)}
\big( \widetilde{F}, \widetilde{\varphi}\big) = \sum \limits_{|q|=k} \int \limits_{\mathbb{R}^n} \widetilde{F}_q(p) \,
D^q \widetilde{\varphi}(p) \, \ud^n p.
\end{equation}
Here $q$ is the miltiindex rangig over all values for which $|q| = k$.
Alternatively the functional $\widetilde{F}$ may also be represented as single (distributional derivative)
of a single continuous function $p \mapsto \widetilde{F}(p)$ of growth not faster than positive integer
power at infinity and not faster than
a negative integer power at zero:
\begin{equation}\label{tilF-representation(tilF)}
\big( \widetilde{F}, \widetilde{\varphi}\big) =
\int \limits_{\mathbb{R}^n} \widetilde{F}(p) \,
D^q \widetilde{\varphi}(p) \, \ud^n p
\end{equation}
for sufficiently large $|q|$.
The same statement holds for $F \in \mathcal{S}^{00}(\mathbb{R}^n)^*$. Any continuous
functional $F \in \mathcal{S}^{00}(\mathbb{R}^n)^*$ can be represented by the formula
\begin{equation}\label{FqF-representation(F)}
\big(F, \varphi\big) =
\sum \limits_{|q|=k} \int \limits_{\mathbb{R}^n} F_q(x) \,
D^q \varphi(x) \, \ud^n x.
\end{equation}
or
\begin{equation}\label{F-representation(F)}
\big(F, \varphi\big) =
\int \limits_{\mathbb{R}^n} F(x) \,
D^q \widetilde{\varphi}(p) \, \ud^n p
\end{equation}
with the corresponding functions $x \mapsto F_q(x)$, and respectively $x \mapsto F(x)$
continuous of growth not faster than positive integer power at infinity and not faster than
a negative integer power at zero.
\end{prop*}
\qedsymbol \, We apply the Theorem of \cite{GelfandII}, Chap. II.4.2
to the nuclear space $\mathcal{S}^{0}(\mathbb{R}^n) = K\{M_m\}$ with the system of functions
$M_m(p)$ defined by (\ref{M}), exactly as Gelfand and Shilov did
for the functional on the space $\mathcal{S}(\mathbb{R}^n) = K\{M_m\}$ defined by
$M_m(x) = \prod_{j=1}^{m} (1 + |x_j|)^m$
in Chap. II.4.3 of their book \cite{GelfandIII}. 

The second part of the statement concerning continuous functionals $F$ on  $\mathcal{S}^{00}(\mathbb{R}^n)
= \widetilde{\mathcal{S}^{0}(\mathbb{R}^n)}$ follows by application of the inverse 
Fourier transform\footnote{In fact it is not that simple and requires some further analysis. But similar result (\ref{tilFqtilF-representation(tilF)}) (or (\ref{F-representation(F)})) in this case may be obtained by using the fact that $\mathcal{S}^{0}(\mathbb{R}^n)$, and thus $\mathcal{S}^{00}(\mathbb{R}^n)$,
is a closed subspace of $\mathcal{S}(\mathbb{R}^n)$. By Hahn-Banach theorem
there exists an extension $f \in \mathcal{S}(\mathbb{R}^n)^*$ of $F$. We apply the Theorem of Chap. II.4.3
classifying continuous functionals on $\mathcal{S}(\mathbb{R}^n)$, to the extension $f$ and obtain the above representation (\ref{FqF-representation(F)}) (or (\ref{F-representation(F)})) of $F$ with $x \mapsto F_q(x), F(x)$ with at most power growth at infinity. That the space 
$\mathcal{S}^{00}(\mathbb{R}^n)^*$ contains also elements $F$ with $x \mapsto F_q(x), F(x)$ with
the inverse power growth at zero follows from application of the Fourier transform 
(understood as a map between $\mathcal{S}^{0}(\mathbb{R}^n)$ and $\mathcal{S}^{00}(\mathbb{R}^n)$) to homogeneous $\widetilde{F}$ (\ref{tilFqtilF-representation(tilF)}) (or (\ref{tilF-representation(tilF)})) 
with homogeneous $p \mapsto \widetilde{F}_q(p), \widetilde{F}(p)$.  So obtained  inverse Fourier transform 
$F$ of $\widetilde{F}$ will be likewise homogeneous in $\mathcal{S}^{00}(\mathbb{R}^n)^*$.
The corresponding $x \mapsto F_q(x), F(x)$ need not be homogeneous, but there are multitude of concrete 
examples in which they indeed do are, compare e.g. \cite{GelfandI} or Section \ref{infra}.}.

\qed

\subsection{{\L}opusza\'nski representation acting on the space 
$E = \mathcal{S}_{A^{(3)}}(\mathbb{R}^3;\mathbb{R}^4)$
and the Pauli-Jordan zero mass function on $\mathcal{S}_{A^{(4)}}(\mathbb{R}^4;\mathbb{R})$}\label{Lop-on-E}

In the previous Subsections we have used the symbol $E$ for general real standard countably Hilbert nuclear
spaces, constructed from standard operators $A$ on real Hilbert spaces $H$, associated with the corresponding Gelfand triples $E \subset H \subset E^*$. But from now on we fix the meaning of $E$ as a concrete real nuclear space:

\begin{defin*}
We put
\[
A = \oplus_{1}^{4} A^{(3)} \,\, \textrm{on} \,\,
L^2(\mathbb{R}^3, \ud^3 \boldsymbol{\p}; \mathbb{R}^4) = \oplus_{1}^{4} L^2(\mathbb{R}^3, \ud^3 \boldsymbol{\p}; \mathbb{R}).
\] 
Let 
\[
E = \mathcal{S}_{A}(\mathbb{R}^3; \mathbb{R}^4) 
= \mathcal{S}_{\oplus A^{(3)}}(\mathbb{R}^3; \mathbb{R}^4) 
= \oplus_{1}^{4} \mathcal{S}_{A^{(3)}}(\mathbb{R}^3; \mathbb{R}),
\]
which may be understood as a subspace
of the Hilbert space $\mathcal{H}'$ of the space of the {\L}opusza\'nski representation and
its conjugation, where the functions $\widetilde{\varphi} \in \mathcal{H}'$
on the orbit $\mathscr{O}_{(1,0,0,1)}$ are treated as functions on $\mathbb{R}^3$ with the three 
momentum components  $\vec{p}$ as the three real coordinates.
\end{defin*}

By the equality $\mathcal{S}_{A^{(n)}}(\mathbb{R}^n) = \mathcal{S}^{0}(\mathbb{R}^n)$
in store of elements and topology (Proposition of the last Subsection \ref{SA=S0})
and the results of Subsection \ref{SA=S0} we can use various equivalent systems of norms
on $\mathcal{S}_{A^{(n)}}(\mathbb{R}^n) = \mathcal{S}^{0}(\mathbb{R}^n)$. Namely, we have for
example the following equivalent systems of norms:
\[
\{|\cdot |_m = |(A^{(n)})^m \cdot|_{{}_{L^2(\mathbb{R}^n)}}\}, \{|| \cdot ||_m \},
\{||| \cdot |||_m \}, \{\rceil \cdot \lceil_{{}_{m}}\}, \{\rceil \cdot \lceil'_{{}_{m}}\}
\]
on $\mathcal{S}_{A^{(n)}}(\mathbb{R}^n) = \mathcal{S}^{0}(\mathbb{R}^n)$
defined in Subsection \ref{SA=S0}.
Various systems of norms are convenient for various continuity questions. In particular using
the system $\{\rceil \cdot \lceil_{{}_{m}}\}$ of $\sup$-norms (\ref{normMm}) with $M_{{}{m}}$
given by (\ref{M}), inherited from
$K\{M_{{}{m}}\} = \mathcal{S}^{0}(\mathbb{R}^n)$ or the system of $\sup$-norms $|| \cdot||_m$
given by (\ref{||.||}), the fact that
the representors of the {\L}opusza\'nski representation and its conjugation
map the space $E$ continuously into itself becomes almost obvious.
Joining this observation with the results of Subsections \ref{dim=1}-\ref{SA=S0}
we obtain in particular the following corollary

\begin{prop*}
If we construct Gelfand triples $E \subset \oplus L^2(\mathbb{R}^3) \subset E^*$ and 
$\mathbb{E} \subset \oplus L^2(\mathbb{R}^3) \subset \mathbb{E}^*$, with the help of
positive self-adjoint operators resp. $A$ and $\mathscr{F}^{-1} A \mathscr{F}$, then the operators
$\sqrt{B}, \sqrt{B}^{-1}$, the operators of multiplication by the functions $r^{-1/2}(\vec{p}) = 
(\vec{p} \cdot \vec{p})^{-1/4}$, $r^{1/2}(\vec{p}) = 
(\vec{p} \cdot \vec{p})^{1/4}$, and the differentiation operator are continuous
as operators $E \to E$, and the ordinary Fourier transform $\mathscr{F}$ and its inverse
$\mathscr{F}^{-1}$, are continuous, respectively, as operators $\mathbb{E} \to E$ and $E \to \mathbb{E}$.
It follows that the operator $\mathcal{F}$ defined by (\ref{F(varphi)}) treated as operator 
$E \ni \widetilde{\varphi} \mapsto \varphi \in \mathbb{E}$ is continuous and onto with the continuous
inverse  $\mathbb{E} \ni \varphi \mapsto \widetilde{\varphi} \in E$, where the functions $\widetilde{\varphi}$
on the orbit $\mathscr{O}_{(1,0,0,1)}$ are treated as functions on $\mathbb{R}^3$ with the three 
momentum components  $\vec{p}$ as the three real coordinates.
The operators $\mathfrak{J}'$, $WU_{{}_{a,\alpha}}^{{}_{(1,0,0,1)}{\L}}W^{-1}$ and 
$\big[WU_{{}_{a,\alpha}}^{{}_{(1,0,0,1)}{\L}}W^{-1}\big]^{*-1}$, $(a,\alpha) \in T_4 \circledS SL(2, \mathbb{C})$
preserve $E$ and are continuous as operators $E \to E$ (resp. $\mathbb{E} \to \mathbb{E} $) 
with respect to the nuclear topology.
The operators $A$ and $\mathscr{F}^{-1} A \mathscr{F}$ preserve conditions A1-A3 of \cite{hida}, \S1  
and the spaces $E$ and $\mathbb{E}$ 
preserve the Kubo-Takenaka conditions  
H1-H3 of \cite{hida}, \S1. 
\end{prop*}

NOTATION. For the simplicity of notation we will frequently use for the operator
\[
A = \oplus_{1}^{4} A^{(3)} \,\, \textrm{on} \,\,
L^2(\mathbb{R}^3, \ud^3 \boldsymbol{\p}; \mathbb{R}^4) = \oplus_{1}^{4} L^2(\mathbb{R}^3, \ud^3 \boldsymbol{\p}; \mathbb{R})
\]
the same symbol $A^{(3)}$ and in general for the operator
\[
B' = \oplus_{1}^{k} B \,\, \textrm{on} \,\,
L^2(\mathbb{R}^3, \ud^3 \boldsymbol{\p}; \mathbb{R}^k) =
\oplus_{1}^{k} L^2(\mathbb{R}^3, \ud^3 \boldsymbol{\p}; \mathbb{R})
\]
the same symbol as for $B$ whenever it is clear that it is equal to the $k$-fold direct sum of the
operator $B$. In particular, we will use the same symbol $\mathscr{F}$ for the Fourier operator
\[
\oplus_{1}^{4} \mathscr{F} \,\, \textrm{on} \,\,
L^2(\mathbb{R}^3, \ud^3 \boldsymbol{\p}; \mathbb{R}^4) = \oplus_{1}^{4} L^2(\mathbb{R}^3, \ud^3 \boldsymbol{\p}; \mathbb{R})
\]
acting on four-component functions as for the Fourier operator acting on one-component scalar functions.

In particular, we will write $\mathcal{S}_{{}{A^{(n)}}}(\mathbb{R}^n)$ for
$\mathcal{S}_{{}{A^{(n)}}}(\mathbb{R}^n; \mathbb{R}^4)$ or even for
$\mathcal{S}_{{}{A^{(n)}}}(\mathbb{R}^n; \mathbb{C}^4)$ whenever it is is clear
that the functions in $\mathcal{S}_{{}{A^{(n)}}}(\mathbb{R}^n)$ are
$\mathbb{R}^4$- or $\mathbb{C}^4$-valued, in order to simplify notation.
Sometimes we omit the complexification sign $(\cdot)_{\mathbb{C}}$ in such expressions like
$E_\mathbb{C} = \mathcal{S}_{A^{(3)}}\big(\mathbb{R}^3;\mathbb{R}^4\big)_{\mathbb{C}} =
\mathcal{S}_{A^{(3)}}\big(\mathbb{R}^3;\mathbb{C}^4\big)$, whenever it is obvious
if we are talking of real or complex valued functions, or whenever a statement
holds for both cases.

We sometimes write $A', A'', A''', \ldots$ for $A^{(1)}, A^{(2)}, A^{(3)}, \ldots$.
\qed

From the results of Subsections \ref{dim=1} -- \ref{SA=S0} it follows 
\begin{prop*}
The following maps
\begin{multline*}
\mathcal{S}^{00}(\mathbb{R}^n) \xrightarrow{\mathscr{F}} \mathcal{S}^{0}(\mathbb{R}^n) \\
= \mathcal{S}_{A^{(n)}}(\mathbb{R}^n)  
\xrightarrow{\textrm{restriction to the cone}} \mathcal{S}_{A^{(n-1)}}(\mathbb{R}^{n-1}) 
= \mathcal{S}^{0}(\mathbb{R}^{n-1}),
\end{multline*}
are continuous.
In particular
\begin{multline*}
\mathcal{S}^{00}(\mathbb{R}^4; \mathbb{C}^4) \xrightarrow{\mathscr{F}} \mathcal{S}^{0}(\mathbb{R}^4; \mathbb{C}^4) \\
= \mathcal{S}_{A^{(4)}}(\mathbb{R}^4; \mathbb{C}^4) 
\xrightarrow{\textrm{restriction to the cone}} \mathcal{S}_{A^{(3)}}(\mathbb{R}^{3}; \mathbb{C}^4) 
= E_\mathbb{C}
\end{multline*}
are continuous.
\end{prop*}
\qedsymbol \,
This is a consequence of the Lemma of Subsection \ref{dim=n} and the equality $\mathcal{S}_{A^{(n)}}(\mathbb{R}^n)
= \mathcal{S}^{0}(\mathbb{R}^n)$ proved in Subsection \ref{SA=S0}.

Let us give another proof. Let $(p_0, p_1,p_2, \ldots , p_{n-1})$ be the Cartesian coordinates of $p$ in $\mathbb{R}^{n}$
and let $\mathscr{O} = \{p: p_{0}^{2} - p_{1}^{2} - \ldots - p_{n-1}^{2} =0, p_0 >0 \}$ or 
$\mathscr{O} = \{p: p_{0}^{2} - p_{1}^{2} - \ldots - p_{n-1}^{2} =0, p_0 <0 \}$ be the (positive or negative)
sheet of the cone in $\mathbb{R}^n$. Let us denote $(p_1, p_2, \ldots, p_{n-1})$ by $\boldsymbol{\p}$
and for the radius function $r(\boldsymbol{\p}) = \sqrt{p_{1}^{2} + p_{2}^{2} + \ldots p_{n-1}^{2}}$ in $\mathbb{R}^{n-1}$
we put $|\boldsymbol{\p}|$.  Then for the radius function 
$r(p)= \sqrt{p_{0}^{2} + p_{1}^{2} + \ldots + p_{n-1}^{2}}$ in $\mathbb{R}^{n}$  we have the following
relation on the (positive or negative) sheet $\mathscr{O}$ of the cone
\[
r(p) = \sqrt{2} |\boldsymbol{\p}|, \,\,\,\,\, 
p = (\pm |\boldsymbol{\p}|, \boldsymbol{\p}) \in \mathscr{O}.
\]

In the proof we will use the system $\{ || \cdot ||_{{}_{m}} \}_{{}_{m\in \mathbb{N}}}$ of norms (\ref{||.||})
in $\mathbb{S}^0(\mathbb{R}^{n-1})= \mathcal{S}_{A^{(n -1)}}(\mathbb{R}^{n-1})$ and in 
$\mathbb{S}^0(\mathbb{R}^n)= \mathcal{S}_{A^{(n)}}(\mathbb{R}^n)$. 

Note that for $\widetilde{\varphi} \in \mathbb{S}^0(\mathbb{R}^n)$
the restriction to the sheet $\mathscr{O}$ of the cone  
is defined by 
\[
\widetilde{\varphi}|_{{}_{\mathscr{O}}}(\boldsymbol{\p}) = \widetilde{\varphi}(\pm |\boldsymbol{\p}|, \boldsymbol{\p}),
\]
so that   
\begin{multline*}
|\boldsymbol{\p}|^{k} \frac{\partial}{\partial p_i} \widetilde{\varphi}|_{{}_{\mathscr{O}}}(\boldsymbol{\p}) 
=
|\boldsymbol{\p}|^{k} \frac{\partial}{\partial p_i} \widetilde{\varphi}(\pm |\boldsymbol{\p}|, \boldsymbol{\p}) \\
=
\Big(\frac{1}{\sqrt{2}}\Big)^k r(p)^k \Big( \frac{\partial}{\partial p_i} \widetilde{\varphi}(\pm |\boldsymbol{\p}|, \boldsymbol{\p})
\pm \frac{\partial}{\partial p_0} \widetilde{\varphi}(\pm |\boldsymbol{\p}|, \boldsymbol{\p}) \frac{p_1}{|\boldsymbol{\p}|} \Big) \,\,\,\, i= 1,2, \ldots n-1.
\end{multline*}
From this the inequality 
\begin{equation}\label{||dim=n-1||<||dim=n||}
|| \widetilde{\varphi}|_{{}_{\mathscr{O}}} ||_{{}_{m}} \leq \sqrt{2}^{m} 3^m || \widetilde{\varphi}||_{{}_{m}}
\end{equation}
follows. Because the system of norms $|| \cdot ||_{{}_{m}}$ 
is equivalent to the system of
norms (\ref{normMm}), i.e. $\rceil \cdot \lceil_{{}_{m}}$, associated with the system $\{M_{{}_{m}}\}$ of functions
(\ref{M}), defining the space $K\{M_{{}_{m}}\} = \mathcal{S}^0(\mathbb{R}^{n-1})$ (compare Subsection \ref{SA=S0}),
then it follows from (\ref{||dim=n-1||<||dim=n||}) and Subsection \ref{SA=S0} (compare also \cite{GelfandII},
Chap. II) that $\widetilde{\varphi}|_{{}_{\mathscr{O}}} \in \mathcal{S}^0(\mathbb{R}^{n-1})$
and that the restriction 
\[
\mathcal{S}^0(\mathbb{R}^{n}) \ni \widetilde{\varphi} \mapsto \widetilde{\varphi}|_{{}_{\mathscr{O}}} \in \mathcal{S}^0(\mathbb{R}^{n-1})
\]
to the (positive or negative sheet $\mathscr{O}$ of the) cone
is continuous as a map from $\mathcal{S}^0(\mathbb{R}^{n})$ into $\mathcal{S}^0(\mathbb{R}^{n-1})$.

\qed

The nuclear test spaces $\mathcal{S}^{0}(\mathbb{R}^4), \mathcal{S}^0(\mathbb{R}^3)$ and
$\mathcal{S}^{00}(\mathbb{R}^4),
\mathcal{S}^{00}(\mathbb{R}^3)$ are fundamental for the correct understanding of the zero mass Pauli-Jordan distribution 
as the commutation function of massless free fields understood
as integral kernel operators with vector-valued kernels in the sense of \cite{obataJFA} in the white noise setup
(when multiplied by the respective smooth invariant factor
correspondingly to the particular zero mass field, e.g.
$g_{\mu \nu}$ in case of the field $A_\mu$). These test spaces (of resp. scalar--, four vector-- e.t.c.
valued functions) compose the indispensable ingredient as the proper domain(s) for commutator function(s).
Although the Pauli-Jordan function extends over to an element of $\mathcal{S}(\mathbb{R}^4)^*$, and moreover this extension is unique if we require preservation of homogeneity and support. This fact has very important consequence for the splitting problem, compare discussion in Subsection \ref{splitting}.
Nonetheless, we describe here the zero mas
Pauli-Jordan commutator function totally within its proper domain $\mathcal{S}^{0}(\mathbb{R}^4), \mathcal{S}^0(\mathbb{R}^3)$.
The reason is that within its proper domain the Pauli-Jordan zero mas commutator function
is more easily manageable, and we avoid indispensable regularizations, necessary when using the improper domain
$\mathcal{S}(\mathbb{R}^4)$ with mathematical rigor.

Namely, the ''singular function''
\begin{equation}\label{tildeD_0(p)}
g_{\mu \nu} \, \widetilde{D_0}(p) =
\frac{\delta(p_0 - |\boldsymbol{\p}|) - \delta(p_0 + |\boldsymbol{\p}|)}{|\boldsymbol{\p}|}
= g_{\mu \nu} \, \textrm{sign} \, p_0 \,\, \delta(p\cdot p),
\end{equation}
with $g_{\mu\nu} \textrm{sign} \, p_0 \,\, \delta(p\cdot p)$ as a (Fourier transformed) commutator function
of the field $A^\mu(x)$, cannot be interpreted as a distribution on $\mathcal{S}(\mathbb{R}^4)$, whenever the
electromagnetic potential field $A^\mu(x)$ is constructed as an integral kernel operator with vector-valued
kernel in the sense of \cite{obataJFA}, within the white noise setup.
This has been already explained in Section \ref{psiBerezin-Hida}, \ref{WightmanField}, and
will be summarized as Theorem \ref{ZeromassTestspace}, Subsection \ref{A=Xi0,1+Xi1,0}.
Nonetheless, the meaning of
the pairing $(\widetilde{D_0},\widetilde{\varphi})$ is clear: the symbol
\[
(\widetilde{D_0},\widetilde{\varphi}) = \int \limits_{\mathbb{R}^4} \widetilde{D_0}(p) \widetilde{\varphi}(p) \, \ud^4 p
\]
stands for the value of the functional on $\widetilde{\varphi}$, defined by the integration of $\widetilde{\varphi}$ along the light cone with respect to the induced invariant measure of the restriction of the test function $\widetilde{\varphi}$ to the cone $\mathscr{O}_{1,0,0,1} \sqcup \mathscr{O}_{-1,0,0,1} = \{p: p\cdot p = 0 \}$ taken with opposite signs on the two sheets of the cone. Namely,
\begin{multline*}
(\widetilde{D_0},\widetilde{\varphi}) = \int \limits_{p\cdot p = 0, p_0 >0}
\, \widetilde{\varphi}|_{{}_{p\cdot p = 0, p_0 >0}}(p) \,
\ud \mu|_{{}_{p\cdot p = 0, p_0 >0}}(p) \\
-
\int \limits_{p\cdot p = 0, p_0 <0}
\, \widetilde{\varphi}|_{{}_{p\cdot p = 0, p_0 <0}}(p) \,
\ud \mu|_{{}_{p\cdot p = 0, p_0 <0}}(p),
\,\,\, \widetilde{\varphi} \in \mathcal{S}^{0}(\mathbb{R}^4) \, \textrm{and} \,
\varphi \in \mathcal{S}^{00}(\mathbb{R}^4),
\end{multline*}
which is a well-defined continuous functional on $\mathcal{S}^{0}(\mathbb{R}^4)$ as a function
of $\widetilde{\varphi} \in \mathcal{S}^{0}(\mathbb{R}^4)$ and a continuous functional
on $\mathcal{S}^{00)}(\mathbb{R}^4)$ as a function
of $\varphi \in \mathcal{S}^{00}(\mathbb{R}^4)$. The functional $(\widetilde{D_0},\widetilde{\varphi})$ as a function of $\varphi \in \mathcal{S}^{00)}(\mathbb{R}^4)$ is by definition equal to $(D_0, \varphi)$, so that $\widetilde{D_0}$
is the Fourier transform of the functional $D_0$, and
\begin{multline*}
(\widetilde{D_0},\widetilde{\varphi}) = i \int \limits_{x\cdot x = 0, x_0 >0}
\, \varphi|_{{}_{x\cdot x = 0, x_0 >0}}(x) \,
\ud \mu|_{{}_{x\cdot x = 0, x_0 >0}}(x) \\
-
i \int \limits_{x\cdot x = 0, x_0 <0}
\, \widetilde{\varphi}|_{{}_{x\cdot x = 0, x_0 <0}}(x) \,
\ud \mu|_{{}_{x\cdot x = 0, x_0 <0}}(x) = (D_0,\varphi), \\
\,\,\, \widetilde{\varphi} \in \mathcal{S}^{0}(\mathbb{R}^4) \, \textrm{and} \,
\varphi \in \mathcal{S}^{00}(\mathbb{R}^4),
\end{multline*}
is a well-defined continuous functional on $\mathcal{S}^{00}(\mathbb{R}^4)$
as a function of $\varphi \in \mathcal{S}^{00}(\mathbb{R}^4)$ so that the value $(D_0, \varphi)$ of the functional
$D_0$ is equal to the integration along the light cone of the restriction of the test function $\varphi$
to the cone with respest to the induced invariant measure on the cone, taken with opposite signs on the two sheets of the cone. Thus, the functionals $D_0$
and $\widetilde{D_0}$ operate identically on the test functions in their domains, which reflects the intuition that the ''singular functions'' $\widetilde{D_0}(p)$
and $D_0(x)$ are equal (replacing the variable $p$ in the first with the variable $x$ we obtain the other and vice versa,
compare \cite{Dirac3rdEd}, pp. 276-277):
\begin{equation}\label{D_0(x)}
D_0 (x) = i \frac{\delta(x_0 - |\boldsymbol{\x}|) - \delta(x_0 + |\boldsymbol{\x}|)}{|\boldsymbol{\x}|}
= = \textrm{sign} \, x_0 \,\, \delta(x\cdot x).
\end{equation}
Namely we have the following
\begin{prop*}
\begin{multline}\label{tildeD_0=D_0}
(\widetilde{D_0},\widetilde{\varphi}) = \int \limits_{p\cdot p = 0, p_0 >0}
\, \widetilde{\varphi}|_{{}_{p\cdot p = 0, p_0 >0}}(p) \,
\ud \mu|_{{}_{p\cdot p = 0, p_0 >0}}(p) \\
-
\int \limits_{p\cdot p = 0, p_0 <0}
\, \widetilde{\varphi}|_{{}_{p\cdot p = 0, p_0 <0}}(p) \,
\ud \mu|_{{}_{p\cdot p = 0, p_0 <0}}(p) \\
= 2\pi i \int \limits_{x\cdot x = 0, x_0 >0}
\, \varphi|_{{}_{x\cdot x = 0, x_0 >0}}(x) \,
\ud \mu|_{{}_{x\cdot x = 0, x_0 >0}}(x) \\
-
2\pi i \int \limits_{x\cdot x = 0, x_0 <0}
\, \varphi|_{{}_{x\cdot x = 0, x_0 <0}}(x) \,
\ud \mu|_{{}_{x\cdot x = 0, x_0 <0}}(x) = (D_0,\varphi),
\end{multline}
is a well-defined continuous functional on $\mathcal{S}^{0}(\mathbb{R}^4)$ as a function
of $\widetilde{\varphi} \in \mathcal{S}^{0}(\mathbb{R}^4)$ and a continuous functional
on $\mathcal{S}^{00)}(\mathbb{R}^4)$ as a function
of $\varphi \in \mathcal{S}^{00}(\mathbb{R}^4)$
\end{prop*}

\qedsymbol \,
Consider the following four maps. 1) The map 
$\widetilde{\varphi} \mapsto \widetilde{\varphi}|_{{}_{\mathscr{O}_{1,0,0,1}}}$
of $\mathcal{S}^0(\mathbb{R}^4)$ into $\mathcal{S}^0(\mathbb{R}^3)$,
2) the map  $\widetilde{\varphi} \mapsto \widetilde{\varphi}|_{{}_{\mathscr{O}_{-1,0,0,1}}}$
of $\mathcal{S}^0(\mathbb{R}^4)$ into $\mathcal{S}^0(\mathbb{R}^3)$,
3) the multiplication by the function $\boldsymbol{\p} \mapsto \frac{1}{|\boldsymbol{\p}|}$
mapping $\mathcal{S}^0(\mathbb{R}^3)$ into itself,
and 4) the map 
\[
\widetilde{\varphi}|_{{}_{\mathscr{O}}} 
\mapsto \int \widetilde{\varphi}|_{{}_{\mathscr{O}}} (\boldsymbol{\p}) \ud^3 \boldsymbol{\p},
\,\,\,
\mathscr{O} = \mathscr{O}_{1,0,0,1}, \mathscr{O}_{-1,0,0,1}
\]
of $\mathcal{S}^0(\mathbb{R}^3)$ into complex numbers.
The functional $\widetilde{\varphi} \mapsto (\widetilde{D_0}, \widetilde{\varphi})$
is equal to the composition of the maps 1), 3) and 4) minus the composition of the maps 2), 3) and 4).
Now the maps 1) and 2) are continuous by the preceding Proposition, the map 3) is continuous
by the results of Subsections \ref{diffSA} and \ref{SA=S0}, and finally continuity of the map 4)
easily follows when using the system of norms (\ref{||.||}).
Continuity of the functional $\widetilde{\varphi} \mapsto (\widetilde{D_0}, \widetilde{\varphi})$
thus follows. Continuity of the functional  $\varphi \mapsto (\widetilde{D_0}, \widetilde{\varphi})$
follows from the continuity of the functional 
$\widetilde{\varphi} \mapsto (\widetilde{D_0}, \widetilde{\varphi})$ and from the continuity 
of the Fourier transform $\mathcal{S}^{00}(\mathbb{R}^3) \rightarrow \mathcal{S}^0(\mathbb{R}^3)$.

Thus, in order to prove the equality (\ref{tildeD_0=D_0}) of the assertion it will be sufficient to prove it
for $\varphi$ ranging over a subspace dense in $\mathcal{S}^{00}(\mathbb{R}^4)$, or what amounts to the same thing
for $\widetilde{\varphi}$ ranging over the subspace dense in $\mathcal{S}^{0}(\mathbb{R}^4)$.
By the results of Subsection \ref{SA=S0} the space of smooth functions with compact support 
 is dense in $\mathcal{S}^{0}(\mathbb{R}^4)$, so it will be sufficient to prove (\ref{tildeD_0=D_0}) for all
$\varphi \in \mathcal{S}^{00}(\mathbb{R}^4)$ for which $\widetilde{\varphi}$ has compact support.

Note that $\mathcal{S}^0(\mathbb{R}^3) \otimes \mathcal{S}^0(\mathbb{R}) \subset \mathcal{S}^0(\mathbb{R}^4)$,
but $\mathcal{S}^0(\mathbb{R}^3) \otimes \mathcal{S}^0(\mathbb{R}) \neq \mathcal{S}^0(\mathbb{R}^4)$,
so that $\mathcal{S}^0(\mathbb{R}^3) \otimes \mathcal{S}^0(\mathbb{R})$ is not dense in the nuclear topology
in $\mathcal{S}^0(\mathbb{R}^4)$. Nonetheless, the restriction to the cone of the elements
$\mathcal{S}^0(\mathbb{R}^3) \otimes \mathcal{S}^0(\mathbb{R}) \subset \mathcal{S}^0(\mathbb{R}^4)$ may approximate
the restriction of any element of $\mathcal{S}^0(\mathbb{R}^4)$ to the cone in the nuclear topology of
$\mathcal{S}^0(\mathbb{R}^3) \oplus \mathcal{S}^0(\mathbb{R}^3)$ on the cone,
which follows easily from the general form
of eigenfunctions of the standard operators $A^{(n)}$ as well as the first Lemma of Subsection
\ref{SA=S0}. Because the left-hand side $(\widetilde{D_0}, \widetilde{\varphi})$ of (\ref{tildeD_0=D_0})
is concentrated on the light cone its value depends only on the restriction
$\widetilde{\varphi}|_{{}_{\mathscr{O}_{1,0,0,1}\sqcup \mathscr{O}_{-1,0,0,1}}}(\boldsymbol{\p}) =
\widetilde{\varphi}(\boldsymbol{\p}, p_0 = \pm |\boldsymbol{\p}|)$ of the Fourier transform $\widetilde{\varphi}$ to the cone. Thus, it will be sufficient to prove (\ref{tildeD_0=D_0}) for
such $\varphi$ that $\widetilde{\varphi}$ has compact support and the
restriction $\widetilde{\varphi} (\boldsymbol{\p}, \pm |\boldsymbol{\p}|)$ has the following form
$\widetilde{\xi} \otimes \widetilde{\eta} (\boldsymbol{\p}, \pm |\boldsymbol{\p}|) =
\widetilde{\xi} (\boldsymbol{\p}) \widetilde{\eta}(\pm |\boldsymbol{\p}|)$, with
$\widetilde{\xi} \in \mathcal{S}^0(\mathbb{R}^3)$, $\widetilde{\eta} \in \mathcal{S}^0(\mathbb{R})$ of compact support.

Thus, let $\varphi$ be any such function belonging to $\mathcal{S}^{00}(\mathbb{R}^4)$ that
$\widetilde{\varphi} \in \mathcal{S}^{0}(\mathbb{R}^4)$ has compact support
and such that 
\begin{multline*}
\widetilde{\varphi} (\boldsymbol{\p}, \pm |\boldsymbol{\p}|) = \widetilde{\xi} (\boldsymbol{\p}) \widetilde{\eta}(\pm |\boldsymbol{\p}|) 
=
\int \limits_{\mathbb{R}^3} \ud^3 \, \boldsymbol{\x} \xi(\boldsymbol{\x}) e^{-i\boldsymbol{\p} \cdot \boldsymbol{\x}}
\int \limits_{\mathbb{R}} \ud x_0 \, \eta(x_0) e^{i \pm |\boldsymbol{\p}|x_0} \\
=
\int \limits_{\mathbb{R}^4} \ud^3 \boldsymbol{\x} \ud x_0 \, \xi(\boldsymbol{\x}) \eta(x_0) e^{-i \boldsymbol{\p} \cdot \boldsymbol{\x} \pm i |\boldsymbol{\p}| x_0} 
= \int \limits_{\mathbb{R}^4} \ud^4 x \, \xi \otimes \eta (x) 
e^{-i \boldsymbol{\p} \cdot \boldsymbol{\x} \pm i |\boldsymbol{\p}| x_0}
\end{multline*} 
with $\widetilde{\xi} \in \mathcal{S}^0(\mathbb{R}^3)$, 
$\widetilde{\eta} \in \mathcal{S}^0(\mathbb{R})$ of compact support and with $\xi \in \mathcal{S}^{00}(\mathbb{R}^3)$,
$\eta \in \mathcal{S}^{00}(\mathbb{R})$. We prove (\ref{tildeD_0=D_0})
for such $\varphi$. By construction 
\begin{equation}\label{(D0,xi-otimes-eta)}
(D_0, \varphi) = (D_0, \xi \otimes \eta),
\end{equation}
because
\[ 
(D_0, \varphi_1) = (\widetilde{D_0}, \widetilde{\varphi_1}) = (\widetilde{D_0}, \widetilde{\varphi_2})
= (D_0, \varphi_2)
\]
whenever the restrictions of $\widetilde{\varphi_1}$ and $\widetilde{\varphi_2}$ coincide
on the light cone in the momentum space.

In this case where $\widetilde{\varphi}$ is of compact support we may apply the Fubini theorem to the 
integral on the left-hand side of (\ref{tildeD_0=D_0}):
\begin{multline}\label{ProofD0eq1}
(\widetilde{D_0},\widetilde{\varphi}) = \int \limits_{\mathscr{O}_{1,0,0,1}} \, \widetilde{\varphi}|_{{}_{\mathscr{O}_{1,0,0,1}}}(p) \, 
\ud \mu|_{{}_{\mathscr{O}_{1,0,0,1}}}(p) \\
-
\int \limits_{\mathscr{O}_{-1,0,0,1}} \, \widetilde{\varphi}|_{{}_{\mathscr{O}_{-1,0,0,1}}}(p) \, 
\ud \mu|_{{}_{\mathscr{O}_{-1,0,0,1}}}(p) \\
= 
\int \limits_{\mathbb{R}^3} \, \frac{1}{|\boldsymbol{\p}|} \,
\widetilde{\varphi}(\boldsymbol{\p}, |\boldsymbol{\p}|) \, 
\ud^3 \boldsymbol{\p}
-
\int \limits_{\mathbb{R}^3} \, \frac{1}{|\boldsymbol{\p}|} \, 
\widetilde{\varphi}(\boldsymbol{\p}, -|\boldsymbol{\p}|) \, 
\ud^3 \boldsymbol{\p} \\
= 
\int \limits_{\mathbb{R}^3} \,\ud^3 \boldsymbol{\p} \, \frac{1}{|\boldsymbol{\p}|} \, 
\int \limits_{\mathbb{R}^3} \ud^3 \boldsymbol{\x} \,
\int \limits_{\mathbb{R}} \ud x_0 
\xi(\boldsymbol{\x}) \eta(x_0) \, e^{-i\boldsymbol{\p} \cdot \boldsymbol{\x} + i|\boldsymbol{\p}|x_0} \\
-
\int \limits_{\mathbb{R}^3} \, \ud^3 \boldsymbol{\p} \, \frac{1}{|\boldsymbol{\p}|} \, \int \limits_{\mathbb{R}^3} \ud^3 \boldsymbol{\x} \,
\int \limits_{\mathbb{R}} \ud x_0 
\xi(\boldsymbol{\x}) \eta(x_0) \, e^{-i\boldsymbol{\p} \cdot \boldsymbol{\x} - i|\boldsymbol{\p}|x_0},
\end{multline} 
where the integrals 
\[
\int \limits_{\mathbb{R}^3} \, \ud^3 \boldsymbol{\p}  \ldots
\]
can be taken over a compact domain, e.g. a ball $\mathbb{B}$ of radius sufficiently large to contain the compact support
of the function $\widetilde{\varphi}$ restricted to the cone.

Now consider the integrand functions
\[
\begin{split}
h_+: \boldsymbol{\p} \times (\boldsymbol{\x} \times x_0) \mapsto
\frac{1}{|\boldsymbol{\p}|} e^{-i\boldsymbol{\p} \cdot \boldsymbol{\x} + i|\boldsymbol{\p}|x_0}
\xi(\boldsymbol{\x}) \eta(x_0), \\
h_-: \boldsymbol{\p} \times (\boldsymbol{\x} \times x_0) \mapsto
\frac{1}{|\boldsymbol{\p}|} e^{-i\boldsymbol{\p} \cdot \boldsymbol{\x} - i|\boldsymbol{\p}|x_0}
\xi(\boldsymbol{\x}) \eta(x_0)
\end{split}
\]
in the above expression. Then
\[
h_+ = (g \otimes (\xi \otimes \eta)) \cdot e_+ \,\,\, \textrm{and} \,\,\,
h_- = (g \otimes (\xi \otimes \eta)) \cdot e_-
\]
where $(g \otimes (\xi \otimes \eta)) (\boldsymbol{\p}, x) = g(\boldsymbol{\p}) \xi \otimes \eta(x)$
and where
\[
\begin{split}
g(\boldsymbol{\p}) = \frac{1}{|\boldsymbol{\p}|}
\,\,\,\textrm{and} \,\,\,
e_{\pm}(\boldsymbol{\p}) = e^{-i\boldsymbol{\p} \cdot \boldsymbol{\x} \pm i|\boldsymbol{\p}|x_0}.
\end{split}
\]
Because (by an easy application of the Scholium 3.9 of \cite{Segal_Kunze}) the functions
$e_+, e_-$ are measurable functions of absolute value equal one on the product measure space $\mathbb{B} \times \mathbb{R}^4$,
and $g, \xi \otimes \eta$ are measurable over the measure spaces $\mathbb{B}$ and $\mathbb{R}^4$ respectively, then again by Scholium 3.9 of \cite{Segal_Kunze},
$h_+$ and $h_-$ are measurable on the product measure space $\mathbb{B} \times \mathbb{R}^4$
and moreover because $g$ is integrable, i.e. belongs to $L^1(\mathbb{B}, \ud^3 \boldsymbol{\p})$
and $\xi \otimes \eta \in L^1(\mathbb{R}^4, \ud^4 x)$, then $h_+, h_-$ are integrable over the product measure
space $\mathbb{B} \times \mathbb{R}^4$ and Fubini theorem (Corollary 3.6.2 of \cite{Segal_Kunze})
is applicable to the integrals (\ref{ProofD0eq1}). Therefore, for the sum of the integrals
(\ref{ProofD0eq1}) we obtain
\begin{multline}\label{ProofD0eq2}
\int \limits_{\mathbb{R}^3} \ud^3 \boldsymbol{\x} \,
\int \limits_{\mathbb{R}^3} \,\ud^3 \boldsymbol{\p} \, \frac{1}{|\boldsymbol{\p}|}
e^{-i\boldsymbol{\p} \cdot \boldsymbol{\x}}
\, \int \limits_{\mathbb{R}} \ud x_0
\xi(\boldsymbol{\x}) \eta(x_0) \, e^{i|\boldsymbol{\p}|x_0} \\
-
\int \limits_{\mathbb{R}^3} \ud^3 \boldsymbol{\x} \,
\int \limits_{\mathbb{R}^3} \, \ud^3 \boldsymbol{\p} \, \frac{1}{|\boldsymbol{\p}|}
e^{-i\boldsymbol{\p} \cdot \boldsymbol{\x}}
\, \int \limits_{\mathbb{R}} \ud x_0
\xi(\boldsymbol{\x}) \eta(x_0) \, e^{- i|\boldsymbol{\p}|x_0} \\
=
\int \limits_{\mathbb{R}^3} \ud^3 \boldsymbol{\x} \,
\int \limits_{0}^{\infty} |\boldsymbol{\p}| \,\ud |\boldsymbol{\p}| \, \int \limits_{0}^{\pi} \int \limits_{0}^{2\pi}\sin \theta \ud \theta \ud \phi
e^{-i|\boldsymbol{\p}| |\boldsymbol{\x}| \cos \theta}
\, \int \limits_{\mathbb{R}} \ud x_0
\xi(\boldsymbol{\x}) \eta(x_0) \, e^{i|\boldsymbol{\p}|x_0} \\
-
\int \limits_{\mathbb{R}^3} \ud^3 \boldsymbol{\x} \,
\int \limits_{0}^{\infty} |\boldsymbol{\p}| \,\ud |\boldsymbol{\p}| \, \int \limits_{0}^{\pi} \int \limits_{0}^{2\pi}\sin \theta \ud \theta \ud \phi
e^{-i|\boldsymbol{\p}| |\boldsymbol{\x}| \cos \theta}
\, \int \limits_{\mathbb{R}} \ud x_0
\xi(\boldsymbol{\x}) \eta(x_0) \, e^{-i|\boldsymbol{\p}|x_0} \\
=
\frac{2\pi}{i}
\int \limits_{\mathbb{R}^3} \ud^3 \boldsymbol{\x} \, \frac{1}{|\boldsymbol{\x}|} \xi(\boldsymbol{\x})
\int \limits_{0}^{\infty} \,\ud |\boldsymbol{\p}| \,
\big\{ e^{i|\boldsymbol{\p}| |\boldsymbol{\x}|} - e^{-i|\boldsymbol{\p}| |\boldsymbol{\x}|} \big\}
\, \int \limits_{\mathbb{R}} \ud x_0
\eta(x_0) \, e^{i|\boldsymbol{\p}|x_0} \\
-
\frac{2\pi}{i}
\int \limits_{\mathbb{R}^3} \ud^3 \boldsymbol{\x} \, \frac{1}{|\boldsymbol{\x}|} \xi(\boldsymbol{\x})
\int \limits_{0}^{\infty} \,\ud |\boldsymbol{\p}| \,
\big\{ e^{i|\boldsymbol{\p}| |\boldsymbol{\x}|} - e^{-i|\boldsymbol{\p}| |\boldsymbol{\x}|} \big\}
\, \int \limits_{\mathbb{R}} \ud x_0
\eta(x_0) \, e^{-i|\boldsymbol{\p}|x_0},
\end{multline}
where, inspired by the hint of Dirac \cite{Dirac3rdEd}, pages 276-277, we have used the polar coordinates
$|\boldsymbol{\p}|, \theta, \phi$, with $\boldsymbol{\x}$ as pointing to the ''North Pole'',
in the integration
\[
\int \limits_{\mathbb{R}^3} \,\ud^3 \boldsymbol{\p} \, \ldots
\]
and where the range of the integration
\[
\int \limits_{0}^{\infty} \,\ud |\boldsymbol{\p}| \, \dots
\]
in the integrals (\ref{ProofD0eq2}) can be taken to be finite and the upper
bound $\infty$ can be replaced with the radius of the ball $\mathbb{B}$.

Because $\widetilde{\eta}$ belongs to $\mathcal{S}^{0}(\mathbb{R})$, then the integrals 
\[
\int \limits_{-\infty}^{+\infty} \ud a  e^{ia |\boldsymbol{\x}|} \widetilde{\eta}(a) 
\]
and 
\[
\int \limits_{-\infty}^{+\infty} \ud a  e^{-ia |\boldsymbol{\x}|} \widetilde{\eta}(a) 
\]
converge absolutely, and as functions of $|\boldsymbol{\x}|$ belong to $\mathcal{S}^{00}(\mathbb{R})$, 
so that the hint of Dirac \cite{Dirac3rdEd}, pages 276-277, becomes legitimate and the 
sum (\ref{ProofD0eq2}) of integrals is equal to 
\begin{multline}\label{ProofD0eq3}
\frac{2\pi}{i}
\int \limits_{\mathbb{R}^3} \ud^3 \boldsymbol{\x} \, \frac{1}{|\boldsymbol{\x}|} \xi(\boldsymbol{\x})
\int \limits_{-\infty}^{+\infty} \,\ud a \, 
e^{ia |\boldsymbol{\x}|}
\, \int \limits_{\mathbb{R}} \ud x_0 
 \eta(x_0) \, e^{iax_0} \\
-
\frac{2\pi}{i}
\int \limits_{\mathbb{R}^3} \ud^3 \boldsymbol{\x} \, \frac{1}{|\boldsymbol{\x}|} \xi(\boldsymbol{\x})
\int \limits_{-\infty}^{+\infty} \,\ud a \, 
 e^{-ia |\boldsymbol{\x}|} 
\, \int \limits_{\mathbb{R}} \ud x_0 
 \eta(x_0) \, e^{iax_0} \\
=
2\pi i
\int \limits_{\mathbb{R}^3} \ud^3 \boldsymbol{\x} \, \frac{1}{|\boldsymbol{\x}|} \xi(\boldsymbol{\x})
\int \limits_{\mathbb{R}} da \, \int \limits_{\mathbb{R}} dx_0 \, 
\eta(x_0)  
\big\{ e^{i(x_0 - |\boldsymbol{\x}|)a} - e^{i(x_0 + |\boldsymbol{\x}|)a}\big\}.
\end{multline} 
Because $\widetilde{\eta}$ belongs to $\mathcal{S}^{0}(\mathbb{R}) \subset \mathcal{S}(\mathbb{R})$, 
then inversion formula for the Fourier transform, \cite{Rudin}, Thm. 7.7, is applicable to the integral (\ref{ProofD0eq3}),
which by the said formula is equal to
\begin{multline*}
2\pi i
\int \limits_{\mathbb{R}^3} \ud^3 \boldsymbol{\x} \, \frac{1}{|\boldsymbol{\x}|} \xi(\boldsymbol{\x})
 \eta(|\boldsymbol{\x}|) 
-
2\pi i
\int \limits_{\mathbb{R}^3} \ud^3 \boldsymbol{\x} \, \frac{1}{|\boldsymbol{\x}|} \xi(\boldsymbol{\x})
 \eta(-|\boldsymbol{\x}|) \\
= 2\pi i \int \limits_{x\cdot x = 0, x_0 >0} 
\, (\xi \otimes \eta)|_{{}_{x\cdot x = 0, x_0 >0}}(x) \, 
\ud \mu|_{{}_{x\cdot x = 0, x_0 >0}}(x) \\
-
2\pi i \int \limits_{x\cdot x = 0, x_0 <0} 
\, (\xi \otimes \eta)|_{{}_{x\cdot x = 0, x_0 <0}}(x) \, 
\ud \mu|_{{}_{x\cdot x = 0, x_0 <0}}(x)
=(D_0, \xi \otimes \eta),
\end{multline*}
and by (\ref{(D0,xi-otimes-eta)}) the last expression is equal to $(D_0, \varphi)$
for all $\varphi \in \mathcal{S}^{00}(\mathbb{R}^4)$ with $\widetilde{\varphi} \in \mathcal{S}^0(\mathbb{R}^4)$ of compact support such that $\widetilde{\varphi} (\boldsymbol{\p}, \pm |\boldsymbol{\p}|) 
= \widetilde{\xi} \otimes \widetilde{\eta} (\boldsymbol{\p}, \pm |\boldsymbol{\p}|) =
\widetilde{\xi} (\boldsymbol{\p}) \widetilde{\eta}(\pm |\boldsymbol{\p}|)$, with
$\widetilde{\xi} \in \mathcal{S}^0(\mathbb{R}^3)$, $\widetilde{\eta} \in \mathcal{S}^0(\mathbb{R})$ of compact support. 
Therefore, (\ref{tildeD_0=D_0}) is proved. 
\qed

Note that because $\widetilde{D_0}$ is concentrated on the light cone in the momentum space then
\[
\boxed{(D_0, \square \varphi) = 0, \,\,\,
\varphi \in \mathcal{S}^{00}(\mathbb{R}^4).}
\]

{\bf REMARK 1}. Note that (\ref{tildeD_0(p)}) understood as a distribution has the interpretation of the operation of integration along the cone of the restriction to the cone taken with the opposite signs on the two sheets of the cone, and as such is a well-defined
functional over $\mathcal{S}^{0}(\mathbb{R}^4)$ . The symbol
$\frac{1}{|\boldsymbol{\p}|} \delta(p_0 - |\boldsymbol{\p}|)$ cannot be interpreted simply as the ordinary multiplication
of the Dirac delta distribution by the function $\boldsymbol{\p} \mapsto \frac{1}{|\boldsymbol{\p}|}$,
even within $\mathcal{S}^{0}(\mathbb{R}^4)^*$ because the function is not any multiplier of the algebra
$\mathcal{S}^{0}(\mathbb{R}^4)$. Indeed, recall that $(\boldsymbol{\p},p_0) \mapsto \frac{1}{r(\boldsymbol{\p},p_0)}
= \frac{1}{\sqrt{p_{0}^{2} + |\boldsymbol{\p}|^2}}$ is a multiplier of $\mathcal{S}^{0}(\mathbb{R}^4)$
but not the function $(\boldsymbol{\p}, p_0) \mapsto \frac{1}{|\boldsymbol{\p}|}$. It is the continuity of the restriction to the (positive or negative sheet of the) cone as a map $\mathcal{S}^0(\mathbb{R}^4) \rightarrow \mathcal{S}^0(\mathbb{R}^3)$ which allows to multiply the delta function $\delta(p_0 - |\boldsymbol{\p}|)$ by the function $\boldsymbol{\p} \mapsto \frac{1}{|\boldsymbol{\p}|}$, because the last function indeed is a multiplier of the
algebra $\mathcal{S}^{0}(\mathbb{R}^3)$. Nonetheless, the ordinary formal rules for differentiation operations are applicable to the symbol (\ref{tildeD_0(p)}). Namely,
the functional (\ref{tildeD_0(p)}), as element of $\mathcal{S}^{0}(\mathbb{R}^4)^*$
defined by the singular cone sub-manifold $\{p , P(p) = p_{0}^{2} - p_{1}^2 - p_{2}^{2} - p_{3}^{2}=0\}$ in
$\mathbb{R}^4$ can be identified with distribution $\textrm{sign} \, p_0 \,\, \delta(P)$, where $\delta(P)$
is the distribution determined by the quadratic form $P$, compare \cite{GelfandI}, Chap.II.2.1, and denoted
there by $\delta(P)$. In case the quadratic form $P$, or generally function $P$, is smooth around the sub-manifold $P=0$ and non-singular
around $P=0$, there is natural and essentially unique construction for
$\delta^{(1)}(P), \delta^{(2)}(P), \ldots $, compare
\cite{GelfandI}, Chap.II.1. If the sub-manifold $\{p, P(p) =0\}$ contains singular points (as is the case for the cone)
the construction of $\delta^{(1)}(P), \delta^{(2)}(P), \ldots $ is less natural and in general singularities appear in the value $\big(\delta^{(k)}(P), \widetilde{\varphi} \big)$, $\widetilde{\varphi} \in \mathcal{S}(\mathbb{R}^4)$, constructed as for the non-singular manifold $P=0$, if the value or the derivatives of
$\widetilde{\varphi}$ do not vanish at singular points, at least up to some order depending on
$k$ and the order of singularity of the manifold $P=0$.
But if the singular points of the manifold $P=0$ compose a discrete separated set, or a finite set, the value
$\big(\delta^{(k)}(P), \widetilde{\varphi} \big)$, computed as for the smooth case, can be regularized. Thus,
definition of the counterparts for
$\delta^{(1)}(P), \delta^{(2)}(P), \ldots $ is still possible, but the regularization is in general non unique. For the subspace $\mathcal{S}^{0}(\mathbb{R}^4)$ these difficulties disappear for
the quadratic form $P(p) = p_{0}^{2} - p_{1}^2 - p_{2}^{2} - p_{3}^{2} = p\cdot p$, and the definition of
$\delta^{(1)}(P), \delta^{(2)}(P), \ldots $ becomes unique with the preservation of all formal rules
for differentiation of distributions applicable to the symbolic function
\[
\frac{\delta(p_0 - |\boldsymbol{\p}|) - \delta(p_0 + |\boldsymbol{\p}|)}{|\boldsymbol{\p}|}
= \textrm{sign} \, p_0 \,\, \delta(p\cdot p).
\]
Unfortunately the values, and the values of derivatives, at zero are in general not equal to zero
for elements $\varphi$ of
$\mathcal{S}^{00}(\mathbb{R}^4)$. In particular for the proof of $\square D_0 =0$
the formal rules for differentiation of distribution functions would be insufficient, and
explicit computation of the Fourier transform $\widetilde{D_0}$ was necessary.

\qed

\subsection{Splitting of homogeneous distributions $\big(\mathcal{S}^{00}(\mathbb{R}^4)^*\big)^{\otimes \, n}$}\label{splitting}

In the proof of the next Proposition we need a connection between the functions $f$ on $\mathbb{R}$ which are Fourier transforms 
of square integrable functions $F$ supported on the positive half $\mathbb{R}_+$
of $\mathbb{R}$, and the Hardy space $H^2(\mathbb{H})$ of functions holomorphic
on the upper half $\mathbb{H} \subset \mathbb{C}$ of the complex plane, $\mathbb{H} =
\{(p + iq) \in \mathbb{C}, q >0\}$. This connection is summarized in the
Paley-Wiener theorem for the functions whose Fourier transforms are boundary values of
holomorphic functions on the upper half complex plane.  In the proof of the next Proposition 
 we use a construction 
of a unitary operator $U$ transforming the Hardy space $H^2(\mathbb{D})$ on the unit disk 
in $\mathbb{C}$ onto the Hardy space $H^2(\mathbb{H})$, which preserves the holomorphic structure 
(resp. of   $\mathbb{D}$ and $\mathbb{H}$) and the smooth 
structure of the boundaries (resp. $\mathbb{S}^1$ and $\mathbb{R}$), and which is naturally 
generated by  conformal equivalence between $\mathbb{D}$ and $\mathbb{H}$. Regularity at the
point at ``infinity'' $\infty$ on the boundary unit circle $\mathbb{S}^1$ is restored
by using the standard operator $A = UH_{(1)}U^{-1}$ on $L^2(\mathbb{S}^1)$, 
defining the nuclear space $\mathcal{S}_{A}(\mathbb{S}^1)$,

and which is the image of the nuclear Schwartz space 
$\mathcal{S}_{H_{(1)}}(\mathbb{R}) = \mathcal{S}(\mathbb{R})$ on the boundary $\mathbb{R}$ of $\mathbb{H}$
under the he unitary operator $U$.
Here $H_{(1)}$ is the Hamiltonian operator of the one dimensional oscillator
and $\mathcal{S}_{A}(\mathbb{S}^1)$ turns out to be equal to the nuclear space of all smooth
functions on the boundary $\mathbb{S}^1$ vanishing together with all derivatives at the pole
of the conformal mapping defining the conformal equivalence of $\mathbb{D}$ and $\mathbb{H}$. 

Let us recall the Paley-Wiener theorem\footnote{Recall that this is a particular case of the whole family known as ``Paley-Wiener theorem(s)'', all characterizing Fourier transforms of functions with a specified support,
as e.g. compact, or compact and convex, e.t.c.. } characterizing Fourier transforms of functions supported on 
the positive real line, as elements of the Hardy space $H^2(\mathbb{H})$, as well as the most
important properties of $H^2(\mathbb{D})$, which we will use below 
(for a proof compare \cite{RudinARZ}, Chap. 17 and 19) 
\begin{twr*}[of PALEY-WIENER for $H^2(\mathbb{H})$]
Let $H(\mathbb{H})$ be the linear space of holomorphic functions on the upper half
$\mathbb{H}$ of the complex plane $\mathbb{C}$.  
\begin{enumerate}
\item[1)]
Suppose that $f \in H(\mathbb{H})$ and 
\begin{equation}\label{PWR+condition}
\sup \limits_{0 <q} \frac{1}{2\pi} \int \limits_{-\infty}^{+\infty} |f(p + iq)|^2 \, \ud p = C < +\infty.
\end{equation}
Let for each $q>0$, $f_q(p) = f(p+iq)$. (\ref{PWR+condition}) means
that $f_q, q>0$ forms a bounded set in $L^2(\mathbb{R})$. 
Then there exists $f^* \in L^2(\mathbb{R})$ such that 
$f_q \rightarrow f^*$ in $L^2(\mathbb{R})$, as
$q \rightarrow 0^+$, and $f_q \rightarrow f^*$ pointwise almost everywhere on the boundary
$\mathbb{R}$ of $\mathbb{H}$, i.e.  
\begin{equation}\label{PWR+Limitcondition}
\lim \limits_{q \rightarrow 0^+} \int \limits_{-\infty}^{+\infty} |f(p +iq) - f^*(p + iq)|^2 \, \ud p = 0,
\end{equation}
the Fourier transform of $f^*$ and of $f_q$ for each $q>0$,  is supported on $\mathbb{R}_+$, i.e. 
\begin{equation}\label{supp(Ff*)inR+}
f^*(p) = \int \limits_{0}^{+\infty} F(x) e^{ixp} \ud x, 
\end{equation}
and
\[
f(z) = \int \limits_{0}^{+\infty} F(x) e^{izp} \ud x \,\,\,(z \in \mathbb{H}),
\]
and finally
\begin{equation}\label{PWR+norm}
||f^*||_2 = \int \limits_{0}^{+\infty} |F(x)|^2 \, \ud x = C.
\end{equation}
\item[2)]
Let $f^* \in L^2(\mathbb{R})$ be such that (\ref{supp(Ff*)inR+}) holds, i.e. its Fourier transform
$F$ is supported on $\mathbb{R}_+$. Let us define 
$f$ on $\mathbb{H}$ by the formula
\[
f(z) = \int \limits_{\mathbb{R}} F(x) e^{izp} \, \ud x.
\]
Then $f \in H(\mathbb{H})$ and satisfies (\ref{PWR+condition}). 

\end{enumerate}
\end{twr*}
     
Note that $H^2(\mathbb{H})$
 consists of all those $f \in H(\mathbb{H})$ for which (\ref{PWR+condition})
holds and the norm $\big|\big| f\big|\big|_{{}_{\mathbb{H}2}}$ of $f$ in $H^2(\mathbb{H})$ 
is precisely equal to the number $C$ in (\ref{PWR+condition}),
so that the above version of Paley-Wiener theorem gives us natural
identification of the Hilbert space of Fourier transforms $f^*$
of functions $F$ in $L^2(\mathbb{R})$ with $\textrm{supp} \, F \subset \mathbb{R}_+$ 
with the Hardy space of $f$ in $H^2(\mathbb{H})$.

Let us remind now definition and fundamental properties of the Hardy space
$H^2(\mathbb{D})$ of Holomorphic functions $g$ on the open unit disc 
$\mathbb{D} = \{z \in \mathbb{C}, |z|<1 \}$ in $\mathbb{C}$.   
For any $g \in H(\mathbb{D})$ we define
\[
V(g;r) = \Bigg(\frac{1}{2\pi} \int \limits_{-\pi}^{\pi} |f(re^{i\alpha})|^2 \, d\alpha \Bigg)^{1/2},
\,\,\, \textrm{and} \,\,\,
\big|\big| g\big|\big|_{{}_{\mathbb{D}2}} = \lim \limits_{r \rightarrow 1} V(g;r).
\]
Then the Hardy space $H^2(\mathbb{D})$ is the space of all those $f \in H(\mathbb{D})$
for which $\big|\big| g\big|\big|_{{}_{\mathbb{D}2}} < \infty$, and the Hilbert space
norm of $g \in H^2(\mathbb{D})$ is given by $\big|\big| g\big|\big|_{{}_{\mathbb{D}2}}$.

Analogously as the space of $f$ in $H^2(\mathbb{H})$ has a natural identification with the closed subspace of $f^*$ in $L^2(\mathbb{R})$
(here with $\mathbb{R}$ regarded as the boundary of $\mathbb{H}$)
consisting of Fourier transforms $f^*$ of functions $F$ in $L^2(\mathbb{R})$ with $\textrm{supp} \, F \subset
\mathbb{R}_+$, we have analogous property for $H^2(\mathbb{D})$. Namely the space of $g$ in $H^2(\mathbb{D})$
can be naturally identified with the closed subspace of these $g^*$ in $L^2(\mathbb{S}^1)$
whose Fourier coefficients $\widehat{g^*}(n)$ vanish for negative integers $n$. 
Here $\mathbb{S}^1$ is regarded as the boundary of $\mathbb{D}$. Compare the Theorem below.  Recall that the norm of 
$g^* \in L^2(\mathbb{S}^1)$ is defined by
\[
||g||_2 = \Bigg( \frac{1}{2\pi} \int \limits_{-\pi}^{\pi} |g^*(e^{i\alpha})|^2 \, d\alpha \Bigg)^{1/2},
\] 
and that each $g^* \in L^2(\mathbb{S}^1)$ has the Fourier coefficients equal
\[
\widehat{g^*}(n) = \frac{1}{2\pi} \int \limits_{-\pi}^{\pi} g^*(e^{i\alpha}) e^{-in\alpha} \, d\alpha, 
\,\,\,\,
n \in \mathbb{Z}.
\]

The most important properties of $H^2(\mathbb{D})$ are collected in the following
theorem (for a proof, compare \cite{RudinARZ}, Chap. 17)
\begin{twr*}
\begin{enumerate}
\item[1)]
A function $g \in H(\mathbb{D})$ of the form
\[
g(z) = \sum \limits_{n=0}^{+\infty} a_n z^n, \,\,\, z \in \mathbb{D},
\]
belongs to $H^2(\mathbb{D})$ if and only if $\sum |a_n|^2 < \infty$; in this case
\[
\big|\big| g\big|\big|_{{}_{\mathbb{D}2}} = \Big( \sum \limits_{n=0}^{+\infty} |a_n|^2 \Big)^{1/2}.
\]
\item[2)]
If $g \in H^2(\mathbb{D})$, then $g$ has the radial limit $g(re^{i\alpha}) \rightarrow g^*(e^{i\alpha})$,
as $r \rightarrow 1$, at almost each point $\alpha$ of the boundary circle $\mathbb{S}^1$ of $\mathbb{D}$
to a function $g^* \in L^2(\mathbb{S}^1)$. 
The $n$-th Fourier coefficient of the function $g^*$ is equal $a_n$, if $n\geq 0$, and is equal zero, if $n <0$.
The following $L^2(\mathbb{S}^1)$-approximation is valid
\[
\lim \limits_{r \rightarrow 1} \frac{1}{2\pi} \int \limits_{-\pi}^{\pi} |g(re^{i\alpha}) - g^*(e^{i\alpha})|^2
\, d\alpha = 0.
\]
The function $g$ is given by the Poisson as well as the Cauchy integral formula over the boundary
$\mathbb{S}^1$ and the boundary function $g^*$ as the integrand: if $z=re^{i\alpha} \in \mathbb{D}$, 
then
\[
g(z) = \frac{1}{2\pi} \int \limits_{-\pi}^{\pi} P_r(\alpha -t) g^*(e^{i\alpha}) \, dt
\] 
and
\[
g(z) = \frac{1}{2\pi i} \int \limits_{\mathbb{S}^1} \frac{g^*(\zeta)}{\zeta - z} \, d \zeta.
\]
\item[3)]
The map $g \to g^*$ is an isometry of the Hardy space $H^2(\mathbb{D})$ onto the closed subspace
of $L^2(\mathbb{S}^1)$ consisting of all those $g^* \in L^2(\mathbb{S}^1)$ for which 
$\widehat{g^*}(n) = 0$ for all $n<0$.
\end{enumerate}

\end{twr*}
 
Now we are ready to give a proof of the following

\begin{prop*}
\begin{enumerate}
\item[1)]
The nuclear space $\mathcal{S}^{00}(\mathbb{R}^n) = \widetilde{\mathcal{S}^{0}(\mathbb{R}^n)}
= \widetilde{\mathcal{S}_{A^{(n)}}(\mathbb{R}^n)}$ contains no function with compact support.
\item[2)]
Let $\mathbb{S}^{n-1} \subset \mathbb{R}^n$ be the unit $(n-1)$-sphere centered at zero. 
For any open set $\Omega \subset \mathbb{S}^{n-1}$ there exists a nonzero function
$\varphi \in \mathcal{S}^{00}(\mathbb{R}^n)$ whose support lies within the cone $C_\Omega$
determined by the open set of directions $\Omega$. The same holds for any translation of this cone.
For such $\varphi$, and any open ball $\mathscr{U} \subset \mathbb{R}^n$ of finite radius, 
the Fourier transform 
$\widetilde{\varphi} \in \mathcal{S}^{0}(\mathbb{R}^n)$ cannot vanish identically on $\mathscr{U}$.
\end{enumerate}
\end{prop*}
\qedsymbol \,
Let $\varphi \in \mathcal{S}^{00}(\mathbb{R}^n)$ be a function of compact support. 
Then by one of the classic versions of the 
Paley-Wiener theorem for Fourier transforms of compactly supported $L^2$ functions 
(\cite{Rudin}, Thm. 7.22), $\widetilde{\varphi} \in \mathcal{S}^0(\mathbb{R}^n)$ 
is the restriction to the boundary $\mathbb{R}^n$ 
of the ''upper half'' of the $\mathbb{C}^n$ complex space of an entire function of exponential type 
of $n$ complex variables. Because all derivatives of $\widetilde{\varphi}$ vanish at zero along
the boundary $\mathbb{R}^n$, then $\widetilde{\varphi} = 0$, and thus $\varphi = 0$.   The first 
assertion thereby follows.

Concerning the second assertion, we start at dimension $n=1$. In this case we show
that there exists $\varphi \in \mathcal{S}^{00}(\mathbb{R})$ with $\textrm{supp} \, \varphi \subset \mathbb{R}_+$. We have to show that there exists a function 
$f^* \in \mathcal{S}^{0}(\mathbb{R}) \subset \mathcal{S}(\mathbb{R})$, whose Fourier
transform $F$ has support in $\mathbb{R}_+$. 

Let us consider first the whole Hilbert space of functions $f^* \in L^2(\mathbb{R})$ whose Fourier 
transforms $F$ are supported in $\mathbb{R}_+$. This is the Paley-Wiener Hilbert space of boundary values 
$f^*$ of analytic functions $f \in H^2(\mathbb{H})$ of the above version of Paley-Wiener theorem. 
It is equal to the closed subspace of 
$L^2(\mathbb{R})$, with $\mathbb{R}$ understood as the boundary of $\mathbb{H}$, naturally isomorphic 
to the Hardy space $H^2(\mathbb{H})$. 

Now we consider a unitary operator $U$ mapping $H^2(\mathbb{D})$ onto $H^2(\mathbb{H})$, which is generated 
by the conformal equivalence 
\[
c: \mathbb{D} \ni z \to z'(z) = \frac{-iz -1}{z+i} \in \mathbb{H}, \,\,\,\,\,\,
c^{-1}: \mathbb{H} \ni z' \to z(z') = \frac{z'-i}{iz'-1} \in \mathbb{D}. 
\]  
Namely for any $f\in H(\mathbb{H})$ and $g \in H(\mathbb{D})$, we define the following operators
\[
Uf(z) = \sqrt{2}(z-i)^{-1} \, f(z'(z)), \,\,\, 
U^{-1} = \frac{1}{\sqrt{2}}(z(z') +i) \, g(z(z')).
\]  
Of course $U$ maps holomorphic functions on $\mathbb{H}$ into holomorphic functions
on $\mathbb{D}$, and vice versa for $U^{-1}$. Both, $U$ and $U^{-1}$, are isometric between the Hardy spaces, because the absolute values of the multipliers in their definitions are precisely the square roots 
of the inverses of the 
Radon-Nikodym derivatives of $c$-- (or resp. $c^{-1}$--) transformed measures with respect to 
non-transformed measures  
(as computed for induced measures on one dimensional curves in $\mathbb{C}$):
\[
\frac{dz'}{dz} = \frac{2}{(z+i)^2}.
\]
The operators are by construction mutually inverse, and possess natural extensions
on the boundary value functions, $g^*$ amd $f^*$, corresponding to the elements $g$ and $f$ of 
$H^2(\mathbb{D})$ and $H^2(\mathbb{H})$ respectively.

Consider now elements $f^* \in \mathcal{S}(\mathbb{R})$ with $\mathbb{R}$ understood as the boundary of 
$\mathbb{H}$, but in general we do not assume that there exist the corresponding $f\in H^2(\mathbb{H})$, 
for which $f^*$ is the boundary value function, equal almost everywhere to the pointwise limit of 
$f$ on the boundary
$\mathbb{R}$ of $\mathbb{H}$. The operators $U$ and $U^{-1}$ still make sense for such functions, and are unitary between the whole Hilbert spaces $L^2(\mathbb{R})$ and $L^2(\mathbb{S}^1)$. 
We now find the images $Uf^* = g^*$ of such elements under
$U$. Of course in general $Uf^* = g^*$ will not be equal to any boundary value function corresponding to 
any element $g$ of $H^2(\mathbb{D})$.  
The space $\mathcal{S}(\mathbb{R})$ can be regarded as the Gelfand-Shilov nuclear space $K\{M''_{{}_{m}}\}$
of smooth functions, with the family of functions $M''_{{}_{m}}$ defined 
by (\ref{M''M'}), compare Subsection \ref{SA=S0} or \cite{GelfandII}, Chap. II.
Now it is easily checked that whenever $f^* \in K\{M''_{{}_{m}}\} = \mathcal{S}(\mathbb{R})$ on 
$\mathbb{R}$, then $Uf^* \in K\{M_{{}_{n}}\}$ of functions on $\mathbb{S}^1$
with
\[
M_{{}_{n}}(\alpha) = |e^{\alpha} - i| \,\, M''_{{}_{m}}\big(c(e^{i\alpha})\big).
\]
Here $c$ is the conformal mapping definig conformal equivalence between $\mathbb{D}$
and $\mathbb{H}$, which streams to infinity at the pole $-i = e^{i3\pi/2}$, i.e. at the 
point $\alpha = 3\pi/2$ of the unit circle $\mathbb{S}^1$.
By the general theory $K\{M_{{}_{n}}\}$ is a nuclear space of smooth functions
on the circle $\mathbb{S}^1$, which vanish together with all derivatives at the pole 
$\alpha = 3\pi/2$ of the conformal map $c$. One can prove this exactly as we did for the 
$\mathcal{S}^0(\mathbb{R}^n)$ in Subsection \ref{SA=S0}, or compare the general theory
in \cite{GelfandII}, Chap. II.  Therefore the boundary value functions $g^{*} = Uf^*
\in L^2(\mathbb{R})$ corresponding to the elements $f^*$, with $f^* \in \mathcal{S}(\mathbb{R})$
(with the corresponding $f \in H^2(\mathbb{H})$ existing or not)
are smooth functions on $\mathbb{S}^1$ vanishing together with derivatives of all orders 
at the pole $-i = e^{-3\pi/2}$ of the map $c$, i.e. at the point of the circle $\mathbb{S}^1$
which corresponds via the map $c$ to the point at infinity on the boundary $\mathbb{R}$
of $\mathbb{H}$. 

Consider now an element $f^* \in \mathcal{S}(\mathbb{R})$ of the Paley-Wiener space for 
which the corresponding $f \in H^2(\mathbb{H})$
exists, or what amounts to the same thing, an element
$f$ of $H^2(\mathbb{H})$ corresponding to the function $F$ which not only belongs to $L^2(\mathbb{R})$
and has support in $\mathbb{R}_+$, but moreover $F \in \mathcal{S}(\mathbb{R})$.
This means that the corresponding boundary value function $f^* \in L^2(\mathbb{R})$
on $\mathbb{R}$, regarded as the boundary of $\mathbb{H}$, is equal to the Fourier transform
of a function $F \in \mathcal{S}(\mathbb{R})$. 
Because Fourier transform maps $\mathcal{S}(\mathbb{R})$ onto $\mathcal{S}(\mathbb{R})$,
then $f^* \in \mathcal{S}(\mathbb{R})$. 

Thus among the elements of the Hardy space $H^2(\mathbb{D})$ there are elements $g= Uf$
(with $f^* \in \mathcal{S}(\mathbb{R})$ such that $f\in H^2(\mathbb{H})$)
whose boundary value functions $g^*$ are smooth
and whose derivatives of all orders vanish at the pole $-i$ of $c$. 

Now we repeat the whole construction of the operator $U$ but with the conformal map
$c$ replaced with another, which differs from $c$ by the factor $e^{i\pi}$, 
i.e. by additional $\pi$-rotation of the unit disk $\mathbb{D}$ and the boundary
circle, moving the pole of $c$ from $-i$ to $i$. The same construction of smooth element
$g^*$ with all derivations vanishing at the pole of $c$ gives a function $g*$
in the Hardy space $H^2(\mathbb{D})$, smooth on $\mathbb{S}^1$ vanishing together with all derivatives
at the new pole $i = e^{i\pi/2}$ of the new conformal map $c$.

Note that the rotation of the disk $\mathbb{D}$ induces a map transforming the Hardy space $H^2(\mathbb{D})$ onto $H^2(\mathbb{D})$. This is 
not the case for reflection or complex conjugation. Recall that according to the above stated Theorem 
on the $H^2(\mathbb{D})$, Fourier coefficients $\widehat{g^*}(n)$ of the boundary valued elements $g^*$ corresponging to the elements $g$ of $H^2(\mathbb{D})$ are zero for all negative integers $n$. 
Therefore complex conjugation and reflection\footnote{This is of importance which in conjuction 
with Stone-Weirestrass theorem explains the considerable loss of flexibility in uniform approximation within the space of continuous $g^*$, with  $g \in H^2(\mathbb{D})$, corrsponding via $U$ to the Paley-Wiener space of Fourier transforms of functions supported on the positive half of the real line. Indeed, recall that by the 
Stone-Weierstrass theorem, a linear algebra $\boldsymbol{A}$ of continuous complex valued functions on a compact space $S$ is uniformly dense in the 
algebra $\mathscr{C}(S)$ of all continuous functions on $S$ if the following two conditions hold. 
I) The algebra $\boldsymbol{A}$ is closed under complex conjugation. II) $\boldsymbol{A}$ 
separates points of $S$: for any two points $s_1, s_2$ there exists a function $g\in \boldsymbol{A}$
such that $g(s_1) \neq g(s_2)$. Now it is easily checked that continuous $g^*$ with $g \in H^2(\mathbb{D})$
are sufficient to separate the points of $\mathbb{S}^1$ (take e.g. the functions $g^*(\alpha) = e^{in\alpha}$,
$n = 1,2, \ldots$). Nonetheless the algebra of continuous $g^*$ with $g\in H^2(\mathbb{D})$ is not closed under the complex conjugation, and the Stone-Weierstrass theorem is not applicable. In fact, we will show that
for any open subset $\mathscr{U} \subset \mathbb{S}^1$ no element $g^* \neq 0$, with $g\in H^2(\mathbb{D})$, 
exists vanishing identically on $\mathscr{U}$.}
 lead us out of the Hardy space $H^2(\mathbb{D})$. Nonetheles pointwise multiplication is allowed. Moreover, the continuous elements $g^*$, with $g \in H^2(\mathbb{D})$, compose a Banach algebra with pointwise multiplication and supremum norm, which is a closed subalgebra 
of the Banach algebra  $\mathscr{C}(\mathbb{S}^1)$ 
of all continuous functions on $\mathbb{S}^1$, endowed with the supremum norm, 
compare \cite{RudinARZ}. In particular the function constructed above by pointwise multiplication of the two 
$g^*$-s with $g \in H^2(\mathbb{D})$,
is therefore justified and gives again a boundary valued element of a function in $H^2(\mathbb{D})$.

Now mutliplying two such constructed $g^*$ (with $g$ in the Hardy space $H(\mathbb{D})$),
both smooth, first vanishing together with all derivatives at $-i$, the second one vanishing together with all
derivatives at $i$, we obtain a third smooth function $g^*$ on $\mathbb{S}^1$ (with $g \in H^2(\mathbb{D})$)
which vanishes together with all derivatives at the two points $-i, i$ of the unit boundary
circle $\mathbb{S}^1$. By construction this $g^*$ is in $K\{M_{{}_{n}}\}$, and $U^{-1}g^*$
is in $K\{M''_{{}_{m}}\} = \mathcal{S}(\mathbb{R})$. The function  $U^{-1}g^*$
vanishes together with derivatives of all orders at the zero point of the boundary
$\mathbb{R}$ of $\mathbb{H}$.  Hence $U^{-1}g^* \in \mathcal{S}^0(\mathbb{R})$. 
Because by construction $U^{-1}g^*$ belongs to $H^2(\mathbb{H})$, then the Fourier transform 
$F$ of $U^{-1}g^*$ is supported on the positive half of the real line. Thus we can take $\varphi = F$.
This gives a proof of the first part of assertion 2) for dimension $n=1$ 
and for the cone which degenerates in this case
to the half space $x>0$. In order to obtain the assertion for the cone (half space)
$x>a$, for some real $a \neq 0$, it is sufficient to take an ordinary translation of $F$ constructed above
with  the corresponding $U^{-1}g^* \in \mathcal{S}^0(\mathbb{R})$ multiplied by the phase $e^{ipa}$.   

Let us prove the second part of the assertion 2) for $n=1$.  Suppose that 
for some nonempty  open ball $\mathscr{U}' \subset \mathbb{R}$ of finite radius 
there exists a nonzero element $f^* \in \mathcal{S}(\mathbb{R})$
with $f \in H^2(\mathbb{H})$, such that $f^*$ vanishes identically on $\mathscr{U}'$. This would imply existence
of a nonzero $g^* = Uf^*$ with $g \in H^2(\mathbb{D})$ which vanishes identically 
on a nonempty open \footnote{Note that $c$ transforms finite open intervals of $\mathbb{R}$ into 
``finite'' open intervals of $\mathbb{S}^1$,
i.e. not containing the pole $-i \in \mathbb{S}^1$.} 
ball $\mathscr{U}= c(\mathscr{U}')$ in 
$\mathbb{S}^1$. By applying a rotation to this $g^*$ (and resp. $g \in H^2(\mathbb{D})$) 
with the rotated $\mathscr{U}$ covering the pole $-i$, we obtain a nonzero function $f^* = U^{-1}g^*$
which necessary has compact support and lies in the Paley-Wiener space, i.e. for which
the corresponding $f \in H^2(\mathbb{H})$ exists.
In this case we obtain nonzero compactly supported function $f^*$
equal to the Fourier transform of function $F$ supported on the positive half of the real line. 
But again by the Paley-Wiener theorem, characterizing Fourier transforms of compactly supported $L^2$
functions (\cite{Rudin}, Thm. 7.22), this $F$ would be equal to the restriction to $\mathbb{R}$
(understood as the boundary of $\mathbb{H}$) of an entire function, and as supported on $\mathbb{R}_+$
would be zero. This contradicts our assumption that $f^* \neq 0$, because $F=0$ forces $f^*=0$.
The second part of assertion 2) for $n=1$ is thereby proved.

Concerning the assertion 2) for $n=2$ and the special cone $C_\Omega$ with the apex at zero
and consisting of the first quarter $\{(x,y) \in \mathbb{R}^2, x>0, y>0\}$ , i.e. 
$\Omega = (0,\pi/2)$, we can use 
\begin{equation}\label{varphi-Quarter}
\varphi(x,y) = \phi \otimes \psi (x,y) = \phi(x) \psi(y)
\end{equation}
with $\phi, \psi \in \mathcal{S}^{00}(\mathbb{R})$
fulfilling the assertion 2) for dimension $n=1$, and the cones (half lines)
$x>0$ and $y>0$. We can do this because by the results of the preceding Subsections
$\mathcal{S}^{00}(\mathbb{R}) \otimes \mathcal{S}^{00}(\mathbb{R}) \subset \mathcal{S}^{00}(\mathbb{R}\times \mathbb{R})
= \mathcal{S}^{00}(\mathbb{R}^2)$. Note however that $\mathcal{S}^{00}(\mathbb{R}) \otimes \mathcal{S}^{00}(\mathbb{R}) 
\neq \mathcal{S}^{00}(\mathbb{R}^2)$, contrary to the case of the ordinary Schwartz space.

Consider now a more general cone $C_\Omega \subset \mathbb{R}^2$ determined by 
$\Omega \subset \mathbb{S}^1$ equal to $(\pi/4 - \epsilon, \pi/4 + \epsilon)$,
still lying in the first quarter of the plane $\mathbb{R}^2$,
symmetrically with respect to the line $x=y$ and with arbitrary small angle diameter 
$|\Omega| = 2\epsilon$. 
We can apply appropriate linear transformation $L$ in $\mathbb{R}^2$,
for example appropriate two-dimensional Lorentz transformation $L$ in $\mathbb{R}^2$,
which changes the support of (\ref{varphi-Quarter}) equal to the first quarter into the cone 
$C_\Omega$, $\Omega = (\pi/4 -\epsilon, \pi/4 + \epsilon)$. 
In this way  we obtain the required function $\varphi' \in \mathcal{S}^{00}(\mathbb{R}^2)$
with $\textrm{supp} \, \varphi \subset C_\Omega$ by applying this linear transformation
$L$ to the function (\ref{varphi-Quarter}):
\begin{equation}\label{varphiCpi/4}
\varphi'(x,y) = \varphi\big(L(x,y)\big).
\end{equation}
In order to obtain the required $\varphi'' \in \mathcal{S}^{00}(\mathbb{R}^2)$
with $\textrm{supp} \, \varphi''' \subset C_\Omega$ with arbitrary small angle diameter $|\Omega|$
and arbitraty direction we can apply euclidean rotation $R$ in $\mathbb{R}^2$ to the function
(\ref{varphiCpi/4}):
\[
\varphi''(x,y) = \varphi'\big(R(x,y)\big).
\]    

Generalization of this proof of 2) to higher dimensions is now obvious, concerning at least the existence
of $\varphi \in \mathcal{S}^{00}(\mathbb{R}^n)$ with $\textrm{supp} \, \varphi \subset C_\Omega$
for arbitrary open subset $\Omega \subset \mathbb{S}^{n-1}$. 

The proof of the second part of 2) for $n>1$, cannot be similarly reduced to the case $n=1$,
because  the $n$-fold projective tensor product 
$\mathcal{S}^{00}(\mathbb{R}) \otimes \ldots \otimes \mathcal{S}^{00}(\mathbb{R})$ is a proper subset of 
$\mathcal{S}^{00}(\mathbb{R}^n)$. Nonetheless, the second part of 2) is likewise true for higher dimensions.

In order to see it, note please, that there exist natural extensions $H^2(\mathbb{D}^n),H^2(\mathbb{H}^n)$, 
of the Hardy space constructions $H^2(\mathbb{D})$, $H^2(\mathbb{H})$,
to higher dimensions in $\mathbb{C}^n$. 
Similarly there exist an analogous Paley-Wiener theorem for $H^2(\mathbb{H}^n)$
characterizing Fourier transforms of functions $F\in L^2(\mathbb{R}^n)$ with the support 
of $F$ concentrated in the half space of $\mathbb{R}^n$. Thus the proof of the whole assertion
2) could have been given for all dimensions with the appropriate construction
of the unitary operator $U: H^2(\mathbb{D}^n) \rightarrow H^2(\mathbb{H}^n)$. 
However, we dispense with detailed presentation, as the idea of the proof 
of 2) (including the second part of the assertion) should be clear now.

\qed

Let for any fixed $\lambda>0$, $S_\lambda$ be the scale transformation $S_\lambda \varphi(x)
= \varphi(\lambda x)$ acting in the respective space of test functions $\varphi$ on $\mathbb{R}^n$.
Let us remind that a functional $F \in \mathcal{S}^{00}(\mathbb{R}^n)^*$ (or in any other test space 
of functions on $\mathbb{R}^n$) is called 
homogeneous of degree $\textrm{deg} \, F$ if for all $\lambda >0$, and all test functions $\varphi$
\begin{equation}\label{homogeneityF}
\big(S_\lambda F, \varphi\big) = \Big(F, S_{\lambda^{-1}} \varphi \Big) 
= \lambda^{\textrm{deg} \, F}\lambda^n (F, \varphi).
\end{equation}
Similarly for any $a \in \mathbb{R}^n$ and the translation $T_a: x \mapsto x-a$ we define the translation
$T_a \varphi(x) = \varphi(x-a)$ of $\varphi$, and dually the translation 
\[
\big(T_a F, \varphi \big) = \big(F, T_{-a} \varphi \big)
\]
of the functional $F$ on the test function space. 

Note that even if the functional $\widetilde{F} \in \mathcal{S}^{0}(\mathbb{R}^n)^*$
(or resp. $F \in \mathcal{S}^{00}(\mathbb{R}^n)^*$) is homogeneous, the corresponding functions
$p \mapsto \widetilde{F}_q(p), p \mapsto \widetilde{F}(x)$ (resp. $F_q, F$) representing this 
functional as in the last Proposition of Subsection \ref{SA=S0}, formulas (\ref{tilFqtilF-representation(tilF)})
or (\ref{tilF-representation(tilF)}) (resp.  (\ref{FqF-representation(F)})
or (\ref{F-representation(F)})), need not be homogeneous.
This is because homogeneity is preserved on the subspaces $\mathcal{S}^{0}(\mathbb{R}^n)$
(resp. $\mathcal{S}^{0}(\mathbb{R}^n)$): 
\begin{multline*}
\big(S_\lambda \widetilde{F}, \widetilde{\varphi} \big) 
= \Big(\widetilde{f}, S_{\lambda^{-1}} \widetilde{\varphi} \Big) 
= \lambda^{\textrm{deg} \, f}\lambda^n (f, \varphi)
\,\,\, \widetilde{\varphi} \in \mathcal{S}^{0}(\mathbb{R}^n) \subset \mathcal{S}(\mathbb{R}^n), \\
\textrm{resp.} \,\,\, 
\big(S_\lambda f, \varphi\big) = \Big(F, S_{\lambda^{-1}} \varphi \Big) 
= \lambda^{\textrm{deg} \, F}\lambda^n (f, \varphi)
\,\,\, \varphi \in \mathcal{S}^{00}(\mathbb{R}^n) \subset \mathcal{S}(\mathbb{R}^n). 
\end{multline*}
A simple inspection will show that there are in general many inhomogeneous functions
$x \mapsto F_q(x), x \mapsto F(x)$ (or functions $x \mapsto F_q(x), x \mapsto F(x)$ of various 
homogeneities not equal $\textrm{deg} \, F$) for which the corresponding functionals, defined as in 
(\ref{tilFqtilF-representation(tilF)})
or (\ref{tilF-representation(tilF)}) (resp.  (\ref{FqF-representation(F)})
or (\ref{F-representation(F)}))) are identically zero on the subspace
$\mathcal{S}^{00}(\mathbb{R}^n)^*$. In general such admixture of non homogeneous (or with various homogeneities)   
degenerating to zero on $\mathcal{S}^{0}(\mathbb{R}^n)$
(resp. $\mathcal{S}^{0}(\mathbb{R}^n)$) cannot in general be clearly separated off. 
In particular existence of a homogeneous extension 
$f \in \mathcal{S}(\mathbb{R}^n)^*$ of a general homogeneous $F \in \mathcal{S}^{00}(\mathbb{R}^n)^*$
is far not obvious. Situation is still less trivial if in addition we will require preservation of the 
support, say of conic-type-shape, during this extension. Nonetheless, situation becomes much better
if we have the functional $F \in \mathcal{S}^{00}(\mathbb{R}^n)^*$ (resp. $\widetilde{F} \in \mathcal{S}^{0}(\mathbb{R}^n)^*$)
 in more explicit form. For example suppose that we know from the outset 
the corresponding continuous
functions $x \mapsto F_q(x), x \mapsto F(x)$ giving the representation (\ref{tilFqtilF-representation(tilF)})
or (\ref{tilF-representation(tilF)}) (resp.  (\ref{FqF-representation(F)})
or (\ref{F-representation(F)})) of the functional. Moreover, suppose that 
that all functions $x \mapsto F_q(x), x \mapsto F(x)$ are homogeneous of degree 
$\textrm{deg} \, F -|q|$. Finally, suppose that $\textrm{deg} \, F$ is an integer (for simplicity). 
In this case the functional $F$ can be extended with preservation of 
homogeneity degree and the support, provided it is of conic-type shape.
The additional complications come when the integer $\textrm{deg} \, F <-n+1$, so that
the functions  $x \mapsto F_q(x), F(x)$ (in the last Proposition of Subsection \ref{SA=S0})
cease to be locally integrable around zero. In this case the integrals in the formula
(\ref{tilFqtilF-representation(tilF)})
or (\ref{tilF-representation(tilF)}) (resp.  (\ref{FqF-representation(F)})
or (\ref{F-representation(F)})) for the functional
 should be understood in the regularized sense 
preserving homogeneity (compare \cite{GelfandI}). In this situation we may extend $F$ with preservation 
of homogeneity and even the support (provided it has a natural conic shape). If the homogeneity degree
is non integer and less than $-n+1$ situation becomes slightly more complicated due mainly to the fact
that the regularization of the integrals (\ref{tilFqtilF-representation(tilF)})
or (\ref{tilF-representation(tilF)}) (resp. (\ref{FqF-representation(F)})
or (\ref{F-representation(F)})) is slightly less easily manageable in computations, compare \cite{GelfandI}.
If $\textrm{deg} \, F \geq -n+1 -|q|$,
where $n$ is the dimension of the space $\mathbb{R}^n$ on which the test functions live,
local integrability of the functions $x \mapsto F_q(x), F(x),
p \mapsto \widetilde{F}_q(p), F\widetilde{F}(p)$
in (\ref{tilFqtilF-representation(tilF)})
or (\ref{tilF-representation(tilF)}) (resp. (\ref{FqF-representation(F)})
or (\ref{F-representation(F)})) is assured $\textrm{deg} \, F_q > -n+1$ and no regularization is needed
there, although in further computations is unavoidable.

\begin{prop*}
\begin{enumerate}
\item[1)]
To each functional $F \in \mathcal{S}^{00}(\mathbb{R}^n)^*$ there exists (in general non-unique),
extension $f \in \mathcal{S}(\mathbb{R}^n)^*$.
For any two possible extensions $f,f'$ the difference $f_{{}_{\Delta}} = f -f'$ runs over the following set of functionals
$f_{{}_{\Delta}}$ equal
\[
\big( f_{{}_{\Delta}}, \varphi \big) = 
\sum \limits_{|q| <N} \,\, \int \limits_{\mathbb{R}^n} c_q x^q \varphi(x) \, \ud^n x,
\]
where $N$ ranges over all natural numbers. Here multi-index notation of Schawartz is unzed with $q$
equal to the multi-index $q = (q_0q_1q_2q_3)$ with $|q| = q_0 + \ldots + q_3$ and 
$x^q = \big(x_0\big)^{q_0}\big(x_1\big)^{q_1}\big(x_2\big)^{q_2}\big(x_3\big)^{q_3}$.
\item[2)] Let $\mathfrak{S}$ be any family of open subsets $\Omega \subset \mathbb{S}^{n-1}$ of
$\mathbb{S}^{n-1}$ centered at zero.
Let $C \subset \mathbb{R}^n$ be the complementary $\mathbb{R}^n \backslash \cup_{\Omega \in \mathfrak{S}} 
C_\Omega$ in $\mathbb{R}^n$, i.e. $C$ is the complementary set
(in the set theoretical sense) of any set theoretic sum of open cones $C_\Omega$ all ceneterd at zero. 
Let $F \in \mathcal{S}^{00}(\mathbb{R}^n)^*$ be any homogeneous of degree $\textrm{deg} \,F$ functional.
Let $\textrm{deg} \, F \in \mathbb{Z}$ and 
$\textrm{supp} \, F \subset C$. Let for $F$ there exists the representation
(\ref{FqF-representation(F)})
or (\ref{F-representation(F)})
 of the last Proposition of Subsection \ref{SA=S0},
with homogeneous of degree $\textrm{deg} \, F - |q|$ functions $x \mapsto F_q(x)$ (or resp.
homogeneous of degree  $\textrm{deg} \, F - |q|$ continuous function $x \mapsto F(x)$).   
In this case there exists unique homogeneous extension
$f \in \mathcal{S}(\mathbb{R}^n)^*$ of $F$ with $\textrm{deg} \, f = \textrm{deg} \, F$ and 

$\textrm{supp} \, f \subset C$. 
\item[3)]
Let $F \in \mathcal{S}^{00}(\mathbb{R}^n)^*$ be homogeneous. Let 
$\widetilde{F} \in \mathcal{S}^{0}(\mathbb{R}^n)^*$ be the Fourier transform of $F$. Then $\widetilde{F}$
is likewise homogeneous. 
Suppose there exist extensions $f \in \mathcal{S}(\mathbb{R}^n)^*$ and $\hat{f} \in \mathcal{S}(\mathbb{R}^n)^*$ 
respectively of $F$ and $\widetilde{F}$,  preserving homogeneities and supports. Then there exist numbers
$c_\alpha$ for multi-indices $\alpha$ with $|\alpha| = \textrm{deg} \, F$, such that
\[
\widetilde{f} - \hat{f} = \sum \limits_{|\alpha| = \textrm{deg} \, F} \,\,\,
c_\alpha D^\alpha \delta(p), \,\,\,\,\,\,\,\,
f - \widetilde{\hat{f}} = \sum \limits_{|\alpha| = \textrm{deg} \, F} \,\,\,
c_\alpha x^\alpha, 
\]
in the notation of Schwartz.
\end{enumerate}
\end{prop*}
 
\qedsymbol,
{\bf Ad 1)}. Because $\mathcal{S}^{00}(\mathbb{R}^n)$ is a closed subspace of $\mathcal{S}(\mathbb{R}^n)$,
then by the Hahn-Banach theorem there exists an extension $f \in \mathcal{S}(\mathbb{R}^n)^*$
of $F \in \mathcal{S}^{00}(\mathbb{R}^n)^*$. Applying Fourier transform we have an extension 
$\widetilde{f} \in \mathcal{S}(\mathbb{R}^n)^*$ (equal to the Fourier transform of $f$)
of the functional $\widetilde{F} \in \mathcal{S}^{0}(\mathbb{R}^n)^*$ (equal to the Fourier transform
of $F$). Therefore, for any two such extensions $f,f'$ we have
\[
\big(\widetilde{f}, \widetilde{\varphi}\big) =  \big(\widetilde{f}', \widetilde{\varphi} \big), 
\,\,\, \textrm{for all} \,\,\ \widetilde{\varphi} \in \mathcal{S}^{0}(\mathbb{R}^n) \subset 
\mathcal{S}(\mathbb{R}^n).
\]
Because $\mathcal{S}^{0}(\mathbb{R}^n)$ contains all smooth functions with compact support
$K \subset \mathbb{R}^n \backslash \{0\}$, then the support of 
$\widetilde{f} - \widetilde{f}' \in \mathcal{S}(\mathbb{R}^n)^*$ is equal to the single zero point set
$\{0\}$. The general functional in $\mathcal{S}(\mathbb{R}^n)^*$ supported on $\{0\}$ has the following form (\cite{GelfandII}, Chap. II.4.5)
\[
\widetilde{f} - \widetilde{f}' = \sum \limits_{|q| \leq N}
c_q D^q \delta(x).
\]
Applying the inverse Fourier transform to $\widetilde{f} - \widetilde{f}'$ we obtain the assertion 1).

{\bf Ad 2)} For simplicity we assume $\textrm{deg} \, F \in \mathbb{Z}$. 
Because by assumption the continuous functions 
$x \mapsto F_q(x)$ (resp. the function $x \mapsto F(x)$)
representing the functional as in (\ref{FqF-representation(F)})
or (\ref{F-representation(F)})
are homogeneous of degree $\textrm{deg} \, F + |q|$, then, by comparison to
Thm. \cite{GelfandII}, Chap. II.4.3 we see that the same formula (\ref{FqF-representation(F)})
or (\ref{F-representation(F)}) defines a continuous functional $f$ in 
$\mathcal{S}(\mathbb{R}^n)^*$. In case $\textrm{deg} \, F + |q| < -n+1$ 
the integrals (\ref{FqF-representation(F)})
or (\ref{F-representation(F)}) are understood in regularized sense, \cite{GelfandI}. 
Because  $x \mapsto F_q(x)$ (resp. the function $x \mapsto F(x)$), are homogeneous 
of degree $\textrm{deg} \, F + |q|$,
the functional $f$ is homogeneous of degree $\textrm{deg} \, F$.  

Let $C_\Omega$ be any open cone (say for $\Omega \in \mathfrak{S}$). 
Suppose that for any function $\varphi \in \mathcal{S}^{00}(\mathbb{R}^n)$ 
with $\textrm{supp} \, \varphi  \subset C_\Omega$ 
(there exists such nontrivial $\varphi$ by the preceding Proposition) 
$\big(F, \varphi \big) = 0$. We will show that for the constructed extension 
$f \in \mathcal{S}(\mathbb{R}^n)^*$ we have
\[
\big(f, \varphi\big) = 0, \,\,\, \textrm{for all} \,\,\, \varphi \in \mathcal{S}(\mathbb{R}^n)
\,\,\, \textrm{with} \,\,\, \textrm{supp} \, \varphi \subset C_\Omega.
\] 

We will proceed as in the proof of the last Proposition starting at dimension
$n=1$. We will use notation from this proof. Racall that in the case $n=1$ the cone $C_\omega$
degenerates to $x>0$ half line.  

Let us assume for simplicity that $\textrm{deg} \, F = \textrm{deg} \, f \in \mathbb{Z}$.
Thus for continuous and homogeneous function $x \mapsto f(x)$ with $\textrm{deg} \big(x \mapsto f(x) \big)
= \textrm{deg} \, f +q \in \mathbb{Z}$ we need to show that 
from
\begin{equation}\label{(f,varphi) = 0onS00R+}
\big(f, \varphi\big) = \int \limits_{\mathbb{R}} \, f(x) \, \frac{d^q\varphi}{dx^q}(x) \, \ud x = 0, 
\,\,\, \textrm{for all} \,\,\, \varphi \in \mathcal{S}^{00}(\mathbb{R})
\,\,\, \textrm{with} \,\,\, \textrm{supp} \, \varphi \subset \mathbb{R}_+,
\end{equation}
it follows that 
\begin{equation}\label{(f,varphi) = 0onSR+}
\big(f, \varphi\big) = \int \limits_{\mathbb{R}} \, f(x) \frac{d^q\varphi}{dx^q}(x) \, \ud x = 0, 
\,\,\, \textrm{for all} \,\,\, \varphi \in \mathcal{S}(\mathbb{R})
\,\,\, \textrm{with} \,\,\, \textrm{supp} \, \varphi \subset \mathbb{R}_+.
\end{equation}
The integral in (\ref{(f,varphi) = 0onS00R+}) is identically zero 
if $q \geq \textrm{deg} \big(x\mapsto f(x)\big)$ irrespectively if 
$\varphi \in \mathcal{S}^{00}(\mathbb{R})$ or $\varphi \in \mathcal{S}(\mathbb{R})$
with $\textrm{supp} \, \varphi \subset \mathbb{R}_+$. Therefore, we may assume that
$q < \textrm{deg} \big(x\mapsto f(x)\big)$. The integral in 
(\ref{(f,varphi) = 0onS00R+}) and (\ref{(f,varphi) = 0onSR+}) is understood in the sense of 
regularization, \cite{GelfandI}, Chap. I.1.7, eq. (3), when $\varphi \in \mathcal{S}(\mathbb{R})$.
Note however that in our case $\varphi(0) = \frac{d\varphi}{dx}(0) = \frac{d^2\varphi}{dx^2}(0) = \ldots = 0$ by 
the assumption that $\textrm{supp} \, \varphi \subset \mathbb{R}_+$. So in our case the integral 
(\ref{(f,varphi) = 0onSR+}) coincides with ordinary 
non regularized integral in both cases $\varphi \in \mathcal{S}^{00}(\mathbb{R})$ or 
$\varphi \in \mathcal{S}(\mathbb{R})$. 

Thus, we need to show that for any fixed integer $m$ and $b \in \mathbb{R}$, from 
\begin{equation}\label{(f,varphi) = 0onS00R+'}
\big(f, \varphi\big) = \int \limits_{\mathbb{R}} \, b \, x^m \varphi(x) \, \ud x = 0, 
\,\,\, \textrm{for all} \,\,\, \varphi \in \mathcal{S}^{00}(\mathbb{R})
\,\,\, \textrm{with} \,\,\, \textrm{supp} \, \varphi \subset \mathbb{R}_+,
\end{equation}
it follows
\begin{equation}\label{(f,varphi) = 0onSR+'}
\big(f, \varphi\big) = \int \limits_{\mathbb{R}} \, b \, x^m \varphi(x) \, \ud x = 0, 
\,\,\, \textrm{for all} \,\,\, \varphi \in \mathcal{S}(\mathbb{R})
\,\,\, \textrm{with} \,\,\, \textrm{supp} \, \varphi \subset \mathbb{R}_+.
\end{equation}
But from (\ref{(f,varphi) = 0onS00R+'}) it follows
\begin{multline}\label{(f,varphi) = 0onS00R+''}
\big(f, \varphi\big) = \int \limits_{\mathbb{R}} \, b x^m \varphi(x -a) \, \ud x = 0, \\ 
\,\,\, \textrm{for all} \,\,\, \varphi \in \mathcal{S}^{00}(\mathbb{R})
\,\,\, \textrm{with} \,\,\, \textrm{supp} \, \varphi \subset \mathbb{R}_+,
\,\,\, \textrm{and all} \,\,\, a>0.
\end{multline}
By applying Fourier transform to (\ref{(f,varphi) = 0onS00R+''}) we obtain
\begin{multline}\label{(f,varphi) = 0onS00R+'''}
\big(f, \varphi\big) = \int \limits_{\mathbb{R}} \, b \frac{d^m\delta}{dx^m}(p -a) 
\,\, \widetilde{\varphi}(p) \, \ud p = 
b \, \frac{d^m \widetilde{\varphi}}{dp^n}(a) = 0, \\
\,\,\, \textrm{for all} \,\,\, \varphi \in \mathcal{S}^{00}(\mathbb{R})
\,\,\, \textrm{with} \,\,\, \textrm{supp} \, \varphi \subset \mathbb{R}_+,
\,\,\, \textrm{and all} \,\,\, a>0,
\end{multline}
if the integer $m \geq 0$, or (up to irrelevant constant)
\begin{multline}\label{(f,varphi) = 0onS00R+''''}
\big(f, \varphi\big) = b \, \int \limits_{\mathbb{R}} \,  \textrm{sign} \, (p - a) \,\, p^{m-1} \widetilde{\varphi}(p) \, \ud p 
= 
b \, \int \limits_{\mathbb{R}} \, \textrm{sign} \, p \,\, \widetilde{\varphi^{(|m|-1)}}(p) \, \ud p
 = 0, \\
\,\,\, \textrm{for all} \,\,\, \varphi \in \mathcal{S}^{00}(\mathbb{R})
\,\,\, \textrm{with} \,\,\, \textrm{supp} \, \varphi \subset \mathbb{R}_+,
\,\,\, \textrm{and all} \,\,\, a>0,
\end{multline}
for the integer $m<0$. Here $\varphi^{(|m|-1)} = \frac{d^{|m|-1}\varphi}{dx^{|m|-1}}$

On the other hand if $\varphi \in \mathcal{S}^{00}(\mathbb{R})$ and 
$\textrm{supp} \, \varphi \subset \mathbb{R}_+$, then for any positive integer $m$,
$x^{m} \varphi, \frac{d^{m-1}\varphi}{d^{m-1}x} \in \mathcal{S}^{00}(\mathbb{R})$
and $\textrm{supp} \, \big(x^{m} \varphi \big) \subset \mathbb{R}_+$,
$\textrm{supp} \, \big(\frac{d^{m-1}\varphi}{d^{m-1}x} \big) \subset \mathbb{R}_+$. Therefore, 
by the preceding Proposition, assertion 2), for each fixed  $\varphi \in \mathcal{S}^{00}(\mathbb{R})$ with
$\textrm{supp} \, \varphi \subset \mathbb{R}_+$ the value 
$\widetilde{x^m \varphi}(a) = \frac{d^n \widetilde{\varphi}}{dp^n}(a) \neq 0$ for some $a>0$. In 
particular for each fixed $\varphi$ in 
(\ref{(f,varphi) = 0onS00R+'''}) the value $\frac{d^n \widetilde{\varphi}}{dp^n}(a) \neq 0$ for
at lest some (in fact almost all) $a>0$. Thus from (\ref{(f,varphi) = 0onS00R+'''})
it follows that $b=0$ and  $\textrm{supp} \,\big(x \mapsto f(x)\big) \subset \mathbb{R}_-$.
Similarly, for each fixed $\varphi$ in (\ref{(f,varphi) = 0onS00R+''''})
$\widetilde{\frac{d^{|m|-1}\varphi}{d^{|m|-1}x}}(a) = a^{|m|-1} \widetilde{\varphi}(a) \neq 0$
for almost all $a>0$. Therefore the integral
\[
\int \limits_{\mathbb{R}} \,  \textrm{sign} \, (p - a) \,\, p^{m-1} \widetilde{\varphi}(p) \, \ud p 
\]

in (\ref{(f,varphi) = 0onS00R+''''}) cannot be zero for all $a>0$. Therefore, 
it follows from (\ref{(f,varphi) = 0onS00R+''''}) that $b=0$, or that 
$\textrm{supp} \big(x \mapsto f(x)\big) \subset \mathbb{R}_-$. Thus, 
from (\ref{(f,varphi) = 0onS00R+'}) it follows (\ref{(f,varphi) = 0onSR+'}).
Therefore for dimension $n=1$ for any open cone $C_\Omega$ (half line),
from $F\big|_{{}_{C_\Omega}} = 0$ it follows for the homogeneous extension
$f$ that $f\big|_{{}_{C_\Omega}} = 0$. 

When the homogeneity degree of $F$
is non integer the integer $m$ will have to be replaced with the corresponding
non integer number $\lambda$. The additional complication comming in is that the Fourier transform 
of the homogeneous distribution function $x^m$ (which we use during the proof), is now replaced with the Fourier
transform of the homogeneous distribution function $(x +i0)^\lambda$. For detailed analysis of 
this distribution
compare \cite{GelfandI}. Its Fourier transform need slightly more sophisticated regularization, 
\cite{GelfandI}.  
 
Similar proof based on the same principle, i.e. assertion 2) of the preceding Proposition,
can be extended to higher dimensions without essential modifications.

{\bf Ad 3)}. $\widetilde{f}$ and $\hat{f}$ coincide on $\mathcal{S}^0(\mathbb{R}^n)$
by assumption. Therefore, $\textrm{supp} \, \big(\widetilde{f} -\widetilde{f}'\big) = \{0\}$
follows as the proof of the assertion 1). Restriction to the subset $\alpha$ with 
$|\alpha| = \textrm{deg} \, F$ follows from homogeneity. 

\qed

In particular for any homogeneous distribution $F$ in $\mathcal{S}^{00}(\mathbb{R}^4)^*$ which is sufficiently regular, i.e. $F$ may be represented by the formula (\ref{F-representation(F)}) with the corresponding
function $x \mapsto F(x)$ in (\ref{F-representation(F)}) which is homogeneous
of degree $\textrm{deg} \, F - |q|$, and with the support equal to the set theoretical sum
$\Gamma^+(0) \cup \Gamma^-(0)$ of the closed forward cone $\Gamma^+(0)$ and the closed backward
cone $\Gamma^-(0)$ of zero (i.e. $\Gamma^+(0) \cup \Gamma^-(0)= \{0\}$), there exists 
unique homogeneous of degree $\textrm{deg} \, F$ extension $f \in \mathcal{S}^{00}(\mathbb{R}^4)^*$ of
$F$, such that $\textrm{supp} \, f = \textrm{supp} \, F = \Gamma^+(0) \cup \Gamma^-(0)$. 
The same holds true if we replace
$\Gamma^+(0), \Gamma^-(0)$ with the forward and backward light cones (boundaries of the convex closed forward and backward cones centered at zero). Of course the same will hold for cones centered at any other point,
provided we replace $f,F$ with the corresponding translations. 

This is of considerable importance and allows to extend the splitting method of Epstein-Glaser,
\cite{Epstein-Glaser}, \S 5,
for causal distributions over to the realm of homogeneous causal distributions (and their translations)
on $\mathcal{S}^{00}(\mathbb{R}^4)$ and more generally on $\mathcal{S}^{00}(\mathbb{R}^4)^{n\otimes}$.
In order to split a homogeneous distribution, with the support say
$\Gamma^+(0) \cup \Gamma^-(0)$ into the difference of homogeneous distributions each supported respectively on
$\Gamma^+(0)$, $\Gamma^-(0)$, we extend the initial distribution with the preservation 
of homogeneity and the support $\Gamma^+(0) \cup \Gamma^-(0)$ over to $\mathcal{S}(\mathbb{R}^4)$
and apply the splitting of Epstein-Glaser, \cite{Epstein-Glaser}, \S 5, to the extended distribution.
This method can, obviously, be extended over the spaces of tensor product
distributions $F\otimes G \otimes \ldots \otimes L$ in $\mathbb{E}^{n\otimes}$,
for homogeneous $F,G, \ldots, L \in \mathbb{E}^* = \mathcal{S}^{00}(\mathbb{R}^4)^*$, 
by extending each factor $F,G, \ldots, L$ over to a distribution over $\mathcal{S}(\mathbb{R}^4)$,
with preservation of the support and homogeneity degree. This makes sense if the supports 
in $\big(\mathbb{R}^{4}\big)^{n\times} = \mathbb{R}^{4n}$ are the Cartesian products
of cones $\Gamma^{\pm}(0)$ with fixed vertex $0$. The same holds of course for any other 
fixed vertex with the distributions accordingly translated. Still, one can
extend this method over to tensor products of distributions in which the distributions
$F,G, \ldots L$ are either over $\mathbb{E}_1$ or $\mathbb{E}_2$, where
$\mathbb{E}_1 = \mathcal{S}^{00}(\mathbb{R}^4)$ or $\mathbb{E}_2 = \mathcal{S}(\mathbb{R}^4)$,
provided that each factor distribution in $F\otimes G \otimes \ldots \otimes L$
is homogeneous and regular in the sense defined above whenever it is a factor in 
$\mathbb{E}_{1}^{*}$ acting on $\mathbb{E}_1$. This is in fact sufficient for the splitting 
of causal distributions in the causal perturbative series, in case the theory 
contains (at the free level) massive as well as zero mass fields, such as the electromagnetic
potential field $A$.

The most important example which can be extended and split in this way is the zero mass
Pauli-Jordan function $D_0$. It is nonetheless crucial to understand that as a commutator
function  of the zero mass field, $D_0$ or respectively $g_{\mu \nu}D_0$, is a distribution over 
$\mathcal{S}^{00}(\mathbb{R}^4)$
and not over $\mathcal{S}(\mathbb{R}^4)$. Similarly, its Fourier transform
$\widetilde{D_0}$ (resp. $g_{\mu\nu}\widetilde{D_0}$) is a distribution over 
$\mathcal{S}^{0}(\mathbb{R}^4)$, and not
over $\mathcal{S}(\mathbb{R}^4)$. This follows from the principles of QFT
relating construction of free fields to the representation theory of the double covering
of the Poincar\'e group, and the white noise construction of free fields\footnote{As we have already empasized 
several times before, costruction of free fields accoring to Wightman allows the Schwartz space
$\mathcal{S}(\mathbb{R}^4)$ of scalar-, vector-, e.t.c. -valued functions -- depending on the field -- as the 
test space on the space-time irrespectively if the field is massive or massless. But Wightman's field
is not useful in the perturbative causal approach for physical theories like QED.}. We have
explained this in details for the free electromagnetic field $A_\mu$
in previous Subsections and in the next Subsection. This fact is also 
transparently expressed by the continuity of the restriction map 
$\widetilde{\varphi} \rightarrow \widetilde{\varphi}\big|_{{}_{\mathscr{O}}}$,
where $\mathscr{O}$ is the positive energy sheet of the light cone -- the orbit defining 
the representation pertinent to zero mass field, i.e. electromagnetic potential field, 
as explained in previous Subsections. The restriction is naturally a map $\mathcal{S}^{0}(\mathbb{R}^4)
\rightarrow \mathcal{S}^{0}(\mathbb{R}^3)$, as explained in the previous Subsection. This continuity 
gives the natural and immediate linkage between the elements of single particle Hilbert space $\mathcal{H}'$
of the field and the distributional solutions $F \in \mathcal{S}^{00}(\mathbb{R}^4)^*$
of d'Alembert equation, as it should be for correctly defined free zero mass quantum field.
Indeed, for  any $S \in \mathcal{S}^{0}(\mathbb{R}^3)^* = E^* \supset \mathcal{H}'$, which for regular $S$
is identifiable with ordinary function on the orbit $\mathscr{O}$, there corresponds naturally 
and uniquely the functional
$\widetilde{F} \in \mathcal{S}^{0}(\mathbb{R}^4)$ given by the formula
\begin{equation}\label{QFTstate=d'AlembertSol}
\widetilde{F}(\widetilde{\varphi}) = S(\widetilde{\varphi}\big|_{{}_{\mathscr{O}}}).
\end{equation} 
$\widetilde{F}$ is well defined because of the continuity of the restriction map 
$\widetilde{\varphi} \rightarrow \widetilde{\varphi}\big|_{{}_{\mathscr{O}}}$. The closure 
of the smooth regular elements $S \in E^*$ with respect to the one particle Hilbert 
space\footnote{Here of 
functions on the orbit $\mathscr{O}$
pertinent to the field.} inner product
 does not lead us out of the space $E^*$, and in this case each element of the single particle Hilbert space
is in a natural manner a solution $F \in \mathcal{S}^{00}(\mathbb{R}^4)^*$ of
d'Alembert equation. Namely $F$ is precisely the inverse Fourier transform of the functional
$\widetilde{F}$ defined by the state $S \in \mathcal{S}^{0}(\mathbb{R}^3)^* = E^*$ 
as in the formula (\ref{QFTstate=d'AlembertSol}). $F$ is a solution of d'Alembert equation because by construction 
$\widetilde{F}$ is supported on the light cone in momentum space. 
The fact that the completion with respect to the inner
product of the single particle subspace does not lead us out of the space $E^* = \mathcal{S}_{A^{(3)}}(\mathbb{R}^3)^*$ 
follows easily by the the very construction of 
$\mathcal{S}_{A^{(3)}}(\mathbb{R}^3) =
\mathcal{S}^{0}(\mathbb{R}^3)$, or alternatively by comparison of the inner product with one (of the various 
equivalent computed above) systems of norms definig the nuclear space $\mathcal{S}_{A^{(3)}}(\mathbb{R}^3)
= \mathcal{S}^{0}(\mathbb{R}^3)$.  It is remarkable that the same continuity 
$\widetilde{\varphi} \rightarrow \widetilde{\varphi}\big|_{{}_{\mathscr{O}}}$ requirement forces  
the fundamental nuclear space $E$ in the one particle Hilbert space of the field $A$ to be 
$E =\mathcal{S}^{0}(\mathbb{R}^3)$, as only in this case 
the white noise construction of the field $A$ possible.  As explained
in previous Subsections this would be impossible with $E = \mathcal{S}(\mathbb{R}^3)$.
Moreover, any linkage between the single particle Hilbert space and distributional
solutions of d'Alembert equation would be impossible if $E$ would be replaced by 
$\mathcal{S}(\mathbb{R}^3)$ -- the restrictions to the orbit $\mathscr{O}$ of Fourier transforms 
of space-time test functions in the space 
equal $\mathcal{S}(\mathbb{R}^4)$ is not continuous as a map 
$\mathcal{S}(\mathbb{R}^4) \rightarrow \mathcal{S}(\mathbb{R}^3)$
(singularity at the vertex of th cone $\mathscr{O}$ intervenes here).

This means that space-time test function space for zero mass free fields is not equal
$\mathcal{S}(\mathbb{R}^4)$ but instead $\mathcal{S}^{00}(\mathbb{R}^4)$. Their inverse Fourier
transforms form the space $\mathcal{S}^{0}(\mathbb{R}^4)$, and their restrictions to the orbit
$\mathscr{O}$, the space $E = \mathcal{S}^{0}(\mathbb{R}^3) \subset \mathcal{H}'$. 
In particular thecomutator function of the electromagnetic
field is equal to the Pauli-Jordan zero mass distribution multiplied by the Minkowski metric components 
$g_{\mu\nu}D_0$ \emph{on the nuclear test space}
$\mathcal{S}^{00}(\mathbb{R}^4; \mathbb{C}^4)$. 
Similarly, the Fourier transform of this distribution, understood as a commutator function, 
is equal to the distribution $g_{\mu\nu}\widetilde{D_0} \in \mathcal{S}^{0}(\mathbb{R}^4; \mathbb{C}^4)$.

It is very remarkable, that despite the less flexibility in localization
(compare the assertion 1) of the last but one Proposition), the test space 
$\mathcal{S}^{00}(\mathbb{R}^4)$ of zero mass free fields is nonetheless
sufficient to subsume all relevant causality-type relations, as they are based 
on conic shape supports, (compare assertion 2) of the said Proposition). Moreover,
and from the point of view of causal perturbative method this perhaps most important, 
the propagators of zero mass fields
are homogeneous distributions, which can be uniquely extended over the Schwartz test space
(in space-time picture) $\mathcal{S}(\mathbb{R}^4)$, with the preservation of both: homogeneity
and cone-shaped causal support. In particular the splitting of Epstein-Glaser is still possible
for all scalar-type distributions occurring in the causal construction of the perturbative series.

Let us look more carefully at the most important case -- the Pauli-Jordan function
$D_0 \in \mathcal{S}^{00}(\mathbb{R}^4)^*$. It has natural extension 
$\boldsymbol{D}_0 \in \mathcal{S}(\mathbb{R}^4)^*$. Its meaning as a distribution, 
which makes it a well defined
element of $\mathcal{S}(\mathbb{R}^4)^*$, is exactly the same as stated above
when considering $D_0 \in \mathcal{S}^{00}(\mathbb{R}^4)^*$:
$\boldsymbol{D}_0 \in \mathcal{S}(\mathbb{R}^4)^*$
in action on a test function $\varphi \in \mathcal{S}(\mathbb{R}^4)$ is equal to the integration 
along the whole cone with the measure equal to the natural induced measure, of the restriction
of $\varphi$ to the whole light cone,
and taken with the positive sign on the forward and negative sign on the backward sheet of the cone: 
\[
\big(\boldsymbol{D}_0, \varphi \big) =
2\pi i
\int \limits_{\mathbb{R}^3} \ud^3 \boldsymbol{\x} \, \frac{1}{|\boldsymbol{\x}|} 
\varphi(\boldsymbol{\x},|\boldsymbol{\x}|) 
-
2\pi i
\int \limits_{\mathbb{R}^3} \ud^3 \boldsymbol{\x} \, \frac{1}{|\boldsymbol{\x}|} 
\varphi(\boldsymbol{\x},-|\boldsymbol{\x}|), \,\,\
\varphi \in \mathcal{S}(\mathbb{R}^4). 
\]
That this functional is continuous for $\varphi \in \mathcal{S}(\mathbb{R})$
easily follows by dividing domains the two integrals into two pieces: one consisting
of the ball $|\boldsymbol{\x}|\leq 1$ and the second piece $|\boldsymbol{\x}|> 1$.
Let us denote this extension $\boldsymbol{D}_0 \in \mathcal{S}(\mathbb{R}^4)^*$
of $D_0 \in \mathcal{S}^{0}(\mathbb{R}^4)^*$ by $\boldsymbol{D}_0$. This distinction is not
merely a matter of pedantism as the functionals $D_0$ and $\boldsymbol{D}_0$
are simply different, with $\boldsymbol{D}_0$ being an extension of $D_0$.
It is easy to to see that $\boldsymbol{D}_0 \in \mathcal{S}(\mathbb{R}^4)^*$
is a homogeneous of degree $\textrm{deg} \, \boldsymbol{D}_0 = \textrm{deg} \, D_0 = -2$
functional, and by construction $\textrm{supp} \, \boldsymbol{D}_0 
= \textrm{supp} \, D_0 = \{x \in \mathbb{R}^4, x \cdot x = 0 \}$. 

Exactly as for $D_0 \in \mathcal{S}^{00}(\mathbb{R}^4)^*$, we can construct the extension
$\boldsymbol{\hat{D}}_0 \in \mathcal{S}(\mathbb{R}^4)^*$ of the functional
$\widetilde{D_0} \in \mathcal{S}^{0}(\mathbb{R}^4)^*$, by putting it equal to integration
along the whole light cone in the momentum space of the restriction to this cone, taken with the negative sign
on the negative energy sheet of the cone:
\[
\big(\boldsymbol{\hat{D}}_0 , \widetilde{\varphi} \big) = 
\int \limits_{\mathbb{R}^3} \, \frac{1}{|\boldsymbol{\p}|} \,
\widetilde{\varphi}(\boldsymbol{\p}, |\boldsymbol{\p}|) \, 
\ud^3 \boldsymbol{\p}
-
\int \limits_{\mathbb{R}^3} \, \frac{1}{|\boldsymbol{\p}|} \, 
\widetilde{\varphi}(\boldsymbol{\p}, -|\boldsymbol{\p}|) \, 
\ud^3 \boldsymbol{\p}, \,\,\,
\widetilde{\varphi} \in \mathcal{S}(\mathbb{R}^4).
\] 
Its continuity follows exactly as for $\boldsymbol{D}_0 \in \mathcal{S}(\mathbb{R}^4)^*$.
Similarly it is readily seen that $\widetilde{D_0} \in \mathcal{S}^{0}(\mathbb{R}^4)^*$
extends $\widetilde{D_0} \in \mathcal{S}^{0}(\mathbb{R}^4)^*$ with preservation of homogeneity,
its degree and support:
\[
\begin{split}
\textrm{deg} \, \boldsymbol{\hat{D}}_0 = \textrm{deg} \, \widetilde{D_0} = -2, \\
\textrm{supp} \, \boldsymbol{\hat{D}}_0 
= \textrm{supp} \, \widetilde{D_0} = \{p \in \mathbb{R}^4, p \cdot p = 0 \}.
\end{split}
\]

Nonetheless, the Fourier transform $\widetilde{\boldsymbol{D}_0} \in \mathcal{S}(\mathbb{R}^4)^*$
need not be equal $\boldsymbol{\hat{D}}_0$. By our previous result,
stating that $\widetilde{D_0} \in \mathcal{S}^{0}(\mathbb{R}^4)$ is indeed equal
to the Fourier transform of $D_0 \in \mathcal{S}^{00}(\mathbb{R}^4)^*$, we only know, that
$\widetilde{\boldsymbol{D}_0} \in \mathcal{S}(\mathbb{R}^4)^*$ is equal to some extension 
of $\widetilde{D_0} \in \mathcal{S}^{0}(\mathbb{R}^4)^*$, to a homogeneous of degree $-2$
distribution, but this distribution does not have to be \emph{a priori} 
equal $\boldsymbol{\hat{D}}_0$. 
Similarly, we know that the the inverse Fourier transform $\mathscr{F}^{-1} \boldsymbol{\hat{D}}_0
\in \mathcal{S}(\mathbb{R}^4)^*$ is equal to a homogeneous of degree $-2$ 
extension $\boldsymbol{\hat{\hat{D}}}_0$ of $D_0 \in \mathcal{S}^{00}(\mathbb{R}^4)^*$, 
but is far not obvious if $\boldsymbol{\hat{\hat{D}}}_0$ is equal 
to $\boldsymbol{D}_0 \in \mathcal{S}(\mathbb{R}^4)^*$. The proof that indeed $D_0, \widetilde{D_0}$
are mutually inverse images under Fourier transform, we have given in the previous Subsection, used strongly the fact that
these distributions are elements resp. of $\mathcal{S}^{00}(\mathbb{R}^4)^*$ and
$\mathcal{S}^{0}(\mathbb{R}^4)^*$, and
heavily rests on the properties pertinent to these nuclear spaces, not shared by the
Schwartz space $\mathcal{S}(\mathbb{R}^4)$.

However, from the last Proposition
it follows that $\boldsymbol{D}_0 \in \mathcal{S}(\mathbb{R}^4)^*$ indeed fulfills d'Alembert equation
as a continuous functional on $\mathcal{S}(\mathbb{R}^4)$. 
Indeed, from the last Proposition we can only infer
the following corollary (but sufficint to infer $\square \boldsymbol{D}_0 = 0$): 
\begin{cor*}
There exists multi-index sequence $c_q$ with $|q| = 2$, such that
\begin{multline*}
\widetilde{\boldsymbol{D}_0} = \boldsymbol{\hat{D}}_0 \,
+ \, \sum \limits_{|q| =2} \, c_q D^q \delta(p)
\\
\boldsymbol{D}_0 = \widetilde{\boldsymbol{\hat{D}}_0} \,
+ \, \sum \limits_{|q| =2} \, c_q x^q 
\,\,\, \textrm{on} \,\,\, \mathcal{S}(\mathbb{R}^4).
\end{multline*}
\end{cor*}

In fact the equality
$\widetilde{\boldsymbol{D}_0} = \boldsymbol{\hat{D}}_0$ in $\mathcal{S}(\mathbb{R}^4)$ is not needed for the theory, in particular for 
the splitting. It is nonetheless remarkable that it indeed holds true, and in particular it follows that 
$\square \boldsymbol{D}_0 = 0$. Namely, we have the following
\begin{prop*}
\begin{equation}
\widetilde{\boldsymbol{D}_0} = \boldsymbol{\hat{D}}_0 \,\,\, \textrm{on} \,\,\, \mathcal{S}(\mathbb{R}^4).
\end{equation}
\end{prop*}
\qedsymbol \, Consider the Fourier transform $\widetilde{D_m} \in \mathcal{S}(\mathbb{R}^4)^*$
of the massive Pauli-Jordan commutator distribution 
$D_m \in \mathcal{S}(\mathbb{R}^4)^*$. As we know, correspondingly to the massive two-sheet orbit
$\mathscr{O}_{m,0,0,0}$ pertinent to massive fields, it is equal to the integral along the 
the whole disjoint sum $\mathscr{O}_{m,0,0,0} \sqcup \mathscr{O}_{m,0,0,0}$ of the two-sheet mass $m$
hyperboloid of the restriction of the test function to $\mathscr{O}_{m,0,0,0} \sqcup \mathscr{O}_{m,0,0,0}$,
with respect to the induced invariant measure, taken with the negative sign on the negative energy sheet of the 
hyperboloid:
\begin{multline*}
\big(D_m , \widetilde{\varphi} \big) = 
\int \limits_{\mathbb{R}^3} \, \frac{1}{\sqrt{\boldsymbol{\p}^2 + m^2}} \,
\widetilde{\varphi}(\boldsymbol{\p}, \sqrt{\boldsymbol{\p}^2 + m^2}) \, 
\ud^3 \boldsymbol{\p} \\
-
\int \limits_{\mathbb{R}^3} \, \frac{1}{\sqrt{\boldsymbol{\p}^2 + m^2}} \, 
\widetilde{\varphi}(\boldsymbol{\p}, -\sqrt{\boldsymbol{\p}^2 + m^2}) \, 
\ud^3 \boldsymbol{\p}, \,\,\,
\widetilde{\varphi} \in \mathcal{S}(\mathbb{R}^4).
\end{multline*}
Now it easily to estimate the value $\big|\big(\boldsymbol{\hat{D}}_0 - \widetilde{D_m}, \widetilde{\varphi} \big)|$
for any fixed element $\widetilde{\varphi} \in \mathcal{S}(\mathbb{R}^4)$, by dividing the 
domain of integration into two pieces: the unit ball $|\boldsymbol{\p}| \leq 1$ and its complementary. By using the system of norms (\cite{Reed_Simon}, Appendix to V.3, here the Schwartz notation is used with
multiindeces $\alpha, \beta$)
\[
\big|\widetilde{\varphi}\big|_n = \sup \limits_{|\alpha|, |\beta| \leq n}
\big|\big| p^\alpha D^\beta \widetilde{\varphi} \big|\big|_{{}_{L^2(\mathbb{R^4})}}
\]
on $\mathcal{S}(\mathbb{R}^4)$ we can in this way easily obtain the estimation
\[
\big|\big(\boldsymbol{\hat{D}}_0 - \widetilde{D_m}, \widetilde{\varphi} \big)| \leq m \, C \, 
\big|\widetilde{\varphi}\big|_2.
\]
This means that there exists a limit $\widetilde{D_m} \rightarrow \boldsymbol{\hat{\hat{D}}}$
in $\mathcal{S}(\mathbb{R}^4)^*$, when $m \rightarrow 0$, and moreover that 
this limit distribution $\boldsymbol{\hat{\hat{D}}}$ must be equal $\boldsymbol{{\hat{D}}}_0$,
compare e.g. \cite{GelfandII}, Chap. II.3. 

On the other hand the inverse Fourier transform $D_m$ of $\widetilde{D_m}$ can be explicitly computed.
This point is the most tricky point of the proof. Here one can proceed in at least two different ways. First
way consists in giving the proof that (we have omitted the factors $2\pi$ ) 
\[
D_m(x) = \boldsymbol{D}_0(x) 
- \Theta(x\cdot x) \frac{m}{2\sqrt{x\cdot x}} \, J_1 \big(m \sqrt{x\cdot x}\big),
\] 
or explicitly
\begin{multline*}
\big(D_m, \varphi \big) =
\int \limits_{\mathbb{R}^3} \ud^3 \boldsymbol{\x} \, \frac{1}{|\boldsymbol{\x}|} 
\varphi(\boldsymbol{\x},|\boldsymbol{\x}|) 
-
\int \limits_{\mathbb{R}^3} \ud^3 \boldsymbol{\x} \, \frac{1}{|\boldsymbol{\x}|} 
\varphi(\boldsymbol{\x},-|\boldsymbol{\x}|) \\
- m \, \int \limits_{\mathbb{R}^4}  \frac{\Theta(x\cdot x)}{2\sqrt{x\cdot x}} \, J_1 \big(m \sqrt{x\cdot x}\big)
\,\varphi(x) \, \ud^4 x, 
\,\,\
\varphi \in \mathcal{S}(\mathbb{R}^4). 
\end{multline*}
Here $J_1$ is the Bessel function of first order. One can prove that for such $D_m$, the distributions
$\widetilde{D_m}, D_m$ are indeed mutual images under Fourier transform, carrying out the integration
with respect to $p_0$ and then in $\boldsymbol{\p}$ as Bogolibov and Shirkov in 
\cite{Bogoliubov_Shirkov}, Chap. 3.15.3. This proof is in fact fully analogous to that given 
in the previous Subsection
for the equality $\widetilde{D_0} = D_0$, only the integration trick of Dirac, \cite{Dirac3rdEd}, Chap XII,
pp. 276-277, is replaced with the integration trick of Bogoliubov-Shirkov. Therefore,
we dispense with presentation of further details as now they should be clear. 

Alternative way consists in explicit computation of the Fourier inverse transform of $\widetilde{D_m}$
in the same way 
as Gelfand and Shilov computed the Fourier transform of $\delta(m^2- p\cdot p)$
in \cite{GelfandI}, Chap. III.2.10. In their computation 
we have to change $\delta(c^2-p\cdot p) = \delta(c^2 +P)$ into $\textrm{sign} \, p_0 \,\, \delta(m^2 -p\cdot p)$ (note that this is well defined distribution to which the Gelfand-Shilov method works pretty well because the two sheets of the massive hyperboloid are regularly separated). 

Having given the Pauli-Jordan distribution $D_m$ computed explicitly we note that, 
similarly as for $\widetilde{D_m} \rightarrow \boldsymbol{\hat{D}}_0$, we can easily
show that $D_m \rightarrow \boldsymbol{D}_0$ whenever $m \rightarrow 0$. 
Indeed $J_1$ stays bounded over all real line,
so that the needed norm estimation easily follows. 

Now because  $D_m \rightarrow \boldsymbol{D}_0$ and  $\widetilde{D_m} \rightarrow \boldsymbol{\hat{D}}_0$
whenever $m \rightarrow 0$, then by isomorphism property of the Fourier transform,
assertion of our Proposition is proved. 
\qed

Now we are ready to resolve the splitting problem for the most important distributions:
$\widetilde{D_0}$ and $D_0$.
Note that the corresponding unique extensions $\boldsymbol{\hat{D}}_0$
and $\boldsymbol{D}_0$ in $\mathcal{S}(\mathbb{R}^n)$, as homogeneous of degree $-2$
coincide with their quasi-asymptotic distributions, and have singularity
degree equal $-2$, necessary coinciding in the case of homogeneous distributions with their 
homogeneity degree. For definition of the quasi-asymptotic distribution and the singularity degree, 
compare \cite{Epstein-Glaser}, \S 5, or \cite{Scharf}, \S 3.2.
Thereofore, by the general splitting construction both,  $\boldsymbol{\hat{D}}_0$
and $\boldsymbol{D}_0$,  
can be uniquelly splitted, thus generating
unique splitting of $\widetilde{D_0}$ ,$D_0$. Namely $\widetilde{D_0}$ can be uniquely splitted 
into positive and negative frequency parts, i.e. $\widetilde{D_0}$ can be uniquely written 
as a sum of two homogeneous distributions first supported on the positive, 
the second on the negative energy sheet of the cone. The Pauli-Jordan distribution $D_0$
can be uniquely written as a difference of the so-called advanced and retarded parts, supported respectively
on the forward and respectively backward sheet of the light cone. 
In fact these are given explicitly in the very definition of the two distributions,
as respectively the two integrals along the positive (or forward) and negative (or backward)
sheet of the light cone.

{\bf REMARK 1}.
In particular $\mathcal{S}^{00}(\mathbb{R}^4)$
contains no functions of compact support, so that the localization within this space is much weaker than
within the ordinary Schwartz space $\mathcal{S}(\mathbb{R}^4)$. 
But although we do not have functions of compact support in $\mathcal{S}^{00}(\mathbb{R}^4)$, 
we have sufficiently many elements
in $\mathcal{S}^{00}(\mathbb{R}^4)$ to distinguish any (arbitrary small) conical shapes. In
particular to any open angular $\Omega$ set of directions on $\mathbb{S}^3 \subset \mathbb{R}^4$
centered at any point in $\mathbb{R}^4$,
we can find an element in $\mathcal{S}^{00}(\mathbb{R}^4)$ with the support lying totally within the
the cone $C_\Omega$ determined by this angular open set $\Omega$ of directions determined by the 
points of $\mathbb{S}^3$. 
In particular the space $\mathcal{S}^{00}(\mathbb{R}^4)$ is sufficient 
to distinguish causally independent regions in space-time from those which are causally related. 
Similarly, this space is sufficient to check if a homogeneous distribution in $\mathcal{S}^{00}(\mathbb{R}^4)^*$ 
fulfills d'Alembert equation, say outside the light cone, or to check if two (homogeneous) elements 
of $\mathcal{S}^{00}(\mathbb{R}^4)^*$ coincide on the conic-shape domain.

\subsection{White noise construction of the free electromagnetic potential field}\label{WhiteNoiseA}

Having given the Gelfand triples $E \subset \oplus L^2(\mathbb{R}^3) \subset E^*$ and 
$\mathbb{E} \subset \oplus L^2(\mathbb{R}^3) \subset \mathbb{E}^*$ constructed by means of the corresponding
positive definite self-adjoint operators $A$ and $\mathscr{F}^{-1} A \mathscr{F}$
interconnected as in the diagram (\ref{2-Gelfand-triples}) we obtain the Gelfand triple
$E \subset \oplus L^2(\mathbb{R}^3) \subset E^*$ of function spaces on $\mathbb{R}^3$, the Gelfand triple
of function spaces on the orbit $\mathscr{O}_{(1,0,0,1)}$ (where the functions on the orbit $\mathscr{O}_{(1,0,0,1)}$
are naturally regarded as functions of the space-like momentum components $\vec{p}$):
$E \subset \mathcal{H}' \subset E^*$
and the Gelfand triple $\mathcal{E} \subset \mathcal{H''} \subset \mathcal{E}^*$ in the position picture 
interconnected in the following way
\begin{equation}\label{3-Gelfand-triples}
\left. \begin{array}{ccccc}              E         & \subset & \oplus L^2(\mathbb{R}^3) & \subset & E^*        \\
                               \downarrow \uparrow &         & \downarrow \uparrow      &         & \downarrow \uparrow  \\
                                         E         & \subset &  \mathcal{H'} & \subset & E^*        \\
                               \downarrow \uparrow &         & \downarrow \uparrow      &         & \downarrow \uparrow  \\
                        \mathcal{E} & \subset & \mathcal{H''} & \subset & \mathcal{E}^* \end{array}\right.,
\end{equation}  
where the vertical arrows represent unitary maps which are continuous between the indicated spaces 
and equal to the operator $\sqrt{B}^{\, -1}$ and its inverse (in the first row
of maps) and by the Fourier transform $\mathcal{F}^{-1}$ defined by (\ref{F(varphi)}) and its inverse (in the second
row of maps), compare Subsection \ref{psiBerezin-Hida}. 

We apply to the diagram (\ref{3-Gelfand-triples})
the functor of second quantization $\Gamma$ exactly as in \cite{hida} (compare also \cite{HKPS} or \cite{luo})
putting for the operator $A$ in \cite{hida} our operator $A$, (compare also
\cite{HKPS}, where the oscillator Hamiltonian operator is used instead of our
$A$, or resp. instead of our  $\sqrt{B}^{\, -1} \, A \, \sqrt{B}$ or 
$\mathcal{F}^{-1} A \mathcal{F}$). Because our operator $A$ (and similarly the operators:
$\sqrt{B}^{\, -1} \, A \, \sqrt{B}$, $\mathscr{F}^{-1} A \mathscr{F}$
and in consequence the operator\footnote{Recall that the transform $\mathcal{F}^{-1}: \widetilde{\varphi} \mapsto 
\mathcal{F}^{-1} \widetilde{\varphi} =\varphi $
is defined by (\ref{F(varphi)}). Note that the operator $\mathcal{F}$ may be defined as the ordinary 
four-dimensional Fourier transform $\mathscr{F}: \varphi \mapsto \mathscr{F} \varphi = \widetilde{\varphi}$, 
with the sign at $ip^0 x_0$ in the exponent changed, followed by the restriction to
the orbit $\mathscr{O}_{(1,0,0,1)}$. In fact the ordinary four dimensional Fourier transform of the elements 
$\varphi$ of $\mathcal{H''}$ are concentrated on the orbit $\mathscr{O}_{(1,0,0,1)}$, and thus are distributions fulfilling d'Alembert equation.}
 $\mathcal{F}^{-1} A \mathcal{F}$) fulfils the conditions
(A1)-(A3) of \S 1 of \cite{hida} (after eventually the trivial modification by adding the unit operator) 
as well as our space $E$ (and similarly the spaces  $\mathbb{E}$ and $\mathcal{E}$) preserves the conditions (H1)-(H3)
of Kubo and Takenaka of \S1 of \cite{hida} and \cite{obata}, we can apply the results of the white noise
calculus of Hida, Obata and Sait\^o \cite{hida}, \cite{obata}, in constructing our quantum electromagnetic 
four-potential field as a generalized operator 
(operator valued distribution) and their theory of operators which can be represented as integrals
of generalized creation and annihilation operators $a^+(\vec{p})$ and $a(\vec{p})$, compare Subsection
\ref{psiBerezin-Hida}. 

Because the operator $A = \oplus A^{(3)}$ leaves invariant each Hilbert subspace 
$L^2(\mathbb{R}^3) \subset \oplus L^2(\mathbb{R}^3)$ spanned by each of the four components of the functions 
$\widetilde{\varphi} \in \oplus L^2(\mathbb{R}^3)$ then by the mentioned property of Gelfand triples
it follows that $E \subset \oplus L^2(\mathbb{R}^3) \subset E^*$ is isomorphic to the direct sum 
$\oplus_{\nu = 0}^{3} \big( E^\nu \subset  L^2(\mathbb{R}^3) \subset E^{\nu \, *} \big)$ of Gelfand triples
$E^\nu \subset  L^2(\mathbb{R}^3) \subset E^{\nu \, *}$ (and similarly for
$E \subset  \mathcal{H}' \subset E^*$,
$\mathbb{E} \subset \oplus L^2(\mathbb{R}^3) \subset \mathbb{E}^*$
and $\mathcal{E} \subset \mathcal{H''} \subset \mathcal{E}^*$) so that we can basically
apply the construction of \cite{hida} for scalar operator $A^{(3)}$ instead of the operator
$A$, separately to each component. 

Let us recapitulate shortly the white noise method after \cite{hida} and \cite{obata}. 
Let
\[
\Gamma(A) = \bigoplus \limits_{n=0}^{\infty} A^{\otimes n}
\]
be the second quantized operator acting in $\Gamma\Big( \oplus L^2(\mathbb{R}^3) \Big)$
with the inner product space norm denoted by $\| \cdot \|_0$. It follows that the
operator $\Gamma(A)$ is standard in the sense of \cite{obata}, i.e. it fulfills the conditions
(A1)-(A3) of of \S 2 of \cite{obata} or of \cite{hida}, \S 1, for the proof compare e.g. \cite{obata}
and references cited there.  

For $k \in \mathbb{N}$ let $(E_k)$ be the closure of the domain $\Dom \, \big( \Gamma(A^k)\big)$
of the operator $\Gamma(A^k)$ with respect to the norm $\| \Gamma(A^k) \cdot \|_{0}$,
and let $(E_{-k})$ be the dual space of $(E_k)$ with the dual norm $\| \cdot \|_{-k}$.
The projective limit $(E) = \cap_k (E_k)$ is a nuclear Frechet reflexive space and its topological dual is equal to the
inductive limit $(E)^* = \cup_k (E_{-k})$. In this way we obtain another Gelfand triple (lifting of the Gelfand triple
$E \subset \oplus L^2(\mathbb{R}^3) \subset E^*$ to the Fock space)
\[
(E) \subset \Gamma\Big( \oplus L^2(\mathbb{R}^3; \mathbb{C}) \Big) \subset (E^*)
\]
with the canonical bilinear form $\langle \langle \cdot , \cdot \rangle \rangle : (E)^* \times (E) \to \mathbb{C}$,
i.e. dual pairing between $(E)^*$ and $(E)$. Similarly we obtain the lifting of the Gelfand triple
$\mathbb{E} \subset \oplus L^2(\mathbb{R}) \subset \mathbb{E}^*$ (compare \cite{obata}, \S2).  

It turns out that the Fock space 
\[
\Gamma\Big( \oplus L^2(\mathbb{R}^3; \mathbb{C}) \Big) = \Gamma\Big( L^2(\mathbb{R}^3; \mathbb{C}^4) \Big)
=  \bigoplus \limits_{n=0}^{\infty} L^2(\mathbb{R}^3; \mathbb{C}^4)^{\widehat{\otimes} n}
\]
(we have used $\widehat{\otimes}$ for the symmetrized tensor product) may be 
naturally realized as the function space
of square integrable (equivalence classes of) functions on a measure space, and moreover the measure space
has natural linear structure being a dual space to a nuclear space allowing to build an effective calculus on it
(including integration, Fr\'echet differentiantion, Taylor formula, e. t. c. for elements of $(E)$).

In the  construction of the realization the real Gelfand triple 
$E_\mathbb{R} \subset \oplus L^2(\mathbb{R}^3; \mathbb{R}) \subset {E_\mathbb{R}}^*$ (or equivalently
$\mathbb{E}_\mathbb{R} \subset \oplus L^2(\mathbb{R}; \mathbb{R}) \subset {\mathbb{E}_\mathbb{R}}^*$)
and Gaussian measures are used. By Minlos theorem, \cite{GelfandIV}, Ch. IV.2.3, Theorem 3 and Prokhorov-Sazanov 
theorem, \cite{GelfandIV}, Ch. IV.4.2,
Theorem 1 (i.e. Bochner's theorem for nuclear spaces), there exists a unique Gaussian measure $\mu$ on $E_{\mathbb{R}}^*$ associated to the Gelfand triple $E_\mathbb{R} \subset \oplus L^2(\mathbb{R}^3; \mathbb{R}) \subset E_{\mathbb{R}}^*$ 
(and similarly for 
$\mathbb{E}_\mathbb{R} \subset \oplus L^2(\mathbb{R}^3; \mathbb{R}) \subset \mathbb{E}_{\mathbb{R}}^*$) such that 
\[
\int \limits_{E_{\mathbb{R}}^{*}} e^{i\langle \bold{\zeta}, \widetilde{\varphi} \rangle} \, \ud \mu(\bold{\zeta}) 
= e^{- \frac{1}{2} |\widetilde{\varphi} |_{0}^{2}},
\,\,\, \widetilde{\varphi} \in E_\mathbb{R},
\] 
where $\langle \cdot , \cdot \rangle$ is the dual pairing between $E_{\mathbb{R}}^*$ and $E_\mathbb{R}$, 
and where $| \cdot |_0$ is the Hilbert space norm in $\oplus L^2(\mathbb{R}^3; \mathbb{R})
= L^2(\mathbb{R}^3; \mathbb{R}^4)$.  

After \cite{hida}, \cite{obata-book} we likewise denote by $| \cdot |_0$ the Hilbert space norm
on the Hilbert space tensor product  
\[
L^2(\mathbb{R}^3; \mathbb{R}^4)^{\otimes n}
\]
and its restriction to the symmetrized tensor product subspace
\[
L^2(\mathbb{R}^3; \mathbb{R}^4)^{\widehat{\otimes} n}.
\]

The measure space $(E_{\mathbb{R}}^*, \mu)$ is the fundamental probability space in the white noise calculus and is called

\emph{white noise space}.

Let us define after \cite{hida} and \cite{obata} the Hilbert space $L^2 (E_{\mathbb{R}}^* , \mu; \mathbb{R}) = 
\oplus_{\nu=0}^{3}  L^2 (E_{\mathbb{R}}^{\nu \, *} , \mu, \mathbb{R})$  
of square summable functions on $E_{\mathbb{R}}^*$, and denote the Hilbert space $L^2$ norm of 
$L^2 (E_{\mathbb{R}}^* , \mu; \mathbb{R})$ by 
$\| \cdot \|_{0}$,  and consider the real Bose-Fock space
\[
\Gamma\Big( \oplus L^2(\mathbb{R}^3; \mathbb{R}) \Big) 
= \bigoplus \limits_{n=0}^{\infty} \big[ L^2(\mathbb{R}^3; \mathbb{R}^4) \big]_{S}^{\otimes n}
= \bigoplus \limits_{n=0}^{\infty} L^2(\mathbb{R}^3; \mathbb{R}^4)^{\widehat{\otimes} n}
\]
over $\oplus L^2(\mathbb{R}^3; \mathbb{R})$. By the well known Wiener-It\^o-Segal chaos decomposition of $L^2 (E_{\mathbb{R}}^* , \mu; \mathbb{R})$
(\cite{obata}, Proposition 2.1) the Hilbert space $L^2 (E_{\mathbb{R}}^* , \mu; \mathbb{R})$ is naturally unitary equivalent (isometrically isomorphic) to
the real Fock space $\Gamma\Big( \oplus L^2(\mathbb{R}^3; \mathbb{R}) \Big)$, compare e.g. \cite{hida} or \cite{HKPS}.
Let us remind shortly the construction of chaos decomposition after \cite{obata}. For this purpose
for each $\zeta \in E_{\mathbb{R}}^*$ and $n \in \mathbb{N}$ we define 
$\boldsymbol{:} \zeta^{\otimes n} \boldsymbol{:} \in (E_{\mathbb{R}}^{\otimes n})^*$ inductively as follows:
\[
\left\{ \begin{array}{ll}
\boldsymbol{:} \zeta^{\otimes 0} \boldsymbol{:} = 1& \\
\boldsymbol{:} \zeta^{\otimes 1} \boldsymbol{:} = \zeta & \\
\boldsymbol{:} \zeta^{\otimes n} \boldsymbol{:} 
= \zeta \widehat{\otimes} \boldsymbol{:} \zeta^{\otimes (n-1)} \boldsymbol{:}
- (n-1) \tau \widehat{\otimes} \boldsymbol{:} \zeta^{\otimes (n-2)} \boldsymbol{:} & \, n \geq 2,
\end{array} \right.
\]
where $\tau \in (E_\mathbb{R} \otimes E_{\mathbb{R}})^*$ is defined by the formula
\[
\langle \tau, \xi \otimes \vartheta \rangle = \langle \xi , \vartheta \rangle , \,\,\, \xi , \vartheta \in E_\mathbb{R}.
\]
A variant of Wiener-It\^o-Segal chaos decomposition of $L^2 (E_{\mathbb{R}}^* , \mu; \mathbb{R})$ 
may be formulated in the following manner.
\begin{wiener-ito*}
For each $\Phi \in L^2 (E_{\mathbb{R}}^* , \mu; \mathbb{R})$ there exists a sequence $f_n
\in L^2(\mathbb{R}^3; \mathbb{R}^4)^{\widehat{\otimes} n}$, $n= 0, 1, 2, \ldots$ such that 
\[
\Phi(\zeta) = \sum \limits_{n=0}^{\infty} \langle \, \boldsymbol{:} \zeta^{\otimes n} \boldsymbol{:} \,
, \,\, f_n \, \rangle , \,\,\,\, \zeta \in E_{\mathbb{R}}^*,
\] 
and  on the right-hand side there is an orthogonal direct sum of functions in 
 $L^2 (E_{\mathbb{R}}^* , \mu; \mathbb{R})$.  Moreover,
\[
\| \Phi \|_0^2 = \sum \limits_{n=0}^{\infty} n! | f_n |_0^2 .
\]
\end{wiener-ito*}

The second quantized operator
\[
\Gamma(A) = \bigoplus \limits_{n=0}^{\infty} A^{\otimes n}
\]
acting in $\Gamma\Big( \oplus L^2(\mathbb{R}^3; \mathbb{R}) \Big)$  
can be naturally lifted to an operator acting on 
$L^2 (E_{\mathbb{R}}^* , \mu; \mathbb{R}) \cong \Gamma\Big( L^2(\mathbb{R}^3; \mathbb{R}^4) \Big)$.
Application of the operator $\Gamma(A)$ (which as we already know respects the conditions (A1)-(3A) 
of \cite{hida}, \S 1 or \cite{obata}, \S 1, allowing the construction) gives the standard construction 
of the real Gelfand triple 
\[
\mathcal{S}_{\Gamma(A)}(E_{\mathbb{R}}^*; \mathbb{R}) = (E_\mathbb{R}) \subset
L^2 (E_{\mathbb{R}}^* , \mu; \mathbb{R}) \subset \mathcal{S}_{\Gamma(A)}(E_{\mathbb{R}}^*; \mathbb{R})
= (E_\mathbb{R})^*.
\]
Its complexification is equal
\[
(E) \subset (L^2) \subset (E)^*
\]
with 
\[
(L^2) =  L^2 (E_{\mathbb{R}}^* , \mu; \mathbb{C}) 
\]
naturally isomorphic (via the chaos decomposition) to the Fock space 
$\Gamma\Big( L^2(\mathbb{R}^3 ; \mathbb{C}^4) \Big)$. In particular for each $\Phi \in (L^2)$
there exists a sequence of $f_n \in L^2(\mathbb{R}^3; \mathbb{C}^4)^{\widehat{\otimes} n}$
such that 
\begin{equation}\label{Phi=white-expansion}
\Phi(\zeta) = \sum \limits_{n=0}^{\infty} \langle \, \boldsymbol{:} \zeta^{\otimes n} \boldsymbol{:} \,
, \,\, f_n \, \rangle , \,\,\,\, \zeta \in E_{\mathbb{R}}^*;
\end{equation}
and $\Phi \in (E)$ if and only if $f_n \in E^{\widehat{\otimes} n}$ for all $n = 0, 1, 2, \ldots$,
and 
\[
\sum \limits_{n=0}^{\infty} n! | f_n |_k^2 < \infty, \,\,\, \textrm{for all} \, k \geq 0.
\]
In that case 
\[
\| \Phi \|_k^2 = \sum \limits_{n=0}^{\infty} n! | f_n |_k^2 < \infty, \,\,\,
\textrm{for all} \, k \geq 0.
\]

Concerning the last two formulas recall that by definition (\cite{hida}, \cite{obata-book}) 
\[
\| \Phi \|_{k} = \| \Gamma(A)^k \Phi \|_0, \,\,\, \textrm{and} \,\,\,
|f_n|_k = |(A^{\otimes n})^k f_n|_0.
\]

Moreover, for each $\zeta \in E_{\mathbb{R}}^*$ the right-hand side of (\ref{Phi=white-expansion})
converges absolutely and defines a continuous function which coincides with $\Phi$
$\mu$-a.e.. In particular $\mathcal{S}_{\Gamma(A)}(E_{\mathbb{R}}; \mathbb{C}) = (E)$
respects the Kubo-Takenaka conditions (H1)-(H3) of \cite{hida},\S 1 or of 
\cite{obata}, \S 1.

As proven in \cite{hida} and \cite{obata} the \emph{exponential vectors} are useful
in computations. Namely, for each $\xi \in E$ we define $\Phi_\xi \in (E)$ after Hida, Obata and Sait\^o
\[
\Phi_\xi (\zeta) = \sum \limits_{n=0}^{\infty} \frac{1}{n!}\langle \, \boldsymbol{:} \zeta^{\otimes n} \boldsymbol{:} \,
, \,\, \xi^{\otimes n} \, \rangle.
\] 
Among other reasons the usefulness of the set $\{\Phi_\xi , \xi \in E \}$ of exponential vectors 
comes from the fact that they span a dense subspace of $(E)$. In particular for $\xi, \zeta \in E$,
$y \in E^*$ we have (for the Hida derivation operator $D_y$ -- extension of the annihilation operator --
defined in (\ref{D-xi-white-dec}))
\[
\langle \langle \Phi_\xi , \Phi_\zeta  \rangle \rangle = e^{\langle \xi , \zeta \rangle},
\,\,\, D_y \Phi_\xi = \langle y, \xi \rangle \Phi_\xi.
\]

Similarly, using the operator $\mathscr{F}^{-1} A \mathscr{F}$ instead of $A$ we construct the
Gelfand triple $\mathbb{E} \subset L^2 (\mathbb{R}^3, \mathbb{C}^4) \subset (\mathbb{E})^*$ 
and its lifting (this time with the respective Gaussian measure $\mu$ on $\mathbb{E}_{\mathbb{R}}^*$)
\[
(\mathbb{E}) \subset L^2(\mathbb{E}_{\mathbb{R}}^*, \mu ; \mathbb{C}) \subset (\mathbb{E})^*,
\]
with the Hilbert space $L^2(\mathbb{E}_{\mathbb{R}}^*, \mu ; \mathbb{C})$ naturally isomorphic (via the chaos decomposition)
with the Fock space $\Gamma\Big( L^2(\mathbb{R}^3 ; \mathbb{C}^4) \Big)$, but this time
in the position picture. The same construction with the operator $\sqrt{B}^{\, -1} \, A \, \sqrt{B}$ 
used instead of $A$ gives the triple $E \subset \mathcal{H}'' \subset E^*$ and its lifting to the Fock space
$(E) \subset \Gamma(\mathcal{H}') \subset (E)^*$. As we have already remarked the construction is even possible
with the operator $A$ replaced with $\mathcal{F}^{-1} A \mathcal{F}$ acting in the one particle Hilbert space
$\mathcal{H}''$, and leads to the triple $\mathcal{E} \subset \mathcal{H}'' \subset (\mathcal{E})^*$ 
and its lifting $(\mathcal{E}) \subset \Gamma(\mathcal{H}'') \subset (\mathcal{E})^*$,  
but the respective pairings $\langle \cdot, \cdot \rangle$, 
$\langle\langle \cdot, \cdot \rangle\rangle$ are substantially more complicated in this case.

\begin{rem*}

Note that by the Wiener-It\^o-Segal decomposition (\ref{Phi=white-expansion}) each element $\Phi$ of $(E)$ or of 
$L^2 (E_{\mathbb{R}}^* , \mu; \mathbb{R})$, regarded as a function on $E_{\mathbb{R}}^*$, 
defines in a unique natural manner a function on the complexification $E^*$
of $E_{\mathbb{R}}^*$. Indeed $E$ (and the same holds for $(E)$) is a nuclear involutive algebra 
with the involution given by complex conjugation, and the algebra $E$ is equal to the complexification of 
the real nuclear algebra $E_\mathbb{R}$ of all real elements of $E$. Every regular functional 
(function distribution) $\zeta \in E^*$ is canonically given by the integral and the same holds
for regular $\zeta' \in E_{\mathbb{R}}^*$. Every such functional $\zeta' \in E_{\mathbb{R}}^*$
defines naturally a real functional $\zeta \in E^*$ in the sense that $\zeta$ takes on real values
on real elements of $E$ (this is actually the meaning of the coupling on the right-hand side of the formula 
(\ref{Phi=white-expansion})). On the other hand every functional $\zeta \in E^*$ is canonically a sum
$\zeta = \zeta' + i \zeta''$, where $\zeta', \zeta'' \in E^*$ are real. Because every element of 
$E^*$ and $E_{\mathbb{R}}^*$ is a limit of regular functionals, then our assertion follows. 
From now on we will regard the elements $\Phi$
of $(E)$ or of $L^2(E_{\mathbb{R}}^*, \mu ; \mathbb{C})$, as functions on
 $E^*$ (resp. on $\mathbb{E}^*$), although the inner product in $L^2(E_{\mathbb{R}}^*, \mu ; \mathbb{C})$
(or in $L^2(\mathbb{E}_{\mathbb{R}}^*, \mu ; \mathbb{C})$)  is defined by the integral
of their restrictions to the real subspace $E_{\mathbb{R}}^*$ (resp. $\mathbb{E}_{\mathbb{R}}^*$)
with respect to the Gaussian measure. This is important regarding the action of 
the double cover of the Poincar\'e group in the momentum picture, 
which does not transform the real part $E_{\mathbb{R}}$, $E_{\mathbb{R}}^*$, $(E_{\mathbb{R}})$ 
and $(E_{\mathbb{R}})^*$ into itself. Althogh in the position picture
the representation transforms the real functions of $\mathcal{E}$ into real functions
this is not the case in passing to the dual space as the inner product in $\mathcal{E}$ is in general 
complex valued for real elements; for the same reason the argument based on complexification does not work for $\mathcal{E}$
and $(\mathcal{E})$, and application of Gaussian measures on duals to real nuclear spaces 
must proceed differently and must be based on the construction for $E$ and $(E)$ and the functoriality
of $\Gamma$ applied to the diagram (\ref{3-Gelfand-triples}). For example, we may use the construction of the 
measure induced by the Gaussian measure on the subspace $E_{\mathbb{R}}^* \subset E^*$ and by the respective map in
the diagram (\ref{3-Gelfand-triples}). In this respect it is convenient to regard the elements
of $(E)$ as functions on the whole space $E^*$. Although $\mathcal{E}$ is not an algebra we construct in this way
appropriate involution ${}^*$ on $\mathcal{E}$, given by the total reflection operation followed by complex conjugation:
$\varphi^*(x) = \overline{\varphi(-x)}$ with the hermitean elements: $\varphi^* = \varphi$ plying the role
of real elements.   
\end{rem*} 

Pointwise multiplication defines a (jointly) continuous map $(E)\times (E) \to (E)$,
compare \cite{kubo-takenaka}, which makes $(E)$ a nuclear algebra (the same holds of course 
for $(\mathbb{E})$ and $(\mathcal{E})$, and everything which will be said of $(E)$ also holds for $(\mathbb{E})$
and $(\mathcal{E})$ as the constructions of the spaces are based on the same principles, the only difference 
in the explicit formulas will come from different pairings $\langle \cdot, \cdot \rangle$, 
$\langle\langle \cdot, \cdot \rangle\rangle$ induced by different inner products of one particle Hilbert spaces
which are different in the constructions of $(E)$, $(\mathbb{E})$ and $(\mathcal{E})$).

Now let us back to the concrete triple $E \subset \oplus L^2(\mathbb{R}^3) \subset  E^* $ and its lifting
\[
(E) \,\,\,\, \subset \,\,\,\, \Gamma(\oplus L^2(\mathbb{R}^3)) \cong L^2(E_{\mathbb{R}}^*, \mu; \mathbb{C}) 
\,\,\,\, \subset \,\,\,\,  (E)^*. 
\]

The space $(E)$ is the \emph{Hida's testing functional space} and the space $(E)^*$
is known under the name of \emph{Hida's generalized functional space}.

Let $\delta_{\vec{p}}^{\nu} = (0 ,\ldots,  \delta_{\vec{p}}, 0, \ldots, 0 )$ be equal to the Dirac delta 
functional $\delta_{\vec{p}}^{\nu}$, equal to the zero functional on each of the four components 
$E^0 ,E^1, E^2, E^3$ of $E$ except 
the $\nu$-th component $E^\nu$ where it is equal to the ordinary scalar Dirac delta functional. 

Let us introduce after \cite{hida} and \cite{obata} the symmetrized contraction $\widehat{\otimes}_m$ of 
symmetrized tensor products
determined (through the polarization formula for symmetric tensors, \cite{obata-book}, Appendix A) as a unique extension of 
\[
\zeta^{\otimes (l+m)} \, \widehat{\otimes}_m \, \xi^{\otimes m}= \langle \zeta, \xi \rangle^m \zeta^{\otimes l},
\,\,\, \xi, \zeta \in E.
\]
In particular for $\boldsymbol{\p} \in \mathbb{R}^3$ (we use $\vec{p}$ and $\boldsymbol{\p}$
interchangeably) and $f \in E^{\widehat{\otimes} (n+1)}$
we have
\[
\delta_{\boldsymbol{\p}}^{\nu} \widehat{\otimes}_1 f \, \big( \underset{1}{\boldsymbol{\p}}, \ldots, 
\underset{n}{\boldsymbol{\p}} \, \big) = 
f^\nu \, \big( \boldsymbol{\p},  \underset{1}{\boldsymbol{\p}}, \ldots, 
\underset{n}{\boldsymbol{\p}} \, \big), \,\,\,
\underset{1}{\boldsymbol{\p}}, \ldots, 
\underset{n}{\boldsymbol{\p}} \in \mathbb{R}^3.
\]

Let $\Phi$ be any element of $(E)$. According to the Wiener-It\^o-Segal decomposition,
$\Phi$ is given by the corresponding function (\ref{Phi=white-expansion}). For
$\xi \in E^*$ we put 
\begin{equation}\label{D-xi-white-dec}
D_\xi \Phi(\zeta) = \sum \limits_{n=0}^{\infty} n \langle \, \boldsymbol{:} \zeta^{\otimes (n-1)} \boldsymbol{:} \,
, \,\, \xi \widehat{\otimes}_1 f_n \, \rangle , \,\,\,\, \zeta \in E^*.
\end{equation}  
It follows that $D_\xi \Phi \in (E)$ and $D_\xi$ is a continuous operator $(E) \to (E)$.

Note that if $\xi \in  L^2(\mathbb{R}^3; \mathbb{C}^4)$ then $D_\xi$ defined by the formula 
(\ref{D-xi-white-dec}) can be identified with the ordinary annihilation operator
$a(\overline{\xi})$ of the Fock space $\Gamma( L^2(\mathbb{R}^3; \mathbb{C}^4))$, but in the representation
which we described shortly in Subsection \ref{Gamma-applied-to-(H',J')}), Remark \ref{TwoRepOfaa^+InBoseFock},
and called there the representation of Hida, Obata, Sait\^o (although it is quite popular in mathematical literature).
When using the representation and norm of the Fock space more popular in physical literature
the formula (\ref{D-xi-white-dec}) would have to be appropriately modified. 
Because the modifications of the formulas are trivial, we prefer here
to use the same representation as Hida, Obata and Sait\^o \cite{hida}, \cite{obata-book}.

Therefore, the operator $D_{\xi}^*$ dual to $D_\xi$ transforms $(E)^* \to (E)^*$
continuously for the strong dual topology on $(E)^*$, and if (in the formal notation
of \cite{obata-book}, \S 3.1)
\[
\Phi(\zeta) = \sum \limits_{n=0}^{\infty} \big\langle \boldsymbol{:} \zeta^{\otimes n}
\boldsymbol{:}, F_n \big\rangle, \,\,\, 
F_n \in (E^{\widehat{\otimes} \,n})^*= (E^*)^{\widehat{\otimes} \, n}
\]
represents the Wiener-It\^o expansion of  $\Phi \in (E)^*$, then for $\xi \in E^*$
\[
D_{\xi}^*\Phi(\zeta) = \sum \limits_{n=0}^{\infty} \big\langle \boldsymbol{:} \zeta^{\otimes (n+1)}
\boldsymbol{:}, \, \xi \widehat{\otimes} \, F_n \big\rangle,
\]
for a proof compare \cite{obata-book}.

On the other hand every element of $(E)$ regarded as a function naturally extended all over the space
$E^*$ is Fr\'echet differentiable up to all orders, compare \cite{hida} of \cite{HKPS}. 
In particular for any $\xi \in E^*$ and any $\Phi \in (E)$ there exists G\^atoux derivative of $\Phi$
at $\zeta \in E^*$ in the direction of $\xi$ and is equal to $D_\xi \Phi(\zeta)$:
\[
D_{\xi} \Phi(\zeta) = \frac{d}{dt} \Phi(\zeta + t \xi)|_{t=0}
= \lim \limits_{t \to 0} \frac{1}{t} \Big[ \Phi(\zeta + t \xi) - \Phi(\zeta) \Big]. 
\]
It follows that for $\xi \in E^*$, $D_{\xi}$ is a continuous derivation on $(E)$, and 
$\{D_\xi , \xi \in E^* \}$ is a commuting family of operators. Moreover, for $\xi \in E$,
$D_\xi$ can be extended to a continuous linear operator $(E)^* \to (E)^*$. And dually:
for any $\xi \in E^*$, the adjoint operator $D_{\xi}^{*}$ is continuous from $(E)^*$
to $(E)^*$ and $\{D_{\xi}^{*} , \xi \in E^* \}$ is a commuting family. The operator
$D_{\xi}^{*}$ restricts to a continuous linear operator from $(E)$ to $(E)$, whenever
$\xi \in E$.
 
It is customary to write $\partial_{\vec{p}}^{\nu}$, $\partial_{{\vec{p}}}^{\nu *}$ for 
$D_{{}_{\delta_{{}_{\vec{p}}}^{\nu}}}$,
$D_{{}_{\delta_{{}_{\vec{p}}}^{\nu}}}^*$ respectively,
when $\zeta = \delta_{\vec{p}}^{\nu} = (0 ,\ldots,  \delta_{\vec{p}}, 0, \ldots, 0 )$ is equal to the Dirac delta functional $\delta_{\vec{p}}^{\nu}$. 

It follows that $\partial_{\boldsymbol{\p}}^{\nu}$ and $\partial_{{\boldsymbol{\p}}}^{\mu *}$
are well defined and continuous if regarded as operators $(E) \to (E)$ and $(E)^* \mapsto (E)^*$
respectively and in particular both are
continuous as operators $(E) \mapsto (E)^*$, but  $\partial_{{\boldsymbol{\p}}}^{\nu *}$
treated as operator on $(L^2)$ has just the zero vector as the only element of its domain, which motivates
introducing the generalized operators. Similarly,  
$\partial_{{\boldsymbol{\p}}}^{\mu *}\partial_{\boldsymbol{\p}'}^{\nu}$ is a well defined continuous operator
$(E) \to (E)^*$ but $\partial_{\boldsymbol{\p}'}^{\nu} \partial_{{\boldsymbol{\p}}}^{\mu *}$ is not well defined
as an operator $(E) \to (E)^*$ (or as an operator $(E)^* \mapsto (E)^*$), which is the mathematical counterpart 
for the necessity of normal Wick's ordering.

\begin{rem*}
Note that in order to reduce the construction of the Hida test space $(E)$ in the Fock space
\begin{multline*}
\Gamma\Big( \oplus_{0}^{3} L^2(\mathbb{R}^3; \mathbb{C}) \Big) 
= \Gamma\Big( L^2(\mathbb{R}^3; \mathbb{C}^4) \Big) 
= \Gamma\Big(L^2(\mathbb{R}^3; \mathbb{R}^4)_{\mathbb{C}} \Big)
= \Gamma\Big(L^2(\mathbb{R}^3; \mathbb{R}^4) \Big)_{\mathbb{C}},
\end{multline*}
together with the corresponding Hida
operators  $\partial_{\boldsymbol{\p}}^{\nu}$, $\partial_{{\boldsymbol{\p}}}^{\nu *}$ (i.e. 
$D_{{}_{\delta_{{}_{\boldsymbol{\p}}}^{\nu}}}, D_{{}_{\delta_{{}_{\boldsymbol{\p}}}^{\nu}}}^*$) 
to the standard general setup, as summarized e.g. in 
\cite{hida} or \cite{obata-book}, we regard the Hilbet space of (equivalence classes)
of $\mathbb{R}^4$- or $\mathbb{C}^4$-valued square summable functions 
\[
\oplus L^2(\mathbb{R}^3; \mathbb{R}) = L^2(\mathbb{R}^3; \mathbb{R}^4)
\,\,\, \textrm{or} \,\,\, 
\oplus L^2(\mathbb{R}^3; \mathbb{C}) = L^2(\mathbb{R}^3; \mathbb{C}^4)
\]
as the Hilbert space of (equivalence classes) of real or complex valued functions
on the disjoint sum 
\[
\mathscr{O} = \mathbb{R}^3 \sqcup \mathbb{R}^3 \sqcup \mathbb{R}^3 \sqcup \mathbb{R}^3
\]
of four copies of the space $\mathbb{R}^3$, with the direct sum measure
on $\mathscr{O}$ coinciding with the ordinary invariant 
(for the ordinary euclidean metric on $\mathbb{R}^3$) Lebesgue measure 
$\ud^3 \boldsymbol{\p}$ on each copy.

Note however that we have the canonical identifications (which behave naturally
under complexification)
\[
\begin{split}
\Gamma\Big( \oplus_{0}^{3} L^2(\mathbb{R}^3; \mathbb{C}) \Big) = 
\Gamma\Big(L^2(\mathbb{R}^3; \mathbb{C}) \Big) \otimes 
\Gamma\Big( L^2(\mathbb{R}^3; \mathbb{C}) \Big) \otimes
\Gamma\Big( L^2(\mathbb{R}^3; \mathbb{C}) \Big) \otimes
\Gamma\Big( L^2(\mathbb{R}^3; \mathbb{C}) \Big), \\
\Gamma\Big( \oplus_{0}^{3} L^2(\mathbb{R}^3; \mathbb{R}) \Big) = 
\Gamma\Big(L^2(\mathbb{R}^3; \mathbb{R}) \Big) \otimes 
\Gamma\Big( L^2(\mathbb{R}^3; \mathbb{R}) \Big) \otimes
\Gamma\Big( L^2(\mathbb{R}^3; \mathbb{R}) \Big) \otimes
\Gamma\Big( L^2(\mathbb{R}^3; \mathbb{R}) \Big), 
\end{split}
\]  
under which the following equalities hold
\begin{multline*}
a(\xi_0 \oplus \xi_1 \oplus \xi_2 \oplus \xi_3) =
a_{{}_{0}}(\xi_0) \otimes \boldsymbol{1} \otimes \boldsymbol{1} \otimes \boldsymbol{1} +
\boldsymbol{1} \otimes a_{{}_{1}}(\xi_1) \otimes  \boldsymbol{1} \otimes \boldsymbol{1} \\ +
\boldsymbol{1} \otimes \boldsymbol{1} \otimes a_{{}_{2}}(\xi_2) \otimes \boldsymbol{1} +
\boldsymbol{1}  \otimes \boldsymbol{1} \otimes \boldsymbol{1} \otimes  a_{{}_{3}}(\xi_3),
\end{multline*}
where $a(\xi_0 \oplus \xi_1 \oplus \xi_2 \oplus \xi_3)$ stands  for the ordinary
annihilation operators on $\Gamma\Big( \oplus_{0}^{3} L^2(\mathbb{R}^3; \mathbb{C}) \Big)$
(or respectively on  $\Gamma\Big( \oplus_{0}^{3} L^2(\mathbb{R}^3; \mathbb{R}) \Big)$) and where
$a_{{}_{\nu}}(\xi_\nu)$ stand for the ordinary annihilation operators acting in the
Fock space $\Gamma\Big(L^2(\mathbb{R}^3; \mathbb{C}) \Big)$) over the $\nu$-th copy
of $L^2(\mathbb{R}^3; \mathbb{C})$ (resp. on the Fock space $\Gamma\Big(L^2(\mathbb{R}^3; \mathbb{R}) \Big)$
over the $\nu$-th copy of $L^2(\mathbb{R}^3; \mathbb{R})$). 
In this manner we obtain the Gelfand triple 
\[
(E) \,\,\,\, \subset \,\,\,\, \Gamma(\oplus_{0}^{3} L^2(\mathbb{R}^3)) \cong 
L^2(E_{\mathbb{R}}^*, \mu; \mathbb{C}) 
\,\,\,\, \subset \,\,\,\,  (E)^*, 
\]
which has the tensor product structure
\[
(E) = (E^0) \otimes (E^1) \otimes (E^2) \otimes (E^3) \,\,\, \textrm{and} \,\,\,
(E)^* = (E^0)^* \otimes (E^1)^* \otimes (E^2)^* \otimes (E^3)^*
\]
with the scalar continuous Hida operators (we denote them as the vector-valued Hida operators 
$\partial_{\boldsymbol{\p}}^{\nu}$, $\partial_{{\boldsymbol{\p}}}^{\nu *}$ constructed above and 
acting on $(E)$ or on $(E)^*$)
\[
\partial_{\boldsymbol{\p}}^{\nu}: (E^\nu) \rightarrow (E^\nu) \,\,\, \textrm{and} \,\,\,
\partial_{{\boldsymbol{\p}}}^{\nu *}: (E^\nu)^* \rightarrow (E^\nu)^*
\]
acting on the ``scalar'' Hida spaces $(E^\nu)$, which compose Gelfand triples
\[
(E^\nu) \,\,\,\, \subset \,\,\,\, \Gamma(L^2(\mathbb{R}^3; \mathbb{C})) \cong 
L^2({E^{\nu}_{\mathbb{R}}}^*, \mu; \mathbb{C}) 
\,\,\,\, \subset \,\,\,\,  (E^\nu)^*
\]
with the Fock spaces
\[
\Gamma(L^2(\mathbb{R}^3; \mathbb{C}))
\] 
over the $\nu$-th copy of $L^2(\mathbb{R}^3; \mathbb{C})$.

Note that in our case we have
\[
E^\nu = \mathcal{S}^{0}(\mathbb{R}^3; \mathbb{C}) = \mathcal{S}_{A^{(3)}}(\mathbb{R}^3; \mathbb{C})
= \mathcal{S}_{A^{(3)}}(\mathbb{R}^3; \mathbb{R})_{\mathbb{C}},
\]
for each $\nu \in \{0,1,2,3\}$, composing the Gelfand triple
\[
\mathcal{S}_{A^{(3)}}(\mathbb{R}^3; \mathbb{R})_{\mathbb{C}} \,\,\, \subset \,\,\,
L^2(\mathbb{R}^3; \mathbb{C}) \,\,\, \subset \,\,\,
\mathcal{S}_{A^{(3)}}(\mathbb{R}^3; \mathbb{C})^*.
\]

Thus, we could have been working exclusively with scalar valued functions in the single particle spaces
and construct four copies of Hida operators using as the single particle space 
$L^2(\mathbb{R}^3; \mathbb{R})$. We should note however that having given the scalar Hida operators
$\partial_{\boldsymbol{\p}}^{\nu}$ acting respectively on $(E^\nu)$, we construct the 
vector valued $\partial_{\boldsymbol{\p}}^{\nu}$, acting on $(E)$ (which we need), in the following manner
\[
\begin{split}
\partial_{\boldsymbol{\p}}^{\nu=0} =
\partial_{\boldsymbol{\p}}^{\nu=0} \otimes \boldsymbol{1} \otimes \boldsymbol{1} \otimes \boldsymbol{1}, \\
\partial_{\boldsymbol{\p}}^{\nu=1} =
\boldsymbol{1} \otimes \partial_{\boldsymbol{\p}}^{\nu=1} \otimes  \boldsymbol{1} \otimes \boldsymbol{1}, \\ 
\partial_{\boldsymbol{\p}}^{\nu=2} =
\boldsymbol{1} \otimes \boldsymbol{1} \otimes \partial_{\boldsymbol{\p}}^{\nu=2} \otimes \boldsymbol{1}, \\
\partial_{\boldsymbol{\p}}^{\nu=3} =
\boldsymbol{1}  \otimes \boldsymbol{1} \otimes \boldsymbol{1} \otimes  \partial_{\boldsymbol{\p}}^{\nu=3},
\end{split}
\]
where on the right-hand side we have the scalar Hida operators 
$\partial_{\boldsymbol{\p}}^{\nu}$, acting  on $(E^\nu)$ and on the left-hand side 
 have the vector-valued Hida operators acting on $(E) = (E^0) \otimes (E^1) \otimes (E^2) \otimes (E^3)$. 
\end{rem*}

Now in \cite{hida},
there has been developed an effective calculus of continuous operators
$(E) \to (E)^*$ which can be expressed as the integrals of the operators
$\partial_{\vec{p}}^{\nu}$, $\partial_{{\vec{p}}}^{\nu *}$. Indeed, it follows that for any
$\Phi, \Psi \in (E)$ and $l,m \in \mathbb{N}$ the function (where we write $\boldsymbol{\p}$ for $\vec{p}$)
\[
\eta_{{}_{\Phi, \Psi}}: 
\,\big(\nu_1, \underset{1}{\boldsymbol{\p}'}, \ldots, \nu_l, \underset{l}{\boldsymbol{\p}'}, 
\mu_1, \underset{1}{\boldsymbol{\p}}, \ldots, \mu_m, \underset{m}{\boldsymbol{\p}} \big) \,
\mapsto \langle \langle 
\,\, \partial_{{\underset{1}{\boldsymbol{\p}'}}}^{\nu_1 *} \ldots \partial_{{\underset{l}{\boldsymbol{\p}'}}}^{\nu_l *}
\partial_{{\underset{1}{\boldsymbol{\p}}}}^{\mu_1 } \ldots \partial_{{\underset{m}{\boldsymbol{\p}}}}^{\mu_m } \,\, \Phi,
\Psi \rangle \rangle
\] 
on $\big(\mathbb{R}^3 \sqcup \mathbb{R}^3 \sqcup \mathbb{R}^3 \sqcup \mathbb{R}^3  \big)^{\otimes(l+m)}$ 
belongs to $E^{\otimes(l+m)}$, compare \cite{hida}, Lemma 2.1. Thus, for any 
$\kappa_{l,m} \in (E^{\otimes(l+m)})^*$ there exists a unique continuous operator (Theorem 2.2 of \cite{hida})
\[
\Xi_{l,m}(\kappa_{l,m}): (E) \to (E)^*
\]
such that 
\[
\langle \langle \Xi_{l,m}(\kappa_{l,m}) \Phi , \, \Psi \rangle \rangle 
= \langle \kappa_{l,m} , \, \eta_{{}_{\Phi, \Psi}} \rangle, \,\,\, \Phi, \Psi \in (E)
\]
Because it is customary to write the dual pairing $\langle \cdot, \cdot \rangle$ between 
$E^*$ and $E$ using formal integral expressions and the formal integral distributional kernels
$\kappa_{\nu_1 \ldots \nu_l \mu_1 \ldots \mu_m}  
\,\big(\underset{1}{\boldsymbol{\p}'}, \ldots, \underset{l}{\boldsymbol{\p}'}, 
\underset{1}{\boldsymbol{\p}}, \ldots, \underset{m}{\boldsymbol{\p}} \big) \,$ corresponding to
$\kappa_{l,m} \in (E^{\otimes(l+m)})^*$, then the operator $\Xi_{l,m}(\kappa_{l,m})$ can be 
formally written 
as the following Berezin-type integral  
\begin{multline}\label{Berezin-integral-p}
\Xi_{l,m}(\kappa_{l,m}) = \\ \sum \limits_{\mu_1, \ldots \nu_l, \mu_1, \ldots \mu_m = 0}^{3} \, 
\int \limits_{(\mathbb{R}^3)^{(l+m)}} 
\kappa_{\mu_1 \ldots \nu_l \mu_1 \ldots \mu_m}  
\,\big(\underset{1}{\boldsymbol{\p}'}, \ldots, \underset{l}{\boldsymbol{\p}'}, 
\underset{1}{\boldsymbol{\p}}, \ldots, \underset{m}{\boldsymbol{\p}} \big) \, \times \\
\times \, 
\partial_{{\underset{1}{\boldsymbol{\p}'}}}^{\nu_1 *} \ldots \partial_{{\underset{l}{\boldsymbol{\p}'}}}^{\nu_l *}
\partial_{{\underset{1}{\boldsymbol{\p}}}}^{\mu_1 } \ldots \partial_{{\underset{m}{\boldsymbol{\p}}}}^{\mu_m } \,\,
\ud^3 \underset{1}{\boldsymbol{\p}'} \ldots \ud^3 \underset{l}{\boldsymbol{\p}'}
\ud^3 \underset{1}{\boldsymbol{\p}} \ldots \ud^3 \underset{m}{\boldsymbol{\p}}.
\end{multline}
In particular the integral (\ref{Berezin-integral-p}) represents a continuous operator $(E) \to (E)^*$
iff $\kappa_{l,m} \in (E^{\otimes(l+m)})^*$, and similarly 
\begin{multline}\label{Berezin-integral-x}
\Xi_{l,m}(\kappa_{l,m}) = \\ 
\sum \limits_{\mu_1, \ldots \nu_l, \mu_1, \ldots \mu_m = 0}^{3} \, 
\int \limits_{(\mathbb{R}^3)^{(l+m)}} 
\kappa_{\mu_1 \ldots \nu_l \mu_1 \ldots \mu_m}  
\,\big(\underset{1}{\boldsymbol{\x}'}, \ldots, \underset{l}{\boldsymbol{\x}'}, 
\underset{1}{\boldsymbol{\x}}, \ldots, \underset{m}{\boldsymbol{\x}} \big) \, \times \\
\times \,
\partial_{{\underset{1}{\boldsymbol{\x}'}}}^{\nu_1 *} \ldots \partial_{{\underset{l}{\boldsymbol{\x}'}}}^{\nu_l *}
\partial_{{\underset{1}{\boldsymbol{\x}}}}^{\mu_1 } \ldots \partial_{{\underset{m}{\boldsymbol{\x}}}}^{\mu_m } \,\,
\ud^3 \underset{1}{\boldsymbol{\x}'} \ldots \ud^3 \underset{l}{\boldsymbol{\x}'}
\ud^3 \underset{1}{\boldsymbol{\x}} \ldots \ud^3 \underset{m}{\boldsymbol{\x}},
\end{multline}
represents a continuous operator\footnote{In particular the results of \cite{hida} extend substantially the idea of Berezin, who proved that every bounded 
operator in the Fock space have the normal integral representation of the form of sum of the 
operators (\ref{Berezin-integral-p}) or equivalently (\ref{Berezin-integral-x}).} $(\mathbb{E}) \to (\mathbb{E})^*$ if $\kappa_{l,m} \in (\mathbb{E}^{\otimes(l+m)})^*$. Since the most important and usually unbounded operators on $(L^2)$ which we encounter in QFT are expressible in this form, theory presented in \cite{hida} is very useful for us.

Obata and Huang, \cite{obata} and \cite{huang}, have then extended this result proving that any continuous linear operator $\Xi: (E) \to (E)^*$ can be represented as a series 
\[
\Xi = \sum_{l,m} \Xi_{l,m}(\kappa_{l,m})
\]
of operators in the normal form $\Xi_{l,m}(\kappa_{l,m})$ as in (\ref{Berezin-integral-x}) or (\ref{Berezin-integral-p}),
in the weak sense that for each $\Phi$ and $\Psi$ which are exponential (``coherent'') over $E$ we have
\[
\langle \langle \Xi \Phi, \Psi \rangle \rangle = \sum \limits_{l,m=0}^{\infty} 
\langle \langle \Xi_{l,m}(\kappa_{l,m}) \Phi, \Psi \rangle \rangle;
\] 
i.e.  every continuous operator $\Xi: (E) \to (E)^*$ admits unique \emph{Fock expansion}
into the series of continuous integral kernel operators $ \Xi_{l,m}(\kappa_{l,m}): (E) \to (E)^*$ (\cite{obata},
Theorem 6.1 or \cite{huang}, Theorem 3.3).

Although we would like to be mathematically rigorous, we should not be too much pedestrian in killing useful
physical ideas concerning the integral kernel operators of Bogoliubov-Berezin type, such as 
(\ref{Berezin-integral-p}) or (\ref{Berezin-integral-x}). The integral expressions 
(\ref{Berezin-integral-p}) or (\ref{Berezin-integral-x}) are much more than merely formal symbols
for the continuous operators $\Xi_{l,m}(\kappa_{l,m}): (E) \to (E)^*$ of Hida, Obata and Sait\'o \cite{hida}.
$E$ (or the larger function spaces $E_k$) may be naturally regarded as subspaces of the dual space $E^*$, 
the so called function (or regular) distributions, with the pairing of this special distributions with
the elements of $E$ given by the ordinary (not merely symbolic) integral, and every element
of $E^{*}$ is a limit in $E^*$ of function distributions. In consequence a wide subclass of the integral kernel operators
are pointwisely actual Pettis integrals\footnote{For definition compare \cite{HKPS}, Chap. 8.A.} (and sometimes even Bochner integrals), 
and every formal integral kernel operator is a limit of kernel
operators given pointwisely by Pettis integral kernel operators. In fact every important operator valued distribution
in QFT is introduced by this limiting process of integral kernel operators. The calculus of integral kernel operators of this type is therefore of much importance and cannot be obscured by formal pedantism.

 For 
$\widetilde{\varphi}', \widetilde{\varphi} \in E$ we have (with abbreviation $\sqcup \mathbb{R}^3$
for $\mathbb{R}^3 \sqcup \mathbb{R}^3 \sqcup \mathbb{R}^3 \sqcup \mathbb{R}^3$)
\[
D_{\widetilde{\varphi}'} = \Xi_{0,1}(\widetilde{\varphi}') 
= \int \limits_{\sqcup \mathbb{R}^3} \widetilde{\varphi}'(\boldsymbol{\p}) 
\partial_{\boldsymbol{\boldsymbol{\p}}}
\, \ud^3 p =  \sum \limits_{\nu} \int \limits_{\mathbb{R}^3} \widetilde{\varphi}'^{\nu}(\boldsymbol{\p}) 
\partial_{\boldsymbol{\p}}^{\nu}
\, \ud^3 p
\]
and 
\[
D_{\widetilde{\varphi}}^{*} = \Xi_{1,0}(\widetilde{\varphi}) = \int \limits_{\sqcup \mathbb{R}^3} 
\widetilde{\varphi}(\boldsymbol{\p}) 
\partial_{\boldsymbol{\p}}^{*}
\, \ud^3 p =  \sum \limits_{\nu} \int \limits_{\mathbb{R}^3} 
\widetilde{\varphi}^{\nu}(\boldsymbol{\p}) 
\partial_{\boldsymbol{\p}}^{\nu *}
\, \ud^3 p,
\]
where for each element $\Phi$ of the Hida's testing functional space $(E)$ the integral 
\[
 \int \limits_{\sqcup \mathbb{R}^3} \widetilde{\varphi}'(\boldsymbol{\p}) 
\partial_{\boldsymbol{\boldsymbol{\p}}} \, \Phi
\, \ud^3 p =  \sum \limits_{\nu} \int \limits_{\mathbb{R}^3} \widetilde{\varphi}'^{\nu}(\boldsymbol{\p}) 
\partial_{\boldsymbol{\p}}^{\nu} \Phi
\, \ud^3 p
\]
and 
\[
 \int \limits_{\sqcup \mathbb{R}^3} \widetilde{\varphi}(\boldsymbol{\p}) 
\partial_{\boldsymbol{\p}}^{*} \, \Phi
\, \ud^3 p =  \sum \limits_{\nu} \int \limits_{\mathbb{R}^3} 
\widetilde{\varphi}^{\nu}(\boldsymbol{\p}) 
\partial_{\boldsymbol{\p}}^{\nu *} \, \Phi
\, \ud^3 p,
\]
exist as Pettis integrals (and the first integral exists even in the Bochner sense for any
$\widetilde{\varphi}' \in L^2(\mathbb{R}^3)$, as an element of one of the Hilbert spaces 
$(E_{k})$ for some $k$ depending on $\widetilde{\varphi}'$ and $\Phi$ and is independent of the choice of 
all possible greater values of $k$, compare \cite{HKPS}, Chap. V.B). In this case it turns out that
the integral belongs to $(E) \subset (E_{k}) \subset (E)^*$.
By Theorem 2.2 of \cite{hida} we can interpret $\partial_{\boldsymbol{\p}}^{*}$ and $\partial_{\boldsymbol{\p'}}$
as operator valued distributions, with the test function space equal $E$ and the domain $\mathscr{D}$ equal to the Hida's
testing functional space $(E)$ and with the nuclear topology of uniform convergence on bounded sets on the linear space 
$\mathscr{L}\big(E), (E)\big)$ of continuous linear operators $(E) \to (E)$.

The oparators $D_{\widetilde{\varphi}'}$ and $D_{\widetilde{\varphi}}^{*}$ are continuous
when regarded as operators $(E) \to (E)$ in this case when $\widetilde{\varphi}', \widetilde{\varphi} \in E$, 
with the compositions 
$D_{\widetilde{\varphi}'}D_{\widetilde{\varphi}}^{*}$ and $D_{\widetilde{\varphi}}^{*}D_{\widetilde{\varphi}'}$
continuous as operators $(E) \to (E)$, and with the composition $D_{\widetilde{\varphi}}^{*}D_{\widetilde{\varphi}'}$ 
equal 

\begin{multline*}
D_{\widetilde{\varphi}}^{*} D_{\widetilde{\varphi}'} 
= \Xi_{1,1}(\widetilde{\varphi} \otimes \widetilde{\varphi}') =
\int \limits_{( \sqcup \mathbb{R}^3) \times (\sqcup \mathbb{R}^3)}  
\widetilde{\varphi}(\boldsymbol{\p}) \widetilde{\varphi}'(\boldsymbol{\p}')
\,\, \partial_{\boldsymbol{\p}}^{*} \partial_{\boldsymbol{\p'}}
\, \ud^3 p \ud^3 p'  \\
=  \sum \limits_{\nu \mu} \int \limits_{\mathbb{R}^3 \times \mathbb{R}^3} 
\widetilde{\varphi}^{\nu }(\boldsymbol{\p}) \widetilde{\varphi}'^{\mu}(\boldsymbol{\p}') 
\partial_{\boldsymbol{\p}}^{\nu *} \partial_{\boldsymbol{\p'}}^{\mu} 
\, \ud^3 p \ud^3 p',
\end{multline*}
and where again the integral operator exist pointwise on $(E)$ as Pettis integral, 
i.e. for each $\Phi, \Psi \in (E)$ the function 
\[
\,\big(\boldsymbol{\p}, \boldsymbol{\p}' \big)  
\mapsto \langle \langle 
\widetilde{\varphi}^{\nu }(\boldsymbol{\p}) \widetilde{\varphi}'^{\mu}(\boldsymbol{\p}') 
\,\, \partial_{\boldsymbol{\p}}^{\nu *} \partial_{\boldsymbol{\p}'}^{\mu}
 \,\, \Phi,
\Psi \rangle \rangle =
\widetilde{\varphi}^{\nu }(\boldsymbol{\p}) \widetilde{\varphi}'^{\mu}(\boldsymbol{\p}') 
\langle \langle 
\,\, \partial_{\boldsymbol{\p}}^{\nu *} \partial_{\boldsymbol{\p}'}^{\mu}
 \,\, \Phi,
\Psi \rangle \rangle
\] 
on $\mathbb{R}^{3} \times \mathbb{R}^{3}$ is measurable and belongs\footnote{In fact the function belongs
to $E \otimes E$ in this case, because  $\widetilde{\varphi}', \widetilde{\varphi} \in E$.} 
to $L^1 (\mathbb{R}^3 \times \mathbb{R}^3)$, so that there 
exists\footnote{Compare the proof of Thm. 2.2 of \cite{hida} and recall that the pairing 
$\langle \cdot , \cdot \rangle$ of 
$\widetilde{\varphi} \in E \subset E^*$
regarded as an element of $E^*$ with an element $\widetilde{\varphi}' \in E$ is given by the inner product:
$\langle \widetilde{\varphi} , \widetilde{\varphi}' \rangle  
= \big(\overline{\widetilde{\varphi}} , \widetilde{\varphi}' \big)_{{}_{\oplus L^2(\mathbb{R}^3)}}$.} 
an element of $(E)^*$, denoted by 
\[
\sum \limits_{\nu \mu} \int \limits_{\mathbb{R}^3 \times \mathbb{R}^3} 
\widetilde{\varphi}^{\nu }(\boldsymbol{\p}) \widetilde{\varphi}'^{\mu}(\boldsymbol{\p}') 
\partial_{\boldsymbol{\p}}^{\nu *} \partial_{\boldsymbol{\p'}}^{\mu} \, \Phi \, 
\, \ud^3 p \ud^3 p',
\]
such that 
\begin{multline*}
\bigg\langle \bigg\langle
\sum \limits_{\nu \mu} \int \limits_{\mathbb{R}^3 \times \mathbb{R}^3} 
\widetilde{\varphi}^{\nu }(\boldsymbol{\p}) \widetilde{\varphi}'^{\mu}(\boldsymbol{\p}') 
\partial_{\boldsymbol{\p}}^{\nu *} \partial_{\boldsymbol{\p'}}^{\mu} \, \Phi \, 
\, \ud^3 p \ud^3 p' \,\, , \,\,\, \Psi \, 
\bigg\rangle \bigg\rangle \\ =
\sum \limits_{\nu \mu} \int \limits_{\mathbb{R}^3 \times \mathbb{R}^3} 
\big\langle \big\langle
\widetilde{\varphi}^{\nu }(\boldsymbol{\p}) \widetilde{\varphi}'^{\mu}(\boldsymbol{\p}') 
\partial_{\boldsymbol{\p}}^{\nu *} \partial_{\boldsymbol{\p'}}^{\mu} \, \Phi \, , \,
\Psi
\big\rangle \big\rangle 
\, \ud^3 p \ud^3 p',
\end{multline*}
for all $\Phi, \Psi \in (E)$.
And thus by Theorem 2.2 of \cite{hida} we can interpret $\partial_{\boldsymbol{\p}}^{*} \partial_{\boldsymbol{\p'}}$
as operator valued distribution, with the test function space equal $E\otimes E$ and the domain $\mathscr{D}$ equal to the Hida's testing functional space $(E)$ and with the nuclear topology on $\mathscr{L}\big(E), (E)\big)$ defined as above. Similarly,  the continuous operator 
$(E) \mapsto (E)^*$
\[
\partial_{{\underset{1}{\boldsymbol{\p}'}}}^{\nu_1 *} \ldots \partial_{{\underset{l}{\boldsymbol{\p}'}}}^{\nu_l *}
\partial_{{\underset{1}{\boldsymbol{\p}}}}^{\mu_1 } \ldots \partial_{{\underset{m}{\boldsymbol{\p}}}}^{\mu_m }, \,\,
\]
may be regarded as operator valued distribution with the test function space equal $E^{\otimes (l+m)}$ and the domain
$\mathscr{D}$ equal to the Hida's
testing functional space $(E)$ and with the nuclear topology on $\mathscr{L}\big(E), (E)\big)$
defined as above.

Because for $\widetilde{\varphi}', \widetilde{\varphi} \in E$ the operator 
$D_{\widetilde{\varphi}'}D_{\widetilde{\varphi}}^{*}$ also is a continuous operator $(E) \to (E)$, then by the general theory of \cite{hida} it follows that this operator likewise 
has (finite) expansion into normal integral Berezin-type kernel operators  (\ref{Berezin-integral-p}) or (\ref{Berezin-integral-x}),  but in order to compute them explicitly we use the well known fact that   
\begin{equation}\label{[D',D]}
[D_{\overline{\widetilde{\varphi}'}}, D_{\widetilde{\varphi}}^{*}] 
= \langle \overline{\widetilde{\varphi}'}, \widetilde{\varphi} \rangle \bold{1} 
= (\widetilde{\varphi}', \widetilde{\varphi})_{{}_{\oplus L^2(\mathbb{R}^3)}} \bold{1}
\end{equation}
for all $\widetilde{\varphi}', \widetilde{\varphi} \in \oplus L^2(\mathbb{R}^3)$ 
and in particular for all $\widetilde{\varphi}', \widetilde{\varphi} \in E$, which after simple
computations follows from the formula (\ref{D-xi-white-dec}). 
Using the continuity of the scalar product $(\cdot, \cdot)_{{}_{\oplus L^2(\mathbb{R}^3)}}$
in the nuclear topology of $E$ (compare \cite{GelfandIV}, Ch. I.4.2) and nuclearity of $E$,
it follows that the bilinear map $\widetilde{\varphi}'\times \widetilde{\varphi} \mapsto 
(\overline{\widetilde{\varphi}'}, \widetilde{\varphi})_{{}_{\oplus L^2(\mathbb{R}^3)}} \bold{1}$
defines an operator valued distribution:
\[
E \otimes E \ni \zeta \mapsto \Xi_{0,0}(\zeta) 
= \int \limits_{(\sqcup \mathbb{R}^{3}) \times (\sqcup \mathbb{R}^{3})}  
\zeta (\boldsymbol{\p}', \boldsymbol{\p})
\tau(\boldsymbol{\p}', \boldsymbol{\p})\, \bold{1} \, \ud^3 p' \ud^3 p  = \tau(\zeta) \bold{1}
\]
where $\tau \in (E \otimes E)^*$ is defined by
\[
\langle \tau, \widetilde{\varphi}' \otimes \widetilde{\varphi} \rangle 
= \langle \widetilde{\varphi}', \widetilde{\varphi} \rangle, \,\,\, \widetilde{\varphi}', \widetilde{\varphi} \in E, 
\]
therefore, we write symbolically 
\begin{multline*}
\Xi_{0,0}(\widetilde{\varphi} \otimes \widetilde{\varphi}') 
= [D_{\widetilde{\varphi}'}, D_{\widetilde{\varphi}}^{*}] 
= \int \limits_{(\sqcup \mathbb{R}^{3}) \times (\sqcup \mathbb{R}^{3})}  
\widetilde{\varphi} \otimes \widetilde{\varphi}' (\boldsymbol{\p}, \boldsymbol{\p}') \,\,
\delta(\boldsymbol{\p}'- \boldsymbol{\p}) \, \bold{1} \, \ud^3 p \ud^3 p'\\
= \sum \limits_{\mu,\nu} \int \limits_{\mathbb{R}^{3} \times \mathbb{R}^{3}}  
\widetilde{\varphi}^\mu \otimes \widetilde{\varphi}'^\nu (\boldsymbol{\p}, \boldsymbol{\p}') \,\,
\delta^{\mu \nu} \, \delta(\boldsymbol{\p}'- \boldsymbol{\p}) \, \bold{1} \, \ud^3 p \ud^3 p' \\
\sum \limits_{\mu,\nu} \int \limits_{\mathbb{R}^{3} \times \mathbb{R}^{3}}  
\widetilde{\varphi}^\mu(\boldsymbol{\p}) \, \widetilde{\varphi}'^\nu(\boldsymbol{\p}') \,\,
\delta^{\mu \nu} \, \delta(\boldsymbol{\p}'- \boldsymbol{\p}) \, \bold{1} \, \ud^3 p \ud^3 p',
\end{multline*}
and
\begin{equation}\label{CCR-distribution}
[\partial_{\boldsymbol{\p'}}^{\mu},  \partial_{\boldsymbol{\p}}^{\nu *}] 
= \delta^{\mu \nu} \delta(\boldsymbol{\p}' - \boldsymbol{\p}) \bold{1}.
\end{equation}

Recall that we treat $\widetilde{\varphi} \in E$ as a function
\[
\sqcup \mathbb{R}^3 \ni (\mu, \boldsymbol{\p}) \mapsto \widetilde{\varphi}(\boldsymbol{\p})
= \widetilde{\varphi}^\mu(\boldsymbol{\p})
\]
and respectively $\tau \in (E \otimes E)^*$ as a ``generalized function'' 
\[
(\sqcup \mathbb{R}^3)  \times (\sqcup \mathbb{R}^3) \ni 
(\mu, \boldsymbol{\p}) \times (\nu, \boldsymbol{\p}') \mapsto \tau(\boldsymbol{\p}, \boldsymbol{\p}')
= \tau_{\mu \nu}(\boldsymbol{\p}, \boldsymbol{\p}').
\]

Thus, the bilinear operator valued map $\widetilde{\varphi}' \times \widetilde{\varphi} \mapsto
D_{\widetilde{\varphi}}^{*}D_{\widetilde{\varphi}'} +  [D_{\widetilde{\varphi}'}, 
D_{\widetilde{\varphi}}^{*}]  
= D_{\widetilde{\varphi}}^{*}D_{\widetilde{\varphi}'} + \tau(\widetilde{\varphi}' \otimes \widetilde{\varphi}) \bold{1} $ defines the operator valued distribution with the following distributional integral kernel 
\[
\partial_{\boldsymbol{\p}'}^{\nu}\partial_{{\boldsymbol{\p}}}^{\mu *} = 
\partial_{{\boldsymbol{\p}}}^{\mu *}\partial_{\boldsymbol{\p}'}^{\nu} 
+ \delta^{\mu \nu} \delta(\boldsymbol{\p}' - \boldsymbol{\p}) \bold{1},
\]
so that 
\begin{multline}\label{Dvarphi'D*varphi}
D_{\widetilde{\varphi}'}D_{\widetilde{\varphi}}^{*} = 
\Xi_{1,1}(\widetilde{\varphi} \otimes \widetilde{\varphi}')
+ \Xi_{0,0}(\widetilde{\varphi} \otimes \widetilde{\varphi}') \\
= \sum \limits_{\mu, \nu} \int \limits_{\mathbb{R}^3 \times \mathbb{R}^3} 
\widetilde{\varphi}^{\nu } \otimes \widetilde{\varphi}'^{\mu}(\boldsymbol{\p},\boldsymbol{\p}') 
\,\, \partial_{\boldsymbol{\p}}^{\nu *} \partial_{\boldsymbol{\p'}}^{\mu} 
\, \ud^3 p \ud^3 p' \\
+ \sum \limits_{\mu,\nu} \int \limits_{\mathbb{R}^{3} \times \mathbb{R}^{3}}  
\widetilde{\varphi}^\mu \otimes \widetilde{\varphi}'^\nu (\boldsymbol{\p}, \boldsymbol{\p}') \,\,
\delta^{\mu \nu} \, \delta(\boldsymbol{\p}'- \boldsymbol{\p}) \, \bold{1} \, \ud^3 p \ud^3 p',
\end{multline}
where the second symbolic integral may also be defined pointwisely on $(E)$ as a limit of actual Pettis, or even Bochner, integral operators (which is termed \emph{regularization process} in physicists parlance).

And although $\partial_{\boldsymbol{\p}}^{\nu}\partial_{{\boldsymbol{\p}}}^{\mu *}$ is not well defined
as operator $(E) \to (E)^*$ it is well defined as operator valued distribution. And similarly,
\[
\partial_{{\underset{1}{\boldsymbol{\p}'}}}^{\nu_1 *} \ldots \partial_{{\underset{l}{\boldsymbol{\p}'}}}^{\nu_l *}
\partial_{{\underset{1}{\boldsymbol{\p}}}}^{\mu_1 } \ldots \partial_{{\underset{m}{\boldsymbol{\p}}}}^{\mu_m }, \,\,
\]
is not only well defined continuous operator $(E) \to (E)^*$, but a well defined operator valued distribution,
and reordering the operators $\partial_{{\underset{k}{\boldsymbol{\p}'}}}^{\nu_k *}$ and 
$\partial_{{\underset{q}{\boldsymbol{\p}}}}^{\mu_q }$ 
in this expression we similarly obtain well defined operator valued distribution 
(although not well defined operator $(E) \to (E)^*$).

Because of (\ref{[D',D]}) and (\ref{CCR-distribution}), 
the operators\footnote{Note that the additional 
complex conjugation 
$\overline{\widetilde{\varphi}}$ in $a(\overline{\widetilde{\varphi}})$
 is due to the physicist's convention, which we adopt here, that the inner product
is conjugate linear in the first argument.}
 $a(\overline{\widetilde{\varphi}})$ and $a(\widetilde{\varphi})^+$ may be identified 
respectively with $D_{\widetilde{\varphi}}$ and $ D_{\widetilde{\varphi}}^{*}$,
and operator valued distributions $a^\nu(\boldsymbol{\p}')$ and $a^{\mu}(\boldsymbol{\p})^{+}$ with
$\partial_{\boldsymbol{\p'}}^{\mu},  \partial_{\boldsymbol{\p}}^{\nu *}$;
where the identification is defined by the naural unitary equivalence between the Fock space
$\Gamma\big( \oplus L^2(\mathbb{R}^3) \big)$ and $(L^2)$. 
 
Using the operator $\sqrt{B}$ of pointwise multiplication by the 
matrix\footnote{Where $\sqrt{B(\boldsymbol{\p}, p^0(\boldsymbol{\p}))}$ is the square root (\ref{sqrtB}) of the positive matrix (\ref{Bmatrix})
$B(\boldsymbol{\p}, p^0(\boldsymbol{\p})) = V(\beta(\boldsymbol{\p}, p^0(\boldsymbol{\p})))^* 
V(\beta(\boldsymbol{\p}, p^0(\boldsymbol{\p})))$, (in the coordinates
$\boldsymbol{\p}$ on the orbit $\mathscr{O}_{1,0,0,1}$).} 
\[
\frac{1}{\sqrt{2 p^0(\boldsymbol{\p})}}\sqrt{B(\boldsymbol{\p}, p^0(\boldsymbol{\p}))}, 
\]
we obtain from (\ref{[D',D]})
\[
[D_{\sqrt{B} \, \overline{\widetilde{\varphi}'}} \, , \, D_{\sqrt{B} \widetilde{\varphi}}^{*}] 
= (\sqrt{B}\widetilde{\varphi}', \sqrt{B}\widetilde{\varphi})_{{}_{\oplus L^2(\mathbb{R}^3)}}
= (\widetilde{\varphi}', B\widetilde{\varphi})_{{}_{\oplus L^2(\mathbb{R}^3)}},
\]
for all $\widetilde{\varphi}', \widetilde{\varphi}$
such that $\sqrt{B}\widetilde{\varphi}', \sqrt{B}\widetilde{\varphi} \in \oplus L^2(\mathbb{R}^3)$, and because
$D_{\sqrt{B} \, \overline{\widetilde{\varphi}'}}, D_{\sqrt{B} \widetilde{\varphi}}^{*}$  
are to be identified with $a(\sqrt{B} \, \widetilde{\varphi})$ and $a(\sqrt{B} \, \widetilde{\varphi})^+$
and by the definition of the inner product in $\mathcal{H}'$, compare (\ref{inn-Lop-1-space}),
the equality (\ref{proof-first}) follows. 

Into the Fock space $\Gamma\big( \oplus L^2(\mathbb{R}^3) \big)$ we introduce the Gupta-Bleuler operator
$\eta$, in the following manner. In order to give the definition we need to distinguish separate
orthogonal components $L^2(\mathbb{R}^3)$ in the one particle Hilbert space $\oplus L^2(\mathbb{R}^3)$
respectively $L^2(\mathbb{R}^3)_\mu$ by the corresponding index $\mu = 0,1,2,3$, with the zero index $0$ 
corresponding to the so called scalar photons. Every element
\[ 
\Phi \in \Gamma\big( \oplus L^2(\mathbb{R}^3) \big) \cong_{U}
\Gamma\big(L^2(\mathbb{R}^3)_0 \big) \otimes \Gamma\big(L^2(\mathbb{R}^3)_1 \big)
\otimes \Gamma\big(L^2(\mathbb{R}^3)_2 \big) \otimes \Gamma\big(L^2(\mathbb{R}^3)_3 \big) 
\]
may be represented by the following decomposition
\[
\Phi = \sum \limits_{n}^{\infty} \Phi^{(n)} \cong_U \sum \limits_{n_0 + n_1 + n_2 + n_3 = 0}^{\infty} 
\Phi^{(n_0)}\otimes \Phi^{(n_1)} \otimes \Phi^{(n_2)} \otimes \Phi^{(n_3)},
\] 
into orthogonal components $\Phi^{(n)} \in \big[\oplus L^2(\mathbb{R}^3) \big]^{\otimes n}_{S}$, but this time every component $\Phi^{(n)}$ may be naturally regarded as an element of
\[ 
\bigoplus \limits_{n_0 + n_1 + n_2 + n_3 = n}
\big[ L^2(\mathbb{R}^3)_0 \big]^{\otimes n_0}_{S} \otimes \big[ L^2(\mathbb{R}^3)_1 \big]^{\otimes n_1}_{S}
\otimes \big[ L^2(\mathbb{R}^3)_2 \big]^{\otimes n_2}_{S}
\otimes \big[ L^2(\mathbb{R}^3)_3 \big]^{\otimes n_3}_{S},
\]

with $n = n_0 + n_1 + n_2 + n_3$.  We define 
\[
\eta \Phi = \sum \limits_{n_0+ n_1+ n_2 + n_3 = 0}^{\infty} 
(-1)^{n_0} \Phi^{(n_0)}\otimes \Phi^{(n_1)} \otimes \Phi^{(n_2)} \otimes \Phi^{(n_3)},
\] 
i.e. $\eta$ is a multiplication operator by a bounded measurable function on a direct sum measure space
and thus it is self-adjoint and bounded operator fulfilling $\eta \eta = \bold{1}$,
with the commutation rules (\ref{[eta,a]}). Note that $\eta$ being defined on the dense subspace
$(E)$ of $\Gamma(\mathcal{H}')$ has unique extension to a unitary and selfadjoint operator
on $\Gamma(\mathcal{H}')$, which we likewise denote by $\eta$.   

In order to show (\ref{proof-second}) note that for any $\widetilde{\varphi}$
such that $\sqrt{B}\widetilde{\varphi} \in \oplus L^2(\mathbb{R}^3)
= L^2(\mathbb{R}^3)_0 \oplus L^2(\mathbb{R}^3)_1 \oplus L^2(\mathbb{R}^3)_2 
\oplus L^2(\mathbb{R}^3)_3$,  the operator $a^+(\sqrt{B}\widetilde{\varphi})$
is equal to the sum 
\[
a^+ (\sqrt{B}\widetilde{\varphi})= a^+\big( (\sqrt{B}\widetilde{\varphi})^0 \big)
+ a^+ \big( (\sqrt{B}\widetilde{\varphi})^1 \big)
+ a^+\big( (\sqrt{B}\widetilde{\varphi})^2 \big)
+ a^+\big( (\sqrt{B}\widetilde{\varphi})^3 \big),
\]
of four commuting operators, where $(\sqrt{B}\widetilde{\varphi})^\mu$ is the function having 
all components zero with the exception
of the $\mu$-th component equal to the $\mu$-th component of $\sqrt{B}\widetilde{\varphi}$.
By the commutation rules (\ref{[eta,a]}) it follows that
\[
\eta a^+\big( (\sqrt{B}\widetilde{\varphi})^0 \big) = - a^+\big( (\sqrt{B}\widetilde{\varphi})^0 \big) \eta
= a^+\big( - (\sqrt{B}\widetilde{\varphi})^0 \big) \eta, 
\]
and
\[
\eta a^+\big( (\sqrt{B}\widetilde{\varphi})^k \big) =  a^+ \big( (\sqrt{B}\widetilde{\varphi})^k \big)\eta,
\,\,\, k = 1,2,3;
\]
and thus, 
\begin{equation}\label{[eta,a(j')]}
\eta a^+(\sqrt{B}\widetilde{\varphi}) = a^+(\mathfrak{J}_{\bar{p}} \sqrt{B} \widetilde{\varphi}) \eta,
\end{equation}
where $\mathfrak{J}_{\bar{p}}$ is the operator of multiplication by the constant matrix (\ref{J-barp}).
On the other hand for any $\underset{1}{\widetilde{\varphi}}, \ldots \underset{n}{\widetilde{\varphi}}$ 
such that $\sqrt{B}\underset{1}{\widetilde{\varphi}}, 
\ldots \sqrt{B}\underset{n}{\widetilde{\varphi}} \, \in \, \oplus L^2(\mathbb{R}^3)$ we have 
\begin{multline}\label{(aphi...omega,aphi...omega)}
\frac{1}{n!}\Bigg( a^+(\sqrt{B} \underset{1}{\widetilde{\varphi}'}) \, 
a^+(\sqrt{B}\underset{2}{\widetilde{\varphi}'})
\, \dots \, 
a^+(\sqrt{B}\underset{n}{\widetilde{\varphi}'}) \Omega , \,\,
a^+(\sqrt{B}\underset{1}{\widetilde{\varphi}}) \, 
a^+(\sqrt{B}\underset{2}{\widetilde{\varphi}})
\, \dots \, 
a^+(\sqrt{B}\underset{n}{\widetilde{\varphi}}) \Omega \Bigg) \\ =
\Bigg( \bigg[\sqrt{B}\underset{1}{\widetilde{\varphi}'} \, \otimes \ldots 
\otimes \, \sqrt{B}\underset{n}{\widetilde{\varphi}'} \bigg]_S,
 \bigg[\sqrt{B}\underset{1}{\widetilde{\varphi}} \, \otimes \ldots 
\otimes \, \sqrt{B}\underset{n}{\widetilde{\varphi}} \bigg]_S \Bigg) \\ =
\frac{1}{n!} \sum \limits_{\pi} 
\big(\sqrt{B}\underset{1}{\widetilde{\varphi}'}, 
\sqrt{B}\underset{\pi(1)}{\widetilde{\varphi}} \big)_{{}_{\oplus L^2(\mathbb{R}^3)}}
\cdot \ldots \cdot
\big(\sqrt{B} \underset{n}{\widetilde{\varphi}'}, 
\sqrt{B} \underset{\pi(n)}{\widetilde{\varphi}} \big)_{{}_{\oplus L^2(\mathbb{R}^3)}},
\end{multline}
where the sum is over all permutations $\pi$ of the first $n$ natural numbers. 
Joining this with (\ref{[eta,a(j')]}) we obtain for any 
$\underset{1}{\widetilde{\varphi}}, \ldots \underset{n}{\widetilde{\varphi}}$ 
such that $\sqrt{B}\underset{1}{\widetilde{\varphi}}, 
\ldots \sqrt{B}\underset{n}{\widetilde{\varphi}} \, \in \, \oplus L^2(\mathbb{R}^3)$
the following equality 
\begin{multline*}
\frac{1}{n!}\Bigg( a^+(\sqrt{B} \underset{1}{\widetilde{\varphi}'}) \, 
a^+(\sqrt{B}\underset{2}{\widetilde{\varphi}'})
\, \dots \, 
a^+(\sqrt{B}\underset{n}{\widetilde{\varphi}'}) \Omega , \,\, \eta \,
a^+(\sqrt{B}\underset{1}{\widetilde{\varphi}}) \, 
a^+(\sqrt{B}\underset{2}{\widetilde{\varphi}})
\, \dots \, 
a^+(\sqrt{B}\underset{n}{\widetilde{\varphi}}) \Omega \Bigg) \\ =
\frac{1}{n!}\Bigg( a^+(\sqrt{B} \underset{1}{\widetilde{\varphi}'}) \, 
a^+(\sqrt{B}\underset{2}{\widetilde{\varphi}'})
\, \dots \, 
a^+(\sqrt{B}\underset{n}{\widetilde{\varphi}'}) \Omega , \,\, 
a^+(\mathfrak{J}_{\bar{p}}\sqrt{B}  \underset{1}{\widetilde{\varphi}}) \, 
a^+(\mathfrak{J}_{\bar{p}}\sqrt{B}  \underset{2}{\widetilde{\varphi}})
\, \dots \, 
a^+(\sqrt{B} \mathfrak{J}_{\bar{p}} \underset{n}{\widetilde{\varphi}}) \Omega \Bigg) \\ =
\Bigg( \bigg[\sqrt{B}\underset{1}{\widetilde{\varphi}'} \, \otimes \ldots 
\otimes \, \sqrt{B}\underset{n}{\widetilde{\varphi}'} \bigg]_S,
 \bigg[\mathfrak{J}_{\bar{p}}\sqrt{B}   \underset{1}{\widetilde{\varphi}} \, \otimes \ldots 
\otimes \, \mathfrak{J}_{\bar{p}}\sqrt{B}  \underset{n}{\widetilde{\varphi}} \bigg]_S \Bigg) \\ =
\frac{1}{n!} \sum \limits_{\pi} 
\big(\sqrt{B}\underset{1}{\widetilde{\varphi}'}, 
\mathfrak{J}_{\bar{p}}\sqrt{B}  \underset{\pi(1)}{\widetilde{\varphi}} \big)_{{}_{\oplus L^2(\mathbb{R}^3)}}
\cdot \ldots \cdot
\big(\sqrt{B} \underset{n}{\widetilde{\varphi}'}, 
\mathfrak{J}_{\bar{p}}\sqrt{B}  \underset{\pi(n)}{\widetilde{\varphi}} \big)_{{}_{\oplus L^2(\mathbb{R}^3)}} \\ =
\Bigg( \bigg[\sqrt{B}\underset{1}{\widetilde{\varphi}'} \, \otimes \ldots 
\otimes \, \sqrt{B}\underset{n}{\widetilde{\varphi}'} \bigg]_S,
 \bigg[\mathfrak{J}_{\bar{p}}\sqrt{B}   \underset{1}{\widetilde{\varphi}} \, \otimes \ldots 
\otimes \, \mathfrak{J}_{\bar{p}}\sqrt{B}  \underset{n}{\widetilde{\varphi}} \bigg]_S \Bigg) \\ =
\frac{1}{n!} \sum \limits_{\pi} 
\big(\sqrt{B}\underset{1}{\widetilde{\varphi}'}, 
\mathfrak{J}_{\bar{p}}\sqrt{B}  \underset{\pi(1)}{\widetilde{\varphi}} \big)_{{}_{\oplus L^2(\mathbb{R}^3)}}
\cdot \ldots \cdot
\big(\sqrt{B} \underset{n}{\widetilde{\varphi}'}, 
\mathfrak{J}_{\bar{p}}\sqrt{B}  \underset{\pi(n)}{\widetilde{\varphi}} \big)_{{}_{\oplus L^2(\mathbb{R}^3)}} \\ =
\frac{1}{n!} \sum \limits_{\pi} 
\big(\underset{1}{\widetilde{\varphi}'}, 
\sqrt{B}\mathfrak{J}_{\bar{p}}\sqrt{B}  \underset{\pi(1)}{\widetilde{\varphi}} \big)_{{}_{\oplus L^2(\mathbb{R}^3)}}
\cdot \ldots \cdot
\big( \underset{n}{\widetilde{\varphi}'}, 
\sqrt{B}\mathfrak{J}_{\bar{p}}\sqrt{B}  \underset{\pi(n)}{\widetilde{\varphi}} \big)_{{}_{\oplus L^2(\mathbb{R}^3)}} \\ =
\frac{1}{n!} \sum \limits_{\pi} 
\big(\underset{1}{\widetilde{\varphi}'}, 
\mathfrak{J}_{\bar{p}}  \underset{\pi(1)}{\widetilde{\varphi}} 
\big)_{{}_{\oplus L^2(\mathbb{R}^3, \ud \mu |_{{}_{\mathscr{O}_{\bar{p}}}})}}
\cdot \ldots \cdot
\big( \underset{n}{\widetilde{\varphi}'}, 
\mathfrak{J}_{\bar{p}} \underset{\pi(n)}{\widetilde{\varphi}} 
\big)_{{}_{\oplus L^2(\mathbb{R}^3, \ud \mu |_{{}_{\mathscr{O}_{\bar{p}}}})}},
\end{multline*}
where the last equality follows from (\ref{BJBJ=1}) (where $\ud \mu |_{{}_{\mathscr{O}_{\bar{p}}}}$
stands for the measure (\ref{inn-Lop-1-space}) on the orbit $\mathscr{O}_{\bar{p}} = \mathscr{O}_{(1,0,0,1)}$
in the coordinates $\boldsymbol{\p}$). By definition and properties (\ref{Kr-inn-Lop-1-space})
and (\ref{Kr-inn-Lop-1-space'}) of the Krein product in $\mathcal{H'}$ the last expression
(after the last equality sign) is equal to
\[
\frac{1}{n!} \sum \limits_{\pi} 
\big(\sqrt{B}\underset{1}{\widetilde{\varphi}'}, 
\sqrt{B} \mathfrak{J}' \underset{\pi(1)}{\widetilde{\varphi}} \big)_{{}_{\oplus L^2(\mathbb{R}^3)}}
\cdot \ldots \cdot
\big(\sqrt{B} \underset{n}{\widetilde{\varphi}'}, 
\sqrt{B} \mathfrak{J}' \underset{\pi(n)}{\widetilde{\varphi}} \big)_{{}_{\oplus L^2(\mathbb{R}^3)}}.
\] 
Comparing this with (\ref{(aphi...omega,aphi...omega)}) we see that 
\begin{multline*}
\frac{1}{n!}\Bigg( a^+(\sqrt{B} \underset{1}{\widetilde{\varphi}'}) \, 
a^+(\sqrt{B}\underset{2}{\widetilde{\varphi}'})
\, \dots \, 
a^+(\sqrt{B}\underset{n}{\widetilde{\varphi}'}) \Omega , \,\, \eta \,
a^+(\sqrt{B}\underset{1}{\widetilde{\varphi}}) \, 
a^+(\sqrt{B}\underset{2}{\widetilde{\varphi}})
\, \dots \, 
a^+(\sqrt{B}\underset{n}{\widetilde{\varphi}}) \Omega \Bigg) \\ =
\frac{1}{n!}\Bigg( a^+(\sqrt{B} \underset{1}{\widetilde{\varphi}'}) \, 
a^+(\sqrt{B}\underset{2}{\widetilde{\varphi}'})
\, \dots \, 
a^+(\sqrt{B}\underset{n}{\widetilde{\varphi}'}) \Omega , \,\, 
a^+(\sqrt{B} \mathfrak{J}' \underset{1}{\widetilde{\varphi}}) \, 
a^+(\sqrt{B} \mathfrak{J}' \underset{2}{\widetilde{\varphi}})
\, \dots \, 
a^+(\sqrt{B} \mathfrak{J}' \underset{n}{\widetilde{\varphi}}) \Omega \Bigg), 
\end{multline*}
for all 
$\underset{1}{\widetilde{\varphi}}, \underset{1}{\widetilde{\varphi}}', 
\ldots \underset{n}{\widetilde{\varphi}}, \underset{n}{\widetilde{\varphi}'}$ 
such that $\sqrt{B}\underset{1}{\widetilde{\varphi}}, \sqrt{B}\underset{1}{\widetilde{\varphi}}'
\ldots \sqrt{B}\underset{n}{\widetilde{\varphi}}, \sqrt{B} \underset{n}{\widetilde{\varphi}}' 
\, \in \, \oplus L^2(\mathbb{R}^3)$.
Because the linear span of vectors of the form
\[
a^+( \underset{1}{\widetilde{\varphi}'}) \, 
a^+(\underset{2}{\widetilde{\varphi}'})
\, \dots \, 
a^+(\underset{n}{\widetilde{\varphi}'}) \Omega, \,\,\,
\underset{k}{\widetilde{\varphi}}' \in \mathcal{H'}
\]
is dense in $\Gamma(\mathcal{H'})$, then the equality (\ref{proof-second}) is thereby proved. 

Having obtained this we proceed further in computing the integral kernel operator representation
of the operator valued distribution (\ref{Pauli-Jordan}) exactly as in the process of computing
 (\ref{Dvarphi'D*varphi}) and we show that (\ref{Pauli-Jordan}) defines an operator valued distribution 
which can be represented as an integral\footnote{Understood formally as a pointwise limit of actual Pettis (or even 
Bochner in this case) integral 
operators.} with the distributional kernel $i g^{\mu \nu} D_0 (x-y)$. In fact we show a slightly stronger result
that this integral representation holds for 
$\underset{1}{\varphi}, \underset{2}{\varphi} \in \widetilde{\mathcal{S}_{{}_{A^{(4)}}}(\mathbb{R}^4)}
= \mathcal{S}^{00}(\mathbb{R}^4)$ in 
(\ref{Pauli-Jordan})
and the distribution defined by (\ref{Pauli-Jordan}) understood over the test function space 
$\widetilde{\mathcal{S}_{{}_{A^{(4)}}}(\mathbb{R}^4)} \otimes \widetilde{\mathcal{S}_{{}_{A^{(4)}}}(\mathbb{R}^4)} = \mathcal{S}^{00}(\mathbb{R}^4) \otimes \mathcal{S}^{00}(\mathbb{R}^4)$ with the domain $\mathscr{D} = (E)$ and nuclear topology of uniform convergence on $\mathscr{L}\big((E), (E)\big)$. To this end note that for any elements $\underset{1}{\varphi}, \underset{2}{\varphi} \in \mathcal{S}^{00}(\mathbb{R}^4)$ the ordinary Fourier transform is defined as follows
\[
\widetilde{\underset{k}{\varphi}}(p) = \int \limits_{\mathbb{R}^4} \underset{k}{\varphi}(x) e^{ip \cdot x} \, \ud^4 x,
\,\,\, k = 1,2,
\]
(for distributional solutions $\underset{k}{\varphi} \in \mathcal{E}$
of d'Alembert equation the Fourier transform $\widetilde{\underset{k}{\varphi}}$ 
is concentrated on the orbit $\mathscr{O}_{1,0,0,1}$ and induce in a canonical way 
ordinary functions on the orbit which belong to
$E = \mathcal{S}^{0}(\mathbb{R}^3) = \mathcal{S}_{A^{(3)}}(\mathbb{R}^3)$). 

Using (\ref{[eta,a(j')]}) the commutation relations
(\ref{[D',D]}) (equivalently the commutation relations for $a(\widetilde{\varphi}')$ and 
$a(\widetilde{\varphi})^+$), the properties (\ref{Kr-inn-Lop-1-space})
and (\ref{Kr-inn-Lop-1-space'}) of the Krein product in $\mathcal{H'}$ and the formula
(\ref{BJBJ=1})  we easily compute (where on the right-hand side the sign of restriction to the cone
$\mathscr{O}_{1,0,0,1}$ at the arguments $\underset{i}{\widetilde{\varphi}}$ has been 
omitted for simplicity)\footnote{Note also that we have used real $\underset{i}{\varphi}$, 
but this restriction is irrelevant
and is introduced only for the simplicity of notation, which otherwise will have to use the additional
superscript $\check{(\cdot)}$ in the argument of the annihilation operators.}
\begin{multline*}
\Big[ A\big(\underset{1}{\varphi}\big) \, , \, A\big(\underset{2}{\varphi} \big) \Big]
=  \Big[a(\sqrt{B}\underset{1}{\widetilde{\varphi}}) 
+ \eta a(\sqrt{B}\underset{1}{\widetilde{\varphi}})^+\eta \, , 
\, a(\sqrt{B}\underset{2}{\widetilde{\varphi}}) 
+ \eta a(\sqrt{B}\underset{2}{\widetilde{\varphi}})^+\eta \big) \Big] \\ =
\Big[a(\sqrt{B}\underset{1}{\widetilde{\varphi}})\, , 
 \eta a(\sqrt{B}\underset{2}{\widetilde{\varphi}})^+\eta \big) \Big] +
\Big[ \eta a(\sqrt{B}\underset{1}{\widetilde{\varphi}})^+\eta \, , 
\, a(\sqrt{B}\underset{2}{\widetilde{\varphi}}) \Big]  \\ =
\Big[a(\sqrt{B} \,\underset{1}{\widetilde{\varphi}})\, , 
 a(\mathfrak{J}_{\bar{p}}\sqrt{B} \, \underset{2}{\widetilde{\varphi}})^+\big) \Big] +
\Big[ a(\mathfrak{J}_{\bar{p}}\sqrt{B} \, \underset{1}{\widetilde{\varphi}})^+ \, , 
\, a(\sqrt{B} \, \underset{2}{\widetilde{\varphi}}) \Big]  \\
= \bigg\{ \big(\underset{1}{\widetilde{\varphi}}, 
\mathfrak{J}_{\bar{p}}  \underset{2}{\widetilde{\varphi}} 
\big)_{{}_{\oplus L^2(\mathbb{R}^3, \ud \mu |_{{}_{\mathscr{O}_{\bar{p}}}})}}
- \big(\underset{2}{\widetilde{\varphi}}, 
\mathfrak{J}_{\bar{p}}  \underset{1}{\widetilde{\varphi}} 
\big)_{{}_{\oplus L^2(\mathbb{R}^3, \ud \mu |_{{}_{\mathscr{O}_{\bar{p}}}})}} \bigg\} \, \bold{1} \\
= \bigg\{ \big(\underset{1}{\widetilde{\varphi}}, 
\mathfrak{J}'  \underset{2}{\widetilde{\varphi}} \big)
- \big(\underset{2}{\widetilde{\varphi}}, 
\mathfrak{J}'  \underset{1}{\widetilde{\varphi}} \big) \bigg\} \, \bold{1},
\end{multline*}
where $(\cdot , \cdot )$ in the last expression is the inner product in $\mathcal{H'}$ and thus, 
with $(\cdot , \mathfrak{J}' \cdot)$ in this expression equal to the Krein-product in $\mathcal{H'}$.

On the other hand we have
\begin{multline*}
\int \limits_{\mathbb{R}^4 \times \mathbb{R}^4} \underset{1}{\varphi_\mu} \otimes \underset{2}{\varphi_\nu}
\big(x,y\big) ig^{\mu \nu} D_0 (x-y) \, \ud^4 x \ud^4 y \\ =
- \int \limits_{\mathbb{R}^4}\, \ud^4 x \, \int \limits_{\mathbb{R}^4} \, \ud^4 y \,
\underset{1}{\varphi_\mu} \big(x \big) \underset{2}{\varphi_\nu} \big(y\big) ig^{\mu \nu} \,
\int \limits_{\mathbb{R}^3} \frac{\ud^3 p}{2 p^0 (\boldsymbol{\p})} e^{-ip \cdot (x-y)} \\ +
\int \limits_{\mathbb{R}^4}\, \ud^4 x \, \int \limits_{\mathbb{R}^4} \, \ud^4 y \,
\underset{1}{\varphi_\mu} \big(x \big) \underset{2}{\varphi_\nu} \big(y\big) ig^{\mu \nu} \,
\int \limits_{\mathbb{R}^3} \frac{\ud^3 p}{2 p^0 (\boldsymbol{\p})} e^{ip \cdot (x-y)} \\ =
\big(\underset{1}{\widetilde{\varphi}}, 
\mathfrak{J}'  \underset{2}{\widetilde{\varphi}} \big)
- \big(\underset{2}{\widetilde{\varphi}}, 
\mathfrak{J}'  \underset{1}{\widetilde{\varphi}} \big)
\end{multline*}
where $(\cdot , \cdot )$ is the inner product in $\mathcal{H'}$ and 
 $(\cdot , \mathfrak{J}' \cdot)$  is the Krein-product in $\mathcal{H'}$, and where in the last expression
the sign of restriction to the cone $\mathscr{O}$ at $\underset{i}{\widetilde{\varphi}}$
has been omitted for simplicity.
Therefore, 
\begin{equation}\label{[A,A']=Pauli-Jordan}
\Big[ A\big(\underset{1}{\varphi}\big) \, , \, A\big(\underset{2}{\varphi} \big) \Big]  =
\bigg\{ \int \limits_{\mathbb{R}^4 \times \mathbb{R}^4} \underset{1}{\varphi_\mu} \otimes \underset{2}{\varphi_\nu}
\big(x,y\big) ig^{\mu \nu} D_0 (x-y) \, \ud^4 x \ud^4 y  \bigg\} \bold{1},
\end{equation}
which was to be shown. The continuity assertions follow from general theory of integral kernel operators
\cite{hida}, \cite{obata-book}, and from the continuity of the restriction map
\[
\mathcal{S}^{0}(\mathbb{R}^4) \ni \widetilde{\varphi} \,\,\, \longrightarrow \,\,\,
\widetilde{\varphi}|_{{}_{\mathscr{O}_{1,0,0,1}}} \in
\mathcal{S}^{0}(\mathbb{R}^3)
\]
proved in Subsection \ref{Lop-on-E}.

The derivation $\frac{\partial A}{\partial x^\nu}$ of the operator valued distribution (in the white noise sense) $\mathcal{S}^{00}(\mathbb{R}^4) \ni \varphi \mapsto A(\varphi) \in 
\mathscr{L}\big((E), (E)\big)$ is defined in the ordinary distributional manner 
\[
\mathcal{S}^{00}(\mathbb{R}^4) \ni \varphi \mapsto \frac{\partial A}{\partial x^\nu}\Big(\varphi \Big)
= A \Big( - \frac{\partial \varphi}{\partial x^\nu} \Big).
\]
Because
\[
\varphi(x) = \int \limits_{\mathbb{R}^4} \widetilde{\varphi}(p) e^{-ip \cdot x} \,\,
\ud^4 p,
\]
then the Fourier transform of 
\[
g^{\nu \mu}\frac{\partial^2 \varphi}{\partial x^\nu \partial x^\mu} 
\]
is equal 
\[
p^\mu p_{\mu} \widetilde{\varphi} = p \cdot p \widetilde{\varphi}.
\]
Therefore, 
\[
g^{\mu \nu} \frac{\partial^2 A}{\partial x^\mu \partial x^\nu}\Big(\varphi \Big) =
A\Big(g^{\mu \nu} \frac{\partial^2 \varphi}{\partial x^\mu \partial x^\nu} \Big) = \,
0, \,\,\, 
\varphi \in \mathcal{S}^{00}(\mathbb{R}^4),
\]
because for $\varphi \in \mathcal{S}^{00}(\mathbb{R}^4)$
\[
g^{\mu \nu} \frac{\partial^2 A}{\partial^\nu \partial x^\nu}\Big(\varphi \Big)
= a(\sqrt{B} \, (p \cdot p \, \check{\widetilde{\varphi}'})|_{{}_{\mathscr{O}}}) + 
\eta a(\sqrt{B} \, (p \cdot p \, \widetilde{\varphi}')|_{{}_{\mathscr{O}}})^+ \eta
\]
and $(p \cdot p \, \overline{\check{\widetilde{\varphi}'}})|_{{}_{\mathscr{O}}}$,
$(p \cdot p \, \widetilde{\varphi}')|_{{}_{\mathscr{O}}}$ are identically equal to zero;
or equivalently because the Fourier transform $\widetilde{g^{\mu \nu} \partial_\mu \partial_\nu\varphi}$ is 
identically equal to zero on the orbit $\mathscr{O}_{(1,0,0,1)} = \{p, p^\nu p_\nu = p\cdot p = 0\}$. 

All the above statements could have been formulated in the position picture,
with the test spaces $\mathbb{E}$ and $(\mathbb{E})$ instead of $E$ and $(E)$,
using the property 
\[
\int \limits_{\mathbb{R}^3} \widetilde{\varphi}_\mu(\boldsymbol{\p}) \,\, \partial_{\boldsymbol{\p}}^{\mu}
\,\,\, \ud^3 \boldsymbol{\p}
= \int \limits_{\mathbb{R}^3} \big(\mathscr{F}^{-1}\widetilde{\varphi}_\mu \big)(\boldsymbol{\x}) \,\,
 \partial_{\boldsymbol{\x}}^{\mu}
\,\,\, \ud^3 \boldsymbol{\x}, \,\,\,
\widetilde{\varphi} \in E.
\]
It follows from the above Theorem of this Subsection that the multiplication operator $M_{2p^0}$ by the function
$2p^0: \boldsymbol{\p} \mapsto 2 (\boldsymbol{\p} \cdot \boldsymbol{\p})^{1/2}$ and the ordinary three-dimensional
Fourier transform $\mathscr{F}$ and their inverses are continuous as operators $E \to E$
and $\mathbb{E} \to E$ respectively and that in particular the space of functions 
$\boldsymbol{\x} \mapsto \varphi(t,\boldsymbol{\x})$ with $\varphi \in \mathcal{E}$ and fixed $t \in \mathbb{R}$ is naturally isomorphic to the space 
$\mathbb{E}$ with every $\varphi \in \mathcal{E}$ which may be treated as one parameter family of elements of $\mathbb{E}$
and having the property that the derivation with respect to the parameter is another family of elements $\mathbb{E}$.  
The operator valued distributions over the test function space
$\mathcal{E}$ may be treated as distributions  over the  space $\mathbb{E}$ of functions of three variables. 
In particular for each fixed $t \in \mathbb{R}$
\begin{multline*}
D_{\sqrt{B} \widetilde{\varphi}} = 
\int \limits_{\mathbb{R}^3} \widetilde{\varphi}_\mu(\boldsymbol{\p}) \,\, \partial_{\boldsymbol{\p}}^{\mu}
\,\,\, \ud^3 \boldsymbol{\p} 
= \int \limits_{\mathbb{R}^3} \mathscr{F}^{-1}\big(\sqrt{B} M_{2p^0} M_{e^{itp^0}} \big) \mathscr{F} \,
\varphi_\mu (t, \boldsymbol{\x}) \,\, \partial_{\boldsymbol{\x}}^{\mu}
\,\,\, \ud^3 \boldsymbol{\x},
\end{multline*}
with the operator 
\[
 \mathscr{F}^{-1}\big(\sqrt{B} M_{2p^0} M_{e^{itp^0}} \big) \mathscr{F} 
\] 
acting on the function $\boldsymbol{\x} \mapsto \varphi (t, \boldsymbol{\x}) \in \mathbb{E}$.
The just mentioned one-parameter families of elements of $\mathbb{E}$  (with the time as the parameter) 
would be sufficient e.g. for the treatment of the translation subgroup, 
Let $\mathbb{E}$ be realized as the space of functions 
$\boldsymbol{\x} \mapsto \varphi (t=0, \boldsymbol{\x})$ for $\varphi \in \mathcal{E}$,
i.e. by the restrictions to $t = 0$ of $\varphi \in \mathcal{E}$.
In particular 
let $\varphi|_{t=0} \in \mathbb{E}$, then
writing $T_a$, $a \in \mathbb{R}$, for the representors of time translations in the {\L}opusza\'nski representation 
and in the conjugate {\L}opusza\'nski representation, we have
\[
T_a \varphi|_{t=0}  \in \mathbb{E}, \,\,\, a \in \mathbb{R},
\]
which follows because $\boldsymbol{\x} \mapsto T_a \varphi(0,\boldsymbol{\x})
= \varphi(t=a, \boldsymbol{\x}) \in \mathbb{E}$. It is easily seen that $T_a$
induces unitary transform in $\oplus L^2(\mathbb{R}^3, \mathbb{C}) = L^2(\mathbb{R}^3, \mathbb{C}^4)$
therefore, the investigation of the translation subgroup may be performed within the
Gelfand triple $\mathbb{E} \subset \oplus L^2(\mathbb{R}^3, \mathbb{C}) \subset \mathbb{E}^*$
and its lifting to the Fock space.
But in the investigation of the full double cover of the Poincar\'e group parametric
families of elements in $\mathbb{E}$ with the values of the parameter in the group would be 
necessary and over the Hilbert spaces with more complicated inner products, therefore we prefer
using the momentum picture. 

Note however that the operator valued distribution
\[
\varphi \mapsto A(\varphi) = a(\sqrt{B} \, \overline{\check{\widetilde{\varphi}}}) 
+ \eta a(\sqrt{B} \, \widetilde{\varphi})^+ \eta,
\]
in the white noise sense, with $\varphi$ ranging over the space
$\mathcal{E} \subset \mathcal{H}''$ (equivalently with the distributional Fourier transforms
of $\varphi \in \mathcal{E}$ concentrated on the orbit $\mathscr{O}_{1,0,0,1}$
and determining uniquely ordinary functions $\widetilde{\varphi}$ on the orbit $\mathscr{O}_{1,0,0,1}$,
which belong to $E$) is not yet equal to the local field in Wightman sense.
Indeed, the elements of $\mathcal{H}''$ compose a space of specific distributional solutions of the 
massless wave equation
which forms an indecomposable representation Krein space of the double covering
of the Poincar\'e group and are far not flexible enough to contain e.g. smooth functions
of compact support in $\mathbb{R}^4$ (regarded as Minkowski spacetime). 
We may nonetheless consider a space of space-time test functions
$\varphi$ on $\mathbb{R}^4$ whose ordinary Fourier transform 
\[
\widetilde{\varphi}(p) = \int \limits_{\mathbb{R}^4} \varphi(x) e^{ip \cdot x} \, \ud^4 x
\]
after restriction to the orbit $\mathscr{O}_{(1,0,0,1)}$ belongs to $E$. From what we have shown
above it follows that we can choose  the elements $\varphi$ from the nuclear space
$\mathcal{S}^{00}(\mathbb{R}^4)$. From what we have already proved it follows that 
the map $\varphi \mapsto \widetilde{\varphi}|_{{}_{\mathscr{O}_{1,0,0,1}}}$ with 
$\widetilde{\varphi}|_{{}_{\mathscr{O}_{1,0,0,1}}}$ equal to the restriction to the orbit 
$\mathscr{O}_{1,0,0,1}$ of the Fourier transform $\widetilde{\varphi}$, is continuous
as the operator $\mathcal{S}^{00}(\mathbb{R}^4) \to E$.
In this case we may define
\[
\varphi \mapsto A(\varphi) = a(\sqrt{B} \, \overline{\check{\widetilde{\varphi}}}|_{{}_{\mathscr{O}_{1,0,0,1}}}) 
+ \eta a(\sqrt{B} \, \widetilde{\varphi}|_{{}_{\mathscr{O}_{1,0,0,1}}})^+ \eta,
\]
as the four-potential field  -- operator valued distribution in the white noise sense of Berezin-Hida.

Moreover, from what we have already shown it easily follows that 
the local field fulfils the Wightman axioms of \cite{wig}, Chap. 3, with the obvious modifications
that our representation of the double covering of the Poincar\'e group is replaced with
a Krein-isometric representation (although unitarity of the representation of the translation subgroup is preserved)
and with the test function space equal $\mathcal{S}^{00}(\mathbb{R}^4)$  in this massless gauge field case
instead of the ordinary Schwartz space $\mathcal{S}(\mathbb{R}^4)$ (correct only for massive nongauge 
fields if the fields are required to be build within the Berezin-Hida white noise formalism) 
and domain $\mathscr{D} = (E)$. But in fact the field $A$ we have constructed using white noise is a much more
subtle object than (so modified) Wightman zero mass field $A$, and in particular is useful in the perturbative
causal approach -- contrary to Wightman fields. 

Although the space  $\mathcal{S}^{00}(\mathbb{R}^4)$ is not flexible enough
to contain any smooth function on $\mathbb{R}^4$ with compact support (except the trivial zero function),
it nonetheless is sufficient for the splitting of causal homogeneous distributions,
which is sufficient for the causal perturbative method, compare Subsection \ref{splitting}. Indeed,
note that pairing and commutation singular functions corresponding to zero mass fields 
(which require the test space to be the space $\mathcal{S}^{00}(\mathbb{R}^4)$ of scalar, vector, 
e.t.c. valued functions depending on the field) are always homogeneous, and for splitting of homogeneous 
distribitions (and their tensor products) the space $\mathcal{S}^{00}(\mathbb{R}^4)$ (and its tensor products)
is pretty sufficient.

Although usefulness of the white noise construction of free fields for the causal perturbative approach
is the main motivation for us, we also mention that it also allows rigorous formulation and proof 
of the generalization of the first Noether
theorem in the realm of free quantum fields. Wightman approach is not effective
for this task. The main trouble comes from the unclear averaging of Wightman-Garding 
``Wick product fields'' over Cauchy surfaces in construction of the conserved currents.
Some (not entirely mathematically controllable) constructions for the massive fields 
have been undertaken with a restricted success,
compare e.g. \cite{Maison-Reeh-1}, \cite{Maison-Reeh-2}, \cite{Requardt},
but the zero mass gauge fields seem to be intractable within the Wightman-G{\aa}rding approach.  
In the next Subsetion we show how the white noise approach allows to solve this problem
even for gauge zero mass field such as the electromagnetic potential field.

\subsection{Bogoliubov-Shirkov quantization postulate for free fields. 
The case of the electromagnetic quantum four-vector potential field}\label{BSH}

Let us give the heuristic formulation of the Postulate in the original form 
as stated in \cite{Bogoliubov_Shirkov}, 
Chap. 2, \S 9.4 (in 1980 Ed.): \emph{The operators for the energy-momentum four-vector
$\boldsymbol{P}$, and the angular momentum tensor $\boldsymbol{M}$, the charge
$\boldsymbol{Q}$, and so on, which are the generators of the corresponding 
symmetry transformations of state vectors, can be expressed in terms of the operator functions 
of the fields by the same relations as in classical field theory with the operators
arranged in the normal order}.

Here we confine ourselves to the case of the free electromagnetic field and to the case of 
translation subgroup with the generators expressed (via Emmy Noether theorem) by the 
spatial integrals of the energy
momentum tensor components $T^{0 \mu}$. The case of massive fields has been proved even in a 
slightly more general context of general Wightman fields fulfilling the Wightmann axioms of \cite{wig} Chap. 3.3.1,
compare eg. \cite{Requardt}. 

Let $T^{0 \mu}$ be the components of energy-momentum tensor for the free classical electromagnetic field
$A^{\mu}$ corresponding to translations via Emmy Noether theorem (compare \cite{Bogoliubov_Shirkov})
expressed in terms of derivatives $\partial_\nu A^\mu$. According to this theorem the spatial 
(or more general integral over any space-like surface) 
\[
\int T^{00} \, \ud^3 \boldsymbol{\x}= - \frac{1}{2}  \int g_{\mu \nu} \sum \limits_{\rho} 
\partial_\rho A^\mu \partial_\rho A^\nu  \, \ud^3 \boldsymbol{\x}, \,\,\,
\int T^{0k} \, \ud^3 \boldsymbol{\x} =   \int g_{\mu \nu} 
\partial_0 A^\mu \partial_k A^\nu  \, \ud^3 \boldsymbol{\x}, \,\,\,
\] 
is equal to the conserved integral corresponding to the translational symmetry, i.e. energy-momentum components
of the field. We replace the classical field in the above integral formally by the quantum fields and arrange them in the normal order.  
Thus, we are going to show that 
\[
\int \boldsymbol{:} T^{0\mu} \boldsymbol{:} \, \ud^3 \boldsymbol{\x} = \boldsymbol{P}^\mu = d\Gamma(P^\mu),
\] 
where $P^\mu$, $\mu = 0,1,2,3$, are the translation generators of the conjugate {\L}opusza\'nski represenation  
$\big[WU^{{}_{(1,0,0,1)}{\L}}W^{-1}\big]^{*-1} = \mathfrak{J'} \big[WU^{{}_{(1,0,0,1)}{\L}}W^{-1}\big]
\mathfrak{J'}$ and thus with 
$\boldsymbol{P}^\mu = d\Gamma(P^\mu)$, $\mu = 0,1,2,3$, equal to the generators of translations of the representation 
\[
\Gamma\Big(\big[WU^{{}_{(1,0,0,1)}{\L}}W^{-1}\big]^{*-1}\Big) 
= \Gamma(\mathfrak{J'}) \, \Gamma\Big( WU^{{}_{(1,0,0,1)}{\L}}W^{-1} \Big) \, \Gamma (\mathfrak{J'}),
\]
of $T_4 \circledS SL(2, \mathbb{C})$ in the Fock space $\Gamma(\mathcal{H}')$.
Because $P^\mu$ commute with $\mathfrak{J}'$ and thus $d\Gamma(P^\mu)$ commute with $\Gamma(\mathfrak{J'}) = \eta$,
then $P^\mu$ are at the same time the translation generators of the {\L}opusza\'nski representation
$WU^{{}_{(1,0,0,1)}{\L}}W^{-1}$ and $d\Gamma(P^\mu)$ are also the translation generators 
of the representation 
\[
\Gamma\Big(WU^{{}_{(1,0,0,1)}{\L}}W^{-1}\Big). 
\]
Equivalently we will show that 
\begin{equation}\label{Bogoliubov-Postulate-1}
\boxed{- \frac{1}{2}  \int \boldsymbol{:} g_{\mu \nu} \sum \limits_{\rho} 
\partial_\rho A^\mu \partial_\rho A^\nu \boldsymbol{:}  \, \ud^3 \boldsymbol{\x}  = d\Gamma(P^0)},
\end{equation} 
\begin{equation}\label{Bogoliubov-Postulate-2}
\boxed{\int g_{\mu \nu} 
\boldsymbol{:} \partial_0 A^\mu \partial_k A^\nu \, \boldsymbol{:}  \, \ud^3 \boldsymbol{\x} = d\Gamma(P_k)},
\end{equation}
where
\[
d\Gamma(P_k) = g_{k \nu} d\Gamma(P^\nu) = - d\Gamma(P^k).
\]

We have to give a rigorous meaning to the integral on the left-hand side of (\ref{Bogoliubov-Postulate-1})
and (\ref{Bogoliubov-Postulate-2}) as well defined continuous operator
$(E) \to (E)$ equal to the translation generator $d\Gamma(P^\mu)$ on $(E)$, i.e. on the core domain of
a self-adjoint operator $d\Gamma(P^\mu)$. Thus, the operator on the left will have a selfadjoint extension equal to $ d\Gamma(P^\mu)$.  
The whole point about the Postulate is that the operators $\boldsymbol{P}^\mu = d\Gamma(P^\mu)$ may be computed in 
therms of Wick polynomials in free fields -- operator valued distributions to which we know how to apply the 
perturbative series in the sense of Bogoliubov-Epstein-Glaser. 
In  the course of the proof of the Postulate the white noise calculus is havily used. We proceed in two steps. 
In the first step we show that for each $\mu = 0,1,2,3$, 
there exist a distribution $\kappa^\mu \in E \otimes E^*$ 
such that the corresponding integral kernel operator  $\Xi_{1,1}(\kappa^\mu)$
 (eq. (\ref{Berezin-integral-p})) is equal to $\boldsymbol{P}^\mu = d\Gamma(P^\mu)$. Then we give the definition 
of the integral on the left-hand side of (\ref{Bogoliubov-Postulate-1}) and (\ref{Bogoliubov-Postulate-2})
and show that it is equal to $\Xi_{1,1}(\kappa^\mu)$.

In the investigation of the representation of the double covering of the Poincar\'e group we could restrict ourselves 
to the following Gelfand triples: 
\begin{equation}\label{E-L2-E*}
\left. \begin{array}{ccccc}              E         & \subset & \oplus L^2(\mathbb{R}^3) & \subset & E^*        \\
                               \downarrow \uparrow &         & \downarrow \uparrow      &         & \downarrow \uparrow  \\
                        \mathbb{E} & \subset & \oplus L^2(\mathbb{R}^3) & \subset & \mathbb{E}^* \end{array}\right.,
\end{equation}
and their liftings
\begin{equation}\label{(E)-(L2)-(E)*}
\left. \begin{array}{ccccc}              (E)  &    \subset    & L^2(E_{\mathbb{R}}^*, \mu; \mathbb{C})
 \cong \Gamma\Big( \oplus L^2(\mathbb{R}^3)\Big) & 
\subset & (E)^*        \\
                               \downarrow \uparrow &         & \downarrow \uparrow      &         & \downarrow \uparrow  \\
                        (\mathbb{E}) & \subset & L^2(\mathbb{E}_{\mathbb{R}}^*, \mu; \mathbb{C})
\cong \Gamma\Big(\oplus L^2(\mathbb{R}^3) \Big) & \subset & (\mathbb{E})^*. 
\end{array}\right.
\end{equation}
But the triple $\mathbb{E} \subset \oplus L^2(\mathbb{R}^3) \subset \mathbb{E}^*$ works
smoothly only for the translation subgroups, the analysis of the other subgroups with the use of this triple
is not very elegant. The triple $E \subset \oplus L^2(\mathbb{R}^3) \subset E^*$ works well and moreover
produces simple formulas due to simple expressions for the pairings induced by the simple inner product formula
in $\oplus L^2(\mathbb{R}^3, \mathbb{C}) = L^2(\mathbb{R}^3, \mathbb{C}^4)$ -- this is why we are using it. 
 Although the inner product (\ref{inn-Lop-1-space}) in $\mathcal{H}'$ have the additional ``weight'' operator 
$B$  and thus the pairings $\langle \cdot, \cdot \rangle$ and $\langle \langle \cdot, \cdot \rangle \rangle$ 
which it induces are given by slightly more complicated formulas, the Gelfand triple 
(constructed in the standard way with the help of the operator 
$\sqrt{B}^{\, -1} \, A \, \sqrt{B}$ in $\mathcal{H}'$)
\begin{equation}\label{Gelfand-triple-H'}
E \subset  \mathcal{H'}  \subset  E^*        
\end{equation}
and its lifting
\begin{equation}\label{Gelfand-triple-H'-lifting}
(E) \,\,\,\,\,\,\,\,\,\,\,\,\,\,\,\, \subset \,\,\,\,\,\,\,\,\,\,\,\,\,\,\,\,  L^2(E_{\mathbb{R}}^*, \mu; \mathbb{C})
\cong \Gamma(\mathcal{H'}) \,\,\,\,\,\,\,\,\,\,\,\,\,\,\,\, \subset \,\,\,\,\,\,\,\,\,\,\,\, (E)^* 
\end{equation}
seems conceptually better suited for the investigation of the action of the double covering of the Poincar\'e group
in the Fock space $\Gamma(\mathcal{H}')$ (of course the Gaussian measures 
${\mu}$ in (\ref{(E)-(L2)-(E)*}) and (\ref{Gelfand-triple-H'-lifting})
depend on the inner product in the respective one particle Hilbert spaces). 
We use it in the proof of the Bogoliubov Postulate to illustrate the interconnection between the formalisms based on differend Gelfand triples. 

Note that although the nuclear space $E$ and $(E)$
are common for the Gelfand triples $E \subset \oplus L^2(\mathbb{R}^3) \subset (E)^*$
and $E \subset \mathcal{H'} \subset (E)^*$
(and their liftings to different Fock spaces) the element $\Phi \in (E)$ 
common for the two Fock spaces $\Gamma (\mathcal{H}')$
and $\Gamma\Big(\oplus L^2(\mathbb{R}^3)\Big)$ has different representations as two different functions 
given by the Wiener-It\^o-Segal decomposition (\ref{Phi=white-expansion}), because the representation as a function
on $E^*$ depends on the pairing $\langle \cdot , \cdot  \rangle$ induced by the inner product
in the one particle Hilbert space. And in the two cases of the Gelfand triples and their liftings
the respective one particle Hilbert spaces are different, so that the operators
$D_{\widetilde{\varphi}}$, $D_{\widetilde{\varphi}}^*$  (whenever well defined as operators
$(E) \to (E)$) regarded as operators in the Fock spaces are different. 
Likewise, the generalized operators $D_{\delta^\nu_{\boldsymbol{\p}}}$,
$D_{\delta^\nu_{\boldsymbol{\p}}}^*$, induce different operator valued distributions 
in the two indicated cases of Gelfand triples. (Although we express the final formulas 
in terms of the canonical set of generalized operators with the canonical commutation relations.)  
Because the only difference in the application of the two mentioned Gelfand triples is 
of technical character and reduces to
the replacement of the pairings in the formulas of \cite{hida} or in the above formulas 
by the pairings $\langle \cdot, \cdot \rangle$, $\langle \langle \cdot, \cdot \rangle \rangle$
induced by the inner product (\ref{inn-Lop-1-space}) in $\mathcal{H}'$, we only list
here the final formulas leaving all details as an exercise. In order to simplify notation we write
$\partial_{\boldsymbol{\p}} = D_{\delta_{\boldsymbol{\p}}}$ for the tuple 
$(\partial_{\boldsymbol{\p}}^0, \ldots , \partial_{\boldsymbol{\p}}^3) 
= (D_{\delta_{\boldsymbol{\p}}^0}, \ldots , D_{\delta_{\boldsymbol{\p}}^3})$ of operators, and the dependence of these
operators on the inner product (\ref{inn-Lop-1-space}) or on the ``weight'' operator $B$ in 
(\ref{inn-Lop-1-space}) will be reflected by the overset character $B$: $\overset{B}{\partial_{\boldsymbol{\p}}}$.
The notation $B \overset{B}{\partial_{\boldsymbol{\p}}}$, $\sqrt{B} \overset{B}{\partial_{\boldsymbol{\p}}}$,
e.t.c. is self-evident. If the overset character is absent then the symbol refers to the respective generalized
operator obtained with the help of the triple $ E \subset \oplus L^2(\mathbb{R}^3) \subset E^* $
and its lifting.

 We have the following formulas when the pairing $\langle \cdot, \cdot \rangle$ induced
by the inner product (\ref{inn-Lop-1-space}): $(\cdot, B \cdot)_{{}_{\oplus L^2(\mathbb{R}^3)}}$ and
when the Gelfand triple (\ref{Gelfand-triple-H'}) and its lifting (\ref{Gelfand-triple-H'-lifting}) are used :
\[
\begin{split}
\overset{B}{D_{\zeta}} 
= \int \limits_{\mathbb{R}^3} \zeta(\boldsymbol{\p}) 
 \overset{B}{\partial_{\boldsymbol{\boldsymbol{\p}}}}
\, \ud^3 \boldsymbol{\p}, \,\,\,
\overset{B}{D_{\zeta}^*} 
= \int \limits_{\mathbb{R}^3} \zeta(\boldsymbol{\p}) 
 \overset{B}{\partial_{\boldsymbol{\boldsymbol{\p}}}^*}
\, \ud^3 \boldsymbol{\p}, \,\,\, \zeta \in E^* \\
\big[ \overset{B}{D_{\zeta}^*} \, , \,\,  \overset{B}{D_{\xi}}  \Big] = \langle \zeta, \xi \rangle 
= (\overline{\zeta}, B \xi)_{{}_{\oplus L^2(\mathbb{R}^3)}}, \,\,\, \zeta, \xi \in \mathcal{H}', \\
\Big[ \overset{B}{\partial_{\boldsymbol{\boldsymbol{\p}}}^*} \, , \,\, 
\overset{B}{\partial_{\boldsymbol{\boldsymbol{\p}'}}} \Big] = B \, \delta(\boldsymbol{\p} - \boldsymbol{\p}'),
\,\,\, \overset{B}{\partial_{\boldsymbol{\boldsymbol{\p}}}} = \sqrt{B} \, \partial_{\boldsymbol{\boldsymbol{\p}}},
\,\,\, \overset{B}{\partial_{\boldsymbol{\boldsymbol{\p}}}^*} 
= \sqrt{B} \, \partial_{\boldsymbol{\boldsymbol{\p}}}^*.  
\end{split}
\]
In particular 
\[
\begin{split}
\overset{B}{D_{\zeta}} = \int \limits_{\mathbb{R}^3} \zeta(\boldsymbol{\p}) 
\sqrt{B} \, \partial_{\boldsymbol{\boldsymbol{\p}}}
\, \ud^3 \boldsymbol{\p}, \,\,\,
\overset{B}{D_{\zeta}^*} 
= \int \limits_{\mathbb{R}^3} \zeta(\boldsymbol{\p}) 
\sqrt{B} \, \partial_{\boldsymbol{\boldsymbol{\p}}}^*
\, \ud^3 \boldsymbol{\p}, \,\,\, \zeta \in E^*, \\
\overset{B}{D_{\widetilde{\varphi}}} = D_{\sqrt{B} \, \widetilde{\varphi}} = a'(\widetilde{\varphi}) 
\,\,\,\,\,\,\,\,\,\,\,\,\,\,\,
\overset{B}{D_{\widetilde{\varphi}}^*} = D_{\sqrt{B} \, \widetilde{\varphi}}^* = a'(\widetilde{\varphi})^+,
\,\,\,  \widetilde{\varphi} \in E.
\end{split} 
\]

By the above theorem of this Subsection the representors of the {\L}opusza\'nski representation 
$WU^{{}_{(1,0,0,1)}{\L}}W^{-1}$ are continuous as operators $E \to E$. In particular this holds
for the translation subgroup representors of this representation equal to the translation subgroup representors
of the conjugate {\L}opusza\'nski representation $\big[WU^{{}_{(1,0,0,1)}{\L}}W^{-1}\big]^{*-1}$.
And because the translation representors in both of the representations commute with the fundamental symmetry
$\mathfrak{J}'$, then in both representations the translation subgroup is unitary and not only Krein-isometric.
Therefore, the translation subgroup in the {\L}opusza\'nski representation and in the conjugate {\L}opusza\'nski 
representation compose the subgroup of the Yoshizawa group $U(E;\mathcal{H}')$.  
The Yoshizawa group $U(E;\mathcal{H}')$ is the group of unitary operators on $\mathcal{H}'$ which induce homeomorphisms
of the test function space $E$ with respect to the nuclear topology of $E$. In other wards the translation representors in the {\L}opusza\'nski and conjugate {\L}opusza\'nski  representation compose automorphisms of the Gelfand triple 
$E \subset \mathcal{H}' \subset E^*$. Moreover,
any one parameter subgroup $\{T_\theta\}_{\theta \in \mathbb{R}}$ of translations in the 
{\L}opusza\'nski representation and in the conjugate {\L}opusza\'nski representation is differentiable,
i.e. $\lim_{\theta \to 0} (T_{\theta}\xi - \xi)/\theta = X\xi$ converges in $E$.  Let us consider the one parameter 
subgroup of translations along the $\mu$-th axis and write in this case $X^\mu$ for $X$,
where $X^\mu$ is the operator $M_{ip^\mu}$ of multiplication by the function 
$\boldsymbol{\p} \to ip^\mu(\boldsymbol{\p})$, and where 
$(p^0(\boldsymbol{\p}), \ldots p^3(\boldsymbol{\p}))
 = (\sqrt{\boldsymbol{\p} \cdot \boldsymbol{\p}}, \boldsymbol{\p}) \in \mathscr{O}_{(1,0,0,1)}$. 
Existence of the limit is equivalent to 
\begin{multline}\label{T-theta-differentiability}
\lim \limits_{\theta \to 0} \bigg| \frac{T_{\theta}\xi - \xi}{\theta} - X^\mu \xi  \bigg|_k^2 \\ =
\lim \limits_{\theta \to 0}
\int \bigg( \frac{\boldsymbol{A}^k \Big(e^{i\theta p^\mu} -1 
- i \theta p^\mu \Big)\xi(\boldsymbol{\p})}{\theta} \, , \,\, 
\frac{B \, \boldsymbol{A}^k \Big(e^{i\theta p^\mu} -1 
- i \theta p^\mu \Big)\xi(\boldsymbol{\p})}{\theta}
\bigg)_{{}_{\mathbb{C}^4}} \,\, 
\ud^3 \boldsymbol{\p} \,\, = 0, \\
 k = 0, 1, 2, \ldots, \,\,\, \xi \in E,
\end{multline}
where $p^\mu$, $\mu = 0,1,2,3$, in the exponent are the functions 
$\boldsymbol{\p} \mapsto (p^\mu(\boldsymbol{\p})) = (\sqrt{\boldsymbol{\p} \cdot \boldsymbol{\p}}, \boldsymbol{\p})$
and where $\boldsymbol{A}$ is the operator $\sqrt{B}^{\, -1} \, A \, \sqrt{B}$ and $A$
is the operator $A$ used in the construction of the Gelfand triple $E \subset \oplus L^2(\mathbb{R}^3)
\subset E^*$ and has been constructed above. Explicit calculation shows that (\ref{T-theta-differentiability}) is fulfilled. 
Therefore, $\{T_\theta\}_{\theta \in \mathbb{R}}$ is differentiable subgroup and by the Banach-Steinhaus
theorem the linear operators $X^\mu$, $\mu = 0,1,2,3$, 
are continuous as operators $E \to E$ and finally by Proposition 3.1 of \cite{hida}
every such subgroup is regular in the sense of \cite{hida}, \S  3.

For every operator $X$ which is continuous as the operator $E \to E$  we define $\Gamma(X)$
and $d \Gamma(X)$ on $(E)$. Let $\Phi \in (E)$ be represented as a function by the 
 Wiener-It\^o-Segal decomposition (\ref{Phi=white-expansion}) corresponding to the Gelfand triples
(\ref{Gelfand-triple-H'}) and (\ref{Gelfand-triple-H'-lifting}), i.e. with the pairing 
$\langle \cdot, \cdot \rangle$ in (\ref{Phi=white-expansion}) induced by the inner product
$(\cdot, B \cdot)_{{}_{\oplus L^2(\mathbb{R}^3)}}$ in $\mathcal{H}'$. Then we define  
\[
(\Gamma(X)\Phi)(\zeta) 
= \sum \limits_{n=0}^{\infty} \langle \, \boldsymbol{:} \zeta^{\otimes n} \boldsymbol{:} \,
, \,\, X^{\otimes n} f_n \, \rangle , \,\,\,\, \zeta \in E^*;
\]
\[
(d\Gamma(X)\Phi)(\zeta) 
= \sum \limits_{n=0}^{\infty} n \langle \, \boldsymbol{:} \zeta^{\otimes n} \boldsymbol{:} \,
, \,\, (X \otimes I^{\otimes (n-1)}) \, f_n \, \rangle , \,\,\,\, \zeta \in E^*.
\]
In this case it is easily seen that the Theorem 4.1 of \cite{hida} is applicable
and that $\{\Gamma(T_\theta ) \}_{\theta \in \mathbb{R}}$, with the generator $X^\mu$, is a regular 
one parameter subgroup with the generator $d\Gamma(X^\mu)$ which continuously maps $(E)$
into itself. 

In this situation it is not difficult to see that  for each $\mu =0,1,2,3$, the proof of Proposition 4.2 
and Theorem 4.3 of \cite{hida} is applicable to any of the one parameter translation subgroups
in the {\L}opusza\'nski representation and in the conjugate {\L}opusza\'nski representation, in particular
for any of the translation subgroup along the direction of the 
$\mu$-th axis, $\mu =0,1,2,3$, there exists  
a symmetric distribution $\kappa^\mu \in E \otimes E^*$ such that 
\begin{equation}\label{dGammaX=Berexin-int}
 d \Gamma(X^\mu) = \Xi_{1,1}(\kappa^\mu) = \int \limits_{\mathbb{R}^3 \times \mathbb{R}^3}
\kappa^\mu(\boldsymbol{\p}', \boldsymbol{\p}) \,\, 
 \overset{B}{\partial_{\boldsymbol{\boldsymbol{\p}'}}^*} 
\overset{B}{\partial_{\boldsymbol{\boldsymbol{\p}}}}  
\,\, \ud^3 \boldsymbol{\p}' \ud^3 \boldsymbol{\p},
\end{equation}
and $\kappa^\mu \in E \otimes E^*$ fulfills
\begin{equation}\label{kappa-distribution-P}
\langle \kappa^\mu, \zeta \otimes \xi \rangle =  \langle \zeta, X^\mu \xi \rangle,
\,\,\, \zeta, \xi \in E. 
\end{equation}
Because the pairings $\langle \cdot, \cdot \rangle$ in the formula are induced by 
the inner product $(\cdot, B \cdot)_{{}_{\oplus L^2(\mathbb{R}^3)}}$ in $\mathcal{H}'$
and the operator $B$ is equal to pointwise multiplication by real symmetric matrix,
and because $X^\mu$ is the operator of multiplication by $ip^\mu$, we have
\[
(\overline{\zeta}, B \, X^\mu \xi)_{{}_{\oplus L^2(\mathbb{R}^3)}} = \langle \zeta, X^\mu \xi \rangle
= \langle X^\mu \xi , \zeta \rangle = \langle \xi , X^\mu \zeta \rangle,
\,\,\, \zeta, \xi \in E, 
\]
so that
\[
\langle \kappa^\mu, \zeta \otimes \xi \rangle = \langle \kappa^\mu, \xi \otimes \zeta \rangle,
\,\,\, \zeta, \xi \in E, 
\] 
and $\kappa^\mu$ is indeed symmetric. Therefore, the right-hand side of (\ref{dGammaX=Berexin-int})
is equal
\[
\int \limits_{\mathbb{R}^3 \times \mathbb{R}^3}
\kappa^\mu(\boldsymbol{\p}', \boldsymbol{\p}) \,\, 
 \overset{B}{\partial_{\boldsymbol{\boldsymbol{\p}'}}^*} 
\overset{B}{\partial_{\boldsymbol{\boldsymbol{\p}}}}  
\,\, \ud^3 \boldsymbol{\p}' \ud^3 \boldsymbol{\p}
= \int \limits_{\mathbb{R}^3 \times \mathbb{R}^3}
\kappa^\mu(\boldsymbol{\p}', \boldsymbol{\p}) B \,\,
\partial_{\boldsymbol{\boldsymbol{\p}'}}^*  
\partial_{\boldsymbol{\boldsymbol{\p}}}  
\,\, \ud^3 \boldsymbol{\p}' \ud^3 \boldsymbol{\p}.
\]
On the other hand the pairing $\langle \cdot, \cdot \rangle$
on left hand side of (\ref{kappa-distribution-P}) expressed in terms
of the kernel $\kappa^\mu (\boldsymbol{\p}', \boldsymbol{\p})$ is likewise induced by the inner
product $(\cdot, B \cdot)_{{}_{\oplus L^2(\mathbb{R}^3)}}$ in $\mathcal{H}'$. Therefore,
we have
\[
\langle \kappa^\mu, \zeta \otimes \xi \rangle = 
\int \limits_{\mathbb{R}^3 \times \mathbb{R}^3}
\kappa^\mu(\boldsymbol{\p}', \boldsymbol{\p}) \,\, 
B\zeta(\boldsymbol{\p}') B \xi( \boldsymbol{\p})
\,\, \ud^3 \boldsymbol{\p}' \ud^3 \boldsymbol{\p}.
\]  
Joining this with (\ref{kappa-distribution-P}) we obtain
\[
\big( \kappa^\mu(\boldsymbol{\p}', \boldsymbol{\p}) B\big)_{\nu \, \lambda} 
= i p^\mu(\boldsymbol{\p}) \delta_{\mu \lambda} \delta(\boldsymbol{\p}'- \boldsymbol{\p}).
\]

Because the operator $P^\mu = -iX^\mu$, $\mu = 0,1,2,3$, acts in the Hilbert space
$\mathcal{H}'$ as operator $M_{{}_{p^\mu}}$ of multiplication  by $p^\mu$ (with $p = (p^0, p^1, p^2, p^3) \in
\mathscr{O}_{(1,0,0,1)}$), then  
\begin{equation}\label{d(Gamma(P)}
\boldsymbol{P}^\mu = d \Gamma(P^\mu) = \int \limits_{\mathbb{R}^3 \times \mathbb{R}^3}
p^\mu(\boldsymbol{\p}) \,\, \delta_{\nu \lambda} \delta(\boldsymbol{\p}'- \boldsymbol{\p}) \,\,\, 
\partial_{\boldsymbol{\p}'}^{\nu \, *} \partial_{\boldsymbol{\p}}^{\lambda}
\,\,\, \ud^3 \boldsymbol{\p}' \ud^3 \boldsymbol{\p},
\end{equation}
which is customary to be written as
\[
\boldsymbol{P}^\mu = d \Gamma(P^\mu) = \sum \limits_\nu \int \limits_{\mathbb{R}^3}
p^\mu(\boldsymbol{\p}) \,\, 
\partial_{\boldsymbol{\p}}^{\nu \, *} \partial_{\boldsymbol{\p}}^{\nu}
\,\,\, \ud^3 \boldsymbol{\p}.
\]
Both operators $d \Gamma(P^\mu)$ and $\Xi_{1,1}(-i\kappa^\mu)$ transform (continuously)
the nuclear, and thus perfect, space $(E)$ into itself and both being equal and symmetric 
on $(E)$ have self-adjoint extension to self-adjoint operator
in the Fock space $\Gamma(\mathcal{H}')$, again by the classical criterion of \cite{Riesz-Szokefalvy}
(p. 120 in Russian Ed. 1954). In general the criterion of Riesz-Sz\"okefalvy-Nagy
does not exclude existence of more than just one self-adjoint extension, but
in our case it is unique. Indeed, because for each $\mu=0,1,2,3$, the one-parameter unitary
group generated by $d \Gamma(P^\mu)$ leaves invariant the dense nuclear space $(E)$, then by general
theory, e.g. Chap. 10.3., it follows that  $d \Gamma(P^\mu)$ with domain $(E)$ is essentially self adjoint
(admits unique self adjoint extension).

It is not difficult to see that the method of Hida, Obata and Sait\^o with the Gelfand triples
(\ref{Gelfand-triple-H'-lifting}) and (\ref{Gelfand-triple-H'-lifting}) is applicable to the representors
of any one parameter subgroup of $T_4 \circledS SL(2, \mathbb{C})$, and the result analogous to 
(\ref{dGammaX=Berexin-int}) may be obtained with  $d \Gamma(X) = \Xi_{1,1}(\kappa)$, $\kappa \in E \otimes E^*$,
transforming $(E)$ continuously into itself. The additional work is required in proving
existence of Krein-self-adjoint extension of $d \Gamma(-iX) = \Xi_{1,1}(-i\kappa)$,
which requires a generalization of the Riesz-Sz\"okefalvy-Nagy criterion for existence of the ordinary
self adjoint extension. 

Now let us back to the Gelfand triple
$E \subset  \oplus L^2(\mathbb{R}^3) \subset  E^*$ in the momentum picture and its lifting, as the pairings 
$\langle \cdot, \cdot \rangle$ and $\langle\langle \cdot, \cdot \rangle\rangle$ 
and the corresponding formulas are simpler in this case.

Now we give a rigorous definition of the spatial integral of the local conserved current equal to the 
energy momentum tensor components regarded as Wick ordered polynomials of free fields on the left hand side of 
the formulas (\ref{Bogoliubov-Postulate-1}) and (\ref{Bogoliubov-Postulate-2}).
Note that the Wick integrands in (\ref{Bogoliubov-Postulate-1}) and (\ref{Bogoliubov-Postulate-2}), evaluated at fixed space-time points, 
are well defined integral kernel operators transforming
continuously the Hida space into its strong dual, by the results of Subsection \ref{OperationsOnXi}. Nonetheless
here we proceed largely independently of the results of Subsection \ref{OperationsOnXi}, and treat the free fields and their Wick products
only as evaluated at fixed space-time points. The definition of the free electromagnetic potential
field understood as an integral kernel operator with vector valued kernel in the sense of \cite{obataJFA}, we postpone to the following Subsections.
We proceed in this manner because we would like to provide not only a proof of the Bogoliubov-Shirkov Quantization Postulate
but at the same we would like to provide a justification of the Rules of Subsection \ref{OperationsOnXi}, so our method will be based on the 
Pettis integrals of the Hida operators, with the free fields and their Wick polynomials understood as evaluated at fixed space-time
points and well defined as generalized operators transforming the Hida space into its strong dual. 
This at the same time gives the connection of the quantum electromagnetic four-potential free field $A$ constructed 
here with the one
used in the standard physical literature.
To this end let $\varphi$ be any real valued element of $\mathcal{S}^{00}(\mathbb{R}^4)$ 
and let $\Phi$ be any element of $(E)$. We consider the Fourier transform $\widetilde{\varphi}$ in $\mathbb{R}^4$
\[
\widetilde{\varphi}(p) = \int \limits_{\mathbb{R}^4} \varphi(x) e^{ip \cdot x} \, \ud^4 x
\]
of $\varphi \in \mathcal{S}^{00}(\mathbb{R}^4; \mathbb{C}^4) = \widetilde{\mathcal{S}^{0}(\mathbb{R}^4; \mathbb{C}^4)}$ and note that 
$\eta: (E) \mapsto (E)$ is continuous as the second quantization of a continuous operator: $E \rightarrow E$,
compare \cite{hida}. 
It easily follows that the 
function\footnote{The map $\boldsymbol{\p} \mapsto \partial_{\boldsymbol{\p}}^{\lambda} \Phi$
is even Bochner strongly measurable (for definition compare \cite{Yosida}, Chap. V.5) being continuous with respect to each fixed $\| \cdot \|_{k}$ in $(E)_{k}$ and separably valued as $(E)_{k}$ is separable. Note that because $\Phi \in (E)$, and $(E) \subset (E)_k$ for each $k>0$, then
$\Phi \in (E)_k$ for each $k>0$. Therefore, the firs summand belongs to $(E)_k$ for each $k>0$,
thus it belongs to the intersection $\cap_{k>0}(E)_k = (E)$.}
 (summation
with respect to $\mu,\nu$) 
\begin{multline}\label{Bochner-integrable}
\mathbb{R}^3 \times \mathbb{R}^4 \ni (\boldsymbol{\p}, x) \mapsto 
\bigg\{
\frac{1}{\sqrt{2 p^0(\boldsymbol{\p})}}\sqrt{B(\boldsymbol{\p}, p^0(\boldsymbol{\p}))}^{\mu}_{\lambda}
 \varphi_{\mu}(x) e^{-ip \cdot x} 
\partial_{\boldsymbol{\p}}^{\lambda} \Phi  \\
+ \frac{1}{\sqrt{2 p^0(\boldsymbol{\p})}}\sqrt{B(\boldsymbol{\p}, p^0(\boldsymbol{\p}))}^{\mu}_{\lambda}
 \varphi_{\mu}(x) e^{ip \cdot x} 
\eta \partial_{\boldsymbol{\p}}^{* \, \lambda} \eta \Phi \bigg\} \in (E)^*,
\end{multline} 
where $p$ in the exponent is equal $\big(p^0(\boldsymbol{\p}),\boldsymbol{\p} \big) 
= \big((\boldsymbol{\p} \cdot \boldsymbol{\p})^{1/2}, \boldsymbol{\p}\big)$, 
is Pettis-inegrable\footnote{The integrand in the first summand is even Bochner strongly measurable on the product measure space $\mathbb{R}^3 \times \mathbb{R}^4$ as a function $\mathbb{R}^3 \times \mathbb{R}^4 \to (E)_k$, for each fixed $k>0$ with $\Phi \in (E)_k \subset (E)$ and bounded measurable $\varphi$.}. 
The following iterated integral 
\begin{multline*}
\int \limits_{\mathbb{R}^3} \, \ud^3 p \, \int \limits_{\mathbb{R}^4} \, \ud^4 x \, \bigg\{
\frac{1}{\sqrt{2 p^0(\boldsymbol{\p})}}\sqrt{B(\boldsymbol{\p}, p^0(\boldsymbol{\p}))}^{\mu}_{\lambda}
\varphi_{\mu}(x) e^{-ip \cdot x} 
\partial_{\boldsymbol{\p}}^{\lambda} \Phi \bigg\} \\
+ \eta \, \int \limits_{\mathbb{R}^3} \, \ud^3 p \, \int \limits_{\mathbb{R}^4} \, \ud^4 x \, \bigg\{
\frac{1}{\sqrt{2 p^0(\boldsymbol{\p})}}\sqrt{B(\boldsymbol{\p}, p^0(\boldsymbol{\p}))}^{\mu}_{\lambda}
\varphi_{\mu}(x) e^{ip \cdot x} 
\partial_{\boldsymbol{\p}}^{\lambda \,*} \, \eta \,  \Phi \bigg\} \\
= \int \limits_{\mathbb{R}^3} \, \ud^3 p \, \int \limits_{\mathbb{R}^4} \, \ud^4 x \, \bigg\{
\frac{1}{\sqrt{2 p^0(\boldsymbol{\p})}}\sqrt{B(\boldsymbol{\p}, p^0(\boldsymbol{\p}))}^{\mu}_{\lambda}
\varphi_{\mu}(x) e^{-ip \cdot x} 
\partial_{\boldsymbol{\p}}^{\lambda} \Phi  \\
+ \frac{1}{\sqrt{2 p^0(\boldsymbol{\p})}}\sqrt{B(\boldsymbol{\p}, p^0(\boldsymbol{\p}))}^{\mu}_{\lambda}
\varphi_{\mu}(x) e^{ip \cdot x} 
\eta \partial_{\boldsymbol{\p}}^{\lambda \, *} \eta \Phi \bigg\} \\
= \int \limits_{\mathbb{R}^3} \, \ud^3 p \, \bigg\{
\frac{1}{\sqrt{2 p^0(\boldsymbol{\p})}}\sqrt{B(\boldsymbol{\p}, p^0(\boldsymbol{\p}))}^{\mu}_{\lambda}
\check{\widetilde{\varphi}}_{\mu}|_{{}_{\mathscr{O}}}(\boldsymbol{\p})
\partial_{\boldsymbol{\p}}^{\lambda} \Phi \\
+ \frac{1}{\sqrt{2 p^0(\boldsymbol{\p})}}\sqrt{B(\boldsymbol{\p}, p^0(\boldsymbol{\p}))}^{\mu}_{\lambda} \,
\widetilde{\varphi}_{\mu}|_{{}_{\mathscr{O}}}(\boldsymbol{\p})
\eta \partial_{\boldsymbol{\p}}^{\lambda \, *} \eta \Phi \bigg\} \\
= D_{{}_{\sqrt{B}\check{\widetilde{\varphi}}|_{{}_{\mathscr{O}}}}} \Phi + 
\eta D_{{}_{\sqrt{B} \,\, \widetilde{\varphi}|_{{}_{\mathscr{O}}}}}^* \eta \Phi 
= a(\sqrt{B} \,\, \overline{\check{\widetilde{\varphi}}|_{{}_{\mathscr{O}}}}) \Phi 
+ \eta a^+ (\sqrt{B} \,\, \widetilde{\varphi}|_{{}_{\mathscr{O}}}) \eta \Phi = A(\varphi) \Phi
\end{multline*}
exists as Pettis integral (the first summand exists even in the Bochner sense as an element of $(E) \subset (E_{k})$ in the Hilbert space 
$(E_{k})$ for some $k$). In the above formula $\widetilde{\varphi}_{\mu}|_{{}_{\mathscr{O}}}$ denotes the restriction
of the Fourier transform $\widetilde{\varphi}_{\mu}$ in $\mathbb{R}^4$ of an element $\varphi_{\mu} \in
\mathcal{S}^{00}(\mathbb{R}^4; \mathbb{C})$ to the light cone $\mathscr{O}_{(1,0,0,1)}$. We have inserted $\eta$
under the integral sign
because it is continuous as an operator $(E) \to (E)$. 
By Corollary 3.9 of \cite{Segal_Kunze} it follows that for each $\Psi \in (E)$ the function 
\begin{multline*}
\mathbb{R}^3 \times \mathbb{R}^4 \ni (\boldsymbol{\p}, x) \mapsto 
\bigg\langle \bigg\langle
\frac{1}{\sqrt{2 p^0(\boldsymbol{\p})}}\sqrt{B(\boldsymbol{\p}, p^0(\boldsymbol{\p}))}^{\mu}_{\lambda}
 \varphi_{\mu}(x) e^{-ip \cdot x} 
\partial_{\boldsymbol{\p}}^{\lambda} \Phi  \\
+ \frac{1}{\sqrt{2 p^0(\boldsymbol{\p})}}\sqrt{B(\boldsymbol{\p}, p^0(\boldsymbol{\p}))}^{\mu}_{\lambda}
 \varphi_{\mu}(x) e^{ip \cdot x} 
\eta \partial_{\boldsymbol{\p}}^{\lambda \, *} \eta \Phi, \, \Psi \bigg\rangle \bigg\rangle  \\ =
\frac{1}{\sqrt{2 p^0(\boldsymbol{\p})}}\sqrt{B(\boldsymbol{\p}, p^0(\boldsymbol{\p}))}^{\mu}_{\lambda}
 \varphi_{\mu}(x) e^{-ip \cdot x} \,
\big\langle \big\langle
\partial_{\boldsymbol{\p}}^{\lambda} \Phi, \Psi \big\rangle \big\rangle  \\
+ \frac{1}{\sqrt{2 p^0(\boldsymbol{\p})}}\sqrt{B(\boldsymbol{\p}, p^0(\boldsymbol{\p}))}^{\mu}_{\lambda}
 \varphi_{\mu}(x) e^{ip \cdot x} \,
\big\langle \big\langle
\eta \partial_{\boldsymbol{\p}}^{\lambda \, *} \eta \Phi, \, \Psi \big\rangle \big\rangle \in \mathbb{R}
\end{multline*} 
is measurable and absolutely integrable on the product measure space $\mathbb{R}^3 \times \mathbb{R}^4$
(note that by Lemma 2.1 of \cite{hida} and the continuity of $\eta = \Gamma(\mathfrak{J}'): (E) \rightarrow (E)$, the  functions
\[ 
\boldsymbol{\p} \mapsto \big\langle \big\langle
\partial_{\boldsymbol{\p}}^{\lambda} \Phi, \Psi \big\rangle \big\rangle \,\,\, \textrm{and} \,\,\,
\boldsymbol{\p} \mapsto \big\langle \big\langle
\eta \partial_{\boldsymbol{\p}}^{\lambda \, *} \eta \Phi, \, \Psi \big\rangle \big\rangle
\]
belong to $E = \mathcal{S}_{A'''}(\mathbb{R}^3) = \mathcal{S}^0 (\mathbb{R}^3)$, and thus because the operator of multiplication by any integer power of $r(\boldsymbol{\p}) = p^0(\boldsymbol{\p})$ is, by the first Lemma 
of Subsection \ref{diffSA}, continuous as operator
$\mathcal{S}^0 (\mathbb{R}^3) = \mathcal{S}_{A'''}(\mathbb{R}^3) \rightarrow \mathcal{S}_{A'''}(\mathbb{R}^3) = 
\mathcal{S}^0 (\mathbb{R}^3)$, the functions belong to 
$\mathcal{S}^0 (\mathbb{R}^3) = \mathcal{S}_{A'''}(\mathbb{R}^3)$)
By the classical Fubini theorem (\cite{Segal_Kunze}, Chap. 3.6, Corollary 3.6.2) 
\begin{multline*}
\int \limits_{\mathbb{R}^4} \, A^\mu(x) \, \varphi_\mu (x) \, \Phi \, \ud^4 x  \\
= \int \limits_{\mathbb{R}^4} \, \ud^4 x \, \int \limits_{\mathbb{R}^3} \, \ud^3 p \,  \bigg\{
\frac{1}{\sqrt{2 p^0(\boldsymbol{\p})}}\sqrt{B(\boldsymbol{\p}, p^0(\boldsymbol{\p}))}^{\mu}_{\lambda}
 \varphi_{\mu}(x) e^{-ip \cdot x} 
\partial_{\boldsymbol{\p}}^{\lambda} \Phi \bigg\} \\
+  \int \limits_{\mathbb{R}^4} \, \ud^4 x \, \int \limits_{\mathbb{R}^3} \, \ud^3 p \, \bigg\{
 \frac{1}{\sqrt{2 p^0(\boldsymbol{\p})}}\sqrt{B(\boldsymbol{\p}, p^0(\boldsymbol{\p}))}^{\mu}_{\lambda}
 \varphi_{\mu}(x) e^{ip \cdot x} 
\eta \, \partial_{\boldsymbol{\p}}^{\lambda \, *} \, \eta \,  \Phi \bigg\} \\
= a(\sqrt{B}\overline{\check{\widetilde{\varphi}}|_{{}_{\mathscr{O}}}}) \Phi 
+ \eta a^+ (\sqrt{B} \widetilde{\varphi}|_{{}_{\mathscr{O}}}) \eta \Phi = A(\varphi) \Phi,
\end{multline*}
where 
$p = \big((\boldsymbol{\p} \cdot \boldsymbol{\p})^{1/2}, \boldsymbol{\p}  \big) \in \mathscr{O}_{(1,0,0,1)}$
and 
\begin{multline*}
A^\mu(x)\Phi = \int \limits_{\mathbb{R}^3} \, \ud^3 p \, \bigg\{
\frac{e^{-ip\cdot x}}{\sqrt{2 p^0(\boldsymbol{\p})}}\sqrt{B(\boldsymbol{\p}, p^0(\boldsymbol{\p}))}^{\mu}_{\lambda}
\partial_{\boldsymbol{\p}}^{\lambda} \Phi \bigg\} \\
+  \int \limits_{\mathbb{R}^3} \, \ud^3 p \, \bigg\{
 \frac{e^{ip\cdot x}}{\sqrt{2 p^0(\boldsymbol{\p})}}\sqrt{B(\boldsymbol{\p}, p^0(\boldsymbol{\p}))}^{\mu}_{\lambda}
\eta \, \partial_{\boldsymbol{\p}}^{\lambda \, *} \, \eta \,  \Phi \bigg\} \\
\end{multline*}  
and where the integrals exist as Pettis 
integrals\footnote{The first summand in the above integrals exists even in the Bochner sense. Indeed, by 
the classical Bochner measurability 
criterion (Theorem 1 of Chap. V.5 of \cite{Yosida}) it follows that 
\begin{multline*}
\mathbb{R}^3 \times \mathbb{R}^4 \ni (\boldsymbol{\p}, x) \mapsto 
\bigg\|
\frac{1}{\sqrt{2 p^0(\boldsymbol{\p})}}\sqrt{B(\boldsymbol{\p}, p^0(\boldsymbol{\p}))}^{\mu}_{\lambda}
 \varphi_{\mu}(x) e^{-ip \cdot x} 
\partial_{\boldsymbol{\p}}^{\lambda} \Phi \bigg\|_{k} \in \mathbb{R}
\end{multline*} 
is measurable on the product measure space $\mathbb{R}^3 \times \mathbb{R}^4$. By the classical Fubini
theorem for scalar functions (\cite{Segal_Kunze}, Chap. 3.6, Corollary 3.6.2) it follows 
again by \cite{Yosida}, Chap. V.5 Theorem 1, that the first summand of the function (\ref{Bochner-integrable})
is Bochner integrable on the product measure space $\mathbb{R}^3 \times \mathbb{R}^4$. We can
therefore apply the Fubini theorem for Bochner integrable functions to the first summand of the function
(\ref{Bochner-integrable}), compare \cite{Dunford-Schwartz} or \cite{Bogdanowicz},
to obtain the above equality for the first summand in the sense of Bochner integrals.}
on the basis of Pettis theorem, compare Proposition 8.1 of \cite{HKPS}. 
Note however that although the first integral is an element
of $(E) \subset \Gamma(\mathcal{H'}) \subset (E)^*$ it is not the case for the last integral which 
in general is an element of $(E_{-k}) \subset (E)^*$ but not of the Fock space $\Gamma(\mathcal{H'})$. 
Therefore, we can write
\begin{multline}\label{q-A-B}
A^\mu(x) = \int \limits_{\mathbb{R}^3} \, \ud^3 p \, \bigg\{
\frac{1}{\sqrt{2 p^0(\boldsymbol{\p})}}\sqrt{B(\boldsymbol{\p}, p^0(\boldsymbol{\p}))}^{\mu}_{\lambda}
a^{\lambda} (\boldsymbol{\p}) e^{-ip\cdot x} \\
+  \frac{1}{\sqrt{2 p^0(\boldsymbol{\p})}}\sqrt{B(\boldsymbol{\p}, p^0(\boldsymbol{\p}))}^{\mu}_{\lambda} \,
\eta \, a^{\lambda}(\boldsymbol{\p})^+ \, \eta \, e^{ip\cdot x}  \bigg\} 
\end{multline}  
where 
$p = \big((\boldsymbol{\p} \cdot \boldsymbol{\p})^{1/2}, \boldsymbol{\p}  \big) \in \mathscr{O}_{(1,0,0,1)}$
and where the integral is understood as pointwisely defined on $(E)$ as Pettis integral 
(and the first summand even in Bochner sense) and thus defines a well defined operator $(E) \rightarrow (E)^{*}$.
But note that $A^{\mu}(x)\Phi$
is not an element of the Fock space (except for $\Phi = 0$) and $A^\mu(x)$ is not well defined as operator
in the Fock space.
Similarly, using the Lemma 2.1 of \cite{hida} and our first Lemma of Subsection \ref{diffSA}  on the continuity of multiplication operators
by the (integer or fractional) power of $r = \sqrt{\boldsymbol{\p}\cdot \boldsymbol{\p}}$: 
$S^0 (\mathbb{R}^3) = \mathcal{S}_{A'''}(\mathbb{R}^3) \rightarrow \mathcal{S}_{A'''}(\mathbb{R}^3) = S^0 (\mathbb{R}^3))$,
and our 5-th Lemma of Subsection \ref{SA=S0} on the equivalence of norms, we easily show that the following operators $(E) \rightarrow (E)^*$ are well defined pointwisely on $(E)$ as Pettis integrals:
\begin{multline*}
\partial_{0}A^\mu(x) = \int \limits_{\mathbb{R}^3} \, \ud^3 p \, \bigg\{
\frac{-ip^0(\boldsymbol{\p})}{\sqrt{2 p^0(\boldsymbol{\p})}}\sqrt{B(\boldsymbol{\p}, p^0(\boldsymbol{\p}))}^{\mu}_{\lambda}
a^{\lambda} (\boldsymbol{\p}) e^{-ip\cdot x} \\
+  \frac{ip^0(\boldsymbol{\p})}{\sqrt{2 p^0(\boldsymbol{\p})}}\sqrt{B(\boldsymbol{\p}, p^0(\boldsymbol{\p}))}^{\mu}_{\lambda} \,
\eta \, a^{\lambda}(\boldsymbol{\p})^+ \, \eta \, e^{ip\cdot x}  \bigg\} 
\end{multline*}  
and
\begin{multline*}
\partial_k A^\mu(x) = \int \limits_{\mathbb{R}^3} \, \ud^3 p \, \bigg\{
\frac{ip^k}{\sqrt{2 p^0(\boldsymbol{\p})}}\sqrt{B(\boldsymbol{\p}, p^0(\boldsymbol{\p}))}^{\mu}_{\lambda}
a^{\lambda} (\boldsymbol{\p}) e^{-ip\cdot x} \\
+  \frac{-ip^k}{\sqrt{2 p^0(\boldsymbol{\p})}}\sqrt{B(\boldsymbol{\p}, p^0(\boldsymbol{\p}))}^{\mu}_{\lambda} \,
\eta \, a^{\lambda}(\boldsymbol{\p})^+ \, \eta \, e^{ip\cdot x}  \bigg\}; 
\end{multline*} 
or more precisely, for each $\Phi \in (E)$ the inegrals
\begin{multline*}
\partial_{0}A^\mu(x) \Phi = \int \limits_{\mathbb{R}^3} \, \ud^3 p \, \bigg\{
\frac{-ip^0(\boldsymbol{\p})}{\sqrt{2 p^0(\boldsymbol{\p})}}\sqrt{B(\boldsymbol{\p}, p^0(\boldsymbol{\p}))}^{\mu}_{\lambda}
a^{\lambda} (\boldsymbol{\p}) \, \Phi \, e^{-ip\cdot x} \\
+  \frac{ip^0(\boldsymbol{\p})}{\sqrt{2 p^0(\boldsymbol{\p})}}\sqrt{B(\boldsymbol{\p}, p^0(\boldsymbol{\p}))}^{\mu}_{\lambda} \,
\eta \, a^{\lambda}(\boldsymbol{\p})^+ \, \eta \, \Phi \, e^{ip\cdot x}  \bigg\} 
\end{multline*}  
and
\begin{multline*}
\partial_k A^\mu(x) \Phi = \int \limits_{\mathbb{R}^3} \, \ud^3 p \, \bigg\{
\frac{ip^k}{\sqrt{2 p^0(\boldsymbol{\p})}}\sqrt{B(\boldsymbol{\p}, p^0(\boldsymbol{\p}))}^{\mu}_{\lambda}
a^{\lambda} (\boldsymbol{\p}) \, \Phi \, e^{-ip\cdot x} \\
+  \frac{-ip^k}{\sqrt{2 p^0(\boldsymbol{\p})}}\sqrt{B(\boldsymbol{\p}, p^0(\boldsymbol{\p}))}^{\mu}_{\lambda} \,
\eta \, a^{\lambda}(\boldsymbol{\p})^+ \, \eta \, \Phi \, e^{ip\cdot x}  \bigg\}. 
\end{multline*}
exist as Pettis integrals and belong to $(E)^*$; which means that for any fixed $\Phi \in (E)$
and any $\Psi \in (E)$ the functionals
\begin{multline*}
\Psi \, \mapsto \, \langle \langle \partial_{0}A^\mu(x) \Phi, \Psi \rangle \rangle 
= \int \limits_{\mathbb{R}^3} \, \ud^3 p \,  \bigg\{ \bigg \langle \bigg \langle
\frac{-ip^0(\boldsymbol{\p})}{\sqrt{2 p^0(\boldsymbol{\p})}}\sqrt{B(\boldsymbol{\p}, p^0(\boldsymbol{\p}))}^{\mu}_{\lambda}
\, e^{-ip\cdot x} \, a^{\lambda} (\boldsymbol{\p}) \, \Phi, \,  \Psi \bigg \rangle \bigg \rangle  \\
+ \bigg \langle \bigg \langle  \frac{ip^0(\boldsymbol{\p})}{\sqrt{2 p^0(\boldsymbol{\p})}}\sqrt{B(\boldsymbol{\p}, p^0(\boldsymbol{\p}))}^{\mu}_{\lambda} \, e^{ip\cdot x} \,
\eta \, a^{\lambda}(\boldsymbol{\p})^+ \, \eta \, \Phi , \Psi \bigg \rangle \bigg \rangle  \bigg\} 
\end{multline*}  
and
\begin{multline*}
\partial_k A^\mu(x) \Phi = \int \limits_{\mathbb{R}^3} \, \ud^3 p \, \bigg\{ \bigg \langle \bigg \langle
\frac{ip^k}{\sqrt{2 p^0(\boldsymbol{\p})}}\sqrt{B(\boldsymbol{\p}, p^0(\boldsymbol{\p}))}^{\mu}_{\lambda}
\, e^{-ip\cdot x} \, a^{\lambda} (\boldsymbol{\p}) \, \Phi, \, \Psi \bigg \rangle \bigg \rangle  \\
+ \bigg \langle \bigg \langle \frac{-ip^k}{\sqrt{2 p^0(\boldsymbol{\p})}}\sqrt{B(\boldsymbol{\p}, p^0(\boldsymbol{\p}))}^{\mu}_{\lambda} \, e^{ip\cdot x} \,
\eta \, a^{\lambda}(\boldsymbol{\p})^+ \, \eta \, \Phi, \, \Psi \bigg \rangle \bigg \rangle \bigg\}. 
\end{multline*}
are continuous functionals on $(E)$, i.e. belong to $(E)^*$.   

 Now using the inequality (2-2) of Lemma 2.1 of \cite{hida} we will prove more, i.e.
\begin{lem*}
For any $x \in \mathbb{R}^4$ the operators $A^\mu(x), \partial_{0}A^\mu(x), \partial_k A^\mu(x): (E) \rightarrow
(E)^*$ are continuous, where $(E)^*$ is equipped with the strong topology (although in this case the linear
spaces $\mathscr{L}((E), (E)^{*}_{\sigma})$ and $\mathscr{L}((E), (E)^{*}_{b})$  
of continuous operators $(E) \rightarrow (E)^*$ for the weak and strong topology on $(E)^*$  
are identical and denoted simply by $\mathscr{L}((E), (E)^{*})$). 
\end{lem*}

\qedsymbol \, 

Let $\Phi, \Psi$ be any elements of $(E)$ and $x = (\boldsymbol{\x}, t)$ be any point in $\mathbb{R}^4$. 
Then we have
\begin{multline*}
\langle \langle A^\mu (\boldsymbol{\x}, t) \Phi, \Psi \rangle \rangle = 
\int \limits_{\mathbb{R}^3} \, \ud^3 p \, e^{i\boldsymbol{\p} \cdot \boldsymbol{\x}}\bigg\{
e^{-ip^0(\boldsymbol{\p})t} \frac{1}{\sqrt{2 p^0(\boldsymbol{\p})}}\sqrt{B(\boldsymbol{\p}, p^0(\boldsymbol{\p}))}^{\mu}_{\lambda}  \,
\big\langle \big\langle
\partial_{\boldsymbol{\p}}^{\lambda} \Phi, \Psi \big\rangle \big\rangle  \bigg\} \\
+ \int \limits_{\mathbb{R}^3} \, \ud^3 p \, e^{i\boldsymbol{\p} \cdot (-\boldsymbol{\x}})\bigg\{
e^{ip^0(\boldsymbol{\p})t}\frac{1}{\sqrt{2 p^0(\boldsymbol{\p})}}\sqrt{B(\boldsymbol{\p}, p^0(\boldsymbol{\p}))}^{\mu}_{\lambda}  \,
\big\langle \big\langle
\eta \partial_{\boldsymbol{\p}}^{\lambda \, *} \eta \Phi, \, \Psi \big\rangle \big\rangle \bigg\};
\end{multline*} 
\begin{multline*}
\langle \langle \partial_0 A^\mu (\boldsymbol{\x}, t) \Phi, \Psi \rangle \rangle = 
\int \limits_{\mathbb{R}^3} \, \ud^3 p \, e^{i\boldsymbol{\p} \cdot \boldsymbol{\x}}\bigg\{
e^{-ip^0(\boldsymbol{\p})t} \frac{-i\sqrt{p^0(\boldsymbol{\p})}}{\sqrt{2}}\sqrt{B(\boldsymbol{\p}, p^0(\boldsymbol{\p}))}^{\mu}_{\lambda}  \,
\big\langle \big\langle
\partial_{\boldsymbol{\p}}^{\lambda} \Phi, \Psi \big\rangle \big\rangle  \bigg\} \\
+ \int \limits_{\mathbb{R}^3} \, \ud^3 p \, e^{i\boldsymbol{\p} \cdot (-\boldsymbol{\x}})\bigg\{
e^{ip^0(\boldsymbol{\p})t}\frac{i\sqrt{p^0(\boldsymbol{\p})}}{\sqrt{2}}\sqrt{B(\boldsymbol{\p}, p^0(\boldsymbol{\p}))}^{\mu}_{\lambda}  \,
\big\langle \big\langle
\eta \partial_{\boldsymbol{\p}}^{\lambda \, *} \eta \Phi, \, \Psi \big\rangle \big\rangle \bigg\};
\end{multline*} 
\begin{multline*}
\langle \langle \partial_k A^\mu (\boldsymbol{\x}, t) \Phi, \Psi \rangle \rangle = 
\int \limits_{\mathbb{R}^3} \, \ud^3 p \, e^{i\boldsymbol{\p} \cdot \boldsymbol{\x}}\bigg\{
e^{-ip^0(\boldsymbol{\p})t} \frac{ip^k}{\sqrt{2 p^0(\boldsymbol{\p})}}\sqrt{B(\boldsymbol{\p}, p^0(\boldsymbol{\p}))}^{\mu}_{\lambda}  \,
\big\langle \big\langle
\partial_{\boldsymbol{\p}}^{\lambda} \Phi, \Psi \big\rangle \big\rangle  \bigg\} \\
+ \int \limits_{\mathbb{R}^3} \, \ud^3 p \, e^{i\boldsymbol{\p} \cdot (-\boldsymbol{\x}})\bigg\{
e^{ip^0(\boldsymbol{\p})t}\frac{-ip^k}{\sqrt{2 p^0(\boldsymbol{\p})}}\sqrt{B(\boldsymbol{\p}, p^0(\boldsymbol{\p}))}^{\mu}_{\lambda}  \,
\big\langle \big\langle
\eta \partial_{\boldsymbol{\p}}^{\lambda \, *} \eta \Phi, \, \Psi \big\rangle \big\rangle \bigg\}.
\end{multline*} 
From Lemma 2.1 of \cite{hida} and continuity of $\eta = \Gamma(\mathfrak{J}'): (E) \rightarrow (E)$ 
it follows that the functions 
\begin{multline*}
\boldsymbol{\p} \mapsto \big\langle \big\langle
\partial_{\boldsymbol{\p}}^{\lambda} \Phi, \Psi \big\rangle \big\rangle  
= \eta_{\Phi, \Psi}^{\lambda}(\boldsymbol{\p}), \\
\boldsymbol{\p} \mapsto \big\langle \big\langle
\eta \partial_{\boldsymbol{\p}}^{\lambda \, *} \eta \Phi, \, \Psi \big\rangle \big\rangle
= {\eta^*}_{\eta \Phi, \eta \Psi}^{\lambda}(\boldsymbol{\p})
\end{multline*}
belong to the nuclear space $E = \mathcal{S}_{A'''}(\mathbb{R}^3) = \mathcal{S}^0 (\mathbb{R}^3)$.

Now for any fixed $t \in  \mathbb{R}$ and $\mu, \lambda \in \{0,1,2,3\}$, consider the operators 
${Op_t}^{\mu}_{\lambda}: E= \mathcal{S}_{A'''}(\mathbb{R}^3)
\rightarrow E = \mathcal{S}_{A'''}(\mathbb{R}^3)$, where for $\xi^\lambda \in E = \mathcal{S}_{A'''}(\mathbb{R}^3)$
$\big(Op_t \xi \big)^\mu (\boldsymbol{\p}) = {Op_t}^{\mu}_{\lambda} \xi^\lambda (\boldsymbol{\p})$ (summation over $\lambda= 0,1,2,3$) is given by one of the following formulas
\begin{multline*}
e^{-ip^0(\boldsymbol{\p})t} \frac{1}{\sqrt{2 p^0(\boldsymbol{\p})}}\sqrt{B(\boldsymbol{\p}, p^0(\boldsymbol{\p}))}^{\mu}_{\lambda}  \, \xi^\lambda (\boldsymbol{\p}) \,\,\, \textrm{or} \,\,\, \\
e^{ip^0(\boldsymbol{\p})t} \frac{1}{\sqrt{2 p^0(\boldsymbol{\p})}}\sqrt{B(\boldsymbol{\p}, p^0(\boldsymbol{\p}))}^{\mu}_{\lambda}  \, \xi^\lambda (\boldsymbol{\p}) \,\,\, \textrm{or} \,\,\, \\
e^{-ip^0(\boldsymbol{\p})t} \frac{-i\sqrt{p^0(\boldsymbol{\p})}}{\sqrt{2}}\sqrt{B(\boldsymbol{\p}, p^0(\boldsymbol{\p}))}^{\mu}_{\lambda}  \, \xi^\lambda (\boldsymbol{\p}) \,\,\, \textrm{or} \,\,\, \\
e^{ip^0(\boldsymbol{\p})t}\frac{i\sqrt{p^0(\boldsymbol{\p})}}{\sqrt{2}}\sqrt{B(\boldsymbol{\p}, p^0(\boldsymbol{\p}))}^{\mu}_{\lambda}  \, \xi^\lambda (\boldsymbol{\p}) \,\,\, \textrm{or} \,\,\, \\
e^{-ip^0(\boldsymbol{\p})t} \frac{ip^k}{\sqrt{2 p^0(\boldsymbol{\p})}}\sqrt{B(\boldsymbol{\p}, p^0(\boldsymbol{\p}))}^{\mu}_{\lambda}  \, \xi^\lambda (\boldsymbol{\p}) \,\,\, \textrm{or} \,\,\, \\
e^{ip^0(\boldsymbol{\p})t}\frac{-ip^k}{\sqrt{2 p^0(\boldsymbol{\p})}}\sqrt{B(\boldsymbol{\p}, p^0(\boldsymbol{\p}))}^{\mu}_{\lambda}  \, \xi^\lambda (\boldsymbol{\p}).
\end{multline*} 
By the lemmas of Subsection \ref{SA=S0} all the operators  ${Op_t}^{\mu}_{\lambda}$ and thus the operators $Op_t$ defined above are continuous linear operators from $E = \mathcal{S}_{A'''}(\mathbb{R}^3) = \mathcal{S}^0 (\mathbb{R}^3)$ into 
$E = \mathcal{S}_{A'''}(\mathbb{R}^3) = \mathcal{S}^0 (\mathbb{R}^3)$. It follows that all functions
\[
\begin{split}
\boldsymbol{\x} \mapsto \langle \langle 
A^\mu (\boldsymbol{\x},t) \Phi, \Psi \rangle \rangle, \\
\boldsymbol{\x} \mapsto \langle \langle 
\partial_0 A^\mu (\boldsymbol{\x}, t) \Phi, \Psi \rangle \rangle, \\
\boldsymbol{\x} \mapsto \langle \langle 
\partial_k A^\mu (\boldsymbol{\x}, t) \Phi, \Psi \rangle \rangle,
\end{split}
\]
belong to $\widetilde{\mathcal{S}^0 (\mathbb{R}^3)} = \widetilde{\mathcal{S}_{A'''}(\mathbb{R}^3)}
= \mathcal{S}^{00}(\mathbb{R}^3)$. i.e. they are equal to the Fourier transforms $\widetilde{\xi}$ of some elements
$\xi$ of $E = \mathcal{S}_{A'''}(\mathbb{R}^3) = \mathcal{S}^0 (\mathbb{R}^3)$. In particular
\[
\langle \langle A^\mu (\boldsymbol{\x}, t) \Phi, \Psi \rangle \rangle
= \big({Op_t}^{\mu}_{\lambda} \eta_{{}_{\Phi, \Psi}}^{\lambda}\big)^{\thicksim}(\boldsymbol{\x}) + 
\big({Op_{-t}}^{\mu}_{\lambda} {\eta^*}_{{}_{\eta \Phi, \eta \Psi}}^{\lambda}\big)^{\thicksim}(\boldsymbol{-\x})
\] 
and similarly for the operators $\partial_0 A^\mu (\boldsymbol{\x}, t)$ and 
$\partial_k A^\mu (\boldsymbol{\x}, t)$ with the corresponding operators ${Op_t}^{\mu}_{\lambda}$
inserted. Therefore,  
\begin{multline}\label{ineq-1}
\big| \langle \langle A^\mu (\boldsymbol{\x}, t) \Phi, \Psi \rangle \rangle \big |^2 \leq
 \Big| \big({Op_t}^{\mu}_{\lambda} \eta_{{}_{\Phi, \Psi}}^{\lambda}\big)^{\thicksim}(\boldsymbol{\x}) \Big|^2  \\ + 
2 \Big| \big({Op_t}^{\mu}_{\lambda} \eta_{{}_{\Phi, \Psi}}^{\lambda}\big)^{\thicksim}(\boldsymbol{\x}) \Big|
\Big|\big({Op_{-t}}^{\mu}_{\lambda} {\eta^*}_{{}_{\eta \Phi, \eta \Psi}}^{\lambda}\big)^{\thicksim}(\boldsymbol{-\x})\Big|
\\ +
\Big|\big({Op_{-t}}^{\mu}_{\lambda} {\eta^*}_{{}_{\eta \Phi, \eta \Psi}}^{\lambda}\big)^{\thicksim}(\boldsymbol{-\x})\Big|^2
\\ \leq 2 \Big| \big({Op_t}^{\mu}_{\lambda} \eta_{{}_{\Phi, \Psi}}^{\lambda}\big)^{\thicksim}(\boldsymbol{\x}) \Big|^2 
+ 2 \Big|\big({Op_{-t}}^{\mu}_{\lambda} {\eta^*}_{{}_{\eta \Phi, \eta \Psi}}^{\lambda}\big)^{\thicksim}(\boldsymbol{-\x})\Big|^2
\end{multline} 
and similarly for $\langle \langle \partial_0 A^\mu (\boldsymbol{\x}, t) \Phi, \Psi \rangle \rangle$
and $\langle \langle \partial_k A^\mu (\boldsymbol{\x}, t) \Phi, \Psi \rangle \rangle$. 
On the basis of the 5-th Lemma of Subsection \ref{SA=S0}, for each $p \in \mathbb{N}$ there exists a finite constant $C_p < + \infty$
and a natural number (depending on $p$) $q(p) \in \mathbb{N}$ such that 
\[
\sup \limits_{0 \leq |k|, |m| \leq p} \int \limits_{\mathbb{R}^3} \big| r^k \varphi^{(m)}(\boldsymbol{\p}) \big|^2 \, \ud^3 \boldsymbol{\p} \, \leq \, C_p \, || (A''')^{q} \varphi ||_{{}_{L^2(\mathbb{R}^3)}}^{2}, \,\,\,
\varphi \in \mathcal{S}_{A'''}(\mathbb{R}^3) = \mathcal{S}^0 (\mathbb{R}^3),
\]
where $m \in \mathbb{N}^3$ denotes a multi-idex and $|m|$ its standard modulus and  $\varphi^{(m)}$
is the derivative of $\varphi$ of degree $|m|$ corresponding to the multi-index $m$. In particular
\begin{equation}\label{ineq-2}
\begin{split}
\int \limits_{\mathbb{R}^3} \big| \varphi(\boldsymbol{\p}) \big|^2 \, \ud^3 \boldsymbol{\p} \, \leq \, C \, || (A''')^{q} \varphi ||_{{}_{L^2(\mathbb{R}^3)}}^{2}, \\
\int \limits_{\mathbb{R}^3} \big| p^i \varphi(\boldsymbol{\p}) \big|^2 \, \ud^3 \boldsymbol{\p} \, \leq \, C \, || (A''')^{q} \varphi ||_{{}_{L^2(\mathbb{R}^3)}}^{2}, \\
\int \limits_{\mathbb{R}^3} \big| p^i p^j \varphi(\boldsymbol{\p}) \big|^2 \, \ud^3 \boldsymbol{\p} \, \leq \, C \, || (A''')^{q} \varphi ||_{{}_{L^2(\mathbb{R}^3)}}^{2}, \\
\int \limits_{\mathbb{R}^3} \big| p^1 p^2 p^3 \varphi(\boldsymbol{\p}) \big|^2 \, \ud^3 \boldsymbol{\p} \, \leq \, C \, || (A''')^{q} \varphi ||_{{}_{L^2(\mathbb{R}^3)}}^{2}, \,\,\,
\varphi \in \mathcal{S}_{A'''}(\mathbb{R}^3) = \mathcal{S}^0 (\mathbb{R}^3),
\end{split}
\end{equation}
for some $q \in \mathbb{N}$ and $C < + \infty$ independent of 
$\varphi \in \mathcal{S}_{A'''}(\mathbb{R}^3) = \mathcal{S}^0 (\mathbb{R}^3)$. After performing the Fourier transformation
$\widetilde{(\cdot)}: \mathcal{S}^0 (\mathbb{R}^3) \rightarrow \widetilde{\mathcal{S}^0 (\mathbb{R}^3)}$
we obtain from (\ref{ineq-2}) the following inequalities
\begin{equation}\label{ineq-3}
\begin{split}
\int \limits_{\mathbb{R}^3} \big| \widetilde{\varphi}(\boldsymbol{\x}) \big|^2 \, \ud^3 \boldsymbol{\x} \, \leq \, C \, || (\widetilde{A'''})^{q} \widetilde{\varphi} ||_{{}_{L^2(\mathbb{R}^3)}}^{2}, \\
\int \limits_{\mathbb{R}^3} \big| \partial_i \widetilde{\varphi}(\boldsymbol{\x}) \big|^2 \, \ud^3 \boldsymbol{\x} \, \leq \, C \, || (\widetilde{A'''})^{q} \varphi ||_{{}_{L^2(\mathbb{R}^3)}}^{2}, \\
\int \limits_{\mathbb{R}^3} \big| \partial_i \partial_j \widetilde{\varphi}(\boldsymbol{\x}) \big|^2 \, \ud^3 \boldsymbol{\x} \, \leq \, C \, || (\widetilde{A'''})^{q} \widetilde{\varphi} ||_{{}_{L^2(\mathbb{R}^3)}}^{2}, \\
\int \limits_{\mathbb{R}^3} \big|\partial_1 \partial_2 \partial_3 \widetilde{\varphi}(\boldsymbol{\x}) \big|^2 \, \ud^3 \boldsymbol{\x} \, \leq \, C \, || (\widetilde{A'''})^{q} \widetilde{\varphi} ||_{{}_{L^2(\mathbb{R}^3)}}^{2}, \,\,\,
\varphi \in  \mathcal{S}^0 (\mathbb{R}^3).
\end{split}
\end{equation}
On the other hand (3-rd Lemma of Subsect. \ref{SA=S0})
\begin{multline}\label{ineq-4}
|\widetilde{\varphi}(\boldsymbol{\x})|^2 \, \leq \, \int \limits_{\mathbb{R}^3} \big| \widetilde{\varphi}(\boldsymbol{\x}) \big|^2 \, \ud^3 \boldsymbol{\x} + \int \limits_{\mathbb{R}^3} \big|\partial_1 \widetilde{\varphi}(\boldsymbol{\x}) \big|^2 \, \ud^3 \boldsymbol{\x} + \int \limits_{\mathbb{R}^3} \big|\partial_2 \widetilde{\varphi}(\boldsymbol{\x}) \big|^2 \, \ud^3 \boldsymbol{\x} + \int \limits_{\mathbb{R}^3} \big|\partial_3 \widetilde{\varphi}(\boldsymbol{\x}) \big|^2 \, \ud^3 \boldsymbol{\x} \\ + \int \limits_{\mathbb{R}^3} \big|\partial_1 \partial_2 \widetilde{\varphi}(\boldsymbol{\x}) \big|^2 \, \ud^3 \boldsymbol{\x} + \int \limits_{\mathbb{R}^3} \big|\partial_1 \partial_3 \widetilde{\varphi}(\boldsymbol{\x}) \big|^2 \, \ud^3 \boldsymbol{\x} + \int \limits_{\mathbb{R}^3} \big|\partial_2 \partial_3 \widetilde{\varphi}(\boldsymbol{\x}) \big|^2 \, \ud^3 \boldsymbol{\x} \\ +
\int \limits_{\mathbb{R}^3} \big|\partial_1 \partial_2 \partial_3 \widetilde{\varphi}(\boldsymbol{\x}) \big|^2 \, \ud^3 \boldsymbol{\x}.
\end{multline}
From (\ref{ineq-3}) and (\ref{ineq-4}) we obtain 
\begin{equation}\label{ineq-5}
|\widetilde{\varphi}(\boldsymbol{\x})|^2 \,
\leq \, C \, || (\widetilde{A'''})^{q} \widetilde{\varphi} ||_{{}_{L^2(\mathbb{R}^3)}}^{2}, 
\end{equation}
with $C<+\infty$ independent of $\varphi \in \mathcal{S}^0 (\mathbb{R}^3)= \mathcal{S}_{A'''}(\mathbb{R}^3)$.
Because (\ref{ineq-5}) is valid for all $\widetilde{\varphi}$, with $\varphi \in \mathcal{S}^0 (\mathbb{R}^3)= \mathcal{S}_{A'''}(\mathbb{R}^3)$, where $\widetilde{\varphi}$ is the Fourier transform of $\varphi$, then from 
(\ref{ineq-1}) and (\ref{ineq-5}) we obtain
\begin{multline}\label{ineq-5'}
\big| \langle \langle A^\mu (\boldsymbol{\x}, t) \Phi, \Psi \rangle \rangle \big|^2 \, \leq \,
C \Big\{
\Big\| (\widetilde{A'''})^{q} \big({Op_t}^{\mu}_{\lambda} \eta_{{}_{\Phi, \Psi}}^{\lambda}\big)^{\thicksim} \Big\|_{{}_{L^2(\mathbb{R}^3)}}^{2} \\
+ \Big\| (\widetilde{A'''})^{q} \big({Op_{-t}}^{\mu}_{\lambda} {\eta^*}_{{}_{\eta \Phi, \eta \Psi}}^{\lambda}\big)^{\thicksim} \Big\|_{{}_{L^2(\mathbb{R}^3)}}^{2} \Big\} \\
= C \Big\{
\Big\| (A''')^{q} {Op_t}^{\mu}_{\lambda} \eta_{{}_{\Phi, \Psi}}^{\lambda} \Big\|_{{}_{L^2(\mathbb{R}^3)}}^{2} 
+ \Big\| (A''')^{q} {Op_{-t}}^{\mu}_{\lambda} {\eta^*}_{{}_{\eta \Phi, \eta \Psi}}^{\lambda} \Big\|_{{}_{L^2(\mathbb{R}^3)}}^{2} \Big\} \\
= C \Big\{
{\Big| {Op_t}^{\mu}_{\lambda} \eta_{{}_{\Phi, \Psi}}^{\lambda} \Big|_q}^2
+ {\Big| {Op_{-t}}^{\mu}_{\lambda} {\eta^*}_{{}_{\eta \Phi, \eta \Psi}}^{\lambda} \Big|_q}^2 \Big\}.
\end{multline}

But from the inequality (2-2) of Lemma 2.1 of \cite{hida} and from the continuity
of $\eta = \Gamma(\mathfrak{J}'): (E) \rightarrow (E)$ for each $p\in \mathbb{N}$ there exist
a positive constant $C_p$ and $q \in \mathbb{N}$ (depending on $p$) such that
\begin{multline}\label{ineq-(*)}
\Big| \eta_{{}_{\Phi, \Psi}}^{\lambda} \Big|_p \leq
\rho^{-p} \Big(\frac{\rho^{-p}}{-2pe \ln \rho}\Big)^{1/2} \| \Phi \|_p \| \Psi \|_p , \\
\Big| {\eta^*}_{{}_{\eta \Phi, \eta \Psi}}^{\lambda} \Big|_p \leq
\rho^{-p} \Big(\frac{\rho^{-p}}{-2pe \ln \rho}\Big)^{1/2} \| \eta \Phi \|_p \| \eta \Psi \|_p  \\ \leq
\rho^{-p} \Big(\frac{\rho^{-p}}{-2pe \ln \rho}\Big)^{1/2} \, C_p \, \| \Phi \|_q \| \Psi \|_q , \,\,\,
\Phi, \Psi \in (E) 
\end{multline} 
where
\[
\rho = \| (A''')^{-1}\|_{OP} = {\lambda_1}^{-1} < 1 
\]
is the operator norm of $(A''')^{-1}$, and $\lambda_1 = \inf \Sp A'''$. Joining (\ref{ineq-5'})
and (\ref{ineq-(*)}) we obtain from the continuity of the operators ${Op_t}^{\mu}_{\lambda}: E \rightarrow E = 
\mathcal{S}_{A'''}(\mathbb{R}^3) = \mathcal{S}^0 (\mathbb{R}^3)$ the following inequalities
\begin{multline}\label{ineq-6}
\big| \langle \langle A^\mu (\boldsymbol{\x}, t) \Phi, \Psi \rangle \rangle \big|^2 \, \leq \,
\rho^{-p} \Big(\frac{\rho^{-p}}{- 2pe \ln \rho}\Big)^{1/2} \, C(t) \, \| \Phi \|_p \| \Psi \|_p , \,\,\,
\Phi, \Psi \in (E), 
\end{multline}
for all $p \in \mathbb{N}$ greater than some fixed $q_0 \in \mathbb{N}$; and 
where $t \mapsto C(t)$ is a positive finite function which can be chosen continuous, as is easily checked.

Analogous inequality (\ref{ineq-6}) holds true for the operators $\partial_0 A^\mu (\boldsymbol{\x}, t)$ and 
$\partial_k A^\mu (\boldsymbol{\x}, t)$.

Now let $V(B, \epsilon)$ be a strong neighbourhood of zero in $(E)^*$ determined by a bounded subset
$B$ of $(E)$ and a positive number $\epsilon$, i.e. $V(B, \epsilon)$ is the set of all those 
functionals $F \in (E)^*$ for which
\[
| \langle \langle F, \Psi \rangle \rangle | < \epsilon, \Psi \in B.
\]
Because $B$ is bounded in $(E)$ there exists for each $p \in \mathbb{N}$ a positive finite number
$C_p$ such that 
\[
\| \Psi \|_p < C_p, \Psi \in B.
\]

Let $\mathscr{U}$ be the zero neighbourhood in $(E)$ equal to the open ball determined by the $p$-th norm
$\| \cdot \|_p$ in $(E)$ of radius 
\[
\frac{\epsilon}{\rho^{-p} \Big(\frac{\rho^{-p}}{-2pe \ln \rho}\Big)^{1/2} \, C(t)},
\]
i.e.
\[
\Phi \in \mathscr{U} \Longleftrightarrow \| \Phi \|_p < 
\frac{\epsilon}{\rho^{-p} \Big(\frac{\rho^{-p}}{-2pe \ln \rho}\Big)^{1/2} \, C(t)}.
\] 
Then from (\ref{ineq-6}) it follows that for $\Phi \in \mathscr{U}$ the value 

$A^\mu (\boldsymbol{\x},t) \Phi \in V(B, \epsilon)$, and thus continuity of the 
operator $A^\mu (\boldsymbol{\x},t): (E) \rightarrow (E)^*$ follows for the strong topology
$(E)^{*}_{b}$ on $(E)^*$. 

Similarly, we obtain the continuity of the operators $\partial_0 A^\mu (\boldsymbol{\x}, t)$ and 
$\partial_k A^\mu (\boldsymbol{\x}, t)$.

\qed

\begin{defin*}
Let for any continuous linear operator $\Xi: (E) \rightarrow (E)^*$ and $\xi, \zeta \in E =
\mathcal{S}_{A'''}(\mathbb{R}^3) = \mathcal{S}^0 (\mathbb{R}^3)$, $\widehat{\Xi}(\xi, \zeta)$
denotes its symbol, i.e.
\[
\widehat{\Xi}(\xi, \zeta) = \langle \langle \Xi \Phi_\xi , \Phi_\zeta \rangle \rangle,
\] 
where $\Phi_\xi$ is the exponential (coherent) vector in $(E)$ corresponding to $\xi \in E$.
Note that we are using the convention of Obata, and our symbol differs from the Wick symbol
introduced by Berezin by the additional factor $e^{\langle \xi, \zeta \rangle}$ so that our symbol is not 
multiplicative under the Wick product of generalized operators but gets additional factor:
\[
\widehat{: \Xi_1 \Xi_2 :} (\xi, \zeta) = e^{-\langle \xi, \zeta \rangle} 
\widehat{\Xi_1} (\xi, \zeta) \widehat{\Xi_2}(\xi, \zeta).
\]  
\end{defin*}

Because for each $x = (\boldsymbol{\x, t}) \in \mathbb{R}^4$ the operators $A^\mu (\boldsymbol{\x, t})$, $\partial_0 A^\mu (\boldsymbol{\x, t})$ and $\partial_k A^\mu (\boldsymbol{\x}, t)$,  belong to $\mathscr{L}((E), (E)^*)$, i.e.
are continuous, then their Wick product
\[
: \partial_0 A^\mu (\boldsymbol{\x, t}) \partial_k A^\nu (\boldsymbol{\x}, t):
\] 
is a well defined and continuous operator $(E) \rightarrow (E)^*$,
i.e. belongs to $\mathscr{L}((E), (E)^*)$, for the proof compare e.g. Lemma 2.1 of \cite{obata-wick}
or \cite{obata-book}. Concerning the symbols of the mentioned operators we have the following and simple
\begin{lem*}
Let $\xi, \zeta \in E = \mathcal{S}_{A'''}(\mathbb{R}^3) = \mathcal{S}^0 (\mathbb{R}^3)$. Then
\begin{multline*}
\widehat{A^\mu(\boldsymbol{\x}, t)}(\xi, \zeta) = e^{\langle \xi, \zeta \rangle} 
\int \limits_{\mathbb{R}^3} \, \ud^3 \boldsymbol{\p} \, 
e^{i \boldsymbol{\p}\cdot \boldsymbol{\x}} \, \bigg\{
\frac{1}{\sqrt{2 p^0(\boldsymbol{\p})}}\sqrt{B(\boldsymbol{\p}, p^0(\boldsymbol{\p}))}^{\mu}_{\lambda}
\xi^{\lambda} (\boldsymbol{\p}) e^{-ip^0(\boldsymbol{\p})t}  \bigg\} \\
+ e^{\langle \xi, \zeta \rangle}\int \limits_{\mathbb{R}^3} \, \ud^3 \boldsymbol{\p} \, 
e^{-i \boldsymbol{\p}\cdot \boldsymbol{\x}} \, \bigg\{
\frac{1}{\sqrt{2 p^0(\boldsymbol{\p})}}\sqrt{B(\boldsymbol{\p}, p^0(\boldsymbol{\p}))}^{\mu}_{\lambda} \,
(\mathfrak{J}_{\bar{p}}\zeta)^{\lambda}(\boldsymbol{\p}) \, e^{ip^0(\boldsymbol{\p})t}  \bigg\} ;
\end{multline*}  
\begin{multline*}
\Big(\partial_0 A^\mu(\boldsymbol{\x}, t)\Big)^\wedge (\xi, \zeta) = e^{\langle \xi, \zeta \rangle}\int \limits_{\mathbb{R}^3} \, \ud^3 \boldsymbol{\p} \, e^{i \boldsymbol{\p}\cdot \boldsymbol{\x}} \, \bigg\{
\frac{-i\sqrt{p^0(\boldsymbol{\p})}}{\sqrt{2}}\sqrt{B(\boldsymbol{\p}, p^0(\boldsymbol{\p}))}^{\mu}_{\lambda}
\xi^{\lambda} (\boldsymbol{\p}) e^{-ip^0(\boldsymbol{\p})\cdot t}  \bigg\} \\
+ e^{\langle \xi, \zeta \rangle}\int \limits_{\mathbb{R}^3} \, \ud^3 \boldsymbol{\p} \, 
e^{-i \boldsymbol{\p}\cdot \boldsymbol{\x}} \, \bigg\{
\frac{i\sqrt{p^0(\boldsymbol{\p})}}{\sqrt{2}}\sqrt{B(\boldsymbol{\p}, p^0(\boldsymbol{\p}))}^{\mu}_{\lambda} \,
(\mathfrak{J}_{\bar{p}}\zeta)^{\lambda}(\boldsymbol{\p}) \, e^{ip^0(\boldsymbol{\p})t}  \bigg\};
\end{multline*}
\begin{multline*}
\Big(\partial_k A^\mu(\boldsymbol{\x}, t)\Big)^\wedge (\xi, \zeta) = e^{\langle \xi, \zeta \rangle}\int \limits_{\mathbb{R}^3} \, \ud^3 \boldsymbol{\p} \, e^{i \boldsymbol{\p}\cdot \boldsymbol{\x}} \,\bigg\{
\frac{ip^k}{\sqrt{2 p^0(\boldsymbol{\p})}}\sqrt{B(\boldsymbol{\p}, p^0(\boldsymbol{\p}))}^{\mu}_{\lambda}
\xi^{\lambda} (\boldsymbol{\p}) e^{-ip^0(\boldsymbol{\p})\cdot t}  \bigg\} \\
+ e^{\langle \xi, \zeta \rangle}\int \limits_{\mathbb{R}^3} \, \ud^3 \boldsymbol{\p} \, 
e^{-i \boldsymbol{\p}\cdot \boldsymbol{\x}} \, \bigg\{
\frac{-i p^k}{\sqrt{2 p^0(\boldsymbol{\p})}}\sqrt{B(\boldsymbol{\p}, p^0(\boldsymbol{\p}))}^{\mu}_{\lambda} \,
(\mathfrak{J}_{\bar{p}}\zeta)^{\lambda}(\boldsymbol{\p}) \, e^{ip^0(\boldsymbol{\p})t}  \bigg\}.
\end{multline*}    
\end{lem*}

\qedsymbol \, 
This Lemma is a simple consequence of the following formulas
\begin{equation}\label{eta-hida-partial-exp}
\eta \Phi_\xi = \Phi_{\mathfrak{J}_{\bar{p}}\xi}, \,\,\,
\partial_{\boldsymbol{\p}}^{\mu} \Phi_\xi = \xi^\mu (\boldsymbol{\p}) \Phi_\xi, \,\,\,
\langle \langle \Phi_\xi , \Phi_\zeta \rangle \rangle = e^{\langle \xi, \zeta \rangle},
\end{equation}
for the Hida derivation operator $\partial_{\boldsymbol{\p}}^{\mu}$ which through the Wiener-It\^o-Segal
decomposition corresponds to the annihilation (generalized) operator $a^\mu (\boldsymbol{\p})$ and from 
the fact that the indicated operators are well defined pointwisely as Pettis integrals.

The first formula in (\ref{eta-hida-partial-exp}) has been shown above, for the proof of the 
second and the third formula in (\ref{eta-hida-partial-exp}) compare e.g. \cite{hida} or 
\cite{obata-book} or \cite{HKPS}.
\qed

\begin{lem*}
For each $x = (\boldsymbol{\x}, t), y = (\boldsymbol{\y}, t) \in \mathbb{R}^4$ and each $\Phi \in (E)$ the integral
\begin{multline*}
\Xi(\boldsymbol{\x}, \boldsymbol{\y}, t) \Phi  = \int \limits_{\mathbb{R}^3} \, \ud^3 \boldsymbol{\p} \, 
\int \limits_{\mathbb{R}^3} \, \ud^3 \boldsymbol{\p}' \, \bigg\{
\frac{-i p^0(\boldsymbol{\p})}{\sqrt{2 p^0(\boldsymbol{\p})}}\sqrt{B(\boldsymbol{\p}, p^0(\boldsymbol{\p}))}^{\mu}_{\lambda}
\, \times \\ 
\times \, \frac{i p'^k}{\sqrt{2 p^0(\boldsymbol{\p}')}}\sqrt{B(\boldsymbol{\p}', p^0(\boldsymbol{\p}'))}^{\nu}_{\gamma}
 e^{-i(p\cdot x + p \cdot y)} \bigg\} \, \partial_{\boldsymbol{\p}}^{\lambda} \, \partial_{\boldsymbol{\p}'}^{\gamma}
\, \Phi 
\end{multline*}
\begin{multline*}
+ \, \int \limits_{\mathbb{R}^3} \, \ud^3 \boldsymbol{\p} \, \int \limits_{\mathbb{R}^3} \, \ud^3 \boldsymbol{\p}' 
\, \bigg\{
\frac{-i p^0(\boldsymbol{\p})}{\sqrt{2 p^0(\boldsymbol{\p})}}\sqrt{B(\boldsymbol{\p}, p^0(\boldsymbol{\p}))}^{\mu}_{\lambda}
\, \times \\ 
\times \, \frac{-i p'^k}{\sqrt{2 p^0(\boldsymbol{\p}')}}\sqrt{B(\boldsymbol{\p}', p^0(\boldsymbol{\p}'))}^{\nu}_{\gamma}
 e^{-i(p\cdot x - p' \cdot y)} \bigg\} \, \eta \, \partial_{\boldsymbol{\p}'}^{\gamma *} \, \eta \,\partial_{\boldsymbol{\p}}^{\lambda} \,  \Phi 
\end{multline*}
\begin{multline*}
+ \, \int \limits_{\mathbb{R}^3} \, \ud^3 \boldsymbol{\p} \, \int \limits_{\mathbb{R}^3} \, \ud^3 \boldsymbol{\p}' 
\, \bigg\{
\frac{i p^0(\boldsymbol{\p})}{\sqrt{2 p^0(\boldsymbol{\p})}}\sqrt{B(\boldsymbol{\p}, p^0(\boldsymbol{\p}))}^{\mu}_{\lambda}
\, \times \\ 
\times \, \frac{i p'^k}{\sqrt{2 p^0(\boldsymbol{\p}')}}\sqrt{B(\boldsymbol{\p}', p^0(\boldsymbol{\p}'))}^{\nu}_{\gamma}
 e^{i(p\cdot x - p' \cdot y)} \bigg\} \, \eta \, \partial_{\boldsymbol{\p}}^{\lambda *} \, \eta \,\partial_{\boldsymbol{\p}'}^{\gamma} \,  \Phi 
\end{multline*}
\begin{multline*}
+ \, \int \limits_{\mathbb{R}^3} \, \ud^3 \boldsymbol{\p} \, \int \limits_{\mathbb{R}^3} \, \ud^3 \boldsymbol{\p}' 
\, \bigg\{
\frac{i p^0(\boldsymbol{\p})}{\sqrt{2 p^0(\boldsymbol{\p})}}\sqrt{B(\boldsymbol{\p}, p^0(\boldsymbol{\p}))}^{\mu}_{\lambda}
\, \times \\ 
\times \, \frac{-i p'^k}{\sqrt{2 p^0(\boldsymbol{\p}')}}\sqrt{B(\boldsymbol{\p}', p^0(\boldsymbol{\p}'))}^{\nu}_{\gamma}
 e^{i(p\cdot x + p' \cdot y)} \bigg\} \, \eta \, \partial_{\boldsymbol{\p}}^{\lambda *} \, \partial_{\boldsymbol{\p}'}^{\gamma *} \,  \eta \, \Phi, 
\end{multline*} 
exists as Pettis integral, i.e. belongs to $(E)^*$, which means that for $\Psi \in (E)$
\begin{multline*}
\Psi \,\, \mapsto \,\, \langle \langle \Xi(\boldsymbol{\x}, \boldsymbol{\y}, t) \Phi, \Psi \rangle \rangle  = \int \limits_{\mathbb{R}^3} \, \ud^3 \boldsymbol{\p} \, 
\int \limits_{\mathbb{R}^3} \, \ud^3 \boldsymbol{\p}' \, \bigg \langle \bigg \langle \bigg\{
\frac{-i p^0(\boldsymbol{\p})}{\sqrt{2 p^0(\boldsymbol{\p})}}\sqrt{B(\boldsymbol{\p}, p^0(\boldsymbol{\p}))}^{\mu}_{\lambda}
\, \times \\ 
\times \, \frac{i p'^k}{\sqrt{2 p^0(\boldsymbol{\p}')}}\sqrt{B(\boldsymbol{\p}', p^0(\boldsymbol{\p}'))}^{\nu}_{\gamma}
 e^{-i(p\cdot x + p \cdot y)} \bigg\} \, \partial_{\boldsymbol{\p}}^{\lambda} \, \partial_{\boldsymbol{\p}'}^{\gamma}
\, \Phi, \Psi \bigg \rangle \bigg \rangle
\end{multline*} 
\begin{multline*}
+ \, \int \limits_{\mathbb{R}^3} \, \ud^3 \boldsymbol{\p} \, \int \limits_{\mathbb{R}^3} \, \ud^3 \boldsymbol{\p}' 
\, \bigg \langle \bigg \langle \bigg\{
\frac{-i p^0(\boldsymbol{\p})}{\sqrt{2 p^0(\boldsymbol{\p})}}\sqrt{B(\boldsymbol{\p}, p^0(\boldsymbol{\p}))}^{\mu}_{\lambda}
\, \times \\ 
\times \, \frac{-i p'^k}{\sqrt{2 p^0(\boldsymbol{\p}')}}\sqrt{B(\boldsymbol{\p}', p^0(\boldsymbol{\p}'))}^{\nu}_{\gamma}
 e^{-i(p\cdot x - p' \cdot y)} \bigg\} \, \eta \, \partial_{\boldsymbol{\p}'}^{\gamma *} \, \eta \,\partial_{\boldsymbol{\p}}^{\lambda} \,  \Phi, \Psi \bigg \rangle \bigg \rangle 
\end{multline*}
\begin{multline*}
+ \, \int \limits_{\mathbb{R}^3} \, \ud^3 \boldsymbol{\p} \, \int \limits_{\mathbb{R}^3} \, \ud^3 \boldsymbol{\p}' 
\, \bigg \langle \bigg \langle \bigg\{
\frac{i p^0(\boldsymbol{\p})}{\sqrt{2 p^0(\boldsymbol{\p})}}\sqrt{B(\boldsymbol{\p}, p^0(\boldsymbol{\p}))}^{\mu}_{\lambda}
\, \times \\ 
\times \, \frac{i p'^k}{\sqrt{2 p^0(\boldsymbol{\p}')}}\sqrt{B(\boldsymbol{\p}', p^0(\boldsymbol{\p}'))}^{\nu}_{\gamma}
 e^{i(p\cdot x - p' \cdot y)} \bigg\} \, \eta \, \partial_{\boldsymbol{\p}}^{\lambda *} \, \eta \,\partial_{\boldsymbol{\p}'}^{\gamma} \,  \Phi, \Psi \bigg \rangle \bigg \rangle 
\end{multline*}
\begin{multline*}
+ \, \int \limits_{\mathbb{R}^3} \, \ud^3 \boldsymbol{\p} \, \int \limits_{\mathbb{R}^3} \, \ud^3 \boldsymbol{\p}' 
\, \bigg \langle \bigg \langle \bigg\{
\frac{i p^0(\boldsymbol{\p})}{\sqrt{2 p^0(\boldsymbol{\p})}}\sqrt{B(\boldsymbol{\p}, p^0(\boldsymbol{\p}))}^{\mu}_{\lambda}
\, \times \\ 
\times \, \frac{-i p'^k}{\sqrt{2 p^0(\boldsymbol{\p}')}}\sqrt{B(\boldsymbol{\p}', p^0(\boldsymbol{\p}'))}^{\nu}_{\gamma}
 e^{i(p\cdot x + p' \cdot y)} \bigg\} \, \eta \, \partial_{\boldsymbol{\p}}^{\lambda *} \, \partial_{\boldsymbol{\p}'}^{\gamma *} \,  \eta \, \Phi, \Psi \bigg \rangle \bigg \rangle 
\end{multline*} 
is a continuous functional on $(E)$, i.e. belongs to $(E)^*$;
and thus the Pettis integral defines a linear operator $\Xi(\boldsymbol{\x}, \boldsymbol{\y}, t): (E) \rightarrow (E)^*$ 
which turns out to be continuous, i.e. belongs to $\mathscr{L}((E),(E)^*)$. 
\end{lem*}

\qedsymbol \, 

Again by Lemma 2.1 of \cite{hida} and continuity of the operator $\eta$ (consider also the commutation rules for $\eta$
and $\partial_{\boldsymbol{\p}}^{\lambda}$) the functions 
\[
\begin{split}
\boldsymbol{\p} \times \boldsymbol{\p}' \mapsto \langle \langle  
\partial_{\boldsymbol{\p}}^{\lambda} \, \partial_{\boldsymbol{\p}'}^{\gamma}
\, \Phi, \,\, \Psi
\rangle \rangle = {\eta_{{}_{\Phi, \Psi}}}^{\lambda \gamma} (\boldsymbol{\p} \times \boldsymbol{\p}'), \\
\boldsymbol{\p} \times \boldsymbol{\p}' \mapsto \langle \langle   
\eta \, \partial_{\boldsymbol{\p}'}^{\gamma*} \, \eta \,\partial_{\boldsymbol{\p}}^{\lambda} 
\, \Phi, \,\, \Psi
\rangle \rangle = {{\eta^{*}}_{{}_{\Phi, \Psi}}}^{\lambda \gamma} (\boldsymbol{\p} \times \boldsymbol{\p}'), \\
\boldsymbol{\p} \times \boldsymbol{\p}' \mapsto \langle \langle  
\eta \, \partial_{\boldsymbol{\p}}^{\lambda *} \, \eta \,\partial_{\boldsymbol{\p}'}^{\gamma}
\, \Phi, \,\, \Psi
\rangle \rangle = {{{}^{*}\eta^{*}}_{{}_{\Phi, \Psi}}}^{\lambda \gamma} (\boldsymbol{\p} \times \boldsymbol{\p}'), \\
\boldsymbol{\p} \times \boldsymbol{\p}' \mapsto \langle \langle  
\eta \, \partial_{\boldsymbol{\p}}^{\lambda *} \, \partial_{\boldsymbol{\p}'}^{\gamma *} \,  \eta 
\, \Phi, \,\, \Psi
\rangle \rangle = {{\eta^{**}}_{{}_{\Phi, \Psi}}}^{\lambda \gamma} (\boldsymbol{\p} \times \boldsymbol{\p}'), 
\end{split}
\]
belong to the nuclear space $E \otimes E = \mathcal{S}_{A'''}(\mathbb{R}^3) \otimes \mathcal{S}_{A'''}(\mathbb{R}^3)
= \mathcal{S}_{A''' \otimes A'''}(\mathbb{R}^3 \times \mathbb{R}^3) = \mathcal{S}^0 (\mathbb{R}^3) \otimes 
\mathcal{S}^0 (\mathbb{R}^3) \subset \mathcal{S}(\mathbb{R}^6)$. Moreover, because the operators $Op_t:
E \rightarrow E = \mathcal{S}_{A'''}(\mathbb{R}^3) = \mathcal{S}^0 (\mathbb{R}^3)$  defined above are continuous,
the same holds for the extensions 
${Op_t}^{\mu}_{\lambda} \otimes {Op_t}^{\nu}_{\gamma}: E \otimes E \rightarrow E\otimes E$ of their algebraic tensor products. 
Thus, for each fixed $x = (\boldsymbol{\x}, t) \in \mathbb{R}^4$ and each $\Phi \in (E)$ 
the functions under the double integration sign belong to 
$L^1(\mathbb{R}^3 \times \mathbb{R}^3) \cap L^2(\mathbb{R}^3 \times \mathbb{R}^3)$, and thus 
the integral $\Xi(\boldsymbol{\x}, t) \Phi$ in the assertion of the Lemma does exist as the Pettis
integral (Prop. 8.1 in \cite{HKPS}, thus defining a linear operator 
$\Xi(\boldsymbol{\x}, \boldsymbol{\y}, t): (E) \rightarrow (E)^*$. 
Moreover, we see that 
for any fixed $\Phi, \Psi \in (E)$ and any fixed $x = (\boldsymbol{\x}, t), y = (\boldsymbol{\y}, t) \in \mathbb{R}^4 \in \mathbb{R}^4$,
$\langle \langle \Xi(\boldsymbol{\x}, \boldsymbol{\y}, t) \Phi, \Psi \rangle \rangle$ is equal 
\begin{multline*}
\widetilde{\xi_{1t}}\big(\boldsymbol{\x} \times \boldsymbol{\y} \big) + 
\widetilde{\xi_{2t}}\big((-\boldsymbol{\x}) \times \boldsymbol{\y}\big) 
+ \widetilde{\xi_{3t}}\big(\boldsymbol{\x} \times (-\boldsymbol{\y})\big) 
+ \widetilde{\xi_{4t}}\big((-\boldsymbol{\x}) \times (-\boldsymbol{\y})\big)
\end{multline*}
where 
$\widetilde{\xi_{1t}}, \ldots \widetilde{\xi_{4t}}$ are Fourier transforms of some elements $\xi_{1t}, \ldots \xi_{4t}$ of 
the nuclear space $E \otimes E = \mathcal{S}_{A'''}(\mathbb{R}^3) \otimes \mathcal{S}_{A'''}(\mathbb{R}^3)
= \mathcal{S}_{A''' \otimes A'''}(\mathbb{R}^3 \times \mathbb{R}^3) = \mathcal{S}^0 (\mathbb{R}^3) \otimes 
\mathcal{S}^0 (\mathbb{R}^3) \subset \mathcal{S}(\mathbb{R}^6)$. Thus 
proceeding similarly as in the proof of the continuity of the operator $A^\mu (\boldsymbol{\x}, t)$ we show
using the inequalities (2-2) of Lemma 2.1 of \cite{hida} that the operator 
$\Xi(\boldsymbol{\x}, \boldsymbol{\y},  t): (E) \rightarrow (E)^*$ is continuous for the strong topology on $(E)^*$. 
Indeed: note that our 5-th Lemma of Subsection \ref{SA=S0} for $\mathbb{R}^3$ with the corresponding nuclear spce $\mathcal{S}_{A'''}(\mathbb{R}^3)$ is likewise valid for 
$\mathbb{R}^3 \times \mathbb{R}^3 = \mathbb{R}^6$ with the corresponding nuclear space $\mathcal{S}_{A''''''}(\mathbb{R}^6)
= \mathcal{S}_{A^{(6)}}(\mathbb{R}^6)$. And on the other hand it is easily checked that
$\mathcal{S}_{A''' \otimes A'''}(\mathbb{R}^3 \times \mathbb{R}^3) \subset \mathcal{S}_{A^{(6)}}(\mathbb{R}^6)$ 
with the system of norms 
\[
\Big\{\|(A''' \otimes A''')^p  \cdot \|_{{}_{L^2(\mathbb{R}^3 \times \mathbb{R}^3)}}\Big\}_{p \in \mathbb{N}}
\]
on $\mathcal{S}^0 (\mathbb{R}^3) \otimes 
\mathcal{S}^0 (\mathbb{R}^3) = \mathcal{S}_{A'''}(\mathbb{R}^3) \otimes \mathcal{S}_{A'''}(\mathbb{R}^3) =
\mathcal{S}_{A''' \otimes A'''}(\mathbb{R}^3 \times \mathbb{R}^3)$
stronger than the system 
\[
\Big\{\|(A^{(6)})^p  \cdot \|_{{}_{L^2(\mathbb{R}^3 \times \mathbb{R}^3)}}\Big\}_{p \in \mathbb{N}}
\]
of norms on $\mathcal{S}^0 (\mathbb{R}^3) \otimes \mathcal{S}^0 (\mathbb{R}^3)$. 

\qed

\begin{lem*} 
For the operator $\Xi(\boldsymbol{\x}, \boldsymbol{\y}, t)$ of the preceding Lemma we have
\[
 \Xi(\boldsymbol{\x}, \boldsymbol{\y}, t) \,\, = \,\,\,\, 
:\partial_0 A^\mu (\boldsymbol{\x}, t) \partial_k A^\nu (\boldsymbol{\y}, t): \,\,.
\]
Moreover, the functions
\begin{equation}\label{int-funct-T_0k}
\mathbb{R}^3 \ni \boldsymbol{\x} \mapsto \langle \langle  
: \partial_0 A^\mu (\boldsymbol{\x}, t) \partial_k A^\nu (\boldsymbol{\x}, t): \Phi, \Psi
\rangle \rangle,
\end{equation}
for $\Phi, \Psi \in (E)$, $t \in \mathbb{R}$, $\mu, \nu \in \{0,1,2,3\}$, $k \in \{1,2,3\}$, 
are continuous, belong to $L^1 (\mathbb{R}^3) \cap L^2(\mathbb{R}^3)$ and even to 
$\mathcal{S}^{00}(\mathbb{R}^3) = \widetilde{\mathcal{S}_{A'''}(\mathbb{R}^3)} = 
\widetilde{\mathcal{S}^0(\mathbb{R}^3)}$ -- the Fourier image of the nuclear space 
$\mathcal{S}^0(\mathbb{R}^3) = \mathcal{S}_{A'''}(\mathbb{R}^3) \subset \mathcal{S}(\mathbb{R}^3)$. 
\end{lem*}

\qedsymbol \, 
In the course of the proof of the preceding Lemma we have shown that the function
\[
\mathbb{R}^3 \times \mathbb{R}^3 \ni \boldsymbol{\x} \times \boldsymbol{\y} \mapsto \langle \langle  
: \partial_0 A^\mu (\boldsymbol{\x}, t) \partial_k A^\nu (\boldsymbol{\y}, t): \Phi, \Psi
\rangle \rangle
\]
is a Fourier transform of an element of $E \otimes E = \mathcal{S}_{A'''}(\mathbb{R}^3) \otimes \mathcal{S}_{A'''}(\mathbb{R}^3)= \mathcal{S}_{A''' \otimes A'''}(\mathbb{R}^3 \times \mathbb{R}^3) = \mathcal{S}^0 (\mathbb{R}^3) \otimes 
\mathcal{S}^0 (\mathbb{R}^3) \subset \mathcal{S}(\mathbb{R}^6)$. Because 
$\mathcal{S}_{A'''}(\mathbb{R}^3) = \mathcal{S}^0(\mathbb{R}^3)$ is an algebra under pointwise multiplication, then
the function indicated in the assertion of the Lemma belongs to $\widetilde{\mathcal{S}^0(\mathbb{R}^3)} \subset \mathcal{S}(\mathbb{R}^3)$; in particular it belongs to $L^1 (\mathbb{R}^3) \cap L^2(\mathbb{R}^3)$. 

From (\ref{eta-hida-partial-exp}) we obtain 
\begin{equation}\label{circ1-circ4}
\begin{split}
\langle \langle  
\partial_{\boldsymbol{\p}}^{\lambda} \, \partial_{\boldsymbol{\p}'}^{\gamma}
\, \Phi_\xi, \,\, \Phi_\zeta
\rangle \rangle = \xi^\lambda(\boldsymbol{\p}) \xi^{\gamma}(\boldsymbol{\p}') e^{\langle \xi, \zeta \rangle}, \\
\langle \langle   
\eta \, \partial_{\boldsymbol{\p}'}^{\gamma*} \, \eta \,\partial_{\boldsymbol{\p}}^{\lambda} 
\, \Phi_\xi , \,\, \Phi_\zeta
\rangle \rangle = \xi^\lambda(\boldsymbol{\p}) (\mathfrak{J}_{\bar{p}}\zeta)^{\gamma}(\boldsymbol{\p}') e^{\langle \xi, \zeta \rangle}, \\ 
\langle \langle  
\eta \, \partial_{\boldsymbol{\p}}^{\lambda *} \, \eta \,\partial_{\boldsymbol{\p}'}^{\gamma}
\, \Phi_\xi , \,\, \Phi_\zeta
\rangle \rangle = \xi^\gamma(\boldsymbol{\p}') (\mathfrak{J}_{\bar{p}}\zeta)^{\lambda}(\boldsymbol{\p}) e^{\langle \xi, \zeta \rangle}, \\ 
\langle \langle  
\eta \, \partial_{\boldsymbol{\p}}^{\lambda *} \, \partial_{\boldsymbol{\p}'}^{\gamma *} \,  \eta 
\, \Phi_\xi, \,\, \Phi_\zeta
\rangle \rangle = (\mathfrak{J}_{\bar{p}}\zeta)^{\lambda}(\boldsymbol{\p}) (\mathfrak{J}_{\bar{p}}\zeta)^{\gamma}(\boldsymbol{\p}') e^{\langle \xi, \zeta \rangle}.
\end{split}
\end{equation} 
From (\ref{circ1-circ4}) almost immediately follows that
\[
\widehat{\Xi(\boldsymbol{\x}, \boldsymbol{\y}, t)} = e^{-\langle \xi, \zeta \rangle}
\big( \partial_0 A^\mu (\boldsymbol{\x}, t) \big)^\wedge 
(\xi, \zeta) \, \big( \partial_k A^\nu (\boldsymbol{\y}, t) \big)^\wedge (\xi, \zeta),
\]
for all $\xi, \zeta \in E = \mathcal{S}_{A'''}(\mathbb{R}^3)$. Because the symbol of the operator
in $\mathscr{L}((E), (E)^*)$ uniquely characterizes the operator itself, compare
e.g. Lemma 4.2 of \cite{obata}, then 
\[
 : \partial_0 A^\mu (\boldsymbol{\x}, t) \partial_k A^\nu (\boldsymbol{\y}, t): \,\,\,\,
 = \Xi(\boldsymbol{\x}, \boldsymbol{\y}, t).
\]

\qed

\begin{lem*}
For $t \in \mathbb{R}$ and $\Phi \in (E)$ the integral
\[
\int \limits_{\mathbb{R}^3} : \partial_0 A^\mu (\boldsymbol{\x}, t) \partial_k A^\nu (\boldsymbol{\x}, t):
\, \Phi \, \ud^3 \boldsymbol{\x}  
\]
is well defined as the Pettis integral and represents an element of $(E)^*$, thus giving a well defined
linear operator, denoted by
\[
\int \limits_{\mathbb{R}^3} : \partial_0 A^\mu (\boldsymbol{\x}, t) \partial_k A^\nu (\boldsymbol{\x}, t):
\, \ud^3 \boldsymbol{\x},
\]
from $(E)$ into $(E)^*$.
\end{lem*}

\qedsymbol \, 
This Lemma is a simple corollary of the preceding Lemma and Prop. 8.1 of \cite{HKPS}.
\qed

\begin{lem*}
The operator 
\[
\int \limits_{\mathbb{R}^3} : \partial_0 A^\mu (\boldsymbol{\x}, t) \partial_k A^\nu (\boldsymbol{\x}, t):
\, \ud^3 \boldsymbol{\x}
\]
$(E) \rightarrow (E)^*$ is continuous for the strong topology on $(E)^*$, i.e. it belongs to 
$\mathscr{L}((E), (E)^*)$. 
\end{lem*}

\qedsymbol \, 
For any $\Phi, \Psi \in (E)$ the quantity
\begin{multline*}
\Bigg| \bigg \langle \bigg \langle  
\int \limits_{\mathbb{R}^3} : \partial_0 A^\mu (\boldsymbol{\x}, t) \partial_k A^\nu (\boldsymbol{\x}, t):
\, \ud^3 \boldsymbol{\x} \, \Phi, \, \Psi
 \bigg \rangle \bigg \rangle \Bigg|^2 \\  = 
\Bigg| \bigg \langle \bigg \langle  
\int \limits_{\mathbb{R}^3} : \partial_0 A^\mu (\boldsymbol{\x}, t) \partial_k A^\nu (\boldsymbol{\x}, t):
\, \Phi \,\,\, \ud^3 \boldsymbol{\x} \, , \, \Psi
 \bigg \rangle \bigg \rangle \Bigg|^2
\end{multline*}
is easily seen to be majorized by four times the sum of four summands of the form
\[
\bigg| \int \limits_{\mathbb{R}^3} \big| \big({Op_t \otimes Op_t \, \eta_{{}_{\Phi, \Psi}}}\big)^\sim
(\boldsymbol{\x} \times \boldsymbol{\x}) \bigg|^2
\, \ud^3 \boldsymbol{\x}
\]
one for each of the functions ${\eta_{{}_{\Phi, \Psi}}}, {\eta^{*}}_{{}_{\Phi, \Psi}}, 
{{}^{*}\eta^{*}}_{{}_{\Phi, \Psi}}, {\eta^{**}}_{{}_{\Phi, \Psi}} \in E \otimes E$ defined above with the continuous operators $Op_t: E \rightarrow E$,
$Op_t \otimes Op_t: E \otimes E \rightarrow E \otimes E$ defined as above. Because the Fourier transform is
unitary for the $L^2$-norm $\| \cdot \|_{{}_{L^2(\mathbb{R}^m)}}$, then we obtain\footnote{Exactly as in the proof of the continuity of the operator $A^\mu(\boldsymbol{\x},t)$.} from the continuity
of $Op_t \otimes Op_t: E \otimes E \rightarrow E \otimes E$ and from the inequality (2-2) of the Lemma 2.1
of \cite{hida} the inequality
\[
\Bigg| \bigg \langle \bigg \langle  
\int \limits_{\mathbb{R}^3} : \partial_0 A^\mu (\boldsymbol{\x}, t) \partial_k A^\nu (\boldsymbol{\x}, t):
\, \ud^3 \boldsymbol{\x} \, \Phi, \, \Psi
 \bigg \rangle \bigg \rangle \Bigg|^2 \leq C(t) \| \Phi \|_p \, \| \Psi \|_p,
\]
for all $p \in \mathbb{N}$ greater than some fixed $q_0 \in \mathbb{N}$; from which the continuity
of the operator of the assertion of the Lemma follows as the continuity of $A^\mu(\boldsymbol{\x},t)$
from (\ref{ineq-6}).

\qed

\begin{lem*}
Symbols of the operators 
\begin{equation}\label{noether-int-T-0k}
g_{\mu \nu} \, \int \limits_{\mathbb{R}^3} : \partial_0 A^\mu (\boldsymbol{\x}, t) \partial_k A^\nu (\boldsymbol{\x}, t):
\, \ud^3 \boldsymbol{\x}
\end{equation}
(summation with respect to $\mu$ and $\nu$) and 
\begin{equation}\label{generator-P-k}
\sum \limits_{\mu, \nu} \, g_{\mu k} \int \limits_{\mathbb{R}^3}
p^\mu(\boldsymbol{\p}) \,\, 
\partial_{\boldsymbol{\p}}^{\nu \, *} \partial_{\boldsymbol{\p}}^{\nu}
\,\,\, \ud^3 \boldsymbol{\p} = d\Gamma(P_k)
\end{equation}
are equal, and thus
\[
g_{\mu \nu} \, \int \limits_{\mathbb{R}^3} : \partial_0 A^\mu (\boldsymbol{\x}, t) \partial_k A^\nu (\boldsymbol{\x}, t):
\, \ud^3 \boldsymbol{\x} =
\sum \limits_{\mu, \nu} \, g_{\mu k} \int \limits_{\mathbb{R}^3}
p^\mu(\boldsymbol{\p}) \,\, 
\partial_{\boldsymbol{\p}}^{\nu \, *} \partial_{\boldsymbol{\p}}^{\nu}
\,\,\, \ud^3 \boldsymbol{\p} = d\Gamma(P_k)
\]
as elements of $\mathscr{L}((E), (E)^*)$.
\end{lem*}

\qedsymbol \,

Because the operator (\ref{noether-int-T-0k}) is pointwisely well defined as Pettis integral,
then for any $\xi, \zeta \in E$ we obtain the following formula for its symbol
\begin{multline}\label{symbol-T-ok}
\bigg( g_{\mu \nu} \, \int \limits_{\mathbb{R}^3} : \partial_0 A^\mu (\boldsymbol{\x}, t) \partial_k 
A^\nu (\boldsymbol{\x}, t): \, \ud^3 \boldsymbol{\x} \bigg)^\wedge (\xi, \zeta) \\ =
 g_{\mu \nu} \, \int \limits_{\mathbb{R}^3} \Big(: \partial_0 A^\mu (\boldsymbol{\x}, t) \partial_k A^\nu (\boldsymbol{\x}, t):\Big)^\wedge (\xi, \zeta) \, \ud^3 \boldsymbol{\x}  \\ = e^{-\langle \xi, \zeta \rangle} \,
g_{\mu \nu} \, \int \limits_{\mathbb{R}^3} \, \Big(\partial_0 A^\mu (\boldsymbol{\x}, t) \Big)^\wedge (\xi, \zeta) \,\,\, 
\Big( \partial_k A^\nu (\boldsymbol{\x}, t)\Big)^\wedge (\xi, \zeta) \,\, \ud^3 \boldsymbol{\x}.
\end{multline}

Now each of the factors under the integral sign, i.e. each of the symbols 
$\Big(\partial_0 A^\mu (\boldsymbol{\x}, t) \Big)^\wedge (\xi, \zeta)$ and 
$\Big( \partial_k A^\nu (\boldsymbol{\x}, t)\Big)^\wedge (\xi, \zeta)$, is the sum of two integrals 
-- the components corresponding to positive and negative sign of the energy frequencies. 
Now we show that the contributions to the above expression (\ref{symbol-T-ok}) 
coming from the product of the components corresponding to the same energy sign is equal to zero.

Namely consider the contribution coming from the product of the components both
containing the factor $e^{-ip\cdot x}$:
\begin{multline*}
g_{\mu \nu} \, \int \limits_{\mathbb{R}^3} \, \ud^3 \boldsymbol{\x} \, \bigg(
\int \limits_{\mathbb{R}^3} \,  
\frac{\ud^3 \boldsymbol{\p}}{\sqrt{2 p^0(\boldsymbol{\p})}} (-ip^0(\boldsymbol{\p}))
\sqrt{B(\boldsymbol{\p}, p^0(\boldsymbol{\p}))}^{\mu}_{\lambda} \xi^\lambda (\boldsymbol{\p}) e^{-ip \cdot x}
\bigg) \, \times \\
\times \, \bigg(
\int \limits_{\mathbb{R}^3} \,  
\frac{\ud^3 \boldsymbol{\p}'}{\sqrt{2 p^0(\boldsymbol{\p}')}} (ip'^k))
\sqrt{B(\boldsymbol{\p}', p^0(\boldsymbol{\p}'))}^{\nu}_{\gamma} \xi^\gamma (\boldsymbol{\p}') e^{-ip' \cdot x}
\bigg) 
\end{multline*}
\begin{multline*}
= g_{\mu \nu} \,
\int \limits_{\mathbb{R}^3}  \,  
\frac{\ud^3 \boldsymbol{\p}}{\sqrt{2 p^0(\boldsymbol{\p})}} 
e^{-ip^0(\boldsymbol{\p})} p^0(\boldsymbol{\p})
\sqrt{B(\boldsymbol{\p}, p^0(\boldsymbol{\p}))}^{\mu}_{\lambda}
 \xi^\lambda (\boldsymbol{\p}) \, \times \\ \times \,
\int \limits_{\mathbb{R}^3} \, \ud^3 \boldsymbol{\x}
e^{i\boldsymbol{\p} \cdot \boldsymbol{\x}}
\int \limits_{\mathbb{R}^3} \,  
\frac{\ud^3 \boldsymbol{\p}'}{\sqrt{2 p^0(\boldsymbol{\p}')}} e^{-ip^0(\boldsymbol{\p}')t} p'^k
\sqrt{B(\boldsymbol{\p}', p^0(\boldsymbol{\p}'))}^{\nu}_{\gamma} 
\xi^\gamma (\boldsymbol{\p}') e^{i\boldsymbol{\p}' \cdot \boldsymbol{\x}}
\end{multline*}
\begin{multline}\label{(+component)times(+component)-0}
= g_{\mu \nu} \,
\int \limits_{\mathbb{R}^3}  \,  
\frac{\ud^3 \boldsymbol{\p}}{\sqrt{2 p^0(\boldsymbol{\p})}} 
e^{-ip^0(\boldsymbol{\p})} p^0(\boldsymbol{\p})
\sqrt{B(\boldsymbol{\p}, p^0(\boldsymbol{\p}))}^{\mu}_{\lambda}
 \xi^\lambda (\boldsymbol{\p}) \, \times \\ \times \, 
\frac{1}{\sqrt{2 p^0(-\boldsymbol{\p})}} e^{-ip^0(-\boldsymbol{\p})t} (-p^k)
\sqrt{B(-\boldsymbol{\p}, p^0(-\boldsymbol{\p}))}^{\nu}_{\gamma} 
\xi^\gamma (-\boldsymbol{\p}), 
\end{multline}
where the first equality follows from the Fubini theorem and the second follows on the basis of 
the Fourier inversion formula \cite{gelfand-comm-norm-rings}, Ch. IV.25.2, 
Thm. 1\footnote{Note that the Fourier transform maps 
continuously $\mathcal{S}(\mathbb{R}^n)$ onto itself, and the Fourier inversion formula holds on 
$\mathcal{S}(\mathbb{R}^n)$, \cite{Rudin} Thm 7.7. In the physical literature the argument is expressed by 
using a formal formula for the Dirac delta function 
$\delta(\boldsymbol{\p})$, namely: the action of the Dirac delta functional 
$\delta(\boldsymbol{\p})$ on 
$\mathcal{S}^0 (\mathbb{R}^3)$ or on $\mathcal{S} (\mathbb{R}^3)$ may be expressed by the integration over 
$\boldsymbol{\p}'$ with the ''function'' $\int \limits_{\mathbb{R}^3} \, \ud^3 \boldsymbol{\x} \, e^{i(\boldsymbol{\p} - \boldsymbol{\p}') \cdot \boldsymbol{\x}}$.}, which is justified because
$\xi^\lambda \in \mathcal{S}^0 (\mathbb{R}^3) = \mathcal{S}_{A'''}(\mathbb{R}^3) 
\subset \mathcal{S}(\mathbb{R}^3) \subset L^1(\mathbb{R}^3) \cap L^2(\mathbb{R}^3)$, and (the first
Lemma of Subsection \ref{diffSA})
the functions 
\[
\begin{split}
\bigg( \boldsymbol{\p} \mapsto \frac{p^k}{\sqrt{2 p^0(\boldsymbol{\p})}} e^{-ip^0(\boldsymbol{\p})t} 
\sqrt{B(\boldsymbol{\p}, p^0(\boldsymbol{\p}))}^{\nu}_{\gamma} 
\xi^\gamma (\boldsymbol{\p}) \bigg) \,\,\, \textrm{and} \\
\bigg( \boldsymbol{\p} \mapsto 
e^{-ip^0(\boldsymbol{\p})} \sqrt{p^0(\boldsymbol{\p})} 
\sqrt{B(\boldsymbol{\p}, p^0(\boldsymbol{\p}))}^{\mu}_{\lambda}
 \xi^\lambda (\boldsymbol{\p}) \bigg), \\
 \xi^\lambda \in \mathcal{S}_{A'''}(\mathbb{R}^3) = \mathcal{S}^0 (\mathbb{R}^3), 
\end{split}
\]
belong to $\mathcal{S}^0 (\mathbb{R}^3) = \mathcal{S}_{A'''}(\mathbb{R}^3) 
\subset \mathcal{S}(\mathbb{R}^3) \subset L^1(\mathbb{R}^3) \cap L^2(\mathbb{R}^3)$.
Therefore, the expression (\ref{(+component)times(+component)-0}) is equal to 
\begin{multline}\label{(+component)times(+component)-1}
- \frac{1}{2} 
\int \limits_{\mathbb{R}^3}  \,  
\ud^3 \boldsymbol{\p} \,
 \xi^\lambda (\boldsymbol{\p}) \xi^\gamma (-\boldsymbol{\p}) 
\sqrt{B(\boldsymbol{\p}, p^0(\boldsymbol{\p}))}^{\mu}_{\lambda}
\sqrt{B(-\boldsymbol{\p}, p^0(\boldsymbol{\p}))}^{\nu}_{\gamma} \,\, g_{\mu \nu} \, 
e^{-2ip^0(\boldsymbol{\p})t} 
\,  p^k \\ =
- \frac{1}{2} 
\int \limits_{\mathbb{R}^3}  \,  
\ud^3 \boldsymbol{\p} \,
 \xi^\lambda (\boldsymbol{\p}) \xi^\gamma (-\boldsymbol{\p}) 
C(\boldsymbol{\p})_{\lambda \gamma} \, 
e^{-2ip^0(\boldsymbol{\p})t} 
\,  p^k
\end{multline}
where we have introduced (summation with respect to $\mu, \nu$)
\[
C(\boldsymbol{\p})_{\lambda \gamma} = 
\sqrt{B(\boldsymbol{\p}, p^0(\boldsymbol{\p}))}^{\mu}_{\lambda}
\sqrt{B(-\boldsymbol{\p}, p^0(\boldsymbol{\p}))}^{\nu}_{\gamma} \,\, g_{\mu \nu}
\]
in order to simplify the notation. Using the explicit formula (\ref{sqrtB})
for the matrix $\sqrt{B(\boldsymbol{\p}, p^0(\boldsymbol{\p}))}$ we can 
show that\footnote{Note in passing that checking of the formulae (\ref{C(p)-mu-nu=C(-p)-nu-mu}) 
is almost immediate when using (\ref{BJBJ=1})
and the formulae
\[
\sqrt{B(-\boldsymbol{\p}, p^0(\boldsymbol{\p}))} = \sqrt{B(\boldsymbol{\p}, p^0(\boldsymbol{\p}))} +
(r^{-2} -1)  \left( \begin{array}{cccc} 0 & p^1 & p^2 & p^3 \\
                                p^1 & 0 & 0 & 0     \\
                                p^2 & 0 & 0 & 0  \\
                                     p^3 & 0 & 0 & 0  \end{array}\right).
\]} 
\begin{equation}\label{C(p)-mu-nu=C(-p)-nu-mu}
C(\boldsymbol{\p})_{\lambda \gamma} = C(-\boldsymbol{\p})_{\gamma \lambda}.
\end{equation}
It is easily seen that from (\ref{C(p)-mu-nu=C(-p)-nu-mu}) it follows that the function
\[
\boldsymbol{\p} \mapsto f(\boldsymbol{\p}) = \xi^\lambda (\boldsymbol{\p}) \xi^\gamma (-\boldsymbol{\p}) 
C(\boldsymbol{\p})_{\lambda \gamma} \, 
e^{-2ip^0(\boldsymbol{\p})t} 
\,  p^k
\]
under the integral sign in the expression (\ref{(+component)times(+component)-1})
is an odd function: $f(-\boldsymbol{\p}) = - f(\boldsymbol{\p})$, and thus the expression
(\ref{(+component)times(+component)-1}), equal to (\ref{(+component)times(+component)-0}),
is equal to zero, i.e. for all $\xi \in E$
\begin{multline}\label{(+component)times(+component)-I}
g_{\mu \nu} \, \int \limits_{\mathbb{R}^3} \, \ud^3 \boldsymbol{\x} \, \Bigg(
\int \limits_{\mathbb{R}^3} \,  
\frac{\ud^3 \boldsymbol{\p}}{\sqrt{2 p^0(\boldsymbol{\p})}} (-ip^0(\boldsymbol{\p}))
\sqrt{B(\boldsymbol{\p}, p^0(\boldsymbol{\p}))}^{\mu}_{\lambda} \xi^\lambda (\boldsymbol{\p}) e^{-ip \cdot x}  
\Bigg) \, \times \\
\times \, \Bigg(
\int \limits_{\mathbb{R}^3} \,  
\frac{\ud^3 \boldsymbol{\p}'}{\sqrt{2 p^0(\boldsymbol{\p}')}} (ip'^k))
\sqrt{B(\boldsymbol{\p}', p^0(\boldsymbol{\p}'))}^{\nu}_{\gamma} \xi^\gamma (\boldsymbol{\p}') e^{-ip' \cdot x}
\Bigg) = 0;
\end{multline}
and similarly for all $\zeta \in E$ we have\footnote{Note that the identities 
(\ref{(+component)times(+component)-I}) and (\ref{(+component)times(+component)-II})
correspond to the fact that the contributions to the conserved invariant $\int :T^{0k}: \ud \boldsymbol{\x}$, 
with the energy-momentum tensor $T^{\mu \nu}$,
coming from the product of the creation part with the creation part as well as from the
product of the annihilation part with the annihilation part is zero, compare e. g. pages 28-30 of the
1980 Ed. of \cite{Bogoliubov_Shirkov}.}
\begin{multline}\label{(+component)times(+component)-II}
g_{\mu \nu} \, \int \limits_{\mathbb{R}^3} \, \ud^3 \boldsymbol{\x} \, \Bigg(
\int \limits_{\mathbb{R}^3} \,  
\frac{\ud^3 \boldsymbol{\p}}{\sqrt{2 p^0(\boldsymbol{\p})}} (ip^0(\boldsymbol{\p}))
\sqrt{B(\boldsymbol{\p}, p^0(\boldsymbol{\p}))}^{\mu}_{\lambda} 
(\mathfrak{J}_{\bar{p}}\zeta)^\lambda (\boldsymbol{\p}) e^{ip \cdot x}  
\Bigg) \, \times \\
\times \, \Bigg(
\int \limits_{\mathbb{R}^3} \,  
\frac{\ud^3 \boldsymbol{\p}'}{\sqrt{2 p^0(\boldsymbol{\p}')}} (-ip'^k))
\sqrt{B(\boldsymbol{\p}', p^0(\boldsymbol{\p}'))}^{\nu}_{\gamma} 
(\mathfrak{J}_{\bar{p}}\zeta)^\gamma (\boldsymbol{\p}') e^{ip' \cdot x}
\Bigg) = 0;
\end{multline}

Therefore, for each $\xi, \zeta \in E$ we have the following formula for the
symbol 
\begin{multline*}
\bigg( g_{\mu \nu} \, \int \limits_{\mathbb{R}^3} : \partial_0 A^\mu (\boldsymbol{\x}, t) \partial_k A^\nu (\boldsymbol{\x}, t): \, \ud^3 \boldsymbol{\x} \bigg)^\wedge (\xi, \zeta) \\ 
 = e^{-\langle \xi, \zeta \rangle} \,
g_{\mu \nu} \, \int \limits_{\mathbb{R}^3} \, \Big(\partial_0 A^\mu (\boldsymbol{\x}, t) \Big)^\wedge (\xi, \zeta) \,\,\, 
\Big( \partial_k A^\nu (\boldsymbol{\x}, t)\Big)^\wedge (\xi, \zeta) \,\, \ud^3 \boldsymbol{\x} \\ = \,
e^{\langle \xi, \zeta \rangle} \,
g_{\mu \nu} \, \int \limits_{\mathbb{R}^3} \, \ud^3 \boldsymbol{\x} \, \Bigg(
\int \limits_{\mathbb{R}^3} \,  
\frac{\ud^3 \boldsymbol{\p}}{\sqrt{2 p^0(\boldsymbol{\p})}} (-ip^0(\boldsymbol{\p}))
\sqrt{B(\boldsymbol{\p}, p^0(\boldsymbol{\p}))}^{\mu}_{\lambda} 
\xi^\lambda (\boldsymbol{\p}) e^{-ip \cdot x}  
\Bigg) \, \times \\
\times \, \Bigg(
\int \limits_{\mathbb{R}^3} \,  
\frac{\ud^3 \boldsymbol{\p}'}{\sqrt{2 p^0(\boldsymbol{\p}')}} (-ip'^k))
\sqrt{B(\boldsymbol{\p}', p^0(\boldsymbol{\p}'))}^{\nu}_{\gamma} 
(\mathfrak{J}_{\bar{p}}\zeta)^\gamma (\boldsymbol{\p}') e^{ip' \cdot x} 
\Bigg) \\ +
e^{\langle \xi, \zeta \rangle} \,
g_{\mu \nu} \, \int \limits_{\mathbb{R}^3} \, \ud^3 \boldsymbol{\x} \, \Bigg(
\int \limits_{\mathbb{R}^3} \,  
\frac{\ud^3 \boldsymbol{\p}}{\sqrt{2 p^0(\boldsymbol{\p})}} (ip^0(\boldsymbol{\p}))
\sqrt{B(\boldsymbol{\p}, p^0(\boldsymbol{\p}))}^{\mu}_{\lambda} 
(\mathfrak{J}_{\bar{p}}\zeta)^\lambda (\boldsymbol{\p}) e^{ip \cdot x}  
\Bigg) \, \times \\
\times \, \Bigg(
\int \limits_{\mathbb{R}^3} \,  
\frac{\ud^3 \boldsymbol{\p}'}{\sqrt{2 p^0(\boldsymbol{\p}')}} (ip'^k))
\sqrt{B(\boldsymbol{\p}', p^0(\boldsymbol{\p}'))}^{\nu}_{\gamma} 
\xi^\gamma (\boldsymbol{\p}') e^{-ip' \cdot x} 
\Bigg)
\end{multline*}
\begin{multline*}
= e^{\langle \xi, \zeta \rangle} \, g_{\mu \nu} \,
\int \limits_{\mathbb{R}^3}  \,  
\frac{\ud^3 \boldsymbol{\p}}{\sqrt{2 p^0(\boldsymbol{\p})}} 
e^{ip^0(\boldsymbol{\p})} (-p^0(\boldsymbol{\p}))
\sqrt{B(\boldsymbol{\p}, p^0(\boldsymbol{\p}))}^{\mu}_{\lambda}
 \xi^\lambda (\boldsymbol{\p}) \, \times \\ \times \,
\int \limits_{\mathbb{R}^3} \, \ud^3 \boldsymbol{\x}
e^{i\boldsymbol{\p} \cdot \boldsymbol{\x}}
\int \limits_{\mathbb{R}^3} \,  
\frac{\ud^3 \boldsymbol{\p}'}{\sqrt{2 p^0(\boldsymbol{\p}')}} e^{-ip^0(\boldsymbol{\p}')t} p'^k
\sqrt{B(\boldsymbol{\p}', p^0(\boldsymbol{\p}'))}^{\nu}_{\gamma} 
[\mathfrak{J}_{\bar{p}}]^{\gamma \sigma} \zeta^\sigma (\boldsymbol{\p}') e^{-i\boldsymbol{\p}' \cdot \boldsymbol{\x}}
\end{multline*}

\begin{multline*}
+ \, e^{\langle \xi, \zeta \rangle} \, g_{\mu \nu} \,
\int \limits_{\mathbb{R}^3}  \,  
\frac{\ud^3 \boldsymbol{\p}}{\sqrt{2 p^0(\boldsymbol{\p})}} 
e^{-ip^0(\boldsymbol{\p})} (-p^0(\boldsymbol{\p}))
\sqrt{B(\boldsymbol{\p}, p^0(\boldsymbol{\p}))}^{\mu}_{\lambda}
 [\mathfrak{J}_{\bar{p}}]^{\lambda \sigma}\zeta^\sigma (\boldsymbol{\p}) \, \times \\ \times \,
\int \limits_{\mathbb{R}^3} \, \ud^3 \boldsymbol{\x}
e^{-i\boldsymbol{\p} \cdot \boldsymbol{\x}}
\int \limits_{\mathbb{R}^3} \,  
\frac{\ud^3 \boldsymbol{\p}'}{\sqrt{2 p^0(\boldsymbol{\p}')}} e^{ip^0(\boldsymbol{\p}')t} p'^k
\sqrt{B(\boldsymbol{\p}', p^0(\boldsymbol{\p}'))}^{\nu}_{\gamma} 
\xi^\gamma (\boldsymbol{\p}') e^{i\boldsymbol{\p}' \cdot \boldsymbol{\x}}
\end{multline*}
\begin{multline*}
= e^{\langle \xi, \zeta \rangle} \,
\int \limits_{\mathbb{R}^3}  \,  
\frac{\ud^3 \boldsymbol{\p}}{2 p^0(\boldsymbol{\p})}
 \, (- p^0(\boldsymbol{\p}) p^k) \times \\ \times \,
\sqrt{B(\boldsymbol{\p}, p^0(\boldsymbol{\p}))}^{\mu}_{\lambda}
\sqrt{B(\boldsymbol{\p}, p^0(\boldsymbol{\p}))}^{\nu}_{\gamma} \,\, g_{\mu \nu} \,
\xi^\lambda (\boldsymbol{\p}) \, g^{\gamma \sigma} \zeta^\sigma (\boldsymbol{\p}) 
\end{multline*}
\begin{multline*}
+ e^{\langle \xi, \zeta \rangle} \,
\int \limits_{\mathbb{R}^3 \times \mathbb{R}^3}  \,  
\frac{\ud^3 \boldsymbol{\p}}{2 p^0(\boldsymbol{\p})} \, (-p^0(\boldsymbol{\p}) p^k) \, \times \\ \times \,
\sqrt{B(\boldsymbol{\p}, p^0(\boldsymbol{\p}))}^{\mu}_{\lambda}
\sqrt{B(\boldsymbol{\p}, p^0(\boldsymbol{\p}))}^{\nu}_{\gamma} \,\, g_{\mu \nu} \,
\ g^{\lambda \sigma} \zeta^\sigma(\boldsymbol{\p}) \, \xi^\gamma(\boldsymbol{\p}) 
\end{multline*}
where the third equality follows from the Fubini theorem and the last equality follows from the 
application of the inversion formula for the Fourier transform,
\cite{gelfand-comm-norm-rings}, Ch. IV.25.2, Theorem 1, justified because the functions 
$\xi^\lambda, \zeta^\gamma$ and (the first Lemma of Subsection \ref{diffSA}) the functions
\[
\begin{split}
\bigg( \boldsymbol{\p} \mapsto \frac{p^k}{\sqrt{2 p^0(\boldsymbol{\p})}} e^{-ip^0(\boldsymbol{\p})t} 
\sqrt{B(\boldsymbol{\p}, p^0(\boldsymbol{\p}))}^{\nu}_{\gamma} 
\xi^\gamma (\boldsymbol{\p}) \bigg) \,\,\, \textrm{and} \\
\bigg( \boldsymbol{\p} \mapsto 
e^{-ip^0(\boldsymbol{\p})} \sqrt{p^0(\boldsymbol{\p})} 
\sqrt{B(\boldsymbol{\p}, p^0(\boldsymbol{\p}))}^{\mu}_{\lambda}
 \xi^\lambda (\boldsymbol{\p}) \bigg), \,\,\, \xi^\lambda \in \mathcal{S}_{A'''}(\mathbb{R}^3),
\end{split}
\]
belong to $\mathcal{S}^0 (\mathbb{R}^3) = \mathcal{S}_{A'''}(\mathbb{R}^3) 
\subset \mathcal{S}(\mathbb{R}^3) \subset L^1(\mathbb{R}^3) \cap L^2(\mathbb{R}^3)$;
and where we have used also the fact that
\[
[\mathfrak{J}_{\bar{p}}]^{\lambda \sigma} = g^{\lambda \sigma}.
\]
Because from (\ref{BJBJ=1}) it follows that
\[
\sqrt{B(\boldsymbol{\p}, p^0(\boldsymbol{\p}))}^{\mu}_{\lambda}
\sqrt{B(\boldsymbol{\p}, p^0(\boldsymbol{\p}))}^{\nu}_{\gamma} \,\, g_{\mu \nu} = g_{\lambda \gamma}
\] 
and because 
\[
g_{\lambda \gamma} g^{\lambda \sigma} = \delta^{\sigma}_{\gamma},
\]
then we obtain the following formula for the symbol
\begin{multline*}
\bigg( g_{\mu \nu} \, \int \limits_{\mathbb{R}^3} : \partial_0 A^\mu (\boldsymbol{\x}, t) \partial_k A^\nu (\boldsymbol{\x}, t): \, \ud^3 \boldsymbol{\x} \bigg)^\wedge (\xi, \zeta) \\ 
 =
- e^{\langle \xi, \zeta \rangle} \,
\int \limits_{\mathbb{R}^3}  \,  
\frac{\ud^3 \boldsymbol{\p}}{2} \, p^k 
 g_{\lambda \gamma} g^{\gamma \sigma} \, \zeta^\sigma (\boldsymbol{\p}) \xi^\lambda (\boldsymbol{\p}) \\
- e^{\langle \xi, \zeta \rangle} \,
\int \limits_{\mathbb{R}^3}  \,  
\frac{\ud^3 \boldsymbol{\p}}{2} \, p^k 
 g_{\lambda \gamma} g^{\lambda \sigma} \, \zeta^\sigma (\boldsymbol{\p}) \xi^\gamma (\boldsymbol{\p})
\end{multline*}
\begin{equation}\label{symbolIntT0k}
= \, - e^{\langle \xi, \zeta \rangle} \,\sum \limits_{\nu}
\int \limits_{\mathbb{R}^3}  \,  
\ud^3 \boldsymbol{\p} \, p^k 
 \, \zeta^\nu (\boldsymbol{\p}) \xi^\nu (\boldsymbol{\p}) \,\,\,\,\, \xi, \zeta \in E.
\end{equation}

On the other hand from Lemma 2.1 of \cite{hida}
it follows that the functions
\[
\boldsymbol{\p} \mapsto 
\langle \langle \partial_{\boldsymbol{\p}}^{\nu \, *} \partial_{\boldsymbol{\p}}^{\nu} \Phi, \Psi
\rangle \rangle, \,\,\, \Phi, \Psi \in (E), \nu = 0, \ldots 3
\]
belong to $\mathcal{S}_{A'''}(\mathbb{R}^3) = \mathcal{S}^0(\mathbb{R}^3) 
\subset \mathcal{S}(\mathbb{R}^3) \subset L^1(\mathbb{R}^3)$, so that for each $\Phi \in (E)$ 
\[
\sum \limits_{\mu, \nu} \, g_{\mu k} \int \limits_{\mathbb{R}^3}
p^\mu(\boldsymbol{\p}) \,\, 
\partial_{\boldsymbol{\p}}^{\nu \, *} \partial_{\boldsymbol{\p}}^{\nu}
\, \Phi \,\, \ud^3 \boldsymbol{\p} =
- \sum \limits_{\nu} \, \int \limits_{\mathbb{R}^3}
p^k \,\, 
\partial_{\boldsymbol{\p}}^{\nu \, *} \partial_{\boldsymbol{\p}}^{\nu}
\,\,\, \Phi \,\, \ud^3 \boldsymbol{\p} 
\]
exists as the Pettis integral and belongs to $(E)^*$, and thus defines an operator
$\Xi: (E) \rightarrow (E)^*$. In fact 
\[
\Xi = - \sum \limits_{\nu} \, \int \limits_{\mathbb{R}^3}
p^k \,\, 
\partial_{\boldsymbol{\p}}^{\nu \, *} \partial_{\boldsymbol{\p}}^{\nu}
\,\,\, \ud^3 \boldsymbol{\p}
\]
because for all $\Phi, \Psi \in (E)$
\[
\langle \langle \Xi \Phi, \Psi \rangle \rangle = 
- \sum \limits_{\nu} \, \int \limits_{\mathbb{R}^3}
p^k \,\, 
\langle \langle \partial_{\boldsymbol{\p}}^{\nu \, *} \partial_{\boldsymbol{\p}}^{\nu}
\, \Phi, \, \Psi \rangle \rangle \, \ud^3 \boldsymbol{\p} = 
\langle \langle d \Gamma(P_k) \Phi, \Psi \rangle \rangle 
\]
by the very definition of the two operators, and in view of Thm. 2.2 of \cite{hida}.
Now because 
\[d \Gamma(P_k) = - \sum \limits_{\nu} \, \int \limits_{\mathbb{R}^3}
p^k \,\, 
\partial_{\boldsymbol{\p}}^{\nu \, *} \partial_{\boldsymbol{\p}}^{\nu}
\,\,\, \ud^3 \boldsymbol{\p} 
\] 
is well defined pointwisely as the Pettis integral, then
using (\ref{eta-hida-partial-exp}) we obtain 
\begin{multline*}
\bigg( \sum \limits_{\mu, \nu} \, g_{\mu k} \int \limits_{\mathbb{R}^3}
p^\mu(\boldsymbol{\p}) \,\, 
\partial_{\boldsymbol{\p}}^{\nu \, *} \partial_{\boldsymbol{\p}}^{\nu}
\,\,\, \ud^3 \boldsymbol{\p} \bigg)^\wedge (\xi, \zeta)  =
-\bigg( \sum \limits_{\nu} \, \int \limits_{\mathbb{R}^3}
p^k \,\, 
\partial_{\boldsymbol{\p}}^{\nu \, *} \partial_{\boldsymbol{\p}}^{\nu}
\,\,\, \ud^3 \boldsymbol{\p} \bigg)^\wedge (\xi, \zeta) \\ =
- \bigg\langle \bigg \langle 
\sum \limits_{\nu} \, \int \limits_{\mathbb{R}^3}
p^k \,\, 
\partial_{\boldsymbol{\p}}^{\nu \, *} \partial_{\boldsymbol{\p}}^{\nu}
\,\,\, \ud^3 \boldsymbol{\p} \, \Phi_\xi \, , \,\, \Phi_\zeta
\bigg \rangle \bigg \rangle  
=
- \bigg\langle \bigg \langle 
\sum \limits_{\nu} \, \int \limits_{\mathbb{R}^3}
p^k \,\, 
\partial_{\boldsymbol{\p}}^{\nu \, *} \partial_{\boldsymbol{\p}}^{\nu}
\,\, \Phi_\xi \,\, \ud^3 \boldsymbol{\p}  , \,\, \Phi_\zeta
\bigg \rangle \bigg \rangle  
\\ =
-  \sum \limits_{\nu} \, \int \limits_{\mathbb{R}^3}
p^k \, \big \langle \big \langle
\partial_{\boldsymbol{\p}}^{\nu \, *} \partial_{\boldsymbol{\p}}^{\nu} \Phi_\xi,
\Phi_\zeta \big \rangle \big \rangle 
\,\,\, \ud^3 \boldsymbol{\p}  
=
-  \sum \limits_{\nu} \, \int \limits_{\mathbb{R}^3}
p^k \, 
\xi^\nu (\boldsymbol{\p}) \zeta^\nu (\boldsymbol{\p})
\, \ud^3 \boldsymbol{\p} \, \cdot 
\big \langle \big \langle \Phi_\xi, \Phi_\zeta 
\big \rangle \big \rangle 
\\ =
-  e^{\langle \xi, \zeta \rangle} \, \sum \limits_{\nu} \, \int \limits_{\mathbb{R}^3}
p^k \, \xi^\nu (\boldsymbol{\p}) \zeta^\nu (\boldsymbol{\p})
\, \ud^3 \boldsymbol{\p}, 
\end{multline*}
for all $\xi, \zeta \in E$ (the last formula may also be inferred immediately from the formula for the symbol
of the integral kernel operator, \cite{obata} or \cite{huang}). 
Comparing this result with (\ref{symbolIntT0k}) we obtain
\begin{multline*}
\bigg( \sum \limits_{\mu, \nu} \, g_{\mu k} \int \limits_{\mathbb{R}^3}
p^\mu(\boldsymbol{\p}) \,\, 
\partial_{\boldsymbol{\p}}^{\nu \, *} \partial_{\boldsymbol{\p}}^{\nu}
\,\,\, \ud^3 \boldsymbol{\p} \bigg)^\wedge (\xi, \zeta) \\ =
\bigg( g_{\mu \nu} \, \int \limits_{\mathbb{R}^3} : \partial_0 A^\mu (\boldsymbol{\x}, t) \partial_k A^\nu (\boldsymbol{\x}, t): \, \ud^3 \boldsymbol{\x} \bigg)^\wedge (\xi, \zeta), \,\,\,\,\, \xi, \zeta \in E.
\end{multline*}
Because each operator in $\mathscr{L}((E), (E)^*)$ is uniquely determined by its symbol,
Lemma 4.2 of \cite{obata}, then we obtain
\[
\sum \limits_{\mu, \nu} \, g_{\mu k} \int \limits_{\mathbb{R}^3}
p^\mu(\boldsymbol{\p}) \,\, 
\partial_{\boldsymbol{\p}}^{\nu \, *} \partial_{\boldsymbol{\p}}^{\nu}
\,\,\, \ud^3 \boldsymbol{\p}   =
 g_{\mu \nu} \, \int \limits_{\mathbb{R}^3} : \partial_0 A^\mu (\boldsymbol{\x}, t) \partial_k A^\nu (\boldsymbol{\x}, t): \, \ud^3 \boldsymbol{\x}.
\]
\qed

\begin{lem*}
Symbols of the operators 
\begin{equation}\label{noether-int-T-00}
-\frac{1}{2} \,  \int \limits_{\mathbb{R}^3} : g_{\mu \nu} \, \sum \limits_{\rho} \partial_\rho A^\mu (\boldsymbol{\x}, t) \partial_\rho A^\nu (\boldsymbol{\x}, t):
\, \ud^3 \boldsymbol{\x}
\end{equation}
(summation with respect to $\mu$ and $\nu$) and 
\begin{equation}\label{generator-P-0}
d\Gamma(P^0) =
\sum \limits_{\nu} \int \limits_{\mathbb{R}^3}
 p^0(\boldsymbol{\p}) \,\, 
\partial_{\boldsymbol{\p}}^{\nu \, *} \partial_{\boldsymbol{\p}}^{\nu}
\,\,\, \ud^3 \boldsymbol{\p},
\end{equation}
are equal, and thus
\[
-\frac{1}{2} \,  \int \limits_{\mathbb{R}^3} : g_{\mu \nu} \, \sum \limits_{\rho} \partial_\rho A^\mu (\boldsymbol{\x}, t) \partial_\rho A^\nu (\boldsymbol{\x}, t):
\, \ud^3 \boldsymbol{\x} =
\sum \limits_{\nu} \int \limits_{\mathbb{R}^3}
 p^0(\boldsymbol{\p}) \,\, 
\partial_{\boldsymbol{\p}}^{\nu \, *} \partial_{\boldsymbol{\p}}^{\nu}
\,\,\, \ud^3 \boldsymbol{\p} =
d\Gamma(P^0),
\]
as elements of $\mathscr{L}((E), (E)^*)$.
\end{lem*}

\qedsymbol \, 
The proof is similar to that of the preceding Lemma.

\qed

The last two Lemmas finish the proof of the Bogoliubov-Shirkov quantization postulate:
\[
\boxed{\int \boldsymbol{:} T^{0\mu} \boldsymbol{:} \, \ud^3 \boldsymbol{\x} = \boldsymbol{P}^\mu = d\Gamma(P^\mu),}
\] 
for the free quantum electromagnetic potential $A^\mu$-field.

We end this Subsection noting that the first summand 
\[
A^{(-) \, \mu}(x) = \int \limits_{\mathbb{R}^3} \, \ud^3 p \, 
\frac{1}{\sqrt{2 p^0(\boldsymbol{\p})}}\sqrt{B(\boldsymbol{\p}, p^0(\boldsymbol{\p}))}^{\mu}_{\lambda}
a^{\lambda} (\boldsymbol{\p}) e^{-ip\cdot x} 
\] 
in
\begin{multline*} 
A^{\mu}(x)  = \int \limits_{\mathbb{R}^3} \, \ud^3 p \, 
\frac{1}{\sqrt{2 p^0(\boldsymbol{\p})}}\sqrt{B(\boldsymbol{\p}, p^0(\boldsymbol{\p}))}^{\mu}_{\lambda}
a^{\lambda} (\boldsymbol{\p}) e^{-ip\cdot x} \\
+
\int \limits_{\mathbb{R}^3} \, \ud^3 p \,
\frac{1}{\sqrt{2 p^0(\boldsymbol{\p})}}\sqrt{B(\boldsymbol{\p}, p^0(\boldsymbol{\p}))}^{\mu}_{\lambda}
\, \eta \, a^{\lambda} (\boldsymbol{\p})^+ \eta \, e^{ip\cdot x},
\end{multline*}
denoted by physicists $A^{(-) \, \mu}(x)$,
is a well defined operator on the Fock space, which transforms
continuously the Hida space $(E)$ into itself. Again this follows from
our previous observation that the integrand in the first summand 
\begin{equation}\label{A^(-)mu(x)}
A^{(-) \, \mu}(x)\Phi = \int \limits_{\mathbb{R}^3} \, \ud^3 p \, 
\frac{e^{-ip\cdot x}}{\sqrt{2 p^0(\boldsymbol{\p})}}\sqrt{B(\boldsymbol{\p}, p^0(\boldsymbol{\p}))}^{\mu}_{\lambda} \, a^{\lambda} (\boldsymbol{\p}) \Phi,
\,\,\, \Phi \in (E)
\end{equation}
in 
\begin{multline*} 
A^{\mu}(x) \Phi = \int \limits_{\mathbb{R}^3} \, \ud^3 p \, 
\frac{e^{-ip\cdot x}}{\sqrt{2 p^0(\boldsymbol{\p})}}\sqrt{B(\boldsymbol{\p}, p^0(\boldsymbol{\p}))}^{\mu}_{\lambda}
a^{\lambda} (\boldsymbol{\p}) \Phi \\
+
 \int \limits_{\mathbb{R}^3} \, \ud^3 p \, 
\frac{ e^{ip\cdot x}}{\sqrt{2 p^0(\boldsymbol{\p})}}\sqrt{B(\boldsymbol{\p}, p^0(\boldsymbol{\p}))}^{\mu}_{\lambda}
\, \eta \, a^{\lambda} (\boldsymbol{\p})^+ \eta \, \Phi,
\,\,\,
\Phi \in (E),
\end{multline*}
is Bochner integrable and the integral (\ref{A^(-)mu(x)}) exists in Bochner's sense and 
belongs to each $(E)_k$, $k>0$. This means that for each $\Phi \in (E)$ the integral 
(\ref{A^(-)mu(x)}) belongs to $\cap_{k>0}(E)_k = (E)$. Joining this observation again
with Lemma 2.1 of \cite{hida} we easily show that $A^{(-) \, \mu}(x): (E) \rightarrow (E)$
is continuous:
\begin{prop*}
The negative frequency part
\[
A^{(-) \, \mu}(x) = \int \limits_{\mathbb{R}^3} \, \ud^3 p \, 
\frac{1}{\sqrt{2 p^0(\boldsymbol{\p})}}\sqrt{B(\boldsymbol{\p}, p^0(\boldsymbol{\p}))}^{\mu}_{\lambda} \partial^{\lambda}_{\boldsymbol{\p}} \, e^{-ip\cdot x}
\] 
and all its space-time derivatives are generalized operators transforming continuously
the Hida test space $(E)$ into $(E)$, and thus represent ordinary (unbounded)
operators on the Fock space. 
\end{prop*}
 \qedsymbol \, 
Because for each $\Phi \in (E)$, $A^{(-) \, \mu}(x)\Phi \in (E) \subset (E)^*$,
then the canonical pairing $\langle \langle A^{(-) \, \mu}(x)\Phi, A^{(-) \, \mu}(x)\Phi
\rangle \rangle$ coincides with the ordinary inner product. Recall that the second quantized 
$\Gamma(A)$ version of the standard operator 
$A$ defininig the single particle Gelfand triple of the field $A^\mu$, transforms continuously
the Hida test space $(E)$ into itself. In particular $\Gamma(A)^r\Phi \in (E)$, $r>0$.
Therefore, we can form the canonical pairing  $\langle \langle A^{(-) \, \mu}(x)
\Gamma(A)^r\Phi, A^{(-) \, \mu}(x)\Gamma(A)^r\Phi \rangle \rangle$, which coincides with 
the inner product and is equal $\| A^{(-) \, \mu}(x)\Phi\|_{r}^{2}$. Estimation of this
$r$-th squared norm with the help of the inequality (2-2) of Lemma 2.1 of \cite{hida},
will give us continuity of the operator $A^{(-) \, \mu}(x)$. Namely, we have the following
estimation
\begin{multline*}
\| A^{(-) \, \mu}(x)\Phi\|_{r}^{2} 
= \langle \langle A^{(-) \, \mu}(x)
\Gamma(A)^r\Phi, A^{(-) \, \mu}(x)\Gamma(A)^r\Phi \rangle \rangle 
\\
= \int \limits_{\mathbb{R}^3} \, 
{\textstyle\frac{\ud^3 p'}{\sqrt{2 p^0(\boldsymbol{\p}')}}}
{\textstyle\frac{\ud^3 p}{\sqrt{2 p^0(\boldsymbol{\p})}}} \, 
\sqrt{B(\boldsymbol{\p}, p^0(\boldsymbol{\p}))}^{\mu}_{\lambda}
\sqrt{B(\boldsymbol{\p}, p^0(\boldsymbol{\p}'))}^{\mu}_{\lambda'} \,
\langle \langle \partial^{\lambda' \,\, *}_{\boldsymbol{\p}'} \partial^{\lambda}_{\boldsymbol{\p}} \Gamma(A)^r \Phi, \Gamma(A)^r \Phi  \rangle \rangle \\
= 
\int \limits_{\mathbb{R}^3} \, 
{\textstyle\frac{\ud^3 p'}{\sqrt{2 p^0(\boldsymbol{\p}')}}}
{\textstyle\frac{\ud^3 p}{\sqrt{2 p^0(\boldsymbol{\p})}}} \,
\sqrt{B(\boldsymbol{\p}, p^0(\boldsymbol{\p}))}^{\mu}_{\lambda}
\sqrt{B(\boldsymbol{\p}, p^0(\boldsymbol{\p}'))}^{\mu}_{\lambda'} \,
\eta_{{}_{\Gamma(A)^r \Phi, \Gamma(A)^r \Phi}}(\boldsymbol{\p}', \boldsymbol{\p}).
\end{multline*}

Because the functional 
\begin{multline*}
E \otimes E \ni \eta_{{}_{\Gamma(A)^r \Phi, \Gamma(A)^r \Phi}}
\longrightarrow \\
\longrightarrow 
\int \limits_{\mathbb{R}^3} \, 
{\textstyle\frac{\ud^3 p'}{\sqrt{2 p^0(\boldsymbol{\p}')}}}
{\textstyle\frac{\ud^3 p}{\sqrt{2 p^0(\boldsymbol{\p})}}} \,
\sqrt{B(\boldsymbol{\p}, p^0(\boldsymbol{\p}))}^{\mu}_{\lambda}
\sqrt{B(\boldsymbol{\p}, p^0(\boldsymbol{\p}'))}^{\mu}_{\lambda'} \,
\eta_{{}_{\Gamma(A)^r \Phi, \Gamma(A)^r \Phi}}(\boldsymbol{\p}', \boldsymbol{\p})
\end{multline*}
is continuous (by the first Lemma of Subsection \ref{diffSA}), then there exists a positive constant 
$C_r$ and a positive integer $q$ (depending on $r$) such that 
\[
\| A^{(-) \, \mu}(x)\Phi\|_{r}^{2} 
\leq C_r \big| \eta_{{}_{\Gamma(A)^r \Phi, \Gamma(A)^r \Phi}} \big|_{q}.
\]
Now from the inequality (2-2) of Lemma 2.1 of \cite{hida} it follows the following inequality
\begin{multline*}
\| A^{(-) \, \mu}(x)\Phi\|_{r}^{2} 
\leq C_r \big| \eta_{{}_{\Gamma(A)^r \Phi, \Gamma(A)^r \Phi}} \big|_{q} \\
\leq C_r 
{\textstyle\frac{\rho^{-2q}}{-2q\, e \, \textrm{ln}\, \rho}}
\|\Gamma(A)^r \Phi\|_q \, \| \Gamma(A)^r \Phi\|_q.
\end{multline*}
Finally, from the continuity of the operator $\Gamma(A)^r: (E) \rightarrow (E)$ it follows
existence (for each natural $q$) of a natural $k = k(q,r)$ (depending on $q$ and $r$) and a finite positive constant $C_q$  such that
\begin{multline*}
\| A^{(-) \, \mu}(x)\Phi\|_{r}^{2} 
\leq C_r \big| \eta_{{}_{\Gamma(A)^r \Phi, \Gamma(A)^r \Phi}} \big|_{q} \\
\leq C_r 
{\textstyle\frac{\rho^{-2q}}{-2q\, e \, \textrm{ln}\, \rho}}
\|\Gamma(A)^r \Phi\|_q \, \| \Gamma(A)^r \Phi\|_q
\\
\leq C_r C_q
{\textstyle\frac{\rho^{-2q}}{-2q\, e \, \textrm{ln}\, \rho}}
\|\Phi\|_{k(q,r)}^{2},
\end{multline*}
which is equivalent to the continuity of the operator $A^{(-) \, \mu}(x): (E) \rightarrow (E)$.
The proof of the continuity of space-time derivatives $D^{\alpha}A^{(-) \, \mu}(x): (E) \rightarrow (E)$ is identical. 
Here $D^\alpha$ is the Schwartz' notation for higher order space-time derivation
with multi-idex $\alpha$. 
 \qed

{\bf REMARK}. Note that we can construct in the same way the negative energy local electromagnetic potential quantum field, 
together with the proof of the Bogoliubov-Shirkov Postulate valid for it. Indeed, it is sufficient to replace the orbit 
$\mathscr{O}_{1,0,0,1}$ with the fixed point $\bar{p} = (1,0,0,1) \in \mathscr{O}_{1,0,0,1}$ with the orbit 
$\mathscr{O}_{-1,0,0,1}$ and the fixed point $\bar{p} = (-1,0,0,1) \in \mathscr{O}_{-1,0,0,1}$.
This replacement is accompanied by the corresponding replacement of the stability subgroup $G_{{}_{(1,0,0,1)}}$
by the corresponding stability subgroup $G_{{}_{(-1,0,0,1)}}$ of the fixed point 
$\bar{p} = \mathscr{O}_{-1,0,0,1}$, and the replacement of the function $\beta(p)$ with the one corresponding to the orbit 
$\mathscr{O}_{-1,0,0,1}$ and to the fixed point $\bar{p} = \mathscr{O}_{-1,0,0,1}$ in it. In this way we obtain the matrix 
$B(p)$ (compare (\ref{Bmatrix})) and the operators $B$, $\mathfrak{J}'$ of multiplication by (\ref{operatorB}) and 
(\ref{operatorJ'}) respectively, corresponding to the orbit $\mathscr{O}_{-1,0,0,1}$. All the results of Sections  
\ref{free-gamma} and \ref{white-noise-proofs} stay valid and all the proofs of \ref{white-noise-proofs}
remain unchanged.

\subsection{The quantum electromagnetic potential field $A$ as an integral kernel operator
with vector-valued distributional kernel}\label{A=Xi0,1+Xi1,0}

Recall that the formula (\ref{q-A-B}): 
\begin{multline}\label{q-A-B'}
A^\mu(x) = \int \limits_{\mathbb{R}^3} \, \ud^3 p \, \bigg\{
\frac{1}{\sqrt{2 p^0(\boldsymbol{\p})}}\sqrt{B(\boldsymbol{\p}, p^0(\boldsymbol{\p}))}^{\mu}_{\lambda}
a^{\lambda} (\boldsymbol{\p}) e^{-ip\cdot x} \\
+  \frac{1}{\sqrt{2 p^0(\boldsymbol{\p})}}\sqrt{B(\boldsymbol{\p}, p^0(\boldsymbol{\p}))}^{\mu}_{\lambda} \,
\eta \, a^{\lambda}(\boldsymbol{\p})^+ \, \eta \, e^{ip\cdot x}  \bigg\} 
\end{multline} 
gives a well defined generalized operator transforming continuously the Hida
space $(E)$ into its strong dual $(E)^*$, where $(E)$ is the Hida space of the Gelfand triple 
$(E) \subset \Gamma(\mathcal{H}') \subset (E)^*$
defining the electromagnetic potential field $A$ within the white noise setup. 
Recall that  $E = \mathcal{S}_{A}(\mathbb{R}^3; \mathbb{C}^4) = 
\mathcal{S}_{\oplus A^{(3)}}(\mathbb{R}^3; \mathbb{C}^4)$ is defined by the standard operator 
$A = \oplus_{0}^{3} A^{(3)}$ on the standard Hilbert space $L^2(\mathbb{R}^3; \mathbb{C}^4)$,
with the operator $A^{(3)}$ defined as in Subsection \ref{dim=n}. 
Recall that  the integral (\ref{q-A-B'}) exists pointwisely as the Pettis integral, compare 
(\ref{q-A-B}), Subsection \ref{BSH}. Nonetheless, the potential field $A$ is naturally
a sum of two integral kernel operators 
\[
A = \Xi_{0,1}(\kappa_{0,1}) + \Xi_{1,0}(\kappa_{1,0})
 \in
\mathscr{L}\big( (E) \otimes \mathscr{E}, \, (E)^* \big) \cong
\mathscr{L}\big( \mathscr{E}, \,\, \mathscr{L}( (E), (E)^*) \big) 
\]
with vector valued kernels $\kappa_{0,1}, \kappa_{1,0} \in \mathscr{L}\big(E, \mathscr{E}^* \big)$
for
\[
\mathscr{E} = 
\mathcal{S}_{\mathscr{F}\oplus A^{(4)}\mathscr{F}^{-1}}(\mathbb{R}^4; \mathbb{C}^4) 
= \mathscr{F} \Big[\mathcal{S}_{\oplus A^{(4)}}(\mathbb{R}^4; \mathbb{C}^4) \Big] 
= \mathcal{S}^{00}(\mathbb{R}^4; \mathbb{C}^4),
\]
in the sense of Obata \cite{obataJFA} explained in Subsection \ref{psiBerezin-Hida}.
The vector valued distributions $\kappa_{0,1}, \kappa_{1,0}$ are defined by the following plane waves 
\[
\begin{split}
\kappa_{0,1}(\nu, \boldsymbol{\p}; \mu, x) =
\frac{\sqrt{B(\boldsymbol{\p}, p^0(\boldsymbol{\p}))}^{\mu}_{\nu}}{\sqrt{2 p^0(\boldsymbol{\p})}}
e^{-ip\cdot x}, \,\,\,\,\,\,
p = (| p^0(\boldsymbol{\p})|, \boldsymbol{\p}) 
\in \mathscr{O}_{1,0,0,1}, \\
\kappa_{1,0}(\nu, \boldsymbol{\p}; \mu, x) = (-1)^{(\mu)}
\frac{\sqrt{B(\boldsymbol{\p}, p^0(\boldsymbol{\p}))}^{\mu}_{\nu}}{\sqrt{2 p^0(\boldsymbol{\p})}}
e^{ip\cdot x},
\,\,\,\,\,\,
p = (| p^0(\boldsymbol{\p})|, \boldsymbol{\p}) 
\in \mathscr{O}_{1,0,0,1},
\end{split}
\]  
with
\[
(-1)^{(\mu)} \overset{\textrm{df}}{=}
\left\{ \begin{array}{ll}
-1 & \textrm{if $\mu = 0$}, \\
1 & \textrm{if $\mu = 1, 2, 3$}.
\end{array} \right., \,\,\,\,
p^0(\boldsymbol{\p}) = |\boldsymbol{\p}|.
\]
The above stated formulas for $\kappa_{0,1}, \kappa_{1,0}$ can be immediately read off from the
formula (\ref{q-A-B}) and the commutation rules (\ref{[eta,a]}) of the Gupta-Bleuler operator $\eta$
and the Hida operators $\partial_{\mu, \boldsymbol{\p}} = a_{\mu}(\boldsymbol{\p})$:
\[
a_{0}(\boldsymbol{\p}) \eta = - \eta a_{0}(\boldsymbol{\p}), \,\,\,\,
a_{i}(\boldsymbol{\p}) \eta = \eta a_{i}(\boldsymbol{\p}), \,\,\,\, i= 1, 2, 3, \,\,\,\,
\eta^2 = \boldsymbol{1}.
\] 
Here we are using the standard convention of Subsection \ref{psiBerezin-Hida} that in the general integral kernel
operator (\ref{electron-positron-photon-Xi}) in the tensor product of the Fock space of the Dirac field 
$\boldsymbol{\psi}$ and of the electromagnetic potential field $A$ we have the ordinary Hida operators 
in the normal order with the ordinary adjoint (linear transpose)
$\partial_{\mu, \boldsymbol{\p}}^{*} = a_{\mu}(\boldsymbol{\p})^{+}$ corresponding to photon variables
$\mu, \boldsymbol{\p}$. This is the convention assumed in mathematical literature concerning integral 
kernel operators. But physicists never use the ordinary
adjoint $\partial_{\mu, \boldsymbol{\p}}^{*} = a_{\mu}(\boldsymbol{\p})^{+}$ whenever usng expansions into normally ordered
creation-annihilation operators for the variables corresponding to the electromagnetic field, but instead they are using
the ``Krein-adjoined'' operators $\eta\partial_{\mu, \boldsymbol{\p}}^{*} \eta = \eta a_{\mu}(\boldsymbol{\p})^{+} \eta$
instead,  as in the formula  (\ref{q-A-B'}). Therefore, it is more convenient, when adopting the integral kernel operators to QED (in Gupta-Bleuler gauge), to change slightly the convention of Subsection
\ref{psiBerezin-Hida} and use for $\partial_{w}^*$ in the general integral kernel operator 
(\ref{electron-positron-photon-Xi}),
on the tensor product of Fock spaces of the Dirac field $\boldsymbol{\psi}$ and the electromagnetic potential field $A$,
the operators  $\eta\partial_{\mu, \boldsymbol{\p}}^{*} \eta$ whenever 
$w = (\mu, \boldsymbol{\p})$ corresponds to the photon variables $\mu, \boldsymbol{\p}$ 
in (\ref{electron-positron-photon-Xi}), 
instead of the ordinary transposed operators $\partial_{\mu, \boldsymbol{\p}}^{*}$. 
With this convention of physicists we will have the following formulas 
\begin{equation}\label{kappa_0,1kappa_1,0A}
\boxed{
\begin{split}
\kappa_{0,1}(\nu, \boldsymbol{\p}; \mu, x) =
\frac{\sqrt{B(\boldsymbol{\p}, p^0(\boldsymbol{\p}))}^{\mu}_{\nu}}{\sqrt{2 p^0(\boldsymbol{\p})}}
e^{-ip\cdot x}, \,\,\,\,\,\,
p \in \mathscr{O}_{1,0,0,1}, \\
\kappa_{1,0}(\nu, \boldsymbol{\p}; \mu, x) = 
\frac{\sqrt{B(\boldsymbol{\p}, p^0(\boldsymbol{\p}))}^{\mu}_{\nu}}{\sqrt{2 p^0(\boldsymbol{\p})}}
e^{ip\cdot x},
\,\,\,\,\,\,
p \in \mathscr{O}_{1,0,0,1},
\end{split}
}
\end{equation}
without the additional factor $(-1)^{(\mu)}$. In fact presence of the factors
\[ 
(-1)^{(\mu_1)} \cdots (-1)^{(\mu_l)}
\]
for the kernels of the corresponding integral kernel operators
is the only difference between the two conventions, and which are absorbed concisely by the 
Gupta-Bleuler operator $\eta$.  

In other words: we will show that for the plane wave kernels (\ref{kappa_0,1kappa_1,0A}) we have
\begin{multline}\label{A=IntKerOpVectValKer}
A(\varphi) = a'( \overline{\check{\widetilde{\varphi}}}|_{{}_{\mathscr{O}_{1,0,0,1}}}) 
+ \eta a'(\widetilde{\varphi}|_{{}_{\mathscr{O}_{1,0,0,1}}})^+ \eta \\ =
a\Big( U\big( \overline{\check{\widetilde{\varphi}}}|_{{}_{\mathscr{O}_{1,0,0,1}}}\big) \Big) 
+ \eta a\Big( U\big(\widetilde{\varphi}|_{{}_{\mathscr{O}_{1,0,0,1}}} \big) \Big)^+ \eta \\ =
a(\sqrt{B} \, \overline{\check{\widetilde{\varphi}}}|_{{}_{\mathscr{O}_{1,0,0,1}}}) 
+ \eta a(\sqrt{B} \, \widetilde{\varphi}|_{{}_{\mathscr{O}_{1,0,0,1}}})^+ \eta \\ =
\sum \limits_{\nu=0}^{3} \int \kappa_{0,1}(\varphi)(\nu, \boldsymbol{\p}) \partial_{\nu, \boldsymbol{\p}} \, 
\ud^3 \boldsymbol{\p}
+ \sum \limits_{\nu=0}^{3} \int \kappa_{1,0}(\varphi)(\nu, \boldsymbol{\p}) \eta \partial_{\nu, \boldsymbol{\p}}^{*}\eta
\, \ud^3 \boldsymbol{\p} \\ =
\Xi_{0,1}\big(\kappa_{0,1}(\varphi)\big) + \Xi_{1,0}\big(\kappa_{1,0}(\varphi)\big),
\,\,\,\,
\varphi \in \mathscr{E}= \mathcal{S}^{00}(\mathbb{R}^4; \mathbb{C}^4).
\end{multline}
Moreover, we will show that the kernels $\kappa_{0,1},\kappa_{1,0}$ defined 
by (\ref{kappa_0,1kappa_1,0A}) can be (uniquely) extended to the elements 
(and denoted by the same $\kappa_{0,1},\kappa_{1,0}$)
\[
\kappa_{0,1},\kappa_{1,0} \in \mathscr{L}(E^*, \mathscr{E}^*),
\]
so that by Thm 3.13 of \cite{obataJFA} (or Thm. \ref{obataJFA.Thm.3.13} of Subsection)
\ref{psiBerezin-Hida}
\[
A = \Xi_{0,1}(\kappa_{0,1}) + \Xi_{1,0}(\kappa_{1,0})
 \in
\mathscr{L}\big( (E) \otimes \mathscr{E}, \, (E) \big) \cong
\mathscr{L}\big( \mathscr{E}, \,\, \mathscr{L}( (E), (E)) \big) 
\] 
and $A$, understood as an integral kernel operator with vector-valued distributional kernels
(\ref{kappa_0,1kappa_1,0A}), determines a well defined operator-valued distribution
on the space-time nuclear test space 
\[
\mathscr{E} = \mathscr{F} \Big[\mathcal{S}_{\oplus A^{(4)}}(\mathbb{R}^4; \mathbb{C}^4)\Big] 
= \mathcal{S}^{00}(\mathbb{R}^4; \mathbb{C}^4).
\]
In the formula (\ref{A=IntKerOpVectValKer}) $\kappa_{0,1}(\phi), \kappa_{1,0}(\phi)$  denote the kernels 
representing distributions
in $E^* = \mathcal{S}_{A}(\mathbb{R}^3, \,\, \mathbb{C}^4)^*$ which are defined in the standard manner
\[
\kappa_{0,1}(\varphi)(\nu, \boldsymbol{\p})
=  \sum_{\mu=0}^{3} \int \limits_{\mathbb{R}^3}
\kappa_{0,1}(\nu, \boldsymbol{\p}; \mu,x) \varphi^{\mu}(x) \, \ud^4 x
\]
and analogously for $\kappa_{1,0}(\phi)$, where $\kappa_{0,1}, \kappa_{1,0}$
are understood as elements of 
\[
\mathscr{L}(\mathscr{E}, E^*) \cong \mathscr{L}\big( E, \, \mathscr{L}(\mathscr{E},\mathbb{C}) \big)
\cong \mathscr{L}\big( E, \, \mathscr{E}^* \big). 
\]
Similarly we have
\[
\kappa_{0,1}(\xi)(\mu, x) 
= \sum_{\nu=0}^{4}  \, \int \limits_{\mathbb{R}^3} 
\kappa_{0,1}(\nu, \boldsymbol{p}; a, x) \, \xi(\nu, \boldsymbol{\p}) \, \ud^3 \boldsymbol{\p},
\,\,\, \xi \in E,
\]
and analogously for $\kappa_{1,0}(\xi)(\mu, x)$, with
$\kappa_{0,1}, \kappa_{1,0}$
understood as elements of 
\[
\mathscr{L}\big( E, \, \mathscr{L}(\mathscr{E},\mathbb{C}) \big)
\cong \mathscr{L}\big( E, \, \mathscr{E}^* \big) \cong \mathscr{L}(\mathscr{E}, E^*); 
\]
with pairings
\begin{multline*}
\langle \kappa_{0,1}(\phi), \xi \rangle 
= \sum_{\mu=0}^{3} \, \int \limits_{ \mathbb{R}^4 \times \mathbb{R}^3} 
\kappa_{0,1}(\phi)(\mu, \boldsymbol{p}) \,\, \xi(s, \boldsymbol{\p}) \, \ud^3 \boldsymbol{\p} \\ 
= \sum_{s=1}^{4} \, \sum_{a=1}^{4} \, \int \limits_{\mathbb{R}^3} 
\kappa_{0,1}(s, \boldsymbol{p}; a, x) \, \phi^{a}(x) \,\, \xi(s, \boldsymbol{\p}) \, \ud^4 x \, \ud^3 \boldsymbol{\p}
= \langle \kappa_{0,1}(\xi), \phi \rangle,
\,\,\, \xi \in E, \phi \in  \mathscr{E},
\end{multline*}
defined through the ordinary Lebesgue integrals.

$U$ is the unitary isomorphism (and its inverse $U^{-1}$)
\[
\begin{split}
U: \mathcal{H}' \ni \xi \mapsto  \sqrt{B} \xi \in L^2(\mathbb{R}^3; \mathbb{C}^4), \\
U^{-1}: L^2(\mathbb{R}^3; \mathbb{C}^4) \ni \zeta \mapsto \sqrt{B}^{-1} \zeta \in \mathcal{H}',
\end{split}
\]
joining the Gelfand triples (\ref{3-Gelfand-triples}) defining the field $A$ through its Fock lifting,
and is defined as pointwise multiplication
\begin{multline*}
\sqrt{B}\xi(\boldsymbol{\p}) \overset{\textrm{df}}{=}
\frac{1}{\sqrt{2 p^0(\boldsymbol{\p})}}\sqrt{B(\boldsymbol{\p}, p^0(\boldsymbol{\p}))} \xi(\boldsymbol{\p}),
\\
\sqrt{B}^{-1}\zeta(\boldsymbol{\p}) \overset{\textrm{df}}{=}
\sqrt{2 p^0(\boldsymbol{\p})}\sqrt{B(\boldsymbol{\p}, p^0(\boldsymbol{\p}))}^{{}^{-1}} \zeta(\boldsymbol{\p})
\end{multline*}
by the matrix (and respectively its inverse)
\begin{equation}\label{sqrtB/sqrtp0}
\frac{1}{\sqrt{2 p^0(\boldsymbol{\p})}}\sqrt{B(\boldsymbol{\p}, p^0(\boldsymbol{\p}))},
\end{equation}
the same which is present in the formula (\ref{q-A-B'}), with the matrix $\sqrt{B(p)}$, 
$p \in \mathscr{O}_{1,0,0,1}$ defined by (\ref{sqrtB}) in Subsection \ref{DefLopRep}.

Note here that the Gelfand triples (\ref{3-Gelfand-triples}) with the joining unitary isomorphism
$U$ plays the same role in the construction of the field $A$ in Subsection \ref{WhiteNoiseA} as does 
the triples (\ref{SinglePartGelfandTriplesForPsi})
joined by the unitary isomorphism (\ref{isomorphismU}) in the construction of the Dirac field
$\boldsymbol{\psi}$, Subsection \ref{psiBerezin-Hida}. 

Concerning the equality (\ref{A=IntKerOpVectValKer}) note that the first equality in 
(\ref{A=IntKerOpVectValKer}) follows by definition, second by the fact that $U$ is the unitary isomorphism
joining the standard Gelfand triple 
\[
E = \mathcal{S}_{A}(\mathbb{R}^3; \mathbb{C}^4)
\subset L^2(\mathbb{R}^3; \mathbb{C}^4) \subset \mathcal{S}_{A}(\mathbb{R}^3; \mathbb{C}^4)^*
\]
with the triple 
\[
E \subset \mathcal{H}' \subset E^*
\]
over the single particle Hilbert space of the field $A$ (the analogue of the unitary isomorphism 
(\ref{isomorphismU}) of Subsection \ref{psiBerezin-Hida}) . The Fock lifting of the standard triple
serves to construct the standard Hida operators $a(\zeta)$, and the Fock lifting of the second triple serves to
construct the Hida operators $a'(\xi)$. Therefore we obtain the second equality (the analogue of the isomorphism
(\ref{a(U(u+v))=a'(u+v)})), compare also Subsection \ref{WhiteNoiseA}. 
Third equality in (\ref{A=IntKerOpVectValKer}) follows by definition of the isomorphism
$U$. Finally note that it follows almost immediately from definition (\ref{kappa_0,1kappa_1,0A}) 
of $\kappa_{0,1}, \kappa_{1,0}$ that
\begin{equation}\label{kappa_0,1(varphi),kappa_1,0(varphi)}
\kappa_{0,1}(\varphi) = \sqrt{B} \check{\widetilde{\varphi}}|_{{}_{\mathscr{O}_{1,0,0,1}}}, \,\,\,\,\,\,\,
\kappa_{1,0}(\varphi) = \sqrt{B} \widetilde{\varphi}|_{{}_{\mathscr{O}_{1,0,0,1}}}.
\end{equation}
Thus, the fourth equality in (\ref{A=IntKerOpVectValKer}) follows by 
Prop. 4.3.10 of \cite{obata-book} (compare also the fermi analogue of Prop. 4.3. 10 of  \cite{obata-book} -- 
the Corollary \ref{D_xi=int(xiPartial)}
of Subsection \ref{psiBerezin-Hida}).

Let $\mathcal{O}'_C, \mathcal{O}_M$ be the algebras of convolutors and multipliers of the ordinary Schwartz
algebra $\mathcal{S}(\mathbb{R}^4; \mathbb{C}^4)$, defined by Schwartz \cite{Schwartz},
compare also Appendix \ref{convolutorsO'_C}. 
If the elements of $\mathcal{O}'_C$ (resp.  of $\mathcal{O}_M$) are understood as continuous linear operators
$\mathcal{S} \rightarrow \mathcal{S}$ of convolution
with distributions in $\mathcal{O}'_C$ (or respectively as continuous operators of multiplication by an element of 
$\mathcal{O}_M$) then we can endow $\mathcal{O}'_C, \mathcal{O}_M$ with the operator topology of 
uniform convergence on bounded sets (after Schwartz). The Fourier exchange theorem of Schwartz then says
that the Fourier transform becomes a topological isomorphism of $\mathcal{O}_M$ onto $\mathcal{O}'_C$,
which exchanges pointwise multiplication product defined by pointwise multiplication of functions in $\mathcal{O}_M$ (representing the corresponding tempered distributions) with the convolution product, defined through the composition
of the corresponding convolution operators in $\mathscr{L}(\mathcal{S, \mathcal{S}})$, compare \cite{Schwartz},
or Appendix \ref{convolutorsO'_C}.

Let $\mathcal{O}_C$ be the predual (a smooth function space determined explicitly by Horv\'ath) of 
the Schwartz convolution algebra $\mathcal{O}'_C$ endowed with the above 
Schwartz operator topology of uniform convergence on bounded sets on $\mathcal{O}'_C$ (strictly stronger 
than the topology inherited from the strong dual space $\mathcal{S}^*$ of tempered distributions), 
compare Appendix \ref{convolutorsO'_C}.  
 
Let $\mathcal{O}'_{CB_2}$ be the algebra of convolutors
of the algebra 
\[
\mathscr{E} = \mathcal{S}^{00}(\mathbb{R}^4; \mathbb{C}^4) = \mathscr{F}\Big[\mathcal{S}^{0}(\mathbb{R}^4; \mathbb{C}^4) \Big]
= \mathscr{F}\Big[\mathcal{S}_{\oplus A^{(4)}}(\mathbb{R}^4; \mathbb{C}^4) \Big]
= \mathcal{S}_{B_2}(\mathbb{R}^4; \mathbb{C}^4),
\]
where we have used the standard operator
\[
B_2 = \mathscr{F} \oplus_{0}^{3} A^{(4)} \mathscr{F}^{-1} \,\,\, \textrm{on} \,\,\,
\oplus_{0}^{3}L^2(\mathbb{R}^4; \mathbb{C}) = L^2(\mathbb{R}^4; \mathbb{C}^4),
\]
introduced in Subsection \ref{psiBerezin-Hida}, and further used in Subsection \ref{OperationsOnXi}.
Recall that the standard operators $A^{(n)}$ on $L^2(\mathbb{R}^n; \mathbb{C})$ have been constructed
in Subsection \ref{dim=n}.

Let $\mathcal{O}'_{MB_2}$ be the algebra of multipliers of the nuclear algebra
\[
\mathcal{S}^{0}(\mathbb{R}^4; \mathbb{C}^4) = \mathcal{S}_{\oplus A^{(4)}}(\mathbb{R}^4; \mathbb{C}^4)
= \mathcal{S}_{B_2}(\mathbb{R}^4; \mathbb{C}^4).
\]
All the spaces $\mathcal{O}_C, \mathcal{O}_M, \mathcal{O}_{MB_2}$ equipped with the Horv\'ath inductive limit or respectively Schwartz operator 
topology of uniform convergence on bounded sets, and their strong duals
$\mathcal{O}'_C, \mathcal{O}'_M, \mathcal{O}'_{MB_2}$, equipped with the Schwartz operator topology of uniform convergence
on bounded sets, are nuclear. 

We have:
\begin{equation}\label{O_C<O_CA,O_M<O_MA}
\begin{split}
\mathcal{O}_M \subset \mathcal{O}_{MB_2}, \\
\mathcal{O}'_C  \subset \mathcal{O}'_{CB_2}, \\
\mathcal{O}'_C  \subset \mathcal{O}'_{CB_2},
\end{split}
\end{equation}
by the results of Subsections \ref{dim=1}-\ref{SA=S0}.

Recall that here $\mathcal{O}_M(\mathbb{R}^m; \mathbb{C}^n)$ is understood as 
the pointwise multiplication algebra of $\mathbb{C}^n$-valued functions on
$\mathbb{R}^3$ in $\mathcal{O}_M(\mathbb{R}^m; \mathbb{C}^n)$, with the elements of 
$\mathcal{O}_M(\mathbb{R}^m; \mathbb{C}^n)$, $\mathcal{S}(\mathbb{R}^m; \mathbb{C}^n)$
understood as $\mathbb{C}$-valued functions on the disjoint sum $\sqcup \mathbb{R}^m$ of $n$ 
copies of $\mathbb{R}^m$, compare Subsection \ref{psiBerezin-Hida}. The 
translation $T_b, b \in \mathbb{R}^m$ is understood as acting on $(a,x) \in \sqcup \mathbb{R}^m$, $a \in \{1,2, \ldots n\}$,
in the following manner $T_b(a,x) = (a, x+b)$. Equivalently $f \in \mathcal{O}_M(\mathbb{R}^m; \mathbb{C}^n)$
(or $f \in \mathcal{O}_C(\mathbb{R}^m; \mathbb{C}^n)$)
means that each component of $f$ belongs to $\mathcal{O}_M(\mathbb{R}^m; \mathbb{C})$
(or resp. to $\mathcal{O}_C(\mathbb{R}^m; \mathbb{C})$).

We need the following Lemma (analogously as in Subsection \ref{psiBerezin-Hida} for the Dirac field).
\begin{lem}\label{kappa0,1,kappa1,0ForA}
For the $\mathscr{L}(\mathscr{E},\mathbb{C})$-valued (or $\mathscr{E}^*$ -valued) distributions 
$\kappa_{0,1}, \kappa_{1,0}$, given by (\ref{kappa_0,1kappa_1,0A}),
in the equality (\ref{A=IntKerOpVectValKer}) defining the electromagnetic potential field $A$,
for the the pairings $\kappa_{0,1}(\xi)$, $\kappa_{1,0}(\xi)$ of
$\kappa_{0,1}, \kappa_{1,0}$ with single particle test functions $\xi$,
we have
\begin{multline*}
\Bigg( \, (\mu,x) \mapsto \sum_{\nu} \, \int \limits_{\mathbb{R}^3}
\kappa_{0,1}(\nu, \boldsymbol{\p}; \mu, x)\, \xi(\nu,\boldsymbol{\p}) \,\ud^3\boldsymbol{\p} = \kappa_{0,1}(\xi)(\mu,x) \,\, \Bigg) 
 \\ 
\in \mathcal{O}_C \subset \mathcal{O}_M \subset \mathscr{E}^*, 
\xi \in \mathcal{S}_{A}(\mathbb{R}^3, \mathbb{C}^4), 
\\
\Bigg( \, (\mu, x) \mapsto \sum_{\nu} \, \int \limits_{\mathbb{R}^3}
\kappa_{1,0}(\nu, \boldsymbol{\p}; \mu, x) \, \xi(s, \boldsymbol{\p}) \,\ud^3\boldsymbol{\p} = \kappa_{1,0}(\xi)(\mu,x) \,\, \Bigg)
\\  
\in \mathcal{O}_C \subset \mathcal{O}_M \subset \mathscr{E}^*, 
 \,\, \xi \in \mathcal{S}_{A}(\mathbb{R}^3, \mathbb{C}^4), 
\end{multline*}
are smooth, with their Fourier transforms $\widetilde{\kappa_{0,1}(\xi)}$, $\widetilde{\kappa_{1,0}(\xi)}$  
concentrated, respectively, on the negative or positive energy orbit 
$\mathscr{O}_{{}_{\pm1, 0,0,1}}$, and equal there 
\[
\begin{split}
\widetilde{\kappa_{0,1}(\xi)}(p_0(\boldsymbol{\p}), \boldsymbol{\p})
=\sum_{\nu =0}^{3} \, \sqrt{B(\boldsymbol{\p}, p^0(\boldsymbol{\p}))}^{\mu}_{\nu}
 \, \xi(\nu,\boldsymbol{\p}),
\\
\widetilde{\kappa_{1,0}(\xi)}(-p_0(\boldsymbol{\p}), -\boldsymbol{\p})
=\sum_{\nu =0}^{3} \,\sqrt{B(\boldsymbol{\p}, p^0(\boldsymbol{\p}))}^{\mu}_{\nu}
 \, \xi(\nu,\boldsymbol{\p}),
\end{split}
\]
\[
\,\,\,\, p_0(\boldsymbol{\p}) = \sqrt{|\boldsymbol{\p}|^2} = |\boldsymbol{\p}|, 
\] 
to elements of $\mathcal{S}_{A}(\mathbb{R}^3; \mathbb{C}^4)$.
In particular $\kappa_{0,1}(\xi)$, $\kappa_{1,0}(\xi)$ have all space-time derivatives bounded. For the pairings of
$\kappa_{0,1}, \kappa_{1,0}$ with space-time test functions $\varphi$ we have
\begin{multline*}
\Bigg( \, (\nu,\boldsymbol{\p}) \mapsto \sum_{\mu} \, \int \limits_{\mathbb{R}^4}
\kappa_{0,1}(\nu, \boldsymbol{\p}; \mu ,x) \, \varphi^\mu(x) \,\ud^4x = \kappa_{0,1}(\varphi)(\nu,\boldsymbol{\p}) \,\, \Bigg) \in 
\mathcal{S}_{A}(\mathbb{R}^3, \mathbb{C}^4), \,\, \varphi \in \mathscr{E}, \\
\Bigg( \, (\nu ,\boldsymbol{\p}) \mapsto \sum_{\mu} \, \int \limits_{\mathbb{R}^4}
\kappa_{1,0}(\nu, \boldsymbol{\p}; \mu, x) \, \varphi^\mu(x) \,\ud^4x = \kappa_{1,0}(\varphi)(\nu,\boldsymbol{\p}) \,\, \Bigg) \in 
\mathcal{S}_{A}(\mathbb{R}^3, \mathbb{C}^4), 
 \,\, \varphi \in \mathscr{E}. 
\end{multline*}
The maps
\[
\begin{split}
\kappa_{0,1}: \mathcal{S}_{A}(\mathbb{R}^3, \mathbb{C}^4) \ni \xi \longmapsto \kappa_{0,1}(\xi) 
\in \mathcal{O}_M, \\
\kappa_{1,0}: \mathcal{S}_{A}(\mathbb{R}^3, \mathbb{C}^4) \ni \xi \longmapsto \kappa_{1,0}(\xi) 
\in \mathcal{O}_M,
\end{split}
\]
(for $\kappa_{0,1}, \kappa_{1,0}$ understood as maps 
$\mathscr{L}\big( \mathcal{S}_{A}(\mathbb{R}^3, \mathbb{C}^4), \,\,
\mathscr{L}(\mathscr{E}, \mathbb{C})  \big) \cong 
\mathscr{L}\big( \mathcal{S}_{A}(\mathbb{R}^3, \mathbb{C}^4), \,\,
\mathscr{E}^*  \big)$) are continuous with the Schwartz operator topology on $\mathcal{O}_M$.

Moreover, the maps 
\[
\begin{split}
\kappa_{0,1}: \mathscr{E} \ni \varphi \longmapsto \kappa_{0,1}(\varphi) 
\in \mathcal{S}_{A}(\mathbb{R}^3, \,\, \mathbb{C}^4), \\
\kappa_{1,0}: \mathscr{E} \ni \varphi \longmapsto \kappa_{1,0}(\varphi) 
\in \mathcal{S}_{A}(\mathbb{R}^3, \,\, \mathbb{C}^4)
\end{split}
\]
are continuous, with $\kappa_{0,1}, \kappa_{1,0}$ understood as maps 
in 
\[
\mathscr{L}\big( \mathscr{E}, \,\, \big(\mathcal{S}_{A}(\mathbb{R}^3, \mathbb{C}^4)^* 
\big) \cong \mathscr{L}\big( \mathcal{S}_{A}(\mathbb{R}^3, \mathbb{C}^4), \,\,
\mathscr{L}(\mathscr{E}, \mathbb{C})  \big)
\] 
and, equivalently,
the maps $\xi \longmapsto \kappa_{0,1}(\xi)$, $\xi \longmapsto \kappa_{1,0}(\xi)$ can be extended to
continuous maps
\[
\begin{split}
\kappa_{0,1}: \mathcal{S}_{A}(\mathbb{R}^3, \mathbb{C}^4)^* \ni \xi \longmapsto \kappa_{0,1}(\xi) 
\in \mathscr{E}^*, \\
\kappa_{1,0}: \mathcal{S}_{A}(\mathbb{R}^3, \mathbb{C}^4)^* \ni \xi \longmapsto \kappa_{1,0}(\xi) 
\in \mathscr{E}^*,
\end{split}
\]
(for $\kappa_{0,1}, \kappa_{1,0}$ understood as maps 
$\mathscr{L}\big( \mathcal{S}_{A}(\mathbb{R}^3, \mathbb{C}^4), \,\,
\mathscr{L}(\mathscr{E}, \mathbb{C})  \big) \cong 
\mathscr{L}\big( \mathcal{S}_{A}(\mathbb{R}^3, \mathbb{C}^4), \,\,
\mathscr{E}^*  \big)$). Therefore, 
not only $\kappa_{0,1}, \kappa_{1,0}
\in \mathscr{L}\big( \mathcal{S}_{A}(\mathbb{R}^3, \mathbb{C}^4), \,\,
\mathscr{L}(\mathscr{E}, \mathbb{C})  \big)$, but both $\kappa_{0,1}, \kappa_{1,0}$
can be (uniquely) extended to elements of 
\[
\mathscr{L}\big( \mathcal{S}_{A}(\mathbb{R}^3, \mathbb{C}^4)^*, \,\,
\mathscr{L}(\mathscr{E}, \mathbb{C})  \big) \cong 
\mathscr{L}\big( \mathcal{S}_{A}(\mathbb{R}^3, \mathbb{C}^4)^*, \,\,
\mathscr{E}^*  \big)  \cong
\mathscr{L}\big( \mathscr{E}, \,\, 
\mathcal{S}_{A}(\mathbb{R}^3, \mathbb{C}^4)  \big).
\]
\end{lem}
\qedsymbol \,
That for each $\xi \in \mathcal{S}_{A}(\mathbb{R}^3, \mathbb{C}^4)$ and each fixed component $\mu$ 
the functions
$\kappa_{0,1}(\xi), \kappa_{1,0}(\xi)$ given by (here $x = (x_0, \boldsymbol{\x})$)
\begin{multline*}
x \mapsto \sum_{\nu=0}^{3} \, \int \limits_{\mathbb{R}^3}
\kappa_{0,1}(\nu, \boldsymbol{\p}; \mu,x)\, \xi(\nu,\boldsymbol{\p}) \,\ud^3\boldsymbol{\p} \\ =
\sum_{\nu =0}^{3} \, \int \limits_{\mathbb{R}^3}
{\textstyle\frac{\sqrt{B(\boldsymbol{\p}, p^0(\boldsymbol{\p}))}^{\mu}_{\nu}}{\sqrt{2p_0(\boldsymbol{\p})}}}
\, 
\xi(\nu,\boldsymbol{\p}) e^{-ip_0(\boldsymbol{\p})x_0 + i\boldsymbol{\p} \cdot \boldsymbol{\x}} \, \ud^3 \boldsymbol{\p} 
\\
= \int \delta(p^2-m^2)\theta(p_0) \zeta^a(p) e^{-ip\cdot x} \, \ud^4p, \,\,\,\, \zeta^a(p_0,\boldsymbol{\p}) =  
\sum_{\nu =0}^{3} \, \sqrt{B(\boldsymbol{\p}, p^0)}^{\mu}_{\nu}
 \, \xi(\nu,\boldsymbol{\p}),
\end{multline*}
\begin{multline*}
x \mapsto \sum_{\nu=0}^{3} \, \int \limits_{\mathbb{R}^3}
\kappa_{1,0}(\nu, \boldsymbol{\p}; \mu, x)\, \xi(\nu,\boldsymbol{\p}) \,\ud^3\boldsymbol{\p} \\ =
\sum_{\nu=0}^{3} \, \int \limits_{\mathbb{R}^3}
{\textstyle\frac{\sqrt{B(\boldsymbol{\p}, p^0(\boldsymbol{\p}))}^{\mu}_{\nu}}{\sqrt{2p_0(\boldsymbol{\p})}}}
\, 
\xi(\nu, \boldsymbol{\p}) e^{i|p_0(\boldsymbol{\p})|x_0 - i\boldsymbol{\p} \cdot \boldsymbol{\x}} \, \ud^3 \boldsymbol{\p}
\\
= \int \delta(p^2-m^2)\theta(-p_0) \zeta^a(-p)e^{-ip\cdot x} \, \ud^4p, \,\,\,\,\, \zeta^a(p_0,\boldsymbol{\p}) =  
\sum_{\nu =0}^{3} \, \sqrt{B(\boldsymbol{\p}, p^0)}^{\mu}_{\nu}
 \, \xi(\nu,\boldsymbol{\p}),
\end{multline*}
have Fourier transforms concentrated on $\mathscr{O}_{{}_{\pm1,0,0,1}}$ and are smooth with all derivatives bounded is immediate,
so in particular they belong to 
$\mathcal{O}_C(\mathbb{R}^4; \mathbb{C}) \subset \mathcal{O}_M(\mathbb{R}^4; \mathbb{C}) \subset \mathscr{E}^*$. 
Indeed, that they are smooth
is obvious, similarly as it is obvious the existence of such a natural $N$ (it is sufficient to take here $N=0$)
that for each multi-index $\alpha \in \mathbb{N}^4$ and each fixed $\mu$ the functions
\[
x \mapsto (1 + |x|^2)^{-N} |\partial_{x^{\alpha}}^{\alpha}\kappa_{0,1}(\xi)(\mu,x)|, \,\,\,
x \mapsto (1 + |x|^2)^{-N} |\partial_{x^{\alpha}}^{\alpha}\kappa_{1,0}(\xi)(\mu,x)|
\]
are bounded (of course for fixed $\xi$). Here $\partial_{x^{\alpha}}^{\alpha}\kappa_{l,m}(\xi)$ denotes the ordinary derivative of 
the function $\kappa_{l,m}(\xi)$
of $|\alpha| = \alpha_0 + \alpha_1+\alpha_2+ \alpha_3$ order with respect to space-time 
coordinates $x= (x_0, x_1, x_2, x_3)$; and here 
$|x|^2= (x_{0})^2 + (x_{1})^2 + (x_{2})^2+ (x_{3})^2$. Recall that by the results of Subsections \ref{diffSA}
and \ref{SA=S0}, the operation of pointwise multiplication by the matrix (\ref{sqrtB/sqrtp0})
is a multiplier of the algebra  
$\mathcal{S}_{A}(\mathbb{R}^3, \mathbb{C}^4) = \mathcal{S}^{0}(\mathbb{R}^3; \mathbb{C}^4)$, similarly
multiplication by the function $|p_0(\boldsymbol{\p})|^{k} = |\boldsymbol{\p}|^k$, $k \in \mathbb{Z}$,
is a multiplier of this algebra, by the same Subsections. Thus, the said integrals defining 
$\kappa_{0,1}(\xi), \kappa_{1,0}(\xi)$ are convergent, similarly as the integrals defining their space-time derivatives
with the obviously preserved mentioned above boundedness. 

That $\xi \mapsto \kappa_{0,1}(\xi), \kappa_{1,0}(\xi) \in \mathcal{O}_M(\mathbb{R}^4)$ are continuous
follows easily if we use the system of norms $p_{{}_{m, \omega}}$, $\omega \in \mathcal{S}(\mathbb{R}^4)$, $m= 0,1,2, \ldots$ 
\[
p_{{}_{m, \omega}}(f) = \underset{|\alpha| \leq m}{\textrm{sup}} \big|\omega. \partial^\alpha f \big|_{{}_{L^\infty}}, \,\,\, 
f \in \mathcal{O}_M(\mathbb{R}^4),
\]
defining the topology in $\mathcal{O}_M(\mathbb{R}^4)$, compare \cite{Larcher}, \cite{Schwartz}.

Then 
\begin{multline*}
p_{{}_{m, \omega}}\big(\kappa_{0,1}(\xi)\big) 
\\
\leq 
\underset{|\alpha|\leq m}{\textrm{sup}} \,\,
\big|\omega\big|_{{}_{L^\infty}} \,
\sum_{\nu=0}^{3} \, \int \limits_{\mathbb{R}^3}
\Big|
{\textstyle\frac{|\boldsymbol{\p}|^{\alpha_0}\big(p_1\big)^{\alpha_1}\big(p_2\big)^{\alpha_2}\big(p_3\big)^{\alpha_3}
\, \sqrt{B(\boldsymbol{\p}, |\boldsymbol{\p}|)}^{\mu}_{\nu}
\xi(\nu, \boldsymbol{\p}) e^{i|\boldsymbol{\p}|x_0 - i\boldsymbol{\p} \cdot \boldsymbol{\x}} }{\sqrt{2}|\boldsymbol{\p}|}}
\Big|
\, \ud^3 \boldsymbol{\p}
\end{multline*}
\begin{multline*}
=
\underset{|\alpha|\leq m}{\textrm{sup}} \,\, 
\big|\omega\big|_{{}_{L^\infty}} \,
\sum_{\nu=0}^{3} \, \int \limits_{\mathbb{R}^3}
\Big|
\,
{\textstyle\frac{(1+|\boldsymbol{\p}|)^n|\boldsymbol{\p}|^{\alpha_0}|p_1|^{\alpha_1}|p_2|^{\alpha_2}|p_3|^{\alpha_3}
|\boldsymbol{\p}|^2 \sqrt{B(\boldsymbol{\p}, |\boldsymbol{\p}|)}^{\mu}_{\nu}}
{\sqrt{2}|\boldsymbol{\p}|^3 \big(1+|\boldsymbol{\p}|\big)^n}} \,\, \times
\\
\times
\, 
\xi(\nu, \boldsymbol{\p}) e^{i|\boldsymbol{\p}|x_0 - i\boldsymbol{\p} \cdot \boldsymbol{\x}} 
\Big|
\, \ud^3 \boldsymbol{\p}
\end{multline*}
\begin{multline*}
=
\underset{|\alpha|\leq m}{\textrm{sup}} \,\, 
\big|\omega\big|_{{}_{L^\infty}} \,
\sum_{\nu=0}^{3} \, \int \limits_{\mathbb{R}^3}
\, 
{\textstyle\frac{\big(1+|\boldsymbol{\p}|\big)^n|\boldsymbol{\p}|^{\alpha_0}|p_1|^{\alpha_1}|p_2|^{\alpha_2}|p_3|^{\alpha_3}
|\boldsymbol{\p}|^2\sqrt{B(\boldsymbol{\p}, |\boldsymbol{\p}|)}^{\mu}_{\nu}}
{\sqrt{2}|\boldsymbol{\p}|^3\big(1+|\boldsymbol{\p}|\big)^n}}
\, 
|\xi(\nu, \boldsymbol{\p})|
\, \ud^3 \boldsymbol{\p}
\\
\leq 
\big|\omega\big|_{{}_{L^\infty}} \,
\underset{0\leq \nu \leq 3, \boldsymbol{\p} \in \mathbb{R}^3, |\alpha| \leq m}{\textrm{sup}} 
\Big\{ 
{\textstyle\frac{|\boldsymbol{\p}|^{\alpha_0}\big||p_1|^{\alpha_1}|p_2|^{\alpha_2}|p_3|^{\alpha_3}\big|
\big|\big(1+|\boldsymbol{\p}|\big)^n}{\sqrt{2}|\boldsymbol{\p}|^3}}
 | \xi(\nu, \boldsymbol{\p})|
\Big\}
\, \times
\\
\times \,
\sum_{\nu=0}^{3} \, \int \limits_{\mathbb{R}^3}
\Bigg|
{\textstyle\frac{|\boldsymbol{\p}|^2|\sqrt{B(\boldsymbol{\p}, |\boldsymbol{\p}|)}^{\mu}_{\nu}}{\big(1+|\boldsymbol{\p}|\big)^n}}
\Bigg|
\, \ud^3 \boldsymbol{\p}.
\end{multline*}

By the Lemmas of Subsection \ref{SA=S0} there exists natural $q(m)$, depending on $m$ and $n$, such that
\[
\underset{0\leq\nu\leq 3, \boldsymbol{\p} \in \mathbb{R}^3, |\alpha| \leq m}{\textrm{sup}} 
\Big\{ 
{\textstyle\frac{|\boldsymbol{\p}|^{\alpha_0}\big|\big(p_1\big)^{\alpha_1}\big(p_2\big)^{\alpha_2}\big(p_3\big)^{\alpha_3}\big|
(1+|\boldsymbol{\p}|)^n}{\sqrt{2}|\boldsymbol{\p}|^3}}
 | \xi(\nu, \boldsymbol{\p})|  
\Big\}
\leq 
c_{m} |\xi|_{q(m)}
\]
where $|\cdot|_q = | A^q \cdot|_{{}_{L^2}}$ is one of the defining norms of the standard nuclear space
$\mathcal{S}_{A}(\mathbb{R}^3, \mathbb{C}^4) = \mathcal{S}^{0}(\mathbb{R}^3; \mathbb{C}^4)$, with $A = \oplus A^{(3)}$.
Therefore
\[
p_{{}_{m, \omega}}\big(\kappa_{0,1}(\xi)\big) 
\leq 
c_{m}
C
\big|\omega\big|_{{}_{L^\infty}} |\xi|_{q(m)}
\]
where
\[
C = \sum\limits_{\nu=0}^{3} C_{\nu} 
\]
and $C_{\nu}$ is equal to the $|\cdot|_{{}_{L^1}}$-norm of the function (the component $\mu$ being fixed)
\[
\boldsymbol{\p} \longmapsto {\textstyle\frac{\sqrt{B(\boldsymbol{\p}, |\boldsymbol{\p}|)}^{\mu}_{\nu}}{\sqrt{2|\boldsymbol{\p}|}\big(1+|\boldsymbol{\p}|\big)^n}}
\]
which for sufficiently large $n$ is absolutely integrable. 
Thus, continuity of the function $\xi \mapsto \kappa_{0,1}(\xi) \in \mathcal{O}_M(\mathbb{R}^4)$ thereby follows.
Proof of the continuity of  $\xi \mapsto \kappa_{1,0}(\xi) \in \mathcal{O}_M(\mathbb{R}^4)$ is identical.

Consider now the functions 
\[
\begin{split}
\varphi \mapsto \kappa_{0,1}(\varphi)
=  \sqrt{B} \check{\widetilde{\varphi}}|_{{}_{\mathscr{O}_{1,0,0,1}}}, \\
\varphi \mapsto \kappa_{1,0}(\varphi)
= \sqrt{B} \widetilde{\varphi}|_{{}_{\mathscr{O}_{1,0,0,1}}},
\end{split}
\] 
with $\varphi \in \mathcal{S}^{00}(\mathbb{R}^4; \mathbb{C}^4)$. 
It is obvious that both functions
$\kappa_{0,1}(\varphi), \kappa_{1,0}(\varphi)$ belong to 
$\mathcal{S}_{A}(\mathbb{R}^3, \mathbb{C}^4) = \mathcal{S}^{0}(\mathbb{R}^3; \mathbb{C}^4)$
whenever $\varphi \in \mathcal{S}^{00}(\mathbb{R}^4; \mathbb{C}^4)$, by the results of Subsections
\ref{diffSA} and \ref{SA=S0}.
That both functions
$\kappa_{0,1}(\varphi), \kappa_{1,0}(\varphi)$ depend continuously on $\varphi$ as maps 
\[
\mathscr{E} = \mathcal{S}^{00}(\mathbb{R}^4; \mathbb{C}^4)
\longrightarrow \mathcal{S}_{A}(\mathbb{R}^3, \,\, \mathbb{C}^4) = \mathcal{S}^{0}(\mathbb{R}^3, \,\, \mathbb{C}^4)
\]
follows from:  1) the results of Subsection \ref{SA=S0} and continuity of the Fourier transform as a map on the Schwartz space, 2) from the continuity of the restriction to the orbits $\mathscr{O}_{1,0,0,1}$ and $\mathscr{O}_{-1,0,0,1}$
 regarded as a map from 
\[
\mathcal{S}^{0}(\mathbb{R}^4;\mathbb{C}) =
\mathcal{S}_{\oplus A^{(4)}}(\mathbb{R}^4, \,\, \mathbb{C}^4)
\]
 into 
\[
\mathcal{S}^{0}(\mathbb{R}^3;\mathbb{C}) = \mathcal{S}_{\oplus A^{(3)}}(\mathbb{R}^3, \,\, \mathbb{C}^4),
\]
compare the second Proposition of Subsection \ref{Lop-on-E}, and finally 3) from the fact 
that the operators of pointwise multiplication by the matrix (\ref{sqrtB/sqrtp0})
are multipliers of the nuclear algebra 
\[
\mathcal{S}_{A}(\mathbb{R}^3, \,\, \mathbb{C}^4) = 
\mathcal{S}_{\oplus A^{(3)}}(\mathbb{R}^3, \,\, \mathbb{C}^4) 
= \mathcal{S}^{0}(\mathbb{R}^3;\mathbb{C}),
\]
compare Subsections \ref{diffSA} and \ref{SA=S0}. 
\qed

From the last Lemma \ref{kappa0,1,kappa1,0ForA} and from Thm. 3.13 of \cite{obataJFA} 
(or equivalently from Theorem \ref{obataJFA.Thm.3.13} of Subsection \ref{psiBerezin-Hida}) 
we obtain the following
\begin{cor}\label{A=intKerOpVectVal=OpValDistr}
Let $E = \mathcal{S}_{A}(\mathbb{R}^3; \mathbb{C}^4) = 
\mathcal{S}_{\oplus A^{(3)}}(\mathbb{R}^3; \mathbb{C}^4)$. 
Let
\[
A = \Xi_{0,1}(\kappa_{0,1}) + \Xi_{1,0}(\kappa_{1,0}) \in
\mathscr{L}\big( (E) \otimes \mathscr{E}, \, (E)^* \big) \cong
\mathscr{L}\big( \mathscr{E}, \,\, \mathscr{L}( (E), (E)^*) \big)
\]
be the free quantum electromagnetic potential field understood as an integral kernel 
operator with vector-valued kernels
\[
\kappa_{0,1}, \kappa_{1,0} \in \mathscr{L}\big( \mathcal{S}_{A}(\mathbb{R}^3, \mathbb{C}^4), \,\,
\mathscr{E}^*  \big) \cong \mathcal{S}_{A}(\mathbb{R}^3, \mathbb{C}^4)^* \otimes \mathscr{E}^*
= E^* \otimes \mathscr{E}^*,
\]
defined by (\ref{kappa_0,1kappa_1,0A}). Then the electromagnetic potential field operator
\[
A  = A^{(-)} + A^{(+)} = \Xi_{0,1}(\kappa_{0,1}) + \Xi_{1,0}(\kappa_{1,0}),
\]
belongs to $\mathscr{L}\big( (E) \otimes \mathscr{E}, \, (E) \big) \cong
\mathscr{L}\Big( \mathscr{E}, \,\, \mathscr{L}\big( (E), (E)\big) \, \Big)$, i.e.
\[
A = \Xi_{0,1}(\kappa_{0,1}) + \Xi_{1,0}(\kappa_{1,0}) \in
\mathscr{L}\big( (E) \otimes \mathscr{E}, \, (E) \big) \cong
\mathscr{L}\Big( \mathscr{E}, \,\, \mathscr{L}\big( (E), (E)\big) \, \Big),
\]
which means in particular that the electromagnetic potential field $A$, understood as a sum 
$A = \Xi_{0,1}(\kappa_{0,1}) + \Xi_{1,0}(\kappa_{1,0})$ of 
two integral kernel operators with vector-valued kernels, 
defines an operator valued distribution through the continuous map
\[
\mathscr{E} \ni \varphi \longmapsto
\Xi_{0,1}\big(\kappa_{0,1}(\varphi)\big) + \Xi_{1,0}\big(\kappa_{1,0}(\varphi)\big)
\in \mathscr{L}\big( (E), (E)\big).
\]
\end{cor}

Note that the last Corollary likewise follows from: 
\begin{enumerate}
\item[1)]
the equality (\ref{A=IntKerOpVectValKer}),
\item[2)]
from Thm. 2.2 and 2.6 of \cite{hida}, 
\item[3)]
continuity of the Fourier transform as a map
on the Schwartz space, 
\item[4)]
continuity of the restriction to the orbit $\mathscr{O}_{1,0,0,1}$
regarded as a map $\mathcal{S}^{0}(\mathbb{R}^4) \longrightarrow \mathcal{S}^{0}(\mathbb{R}^3)$ and 
finally 
\item[5)]
from continuity of the multiplication by the matrix (\ref{sqrtB/sqrtp0}), regarded as a map
$\mathcal{S}^{0}(\mathbb{R}^3; \mathbb{C}^4) \longrightarrow \mathcal{S}^{0}(\mathbb{R}^3; \mathbb{C}^4)$.
\end{enumerate}

It is important to emphasize here that by the Thm. 3.13 of \cite{obataJFA}
(or Thm. \ref{obataJFA.Thm.3.13} of Subsection \ref{psiBerezin-Hida}) the continuity of the map
$\varphi \longmapsto \kappa_{1,0}(\varphi)$, regarded as a map
$\mathscr{E} \longrightarrow E = \mathcal{S}_{A}(\mathbb{R}^3; \mathbb{C}^4)$,
equivalent to the continuous unique extendibility of $\kappa_{1,0}$
to an element of $\mathscr{L}(E^*, \mathscr{E}^*)$, is a necessary and sufficient
condition for the operator $A = \Xi_{0,1}(\kappa_{0,1}) + \Xi_{1,0}(\kappa_{1,0})$
to be an element of 
\[
\mathscr{L}\big( (E) \otimes \mathscr{E}, \, (E) \big) \cong
\mathscr{L}\Big( \mathscr{E}, \,\, \mathscr{L}\big( (E), (E)\big) \, \Big),
\]
i.e. for $A$ being a sum of integral kernel operators with vector-valued kernels which defines 
an operator-valued distribution on $\mathscr{E}$.
On the other hand the continuity of the map
\[
\mathscr{E} \ni \varphi \longmapsto \kappa_{1,0}(\varphi) \in E = \mathcal{S}_{A}(\mathbb{R}^3; \mathbb{C}^4) 
\]
is equivalent, as we have seen, to the continuity of the restriction to the cone $\mathscr{O}_{1,0,0,1}$,
regarded as a map
\[
\widetilde{\mathscr{E}} \longrightarrow E = \mathcal{S}_{A}(\mathbb{R}^3; \mathbb{C}^4),
\]
followed by the multiplication by the matrix 
(\ref{sqrtB/sqrtp0}), and regarded as a map $E \rightarrow E$. 
From this it follows that 
\[
\widetilde{\mathscr{E}} \neq \mathcal{S}(\mathbb{R}^4), \,\,\, E \neq \mathcal{S}(\mathbb{R}^3)
\]
for the space-time test space of the zero mass field $A$ determined by a representation pertinent
to the cone orbit $\mathscr{O}_{1,0,0,1}$, because restriction to the cone
$\mathscr{O}_{1,0,0,1}$ is not continuous as a map $\mathcal{S}(\mathbb{R}^4) \rightarrow \mathcal{S}(\mathbb{R}^3)$,
nor the multiplication by the matrix (\ref{sqrtB/sqrtp0}) regarded as a map
$\mathcal{S}(\mathbb{R}^3) \rightarrow \mathcal{S}(\mathbb{R}^3)$.
This is in general the case for any zero mass (free) field. Namely, we have the following
\begin{twr}\label{ZeromassTestspace}
For any zero mass field, pertinent to the cone orbit $\mathscr{O}_{1,0,0,1}$, such as the electromagnetic potential field, 
which can be regarded as an integral kernel operator
\[
\Xi_{0,1}(\kappa_{0,1}) + \Xi_{1,0}(\kappa_{1,0})
\]
with vector-valued kernels 
\[
\kappa_{0,1}, \kappa_{1,0} \in \mathscr{L}\big( \mathcal{S}_{A}(\mathbb{R}^3, \mathbb{C}^4), \,\,
\mathscr{E}^*  \big) \cong \mathcal{S}_{A}(\mathbb{R}^3, \mathbb{C}^4)^* \otimes \mathscr{E}^*
= E^* \otimes \mathscr{E}^*,
\]
extendible to 
\[
\kappa_{0,1}, \kappa_{1,0} \in \mathscr{L}\big( \mathcal{S}_{A}(\mathbb{R}^3, \mathbb{C}^4)^*, \,\,
\mathscr{E}^*  \big) \cong \mathcal{S}_{A}(\mathbb{R}^3, \mathbb{C}^4) \otimes \mathscr{E}^*
= E \otimes \mathscr{E}^*,
\]
and defined by plane waves
\[
\begin{split}
\kappa_{0,1}(s, \boldsymbol{\p}; a, x) = u^a(s, \boldsymbol{\p})\,e^{-ip\cdot x}, \,\,\,\,
p = (p_0(\boldsymbol{\p}), \boldsymbol{\p}) \in \mathscr{O}_{1,0,0,1}, \\
\kappa_{0,1}(s, \boldsymbol{\p}; a, x) = v^a(s, \boldsymbol{\p})\,e^{ip\cdot x}, \,\,\,\,
p = (p_0(\boldsymbol{\p}), \boldsymbol{\p}) \in \mathscr{O}_{1,0,0,1}, \\
s,a = 1,2, \ldots N
\end{split}
\]
the space-time test space $\mathscr{E}$ cannot be equal to the ordinary Schwartz space
 $\mathcal{S}(\mathbb{R}^4; \mathbb{C}^N)$ but instead it has to be equal
\[
\mathscr{E} = \mathcal{S}^{00}(\mathbb{R}^4; \mathbb{C}^N)
=  \mathscr{F} \Big[ \mathcal{S}^{0}(\mathbb{R}^4; \mathbb{C}^N) \Big]
= \mathscr{F} \Big[ \mathcal{S}_{\oplus A^{(4)}}(\mathbb{R}^4; \mathbb{C}^N) \Big],
\]
where $A^{(4)}$ is the standard operator on $L^2(\mathbb{R}^4; \mathbb{C})$
constructed in Subsection \ref{dim=n}, and $\oplus A^{(4)}$ denotes direct sum of
$N$ copies of the operator $A^{(4)}$ acting on 
\[
L^2(\mathbb{R}^4; \mathbb{C}^N) = \oplus_{1}^{N} L^2(\mathbb{R}^4; \mathbb{C}).
\]
\end{twr}
 
In particular this Theorem holds for all zero mass free gauge fields $A$ of the Standard Model. 

Let us stress once more that the conclusion of the last Theorem is inapplicable to 
zero-mass fields in the sense of Wightman, which allows the ordinary 
Schwartz space as the space-time test space. This follows immediately from the fact
that the integration of the restriction of the test function to the cone orbit 
$\mathscr{O}_{1,0,0,1}$ along $\mathscr{O}_{1,0,0,1}$ with respect to the measure 
induced by the ordinary measure of the ambient space $\mathbb{R}^4$, is a well defined
continuous functional on the ordinary Schwartz space $\mathcal{S}(\mathbb{R}^4; \mathbb{C})$.
We have also used this fact in extending the zero mass Pauli-Jordan function from
$\mathcal{S}^{00}(\mathbb{R}^4)$ over to a functional on $\mathcal{S}(\mathbb{R}^4)$,
with preservation of the homogeneity and its degree, compare Subsection \ref{Lop-on-E}.

\subsection{Justification of the rules of Subsection \ref{psiBerezin-Hida}}

During the proof of the Bogoliubov-Shirkov Postulate for the 
free electromagnetic potential field $A$, Subsection \ref{BSH}, we have not 
used the fact that the field $A$ is a well defined integral kernel
operator 
\[
A = \Xi_{0,1}(\kappa_{0,1}) + \Xi_{1,0}(\kappa_{1,0})
\]
with vector valued distributional kernels $\kappa_{0,1}, \kappa_{0,1}$ 
defined by the plane waves (\ref{kappa_0,1kappa_1,0A}), having the extendibility properties 
of Lemma \ref{kappa0,1,kappa1,0ForA}. During this proof
we have used the fact that for each fixed spacetime point $x$
the integral (\ref{q-A-B}), Subsect. \ref{BSH}, exists pointwisely as the Pettis
integral and defines continuous operator
$(E) \rightarrow (E)^*$. Similarly, using the Pettis integration we have defined pointwisely
the operators
\[
\partial_\nu A^\mu(x), \,\,\, {:}\partial_0 A^\mu(x) \partial_{k}A^\nu(x){:}, \,\,\,
\int {:}\partial_0 A^\mu(x_0, \boldsymbol{\x}) \partial_{k}A^\nu(x_0, \boldsymbol{\x}){:} \, \ud^3 \boldsymbol{\x}.
\] 
During this proof we have obtained a justification for the rules of differentiation, Wick product, 
and integration, understood as the operations performed upon integral kernel operators
\[
\Xi_{0,1}(\kappa_{0,1}) = A^{(-)}, \,\,\,\,\,\,\,\, \Xi_{1,0}(\kappa_{1,0}) = A^{(+)}
\]
defined by the free field $A = \Xi_{0,1}(\kappa_{0,1}) + \Xi_{1,0}(\kappa_{1,0}) = A^{(-)} + A^{(+)}$. 
According to these rules the said operations are determined
by the corresponding operations performed upon the kernel distributions $\kappa_{l,m}$
corresponding to the involved integral kernel operators.

The presented proof of the rules performed upon 
$\Xi_{0,1}(\kappa_{0,1}) = A^{(-)}, \Xi_{1,0}(\kappa_{1,0}) = A^{(+)}$,
is easily applicable to the proof of general rules as stated
in Subsection \ref{psiBerezin-Hida}, and involving Wick product
of any $\Xi_{0,1}(\kappa_{0,1}), \Xi_{1,0}(\kappa_{1,0})$, coming from any (finite) 
number of free fields,
provided they can be constructed as integral kernel operators
$\Xi_{0,1}(\kappa_{0,1}) + \Xi_{1,0}(\kappa_{1,0})$ with vector-valued 
kernels $\kappa_{0,1}, \kappa_{1,0}$ having the properties 
expressed in Lemma \ref{kappa0,1,kappa1,0ForA}, Subsection \ref{A=Xi0,1+Xi1,0}
or Lemma  \ref{kappa0,1,kappa1,0psi}, Subsection \ref{psiBerezin-Hida}.
In fact the integration of the Wick polynomials of free fields over the whole space-time
(or over its cartresian product of the respective integral kernel operators with kernels
having values over the respective tensor product of the space-time test spaces)
has not been analysed in Subsection \ref{BSH}, but it can be analysed in the same manner.

\subsection{Comparison of two realizations of the free local electromagnetic
potential quantum field. Decomposition of the representation of $SL(2,\mathbb{C})$ pertinent to the standard realization,
and the associated decomposition of the standard free electromagnetic potential field}\label{equivalentA-s}

Let $U^{*-1} = WU^{{}_{(1,0,0,1)}{\L}}W^{-1}$ and  $U = \big[WU^{{}_{(1,0,0,1)}{\L}}W^{-1}\big]^{*-1}$
be the {\L}opusza\'nski representation and its conjugation $U$ acting in the single
particle space of the quantum field $A$ realization of Sections \ref{free-gamma} and  \ref{white-noise-proofs}.
Both $U^{*-1}$, and $U$ transform continuously the nuclear space $E_\mathbb{C}$ into itself (let us write
simply $E$ instead $E_\mathbb{C}$ for simplicity). Similarly, the lifting $\Gamma(U)$ of $U$
acting in the Krein-Fock space $(\Gamma(\mathcal{H}'), \Gamma(\mathfrak{J}'))$ transforms
continuously the nuclear Hida's test space $(E)$ onto itself, and is Krein isometric in the Krein-Fock
space of the field $A$. 

We can consider different such realizations of $A$, with the representations $U$ and $\Gamma(U)$ restricted to the translation subgroup commuting with the Krein fundamental symmetry $\mathfrak{J}'$, and resp. 
$\Gamma(U)$ commuting with the Gupta-Bleuler operator $\Gamma(\mathfrak{J}')$, and thus with translations being represented by unitary and Krein-unitary operators. The natural equivalence for such realizations is the existence of Krein isometric mapping transforming bi-uniquely and bi-continuously $E$, resp. $(E)$,
onto itself, and which intertwines the representations. It is easily seen that in case of ordinary non gauge fields with unitary representations, this equivalence reduces to the ordinary unitary equivalence
of the realizations of the fields. In case of gauge massless fields, such as electromagnetic potential field 
$A$, where $U$ and $\Gamma(U)$ are unbounded (and Krein-isometric) the equivalence is weaker,
although preserves the pairing functions of the field, the linear equation it fulfills and its local transformation formula. 
Nonetheless, the analytic properties of the representation may be substantially different
for equivalent realizations of the field $A$, especially the behavior of the restriction of the representation
$U$ or $\Gamma(U)$ of $T_4 \circledS SL(2, \mathbb{C})$ to the subgroup $SL(2, \mathbb{C})$, 
as is no very surprising as the representors of the Loretz hyperbolic rotations
are unbounded, contrary to the representors of translations, which are bounded (even unitary and Krein-unitary).

We illustrate this phenomena on a concrete example of different equivalent realizations of the free field $A$. 
Although the example is concrete it can be shown 
that the construction encountered is generic, and that the general class of equivalent realizations
may be constructed
without any substantial modification. The general construction of a realization of the free field $A$
is equivalent to the construction of the most general intertwining operator bi-uniquelly and bi-continuously mapping the 
nuclear spaces, where the initial spaces and representations are these given in 
Sections \ref{free-gamma} and  \ref{white-noise-proofs} for the realization of $A$ given there.  
We give a concrete example of such an intertwining operator, in case where the nuclear spaces corresponding to different realizations are identical. Because this assumption is not relevant, and because the construction
of the general intertwining operator is general for the case where the nuclear spaces are identical, 
we prefer to give the concrete example
instead of going immediately into a general situation, which would be less transparent.

On the single particle space $(\mathcal{H}', \mathfrak{J}')$ of the realization of $A$ of Sect. 
\ref{free-gamma} and  \ref{white-noise-proofs} there exists, besides $U, U^{*-1}$, the Krein-isometric
representation (compare Proposition of introduction to Section \ref{constr-of-VF})
\begin{equation}\label{Lop-rep--on-tildevarphi-ass}
\begin{split}
{}^{{}^{\textrm{acc}}}U(0,\alpha) \widetilde{\varphi} (p) 
= \sqrt{B(p)}^{-1}V(\alpha)\sqrt{B(\Lambda(\alpha)p)} \widetilde{\varphi} (\Lambda(\alpha)p) \\
= \sqrt{B(p)}^{-1}\Lambda(\alpha^{-1})\sqrt{B(\Lambda(\alpha)p)} \widetilde{\varphi} (\Lambda(\alpha)p), \\
{}^{{}^{\textrm{acc}}}U(a,1) \widetilde{\varphi} (p) =  \sqrt{B(p)}^{-1}T(a) \sqrt{B(p)} \widetilde{\varphi}(p) 
= e^{i a \cdot p}\widetilde{\varphi}(p). 
\end{split}
\end{equation}
accompanying the {\L}opusza\'nski representation $U^{*-1} = WU^{{}_{(1,0,0,1)}{\L}}W^{-1}$,
where $\sqrt{B(p)}$ is the (positive) square root of the (positive) matrix $B(p), p \in \mathscr{O}_{{}_{1,0,0,1}}$ (\ref{Bmatrix}), equal  (\ref{sqrtB}). Recall that for each fixed point 
$p \in \mathscr{O}_{1,0,0,1}$, the matrices $\sqrt{B(p)}$, $B(p)$, $\mathfrak{J}'_{{}_{\bar{p}}}
=V(\beta(p))^{-1} \mathfrak{J}_{\bar{p}} V(\beta(p)) = 
\mathfrak{J}_{\bar{p}} B(p)$, are all Krein-unitary in the Krein space $(\mathbb{C}^4, \mathfrak{J}_{\bar{p}})$,
where $\mathfrak{J}_{\bar{p}}$ is the constant matrix (\ref{J-barp}). In other words all the matrices 
$\sqrt{B(p)}$, $B(p)$, $\mathfrak{J}'_{{}_{\bar{p}}}$ are Lorentz matrices preserving the the Lorentz
metric $g^{\mu \nu} = \textrm{diag}(-1,1,1,1)$.

This representation is Krein-isometrically equivalent to the 
{\L}opusza\'nski representation $U^{*-1} = WU^{{}_{(1,0,0,1)}{\L}}W^{-1}$ (\ref{Lop-rep--on-tildevarphi}).
(Analogously its conjugation is equivalent to the conjugation $U$ of the {\L}opusza\'nski representation
$U^{*-1}$). Indeed, the intertwining operator $C$, understood as an operator
$(\mathcal{H}', \mathfrak{J}') \rightarrow (\mathcal{H}', \mathfrak{J}')$, 
acting in the single particle space is equal
\[
C \widetilde{\varphi}(p) = \sqrt{B(p)}^{-1} \widetilde{\varphi}(p), \,\,\,\,
C^{-1} \widetilde{\varphi}(p) = \sqrt{B(p)} \widetilde{\varphi}(p),
\]
and $C$ transforms bi-uniquely and bi-continuously the nuclear space $E$ onto itself (compare the first Proposition of 
Subsection \ref{Lop-on-E}) and the intertwining operator $\Gamma(C)$ transforms bi-uniquely and bi-continuously $(E)$ 
onto itself $(E)$, \cite{hida}, \cite{obata-book}. One easily checks that $C$ indeed intertwines $U^{*-1}$ and ${}^{{}^{\textrm{acc}}}U$:
\[
C \, U^{*-1} \, C^{-1} = {}^{{}^{\textrm{acc}}}U 
\] 
and thus that $\Gamma(C)$ intertwines $\Gamma(U^{*-1})$ and $\Gamma({}^{{}^{\textrm{ac}}}U)$. 

Let us introduce another operator $K$:
\[
K \widetilde{\varphi}(p) = \sqrt{B(p)} \widetilde{\varphi}(p), \,\,\,
K^{-1} \widetilde{\varphi}(p) = \sqrt{B(p)}^{-1} \widetilde{\varphi}(p),
\]
understood as a Krein-unitary operator mapping the Krein space $(\mathcal{H}', \mathfrak{J}')$
onto the Krein space $(K\mathcal{H}', K\mathfrak{J}'K^{-1}) 
= (K\mathcal{H}', \mathfrak{J}_{{}_{\bar{p}}})$, where the Krein fundamental symmetry
in the Krein space $(K \mathcal{H}', \mathfrak{J}_{{}_{\bar{p}}})$ is equal to the operator
of multiplication by the constant matrix $\mathfrak{J}_{{}_{\bar{p}}}$ equal (\ref{J-barp}).
Recall that the Krein fundamental symmetry operator $\mathfrak{J}'$ in the single particle
Krein space $(\mathcal{H}', \mathfrak{J}')$ is equal to the operator of multiplication
by the matrix (\ref{operatorJ'})
\[
\mathfrak{J'}_{p} = V(\beta(p))^{-1} \mathfrak{J}_{\bar{p}} V(\beta(p)) = 
\mathfrak{J}_{\bar{p}} B(p),
\]
where $B(p)$ is equal to the matrix (\ref{Bmatrix}). The operator $K$ gives a Krein-unitary equivalence
between the representation ${}^{{}^{\textrm{acc}}}U$ acting on the Krein space $(\mathcal{H}', \mathfrak{J}')$
and defined by the formula (\ref{Lop-rep--on-tildevarphi-ass}) with the dense nuclear domain
$E$, and the Krein-isometric representation given by formula (\ref{Lop-rep--on-tildevarphi}) identical as for the
{\L}opusza\'nski representation $U^{*-1}$ on $E$, but on the Krein space  
$(K \mathcal{H}', \mathfrak{J}_{{}_{\bar{p}}})$ and with the nuclear domain $E$, which differs from 
the Krein space of Sections
\ref{free-gamma} and  \ref{white-noise-proofs} by the replacement of the Lorentz matrices
$\sqrt{B(p)}$ and $B(p)$ everywhere with the constant unit matrix $\boldsymbol{1}$. Because on the other hand 
the {\L}opusza\'nski representation $U^{*-1}$, defined by (\ref{Lop-rep--on-tildevarphi}),
and the representation ${}^{{}^{\textrm{acc}}}U$, both acting on the Krein space
$(\mathcal{H}', \mathfrak{J}')$ are Krein isometric equivalent (with $C$ defining the equivalence),
then it follows that the {\L}opusza\'nski representation, defined by (\ref{Lop-rep--on-tildevarphi}),
with the nuclear domain $E$, on the Krein space $(\mathcal{H}', \mathfrak{J}')$ (with
the matrix $B(p) \neq \boldsymbol{1}$ and equal (\ref{Bmatrix})) is equivalent to the Krein isometric
representation defined by the same formula (\ref{Lop-rep--on-tildevarphi}) and the same nuclear domain $E$,
but on the Krein space in which the operators $B(p)$ and $\sqrt{B(p)}$ are everywhere replaced
by the constant unit matrices $\boldsymbol{1}$.
      
In this way we have obtained two equivalent realizations of the free quantum field $A$. 
The one is obtained as in Sections \ref{free-gamma} and  \ref{white-noise-proofs}. The other is obtained exactly as in Sections
\ref{free-gamma} and  \ref{white-noise-proofs} by the replacement everywhere in the formulas of the 
positive Lorentz matrices $B(p)$ and $\sqrt{B(p)}$ by the unit $4 \times 4$-matrix. A simple inspection shows 
that all proofs remain valid if we replace $B(p), \sqrt{B(p)}$ by $\boldsymbol{1}$ in Sections 
\ref{free-gamma} and  \ref{white-noise-proofs}. In particular, we obtain in this way a local
mass-less quantum four-vector field $A$, fulfilling d'Alembert equation with the pairing equal to the
zero mass Pauli-Jordan distribution function multiplied by the Minkowski metric components.
In particular this realization should be identified with the one used e.g. in \cite{Scharf},
\cite{DKS1}-\cite{DKS4}. In particular replacement of the matrix 
\[
\sqrt{B(\boldsymbol{\p}, p^0(\boldsymbol{\p}))}^{\mu}_{\lambda}
\]
by the unit $4\times 4$ matrix in the formula (\ref{q-A-B}): 
\begin{multline*}
A^\mu(x) = \int \limits_{\mathbb{R}^3} \, \ud^3 p \, \bigg\{
\frac{1}{\sqrt{2 p^0(\boldsymbol{\p})}}\sqrt{B(\boldsymbol{\p}, p^0(\boldsymbol{\p}))}^{\mu}_{\lambda}
a^{\lambda} (\boldsymbol{\p}) e^{-ip\cdot x} \\
+  \frac{1}{\sqrt{2 p^0(\boldsymbol{\p})}}\sqrt{B(\boldsymbol{\p}, p^0(\boldsymbol{\p}))}^{\mu}_{\lambda} \,
\eta \, a^{\lambda}(\boldsymbol{\p})^+ \, \eta \, e^{ip\cdot x}  \bigg\} 
\end{multline*} 
gives exactly the formula (2.11.45):
\begin{equation}\label{q-A'-B}
A^\mu(x) = \int \limits_{\mathbb{R}^3} \, \ud^3 p \, \bigg\{
\frac{1}{\sqrt{2 p^0(\boldsymbol{\p})}}
a^{\mu} (\boldsymbol{\p}) e^{-ip\cdot x} \\
+  \frac{1}{\sqrt{2 p^0(\boldsymbol{\p})}} \,
\eta \, a^{\mu}(\boldsymbol{\p})^+ \, \eta \, e^{ip\cdot x}  \bigg\} 
\end{equation} 
 of \cite{Scharf} (the lack of the additional constant factor $(2\pi)^{-3/2}$
in our formula comes from the fact that we have discarded the normalization factor 
for the measures in the Fourier transforms, in order to simplify notation). 
Similarly, for other operator-valued distributions,
or ordinary operators, which we obtain by inserting the unit matrix for $\sqrt{B(p)}$. 

However, the explicit formula for the Krein-isometric representation of
$T_4 \circledS SL(2, \mathbb{C})$ is lacking in the cited works as well as in other
works (as to the knowledge of the author) using the Gupta-Bleuler or BRST method. 
Moreover, any analysis of the electromagnetic potential field in the Gupta-Bleuler approach, 
giving the linkage to 
the (generalized) induced representation theory of Mackey necessary uses the operator $\sqrt{B(p)}
\neq \boldsymbol{1}$. In particular no explicit construction of the representation of 
$T_4 \circledS SL(2, \mathbb{C})$ would be possible and its immediate linkage to the induced 
{\L}opusza\'nski representation, without the analysis using explicitly the
realization of the field $A$ with the matrix $B(p)$ equal (\ref{Bmatrix}). 
We can pass to the (apparently)
simpler formulas only after using the intertwining operators, $C, K$, defined again with the hepl of 
$\sqrt{B(p)}$, and starting with the realization of $A$ presented in Sections \ref{free-gamma} and  
\ref{white-noise-proofs}.

Perhaps we should emphasize that the two realizations of the free electromagnetic potential 
quantum field $A$: 1) the one with with $\sqrt{B(p)} \neq \boldsymbol{1}$ equal (\ref{sqrtB})
and presented in Section \ref{free-gamma} and 2) the one with 
$\sqrt{B(p)} = \boldsymbol{1}$, differ substantially. In particular, we have the following
\begin{prop*}
 Consider the restriction
of the Krein-isometric representations of $T_4 \circledS SL(2, \mathbb{C})$ to the subgroup 
$SL(2, \mathbb{C})$, acting in the single particle Krein-Hilbert spaces in the two realizations, 1) and 2).
Then for the second realization 2)  (with $\sqrt{B(p)} = \boldsymbol{1}$) the restriction can be 
decomposed into
ordinary Hilbert space direct integral of subrepresentations $U_{{}_{\chi}}$ each acting in the
Hilbert space of generalized homogeneous of degree $\chi$ eigenstates $\in E^*$ (distributions) 
of the scaling operator $\widetilde{S_{\lambda}}$:
\[
\widetilde{S_\lambda} \widetilde{\varphi}(p) = \widetilde{\varphi}(\lambda p), \widetilde{\varphi} \in E,
\]       
where $\lambda$ is a fixed positive real number. 

No such decomposition is possible for the 1) realization of $A$ (with $\sqrt{B(p)} \neq \boldsymbol{1}$ 
and equal (\ref{sqrtB})).
\end{prop*}
{\bf REMARK}.
The statement of the last Proposition can be easily lifted to the Fock-Krein spaces of the realizations
1) and 2) of the 
field $A$, therefore we consider the statement and the proof only for the single particle 
Krein-Hilbert spaces. \qed

\qedsymbol \, 
We consider the two versions of the {\L}opusza\'nski representation $U^{*-1}$ 
with $\sqrt{B(p)}$ equal respectively (\ref{sqrtB}) or  $\boldsymbol{1}$ in case 1) or 2).
The results for its conjugation $U$ actually acting in the single particle space will follow
as a consequence from the result for the {\L}opusza\'nski representation $U^{*-1}$ itself.  

Note that in both realizations the operator $\widetilde{S_\lambda}$ (checking of which we leave as an easy exercise) 
has (unique) bounded extension to a normal operator, i.e. commuting with its adjoint $\widetilde{S_{\lambda}}^{*}$ 
(with respect
to the ordinary Hilbert space inner product $(\cdot, \cdot)$, and not with respect to the Krein-inner product
$(\cdot, \mathfrak{J}' \cdot)$). 

The point is that the operators $\widetilde{S_\lambda}, \widetilde{S_{\lambda}}^{*}$, both commute 
with the {\L}opusza\'nski representation $U^{*-1}$ in the second realization 2) (with $\sqrt{B(p)} = \boldsymbol{1}$) 
and with the operator $\mathfrak{J}'$ (which in the realization 2) with $\sqrt{B(p)} = \boldsymbol{1}$
reduces to the constant matrix operator $\mathfrak{J}_{\bar{p}}$ equal to (\ref{J-barp})). 
But in the first realization 1) (with $\sqrt{B(p)}$ equal (\ref{sqrtB})), although
$\widetilde{S_\lambda}$ commutes with the {\L}opusza\'nski representation $U^{*-1}$, the adjoint
operator $\widetilde{S_{\lambda}}^{*}$ does not commute with the {\L}opusza\'nski representation $U^{*-1}$,
nor with the operator $\mathfrak{J}'$. Checking the commutation rules we again leave as an easy 
exercise to the reader. 

The proof of the statement of the Proposition can now be essentially reduced to the application of Theorems 1 and 2, \cite{Segal_dec_I}, with the commutative decomposition $*$-algebra $C$ of Thm. 2 in 
\cite{Segal_dec_I} equal to the one generated by the 
commuting operators $\widetilde{S_\lambda}, \widetilde{S_{\lambda}}^{*}$.

In both realizations, 1) and 2), the operators $\widetilde{S_\lambda}, \widetilde{S_{\lambda}}^{*}$ transform continuously 
the nuclear space $E$ into itself, which follows easily by the results of Section \ref{white-noise-proofs} 
(compare the proof of the first Proposition of Subsection \ref{Lop-on-E}). On the other hand $E$, 
the single particle Krein-Hilbert space $\mathcal{H}'$
and $E^*$, compose the Gefand triple $E \subset \mathcal{H}' \subset E^*$ (or a rigged Hilbert space). Thus,  
the decomposition
\[
\mathcal{H}'= \int\limits_{\textrm{Sp} \, \widetilde{S_\lambda}} \mathcal{H}'_{{}_{\chi}} \, \ud \chi,
\,\,\,\,
\int\limits_{\textrm{Sp} \, \widetilde{S_\lambda}} U_{{}_{\chi}} \, \ud \chi,
\,\,\,\,
\mathfrak{J}'= \int\limits_{\textrm{Sp} \, \widetilde{S_\lambda}} \mathfrak{J}'_{{}_{\chi}} \, \ud \chi
\]
of $U$ (restricted to $SL(2, \mathbb{C})$) and of $\mathfrak{J}'$ in the realization 2), is precisely the decomposition 
corresponding to the decomposition 
\[
\mathcal{H}'= \int\limits_{\textrm{Sp} \, \widetilde{S_\lambda}} \mathcal{H}'_{{}_{\chi}} \, \ud \chi,
\,\,\,\,
\int\limits_{\textrm{Sp} \, \widetilde{S_\lambda}} \big(\widetilde{S_\lambda}\big)_{{}_{\chi}} \, \ud \chi
\]
associated to the normal operator $\widetilde{S_\lambda}$,
into the direct integral of (generalized) Hilbert subspaces $\mathcal{H}'_{{}_{\chi}}$ of generalized eigenvectors in $E^*
= \mathcal{S}^{0}(\mathbb{R}^3; \mathbb{C}^4)^*$ of $\widetilde{S_\lambda}$, constructed as in Chap. I.4. of \cite{GelfandIV}.
\qed

Here we give this decomposition  (using the method of Chap. I.4. of \cite{GelfandIV}) in explicit form,
\emph{i.e.} we give the complete system of homogeneous states 
\begin{equation}\label{Fchi,l,m,s}
\begin{split}
F_{{}_{\chi, l,m,s}} \in E^* =  \mathcal{S}^{0}(\mathbb{R}^3, \mathbb{C}^4)^*, 
\,\,\,\,\,
\lambda^{{}^{\chi}} \in \textrm{Sp} \, \widetilde{S_\lambda}, 
\\
 l \in \{0, 1, 2, \ldots \}, \,\, -l \leq m \leq l, \,\,
 s \in \{+,-,0+,0-\},
\end{split}
\end{equation}
of homogeneity degree $\chi$ such that for each fixed $\chi$ with $\lambda^{{}^{\chi}} \in \textrm{Sp} \, \widetilde{S_\lambda}$,
it composes a complete orthonormal system of states in the Hilbert-Krein space direct summand $(\mathcal{H}'_{{}_{\chi}}, \mathfrak{J}'_{{}_{\chi}})$
of the decomposition, together with the Plancherel formula associated to this decomposition. 

Please note, that $\widetilde{S_\lambda}^* = \lambda^{-2}\widetilde{S_{1/\lambda}}$ (of course in the decomposable case with $B(p) = \boldsymbol{1}$), so that
\[
\widetilde{S_\lambda} \widetilde{S_\lambda}^* = \lambda^{-2} \boldsymbol{1}.
\]
Therefore, each spectral value $\sigma \in \textrm{Sp} \, \widetilde{S_\lambda}$ of the scaling operator $\widetilde{S_\lambda}$
respects the condition
\begin{equation}\label{sigmasigmabar=lambdato-2ForA}
\sigma \overline{\sigma} = \lambda^{-2}.
\end{equation}
Thus, writing the spectral values $\sigma = \lambda^{{}^{\chi}}$ in terms of the homogeneity degree $\chi = \mu + i \nu$, $\mu,\nu \in \mathbb{R}$,
the condition (\ref{sigmasigmabar=lambdato-2ForA}) implies that
\[
\chi = -1 + i\nu, \,\,\, \nu \in \mathbb{R}
\]
are the only allowed values for the homogeneities of the decomposition and that
\[
\textrm{Sp} \, \widetilde{S_\lambda} \subset \{\lambda^{{}^{-1 + i\nu}}, \nu \in \mathbb{R} \}.
\]
From the Plancherel formula, which we give below, it follows that the equality holds
\[
\textrm{Sp} \, \widetilde{S_\lambda} = \{\lambda^{{}^{-1 + i\nu}}, \nu \in \mathbb{R} \}.
\]

In constructing the complete system of generalized states (\ref{Fchi,l,m,s}), and in the proof of the Plancherel
formula we use the canonical factorization 
\[
E = \mathcal{S}^{0}(\mathbb{R}^3; \mathbb{C}^4) = \mathcal{S}^{0}(\mathbb{R}_+;\mathbb{C}) \otimes \mathscr{C}^\infty(\mathbb{S}^2; \mathbb{C}^4)
= \mathcal{S}_{{}_{A^{(1)}}}(\mathbb{R}_+;\mathbb{C}) \otimes \mathcal{S}_{{}_{\Delta_{\mathbb{S}^2}}}(\mathbb{S}^2; \mathbb{C}^4)
\]
of the nuclear test space $E \subset \mathcal{H}'$ (compare Subsections \ref{dim=1}-\ref{SA=S0})
and the system (\ref{eigen-vectors-matrixB}) of eigenvectors
\begin{equation}\label{eigenvectorsOfB(p)}
w^{{}^{+}}, w^{{}^{-}}, w_{{}_{r^{-2}}}, w_{{}_{r^{-2}}}
\end{equation}
of the matrix $B(p)$, given by (\ref{Bmatrix}) of Subsection \ref{DefLopRep}. 

Next we consider the Plancherel formula
for simple tensors $\widetilde{\varphi} = \xi \otimes \zeta \in E$, with $\xi \in \mathcal{S}^{0}(\mathbb{R}_+;\mathbb{C})$ 
and $\zeta \in \mathscr{C}^\infty(\mathbb{S}^2; \mathbb{C}^4) = \mathcal{S}_{{}_{\Delta_{\mathbb{S}^2}}}(\mathbb{S}^2; \mathbb{C}^4)$,
using the spherical coordinates $(r, \theta, \vartheta) = (r, \omega)$ on the cone, with the invariant measure 
\[
\begin{split}
\ud \mu_{{}_{\mathscr{O}_{1,0,0,1}}}(\boldsymbol{\p}) =  \ud^3 \boldsymbol{p}/|\boldsymbol{\p}| = r dr \, \ud \mu_{{}_{\mathbb{S}^2}} = rdr d\omega,
\\
r^2= |\boldsymbol{\p}|^2 = p_0(\boldsymbol{\p})^2 = \big(p_1\big)^2 + \big(p_2\big)^2  +\big(p_3\big)^2, 
\,\,\,\, \ud \mu_{{}_{\mathbb{S}^2}} = d\omega = \sin \theta \, d\theta d \vartheta,
\end{split}
\]
on the cone, compare Subsection \ref{DefLopRep}. 

Note that the orthonormal (at each point $p = (|\boldsymbol{\p}|, \boldsymbol{\p})$ of the cone $\mathscr{O}_{1,0,0,1}$)
eigenvectors (\ref{eigenvectorsOfB(p)}) are homogeneous of degree zero $\mathbb{R}^4$-valued functions on the cone, uniquely
determined by their values on the unit sphere $\mathbb{S}^2$ in the cone $\mathscr{O}_{1,0,0,1}$. We will regard (\ref{eigenvectorsOfB(p)}) as 
functions on the unit sphere $\mathbb{S}^2 \subset \mathscr{O}_{1,0,0,1}$. Let $Y_{lm}$ be the ordinary spherical functions
on $\mathbb{S}^2$ -- the complete system of eigenvectors of the Laplace operator $\Delta_{\mathbb{S}^2}$ on the sphere. 
Finally, consider the pair of Lie Abelian groups -- the additive group $\mathbb{R}$ of real numbers $\nu$ and the multiplicative group
$\mathbb{R}_+$ of strictly positive real numbers $r$ with, respectively, the ordinary additively invariant measure $d\nu$ on the additive group
$\mathbb{R}$ and with the multiplicatively invariant measure $r^{-1}dr$ on the multiplicative group $\mathbb{R}_+$. Recall that  
\[
r \longrightarrow r^{i\nu}, \,\,\,\, \nu \in \mathbb{R}, 
\,\,\,\,\,\,\,\,\,\,\,\,\,\,
\nu \longrightarrow r^{i\nu}, \,\,\,\, r \in \mathbb{R}_+,
\]
compose the complete systems of unitary characters, respectively, of $\mathbb{R}_+$ and of $\mathbb{R}$, 
which make the  Abelian groups $\mathbb{R}$ and $\mathbb{R}_+$ mutually dual in the Pontrjagin sense. 
In particular by the generalized Plancherel Theorem (compare e.g. \cite{gelfand-comm-norm-rings}, p. 154)
\begin{equation}\label{PlancherelForR+andR}
\int\limits_{\mathbb{R}_+} |\xi(r)|^2 \, r^{-1}dr = \int \limits_{\mathbb{R}} \Bigg| \int\limits_{\mathbb{R}_+} r^{i\nu} \xi(r)  \, r^{-1}dr \Bigg|^2 \, d\nu.
\end{equation}

We have the similar Plancherel formula for any complex valued square integrable function $f$ on the unit sphere
\begin{equation}\label{PlancherelForSphere}
\int\limits_{\mathbb{S}^2} |f(\omega)|^2 \, d\omega 
= \sum \limits_{l,m} \Bigg| \int\limits_{\mathbb{S}^2} \overline{Y_{lm}(\omega)} f(\omega)  \, d\omega \Bigg|^2
\end{equation}

Let us define the complete system (\ref{Fchi,l,m,s}) of homogeneous of degree $\chi = -1+i\nu$ functionals on the cone,
by the following corresponding homogeneous functions on the cone
\begin{align*}
F^{{}^{+}}_{{}_{\nu, lm}}(r,\omega)  & = r^{-1+i\nu}Y_{lm}(\omega) w^{{}^{+}}(\omega),
\\
F^{{}^{-}}_{{}_{\nu, lm}}(r,\omega)  & = r^{-1+i\nu}Y_{lm}(\omega)w^{{}^{-}}(\omega),
\\
F^{{}^{0+}}_{{}_{\nu, lm}}(r,\omega) & = r^{-1+i\nu}Y_{lm}(\omega) w_{{}_{r^{-2}}}(\omega),
\\
F^{{}^{0+}}_{{}_{\nu, lm}}(r,\omega) & = r^{-1+i\nu}Y_{lm}(\omega) w_{{}_{r^{2}}}(\omega),
\end{align*}
\[
l = 0, 1, 2, \ldots, \,\,\,\, -l \leq m \leq l, \,\,\,\, \chi = -1 + i\nu.
\]

Valuation $F^{{}^{+}}_{{}_{\nu, lm}}(\widetilde{\varphi})$
of the functional $F^{{}^{+}}_{{}_{\nu, lm}}$
on $\widetilde{\varphi}\in E$, or the invariant pairing
\[
\Big\langle F^{{}^{s}}_{{}_{\nu,lm}}, \widetilde{\varphi} \Big\rangle_{{}_{\textrm{pairing}}},
\]
of $F^{{}^{s}}_{{}_{\nu,lm}}$ and $\widetilde{\varphi}$ (we consider simple tensors $\widetilde{\varphi}(r,\omega) = \xi(r)\zeta(\omega)$)
is defined, as usual, through the invariant Krein inner product of the complex conjugation of the corresponding function $F^{{}^{+}}_{{}_{\nu, lm}}$
with the test function $\widetilde{\varphi} \in E$, \emph{i.e.} integral of the Krein product
\[
\big(\overline{F^{{}^{s}}_{{}_{\nu,lm}}(p)}, \mathfrak{J}_{{}_{\bar{p}}} \widetilde{\varphi}(p) \big)_{{}_{\mathbb{C}^4}}
= \big(\overline{F^{{}^{s}}_{{}_{\nu,lm}}(r,\omega)}, \mathfrak{J}_{{}_{\bar{p}}} \widetilde{\varphi}(r,\omega) \big)_{{}_{\mathbb{C}^4}}
\]
over the cone orbit $\mathscr{O}_{1,0,0,1}$ with respect to the invariant measure $\ud \mu_{{}_{\mathscr{O}_{1,0,0,1}}}(r,\omega) = rd\omega$
on the cone $\mathscr{O}_{1,0,0,1}$, \emph{i.e.} by the extension
of the ordinary rule for the Krein inner product in the single particle space $\mathcal{H}'$ over to the pairing of a functional
$F^{{}^{+}}_{{}_{\nu, lm}} \in E^*$ (represented by the ordinary function $F^{{}^{+}}_{{}_{\nu, lm}}$) with the
test function $\widetilde{\varphi}$. Of course, we have the analogue formula for the valuation
$F^{{}^{-}}_{{}_{\nu, lm}}(\widetilde{\varphi}), F^{{}^{0+}}_{{}_{\nu, lm}}(\widetilde{\varphi}),
F^{{}^{0-}}_{{}_{\nu, lm}}(\widetilde{\varphi})$ of any other functional
$F^{{}^{-}}_{{}_{\nu, lm}}, F^{{}^{0+}}_{{}_{\nu, lm}}, F^{{}^{0-}}_{{}_{\nu, lm}}$:
\[
\begin{split}
F^{{}^{+}}_{{}_{\nu,lm}}\big(\widetilde{\varphi}\big)
= \Big\langle F^{{}^{s}}_{{}_{\nu,lm}}, \widetilde{\varphi} \Big\rangle_{{}_{\textrm{pairing}}}
\,\,\,\,\,\,\,\,\,\,\,\,\,\,\,\,\,\,\,\,\,\,\,\,\,\,\,\,\,\,\,\,\,\,\,\,\,\,\,\,\,\,\,\,\,\,\,\,\,\,\,\,\,\,\,\,\,\,\,\,\,\,\,\,\,\,\,\,\,\,
\,\,\,\,\,\,\,\,\,\,\,\,\,\,\,\,\,\,\,\,\,\,\,\,\,\,\,\,\,\,\,\,\,\,\,\,\,\,\,\,\,\,\,\,\,\,\,\,\,\,\,\,\,\,\,\,\,\,\,\,\,\,\,\,\,\,\,\,\,\,
\\
= \int \limits_{\mathscr{O}_{1,0,0,1}} \big(\overline{F^{{}^{+}}_{{}_{\nu,lm}}(p)}, \mathfrak{J}_{{}_{\bar{p}}} \widetilde{\varphi}(p) \big)_{{}_{\mathbb{C}^4}}
\, \ud \mu_{{}_{\mathscr{O}_{1,0,0,1}}}(p)
=\int \limits_{\mathscr{O}_{1,0,0,1}} \big(\overline{(F^{{}^{+}}_{{}_{\nu,lm}}(p)}, \widetilde{\varphi}(p) \big)_{{}_{\mathbb{C}^4}}
\, \ud \mu_{{}_{\mathscr{O}_{1,0,0,1}}}(p)
\\
F^{{}^{-}}_{{}_{\nu,lm}}\big(\widetilde{\varphi}\big)
= \Big\langle F^{{}^{s}}_{{}_{\nu,lm}}, \widetilde{\varphi} \Big\rangle_{{}_{\textrm{pairing}}}
\,\,\,\,\,\,\,\,\,\,\,\,\,\,\,\,\,\,\,\,\,\,\,\,\,\,\,\,\,\,\,\,\,\,\,\,\,\,\,\,\,\,\,\,\,\,\,\,\,\,\,\,\,\,\,\,\,\,\,\,\,\,\,\,\,\,\,\,\,\,
\,\,\,\,\,\,\,\,\,\,\,\,\,\,\,\,\,\,\,\,\,\,\,\,\,\,\,\,\,\,\,\,\,\,\,\,\,\,\,\,\,\,\,\,\,\,\,\,\,\,\,\,\,\,\,\,\,\,\,\,\,\,\,\,\,\,\,\,\,\,
\\
= \int \limits_{\mathscr{O}_{1,0,0,1}} \big(\overline{F^{{}^{-}}_{{}_{\nu,lm}}(p)}, \mathfrak{J}_{{}_{\bar{p}}} \widetilde{\varphi}(p) \big)_{{}_{\mathbb{C}^4}}
\, \ud \mu_{{}_{\mathscr{O}_{1,0,0,1}}}(p)
=\int \limits_{\mathscr{O}_{1,0,0,1}} \big(\overline{F^{{}^{-}}_{{}_{\nu,lm}}(p)}, \widetilde{\varphi}(p) \big)_{{}_{\mathbb{C}^4}}
\, \ud \mu_{{}_{\mathscr{O}_{1,0,0,1}}}(p)
\\
F^{{}^{0+}}_{{}_{\nu,lm}}\big(\widetilde{\varphi}\big)
= \Big\langle F^{{}^{s}}_{{}_{\nu,lm}}, \widetilde{\varphi} \Big\rangle_{{}_{\textrm{pairing}}}
\,\,\,\,\,\,\,\,\,\,\,\,\,\,\,\,\,\,\,\,\,\,\,\,\,\,\,\,\,\,\,\,\,\,\,\,\,\,\,\,\,\,\,\,\,\,\,\,\,\,\,\,\,\,\,\,\,\,\,\,\,\,\,\,\,\,\,\,\,\,
\,\,\,\,\,\,\,\,\,\,\,\,\,\,\,\,\,\,\,\,\,\,\,\,\,\,\,\,\,\,\,\,\,\,\,\,\,\,\,\,\,\,\,\,\,\,\,\,\,\,\,\,\,\,\,\,\,\,\,\,\,\,\,\,\,\,\,\,\,\,
\\
= \int \limits_{\mathscr{O}_{1,0,0,1}} \big(\overline{(F^{{}^{0+}}_{{}_{\nu,lm}}(p)}, \mathfrak{J}_{{}_{\bar{p}}} \widetilde{\varphi}(p) \big)_{{}_{\mathbb{C}^4}}
\, \ud \mu_{{}_{\mathscr{O}_{1,0,0,1}}}(p)
=-\int \limits_{\mathscr{O}_{1,0,0,1}} \big(\overline{F^{{}^{0-}}_{{}_{\nu,lm}}(p)}, \widetilde{\varphi}(p) \big)_{{}_{\mathbb{C}^4}}
\, \ud \mu_{{}_{\mathscr{O}_{1,0,0,1}}}(p)
\\
F^{{}^{0-}}_{{}_{\nu,lm}}\big(\widetilde{\varphi}\big)
= \Big\langle F^{{}^{s}}_{{}_{\nu,lm}}, \widetilde{\varphi} \Big\rangle_{{}_{\textrm{pairing}}}
\,\,\,\,\,\,\,\,\,\,\,\,\,\,\,\,\,\,\,\,\,\,\,\,\,\,\,\,\,\,\,\,\,\,\,\,\,\,\,\,\,\,\,\,\,\,\,\,\,\,\,\,\,\,\,\,\,\,\,\,\,\,\,\,\,\,\,\,\,\,
\,\,\,\,\,\,\,\,\,\,\,\,\,\,\,\,\,\,\,\,\,\,\,\,\,\,\,\,\,\,\,\,\,\,\,\,\,\,\,\,\,\,\,\,\,\,\,\,\,\,\,\,\,\,\,\,\,\,\,\,\,\,\,\,\,\,\,\,\,\,
\\
= \int \limits_{\mathscr{O}_{1,0,0,1}} \big(\overline{F^{{}^{0-}}_{{}_{\nu,lm}}(p)}, \mathfrak{J}_{{}_{\bar{p}}} \widetilde{\varphi}(p) \big)_{{}_{\mathbb{C}^4}}
\, \ud \mu_{{}_{\mathscr{O}_{1,0,0,1}}}(p)
=-\int \limits_{\mathscr{O}_{1,0,0,1}} \big(\overline{F^{{}^{0+}}_{{}_{\nu,lm}}(p)}, \widetilde{\varphi}(p) \big)_{{}_{\mathbb{C}^4}}
\, \ud \mu_{{}_{\mathscr{O}_{1,0,0,1}}}(p).
\end{split}
\]
Here $( \,\cdot \, , \, \cdot \, )_{{}_{\mathbb{C}^4}}$ stands for the ordinary Hilbert space inner product in $\mathbb{C}^4$
and $\mathfrak{J}_{{}_{\bar{p}}}$ is the constant fundamental symmetry matrix (\ref{J-barp}) of Subsection \ref{DefLopRep}
associated to the {\L}opusza\'nski representation.
Therefore, the invariant valuations $F^{{}^{+}}_{{}_{\nu,lm}}(\widetilde{\varphi})$ and $F^{{}^{-}}_{{}_{\nu,lm}}(\widetilde{\varphi})$, or pairings,
coincide with the ordinary inner product (complex-conjugated in the first variable), the invariant valuation (pairing)
$F^{{}^{0+}}_{{}_{\nu,lm}}(\widetilde{\varphi})$ is given by the ordinary inner product of the complex conjugated
function $-F^{{}^{0-}}_{{}_{\nu,lm}}$ with $\widetilde{\varphi}$ and \emph{vice versa} for the invariant valuation
$F^{{}^{0-}}_{{}_{\nu,lm}}(\widetilde{\varphi})$.
Let us emphasize once again that the correct valuations should be invariant in order to make the distribution theory
easily applicable to the analysis of the associated representation of the corresponding group. In particular, only for invariant
valuations the rule for the transformation formula of the functionals (generalized states), given through the
dual formulas by the action on the normalizable states in $E$, the action
of the group in question does indeed coincide with the ordinary rule acting on functions whenever the functional (generalized state)
coincides with an ordinary function (as is the case for our functionals $F^{{}^{s}}_{{}_{\nu,lm}}$ regarded as elements of $E^*$).
Thus, working with invariant pairings simplifies matters, and allows to represent the corresponding invariant subspaces
of generalized states with simple local formulas for the representation acting as on the ordinary functions.

Note that the square $\| \widetilde{\varphi} \|^2$ of the single particle Hilbert space norm of $\widetilde{\varphi} \in E \subset \mathcal{H}'$
is equal
\begin{equation}\label{w-decompositionOfTheNormOfWidetildevarphi}
\| \widetilde{\varphi} \|^2 = 
\int \big|\widetilde{\varphi}(r,\omega)\big|_{{}_{\mathbb{C}^4}}^{2} \, r dr d\omega =
\| \widetilde{\varphi}_{{}_{+}} \|^2 + \| \widetilde{\varphi}_{{}_{-}} \|^2 +
 \| \widetilde{\varphi}_{{}_{0+}} \|^2 + \| \widetilde{\varphi}_{{}_{0+}} \|^2,
\end{equation}
where 
\[
\begin{split}
\| \widetilde{\varphi}_{{}_{+}} \|^2 = \int |\widetilde{\varphi}_{{}_{+}}(r,\omega)|^2 \, r dr d\omega, 
\,\,\,\,\,
\| \widetilde{\varphi}_{{}_{-}} \|^2 = \int |\widetilde{\varphi}_{{}_{-}}(r,\omega)|^2 \, r dr d\omega,
\\
\| \widetilde{\varphi}_{{}_{0+}} \|^2 = \int |\widetilde{\varphi}_{{}_{0+}}(r,\omega)|^2 \, r dr d\omega,
\,\,\,\,\,
\| \widetilde{\varphi}_{{}_{0-}} \|^2 = \int |\widetilde{\varphi}_{{}_{0-}}(r,\omega)|^2 \, r dr d\omega,
\end{split}
\]
and where
\[
\begin{split}
\widetilde{\varphi}_{{}_{+}}(r,\omega) = \Big( w^{{}^{+}}(\omega), \widetilde{\varphi}(r,\omega) \Big)_{{}_{\mathbb{C}^4}},
\,\,\,\,
\widetilde{\varphi}_{{}_{-}}(r,\omega) = \Big( w^{{}^{-}}(\omega), \widetilde{\varphi}(r,\omega) \Big)_{{}_{\mathbb{C}^4}},
\\
\widetilde{\varphi}_{{}_{0+}}(r,\omega) = \Big(w_{{}_{r^{-2}}}(\omega), \widetilde{\varphi}(r,\omega) \Big)_{{}_{\mathbb{C}^4}},
\,\,\,\,
\widetilde{\varphi}_{{}_{0+}}(r,\omega) = \Big(w_{{}_{r^{2}}}(\omega), \widetilde{\varphi}(r,\omega) \Big)_{{}_{\mathbb{C}^4}},
\end{split}
\]
are the projections of $\widetilde{\varphi}(r,\omega)$ on the respective orthonormal eigenvectors (\ref{eigenvectorsOfB(p)})
with respect to the Hilbert space inner product in $\mathbb{C}^4$.

\begin{prop*}
Let $\widetilde{\varphi} \in E \subset \mathcal{H}'$.
The following Plancherel formula holds
\begin{equation}\label{PlancherelForSingleHforA}
\|\widetilde{\varphi} \|^2 =  
\int \limits_{\mathscr{O}_{1,0,0,1}} \big|\widetilde{\varphi}(p)\big|_{{}_{\mathbb{C}^4}}^{2} \, \ud \mu_{{}_{\mathscr{O}_{1,0,0,1}}}(p) 
= \sum \limits_{l,m,s} \,\, \int\limits_{\mathbb{R}} \big| F^{{}^{s}}_{{}_{\nu,lm}}(\widetilde{\varphi}) \big|^2 \, d \nu  
\end{equation}
\[
l = 0, 1, 2,  \ldots, \,\,\, -l \leq m \leq l, \,\,\,\, s = +, -, 0+, 0-.
\]
The indicated generalized states 
$F^{{}^{+}}_{{}_{\nu, lm}}, F^{{}^{-}}_{{}_{\nu, lm}}, F^{{}^{0+}}_{{}_{\nu, lm}}, F^{{}^{0-}}_{{}_{\nu, lm}} \in E^*$
compose, for each fixed $\nu \in \mathbb{R}$, a complete orthonormal system in $\mathcal{H}'_{{}_{\chi}} = \mathcal{H}'_{{}_{-1+i\nu}}$.
\end{prop*}

\qedsymbol \,
It is sufficient to show this formula for the simple tensors $\widetilde{\varphi}(r,\omega) = \xi(r) \zeta(\omega)$, 
with $\xi \in \mathcal{S}^{0}(\mathbb{R}_+;\mathbb{C})$ 
and $\zeta \in \mathscr{C}^\infty(\mathbb{S}^2; \mathbb{C}^4) = \mathcal{S}_{{}_{\Delta_{\mathbb{S}^2}}}(\mathbb{S}^2; \mathbb{C}^4)$.
Moreover, it is sufficient to consider each of the four direct summands in (\ref{w-decompositionOfTheNormOfWidetildevarphi}) separately.
But the equalities 
\[
\begin{split}
\int \limits_{\mathscr{O}_{1,0,0,1}} \big|\widetilde{\varphi}_{{}_{+}}\big(p\big)\big|^2 \, \ud \mu_{{}_{\mathscr{O}_{1,0,0,1}}}(p) 
= \sum \limits_{l,m,s} \,\, \int\limits_{\mathbb{R}} \Big| F^{{}^{+}}_{{}_{\nu,lm}}\big(\mathfrak{J}'\widetilde{\varphi}\big) \Big|^2 \, d \nu 
= \sum \limits_{l,m,s} \,\, \int\limits_{\mathbb{R}} \Big| F^{{}^{+}}_{{}_{\nu,lm}}\big(\widetilde{\varphi}\big) \Big|^2 \, d \nu ,
\\
\int \limits_{\mathscr{O}_{1,0,0,1}} \big|\widetilde{\varphi}_{{}_{-}}\big(p\big)\big|^2 \, \ud \mu_{{}_{\mathscr{O}_{1,0,0,1}}}(p)
= \sum \limits_{l,m,s} \,\, \int\limits_{\mathbb{R}} \Big| F^{{}^{-}}_{{}_{\nu,lm}}\big(\mathfrak{J}'\widetilde{\varphi}\big) \Big|^2 \, d \nu
= \sum \limits_{l,m,s} \,\, \int\limits_{\mathbb{R}} \Big| F^{{}^{-}}_{{}_{\nu,lm}}\big(\widetilde{\varphi}\big) \Big|^2 \, d \nu,
\\
\int \limits_{\mathscr{O}_{1,0,0,1}} \big|\widetilde{\varphi}_{{}_{0+}}\big(p\big)\big|^2 \, \ud \mu_{{}_{\mathscr{O}_{1,0,0,1}}}(p) 
= \sum \limits_{l,m,s} \,\, \int\limits_{\mathbb{R}} \Big| F^{{}^{0+}}_{{}_{\nu,lm}}\big(\mathfrak{J}'\widetilde{\varphi}\big) \Big|^2 \, d \nu 
= \sum \limits_{l,m,s} \,\, \int\limits_{\mathbb{R}} \Big| F^{{}^{0-}}_{{}_{\nu,lm}}\big(\widetilde{\varphi}\big) \Big|^2 \, d \nu ,
\\
\int \limits_{\mathscr{O}_{1,0,0,1}} \big|\widetilde{\varphi}_{{}_{0-}}\big(p\big)\big|^2 \, \ud \mu_{{}_{\mathscr{O}_{1,0,0,1}}}(p) 
= \sum \limits_{l,m,s} \,\, \int\limits_{\mathbb{R}} \Big| F^{{}^{0-}}_{{}_{\nu,lm}}\big(\mathfrak{J}' \widetilde{\varphi}\big) \Big|^2 \, d \nu
= \sum \limits_{l,m,s} \,\, \int\limits_{\mathbb{R}} \Big| F^{{}^{0+}}_{{}_{\nu,lm}}\big(\widetilde{\varphi}\big) \Big|^2 \, d \nu ,
\end{split}
\]
for simple tensors $\widetilde{\varphi}(r,\omega) = \xi(r) \zeta(\omega)$, written in spherical coordinates,
factorize, and they follow easily from the Plancherel formulas (\ref{PlancherelForR+andR}) and (\ref{PlancherelForSphere}).

Note that to the component $\|\widetilde{\varphi}_{{}_{0+}}\|^2$ on the left-hand side correspond the integral of the squared absolute value 
$\big|F^{{}^{0-}}_{{}_{\nu,lm}}\big(\widetilde{\varphi}\big)\big|^2$ of the valuation
$F^{{}^{0-}}_{{}_{\nu,lm}}\big(\widetilde{\varphi}\big)$, and \emph{vice versa} for $\|\widetilde{\varphi}_{{}_{0-}}\|^2$,
due to the invariance of the valuation involving the Krein involutive symmetry $\mathfrak{J}'$. 
\qed

Having given the Hilbert space $\mathcal{H}'_{{}_{\chi}} =\mathcal{H}'_{{}_{-1+i\nu}} \subset E^*$ of functionals (generalized states) of the normal operator
$\widetilde{S_\lambda}$, corresponding to the eigenvalue $\lambda^{{}^{\chi}} = \lambda^{{}^{-1 +i\nu}}$, and the above stated Plancherel
formula, we can construct, after \cite{GelfandIV}, Chap. I.4, in explicit form decomposition
\[
\widetilde{\varphi} = \int\limits_{\textrm{Sp} \, \widetilde{S_\lambda}} \widetilde{\varphi}_{{}_{\nu}} \, \ud \nu,
\]
of
\[
\widetilde{\varphi} \in E \subset \mathcal{H}'= \int\limits_{\textrm{Sp} \, \widetilde{S_\lambda}} \mathcal{H}'_{{}_{-1+i\nu}} \, \ud \nu,
\]
corresponding to the spectral decomposition of $\widetilde{S_\lambda}$.
Each component $\widetilde{\varphi}_{{}_{\nu}}$ of this decomposition is equal to $T_{{}_{\nu}} (\widetilde{\varphi})$, where
\begin{equation}\label{(Tnu(varphi), Fk)nu}
T_{{}_{\nu}}(\widetilde{\varphi}) = \sum \limits_{k} \lambda_{k}^{-n}F'_{{}_{k}}(\widetilde{\varphi})h_{{}_{k \nu}}
= \sum \limits_{k} \lambda_{k}^{-n} \lambda_{k}^{n} F_{{}_{k}}(\widetilde{\varphi})h_{{}_{k \nu}}
= \sum \limits_{k} F_{{}_{k}}(\widetilde{\varphi})h_{{}_{k \nu}}
\end{equation}
with the nuclear operator $T_{{}_{\nu}}$ mapping $E$ into $\mathcal{H}'_{{}_{-1+i\nu}}$ of Theorem 1' of \cite{GelfandIV},
Chap. I.4.4, p. 117, associated to the above decomposition of the rigged Hilbert space $E \subset \mathcal{H}' \subset E^*$.
Here
\[
\begin{split}
F'_{{}_{k}} = \lambda_{k}^{n}F_{{}_{k}} \in E_{n}^{*} = E_{{}_{-n}} \subset E^*, \,\,\,\, F_{{}_{k}} \in E_{0}^* \cong \mathcal{H}' \cong E_0,
\\
E_n = \overline{\textrm{Dom} \, A^n} \,\,\, \textrm{with} \,\,\,
(\cdot,\cdot)_{{}_{n}} = \big({A}^n \, \cdot \,\, , {A}^n \, \cdot \, \big)_{{}_{\mathcal{H}'}},
\end{split}
\]
\[
k=0,1,2, \ldots
\]
being equal to orthonormal complete systems, respectively, in $E_{n}^{*}$ and in $E_{0}^* \cong \mathcal{H}'$,
with $\lambda_{k}$ being the spectral values of the standard operator $A = \oplus_{0}^{3} A^{(3)}$ 
(with the operator $A^{(3)}$ constructed in Subsection \ref{dim=n}) 
defining the standard nuclear space
\[
E = \cap_{{}_{n}} E_{{}_{n}}.
\]
$h_{{}_{k}}$, $k=1,2, \ldots$, is the orthonormal system in $\mathcal{H}'$ dual to $F_{{}_{k}}$, $k=1,2, \ldots$, in 
$E_{0}^* = \mathcal{H}'^{*} \cong \mathcal{H}'$.
$h_{{}_{k \nu}}$ are decomposition components of $h_{{}_{k}}$ which, for each fixed $\nu$,
provide a complete orthonormal system in $\mathcal{H}'_{{}_{-1+i\nu}}$.

Thus, from (\ref{(Tnu(varphi), Fk)nu}) we get
\[
\big(\overline{T_{{}_{\nu}}(\widetilde{\varphi})}, h_{{}_{k}})_{{}_{-1+i\nu}} = F_{{}_{k}}(\widetilde{\varphi}),
\]
where the inner product $(\, \cdot \, , \, \cdot \, )_{{}_{-1+i\nu}}$ is taken in ${\mathcal{H}'}_{{}_{-1+i\nu}}$.

Each component $\widetilde{\varphi}_{{}_{\nu}}$
of this decomposition is understood as a functional on $\mathcal{H}'_{{}_{\chi}} =\mathcal{H}'_{{}_{-1+i\nu}} \subset E^*$.
$\mathcal{H}'_{{}_{-1+i\nu}} \cong {\mathcal{H}'}_{{}_{-1+i\nu}}^*$ is understood as the space dual to the Hilbert space of functionals
$\mathcal{H}'_{{}_{-1+i\nu}}$ endowed with the Hilbert space inner product in which the functionals $F^{{}^{s}}_{{}_{\nu, lm}}$
compose a complete orthonormal 
system\footnote{Provided there are used invariant pairings $F(\widetilde{\varphi})$ in
Theorem 1' of \cite{GelfandIV}, p. 117, and thus $T_{{}_{\nu}}$
commuting with the representation of $SL(2, \mathbb{C})$, respectively, equal to the restriction of the {\L}opusza\'nski representation to the subgroup
$SL(2,\mathbb{C})$ in $E \subset \mathcal{H}'$ and acting in the image $\mathcal{H}'_{{}_{-1+i\nu}}$ of $T_{{}_{\nu}}$ as on the ordinary four-vector functions. In case the pairings in Theorem 1' of \cite{GelfandIV}, p. 117 are not invariant and equal to the extension of the ordinary inner product (with the first variable complex conjugate) the formula (\ref{GelfandeDecompositionFunction}) will have to be changed into:
\[
\widetilde{\varphi}_{{}_{\nu}}: \, \mathcal{H}'_{{}_{-1+i\nu}} \ni
F_{{}_{\nu}} \longmapsto \big(\overline{\widetilde{\varphi}_{{}_{\nu}}}, F_{{}_{\nu}} \big)_{{}_{-1+i\nu}}
= \big(\overline{T_{{}_{\nu}} \widetilde{\varphi}}, F_{{}_{\nu}} \big)_{{}_{-1+i\nu}}
= F_{{}_{\nu}}\big(\mathfrak{J}' \widetilde{\varphi} \big).
\]
}
\begin{equation}\label{GelfandeDecompositionFunction}
\widetilde{\varphi}_{{}_{\nu}}: \, \mathcal{H}'_{{}_{-1+i\nu}} \ni
F_{{}_{\nu}} \longmapsto \big(\overline{\widetilde{\varphi}_{{}_{\nu}}}, F_{{}_{\nu}} \big)_{{}_{-1+i\nu}}
= \big(\overline{T_{{}_{\nu}} \widetilde{\varphi}}, F_{{}_{\nu}} \big)_{{}_{-1+i\nu}}
= F_{{}_{\nu}}\big(\widetilde{\varphi} \big),
\end{equation}
where the inner product $(\, \cdot \, , \, \cdot \, )_{{}_{-1+i\nu}}$ is taken in ${\mathcal{H}'}_{{}_{-1+i\nu}}^{*}
 \cong \mathcal{H}'_{{}_{-1+i\nu}}$. 
In fact (\ref{GelfandeDecompositionFunction}), \emph{i.e.} existence of $\widetilde{\varphi}_{{}_{\nu}} \in \mathcal{H}'_{{}_{-1+i\nu}}$ is a consequence
of the Riesz representation theorem.
Recall that the elements of ${\mathcal{H}'}_{{}_{-1+i\nu}}^{*}
\cong \mathcal{H}'_{{}_{-1+i\nu}}$ are represented by complex valued functions on the cone and the complex conjugation
$\overline{(\, \cdot \,)}$ is well-defined on ${\mathcal{H}'}_{{}_{-1+i\nu}}^{*}
\cong \mathcal{H}'_{{}_{-1+i\nu}}$.
Because
\[
\big\| \widetilde{\varphi}_{{}_{\nu}} \big\|_{{}_{-1+i\nu}}^{2} = \sum \limits_{l,m,s} \big| \big(\overline{\widetilde{\varphi}_{{}_{\nu}}}, F^{{}^{s}}_{{}_{\nu, lm}} \big)_{{}_{-1+i\nu}} \big|^2
= \sum \limits_{l,m,s} \big| F^{{}^{s}}_{{}_{\nu, lm}}\big(\widetilde{\varphi} \big) \big|^2,
\]
as the functionals $F^{{}^{s}}_{{}_{\nu, lm}}$
compose a complete orthonormal system in ${\mathcal{H}'}_{{}_{-1+i\nu}}^{*} \cong \mathcal{H}'_{{}_{-1+i\nu}}$, we can rewrite the Plancherel
formula in terms of decomposition components $\widetilde{\varphi}_{{}_{\nu}} \in \mathcal{H}'_{{}_{-1+i\nu}}$ 
in the following form, and also call \emph{Plancherel formula}:
\[
\boxed{
\|\widetilde{\varphi} \|^2 = \int\limits_{\mathbb{R}} \big\| \widetilde{\varphi}_{{}_{\nu}} \big\|_{{}_{-1+i\nu}}^{2}
\, d\nu,
}
\]
where
\[
\|\widetilde{\varphi} \|^2
= \int \limits_{\mathscr{O}_{1,0,0,1}} \big|\widetilde{\varphi}(p)\big|_{{}_{\mathbb{C}^4}}^{2} \, \ud \mu_{{}_{\mathscr{O}_{1,0,0,1}}}(p).
\]

The formulas for the decomposition components $\mathfrak{J}'_{{}_{\chi=-1+i\nu}}$, written $  \mathfrak{J}'_{{}_{\nu}}$ with the subscript $\nu$,
of the fundamental symmetry $\mathfrak{J}' = {\mathfrak{J}'}^{-1} $ are easily seen to be equal
\[
\mathfrak{J}'_{{}_{\nu}}F^{{}^{+}}_{{}_{\nu,lm}} = F^{{}^{+}}_{{}_{\nu,lm}},
\,\,\,\,
\mathfrak{J}'_{{}_{\nu}}F^{{}^{-}}_{{}_{\nu,lm}} = F^{{}^{-}}_{{}_{\nu,lm}},
\,\,\,\,
\mathfrak{J}'_{{}_{\nu}}F^{{}^{0+}}_{{}_{\nu,lm}} = -F^{{}^{0-}}_{{}_{\nu,lm}},
\,\,\,\,
\mathfrak{J}'_{{}_{\nu}}F^{{}^{0-}}_{{}_{\nu,lm}} = -F^{{}^{0+}}_{{}_{\nu,lm}},
\]
by the very definition
\[
\begin{split}
\mathfrak{J}'_{{}_{\nu}}F^{{}^{+}}_{{}_{\nu,lm}}\big(\widetilde{\varphi}\big) 
= F^{{}^{+}}_{{}_{\nu,lm}}\big({\mathfrak{J}'}^{-1}\widetilde{\varphi}\big)
= F^{{}^{+}}_{{}_{\nu,lm}}\big(\mathfrak{J}'\widetilde{\varphi}\big)
=F^{{}^{+}}_{{}_{\nu,lm}}\big(\widetilde{\varphi}\big),
\\
 \mathfrak{J}'_{{}_{\nu}}F^{{}^{-}}_{{}_{\nu,lm}}\big(\widetilde{\varphi}\big) 
= F^{{}^{-}}_{{}_{\nu,lm}}\big({\mathfrak{J}'}^{-1}\widetilde{\varphi}\big)
= F^{{}^{-}}_{{}_{\nu,lm}}\big(\mathfrak{J}'\widetilde{\varphi}\big)
= F^{{}^{-}}_{{}_{\nu,lm}}\big(\widetilde{\varphi}\big),
\\
\mathfrak{J}'_{{}_{\nu}}F^{{}^{0+}}_{{}_{\nu,lm}}\big(\widetilde{\varphi}\big) 
= F^{{}^{0+}}_{{}_{\nu,lm}}\big({\mathfrak{J}'}^{-1}\widetilde{\varphi}\big) 
= F^{{}^{0+}}_{{}_{\nu,lm}}\big(\mathfrak{J}'\widetilde{\varphi}\big) 
=  - F^{{}^{0-}}_{{}_{\nu,lm}}\big(\widetilde{\varphi}\big),
\\
\mathfrak{J}'_{{}_{\nu}}F^{{}^{0-}}_{{}_{\nu,lm}}\big(\widetilde{\varphi}\big) 
= F^{{}^{0-}}_{{}_{\nu,lm}}\big({\mathfrak{J}'}^{-1}\widetilde{\varphi}\big)
= F^{{}^{0-}}_{{}_{\nu,lm}}\big(\mathfrak{J}'\widetilde{\varphi}\big)
= - F^{{}^{0+}}_{{}_{\nu,lm}}\big(\widetilde{\varphi}\big).
\end{split}
\]

From the stated above Plancherel formula we can easily read-off the
(non-invariant, because {\L}opusza\'nski representation is not unitary) inner product formula 
\[
\big( \, \cdot \, , \, \cdot \, \big)_{{}_{-1+i\nu}}
\,\,\,\, \textrm{in} \,\,\,\,
\mathcal{H}'_{{}_{-1+i\nu}}.
\]
Namely 
\[
\Big( F^{{}^{s}}_{{}_{\nu,lm}} , F^{{}^{s'}}_{{}_{\nu,l'm'}} \Big)_{{}_{-1+i\nu}} = \delta^{{}^{ss'}}\delta_{{}_{ll'}}\delta_{{}_{mm'}}.
\]
For the invariant Krein inner product
\[
\Big( \, \cdot \, , \mathfrak{J}'_{{}_{\nu}} \, \cdot \, \Big)_{{}_{-1+i\nu}}
\,\,\,\, \textrm{in} \,\,\,\,
\mathcal{H}'_{{}_{-1+i\nu}}
\]
we have the analogous formula immediately following from the formulas for the inner product 
$(\, \cdot \, , \, \cdot \, )_{-1+i\nu}$ 
and the Krein involutive symmetry
operator $\mathfrak{J}'_{{}_{\nu}}$ in $\mathcal{H}'_{{}_{-1+i\nu}}$.

Similarly, we have for the decomposition components $U_{{}_{\nu}}$  of the {\L}opusza\'nski representation
$U$, given by the formula (\ref{Lop-rep--on-tildevarphi}) of Subsection \ref{DefLopRep}, the following formula:
\[
U_{{}_{\nu}}F^{{}^{s}}_{{}_{\nu,lm}}\big(\widetilde{\varphi}\big) 
= F^{{}^{s}}_{{}_{\nu,lm}}\big(U^{-1}\widetilde{\varphi}\big).
\]
By invariance of the pairing it is easily seen that $U_{{}_{\nu}}$ act by the same formula on the functions $F^{{}^{s}}_{{}_{\nu,lm}}$, representing
the corresponding functionals $F^{{}^{s}}_{{}_{\nu,lm}}$, as $U$ on the functions $\widetilde{\varphi}$
in (\ref{Lop-rep--on-tildevarphi}), Subsection \ref{DefLopRep}. Decomposition components of the representation 
$\mathfrak{J}'U\mathfrak{J}' = U^{* \, -1}$, given by (\ref{conjugated-Lop-rep}), which is conjugated to the {\L}opusza\'nski representation
$U$, are equal $\mathfrak{J}'_{{}_{\nu}}U_{{}_{\nu}}\mathfrak{J}'_{{}_{\nu}}$, and act on the functions 
$F^{{}^{s}}_{{}_{\nu,lm}}$, representing the corresponding functionals $F^{{}^{s}}_{{}_{\nu,lm}}$, exactly as the representation 
$\mathfrak{J}'U\mathfrak{J}' = U^{* \, -1}$ on $\widetilde{\varphi}$ in  (\ref{conjugated-Lop-rep}).

Note also, that the Lorentz invariant generalized subspace $\mathcal{H}'_{{}_{-1+i\nu}}$, has the closed Lorentz invariant subspace
$\mathcal{H}'_{{}_{-1+i\nu \,\, \textrm{tr}}}$ consisting of transversal generalized homogeneous states,  spanned
by the functionals $F^{{}^{+}}_{{}_{\nu,lm}}, F^{{}^{-}}_{{}_{\nu,lm}}, F^{{}^{0+}}_{{}_{\nu,lm}}$, $l=0,1,2, \ldots$, $-l \leq m \leq l$,
although it has no complementary and Lorentz invariant subspace in $\mathcal{H}'_{{}_{-1+i\nu}}$. 
This follows from the transformation formulas of Subsection \ref{SingleKreinLopRep} for the eigenvectors (\ref{eigen-vectors-matrixB})
of Subsection \ref{DefLopRep}. For a unitary representation such a situation would be impossible.
This is possible for the {\L}opusza\'nski representation because
it is Krein isometric but not unitary.

As we have already explained, each element $F \in E^* = \mathcal{S}^{0}(\mathbb{R}^3; \mathbb{C}^4)^*$ defines
uniquely a functional $S \in \mathcal{S}^{0}(\mathbb{R}^4; \mathbb{C}^4)^*$ concentrated on the 
cone $\mathscr{O}_{1,0,0,1}$ by the rule
\[
S(\widetilde{\varphi}) =  F\big(\widetilde{\varphi}\big|_{{}_{\mathscr{O}_{1,0,0,1}}}\big), 
\,\,\,\,\, \widetilde{\varphi} \in \mathcal{S}^{0}(\mathbb{R}^4; \mathbb{C}^4),
\]
well-defined by the continuity of the map
\[
\mathcal{S}^{0}(\mathbb{R}^4; \mathbb{C}^4) \ni \widetilde{\varphi} \longrightarrow \widetilde{\varphi}\big|_{{}_{\mathscr{O}_{1,0,0,1}}}
\in \mathcal{S}^{0}(\mathbb{R}^3; \mathbb{C}^4),
\]
proved in Subsection \ref{Lop-on-E}. Inverse Fourier transforms of the functionals $S$ (associated to regular $F \in \mathcal{H}' \subset E^*$)
represent the single particle states as the solutions of the d'Alembert equation in the ``position picture'' or ``space-time picture''. 
Note that the inverse Fourier transforms $S_\lambda, S_{\lambda}^{*}$ of the scaling operator $\widetilde{S_\lambda}$ 
and of its adjoint $\widetilde{S_\lambda}^*$ (acting in the single particle space
of solutions $\varphi$ of d'Alembert equation in the ``space-time picture'') are given by
\[
S_\lambda \varphi(x) = \lambda^{-2} \varphi(\lambda^{-1}x), \,\,\,\, S_{\lambda}^{*} \varphi(x) = \varphi(\lambda x)
\]
From this it follows that indeed the Hilbert space $\mathcal{H}'_{{}_{\chi}} = \mathcal{H}'_{{}_{-1+i\nu}}$ of generalized eigenstates 
of $S_\lambda$, viewed as elements of $\mathcal{S}^{00}(\mathbb{R}^4; \mathbb{C}^4)^*$, are homogeneous of degree $\chi = -1 + i\nu$
solutions of d'Alembert equation.

Using the formula (\ref{q-A'-B}) for the electromagnetic potential field operator, regarded as 
the sum of integral kernel operators 
\[
A = \Xi_{0,1}(\kappa_{0,1}) + \Xi_{1,0}(\kappa_{1,0})
\]
with vector-valued distributional plane wave kernels 
\[
\kappa_{0,1}, \kappa_{1,0} \in \mathscr{L}\big( \mathcal{S}_{A}(\mathbb{R}^3, \mathbb{C}^4), \,\,
\mathscr{E}^*  \big) \cong \mathcal{S}_{A}(\mathbb{R}^3, \mathbb{C}^4)^* \otimes \mathscr{E}^*
= E^* \otimes \mathscr{E}^*,
\]
we will have the following formula for the plane wave kernels:
\begin{equation}\label{kappa_0,1kappa_1,0A'}
\boxed{
\begin{split}
\kappa_{0,1}(\nu, \boldsymbol{\p}; \mu, x) =
\frac{g_{\nu \mu}}{\sqrt{2 p^0(\boldsymbol{\p})}}
e^{-ip\cdot x}, \,\,\,\,\,\,
p \in \mathscr{O}_{1,0,0,1}, \\
\kappa_{1,0}(\nu, \boldsymbol{\p}; \mu, x) = 
\frac{-g_{\nu \mu}}{\sqrt{2 p^0(\boldsymbol{\p})}}
e^{ip\cdot x},
\,\,\,\,\,\,
p \in \mathscr{O}_{1,0,0,1},
\end{split}
}
\end{equation}
defining the distributions $\kappa_{0,1}, \kappa_{1,0}$
instead of (\ref{kappa_0,1kappa_1,0A}), if the Krein adjont Hida annihilation operators are used
in the formula for $A$.
If the the adjoint of the  Hida annihilation operators are used
in the formula for $A$, then
\[
\boxed{
\begin{split}
\kappa_{0,1}(\nu, \boldsymbol{\p}; \mu, x) =
\frac{\delta_{\nu \mu}}{\sqrt{2 p^0(\boldsymbol{\p})}}
e^{-ip\cdot x}, \,\,\,\,\,\,
p \in \mathscr{O}_{1,0,0,1}, \\
\kappa_{1,0}(\nu, \boldsymbol{\p}; \mu, x) = 
\frac{-g_{\nu \mu}}{\sqrt{2 p^0(\boldsymbol{\p})}}
e^{ip\cdot x},
\,\,\,\,\,\,
p \in \mathscr{O}_{1,0,0,1}.
\end{split}
}
\]
Here $g_{\nu\mu}$ is the Minkowski space-time metric. 
Proof that they can be (uniquely) extended
to elements
\[
\kappa_{0,1}, \kappa_{1,0} \in \mathscr{L}\big( \mathcal{S}_{A}(\mathbb{R}^3, \mathbb{C}^4)^*, \,\,
\mathscr{E}^*  \big) \cong \mathcal{S}_{A}(\mathbb{R}^3, \mathbb{C}^4) \otimes \mathscr{E}^*,
\] 
remains the same as for the kernels (\ref{kappa_0,1kappa_1,0A}) in Lemma \ref{kappa0,1,kappa1,0ForA}, 
Subsection \ref{A=Xi0,1+Xi1,0}.
Thus, by Thm. 3.13 of \cite{obataJFA} (or Thm. \ref{obataJFA.Thm.3.13} 
of Subsection \ref{psiBerezin-Hida}) we obtain the corollary that
\[
A = \Xi_{0,1}(\kappa_{0,1}) + \Xi_{1,0}(\kappa_{1,0})
\in
\mathscr{L}\big( (E) \otimes \mathscr{E}, \, (E) \big) \cong
\mathscr{L}\Big( \mathscr{E}, \,\, \mathscr{L}\big( (E), (E)\big) \, \Big),
\]
with $\kappa_{0,1}, \kappa_{1,0}$ defined by (\ref{kappa_0,1kappa_1,0A'}).
Thus, the field $A = \Xi_{0,1}(\kappa_{0,1}) + \Xi_{1,0}(\kappa_{1,0})$, understood as integral kernel operator
defines an operator-valued distribution through the continuous map
\[
\mathscr{E} \ni \varphi \longmapsto
A(\varphi) =
\Xi_{0,1}\big(\kappa_{0,1}(\varphi)\big) + \Xi_{1,0}\big(\kappa_{1,0}(\varphi)\big)
\in \mathscr{L}\big( (E), (E)\big),
\]
\[
\mathscr{E} = S^{00}(\mathbb{R}^4; \mathbb{C}^4).
\]

Using the Hida operators $a', a'^+$ associated to the free four-potential field $A$ in the second realization 2)
and the Gupta-Bleuler operator $\eta$ we can write 
\[
A(\varphi) = a'\big(\overline{\check{\widetilde{\varphi}}|_{{}_{\mathscr{O}_{1,0,0,1}}}}\big)
+
\eta a'\big(\widetilde{\varphi}|_{{}_{\mathscr{O}_{1,0,0,1}}}\big)^{+} \eta,
\]
\[
\varphi \in \mathscr{E} = S^{00}(\mathbb{R}^4; \mathbb{C}^4).
\]
Note please, that the restriction $\widetilde{\varphi}|_{{}_{\mathscr{O}_{1,0,0,1}}} \in E$ to the cone $\mathscr{O}_{1,0,0,1}$
of the Fourier transform of an element $\varphi \in \mathscr{E} = S^{00}(\mathbb{R}^4; \mathbb{C}^4)$, should be identified 
with $\widetilde{\varphi} \in E$ of the last two Propositions. Therefore, the direct integral decomposition
of the single particle Hilbert-Krein space $\mathcal{H}'$, described in the last two Propositions, allows to write
the following direct integral decomposition of the field operator
\begin{multline}\label{potentialFieldA-decomposition}
A(\varphi) = \int \limits_{\textrm{Sp} \, S_\lambda} A(\varphi)_{{}_{\chi}} \ud \chi
= \int \limits_{\textrm{Sp} \, S_\lambda} \Bigg\{ 
a'_{{}_{\chi}}\big(\overline{\check{\widetilde{\varphi}}\big|_{{}_{\mathscr{O}_{1,0,0,1} \, \chi}}}\big)
\, + \,\,
\eta_{{}_{\chi}} a'_{{}_{\chi}}\big(\widetilde{\varphi}\big|_{{}_{\mathscr{O}_{1,0,0,1} \, \chi}}\big)^{+} \eta_{{}_{\chi}}
\Bigg\} \, \ud \chi.
\end{multline}
Here 
\[
\begin{split}
\widetilde{\varphi}|_{{}_{\mathscr{O}_{1,0,0,1}}} 
= \int \limits_{\textrm{Sp} \, S_\lambda} \widetilde{\varphi}\big|_{{}_{\mathscr{O}_{1,0,0,1} \, \chi}} \ud \chi,
\\
\eta =  \int \limits_{\textrm{Sp} \, S_\lambda} \eta_{{}_{\chi}} \ud \chi 
= \int \limits_{\textrm{Sp} \, S_\lambda} \Gamma\big(\mathfrak{J}_{{}_{\chi}}\big) \, \ud \chi,
\end{split}
\]
are the direct integral decompositions of the elements $\widetilde{\varphi}\big|_{{}_{\mathscr{O}_{1,0,0,1}}} \in E \subset \mathcal{H}'$
and of $\eta = \Gamma(\mathfrak{J})$, constructed in the last two Propositions. To each direct integral summand $\mathcal{H}'_{{}_{\chi}}
= \mathcal{H}'_{{}_{-1+i\nu}}$ we associate the corresponding Hida annihilation-creation operators
$a'_{{}_{\chi}}(\cdot), a'_{{}_{\chi}}(\cdot)^+$
and construct the Fock space over each direct integral summand $\mathcal{H}'_{{}_{\chi}}$ and the associated
Gelfand tripe 
\[
E_{{}_{\chi}} \subset \mathcal{H}'_{{}_{\chi}} \subset E_{{}_{\chi}}^*,
\]
where each element of $E_{{}_{\chi}}$ is a smooth $\mathbb{C}^4$-valued function on the cone 
of homogeneity degree $\chi$ (with $\lambda^{{}^{\chi}} \in \textrm{Sp} \, S_\lambda$) 
identified with its restriction to the unit two-sphere on the cone, e.g. with a smooth $\mathbb{C}^4$-valued function on the two-sphere:
\[
E_{{}_{\chi}} = \mathscr{C}^\infty(\mathbb{S}^2; \mathbb{C}^4) = \mathcal{S}_{{}_{\Delta_{\mathbb{S}^2}}}(\mathbb{S}^2; \mathbb{C}^4).
\]
Accordingly, the Hilbert space $\mathcal{H}'_{{}_{\chi}}$ is identified with the square summable $\mathbb{C}^4$-valued functions on the two-sphere in the cone:
\[
\mathcal{H}'_{{}_{\chi}} = L^2(\mathbb{S}^2; \mathbb{C}^4),
\]
which, by homogeneity, are uniquely extended on the whole cone. Similarly, the elements of $E_{{}_{\chi}}^*$ become identifiable
with the space of distributions on the two-sphere:
\[
E_{{}_{\chi}}^* = \mathscr{C}^\infty(\mathbb{S}^2; \mathbb{C}^4)^* = \mathcal{S}_{{}_{\Delta_{\mathbb{S}^2}}}(\mathbb{S}^2; \mathbb{C}^4)^*.
\]

We want to rewrite the decomposition (\ref{potentialFieldA-decomposition}) in the form
\begin{multline*}
A(\varphi) = \int \limits_{\textrm{Sp} \, S_\lambda} A_{{}_{\chi}}(\varphi_{{}_{\chi}}) \ud \chi
= \int \limits_{\textrm{Sp} \, S_\lambda} \Bigg\{ 
a'_{{}_{\chi}}\big(\overline{\check{\widetilde{\varphi}}\big|_{{}_{\mathscr{O}_{1,0,0,1} \, \chi}}}\big)
\, + \,\,
\eta_{{}_{\chi}} a'_{{}_{\chi}}\big(\widetilde{\varphi}\big|_{{}_{\mathscr{O}_{1,0,0,1} \, \chi}}\big)^{+} \eta_{{}_{\chi}}
\Bigg\} \, \ud \chi,
\end{multline*}
as a decomposition into actual fields $A_{{}_{\chi}}$ over the single particle Hilbert spaces $\mathcal{H}'_{{}_{\chi}}$ --
\emph{the homogeneous parts of degree $\chi$ of the free electromagnetic four-potential field} $A$. 
This can be possible if, instead of general $\varphi \in \mathcal{S}^{00}(\mathbb{R}^4; \mathbb{C}^4)$
we will use the elements $\varphi^*$ of $\mathcal{S}^{00}(\mathbb{R}^4; \mathbb{C}^4)^*$ whose Fourier transforms are
concentrated on the positive energy cone $\mathscr{O}_{1,0,0,1}$ and coincide there with
$\widetilde{\varphi}\big|_{{}_{\mathscr{O}_{1,0,0,1}}}$ for the corresponding $\varphi \in \mathcal{S}^{00}(\mathbb{R}^4; \mathbb{C}^4)$. 
Such $\varphi^*$ are identifiable with the elements
of $E \subset \mathcal{H}'$ and, as such, undergo the decomposition
\[
 \widetilde{\varphi}^{*}
= \int \limits_{\textrm{Sp} \, S_\lambda} \widetilde{\varphi}^{*}_{{}_{\chi}} \, \ud \chi.
\]   
Therefore we can write the following direct integral decomposition of the free potential field $A$ into actual
homogeneous of degree $\chi = -1 +i\nu$, $\nu \in \mathbb{R}$, parts $A_{{}_{\chi}}$, which are actually fields $A_{{}_{\chi}}$
on the Fock spaces $\Gamma\big( \mathcal{H}'_{{}_{\chi}}\big)$ over the Hilbert spaces $\mathcal{H}'_{{}_{\chi}}$ 
of generalized homogeneous of degree $\chi = -1 +i\nu$ states:
\begin{multline}\label{potentialFieldA-decompositionProper}
A(\varphi) = \int \limits_{\textrm{Sp} \, S_\lambda} A_{{}_{\chi}}(\varphi^{*}_{{}_{\chi}}) \ud \chi
= \int \limits_{\textrm{Sp} \, S_\lambda} \Bigg\{ 
a'_{{}_{\chi}}\big(\overline{\check{\widetilde{\varphi^{*}}}\big|_{{}_{\mathscr{O}_{1,0,0,1} \, \chi}}}\big)
\, + \,\,
\eta_{{}_{\chi}} a'_{{}_{\chi}}\big(\widetilde{\varphi^{*}}\big|_{{}_{\mathscr{O}_{1,0,0,1} \, \chi}}\big)^{+} \eta_{{}_{\chi}}
\Bigg\} \, \ud \chi
\\
\int \limits_{\textrm{Sp} \, S_\lambda} \Bigg\{ 
a'_{{}_{\chi}}\big(\overline{\check{\widetilde{\varphi}}\big|_{{}_{\mathscr{O}_{1,0,0,1} \, \chi}}}\big)
\, + \,\,
\eta_{{}_{\chi}} a'_{{}_{\chi}}\big(\widetilde{\varphi}\big|_{{}_{\mathscr{O}_{1,0,0,1} \, \chi}}\big)^{+} \eta_{{}_{\chi}}
\Bigg\} \, \ud \chi.
\end{multline}

We regard
\[
\mathscr{E} \ni \varphi \longmapsto A_{{}_{\chi}}(\varphi^{*}_{{}_{\chi}})  = A_{{}_{\chi}}(\varphi) 
= a'_{{}_{\chi}}\big(\overline{\check{\widetilde{\varphi}}\big|_{{}_{\mathscr{O}_{1,0,0,1} \, \chi}}}\big)
\, + \,\,
\eta_{{}_{\chi}} a'_{{}_{\chi}}\big(\widetilde{\varphi}\big|_{{}_{\mathscr{O}_{1,0,0,1} \, \chi}}\big)^{+} \eta_{{}_{\chi}}
\]
as a field over the Minkowski space-time, with the same space-time test space $\mathscr{E} = \mathcal{S}^{00}(\mathbb{R}^4; \mathbb{C}^4)$ 
as for the electromagnetic potential field $A$.  By construction, the Hida annihilation-creation operators fulfill the following
canonical commutation relations (where for $\chi=-1+i\nu$ we have put $F^{{}^{s}}_{{}_{\chi,lm}} = F^{{}^{s}}_{{}_{\nu,lm}}$
and $\widetilde{\varphi}_{{}_{\chi}} = \widetilde{\varphi}_{{}_{\nu}}$)
\[
\Big[ a'_{{}_{\chi}}\big(F^{{}^{s}}_{{}_{\chi,lm}}\big),   a'_{{}_{\chi}}\big(F^{{}^{s}}_{{}_{\nu,l'm'}}\big)^+\Big]
= \big(F^{{}^{s}}_{{}_{\chi,lm}} , F^{{}^{s'}}_{{}_{\chi,l'm'}}\Big)_{{}_{\chi}} = \delta_{{}_{l \,l'}}\delta_{{}_{m \,m'}} \delta^{{}^{s \,s'}}. 
\]
Let us introduce
\[
{a'}_{{}_{\chi, lms}} \coloneqq {a'}_{{}_{\chi}}\big(F^{{}^{s}}_{{}_{\chi,lm}}\big),
\]
with the canonical commutation relations
\[
\big[{a'}_{{}_{\chi, lms}}, {a'}_{{}_{\chi, l'm's'}}^{+} \big] = \delta_{{}_{l \,l'}}\delta_{{}_{m \,m'}}\delta_{{}_{s \,s'}},
\,\,\, \big[{a'}_{{}_{\chi, lms}}, {a'}_{{}_{\chi, l'm's'}} \big] = \big[{a'}_{{}_{\chi, lms}}^{+}, {a'}_{{}_{\chi, l'm's'}}^{+} \big] = 0.
\]

For each $\varphi \in \mathscr{E}$
\[
\widetilde{\varphi}\big|_{{}_{\mathscr{O}_{1,0,0,1} \, \chi}} = \sum\limits_{l,m,s} c_{{}^{lm}}^{{}^{s}} F^{{}^{s}}_{{}_{\chi,lm}}
\] 
where 
\begin{equation}\label{clms}
c_{{}^{lm}}^{{}^{s}} = \Big(\widetilde{\varphi}\big|_{{}_{\mathscr{O}_{1,0,0,1} \, \chi}}, F^{{}^{s}}_{{}_{\chi,lm}}  \Big)_{{}_{\chi}}
=
F^{{}^{s}}_{{}_{\chi,lm}} \big( \overline{\widetilde{\varphi}\big|_{{}_{\mathscr{O}_{1,0,0,1}}}} \Big)
\end{equation}
by (\ref{GelfandeDecompositionFunction}). We have analogous expansion 
\[
\overline{\check{\widetilde{\varphi}}\big|_{{}_{\mathscr{O}_{1,0,0,1} \, \chi}}} = \sum\limits_{l,m,s} c_{{}^{lm}}^{{}^{s}} F^{{}^{s}}_{{}_{\chi,lm}}
\] 
where 
\begin{equation}\label{checkclms}
\check{c}_{{}^{lm}}^{{}^{s}} = \Big(\overline{\check{\widetilde{\varphi}}\big|_{{}_{\mathscr{O}_{1,0,0,1} \, \chi}}},  F^{{}^{s}}_{{}_{\chi,lm}} \Big)_{{}_{\chi}}
=
F^{{}^{s}}_{{}_{\chi,lm}} \big( \check{\widetilde{\varphi}}\big|_{{}_{\mathscr{O}_{1,0,0,1}}} \Big).
\end{equation}

Here have not only the convergence
\[
\sum\limits_{l,m,s} \big| c_{{}^{lm}}^{{}^{s}} \big|^2 = \big\| \widetilde{\varphi}\big|_{{}_{\mathscr{O}_{1,0,0,1} \, \chi}} \big\|_{{}_{\chi}}^{2} < +\infty,
\,\,\,\,
\sum\limits_{l,m,s} \big| \check{c}_{{}^{lm}}^{{}^{s}} \big|^2 = \big\| \overline{\check{\widetilde{\varphi}}\big|_{{}_{\mathscr{O}_{1,0,0,1} \, \chi}}} \big\|_{{}_{\chi}}^{2} < +\infty
\]
in the Hilbert space $\mathcal{H}'_{{}_{\chi}} = \mathcal{H}'_{{}_{-1+i\nu}}$, but a rapid convergence
with rapidly decreasing $\{ c_{{}^{lm}}^{{}^{s}} \}$, $\{ \check{c}_{{}^{lm}}^{{}^{s}} \}$:
\[
\sum\limits_{l,m,s} \big[l(l+1)\big]^n \,\, \big| c_{{}^{lm}}^{{}^{s}} \big|^2 < +\infty, \,\,\,
\,\,\,
\sum\limits_{l,m,s} \big[l(l+1)\big]^n \,\, \big| \check{c}_{{}^{lm}}^{{}^{s}} \big|^2 < +\infty,
\,\,\,
n \in \mathbb{N}.
\]
This is the case because
$\overline{\check{\widetilde{\varphi}}\big|_{{}_{\mathscr{O}_{1,0,0,1} }}}, \widetilde{\varphi}\big|_{{}_{\mathscr{O}_{1,0,0,1} }} \in E$
and $\overline{\check{\widetilde{\varphi}}\big|_{{}_{\mathscr{O}_{1,0,0,1}}}}, \widetilde{\varphi}\big|_{{}_{\mathscr{O}_{1,0,0,1} }}$
lie in $\textrm{Dom} \, A^n$, for each $n \in \mathbb{N}$, of the standard operator $A$ defining the standard nuclear space
$E$. For the Hida operator $a'_{{}_{\chi}}$ the maps
\[
\begin{split}
E_{{}_{\chi}} \ni \xi \mapsto a'_{{}_{\chi}}(\xi) \in \mathscr{L}\big((E_{{}_{\chi}}), (E_{{}_{\chi}})\big),
\\
E_{{}_{\chi}} \ni \xi \mapsto a'_{{}_{\chi}}(\xi)^+ \in \mathscr{L}\big((E_{{}_{\chi}}), (E_{{}_{\chi}})^* \big),
\end{split}
\]
are continuous. Therefore, the sum operation, representing rapidly decreasing expansions 
\[
\overline{\check{\widetilde{\varphi}}\big|_{{}_{\mathscr{O}_{1,0,0,1} \, \chi}}} = \sum\limits_{l,m,s} c_{{}^{lm}}^{{}^{s}} F^{{}^{s}}_{{}_{\chi,lm}},
\,\,\,\,\,\,\,\,\,
\widetilde{\varphi}\big|_{{}_{\mathscr{O}_{1,0,0,1} \, \chi}} = \sum\limits_{l,m,s} c_{{}^{lm}}^{{}^{s}} F^{{}^{s}}_{{}_{\chi,lm}}
\] 
of
\[
\widetilde{\varphi}\big|_{{}_{\mathscr{O}_{1,0,0,1} \, \chi}}, \overline{\check{\widetilde{\varphi}}\big|_{{}_{\mathscr{O}_{1,0,0,1} \, \chi}}} \in E_{{}_{\chi}},
\]
can be pull-out of the arguments of $a'_{{}_{\chi}}\big(\cdot \big), a'_{{}_{\chi}}\big(\cdot \big)^+$:
\begin{multline*}
A_{{}_{\chi}}(\varphi^{*}_{{}_{\chi}}) = A_{{}_{\chi}}(\varphi)
= a'_{{}_{\chi}}\big(\overline{\check{\widetilde{\varphi}}\big|_{{}_{\mathscr{O}_{1,0,0,1} \, \chi}}}\big)
\, + \,\,
\eta_{{}_{\chi}} a'_{{}_{\chi}}\big(\widetilde{\varphi}\big|_{{}_{\mathscr{O}_{1,0,0,1} \, \chi}}\big)^{+} \eta_{{}_{\chi}}
\\
=
\sum\limits_{l,m,s}
\overline{\check{c}_{{}^{lm}}^{{}^{s}}}
\,\, {a'}_{{}_{\chi, lms}}
+ \sum\limits_{l,m,s} c_{{}^{lm}}^{{}^{s}} \,\, \eta_{{}_{\chi}} \, {a'}_{{}_{\chi, lms}}^{+} \, \eta_{{}_{\chi}}
\end{multline*}
\begin{multline*}
=
\sum\limits_{l,m,s}
\overline{F^{{}^{s}}_{{}_{\chi,lm}} \big( \check{\widetilde{\varphi}}\big|_{{}_{\mathscr{O}_{1,0,0,1}}} \Big)}
\,\, {a'}_{{}_{\chi, lms}}
+ \sum\limits_{l,m,s}
F^{{}^{s}}_{{}_{\chi,lm}} \big( \overline{\widetilde{\varphi}\big|_{{}_{\mathscr{O}_{1,0,0,1}}}} \Big)
\,\, \eta_{{}_{\chi}} \, {a'}_{{}_{\chi, lms}}^{+} \, \eta_{{}_{\chi}}
\\
=
\sum\limits_{l,m,s}
\overline{\check{F}^{{}^{s}}_{{}_{\chi,lm}} \big( \widetilde{\varphi}\big|_{{}_{\mathscr{O}_{1,0,0,1}}} \Big)}
\,\, {a'}_{{}_{\chi, lms}}
+ \sum\limits_{l,m,s}
F^{{}^{s}}_{{}_{\chi,lm}} \Big( \overline{\widetilde{\varphi}\big|_{{}_{\mathscr{O}_{1,0,0,1}}}} \Big)
\,\, \eta_{{}_{\chi}} \, {a'}_{{}_{\chi, lms}} \, \eta_{{}_{\chi}}
\end{multline*}
\begin{multline}\label{Achi(varphi)}
=
\sum\limits_{l,m,s}
\overline{\check{F}^{{}^{s}}_{{}_{\chi,lm}} \big( \widetilde{\varphi}\big|_{{}_{\mathscr{O}_{1,0,0,1}}} \Big)}
\,\, {a'}_{{}_{\chi, lms}}
+ \sum\limits_{l,m,s}
\overline{F^{{}^{s}}_{{}_{\chi,lm}} \Big( \widetilde{\varphi}\big|_{{}_{\mathscr{O}_{1,0,0,1}}} \Big)}
\,\, \eta_{{}_{\chi}} \, {a'}_{{}_{\chi, lms}}^{+} \, \eta_{{}_{\chi}}
\\
=
\sum\limits_{l,m,s}
\overline{\check{f}^{{}^{s}}_{{}_{\chi,lm}}( \varphi)}
\,\, {a'}_{{}_{\chi, lms}}
+ \sum\limits_{l,m,s}
\overline{f^{{}^{s}}_{{}_{\chi,lm}} (\varphi)}
\,\, \eta_{{}_{\chi}} \, {a'}_{{}_{\chi, lms}}^{+} \, \eta_{{}_{\chi}},
\end{multline}
where $f^{{}^{s}}_{{}_{\chi,lm}} \in \mathscr{E}^*$ is the inverse Fourier transform of the distribution
\[
\widetilde{\mathscr{E}} = \mathcal{S}^{0}(\mathbb{R}^4; \mathbb{C}^4) \ni \widetilde{\varphi}
\longmapsto F^{{}^{s}}_{{}_{\chi,lm}} \Big( \widetilde{\varphi}\big|_{{}_{\mathscr{O}_{1,0,0,1} \, \chi}} \Big),
\]
concentrated on $\mathscr{O}_{{}_{1,0,0,1}}$, and equal to the invariant integration along $\mathscr{O}_{{}_{1,0,0,1}}$
of the test function $\widetilde{\varphi}$, Krein projected on the real four-vector function $F^{{}^{s}}_{{}_{\chi,lm}}$, in accordance to the definition
given above. The distributions $f^{{}^{s}}_{{}_{\chi,lm}} \in \mathscr{E}^*$ are regular, with the valuations
$f^{{}^{s}}_{{}_{\chi,lm}} (\varphi)$ representable by integration
with ordinary four-vector space-time functions $x \mapsto f^{{}^{s}}_{{}_{\chi,lm}}(x)$, smooth everywhere except the light cone, where they have discontinuity along the light cone. Therefore, using the inverse Fourier transformed
functions $x \mapsto f^{{}^{s}}_{{}_{\chi,lm}}(x)$ we can explicitly compute the kernel
$A_{{}_{\chi}}(x)$ of the homogeneous part $A_{{}_{\chi}}$ of the electromagnetic potential field
$A$ of homogeneity degree $\chi \in \{-1 + i\nu, \nu \in \mathbb{R} \}$. Namely, using the pairing
\[
f^{{}^{s}}_{{}_{\chi,lm}} (\varphi) = \int\limits_{\mathbb{R}^4} f^{{}^{s}}_{{}_{\chi,lm}}(x) \varphi(x) \, \ud^4 x
= \int\limits_{\mathbb{R}^4} f^{{}^{s \,\, \mu}}_{{}_{\chi,lm}}(x) \varphi_{{}_{\mu}}(x) \, \ud^4 x
\]
and (\ref{Achi(varphi)})  we can write
\[
A_{{}_{\chi}}(\varphi) = \int\limits_{\mathbb{R}^4} A_{{}_{\chi}}(x) \varphi(x) \, \ud^4 x
= \int\limits_{\mathbb{R}^4} A_{{}_{\chi}}^{{}^{\mu}}(x) \varphi_{{}_{\mu}}(x) \, \ud^4 x,
\]
where
\begin{multline*}
A_{{}_{\chi}}(x)
=
\sum\limits_{l,m,s}
\overline{\check{f}^{{}^{s}}_{{}_{\chi,lm}}(x)}
\,\, {a'}_{{}_{\chi, lms}}
+ \sum\limits_{l,m,s}
\overline{f^{{}^{s}}_{{}_{\chi,lm}} (x)}
\,\, \eta_{{}_{\chi}} \, {a'}_{{}_{\chi, lms}} \, \eta_{{}_{\chi}}
\\
=
\sum\limits_{l,m,s}
\overline{f^{{}^{s}}_{{}_{\chi,lm}}(-x)}
\,\, {a'}_{{}_{\chi, lms}}
+ \sum\limits_{l,m,s}
\overline{f^{{}^{s}}_{{}_{\chi,lm}} (x)}
\,\, \eta_{{}_{\chi}} \, {a'}_{{}_{\chi, lms}}^{+} \, \eta_{{}_{\chi}}.
\end{multline*}
In fact for $f^{{}^{s}}_{{}_{\chi,lm}}$ we have $f^{{}^{s}}_{{}_{\chi,lm}} (-x) = \overline{f^{{}^{s}}_{{}_{\chi,lm}} (x)}$, so  that  
\begin{equation}\label{Achi(x)}
A_{{}_{\chi}}(x)
=
\sum\limits_{l,m,s}
f^{{}^{s}}_{{}_{\chi,lm}}(x)
\,\, {a'}_{{}_{\chi, lms}}
+ \sum\limits_{l,m,s}
\overline{f^{{}^{s}}_{{}_{\chi,lm}} (x)}
\,\, \eta_{{}_{\chi}} \, {a'}_{{}_{\chi, lms}}^{+} \, \eta_{{}_{\chi}},
\end{equation}
with distributional solutions $f^{{}^{s}}_{{}_{\chi,lm}}$ of d'Alembert equation which are homogeneous of homogeneity
degree $\chi = -1 - i\nu$, $\nu \in \mathbb{R}$.

Note that $\varphi^{*}_{{}_{\chi}}$ in the decomposition (\ref{potentialFieldA-decompositionProper}) are homogeneous
functions of homogeneity degree $\chi \in \{ -1 -i\nu, \nu \in \mathbb{R}\}$ and, similarly, $x \mapsto f^{{}^{s}}_{{}_{\chi,lm}}(x)$,
are homogeneous of homogeneity degree $\chi \in \{ -1 -i\nu, \nu \in \mathbb{R}\}$, so that each $A_{{}_{\chi}}$
induces naturally a field on the de Sitter hyperboloid. 

In Subsection \ref{AS} we will give a detailed analysis of the Hilbert space $\mathcal{H}'_{{}_{\chi}} = \mathcal{H}'_{{}_{-1+i\nu}}$
of generalized homogeneous of degree $\chi = -1 +i\nu$ states, together with the representation
of $SL(2, \mathbb{C})$ group acting in $\mathcal{H}'_{{}_{-1+i\nu}}$, for the special
case $\chi = -1$.
We will see there that $\mathcal{H}'_{{}_{-1+i\nu}}$ decomposes into the direct sum 
\[
\mathcal{H}'_{{}_{-1+i\nu}} = {\mathcal{H}'}_{{}_{-1+i\nu \,\,\, \textrm{tr}}}^{\textrm{e}} \oplus {\mathcal{H}'}_{{}_{-1+i\nu}}^{\textrm{m}}
\]
of two orthogonal closed Lorentz invariant subspaces which, in case $\nu=0$, we have denoted there
by ${\mathcal{H}'}_{{}_{\chi=-1}}^{e}$ and ${\mathcal{H}'}_{{}_{\chi=-1}}^{m}$.
On the closed subspace ${\mathcal{H}'}_{{}_{-1+i\nu}}^{e}$ the invariant Krein inner product 
\[
\Big( \, \cdot \, , \mathfrak{J}'_{{}_{-1+i\nu}} \, \cdot \, \Big)_{{}_{-1+i\nu}}
\]
coincides with the inner product 
\[
\Big( \, \cdot \, , \, \cdot \, \Big)_{{}_{-1+i\nu}}
\]
and the Krein operator $ \mathfrak{J}'_{{}_{-1+i\nu}}$ degenerates to  the unit operator $\boldsymbol{1}$. 
The invariant Krein inner product is degenerate but positive definite  on the transversal subspace
${\mathcal{H}'}_{{}_{-1+i\nu \,\,\textrm{tr}}}^{m} \subset {\mathcal{H}'}_{{}_{-1+i\nu}}^{m}$ and the quotient space 
${\mathcal{H}'}_{{}_{-1+i\nu \,\,\textrm{tr}}}^{m} \big{/} N_{{}_{-1+i\nu \,\,\textrm{tr}}}$ 
is endowed with the strictly positive invariant Hilbert space inner product induced by
\[
\Big( \, \cdot \, , \mathfrak{J}'_{{}_{-1+i\nu}} \, \cdot \, \Big)_{{}_{-1+i\nu}} 
\,\,\,\textrm{with} \,\,  N_{{}_{-1+i\nu}} 
\coloneqq \textrm{Ker} \, \Big( \, \cdot \, , \mathfrak{J}'_{{}_{-1+i\nu}} \, \cdot \, \Big)_{{}_{-1+i\nu=-1}}.
\]
The action of $SL(2, \mathbb{C})$ on ${\mathcal{H}'}_{{}_{\chi=-1}}^{e}$ is
equivalent to the unitary irreducible representation, which in \cite{Geland-Minlos-Shapiro} is denoted
by the pair $(l_0, l_1) = (1,0)$.

\section{Higher order contributions $A_{\textrm{int}}^{(n)}(g=1,x)$ and $\psi_{\textrm{int}}^{(n)}(g=1,x)$
to the interacting fields $A^{\mu}_{\textrm{int}}(g=1,x)$ and 
$\psi_{\textrm{int}}(g=1,x)$. Causal method and space-time geometry}\label{A(1)psi(1)}

The only modification which we introduce into the causal perturbative approach
to spinor QED, which goes back to St\"uckelberg and Bogoliubov is that we are using
the white noise construction of free fields of the theory.

This allows us to treat each free field at specified space-time point 
as a well defined generalized Hida operator, 
but moreover each free field gains the  mathematical interpretation
of an integral kernel operator with vector-valued kernel
in the sense of Obata \cite{obataJFA}. We have constructed the free Dirac and electromagnetic potential
fields as integral kernel operators with vector-valued kernels in the sense of Obata, respectively, 
in Subsections \ref{psiBerezin-Hida} and \ref{A=Xi0,1+Xi1,0}. 
The operations of Wick product, differentiation, integration, convolution with tempered distributions, 
which can be performed upon field operators understood
as integral kernel operators in the sense of Obata, have been described 
in Subsection \ref{OperationsOnXi}. Construction of the free fields as integral kernel 
operators opens us to 
the general and effective theory of integral kernel operators due to 
Hida-Obata-Sait\^o. In particular we can
treat the Wick product  (compare the so called ``Wick theorem''
in the book \cite{Bogoliubov_Shirkov} and in Subsection \ref{WickForProduct}) in the rigorous mathematically controllable fashion, 
necessary for the needs of the causal method (note here that in particular Wightman's definition is 
not effective here). The whole causal method is left completely untouched. We just put the free fields,
understood as integral kernel operators, into the formulas for the causal perturbative series
using the computational Rules for the Wick product, integration and convolution with tempered
distributions, which are given in Subsection \ref{OperationsOnXi}. The only nontrivial point is
the splitting of the causal distributions. Namely, (if the free fields are understood as integral kernel operators) 
each contribution to the causal scattering matrix
is a finite sum 
\[
\sum \limits_{l,m} \Xi_{l,m}(\kappa_{l,m})
\]
of well defined integral kernel operators (which almost immediately follows from the 
results summarized in Subsection \ref{OperationsOnXi})
\[
\Xi_{l,m}(\kappa_{l,m}) \in \mathscr{L}\big( (\boldsymbol{E}) \otimes \mathscr{E}, \, (\boldsymbol{E}^*) \big)
\cong \mathscr{L}\big(\mathscr{E}, \, \mathscr{L}((\boldsymbol{E}), (\boldsymbol{E})) \big)
\]
with vector-valued kernels
\[
\kappa_{l,m} \in \mathscr{L} \big(E_{i_1} \otimes \cdots \otimes E_{i_{l+m}}, \,\, \mathscr{E}^* \big) 
\cong E_{i_1}^{*} \otimes \cdots \otimes E_{i_{l+m}}^{*} \otimes \mathscr{E}^* 
\] 
in the sense of Obata, compare Subsections \ref{psiBerezin-Hida}, \ref{OperationsOnXi}, \ref{WickForProduct} and \ref{WickForChronological},
where the Hida subspace $(\boldsymbol{E})$
in the tensor product of the Fock spaces of the Dirac fied and the electromagnetic potential field is constructed
and the scatterig operator. 

Here 
\[
\mathscr{E} = \mathscr{E}_{1} \otimes \cdots \otimes \mathscr{E}_{1},
\]
is equal to the tensor product of several space-time test function spaces
\[
\mathscr{E}_{1} = \mathcal{S}(\mathbb{R}^4; \mathbb{C}).
\]
The nontrivial task in construction is the splitting of vector-valued
causal distribution kernels $\kappa_{l,m}$ into retarded and advanced parts, which in practical 
computation reduces to the slitting of causal distributions in
\[
\mathscr{E} = \mathscr{E}_{1}^{*} \otimes \cdots \otimes \mathscr{E}_{1}^{*}, 
\]
causally supported into retarded and advanced parts, compare Subsection \ref{WickForChronological}. 
This problem has been solved by Epstein and 
Glaser \cite{Epstein-Glaser}, but their method can be simplified and used in practical
computations as in Subsection \ref{WickForChronological}.

Summing up we can insert the free fields, understood as integral kernel operators in the sense of Obata,
into the formulas for the causal perturbative series for interacting fields. The necessary
operations of Wick product, splitting, integrations, have a rigorous meaning as operations
performed upon integral kernel operations explained in  Subsections \ref{OperationsOnXi} and  \ref{WickForProduct}. 
The formulas for the contributions are exactly the same as in the standard perturbative causal spinor QED, 
compare e.g. \cite{DKS1} or \cite{Scharf}, but with the Wick product and integration
in these formulas rigorously understood as performed upon integral kernel operators and expressed
by the Rules of  Subsection \ref{OperationsOnXi} and \ref{WickForProduct}. Recall that in order to define (the higher order contributions to) 
the local interacting field $\mathbb{A}_{{}_{\textrm{int}}}(x)$ corresponding to a free field $\mathbb{A}(x)$ 
we are using the Bogoliubov \cite{Bogoliubov_Shirkov} definition of the interacting field as the limit 
\[
\mathbb{A}_{{}_{\textrm{int}}}(x) = \underset{g\rightarrow 1}{\textrm{lim}} \,\, \mathbb{A}_{{}_{\textrm{int}}}(g,x)
\]
of the  variational derivative
\begin{multline*}
\mathbb{A}_{{}_{\textrm{int}}}(g,x) =  S^{-1}(g\mathcal{L})
\frac{\delta S(g\mathcal{L}+h\mathbb{A})}{\delta h(x)}\Bigg{|}_{{}_{h=0}}
\\
= \mathbb{A}(x) + \sum \limits_{n=1}^{\infty} {\textstyle\frac{1}{n!}} 
\int \ud^4 x_1 \cdots \ud^4 x_n \mathbb{A}^{(n)}(x_1, \ldots, x_n,x) \,
g(x_1) \cdots g(x_n),
\end{multline*}
which has the form of the formal functional power series in $g$. Here we are using the scattering
operator $S(g\mathcal{L})$, denoted in Subsection \ref{MotivationForHida} and \ref{WickForProduct}
shortly by $S(g)$, where we have indicated only the functional dependence of the scattering operator $S$
on the ``switching-of-interaction-function'' $g$. Existence of the above limit $g\rightarrow 1$ for ech higher order contribution, 
as the finite sum of integral kernel operators with vector-valued kernels (in the sense of \cite{obataJFA}),
has been proved for QED in Subsection \ref{OperationsOnXi}.
  However, it will be more appropriate to 
indicate dependence of the scattering operator $S$ on the whole interaction Lagrange density 
$g\mathcal{L}$.
The operator functional $S(g)=S(g\mathcal{L})$ with the Lagrange density
interaction of the theory (here the Lagrange density interaction $\mathcal{L}$ of QED)
is constructed as the formal functional power series
\[
\begin{split}
S(g\mathcal{L}) = \boldsymbol{1}+
\sum \limits_{n=1}^{\infty} {\textstyle\frac{1}{n!}} 
\int \ud^4 x_1 \cdots \ud^4 x_n S(x_1, \ldots, x_n) \,
g(x_1) \cdots g(x_n),
\\
S^{-1}(g\mathcal{L}) = \boldsymbol{1} +
\sum \limits_{n=1}^{\infty} {\textstyle\frac{1}{n!}} 
\int \ud^4 x_1 \cdots \ud^4 x_n \overline{S}(x_1, \ldots, x_n) \,
g(x_1) \cdots g(x_n),
\end{split}
\]
inductively within the causal method  explained in Subsection
\ref{MotivationForHida}, based on the axioms (I)-(V) of Subsection \ref{MotivationForHida} and \ref{WickForProduct}.
Physical motivation for the axioms the reader will find in \cite{Bogoliubov_Shirkov}.
Here we recall shortly the motivation for the causality axiom (I).
For this purpose let $G_1$ and $G_2$ be two regions in space-time such that
in some Lorentz frame all points of $G_1$ precede an equal-time surface $t=\tau$  
and all points of $G_2$ have time coordinate greater than $\tau$,
which we denote $G_1 \prec G_2$. 
\begin{center}
\begin{tikzpicture}[yscale=1]
    \draw[thin, dashed] (-2.5,0) -- (2.5,0);
    \draw[thin, ->] (0,1.5) -- (0,2);

\fill[color=gray!, fill opacity=0.1] (0,0.9) ellipse (0.5 and 0.3);
\fill[color=gray!, fill opacity=0.1] (0,-0.8) ellipse (1 and 0.3);
\node [right] at (2.6,0) {$t=\textrm{const}$};
\node [right] at (-4,1.5) {$G_1 \prec G_2$};
\node [left] at (0,2) {$t$};
\node [above] at (0,0.6) {$G_2$};
\node [below] at (0,-0.6) {$G_1$};
\end{tikzpicture}
\end{center}

Let $g_1$ and $g_2$ be two smooth functions
``switching-on-off'' the intensity of interaction with the property
$\textrm{supp} \, g_1 \subset G_1$ and $\textrm{supp} \, g_2 \subset G_2$.
Then consider the ``switching off'' the intensity of interaction 
$g = g_1 + g_2$. Then the interaction $g\mathcal{L}$ is switched off completely
outside $G_1 \cup G_2$, and the state of the system $\Phi$ in times preceding 
both regions $G_i$ evolves as a state of a free system, and  it is a constant state in
the ``interaction picture'' in the times preceding both $G_i$, which coincides there also with the constant state in the Heisenberg picture with the free field operators evolving accordingly to the evolution of the system of free fields. Similarly, the state 
of the system in the ``interaction picture'' is constant in time for all times later than both $G_i$. 
The scattering operator $S(g) = S(g\mathcal{L})$ transforms the initial state $\Phi$ into the final  $S(g)\Phi
= S(g\mathcal{L})\Phi$. This final state can be obtained from the
intermediate state $\Phi_{{}_{\tau}}$ of the system which it assumes at time $t=\tau$.
Now the causality requirement says that the state  $\Phi_{{}_{\tau}}$ should not depend
on the interaction in the region $G_2$ lying in the future of $t=\tau$, \emph{i.e.}
\[
\Phi_{{}_{\tau}} = S(g_1)\Phi = S(g_1\mathcal{L})\Phi,
\]
where $S(g_1\mathcal{L})$ is the scattering operator in which the interaction is switched on
with the intensity $g_1$; and moreover the final state $S(g\mathcal{L})\Phi$
should be obtainable from the intermediate one $\Phi_{{}_{\tau}}$
by the application of the scattering operator $S(g_2\mathcal{L})$
with the intensity of interaction equal $g_2$:
\[
S(g\mathcal{L})\Phi = S(g_2\mathcal{L})\Phi_{{}_{\tau}}.
\]
We therefore arrive at the causality condition
\[
S\big((g_1 + g_2)\mathcal{L}\big) = S(g_2\mathcal{L}) S(g_1\mathcal{L})
\,\,\,\,\,\,
\textrm{if}
\,\, G_1 \prec G_2.
\]
In case the regions $G_1$ and $G_2$ are mutually space-like, which we write $G_1 \sim G_2$,
\begin{center}
\begin{tikzpicture}[yscale=1]
   \draw[thin, ->] (-2,-1) -- (-2.25,-0.5);
    \draw[thin, ->] (-2,1) -- (-1.75,1.5);
    \draw[ultra thin, domain=-1.5:1.5] plot(\x, {\x});
    \draw[ultra thin, domain=-1.5:1.5] plot(\x, -\x);
    \draw[thin,dashed] (-2,-1) -- (2,1);
     \draw[thin,dashed] (-2,1) -- (2,-1);
    \fill[color=gray!, fill opacity=0.1] (1.1,0) ellipse (0.5 and 0.3);
\fill[color=gray!, fill opacity=0.1] (-1.6,0) ellipse (1 and 0.3);
\node [above] at (1.5,1.6) {\textrm{\tiny $(ct)^2 = \boldsymbol{\x}^2$}};
\node [right] at (2,1) {$t=\textrm{const}$};
\node [right] at (2,-1) {$t'=\textrm{const}$};
\node [right] at (0.8,0) {$G_2$};
\node [left] at (-1,0) {$G_1$};
\node [right] at (-4,1.5) {$G_1 \sim G_2$};
\node [left] at (-2.25,-0.5) {\textrm{\tiny $t$}};
\node [left] at (-1.75,1.5) {\textrm{\tiny $t'$}};
\end{tikzpicture}
\end{center}
there always exists a Lorentz
frame in which $G_1 \prec G_2$, as well as  a Lorentz frame in which $G_2 \prec G_1$,
so that repeating the above causality argument we arrive at the conclusion
\[
S\big((g_1 + g_2)\mathcal{L}\big) = S(g_2\mathcal{L}) S(g_1\mathcal{L})
\,\,\,\,\,\,
\textrm{if}
\,\, G_1 \sim G_2.
\]
Bogoliubov was the first, who converted this form of causality condition (due to
St\"uckelberg) into the practical form in terms of the kernels of the
functional expansion of the scattering operator functional $S$, compare
\cite{Bogoliubov_Shirkov} and Subsection \ref{MotivationForHida}.
Using his form of causality (I), Poincar\'e covariance (II), unitarity (III),
correspondence principle (IV) identifying the kernel of the first order contribution
with the Lagrange density of interaction, and (V) preservation of the degree of distribution when computing its splitting
into retarded and advanced part, Epstein and Glaser were able to prove
that all higher order kernels are essentially determined by the axioms (I)--(V).
However, Epstein and Glaser, using the free fields as Wigthman operator distributions,
have not been able to restore the physical value $g=1$ of the intensity $g$ of interaction,
particularly in case of theories containing among the underlying free fields a massless
field, e.g in case of QED.
We reinterpret the free fields as integral kernel operators with vector-valued kernels
in the sense of \cite{obataJFA},
which allows us to put $g=1$ in the final formulas, for the proof compare 
Subsection \ref{OperationsOnXiIF}. Although we should emphasize that the resulting theory
cannot be applied to normalizable states but only to generalized states,
such as the many particle plane wave states, which nonetheless is sufficient
for the computation of the effective cross sections involving many particle plane wave states,
as we have already said in Introduction. 

Similarly, using the causal perturbative 
method of Subsections \ref{MotivationForHida} and \ref{WickForProduct}, we construct inductively 
the scattering operator functional
\begingroup\makeatletter\def\f@size{5}\check@mathfonts
\def\maketag@@@#1{\hbox{\m@th\large\normalfont#1}}%
\begin{multline*}
S(g\mathcal{L} + h\mathbb{A}) = \boldsymbol{1}+
\\
\sum \limits_{n,m=0}^{\infty} {\textstyle\frac{1}{(n+m)!}} 
\int \ud^4 x_1 \cdots \ud^4 x_n y_1 \cdots \ud^4 y_m S(x_1, \ldots, x_n,y_1, \ldots, y_m) \,
g(x_1) \cdots g(x_n) h(y_1) \cdots h(y_m),
\end{multline*}
\begin{multline*}
S^{-1}(g\mathcal{L}+h\mathbb{A}) = \boldsymbol{1} + \\
\sum \limits_{n,m=0}^{\infty} {\textstyle\frac{1}{(n+m)!}} 
\int \ud^4 x_1 \cdots \ud^4 x_n\ud^4 y_1 \cdots \ud^4 y_m \overline{S}(x_1, \ldots, x_n,y_1, \ldots, y_m) \,
g(x_1) \cdots g(x_n)h(y_1) \cdots h(y_m),
\end{multline*}
\[
m+n\neq 0,
\]
\endgroup
on using the interaction Lagrange density $g\mathcal{L} + h\mathbb{A}$ with two-component
$(g,h)$ ``intensity-of-interaction'' function $(g,h)$ serving as a tool for implementing the causality condition
\begin{multline*}
S\Big((g_1+g_2)\mathcal{L} + (h_1+h_2)\mathbb{A}\Big)
= S\big(g_2\mathcal{L} + h_2\mathbb{A}\big)S\big(g_1\mathcal{L} + h_1\mathbb{A}\big)
\\
\textrm{\tiny whenever $\textrm{supp} \, (g_1,h_1) \preceq \textrm{supp} \, (g_2,h_2)$},
\end{multline*}
compare Subsection \ref{MotivationForHida}, \ref{WickForProduct} or \cite{DKS1}, \cite{Scharf}, \cite{DutFred}.
The kernel $S(x_1, \ldots, x_n,y)$ of the $n+1$-order term
\[ 
{\textstyle\frac{1}{(n+1)!}} 
\int \ud^4 x_1 \cdots \ud^4 x_n\ud^4 y S(x_1, \ldots, x_n,y) \,
g(x_1) \cdots g(x_n)h(y) 
\]
contribution to $S(g\mathcal{L} + h\mathbb{A})$ into which $h$ enters linearly,
we denote shortly by $S(Z,y)$, using the abbreviated notation $Z$ of Epstein-Glaser 
for the set $\{x_1, \ldots, x_n\}$ of space-time variables. The kernels of the $n$-th order contributions 
\begin{multline*}
{\textstyle\frac{1}{(n)!}} 
\int \ud^4 x_1 \cdots \ud^4 x_n S(x_1, \ldots, x_n) \,
g(x_1) \cdots g(x_n)
\\
\textrm{and} \,\,
{\textstyle\frac{1}{(n)!}} 
\int \ud^4 x_1 \cdots \ud^4 x_n \overline{S}(x_1, \ldots, x_n) \,
g(x_1) \cdots g(x_n)
\end{multline*}
to $S(g\mathcal{L})$, and respectively to $S^{-1}(g\mathcal{L})$,
we denote simply by $S(x_1, \ldots, x_n) = S(Z)$ and $\overline{S}(x_1, \ldots, x_n)
= \overline{S}(Z)$. 

The variational derivative at $h=0$ in definition of interacting fields is understood,
as nothing else but taking the sum 
\[
\int \Sigma(g,x)h(x) dx
\]
of all terms which are linear in $h$ in the formal expansion for 
$S(g\mathcal{L} +h\mathbb{A})$ and putting the kernel $\Sigma(g,x)$ of this term as the derivative. 
For a rigorous definition of the higher order contributions to $S(g,h) = S(g\mathcal{L}+h\mathbb{A})$,
with the creation-annihilation operators understood as the Hida operators, including the case of the Grassmann-valued
test functions $h$, compare Subsection \ref{WickForProduct}. For a rigorous treatement of the variational derivatives with 
respect to Grassmann-valued test functions $h$, evaluated at $h\neq 0$, where we have to distinguish the left-hand-side
and right-hand-side derivatives, compare \cite{Berezin}.
 
Note here that $\mathbb{A}$
need not be equal to any one of the elementary free fields of the theory, but it can be equal to any 
local field expressed as a Wick polynomial of free fields. We apply this
construction to the physical observables $\mathbb{A}$ -- the Noether conserved currents, which are Hermitian symmetric
(equal to their Hermitian adjoints $(\, \cdot \, )^+$, \emph{i.e.} the linear transpositions preceded and followed by the complex conjugation operation,
eventually with the additional application of the Gupta-Bleuler operator in case of the theory with gauge fields,
compare Subsection \ref{psiBerezin-Hida}). 

Performing the formal functional variation we can easily see that
$\mathbb{A}^{(n)}(x_1, \ldots, x_n,x) = \mathbb{A}^{(n)}(Z,x)$
is eqaual
\begin{multline*}
\mathbb{A}^{(n)}(Z,x) =  {\textstyle\frac{1}{i}} \sum \limits_{X \sqcup Y =Z}
\overline{S}(X)S(Y,x),
\\ 
\textrm{\tiny sum over all partitions $X \sqcup Y$ of $Z$ including $X = \emptyset$}
\\
= {\textstyle\frac{1}{i}} \sum' \limits_{X \sqcup Y=Z}
\overline{S}(X)S(Y,x) + S(Z,x),
\\
\textrm{\tiny sum over all partitions $X \sqcup Y$ of $Z$ with $X \neq \emptyset$},
\end{multline*}
compare \cite{DKS1} and \cite{Scharf}.
This is precisely the kernel $A_{(n+1)}(Z,x)$ of Epstein-Glaser introduced 
in Subsection \ref{MotivationForHida} and \ref{WickForProduct} (computed for the scattering 
operator $S(g,h) = S(g\mathcal{L}+h\mathbb{A})$) with the support properties 
summarized in Subsection \ref{MotivationForHida}. In particular, it follows
that $\mathbb{A}^{(n)}(Z,x)$ is equal to the advanced part of the kernel 
$D_{(n+1)}(Z,x)$ in the inductive constrution the $n+1$-order contribution (defined above)
to the scattering operator $S(g\mathcal{L}+h\mathbb{A})$ , multiplied
by $-i$. More precisely let us introduce after Epstein and Glaser, besides
\[
A'_{(n+1)}(Z,x) = \sum' \limits_{X \sqcup Y=Z}
\overline{S}(X)S(Y,x)
\]
the kernel
\[
R'_{(n+1)}(Z,x) = \sum' \limits_{X \sqcup Y=Z} S(Y,x) \overline{S}(X)
\]
(note that the primed sums run over those partitions which do not include the 
empty set $X=\emptyset$).
Introducing further, after Epstein and Glaser, the kernel
\[
D_{(n+1)}(Z,x) = R'_{(n+1)}(Z,x) - A'_{(n+1)}(Z,x)
\]
we see (compare Subsection \ref{MotivationForHida} and \ref{WickForProduct}) that
\[
\mathbb{A}^{(n)}(Z,x) = {\textstyle\frac{1}{i}} A_{(n+1)}(Z,x)
\]
where
\[
A_{(n+1)}(Z,x) = \textrm{advanced part} \big[D_{(n+1)}(Z,x)  \big]
\]
in the decomposition 
\[
D_{(n+1)}(Z,x) = R_{(n+1)}(Z,x)-A_{(n+1)}(Z,x)
\]
of $D_{(n+1)}$ into the advanced $A_{(n+1)}$ and retarded part $R_{(n+1)}$:
\begin{multline*}
A_{(n+1)}(Z,x) =  \sum \limits_{Z = X \sqcup Y}
\overline{S}(X)S(Y,x),
\\ 
\textrm{\tiny sum over all partitions $X \sqcup Y$ of $Z$ including $X = \emptyset$}
\\
= {\textstyle\frac{1}{i}} \sum' \limits_{Z = X \sqcup Y}
\overline{S}(X)S(Y,x) + S(Z,x),
\\
\textrm{\tiny sum over all partitions $X \sqcup Y$ of $Z$ with $X \neq \emptyset$},
\end{multline*}\begin{multline*}
R_{(n+1)}(Z,x) =  \sum \limits_{X \sqcup Y=Z}
S(Y,x)\overline{S}(X),
\\ 
\textrm{\tiny sum over all partitions $X \sqcup Y$ of $Z$ including $X = \emptyset$}
\\
= {\textstyle\frac{1}{i}} \sum' \limits_{X \sqcup Y=Z}
S(Y,x)\overline{S}(X) + S(Z,x),
\\
\textrm{\tiny sum over all partitions $X \sqcup Y$ of $Z$ with $X \neq \emptyset$}.
\end{multline*}

In particular for $\mathbb{A}(x) = A^\mu(x)$ equal to the $\mu$-th component 
of electromagnetic potential field we have the following formula
\[
A^{\mu \, (1)}(x_1,x) = {\textstyle\frac{1}{i}} \textrm{advanced part} \big[D_{(2)}(x_1,x)  \big]
\]
for the kernel of the first order contribution to the interacting
field $A_{{}_{\textrm{int}}}^{\mu}(x)$,
where ($\eta$ is the Gupta-Bleuler operator in the total Fock space of the free
$A$ and $\boldsymbol{\psi}$ fields acting trivially in the factor Fock space
of the field $\boldsymbol{\psi}$)
\begin{multline*}
D_{(2)}(x_1,x) = \eta \big(i\mathcal{L}(x_1)\big)^{+} \eta \,\, iA^\mu(x) 
\,\,\,
-
\,\,\,
iA^\mu(x)   \,\,\, \eta \big(i\mathcal{L}(x_1)\big)^{+} \eta 
\\
=
-eA^\mu(x) \,  \boldsymbol{{:}} \psi^\sharp(x_1) \gamma_\nu A^\nu(x_1) \psi(x_1) \boldsymbol{{:}}
\,\,\,\,
+ 
\,\,\,\,
e \boldsymbol{{:}} \psi^\sharp(x_1) \gamma_\nu  A^\nu(x_1) \psi(x_1) \boldsymbol{{:}} \,  A^\mu(x) 
\\
=
-e[A^\nu(x_1), A^\mu(x) ] \,
\boldsymbol{{:}} \psi^\sharp(x_1) \gamma_\nu \psi(x_1) \boldsymbol{{:}} 
\,\,\,\,\,\,
=
-ie D_{0}(x_1-x) \,
\boldsymbol{{:}} \psi^\sharp(x_1) \gamma_\mu \psi(x_1) \boldsymbol{{:}}. 
\end{multline*}

For $\mathbb{A}(x) = \boldsymbol{\psi}^a(x)$ equal to the $a$-th component 
of Dirac's spinor field we have the following formula
\[
\boldsymbol{\psi}^{a \, (1)}(x_1,x) = {\textstyle\frac{1}{i}} \textrm{advanced part} \big[D_{(2)}(x_1,x)  \big]
\]
for the kernel of the first order contribution to the interacting
field $\boldsymbol{\psi}_{{}_{\textrm{int}}}^{a}(x)$, with
\begin{multline*}
D_{(2)}(x_1,x) = \eta \big(i\mathcal{L}(x_1)\big)^{+} \eta \,\, i\boldsymbol{\psi}^a(x) 
\,\,\,
-
\,\,\,
i\boldsymbol{\psi}^a(x)  \,\,\, \eta \big(i\mathcal{L}(x_1)\big)^{+} \eta 
\\
=
ie S(x-x_1)\gamma_\nu \boldsymbol{\psi}(x_1)A^\nu(x_1).
\end{multline*}

The computation, with the only nontrivial step lying in the splitting which can essentially be borrowed from  Subsection
\ref{WickForChronological} (compare \cite{Scharf}, \cite{DKS1}, \cite{DutFred}), can therefore be omitted. 
We give only the final formulas for the contributions of the first and second order to interacting fields 
(with summation convention and Dirac adjoined bispinor 
$\boldsymbol{\psi}^\sharp(x) = \boldsymbol{\psi}(x)^\sharp = \boldsymbol{\psi}(x)^+\gamma_0$, 
compare \cite{DKS1}, \cite{Scharf}, \cite{DutFred} where $\boldsymbol{\psi}^\sharp(x)$
is denoted by $\overline{\boldsymbol{\psi}}(x)$, and with the 
sign $\{x_1 \leftrightarrow x_2 \}$ denoting expression immediately preceding it with the variables
$x_1$ and $x_2$ interchanged) 
\[
\boldsymbol{\psi}_{{}_{\textrm{int}}}^{a}(g, x) =
\boldsymbol{\psi}^{a}(x) + \sum \limits_{n=1}^{\infty} \frac{1}{n!}
\int \limits_{\mathbb{R}^{4n}} \ud^4x_1 \cdots \ud^4 x_n \boldsymbol{\psi}^{a \, (n)}(x_1, \ldots, x_n; x)
g(x_1) \cdots g(x_n), 
\]
with
\[
\boldsymbol{\psi}^{a \, (1)}(x_1; x) = 
e S_{{}_{\textrm{ret}}}^{aa_1}(x-x_1) \gamma^{\nu_1 \, a_1a_2} \boldsymbol{\psi}^{a_2}(x_1)A_{\nu_1}(x_1), 
\]
\begin{multline*}
\boldsymbol{\psi}^{a \, (2)}(x_1, x_2; x) = \\
e^2 \Bigg\{ S_{{}_{\textrm{ret}}}^{aa_1}(x-x_1) \gamma^{\nu_1 \, a_1a_2}S_{{}_{\textrm{ret}}}^{a_2a_3}(x_1-x_2)
\gamma^{\nu_2 \, a_3a_4} \,  {:}\boldsymbol{\psi}^{a_4}(x_2)  A_{\nu_1}(x_1)A_{\nu_2}(x_2) {:} \\ 
- S_{{}_{\textrm{ret}}}^{aa_1}(x-x_1) \gamma^{\nu_1 \, a_1a_2} \,
{:} \boldsymbol{\psi}^{a_2}(x_1) 
\boldsymbol{\psi}^{\sharp \, a_3}(x_2) \gamma_{\nu_1}^{a_3a_4} \boldsymbol{\psi}^{a_4}(x_2){:} \,
D^{{}^{\textrm{ret}}}_{0}(x_1-x_2) \\
+S_{{}_{\textrm{ret}}}^{aa_1}(x-x_1) \Sigma_{{}_{\textrm{ret}}}^{a_1a_2}(x_1-x_2)\boldsymbol{\psi}^{a_2}(x_2)
\Bigg\} \,\,\, +  \,\,\, \Bigg\{ x_1 \longleftrightarrow x_2 \Bigg\},
\end{multline*}
\[
\textrm{e. t. c.}
\]
and  
\[
{A_{{}_{\textrm{int}}}}_{\mu}(g, x) =
A_{\mu}(x) + \sum \limits_{n=1}^{\infty} \frac{1}{n!}
\int \limits_{\mathbb{R}^{4n}} \ud^4x_1 \cdots \ud^4 x_n A_{\mu}^{\, (n)}(x_1, \ldots, x_n; x)
g(x_1) \cdots g(x_n),
\]
with
\[
A_{\mu}^{\, (1)}(x_1;x) = -e D^{{}^{\textrm{av}}}_{0}(x_1-x) \,
{:}\boldsymbol{\psi}^{\sharp \, a_1}(x_1) \gamma_{\mu}^{a_1a_2} \boldsymbol{\psi}^{a_2}(x_1){:},
\]
\begin{multline*}
A_{\mu}^{\, (2)}(x_1, x_2; x) = 
e^2 \Bigg\{ 
{:}\boldsymbol{\psi}^{\sharp \, a_1}(x_1) 
\Big( 
 \gamma_{\mu}^{a_1a_2} S_{{}_{\textrm{ret}}}^{a_2a_3}(x_1-x_2) \gamma^{\nu_1 \, a_3a_4}
D^{{}^{\textrm{av}}}_{0}(x_1-x) A_{\nu_1}(x_2) \\
+ \gamma^{\nu_1 \, a_1a_2}S_{{}_{\textrm{av}}}^{a_2a_3}(x_1-x_2) \gamma_{\mu}^{a_3a_4}
D^{{}^{\textrm{av}}}_{0}(x_2-x)A_{\nu_1}(x_1)
\Big)  \boldsymbol{\psi}^{a_4}(x_2){:} \\
+ D^{{}^{\textrm{av}}}_{0}(x_1-x) {\Pi^{{}^{\textrm{av}}}}_{\mu}^{\nu_1}(x_2-x_1)A_{\nu_1}(x_2)
\Bigg\} \,\,\, + \,\,\, \Bigg\{ x_1 \longleftrightarrow x_2 \Bigg\}
\end{multline*}
\[
\textrm{e. t. c.}
\]
where $g$ is the intensity-of-interaction function over space-time which is assumed to be an element of
the ordinary Schwartz space $\mathcal{S}(\mathbb{R}^4; \mathbb{C})$, and which plays a technical role in 
realizing the causality condition in the form we have learned from Bogoliubov and Shirkov
\cite{Bogoliubov_Shirkov}, compare \cite{DKS1}, \cite{Scharf}, \cite{DutFred}.
This intensity function $g$ modifies the interaction into unphysical in the regions
which lie outside the domain on which $g$ is constant and equal to $1$. It is therefore important
problem to pass to a ``limit'' case of physical interaction with $g=1$ everywhere over the space-time.  

\[
\begin{split}
\boldsymbol{\psi}_{{}_{\textrm{int}}}^{a \,(n)}(g, x) =
\frac{1}{n!}
\int \limits_{\mathbb{R}^{4n}} \ud^4x_1 \cdots \ud^4 x_n \boldsymbol{\psi}^{a \, (n)}(x_1, \ldots, x_n; x), \\
{A_{{}_{\textrm{int}}}}_{\mu}^{\, (n)}(g, x) = 
\frac{1}{n!}
\int \limits_{\mathbb{R}^{4n}} \ud^4x_1 \cdots \ud^4 x_n A_{\mu}^{\, (n)}(x_1, \ldots, x_n; x),
\end{split}
\]  
are the respective $n$-th order contributions to the interacting Dirac and electromagnetic potential fields. 

Here in the above formulas for the $n$-th order contributions to interacting fields the free Dirac and electromagnetic 
fields $\boldsymbol{\psi}$ and $A$  we understood as integral kernel operators with vector-valued kernels as explained
in \ref{psiBerezin-Hida} and \ref{A=Xi0,1+Xi1,0}. Correspondingly the Wick product
and the integrations in these formulas are understood in a rigorous sense as operations performed 
upon integral kernel operators, and summarized in the Rules of Subsection \ref{OperationsOnXi} and  \ref{WickForProduct}.
It turns out that each order contribution is equal
\[
\begin{split}
\boldsymbol{\psi}_{{}_{\textrm{int}}}^{a, \,(n)}(g) \,\, = \,\,
 \sum \limits_{l,m} \Xi(\kappa_{l,m}), \\
{A_{{}_{\textrm{int}}}}_{\mu}^{\, (n)}(g) \,\, = \,\,
\sum \limits_{l,m} \Xi(\kappa'_{l,m}),
\end{split}
\] 
to a finite sum of well defined integral kernel operators $\Xi(\kappa_{l,m}), \Xi(\kappa'_{l,m})$ 
with vector-valued distributional kernels $\kappa_{l,m}, \kappa'_{l,m}$ in the sense of Obata 
\cite{obataJFA} (compare Subsection \ref{OperationsOnXi}).

But the main and the whole point is that if the free fields are 
understood as integral kernel operators in the sense of Obata, then the above formulas for each 
$n$-th order contribution to interacting fields, preserve their rigorous mathematical meaning
even if we put $g=1$ everywhere:
namely for $g$ put everywhere equal to $1$ the formulas for each order contributions
to interacting fields represent well defined integral kernel operators in the sense of Obata. 
This we have proved as Theorem \ref{g=1InteractingFieldsQED}, Subsection \ref{OperationsOnXi}. 
Free fields are of course understood as integral kernel operators in the formulas
for contributions to interacting fields, 
and the respective operations of Wick product and integrations with pairing 
functions are understood as performed upon integral kernel operators according to the Rules
of Subsection \ref{OperationsOnXi} and \ref{WickForProduct}.   

Thus, each order contribution to interacting fields in the adiabatic limit $g=1$ of physical
interaction is well defined integral kernel operator and belongs to the same general class of integral 
kernel operators as the Wick product at the same space-time point of free massless fields
(such as the free electromagnetic potential field). Thus, the construction of the free fields within the white 
noise setup as integral kernel operators allows us to solve the adiabatic limit problem in the 
causal perturbative and spinor QED.

Presented method of solution of this problem is general enough to be applicable to other more general and realistic
QFT, provided they can be formulated within the causal perturbative approach, which is for example the case for the Standard Model with the Higgs field \cite{DKS2}, \cite{DKS3}.  

Moreover, the interacting fields $\mathbb{A}_{{}_{\textrm{int}}}$ are given through Fock expansions
\begin{equation}\label{generalAint}
\mathbb{A}_{{}_{\textrm{int}}} = \sum \limits_{l,m} \Xi(\kappa_{l,m})
\end{equation}
into integral kernel operators in the sense of \cite{obataJFA} which can be subject to a precise and computable
convergence criteria, which utilize the symbol calculus of Obata, compare \cite{obataJFA},
\cite{obata}, \cite{obata-book}. This allows us to verify the convergence of the perturbative series 
with the tools which were beyond our reach before. 

In particular the series (\ref{generalAint}) for interacting field 
$\mathbb{A}_{{}_{\textrm{int}}}$ converges as the series of integral kernel operators with vector-valued kernels $\kappa_{l,m}$ in the sense of Thm 4.8 of \cite{obataJFA}
if and only if it represents an element $\mathbb{A}_{{}_{\textrm{int}}}$ of
\[
\mathscr{L}\big( (\boldsymbol{E}) \otimes \mathscr{E}, \, (\boldsymbol{E})^* \big) \cong
\mathscr{L}\Big( \mathscr{E}, \,\, \mathscr{L}\big( (\boldsymbol{E}), (\boldsymbol{E})^*\big) \, \Big),
\]
with continuous map
\[
\mathscr{E} \ni \phi \longmapsto
\mathbb{A}_{{}_{\textrm{int}}}(\phi) = \sum \limits_{l,m} \Xi\big(\kappa_{l,m}(\phi)\big)
\in \mathscr{L}\big( (\boldsymbol{E}), (\boldsymbol{E})^*\big).
\]
In this case the variational derivative in the Bogoliubov definition of the interactig
field $\mathbb{A}_{{}_{\textrm{int}}}$ is not only formal but can be given a sense
of the limit
\[
\mathbb{A}_{{}_{\textrm{int}}}(\phi)\Phi
= \lim \limits_{\epsilon \rightarrow 0}
S^{-1}(\mathcal{L})\frac{S\big(\mathcal{L} +\epsilon\phi \mathbb{A}\big) - S(\mathcal{L})}{\epsilon}
\Phi, \,\,\,\, \Phi \in (\boldsymbol{E}), \,\,\,
\phi \in \mathscr{E},
\]
in the strong dual topology of the space $(\boldsymbol{E})^*$ strongly dual
to the Hida test space $(\boldsymbol{E})$.

Note, please, that the above causal inductive construction of the chronological and anti-chronological
products $S_n(x_1, \ldots, x_n)$ and $\overline{S_n}(x_1, \ldots, x_n)$ could have been simplified without any
need for computation of the quasi asymptotics during the splitting of the causally supported tempered distributions 
into advanced and retarded parts.
Indeed, we could have used the ``natural'' construction of $S_n(x_1, \ldots, x_n)$ and $\overline{S_n}(x_1, \ldots, x_n)$,
given in Subsection \ref{WickForChronological}, which is equivalent with making specific choices in the Epstein-Glaser
splittings. Having given the ``natural'' chronological products $S_n(x_1, \ldots, x_n)$ and $\overline{S_n}(x_1, \ldots, x_n)$,
we could also have based on it the construction of the interacting fields. We nonetheless have chosen the general causal construction
of $S_n(x_1, \ldots, x_n)$ and $\overline{S_n}(x_1, \ldots, x_n)$, in order to keep our exposition at the most general level.

\subsection{Effective cross-section}\label{EffCrossSection}

Each higher order contribution
\[
S_n(g\mathcal{L}) = 
\sum \limits_{n=1}^{\infty} {\textstyle\frac{1}{n!}} 
\int \ud^4 x_1 \cdots \ud^4 x_n S(x_1, \ldots, x_n) \,
g(x_1) \cdots g(x_n), \,\,\,\,\,\, g \in \mathscr{E},
\]
to the scattering operator belongs to $\mathscr{L}((\boldsymbol{E}), \, (\boldsymbol{E})^*)$ for each fixed
space-time test function $g \in \mathscr{E}$, and defines a continuous map $\mathscr{E} \ni g \longmapsto S_n(g\mathcal{L}) \in \mathscr{L}((\boldsymbol{E}), \, (\boldsymbol{E})^*)$. For the proof compare Corollary to Thm. \ref{g=1InteractingFieldsQED} of Subsection \ref{OperationsOnXi}
or Subsection \ref{WickForChronological}. Contrary to the higher order contributions to interacting fields, we have not unique choice for
$S(g\mathcal{L})$ with $g$ put equal to the constant function $1_{{}_{\mathbb{R}^4}}$ equal $1$ everywhere on the space-time $\mathbb{R}^4$, and preserve 
the meaning of $S(g\mathcal{L})$ to be a well defined generalized operator transforming
continuously the Hida space into its dual. This is because $1_{{}_{\mathbb{R}^4}} \notin \mathscr{E}$. Recall that here, on the Minkowski space-time, 
$\mathscr{E}$ is equal $\mathcal{S}(\mathbb{R}^4; \mathbb{C})$.  
At least this choice is not uniquely and automatically determined by the presented theory based on the white noise analysis, solely by the axioms
(I)-(V). However, the physical circumstances accompanying the scattering process will allow us to use some further conditions, 
which allow us to eliminate dependence on the possible choices which can appear here. 
It turns out that the fact that the map 
\begin{equation}\label{SnInL(EE, L(E,E*))}
S_n: \mathscr{E} \ni g \longmapsto S_n(g\mathcal{L}) \in \mathscr{L}\big( (\boldsymbol{E}), \, (\boldsymbol{E})^* \big) \cong
(\boldsymbol{E})^* \otimes (\boldsymbol{E})^* \cong \big[(\boldsymbol{E}) \otimes (\boldsymbol{E}) \big]^*
\end{equation}
is continuous is sufficient for the computation of the effectve cross-section in the adiabatic limit $g=1$ for the scattering with the 
many particle plane wave states of the elementary free fields of the theory, as explained in Introduction. 
Indeed, because of the continuity of (\ref{SnInL(EE, L(E,E*))}), then the operator $S_n$ defines the corresponding continuous map 
\[
(\boldsymbol{E}) \times (\boldsymbol{E}) \ni \Phi \times \Psi \longmapsto 
\big\langle \big\langle S_n(g\mathcal{L}) \Phi, \Psi \big\rangle \big\rangle \in \mathbb{C},
\]
\emph{i.e.} a distribution, whose distribution kernel can be naturally identified with, say, ``matrix elements'':
\begin{equation}\label{<SnPhi...s,p..., Psi...s',p'...>}
\big\langle S_n(g\mathcal{L}) \Phi_{{}_{\ldots s,p \ldots}}, \Psi_{{}_{\ldots s',p' \ldots}} \big\rangle
\end{equation}
of the generalized operator $S_n(g\mathcal{L})$ in the non-normalizable many-particle plane wave states
\[
\Phi_{{}_{\ldots s,\boldsymbol{\p} \ldots }} =
\cdots a_{s}(\boldsymbol{\p})^{+} \cdots |0\rangle, 
\,\,\,\,
\Psi_{{}_{\ldots s',\boldsymbol{\p}' \ldots }} =
\cdots a_{s'}(\boldsymbol{\p}')^{+} \cdots |0\rangle,
\]
with the creation (Hida) operators $a_{s}(\boldsymbol{\p})^{+}$ in the momentum picture and with 
$|0\rangle = \Psi_{{}_{0}}$ being the vacuum in the Fock space of free fields of the theory. 

The distribution kernel  (\ref{<SnPhi...s,p..., Psi...s',p'...>}) with $g \in \mathscr{E}$
is regular enough to be represented by ordinary function, and has the form of convolution of the Fourier transform
$\widetilde{g^{\otimes \, n}} = \widetilde{g}^{\otimes \, n}$ with an element which is a convolutor\footnote{In fact a smooth very slowly increasing function -- element of the predual of the convolutor algebra of $E_{{}_{n_1}}^{*} \otimes 
\ldots E_{{}_{n_1}}^{*}$.} of $E_{{}_{n_1}}^{*} \otimes 
\ldots E_{{}_{n_1}}^{*}$ which allows the operation of multiplication
\begin{multline}\label{<PlaneWaveS(gnot1),PlaneWave>}
\overline{\big \langle S_{n}(g\mathcal{L}) \Phi_{{}_{\ldots s,\boldsymbol{\p} \ldots }}, \Psi_{{}_{\ldots s',\boldsymbol{\p}' \ldots }}\big\rangle}
\big \langle S_{n}(g\mathcal{L}) \Phi_{{}_{\ldots s,\boldsymbol{\p} \ldots }}, \Psi_{{}_{\ldots s',\boldsymbol{\p}' \ldots }}\big\rangle = \\ =
\big| \big \langle S_{n}(g\mathcal{L}) \Phi_{{}_{\ldots s,\boldsymbol{\p} \ldots }}, \Psi_{{}_{\ldots s',\boldsymbol{\p}' \ldots }}\big\rangle \big|^2,
\,\,\,\, g \in \mathscr{E}.
\end{multline} 

These circumstances, in fact noticed by Bogoliubov and Shirkov in \cite{Bogoliubov_Shirkov}, \S 24.5, allows us to compute
\[
\Big| \Big \langle S_{n}(g\mathcal{L}) \Phi_{{}_{\ldots s,\boldsymbol{\p} \ldots }}, \Psi_{{}_{\ldots s',\boldsymbol{\p}' \ldots }}\Big\rangle \Big|^2,
\,\,\, \textrm{for} \, g=1,
\]
and to extract the ``residual part''
\begin{equation}\label{|<Sn(g=1)Phi...s,p..., Psi...s',p'...>|doPotegi2}
\Big| \Big \langle S_{n}\big((g=1)\mathcal{L}\big) \Phi_{{}_{\ldots s,\boldsymbol{\p} \ldots }}, 
\Psi_{{}_{\ldots s',\boldsymbol{\p}' \ldots }}\Big\rangle \Big|^2,
\end{equation}
in the adiabatic limit $g \rightarrow g=1$, and the computation of the effective cross section in the adiabatic limit, even if  
(\ref{<SnPhi...s,p..., Psi...s',p'...>}) was initially given for $g \in \mathscr{E}$. We encourage the reader to consult \cite{Bogoliubov_Shirkov}, Chap. IV, \S\S 24.5 and 25, where it is also noticed that the general convolution-dependence of (\ref{<SnPhi...s,p..., Psi...s',p'...>}) on $\widetilde{g}$
can be simplified with a simplifying approximation satisfactory for the computation of the effective cross-section. 
The extraction is possible because we are interested not with the amplitudes of the absolute
probabilities, but only with the probability for the registration of a particle with given quantum numbers \emph{per unit volume and per unit time}.
Note that the \emph{in} and \emph{out} many particle plane-wave states are non-normalizable, hence the distributional character of the ``matrix elements''
of the scattering operator in such states is what should be expected. Nonetheless, there is natural and essentially unique manner for normalization 
of the probability of registration of free quanta in such states counted per unit time and unit volume, and this allows us to make practical use
of the distributional ``matrix elements''.  
However, any need for handling infinite infra-red or ultra-violet infinities has in this way been eliminated. In the limit $g \rightarrow g=1$ the
Dirac delta function in the limiting formula (\ref{|<Sn(g=1)Phi...s,p..., Psi...s',p'...>|doPotegi2}) appears, but the difficulty in 
computing its square can be avoided by considering the expression (\ref{<PlaneWaveS(gnot1),PlaneWave>}), and using the approximation
\[
\widetilde{g}(p)\widetilde{g}(p) \approx \delta(p){\textstyle\frac{VT}{(2\pi)^4}}, 
\]
in it when passing to the adiabatic limit and counting registration of free quanta 
per unit of time and per unit of volume. Here $V$, $T$ stand for the volume and time period in which the scattering takes place,
and where $g=1$. This essentially was noted in \cite{Bogoliubov_Shirkov}, Chap. IV, \S\S 24.5 and 25, and from the physical point of
view is the basis for the solution of the adiabatic limit problem in the computation of the cross-section involving many particle plane-wave states of the
elementary fields in the manner shown in  \cite{Bogoliubov_Shirkov}, Chap. IV, \S\S 24.5 and 25.

Finally, let us look at ``extraction'' of the residual part (\ref{|<Sn(g=1)Phi...s,p..., Psi...s',p'...>|doPotegi2}) from the point of view, which is
based on the ``locality'' properties of the scattering operator which follow from the axioms (I)-(V), 
as well as from the ``adiabatic'' character of passing to the limit
$g \rightarrow 1_{{}_{\textrm{R}^4}}$, where   $1_{{}_{\textrm{R}^4}}$ is the constant function on space-time equal  everywhere to $1$.
Let us write the Fock expansion
\[
S_n(g\mathcal{L}) = \sum_{\substack{l, m\\
                  l+m \leq n }}
 \Xi\big(\kappa_{l,m}(g)\big)
\]
for the operator $S_n(g\mathcal{L})$. Because it defines the continuous map (\ref{SnInL(EE, L(E,E*))}), \emph{i.e.} the continuous map
\[
\mathscr{E} \ni g \longmapsto
S_n(g\mathcal{L}) = \sum_{\substack{l, m\\
                  l+m =n }}
\Xi\big(\kappa_{l,m}(g)\big)
\in \mathscr{L}\big( (\boldsymbol{E}), (\boldsymbol{E})^*\big),
\]
then the kernels
\[
\kappa_{l,m}(g) \in \mathscr{L}(E_{{}_{n_1}} \otimes \ldots \otimes  E_{{}_{n_n}}, \mathbb{C}) \cong 
E_{{}_{n_1}}^* \otimes \ldots \otimes  E_{{}_{n_n}}^{*} \cong \big[ E_{{}_{n_1}} \otimes \ldots \otimes  E_{{}_{n_n}}\big]^*
\]
define continuous maps
\begin{equation}\label{continuous:g->kappa_lm(g)}
\mathscr{E} \ni g \longmapsto \kappa_{l,m}(g) \in \big[ E_{{}_{n_1}} \otimes \ldots \otimes  E_{{}_{n_n}}\big]^*,
\end{equation}
by Thm. 3.13 of \cite{obataJFA} (or its fermionic analogue, Thm. \ref{obataJFA.Thm.3.13} of Subsection \ref{psiBerezin-Hida}).
Recall that here
\[
E_{{}_{n_k}} = \mathcal{S}(\mathbb{R}^3) \,\,\, \textrm{or} \,\,\,
E_{{}_{n_k}} = \mathcal{S}^{0}(\mathbb{R}^3).
\]
This fact can be understood as a consequence of (I)-(V) with the rigorous interpretation of the canonical commutation rules for free
fields realized through the Hida operators. Now we consider a sequence $g_j \in \mathscr{E}$, $j \in \mathbb{N}$, which converges to 
$1_{{}_{\textrm{R}^4}}$ in $\mathscr{E}^*$. This sequence cannot, of course, converge in $\mathscr{E}$, but can always be chosen 
in such a manner that it is bounded in $\mathscr{E}$. This is the first ``adiabatic'' condition put on the  limit 
$g_j \overset{j \rightarrow \infty}{\longrightarrow} 1_{{}_{\textrm{R}^4}}$ in $\mathscr{E}^*$. Let
$\big| \cdot \big|_{{}_{N}}$ be the Hilbertian norms defining the topology of $\mathscr{E}$, 
the boundedness of the sequence $\{g_j \}_{j \in \mathbb{N}}$
in $\mathscr{E}$ means that there exists
a sequence of finite numbers $C_{{}_{N}} < \infty$, such that
\[
\big| g_j \big|_{{}_{N}} < C_{{}_{N}}, \,\,\, j,N \in \mathbb{N}.
\]
Taking into account the form of
Hilbertian norms $\big| \cdot \big|_{{}_{N}}$ this boundedness, or the `adiabatic'' condition, means in fact that each higher order derivative
of $g_j$ is uniformly bounded in space-time variables and in the index $j$. Next, because each of the maps (\ref{continuous:g->kappa_lm(g)})
defined by the kernels $\kappa_{l,m}$ determining Fock expansion of $S_n$ is continuous, then the set of functionals
\[
\kappa_{l,m}(g_j) \in \big[ E_{{}_{n_1}} \otimes \ldots \otimes  E_{{}_{n_n}}\big]^*, \,\,\, j \in \mathbb{N},
\] 
is bounded in the strong dual topology of
\[
\big[ E_{{}_{n_1}} \otimes \ldots \otimes  E_{{}_{n_n}}\big]^*.
\]
Finally, because 
\[
E_{{}_{n_1}} \otimes \ldots \otimes  E_{{}_{n_n}}
\]
is countably Hilbert and nuclear Fr\'echet space, in particular 
\[
E_{{}_{n_1}} \otimes \ldots \otimes  E_{{}_{n_n}}
\]
is perfect (in the sense of \cite{GelfandII}), then the sequence of functionals
\[
\kappa_{l,m}(g_j) \in \big[ E_{{}_{n_1}} \otimes \ldots \otimes  E_{{}_{n_n}}\big]^*, \,\,\, j \in \mathbb{N},
\] 
possesses a subsequence
\[
\kappa_{l,m}(g_{j_k}) \in \big[ E_{{}_{n_1}} \otimes \ldots \otimes  E_{{}_{n_n}}\big]^*, \,\,\, k \in \mathbb{N},
\] 
which is strongly convergent to an element, say
\[
\kappa_{l,m}\big(1_{{}_{\textrm{R}^4}}\big) \in \big[ E_{{}_{n_1}} \otimes \ldots \otimes  E_{{}_{n_n}}\big]^*,
\]
compare  \cite{GelfandII}. By Proposition 3.9 of \cite{obataJFA} (or its fermionic analogue, compare Subsection \ref{psiBerezin-Hida}),
the operators 
\[
S_n\big(g_{j_k}\mathcal{L}\big) = \sum_{\substack{l, m\\
                  l+m =n }}
\Xi\big(\kappa_{l,m}(g_{j_k})\big)
\in \mathscr{L}\big( (\boldsymbol{E}), (\boldsymbol{E})^*\big), \,\,\, k \in \mathbb{N},
\]
converge in $\mathscr{L}\big( (\boldsymbol{E}), (\boldsymbol{E})^*\big)$ to an operator
\[
S_n\big((g=1)\mathcal{L}\big) = \sum_{\substack{l, m\\
                  l+m =n }}
\Xi\Big(\kappa_{l,m}\big(1_{{}_{\textrm{R}^4}}\big)\Big)
\in \mathscr{L}\big( (\boldsymbol{E}), (\boldsymbol{E})^*\big).
\]

Now, because this operator $S_n\big((g=1)\mathcal{L}\big)$ belongs to $\mathscr{L}\big( (\boldsymbol{E}), (\boldsymbol{E})^*\big)$,
then it defines distributional ``matrix elements''  which, by translational invariance of the causal distributions 
(and their retarded and advanced parts in the causal higher order perturbatie contribution $S_n$) have the general form
\[
\Big \langle S_{n}\big((g=1)\mathcal{L}\big) \Phi_{{}_{\ldots s,\boldsymbol{\p} \ldots }}, 
\Psi_{{}_{\ldots s',\boldsymbol{\p}' \ldots }}\Big\rangle = 
\delta(\Sigma p - \Sigma p') F(\ldots sp \ldots, \,\,\, \ldots s'p' \ldots).
\]

Of course, it is \emph{a priori} possible that using other subsequence, we will obtain another limit operator
$S_n\big((g=1)\mathcal{L}\big) \in \mathscr{L}\big( (\boldsymbol{E}), (\boldsymbol{E})^*\big)$. But the point is that 
keeping the adiabatic character of the limit, \emph{i.e.} with the bounds $C_{{}_{N}}$ for the derivatives
of the limiting sequence $g_j$ as small as possible, we should get ``matrix elements''
\[
\Big \langle S_{n}\big((g=1)\mathcal{L}\big) \Phi_{{}_{\ldots s,\boldsymbol{\p} \ldots }}, 
\Psi_{{}_{\ldots s',\boldsymbol{\p}' \ldots }}\Big\rangle = 
\delta(\Sigma  p -   \Sigma p') \, F'(\ldots sp \ldots, \,\,\, \ldots s'p' \ldots).
\]
with the functions $F'$ which give essentially the same cross-sections.
 This is the case for QED as we know by the general
form of the ``matrix elements'' (\ref{<PlaneWaveS(gnot1),PlaneWave>}) and their dependence on $g \in \mathscr{E}$. Namely,
it is not exactly the function $F'$ of the last formula which plays the role in practical computation of the effective-cross
section, but the function $F$ in the formula
\[
\Big \langle S_{n}\big(g\mathcal{L}\big) \Phi_{{}_{\ldots s,\boldsymbol{\p} \ldots }}, 
\Psi_{{}_{\ldots s',\boldsymbol{\p}' \ldots }}\Big\rangle \approx 
\widetilde{g}(\Sigma  p -   \Sigma p') \,  F(\ldots sp \ldots, \,\,\, \ldots s'p' \ldots),
\,\,\, g \in \mathscr{E},
\]
which is only approximately valid for $\widetilde{g}$ tending to  the Dirac $\delta$-function in $\mathscr{E}^*$, which is used
in the computation of the cross-section. We extract the function $F$ before passing to the adiabatic limit $\widetilde{g} \rightarrow
\delta$ (or $g\rightarrow 1_{{}_{\textrm{R}^4}}$) in $\mathscr{E}^*$.
It is therefore not the (in principle non-unique) limit operator $S_n\big((g=1)\mathcal{L}\big) \in \mathscr{L}\big( (\boldsymbol{E}), (\boldsymbol{E})^*\big)$,
which matters in the computation of the effective cross-section in the adiabatic limit, but the dependence of the expression (\ref{<PlaneWaveS(gnot1),PlaneWave>})
on $g \in \mathscr{E}$. Having given the last approximate expression for the ``matrix elements'' we use the following approximation for
\[
\Big|\Big \langle S_{n}\big(g\mathcal{L}\big) \Phi_{{}_{\ldots s,\boldsymbol{\p} \ldots }}, 
\Psi_{{}_{\ldots s',\boldsymbol{\p}' \ldots }}\Big\rangle \Big|^2 \approx 
{\textstyle\frac{VT}{(2\pi)^4}}
\delta(\Sigma  p -   \Sigma p') \, |F(\ldots sp \ldots, \,\,\, \ldots s'p' \ldots)|^2,
\] 
in the formula for the effective cross-section, compare \cite{Bogoliubov_Shirkov}, \S\S 24.5 and 25. 

Concerning the scattering process with many-particle plane-wave states of elementary fields, we expect that the space-time domain $V \times T$ 
on which the behavior of $g$ in (\ref{<PlaneWaveS(gnot1),PlaneWave>}) is crucial, depends on the interaction 
Lagrange density of the theory. The domain the bahavior of $g$ is expected to be crucial, should correspond to the domain of the asymptotic freedom
of the theory in question, if we are about to give cross-sections which make physical sense. 
For QED we expect that the limit $g=1$ is more essential in comparison with strong interactions in the perturbative approach. For strong interactions we expect that
the perturbative approach is justified at high energy limit, and for the scattering at the level of the many-particle plane wave states of the elementary fields
these results which have physical sense are pertinent to small space-time domains on which the behavior of $g$ is crucial in this case.
In this situation the limit $g=1$ (in perturbative approach) has only abstract character without any immediate physical meaning. 
When considering the problem of the adiabatic 
limit for strong interactions on the Minkowski space-time, we should remember that only the small domains are crucial there on which 
$g$ should be put equal $1$.  Trying to put $g=1$ (within perturbative approach) at the regions lying far apart from the scattering region 
seems to have no immediate physical sense in this case.

\subsection{Bound states. Stable and meta-stable particles}\label{BoundStates}

The bound state problems cannot be entirely reduced to the properties of the generalized scattering operator in action on the 
generalized un-normalizable plane wave states of elementary fields. Here only the essentially ``local part'' of the problem can be treated on Minkowski space-time 
within the causal method using auxiliary switching off function $g$ in the scattering operator
functional $g \rightarrow  S(g\mathcal{L})$, where it is sufficient only in the construction of the so-called many-particle (or higher order) Green functions.
On the Minkowski spacetime the many-particle wave functions (describing the corresponding many-particle bound states), canonically related to these Green functions, 
are only heuristically 
related to actual well-defined states of the Fock space in which the interacting fields act as generalized operators.  

In fact the dependence
on $g$ of the scattering operator and of interacting fields, $\boldsymbol{\psi}_{{}_{\textrm{int}}}(g,x)$, 
$A_{{}_{\textrm{int}}}^{\mu}(g,x)$, and more generally for other observables $\mathbb{A}_{{}_{\textrm{int}}}(g,x)$  with $\mathbb{A}$ being equal to a Wick
polynomial in free fields (say Noether conserved currents in free fields),
is too much singular and cannot be eliminated to the same extent as in the computation of the effective cross section for the many particle plane
wave states of the free elementary fields underlying the particular QFT in question. On the Minkowski spacetime we are able only to compute 
the higher order Green functions, without the need for renormalization or infrared infinities, using the causal perturbative construction of the scattering operator
and Hida white noise operators. But the relation of the corresponding many-particle wave functions to well-defined states of the Fock
space is lost. This is related to the fact that the interacting fields 
$\boldsymbol{\psi}_{{}_{\textrm{int}}}(g=1,x)$, 
$A_{{}_{\textrm{int}}}^{\mu}(g=1,x)$, or more generally  $\mathbb{A}_{{}_{\textrm{int}}}(g=1,x)$, on the Minkowski space-time, 
are quite singular generalized operators, such that
\[
\boldsymbol{\psi}_{{}_{\textrm{int}}}(g=1), A_{{}_{\textrm{int}}}^{\mu}(g=1), 
\mathbb{A}_{{}_{\textrm{int}}}(g=1) \in \mathscr{L}\big( (\boldsymbol{E}) \otimes \mathscr{E}, (\boldsymbol{E})^* \big).
\]
Similarly, the scattering operator $S(g=1)$ on the Minkowski space-time, although being essentially uniquely determined (as we have explained in Subsection \ref{EffCrossSection}), is quite singular and
\[
S(g=1) \in \mathscr{L}\big( (\boldsymbol{E}), (\boldsymbol{E})^* \big),
\,\,\,\,
S(g=1) \notin \mathscr{L}\big( (\boldsymbol{E}), (\boldsymbol{E}) \big)
\]
and it is not an ordinary operator in the Fock space (this is the case for the separate higher order 
contributions $S_n(g=1)$ to $S(g=1)$).  The stable particles should correspond to the normalizable -- \emph{i.e.} bound -- states of the temporal 
translation Noether generator computed for the interacting fields.
The meta-stable particles, which decay with the decay process understood as caused by the interaction, 
cannot be exactly equal to bound eigenstates of the
temporal translation Noether generator computed for the interacting fields (say of the hamiltonian including interaction), but rather
of some other intermediate operator (\emph{i.e.} of the Noether translation generator computed for free fields, which exists on the Minkowski 
space-time, but strictly speaking it is more likely to correspond to the local time symmetry flow which can be identified
with an actual space-time symmetry of a static space-time -- the case in which the bound state problem can be treated perturbatively, 
compare the next Subsection). 

Here we should emphasize that the higher order Green functions can be computed, and this can be done without infrared and ultraviolet infinities, 
using the presented above causal
construction of the scattering generalized operator.  In particular the radiative corrections, say to the bound state
of the positronium can be computed, provided we can use the positron-electron Green function as defining the corresponding wave function of the positron-electron
interacting system from the Cauchy data on the Cauchy surface. But a serious problem arises in giving mathematical justification for 
such electron-positron wave function as corresponding to a superposition of well defined bound eigenstates of the  
temporal translation Noether generator computed for the interacting fields. This is because the last operator is not well-defined 
and is not ordinary self adjoint operator acting in the Fock space.

The problem of explanation of the very existence of stable (or meta-stable) states is beyond the QFT theory on the Minkowski space-time
with the realistic Lagrange density interactions. 
In particular the very existence of the meta-stable bound state corresponding to the positronium, lies beyond the QED on the Minkowski space-time,
with the ordinary gauge Lagrange interaction.

The same situation we have for the radiative corrections to bound states, in particular the Lamb shift, or vacuum  polarization, computed by Schwinger.
Here the radiative corrections weakly depend on the concrete form of the switch-off function $g$ at the remote parts of space-time, but no sensible finite value to the correction could have been reached without putting the adiabatic switch-off $g=0$  at the remote past and remote future (at infinity)
in the intermediate stages of the computation. Nonetheless, the concrete values of the corrections
do not depend on the  concrete form of the adiabatic switch-off $g$ at infinity and, similarly as for the cross-section, dependence on $g$
in the final formulas for the Green functions can be eliminated. 
The crucial point, however, lies in giving a rigorous identification of the many-particle wave function determined by the higher order Green function
with a well-defined superposition of bounded eigenstates in the Fock space of the temporal translation Noether generator computed for the interacting fields.
This identification cannot be reached within the perturbative causal method on the flat Minkowski space-time and with Hida operators as the creation-annihilation operators. The radiative corrections to bound states we discuss in the next Subsection.

In this Subsection we explain why the operator of the  
temporal translation Noether generator computed for the interacting fields is not equal to any ordinary self-adjoint operator in the Fock space,
which is essential if one wants to account for the very existence of meta-stable
and stable particles.

Passing to the bound state problem, we need to consider the Noether conserved currents corresponding to translations,
and the Noether generators -- the equal time integrals of the corresponding conserved currents -- the translation generators
$\boldsymbol{P}^\mu$, $\mu = 0,1,2,3$, \emph{i.e.} the conserved integrals. In case of free fields we have already shown
that indeed the translation generators $\boldsymbol{P}^\mu$ of the representation of the double cover of the Pincar\'e group acting
in the Fock space is indeed equal to the integral over the Cauchy equal-time surface of the currents
$\mathbb{A} = \,\,\, {:}T^{0\mu}{:}$, expressed as the Wick polynomials of the free fields, which are the quantum counterparts of the classic conserved currents 
$T^{0\mu}$ of free fields. 
In fact this equality can also serve as the Quantization Postulate for free quantum fields, compare \cite{Bogoliubov_Shirkov}.
In passing to the analogue of this equality for interacting fields we need to have the quantum interacting analogue of the conserved current
$\mathbb{A} = \,\, \boldsymbol{{:}} T^{0 \mu} \boldsymbol{{:}}$.  We have learned after \cite{Bogoliubov_Shirkov}
how the interacting counterpart $\mathbb{A}_{\textrm{int}}$ can be computed for any Wick polynomial $\mathbb{A}$ of free fields.
In the previous Subsection we have given the general construction of $\mathbb{A}_{\textrm{int}}$, corresponding to any 
Wick polynomial in free fields. Here we apply it to the Wick polynomial $\mathbb{A}$ of free fields which is the 
analogue of the classic conserved current $T^{0\mu}$ for the classic \emph{interacting} fields, which necessary includes
the interaction Lagrange density $\mathcal{L}$.
We have given the general formula for $\mathbb{A} = \,\,\, {:}T^{0\mu}{:}$ including such $\mathbb{A} = \,\, \boldsymbol{{:}} T^{0 \mu} \boldsymbol{{:}}$
in the previous Subsection. Having given $\mathbb{A}_{\textrm{int}} = \big(\boldsymbol{{:}} T^{0 \mu} \boldsymbol{{:}} \big)_{{}_{\textrm{int}}}$ 
we compute the generators
\begin{equation}\label{PmuInteractingGeneral}
\boldsymbol{P}_{\textrm{int}}^{\mu} = \int \big(\boldsymbol{{:}} T^{0 \mu} \boldsymbol{{:}} \big)_{{}_{\textrm{int}}}(g=1; t, \boldsymbol{\x}) 
\, \ud^3 \boldsymbol{\x}
\end{equation}
corresponding to the system of interacting fields.
Here the interaction Lagrangian term $\mathcal{L}$ understood as a Wick polynomial in
free fields is included into the full Lagrangian $\mathcal{L}_{{}_{\textrm{Full}}}$ (expressed as a Wick polynomial in free fields) 
when computing Wick polynomial ${:}T^{0\mu}{:}$. Thus ${:}T^{0\mu}{:}$ is put equal 
to the Wick formal counterpart of the classical expression for $T^{0\mu}$ in the Noether theorem for classical interacting fields, in which the classical fields are replaced by free fields and their product replaced formally by the Wick product of free fields. 
Such Wick polynomial ${:}T^{0\mu}{:}$ including the Wick interaction term is then inserted into 
the Bogoliubov-Shirkov-Stepanov perturbative formula \cite{Bogoliubov_Shirkov}, \S\S 40.2 and 40.5:
\begin{multline}\label{TmunuInteractingGeneral}
\big(\boldsymbol{{:}} T^{0 \mu} \boldsymbol{{:}} \big)_{{}_{\textrm{int}}}(g; x) =  S^{-1}(g\mathcal{L})
\frac{\delta S(g\mathcal{L} \,\, + \,\, h\boldsymbol{{:}} T^{0 \mu} \boldsymbol{{:}} \,\,)}{\delta h(x)}\Bigg{|}_{{}_{h=0}}
= \\
= \,\,\,\,\, \boldsymbol{{:}} T^{0 \mu} \boldsymbol{{:}}(x) + \sum \limits_{n=1}^{\infty} {\textstyle\frac{1}{n!}} 
\int \ud^4 x_1 \cdots \ud^4 x_n \big(\boldsymbol{{:}} T^{0 \mu} \boldsymbol{{:}} \big)^{(n)}(x_1, \ldots, x_n,x) \,
g(x_1) \cdots g(x_n).
\end{multline}
Note that the conserved current $\boldsymbol{{:}} T^{0 \mu} \boldsymbol{{:}}(x)$ for $\mu=0$ includes the 
interaction Lagrange density $\mathcal{L}(x) = \boldsymbol{{:}} \boldsymbol{\psi}^{\sharp}(x) \gamma_\mu \boldsymbol{\psi}(x) A^\mu(x)\boldsymbol{{:}}$
(in case of spinor QED). 
Here we only mention that the above Bogoliubov-Shirkov-Stepanov construction  is an immediate generalization 
of the local quantities introduced by Schwinger \cite{Schwinger}.

The stable and meta-stable particles correspond to the bound, and thus normalizable, eigensates of the operator 
$\boldsymbol{P}_{\textrm{int}}^{0}$. Now for free fields the operators $\boldsymbol{P}^{\mu}$ (including
$\boldsymbol{P}^{0}$) are indeed ordinary self adjoint operators on the Fock space of free fields, and even operators 
which transform continuously the test Hida dense space into itself, as we have proved in Subsection \ref{BSH} (for the free electromagnetic
potential field) or in Subsection \ref{StandardDiracPsiField} (for the free Dirac field), from which it follows 
for the joint system of the free electromagnetic potential field and the free Dirac field, as we have already mentioned in 
Subsection \ref{G} of Introduction. Now because the (higher order contributions to the) interacting electromagnetic potential and Dirac fields 
preserve their meaning of (finite sums  of) integral kernel operators with vector valued kernels (in the sense of \cite{obataJFA})
even with $g=1$ (for the proof compare Subsection \ref{OperationsOnXi} and Introduction to this Section) and are no more singular 
than the Wick products of free massless fields, the reader could have a hope that
the interacting integral conserved currents will preserve the meaning of ordinary (unbounded) operators on the 
Fock space when integrated over the equal time surface. This hope could also come from the fact that this is so 
for free fields. But this not the case and the higher order contributions to 
$\boldsymbol{P}_{\textrm{int}}^{0}$ are not ordinary (densely) defined operators on the Fock space if $g=1$,
in particular 
\[
\textrm{higher order contributions to} \,\, \boldsymbol{P}_{\textrm{int}}^{0} \notin \mathscr{L}\big( (\boldsymbol{E}), (\boldsymbol{E})\big)
\,\,\, \textrm{if} \,\, g=1.
\]
The higher order contributions to $\boldsymbol{P}_{\textrm{int}}^{0}$ preserve only the meaning of generalized 
operators which transform the test Hida space into its strong dual, namely
\[
\textrm{higher order contributions to} \,\, \boldsymbol{P}_{\textrm{int}}^{0} \in \mathscr{L}\big( (\boldsymbol{E}), (\boldsymbol{E})^* \big)
\,\,\, \textrm{if} \, g=1.
\]

The first assertion is an immediate consequence of the following 
\begin{lem*}
The operator 
\[
\int \mathcal{L} (t, \boldsymbol{\x}) 
\, \ud^3 \boldsymbol{\x} =
\int \boldsymbol{{:}} \boldsymbol{\psi}^{\sharp}(t, \boldsymbol{\x}) \gamma_\mu \boldsymbol{\psi}(t, \boldsymbol{\x}) A^\mu(t, \boldsymbol{\x})\boldsymbol{{:}}
\, \ud^3 \boldsymbol{\x}
\]
does not belong to 
\[
\mathscr{L}\big( (\boldsymbol{E}), (\boldsymbol{E})\big),
\]
but it is a finite sum of  well defined integral kernel operators belonging to
\[
\mathscr{L}\big( (\boldsymbol{E}), (\boldsymbol{E})^*\big).
\]
Summing up
\[
\int \mathcal{L} (t, \boldsymbol{\x}) 
\, \ud^3 \boldsymbol{\x}  \in \mathscr{L}\big( (\boldsymbol{E}), (\boldsymbol{E})^*\big)
\,\,\,
\textrm{but}
\,\,\,
\int \mathcal{L} (t, \boldsymbol{\x}) 
\, \ud^3 \boldsymbol{\x}  \notin \mathscr{L}\big( (\boldsymbol{E}), (\boldsymbol{E})\big).
\]
\end{lem*}
\qedsymbol \,
Indeed, using the Rules of Subsection \ref{OperationsOnXi} we immediately compute
the integral scalar-valued kernels $\kappa_{\ell,m}$, $\ell, m \in \{0,1,2,3 \}$, $\ell+m = 3$, 
of this integral kernel operator. It is immediately seen from the explicit form of the kernel
$\kappa_{3,0}$ expressed through the Dirac delta distribution, that
\[
\kappa_{3,0} \notin \mathscr{L}(E_{1}^{*} \otimes E_{1}^{*} \otimes E_{2}^{*}, \mathbb{C})
\cong E_{1} \otimes E_{1} \otimes E_{2},
\]
but 
\begin{multline*}
\kappa_{\ell,m} \in \mathscr{L}( E_{n_1} \otimes \ldots \otimes E_{n_\ell} \otimes E_{n_\ell+1}  \otimes \ldots E_{n_{\ell+m}}, \mathbb{C})
\\
\cong E_{n_1}^{*} \otimes \ldots \otimes E_{n_\ell}^{*} \otimes E_{n_\ell+1}^{*}  \otimes \ldots E_{n_{\ell+m}}^{*},
\,\,\, \ell+m = 3,
\end{multline*}
\[
 E_{1} = \mathcal{S}(\mathbb{R}^3), \,\,\,\, E_{2} = \mathcal{S}^{0}(\mathbb{R}^3).
\,\,\, n_i \in \{1,2\}
\]
Whence by Thm. 2.2 and 2.6 of \cite{hida}, or their generalization -- Thm. \ref{Xi_l,m} and \ref{Xi_l,m:Hida->Hida} 
of Subsection \ref{psiBerezin-Hida}, we obtain the assertion.
\qed

Because already the  contribution $\mathcal{L}$ enters additively into $\boldsymbol{{:}} T^{0 0} \boldsymbol{{:}}$,
which in turn enters into $\big(\boldsymbol{{:}} T^{0 0} \boldsymbol{{:}} \big)_{{}_{\textrm{int}}}$
(\emph{i.e.} the first order contribution in the coupling constant, multiplying the Lagrange interaction density function $\mathcal{L}$)
as the zero order contribution, then the first order contribution to $\boldsymbol{P}_{\textrm{int}}^{0}$ 
includes additively the operator 
\[
\int \mathcal{L} (t, \boldsymbol{\x}) 
\, \ud^3 \boldsymbol{\x} =
\int \boldsymbol{{:}} \boldsymbol{\psi}^{\sharp}(t, \boldsymbol{\x}) \gamma_\mu \boldsymbol{\psi}(t, \boldsymbol{\x}) A^\mu(t, \boldsymbol{\x})\boldsymbol{{:}}
\, \ud^3 \boldsymbol{\x}.
\]
Thus, the assertion that the first order contribution to
$\boldsymbol{P}_{\textrm{int}}^{0}$ cannot be equal to an operator on the Fock space transforming continuously the Hida space into itself,
but only the Hida space into its strong dual, follows. Similar assertion holds true for the higher order contributions, as it is easily seen
that the kernels of the higher orders always belong to the distribution spaces
\[
\kappa_{\ell, m} \in E_{n_1}^{*} \otimes \ldots \otimes E_{n_\ell}^{*} \otimes E_{n_\ell+1}^{*}  \otimes \ldots E_{n_{\ell+m}}^{*}
\]
but in general
\[
\kappa_{\ell, m} \notin E_{n_1} \otimes \ldots \otimes E_{n_\ell} \otimes E_{n_\ell+1}^{*}  \otimes \ldots E_{n_{\ell+m}}^{*}.
\]

Therefore, we have the following
\begin{prop*}
\[
\textrm{higher order contributions to} \,\, \boldsymbol{P}_{\textrm{int}}^{0} \notin \mathscr{L}\big( (\boldsymbol{E}), (\boldsymbol{E})\big)
\,\,\, \textrm{if} \,\, g=1
\]
The higher order contributions to $\boldsymbol{P}_{\textrm{int}}^{0}$ preserve only the meaning of generalized 
operators which transform the test Hida space into its strong dual, namely
\[
\textrm{higher order contributions to} \,\, \boldsymbol{P}_{\textrm{int}}^{0} \in \mathscr{L}\big( (\boldsymbol{E}), (\boldsymbol{E})^* \big)
\,\,\, \textrm{if} \, g=1.
\]
\end{prop*}

In particular, even if we restrict our consideration only to a finite number of higher order contributions
to the generator  $\boldsymbol{P}_{\textrm{int}}^{0}$, still it remains to be an operator transforming continuously the Hida space into
its strong dual. The notion of bounded (normalizable) eigenstate for such generalized operator does not make any sense.
Here we cannot repeat the approach used in computation of the effecitve cross-section, and compute first the higher order contributions
keeping $g$ within the test space $\mathscr{E}$, and pass to the ``adiabatic limit'' at the end, because here the limits
of the higher order contributions do not make sense of ordinary operators in the limit $g=1$, but only generalized operators
which transform the Hida space into the space strongly dual to it. 

Possibly one could hope that keeping the switching off function  $g$ within the test space $\mathscr{E}$, we can somehow save
the ordinary operator meaning of the total energy of the system, appropriately modified due to the presence of the 
switching off factor $g$ modifying the Poincar\'e invariance of the theory.
Namely, after Bogoliubov, we can consider a class of allowed $g$ which does not include the constant function. This modifies
the Poincar\'e invariance of the theory, destroyed by the switching off function $g$, which modifies the invariant Lagrange density of interaction, 
with the corresponding and natural modification of the generators
\begin{equation}\label{PintOng-ModifiedSpacetime}
\boldsymbol{P}_{\textrm{int}}^{\mu}(g)
=
\boldsymbol{P}^{\mu} -
\int H(g,x) \,
{\textstyle\frac{\partial g}{\partial x_\mu}}
\, \ud^4 x.
\end{equation}
Here 
\[
H(g,x) = S^{-1}(g\mathcal{L})
\frac{\delta S(g\mathcal{L})}{\delta g(x)}, \,\,\,\,\, g \in \mathscr{E},
\]
is the generalized hamiltonian density function, introduced in \cite{Bogoliubov_Shirkov}, \S 20.5 and 39.
In $\boldsymbol{P}_{\textrm{int}}^{\mu}(g)$ the switching off function $g\neq 1$, 
compare \cite{Bogoliubov_Shirkov}, \S 40.1. Next we notice high insensitivity 
of the observed quantities on the manner in which the function $g$ tends to zero in the remote past and remote future, 
\cite{Bogoliubov_Shirkov}, \S 40.1. 

Accordingly, the equations of motion for the interacting fields $A_{\textrm{int}}^{\mu}(g,x)$, $\boldsymbol{\psi}_{\textrm{int}}^{a}(g,x)$
with $g \in  \mathscr{E}$ are modified with the presence of the factor $g \neq 1$ replacing the coupling constant, which reflects
the modification of the Poincar\'e invariance, modified by the factor $g \neq 1$, compare \cite{DKS1}, \cite{DutFred}. The formula (\ref{PintOng-ModifiedSpacetime}) 
for the generators is also modified by $g$, which for $g\neq 1$ modifies the Poincar\'e invariance, \cite{Bogoliubov_Shirkov}, \S 40.1. 
Nonetheless, there is a natural relation between
the generators  (\ref{PmuInteractingGeneral}) and the modified generators (\ref{PintOng-ModifiedSpacetime}), where the formula
(\ref{PintOng-ModifiedSpacetime}) with $g$ converging to the step function in $\mathscr{E}^*$ should converge 
to the formula (\ref{PmuInteractingGeneral}), but although each $\boldsymbol{P}_{\textrm{int}}^{\mu}(g) \in 
\mathscr{L}\big( (\boldsymbol{E}), (\boldsymbol{E})\big)$ for $g \in  \mathscr{E}$, the limit cannot converge in 
$\mathscr{L}\big( (\boldsymbol{E}), (\boldsymbol{E})\big)$
but only in $\mathscr{L}\big( (\boldsymbol{E}), (\boldsymbol{E})^*\big)$.

However, this modification of the ``generators'' (\ref{PintOng-ModifiedSpacetime}) does not help much, because the higher order
contributions to the operator
$\boldsymbol{P}_{\textrm{int}}^{0}(g)$ in (\ref{PintOng-ModifiedSpacetime}) still behave singularly even if we keep  $g \in \mathscr{E}$ fixed.
This can be immediately seen by noticing that the first order contribution to $\boldsymbol{P}_{\textrm{int}}^{0}(g)$
is equal
\[
S_1\big(g{\textstyle\frac{\partial g}{\partial x_\mu}}\big)
\]
which, by the results of Subsections \ref{OperationsOnXi} and \ref{WickForChronological}, is equal to a generalized operator belonging to
\[
\mathscr{L}\big( (\boldsymbol{E}), (\boldsymbol{E})^*\big),
\]
but not to
\[
\mathscr{L}\big( (\boldsymbol{E}), (\boldsymbol{E})\big).
\]
Thus even if we keep  $g \in \mathscr{E}$ fixed the higher order contributions to the modified generators  $\boldsymbol{P}_{\textrm{int}}^{\mu}(g)$
remain to be singular integral kernel operators, and transform continuously the Hida space into its strong dual, namely
\[
\textrm{higher order contributions to} \,\, \boldsymbol{P}_{\textrm{int}}^{\mu}(g) \notin \mathscr{L}\big( (\boldsymbol{E}), (\boldsymbol{E})\big)
\,\,\, \textrm{even if} \,\, g \in \mathscr{E},
\]
but it holds true only the following assertion:
\[
\textrm{higher order contributions to} \,\, \boldsymbol{P}_{\textrm{int}}^{\mu}(g) \in \mathscr{L}\big( (\boldsymbol{E}), (\boldsymbol{E})^*\big)
\,\,\, \textrm{if} \,\, g \in \mathscr{E}.
\]
In particular $\boldsymbol{P}_{\textrm{int}}^{0}(g)$, $g \in \mathscr{E}$, cannot serve as a basis for the qualitative
analysis of the bound state problem, at least in the theories with interactions with relatively large domains of asymptotic freedom,
where $g=1$ can sensibly be put equal $1$ on large space-time domains, e.g. QED. But this is the case also because the whole perturbative 
series for $\boldsymbol{P}_{\textrm{int}}^{0}(g)$ and for $S(g\mathcal{L})$, with fixed $g \in \mathscr{E}$, 
cannot be convergent in the stronger sense, which allows to keep the meaning of the limit operators $\boldsymbol{P}_{\textrm{int}}^{0}(g)$ and $S(g\mathcal{L})$
 to be  elements of the space $\mathscr{L}\big( (\boldsymbol{E}), (\boldsymbol{E})\big)$. To the contrary the  higher order contributions
to the scattering operator  $S\big((g=1)\mathcal{L}\big)$ and to interacting fields in the adiabatic limit are well defined finite sums of integral kernel operators
with vector valued kernels, and we can rather expect their convergence to the limit generalized operators with vector valued kernels
in the sense explained in the introduction to this Section.  Convergence of the perturbative series for the scattering operator 
$S(g\mathcal{L})$ even with fixed $g \in \mathscr{E}$  to an operator in $\mathscr{L}\big( (\boldsymbol{E}), (\boldsymbol{E})\big)$
is impossible simply because each higher order contribution already does not belong to $\mathscr{L}\big( (\boldsymbol{E}), (\boldsymbol{E})\big)$
but only to $\mathscr{L}\big( (\boldsymbol{E}), (\boldsymbol{E})^*\big)$, compare Subsections \ref{OperationsOnXi} and \ref{WickForChronological}.
This is what has been suspected by physicists for a long time, compare e.g. \cite{Kazakov_Shirkov}.
Similarly convergence of the perturbative series for 
$\boldsymbol{P}_{\textrm{int}}^{0}(g)$ with fixed $g \in \mathscr{E}$  to an element of $\mathscr{L}\big( (\boldsymbol{E}), (\boldsymbol{E})\big)$
is impossible because the separate higher order contributions to $\boldsymbol{P}_{\textrm{int}}^{0}(g)$  do not belong to  
$\mathscr{L}\big( (\boldsymbol{E}), (\boldsymbol{E})\big)$
but only to  $\mathscr{L}\big( (\boldsymbol{E}), (\boldsymbol{E})^*\big)$. Because of these singular behavior of the scattering
operator, of the interacting fields and of the conserved currents on the Minkowski space-time, we cannot give rigorous meaning to the many-particle wave function 
(determined by the many-particle Green function-propagator) as corresponding to a well defined state of the Fock space, which is heuristically outlined 
in  \cite{Bogoliubov_Shirkov}, Chap. VII, \S 42.  

Thus, the only role of the auxiliary switching-off function $g \in \mathscr{E}$ is to implement the causality condition of the axioms
(I)-(V) for the construction of the scattering generalized functional $g \mapsto S(g \mathcal{L})$ set up in Subsection \ref{WickForProduct} 
and in the introduction to this Section. The Green functions, similarly as the effective cross sections in Subsection \ref{EffCrossSection}, 
can be computed with the help of this
scattering functional, and the dependence of the Green functions on the auxiliary function $g$ can be eliminated, similarly as for the 
cross section, which we explain in more details in the next Subsection. However rigorous relation of the ``many-particle'' Green function
to a well defined state in the Fock space is untenable on the Minkowski space-time, because of the singular behavior of the Noether 
conserved current integrals for interacting fields and of the scattering operator itself, which (even for fixed $g \in \mathscr{E}$)
is not an ordinary operator on the Fock space. 

Nonetheless, because the radiative corrections to the bound states computed from the higher order (\emph{i.e.} ``many-particle'') 
Green functions are essentially insensitive to the manner in which the function $g$ tends to zero at infinity 
shows that the results obtained with the help of Green functions have some deeper physical meaning. We think that this is so, because the results which we achieve in this manner for the radiative corrections to the bound states, vacuum polarization, anomalous magnetic moment of the electron, radiative corrections
to the meta-stable bound states of the positronium (in case of spinor QED), \emph{e.t.c.},
agree with experiment to a very high level of accuracy (in case in which we can safely control and separate contributions coming from
electromagnetic interactions). We should emphasize here again that these experiments depend essentially on the local
properties of the many-particle wave functions of bound states determined by the corresponding Green functions (propagators). This is because they essentially are 
concerned with the corrections to the lowest bound states close to the ground states, insensitive to the behavior of the wave function at the remote 
parts of space-time. Thus, similarly as in the high energy scattering process, we again probe experimentally essentially only the local 
behaviour of the many particle wave function determined by the corresponding Green function. Thus the results obtained with the help of Green functions 
are again sensitive to the local properties of the wave many-particle functions determined by them.  

We would like to understand agreement of radiative corrections to bound states computed with the help of these Green functions with experiments without 
introducing any extra laws and keep only the canonical commutation relations
for free fields of the theory and the causal perturbative method. In order to achieve this we observe that: 1) the causal axioms for the scattering operator 
make sense on more general globally causal space-time, not necessary the Minkowski space-time, 2) the canonical commutation laws for free fields can be applied for more general space-time (at least for a wide class of space-times -- including those with compact Cauchy surfaces -- where we know how to solve the problem 
of division of solutions into ``positive and negative frequency solutions'') and 3) the scattering operator on space-times with compact Cauchy surfaces 
and non-zero curvature behaves much more regularly and becomes an ordinay operator in the Fock space, 4) the Noether integral for interacting fields
(if the space-time possesses a one-parameter symmetry group) becomes ordinary self-adjoint operator, 5) the said experiments probe only the local properties
of the space-time.  These circumstances allow  us to  recover rigorously relation of many-particle Green functions and the corresponding wave functions to
well defined states of the Fock space and to justify Schwinger's method of Green functions within QED, with the only proviso that we have to change the space-time into some other with compact Cauchy surfaces and use the Hida operators as the creation-annihilation operators of free fields underlying the theory. We do it explicitly for the concrete example of the Einstein static Universe space-time in Section \ref{EUandG}. 
We propose to look at the insensitivity of the results obtained with Green functions on the behavior of the switching off factor $g \neq 1$ at the remote parts of space-time, as indication that the geometry of the remote part of space-time itself is irrelevant for the structure
of Green functions computed from the scattering operator, at least for the actual experiments and bound state problems treated with Green functions method.
In fact in the experiments probing the radiative correction to the fine and hyperfine structure of the lowest bound states of atoms only the local behavior of the many particle wave functions determined by the Green functions is essential. But on the other hand the interacting fields on the space-times with compact Cauchy surface (e.g. Einstein Universe) behave much more regularly
with the Noether conserved current (\ref{PmuInteractingGeneral}) corresponding to time translation (if the space-time is static, like Einstein Universe) 
being equal to ordinary self-adjoint operator in the Fock space of free fields, and with the scattering operator $S\big((g=1) \mathcal{L}\big)$ equal
to an ordinary operator in the Fock space, which allows us to convert the heuristic relation\footnote{In \cite{Bogoliubov_Shirkov}, Chap. VII, \S 42.} 
of the many particle bound state determined by the
many-particle Green function with a well defined state in the Fock space, into a mathematically rigorous construction. 
Indeed, the scattering operator and the axioms (I)-(V)
of introduction to this Section (compare also Subsections \ref{MotivationForHida} and \ref{WickForProduct}) 
make sense on more general globally causal space-times, and we present detailed construction of the 
scattering operator and of the interacting fields on the Einstein Universe in Section \ref{EUandG}.
 
The theory determined by the axioms (I)-(V) on the Minkowski space-time given in the  introduction to this Section 
and Subsection \ref{WickForProduct} is applicable to the scattering phenomena
with the many particle plane wave states of the elementary fields as the initial and final states. This theory can also be applied to the other
class of generalized states -- the infrared states as will be explained in Section \ref{infra}.  Now the scattering phenomena
involving the many particle plane wave states are insensitive to the global geometry of space-time, with the geometry being relevant within the rather small domains on which $g=1$ and where the theory is asymptotically free, where it is assumed to be flat 
Minkowski space-time. Similarly, the Coulomb law, valid for the Maxwell equations on flat Minkowski space-time, remains to
be applicable to infrared states which make sense in situations where the space-time can indeed be assumed flat on large parts of the space-time,
sufficient to the experimental identification of the infrared states in the Bremsstrahlung phenomena, so that  the homogeneous parts
of homogeneity $-1$ in the radiation can sensibly be indentified. Thus we obtain theory with well defined range of applicability which 
works well within this range. The same axioms (I)-(V) lead us to the conclusion that the interacting fields with $g=1$ are to much singular on the flat
Minkowski space-time to serve as a basis for explanation of the very existence of the stable and meta-stable particles. 
In particular the Noether conserved energy integral (\ref{PmuInteractingGeneral}) becomes a singular operator which cannot serve as a basis for the analysis of its 
bound eigenstates. 

On the other hand the axioms (I)-(V) given in the  introduction to this Section, with the Hida operators as the mathematical realization
of the canonical commutation relations for free fields (and, in expanded form, in Subsection \ref{WickForProduct}), 
make sense on more general globally causal space-times.
It is true at least for the space-time which -- as a smooth manifold -- is a Lie group with four distinguished one parameter right invariant
vector fields preserving causality and plying the role of translations, and which causally and periodically covers the Minkowski space-time. 
The simplest case we obtain when these distinguished one-parameter groups-vector fields are actual space-time symmetries.
The Einstein Universe is an example of such a globally causal space-time. Moreover, for such space-times with compact Cauchy surface  the
interacting fields and the scattering operator behave much more regularly, with the scattering operator $S(g=1)$ being an ordinary operator in the Fock space. 
Because the axioms (I)-(V) make sense on such more general globally causal  space-times we arrive at the conclusion
that the global geometry of the flat Minkowski space-time need to be modified in order to account for the very existence
of the stable and meta-stable particles numbered by discrete quantum numbers.
As we will see in Section \ref{EUandG} the conserved Noether integrals for interacting fields on the Einstein Universe space-time
behave so regularly that they can account for existence of stable and meta-stable particles, or bound states of the Noether integrals
for interacting fields,  where the conserved integrals become ordinary self-adjoint operators on the Fock space 
without any necessity of introducing any modifications with the extra switching off function $g \neq 1$ in the formula for these integrals. 
Similarly, the interacting fields with $g=1$ behave much more regularly on the Einstein Static Universe, in particular all massive
interacting fields become ordinary (densely) defined operators on the Fock space transforming continuously the Hida space into itself,
even when evaluated at fixed space-time point.

This conclusion should be treated seriously also for one more and important reason. The perturbative series for
(\ref{PmuInteractingGeneral}) cannot converge on the Minkowski space-time to an operator in $\mathscr{L}\big( (\boldsymbol{E}), (\boldsymbol{E})\big)$ 
even in theories with relatively weak interactions, like QED. In case of strong interactions this approach to the bound state problem based on 
Green functions computed from the scattering operator is 
not effective on  the Minkowski space-time because here the asymptotic freedom restricts
the essential support of $g$ to very small domains lying `` deeply inside nucleus'' and consideration of bound states within the perturbative approach with strong interactions and with $g \in \mathscr{E}$ is physically meaningless. The additional difficulty with strong interactions comes from the fact that the higher order contributions to Green functions are comparable to the lower order terms, so even qualitative results which count for only the first order terms are physically useless. 
On globally causal space-times with compact Cauchy surfaces situation becomes
much more better. First of all the perturbative series for (\ref{PmuInteractingGeneral}) becomes much more likely to be convergent 
to ordinary operator in the Fock space, and this is so even with strong interactions, at least for the Einstein Universe. 
This is mainly because on the Einstein Universe all free massive fields have finite-dimensional single particle Hilbert spaces and even Fock spaces
if the spin is odd, compare \cite{SegalZhouQED}, \cite{PaneitzSegalI}-\cite{PaneitzSegalIII}, or Section \ref{EUandG}
or Section \ref{EUandG}. 
This reduces tremendously the number of states which need to be summed up in the perturbative series
due to the form of gauge couplings, which are sesquilinear forms in the massive fields, and kills most of the terms due to the orthogonality
of the fundamental solutions composing the single particle Hilbert spaces of the corresponding free fields underlying the theory in question.  
It is not the naive power counting in the (comparable to unity for strong interactions) coupling constants which is essential here, 
but rather it is the effect of cancelling out most of the states, which greatly improves the convergence of the perturbative series on 
space-times with compact Cauchy surfaces. This opens up a new perspective to the investigation of stable and meta-stable particles within QFT
with realistic interactions, which is impossible on the flat Minkowski space-time, and opens up the way for converting the heuristic 
interpretation of the many-particle wave function determined by the many particle Green function (propagator) as a well defined state in the Fock space
of free fields, also acted on by the interacting fields, compare the next Subsection and Section \ref{EUandG}.

\subsection{Higher order Green functions--propagators. Radiative corrections 
\\ to many-particle Green functions. Lamb shift}\label{Green}

For free fields understood as integral kernel operators with vector-valued kernels, say $u_1, \ldots u_k$, 
there is an important object, namely the ``vacuum expectation''
\begin{equation}\label{FreeGreen}
G_{0}(x_1, \ldots, x_k) = \Big\langle \Big\langle \Phi_{{}_{0}}, T\big[u_1(x_1)u_2(x_2) \ldots u_k(x_k) \big] \Phi_{{}_{0}} \Big\rangle \Big\rangle
\end{equation}
of the chronological product 
\[
T\big[u_1(x_1)u_2(x_2) \ldots u_k(x_k)\big],
\] 
with
\[
T\big[u_1(x_1)u_2(x_2) \ldots u_k(x_k)\big]
\] 
understood rigorously in the sense 
explained in Subsection \ref{WickForChronological}.
As noted first by Schwinger the numerical distribution $G_{0}(x_1, \ldots, x_k)$ determines the $k-1$-particle wave function as its Green function.
In particular for $k=2$ and for $u_1$ equal to the free Dirac field $\boldsymbol{\psi}$ and $u_2$ equal to Dirac-conjugated field $\boldsymbol{\psi}^\sharp$ we obtain the Green function $G_{0}(x_1, x_2)=S^{c}(x_1,x_2)$ of the wave function of a single-electron state of the free Dirac field, which is equal to the fundamental solution-propagator of the free Dirac equation -- corresponding to the free
Dirac field
\[
\big[\gamma^{\mu} i{\textstyle\frac{\partial}{\partial x_\mu}} -m \big]G_0(x,y) = - \delta(x-y).
\]
Similarly, we have for other free fields and the corresponding Green functions of the differential equations they respect.

The ``$k-1$-particle wave function'' corresponding to the $k$-th order Green function is determined through the Huygens rule by the integral of the initial data on a Cauchy surface multiplied with the Green function. 

Passing to interacting fields we are still able to compute (perturbatively) the Green functions of interacting fields -- the formal counterparts of the free Green functions, which account for the interaction terms. In this computation we again start with the causal perturbative construction of the scattering generalized operator
$S$ and with the Hida operators as the creation-annihilation operators.  The free fields underlying the theory (free Dirac and electromagnetic potential fields in case of spinor QED) are understood as integral kernel operators with vector-valued kernels as explained in previous Sections. 
No renormalization is needed and no infrared infinities we encounter in this computation.
However matters concerning indentification of the
the ``$k-1$-particle wave function'' corresponding to the $k$-th order Green function for interacting fields as a well defined state in the Fock space acted on 
by the creation-annihilation operators of the free fields, are much worse, and even impossible for the realization on the flat Minkowski space-time.

We now pass to the computation of the complete higher order Green functions, but in terms of the scattering generalized operator,
and with free fields understood as integral kernel operators with vector valued kernels in the sense \cite{obataJFA} explained in previous Sections.
This allows us to compute all higher order contributions to complete Green functions in well defined mathematical terms without the need for renormalization prescription. In this process the linkage of the wave functions defined by the complete Green functions to well defined states of the Fock space of free fields is
lost in case of Minkowski space-time.
 We do it in the form which allows us to understand to some extend the role of the global structure of space-time itself in keeping the linkage of
the many-particle wave function defined by complete higher order Green function with a well defined state in the Fock space. 
Namely, we should keep in mind at the beginning of the computation that the causal axioms (I)-(V) on the Minkowski space-time for the construction of the scattering operator have their natural counterparts on more general
globally causal space-time. For example, they can be formulated on the Einstein Universe, with the Abelian group of translations replaced with 
the corresponding (non-Abelian) subgroup $\mathbb{R}\times SU(2, \mathbb{C})$ of the isometry group of the Einstein Universe. 
We have in view for example space-times which moreover are Lie groups, which are equal to causal isomorphisms images of the Einstein Universe, and in particular are 
Lie groups $\mathbb{R}\times SU(2, \mathbb{C})$ of causal diffeomorphisms, when regarded as manifolds. This allows us to avoid the problem of division into positive and negative frequency solutions composing the single particle Hilbert spaces of free fields. The simplest case of course we obtain in case all the four corresponding right invariant vector fields generate actual space-time symmetries (as is the case for the Einstein Universe), but it may also happen that no one of them is an 
actual space-time symmetry. For the consideration of the bound state problem we assume however that at least one of the right invariant vector fields 
represents a time-like symmetry of the space-time.
In Section \ref{EUandG} we have presented detailed construction of the scattering 
generalized operator based on the analogues of the causal axioms (I)-(V), with the white noise construction of free fields understood as (finite sums of) integral kernel operators with vector valued kernels in the sense of \cite{obataJFA}. Having this in mind the reader should keep in mind that scattering operator $S(g)$,
the Noether generators $\boldsymbol{P}^\mu$ have their counterparts on the Einstein Universe (or at least the time-translation $\boldsymbol{P}^0$  generator in case of static space-times of the indicated class with compact Cauchy surfaces) with the  general property $[S(g=1), \boldsymbol{P}^\mu]=0$, \emph{i.e.}
$S(g=1)$ commutes\footnote{Of course in case the scattering operator $S(g=1)$ is generalized and transforms
the Hida space into its strong dual (as for spinor QED on the Minkowski space-time) we use the fact 
that the operators $\boldsymbol{P}^\mu$ always transform continuously
the Hida space into itself and thus their linear traspositions transform continuously
the strong dual of the Hida space into itself. Thus, in this case 
the commutativity condition reads: the operator $S(g=1)$ composed with the transposition 
of $\boldsymbol{P}^\mu$ is equal to the operator  $\boldsymbol{P}^\mu$
composed with $S(g=1)$.} with $\boldsymbol{P}^\mu$.  
In passing to interacting fields we replace the free Green function (\ref{FreeGreen}) with its analogue
\begin{multline}\label{CompleteGreen}
G(g; x_1, \ldots, x_k) = {\textstyle\frac{1}{S_0}}
\Big\langle \Big\langle \Phi_{{}_{0}}, T\big[u_1(x_1)u_2(x_2) \ldots u_k(x_k) S(g) \big] \Phi_{{}_{0}} \Big\rangle \Big\rangle,
\\
S_0 = \langle \langle \Phi_{{}_{0}}, S(g) \Phi_{{}_{0}} \rangle \rangle,
\end{multline}
and define (\ref{CompleteGreen}) as the Green function-propagator for the interacting fields, the so called \emph{complete Green function}.
That this expression is indeed the analogue of the free field propagator (\ref{FreeGreen}) can be seen at the heuristic level by noticing the fact that 
(\ref{CompleteGreen}) arises formally from (\ref{FreeGreen}) by adding all intermediate ``connected Feynman graphs with fixed number $k$ of external lines'',
and thus represents the propagator counting for the interaction, compare e.g. \cite{Bogoliubov_Shirkov}, \S 37. In fact this is so only for the realistic
interaction with $g=1$, thus we need to consider the ``limit'' value of the expression (\ref{CompleteGreen}) for $g=1$.
Of course this heuristic argumentation cannot 
be regarded as a final argument. Below we give, after \cite{Bogoliubov_Shirkov}, a construction of the many-particle wave function
which respects the Schwinger equation (associated to the complete Green function) which opens up a relation of the wave function
to a state of the Fock space. On the Minkowski space-time the corresponding state in the Fock space is not well defined, because the generalized scattering
operator cannot be represented by a perturbative series converging to an ordinary operator in the Fock space, at least on the Minkowski space-time
with realistic interactions (for example spinor QED). Nonetheless, the complete Green function itself and its wave function can be computed as a well defined mathematical object, even for QED on the Minkowski space-time, with the scattering operator understood as the generalized operator with the free fields understood as integral kernel operators with vector-valued kernels. Moreover, on the space-times with compact Cauchy surfaces the state in the Fock space corresponding 
to the wave function becomes well defined, and the Schwinger's method acquires a profound justification on such space-times.

Here we should remind that, depending on the specific application, we construct after Schwinger and Bogoliubov the scattering operator with the interaction Lagrange density $g(x) \mathcal{L}(x)$ to which an auxiliary additional terms are added
\[
\eta^\sharp(x)\boldsymbol{\psi}(x) +  \boldsymbol{\psi}^\sharp(x) \eta(x) + J(x)A(x)
\]
with the free Dirac field and its conjugation $\boldsymbol{\psi}, \boldsymbol{\psi}^\sharp$ and with the free electromagnetic potential field $A$, 
together with the many component
switching off function $(g,\eta, \eta^\sharp, J)$, and constructed as explained in the
introduction to the current Section \ref{A(1)psi(1)} and Subsection \ref{WickForProduct}. 
In the last formula the summed up spinor and Lorentz indices are not explicitly written 
in order to simplify notation.
Then the corresponding interacting fields are constructed through the variational derivatives
of the scattering operator with respect to $\eta^{\sharp \,\, a}$ (for the interacting $\boldsymbol{\psi}^{a}$) or with respect to
$\eta^{a}$ (for the interacting $\boldsymbol{\psi}^{\sharp \,\, a}$) or finally with respect to $J^\mu$ (for the interacting field $A_\mu$),
as explained in introduction to the current Section \ref{A(1)psi(1)}. All those variational derivatives are taken at $J=0, \eta=0, \eta^\sharp=0$
and with intensity of interaction $g=1$ in the final formulas in case there is no external field present. In case the there is present an external 
classical field, represented by the electric classical current field $J_{{}_{0}}$ the said variational derivatives are taken
at $J = J_{{}_{0}}$, $\eta=0, \eta^\sharp=0, g=1$. For the bound state problem we consider, after Schwinger the case when 
$J_{{}_{0}}$ is time independent (invariant under the time-like space-time symmetry). Accordingly, we denote shortly by $S(g)$
the scattering generalized operator functional evaluated at $J = J_{{}_{0}}$, $\eta=0, \eta^\sharp=0, g$ (in case an external classic
source is present represented by the classic current field $J_{{}_{0}}$) or at $J = 0$, $\eta=0, \eta^\sharp=0, g$
(in case there are no external fields present). Thus we are dealing with the generated functional method initiated by Schwinger
and then developed by Bogoliubov and his school, compare \cite{Bogoliubov_Shirkov}.

First of all we should emphasize that the expression (\ref{CompleteGreen}) is symbolic and in fact its numerator should be understood as the 
following perturbative series -- coming from the causal perturbative construction of the scattering generalized operator according to the
axioms (I)-(V), or better according to the ``natural'' chronological product of Subsection \ref{WickForChronological}:
\begin{multline}\label{SeriesForGreen}
\Big\langle \Big\langle \Phi_{{}_{0}}, T\big[u_1(x_1)u_2(x_2) \ldots u_k(x_k) S(g) \big] \Phi_{{}_{0}} \Big\rangle \Big\rangle =
\\
= \big\langle \big\langle \Phi_{{}_{0}}, T\big[ u_1(x_1)u_2(x_2) \ldots u_k(x_k)\big] \Phi_{{}_{0}} \big\rangle \big\rangle 
\,\,\,\,\,\,\,\,\,\,\,\,\,\,\,\,\,\,\,\,\,\,\,\,\,\,\,\,\,\,\,\,\,\,\,\,\,\,\,\,\,\,\,\,\,\,\,\,\,\,\,\,\,\,\,\,\,\,\,\,\,\,\,\,\,
\\
+ i \int \, \ud^4 y_1 \, \Big\langle \Big\langle \Phi_{{}_{0}}, T\big[u_1(x_1)u_2(x_2) \ldots u_k(x_k) g(y_1)\mathcal{L}(y_1) \big] \Phi_{{}_{0}} \Big\rangle
\Big\rangle  
\\
+ {\textstyle\frac{i^2}{2!}} \int \, \ud^4 y_1 \ud^4 y_2  \, \Big\langle \Big\langle \Phi_{{}_{0}}, T\big[u_1(x_1)u_2(x_2) \ldots u_k(x_k) g(y_1)\mathcal{L}(y_1) 
g(y_2)\mathcal{L}(y_2) \big] \Phi_{{}_{0}} \Big\rangle \Big\rangle 
\\
+ {\textstyle\frac{i^3}{3!}} \int \, \ud^4 y_1 \ud^4 y_2 \ud^4 y_3  \, \Big\langle \Big\langle \Phi_{{}_{0}}, T\big[u_1(x_1)u_2(x_2) \ldots u_k(x_k) g(y_1) \mathcal{L}(y_1) g(y_2)\mathcal{L}(y_2) g(y_3)\mathcal{L}(y_3) \big] \Phi_{{}_{0}} \Big\rangle \Big\rangle
\\
+ \ldots \,\,\,\,\,\,\,\,\,\,\,\,\,\,\,\,\,\,\,\,\,\,\,\,\,\,\,\,\,\,\,\,\,\,\,\,\,\,\,\,\,\,\,\,\,\,\,\,\,\,\,\,\,\,\,\,\,\,\,\,\,\,\,\,\,\,\,\,\,\,\,
\end{multline}
The causal axioms (I)-(V) can in principle be understood as defining the chronological products $ T\big[u_1(x_1)u_2(x_2) \ldots \big]$ present in this formula,
and analysed in Subsection \ref{WickForProduct}. 
Here we should however proceed
as in Subsection \ref{WickForChronological} and use  the ``natural'' chronological product $T[\ldots]$ defined as in Subsection \ref{WickForChronological}
and the method of computation of $S_n$ which reduces the splitting to the splitting of only several scalar ``basic distributions'' which enter the Wick
decomposition of the product $\mathcal{L}(x)\mathcal{L}(y)$.  
The non uniqueness in the splitting is completely and naturally eliminated, and the scattering operator is fully determined there solely by the requirement that
the higher order contributions to interacting fields exist in the adiabatic limit  as  well-defined finite sums of integral kernel operators,
compare Subsection \ref{OperationsOnXi}.

Note that in the perturbative formula (\ref{SeriesForGreen}) we fix the switching off function $g \in \mathscr{E}$. Similarly as in the 
construction of the generalized operator $S(g=1)$ in Subsection \ref{EffCrossSection}, we construct 
\[
\Big\langle \Big\langle \Phi_{{}_{0}}, T\big[u_1(x_1)u_2(x_2) \ldots u_k(x_k) S(g=1) \big] \Phi_{{}_{0}} \Big\rangle \Big\rangle
\]
or
\begin{multline}\label{g=1SeriesForGreen}
\Big\langle \Big\langle \Phi_{{}_{0}}, T\big[u_1(x_1)u_2(x_2) \ldots u_k(x_k) S(g=1) \big] \Phi_{{}_{0}} \Big\rangle \Big\rangle =
\\
= \big\langle \big\langle \Phi_{{}_{0}}, T\big[ u_1(x_1)u_2(x_2) \ldots u_k(x_k)\big] \Phi_{{}_{0}} \big\rangle \big\rangle 
\,\,\,\,\,\,\,\,\,\,\,\,\,\,\,\,\,\,\,\,\,\,\,\,\,\,\,\,\,\,\,\,\,\,\,\,\,\,\,\,\,\,\,\,\,\,\,\,\,\,\,\,\,\,\,\,\,\,\,\,\,\,\,\,\,
\\
+ i \int \, \ud^4 y_1 \, \Big\langle \Big\langle \Phi_{{}_{0}}, T\big[u_1(x_1)u_2(x_2) \ldots u_k(x_k) \mathcal{L}(y_1) \big] \Phi_{{}_{0}} \Big\rangle
\Big\rangle  
\\
+ {\textstyle\frac{i^2}{2!}} \int \, \ud^4 y_1 \ud^4 y_2  \, \Big\langle \Big\langle \Phi_{{}_{0}}, T\big[u_1(x_1)u_2(x_2) \ldots u_k(x_k) \mathcal{L}(y_1) 
\mathcal{L}(y_2) \big] \Phi_{{}_{0}} \Big\rangle \Big\rangle 
\\
+ {\textstyle\frac{i^3}{3!}} \int \, \ud^4 y_1 \ud^4 y_2 \ud^4 y_3  \, \Big\langle \Big\langle \Phi_{{}_{0}}, T\big[u_1(x_1)u_2(x_2) \ldots u_k(x_k)  \mathcal{L}(y_1) \mathcal{L}(y_2) \mathcal{L}(y_3) \big] \Phi_{{}_{0}} \Big\rangle \Big\rangle
\\
+ \ldots \,\,\,\,\,\,\,\,\,\,\,\,\,\,\,\,\,\,\,\,\,\,\,\,\,\,\,\,\,\,\,\,\,\,\,\,\,\,\,\,\,\,\,\,\,\,\,\,\,\,\,\,\,\,\,\,\,\,\,\,\,\,\,\,\,\,\,\,\,\,\,
\end{multline}
in each order separately, by the ``adiabatic'' limit construction presented there, and thus the complete Green functions
\begin{multline}\label{GreenFunctions}
G(x_1, x_2)= i {\textstyle\frac{1}{S_0}}
 \Big\langle \Big\langle \Phi_{{}_{0}}, T\big[\boldsymbol{\psi}(x_1)\boldsymbol{\psi}^\sharp(x_2) S(g=1) \big] \Phi_{{}_{0}} \Big\rangle \Big\rangle, 
\,\,\, S_{{}_{0}} = \big\langle \big\langle \Phi_{{}_{0}}, S(g=1) \Phi_{{}_{0}} \big\rangle \big\rangle,
\\
G(x_1, x_2, x_3, x_4)=  {\textstyle\frac{1}{S_0}}
\Big\langle \Big\langle \Phi_{{}_{0}}, T\big[\boldsymbol{\psi}(x_1)\boldsymbol{\psi}(x_2)\boldsymbol{\psi}^\sharp(x_3) 
\boldsymbol{\psi}^\sharp(x_4) S(g=1) \big] \Phi_{{}_{0}} \Big\rangle \Big\rangle, 
\\
\ldots
\end{multline}
can be computed in the limit. Similarly as in Subsection \ref{EffCrossSection}, the limit value, in principle, may depend on the choice of the 
strongly bounded sequence $\{g_j \} \subset \mathscr{E}$ converging to $g=1$ in the strong dual topology of $\mathscr{E}^* \supset \mathscr{E}$.
The point however is that the final results, such as the radiative energy shifts of the atomic bound states computed
from the wave functions determined by the Green functions are independent of the behavior of the switching off function
$g$ at the remote parts of the space-time, and thus the specific choice of the limit is irrelevant for these computations.
This has been already noted in \cite{Bogoliubov_Shirkov}. Thus from the mathematical point of view the ``adiabatic
problem'' remains in that we have no unique choice for the limit $g=1$ in the formula for the complete Green functions, but the practical
application restricts the essential domain of the wave function of nearly ground states of bounded systems (atoms) with local 
dependence on $g$ restricted to this domain. It can be shown, however, that the limit $g=1$ of the Green functions is essentially
unique, compare also \cite{duch}.

Using the Wick theorem for the chronological product of Subsection \ref{WickForChronological} (in case we are using 
the ``natural'' chronological product of Subsection \ref{WickForChronological}) applied to (\ref{g=1SeriesForGreen}) we obtain
the Schwinger's integral equation for the $k$-th order complete Green function $G$, at least in the perturbative form, compare
\cite{Bogoliubov_Shirkov}, \S 37-38, for the motivation. On static space-time with compact Cauchy surfaces the splitting problem trivializes,
and we obtain simple recurrence rule for the chronological product and thus for Green functions, at least for realistic theories like QED,
compare Subsection \ref{WickForChronologicalEU}.

Now let $\Phi_1$ be a single particle (say single electron) state, smooth enough to be an element of the Hida space (recall
that the vacuum by construction belongs to the Hida space). We can then consider, again after \cite{Bogoliubov_Shirkov}, \S 42,
the following ordinary numerical spinor
\begin{multline}\label{psi(x1)}
\psi(x_1) = \Big\langle \Big\langle S(g=1)\Phi_{{}_{0}}, \boldsymbol{\psi}_{{}_{\textrm{int}}}(g=1, x_1)S(g=1)\Phi_{{}_{1}} \Big\rangle \Big\rangle
\\
=
{\textstyle\frac{1}{i}} \Big\langle \Big\langle \Phi_{{}_{0}}, S(g=1)^{+}  {\textstyle\frac{\delta S}{\delta \eta^\sharp(x_1)}} \Phi_{{}_{2}} \Big\rangle \Big\rangle
=
\Big\langle \Big\langle \Phi_{{}_{0}}, S(g=1)^{+}T\big[\boldsymbol{\psi}(x_1)S(g=1) \big] \Phi_{{}_{1}} \Big\rangle \Big\rangle
\\
=
{\textstyle\frac{1}{S_0}}
\Big\langle \Big\langle \Phi_{{}_{0}}, T\big[\boldsymbol{\psi}(x_1) S(g=1) \big] \Phi_{{}_{1}} \Big\rangle \Big\rangle
\end{multline}
which is an immediate analogue of the numerical spinor $\varphi(x_1) = \langle \langle \Phi_0, \boldsymbol{\psi}(x_1) \Phi_1 \rangle \rangle$ for the free 
Dirac field $\boldsymbol{\psi}$. The spinor (\ref{psi(x1)}) is the candidate for the single electron wave function which counts for the interaction,
and which should be identified with the wave function of the single electron complete Green function $G(x_1,x_2)$, compare \cite{Bogoliubov_Shirkov}, \S 42.
Thus we need to show that indeed the spinor (\ref{psi(x1)}) respects the Schwinger's equation for the single electron wave function determined
by the Green function $G(x_1, x_2)$ with radiative corrections.
Doing this we follow Bogoliubov and Shirkov \cite{Bogoliubov-Shirkov}, \S 42.

Utilizing the equality
\[
i\Big\langle \Big\langle \Phi_{{}_{0}}, T\big[\boldsymbol{\psi}(x_1)\boldsymbol{\psi}^\sharp(x_2) \big] \Phi_{{}_{0}} \Big\rangle \Big\rangle
= S^c(x_1-x_2)
\]
in each order term in the expansion (\ref{g=1SeriesForGreen}) we arrive at the relation between $\psi(x_1)$ and $G(x_1,x_2)$ which can be written
in the following general form (compare \cite{Bogoliubov_Shirkov}, \S 42)
\[
\psi(x_1) = \int G(x_1, x_2) \big[S^c(x_2-x_3)\big]^{-1} \varphi(x_3) \ud x_2 \ud x_3,
\]
which, joined with the Schwinger integral equation for $G$, gives the Schwinger equation for the single electron
wave function $\psi$ with radiative corrections, compare \cite{Bogoliubov_Shirkov}, \S 42.

Analogously having given a single positron state $\Phi_1$ in the Hida subspace of the Fock space, and replacing the free Dirac field with
Dirac-conjugated field $\boldsymbol{\psi}^\sharp$, we can construct the the single-positron wave function out of the state $\Phi_1$
and using the scattering generalized operator and its expansion into chronological products:
\begin{multline}\label{psi^sharp(x1)}
\psi^\sharp(x_1) = \Big\langle \Big\langle S(g=1)\Phi_{{}_{0}}, \boldsymbol{\psi}^{\sharp}_{{}_{\textrm{int}}}(g=1, x_1)S(g=1)\Phi_{{}_{1}} \Big\rangle \Big\rangle
\\
=
{\textstyle\frac{1}{i}} \Big\langle \Big\langle \Phi_{{}_{0}}, S(g=1)^{+}  {\textstyle\frac{\delta S}{\delta \eta(x_1)}} \Phi_{{}_{2}} \Big\rangle \Big\rangle
=
\Big\langle \Big\langle \Phi_{{}_{0}}, S(g=1)^{+}T\big[\boldsymbol{\psi}^\sharp(x_1)S(g=1) \big] \Phi_{{}_{1}} \Big\rangle \Big\rangle
\\
=
{\textstyle\frac{1}{S_0}}
\Big\langle \Big\langle \Phi_{{}_{0}}, T\big[\boldsymbol{\psi}^\sharp(x_1) S(g=1) \big] \Phi_{{}_{1}} \Big\rangle \Big\rangle
\end{multline}

Analogously, having given a two-particle (say positron-electron) state $\Phi_2$ in the Fock space which belongs to the Hida space, we can form
the positron-electron wave function
\begin{multline}\label{psi(x1,x2)}
\psi(x_1, x_2) =  
{\textstyle\frac{1}{i^2}} \Big\langle \Big\langle \Phi_{{}_{0}}, S(g=1)^{+}  {\textstyle\frac{\delta^2 S}{\delta\eta^\sharp(x_1)\delta \eta(x_2)}} \Phi_{{}_{2}} \Big\rangle \Big\rangle
\\
=
{\textstyle\frac{1}{S_0}}
\Big\langle \Big\langle \Phi_{{}_{0}}, T\big[\boldsymbol{\psi}(x_1) \boldsymbol{\psi}^\sharp(x_2) S(g=1) \big] \Phi_{{}_{2}} \Big\rangle \Big\rangle
\end{multline}
which, as a consequence of the integral Schwinger's equation for $G(x_1,x_2,x_3,x_4)$, respects the Schwinger equation with radiative corrections.

Analogously, repeating the operation of functional derivation 
\[
{\textstyle\frac{\delta^k S}{\delta \eta^\sharp(x_1) \ldots \delta \eta(x_k)}}
\]
with respect to $\eta^\sharp$ or $\eta$, we can construct wave functions for many-electron or many electron-positron systems,
respecting the wave equation with radiative corrections coming from the Schwinger equation for the
corresponding higher order complete Green functions.

In considering the positronium we assume that the external field is zero, and put $J=0$ in the final formulas.
But considering the radiative contribution to the bound state of atoms we put, in the final formulas, 
$J=J_{{}_{0}}$ and assume this classic current to be time independent and coming from the Coulomb field of the nucleus.

In order to find the energy levels we insert into the Schwinger equation the wave function $\psi$ of the general form
\[
\psi(x_1) = e^{-iE x_{01}} \chi(\boldsymbol{\x}_{1}), \,\,\, \textrm{with} \,\, x_1 = (x_{10}, \boldsymbol{\x}_{1})
\]
in case of the single-electron wave function with the external time-independent current put equal $J=J_{{}_{0}}$ 
in the scattering generalized operator (coming from the nucleus
Coulomb field, e.g. in case of the computation of the radiative correction to the energy shift of the hydrogen atom),
so that
\begin{equation}\label{dx10psi(x1)=Epsi(x1)}
i {\textstyle\frac{\partial}{\partial x_{1 0}}} \psi(x_1) = E \psi(x_1);
\end{equation}
or 
\[
\psi(x_1, x_2) = \int \limits_{(p_1+p_2)^2 = m^2} e^{-ip_1 \cdot x_{1} -i p_2 \cdot x_2} \chi(p_1, p_2) \ud^4 p_1 \ud^4 p_2, 
\]
\[
\textrm{with fixed}  \, \,  (p_1+p_2)\cdot (p_1+p_2) = m^2,
\]
in case of the positron-electron wave function (\ref{psi(x1,x2)}) in computation of the radiative corrections to the bound states of the positronium,
so that
\begin{equation}\label{X.Xpsi=m2psi}
\Big[i{\textstyle\frac{\partial}{\partial x_{1 \mu}}} + i {\textstyle\frac{\partial}{\partial x_{2 \mu}}}\Big]\Big[ i\textstyle\frac{\partial}{\partial x_{1}^{\mu}}
+i\textstyle\frac{\partial}{\partial x_{2}^{\mu}} \Big] \psi(x_1,x_2)
= m^2 \psi(x_1, x_2)
\end{equation}
with $J=0$ put equal zero in the scattering operator in the final formulas. In case an external time-independent classic bounding potential field is present
in a, say two-electron system, we insert into the Schwinger's equation the wave function (given by the formula analogous to 
(\ref{psi(x1,x2)}) with $\boldsymbol{\psi}^\sharp$ replaced by $\boldsymbol{\psi}$ in it) of the following general form
\[
\psi(x_1, x_2) = \int \limits_{p_{10}+p_{20} =E} e^{-ip_1 \cdot x_{1} -i p_2 \cdot x_2} \chi(p_1, p_2) \ud^4 p_1 \ud^4 p_2, 
\]
\[
\textrm{with fixed}  \, \,  p_{10}+p_{20} =E 
\]
in case of the two-electron wave function in computation of the radiative corrections to the bound states of the system,
so that
\begin{equation}\label{X0psi=Epsi}
\Big[i {\textstyle\frac{\partial}{\partial x_{1 0}}} + i {\textstyle\frac{\partial}{\partial x_{2 0}}} \Big] \psi(x_1,x_2)
= E \psi(x_1, x_2).
\end{equation}
Analogously we have for multi-electron or multi-electron-positron systems. 
In any case we always insert into the Schwinger equation the wave function $\psi$, defined respectively by (\ref{psi(x1)}), (\ref{psi(x1,x2)}), \ldots, 
which is a proper eigenfunction of the 
differential operator $\boldsymbol{X}^\mu\boldsymbol{X}_\mu$ or $\boldsymbol{X}^0$, with differential operator 
\[
\boldsymbol{X}^\mu = i{\textstyle\frac{\partial}{\partial x_{1 \mu}}} 
\,\,\,\,\, \textrm{or}
\,\,\,\,\,
i{\textstyle\frac{\partial}{\partial x_{1 \mu}}} + i{\textstyle\frac{\partial}{\partial x_{2 \mu}}} 
\,\,\,\,\,
\textrm{or} \,\,\,\,\,
i{\textstyle\frac{\partial}{\partial x_{1 \mu}}} + i{\textstyle\frac{\partial}{\partial x_{2 \mu}}} 
+i{\textstyle\frac{\partial}{\partial x_{3 \mu}}} 
\,\,\,\,\,\,
\textrm{or}
\,\,\,
\ldots
\]
representing the symmetry of the system
(in the case of Minkowski space-time the symmetries considered correspond to all translations in case when no external fields
are present or to the time-translation symmetry generator $\boldsymbol{X}^0$ in case an external and classic time-independent field is present).

Now we look for a deeper justification of this Schwinger-Bogoliubov method, so successful in all its applications, at least within QED.

First we observe (which holds on the Minkowski space-time but the analogous statement holds on other globally causal symmetric space-times) 
that by the translational invariance of the scattering
operator or the translational covariance axiom (II) for the chronological product, we have the following
formulas for the derivatives of the wave functions $\psi$ defined respectively by (\ref{psi(x1)}), (\ref{psi(x1,x2)}), \ldots:  
\begin{equation}\label{partialpsi(x1)}
i {\textstyle\frac{\partial}{\partial x_{1 \mu}}} \psi(x_1) 
=
{\textstyle\frac{1}{S_0}}
\Big\langle \Big\langle \Phi_{{}_{0}}, T\big[\boldsymbol{\psi}(x_1) S(g=1) \big] \boldsymbol{P}^\mu \Phi_{{}_{1}} \Big\rangle \Big\rangle
\end{equation}
or 
\begin{equation}\label{timepartialpsi(x1)}
i {\textstyle\frac{\partial}{\partial x_{1 0}}} \psi(x_1)  
=
{\textstyle\frac{1}{S_0}}
\Big\langle \Big\langle \Phi_{{}_{0}}, T\big[\boldsymbol{\psi}(x_1) S(g=1) \big] \boldsymbol{P}^0 \Phi_{{}_{1}} \Big\rangle \Big\rangle
\end{equation}
when an external time-independent classical field $J=J_{{}_{0}}$ is present. Similarly
\begin{equation}\label{partialpsi(x1,x2)}
\Big[i{\textstyle\frac{\partial}{\partial x_{1 \mu}}} + i {\textstyle\frac{\partial}{\partial x_{2 \mu}}}\Big] \psi(x_1,x_2)
= 
{\textstyle\frac{1}{S_0}}
\Big\langle \Big\langle \Phi_{{}_{0}}, T\big[\boldsymbol{\psi}(x_1) \boldsymbol{\psi}^\sharp(x_2) S(g=1) \big] \boldsymbol{P}^\mu \Phi_{{}_{2}} \Big\rangle \Big\rangle
\end{equation}
or 
\begin{equation}\label{timepartialpsi(x1,x2)}
\Big[i{\textstyle\frac{\partial}{\partial x_{1 0}}} + i {\textstyle\frac{\partial}{\partial x_{2 0}}}\Big] \psi(x_1,x_2)
= 
{\textstyle\frac{1}{S_0}}
\Big\langle \Big\langle \Phi_{{}_{0}}, T\big[\boldsymbol{\psi}(x_1) \boldsymbol{\psi}^\sharp(x_2) S(g=1) \big] \boldsymbol{P}^0 \Phi_{{}_{2}} \Big\rangle \Big\rangle.
\end{equation}
Analogous statements hold true on other globally causal space-times, with the four groups of space-time symmetries generated by 
the respective $\boldsymbol{X}^\mu$, as for example on the Einstein Universe (compare Section \ref{EUandG}),
but in general $\boldsymbol{X}^\mu$ and $\boldsymbol{X}^\nu$ do not commute on the Einstein Universe
(with the generators $\boldsymbol{X}^\nu$ of the space-time symmetry subgroup $\mathbb{R} \times SU(2, \mathbb{C})$ plying the role of generators of translations, 
in case of a more general symmetric space-time
without external fields
or time-like symmetry generator $\boldsymbol{X}^0$ in case of a still more general static space-time with an external time-indepent and classic field,
which is a causal isomorphism image of the Einstein Universe). 

We see that for any 
state $\Phi_1, \Phi_2, \ldots \Phi_k, \ldots$ of the Fock space, belonging to the Hida space (thus normalizable or bound state) and such that 
\[
\boldsymbol{P}^\mu\boldsymbol{P}_\mu \Phi_k = p^\mu p_\mu \Phi_k = m^2 \Phi_k
\,\,\,
\textrm{or} \,\,\,
\boldsymbol{P}^0 \Phi_k = E \Phi_k
\]
the corresponding wave function $\psi$, given respectively by (\ref{psi(x1)}), (\ref{psi(x1,x2)}), \ldots 
respects the Schwinger equation, including the radiative corrections, and corresponds to the eigenvalue
$m^2$ or $E$ of the differential operator $\boldsymbol{X}^\mu\boldsymbol{X}_\mu$ or $\boldsymbol{X}^0$.
Such a state $\Phi_k$ with $k>1$ cannot exist  in the Fock space of free fields on the Minkowski space-time. This is because for $k>1$
the spectrum of $\boldsymbol{P}^\mu\boldsymbol{P}_\mu$ on the $k$-th particle states is purely continuous in this case (with no normalizable eigenstates), 
and the operator $\boldsymbol{P}^0$ has purely continuos spectrum on the Fock space of free fields on the Minkowski space-time. For $k=1$,
of course, all normalizable single electron or single positron states which belong to the Hida space are at the same time the normalizable
eigenstates of the operator $\boldsymbol{P}^\mu\boldsymbol{P}_\mu$ corresponding to the eigen-value $m^2$ with $m$ equal to the electron mass. 

The last class of states $\Phi_1$ is far too small for the justification of the Schwinger-Bogoliubov method for computation 
of the radiative corrections to the energy levels in the bound state problem. This can be seen by the comparison
of the formula  (\ref{timepartialpsi(x1)}) with the formula 
(\ref{dx10psi(x1)=Epsi(x1)}) (in case an external $J\neq 0$ time-independent classical field is present) or respectively
the formulas (\ref{X.Xpsi=m2psi}) and (\ref{partialpsi(x1,x2)}), noting that the repeated application of the operator
$\boldsymbol{P}_\mu$ to the state $\Phi_2$ gives
\begin{multline*}
\Big[i{\textstyle\frac{\partial}{\partial x_{1 \mu}}} + i {\textstyle\frac{\partial}{\partial x_{2 \mu}}}\Big] 
\Big[ i\textstyle\frac{\partial}{\partial x_{1}^{\mu}}
+i\textstyle\frac{\partial}{\partial x_{2}^{\mu}} \Big]\psi(x_1,x_2)
\\
= 
{\textstyle\frac{1}{S_0}}
\Big\langle \Big\langle \Phi_{{}_{0}}, T\big[\boldsymbol{\psi}(x_1) \boldsymbol{\psi}^\sharp(x_2) S(g=1) \big] \boldsymbol{P}^\mu
\boldsymbol{P}_\mu \Phi_{{}_{2}} \Big\rangle \Big\rangle,
\end{multline*}
or finally, in case $J\neq 0$, the formulas
(\ref{X0psi=Epsi}) and  (\ref{timepartialpsi(x1,x2)}).

In particular in order to construct the single-electron wave functions $\psi$, given by (\ref{psi(x1)}), in the classical field of a nucleus 
(e.g. Lamb shift calculation for hydrogen) we have so seek for the appropriate state $\Phi_1$ as a direct integral of states corresponding to various eigenvalues
of the operator $\boldsymbol{P}^0$, because there are no elements $\Phi_1$ of the Hida space (thus normalizable) which are normalizable eigenstates
of $\boldsymbol{P}^0$. But the comparison of the formulas  (\ref{timepartialpsi(x1)}) and
(\ref{dx10psi(x1)=Epsi(x1)}) shows that this is hardly possible (in fact impossible) to find such $\Phi_1$, on account of the form 
(\ref{psi(x1)}) of the wave function $\psi$. Similarly, in order to construct an element $\Phi_2$ in the Hida subspace (thus normalizable) of the Fock space
defining the wave function (\ref{psi(x1,x2)}) we cannot use $\Phi_2$ which would be a normalizable eigenstate of the operator 
$\boldsymbol{P}^\mu\boldsymbol{P}_\mu$ because there are no such normalizable states on the Minkowski space-time, 
and instead we have to use $\Phi_2$ as a direct integral
of eigenstates of $\boldsymbol{P}^\mu\boldsymbol{P}_\mu$ corresponding to various eigenvalues in order to achieve a normalizable $\Phi_2$.
Again, the general form (\ref{psi(x1,x2)}) of $\psi$, and comparison of the formulas (\ref{X.Xpsi=m2psi}) and (\ref{partialpsi(x1,x2)})
shows that such $\Phi_2$ hardly exist (in fact does not exists in the Fock space on the Minkowski space-time). 

Therefore we arrive at the conclusion that the Schwinger-Bogoliubov method would be justified if the operators 
$\boldsymbol{P}^\mu\boldsymbol{P}_\mu$ and $\boldsymbol{P}^0$ had a reach class of normalizable (lying in the Hida subspace)
$k$-th particle states $\Phi_k$, $k>0$, in the Fock space of free fields. Even more, all those states would have to form a discrete parts of the spectra
of the said operators $\boldsymbol{P}^\mu\boldsymbol{P}_\mu$ and $\boldsymbol{P}^0$ with an extraordinary stability: 
their discrete eigenvalues $m^2$ and respectively $E$ should coincide with the eigenvalues of the corresponding differential operators
$\boldsymbol{X}^\mu\boldsymbol{X}_\mu$ and $\boldsymbol{X}^0$ with the wave functions $\psi$,
 given respectively by (\ref{psi(x1)}), (\ref{psi(x1,x2)}), \ldots, as the proper eigenfunctions of
$\boldsymbol{X}^\mu\boldsymbol{X}_\mu$ and $\boldsymbol{X}^0$. Thus the spectra should coincide with the perturbed spectra 
which include radiative corrections. This, as we have seen, is impossible on the Minkowski space-time.

In order to understand this paradoxical situation without introducing any \emph{ad hoc} laws, we observe that this conclusion
would be in fact true, although not on the flat Minkowski space-time, but on another globally causal space-times with compact Cauchy surfaces
and nonzero curvature.
In particular it is true on the Einstein Universe, where the operators $\boldsymbol{P}^\mu$ and $\boldsymbol{P}^\mu\boldsymbol{P}_\mu$ 
on the Fock space (corresponding to the one parameter subgroups of space-time symmetry subgroup $\mathbb{R}\times SU(2, \mathbb{C})$
generated by $\boldsymbol{X}^\mu$ -- the formal analogue of the translation subgroup on the Minkowski space-time) have purely
discrete spectra, for the proof compare Section \ref{EUandG}. By the invariance of Schwinger equation it follows the required stability
property of the spectra. Thus in order to justify the Schwinger-Bogoliubov method, confirmed by experiments, we propose to
assume that the space-time (which is compatible with this method) is not flat, but some other with compact Cauchy surfaces,
e.g. a space-time which can be obtained as the image of the Einstein Universe through a diffeomorphism preserving the causal structure
(``conformal''), in particular Einstein Universe itself, but not necessarily. 

This conclusion is justified by the fact that only the local properties of space-time are crucial in considering bound states
which are nearly ground states of sufficiently strongly bounded systems (bound states of electrons in atoms), as we have already mentioned
in the previous Subsection. This is because the Green functions computed for the flat space-time are pretty
good for the practical computation of the energy levels of nearly ground states of atoms. 
It is only the relation of the wave function determined by the Green function to a well defined state 
$\Phi_k$ of the Fock space where the global structure of space-time intervenes dramatically. But staying at the level of Green function
does not require to account for the global structure especially in application to the energy levels of sufficiently strongly bound systems.
Moreover, with the theory based on the causal perturbative construction of the scattering
generalized operator, based on the axioms (I)-(V) (or their immediate generalization), we construct the scattering
operator which is invariant whenever the space-time and the classical sources have a common symmetry. In particular,
using the Schwinger-Bogoliubov method, we arrive at the conclusion that the atomic clock shows the ``time'' 
(up to an irrelevant constant factor) -- \emph{i.e.} the 
parameter of the time symmetry flow of space-time (if the space-time do have such a symmetry). This theoretical conclusion is confirmed experimentally 
by the timing used in Global Positioning System (GPS). This timing, based on the hyperfine transitions in the ground state of caesium ${}^{133}\textrm{Cs}$ works correctly if (among other effects) the correction of the atomic clock rate coming from the Earth's gravitational field is accounted for, 
and moreover if it is done in such a manner which assumes that the time
(in the frequency of the hyperfine transitions of caesium atomic clocks boarded on the satellites and ground stations) actually coincides with 
the local time symmetry parameter of the static space-time geometry around the Earth. This is non-trivial and in fact we do not have any other
explanation of this experimental fact. We should emphasize here that this experimental fact is not obvious, and assertion that time shown 
by atomic clock coincides with the actual parameter
of the one-parametrer group of time symmetry diffeomorphisms (up to an irrelevant overall constant factor and in case the space-time does have such a symmetry) 
at the place where it is actually placed, does not have to be \emph{a priori} true. This has been emphasized by physicists, especially some 40 years ago before the era of easily accessible and accurate atomic clocks has come. For example Dirac \cite{DiracHund} was one among the physicists who emphasized that the time shown by an atomic clock need not coincide with the time understood as time symmetry parameter of a static space-time in the Einstein's Theory of General Relativity.
Of course this experimental fact will remain to be non-trivial forever, but nowadays with atomic clocks all around us we are more inclined to
succumb to the power of habit, and forget about it. 
 
Perhaps we should emphasize that the ordinary methods, based on the perturbative prescription, are far too weak to account for
the radiative corrections that would be required in order to achieve the required accuracy comparable with the frequency shifts
in the hyperfine structure transitions of the atomic clocks in GPS caused by the Earth's gravitational field. Indeed 
the relative shifts (difference between the ratio of the shifted frequencies and the unity) caused by the gravitational field in GPS atomic clocks 
are at the level of $10^{-10}$.
On the other hand, the relative accuracy of the radiative corrections to bound state energy levels is at the level
of $10^{-5}$ (for the situation where only the QED effects are relevant, computed for the Lamb shift of $2S_{{}_{1/2}}$ and
$1P_{{}_{1/2}}$ of the hydrogen atom) or by one order better at the level $10^{-6}$ (for the same states of hydrogen atom 
including effects outside QED interactions). We hope that by eliminating mathematically ill-defined quantities and thus by eliminating handling with them
in the renormalization prescription, we have improved the computational possibilities of the theory by working with the chronological
product well-defined mathematically, using the computational method presented in Subsection \ref{WickForChronological}.
With this method, the only operations we need to perform is to form the simple products of kernels $\kappa_{l,m}$,
with some products taken at the same space-time point,
next integrations (contractions)  and  dispersion integrations (computation of ret and av parts), 
and finally, symmetrizations and antisymmetrizations in the respective momentum variables. 
These operations are, in principle, subjectable to algorithmic computations. 

On the static space-times with compact Cauchy surface, 
with additional symmetries, e.g. the Einstein Universe, which can be used to the automatic computation of the ``plane wave'' kernels
$\kappa_{0,1}, \kappa_{1,0}$ defining free fields, the computational method based on the white noise analysis
is still much more effective for QFT theories in which each of the Wick monomials in the interaction density $\mathcal{L}$
has at most one massless field. In this case the additional problem with non-unique splitting disappears
completely and no idempotential projection operators of  Subsection \ref{WickForChronological} are needed,
compare the first Wick Theorem for chronological product of Subsection \ref{WickForChronologicalEU}, case 1).

\subsection{Causal method and space-time geometry}\label{CausalSandSpace-time}

The physical motivation for the construction of all relevant dynamical quantities,
including interacting fields, through the scattering operator goes in fact to Heisenberg, and 
tries to avoid usage of the interaction hamiltonian, a very singular quantity in relativistic
quantum mechanical systems, \emph{i.e.} for the systems of relativistic quantum fields. 
The motivation is clear:
first we define free quantum particles and the Fock space of symmetrized or anti-summetrized states
with arbitrary many quantum particles.   By the classic result of Dirac, Wigner and Jordan this 
is equivalent to the quantization of the classical linear system of non-interacting wave fields
(recall, please, their wonderful explanation of the duality $\tiny{\ll}$wave-particle$\tiny \gg$: ``quantization of the classical wave = system with symmetrized/antisymmetrized states of 
arbitrary many quantum particles'' \cite{TomonagaII}).
The point lies in extension of the result shown by Dirac, Wigner and Jordan:
\begin{enumerate}
\item[]
\emph{quantization of the 
classical wave = system with symmetrized or antisymmetrized states of arbitrary many quantum particles}, 
\end{enumerate}
proven strictly for a class of classical interacting non-relativistic wave fields in the external binding potential which leads to the eigenvalue problem with purely discrete spectra (or for interacting wave fields for wave fields enclosed in a box), compare \cite{TomonagaII}, Chap. 10.74 or \cite{Heisenberg}, Appendix, \S 11, 
over to relativistic interacting classical wave fields  with realistic interactions. The interaction Hamitlonian is not the only possibility of introduction of the interaction between the quanta. Indeed inspired by the experiment we see that the scattering process is the fundamental device which allows us to read off the interaction between quanta, which at the same time allows us to avoid the whole complexity of bound states they may compose with concrete interactions. By the very nature of the scattering the particles in the initial and final state are asymptotically free, therefore the scattering operator may serve as the correct device of introduction interaction between the initially free quanta, at least for interactions which allow existence of the regime of asymptotic freedom during the scattering process.
In case of QED existence of the regime of the energy of scattered particles such that the particles become asymptotically free before and after the scattering is obvious and is achieved for the relatively low energy, but for other interactions of the SM existence of the asymptotic freedom regime for high energy scattering is much less evident and it is much more difficult to be shown its very existence.
But as far as we know, the introduction of the interaction through the scattering operator is justified
for the fields interacting as in the Standard Model.
Nonetheless, we confine attention only to such interactions. This was also the case for the milestone non relativistic quantum mechanics developed in the investigation of the atom structure: it was the scattering of alpha particles by the nucleon which allowed us to infer how the charged quantum particle interacts with the nuclei. We have similar situation for other interacting quanta:
we assume that the scattering operator is the correct tool for introduction of interaction
between the initially free quanta. 
Because the scattering operator defines interaction between quanta, we hope that all the rest
should be determined by the quantum mechanics of many quantum particles, as in the mentioned wonderful classic works of Dirac, Wigner and Jordan. But it was St\"uckelberg and then Bogoliubov, who developed this idea into a computationally effective theory. First they discovered the causality, which indeed may replace the evolution principle in construction of the scattering operator and avoids at the very start much of the difficulties pertinent to the Hamiltonian method. The switching-of-interaction
function $g$ is the tool for implementing the causality in Bogoliubov approach \cite{Bogoliubov_Shirkov}. By the comparison to the Schwinger-Tomonaga evolution equation,
Bogoliubov and his students made a guess how the other local dynamic variables should be constructed from the scattering operator \cite{Bogoliubov_Shirkov}, in particular the interacting fields.
Namely they discovered that the local interacting fields should be defined through the functional variation of the scattering operator functional, according
to the definition stated above. That they indeed deserve the name of interacting fields (e.g. in case of QED) it follows from the fact that they respect the correct operator equations
of motion. In case of QED they respect the following operator equations
\begin{align*}
\big(D - m\big) \boldsymbol{\psi} = -e \big(\gamma_\mu A^\mu \boldsymbol{\psi}\big)_{{}_{\textrm{int}}},
\\
\square A^{\mu}_{{}_{\textrm{int}}} = -e \big(\boldsymbol{{:}}\boldsymbol{\psi}^\sharp \gamma^\mu \boldsymbol{\psi} \boldsymbol{{:}}\big)_{{}_{\textrm{int}}},
\end{align*}
compare \cite{DKS1} or \cite{DutFred}. 
From this it follows that the inhomogeneous Maxell-Dirac equations 
\begin{align*}
\big(D - m\big) 
\langle\langle \Phi_{{}_{\textrm{phys}}}, \boldsymbol{\psi}_{{}_{\textrm{int}}} 
\Phi_{{}_{\textrm{phys}}} \rangle\rangle
= -e \Big\langle\Big\langle \Phi_{{}_{\textrm{phys}}},
\big(\gamma_\mu A^{\mu}_{{}_{\textrm{int}}}  \boldsymbol{\psi}_{{}_{\textrm{int}}} \big)_{{}_{\textrm{int}}}
\Phi_{{}_{\textrm{phys}}} \Big\rangle\Big\rangle, \\
d\star d \,\, \langle\langle \Phi_{{}_{\textrm{phys}}}, A_{{}_{\textrm{int}}} 
\Phi_{{}_{\textrm{phys}}} \rangle\rangle
= - e \langle\langle \Phi_{{}_{\textrm{phys}}}, j_{{}_{\textrm{int}}}
\Phi_{{}_{\textrm{phys}}} \rangle\rangle
\end{align*}
\[
\textrm{with} \,\,
j_{{}_{\textrm{int}}}^{\mu}(x)
\overset{\textrm{df}}{=}
\Big(\boldsymbol{{:}}\boldsymbol{\psi}^\sharp \gamma^\mu
\boldsymbol{\psi} \boldsymbol{{:}}\Big)_{{}_{\textrm{int}}}
\]
of the interacting
Maxwell-Dirac system are fulfilled by the ``averages'' 
\[
\langle\langle \Phi_{{}_{\textrm{phys}}},
(\, \cdot \, ) \Phi_{{}_{\textrm{phys}}} \rangle\rangle
\]
of the interacting Dirac and electromagnetic fields, in the physical states 
$\Phi_{{}_{\textrm{phys}}} \in (\boldsymbol{E})$ belonging to the Hida test space which respect
the Lorentz condition
\[
\partial_\mu  A^{\mu(-)}_{{}_{\textrm{int}}} 
\Phi_{{}_{\textrm{phys}}}
=0.
\]
But we should emphasize that $\langle\langle \cdot, \cdot \rangle\rangle$
is the canonical pairing on $(\boldsymbol{E}) \times (\boldsymbol{E})^*$
given by the evaluation $F(\Phi)$ of the functional $F \in (\boldsymbol{E})^*$ on the elements 
$\Phi \in (\boldsymbol{E})$, which coincides with the ordinary inner product only if the functional
$F$ belongs to $(\boldsymbol{E}) \subset(\boldsymbol{E})^*$. 

Recall that here $D$ is the Dirac operator, $d$ is the external differentiation
and $\star$ the Hodge star operator defined by the Minkowski metric, and 
$d\star d$ is the differential operator which transforms one-forms into one forms, 
here understood as ordinary differential operator acting on four component smooth functions -- components of one-forms.

We should also emphasize that in case of the Minkowski space-time the interacting fields
$\mathbb{A}_{{}_{\textrm{int}}}$, in particular $A_{{}_{\textrm{int}}}$, 
$\boldsymbol{\psi}_{{}_{\textrm{int}}}$ or
$\big(\boldsymbol{{:}}\boldsymbol{\psi}^\sharp \gamma^\mu
\boldsymbol{\psi} \boldsymbol{{:}}\big)_{{}_{\textrm{int}}}$, up to any arbitrary high order,
are well defined integral kernel operators, but they belong to the general class
\[
\mathscr{L}\big( (\boldsymbol{E}) \otimes \mathscr{E}, \, (\boldsymbol{E})^* \big) \cong
\mathscr{L}\Big( \mathscr{E}, \,\, \mathscr{L}\big( (\boldsymbol{E}), (\boldsymbol{E})^*\big) \, \Big),
\]
and even if the series (\ref{generalAint}) converges in the sense of Thm 4.8 of \cite{obataJFA},
they can at most represent elements of
\[
\mathscr{L}\big( (\boldsymbol{E}) \otimes \mathscr{E}, \, (\boldsymbol{E})^* \big) \cong
\mathscr{L}\Big( \mathscr{E}, \,\, \mathscr{L}\big( (\boldsymbol{E}), (\boldsymbol{E})^*\big) \, \Big),
\]
and this is the case if an only if the series
\[
\mathbb{A}_{{}_{\textrm{int}}} \Phi = \sum \limits_{l,m} \Xi(\kappa_{l,m})\Phi,
\,\,\,
\Phi \in (\boldsymbol{E}),
\]
converges in the strong dual topology of $(\boldsymbol{E})^*$. This is so because the 
finite order contributions belong in general to
\[
\mathscr{L}\big( (\boldsymbol{E}) \otimes \mathscr{E}, \, (\boldsymbol{E})^* \big) \cong
\mathscr{L}\Big( \mathscr{E}, \,\, \mathscr{L}\big( (\boldsymbol{E}), (\boldsymbol{E})^*\big) \, \Big),
\]
but do not belong to 
\[
\mathscr{L}\big( (\boldsymbol{E}) \otimes \mathscr{E}, \, (\boldsymbol{E}) \big) \cong
\mathscr{L}\Big( \mathscr{E}, \,\, \mathscr{L}\big( (\boldsymbol{E}), (\boldsymbol{E})\big) \, \Big).
\]
In particular constructed here QFT theory on the Minkowski space-time can only be applied to the generalized states, such as many-particle plane wave states of the free fields underlying the theory in question, but not to the normalizable states of the Fock space. The constructed scattering operator functional can be used for the computation of the effective cross section for the said generalized states, as we have already said in the beginning of Introduction. Another class of generalized states, \emph{i.e.} homogeneous states, pertinent to the infrared phenomena, will be discussed in the next Section \ref{infra}. 

Unfortunately the interacting counterparts $\mathbb{A}_{{}_{\textrm{int}}}$ of the conserved current densities corresponding to free fields conserved currents $\mathbb{A}$ of the the Emmy Noether integrals
$\boldsymbol{P}^{{}^{\mu}}$, $\boldsymbol{Q}_{{}_{\textrm{int}}}$  for the symmetry groups
of translations or global gauge transformations (the total charges) are singular operators. 
In particular the Emmy Noether integrals 
$\boldsymbol{P}^{{}^{\mu}}_{{}_{\textrm{int}}}$, $\boldsymbol{Q}_{{}_{\textrm{int}}}$
for the interacting fields are singular and do not represent ordinary operators
in the Fock space, and even the particular higher order contributions to
$\boldsymbol{P}^{{}^{\mu}}_{{}_{\textrm{int}}}$, $\boldsymbol{Q}_{{}_{\textrm{int}}}$
represent generalized operators which belong to 
\[
\mathscr{L}\big( (\boldsymbol{E}), (\boldsymbol{E})^*\big) 
\]
and are not well defined as operators on the Fock space. In particular the investigation
of the stationary states (say stable particles) of the operator
$\boldsymbol{P}^{{}^{0}}_{{}_{\textrm{int}}}$ is impossible, as it is not a well defined
(densely defined) self-adjoint operator on the Fock space. This is why we are forced
to stay at the level of generalized many particle plane-wave states of the free fields
and compute the cross sections only for such generalized states. Similarly in the
investigation of the global QED gauge group and the corresponding electric charge, we are
forced to investigate an idealization of the scattering process, and look at the scattering
from the sufficiently high distance when the radiation becomes practically homogeneous,
and confine attention to the generalized states which are homogeneous (compare Section \ref{infra}). 

With this theory at hand on the Minkowski space-time we are not able to 
built the operator $\boldsymbol{P}^{{}^{0}}_{{}_{\textrm{int}}}$ (and $\boldsymbol{P}^{{}^{\mu}}_{{}_{\textrm{int}}}$, $\boldsymbol{Q}_{{}_{\textrm{int}}}$) as well defined self-adjoint operator
and account for the very existence of stable (and meta-stable) particles as bound eigenstates
(or their superpositions) of the operator $\boldsymbol{P}^{{}^{0}}_{{}_{\textrm{int}}}$.
This is impossible even if we account for only the first arbitrarily high but finite number of higher order contributions. 
But we should not think that all the fundamental laws we are using are in conflict with the very existence of such particles. 
In fact the computation of the effective  cross sections
for the many particle plane wave states of the free fields, is confirmed by the experiments
in deeply inelastic scattering of the electron by the nucleon. Presented theory, which
accounts for correctly the effective cross section for the said generalized states, is based 
solely on the construction of free fields and the interaction introduced through the
scattering operator functional. Therefore it is based on the validity of the classical
free fields and the canonical commutation rules (realized through the Hida operators),
as well as on the validity of introducing interactions between the free field
quanta through the causal construction of the scattering operator functional. 
This procedure of introducing
interaction is, as we have seen, essentially local and can be repeated without any essential obstacles 
on more general globally causal space-time not necessary the flat Minkowski space-time. Moreover
the interaction Lagrange density $\mathcal{L}$, giving the first order contribution,
is introduced on the basis of Bohr's correspondence principle and validity of the interaction
of the classical fields, in case of QED the interaction in the classical Maxwell-Dirac system.
(Concerning the status of the Dirac field as a well defined classical de Broglie wave field with polarization, 
compare \cite{TomonagaII} and Introduction, Subsection \ref{G}.)
Introduction of interaction is therefore, in case of QED, based on the established validity of the Bohr's correspondence priciple and the validity of the classical electrodynamics. All said laws make sense on more general globally causal space-times. There will arise the only difficulty of constructing free fields
on more general globally causal space-time, which is related to the correct division of the fundamental ``plane wave'' solutions into positive and negative-energy solutions, but this difficulty has rather technical character. In fact it is related to a physical problem, solution of which will at the same time provide a natural solution of the division into positive and negative energy solutions.   
Namely, in changing the Minkowski space-time into some other globally causal we should explain
why the cross section computed on the Minkowski space-time for the many particle plane wave states
are correct. This will require existence of a natural identification of the generalized states
which correspond to the plane wave states on the Minkowski space-time and comparison of the 
effective cross section, which computed for the new scattering operator on the changed space-time
should coincide with that computed for the plane waves on the Minkowski space-time from the 
scattering functional on the Minkowski space-time. Inspired by the Einstein Universe
example worked out in \cite{PaneitzSegalI}-\cite{PaneitzSegalIII}, \cite{SegalZhouQED},
we confine attention to space-times into which the causal compactification of the Minkowski
space-time is causally (``conformally'', as termed by some authors) periodically embedded.
Speaking otherwise: we confine attention to space-times which are naturally realized as causal periodic coverings 
of the causal compactification of the Minkowski space-time (with appropriate boundary identifications).
This provides natural identifications of the Minkowski wave packets with the 
corresponding packets on the chosen space-time by the requirement that the only allowed packets 
on the space-time are those which arise as the extensions of the Minkowski packets (regarded as embedded
into the new space-time) and, among other things, provides the natural frequency division as inherited from 
the Minkowski space-time through the said embedding. 

While keeping the effective cross sections for the states corresponding 
to the Minkowski many particle plane-wave states essentially unchanged, and all the laws, except the 
global geometry of Minkowski space-time, we can change the global space-time geometry into
some other. The point is that on the (globally causal) space-time with compact Cauchy surfaces
the scattering operator functional and the interacting fields are much more regular.
First of all the higher order contributions to the scattering operator functional
$S(g=1) = S(\mathcal{L})$ belong to 
\[
S(g\mathcal{L}) \in \mathscr{L}\big( (\boldsymbol{E}), (\boldsymbol{E})\big),
\]
and in particular represent ordinary operators in the Fock space. Each free field operator of a massive field 
evaluated at at space-time point becomes a well defined operator in the Fock space transforming continuously the Hida space into itself. 
Moreover all higher order contributions to interacting fields (\ref{generalAint}) belong to 
\[
\mathscr{L}\big( (\boldsymbol{E}) \otimes \mathscr{E}, \, (\boldsymbol{E}) \big) \cong
\mathscr{L}\Big( \mathscr{E}, \,\, \mathscr{L}\big( (\boldsymbol{E}), (\boldsymbol{E})\big) \, \Big).
\]
Even all higher order contributions to the interacting fields evaluated at space-time 
point $x$
\[
\mathbb{A}^{(n)}_{{}_{\textrm{int}}}(g=1,x) = {\textstyle\frac{1}{n!}} \int \ud^4 x_1 \cdots \ud^4 x_n
\, \mathbb{A}^{(n)}(x_1, \ldots, x_n, x)
\]
are well defined operators on the Fock space transforming continuously the Hida space into itself
and this is so for any Wick polynomial $\mathbb{A}(x)$ in free fields of the theory.
In particular all higher order contributions to Noether integrals 
$\boldsymbol{P}^{{}^{\mu}}_{{}_{\textrm{int}}}$, $\boldsymbol{Q}_{{}_{\textrm{int}}}$
(if the space-time admits the corresponding symmetry) are well defined self adjoint
operators on the Fock space and moreover transform continuously the Hida space into itself.
We will prove those statements for the example of QED on the Einstein Universe in Section 
\ref{EUandG}. 

These results show that the singular character of interacting (and even of the free fields)
on the flat Minkowski space-time has deeper physical reason.

\subsection{Example 1: kernels $\kappa_{l,m}$ 
corresponding to $A_{{}_{\textrm{int}}}^{(1)}(g=1,x)$}\label{analysis-of-klm-A(1)}

Here we give explicit formula for the (finite set of) kernels $\kappa'_{l,m}$ for which
\[
A_{{}_{\textrm{int}}}^{(1)}(g=1) \,\, = \,\,
\sum \limits_{l,m} \Xi(\kappa'_{l,m}),
\]
\emph{i. e.} which define (finite set of) integral kernel operators, (finite) 
sum of which gives the first order contribution to the interacting electromagnetic
potential field in the adiabatic limit $g=1$. 
More explicitly (using the notation of Subsections \ref{psiBerezin-Hida}
and \ref{A=Xi0,1+Xi1,0})
\begin{multline*}
{A_{{}_{\textrm{int}}}}_{\mu}^{\, (1)}(g=1, x) = \\
= \sum \limits_{s,s'=1}^{4} \int \limits_{\mathbb{R}^3\times \mathbb{R}^3}
\kappa'_{2,0}(\boldsymbol{\p}', s', \boldsymbol{\p}, s; \mu, x) \,
\partial_{s',\boldsymbol{\p}'}^{*} \partial_{s, \boldsymbol{\p}}^{*} \, \ud^3 \boldsymbol{\p}' \ud^3 \boldsymbol{\p} \\
+ \sum \limits_{s,s'=1}^{4} \int \limits_{\mathbb{R}^3\times \mathbb{R}^3}
\kappa'_{1,1}(\boldsymbol{\p}', s', \boldsymbol{\p}, s; \mu, x) \,
\partial_{s', \boldsymbol{\p}'}^{*} \partial_{s, \boldsymbol{\p}} \, \ud^3 \boldsymbol{\p}' \ud^3 \boldsymbol{\p} \\
+\sum \limits_{s,s'=1}^{4} \int \limits_{\mathbb{R}^3\times \mathbb{R}^3}
\kappa'_{0,2}(\boldsymbol{\p}', s', \boldsymbol{\p}, s; \mu, x) \,
\partial_{s', \boldsymbol{\p}'} \partial_{s, \boldsymbol{\p}} \, \ud^3 \boldsymbol{\p}' \ud^3 \boldsymbol{\p}
\end{multline*}
or otherwise (according to the notation for the Hida operators $\partial_{s, \boldsymbol{\p}}, 
\partial_{\nu, \boldsymbol{\p}}$
\emph{i. e.} the annihilation operators $a_{s}(\boldsymbol{\p}), a_{\mu}(\boldsymbol{\p})$
introduced in Subsection \ref{psiBerezin-Hida})
\begin{multline*}
{A_{{}_{\textrm{int}}}}_{\mu}^{\, (1)}(g=1, x) = \\
= \sum \limits_{s,s'=1}^{4} \int \limits_{\mathbb{R}^3\times \mathbb{R}^3}
\kappa'_{2,0}(\boldsymbol{\p}', s', \boldsymbol{\p}, s; \mu, x) \,
a_{s'}(\boldsymbol{\p}')^{+}a_{s}(\boldsymbol{\p})^{+} \, \ud^3 \boldsymbol{\p}' \ud^3 \boldsymbol{\p} \\
+ \sum \limits_{s,s'=1}^{4} \int \limits_{\mathbb{R}^3\times \mathbb{R}^3}
\kappa'_{1,1}(\boldsymbol{\p}', s', \boldsymbol{\p}, s; \mu, x) \,
a_{s'}(\boldsymbol{\p}')^{+}a_{s}(\boldsymbol{\p}) \, \ud^3 \boldsymbol{\p}' \ud^3 \boldsymbol{\p} \\
+\sum \limits_{s,s'=1}^{4} \int \limits_{\mathbb{R}^3\times \mathbb{R}^3}
\kappa'_{0,2}(\boldsymbol{\p}', s', \boldsymbol{\p}, s; \mu, x) \,
a_{s'}(\boldsymbol{\p}') a_{s}(\boldsymbol{\p}) \, \ud^3 \boldsymbol{\p}' \ud^3 \boldsymbol{\p}
\end{multline*}
or using still another notation for the annihilation and creation operators 
(used e.g. in \cite{Scharf}, compare Subsection \ref{psiBerezin-Hida})
\begin{multline*}
{A_{{}_{\textrm{int}}}}_{\mu}^{\, (1)}(g=1, x) = \\
= \sum \limits_{s,s'=1}^{2} \int \limits_{\mathbb{R}^3\times \mathbb{R}^3}
{\kappa'}_{2,0}^{+-}(\boldsymbol{\p}', s', \boldsymbol{\p}, s; \mu, x) \,
b_{s'}(\boldsymbol{\p}')^{+}d_{s}(\boldsymbol{\p})^{+} \, \ud^3 \boldsymbol{\p}' \ud^3 \boldsymbol{\p} \\
+ \sum \limits_{s,s'=1}^{2} \int \limits_{\mathbb{R}^3\times \mathbb{R}^3}
{\kappa'}_{1,1}^{++}(\boldsymbol{\p}', s', \boldsymbol{\p}, s; \mu, x) \,
b_{s'}(\boldsymbol{\p}')^{+}b_{s}(\boldsymbol{\p}) \, \ud^3 \boldsymbol{\p}' \ud^3 \boldsymbol{\p} \\
+\sum \limits_{s,s'=1}^{2} \int \limits_{\mathbb{R}^3\times \mathbb{R}^3}
{\kappa'}_{1,1}^{--}(\boldsymbol{\p}', s', \boldsymbol{\p}, s; \mu, x) \,
d_{s'}(\boldsymbol{\p}')^{+}d_{s}(\boldsymbol{\p}) \, \ud^3 \boldsymbol{\p}' \ud^3 \boldsymbol{\p} \\
\sum \limits_{s,s'=1}^{2} \int \limits_{\mathbb{R}^3\times \mathbb{R}^3}
{\kappa'}_{0,2}^{-+}(\boldsymbol{\p}', s', \boldsymbol{\p}, s; \mu, x) \,
d_{s'}(\boldsymbol{\p}')b_{s}(\boldsymbol{\p}) \, \ud^3 \boldsymbol{\p}' \ud^3 \boldsymbol{\p}
\end{multline*} 
where we have put
\[
\kappa'_{2,0}(\boldsymbol{\p}', s', \boldsymbol{\p}, s; \mu, x)
= 
\left\{\begin{array}{ll}
{\kappa'}_{2,0}^{+-}(\boldsymbol{\p}', s', \boldsymbol{\p}, s-2; \mu, x) & 
s' = 1, 2, s = 3,4 \\
0 & \textrm{otherwise} 
\end{array} \right.,
\]
\[
\kappa'_{1,1}(\boldsymbol{\p}', s', \boldsymbol{\p}, s; \mu, x)
= 
\left\{\begin{array}{ll}
{\kappa'}_{1,1}^{++}(\boldsymbol{\p}', s', \boldsymbol{\p}, s; \mu, x) & 
s' = 1, 2, s = 1,2 \\
{\kappa'}_{1,1}^{--}(\boldsymbol{\p}', s'-2, \boldsymbol{\p}, s-2; \mu, x) & 
s' = 3,4, s = 3,4 \\
0 & \textrm{otherwise} 
\end{array} \right.,
\]
\[
\kappa'_{0,2}(\boldsymbol{\p}', s', \boldsymbol{\p}, s; \mu, x)
= 
\left\{\begin{array}{ll}
{\kappa'}_{0,2}^{-+}(\boldsymbol{\p}', s'-2, \boldsymbol{\p}, s; \mu, x) & 
s' = 3, 4, s = 1,2 \\
0 & \textrm{otherwise} 
\end{array} \right..
\]

Let us assume the standard plane wave distribution kernels, $\kappa_{0,1}$ and $\kappa_{1,0}$, 
namely (\ref{kappa_0,1}), (\ref{kappa_1,0}),
Subsection \ref{psiBerezin-Hida} and (\ref{kappa_0,1kappa_1,0A'}),
Subsection \ref{equivalentA-s}, as the plane wave distribution kernels 
 defining the free fields $\boldsymbol{\psi}$, $A$ of the 
theory. In fact, we have ignored the constant $\pm 1/2$ in (\ref{kappa_0,1}), (\ref{kappa_1,0}).

Application of the Rules II, IV and VI immediately gives the following result
\begin{multline*}
\langle {\kappa'}_{2,0}^{+-}(\zeta, \chi), \varphi \rangle \overset{\textrm{df}}{=} \\
\overset{\textrm{df}}{=} 
\sum \limits_{s,s',\mu} \int \limits_{\mathbb{R}^3 \times \mathbb{R}^3 \times \mathbb{R}^4} 
{\kappa'}_{2,0}^{+-}(\boldsymbol{\p}', s', \boldsymbol{\p}, s; \mu, x) \, 
\zeta(s', \boldsymbol{\p}') \chi(s, \boldsymbol{\p})\varphi^\mu(x)
\ud^3 \boldsymbol{\p}' \ud^3 \boldsymbol{\p} \ud^4 x \\ =
-e \sum \limits_{s,s',\mu} \int \limits_{\mathbb{R}^3 \times \mathbb{R}^3} 
\ud^3 \boldsymbol{\p}' \ud^3 \boldsymbol{\p} u_{s'}(\boldsymbol{\p}')^{+}\gamma_0\gamma_\mu v_{s}(\boldsymbol{\p}) 
\frac{\widetilde{\varphi}^\mu(\boldsymbol{\p} + \boldsymbol{\p}', E(\boldsymbol{\p}) + E(\boldsymbol{\p}'))
\, \zeta(s', \boldsymbol{\p}') \chi(s, \boldsymbol{\p})}
{|\boldsymbol{\p} + \boldsymbol{\p}'|^2 - (E(\boldsymbol{\p}) + E(\boldsymbol{\p}'))^2}
\end{multline*}
\begin{multline*}
\langle {\kappa'}_{1,1}^{++}(\zeta, \chi), \varphi \rangle = \\ =
-e \sum \limits_{s,s',\mu} \int \limits_{\mathbb{R}^3 \times \mathbb{R}^3} 
\ud^3 \boldsymbol{\p}' \ud^3 \boldsymbol{\p} u_{s'}(\boldsymbol{\p}')^{+}\gamma_0\gamma_\mu u_{s}(\boldsymbol{\p}) 
\frac{\widetilde{\varphi}^\mu(\boldsymbol{\p}' - \boldsymbol{\p}, E(\boldsymbol{\p}') - E(\boldsymbol{\p}))
\, \zeta(s', \boldsymbol{\p}') \chi(s, \boldsymbol{\p})}
{|\boldsymbol{\p}' - \boldsymbol{\p}|^2 - (E(\boldsymbol{\p}') - E(\boldsymbol{\p}))^2}
\end{multline*}
\begin{multline*}
\langle {\kappa'}_{1,1}^{--}(\zeta, \chi), \varphi \rangle = \\ =
-e \sum \limits_{s,s',\mu} \int \limits_{\mathbb{R}^3 \times \mathbb{R}^3} 
\ud^3 \boldsymbol{\p}' \ud^3 \boldsymbol{\p} v_{s'}(\boldsymbol{\p}')^{+}\gamma_0\gamma_\mu v_{s}(\boldsymbol{\p}) 
\frac{\widetilde{\varphi}^\mu(\boldsymbol{\p} - \boldsymbol{\p}', E(\boldsymbol{\p}) - E(\boldsymbol{\p}'))
\, \zeta(s', \boldsymbol{\p}') \chi(s, \boldsymbol{\p})}
{|\boldsymbol{\p} - \boldsymbol{\p}'|^2 - (E(\boldsymbol{\p}) - E(\boldsymbol{\p}'))^2}
\end{multline*}
\begin{multline*}
\langle {\kappa'}_{0,2}^{-+}(\zeta, \chi), \varphi \rangle = \\ =
-e \sum \limits_{s,s',\mu} \int \limits_{\mathbb{R}^3 \times \mathbb{R}^3} 
\ud^3 \boldsymbol{\p}' \ud^3 \boldsymbol{\p} v_{s'}(\boldsymbol{\p}')^{+}\gamma_0\gamma_\mu u_{s}(\boldsymbol{\p}) 
\frac{\widetilde{\varphi}^\mu\big(-(\boldsymbol{\p} + \boldsymbol{\p}'), -(E(\boldsymbol{\p}) + E(\boldsymbol{\p}'))\big)
\, \zeta(s', \boldsymbol{\p}') \chi(s, \boldsymbol{\p})}
{|\boldsymbol{\p} + \boldsymbol{\p}'|^2 - (E(\boldsymbol{\p}) + E(\boldsymbol{\p}'))^2}
\end{multline*}
with
\[
\zeta, \chi \in  \mathcal{S}(\mathbb{R}^3; \mathbb{C}^2), \,\,\,
\varphi \in \mathscr{E}_2 = \mathcal{S}^{00}(\mathbb{R}^4; \mathbb{C}^4), \,\,\,
\widetilde{\varphi} \in  \mathscr{F}\mathscr{E}_2 = \mathcal{S}^{0}(\mathbb{R}^4; \mathbb{C}^4),
\]
and with the convention that $\mathcal{S}(\mathbb{R}^3; \mathbb{C}^2) \subset \mathcal{S}(\mathbb{R}^3; \mathbb{C}^4)
= E_1$ with the convention that only two components of $\zeta$ or $\chi$ are non zero when 
$\xi, \chi$ are regarded as elements of $E_1$. Here
\[
E(\boldsymbol{\p}) = \sqrt{|\boldsymbol{\p}|^2+m^2}, \,\,\, E(\boldsymbol{\p}') = \sqrt{|\boldsymbol{\p}'|^2+m^2}.
\]

It follows from the general Theorem \ref{g=1InteractingFieldsQED} of 
Subsection \ref{OperationsOnXi} that    
\begin{equation}\label{kappaA^(1)}
\kappa'_{2,0}, \kappa'_{1,1}, \kappa'_{0,2} \in \mathscr{L} \big(E_1 \otimes E_2, \,\, \mathscr{E}_{2}^* \big),
\end{equation}
so that (compare generalization of Thm 3.9 of \cite{obataJFA}, and Subsection \ref{psiBerezin-Hida}) 
\[
\Xi_{l,m}(\kappa'_{l,m}) \in 
\mathscr{L}\big((\boldsymbol{E}) \otimes \mathscr{E}, (\boldsymbol{E})^*\big) \cong 
\mathscr{L}\Big(\mathscr{E}, \,\, \mathscr{L}\big((\boldsymbol{E}), \, (\boldsymbol{E})^*\big) \Big).
\]

But (\ref{kappaA^(1)}) can also be shown with the help of the explicit formulas for the kernels
$\kappa'_{l,m}$ by repeating the proof of Lemma \ref{Cont.Ofkappa.kappa}, 
Subsection \ref{OperationsOnXi}. 

Moreover we have the following 
\begin{prop*}
\begin{enumerate}
\item[1)]
The bilinear map
\[
\xi \times \eta \mapsto \kappa'_{1,1}(\xi \otimes \eta), 
\,\,\,\,
\xi , \eta \in E_1, 
\]
can be extended to a separately continuous bilinear map from
\[
E_{1}^{*} \times E_{1} \,\,\, \textrm{into} \,\,\,\mathscr{L}(\mathscr{E}, \mathbb{C}) = \mathscr{E}^*.
\]
\item[2)]
The bilinear maps
\[
\xi \times \eta \mapsto \kappa'_{2,0}(\xi \otimes \eta),
\,\,\,\,
\xi \times \eta \mapsto \kappa'_{0,2}(\xi \otimes \eta), 
\,\,\,\,\,\,\,\,\,\,\,\,
\xi, \eta \in E_1, 
\]
can be extended to continuous bilinear maps from
\[
E_{1}^{*} \times E_{1}^{*} \,\,\, \textrm{into} \,\,\,\mathscr{L}(\mathscr{E}, \mathbb{C}) = \mathscr{E}^*.
\]
Therefore 
\[
\Xi_{l,m}(\kappa'_{l,m}) \in 
\mathscr{L}\big((\boldsymbol{E}) \otimes \mathscr{E}, (\boldsymbol{E})\big) \cong 
\mathscr{L}\Big(\mathscr{E}, \,\, \mathscr{L}\big((\boldsymbol{E}), \, (\boldsymbol{E})\big) \Big)
\]
and
\[
A_{{}_{\textrm{int}}}^{(1)}(g=1) \,\, = \,\,
\sum \limits_{l,m} \Xi(\kappa'_{l,m}) \in 
\mathscr{L}\big((\boldsymbol{E}) \otimes \mathscr{E}, (\boldsymbol{E})\big) \cong 
\mathscr{L}\Big(\mathscr{E}, \,\, \mathscr{L}\big((\boldsymbol{E}), \, (\boldsymbol{E})\big) \Big),
\]
by Thm. \ref{obataJFA.Thm.3.13}, Subsection \ref{psiBerezin-Hida}. 
\end{enumerate}
\end{prop*}

The same holds for the other (nonstandard) choices, 
(\ref{skappa_0,1}),
(\ref{skappa_1,0}), Subsection \ref{StandardDiracPsiField} and (\ref{kappa_0,1kappa_1,0A}), 
Subsection \ref{A=Xi0,1+Xi1,0},  of $\kappa_{0,1}$, $\kappa_{1,0}$, which define, respectively, the (nonstandard) free Dirac (\ref{standardpsi(x)}) 
and the (nonstandard) electromagnetic potential (\ref{q-A'-B}) 
fields as sums of two integral kernel operators with vector valued kernels $\kappa_{0,1}$ and $\kappa_{1,0}$.

\subsection{Example 2: kernels $\kappa_{l,m}$ 
corresponding to $\boldsymbol{\psi}_{{}_{\textrm{int}}}^{\,(1)}(g=1,x)$}\label{analysis-of-klm-Psi(1)}

Here we give explicit formula for the (finite set of) kernels $\kappa_{l,m}$ for which
\[
\boldsymbol{\psi}_{{}_{\textrm{int}}}^{(1)}(g=1) \,\, = \,\,
 \sum \limits_{l,m} \Xi(\kappa_{l,m}).
\]
\emph{i. e.} which define (finite set of) integral kernel operators, (finite) 
sum of which gives the first order contribution to the interacting Dirac field in the adiabatic 
limit $g=1$. 
More explicitly (using the notation of Subsections \ref{psiBerezin-Hida}
and \ref{A=Xi0,1+Xi1,0})
\begin{multline*}
\boldsymbol{\psi}_{{}_{\textrm{int}}}^{a \,(1)}(g=1,x) = \\
= \sum \limits_{\nu'=0}^{3} \sum \limits_{s=1}^{4} \int \limits_{\mathbb{R}^3\times \mathbb{R}^3}
\kappa_{2,0}(\boldsymbol{\p}', \nu', \boldsymbol{\p}, s; a, x) \,
\eta\partial_{\nu',\boldsymbol{\p}'}^{*}\eta \partial_{s, \boldsymbol{\p}}^{*} \, 
\ud^3 \boldsymbol{\p}' \ud^3 \boldsymbol{\p} \\
+ \sum \limits_{\nu'=0}^{3} \sum \limits_{s=1}^{4} \int \limits_{\mathbb{R}^3\times \mathbb{R}^3}
\kappa_{1,1}(\boldsymbol{\p}', \nu', \boldsymbol{\p}, s; a, x) \,
\eta \partial_{\nu', \boldsymbol{\p}'}^{*} \eta \partial_{s, \boldsymbol{\p}} \, 
\ud^3 \boldsymbol{\p}' \ud^3 \boldsymbol{\p} \\
+ \sum \limits_{\nu=0}^{3} \sum \limits_{s'=1}^{4} \int \limits_{\mathbb{R}^3\times \mathbb{R}^3}
\kappa_{1,1}(\boldsymbol{\p}', s', \boldsymbol{\p}, \nu; a, x) \,
\partial_{s', \boldsymbol{\p}'}^{*} \partial_{\nu, \boldsymbol{\p}} \, 
\ud^3 \boldsymbol{\p}' \ud^3 \boldsymbol{\p} \\
+\sum \limits_{\nu'=0}^{3} \sum \limits_{s=1}^{4} \int \limits_{\mathbb{R}^3\times \mathbb{R}^3}
\kappa_{0,2}(\boldsymbol{\p}', \nu', \boldsymbol{\p}, s; a, x) \,
\partial_{\nu', \boldsymbol{\p}'} \partial_{s, \boldsymbol{\p}} \, \ud^3 \boldsymbol{\p}' \ud^3 \boldsymbol{\p}
\end{multline*}
or otherwise (according to the notation for the Hida operators $\partial_{s, \boldsymbol{\p}}, 
\partial_{\nu, \boldsymbol{\p}}$
\emph{i. e.} the annihilation operators $a_{s}(\boldsymbol{\p}), a_{\mu}(\boldsymbol{\p})$
introduced in Subsection \ref{psiBerezin-Hida})
\begin{multline*}
\boldsymbol{\psi}_{{}_{\textrm{int}}}^{a \,(1)}(g=1,x) = \\
= \sum \limits_{\nu'=0}^{3} \sum \limits_{s=1}^{4} \int \limits_{\mathbb{R}^3\times \mathbb{R}^3}
\kappa_{2,0}(\boldsymbol{\p}', \nu', \boldsymbol{\p}, s; a, x) \,
\eta a_{\nu'}(\boldsymbol{\p}')^{+}\eta a_{s}(\boldsymbol{\p})^{+} \, 
\ud^3 \boldsymbol{\p}' \ud^3 \boldsymbol{\p} \\
+ \sum \limits_{\nu'=0}^{3} \sum \limits_{s=1}^{4} \int \limits_{\mathbb{R}^3\times \mathbb{R}^3}
\kappa_{1,1}(\boldsymbol{\p}', \nu', \boldsymbol{\p}, s; a, x) \,
\eta a_{\nu'}(\boldsymbol{\p}')^{+} \eta a_{s}(\boldsymbol{\p}) \, 
\ud^3 \boldsymbol{\p}' \ud^3 \boldsymbol{\p} \\
+ \sum \limits_{\nu=0}^{3} \sum \limits_{s'=1}^{4} \int \limits_{\mathbb{R}^3\times \mathbb{R}^3}
\kappa_{1,1}(\boldsymbol{\p}', s', \boldsymbol{\p}, \nu; a, x) \,
a_{s'}(\boldsymbol{\p}')^{+} a_{\nu}(\boldsymbol{\p}) \, 
\ud^3 \boldsymbol{\p}' \ud^3 \boldsymbol{\p} \\
+\sum \limits_{\nu'=0}^{3} \sum \limits_{s=1}^{4} \int \limits_{\mathbb{R}^3\times \mathbb{R}^3}
\kappa_{0,2}(\boldsymbol{\p}', \nu', \boldsymbol{\p}, s; a, x) \,
a_{\nu'}(\boldsymbol{\p}') a_{s}(\boldsymbol{\p}) \, \ud^3 \boldsymbol{\p}' \ud^3 \boldsymbol{\p}
\end{multline*}
or using still another notation for the annihilation and creation operators 
(used e.g. in \cite{Scharf}, compare Subsection \ref{psiBerezin-Hida})
\begin{multline*}
\boldsymbol{\psi}_{{}_{\textrm{int}}}^{a \,(1)}(g=1,x) = \\
= \sum \limits_{\nu'=0}^{3} \sum \limits_{s=1}^{2} \int \limits_{\mathbb{R}^3\times \mathbb{R}^3}
\kappa_{2,0}^{+-}(\boldsymbol{\p}', \nu', \boldsymbol{\p}, s; a, x) \,
\eta a_{\nu'}(\boldsymbol{\p}')^{+}\eta d_{s}(\boldsymbol{\p})^{+} \, 
\ud^3 \boldsymbol{\p}' \ud^3 \boldsymbol{\p} \\
+ \sum \limits_{\nu'=0}^{3} \sum \limits_{s=1}^{2} \int \limits_{\mathbb{R}^3\times \mathbb{R}^3}
\kappa_{1,1}^{++}(\boldsymbol{\p}', \nu', \boldsymbol{\p}, s; a, x) \,
\eta a_{\nu'}(\boldsymbol{\p}')^{+} \eta b_{s}(\boldsymbol{\p}) \, 
\ud^3 \boldsymbol{\p}' \ud^3 \boldsymbol{\p} \\
+ \sum \limits_{\nu=0}^{3} \sum \limits_{s'=1}^{2} \int \limits_{\mathbb{R}^3\times \mathbb{R}^3}
\kappa_{1,1}^{-+}(\boldsymbol{\p}', s', \boldsymbol{\p}, \nu; a, x) \,
d_{s'}(\boldsymbol{\p}')^{+} a_{\nu}(\boldsymbol{\p}) \, 
\ud^3 \boldsymbol{\p}' \ud^3 \boldsymbol{\p} \\
+\sum \limits_{\nu'=0}^{3} \sum \limits_{s=1}^{2} \int \limits_{\mathbb{R}^3\times \mathbb{R}^3}
\kappa_{0,2}^{++}(\boldsymbol{\p}', \nu', \boldsymbol{\p}, s; a, x) \,
a_{\nu'}(\boldsymbol{\p}') b_{s}(\boldsymbol{\p}) \, \ud^3 \boldsymbol{\p}' \ud^3 \boldsymbol{\p}
\end{multline*}
where we have put
\[
\kappa_{2,0}(\boldsymbol{\p}', \nu', \boldsymbol{\p}, s; a, x)
= 
\left\{\begin{array}{ll}
\kappa_{2,0}^{+-}(\boldsymbol{\p}', \nu', \boldsymbol{\p}, s-2; a, x) & 
s = 3,4 \\
0 & \textrm{otherwise} 
\end{array} \right.,
\]
\[
\kappa_{1,1}(\boldsymbol{\p}', \nu', \boldsymbol{\p}, s; a, x) 
= 
\left\{\begin{array}{ll}
\kappa_{1,1}^{++}(\boldsymbol{\p}', \nu', \boldsymbol{\p}, s; a, x)  & 
s = 1,2 \\
0 & \textrm{otherwise} 
\end{array} \right.,
\]
\[
\kappa_{1,1}(\boldsymbol{\p}', s', \boldsymbol{\p}, \nu; a, x) 
= 
\left\{\begin{array}{ll}
\kappa_{1,1}^{-+}(\boldsymbol{\p}', s'-2, \boldsymbol{\p}, \nu; a, x) & 
s' = 3,4 \\
0 & \textrm{otherwise} 
\end{array} \right.,
\]
\[
\kappa_{0,2}(\boldsymbol{\p}', \nu', \boldsymbol{\p}, s; a, x) 
= 
\left\{\begin{array}{ll}
\kappa_{0,2}^{++}(\boldsymbol{\p}', \nu', \boldsymbol{\p}, s; a, x) & 
s = 1,2 \\
0 & \textrm{otherwise} \\
\end{array} \right..
\]

Let us assume the standard plane wave distribution kernels, $\kappa_{0,1}$ and $\kappa_{1,0}$, 
namely (\ref{kappa_0,1}), (\ref{kappa_1,0}),
Subsection \ref{psiBerezin-Hida} and (\ref{kappa_0,1kappa_1,0A'}),
Subsection \ref{equivalentA-s}, as the plane wave distribution kernels 
 defining the free fields $\boldsymbol{\psi}$, $A$ of the 
theory. In fact, we have ignored the constant $\pm 1/2$ in (\ref{kappa_0,1}), (\ref{kappa_1,0}).

Application of the Rules II, IV and VI immediately gives the following result
\begin{multline*}
\langle \kappa_{2,0}^{+-}(\zeta, \chi), \phi \rangle \overset{\textrm{df}}{=} \\
\overset{\textrm{df}}{=} 
\sum \limits_{\nu',s,a} \int \limits_{\mathbb{R}^3 \times \mathbb{R}^3 \times \mathbb{R}^4} 
\kappa_{2,0}^{+-}(\boldsymbol{\p}', \nu', \boldsymbol{\p}, s; a, x) \, 
\zeta(s', \boldsymbol{\p}') \chi(s, \boldsymbol{\p})\phi^a(x)
\ud^3 \boldsymbol{\p}' \ud^3 \boldsymbol{\p} \ud^4 x \\ =
e \sum \limits_{\nu',s,b,c,a} \int \limits_{\mathbb{R}^3 \times \mathbb{R}^3} 
\ud^3 \boldsymbol{\p}' \ud^3 \boldsymbol{\p} v_{s}^{c}(\boldsymbol{\p}) 
\big(
-(\boldsymbol{p}'+ \boldsymbol{p}) \cdot \vec{\boldsymbol{\gamma}}_{ab} 
+(E'(\boldsymbol{p}')+E(\boldsymbol{p}) \gamma^0 + \boldsymbol{1}_{ab}m 
\big)
\gamma^{\nu'}_{bc} \,\, \times \\ \times \,\, 
\frac{\widetilde{\phi}^a(\boldsymbol{\p} + \boldsymbol{\p}', E(\boldsymbol{\p}) + E'(\boldsymbol{\p}'))
\, \zeta(\nu', \boldsymbol{\p}') \chi(s, \boldsymbol{\p})}
{2|\boldsymbol{\p}'|\big( |\boldsymbol{\p}'| E(\boldsymbol{\p}) - \langle \boldsymbol{\p}'|\boldsymbol{\p}\rangle\big)}
\end{multline*}

\begin{multline*}
\langle \kappa_{1,1}^{++}(\zeta, \chi), \phi \rangle = \\ =
e \sum \limits_{\nu',s,b,c,a} \int \limits_{\mathbb{R}^3 \times \mathbb{R}^3} 
\ud^3 \boldsymbol{\p}' \ud^3 \boldsymbol{\p} u_{s}^{c}(\boldsymbol{\p}) 
\big(
-(\boldsymbol{p}'- \boldsymbol{p}) \cdot \vec{\boldsymbol{\gamma}}_{ab} 
+(E'(\boldsymbol{p}')-E(\boldsymbol{p}) \gamma^0 + \boldsymbol{1}_{ab}m 
\big)
\gamma^{\nu'}_{bc} \,\, \times \\ \times \,\, 
\frac{\widetilde{\phi}^a(\boldsymbol{\p}'- \boldsymbol{\p}, E'(\boldsymbol{\p}') - E(\boldsymbol{\p}))
\, \zeta(\nu', \boldsymbol{\p}') \chi(s, \boldsymbol{\p})}
{2|\boldsymbol{\p}'|\big(\langle \boldsymbol{\p}'|\boldsymbol{\p}\rangle - |\boldsymbol{\p}'| E(\boldsymbol{\p}) \big)}
\end{multline*}

\begin{multline*}
\langle \kappa_{1,1}^{-+}(\zeta, \chi), \phi \rangle = \\ =
e \sum \limits_{\nu',s,b,c,a} \int \limits_{\mathbb{R}^3 \times \mathbb{R}^3} 
\ud^3 \boldsymbol{\p}' \ud^3 \boldsymbol{\p} v_{s}^{c}(\boldsymbol{\p}) 
\big(
(\boldsymbol{p}'- \boldsymbol{p}) \cdot \vec{\boldsymbol{\gamma}}_{ab} 
+(E'(\boldsymbol{p}')-E(\boldsymbol{p}) \gamma^0 + \boldsymbol{1}_{ab}m 
\big)
\gamma^{\nu'}_{bc} \,\, \times \\ \times \,\, 
\frac{\widetilde{\phi}^a(\boldsymbol{\p} - \boldsymbol{\p}', E(\boldsymbol{\p}) - E'(\boldsymbol{\p}'))
\, \zeta(\nu', \boldsymbol{\p}') \chi(s, \boldsymbol{\p})}
{2|\boldsymbol{\p}'|\big( \langle \boldsymbol{\p}'|\boldsymbol{\p}\rangle - |\boldsymbol{\p}'| E(\boldsymbol{\p})\big)}
\end{multline*}

\begin{multline*}
\langle \kappa_{0,2}^{++}(\zeta, \chi), \phi \rangle = \\ =
e \sum \limits_{\nu',b,c,a} \int \limits_{\mathbb{R}^3 \times \mathbb{R}^3} 
\ud^3 \boldsymbol{\p}' \ud^3 \boldsymbol{\p} u_{s}^{c}(\boldsymbol{\p}) 
\big(
(\boldsymbol{p}'+ \boldsymbol{p}) \cdot \vec{\boldsymbol{\gamma}}_{ab} 
-(E'(\boldsymbol{p}')+E(\boldsymbol{p}) \gamma^0 + \boldsymbol{1}_{ab}m 
\big)
\gamma^{\nu'}_{bc} \,\, \times \\ \times \,\, 
\frac{\widetilde{\phi}^a\big(-(\boldsymbol{\p} + \boldsymbol{\p}'), -(E(\boldsymbol{\p}) + E'(\boldsymbol{\p}'))\big)
\, \zeta(\nu', \boldsymbol{\p}') \chi(s, \boldsymbol{\p})}
{2|\boldsymbol{\p}'|\big( |\boldsymbol{\p}'| E(\boldsymbol{\p}) - \langle \boldsymbol{\p}'|\boldsymbol{\p}\rangle\big)}
\end{multline*}
with summation over repeated spinor indices $b,c \in \{1,2,3,4\}$ and with
\[
\zeta \in  \mathcal{S}^{0}(\mathbb{R}^3; \mathbb{C}^4) = E_2, \,\,\,
\chi \in \mathcal{S}(\mathbb{R}^3; \mathbb{C}^2),
\phi \in \mathscr{E}_1 = \mathcal{S}(\mathbb{R}^4; \mathbb{C}^4), \,\,\,
\]
and with the convention that $\mathcal{S}(\mathbb{R}^3; \mathbb{C}^2) \subset \mathcal{S}(\mathbb{R}^3; \mathbb{C}^4)
= E_1$ with the convention that only two components of $\chi$ are non-zero when 
$\chi$ is regarded as an element of $E_1$. Here
\[
E(\boldsymbol{\p}) = \sqrt{|\boldsymbol{\p}|^2 + m^2}, \,\,\,\,\,\,\, E'(\boldsymbol{\p}') = |\boldsymbol{\p}'|.
\]

It follows from the general Theorem \ref{g=1InteractingFieldsQED} of 
Subsection \ref{OperationsOnXi} that    
\begin{equation}\label{kappapsi^(1)}
\kappa_{2,0}, \kappa_{1,1}, \kappa_{0,2} \in \mathscr{L} \big(E_1 \otimes E_2, \,\, \mathscr{E}_{1}^* \big),
\end{equation}
so that (compare generalization of Thm 3.9 of \cite{obataJFA}, and Subsection \ref{psiBerezin-Hida}) 
\[
\Xi_{l,m}(\kappa'_{l,m}) \in 
\mathscr{L}\big((\boldsymbol{E}) \otimes \mathscr{E}, (\boldsymbol{E})^*\big) \cong 
\mathscr{L}\Big(\mathscr{E}, \,\, \mathscr{L}\big((\boldsymbol{E}), \, (\boldsymbol{E})^*\big) \Big).
\]

But (\ref{kappapsi^(1)}) can also be shown with the help of the explicit formulas for the kernels
$\kappa_{l,m}$ by repeating the proof of Lemma \ref{Cont.Ofkappa.kappa}, 
Subsection \ref{OperationsOnXi}. 

Thus the first order contribution to the interacting Dirac field is equal to a finite sum
\[
\boldsymbol{\psi}_{{}_{\textrm{int}}}^{(1)}(g=1) \,\, = \,\,
 \sum \limits_{l,m} \Xi(\kappa_{l,m}) \in 
\mathscr{L}\big((\boldsymbol{E}) \otimes \mathscr{E}, (\boldsymbol{E})^*\big) \cong 
\mathscr{L}\Big(\mathscr{E}, \,\, \mathscr{L}\big((\boldsymbol{E}), \, (\boldsymbol{E})^*\big) \Big)
\]
of well defined integral kernel operators $\Xi(\kappa_{l,m})$ with vector-vaued distributional kernels
in the sense of Obata, compare \cite{obataJFA} or Subsections \ref{psiBerezin-Hida} and 
\ref{OperationsOnXi}.

However 
\[
\boldsymbol{\psi}_{{}_{\textrm{int}}}^{(1)}(g=1) \,\, = \,\,
 \sum \limits_{l,m} \Xi(\kappa_{l,m}) 
\notin \mathscr{L}\big((\boldsymbol{E}) \otimes \mathscr{E}, (\boldsymbol{E})\big) \cong 
\mathscr{L}\Big(\mathscr{E}, \,\, \mathscr{L}\big((\boldsymbol{E}), \, (\boldsymbol{E})\big) \Big)
\]
similarly as for Wick products of free massless fields (such as $A_\mu(x)$) at the same space-time point
$x$ which do belong to
\[
\mathscr{L}\big((\boldsymbol{E}) \otimes \mathscr{E}, (\boldsymbol{E})^*\big) \cong 
\mathscr{L}\Big(\mathscr{E}, \,\, \mathscr{L}\big((\boldsymbol{E}), \, (\boldsymbol{E})^*\big) \Big),
\] 
but do not belong to
\[
\mathscr{L}\big((\boldsymbol{E}) \otimes \mathscr{E}, (\boldsymbol{E})\big) \cong 
\mathscr{L}\Big(\mathscr{E}, \,\, \mathscr{L}\big((\boldsymbol{E}), \, (\boldsymbol{E})\big) \Big).
\] 

The same holds for the other(nonstandard) choices, 
(\ref{skappa_0,1}),
(\ref{skappa_1,0}), Subsection \ref{StandardDiracPsiField} and (\ref{kappa_0,1kappa_1,0A}), 
Subsection \ref{A=Xi0,1+Xi1,0},  of $\kappa_{0,1}$, $\kappa_{1,0}$, which define, respectively, the (nonstandard) free Dirac (\ref{standardpsi(x)}) 
and the (nonstandard) electromagnetic potential (\ref{q-A'-B}) 
fields as sums of two integral kernel operators with vector valued kernels $\kappa_{0,1}$ and $\kappa_{1,0}$.

\pagebreak 

\vspace*{5cm}

\section{Infrared fields and the Theory of Staruszkiewicz}\label{infra}

As we have already shown it is very important for the construction of the free quantum electromagnetic four-potential field
what kind of the test function space is used. This is at least the case for the white noise construction of this field,
understood as integral kernel operator with vector-valued kernels, 
useful in the causal perturbative approach to QED (and more generally QFT), compare e.g. Thm \ref{ZeromassTestspace},
Subsect. \ref{A=Xi0,1+Xi1,0}. Construction of the free field due to Wightman is not so much sensitive to the choice
of the test space, allowing both $\mathcal{S}(\mathbb{R}^4)$ and $\mathcal{S}^{00}(\mathbb{R}^4)$, 
but at the same time Wightman's definition of quantum field is not useful in causal perturbative approach 
to physical quantum field theories like QED. The field $A$ is an integral kernel operator with vector-valued 
kernel (defining operator-valued distribution within the white noise formalism)
and its construction within the pure Hilbert space structure is impossible. We have also shown that the 
Schwartz space is not the correct space for the construction of the free field $A$ as integral kernel operator with vector-valued kernel (and generally
white noise construction of a massless free field), but instead we have to use in this case 
$\mathcal{S}^{00}(\mathbb{R}^4;\mathbb{C}^4)$
as the test space over the spacetime and $\mathcal{S}^0(\mathbb{R}^4; \mathbb{C}^4)$ for the test function space 
in the momentum picture, compare e.g. Thm. \ref{ZeromassTestspace}, Subsection \ref{A=Xi0,1+Xi1,0}. 
Moreover construction of the free fields as integral kernel operators allows us to give a rigorous meaning 
to all higher order contributions to interacting fields in the adiabatic limit $g=1$, as well defined integral kernel 
operators (Subsection \ref{OperationsOnXi} and Section \ref{A(1)psi(1)}). This construction requires the test 
function space for the electromagnetic field to be equal  $\mathscr{E} = \mathcal{S}^{00}(\mathbb{R}^4;\mathbb{C}^4)$ if one wants to have
the zero order and the first order contributions to $A_{{}_{\textrm{int}}}$ to be operators of the class
\[
\mathscr{L}\big(\mathscr{E}, \, \mathscr{L}((\boldsymbol{E}), (\boldsymbol{E}))\big).
\]
In case  $\mathscr{E} = \mathcal{S}(\mathbb{R}^4;\mathbb{C}^4)$ the higher order contributions, within the white noise setup for free fields
understood as integral kernel operators, belong to a slightly more general class
\[
\mathscr{L}\big(\mathscr{E}, \, \mathscr{L}((\boldsymbol{E}), (\boldsymbol{E})^*)\big),
\]
still allowing the operator product defined through
a limit operation, compare Subsection \ref{WickForProduct}.
 
But, and which is more important, it is only within the white noise construction of free fields,
that the higher order contributions to interacting fields exist in the adiabatic limit $g\rightarrow 1$, 
and are equal to finite sums of the generalized integral kernel operators
(with $\mathscr{E} = \mathcal{S}(\mathbb{R}^4;\mathbb{C}^4)$ as the test space) in the sense of \cite{obataJFA}
(Thm. \ref{g=1InteractingFieldsQED},
Subsection \ref{OperationsOnXi}),
which are decomposable with respect to the (continuous) decomposition of the representation
of the $SL(2, \mathbb{C})$ group acting in the Fock space of free fields (Subsections \ref{psichi} and \ref{equivalentA-s}). Moreover,
higher order contributions to interacting fields exist in the adiabatic limit $g\rightarrow 1$ as finite
sums of well-defined integral kernel operators
in the sense of \cite{obataJFA} if and only if the charged fields are massive (Thms. \ref{ExistenceIntFields.g=1.m>0}
and \ref{NonExistenceIntFields.g=1.m=0}, Subsection \ref{OperationsOnXi}).
This decomposition determines naturally decomposition of the higher order
contributions into the (asymptotically) homogeneous parts of the interacting field $A_{{}_{\textrm{int}}}$.
In this Section we provide the comparison of the (asymptotically) homogeneous part of degree $-1$ of
$A_{{}_{\textrm{int}}}$ with the theory of Staruszkiewicz.

It is remarkable that using the  test space $\mathscr{E} = \mathcal{S}^{00}(\mathbb{R}^4;\mathbb{C}^4)$ in the white noise construction of the quantum free
four-potential $A^\mu$, we obtain  
\[
A \in \mathscr{L}\big(\mathscr{E}, \, \mathscr{L}((\boldsymbol{E}), (\boldsymbol{E}))\big),
\]
and similarly for the first order contribution, but for the test space $\mathscr{E} = \mathcal{S}(\mathbb{R}^4;\mathbb{C}^4)$ we obtain 
\[
A \in \mathscr{L}\big(\mathscr{E}, \, \mathscr{L}((\boldsymbol{E}), (\boldsymbol{E})^*)\big),
\]
and similarly for the first order, and higher order, contributions. 
Moreover, usage of the  test space $\mathscr{E} = \mathcal{S}^{00}(\mathbb{R}^4;\mathbb{C}^4)$,
simplifies the treatment of the zero mas Pauli-Jordan function (in avoiding regularizations of its derivatives), and simplifies inclusion  (as generalized states) of the single particle state space of the quantum electromagnetic potential field as the states of the  homogeneous electromagnetic potential fields. 
Among them there are homogeneous of degree $-1$ ``electric type'' fields, i.e. the space of homogeneous of degree $-1$ solutions of d'Alembert equation:
\begin{equation}\label{fmuHomogeneity=-1}
\square f^\mu = 0, \,\,\, f^\mu(\lambda x) = \lambda^{-1} f^\mu(x), \,\,\, \lambda >0, 
\end{equation}
which are spanned by  Lorentz (proper ortochronous\footnote{Or of its double covering $SL(2, \mathbb{C})$, through the natural double covering homomorphism 
between these Lorentz transformations and the $SL(2, \mathbb{C})$ group.}, so reflections are excluded here) transformations 
 of the \emph{Dirac homogeneous of degree $-1$ solution}, defined by the formula 
(\ref{f^muCorrespondingTo-th(psi)}), Subsect. \ref{infra-electric-transversal-generalized-states}. Because the Dirac homogeneous of degree $-1$
solution (\ref{f^muCorrespondingTo-th(psi)}), Subsect. \ref{infra-electric-transversal-generalized-states}, is even, and reflections are excluded,
the considered here homogeneous of degree $-1$  ``electric type'' fields are even. 
These are well-defined distributions on $\mathcal{S}^{00}(\mathbb{R}^4;\mathbb{C}^4)$ and 
$\mathcal{S}^0(\mathbb{R}^4; \mathbb{C}^4)$ in the position (over space-time) and in the momentum pictures respectively. 
Moreover, the Fourier transforms $\widetilde{F}$ of infrared ``electric type'' 
solutions $F$ are not only concentrated on the light cone in the momentum picture but determine in a unique and natural fashion a regular, i.e. function-like,
distributions $\widetilde{S}$ over the disjoint sum $\mathscr{O}_{1,0,0,1} \sqcup \mathscr{O}_{-1,0,0,1}$ of the positive
$\mathscr{O}_{1,0,0,1}$ and the negative energy light cone $\mathscr{O}_{-1,0,0,1}$ in the momentum picture, 
i.e. they determine a unique continuous and regular functional $\widetilde{S}$ on
\begin{multline*}
\mathcal{S}^0(\mathscr{O}_{1,0,0,1} \sqcup \mathscr{O}_{-1,0,0,1};\mathbb{C}^4) =
\mathcal{S}^0(\mathscr{O}_{1,0,0,1}; \mathbb{C}^4) \oplus \mathcal{S}^0(\mathscr{O}_{-1,0,0,1}; \mathbb{C}^4) \\
= \mathcal{S}^0(\mathbb{R}^3; \mathbb{C}^4) \oplus \mathcal{S}^0(\mathbb{R}^3; \mathbb{C}^4).
\end{multline*}
The unique relation between $F$ and $\widetilde{S}$ is the following
\begin{equation}\label{S<->F}
\widetilde{F} (\widetilde{\varphi}) =  \widetilde{S}
\Big(\widetilde{\varphi}\Big|_{{}_{\mathscr{O}_{1,0,0,1} \sqcup \mathscr{O}_{-1,0,0,1}}} \Big).
\end{equation}
This makes sense because of continuity of the map induced by the restriction to the cone, compare 
second Proposition of Subsection \ref{Lop-on-E}. In particular to the homogeneous of degree
$-1$ Dirac solution $F \in \mathcal{S}^{00}(\mathbb{R}^4)^*$ defined
by the function (\ref{f^muCorrespondingTo-th(psi)}), there corresponds the restriction 
$\widetilde{S}$ defined by the four-vector function (\ref{thpsi-state})
on the cone.

The infrared electric type fields (\ref{fmuHomogeneity=-1})  are in general not transversal
and not fulfill vacous Maxwell equations (in the distributional sense,  understood as 
electromagnetic potentials) as they include the Coulomb potential field. Indeed the very solution 
(\ref{f^muCorrespondingTo-th(psi)}) (resp. (\ref{thpsi-state})) of Dirac is not 
transversal and defines, outside the light cone, just the Coulomb 
potential field. The transversal electric type solutions (\ref{fmuHomogeneity=-1}) are those which 
can be obtained by subtraction of the initial untransformed Dirac solution (\ref{f^muCorrespondingTo-th(psi)})
(resp. (\ref{thpsi-state})) from the Lorentz transformed Dirac solution. They generate the transversal electric type, 
and homogeneous of degree $-1$ solutions of vacous Maxwell equations.
In any case the solutions (\ref{fmuHomogeneity=-1}) represent solutions of vacous Maxwell equations
but only outside the light cone part of space-time.

The infrared solutions (\ref{fmuHomogeneity=-1}) which are generated by $SL(2, \mathbb{C})$ transforms of 
(\ref{f^muCorrespondingTo-th(psi)}) 
(resp. (\ref{thpsi-state})), and thus which are even,
have a remarkable property that they are spatial-like supported, i.e. in that part of space-time which lies 
outside the light cone. These are precisely the fields which are produced in the Bremsstrahlung radiation. 
Let us emphasize that this is true only for the
homogeneous of degree $-1$ ``electric type'' solutions, which are generated by the Dirac homogeneous of degree
$-1$ solution through Lorentz (or rather $SL(2, \mathbb{C})$) transforms, and which are even. There exist also odd transversal,
homogeneous of degree $-1$ ``electric type'' solutions which are not supported outside the light cone, which are not
produced in the Bremsstrahlung radiation.  
This makes rigorous sense for (\ref{fmuHomogeneity=-1}) or for 
(\ref{thpsi-state}) treated as distributions on $\mathcal{S}^{00}(\mathbb{R}^4)$
although this test space is much less flexible for testing locality in comparison with  $\mathcal{S}(\mathbb{R}^4)$. 
This is because this test space
contains enough elements to find for any open cone an element supported on this cone. This is enough
for example to distinguish homogeneous distributions and to check if they vanish, say,  
outside the light cone. For the proof compare Subsection \ref{splitting}. This fact that these infrared solutions (\ref{fmuHomogeneity=-1}) which are 
generated by the $SL(2, \mathbb{C})$ transforms of the Dirac homogeneous of degree $-1$ solution, \emph{i.e.} even ``electric type'' solutions, 
are supported outside the lightcone, has important physical ramifications, which will be explained below. 
The statement that the solutions (\ref{fmuHomogeneity=-1}) which are generated by $SL(2, \mathbb{C})$ transforms of (\ref{f^muCorrespondingTo-th(psi)}) 
are supported outside the light cone becomes a theorem if they are treated as elements of 
$\mathcal{S}^{00}(\mathbb{R}^4)^*$. These solutions have
also unique extensions to elements of $\mathcal{S}(\mathbb{R}^4)^*$, with the extensions determined by the condition of 
preservation of homogeneity degree $-1$, and are supported outside the light cone also as elements of $\mathcal{S}(\mathbb{R}^4)^*$. 
In accordance to the second Proposition of Subsection \ref{splitting},
we can extend the homogeneous solutions (\ref{fmuHomogeneity=-1}) generated by
(\ref{f^muCorrespondingTo-th(psi)}) over the ordinary Schwatrz test space, 
with the preservation of the homogeneity and the property that they fulfill d'Alembert equation. 
Accordingly to this Proposition, such extension is in general far not unique
and during the extension the space-time support will not in general be preserved and in general will be prolonged into the
inside part of the light cone. However in case of  solutions (\ref{fmuHomogeneity=-1}) generated by (\ref{f^muCorrespondingTo-th(psi)}),
their extension, by the same Proposition, becomes uniquely determined by the preservation of homogeneity $-1$ condition and,
treated as elements of $\mathcal{S}(\mathbb{R}^4)^*$, they are supported outside the light cone and respect d'Alembert equation.
However the (even) solutions (\ref{fmuHomogeneity=-1}) which are generated by $SL(2, \mathbb{C})$ transforms of (\ref{f^muCorrespondingTo-th(psi)}) 
also should be primary understood 
as distributions on $\mathcal{S}^{00}(\mathbb{R}^4)$. In consequence
we get the theorem that all these solutions are supported outside the light cone as distributions
in $\mathcal{S}^{00}(\mathbb{R}^4)^*$ and as distributions in  $\mathcal{S}(\mathbb{R}^4)^*$. This has important physical consequence, recognized first
by Staruszkiewicz \cite{Staruszkiewicz1987}, \cite{Staruszkiewicz}, and which we explain below.

The representation of $SL(2, \mathbb{C})$ spanned by the homogeneous of degree $-1$
solutions (\ref{fmuHomogeneity=-1}) is closely related to the ''pathological'' representation
of Dirac, which he constructed in his last research paper \cite{DiracLastPater}.

The representation of Dirac can be characterized as the representation spanned by the
scale-invariant scalar solutions of d'Alembert equation incuding Lorentz transformations 
of the scale-invariant \emph{Dirac homogeneous of degree zero solution} $f$ of d'Alembert equation
\begin{equation}\label{DiracHomogeneity=0}
\square f = 0, \,\,\, f(\lambda x) = f(x), \,\,\, \lambda >0
\end{equation} 
in the (ordinary four dimensional) Minkowski spacetime defined by (\ref{f=homogeneity=0}). Namely let 
$\boldsymbol{\x} = (x_{1}, x_{2}, x_{3})$ and
$|\boldsymbol{\x}| = \sqrt{(x_{1})^2 + (x_{2})^2 + (x_{3})^2}$. Then the regular (function-like) distribution
in $\mathcal{S}^{00}(\mathbb{R}^4)^*$ determined by the function
\begin{equation}\label{f=homogeneity=0}
f(x_0, \boldsymbol{\x}) = \left \{ \begin{array}{ll}
1 & \textrm{for $x_0 > |\boldsymbol{\x}|$} \\
\frac{x_0}{|\boldsymbol{\x}|}  & \textrm{for $-|\boldsymbol{\x}| < x_0 < |\boldsymbol{\x}|$} \\
-1 & \textrm{for $x_0 < - |\boldsymbol{\x}|$} \\
\end{array} \right.
\end{equation}  
\begin{center}
\begin{tikzpicture}[yscale=1]
    \draw[thin, ->] (2,0) -- (2.5,0);
    \draw[thin, ->] (0,1.5) -- (0,2);
    \draw[thin, domain=-1.5:1.5] plot(\x, {\x});
    \draw[thin, domain=-1.5:1.5] plot(\x, -\x);
\fill[color=gray, opacity=0.2] (0,0) -- (2,2) -- (-2,2) -- (0,0) -- cycle;
\fill[color=gray, opacity=0.1] (0,0) -- (1.5,-1.5) -- (-1.5,-1.5) -- (0,0) -- cycle;
\node [right] at (2.2,-0.2) {$\boldsymbol{\x}$};
\node [left] at (0,2) {$x_0$};
\node [above] at (0,0.6) {$f=1$};
\node [below] at (0,-0.6) {$f=-1$};
\node [right] at (0.2,0) {$f= \frac{x_0}{|\boldsymbol{\x}|}$};
\end{tikzpicture}
\end{center}
is an example of such a solution. We call it \emph{Dirac homogeneous of degree zero solution} because it was Dirac (\cite{Dirac3rdEd}, pages 303-304) who discovered it (compare also \cite{DiracLastPater}). It seems that Staruszkiewicz was the first who recognized its role for infrared fields in QED, \cite{Staruszkiewicz1987}, \cite{Staruszkiewicz}. The intimate relation
between the homogeneous of degree $-1$ electric type fields $f_\mu$ including those
spanned by the Dirac homogeneous of degree $-1$ solution (\ref{fmuHomogeneity=-1}), 
and the homogeneous of degree zero scalar solutions $f$ of d'Alembert equation,
including those spanned by the Dirac homogeneous of degree zero solution is the following. To each homogeneous of degree
$-1$ solution $f^\mu$ there correspond bi-uniquely the homogeneous of degree zero solution $f$
uniquely determined by the condition that $f = x^\mu f_\mu$ outside the light cone. In case of transversal
$f^\mu$ (i.e. such that $\partial_\mu f^\mu = 0$) solution the equality $f= x^\mu f_\mu$ 
holds in the whole space-time, but in case of the non-transversal $f_\mu$ (e.g. Dirac homogeneous of degree $-1$ solution
is not transversal) the equality $f= x^\mu f_\mu$ holds only outside the light cone. 
This relation, between the two
representations was discovered by Staruszkiewicz. We explain below the physical motivation which lead
Staruszkiewicz to this relation.

Fourier transform $\widetilde{F}$ of the \emph{Dirac homogeneous of degree zero solution} is not only concentrated on the disjoint sum 
$\mathscr{O}_{1,0,0,1} \sqcup \mathscr{O}_{-1,0,0,1}$ of the positive and negative energy light cones in the momentum space, but determines a regular distribution 
$\widetilde{S}$ on $\mathscr{O}_{1,0,0,1} \sqcup \mathscr{O}_{-1,0,0,1}$, by the formula (\ref{S<->F}), i.e. determines
a regular functional $\widetilde{S}$ on
\begin{multline*}
\mathcal{S}^0(\mathscr{O}_{1,0,0,1} \sqcup \mathscr{O}_{-1,0,0,1};\mathbb{C}^4) =
\mathcal{S}^0(\mathscr{O}_{1,0,0,1}; \mathbb{C}^4) \oplus \mathcal{S}^0(\mathscr{O}_{-1,0,0,1}; \mathbb{C}^4) \\
= \mathcal{S}^0(\mathbb{R}^3; \mathbb{C}^4) \oplus \mathcal{S}^0(\mathbb{R}^3; \mathbb{C}^4).
\end{multline*}

Staruszkiewicz recognized that the infrared fields, generated by the $SL(2, \mathbb{C})$ transforms 
of the Dirac homogeneous of degree
$-1$ solution, including the Dirac homogeneous of degree
$-1$ solution (\ref{thpsi-state}) $f_\mu(x)$ (here we will write more suggestively $A_\mu(x)$
for this solution $f_\mu(x)$) 
coinciding with the Coulomb potential outside the cone, are exceptional if subjected 
to the criterion, pertinent to the old quantum mechanics, for the electromagnetic 
field $F_{\mu\nu}(x)$ to be approximately classical. It is concisely formulated
by Berestecky, Lifshitz, and Pitaevsky, in the following form:  the field $F_{\mu\nu}(x)$
is approximately classical if ($\hslash =1=c$)
\[
\big( \Delta x^0 \big)^2 \, \sqrt{F_{01}^2 + F_{02}^2 + F_{03}^2} \gg 1.
\] 
Here $\Delta x^0$ is the observation time over which the field can be averaged without being 
significantly changed. Now in case of the ordinary Coulomb electric field
$(F_{01}, F_{02}, F_{03})$ corresponding to the ordinary Coulomb potential
we have at our disposal arbitrary long time $\Delta x^0$, in fact the whole eternity in this case.
This is in particular the case of the ordinary Coulomb field of atomic nuclei when considering 
the bound state problem of electron in the atom. Therefore by the said Berestecky-Lifshitz-Pitaevsky inequality
the Coulomb field (in particular the Coulomb field of atomic nuclei) is ``exactly''
classical. But this is not the case for the homogeneous of degree $-2$
fields $F_{\mu\nu}(x)$ corresponding to the electromagnetic infrared potential fields generated
by the $SL(2, \mathbb{C})$ transforms of the Dirac solution. Indeed for these fields, which are zero inside the light cone
$\Delta x^0$ no longer extends to the whole eternity but is confined to the outside
part of the light cone: 
$|\Delta x^0| < 2r$, $r = \sqrt{\big(x^1\big)^2 + \big(x^2\big)^2 + \big(x^3\big)^2}$.
In particular for the Dirac homogeneous of degree $-1$ solution (\ref{thpsi-state})
we obtain the ordinary Coulomb electric field $(F_{01} = x^1Q/r^3, F_{02} = x^2Q/r^3, F_{03}= x^3Q/r^3)$
confined to the outside part of the light cone
for which the Berestecky-Lifshitz-Pitaevsky inequality takes the following form
(in the natural units $\hslash=c=1$)
\[
\big( 2r\big)^2 \frac{|Q|}{r^2} \gg 1 \,\,\, \textrm{or} \,\,\,
|Q| \gg \frac{1}{4}.
\]
In the natural units, in which $\hslash=c=1$, the elementary charge $e = \tfrac{1}{\sqrt{137}}$, so that
the inequality
\[
|Q| \gg {\textstyle\frac{1}{4}}\sqrt{137} \, e \approx 2.93 \, e,
\]
must be satisfied for the total electric charge $Q$ to be approximately classical. Otherwise:
the total electric charge $Q$ is approximately classical if it is substantially larger than three
elementary charges. The last inequality condition follows from the observed value of the fine structure 
constant and from the limitation of the available observation
time pertinent to the infrared solutions (\ref{fmuHomogeneity=-1}) which are generated
by $SL(2, \mathbb{C})$ transforms of the Dirac homogeneous of degree $-1$ solution.
This inequality cannot hold for the charge $Q=e$ of a single electron.
So that the infrared field at spatial infinity, i.e. of the form (\ref{fmuHomogeneity=-1})
which accompany particles with the charge of the order of magnitude of the electron charge,
cannot be, even approximately, classical. It is important to note that scattered charged
particles produce infrared fields (Bremsstrahlung) (\ref{fmuHomogeneity=-1}), which have the form of linear combinations
of $SL(2, \mathbb{C})$ transforms of the Dirac homogeneous of degree $-1$ solution, compare
\cite{Staruszkiewicz1981}, so these fields are real.

Next Staruszkiewicz has shown in \cite{Staruszkiewicz1981}  that if a test charged particle
moves through an infrared field (\ref{fmuHomogeneity=-1}) then the phase of each plane 
wave component of the whole packet receives a finite phase shift, which changes non trivially 
the whole packet, in particular changes the norm of the whole packet. In \cite{Staruszkiewicz1981} 
the quasi classical approximation was used for the quantum particle treated non-relativistically, which is legitimate.
This supports existence of a nontrivial back-reaction of the infrared field on the quantum charged particle,
which can be measured, and confirms real character of infrared fields which have the form of linear combinations of
$SL(2, \mathbb{C})$ transforms of the Dirac homogeneous of degree $-1$ solution. 

A theory of quantum homogenous of degree $-1$
electric type fields was proposed in \cite{Staruszkiewicz1987} and \cite{Staruszkiewicz} 
where it was based on the fact that every system containing charged particles, possess
a field $S(x)$, say a phase, intimately related to the electromagnetic potential field
$A_\mu$ by the condition that $eA_\mu + \partial_\mu S$ is a gauge invariant quantity. 
As shown in \cite{Staruszkiewicz1987} and \cite{Staruszkiewicz} this condition determines 
bi-uiquely the phase $S$ corresponding to homogeneous of degree $-1$ electric type
field $f_\mu$ (\ref{fmuHomogeneity=-1}) (here written more suggestively by $A_\mu$)
from the space of solutions generated by the Dirac homogeneous of degree $-1$
solution  (\ref{thpsi-state}). This relation is precisely that indicated above
between the solutions (\ref{fmuHomogeneity=-1}) (now written $A_\mu$)
and the scalar homogeneous of degree zero solutions $f$ (\ref{DiracHomogeneity=0}) 
(now written as $S$).

As pointed in the said papers \cite{Staruszkiewicz1987} and \cite{Staruszkiewicz}, 
the zero component $\frac{1}{e}j_0$ of the current density is the momentum canonically 
conjugated to the phase $S$, thus the commutation relation naturally follows 
\[
\Big[\frac{1}{e} j_0(x), S(y)\Big]_{{}_{x_0=y_0}} = i \delta(\boldsymbol{\x} - \boldsymbol{\y}),
\]
between the phase field $S(x)$ and the zero component $j_0(x)$ of the electric current density. 
After integration over the hyperplane $x_0=y_0$ 
\[
[Q, S(x)] = ie, \,\,\,\,\,Q = \int \ud^3 \x \, j_0. 
\]
On this canonical commutation relation Staruszkiewicz's theory of quantum homogeneous of degree $-1$ electric type field 
has been based, as a theory of a quantum homogeneous of degree zero phase field $S(x)$, i.e. a quantum field 
on de Sitter $3$-hyperboloid space-time,
compare \cite{Staruszkiewicz1987} and \cite{Staruszkiewicz}. It subsumes 
the total charge operator $Q$ and the quantum Coulomb field (at spatial infinity) 
-- quantum counterpart of the Dirac homogeneous of degree 
$-1$ solution (\ref{f^muCorrespondingTo-th(psi)}).

In this Section we provide a mathematical analysis of the theory of Staruszkiewicz \cite{Staruszkiewicz}.
Namely we give a proof that the Dirac homogeneous of degree $-1$ solution as well as the homogeneous of degree
zero, and all remaining solutions (\ref{DiracHomogeneity=0}) and (\ref{fmuHomogeneity=-1})
belong to the space of continuous functionals on the space $\mathcal{S}^{00}(\mathbb{R}^4)$ 
(of $\mathbb{C}$-valued or $\mathbb{C}^4$-valued functions). We prove that their Fourier transforms belong to $\mathcal{S}^{0}(\mathbb{R}^4)^*$ 
and have support concentrated on the light cone (disjoint sum of the positive and the negative energy sheet
of the cone), and corresponding to the two sheets can be uniquely split into sums with the supports respectively equal to the positive and the negative energy sheets. Each such component is a regular functional on each sheet separately and is an element of $\mathcal{S}^0(\mathbb{R}^3)^*$ as an functional on the sheet of the cone.
These results are contained in Subsection \ref{DiracHom=-1Sol}. Next we give a more detailed description
of the \emph{standard representation} of the commutation relations proposed in \cite{Staruszkiewicz}
and a consistency proof of the axioms of the theory proposed in \cite{Staruszkiewicz}
in the standard representation.
The consistency proof is essentially based on the three pillars 1) positive definiteness of an invariant 
kernel on the Lobachevsky space proved by using Schoenberg theorem on conditionally negative definite functions, 
2) the generalized Bochner theorem for spherical-type representations of the $SL(2,\mathbb{C})$
group and finally  3) explicit construction of the representation of the $SL(2, \mathbb{C})$ group
acting in the Hilbert space of the quantum phase field $S(x)$,
together with the explicit construction of the operators $S_0, Q, c_{lm}^{+}, c_{lm}$ 
in this Hilbert space (in the notation of \cite{Staruszkiewicz}).
Proof of the first part 1) (positive definiteness of an invariant kernel) 
is presented at the end of Subsection \ref{AS}. The third part 3), i.e. 
explicit construction of $U, S_0, Q, c_{lm}^{+}, c_{lm}$ and the Hilbert space in which they act, 
is given in Subsections \ref{infra-electric-transversal-generalized-states}, \ref{globalU(1)} and \ref{Ustructure}.  
The proof of consistency
is given in Subsection \ref{Consistency}.
We give full classification of all, say  nonstandard, representations of the commutation 
relations of Staruszkiewicz theory  in Subsection \ref{globalU(1)} and characterize the standard representation 
in terms of its relation to the spectral construction of the global $U(1)$ group 
by the operators $e^{iS(u)}, (1/e) Q$ in this representation.
Finally in Subsection \ref{infra-electric-transversal-generalized-states} 
we show that the subspace of transversal infrared states
and the operators $c_{lm}, c_{lm}^{+}$ of Staruszkiewicz theory can be identified respectively  
with the transversal states of the homogeneous of degree zero part of the field $-ex_\mu A_{\textrm{int}}^{\mu}(g=1, x)$
(defined as in the Subsection \ref{IntAspatialInfty} of Introduction and more rigorously in Subsect. 
\ref{AS}) and the annihilation 
and creation operators of this homogeneous of degree zero part of the field
$-ex_\mu A_{\textrm{int}}^{\mu}(g=1, x)$. 

But this is not the whole story. Because we have learn how to compute the  
perturbative corrections to the interacting field in the adiabatic limit $g=1$
(in the causal formulation as given in \cite{Scharf}, \cite{DKS1}),
compare Subsection \ref{OperationsOnXi}, Subsection \ref{BSH}, Section \ref{A(1)psi(1)}, and Introduction, 
then we can compute the homogeneous of degree
zero part $\big(x_\mu A_{\textrm{int}}^{\mu}(g=1, x))_{{}_{\chi=0}}$  
of the interacting field $x_\mu A_{\textrm{int}}^{\mu}(g=1, x)$
in the adiabatic limit $g(x) \rightarrow 1$, and compute also the operator $Q$
of Staruszkiewicz theory by comparing $-e \big(x_\mu A_{\textrm{int}}^{\mu}(g=1, x))_{{}_{\chi=0}}$ 
to $S(x)$ of Staruszkiewicz theory, compare Introduction and Subsection \ref{IntAspatialInfty}. 
Unfortunately we have not lead the computation to an end (compare, however, Subsection \ref{A(1)chi}), 
and have not proved completely that 
$-e\big(x_\mu A_{\textrm{int}}^{\mu}(g=1, x))_{{}_{\chi=0}}$ is equal to that part of $S(x)$
which does not contain $S_0$ (compare Subsection \ref{A(1)chi}, where for the scalar QED, our proof is almost completed).

Nonetheless, we hope that we have made a step forward on the way in giving a 
proof of universality of the electric charge, emerging from comparison of $-e\big(x_\mu A_{\textrm{int}}^{\mu}(g=1, x))_{{}_{\chi=0}}$
to that part of Staruszkiewicz's phase, which does not contain $S_0$,
 compare Introduction, Subsection \ref{IntAspatialInfty} and this Section. 
Speaking more precisely:  the universality of the scale of electric charge
arises by the equality of the part of the quantum phase field $S(x)$ of Staruszkiewicz theory
 not containing $S_0$ (resp. $S(u)$) with the homogeneous of degree zero part of the interacting
field $-e\big(x^\mu A_\mu \big)_{{}_{\textrm{int}}}$. This equality guaranties existence
of $S_0$ (resp. $S(u)$)
such that $\big(e^{iS_0}, (1/e) Q\big)$ provide a spectral description of the global gauge group
$U(1)$, because this is the case for Staruszkiewicz theory, compare 
Subsection \ref{globalU(1)}. On the other hand this is possible only if the particular contributions 
to the total charge operator coming from different fields coupled to $A$ 
have common spectrum $e\mathbb{Z}$. 

Thus the universality for electric charge can be understood as arising
from the existence of the spectral realization of the global gauge $U(1)$ group 
in the Hilbert space of the homogeneous of degree zero part of the interacting field
$\big(x^\mu A_\mu \big)_{{}_{\textrm{int}}}$ of the whole system of charged fields 
coupled to the electromagnetic potential $A$.

Here we emphasize that the argument for the universality of the charge scale due to
Staruszkiewicz himself is simpler, compare the end of Subsection \ref{IntAspatialInfty}
of Introduction.

\subsection{Dirac's homogeneous of degree $-1$ solution}\label{DiracHom=-1Sol}

In the proof that  the Dirac homogeneous of degree zero  spherically symmetric function (\ref{f=homogeneity=0}) is a distributional solution of the d'Alembert equation we proceed along the general lines given by Dirac
himself in the third edition of his ``Principles'' \cite{Dirac3rdEd}, pages 276-277 and 303-304: namely we will show that the Fourier transformed distribution has the support concentrated on the light cone in the momentum space.
In his ``heuristic proof'' Dirac treats not only the scalar homogeneous solution (\ref{f=homogeneity=0})
but he simultaneously gives the proof that the following associated four-vector function 
\begin{equation}\label{fmu=homogeneity=-1}
\begin{split}
f_0(x_0, \boldsymbol{\x}) = \left \{ \begin{array}{ll}
0 & \textrm{for $x_0 > |\boldsymbol{\x}|$} \\
\frac{1}{|\boldsymbol{\x}|}  & \textrm{for $-|\boldsymbol{\x}| < x_0 < |\boldsymbol{\x}|$} \\
0 & \textrm{for $x_0 < - |\boldsymbol{\x}|$}, \\
\end{array} \right. \\
f_i(x_0, \boldsymbol{\x}) = \left \{ \begin{array}{ll}
0 & \textrm{for $x_0 > |\boldsymbol{\x}|$} \\
\frac{x_0 x_i}{|\boldsymbol{\x}|^3}  & \textrm{for $-|\boldsymbol{\x}| < x_0 < |\boldsymbol{\x}|$} \\
0 & \textrm{for $x_0 < - |\boldsymbol{\x}|$}, \\
\end{array} \right.
\end{split}
\end{equation}  
\begin{center}
\begin{tikzpicture}[yscale=1]
    \draw[thin, ->] (2,0) -- (2.5,0);
    \draw[thin, ->] (0,1.5) -- (0,2);
    \draw[thin, domain=-1.5:1.5] plot(\x, {\x});
    \draw[thin, domain=-1.5:1.5] plot(\x, -\x);
\fill[color=gray!, fill opacity=0.1] (0,0) -- (2,2) -- (-2,2) -- (0,0) -- cycle;
\fill[color=gray!, fill opacity=0.1] (0,0) -- (1.5,-1.5) -- (-1.5,-1.5) -- (0,0) -- cycle;
\node [right] at (2.2,-0.2) {$\boldsymbol{\x}$};
\node [left] at (0,2) {$x_0$};
\node [above] at (0,0.6) {$f_0=0$};
\node [below] at (0,-0.6) {$f_0=0$};
\node [right] at (0.2,0) {$f_0= \frac{1}{|\boldsymbol{\x}|}$};
\end{tikzpicture}
\end{center}
\begin{center}
\begin{tikzpicture}[yscale=1]
    \draw[thin, ->] (2,0) -- (2.5,0);
    \draw[thin, ->] (0,1.5) -- (0,2);
    \draw[thin, domain=-1.5:1.5] plot(\x, {\x});
    \draw[thin, domain=-1.5:1.5] plot(\x, -\x);
\fill[color=gray, opacity=0.1] (0,0) -- (2,2) -- (-2,2) -- (0,0) -- cycle;
\fill[color=gray, opacity=0.1] (0,0) -- (1.5,-1.5) -- (-1.5,-1.5) -- (0,0) -- cycle;
\node [right] at (2.2,-0.2) {$\boldsymbol{\x}$};
\node [left] at (0,2) {$x_0$};
\node [above] at (0,0.6) {$f_i=0$};
\node [below] at (0,-0.6) {$f_i=0$};
\node [right] at (0.2,0) {$f_i= \frac{x_0 x_i}{|\boldsymbol{\x}|^3}$};
\end{tikzpicture}
\end{center}
is a transversal homogeneous of degree $-1$ solution of d'Alemebert equation. In fact the $\mu$-component of
the last distributional solution can be obtained 
as the distributional derivative $\frac{\partial}{\partial x_\mu}$ applied to the distribution determined by
(\ref{f=homogeneity=0}). In fact the functions (\ref{f=homogeneity=0}) as well as all the component functions
(\ref{fmu=homogeneity=-1}) $f_\mu$ are locally integrable in $L^2(\mathbb{R}^4, \ud^4 x)$ so that they determine
well defined regular (function like) distributions over $\mathcal{S}(\mathbb{R}^4)$. But in fact the 
hint of Dirac suggests much more than just merely the fact that (\ref{f=homogeneity=0})
and (\ref{fmu=homogeneity=-1}) understood as distributions have supports, after Fourier transformation, 
concentrated on the light cone in the momentum space.  In fact the hint of Dirac suggests that their 
Fourier transforms should determine regular, i.e. function like, distributions
over the light cone in the momentum space. However, this is impossible if we understand the functions
(\ref{f=homogeneity=0}) and (\ref{fmu=homogeneity=-1}) as distributions over the ordinary 
Schwartz test space $\mathcal{S}(\mathbb{R}^4)$.

But if we understand (\ref{f=homogeneity=0})
and (\ref{fmu=homogeneity=-1}) as functions defining regular distributions over the test space 
$\mathcal{S}^{00}(\mathbb{R}^4)$ then the intuitive argument of Dirac, as placed in
\cite{Dirac3rdEd}, pages 276-277 and 303-304, regains its full mathematical justification:
the Fourier transforms of the distributions on $\mathcal{S}^{00}(\mathbb{R}^4)$ defined by (\ref{f=homogeneity=0})
and (\ref{fmu=homogeneity=-1}) are well defined distributions on $\mathcal{S}^{0}(\mathbb{R}^4)$
in the momentum space, and because the restriction to the cone is a continuous map 
$\mathcal{S}^{0}(\mathbb{R}^4) \rightarrow \mathcal{S}^{0}(\mathbb{R}^3) = \mathcal{S}^{0}(\mathscr{O})$
(for the positive $\mathscr{O} = \mathscr{O}_{1,0,0,1}$ as well as negative energy light cone 
$\mathscr{O} = \mathscr{O}_{-1,0,0,1}$) then indeed the Fourier transforms
of the said distributions determine unique regular, i.e. function like, distribution on the light cone
i.e. continuous, function like, functionals on
\[
\mathcal{S}^{0}(\mathbb{R}^3) \oplus \mathcal{S}^{0}(\mathbb{R}^3) 
= \mathcal{S}^{0}(\mathscr{O}_{1,0,0,1}) \oplus \mathcal{S}^{0}(\mathscr{O}_{-1,0,0,1}),
\]
with the functions on the light cone determining them precisely the same as that in the hint of Dirac.
This is exactly what we are going to show in this Subsection.

We need the following technical Lemmas.

\begin{lem}
Let $\eta$ be a function of one real variable $x_0$ belonging to $\mathcal{S}(\mathbb{R})$
and let $\eta$ be such that for some other function $u \in \mathcal{S}(\mathbb{R})$ (or a differentiable function 
$u \in L^2(\mathbb{R})$) we have
\[
\eta = u' = \frac{d u}{dx_0}.
\]
Then the following identity holds
\[
\int \limits_{\mathbb{R}} da \, \int \limits_{\mathbb{R}} dx_0 \, 
\eta(x_0) \frac{i}{a} 
\big\{ e^{i(x_0 - |\boldsymbol{\x}|)a} - e^{i(x_0 + |\boldsymbol{\x}|)a}\big\}
=
\int \limits_{-|\boldsymbol{\x}|}^{|\boldsymbol{\x}|} \eta(x_0) \, dx_0.
\]
\label{lem1-DiracH=0,-1}
\end{lem}

\qedsymbol \, 
\begin{multline*}
\int \limits_{\mathbb{R}} da \, \int \limits_{\mathbb{R}} dx_0 \, 
\eta(x_0) \frac{i}{a} 
\big\{ e^{i(x_0 - |\boldsymbol{\x}|)a} - e^{i(x_0 + |\boldsymbol{\x}|)a}\big\} \\
=
\int \limits_{\mathbb{R}} da \, \frac{i}{a} \big\{ e^{-i|\boldsymbol{\x}|a} - e^{i|\boldsymbol{\x}|a}\big\} \, 
\underbrace{\int \limits_{\mathbb{R}} dx_0 \, \eta(x_0) e^{ix_0 a}}_{\widetilde{\eta}(a)}  \\
=\int \limits_{\mathbb{R}} da \, \frac{i}{a} \widetilde{\eta}(a) e^{-i |\boldsymbol{\x}|a}
- \int \limits_{\mathbb{R}} da \, \frac{i}{a} \widetilde{\eta}(a) e^{i |\boldsymbol{\x}|a}.
\end{multline*}
Let $u$ be the function from the assumption of the Lemma, then the last expression equals to
\begin{multline*}
\int \limits_{\mathbb{R}} da \, \frac{i}{a} \widetilde{\frac{d u}{dx_0}}(a) e^{-i |\boldsymbol{\x}|a}
- \int \limits_{\mathbb{R}} da \, \frac{i}{a} \widetilde{\frac{d u}{dx_0}}(a) e^{i |\boldsymbol{\x}|a} \\
=
\int \limits_{\mathbb{R}} da \, \frac{i}{a}(-ia) \widetilde{u}(a) e^{-i |\boldsymbol{\x}|a}
- \int \limits_{\mathbb{R}} da \, \frac{i}{a}(-ia) \widetilde{u}(a) e^{i |\boldsymbol{\x}|a} \\
=
\int \limits_{\mathbb{R}} da \,  \widetilde{u}(a) e^{-i |\boldsymbol{\x}|a}
- \int \limits_{\mathbb{R}} da \, \widetilde{u}(a) e^{i |\boldsymbol{\x}|a} 
= u(|\boldsymbol{\x}|) - u(-|\boldsymbol{\x}|).
\end{multline*}

Because 
\[
\int \limits_{-\infty}^{|\boldsymbol{\x}|} \eta(x_0) \, dx_0 
= 
\int \limits_{-\infty}^{|\boldsymbol{\x}|} \frac{d u}{dx_0}(x_0) \, dx_0 
= u(|\boldsymbol{\x}|),
\]
then
\[
u(|\boldsymbol{\x}|) - u(-|\boldsymbol{\x}|) 
=
\int \limits_{-|\boldsymbol{\x}|}^{|\boldsymbol{\x}|} \eta(x_0) \, dx_0, 
\]
and our Lemma is proved.
\qed

\begin{lem}
The functions (\ref{f=homogeneity=0}) and (\ref{fmu=homogeneity=-1}) regarded as distributions
\[
\begin{split}
\mathcal{S}(\mathbb{R}^4) \ni  \varphi \mapsto (f_\mu , \varphi) =
\int \limits_{\mathbb{R}^4} f_\mu(x) \, \varphi(x) \ud^4 x \in \mathbb{C}, \\
\mathcal{S}(\mathbb{R}^4) \ni  \varphi \mapsto (f , \varphi) =
\int \limits_{\mathbb{R}^4} f(x) \, \varphi(x) \ud^4 x \in \mathbb{C},
\end{split}
\] 
over the Schwartz test function space $\mathcal{S}(\mathbb{R}^4) = \mathcal{S}_{H_{(4)}}(\mathbb{R}^4)$
are distributional solutions of the d'Alembert (i.e. wave) equation:
\[
(f, \square \varphi) = 0, (f_\mu, \square \varphi) =0, \,\,\, \varphi \in \mathcal{S}(\mathbb{R}^4);
\]
which means that their Fourier transforms regarded as distributions are concentrated on the light cone in the momentum 
space:
\[
\begin{split}
(f, \widetilde{\varphi}_1) = (f,  \widetilde{\varphi}_2) \,\,\, \textrm{and} \\
(f_\mu, \widetilde{\varphi}_1) = (f_\mu, \widetilde{\varphi}_2), \,\,\, \mu =0,1,2,3;
\end{split}
\]
whenever $\widetilde{\varphi}_1 = \widetilde{\varphi}_2$ on the light cone in the momentum space.
\label{lem2-DiracH=0,-1}
\end{lem}
\qedsymbol \,
It is easily seen that the functions (\ref{f=homogeneity=0}) and (\ref{fmu=homogeneity=-1})
are locally integrable, and thus can be regarded as regular functionals on the test function space
$\mathcal{S}(\mathbb{R}^4)$ of Schwartz.
Because $\mathcal{S}(\mathbb{R}^4) = \mathcal{S}_{H_{(4)}}(\mathbb{R}^4) = \mathcal{S}_{\Gamma_4(H_{(1))}}(\mathbb{R}^4)
= \mathcal{S}_{H_{(3)}}(\mathbb{R}^3) \otimes \mathcal{S}_{H_{(1)}}(\mathbb{R}) 
= \mathcal{S}(\mathbb{R}^3) \otimes \mathcal{S}(\mathbb{R})$, then it is sufficient to prove the Lemma
for $\varphi = \xi \otimes \eta$, $\xi \in \mathbb{S}(\mathbb{R}^3)$, $\eta \in \mathcal{S}(\mathbb{R})$, where
$\varphi(\boldsymbol{\x}, x_0) = (\xi \otimes \eta)(\boldsymbol{\x}, x_0) = \xi(\boldsymbol{\x}) \eta(x_0)$. 
Moreover, because $f$ defined by (\ref{f=homogeneity=0}) is odd: $f(- \boldsymbol{\x}, -x_0) = - f(\boldsymbol{\x},x_0)$ 
and all $f_\mu$ defined by (\ref{fmu=homogeneity=-1}) are even, for example if we put $\theta$ (as usually) for the Heaviside 
step theta function then we have:
\[
\begin{split}
f(x) = \frac{\big|x_0 + |\boldsymbol{\x}| \big| - \big|x_0 - |\boldsymbol{\x}| \big|}{2 |\boldsymbol{\x}|}, \\
f_0(x) = \frac{\theta (x_0 + |\boldsymbol{\x}|) - \theta(x_0 - |\boldsymbol{\x}|)}{|\boldsymbol{\x}|}, \\
f_i(x) = \frac{\theta (x_0 + |\boldsymbol{\x}|) - \theta(x_0 - |\boldsymbol{\x}|)}{|\boldsymbol{\x}|^3} x_0 x_i, \,\, i=1,2,3,
\end{split}
\]
then we can assume respectively that $\xi, \eta$ are odd (in analysing $f$) and even (whenever we consider
$f_\mu$). Moreover because $\mathcal{S}(\mathbb{R}^3) = \mathcal{S}_{H_{(3)}}(\mathbb{R}^3)$, where $H_{(3)}$ is the 
$3$-dimensional quantum oscillator Hamiltonian operator, which splits in the spherical coordinates, so that 
in these coordinates the eigenfunctions of $H_{(3)}$ have the general form $\xi(r, \theta, \phi) = \rho(r)\omega(\theta, \phi)$,
and because by the first Lemma of Subsection \ref{dim=1} or the first Lemma of Subsection \ref{SA=S0} valid
for any standard operator $A$, in particular for $A = H_{(3)}$ (compare also \cite{Reed_Simon}, Appendix to Ch V.3, pp. 141-145) the eigenfunctions of $H_{(3)}$ are dense in the nuclear topology of $\mathcal{S}(\mathbb{R}^3) = \mathcal{S}_{H_{(3)}}(\mathbb{R}^3)$, then we can restrict ourselves to the case when $\xi(r,\theta, \phi) = \rho(r) \omega(\theta, \phi)$
in the spherical coordinates. 

Consider for example the distribution $f_0$ (the treatment of the remaining distributions $f_1, f_2, f_3$ and $f$
is analogous). Thus we can assume $\xi, \eta$ to be even: $\xi(-\boldsymbol{\x}) = \xi(\boldsymbol{\x})$,
$\eta(-x_0) = \eta(x_0)$.  In the proof of the equality $(f_0, \square \varphi) = 0$, for all $\varphi = 
\xi \otimes \eta$, for even $\xi \in \mathcal{S}(\mathbb{R}^3)$, and even $\eta \in \mathcal{S}(\mathbb{R})$, we need the following equality
\begin{equation}\label{int1/rDeltau}
\int \limits_{\mathbb{R}^3} \ud^3 \, \boldsymbol{\x} \frac{1}{|\boldsymbol{\x}|} \Delta_{\mathbb{R}^3} u(\boldsymbol{\x}) = -4\pi u(0) = 0 
\end{equation}
for 
\[
u(\boldsymbol{\x}) = \xi(\boldsymbol{\x}) \int \limits_{-|\boldsymbol{\x}|}^{|\boldsymbol{\x}|} \eta(x_0) \ud x_0
= \xi(\boldsymbol{\x}) 2 \int \limits_{0}^{|\boldsymbol{\x}|} \eta(x_0) \ud x_0, \,\,\, \eta \in \mathcal{S}(\mathbb{R}),
\xi \in \mathcal{S}(\mathbb{R}^3).
\]
Indeed, note that $u(0) = 0$, and $u$ is continuous everywhere, locally integrable, and smooth everywhere
except the zero point, and derivative of any order of $u$ multiplied by any polynomial in $|\boldsymbol{\x}|$ tends to zero at infinity and is integrable; and in particular $\boldsymbol{x} \mapsto  \frac{\Delta_{\mathbb{R}^3} u(\boldsymbol{\x})}{|\boldsymbol{\x}|}$ is integrable. For such function 
\[
\int \limits_{\mathbb{R}^3} \frac{\Delta_{\mathbb{R}^3} u(\boldsymbol{\x})}{|\boldsymbol{\x}|} \, \ud^3 \boldsymbol{\x} 
= \lim \limits_{\epsilon \rightarrow 0} \int \limits_{|\boldsymbol{\x}| \geq \epsilon} \frac{\Delta_{\mathbb{R}^3} u(\boldsymbol{\x})}{|\boldsymbol{\x}|} \, \ud^3 \boldsymbol{\x}.
\]
Integration by parts yields (where $r = |\boldsymbol{\x}|$ and $\mathbb{S}_{r}^{2}$ stands for the $2$-sphere of radius $r$
with the invariant measure $\ud {\mu}_{\mathbb{S}_{r}^{2}}$ inherited from the euclidean $3$-space)
\begin{multline*}
\lim \limits_{\epsilon \rightarrow 0} \int \limits_{r \geq \epsilon} \frac{\Delta_{\mathbb{R}^3} u(\boldsymbol{\x})}{r} \, \ud^3 \boldsymbol{\x}
=
\lim \limits_{\epsilon \rightarrow 0} \int \limits_{r \geq \epsilon} \Delta_{\mathbb{R}^3} 
\bigg( \frac{1}{r} \bigg) u(\boldsymbol{\x})\, \ud^3 \boldsymbol{\x} \\
- \int \limits_{r =\epsilon} \frac{\partial u}{\partial r} \frac{1}{r} \, \ud {\mu}_{\mathbb{S}_{r}^{2}}
+ \int \limits_{r = \epsilon} u \frac{\partial}{\partial r} \bigg(\frac{1}{r}\bigg) \, 
\ud {\mu}_{\mathbb{S}_{r}^{2}},
\end{multline*}
where the first term vanishes, the second is of order $\epsilon$, and the third is equal to
\[
- \epsilon^{-2} \int \limits_{r = \epsilon} u \, \ud {\mu}_{\mathbb{S}_{r}^{2}} 
\]
i.e. $\epsilon^{-2}$ times the average of $u$ on the sphere of radius $\epsilon$. Thus letting $\epsilon \rightarrow 0$
we get (\ref{int1/rDeltau}). 

Thus for any $\varphi = \xi \otimes \eta$ with even $\xi \in \mathcal{S}(\mathbb{R}^3)$ of the form
$\xi(r,\theta, \phi) = \rho(r) \omega(\theta,\phi)$ in the spherical coordinates and even 
$\eta \in \mathcal{S}(\mathbb{R})$, the equality (\ref{int1/rDeltau}) yields
on introducing $g(r) = \int_{0}^{r} \eta(x_0) \ud x_0$:
\begin{multline*}
\int \limits_{\mathbb{R}^4} \ud^3 \boldsymbol{\x} \ud x_0 \, f_0(\boldsymbol{\x}, x_0) 
\Delta_{{}_{\boldsymbol{\x}}} \varphi(\boldsymbol{\x}, x_0) 
=
\int \limits_{\mathbb{R}^3} \ud^3 \boldsymbol{\x} \, \frac{1}{|\boldsymbol{\x}|} \Delta_{{}_{\boldsymbol{\x}}} \xi(\boldsymbol{\x}) 2 \int \limits_{0}^{|\boldsymbol{\x}|} \eta(x_0) \ud x_0 \\
= \underbrace{2 \int r^2 dr \sin \theta d \theta d \phi \frac{1}{r} 
\bigg( \frac{\partial^2}{\partial r^2} +  \frac{2}{r} \frac{\partial}{\partial r} 
+ \frac{1}{r^2} \Delta_{{}_{\mathbb{S}^2}}\bigg)
\big(\rho(r) \omega(\theta, \phi) g(r) \big)}_{\int \limits_{\mathbb{R}^3} \ud^3 \boldsymbol{\x} \, \frac{1}{|\boldsymbol{\x}|} \Delta_{{}_{\boldsymbol{\x}}} \Big( \xi(\boldsymbol{\x}) 2 \int \limits_{0}^{|\boldsymbol{\x}|} 
\eta(x_0) \ud x_0 \Big) = 0 \,\, \textrm{by (\ref{int1/rDeltau})}}  \\
-
2 \int r^2 dr \sin \theta d \theta d \phi \omega(\theta, \phi) \frac{1}{r}
\bigg(2 \frac{\partial \rho(r)}{\partial r} \underbrace{\frac{\partial g(r)}{\partial r}}_{\eta(r)} 
+  2 \frac{ \rho(r)}{r} \underbrace{\frac{\partial g(r)}{\partial r}}_{\eta(r)} 
+ \rho(r) \underbrace{\frac{\partial^2 g(r)}{\partial r^2}}_{\frac{d\eta}{dx_0}(r)} \bigg).
\end{multline*}
Integration by parts yields further
\begin{multline}\label{int1/rDelta(xi)inteta}
- 2 \int \sin \theta dr \theta d \phi \omega(\theta, \phi) r \rho(r) \frac{d \eta}{d r}(r)
- 2 \int \sin \theta dr \theta d \phi \omega(\theta, \phi) 2 \rho(r) \eta(r) \\
- 2 \int \sin \theta dr \theta d \phi \omega(\theta, \phi) 2r \frac{d \rho}{dr}(r) \eta(r) \\
=
- 2 \int \sin \theta dr \theta d \phi \omega(\theta, \phi) r \rho(r) \frac{d \eta}{d r}(r)
- 2 \int \sin \theta dr \theta d \phi \omega(\theta, \phi) 2 \frac{d }{dr} \big( r \rho(r) \big) \eta(r) \\
= 2 \int \sin \theta dr \theta d \phi \omega(\theta, \phi) r \rho(r) \frac{d \eta}{d r}(r).
\end{multline}

On the other hand for any $\varphi = \xi \otimes \eta$ with even $\xi \in \mathcal{S}(\mathbb{R}^3)$ of the form
$\xi(r,\theta, \phi) = \rho(r) \omega(\theta,\phi)$ in the spherical coordinates and even 
$\eta \in \mathcal{S}(\mathbb{R})$ we have (note that $\frac{d \eta}{dx_0}$ is odd)
\begin{multline}\label{int1/r(xi)int partial_x0^2eta}
\int \limits_{\mathbb{R}^4} \ud^3 \boldsymbol{\x} \ud x_0 \, f_0(\boldsymbol{\x}, x_0) 
\frac{\partial^2}{\partial x_{0}^{2}} \varphi(\boldsymbol{\x}, x_0) 
=
\int \limits_{\mathbb{R}^3} \ud^3 \boldsymbol{\x} \, \frac{1}{|\boldsymbol{\x}|} 
\xi(\boldsymbol{\x})  \int \limits_{-|\boldsymbol{\x}|}^{|\boldsymbol{\x}|} 
\frac{d^2 \eta}{d x_{0}^{2}}(x_0) \ud x_0 \\
= 
\int \limits_{\mathbb{R}^3} \ud^3 \boldsymbol{\x} \, \frac{1}{|\boldsymbol{\x}|} 
\xi(\boldsymbol{\x}) \bigg\{ 
\frac{d \eta}{d x_{0}}(|\boldsymbol{\x}|) 
-\frac{d \eta}{d x_{0}}(-|\boldsymbol{\x}|)\bigg\}\\
=
2\int \limits_{\mathbb{R}^3} \ud^3 \boldsymbol{\x} \, \frac{1}{|\boldsymbol{\x}|} 
\xi(\boldsymbol{\x}) 
\frac{d \eta}{d x_{0}}(|\boldsymbol{\x}|) \\
= 2 \int \sin \theta dr \theta d \phi \omega(\theta, \phi) r \rho(r) \frac{d \eta}{d r}(r).
\end{multline}

Comparing (\ref{int1/rDelta(xi)inteta}) with (\ref{int1/r(xi)int partial_x0^2eta}) we get
\[
\int \limits_{\mathbb{R}^4} \ud^3 \boldsymbol{\x} \ud x_0 \, f_0(\boldsymbol{\x}, x_0) 
\Big( -\Delta_{{}_{\boldsymbol{\x}}} + \frac{\partial^2}{\partial x_{0}^{2}} \Big) \varphi(\boldsymbol{\x}, x_0) =0
\]
for any $\varphi = \xi \otimes \eta$ with even $\xi \in \mathcal{S}(\mathbb{R}^3)$ of the form
$\xi(r,\theta, \phi) = \rho(r) \omega(\theta,\phi)$ in the spherical coordinates and even 
$\eta \in \mathcal{S}(\mathbb{R})$. Therefore 
\[
(f_0, \square \varphi) = 0, \,\,\, \varphi \in \mathcal{S}(\mathbb{R}^4).
\]

The same assertion follows similarly for $f_1,f_2, f_3$ and $f$. 

\qed

The  Fourier trasforms of the regular distributions, as functionals on $\mathcal{S}(\mathbb{R}^4)$, 
defined by the locally integrable functions
$f$ and $f_\mu$ given resp. by (\ref{f=homogeneity=0}) and (\ref{fmu=homogeneity=-1}), are of course not regular  
-- i.e. not function like -- distributions. The mentioned Ansatz of Dirac suggests much more: the Fourier transform
of the distributions (\ref{f=homogeneity=0}) and (\ref{fmu=homogeneity=-1}) should define again regular, i.e. function like, distributions
on the light cone in the momentum space. The hint of Dirac would be however impossible for the realization within the ordinary Schwartz test function space.
The test function space in the momentum space should be $\mathcal{S}^{0}(\mathbb{R}^4)$ and in the position (space time)
coordinates it should be $\mathcal{S}^{00}(\mathbb{R}^4)$ in order to save the initial intuition of Dirac.

Namely we now prove that the Fourier transform $\frac{p_\mu}{p_{0}^{2}} \widetilde{ D_0}(p) = \widetilde{f_\mu}$
is not only well defined distribution on $\mathcal{S}^0(\mathbb{R}^4)$ concentrated on the light cone, 
but determines a regular (i.e. function like)  distribution on the light cone, i.e. regular functional on
\[
\mathcal{S}^{0}(\mathbb{R}^3) \oplus \mathcal{S}^{0}(\mathbb{R}^3) 
= \mathcal{S}^{0}(\mathscr{O}_{1,0,0,1}) \oplus \mathcal{S}^{0}(\mathscr{O}_{-1,0,0,1}),
\]
which in turn is associated to the function $\boldsymbol{\p} \mapsto 
\frac{\textrm{sgn} \, (p_0) p_\mu}{p_0(\boldsymbol{\p})^2} = \frac{\textrm{sgn} \, (p_0) p_\mu}{|\boldsymbol{\p}|^2}$ on the cone in the momentum space.
Or more precisely
\begin{prop*}
\begin{enumerate}
\item[1)]
$\mathcal{S}^{0}(\mathbb{R}^4) \ni \widetilde{\varphi} \mapsto \widetilde{\varphi}|_{{}_{\mathscr{O}}}
\in \mathcal{S}^{0}(\mathbb{R}^3)$ is contunuous for $\mathscr{O} = \mathscr{O}_{1,0,0,1}$ or $\mathscr{O} = \mathscr{O}_{-1,0,0,1}$.

\item[2)]
The functional 
\begin{multline}\label{fmuDistr}
\widetilde{\varphi} \mapsto (\widetilde{f_\mu},\widetilde{\varphi}) \\
= \int \limits_{\mathscr{O}_{1,0,0,1}} \, \frac{p_\mu}{p_{0}^{2}} \, \widetilde{\varphi}|_{{}_{\mathscr{O}_{1,0,0,1}}}(p) \, 
\ud \mu|_{{}_{\mathscr{O}_{1,0,0,1}}}(p)
-
\int \limits_{\mathscr{O}_{-1,0,0,1}} \, \frac{p_\mu}{p_{0}^{2}} \, \widetilde{\varphi}|_{{}_{\mathscr{O}_{-1,0,0,1}}}(p) \, 
\ud \mu|_{{}_{\mathscr{O}_{-1,0,0,1}}}(p), \\
\,\,\, \widetilde{\varphi}  \in \mathcal{S}^{0}(\mathbb{R}^4) \, \textrm{and} \,
\varphi \in \mathcal{S}^{00}(\mathbb{R}^4)
\end{multline}
belongs to $\mathcal{S}^0(\mathbb{R}^4)^*$.

\item[3)]
The functional
\begin{equation}\label{tildefmuDistr}
\varphi \mapsto (\widetilde{f_\mu},\widetilde{\varphi}) \\
\,\,\, \widetilde{\varphi}  \in \mathcal{S}^{0}(\mathbb{R}^4) \, \textrm{and} \,
\varphi \in \mathcal{S}^{00}(\mathbb{R}^4)
\end{equation}
belongs to $\mathcal{S}^{00}(\mathbb{R}^4)^*$. 

\item[4)]
The functional
\begin{multline}\label{fmuRestCone}
\mathcal{S}^{0}(\mathbb{R}^3) \oplus \mathcal{S}^{0}(\mathbb{R}^3) \ni \widetilde{\varphi}|_{{}_{\mathscr{O}_{1,0,0,1} \sqcup \mathscr{O}_{-1,0,0,1}}} 
\mapsto (\widetilde{f_\mu}|_{{}_{\mathscr{O}_{1,0,0,1} \sqcup \mathscr{O}_{-1,0,0,1}}},
\widetilde{\varphi}|_{{}_{\mathscr{O}_{1,0,0,1} \sqcup \mathscr{O}_{-1,0,0,1}}}) \\ = 
\int \limits_{\mathscr{O}_{1,0,0,1}} \, \frac{p_\mu}{p_{0}^{2}} \, \widetilde{\varphi}|_{{}_{\mathscr{O}_{1,0,0,1}}}(p) \, 
\ud \mu|_{{}_{\mathscr{O}_{1,0,0,1}}}(p) \\
-
\int \limits_{\mathscr{O}_{-1,0,0,1}} \, \frac{p_\mu}{p_{0}^{2}} \, \widetilde{\varphi}|_{{}_{\mathscr{O}_{-1,0,0,1}}}(p) \, 
\ud \mu|_{{}_{\mathscr{O}_{-1,0,0,1}}}(p),
\end{multline}
belongs to $\mathcal{S}^{0}(\mathbb{R}^3)^* \oplus \mathcal{S}^{0}(\mathbb{R}^3)^*$ and by construction its 
composition with the restriction to the cone gives the functional from 2)
\[
(\widetilde{f_\mu}|_{{}_{\mathscr{O}_{1,0,0,1} \sqcup \mathscr{O}_{-1,0,0,1}}},\widetilde{\varphi}|_{{}_{\mathscr{O}_{1,0,0,1} \sqcup \mathscr{O}_{-1,0,0,1}}}) = (\widetilde{f_\mu},\widetilde{\varphi}).
\]
\item[5)]
And
\begin{multline*}
(\widetilde{f_\mu},\widetilde{\varphi}) = \int \limits_{\mathscr{O}_{1,0,0,1}} \, \frac{p_\mu}{p_{0}^{2}} \, \widetilde{\varphi}|_{{}_{\mathscr{O}_{1,0,0,1}}}(p) \, 
\ud \mu|_{{}_{\mathscr{O}_{1,0,0,1}}}(p)
-
\int \limits_{\mathscr{O}_{-1,0,0,1}} \, \frac{p_\mu}{p_{0}^{2}} \, \widetilde{\varphi}|_{{}_{\mathscr{O}_{-1,0,0,1}}}(p) \, 
\ud \mu|_{{}_{\mathscr{O}_{-1,0,0,1}}}(p) \\
= 
 \left \{ \begin{array}{ll}
2\pi^2 \int \limits_{x \cdot x < 0} \, \frac{1}{|\boldsymbol{\x}|} \, \varphi(x) \, \ud^4 x & \mu = 0 \\
2\pi^2 \int \limits_{x \cdot x < 0} \, \frac{x_0 x_i}{|\boldsymbol{\x}|^3} \, \varphi(x) \, \ud^4 x & \mu = 1,2,3 
\end{array} \right. \\
= 2\pi^2 \int \limits_{x \cdot x < 0} \, f_\mu(x) \, \varphi(x) \, \ud^4 x 
= (f_\mu, \varphi), \,\,\,\,\, \varphi \in \mathcal{S}^{00}(\mathbb{R}^4).
\end{multline*} 
where the functions $f_\mu$ are given by (\ref{fmu=homogeneity=-1}).
\end{enumerate}
\end{prop*} 

\qedsymbol \,
Ad. 1). The assertion 3) follows from the Lemma of Subsection \ref{dim=n} and from the Proposition of 
Subsection \ref{SA=S0}, compare also one of the Propositions of Subsection \ref{Lop-on-E}. 

Ad. 2). Continuity of the functional $\widetilde{\varphi} \mapsto (\widetilde{f_\mu},\widetilde{\varphi})$ 
follows from 1) similarly as the continuity of the Pauli-Jordan functional 
$\widetilde{\varphi} \mapsto (\widetilde{D_0},\widetilde{\varphi})$ in Subsection \ref{Lop-on-E}. 
But we prefer to give here another more explicit proof, which could have been applied also in showing 
the continuity of the Fourier transform of the Pauli-Jordan distribution 
$\widetilde{D_0} \in \mathcal{S}^{0}(\mathbb{R}^4)^*$.
 
By the results of Subsection \ref{SA=S0} we may use the system $\{||\cdot ||_m\}_{{}_{m \in \mathbb{N}}}$
of norms defined by (\ref{||.||}) on the nuclear space $\mathcal{S}^0(\mathbb{R}^4) = \mathcal{S}_{A^{(4)}}(\mathbb{R}^4)$.
Note that for the radius $r(p) = \sqrt{(p_0)^2 + (p_1)^2 + (p_2)^2 + (p_3)^2}$ we have the following relation on the cone
\[
r(p) = \sqrt{2} |\boldsymbol{\p}|, \,\,\, p = (\pm |\boldsymbol{\p}|, \boldsymbol{\p}) \in \mathscr{O}_{1,0,0,1} \sqcup \mathscr{O}_{-1,0,0,1}.
\]
We have
\begin{multline*}
|(\widetilde{f_\mu},\widetilde{\varphi})| \leq \\
=  \int \limits_{\mathscr{O}_{1,0,0,1}} \, \bigg|\frac{p_\mu}{p_{0}^{2}} \, \widetilde{\varphi}|_{{}_{\mathscr{O}_{1,0,0,1}}}(p) \bigg| \, 
\ud \mu|_{{}_{\mathscr{O}_{1,0,0,1}}}(p)
+
\int \limits_{\mathscr{O}_{-1,0,0,1}} \, \bigg| \frac{p_\mu}{p_{0}^{2}} \, \widetilde{\varphi}|_{{}_{\mathscr{O}_{-1,0,0,1}}}(p) \bigg| \, 
\ud \mu|_{{}_{\mathscr{O}_{-1,0,0,1}}}(p).
\end{multline*}

On the other hand the function 
\[
\boldsymbol{\p} \mapsto (1 + |\boldsymbol{\p}|^2)^{-2}
\] 
belongs to $L^1(\mathbb{R}^3, \ud^3 \boldsymbol{\p}) \cap L^2(\mathbb{R}^3, \ud^3 \boldsymbol{\p})$ 
and let $C$ be the $L^2$ squared norm of it.
By the assertion 1) the functions 
\[
\begin{split}
\boldsymbol{\p} \mapsto \widetilde{\varphi}(\boldsymbol{\p}, \pm |\boldsymbol{\p}|),
\boldsymbol{\p} \mapsto |\boldsymbol{\p}|^{-k} \widetilde{\varphi}(\boldsymbol{\p}, \pm |\boldsymbol{\p}|),
\,\,\,
 k = 1,2,3, \ldots, \\ 
\boldsymbol{\p} \mapsto (1 + |\boldsymbol{\p}|^2)^{2} 
\frac{1}{|\boldsymbol{\p}|^2} \widetilde{\varphi}(\boldsymbol{\p}, \pm |\boldsymbol{\p}|),
\end{split}
\]
belong to $\mathcal{S}^0(\mathbb{R}^3) \subset L^1(\mathbb{R}^3, \ud^3 \boldsymbol{\p}) \cap L^2(\mathbb{R}^3, \ud^3 \boldsymbol{\p})$, because the functions
\[
\begin{split}
\boldsymbol{\p} \mapsto |\boldsymbol{\p}|^{-k},
\,\,\,
 k = 1,2,3, \ldots, \\
\boldsymbol{\p} \mapsto (1 + |\boldsymbol{\p}|^2)^{2} 
\frac{1}{|\boldsymbol{\p}|^2}
\end{split}
\]
are multipliers of the algebra $\mathcal{S}^{0}(\mathbb{R}^3)$ (compare Subsect. \ref{diffSA} and \ref{SA=S0}).
Therefore, for the case of $\mu =0$  we have 
\begin{multline*}
|(\widetilde{f_0},\widetilde{\varphi})| \leq \\
=  \int \limits_{\mathbb{R}^3} \, \bigg|\frac{1}{|\boldsymbol{\p}|^2} \,
\widetilde{\varphi}(\boldsymbol{\p}, |\boldsymbol{\p}|) \bigg| \, 
\ud^3 \boldsymbol{\p}
+
\int \limits_{\mathbb{R}^3} \, \bigg| \frac{1}{|\boldsymbol{\p}|^{2}} \, 
\widetilde{\varphi}(\boldsymbol{\p}, -|\boldsymbol{\p}|) \bigg| \, 
\ud^3 \boldsymbol{\p} \\
=
\int \limits_{\mathbb{R}^3} \, (1 + |\boldsymbol{\p}|^2)^{-2} (1 + |\boldsymbol{\p}|^2)^{2} 
\frac{1}{|\boldsymbol{\p}|^2} \,
|\widetilde{\varphi}(\boldsymbol{\p}, |\boldsymbol{\p}|)| \, 
\ud^3 \boldsymbol{\p} \\
+
\int \limits_{\mathbb{R}^3} \, 
(1 + |\boldsymbol{\p}|^2)^{-2}(1 + |\boldsymbol{\p}|^2)^{2}
\frac{1}{|\boldsymbol{\p}|^{2}} \, 
|\widetilde{\varphi}(\boldsymbol{\p}, -|\boldsymbol{\p}|)| \, 
\ud^3 \boldsymbol{\p} \\
\leq C \sup \limits_{\boldsymbol{\p} \in \mathbb{R}^3} \frac{(1 + |\boldsymbol{\p}|^2)^{2} }{|\boldsymbol{\p}|^2}
|\widetilde{\varphi}(\boldsymbol{\p}, |\boldsymbol{\p}|)|
+
C \sup \limits_{\boldsymbol{\p} \in \mathbb{R}^3} \frac{(1 + |\boldsymbol{\p}|^2)^{2} }{|\boldsymbol{\p}|^2}
|\widetilde{\varphi}(\boldsymbol{\p}, -|\boldsymbol{\p}|)| \\
\leq C \sup \limits_{p \in \mathscr{O}_{1,0,0,1} \sqcup \mathscr{O}_{-1,0,0,1}} 
\frac{1}{2}\bigg(\frac{1}{r(p)^2} + 2 + r(p)^2 \bigg) 
|\widetilde{\varphi}(p)| \\
\leq C \sup \limits_{p \in \mathbb{R}^4} 
\frac{1}{2}\bigg(\frac{1}{r(p)^2} + 2 + r(p)^2 \bigg) 
|\widetilde{\varphi}(p)| 
\leq 6 C || \widetilde{\varphi} ||_2;
\end{multline*}
and thus the continuity of the functional 
\[
\mathcal{S}^0(\mathbb{R}^4) \ni \widetilde{\varphi} \mapsto (\widetilde{f_0},\widetilde{\varphi}) 
\]
follows.
Similarly for $\mu = i = 1,2,3$ we have
\begin{multline*}
|(\widetilde{f_i},\widetilde{\varphi})| \leq \\
=  \int \limits_{\mathbb{R}^3} \, \bigg|\frac{p_i}{|\boldsymbol{\p}|^3} \,
\widetilde{\varphi}(\boldsymbol{\p}, |\boldsymbol{\p}|) \bigg| \, 
\ud^3 \boldsymbol{\p}
+
\int \limits_{\mathbb{R}^3} \, \bigg| \frac{p_i}{|\boldsymbol{\p}|^{3}} \, 
\widetilde{\varphi}(\boldsymbol{\p}, -|\boldsymbol{\p}|) \bigg| \, 
\ud^3 \boldsymbol{\p} \\
=
\int \limits_{\mathbb{R}^3} \, (1 + |\boldsymbol{\p}|^2)^{-2} (1 + |\boldsymbol{\p}|^2)^{2} 
\frac{1}{|\boldsymbol{\p}|^2} \,
|\widetilde{\varphi}(\boldsymbol{\p}, |\boldsymbol{\p}|)| \, 
\ud^3 \boldsymbol{\p} \\
+
\int \limits_{\mathbb{R}^3} \, 
(1 + |\boldsymbol{\p}|^2)^{-2}(1 + |\boldsymbol{\p}|^2)^{2}
\frac{1}{|\boldsymbol{\p}|^{2}} \, 
|\widetilde{\varphi}(\boldsymbol{\p}, -|\boldsymbol{\p}|)| \, 
\ud^3 \boldsymbol{\p} \\
\leq C \sup \limits_{\boldsymbol{\p} \in \mathbb{R}^3} \frac{(1 + |\boldsymbol{\p}|^2)^{2} }{|\boldsymbol{\p}|^2}
|\widetilde{\varphi}(\boldsymbol{\p}, |\boldsymbol{\p}|)|
+
C \sup \limits_{\boldsymbol{\p} \in \mathbb{R}^3} \frac{(1 + |\boldsymbol{\p}|^2)^{2} }{|\boldsymbol{\p}|^2}
|\widetilde{\varphi}(\boldsymbol{\p}, -|\boldsymbol{\p}|)| \\
\leq C \sup \limits_{p \in \mathscr{O}_{1,0,0,1} \sqcup \mathscr{O}_{-1,0,0,1}} 
\frac{1}{2}\bigg(\frac{1}{r(p)^2} + 2 + r(p)^2 \bigg) 
|\widetilde{\varphi}(p)| \\
\leq C \sup \limits_{p \in \mathbb{R}^4} 
\frac{1}{2}\bigg(\frac{1}{r(p)^2} + 2 + r(p)^2 \bigg) 
|\widetilde{\varphi}(p)| 
\leq 6 C || \widetilde{\varphi} ||_2;
\end{multline*}
and the continuity of the functionals 
\[
\mathcal{S}^0(\mathbb{R}^4) \ni \widetilde{\varphi} \mapsto (\widetilde{f_i},\widetilde{\varphi}) \,\,\,
i = 1,2,3,
\]
follows.

Ad. 3). By the continuity of the Fourier transform and its inverse of the Schwartz space onto itself
and by the Proposition of Subsection \ref{SA=S0} the assertion 3) follows from the assertion 2).

Ad. 4). Because for $dim =3$  we have $r(\boldsymbol{\p}) = \sqrt{(p_1)^2 + (p_2)^2 + (p_3)^2} = |\boldsymbol{\p}|$
and by the results of Subsection \ref{diffSA} multiplication by the functions $r^{-k}$, $k \in \mathbb{N}$,
in particular multiplication by the functions $r^{-2}$ or $r^{-3}$, as well as as multliplication by the 
Cartesian coordinates $p_i$, $i = 1,2,3$, are continuous maps of $\mathcal{S}^0(\mathbb{R}^3) = \mathcal{S}_{A^{(3)}}(\mathbb{R}^3)$ into itself, then 3) follows immediately when using the norms $|| \cdot ||_m$ defined
by (\ref{||.||}) on $\mathcal{S}^0(\mathbb{R}^3) = \mathcal{S}_{A^{(3)}}(\mathbb{R}^3)$.

Ad. 5). Because the functions $f_\mu$ are locally integrable in $L^2(\mathbb{R}^4, \ud^4 p)$ then 
the right hand side of 5) is a continuous functional on $\mathcal{S}^{00}(\mathbb{R}^4)$
as a function of $\varphi$.
By 2) the left hand side of 5) is likewise a continuous functional on $\mathcal{S}^{00}(\mathbb{R}^4)$
 as a function of $\varphi$. Thus in order to prove 5) it will be sufficient to prove it
for $\varphi$ ranging over a subspace dense in $\mathcal{S}^{00}(\mathbb{R}^4)$, or what amounts to the same thing
for $\widetilde{\varphi}$ ranging over the subspace dense in $\mathcal{S}^{0}(\mathbb{R}^4)$.
Because by the results of Subsection \ref{SA=S0} the space of smooth functions with compact support (not containing the zero point) is dense in $\mathcal{S}^{0}(\mathbb{R}^4)$, it will be sufficient to prove 5) for all
$\varphi \in \mathcal{S}^{00}(\mathbb{R}^4)$ for which $\widetilde{\varphi}$ has compact support.

Note that $\mathcal{S}^0(\mathbb{R}^3) \otimes \mathcal{S}^0(\mathbb{R}) \subset \mathcal{S}^0(\mathbb{R}^4)$,
but $\mathcal{S}^0(\mathbb{R}^3) \otimes \mathcal{S}^0(\mathbb{R}) \neq \mathcal{S}^0(\mathbb{R}^4)$,
so that $\mathcal{S}^0(\mathbb{R}^3) \otimes \mathcal{S}^0(\mathbb{R})$ is not dense in the nuclear topology
in $\mathcal{S}^0(\mathbb{R}^4)$. Nonetheless the restriction to the cone of the elements
$\mathcal{S}^0(\mathbb{R}^3) \otimes \mathcal{S}^0(\mathbb{R}) \subset \mathcal{S}^0(\mathbb{R}^4)$ may approximate 
the restriction of any element of $\mathcal{S}^0(\mathbb{R}^4)$ to the cone in the nuclear topology of 
$\mathcal{S}^0(\mathbb{R}^3) \oplus \mathcal{S}^0(\mathbb{R}^3)$ on the cone,
which follows easily from the general form 
of eigenfunctions of the standard operators $A^{(n)}$ as well as the first Lemma of Subsection 
\ref{SA=S0}. By Lemma   
\ref{lem2-DiracH=0,-1} of this Subsection the value of the right hand side of 5) 
is fully determined by the restriction 
$\widetilde{\varphi}|_{{}_{\mathscr{O}_{1,0,0,1}\sqcup \mathscr{O}_{-1,0,0,1}}}(\boldsymbol{\p}) =
\widetilde{\varphi}(\boldsymbol{\p}, p_0 = \pm |\boldsymbol{\p}|)$ of the Fourier transform $\widetilde{\varphi}$ to 
the cone, and the same is obvious for the left hand side of 5). Thus it will be sufficient to prove 5) for
such $\varphi$ that $\widetilde{\varphi}$ has compact support and the
restriction $\widetilde{\varphi} (\boldsymbol{\p}, \pm |\boldsymbol{\p}|)$ has the following form 
$\widetilde{\xi} \otimes \widetilde{\eta} (\boldsymbol{\p}, \pm |\boldsymbol{\p}|) =
\widetilde{\xi} (\boldsymbol{\p}) \widetilde{\eta}(\pm |\boldsymbol{\p}|)$, with
$\widetilde{\xi} \in \mathcal{S}^0(\mathbb{R}^3)$, $\widetilde{\eta} \in \mathcal{S}^0(\mathbb{R})$ of compact support.

Thus let $\varphi$ be any such function belonging to $\mathcal{S}^{00}(\mathbb{R}^4)$ that
$\widetilde{\varphi} \in \mathcal{S}^{0}(\mathbb{R}^4)$ has compact support
and such that 
\begin{multline*}
\widetilde{\varphi} (\boldsymbol{\p}, \pm |\boldsymbol{\p}|) = \widetilde{\xi} (\boldsymbol{\p}) \widetilde{\eta}(\pm |\boldsymbol{\p}|) 
=
\int \limits_{\mathbb{R}^3} \ud^3 \, \boldsymbol{\x} \xi(\boldsymbol{\x}) e^{-i\boldsymbol{\p} \cdot \boldsymbol{\x}}
\int \limits_{\mathbb{R}} \ud x_0 \, \eta(x_0) e^{i \pm |\boldsymbol{\p}|x_0} \\
=
\int \limits_{\mathbb{R}^4} \ud^3 \boldsymbol{\x} \ud x_0 \, \xi(\boldsymbol{\x}) \eta(x_0) e^{-i \boldsymbol{\p} \cdot \boldsymbol{\x} \pm i |\boldsymbol{\p}| x_0} 
= \int \limits_{\mathbb{R}^4} \ud^4 x \, \xi \otimes \eta (x) 
e^{-i \boldsymbol{\p} \cdot \boldsymbol{\x} \pm i |\boldsymbol{\p}| x_0}
\end{multline*} 
with $\widetilde{\xi} \in \mathcal{S}^0(\mathbb{R}^3)$, 
$\widetilde{\eta} \in \mathcal{S}^0(\mathbb{R})$ of compact support and with $\xi \in \mathcal{S}^{00}(\mathbb{R}^3)$,
$\eta \in \mathcal{S}^{00}(\mathbb{R})$. We prove 5)
for such $\varphi$. By Lemma \ref{lem2-DiracH=0,-1} of this Subsection 
\begin{equation}\label{(fmu,xi-otimes-eta)}
(f_\mu, \varphi) = (f_\mu, \xi \otimes \eta).
\end{equation}
In this case where $\widetilde{\varphi}$ is of compact support we may apply the Fubini theorem to the 
integral on the left hand side of 5).

Consider for example the case $\mu = 0$. Then
\begin{multline}\label{Prooff0eq1}
(\widetilde{f_0},\widetilde{\varphi}) = \int \limits_{\mathscr{O}_{1,0,0,1}} \, \frac{p_0}{p_{0}^{2}} \, \widetilde{\varphi}|_{{}_{\mathscr{O}_{1,0,0,1}}}(p) \, 
\ud \mu|_{{}_{\mathscr{O}_{1,0,0,1}}}(p) \\
-
\int \limits_{\mathscr{O}_{-1,0,0,1}} \, \frac{p_0}{p_{0}^{2}} \, \widetilde{\varphi}|_{{}_{\mathscr{O}_{-1,0,0,1}}}(p) \, 
\ud \mu|_{{}_{\mathscr{O}_{-1,0,0,1}}}(p) \\
= 
\int \limits_{\mathbb{R}^3} \, \frac{1}{|\boldsymbol{\p}|^2} \,
\widetilde{\varphi}(\boldsymbol{\p}, |\boldsymbol{\p}|) \, 
\ud^3 \boldsymbol{\p}
+
\int \limits_{\mathbb{R}^3} \, \frac{1}{|\boldsymbol{\p}|^{2}} \, 
\widetilde{\varphi}(\boldsymbol{\p}, -|\boldsymbol{\p}|) \, 
\ud^3 \boldsymbol{\p} \\
= 
\int \limits_{\mathbb{R}^3} \,\ud^3 \boldsymbol{\p} \, \frac{1}{|\boldsymbol{\p}|^2} \, 
\int \limits_{\mathbb{R}^3} \ud^3 \boldsymbol{\x} \,
\int \limits_{\mathbb{R}} \ud x_0 
\xi(\boldsymbol{\x}) \eta(x_0) \, e^{-i\boldsymbol{\p} \cdot \boldsymbol{\x} + i|\boldsymbol{\p}|x_0} \\
+
\int \limits_{\mathbb{R}^3} \, \ud^3 \boldsymbol{\p} \, \frac{1}{|\boldsymbol{\p}|^2} \, \int \limits_{\mathbb{R}^3} \ud^3 \boldsymbol{\x} \,
\int \limits_{\mathbb{R}} \ud x_0 
\xi(\boldsymbol{\x}) \eta(x_0) \, e^{-i\boldsymbol{\p} \cdot \boldsymbol{\x} - i|\boldsymbol{\p}|x_0},
\end{multline} 
where the integrals 
\[
\int \limits_{\mathbb{R}^3} \, \ud^3 \boldsymbol{\p}  \ldots
\]
can be taken over a compact domain, e.g. a ball $\mathbb{B}$ of radius sufficiently large to contain the compact support
of the function $\widetilde{\varphi}$ restricted to the cone. 

Now consider the integrand functions 
\[
\begin{split}
h_+: \boldsymbol{\p} \times (\boldsymbol{\x} \times x_0) \mapsto 
\frac{1}{|\boldsymbol{\p}|^2} e^{-i\boldsymbol{\p} \cdot \boldsymbol{\x} + i|\boldsymbol{\p}|x_0}
\xi(\boldsymbol{\x}) \eta(x_0), \\
h_-: \boldsymbol{\p} \times (\boldsymbol{\x} \times x_0) \mapsto 
\frac{1}{|\boldsymbol{\p}|^2} e^{-i\boldsymbol{\p} \cdot \boldsymbol{\x} - i|\boldsymbol{\p}|x_0}
\xi(\boldsymbol{\x}) \eta(x_0)
\end{split}
\]
in the above expression (\ref{Prooff0eq1}). Then 

\[
h_+ = (g \otimes (\xi \otimes \eta)) \cdot e_+ \,\,\, \textrm{and} \,\,\,
h_- = (g \otimes (\xi \otimes \eta)) \cdot e_- 
\]
where $(g \otimes (\xi \otimes \eta)) (\boldsymbol{\p}, x) = g(\boldsymbol{\p}) \xi \otimes \eta(x)$
and where 
\[
\begin{split}
g(\boldsymbol{\p}) = \frac{1}{|\boldsymbol{\p}|^2} 
\,\,\,\textrm{and} \,\,\,
e_{\pm}(\boldsymbol{\p}) = e^{-i\boldsymbol{\p} \cdot \boldsymbol{\x} \pm i|\boldsymbol{\p}|x_0}.
\end{split}
\]
Because (by an easy application of the Scholium 3.9 of \cite{Segal_Kunze}) the functions 
$e_+, e_-$ are measurable of modulus one functions on the product measure space $\mathbb{B} \times \mathbb{R}^4$,
and $g, \xi \otimes \eta$ are measurable over the measure spaces $\mathbb{B}$ and $\mathbb{R}^4$ respectively, then again by Scholium 3.9 of \cite{Segal_Kunze},
$h_+$ and $h_-$ are measurable on the product measure space $\mathbb{B} \times \mathbb{R}^4$
and moreover because $g$ is integrable, i.e. belongs to $L^1(\mathbb{B}, \ud^3 \boldsymbol{\p})$
and $\xi \otimes \eta \in L^1(\mathbb{R}^4, \ud^4 x)$, then $h_+, h_-$ are integrable over the product measure
space $\mathbb{B} \times \mathbb{R}^4$ and Fubini theorem (Corollary 3.6.2 of \cite{Segal_Kunze}) 
is applicable to the integrals (\ref{Prooff0eq1}). Therefore, for the sum  of the integrals 
(\ref{Prooff0eq1}) we obtain
\begin{multline}\label{Prooff0eq2}
\int \limits_{\mathbb{R}^3} \ud^3 \boldsymbol{\x} \,
\int \limits_{\mathbb{R}^3} \,\ud^3 \boldsymbol{\p} \, \frac{1}{|\boldsymbol{\p}|^2} 
e^{-i\boldsymbol{\p} \cdot \boldsymbol{\x}}
\, \int \limits_{\mathbb{R}} \ud x_0 
\xi(\boldsymbol{\x}) \eta(x_0) \, e^{i|\boldsymbol{\p}|x_0} \\
+
\int \limits_{\mathbb{R}^3} \ud^3 \boldsymbol{\x} \,
\int \limits_{\mathbb{R}^3} \, \ud^3 \boldsymbol{\p} \, \frac{1}{|\boldsymbol{\p}|^2} 
e^{-i\boldsymbol{\p} \cdot \boldsymbol{\x}}
\, \int \limits_{\mathbb{R}} \ud x_0 
\xi(\boldsymbol{\x}) \eta(x_0) \, e^{- i|\boldsymbol{\p}|x_0} \\
=
\int \limits_{\mathbb{R}^3} \ud^3 \boldsymbol{\x} \,
\int \limits_{0}^{\infty} \,\ud |\boldsymbol{\p}| \, \int \limits_{0}^{\pi} \int \limits_{0}^{2\pi}\sin \theta \ud \theta \ud \phi 
e^{-i|\boldsymbol{\p}| |\boldsymbol{\x}| \cos \theta}
\, \int \limits_{\mathbb{R}} \ud x_0 
\xi(\boldsymbol{\x}) \eta(x_0) \, e^{i|\boldsymbol{\p}|x_0} \\
+
\int \limits_{\mathbb{R}^3} \ud^3 \boldsymbol{\x} \,
\int \limits_{0}^{\infty} \,\ud |\boldsymbol{\p}| \, \int \limits_{0}^{\pi} \int \limits_{0}^{2\pi}\sin \theta \ud \theta \ud \phi 
e^{-i|\boldsymbol{\p}| |\boldsymbol{\x}| \cos \theta}
\, \int \limits_{\mathbb{R}} \ud x_0 
\xi(\boldsymbol{\x}) \eta(x_0) \, e^{-i|\boldsymbol{\p}|x_0} \\
=
\frac{2\pi}{i}
\int \limits_{\mathbb{R}^3} \ud^3 \boldsymbol{\x} \, \frac{1}{|\boldsymbol{\x}|} \xi(\boldsymbol{\x})
\int \limits_{0}^{\infty} \,\ud |\boldsymbol{\p}| \, \frac{1}{|\boldsymbol{\p}|}
\big\{ e^{i|\boldsymbol{\p}| |\boldsymbol{\x}|} - e^{-i|\boldsymbol{\p}| |\boldsymbol{\x}|} \big\}
\, \int \limits_{\mathbb{R}} \ud x_0 
 \eta(x_0) \, e^{i|\boldsymbol{\p}|x_0} \\
+
\frac{2\pi}{i}
\int \limits_{\mathbb{R}^3} \ud^3 \boldsymbol{\x} \, \frac{1}{|\boldsymbol{\x}|} \xi(\boldsymbol{\x})
\int \limits_{0}^{\infty} \,\ud |\boldsymbol{\p}| \, \frac{1}{|\boldsymbol{\p}|}
\big\{ e^{i|\boldsymbol{\p}| |\boldsymbol{\x}|} - e^{-i|\boldsymbol{\p}| |\boldsymbol{\x}|} \big\}
\, \int \limits_{\mathbb{R}} \ud x_0 
 \eta(x_0) \, e^{-i|\boldsymbol{\p}|x_0},
\end{multline}  
where, inspired by the hint of Dirac \cite{Dirac3rdEd}, pages 276-277, we have used the polar coordinates 
$|\boldsymbol{\p}|, \theta, \phi$, with $\boldsymbol{\x}$ as pointing to the ``north pole'', 
in the integration
\[
\int \limits_{\mathbb{R}^3} \,\ud^3 \boldsymbol{\p} \, \ldots
\]
and where the range of the integration 
\[
\int \limits_{0}^{\infty} \,\ud |\boldsymbol{\p}| \, \dots
\]
in the last integrals (\ref{Prooff0eq2}) can be taken to be finite and the upper 
bound $\infty$ can be replaced with the radius of the ball $\mathbb{B}$. 

Because $\widetilde{\eta}$ belongs to $\mathcal{S}^{0}(\mathbb{R})$, then the functions defined on $\mathbb{R}$ by
\[ 
|\boldsymbol{\p}| \mapsto \frac{1}{|\boldsymbol{\p}|} \widetilde{\eta}(|\boldsymbol{\p}|), \,\,\,
-|\boldsymbol{\p}| \mapsto -\frac{1}{|\boldsymbol{\p}|} \widetilde{\eta}(-|\boldsymbol{\p}|)
\]
and 
\[ 
|\boldsymbol{\p}| \mapsto \frac{1}{|\boldsymbol{\p}|} \widetilde{\eta}(|\boldsymbol{\p}|), \,\,\,
-|\boldsymbol{\p}| \mapsto \frac{1}{|\boldsymbol{\p}|} \widetilde{\eta}(-|\boldsymbol{\p}|)
\]
belong to $\mathcal{S}^{0}(\mathbb{R}) \subset \mathcal{S}(\mathbb{R})$ by the results of 
Subsection \ref{dim=1} and \ref{SA=S0}. In particular the integrals 
\[
\int \limits_{-\infty}^{+\infty} \ud a \frac{1}{ia} e^{ia |\boldsymbol{\x}|} \widetilde{\eta}(a) 
\]
and 
\[
\int \limits_{-\infty}^{+\infty} \ud a \frac{1}{ia} e^{-ia |\boldsymbol{\x}|} \widetilde{\eta}(a) 
\]
converge absolutely, and the hint of Dirac \cite{Dirac3rdEd}, pages 276-277, becomes legitimate so that the 
sum (\ref{Prooff0eq2}) of integrals is equal to 
\begin{multline}\label{Prooff0eq3}
2\pi
\int \limits_{\mathbb{R}^3} \ud^3 \boldsymbol{\x} \, \frac{1}{|\boldsymbol{\x}|} \xi(\boldsymbol{\x})
\int \limits_{-\infty}^{+\infty} \,\ud a \, \frac{1}{ia}
e^{ia |\boldsymbol{\x}|}
\, \int \limits_{\mathbb{R}} \ud x_0 
 \eta(x_0) \, e^{iax_0} \\
-
2\pi
\int \limits_{\mathbb{R}^3} \ud^3 \boldsymbol{\x} \, \frac{1}{|\boldsymbol{\x}|} \xi(\boldsymbol{\x})
\int \limits_{-\infty}^{+\infty} \,\ud a \, \frac{1}{ia}
 e^{-ia |\boldsymbol{\x}|} 
\, \int \limits_{\mathbb{R}} \ud x_0 
 \eta(x_0) \, e^{iax_0} \\
=
2\pi
\int \limits_{\mathbb{R}^3} \ud^3 \boldsymbol{\x} \, \frac{1}{|\boldsymbol{\x}|} \xi(\boldsymbol{\x})
\int \limits_{\mathbb{R}} da \, \int \limits_{\mathbb{R}} dx_0 \, 
\eta(x_0) \frac{i}{a} 
\big\{ e^{i(x_0 - |\boldsymbol{\x}|)a} - e^{i(x_0 + |\boldsymbol{\x}|)a}\big\}.
\end{multline} 
Because $\widetilde{\eta}$ belongs to $\mathcal{S}^{0}(\mathbb{R})$, then the function defined on $\mathbb{R}$ by
\[ 
a \mapsto \frac{1}{ia} \widetilde{\eta}(a) \overset{\textrm{df}}{=} \widetilde{u}(a) 
\]
again belongs to $\mathcal{S}^{0}(\mathbb{R}) \subset \mathcal{S}(\mathbb{R})$ by the results of 
Subsection \ref{dim=1} and \ref{SA=S0}. Therefore, there exists such a function $u \in \mathcal{S}^{00}(\mathbb{R})
\subset \mathcal{S}(\mathbb{R})$  that 
\[
\eta = u' = \frac{d u}{dx_0},
\]
and the Lemma \ref{lem1-DiracH=0,-1} of this Subsection is applicable to the integral (\ref{Prooff0eq3}),
which by the said Lemma is equal to
\[
2\pi
\int \limits_{\mathbb{R}^3} \ud^3 \boldsymbol{\x} \, \frac{1}{|\boldsymbol{\x}|} \xi(\boldsymbol{\x})
\int \limits_{-|\boldsymbol{\x}|}^{|\boldsymbol{\x}|} \eta(x_0) \, \ud x_0
= 2 \pi 
\int \limits_{x \cdot x <0} \frac{1}{|\boldsymbol{\x}|} \xi \otimes \eta (x) \, \ud^4 x =
(f_0, \xi \otimes \eta),
\]
and by (\ref{(fmu,xi-otimes-eta)}) the last expression is equal to $(f_0, \varphi)$
for all $\varphi \in \mathcal{S}^{00}(\mathbb{R}^4)$ with $\widetilde{\varphi} \in \mathcal{S}^0(\mathbb{R}^4)$ of
compact support such that $\widetilde{\varphi} (\boldsymbol{\p}, \pm |\boldsymbol{\p}|) 
= \widetilde{\xi} \otimes \widetilde{\eta} (\boldsymbol{\p}, \pm |\boldsymbol{\p}|) =
\widetilde{\xi} (\boldsymbol{\p}) \widetilde{\eta}(\pm |\boldsymbol{\p}|)$, with
$\widetilde{\xi} \in \mathcal{S}^0(\mathbb{R}^3)$, $\widetilde{\eta} \in \mathcal{S}^0(\mathbb{R})$ of compact support. 
Therefore 5) is proved for $f_0$. The proof of 5) for $f_1, f_2,f_3$ is similar. 

\qed

Let $\widetilde{f}_\mu$, $\mu= 0,1,2,3$, be homogeneous of degree $-1$ measurable functions on the light cone in the momentum space whose restrictions to the unit two-sphere  $\mathbb{S}^2$ belong to $L^{1}(\mathbb{S}^2; \ud \mu_{{}_{\mathbb{S}^2}})$. For any such $\widetilde{f}_\mu$ there correspond
a regular distribution on the light cone, i.e. a continuous functional on $\mathcal{S}^0(\mathscr{O}_{1,0,0,1})
\oplus \mathcal{S}^0(\mathscr{O}_{-1,0,0,1}) = \mathcal{S}^0(\mathbb{R}^3) \oplus \mathcal{S}^0(\mathbb{R}^3)$, and 
the corresponding distribution on $\mathcal{S}^{0}(\mathbb{R}^4)$ given by 
\begin{multline}\label{Hom=-1Distr}
(\widetilde{f}_\mu,\widetilde{\varphi}) = \int \limits_{\mathscr{O}_{1,0,0,1}} \, \widetilde{f}_\mu(p)
 \, \widetilde{\varphi}|_{{}_{\mathscr{O}_{1,0,0,1}}}(p) \, 
\ud \mu|_{{}_{\mathscr{O}_{1,0,0,1}}}(p) \\
-
\int \limits_{\mathscr{O}_{-1,0,0,1}} \, \widetilde{f}_\mu(p) \, \widetilde{\varphi}|_{{}_{\mathscr{O}_{-1,0,0,1}}}(p) \, 
\ud \mu|_{{}_{\mathscr{O}_{-1,0,0,1}}}(p).
\end{multline}
Namely we have
\begin{prop*}
Let $\widetilde{f}_\mu$, $\mu= 0,1,2,3$, be homogeneous of degree $-1$ measurable functions on the light cone in the momentum space whose restrictions to the unit two-sphere  
$\mathbb{S}^2$ belong to $L^{1}(\mathbb{S}^2; \ud \mu_{{}_{\mathbb{S}^2}})$.
The functionals $\widetilde{\varphi} \mapsto (\widetilde{f}_\mu,\widetilde{\varphi}) = (f_\mu, \varphi)$,
defined by (\ref{Hom=-1Distr}), are continuous on 
$\mathcal{S}^0(\mathbb{R}^4)$ as the functions of $\widetilde{\varphi} \in \mathcal{S}^0(\mathbb{R}^4)$
as well as $\varphi \mapsto (\widetilde{f}_\mu,\widetilde{\varphi}) = (f_\mu, \varphi)$ is continuous on $\mathcal{S}^{00}(\mathbb{R}^4)$ as the function of $\varphi \in \mathcal{S}^{00}(\mathbb{R}^4)$. 
The functionals 
\[
\mathcal{S}^{0}(\mathbb{R}^3) \oplus \mathcal{S}^{0}(\mathbb{R}^3) \ni \widetilde{\varphi}|_{{}_{\mathscr{O}_{1,0,0,1} \sqcup \mathscr{O}_{-1,0,0,1}}} 
\mapsto
(\widetilde{f}_\mu,\widetilde{\varphi}), \,\,\, \mu =0,1,2,3,
\]
are continuous.
Because the support of the distribution $\widetilde{\varphi} \mapsto (\widetilde{f}_\mu,\widetilde{\varphi})$ is concetrated on the light cone, then
\[
(f_\mu, \square \varphi) = 0, 
\,\,\,\, \varphi \in \mathcal{S}^{00}(\mathbb{R}^4).
\]
\end{prop*}

\qedsymbol \,
There exists a $c_0 >0$ such that
\[
\textrm{ess sup} 
|\widetilde{f}_\mu (p)| \leq c_0 \frac{1}{|\boldsymbol{\p}|}, 
\]
with $\textrm{ess sup}$ taken over all those 
\[
p = (\pm |\boldsymbol{\p}|, \boldsymbol{\p}) \in \mathscr{O}_{1,0,0,1} \sqcup \mathscr{O}_{-1,0,0,1}
\]
which have fixed $|\boldsymbol{\p}|$, i.e. over the disjoint sum of the spheres of radius $|\boldsymbol{\p}|$, 
one in the positive, and the other one in the negative energy sheet of the cone.
Therefore the continuity follows like in the proof of 2), 3) and 4) of the preceding Proposition.
\qed

In particular, let $\widetilde{f}_\mu$ be defined on the light cone
$\mathscr{O}_{1,0,0,1}$ $\sqcup \mathscr{O}_{-1,0,0,1}$
 by the following formula
\begin{equation}\label{tildefmuHom=-1}
\widetilde{f}_\mu(p) = \textrm{sgn} (p_0) \sum \limits_{s} \, e_s \frac{u_{s\mu}}{u_s \cdot p}, \,\,\,
\end{equation}
and let 
\begin{equation}\label{fmuHom=-1}
f_\mu(x) = \theta(-x \cdot x) \sum \limits_{s} \, e_s \frac{u_{s\mu}}{\boldsymbol{r}(u_s)}, 
\end{equation}
where 
\[
\boldsymbol{r}(u)^2 = (u\cdot x)^2 - (u \cdot u)(x\cdot x), \,\,\,
u\cdot x = g_{\mu\nu}u^\mu x^\nu.
\]

Then, similarly as the last two Propositions, we show validity of the following

\begin{prop*}
The functions $\widetilde{f}_\mu$ on the cone, given by (\ref{tildefmuHom=-1}), define via the formula
(\ref{Hom=-1Distr}) a continuous functional 
$\widetilde{\varphi} \mapsto (\widetilde{f}_\mu,\widetilde{\varphi})$ 
on $\mathcal{S}^{0}(\mathbb{R}^4)$ which as a function of $\varphi \in \mathcal{S}^{00}(\mathbb{R}^4)$
is continuous on $\mathcal{S}^{00}(\mathbb{R}^4)$ and equal to
\[
(\widetilde{f}_\mu,\widetilde{\varphi}) = \int \limits_{\mathbb{R}^4} f_\mu(x) \varphi(x) \, \ud^4 x
= (f_\mu, \varphi),
\]
where the functions $f_\mu$ in the last formula are equal (\ref{fmuHom=-1}). 
Because $\widetilde{f}_\mu$ as a distribution $\widetilde{\varphi} \mapsto (\widetilde{f}_\mu,\widetilde{\varphi})$
is concentrated on the light cone, then
\[
(f_\mu, \square \varphi) = 0, \,\,\,
\varphi \in \mathcal{S}^{00}(\mathbb{R}^4),
\]
and is transversal
\[
(f_\mu, \partial^\mu \varphi) = 0, \,\,\,
\varphi \in \mathcal{S}^{00}(\mathbb{R}^4),
\]
if and only if 
\[
Q = \sum \limits_{s} e_s = 0.
\]
\end{prop*}

Proceeding similarly as in the proof of the first Proposition of this Subsection we can show the following
\begin{prop*}
\begin{enumerate}
\item[1)]
The functional 
\begin{multline}\label{fDistr}
\widetilde{\varphi} \mapsto (\widetilde{f},\widetilde{\varphi}) \\
= \int \limits_{\mathscr{O}_{1,0,0,1}} \, \frac{1}{p_{0}^{2}} \, \widetilde{\varphi}|_{{}_{\mathscr{O}_{1,0,0,1}}}(p) \, 
\ud \mu|_{{}_{\mathscr{O}_{1,0,0,1}}}(p)
-
\int \limits_{\mathscr{O}_{-1,0,0,1}} \, \frac{1}{p_{0}^{2}} \, \widetilde{\varphi}|_{{}_{\mathscr{O}_{-1,0,0,1}}}(p) \, 
\ud \mu|_{{}_{\mathscr{O}_{-1,0,0,1}}}(p), \\
\,\,\, \widetilde{\varphi}  \in \mathcal{S}^{0}(\mathbb{R}^4) \, \textrm{and} \,
\varphi \in \mathcal{S}^{00}(\mathbb{R}^4)
\end{multline}
belongs to $\mathcal{S}^0(\mathbb{R}^4)^*$.

\item[2)]
The functional
\begin{equation}\label{tildefDistr}
\varphi \mapsto (\widetilde{f},\widetilde{\varphi}) \\
\,\,\, \widetilde{\varphi}  \in \mathcal{S}^{0}(\mathbb{R}^4) \, \textrm{and} \,
\varphi \in \mathcal{S}^{00}(\mathbb{R}^4)
\end{equation}
belongs to $\mathcal{S}^{00}(\mathbb{R}^4)^*$. 

\item[3)]
The functional
\begin{multline}\label{fRestCone}
\mathcal{S}^{0}(\mathbb{R}^3) \oplus \mathcal{S}^{0}(\mathbb{R}^3) \ni \widetilde{\varphi}|_{{}_{\mathscr{O}_{1,0,0,1} \sqcup \mathscr{O}_{-1,0,0,1}}} 
\mapsto (\widetilde{f}|_{{}_{\mathscr{O}_{1,0,0,1} \sqcup \mathscr{O}_{-1,0,0,1}}},
\widetilde{\varphi}|_{{}_{\mathscr{O}_{1,0,0,1} \sqcup \mathscr{O}_{-1,0,0,1}}}) \\ = 
\int \limits_{\mathscr{O}_{1,0,0,1}} \, \frac{1}{p_{0}^{2}} \, \widetilde{\varphi}|_{{}_{\mathscr{O}_{1,0,0,1}}}(p) \, 
\ud \mu|_{{}_{\mathscr{O}_{1,0,0,1}}}(p) \\
-
\int \limits_{\mathscr{O}_{-1,0,0,1}} \, \frac{1}{p_{0}^{2}} \, \widetilde{\varphi}|_{{}_{\mathscr{O}_{-1,0,0,1}}}(p) \, 
\ud \mu|_{{}_{\mathscr{O}_{-1,0,0,1}}}(p),
\end{multline}
belongs to $\mathcal{S}^{0}(\mathbb{R}^3)^* \oplus \mathcal{S}^{0}(\mathbb{R}^3)^*$ and by construction its 
composition with the restriction to the cone gives the functional from 1)
\[
(\widetilde{f}|_{{}_{\mathscr{O}_{1,0,0,1} \sqcup \mathscr{O}_{-1,0,0,1}}},\widetilde{\varphi}|_{{}_{\mathscr{O}_{1,0,0,1} \sqcup \mathscr{O}_{-1,0,0,1}}}) = (\widetilde{f},\widetilde{\varphi}).
\]
\item[4)]
And
\begin{multline*}
(\widetilde{f},\widetilde{\varphi}) = \int \limits_{\mathscr{O}_{1,0,0,1}} \, \frac{1}{p_{0}^{2}} \, \widetilde{\varphi}|_{{}_{\mathscr{O}_{1,0,0,1}}}(p) \, 
\ud \mu|_{{}_{\mathscr{O}_{1,0,0,1}}}(p)
-
\int \limits_{\mathscr{O}_{-1,0,0,1}} \, \frac{1}{p_{0}^{2}} \, \widetilde{\varphi}|_{{}_{\mathscr{O}_{-1,0,0,1}}}(p) \, 
\ud \mu|_{{}_{\mathscr{O}_{-1,0,0,1}}}(p) \\
=
2\pi^2 \int \limits_{x \cdot x < 0} \, \frac{x_0}{|\boldsymbol{\x}|} \, \varphi(x) \, \ud^4 x  \,\,\,
+ \,\,\, 2\pi^2 \int \limits_{x \cdot x > 0, x_0>0} \, \varphi(x) \, \ud^4 x \,\,\, \\ 
-2\pi^2 \int \limits_{x \cdot x > 0, x_0<0} \, \varphi(x) \, \ud^4 x \\
= 2\pi^2 \int \limits_{\mathbb{R}^4} \, f(x) \, \varphi(x) \, \ud^4 x 
= (f, \varphi), \,\,\,\,\, \varphi \in \mathcal{S}^{00}(\mathbb{R}^4).
\end{multline*} 
where the function $f$ is the Dirac's homogeneous of degree zero function 
given by (\ref{f=homogeneity=0}).
\end{enumerate}

\end{prop*}

\subsection{Hilbert space of the supplementary series representation
of $SL(2, \mathbb{C})$ as a space of homogeneous solutions of d'Alembert equation
belonging to $\mathcal{S}^{00}(\mathbb{R}^4;\mathbb{C})^*$}\label{SupplementaryDistributions}

In this Subsection we give a proof of the following

\begin{prop*}
Consider the Hilbert space $\mathcal{H}_{z-2}$ 
generated by functions $\widetilde{f}$ on the (positive sheet of the) cone homogeneous of degree $z-2$, where 
$z \in (0,1)$ with the invariant inner product \cite{Staruszkiewicz1992ERRATUM} 
\begin{equation}\label{suppInnnerProd}
(\widetilde{f},\widetilde{g}) =   \int \limits_{\mathbb{S}^2 \times \mathbb{S}^2} 
\frac{d^2 k d^2 l}{(k \cdot l)^{{}^{z}}} \overline{\widetilde{f}(k)} \,\widetilde{g}(l),
\end{equation} 
where $d^2k$ (resp. $d^2l$) is the invariant measure (\cite{GelfandV}) on the space of rays 
(which can be identified with the unit 2-sphere $\mathbb{S}^2$) on the cone  $k \cdot k = 0, k_0 >0$ 
(resp. $l \cdot l = 0, l_0 >0$). The Lorentz group acts naturally in this space, 
and it has been recognized in \cite{Staruszkiewicz} that  $\mathcal{H}_{z-2}$ 
with the inner product (\ref{suppInnnerProd}) gives the irreducible unitary representation 
of the $SL(2, \mathbb{C})$ group of the supplementary series with parameter of the series 
equal $1-z$).  
The homogeneous of degree $z-2$  functions $\widetilde{f}$ on the cone, whose restrictions to
the unit two-sphere $\mathbb{S}^2$ on the cone belong to $L^2(\mathbb{S}^2, \ud \mu_{{}_{\mathbb{S}^2}})$ 
can be naturally regarded as continuous functionals on $\mathcal{S}^{0}(\mathbb{R}^3; \mathbb{C}) = 
\mathcal{S}_{A^{(3)}}(\mathbb{R}^3; \mathbb{C})$ (with the spatial momentum components as coordinates on the cone), 
by the second Proposition of Subsection \ref{DiracHom=-1Sol}. Any element $S$ of 
$\mathcal{S}^{0}(\mathbb{R}^3; \mathbb{C})^* = 
\mathcal{S}_{A^{(3)}}(\mathbb{R}^3; \mathbb{C})^*$, as a distribution on the cone $\mathscr{O}$,
determines uniquely and canonically an element $\widetilde{F}$ of 
$\mathcal{S}^{0}(\mathbb{R}^4; \mathbb{C})^* = 
\mathcal{S}_{A^{(4)}}(\mathbb{R}^3; \mathbb{C})^*$ by the condition that 
for each $\widetilde{\varphi} \in \mathcal{S}^{0}(\mathbb{R}^4; \mathbb{C}) = 
\mathcal{S}_{A^{(4)}}(\mathbb{R}^3; \mathbb{C})$
\[
\widetilde{F}(\widetilde{\varphi}) = S(\widetilde{\varphi}|_{{}_{\mathscr{O}}}).
\]
This is correct definition, because by the second Proposition of Subsection \ref{Lop-on-E}
the restriction $\widetilde{\varphi} \rightarrow \widetilde{\varphi}|_{{}_{\mathscr{O}}}$ 
maps continuously $\mathcal{S}^{0}(\mathbb{R}^4; \mathbb{C})$ into 
$\mathcal{S}^{0}(\mathbb{R}^3; \mathbb{C})$. Inverse Fourier transform $F$ of such $\widetilde{F}$ is, by construction,
a continuous functional
on $\mathcal{S}^{00}(\mathbb{R}^4; \mathbb{C})$ fulfilling d'Alembert equation, as $\widetilde{F}$ is concentrated on the cone
$\mathscr{O}$. It follows that the Hilbert space closure $\mathcal{H}_{z-2}$ of the space of homogeneous of degree $z-2$ functions with respect to the inner product (\ref{suppInnnerProd}) leads us out of the space of (equivalence classes of) ordinary homogeneous functions on the cone (up to almost everywhere equality).

But we claim that the closure of the space of homogeneous of degree $z-2$ functions 
with respect to (\ref{suppInnnerProd}) does not leads us out of the space 
$\mathcal{S}^{0}(\mathbb{R}^3; \mathbb{C})^* = 
\mathcal{S}_{A^{(3)}}(\mathbb{R}^3; \mathbb{C})^*$, i.e.
\[
\mathcal{H}_{z-2} \subset \mathcal{S}^{0}(\mathbb{R}^3; \mathbb{C})^* = 
\mathcal{S}_{A^{(3)}}(\mathbb{R}^3; \mathbb{C})^*.
\]
This means that the elements of the supplementary series Hilbert space $\mathcal{H}_{z-2}$
can be regarded as homogeneous of degree $2-z$ distributions  
$S \in \mathcal{S}^{0}(\mathbb{R}^3; \mathbb{C})^* = 
\mathcal{S}_{A^{(3)}}(\mathbb{R}^3; \mathbb{C})^*$, and determine canonically
homogeneous of degree $z-2$ distribution $\widetilde{F} \in \mathcal{S}^{0}(\mathbb{R}^4; \mathbb{C})^* = 
\mathcal{S}_{A^{(4)}}(\mathbb{R}^3; \mathbb{C})^*$, whose inverse Fourier transforms 
$F \in \mathcal{S}^{00}(\mathbb{R}^4; \mathbb{C})$
are homogeneous of degree $-z$ solutions of d'Alembert equation. Thus the Hilbert space
$\mathcal{H}_{z-2}$ can be regarded as a linear space of homogeneous of degree $-z$
solutions $F \in \mathcal{S}^{00}(\mathbb{R}^4; \mathbb{C})$ of d'Alembert equation. 
\end{prop*}

\qedsymbol \,

Note that the nuclear space 
$\mathcal{S}_{A^{(3)}}(\mathbb{R}^3; \mathbb{C}) = \mathcal{S}^{0}(\mathbb{R}^3; \mathbb{C})$ is the complexification
of a standard countably Hilbert nuclear space $\mathcal{S}_{A^{(3)}}(\mathbb{R}^3; \mathbb{R})$ constructed 
on the standard pair $(A, H) = (A^{(3)}, H= L^2(\mathbb{R}^3; \mathbb{R}))$, compare Subsection \ref{white-setup},
where $A^{(3)}$ is the standard self adjoint operator on $H= L^2(\mathbb{R}^3; \mathbb{R}))$
constructed in Subsection \ref{dim=n}. Recall that $\mathcal{S}_{A^{(3)}}(\mathbb{R}^3; \mathbb{R})
= \cap_k E_k$ is the inductive limit of Hilbert spaces $E_k$, $k \in \mathbb{Z}$ -- the completions 
of $\textrm{Dom} A^k$ with respect to the Hilbertian norms $|A^k \cdot |_{{}_{L^2(\mathbb{R}^3;\mathbb{R})}}$
joined by the topological inclusions ($k_2 >k_1$ implies $E_{k_2} \subset E_{k_1}$):
\[
\mathcal{S}_{A}(\mathbb{R}^3; \mathbb{R}) \subset \ldots \subset E_{k} \ldots \subset E_0 = H \subset \ldots 
\subset E_{-k} \subset  \ldots 
\subset \mathcal{S}_{A}(\mathbb{R}^3; \mathbb{R})^*.
\] 
$\mathcal{S}_{A}(\mathbb{R}^3; \mathbb{R})^*$ with its strong dual topology (coinciding with its weak dual topology)
is the inductive limit $\cup_k E_k$ of the Hilbert spaces $E_k$.
Recall that $A^{(3)}$ is unitarily equivalent to  
\[
H_{{}_{(1)}} \otimes \boldsymbol{1} + \boldsymbol{1} \otimes \Delta_{{}_{\mathbb{S}^2}} 
\]
where $H_{{}_{(1)}}$ is the hamiltonian operator of one dimensional oscillator, and $\Delta_{{}_{\mathbb{S}^2}}$
is the Laplace operator on the two-sphere. It is well known (compare the second Proposition of
Subsection \ref{white-setup}) that in the standard construction of $\mathcal{S}_{A^{(3)}}(\mathbb{R}^3; \mathbb{R})$
and its dual $\mathcal{S}_{A^{(3)}}(\mathbb{R}^3; \mathbb{R})^*$ we can replace the standard operator
\[
H_{{}_{(1)}} \otimes \boldsymbol{1} + \boldsymbol{1} \otimes \Delta_{{}_{\mathbb{S}^2}} 
\]
by 
\[
H_{{}_{(1)}} \otimes \Delta_{{}_{\mathbb{S}^2}} 
\]
so that instead of the operator 
\[A^{(3)} = U\big(H_{{}_{(1)}} \otimes \boldsymbol{1} + \boldsymbol{1} \otimes \Delta_{{}_{\mathbb{S}^2}}\big) U^{-1}
\]
where $U$ is the unitary operator of Subsect. \ref{dim=n}, we will use 
the operator 
\begin{equation}\label{operatorA}
A = U \big( H_{{}_{(1)}} \otimes \Delta_{{}_{\mathbb{S}^2}} \big)  U^{-1}.
\end{equation}
Despite this changing we have
\[
\mathcal{S}_{A^{(3)}}(\mathbb{R}^3; \mathbb{R}) = \mathcal{S}_{A}(\mathbb{R}^3; \mathbb{R})
\]
in store of elements and in topology.
We make this modification of the standard operator for computational convenience, because the eigenvalues
of the operator $A$ are equal 
\[
\lambda_{{}_{n,l,m}} = (2n + 2)l(l+1), n,l = 0,1,2, \ldots, -l \leq m \leq l,
\]
each with multiplicity one and thus each eigenvalue
\[
(2n + 2)l(l+1), n,l = 0,1,2, \ldots
\]
enters with multiplicity $2l +1$. This formula for eigenvalues will simplify slightly 
the computations which are to follow.

We have to show that the  inequalities  $0 < z < 1$  
assure that the convergence in the norm defined by (\ref{suppInnnerProd}) 
of a sequence of functionals in $\mathcal{S}_{A^{(3)}}(\mathbb{R}^3; \mathbb{C})^*$ regarded as functionals on 
$\mathcal{S}_{A^{(3)}}(\mathbb{R}^3; \mathbb{R}) = \mathcal{S}_{A}(\mathbb{R}^3; \mathbb{R})$
will have as a consequence convergence of that sequence in the weak topology of 
$\mathcal{S}_{A}(\mathbb{R}^3; \mathbb{R})^*$. Now for the validity of this implication it will be 
sufficient that for some positive integer
$k$ the norm $| A^{-k} \cdot |_{{}_{L^2(\mathbb{R})}}$ will be weaker than the norm defined by
(\ref{suppInnnerProd}) for all homogeneous of degree $z-2$ functions $\widetilde{f}$ on the cone,
smooth when restricted to the unit $2$-sphere $\mathbb{S}^2$, compare \cite{GelfandII}, Chap. I \S 5.6, p. 50,
or \cite{GelfandIV}. In other words we are going to show now that this is indeed the case, i.e. we are going
to show that if $0 < z < 1$, then there exists a constant $c < \infty$ 
such that 
\begin{equation}\label{(-k)norm-weaker-long-norm}
| A^{-k} \widetilde{f} |_{{}_{L^2(\mathbb{R}^3)}}^{2} 
\leq c \, (\widetilde{f}, \widetilde{f})
\end{equation}
for any (expressed in spheraical coordinates $r,\theta, \phi$, on the cone) 
\begin{equation}\label{tildef=r^(z-2)s(teta,phi)}
\widetilde{f}(r,\theta, \phi) = r^{z-2} \, s(\theta, \phi), \,\,\, s \in \mathscr{C}^\infty (\mathbb{S}^2; \mathbb{R})
\end{equation}  
and where $k$ is a natural number greather than some fixed natural $k_0$.

In order to achieve this we have to compute the operator $A^{-k}$ more explicitly. 
Note that the eigenfunctions $u_{{}_{n,l,m}}$ corresponding to the eigenvalues 
$\lambda_{{}_{n,l,m}}$ of the operator $A$ have the following form (in spherical coordinates)
\[
u_{{}_{n,l,m}}(r, \theta, \phi) = \frac{\sqrt{1 +r^2}}{r^2}h_{{}_{n}}(t(r)) \, Y_{{}_{l,m}}(\theta, \phi)
\] 
where $t(r)= r-r^{-1}$, $h_{{}_{n}}$ are the Hermite functions and $Y_{{}_{l,m}}$ are the spherical functions,
and thus we can write
\[
u_{{}_{n,l,m}}(r, \theta, \phi) = q_{{}_{n}} \otimes Y_{{}_{l,m}}(r, \theta, \phi)
\] 
where 
\[
q_{{}_{n}}(r) = \frac{\sqrt{1 +r^2}}{r^2}h_{{}_{n}}(t(r)) = \sqrt{2}\frac{r^2 +1}{r^3} u_n(r)
\]
where $u_n \in \mathcal{S}^0(\mathbb{R};\mathbb{R})$ are the eigenfunctions of the self adjoint standard operator 
$A^{(1)}$ on $L^2(\mathbb{R}; \mathbb{R})$ constructed in Subsection \ref{dim=1}. Because by the results
of Subsection \ref{dim=1} the function 
\[
\mathbb{R} \ni p \mapsto \frac{|p|^2 + 1}{|p|^3}
\]
is a multiplier of $\mathcal{S}^0(\mathbb{R};\mathbb{R}) = \mathcal{S}_{A^{(1)}}(\mathbb{R};\mathbb{R})$
then it follows that the functions $q_{{}_{n}}$ belong to the nuclear space 
$\mathcal{S}^{0}(\mathbb{R};\mathbb{R}) = \mathcal{S}_{A^{(1)}}(\mathbb{R};\mathbb{R})$ and similarly 
$Y_{{}_{l,m}}$ belongs to the nuclear space $\mathscr{C}^{\infty}(\mathbb{S}^2;\mathbb{R}) = \mathcal{S}_{{}_{\Delta_{\mathbb{S}^2}}}(\mathbb{S}^2; \mathbb{R})$.

Now it follows that for any positive integer $k$ and any 
$\widetilde{f} \in \mathcal{S}_{A^{(3)}}(\mathbb{R}^3; \mathbb{R})
= \mathcal{S}_{A}(\mathbb{R}^3; \mathbb{C})$ and in fact for any $\widetilde{f} \in H = L^2(\mathbb{R}^3; \mathbb{R})$ the series 
\begin{multline}\label{A^(-k)}
A^{-k} \tilde{f}(r,\theta,\phi)  \\ =
\sum \limits_{n,l,m} \lambda_{{}_{n,l,m}}^{-k}  u_{{}_{n,l,m}}(r,\theta,\phi) \, 
\int u_{{}_{n,l,m}}(r', \theta', \phi') \tilde{f}(r',\theta',\phi') r'^{2} \sin \theta' dr' d\theta' d\phi'
\end{multline}
converges in the $L^2$- norm of the Hilbert space $H = L^2(\mathbb{R}^3; \mathbb{R})$.  
In fact for the integer $k$ greater than a fixed positive integer $k_0$ this series converges in the $L^2$
norm of $H = L^2(\mathbb{R}^3; \mathbb{R})$ for any $\widetilde{f}^\mu$ of the form
(\ref{tildef=r^(z-2)s(teta,phi)}), expressed in spherical coordinates on the cone.
Indeed for any $\widetilde{f}$ of the form $\tilde{f}(r,\theta, \phi) = \tilde{g}(r) s(\theta, \phi)$
the series (\ref{A^(-k)}) takes on the following form
\begin{multline*}
A^{-k} \tilde{f}(r,\theta,\phi)  =
\sum \limits_{n,l,m} \lambda_{{}_{n,l,m}}^{-k}  u_{{}_{n,l,m}}(r,\theta,\phi) \, \times \\ 
\times \Bigg[ \int \limits_{\mathbb{R}_{+}} \ud r' r'^{2} \, q_{{}_{n}}(r') \tilde{g}(r') \,
\int \limits_{\mathbb{S}^2} Y_{{}_{l,m}}(r', \theta', \phi') s(\theta',\phi') \sin \theta' d\theta' d\phi'
\, \Bigg].
\end{multline*}
But for the homogeneous $\widetilde{f}$ of degree $-2 + z$ of the form
\[
\widetilde{f}(r,\theta,\phi) = r^{-2 +z} s(\theta,\phi)
\]
with $s \in \mathscr{C}^\infty(\mathbb{S}^2;\mathbb{R})$
the modulus of the integral 
\begin{equation}\label{bound-on-s_(lm)}
\Bigg| \int \limits_{\mathbb{S}^2} Y_{{}_{l,m}}(r', \theta', \phi') s(\theta',\phi') \sin \theta' d\theta' d\phi' \Bigg|
\leq \textrm{const.} \,\,\,\textrm{for all $l$ and $-l \leq m \leq l$}
\end{equation}
remains bounded uniformly with respect to $l,m$. 
Note please that the function $\tilde{g}(r) = r^{-2 + z}$  remains the same
for all $\widetilde{f}$ of the form (\ref{tildef=r^(z-2)s(teta,phi)}).
On the other hand for $0 < z <1$ the function
\[
\mathbb{R} \ni p \mapsto \frac{|p|^{z}}{1+|p|^2} 
\]
belongs to $L^2(\mathbb{R}; \mathbb{R})$ and the function 
\[
\mathbb{R} \ni p \mapsto \frac{(1 + |p|^2)^2}{|p|^3}
\]
again by Subsection \ref{dim=1} is a multiplier of the algebra 
$\mathcal{S}^{0}(\mathbb{R};\mathbb{R}) = \mathcal{S}_{A^{(1)}}(\mathbb{R};\mathbb{R})$, therefore
there exists a constant $c_{00}$ and a positive integer $k_0$ such that
\begin{multline}\label{c_(n)-estimation}
|c_n| = \Bigg| \int \limits_{\mathbb{R}_{+}} q_{{}_{n}}(r) r^{-2 + z} \, r^2 dr \Bigg|
=
\Bigg| \int \limits_{\mathbb{R}_{+}} \frac{r^{z}}{1 + r^2} (1 + r^2) q_{{}_{n}}(r)  dr \Bigg| \\
\leq
\Bigg| \frac{r^{z}}{1 + r^2} \Bigg|_{{}_{L^2(\mathbb{R}; \mathbb{R})}} \,
\Bigg|(1 + r^2) q_{{}_{n}} \Bigg|_{{}_{L^2(\mathbb{R}; \mathbb{R})}} 
=
\Bigg| \frac{r^{z}}{1 + r^2} \Bigg|_{{}_{L^2(\mathbb{R}; \mathbb{R})}} \,
\Bigg| \frac{(1 + r^2)^2}{r^3} u_{{}_{n}} \Bigg|_{{}_{L^2(\mathbb{R}; \mathbb{R})}} \\
\leq  \Bigg| \frac{r^{z}}{1 + r^2} \Bigg|_{{}_{L^2(\mathbb{R}; \mathbb{R})}} \,
c_{00} \, \Big| \big(A^{(1)}\big)^{k_0} u_{{}_{n}} \Big|_{{}_{L^2(\mathbb{R}; \mathbb{R})}} 
= c_{0} (2n +1)^{{}^{k_0}}.
\end{multline} 
Although the integration in the estimated quantity we start with is extended only over the positive 
half of the reals the inequalities remain legitimate, because the function defined as the zero function for negative 
real numbers and for positive $r \in \mathbb{R}$ defined to be equal $u_{{}_{n}}(r)$ still belongs to
the nuclear space $\mathcal{S}^{0}(\mathbb{R};\mathbb{R}) = \mathcal{S}_{A^{(1)}}(\mathbb{R};\mathbb{R})$; 
and the last equality is legitimate because such a function acted on by the differential operator $\big(A^{(1)} \big)^{k_0}$ is equal to the zero
function for negative real numbers and is equal $(2n+1)^{{}^{k_0}}u_{{}_{n}}(r)$ for $r\geq 0$, compare
Subsection \ref{dim=1}.
From the estimations (\ref{c_(n)-estimation}) and (\ref{bound-on-s_(lm)}) our assertion that the series 
(\ref{A^(-k)}) converges in the norm
of $L^2(\mathbb{R}^3; \mathbb{C})$ for each element $\widetilde{f}$ of the form 
(\ref{tildef=r^(z-2)s(teta,phi)}) now easily follows
if $k > k_0$, where $k_0$ is the positive integer independent of the choice
of $\widetilde{f}$. 

Having obtained this we can easily compute 
$(A^{-k} \widetilde{f}, A^{-k} \widetilde{f}')_{{}_{L^2(\mathbb{R})}}$
for $k > k_0$ and for $\widetilde{f}$ of the form (\ref{tildef=r^(z-2)s(teta,phi)}). 
If we put $\widetilde{f}(r,\theta, \phi) = r^{-2 +z} s(\theta,\phi)$ and
$\widetilde{f}'(r,\theta, \phi) = r^{-2 +z} s'(\theta,\phi)$ 
for these homogeneous functions $\widetilde{f}, \widetilde{f}'$, then we will get
\begin{multline}\label{scal-prod-(-k)-tildeftildef'}
\big(A^{-k} \widetilde{f}, A^{-k} \widetilde{f}' \big)_{{}_{L^2(\mathbb{R})}}
=
\sum \limits_{n,l,m} \lambda_{{}_{n,l,m}}^{-2k} c_{{}_{n}}^{2} \overline{s_{lm}} s'_{lm} \\
=
\sum \limits_{n,l,m} \frac{1}{2^{{}^{2k}}(n+1)^{{}^{2k}} l^{{}^{2k}}(l+1)^{{}^{2k}}} c_{{}_{n}}^{2} 
\overline{s_{lm}} s'_{lm} \\
= \sum \limits_{l,m} b^{{}^{2}}\, \frac{1}{l^{{}^{2k}}(l+1)^{{}^{2k}}}  
\overline{s_{lm}} s'_{lm}
\end{multline}
where 
\[
s_{lm} = \int \limits_{\mathbb{S}^2} Y_{{}_{lm}}(\theta, \phi) s(\theta,\phi) \, \sin \theta d\theta d \phi
\,\, \textrm{and} \,\,
s'_{lm} = \int \limits_{\mathbb{S}^2} Y_{{}_{lm}}(\theta, \phi) s'(\theta,\phi) \, \sin \theta d\theta d \phi
\]
and  where
\[
c_{{}_{n}} = \int \limits_{\mathbb{R}_{+}} q_{{}_{n}}(r) r^{-2 + z} \,r^2 \ud r, \,\,\,
b^{{}^{2}} = \sum \limits_{n} \frac{1}{2^{{}^{2k}}(n+1)^{{}^{2k}}} c_{{}_{n}}^{2} < \infty
\]
because the system $u_{{}_{n,l,m}}$ is orthonormal and complete in $L^2(\mathbb{R}^3; \mathbb{C})$.

The series defining the positive constant $b^{{}^{2}}$ is convergent for all positive integers $k > k_0$,
which easily follows from the estimation (\ref{c_(n)-estimation}).

Now we are going to estimate the inner product (\ref{suppInnnerProd}) norm 
$(\widetilde{f}, \widetilde{f})$ on the right hand side of the inequality 
(\ref{(-k)norm-weaker-long-norm}) comparable with the expression 
(\ref{scal-prod-(-k)-tildeftildef'}) obtained for the left hand side of the inequality
(\ref{(-k)norm-weaker-long-norm}). In doing this we use a reproducing property
(\ref{reproducing-property-of-(kl)^z}) of the kernel
$k \times l \rightarrow 1/(k \cdot l)^z$. This reproducing property has been noticed 
in \cite{Staruszkiewicz1998}, and is quite useful in the investigation of the supplementary series
representation, e.g. for the proof of positive definiteness of the supplementary series inner product
 (\ref{suppInnnerProd}) for $0<z<1$ independent of the proof given by Gelfand and Neumark
\cite{GelfandNeumark}, compare \cite{Staruszkiewicz1998}.

Now consider the homogeneous of degree $z-2$ functions 
$\widetilde{f}, \tilde{f}'$ the same as in the formula (\ref{scal-prod-(-k)-tildeftildef'}).
For this pair homogeneous functions $\widetilde{f}(r,\theta,\phi)
= r^{-2 + z} s(\theta,\phi)$, $\widetilde{f}'(r,\theta,\phi)$
$= r^{-2 + z} s'(\theta,\phi)$ we give the estimation of their inner product
$(\widetilde{f}, \widetilde{f}')$ defined by (\ref{suppInnnerProd}). 

In order to achieve this we use the fact that the function
\begin{equation}\label{SingularKernel}
\mathbb{S}^2 \times \mathbb{S}^2 \ni p\times q \mapsto 
\frac{1}{(p\cdot q)^{z}},
\end{equation} 
regarded as a function on $\mathbb{S}^2 \times \mathbb{S}^2$ reproduces the spherical functions, 
or more precisely
\begin{equation}\label{reproducing-property-of-(kl)^z}
\int \limits_{\mathbb{S}^2} 
\frac{\ud^2 q}{(p\cdot q)^{z}} Y_{{}_{lm}}(q) =
2\pi \frac{2^{{}^{1-z}}\Gamma(1-z)\Gamma(l+ z)}{\Gamma(z)\Gamma(l+2 -z)}
Y_{{}_{lm}}(p),
\end{equation}
where $\Gamma$ is the Euler gamma function. Using this fact we rewritte 
$(\widetilde{f}, \widetilde{f}')$ defined by (\ref{suppInnnerProd}) as follows 
\begin{equation}\label{sc-prod-long-tildeftildef'I}
(\widetilde{f}, \widetilde{f}')
=
 \,\,
\sum \limits_{l,m}  
2\pi \frac{2^{{}^{1-z}}\Gamma(1-z)\Gamma(l+ z)}{\Gamma(z)\Gamma(l+2 -z)} 
\, \overline{s_{lm}} \, s'_{lm}.
\end{equation}
Observe now that 
\[
2\pi \frac{2^{{}^{1-z}}\Gamma(1-z)\Gamma(l+ z)}{\Gamma(z)\Gamma(l+2 -z)} 
\]
decreases monotonically as a function of $l$, and on using Stirling's formula we see that asymptotically it behaves like
\[
2\pi \frac{2^{{}^{1-z}}\Gamma(1-z)\Gamma(l+ z)}{\Gamma(z)\Gamma(l+2 -z)} 
\sim 
2\pi \frac{2^{{}^{1-z}}\Gamma(1-z)}{\Gamma(z)} \, 
\frac{1}{l^{{}^{2(1-z)}}}
\]
for large $l$. Using this fact as well as the fact that
\[
0 < \frac{\Gamma(1-z)}{\Gamma(z)} < \infty,
\]
and comparing (\ref{sc-prod-long-tildeftildef'I}) with
(\ref{scal-prod-(-k)-tildeftildef'}) we see that there exists a positive finite number $c$ such that
(\ref{(-k)norm-weaker-long-norm}) is preserved for all $\widetilde{f}$ of the form
(\ref{tildef=r^(z-2)s(teta,phi)}).

Thus the the assertion of our Proposition is thereby proved.
\qed

{\bf REMARK 1}. Note that the integer number $k$ in the inequality (\ref{(-k)norm-weaker-long-norm})
valid for all $\widetilde{f} \in \mathcal{H}_{z-2}$ is greater than zero. Therefore
the elements of $\mathcal{H}_{z-2}$ belong to the Hilbert space $E_{-k} \subset 
\mathcal{S}^{0}(\mathbb{R}^3; \mathbb{C})^* = 
\mathcal{S}_{A^{(3)}}(\mathbb{R}^3; \mathbb{C})^*$ and are distributions of
of order $k>0$. In general they are not equal to distributions canonically identifiable with 
the elements of $L^{2}(\mathbb{R}^3) \subset E_{-k} \subset \mathcal{S}^{0}(\mathbb{R}^3; \mathbb{C})^*$.

That the Hilbert space completion $\mathcal{H}_{z-2}$  of the linear space of smooth 
homogeneous functions of degree $z-2$ on the cone with respect to
(\ref{suppInnnerProd}) contains elements which are not identifiable with (equivalence classes modulo
equality almost everywhere of) ordinary homogeneous of degree $z-2$ functions on the cone
has been already noted by Gelfand and Neumark, compare \cite{GelfandNeumark}, \S 6, \cite{NeumarkLorentzBook}, \S 12.2-12.3, \cite{nai2}, \S 2. In particular restrictions to the unit sphere $\mathbb{S}^2$
of sequences of homogeneous of degree $z-2$ functions converging with respect to  
(\ref{suppInnnerProd}), do not in general converge in $L^2(\mathbb{S}^2)$.

{\bf REMARK 2}.
The elements $\widetilde{F} \in \mathcal{S}^{0}(\mathbb{R}^4; \mathbb{C})^*$
naturally and bi-uniquely corresponding to 
$\widetilde{f} \in \mathcal{H}_{z-2} \subset \mathcal{S}^{0}(\mathbb{R}^3; \mathbb{C})^*$
are all the more not identifiable with function like distributions in 
$\mathcal{S}^{0}(\mathbb{R}^4; \mathbb{C})^*$.
This does not yet mean that the inverse Fourier transforms $F \in \mathcal{S}^{00}(\mathbb{R}^4)$ 
of these $\widetilde{F}$ are not identifiable with ordinary functions, or as ordinary functions
only outside the cone. Thus the problem if the elements $F$ corresponding
to  $\widetilde{f} \in \mathcal{H}_{z-2}$ are indeed regular or indetifiable with ordinary
functions outside the ligh cone (i.e. with ordinary scalar solutions of the inhomogeneous-- say massive -- wave 
equation on de Sitter $3$-hyperboloid) is still open. But note that by 
(\ref{scal-prod-(-k)-tildeftildef'}) any sequence of homogeneous (with fixed homogeneity)
regular function like elements $\widetilde{f}$ of
$E_{-k} \subset \mathcal{S}^{0}(\mathbb{R}^3; \mathbb{C})^*$ with bounded 
norm $\| \cdot \|_{-k} = | A^{-k} \cdot |_{{}_{L^2(\mathbb{R})}}$ has also, when restricted to 
$\mathbb{S}^2$, finite harmonic Fourier coefficients $s_{lm}$ with
\begin{equation}\label{NormHomogStates}
\sum \limits_{l,m} \, \frac{1}{l^{{}^{2k}}(l+1)^{{}^{2k}}}  \,
\overline{s_{lm}} s'_{lm} 
= \int \limits_{\mathbb{S}^2} \overline{s(\theta, \phi)} \,\,
\Delta_{{}_{\mathbb{S}^2}}^{-2k} s(\theta, \phi) \, \, \ud \mu_{{}_{\mathbb{S}^2}} \, < \,  \infty, \,\,\,
s = \widetilde{f}\big|_{{}_{\mathbb{S}^2}} .
\end{equation}
Thus in general their restrictions $s$ to $\mathbb{S}^2$ need not belong to $L^2(\mathbb{S}^2)$.
In other words all restrictions to $\mathbb{S}^2$, of a sequence of regular homogeneous (with 
fixed homogeneity degree)
elements of $E_{\mathbb{C}}^{*} = \mathcal{S}^{0}(\mathbb{R}^3; \mathbb{C})^*$
converging in $E_{\mathbb{C}}^{*}$, converge also in 
$\mathcal{S}_{\Delta_{{}_{\mathbb{S}^2}}}(\mathbb{S}^2)^* = \mathscr{C}^\infty(\mathbb{S}^2)^*$,
which is what one should expect by the already proved isomorphism
$E_{\mathbb{C}}^{*} = \mathcal{S}^{0}(\mathbb{R}^3; \mathbb{C})^* \cong 
\mathscr{C}^\infty(\mathbb{S}^2)^* \otimes \mathcal{S}^{0}(\mathbb{R}_+)^*$.

Thus the problem is essentially reduced to the characterization
of the space of distributions on the $2$-sphere, and its relation to the space of restrictions to the cone
of Fourier transforms $\widetilde{F}$
of homogeneous of degree $-z \in (-1,0)$ solutions $F \in \mathcal{S}^{00}(\mathbb{R}^4)^*$
of d'Alembert equation.  

We should start with the homogeneous of degree 
$\lambda = -z$ solution $P_{-}^{\lambda}$ of Gelfand-Shilov \cite{GelfandI}, Chap. III,
which indeed coincides with ordinary function outside the light cone and vanishes inside the light cone.
Then we should consider the space generated by all Lorentz transforms of $P_{-}^{\lambda}$.
The distributional solution $P_{-}^{\lambda}$ of Gelfand-Shilov is defined as follows. 
The following integral
\begin{multline*}
\big(P_{-}^{\lambda}, \varphi \big) =
\int \limits_{P<0} \, P^\lambda (x) \, \varphi(x) \, \ud^4 x, \\
\,\,\, P(x)= \big(x_0\big)^2 - \big(x_1\big)^2 -\big(x_2\big)^2 - \big(x_3\big)^2,
\,\,\, \varphi \in \mathcal{S}(\mathbb{R}^4),
\end{multline*}
converges and represents an analytic function of $\lambda \in \mathbb{C}$ when 
$\textrm{Re} \, \lambda \geq 0$.
Its analytic continuation defines a well defined functional $ P_{-}^{\lambda}$ on $\mathcal{S}(\mathbb{R}^4)$
(and \emph{a fortiori} on $\mathcal{S}^{00}(\mathbb{R}^4)$)
also for $\lambda$ with $\textrm{Re} \, \lambda <0$, except for the integer points $\lambda = -1, -2, \ldots$,
where it has poles of order $2$ at $\lambda = -2,-3, -4, \ldots$ and a simple pole 
at $\lambda = -1$, compare \cite{GelfandI}, Chap. III.2.2. A corresponding functional can still
be defined  at the pole  $\lambda = -1$ through the residue
\[
\textrm{res}_{\lambda = - 1} P_{-}^{\lambda}
\]
and at the poles $\lambda = -2, -3, \ldots$ of order two, by the Laurent expansion
around the corresponding pole. The functional at the singular points becomes more singular. In particular
the functional defined by the residue of $P_{-}^{\lambda}$ at $\lambda = -1$ is concentrated on the 
light cone $P=0$, and cannot induce any nontrivial solution of the (inhomogeneous) wave equation on the de Sitter hyperboloid. 
Fortunately the domain covered by real 
$\lambda \in (-1,0)$ and relevant for us is regular. We expect that the Lorentz transforms
of the Gelfand-Shilov homogeneous of degree $\lambda$ solution $P_{-}^{\lambda}$ of d'Alembert equation, 
with $\lambda = -z$, 
generate an irreducible representation of $SL(2, \mathbb{C})$ equivalent to the supplementary
series representation acting on the homogeneous of degree $z-2$ functions on the cone, described in the  Proposition of this Section.      
Unfortunately we have not checked it through explicit 
computation of the wave functions on de Sitter hyperboloid induced by the distribution
$P_{-}^{\lambda}$ and its Lorentz transforms. 
This method would for sure be more comfortable for a physicist, who in particular will be interested in 
the locality of the transformation rule. But locality of the transformation on de Sitter
hyperboloid will follow from the
transformation rule of the Fourier transforms which are  homogeneous of degree $z-2$
functions, i.e. from the absence of any multiplier depending on the momentum
in the transformation law in momentum space. This is the case even for 
non zero homogeneity order $\lambda = -z$ of $P_{-}^{\lambda}$. Indeed, in passing
from homogeneous functions in full Minkowski space-time with non zero homogeneity, to their restrictions
to de Sitter hyperboloid eventual additional space-time dependent multiplier will cease to come in during transformation. 
If any such were present, they would have to come from the
eventual change of the ''radius'' of the hyperboloid. But it is impossible because
de Sitter hyperboloid is invariant for the Lorentz transformation. Of course translation will produce non-local,
space-time dependent multipliers, but this does not bother us, because here we are not interesting in translations.

{\bf REMARK 3}.
Note that $F \in \mathcal{S}^{00}(\mathbb{R})^*$, regular outside the cone, 
i.e. identifiable there with ordinary functions, and homogeneous with nonzero homogeneity degree, 
and which are solutions of d'Alembert equation in the full Minkowski space-time define, by restriction to de Sitter hyperboloid, wave functions
fulfilling \emph{inhomogeneous} wave equation on de Sitter hyperboloid, say ``massive'' waves, with 
the constant ``mass'' term coming from nonzero homogeneity degree. Thus the Hilbert space $\mathcal{H}_{z-2}$
is identifiable with the single particle Hilbert space of a homogeneous of dgree $-z$ field in Minkowski
spacetime, which induces on de Sitter $3$-hyperboloid spacetime a free massive field. This is very non trivial
fact and still does not follow yet from the above Proposition, 
and even not yet from the previous Remark asserting
that the states $\widetilde{F} \in \mathcal{H}_{z-2}$, have the property that their inverse
Fourier transforms $F$ are indetifiable with ordinary (say wave) functions on de Sitter $3$-hyperboloid.
It is important that the supplementary inner product $(s,s')_{{}_{z-2}}$, defined by (\ref{suppInnnerProd})
 (on states $\widetilde{f} = \widetilde{F}\big|_{{}_{\mathscr{O}}}$ regarded as ordinary functions
$s = \widetilde{f}\big|_{{}_{\mathbb{S}^2}}, s' = \widetilde{f}'\big|_{{}_{\mathbb{S}^2}}$ on the unit
sphere $\mathbb{S}^2$), is continuous on $\mathscr{C}^\infty(\mathbb{S}^2)$
with respect to the nuclear topology of 
$\mathscr{C}^\infty(\mathbb{S}^2) = \mathcal{S}_{\Delta_{{}_{\mathbb{S}^2}}}(\mathbb{S}^2)$,
and moreover that $\mathcal{H}_{z-2}$ composes with $\mathcal{S}_{\Delta_{{}_{\mathbb{S}^2}}}(\mathbb{S}^2)$
and $\mathcal{S}_{\Delta_{{}_{\mathbb{S}^2}}}(\mathbb{S}^2)^*$ a Gelfand triple:
\[
\mathcal{S}_{\Delta_{{}_{\mathbb{S}^2}}}(\mathbb{S}^2)
\subset \mathcal{H}_{z-2}
\subset \mathcal{S}_{\Delta_{{}_{\mathbb{S}^2}}}(\mathbb{S}^2)^*.
\]

It follows from the existence of a nonzero finite constant $M$ and a positive integer $k$
for which (here $s = \widetilde{f}\big|_{{}_{\mathbb{S}^2}}, s' = \widetilde{f}'\big|_{{}_{\mathbb{S}^2}}$
and $s_{lm}, {s'}_{lm}$ are the spherical harmonic coefficients, respectively, of $s, s'$)
\begin{multline*}
M^{-1} \, \Big(\Delta_{{}_{\mathbb{S}^2}}^{-k}s, \,\,\,\,
 \Delta_{{}_{\mathbb{S}^2}}^{-k} s'\Big)_{{}_{L^2(\mathbb{S}^2)}} = 
\, M^{-1} \, \sum \limits_{l,m} \, \frac{1}{l^{{}^{2k}}(l+1)^{{}^{2k}}} \,  
\overline{s_{lm}} s'_{lm} \\ 
\leq \, (s,s')_{{}_{z-2}}  = \, \sum \limits_{l,m}  
2\sqrt{2}\pi^{3/2} \frac{2^{{}^{1-z}}\Gamma(1-z)\Gamma(l+ z)}{\Gamma(z)\Gamma(l+2 -z)} 
\, \overline{s_{lm}} \, s'_{lm} \\ \leq  
\, M \, \sum \limits_{l,m} \, l^{{}^{2k}}(l+1)^{{}^{2k}} \, 
\overline{s_{lm}} s'_{lm}
= 
\, M \,
\Big(\Delta_{{}_{\mathbb{S}^2}}^{k}s, \,\,\,\,\,
\Delta_{{}_{\mathbb{S}^2}}^{k} s'\Big)_{{}_{L^2(\mathbb{S}^2)}}
, \,\,\, \\ \textrm{for all} \,\,\,
s,s' \in \mathcal{S}_{\Delta_{{}_{\mathbb{S}^2}}}(\mathbb{S}^2) = \mathscr{C}^\infty(\mathbb{S}^2)
\end{multline*}
or in short 
\begin{multline}\label{SupplementaryGelfandTripleIneq}
M^{-1} \, \big(\Delta_{{}_{\mathbb{S}^2}}^{-k} \cdot, \,
\Delta_{{}_{\mathbb{S}^2}}^{-k} \cdot \big)_{{}_{L^2(\mathbb{S}^2)}} \,\,\,\,
\leq  \, (\cdot, \cdot)_{{}_{z-2}} \leq \,\,\,\,
M \, \big(\Delta_{{}_{\mathbb{S}^2}}^{k} \cdot,  \,
\Delta_{{}_{\mathbb{S}^2}}^{k} \cdot \big)_{{}_{L^2(\mathbb{S}^2)}} \\
\textrm{on} \,\,\, \mathcal{S}_{\Delta_{{}_{\mathbb{S}^2}}}(\mathbb{S}^2).
\end{multline}
Indeed only in this case we can construct the free homogeneous field, exactly as we did for the free
field $A_\mu$.

Note that the first inequality in (\ref{SupplementaryGelfandTripleIneq}) is essentially equivalent to
the inequality proved in the proof of the last Proposition, and it assures the (topological)
inclusion $\mathcal{H}_{z-2} \subset E^* = \mathcal{S}_{A^{(3)}}(\mathbb{R}^3)^* 
= \mathcal{S}^{0}(\mathbb{R}^3)^*$. \qed

{\bf REMARK 4}. By the above Proposition and the preceding Remark, the Hilbert space
$\mathcal{H}_{z-2}$ can, in principle, be used as a single particle Hilbert space of a homogeneous of degree
$-z \in (-1,0)$ part of the scalar mass less field. Therefore for these fields 
homogeneities lying in the open interval $(-1, 0)$ are, in priciple at least, possible  (for a natural definition
of \emph{allowed (asymptotic) homogeneity}, which in fact excludes the interval  $(-1, 0)$,  
compare Subsections \ref{psichi}, \ref{equivalentA-s}). 
Moreover, by the Gelfand-Graev-Vilenkin
\cite{GelfandV}, it follows that homogeneities $-1 + i\nu$, $\nu \in \mathbb{R}$ are likewise possible. 
Indeed, also for them one can construct homogeneous of degree 
$-1 +i\nu$ parts of the (mass less) scalar fields. Indeed in this case we construct the
corresponding Hilbert space $\mathcal{H}_{-1-i\nu}$ as the Hilbert space completion
of homogeneous of degree $-1-i\nu$ functions $\widetilde{f}$ on the positive energy sheet of the cone
with  the inner product (with the ordinary identifications 
$s = \widetilde{f}\big|_{{}_{\mathbb{S}^2}}, s' = \widetilde{f}'\big|_{{}_{\mathbb{S}^2}}$):
\[
(\widetilde{f},\widetilde{g})_{{}_{-1-i\nu}} =   \int \limits_{\mathbb{S}^2} 
d^2 p \overline{\widetilde{f}(p)} \,\widetilde{g}(p) = 
\int \limits_{\mathbb{S}^2} \ud \mu_{{}_{\mathbb{S}^2}} \, \overline{s} s'.
\]
Verification of the continuity of $(\cdot,\cdot)_{{}_{-1-i\nu}}$ on $\mathscr{C}^\infty(\mathbb{S}^2)$ 
with respect to the
nuclear topology or the inequalities
analogous to (\ref{SupplementaryGelfandTripleIneq}) is immediate, in fact they follow immediately 
from the construction of
$\mathscr{C}^\infty(\mathbb{S}^2)$ as equal to $\mathcal{S}_{\Delta_{{}_{\mathbb{S}^2}}}(\mathbb{S}^2)$. Therefore, 
we likewise have the required Gelfand triple over $\mathcal{H}_{-1-i\nu}$:
\[
\mathcal{S}_{\Delta_{{}_{\mathbb{S}^2}}}(\mathbb{S}^2) \subset
\mathcal{H}_{-1-i\nu}
\subset \mathcal{S}_{\Delta_{{}_{\mathbb{S}^2}}}(\mathbb{S}^2)^*.
\]
Invariant positive definite inner products on spaces of functions $\widetilde{f}$
homogeneous of degre $\chi \notin (-1,0)$ and $\chi \neq -1 -i\nu$, $\nu \in \mathbb{R}$,
which are continuous on $\mathscr{C}^\infty(\mathbb{S}^2)$ do not exist. 
This follows from the classification
of such invariant inner products, i.e. positive definite invariant Hermitian bilinear forms, 
on $\mathscr{C}^\infty(\mathbb{S}^2)$
due to Gelfand, Graev and Vilenkin \cite{GelfandV}, Chap.III.4 (note that the authors are using 
stereographic coordinates on $\mathbb{S}^2$ much better for analysis of invariant bilinear forms).  
Thus the computation of all, in principle allowed, homogeneities for the (mass less) scalar field 
is now complete. Thus passing to the inverse Fourier transforms, we arrive at the conclusion
that the only, in principle possible, homogeneities $\chi$ of the homogeneous parts $\boldsymbol{\psi}_{\chi}$ of the free 
mass less scalar field $\boldsymbol{\psi}$ assume the values $-1 <\chi<0$ and  
$\chi = -1 + i\nu$, $\nu \in \mathbb{R}$. 

This follows by general considerations. But in fact the (asymtotically) 
homogeneous parts $\boldsymbol{\psi}_{\chi}$ of a free (massive)
field $\boldsymbol{\psi}$ are canonically determined by the
decomposition of the representation of the subgroup $SL(2, \mathbb{C})$ acting in the single particle
Hilbert space, determined by the spectral decomposition of the Casimir operators (in the massive case)
or by the scaling operator (in the mass less case). The (asymptotic) homogeneity $\chi$ is the invariant,
which in each case parametrizes the components of this decomposition, and determinates the
(asymptotically) homogeneous parts $\boldsymbol{\psi}_{\chi}$ of (asymptotic) homogeneity $\chi$.
The allowed values of $\chi$ in this decomposition are determined by the joint spectrum
of the Casimir operators (in the massive case, Subsection \ref{psichi})
or by the spectrum of the scaling operator (in the mass less case, Subsection \ref{equivalentA-s}).
For the scalar field (massive or mass less) it turns out that the spectrum is always equal $\chi = -1 + i\nu$, $\nu \in \mathbb{R}$,
so that the interval $-1 <\chi<0$, which could in principle be possible for mass less scalar fields, is in fact excluded
by this decomposition.

\qed

\subsection{Several spaces of homogeneous states in $E^* = \mathcal{S}^{0}(\mathbb{R}^3)^*$}\label{AS}

As explained in Subsection \ref{IntAspatialInfty} of Introduction, we need to classify all invariant
Hilbert space inner products on the subspace $E_{\chi}^{*}$ of homogeneous of degree $\chi$
states in $E^* = \mathcal{S}^{0}(\mathbb{R}^3; \mathbb{C}^4)^*$ in order to construct the single
particle Hilbert space of the homogeneous of degree $-2 -\chi$ part of the free field $A_\mu$.
Here $E = \mathcal{S}^{0}(\mathbb{R}^3; \mathbb{C}^4)$ is the nuclear space which together with 
the single particle Hilbert-Krein space $\mathcal{H}'$ of the free field $A_\mu$ composes the Gelfand
triple  $E \subset \mathcal{H}' \subset E^*$ used in the white noise construction of the field $A_\mu$.

In this Section we give the construction of the homogeneous 
of degree $-1$ electric part of the free field $A_\mu$ and determine the invariant subspace 
\[
(E^{*})_{tr}^{e} \subset E_{\chi=-1}^{*} \subset E^*
\]
 of electric type 
homogeneous of degree $-1$ states together with a natural invariant inner product on 
$(E^{*})_{tr}^{e}$ in  $E_{\chi=-1}^{*}$. The positive definite inner product which serves to define single particle 
space of a homogeneous part of the field $A_\mu$, is determined by the general direct integral decomposition 
of the field $A$ into the homogeneous of degree parts $A_{{}_{\chi}}$, determined by the decomposition
of the representation of $SL(2, \mathbb{C})$ acting in the single particle space of the field, and presented in Subsection
\ref{equivalentA-s}. This inner product has, as expected, the property that the completion of 
$(E)_{tr}^{e} \subset E_{\chi=-1}^{*}$ with respect to it cannot lead us 
out of the space $E^*$. For justification
see Subsection \ref{equivalentA-s} and Subsection \ref{IntAspatialInfty} of Introduction, and this Subsection.  
But here we confine attention to  transversal homogeneous states in
$E_{\chi=-1}^{*}$. 

To get insight into the situation we investigate spaces of ordinary homogeneous of degree $\chi = -1$ (or more generally $\chi = -1 + i \nu$, $\nu
\in \mathbb{R}$, as only these homogeneities enter into the decomposition of Subsection \ref{equivalentA-s})
transversal four-vector functions on the cone,
not interpreting them as elements of $E^*$, at least at the initial stage of investigation. 
There remain two invariant subspaces of such functions, ``electric'' and respectively ``magnetic type'', 
transversal functions. This provides a useful hint for the investigation 
of the space $E_{\chi=-1}^{*}$. Namely the linear space of ``electric-type'' homogeneous 
of degree $-1$ functions gives rise to the construction of a well defined 
invariant subspace $(E^{*})_{tr}^{e}$ of $E_{\chi=-1}^{*} \subset E^*$ of ``electric type'' 
transversal states $(E^{*})_{tr}^{e}$ in $E_{\chi=-1}^{*}$. 

There exists essentially only one  invariant Hilbert space inner 
product on the space $(E^{*})_{tr}^{e}$ of electric type 
transversal states with the property that the operation of completion of this space of states with respect to  this invariant inner product is contained in $E^*$. 
This inner product is uniquely determined by the decomposition of Subsection \ref{equivalentA-s}.
But also, this is a consequence of the classification of invariant positive definite Hermitian bilinear 
forms on the nuclear space of smooth homogeneous of degree zero functions 
on the cone due to \cite{GelfandIV}, because the electric type homogeneous of degree zero smooth transversal states are identifiable as homogeneous of degree zero smooth functions on the cone. 
Completion of this space with respect to the corresponding invariant inner product will serve as the single particle Hilbert space
of the homogeneous of degree $-1$ electric part of the general $\big(A_\mu\big)_{{}_{\chi=-1}}$ homogeneous of degree $\chi=-1$
part of the free field $A_\mu$, constructed in Subsection \ref{equivalentA-s}.

In Subsections  \ref{infra-electric-transversal-generalized-states} and \ref{Comparison2}, we compare
the electric part of $\big(A_\mu\big)_{{}_{\chi=-1}}$, or rather its contraction with $x^\mu$: $x^\mu\big(A_\mu\big)_{{}_{\chi=-1}}$,  
with the transversal part of the phase field $S(x)$ of Staruszkiewicz theory.

The additional condition that the states in $E_{\chi=-1}^{*}$, which serve as the single particle states of the homogeneous of degree $-1$
electric part of the field $A$ should be transversal means that 
the corresponding  distributional solutions of d'Alembert equation, and belonging to 
$\mathcal{S}^{00}(\mathbb{R}^4; \mathbb{C}^4)^*$,  are transversal.
Here the relation between $\widetilde{S} \in E^* = \mathcal{S}^{0}(\mathbb{R}^3)^*$
and the solutions $F \in \mathcal{S}^{00}(\mathbb{R}^4; \mathbb{C}^4)^*$ of d'Alembert equation
is given by the general rule  (\ref{S<->F}), restricted to the positive energy sheet
$\mathscr{O}_{1,0,0,1}$ of the cone $p\cdot p=0$, equal to the orbit pertinent to the zero mas field 
$A_\mu$. The allowed homogeneities $\chi$ are determined by the spectrum of the scaling operator and the decomposition of $SL(2, \mathbb{C})$
acting in the single particle space of the field $A$, given in
Subsection \ref{equivalentA-s}.  The allowed homogeneities (which enter into this decomposition) are equal $\chi = -1 +i\nu$, $\nu \in \mathbb{R}$. Each space
of homogeneous states-functions of homogeneity $\chi$ belonging to this set, square integrable, when restricted to
the unit two-sphere, serve as the single particle Hilbert space $E^{*}_{{}_{\chi}} = \mathcal{H}'_{{}_{\chi=-1+i\nu}}$ of the homogeneous of degree $\chi$
part $A_{{}_{\chi}}$ of the field, compare Subsection \ref{equivalentA-s}. In this Subsection we pay a particular
attention to the part of homogeneity $\chi=-1$ with $\nu=0$, and decompose $E^{*}_{{}_{\chi}} = \mathcal{H}'_{{}_{\chi=-1+i\nu}}$
into invariant orthogonal subspaces. Correspondingly to this decomposition, the homogeneous
of degree $\chi=-1$ part $A_{{}_{\chi}}$ of the field decomposes into the sum of two fields, say ``electric type''
and ``magnetic type'' homogeneous of degree $-1$ part of $A$.   
In fact in case $\chi=-1$ there will remain the transversal homogeneous of degree $-1$  ``magnetic type'' states composing the invariant orthogonal complementary
of the space of electric type transversal homogeneous of degree $-1$ states.

Finally we give in this Subsection an example of a linear subspace $L[\mathfrak{F}_{\chi=1}]$ of 
$E_{\chi=-1}^{*}$ consisting  of states
which in general are not necessary transversal. We then construct invariant Hilbert space inner product on this 
subspace $L[\mathfrak{F}_{\chi=1}]$ by the kernel method. A property of the indefinite inner product, induced by the Krein-inner product of the single particle Hilbert space of the field $A_\mu$
on the space of homogeneous of degree $-1$ states of $L[\mathfrak{F}_{\chi=1}]$, will allow application of a theorem of Schoenberg 
in proving positivity of this kernel. By construction
the kernel is an invariant kernel on the Lobachevsky space. The closure of $L[\mathfrak{F}_{\chi=1}]$ with respect to 
the inner product defined by the kernel is not contained within
$E^*$. Therefore the constructed Hilbert space cannot serve as a single particle Hilbert space of any homogeneous of 
degree $\chi=-1$ part of the free field $A_\mu$.  
But the constructed kernel
coincides with a kernel which is fundamental in Staruszkiewicz theory, and having proven its positivity
independently of Staruszkiweicz theory, will allow us to prove (relative) consistency of
his theory.

In general there is no natural way of extension of the inner product (or the Krein-product)
from the Hilbert space $\mathcal{H}'$ of a Gelfand triple 
$E \subset \mathcal{H}' \subset E^{*}$ 
over to the dual space $E^{*}$ of generalized states. However, we are in a privileged situation that we are interested only in a closed subspace consisting of generalized states of the very specific character in $E^{*}$ (and generally their tensor products in 
$\big[(E)_{\textrm{Sym}}^{\otimes n}\big]^* = (E)_{\textrm{Sym}}^{*\otimes n}$ by the kernel theorem), namely
the distributions which are homogeneous of degree $\chi =-1+i\nu$ with the very specific geometry of the light cone which allows to pull back
the Lorentz invariant bilinear forms on $E \subset \mathcal{H}'$ over to the spaces of homogeneous 
 distributions in $E^*$. And on the other hand our orbit $\mathscr{O}_{1,0,0,1}$ 
(in case of the positive energy field and the negative energy cone $\mathscr{O}_{-1,0,0,1}$ in case of negative energy field) defining the {\L}opusza\'nski representation is the light cone which possess an extra internal Lorentz invariant structure in comparison to the remaining one-sheet hyperboloid orbits defining the representations which serve to compose massive fields. Namely the metric induced on the light cone by the Minkowski metric of the surrounding Minkowski momentum space is degenerate and selects the zero direction on the cone. The lines along the zero direction are called rays
or linear generators of the cone. There is a Lorenz invariant metric and measure on the manifold of rays, and separately along the rays. In particular
the metric along the rays, which thus may be coordinated by the one $0$-coordinate $p_0$ of the momentum, and thus gives
a Lorentz invariant measure along the rays, is equal
\[
\frac{\ud p_0}{p_0}.
\] 
The invariant measure on the space of rays, which may be coordinated by the spatial $p_1, p_2, p_3$
coordinates (the coordinates $p_1, p_2, p_3$ correspond to one and the same ray whenever $p_1:p_2:p_3 = const$) is equal 
\begin{equation}\label{d2p}
\ud^2 p  = \frac{p_1 dp_2 \wedge dp_3 + p_2 dp_3 \wedge dp_1 + p_3 dp_1 \wedge dp_2}{p_0}.
\end{equation}
In particular the invariant measure $\ud \mu_{{}_{\mathscr{O}_1,0,0,1}}$ on the cone is equal to the product measure
\[
\ud \mu_{{}_{\mathscr{O}_1,0,0,1}} = \frac{\ud^3 \boldsymbol{\p}}{p_0} =
\frac{dp_1 \wedge dp_2 \wedge dp_3}{p_0} = \frac{dp_0}{p_0} \wedge \ud^2 p.
\] 
The invariant measure induces in a natural way a measure with respect to which the functions which ``lives'' effectively 
on the space of rays (e.g. on the space of functions $\tilde{f}^\mu$ homogeneous of degree $\chi= -1+i\nu$, $\nu \in \mathbb{R}$,
which may be treated as distributions i.e. continuous functionals on 
\[
\mathcal{S}^0(\mathscr{O}_{1,0,0,1};\mathbb{C}^4) = \mathcal{S}^0(\mathbb{R}^3;\mathbb{C}^4) 
= \mathcal{S}^0(\mathbb{R}; \mathbb{C}) \otimes \mathscr{C}^\infty(\mathbb{S}^2; \mathbb{C}^4) =  E
\]
may be integrated, and in the case when the integrand is homogeneous of degree $-2$,
the integral is Lorentz invariant. In particular in the case when $\tilde{g}$ is a scalar homogeneous of degree $-2$, e.g. for 
$\tilde{g} = \overline{\tilde{f}^\mu} \tilde{f}_\mu$,
it can be integrated with respect to $\ud^2 p$ and the integral is Lorentz invariant, which allows to introduce natural inner products in the space of homogeneous of degree $\chi =-1+i\nu$,  $\nu \in \mathbb{R}$, four-vector solutions of the wave equation, or Maxwell equations. 
It is important to understand it correctly. Namely in the spherical coordinates $(r,\theta, \vartheta)$ 
on the light cone $\mathscr{O}_{1,0,0,1}$ defined as in the formula (\ref{cone-r,theta,phi-coordinates}) (and respectively 
on the negative energy sheet $\mathscr{O}_{-1,0,0,1}$ of the cone with the opposite sign of the zero coordinate $p^0 = -r$) we have
(here we are using the symbol $\phi$ for angle $\vartheta$)
\[
\begin{split}
\ud^2 p = r^2 \sin \theta d \phi \wedge d \theta,
\\
\ud^2 p\big|_{{}_{\mathbb{S}^2}} = \sin \theta d \phi \wedge d \theta = \ud \mu_{{}_{\mathbb{S}^2}}
\end{split}
\]
Consider now a sufficiently regular (say measurable) set of rays, namely a pencil of rays cutting an angular
set $\Omega$ on the unit $r = 1$ sphere $\mathbb{S}^2$ in the cone whose area is equal $\mu_{{}_{\mathbb{S}^2}}(\Omega)$ if measured with respect to the ordinary spherical volume form $\ud \mu_{{}_{\mathbb{S}^2}}$.
Then consider the image  of the pencil of rays under a hyperbolic transformation (a Lorentz transformation with hyperbolic angle $\lambda$) $\Lambda(\lambda)$ or more precisely its intersection $\Lambda(\lambda)\Omega$ with the unit sphere 
$\mathbb{S}^2$ of rays on the positive or negative
energy sheet of the cone. The Lorentz invariance of the one-form $\frac{\ud p_0}{p_0}$, the Lorentz invariance of the two-form $\ud^2 p$ and the Lorenz invariance of the three-form $\ud \mu_{{}_{\mathscr{O}_1,0,0,1}}$ imply among other things
that the Radon-Nikodym derivative of the transformed measure on $\mathbb{S}^2$ with respect to the initial one is equal
\begin{equation}\label{RadonNkodym-on-S2rays}
\frac{d\mu_{{}_{\mathbb{S}^2}}(\Lambda(\lambda)(\theta,\phi))}{d\mu_{{}_{\mathbb{S}^2}}(\theta,\phi)} =
\Big(\frac{p^0}{(\Lambda(\lambda)p)^0}\Big)^2 = \Big(\frac{p^0}{p'^0}\Big)^2
\end{equation}
where $p'^0$ is the zero component of the transformed four-momentum $p' = \Lambda(\lambda)p$. Thus if the integrand function 
$\tilde{g}$ is homogeneous of degree $-2$ then the non invariance of the ordinary angular measure $\ud \mu_{{}_{\mathbb{S}^2}}$ on the unit sphere $\mathbb{S}^2$ of rays is just compensated for by the homogeneity factor of the integrand function $\tilde{g}$ so that the integral 
\[
\int \limits_{\mathbb{S}^2} \tilde{g}(p) \, \ud^2 p
\]
is Lorentz invariant. Indeed, the Lorentz transformed function $p \mapsto \widetilde{g}(\Lambda p)$ when expressed
in terms of $\widetilde{g}$ restricted to the sphere $\mathbb{S}^2$  gives a multiplier representation (in the sense of 
\cite{Bargmann}, Definition 1, p. 579) of the Lorentz group:
\[
\widetilde{g}(\Lambda p) = \Big(\frac{p^0}{(\Lambda p)^0}\Big)^2 \widetilde{g}(\Lambda (\theta,\phi))
= \frac{d\mu_{{}_{\mathbb{S}^2}}(\Lambda (\theta,\phi))}{d\mu_{{}_{\mathbb{S}^2}}(\theta,\phi)}
 \widetilde{g}(\Lambda (\theta,\phi))
\]
with the multiplier which just compensates for the non-invariance of the measure $d\mu_{{}_{\mathbb{S}^2}}$
under the Lorentz transform so that
\begin{multline*}
\int \limits_{\mathbb{S}^2} \widetilde{g}(\Lambda p) \, \ud^2 p 
= \int \limits_{\mathbb{S}^2} \Big(\frac{p^0}{(\Lambda p)^0}\Big)^2 \widetilde{g}(\Lambda (\theta,\phi))
  \, \ud \mu_{{}_{\mathbb{S}^2}}(\theta,\phi) \\
=
\int \limits_{\mathbb{S}^2} \frac{d\mu_{{}_{\mathbb{S}^2}}(\Lambda (\theta,\phi))}{d\mu_{{}_{\mathbb{S}^2}}(\theta,\phi)}
 \tilde{g}(\Lambda (\theta,\phi)) \, \ud \mu_{{}_{\mathbb{S}^2}}(\theta,\phi) \\ 
=
\int \limits_{\mathbb{S}^2} \tilde{g}(\Lambda (\theta,\phi)) \, \ud \mu_{{}_{\mathbb{S}^2}}(\Lambda (\theta,\phi))
=
\int \limits_{\mathbb{S}^2} \tilde{g}(\theta,\phi) \, \ud \mu_{{}_{\mathbb{S}^2}}(\theta,\phi)
=
\int \limits_{\mathbb{S}^2} \tilde{g}(p) \, \ud^2 p 
\end{multline*}
This fact will be useful here and in Subsections \ref{infra-electric-transversal-generalized-states} 
and \ref{Consistency}.

In fact this was already noticed in \cite{Staruszkiewicz1981}, where it was shown that the indefinite  Krein-inner product (\ref{Kr-inn-Lop-1-space}) (or (\ref{Kr-inn-Lop-1-space'}) or in the position picture 
the Krein-inner product (\ref{krein-prod-1-photon})) induces naturally (degenerate, indefinite) Hermitian bilinear form
\begin{multline}\label{krein-prod-infra-red}
(\tilde{f},\tilde{f}')_{{}_{\mathfrak{J}}}^{\textrm{tr}} = -\frac{1}{8\pi}\int \limits_{\mathbb{S}^2 \times \mathbb{S}^2} 
\tilde{f}^\nu(q)\Big( g_{\mu \nu} + p_\mu \frac{\partial}{\partial p^\nu} - p_\nu \frac{\partial}{\partial p^\mu} \Big) 
\tilde{f}'^\mu(p)
\, \frac{\ud^2 p \, \ud^2 q}{p \cdot q} \\
- \frac{1}{8\pi}\int \limits_{\mathbb{S}^2 \times \mathbb{S}^2} 
\tilde{f}'^\nu(q)\Big( g_{\mu \nu} + p_\mu \frac{\partial}{\partial p^\nu} - p_\nu \frac{\partial}{\partial p^\mu} \Big) 
\tilde{f}^\mu(p)
\, \frac{\ud^2 p \, \ud^2 q}{p \cdot q},
\end{multline}
on the space of transversal, i.e. fulfilling $p_\mu \tilde{f}^\mu =0$, four-vector functions $\tilde{f}^\mu$ on the cone, homogeneous of degree $-1$.
Here $p \cdot q = g_{\mu \nu} p^\mu q^\nu$, and $g_{\mu \nu}$ are the Minkowski metric components.
 
 The bilinear form (\ref{krein-prod-infra-red})
has three important properties: 1) it is gauge invariant, which means that after addition of a gauge 
term $\delta \tilde{f}^\mu(p)$ to $\tilde{f}^\mu(p)$ (preserving homogeneity), which in the momentum picture has the general form $\delta \tilde{f}^\mu(p) = p^\mu \tilde{g}(p)$, 
with the function $\tilde{g}$ homogeneous of degree $-2$ on the light cone, the value of $(\tilde{f},\tilde{f})_{{}_{\mathfrak{J}}}^{\textrm{tr}}$
will stay unchanged; 2) the inner product (\ref{krein-prod-infra-red}) does exist for transversal 
$\tilde{f}^\mu, \tilde{f}'^\mu$, i.e. fulfilling $p^\mu \tilde{f}_\mu = 0$, homogeneous of degree $-1$ electric type states defined in (\ref{SolMaxwellHom=-1}); 3) (\ref{krein-prod-infra-red}) vanishes for a homogeneous of degree $-1$ gradient field $\tilde{f}^\mu(p) = \tilde{f}'^\mu(p) = p^\mu \tilde{g}(p)$ with 
$\tilde{g}$ homogeneous of degree $-2$ (which in fact follows from the property 1)).

We define the linear space $(E^{*})_{tr}^{e}$ of electric type transversal homogeneous of degree $\chi=-1$ states 
as the space of states spanned (over $\mathbb{C}$) by the following states 
\begin{equation}\label{SolMaxwellHom=-1}
\widetilde{f}_\mu(p) = \sum \limits_{i}^{N} \, \alpha_i \frac{u_{i\mu}}{u_i \cdot p}, \,\,\,
\sum \limits_{i}^{N} \alpha_i = 0,
\end{equation}
where $u_i$ runs over a finite set of time like unit ($u_i \cdot u_i = 1$) four-vectors, 
and $p$ runs over the positive energy sheet of the light cone in momentum space. 
Note that if we allow in this definition only real $\widetilde{f}$ and $\alpha_i$ and both energy sheets of the 
light cone in the momentum space, and finally discard the condition $\sum \alpha_i = 0$ ,
then we obtain the space of solutions $F \in \mathcal{S}^{00}(\mathbb{R}^4)^*$ 
(\ref{fmuHomogeneity=-1}) generated by the Dirac solution (\ref{thpsi-state}) 
(resp. (\ref{f^muCorrespondingTo-th(psi)})), defined in Subsection 
\ref{infra-electric-transversal-generalized-states}. 
In this identification we regard the states (\ref{SolMaxwellHom=-1}) of course as the restrictions 
$\widetilde{S} \in \mathcal{S}^{0}(\mathscr{O})$ of the Fourier transforms
$\widetilde{F} \in \mathcal{S}^{0}(\mathbb{R}^4)^*$ to the cone $\mathscr{O}$,
according to the general rule (\ref{S<->F}). 
Note that the condition
\[
\sum \limits_{i} \alpha_i = 0
\]
is equivalent to the transversality condition
\[
p^\mu \tilde{f}_\mu = 0,
\]
which, together with the assumption that $\supp \tilde{f}_\mu \subset \{p; p \cdot p = 0\}$) assures the Fourier transform 
of $\tilde{f}_\mu$ regarded as distribution on $\mathcal{S}^{0}(\mathbb{R}^4)$ concentrated on the light cone, 
to be a solution of the vacuous Maxwell equations, compare Subsection \ref{DiracHom=-1Sol}.

For the infrared fields having the form (\ref{SolMaxwellHom=-1}) the invariant inner product (\ref{krein-prod-infra-red}), extended over complex-valued 
(\ref{SolMaxwellHom=-1}), is equal 
\begin{equation}\label{krein-prod-infra-red-1}
(\tilde{f},\tilde{f})_{{}_{\mathfrak{J}}}= \int \limits_{\mathbb{S}^2} 
\Big( \tilde{f}(p), \mathfrak{J}_{\bar{p}} \tilde{f}(p) \Big)_{\mathbb{C}^4} 
\, \ud^2 p 
\,\,\,
= 
\,\,\,
- \, \int \limits_{\mathbb{S}^2}
\overline{ 
\tilde{f}_\mu(p)
}
\tilde{f}^\mu(p)
\, \ud^2 p,
\end{equation}
and, as we will soon see, on the states of the specific form (\ref{SolMaxwellHom=-1}) it coincides with 
\begin{equation}\label{inn-prod-infra-red-1}
(\tilde{f},\tilde{f})= \int \limits_{\mathbb{S}^2} 
\Big( \tilde{f}(p), \tilde{f}(p) \Big)_{\mathbb{C}^4} 
\, \ud^2 p.
\end{equation}

In case when both sheets of the light cone in momentum space are allowed, and $\widetilde{f}$ as well as 
$\alpha_i$ are real then the sum (\ref{SolMaxwellHom=-1}) can be realized physically as the electromagnetic 
potential of the infrared radiation field produced in the scattering process of point charges
$\alpha_i$, with some of the four-velocities $p_{i}$ coming in  (which have, say, the corresponding $\alpha_i$ positive) and with the 
four-velocities $p_i$ coming out which have  the corresponding $\alpha_i$ with the opposite sign, 
compare \cite{Staruszkiewicz1981}.

In particular for the potential 
\[
\tilde{f}_\mu(p) = \frac{e}{2\pi} \bigg( \frac{u_\mu}{u \cdot p} - \frac{v_\mu}{v \cdot p} \bigg)
\]
corresponding to the infrared field produced by a point charge $e$ scattered at the origin such that 
$u^\mu, v^\mu$ are the time like four-velocities of the point charge before and after the scattering respectively, 
the inner product (\ref{krein-prod-infra-red}) is equal
\[
(\tilde{f},\tilde{f})_{{}_{\mathfrak{J}}} = 2 \frac{e^2}{\pi} \bigg( \lambda \textrm{coth} \lambda - 1\bigg),
\] 
where $\lambda$ is the hyperbolic angle between $u$ and $v$, i.e. $\textrm{cosh} \, \lambda
= u\cdot v$, compare \cite{Staruszkiewicz1981}.

In the investigation of the Hermitian form (\ref{krein-prod-infra-red-1}) the operator $B$ standing
in the formula (\ref{inn-Lop-1-space}) or in the formula (\ref{Kr-inn-Lop-1-space'})
will be useful. There is a canonical decomposition of the one particle Krein-Hilbert space 
$\mathcal{H}'$ of the field $A_\mu$ associated to the operator $B$, which allows the construction 
of the subspace $\mathcal{H}'_{tr} \subset \mathcal{H}'$ of physical transversal states. 
The decomposition of $\mathcal{H}'$ associated to $B$ can in principle be extended over the space
of homogeneous functions on the cone. If in addition restrictions of these functions to the unit sphere
$\mathbb{S}^2$ belong to $L^2(\mathbb{S}^2)$, then this decomposition will allow us to make some 
statements concerning positivity of the form (\ref{krein-prod-infra-red-1}) 
as defined on homogeneous of degree $-1$
four-vector functions summable on $\mathbb{S}^2$.   

Namely recall that the ordinary one particle state, i.e. a four-component function $\widetilde{\varphi}^\mu$ 
on the cone -- an element of the Hilbert space 
$\mathcal{H}'$, has the unique decomposition
\[
\widetilde{\varphi} = 
{w_1}^+ \tilde{f}_+ + {w_1}^- \tilde{f}_- + w_{r^{-2}} \tilde{f}_{0+} + w_{r^{2}} \tilde{f}_{0-},
\] 
where the four-component functions $w$, given by the formula (\ref{eigen-vectors-matrixB}) of Subsection \ref{DefLopRep}, 
are at each point $p$ of the cone $\mathscr{O}_{1,0,0,1}$ equal to the 
eigenvectors of the $4 \times 4$ matrix $B(p)$ given by (\ref{Bmatrix}), corresponding respectively to the eigenvalues
$1,1,r^{-1}, r^2$; and where the complex valued functions $\tilde{f}_+, \tilde{f}_-$ are square integrable on the cone with respect to the invariant measure on the cone, and the scalar function $\tilde{f}_{0+}$ is square integrable with respect to the measure $\frac{\ud^3 \boldsymbol{\p}}{|\boldsymbol{\p}|^3}$; and finally with the complex valued function $\tilde{f}_{0-}$ square integrable on the cone with respect to the measure $|\boldsymbol{\p}| \, \ud^3 \boldsymbol{\p}$. 
The subspace $\mathcal{H}'_{tr}$ of physical (one particle) states 
consists precisely of all those functions $\widetilde{\varphi}$ which have the decomposition
\[
\widetilde{\varphi} = {w_1}^+ \tilde{f}_+ + {w_1}^- \tilde{f}_-.
\] 
Note in particular that the elements of $\mathcal{H}'_{tr}$ are transversal in the stronger sense, i.e. not only 
$p^\mu \widetilde{\varphi}_\mu = 0$ but $p_1\widetilde{\varphi}_1 + p_1 \widetilde{\varphi}_2 + p_3 \widetilde{\varphi}_3 =0$. Let now the (four-component) function  
$\widetilde{\varphi}$ be replaced with a function $\widetilde{f}$ on the cone, homogeneous of degree $\chi =-1+i\nu$,
$\nu \in \mathbb{R}$.  In this case $\widetilde{f}$ likewise has the unique decomposition 
\[
\widetilde{f} = {w_1}^+ \tilde{f}_+ + {w_1}^- \tilde{f}_- + w_{r^{-2}} \tilde{f}_{0+} + w_{r^{2}} \tilde{f}_{0-},
\] 
where in this decomposition the functions $\tilde{f}_+, \tilde{f}_-, \tilde{f}_{0+}, \tilde{f}_{0-}$, 
are homogeneous of degree $\chi =-1+i\nu$, as the functions
${w_1}^+, {w_1}^-,w_{r^{-2}}, w_{r^{2}}$ are homogeneous of degree zero functions on the light cone. 
Recall that the homogeneous elements of $E^{*}$ can always be represented by ordinary functions 
$\widetilde{f}$ whose restrictions to $\mathbb{S}^2$ fulfills the condition (\ref{NormHomogStates}). 
This is the case for the elements of the decomposition Hilbert space $\mathcal{H}'_{{}+{\chi}}$ of Subsection \ref{equivalentA-s},
which are representable by ordinary functions homogeneous of degree $\chi = -1+i\nu$, with square summable restrictions
to the unit two sphere $\mathbb{S}^2$.
We can therefore assume that the function $\widetilde{f}$ is 
regular enough in having the restriction to the unit sphere $\mathbb{S}^2$ which 
belongs to $L^2(\mathbb{S}^2)$, compare also Subsection \ref{DiracHom=-1Sol}. In this case the decomposition of 
$\widetilde{f}$ can be used to the analysis of the positivity of (\ref{krein-prod-infra-red-1}) on the linear space of homogeneous of degree
$\chi=-1+i\nu$ states, or in particular on homogeneous of degree $-1$ states of the form (\ref{SolMaxwellHom=-1}) which can be transversal, i.e.
\[
\sum \limits_{i}^{N} \alpha_i = 0
\]
or not necessary transversal, i.e. 
\[
\sum \limits_{i}^{N} \alpha_i \neq 0.
\]
Note that the functions 
homogeneous of degree $-1$, summable on $\mathbb{S}^2$ are, by the results of the said Subsections
and the Subsection \ref{DiracHom=-1Sol}, well defined continuous functionals on 
$E$ (compare also the Subsection \ref{DiracHom=-1Sol}).  In particular
we can consistently define the physical subspace $(E^{*})_{tr}$ of homogeneous states 
as the space of all those functions on the cone
which can be represented as the linear combination
\[
{w_1}^+ \tilde{f}_+ + {w_1}^- \tilde{f}_- + w_{r^{-2}} \tilde{f}_{0+}
\]
with $\tilde{f}_+,\tilde{f}_-, \tilde{f}_{0+}$ homogeneous of degree $\chi = -1+i\nu$,
with restrictions to $\mathbb{S}^2$ belonging to $L^2(\mathbb{S}^2)$. 
Note in particular that the elements of 
$(E^{*})_{tr}$ are transversal: $p^\mu \tilde{f}_\mu = 0$.

Observe that for homogeneous transversal states $\tilde{f}_\mu$ of homogeneity degree $\chi=-1+i\nu$, or
for any $\tilde{f}_\mu \in (E^{*})_{tr}^{e}$ of the general form (\ref{SolMaxwellHom=-1})
with
\[
\sum \limits_{i}^{N} \alpha_i = 0,
\]
the bilinear form (\ref{krein-prod-infra-red-1}) is non negatively defined. Indeed 
any such element can be decomposed into the three components
\[
\tilde{f}_\mu = {{w_1}^+}_\mu \tilde{f}_+ + {{w_1}^-}_\mu \tilde{f}_- + {w_{r^{-2}}}_\mu \tilde{f}_{0+},
\]
(the fourth component of the general decomposition is lacking because of the transversality).
On the other hand the components ${{w_1}^+}_\mu \tilde{f}_+$, ${{w_1}^-}_\mu \tilde{f}_-$, 
${w_{r^{-2}}}_\mu \tilde{f}_{0+}$ are orthogonal with respect to (\ref{krein-prod-infra-red-1}),
the bilinear form (\ref{krein-prod-infra-red-1}) is positive (for the first two components 
${{w_1}^+}_\mu \tilde{f}_+$, ${{w_1}^-}_\mu \tilde{f}_-$)
or zero (for the last ${w_{r^{-2}}}_\mu \tilde{f}_{0+}$). Thus non negativity on transversal
homogeneous of degree $\chi=-1+i\nu$ as well as positivity
on the states (\ref{SolMaxwellHom=-1})
follows whenever 
\[
\sum \limits_{i}^{N} \alpha_i = 0.
\]

Thus we may summarize the results in the following
\begin{prop*}
The invariant Hermitian bilinear form (\ref{krein-prod-infra-red-1})  
\[
(\tilde{f},\tilde{f})_{{}_{\mathfrak{J}}}= \int \limits_{\mathbb{S}^2} 
\Big( \tilde{f}(p), \mathfrak{J}_{\bar{p}} \tilde{f}(p) \Big)_{\mathbb{C}^4} 
\, \ud^2 p 
\,\,\,
= 
\,\,\,
- \, \int \limits_{\mathbb{S}^2} 
\overline{
\tilde{f}_\mu(p)
}
\tilde{f}^\mu(p)
\, \ud^2 p
\]
is non-negatively definite on the linear space  $(E^{*})_{tr}$ of transversal homogeneous of degree
$\chi = -1+i\nu$ states as well as on the space $(E^{*})_{tr}^{e}$ of transversal electric-type 
states
\[
\tilde{f}_\mu(p) = \sum \limits_{i}^{N} \, \alpha_i \frac{u_{i\mu}}{u_i \cdot p}, \,\,\,\,\,\,\,
\sum \limits_{i}^{N} \alpha_i = 0, \,\, u_i \cdot u_i =1, \, i=1, \ldots, N.
\]
Each element (\ref{SolMaxwellHom=-1}) of $(E^{*})_{tr}^{e}$ is a gradient of a homogeneous of degree
zero function $\widetilde{f}$:
\begin{multline*}
\widetilde{f}_\mu(p) = \frac{\partial \widetilde{f}}{\partial p^\mu}, \,\,\,
\widetilde{f} = \ln \,\Big(\big(u_1 \cdot p\big)^{\textrm{Re} \,\alpha_1} \ldots 
\big(u_N \cdot p\big)^{\textrm{Re} \, \alpha_N} \Big) \\
+ i \ln \,\Big(\big(u_1 \cdot p\big)^{\textrm{Im} \,\alpha_1} \ldots 
\big(u_N \cdot p\big)^{\textrm{Im} \, \alpha_N} \Big). 
\end{multline*}
\end{prop*}
Note that $u\cdot p >0$ for all $u$ ranging over the Lobachevsky space $\mathscr{L}_3 = \{u, u\cdot u =1 \}$
and $p$ ranging over the positive energy sheet of the cone. Note also that in the above Proposition
$\textrm{Re} \,\alpha_1 + \ldots + \textrm{Re} \,\alpha_N = \textrm{Im} \,\alpha_1 + \ldots +
\textrm{Im} \,\alpha_N = 0$.

Note, please, that the respective component $\widetilde{f}_+(p), \widetilde{f}_-(p), \widetilde{f}_{0+}(p)$,
of the four-vector state $\widetilde{f} = \big(\widetilde{f}_\mu\big)$ is equal to the projection
of the four-vector state $\widetilde{f}(p) = (\widetilde{f}_\mu(p))$
on the corresponding eigen-four-vector ${{w_1}^+}(p)$, ${{w_1}^-}(p)$, ${w_{r^{-2}}}(p)$:
\[
\widetilde{f}_+(p) = \big({{w_1}^+}(p), \widetilde{f}(p)\big)_{{}_{\mathbb{C}^4}},
\,\,\,\,
\widetilde{f}_-(p) = \big({{w_1}^-}(p), \widetilde{f}(p)\big)_{{}_{\mathbb{C}^4}},
\,\,\,\,
\widetilde{f}_{0+}(p) = \big({w_{r^{-2}}}(p), \widetilde{f}(p)\big)_{{}_{\mathbb{C}^4}}.
\]
Let $\widetilde{f} \in (E^{*})_{tr}^{e}$. Then, by the above Proposition,
$\widetilde{f} = \big(\widetilde{f}_\mu\big) = \big(\tfrac{\partial \widetilde{f}}{\partial p^\mu}\big)$
and is homogeneous of degree $\chi =-1$, with the corresponding scalar function $\widetilde{f}$ homogeneous
of degree zero. Using the spherical
coordinates $(r,\theta, \vartheta)$ on the cone (defined in (\ref{cone-r,theta,phi-coordinates}), but here
we denote the ange $\vartheta$ by $\phi$) we easily see that
\begin{equation}\label{f+=1/sintheta.partialf/partialphi,f-=partialf/partialtheta}
\widetilde{f}_+ =
{\textstyle\frac{1}{r \sin \theta}} {\textstyle\frac{\partial \widetilde{f}}{\partial \phi}},
\,\,\,\,
\widetilde{f}_- =
{\textstyle\frac{1}{r}}
{\textstyle\frac{\partial \widetilde{f}}{\partial \theta}},
\,\,\,\, \widetilde{f}_{0+} = {\textstyle\frac{1}{\sqrt{2} r}} {\textstyle\frac{\partial \widetilde{f}}{\partial r}} = 0,
\end{equation}
where the equality $\widetilde{f}_{0+}=0$ follows because the scalar function $\widetilde{f}$ is homogeneous of degree zero.
Thus,
\begin{equation}\label{Eetr=f+w++f-w-}
\boxed{
\textrm{if} \,\, \widetilde{f} \in (E^{*})_{tr}^{e} \,\,\,\,\,\, \textrm{then} \,\, \widetilde{f}_{0+} =0, \widetilde{f}_{0-} =0,
}
\end{equation}
and
\[
\widetilde{f} = \widetilde{f}_{+} {{w_1}^+} + \widetilde{f}_{-} {{w_1}^-}.
\]
Thus the space $(E^{*})_{tr}^{e}$ of ''electric type'' states is strongly transversal and is spanned
solely by the first two eigenvectors ${{w_1}^+}$ and ${{w_1}^-}$ and the equality
between (\ref{krein-prod-infra-red-1}) and (\ref{inn-prod-infra-red-1}) on the subspace
$(E^{*})_{tr}^{e}$ immediately follows from (\ref{Eetr=f+w++f-w-}).
The formulas (\ref{f+=1/sintheta.partialf/partialphi,f-=partialf/partialtheta}) and the elementary ''by parts integration'' gives
\begin{multline}\label{e-type-invariat-inner-prod}
\big(\tilde{f}, \tilde{f})_{{}_{\mathfrak{J}}}^{\textrm{tr}}= \int \limits_{\mathbb{S}^2}
\Big( \tilde{f}(p), \mathfrak{J}_{\bar{p}} \tilde{f}(p) \Big)_{\mathbb{C}^4}
\, \ud^2 p =
\int\limits_{\mathbb{S}^2} \big\{|\widetilde{f}_+(\phi, \theta)|^2 + |\widetilde{f}_-(\phi,\theta)|^2 \big\} \, \sin \theta \ud \theta \ud \phi
\\
=
\int\limits_{\mathbb{S}^2} \overline{\widetilde{f}} \big[-\partial_{\theta}^{2} - \textrm{ctg} \, \theta \, \partial_\theta
-{\textstyle\frac{1}{\sin^2 \theta}} \partial_{\phi}^{2} \big] \widetilde{f} \,\, \sin \theta \ud \theta \ud \phi
\\
= \int\limits_{\mathbb{S}^2} \overline{\widetilde{f}} \,\, \big[-\Delta_{{}_{\mathbb{S}^2}} \big] \widetilde{f} \, \ud \mu_{{}_{\mathbb{S}^2}},
\end{multline}
for the state of the form
\[
\widetilde{f} = (\widetilde{f}_\mu) = \big({\textstyle\frac{\partial\widetilde{f}}{\partial p^\mu}}\big).
\]
Here $\Delta_{{}_{\mathbb{S}^2}}$ is the standard Laplace operator on the unit two-sphere and in the last two integrals we have the scalar
function $\widetilde{f}$ corresponding to the state of the form $\widetilde{f}_\mu = \tfrac{\partial \widetilde{f}}{\partial p^\mu}$.
In the first integral we have also used the abbreviated notation $\widetilde{f}$ for the four-vector state
$\widetilde{f} = (\widetilde{f}_\mu) = \big(\tfrac{\partial \widetilde{f}}{\partial p^\mu}\big)$, hoping that it will not be confused with
the corresponding scalar $\widetilde{f}$.
It follows from (\ref{e-type-invariat-inner-prod}) that for such a state $\widetilde{f} \in (E^{*})_{tr}^{e}$
\[
\big(\tilde{f},\tilde{f}\big)_{{}_{\mathfrak{J}}}^{\textrm{tr}} = 0,
\]
if and only if the corresponding scalar function $\widetilde{f}$ is constant, when restricted to the unit sphere $\mathbb{S}^2$.
Because the scalar function $\widetilde{f}$ is homogeneous of degree zero, then the scalar $\widetilde{f}$ must be constant on the cone.
Thus, we have proved that
the state $\widetilde{f}_\mu = \partial \widetilde{f}/\partial p^\mu$ of the space
$(E^{*})_{tr}^{e}$ belongs to the zero subspace
$N \subset (E^{*})_{tr}^{e}$ of the hermitian form (\ref{krein-prod-infra-red-1})
if and only if the scalar function $\widetilde{f}$ is constant, and thus if and only if the four vector state
$\widetilde{f}_\mu = \partial \widetilde{f}/\partial p^\mu$ is zero. Thus, the kernel $N$ of the positive definite
inner product on the space $(E^{*})_{tr}^{e}$ is equal zero $N = \{ 0 \}$.
The completion of the
quotient $(E^{*})_{tr}^{e}/N = (E^{*})_{tr}^{e}$, with respect to the inner product
(\ref{krein-prod-infra-red-1})
is equal to the space of (equivalence classes modulo equality everywhere) functions of the form
$\widetilde{f}_\mu = \partial \widetilde{f}/\partial p^\mu$ with $\widetilde{f}$ equal
almost everywhere to a homogeneous of degree zero function, with summable
restriction to the unit sphere $\mathbb{S}^2$.
By the second Proposition of Subsection \ref{DiracHom=-1Sol}, it then follows that
the elements of the closure of $(E^{*})_{tr}^{e}/N = (E^{*})_{tr}^{e}$ with respect to
the inner product (\ref{krein-prod-infra-red-1}) belong to $E^*$, and can serve as states
of the single particle space of a homogeneous of degree $-1$ part of the free field $A_\mu$.

In general the same formulas (\ref{inn-prod-infra-red-1}) and (\ref{krein-prod-infra-red-1}) give, respectively, the inner
product and the Krein inner product on the homogeneous of degree $\chi=-1+i\nu$ states (functions) $\widetilde{f}$ of  
\[
 \mathcal{H}'_{{}_{-1+i\nu}}
\]
in the decomposition Hilbert space $\mathcal{H}'_{{}_{\chi =-1+i\nu}}$ of Subsection \ref{equivalentA-s}. This follows
by insertion of the homogeneous functions $\widetilde{f} = F^{{}^{s}}_{{}_{\chi, lm}}$ of Subsection \ref{equivalentA-s} into the formulas 
(\ref{inn-prod-infra-red-1}) and (\ref{krein-prod-infra-red-1}). Namely,
\begin{eqnarray}\label{Krein-and-inner-prod-chi}
(\tilde{f},\tilde{f})=(\tilde{f}, \tilde{f})_{{}_{\chi}}= \int \limits_{\mathbb{S}^2} 
\Big( \tilde{f}(p), \mathfrak{J}_{\bar{p}} \tilde{f}(p) \Big)_{\mathbb{C}^4} 
\, \ud^2 p,
\label{inner-prod-chi}
\\
(\tilde{f},\tilde{f})_{{}_{\mathfrak{J}}} =(\tilde{f},\mathfrak{J}_{{}_{\chi}}\tilde{f})_{{}_{\chi}} = \int \limits_{\mathbb{S}^2} 
\Big( \tilde{f}(p), \mathfrak{J}_{\bar{p}} \tilde{f}(p) \Big)_{\mathbb{C}^4} 
\, \ud^2 p 
\,\,\,
= 
\,\,\,
- \, \int \limits_{\mathbb{S}^2}
\overline{ 
\tilde{f}_\mu(p)
}
\tilde{f}^\mu(p)
\, \ud^2 p,
\label{Krein-inner-prod-chi}
\end{eqnarray}
\[
\textrm{for} \, \widetilde{f} \in  \mathcal{H}'_{{}_{\chi}}, \,\, \chi =-1+i\nu,
\]
with any fixed $\nu$ ranging over real numbers. 

The Hilbert spaces $\mathcal{H}'_{{}_{\chi = -1+i\nu}}$ of homogeneous of degree $\chi = -1+i\nu$ states are reducible
and decompose further into two orthogonal (with respect to the inner product (\ref{inner-prod-chi}) and with respect to the invariant
inner product (\ref{Krein-and-inner-prod-chi}) in $\mathcal{H}'_{{}_{\chi = -1+i\nu}}$) and invariant subspaces. 
Because we are primarily interested in the case $\chi =-1$, we will show it in details for the case $\chi=-1$ 
and give only the proof for existence of these invariant subspaces for the general case with $\nu\neq 0$ (explicit computation 
of the complete system of solutions is slightly more complicated for the case $\nu\neq 0$).

Let us concentrate our attention on $\mathcal{H}'_{{}_{\chi = -1+i\nu}}$, with $\nu=0$, containing the closed and invariant
subspace $(E^{*})_{tr}^{e}$ (to see invariance of $(E^{*})_{tr}^{e}$, recall please that $SL(2, \mathbb{C})$ acts on the homogeneous of degree
$-1$ state $\widetilde{f} = (\widetilde{f}_\mu) = (\partial \widetilde{f}/\partial p^\mu)$ as ordinary Lorentz transformation
on the four-vector, and thus on the corresponding scalar $\widetilde{f}$ as the ordinary Lorentz transformation acting on scalars,
leaving the gradient form and homogeneity unchanged). Let us define now the orthogonal complementary
$(E^{*})^{\mathfrak{m}}={\mathcal{H}'}_{{}_{\chi = -1}}^{\mathfrak{m}}$ of $(E^{*})_{tr}^{e} = {\mathcal{H}'}^{e}_{{}_{\chi = -1}}$ in
$\mathcal{H}'_{{}_{\chi = -1}}$:
\begin{equation}\label{H-1=He-1+Hm-1}
\mathcal{H}'_{{}_{\chi = -1}} = {\mathcal{H}'}^{e}_{{}_{\chi = -1}} \oplus
{\mathcal{H}'}_{{}_{\chi = -1}}^{\mathfrak{m}}
\end{equation}
with respect to the inner product (\ref{inner-prod-chi}) in $\mathcal{H}'_{{}_{\chi = -1}}$. It is not obvious that this orthogonal
complementary $(E^{*})^{\mathfrak{m}}={\mathcal{H}'}_{{}_{\chi = -1}}^{\mathfrak{m}}$ is Lorentz invariant because
the inner product (\ref{inner-prod-chi}) is not invariant and the {\L}opusza\'nski representation is not unitary.
But the orthogonality condition of a state $\widetilde{g}$ to any state
$\widetilde{f}\in (E^{*})_{tr}^{e} = {\mathcal{H}'}^{e}_{{}_{\chi = -1}}$ with respect to (\ref{inner-prod-chi})
coincides with the orthogonality with respect to the invariat Krein inner product (\ref{Krein-inner-prod-chi}), because the components
$\widetilde{f}_{0+}$ and $\widetilde{f}_{0-}$ of the electric type state $\widetilde{f}$ are both equal zero, whence
the invariance of the orthogonality condition with respect to (\ref{inner-prod-chi}). Therefore,
$(E^{*})^{m}={\mathcal{H}'}_{{}_{\chi = -1}}^{m}$ is indeed invariant.
Let $\widetilde{g}$ be any homogeneous of degree $-1$ four-vector state $\widetilde{g}$ with smooth
components $\widetilde{g}_{+}$, $\widetilde{g}_{-}$, $\widetilde{g}_{0+}$, $\widetilde{g}_{0-}$.
Let us write the orthogonality
condition for such a state $\widetilde{g}$ to all smooth $\widetilde{f}\in (E^{*})_{tr}^{e}
= {\mathcal{H}'}^{e}_{{}_{\chi = -1}}$ in more explicit form. To this end we again use the
spherical coordinates. By assumption the restrictions to $\mathbb{S}^2$ of the scalar functions $\widetilde{f}$
corresponding to $\widetilde{f} = \partial\widetilde{f}/\partial p^\mu \in (E^{*})_{tr}^{e}
= {\mathcal{H}'}^{e}_{{}_{\chi = -1}}$
and of the component functions $\widetilde{g}_{+}$, $\widetilde{g}_{-}$, $\widetilde{g}_{0+}$, $\widetilde{g}_{0-}$
belong to $\mathcal{S}_{\Delta_{\mathbb{S}^2}}(\mathbb{S}^2;\mathbb{C}) = \mathscr{C}^\infty(\mathbb{S}^2;\mathbb{C})$.
The formulas (\ref{f+=1/sintheta.partialf/partialphi,f-=partialf/partialtheta})
for $\widetilde{f}_{+}$, $\widetilde{f}_{-}$ and the elementary ''by parts integration'' gives:
\begin{multline*}
\big(\widetilde{f}, \widetilde{g} \big)_{{}_{\chi=-1}}
=\big(\widetilde{f}_{+}{w_1}^+ + \widetilde{f}_{-}{w_1}^-, \,\, \widetilde{g}_{+}{w_1}^+ + \widetilde{g}_{-}{w_1}^-
\widetilde{g}_{0+}{w_{r^{-2}}} + \widetilde{g}_{0-}{w_{r^{2}}}\big)
\\
=
\big(\widetilde{f}_{+}{w_1}^+ + \widetilde{f}_{-}{w_1}^-, \,\, \widetilde{g}_{+}{w_1}^+ + \widetilde{g}_{-}{w_1}^-
\widetilde{g}_{0+}{w_{r^{-2}}} + \widetilde{g}_{0-}{w_{r^{2}}}\big)_{{}_{\mathfrak{J}}}
=\big(\widetilde{f}, \widetilde{g})_{{}_{\mathfrak{J}}}
\\
= \int \limits_{\mathbb{S}^2}
\Big( \widetilde{f}(p), \mathfrak{J}_{\bar{p}} \widetilde{f}(p) \Big)_{\mathbb{C}^4}
\, \ud^2 p =
\int\limits_{\mathbb{S}^2} \big\{\overline{\widetilde{f}_+(\phi, \theta)}\widetilde{g}_+(\phi, \theta) +
\overline{\widetilde{f}_-(\phi, \theta)}\widetilde{g}_-(\phi, \theta) \big\} \, \sin \theta \ud \theta \ud \phi
\\
=
-\int\limits_{\mathbb{S}^2} \overline{\widetilde{f}(\phi,\theta) } \big[
{\textstyle\frac{1}{\sin \theta}} \partial_{\phi}\widetilde{g}_{+}(\phi,\theta)
+\partial_{\theta}\widetilde{g}_{-}(\phi,\theta)
+ \textrm{ctg} \, \theta \, \widetilde{g}_{-}(\phi,\theta)
\big] \,\, \sin \theta \ud \theta \ud \phi =0
\end{multline*}
\[
\textrm{for each smooth} \, \widetilde{f} \in L^2(\mathbb{S}^2; \mathbb{C}).
\]
In the last integral we have the scalar functions $\widetilde{f}$ corresponding to the smooth electric type states
\[
\widetilde{f} = (\widetilde{f}_\mu) = (\partial \widetilde{f}/\partial p^\mu) \in (E^{*})_{tr}^{e}
= {\mathcal{H}'}_{{}_{\chi = -1}}^{e},
\]
restricted to the unit two sphere and expressed in spherical coordinates.
Thus, $(E^{*})^{\mathfrak{m}}={\mathcal{H}'}_{{}_{\chi = -1}}^{\mathfrak{m}}$ is equal to the closure,
with respect to (\ref{inner-prod-chi}),
of the space of all homogeneous of degree $-1$ states $\widetilde{g}$ with smooth
$\widetilde{g}_+$, $\widetilde{g}_-$, $\widetilde{g}_{0+}$, $\widetilde{g}_{0-}$, and with
the components $\widetilde{g}_+, \widetilde{g}_-$ which respect the following differential equation
\begin{equation}\label{DiffEqForg}
{\textstyle\frac{1}{\sin \theta}} \partial_{\phi}\widetilde{g}_{+}
+\partial_{\theta}\widetilde{g}_{-}
+ \textrm{ctg} \, \theta \, \widetilde{g}_{-} =0.
\end{equation}
In particular the components $\widetilde{g}_{0+}, \widetilde{g}_{0-}\in L^2(\mathbb{S}^2; \mathbb{C})$ can be chosen arbitrarily.
Equivalently, $(E^{*})^{\mathfrak{m}}={\mathcal{H}'}_{{}_{\chi = -1}}^{\mathfrak{m}}$ is equal to the closure of all homogeneous of degree $-1$
four vector functions $\widetilde{g}$ with smooth $\widetilde{g}_+$, $\widetilde{g}_-$, $\widetilde{g}_{0+}$, $\widetilde{g}_{0-}$
fulfilling (\ref{DiffEqForg}),
with respect to the inner product
\[
\int\limits_{\mathbb{S}^2} \big\{ |\widetilde{g}_{+}(p)|^2 + |\widetilde{g}_{-}(p)|^2
+ |\widetilde{g}_{0+}(p)|^2 + |\widetilde{g}_{0-}(p)|^2 \big\} \, d^2p.
\]
Note that the elements
$\widetilde{f}\in (E^{*})_{tr}^{e} = {\mathcal{H}'}_{{}_{\chi = -1}}^{e}$
and $\widetilde{g} \in (E^{*})^{\mathfrak{m}}={\mathcal{H}'}_{{}_{\chi = -1}}^{\mathfrak{m}}$ are completely independent.
By the last Proposition the invariant Krein inner product (\ref{Krein-inner-prod-chi}) defines a non-negatively
definite, degenerate, and invariant Hermitian form on the transversal subspace
$(E^{*})^{\mathfrak{m}}_\textrm{tr}={\mathcal{H}'}_{{}_{\chi = -1 \,\,\textrm{tr}}}^{\mathfrak{m}}$
of $(E^{*})^{\mathfrak{m}}={\mathcal{H}'}_{{}_{\chi = -1}}^{\mathfrak{m}}$ spanned by the first three eigenvectors ${w_1}^+, {w_1}^-, {w_{r^{-2}}}$. The kernel
\[
N_{{}_{\chi=-1 \, \textrm{tr}}} = N_{{}_{\textrm{tr}}} = \big\{\widetilde{g} = \widetilde{g}_{0+}{w_{r^{-2}}},
\,\,\, \widetilde{g}_{0+} \in L^2(\mathbb{S}^2; \mathbb{C}) \big\}
\]
of this invariant Hermitian form is a nonzero and invariant subspace of
$(E^{*})^{\mathfrak{m}}_\textrm{tr}={\mathcal{H}'}_{{}_{\chi = -1 \,\,\textrm{tr}}}^{\mathfrak{m}}$.
The quotient space $(E^{*})^{\mathfrak{m}}_\textrm{tr}/N_\textrm{tr}
={\mathcal{H}'}_{{}_{\chi = -1 \,\,\textrm{tr}}}^{\mathfrak{m}}/N_\textrm{tr}$
is naturally a pre-Hilbert space with the well-defined positive definite and invariant inner product on the equivalence classes,
with the well-defined action of $SL(2, \mathbb{C})$ which is unitary. We can also proceed as in Subsection \ref{SingleKreinLopRep}
and compute the action of $SL(2, \mathbb{C})$ induced on the strongly transversal subspace of ${\mathcal{H}'}_{{}_{\chi = -1}}^{\mathfrak{m}}$,
spanned by the first two ${w_1}^+, {w_1}^-$, and compute the action of $SL(2, \mathbb{C})$ modulo the ''non-physical'' longitudinal
states of Krein norm zero. The formula for the action of $SL(2, \mathbb{C})$, expressed in terms of the coefficient functions
$\widetilde{g}_+,\widetilde{g}_-$ coincides with the analogue formula given in Subsection \ref{SingleKreinLopRep} with the same
phase function $\Theta(\alpha,p)$ (where we have replaced $f_+,f_-$ with $\widetilde{g}_+,\widetilde{g}_-$), and the difference is that
now the strongly transversal Hilbert space of $\widetilde{g}_+,\widetilde{g}_-$ is the closure with respect to the inner product
\[
\int\limits_{\mathbb{S}^2} \big\{ |\widetilde{g}_+(p)|^2 + |\widetilde{g}_-(p)|^2 \big\} \, d^2p
\]
of smooth homogeneous of degree $-1$ functions $\widetilde{g}_+,\widetilde{g}_-$ which respect the differential equation (\ref{DiffEqForg}).

We note here that, similarly as for the electric type states, also each state $\widetilde{g}$ of ${\mathcal{H}'}_{{}_{\chi = -1}}^{\mathfrak{m}}$
can essentially, \emph{i.e.} up to elements of Krein-norm zero, be characterized by the corresponding scalar function
\[
\widetilde{m}_{{}_{\widetilde{g}}} =
-2\frac{p_1 \partial_{{}_{p^{[2}}} \widetilde{g}_{{}_{3]}} + p_2 \partial_{{}_{p^{[3}}} \widetilde{g}_{{}_{1]}}
+ p_3 \partial_{{}_{p^{[1}}} \widetilde{g}_{{}_{2]}}}{p_0},
\]
where the bracket $[ik]$ stands for antisymmetrization of the indices $ik$. That $\widetilde{m}_{{}_{\widetilde{g}}}$
is a scalar under Lorentz transformation of the corresponding state $\widetilde{g}$ follows analogously
as the invariance of the measure (\ref{d2p}). It is easily seen that the Krein zero norm longitudinal
components $\widetilde{g}_{0+}{w_{r^{-2}}}$ and $\widetilde{g}_{0-}{w_{r^{2}}}$ have zero contribution
to the corresponding scalar $\widetilde{m}_{{}_{\widetilde{g}}}$.

In Subsection \ref{equivalentA-s} we have determined the complete orthonormal system
of states $F^{{}^{s}}_{{}_{\chi, lm}}$, $l=0,1, \ldots$, $-l\leq m \leq l$, $s=+,-,0+,0-$,
in $\mathcal{H}'_{{}_{\chi = -1+i\nu}}$ for each $\nu \in \mathbb{R}$. Now we determine complete orthonormal system
$F_{{}_{\chi=-1 \, lm}}^{e}$, $l=1, \ldots$, $-l\leq m \leq l$, in  ${\mathcal{H}'}^{e}_{{}_{\chi = -1}}$ and, respectively,
a complete orthonormal system $\big\{F_{{}_{\chi=-1 \, lm}}^{\mathfrak{m} \,\, s}, F_{{}_{\chi=-1 \, 00}}^{\mathfrak{m} \,\, 0}\big\}$
in ${\mathcal{H}'}^{\mathfrak{m}}_{{}_{\chi = -1}}$. In the last system $s$ ranges over the three element set $0,0+,0-$,
with the range of $l,m$ depending on $s$. For $s=0$,
$l=1, 2, \ldots$, $-l\leq m \leq l$, $m\neq 0$, and for $s=0+$ or $s=0-$, 
$l=0,1, \ldots$, $-l\leq m \leq l$.

From the formulas (\ref{f+=1/sintheta.partialf/partialphi,f-=partialf/partialtheta}) and from the formula (\ref{e-type-invariat-inner-prod})
for the inner product of the electric type states it immediately follows that 
\begin{equation}\label{Felectric}
F_{{}_{\chi=-1, lm} \, \mu}^{e}(p) = {\textstyle\frac{1}{\sqrt{l(l+1)}}}\frac{\partial Y_{{}_{lm}}}{\partial p^\mu}(p), 
\end{equation}
\[
l=1, 2, \ldots,  \,\,\, -l\leq m \leq l,
\]
composes a complete orthonormal system of states in $(E^{*})_{tr}^{e} = {\mathcal{H}'}^{e}_{{}_{\chi = -1}}$. 
Here $ Y_{{}_{lm}}$ is understood as the homogeneous of degree zero function coinciding with the spherical
harmonics $Y_{lm}$ on the unit two-sphere $\mathbb{S}^2$. Using spherical coordinates on the cone we can write this system in the
form
\[
F_{{}_{\chi=-1 \, lm}}^{e} = 
{\textstyle\frac{1}{\sqrt{l(l+1)} \,\, r \sin \theta}} {\textstyle\frac{\partial Y_{{}_{lm}}}{\partial \phi}} {w_1}^+
+
{\textstyle\frac{1}{\sqrt{l(l+1)} \,\, r}}
{\textstyle\frac{\partial Y_{{}_{lm}}}{\partial \theta}} {w_1}^-
\]
\[
l=1, 2, \ldots,  \,\,\, -l\leq m \leq l.
\]
In the last formula $Y_{{}_{lm}}$ are the ordinary spherical functions on $\mathbb{S}^2$.

In order to determine the complete orthonormal system of states $F_{{}_{\chi=-1 \, lm}}^{\mathfrak{m} \,\, s}$ in ${\mathcal{H}'}^{\mathfrak{m}}_{{}_{\chi = -1}}$
we determine first a complete system of smooth solutions of the differential equation (\ref{DiffEqForg}). The equation
(\ref{DiffEqForg}) can be rewritten in the following form
\begin{equation}\label{DiffEqForgCanonical}
\big[\sin \theta \, \partial_{\theta}
+ \textrm{cos} \, \theta \big] \, \widetilde{g}_{-} = -\partial_{\phi}\widetilde{g}_{+},
\end{equation}
\[
\textrm{for} \,\,\, \widetilde{g}_{-} ,\widetilde{g}_{+} \in \mathcal{S}_{\Delta_{\mathbb{S}^2}}(\mathbb{S}^2;\mathbb{C})
= \mathscr{C}^\infty (\mathbb{S}^2;\mathbb{C}).
\]
Here we understood the space $\mathscr{C}^\infty (\mathbb{S}^2;\mathbb{C})$ of smooth $\mathbb{C}$-valued functions
on the unit two-sphere $\mathbb{S}^2$ as the standard nuclear space $\mathcal{S}_{\Delta_{\mathbb{S}^2}}(\mathbb{S}^2;\mathbb{C})$,
determined by the standard operator $\Delta_{\mathbb{S}^2}$
-- the Laplace operator on $\mathbb{S}^2$. Each element $g \in \mathcal{S}_{\Delta_{\mathbb{S}^2}}(\mathbb{S}^2;\mathbb{C})$
has rapidly decreasing development
\[
g = \sum\limits_{lm} g_{{}_{lm}} Y_{{}_{lm}}
\]
with rapidly decreasing sequence $\{g_{{}_{lm}}\}$, \emph{i.e.} with the series
\[
\sum\limits_{lm} \big|\lambda_{{}_{lm}}^{n} g_{{}_{lm}}\big|^2 <\infty
\]
convergent for each natural $n$, where $\lambda_{{}_{lm}} = l(l+1)$, $l=0,1,\ldots$, $-l\leq m \leq l$,
are the spectral values of $\Delta_{\mathbb{S}^2}$. From the second and the fourth Lemma of Subsection
\ref{diffSA} it follows that the differential operators
$\sin \theta \, \partial_{\theta} + \cos \theta$ and $\partial_{\phi}$, map
continuously $\mathcal{S}_{\Delta_{\mathbb{S}^2}}(\mathbb{S}^2;\mathbb{C})$ into itself. Because
$Y_{{}_{lm}}$ are the eigenfunctions of the operator $\partial_\phi$ to the eigenvalue $im$, then
any smooth $g$ lying in the kernel of $\partial_\phi$ is representable by rapidly decreasing development
into $Y_{{}_{l0}}$ with $m=0$ and no smooth $g$ can lie in the image of $\partial_\phi$ which has
in its rapidly decreasing development nonzero components $Y_{{}_{l0}}$ with $m=0$. On the other hand easy
computation using the general properties of the spherical harmonics shows that the result of the action
of the operator $\sin \theta \, \partial_{\theta} + \cos \theta$ on $Y_{{}_{lm}}$ is equal
\[
\big[\sin \theta \, \partial_{\theta} + \cos \theta \big] Y_{{}_{l \,\, m}}
= (l+1)
\sqrt{{\textstyle\frac{(l+1)^2 - m^2}{(2l+1)(2l+3)}}} Y_{{}_{l+1 \,\, m}}
-
l
\sqrt{{\textstyle\frac{l^2 - m^2}{4l^2-1}}} Y_{{}_{l-1 \,\, m}}.
\]
This function lies in the image of the operator $-\partial_\phi$ if and only if $m \neq 0$, which also implies $l\neq 0$.
In this case, \emph{i.e.} $m\neq 0, l\neq 0$, the last formula can be written
\[
\big[\sin \theta \, \partial_{\theta} + \cos \theta \big] Y_{{}_{l \,\, m}}
= -\partial_\phi \Bigg[
{\textstyle\frac{i(l+1)}{m}}
\sqrt{{\textstyle\frac{(l+1)^2 - m^2}{(2l+1)(2l+3)}}} Y_{{}_{l+1 \,\, m}}
-
{\textstyle\frac{il}{m}}
\sqrt{{\textstyle\frac{l^2 - m^2}{4l^2-1}}} Y_{{}_{l-1 \,\, m}}
\Bigg].
\]
It follows that $\widetilde{g}_{+}, \widetilde{g}_{-}$,
given by
\[
\widetilde{g}_{+} = {\textstyle\frac{i(l+1)}{m}}
\sqrt{{\textstyle\frac{(l+1)^2 - m^2}{(2l+1)(2l+3)}}} Y_{{}_{l+1 \,\, m}}
-
{\textstyle\frac{il}{m}}
\sqrt{{\textstyle\frac{l^2 - m^2}{4l^2-1}}} Y_{{}_{l-1 \,\, m}},
\,\,\,\,\,\,\,
\widetilde{g}_{-} = Y_{{}_{l \,\, m}}
\]
\[
m \neq 0, l=1,2, \ldots, -l \leq m \leq l,
\]
together with the constant solution
\[
\widetilde{g}_{+} = Y_{{}_{0 \,\, 0}},
\,\,\,\,\,\,\,
\widetilde{g}_{-} = 0,
\]
composes a complete system of smooth solutions of the equation (\ref{DiffEqForgCanonical}) or equivalently of the equation
(\ref{DiffEqForg}). Its completeness can also be checked directly by checking that any state $\widetilde{h}$
with components
\[
\widetilde{h}_{+} = \sum\limits_{lm} \widetilde{h}_{{}_{+lm}} Y_{{}_{l \,\, m}} , \,\,\,\,\, \widetilde{h}_{-} = 0,
\]
or
\[
\widetilde{h}_{+} = 0, \,\,\,\,\, \widetilde{h}_{-} = \sum\limits_{lm} \widetilde{h}_{{}_{-lm}} Y_{{}_{l \,\, m}}
\]
orthogonal to all $\widetilde{g}_{+}, \widetilde{g}_{-}$ of the above complete system of solutions, with respect to the inner product
\begin{equation}\label{InnerProductg+g-}
\int\limits_{\mathbb{S}^2} \big\{ \overline{\widetilde{g}_+(p)}\widetilde{h}_{+}(p) + \overline{\widetilde{g}_-(p)}\widetilde{h}_{-}(p) \big\} \, d^2p,
\end{equation}
is equal zero. This, again, can be checked, using the standard properties of the spherical harmonics.
However, the complete system of smooth solutions $\widetilde{g}_{+}, \widetilde{g}_{-}$ given above is not orthonormal with respect to the inner product
(\ref{InnerProductg+g-}). Equivalently, the corresponding states
\[
\begin{split}
G_{{}_{\chi=-1 \,\,\, lm}}^{\mathfrak{m}} = \Bigg[{\textstyle\frac{i(l+1)}{m}}
\sqrt{{\textstyle\frac{(l+1)^2 - m^2}{(2l+1)(2l+3)}}} {\textstyle\frac{1}{r}}Y_{{}_{l+1 \,\, m}}
-
{\textstyle\frac{il}{m}}
\sqrt{{\textstyle\frac{l^2 - m^2}{4l^2-1}}} {\textstyle\frac{1}{r}}Y_{{}_{l-1 \,\, m}} \Bigg]\, {w_1}^+ + {\textstyle\frac{1}{r}}Y_{{}_{l \,\, m}} \, {w_1}^-,
\\
G_{{}_{\chi=-1 \,\,\, 00}}^{\mathfrak{m}} = {\textstyle\frac{1}{r}}Y_{{}_{0 \,\, 0}}{w_1}^+,
\,\,\,\,\,\,\,\,\,\,\,\,\,\,\,\,\,\,\,\,\,\,\,\,\,\,\,\,\,\,\,\,\,\,\,\,\,\,\,\,\,\,\,\,\,\,\,\,\,\,\,\,\,\,\,\,
\,\,\,\,\,\,\,\,\,\,\,\,\,\,\,\,\,\,\,\,\,\,\,\,\,\,\,\,\,\,\,\,\,\,\,\,\,\,\,\,\,\,\,\,\,\,\,\,\,\,\,\,\,\,\,\,
\,\,\,\,\,\,\,\,\,\,\,\,\,\,\,\,\,\,\,\,\,\,\,\,\,\,\,\,\,\,
\\
m \neq 0, l=1,2, \ldots, -l \leq m \leq l,
\,\,\,\,\,\,\,\,\,\,\,\,\,\,\,\,\,\,\,\,\,\,\,\,\,\,\,\,\,\,\,\,\,\,\,\,\,\,\,\,\,\,\,\,\,\,\,\,\,\,\,\,\,\,\,\,
\,\,\,\,\,\,\,\,\,\,\,\,\,
\end{split}
\]
are not orthonormal with respect to the inner product (\ref{inner-prod-chi}) for $\chi=-1$. But the orthonormalization
of this system $\{ G_{{}_{\chi=-1 \,\,\, lm}}^{\mathfrak{m}},G_{{}_{\chi=-1 \,\,\, 00}}^{\mathfrak{m}}\}$,
$m \neq 0$, $l=1,2, \ldots$, $-l \leq m \leq l$, gives the orthonormal system
$\{ F_{{}_{\chi=-1 \,\,\, lm}}^{\mathfrak{m} \,\, 0}, F_{{}_{\chi=-1 \,\,\, 00}}^{\mathfrak{m} \,\, 0}\}$,
$m \neq 0$, $l=1,2, \ldots$, $-l \leq m \leq l$,
which together with
the longitudinal states $F_{{}_{\chi \,lm}}^{{}^{s}}$, $s=0+,0-$, $l=0,1,\ldots$, $-l\leq m\leq l$, of Subsection \ref{equivalentA-s},
composes the required complete orthonormal system
$\{ F_{{}_{\chi=-1 \,\,\, lm}}^{\mathfrak{m} \,\, s}\}$, $s=0,0+,0-$, of states in ${\mathcal{H}'}^{m}_{{}_{\chi = -1}}$.
Of course, in the last system the range of $l,m$ depends on the value of the index $s$, and for $s=0$, the range is equal
$m \neq 0$, $l=1,2, \ldots$, $-l \leq m \leq l$. For $s=0+$ and $s=0-$, the range of $l,m$ is equal
$l=0,1,\ldots$, $-l\leq m\leq l$.
We are not giving the explicit formula for $\{ F_{{}_{\chi=-1 \,\,\, lm}}^{\mathfrak{m} \,\, s}\}$
because in this particular case it follows almost immediately from the formula for
$G_{{}_{\chi=-1 \,\,\, lm}}^{\mathfrak{m}}$. First, we put the normalized $G_{{}_{\chi=-1 \,\,\, 00}}^{\mathfrak{m}}$ for
$F_{{}_{\chi=-1 \,\,\, 00}}^{\mathfrak{m} \,\, 0}$, the normalized $G_{{}_{\chi=-1 \,\,\, lm}}^{\mathfrak{m}}$, $l=1,2$,
$-l\leq m \leq l$, for $F_{{}_{\chi=-1 \,\,\, lm}}^{\mathfrak{m} \,\, 0}$, $l=1,2$,
$-l\leq m \leq l$. Next, starting with $l=3$, we normalize each element of $\{ G_{{}_{\chi=-1 \,\,\, lm}}^{\mathfrak{m}}\}$,
and subtract from the normalized $G_{{}_{\chi=-1 \,\,\, l \,\, m}}^{\mathfrak{m}}$ its projection
on the normalized $G_{{}_{\chi=-1 \,\,\, l-2 \,\, m}}^{\mathfrak{m}}$, and normalize the result, which gives
$F_{{}_{\chi=-1 \,\,\, lm}}^{\mathfrak{m} \,\, 0}$. Finally, we adjoint the longitudinal states
$F_{{}_{\chi=-1 \,lm}}^{{}^{s}}$, $s=0+,0-$, $l=0,1,\ldots$, $-l\leq m\leq l$, of Subsection \ref{equivalentA-s}.

Similar decomposition 
\begin{equation}\label{Hchi=Hechi+Hmchi}
\mathcal{H}'_{{}_{\chi = -1+i\nu}} = {\mathcal{H}'}^{e}_{{}_{\chi = -1+i\nu}} \oplus
{\mathcal{H}'}_{{}_{\chi = -1+i\nu}}^{\mathfrak{m}}
\end{equation}
is possible for $\mathcal{H}'_{{}_{\chi = -1+i\nu}}$, $\nu \neq 0$. We replace in this decomposition
the invariant strongly transversal subspace $(E^{*})_{tr}^{e}$ of homogeneous of degree $-1$ electric type states by 
an invariant strongly transversal subspace ${\mathcal{H}'}_{{}_{\chi = -1+i\nu}}^{e} 
\subset \mathcal{H}'_{{}_{\chi = -1+i\nu}}$ spanned by the homogeneous of degree $\chi = -1+i\nu$ states
of the form
\[
\widetilde{f}_\mu(p) = \sum \limits_{i}^{N} \, \alpha_i \frac{u_{i\mu}}{u_i \cdot p} \widetilde{f}(p), \,\,\,\,\,\,\,
\sum \limits_{i}^{N} \alpha_i = 0, \,\, u_i \cdot u_i =1, \, i=1, \ldots, N,
\]
where $\widetilde{f}$ is any scalar homogeneous of degree $i\nu$ function on the cone, with the restriction to the unit two-sphere
$\mathbb{S}^2$ square summable. Explicit construction of complete orthonormal systems of states, respectively, in both invariant orthogonal 
subspaces ${\mathcal{H}'}^{e}_{{}_{\chi = -1=i\nu}}$ and
 ${\mathcal{H}'}_{{}_{\chi = -1+i\nu}}^{\mathfrak{m}} 
\subset \mathcal{H}'_{{}_{\chi = -1+i\nu}}$ is slightly more complicated in comparison to the case $\chi=-1$. 
In case $\nu\neq 0$ the states of these invariant subspaces cannot be characterized by simple scalars, and have more involved structure. 
Because we are not interested here in the case $\chi=-1+i\nu$ with $\nu\neq 0$, we do not enter into details of this construction. 

In each case the Casimir operators of the $SL(2, \mathbb{C})$ group can be easily computed together with the representation structure
of the compact subgroup $SU(2, \mathbb{C})$, and the minimal weight $l$ of the corresponding irreducible component
$L_l$ with weight $l$, so that using \cite{Geland-Minlos-Shapiro} or \cite{NeumarkLorentzBook}, we can easily identify the 
type of the representation of $SL(2, \mathbb{C})$ acting in ${\mathcal{H}'}_{{}_{\chi = -1}}^{e}$,
${\mathcal{H}'}_{{}_{\chi = -1 \,\, \textrm{tr}}}^{m}/N_\textrm{tr}$, or more generally in 
 ${\mathcal{H}'}_{{}_{\chi = -1+i\nu \,\,\, \textrm{tr}}}^{e}$ and finally in 
${\mathcal{H}'}_{{}_{\chi = -1+i\nu}}^{m}$ modulo the states of Krein norm zero. 
In particular in the next Section we will show that the unitary representation of 
$SL(2, \mathbb{C})$ acting in  ${\mathcal{H}'}_{{}_{\chi = -1}}^{e}$ is irreducible and
is characterized by the pair $(l_0=1, l_1=0)$ in the classification scheme of \cite{Geland-Minlos-Shapiro}.

\begin{center}
{\small DECOMPOSITION OF $A_{{}_{\chi}}$}
\end{center}

To the direct sum decomposition (\ref{Hchi=Hechi+Hmchi}) there correspond the decomposition
of the annihilation-creation operators $a'_{{}_{\chi}},{a'}_{{}_{\chi}}^{+}$ of the homogeneous of degree $\chi=-1+i\nu$
part $A_{{}_{\chi}}$ of the electromagnetic potential field $A$, constructed in Subsection \ref{equivalentA-s}.
Recall that $a'_{{}_{\chi}}(\widetilde{h}),{a'}^{+}_{{}_{\chi}}(\widetilde{h}) = {a'}_{{}_{\chi}}(\widetilde{h})^{+}$
are the annihilation-creation operators associated to the Fock space $\Gamma\big(\mathcal{H}'_{{}_{\chi = -1+i\nu}} \big)$
over the single particle Hilbert space $\mathcal{H}'_{{}_{\chi = -1+i\nu}}$ of homogeneous of degree $\chi$
states $\widetilde{h}$ (functions on the cone).
Recall please, that $\mathcal{H}'_{{}_{\chi = -1+i\nu}}$ are the direct integral decomposition components of the single particle
Hilbert space of the free electromagnetic potential field $A$, determined by the spectral decomposition of the scaling operator,
which determines uniquely the spectrum $\chi=-1=i\nu$ of this decomposition (all allowable homogeneity degrees $\chi$)
and the homogeneous parts $A_{{}_{\chi}}$.
Recall please, that to the decomposition (\ref{Hchi=Hechi+Hmchi})
there corresponds canonically the factorization
\[
\Gamma\big(\mathcal{H}'_{{}_{\chi = -1+i\nu}} \big) = \Gamma\big({\mathcal{H}'}^{e}_{{}_{\chi = -1+i\nu}} \oplus
{\mathcal{H}'}_{{}_{\chi = -1+i\nu}}^{\mathfrak{m}} \big)
= \Gamma\big({\mathcal{H}'}^{e}_{{}_{\chi = -1+i\nu}} \big) \otimes \Gamma\big({\mathcal{H}'}_{{}_{\chi = -1+i\nu}}^{\mathfrak{m}} \big)
\]
of the Fock space, and the canonical decomposition of the annihilation-creation operators. Namely,
if $\widetilde{h} = \widetilde{f} \oplus \widetilde{g}$ is the orthogonal decomposition of
$\widetilde{h} \in \mathcal{H}'_{{}_{\chi = -1+i\nu}}$ into the components $\widetilde{f} \in {\mathcal{H}'}^{e}_{{}_{\chi = -1+i\nu}}$
and $\widetilde{g} \in {\mathcal{H}'}_{{}_{\chi = -1+i\nu}}^{\mathfrak{m}}$, then
\[
\begin{split}
a'_{{}_{\chi}}(\widetilde{h}) = a'_{{}_{\chi}}(\widetilde{f} \oplus \widetilde{g})
= a^{e}_{{}_{\chi}}(\widetilde{f}) \otimes \boldsymbol{1} + \boldsymbol{1} \otimes a^{\mathfrak{m}}_{{}_{\chi}}(\widetilde{g}),
\\
a'_{{}_{\chi}}(\widetilde{h})^+ = a^{e}_{{}_{\chi}}(\widetilde{f} \oplus \widetilde{g})^+
= a^{e}_{{}_{\chi}}(\widetilde{f})^+ \otimes \boldsymbol{1} + \boldsymbol{1} \otimes a^{\mathfrak{m}}_{{}_{\chi}}(\widetilde{g})^+
\end{split}
\]
on
\[
\Gamma\big({\mathcal{H}'}^{e}_{{}_{\chi = -1+i\nu}} \big) \otimes \Gamma\big({\mathcal{H}'}_{{}_{\chi = -1+i\nu}}^{\mathfrak{m}} \big).
\]
Here $a^{e}_{{}_{\chi}}(\widetilde{f}),a^{e}_{{}_{\chi}}(\widetilde{f})^+$ are the annihilation-creation operators
of the Fock space $\Gamma\big({\mathcal{H}'}^{e}_{{}_{\chi = -1+i\nu}} \big)$, and, analogously,
$a^{\mathfrak{m}}_{{}_{\chi}}(\widetilde{g}),a^{\mathfrak{m}}_{{}_{\chi}}(\widetilde{g})^+$ are the annihilation-creation
operators of the Fock space $\Gamma\big({\mathcal{H}'}_{{}_{\chi = -1+i\nu}}^{\mathfrak{m}} \big)$.
We write this decomposition shortly
\[
\begin{split}
a'_{{}_{\chi}}(\widetilde{f} \oplus \widetilde{g}) = a^{e}_{{}_{\chi}}(\widetilde{f}) + a^{\mathfrak{m}}_{{}_{\chi}}(\widetilde{g}),
\\
a'_{{}_{\chi}}(\widetilde{f} \oplus \widetilde{g})^+ = a^{e}_{{}_{\chi}}(\widetilde{f})^+ + a^{\mathfrak{m}}_{{}_{\chi}}(\widetilde{g})^+,
\end{split}
\]
but in the last equality the operators $a^{e}_{{}_{\chi}}(\widetilde{f})$, $a^{e}_{{}_{\chi}}(\widetilde{f})^+$,
$a^{\mathfrak{m}}_{{}_{\chi}}(\widetilde{g})$, $a^{\mathfrak{m}}_{{}_{\chi}}(\widetilde{g})^+$, are understood as acting
in the full Fock space
\[
\Gamma\big(\mathcal{H}'_{{}_{\chi = -1+i\nu}} \big) = \Gamma\big({\mathcal{H}'}^{e}_{{}_{\chi = -1+i\nu}} \oplus
{\mathcal{H}'}_{{}_{\chi = -1+i\nu}}^{\mathfrak{m}} \big)
= \Gamma\big({\mathcal{H}'}^{e}_{{}_{\chi = -1+i\nu}} \big) \otimes \Gamma\big({\mathcal{H}'}_{{}_{\chi = -1+i\nu}}^{\mathfrak{m}} \big).
\]
over all homogeneous of degree $\chi=-1+i\nu$ states.

This decomposition of the annihilation-creation operators
$a'_{{}_{\chi}}(\widetilde{f} \oplus \widetilde{g}), a'_{{}_{\chi}}(\widetilde{f} \oplus \widetilde{g})^+$,
defines canonically the corresponding decomposition
\[
A_{{}_{\chi}} = A_{{}_{\chi}}^{e} + A_{{}_{\chi}}^{\mathfrak{m}}
\]
of the homogeneous of degree $\chi$ part
$A_{{}_{\chi}}$ of the free electromagnetic potential field $A$, into two different types of homogeneous of degree $\chi=-1+i\nu$
electromagnetic potential fields. Using the complete orthonormal systems, respectively, $\{ F_{{}_{\chi=-1+i\nu \,\,\, n}}^{e}\}$
and $\{ F_{{}_{\chi=-1 +i\nu\,\,\,n}}^{\mathfrak{m}}\}$, in ${\mathcal{H}'}^{e}_{{}_{\chi = -1+i\nu}}$
and ${\mathcal{H}'}_{{}_{\chi = -1+i\nu}}^{\mathfrak{m}}$, exactly as we did with the complete orthonormal systems
$\{ F_{{}_{\chi=-1+i\nu \,\,\, lm}}^{{}^{s}}\}$ in Subsection \ref{equivalentA-s}, we can determine the kernels
$A_{{}_{\chi}}^{e}(x)$ and $A_{{}_{\chi}}^{\mathfrak{m}}(x)$ of the fields $ A_{{}_{\chi}}^{e}$
and $A_{{}_{\chi}}^{\mathfrak{m}}$. In fact in order to do it explicitly we need to construct explicitly the complete
orthonormal systems $\{ F_{{}_{\chi=-1+i\nu \,\,\, n}}^{e}\}$
and $\{ F_{{}_{\chi=-1+i\nu \,\,\, n}}^{\mathfrak{m}}\}$, which we have done only for the special case
$\chi=-1$. Thus, we can do it for the special case $\chi=-1$ using the complete orthonormal systems
$\{ F_{{}_{\chi=-1 \,\,\, lm}}^{e}\}$
and $\{ F_{{}_{\chi=-1 \,\,\, lm}}^{\mathfrak{m} \,\, s}\}$, respectively,
in ${\mathcal{H}'}^{e}_{{}_{\chi = -1}}$
and ${\mathcal{H}'}_{{}_{\chi = -1}}^{\mathfrak{m}}$.

We proceed generally after Subsection \ref{equivalentA-s}, including all $\chi = -1 +i\nu$, $\nu \in \mathbb{R}$,
assuming that we have given explicitly
the complete systems $\{ F_{{}_{\chi=-1+i\nu \,\,\, n}}^{e}\}$
and $\{ F_{{}_{\chi=-1 +i\nu\,\,\,n}}^{\mathfrak{m}}\}$, respectively, in ${\mathcal{H}'}^{e}_{{}_{\chi = -1+i\nu}}$
and in ${\mathcal{H}'}_{{}_{\chi = -1+i\nu}}^{\mathfrak{m}}$. At the end we restrict ourselves to
the special case $\chi = -1$, and insert the complete orthonormal systems
$\{ F_{{}_{\chi=-1 \,\,\, lm}}^{e}\}$
and $\{ F_{{}_{\chi=-1 \,\,\, lm}}^{\mathfrak{m} \,\, s}\}$, respectively,
in ${\mathcal{H}'}^{e}_{{}_{\chi = -1}}$
and ${\mathcal{H}'}_{{}_{\chi = -1}}^{\mathfrak{m}}$ for
$\{ F_{{}_{\chi=-1+i\nu \,\,\, n}}^{e}\}$
and $\{ F_{{}_{\chi=-1 +i\nu\,\,\,n}}^{\mathfrak{m}}\}$,
which we have constructed explicitly for the case $\chi=-1$.

Namely, let $\varphi \in \mathcal{S}^{00}(\mathbb{R}^4;\mathbb{C}^4)=\mathscr{E}$ be any space-time
test function of the electromagnetic potential field $A$. The restriction
$\widetilde{\varphi}\big|_{{}_{\mathscr{O}_{{}_{1,0,0,1}}}}$ of its Fourier transform $\widetilde{\varphi}$ 
to the positive energy sheet $\mathscr{O}_{{}_{1,0,01}}$
of the cone  is an element of the nuclear space $E\subset\mathcal{H}'$ in the single particle Hilbert space of the field
$A$, which composes the single particle Gelfand triple $E\subset\mathcal{H}'\subset E^*$
of the field $A$. $\widetilde{\varphi}\big|_{{}_{\mathscr{O}_{{}_{1,0,0,1}}}}$, as an element of $E\subset\mathcal{H}'$
has the direct integral decomposition
\[
\widetilde{\varphi}\big|_{{}_{\mathscr{O}_{{}_{1,0,0,1}}}} =
\int \widetilde{\varphi}\big|_{{}_{\mathscr{O}_{{}_{1,0,0,1}} \,\, \chi}} \, \ud \chi
\]
determined by the spectral decomposition of the scaling operator $\widetilde{S_\lambda}$, compare Subsection \ref{equivalentA-s}, with the
decomposition components
\[
\widetilde{\varphi}\big|_{{}_{\mathscr{O}_{{}_{1,0,0,1}} \,\, \chi}} \in \mathcal{H}'_{{}_{\chi = -1+i\nu}},
\,\,\,\,\,\, \chi \in \{ -1+i\nu, \,\, \nu \in \mathbb{R}\}, \,\, \lambda^{{}^{\chi}} \in \textrm{Spec} \, \widetilde{S_\lambda}.
\]
Accordingly to the general decomposition theory of \cite{GelfandIV}, based on the Gelfand triples (rigged Hilbert spaces)
and Subsection \ref{equivalentA-s}, decomposition component $\widetilde{\varphi}\big|_{{}_{\mathscr{O}_{{}_{1,0,0,1}} \,\, \chi}}$
has the following rapidly decreasing development
\[
\widetilde{\varphi}\big|_{{}_{\mathscr{O}_{{}_{1,0,0,1}} \,\, \chi}} =
\sum\limits_{n} c^{e}_{{}_{n}} F_{{}_{\chi=-1 \,\,\, n}}^{e}
+
\sum\limits_{n} c^{\mathfrak{m}}_{{}_{n}} F_{{}_{\chi=-1 \,\,\, n}}^{\mathfrak{m}},
\]
where 
\[
\begin{split}
c^{e}_{{}_{n}}=
\Big(\widetilde{\varphi}\big|_{{}_{\mathscr{O}_{{}_{1,0,0,1}} \,\, \chi}}, F_{{}_{\chi=-1 \,\,\, n}}^{e}\Big)_{{}_{\chi}}
= F_{{}_{\chi=-1 \,\,\, n}}^{e} \big( \overline{\widetilde{\varphi}\big|_{{}_{\mathscr{O}_{{}_{1,0,0,1}}}}}\big)
\\
c^{\mathfrak{m}}_{{}_{n}} =
\Big(\widetilde{\varphi}\big|_{{}_{\mathscr{O}_{{}_{1,0,0,1}} \,\, \chi}}, F_{{}_{\chi=-1 \,\,\, n}}^{\mathfrak{m}}\Big)_{{}_{\chi}}
= F_{{}_{\chi=-1 \,\,\, n}}^{\mathfrak{m}} \big( \overline{\widetilde{\varphi}\big|_{{}_{\mathscr{O}_{{}_{1,0,0,1}}}}}\big).
\end{split}
\]
Thus, using the orthogonal projectors $P_{{}_{e}}$ and $P_{{}_{\mathfrak{m}}}$ in $\mathcal{H}'_{{}_{\chi = -1+i\nu}}$
on the orthogonal subspaces ${\mathcal{H}'}^{e}_{{}_{\chi = -1+i\nu}}$ and ${\mathcal{H}'}_{{}_{\chi = -1+i\nu}}^{\mathfrak{m}}$,
defined by the complete orthonormal systems,
we can write
\[
\widetilde{\varphi}\big|_{{}_{\mathscr{O}_{{}_{1,0,0,1}} \,\, \chi}} = 
\sum\limits_{n} c^{e}_{{}_{n}} F_{{}_{\chi=-1 \,\,\, n}}^{e}
+
\sum\limits_{n} c^{\mathfrak{m}}_{{}_{n}} F_{{}_{\chi=-1 \,\,\, n}}^{\mathfrak{m}}
=
P_{{}_{e}} \widetilde{\varphi}\big|_{{}_{\mathscr{O}_{{}_{1,0,0,1}} \,\, \chi}} + 
P_{{}_{\mathfrak{m}}}\widetilde{\varphi}\big|_{{}_{\mathscr{O}_{{}_{1,0,0,1}} \,\, \chi}},
\]
and analogously
\[
\overline{\check{\widetilde{\varphi}}\big|_{{}_{\mathscr{O}_{{}_{1,0,0,1}} \,\, \chi}}} = 
\sum\limits_{n} \check{c}^{e}_{{}_{n}} F_{{}_{\chi=-1 \,\,\, n}}^{e}
+
\sum\limits_{n} \check{c}^{\mathfrak{m}}_{{}_{n}} F_{{}_{\chi=-1 \,\,\, n}}^{\mathfrak{m}}
=
P_{{}_{e}} \overline{\check{\widetilde{\varphi}}\big|_{{}_{\mathscr{O}_{{}_{1,0,0,1}} \,\, \chi}}}  + 
P_{{}_{\mathfrak{m}}}\overline{\check{\widetilde{\varphi}}\big|_{{}_{\mathscr{O}_{{}_{1,0,0,1}} \,\, \chi}}} ,
\]
with
\[
\begin{split}
\check{c}^{e}_{{}_{n}}=
\Big(\overline{\check{\widetilde{\varphi}}\big|_{{}_{\mathscr{O}_{{}_{1,0,0,1}} \,\, \chi}}}, F_{{}_{\chi=-1 \,\,\, n}}^{e}\Big)_{{}_{\chi}}
= F_{{}_{\chi=-1 \,\,\, n}}^{e} \big( \check{\widetilde{\varphi}}\big|_{{}_{\mathscr{O}_{{}_{1,0,0,1}}}}\big)
\\
\check{c}^{\mathfrak{m}}_{{}_{n}} =
\Big(\overline{\check{\widetilde{\varphi}}\big|_{{}_{\mathscr{O}_{{}_{1,0,0,1}} \,\, \chi}}}, F_{{}_{\chi=-1 \,\,\, n}}^{\mathfrak{m}}\Big)_{{}_{\chi}}
= F_{{}_{\chi=-1 \,\,\, n}}^{\mathfrak{m}} \big( \check{\widetilde{\varphi}}\big|_{{}_{\mathscr{O}_{{}_{1,0,0,1}}}}\big).
\end{split}
\]
We, thus, get
\[
A_{{}_{\chi}}(\varphi) 
=
a'_{{}_{\chi}}\Big(\overline{\check{\widetilde{\varphi}}\big|_{{}_{\mathscr{O}_{{}_{1,0,0,1}} \,\, \chi}}}\Big) 
+ \, \eta_{{}_{\chi}} a'_{{}_{\chi}}\Big(\widetilde{\varphi}\big|_{{}_{\mathscr{O}_{{}_{1,0,0,1}} \,\, \chi}}\Big)^+ \eta_{{}_{\chi}} =
\,\,\,\,\,\,\,\,\,\,\,\,\,\,\,\,\,\,\,\,\,\,\,\,\,\,\,\,\,\,\,\,\,\,\,\,\,\,\,\,\,\,\,\,\,\,\,\,\,\,\,\,\,\,\,\,\,\,\,\,
\,\,\,\,\,\,\,\,\,\,\,\,\,\,\,\,\,\,\,\,\,\,\,\,\,\,\,
\]
\begin{multline*}
= a^{e}_{{}_{\chi}}\Big(P_{{}_{e}}\overline{\check{\widetilde{\varphi}}\big|_{{}_{\mathscr{O}_{{}_{1,0,0,1}} \,\, \chi}}}\Big) 
+a^{e}_{{}_{\chi}}\Big(P_{{}_{e}}\widetilde{\varphi}\big|_{{}_{\mathscr{O}_{{}_{1,0,0,1}} \,\, \chi}}\Big)^+
\\
+ 
a^{\mathfrak{m}}_{{}_{\chi}}\Big(P_{{}_{\mathfrak{m}}}\overline{\check{\widetilde{\varphi}}\big|_{{}_{\mathscr{O}_{{}_{1,0,0,1}} \,\, \chi}}}\Big) 
+\eta_{{}_{\chi}} a^{\mathfrak{m}}_{{}_{\chi}}\Big(P_{{}_{\mathfrak{m}}}\widetilde{\varphi}\big|_{{}_{\mathscr{O}_{{}_{1,0,0,1}} \,\, \chi}}\Big)^+ \eta_{{}_{\chi}}
\\
= A_{{}_{\chi}}^{e}(\varphi) + A_{{}_{\chi}}^{\mathfrak{m}}(\varphi),
\end{multline*}
where in the contribution $A_{{}_{\chi}}^{e}(\widetilde{\varphi})$ the decomposition component 
$\eta_{{}_{\chi}}= \Gamma(\mathfrak{J}_{{}_{\chi}})$ of the Gupta-Bleuler operator
is equal $\boldsymbol{1}$, in accordance with the fact that $\mathfrak{J}_{{}_{\chi}} = \boldsymbol{1}$ on the subspace
${\mathcal{H}'}^{e}_{{}_{\chi = -1+i\nu}}$ of strongly transversal states of homogeneity degree $\chi = -1+i\nu$.

Proceeding further after Subsection \ref{equivalentA-s}, we observe that for each space-time test function $\varphi \in \mathscr{E}$ the components
\[
P_{{}_{e}} \widetilde{\varphi}\big|_{{}_{\mathscr{O}_{{}_{1,0,0,1}} \,\, \chi}}, 
P_{{}_{\mathfrak{m}}}\widetilde{\varphi}\big|_{{}_{\mathscr{O}_{{}_{1,0,0,1}} \,\, \chi}},
\]
belong to the single particle nuclear spaces $E_{{}_{\chi}}^{e},E_{{}_{\chi}}^{\mathfrak{m}}$
in the Hilbert spaces ${\mathcal{H}'}^{e}_{{}_{\chi = -1+i\nu}}$, ${\mathcal{H}'}^{\mathfrak{m}}_{{}_{\chi = -1+i\nu}}$,
and that the annihilation-creation operators $a^{e}_{{}_{\chi}}\big( \, \cdot \, \big), a^{e \, +}_{{}_{\chi}}\big( \, \cdot \, \big)$ 
and $a^{\mathfrak{m}}_{{}_{\chi}}\big( \, \cdot \, \big), a^{\mathfrak{m}\, +}_{{}_{\chi}}\big( \, \cdot \, \big)$ 
are understood as the Hida operators over the corresponding
single particle Gelfand triples 
\[
\begin{split}
E_{{}_{\chi}}^{e} \subset {\mathcal{H}'}^{e}_{{}_{\chi}} \subset E_{{}_{\chi}}^{e \, *},
\\
E_{{}_{\chi}}^{\mathfrak{m}} \subset {\mathcal{H}'}^{\mathfrak{m}}_{{}_{\chi}}  \subset E_{{}_{\chi}}^{\mathfrak{m} \, *}.
\end{split}
\]
In particular the sum operation, defining the projections
\[
P_{{}_{e}} \widetilde{\varphi}\big|_{{}_{\mathscr{O}_{{}_{1,0,0,1}} \,\, \chi}}, 
P_{{}_{\mathfrak{m}}}\widetilde{\varphi}\big|_{{}_{\mathscr{O}_{{}_{1,0,0,1}} \,\, \chi}},
P_{{}_{e}} \overline{\check{\widetilde{\varphi}}\big|_{{}_{\mathscr{O}_{{}_{1,0,0,1}} \,\, \chi}}}, 
P_{{}_{\mathfrak{m}}}\overline{\check{\widetilde{\varphi}}\big|_{{}_{\mathscr{O}_{{}_{1,0,0,1}} \,\, \chi}}} ,
\]
can be pull out of the arguments of the operators $a^{e}_{{}_{\chi}}\big( \, \cdot \, \big)$, $a^{e \, +}_{{}_{\chi}}\big( \, \cdot \, \big)$ 
and $a^{\mathfrak{m}}_{{}_{\chi}}\big( \, \cdot \, \big)$, $a^{\mathfrak{m}\, +}_{{}_{\chi}}\big( \, \cdot \, \big)$.
Therefore introducing the operators
\[
\begin{split}
{a}_{{}_{\chi, n}}^{e} \coloneqq a'_{{}_{\chi}}\big(F^{e}_{{}_{\chi,n}} \big) = a^{e}_{{}_{\chi}}\big(F^{e}_{{}_{\chi,n}} \big) \otimes \boldsymbol{1}, 
\,\,\,\,\,\,\,\,
{a}_{{}_{\chi, n}}^{e \, +} \coloneqq a'_{{}_{\chi}}\big(F^{e}_{{}_{\chi,n}} \big)^+ = a^{e}_{{}_{\chi}}\big(F^{e}_{{}_{\chi,n}} \big)^+ \otimes \boldsymbol{1}
\\
{a}_{{}_{\chi, n}}^{\mathfrak{m}} \coloneqq a'_{{}_{\chi}}\big(F^{\mathfrak{m}}_{{}_{\chi,n}} \big)  = \boldsymbol{1} \otimes a^{\mathfrak{m}}_{{}_{\chi}}\big(F^{\mathfrak{m}}_{{}_{\chi,n}} \big),
\,\,\,\,\,\,\,\,
{a}_{{}_{\chi, n}}^{\mathfrak{m} \, +} \coloneqq a'_{{}_{\chi}}\big(F^{\mathfrak{m}}_{{}_{\chi,n}} \big)^+ = \boldsymbol{1} \otimes a^{\mathfrak{m}}_{{}_{\chi}}\big(F^{\mathfrak{m}}_{{}_{\chi,n}} \big)^+,
\end{split}
\]
\[
\textrm{on}
\,\, \Gamma\big({\mathcal{H}'}^{e}_{{}_{\chi = -1+i\nu}}\big) \otimes \Gamma\big({\mathcal{H}'}^{\mathfrak{m}}_{{}_{\chi = -1+i\nu}}\big)
= \Gamma(\mathcal{H}'_{{}_{\chi = -1+i\nu}})
\]
which respect the canonical commutation relations
\[
\begin{split}
\big[{a}_{{}_{\chi, n}}^{e} , {a}_{{}_{\chi, n'}}^{e \, +}  \big]_{{}_{-}} = \delta_{{}_{nn'}},
\,\, \big[{a}_{{}_{\chi, n}}^{e} , {a}_{{}_{\chi, n'}}^{e}  \big]_{{}_{-}} 
=  \big[{a}_{{}_{\chi, n}}^{e \, +} , {a}_{{}_{\chi, n'}}^{e \, +}  \big]_{{}_{-}} = 0,
\\
\big[{a}_{{}_{\chi, n}}^{\mathfrak{m}} , {a}_{{}_{\chi, n'}}^{\mathfrak{m} \, +}  \big]_{{}_{-}} = \delta_{{}_{nn'}},
\,\, \big[{a}_{{}_{\chi, n}}^{\mathfrak{m}} , {a}_{{}_{\chi, n'}}^{\mathfrak{m}}  \big]_{{}_{-}} 
=  \big[{a}_{{}_{\chi, n}}^{\mathfrak{m} \, +} , {a}_{{}_{\chi, n'}}^{\mathfrak{m} \, +}  \big]_{{}_{-}} = 0,
\end{split}
\]
with all the remaining pairs of these operators commuting, we obtain
\[
A_{{}_{\chi}}(\varphi) =
\sum\limits_{n}
\overline{\check{c}_{{}_{n}}^{e}}
\,\, {a}_{{}_{\chi, n}}^{e}
+ \sum\limits_{n} c_{{}_{n}}^{e} \,\,  \, {a}_{{}_{\chi, n}}^{e \, +} 
+ \sum\limits_{n}
\overline{\check{c}_{{}_{n}}^{\mathfrak{m}}}
\,\, {a}_{{}_{\chi, n}}^{\mathfrak{m}}
+ \sum\limits_{n} c_{{}_{n}}^{\mathfrak{m}} \,\,  \, \eta_{{}_{\chi}} \, {a}_{{}_{\chi, n}}^{\mathfrak{m} \, +} \, \eta_{{}_{\chi}} 
\]
\begin{multline*}
=
\sum\limits_{n}
\overline{F^{\mathfrak{m}}_{{}_{\chi, n}} \big( \check{\widetilde{\varphi}}\big|_{{}_{\mathscr{O}_{1,0,0,1}}} \Big)}
\,\, a_{{}_{\chi, n}}^{\mathfrak{m}}
+ \sum\limits_{n}
F^{\mathfrak{m}}_{{}_{\chi,n}} \big( \overline{\widetilde{\varphi}\big|_{{}_{\mathscr{O}_{1,0,0,1}}}} \Big)
\,\,  a_{{}_{\chi, n}}^{\mathfrak{m} \, +} 
\\
+
\sum\limits_{n}
\overline{F^{\mathfrak{m}}_{{}_{\chi, n}} \big( \check{\widetilde{\varphi}}\big|_{{}_{\mathscr{O}_{1,0,0,1}}} \Big)}
\,\, a_{{}_{\chi, n}}^{\mathfrak{m}}
+ \sum\limits_{n}
F^{\mathfrak{m}}_{{}_{\chi,n}} \big( \overline{\widetilde{\varphi}\big|_{{}_{\mathscr{O}_{1,0,0,1}}}} \Big)
\,\,  \eta_{{}_{\chi}} a_{{}_{\chi, n}}^{\mathfrak{m} \, +} \eta_{{}_{\chi}}
\end{multline*}
\begin{multline*}
=
\sum\limits_{n}
\overline{\check{F}^{e}_{{}_{\chi,n}} \big( \widetilde{\varphi}\big|_{{}_{\mathscr{O}_{1,0,0,1}}} \Big)}
\,\, a_{{}_{\chi, n}}^{e}
+ \sum\limits_{n}
F^{e}_{{}_{\chi,n}} \Big( \overline{\widetilde{\varphi}\big|_{{}_{\mathscr{O}_{1,0,0,1}}}} \Big)
\,\,  a_{{}_{\chi, n}}^{e \, +}
\\
+
\sum\limits_{n}
\overline{\check{F}^{\mathfrak{m}}_{{}_{\chi,n}} \big( \widetilde{\varphi}\big|_{{}_{\mathscr{O}_{1,0,0,1}}} \Big)}
\,\, a_{{}_{\chi, n}}^{\mathfrak{m}}
+ \sum\limits_{n}
F^{\mathfrak{m}}_{{}_{\chi,n}} \Big( \overline{\widetilde{\varphi}\big|_{{}_{\mathscr{O}_{1,0,0,1}}}} \Big)
\,\,  \eta_{{}_{\chi}} \, a_{{}_{\chi, n}}^{\mathfrak{m} \, +} \, \eta_{{}_{\chi}}
\end{multline*}
\begin{multline*}
=
\sum\limits_{n}
\overline{\check{F}^{e}_{{}_{\chi,n}} \big( \widetilde{\varphi}\big|_{{}_{\mathscr{O}_{1,0,0,1}}} \Big)}
\,\, a_{{}_{\chi, n}}^{e}
+ \sum\limits_{n}
\overline{F^{e}_{{}_{\chi,n}} \Big( \widetilde{\varphi}\big|_{{}_{\mathscr{O}_{1,0,0,1}}} \Big)}
\,\, a_{{}_{\chi, n}}^{e \, +} 
\\
+
\sum\limits_{n}
\overline{\check{F}^{\mathfrak{m}}_{{}_{\chi,n}} \big( \widetilde{\varphi}\big|_{{}_{\mathscr{O}_{1,0,0,1}}} \Big)}
\,\, a_{{}_{\chi, n}}^{\mathfrak{m}}
+ \sum\limits_{n}
\overline{F^{\mathfrak{m}}_{{}_{\chi,n}} \Big( \widetilde{\varphi}\big|_{{}_{\mathscr{O}_{1,0,0,1}}} \Big)}
\,\, \eta_{{}_{\chi}} \, a_{{}_{\chi, n}}^{\mathfrak{m} \, +} \, \eta_{{}_{\chi}} 
\end{multline*}
\begin{multline}\label{Aechi(varphi)+Amchi(varphi)}
=
\sum\limits_{n}
\overline{\check{f}^{e}_{{}_{\chi,n}}( \varphi)}
\,\, a_{{}_{\chi,n}}^{e}
+ \sum\limits_{n}
\overline{f^{e}_{{}_{\chi,n}} (\varphi)}
\,\, a_{{}_{\chi, lms}}^{e \, +}
\\
+
\sum\limits_{n}
\overline{\check{f}^{\mathfrak{m}}_{{}_{\chi,n}}( \varphi)}
\,\, a_{{}_{\chi,n}}^{\mathfrak{m}}
+ \sum\limits_{n}
\overline{f^{\mathfrak{m}}_{{}_{\chi,n}} (\varphi)}
\,\, \eta_{{}_{\chi}} \, a_{{}_{\chi, n}}^{\mathfrak{m} \, +} \, \eta_{{}_{\chi}}
= A_{{}_{\chi}}^{e}(\varphi) + A_{{}_{\chi}}^{\mathfrak{m}}(\varphi),
\end{multline}
where $f^{e}_{{}_{\chi,n}}$, $\check{f}^{e}_{{}_{\chi,n}}$ 
$f^{\mathfrak{m}}_{{}_{\chi,n}}$, $\check{f}^{\mathfrak{m}}_{{}_{\chi,n}}$,
are the inverse Fourier transforms, respectively, of the distributions 
\[
\begin{split}
\widetilde{\varphi} \longmapsto F_{{}_{\chi, n}}^{e} \big(\widetilde{\varphi}\big|_{{}_{\mathscr{O}_{{}_{1,0,0,1}}}}\big),
\,\,\,\,\,\,
\widetilde{\varphi} \longmapsto \check{F}^{e}_{{}_{\chi,n}}  \big(\widetilde{\varphi}\big|_{{}_{\mathscr{O}_{{}_{1,0,0,1}}}}\big),
\\
\widetilde{\varphi} \longmapsto F_{{}_{\chi, n}}^{\mathfrak{m}} \big(\widetilde{\varphi}\big|_{{}_{\mathscr{O}_{{}_{1,0,0,1}}}}\big),
\,\,\,\,\,\,
\widetilde{\varphi} \longmapsto \check{F}^{\mathfrak{m}}_{{}_{\chi,n}} \big(\widetilde{\varphi}\big|_{{}_{\mathscr{O}_{{}_{1,0,0,1}}}}\big).
\end{split}
\]
Because the distributions 
\[
\varphi \longmapsto f^{e}_{{}_{\chi,n}}(\varphi),  
\,\,\,\ 
\varphi \longmapsto \check{f}^{e}_{{}_{\chi,n}}(\varphi), 
\,\,\,\,
\varphi \longmapsto f^{\mathfrak{m}}_{{}_{\chi,n}}(\varphi)
\,\,\,\,
\varphi \longmapsto \check{f}^{\mathfrak{m}}_{{}_{\chi,n}}(\varphi),  
\]
are regular with the values on the test functions $\varphi$ representable by integration of the contraction
$f^{e \, \mu}_{{}_{\chi,n}}\varphi_\mu, \ldots $ with the ordinary homogeneous function $f^{e \, \mu}_{{}_{\chi,n}}, \ldots$,
smooth everywhere except the light cone, then the kernels $A_{{}_{\chi}}^{e}(x), A_{{}_{\chi}}^{\mathfrak{m}}(x)$
of the fields $A_{{}_{\chi}}^{e}, A_{{}_{\chi}}^{\mathfrak{m}}$ are determined
by (\ref{Aechi(varphi)+Amchi(varphi)}) and by the kernels $f^{e}_{{}_{\chi,n}}(x), \ldots$, of the said homogeneous
distributions $f^{e}_{{}_{\chi,n}}, \ldots$. Namely, (compare Subsection \ref{equivalentA-s})
\begin{multline*}\label{Achi(x)=Aechi(x)+Amchi(x)}
A_{{}_{\chi}}(x)
=
\sum\limits_{n}
f^{e}_{{}_{\chi,n}}(x)
\,\, a_{{}_{\chi, n}}^{e}
+ \sum\limits_{n}
\overline{f^{e}_{{}_{\chi,n}} (x)}
\,\, a_{{}_{\chi, n}}^{e \, +} 
\\
+
\sum\limits_{n}
f^{\mathfrak{m}}_{{}_{\chi,n}}(x)
\,\, a_{{}_{\chi, n}}^{\mathfrak{m}}
+ \sum\limits_{n}
\overline{f^{\mathfrak{m}}_{{}_{\chi,n}} (x)}
\,\, \eta_{{}_{\chi}} \, a_{{}_{\chi, n}}^{\mathfrak{m} \, +} 
\, \eta_{{}_{\chi}}
= A_{{}_{\chi}}^{e}(x) + A_{{}_{\chi}}^{\mathfrak{m}}(x),
\end{multline*}

Now we restrict $\chi = -1+i\nu$ to
the special case $\chi = -1$, and insert the complete orthonormal systems
$\{ F_{{}_{\chi=-1 \,\,\, lm}}^{e}\}$
and $\{ F_{{}_{\chi=-1 \,\,\, lm}}^{\mathfrak{m} \,\, s}\}$, respectively,
in ${\mathcal{H}'}^{e}_{{}_{\chi = -1}}$
and ${\mathcal{H}'}_{{}_{\chi = -1}}^{\mathfrak{m}}$ for
$\{ F_{{}_{\chi=-1+i\nu \,\,\, n}}^{e}\}$
and $\{ F_{{}_{\chi=-1 +i\nu\,\,\,n}}^{\mathfrak{m}}\}$,
which we have constructed explicitly for the case $\chi=-1$.
We obtain the following formula, with the self-understood notation:
\begin{multline*}
A_{{}_{\chi=-1}}(\varphi)=
\sum\limits_{lm}
\overline{\check{F}^{e}_{{}_{\chi=-1,lm}} \big( \widetilde{\varphi}\big|_{{}_{\mathscr{O}_{1,0,0,1}}} \Big)}
\,\, a_{{}_{\chi=-1, lm}}^{e}
+ \sum\limits_{lm}
\overline{F^{e}_{{}_{\chi=-1,lm}} \Big( \widetilde{\varphi}\big|_{{}_{\mathscr{O}_{1,0,0,1}}} \Big)}
\,\, a_{{}_{\chi=-1, lm}}^{e \, +} 
\\
+
\sum\limits_{lms}
\overline{\check{F}^{\mathfrak{m} \, s}_{{}_{\chi=-1,lm}} \big( \widetilde{\varphi}\big|_{{}_{\mathscr{O}_{1,0,0,1}}} \Big)}
\,\, a_{{}_{\chi=-1, lms}}^{\mathfrak{m}}
+ \sum\limits_{lms}
\overline{F^{\mathfrak{m} \, s}_{{}_{\chi=-1,lm}} \Big( \widetilde{\varphi}\big|_{{}_{\mathscr{O}_{1,0,0,1}}} \Big)}
\,\, \eta_{{}_{\chi=-1}} \, a_{{}_{\chi=-1, lms}}^{\mathfrak{m} \, +} \, \eta_{{}_{\chi=-1}} 
\end{multline*}
\[
= A_{{}_{\chi}}^{e}(\varphi) + A_{{}_{\chi}}^{\mathfrak{m}}(\varphi),
\]
with the following kernel
\begin{multline*}\label{Achi=-1(x)=Aechi(x)+Amchi=-1(x)}
A_{{}_{\chi=-1}}(x)
=
\sum\limits_{lm}
f^{e}_{{}_{\chi=-1, lm}}(x)
\,\, a_{{}_{\chi=-1, lm}}^{e}
+ \sum\limits_{lm}
\overline{f^{e}_{{}_{\chi=-1,lm}} (x)}
\,\, a_{{}_{\chi=-1, lm}}^{e \, +} 
\\
+
\sum\limits_{lms}
f^{\mathfrak{m} \, s}_{{}_{\chi,lm}}(x)
\,\, a_{{}_{\chi=-1, lms}}^{\mathfrak{m}}
+ \sum\limits_{lms}
\overline{f^{\mathfrak{m} \, s}_{{}_{\chi=-1,lms}} (x)}
\,\, \eta_{{}_{\chi=-1}} \, a_{{}_{\chi=-1, lms}}^{\mathfrak{m} \, +} 
\, \eta_{{}_{\chi=-1}}
= A_{{}_{\chi=-1}}^{e}(x) + A_{{}_{\chi=-1}}^{\mathfrak{m}}(x),
\end{multline*}

Although the notation is self-evident, let us recall explicitly that the operators used in these formulas are equal
\[
\begin{split}
{a}_{{}_{\chi=-1, lm}}^{e} \coloneqq a'_{{}_{\chi=-1}}\big(F^{e}_{{}_{\chi,lm}} \big) = a^{e}_{{}_{\chi=-1}}\big(F^{e}_{{}_{\chi=-1,lm}} \big) \otimes \boldsymbol{1}, 
\\
{a}_{{}_{\chi=-1, lm}}^{e \, +} \coloneqq a'_{{}_{\chi=-1}}\big(F^{e}_{{}_{\chi=-1,lm}} \big)^+ = a^{e}_{{}_{\chi=-1}}\big(F^{e}_{{}_{\chi=-1,lm}} \big)^+ \otimes \boldsymbol{1}
\\
{a}_{{}_{\chi=-1, lms}}^{\mathfrak{m}} \coloneqq a'_{{}_{\chi=-1}}\big(F^{\mathfrak{m} \, s}_{{}_{\chi=-1,lm}} \big)  
= \boldsymbol{1} \otimes a^{\mathfrak{m}}_{{}_{\chi=-1}}\big(F^{\mathfrak{m} \, s}_{{}_{\chi=-1,lm}} \big),
\\
{a}_{{}_{\chi=-1, lms}}^{\mathfrak{m} \, +} \coloneqq a'_{{}_{\chi=-1}}\big(F^{\mathfrak{m} \, s}_{{}_{\chi=-1,lm}} \big)^+ 
= \boldsymbol{1} \otimes a^{\mathfrak{m}}_{{}_{\chi=-1}}\big(F^{\mathfrak{m} \, s}_{{}_{\chi=-1,lm}} \big)^+,
\end{split}
\]
\[
\textrm{on}
\,\, \Gamma\big({\mathcal{H}'}^{e}_{{}_{\chi = -1}}\big) \otimes \Gamma\big({\mathcal{H}'}^{\mathfrak{m}}_{{}_{\chi = -1}}\big)
= \Gamma(\mathcal{H}'_{{}_{\chi = -1}})
\]
and respect the canonical commutation relations
\begin{equation}\label{[achi=-1,achi=-1+]}
\begin{split}
\big[{a}_{{}_{\chi=-1, lm}}^{e} , {a}_{{}_{\chi=-1, l'm'}}^{e \, +}  \big]_{{}_{-}} = \delta_{{}_{l \, l'}} \delta_{{}_{m \, m'}},
\,\,\,\,\,\,\,\,\,\, \big[{a}_{{}_{\chi=-1, lm}}^{e} , {a}_{{}_{\chi=-1, l'm'}}^{e}  \big]_{{}_{-}} 
\\
=  \big[{a}_{{}_{\chi=-1, lm}}^{e \, +} , {a}_{{}_{\chi=-1, l'm'}}^{e \, +}  \big]_{{}_{-}} = 0,
\\
\big[{a}_{{}_{\chi=-1, lms}}^{\mathfrak{m}} , {a}_{{}_{\chi=-1, l'm's'}}^{\mathfrak{m} \, +}  \big]_{{}_{-}} 
= \delta_{{}_{l \, l'}} \delta_{{}_{m \, m'}} \delta_{{}_{s \, s'}},
\,\,\,\,\,\,\,\,\,\, \big[{a}_{{}_{\chi=-1, lms}}^{\mathfrak{m}} , {a}_{{}_{\chi=-1, l'm's'}}^{\mathfrak{m}}  \big]_{{}_{-}} 
\\
=  \big[{a}_{{}_{\chi=-1, lms}}^{\mathfrak{m} \, +} , {a}_{{}_{\chi=-1, l'm's'}}^{\mathfrak{m} \, +}  \big]_{{}_{-}} = 0,
\end{split}
\end{equation}
with all the remaining pairs of these operators commuting.

Here the distributions 
\[
\begin{split}
\varphi \longmapsto f^{e}_{{}_{\chi=-1,lm}}(\varphi) = \int f^{e \, \mu}_{{}_{\chi=-1,lm}}(x)\varphi_\mu(x) \, \ud^4 x  ,  
\\ 
\varphi \longmapsto f^{\mathfrak{m} \, s}_{{}_{\chi=-1,lm}}(\varphi) = \int f^{\mathfrak{m} \, s \, \mu}_{{}_{\chi=-1,lm}}(x)\varphi_\mu(x) \, \ud^4 x,
\end{split}
\]
(representable by ordinary functions $x\mapsto f^{e}_{{}_{\chi=-1,lm}}(x)$, $x\mapsto f^{\mathfrak{m} \, s}_{{}_{\chi=-1,lm}}(x)$)
are equal to the inverse Fourier transforms of the distributions
\[
\begin{split}
\widetilde{\varphi} \longmapsto F_{{}_{\chi=-1, lm}}^{e} \big( \widetilde{\varphi}\big|_{{}_{\mathscr{O}_{{}_{1,0,0,1}}}}\big)
=\int \limits_{\mathscr{O}_{{}_{1,0,0,1}}} F_{{}_{\chi=-1, lm}}^{e \, \mu} \widetilde{\varphi}_\mu\big|_{{}_{\mathscr{O}_{{}_{1,0,0,1}}}}
\,\, \ud \mu_{{}_{\mathscr{O}_{{}_{1,0,0,1}}}},
\\
\widetilde{\varphi} \longmapsto F_{{}_{\chi=-1, lm}}^{\mathfrak{m}} \big( \widetilde{\varphi}\big|_{{}_{\mathscr{O}_{{}_{1,0,0,1}}}}\big)
= \int \limits_{\mathscr{O}_{{}_{1,0,0,1}}} F_{{}_{\chi=-1, lm}}^{\mathfrak{m} \, s \, \mu} \widetilde{\varphi}_\mu\big|_{{}_{\mathscr{O}_{{}_{1,0,0,1}}}}
\,\, \ud \mu_{{}_{\mathscr{O}_{{}_{1,0,0,1}}}},
\end{split}
\] 
concentrated on the positive energy sheet $\mathscr{O}_{{}_{1,0,0,1}}$ of the cone in the momentum space.

\begin{center}
{\small $A_{{}_{\chi=-1}}$ AND THE PHASE $S$}
\end{center}

In particular, the electric type homogeneous of degree $-1$ part of the field $x^\mu A_\mu$ is equal 
\[
x^\mu A_{{}_{\chi=-1} \, \mu}^{e}(x)=
\sum\limits_{lm}
x^\mu f^{e}_{{}_{\chi=-1, lm} \, \mu}(x)
\,\, a_{{}_{\chi=-1, lm}}^{e}
+ \sum\limits_{lm}
\overline{x^\mu f^{e}_{{}_{\chi=-1,lm} \, \mu} (x)}
\,\, a_{{}_{\chi=-1, lm}}^{e \, +}. 
\]

The Fourier transform $F_{{}_{\chi=-1, lm} \, \mu}^{e}$ of  
$f^{e}_{{}_{\chi=-1,lm} \, \mu}$ is, by the formula (\ref{Felectric}), equal
\begin{equation}\label{F(felm)}
\mathscr{F}\big(f^{e}_{{}_{\chi=-1,lm} \, \mu}\big)(p) 
= F_{{}_{\chi=-1, lm} \, \mu}^{e}(p) = {\textstyle\frac{1}{\sqrt{l(l+1)}}} \frac{\partial Y_{lm}(p)}{\partial p^\mu}, 
\end{equation}
on $\mathscr{O}_{1,0,0,1}$ and is concentrated on the positive energy sheet 
\[
\mathscr{O}_{1,0,0,1} = \{p: p \cdot p = 0, p^0 >0\} 
\]
of the cone. Therefore, the Fourier transform   
of $x^\mu f^{e}_{{}_{\chi=-1,lm} \, \mu}$ is equal 
\begin{multline}\label{F(xfelm)}
\mathscr{F}\big(x^\mu f^{e}_{{}_{\chi=-1,lm} \, \mu}\big)(p)
= -i \frac{\partial}{\partial p_\mu} F_{{}_{\chi=-1, lm} \, \mu}^{e}(p) 
\\
= -i {\textstyle\frac{1}{\sqrt{l(l+1)}}} \frac{\partial}{\partial p_\mu}\frac{\partial Y_{lm}}{\partial p^\mu}(p)
 = -i {\textstyle\frac{1}{\sqrt{l(l+1)}}} \frac{\partial^2 Y_{lm}}{\partial p_\mu \partial p^\mu}(p). 
\end{multline}
on $\mathscr{O}_{1,0,0,1}$ and is concentrated on $\mathscr{O}_{1,0,0,1}$.

Comparing the formulas (\ref{F(felm)}) and (\ref{F(felm)}) with the formulas
(\ref{a(p)ASsolution}) and (\ref{F(f(+)lm)}) of the next Subsection we see that 
\[
x^\mu f^{e}_{{}_{\chi=-1,lm} \, \mu}(x) = 2\sqrt{\pi} f^{(+)}_{lm}(x), \,\,\, l=1,2, \ldots, \,\, -l \leq m \leq l,
\]
where $f^{(+)}_{lm}$ are the partial waves in the phase field $S$ of Staruszkiewicz theory, extended by homogeneity
all over the whole space-time, and which are present in the formula of axiom (II) of the next Subsection. 

Thus, we have the equality
\begin{equation}\label{xmuAchi=-1mu(x)}
x^\mu A_{{}_{\chi=-1} \, \mu}^{e}(x)=
\sum\limits_{lm}
f^{(+)}_{lm}(x) \,\, 2\sqrt{\pi}
\,\, a_{{}_{\chi=-1, lm}}^{e}
+ \sum\limits_{lm}
\overline{f^{(+)}_{lm}(x)} \,\,2\sqrt{\pi}
\,\, a_{{}_{\chi=-1, lm}}^{e \, +}.
\end{equation}
Recall that 
\[
x^\mu f^{e}_{{}_{\chi=-1,lm} \, \mu}(x) = 2\sqrt{\pi} \, f^{(+)}_{lm}(x), \,\,\, l=1,2, \ldots, \,\, -l \leq m \leq l,
\]
transform as scalars under the Lorentz group action of $SL(2, \mathbb{C})$, which, when restricted to the
single particle space ${\mathcal{H}'}^{e}_{{}_{\chi = -1}}$ of electric type states acts on the homogeneous of degree zero scalars representing the states
of ${\mathcal{H}'}^{e}_{{}_{\chi = -1}}$ in the ordinary fashion, \emph{i.e.} as the ordinary Lorentz group action on the scalars
on the cone with the inner product (\ref{e-type-invariat-inner-prod}). Easy computation (compare Subsection 
\ref{infra-electric-transversal-generalized-states}) shows that this representation is irreducible and equivalent to the representation $(l_0,l_1) = (1,0)$
of \cite{Geland-Minlos-Shapiro}, with the normalized states $F_{{}_{\chi=-1, lm}}^{e}$ given by (\ref{Felectric})
corresponding to the normalized states $\xi_{lm}$ of the representation $(l_0,l_1) = (1,0)$ in \cite{Geland-Minlos-Shapiro}. 
Recall, please, that the single particle normalized states $F_{{}_{\chi=-1, lm}}^{e} \in {\mathcal{H}'}^{e}_{{}_{\chi = -1}}$, regarded as the elements
of the Fock space
\begin{equation}\label{Gamma(Hch=-1)}
\Gamma\big({\mathcal{H}'}^{e}_{{}_{\chi = -1}}\big) \otimes \Gamma\big({\mathcal{H}'}^{\mathfrak{m}}_{{}_{\chi = -1}}\big)
= \Gamma(\mathcal{H}'_{{}_{\chi = -1}})
\end{equation}
are by construction equal
\[
F_{{}_{\chi=-1, lm}}^{e} = {a}_{{}_{\chi=-1, lm}}^{e \, +} |0\rangle = {a}_{{}_{\chi=-1, lm}}^{e \, +} |0\rangle^{e} \otimes |0\rangle^{\mathfrak{m}}, 
\]
where $|0\rangle$ is the (normalized) vacuum state in the Fock space (\ref{Gamma(Hch=-1)}),
and $|0\rangle^{e}$, $|0\rangle^{\mathfrak{m}}$, are, respectively, the vacuum states in the Fock spaces
\[
\Gamma\big({\mathcal{H}'}^{e}_{{}_{\chi = -1}}\big) \,\,\,\, \textrm{and} \,\,\,\, \Gamma\big({\mathcal{H}'}^{\mathfrak{m}}_{{}_{\chi = -1}}\big).
\]
Comparing the commutation rules (\ref{[achi=-1,achi=-1+]}) with the commutation rules
in the axiom (IV) of the next Subsection for the operators $c_{lm}, c_{lm}^{+}$ of the Staruszkiewicz theory
we see that the operators 
\[
c_{lm}, c_{lm}^{+}  \,\,\,\,\, \textrm{on the subspace} \, Q=0
\]
of the Staruszkiewicz theory can be equated with the operators
\[
\pm e 2\sqrt{\pi} \, {a}_{{}_{\chi=-1, lm}}^{e}, \,\, \pm e 2\sqrt{\pi} \, {a}_{{}_{\chi=-1, lm}}^{e \, +}
\]
\[
\textrm{on}
\,\, \Gamma\big({\mathcal{H}'}^{e}_{{}_{\chi = -1}}\big) \otimes |0\rangle^{\mathfrak{m}}
\subset \Gamma(\mathcal{H}'_{{}_{\chi = -1}}),
\]
where $|0\rangle^{\mathfrak{m}}$ is the vacuum state in the Fock space 
\[
\Gamma\big({\mathcal{H}'}^{\mathfrak{m}}_{{}_{\chi = -1}}\big)
\]
of the ``magnetic type'' states. Here $e$ is the constant, which is present in the commutation
relations in the axiom (IV) of Staruszkiewicz theory, Subsection \ref{infra-electric-transversal-generalized-states}.  
Comparing this with the results of \cite{Staruszkiewicz1995}, we see that the representation of $SL(2, \mathbb{C})$
acting on the single particle states spanned by 
\begin{equation}\label{e2sqrtpiae+|0>}
\pm e 2\sqrt{\pi} \, F_{{}_{\chi=-1, lm}}^{e} = 
\pm e 2\sqrt{\pi} \, {a}_{{}_{\chi=-1, lm}}^{e \, +} |0\rangle = \pm e 2\sqrt{\pi} \, {a}_{{}_{\chi=-1, lm}}^{e \, +} 
|0\rangle^{e} \otimes |0\rangle^{\mathfrak{m}}, 
\end{equation}
and on the subspace spanned by the states $c_{lm}^{+}|0\rangle$ are identical, or unitarily equivalent, with the unitary equivalence 
$\mathcal{U}$ which maps each state (\ref{e2sqrtpiae+|0>}) into the state $c_{lm}^{+}|0\rangle$ of Staruszkiewicz theory, and thus we arrive
at the equality
\begin{equation}\label{xAechi=-1(x)=transversalS}
\pm e  \,  x^\mu A_{{}_{\chi=-1} \, \mu}^{e}(x)\Bigg|_{{}_{\Gamma({\mathcal{H}'}^{e}_{{}_{\chi = -1}}) \otimes |0\rangle^{\mathfrak{m}}}}
= 
\Bigg[\sum\limits_{lm}
f^{(+)}_{lm}(x)
\,\, c_{lm}
+ \sum\limits_{lm}
\overline{f^{(+)}_{lm}(x)} 
\,\, c_{lm}^{+}\Bigg]_{{}_{Q=0}},
\end{equation}
which could also be understood as unitary equivalence, given by the unitary Fock lifting $V=\Gamma(\mathcal{U}) \otimes \boldsymbol{1}$.

Let $\phi \in \mathcal{S}^{00}(\mathbb{R}; \mathbb{C})$. Consider now the evaluation $x_\mu A_{{}_{\chi=-1}}^{\mu}(\phi)$ of the
field with the kernel $x_\mu A_{{}_{\chi=-1}}(x)$ on the test function $\phi$ (here we have contraction with respect to the
Lorentz index $\mu$, \emph{i.e.} with the Einstein summation convention with respect to the repeated index $\mu$):
\[
x_\mu A_{{}_{\chi=-1}}^{\mu}(\phi) \coloneqq 
\int A_{{}_{\chi=-1}}^{\mu}(x)x_\mu \phi(x) \, \ud^4 x.
\]
It follows from this formula, that 
\[
x_\mu A_{{}_{\chi=-1}}^{\mu}(\phi) = A_{{}_{\chi=-1}}(\varphi), \,\,\, \varphi_\mu = x_\mu \phi.
\]
Therefore 
\[
x_\mu A_{{}_{\chi=-1}}^{\mu}(\phi) =
a'_{{}_{\chi}}\Big(\overline{\check{\partial \widetilde{\phi}}\big|_{{}_{\mathscr{O}_{{}_{1,0,0,1}} \,\, \chi}}}\Big) 
+ \, \eta_{{}_{\chi}} a'_{{}_{\chi}}\Big(\partial \widetilde{\phi}\big|_{{}_{\mathscr{O}_{{}_{1,0,0,1}} \,\, \chi}}\Big)^+ \eta_{{}_{\chi}},
\]
where $\partial \widetilde{\phi}$ denotes the following four-vector test function
\[
(\partial \widetilde{\phi})_\mu \coloneqq \widetilde{\varphi}_\mu = \widetilde{x_\mu \phi} = -i \partial_{{}_{p^\mu}} \widetilde{\phi}.
\]
It, thus, follows the following formula
\begin{multline*}
x_\mu A_{{}_{\chi=-1}}^{\mu}(\phi) =
\sum\limits_{lm}
\overline{\check{F}^{e}_{{}_{\chi=-1,lm}} \big( \partial \widetilde{\phi}\big|_{{}_{\mathscr{O}_{1,0,0,1}}} \Big)}
\,\, a_{{}_{\chi=-1, lm}}^{e}
+ \sum\limits_{lm}
\overline{F^{e}_{{}_{\chi=-1,lm}} \Big( \partial \widetilde{\phi}\big|_{{}_{\mathscr{O}_{1,0,0,1}}} \Big)}
\,\, a_{{}_{\chi=-1, lm}}^{e \, +} 
\\
+
\sum\limits_{lms}
\overline{\check{F}^{\mathfrak{m} \, s}_{{}_{\chi=-1,lm}} \big( \partial \widetilde{\phi}\big|_{{}_{\mathscr{O}_{1,0,0,1}}} \Big)}
\,\, a_{{}_{\chi=-1, lms}}^{\mathfrak{m}}
+ \sum\limits_{lms}
\overline{F^{\mathfrak{m} \, s}_{{}_{\chi=-1,lm}} \Big( \partial \widetilde{\phi}\big|_{{}_{\mathscr{O}_{1,0,0,1}}} \Big)}
\,\, \eta_{{}_{\chi=-1}} \, a_{{}_{\chi=-1, lms}}^{\mathfrak{m} \, +} \, \eta_{{}_{\chi=-1}} 
\end{multline*}
\begin{multline*}
= a^{e}_{{}_{\chi}}\Big(P_{{}_{e}}\overline{\check{\partial\widetilde{\phi}}\big|_{{}_{\mathscr{O}_{{}_{1,0,0,1}} \,\, \chi}}}\Big) 
+a^{e}_{{}_{\chi}}\Big(P_{{}_{e}}\partial\widetilde{\phi}\big|_{{}_{\mathscr{O}_{{}_{1,0,0,1}} \,\, \chi}}\Big)^+
\\
+
a^{\mathfrak{m}}_{{}_{\chi}}\Big(P_{{}_{\mathfrak{m}}}\overline{\check{\partial\widetilde{\phi}}\big|_{{}_{\mathscr{O}_{{}_{1,0,0,1}} \,\, \chi}}}\Big) 
+\eta_{{}_{\chi=-1}} \, 
a^{\mathfrak{m}}_{{}_{\chi}}\Big(P_{{}_{\mathfrak{m}}}\partial\widetilde{\phi}\big|_{{}_{\mathscr{O}_{{}_{1,0,0,1}} \,\, \chi}}\Big)^+
\, \eta_{{}_{\chi=-1}},
\end{multline*}
\[
= x_\mu A_{{}_{\chi=-1}}^{e \, \mu}(\phi) + x_\mu A_{{}_{\chi=-1}}^{\mathfrak{m} \, \mu}(\phi).
\]
On the other hand, by the very construction of the states $F^{e}_{{}_{\chi=-1,lm}}$ $\in {\mathcal{H}'}^{e}_{{}_{\chi = -1}}$
and $F^{\mathfrak{m} \, s}_{{}_{\chi=-1,lm}}$ $\in {\mathcal{H}'}^{\mathfrak{m}}_{{}_{\chi = -1}}$, 
it follows that for any test function 
of gradient type $\partial \widetilde{\phi}$ 
\begin{equation}\label{Pm(partialtildephi)=0}
\begin{split}
P_{{}_{e}}  \partial \widetilde{\phi}\big|_{{}_{\mathscr{O}_{1,0,0,1} \, \chi=-1 }} 
=  \partial \widetilde{\phi}\big|_{{}_{\mathscr{O}_{1,0,0,1} \, \chi=-1}} + \textrm{longitudinal admixture},
\\
P_{{}_{\mathfrak{m}}} \partial \widetilde{\phi}\big|_{{}_{\mathscr{O}_{1,0,0,1} \, \chi=-1}} 
= 0 + \textrm{longitudinal admixture}.
\end{split}
\end{equation}
Here ``longitudinal admixture'' denotes a state which is a linear combination of the two eigenvectors $w_{{}_{r^{-2}}}, w_{{}_{r^{2}}}$,
of zero Krein norm. 
We arrive, thus, at the following formula
\begin{equation}\label{xA=xAemodL0}
x_\mu A_{{}_{\chi=-1}}^{\mu}(\phi)\Bigg|_{{}_{\Gamma({\mathcal{H}'}^{e}_{{}_{\chi = -1}}) \otimes |0\rangle^{\mathfrak{m}} }}
= x_\mu A_{{}_{\chi=-1}}^{e \, \mu}(\phi)\Bigg|_{{}_{\Gamma({\mathcal{H}'}^{e}_{{}_{\chi = -1}}) \otimes |0\rangle^{\mathfrak{m}} }}
\,\, \textrm{mod} \, L_{0}. 
\end{equation}
Here $L_{0} \subset \Gamma({\mathcal{H}'}_{{}_{\chi = -1}})$ is the Lorenz invariant subspaces
spanned by the longitudinal states of zero Krein norm. 
This formula tells us that the part $A_{{}_{\chi=-1}}^{\mathfrak{m}}$ of the field $A_{{}_{\chi=-1}}$, in action on the Lorentz 
invariant subspace  $\Gamma({\mathcal{H}'}^{e}_{{}_{\chi = -1}}) \otimes|0\rangle^{\mathfrak{m}}$,
has the purely longitudinal contribution $x_\mu A_{{}_{\chi=-1}}^{\mathfrak{m} \, \mu}$ to  the Lorentz contraction  
$x_\mu A_{{}_{\chi=-1}}^{\mu}$. It follows in particular that the equality (\ref{xAechi=-1(x)=transversalS}) can be rewritten
in the form
\begin{equation}\label{xAchi=-1(x)=transversalS}
\pm e  \,  x^\mu A_{{}_{\chi=-1} \, \mu}(x)\Bigg|_{{}_{\Gamma({\mathcal{H}'}^{e}_{{}_{\chi = -1}}) \otimes |0\rangle^{\mathfrak{m}} }}
\,\,\, \textrm{mod} \, L_{0}
= 
\Bigg[\sum\limits_{lm}
f^{(+)}_{lm}(x)
\,\, c_{lm}
+ \sum\limits_{lm}
\overline{f^{(+)}_{lm}(x)} 
\,\, c_{lm}^{+}\Bigg]_{{}_{Q=0}},
\end{equation}

Let us consider the homogeneous of degree $-2$ part of the free electromagnetic tensor field $F$ and its dual $F^{*}$, 
computed for the homogeneous of degree $-1$ part $A_{{}_{\chi=-1}}^{e}$ or $A_{{}_{\chi=-1} \, \mu}^{\mathfrak{m}}$ 
of the free potential $A$. This allows us to express the invariant decomposition
$A_{{}_{\chi=-1}} = A_{{}_{\chi=-1}}^{e}+ A_{{}_{\chi=-1} \, \mu}^{\mathfrak{m}}$  in therms 
of the tensors $F$ and $F^{*}$, in the form analogous to that of Alexander and Bergmann \cite{Bergmann}
for the classical homogeneous of degree $-2$ electromagnetic tensor field $F$. 
Namely, using homogeneity it follows from (\ref{Pm(partialtildephi)=0}), or equivalently from (\ref{xA=xAemodL0}),
that
\[
\begin{split}
x^\mu\big[ 
\partial_\mu A_{{}_{\chi=-1} \, \nu} -\partial_\nu A_{{}_{\chi=-1} \, \mu} 
\big]\Bigg|_{{}_{\Gamma({\mathcal{H}'}^{e}_{{}_{\chi = -1}}) \otimes |0\rangle^{\mathfrak{m}} }}
= 
\partial_\nu\big[
x^\mu A_{{}_{\chi=-1} \, \mu}^{e} 
\big]\Bigg|_{{}_{\Gamma({\mathcal{H}'}^{e}_{{}_{\chi = -1}}) \otimes |0\rangle^{\mathfrak{m}} }}
\,\,\, \textrm{mod} \, L_{0},
\\
x^\mu\big[ 
\partial_\mu A_{{}_{\chi=-1} \, \nu}^{\mathfrak{m}} -\partial_\nu A_{{}_{\chi=-1} \, \mu}^{\mathfrak{m}}  
\big]\Bigg|_{{}_{\Gamma({\mathcal{H}'}^{e}_{{}_{\chi = -1}}) \otimes |0\rangle^{\mathfrak{m}} }}
= 
0
\,\,\, \textrm{mod} \, L_{0},
\\
x^\mu\big[ 
\partial_\mu A_{{}_{\chi=-1} \, \nu}^{e} -\partial_\mu A_{{}_{\chi=-1} \, \nu}^{e} 
\big]
=
\partial_\nu\big[
x^\mu A_{{}_{\chi=-1} \, \mu}^{e} 
\big],
\end{split}
\]
and 
\[
\begin{split}
{\textstyle\frac{1}{2}} \epsilon^{\mu\nu\rho\sigma}
x_\mu\big[ 
\partial_\rho A_{{}_{\chi=-1} \, \sigma}^{\mathfrak{m}} -\partial_\sigma A_{{}_{\chi=-1} \, \rho}^{\mathfrak{m}}  
\big] \neq 0,
\\
{\textstyle\frac{1}{2}} \epsilon^{\mu\nu\rho\sigma}
x_\mu\big[ 
\partial_\rho A_{{}_{\chi=-1} \, \sigma}^{e} -\partial_\sigma A_{{}_{\chi=-1} \, \rho}^{e}  
\big] = 0.
\end{split}
\]

Finally, let us note, that the scalar (homogeneous of degree zero on the space-time test functions $\phi\in \mathcal{S}^{00}(\mathbb{R}^4;\mathbb{C})$) 
field 
\[
x_\mu A_{{}_{\chi}}^{e \, \mu}(\phi)  
= a^{e}_{{}_{\chi}}\Big(P_{{}_{e}}\overline{\check{\partial\widetilde{\phi}}\big|_{{}_{\mathscr{O}_{{}_{1,0,0,1}} \,\, \chi}}}\Big) 
+a^{e}_{{}_{\chi}}\Big(P_{{}_{e}}\partial\widetilde{\phi}\big|_{{}_{\mathscr{O}_{{}_{1,0,0,1}} \,\, \chi}}\Big)^+,
\]
regarded as acting in the Fock space 
\[
\Gamma\big({\mathcal{H}'}^{e}_{{}_{\chi = -1}}\big) \otimes |0\rangle^{\mathfrak{m}},
\]
is equal to the free field over the single particle Hilbert space of homogeneous states (distributions)
which are spanned by the following orthonormal system of states (distributions)
\begin{multline*}
\widetilde{\phi}
\longmapsto  F^{e}_{{}_{\chi=-1,lm}} \Big( \partial \widetilde{\phi}\big|_{{}_{\mathscr{O}_{1,0,0,1}}} \Big)
= \int\limits_{\mathscr{O}_{{}_{1,0,0,1}}} F^{e \, \mu}_{{}_{\chi=-1,lm}}(p) 
(-i)\partial_{{}_{p^\mu}}\widetilde{\phi}\big|_{{}_{\mathscr{O}_{{}_{1,0,0,1}}}}
\\
=
\int\limits_{\mathscr{O}_{{}_{1,0,0,1}}} (-i)\partial_{{}_{p^\mu}}F^{e \, \mu}_{{}_{\chi=-1,lm}}(p) 
\widetilde{\phi}\big|_{{}_{\mathscr{O}_{{}_{1,0,0,1}}}}
\end{multline*}
concentrated on the cone $\mathscr{O}_{{}_{1,0,0,1}}$, and represented by the functions
\[
-i\partial_{{}_{p^\mu}}F^{e \, \mu}_{{}_{\chi=-1,lm}} = -i {\textstyle\frac{1}{\sqrt{l(l+1)}}} 
{\textstyle\frac{\partial^2}{\partial p_\mu \partial p^\mu}} Y_{{}_{lm}}
\]
on the cone which are homogeneous of degree $-2$. 
Therefore, the scalar field $\phi \mapsto x_\mu A_{{}_{\chi}}^{e \, \mu}(\phi)$, regarded as a field
on the Fock space regarded as acting in the Fock space 
\[
\Gamma\big({\mathcal{H}'}^{e}_{{}_{\chi = -1}}\big) \otimes |0\rangle^{\mathfrak{m}},
\]
is equal to the free field over the single particle Hilbert space of homogeneous
of degree $-2$ states of the form
\[
{\textstyle\frac{\partial^2}{\partial p_\mu \partial p^\mu}} \widetilde{f}
\]
uniquely determined by the corresponding homogeneous of degree zero scalar functions $\widetilde{f}$, 
with the invariant scalar product (\ref{e-type-invariat-inner-prod}) expressed in therms of the 
corresponding homogeneous of degree zero scalar functions $\widetilde{f}$. The inverse Fourier transform
of the states (distributions) $-i\partial_{{}_{p^\mu}}F^{e \, \mu}_{{}_{\chi=-1,lm}}$
coincide (up to an irrelevant constant) with the partial waves $f^{(+)}_{lm}$ of Staruszkiewicz theory and span the irreducible
representation $(l_0,l_1) = (1,0)$ of the $SL(2, \mathbb{C})$ group. 
The transversal part of the phase field $S$ of Staruszkiewicz theory, restricted to
the invariant zero charge subspace $Q=0$ is, accordingly to \cite{Staruszkiewicz1995}, equal to the free
field over the Fock space constructed over the same single particle Hilbert space. Thus, we again
arrive at the equality (\ref{xAechi=-1(x)=transversalS}), or equivalently, 
the equality (\ref{xAchi=-1(x)=transversalS}).

\begin{center}
{\small POSITIVITY OF AN INVARIANT KERNEL}
\end{center}

Now we construct a continuous and invariant kernel $\langle \cdot|\cdot\rangle$ 
on the Lobachevsky space $\mathscr{L}_3$,
which is of considerable importance for Staruszkiewicz theory. We give a proof of its positivity. 
In order to achieve this result we use the linear subspace $L[\mathfrak{F}_{\chi=1}]$ 
of states homogeneous of degree $-1$. This subspace $L[\mathfrak{F}_{\chi=1}]$ is useless in construction of single particle space of any homogeneous part 
of the free field $A_\mu$, but useful for giving very simple proof of the positivity of the said kernel.
 
For this reason consider now the specific homogeneous of degree $-1$ state (Fourier transform
of the Dirac homogeneous of degree $-1$ solution restricted to the positive energy sheet of the cone)
of the form
\begin{equation}\label{tildefu}
\widetilde{f}^{|u\rangle}_{\mu}(p) = \frac{u_\mu}{u \cdot p}
\end{equation}
with a fixed unit time like vector $u$ in the Lobachevsky space. 
Then construct the linear span $L[\mathfrak{F}_{\chi=1}]$ of all such (\ref{tildefu}) with $u$
ranging over the Lobachevsky space $\mathscr{L}_3$ of unit time like vectors $u$, \emph{i.e} such that $u\cdot u =1$.
In other words we consider the space $L[\mathfrak{F}_{\chi=1}]$
spanned by all Lorentz   
transforms
\[
\begin{array}{cc}
\Lambda(g)^{-1} \tilde{f}^{|u\rangle}(\Lambda(g)p) =
\tilde{f}^{|u'\rangle}(p) = \frac{u'}{u' \cdot p}, \,\,\,
u' = \Lambda(g)^{-1} u,
\\
g \in SL(2, \mathbb{C}),
\end{array}
\]
of one single  state of the form $\tilde{f}^{|u\rangle}$. Here, similarly as in Sections \ref{constr-of-VF}, \ref{e+e-}, \ref{free-gamma}, 
we have used the natural antihomomorphism $g \rightarrow \Lambda(g)$ from the $SL(2, \mathbb{C})$ into the group of Lorentz transformations. 

Note that this Lorentz transformation is induced by the linear dual of the (conjugated) {\L}opusza\'nski transformation acting in the Fock space of the quantum field $A_\mu$. Indeed it easily follows by the formula for the pairing
between the test space $E_\mathbb{C} = \mathcal{S}^0(\mathscr{O}_{\pm 1,0,0,1};\mathbb{C}^4) = \mathcal{S}_{A^{(3)}}(\mathbb{R}^3; \mathbb{C}^4)$ and its dual $E_{\mathbb{C}}^{*}$. Namely we put the natural formula 
(\ref{Kr-inn-Lop-1-space}) for the invariant pairing
\begin{multline}\label{pairing-formula}
(\tilde{f}, \widetilde{\varphi})_{{}_{\textrm{pairing}}} = (\tilde{f}, \mathfrak{J}'\widetilde{\varphi})
= - \int \limits_{\mathscr{O}_{\pm 1,0,0,1}} \tilde{f}^\mu (p) \widetilde{\varphi}_\mu (p) \, \ud \mu_{{}_{\mathscr{O}_{\pm 1,0,0,1}}}(p) \\
= \int \limits_{\mathscr{O}_{\pm 1,0,0,1}} \big( \tilde{f} (p), \mathfrak{J}_{{}_{\bar{p}}} 
\widetilde{\varphi} (p) \big)_{{}_{\mathbb{C}^4}} \, \ud \mu_{{}_{\mathscr{O}_{\pm 1,0,0,1}}}(p).
\end{multline}

The space $L[\mathfrak{F}_{\chi=1}]$ contains the transversal electric type homogeneous states (respectively homogeneous of degree $-1$
solutions of d'Alembert equation) of the form
(\ref{SolMaxwellHom=-1}) with
\[
\sum \limits_{i}^{N} \alpha_i = 0
\]
along with the longitudinal solutions of the form (\ref{SolMaxwellHom=-1}) with
\[
\sum \limits_{i} \alpha_j \neq 0.
\]

The space of invariant kernels $\langle \cdot | \cdot \rangle$ on $\mathscr{L}_3$ is rather reach. 
However, in this particular case of positive definite kernels on the Lobachevsky space 
$\mathscr{L}_3 \cong SL(2, \mathbb{C})/SU(2, \mathbb{C})$ acted on by $SL(2, \mathbb{C})$
the invariant kernels are fully classified, compare e.g. \cite{Gangolli}.
The manifold $\mathscr{L}_3$ can also be realized by $2\times 2$ Hermitian complex matrices $\widehat{u} = \sigma_\mu u^\mu$, where 
$\sigma_0 = \boldsymbol{1}_{{}_{2}}$ and $\sigma_i$, $i=1$,$2$,$3$, are the Pauli matrices and $u\cdot u =1$. Next we consider the smooth 
left action $SL(2, \mathbb{C}) \times \mathscr{L}_3 \ni g \times \widehat{u} \rightarrow g \cdot \widehat{u} \in \mathscr{L}_3$
of  $SL(2, \mathbb{C})$ on $\mathscr{L}_3$ defined by the formula
\[
g \cdot \widehat{u} = g \widehat{u} g^* = \widehat{\Lambda(g^{-1})u}, \,\,\,\,\,\,\,\,\,
g \in SL(2, \mathbb{C}), \widehat{u} \in \mathscr{L}_3.
\]
Then  $\mathscr{L}_3$ is equal to the orbit $\mathscr{O}_{{}_{\widehat{u}}} = \{g \cdot \widehat{u}, g \in  SL(2, \mathbb{C}) \}$ of
the point $\widehat{u}$ with $u = (1,0,0,0)$. The isotropy group at the point $\widehat{u}$ with $u = (1,0,0,0)$, is equal
to the maximal compact subgroup  $K = SU(2, \mathbb{C})$ of $SL(2, \mathbb{C})$. Therefore, $\mathscr{L}_3 = \mathscr{O}_{{}_{\widehat{u}}}$,
$u = (1,0,0,0)$, is diffeomorphic to the space $\mathscr{L}_3 \cong SL(2, \mathbb{C})/SU(2, \mathbb{C})$ of left cosets
$qK$, $q \in SL(2, \mathbb{C})$, with the above left action, which coincides, under this diffeomorphism, with the ordinary left action
$g\times qK \rightarrow gqK$ on the left cosets $qK$. To the coset $eK$ of the identity element $e \in SL(2, \mathbb{C})$
there corresponds the point $\widehat{u}$, $u = (1,0,0,0)$, invariant under $SU(2, \mathbb{C})$. This left action of $G$
on $G/K$ is transitive.
Choosing various invariant positive definite kernels $\langle \cdot | \cdot \rangle$ on $\mathscr{L}_3$ we achieve in this way various cyclic spherical unitary 
representations $\mathbb{U}$ of the $SL(2, \mathbb{C})$ group
on the completion of $L[\mathfrak{F}_{\chi=1}]$ with respect to the inner product defined by the invariant kernel and with the unit cyclic vector
which is invariant under the action of the maximal compact subgroup $K$.  
Let $K= SU(2, \mathbb{C})$ be the maximal compact subgroup of 
$SL(2, \mathbb{C})$. Recall, that a unitary representation $U$ of $SL(2, \mathbb{C})$ is called $K$-spherical (or merely spherical) if the decomposition
of the restriction of $U$ to $K$ contains the trivial representation $k \rightarrow 1$ of $K$. Equivalently $U$
is spherical whenever there is a unit vector $v \in H_{U}$ such that $U_k v = v$ for all $k \in K$.
Then in particular it follows by the classification results (or the theory of spherical functions
and generalization of Bochner's theorem for semi-simple Lie groups, in particular for $SL(2, \mathbb{C})$ group, 
worked out by Krein, Naimark and Gelfand, compare \S 31.10 of \cite{Neumark_dec} or \cite{Gangolli}), that each unitary cyclic and spherical representation $\mathbb{U}$ of $SL(2, \mathbb{C})$, with unit cyclic $v$ such that
$\mathbb{U}(k)v=v$ for all $k\in K$, can be reached by the respective choice of the invariant kernel on the Lobachevksy space, or to each such representation there exists the corresponding invariant kernel and this correspondence 
is bi-unique in the sense that two such representations with the same kernel are unitary equivalent if and only if the corresponding kernels are equal. 

It follows that the most general representation $\mathbb{U}$ which can be achieved in this way
has the general form (\cite{Gangolli})
\begin{equation}\label{BochnerKernelDecomposition}
\mathbb{U} = \int \limits_{\mathbb{R}_+} \mathfrak{S}(m = 0, \rho) \, \ud \rho \oplus \int \limits_{[0,1] \subset \mathbb{R}} \mathfrak{D}(\nu) \, \ud \nu
\end{equation}
where $\mathfrak{S}(m , \rho)$ is the irreducible representation of the principal series denoted by the pair
$(l_0 = \frac{m}{2}, l_1 = \frac{i\rho}{2})$, with $m \in \mathbb{Z}$
and $\rho \in \mathbb{R}$ in the notation of the book \cite{Geland-Minlos-Shapiro}, and correspond to the characters
$\chi = (n_1, n_2) = \big(\frac{m}{2} + \frac{i\rho}{2}, - \frac{m}{2} + \frac{i\rho}{2}\big)$ in the notation
of the book \cite{GelfandV}. Here $\mathfrak{D}(\nu)$
are the irreducible unitary representations of the supplementary series denoted by the pair
$(l_0 = 0, l_1 = \nu)$ in the notation of the book \cite{Geland-Minlos-Shapiro}, and correspond to the character
$\chi = (n_1, n_2) = \big(\nu, \nu)$ in the notation
of the book \cite{GelfandV} with \footnote{In the notation 
of \cite{nai1}-\cite{nai3} the parameter $\nu$ numbering the supplementary sries $\mathfrak{D}(\nu)$
is twice as ours $\nu$ and ranges over the interval $(0,2)$.} the real parameter $\nu \in (0,1)$. Finally 
$\ud \rho$ and $\ud \nu$ are arbitrary nonnegative $\sigma$-measures 
on the positive reals $\mathbb{R}_+$ and on the interval $[0,1] \subset \mathbb{R}$ respectively. 

Therefore the classification of positive definite invariant kernels on the 
Lobachevsky 
space $\mathscr{L}_3$
 is reduced to the
classification of the equivalence classes (with equivalence meaning the mutual absolute continuity) 
of nonnegative $\sigma$-measures on $\mathbb{R}_+$ and on the interval $[0,1]$, compare
e.g. in \cite{Gangolli}, and it is difficult in each particular case, to 
give explicit form to the possible kernels, compare e.g. the example of positive definite kernels 
on the Lobachevsky plane $\mathscr{L}_2$ $=  SL(2, \mathbb{R})/SO(2)$ invariant under $SL(2, \mathbb{R})$
in \cite{Gangolli}. The problem is that the set of equivalence classes of such $\sigma$-measures
is enormous, and the invariant kernel, corresponding to a particular  $\sigma$-measure is in general difficult to handle in explicit
form. In particular the continuous singular $\sigma$-measures with effective support $[0,1]$ are well known, and there are at least continuum many
equivalence classes of such $\sigma$-measures.
In particular the classification results, as given
in \cite{Gangolli},  when applied to the case $\mathscr{L}_3 =  SL(2, \mathbb{C})/SU(2)$, are difficult to handle in obtaining explicit formulas 
for the invariant kernels on $\mathscr{L}_3$. 
Therefore, we prefer to construct the required invariant kernel on $\mathscr{L}_3$, which is of particular importance, 
with the help of the Hermitian form (\ref{krein-prod-infra-red-1}), and prove its positivity with the help of the Schoenberg theorem.

Recall that for
two points $u, v$ of the Lobachevsky space we have
\[
(\tilde{f}^{|u\rangle},\tilde{f}^{|v\rangle})_{{}_{\mathfrak{J}}} = -4\pi \lambda \textrm{coth} \lambda,
\]
where $\lambda$ is the hyperbolic angle between $u$ and $v$: $\textrm{cosh} \, \lambda = u\cdot v$, 
compare \cite{Staruszkiewicz1981}. The Hermitian bilinear invariant form (\ref{krein-prod-infra-red-1}) is not positive definite on the linear space $L[\mathfrak{F}_{\chi=1}]$ of states spanned by the states 
$\tilde{f}^{|u'\rangle}$ of the form (\ref{tildefu}) with $u'$
ranging over the Lobachevsky space $\mathscr{L}_3$. Nonetheless it defines (after addition of the constant term
$4\pi$ and changing the sign) the ``polarization'' of a L\'evy-Schoenberg kernel on the Lobachevsky space
$\mathscr{L}_3 =  SL(2, \mathbb{C})/K = SL(2, \mathbb{C})/SU(2, \mathbb{C})$ (we are using the terminology of 
\cite{Gangolli}).
Namely the kernel 
\[
u \times v \mapsto -((\tilde{f}^{|u\rangle},\tilde{f}^{|v\rangle})_{{}_{\mathfrak{J}}} + 4\pi)
\]
on $\mathscr{L}_3 = SL(2, \mathbb{C})/SU(2, \mathbb{C})$ preserves the conditions 
(2.16)-(2.19) of \cite{Gangolli}. 
In particular (2.19) of \cite{Gangolli} means in our case that for each positive real number $t$ 
\begin{equation}\label{kernel-t}
u \times v \mapsto \langle u|v\rangle_t = e^{t((\tilde{f}^{|u\rangle},\tilde{f}^{|v\rangle})_{{}_{\mathfrak{J}}} + 4\pi)}
=
e^{-t4\pi(\lambda \textrm{coth}\lambda -1)}
\end{equation} 
is an invariant positive definite kernel 
on the Lobachevsky space  and thus defines positive definite 
and invariant inner product on the linear space $S$ spanned by $\tilde{f}^{|u\rangle}$
and all its Lorentz transforms $\tilde{f}^{|u'\rangle}$ defined by (\ref{tildefu}) with $u'$ ranging over 
the Lobachevsky space. Here $\lambda$ is the hyperbolic angle between $u$ and $v$.

Indeed that the conditions (2.16)-(2.18) of \cite{Gangolli} are preserved is immediate. We need only show that 
(2.19) of \cite{Gangolli} is preserved, i.e. that the kernel (\ref{kernel-t})
is positive definite. But in order to see this note that
\[
\Big( \sum \limits_{i} \alpha_i \tilde{f}^{|u_i \rangle}, \sum \limits_{j} \alpha_j \tilde{f}^{|u_j\rangle} 
\Big)_{{}_{\mathfrak{J}}} \geq 0
\]
whenever 
\[
\sum \limits_{i} \alpha_i =0
\]
for $\tilde{f}^{|u\rangle}$ defined by (\ref{tildefu}), as we have already shown that the bilinear form 
$(\cdot,\cdot)_{{}_{\mathfrak{J}}}$ is positive definite on the linear space of electric type transversal states
(\ref{SolMaxwellHom=-1}), compare the preceding Proposition.
This means that the function
\[
u \times v \mapsto -((\tilde{f}^{|u\rangle},\tilde{f}^{|v\rangle})_{{}_{\mathfrak{J}}} + 4\pi)
\]
is a conditionally negative definite kernel on the Lobachevsky space in the sense of Schoenberg
\cite{Schoenberg}, compare also \cite{PaulsenRaghupathi} \S 9.1. Thus by the classical result 
of Schoenberg \cite{Schoenberg}  (compare e. g. also
\cite{PaulsenRaghupathi} \S 9.1, Theorem 9.7) 
\[
u \times v \mapsto \langle u|v\rangle_t = e^{t((\tilde{f}^{|u\rangle},\tilde{f}^{|v\rangle})_{{}_{\mathfrak{J}}} + 4\pi)}
=
e^{-t4\pi(\lambda \textrm{coth}\lambda -1)}
\]
is a positive definite kernel on the Lobachevsky space for all positive $t$. Its invariance follows from the invariance of the bilinear form $(\cdot,\cdot)_{{}_{\mathfrak{J}}}$ and the transformation rule for 
$\tilde{f}^{|u\rangle}$ defined by (\ref{tildefu}). 

This positivity result is of particular importance in the theory of Staruszkiewicz, so we state it as 
a separate
\begin{prop*}
For each positive real number $t$ the function 
\[
u \times v \mapsto \langle u|v\rangle_t = e^{t((\tilde{f}^{|u\rangle},\tilde{f}^{|v\rangle})_{{}_{\mathfrak{J}}} + 4\pi)}
=
e^{-t4\pi(\lambda \textrm{coth}\lambda -1)}
\]
defines a positive definite invariant kernel on the Lobachevsky space $\mathscr{L}_3$
of unit time like vectors $u$. Here $\lambda$ is the hyperbolic angle between $u$ and $v$
in $\mathscr{L}_3$. 
\end{prop*}

We choose $u \times v \mapsto \langle u|v\rangle_t$ as the invariant kernel defining the inner product
$\langle \cdot | \cdot \rangle_{{}_{\mathfrak{J}, t}}$ on the 
linear space $L[\mathfrak{F}_{\chi=1}]$ of states spanned by  $\tilde{f}^{|u'\rangle}$
defined by (\ref{tildefu}), with $u'$ ranging over the Lobachevsky space, by the formula
\begin{equation}\label{krein-prod-infra-red-long}
\Big\langle \sum \limits_{i=1}^{m} \alpha_i \tilde{f}^{|u_i\rangle} \Big| 
\sum \limits_{i=1}^{m} \beta_j \tilde{f}^{|v_j\rangle} \Big\rangle_{{}_{\mathfrak{J}, t}}
= \sum \limits_{i,j=1}^{m} \overline{\alpha_i} \beta_j \langle u_i | v_j \rangle_t,
\end{equation} 
and let us define the Hilbert space completion $\mathcal{H}_{t} \nsubseteq E^{*}$ of it.
Then we recover the unitary representation $\mathbb{U}^{t}$ of the $SL(2, \mathbb{C})$ group which the 
action of the dual of the (conjugate) of the {\L}opusza\'nski representation induces on the  linear
space  $L[\mathfrak{F}_{\chi=1}]$  of states and its Hilbert space completion $\mathcal{H}_{t}$.

Indeed by comparing this construction with the result  of \cite{Staruszkiewicz1992ERRATUM} 
and \cite{Staruszkiewicz2009} we obtain the following formula
\begin{equation}\label{decASlong}
\mathbb{U}^{t} = 
\left\{ \begin{array}{lll}
\mathfrak{D}(\nu_0) \bigoplus \int \limits_{\rho>0} \mathfrak{S}(m=0, \rho) \ud \rho,
& \nu_0 = 1 - 4 \pi t, & \textrm{if} \, 0 < 4 \pi t <1  \\
\int \limits_{\rho>0} \mathfrak{S}(m=0, \rho) \, \ud \rho, & &
\textrm{if} \, 1 < 4 \pi t, 
\end{array} \right.
\end{equation}
where $\ud \rho$ is the ordinary Lebesgue measure on $\mathbb{R}_+$, compare
(\ref{decompositionAS}).

\begin{rem*}
The Hilbert space completion $\mathcal{H}_{t}$ of the linear space $L[\mathfrak{F}_{\chi=1}]$
 generated by states
of the form (\ref{tildefu}) with $u \in \mathscr{L}_3$, with respect to the 
inner product (\ref{krein-prod-infra-red-long}) 
$\langle \cdot | \cdot \rangle_{{}_{\mathfrak{J}, t}}$,
generated by the kernel (\ref{kernel-t}), is not contained in  
$E^{*}$ nor in any of its natural quotient spaces. 
Therefore, $\mathcal{H}_{t}$ cannot serve as a single particle Hilbert space of any homogeneous part of the 
free electromagnetic potential field $A_\mu$.  
\end{rem*}

\qedsymbol \, 
We find a sequence in
$L[\mathfrak{F}_{\chi=1}]$
the elements of which regarded as linear functionals in 
$E^{*}$
do not converge weakly in $E^{*}$
although they converge in the norm of the inner product (\ref{krein-prod-infra-red-long}). 

Namely consider the sequence of partial sums of the series
\begin{equation}\label{a-counter-example-sequence}
\sum \limits_{k \in \mathbb{N}} \frac{1}{k} \, \tilde{f}^{|u_k \rangle}
\end{equation}
with $\tilde{f}^{|u \rangle}$ of the form (\ref{tildefu}).
We choose the unit time-like vectors $u_i \in \mathscr{L}_3$ as images of a fixed $u_1$
under the Lorentz transforms in a fixed plane (say $0-3$ plane) with the hyperbolic angles
$\lambda_{{}_{k+1 \,\,\, k}}$, where $u_{k+1} \cdot u_k = \cosh \lambda_{{}_{k+1 \,\,\, k}}$ 
between neighboring $u_{k}$ and $u_{k+1}$ growing sufficiently fast
as $k \in \mathbb{N}$ tends to infinity. 
Note that for the inner product 
$\big\langle \tilde{f}^{|u\rangle}, \tilde{f}^{|v\rangle} \big\rangle_{{}_{\mathfrak{J}, t}}$ we have the following formula
\[
\big\langle \tilde{f}^{|u\rangle}, \tilde{f}^{|v\rangle} \big\rangle_{{}_{\mathfrak{J}, t}}
= e^{-4\pi t(\lambda_{{}_{u \,\,\, v}} \textrm{coth} \lambda_{{}_{u \,\,\, v}} -1)}
\]
so that asymptotically, i.e. for large $\lambda_{{}_{u \,\,\, v}}$ hyperbolic angle between
$u$ and $v$ in $\mathscr{L}_3$ it decreases exponentially together with 
$\lambda_{{}_{u \,\,\, v}}$ going to infinity
\[
\big\langle \tilde{f}^{|u\rangle}, \tilde{f}^{|v\rangle} \big\rangle_{{}_{\mathfrak{J}, t}}
\sim 
e^{-4\pi t \, \lambda_{{}_{u \,\,\, v}}}.
\]
Therefore, it is easily seen that the sequence $u_i = (\sqrt{1 + |\boldsymbol{u}_i|^2}, \boldsymbol{u}_i) 
\in \mathscr{L}_3$ with all $\boldsymbol{u}_i$ having the same direction may be so chosen
that the series (\ref{a-counter-example-sequence}) is convergent with respect to the norm
$\| \cdot \|_{{}_{\mathfrak{J}, t}}$ defined by (\ref{krein-prod-infra-red-long}).

On the other hand consider $\tilde{f}^{|u \rangle}$ defined by (\ref{tildefu})
as a functional on  $E = \mathcal{S}_{A^{(3)}}(\mathbb{R}^3; \mathbb{C}^4)$ 
evaluated on the test function
$\widetilde{\varphi} \in E$
of the form $\widetilde{\varphi}(r,\theta,\phi) = v \, q(r)$,
with some fixed $v \in \mathscr{L}_3$ and $q \in \mathcal{S}^{0}(\mathbb{R})$. 
It is easy to see that in this case 
for the values
$\big(\tilde{f}^{|u_k \rangle}, \widetilde{\varphi} \big)_{{}_{\textrm{pairing}}}$ 
of the functional $\tilde{f}^{|u_k \rangle}$ for the unit four vectors $u = u_k$ which are present in the series
(\ref{a-counter-example-sequence}) the absolute value of 
\[
\big(\tilde{f}^{|u_k \rangle}, \widetilde{\varphi} \big)_{{}_{\textrm{pairing}}}
\sim
4\pi d_{{}_{n}} \frac{1}{-1 + 4\pi t} |\boldsymbol{u}_k|^{-2 + 4\pi t}. 
\]
is bounded from below (even streams to infinity).
Therefore  the series 
\[
\sum \limits_{k \in \mathbb{N}} \frac{1}{k}
\big(\tilde{f}^{|u_k \rangle}, \widetilde{\varphi} \big)_{{}_{\textrm{pairing}}}
\]
is divergent. 

For the justification of the rule that the closure with respect to the invariant Hilbert space inner product 
of  a homogeneous part of the free field $A_\mu$ must be contained in 
$E^{*}$, compare Subsection \ref{IntAspatialInfty} of Introduction and Subsections \ref{psichi}, \ref{equivalentA-s}.
It follows from the rule that Fourier transforms of all elements of that space should be 
homogeneous (of fixed degree) solutions of d'Alembert equation. In going outside  $E^{*}$
we lose any natural way in forming the Fourier transform and all the more
in giving any strict sense in which the elements of the Hilbert space $\mathcal{H}_{t}$ are solutions 
of d'Alembert equation. 
\qed

\subsection{Comparison with the 
theory of Staruszkiewicz. The case of infrared electric-type and transversal 
generalized states}\label{infra-electric-transversal-generalized-states}

In the first part of this Subsection we concentrate our attention on the action of the
$SL(2, \mathbb{C})$ group on the decomposition component ${\mathcal{H}'}_{{}_{\chi = -1}}^{e}$
the homogeneous of degree $\chi = -1$ electric type states. This action gives the decomposition component
$\boldsymbol{U}_{{}_{\chi=-1, \,e}}$ of the decomposition of {\L}opusza\'nski representation
constructed in Subsections \ref{equivalentA-s} and \ref{AS}, and determined by the spectral decomposition of the scaling operator
and then by the decomposition of the homogeneous of degree $-1$ states into the the states of electric type and its invariat orthogonal
complementary: 
\[
{\mathcal{H}'}_{{}_{\chi = -1}}= {\mathcal{H}'}_{{}_{\chi = -1}}^{e} \oplus {\mathcal{H}'}_{{}_{\chi = -1}}^{\mathfrak{m}}.
\]
We establish the type of this representation
within the classification scheme of Gelfand-Neumark. Next we establish the type of the representation 
of $SL(2,\mathbb{C})$ acting in the single particle Hilbert space ${\mathcal{H}''}_{{}_{\chi = -1}}^{e}$
of the scalar homogeneous of degree zero field  $\phi \mapsto -ex_\mu A_{{}_{\chi}}^{e \, \mu}(\phi)$,
constructed in Subsection \ref{AS}, 
and compare it with the representation of
$SL(2, \mathbb{C})$ acting in the single particle subspace $\mathcal{H}_{0}^{1}$ spanned by the states $c_{lm}|0\rangle$
of Staruszkiewicz theory. It turns out that they are identical. From this we obtain another proof
of the equality (\ref{xAechi=-1(x)=transversalS}), established in Subsection \ref{AS}.
Finally, we recapitulate the theory of Staruszkiewicz \cite{Staruszkiewicz}.

By the first Proposition of Subsection \ref{Lop-on-E} representors of 
the {\L}opusza\'nski representation  (eq. (\ref{Lop-rep--on-tildevarphi}) ) of 
$T_4 \circledS SL(2,\mathbb{C})$ and its conjugation (eq. (\ref{conjugated-Lop-rep})) transform the 
nuclear space $E \subset \mathcal{H}'$
into itself  continuously with respect to the nuclear topology
of $E = \mathcal{S}_{A^{(3)}}(\mathbb{R}^3; \mathbb{C}^4)$. 
Note that $T_4 \circledS SL(2,\mathbb{C})$ acts in the Fock-Krein space of the quantum four-vector potential field
through the second quantization functor $\Gamma$ of the conjugated {\L}opusza\'nski representation (\ref{conjugated-Lop-rep})
and not through the second quantized functor of the {\L}opusza\'nski represetation itself (\ref{Lop-rep--on-tildevarphi}).
By the standard duality theorem
(compare e.g. \cite{treves}, Proposition 19.5 and its Corollary) the linear dual (conjugation) of the conjugated 
{\L}opusza\'nski representation (\ref{conjugated-Lop-rep}) gives a representation
on the dual space $E^{*}$,
whose representors act continuously on $E^{*}$. Because the generalized infrared 
(homogeneous of degree $-1$) electric-type
and transversal states concentrated on the positive energy sheet of the cone
compose a closed subspace  $(E^{*})_{tr}^{e} = {\mathcal{H}'}_{{}_{\chi = -1}}^{e}$ of $E^{*}$, 
then the dual (conjugation) of the conjugated
{\L}opusza\'nski representation restricted to the $SL(2, \mathbb{C})$ subgroup\footnote{Regarded as a  
representation on $E$.} acts naturally and continuously on the invariant subspace $(E^{*})_{tr}^{e}= {\mathcal{H}'}_{{}_{\chi = -1}}^{e}$. 
Let us denote it by $\boldsymbol{U}_{{}_{\chi=-1, \,e}}$. 
By the results of the Subsections \ref{DiracHom=-1Sol} and \ref{AS}, the elements of 
$(E_{\mathbb{C}}^{*})_{tr}^{e}$ are regular, and can be identified with ordinary functions $\tilde{f}_\mu$
on (the positive energy sheet of) the light cone, and the representation 
$\boldsymbol{U}_{{}_{\chi=-1, \,e}}$ acts on 
the corresponding four component functions $\tilde{f}$ exactly as the {\L}opusza\'nski representation (\ref{Lop-rep--on-tildevarphi}) on the ordinary states $\widetilde{\varphi}
\in \mathcal{H}'$ with $\widetilde{\varphi}$ in the formula (\ref{Lop-rep--on-tildevarphi}) 
replaced with $\tilde{f}$. Moreover the analogous continuity statements of the second quantized version 
of the representation $\Gamma(\boldsymbol{U}_{{}_{\chi=-1, \,e}})$ on $\Gamma(E)$ hold as well, compare \cite{hida},
\cite{obata-book}. In particular $(E \otimes E)^* = E^{*} \otimes E^{*}$
by the celebrated kernel theorem for nuclear spaces. 

Consider the one-particle space $(E^{*})_{tr}^{e} = {\mathcal{H}'}_{{}_{\chi = -1}}^{e}$ of generalized transversal electric-type positive energy states, and the representation $\boldsymbol{U}_{{}_{\chi=-1, \,e}}$ acting on it. If $\widetilde{f}, \widetilde{f}'$ are measurable homogeneous of degree $-1$ functions on the cone, representing the corresponding distributions in 
$(E^{*})_{tr}^{e} = {\mathcal{H}'}_{{}_{\chi = -1}}^{e}$, 
then we have the inner product $(\tilde{f},\tilde{f})_{{}_{\mathfrak{J}}}$ on $(E^{*})_{tr}^{e}$ 
defined by the formula (\ref{krein-prod-infra-red}) or (\ref{krein-prod-infra-red-1}),
compare the first Proposition of Subsection \ref{AS}. The inner product 
$(\cdot,\cdot)_{{}_{\mathfrak{J}}}$ 
is positive definite on $(E^{*})_{tr}^{e} = {\mathcal{H}'}_{{}_{\chi = -1}}^{e}$ and non-degenerate. By the Cauchy-Schwarz inequality
for non-negatively definite bilinear Hermitian forms the subspace $N$ of generalized states with 
zero $(\cdot,\cdot)_{{}_{\mathfrak{J}}}^{\textrm{tr}}$-norm is a linear subspace of 
$(E^{*})_{tr}^{e} = {\mathcal{H}'}_{{}_{\chi = -1}}^{e}$
and the quotient linear space  $(E_{\mathbb{C}}^{*})_{tr}^{e}/N$ is naturally a pre-Hilbert space with the well defined inner product
on the equivalence classes given by the inner product $(\cdot,\cdot)_{{}_{\mathfrak{J}}}$
of the corresponding representative elements. As we have seen in the previous Subsection, $N = \{ 0\}$, so that
(\ref{krein-prod-infra-red-1}) is non-degenerate on the electric-type states $(E^{*})_{tr}^{e}= {\mathcal{H}'}_{{}_{\chi = -1}}^{e}$, and
$(E^{*})_{tr}^{e}/N = (E_{\mathbb{C}}^{*})_{tr}^{e} = {\mathcal{H}'}_{{}_{\chi = -1}}^{e}$.  Its closure --  the corresponding Hilbert space --
is equal ${\mathcal{H}'}_{{}_{\chi = -1}}^{e}$.

Because the bilinear Hermitian form $(\cdot,\cdot)_{{}_{\mathfrak{J}}}$ 
on $(E_{\mathbb{C}}^{*})_{tr}^{e} = {\mathcal{H}'}_{{}_{\chi = -1}}^{e}$ is invariant for the decomposition component $\boldsymbol{U}_{{}_{\chi=-1, \,e}}$
of the {\L}opusza\'nski representation $\boldsymbol{U}$  
then it follows that the {\L}opusza\'nski representation induces a unitary representation 
$\boldsymbol{U}_{{}_{\chi=-1, \,e}}$ on the Hilbert space
${\mathcal{H}'}_{{}_{\chi = -1}}^{e}$ of transversal homgeneous of degree $\chi=-1$ electric type states -- the electric type component of the  $\chi=-1$
component of the decomposition constructed in Subsection \ref{equivalentA-s}.

In order to establish the irreducibility of the representation $\boldsymbol{U}_{{}_{\chi=-1, \,e}}$
and its type within the classification
scheme of Gelfand-Neumark, and in order to establish the explicit formula for $\boldsymbol{U}_{{}_{\chi=-1, \,e}}$, 
we use the first Proposition of Subsection \ref{AS}. 
Then we refer to the general theory of Gelfand-Neumark and to the general properties 
of infrared transversal and positive energy solutions of the wave equation (i.e. positive energy infrared solutions of the vacous Maxwell equations), summarised in the said Proposition. 

Namely, we use the general property of states in $(E^{*})_{tr}^{e} = {\mathcal{H}'}_{{}_{\chi = -1}}^{e}$.  In particular (compare the first Proposition of Subsect. \ref{AS}) the electric type generalized states are determined by the 
homogeneous of degree $-1$ functions $\tilde{f}_{\mu}$, $\mu=0,1,2,3$ on the positive energy sheet of the cone,
which are of the form
\begin{equation}\label{tildefmu=prtialmutildef}
\tilde{f}_{\mu} = \frac{\partial \tilde{f}}{\partial p^\mu} 
\end{equation}     
where $\tilde{f}$ is a restriction to the cone of a smooth (except zero) homogeneous of degree zero scalar function. Equivalence classes
of such states compose a dense subspace 
in ${\mathcal{H}'}_{{}_{\chi = -1}}^{e}$, compare Lemma of Subsect  \ref{AS}. 
The {\L}opusza\'nski representation
$\boldsymbol{U}_{{}_{\chi=-1, \,e}}$ acts on $\tilde{f}_{\mu}$ as on the ordinary state according to the formula
(\ref{Lop-rep--on-tildevarphi}) with $\widetilde{\varphi}$ replaced by $\tilde{f}$, which is equivalent to the ordinary action on the scalar $\tilde{f}$ 
\begin{equation}\label{Utr,e(f)}
{\boldsymbol{U}_{{}_{\chi=-1, \,e}}}_{\alpha} \tilde{f}(p)= \tilde{f}(\Lambda(\alpha)p),
\,\,\,
\alpha \in SL(2, \mathbb{C}), p \in \mathscr{O}_{1,0,0,1}.
\end{equation}

Because the smooth $\tilde{f}$ in the formula (\ref{tildefmu=prtialmutildef}) for the electric type generalized state
is homogeneous of degree zero, then it ``lives effectively'' on the unit $2$-sphere $\mathbb{S}^2$ 
of rays of the cone in the momentum space and the representation $\boldsymbol{U}_{{}_{\chi=-1, \,e}}$
may be considered as acting on the Hilbert space of functions on the unit $2$-sphere
modulo the constant functions on the unit $2$-sphere, and is induced by (\ref{Utr,e(f)}).
In particular the restriction of the representation $\boldsymbol{U}_{{}_{\chi=-1, \,e}}$ to the
subgroup $SU(2, \mathbb{C})$ (double covering of the rotation group $SO(3)$) induced by (\ref{Utr,e(f)}) 
on the functions on the $2$-sphere modulo the constant functions coincides with the 
ordinary representation of the rotation group on the subspace of scalar functions on the $2$-sphere 
$\mathbb{S}^2$
orthogonal to the one dimensional subspace of constant functions on $\mathbb{S}^2$.
Indeed let $\tilde{f}_\mu$ and $\tilde{f}'_\mu$ be two electric type transversal homogeneous of degree zero
solutions of the form (\ref{tildefmu=prtialmutildef}) with the corresponding homogeneous of degree zero functions 
$\tilde{f}$ and $\tilde{f}'$ on the cone. Then their inner product (\ref{krein-prod-infra-red}), equal to
(\ref{krein-prod-infra-red-1}) expressed in terms of the scalar homogeneous of degree zero functions $\tilde{f}$
and $\tilde{f}'$ has the following form
\begin{multline}\label{krein-prod-infra-red-3}
(\tilde{f},\tilde{f}')_{{}_{\mathfrak{J}}} = \,\,\, 
- \, \int \limits_{\mathbb{S}^2} \overline{\tilde{f}_\mu(p)}\tilde{f}'^\mu(p)
\, \ud \mu_{{}_{\mathbb{S}^2}} \\
=
- \, \int \limits_{\mathbb{S}^2} \overline{\frac{\partial \tilde{f}}{\partial p^\mu}}
\frac{\partial \tilde{f}'}{\partial p_\mu}
\, \ud \mu_{{}_{\mathbb{S}^2}} 
= \, \int \limits_{\mathbb{S}^2} \overline{\tilde{f}}
\big[- \Delta_{{}_{\mathbb{S}^2}}\tilde{f}'\big]
\, \ud \mu_{{}_{\mathbb{S}^2}},
\end{multline}
where $\Delta_{{}_{\mathbb{S}^2}}$ is the standard Laplace operator on the unit $2$-sphere $\mathbb{S}^2$ 
and where in the last equality we have used the spherical coordinates and the integration by parts, as in the previous Subsection. 
Thus our Hilbert space ${\mathcal{H}'}_{{}_{\chi = -1}}^{e}$ consist
of all functions on the unit sphere for which 
\begin{equation}\label{krein-prod-infra-red-4}
(\tilde{f},\tilde{f})_{{}_{\mathfrak{J}}}
=\int \limits_{\mathbb{S}^2} \overline{\tilde{f}}
\big[- \Delta_{{}_{\mathbb{S}^2}}\tilde{f} \big]
\, \ud \mu_{{}_{\mathbb{S}^2}}
\end{equation}
is finite, where $\Delta_{{}_{\mathbb{S}^2}}$ is understood as the self-adjoint operator on the Hilbert space of all
square integrable functions on $\mathbb{S}^2$ with respect to the invariant measure $\ud \mu_{{}_{\mathbb{S}^2}}$. 
More precisely ${\mathcal{H}'}_{{}_{\chi = -1}}^{e}$ is the closure of the domain of the self adjoint 
operator $\sqrt{\Delta_{{}_{\mathbb{S}^2}}}$ in $L^2(\mathbb{S}^2; \ud \mu_{{}_{\mathbb{S}^2}})$ 
with respect to the inner product (\ref{krein-prod-infra-red-4}).   In particular 
the constant functions compose just the whole linear subspace of zero 
$(\cdot,\cdot)_{{}_{\mathfrak{J}}}$-norm in agreement 
with our assertion formulated above. Moreover our representation is spanned by the system of functions  
\[
{\textstyle\frac{1}{\sqrt{l(l+1)}}} Y_{lm}, \,\,\, -l \leq m \leq l, l = 1, 2, 3, \ldots
\]  
orthonormal with respect to the norm $(\cdot,\cdot)_{{}_{\mathfrak{J}}}$,
computed as in (\ref{krein-prod-infra-red-3}).
This system is complete in ${\mathcal{H}'}_{{}_{\chi = -1}}^{e}$ which easily follows from the completeness 
of the system $\{Y_{lm} \}$, $l= 0,1,2, \ldots$,
$-l \leq m \leq l$, of spherical functions in the Hilbert space 
$L^2(\mathbb{S}^2, \ud \mu_{{}_{\mathbb{S}^2}})$ of square integrable
functions on $\mathbb{S}^2$. Note also that the inner product (\ref{krein-prod-infra-red-4})
is indeed not only rotationally but likewise Lorentz invariant, although it is not immediately visible,
so that $\boldsymbol{U}_{{}_{\chi=-1, \,e}}$ is a unitary representation of $SL(2, \mathbb{C})$
on ${\mathcal{H}'}_{{}_{\chi = -1}}^{e}$ with the inner
product defined by (\ref{krein-prod-infra-red-4}). Let us explain this assertion.
Any element of the form (\ref{tildefmu=prtialmutildef}) is identified with the corresponding homogeneous 
of degree zero function $\tilde{f}$ in (\ref{tildefmu=prtialmutildef}). Any such function $\tilde{f}$ 
is uniquely determined by its restriction to the unit $2$-sphere $\mathbb{S}^2$. The action 
of the Lorentz (or rotation) transformation on $\tilde{f}$ 
can be understood as the action on $\tilde{f}$ understood as a function on $\mathbb{S}^2$. Namely we  
act on $\tilde{f}$ regarded as a homogeneous of degree zero function on the cone accordingly to the formula
(\ref{Utr,e(f)}), and then restrict the result of the action to the sphere $\mathbb{S}^2$. The action
$(\theta, \phi) \mapsto \Lambda(\theta,\phi)$ of the Lorentz group on $\mathbb{S}^2$ is defined through the 
natural action of the Lorentz group on the rays (i.e. linear generators) of the cone. This furnishes 
a ``standard representation'' ${\boldsymbol{U}_{{}_{\chi=-1, \,e}}}_{\Lambda}\tilde{f}(\theta, \phi) = \tilde{f}(\Lambda(\theta, \phi))$ of the Lorentz group
in the terminology of \cite{Bargmann}, p. 577, induced by the action $(\theta,\phi) \mapsto \Lambda(\theta,\phi)$
on the manifold $\mathbb{S}^2$ with the trivial multiplier equal $1$ because the functions $\tilde{f}$ are assumed to be homogeneous of degree zero.
The measure $\ud \mu_{{}_{\mathbb{S}^2}}$ on $\mathbb{S}^2$ is rotationally invariant but it is not Lorentz
invariant. Nonetheless the inner product (\ref{krein-prod-infra-red-4}) is Lorentz 
invariant because the non-invariance of the measure $\mu_{{}_{\mathbb{S}^2}}$ under 
the hyperbolic rotation $\Lambda(\lambda)$, i.e. Lorentz transformation, 
is compensated for by the non-invariance of the Laplace operator $\Delta_{{}_{\mathbb{S}^2}}$ 
under the Lorentz transformation. In other words the nontrivial
Radon-Nikodym derivative  (\ref{RadonNkodym-on-S2rays}) of the measure $\mu_{{}_{\mathbb{S}^2}}$ transformed 
by $\Lambda(\lambda)$ with respect to the non-transformed measure $\mu_{{}_{\mathbb{S}^2}}$ is just compensated 
for by the non-invariance of the Laplace operator
$\Delta_{{}_{\mathbb{S}^2}}$ on $\mathbb{S}^2$ under the action of the Lorentz transformation 
$\Lambda(\lambda)$:
\begin{multline*}
{\boldsymbol{U}_{{}_{\chi=-1, \,e}}}_{{}_{\Lambda(\lambda)}} \Delta_{{}_{\mathbb{S}^2}} 
\Big({\boldsymbol{U}_{{}_{\chi=-1, \,e}}}_{{}_{\Lambda(\lambda)}}\Big)^{-1} 
=
\frac{d\mu_{{}_{\mathbb{S}^2}}(\theta,\phi)}{d\mu_{{}_{\mathbb{S}^2}}(\Lambda(\lambda)(\theta,\phi))}
\Delta_{{}_{\mathbb{S}^2}} 
\\
= 
\Big(\frac{(\Lambda(\lambda)p)^0}{p^0}\Big)^2 \Delta_{{}_{\mathbb{S}^2}} = \Big(\frac{p'^0}{p^0}\Big)^2
\Delta_{{}_{\mathbb{S}^2}}, 
\end{multline*}
or equivalently
\[
\frac{d\mu_{{}_{\mathbb{S}^2}}(\Lambda(\lambda)(\theta,\phi))}{d\mu_{{}_{\mathbb{S}^2}}(\theta,\phi)} 
{\boldsymbol{U}_{{}_{\chi=-1, \,e}}}_{{}_{\Lambda(\lambda)}} \Delta_{{}_{\mathbb{S}^2}} 
=
\Delta_{{}_{\mathbb{S}^2}}
{\boldsymbol{U}_{{}_{\chi=-1, \,e}}}_{{}_{\Lambda(\lambda)}}
\]
where ${\boldsymbol{U}_{{}_{\chi=-1, \,e}}}_{{}_{\Lambda(\lambda)}}$, and $\Lambda(\lambda)$, is understood as acting on functions on $\mathbb{S}^2$, 
or, respectively, on points of $\mathbb{S}^2$, as explained above. Unitarity of the Lorentz transformation immediately 
thus follows:
\begin{multline*}
\int \limits_{\mathbb{S}^2} {\boldsymbol{U}_{{}_{\chi=-1, \,e}}}_{{}_{\Lambda(\lambda)}}\tilde{f}(\theta, \phi) 
\Delta_{{}_{\mathbb{S}^2}} {\boldsymbol{U}_{{}_{\chi=-1, \,e}}}_{{}_{\Lambda(\lambda)}} \tilde{g}(\theta, \phi) 
\, \ud \mu_{{}_{\mathbb{S}^2}}(\theta, \phi)  \\
=
\int \limits_{\mathbb{S}^2} {\boldsymbol{U}_{{}_{\chi=-1, \,e}}}_{{}_{\Lambda(\lambda)}}\tilde{f}(\theta, \phi) 
{\boldsymbol{U}_{{}_{\chi=-1, \,e}}}_{{}_{\Lambda(\lambda)}} \big(\Delta_{{}_{\mathbb{S}^2}} \tilde{g} \big) (\theta, \phi) \,
\frac{d\mu_{{}_{\mathbb{S}^2}}(\Lambda(\lambda)(\theta,\phi))}{d\mu_{{}_{\mathbb{S}^2}}(\theta,\phi)}
\, \ud \mu_{{}_{\mathbb{S}^2}} (\theta, \phi)  
\end{multline*}
\begin{multline*}
=
\int \limits_{\mathbb{S}^2} \tilde{f}(\Lambda(\lambda)(\theta,\phi))
\big(\Delta_{{}_{\mathbb{S}^2}} \tilde{g} \big) (\Lambda(\lambda)(\theta,\phi)) \,
\frac{d\mu_{{}_{\mathbb{S}^2}}(\Lambda(\lambda)(\theta,\phi))}{d\mu_{{}_{\mathbb{S}^2}}(\theta,\phi)}
\, \ud \mu_{{}_{\mathbb{S}^2}} (\theta, \phi)  \\
=
\int \limits_{\mathbb{S}^2} \tilde{f}(\Lambda(\lambda)(\theta,\phi))
\big(\Delta_{{}_{\mathbb{S}^2}} \tilde{g} \big) (\Lambda(\lambda)(\theta,\phi)) \,
d\mu_{{}_{\mathbb{S}^2}}(\Lambda(\lambda)(\theta,\phi)) \\
=
\int \limits_{\mathbb{S}^2} \tilde{f}(\theta, \phi) 
\Delta_{{}_{\mathbb{S}^2}} \tilde{g}(\theta, \phi) 
\, \ud \mu_{{}_{\mathbb{S}^2}} (\theta, \phi) .
\end{multline*}

Therefore, we have just shown (compare e.g. \cite{Geland-Minlos-Shapiro} or 
\cite{NeumarkLorentzBook}
where the representation of the rotation group on $L^2(\mathbb{S}^2, \ud \mu_{{}_{\mathbb{S}^2}})$
is analysed systematically) that the restriction of the representation
$\boldsymbol{U}_{{}_{\chi=-1, \,e}}$ to the subgroup $SU(2, \mathbb{C})$ (doubly covering the group of rotations) is unitary equivalent with
\[
L_1 \oplus L_2 \oplus L_3 \oplus \ldots,
\]
where $L_l$ is the standard irreducible unitary representation of $SU(2, \mathbb{C})$  
corresponding to the weight $l$ (with the ``angular momentum quantum number'' equal $l = 1,2,3, \ldots$)
so that the representation with $l=0$ does not enter into the decomposition.

Therefore, the results of \cite{NeumarkLorentzBook} Chap III (or the results of \cite{Geland-Minlos-Shapiro}, 
Part II, Chap. I) are applicable to the representation $\boldsymbol{U}_{{}_{\chi=-1, \,e}}$. 
In particular the unitary representation $\boldsymbol{U}_{{}_{\chi=-1, \,e}}$ is irreducible.

Moreover easy computation shows that the second Casimir operator, which Neumark denotes $\Delta'$,
eq. (2), page 167 of his book \cite{NeumarkLorentzBook},  corresponding to the representation  
$\boldsymbol{U}_{{}_{\chi=-1, \,e}}$  of the $SL(2, \mathbb{C})$ group is identically zero: $\Delta' =0$. 
In particular by Theorem 2 of \S 8.3 and Theorem 3 of \S 8.4 of \cite{NeumarkLorentzBook} we see that the unitary representation
 $\boldsymbol{U}_{{}_{\chi=-1, \,e}}$  of $SL(2, \mathbb{C})$ is unitary equivalent to the representation
of the principal series, which is denoted by the pair of numbers $(k_0=1, c =0)$  in the notation of \cite{NeumarkLorentzBook} (or $(l_0 = 1, l_1 = 0)$
in the notation of \cite{Geland-Minlos-Shapiro}). 
The normalized functions 
\[
{\textstyle\frac{1}{\sqrt{l(l+1)}}} Y_{lm}, \,\,\, -l \leq m \leq l, l = 1, 2, 3, \ldots
\] 
correspond to the normalized states $\xi_{lm}$ of the representation $(l_0,l_1) = (1,0)$ in \cite{Geland-Minlos-Shapiro}.
Thus we have proved the following

\begin{prop*}
The representation $\boldsymbol{U}_{{}_{\chi=-1, \,e}}$ of $SL(2, \mathbb{C})$ acting on the Hilbert space of electric-type infrared transversal generalized states 
${\mathcal{H}'}_{{}_{\chi = -1}}^{e}$ is
unitary equivalent with the irreducible unitary  representation of $SL(2, \mathbb{C})$, which in the classification scheme of Gelfand-Neumark is the representation of the principal series denoted by the pair of numbers $(k_0=1,c=0) = \mathfrak{S}(m=2, \rho = 0)$ (in the book \cite{NeumarkLorentzBook}) and by the pair of numbers $(l_0=1, l_1 = 0) $ (in the book \cite{Geland-Minlos-Shapiro}). The normalized functions 
\[
{\textstyle\frac{1}{\sqrt{l(l+1)}}} Y_{lm}, \,\,\, -l \leq m \leq l, l = 1, 2, 3, \ldots
\] 
correspond to the states $\xi_{lm}$ of the representation $(l_0,l_1) = (1,0)$ in \cite{Geland-Minlos-Shapiro}.
\end{prop*}

Consider now the single particle Hilbert space ${\mathcal{H}''}_{{}_{\chi = -1}}^{e}$ of homogeneous
of degree $-2$ states of the form
\[
{\textstyle\frac{\partial^2}{\partial p_\mu \partial p^\mu}} \widetilde{f},
\]
where $\widetilde{f}$ are homogeneous of degree zero functions on the cone, with the invariant
inner product (\ref{krein-prod-infra-red-4}) expressed in therms of the corresponding
homogeneous of degree zero scalars $\widetilde{f}, \widetilde{f}'$. Let the $SL(2, \mathbb{C})$
acts on the states through the ordinary Lorentz group action on the corresponding
scalars $\widetilde{f}$ homogeneous of degree zero.
The following states
\begin{equation}\label{partialFe}
-i\partial_{{}_{p^\mu}}F^{e \, \mu}_{{}_{\chi=-1,lm}} = -i {\textstyle\frac{1}{\sqrt{l(l+1)}}}
{\textstyle\frac{\partial^2}{\partial p_\mu \partial p^\mu}} Y_{{}_{lm}}
\end{equation}
on the cone which are homogeneous of degree $-2$, and with the harmonics $Y_{{}_{lm}}$ understood as homogeneous
of degree zero functions coinciding with the ordinary harmonics $Y_{{}_{lm}}$ on the unit two-sphere $\mathbb{S}^2$,
compose the complete orthonormal system of the Hilbert space ${\mathcal{H}''}_{{}_{\chi = -1}}^{e}$
of homogeneous of degree $-2$ states of the indicated form.
It is obvious, by the above Proposition and its proof, that $SL(2,\mathbb{C})$ acts on
${\mathcal{H}''}_{{}_{\chi = -1}}^{e}$ irreducibly through the representation $(l_0,l_1) = (1,0)$
with the normalized states (\ref{partialFe}) corresponding to the states
$\xi_{lm}$ of the representation $(l_0,l_1) = (1,0)$ in \cite{Geland-Minlos-Shapiro}.
Because the inverse Fourier transform of the states (\ref{partialFe}) coincides with
the partial waves $f^{(+)}_{lm}$ of Staruszkiewicz theory, then ${\mathcal{H}''}_{{}_{\chi = -1}}^{e}$
coincides with the single particle Hilbert space $\mathcal{H}_{0}^{1}$ of Staruszkiewicz theory, spanned by $c_{lm}^{+}|0\rangle$.

Let $A_{{}_{\chi=-1}}^{e}$ be the electric part of the homogeneous of degree $-1$ part $A_{{}_{\chi=-1}}$
of the free electromagnetic potential field $A$, which we have constructed in Subsection \ref{AS}.
In Subsection \ref{AS} we have seen that the scalar homogeneous of degree zero field $\phi \mapsto -ex_\mu A_{{}_{\chi}}^{e \, \mu}(\phi)$,
with the kernel
\begin{equation}\label{xAhom=-1}
-e x_\mu A_{{}{\chi=-1}}^{e \, \mu} (x) =
\sum \limits_{l=1}^{\infty} \sum \limits_{m = -l}^{m= +l}
\Big\{ {c'}_{lm} f_{lm}^{(+)}(x) + {c'}_{lm}^{+} \overline{f_{lm}^{(+)}(x)}
\Big\},
\end{equation}
regarded as acting in the Fock space
\[
\Gamma\big({\mathcal{H}'}^{e}_{{}_{\chi = -1}}\big) \otimes |0\rangle^{\mathfrak{m}},
\]
is equal to the free field over the single particle Hilbert space
${\mathcal{H}''}_{{}_{\chi = -1}}^{e} \cong_{{}_{U}} {\mathcal{H}'}_{{}_{\chi = -1}}^{e}$.
But from the results of \cite{Staruszkiewicz1995} it follows that the transversal part
of the phase field $S$, regarded as a field on the invariant zero charge subspace $\mathcal{H}_0$
$=\Gamma(\mathcal{H}_{0}^{1})$ $\subset \mathcal{H}$,
is equal to the free field over the Fock space $\Gamma(\mathcal{H}_{0}^{1})$ over the single particle Hilbert space
$\mathcal{H}_{0}^{1}$ spanned by $c_{lm}^{+}|0\rangle$. We therefore obtain the equality (\ref{xAchi=-1(x)=transversalS}),
compare Subsection \ref{AS} for details. In Subsection \ref{Comparison2}
we extend the equality (\ref{xAchi=-1(x)=transversalS}) all over the whole Hilbert space $\mathcal{H}$
of the phase field $S$, using a non-invariant tensor product factorization of the
Hilbert space $\mathcal{H}$ of the phase field $S$. Note here that this factorization depends on the reference frame,
and is associated to the frame in which the partial waves $f^{(+)}_{lm}$ were computed. This must be the case
because the division of the phase $S$ into the transversal part, present in the equality
(\ref{xAechi=-1(x)=transversalS}), and the rest $S_0 -e Q \textrm{th}\psi$, is not invariant and depends
on the reference frame.

Now let us recapitulate, for the completeness of our presentation, the quantum theory of infrared fields of Staruszkiewicz. 
For the original account the reader is encouraged to consult the works, \cite{Staruszkiewicz1987},
\cite{Staruszkiewicz} , \cite{Staruszkiewicz1992ERRATUM} and \cite{Staruszkiewicz1992}. 

We start at the classical level.
Here we consider only the electric type homogeneous of degree $-1$
solutions of d'Alembert equation generated by the Lorentz transformations of the Dirac homogeneous of degree 
$-1$ solution (\ref{thpsi-state}) the same as those in (\ref{fmuHomogeneity=-1}).
Note that by subtraction of the untransformed Dirac solution (\ref{thpsi-state}) 
from the transformed Dirac solution we get a transversal electric type solution 
entering the set of solution generated by 
Dirac solution (\ref{thpsi-state}).  
Here we add to them also the odd solutions $f_\mu(-x) = -f_\mu(x)$
although in the real Bremsstrahlung infrared radiation there are present only the even solutions
$f_\mu(-x) = f_\mu(x)$ (\ref{fmuHomogeneity=-1}). 
We do this after \cite{Staruszkiewicz1987},
\cite{Staruszkiewicz} because we need among their Fourier transforms such which are 
complex valued on the cone, in order to construct a positive and negative energy solutions
which then serve as a basis of quantization of the phase field in the Hilbert space which is over $\mathbb{C}$
and not  over $\mathbb{R}$. But compare the second Remark ending this Subsection. 
Their Fourier transforms are concentrated on both sheets of the cone.
Among them there are transversal solutions. Each such solution has unique decomposition 
into the sum of two solutions -- Fourier transform of the first one is concentrated on the positive energy sheet of the cone
and Fourier transform of the second one is concentrated on the negative energy sheet of the cone.
All of them are regular in the position picture, i.e. zero order or function type in 
$\mathcal{S}^{00}(\mathbb{R}^4)^*$
and their Fourier transform (concentrated on the cone) determine regular (of order zero) distributions on the cone, 
whenever treated as elements in $E^* = \mathcal{S}^{0}(\mathbb{R}^3;\mathbb{C}^4)^*$ (no longer 
regular as elements of $\mathcal{S}^{0}(\mathbb{R}^4; \mathbb{C}^4)^*$).
Note that the splitting of the said distributions in $\mathcal{S}^{00}(\mathbb{R}^4)^*$ into positive 
and negative energy solutions is unique (compare Subsect. \ref{Lop-on-E}).

Essentially each global, positive frequency electric-type and homogeneous of degree $-1$ solution
of Maxwell equations (i.e. transversal electric-type solution of the wave equation)
can be written in the form
\begin{equation}\label{TrElSolWaveEq}
A_\mu(x) = \int \limits_{\mathscr{O}_{1,0,0,1}} \ud \mu_{{}_{\mathscr{O}_{1,0,0,1} }} \frac{\partial a(p)}{\partial p^\mu}
e^{-ip\cdot x}
\end{equation}   
and where $a$ is a differentiable (except zero) and homogeneous of degree zero function on the cone, and can be interpreted as a distribution
(in $\mathcal{S}^{00}(\mathbb{R}^4)^*$ as a function of $\varphi$) defined
by the formula (\ref{Hom=-1Distr}) (with the second term -- the integral over $\mathscr{O}_{-1,0,0,1}$ -- equal zero) and with 
\[
\widetilde{f}_\mu(p) = \frac{\partial a(p)}{\partial p^\mu},
\] 
which fulfills the d'Alembert and transversality equations, and thus is a distributional solution of the Maxwell equations, 
compare Subsections \ref{DiracHom=-1Sol} and \ref{AS}. It also defines via the formula (\ref{Hom=-1Distr}) a regular distribution
on the light cone $\mathscr{O}_{1,0,0,1} = \{p: p \cdot p = 0, p^0 >0\}$, as a function of  
$\widetilde{\varphi}|_{{}_{\mathscr{O}_{1,0,0,1}}}$ and a distribution in $\mathcal{S}^{0}(\mathbb{R}^4)$
as function of $\widetilde{\varphi}$, compare Subsection \ref{DiracHom=-1Sol}. 
To each such solution there corresponds 
the classical scalar field $S(x) = -e x^\mu A_\mu(x)$ which  
is homogeneous of degree zero, and thus ``lives effectively'' on the $3$-dim de Sitter hyperboloid,
and fulfills d'Alembert equation (correspondingly the wave equation on the de Sitter hyperboloid).

On the other hand the wave equation on the de Sitter $3$-hyperboloid has the general real solution as a 
function on the de Sitter hyperboloid (we are using the spherical coordinates 
with the hyperbolic angle $\psi$ ranging over $\mathbb{R}$) 
\begin{equation}\label{classicalS(x)}
S(\psi, \theta, \phi) = S_0 -e Q \textrm{th} \psi +  \sum \limits_{l=1}^{\infty} \sum \limits_{m = -l}^{m= +l}
\big\{ c_{lm} f_{lm}^{(+)}(\psi, \theta, \phi) 
+ \overline{c_{lm}} \overline{f_{lm}^{(+)}(\psi, \theta, \phi)} \big\} 
\end{equation}
where 
\[
f_{lm}^{(+)}(\psi, \theta, \phi) 
\]
\emph{corresponds} to the classical transversal solution (\ref{Hom=-1Distr}) of the wave equation with
\begin{equation}\label{a(p)ASsolution}
\widetilde{f}_\mu(p) = \frac{\partial a(p)}{\partial p^\mu}, \,\,\, a(p) =\, {\textstyle\frac{1}{2\sqrt{\pi} \sqrt{l(l+1)}}} \, Y_{lm}(p), \,\,\, p \in \mathbb{S}^2
\end{equation}
and concentrated on the positive energy sheet 
\[
\mathscr{O}_{1,0,0,1} = \{p: p \cdot p = 0, p^0 >0\} 
\]
of the cone. Here \emph{correspondence} means that the corresponding of degree zero solution is equal to 
$x\cdot x \, f$, where $f$ is the scalar homogeneous of degree $-2$ solution whose Fourier transform is equal to the homogeneous of degree zero $a(p)$ in (\ref{a(p)ASsolution}). Therefore, the Fourier transform of $f_{lm}^{(+)}(x)$ -- the unique extension of
$f_{lm}^{(+)}(\psi, \theta, \phi)$ over the whole space-time, determined by the homogeneity, is equal 
\begin{equation}\label{F(f(+)lm)}
\widetilde{f_{lm}^{(+)}}(p) = -i {\textstyle\frac{1}{2\sqrt{\pi} \sqrt{l(l+1)}}} \frac{\partial^2 Y_{lm}(p)}{\partial p_\mu \partial p^\mu}, 
\end{equation}
where $Y_{lm}$ is the homogeneous of degree zero function coinciding with the ordinary harmonics $Y_{lm}$ on the unit two-sphere
$\mathbb{S}^2$. Compare \cite{Staruszkiewicz}, \S 4
and \cite{Staruszkiewicz1995}, \S 3,
for more details on $f_{lm}^{(+)}(\psi, \theta, \phi)$, their normalization, as well as explicit formulas
in terms of the hypergeometric function ${}_2 F_1$.
Note that the function $a$ being homogeneous of degree zero is uniquely determined by its restriction to the
unit $2$-sphere and is a function of the angles only when expressed in the spherical coordinates. 

The constant solution 
\[
S(\psi, \theta, \phi) = S_0
\]
and the solution
\[
S(\psi, \theta, \phi) = \textrm{th} \psi
\]
have no counterpart among the transversal solutions (\ref{Hom=-1Distr}) or, respectively (\ref{TrElSolWaveEq}) 
with $f_\mu$ denoted by $A_\mu$, of the wave equation.

But there is the solution (\ref{Hom=-1Distr}) or, respectively (\ref{TrElSolWaveEq}) with $f_\mu$ denoted by $A_\mu$ 
(regarded as a distribution $\mathcal{S}^{00}(\mathbb{R}^4)
\ni \varphi \mapsto (\widetilde{f_\mu}, \widetilde{\varphi})$) which defines the corresponding 
regular distribution on the cone
$\mathscr{O}_{1,0,0,1} \sqcup \mathscr{O}_{-1,0,0,1} = \{p: p \cdot p = 0\}$ defined by the functions
$\tilde{f}_\mu$ homogeneous of deegree $-1$ on the cone which are not transversal and correspond
to the the solution $S(\psi, \theta, \phi) = \textrm{th}\psi$
on the de Sitter hyperboloid. Namely by the results of Subsection \ref{DiracHom=-1Sol} it follows that the 
solution (\ref{Hom=-1Distr}) (as a distribution in
$\mathcal{S}^{00}(\mathbb{R}^4)^*$) of the wave equation defined by
\begin{equation}\label{thpsi-state}
(\tilde{f}_0 = \frac{\textrm{sgn}(p_0)}{p_0}, \tilde{f}_1 =0, \tilde{f}_2 =0,\tilde{f}_3 =0) 
=  \frac{\textrm{sgn}(p_0) \, u}{u \cdot p}, \,\,\textrm{where} \, (u_0 = 1, u_1=u_2= u_3=0),
\end{equation}
on the cone
$\{p: p\cdot p =0\}$ with 
\begin{equation}\label{f^muCorrespondingTo-th(psi)}
f_0(x) = \theta(-x \cdot x) \frac{1}{|\boldsymbol{\x}|}, \,\,\, f_1 = f_2 = f_3 = 0,
\end{equation}
corresponds to the solution $S(\psi, \theta, \phi) = \textrm{th} \psi$ of the wave equation 
on the de Sitter $3$-hyperboloid. We call this solution the \emph{Dirac homogeneous of degree $-1$ solution}.

Indeed, the general rule giving the correspondence between the classical electromagnetic field $f_\mu$
and the scalar field solution $S$ of the wave equation on de Sitter $3$-hyperbolid, called ``phase''
in \cite{Staruszkiewicz},
goes through the construction of a homogeneous of degree zero function $S$ in the Minkowski spacetime
which thus ``lives'' on the $3$-hyperboloid, and has the property that the quantity
\[
e f_\mu + \partial_\mu S
\]
is gauge invariant. One can associate to each such $f_\mu$ the corresponding phase $S$ in the Poincar\'e
invariant way, as in \cite{Staruszkiewicz}, \S 3. The method of \cite{Staruszkiewicz}, \S 3, has the justification
within distribution theory, as follows from the results of Subsection \ref{DiracHom=-1Sol}. 
In particular the correspondence between $f_\mu$ and $S$ can be prolonged on $f_\mu$ which are not transversal,
e.g. on the homogeneous of degree $-1$ solutions of the wave equations defined by (\ref{tildefmuHom=-1})
with
\[
\sum \limits_{i} \alpha_i \neq 0,
\]  
which is the case e.g. for (\ref{thpsi-state}). In this case, i.e. for (\ref{thpsi-state}) the homogeneous 
of degree zero function $S$ has the following form
\begin{equation}\label{distribution-xx(x0/|x|)}
S(x) = -e x^\mu f_\mu(x) = -e \theta(-x \cdot x) \frac{x_0}{|\boldsymbol{\x}|}.
\end{equation} 
This scale invariant $S$ does not fulfil d'Alembert equation in the whole Minkowski space, but fulfils 
d'Alembert equation in space-like region $x\cdot x <0$ outside the light cone, and defines 
the function $S(\psi, \theta, \phi) = -e\textrm{th} \psi$ on de Sitter hyperboloid. 
In fact any non-transversal homogeneous of degree $-1$ solution $f_\mu$ of d'Alembert equation 
determines a unique homogeneous of degree zero solution $S$ of d'Alembert equation 
in the whole Minkowski space by the rule that the solution 
$S$ coincides with $-ex^\mu f_\mu$ outside the light cone. In case of 
(\ref{thpsi-state}) the homogeneous of degree zero solution $S$  of d'Alembert equation coinciding with
$-ex^\mu f_\mu$ outside the light cone is given by the Dirac homogeneous of degree zero solution (\ref{f=homogeneity=0}),
by the results of Subsection \ref{DiracHom=-1Sol}. 

Note that this makes sense although (by the Paley-Wiener theorem) the nuclear space
$\mathcal{S}^{00}(\mathbb{R}^4)$
contains no functions of compact support, so that the localization within this space is much weaker than
within the ordinary Schwartz space $\mathcal{S}(\mathbb{R}^4)$. The above statement that $S$
and $-ex^\mu f_\mu$ coincide outside the cone, and that $-ex^\mu f_\mu$ fulfills d'Alemebert equation
outside the cone (as elements of $\mathcal{S}^{00}(\mathbb{R}^4)^*$) makes sense not because
the distributions $S$ and resp. $-e x^\mu f_\mu$ are regular and are defined by ordinary functions
(\ref{f=homogeneity=0}) and respectively (\ref{distribution-xx(x0/|x|)}) and coincide outside the cone. 
This would be insufficient. 
Indeed, that this assertion makes sense follows from the said regularity and equality,  together with the two
Propositions of Subsection \ref{splitting} (compare also Remark 1 of Subsect. \ref{splitting}),
which assert, among other things, that for any cone determined by any open set 
$\Omega \subset \mathbb{S}^3 \subset \mathbb{R}^4$ there exists an element 
$\varphi \in \mathcal{S}^{00}(\mathbb{R}^4)$ with the support contained in the cone of 
directions $\Omega$.

Summing up the space of classical homogeneous of degree zero scalar-type solutions $S$ of the wave 
equation, i.e. scalar-type solutions of the wave equations on the de Sitter $3$-hyperboloid subsumes all classical electric type homogeneous
of degree $-1$ solutions of the Maxwell equations (i.e. electric-type transversal solutions of the wave equation)
as well as the Coulomb field (at least the Coulomb field solution in spatial region outside the light cone,
sufficient for the determinantion of the charge by the Gauss law).

As shown in \cite{Staruszkiewicz1987} and \cite{Staruszkiewicz}, the constant $Q$ in the classical
phase $S$ solution (\ref{classicalS(x)}) is equal to the total charge computed for
the solution $f^\mu(x) = A^\mu(x)$ of Maxwell equations, for which $-ex_\mu f^\mu$ coincides with  
$S$ outside the light cone or on de Sitter hyperboloid.

The quantum theory
 of the real scalar field $S$ fulfilling the wave equation on de Sitter $3$-hyperboloid
is summarized in axioms supported by the canonical commutation relation
(derived in \cite{Staruszkiewicz1987} and \cite{Staruszkiewicz})
\[
\Big[\frac{1}{e} j_0(x), S(y)\Big]_{{}_{x_0=y_0}} = i \delta(\boldsymbol{\x} - \boldsymbol{\y}),
\]
between the phase field $S(x)$ and the zero component $j_0(x)$ of the electric current density, 
or after integration over the hyperplane $x_0=y_0$ 
\[
[Q, S(x)] = ie, \,\,\,\,\,Q = \int \ud^3 \x \, j_0. 
\]
The axioms are (compare \cite{Staruszkiewicz1987} and \cite{Staruszkiewicz}):

\begin{enumerate}
\item[(I)]
In the Hilbert space $\mathcal{H}$ of the quantum field $S$ there acts a unitary representation $U$ of the 
$SL(2, \mathbb{C})$ group.
\item[(II)]
\[
S(\psi, \theta, \phi) = S_0 -e Q \textrm{th} \psi + \sum \limits_{l=1}^{\infty} \sum \limits_{m = -l}^{m= +l}
\big\{ c_{lm} f_{lm}^{(+)}(\psi, \theta, \phi) + \textrm{h.c.} \big\}, 
\]
is a quantum field, transforming as a scalar field under the action of $U$. 
\item[(III)]
If $M_{\mu\nu}$ stand for the corresponding generators of the unitary representation then
there exists a unique normalized Lorentz invariant vacuum state $|0 \rangle$ in 
$\mathcal{H}$:
\[
M_{\mu \nu}|0\rangle = 0, \,\,\,\langle 0| M_{\mu \nu} = 0,
\]
such that
\[
c_{lm}|0\rangle = 0, \,\,\, \langle 0| c_{lm}^{+} =0, \,\,\, Q|0\rangle = 0, \,\,\,\langle 0 |Q = 0.
\]

\item[(IV)]
\[
[Q, S_0] = ie, \,\,\, [Q, c_{lm}] = 0, \,\,\, [S_0, c_{lm}] = 0,
\]
\[
[c_{lm}, c_{l'm'}^{+}] = 4\pi e^2 \, \delta_{ll'} \delta_{mm'}, \,\,\,
[c_{lm}, c_{l'm'}] = 0.
\]
\item[(V)]
The state $|0\rangle$ is such that the vectors 
\begin{multline*}
(c_{l_1m_1}^+)^{\alpha_1}  \ldots (c_{l_k m_k}^+)^{\alpha_k} e^{ imS_0} |0\rangle, \\
k = 1,2, \ldots \,\, l_i = 1,2, \ldots,  -l_{i} \leq m_i \leq l_i, \alpha_i = 0,1, \ldots, \,\, m \in \mathbb{Z} 
\end{multline*}
span a dense subspace of the Hilbert space of the quantum field $S$. 
\end{enumerate}

Note that the equality 
\begin{equation}\label{[Q,S_0]}
[Q, S_0] = ie
\end{equation}
really means that $S_0$ and $Q$ are supposed to be self adjoint operators such that 
for each smooth function on the spectrum of $S_0$
\[
[Q, f(S_0)] = ie f'(S_0), \,\,\, \textrm{where} \, f'(t) = \frac{df(t)}{dt}
\] 
and $|0\rangle$ is such that 
\[
e^{-inS_0} |0\rangle, \,\,\, n \in \mathbb{Z}
\]
span a complete set of states in the subspace orthogonal to all vectors of the form
\[
(c_{l_1m_1}^+)^{\alpha_1} \ldots  (c_{l_nm_n}^+)^{\alpha_n} e^{-imS_0} 
|0\rangle, \,\,\, n = 1,2,3, \ldots, \,\,
\alpha_i = 1,2,3, \ldots, \, m \in \mathbb{Z},
\]
(note that $\alpha_i$ have to be non zero here, so that the states span the subspace with $l>0$).

As already noted in \cite{Staruszkiewicz} the consistency of the axioms (I)-(V) can be shown 
by noting that there exist a model which realizes the corresponding operators $S_0, Q, c_{lm}, c_{lm}^{+}$
respecting the axioms. Namely we can use the ordinary discrete set of oscillators acting in the ordinary Fock space
$\mathcal{H}_{\textrm{Fock}}$ together with the corresponding annihilation-creation operators (not distributions) ${c'}_{lm}, {c'}_{lm}^{+}$,
the self adjoint and bounded operator $S'_0$  of multiplication by the periodic function 
\[
S'_0(\alpha)= \alpha -n2\pi, \,\,\, n2\pi \leq \alpha < (n+1)2\pi, n \in \mathbb{Z},
\]
\begin{center}
\begin{tikzpicture}[yscale=1]
    \draw[thin, ->] (-2.2,0) -- (2.5,0);
    \draw[thin, ->] (0,-0.5) -- (0,1.5);
    \draw[thin, domain=-2.2:-2] plot(\x, {\x + 3});
    \draw[thin, domain=-2:-1] plot(\x, {\x + 2});
    \draw[thin, domain=-1:0] plot(\x, {\x + 1});
    \draw[thin, domain=0:1] plot(\x, {\x});
    \draw[thin, domain=1:2] plot(\x, {\x - 1});
    \draw[thin, domain=2:2.5] plot(\x, {\x - 2});
\draw[dotted, thin] (-2,0) -- (-2,1);
\draw[dotted, thin] (-1,0) -- (-1,1);
\draw[dotted, thin] (1,0) -- (1,1);
\draw[dotted, thin] (2,0) -- (2,1);

\node [below] at (-2,0) {$-4\pi$};
\node [below] at (-1,0) {$-2\pi$};
\node [below] at (1,0) {$2\pi$};
\node [below] at (2,0) {$4\pi$};
\node [right] at (0,1) {$2\pi$};
\node [right] at (2.6,-0.2) {$\alpha$};
\node [left] at (0,1.5) {$S'_0(\alpha)$};
\end{tikzpicture}
\end{center} 
on the space of  periodic functions square integrable on $\mathbb{S}^1$ (with respect to the invariant Lebesgue measure $\ud \alpha$ on $\mathbb{S}^1$), i.e.
on $L^2(\mathbb{S}^1, \ud \alpha)$; and $Q'$ defined by the extension of the operator  $ie \frac{d}{d \alpha}$ on the 
domain equal to the perfect space $\mathscr{C}^{\infty}(\mathbb{S}^1)$ to a self adjoint operator on  
$L^2(\mathbb{S}^1, \ud \alpha)$.  It is well known that the operator
$ie \frac{d}{d \alpha}$ on the domain $\textrm{Dom} \, \big( ie \frac{d}{d \alpha} \big) 
= \mathscr{C}^{\infty}(\mathbb{S}^1)$ is essentially self adjoint (indeed it is unitarily equivalent to multiplication
operator on a standard, here discrete, measure space -- just apply the Fourier transform on $\mathbb{S}^1$).
Thus the self adjoint operator $Q'$ with $\textrm{Dom} \, Q'$ is uniquely determined.
On application of the Fourier transform on $\mathbb{S}^1$, which defines unitary operator converting
$ie \frac{d}{d \alpha}$ into a multiplication operator on the discrete measure space, one can easily 
see that $\textrm{Dom} \, Q'$ consists of all absolutely continuous functions $f$ on $\mathbb{S}^1$
(i.e. absolutely continuous on $(0, 2\pi)$ and such that $f(0) = f(2\pi)$) and such that
the derivative $f'$ (which exits almost everywhere for absolutely continuous function $f$) 
is square integrable on $\mathbb{S}^1$ with respect to $\ud \alpha$. 
Moreover the operator $ie \frac{d}{d \alpha}$ is essentally self adjoint on many dense subdomains of
$\textrm{Dom} \, Q'$ other than $\mathscr{C}^{\infty}(\mathbb{S}^1)$. For example
$ie \frac{d}{d \alpha}$ is essentially self adjoint on the domain
$\mathscr{C}_{0}^{\infty}(\mathbb{S}^1)$ defined in the Remark below.  

 We can then put
\begin{equation}\label{Hilb-space-rep-clm}
\mathcal{H}_{\textrm{Fock}} \otimes  L^2(\mathbb{S}^1, \ud \alpha) 
\end{equation} 
for the Hilbert space of the quantum field $S$ and 
\begin{equation}\label{rep-clm}
c_{lm} = {c'}_{lm} \otimes \boldsymbol{1}, , \,\,\, 
{c}_{lm}^{+} = {c'}_{lm}^{+} \otimes \boldsymbol{1} \,\,\,
S_0 = \boldsymbol{1} \otimes S'_{0}, \,\,\,
Q =  \boldsymbol{1} \otimes Q'.
\end{equation}

In particular we can apply the white-noise method of Hida-Obata-Sait\^o \cite{hida}
to construct ${c'}_{lm}, {c'}_{lm}^{+}$ and prove the statement that the average of the quantum field $S$
over any smooth Cauchy surface (which is compact) is a self adjoint operator in 
$\mathcal{H}_{\textrm{Fock}} \otimes L^2(\mathbb{S}^1, \ud \alpha)$. Indeed in this case we may put for $\mathscr{O}$ in the standard white noise 
setup of Subsection  \ref{white-setup} the discrete measure space 
$\{(l,m)\}_{{}_{l \in \mathbb{N}, -m \leq m \leq l}}$ with the discrete topology in which every point
point $(l,m) \in \mathscr{O}$ is open, closed and compact as a one element set. We consider a discrete measure
on $\mathscr{O} = \{(l,m)\}_{{}_{l \in \mathbb{N}, -m \leq m \leq l}}$ with the measure of the one point set
$\{ (l,m)\}$ equal $4\pi e^2$. Let $\chi_{{}_{lm}}$ be the characteristic function of the one point set
$\{ (l,m)\}$. We may define the standard operator $A$ by choosing the set
of functions $\frac{1}{2\sqrt{\pi}e}\chi_{{}_{lm}}$ as the complete set 
of its eigenfunctions corresponding to the eigenvalues $l +1$. 
Because $\mathscr{O}$ is discrete topological, then the Kubo-Takenaka conditions 
are trivially preserved by the corresponding nuclear space $\mathcal{S}_{A}(\mathscr{O})$. 
Then we consider the Gelfand triple
\[
\mathcal{S}_{A}(\mathscr{O}) \subset L^2(\mathscr{O}) \subset \mathcal{S}_{A}(\mathscr{O})^*
\] 
with the nuclear space $\mathcal{S}_{A}(\mathscr{O})$ equal to the space of rapidly decreasing sequences
and with the correspoding amplification of the Gelfand triple to the Fock space
\[
\big( \mathcal{S}_{A}(\mathscr{O})\big)  \subset \Gamma\big(L^2(\mathscr{O})\big) = \mathcal{H}_{\textrm{Fock}}
\subset \big( \mathcal{S}_{A}(\mathscr{O})\big)^*.
\]
In this case (compare \cite{hida}) not only ${c'}_{lm}$, $l \in \mathbb{N}$, $-l \leq m \leq l$, transform continuously 
the nuclear space $\big( \mathcal{S}_{A}(\mathscr{O})\big)$ into itself but also
${c'}_{lm}^{+}$ transform  continuously the nuclear space $\big( \mathcal{S}_{A}(\mathscr{O})\big)$
into itself (as $\mathscr{O}$ is a discrete topological space), and in particular by \cite{hida} the integral
\begin{equation}\label{oparatorS'}
S' = \frac{1}{4\pi}\int \limits_{\textrm{C.S.}} S_1 \ud \mu_{{}{\textrm{C.S.}}}
\end{equation}
over a Cauchy surface C.S. of the operator 
\[
S_1 = \sum \limits_{l=1}^{\infty} \sum \limits_{m = -l}^{m= +l}
\{ {c'}_{lm} f_{lm}^{(+)}(\psi, \theta, \phi) +  {c'}_{lm}^{+} \overline{f_{lm}^{(+)}(\psi, \theta, \phi)}\}
\]
is a well defined operator transforming continuously the nuclear space
$\big( \mathcal{S}_{A}(\mathscr{O})\big)$ into itself. 
This follows from Thm. 2.6 of \cite{hida}. Indeed, for rapidly decreasing sequances 
$\{s_{lm}\} \in \mathcal{S}_{A}(\mathscr{O})$, and for 
\[
\begin{split}
F^{(+)}_{lm} = \frac{1}{4\pi}\int \limits_{\textrm{C.S.}} \,
 f_{lm}^{(+)}(\psi(\theta, \phi), \theta, \phi) \,
 \ud \mu_{{}{\textrm{C.S.}}}(\theta, \phi), \\
\overline{F^{(+)}_{lm}} = \frac{1}{4\pi}\int \limits_{\textrm{C.S.}} \,
 \overline{f_{lm}^{(+)}(\psi(\theta, \phi), \theta, \phi)} \,
 \ud \mu_{{}{\textrm{C.S.}}}(\theta, \phi),
\end{split}
\]
the functionals 
\[
\begin{split}
\{s_{lm}\} \rightarrow F^{(+)}(\{s_{lm}\}) = \sum \limits_{l=1}^{\infty} \sum \limits_{m = -l}^{m= +l}
F^{(+)}_{lm} s_{lm}, \\
\{s_{lm}\} \rightarrow \overline{F^{(+)}}(\{s_{lm}\}) = \sum \limits_{l=1}^{\infty} \sum \limits_{m = -l}^{m= +l}
\overline{F^{(+)}_{lm}} s_{lm}
\end{split}
\]
belong to $\mathcal{S}_{A}(\mathscr{O})^*$. 
Because the operator $S'$ is symmetric and because every nuclear 
space is perfect, then by the Riesz and Sz\"okefalvy-Nagy criterion \cite{Riesz-Szokefalvy} (p. 120 in Russian 1954 Ed.)
the said operator (\ref{oparatorS'}) on the domain $\big( \mathcal{S}_{A}(\mathscr{O})\big)
\subset \Gamma\big(L^2(\mathscr{O})\big)$ has a self adjoint extension to (an unbounded) self adjoint operator in 
$\Gamma\big(L^2(\mathscr{O})\big) = \mathcal{H}_{\textrm{Fock}}$. In general the
Riesz and Sz\"okefalvy-Nagy criterion does not exclude existence of more than just one self-adjoint extension.
But because the one-parameter unitary group generated by $S'$ leaves invariant the dense nuclear
Hida's test space $\big( \mathcal{S}_{A}(\mathscr{O})\big)$, then by general theory, e.g. \cite{Segal_Kunze},
p. 289, $S'$ is essentially self-adjoint on $\big( \mathcal{S}_{A}(\mathscr{O})\big)$, i.e. admits
just one self adjoint extension. Thus it follows in particular the following
lemma (on application of the general theorem on tensor products
of essentially self adjoint operators \cite{Reed_Simon}, Ch. VIII.10, and self adjointness of the operator
$S'_{0} - eQ'$, regarded as an operator on the domain $\textrm{Dom} \, Q'$, and essential self adjointness of
$S'$). 
\begin{lem*}
For any smooth Cauchy surface \emph{(C.S.)} on the de Sitter $3$-hyperboloid, the integral
\begin{multline*}
S(\textrm{\emph{C.S.}}) = \frac{1}{4\pi} \int \limits_{\textrm{C.S.}} 
S(x) \, \ud \mu_{{}{\textrm{C.S.}}}(x) = c_1 S_0 - c_2 eQ +  S' \otimes \boldsymbol{1} \\
= c_1 \boldsymbol{1} \otimes S'_{0}  -  c_2 \boldsymbol{1} \otimes eQ' +  S' \otimes \boldsymbol{1}, 
\end{multline*}
with the constants $c_1,c_2$ depending on the Cauchy surface,
is essentially self adjoint operator in the Hilbert space of the quantum scalar field 
$S$ on the $3$-hyperboloid $\{x, x \cdot x = -1\}$, on the domain
$\big( \mathcal{S}_{A}(\mathscr{O})\big) \otimes \textrm{Dom} \, Q'$. It is essentially self adjoint 
also on the invariant domain $\big( \mathcal{S}_{A}(\mathscr{O})\big) \otimes \mathscr{C}^{\infty}_{0} (\mathbb{S}^1)$, 
defined in the Remark below.  In the above formula 
$\ud \mu_{{}{\textrm{C.S.}}}(x)$ is the induced measure on the Cauchy surface. In particular for any measurable and periodic function
$f$ the operator $f(S(\textrm{C.S.}))$ is a well defined normal (self adjoint if $f$ is real valued) operator. 
In particular $S(\textrm{\emph{C.S.}})$ can be exponentiated
and 
\[
e^{i S(\textrm{\emph{C.S.}})}
\]
is a unitary operator\footnote{It belongs to the ``folklor knowledge'' that the free real quantum field on a globally hyperbolic spacetime, integrated over a compact subset of a Cauchy surface (in the case of de Sitter 
$3$-hyperboloid the whole of Cauchy surface is compact), is a densely defined self adjoint operator. We have to be careful however because Cauchy surface has one dimension less then the space-time itself. Integral with a test function of compact support (of full space-time dimension) would be a well defined operator by construction of the field, but this is not this simple situation.  The massless fields on the flat Minkowski space time still behave much worse than on space-times of constant curvature with compact Cauchy surfaces. Compare our proof of Bogoliubov-Shirkov Hypothesis, where the proof of a similar ``folklor knowledge'' statement requires a much work. Lacking of a precise mathematical status in construction of such operators for fields on Minkowski space-time, e.g. in the works of A. Jaffe and J. Glimm (e.g. ``Wick polynomials at fixed times'', \emph{J. Math. Phys.} {\bf 7} (1966) 1250-1255), was noticed by Segal \cite{Segal-NFWP.I}.  
It is the compactness of the Cauchy surface on the 
$3$-hyperboloid and its non zero curvature which saved the quantum theory of the scalar field on the $3$-hyperboloid from the distribution-type- subtleties.  One should note that the proper treatment of the propagator distributions of massless fields on the flat Minkowski spacetime require much more care and the correct manipulations with them is much more difficult to control in comparison with the propagators of the massless fields on globally hyperbolic symmetric space-times of constant non zero (negative or positive) curvature with compact Cauchy surfaces. It was already noticed by many authors. For example Segal, Zhou and Paneitz, \cite{SegalZhouPhi4}, \cite{SegalZhouQED}, 
\cite{PaneitzSegalI}-\cite{PaneitzSegalIII}, have worked out in details the case of the $: \varphi^4 :$ scalar theory as well as the QED on the (static) Einstein Universe space-time, compare also Section
\ref{EUandG}. }. 
\end{lem*}

In particular for the Cauchy surface on the de Sitter $3$-hyperboloid determined by the intersection of the space like 
hyperplane $u \cdot x = g_{\mu \nu}u^\mu x^\nu = 0$ with the hyperboloid $x \cdot x = -1$, where 
$u$ is any unit (i.e. $u \cdot u = 1$) time like vector, the integral
\begin{multline*}
S(u) = \frac{1}{4\pi} \int \limits_{\{u \cdot x =0 \} \cap \{x \cdot x = -1 \}} 
S(x) \, \ud \mu_{{}_{\{u \cdot x =0 \} \cap \{x \cdot x = -1 \}}}(x) = c_1 S_0 - c_2 eQ +  S' \otimes \boldsymbol{1} \\
= S_0 +  S' \otimes \boldsymbol{1} = \boldsymbol{1} \otimes S'_{0} +  S' \otimes \boldsymbol{1}, 
\end{multline*}
because for this particular Cauchy surface the integration constants $c_1=1, c_2=0$.
In particular if the partial waves $f_{lm}^{(+)}$ on de Sitter hyperboloid are computed in the reference frame in which $u$ is the unit time like vector along the time like axis, then 
\begin{equation}\label{S(u=(1,0,0,0))}
S(u) = S_0 +  S' \otimes \boldsymbol{1},
\end{equation}
on the dense, invariant for $S_0$ and $Q$, essentially self adjoint on the nuclear subspace
$\big(\mathcal{S}_{A}(\mathscr{O})\big) \otimes \mathscr{C}^{\infty}_{0} (\mathbb{S}^1)$ of the Hilbert space of the field $S$, defined in the Remark closing
this Subsection, or 
on $\big(\mathcal{S}_{A}(\mathscr{O})\big) \otimes \textrm{Dom} \, Q'$. 
By the the axioms (I)-(V) (in particular by (\ref{[Q,S_0]}) and the Baker-Hausdorff-Campbell
formula) it follows that in this reference frame 
\begin{equation}\label{|u>|u= (1000)}
|u \rangle = e^{-iS(u)} |0\rangle = e^{-i S_0} |0\rangle,  
\end{equation}
up to an irrelevant constant phase factor\footnote{Here $e$ in the exponent index of $e^{ie/2}$ is the constant
present in the commutation rules (IV), and not the basis of the natural logarithms.} $e^{ie/2}$. 

One can check by explicit computations, compare \cite{Staruszkiewicz}, \cite{Staruszkiewicz1992}, 
\cite{Staruszkiewicz1995}, that the quantum field $S$ on the de Sitter $3$-hyperboloid respecting
(I)-(V) in the concrete representation (\ref{Hilb-space-rep-clm})
and (\ref{Hilb-space-rep-clm}) is indeed a quantum scalar field with the transformation rule of scalar field
under the representation $U$, and moreover one can compute the representation $U$ explicitly. It likewise follows from (I)-(V) the following relation
\[
[Q, S(x)] = ie \boldsymbol{1}
\]
on the dense nuclear subspace 
\[
\big(\mathcal{S}_{A}(\mathscr{O})\big) \otimes \mathscr{C}^{\infty}_{0} (\mathbb{S}^1)
\subset \mathcal{H}_{\textrm{Fock}}  \otimes L^2(\mathbb{S}^1)
\]
of the Hilbert space of the field $S$ defined in the Remark closing
this Subsection.

Note that the essential self adjointness of $S(u)$, for any unit time-like $u$, easily follows from the consistency of the relations (I)-(IV)
in the specified above concrete representation (\ref{Hilb-space-rep-clm})
and (\ref{Hilb-space-rep-clm}).
In particular for the indicated specific representation of (I)-(IV), existence
of the unitary representation $U$ for which $S(x)$ is a scalar field respecting (I)-(IV), implies easily 
this assertion. Indeed  $S(u)$, when computed in the reference frame 
in which the partial waves $f_{lm}^{(+)}$ on de Sitter hyperboloid are computed (i.e. $u = (1,0,0,0)$),
becomes equal (\ref{S(u=(1,0,0,0))}) on the invariant dense domain
$\big(\mathcal{S}_{A}(\mathscr{O})\big) \otimes \mathscr{C}^{\infty}_{0} (\mathbb{S}^1)$,
on which (\ref{S(u=(1,0,0,0))}) is essentially self adjoint (compare Remark below). But in order to compute
$S(u)$ for $u \neq(1,0,0,0)$, it is sufficient to apply the unitary transformation $U_\Lambda$
for the Lorentz transformation $\Lambda$ which transforms $(1,0,0,0)$ into $u$, and the hyperplane
$x_0=0$ into the hyperplane $u \cdot x = 0$. Thus we prove in this way that $S(u)$ for each $u$
in the Lobachevsky hyperboloid, is unitarily equivalent to an essentially self adjoint operator
(\ref{S(u=(1,0,0,0))}) on the invariant domain 
$\big(\mathcal{S}_{A}(\mathscr{O})\big) \otimes \mathscr{C}^{\infty}_{0} (\mathbb{S}^1)$
(likewise invariant for the unitary representation $U$). Nonetheless we have
indicated the relation of the Lemma to the white noise calculus, because the  proof indicated above and 
using white noise calculus is more general, and can be applied for other fields on the
de Sitter $3$-hyperboloid space-time.

In particular by the results of \cite{Staruszkiewicz1995} the unitary representation of $SL(2, \mathbb{C})$
acting on the invariant subspace spanned by the vectors $c_{lm}^+ |0\rangle$ is exactly equal to the irreducible unitary representation which Gelfand, Minlos and Shapiro denoted by $(l_0 = 1, l_1 =0) = \mathfrak{S}(m=2, \rho = 0)$ in their book
\cite{Geland-Minlos-Shapiro} with the vectors $c_{lm}^+ |0\rangle$ corresponding to the vectors
$\xi_{lm}$ of Gelfand-Minlos-Shapiro book, pages 188-189. 

Therefore the subspace of states spanned by the vectors $c_{lm}^+ |0\rangle$ should be identified
with the Hilbert space of electric-type infrared transversal states 
${\mathcal{H}''}_{{}_{\chi = -1}}^{e} \cong_U {\mathcal{H}'}_{{}_{\chi = -1}}^{e}$ understood as the Fourier transforms of scalar
homogeneous of degree zero (or resp. $-2$) solutions of d'Alembert equation belonging 
to $\mathcal{S}^{00}(\mathbb{R}^4; \mathbb{C})$, described above. The identification can be realized in such
a manner that the action of $SL(2, \mathbb{C})$  through $\boldsymbol{U}_{{}_{\chi=-1, \,e}}$ will coincide
with the action of the representation $(l_0 = 1, l_1=0) = \mathfrak{S}(m=2, \rho = 0)$ on the subspace 
spanned by $c_{lm}^+ |0\rangle$.
In this way the invariant subspace of transversal states spanned by
\begin{equation}\label{Q=0-states}
(c_{l_1m_1}^+)^{\alpha_1} \ldots  (c_{l_nm_n}^+)^{\alpha_n}|0\rangle, \,\,\, n = 1,2,3, \ldots, \,\,
\alpha_i = 0,1,2,3, \ldots.
\end{equation}  
(note that (\ref{Q=0-states}) include $|0\rangle$), acted on by $\Gamma(l_0=1,l_1=0) = \Gamma(\mathfrak{S}(m=2, \rho = 0))$ will be identical with the subspace
\[
\Gamma({\mathcal{H}'}_{{}_{\chi = -1}}^{e})
\]
acted on by 
\begin{equation}\label{Utre}
\Gamma\big(\boldsymbol{U}_{{}_{\chi=-1, \,e}}\big)
= \Gamma\big( \,(l_0=1, l_1=0)  \,  \big)
= \Gamma\big( \, \mathfrak{S}(m=2, \rho = 0) \, \big).  
\end{equation}
By the results of Neumark \cite{nai1}-\cite{nai2}
\[
\Gamma\big(\boldsymbol{U}_{{}_{\chi=-1, \,e}}\big) \cong_U \Gamma(\mathfrak{S}(m=2, \rho = 0))
\cong_U [\infty] \underset{m\in 2\mathbb{Z}}{\oplus} \int\limits_{\mathbb{R}}\mathfrak{S}(m=2, \rho) \, \ud \rho \oplus \boldsymbol{1}.
\]
Here $\ud \rho$ is the Lebesgue measure on $\mathbb{R}$, and $[\infty]$ means that for each $m$ the representation
\[
 \int\limits_{\mathbb{R}}\mathfrak{S}(m=2, \rho) \, \ud \rho 
\]
enters into the decomposition with uniform infnite multiplicity, and finally $\boldsymbol{1}$
denotes the trivial representation on $\mathbb{C}$

The wave functions $f_{lm}^{(+)}$ of Staruszkiewicz theory, when uniquely extended byhomogeneity over the whole space-time,  
become identical
with the restrictions
of homogeneous of degree zero solutions $f_{lm}^{(+)}$ in  (\ref{xAhom=-1}) of d'Alembert equation
to the de Sitter hyperboloid. Finally, the operators $c_{lm}, c_{lm}^{+}$ of Staruszkiewicz theory in (II)
become identical with the operators ${c'}_{lm}, {c'}_{lm}^{+}$ in (\ref{xAhom=-1}).
More precisely the operators  ${c'}_{lm}, {c'}_{lm}^{+}$ in (\ref{rep-clm})
should be identified with the operators ${c'}_{lm}, {c'}_{lm}^{+}$ in (\ref{xAhom=-1}). 

In particular in the degenerate case of the Staruszkiewicz theory, restricted to the zero charge eigenspace $\mathcal{H}_{0}$, 
we can use the representation (\ref{Hilb-space-rep-clm}),
(\ref{rep-clm}) with $L^2(\mathbb{S}^1)$ replaced by $\mathbb{C}$, and with the operator 
$Q'$ acting on the first factor $\mathbb{C}$ (replacing $L^2(\mathbb{S}^1)$ in (\ref{Hilb-space-rep-clm}))
as the zero operator. We thus obtain the equality (\ref{xAechi=-1(x)=transversalS}) of Subsection \ref{AS},
or equality up to unitary equivalence $V$ with the unitary operator $V$ defined as in Subsection \ref{AS}.

But it turns out that also in case of the full non degenerate case 
of Staruszkiewicz theory, the Hilbert space $\mathcal{H}$ of the quantum phase $S(x)$
has the tensor product structure (\ref{Hilb-space-rep-clm})) 
 on which the operators $c_{lm}, c_{lm}^{+}$
have the form (\ref{rep-clm}), and that the operators 
${c'}_{lm}, {c'}_{lm}^{+}$ in (\ref{rep-clm})
coincide with the operators ${c'}_{lm}, {c'}_{lm}^{+}$ in (\ref{xAhom=-1})
defining the field (\ref{xAhom=-1}) with $A_{{}{\chi=-1}}^{e \, \mu} (x)$ equal to the 
homogeneous of degree $\chi = -1$ electric part of the free electromagnetic potential field $A^\mu$, 
and the equality (\ref{xAechi=-1(x)=transversalS}) can be extended over the whole Hibert space 
$\mathcal{H}$ of the phase field $S$, for the proof compare Subsection
\ref{Comparison2}. 

Now let us go back to the consistency of the axioms (I)-(V) in the concrete representation
(\ref{Hilb-space-rep-clm}) and (\ref{rep-clm}).
This consistency is equivalent to the construction of the unitary representation $U$
of the $SL(2, \mathbb{C})$ which makes the field $S(x)$ a scalar field. This representation
has been constructed almost explicitly in 
\cite{Staruszkiewicz1995} and \cite{Staruszkiewicz1992ERRATUM}, and in fact it has been proved constructively 
in \cite{Staruszkiewicz1995} that the representation acting on the subspace of states generated by 
(\ref{Q=0-states}) is indeed unitary and equal to 
the representation $\Gamma\big((l_0 = 1, l_1=0) \big)$. In the Subsection \ref{Ustructure}
using the results of Staruszkiewicz obtained in \cite{Staruszkiewicz1992ERRATUM}, \cite{Staruszkiewicz1992}
and \cite{Staruszkiewicz1995} we construct explicitly the representation 
$U$. Consistency (compare also \cite{Staruszkiewicz1992ERRATUM}) imlies in particular that the mapping 
\begin{equation}\label{kernelLobachevskyAS}
u \times v \mapsto \langle 0| e^{iS(u)}e^{-iS(v)} |0\rangle = 
\langle u| v \rangle = \exp\Big\{ - \frac{e^2}{\pi}(\lambda \textrm{coth} \lambda - 1) \Big\},
\end{equation}
for $u,v$ ranging over the Lobachevsky space $\mathscr{L}_3$: $u\cdot u = 1$, $v \cdot v = 1$, is equal to an invariant 
positive definite kernel on the Lobachevsky space. Here $\lambda$ is the hyperbolic angle
between $u$ and $v$: $\textrm{cosh} \, \lambda = u \cdot v$. This assertion of course would immediately follow 
from the consistency of (I)-(V) in the representation (\ref{Hilb-space-rep-clm}) and (\ref{rep-clm}),
but theorem that (\ref{kernelLobachevskyAS}) 
is an invariant positive definite kernel is not yet sufficient for the consistency of (I)-(V).
We have already shown in the second Proposition of Subsection \ref{AS}, that the function (\ref{kernelLobachevskyAS}) 
defines indeed an invariant positive definite kernel on the Lobachevsky space, 
using the Schoenberg's theorem on conditionally negative functions. The  full and costructive consistency
proof of the axioms (I)-(V) in the concrete representation (\ref{Hilb-space-rep-clm}) and (\ref{rep-clm}) (compare
also the Remark), is given in Subsection \ref{Consistency}.

Note also that in the work \cite{Staruszkiewicz} assumption 
that (\ref{kernelLobachevskyAS}) is positive definite is implicitly used. In 
\cite{Staruszkiewicz1992ERRATUM} it was obtained the Fourier transform of 
(\ref{kernelLobachevskyAS}), regarded as a distribution on $\mathscr{L}_3 \times \mathscr{L}_3$, 
and thus its decomposition into the Fourier integral (with the Fourier transform of Gelfand-Graev-Vilenkin
on the Lobachevsky space). In particular in
\cite{Staruszkiewicz1992ERRATUM} one finds the following decomposition\footnote{In the abstract of the 
paper \cite{Staruszkiewicz} it was placed a  statement which might mislead the reader into thinking that 
$0 < e^2/\pi < 1$ is a necessary condition for the positive definitenes of the kernel 
(\ref{kernelLobachevskyAS}),  i.e. for the positive definitenes of the inner product of the theory. 
It was clearly stated in the following paper \cite{Staruszkiewicz1992ERRATUM}, that the value $e^2/\pi = 1$ is critical 
in the sense that it separates two domains in which the kernel (\ref{kernelLobachevskyAS}) 
is positive definite but behaves differently in them. For $e^2/\pi<1$ there is present the discrete 
supplementary series representation (corresponding to the second term in (\ref{decompositionAS})) 
in the decomposition of the representation of  $SL(2, \mathbb{C})$ in the reproducing kernel Hilbert space defined by the kernel 
(\ref{kernelLobachevskyAS}). For $e^2/\pi >1$ it is absent.} 
(here $z = e^2/\pi$ with the universal constant $e$ entering the above stated axioms (I)-(V),
 and the second term below is absent for $z>1$)\footnote{In the formula 
(\ref{decompositionAS}) the constant $e$ stands for the basis of natural logarithms.}
\begin{multline}\label{decompositionAS}
\langle f | f\rangle = {\textstyle\frac{1}{(2\pi)^3}} \int \limits_{0}^{\infty} \,
d \nu \, \nu^2 \, K(\nu; z) \, \int \limits_{\mathbb{S}^2} \, d^2 p \,
|\mathcal{F}f(p;\nu)|^2 \\ +
{\textstyle\frac{(1-z)^2(2e)^z}{16\pi^2}} \int \limits_{\mathbb{S}^2 \times \mathbb{S}^2} 
\frac{d^2p \, d^2 k}{(p \cdot k)^z} \,\, \overline{\mathcal{F}f(p;i(1-z))} \,\,
\mathcal{F}f(k;i(1-z))
\end{multline}
for $|f\rangle = \int du f(u) |u\rangle$ with smooth $f$ of compact support on the Lobachevsky
space $u \cdot u = 1$, with the invariant measure $du$ on the Lobachevsky space. The second term 
in (\ref{decompositionAS}) is present if and only if $0 < z <1$. The formula (\ref{decompositionAS}) 
for the kernel (\ref{kernelLobachevskyAS}), evaluated for arbitrary (but fixed) compactly supported test 
functions $f, g=f$ on the Lobachevsky space, 
is computed first, using invariance of (\ref{kernelLobachevskyAS}), in the domain $z>1$
 (it does not contain the second term). Next using analyticity in $z,\nu$
it is computed, by analytic continuation, in the domain $0<z<1$ (in \cite{GelfandI} 
the reader will find general theory underlying the method used in \cite{Staruszkiewicz1992ERRATUM}).  
Here the Gelfand-Graev-Vilenkin inverse Fourier transform of $f$ on the Lobachevsky space is used
\[
f(u) = {\textstyle\frac{1}{(2\pi)^3}} \int \limits_{0}^{\infty} \,
d \nu \, \nu^2 \, \int \limits_{\mathbb{S}^2} \, d^2 p \,
\mathcal{F}f(p;\nu) \, (p \cdot u)^{-i\nu-1}
\] 
together with the Gelfand-Graev-Vilenkin Fourier transform  $\mathcal{F}f$ of $f$ 
on the Lobachevsky space, equal
\[
\mathcal{F}f(p;\nu) = \int du \, f(u) \, (p \cdot u)^{i\nu-1},
\]
which is  a homogeneous of degree $i\nu-1$ function of $p$ on the positive sheet of the cone
(and thus with $\mathcal{F}f(p;i(1-z)$ homogeneous of degree $z-2$ in $p$), and which, in principle, can be understood 
as a distribution in $\mathcal{S}^{0}(\mathbb{R}^3)^*$ (and canonically as a distribution
in $\mathcal{S}^{0}(\mathbb{R}^4)^*$ with the support equal to the positive sheet of the cone, 
whose Fourier transform belongs to $\mathcal{S}^{0}(\mathbb{R}^4)^*$ and fulfils d'Alembert equation, 
by the results of Subsection \ref{DiracHom=-1Sol}). 
But the Fourier decomposition (\ref{decompositionAS}) can be computed as in \cite{Staruszkiewicz1992ERRATUM}
without the assumption of positive definitenes of (\ref{kernelLobachevskyAS}) (as the invariance of 
(\ref{kernelLobachevskyAS}) is evident).
Positive definiteness of (\ref{kernelLobachevskyAS}) is equivalent to the positivity of the weight function $K(\nu;z=e^2/\pi)$:
\[
K(\nu;z) = -{\textstyle\frac{4\pi}{\nu}}z^2 e^{z} 
\sum\limits_{n=-\infty}^{+\infty}{\textstyle\frac{[\nu+i(2n+1-z)]^{n-1}}{[\nu+i(2n+1+z)]^{n+2}}}
\]
in (\ref{decompositionAS}) for
each positive real $\nu$. However, the positivity of the weight function 
$K(\nu;e^2/\pi)$ is not evident, compare \cite{Staruszkiewicz1992ERRATUM}.

Nonetheless, positivity of $K(\nu;e^2/\pi)$ follows from the positive definiteness of the 
invariant kernel (\ref{kernelLobachevskyAS}) by the generalization of the Bochner's theorem (due to Krein-Naimark-Gelfand \cite{Neumark_dec}, \S 31.10)
extended on the relation between positive measures on the set of irreducible $K$-spherical unitary representations 
of semi-simple Lie groups $G$ and the corresponding positive definite kernels on $G/K \times G/K$
(or positive definite functions
on $G$, corresponding to positive definite kernels on $G/K \times G/K$), compare e.g. \cite{Gangolli}, 
with $G= SL(2, \mathbb{C})$, $K = SU(2, \mathbb{C})$ and with the Lobachevsky space $G/K = \mathscr{L}_3$
as the homogeneous Riemannian manifold. Recall, please, that a continuous function $\varphi$ on $G$ is called positive definite iff
\[
\sum_{i,j} \varphi(g_{i}^{-1}g_{j}) \alpha_i \overline{\alpha_j} \geq 0
\]
for any finte set of complex numbers $\alpha_i$, and any finite set of $g_i \in G$; and that $\varphi$ is $K$-spherical
iff $\varphi(k_1gk_2) = \varphi(g)$, for all $g \in G$, $k_1,k_2 \in K\subset G$. $\varphi$ is called normalized iff
$\varphi(e)=1$. Recall, that a unitary representation $U$ of $G$ is called $K$-spherical if there exists a unit vecor $v\in H(U)$,
such that $U(k)v= v$ for each $k\in K$. Recall further that to any $K$-spherical
unitary representation $U$ of $G$ there corresponds the $K$-spherical continuous normalized and positive definite
function $\varphi(g) = (U(g)v,v)$, where $(\cdot, \cdot)$ is the inner product in the Hilbert space $H(U)$ of the reprsentation $U$. 
If the $K$-spherical representation $U$ is cyclic, with the cyclic vector $v$ which is invariant under $U(k)$, $k\in K$,
then the correspondence between the unitary equivalence class of $U$ and the spherical function $\varphi(g) = (U(g)v,v)$
is bi-unique. This is in paricular the case for irreducible $K$-spherical $U$. 
If the $K$-spherical representation $U$ is irreducible, then
the corresponding  $K$-spherical normalized and positive definite
function $\varphi$ is called \emph{elementary}. Recall that each positive definite kernel $\kappa$ on $G/K \times G/K$
can be lifted to the bi-uniquelly corresponding positive definite function $\varphi$ on $G$ through the formula
$\varphi(g)= \kappa(gK, eK)$ (compare e.g. \cite{Gangolli}). Let $K$ be maximal compact subgroup of a semisimple $G$. 
Finally, let $\nu \in \mathscr{M}$ be all equivalence classes of 
all irreducible $K$-spherical representations $U_{{}_{\nu}}$ of a semisimple Lie group $G$
and let $\varphi_{{}_{\nu}}$ be the elementary normalized positive definite $K$-spherical
functions correspoding to the irreducible representants $U_{{}_{\nu}}$, one for each unitary equivalence class. 
Then we have the following generalization of Bochner theorem
(compare \S 31.10, eq. (4), p. 426 of \cite{Neumark_dec}, or Theorem 3.23 in the review article \cite{Gangolli}):
\begin{twr*}[{\bf Generalized Bochner theorem}]
Let $\varphi$ be a continuous positive definite $K$-spherical function on $G$. Then
there exists a unique nonnegative measure $\mu$ on $\mathscr{M}$ such that
\[
\varphi(g) = \int\limits_{\mathscr{M}} \varphi_{{}_{\nu}}(g) \, d\mu(\nu),
\,\,\,\,\,\,\,\,\,\,\,\,\,\,\,\, g \in G.
\]
\end{twr*}
The Fourier transform of the kernel (\ref{kernelLobachevskyAS}) found in \cite{Staruszkiewicz1992ERRATUM} 
(or decomposition  (\ref{BochnerKernelDecomposition})
of the representation corresponding to the kernel (\ref{kernelLobachevskyAS}) or equivalently to the kernel (\ref{kernel-t}) with $4\pi t$ put equal $e^2/\pi$) 
applied to the spherical positive definite function $\varphi$ corresponding to the kernel (\ref{kernelLobachevskyAS}) gives decomposition 
of $\varphi$ into the elementary spherical functions $\varphi_{{}_{\nu}}$ with the same measure $\mu$ on the set of equivalence classes of spherical 
unitary irreducible representations of $G=SL(2, \mathbb{C})$ as in (\ref{decompositionAS}). 
On the set of spherical representations
of the principal series
it is given by the Lebesgue mesure $d\nu$ on $\mathbb{R}_+$ with the weight function equal 
$(2\pi^2)^{-1}\nu^2 K(\nu;e^2/\pi)$: $d\mu(\nu) = (2\pi^2)^{-1} \nu^2 K(\nu;e^2/\pi) d\nu$. 
Therefore, the generalized Bochner theorem  (eq. (4), p. 426 of \cite{Neumark_dec} or Theorem 3.23 of \cite{Gangolli}) implies
positivity of the measure $d\mu(\nu) = (2\pi^2)^{-1} \nu^2 K(\nu;e^2/\pi) d\nu$. 
Thus, positivity of the weight function
$K(\nu;e^2/\pi)$ in (\ref{decompositionAS}) follows for almost all $\nu$. By the analyticity of
$K(\nu;e^2/\pi)$ in both arguments (compare \cite{Staruszkiewicz1992ERRATUM}) positivity of  
$K(\nu;e^2/\pi)$ in ordinary sense follows. Let us explain it in more detail. 
Recall, please, that the unitary irreducible representations of the principal series $\mathfrak{S}(m=0, \nu)$, $\nu \in \mathbb{R}$ and 
of the supplementary series $\mathfrak{D}(\nu_0)$, $\nu_0 \in [0,1]$, exhaust all irreducible equivalence classes of all spherical unitary irreducible
representations of $G= SL(2, \mathbb{C})$. Next, recall that the Fourier decomposition
of the kernel $\kappa(u,v) = \langle u | v \rangle$, given by (\ref{kernelLobachevskyAS}), and found in \cite{Staruszkiewicz1992ERRATUM},
is equal
\begin{multline}\label{kerneldecompositionAS}
\langle u | v \rangle = {\textstyle\frac{1}{(2\pi)^3}} \int \limits_{0}^{\infty} \,
d \nu \, \nu^2 \, K(\nu; z) \, \int \limits_{\mathbb{S}^2} \, d^2 p \,
\overline{(p\cdot u)^{i\nu-1}}(p\cdot v)^{i\nu-1} \\ +
{\textstyle\frac{(1-z)^2(2e)^z}{16\pi^2}} \int \limits_{\mathbb{S}^2 \times \mathbb{S}^2} 
\frac{d^2p \, d^2 k}{(p \cdot k)^z} \,\, \overline{(p\cdot u)^{z-2}} \,\,
(k\cdot v)^{z-2} 
\\
= \kappa(u,v) = \kappa(gK, eK) =  \varphi(g),
\\
u = gv = \Lambda(g^{-1})v, \,\,\, v= (1,0,0,0), \,\,\, g \in SL(2, \mathbb{C}), 
\end{multline}
here with $e$ equal to the basis of natural logarithms.
Recall that the $d^2 p$ integral over $\mathbb{S}^2$ in the first summand is equal to the inner product of two functions
\[
p \rightarrow (p\cdot u)^{i\nu-1}, \,\,\,\, p \rightarrow (p\cdot v)^{i\nu-1}
\]
homogeneous of degree $i\nu -1$ on the positive cone $p\cdot p =0$, $p_0>0$, in the Hilbert 
space of the unitary spherical representation of the principal series $\mathfrak{S}(m=0, \nu)$, realized as the closure
with this inner product of all homogeneous of degree $i\nu -1$ continuous functions on the cone.   
The double integral $d^2 p \times d^2 k$ over $\mathbb{S}^2 \times \mathbb{S}^2$ of the second summand in (\ref{kerneldecompositionAS})
is equal  to the inner product of two functions
\[
p \rightarrow (p\cdot u)^{z-2}, \,\,\,\, p \rightarrow (p\cdot v)^{z-2}
\]
homogeneous of degree $z-2$ on the positive cone $p\cdot p =0$, $p_0>0$, in the Hilbert 
space of the unitary spherical representation of the supplementary series $\mathfrak{D}(\nu_0 = 1-z)$, realized as the closure
with this inner product of all homogeneous of degree $2-z$ continuous functions on the cone. It is easily seen that
the function  
\[
p \rightarrow {\textstyle\frac{1}{\sqrt{4\pi}}} (p\cdot v)^{i\nu-1},  \,\,\,\, v= (1,0,0,0),
\]
represents the unit state invariant under the action of all elements of $K=SU(2, \mathbb{C})$ in the Hilbert space of 
the $K$-spherical irreducible representation $\mathfrak{S}(m=0, \nu)$ of the principal series, in the realization stated above. Similarly, 
\[
p \rightarrow \left({\textstyle\frac{(1-z)2^z}{16 \sqrt{2} \pi^{5/2}}}\right)^{1/2} (p\cdot v)^{z-2},  \,\,\,\, v= (1,0,0,0),
\]
is the function representing the unit state invariant under the action of all elements of  $K=SU(2, \mathbb{C})$ 
in the Hilbert space of the $K$-spherical representation
$\mathfrak{D}(\nu_0 = 1-z)$ of the supplementary series, in the realization stated above. Therefore the elementary positive definite 
and normalized spherical functions $\varphi_{{}_{\nu}}$ and $\varphi_{{}_{\nu_0}}$ corresponding, respectively, to the spherical representations
of the classes $\mathfrak{S}(m=0, \nu)$ and $\mathfrak{D}(\nu_0 = 1-z)$ are, respectively, equal
\[
\varphi_{{}_{\nu}}(g) = 
{\textstyle\frac{1}{4\pi}}
\int \limits_{\mathbb{S}^2} \, d^2 p \,
\overline{(p\cdot u)^{i\nu-1}}(p\cdot v)^{i\nu-1}, \,\,\, v=(1,0,0,0), \, u = gv = \Lambda(g^{-1})v
\]
\begin{multline*}
\varphi_{{}_{\nu_0}}(g) =
{\textstyle\frac{(1-z)2^z}{16 \sqrt{2} \pi^{5/2}}} \int \limits_{\mathbb{S}^2 \times \mathbb{S}^2} 
\frac{d^2p \, d^2 k}{(p \cdot k)^z} \,\, \overline{(p\cdot u)^{2-z}}(k\cdot u)^{z-2},
\\
\,\,\, v=(1,0,0,0), \, u = gv = \Lambda(g^{-1})v.
\end{multline*}
Inserting this formulas into (\ref{kerneldecompositionAS})
we obtain
\begin{multline}\label{decompositionOfvarphi}
\varphi(g)
= \int\limits_{\mathbb{R_+}} \varphi_{{}_{\nu}}(g) \ud \mu(\nu) + \int\limits_{[0,1]} \varphi_{{}_{\nu_0}}(g) d\mu(\nu_0)
\\
= {\textstyle\frac{1}{2\pi^2}} \int \limits_{0}^{\infty} \, \varphi_{{}_{\nu}}(g) \,
\, \nu^2 \, K(\nu; z) d\nu  +
{\textstyle\frac{(1-z)e^z}{\sqrt{2\pi}}} \, \varphi_{{}_{\nu_0}}(g),
\end{multline}
with the restriction of the measure $d\mu(\nu)$ to the set $\mathbb{R}_+$ of the equivalence classes $\nu\in \mathbb{R}_+$ 
of the spherical representations of the principal series 
$\mathfrak{S}(m=0, \nu)$ equal 
\[
d\mu(\nu) = {\textstyle\frac{1}{2\pi}} \nu^2 \, K(\nu; z) d\nu 
\]
and with the restriction of the measure $d\mu(\nu_0)$ to the set $[0,1]$ of the equivalence classes $\nu_0\in[0,1]$ 
of the spherical representations of the supplementary series 
$\mathfrak{D}(\nu_0)$ equal to the discrete measure concentrated at the single point $\nu_0 = 1-z$, and giving
the measure 
\[
{\textstyle\frac{(1-z)e^z}{\sqrt{2\pi}}}
\] 
to the single point set $\{ \nu_0 = 1-z \}$. Recall, that the second discrete term 
in (\ref{decompositionOfvarphi}) is present if and only if $0< z <1$. In (\ref{decompositionOfvarphi}) and in the last formula, 
$e$ is equal to the basis of natural logarithms.
Thus, from the generalized Bochner theorem positivity of the weight function 
$\nu \rightarrow  K(\nu; z)$ follows for all $z, \nu >0$.

One can find out that there exists a small positive $\epsilon$ such that
$\nu \rightarrow K(\nu; z=e^2/\pi)$ is a nonnegative function for $e^2/\pi$ in the interval $0< e^2/\pi < \epsilon$, 
on using explicit inspection, compare \cite{Staruszkiewicz2020}. Thus, in this asymptotic case  $e^2/\pi \ll 1$,
positivity of the weight $K$ follows by explicit inspection, and then positivity
of the kernel (\ref{kernelLobachevskyAS}) for $e^2/\pi \ll 1$, by the generalized Bochner theorem. Note here that 
the experimental value of $e^2/\pi$ is $\approx 0.0023$. Numerical calculations, which confirm positivity of the weight $K(\nu; e^2/\pi)$, 
were also carried out \cite{Staruszkiewicz2020}.

\begin{defin*}
Let us call the specific representation (\ref{Hilb-space-rep-clm}) and (\ref{rep-clm})  of \emph{(I)-(V)}, the standard representation of \emph{(I)-(V)}. 
\end{defin*}
The theory was further developed in \cite{Staruszkiewicz1987}-\cite{Staruszkiewicz2009} in an elegant 
fashion, free of any concrete particularities pertinent to any concrete representation of (I)-(V), 
and based solely on the abstract assumptions (I)-(V). Nonetheless, an implicit assumption is made:

{\bf ASSUMPTION -- VERSION I}. \emph{The representation of} (I)-(V) \emph{is unitarily equivalent
to the standard representation}.

In fact, we should have in view also the possibility of discarding the uniqueness and cyclicity assumption
(V) of the vacuum $|0\rangle$. In this case representations may appear which are unitarily equivalent
to the standard representation, but only up to possible uniform multiplicity, which may be infinite. 
In fact the uniqueness and cyclicity (V) of the vacuum $|0\rangle$ seems to have a profound meaning.

Nonetheless, in passing from ordinary states to the generalized infrared states 
the physical reason for keeping the uniqueness and cyclicity (V) of the vacuum in the space 
of generalized infrared states is not yet fully understood. Although we prefer the 
Version I of our Assumption we should be careful at the present stage of the theory, 
and we should have in view the following

{\bf ASSUMPTION -- VERSION II} \emph{We keep only the axioms} (I)-(IV) \emph{and discard uniqueness of $|0\rangle$, 
and assume that the representation of} (I)-(IV) \emph{is, up to uniform multiplicity, 
unitarily equivalent to the standard representation of} (I)-(V). 

In Section \ref{globalU(1)} a justification for Assumption in Version I, or eventually Assumtion -- Version II, will be given.

In the Version II case 
the Hilbert space $L^2(\mathbb{S}^1)$ is replaced with direct sum $\oplus L^2(\mathbb{S}^1)$
and the respective operators ${S'}_0, Q'$ on $L^2(\mathbb{S}^1)$, are replaced with direct
sums $\oplus {S'}_0, \oplus Q'$ of the copies of ${S'}_0, Q'$ on  $\oplus L^2(\mathbb{S}^1)$. 
The corresponding canonical vacuum states $|0\rangle_k$ generating any state fulfilling the 
conditions put on the vacuum are equal 
\[
\big(\oplus 0 \oplus \ldots \oplus 1_{{}_{\mathbb{S}^1}} \oplus 0 \oplus \dots \big) \otimes 1.
\]

Here $1_{{}_{\mathbb{S}^1}}$
is the constant function on $\mathbb{S}^1$, everywhere equal to $1$ put as the $k$-th
term of the direct sum. Remaining summands are all equal zero. The second factor $1 \in \mathbb{C}$
represents the vacuum in $\mathcal{H}_{\textrm{Fock}}$.
In other words, the representation is unitarily equivalent to a (denumerable at most) 
set of copies of the standard representation.
 
This Assumption, in both possible versions, I and II, we would like to make explicit.

Note that Assumption-Version II, introduces only trivial modification from the computational
point of view -- just we apply the results developed by Staruszkiewicz to each cyclic
subspace of the $k$-th canonical vacuum $|0\rangle_k$ state, and simply replacing $|0\rangle$ with $|0\rangle_k$
in his theorems in order to obtain corresponding theorems valid in this cyclic subspace of $\mathcal{H}$.

{\bf REMARK 1.} 
The domain of $Q'$ is not invariant under $S'_{0}$. For example the 
constant function $1_{{}_{\mathbb{S}^1}}$ on $\mathbb{S}^1$
belongs to $\textrm{Dom} \, Q'$, but the image 
$S'_{0} \, 1_{{}_{\mathbb{S}^1}}$ does not belong
to $\textrm{Dom} \, Q'$. This is of course an elementary observation,
because $\alpha \rightarrow S'_{0} \, 1_{{}_{\mathbb{S}^1}}(\alpha) = \alpha$ is not absolutely continuous on $\mathbb{S}^1$.
But in this simple case it also easily follows from the very definition of the adjoint operator. 
Indeed, note that there is no finite constant $C$, such that for all finite 
sequences\footnote{Recall in particular that $l^2(\mathbb{Z}) \nsubseteq 
l^1(\mathbb{Z})$.} 
$\{a_m\}_{{}_{m \in \mathbb{Z}}}$
\begin{multline*}
2 \pi e \Big| \sum \limits_{m \in \mathbb{Z}} \overline{a_{m}} \Big| =
\Bigg| \sum \limits_{m \in \mathbb{Z}} \overline{a_{m}} 
\int \limits_{0}^{2\pi} \overline{ie\frac{d e^{im\alpha}}{d \alpha}} \,\,\, \alpha \,\, \ud \alpha  \Bigg| \\
\leq \, C \, \Bigg( \sum \limits_{m,n \in \mathbb{Z}} \overline{a_{m}} a_n 
\int \limits_{0}^{2\pi} \overline{e^{im\alpha}} \,\, e^{in\alpha} \,\, \ud \alpha \Bigg)^{1/2}
= (2\pi)^{1/2} C \Big( \sum_{m \in \mathbb{Z}} |a_{m}|^{2} \Big)^{1/2}.
\end{multline*}
It is sufficient to consider only a special sequence of finite
sequences $\{a_m\}$. Namely, consider the following infinite sequence  of finite sequences
$\{ a_m\}$:
\[
\begin{split}
\ldots, 0, 1, 0, \ldots, \\
\ldots, 0, 1, \frac{1}{2}, 0, \ldots, \\
\ldots, 0, 1, \frac{1}{2},\frac{1}{3}, 0, \ldots, \\
\textrm{e.t.c.}
\end{split}
\]
Putting this finite sequences into the above inequality, we easily see that the left-hand side will
be growing to infinity (because the harmonic series is divergent), but the right-hand side
will stay bounded, because 
\[
1 + \frac{1}{4} + \frac{1}{9} + \frac{1}{16} + \ldots < \infty.
\]
This means that $\big(Q'f,S'_{0} \, 1_{{}_{\mathbb{S}^1}}\big)$ is not bounded
as a linear functional of 
\[
f(\alpha) = \sum \limits_{m \in \mathbb{Z}} a_{m} e^{im\alpha} \,\, 
\in \textrm{Dom} \, Q'.
\] 
By the Riesz 
representation theorem, no element $g \in L^2(\mathbb{S}^1, \ud \alpha)$ exists such that
$\big(Q'f,S'_{0} \, 1_{{}_{\mathbb{S}^1}}\big) = \big(f,g\big)$ for all
$f \in \textrm{Dom} \, Q'$, so that
$S'_{0} \, 1_{{}_{\mathbb{S}^1}} \notin \textrm{Dom} \, Q'$.

The fact that the domain of the self adjoint (unbounded) operator $Q'$
is not invariant under the action of the (self adjoint and bounded) operator $S'_{0}$ 
complicates slightly the computations.  Respective care has to be paid 
in order to control the domains of the respective expressions containing several factors $S$
and $Q$. But note that there exists a nuclear (and thus perfect) space 
$\mathscr{C}^{\infty}_{0}(\mathbb{S}^1) \subset L^2(\mathbb{S}^1)$ dense in $L^2(\mathbb{S}^1)$ lying in the domain
of $Q'$ (and of course in the domain of $S'_{0}$, as $S'_{0}$ is bounded) which is invariant under both
operators $S'_{0}$ and $Q'$. Namely, $\mathscr{C}^{\infty}_{0}(\mathbb{S}^1)$ consists of the periodic smooth functions $f$ such that
\[
\frac{d^kf(\alpha)}{d \alpha^k}(m2\pi) = 0, \,\,\,
k =0,1,2 \ldots, m \in \mathbb{Z},
\] 
i.e. of all functions $f$ whose derivatives of all orders vanish at 
$0, \pm2\pi$, $\pm2(2\pi)$, $\pm 3(2\pi), \ldots$.
\begin{center}
\begin{tikzpicture}[yscale=1]
    \draw[thin, ->] (-4.4,0) -- (5,0);
    \draw[thin, ->] (0,-0.5) -- (0,1);
    \draw[very thick]  (-4.1,0) -- (-3.9,0);
    \draw[very thick] (-3.9,0) to [out=0, in=180] (-3.5,0.5);
    \draw[very thick] (-3.5,0.5) to [out=0, in=180] (-2.1,0);
    \draw[very thick]  (-2.1,0) -- (-1.9,0);
    \draw[very thick] (-1.9,0) to [out=0, in=180] (-1.5,0.5);
    \draw[very thick] (-1.5,0.5) to [out=0, in=180] (-0.1,0);
    \draw[very thick]  (-0.1,0) -- (0.1,0);
    \draw[very thick] (0.1,0) to [out=0, in=180] (0.5,0.5);
    \draw[very thick] (0.5,0.5) to [out=0, in=180] (1.9,0);
    \draw[very thick]  (1.9,0) -- (2.1,0);
    \draw[very thick] (2.1,0) to [out=0, in=180] (2.5,0.5);
    \draw[very thick] (2.5,0.5) to [out=0, in=180] (3.9,0);
    \draw[very thick]  (3.9,0) -- (4.1,0);

\node [below] at (-4,0) {$-4\pi$};
\node [below] at (-2,0) {$-2\pi$};
\node [below] at (2,0) {$2\pi$};
\node [below] at (4,0) {$4\pi$};
\node [right] at (5.2,-0.2) {$\alpha$};
\node [left] at (0,1.2) {$f(\alpha)$};
\end{tikzpicture}
\end{center} 
In particular
\[
[Q', S'_{0}] = ie \boldsymbol{1} \,\,\,\textrm{on} \,\,\,
\mathscr{C}^{\infty}_{0}(\mathbb{S}^1) \subset L^2(\mathbb{S}^1, \ud \alpha)
\]
and 
\[
[Q, S_{0}] = ie \boldsymbol{1} \,\,\,\textrm{on} \,\,\,
\mathcal{H}_{\textrm{Fock}}  \otimes \mathscr{C}^{\infty}_{0}(\mathbb{S}^1)
\subset \mathcal{H}_{\textrm{Fock}} \otimes  L^2(\mathbb{S}^1).
\] 
Nonetheless, the constant functions do not belong to $\mathscr{C}^{\infty}_{0}(\mathbb{S}^1)$, and moreover
for any constant function $c 1_{{}_{\mathbb{S}^1}}$ on $\mathbb{S}^1$, the image 
$S'_{0}\, c 1_{{}_{\mathbb{S}^1}}$ does not belong
to the domain of the operator $Q'$ on $L^2(\mathbb{S}^1, \ud \alpha)$. In particular, it follows that 
$S_0 |0\rangle$ does not belong to the domain of the operator $Q$ on the Hilbert space 
$L^2(\mathbb{S}^1,\ud \alpha) \otimes \mathcal{H}_{\textrm{Fock}}$ of the quantum field $S$. Thus special care has to be paid in computation of correlation functions which involve the vacuum state $|0\rangle$. However, the transformation rule of the quantum field $S$ is such that the total charge operator $Q$ and the operator $S_0$ cancel out in differences $S(u) - S(v)$
(mainly because of the Lorentz invariance of the charge operator $Q$, compare \cite{Staruszkiewicz1992}) 
so that $(S(u) - S(v))|0\rangle$
again lies in the domain of $Q$ and in the domain of the quantum field $S$ (which may be understood as an operator
transforming continuously the nuclear space 
$  \big(\mathcal{S}_{A}(\mathscr{O})\big) \otimes \mathscr{C}^{\infty}_{0} (\mathbb{S}^1)
\subset   \mathcal{H}_{\textrm{Fock}} \otimes L^2(\mathbb{S}^1,\ud \alpha)$ into itself).
One practical rule is however useful: the operator $S$ should be think of as of a ``phase'', a quantity determined up to a multiple of $2\pi$. Thus such expressions as correlation functions of differences of phase operators $S$ and their powers, as well as of the exponentiation $e^{i(\cdot)}$ of the phase or more generally of any smooth periodic function of the phase should be well defined and should behave much better than the phases themselves,  
compare also \cite{Staruszkiewicz2002} or \cite{Staruszkiewicz1992}. \qed

{\bf REMARK 2}. Note that among the solutions $f^\mu(x) = A^\mu(x)$ of Maxwell
equations corresponding to the general classical field $S(x)$ solution (\ref{classicalS(x)})
there are on equal footing the solutions of even: $A^\mu(-x) = A^\mu(x)$ as well as of
odd $A^\mu(-x) = -A^\mu(x)$ parity. Correspondingly the Fourier transforms
of these solutions (concentrated on the positive energy cone, except the solution corresponding to the Coulomb
field which is concentrated on both sheets of the cone in momentum space) are respectively real 
or pure imaginary valued. 
The transversal solutions,
when treated as generalized homogenous of degree $-1$ states of the electromagnetic potential field, as in 
the first part of this Subsection, explain occurrence of both 
parities. Indeed, if the Hilbert space of states is over $\mathbb{C}$
and not over $\mathbb{R}$, both parities naturally occur as multiplication by imaginary
unit $i$ is unavoidable in the Hilbert space of states. 
Nonetheless, one should emphasize that the parity (reality) is preserved by the representation of the Lorentz group 
acting naturally in the space of states. Thus, we can in principle restrict the Hilbert space of the quantized scalar field $S$ to the subspace of fixed, say
even, parity. Moreover, the homogeneous of degree $-1$ solutions $A^\mu(x)$ which are present in Bremsstrahlung radiation are always of even parity. Of course the solutions $f^\mu = A^\mu$
are, regarded either as generalized homogeneous of degree $-1$ states of the single particle quantum potential field (as above in this Subsection), or regarded as unquantized classical fields, need not have any physical interpretation as classical fields. 
It is difficult (if possible at all) to find a physical process in which the odd solutions
are produced. Possibly the odd (in potential) solutions are unphysical and do not have any interpretation as 
physical fields at the classical non quantized level (treated as states obviously need not have
any such interpretation). This however is what we expect by the very nature of the 
``phase field $S(x)$'' as it is unphysical field also at the classical unquantized level. 
We do not bother about it as the most important fields encountered in QFT are unphysical as unquantized 
fields, e.g. the electromagnetic potential field itself or the Dirac spinor field. Moreover, the constant even part (even as classical phase, odd in potential) $S_0$, thus corresponding to the odd $f^\mu = A^\mu(x)$ (at the classical level gauge equivalent to the trivial zero solution) together with the remaining odd (in potential, even in phase) solutions, are fundamental, at the quantum level, for the reconstruction of the global gauge group and explanation of the universality of the scale of electric charges. Therefore, we do not go into details of the possibility of the restriction to the subspace of even parity -- a direction undertaken in \cite{HerdegenMemory}. 
\qed

\subsection{A characterization of the standard representation of the relations (I)-(V)
of Staruszkiewicz theory and its connection to the global gauge $U(1)$ group}\label{globalU(1)}

In this Subsection we will characterize the standard representation 
(\ref{Hilb-space-rep-clm}) and (\ref{rep-clm}) of the relations
(III)-(V) of Staruszkiewicz theory (given in Subsect. 
\ref{infra-electric-transversal-generalized-states}) in terms of the global gauge $U(1)$-group
structure involved spectrally into this representation.

Namely, this representation is uniquely determined by the condition, that in each (fixed) reference frame 
with unit timelike axis $u$, the operators $e^{iS(u)}$ and $Q$ define spectrally the group $U(1)$
in the Hilbert space $\mathcal{H}$ of the field $S(x)$, i.e. the group of the circle $\mathbb{S}^1$ (in the Connes sense, up to infinite uniform multiplicity). The manifold structure, the natural invariant metric, volume form and group structure of $\mathbb{S}^1$ are defined canonically by the operators $e^{iS(u)}$ and $Q$ in $\mathcal{H}$. 

Before we formulate this assertion in details, let us recapitulate some rudiments on (here we mean compact)
finite dimensional unimodular Lie groups $G$. 
The manifold structure of $G$ is characterized by the pre-$C^*$-algebra 
$\mathcal{A}$ of smooth complex valued (Krein-representative) functions  
on $G$ with the corresponding involution defined by complex conjugation. The manifold and the invariant metric structure on the Gelfand spectrum\footnote{Here the space of involutive characters for the first (i.e. commutative pointwise) multiplication of functions and the corresponding involution defined by complex conjugation.} 
$\textrm{Spec} \, \mathcal{A}$ as well as the Haar measure $dg$ can be defined spectrally by
the algebra $\mathcal{A}$, understood as algebra of operators of pointwise multiplication 
on the Hilbert space $\mathscr{H}$ of sections of 
some (naturally defined) Clifford module over 
$\mathscr{C}^{\infty}(G)$ on $G$, and by the appropriate (invariant) Dirac operator $D$ 
acting on the Hilbert space $\mathscr{H}$ (compare \cite{Connes_spectral}). The group structure
is described by the convolution-product $\ast$ and the corresponding involution 
$(\cdot)^{\circledast}$ defined
by $f^{\circledast}(g) = \overline{f(g^{-1})}$ for $f \in \mathcal{A}$. 
Note that to each unitary irreducible representation of the group $G$ there corresponds uniquely
the involutive representation of the algebra $(\mathcal{A}, \ast, (\cdot)^{\circledast})$, 
understood as the involutive algebra with the product
defined by the convolution $\ast$ and with the corresponding to the product $\ast$ involution 
$(\cdot)^{\circledast}$.
Similarly to each irreducible involutive representation (Gelfand character $h_{g}$) of 
$(\mathcal{A}, \cdot, (\cdot)^*)$
understood as the algebra of pointwise multiplication operators with commutative pointwise multiplication
$\cdot$ of functions with the corresponding 
involution $(\cdot)^*$ defined by complex conjugation, there corresponds a unique point 
$g \in G$.  
The two structures of involutive algebras on $\mathcal{A}$ are thus not accidental, 
and are uniquely interrelated. This interrelation is determined by the harmonic analysis
on the group, which can be reduced to the properties of the decomposition of the regular representation.
In case of (compact) Lie group $G$ the commutative involutive Krein-representative algebra 
(with the first pair of multiplication and involution connected to the manifold $G$) $\mathcal{A}$
has the additional \emph{square-block-algebra} structure of Krein corresponding to the (finite dimensional)
irreducible components $U(g,\gamma)$ of the regular representation, compare \cite{Neumark_dec}, Chap. VI.32
($\gamma$ is the parameter counting irreducible components of the regular representation,
the totality of which will be denoted by $\widehat{G}$).
Thus, using the Gelfand-Neumark generalized Fourier transform $\widetilde{f}$ (written also  
$\mathcal{F} f$) of $f \in \mathcal{A}$, corresponding to the decomposition
of the regular representation, we have
\[
\begin{split}
\mathcal{F} f(\gamma) = \widetilde{f}(\gamma) = \int \limits_{G} \, f(g) U^*(g,\gamma) \, dg  \\ 
f(g) = \int \limits_{\widehat{G}} 
\textrm{Tr} \big[ \widetilde{f}(\gamma) U(g,\gamma) \big] \, d\gamma,
\end{split}
\]   
where $d\gamma$ is the Plancherel measure on $\widehat{G}$. From this the interrelation between 
the two involutive algebra structures on $\mathcal{A}$, in principle at least, can be deduced.
The case of the Abelian compact $G= \mathbb{S}^1$ is very simple, so we give its final spectral description
without going into the details of its extraction.
Let us define the commutative spectral triple $(\mathcal{A}, \mathscr{H}, D)$ corresponding to the manifold
$\mathbb{S}^1$ with the commutative involutive algebra $\mathcal{A}$ of operators on a separable Hilbert space 
$\mathscr{H}$ with the commutative multiplication $\cdot$ represented by operator product and with the involution $(\cdot)^{*}$
represented by operator-adjoint, together with self-adjoint operator
$D$ on $\mathscr{H}$ respecting the conditions (1)-(5) of
\cite{Connes_spectral} (with assumption of uniform multiplicity of $\mathcal{A}''$
equal one), and together with the additional convolution multiplication $\ast$ and the corresponding involution
$(\cdot)^{\circledast}$ determined by the block-algebra structure of $\mathcal{A}$. Then 
conditions (1)-(5) of \cite{Connes_spectral}  together with the concrete $\ast$ 
and $(\cdot)^{\circledast}$ determined by the Krein-block-algebra structure will imply: 
\begin{enumerate}
\item[1)]
The set of characters $\textrm{Spec} \, \mathcal{A} = \mathbb{S}^1$ for $\mathcal{A}$ 
understood as the involutive 
commutative algebra of operators with operator-adjoint as the involution. 
\item[2)]
$\mathcal{A}$ can be identified with the algebra of pointwise multiplications
by smooth functions on $\mathscr{H} = L^2(\mathbb{S}^1, \ud \alpha)$ with the invariant
Lebesgue measure $\ud \alpha$ on $\textrm{Spec} \, \mathcal{A} = \mathbb{S}^1$.
\item[3)]
Involutive characters for the algebra
$\big( \mathcal{A}, \ast, (\cdot)^{\circledast}\big)$ with the convolution multiplication $\ast$ and the corresponding involution $(\cdot)^{\circledast}$,  
bi-uniquelly correspond to the character of the group $\mathbb{S}^1 = \textrm{Spec} \, \mathcal{A}$. 
\end{enumerate}

The block-algebra structure of $\mathcal{A}$ in the case of $G = \mathbb{S}^1$ reduces to the 
fact that there is 
a specified unitary operator $V \in \mathcal{A}$. Namely,
the spectral triple is defined by the commutative algebra $\mathcal{A}$ of operators
\[
\sum \limits_{m \in \mathbb{Z}} \tilde{f}_m V^{m}
\]     
where $\{ \tilde{f}_m\} = \{ \mathcal{F}f_m \}\in \mathcal{S}_{\mathcal{F}\Delta_{\mathbb{S}^1}\mathcal{F}^{-1}}(\mathbb{Z})
\subset \mathcal{F} \big[ L^2(\mathbb{S}^1)\big] = L^2(\mathbb{Z})$, i.e. $\{\tilde{f}_m \}$ is the Fourier transform image 
\[
\mathcal{F}f \in \mathcal{F} \big[ \mathcal{S}_{\Delta_{\mathbb{S}^1}}(\mathbb{S}^1)\big]
\]
of an element $f$ lying in the standard countably Hilbert nuclear space
\[
\mathcal{S}_{\Delta_{\mathbb{S}^1}}(\mathbb{S}^1) 
= \mathscr{C}^\infty(\mathbb{S}^1) \subset L^2(\mathbb{S}^1)
\]
determined by the standard operator
\[
\Delta_{\mathbb{S}^1} = -{\textstyle\frac{d^2}{d\alpha^2}} + 1
\,\,\,
\textrm{on}
\,\,\,
 L^2(\mathbb{S}^1).
\]
Therefore, $\mathcal{F} \big[ \mathcal{S}_{\Delta_{\mathbb{S}^1}}(\mathbb{S}^1)\big]
= \mathcal{S}_{\mathcal{F}\Delta_{\mathbb{S}^1}\mathcal{F}^{-1}}(\mathbb{Z})$
is determined by the standard operator
\[
\mathcal{F} \Delta_{\mathbb{S}^1} \mathcal{F}^{-1}
\,\,\,
\textrm{on}
\,\,\,
L^2(\mathbb{Z}) = \mathcal{F} \big[ L^2(\mathbb{S}^1)\big].
\]
Thus, $m \mapsto \tilde{f}_m$ belongs to 
\[
\mathcal{S}_{\mathcal{F}\Delta_{\mathbb{S}^1}\mathcal{F}^{-1}}(\mathbb{Z})
\]
if and only if 
\[
\Big( \, \big[\mathcal{F} \Delta_{\mathbb{S}^1} \mathcal{F}^{-1}\big]^k \tilde{f} \, , 
\, \big[\mathcal{F} \Delta_{\mathbb{S}^1} \mathcal{F}^{-1}\big]^k \tilde{f} \, \Big)_{{}_{L^2(\mathbb{Z})}}
= \sum \limits_{m \in \mathbb{Z}} (1 + m^2)^{2k}\big| \tilde{f}_m\big|^2 < \infty
\]
for each $k \in \mathbb{N}$; or equivalently iff
\[
\sup \limits_{m \in \mathbb{Z}} (1 + m^2)^k |\tilde{f}_m|^2 < \infty \,\,\,
\textrm{for all $k \in \mathbb{N}$}.
\] 

Recall please that in this simple case $G = \mathbb{S}^1 = \mathbb{R} \,  \textrm{mod} \, 2\pi$, 
all irreducible representation $U(g,\gamma)$ are one dimensional and equal to the characters
$\alpha \mapsto U(\alpha, m) = e^{i m \alpha}$ of $\mathbb{S}^1$, with $g= \alpha \in [0,2\pi) \cong \mathbb{S}^1$, $\gamma = m \in  \mathbb{Z}$, and the Fourier transform and its inverse reduces to
\[
\begin{split}
\mathcal{F} f_m = \widetilde{f}_m = {\textstyle\frac{1}{2\pi}} 
\int \limits_{G} \, f(\alpha) U(\alpha, -m) \, d\alpha  \\ 
f(\alpha) = \sum \limits_{m\in \widehat{\mathbb{S}^1}= \mathbb{Z}}
 \widetilde{f}_m U(\alpha, m).
\end{split}
\]
In order to simplify notation the Fourier transform image
\[
\mathcal{F} \big[ \mathcal{S}_{\Delta_{\mathbb{S}^1}}(\mathbb{S}^1)\big]
= \mathcal{S}_{\mathcal{F}\Delta_{\mathbb{S}^1}\mathcal{F}^{-1}}(\mathbb{Z})
\]
of the nuclear space
\[
\mathcal{S}_{\Delta_{\mathbb{S}^1}}(\mathbb{S}^1) 
= \mathscr{C}^\infty(\mathbb{S}^1)
\] 
will be simply denoted by
\[
\mathcal{S}(\mathbb{Z})
\]
omitting the standard operator $\mathcal{F}\Delta_{\mathbb{S}^1}\mathcal{F}^{-1}$
which defines it as a standard countably Hilbert nuclear space.
       
$D$ is self adjoint on $\mathscr{H}$ fulfilling $[D, V] = - \boldsymbol{1}$,
and with $(\mathcal{A}, \mathscr{H}, D)$ respecting all remaining conditions of 
\cite{Connes_spectral}. The commutative multiplication $\cdot$ and the corresponding involution
$(\cdot)^*$ on $\mathcal{A}$ is given by the composition of operators and the operator-adjoint operation. 
The convolution multiplication $\ast$ is defined as follows
\[
\big( \sum \limits_{m \in \mathbb{Z}} \tilde{f}_m V^{m} \big) \ast 
\big( \sum \limits_{m \in \mathbb{Z}} \tilde{g}_m V^{m} \big) = 
\sum \limits_{m \in \mathbb{Z}} \tilde{f}_m \tilde{g}_m V^{m}.
\]
The corresponding involution $(\cdot)^{\circledast}$ has the following definition
\[
\sum \limits_{m \in \mathbb{Z}} \tilde{f}_m V^{m} \rightarrow
\sum \limits_{m \in \mathbb{Z}} \overline{\tilde{f}_m} V^{m}.
\]
One easily checks that ${}^{\circledast}$-involutive characters $\chi_{m}$, $m \in \mathbb{Z}$ of the algebra
$(\mathcal{A}, \ast, (\cdot)^{\circledast})$ are given by 
\[
\chi_{m}\Bigg( \sum \limits_{n \in \mathbb{Z}} \tilde{f}_n V^{n} \Bigg) = \tilde{f}_m
\]   
and that the characters $h_\alpha \in \textrm{Spec} \, \mathcal{A}$ of 
$\big(\mathcal{A}, \cdot, (\cdot)^{*}\big)$
correspond to the spectral points $\alpha \in \mathbb{S}^1$ of the operator $V$. 
The characters  $\chi_{m}$ of $(\mathcal{A}, \ast, (\cdot)^{\circledast})$
correspond bi-uniquely to the irreducible unitary representations (characters)
$\mathbb{S}^1 \ni \alpha \rightarrow U(\alpha, m)$ of the group $\mathbb{S}^1$, 
with the correspondence determined by the formula
\[
\chi_{m}(f) = \int h_\alpha(V^m) \, h_\alpha(f) \, \ud \alpha
\]
for 
\[
f = \sum \limits_{n \in \mathbb{Z}} \tilde{f}_n V^{n} \, \in \mathcal{A}
\]
and 
\[
 h_\alpha(V^m) = U(\alpha,m), \,\,\, h_\alpha \in \textrm{Spec} \, \mathcal{A}.
\]
 
Note that the 
conditions of \cite{Connes_spectral}, put on the triple 
$(\mathcal{A}, \mathscr{H}, D)$, imply (among other things) that $V$ is cyclic, has the spectrum equal to 
$\mathbb{S}^1$ of uniform multiplicity equal one, and that moreover the spectral measure of $V$
on $\mathbb{S}^1$ is absolutely continuous (Lebesgue) and in fact equal to the ordinary invariant
measure $\ud \alpha$ on $\mathbb{S}^1 = \textrm{Spec} \, V = \textrm{Spec} \, \mathcal{A}$. 

Now let us go back to the characterization of the standard representation (\ref{Hilb-space-rep-clm}) and (\ref{rep-clm}),
and fix (for simplicity) the reference frame in which $S(u) = S_0$.
Note that if we put for $\mathscr{H}$ the closure of the linear span of all
elements
\[
e^{imS_0} |0 \rangle, \,\,\, m \in \mathbb{Z},
\]  
then the operators $e^{iS_0}$ and $(1/e) \, Q$, in their action on $\mathscr{H}$, can be identified with
the operators $e^{iS'_{0}}$ and $(1/e) \,Q'$ of (\ref{Hilb-space-rep-clm}) and (\ref{rep-clm}), in their action
on $L^2(\mathbb{S}^1, \ud \alpha)$. One easily checks that $V = e^{iS'_{0}}$ and $D = (1/e) \, Q'$
in their action on $\mathscr{H} = L^2(\mathbb{S}^1, \ud \alpha)$, respect all the conditions 
of the Connes spectral triple of the group $\mathbb{S}^1$, described above.

Moreover, by specific form of the representation 
(\ref{Hilb-space-rep-clm}) and (\ref{rep-clm}), it follows that the algebra of operators 
\begin{equation}\label{bigS}
\sum \limits_{m \in \mathbb{Z}} \tilde{f}_m e^{imS_0}, \,\,\, 
\{\tilde{f}_m\} \in \mathcal{S}(\mathbb{Z}) 
\end{equation}
and the operator
\[
\frac{1}{e}Q
\]
act with infinite uniform multiplicity on $\mathcal{H}$, so that $\mathcal{H}$
is a direct sum $\oplus_{j=0,1,2, \ldots} \mathscr{H}_j$ of orthogonal subspaces 
$\mathscr{H}_j$ on each of which the algebra
(\ref{bigS}) acts with uniform multiplicity one, and $(1/e) \, Q$ has simple discrete spectrum
equal $\mathbb{Z}$ on each $\mathscr{H}_j$, 
and 
\begin{equation}\label{scircleSj}
\Bigg( \,\,\,\,\, \Big(\sum \limits_{m \in \mathbb{Z}} \tilde{f}_m e^{imS_0}, \,
\{\tilde{f}_m\} \in \mathcal{S}(\mathbb{Z})  \Big)\Big|_{{}_{\mathscr{H}_j}} \,\,\,, 
\,\,\,\,\,\,\,\,\,\,\,\,\,\,
\mathscr{H}_j \,\,\,\,\,,
\,\,\,\,\,\,\,\,\,\,\,\,\, \frac{1}{e}Q\Big|{{}_{\mathscr{H}_j}} \,\,\,\,\,\, \Bigg)
\end{equation} 
is a Connes spectral triple of the group $\mathbb{S}^1$. In particular for 
$\mathscr{H}_0$ we can take the Hilbert space spanned by $e^{imS_0}|0\rangle$. Then in the subspace orthogonal
to $\mathscr{H}_0$ we choose a next $\mathscr{H}_1$ on which (\ref{scircleSj}) (with $j=1$)
is a spectral triple of the group $\mathbb{S}^1$, which is possible by the form  
(\ref{Hilb-space-rep-clm}) and (\ref{rep-clm}) of the operators $S_0, Q$ in the standard representation.
Indeed, by choosing an orthonormal basis $b_j$ in the second factor 
$\mathcal{H}_{\textrm{Fock}}$ of
$\mathcal{H} =  L^2(\mathbb{S}^1,\ud \alpha) \otimes \mathcal{H}_{\textrm{Fock}}$
we obtain $\mathscr{H}_j$ by replacing the state\footnote{Recall
that here $1_{{}_{\mathbb{S}^1}}$ is the constant function on $\mathbb{S}^1$
equal everywhere to one, and that $1$ in the second factor is the ordinary unit in $\mathbb{C}$
representing the vacuum in $\mathcal{H}_{\textrm{Fock}}$.} 
$|0\rangle = 1_{{}_{\mathbb{S}^1}} \otimes 1 = 1_{{}_{\mathbb{S}^1}}$ in the construction of 
$\mathscr{H}_0$
with the state $1_{{}_{\mathbb{S}^1}}\otimes b_j$. Of course  
\[
\mathcal{H} = \oplus_{j=0,1,2, \ldots} \mathscr{H}_j
\]
for the whole Hilbert space $\mathcal{H}$ of the quantum phase field $S(x)$.

Thus, we can consider the triple 
\[
\Bigg( \,\,\,\,\, \Big( \sum \limits_{m \in \mathbb{Z}} \tilde{f}_m e^{imS_0}, \, 
\{\tilde{f}_m\} \in \mathcal{S}(\mathbb{Z})  \Big) \,\,\,, 
\,\,\,\,\,\,\,\,\,\,\,\,\,\, 
\mathcal{H} \,\,\,\,\,,
\,\,\,\,\,\,\,\,\,\,\,\,\, \frac{1}{e}Q \,\,\,\,\,\, \Bigg)
\]
as a spectral triple of the group $\mathbb{S}^1 = U(1)$, but acting with infinite uniform multiplicity,
and respecting all conditions of the ordinary spectral triple, when restricted to each direct sum subspace
$\mathscr{H}_j$. In each subspace $\mathscr{H}_j \subset \mathcal{H}$ there exists 
(for each possible value $em$, $m \in \mathbb{Z}$) exactly one eigenstate $|m, j\rangle$ of the charge operator $Q$, with the eigenvalue $em$. For instance in 
\[
\mathscr{H}_0  = \overline{\textrm{Linear span} \, \big(e^{imS_0}|0\rangle, m \in \mathbb{Z} \big)},
\]
$|m,0 \rangle = e^{imS_0}|0\rangle$. Each such eigenvector $|m, j\rangle \in
\mathscr{H}_j$ of the charge operator $Q$ determines in a natural manner
the corresponding irreducible unitary representation $\mathbb{S}^1 \ni \alpha \rightarrow 
U(\alpha, m)$ (character)
of the group $\mathbb{S}^1$, or equivalently the $(\cdot)^{\circledast}$-involutive character $\chi_m$
of the algebra
\[
\Bigg( \Big( \sum \limits_{m \in \mathbb{Z}} \tilde{f}_m e^{imS_0}, \,\,\, 
\{\tilde{f}_m\} \in \mathcal{S}(\mathbb{Z})  \Big), \,\,\,\,\,\,\ast \,\,\,, \,\,\,\,\,\,(\cdot)^{\circledast}  \Bigg)
\]
by the following formula:
\[
\chi_m \Bigg(\sum \limits_{n \in \mathbb{Z}} \tilde{f}_n e^{inS_0} \Bigg) = 
\Bigg\langle 0,j \Bigg| \sum \limits_{n \in \mathbb{Z}} \tilde{f}_n e^{inS_0} \Bigg|m,j \Bigg\rangle.
\]
Therefore, the irreducible unitary representations or unitary characters
of the gauge group $\mathbb{S}^1 = U(1)$ correspond to eigenstates of charge
operator $Q$ and bi-uniquelly to spectral values of $Q$. 
Moreover if we fix reference frame, then the specific eigenstates of $Q$ corresponding bi-uniquely to
the unitary characters of $U(1)$ are naturally determined by the fact that in each $\mathscr{H}_j$ 
to each spectral value of $Q$ there exists exactly one eigenstate of $Q$.

It is easily seen that 
the conjugation of a character 
(the conjugated irreducible representation of the group $\mathbb{S}^1$)
corresponds to the opposite charge. To the tensor product $U(m_1) \otimes U(m_2)$ of 
irreducible unitary representations $\alpha \rightarrow U(\alpha, m_i)$ (unitary characters)
of $\mathbb{S}^1$, equal in this case to the ordinary multiplication
$U(m_1)U(m_2)= U(m_1+m_2)$ (resp. $\chi_{m_1+m_2}$) of unitary characters, 
corresponds to the composition $em_1 + em_2$ of the corresponding charges. 

This can be compared to the Doplicher-Haag-Roberts theory of global gauge
groups and the corresponding generalized charges, \cite{Haag}, Ch. IV.

Of course changing of the reference frame leaves unchanged the whole structure, as it 
is reduced to the application of the unitary operator $U_\Lambda$, and the spectral
triple description of $U(1)$ is unitary invariant, and Lorentz invariant. 
Indeed, the operators $e^{iS_0}$ and $Q$, in passing to another reference frame,
are replaced with the 
corresponding ones $U_\Lambda e^{iS_0} U_{\Lambda}^{-1}$, 
$U_\Lambda Q U_{\Lambda}^{-1} = Q$ (by Lorentz invariance of $Q$) and the decomposition
$\mathcal{H} = \oplus_{j=0,1,2, \ldots} \mathscr{H}_j$ is replaced with
$\mathcal{H} = \oplus_{j=0,1,2, \ldots} U_\Lambda \mathscr{H}_j U_{\Lambda}^{-1}$.

The whole point lies of course in the fact that in the standard representation
(\ref{Hilb-space-rep-clm}) and (\ref{rep-clm})
the unitary operator $e^{iS_0}$,  restricted to the subspace 
\[
\mathscr{H}_0  = \overline{\textrm{Linear span} \, \big(e^{imS_0}|0\rangle, m \in \mathbb{Z} \big)},
\]
has simple absolutely continuous Lebesgue spectrum
equal to the whole circle $\mathbb{S}^1$, or what amounts to the same think, the operator
$e^{iS'_{0}}$ has by construction this property on the Hilbert space
$L^2(\mathbb{S}^1; \ud \alpha)$ (of course with the invariant Lebesgue measure $\ud \alpha$ on $\mathbb{S}^1$).

There are (infinitely) many different representations of the relations
(IV)-(V) of Subsection \ref{infra-electric-transversal-generalized-states}, 
not equivalent with the standard representation 
(\ref{Hilb-space-rep-clm}) and (\ref{rep-clm}).
Even if we add the axiom (III) concerning existence and uniqueness of the vacuum, 
there will remain infinitely many non equivalent representations.
 But for them the spectral construction, as presented above,
of the gauge group $\mathbb{S}^1$ is impossible, at least with $V = e^{iS(u)}, D = (1/e) \, Q$. 
For example in order to construct
other representations, we replace the invariant Lebesgue measure $\ud \alpha$ in the 
Hilbert space
$L^2(\mathbb{S}^1,\ud \alpha)$, with the Lebesgue  measure $\ud|_{{}_{[-\alpha_0,\alpha_0]}} \alpha$ on 
$\mathbb{S}^1$, but concentrated
on some interval $-\alpha_0/2 < \alpha < \alpha_0/2$ (where $0<\alpha_0/2< \pi)$
of the circle $\mathbb{S}^1 = \mathbb{R} \, \textrm{mod} \, 2\pi$. The operator $S'_{0}$ is defined
as before as multiplication by $\alpha$. Then we consider the differential operator 
$L_0 = i d/d\alpha$ on $L^2(\mathbb{S}^1, \ud|_{{}_{[-\alpha_0/2,\alpha_0/2]}} \alpha)$, with $\textrm{Dom} \, L_0$
equal to all functions $y$ on the said interval $(-\alpha_0/2, \alpha_0/2)$ of the circle, 
which are absolutely continuous on $(-\alpha_0/2, \alpha_0/2)$, and fulfill the conditions
\[
\int \limits_{-\alpha_0/2}^{\alpha_0/2} \Big| \frac{dy}{d\alpha} \Big|^2 \ud \alpha < \infty, \,\,\,
y|_{{}_{-\alpha_0/2}} = y|_{{}_{\alpha_0/2}} = 0.
\]  
We define as the operator $(1/e) Q'$, as one of the (infinitely many) possible self adjoint extensions
of the operator $L_0$. Because this operator\footnote{This was one of the first operators 
investigated by von Neumann
when he was discovering his theory of self adjoint operators.} is one of the simplest, 
for which the self adjoint extensions have been completely classified (compare e.g. the general Krein's method of directional functionals,
as presented in \cite{NeumarkLinOp}), then we will not go into details here
and present only the final result. Namely, the self adjoint extensions $(1/e) \, Q'$ of $L_0$ are parametrized
by $\beta \in [0,2\pi]$. Each such extension $(1/e) \, Q'$ corresponding to $\beta$ has the following
domain
\[
\textrm{Dom} \, (1/e) \, Q' = \big\{y + a(\textrm{exp}^{-1} + e^{i\beta}\textrm{exp}), 
y \in \textrm{Dom} \, L_0, \, a \in \mathbb{C} \big\}
\]
and the action of $(1/e) \, Q'$ on the elements $y + a(\textrm{exp}^{-1} + e^{i\beta}\textrm{exp})$ of $\textrm{Dom} \, Q'$ is equal
\begin{multline*}
(1/e) \, Q'\Big(y + a(\textrm{exp}^{-1} + e^{i\beta}\textrm{exp})\Big)(\alpha) \\
= L_0 y(\alpha) - ia \textrm{exp}^{-1}(\alpha) +ia\textrm{exp}(\alpha)
= i \frac{dy}{d\alpha}(\alpha) - iae^{-\alpha} + ia e^{i\beta}e^{\alpha}. 
\end{multline*}
In particular\footnote{Note that for any $\beta \in [0,2\pi]$
\[
\Bigg| \frac{e^{-\alpha_0/2 + e^{i\beta}e^{\alpha_0/2}}}{e^{\alpha_0/2 + e^{i\beta}e^{-\alpha_0/2}}} \Bigg| =1.
\]
} 
\[
f(\alpha_0/2) 
= \frac{e^{-\alpha_0/2 + e^{i\beta}e^{\alpha_0/2}}}{e^{\alpha_0/2 + e^{i\beta}e^{-\alpha_0/2}}} \, 
f(-\alpha_0/2),
\,\,\,
f \in \textrm{Dom} \, Q'.
\]
All self adjoint extensions $(1/e) \, Q'$ of $L_0$ have purely point spectrum (discrete) 
of uniform multiplicity one (i.e. simple), which is equal to $\mathbb{Z}$ multiplied by a nonzero real constant 
(plus eventually some shift, in which case zero does not enter the spectrum
of $Q'$). But among the possible self adjoint extensions $(1/e) \, Q'$ of $L_0$ there are (infinitely many) 
such which have
simple spectrum equal $\mathbb{Z}$ multiplied by a constant, and thus including zero, for which moreover
the relation $[Q',S'_{0}] = ie$ makes sense on $\textrm{Dom} \, L_0$ (although in general
$S'_{0} (\textrm{Dom} \, Q') \nsubseteq \textrm{Dom} \, Q'$ and even 
$e^{iS'_{0}} (\textrm{Dom} \, Q') \nsubseteq \textrm{Dom} \, Q'$). With these self adjoint
extensions $Q'$ and the operator $S'_{0}$ of multiplication by $\alpha$
on $L^2(\mathbb{S}^1, \ud|_{{}_{[-\alpha_0/2,\alpha_0/2]}} \alpha)$, substituted for the operators
$Q'$ and $S'_{0}$ on $L^2(\mathbb{S}^1, \ud \alpha)$ in the Subsection 
\ref{infra-electric-transversal-generalized-states}, we obtain another non-standard representations of
(III)-(V) not equivalent to the standard representation. In particular for such
non-standard representations the phase operator $e^{iS_0}$ and the 
charge operator $Q$ behave differently in comparison to the standard representation. 
In particular the operator
\[
e^{ikS_0}
\] 
transforms eigenstate of the operator $Q$ onto another eigenstate of the operator
$Q$ only if $k$ assumes some particular integer values. In some cases it may even happen
that for no integer $k$ the operator $e^{ikS_0}$ transforms eigenstates into eigenstates of $Q$. 
In general, even in the non-standard
representation, the operator $S(u)$ is essentially self adjoint on\footnote{For definition of 
$(\mathcal{S}_{A}(\mathscr{O}))$
compare Subsection \ref{infra-electric-transversal-generalized-states}.} 
$\textrm{Dom} \, Q' \otimes (\mathcal{S}_{A}(\mathscr{O}))$, so that the unitary operator
$e^{-iS(u)}$ and the state
$|u\rangle = e^{-iS(u)} |0\rangle$
are well defined. But this time $|u\rangle$ is in general not equal to any eigenstate of $Q$.
 In this non-standard case the representation
$U$ of $SL(2, \mathbb{C})$ in the Hilbert space $\mathcal{H}$ of the quantum field
$S(x)$ in (II) is different from that representation $U$
corresponding to the standard representation of (IV)-(V). In particular only some non integer powers
of the operator $e^{iS(u)}$ transform eigenstate of $Q$ into another eigenstate. In consequence the explicit
construction of the unitary representation $U$ of $SL(2, \mathbb{C})$, which meets the requirements
(I)-(II) is different from the standard case. It is not evident if for non-standard representation
of (IV)-(V) not only (III)-(V) are consistent, but moreover that the representation $U$ do actually exists
and makes all axioms (I)-(V) consistent. Nonetheless, one cannot exclude, that the representation $U$
do exist, together with the spectral realization of the gauge group, but with the 
operator $V = e^{iS_0}$ in this realization replaced with some (in general non integer) 
$c$ power $e^{icS_0}$ of
$e^{iS_0}$. In any way, we have the following three possibilities: 

\begin{enumerate}
\item[1)]
In the non-standard representation of (IV)-(V) the axioms
(I)-(V) are consistent and admits the spectral realization of the gauge group,
with $V = e^{icS_0}, D = (1/e) \, Q$
but with $\textrm{Spec} \, Q = ce\mathbb{Z}$ with a constant $ce$ not equal
to the constant $e$ in (I)-(V). 
\item[2)] 
In the non-standard representation of (IV)-(V) the axioms

(I)-(V) are consistent but the spectral realization of the gauge group
is impossible, and $\textrm{Spec} \, Q = ce\mathbb{Z}$ with a constant $ce$ not equal
to the constant $e$ in (I)-(V). 
\item[3)] 
In the non-standard representation of (IV)-(V) the axioms
(I)-(V) are inconsistent, so that $|0\rangle$ and $U$ which meet (I)-(V) do not exist.
\end{enumerate}

The first possibility 1) takes place for the self adjoint extension $(1/e) \, Q'$ corresponding
to the parameter $\beta=0$. In this case 
\[
f(\alpha_0/2) = f(-\alpha_0/2), \,\,\, f \in \textrm{Dom} \, (1/e) \, Q',
\]
and $\textrm{Dom} \, (1/e) \, Q'$ can be identified with the linear space of all absolutely continuous 
functions on the circle $\mathbb{R} \, \textrm{mod} \, \alpha_0$, with square integrable
derivative on this circle, with the eigen-functions of $(1/e) \, Q'$, which are smooth on
the circle $\mathbb{R} \, \textrm{mod} \, \alpha_0$. In this case $c = 2\pi/\alpha_0$.
The pair $V= e^{icS'_{0}}, D = (1/e) \, Q'$ can serve as the spectral realization of the group
$\mathbb{R} \, \textrm{mod} \, \alpha_0$.
In this case the state $e^{-imcS(u)} |0\rangle$ is the eigenstate of $Q$ corresponding to the
eigenvalue $ecm$, $m \in \mathbb{Z}$. The state $e^{-icS(u)} |0\rangle$ plays the same role in 
the theory, as $|u\rangle$ does in the standard representation.

We obtain the second possibility 2) when using the self adjoint extension $(1/e) \, Q'$
corresponding to nonzero parameter $\beta \in [0,2\pi]$, which has to be particularly chosen
in order to achieve $\textrm{Spec} \, Q' = ce\mathbb{Z}$. We have countably many possibilities 
for achieving this for each fixed $\alpha_0$. In this case likewise $c = 2\pi/\alpha_0$,
but this time the eigenfunctions of $(1/e) \, Q'$ (contained in $\textrm{Dom} \, (1/e) \, Q'$) 
are not smooth on the circle $\mathbb{R} \, \textrm{mod} \, \alpha_0$. Indeed,
this time
\[
f(\alpha_0/2) = e^{i\theta_0} f(-\alpha_0/2), \,\,\,
f \in \textrm{Dom} \, Q',
\]
with some $\theta_0 \neq 0 \, \textrm{mod} \, 2\pi$,
so that these $f$ do not glue to any smooth functions on the circle 
$\mathbb{R} \, \textrm{mod} \, \alpha_0$. In particular, the pair $V= e^{icS'_{0}}, D = (1/e) \, Q'$
cannot serve as the operators defining the group $\mathbb{R} \, \textrm{mod} \, \alpha_0$
spectrally. In this case the state $e^{-imcS(u)} |0\rangle$ is the eigenstate of $Q$ corresponding to the
eigenvalue $ecm$, $m \in \mathbb{Z}$. The state $e^{-icS(u)} |0\rangle$ plays the same role in 
the theory, as $|u\rangle$ does in the standard representation.

The third possibility 3) takes place when the self adjoint extension $(1/e) \, Q'$ 
corresponds to nonzero parameter $\beta$, but chosen in such a manner that
we get additional shift $\lambda$ in the spectrum $\textrm{Spec} \, Q' = ce\mathbb{Z} + \lambda$  
of $Q'$. We have uncountably many possibilities to achieve this situation for each 
fixed $\alpha_0$. In this case zero does not enter the spectrum of $Q'$, so that
no vacuum state $|0\rangle$ respecting (III) can exist.  

All self adjoint extensions $(1/e) \, Q'$ of $L_0$ are in this way exhausted. In this way 
essentially\footnote{As the case of the spectrum of $S'_{0}$ equal to a (countable) disjoint sum of 
intervals of the circle $\mathbb{S}^1$ is excluded 
by the uniqueness of the vacuum $|0\rangle$.}
all possible pairs of operators $Q, S_{0}$ on the cyclic subspace $\mathscr{H}_0$ spanned by 
$e^{imS_0}|0\rangle$, $m \in \mathbb{Z}$, are exhausted,
in which $e^{iS_{0}}\big|_{{}_{\mathscr{H}_0}}$ has purely absolutely continuous spectrum, i.e. with the spectral measure absolutely continuous on 
$\textrm{Spec} \, e^{iS_{0}}\big|_{{}_{\mathscr{H}_0}} \subset \mathbb{S}^1$. 
The standard representation corresponds to the case  
$\textrm{Spec} \, e^{iS_{0}}\big|_{{}_{\mathscr{H}_0}} = \mathbb{S}^1$ with absolutely continuous spectral measure
of $e^{iS_{0}}\big|_{{}_{\mathscr{H}_0}}$. The case in which 
the spectral measure of the operator $e^{iS_{0}}\big|_{{}_{\mathscr{H}_0}}$ contains pure point component (existing in addition to the absolutely continuous
component of the spectral measure of 
$e^{iS_{0}}\big|_{{}_{\mathscr{H}_0}}$) is excluded by the uniqueness of the vacuum $|0\rangle$. 
The case in which $e^{iS_{0}}\big|_{{}_{\mathscr{H}_0}}$ has pure point (i.e. discrete) spectrum 
is a priori possible. The simplest case comes from
$\textrm{Spec} \, e^{iS_{0}}\big|_{{}_{\mathscr{H}_0}}$ consisting of single point. But in this case the charge operator degenerates to a zero operator-- again by the uniqueness
of the vacuum $|0\rangle$.
Thus, we obtain the degenerate case of Staruszkiewicz theory with the constant $e$ in (I)-(V)
equal zero.
We do not continue the analysis the discrete case any further because in this case  
(even if it admits at all any other 
nontrival cases) the spectral construction of the
gauge group $\mathbb{S}^1$ is impossible.  Indeed, in this case 
$\textrm{Spec} \, e^{iS_{0}}\big|_{{}_{\mathscr{H}_0}}$
by construction does not contain any open interval of $\mathbb{S}^1$. 
 There remains only the case in which the spectrum 
$\textrm{Spec} \, e^{iS_{0}}\big|_{{}_{\mathscr{H}_0}}$
is purely singular, i.e. the case in which the Lebesgue measure $\ud \alpha$
on $\mathbb{S}^1$ in (\ref{Hilb-space-rep-clm}) and (\ref{rep-clm}) is replaced by a purely singular
measure on $\textrm{Spec} \, e^{iS_{0}}\big|_{{}_{\mathscr{H}_0}} \subset \mathbb{S}^1$, i.e. continuous but not absolutely continuous. 
An example comes from the singular measure concentrated on the Cantor set 
(regarded as a subset of $\mathbb{S}^1$) and determined by the Cantor singular function. We should not 
\emph{a priori} exclude
existence of the corresponding self adjoint operator $(1/e) Q\big|_{{}_{\mathscr{H}_0}}$
on $\mathscr{H}_0$ in this case (compare the spectral differential calculus of Connes on fractal sets
\cite{Connes}, Chap. IV.3), which would provide a representation of (III)-(V). But in this case the spectral
realization of the gauge group is impossible because  $\textrm{Spec} \, e^{iS_{0}}\big|_{{}_{\mathscr{H}_0}}$
covers no open interval of the circle $\mathbb{S}^1$. 

Note that any representation of (I)-(V) is equivalent to the one
which have the general tensor product form (\ref{Hilb-space-rep-clm}) and (\ref{rep-clm})
with $\mathbb{S}^1$ and the measure $\ud \alpha$ replaced with 
 $\textrm{Spec} \, e^{iS_{0}}\big|_{{}_{\mathscr{H}_0}}$ and the spectral measure
of the operator $e^{iS_{0}}\big|_{{}_{\mathscr{H}_0}}$, compare 
the spectral theorem for cyclic unitary operator in \cite{GelfandIV}, Chap.I. 4.5, Thm 2 or \cite{Segal_Kunze},
Chap. IX.2, Scholium 9.2.

Thus, in any case we have the following
\begin{twr*}
 The standard representation of (I)-(V) is uniquely (up to unitary equivalence) 
characterized by the two 
conditions
\begin{enumerate}
\item[1)]
The gauge group has spectral realization (\ref{scircleSj}) in this representation.
\item[2)] 
$\textrm{Spec} \, Q = e\mathbb{Z}$ with the constant $e$ the same as in (I)-(V);
\end{enumerate}
or by the following single condition 
\begin{enumerate}
\item[3)]
In each reference frame (with the unit vector along the time like axis
equal $u$) the gauge group has the spectral realization (\ref{scircleSj}) 

with $V= e^{iS(u)}, D = (1/e) \, Q$. 
\end{enumerate}

For each real number $c > 1$ there exists one (up to unitary equivalence) 
non-standard representation of (I)-(V) such that 
\begin{enumerate}
\item[1)]
In each reference frame with the unit vector along the time like axis
equal $u$, the gauge group has spectral realization (\ref{scircleSj}) in this representation with the 
operators $V= e^{iS(u)}, D= (1/e) Q$ replaced with  $V= e^{icS(u)}, D = (1/e) Q$.
\item[2)] 
$\textrm{Spec} \, Q = ec \mathbb{Z}$ with the constant $e$ equal to that in (I)-(V).
\end{enumerate}
\end{twr*}

Note in particular that for the standard representation of (I)-(V)
\[
\Bigg( \,\,\,\,\, \Big(\sum \limits_{m \in \mathbb{Z}} \tilde{f}_m e^{imS_0}, \,
\{\tilde{f}_m\} \in \mathcal{S}(\mathbb{Z})  \Big)\Big|_{{}_{\mathscr{H}_0}} \,\,\,, 
\,\,\,\,\,\,\,\,\,\,\,\,\,\,
\mathscr{H}_0 \,\,\,\,\,,
\,\,\,\,\,\,\,\,\,\,\,\,\, \frac{1}{e}Q\Big|{{}_{\mathscr{H}_0}} \,\,\,\,\,\, \Bigg)
\]
composes a one dimensional spectral triple in the sense of Connes. This in particular 
means that the operator $e^{-iS_0}$ has simple spectrum (of multiplicity one)
equal to $\mathbb{S}^1$ on the subspace $\mathscr{H}_0$ generated by $e^{-imS_0}|0\rangle$,
$m \in \mathbb{Z}$. Let us look more closely at this condition.

Consider the closed subspace $\mathcal{H}_0 \in \mathcal{H}$ spanned by the vectors (\ref{Q=0-states})
of the form
\[
\big(c_{\alpha_1}^{+}\big)^{\beta_1} \ldots \big(c_{\alpha_n}^{+}\big)^{\beta_n}|0\rangle, \,\,\,n=0,1, \ldots,
\,\,\beta_i = 0,1, \ldots.
\] 
Here we have put $\alpha_i$ for $(l_i,m_i)$. 
It is easy to see that the bilinear map $\otimes$
\[
\Big( \big(c_{\alpha_1}^{+}\big)^{\beta_1} \ldots \big(c_{\alpha_n}^{+}\big)^{\beta_n}|0\rangle \Big) 
\otimes \Big( e^{-imS_0}|0\rangle\Big) = \big(c_{\alpha_1}^{+}\big)^{\beta_1} \ldots 
\big(c_{\alpha_n}^{+}\big)^{\beta_n}e^{-imS_0}|0\rangle
\]
defines a bilinear map $\mathcal{H}_0 \times \mathscr{H}_0 \rightarrow \mathcal{H}$,
whose image is dense in $\mathcal{H}$, and under which  
$\mathcal{H}_0$ and  $\mathscr{H}_0$ are $\otimes$-disjoint (\cite{treves},
Part III, Definition 39.1). Thus, $\otimes$ can serve to define the algebraic 
tensor product $\mathcal{H}_0 \otimes_{{}_{\textrm{alg}}} \mathscr{H}_0$
densely included into $\mathcal{H}$. Moreover the Hilbert space inner product
of $\mathcal{H}$, coincides on simple tensors with the Hilbert space tensor product.
Thus, $\mathcal{H}$ is canonically equal to the following Hilbert space tensor product
$\mathcal{H}_0 \otimes \mathscr{H}_0$ of its own subspaces:
\[
\mathcal{H} = \mathcal{H}_0 \otimes  \mathscr{H}_0.
\]

Now we can back to the condition that $e^{iS_0}$ has simple spectrum on $\mathscr{H}_0$.
This means that $e^{imS_0}|0\rangle$ is dense in $\mathscr{H}_0$, or that
$|0\rangle$ is cyclic on the space $\mathscr{H}_0$, which tensored  with $\mathcal{H}_0$
gives the whole Hilbert space $\mathcal{H} = \mathcal{H}_0 \otimes \mathscr{H}_0$.

The same holds true if we replace the subspace $\mathscr{H}_0$ with any other
$\mathscr{H}_j$ in (\ref{scircleSj}).

In the sequel we consider only representations of (I)-(IV), which fulfil one of the following two assumptions

{\bf ASSUMPTION -- VERSION I}. \emph{The representation of (I)-(V) is unitarily equivalent
to the standard representation}. 

{\bf ASSUMPTION -- VERSION II}. \emph{We keep only the axioms (I)-(IV) and discard uniqueness of $|0\rangle$, 
and assume that the representation of (I)-(IV) is, up to uniform multiplicity, 
unitarily equivalent to the standard representation of (I)-(V)}. 

 Of course for the standard representation
acting with uniform multiplicity the spectral construction of the global gauge group $U(1)$
will be preserved in an obvius manner. In this case 
\[
\mathcal{H} = \mathcal{H}_0 \otimes \Big(\oplus_k \mathscr{H}_{0k}\Big)
\]
where $\mathscr{H}_{0k}$ is defined as $\mathscr{H}_{0}$ by replacing 
the unique vacuum $|0\rangle$ with one of the canonical vacuum states $|0\rangle_{k}$
in the direct sum of the standard representation, compare the end of 
Subsection \ref{infra-electric-transversal-generalized-states}. 
Similarly, the Hilbert subspaces $\mathscr{H}_{j}$ of
spectral realization (\ref{scircleSj}) of $U(1)$ will have to be replaced by the
corresponding $\mathscr{H}_{jk}$.

\subsection{A constructive consistency proof of the axioms (I) -- (V)}\label{Consistency}

We give a constructive consistency proof
of the axioms (I - (V).

We do not use the concrete form of the \emph{standard representation} of 
the axioms (I)-(V), Subsection \ref{infra-electric-transversal-generalized-states},
but nonetheless we made the following

{\bf ASSUMPTION -- VERSION I}. \emph{The representation of (I)-(V) is unitarily equivalent
to the standard representation}.  \qed

{\bf DEFINITION}. \emph{Let the representation of (I)-(V) we are using be called
abstract representation}. \qed

All these distinctions may seem pedantic at first sight, but in fact are essential
to understand the theory. In particular without making these distinctions
the relation of the phase field $S(x)$ of Staruszkiewicz theory to the 
homogeneous of degree zero part of the 
interacting field $x_\mu A^{\mu}_{\textrm{int}}(x)$,
as outlined in the Subsection \ref{IntAspatialInfty} of Introduction, would be difficult to understand.
In Subsection \ref{infra-electric-transversal-generalized-states} we have made the first step
in this direction, by explaining this relation at the free theory level, and the degenerate form
of Staruszkiewicz theory with the constant $e$ in (I)-(V) put equal zero. Doing this we have been using the standard representation. 
In this degenerate case the standard representation likewise degenerates,
by replacing of the circle $\mathbb{S}^1$ in its construction, 
with just a single point set. The mentioned relation is most easily seen in the \emph{standard representation}, 
or a finite or countable number of copies of the standard representation. In order to make the results obtained
in \ref{infra-electric-transversal-generalized-states} applicable 
to the problem of comparison of the phase field $S(x)$ of Staruszkiewicz
theory with the homogeneous part of the interacting field $x_\mu A^{\mu}_{\textrm{int}}(x)$, 
we have to make explicit the relation
of the operators of Staruszkiewicz theory in the abstract representation to the 
corresponding operators in the standard representation (or direct sum of copies of the standard 
representation). 

Having this point in mind, we choose the following plan for Subsections \ref{Consistency} -- \ref{Comparison2}.
First we reconstruct the operators $S_0, Q, c_{lm}, c_{lm^{+}}$ and the representation 
$U$ of $SL(2, \mathbb{C})$ in the abstract Hilbert space $\mathcal{H}$, 
using the abstract representation of (I)-(V). In fact, we compute them in 
explicit form. Let the corresponding operators in the standard representation acting with uniform multiplicity
be denoted with the bold fonts $\boldsymbol{S_0}$, $\boldsymbol{Q}$, $\boldsymbol{c_{lm}}$, 
$\boldsymbol{c_{lm}}$, $\boldsymbol{c_{lm}^{+}}$ and $\boldsymbol{U}$ acting in $\boldsymbol{\mathcal{H}}$, 
respectively. 
Then we construct unitary operator 
$V: \boldsymbol{\mathcal{H}} \rightarrow \mathcal{H}$, having the property
that the operators $V \boldsymbol{S_0} V^{-1} = S_0$, $V \boldsymbol{Q} V^{-1}=Q$, 
$V \boldsymbol{c_{lm}} V^{-1}= c_{lm}$, $V \boldsymbol{c_{lm}^{+}}V^{-1}=c_{lm}^{+}$ and 
$V\boldsymbol{U}V^{-1} = U$ are equal to the operators $S_0, Q, c_{lm}, c_{lm}^{+}, U$ 
of the abstract representation acting in the abstract Hilbert space $\mathcal{H}$. Then we make explicit
the relation between the quantum phase operator $V \boldsymbol{S(x)} V^{-1} = S(x)$
acting in the abstract Hilbert space $\mathcal{H}$, and the zero order contribution to the 
homogeneous of degree zero part of the interacting field $x_\mu A^{\mu}_{\textrm{int}}(x)$.  

Finally, we introduce the assumption that our abstract representation
fulfills the following

{\bf ASSUMPTION -- VERSION II} \emph{We keep only the axioms (I)-(IV) and discard uniqueness of 
$|0\rangle$, and assume that the representation of (I)-(IV) is, up to uniform multiplicity, 
unitarily equivalent to the standard representation of (I)-(V)}. 

In order to avoid unnecessary repetitions, we construct in fact the operator $V$ at once
in this more general case under the Version II of our Assumption, as it simply degenerates to the
particular case in which the multiplicity of the standard representation is equal one.

Under this assumption we construct the unitary operator $V$ as before,
$V: \boldsymbol{\mathcal{H}} \rightarrow \mathcal{H}$, having the property
that the operators $V \boldsymbol{S_0} V^{-1} = S_0$, $V \boldsymbol{Q} V^{-1}=Q$, 
$V \boldsymbol{c_{lm}} V^{-1}= c_{lm}$, $V \boldsymbol{c_{lm}^{+}}V^{-1}=c_{lm}^{+}$ and 
$V\boldsymbol{U}V^{-1} = U$ are equal to the 
operators $S_0, Q, c_{lm}, c_{lm}^{+}, U$ of the abstract representation acting in
the abstract Hilbert space $\mathcal{H}$. But here 
$\boldsymbol{S_0}$, $\boldsymbol{Q}$, $\boldsymbol{c_{lm}}$, 
$\boldsymbol{c_{lm}}$, $\boldsymbol{c_{lm}^{+}}$ and $\boldsymbol{U}$ acting in $\boldsymbol{\mathcal{H}}$, 
respectively denote the operators in the direct sum of copies of the standard representation. 
This means that each bolded operator here is equal to the direct sum of copies of the corresponding operator of the standard representation.

It is sufficient to show consistency with the operators $S_0, Q, c_{lm}, c_{lm^{+}}, U$ in 
the standard representation of  (I)-(V), acting in the Hilbert space $\mathcal{H}$ 
of the standard representation, acting with uniform multiplicity one.

The Hilbert space $\mathcal{H}$ of the quantum field $S$ determined by the axioms (I)-(V)  of the 
Subsection \ref{infra-electric-transversal-generalized-states}, is a direct sum 
$\oplus_{{}_{n \in \mathbb{Z}}} \mathcal{H}_{{}_{n}}$ of orthogonal subspaces $\mathcal{H}_{{}_{n}}$ invariant under the 
unitary representation $U$ of $G= SL(2, \mathbb{C})$,  each of which correspond respectively to the eigenvalue 
$ne$ of the total charge operator $Q$. In particular the subspace $\mathcal{H}_{{}_{n=0}}$ corresponding to the zero eigenspace
of the charge operator $Q$ and is spanned by the vectors (\ref{Q=0-states}). 

The eigenspace $\mathcal{H}_{{}_{n}}$ corresponding
to the eigenvalue $ne$, $n \in \mathbb{Z}$, is spanned by the following vectors  
\begin{equation}\label{x}
x = c_{\alpha_1}^+ \ldots c_{\alpha_\mathfrak{q}}^+ e^{-inS(u)}|0\rangle, 
\end{equation}
or $x = e^{-inS(u)}|0\rangle = |u\rangle$,
with $\mathfrak{q} = 1, 2, \ldots$, $\alpha_i = (l_i,m_i)$, $i,l=1,2, \ldots$, $-l_i \leq m_i \leq l_i$.
With this convention $x$ will sometimes be denoted by $x= U(e)x = |e\rangle$. 
In what follows $e$ stands for the basis of natural logarithms or the unit element in $G$, 
and $\mathfrak{e}^2$ in $z = n^2\mathfrak{e}^2/\pi$ stands for the square of the elementary charge, 
experimental value of which is approximately equal to $1/137$ in units in which $\hbar = c = 1$. 

We use the following
coordinate system $\left(\theta_{{}_{1}}, \varphi_{{}_{1}}, \vartheta_{{}_{1}}, \vartheta, \varphi, \lambda \right)$ on $G$. 
Let $g\in G$. Let $g_{{}_{03}}(\lambda) \in G$ be the (matrix representing) hyperbolic rotation along the $03$ 
plane with hyperbolic angle $\lambda$. Let $g_{{}_{ik}}(\theta)$, for $i,k\in\{1,2,3\}$, be the (matrix representing) spatial rotation along the $ik$ plane 
with the rotation angle $\theta$. Then each $g\in G$ can be uniquely decomposed as follows
\begin{multline}\label{g}
g=g_{{}_{12}}(\theta_{{}_{1}})g_{{}_{13}}(\varphi_{{}_{1}})g_{{}_{12}}(\vartheta_{{}_{1}})g_{{}_{12}}(-\vartheta)
g_{{}_{13}}(-\varphi)g_{{}_{03}}(\lambda)g_{{}_{13}}(\varphi)g_{{}_{12}}(\vartheta) 
\\
= a_1(\theta_{{}_{1}},\varphi_{{}_{1}},\vartheta_{{}_{1}})^*a_2(\vartheta,\varphi)^*g_{{}_{03}}(\lambda)a_2(\vartheta,\varphi)
\end{multline}     
with the invariant measure on $G$
\begin{multline*}
dg = {\textstyle\frac{1}{8\pi^2} \sin \varphi_{{}_{1}}}d\theta_{{}_{1}}d\varphi_{{}_{1}}d\vartheta_{{}_{1}} \pi^2 \textrm{sinh}^2 \lambda \sin \varphi
d\lambda d\vartheta d\varphi 
\\
= da_{{}_{1}} \pi^2 \textrm{sinh}^2 \lambda \sin  \varphi
d\lambda d\vartheta d\varphi
\end{multline*}
normalized as in \cite{NeumarkLorentzBook}, where $da_{{}_{1}} $ is the invariant measure on $SU(2,\mathbb{C})$ normalized to unity.  
Here
\[
0\leq \varphi_{{}_{1}},\varphi \leq \pi, \,\,\, 0\leq \vartheta_{{}_{1}},\vartheta \leq 2\pi, \,\,\, 0 \leq \lambda < \infty.
\]

Consistency, for each $\mathfrak{e}^2\geq 0$, has in principle 
been almost completely shown already in the works \cite{Staruszkiewicz}, 
\cite{Staruszkiewicz1992}, \cite{Staruszkiewicz1995}. We finish the proof 
initiated there. As already remarked, in a concrete Lorentz frame, the operators $S_0 = S(u),Q,c_\alpha, c_\alpha^+$, where $u=(1,0,0,0)$ is the time-like unit versor of the reference frame, can be constructed on the tensor product of the Hilbert space of square summable functions on the unit circle with the bosonic Fock space over the single particle infrared electric-type transversal states -- the standard representation. 
As shown in \cite{Staruszkiewicz1992} for the standard representation, any representation $U$ of $G$, giving transformation rule 
$S(u)'= US(u)U^{-1}, Q' = UQU^{-1} = Q, \ldots, {c'}^{+}_{\alpha}=Uc_{\alpha}U^{-1}$, of the operators $S(u),Q,c_\alpha, c_\alpha^+$,
preserving the commutation rules 
\begin{align}
[Q,S(u)] = i \mathfrak{e}, \,\, \left[Q,c_\alpha\right] = \left[S(u),c_\alpha\right] =0, &
\left[c_\alpha, c_\beta^+ \right] = 4\pi \mathfrak{e}^2 \delta_{{}_{\alpha \,\,\, \beta}},
\label{CommutationRules}
\\
c_\alpha |0\rangle = \langle 0|c_\alpha^+ = Q|0\rangle = \langle 0| Q =0, &
\label{Vacuum}
\end{align}
of the theory \cite{Staruszkiewicz}, necessary has the general form 
\begin{equation}\label{c'S'}
c'_{\alpha} = 
\sum\limits_{\beta} c_{{}_{\beta}}A_{{}_{\beta \,\, \alpha}} + B_{{}_{\alpha}}Q, \,\,\,\,\,
S(u)' = S+ 
{\textstyle\frac{1}{4\pi i\mathfrak{e}}}
\sum\limits_{\alpha,\beta}[c_{{}_{\alpha}} A_{{}_{\alpha \,\, \beta}} \overline{B_{{}_{\beta}}}
- c^+_{{}_{\alpha}} \overline{A_{{}_{\alpha \,\, \beta}}} B_{{}_{\beta}}],
\end{equation}
where  
$A$ is a unitary matrix-valued function, and $B$ is a vector-valued function on $G$ preserving the conditions
\begin{enumerate}
\item[(i)] 
$A(gh)=A(g)A(h)$, 
\item[(ii)]
$B(gh) = B(g)A(h)+B(h)$, $g,h \in G$, 
\item[(iii)] 
$B(g) = 0$, $g\in SU(2,\mathbb{C}) \subset G$, 
\end{enumerate}
with the products understood as the ordinary matrix products of a matrix $A$ with a vector $B$. Conditions (i)-(ii) immediately
follow from the assumed representation (or homomorphism) property of $U$, for (iii), compare \cite{Staruszkiewicz1995}.
(Here we are using the convention with right multiplication by $A$ in (ii), or with left multiplication by $A$ transposed in (ii), 
in order to keep ordinary representation property of $A$. In the convention used in \cite{Staruszkiewicz1992}, 
$A$ is a representatation of the group opposite to $G$). 
The second formula in (\ref{c'S'}) holds for $S(u)' = U(g)S(u)U(g)^{-1}$ with $g\in G$ only if $g \notin SU(2,\mathbb{C})$.
Next, we observe that preservation of the commutation rules by $U$ implies
preservation of the orthogonality of the complete system of vectors (\ref{x}) and of their norms, and thus implies unitarity of $U$.
It was shown in \cite{Staruszkiewicz1995} that $A=U^{{}^{(l_0=1,l_1=0)}}$ is the matrix of the irreducible unitary 
representation $(l_0=1, l_1=0)$ of \cite{Geland-Minlos-Shapiro}. 
It remains to determine $B$. Consistency will be proved if we construct explicitly  
$B$ on $G$, which together with $A$ preserves the conditions (i)-(iii). 
The equation
\begin{equation}\label{infinitesimalB}
{\textstyle\frac{d}{d\lambda}}B_{{}_{l,m}}(\lambda=0) = \mathfrak{e} {\textstyle\sqrt{\frac{8}{3}}} \delta_{{}_{l \, 1}} \delta_{{}_{m \, 0}}
\end{equation}
for $B(\lambda)=B(g_{{}_{03}}(\lambda))$, was found in \cite{Staruszkiewicz1995}, 
with the initial conditions
\begin{equation}\label{B(0)}
B_{{}_{l,m}}(\lambda=0)=0, \,\,\, l=1,2, \ldots, \,\,\, -l \leq m \leq l,
\end{equation}
and the square 
\begin{equation}\label{||B||}
\|B(g)\|^2 = \sum_\alpha |B(g)_{{}_{\alpha}}|^2 = 8 \mathfrak{e}^2(\lambda \textrm{coth}\lambda - 1)
\end{equation}
of the norm of $B$, computed in \cite{Staruszkiewicz}, for hyperbolic rotation $g$ with hyperbolic angle $\lambda$. 
Let us recall the formulas
\begin{multline}\label{A(g)}
A_{{}_{lm \,\, l'm'}}\left(g_{{}_{03}}(\lambda)\right)= A_{{}_{lm \,\, l'm'}}(\lambda) 
= U_{{}_{lm \,\, l'm'}}^{{}^{(l_0=1,l_1=0)}}\left(g_{{}_{03}}(\lambda)\right) 
\\
= \delta_{{}_{m,m'}}
\sqrt{\textstyle{\frac{l(l+1)(2l+1)(l-m)!(2l'+1)(l'-m')!}{l'(l'+1)2(l+m)!2(l'+m')!}}}
\,
\int\limits_{-1}^{1} P_{{}_{l,m}}(y) P_{{}_{l',m'}}\big(\textstyle{\frac{\textrm{tanh}(\lambda)+y}{1+\textrm{tanh}(\lambda)y}}\big)
\, dy
\end{multline}
and 
\[
U^{{}^{(l_0,l_1)}}_{{}_{lm \,\, l'm'}}\left(a\right) = \delta_{{}_{l \, l'}} \, T^{{}^{l}}_{{}_{m \,\,m'}}(a) 
\,\,\, \textrm{for} \,\,\, a \in SU(2, \mathbb{C}).
\]
Here
\begin{multline*} 
T^{{}^{l}}_{{}_{m \,\, m'}}(a) = (-1)^{2l-m-m'}{\textstyle\sqrt{\frac{(l-m)!(l+m)!}{(l-m')!(l+m')!}}}
\sum\limits_{\alpha = \textrm{max}\{0,-m-m'\}}^{\textrm{min}\{l-m,l-m'\}} \Big[
\\
{\textstyle\binom{l-m'}{\alpha}\binom{l+m'}{l-m-\alpha}
(a_{11})^{\alpha}(a_{12})^{l-m-\alpha}(a_{21})^{l-m'-\alpha}(a_{22})^{m+m'+\alpha}} \Big],
\end{multline*} 
for each fixed $l$, are the standard matrices of irreducible representations of weight $l$ of $SU(2, \mathbb{C})$,
\cite{Geland-Minlos-Shapiro}, \cite{NeumarkLorentzBook}. The formula (\ref{A(g)}) uses the realization of the representation
$(l_0=1,l_1=0)$ in the space of homogeneous of degree zero functions $f$ on the cone, thus living effectively on the unit $\mathbb{S}^2$ 
sphere in the cone, with the invariant inner product
\[
(f,g) = \int\limits_{{}_{\mathbb{S}^2}} \overline{f} \Delta_{{}_{\mathbb{S}^2}} g \,\,\, d\mu_{{}_{\mathbb{S}^2}}
\]
and with ordinary Laplace operator and invariant measure $\Delta_{{}_{\mathbb{S}^2}}, \mu_{{}_{\mathbb{S}^2}}$ on $\mathbb{S}^2$, 
with the matrix $A$ coinciding with the standard matrix of the representation $(l_0=1,l_1=0)$
\cite{Geland-Minlos-Shapiro},  \cite{NeumarkLorentzBook}, compare Subsection \ref{infra-electric-transversal-generalized-states}.
Integration of the condition (ii) along the one parameter subgroup $g_{{}_{03}}(\lambda)$ of hyperbolic rotations, 
using (\ref{infinitesimalB}) and
the initial condition (\ref{B(0)}), gives explicit formula 
for $B$ along this subgroup. Namely, writing $B\big(g_{{}_{03}}(\lambda)\big) = B(\lambda)$, condition (ii) gives
\[
B_{{}_{l,m}}(\lambda'+\lambda) = \sum\limits_{l',m'}B_{{}_{l',m'}}(\lambda')A_{{}_{l',m' \,\,\, l,m}}(\lambda) + B_{{}_{l,m}}(\lambda).
\]
Differentiating this equation with respect to $\lambda'$ at $\lambda'=0$ and using (\ref{infinitesimalB}) we obtain
\[
{\textstyle\frac{dB_{{}_{l,m}}(\lambda)}{d\lambda}} = i\mathfrak{e}{\textstyle\sqrt{\frac{8}{3}}} A_{{}_{1,0 \,\,\, l,m}}(\lambda),
\]
which, together with the initial conditions (\ref{B(0)}), determines $B(\lambda)$ uniquely.
Similarly, using (iii), we have trivial 
zero for $B$ along any one parameter subgroup of unitary elements of $G$.
Using decomposition (\ref{g}), we immediately see that conditions (i)-(iii) imply the value of $B$ at a general element (\ref{g}) to be equal
\begin{equation}\label{B(g)1}
B_{{}_{l,m}}(g) = 
B_{{}_{l,0}}(\lambda)A_{{}_{l,0 \,\,\, l,m}}\Big(g_{{}_{13}}(\varphi)g_{{}_{12}}(\vartheta) \Big) = \left(B(\lambda)A\left(a_2\right)\right)_{{}_{lm}},
\end{equation}
\begin{multline*}
B_{{}_{l,m}}\left(g^{-1}\right) = 
(-1)^l B_{{}_{l,0}}(\lambda) A_{{}_{l0 \,\,\, l,m}}\Big(g_{{}_{13}}(\varphi)g_{{}_{12}}(\vartheta)g_{{}_{12}}(-\vartheta_1)g_{{}_{13}}(-\varphi_1)g_{{}_{12}}(-\theta_1) \Big)
\\
 = \left(B(-\lambda)A\left(a_2a_{1}^{*}\right)\right)_{{}_{lm}},
\end{multline*}
\begin{equation}
 B_{{}_{l,m}}\left(g_{{}_{03}}(\lambda)\right) = B_{{}_{l,m}}(\lambda) = (-1)^l B_{{}_{l,m}}(-\lambda) 
= i \mathfrak{e} \sqrt{\textstyle{\frac{8}{3}}} \, \int\limits_{0}^{\lambda} A_{{}_{1,0 \,\, l,m}}(\lambda') d\lambda'.
\label{B(g)} 
\end{equation}
Note that $B_{{}_{l,m}}(\lambda) = 0$ for $m\neq 0$, by the property 
\[
A_{{}_{l',0 \,\, l,m}}(\lambda) = \delta_{{}_{m \, 0}} \, A_{{}_{l',0 \,\, l,m}}(\lambda)
\]
of the matrix $A(\lambda)$. 
Here $P_{{}_{l,m}}$ are the associated Legendre ``polynomials''
\[
P_{{}_{l,m}}(y)= (-1)^m 2^l (1-s^2)^{m/2} 
\displaystyle\sum_{k=m}^{l} \textstyle{\frac{k!}{(k-m)!}}\binom{l}{k}\binom{\frac{l+k-1}{2}}{l}y^{k-m}
\] 
with the generalized binomial symbol
\[
{\textstyle\binom{w}{k} = \frac{w(w-1)(w-2)\ldots (w-k+1)}{k(k-1)\ldots 1}}, \,\,\,\, w\in \mathbb{R}, k\in \mathbb{N}.
\]
To finish the proof we have to show that $A,B$ given by the formulas (\ref{A(g)}), (\ref{B(g)1}), (\ref{B(g)}), 
respect conditions (i)-(iii). (i) and (iii) are trivially fulfilled. In order to show (ii) we note the parity property 
$A_{{}_{l,0 \,\,\, l',0}}(-\lambda) = (-1)^{l+l'}A_{{}_{l,0 \,\,\, l',0}}(\lambda)$
of the matrix $A(\lambda)$ for the hyperbolic rotations parallel to the $03$ hyperplane. This parity property is not immediately seen from the 
formula for $A = U^{{}^{(l_0=1, l_1=0)}}$ given above, but can be easily seen by the exponentiation of the generator $M_{{}_{03}}$ of the representation
$(l_0=1,l_1=0)$ given explicitly in \cite{Geland-Minlos-Shapiro},  \cite{NeumarkLorentzBook}, and which immediately gives the 
series expansion at $\lambda=0$ of the general form
\begin{equation}\label{Aexpansion}
A_{{}_{l',0 \,\,\, l,0}}(\lambda) = 
a_{{}_{l', \, l, \, 0}} \lambda^{|l-l'|}+a_{{}_{l', \, l, \, 2}} \lambda^{|l-l'|+2} 
+a_{{}_{l', \, l, \, 4}} \lambda^{|l-l'|+4}
+ \ldots,
\end{equation}
\begin{equation}\label{Aexpansion'}
\\
a_{{}_{l', \, l, \, 0}} = {\textstyle\frac{(-1)^{l-l'}}{(l-l')!}}
\prod\limits_{j=0}^{l-l'-1}\Bigg[(l-j) \sqrt{{\textstyle\frac{(l-j)^2-1}{4(l-j)^2-1}}}\Bigg],
\,\,\, l'<l,
\end{equation}
\[
a_{{}_{l, \, l', \, r}} = (-1)^{l+l'}a_{{}_{l, \, l', \, r}}.  
\]   
We are not using here the explicit values of the coefficients $a_{{}_{l', \, l, \, 0}}, a_{{}_{l', \, l, \, 2}} \ldots$.
Using the parity property of $A(\lambda)$, and representation property of $A$, we can easily see that for any $g\in G$ of the form (\ref{g})
and for $B(g)$ given by (\ref{B(g)1}) and (\ref{B(g)})
\begin{equation}\label{*}
B(g)A\left(g^{-1}\right) = - B(-\lambda)A\left(a_2a_{1}^{*}\right) = - B(g^{-1}).
\end{equation}
Indeed, using (\ref{B(g)1}), (\ref{B(g)}), the representation property of $A$, and the said parity property of $A$, we have
\begin{multline*}
\Big(B(g)A\left(g^{-1}\right)\Big)_{{}_{lm}} = \Big(B(\lambda)A(g_{{}_{03}}(-\lambda)a_{2}a_{1}^*)\Big)_{{}_{lm}} 
\\
=
i \mathfrak{e} \sqrt{\textstyle{\frac{8}{3}}} \, \sum\limits_{l',l'',m''}\int\limits_{0}^{\lambda}
A_{{}_{1,0 \,\,\,l',0}}(\lambda')A_{{}_{l',0 \,\,\,l'',m''}}(-\lambda)
A_{{}_{l'',m'' \,\,\,l,m}}(a_{2}a_{1}^*) \, d\lambda'
\end{multline*}
\begin{multline*}
=
i \mathfrak{e} \sqrt{\textstyle{\frac{8}{3}}} \, \sum\limits_{l'',m''}\int\limits_{0}^{\lambda}A_{{}_{1,0 \,\,\,l'',m''}}(\lambda'-\lambda)
A_{{}_{l'',m'' \,\,\,l,m}}(a_{2}a_{1}^*) \, d\lambda'
\\
=
i \mathfrak{e} \sqrt{\textstyle{\frac{8}{3}}} \, \sum\limits_{l''}\int\limits_{-\lambda}^{0}A_{{}_{1,0 \,\,\,l'',0}}(\lambda'')
A_{{}_{l'',0 \,\,\,l,m}}(a_{2}a_{1}^*) \, d\lambda''
\end{multline*}
\begin{multline*}
=
i \mathfrak{e} \sqrt{\textstyle{\frac{8}{3}}} \, \sum\limits_{l''}(-1)^{l''+1}\int\limits_{0}^{\lambda}A_{{}_{1,0 \,\,\,l'',0}}(\lambda')
A_{{}_{l'',0 \,\,\,l,m}}(a_{2}a_{1}^*) \, d\lambda'
\\
=
i \mathfrak{e} \sqrt{\textstyle{\frac{8}{3}}} \, \sum\limits_{l''}\int\limits_{0}^{\lambda}A_{{}_{1,0 \,\,\,l'',0}}(-\lambda')
A_{{}_{l'',0 \,\,\,l,m}}(a_{2}a_{1}^*) \, d\lambda'
\end{multline*}
\begin{multline*}
=
-i \mathfrak{e} \sqrt{\textstyle{\frac{8}{3}}} \, \sum\limits_{l''}\int\limits_{0}^{-\lambda}A_{{}_{1,0 \,\,\,l'',0}}(\lambda'')\, d\lambda''\,\,
A_{{}_{l'',0 \,\,\,l,m}}(a_{2}a_{1}^*) 
\\
- \Big(B(-\lambda)A\left(a_2a_{1}^{*}\right)\Big)_{{}_{l,m}} = - B_{{}_{l,m}}(g^{-1}),
\end{multline*}
which proves (\ref{*}).
From (\ref{*}) it follows that the condition (ii) is fulfilled with $h$ of the form $g^{-1}k$, for
any $g,k \in G$, and for $B(g)$ defined by (\ref{B(g)1}), (\ref{B(g)}), 
because the right-hand side of (ii) for $h=g^{-1}k$ is equal
\[
B(g) A\left(g^{-1}k\right) + B\left(g^{-1}\right)A(k) + B(k) = B(k) 
\]
by (\ref{*}), and thus equal to the left-hand-side of (ii). Putting $k=gh$, we get (ii) for all $g,h \in G$, 
which proves consistency of the theory. 

From the general formulas (\ref{A(g)}), (\ref{B(g)}), it follows the formula (\ref{||B||}) and
\begin{equation}\label{<u,ugh-1>}
\langle u|u'\rangle = \langle u| gu\rangle  = \langle 0| e^{inS(u)}e^{-inS(gu)}|0\rangle = e^{-z(\lambda \textrm{coth} \lambda -1)},
\end{equation}    
for general $g$ of the form (\ref{g}). Consistency condition $\mathfrak{e}^2\geq 0$ follows from (\ref{||B||}). 
In Subsection \ref{AS}, compare also \cite{{wawrzyckiKernel}}, we have already proved that the invariant kernel (\ref{<u,ugh-1>})
is positive definite on the Lobachevsky space $u\cdot u=1$, for all $\mathfrak{e}^2\geq 0$, 
independently of the axioms of \cite{Staruszkiewicz}. Of course, consistency of the theory, implying unitarity of $U$,
implies also positive definiteness of the kernel 
\begin{equation}\label{kernel<g|h>ofHx}
\langle g|h\rangle, \,\,\, g,h \in G, 
\,\,\,\,\,\,\,\,\,\,\,\,\,\,\,\,\,\,\,\,
|g\rangle = U(g)x
\end{equation}
associated to the cyclic representation with any cyclic vector $x$, 
in particular with the cyclic vector of the form (\ref{x}).

\subsection{Structure of the representation $U$ of $SL(2,\mathbb{C})$  
acting in the Hilbert space $\mathcal{H}$
of the quantum phase field $S(x)$}\label{Ustructure}

Here we present the structure of the representation
$U$ of $G = SL(2, \mathbb{C})$ acting in the Hilbert space of the theory
subsumed by the axioms (I) -- (V) in the standard representation 
acting with uniform multiplicity one.

The representation $U$ and the inner product in the Hilbert space of the quantum ``phase'' field $S$ are fixed by the 
axioms (I)-(V), compare Subsection \ref{Consistency}. It turns out that they strongly 
depend on the value of the fine structure constant $e^2$, compare \cite{Staruszkiewicz1992}, 
\cite{Staruszkiewicz1992ERRATUM}, but this dependence shows up for $U$ restricted 
to the invariant subspace $\mathcal{H}_{{}_{n}}$ spanned by 
(\ref{x}) corresponding to the proper value $ne$ of $Q$ on which 
$Q = ne\boldsymbol{1}$, $n \neq 0$, and which is orthogonal to the subspace spanned by 
the vectors (\ref{Q=0-states}) and $|0\rangle$. However, the computation of the  explicit formula 
for $U$ restricted to the eigenspace on which $Q = ne\boldsymbol{1}$, $n \neq 0$ is more
difficult in comparison to the formula for $U$ on the subspace $\mathcal{H}_{{}_{n=0}}$ 
on which $Q=0$ spanned by (\ref{x}) with $n=0$ and by $|0\rangle$.

Now let $\textrm{Int} \, k$ for any positive real number $k$ be the least natural number among
all natural numbers $n$ for which $k \leq n$, say the ``integer part of $k$''.
We have the following
\begin{twr}
Let $U|_{{}_{\mathcal{H}_{{}_{n}}}}$ be the restriction of the unitary representation $U$ of $G= SL(2, \mathbb{C})$
in the Hilbert space of the quantum phase field $S$ to the invariant eigenspace ${\mathcal{H}_{{}_{n}}}$
of the total charge operator $Q$ corresponding to the eigenvalue $ne$ for some integer $n$. Then 
for all $n$ such that
\[
|n| > \textrm{Int} \Big( {\textstyle\sqrt{\frac{\pi}{e^2}}} \Big)
\]
the representations $U|_{{}_{\mathcal{H}_{{}_{n}}}}$ do not contain any supplementary series components 
and are unitarily equivalent:
\[
U|_{{}_{\mathcal{H}_{{}_{n}}}} \cong_{{}_{U}}
U|_{{}_{\mathcal{H}_{{}_{n'}}}}
\]
whenever 
\[
|n| \geq \textrm{Int} \Big( {\textstyle\sqrt{\frac{\pi}{e^2}}} \Big), \,\,\,
|n'| \geq \textrm{Int} \Big( {\textstyle\sqrt{\frac{\pi}{e^2}}} \Big).
\]
 
On the other hand if the two 
integers $n,n'$ have different absolute values $|n| \neq |n'|$ and are such that 
\[
|n| \leq  {\textstyle\sqrt{\frac{\pi}{e^2}}}, \,\,\,
|n'| \leq {\textstyle\sqrt{\frac{\pi}{e^2}}},
\]
then the representations $U|_{{}_{\mathcal{H}_{{}_{n}}}}$ and $U|_{{}_{\mathcal{H}_{{}_{n'}}}}$
are inequivalent and contain the supplementary series component.
\label{valueOfalpha}
\end{twr}

This theorem was announced in \cite{wawrzycki-alfa}. However, in the proof of the lemma $U|_{{}_{\mathcal{H}_{{}_{n}}}} = U|_{{}_{\mathcal{H}_{{}_{0}}}}
\otimes U|_{{}_{\mathcal{H}_{{}_{|u\rangle}}}}$ of \cite{wawrzycki-alfa}, 
a computational error was found \cite{Herdegen}. It trurned out not only that $U|_{{}_{\mathcal{H}_{{}_{n}}}}$ is not equal, but 
it is not unitarily equivalent to the Hilbert-space-tensor product $U|_{{}_{\mathcal{H}_{{}_{0}}}}
\otimes U|_{{}_{\mathcal{H}_{{}_{|u\rangle}}}}$, where $\mathcal{H}_{{}_{|u\rangle}}$ is the invariant subspace spanned by the transforms
of the spherically symmetric state $|u\rangle$.  But this error and invalidity of this lemma were not essential and
in \cite{wawrzyckiJT} we presented a corrected proof removing this error. We outline the proof. First, we compute explicitly
the components of $A,B$, and matrix elements of the irreducible unitary representations $(l_0,l_1 = i\rho)$ of \cite{Geland-Minlos-Shapiro}, 
as rational combinations of the exponent function. Next, we compute decomposition of each cyclic representation with
cyclic vector of the general form (\ref{x}). We do it by computing the Fourier transform (and its analytic continuation) on $G$
of the positive definite function $\varphi(g) = \langle e|g\rangle$, determined by the kernel (\ref{kernel<g|h>ofHx}) and
defined by the cyclic vector. Finally, we compare the cyclic subspaces, {i.e.} recover inclusion relations between them, 
using the orthogonality relations of the matrix elements of the representations $(l_0,l_1 = i\rho)$ and the expansions
of $A,B$ at $e \in G$. In fact we obtained in \cite{wawrzyckiJT} substantially more than is stated in theorem \ref{valueOfalpha}.
In particluar we have shown that 
\[
U|_{{}_{\mathcal{H}_{{}_{n}}}} =
\underset{l_0 \in \mathbb{Z}}{\bigoplus} \int\limits_{0}^{+\infty} (l_0,i\rho) \, \nu(\rho,z) d\rho \, \bigoplus \nu(z) \, (l_0=0, 1-z),
\,\,\,\,\, z = {\textstyle\frac{n^2\mathfrak{e}^2}{\pi}},
\]
where for each $z>0$ the weight $\nu(\rho,z)$ is almost everywhere $>0$ and with a positive weight $\nu(z)$ of the supplementary component $(l_0,l_1) = (0,1-z)$,
nonzero  if and only if $0<z<1$. From this decomposithion it follows, in particular, theorem \ref{valueOfalpha}.

\subsection{Comparison with Staruszkiewicz's theory 
continued}\label{Comparison2}

In this Subsection we give a comparison of the quantum phase field $S(x)$ of Staruszkiewicz 
theory with the homogeneous of degree zero part of the field
with the zero order contribution to the homogeneous of degree zero part
of the interacting field  $x_\mu A^{\mu}_{\textrm{int}}(x)$, \emph{i.e.}
we extend the equality (\ref{xAechi=-1(x)=transversalS}) of Subsecton \ref{AS}
all over the whole Hilbert space of the phase field $S$. 
Although our comparison presents all 
details only for the zero order contribution, it nonetheless provides the basis for comparison with the full
homogeneous of degree zero part of the interacting field
$x_\mu A^{\mu}_{\textrm{int}}(x)$, outlined in the Subsection \ref{IntAspatialInfty} 
of Introduction.


We use the factorization of $\mathcal{H}$, not very much 
convenient for the analysis of $U$, but 
it is intimately related to the standard representation, and factorizes
$S_0, Q, c_{lm}, c_{lm}, c_{lm}^{+}$, compare the end of Subsection \ref{globalU(1)}.

We do it at once under the more general Assumption-Version II, as going to the simpler
case of Assumption -- Version I, with the standard representation acting with multiplicity one (correspondingly with cyclic vacuum $|0\rangle$) is trivial.

Namely we consider, for each canonical vacuum $|0\rangle_k$ in the direct sum of standard representation, 
exactly as at the end of Subsection \ref{globalU(1)}, the
subspace $\mathscr{H}_{0k}$ spanned by the vectors
\[
e^{-inS_0}|0\rangle_k = e^{-inS(u)}|0\rangle_k, \,\,\, n \in \mathbb{Z}.
\] 
The formula 
\[
\Big( \big(c_{\alpha_1}^{+}\big)^{\beta_1} \ldots \big(c_{\alpha_\mathfrak{q}}^{+}\big)^{\beta_\mathfrak{q}}|0\rangle \Big) 
\otimes \Big( e^{-inS_0}|0\rangle\Big) = \big(c_{\alpha_1}^{+}\big)^{\beta_1} \ldots 
\big(c_{\alpha_\mathfrak{q}}^{+}\big)^{\beta_\mathfrak{q}}e^{-inS_0}|0\rangle
\]
induces a well defined Hilbert space tensor product, because 
$c_{\alpha_1}$ and $S_0 = S(u)$ are computed in one and the same reference frame with time-like 
unit versor $u=(1,0,0,0)$.
With this bilinear map the Hilbert cyclic subspace $\mathcal{H}_{|0\rangle_k}$, with the cyclic
$|0\rangle_k$ is equal to the Hilbert space tensor product
\begin{equation}\label{H=H0x(SumL2(S))}
\mathcal{H}_{|0\rangle_k} = \mathcal{H}_0 \otimes \mathscr{H}_{0k}
\end{equation}
of its subspaces $\mathcal{H}_{0k}$ and $\mathscr{H}_{0k}$.

By Assumption--Version II the operators $c_{\alpha}$, $c_{\alpha}^{+}$, $S_0$ and $Q$
act on the direct sum Hilbert space
\[
\mathcal{H} = \oplus_k \mathcal{H}_{|0\rangle_k} = \mathcal{H}_0 \otimes \Big(\oplus_k \mathscr{H}_{0k} \Big)
\]
through the direct sum of copies of the corresponding operators in the standard representation.
It is easily seen that now the operators $c_{\alpha}$, $c_{\alpha}^{+}$, $S_0$ and $Q$
(let us denote them with the same symbols) do factorize with respect to (\ref{H=H0x(SumL2(S))}) in the following manner
\begin{equation}\label{abstractS0Qclm}
\begin{split}
c_\alpha = {c'}_{\alpha} \otimes \boldsymbol{1}, \,\,\,
{c'}_{\alpha}^{+} = c_{\alpha}^{+}\big|_{\mathcal{H}_{0}}, \\
c_{\alpha}^{+} = {c'}_{\alpha}^{+} \otimes \boldsymbol{1}, \,\,\,
{c'}_{\alpha}^{+} = c_{\alpha}^{+}\big|_{\mathcal{H}_{0}}, \\
S(u) = S_0 =  \boldsymbol{1} \otimes {S'}_{0}, \,\,\,
{S'}_{0} = S_0\big|_{\oplus_k\mathscr{H}_{0k}}, \,\,\,
Q = \boldsymbol{1} \otimes Q' \,\,\,
Q' = Q\big|_{\oplus_k\mathscr{H}_{0k}}.
\end{split}
\end{equation}

By our Assumption, placed at the beginning of this Subsection, $\oplus_k\mathscr{H}_{0k}$
and the operators ${S'}_0$ and  $Q'$ on it, present in the formulas (\ref{abstractS0Qclm}), 
are identifiable by a unitary operator $U_2$ 
respectively with
\[
\begin{split}
U_{2} \mathscr{H}_0 = \oplus L^2(\mathbb{S}^1), \\
U_{2} {S'}_0 U_{2}^{-1} = \oplus {S'_0} =  \boldsymbol{{S'}_0} \,\,\,\textrm{on} \,\,\,
\oplus L^2(\mathbb{S}^1), \\
U_{2} Q' U_{2}^{-1} = \oplus {Q'} =  \boldsymbol{Q'} \,\,\,\textrm{on} \,\,\,
\oplus L^2(\mathbb{S}^1),
\end{split}
\]
where on the left-hand sides the operators $Q', {S'}_0$ refers to the operators
defined in (\ref{abstractS0Qclm}) in the abstract representation, and the same symbols
$Q', {S'}_0$ on the right denote the operators in the standard representation 
(\ref{Hilb-space-rep-clm}) and (\ref{rep-clm})
(acting with multiplicity one) and the bolded $\boldsymbol{Q'}$, $\boldsymbol{{S'}_0}$
the corresponding operators in the standard representation acting with multiplicity.

Now it is easily seen that putting for $U_1$ the Fock lifting of the unitary operator
which the single particle space basis vector ${c'}_{\alpha}^{+}|0\rangle$
of the Fock space $\mathcal{H}_{\textrm{Fock}}$ (in (\ref{rep-clm}))
puts into correspondence with ${c'}_{\alpha}^{+}|0\rangle = {c}_{\alpha}^{+}|0\rangle$
of $\mathcal{H}_0$
we obtain the equalities 
\[
\begin{split}
U_{1}  \mathcal{H}_0  = \mathcal{H}_{\textrm{Fock}}, \\
U_{1} {c'}_{\alpha}U_{1}^{-1} = {c'}_{\alpha} \,\,\,\textrm{on} \,\,\,
 \mathcal{H}_{\textrm{Fock}}, \\
U_{1} {c'}_{\alpha}^{+}U_{1}^{-1} = {c'}_{\alpha}^{+} \,\,\,\textrm{on} \,\,\,
 \mathcal{H}_{\textrm{Fock}},
\end{split}
\]
where on the left-hand side there are operators ${c'}_{\alpha},{c'}_{\alpha}^{+}$
of the standard representation (\ref{Hilb-space-rep-clm}) and (\ref{rep-clm})
and on the right-hand side there are the operators  ${c'}_{\alpha},{c'}_{\alpha}^{+}$ standing 
in (\ref{abstractS0Qclm}) in the abstract representation.

Denoting the operators $S_0$,$Q$, $c_{lm}$, $c_{lm}^{+}$, $U$ in the standard representation
acting with uniform multiplicity with the bold fonts, $\boldsymbol{S_0}$,$\boldsymbol{Q}$, 
$\boldsymbol{c_{lm}}$, $\boldsymbol{c_{lm}^{+}}$, $\boldsymbol{U}$,
and putting $V = U_{1}^{-1} \otimes U_{2}^{-1}$ we thus obtain
$V \boldsymbol{S_0} V^{-1} = S_0$, $V \boldsymbol{Q} V^{-1}=Q$, 
$V \boldsymbol{c_{lm}} V^{-1}= c_{lm}$, $V \boldsymbol{c_{lm}^{+}}V^{-1}=c_{lm}^{+}$ and 
$V\boldsymbol{U}V^{-1} = U$.

Now let 
\[
S(x) = S_0 -e \, x_0/r \, Q + \sum \limits_{l=1}^{\infty} \sum \limits_{m = -l}^{m= +l}
\{ c_{lm} f_{lm}^{(+)}(x) + \textrm{h.c.} \}, \,\,\,\, x\cdot x <0,
\]
be the expansion (II) of the quantum phase operator $S(x)$. Then 
joining this results with the results obtained in Subsections \ref{AS} and
\ref{infra-electric-transversal-generalized-states} we have shown that 
\[
-e x_\mu A_{{}_{\chi=-1}}^{e \, \mu}(x)\Bigg|_{{}_{\Gamma({\mathcal{H}'}^{e}_{{}_{\chi = -1}}) \otimes |0\rangle^{\mathfrak{m}}}} = 
\sum \limits_{l=1}^{\infty} \sum \limits_{m = -l}^{m= +l}
\{ c_{lm} f_{lm}^{(+)}(x) + \textrm{h.c.} \}, 
\]
where $A_{{}_{\chi =-1}}^{e \, \mu}(x)$ is the homogeneous of degree
$\chi=-1$ electric part of the free electromagnetic potential $A^{\mu}(x)$ field,
constructed in Subsection \ref{AS}.
Here $f_{lm}^{(+)}$ coincide with the partial waves of the expansion of the phase in the domain $x\cdot x<0$,
and can be understood there as extensions of the partial waves from de Sitter $3$-hyperboloid over the whole spacetime by keeping
homogeneity zero.

\subsection{The (asymptotically) homogeneous of degree zero part of the first order contribution 
$x_\mu A^{(1) \, \mu}_{{}_{\textrm{int}}}(x)$.
Comparison with Staruszkiewicz theory continued}\label{A(1)chi}

In this Subsetion we compute the asymptotically (infrared/ultraviolet) homogeneous of degree zero part
\[
\underset{\lambda\rightarrow +\infty}{\textrm{lim}} \,
\big[-e x_\mu \lambda A^{(1) \, \mu}_{{}_{\textrm{int} \, \chi}}\big(\lambda x\big)\big]
\,\,\, (\textrm{infrared})
\] 
or, respectively,
\[
\underset{\lambda\rightarrow +\infty}{\textrm{lim}} \,
\big[-e x_\mu {\textstyle\frac{1}{\lambda}} A^{(1) \, \mu}_{{}_{\textrm{int} \, \chi}}\big({\textstyle\frac{1}{\lambda}}x\big)\big]
\,\,\, (\textrm{ultraviolet}),
\] 
of the first order contribution
$-e x_\mu A^{(1) \, \mu}_{{}_{\textrm{int}}}(x)$.

We compare the asymptotically (infrared) homogeneous of degree zero part
\[
\underset{\lambda\rightarrow +\infty}{\textrm{lim}} \,
\big[-e x_\mu \lambda A^{(1) \, \mu}_{{}_{\textrm{int} \, \chi}}\big(\lambda x\big)\big]
\] 
of the first order contribution
$-e x_\mu A^{(1) \, \mu}_{{}_{\textrm{int}}}(x)$ with the part $-e {\textstyle\frac{x_0}{|\boldsymbol{x}|}}Q$
of the phase $S(x)$ of Staruszkiewicz theory. This comparison depends on the reference frame, as the division
of the phase $S(x)$ into the ``transversal'' part
\[
\sum \limits_{l=1}^{\infty} \sum \limits_{m = -l}^{m= +l}
\{ c_{lm} f_{lm}^{(+)}(x) + \textrm{h.c.} \} 
\]
and the ``longitudinal'' part
\[
S_0 -e \, {\textstyle\frac{x_0}{|\boldsymbol{x}|}} \, Q 
\]
is not Lorentz invariant. Correspondingly the equality 
\[
-e x_\mu A_{{}_{\chi=-1}}^{e \, \mu}(x)\Bigg|_{{}_{\Gamma({\mathcal{H}'}^{e}_{{}_{\chi = -1}}) \otimes \mathcal{H}'_1}} = 
\sum \limits_{l=1}^{\infty} \sum \limits_{m = -l}^{m= +l}
\{ c_{lm} f_{lm}^{(+)}(x) + \textrm{h.c.} \}\Bigg|_{{}_{\mathcal{H}_{0}\otimes\mathscr{H}_{0}}}
\]
holds in the factorization 
\[
\mathcal{H} = \mathcal{H}_{0}\otimes\mathscr{H}_{0}
\]
of the Hilbert space $\mathcal{H}$ of the phase $S$, which depends on the reference frame. Namely, the factor
$\mathscr{H}_{0}$ is a closed subspace of $\mathcal{H}$, spanned by the following spherically symmetric
eigenstates of the total charge operator $Q$:
\[
|n,u\rangle = e^{-inS(u)}|0\rangle = e^{-inS_0}|0\rangle, \,\,\,\, n \in \mathbb{Z},
\]
where $u$ is the time like unit versor of the reference frame in which the partial waves $f_{lm}^{(+)}$ were computed.

In order to show that the homogeneous of degree zero electric part
\[
-e x_\mu A_{{}_{\chi=-1}}^{e \, \mu}(x)
\] 
of the free electromagnetic potential
field $A$, together with the homogeneous of degree zero part 
\[
\underset{\lambda\rightarrow +\infty}{\textrm{lim}} \,
\big[-e x_\mu \lambda A^{(1) \, \mu}_{{}_{\textrm{int} \, \chi}}\big(\lambda x\big)\big]
\] 
of $-e x_\mu A^{(1) \, \mu}_{{}_{\textrm{int}}}(x)$ is equal to the whole phase $S(x)$, except the therm $S_0$,
we have to prove existence of the spherically symmetric states $|S_{m,u}\rangle$ of the field
$-e x_\mu A^{(1) \, \mu}_{{}_{\textrm{int} \, \chi}}(x)$, which correspond to the spherically symmetric
states $|m,u\rangle$, and which have the following property
\[
\underset{\lambda\rightarrow +\infty}{\textrm{lim}} \,
\big[-ex_\mu \lambda A^{(1) \, \mu}_{{}_{\textrm{int} \, \chi}}\big(\lambda x\big)\big]|S_{m,u}\rangle = 
-e {\textstyle\frac{x_0}{|\boldsymbol{x}|}}Q|S_{m,u}\rangle, \,\,\,\,\, m\in \mathbb{Z}, \,\, x\cdot x < 0
\]
at least outside the light cone. Equivalently, we need to show existence of spherically symmetric states $|S_{m,u}\rangle$, 
$m\in \mathbb{Z}$, such that
\begin{equation}\label{-exA(1)intchiSmu=-e2mx0/rSmu}
\underset{\lambda\rightarrow +\infty}{\textrm{lim}} \,
\big[-e x_\mu \lambda A^{(1) \, \mu}_{{}_{\textrm{int} \, \chi}}\big(\lambda x\big)\big]|S_{m,u}\rangle = 
-me^2 {\textstyle\frac{x_0}{|\boldsymbol{x}|}} |S_{m,u}\rangle, \,\,\,\,\, m\in \mathbb{Z}, \,\, x\cdot x < 0.
\end{equation}
This Subsection is devoted to the solution of this problem. Here, in the natural units, in which $\hslash = c= 1$, the elementary
charge $e=\frac{1}{\sqrt{137}}$ for $e$ in the left-hand side of (\ref{-exA(1)intchiSmu=-e2mx0/rSmu}) if the elementary charge of the field coupled to the
potential is equal to the electron's charge. The constatnt $e$ on the right-hand side of (\ref{-exA(1)intchiSmu=-e2mx0/rSmu}) is the elementary charge, 
the universal constant of the Staruszkiewicz theory, which we have \emph{a priori} assumed to be the same, but in fact the actual relation between
the two constants should be determined by the equality (\ref{-exA(1)intchiSmu=-e2mx0/rSmu}), which is supposed to hold for spinor, scalar, \emph{e.t.c.}
QED's, if among the charged fields of the whole system there is present the field with the charge equal to the electron's charge.   

The first order contribution to the interacting potential field $A_{{}_{\textrm{int}}}$ (compare the formula (\ref{1-ord-A-g=1}) of Introduction) 
\[
A_{{}_{\textrm{int}}}^{\mu \,(1)}(g=1,x) =
-{\textstyle\frac{e}{4\pi}} \int \ud^3 \boldsymbol{x_{1}}
{\textstyle\frac{1}{|\boldsymbol{x_1} - \boldsymbol{x}|}}
\,
{:} \boldsymbol{\psi}^+ \gamma^0 \gamma^\mu \boldsymbol{\psi} {:} (x_0 - |\boldsymbol{x_1} - \boldsymbol{x}|, \boldsymbol{x_1})
\]
in spinor causal perturbative QED has been analyzed in details in Subsection \ref{analysis-of-klm-A(1)}. By the general results
of Subsection \ref{OperationsOnXi}, \ref{WickForProduct}, \ref{WickForChronological}, the first order contribution
$A_{{}_{\textrm{int}}}^{\mu \,(1)}(g=1)$, together with all higher order contributions, 
is a generalized integral kernel operator 
\begin{multline*}
{A_{{}_{\textrm{int}}}}_{\mu}^{\, (1)}(g=1, x) = \\
= \sum \limits_{s,s'=1}^{2} \int \limits_{\mathbb{R}^3\times \mathbb{R}^3}
{\kappa'}_{2,0}^{+-}(\boldsymbol{\p}', s', \boldsymbol{\p}, s; \mu, x) \,
b_{s'}(\boldsymbol{\p}')^{+}d_{s}(\boldsymbol{\p})^{+} \, \ud^3 \boldsymbol{\p}' \ud^3 \boldsymbol{\p} \\
+ \sum \limits_{s,s'=1}^{2} \int \limits_{\mathbb{R}^3\times \mathbb{R}^3}
{\kappa'}_{1,1}^{++}(\boldsymbol{\p}', s', \boldsymbol{\p}, s; \mu, x) \,
b_{s'}(\boldsymbol{\p}')^{+}b_{s}(\boldsymbol{\p}) \, \ud^3 \boldsymbol{\p}' \ud^3 \boldsymbol{\p} \\
+\sum \limits_{s,s'=1}^{2} \int \limits_{\mathbb{R}^3\times \mathbb{R}^3}
{\kappa'}_{1,1}^{--}(\boldsymbol{\p}', s', \boldsymbol{\p}, s; \mu, x) \,
d_{s'}(\boldsymbol{\p}')^{+}d_{s}(\boldsymbol{\p}) \, \ud^3 \boldsymbol{\p}' \ud^3 \boldsymbol{\p} \\
\sum \limits_{s,s'=1}^{2} \int \limits_{\mathbb{R}^3\times \mathbb{R}^3}
{\kappa'}_{0,2}^{-+}(\boldsymbol{\p}', s', \boldsymbol{\p}, s; \mu, x) \,
d_{s'}(\boldsymbol{\p}')b_{s}(\boldsymbol{\p}) \, \ud^3 \boldsymbol{\p}' \ud^3 \boldsymbol{\p}
\end{multline*} 
with vector valued kernels in the sense of \cite{obataJFA}. But, accordingly
to the Proposition of Subsection \ref{analysis-of-klm-A(1)}, $A_{{}_{\textrm{int}}}^{\mu \,(1)}(g=1)$ 
is a regular integral kernel operator which transforms continuously
the test space into the nuclear space of continuous operators from the test Hida space into itself.
In particular the first order contribution $A_{{}_{\textrm{int}}}^{\mu \,(1)}(g=1)$ is a well-defined operator valued distribution.
The kernels ${\kappa'}_{\mathpzc{l},\mathpzc{m}}^{\pm\pm}$, $(\mathpzc{l},\mathpzc{m})$  $=(2,0)$, $(0,2)$, $(1,1)$,
of $A_{{}_{\textrm{int}}}^{\mu \,(1)}(g=1)$ are given in Subsection \ref{analysis-of-klm-A(1)}. 

Of course the form of the first order contribution $A_{{}_{\textrm{int}}}^{\mu \,(1)}(g=1)$ depends on the type of QED in question. 
For example for the scalar causal perturbative QED, where we have the scalar complex (massive) field $\boldsymbol{\psi}$
coupled minimally to the electromagnetic potential, the electromagnetic current field has different form
\[
{:} \boldsymbol{\psi}^+ \overset{\leftrightarrow}{\partial}{}^\mu \boldsymbol{\psi} {:}(x)
\overset{\textrm{df}}{=} \,\, {:} \big[\boldsymbol{\psi}(x)^+ \partial^\mu \boldsymbol{\psi}(x) - \partial^\mu\boldsymbol{\psi}(x)^+ \boldsymbol{\psi}(x)\big]{:}, 
\]
and the first order contribution is equal 
\[
A_{{}_{\textrm{int}}}^{\mu \,(1)}(g=1,x) =
-{\textstyle\frac{e}{4\pi}} \int \ud^3 \boldsymbol{x_{1}}
{\textstyle\frac{1}{|\boldsymbol{x_1} - \boldsymbol{x}|}}
\,
{:} \boldsymbol{\psi}^+ \overset{\leftrightarrow}{\partial}{}^\mu \boldsymbol{\psi} {:} (x_0 - |\boldsymbol{x_1} - \boldsymbol{x}|, \boldsymbol{x_1}).
\]
Analogue methods, which we have used for spinor QED, show that also in scalar QED the first order contribution $A_{{}_{\textrm{int}}}^{\mu \,(1)}(g=1)$ 
to the interacting field $A_{{}_{\textrm{int}}}$ is a regular integral kernel operator, which is equal to an ordinary operator valued distribution.

The first order approximation in perturbative QED is in perfect analogy with the first order approximation
in classical electrodynamics, in which the back reaction of the field on the electric current distribution
is neglected. This approximation leads to the well-known construction of the so called
retarded potential, associated to the electric current distribution. Indeed, it is easily seen that also
in perturbative QED the first order contribution respects the equation
\[
\partial^\nu\partial_\nu A_{{}_{\textrm{int}}}^{\mu \,(1)}(g=1,x) = -e \,\, {:} \boldsymbol{\psi}^+ \gamma^0 \gamma^\mu \boldsymbol{\psi} {:} (x)
\]
in spinor QED, and analogously 
\[
\partial^\nu\partial_\nu A_{{}_{\textrm{int}}}^{\mu \,(1)}(g=1,x) = 
-e \,\, {:} \boldsymbol{\psi}^+ \overset{\leftrightarrow}{\partial}{}^\mu \boldsymbol{\psi} {:} (x)
\]
in scalar QED, with the unperturbed (zero order approximation of the) electric current field on the right-hand side of the equation. 
Note also, that the formula for $A_{{}_{\textrm{int}}}^{\mu \,(1)}(g=1)$  is the immediate quantum field analogue of the formula
for the classical retarded potential.

In order to determine the homogeneous of degree $-1$ part of the first order contribution $A_{{}_{\textrm{int}}}^{\mu \,(1)}(g=1)$
we compute first the \emph{asymptotically} homogeneous part $A_{{}_{\textrm{int} \, \chi}}^{\mu \,(1)}(g=1)$ 
of the first order contribution $A_{{}_{\textrm{int}}}^{\mu \,(1)}(g=1)$, determined through the decomposition of the
$SL(2, \mathbb{C})$ acting in the two particle Hilbert space of the free Dirac field. The decomposition component (or asymptotically
homogeneous part) $A_{{}_{\textrm{int} \, \chi}}^{\mu \,(1)}(g=1)$ is naturally determined by the  
decomposition of $A_{{}_{\textrm{int}}}^{\mu \,(1)}(g=1)$, associated to the decomposition of $SL(2, \mathbb{C})$
acting  in $E^{\otimes \, 2}$ into the direct integral component fields $A_{{}_{\textrm{int} \, \chi}}^{\mu \,(1)}(g=1)$,
each having (ultraviolet asymptotic) homogeneity $-1$. 
Indeed, $A_{{}_{\textrm{int}}}^{\mu \,(1)}(g=1; \varphi)$, as a finite sum of integral kernel operators $\Xi(\kappa_{\mathpzc{l},\mathpzc{m}}(\varphi))$,
with $\kappa_{\mathpzc{l},\mathpzc{m}} \in E^{\otimes \, (\mathpzc{l}+\mathpzc{m})}$, and $\mathpzc{l}+\mathpzc{m} = 2$,
can be decomposed into a direct integral (compare Subsection \ref{psichi})
\[
\sum\limits_{\mathpzc{l}+\mathpzc{m} = 2}  \Xi(\kappa_{\mathpzc{l},\mathpzc{m}}(\varphi)) 
= \sum\limits_{\mathpzc{l}+\mathpzc{m} = 2} \int \Xi_{{}_{\chi}}(\kappa_{\chi, \, \mathpzc{l},\mathpzc{m}}(\varphi))
\, \ud\chi
\]
with the kernels $\kappa_{\chi, \, \mathpzc{l},\mathpzc{m}}(\varphi)$ of the decomposition component operators 
$\Xi_{{}_{\chi}}(\kappa_{\chi, \, \mathpzc{l},\mathpzc{m}}(\phi))$ equal to the Fourier
transforms $\mathcal{F}[\kappa_{\mathpzc{l},\mathpzc{m}}(\varphi)](\chi,\chi)$ of the distributions 
$\kappa_{\mathpzc{l},\mathpzc{m}}(\varphi)\in E^{\otimes \, 2}$, restricted to the diagonal, compare Subsection \ref{psichi}.
Let us emphasize that $\mathcal{F}$ is the Fourier transform associated to the decomposition of $SL(2, \mathbb{C})$ acting
in  the two particle nuclear space $E^{\otimes \, (\mathpzc{l}+\mathpzc{m})} = E^{\otimes \, 2}$ of the Dirac field,
or in the two particle Hilbert space of the Dirac free field. Thus, in order to determine the kernels
$\kappa_{\chi, \, \mathpzc{l},\mathpzc{m}}^{\pm\pm}(\varphi)$ of
\[
A_{{}_{\textrm{int} \, \chi}}^{\mu \,(1)}(g=1; \varphi_\mu) = 
\sum\limits_{\mathpzc{l}+\mathpzc{m} = 2} \Xi_{{}_{\chi}}(\kappa_{\chi, \, \mathpzc{l},\mathpzc{m}}(\varphi)) 
= \sum\limits_{\mathpzc{l}+\mathpzc{m} = 2, \pm} \Xi_{{}_{\chi}}(\kappa_{\chi, \, \mathpzc{l},\mathpzc{m}}^{\pm\pm}(\varphi)) 
\]
we need to compute the Fourier transforms $\mathcal{F}[\kappa_{\mathpzc{l},\mathpzc{m}}^{\pm\pm}(\varphi)](\chi,\chi)$ accordingly to the
general prescription given in Subsection \ref{psichi}. 

Decomposition of the free field (see Subsection \ref{psichi} for the Dirac free field and Subsection \ref{equivalentA-s}
for the free  electromagnetic potential field) is the special case of the decomposition 
\[
\boldsymbol{\psi}(\varphi)
=
\int \boldsymbol{\psi}_{{}_{\chi}}(\varphi)  \, \ud\chi
= \int \big[ \Xi_{{}_{\chi}}(\kappa_{\chi \, 0,1}(\varphi)) + \Xi(\kappa_{\chi \, 1,0}(\varphi)) \big] \, \ud \chi
\]
of the integral kernel operator
\[
\boldsymbol{\psi}(\varphi)
= \Xi(\kappa_{0,1}(\varphi)) + \Xi(\kappa_{1,0}(\varphi)), \,\,\, \kappa_{0,1}(\varphi), \kappa_{1,0}(\varphi) \in E \subset \mathcal{H}' \subset E^*
\]
with the kernels $\kappa_{\chi \, 0,1}(\varphi)$, $\kappa_{\chi \, 1,0}(\varphi)$ of the decomposition component fields 
(integral kernel operators) equal to the
Fourier transforms $\mathcal{F}[\kappa_{0,1}(\varphi)](\chi)$, $\mathcal{F}[\kappa_{1,0}(\varphi)](\chi)$. But here 
with the Fourier transform $\mathcal{F}$ associated to the decomposition of $SL(2, \mathbb{C})$
acting in the single particle nuclear and singe particle Hilbert space, composing the single particle
Gelfand triple $E \subset \mathcal{H}' \subset E^*$, compare Subsections \ref{psichi}, \ref{equivalentA-s}.

Let $\boldsymbol{\psi}_{{}_{\chi}}$ be the asymptotically (ultraviolet) homogeneous decomposition component
of the free (massive) complex\footnote{Rather non-neutral, \emph{i.e.} charged, if understood as a quantum field.}
field $\boldsymbol{\psi}$ coupled to the electromagnetic potential.
The kernels of these components $\boldsymbol{\psi}_{{}_{\chi}}$ are well-defined distributions, regular outside the
light cone, but quite singular on the cone and inside the light cone. In particular these kernels cannot be multiplied,
as there is no general rule for multiplication of singular distributions. In particular kernels of the
decomposition components of the Wick products representing the current, cannot be naively computed through
the ''Wick products'' ${:}\boldsymbol{\psi}^{\sharp}_{{}_{\chi}}(x) \gamma^0\gamma^\mu \boldsymbol{\psi}_{{}_{\chi}}(x){:}$
of the component fields, because there is no natural rule for pointwise multiplication of the kernels of the component fields $\boldsymbol{\psi}_{{}_{\chi}}$,
which cannot be naturally multiplied, being singular on the cone or inside the light cone. Therefore, the heuristic rules
\begin{equation}\label{HeuriscticA(1)chispinor}
A_{{}_{\textrm{int} \, \chi}}^{\mu \,(1)}(g=1,x) =
-{\textstyle\frac{e}{4\pi}} \int \ud^3 \boldsymbol{x_{1}}
{\textstyle\frac{1}{|\boldsymbol{x_1} - \boldsymbol{x}|}}
\,
{:} \boldsymbol{\psi}_{{}_{\chi}}^+ \gamma^0 \gamma^\mu \boldsymbol{\psi}_{{}_{\chi}} {:} (x_0 - |\boldsymbol{x_1} - \boldsymbol{x}|, \boldsymbol{x_1}),
\end{equation}
in spinor case or, respectively,
\begin{equation}\label{HeuriscticA(1)chiscalar}
A_{{}_{\textrm{int} \, \chi}}^{\mu \,(1)}(g=1,x) =
-{\textstyle\frac{e}{4\pi}} \int \ud^3 \boldsymbol{x_{1}}
{\textstyle\frac{1}{|\boldsymbol{x_1} - \boldsymbol{x}|}}
\,
{:} \boldsymbol{\psi}_{{}_{\chi}}^+ \overset{\leftrightarrow}{\partial}{}^\mu \boldsymbol{\psi}_{{}_{\chi}} {:} (x_0 - |\boldsymbol{x_1} - \boldsymbol{x}|, \boldsymbol{x_1})
\end{equation}
in the scalar complex case, cannot be used. Therefore, we should compute $A_{{}_{\textrm{int} \, \chi}}^{\mu \,(1)}(g=1,x)$
from the beginning, similarly as the decomposition components of the free electromagnetic potential, as naturally
determined by the decomposition of $SL(2, \mathbb{C})$ acting in the two particle Hilbert space,
and should compute the kernels of $A_{{}_{\textrm{int} \, \chi}}^{\mu \,(1)}(g=1,x)$ as the Fourier transform $\mathcal{F}$
of the kernels of $A_{{}_{\textrm{int}}}^{\mu \,(1)}(g=1,x)$. Of course, now with $\mathcal{F}$
determined by the decomposition of $SL(2, \mathbb{C})$ acting in the two particle Hilbert space.
Decomposition of a wide class of integral kernel operators, as naturally arising from decomposition of the representation
of $SL(2, \mathbb{C})$ acting in the Fock space of the free fields, including decomposition of higher order contributions to interacting
fields, has been given in Subsection \ref{psichi}.
Although
for realistic QED the formulas (\ref{HeuriscticA(1)chispinor}) and (\ref{HeuriscticA(1)chiscalar}) cannot be used,
they have a heuristic value, and for more regular abstract integral kernel operators, instead of $\boldsymbol{\psi}$, for which the kernels of decomposition components
$\boldsymbol{\psi}_{{}_{\chi}}$ can be multiplied, these formulas are valid.

Let us remind, that the asymptotically (ultraviolet) homogeneous
parts $\boldsymbol{\psi}_{{}_{\chi}}$ of the free massive field $\boldsymbol{\psi}$ are naturally and uniquely determined
by the direct integral decomposition
\[
\int U(\alpha)_{{}_{\chi}} \, \ud \chi,
\]
into irreducible components $U(\alpha)_{{}_{\chi}}$
of the representation $\alpha \mapsto U(\alpha)$ of $SL(2, \mathbb{C})$ acting in the single particle Hilbert space
\[
\mathcal{H}^{\oplus}_{\boldsymbol{m},0} \oplus \mathcal{H}^{\ominus \, \flat}_{-\boldsymbol{m},0}
= \int \big(\mathcal{H}^{\oplus}_{\boldsymbol{m},0}\big)_{{}_{\chi}} \oplus \big(\mathcal{H}^{\ominus \, \flat}_{-\boldsymbol{m},0}\big)_{{}_{\chi}}
\, \ud \chi
\]
of the free massive field $\boldsymbol{\psi}$. This decomposition is determined by the spectral
decomposition of the two Casimir operators of the group $SL(2, \mathbb{C})$. In fact each decomposition
Hilbert space
\[
\big(\mathcal{H}^{\oplus}_{\boldsymbol{m},0}\big)_{{}_{\chi}} \oplus \big(\mathcal{H}^{\ominus \, \flat}_{-\boldsymbol{m},0}\big)_{{}_{\chi}}
\]
is spanned by the generalized eigenstates of the two Casimir operators, and the decomposition
parameter $\chi$ can be identified as the asymptotic homogeneity for infinite momenta of the
generalized eigenstates of the two Casimir operators which span the decomposition component
\[
\big(\mathcal{H}^{\oplus}_{\boldsymbol{m},0}\big)_{{}_{\chi}} \oplus \big(\mathcal{H}^{\ominus \, \flat}_{-\boldsymbol{m},0}\big)_{{}_{\chi}},
\]
designated by $\chi$.
Indeed, although the homogeneity does not make sense on the massive hyperboloid $\mathscr{O}_{{}_{\boldsymbol{m},0,0,0}}$,
the asymptotic homogeneity degree $\chi$
(for infinitely large momenta $p$ and at zero in space-time coordinates $x$)
makes perfect sense and is equal to the invariant which parametrizes the irreducible components. This means that
(in space-time picture in $x$-variables) for the generalized eigenstate $f\neq 0$
in
\[
\big(\mathcal{H}^{\oplus}_{\boldsymbol{m},0}\big)_{{}_{\chi}} \oplus \big(\mathcal{H}^{\ominus \, \flat}_{-\boldsymbol{m},0}\big)_{{}_{\chi}}
\]
the limit
\begin{equation}\label{functionlim1/lambdaf(x/lambda)}
\underset{\lambda\rightarrow +\infty}{\textrm{lim}} \,
\big({\textstyle\frac{1}{\lambda}}\big)^{-s}f\big({\textstyle\frac{1}{\lambda}}x\big)
\end{equation}
exists for $s=\chi$, and represents a homogeneous of degree $\chi$ function, whereas it is zero or does not exist
for $s\neq \chi$.
The limit (\ref{functionlim1/lambdaf(x/lambda)}) should be rather understood in the distributional sense
\begin{equation}\label{distributionlim1/lambdaf(x/lambda)}
\underset{\lambda\rightarrow +\infty}{\textrm{lim}} \,
\big({\textstyle\frac{1}{\lambda}}\big)^{-s} \lambda^4 f\big(S_\lambda\varphi\big), \,\,\,\,\,
S_\lambda\varphi(x) = \varphi(\lambda x)
\end{equation}
for any test function $\varphi$.
In the mass less case the decomposition is determined by the scaling operator (compare Subsection \ref{equivalentA-s})
commuting with the representation $\alpha \mapsto U(\alpha)$ of $SL(2, \mathbb{C})$, and the invariant $\chi$,
determined by the spectrum of the scaling operator becomes equal to the ordinary homogeneity. In each case, massive or mass less (although
it seems that the mass less charge carrying fields $\boldsymbol{\psi}$ seem to be non-existing in reality), the
decomposition parameter $\chi$ is equal to the (asymptotic) homogeneity determined by the spectrum of the
two Casimir operators (in the massive case) or by the spectrum of the scaling operator (in the mass less case).

The direct integral decomposition of the free charged field $\boldsymbol{\psi}$
into the (asymptotically) homogeneous fields $\boldsymbol{\psi}_{{}_{\chi}}$, associated to the decomposition
of the representation of $SL(2, \mathbb{C})$, we have determined in Subsection \ref{psichi} for the
massive Dirac field. For the massive scalar field $\boldsymbol{\psi}$, we obtain the decomposition using the same method
and the decomposition of the representation of $SL(2, \mathbb{C})$ acting in the single particle Hilbert space
of $\boldsymbol{\psi}$ (\emph{i.e.} on the scalars living on the Lobachevsky space or on the positive energy sheet
$\mathscr{O}_{{}_{\boldsymbol{m},0,0,0}}$) determined in \cite{GelfandV}. The case of mass less $\boldsymbol{\psi}$
is simpler, and the direct integral decomposition of the representation of $SL(2, \mathbb{C})$ acting in the single particle space
together with the associated decomposition of the field can be determined as in Subsection\footnote{In particular for the
mass less spinor field we replace the eigenvectors $w$ of Subsection \ref{equivalentA-s} by the fundamental solutions
$u,v$ of the mass less Dirac equation, which in the mass less case become homogeneous of degree zero functions on the cone.}
\ref{equivalentA-s}.

In general, we obtain the following formula (compare the formula (\ref{psichi(x)}) of Subsection \ref{psichi})
\begin{equation}\label{psichi(x)AS}
\boldsymbol{\psi}_{{}_{\chi}}(x) = \sum\limits_{lm} \overline{ f^{{}^{\oplus}}_{{}_{\chi, \, lm}}(x) } \,\,\, b'_{{}_{\chi, \, lm}}
+
\sum\limits_{lm} \overline{f^{{}^{\ominus}}_{{}_{\chi, \, lm}}(x) } \,\,\, {d'}_{{}_{\chi, \, lm}}^+
\end{equation}
for the (asymptotically) homogeneous fields $\boldsymbol{\psi}_{{}_{\chi}}$,
with distributional positive $f^{{}^{\oplus}}_{{}_{\chi, \, lm}}$ and negative energy solutions
$f^{{}^{\ominus}}_{{}_{\chi, \, lm}}$ of the associated equation, e.g. Dirac equation in case of the spinor field.
Their Fourier transforms, regarded as distributions,
are equal
\[
\begin{split}
\mathcal{S}(\mathbb{R}^4; \mathbb{C}^4) \ni \widetilde{\phi} \longmapsto
F^{{}^{\oplus}}_{{}_{\chi, \, lm}}(\widetilde{\phi}|_{{}_{\mathscr{O}_{\boldsymbol{m},0,0,0}}})
\\
\mathcal{S}(\mathbb{R}^4; \mathbb{C}^4) \ni \widetilde{\phi} \longmapsto
F^{{}^{\ominus}}_{{}_{\chi, \, lm}}(\widetilde{\phi}|_{{}_{\mathscr{O}_{-\boldsymbol{m},0,0,0}}}).
\end{split}
\]
The corresponding functions $F^{{}^{\oplus}}_{{}_{\chi, \, lm}}$ on the massive
hyperboloid $\mathscr{O}_{{}_{\boldsymbol{m},0,0,0}}$ (or on the cone $\mathscr{O}_{{}_{1,0,0,1}}$
for mass less $\boldsymbol{\psi}$), has been determined in Subsection \ref{psichi}
for the spinor massive field $\boldsymbol{\psi}$, and both $\{F^{{}^{\oplus}}_{{}_{\chi=-1/2+i\nu, \, lm}}\}$
and $\{F^{{}^{\ominus}}_{{}_{\chi=-1/2+i\nu, \, lm}}\}$, $\nu \in \mathbb{R}_+$, span the irreducible unitary representation
$(l_0=1/2, l_1=i\nu)$ of \cite{Geland-Minlos-Shapiro}. $F^{{}^{\oplus}}_{{}_{\chi, \, lm}}$ and $F^{{}^{\ominus}}_{{}_{\chi, \, lm}}$ are
the generalized eigen-states of the two Caismir
operators of $SL(2, \mathbb{C})$ acting in the single particle Hilbert space (in case of charged massive fields,
or the generalized eigenstates of the calling operator in case of mass less fields).
For the massive scalar field they immediately
follow from the decomposition of the representation of $SL(2, \mathbb{C})$
acting on the Lobachevsky space, computed in \cite{GelfandV}, and span the irreducible unitary representation $(l_0=0, l_1=i\nu)$,
$\nu \in \mathbb{R}_+$ of \cite{Geland-Minlos-Shapiro}
with $\chi=-1+i\nu$, $\nu \in \mathbb{R}_+$.

Note that the (scalar, bispinor, \emph{e.t.c.}) functions $\overline{ f^{{}^{\oplus}}_{{}_{\chi, \, lm}}(x)}$ and
$\overline{f^{{}^{\ominus}}_{{}_{\chi, \, lm}}(x)}$ are smooth everywhere outside the light cone,
and respect everywhere (as distributions) the same associated equation (Dirac equation in case of the spinor field), 
as the kernel $\boldsymbol{\psi}(x)$ of the original
(scalar, spinor, \emph{e.t.c}) field $\boldsymbol{\psi}$. Therefore, $\boldsymbol{\psi}_{{}_{\chi}}(x) $ respects the same associated equation 
(say Dirac equation) as $\boldsymbol{\psi}(x)$ does. By the invariance of the pairings, the representation component
$U_{{}_{\chi}}(\alpha)$ act on the (scalar, bispinor, \emph{e.t.c} functions $\overline{ f^{{}^{\oplus}}_{{}_{\chi, \, lm}}(x)}$ and
$\overline{f^{{}^{\ominus}}_{{}_{\chi, \, lm}}(x)}$ as the ordinary local (scalar, bispinor, \emph{e.t.c.}) transformation.

Here $b'_{{}_{\chi, \, lm}}$, ${b'}_{{}_{\chi, \, lm}}^{+}$,
$d'_{{}_{\chi, \, lm}}$, ${d'}_{{}_{\chi, \, lm}}^{+}$ are the Hida annihilation-creation operators, constructed over the single
particle Gelfand triples (rigged Hilbert spaces)
\[
E^{\oplus}_{{}_{\chi}} \oplus E^{\ominus \, \flat}_{{}_{\chi}} \,\,\, \subset \,\,\,
\big(\mathcal{H}^{\oplus}_{\boldsymbol{m},0}\big)_{{}_{\chi}} \oplus \big(\mathcal{H}^{\ominus \, \flat}_{-\boldsymbol{m},0}\big)_{{}_{\chi}}
\,\,\, \subset \,\,\,
E^{\oplus \, *}_{{}_{\chi}} \oplus E^{\ominus \, \flat *}_{{}_{\chi}}
\]
of the associated charged (asymptotically) homogeneous fields $\boldsymbol{\psi}_{{}_{\chi}}$, compare Subsections \ref{psichi}
and \ref{equivalentA-s}. The annihilation-creation operators $b'_{{}_{\chi, \, lm}}$, ${b'}_{{}_{\chi, \, lm}}^{+}$,
$d'_{{}_{\chi, \, lm}}$, ${d'}_{{}_{\chi, \, lm}}^{+}$ respect the canonical anticommutation or commutation relations, depending
on the type of the particular field $\boldsymbol{\psi}(x)$. If the representation of the rotation subgroup, acting in the
single particle space of the field $\boldsymbol{\psi}(x)$ decomposes into irreducible constituents with integer weights,
then these operators respect canonical commutation rules. If this decomposition contains only the irreducible summands
with half an odd integer weights, then $b'_{{}_{\chi, \, lm}}$, ${b'}_{{}_{\chi, \, lm}}^{+}$,
$d'_{{}_{\chi, \, lm}}$, ${d'}_{{}_{\chi, \, lm}}^{+}$ respect the canonical anticommutation relations. Compare
Subsection \ref{psichi} for the Dirac massive field, which is a Fermi field with the canonical anticommutation relations.

The partial waves $f^{{}^{\oplus}}_{{}_{\chi, \, lm}}$ are (asymptotically) homogeneous of (asymptotic) homogeneity
$\chi$, \emph{i.e.} the limits
\begin{equation}\label{homogeneousf+f-}
\underset{\lambda\rightarrow +\infty}{\textrm{lim}} \,
\big({\textstyle\frac{1}{\lambda}}\big)^{-s}f^{{}^{\oplus}}_{{}_{\chi, \, lm}}\big({\textstyle\frac{1}{\lambda}}x\big),
\,\,\,\,\,\,\,\,
\underset{\lambda\rightarrow +\infty}{\textrm{lim}} \,
\big({\textstyle\frac{1}{\lambda}}\big)^{-s}f^{{}^{\ominus}}_{{}_{\chi, \, lm}}\big({\textstyle\frac{1}{\lambda}}x\big)
\end{equation}
exist for $s$ equal to the asymptotic homogeneity $\chi$ determined by the spectral decomposition, and is zero
or does not exist for $s\neq \chi$.
It should be stressed that the spectrum of the decomposition depends on the particular type of the charged field
$\boldsymbol{\psi}$. In particular for the spinor field $\boldsymbol{\psi}$, and generally for a more complicated
multiplet multispinor field $\boldsymbol{\psi}$, for which the current has the general form
${:}\boldsymbol{\psi}^+ \Gamma^0 \Gamma^\mu \boldsymbol{\psi}{:}$, with a Clifford
generators $\Gamma^\mu$ associated to the Minkowski metric, and does not contain any derivatives,
the spectrum is equal
\[
\chi = -3/2 -i\nu, \,\, \nu \in \mathbb{R}.
\]
For the scalar charged field $\boldsymbol{\psi}$, with the current of the form
${:} \boldsymbol{\psi}^+ \overset{\leftrightarrow}{\partial}{}^\mu \boldsymbol{\psi} {:}$, the spectrum is equal
\[
\chi = -1 -i\nu, \,\, \nu \in \mathbb{R}.
\]
This is quite remarkable, because in each case the asymptotically (ultraviolet) homogeneous part
\[
A_{{}_{\textrm{int} \, \chi}}^{(1)}(g=1,x)
\]
of the first order contribution,
determined by the asymptotically homogeneous part $\boldsymbol{\psi}_{{}_{\chi}}$ of
$\boldsymbol{\psi}$, has the asymptotic (ultraviolet) homogeneity $-1$. Indeed, in the case of the spinor,
or multispinor (massive) field $\boldsymbol{\psi}$, the current
$j^{\mu}_{{}_{\chi}}(x)=\big({:}\boldsymbol{\psi}^+ \gamma^0 \gamma^\mu \boldsymbol{\psi}{:}\big)_{{}_{\chi}}$
has the asymptotic (ultraviolet) homogeneity degree $-3$, for the proof compare Subsection \ref{psichi}.
This is not obvious as the naive formula
$j^{\mu}_{{}_{\chi}}(x)={:}\boldsymbol{\psi}_{{}_{\chi}}^+(x) \Gamma^0 \Gamma^\mu \boldsymbol{\psi}_{{}_{\chi}}(x){:}$
cannot be used due to the singular character of the kernels of $\boldsymbol{\psi}_{{}_{\chi}}^+(x)$
and $\boldsymbol{\psi}_{{}_{\chi}}(x)$, which cannot be naturally multiplied.
Similarly, for the scalar (massive) case the current
$j^{\mu}_{{}_{\chi}}(x)= \big({:}\boldsymbol{\psi}^+ \overset{\leftrightarrow}{\partial}{}^\mu \boldsymbol{\psi}{:}\big)_{{}_{\chi}}$
has the same asymptotic (ultraviolet) homogeneity degree $-3$. In fact, in each case
the asymptotically (ultraviolet) homogeneous part
\begin{equation}\label{homogeneousA(1)intchi}
A_{{}_{\textrm{int} \, \chi}}^{\mu \,(1)}(g=1,x) =
-{\textstyle\frac{e}{4\pi}} \int \ud^3 \boldsymbol{x_{1}}
{\textstyle\frac{1}{|\boldsymbol{x_1} - \boldsymbol{x}|}}
\,
j^{\mu}_{{}_{\chi}}(x_0 - |\boldsymbol{x_1} - \boldsymbol{x}|, \boldsymbol{x_1})
\end{equation}
of the first order contribution, always has the asymptotic (ultraviolet) homogeneity $-1$ for each
$\chi$ of the spectrum of the decomposition of the field $j^{\mu}$
into the asymptotically homogeneous parts $j^{\mu}_{{}_{\chi}}$.

Let 
\begin{multline*}
E = \mathcal{S}_{H_{(4)}}(\mathbb{R}^4;\mathbb{C}^d) = \mathcal{S}(\mathbb{R}^4;\mathbb{C}^d) = E^{+} \oplus E^{-}
\\
= U^{-1} \big[P^{\oplus} \mathcal{S}(\mathbb{R}^4;\mathbb{C}^d) \oplus [P^\ominus \mathcal{S}(\mathbb{R}^4;\mathbb{C}^d)]^\flat \big]
=
U^{-1}\big[E^\oplus \oplus  E^{\ominus \, \flat}  \big]. 
\end{multline*}
be the standard single particle nuclear space of a charged massive free field $\boldsymbol{\psi}$. For the Dirac free field $\boldsymbol{\psi}$ the space
 $E$ is equal to the standard nuclear single particle space of the Dirac free field of Subsection \ref{psiBerezin-Hida} with $d=4$ and
with the unitary isomorphism $U$ equal
(\ref{isomorphismU}) joining the single particle Gelfand triples of Subsection \ref{psiBerezin-Hida}.
For the (complex) scalar field $d=1$ and $U$ reduces to the multiplication by the function $\boldsymbol{\p} \mapsto \tfrac{1}{\sqrt{p_0(\boldsymbol{\p})}}$ and its inverse $U^{-1}$ to the  multiplication by the function $\boldsymbol{\p} \mapsto \sqrt{p_0(\boldsymbol{\p})}$,
and the projectors $P^\oplus, P^\ominus$ to unit operators.

Let
\[
\mathscr{E} = \mathscr{F}^{-1}\mathcal{S}_{A^{(4)}}(\mathbb{R}^4;\mathbb{C}^4) = \mathcal{S}^{00}(\mathbb{R}^4;\mathbb{C}^4) 
\]
be the nuclear space-time test space of the free electromagnetic potential field $A$.

Due to the decomposition $E=E^+\oplus E^-$, each kernel ${\kappa'}_{\mathpzc{l},\mathpzc{m}}(\varphi) \in E^{* \widehat{\otimes} \, \mathpzc{l}}
\otimes E^{* \widehat{\otimes} \, \mathpzc{m}}$ can be decomposed into the finite sum
of kernels ${\kappa'}_{\mathpzc{l},\mathpzc{m}}^{\pm\pm}(\varphi) \in E^{\pm * \widehat{\otimes} \, \mathpzc{l}}
\otimes E^{\pm * \widehat{\otimes} \, \mathpzc{m}}$.

The kernels ${\kappa'}_{2,0}^{\pm\pm}$, ${\kappa'}_{0,2}^{\pm\pm}$, ${\kappa'}_{1,1}^{\pm\pm}$ 
of the first order contribution $A_{{}_{\textrm{int} \, \chi}}^{\mu \,(1)}(g=1,x)$
for spinor QED are given in Subsection \ref{analysis-of-klm-A(1)}. Although we should be careful here: if we are using
the free Dirac field $\boldsymbol{\psi}$ in the perturbative spinor QED, which we have decomposed 
in \ref{psichi} with the corresponding formula for the unitary isomorphism (\ref{isomorphismU})  of Subsection \ref{psiBerezin-Hida},
in which we have the fundamental solutions $u,v$,
then the fundamental solutions $u,v$ in Subsection \ref{analysis-of-klm-A(1)} should be replaced with the fundamental
solutions $\underset{*}{u} = \gamma^0u$, $\underset{*}{v} = \gamma^0 v$.  

Because (Subsection \ref{analysis-of-klm-A(1)})
\[
{\kappa'}_{2,0}^{+-}(\phi) \in E^+\otimes E^-, \,\,\,
{\kappa'}_{0,2}^{-+}(\phi) \in E^-\otimes E^+, \,\,\, \kappa_{1,1}^{++}(\phi) \in E^+\otimes E^{*+},
\,\,\, \kappa_{1,1}^{--}(\phi) \in E^-\otimes E^{*-},
\]
the kernels ${\kappa'}_{2,0}^{+-}(\phi)$, ${\kappa'}_{2,0}^{+-}(\phi)$
can be evaluated at the basic functionals.
Therefore the $(l'm'lm)$ components 
\[
\begin{split}
\Bigg\langle
\mathcal{F}\big[{\kappa'}_{2,0}^{+-}(\phi)\big](\chi,\chi), \,\,
F^{+}_{{}_{\chi, \, l'm'}} \, \otimes \, 
F^{-}_{{}_{\chi, \, lm}}
\Bigg\rangle
\\
\Bigg\langle
\mathcal{F}\big[{\kappa'}_{0,2}^{-+}(\phi)\big](\chi,\chi), \,\,
F^{-}_{{}_{\chi, \, l'm'}} \, \otimes \, 
F^{+}_{{}_{\chi, \, lm}}
\Bigg\rangle,
\end{split}
\]
and
\[
\begin{split}
\Bigg\langle
\mathcal{F}\big[{\kappa'}_{1,1}^{++}(\phi)\big](\chi,\chi), \,\,
F^{+}_{{}_{\chi, \, l'm'}} \, \otimes \, 
F^{+}_{{}_{\chi, \, lm}}
\Bigg\rangle
\\
\Bigg\langle
\mathcal{F}\big[{\kappa'}_{0,2}^{--}(\phi)\big](\chi,\chi), \,\,
F^{-}_{{}_{\chi, \, l'm'}} \, \otimes \, 
F^{-}_{{}_{\chi, \, lm}}
\Bigg\rangle,
\end{split}
\]
of the Fourier transforms $\mathcal{F}\big[{\kappa'}_{\mathpzc{l},\mathpzc{m}}^{\pm\pm}\big](\chi,\chi)$
(here with $\mathcal{F}$ which is associated to the decomposition of $SL(2, \mathbb{C})$ acting in $E^{\otimes \, 2}$) are equal
\begin{equation}\label{KernelsOfA(1)chi}
\begin{split}
{\kappa'}_{\chi\, 2,0}^{+-}(\varphi)(l'm',lm) 
= \Big\langle {\kappa'}_{2,0}^{+-}(\varphi), \,\, \overline{U F^{{}^{\oplus}}_{{}_{\chi, \, l'm'}}} \otimes 
\overline{UF^{{}^{\ominus \, \flat}}_{{}_{\chi, \, lm}}} \Big\rangle,
\\
{\kappa'}_{\chi \, 0,2}^{-+}(\varphi)(l'm',lm) 
= \Big\langle {\kappa'}_{2,0}^{-+}(\varphi), \,\, U F^{{}^{\ominus \, \flat}}_{{}_{\chi, \, l'm'}} \otimes UF^{{}^{\oplus}}_{{}_{\chi, \, lm}} \Big\rangle,
\\
{\kappa'}_{\chi \, 1,1}^{++}(\varphi)(l'm',lm) 
= \underset{\epsilon \rightarrow 0}{\textrm{lim}}
\Big\langle {\kappa'}_{1,1}^{++}(\varphi)_{{}_{\epsilon}}, \,\, \overline{U F^{{}^{\oplus}}_{{}_{\chi, \, l'm'}}} \, \widehat{\otimes} \, 
 UF^{{}^{\oplus}}_{{}_{\chi, \, lm}} \Big\rangle,
\\
{\kappa'}_{\chi \, 1,1}^{--}(\varphi)(l'm',lm) 
= 
\underset{\epsilon \rightarrow 0}{\textrm{lim}}
\Big\langle {\kappa'}_{1,1}^{--}(\varphi)_{{}_{\epsilon}}, \,\, U F^{{}^{\ominus \, \flat}}_{{}_{\chi, \, l'm'}} \, \widehat{\otimes} \,
 \overline{UF^{{}^{\ominus \, \flat}}_{{}_{\chi, \, lm}}} \Big\rangle.
\end{split}
\end{equation}
Recall that $F^{+}_{{}_{\chi, \, lm}} = UF^{\oplus}_{{}_{\chi, \, lm}}$,
$F^{-}_{{}_{\chi, \, lm}} = UF^{\ominus \flat}_{{}_{\chi, \, lm}}$, compare Subsection \ref{psichi}.
The pairings
\[
\begin{split}
\langle {\kappa'}_{\mathpzc{l},\mathpzc{m}}^{\pm\pm}(\varphi), \zeta \otimes \eta \rangle
= \langle {\kappa'}_{\mathpzc{l},\mathpzc{m}}^{\pm \pm}(\zeta \otimes \eta), \varphi \rangle
\\
\textrm{for} \,\, \zeta,\eta \in E^{\pm}, \varphi \in E =\mathcal{S}^{00}(\mathbb{R}^4; \mathbb{C}^4) \,\,\,\,
(\mathpzc{l},\mathpzc{m}) = (0,2), (2,0) \, \textrm{or} \, (1,1),
\end{split}
\]
are the canonical \emph{bilinear} pairings, given in Subsection \ref{analysis-of-klm-A(1)}. The regularization may be chosen in the following
form
\[
\begin{split}
{\kappa'}_{1,1}^{++}(\varphi)_{{}_{\epsilon}}(s' \boldsymbol{\p}', s, \boldsymbol{\p}) = 
e^{-\epsilon\{|\boldsymbol{\p}'|^2+ |\boldsymbol{\p}|^2 \}}
{\kappa'}_{1,1}^{++}(\varphi)(s' \boldsymbol{\p}', s, \boldsymbol{\p})
\\
{\kappa'}_{1,1}^{--}(\varphi)_{{}_{\epsilon}}(s' \boldsymbol{\p}', s, \boldsymbol{\p}) = 
e^{-\epsilon\{|\boldsymbol{\p}'|^2 + |\boldsymbol{\p}|^2 \}}
{\kappa'}_{1,1}^{--}(\varphi)(s' \boldsymbol{\p}', s, \boldsymbol{\p}),
\end{split}
\]
but the concrete form of the regularizing factor is irrelevant except for the simplification of the
computation of the limit.

Therefore, accordingly to Subsection \ref{psichi}, the first order contribution has the following direct integral decomposition
\[
A_{{}_{\textrm{int}}}^{\mu \,(1)}(g=1,\varphi_\mu) = \int A_{{}_{\textrm{int} \, \chi}}^{\mu \,(1)}(g=1,\varphi_\mu) \,
\ud \chi,
\]
where
\begin{multline}\label{IntKerA(1)chi}
A_{{}_{\textrm{int} \, \chi}}^{\mu \,(1)}(g=1,\varphi_\mu) 
\\
=
\sum\limits_{l',m',l,m} {\kappa'}_{\chi \, 1,1}^{++}(\varphi)(l'm',lm) \,\, {b'}_{{}_{\chi, \, l'm'}}^{+} b'_{{}_{\chi, \, lm}} +
\sum\limits_{l',m',l,m} \kappa_{\chi \, 1,1}^{--}(\varphi)(l'm',lm) \,\, {d'}_{{}_{\chi, \, l'm'}}^{+} d'_{{}_{\chi, \, lm}} 
\\
+
\sum\limits_{l',m',l,m} {\kappa'}_{\chi \, 2,0}^{+-}(\varphi)(l'm',lm) \,\, {b'}_{{}_{\chi, \, l'm'}}^{+} {d'}_{{}_{\chi, \, lm}}^{+} +
\sum\limits_{l',m',l,m} {\kappa'}_{\chi \, 0,2}^{-+}(\varphi)(l'm',lm) \,\, {d'}_{{}_{\chi, \, l'm'}} b'_{{}_{\chi, \, lm}}. 
\end{multline}

Asymptotic (ultraviolet) homogeneity properties of the charged fields $\boldsymbol{\psi}$ follow from 
the asymptotic (ultraviolet) homogeneity of the functions $F^{{}^{\oplus}}_{{}_{\chi, \, lm}}$,
$F^{{}^{\ominus}}_{{}_{\chi, \, lm}}$ on $\mathscr{O}_{{}_{\boldsymbol{m},0,0,0}}$
and, respectively, on $\mathscr{O}_{{}_{-\boldsymbol{m},0,0,0}}$, which have the
asymptotic homogeneity property (at infinity in momentum) with homogeneity degree depending on the massive
complex field $\boldsymbol{\psi}$. Namely, for the Dirac field their asymptotic homogeneity is equal
$-1/2 +i\nu$, $\nu \in \mathbb{R}$ (compare Subsection \ref{psichi}), for the scalar complex massive field
the functions $F^{{}^{\oplus}}_{{}_{\chi, \, lm}}$,
$F^{{}^{\ominus}}_{{}_{\chi, \, lm}}$ are scalars and, by the results of \cite{GelfandV}, Chap. VI.3.3, are equal
\[
F^{{}^{\oplus}}_{{}_{\chi, \, lm}}(p) = \int\limits_{\mathbb{S}^2} (p\cdot k)^{-1+i\nu}\overline{Y_{{}_{lm}}(k)} \, d^2k,
\,\,\,\,\,\,\,\, p\in \mathscr{O}_{{}_{\boldsymbol{m},0,0,0}}, \,\,\, k\in \mathscr{O}_{{}_{1,0,0,1}}
\]
\[
F^{{}^{\ominus}}_{{}_{\chi, \, lm}}(p) = \int\limits_{\mathbb{S}^2} (p\cdot k)^{-1+i\nu}\overline{Y_{{}_{lm}}(k)} \, d^2k,
\,\,\,\,\,\,\,\, p\in \mathscr{O}_{{}_{-\boldsymbol{m},0,0,0}}, \,\,\, k\in \mathscr{O}_{{}_{-1,0,0,1}}
\]
with their asymptotic homogeneity equal $-1 +i\nu$. This means that (using the Cartesian coordinates $\boldsymbol{\p}$ on
$\mathscr{O}_{{}_{\pm \boldsymbol{m},0,0,0}}$ and $\boldsymbol{\q}$ on $\mathscr{O}_{{}_{\pm 1,0,0,1}}$)
\begin{equation}\label{Fplus,Fminus, scalar}
\underset{\lambda\rightarrow +\infty}{\textrm{lim}} \,
\lambda^{1-i\nu}
F^{{}^{\oplus}}_{{}_{\chi=-1+i\nu, \, lm}}\big(\lambda \boldsymbol{\p}\big),
\,\,\,\,\,
\underset{\lambda\rightarrow +\infty}{\textrm{lim}} \,
\lambda^{1-i\nu}
F^{{}^{\ominus}}_{{}_{\chi=-1+i\nu, \, lm}}\big(\lambda \boldsymbol{\p}\big)
\end{equation}
exist for each $\boldsymbol{\p}$ and represent homogeneous of degree $-1+i\nu$ functions of $\boldsymbol{\p}$, and each of the limits
\[
\underset{\lambda\rightarrow +\infty}{\textrm{lim}} \,
{\textstyle\frac{1}{\lambda^{s}}}
F^{{}^{\oplus}}_{{}_{\chi=-1+i\nu, \, lm}}\big(\lambda \boldsymbol{\p}\big),
\,\,\,\,
\underset{\lambda\rightarrow +\infty}{\textrm{lim}} \,
{\textstyle\frac{1}{\lambda^{s}}}
F^{{}^{\ominus}}_{{}_{\chi=-1+i\nu, \, lm}}\big(\lambda\boldsymbol{\p}\big)
\]
is zero, $\infty$ or does not exist for each value of $s \neq -1+i\nu$. In this scalar case the explicit formula for the
homogeneous functions (\ref{Fplus,Fminus, scalar}) follows from the analytic
continuation of the formula (\ref{reproducing-property-of-(kl)^z}) for the values $z = 1-i\nu$
(using the Cartesian coordinates $\boldsymbol{\p}$ on
$\mathscr{O}_{{}_{\pm \boldsymbol{m},0,0,0}}$ and $\boldsymbol{\q}$ on $\mathscr{O}_{{}_{\pm 1,0,0,1}}$):
\begin{multline}\label{implicithomogeneousF+F-}
\underset{\lambda\rightarrow +\infty}{\textrm{lim}} \,
\lambda^{1-i\nu}
F^{{}^{\oplus}}_{{}_{\chi=-1+i\nu, \, lm}}(\lambda \boldsymbol{\p}) =
 \int\limits_{\mathbb{S}^2}(|\boldsymbol{\p}| | \boldsymbol{\q}|- \boldsymbol{\p}\cdot \boldsymbol{\q})^{-1+i\nu}\overline{Y_{{}_{lm}}(k)} \, d^2k
\\
=
\int\limits_{\mathbb{S}^2} (q\cdot k)^{-1+i\nu} \overline{Y_{{}_{lm}}(k)} \, d^2k
\end{multline}
where $q = (|\boldsymbol{\p}|, \boldsymbol{\p})$, $k=(|\boldsymbol{\q}|, \boldsymbol{\q})$ are elements of the positive energy sheet
$\mathscr{O}_{{}_{1,0,0,1}}$ of the cone, so that we can use (\ref{reproducing-property-of-(kl)^z}), prolonged analytically for $z=1-i\nu$.
Similar formula we have for the negative energy homogeneous function (\ref{Fplus,Fminus, scalar})
on the negative energy cone, using the analogue of the reproducing formula
(\ref{reproducing-property-of-(kl)^z}) on the negative energy sheet of the cone.
From (\ref{implicithomogeneousF+F-}) we see that the limits
(\ref{Fplus,Fminus, scalar}) indeed exist for all values of the decomposition (asymptotic homogeneity) parameter
$\chi=-1+i\nu$, with $\nu \in \mathbb{R}$. However, as follows by the elementary properties of the Euler gamma function
the reproducing formula (\ref{reproducing-property-of-(kl)^z})
cannot be used in the case $\nu=0$, to convert the implicit formula (\ref{implicithomogeneousF+F-}) for the limits (\ref{Fplus,Fminus, scalar})
into the more explicit form expressed through ordinary spherical harmonic functions.
In case $\nu=0$, the homogeneous parts (\ref{Fplus,Fminus, scalar}) are obtained by passing to the limit $\nu \rightarrow 0$
in (\ref{implicithomogeneousF+F-}), and (compare \cite{Staruszkiewicz1998}) it follows that in case $\nu=0$
all homogeneous parts (\ref{Fplus,Fminus, scalar}) are zero except for the spherically
symmetric state with $l=m=0$.

Let us emphasize once more that (\ref{KernelsOfA(1)chi})
\emph{i.e.} the kernels defining $A^{(1) \, \mu}_{{}_{\textrm{int} \, \chi}}$, eq. (\ref{IntKerA(1)chi}), 
evaluated at the space-time test function $\varphi$ 
cannot be computed through the naive formulas
\[
-{\textstyle\frac{e}{4\pi}} \int \ud^3 \boldsymbol{\x}\ud x_0 \ud^3 \boldsymbol{x_{1}}
{\textstyle\frac{f^{{}^{\oplus}}_{{}_{\chi, \, l'm'}}(x_0 - |\boldsymbol{x_1} - \boldsymbol{x}|, \boldsymbol{x_1}) \gamma^0\gamma^\mu
\overline{f^{{}^{\ominus}}_{{}_{\chi, \, lm}}(x_0 - |\boldsymbol{x_1} - \boldsymbol{x}|, \boldsymbol{x_1})}}{|\boldsymbol{x_1} - \boldsymbol{x}|}}
\varphi_\mu(\boldsymbol{\x}, x_0),
\]
\[
e.t.c.
\]
or (in scalar case)
\[
-{\textstyle\frac{e}{4\pi}} \int \ud^3 \boldsymbol{\x}\ud x_0 \ud^3 \boldsymbol{x_{1}}
{\textstyle\frac{f^{{}^{\oplus}}_{{}_{\chi, \, l'm'}}(x_0 - |\boldsymbol{x_1} - \boldsymbol{x}|, \boldsymbol{x_1}) \overset{\leftrightarrow}{\partial}{}^\mu 
\overline{f^{{}^{\ominus}}_{{}_{\chi, \, lm}}(x_0 - |\boldsymbol{x_1} - \boldsymbol{x}|, \boldsymbol{x_1})}}{|\boldsymbol{x_1} - \boldsymbol{x}|}}
\varphi_\mu(\boldsymbol{\x}, x_0),
\]
\[
e.t.c.
\]
as the integrands involve products of singular distributions $f^{{}^{\oplus}}_{{}_{\chi, \, lm}}$, $\partial_\mu f^{{}^{\oplus}}_{{}_{\chi, \, lm}}$ 
\emph{e.t.c.} and are not well-defined. It is thus necessary to use the formulas (\ref{KernelsOfA(1)chi})
which are naturally determined by the decomposition of the first order contribution 
$A^{(1) \, \mu}_{{}_{\textrm{int}}}$, determined solely by the decomposition of the representation of $SL(2, \mathbb{C})$, 
exactly as the homogeneous parts of the free electromagnetic potential field. 

In the massive case the kernels
\[
\kappa_{\chi \, \mathpzc{l},\mathpzc{m}}^{\pm \pm}(\varphi)(l'm',lm), \,\,\,\,\,\,\, (\mathpzc{l},\mathpzc{m}) = (1,1), (2,0), (0,2), 
\]
of the homogeneous of degree $-1$ (ultraviolet/infrared) part\footnote{With the first expression 
\[
\underset{\lambda\rightarrow +\infty}{\textrm{lim}} \,\int
{\textstyle\frac{1}{\lambda}}A^{(1) \, \mu}_{{}_{\textrm{int} \, \chi}}\big({\textstyle\frac{1}{\lambda}}x\big)\,
\varphi(x) \, \ud^4x
\,\,\, \textrm{or} \,\,\, 
\underset{\lambda\rightarrow +\infty}{\textrm{lim}} \,\int
\lambda A^{(1) \, \mu}_{{}_{\textrm{int} \, \chi}}\big(\lambda x\big)\,
\varphi(x) \, \ud^4x
\]
having only a formal symbolic and heuristic character.}
\begin{equation}\label{HomPartIntKerA(1)chi}
\underset{\lambda\rightarrow +\infty}{\textrm{lim}} \,\int
{\textstyle\frac{1}{\lambda}}A^{(1) \, \mu}_{{}_{\textrm{int} \, \chi}}\big({\textstyle\frac{1}{\lambda}}x\big)\,
\varphi_\mu(x) \, \ud^4x = 
\underset{\lambda\rightarrow +\infty}{\textrm{lim}} \, \lambda^3 \int
A^{(1) \, \mu}_{{}_{\textrm{int} \, \chi}}\big(x\big)\varphi_\mu(\lambda x) \ud^4x
\,\,\, (\textrm{ultraviolet})
\end{equation}
or
\begin{equation}\label{HomPartIntKerA(1)chiIR}
\underset{\lambda\rightarrow +\infty}{\textrm{lim}} \,\int
\lambda A^{(1) \, \mu}_{{}_{\textrm{int} \, \chi}}\big(\lambda x\big)\,
\varphi_\mu(x) \, \ud^4x = 
\underset{\lambda\rightarrow +\infty}{\textrm{lim}} \, (1/\lambda)^3 \int
A^{(1) \, \mu}_{{}_{\textrm{int} \, \chi}}\big(x\big)\varphi_\mu({\textstyle\frac{1}{\lambda}} x) \ud^4x
\,\,\, (\textrm{infrared})
\end{equation}
\begin{multline*}
=
\sum\limits_{l',m',l,m} \kappa_{\chi \, 1,1}^{++}(\varphi)(l'm',lm) \,\, {b'}_{{}_{\chi, \, l'm'}}^{+} b'_{{}_{\chi, \, lm}} +
\sum\limits_{l',m',l,m} \kappa_{\chi \, 1,1}^{--}(\varphi)(l'm',lm) \,\, {d'}_{{}_{\chi, \, l'm'}}^{+} d'_{{}_{\chi, \, lm}} 
\\
+
\sum\limits_{l',m',l,m} \kappa_{\chi \, 2,0}^{+-}(\varphi)(l'm',lm) \,\, {b'}_{{}_{\chi, \, l'm'}}^{+} {d'}_{{}_{\chi, \, lm}}^{+} +
\sum\limits_{l',m',l,m} \kappa_{\chi \, 0,2}^{-+}(\varphi)(l'm',lm) \,\, {d'}_{{}_{\chi, \, l'm'}} b'_{{}_{\chi, \, lm}}. 
\end{multline*}
of $A^{(1)}_{{}_{\textrm{int} \, \chi}}$ are equal
\[
\kappa_{\chi \, \mathpzc{l},\mathpzc{m}}^{\pm \pm}(\varphi)(l'm',lm) = \underset{\lambda \rightarrow 0}{\textrm{lim}}
\lambda^3
{\kappa'}_{\chi \, \mathpzc{l},\mathpzc{m}}^{\pm\pm}(S_\lambda \varphi)(l'm',lm), \,\,\,\,\,\, \,\,\, (\textrm{ultraviolet})
\]
or, respectively,
\[
\kappa_{\chi \, \mathpzc{l},\mathpzc{m}}^{\pm \pm}(\varphi)(l'm',lm) = \underset{\lambda \rightarrow 0}{\textrm{lim}}
(1/\lambda)^3
{\kappa'}_{\chi \, \mathpzc{l},\mathpzc{m}}^{\pm\pm}(S_{1/\lambda} \varphi)(l'm',lm), \,\,\,\,\,\, \,\,\, (\textrm{infrared})
\]
or 
\begin{equation}\label{KernelsOfHomPartA(1)chi}
\begin{split}
{\kappa}_{\chi\, 2,0}^{+-}(\varphi)(l'm',lm) 
= 
\underset{\lambda \rightarrow +\infty}{\textrm{lim}}
\lambda^3
\Big\langle {\kappa'}_{2,0}^{+-}(S_\lambda\varphi), \,\, \overline{U F^{{}^{\oplus}}_{{}_{\chi, \, l'm'}}} \otimes 
\overline{UF^{{}^{\ominus \, \flat}}_{{}_{\chi, \, lm}}} \Big\rangle,
\\
{\kappa}_{\chi \, 0,2}^{-+}(\varphi)(l'm',lm) 
= 
\underset{\lambda \rightarrow +\infty}{\textrm{lim}}
\lambda^3
\Big\langle {\kappa'}_{2,0}^{-+}(S_\lambda\varphi), \,\, U F^{{}^{\ominus \, \flat}}_{{}_{\chi, \, l'm'}} \otimes UF^{{}^{\oplus}}_{{}_{\chi, \, lm}} \Big\rangle,
\\
{\kappa}_{\chi \, 1,1}^{++}(\varphi)(l'm',lm) 
= 
\underset{\lambda \rightarrow +\infty}{\textrm{lim}}
\lambda^3
\Big\langle {\kappa'}_{1,1}^{++}(S_\lambda\varphi)_{{}_{\epsilon}}, \,\, \overline{U F^{{}^{\oplus}}_{{}_{\chi, \, l'm'}}} \, \widehat{\otimes} \, 
 UF^{{}^{\oplus}}_{{}_{\chi, \, lm}} \Big\rangle,
\\
{\kappa}_{\chi \, 1,1}^{--}(\varphi)(l'm',lm) 
= 
\underset{\lambda \rightarrow +\infty}{\textrm{lim}}
\lambda^3
\Big\langle {\kappa'}_{1,1}^{--}(S_\lambda\varphi)_{{}_{\epsilon}}, \,\, U F^{{}^{\ominus \, \flat}}_{{}_{\chi, \, l'm'}} \, \widehat{\otimes} \,
 \overline{UF^{{}^{\ominus \, \flat}}_{{}_{\chi, \, lm}}} \Big\rangle,
\\
(\textrm{ultraviolet})
\end{split}
\end{equation}
\begin{equation}\label{KernelsOfHomPartA(1)chiIR}
\begin{split}
{\kappa}_{\chi\, 2,0}^{+-}(\varphi)(l'm',lm) 
= 
\underset{\lambda \rightarrow +\infty}{\textrm{lim}}
(1/\lambda)^3
\Big\langle {\kappa'}_{2,0}^{+-}(S_{1/\lambda}\varphi), \,\, \overline{U F^{{}^{\oplus}}_{{}_{\chi, \, l'm'}}} \otimes 
\overline{UF^{{}^{\ominus \, \flat}}_{{}_{\chi, \, lm}}} \Big\rangle,
\\
{\kappa}_{\chi \, 0,2}^{-+}(\varphi)(l'm',lm) 
= 
\underset{\lambda \rightarrow +\infty}{\textrm{lim}}
(1/\lambda)^3
\Big\langle {\kappa'}_{2,0}^{-+}(S_{1/\lambda}\varphi), \,\, U F^{{}^{\ominus \, \flat}}_{{}_{\chi, \, l'm'}} \otimes UF^{{}^{\oplus}}_{{}_{\chi, \, lm}} \Big\rangle,
\\
{\kappa}_{\chi \, 1,1}^{++}(\varphi)(l'm',lm) 
= 
\underset{\lambda \rightarrow +\infty}{\textrm{lim}}
(1/\lambda)^3
\Big\langle {\kappa'}_{1,1}^{++}(S_{1/\lambda}\varphi)_{{}_{\epsilon}}, \,\, \overline{U F^{{}^{\oplus}}_{{}_{\chi, \, l'm'}}} \, \widehat{\otimes} \, 
 UF^{{}^{\oplus}}_{{}_{\chi, \, lm}} \Big\rangle,
\\
{\kappa}_{\chi \, 1,1}^{--}(\varphi)(l'm',lm) 
= 
\underset{\lambda \rightarrow +\infty}{\textrm{lim}}
(1/\lambda)^3
\Big\langle {\kappa'}_{1,1}^{--}(S_{1/\lambda}\varphi)_{{}_{\epsilon}}, \,\, U F^{{}^{\ominus \, \flat}}_{{}_{\chi, \, l'm'}} \, \widehat{\otimes} \,
 \overline{UF^{{}^{\ominus \, \flat}}_{{}_{\chi, \, lm}}} \Big\rangle,
\\
(\textrm{infrared})
\end{split}
\end{equation}
where $S_\lambda\varphi(x) = \varphi(\lambda x)$, and in these (last two in (\ref{KernelsOfHomPartA(1)chi}) 
and (\ref{KernelsOfHomPartA(1)chiIR})) limits, $\epsilon = \epsilon(\lambda)$ is a \emph{slowly
varying function of $\lambda$ at infinity}
tending to zero when $\lambda$ goes to infinity, chosen in such manner that the limits 
(\ref{KernelsOfHomPartA(1)chi}) or (\ref{KernelsOfHomPartA(1)chiIR})) exist, are finite and nonzero. If no choice of this kind
wolud be possible, then by definition the homogeneous of degree $-1$ (ultraviolet or infrared) asymptotics 
(\ref{KernelsOfHomPartA(1)chi}) or, respectively, (\ref{KernelsOfHomPartA(1)chiIR}), of (\ref{IntKerA(1)chi}) 
is regarded as being equal zero. The condition that $\epsilon$ is a \emph{slowly
varying function of $\lambda$ at infinity} means that it is positive for large $\lambda$ and 
\[
\underset{\lambda \rightarrow +\infty}{\textrm{lim}} {\textstyle\frac{\epsilon(a\lambda)}{\epsilon(\lambda})} = 1
\]
for any $a>0$  (for example: $\epsilon(\lambda) = 1/\textrm{ln} \lambda$).

The homogeneous parts (\ref{KernelsOfHomPartA(1)chi}) of the kernels (\ref{KernelsOfA(1)chi}) 
depend only on the homogeneous parts of the asymptotically homogeneous functionals 
$F^{{}^{\oplus}}_{{}_{\chi, \, lm}}$, $F^{{}^{\ominus}}_{{}_{\chi, \, lm}}$. Therefore, in 
(\ref{KernelsOfHomPartA(1)chi}) we can use the homogeneous 
parts (\ref{Fplus,Fminus, scalar}) (with $1-i\nu$ replaced by $1/2-i\nu$ for the spinor case). 
For the  massless charged $\boldsymbol{\psi}$, 
$A^{(1) \, \mu}_{{}_{\textrm{int} \, \chi}}$ does not exists, compare Subsect. \ref{OperationsOnXi}. 
Thus, the massless charged field $\boldsymbol{\psi}$ is unrealistic, as is also shown by the experimental evidence, 
and which is confirmed by our mathematical 
proof of nonexistence of the adiabatic limit for higher order contributions to interacting fields for QED 
in this massless charged case, compare Subsection \ref{OperationsOnXi}.

In order to solve our problem we need to analyze the distribution kernels
(\ref{KernelsOfHomPartA(1)chiIR}) of the field (\ref{HomPartIntKerA(1)chiIR}) (infrared case) or of the field
(\ref{HomPartIntKerA(1)chi}) (ultraviolet case)
\[
\underset{\lambda\rightarrow +\infty}{\textrm{lim}} \,
\big[\lambda A^{(1) \, \mu}_{{}_{\textrm{int} \, \chi}}\big(\lambda x\big)\big]
\,\,\, \textrm{or} \,\,\, 
\underset{\lambda\rightarrow +\infty}{\textrm{lim}} \,
\big[{\textstyle\frac{1}{\lambda}} A^{(1) \, \mu}_{{}_{\textrm{int} \, \chi}}\big({\textstyle\frac{1}{\lambda}}x\big)\big],
\]
regarded as distributions on the space-time test functions $\varphi \in \mathcal{S}^{00}(\mathbb{R}^4; \mathbb{C}^4)$.

In the single particle Hilbert space
\begin{equation}\label{SingleParticleHchi}
\big(\mathcal{H}^{\oplus}_{\boldsymbol{m},0}\big)_{{}_{\chi}} \oplus \big(\mathcal{H}^{\ominus \, \flat}_{-\boldsymbol{m},0}\big)_{{}_{\chi}}
\end{equation}
of the asymptotically homogeneous part (\ref{psichi(x)AS}) of the charged field $\boldsymbol{\psi}$,
coupled to the electromagnetic potential, there is naturally defined the
unitary group $U_t$, $t\in\mathbb{R}$, of one parameter phase transformations, defined through the multiplication by the phase $e^{it}$.
The self adjoint generator of the transformations $\Gamma(U_t)$, raised to the Fock space
\begin{equation}\label{Fockchi}
\Gamma\big(\big(\mathcal{H}^{\oplus}_{\boldsymbol{m},0}\big)_{{}_{\chi}} \oplus \big(\mathcal{H}^{\ominus \, \flat}_{-\boldsymbol{m},0}\big)_{{}_{\chi}}
\Big),
\end{equation}
is equal to the charge operator
\begin{equation}\label{QinHchi}
Q = e \sum\limits_{lm} \big[{b'}_{{}_{\chi, \, lm}}^{+} b'_{{}_{\chi, \, lm}} -{d'}_{{}_{\chi, \, lm}}^{+} d'_{{}_{\chi, \, lm}} \big]
\end{equation}
and it transforms continuously the test Hida space
\[
(E_{{}_{\chi}}) \subset \Gamma\big(\big(\mathcal{H}^{\oplus}_{\boldsymbol{m},0}\big)_{{}_{\chi}} \oplus \big(\mathcal{H}^{\ominus \, \flat}_{-\boldsymbol{m},0}\big)_{{}_{\chi}}
\Big),
\]
into itself. It is easily seen that the operator (\ref{HomPartIntKerA(1)chi}) commutes with $Q$, as each term
in (\ref{HomPartIntKerA(1)chi}) creates as many particles as antiparticles. Similarly, 
\[
\underset{\lambda\rightarrow +\infty}{\textrm{lim}} \,
\big[\lambda A^{(1) \, \mu}_{{}_{\textrm{int} \, \chi}}\big(\lambda x\big)\big]
\,\,\, \textrm{or} \,\,\, 
\underset{\lambda\rightarrow +\infty}{\textrm{lim}} \,
\big[{\textstyle\frac{1}{\lambda}} A^{(1) \, \mu}_{{}_{\textrm{int} \, \chi}}\big({\textstyle\frac{1}{\lambda}}x\big)\big],
\]
is equal to the operator
\begin{multline}\label{HomPartIntKerA(1)chi(x)}
\sum\limits_{l',m',l,m} \kappa_{\chi \, 1,1}^{++}(x;\mu)(l'm',lm) \,\, {b'}_{{}_{\chi, \, l'm'}}^{+} b'_{{}_{\chi, \, lm}} +
\sum\limits_{l',m',l,m} \kappa_{\chi \, 1,1}^{--}(x;\mu)(l'm',lm) \,\, {d'}_{{}_{\chi, \, l'm'}}^{+} d'_{{}_{\chi, \, lm}}
\\
+
\sum\limits_{l',m',l,m} \kappa_{\chi \, 2,0}^{+-}(x;\mu)(l'm',lm) \,\, {b'}_{{}_{\chi, \, l'm'}}^{+} {d'}_{{}_{\chi, \, lm}}^{+} +
\sum\limits_{l',m',l,m} \kappa_{\chi \, 0,2}^{-+}(x;\mu)(l'm',lm) \,\, {d'}_{{}_{\chi, \, l'm'}} b'_{{}_{\chi, \, lm}},
\\
\,\,\,\,\,\,
x\cdot x <0,
\end{multline}
which commutes with $Q$, at least in the domain $x\cdot x<0$, where the distributions
\[
\kappa_{\chi \, 1,1}^{++}(x;\mu)(l'm',lm), \,\, e.t.c.
\]
are represented by ordinary functions and (\ref{HomPartIntKerA(1)chi(x)}) is a (generalized) operator
in (\ref{Fockchi}). However, even on this domain, where these distributions coincide with
ordinary functions, the sums in (\ref{HomPartIntKerA(1)chi(x)}) are \emph{a priori} infinite,
and (\ref{HomPartIntKerA(1)chi(x)}) may, in principle at least, be equal to a generalized
operator transforming $(E_{{}_{\chi}})$ into its dual, which cannot be regarded as an ordinary
operator in the Fock space (\ref{Fockchi}). In fact (\ref{HomPartIntKerA(1)chi(x)}) will be an ordinary operator
transforming continuously $(E_{{}_{\chi}})$ into itself
if the kernel $\kappa_{\chi \, 2,0}^{+-}(x;\mu)(l'm',lm)$, regarded as a function of the indices $(l'm',lm)$, is rapidly
decreasing in $l'm'lm$ and $\kappa_{\chi \, 1,1}^{++}(x;\mu)(l'm',lm)$, $\kappa_{\chi \, 1,1}^{--}(x;\mu)(l'm',lm)$
are rapidly decreasing in $l'm'$ (Thm. \ref{Xi_l,m:Hida->Hida}, Subsection \ref{psiBerezin-Hida}).
A weaker condition is needed for these kernels in order to assure
(\ref{HomPartIntKerA(1)chi(x)}) to be an ordinary densely defined operator in (\ref{Fockchi}): e.g. by requiring them
to be of Hilbert-Schmidt class matrices, regarded as matrices with the indices $l'm' \times lm$. 


From now on we consider only the scalar QED and the infrared asymptotic of the first order contribution.

In order to 
complete our proof of the existence of the states $|S_{m,u}\rangle$ we need to compute
\[
\kappa_{\chi \, \mathpzc{l},\mathpzc{m}}^{\pm\pm}(x;\mu)(l'm',lm), \,\,\,\,\, \,\,\, x \cdot x <0, 
\]
explicitly using (\ref{KernelsOfHomPartA(1)chiIR}).

If we weaken the condition and instead of the infrared (IR)-asymptotic (\ref{HomPartIntKerA(1)chiIR}) of (\ref{IntKerA(1)chi}), 
we use IR-quasi asymtotic \cite{Vladimirov1}, \cite{Vindas},
or rather, the IR-quasi asymptotic of the spherically symmetric parts of the kernels of (\ref{IntKerA(1)chi}), 
situation becomes easier to handle. Namely, when we extract the IR-quasi asymptotic we can replace $\tfrac{1}{\lambda^3}$,
in the formulas (\ref{KernelsOfHomPartA(1)chiIR}), with $\tfrac{1}{\lambda^3L(\lambda)}$, where $L$ is a \emph{slowly
varying function at infinity}. It is immediately seen that the IR-asymtotic (and quasi-asymptotic) 
\[
{\kappa}_{\chi\, 2,0}^{+-}(x;\mu)(l'm',lm), {\kappa}_{\chi\, 0,2}^{-+}(x;\mu)(l'm',lm),
\]
 given by the first two formulas in
(\ref{KernelsOfHomPartA(1)chiIR}) are zero for all $\chi=-1+i\nu$, and all spatial components ($\mu=1,2,3$) of the IR-asymptotics 
(and quasi-asymptotics) 
\[
{\kappa}_{\chi \, 1,1}^{++}(x;\mu)(l'm',lm), {\kappa}_{\chi \, 1,1}^{--}(x;\mu)(l'm',lm),
\]  
given by the last two formulas in (\ref{KernelsOfHomPartA(1)chiIR}), are equal zero. Thus, without
loss of generality, we can use the test functions $\varphi = (\varphi_0,0,0,0)$ with all spatial components equal zero. This is important 
and shows that the IR-asymptotic (and quasi-asymptotic) of the first order contribution to interactng potential is of ``electric type''.
Further, instead of computing the IR-asymtotic or quasi-asymptotic of the kernels of (\ref{IntKerA(1)chi}), we compute asymptotic 
only of the spherically symmetric parts of the kernels of (\ref{IntKerA(1)chi}). 
They are given by the same formulas (\ref{KernelsOfHomPartA(1)chiIR}), eventually with
$1/\lambda^3$ replaced with $1/(\lambda^3L(\lambda))$ (for the quasi-asymptotic case), in which in the r.h.s of these formulas 
the test function $\varphi= (\varphi_0,\varphi_1,\varphi_2,\varphi_3)$ is replaced with the  averaged test function
$\langle\varphi\rangle_{{}_{\textrm{av}}} = (\langle\varphi_0\rangle_{{}_{\textrm{av}}},\varphi_1,\varphi_2,\varphi_3)$
in which the zero component is averaged over the $SU(2,\mathbb{C})$ subgroup:
\[
\langle\varphi_0\rangle_{{}_{\textrm{av}}}(x) = {\textstyle\frac{1}{\textrm{volume of $SU(2,\mathbb{C})$} }}
\int\limits_{{}_{SU(2,\mathbb{C})}} \varphi_0(\Lambda(\alpha)x) d \alpha.
\]
For any rotationally invariant kernel $K \in L^2(\mathbb{S}^2 \times \mathbb{S}^2)$, we have, by the invariance, the following reproducing
property:
\[
\int\limits_{\mathbb{S}^2} K(\boldsymbol{n},\boldsymbol{m}) Y_{{}_{lm}}(\boldsymbol{m})
d^2 \boldsymbol{m} = \mathfrak{c}_{lm} Y_{{}_{lm}}\left(\boldsymbol{n}\right) 
\]
with the constants $\mathfrak{c}_{lm}$ depending on $K$. In particular, we have
\[
\int\limits_{\mathbb{S}^2} {\textstyle\frac{1}{\sqrt{\lambda^2+|\boldsymbol{\p}|^2}- \boldsymbol{\p}\cdot \boldsymbol{\q} }} Y_{{}_{lm}}(\boldsymbol{\q})
d^2 \boldsymbol{\q} = Y_{{}_{lm}}\left({\textstyle\frac{\boldsymbol{\p}}{|\boldsymbol{\p}|}}\right) \,\, {\textstyle\frac{2\pi}{|\boldsymbol{\p}|}} \, 
Q_{{}_{l}}\left( {\textstyle\frac{\sqrt{\lambda^2+|\boldsymbol{\p}|^2}}{|\boldsymbol{\p}|}} \right),
\]
where $Q_{{}_{l}}$ are the Legendre functions of second kind. Using this fact
it is not difficult to see that for $\chi = -1$, the IR-quasi-asymptotic 
of the spherically symmetric parts of the kernels of  (\ref{IntKerA(1)chi}), defined in this way, are equal
\begin{multline*}
{\kappa}_{\chi \, 1,1}^{++}(x;\mu=0)(l'm',lm) = {\textstyle\frac{\mathfrak{c}^{++}_{lm}}{|\boldsymbol{x}|}}\delta_{{}_{l \, l'}}\delta_{{}_{m \, m'}},
\\
{\kappa}_{\chi \, 1,1}^{--}(x;\mu=0)(l'm',lm) = {\textstyle\frac{\mathfrak{c}^{--}_{lm}}{|\boldsymbol{x}|}}\delta_{{}_{l \, l'}}\delta_{{}_{m \, m'}},
\,\,\, \chi=-1,
\end{multline*}    
with IR-quasi-asymptotic of the spherically symmetric parts of all the remaining kernels equal zero. 
To check if the constants $\mathfrak{c}^{++}_{lm}, \mathfrak{c}^{--}_{lm}$ are finite and nonzero  
we can use the saddle-point method (or the Laplace theorem). We now observe,
that if we compute the IR-quasi-asymptotic of each $l'm',lm$-component separately (i.e. chosing the slowly varying $L$ separately to each
component), then in fact the constants $\mathfrak{c}^{++}_{lm}, \mathfrak{c}^{--}_{lm}$ cannot be fixed, as quasi-asymtotic is determined up to a nonzero
constant factor. Therefore, using quasi-asymptotic, we cannot fix them. But we can fix them to be equal
$\mathfrak{c}^{++}_{lm} = -e, \mathfrak{c}^{--}_{lm} = +e$ by comparison with the Staruszkiewicz theory.  

Let the field (\ref{IntKerA(1)chi}), with the kernels ${\kappa'}_{\chi \, \mathpzc{l},\mathpzc{m}}^{\pm\pm}(x;\mu)(l'm',lm)$ replaced 
with the IR-quasi-asymptotic of the spherically symmetric parts 
of these kernels, be denoted symbolically by 
\[
\underset{\lambda\rightarrow +\infty}{\textrm{lim}} \,
\left[\frac{\lambda A^{(1) \, \mu}_{{}_{\textrm{int} \, \chi}}\big(\lambda x\big)}{L(\lambda)}\right].
\] 
Then, using the above choice of the constants (which are free in the computation using quasi-asymptotic), we have
\[
\underset{\lambda\rightarrow +\infty}{\textrm{lim}} \,
\left[\frac{-e x_\mu \lambda A^{(1) \, \mu}_{{}_{\textrm{int} \, \chi}}\big(\lambda x\big)}{L(\lambda)}\right]
= -e{\textstyle\frac{x_0}{|\boldsymbol{x}|}}Q
\]
for $\chi = -1$, where $Q$ is the charge (\ref{QinHchi}).

The problem with uniqueness concerns quasi-asymptotic, not asymptotic. 
However, the calculations show that we cannot avoid the additional function $L$, slowly varying at infinity, 
of the form $L(\lambda) = \textrm{ln}(\lambda^{\varepsilon})$, with any positive constant $\varepsilon$,
to ensure convergence of the considered limits. Thus, quasi-asymptotic seems inevitable. We can reduce (using quasi-asymptotic) 
the freedom in the choice of the constants $\mathfrak{c}^{++}_{lm}, \mathfrak{c}^{--}_{lm}$ up to within one common nonzero constant factor, by requiring
$L$ to be the same in the formulas for quasi-asymptotic of all components of the kernels (and it is indeed possible).
Chosing common $L$, mentioned above, we have only checked that the constants are finite and nonzero, but have not yet checked
if $\mathfrak{c}^{++}_{lm}$ and, separately, $\mathfrak{c}^{--}_{lm}$, are all equal between themselves 
(that $\mathfrak{c}^{--}_{lm} = - \mathfrak{c}^{++}_{lm}$ is almost immediate).

Above we have established the free constants only by comparison with the Staruszkiewicz theory.

We should warn the reader that in the last two formulas in (\ref{KernelsOfA(1)chi}), 
and the last two  in (\ref{KernelsOfHomPartA(1)chiIR}), we have used an implicit assumption. 
In fact, we should compute the last two Fourier transforms in (\ref{KernelsOfA(1)chi}), associated to the decomposition of $SL(2, \mathbb{C})$ 
acting in the Fock space of the charged field, as in Example 3 of Subsection \ref{psichi}. 
Next, to the Fourier transforms calculated in this way, we should apply the usual formulas for the infrared 
asymptotic (quasi-asymptotic), such as that in the first two formulas in (\ref{KernelsOfHomPartA(1)chiIR}). Using 
the last two formulas in (\ref{KernelsOfA(1)chi}), we have implicitly made assumption that the said Fourier transforms
(regarded as functions of the decomposition parameter $\chi$) are absolutely continuous with respect 
to the spectral measure of the decomposition of $SL(2, \mathbb{C})$), compare Subsection \ref{psichi}. 
We arrived at this assumption by noting that the kernels
${\kappa'}_{\chi \, 1,1}^{++}(\varphi),{\kappa'}_{\chi \, 1,1}^{--}(\varphi)$ of the first order contribution 
to interacting potential, regarded as functions of momenta, decrease faster than decomposable kernel $\kappa_{1,1}$ (for $t>1$) 
of Example 3 of Subsection \ref{psichi} ($\widetilde{\varphi}$ is rapidly decreasing).  

Therefore, we have identified
\begin{equation}\label{homogeneous(-1)AintFixedFrame}
-e x_\mu A_{{}_{\chi=-1}}^{e \, \mu}(x)\Bigg|_{{}_{\Gamma({\mathcal{H}'}^{e}_{{}_{\chi = -1}}) \otimes \mathcal{H}'_1}}
+
\underset{\lambda\rightarrow +\infty}{\textrm{lim}} \,
\left[\frac{\lambda A^{(1) \, \mu}_{{}_{\textrm{int} \, \chi}}\big(\lambda x\big)}{L(\lambda)}\right]\Bigg|_{{}_{\Gamma({\mathcal{H}'}^{e}_{{}_{\chi = -1}}) \otimes \mathcal{H}'_1}}
\end{equation}
with the phase operator
\begin{equation}\label{ASphase}
S(x) = S_0 -e \, x_0/r \, Q + \sum \limits_{l=1}^{\infty} \sum \limits_{m = -l}^{m= +l}
\{ c_{lm} f_{lm}^{(+)}(x) + \textrm{h.c.} \}, \,\,\,\, x\cdot x <0,
\end{equation}
of  \cite{Staruszkiewicz} in a fixed reference frame with the time-like unit versor $u= (1,0,0,0)$.
Here $\mathcal{H}'_{1}$ is the closed subspace of the Fock space of the asymptotically homogeneous states of symptotic homogeneity $\chi = -1$ 
of the charged field (in this case scalar complex field $\boldsymbol{\psi}$), 
which is spanned by the spherically symmetric eigenstates $|S_{n,u}\rangle$, $n \in \mathbb{Z}$, of the total charge operator $Q$:
\[
Q|S_{n,u}\rangle = ne |S_{n,u}\rangle, \,\,\,\, n \in \mathbb{Z}.
\]
The operators $c_{l,m}, c_{l,m}^{+}$, which are to be identified with the operators $c_{l,m}, c_{l,m}^{+}$ acting on zero charge states 
of \cite{Staruszkiewicz}, are constructed as in Subsections \ref{AS}, \ref{infra-electric-transversal-generalized-states}, \ref{Comparison2}.
They act in the Fock space $\Gamma({\mathcal{H}'}^{e}_{{}_{\chi = -1}})$ over homogeneous of degree $\chi =-1$ generalized states of electric type.
As we have seen, these states can be regarded as generalized states in the Fock space of the free e.m. potential, and arise in decomposition of the
representation of the restrictin of double covering of the Poincar\'e group to the subgroup $SL(2,\mathbb{C})$
into irreducible components (and their construction is possible only 
with that realization of the free e.m. potential field, in which the action of $SL(2,\mathbb{C})$ is decomposable, 
Subsection \ref{equivalentA-s}). The second term in (\ref{homogeneous(-1)AintFixedFrame}) denotes the exctraction
of the infrared-red spherically symmetric quasi-asymptotic part of homogeneity degree $\chi =-1$ of the homogeneous part of asymptotic ultraviolet 
homogeneity $-1$ of the first order contribution to the interacting e.m. potential, as explained above. 
The phase $S_0 = S(u)$, which is to be identified with $S_0$ in the fixed
reference frame $u=(1,0,0,0)$, is defined by
\[
e^{-inS(u)}|0\rangle = |S_{n,u}\rangle, \,\, n \in \mathbb{Z}. 
\]
Therefore, we construct the operators $S(u), Q, c_{l,m}, c_{l,m}^+$, component-wise, \emph{i.e.} on a subspace of the Fock space over 
a subspace of homogeneous states of fixed homogeneity $\chi =- 1$. In order to extend the identification between (\ref{homogeneous(-1)AintFixedFrame})
and (\ref{ASphase}) over to general reference frame, we observe that the transformation rule of these operators is fixed by the commutation rules 
(\ref{CommutationRules}), (\ref{Vacuum}) and invariance, compare Subsection \ref{Consistency}. The commutation rules (\ref{CommutationRules}), 
(\ref{Vacuum}), in turn follow from the laws of quantum mechanics
(recall the canonical commutation rules for the phase and the charge). But, compare Subsection \ref{Consistency}, 
the transformation rule preserving (\ref{CommutationRules}), (\ref{Vacuum}), must have the general form (\ref{c'S'})
in which $A$ is the matrix of the representation $(l_0=1,l_1=0)$ acting on the states of zero charge, with the matrix and vector valued
functions $A,B$ on $G=SL(2,\mathbb{C})$ preserving (i) -- (iii) of Subsection \ref{Consistency}, with the initial
conditions (\ref{infinitesimalB}) and (\ref{B(0)}). This determines $A,B$ uniquely, as we have already seen in Subsection \ref{Consistency},
and thus the transformation rule for $S(u), c_{l,m}, c_{l,m}^+$, because this rule is fixed by $A,B$.  

In this work we are primarily interested with the identification of the generalized states on which the interacting e.m. potential 
field, contracted with $x^\mu$, can be identified with the phase of \cite{Staruszkiewicz}. As we have seen, this happens for the Fock space over 
a subspace of generalized states of (asymptotic) homogeneity $\chi =-1$. But the above construction of $S(u), Q, c_{l,m}, c_{l,m}^+$ 
can be exteded component-wise over the 
remaining homogeneity componets $\chi$ with $\chi$ determined by the decomposition of the action of $SL(2,\mathbb{C})$. 
Here we mention only that the matrix $A$ should in general be replaced by the matrix $A = U^{{}^{(l_0,l_1)}}$ 
of the irreducible representation $(l_0,l_1)$ of the homogeneous single particle states, of the respective subspaces of fixed homogeneity, constructed
in Subsection \ref{AS}. The corresponding $c_{l,m}, c_{l,m}^+$, $l=l_0,l_0+1, \ldots$, which respect (\ref{CommutationRules}) and (\ref{Vacuum}),
acting in the Fock space over the generalized states of the corresponding homogeneity, are determined analogously, by decomposition 
into irreducible components of the representation of $SL(2,\mathbb{C})$. 
The corresponding $A,B$, determining the transformation rule of the analogous operators $S(u), Q, c_{l,m}, c_{l,m}^+$, acting in the 
Fock space over generalized states of the corresponding homogeneity $\chi$,
fulfil (i) -- (iii) of Subsection \ref{Consistency}, with the initial conditions
\[
{\textstyle\frac{d}{d\lambda}}B_{{}_{l,m}}(\lambda=0) = \mathfrak{e}  \,\, \textrm{const.} \,\,  \delta_{{}_{l \, l_0}} \delta_{{}_{m \, 0}}
\]
and
\[
B_{{}_{l,m}}(\lambda=0)=0, \,\,\, l=l_0,l_0+1,l_0+2, \ldots, \,\,\, -l \leq m \leq l.
\]
The general consistency proof remains the same as that in Subsection \ref{Consistency}, because of the parity
property $U^{{}^{(l_0,l_1)}}(\lambda)_{{}_{l,0 \,\,\, l',0}}(-\lambda) = (-1)^{l+l'} U^{{}^{(l_0,l_1)}}(\lambda)_{{}_{l,0 \,\,\, l',0}}(\lambda)$,
for the hyperbolic angle $\lambda$ of a hyperbolic rotation. Here we should mention an algebraic formulation \cite{Herdegen2005} of the infrared part of QED,
which constructs an abstract $C^*$-algebra, let say of ``abstract operators $S(u), Q, c_{l,m}, c_{l,m}^+$''. It is a generalization
of \cite{Staruszkiewicz} with the property that for a particular $*$-representation of the $C^*$-algebra of \cite{Herdegen2005}, induced by a particular 
positive-definie state on the $C^*$-algebra, we obtain the particular theory \cite{Staruszkiewicz}. It is worth mentioning here that the
$A,B$, chosen as above, give the corresponding various positive-definite states on the $C^*$-algebra of \cite{Herdegen2005}.

\subsection{Towards an extension of the Staruszkiewicz theory}\label{ASTheoryextended}

As we have already  emphasized, these are the scattering processes with the generalized (unnormalizable) many-particle plane wave states
as the \emph{in} and \emph{out} states, which fall into the proper range of applicability of QED on the Minkowski space-time. 
The radiative corrections to the bound state problems
can also be solved, with the method worked out by Schwinger, but cannot be treated purely in the realm of QED, 
understood as a quantum field theory on the Minkowski spacetime. 
We need assumptions and computational rules which cannot be inferred from the general 
principles of quantum field theory in order to compute the radiative corrections to the bound energy levels.
In particular the Schwinger many particle function is associated to the many particle Green function, or propagator,
purely through the Huygens Principle, and we are not able to associate the wave function with a well-defined
state in the Fock space. This is at least the case on the flat Minkowski space-time. We will show in section 
\ref{EUandG} that QED on space-times with compact cauchy surfaces and non-zero curvature can be extended
over other phenomena, in particular including bound states, and the Schwinger method can be justified
on such space-times, within quantum field theory, understood as the causal perturbative quantum field theory.

Similarly, these are the (multiparticle and generalized) homogeneous (infrared) states, which naturally fall into the range of applicability  
of the theory \cite{Staruszkiewicz}. Recall that the infrared fields are radiated by the scattered charged particles (Bremsstrahlung).
Thus again we see that these are the scattering processes, now looked upon from large distances from the
scattering center, and the radiation produced in this process, which can naturally be subsumed by the Staruszkiewicz theory. 

Nonetheless, comparisons of some of the consequences of Staruszkiewicz theory with some of the results 
concerning bound state problems seem to be important.

Recall, please, Theorem \ref{valueOfalpha} of Subsection \ref{Ustructure} of Staruszkiewicz's theory, 
which asserts that there is a strong relation between the eigenvalue $me$
of the total charge $Q$, and the type of representation of $SL(2,\mathbb{C})$ acting in the eigenspace 
of the respective (generalized) homogeneous states
determined by $me$, and which asserts that these representations become the same (concerning unitary class)
for all appropriately large $m$ greater than a fixed constant value $m_0$. 
This remarkable result can be compared to the well known and curious coincidence concerning self adjointness 
of the Hamiltonian of the bounded system composed by a heavy source (say nucleus) of the classical Coulomb
field and a relativistic charged particle in this field.  Namely, it is a well known phenomenon in relativistic 
wave mechanics that whenever the charge of the nuclei is of the order of magnitude comparable to the inverse 
of the fine structure constant or greater, then the Hamiltonian loses the self adjointness property 
(which sometimes is interpreted as an indication that the system, when passing to the quantum field theory level, 
becomes unstable). On the other hand (and this is a coincidence which no one understands) the nuclei of real atoms 
are unstable whenever the charge of the nuclei reaches the value of the same order (inverse of the fine structure 
constant). The mentioned breakdown of self adjointness cannot explain of course this phenomenon because there 
are mostly the strong (and not electromagnetic) forces which govern the stability of nuclei. 

To this coincidence we add another coming from the quantum theory 
of infrared photons of the quantized Coulomb field 
which asserts similar dependence of the state of the infrared field on the total charge due to Staruszkiewicz's
theory. It therefore seems to be important question if Staruszkiewicz's theory, summarized in the 
condition (derived in \cite{Staruszkiewicz1987} and \cite{Staruszkiewicz})
\begin{equation}\label{AS1}
\Big[\frac{1}{e} j_0(x), S(y)\Big]_{{}_{x_0=y_0}} = i \delta(\boldsymbol{\x} - \boldsymbol{\y}),
\end{equation}
joining the phase field $S(x)$ and the zero component $j_0(x)$ of the electric current density, 
or, after integration over the hyperplane $x_0=y_0$ 
\begin{equation}\label{AS2}
[Q, S(x)] = ie, \,\,\,\,\, Q = \int \ud^3 \x \, j_0,
\end{equation}
can be extended over to QED on the space-times on which QED can be consistently applied also to bound state problems.
Before we go into this direction we remind a problem recognized and set by Staruszkiewicz \cite{Star1},
concerning universality of  the electric charge. The electron and proton, being essentially different
particles, have the electric charge equal, up to sign, with absurdly high experimental accuracy.
We know that the difference between the absolute values of the two charges is less than
$10^{-21}$ part of the electron's charge. This must be a mathematical equality.
Now Staruszkiewicz asks the following question: how such a coincidence can be possible at all?
It can hardly be a consequence of any dynamical law, say coming from the assumed interactions of the elementary
fields of the ``Standard Model''. As we have seen this coincidence can naturally be explained within the 
Staruszkiewicz's theory of infrared states, and comes from the fact that the operator 
$S_0$, the integral of the phase quantum field $S(x)$ over a Cauchy surface and the total charge $Q$
provide a spectral description of the global $U(1)$ gauge group, as we explained in Subsection \ref{globalU(1)}.
This is possible only if the charges of the particles of fields coupled minimally to $A$ are equal to multiplicities
of a fixed elementary charge. Thus, the equality of the proton and electron charges and more generally the universal 
scale of electric charge of stable and metastable particles, similarly as the equality of the inertial 
$m_i$ and gravitational $m_g$ masses in Einstein's theory of gravity, comes from the universal geometry of the gauge 
group (in analogy of the space-time geometric character of the gravitational field, which is the simplest explanation 
of the equality $m_i = m_g$).

In passing to the generalization of (\ref{AS1}) or (\ref{AS2}) over to QED on globally causal 
space-times with compact Cauchy surfaces, we note first that, when we stay at the classical field theory level,
the argument of \cite{Staruszkiewicz1987} or \cite{Staruszkiewicz}, showing that the phase $S(x)$ of the complex
field $\psi(x)$ minimally coupled to the electromagnetic potential field $A(x)$, is the generalized momentum conjugated to the zero 
component of the electric current density, is universally valid and the specific geometry of space-time has nothing to do in 
the argument of \cite{Staruszkiewicz1987}. At the first sight the problem arises when passing to quantum field
operators $\boldsymbol{\psi}(x), A(x), j_0(x), S(x)$, because in general, such operators, when evaluated at single space-time
point $x$, are meaningless as ordinary
operators on the Fock space. In particular the phase $S(x)$ of a quantum field operator $\boldsymbol{\psi}(x)$
is in general meaningless. This is indeed the case on the Minkowski space-time. Thus, on the Minkowski space-time
we are using the homogeneous of degree $-1$ part $\big(A^{\mu}_{{}_{\textrm{int}}}(x)\big)_{{}_{\chi=-1}}$ 
of the field $A_{{}_{\textrm{int}}}(x)$, with well-defined ``phase'' only on a subspace of generalized states. 

However, we will show in Subsection \ref{EUandG} that on globally causal space-times with compact Cauchy 
surfaces and non-zero curvature (we study the Einstein Universe as an example), 
where QED is convergent, and can be applied to bound state problems,
e.g. on the Einstein Universe, the interacting quantum fields behave much more regularly. 
In particular, we will show, that on such space-times the complex fields $\psi(x)$ minimally
coupled to the electromagnetic potential must necessary be massive in order to
preserve regular character of the interacting fields, which after smearing with test function
become ordinary operators on the Fock space, compare Subsection \ref{CausalSonEU}. Moreover, we will
show that any massive and interacting field (strictly speaking each higher order contribution to such field)
becomes a well-defined operator on the Fock space, even when evaluated at single space-time point $x$,
and all higher order contributions have common perfect dense domain in the Fock space; 
for the proof compare Subsection \ref{CausalSonEU}.

In particular for a complex massive field (at least for the sum of higher order contributions up to any order)
the polar decomposition of each component of the interacting field $\boldsymbol{\psi}_{{}_{\textrm{int}}}(x)$ 
is well-defined as an operator in the Fock space. The total charge -- the conserved Noether integral corresponding 
to the global gauge symmetry group $U(1)$--
and computed for interacting fields becomes well-defined self adjoint operator 
\[
Q = \int\limits_{\textrm{Cauchy surface}} \ud^3 \x \, j^{0}_{{}_{\textrm{int}}}(x)
\]
on the  Fock space (this is the case at least for the sum including arbitrary large number of higher order contributions
to the interacting field $j^{0}_{{}_{\textrm{int}}}$, compare Subsection \ref{CausalSonEU}). 

This would be impossible on the Minkowski spacetime,
where the last integral is meaningless as an operator on the Fock space, even each separate higher order contribution
to this integral is meaningless as an operator on the Fock space, for QED on the Minkowski space-time,
whence we have to confine the theory to the generalized plane wave states, or homogeneous states,
in case of Minkowski space-time.  

Inspired by the spectral construction of Subsection \ref{globalU(1)} we set the conjecture that 
on globally causal space-times with compact Cauchy surfaces and non-zero curvature (where the interacting
fields behave regularly) the phase operator, say
\[
S_0 = \int\limits_{\textrm{Cauchy surface}} \ud^3 \x \, S(x),
\]
should exist which together with the total charge operator $Q$ for interacting fields, 
should provide spectral description of the global gauge group $U(1)$ in the Fock space,
similarly as the operators $S_0, Q$ of Staruszkiewicz's theory do in the sense 
explained in Subsection \ref{globalU(1)}. Here $e^{iS(x)}$ is understood as the phase operator
acting (through operator composition) on the partial isometry component 
of $\boldsymbol{\psi}_{{}_{\textrm{int}}}(x)$. 
From this universality of the electric charge would follow.
This also gives a non-trivial linkage
of the spectrum of the operator $Q$ to the considered system of fields coupled minimally to the 
electromagnetic potential field, and the assumed interaction Lagrange densities, which \emph{a priori} include 
also the ``Standard Model''.  One should emphasize that even if the interactions are based on more
complicated gauge groups, including $U(1)$ as a subgroup and, at the level of free fields, include free fields with 
charges not equal to the multiplicities of the electron charge, does not exclude the possibility
that indeed the spectrum of $Q$ for interacting fields consists of integer multiplicities of a fixed elementary charge, 
say the charge of the electron.

\pagebreak
    
\vspace*{5cm}

\section{On the  relation  between the space-time geometry and the quantum field}\label{EUandG}

The general principle (\ref{SpacetimeTupleFields}) of Subsection \ref{G} of Introduction, relating the space-time geometry expressed in operator-format with the quantum field, can be extended over less degenerate case of non-flat space-time with causal structure. For simplicity, we assume that the space-time
posses four independent one-parameter groups of symmetries (i.e. the corresponding vector fields on the spacetime everywhere span the tangent space)
for the causal perturbative approach to be easily applicable.
We choose as the example the static Einstein Universe $\mathbb{R} \times \mathbb{S}^{3}$,
which can be identified with the Lie group $\mathbb{R} \times SU(2, \mathbb{C})$ and the canonical right
invariant Riemannian and pseudo-Riemannian metrics on it (with the right invariant pseudo-Riemannian metric coinciding 
with the ordinary space-time  pseudo-Riemannian metric) and with
(besides the right action of the Lie group $\mathbb{R} \times SU(2, \mathbb{C})$
on itself) the natural action of the Lie group $\mathbb{R} \times SO(4)$ of isometries
(of the Einstein Universe) 
on the cylinder $\mathbb{R} \times \mathbb{S}^{3}$ (Einstein Universe) given by the embedding of the cylinder 
into the five dimensional pseudo-euclidean space. It is the twofold cover
$\mathbb{R} \times SU(2,\mathbb{C}) \times SU(2,\mathbb{C})$ of the group 
$\mathbb{R} \times SO(4)$ of isometries of the Einstein Universe, which plays the more fundamental role in QFT on the Einstein Universe than the group $\mathbb{R} \times SO(4)$ itself for the same reason for which it is the twofold cover $T_4 \circledS SL(2, \mathbb{C})$ of the Poincar\'e group\footnote{Here Poincar\'e group is understood as the semidirect product of the translation group and the group of proper ortochronous Lorentz transformations.} which is more important for QFT on the Minkowski spacetime than the Poincar\'e group itself. The right action of a group element $s \times u \times v
\in \mathbb{R} \times SU(2, \mathbb{C}) \times SU(2, \mathbb{C})$ on the space-time point
$t \times x \in \mathbb{R} \times SU(2, \mathbb{C})$ is equal
\[
(t + s) \times v^{-1}xu \in \mathbb{R} \times SU(2, \mathbb{C}),
\]
which after restriction of this action to the (right) action of the subgroup 
\[
\mathbb{R} \times SU(2, \mathbb{C})
\cong \mathbb{R} \times SU(2, \mathbb{C}) \times \boldsymbol{1}_{{}_{2}} \subset 
\mathbb{R} \times SU(2, \mathbb{C}) \times SU(2, \mathbb{C})
\]
coincides with the action of $\mathbb{R} \times SU(2, \mathbb{C})$ by right translations 
$R_{{}_{s \times u}}: t \times x \mapsto (t+s) \times xu$ on $\mathbb{R} \times SU(2, \mathbb{C})$, 
compare \cite{PaneitzSegalI}. 
The subgroup 
$\mathbb{R} \times SU(2, \mathbb{C}) \subset \mathbb{R} \times SU(2, \mathbb{C}) \times SU(2, \mathbb{C})$ plays the role for the Einstein Universe which the translation subgroup 
$T_4 \subset T_4 \circledS SL(2, \mathbb{C})$ does for the flat Minkowski space-time.

\subsection{A generalization of the notion of canonical Schr\"odinger-von Neumann pairs}\label{GeneralizedSchrodinger-VonNeumannPairs}

Before continuing our discussion of the Einstein Universe it is convenient here to introduce 
a natural extension of the notion of the canonical pair
${}^{{}^{R}}U,\mathbb{V}$ for Abelian Lie group\footnote{The construction makes sense for locally compact Abelian $G$, but we need the construction at the level of generators in order to have actually the corresponding Schr\"odinger-von Neumann canonical pair, so that Lie group is needed.} $G$
over to a compact Lie group $G$ (compare the  construction of the canonical pair 
due to Mackey \cite{Mackey2} for the locally compact Abelian case, which in case of the Abelian Lie group of translations $T_n \cong \mathbb{R}^n$ coincides\footnote{With $\mathbb{R}^n$ regarded as the additive group of vector addition, \emph{i.e.} componentwise addition.} with the canonical pair in the form due to Weyl, Schr\"odinger-von Neumann representation is obtained after passing to generators). Namely let $G$ be a compact Lie group, and let 
$\widehat{G}$ be its Tannaka-Krein dual (discrete in case of compact $G$) being equal (as a set)
to the set of concrete irreducible unitary representations $\widehat{x}: x \mapsto \widehat{x}(x)$ of $G$, exactly one for each unitary equivalence class, exhausting all irreducible unitary equivalence classes. For each $\widehat{x} \in \widehat{G}$, $x \in G$, let $\widehat{x}(x)_{ij}$, $i,j = 1, \ldots, \textrm{dim} \, \widehat{x}$ be the matrix of a (here necessary finite dimensional) irreducible
unitary representation $x \mapsto  \widehat{x}(x)$ of $G$. 
For each $x \in G$, $\widehat{x} \in \widehat{G}$ we define the corresponding families
of operators 
\[
{}^{{}^{L}}U_{{}_{x}}f(y) = f(x^{-1}y), \,\,\, 
f \in L^2(G), x,y \in G,
\]
\[
{}^{{}^{R}}U_{{}_{x}}f(y) = f(yx), \,\,\, 
f \in L^2(G), x,y \in G,
\]
\[
\mathbb{V}_{{}_{\widehat{x}, i, j}}f(y) = \widehat{x}(y)_{ij} f(y), \,\,\, 
f \in L^2(G), \widehat{x} \in \widehat{G},
y \in G, 
\]
on the Hilbert space $L^2(G)$ of (equivalence classes of) functions 
square summable with respect to right (and here also left) Haar measue $dy$ on $G$. 
We easily  verify the following ``canonical'' commutation rules for the right representation:
\begin{equation}\label{Rcanonical}
{}^{{}^{R}}U_{{}_{x}}\mathbb{V}_{{}_{\widehat{x}, i, j}} = 
\sum \limits_{k=1}^{\textrm{dim} \, \widehat{x}} 
\widehat{x}(x)_{kj} \mathbb{V}_{{}_{\widehat{x}, i, k}} {}^{{}^{R}}U_{{}_{x}},
\end{equation}
analogously we have  the following ``canonical'' commutation rules for the left representation:
\begin{equation}\label{Lcanonical}
{}^{{}^{L}}U_{{}_{x}}\mathbb{V}_{{}_{\widehat{x}, i, j}} = 
\sum \limits_{k=1}^{\textrm{dim} \, \widehat{x}} 
\overline{\widehat{x}(x)_{ki}} \mathbb{V}_{{}_{\widehat{x}, k, j}} {}^{{}^{L}}U_{{}_{x}}.
\end{equation}
Using the Peter-Weyl theorem we easily verify that there is no non-trivial
closed proper subspace of $L^2(G)$ invariant under the set of ``canonical'' operators
\[
{}^{{}^{R}}U_{{}_{x}}, \mathbb{V}_{{}_{\widehat{x}, i, j}}, \,\,\,
\widehat{x} \in \widehat{G}, i,j =1, \ldots \textrm{dim} \, \widehat{x}, \,\,\,
x \in G, 
\]
i.e. the right regular ''canonical'' pair ${}^{{}^{R}}U,\mathbb{V}$ is irreducible. 
By the same reason, \emph{i. e.} by the Peter-Weyl theorem we easily verify that there is no non-trivial
closed proper subspace of $L^2(G)$ invariant under the set of ``canonical'' operators
\[
{}^{{}^{L}}U_{{}_{x}}, \mathbb{V}_{{}_{\widehat{x}, i, j}}, \,\,\,
\widehat{x} \in \widehat{G}, i,j =1, \ldots \textrm{dim} \, \widehat{x}, \,\,\,
x \in G, 
\]
i.e. the left regular ''canonical'' pair ${}^{{}^{L}}U,\mathbb{V}$ is irreducible. 

Conversely: Let $U'$ be a (strongly continuous) unitary representation of a compact
group $G$ in a Hilbert space $H_{{}_{U'}}$. Let $\mathbb{V}'$ be a $\ast$-representation of 
the $C^\ast$-algebra $\mathscr{C}(G)$ of continuous complex valued functions
on $G$, with pointwise product and involution given by the complex conjugation, in the same Hilbert space
$H_{{}_{\mathbb{V}'}} = H_{{}_{U'}}$. Let moreover $U',\mathbb{V}'$ fulfil the canonical rules
(\ref{Rcanonical}) (or, respectively, (\ref{Lcanonical})) , with $\mathbb{V}'_{{}_{\widehat{x}, i, j}}$
understood as the representors, under 
$\mathbb{V}'$, of the functions $x\mapsto \widehat{x}_{{}_{i \, j}}(x)$. Let, finally, $U',\mathbb{V}'$ be an irreducible pair.
Then the pair $U',\mathbb{V}'$ is unitarily equivalent to the above pair  ${}^{{}^{R}}U, \mathbb{V}$ 
(or, respectively, to the pair ${}^{{}^{L}}U, \mathbb{V}$). For the proof we extend the imprimitivity system argument of Mackey, 
who applied it to the Abelian case. Indeed, by theorem II, \S 17 of  \cite{Neumark_dec} (compare also Corollary 9.2.1, p. 248 
of  \cite{Segal_Kunze}), there exists unique 
spectral measure $P$ on the Baire subsets of $G$, such that 
\[
\mathbb{V}'_{{}_{f}} = \int\limits_{G} f(x) \,\, \ud P(x), \,\,\, f \in \mathscr{C}(G),
\]
and commuting with $\mathbb{V}'_{{}_{f}}$, $f \in \mathscr{C}(G)$, and with all bounded operators which commute with all operators 
$\mathbb{V}'_{{}_{f}}$. 
In particular
\[
\mathbb{V}'_{{}_{\widehat{x}, i, j}} = \int\limits_{G} \widehat{x}_{{}_{i j}}(x) \,\, \ud P(x),
\]
for all $\widehat{x} \in \widehat{G}$.
It is easily seen, using the canonical rules (\ref{Rcanonical}), that 
\[
U'_{{}_{x}}\left[\int\limits_{G} \widehat{x}_{{}_{i j}}(x) \,\, \ud P(x)\right]{U'_{{}_{x}}}^{-1} =  
\int\limits_{G} \widehat{x}_{{}_{i j}}(x) \,\, \ud P(xy^{-1}).
\]
(In case of the left canonical rules (\ref{Lcanonical}) we get analogous identity with the argument $xy^{-1}$ of the last $P$
replaced with $yx$.)
Again using the Peter-Weyl theorem, we easily infer from this that 
\[
U'_{{}_{x}}P(x){U'_{{}_{x}}}^{-1} =  P(xy^{-1}).
\]
Irreducibility of $U',\mathbb{V}'$ implies irreducibility of the pair $U',P$. 
Therefore, by the irreducibility of the pair $U',P$, and Corollary 1, p. 180 of \cite{Mackey2}, the pair $U',\mathbb{V}'$ 
is unitarily equivalent to the above right regular canonical pair ${}^{{}^{R}}U, \mathbb{V}$ 
(or, respectively, to the left regular canonical pair ${}^{{}^{L}}U, \mathbb{V}$). 

In fact the irreducible right ${}^{{}^{R}}U, \mathbb{V}$  and left ${}^{{}^{L}}U, \mathbb{V}$ regular canonical pair 
construction, having the uniqueness (up to unitary equivalence) property,
can be extended over any locally compact group $G$, compare definition in \S 10.4, and theorem 10.4, p. 291 of \cite{Segal_Kunze},
with the $C^\ast$-algebra $\mathscr{C}(G)$ replaced with the $C^\ast$-algebra $\mathscr{C}_{{}_{0}}(G)$
of continuous functions of compact support.
In order to have generators and manifold structure of the spectra (and e.g. the absolute continuity of the spectral measures
of the representors of $\mathbb{V}$) we confine ourselves to locally compact Lie groups $G$.
In order to keep the formulation adopted
here, with the matrix elements of a complete class of unitary representations as generating the algebra of smooth functions on $G$, 
we would need the regular representation of $G$ to be of Type I, with $\widehat{G}$ equal to the set of the equivalence classes of the
irreducible components of the right (or left) regular representation of $G$,
and assume the topology of $G$ to be second countable with $\mathscr{C}(G)$ replaced by the algebra 
$\mathscr{C}_{{}_{0}}(G)$ of continuous functions vanishing at infinity. But here, we need a very special
case of the Lie group $G$ being a product of an Abelian Lie group and a compact Lie group, as the Einstein Universe 
is isomorphic to the Lie group $\mathbb{R} \times SU(2, \mathbb{C})$. We introduce the canonical pair on it gradually
starting with the compact case first. 

For a more general symmetric space-time, which is not a Lie group, acted on by a symmetry Lie group $G$, 
preserving the Lorentz structure, the construction of the canonical pair is still possible in the sense of the cited definition
 of \cite{Segal_Kunze}. 
But in this more general situation, the class  $\widehat{G}$ of unitary irreducible components of the regular representation, good for the canonical pair 
reconstructing spectrally the space-time, is in general not optimal for the class of representations if looked upon from the 
admissible free fields on the space-time, which makes the situation different in comparison to the QFT on Einstein Universe, in that the Fourier transform
theory associated naturally to the space-time, and based on the class $\widehat{G}$ is not enough in the construction and analysis of free quantum fields,
requiring addition of some classes of representations, and not using of a considerable part of $\widehat{G}$.  
As a rule, the class $\widehat{G}$ is not enough in general (and contains many elements which are superfluous) if looked upon 
from the needs of free quantum fields. For it should be remembered that in the case 
of a relativistically invariant theory, the existence of a positively definite and Lorentz-invariant Hermitian form 
in the underlying single particle Hilbert space of a free field may in general be impossible (recall for example the Gupta-Bleuler quantization, or BRST). 
However, there always exists an invariant Krein structure, \emph{i.e.} an invariant Hermitian form compatible (in the analytical sense
defined e.g. in \cite{Bog}) with the underlying Hilbert space structure. 
Therefore, in the more general symmetric space-time to the class $\widehat{G}$, 
it will in general be necessary to add finite-dimensional (or, a priori, even infinite dimensional)
representations that admit only an invariant Krein structure, in order to account for the 
components of representations we can find in the Hilbert spaces acted on by free fields. Even more, in case $G = SL(2, \mathbb{C})$,
we have to add the unitary irreducible representations of the supplementary series, 
which do not enter into decomposition of the regular representation of $G = SL(2, \mathbb{C})$, in order to include irreducible components
which can appear in the invariant subspaces of the Hilbert spaces acted on by some free fields (compare Subsection \ref{Ustructure}).  
As we will see, in case of the Einstein Universe (EU) situation simplifies, and 
there always exists invariant and compatible Krein structure in the single partice space of a free field, 
for which the underlying Hilbert space structure with which it is compatible,
is also invariant, and any unitary irreducible representation of $\mathbb{R} \times SU(2, \mathbb{C})$ is a direct integral \slash sum 
of the irreducible components of the regular representation, so that we can restrict ourselves
to the class  $\widehat{\mathbb{R} \times SU(2, \mathbb{C})}$ consisting solely of the unitary and irreducible components
of the (unitary) regular representation of $\mathbb{R} \times SU(2, \mathbb{C})$. 
In addition in case of more general (globally hyperbolic) 
space-time there is the problem of division of the space of solutions of the classical 
equation which is to be quantized, into the positive and negative frequency solutions, 
which is the basic step in the construction of the quantized free field.  
This is in general nontrivial and technically difficult problem, but which becomes trivial on EU, 
as it possess time symmetry. A class of space-times (including de Sitter space-time and the cosmological models \cite{HawkingEllis}) are conformally equivalent 
to a part of the Einstein Universe (EU) on which the problem has trivial solution. This allows to solve globally the problem of
division of frequences on these space-times at least for the fields which are conformally invariant (e.g. massless fields) and for those wave 
solutions whose images on EU through the conformal equivalence, can be extended all over the whole EU and belong to a set of solutions
of a fixed conformally invariant wave equation on EU. In this situation frequency division is induced
by the frequency division on EU. It may however happen that the linear subspace of the basic solutions of a wave equation for which consistent 
division of frequencies is possible is not the full space of basic wave solutions. For the remaining solutions  there is no consistent division of frequencies.
By construction the subspace with division of freqiencies well-defined is always invariant, as well as the whole space of basic waves.
In case the subspace is the whole space of basic solutions the standard Fock method of quantization gives us the quantum free field on the standard Fock space.
In case there exists wave solutions not admitting division into positive and negative frequency solutions, 
we have several possibilites: either ignore the solutions with undefined divisin of freuencies, 
or construct the Hilbert space of the field as tensor product of the Fock space (over the invariant subspace of positive frequency solutions) 
with the Hilbert space corresponding to the Hilbert space of the
Fourier coefficients of the waves for which division is inapplicable, regarded as the corresponding operators, with the commutation rules determined by the Poisson 
brackets of the Fourier components. This may work in case each element not admitting division can be written as finite combination of 
fixed finite number of elements modulo linear combinantion of elements admitting division, compare \cite{Staruszkiewicz} 
for the particular case of the scalar field on de Sitter hyperplane. 
This is why, among other reasons, we should start the study of QFT on Einstein Universe. 
To this end, we first develop the Fourier analysis on the EU space-time in a form which is useful for the application of the white noise method
of construction of the free fields as generalized integral kernel operators and for the analysis of the generalized integral kernel operators
including Fock expansions, in a form which can be extended over other symmetric space-times.

After these general remarks, let us back first to the compact Lie group $G$ case and the 
right and left canonical pairs upon it. (We have in mind $G=SU(2, \mathbb{C})$.)
Let us consider the one parameter subgroups $\tau \mapsto x_k(\tau)$,
$k=1, \ldots \textrm{dim} \, G$, whose tangent vectors at zero (at the unit of the group $G$) 
span the Lie algebra of $G$.
They define in the canonical manner the right invariant vector fields  $X^1, \ldots, X^{\textrm{dim} \, G}$ 
which are everywhere linearly independent vector fields on $G$
and at each point $x \in G$ span the tangent space $T_{x} G$ at $x$. Similarly, the same one
parameter subgroups $\tau \mapsto x_k(\tau)$, define the corresponding left invariant vector fields
$Y^1, \ldots, Y^{\textrm{dim} \, G}$ which are everywhere linearly independent vector fields on $G$
and at each point $x \in G$ span the tangent space $T_{x} G$ at $x$.

Namely, for the integral curves\footnote{Note that the \emph{right} invariant vector fields $X^k$ generate  one parameter groups of \emph{left} translations $w \mapsto x_k(\tau)^{-1}w$, and thus are related to generators of the \emph{left} representation${}^{{}^{L}}U$; and \emph{vice versa}
for \emph{left} invariant $Y^k$, which generate one parameter groups of \emph{right} translations $w \mapsto wx_k(\tau)$, and thus are related to generators of the \emph{right} representation
${}^{{}^{R}}U$.}  of $X^k$ passing through $w\in G$ we have 
$\tau \mapsto x_k(\tau)^{-1}w = x_k(-\tau)w$. For the integral curves of  $Y^k$, passing through $w\in G$, we have $\tau \mapsto w x_k(\tau)$, $w \in G$.
We thus we have the corresponding left infinitesimal ``Schr\"odinger-von Neumann left representation'' 
\[
P^1, \ldots P^{{}^{\textrm{dim} \, G}},
\]
\[
\mathbb{V}_{{}_{\widehat{x}, i, j}}, \,\,\, \widehat{x} \in \widehat{G},
i,j = 1, \ldots \textrm{dim} \, \widehat{x}, 
\]
where the generators $P^k$ on $L^2(G)$ are defined by
\[
\tau \mapsto {}^{{}^{L}}U_{{}_{x_k(\tau)}} = \textrm{exp}(i\tau P^k)
\]
through the Stone-von Neumann theorem. 
Analogously we have the corresponding infinitesimal ``Schr\"odinger-von Neumann right representation'' 
\[
R^1, \ldots R^{{}^{\textrm{dim} \, G}},
\]
\[
\mathbb{V}_{{}_{\widehat{x}, i, j}}, \,\,\, \widehat{x} \in \widehat{G},
i,j = 1, \ldots \textrm{dim} \, \widehat{x}, 
\]
where the generators $R^k$ on $L^2(G)$ are defined by
\[
\tau \mapsto {}^{{}^{R}}U_{{}_{x_k(\tau)}} = \textrm{exp}(i\tau R^k)
\]
through the Stone-von Neumann theorem. 

Note also that the smooth functions
${\varphi'}^{{}^{k}}: G \rightarrow \mathbb{R}$ defined by the maps 
\[
\varphi =
\big({\varphi'}^{{}^{1}}, \ldots {\varphi'}^{{}^{\textrm{dim}} \, G}\big) \in \textrm{Atl} \, G,
\]
\[ 
\varphi': G \rightarrow \mathbb{R}^{{}^{\textrm{dim} \, G}}
\]
of the smooth atlas $\textrm{Atl} \, G$ of $G$, regarded as a smooth manifold, can each be extended to smooth functions $\varphi^k$ over the whole $G$. 
Then the operators $\mathbb{V}_{{}_{\widehat{x}, i, j}}$ can be regarded as complex valued smooth\footnote{With their real and imaginary parts being smooth real functions.} functions of the operators $Q^1, \ldots, Q^m$ of multiplication by the functions $\varphi^k$, $\varphi' \in \textrm{Atl} \, G$, $k \in \{1, \ldots,\textrm{dim} \, G \}$. Note that in general $m > \textrm{dim} \, G$, as in general we have more than one map in $\textrm{Atl} \, G$. 
In fact the real and imaginary parts of the functions $\widehat{x}_{ij}$, 
$i,j = 1, \ldots, \textrm{dim} \widehat{x}$, $\widehat{x} \in \widehat{G}$, can be rearranged so that they can serve themselves as maps of an atlas of $G$ defining the smooth manifold structure on $G$
compatible with the initial $\textrm{Atl} \, G$. Because the tangent bundle $TG$ of
each Lie group is trivial, then any Clifford module over $TG$ corresponding to 
a right invariant pseudo-Riemannian or Riemannian metric on $G$ can be constructed
as a finite direct sum of copies of the ``canonical'' pair ${}^{{}^{R}}U,\mathbb{V}$  acting with uniform
multiplicity on the direct sum Hilbert space $\mathscr{H} = \bigoplus L^2(G)$, and with the 
tuple
\begin{multline*}
\Bigg( \,\,\,\,\,\,\,\, \mathcal{A} = 
\{f (\boldsymbol{Q}^{1}, \ldots, \boldsymbol{Q}^{m}), 
f \in \mathscr{C}^\infty(\mathbb{R}^m)\} \,\,,
\,\,\,\,\,\,\,\,
\mathscr{H} = \oplus L^2(G) \,, \\
D_{{}_{\textrm{ell}}} \, {} = \Gamma^{{}^{1}} \boldsymbol{P}^{{}^{1}} + 
\ldots + \Gamma^{{}^{\textrm{dim} \, G}} \boldsymbol{P}^{{}^{\textrm{dim} G}} + \Gamma
  \,\,\,\, \Bigg),
\end{multline*}
which spectrally (in the sense of Connes \cite{Connes_spectral}) defines the Lie group $G$ together with 
the corresponding Riemannian right invariant metric, and the corresponding Clifford algebra  matrix generators
$\Gamma^{{}^{1}}, \ldots \Gamma^{{}^{\textrm{dim} \, G}}$, 
necessary constant throughout
$G$ if the (elliptic) Dirac operator $D_{{}_{\textrm{ell}}}$ is to be invariant. Here the operators written with the bold fonts
on $\oplus L^2(G)$ are equal to the uniform multiplicities of the corresponding operators
of the ``canonical'' left representation ${}^{{}^{L}}U,\mathbb{V}$  in $L^2(G)$. 
The right invariant vector fields $X^1, \ldots, X^{\textrm{dim} \, G}$ are assumed to be everywhere
orthonormal with respect to the chosen right invariant Riemannian metric on $G$ and thus the 
Clifford algebra generators $\Gamma^{{}^{1}}, \ldots \Gamma^{{}^{\textrm{dim} \, G}}$ 
are assumed to fulfill
\[
\Gamma^k \Gamma^j + \Gamma^j \Gamma^k = 2 \delta^{kj}.
\]       
The operator $\Gamma$ of multiplication by the element $\Gamma$ of the Clifford algebra generated
by $\Gamma^{{}^{1}}, \ldots \Gamma^{{}^{\textrm{dim} \, G}}$ is a function of 
$\Gamma^{{}^{1}}, \ldots \Gamma^{{}^{\textrm{dim} \, G}}$ whose concrete form depends on the
commutation relations 
\[
[X^r, X^j] = \sum \limits_{k} c_{k}^{rj} X^k 
\,\,\,
\textrm{respectively}
\,\,\,
[P^r, P^j] = i \sum \limits_{k} c_{k}^{rj} P^k, 
\]
\[
\textrm{where}
\,\,\,
P^k = i X^k \,\,\,
\textrm{on}
\,\,\,
L^2(G),
\]
and on the specific form of the element $\Delta$ of the center of the Lie algebra of $G$,
expressed as a function of $X_k$ (resp. $P_k$) on $L^2(G)$ (say a Laplacian\footnote{In general $\Delta$ is equal 
to the Lapalace-Beltrami operator up to a multiplicative constant factor and up an additive constant 
operator $constant \cdot \boldsymbol{1}$,
with the $constant$ proportional to the curvature.} on $G$), which is expected to fulfill
\[
\big(D_{{}_{\textrm{ell}}} \big)^2 = \boldsymbol{1}_{{}_{d}} \Delta  = \oplus \Delta,
\,\,\,
\textrm{on}
\,\,\,
\bigoplus \limits_{1}^{d} L^2(G),
\]
where $d$ is the multiplicity  of the ``canonical'' pair ${}^{{}^{L}}U,\mathbb{V}$ 
equal to the multiplicity of the associated ``canonical'' pair ${}^{{}^{R}}U,\mathbb{V}$,
and where the Clifford algebra generators $\Gamma^{{}^{1}}, \ldots \Gamma^{{}^{\textrm{dim} \, G}}$
are represented as $d \times d$ matrix operators in $\mathbb{C}^d$.

Equivalently one can use the spectral tuple  
\begin{multline*}
\Bigg( \,\,\,\,\,\,\,\, \mathcal{A} = 
\Big\{ \sum \limits_{\widehat{x} \in \widehat{G}, i,j=1, \ldots \textrm{dim} \, \widehat{x}}
\, \textrm{dim} \, \widehat{x} \, \tilde{f}_{{}_{\widehat{x}, j, i}} \boldsymbol{\mathbb{V}}_{{}_{\widehat{x}, i, j}}, \,
\{ \tilde{f}_{{}_{\widehat{x}, j, i}}\} \in \mathcal{F}\big[\mathcal{S}_{\Delta}(G)\big] \Big\} \,\,, \\
\,\,\,\,\,\,\,\,
\mathscr{H} = \oplus L^2(G) \,, \\
D_{{}_{\textrm{ell}}} \, {} = \Gamma^{{}^{1}} \boldsymbol{P}^{{}^{1}} + 
\ldots + \Gamma^{{}^{\textrm{dim} \, G}} \boldsymbol{P}^{{}^{\textrm{dim} \, G}}  + \Gamma 
  \,\,\,\, \Bigg),
\end{multline*}
where $\mathcal{F}$ is the Fourier transform on $G$, and $\mathcal{F}\big[\mathcal{S}_{\Delta}(G)\big]$
is the set of rapidly decreasing sequences. Namely, the sequence 
$\{\tilde{f}_{{}_{\widehat{x}, j, i}} \}_{{}_{\widehat{G}\times \mathbb{N}\times \mathbb{N}}}$ or simply 
a function $\tilde{f}$ on $\widehat{G}\times \mathbb{N}\times \mathbb{N}$ we call rapidly decreasing
iff
\[
\Big( \big(\mathcal{F} \Delta \mathcal{F}^{-1}\big)^k  \tilde{f}, \,
\big(\mathcal{F} \Delta \mathcal{F}^{-1}\big)^k  \tilde{f}
\Big)_{{}_{\mathcal{F} L^2(G)}}
= \sum \limits_{\widehat{x} \times i \times j  \in 
\widehat{G}\times \mathbb{N}\times \mathbb{N}}  
\big(\lambda_{{}_{\widehat{x}}}\big)^{2k} \,\,\, \big|\tilde{f}_{{}_{\widehat{x}, i, j}}\big|^2
 < \infty
\]
for all $k \in \mathbb{N}$;
or equivalently iff
\[
\textrm{sup} \big\{ \big(\lambda_{{}_{\widehat{x}}}\big)^{2k} \,\,\, \big|\tilde{f}_{{}_{\widehat{x}, j, i}}\big|^2, \,\,\, \widehat{x} \times j \times i  \in 
\widehat{G}\times \mathbb{N}\times \mathbb{N} \big\} < \infty
\]
for all $k \in \mathbb{N}$. Here
\[
\Delta \, \widehat{x}_{ij} = \lambda_{{}_{\widehat{x}}} \,
\widehat{x}_{ij}
\,\,\,\,\,
\textrm{on}
\,\,\,
L^2(G),
\]
so that $\lambda_{{}_{\widehat{x}}}$ is the eigenvalue of the Laplacian $\Delta$ corresponding to the eigenfunction
\[
\widehat{x}_{ij}: G \ni x \mapsto \widehat{x}(x)_{ij}, \,\,\,\,\,
\widehat{x} \in \widehat{G},
\] 
which by definition of $\Delta$ depends only on the representation $\widehat{x} \in \widehat{G}$
and remains the same for all representations belonging to the unitary equivalence class 
of $\widehat{x}$.  

Note here that the operator $A=\Delta$ on $L^2(G)$ is standard\footnote{Strictly speaking we need the spectrum of the standard operator $A=\Delta$ to be greater than $1$, compare condition (A2) of \cite{hida}
or of Subsection \ref{white-setup}, in order to keep standard nature of the second quantized operator 
$\Gamma(A) = \Gamma(\Delta +\boldsymbol{1})$ when constructing the Hida test space in the Fock space of free fields. In practice the additive term proportional to the curvature is too small for the realistic values of the curvature, but we compensate this by addition of the constant unit operator $\boldsymbol{1}$ to $\Delta$ in order to preserve all conditions (A1) -- (A3) put on the standard operator. Thus, whenever we construct the nuclear test space
$\mathcal{S}_{\Delta}(G)$ we understood $\Delta$ as the Laplacian to which the constant unit operator has been added.}, \emph{i.e.} self-adjoint positive whose some negative power $A^{-r}$
is of Hilbert-Schmidt class (here it is so for $r > \textrm{dim} \, G$) and can serve to construct 
the nuclear space
\[
\mathcal{S}_\Delta(G) = \mathscr{C}^\infty(G)
\]
as the standard nuclear countably Hilbert space in the sense of Gelfand \cite{GelfandIV} 
and Hida-Obata \cite{obata-book}, \cite{obataJFA}, compare also Subsect. \ref{white-setup}. Note also that
the Fourier transform $\mathcal{F}$ on the group $G$ serves as the unitary map $\mathcal{F}:
 L^2(G) \rightarrow  L^2(\widehat{G}\times \mathbb{N} \times \mathbb{N})$ (compare discussion below),
which gives the operator
\[
\mathcal{F} \Delta \mathcal{F}^{-1} 
\,\,\,
\textrm{on}
\,\,\,
\mathcal{F}\big[L^2(G)\big]=
L^2(\widehat{G}\times \mathbb{N} \times \mathbb{N})
\]
acting as multiplication operator by the eigenvalue $\lambda_{{}_{\widehat{x}}}$ on each
subspace spanned by the Fourier components   
\[
\widetilde{f}_{{}_{ji}}(\widehat{x}) = \big(\mathcal{F}f\big)_{{}_{ji}}(\widehat{x}) = 
\int \limits_{G} f(y) \overline{\widehat{x}_{ij}(y)} \, dy,
\,\,\,
f \in L^2(G), 1 \leq i,j \leq \textrm{dim} \, \widehat{x},
\]
with fixed $\widehat{x}$ (we will use $\widetilde{f}$ and $\mathcal{F}f$ interchangeably).
Thus, we have 
\begin{equation}\label{FS_Delta(G) = S_Delta^(G^)}
\mathcal{S}_{\mathcal{F}\Delta \mathcal{F}^{-1}}(\widehat{G}\times \mathbb{N} \times \mathbb{N})
= \mathcal{F} \mathcal{S}_{\Delta}(G) = \mathcal{F} \mathscr{C}^\infty(G),
\end{equation}
\emph{i. e.} the standard operator
\[
\big(\mathcal{F}\Delta \mathcal{F}^{-1}, \mathcal{F}L^2(G)=
L^2(\widehat{G}\times \mathbb{N} \times \mathbb{N}) \big)
\]
is a unitary image of the standard operator
\[
(\Delta, L^2(G))
\]
under the unitary map $\mathcal{F}$, so that by construction of the standard countably
Hilbert and nuclear space we have the equality (\ref{FS_Delta(G) = S_Delta^(G^)}), compare 
Subsections \ref{white-setup} and \ref{dim=1}. 

Note also that here the components
$\tilde{f}_{{}_{\widehat{x}, j, i}}$ of
the rapidly decreasing sequence  
$\{\tilde{f}_{{}_{\widehat{x}, j, i}} \}_{{}_{\widehat{G}\times \mathbb{N}\times \mathbb{N}}}$
can be interpreted as the $i\times j$ matrix elements of the values
\[
\big(\mathcal{F}f\big)_{{}_{ji}}(\widehat{x})
\]
of the Fourier transform $\mathcal{F}f$ of a function 
\[
f \in \mathcal{S}_{\Delta}(G) \subset L^2(G)
\]
at $\widehat{x} \in \widehat{G}$.

Note, please, that
\[
\Big( \,\,\,\, \cdot \,\,\,\, , \,\,\,\, \cdot \,\,\,\, \Big)_{{}_{\mathcal{F} L^2(G)}}
\] 
is the Hilbert space inner product on
\[
\mathcal{F} \big[ L^2(G) \big] 
= L^2(\widehat{G} \times \mathbb{N} \times \mathbb{N})
\]
determined uniquely by the Fourier transform image of the inner product 
in $L^2(G)$ through the Plancherel formula.

Of course, for noncommutative Lie group $G$
the right invariant vector fields $X^1, \ldots, X^{\textrm{dim} \, G}$ do not commute, and correspondingly the operators $P^{{}^{1}}, \ldots P^{{}^{\textrm{dim} \, G}}$,
and thus also the operators $\boldsymbol{P}^{{}^{1}}, \ldots \boldsymbol{P}^{{}^{\textrm{dim} \, G}}$, do not commute. Also, the number $m$ of operators $Q^1, \ldots, Q^m$ exceeds in general the number  
$\textrm{dim} \, G$ of dimensions of $G$, contrary to the case of the standard Schr\"odinger-von Neumann representation for the additive Abelian group $G = T_n \cong \mathbb{R}^n$ with $\mathbb{R}^n$ regarded as the additive group of translations, where $m = \textrm{dim} \, G = n$ and where 
$P^{{}^{1}}, \ldots P^{{}^{\textrm{dim} \, G}}$,
and thus also with the commuting operators $\boldsymbol{P}^{{}^{1}}, \ldots \boldsymbol{P}^{{}^{\textrm{dim} \, G}}$. 

Recall, please, that for each $f \in L^2(G)$ we have the following Fourier transform
\[
F(\widehat{x}) = \widetilde{f}(\widehat{x}) = \int \limits_{G} f(y) \widehat{x}(y)^* \, dy
\]
or more explicitly
\[
F_{{}_{ji}}(\widehat{x})= \widetilde{f}_{{}_{ji}}(\widehat{x}) = \int \limits_{G} f(y) \overline{\widehat{x}_{ij}(y)} \, dy,
\]
which can be regarded as a matrix-valued function  $F$ on $\widehat{G}$. The inverse Fourier transform
is equal
\[
f(y) = \sum \limits_{\widehat{x} \in \widehat{G}} \textrm{dim} \, \widehat{x} \,\,\,
\textrm{Tr}\,[F(\widehat{x}) \widehat{x}(y)].
\]
From Peter-Weyl theorem it follows the Plancherel formula
\[
\int \limits_{G} f(y) \overline{f(y)} \, dy = 
\sum \limits_{\widehat{x} \in \widehat{G}} \textrm{dim} \, \widehat{x} \,\,\,
\textrm{Tr}\,[F(\widehat{x}) \, F(\widehat{x})^*],
\]
where for the matrix $F(\widehat{x})$ we define $F(\widehat{x})^*$, the ordinary adjoint (transposition and complex conjugation) of the matrix $F(\widehat{x})$.

In the sequel we will also consider vector valued functions $f$ on a group $G$. If the linear space 
of values of $f$ has a structure which allows a particular linear operation $\sharp$ in it, say transposition $T$ (or passing from a column-vector to a row-vector), then we distinguish between the operations $T$ performed first upon $f$ and then $\mathcal{F}$, which can be written
$\mathcal{F}(f^T)$, and the operation $\mathcal{F}$ performed first and the operation
$T$ performed upon $\mathcal{F}f(\widehat{x})$. The second operation will always be denoted
$\mathcal{F}f(\widehat{x})^T$. The value of the function resulting by the first sequence of the operations at $\widehat{x}$ will always be denoted $\mathcal{F}(f^T)(\widehat{x})$. We will also use the notation
$\mathcal{F}f^T$ for $\mathcal{F}(f^T)$. In general for an operation $\sharp$ on the linear space of values of $f$, defining $f^\sharp$ by the formula $f^\sharp(x) = f(x)^\sharp$, we denote
$\mathcal{F}(f^\sharp)$ simply by $\mathcal{F}f^\sharp$, and always understood 
$\mathcal{F}f^\sharp$ as that function which comes by the operation $\sharp$ performed first upon
$f$ and followed then by $\mathcal{F}$. If the analogue operation $\sharp$ makes also sense
in the linear space of values of the function $\mathcal{F}f$, and thus the function $\widehat{x} \mapsto \mathcal{F}f(\widehat{x})^\sharp = \big(\mathcal{F}f(\widehat{x})\big)^\sharp$ make sense, then it is never written as $\mathcal{F}f^\sharp$, or its value at $\widehat{x}$ is never written 
$\mathcal{F}f^\sharp(\widehat{x})$, but always 
$\mathcal{F}f(\widehat{x})^\sharp$, if the context do not indicate the order of the performed operations. This notation properly reflects the difference
between the order of the performed operations only when the resulting functions are evaluated at specific points  $\widehat{x}$. In order to distinguish between the resulting functions obtained by the operations performed in various orders we write  $\mathcal{F}f^\sharp$ for the function 
$\widehat{x} \mapsto \mathcal{F}(f^\sharp)(\widehat{x})$ and $(\mathcal{F}f)^\sharp$
for the function $\widehat{x} \mapsto \mathcal{F}f(\widehat{x})^\sharp$, by using the additional 
bracket in order to emphasize
that in the last case the operation $\sharp$ is performed at the very end after evaluation at 
$\widehat{x}$. In particular for a $\mathbb{C}^d$-valued function $f$ the value $\mathcal{F}f(\widehat{x})$  of the Fourier transform $\mathcal{F}f$ at $\widehat{x}$ can be interpreted as the column 
\[
\left( \begin{array}{c}  
\mathcal{F}f^1(\widehat{x}) \\
 \vdots \\
\mathcal{F}f^d(\widehat{x})
                     \end{array}\right)   
\]
of $d$ matrices 
\[
\left( \begin{array}{c}  
\mathcal{F}(f^{1})_{{}_{ji}}(\widehat{x}) \\
 \vdots \\
\mathcal{F}(f^{d})_{{}_{ji}}(\widehat{x})
                     \end{array}\right), \,\,\,\, 1 \leq i,j \leq  \textrm{dim} \, \widehat{x}
\]
$\mathcal{F}(f^{a})_{{}_{ji}}(\widehat{x})$, $1 \leq a \leq d$, each 
$\textrm{dim} \, \widehat{x} \times \textrm{dim} \, \widehat{x}$, and in this case 
$\mathcal{F}f(\widehat{x})$ is regarded as matrix
\[
F_{{}_{a \,\,\,\,\,ji}} = \mathcal{F}(f^{a})_{{}_{ji}}(\widehat{x})
\]
of a linear operator from the Hilbert space of $\textrm{dim} \, \widehat{x} \times \textrm{dim} \, \widehat{x}$ matrices (regarded as 
the linear space of Hilbert-Schmidt operators with the Hilbert-Schmidt norm multiplied by $\textrm{dim} \, \widehat{x}$) to the conjugation\footnote{In the sense of \cite{Mackey}} 
of the Hilbert space $\mathbb{C}^d$.
 Equivalently $\mathcal{F}f(\widehat{x})$ may be regarded as a matrix
\[
F_{{}_{ji \,\,\,\,\,a}} = \mathcal{F}(f^{a})_{{}_{ji}}(\widehat{x})
\]
of a linear operator which maps the linear Hilbert space $\mathbb{C}^d$ into the conjugation\footnote{In the sense of \cite{Mackey},
\S 5.} of the Hilbert space of $\textrm{dim} \, \widehat{x} \times \textrm{dim} \, \widehat{x}$ matrices (regarded as 
the linear space of Hilbert-Schmidt operators with the Hilbert-Schmidt norm multiplied by $\textrm{dim} \, \widehat{x}$). But, still equivalently,
$\mathcal{F}f(\widehat{x})$ can be regarded as a matrix 
\[
F_{{}_{aj \,\,\,\,\,i}} = \mathcal{F}(f^{a})_{{}_{ji}}(\widehat{x})
\]
of a linear operator which maps the linear Hilbert space of matrices (regarded as Hilbert-Schidt operators
from the Hilbert space $\mathcal{H}_{\widehat{x}}$ of the representation $\widehat{x}$ to the conjugation
of the Hilbert space $\mathbb{C}^d$ with the generalized Hilbert-Schmidt norm\footnote{Compare Subsection \ref{kronecker}, or \cite{Mackey2}, \S 3.1, pp. 125-127.} multiplied by $\textrm{dim} \, \widehat{x}$) into the conjugation of $\mathcal{H}_{\widehat{x}}$. Analogously
$\mathcal{F}f(\widehat{x})$ can be regarded as a matrix 
\[
F_{{}_{j \,\,\,\,\,ai}} = \mathcal{F}(f^{a})_{{}_{ji}}(\widehat{x})
\]
or as a matrix
\[
F_{{}_{j \,\,\,\,\,ia}} = (\mathcal{F}f^{a})_{{}_{ji}}(\widehat{x}).
\]
Correspondingly we have several possible transpositions $T$, which to the operator and its matrix 
assign the linearly transposed operator (between the dual spaces) and its matrix (called transposition
of the initial matrix). In particular for the various interpretations said above, we would have
\[
F^{T}_{{}_{a \,\,\,\,\,ji}} = F_{{}_{ji \,\,\,\,\, a}}, \,\,
F^{T}_{{}_{a \,\,\,\,\, ji}} =  F_{{}_{ji \,\,\,\,\, a}} \,\,
F^{T}_{{}_{aj \,\,\,\,\, i}} = F_{{}_{i \,\,\,\,\,aj}} \,\,\, \textrm{e. t. c.}
\]
in the possible linear spaces which can serve as the spaces of values of $\mathcal{F}f$
for $\mathbb{C}^d$-valued $f$ on $G$. Correspondingly to these various transposition operations $T$
in these spaces we have the functions $\widehat{x} \mapsto \big(\mathcal{F}f\big)^T(\widehat{x}) =  \big(\mathcal{F}f(\widehat{x})\big)^T$. 
Similarly the operation of transposition $T$ can be regarded
as applied to the space $\mathbb{C}^d$ and its dual, and changes the column vectors of 
$\mathbb{C}^d$ into the row-vectors of $\mathbb{C}^d$. In particular this allows us to define
$f^T$ for $\mathbb{C}^d$-valued function, and the function 
$\widehat{x} \mapsto \mathcal{F}f^T(\widehat{x})$. In case of the first interpretation
of  $\mathcal{F}f(\widehat{x})$ as a matrix
\[
F_{{}_{a \,\,\,\,\,ji}} = \mathcal{F}(f^{a})_{{}_{ji}}(\widehat{x}),
\]
\emph{i. e.} column 
\[
\left( \begin{array}{c}  
\mathcal{F}(f^{1})_{{}_{ji}}(\widehat{x}) \\
 \vdots \\
\mathcal{F}(f^{d})_{{}_{ji}}(\widehat{x})
                     \end{array}\right) 
\]
of $d$ matrices  $\mathcal{F}(f^{a})_{{}_{ji}}(\widehat{x})$, $d=1, \ldots, d$, the 
value $\mathcal{F}f^T(\widehat{x})$ is equal to the row
\[
\left( \begin{array}{ccc}  
\mathcal{F}(f^{1})_{{}_{ji}}(\widehat{x}) & 
 \ldots &
\mathcal{F}(f^{d})_{{}_{ji}}(\widehat{x})
                     \end{array}\right)
\]
of $d$ matrices  $\mathcal{F}(f^{a})_{{}_{ji}}(\widehat{x})$,
$1 \leq a \leq d$.
Thus for the first interpretation of the matrix $\mathcal{F}f(\widehat{x})$
we have the equality $\mathcal{F}f^T(\widehat{x}) = \big(\mathcal{F}f\big)^T(\widehat{x})$.
But note that in general for the other interpretations of $\mathcal{F}f(\widehat{x})$
we will in general have $\mathcal{F}f^T(\widehat{x}) \neq \big(\mathcal{F}f\big)^T(\widehat{x})$.
In the sequel whenever we are considering $\mathbb{C}^d$-valued functions $f$ on a group $G$ we choose 
the first interpretation of $\mathcal{F}f(\widehat{x})$, and regard it as a matrix  
\[
F_{{}_{a \,\,\,\,\,ji}} = \mathcal{F}(f^{a})_{{}_{ji}}(\widehat{x})
\]
of a linear operator from the Hilbert space of $\textrm{dim} \, \widehat{x} \times \textrm{dim} \, \widehat{x}$ matrices (regarded as 
the linear space of Hilbert-Schmidt operators with the Hilbert-Schmidt norm multiplied by $\textrm{dim} \, \widehat{x}$) to the conjugation of the Hilbert space $\mathbb{C}^d$, so that
\begin{equation}\label{TranspositionConventionForFourierTr}
\boxed{
\widetilde{f^T}(\widehat{x}) = \mathcal{F}f^T(\widehat{x}) = \big(\mathcal{F}f\big)^T(\widehat{x})
= \big(\widetilde{f} \big)^T(\widehat{x}).}
\end{equation}

The Plancherel formula shows that the Fourier transform gives us a unitary equivalence
of the right regular representation ${}^{{}^{R}}U$ of the ``canonical pair''
${}^{{}^{R}}U, \mathbb{V}$, to the
direct sum $\widehat{{}^{{}^{R}}U}$ of all irreducible representations $\widehat{x} \in \widehat{G}$ into which each
$\widehat{x} \in \widehat{G}$ enters with multiplicity $\textrm{dim} \, \widehat{x}$.
Indeed, for fixed (but arbitrary) $\widehat{x}$
\[
\big(\widehat{{}^{{}^{R}}U}_{{}_{x}}F\big)_{{}_{ji}}(\widehat{x})=
\int \limits_{G} {}^{{}^{R}}U_{{}_{x}}f(y) \overline{\widehat{x}_{ij}(y)} \, dy 
= \int \limits_{G} f(yx) \overline{\widehat{x}_{ij}(y)} \, dy 
= \sum \limits_{k} \widehat{x}_{{}_{jk}}(x) F_{{}_{ki}}(\widehat{x}),
\]
thus each $i$-th column (and we have of course $\textrm{dim} \, \widehat{x}$ columns) 
\[
\left( \begin{array}{c}   F_{{}_{1i}}(\widehat{x}) \\
                                           F_{{}_{2i}}(\widehat{x}) \\
                                           \vdots \\
                   F_{{}_{\textrm{dim} \, \widehat{x} \, i}}(\widehat{x})  \end{array}\right) 
\]
of the matrix $F(\widehat{x}) = \mathcal{F}f(\widehat{x})= \widetilde{f}(\widehat{x})$ 
transforms under the irreducible unitary representation $\widehat{x}$. 
Similarly Fourier transform gives us a unitary equivalence
of the left regular representation ${}^{{}^{L}}U_{{}_{x}}f(y) = f(x^{-1}y)$ to the
direct sum $\widehat{{}^{{}^{L}}U}$ of complex conjugations $\overline{\widehat{x}}$ of all irreducible representations $\widehat{x} \in \widehat{G}$ into which each $\overline{\widehat{x}}$,
$\widehat{x} \in \widehat{G}$, enters with multiplicity $\textrm{dim} \, \widehat{x}$. Indeed,
again for a fixed $\widehat{x}$:
\[
\big(\widehat{{}^{{}^{L}}U}_{{}_{x}}F\big)_{{}_{ji}}(\widehat{x}) =
\int \limits_{G} {}^{{}^{L}}U_{{}_{x}}f(y) \overline{\widehat{x}_{ij}(y)} \, dy 
= \int \limits_{G} f(x^{-1}y) \overline{\widehat{x}_{ij}(y)} \, dy 
= \sum \limits_{m} \overline{\widehat{x}_{{}_{im}}(x)} F_{{}_{jm}}(\widehat{x}),
\]
thus each $j$-th row (and we have of course $\textrm{dim} \, \widehat{x}$ rows) 
\[
\left( \begin{array}{cccc}   F_{{}_{j1}}(\widehat{x}) &
                                           F_{{}_{j2}}(\widehat{x}) &
                                           \ldots &
                   F_{{}_{\textrm{dim} \, \widehat{x}}}(\widehat{x})  \end{array}\right) 
\]
of the matrix $F(\widehat{x})$ transforms under the irreducible unitary representation
$\overline{\widehat{x}}$. Thus the commuting 
\[
{}^{{}^{L}}U \,\,\, {}^{{}^{R}}U = {}^{{}^{R}}U \,\,\, {}^{{}^{L}}U, \,\,\,\,
\textrm{on} \,\,\, L^2(G) 
\]
left and right representations ${}^{{}^{L}}U, {}^{{}^{R}}U$ on $L^2(G)$, regarded as representation
\[
x \times y \mapsto {}^{{}^{R}}U_{{}_{x}} \,\, {}^{{}^{L}}U_{{}_{y}} = {}^{{}^{L}}U_{{}_{y}}  \,\,
 {}^{{}^{R}}U_{{}_{x}}  
\]
of the direct product $G \times G$ on $L^2(G)$,
is unitarily equivalent, under the Fourier transform, to the direct sum
\[
\widehat{{}^{{}^{R}}U \,\, {}^{{}^{L}}U} =
\bigoplus \limits_{\widehat{x} \in \widehat{G}} \, \widehat{x} \times \overline{\widehat{x}}
\]
of \emph{outer Kronecker product} representations 
\[
G \times G \ni x \times y \mapsto \widehat{x} \times \overline{\widehat{x}}(x\times y) = 
\widehat{x}(x) \times \overline{\widehat{x}}(y)
\]
of the direct product $G \times G$ on the corresponding Hilbert spaces of matrices $F_{{}_{\widehat{x}}}$ with the Hilbert space inner product
given by the trace multiplied by $\textrm{dim} \, \widehat{x}$, naturally identifiable with
the Hilbert space tensor product
\[
\mathcal{H}_{\widehat{x}} \otimes \mathcal{H}_{\overline{\widehat{x}}},
\]
of the Hilbert spaces of representations (with the inner products appropriately weighted)
$\widehat{x}$ and $\overline{\widehat{x}}$. Under this identification
\[
\widehat{x}(x) \times \overline{\widehat{x}}(y) = \widehat{x}(x) \otimes \overline{\widehat{x}}(y)
\]
with the ordinary Hilbert space operator tensor product. We have
\[
\big(\widehat{{}^{{}^{R}}U_{{}_{x}}{}^{{}^{L}}U_{{}_{y}}}F\big)_{{}_{ji}}(\widehat{x}) 
= \int \limits_{G} {}^{{}^{R}}U_{{}_{x}}{}^{{}^{L}}U_{{}_{y}}f(y) \overline{\widehat{x}_{{}_{ij}}(y)} \, dy 
= \sum \limits_{m,k} \widehat{x}_{{}_{jk}}(x)\overline{\widehat{x}_{{}_{im}}(y)} 
F_{{}_{km}}(\widehat{x}).
\]
for the Fourier transform image 
\[
\widehat{{}^{{}^{R}}U_{{}_{x}}{}^{{}^{L}}U_{{}_{y}}}
\]
of the representation
\[
{}^{{}^{R}}U_{{}_{x}}{}^{{}^{L}}U_{{}_{y}}
\]
of the direct product $G\times G$, with first factor acting through right and the second through left translations on $L^2(G)$. 

The operator $\mathbb{V}_{{}_{\widehat{z}, m,k}}$ of the ``canonical pair'' ${}^{{}^{R}}U,\mathbb{V}$, is transformed by the Fourier transform into the operator $\widehat{\mathbb{V}_{{}_{\widehat{z}, m,k}}}$:
\[
\big(\widehat{\mathbb{V}_{{}_{\widehat{z}, m,k}}} F\big)_{{}_{ji}}(\widehat{x})
= \sum \limits_{\widehat{y} \in \textrm{dec} (\widehat{z} \otimes \overline{\widehat{x}})} \,\,\,\,\,
\sum \limits_{1 \leq s,q \leq \textrm{dim} \, \widehat{y}} 
\,\,\,\,\,\,
\underset{\widehat{z}, m,k}{M}_{{}_{\widehat{x} \,\, ji}}^{{}^{\widehat{y} \,\, sq}} 
\,\,\,\, F_{{}_{sq}}(\widehat{y}),
\]
where the summation runs over all those $\widehat{y} \in \widehat{G}$ which (up to unitary equivalence)
enter into the direct sum decomposition of the representation $\widehat{z} \otimes \overline{\widehat{x}}$,
which we have denoted by
\[
\widehat{y} \in \textrm{dec} (\widehat{z} \otimes \overline{\widehat{x}}),
\]
and where 
\[
\underset{\widehat{z}, m,k}{M}_{{}_{\widehat{x} \,\, ji}}^{{}^{\widehat{y} \,\, sq}} 
\]
is the matrix determining the pointwise product
\[
\widehat{z}(x)_{mk} \,\, \overline{\widehat{x}(x)_{ij}}=
\sum \limits_{\widehat{y} \in \textrm{dec} (\widehat{z} \otimes \overline{\widehat{x}})} \,\,\,\,\,
\sum \limits_{1 \leq s,q \leq \textrm{dim} \, \widehat{y}} 
\,\,\,\,\,\,\,
\underset{\widehat{z}, m,k}{M}_{{}_{\widehat{x} \,\, ji}}^{{}^{\widehat{y} \,\, sq}} 
\,\,\,\,
\widehat{y}(x)_{sq},
\]
and determined uniquely by the coefficients of the unitary matrix $S$  
\[
\widehat{z} \otimes \overline{\widehat{x}}
= S\Bigg( \bigoplus \limits_{\widehat{y} \in \textrm{dec} (\widehat{z} \otimes \overline{\widehat{x}})} 
\,\,\,\widehat{y}\Bigg)S^{-1}
\]
giving the unitary equivalence between 
\[
\widehat{z} \otimes \overline{\widehat{x}}
\,\,\,\,\,\,\,\,\,\,\,\,\,\, \textrm{and} \,\,\,\,\,\,\,\,\,\,
\bigoplus \limits_{\widehat{y} \in \textrm{dec} (\widehat{z} \otimes \overline{\widehat{x}})} 
\,\,\,\widehat{y},
\]
In case of $G=SU(2,\mathbb{C})$, we can use the Clebsch-Gordan coefficients $C^{{}^{\widehat{y},s}}_{{}_{\widehat{z},m \,\,\, \widehat{x},i}}$ 
of the group $G = SU(2,\mathbb{C})$ to compute the multiplication matrix $M$:
\[
\underset{\widehat{z}, m,k}{M}_{{}_{\widehat{x} \,\, ji}}^{{}^{\widehat{y} \,\, sq}} =
C^{{}^{\widehat{y},s}}_{{}_{\widehat{z},m \,\,\, \widehat{x},-i}}C^{{}^{\widehat{y},q}}_{{}_{\widehat{z},k \,\,\, \widehat{x},-j}},
\] 
which respect the identity
\[
C^{{}^{\widehat{y},s}}_{{}_{\widehat{z},0 \,\,\, \widehat{x},i}} = \delta_{{}_{i}}^{{}^{s}} C^{{}^{\widehat{y},s}}_{{}_{\widehat{z},0 \,\,\, \widehat{x},i}}
\]
if $G= SU(2,\mathbb{C})$. 

Note that the construction of the right ``canonical'' pair ${}^{{}^{R}}U,\mathbb{V}$ 
(and the corresponding left ``canonical'' pair ${}^{{}^{L}}U,\mathbb{V}$) can be easily extended over a more general class of locally compact Lie groups $G$ which are equal to direct (or semi-direct) product 
$G_{{}_{\textrm{Ab}}} \times G$ (or $G_{{}_{\textrm{Ab}}} \circledS G$) of a locally compact Abelian Lie group $G_{{}_{\textrm{Ab}}}$ and a compact Lie group $G$, or still further on a class of Type I Lie groups.  As we will see the compact case (where $G_{{}_{\textrm{Ab}}}$ is also compact) is essentially all we need here. 

Using the Pontrjagin-Weil-Raikov-Shilov\footnote{Compare \cite{Mackey2} for Mackey's form of the ``canonical'' pair on $L^2(G_{{}_{\textrm{Ab}}} )$ with $\mathbb{V}_{{}_{\widehat{x}}}$
equal to the operator of multiplication on $L^2(G_{{}_{\textrm{Ab}}})$ by the unitary character
$\widehat{x}$ of $G_{{}_{\textrm{Ab}}}$.} duality theorem \cite{gelfand-comm-norm-rings} 
(generalization of Plancherel and Fourier transform and its inverse) for the Abelian Lie group 
$G_{{}_{\textrm{Ab}}}$, we can easily extend the above construction
of the ``canonical'' pair ${}^{{}^{R}}U,\mathbb{V}$ from the compact Lie group $G$ over the locally compact
direct (or semi-direct) product Lie group $G_{{}_{\textrm{Ab}}} \times G$ 
(or $G_{{}_{\textrm{Ab}}} \circledS G$).
In particular using the ordinary Fourier transform for the additive Lie group of real numbers $\mathbb{R}$ 
we easily extend the construction of the ``canonical'' pair ${}^{{}^{R}}U,\mathbb{V}$ from $G$ over to the case of the direct product Lie group $\mathbb{R} \times G$. Indeed, the group $G_{{}_{\textrm{Ab}}} = \mathbb{R}$ being Abelian has all irreducible unitary representations (characters) $\widehat{p} \in \widehat{\mathbb{R}} = \mathbb{R}$ one dimensional:
\[
\mathbb{R} \ni t \mapsto \widehat{p}(t) = e^{ipt}, \,\,\,\,
p \in \mathbb{R}= \widehat{\mathbb{R}};
\]
and the set of irreducible unitary representations $\widehat{p} \cdot \widehat{x} \in \widehat{\mathbb{R} \times G}$
\[
\mathbb{R} \times G \ni t \times x \mapsto \widehat{t} \cdot \widehat{x}(s\times x) = \widehat{p}(t) \widehat{x}(x) = e^{ipt} \widehat{x}(x), \,\,\,\,
p \in \mathbb{R}= \widehat{\mathbb{R}}, \,\, 
\widehat{x} \in \widehat{G},
\]
exhaust all, up to unitary equivalence, irreducible representations of $\mathbb{R} \times G$.
For $f \in L^2(\mathbb{R} \times G)$ we have the Fourier transform 
\[
\widetilde{f}(\widehat{p} \cdot \widehat{x}) = F(\widehat{p} \cdot \widehat{x}) = \int \limits_{\mathbb{R} \times G} f(t,y) \big[\widehat{p} \cdot\widehat{x}(t,y)\big]^* \, 
\, \frac{1}{\sqrt{2\pi}} dt dy, 
\]
or more explicitly
\[
\widetilde{f}_{{}_{ji}}(\widehat{p} \cdot \widehat{x}) = F_{{}_{ji}}(\widehat{p} \cdot \widehat{x}) = \int \limits_{\mathbb{R} \times G} f(t,y) e^{-ipt}\overline{\widehat{x}_{{}_{ij}}(y)} \,
\, \frac{1}{\sqrt{2\pi}} \, dt dy,
\]
which can again be regarded as a matrix-valued function  $F$ on $\widehat{\mathbb{R} \times G}$. 
The inverse Fourier transform
is equal
\begin{multline*}
f(t,y) = \sum \limits_{\widehat{x} \in \widehat{G}} \int \limits_{\mathbb{R}} 
\textrm{dim} \,  (\widehat{p} \cdot \widehat{x}) \,\,\,
\textrm{Tr}\,[F(\widehat{p} \cdot \widehat{x}) \,\,\,\,\, \widehat{p} \cdot \widehat{x}(t,y)]
\, d\widehat{p} 
\\
= 
\sum \limits_{\widehat{x} \in \widehat{G}} \int \limits_{\mathbb{R}} 
\textrm{dim} \,  (\widehat{p} \cdot \widehat{x}) \,\,\,
\textrm{Tr}\,[\widetilde{f}(\widehat{p} \cdot \widehat{x}) \,\,\,\,\, \widehat{p} \cdot \widehat{x}(t,y)]
\, d\widehat{p}, 
\end{multline*}
\[
d\widehat{p} = \frac{1}{\sqrt{2\pi}} \, dp, \,\,\,\,
\textrm{dim} \, (\widehat{p} \cdot \widehat{x}) = \textrm{dim} \, \widehat{x}.
\]
From Peter-Weyl theorem and the ordinary Plancherel theorem for the additive $\mathbb{R}$ it follows the Plancherel formula
\begin{multline*}
 \int \limits_{G} f(t,y) \overline{f(t,y)} \, dt dy = 
\sum \limits_{\widehat{x} \in \widehat{G}} 
\int \limits_{\mathbb{R}}
\textrm{dim} \, \widehat{x} \,\,\,
\textrm{Tr}\,\big [ \, F(\widehat{p} \cdot \widehat{x}) \,\, F(\widehat{p} \cdot \widehat{x})^* \, \big ]
\, d\widehat{p}
\\
=
\sum \limits_{\widehat{x} \in \widehat{G}} 
\int \limits_{\mathbb{R}}
\textrm{dim} \, \widehat{x} \,\,\,
\textrm{Tr}\,\big [ \, \widetilde{f}(\widehat{p} \cdot \widehat{x}) \,\, \widetilde{f}(\widehat{p} \cdot \widehat{x})^* \, \big ]
\, d\widehat{p}.
\end{multline*}
Let us introduce the following right regular representation ${}^{{}^{R}}U$
\[
{}^{{}^{R}}U_{{}_{s \times u}}f(t, x) = f\big((t+s) \times xu \big)
\]
of $\mathbb{R} \times G$ on $L^2(\mathbb{R} \times G)$. Togther with
\[
\mathbb{V}_{{}_{\widehat{p} \cdot \widehat{x}, i, j}}f(t, y) = \widehat{p} \cdot \widehat{x}(s, y)_{ij} f(t, y), \,\,\, 
f \in L^2(\mathbb{R} \times G), \widehat{p} \cdot \widehat{x} \in \widehat{\mathbb{R} \times G},
t \times y \in \mathbb{R} \times G, 
\]
${}^{{}^{R}}U$ composes the right canonical pair ${}^{{}^{R}}U, \mathbb{V}$ on the Hilbert space $L^2(\mathbb{R} \times G)$
veryfing the right``canonical'' commutation rules:
\[
{}^{{}^{R}}U_{{}_{s \times u}}\mathbb{V}_{{}_{\widehat{p} \cdot \widehat{x}, i, j}} = 
\sum \limits_{k=1}^{\textrm{dim} \, \widehat{p} \cdot \widehat{x}} 
\widehat{p} \cdot \widehat{x}(s, u)_{kj} \mathbb{V}_{{}_{\widehat{p} \cdot \widehat{x}, i, k}} {}^{{}^{R}}U_{{}_{x}}.
\]
Analogously we have  the ``canonical'' commutation rules for the left regular representation 
of $\mathbb{R} \times G$.

Next, let us consider the left regular representation
${}^{{}^{L}}U$ 
\[
{}^{{}^{L}}U_{{}_{v}}f(t \times x) = f(t \times v^{-1}x)
\]
of $G \subset \mathbb{R} \times G$ on $L^2(\mathbb{R} \times G)$ (which is not the left regular 
representation of the whole group $\mathbb{R} \times G$, 
and should not be mixed with it).

${}^{{}^{R}}U$ and ${}^{{}^{L}}U$ define the representation
\[
\mathbb{R} \times G \times G \ni 
s \times u \times v \mapsto {}^{{}^{R}}U_{{}_{s \times u}}
\,\,\, {}^{{}^{L}}U_{{}_{v}} = {}^{{}^{L}}U_{{}_{v}} \,\,\,
{}^{{}^{R}}U_{{}_{s \times u}}
\]
of the direct product group $\mathbb{R} \times G \times G$ on 
$L^2(\mathbb{R} \times G)$. The Fourier transform defines the unitary equivalence of the ``regular'' representation
$s \times u \times v \mapsto {}^{{}^{R}}U_{{}_{s \times u}}
\,\,\, {}^{{}^{L}}U_{{}_{v}} = {}^{{}^{L}}U_{{}_{v}} \,\,\, {}^{{}^{R}}U_{{}_{s \times u}}$
on $L^2(\mathbb{R} \times G)$ with the direct sum/integral 
\[
\bigoplus \limits_{\widehat{x} \in \widehat{G}} \,\,\,\, \int \limits_{p \in \mathbb{R}}
\big(\widehat{p} \cdot \widehat{x}\big) \times \overline{\widehat{x}} \,\,\,\, d \widehat{p}
\]
of the \emph{outer Kronecker product} representations
\[
\big(\widehat{p} \cdot \widehat{x}\big) \times \overline{\widehat{x}}
= \widehat{p} \times \widehat{x} \times \overline{\widehat{x}},
\,\,\, \widehat{p} \in \widehat{\mathbb{R}} = \mathbb{R}, \widehat{x} \in \widehat{G}
\]
of the direct product group $\mathbb{R} \times G \times G$. This in particular means that  the restriction
of the representation ${}^{{}^{L}}U \,\,\, {}^{{}^{R}}U$ to the subgroup $\mathbb{R} \times G$ 
(with ${}^{{}^{L}}U$ degenerated to $\boldsymbol{1}$), \emph{i. e.} the representation ${}^{{}^{R}}U$,
is equivalent, via the Fourier transform, to the direct sum/integral 
\[
\widehat{{}^{{}^{R}}U} = \bigoplus \limits_{\widehat{x} \in \widehat{G}} \,\,
\textrm{dim} \, \widehat{x} \,
\int  \limits_{\mathbb{R}} \widehat{p} \cdot \widehat{x} \,\, d \widehat{p}
\]
of all representations
\[
\widehat{p} \cdot \widehat{x}
= \widehat{p} \times \widehat{x}, \,\,\,\,\,
\widehat{p} \in \widehat{\mathbb{R}} = \mathbb{R}, \widehat{x} \in \widehat{G}
\]
each entering with multiplicity 
\[
\textrm{dim} \, \overline{\widehat{x}} = \textrm{dim} \, \widehat{x} = \textrm{dim} \, \widehat{p} \cdot \widehat{x}.
\]

We introduce now the square summable (with respect to the Haar measure on $\mathbb{R} \times G$) multispinors\footnote{Recall that the smooth multispinors $\phi$ on $\mathbb{R} \times G$ 
compose the trivial smooth bundle because of the triviality
of the tangent bundle $T(\mathbb{R} \times G)$.}
\[
\phi \in L^2(\mathbb{R} \times G; \mathbb{C}^d) = \bigoplus \limits_{1}^{d} L^2(\mathbb{R} \times G;  \mathbb{C}) = \bigoplus \limits_{1}^{d} L^2(\mathbb{R} \times G)
\]
 with the following local transformation
formula 
\begin{equation}\label{UphiOnRxG}
U_{{}_{s\times u \times v}} \phi(t \times x) = 
V(v) \phi\big((t+s) \times (v^{-1}xu) \big),
\end{equation}
\[
s\times u \times v \in \mathbb{R} \times G \times G,
\]
where $V$ is a representation of $G$ on $\mathbb{C}^d$ with the standard inner product,
which is unitary and Krein-unitary with respect to a natural Krein fundamental symmetry
$\mathfrak{J}$: $\mathfrak{J}^* = \mathfrak{J}$, $\mathfrak{J}^2 = \boldsymbol{1}$, defined in a natural 
way by the generators $\Gamma^{{}^{1}}, \ldots, \Gamma^{{}^{\textrm{dim} \, G}}$ of the Clifford algebra
which in turn correspond to a right invariant Riemannian metric on $G$ and by an additional 
$\Gamma^{{}^{0}}$ anti-commuting with all $\Gamma^{{}^{1}}, \ldots, \Gamma^{{}^{\textrm{dim} \, G}}$
and with $\Gamma$. Below we will give detailed description of the representations $V$ 
standing in the transformation formula (\ref{UphiOnRxG}) and of the generators 
$\Gamma^{{}^{1}}, \ldots, \Gamma^{{}^{\textrm{dim} \, G}}$ and $\Gamma^{{}^{0}}$ of the Clifford algebra
determined by the Riemannian metric naturally associated to a natural right invariant pseudo-Riemannian metric 
on $\mathbb{R} \times G$ and 
the fundamental symmetry operators $\mathfrak{J}$ corresponding to the  pseudo-Riemannian metric 
and to the representation $V$ for the 
special case $\mathbb{R} \times G = \mathbb{R} \times SU(2, \mathbb{C})$ we are interested in.

The Fourier transform on $\mathbb{R} \times G$ gives us the unitary equivalence of the unitary
(and Krein-unitary) representation (\ref{UphiOnRxG}) with the direct sum/integral 
\begin{equation}\label{decUphiOnRxG}
\widehat{U} = V_\mathcal{F} U {V_\mathcal{F}}^{-1}
= \bigoplus \limits_{\widehat{x} \in \widehat{G}} \,\,\,
\int \limits_{\mathbb{R}} \big(\widehat{p} \cdot \widehat{x}\big) \times 
\big(\overline{\widehat{x}} \otimes V \big) \,\,\,\, d\widehat{p}
\end{equation}
of \emph{outer Kronecker products}
\[
\big(\widehat{p} \cdot \widehat{x} \big) \times 
\big(\overline{\widehat{x}} \otimes V \big)
\]
of representations
\[
\widehat{p} \cdot \widehat{x} 
\,\,\,\,
\textrm{and}
\,\,\,\,
\overline{\widehat{x}} \otimes V
\]
respectively of $\mathbb{R} \times G$ and $G$.
Therefore each direct summand  
\[
\textrm{dim} \, \widehat{x} \,\, \big(\widehat{p} \cdot \widehat{x}\big) 
\]
of the decomposition 
\[
\widehat{{}^{{}^{R}}U} = \bigoplus \limits_{\widehat{x} \in \widehat{G}} \,\,
\textrm{dim} \, \widehat{x} \,
\int  \limits_{\mathbb{R}} \widehat{p} \cdot \widehat{x} \,\, d \widehat{p}
\]
enters into the decomposition\footnote{Restricted to the subgroup $\mathbb{R} \times G \cong
\mathbb{R} \times G \times \boldsymbol{}1$  with $V$ put equal $\boldsymbol{1}_{{}_{\textrm{dim} \, V}}$
in (\ref{UphiOnRxG}).} (\ref{decUphiOnRxG}) with multiplicity 
$\textrm{dim} \, V = d$ equal to the dimension $d$ of the representation $V$. This means
that the operator ${}^{{}^{R}}U$ of the ``canonical'' pair ${}^{{}^{R}}U, \mathbb{V}$
determined by $\mathbb{R} \times G$ and equal to (\ref{UphiOnRxG}) restricted to the subgroup 
\[
\mathbb{R} \times G \cong \mathbb{R} \times G \times \boldsymbol{1} \subset \mathbb{R} \times G \times G,
\]
enters into the representation\footnote{With (\ref{UphiOnRxG}) restricted to the subgroup 
$\mathbb{R} \times G \cong \mathbb{R} \times G \times \boldsymbol{}1$.} 
(\ref{UphiOnRxG}) with uniform multiplicity equal $\textrm{dim} \, V = d$. Otherwise: restriction
of (\ref{UphiOnRxG}) to the subgroup 
\[
\mathbb{R} \times G \cong \mathbb{R} \times G \times \boldsymbol{1} \subset \mathbb{R} \times G \times G,
\]
is equal to the action of ${}^{{}^{R}}U$ with uniform multiplicity $\textrm{dim} \, V = d$.

We have correspondingly denoted the Fourier transform on $L^2(\mathbb{R} \times G; \mathbb{C}^d)$
of square summable multispinors -- composing  $\textrm{dim} \, V = d$ copies of ``canonical'' pairs --
equal in fact to the direct sum 
of $\textrm{dim} \, V = d$ copies of the Fourier transform on $L^2(\mathbb{R} \times G; \mathbb{C})
= L^2(\mathbb{R} \times G)$ acting in the space of the ``canonical''pair, by the symbol
$V_\mathcal{F}$. This is because it is intimately connected to the representation $V$ 
in (\ref{UphiOnRxG}) determining
the transformation formula (\ref{UphiOnRxG}) for the space of multispinors on which the Fourier transform 
$V_\mathcal{F}$ is defined.

We consider only the cases in which the Clifford algebra generators 
$\Gamma^{{}^{1}}, \ldots, \Gamma^{{}^{\textrm{dim} \, G}}$ corresponding to the right invariant
Riemannian metric on $G$ can be extended by adjoining one more generator $\Gamma^{{}^{0}}$
anti-commuting with all $\Gamma^{{}^{1}}, \ldots, \Gamma^{{}^{\textrm{dim} \, G}}$ and $\Gamma$,
such that $\Gamma^{{}^{0}}, \Gamma^{{}^{1}}, \ldots, \Gamma^{{}^{\textrm{dim} \, G}}$ 
compose Clifford algebra generators corresponding to a right invariant Riemannian
metric on $\mathbb{R} \times G$,
with the corresponding Dirac operator
\[
D_{{}_{\textrm{ell}}} =  \Gamma^{{}^{0}} \boldsymbol{P}^{{}^{0}} + 
\Gamma^{{}^{1}} \boldsymbol{P}^{{}^{1}} + 
\ldots + \Gamma^{{}^{\textrm{dim} \, G}} \boldsymbol{P}^{{}^{\textrm{dim} G}} + \Gamma
\] 
 and 
\[
\widehat{\Gamma}^{{}^{0}} = \Gamma^{{}^{0}}, \widehat{\Gamma}^{{}^{1}} = -i \Gamma^{{}^{1}}, \ldots, 
\widehat{\Gamma}^{{}^{\textrm{dim} \, G}} = -i\Gamma^{{}^{\textrm{dim} \, G}},
\widehat{\Gamma} = -i \Gamma,
\]
correspond to  right invariant pseudo-Riemannian metric on 
$\mathbb{R} \times G$, with the corresponding Dirac operator
\[
D = \widehat{\Gamma}^{{}^{0}} \boldsymbol{P}^{{}^{0}} + \widehat{\Gamma}^{{}^{1}} \boldsymbol{P}^{{}^{1}} + 
\ldots + \widehat{\Gamma}^{{}^{\textrm{dim} \, G}} \boldsymbol{P}^{{}^{\textrm{dim} G}} + \widehat{\Gamma},
\] 
on the space of multispinors on $\mathbb{R} \times G$ with $\boldsymbol{P}^{{}^{0}}$ being equal to the generator of the one-parameter subgroup 
$\tau \mapsto \tau \times \boldsymbol{1} \subset \mathbb{R} \times G$. 

On the Hilbert space $\oplus L^2(\mathbb{R} \times G)$ there exists a natural fundamental symmetry
operator $\mathfrak{J}$ of multiplication by a constant matrix which can be expressed in terms
of the generators $\Gamma^{{}^{0}}, \Gamma^{{}^{1}}, \ldots, \Gamma^{{}^{\textrm{dim} \, G}}$.
The operators $D_{{}_{\textrm{ell}}}$, $D$ and $\mathfrak{J}$ have the property that
\[
\frac{1}{2} \Big\{(D\mathfrak{J})^2 + (\mathfrak{J}D)^2 \Big\}
= \big(D_{{}_{\textrm{ell}}} \big)^2 = \boldsymbol{1}_d \Delta = \oplus \Delta,
\]
therefore we will write 
\[
D_\mathfrak{J} = D_{{}_{\textrm{ell}}}
\]
for the invariant elliptic Dirac operator $D_{{}_{\textrm{ell}}}$ in the sequel indicating its connection to 
the (non-elliptic) invariant Dirac operator $D$ and to the fundamental symmetry operator $\mathfrak{J}$
determined by the pseudo-Riemannian right invariant metric and naturally associated to the right invariant Riemannian metric and the elliptic Dirac operator $D_\mathfrak{J} = D_{{}_{\textrm{ell}}}$. 
Here  $\Delta$ is the Laplacian on $L^2(\mathbb{R} \times G)$ corresponding to the invariant Riemannian
metric on $\mathbb{R} \times G$, equal to the Laplace-Beltrami operator with, in general, additional and additive constant term proportional 
to the curvature. $d = \textrm{dim} \, V$ is equal to the multiplicity of the 
``canonical'' pair ${}^{{}^{R}}U, \mathbb{V}$, and equal to the dimension of the matrices 
$\Gamma^{{}^{0}}, \Gamma^{{}^{1}}, \ldots, \Gamma^{{}^{\textrm{dim} \, G}}$. 

Here for the direct product $\mathbb{R} \times G$  we have, besides the generators $P^k$ on $L^2(\mathbb{R} \times G)$ defined by
\[
\tau \mapsto {}^{{}^{L}}U_{{}_{x_k(\tau)}} = \textrm{exp}(i\tau P^k)
\]
for the one-parameter subgroups $\tau \mapsto x_k(\tau) =  1 \times g_k(\tau)$, $k = 1, \ldots \textrm{dim} \, G$
(with $\tau \mapsto g_k(\tau)$ being the one parameter subgroups corresponding to \emph{right} invariant vector fields $X_1, \ldots X_{{}_{\textrm{dim} \, G}}$ on $G$, the generators of the \emph{left} ``canonical'' pair), also the ``time translation'' generator $P^0$, defined by
\[
\tau \mapsto {}^{{}^{R}}U_{{}_{x_0(\tau)}} = {}^{{}^{L}}U_{{}_{x_0(\tau)}} = \textrm{exp}(i\tau P^0)
\]
for the one-parameter ``time translation'' subgroup $\tau \mapsto x_0(\tau)$:
\[
\tau \mapsto x_0(\tau) = \tau \times \boldsymbol{1} \in \mathbb{R} \times G.
\]

On the Hilbert space 
\[
\mathscr{H} =
\bigoplus \limits_{1}^{d} L^2(\mathbb{R} \times G)
\]
of multispinors we introduce the following spectral tuple
\begin{multline}\label{1stSpectralTupleForRxG}
\Bigg( \,\,\,\,\,\,\,\, \mathcal{A} = 
\{f (\boldsymbol{Q}^{0}, \boldsymbol{Q}^{1}, \ldots, \boldsymbol{Q}^{m}), 
f \in \mathcal{S}_{A'}(\mathbb{R}^m)\} \,\,,
\,\,\,\,\,\,\,\,
\mathscr{H} = \oplus L^2(\mathbb{R} \times G) \,, \\
D_{{}_{\textrm{ell}}} = D_{{}_{\mathfrak{J}}} \, {} = \Gamma^{{}^{0}} \boldsymbol{P}^{{}^{0}} + \Gamma^{{}^{1}} \boldsymbol{P}^{{}^{1}} + 
\ldots + \Gamma^{{}^{\textrm{dim} \, G}} \boldsymbol{P}^{{}^{\textrm{dim} G}} 
+\Gamma \,\,, 
\,\,\,\,\,\,\,\,\,  \\
D \, {} =  \widehat{\Gamma}^{{}^{0}} \boldsymbol{P}^{{}^{0}} + \widehat{\Gamma}^{{}^{1}} \boldsymbol{P}^{{}^{1}} + 
\ldots + \widehat{\Gamma}^{{}^{\textrm{dim} \, G}} \boldsymbol{P}^{{}^{\textrm{dim} \, G}} 
+ \widehat{\Gamma} \,,
\,\,\,\,\,\,\,\,\,
\mathfrak{J}  \,\,\,\, \Bigg).
\end{multline}

Equivalently one can use the spectral tuple  
\begin{multline}\label{2ndSpectralTupleForRxG}
\Bigg( \,\,\,\,\,\,\,\, \mathcal{A} = 
\Bigg\{ 
\sum_{\substack{\widehat{x} \in \widehat{G} \\
                  i,j=1, \ldots \textrm{dim} \,\widehat{x}}}
 \,\,
\textrm{dim} \, \widehat{x} 
\,\,
\int \limits_{\mathbb{R}}
\tilde{f}_{{}_{\widehat{p} \cdot \widehat{x}, j, i}} \boldsymbol{\mathbb{V}}_{{}_{\widehat{p} \cdot \widehat{x}, i, j}} \, d\widehat{p}, \,
\tilde{f}\in \mathcal{F}\big[\mathcal{S}_{A'}(\mathbb{R} \times G)\big] \Bigg\} \,\,, \\
\,\,\,\,\,\,\,\,
\mathscr{H} = \oplus L^2(\mathbb{R} \times G) \,, \\
D_{{}_{\textrm{ell}}} =  D_{{}_{\mathfrak{J}}} \, {} = \Gamma^{{}^{0}} \boldsymbol{P}^{{}^{0}} + \Gamma^{{}^{1}} \boldsymbol{P}^{{}^{1}} + 
\ldots + \Gamma^{{}^{\textrm{dim} \, G}} \boldsymbol{P}^{{}^{\textrm{dim} \, G}} + \Gamma \,\,, 
\,\,\,\,\,\,\,\,\, \\
D \, {} = \widehat{\Gamma}^{{}^{0}} \boldsymbol{P}^{{}^{0}} + \widehat{\Gamma}^{{}^{1}} \boldsymbol{P}^{{}^{1}} + 
\ldots + \widehat{\Gamma}^{{}^{\textrm{dim} \, G}} \boldsymbol{P}^{{}^{\textrm{dim} \, G}} 
+ \widehat{\Gamma} \,,
\,\,\,\,\,\,\,\,\,
\mathfrak{J}  \,\,\,\, \Bigg),
\end{multline}
where 
\[
\mathcal{F}\big[\mathcal{S}_{A'}(\mathbb{R} \times G)\big] =
\mathcal{S}_{\mathcal{F} A' \mathcal{F}^{-1}}(\mathbb{R} \times \widehat{G} \times \mathbb{N} \times \mathbb{N})
\]
is the standard countably Hilbert nuclear space of rapidly decreasing functions $\tilde{f}$ on $\mathbb{R} \times \widehat{G}\times \mathbb{N}\times \mathbb{N}$, determined by the standard operator
\[
\mathcal{F}A' \mathcal{F}^{-1} 
\,\,\,
\textrm{on}
\,\,\,
\mathcal{F} L^2(\mathbb{R} \times G) 
= L^2(\mathbb{R} \times \widehat{G} \times \mathbb{N} \times \mathbb{N}),
\]
and equal to the Fourier image of the standard nuclear space
\begin{multline*}
\mathcal{S}_{A'}(\mathbb{R} \times G) = 
\mathcal{S}_{(\Delta_\mathbb{R} +t^2 +1 )\otimes \boldsymbol{1}+ \boldsymbol{1} \otimes \Delta_G}(\mathbb{R} \times G) \\ =
\mathcal{S}_{\Delta_\mathbb{R} +t^2 +1}(\mathbb{R}) \otimes \mathcal{S}_{\Delta_G}(G)
= \mathcal{S}(\mathbb{R}) \otimes \mathscr{C}^\infty(G)
\end{multline*} 
determined by the standard nuclear operator ($X^0 = \tfrac{\partial}{\partial t}$):
\begin{multline*}
A' = (\Delta_\mathbb{R} +t^2 +1 )\otimes \boldsymbol{1}+ \boldsymbol{1} \otimes \Delta_G 
= (-{\textstyle\frac{d^2}{dt^2}} +t^2 +1 )\otimes \boldsymbol{1}+ \boldsymbol{1} \otimes \Delta_G \\
= -\big(X^0)^2 - \Delta_G + V'_{A'} = \Delta + V'_{A'}
\end{multline*} 
on
\[
L^2(\mathbb{R}) \otimes L^2(G),
\]
with $V'_{A'}$ equal to the multiplication by the function $\mathbb{R} \times G \ni t \times w
\mapsto 1+ t^2$, compare the first and the second Proposition of Subsection \ref{white-setup}.

Here the function $\tilde{f}$
\[
p \times \widehat{x}\times j \times i \mapsto
\tilde{f}_{{}_{\widehat{p} \cdot \widehat{x}, j, i}} 
= \mathcal{F}(f)_{{}_{ji}}(\widehat{p} \cdot \widehat{x})
= \widetilde{f}_{{}_{ji}}(\widehat{p} \cdot \widehat{x})
\]
on $\mathbb{R} \times \widehat{G}\times \mathbb{N}\times \mathbb{N}$ we call rapidly decreasing
iff
\begin{multline*}
\Big( \big(\mathcal{F} A \mathcal{F}^{-1}\big)^k  \tilde{f}, \,
\big(\mathcal{F} A \mathcal{F}^{-1}\big)^k  \tilde{f}
\Big)_{{}_{\mathcal{F} L^2(\mathbb{R} \times G)}} \\
= \sum \limits_{\widehat{G} \times \mathbb{N} \times \mathbb{N}}
\int \limits_{\mathbb{R}} \big(1 -{\textstyle\frac{d^2}{dp^2}} +p^2 +  
\lambda_{{}_{\widehat{x}}} \big)^{2k} \,\,
\big|\tilde{f}_{{}_{\widehat{p} \cdot \widehat{x}, i, j}}\big|^2
< \infty, \,\,\,
k \in \mathbb{N},
\end{multline*}
or equivalently iff
\[
\textrm{sup} \Big\{ \big(1 + |p|^2 + \lambda_{{}_{\widehat{x}}} \big)^{2k} 
\big|\partial_{p}^{m}\tilde{f}_{{}_{\widehat{p} \cdot \widehat{x}, j, i}}\big|^2, 
\,\,\, p \times \widehat{x} \times j \times i  \in 
\mathbb{R} \times \widehat{G}\times \mathbb{N}\times \mathbb{N} \Big\} < \infty
\]
for all $k,m \in \mathbb{N}$, and denote it by 
$\tilde{f} \in \mathcal{S}_{\mathcal{F}\Delta \mathcal{F}^{-1}}(\mathbb{R} \times \widehat{G} \times \mathbb{N} \times \mathbb{N})$, compare Subsection \ref{white-setup}. 
Here $\lambda_{{}_{\widehat{x}}}$ is the eigenvalue of the Laplace operator 
$\Delta_G$ on $G$ corresponding to the eigenfunctions $x \mapsto \widehat{x}(x)_{ij}$
determined by the matrix elements of the representation $\widehat{x} \in \widehat{G}$,
and common for all representations of the unitary equivalence class of the representation $\widehat{x}$.
Recall please that
\[
\Big( \,\,\,\, \cdot \,\,\,\, , \,\,\,\, \cdot \,\,\,\, \Big)_{{}_{\mathcal{F} L^2(\mathbb{R} \times G)}}
\] 
is the Hilbert space inner product on
\[
\mathcal{F} L^2(\mathbb{R} \times G)  
= L^2(\mathbb{R} \times \widehat{G} \times \mathbb{N} \times \mathbb{N})
\]
determined uniquely by the Fourier transform image of the inner product 
in $L^2(\mathbb{R} \times G)$ through the Plancherel formula, of course with the Fourier transform on the
direct product group $\mathbb{R} \times G$.

The bold-written $\boldsymbol{P}^{{}^{0}}, \boldsymbol{P}^{{}^{1}}, \ldots$ stand for the generators
$\oplus P^{{}^{0}}, \oplus P^{{}^{1}}, \ldots$ of the subgroups
\[
\tau \mapsto \boldsymbol{{}^{{}^{L}}U}_{{}_{x_\mu(\tau)}} = \textrm{exp}(i\tau \boldsymbol{P}^\mu),
\,\,\, \mu = 0, 1, \ldots, \textrm{dim} \, G
\]
in the representation
\[
\boldsymbol{{}^{{}^{L}}U} = \oplus {}^{{}^{L}}U 
\,\,\, \textrm{on}
\,\,\, \oplus L^2(\mathbb{R} \times G)
\]
equal to the $(\textrm{dim} \, V = d)$-fold copy of the left regular representation
\[
{}^{{}^{L}}U 
\,\,\, \textrm{on}
\,\,\, L^2(\mathbb{R} \times G)
\]
of the ``canonical'' pair ${}^{{}^{L}}U, \mathbb{V}$ on $ L^2(\mathbb{R} \times G)$ determined by
the direct product Lie group $\mathbb{R} \times G$.
 In the spectral tuple (\ref{2ndSpectralTupleForRxG}) the operator $\mathbb{V}_{{}_{\widehat{p} \cdot \widehat{x}, i, j}}$
is in fact also understood as $(\textrm{dim} \, V = d)$-fold copy
$\oplus \mathbb{V}_{{}_{\widehat{p} \cdot \widehat{x}, i, j}}$ of the operator
\[
\mathbb{V}_{{}_{\widehat{p} \cdot \widehat{x}, i, j}}
\,\,\, \textrm{on}
\,\,\, L^2(\mathbb{R} \times G)
\]
of the ``canonical'' pair ${}^{{}^{L}}U, \mathbb{V}$ (equivalently of the right ``canonical'' pair 
${}^{{}^{R}}U, \mathbb{V}$) on $ L^2(\mathbb{R} \times G)$ determined by
the direct product Lie group $\mathbb{R} \times G$. Similarly, we have for the operators 
$\boldsymbol{Q}^{{}^{0}}, \boldsymbol{Q}^{{}^{1}}, \ldots$ standing for the $d$-fold copies of the corresponding operators
$\oplus Q^{{}^{0}}, \oplus Q^{{}^{1}}, \ldots$ determined by the the ``canonical''
pair ${}^{{}^{R}}U, \mathbb{V}$ or ${}^{{}^{L}}U, \mathbb{V}$ on $ L^2(\mathbb{R} \times G)$.

We need the Connes spectral reconstruction theorem, i.e. the possibility of reconstruction
of the manifold $\mathbb{R} \times G$ from the spectral tuple (\ref{1stSpectralTupleForRxG}) 
or equivalently (\ref{2ndSpectralTupleForRxG}).
The spectral tuple (\ref{1stSpectralTupleForRxG}) or equivalently
(\ref{2ndSpectralTupleForRxG}) can indeed be used as the spectral description of the Lie group
$\mathbb{R} \times G$ similar to that presented in \cite{Connes_spectral}. 
In constructing (\ref{1stSpectralTupleForRxG}) or equivalently
(\ref{2ndSpectralTupleForRxG}) we used the Lie group structure, in particular the manifold structure
of $\mathbb{R} \times G$. 
However, going the other way round, i.e. construction of the Lie group $\mathbb{R} \times G$
together with its manifold structure from (\ref{1stSpectralTupleForRxG}) or equivalently
from (\ref{2ndSpectralTupleForRxG}), is quite nontrivial and hard problem, solved only for the
compact manifold in \cite{Connes_spectral}. Here we have the additional complication
coming from the non-compact character of $\mathbb{R} \times G$. As we have explained in Appendix \ref{AppendixNonCompMani}, 
this requires an additional step 
in comparison to \cite{Connes_spectral}, with conformal embedding of the manifold $\mathbb{R} \times G$
into the compact group $U(1) \times G \cong \mathbb{S}^1 \times G$, which introduces auxiliary 
``scaling'' $Q_{{}_{\textrm{scaling}}}$ and ``binding potential'' $V_{{}_{\textrm{binding}}}$ operators 
(denoted there by $Q$ and $V$) affiliated with $(\mathcal{A}'', \mathscr{H})$, and which reduces 
the spectral reconstruction of $\mathbb{R} \times G$ to the spectral reconstruction
of the compact group $\mathbb{S}^1 \times G$ worked out in \cite{Connes_spectral}, compare
Appendix \ref{AppendixNonCompMani}. However, the general method of Appendix \ref{AppendixNonCompMani}
requiring introduction of the intermediate auxiliary ``binding'' and ``scaling'' operators
$Q_{{}_{\textrm{scaling}}}$ and $V_{{}_{\textrm{binding}}}$ and the unital algebra $\mathcal{A}^+$
can in this case be considerably simplified 
(which in case of the flat Minkowski space-time was impossible\footnote{In particular the toral compactification of the Minkowski space-time cannot be used and cannot be achieved just by restriction to spatio-temporal periodic plane wave solutions, because the general plane waves on the Minkowski space-time have no common periods in space and time.}), and the construction of the auxiliary
operators can be avoided. Instead, we restrict attention to multispinors $\phi$
which are periodic in time, and which effectively live on $\widetilde{\mathbb{S}^1} \times G$,
with  $\widetilde{\mathbb{S}^1}$ equal to the two-fold covering of the unit circle $\mathbb{S}^1$,
equal to the circle of radius $2$,
and are square integrable on $\widetilde{\mathbb{S}^1} \times G$. 
This assumption is not arbitrary but, as we will see, comes from QFT on 
the Einstein Universe $\mathbb{R} \times G = \mathbb{R} \times SU(2, \mathbb{C})
= \mathbb{R} \times \mathbb{S}^3$. This allows us to replace the non-compact
$\mathbb{R} \times G = \mathbb{R} \times SU(2, \mathbb{C})$ by compact 
$\widetilde{\mathbb{S}^1} \times G = \widetilde{\mathbb{S}^1} \times SU(2, \mathbb{C})$, 
the continuous parameter 
$p\in \mathbb{R}$ by the discrete $\tfrac{n}{2}, n \in \mathbb{Z}$, and the integral 
\[
\int \limits_{\mathbb{R}} \ldots d\widehat{p}
\]
by ordinary sum
\[
\sum \limits_{n \in \mathbb{Z}} \ldots \,\,\,\ ,
\]
and with the characters $\widehat{p}$
\[
\mathbb{R} \ni t \mapsto \widehat{p}(t) = e^{-ipt}
\]
of the additive group $\mathbb{R}$ replaced with the characters $\widehat{n}$
\[
\widetilde{\mathbb{S}^1} \ni t \mapsto \widehat{n}(t) = e^{-i{\textstyle\frac{n}{2}}t},
\,\,\, n \in \mathbb{Z}
\]
of the circle group $\widetilde{\mathbb{S}^1}$ -- twofold cover of $\mathbb{S}^1$.

Thus, below we will need the spectral tuple of $\widetilde{\mathbb{S}^1} \times G 
= \widetilde{\mathbb{S}^1} \times SU(2, \mathbb{C})$:
\begin{multline}\label{1stSpectralTupleForSxG}
\Bigg( \,\,\,\,\,\,\,\, \mathcal{A} = 
\{f (\boldsymbol{Q}^{0}, \boldsymbol{Q}^{1}, \ldots, \boldsymbol{Q}^{m}), 
f \in \mathscr{C}^\infty (\mathbb{R}^m)\} \,\,,
\,\,\,\,\,\,\,\,
\mathscr{H} = \oplus L^2(\widetilde{\mathbb{S}^1} \times G) \,, \\
D_{{}_{\textrm{ell}}} = D_{{}_{\mathfrak{J}}} \, {} = \Gamma^{{}^{0}} \boldsymbol{P}^{{}^{0}} + \Gamma^{{}^{1}} \boldsymbol{P}^{{}^{1}} + 
\ldots + \Gamma^{{}^{3}} \boldsymbol{P}^{{}^{3}} + \Gamma \,\,, 
\,\,\,\,\,\,\,\,\,  \\
D \, {} =  \widehat{\Gamma}^{{}^{0}} \boldsymbol{P}^{{}^{0}} + \widehat{\Gamma}^{{}^{1}} \boldsymbol{P}^{{}^{1}} + 
\ldots + \widehat{\Gamma}^{{}^{3}} \boldsymbol{P}^{{}^{3}} 
+ \widehat{\Gamma}   \,,
\,\,\,\,\,\,\,\,\,
\mathfrak{J}  \,\,\,\, \Bigg),
\end{multline}
\begin{eqnarray}\label{Gammas}
\Gamma^0 = \widehat{\Gamma}^0,
\Gamma^1 = i\widehat{\Gamma}^1, \Gamma^2 = i\widehat{\Gamma}^2,
\Gamma^3 = i\widehat{\Gamma}^3,
\\
\widehat{\Gamma} = - {\textstyle\frac{i}{2}} \widehat{\Gamma}^{{}^{1}} \widehat{\Gamma}^{{}^{2}} \widehat{\Gamma}^{{}^{3}},
\,\,\,\,
\Gamma = i \widehat{\Gamma} = {\textstyle\frac{i}{2}} \Gamma^1 \Gamma^2 \Gamma^3,
\,\,\,\,
\textrm{dim} \, G = \textrm{dim} \, SU(2, \mathbb{C}) = 3,
\\
\big(\widehat{\Gamma}^0 \big)^* =\widehat{\Gamma}^0, 
\big(\widehat{\Gamma}^k \big)^* =-\widehat{\Gamma}^k, \,\,\,
k=1,2,3,
\\
\widehat{\Gamma}^{{}^{\mu}} \widehat{\Gamma}^{{}^{\nu}}
+
\widehat{\Gamma}^{{}^{\nu}} \widehat{\Gamma}^{{}^{\mu}}
=
2 g^{\mu \nu}\\
g^{\mu \nu} = 
\left( \begin{array}{cccc}  1 & 0 & 0 & 0  \\
                              0 & -1 & 0 & 0 \\
                              0 & 0 & -1 & 0 \\
                              0 & 0 & 0 & -1 \end{array}\right) 
\end{eqnarray}
In the Hilbert space $\mathscr{H} = \oplus L^2(\widetilde{\mathbb{S}^1} \times G)$
of the tuple (\ref{1stSpectralTupleForSxG}) there acts unitary and Krein-unitary representation
\begin{equation}\label{UphiOnSxG}
U_{{}_{s\times u \times v}} \phi(t \times x) = 
V(v) \phi\big((t+s) \times (v^{-1}xu) \big), 
\end{equation}
\[
s\times u \times v \in \mathbb{R} \times G \times G,
\,\,\,
\textrm{with} \,\,\,
G = SU(2, \mathbb{C}), 
\]
\[ 
\textrm{and} \,\,\,
\,\,\, \phi \in \oplus L^2(\widetilde{\mathbb{S}^1} \times G) = 
L^2(\widetilde{\mathbb{S}^1} \times G; \mathbb{C}^d)
\] 
of the isometry group $\mathbb{R} \times SU(2, \mathbb{C}) \times SU(2, \mathbb{C})$ of the
Einstein Universe, which in fact coincides with the action of the compactified isometry
group
\[
\widetilde{\mathbb{S}^1} \times SU(2, \mathbb{C}) \times SU(2, \mathbb{C})
\]
and (\ref{UphiOnSxG}) can be identified with a representation of the compactified isometry group
of the Einstein Universe.

The right invariant vector fields $X^1, X^2, X^3$ and the associated left invariant vector fields
$Y^1, Y^2, Y^3$ on $G=SU(2, \mathbb{C})$ we choose to be determined respectively by the one parameter subgroups of rotations around the ``first'', ``second'' and the ``third axis'' in the 
Euclidean space $\mathbb{R}^3$. In terms of the left invariant vector fields (and thus generating one parameter groups of right translations) $A_1, A_2, A_3$ on $SU(2, \mathbb{C})$ used by Gelfand, Minlos and Shapiro in \cite{Geland-Minlos-Shapiro}, Chap. II.7.2, our
$Y^1,Y^2,Y^3$ are equal 
\[
Y^1 = -A_1, \,\,\,  Y^2 = A_2, \,\,\, Y^3 = A_3, 
\] 
\emph{i. e.} with the sign of the angle of rotation around the first axis reversed in comparison to
\cite{Geland-Minlos-Shapiro}, Chap. II.7.2. We have thus the
following commutation rules
\[
[Y^2,Y^1] = Y^3, \,\,\, [Y^3,Y^1] = -Y^2, \,\,\, [Y^3,Y^2] = Y^1, \,\,\, [Y^i, X^k]=0,
\]
and
\begin{align}\label{[X,X]}
[X^2,X^1] = X^3  
& 
\,\,\,\,\,\,\,\,\,\,\,\,\,\,\,\,\,\,\,\,\,\,\,\, \textrm{or}
&
[P^2,P^1] = iP^3 
\\
[X^3,X^1] = -X^2  
&
\,\,\,\,\,\,\,\,\,\,\,\,\,\,\,\,\,\,\,\,\,\,\,\, \textrm{or}
&
[P^3,P^1] = -iP^2 
\\
[X^3,X^2] = X^1  
& 
\,\,\,\,\,\,\,\,\,\,\,\,\,\,\,\,\,\,\,\,\,\,\,\, \textrm{or}
&
[P^3,P^2] = iP^1
\\
[X^0, X^k] = 0, 
&
\,\,\,\,\,\,\,\,\,\,\,\,\,\,\,\,\,\,\,\,\,\,\,\, \textrm{or}
&
[P^0, P^k] = 0, 
\\
P^0 = i X^0  
&
\,\,\,\,\,\,\,\,\,\,\,\,\,\,\,\,\,\,\,\,\,\,\,\, \textrm{and}
&
P^k = iX^k,
\end{align}
on $L^2(\widetilde{\mathbb{S}^1} \times SU(2, \mathbb{C}))$ and with $k=1,2,3$.
In this case the Laplacian $\Delta$, associated to $D_\mathfrak{J}$,
\emph{i. e.} differential operator on $L^2(\widetilde{\mathbb{S}^1} \times SU(2, \mathbb{C}))$ 
such that 
\[
D_\mathfrak{J}^2 = \boldsymbol{1}_{{}_{d}} \, \Delta,
\]
is equal\footnote{Note however that our operators
\[
\Delta =  -\big(X^0\big)^2 - \big(X^1\big)^2 - \big(X^2\big)^2 - \big(X^3\big)^2 
+  {\textstyle\frac{1}{4}} \boldsymbol{1}
\]
\[
- \big(X^1\big)^2 - \big(X^2\big)^2 - \big(X^3\big)^2  + {\textstyle\frac{1}{4}} \boldsymbol{1}
= \Delta_{SU(2,\mathbb{C})}
\]
are respectively equal to $\tfrac{1}{4}\Delta_{\widetilde{\mathbb{S}^1} \times \mathbb{S}^3} + \tfrac{1}{4} \boldsymbol{1}$ and $\tfrac{1}{4}\Delta_{\mathbb{S}^3} + \tfrac{1}{4} \boldsymbol{1}$ under the natural identification 
$SU(2, \mathbb{C}) \cong \mathbb{S}^3$, where $\Delta_{\widetilde{\mathbb{S}^1} \times \mathbb{S}^3}$ and 
$\Delta_{\mathbb{S}^3}$ are the ordinary Laplace-Beltrami 
operators, respectively, on $\widetilde{\mathbb{S}^1} \times \mathbb{S}^3$ and $\mathbb{S}^3 \cong SU(2, \mathbb{C})$. We use this normalization for the vector fields $X^i$
on $SU(2, \mathbb{C})$ and $X^0, X^i$ on $\widetilde{\mathbb{S}^1} \times SU(2,\mathbb{C})$, because the single particle states of Fock spaces of local fields on the Einstein Universe $\mathbb{R} \times SU(2,\mathbb{C})$ live effectively on 
$\widetilde{\mathbb{S}^1} \times SU(2, \mathbb{C})$ and not on $\mathbb{S}^1 \times SU(2, \mathbb{C})$
with the possible eigenvalues of $(P^0)^2 = -(X^0)^2$ belonging to $\tfrac{1}{4}\mathbb{N}$. This allows us application of the common factor $2$ to our differential operators $X^\mu, Y^\mu$ in order to obtain that used in \cite{PaneitzSegalI}-\cite{PaneitzSegalIII}, and similarly application of the factor $4$
to our $\square$ gives that $\square_c$ used in \cite{PaneitzSegalI}-\cite{PaneitzSegalIII}.}
\[
\Delta =  -\big(X^0\big)^2 - \big(X^1\big)^2 - \big(X^2\big)^2 - \big(X^3\big)^2 
+  {\textstyle\frac{1}{4}} \boldsymbol{1}
\]
and thus is equal to the invariant Laplace-Beltrami operator
on the group $\mathbb{S}^1 \times SU(2, \mathbb{C})$ plus the constant
operator $\tfrac{1}{4} \boldsymbol{1}$. Similarly,
\[
D^2 = \boldsymbol{1}_{{}_{d}} \, \square,
\,\,\, 
\square = -\big(X^0\big)^2 + \big(X^1\big)^2 + \big(X^2\big)^2 + \big(X^3\big)^2
- {\textstyle\frac{1}{4}} \boldsymbol{1} 
\]
with the operator $\square$ equal to the ordinary invariant wave operator on the Einstein Universe
minus the constant operator $\tfrac{1}{4} \boldsymbol{1}$. 

The fundamental symmetry operator $\mathfrak{J}$ corresponding to the right invariant pseudo-Riemannian
metric on $\widetilde{\mathbb{S}^1} \times SU(2, \mathbb{C})$ is canonically determined as the operator of multiplication by the Clifford algebra (generated by $\widehat{\Gamma}^{{}^{0}}, \ldots, \widehat{\Gamma}^{{}^{0}}$) element, it depends together with $\widehat{\Gamma}^{{}^{0}}, \ldots, \widehat{\Gamma}^{{}^{0}}$ on the specific representation $V$ in (\ref{UphiOnSxG}) and is such that
the operators $D_{{}_{\textrm{ell}}} = D_\mathfrak{J}$, $D$ and $\mathfrak{J}$ in (\ref{1stSpectralTupleForSxG}), 
defined by (\ref{Gammas}), 
indeed have the property that
\[
{\textstyle\frac{1}{2}} \Big\{(D\mathfrak{J})^2 + (\mathfrak{J}D)^2 \Big\}
= \big(D_\mathfrak{J} \big)^2 = \boldsymbol{1}_d \, \Delta.
\]

Below we give the general class of representations $V$, which can be used in the transformation formula
in (\ref{UphiOnSxG}) of multispinor or of a more general local object, 
the associated with $V$ generators (generalized Dirac
matrices) $\widehat{\Gamma}^{{}^{0}}, \ldots, \widehat{\Gamma}^{{}^{0}}$, and the fundamental
symmetry operators $\mathfrak{J}$ associated to $V$ corresponding to both opposite signatures 
of the space-time pseudo-Riemannian metric on the Einstein universe $\mathbb{R} \times SU(2, \mathbb{C})$ 
or its compactified periodic version $\widetilde{\mathbb{S}^1} \times SU(2, \mathbb{C}) = \widetilde{\mathbb{S}^1} \times \mathbb{S}^3$.

Equivalently one can use the spectral tuple  
\begin{multline}\label{2ndSpectralTupleForSxG}
\Bigg( \,\,\,\,\,\,\,\, \mathcal{A} = 
\Bigg\{ 
\sum_{\substack{\widehat{n} \in \mathbb{Z}\\
                  \widehat{x} \in \widehat{SU(2,\mathbb{C})}\\
i,j=1, \ldots \textrm{dim} \, \widehat{x} }}
 \,\,
\textrm{dim} \, \widehat{x} 
\,\,
\tilde{f}_{{}_{\widehat{n} \cdot \widehat{x}, j, i}} \boldsymbol{\mathbb{V}}_{{}_{\widehat{n} \cdot \widehat{x}, i, j}}, \,
\tilde{f}\in \mathcal{F}\big[\mathcal{S}_{\Delta}(\widetilde{\mathbb{S}^1} \times SU(2,\mathbb{C}))\big] \Bigg\} \,\,, \\
\,\,\,\,\,\,\,\,
\mathscr{H} = \oplus L^2(\widetilde{\mathbb{S}^1} \times SU(2,\mathbb{C})) \,, \\
D_{{}_{\textrm{ell}}} =  D_{{}_{\mathfrak{J}}} \, {} = \Gamma^{{}^{0}} \boldsymbol{P}^{{}^{0}} + \Gamma^{{}^{1}} \boldsymbol{P}^{{}^{1}} + 
\ldots + \Gamma^{{}^{3}} \boldsymbol{P}^{{}^{3}}  + \Gamma \,\,, 
\,\,\,\,\,\,\,\,\, \\
D \, {} = \widehat{\Gamma}^{{}^{0}} \boldsymbol{P}^{{}^{0}} + \widehat{\Gamma}^{{}^{1}} \boldsymbol{P}^{{}^{1}} + 
\ldots + \widehat{\Gamma}^{{}^{3}} \boldsymbol{P}^{{}^{3}} + \widehat{\Gamma} \,,
\,\,\,\,\,\,\,\,\,
\mathfrak{J}  \,\,\,\, \Bigg),
\end{multline}
where 
\begin{multline*}
\mathcal{F}\big[\mathcal{S}_{\Delta}(\widetilde{\mathbb{S}^1} \times SU(2,\mathbb{C}))\big] =
\mathcal{S}_{\mathcal{F} \Delta \mathcal{F}^{-1}}(\mathbb{Z} \times \widehat{SU(2,\mathbb{C})} \times \mathbb{N} \times \mathbb{N}) \\
= \mathcal{S}_{\widetilde{\Delta}}(\mathbb{Z} \times \widehat{SU(2,\mathbb{C})} \times \mathbb{N} \times \mathbb{N})
\end{multline*}
is the standard countably Hilbert nuclear space of rapidly decreasing functions $\tilde{f}$ on $\mathbb{Z} \times \widehat{SU(2,\mathbb{C})}\times \mathbb{N}\times \mathbb{N}$, determined by the standard operator
\[
\mathcal{F}\Delta \mathcal{F}^{-1} = \widetilde{\Delta}
\,\,\,
\textrm{on}
\,\,\,
\mathcal{F} L^2(\widetilde{\mathbb{S}^1} \times SU(2,\mathbb{C})) 
= L^2(\mathbb{Z} \times \widehat{SU(2,\mathbb{C})} \times \mathbb{N} \times \mathbb{N}),
\]
and equal to the Fourier image of the standard nuclear space\footnote{Recall once more that here in the formulas for the standard countably Hilbert spaces 
$\mathcal{S}_{\Delta}$, $\mathcal{S}_{\Delta_{\widetilde{\mathbb{S}^1}}}$, 
$\mathcal{S}_{\Delta_{SU(2, \mathbb{C})}}$
we are using in fact the Laplace operators $\Delta, \Delta_{\widetilde{\mathbb{S}^1}},
\mathcal{S}_{\Delta_{SU(2, \mathbb{C})}}$ 
to which the constant unit operator has been added.}
\begin{multline*}
\mathcal{S}_{\Delta}(\widetilde{\mathbb{S}^1} \times SU(2,\mathbb{C})) = 
\mathcal{S}_{{}_{\Delta_{\widetilde{\mathbb{S}^1}} \otimes \boldsymbol{1}
+ \boldsymbol{1} \otimes \Delta_{SU(2,\mathbb{C})}}}(\widetilde{\mathbb{S}^1} \times SU(2,\mathbb{C})) \\ =
\mathcal{S}_{\Delta_{\widetilde{\mathbb{S}^1}}}(\widetilde{\mathbb{S}^1}) 
\otimes \mathcal{S}_{\Delta_{SU(2,\mathbb{C})}}(SU(2,\mathbb{C}))
= \mathscr{C}^\infty(\widetilde{\mathbb{S}^1}) \otimes \mathscr{C}^\infty(SU(2,\mathbb{C}))
\end{multline*} 
determined by the standard (\emph{i.e.} self-adjoint positive with nuclear power $A^{-r}$,
for $r > \tfrac{1}{2} \textrm{dim} \, [\widetilde{\mathbb{S}^1} \times SU(2,\mathbb{C})]
= 2$) operator (here $X^0 = \tfrac{\partial}{\partial t}$):
\begin{multline*}
A= \Delta_{\widetilde{\mathbb{S}^1}}  \otimes \boldsymbol{1}+ \boldsymbol{1} \otimes \Delta_{SU(2,\mathbb{C})}
= \big(-\big(X^0)^2 \big)\otimes \boldsymbol{1}+ \boldsymbol{1} \otimes \Delta_{SU(2,\mathbb{C})} \\
= -\big(X^0)^2  -\big(X^1)^2 -\big(X^2)^2-\big(X^3)^2 + {\textstyle\frac{1}{4}} \boldsymbol{1}  
= \Delta 
\end{multline*} 
on
\[
L^2(\widetilde{\mathbb{S}^1}) \otimes L^2(SU(2,\mathbb{C})) 
= L^2(\widetilde{\mathbb{S}^1} \times SU(2, \mathbb{C})).
\]

Here the function $\tilde{f}$
\[
n \times \widehat{x}\times j \times i \mapsto
\tilde{f}_{{}_{\widehat{n} \cdot \widehat{x}, j, i}} 
= \mathcal{F}(f)_{{}_{ji}}(\widehat{n} \cdot \widehat{x})
= \widetilde{f}_{{}_{ji}}(\widehat{n} \cdot \widehat{x})
\]
on $\mathbb{Z} \times \widehat{G}\times \mathbb{N}\times \mathbb{N}$ we call rapidly decreasing,
\emph{i. e.} belonging to 
\[
\mathcal{F}\big[\mathcal{S}_{\Delta}(\widetilde{\mathbb{S}^1} \times SU(2,\mathbb{C}))\big] =
\mathcal{S}_{\mathcal{F} \Delta \mathcal{F}^{-1}}(\mathbb{Z} \times \widehat{SU(2,\mathbb{C})} \times \mathbb{N} \times \mathbb{N}),
\]
iff
\begin{multline*}
\Big( \big(\mathcal{F} \Delta \mathcal{F}^{-1}\big)^k  \tilde{f}, \,
\big(\mathcal{F} \Delta \mathcal{F}^{-1}\big)^k  \tilde{f}
\Big)_{{}_{\mathcal{F} L^2(\widetilde{\mathbb{S}^1} \times SU(2, \mathbb{C}))}} \\
= \sum \limits_{n\times \widehat{x} \times i \times j \in \mathbb{Z} \times\widehat{SU(2, \mathbb{C})} \times \mathbb{N} \times \mathbb{N}}
\big(\lambda_{{}_{\widehat{n} \cdot \widehat{x}}} \big)^{2k} \,\,
\big|\tilde{f}_{{}_{\widehat{n} \cdot \widehat{x}, j, i}}\big|^2
< \infty, \,\,\,
k \in \mathbb{N},
\end{multline*}
or equivalently iff
\[
\textrm{sup} \Big\{ \big(\lambda_{{}_{\widehat{n} \cdot \widehat{x}}} \big)^{2k} 
\big|\tilde{f}_{{}_{\widehat{n} \cdot \widehat{x}, j, i}}\big|^2, 
\,\,\, n \times \widehat{x} \times j \times i  \in 
\mathbb{Z} \times \widehat{SU(2, \mathbb{C})}\times \mathbb{N}\times \mathbb{N} \Big\} < \infty
\]
for all $k \in \mathbb{N}$, and denote it by 
$\tilde{f} \in \mathcal{S}_{\mathcal{F}\Delta \mathcal{F}^{-1}}(\mathbb{Z} \times \widehat{G} \times \mathbb{N} \times \mathbb{N})$, compare Subsection \ref{white-setup}. 
Here $\lambda_{{}_{\widehat{n} \cdot \widehat{x}}}$ is the eigenvalue of the Laplace operator 
$\Delta$ on $L^2(\widetilde{\mathbb{S}^1} \times SU(2, \mathbb{C}))$, corresponding to the eigenfunctions 
$t \times x \mapsto \widehat{n}(t) \cdot \widehat{x}(x)_{ij}$
determined by the matrix elements of the representation $\widehat{n} \cdot \widehat{x}$ of the group 
$\widetilde{\mathbb{S}^1} \times SU(2, \mathbb{C})$,
and common for all representations of the unitary equivalence class of the representation 
$\widehat{n} \cdot \widehat{x}$. For the unitary irreducible representation $\widehat{x} = \widehat{l}$
of $SU(2, \mathbb{C})$ of weight $l \in \{0,\tfrac{1}{2}, 1, \tfrac{3}{2}, 2, \ldots\}$
we have
\[
\lambda_{{}_{\widehat{n} \cdot \widehat{l}}} = {\textstyle\frac{n^2}{4}} + l(l+1) +
{\textstyle\frac{1}{4}}, \,\,\, n \in \mathbb{Z}
\]
(so that $\textrm{Sp} \, A > \tfrac{1}{4}$, but after addition of the unit operator to $\Delta$ we obtain a standard  operator $A$
with $\textrm{Sp} \, A > 1$, as we have already mentioned).

Of course whenever working with the group $\widetilde{\mathbb{S}^1} \times G = 
\widetilde{\mathbb{S}^1} \times SU(2, \mathbb{C})$ we are using the Fourier transform
$\mathcal{F}$ on $\widetilde{\mathbb{S}^1} \times SU(2, \mathbb{C})$:
\[
\widetilde{f}_{{}_{ji}} (\widehat{n} \cdot \widehat{x}) 
= \big(\mathcal{F}f\big)_{{}_{ji}}(\widehat{n} \cdot \widehat{x}) 
= \int \limits_{\widetilde{\mathbb{S}^1} \times G} f(t,y) e^{-{\textstyle\frac{i}{2}}nt}
\overline{\widehat{x}_{ij}(y)} \,
\, {\textstyle\frac{1}{\sqrt{4\pi}}} \, dt dy,
\]
which can again be regarded as a matrix-valued function  $\tilde{f} = \mathcal{F}f$ on 
$\widehat{\widetilde{\mathbb{S}^1} \times G}$. 
The inverse Fourier transform on $\widetilde{\mathbb{S}^1} \times SU(2, \mathbb{C})$
is equal
\begin{multline*}
f(t,y) = {\textstyle\frac{1}{\sqrt{4\pi}}} \sum \limits_{\widehat{n} \cdot \widehat{x} \in \widehat{\widetilde{\mathbb{S}^1} \times G}} 
\textrm{dim} \,  (\widehat{n} \cdot \widehat{x}) \,\,\,
\textrm{Tr}\,\big[\, \mathcal{F}f(\widehat{n} \cdot \widehat{x}) \,\,\, 
\widehat{n} \cdot \widehat{x}(t,y)\big]
\\
= 
{\textstyle\frac{1}{\sqrt{4\pi}}}
\sum \limits_{n\in \mathbb{Z}, \widehat{x} \in \widehat{\widetilde{\mathbb{S}^1} \times G}} 
\textrm{dim} \,  (\widehat{n} \cdot \widehat{x}) \,\,\,
\textrm{Tr}\,\big[\, \mathcal{F}f(\widehat{n} \cdot \widehat{x}) \,\,\, 
\widehat{n} \cdot \widehat{x}(t,y) \, \big]
\\
=
{\textstyle\frac{1}{\sqrt{4\pi}}}
\sum \limits_{n\in \mathbb{Z}, \widehat{x} \in \widehat{\widetilde{\mathbb{S}^1} \times G}} 
\textrm{dim} \,  \widehat{x} \,\,\,
\textrm{Tr}\,\big[\, \mathcal{F}f(\widehat{n} \cdot \widehat{x}) \,\,\, 
\widehat{n} \cdot \widehat{x}(t,y) \, \big]
\\
=
{\textstyle\frac{1}{\sqrt{4\pi}}}
\sum \limits_{n\in \mathbb{Z}, \widehat{x} \in \widehat{\widetilde{\mathbb{S}^1} \times G}} 
\textrm{dim} \,  \widehat{x} \,\,\,
\widehat{n}(t) \,\,
\textrm{Tr}\,\big[\, \mathcal{F}f(\widehat{n} \cdot \widehat{x}) \,\, \widehat{x}(y) \, \big]
\\
=
{\textstyle\frac{1}{\sqrt{4\pi}}}
\sum \limits_{n\in \mathbb{Z}, \widehat{x} \in \widehat{\widetilde{\mathbb{S}^1} \times G}} 
\textrm{dim} \,  \widehat{x} \,\,\,
\widehat{n}(t) \,\,
\textrm{Tr}\,\big[\, \widetilde{f}(\widehat{n} \cdot \widehat{x}) \,\, \widehat{x}(y) \, \big]
\end{multline*}
This Fourier transform $\mathcal{F}$ is used in particular in the formula (\ref{2ndSpectralTupleForSxG})
and below, and will be used below in this Subsection whenever we are working with the group 
$\widetilde{\mathbb{S}^1} \times SU(2, \mathbb{C})$.

The spectral tuple (\ref{1stSpectralTupleForSxG}) or (\ref{2ndSpectralTupleForSxG})
is defined on the Hilbert space 
\[
\mathscr{H} =
\bigoplus \limits_{1}^{d} L^2(\widetilde{\mathbb{S}^1} \times G), \,\,\,
d= \textrm{dim} \, V 
\,\,\,
G=SU(2, \mathbb{C}),
\]
of multispinors $\phi$ with local transformation (\ref{UphiOnSxG}), which under the Fourier transform
$V_\mathcal{F}$ on $\bigoplus \limits_{1}^{d} L^2(\widetilde{\mathbb{S}^1} \times G)$, decomposes into the direct sum
\begin{equation}\label{decUphiOnSxG}
\widehat{U} = V_\mathcal{F} U {V_\mathcal{F}}^{-1}
= \bigoplus \limits_{n \in \mathbb{Z}, \widehat{x} \in \widehat{G}}  \,\,
\big[ \, \widehat{n} \cdot \widehat{x} \, \big] \,  \times  \,
\big[ \,\, \overline{\widehat{x}} \otimes V \,\, \big]
\end{equation}
of \emph{outer Kronecker products}
\[
\big[ \, \widehat{n} \widehat{x}  \, \big] \, \times \,
\big[ \,\, \overline{\widehat{x}} \otimes V \,\, \big]
\]
of representations
\[
\widehat{n} \cdot \widehat{x} 
\,\,\,\,
\textrm{and}
\,\,\,\,
\overline{\widehat{x}} \otimes V
\]
respectively of $\widetilde{\mathbb{S}^1}  \times G$ and $G$.

Note that the operator $\mathbb{V}_{{}_{\widehat{n} \cdot \widehat{z}, m,k}}$ of the ``canonical pair'' 
${}^{{}^{R}}U,\mathbb{V}$ on $L^2(\widetilde{\mathbb{S}^1}  \times G)$, is transformed, by the Fourier transform on $\widetilde{\mathbb{S}^1}  \times G$, into the operator $\widehat{\mathbb{V}_{{}_{\widehat{n} \cdot \widehat{z}, m,k}}}$:
\[
\big(\widehat{\mathbb{V}_{{}_{\widehat{n'} \cdot \widehat{z}, m,k}}} F\big)_{{}_{ji}} (\widehat{n} \cdot \widehat{x} )
= 
\sum \limits_{\widehat{y} \in \textrm{dec} (\widehat{z} \otimes \overline{\widehat{x}})} \,\,\,\,\,
\sum \limits_{1 \leq s,q \leq \textrm{dim} \, \widehat{y}} 
\,\,\,\,\,\,
\underset{\widehat{z}, m,k}{M}_{{}_{\widehat{x} \,\, ji}}^{{}^{\widehat{y} \,\, sq}} 
\,\,\,\, F_{{}_{sq}}(\widehat{n'-n} \cdot \widehat{y}),
\]
where the summation runs over all those $\widehat{y} \in \widehat{G}$ which (up to unitary equivalence)
enter into the direct sum decomposition of the representation $\widehat{z} \otimes \overline{\widehat{x}}$,
and where 
\[
\underset{\widehat{z}, m,k}{M}_{{}_{\widehat{x} \,\, ji}}^{{}^{\widehat{y} \,\, sq}} 
\]
is the matrix defined above determining the pointwise product
\[
\widehat{z}(x)_{mk} \,\, \overline{\widehat{x}(x)_{ij}}=
\sum \limits_{\widehat{y} \in \textrm{dec} (\widehat{z} \otimes \overline{\widehat{x}})} \,\,\,\,\,
\sum \limits_{1 \leq s,q \leq \textrm{dim} \, \widehat{y}} 
\,\,\,\,\,\,\,
\underset{\widehat{z}, m,k}{M}_{{}_{\widehat{x} \,\, ji}}^{{}^{\widehat{y} \,\, sq}} 
\,\,\,\,
\widehat{y}(x)_{sq},
\]
and determined by the Clebsch-Gordan coefficients of the group $G = SU(2, \mathbb{C})$.  
Recall that $V_\mathcal{F}$ acting on $\bigoplus \limits_{1}^{d} L^2(\widetilde{\mathbb{S}^1}  \times G)$
is equal to the direct $d$-fold copy of the Fourier transform acting in the space
$L^2(\widetilde{\mathbb{S}^1}  \times G)$ of the ``canonical'' pair 
${}^{{}^{R}}U,\mathbb{V}$ on $L^2(\widetilde{\mathbb{S}^1}  \times G)$. 

Before continuing our discussion let us recall that all the irreducible representations
$\widehat{x} \in \widehat{G} = \widehat{SU(2, \mathbb{C})}$ exhausting all (up to unitary equivalence)
irreducible representations of $SU(2, \mathbb{C})$ are uniquely determined by the (non negative integer or half an odd non negative integer) 
weight $l \in \{0, \frac{1}{2}, 1, \frac{3}{2}, 2, \ldots \}$ (called total angular momentum or spin
depending on the context). It is therefore useful to denote the irreducible representations of weight $l$, 
exhausting all (up to unitary equivalence) irreducible unitary representations of 
$G = SU(2, \mathbb{C})$ by the corresponding $\widehat{l}$,
instead of $\widehat{x}$. For the explicit construction of $\widehat{l}$ compare \cite{Geland-Minlos-Shapiro}
or \cite{NeumarkLorentzBook}\footnote{The matrix elements of the unitary representation $\widehat{l}$
of weight $l$ are denoted by $T_{ij}^{l}$ in \cite{Geland-Minlos-Shapiro} and by $c_{ij}^{l}$ in
\cite{NeumarkLorentzBook}. Note also that $\textrm{dim} \, \widehat{l} = 2l+1$.}. 
In the sequel the concrete irreducible unitary representations $\widehat{l}$ of $SU(2, \mathbb{C})$,
constructed in \cite{Geland-Minlos-Shapiro}
or \cite{NeumarkLorentzBook} will also be denoted by $L^l$. 

Thus using the weight-notation for the canonical class of the mentioned complete class
of unitary irreducible representations of $G = SU(2, \mathbb{C})$, we see that the formula
(\ref{decUphiOnSxG}) for the decomposition of the representation
(\ref{UphiOnSxG}) acting on the Hilbert space square summable time-periodic multispinors
\[
\phi \in \mathscr{H} = \bigoplus \limits_{1}^{\textrm{dim} \, V} L^2(\widetilde{\mathbb{S}^1} \times SU(2, \mathbb{C}))
\]
can be written in the form
\begin{equation}\label{l-decUphiOnSxG}
\widehat{U} = V_\mathcal{F} U {V_\mathcal{F}}^{-1}
= \bigoplus \limits_{n \in \mathbb{Z}, l  \in \{0, \frac{1}{2}, 1, \ldots \}}  
\big[ \, \widehat{n} \cdot \widehat{l}  \, \big] \,  \times  \,
\big[ \,\, \overline{\, \widehat{l} \,} \otimes V \,\, \big]
\end{equation}
where the summation runs over all $n \in \mathbb{Z}$ and all $l \in \{0, \frac{1}{2}, 1, \frac{3}{2}, 2, \ldots \}$ non negative integer 
or half an odd non negative integer $l$. Note also that the representations $\widehat{l}$ and  $\overline{\, \widehat{l} \, }$ 
are unitary equivalent, and we have:
\[
\overline{\, \widehat{l} \,}(v)_{{}_{m \,\,\,\, k}}= \widehat{l}(v)_{{}_{-m \,\,\,\, -k}}, 
\,\,\,\,\,\,\,\,
-l \leq m,k \leq l, \,\,\, v \in SU(2, \mathbb{C}).
\]

The possible representations $V$ of $G= SU(2, \mathbb{C})$ in the transformation formula
(\ref{UphiOnRxG}) and the corresponding
Clifford algebra generators $\Gamma^{{}^{0}}, \Gamma^{{}^{1}}, \ldots, \Gamma^{{}^{3}}$ 
and $\widehat{\Gamma}^{{}^{0}}, \widehat{\Gamma}^{{}^{1}}, \ldots, \widehat{\Gamma}^{{}^{3}}$
and the Krein fundamental symmetry operator $\mathfrak{J}$ 
in (\ref{1stSpectralTupleForSxG}) or in (\ref{2ndSpectralTupleForSxG}) 
need not be separately constructed for the Einstein Universe, but instead we can borrow  them
from our computations for the flat Minkowski space-time. Indeed, the possible
representations $V$ of $G= SU(2, \mathbb{C})$ in (\ref{UphiOnSxG}) can be taken to be 
equal to the representations
$V$ of $SL(2, \mathbb{C})$ which we have constructed in the following Subsections
\ref{e1} -- \ref{DirectIntRepVF}, restricted to the subgroup $SU(2, \mathbb{C})$. 
This in particular means (recall that $v^{* -1} = v$, $v \in SU(2, \mathbb{C})$) 
that for any fixed natural $n$ we can put
\begin{equation}\label{multispinorV}
V(v) = v \otimes \cdots \otimes v \oplus  \ldots \oplus v \otimes \cdots \otimes v
 =  \bigoplus \limits_{1}^{2^n} v^{\otimes n} = 2^n \, v^{\otimes n}
 \cong_U (v \oplus v)^{\otimes n}
\end{equation}
\emph{i. e.} we have here $2^n$ direct summands each equal to the $n$-fold tensor product of the irreducible identity representation
\[
\widehat{{\textstyle\frac{1}{2}}} = L^{1/2}: v \mapsto v, \,\,\, v \in SU(2, \mathbb{C})
\]
\emph{i. e.} we can put
\[
V = 2^n \,\,\, \widehat{{\textstyle\frac{1}{2}}}^{\otimes n} = 2^n \, \big(L^{1/2}\big)^{\otimes n},
\]
for the representation $V$ in (\ref{UphiOnSxG}). 

In particular for $n=1$ we have
\begin{equation}\label{spinorV}
V = 2 \,\,\, \widehat{{\textstyle\frac{1}{2}}} = \widehat{{\textstyle\frac{1}{2}}} \oplus 
\widehat{{\textstyle\frac{1}{2}}}.
\end{equation}
For $n=2$ we have 
\begin{multline*}
V = 4 \,\,\, \widehat{{\textstyle\frac{1}{2}}}^{\otimes 2}
= 4 \,\,\, \widehat{{\textstyle\frac{1}{2}}} \otimes \widehat{{\textstyle\frac{1}{2}}}  
= 4 \,\,\, S_{{}_{{\textstyle\frac{1}{2}},{\textstyle\frac{1}{2}}}}\big( \widehat{0} \oplus \widehat{1} \big)S_{{}_{{\textstyle\frac{1}{2}},{\textstyle\frac{1}{2}}}}^{-1}
=\Bigg(4S_{{}_{{\textstyle\frac{1}{2}},{\textstyle\frac{1}{2}}}}\Bigg) \, 
\big(4 \,\,\, (\widehat{0} \oplus \widehat{1})\big) \,
\Bigg(4S_{{}_{{\textstyle\frac{1}{2}},{\textstyle\frac{1}{2}}}}\Bigg)^{-1}
 \\
=
S_{{}_{{\textstyle\frac{1}{2}},{\textstyle\frac{1}{2}}}}\big( \widehat{0} \oplus \widehat{1} \big)
S_{{}_{{\textstyle\frac{1}{2}},{\textstyle\frac{1}{2}}}}^{-1} \,\,\, \oplus \,\,\,
S_{{}_{{\textstyle\frac{1}{2}},{\textstyle\frac{1}{2}}}}
\big( \widehat{0} \oplus \widehat{1} \big)S_{{}_{{\textstyle\frac{1}{2}},{\textstyle\frac{1}{2}}}}^{-1}
\,\,\, \oplus \,\,\, S_{{}_{{\textstyle\frac{1}{2}},{\textstyle\frac{1}{2}}}}
\big( \widehat{0} \oplus \widehat{1} \big)S_{{}_{{\textstyle\frac{1}{2}},{\textstyle\frac{1}{2}}}}^{-1} 
\,\,\, \oplus \,\,\,
S_{{}_{{\textstyle\frac{1}{2}},{\textstyle\frac{1}{2}}}}
\big( \widehat{0} \oplus \widehat{1} \big)S_{{}_{{\textstyle\frac{1}{2}},{\textstyle\frac{1}{2}}}}^{-1},
\end{multline*}
where $S_{{}_{{\textstyle\frac{1}{2}},{\textstyle\frac{1}{2}}}}$ is the unitary equivalence between
\[
\widehat{{\textstyle\frac{1}{2}}} \otimes \widehat{{\textstyle\frac{1}{2}}} 
\,\,\,
\textrm{and}
\,\,\,
\widehat{0} \oplus \widehat{1}
\]
which follows from the general theorem asserting that the representations
\[
\widehat{l_1} \otimes \widehat{l_2} 
\,\,\,
\textrm{and}
\,\,\,
\widehat{|l_1-l_2|} \oplus \widehat{(|l_1-l_2| + 1)} \oplus \ldots \oplus \widehat{l_1+l_2}
\]
are unitarily equivalent with the unitary operator giving the equivalence 
denoted by $S_{{}_{l_1,l_2}}$:
\begin{equation}\label{l1xl2}
\widehat{l_1} \otimes \widehat{l_2}  = S_{{}_{l_1,l_2}} \,
\big( \widehat{|l_1-l_2|} \, \oplus \,  \widehat{(|l_1-l_2| + 1)} \, \oplus \, \ldots \, \oplus \, \widehat{l_1+l_2}\big) \, S_{{}_{l_1,l_2}}^{-1},
\end{equation}
with the unitary operator $S_{{}_{l_1,l_2}}$ determining (and determined) by the Clebsh-Gordan coefficients of the group $SU(2, \mathbb{C})$, compare \cite{Geland-Minlos-Shapiro}, Chap. I Section 4, \S 3
and Chap. II Section 10. Thus, for the representation $V$ in the case $n=2$ we may shortly 
write
\[
V = 4 \,\,\, \widehat{{\textstyle\frac{1}{2}}}^{\otimes 2}
\cong_{S}
4 \,\,\, (\widehat{0} \oplus \widehat{1})
\]
with $\cong_{S}$ denoting unitarily equivalence given by
\[
4S_{{}_{{\textstyle\frac{1}{2}},{\textstyle\frac{1}{2}}}}.
\]
Similarly, for $V$ in the case $n=3$, $n=4$, \emph{e.t.c.}, respectively, we obtain by repeated 
application of theorem (\ref{l1xl2})
\[
V = 8 \,\,\, \widehat{{\textstyle\frac{1}{2}}}^{\otimes 3}
\cong_{S}
8 \,\,\, (\widehat{{\textstyle\frac{1}{2}}}  \oplus \widehat{{\textstyle\frac{1}{2}}} \oplus \widehat{{\textstyle\frac{3}{2}}})
=
8 \,\,\, (2\widehat{{\textstyle\frac{1}{2}}} \oplus \widehat{{\textstyle\frac{3}{2}}}),
\]
\[
V = 16 \,\,\, \widehat{{\textstyle\frac{1}{2}}}^{\otimes 4}
\cong_{S}
16 \,\,\, (\widehat{0} \oplus \widehat{0} \oplus \widehat{1} \oplus \widehat{1} \oplus \widehat{1}
\oplus \widehat{2})
=
16 \,\,\, (2 \,\widehat{0}  \oplus 3 \, \widehat{1} \oplus \widehat{2}),
\]
\[
\textrm{\emph{e. t. c.}}
\]
with the respective unitary equivalence operator $S$ determined by the respective 
Clebsch-Gordan coefficients. The general algorithm for computation of $V$ in terms
of direct sums of the irreducible representations $\widehat{l} \in \widehat{SU(2, \mathbb{C})}$
can be given, valid for general value of $n \in \mathbb{N}$.

The Clifford algebra generators
$\Gamma^{{}^{0}}, \Gamma^{{}^{1}}, \Gamma^{{}^{2}}, \Gamma^{{}^{3}}$ 
and $\widehat{\Gamma}^{{}^{0}}, \widehat{\Gamma}^{{}^{1}}, \ldots, \widehat{\Gamma}^{{}^{3}}$
and the Krein fundamental symmetry operator $\mathfrak{J}$ can be put equal to the ones
determined in Subsection \ref{DirectIntRepVF} corresponding to the representation
$V$ of $SL(2, \mathbb{C})$, which we here restrict to the subgroup $SU(2, \mathbb{C})$. 
In particular for $n=1$ the generators $\widehat{\Gamma}^{{}^{0}}, \widehat{\Gamma}^{{}^{1}}, \widehat{\Gamma}^{{}^{2}}, \widehat{\Gamma}^{{}^{3}}$ can be put equal to the Dirac gamma matrices $\gamma^0, \gamma^{1},
\gamma^2, \gamma^3$, say in the chiral representation (\ref{chiralgamma}).
The generators $\Gamma^{{}^{0}}, \Gamma^{{}^{1}}, \Gamma^{{}^{2}}, \Gamma^{{}^{3}}$
can be put equal $\gamma^0, i\gamma^{1},i\gamma^2, i\gamma^3$, where $\gamma^0, \gamma^{1},
\gamma^2, \gamma^3$ are  the Dirac gamma matrices, of course in the same chiral representation (\ref{chiralgamma}). The Krein fundamental symmetry operator $\mathfrak{J}$ can be put 
equal to the operator of multiplication by the matrix $\gamma^0$, if the signature of 
the space-time pseudo-metric is chosen to be equal
$(1,-1,-1,-1)$, or by the matrix $\gamma^1 \gamma^2 \gamma^3$, if the signature of the space-time
metric is chosen to be equal $(-1,1,1,1)$. In the second case the matrices 
$\widehat{\Gamma}^{{}^{0}}, \widehat{\Gamma}^{{}^{1}}, \widehat{\Gamma}^{{}^{2}}, \widehat{\Gamma}^{{}^{3}}$ are to be replaced with $i\widehat{\Gamma}^{{}^{0}}, i\widehat{\Gamma}^{{}^{1}}, i\widehat{\Gamma}^{{}^{2}}, i\widehat{\Gamma}^{{}^{3}}$. 

For $n=2$ we obtain $2^{2 \cdot 2} \times 2^{2\cdot 2} = 16 \times 16$ matrices 
$\mathfrak{J}$, $\widehat{\Gamma}^{{}^{0}}, \widehat{\Gamma}^{{}^{1}}, \ldots, \widehat{\Gamma}^{{}^{3}}$
and $\Gamma^{{}^{0}}, \Gamma^{{}^{1}},  \widehat{\Gamma}^{{}^{2}} , \Gamma^{{}^{3}}$:
\[
 \widehat{\Gamma}^{{}^{0}} = \left( \begin{array}{cccc} \bold{0}_4 & \bold{0}_4 & \bold{0}_4 & \bold{1}_4  \\
                              \bold{0}_4  & \bold{0}_4 & \bold{1}_4 & \bold{0}_4 \\
                             \bold{0}_4 & \bold{1}_4 & \bold{0}_4 & \bold{0}_4 \\
               \bold{1}_4 & \bold{0}_4 & \bold{0}_4 &\bold{0}_4  \end{array}\right) \,\,\,
 \widehat{\Gamma}^{{}^{k}} = \left( \begin{array}{cccc} \bold{0}_4 & \bold{0}_4 & \bold{0}_4 & -\bold{1}_2 \otimes \sigma_k  \\
                              \bold{0}_4  & \bold{0}_4 & -\bold{1}_2 \otimes \sigma_k & \bold{0}_4 \\
                             \bold{0}_4 & \bold{1}_2 \otimes \sigma_k & \bold{0}_4 & \bold{0}_4 \\
               \bold{1}_2 \otimes \sigma_k & \bold{0}_4 & \bold{0}_4 &\bold{0}_4  \end{array}\right),
\]
\[
\Big( \mathfrak{J} \, \phi \Big)(t \times x) 
= \left( \begin{array}{cccc}  \bold{0}_4 & \bold{0}_4 & \bold{1}_4 & \bold{0}_4  \\
                              \bold{0}_4 & \bold{0}_4 & \bold{0}_4 & \bold{1}_4 \\
                              \bold{1}_4 & \bold{0}_4 & \bold{0}_4 & \bold{0}_4 \\
                              \bold{0}_4 & \bold{1}_4 & \bold{0}_4 & \bold{0}_4 \end{array}\right) 
\big(\phi(t \times x) \big),
\]
\[
\widehat{\Gamma} =  
- {\textstyle\frac{i}{2}} \,  \widehat{\Gamma}^{{}^{1}} \widehat{\Gamma}^{{}^{2}} \widehat{\Gamma}^{{}^{3}}
= {\textstyle\frac{1}{2}} \, \left( \begin{array}{cccc}  \bold{0}_4 & \bold{0}_4 & \bold{0}_4 & \bold{1}_4  \\
                              \bold{0}_4 & \bold{0}_4 & \bold{1}_4 & \bold{0}_4 \\
                              \bold{0}_4 & -\bold{1}_4 & \bold{0}_4 & \bold{0}_4 \\
                              -\bold{1}_4 & \bold{0}_4 & \bold{0}_4 & \bold{0}_4 \end{array}\right),
\]
\[
\Gamma_0 =\widehat{\Gamma}_0, \,\,\, \Gamma_1 = i\widehat{\Gamma}_1, \ldots
\Gamma_3 = i \widehat{\Gamma}_3, \,\,\, \Gamma = i \widehat{\Gamma},
\]
with $\sigma_k$ equal to the Pauli matrices 
\[
\sigma_1  =
\left( \begin{array}{cc}  0 &  1  \\
                                           
                                                   1            & 0  \end{array}\right), \,\,
\sigma_2  =
\left( \begin{array}{cc}  0 &  -i  \\
                                           
                                                   i            & 0  \end{array}\right), \,\,
                                                   \sigma_3  =
\left( \begin{array}{cc}  1 &  0  \\
                                           
                                                   0            & -1  \end{array}\right).
\]
(compare Subsection \ref{e4} and formula (\ref{JforU^m0004L^1/2otimes2_00})). This is for the signature of space-time pseudo-metric equal $(1,-1,-1,-1)$. For the opposite signature of the space-time 
pseudo-metric we have to replace the Dirac gamma matrices
$\widehat{\Gamma}^\mu$ with $i \widehat{\Gamma}^\mu$ and accordingly we have to choose another
fundamental symmetry $\mathfrak{J}$, compare Subsect. \ref{e1} -- \ref{DirectIntRepVF}.

For higher value of $n$ we obtain $2^{2n} \times 2^{2n}$ matrices 
$\mathfrak{J}$, $\widehat{\Gamma}^{{}^{0}}, \widehat{\Gamma}^{{}^{1}}, \ldots, \widehat{\Gamma}^{{}^{3}}$
and $\Gamma^{{}^{0}}, \Gamma^{{}^{1}}, \ldots, \Gamma^{{}^{3}}$. In particular 
the Krein fundamental symmetry operator is equal to the operator of multiplication by the matrix $U^{(n)}(\gamma^0 \otimes \gamma^0 \otimes \ldots \otimes \gamma^0){U^{(n)}}^{-1}$ or by the matrix
(all tensor products here are $n$-fold and embrace all possibilities with the factors equal $\gamma^0$ or 
$\gamma^1 \gamma^2 \gamma^3$)
$U^{(n)}(\gamma^0 \otimes \ldots \otimes \gamma^0 \otimes \gamma^1 \gamma^2 \gamma^3 ){U^{(n)}}^{-1}$, 
$\ldots$ or by the matrix 
$U^{(n)}(\gamma^1 \gamma^2 \gamma^3 \otimes \gamma^1 \gamma^2 \gamma^3 \otimes \ldots \otimes \gamma^1 \gamma^2 \gamma^3 ){U^{(n)}}^{-1}$. The unitary involutive operator $U^{(n)}$ is computed
in Subsection \ref{DirectIntRepVF}. We also refer to Subsection \ref{DirectIntRepVF} for the 
general formulas for the Clifford algebra generators in this case.  However, one has to be careful because
among the fundamental symmetry operators $\mathfrak{J}$ determined in Subsection \ref{DirectIntRepVF}
some correspond to the space-time pseudo-metric signature $(1,-1,-1,-1)$ and fulfill
\[
\widehat{\Gamma}^0 \mathfrak{J} = \mathfrak{J} \widehat{\Gamma}^0 , \,\,\,
\widehat{\Gamma}^k \mathfrak{J} = -\mathfrak{J} \widehat{\Gamma}^k, \,\,\, k=1,2,3,
\]
and some $\mathfrak{J}$ correspond to the space-time pseudo-metric signature $(-1,1,1,1)$ and fulfill
\[
\widehat{\Gamma}^0 \mathfrak{J} = -\mathfrak{J} \widehat{\Gamma}^0 , \,\,\,
\widehat{\Gamma}^k \mathfrak{J} = \mathfrak{J} \widehat{\Gamma}^k, \,\,\, k=1,2,3.
\]

It is time to tell that the class of representations $V$ given by the formula
(\ref{multispinorV}), which can be used in (\ref{UphiOnSxG}) 
(equivalently in (\ref{UphiOnRxG})), corresponding to them Hilbert spaces
$\mathscr{H} = \oplus L^2(\widetilde{\mathbb{S}^1} \times SU(2, \mathbb{C})) = 
L^2(\widetilde{\mathbb{S}^1} \times SU(2, \mathbb{C}); \mathbb{C}^d)$
of multispinors, the Clifford algebra generators, $\Gamma^0, \ldots$, $\widehat{\Gamma}^0 
\ldots $, the Krein fundamental symmetry $\mathfrak{J}$, and thus the spectral tuple (\ref{1stSpectralTupleForSxG}) or (\ref{2ndSpectralTupleForSxG}), describing 
$\widetilde{\mathbb{S}^1}\times SU(2, \mathbb{C})$ with the right invariant pseudo-Riemannian $g$
and the corresponding Riemannian metric $g_{\mathfrak{J}}$, have been prepared from the 
spectral triple on the bispinor Hilbert space, based on the representation
$V$ equal (\ref{spinorV}). Indeed, in order to prepare $V$ given by the formula
(\ref{multispinorV}), together with the the Clifford generator matrices 
$\Gamma^0, \ldots$, $\widehat{\Gamma}^0\ldots $, and $\mathfrak{J}$ corresponding to 
$V(v) \cong_U  (v\oplus v)^{\otimes n}$, we have applied the operation of tensoring to the
Clifford generators and the fundamental symmetry operator, which corresponds to the bispinor
representation determined by $V(v) = v\oplus v$. But this by no means exhaust all
natural spectral triple constructions of the manifold $\widetilde{\mathbb{S}^1}\times SU(2, \mathbb{C})$ with the right invariant pseudo-Riemannian $g$ and the corresponding Riemannian metric 
$g_{\mathfrak{J}}$, with the uniform multiplicity $d$ of the action of the smooth  
algebra $\mathcal{A} = \mathscr{C}^\infty(\widetilde{\mathbb{S}^1}\times SU(2, \mathbb{C}))$
equal to the power of the number $2$: $d = 2^{2n}$, $n=1, 2, \ldots$. We have also another infinite 
family of spectral tuples decribing $\widetilde{\mathbb{S}^1}\times SU(2, \mathbb{C})$ with 
the right invariant Riemannian $g$ and the corresponding pseudo-Riemannian metric 
$g_{\mathfrak{J}}$, with the uniform multiplicity $d$ of the multiplicity action of the  
algebra $\mathcal{A}'' = \mathscr{C}^\infty(\widetilde{\mathbb{S}^1}\times SU(2, \mathbb{C}))''$
equal to the power of the number $2$: $d = 2^{4n}$, $n=1, 2, \ldots$. 
In order to construct 
them explicitly we proceed exactly as in  construction of the Clifford generators and the operator
$\mathfrak{J}$ also by application of the operation of tensoring,  but we start not with the spectral tuple acting on the square integrable bispinors.
Instead of the initial spectral tuple acting on the square integrable bispinors, with the transformation rule (\ref{UphiOnSxG}) defined by $V(v) = v\oplus v$, and with the Dirac gamma matrices $\widehat{\Gamma}^\mu = \gamma^\mu$ in the chiral representation (\ref{chiralgamma}), 
$\Gamma^0 = \gamma^0, \Gamma^k = i\gamma^k$, $k=1,2,3$, and $\mathfrak{J} = \gamma^0$, 
corresponding to $V(v) = v \oplus v$, we use another natural spectral tuple, namely that acting
on square integrable forms with complex coefficients on $\widetilde{\mathbb{S}^1} \times SU(2, \mathbb{C})$, compare \cite{Connes_spectral}. 

Indeed, we can construct spectral triple on the complexified space of forms
$\omega \in \bigwedge^{*}_{\mathbb{C}} T^*(\widetilde{\mathbb{S}^1}\times SU(2, \mathbb{C}))$ 
with complex coefficients, which are square integrable,
\emph{i.e.} belong to
\[ 
\mathscr{H} = L^2\Big(\widetilde{\mathbb{S}^1}\times SU(2, \mathbb{C}); {\bigwedge}^{*}_{\mathbb{C}}\Big)
\]
with respect to the natural inner product
\[
(\omega, \eta) = \sum \limits_{k=0}^{4} \,\,\,\,\, \int \overline{\omega_k} \,\,\, \wedge 
\,\,\, \star_{{}_{\mathfrak{J}}} \,\, \eta_k
\]
where $\star_{{}_{\mathfrak{J}}}$ is the Hodge star operation determined by the ordinary right invariant Riemannian 
metric $g_{{}_{\mathfrak{J}}}$, and where
\[
\omega = \omega_0 \oplus \omega_1 \oplus \omega_2 \oplus \omega_3 \oplus \omega_4,
\,\,\,
\eta = \eta_0 \oplus \eta_1 \oplus \eta_2 \oplus \eta_3 \oplus \eta_4,
\,\,\,
\omega_k, \eta_k \in {\bigwedge}^{k}_{\mathbb{C}} T^*(\widetilde{\mathbb{S}^1}\times SU(2, \mathbb{C}))
\]
are decompositions of the forms $\omega, \eta \in \bigwedge^{*}_{\mathbb{C}} T^*(\widetilde{\mathbb{S}^1}\times SU(2, \mathbb{C}))$ 
into the direct sum of $k$-form components. 
The ordinary elliptic Dirac operator $D_\mathfrak{J}$ is equal to the signature operator 
\[
D_{{}_{\mathfrak{J}}} = d \,\,\,\, - \,\,\,\, \star_{{}_{\mathfrak{J}}} \,  d \,\,  \star_{{}_{\mathfrak{J}}}
= d + d^*
\]
determined by the Riemannian metric $g_{{}_{\mathfrak{J}}}$. 
Note that our manifold $\widetilde{\mathbb{S}^1} \times SU(2, \mathbb{C})$ has even dimension $4$,
so that the adjoint of the operator $d$ is equal 
$- \star_{{}_{\mathfrak{J}}} \,  d \,\,  \star_{{}_{\mathfrak{J}}}$, compare e. g. \cite{Rosenberg}. 
The Dirac operator corresponding to
the pseudo-Riemannian right invariant (space-time) metric $g$
is equal to the signature operator 
\[
D = d \,\, - \,\, \star \, d \, \star = d + d^\dagger = d + \mathfrak{J} d^* \mathfrak{J}
\]
where this time $\star$ is the Hodge star operation determined by the pseudo-Riemannian space-time metric
$g$ and $d^\dagger = \mathfrak{J} d^* \mathfrak{J}$ stands for the Krein-adjoint of the operator $d$. 
The fundamental Krein-symmetry operator $\mathfrak{J}$ for the spectral tuple acting
on the square integrable complex valued forms is equal to the operator of pointwise multiplication
by a direct sum matrix
\[
\mathfrak{J}_{(1)} = 
1 \oplus \mathfrak{J}_{\bar{p}} \oplus \mathfrak{J}_{\bar{p}}^{\widehat{\otimes} \, 2}
\oplus \mathfrak{J}_{\bar{p}}^{\widehat{\otimes} \, 3} \oplus \mathfrak{J}_{\bar{p}}^{\widehat{\otimes} \, 4}
\]
acting on
\begin{multline*}
\omega = \omega_0 \oplus \omega_1 \oplus \omega_2 \oplus \omega_3 \oplus \omega_4 \in
\\
\in L^2\Big(\widetilde{\mathbb{S}^1}\times SU(2, \mathbb{C}); {\bigwedge}^{0}_{\mathbb{C}}\Big)
\oplus L^2\Big(\widetilde{\mathbb{S}^1}\times SU(2, \mathbb{C}); {\bigwedge}^{1}_{\mathbb{C}}\Big)
\oplus
L^2\Big(\widetilde{\mathbb{S}^1}\times SU(2, \mathbb{C}); {\bigwedge}^{2}_{\mathbb{C}}\Big) \oplus \\
\oplus L^2(\widetilde{\mathbb{S}^1}\times SU(2, \mathbb{C}); {\bigwedge}^{3}_{\mathbb{C}})
\oplus
L^2\Big(\widetilde{\mathbb{S}^1}\times SU(2, \mathbb{C}); {\bigwedge}^{4}_{\mathbb{C}}\Big) 
= L^2\Big(\widetilde{\mathbb{S}^1}\times SU(2, \mathbb{C}); {\bigwedge}^{*}_{\mathbb{C}}\Big)
\end{multline*}
where the matrix $\mathfrak{J}_{\bar{p}}$ is equal to the matrix (\ref{J-barp}) 
of Subsection \ref{DefLopRep}, defining the fundamental symmetry operator associated to 
the {\L}opusza\'nski representation. The algebra 
$\mathcal{A}=\mathscr{C}^\infty(\widetilde{\mathbb{S}^1}\times SU(2, \mathbb{C}))$ acts on the square integrable forms
$\omega \in \mathscr{H} = 
L^2(\widetilde{\mathbb{S}^1}\times SU(2, \mathbb{C}); {\bigwedge}^{*}_{\mathbb{C}})$
through pointwise multiplication, and its weak closure $\mathcal{A}''$, regarded as algebra of 
operators on $L^2(\widetilde{\mathbb{S}^1}\times SU(2, \mathbb{C}); {\bigwedge}^{*}_{\mathbb{C}})$
acts with uniform multiplicity $d=16$.

To this spectral tuple 
\begin{multline*}
\Big( \,\, \mathcal{A} = \mathscr{C}^\infty(\widetilde{\mathbb{S}^1}\times SU(2, \mathbb{C})), 
\,\,\, \mathscr{H} = 
L^2\big(\widetilde{\mathbb{S}^1}\times SU(2, \mathbb{C}); {\bigwedge}^{*}_{\mathbb{C}}\big),
\\
\,\,\,
D_\mathfrak{J} = d + d^*, 
\,\,\,
D = d + d^\dagger, 
\,\,\,
\mathfrak{J} \Big)
\end{multline*}
we can give the form (\ref{1stSpectralTupleForSxG}) or (\ref{2ndSpectralTupleForSxG})
acting on the Hilbert space of square integrable $\mathbb{C}^{16}$-valued ``multispinors'' $\phi
\in L^2(\widetilde{\mathbb{S}^1}\times SU(2, \mathbb{C}); \mathbb{C}^{16})$,
with the unitary and Krein unitary representation having the general local form 
(\ref{UphiOnSxG}), and with the representation $V$ of $SL(2, \mathbb{C})$
in (\ref{UphiOnSxG}) equal to the direct sum 
\begin{equation}\label{V^(1)forms}
V(v) = V^{(1)}(v) = \bigoplus \limits_{k = 0}^{4} \big(S(v \otimes \overline{v})S\big)^{\widehat{\otimes} k}, 
\end{equation}
of antisymmetrized $k$-fold tensor products $\widehat{\otimes}$
\[
\big(S(v \otimes \overline{v})S\big)^{\widehat{\otimes} k}
\]
of the representation (denoted by $V$ in Subsection \ref{DefLopRep})
\begin{equation}\label{LopRepV}
v \mapsto S(v \oplus \overline{v})S, \,\,\, v \in SU(2, \mathbb{C})
\end{equation}
which is precisely the extension of the representation defining the {\L}opusza\'nski
representation ${\L}$ in Subsection \ref{DefLopRep}, here restricted to the subgroup $SU(2, \mathbb{C})$.
Here $S$ the unitary involutive (equal to its inverse) $4\times 4$ matrix over the complex field
defined in Subsection \ref{DefLopRep}. 

Indeed the bundle 
${\bigwedge}^{*}_{\mathbb{C}}T^*(\widetilde{\mathbb{S}^1}\times SU(2, \mathbb{C}))$ is trivial and
its elements 
\[
\omega = \omega_0 \oplus \omega_1 \oplus \omega_2 \oplus \omega_3 \oplus \omega_4
\]
can be regarded as $\mathbb{C}^{16}$-valued functions on the manifold $\widetilde{\mathbb{S}^1}\times SU(2, \mathbb{C})$. 
Introducing the basis of right invariant one-forms $\beta_0, \beta_1, \beta_2, \beta_3$, which are dual do the right invariant and othonormal vector fields\footnote{Recall please, that we have used normalizations in which $2X^0, 2X^1, 2X^2, 2X^3$ are orthonormal with respect to 
$g$ and $g_\mathfrak{J}$.} $2X^0, 2X^1, 2X^2, 2X^3$, we see that
each element  
\[
\omega = \omega_0 \oplus \omega_1 \oplus \omega_2 \oplus \omega_3 \oplus \omega_4
\in \mathscr{H} = 
L^2\Big(\widetilde{\mathbb{S}^1}\times SU(2, \mathbb{C}); {\bigwedge}^{*}_{\mathbb{C}}\Big)
\]
can be uniquely decomposed in the basis
\[
1, \,\,\,\, \beta_\mu, \,\,\,\, \beta_\mu \wedge \beta_\nu, \,\,\,\, 
\beta_{\mu} \wedge \beta_\nu \wedge \beta_\rho,
\,\,\,\, \beta_{\mu} \wedge \beta_\nu \wedge \beta_\rho \wedge \beta_\sigma,
\]
where $1$ is the zero-form (\emph{i. e.} $\mathbb{C}$-valued function) equal everywhere $1$, and where we need only fix the order in the wedge products of the basic forms $\beta_\mu$ giving the basis of the two, three and four-forms respectively, by the requirement that the zero index always comes first (whenever appears), and the remaining three are always in the cyclic order. Thus in the basis wedge products
we choose respectively: $(01)$, $(02)$, $(03)$, $(12)$, $(23)$, $(31)$ for the basic wedge products of two-forms,  $(012)$, $(023)$, $(031)$, $(123)$ for the basic wedge products of three forms,
and $(0123)$ for the four-forms. We call this set of multi-indices \emph{cyclic}.
Thus, for the element $\omega  = \omega_0 \oplus \omega_1 \oplus \omega_2 \oplus \omega_3 \oplus \omega_4 \in \mathscr{H}$ we have unique canonical components
\[
(\phi_{(0)}, \phi_\mu, \phi_{\mu \nu}, \phi_{\mu \nu \rho}, \phi_{\mu \nu \rho \sigma})
\]
with the multi-indices ranging only over the \emph{cyclic} set, and such that
\[
\omega_0 = \phi_{(0)}, \,\,\,\,\,
\omega_1 = \sum \limits_{\mu} \phi_\mu \beta_\mu,
\,\,\,\,\,
\omega_2 = \sum' \limits_{\mu \nu} \phi_{\mu \nu} \beta_\mu \wedge \beta_{\nu},
\]
\[
\omega_3 = \sum' \limits_{\mu \nu \rho} \phi_{\mu \nu \rho} \beta_\mu \wedge \beta_{\nu} \wedge \beta_\rho,
\,\,\,\,\,
\omega_4 = \phi_{0123} \beta_0 \wedge \beta_1 \wedge \beta_2 \wedge \beta_3,
\]
where the prime indicates that the sums run over the \emph{cyclic} set of multi-indices.
The inner product expressed in terms of the $16$ component function $\phi$
is equal 
\begin{multline*}
(\omega, \omega) = 
 \sum \limits_{k=0}^{4} \,\,\,\,\,\, \int \overline{\omega_k} \,\,\, \wedge 
\,\,\, \star_{{}_{\mathfrak{J}}} \,\, \omega_k \\
= \int \overline{\phi_{(o)}(t,u)} \phi_{(o)}(t,u) \, dtdu
+ \sum \limits_{\mu} \int \overline{\phi_{\mu}(t,u)} \phi_{\mu}(t,u) \, dtdu \\
+ \sum' \limits_{\mu \nu} \int \overline{\phi_{\mu \nu}(t,u)} \phi_{\mu \nu}(t,u) \, dtdu 
+ \sum' \limits_{\mu \nu \rho} \int \overline{\phi_{\mu \nu \rho}(t,u)} \phi_{\mu \nu \rho}(t,u) \, dtdu \\
+ \int \overline{\phi_{0123}(t,u)} \phi_{0123}(t,u) \, dtdu
= \int \big(\phi(t,u), \phi(t,u)\big)_{{}_{\mathbb{C}^{16}}} \, dtdu,
\end{multline*} 
where $dtdu = \beta_0 \wedge \beta_1 \wedge \beta_2 \wedge \beta_3$ is the invariant measure
on $\widetilde{\mathbb{S}^1} \times SU(2, \mathbb{C})$ and where 
\[
\big(\cdot, \cdot \big)_{{}_{\mathbb{C}^{16}}}
\]
is the standard inner product on $\mathbb{C}^{16}$. It is easily seen that
the signature operators $D_\mathfrak{J}$ and $D$ are invariant with respect
to the representation (\ref{UphiOnSxG}) acting on the square integrable complex valued
forms $\omega$ through their coefficients $\phi$ defined as above, and with the representation $V$
of $SU(2, \mathbb{C})$  in the formula (\ref{UphiOnSxG})
given by (\ref{V^(1)forms}). Also a simple computation shows that
the action of the signature operators $D_\mathfrak{J}$ and $D$ expressed in terms of the
coefficients $\phi$ is equal to the respective operators $D_\mathfrak{J}$ and $D$ of the
form expressed in (\ref{1stSpectralTupleForSxG})
or equivalently in (\ref{2ndSpectralTupleForSxG}) through the right invariant vector fields
$X^0, X^1, X^2, X^3$ and the corresponding Clifford algebra generators
$\widehat{\Gamma}^{0}_{(1)}, \widehat{\Gamma}^{1}_{(1)}, \dots$, $\Gamma^{0}_{(1)}, \Gamma^{1}_{(1)}, \ldots$ and the fundmental symmetry operator equal to the pointwise multiplication by the
matrix $\mathfrak{J}_{(1)}$. The only difference in comparison to the  
$D_\mathfrak{J}$ and $D$ of the
form expressed in (\ref{1stSpectralTupleForSxG})
or equivalently in (\ref{2ndSpectralTupleForSxG}) is that the signature operators
$D_\mathfrak{J}$ and $D$ in action on the components $\phi$ are equal to that 
in (\ref{1stSpectralTupleForSxG})
or equivalently in (\ref{2ndSpectralTupleForSxG}) up to the trivial factor 2 (coming from our normalization 
of the right invariant vector fields $X^0,X^1,X^2,X^3$) and that the Clifford algebra
additive element $\Gamma$ and thus the element $\widehat{\Gamma}$, definining the constant additive terms
in the signature operators $D_\mathfrak{J}$ and $D$, are not equal $i\Gamma^1\Gamma^2\Gamma^3$
or respectively $-i\widehat{\Gamma}^1\widehat{\Gamma}^2\widehat{\Gamma}^3$, but are chosen differently 
and which are singular but non-zero. This has the
consequence that the (still invariant together with $D_\mathfrak{J}$ and $D$)
signature Laplacian and signature wave operators $D_{\mathfrak{J}}^{2}$ and $D^2$
are no longer scalar operators, but instead 
\begin{multline*}
D_{\mathfrak{J}}^{2} = - \boldsymbol{1}_{{}_{16}} \, 4 \Delta 
-  4 \sum \limits_{(k,j) \, cycl.} \Gamma^{{}^{k}}\Gamma^{{}^{j}} [X^{{}^{k}},X^{{}^{j}}] 
+ 4 \sum \limits_{k} \{\Gamma^{{}^{k}}, \Gamma\} X^k + (\Gamma)^2 \\
=
- \boldsymbol{1}_{{}_{16}} \, 4 \Delta 
-  4 \sum \limits_{(k,j,i) \, cycl.} \Gamma^{{}^{k}}\Gamma^{{}^{j}} X^{{}^{i}}
+ 4 \sum \limits_{k} \{\Gamma^{{}^{k}}, \Gamma\} X^k + (\Gamma)^2, 
\end{multline*}
and\footnote{Note that with our normalization of $X^0,X^1,X^2, X^3$ these are the operators
$4 \Delta$ and $4\square$ which are equal to the ordinary metric Laplace-Beltrami (resp. wave) 
operator corresponding, respectively, to
the invariant Riemannian metric $g_\mathfrak{J}$ and pseudo-Riemann metric $g$.}
\begin{multline*}
D^{2} = - \boldsymbol{1}_{{}_{16}} \, 4 \square 
-  4 \sum \limits_{(k,j) \, cycl.} 
\widehat{\Gamma}^{{}^{k}}\widehat{\Gamma}^{{}^{j}} [X^{{}^{k}},X^{{}^{j}}] 
+ 4 \sum \limits_{k} \{\widehat{\Gamma}^{{}^{k}}, \widehat{\Gamma}\} X^k + (\widehat{\Gamma})^2
\\
=
- \boldsymbol{1}_{{}_{16}} \, 4 \square 
-  4 \sum \limits_{(k,j,i) \, cycl.} \widehat{\Gamma}^{{}^{k}}\widehat{\Gamma}^{{}^{j}} X^{{}^{i}}
+ 4 \sum \limits_{k} \{\widehat{\Gamma}^{{}^{k}}, \widehat{\Gamma}\} X^k + (\widehat{\Gamma})^2,
\end{multline*}
in action on components $\phi$ of the forms. Here the sums run over $(kj) = (12), (23), (31)$, 
$(kji) = (123), (231), (312)$, and $k=1,2,3$ respectively. The curly brackets $\{ \cdot , \cdot \}$
denote anticommutator. Thus, the constant (here singular) and additive terms $\Gamma$ and $\widehat{\Gamma}$
are so chosen that the first order linear contributions do not cancel out in the squares
$D_\mathfrak{J}^{2}$ and $D^2$.  But the principal second order parts of the signature Laplacian (resp. wave) operators $D_\mathfrak{J}^{2}$ and $D^2$ are scalar operators equal to the ordinary scalar Laplace-Belrtami (resp. wave) operator $- \boldsymbol{1}_{{}_{16}} \, 4 \Delta$ 
(resp.$- \boldsymbol{1}_{{}_{16}} \, 4 \square$).  

In order 
to construct the Clifford algebra generators and the fundamental symmetry
corresponding to the representation 
\begin{equation}\label{V^(n)forms}
V(v) = V^{(n)}(v) = \big[V^{(1)}(v)\big]^{\otimes n} = 
\Big[\bigoplus \limits_{k = 0}^{4} \big(S(v \otimes \overline{v})S\big)^{\widehat{\otimes} k}
\Big]^{\otimes n}, 
\end{equation}
in (\ref{UphiOnSxG}) we apply the tensor product operation to the Clifford algebra generators
$\widehat{\Gamma}^{0}_{(1)}, \widehat{\Gamma}^{1}_{(1)}, \dots$, $\Gamma^{0}_{(1)}, 
\Gamma^{1}_{(1)}, \ldots$
corresponding to the representation $V=V^{(1)}$ given by (\ref{V^(1)forms}), 
\emph{i.e.} to the spectral triple acting on the square integrable complex valued forms.
We can write at once the formulas for the Clifford algebra
generators $\widehat{\Gamma}^0 = \widehat{\Gamma}^{0}_{(1)} \otimes \boldsymbol{1}_{16}^{ \otimes \, (n-1)}, \widehat{\Gamma}^1 = \widehat{\Gamma}^{1}_{(1)} \otimes \boldsymbol{1}_{16}^{ \otimes \, (n-1)}, \dots$, 
$\Gamma^0  = \Gamma^{0}_{(1)} \otimes \boldsymbol{1}_{16}^{ \otimes \, (n-1)}, 
\Gamma^1= \Gamma^{1}_{(1)} \otimes \boldsymbol{1}_{16}^{ \otimes \, (n-1)}, \ldots$ corresponding to $V = V^{(n)}$ in (\ref{UphiOnSxG}), and write down the formulas for the spectral triples
acting on square summable  $\mathbb{C}^{16n}$-valued multispinors $\phi$ which transform under
unitary and Krein unitary representation (\ref{UphiOnSxG}) with $V = V^{(n)}$ in it and with
Dirac operators $D_\mathfrak{J}$ and $D$ invariant under this representation.  
The Krein fundamental symmetry operator corresponding to $V=V^{(n)}$ equal (\ref{V^(n)forms}), 
is equal to the operator of pointwise multiplication by the matrix 
$\mathfrak{J}_{(1)} \otimes \boldsymbol{1}_{16}^{ \otimes \, (n-1)}$,
where $\mathfrak{J}_{(1)}$ is the matrix defining the fundamental symmetry of the spectral tuple
acting on square summable forms.

The procedure of tensoring the Clifford algebra generators corresponding to $V=V^{(1)}$
presented here is much simpler in comparison to that performed in Section \ref{DirectIntRepVF} for the representation $V^{(1)}$ of Subsection \ref{DirectIntRepVF} and the corresponding ordinary Dirac gamma matrices corresponding to $V^{(1)}$ of Section \ref{DirectIntRepVF}. This is because we have chosen 
$V(v) = V^{(n)}(v) = \big[V^{(1)}(v)\big]^{\otimes n}$ with the ordinary equality. In
Subsection \ref{DirectIntRepVF} we have $V(v) = V^{(n)}(v) \cong_U \big[V^{(1)}(v)\big]^{\otimes n}$
up to unitary equivalence with the additional task of determination of unitary equivalence matrix $U$
realizing the equivalence $V(v) = V^{(n)}(v) = U \big[V^{(1)}(v)\big]^{\otimes n} U^{-1}$. For our choice here the matrix $U$ degenerates to the unit matrix. Of course the choice of the representations
$V^{(n)}(v) \cong_U \big[V^{(1)}(v)\big]^{\otimes n}$ of Subsection \ref{DirectIntRepVF} was not arbitrary but rather emerged from the form of direct components of the representation of the double covering of the 
Poincar\'e group acting in the Fock space of free fields
on Minkowski space-time. Of course one may simplify the computation of the Clifford algebra matrix generators
and of the fundamental symmetry corresponding to the representations $V(v) = 2^n \, v^{\otimes n}
 \cong_U (v \oplus v)^{\otimes n}$ of the class
(\ref{multispinorV}), by replacing them with the representations unitary and Krein-unitary equivalent
$V(v) = (v \oplus v)^{\otimes n}$. This simplification would be only apparent because 
the representations $2^n \, v^{\otimes n}$ have the form in which they are
already partially decomposed. In the further stage of analysis we will have to decompose the representation
$V$, so that the computation of the corresponding Clebsch-Gordan coefficients is unavoidable.
When we are using  $V(v) = 2^n \, v^{\otimes n}$ instead of $V(v) = (v \oplus v)^{\otimes n}$, we have the process of decomposition partially done.

We have thus constructed spectral tuples (\ref{1stSpectralTupleForSxG}) or
equivalently the spectral tuples (\ref{2ndSpectralTupleForSxG}) which spectrally (in the sense of
\cite{Connes_spectral}) describe the (compactified) Einstein Universe
$\widetilde{\mathbb{S}^1} \times SU(2, \mathbb{C})$ together with the invariant pseudo-Riemann
$g = \beta_0 \otimes \beta_0 - \beta_1 \otimes \beta_1 - \beta_2 \otimes \beta_2 - \beta_3 \otimes \beta_3$ and the corresponding invariant Riemann metric $g_\mathfrak{J} = \beta_0 \otimes \beta_0 + \beta_1 \otimes \beta_1 + \beta_2 \otimes \beta_2 + \beta_3 \otimes \beta_3$.
All these spectral tuples act on the Hilbert spaces $\mathscr{H}$ of square integrable multispinors
$\phi$, and are invariant under the local unitary and Krein-unitary representation (transformation )
(\ref{UphiOnSxG}) uniquely determined by the representation $V$ of $SU(2, \mathbb{C})$ in 
the formula (\ref{UphiOnSxG}). The spectral tuples, and thus the Clifford algebra generators and the Krein fundamental symmetry $\mathfrak{J}$, and thus the Dirac operators $D$ and $D_\mathfrak{J}$
are uniquely determined by the representation $V$ in the formula (\ref{UphiOnSxG}). The Dirac operators 
are invariant under (\ref{UphiOnSxG}), which corresponds to the invariance of the metrics
$g$ and $g_\mathfrak{J}$. The class of possible representations $V$ constructed so far and the 
corresponding spectral tuples (\ref{1stSpectralTupleForSxG}) or (\ref{2ndSpectralTupleForSxG}) is given by 
(\ref{multispinorV}) or by (\ref{V^(n)forms}). The uniform multiplicity of the action of 
algebra $\mathcal{A}''= \mathscr{C}^\infty(\widetilde{\mathbb{S}^1} \times SU(2, \mathbb{C}))''
= L^\infty(\widetilde{\mathbb{S}^1} \times SU(2, \mathbb{C}))$
on $\mathscr{H}$ in (\ref{2ndSpectralTupleForSxG}) is equal $2^{2n}$
(if the spectral tuple (\ref{2ndSpectralTupleForSxG}) corresponds to the representation $V$ of the class (\ref{multispinorV}))
or $2^{4n}$ (if the spectral tuple (\ref{2ndSpectralTupleForSxG}) corresponds to the representation
$V$ of the class (\ref{V^(n)forms})). Recall that the representations $V$ of the class 
(\ref{multispinorV}) and the corresponding Clifford generators are constructed by a tensoring 
from the bispinor representation (\ref{spinorV}) and the corresponding spectral tuple acting on the bispinors. The class (\ref{V^(n)forms}) of representations $V$ and the corresponding spectral
tuples are constructed from the representation (\ref{V^(1)forms}) and from the spectral tuple acting
on square summable forms with complex coefficients. 

We need these spectral realizations of the (compactified) Einstein Universe in further stage. Namely, we consider the invariant subspaces $\mathcal{H}_{\textrm{inv}}$ of the Fock space $\mathcal{H}$
of free fields (in fact in the tensor product of Fock spaces of free fields of both energy signs),
on which there act the corresponding direct summad subrepresentations  $U_{\textrm{inv}}$  of the 
Einstein isometry group $\widetilde{\mathbb{S}^1} \times SU(2, \mathbb{C}) \times SU(2, \mathbb{C})$  
of the (compactified) Einstein Universe.
We will seek the appropriate direct summand subrepresentations $U_{\textrm{inv}}$ which, when computed in the Fock space of Fourier transformed states, are equal to the representations
(\ref{l-decUphiOnSxG}), which are precisely the Fourier transforms 
$V_\mathcal{F}UV_{\mathcal{F}}^{-1}$ of the representations $U$ given by
(\ref{UphiOnSxG}) acting on the Hilbert space $\mathscr{H} = 
V_\mathcal{F}^{-1} \mathcal{H}_{\textrm{inv}}$ of the spectral tuples corresponding to the representations $V$ of the allowed class, for example of the class (\ref{multispinorV}) or  (\ref{V^(n)forms}). We compare them with the invariant subspaces $\mathcal{H}_{\textrm{inv}}$ of the representation acting in the Fock space, and compute the spectral tuples on each respective invariant subspace $\mathcal{H}_{\textrm{inv}}$, that is we compute the generators $P^\mu$ in terms of the generators of the isometry group acting in the Fock space. These generators in turn can be expressed in terms of Noether integrals through Wick products of free quantum fields (generalized operators) integrated over the compact Cauchy surface $\{t\} \times SU(2, \mathbb{C}) = \{t\} \times \mathbb{S}^3$. To this end we therefore need to know a large class of spectral tuples invariant under the unitary and Krein-unitary local representations (\ref{UphiOnSxG}) determined by $V$, with $V$ ranging over a possible large class of representations of $SU(2, \mathbb{C})$. Thus, a natural question arises if the classes (\ref{multispinorV}) and (\ref{V^(n)forms}) exhaust all, up to unitary and Krein-unitary equivalence, possible representations and the corresponding spectral tuples realizing spectrally the (compactified) Einstein Universe. The answer is: no. For example one can take any representation $V = V_1$ of the class (\ref{multispinorV}) or of the class  (\ref{V^(n)forms}), and any other $V = V_2$ coming of these classes, and then one can perform the tensoring operation to the Clifford
algebra generators corresponding to these representations, in order to obtain the spectral tuple  
corresponding to $V_1 \otimes V_2$ as was done in preparation of the Clifford algebra generators
and spectral tuples corresponding to (\ref{multispinorV}) and (\ref{V^(n)forms}). But this is not the only natural operation which can be used to prepare new spectral tuple from two given ones.
We can likewise apply the operation of direct sum $V_1 \oplus V_2$
in order to obtain a new possible representation $V = V_1 \oplus V_2$ in (\ref{UphiOnSxG})
with the corresponding Clifford generators equal do direct sum of Clifford algebra generators
corresponding respectively to $V_1$ and $V_2$. 

A caution is in order. In general the spectral tuple obtained by the operation od direct sum 
applied to spectral tuples respecting the conditions of \cite{Connes_spectral}, especially the strong regularity condition, will not preserve strong regularity condition. Our case is however special, as all final spectral tuples obtained through the ``tensoring'' and direct sum operations are the 
spectral tuples acting on multispinors with $D$ and $D_{{}_{\mathfrak{J}}}$ equal to Dirac operators,
with the elliptic operator $D_\mathfrak{J}$ whose square $D_{{}_{\mathfrak{J}}}^{2}$ is either equal to the scalar Laplacian $\boldsymbol{1}\Delta$ (for the class of spectral tuples prepared from that acting on bispinors) or to the operator whose principal second order contribution is the scalar Laplacian (in case of spectral tuples prepared from that that acting on square summable forms).  Therefore, in each case the principal symbol of the elliptic $D_{{}_{\mathfrak{J}}}^{2}$ is a scalar. Therefore, the proof of Thm. 11.4 of \cite{Connes_spectral} that all our spectral tuples obtained by ``tensoring'' and direct summation respect the five conditions together with the strong regularity condition of \cite{Connes_spectral}, can be repeated after \cite{Connes_spectral}. 

Returning to the question of the classification of all, up to unitary and Krein-unitary equivalence,
representations $V$ of  $SU(2, \mathbb{C})$ in (\ref{UphiOnSxG}), which allows the corresponding 
Clifford algebra generators and fundamental symmetry giving the corresponding 
invariant spectral tuples (\ref{2ndSpectralTupleForSxG}), 
it can be rather easily and explicitly solved by going through the commutation relations
of the generators of the representation $V$ with the Clifford algebra generators,
imposed by the invariance, which relate them to the Clifford algebra
generators. This finite-linear problem can be solved exactly as the problem of classification
of rotationally invariant linear differential equations\footnote{Compare also the Bargmann's solution of the Lie group extension, \cite{BargmannExtension} or \cite{WawrzyckiExtension}.} by Gelfand, Minlos and Shapiro, \cite{Geland-Minlos-Shapiro}, Part I, Section 9, \S 3. But it is not even the explicit solution of the problem which is most important for us, but rather the recognition that all the solutions
are obtained by application of simple operations of tensoring and direct summation to the fundamental solution. This follows from the simple observation that each representation of $SU(2, \mathbb{C})$ can be decomposed into direct sum of irreducible components. Each irreducible component can be obtained in the process of decomposition of tensor product applied to the fundamental spinor representation. Recall that $V$ is a unitary and Krein unitary representation of $SU(2, \mathbb{C})$. On the other hand the matrices $\Gamma^0, \Gamma^1, \ldots$, are not arbitrary, but they are Clifford algebra generators 
corresponding to the forms $\textrm{diag} \, (1,1,1,1)$ and respectively $\textrm{diag} \, (1,-1,-1,-1)$. We know that each representation of such Clifford algebra (thus any corresponding set of
the  algebra generators) 
is a direct sum of the fundamental irreducible identity representation acting on bispinors,
associated to the fundamental representation (\ref{spinorV}). Thus, in principle all allowed
representations $V$ should be reached by application of the tensor product operation and direct sum
operation applied to the representation (\ref{spinorV}). Correspondingly essentially all
invariant spectral space-time tuples are that which can be reached by the operations corresponding to tensoring of $V$ (which we have also called ``tensoring'') and direct summation applied to the Clifford algebra generators of the tuple corresponding to  (\ref{spinorV}). In particular the representation $V$
equal (\ref{V^(1)forms}) and the class (\ref{V^(n)forms}) of representations $V$ is subsumed
under the class (\ref{multispinorV}) or within the class which is closed under direct sum operation 
and tensor product operation and contains all (\ref{multispinorV}). In particular, it is easily seen that
the representation (\ref{V^(1)forms}) is unitary and Krein unitary equivalent to the representation
(\ref{multispinorV}) with $n=2$:
\begin{multline*}
V = V^{(1)} = \\ =
\big(\widehat{0}\big) \oplus 
\Big(S\big( \widehat{{\textstyle\frac{1}{2}}} \otimes 
\overline{\widehat{{\textstyle\frac{1}{2}}}}\big)S\Big)
\oplus
\big(S(\widehat{{\textstyle\frac{1}{2}}} 
\otimes \overline{\widehat{{\textstyle\frac{1}{2}}}}\big)S\Big)^{\widehat{\otimes} 2}
\oplus
\Big(S\big(\widehat{{\textstyle\frac{1}{2}}} 
\otimes \overline{\widehat{{\textstyle\frac{1}{2}}}}\big)S\Big)^{\widehat{\otimes} 3}
\oplus
\Big(S\big(\widehat{{\textstyle\frac{1}{2}}} 
\otimes \overline{\widehat{{\textstyle\frac{1}{2}}}}\big)S\Big)^{\widehat{\otimes} 4} \\
\cong_U
\big(\widehat{0}\big) 
\oplus 
\big(\widehat{0} \oplus \widehat{1}\big)
\oplus
\big(\widehat{1} \oplus \widehat{1}\big)
\oplus
\big(\widehat{0} \oplus \widehat{1}\big)
\oplus
\big(\widehat{0} \big) \cong_U 4 \,\,\,  \big(\,\widehat{0} \oplus \widehat{1} \, \big) \\
\cong_U 4 \,\,\, \widehat{{\textstyle\frac{1}{2}}}^{\otimes 2}.
\end{multline*}
Also the Clifford algebra generators correspondnig to these equivalent $V$-s, are joined by 
the similarity transformation, the same which connects the equivalent $V$-s. Note also that the corresponding
Dirac operators $D_\mathfrak{J}$ and $D$, namely that acting on bispinors $\phi$ and the signature 
operators acting on the components $\phi$  of forms are essentially uniquely determined. 
Indeed, the Clifford generators $\Gamma^0, \widehat{\Gamma}^0, \ldots, \Gamma^3, \widehat{\Gamma}^3$ 
are fixed (and they are necessarily connected by the said similarity transformation). The only freedom 
left by invariance lies in the choice of the additive matrix terms $\Gamma$ and 
$\widehat{\Gamma}$, which still are not entirely arbitrary, as they  commute with $\Gamma^k$
and anticommute with $\Gamma^0$, and commute with the corresponding representation $V$, 
as a consequence of invariance. In case of the signature operators $D_\mathfrak{J}$ and $D$ the constant
additive terms are chosen in a less fortunate manner in which the squares of the 
$D_\mathfrak{J}$ and $D$ are not equal to the ``scalar'' Lapalace-Beltrami (resp. wave) operators,
but to the Laplace (or wave) operators on forms.  The signature operator $D_\mathfrak{J}$
acting on components $\phi$ of forms and associated to the representation $V$ equal (\ref{V^(1)forms}) has slightly different spectrum in comparison to the Dirac operator $D_\mathfrak{J}$ acting on the multispinors $\phi$ associated to the equivalent representation (\ref{multispinorV}) with $n=2$. This difference (in particular spectrum of the signature operator includes zero and the Dirac operator on multispinors defined as above and corresponding to (\ref{multispinorV}) with $n=2$ does not contain zero) is inessential in relation to spectral reconstruction of geometry $g_\mathfrak{J},g$. But in computations
we need to  have the spectral tuples corresponding to respective $V$ with the spectra of the corresponding
Dirac operators $D_\mathfrak{J}$ which behave in as uniform and similar fashion as possible
(if multiplicities are disregarded). The class of Dirac operators based on forms and their tensor product and direct sum generalizations
behaves slightly differently (if multiplicities are disregarded) in comparison to the Dirac operator   
$D_\mathfrak{J}$ on bispinors and  its tensor product and direct sum generalizations. 
 Fortunately the class of spectral tuples based on the signature operator is redundant and subsumed under the class obtained from tensoring and direct sum applied to bispinors. Summing up our discussion of classification of allowed $V$ and of the associated to $V$ invariant spectral triples based on Clifford algebra generators we have: 1) the tuples (\ref{2ndSpectralTupleForSxG}) are essentially uniquely determined by the representations $V$. 2) The class of all (up to unitary and Krein unitary equivalence) representations $V$ which determine the corresponding invariant space-time spectral tuples (\ref{2ndSpectralTupleForSxG}) is that obtained from the representation (\ref{spinorV}) by the operation of tensoring and direct summation. In particular the representation (\ref{V^(1)forms}) and the spectral tuple on forms corresponding to it and the class  obtained from it by tensoring and direct summation is redundant and is subsumed by the class of representations and corresponding spectral tuples which are based on (\ref{spinorV}) or on the representation obtained from it by tensor product and direct summation.

\begin{defin*}
Space-time spectral tuple (\ref{2ndSpectralTupleForSxG}) and the associated representation $V$ we call
\emph{allowed}  if (\ref{2ndSpectralTupleForSxG}) is that acting on multispinors with ordinary ``scalar'' squares of Dirac operators in the associated spectral tuples (\ref{2ndSpectralTupleForSxG}),
which can be obtained from the tuple acting on bispinors based on (\ref{spinorV}) by tensoring and direct summation. 
\end{defin*}
In the sequel when speaking of space-time spectral tuple we confine ourselves to the class of \emph{allowed representations} $V$ and associated with them \emph{allowed spectral tuples}.
This assumption is not arbitrary but in fact exhaust essentially all possibilities, as we have just explained. Note however that the tuples constructed by tensoring and direct summation of the tuple acting on forms are not allowed.

Before returning to QFT on the Einstein Universe we will make three further technical remarks 
concerning our infinite family of space-time spectral tuples (\ref{2ndSpectralTupleForSxG})
corresponding to the respective \emph{allowed representations} $V$ of $SU(2, \mathbb{C})$. 

The first remark is concerned with the process of seeking the proper subrepresentations, and then the reconstruction of space-time spectral tuples in terms of Noether integrals on their invariant subspaces
in the Fock space. This process is divided into two stages.
In the first stage we focus only on the content of irreducible components, up to unitary
and Krein unitary equivalence. In decomposition of the unitary and Krein unitary representation 
of the twofold cover $\widetilde{\mathbb{S}^1} \times SU(2, \mathbb{C}) \times SU(2, \mathbb{C})$ of the isometry group of the (compactified) $\widetilde{\mathbb{S}^1} \times SU(2, \mathbb{C})$ Einstein Universe 
$\mathbb{R} \times SU(2, \mathbb{C})$  acting in the Fock space we encounter the outer Kronecker products
$[\,\widehat{n} \cdot \widehat{l_1} \,] \times [\, \widehat{l_2} \,]$ 
of the irreducible representations $\widehat{n} \cdot \widehat{l_1}$ and $\widehat{l_2}$
of the groups $\widetilde{\mathbb{S}^1} \times SU(2, \mathbb{C})$ and
$SU(2, \mathbb{C})$. In the second stage, all the fun begins with calculating the Clebsch-Gordan coefficients and converting the unitary equivalences into actual equalities of representations. Therefore, in order to simplify the second stage it is convenient to have the class of spectral tuples
(\ref{2ndSpectralTupleForSxG}) corresponding to the allowed representations $V$ in the form as flexible as possible regarding the unitary equivalences. Thus, in particular we need to have the spectral tuples 
(\ref{2ndSpectralTupleForSxG}) corresponding not only to the 
specific allowed $V$ but also corresponding to $SVS^{-1}$, where $S$ is any unitary and Krein unitary (with the Krein structure corresponding to the actual $V$) matrix $S$. Of course in this case
the Clifford algebra generators need to be replaced with the equivalent ones $S\Gamma^\mu S^{-1}$,
$S\widehat{\Gamma}^\mu S^{-1}$. 
 
The second remark is concerned with generators $\boldsymbol{P}^\mu$ defining the Dirac operators in the tuples (\ref{2ndSpectralTupleForSxG}) acting on multispinors, invariant under 
(\ref{UphiOnSxG}), and corresponding to an \emph{allowed representation} $V$
in (\ref{UphiOnSxG}). Namely, recall that $\boldsymbol{P}^k$, $k=1,2,3$, are the generators of the one-parameter subgroups $\tau \mapsto 1\times x_k(\tau)$ in
the representation $\oplus_{1}^{d} \,\,{{}^{L}}U = \boldsymbol{{{}^{L}}U}$ -- the direct sum of $d$ copies of the left regular representation. But the representation $\oplus_{1}^{d} \,\,\, {{}^{L}}U$ is not equal to the representation  (\ref{UphiOnSxG}) corresponding to $V$, and restricted to the subgroup $1 \times 1\times SU(2, \mathbb{C})$. The representation $\oplus_{1}^{d} \,\,{{}^{L}}U = \boldsymbol{{{}^{L}}U}$ 
is equal to (\ref{UphiOnSxG}) restricted to the subgroup $1 \times 1\times SU(2, \mathbb{C})$, but with $V$ put equal $\boldsymbol{1}$. In the further stage when comparing the spectral tuples (\ref{2ndSpectralTupleForSxG}) corresponding to respective $V$ with the actions of the Noether integral generators of the isometry group on the corresponding invariant subspaces $\mathcal{H}_{{}_{\textrm{inv}}}$ of the Fock space, it  will be convenient to note however, that the generators
of the two representations $\oplus_{1}^{d} {{}^{L}}U$  and (\ref{UphiOnSxG}) restricted to $1\times 1 \times SU(2, \mathbb{C})$, both acting on the common Hilbert space $\mathscr{H}$ of the spectral tuple corresponding to $V$, are related by a very simple formula. Namely, if we denote the generators of the subgroups $1 \times 1 \times x_k(\tau) \in 1 \times SU(2, \mathbb{C})$ in the representation 
(\ref{UphiOnSxG}) of the isometry group $\widetilde{\mathbb{S}^1} \times SU(2, \mathbb{C}) \times SU(2, \mathbb{C})$ by 
\[
\mathbb{P}^k, \,\,\,
\textrm{where}
\,\,\,
U_{{}{1\times1\times x_k(\tau)}} = e^{i\tau \mathbb{P}^k},
\,\,\,
\textrm{with $U$ given by (\ref{UphiOnRxG})},
\]
then we have
\[
\mathbb{P}^0 = P^{0} + iA^0 
\,\,\,
\textrm{or}
\,\,\,
P^{k} = \mathbb{P}^k -  iA^k, 
\]
where $A^k$ is the operator of multiplication by the constant matrix
\[
A^k = {\textstyle\frac{d}{d \tau}}|_{{}_{\tau = 0}} V\big(x_k(\tau)\big),
\]
thus equal to the generator of the representation $V$ associated to the considered spectral
tuple (\ref{2ndSpectralTupleForSxG}) invariant under (\ref{UphiOnSxG}). 
When comparing the spectral tuple (\ref{2ndSpectralTupleForSxG})  with that acting 
on the respective invariant subspace of the Fock space, it will be more convenient to use the tuple 
(\ref{2ndSpectralTupleForSxG})  in the following equivalent form
\begin{multline}\label{3rdSpectralTupleForSxG}
\Bigg( \,\,\,\,\,\,\,\, \mathcal{A} = 
\Bigg\{ 
\sum_{\substack{\widehat{n} \in \mathbb{Z}\\
                  \widehat{x} \in \widehat{SU(2,\mathbb{C})}\\
i,j=1, \ldots \textrm{dim} \, \widehat{x} }}
\textrm{dim} \, \widehat{x} 
\,\,
\tilde{f}_{{}_{\widehat{n} \cdot \widehat{x}, j, i}} \boldsymbol{\mathbb{V}}_{{}_{\widehat{n} \cdot \widehat{x}, i, j}}, \,
\tilde{f}\in \mathcal{F}\big[\mathcal{S}_{\Delta}(\widetilde{\mathbb{S}^1} \times SU(2,\mathbb{C}))\big] \Bigg\} \,\,, \\
\,\,\,\,\,\,\,\,
\mathscr{H} = \oplus L^2(\widetilde{\mathbb{S}^1} \times SU(2,\mathbb{C})) \,, \\
D_{{}_{\textrm{ell}}} =  D_{{}_{\mathfrak{J}}} \, {} = \Gamma^{{}^{0}} \boldsymbol{P}^{{}^{0}} + \Gamma^{{}^{1}} \mathbb{P}^{{}^{1}} + 
\ldots + \Gamma^{{}^{3}} \mathbb{P}^{{}^{3}}  + \Gamma 
- i\Gamma^{{}^{1}} A^{{}^{1}} - 
\ldots -i \Gamma^{{}^{3}} A^{{}^{3}} \,\,, 
\,\,\,\,\,\,\,\,\, \\
D \, {} = \widehat{\Gamma}^{{}^{0}} \boldsymbol{P}^{{}^{0}} + \widehat{\Gamma}^{{}^{1}} \mathbb{P}^{{}^{1}} + 
\ldots + \widehat{\Gamma}^{{}^{3}} \mathbb{P}^{{}^{3}} + \widehat{\Gamma}
- i\widehat{\Gamma}^{{}^{1}} A^{{}^{1}} - 
\ldots -i \widehat{\Gamma}^{{}^{3}} A^{{}^{3}} \,,
\,\,\,\,\,\,\,\,\,
\mathfrak{J}  \,\,\,\, \Bigg),
\end{multline}
with the generators $P^{{}^{k}} = \mathbb{P}^{{}^{k}} -  iA^{{}^{k}}$ of the direct sum of copies of the left regular representation expressed in terms of the generators $\mathbb{P}^{{}^{k}}$ of the representation (\ref{UphiOnSxG})  of the isometry group (restricted of course to
the subgroup $1\times 1\times SU(2,\mathbb{C})$)  inserted into the Dirac operators of  
(\ref{2ndSpectralTupleForSxG}). This will simplify comparison because the Noether generators are actually the generators $\boldsymbol{P}^{{}^{0}}, \mathbb{P}^{{}^{k}}$ of the (Fourier transformed $V_\mathcal{F}$) symmetry group (\ref{UphiOnSxG}) acting in the corresponding invariant subspace $\mathcal{H}_{{}_{\textrm{inv}}}$  of the Fock space. Recall that the matrices $A^{{}^{k}}$, generators
of the subgroups $\tau \mapsto x_k(\tau) \in SU(2, \mathbb{C})$ in the representation $V$ corresponding
to the tuple (\ref{3rdSpectralTupleForSxG}) or respectively (\ref{2ndSpectralTupleForSxG})
can be easily computed. In particular for the representation $V$ of the form
(\ref{multispinorV}) we have
\[
A^{{}^{j}} =  {\textstyle\frac{d}{d \tau}}|_{{}_{\tau = 0}} V\big(x_j(\tau)\big) =
{\textstyle\frac{i}{2}} \bigoplus \limits_{1}^{2^n} \, d\Gamma_n (\sigma_j),
\]
where
\begin{multline*}
d\Gamma_n (\sigma_j) = \sigma_j \otimes \boldsymbol{1}^{\otimes (n-1)}
+ \boldsymbol{1} \otimes \sigma_j \otimes \boldsymbol{1}^{\otimes (n-2)} 
+ \ldots + \boldsymbol{1}^{\otimes (n-1)} \otimes \sigma_j 
\\ 
= 
\sum \limits_{m=0}^{n-1} 
\boldsymbol{1}^{\otimes m} \otimes \sigma_k \otimes \boldsymbol{1}^{\otimes (n-m -1)}, 
\end{multline*}
with the ordinary Pauli matrices $\sigma_k$. Here $\boldsymbol{1}$ are $2 \times 2$ unit matrices.
For the representation $V$ of the form\footnote{Although they need not be considered separately any further because (\ref{V^(n)forms}) for any particular $n=n_0$ is unitary and Krein unitary equivalent to (\ref{multispinorV}) with $n=2n_0$.} (\ref{V^(n)forms}) we have
\[
A^{{}^{j}} =  {\textstyle\frac{d}{d \tau}}|_{{}_{\tau = 0}} V\big(x_j(\tau)\big) =
  d\Gamma_n (B_j),
\]
where
\[
B_j = {\textstyle\frac{1}{2}} \bigoplus \limits_{k=0}^{4} d\Gamma_k\big(S(i\sigma_j \otimes\boldsymbol{1}
\oplus \boldsymbol{1} \otimes \overline{i\sigma_j})S\big)
\]
with the general prescription that for any operator $B$
\[
d\Gamma_k(B)
= \sum \limits_{m=0}^{k-1} 
\boldsymbol{1}^{\otimes m} \otimes B \otimes \boldsymbol{1}^{\otimes (k-m -1)},
\,\,\,\,
d\Gamma_0(B) = 0.
\]
This of course follows from the fact that the generators of the subgroups
$\tau \mapsto x_j(\tau)$ in the representation $v \mapsto \widehat{\tfrac{1}{2}}(v) = v$,
$v \in SU(2, \mathbb{C})$ are equal
\[
{\textstyle\frac{d}{d \tau}}|_{{}_{\tau = 0}} \widehat{{\textstyle\frac{1}{2}}}\big(x_j(\tau)\big) =
{\textstyle\frac{i}{2}} \sigma_j,
\]
and from the general formulas for generators of tensor product and direct sum representations
expressed in terms of generators of the component representations.

The third technical remark is concerned with the computation of the spectral values $\lambda$
and the corresponding eigenstates $\phi_{{}_{\lambda}}$
of the elliptic Dirac operator $D_{{}_{\mathfrak{J}}}$ corresponding to the invariant
Riemannian metric $g_{{}_{\mathfrak{J}}}$ (naturally associated to the pseudo-Riemannian
invariant space-time metric $g$). We will need the formula for 
$\lambda, \phi_{{}_{\lambda}}$ for $D_{{}_{\mathfrak{J}}}$ generally
for the Dirac operator $D_{{}_{\mathfrak{J}}}$ of any \emph{allowed space-time spectral tuple} 
(\ref{2ndSpectralTupleForSxG}) or (\ref{3rdSpectralTupleForSxG}) corresponding to
any  \emph{allowed representation} $V$ of $SU(2, \mathbb{C})$ defining the representation (\ref{UphiOnSxG}) 
under which the tuple (\ref{2ndSpectralTupleForSxG}) or (\ref{3rdSpectralTupleForSxG})
is invariant. These eigenstates, identified as states of the Fock space (as elements
of the invariant subspace $\mathcal{H}_{{}_{\textrm{inv}}}$ of the Fock space $\mathcal{H}$
on which there acts precisely the representation identical with that given by (\ref{l-decUphiOnSxG})
and equal to the Fourier transform of the representation acting on the Hilbert space $\mathscr{H}$
of the tuple (\ref{2ndSpectralTupleForSxG}) or (\ref{3rdSpectralTupleForSxG}) corresponding to
$V$) can in particular serve for computation of the Dixmier traces of the (functions of the) Noether generators
restricted to $\mathcal{H}_{{}_{\textrm{inv}}}$ and multiplied by the gamma Clifford generators, and 
for comparison of these traces with the Connes local trace formula 
for the elliptic operator $D_{{}_{\mathfrak{J}}}$ or its functions (e.g. the Dixmier trace of $D_{{}_{\mathfrak{J}}}^{-2}$ is equal to the scalar Ricci curvature of the metric $g_{\mathfrak{J}}$ multiplied by the volume of our manifold $\widetilde{\mathbb{S}^1} \times SU(2, \mathbb{S})$, compare \cite{Lesch} and references therein), as well as in the comparison of this trace with the average values of the quantum field theory counterparts of the local observable (self-adjoint) quantities which enter into the classic formula of Hilbert's energy-momentum tensor. Details will be given in the later part of this Subsection. Therefore, we should have $\lambda, \phi_{{}_{\lambda}}$ in a form flexible enough in computation. The eigenvalues $\lambda$ of $D_{{}_{\mathfrak{J}}}$
in the tuple (\ref{2ndSpectralTupleForSxG}) or (\ref{3rdSpectralTupleForSxG})
corresponding to $V$ do not depend on the concrete $V$ within the class of \emph{allowed}  tuples 
and corresponding to them \emph{allowed} $V$-s, but their respective multiplicities as well as the concrete form of the eigen-states do of course depend on the particular \emph{allowed tuple} and $V$. The eigenvalue and eigenstate are therefore more
adequately written as $\lambda, \phi_{{}_{V,\lambda}}$ indicating dependence of the form of the eigenstates on the representation $V$. Also, the eigenvalues should have the subscript $V$ if they were counted with multiplicities. However, this dependence of multiplicities is trivial within \emph{allowed} $V$ and the multiplicities of the corresponding eigenvalues are in the constant ratio, equal to the ratio of dimensions of the corresponding representations $V$. The eigenvalues $\lambda$
and their multiplicities can be easily computed and are equal
\[
\lambda_{{}_{n,l}} = \pm \sqrt{{\textstyle\frac{n^2}{4}} + l(l+1) + {\textstyle\frac{1}{4}}},
\,\,\,
n \in \mathbb{Z}, l = 0, {\textstyle\frac{1}{2}}, 1, {\textstyle\frac{3}{2}}, \ldots
\] 
with the multiplicity of each eigenvalue $\lambda_{{}_{n,l}}$ equal to 
\[
{\textstyle\frac{1}{2}} \, \textrm{dim} \, \Big[ \mathcal{H}_{\widehat{l}} \otimes 
\mathcal{H}_{\overline{\widehat{l}}} \otimes \mathcal{H}_V  \Big]
= {\textstyle\frac{1}{2}} \, \textrm{dim} \, \mathcal{H}_{\widehat{l}} \cdot 
\textrm{dim} \, \mathcal{H}_{\overline{\widehat{l}}} \cdot
\textrm{dim} \, \mathcal{H}_V = {\textstyle\frac{1}{2}} \, (2l+1)^2 \, \textrm{dim} \, \mathcal{H}_V,
\]
where $\mathcal{H}_{\widehat{l}}, \mathcal{H}_{\overline{\widehat{l}}}$ and $\mathcal{H}_V$ are the Hilbert spaces of the representations $\widehat{l}, \overline{\widehat{l}} \cong_U \widehat{l}$ and
$V$ respectively. 

Note that the self adjoint operators $P^0, \sum_k (\mathbb{P}^k)^2, \Delta$ and $D_\mathfrak{J}$
commute, so there exist a common complete set of common eigenstates in their common
Hilbert space $\mathscr{H}$ in (\ref{2ndSpectralTupleForSxG}) or (\ref{3rdSpectralTupleForSxG}),
which can be chosen to be equal $\{\phi_{{}_{V,\lambda_{{}_{n,l}}}}\}$. Indeed, by invariance
$P^0, \mathbb{P}^1, \mathbb{P}^2, \mathbb{P}^3$ commute with $D_\mathfrak{J}$ and thus with 
$\Delta = (D_\mathfrak{J})^2$. 

Of course similarly the self adjoint operators $P^0, \sum_k (\mathbb{P}^k)^2, \square$ and the operator $D$ also commute. Indeed by invariance
$P^0, \mathbb{P}^1, \mathbb{P}^2, \mathbb{P}^3$ commute with $D$ and thus with 
$\square = D^2$. There exist a common complete set of common eigen-states for them in their common
Hilbert space $\mathscr{H}$ in (\ref{2ndSpectralTupleForSxG}) or in (\ref{3rdSpectralTupleForSxG}),
but cannot be equal $\{\phi_{{}_{V,\lambda_{{}_{n,l}}}}\}$, because the operators 
$D$ and $D_\mathfrak{J}$ do not commute. This assertion however is not obvious because
$D$ is not a normal operator. However, we know that its square $\square$ is normal (even self adjoint). 
Thus, our assertion follows from the observation that $D$ commutes with the self adjoint $D^2$
and thus is decomposable with respect to the $C^*$-decomposition algebra generated by $D^2$
(compare \cite{Segal_dec_I}), and from the fact (essential here) that $\textrm{Spec} \, D^2$ 
is discrete -- which is related to the compactness of our manifold 
$\widetilde{\mathbb{S}^1} \times SU(2, \mathbb{C})$.\footnote{This assertion on existence 
of common set of generalized eigenstates can be generalized to the non-compact case, 
but we need to deal with the generalized eigenspaces of $D^2$ of infinite dimension 
and with direct integral decompositions, using the same decomposition theorem of 
\cite{Segal_dec_I}, as we will see for the example of non-compact Minkowski space-time e. g. in Subsect. \ref{1/2VF}.}

In computation of $\phi_{{}_{V,\lambda_{{}_{n,l}}}}$ we note that the class of allowed
representations $V$ is constructed from the fundamental representation $V_1$
equal (\ref{spinorV}), by the application 
of the two operations: tensoring and direct sum operation. We note that 
$\phi_{{}_{V,\lambda_{{}_{n,l}}}}$ corresponding to the operation of direct sum
$V= V_1 \oplus V_2$ is equal to direct sum 
$\phi_{{}_{V_1,\lambda_{{}_{n,l}}}} \oplus \phi_{{}_{V_2,\lambda_{{}_{n,l}}}}$ of the respective 
$\phi_{{}_{V_1,\lambda_{{}_{n,l}}}}$ and $\phi_{{}_{V_2,\lambda_{{}_{n,l}}}}$. 
Similarly, we have for the tensoring operation, but here we need 
to be careful because for example in the class (\ref{multispinorV}) of representations $V$ we applied 
not the simple tensoring of the fundamental representation (\ref{spinorV}) but in addition 
we have applied a unitary transformation. Thus a two operations corresponding to tensoring and to
the unitary operation will have to be applied to the
initial $\phi_{{}_{V,\lambda_{{}_{n,l}}}}$ in order to obtain the resulting 
$\phi_{{}_{V,\lambda_{{}_{n,l}}}}$. We explain this procedure by examining first 
the situation in which the resulting
$V$ is equal $V = V \otimes V'$ (and not merely only unitary and Krein 
unitary equivalent) to simple tensoring of the initial representations $V$ and $V'$. 

Suppose we have given the eigenstates
\[
\phi_{{}_{{V},\lambda_{{}_{n,l}}}} \in
\mathscr{H} = L^2(\widetilde{\mathbb{S}^1} \times G; \mathbb{C}^{d})
=  L^2(\widetilde{\mathbb{S}^1} \times G; \mathbb{C}) \otimes \mathbb{C}^{d}
= L^2(\widetilde{\mathbb{S}^1} \times G; \mathbb{C}) \otimes \mathcal{H}_{V}.
\]
Using the canonical basis
\[
e_1 = \left( \begin{array}{c}  
1  \\
0  \\
\vdots   \\
0
                     \end{array}\right) , \ldots,
                      e_{d} = \left( \begin{array}{c}  
0  \\
0  \\
\vdots   \\
1                    \end{array}\right) 
\,\,\,\,\,
\textrm{and}
\,\,\,\,\,
e'_{1} = \left( \begin{array}{c}  
1  \\
0  \\
\vdots   \\
0
                     \end{array}\right) , \ldots
                      e'_{d'} = \left( \begin{array}{c}  
0  \\
0  \\
\vdots   \\
1                    \end{array}\right) , \dots,
\]
of the Hilbert spaces $\mathcal{H}_{V}$ and $\mathcal{H}_{V'}$ of the representations
$V$ and $V'$, respectively, we can write the canonical decompositions
\[
\phi_{{}_{{V},\lambda_{{}_{n,l}}}} = \sum \limits_{1 \leq b \leq d} 
f_{{}_{{V},\lambda_{{}_{n,l}}, b}} \otimes e_b,
\]
with some canonical component functions 
\begin{equation}\label{tesorDecOfEigenStates}
f_{{}_{{V},\lambda_{{}_{n,l}}, b}} \in L^2(\widetilde{\mathbb{S}^1} \times G; \mathbb{C}),
\,\,\,  b= 1, \dots d.
\end{equation}
Let $\Gamma^\mu $, $\widehat{\Gamma}^\mu $ be the Clifford algebra generators and fundamental symmetry 
$\mathfrak{J}$ of the tuple  corresponding to $V$.
Let now $D_{\mathfrak{J}, V \otimes V'}$ be the Dirac operator corresponding to 
$V \otimes V'$ with the Clifford matrix generators $\Gamma^\mu \otimes \boldsymbol{1}_{{}_{V'}}$,
$\widehat{\Gamma}^\mu \otimes \boldsymbol{1}{{}_{V'}}$ and fundamental symmetry 
$\mathfrak{J} \otimes \boldsymbol{1}{{}_{V'}}$ or $\mathfrak{J} \otimes \mathfrak{J}'$
where $\mathfrak{J}'$ is the fundamental symmetry of the tuple corresponding to 
$V'$. Then the eigenvalues $\lambda$
(discarding multiplicity) of $D_{\mathfrak{J}, V \otimes V'}$ are
the same as the eigenvalues of the Dirac operator $D_{\mathfrak{J}, V}$
of the tuple corresponding to $V$. The eigenstates
\begin{equation}\label{tesorDecSpanOfEigenStates}
\phi_{{}_{{V},\lambda_{{}_{n,l}}}} \otimes e'_{c}
=\sum \limits_{1 \leq b \leq d} 
f_{{}_{{V},\lambda_{{}_{n,l}}, b}} \otimes e_b \otimes e'_{c}
, \,\,\,\, c = 1, \ldots, d'
\end{equation}
compose the basis of the eigenspace of the Dirac operator  $D_{\mathfrak{J}, V \otimes V'}$,
corresponding to the eigenvalue $\lambda_{{}_{n,l}}$, associated to the tensor product
representation $V\otimes V'$. If in addition we apply a unitary and Krein-unitary transformation
$S$ to $V\otimes V'$, then the span of eigenspace of the Dirac operator
$D_{\mathfrak{J}, S(V \otimes V')S^{-1}}$ corresponding to the eigenvalue $\lambda_{{}_{n,l}}$ will be obtained by the application of the operator
\[
\boldsymbol{1}_{{}_{L^2}} \otimes S
\]
to the state vectors (\ref{tesorDecSpanOfEigenStates}). Here $\boldsymbol{1}_{{}_{L^2}}$
is the unit operator on the first factor $ L^2(\widetilde{\mathbb{S}^1} \times G; \mathbb{C})$,
and of course $S$ acts on $\mathcal{H}_V \otimes \mathcal{H}_{V'}$. Recall that the Clifford algebra matrices in the tuple corresponding to the representation $S(V \otimes V')S^{-1}$ are respectively
equal $S(\Gamma^\mu \otimes \boldsymbol{1}_{{}_{V'}})S^{-1}$,
$S(\widehat{\Gamma}^\mu \otimes \boldsymbol{1}_{{}_{V'}})S^{-1}$ and fundamental symmetry 
$S(\mathfrak{J} \otimes \boldsymbol{1}_{{}_{V'}})S^{-1}$ or $S(\mathfrak{J} \otimes \mathfrak{J}')S^{-1}$.

Thus having given explicitly the eigenstates
\[
\phi_{{}_{{V},\lambda_{{}_{n,l}}}}
\]
of the (elliptic) Dirac operator $D_{\mathfrak{J}, V}$ corresponding to each eigenvalue
$\lambda_{{}_{n,l}}$, and their canonical decompositions  (\ref{tesorDecOfEigenStates}),
only for the fundamental bispinor representation $V$ equal (\ref{spinorV}):
\[
V= \widehat{{\textstyle\frac{1}{2}}} \oplus \widehat{{\textstyle\frac{1}{2}}},
\]
we can give the eigenstatates
$\phi_{{}_{{V},\lambda_{{}_{n,l}}}}$ for all \emph{allowed} $V$ in simple and concise 
form. 

Now we prove the property (\ref{tesorDecSpanOfEigenStates}) and determine the eigenstates $\phi_{{}_{{V},\lambda_{{}_{n,l}}}}$ 
and their canonical decomposition (\ref{tesorDecOfEigenStates})
of the Dirac operator $D_\mathfrak{J}$ of the spectral tuple corresponding to the
fundamental bispinor representation $V$ equal (\ref{spinorV}).

For this purpose we will need to find the Fourier transform $\mathcal{F}X^0\mathcal{F}^{-1}
= \widetilde{X^0}, \mathcal{F}X^1\mathcal{F}^{-1}
= \widetilde{X^1}, \mathcal{F}X^2\mathcal{F}^{-1}
= \widetilde{X^2}, \mathcal{F}X^3\mathcal{F}^{-1}
= \widetilde{X^3}$ of the operators 
$X^0, X^1, X^2, X^3$ understood as differential operators acting on smooth functions
 $\phi \in \mathscr{C}^\infty(\widetilde{\mathbb{S}^1}\times SU(2, \mathbb{C}); \mathbb{C}^d)
\subset L^2(\widetilde{\mathbb{S}^1}\times SU(2, \mathbb{C}); \mathbb{C}^d) =\mathscr{H}$. 
Here $d$ is equal to the dimension of the representation space $\mathcal{H}_V$ of the representation
$V$. The space $\mathscr{C}^\infty(\widetilde{\mathbb{S}^1} \times SU(2, \mathbb{C}); \mathbb{C}^d)$  of smooth $\mathbb{C}^d$-valued functions is treated as the standard nuclear countably
Hilbert space $\mathcal{S}_{\oplus \Delta}(\widetilde{\mathbb{S}^1}\times SU(2, \mathbb{C}); \mathbb{C}^d)=  \mathcal{S}_{D_{\mathfrak{J}}^{2}}(\widetilde{\mathbb{S}^1}\times SU(2, \mathbb{C}); \mathbb{C}^d)$. The Fourier transform $\mathcal{F}$: $\phi \mapsto \mathcal{F}\big(\phi\big)= \widetilde{\phi}$,
where
\[
\widetilde{\phi}_{{}_{ji}}(\widehat{n} \cdot \widehat{l}) = 
\mathcal{F}\big(\phi\big)_{{}_{ji}}(\widehat{n} \cdot \widehat{l}) = 
{\textstyle\frac{1}{\sqrt{4\pi}}} \int \limits_{\widehat{\mathbb{S}^1} \times G} \phi(t,w) 
\, e^{-i{\textstyle\frac{n}{2}}t} \, \overline{\widehat{l}(w)_{{}_{ij}}} \,
 \, dt dw,
\] 
is treated as a map
which is the isomorphism 
\[
\mathcal{F}: \, 
\mathcal{S}_{\oplus \Delta}(\widetilde{\mathbb{S}^1} \times G; \mathbb{C}^d)
\longrightarrow 
\, 
\mathcal{S}_{\oplus \mathcal{F}\Delta\mathcal{F}^{-1}}
(\widehat{\widetilde{\mathbb{S}^1} \times G} \times \mathbb{N} \times \mathbb{N}; \mathbb{C}^d), 
\,\,\,
G= SU(2, \mathbb{C}),
\]
of the indicated standard nuclear countably Hilbert spaces uniquely extending to 
a unitary map (denoted by the same symbol $\mathcal{F}$)
\[
\mathcal{F}: \,
L^2(\widetilde{\mathbb{S}^1}\times G; \mathbb{C}^d)
\longrightarrow 
\, 
L^2(\widehat{\widetilde{\mathbb{S}^1} \times G} \times \mathbb{N} \times \mathbb{N}; \mathbb{C}^d)
\] 
from the Hilbert space $\mathscr{H}$ of square summable $\mathbb{C}^d$-valued functions 
$u\mapsto \phi(u)
= \phi^1(u), \ldots, \phi^d(u)$
on the group $\widetilde{\mathbb{S}^1}\times G = \widetilde{\mathbb{S}^1}\times SU(2, \mathbb{C})$ 
to the Hilbert space of $\mathbb{C}^d$-valued matrix functions 
$\widehat{n}\cdot \widehat{l} \mapsto \mathcal{F}\big(\phi\big)(\widehat{n}\cdot \widehat{l})$ on the space $\widehat{\widetilde{\mathbb{S}^1} \times G} = \{\widehat{n}\cdot \widehat{l}\}$
dual to $\widetilde{\mathbb{S}^1} \times G$, which are ``square trace summable'' in having finite norm given by the Plancherel formula:
\begin{multline*}
\sum \limits_{a=1,\ldots,d}
 \int \limits_{G} \phi^a(t,u) \overline{\phi^a(t,u)} \, dt du  
\\
= 
\sum \limits_{\widehat{n} \cdot \widehat{l} \in \widehat{\widetilde{\mathbb{S}^1} \times G},
a=1, \ldots, d} 
(2l+1) \,\,\,
\textrm{Tr}\,\Big[\, \mathcal{F}\big(\phi^a\big)(\widehat{n}\cdot \widehat{l}) \,\,
\mathcal{F}\big(\phi^a\big)(\widehat{n}\cdot \widehat{l})^{*} \, \Big] \\
= \sum \limits_{\widehat{n} \cdot \widehat{l} \in \widehat{\widetilde{\mathbb{S}^1} \times G},
1 \leq a \leq d, -l \leq i,j \leq l} (2l+1) \,\,\, 
\big| \mathcal{F}\big(\phi^a\big)_{{}_{ji}}(\widehat{n}\cdot \widehat{l})  \big|^2 \\
= 
\sum \limits_{\widehat{n} \cdot \widehat{l} \in \widehat{\widetilde{\mathbb{S}^1} \times G}} (2l+1) \,\,\, 
\textrm{Tr}\,\Big[\mathcal{F}\phi(\widehat{n}\cdot \widehat{l}) \,
\mathcal{F}\phi(\widehat{n}\cdot \widehat{l})^{*}\Big],
\end{multline*}
where in the last expression the value $\mathcal{F}\phi(\widehat{n}\cdot \widehat{l})$ 
of the Fourier transform $\mathcal{F}\phi$
at $\widehat{n}\cdot \widehat{l} \in \widehat{\widetilde{\mathbb{S}^1} \times G}$ is understood 
as the matrix 
\[
\Big[\mathcal{F}\phi(\widehat{n}\cdot \widehat{l})\Big]_{{}_{a \,\,\,\, ji}} =
\mathcal{F}\big(\phi^a\big)_{{}_{ji}}(\widehat{n}\cdot \widehat{l}),
\]
with matrix indices $a$ and $ij$, of an operator from the Hilbert space endowed with Hilbert-Schmidt norm of scalar $(2l + 1) \times (2l+1)$ matrices to the (conjugation of the) Hilbert space $\mathbb{C}^d = \mathcal{H}_V$, with $\textrm{Tr}$ and $(\cdot)^*$ equal to the  ordinary trace and adjoint operation
of an operator between a Hilbert space and a conjugation\footnote{Compare \cite{Mackey}, \S 5 or Subsect. \ref{kronecker}.} of Hilbert space defined as in 
Subsection \ref{kronecker} or \cite{Mackey2}, \S 3.1, \cite{Mackey}, \S 5\footnote{We have used conjugate linear operators between Hilbert spaces in Subsection \ref{kronecker}, as \cite{Mackey2}, but instead one can use linear operators between Hilbert space and conjugation of the other Hilbert space with obvious modification as in \cite{Mackey}, \S 5.}, respectively. $\mathcal{F}\big(\phi^a\big)(\widehat{n}\cdot \widehat{l})^{*}$ denotes the ordinary adjoint (transposition and complex conjugation) of the
$(2l+1)\times (2l+1)$ matrix
\[
\mathcal{F}\big(\phi^a\big)(\widehat{n}\cdot \widehat{l}).
\] 

The Fourier transform $\mathcal{F}$ of course need not be confined to the Hilbert space
\[
\mathscr{H} = L^2(\widetilde{\mathbb{S}^1} \times G; \mathbb{C}^d)
=  L^2(\widetilde{\mathbb{S}^1} \times G; \mathbb{C}) \otimes \mathbb{C}^d
= L^2(\widetilde{\mathbb{S}^1} \times G; \mathbb{C}) \otimes \mathcal{H}_V
\]
of the spectral tuple (\ref{2ndSpectralTupleForSxG}) with the restriction that 
$V$ belongs to the \emph{allowed} class. We will use in the sequel the Fourier transform not only in analysing of the Dirac operators of the spectral tuples, but
on more general Hilbert spaces of square summable $\mathbb{C}^d$ valued functions acted on by general local transformation (\ref{UphiOnSxG}) with $\mathbb{C}^d$
not necessary being the Hilbert space of \emph{allowed} $V$-s, and with any finite dimensional representation $V$ of $SU(2, \mathbb{C})$. Because the differential equations which are fulfilled by local
quantum fields on the Einstein Universe are expressed in terms of the differential operators
$X^0, X^1, X^2, X^3$ acting on Hilbert spaces $\mathscr{H}$  of square summable $\mathbb{C}^d$-valued functions on $\widetilde{\mathbb{S}^1} \times SU(2, \mathbb{C})$ transforming locally 
according to the law (\ref{UphiOnRxG}) with any finite-dimensional representation $V$ of $SU(2, \mathbb{C})$, computation of their Fourier transform will be useful in the sequel when constructing these fields and not only in computation of the eigenvalues and eigenstates of the Dirac operators.

The differential operators $iX^\mu$, $\mu=0,1,2,3$, are symmetric and transform the indicated above 
dense nuclear subspaces $\mathcal{S}_{\oplus \Delta}$ of $\mathscr{H}$ continuously into themselves, thus by the known criterion they possess self adjoint extensions, \cite{Riesz-Szokefalvy} (p. 120 in Russian 1954 Ed.). Because the one-parameter groups of unitary transformations generated by $iX^\mu$, $\mu=0,1,2,3$, live invariant the nuclear subspaces
$\mathcal{S}_{\oplus \Delta}$, then by the known criterion, \cite{Segal_Kunze}, these extensions are unique, and the operators $iX^\mu$ are essentially self adjoint on $\mathcal{S}_{\oplus \Delta}
\subset \mathscr{H}$. The same criterion of essential self-adjointness applies to the Dirac
operators $D_\mathfrak{J}$, and to the squares $D_{\mathfrak{J}}^{2} = \Delta$ and $D^2= \square$, 
which are essentially
self adjoint on the invariat dense nuclear space $\mathcal{S}_{\oplus \Delta} \subset \mathscr{H}$. 
However, these criteria for essential self adjointness are here unnecessary and too sophisticated, as the essential self adjointness is almost trivial in this compact case. Indeed, the application of the Fourier transform will convert the said operators into the operators of multiplication by functions on discrete measure spaces (with an additional exercise of finiding solutions of finite linear systems of equations in case of the Dirac operator $D_\mathfrak{J}$) unitarily equivalent to these operators.

Let us compute the Fourier transforms $\mathcal{F}\big(X^\mu\big)\mathcal{F}^{-1}
= \widetilde{X^\mu}$ of the differential operators $X^\mu$, $\mu=0,1,2,3$. 

By definition of the differential operators $X^\mu$ we have
\begin{align*}
X^k \phi^a(w) & = {\textstyle\frac{d}{d\tau}} \phi^a\big(t \, \times \, x_k (-\tau)w \, \big)\big|_{{}_{\tau=0}}, \,\,\,\, k=1,2,3,
\\
X^0 \phi^a(w) & = {\textstyle\frac{d}{d\tau}} \phi^a\big((t-\tau) \, \times \, w \, \big)\big|_{{}_{\tau=0}}.
\end{align*}
By definition of the Fourier transform, by invariance of the measure $dtdw$ and by  the representation property $\widehat{l}(u)\widehat{l}(w)
= \widehat{l}(uw)$ of $\widehat{l}$, we have the following formulas
\begin{equation}\label{FT(X_0)}
\mathcal{F}\big(X^0 \phi^a \big)_{{}_{ji}}(\widehat{n}\cdot \widehat{l})
= -{\textstyle\frac{in}{2}}  \, 
\mathcal{F}\big(\phi^a \big)_{{}_{ji}}(\widehat{n}\cdot \widehat{l}),
\end{equation}
and
\begin{multline*}
\mathcal{F}\big(X^k \phi^a \big)_{{}_{ji}}(\widehat{n}\cdot \widehat{l}) =
\frac{d}{d\tau}\Bigg|_{{}_{\tau=0}}
{\textstyle\frac{1}{\sqrt{4\pi}}} \int \limits_{\widehat{\mathbb{S}^1} \times G} 
\phi^a(t,x_k (-\tau)w)
 \, \overline{\widehat{l}(w)_{{}_{ij}}} \, e^{-i{\textstyle\frac{n}{2}}t} \, 
 \, dt dw
 \\
= \frac{d}{d\tau}\Bigg|_{{}_{\tau=0}}
{\textstyle\frac{1}{\sqrt{4\pi}}} \int \limits_{\widehat{\mathbb{S}^1} \times G} \phi^a(t,w) 
 \, \overline{\widehat{l}(x_k (\tau)w)_{{}_{mj}}} \, e^{-i{\textstyle\frac{n}{2}}t} \,  
 \, dt dw
\end{multline*}
\begin{multline*}
=  \sum \limits_{m}
 {\textstyle\frac{d}{d\tau}} \overline{\widehat{l}(x_k (\tau))_{{}_{im}}}\big|_{{}_{\tau=0}}
 \,\,
{\textstyle\frac{1}{\sqrt{4\pi}}} \int \limits_{\widehat{\mathbb{S}^1} \times G} \phi^a(t,w) 
 \, \overline{\widehat{l}(w)_{{}_{ij}}} \, e^{-i{\textstyle\frac{n}{2}}t} \,
 \, dt dw
 \\
 \sum \limits_{m}
 {\textstyle\frac{d}{d\tau}} \overline{\widehat{l}(x_k (\tau))_{{}_{im}}}\big|_{{}_{\tau=0}}
 \, \mathcal{F}\big(\phi^a\big)_{{}_{jm}}(\widehat{n} \cdot \widehat{l})
 =  \sum \limits_{m} \overline{\mathbb{A}^{k}_{{}_{\widehat{l}, \,\,\, im}}} \, 
 \mathcal{F}\big(\phi^a\big)_{{}_{jm}}(\widehat{n} \cdot \widehat{l}),
\end{multline*}
or in concise matrix notation
\begin{equation}\label{FT(X_k)}
\mathcal{F}\big(X^k \phi^a \big)(\widehat{n} \cdot \widehat{l}) =
\mathcal{F}\big(\phi^a\big)(\widehat{n} \cdot \widehat{l})
\big(\overline{\mathbb{A}^{k}_{{}_{\widetilde{l}}}}\big)^{T}.
\end{equation}
The generators
\[
\mathbb{A}^{k}_{{}_{\widehat{l}}} = 
{\textstyle\frac{d}{d\tau}} \widehat{l}(x_k(\tau))\big|_{{}_{\tau=0}},
\,\,\,
k=1,2,3,
\]
are equal to the following matrices
\begingroup\makeatletter\def\f@size{5}\check@mathfonts
\def\maketag@@@#1{\hbox{\m@th\large\normalfont#1}}%
\begin{equation}\label{tildeX_1}
\mathbb{A}^{1}_{{}_{\widetilde{l}}} = 
{\textstyle\frac{i}{2}}
\left( \begin{array}{ccccccc}  
0 &\alpha_{{}_{-l+1}}  &  &  &  &  &  0  \\
\alpha_{{}_{-l+1}} & 0  &  &  &  &  &   \\
 &  & 0 & \alpha_{{}_{-l+2}} &  &  &   \\
 &  & \alpha_{{}_{-l+2}} & 0 &  &  &   \\
 &  &  &  & \ddots  &  &   \\
 &  &  &  &  & 0 & \alpha_{{}_{l}}  \\
 0 &  &  &  &  & \alpha_{{}_{l}} & 0  
                     \end{array}\right)   
\end{equation}\endgroup
\begingroup\makeatletter\def\f@size{5}\check@mathfonts
\def\maketag@@@#1{\hbox{\m@th\large\normalfont#1}}%
\begin{equation}\label{tildeX_2}
\mathbb{A}^{2}_{{}_{\widetilde{l}}} = 
{\textstyle\frac{1}{2}} \left( \begin{array}{ccccccc}  
0 &\alpha_{{}_{-l+1}}  &  &  &  &  &  0  \\
-\alpha_{{}_{-l+1}} & 0  &  &  &  &  &   \\
 &  & 0 & \alpha_{{}_{-l+2}} &  &  &   \\
 &  & -\alpha_{{}_{-l+2}} & 0 &  &  &   \\
 &  &  &  & \ddots  &  &   \\
 &  &  &  &  & 0 & \alpha_{{}_{l}}  \\
 0 &  &  &  &  & -\alpha_{{}_{l}} & 0  
                     \end{array}\right)   
\end{equation}\endgroup
\begingroup\makeatletter\def\f@size{5}\check@mathfonts
\def\maketag@@@#1{\hbox{\m@th\large\normalfont#1}}%
\begin{equation}\label{tildeX_3}
\mathbb{A}^{3}_{{}_{\widetilde{l}}} = 
\left( \begin{array}{ccccccc}  
il &  &  &  &  &  &  0  \\
 & i(l-1)  &  &  &  &  &   \\
 &  & i(l-2) &  &  &  &   \\
 &  &  & i(l-3) &  &  &   \\
 &  &  &  & \ddots  &  &   \\
 &  &  &  &  & i(-l+1) &   \\
 0 &  &  &  &  &  & -il  
                     \end{array}\right)   
\end{equation}\endgroup
\[
\alpha_{{}_{m}} = \sqrt{(l+m)(l-m+1)},
\]
compare, \cite{Geland-Minlos-Shapiro}, Part I, Section 2, \S 3, formula (3). Note also that our generator matrices $\mathbb{A}^{k}_{{}_{\widetilde{l}}}$ in the representation $\widehat{l}$ correspond to the
matrices denoted by $A_k$ or by $A_{k}^{(l)}$ in \cite{Geland-Minlos-Shapiro}, Part I, Section 2, and that our sign convention is different so that our $\mathbb{A}^{1}_{{}_{\widetilde{l}}}$ is equal to minus $A_1$ used in \cite{Geland-Minlos-Shapiro}. 

Thus, the Fourier transformed $\mathcal{F} X^k \mathcal{F}^{-1} = \widetilde{X^k}$ operator $X^k$,
$k=1,2,3$, in action on a 
``square trace summable'' $\mathbb{C}^d$-valued matrix function\footnote{Note once again that
$F(\widehat{n}\cdot \widehat{l})$ can also be understood as a matrix 
$\big[F(\widehat{n}\cdot \widehat{l})\big]_{{}_{a \,\,\,ji}}$  over $\mathbb{C}$ equal 
$\big[F(\widehat{n}\cdot \widehat{l})\big]_{{}_{a \,\,\,ji}} = 
F^{a}_{{}_{ji}}(\widehat{n}\cdot \widehat{l})$ with matrix indices $-l \leq i,j \leq l$  and 
$1 \leq a \leq d$ of an operator from the Hilbert space of $(2l+1) \times (2l+1)$ matrices with Hilbert-Schmidt matrix norm multiplied by the Plancherel weight $2l+1$, to the complex conjugation of the Hilbert space $\mathbb{C}^d$, compare  Subsect. \ref{kronecker}, or \cite{Mackey}, \S 5. We will use both these equivalent interpretations of $F(\widehat{n}\cdot \widehat{l})$. 
Then the norm in the Hilbert space of the matrices 
$F(\widehat{n}\cdot \widehat{l})$, regarded as matrices over $\mathbb{C}$ with indices $i,j,a$,
and coming from the Plancherel formula, is equal 
$(2l+1) \, \textrm{Tr} \,\, \big[ F(\widehat{n}\cdot \widehat{l}) \, F(\widehat{n}\cdot \widehat{l})^{*}  \big]$ 
with ordinary trace $\textrm{Tr}$ and adjoint $(\cdot)^*$ of a matrix
of an operator between a Hilbert space and the conjugation of another Hilbert space defined in the natural manner, 
compare Subsection \ref{kronecker}.}
\[
\widehat{n}\cdot \widehat{l} \mapsto
\left( \begin{array}{c}  
F^{1}(\widehat{n}\cdot \widehat{l})  \\
 \vdots \\
F^{d}(\widehat{n}\cdot \widehat{l})
                     \end{array}\right)   = F(\widehat{n}\cdot \widehat{l})
\]
on $\widehat{\widetilde{\mathbb{S}^1} \times G}$
is equal to the pointwise right multiplication by the matrix valued function
\[
\widehat{n}\cdot \widehat{l} \mapsto \widetilde{X^k}(\widehat{n}\cdot \widehat{l})
 = -\mathbb{A}^{k}_{{}_{\widetilde{l}}} = \big(\overline{\mathbb{A}^{k}_{{}_{\widetilde{l}}}}\big)^{T}
 = \big(\mathbb{A}^{k}_{{}_{\widetilde{l}}}\big)^*
\]
(independent of $\widehat{n}$ and the multispinor index $a$), and gives the resulting  
$\mathbb{C}^d$-valued matrix function of the form
\begin{multline*}
\widehat{n}\cdot \widehat{l} \mapsto
\Big(\widetilde{X^k} \, F\Big)(\widehat{n}\cdot \widehat{l}) = 
F(\widehat{n}\cdot \widehat{l}) \widetilde{X^k}(\widehat{n}\cdot \widehat{l}) =
\\ =
\left( \begin{array}{c}  
F^{1}(\widehat{n}\cdot \widehat{l}) \,\, \widetilde{X^k}(\widehat{n}\cdot \widehat{l}) \\
 \vdots \\
F^{d}(\widehat{n}\cdot \widehat{l}) \,\, \widetilde{X^k}(\widehat{n}\cdot \widehat{l})
                     \end{array}\right)   = 
           \left( \begin{array}{c}  
F^{1}(\widehat{n}\cdot \widehat{l}) \big(\overline{\mathbb{A}^{k}_{{}_{\widetilde{l}}}}\big)^{T} \\
 \vdots \\
F^{d}(\widehat{n}\cdot \widehat{l}) \big(\overline{\mathbb{A}^{k}_{{}_{\widetilde{l}}}}\big)^{T}
                     \end{array}\right)             
=                     
\left( \begin{array}{c}  
-F^{1}(\widehat{n}\cdot \widehat{l}) \mathbb{A}^{k}_{{}_{\widetilde{l}}} \\
 \vdots \\
-F^{d}(\widehat{n}\cdot \widehat{l}) \mathbb{A}^{k}_{{}_{\widetilde{l}}}
                     \end{array}\right)   
\end{multline*}
on $\widehat{\widetilde{\mathbb{S}^1} \times G}$.

The Fourier transformed $\mathcal{F} X^0 \mathcal{F}^{-1} =  \widetilde{X^0}$ operator $X^0$ in action on a ``square trace summable'' $\mathbb{C}^d$-valued matrix function
\[
\widehat{n}\cdot \widehat{l} \mapsto
\left( \begin{array}{c}  
F^{1}(\widehat{n}\cdot \widehat{l})  \\
 \vdots \\
F^{d}(\widehat{n}\cdot \widehat{l})
                     \end{array}\right)   = F(\widehat{n}\cdot \widehat{l})
\]
on $\widehat{\widetilde{\mathbb{S}^1} \times G}$
is equal to the pointwise multiplication by the scalar valued function
\[
\widehat{n}\cdot \widehat{l} \mapsto \widetilde{X^0}(\widehat{n}\cdot \widehat{l})
 = {\textstyle\frac{in}{2}} \,\,\ \boldsymbol{1}_{{}_{2l+1}}
\]
(independent of $\widehat{l}$ and the multispinor index $a$), and gives the resulting  
$\mathbb{C}^d$-valued matrix function of the form
\begin{multline*}
\widehat{n}\cdot \widehat{l} \mapsto
\Big(\widetilde{X^0} F\Big)(\widehat{n}\cdot \widehat{l}) = 
F(\widehat{n}\cdot \widehat{l})  \,\,\, \widetilde{X^0}(\widehat{n}\cdot \widehat{l}) =
\\ =
\left( \begin{array}{c}  
F^{1}(\widehat{n}\cdot \widehat{l}) \,\, \widetilde{X^0}(\widehat{n}\cdot \widehat{l}) \\
 \vdots \\
F^{d}(\widehat{n}\cdot \widehat{l}) \,\,  \widetilde{X^0}(\widehat{n}\cdot \widehat{l})
                     \end{array}\right)   = 
           \left( \begin{array}{c}  
F^{1}(\widehat{n}\cdot \widehat{l}) {\textstyle\frac{in}{2}}\\
 \vdots \\
F^{d}(\widehat{n}\cdot \widehat{l}) {\textstyle\frac{in}{2}}
                     \end{array}\right)             
\end{multline*}
on $\widehat{\widetilde{\mathbb{S}^1} \times G}$.

Because the sum of squares of the generators of the subgroups $\tau \mapsto x_k(\tau)$
is in the center of the algebra of $SU(2, \mathbb{C})$ and equal to the scalar $l(l+1)$, 
then in any irreducible representation
$\widehat{l}$
\[
-\big(\mathbb{A}^{1}_{{}_{\widetilde{l}}} \big)^2 -
\big(\mathbb{A}^{2}_{{}_{\widetilde{l}}} \big)^2 -
\big(\mathbb{A}^{3}_{{}_{\widetilde{l}}} \big)^2 =
l(l+1) \, \boldsymbol{1}_{{}_{2l+1}}
\]
is a scalar matrix, equal $l(l+1)$, compare e.g. \cite{Geland-Minlos-Shapiro}, Part I, Section 2.
Therefore, the Fourier transformed 
\begin{multline*}
\mathcal{F} \Delta \mathcal{F}^{-1} = \widetilde{\Delta} = \Big(-\big(\widetilde{X^0}\big)^2 -\big(\widetilde{X^1}\big)^2 -\big(\widetilde{X^2}\big)^2 -\big(\widetilde{X^3}\big)^2 + 
{\textstyle\frac{1}{4}}\Big) =  \\
- \Big(\widetilde{X^0}\Big)^2 - \Big(\widetilde{X^1}\Big)^2
- \Big(\widetilde{X^2}\Big)^2 -\Big(\widetilde{X^3}\Big)^2
+{\textstyle\frac{1}{4}} \boldsymbol{1}
\end{multline*}
operator $\Delta =  -\big(X^0\big)^2 -\big(X^1\big)^2 -\big(X^2\big)^2 -\big(X^3\big) 
+ \tfrac{1}{4} \boldsymbol{1}$ in action on a 
``square trace summable'' $\mathbb{C}^d$-valued matrix function
\[
\widehat{n}\cdot \widehat{l} \mapsto
\left( \begin{array}{c}  
F^{1}(\widehat{n}\cdot \widehat{l})  \\
 \vdots \\
F^{d}(\widehat{n}\cdot \widehat{l})
                     \end{array}\right)   = F(\widehat{n}\cdot \widehat{l})
\]
on $\widehat{\widetilde{\mathbb{S}^1} \times G}$
is equal to the pointwise multiplication by the scalar valued function
\[
\widehat{n}\cdot \widehat{l} \mapsto \widetilde{\Delta}(\widehat{n}\cdot \widehat{l})
 = \big({\textstyle\frac{n^2}{4}} + l(l+1) + {\textstyle\frac{1}{4}}\big) \boldsymbol{1}
\]
(independent of the multispinor index $a$), and gives the resulting  
$\mathbb{C}^d$-valued matrix function of the form
\begin{multline*}
\widehat{n}\cdot \widehat{l} \mapsto
\Big(\widetilde{\Delta} \, F\Big)(\widehat{n}\cdot \widehat{l}) = 
F(\widehat{n}\cdot \widehat{l}) \,\widetilde{\Delta}(\widehat{n}\cdot \widehat{l}) =
\\ =
\left( \begin{array}{c}  
F^{1}(\widehat{n}\cdot \widehat{l}) \,\, \widetilde{\Delta}(\widehat{n}\cdot \widehat{l}) \\
 \vdots \\
F^{d}(\widehat{n}\cdot \widehat{l}) \,\, \widetilde{\Delta}(\widehat{n}\cdot \widehat{l})
                     \end{array}\right)   = 
                     \big( {\textstyle\frac{n^2}{4}} + l(l+1) + {\textstyle\frac{1}{4}} \big) \,
           \left( \begin{array}{c}  
F^{1}(\widehat{n}\cdot \widehat{l}) \\
 \vdots \\
F^{d}(\widehat{n}\cdot \widehat{l}) 
                     \end{array}\right)             
\end{multline*}
on $\widehat{\widetilde{\mathbb{S}^1} \times G}$.

Thus, the Fourier transformed $\mathcal{F}\Delta \mathcal{F}^{-1}  = \widetilde{\Delta}$ operator $\Delta$ 
reduces to the scalar operator of multiplication by the number
\[
{\textstyle\frac{n^2}{4}} + l(l+1) + {\textstyle\frac{1}{4}}
\]
on the Hilbert subspace of all ``square trace summable'' $\mathbb{C}^d$ -valued matrix 
functions $F: \widehat{n}\cdot \widehat{l} \mapsto F(\widehat{n}\cdot \widehat{l})$ which are concentrated on the single point $\widehat{n} \cdot \widehat{l}$
of the dual to our group $\widetilde{\mathbb{S}^1} \times SU(2, \mathbb{C})$.  
Let us fix arbitrarily $\widehat{n} \cdot \widehat{l}$ and the indices
$a \in \{1, \ldots, d\}$, $i,j \in \{-l, \ldots, l\}$. Let $F(\widehat{n}\cdot \widehat{l})$
be a matrix whose only nonzero component is equal
\[
F^{a}_{{}_{ji}}(\widehat{n}\cdot \widehat{l}),
\]
for the fixed value of indices $a,j,i$.
Let us define the function
\begin{equation}\label{special-phi}
\phi^a(t,w) = F^{a}_{{}_{ji}}(\widehat{n}\cdot \widehat{l}) \, \widehat{l}(w)_{{}_{ij}}
\, e^{i{\textstyle\frac{n}{2}}t}
\end{equation}
whose remaining components except the indicated component $a$ are zero.
Then by the formula for the Fourier transform and by the inverse Fourier transform formula
\[
\phi^a(t,w)
=
\sum \limits_{n\in \mathbb{Z}, \widehat{l} \in \widehat{SU(2, \mathbb{C})}} 
(2l+1)\,\,\,
\textrm{Tr}\,\Big[\, \mathcal{F} \big(\phi^a\big)(\widehat{n}\cdot \widehat{l}) \,\, 
\widehat{l}(w) \,\Big] \,\, e^{i{\textstyle\frac{n}{2}}t}
\]
\[
\textrm{with} \,\,\, F^{a}(\widehat{n}\cdot \widehat{l}) 
= \mathcal{F} \big(\phi^a\big)(\widehat{n}\cdot \widehat{l}),
\]
the Fourier image of the function (\ref{special-phi})
is precisely equal to the ``square trace summable'' $\mathbb{C}^d$ -valued matrix 
function $F: \widehat{n}\cdot \widehat{l} \mapsto F(\widehat{n}\cdot \widehat{l})$ which 
is concentrated on the single point $\widehat{n} \cdot \widehat{l}$ with the only nonzero element
of the matrix $F(\widehat{n}\cdot \widehat{l})$ equal $F^{a}_{{}_{ji}}(\widehat{n}\cdot \widehat{l})$. 
Thus, the Fourier transform of all functions $\phi \in \mathscr{H}$
of the type  (\ref{special-phi}) with fixed $\widehat{n} \cdot \widehat{l}$,
and with $a$ ranging over $\{1, \ldots, d\}$ and $i,j$ over $\{-l, \ldots, l\}$, span all the 
``square trace summable'' $\mathbb{C}^d$ -valued matrix 
functions $F: \widehat{n}\cdot \widehat{l} \mapsto F(\widehat{n}\cdot \widehat{l})$  
concentrated on the single point $\widehat{n} \cdot \widehat{l}$, wih their values
$F(\widehat{n}\cdot \widehat{l})$ ranging over the whole linear space of all
$(2l+1)\times (2l+1) \times d$, $\mathbb{C}$-valued matrices 
$F^{a}_{{}_{ji}}(\widehat{n}\cdot \widehat{l})$. They also compose the space of the direct
summand 
\[
[ \, \widehat{n} \cdot \widehat{l} \, ] \, \times \, [ \, \overline{\widehat{l}} \otimes V \, ]
\]
representation in the decomposition (\ref{l-decUphiOnSxG}) of the Fourier transformed
representation (\ref{UphiOnRxG}). This is immediately seen by looking at the Fourier
transform $\mathcal{F}\big(U_{{}_{s\times u \times v}} \phi \big)$ of the transformed 
$U_{{}_{s\times u \times v}} \phi$ function $\phi \in \mathscr{H}$, with the 
transformation marix $\widehat{n} \cdot\widehat{l}(s\times u) = \widehat{n}(s) \widehat{l}(u)$ acting on the index $j$, with the transformation
matrix $\overline{\widehat{l}(v)}$ acting on the index $i$ and finally with the transformation
matrix $V(v)$ acting on the ``spinor index'' $a$ of the matrix
\[
F^{a}_{{}_{ji}}(\widehat{n}\cdot \widehat{l}).
\]
These matrices with the Hilbert-Schmidt square norm (compare also the Plancherel formula)
\[
\langle F(\widehat{n}\cdot \widehat{l}), F(\widehat{n}\cdot \widehat{l}) \rangle 
= l(l+1) \,\, \textrm{Tr} \big[F(\widehat{n}\cdot \widehat{l}) \, 
F(\widehat{n}\cdot \widehat{l})^{*}\big]
= (2l+1) \,\, \sum \limits_{-l \leq i,j \leq l, 1 \leq a \leq d}
\Big|F^{a}_{{}_{ji}}(\widehat{n}\cdot \widehat{l}) \Big|^2
\]
compose the Hilbert space tensor product\footnote{It is well known that also for general (including infinite dimensional) Hilbert spaces the Hilbert space tensor product can be realized through the class of Hilbert-Schmidt operators transforming the first Hilbert space into the conjugation of the second (or through conjugate linear Hilbert-Schmidt operators between the Hilbert spaces in question), as was systematically described for the first time in the classic work of Murray-Von Neumann, compare Subsection \ref{kronecker}.}
\[
\mathcal{H}_{\widehat{l}} \otimes \mathcal{H}_{\overline{\widehat{l}}} \otimes \mathcal{H}_V
\]
acted on by the outer Kronecker product
\[
[ \, \widehat{n} \cdot \widehat{l} \, ] \, \times \, [ \, \overline{\widehat{l}} \otimes V \, ],
\]
with $\widehat{n} \cdot \widehat{l}$ acting on
$\mathcal{H}_{\widehat{l}} \cong \mathbb{C} \otimes \mathcal{H}_{\widehat{l}}$
and with $\overline{\widehat{l}} \otimes V$
acting on $\mathcal{H}_{\overline{\widehat{l}}} \otimes \mathcal{H}_V$.

Using the identification of each subspace of ``square trace summable'' $\mathbb{C}^d$ -valued matrix functions $F: \widehat{n} \cdot \widehat{l} \mapsto F(\widehat{n} \cdot \widehat{l})$, which
are concentrated on single point $\widehat{n} \cdot \widehat{l}$, with the tensor product 
\[
\mathcal{H}_{\widehat{l}} \otimes \mathcal{H}_{\overline{\widehat{l}}} \otimes \mathcal{H}_V
\]
space of the subrepresentation 
\[
[ \, \widehat{n} \cdot \widehat{l} \, ] \, \times \, [ \, \overline{\widehat{l}} \otimes V \, ]
\]
in the decomposition (\ref{l-decUphiOnSxG}), 
the Fourier transform $\mathcal{F}(X^k)$ of the operator can be identified with the operator of multiplication by the matrix function
\[
\widetilde{X^k}: \widehat{n} \cdot \widehat{l} \mapsto 
\widetilde{X^k}(\widehat{n} \cdot \widehat{l})
\]
with the matrix
\[
\widetilde{X^k}(\widehat{n} \cdot \widehat{l}) = \boldsymbol{1}_{\widehat{l}} \otimes 
\overline{\mathbb{A}^{k}_{{}_{\widetilde{l}}}} \otimes \boldsymbol{1}_V
\]
as a matrix of an operator in the Hilbert space
\begin{equation}\label{Fnl=HoHoHV}
\mathcal{H}_{\widehat{l}} \otimes \mathcal{H}_{\overline{\widehat{l}}} \otimes \mathcal{H}_V
\cong 
\,\,\,
\textrm{marix functions $F$ concentrated on single point $\widehat{n} \cdot \widehat{l}$}.
\end{equation}

Now let us return to \emph{allowed} $V$ not necessary equal to the fundamental (\ref{spinorV}),
and to the Hilbert space $\mathscr{H} = L^2(\widetilde{\mathbb{S}^1} \times SU(2, \mathbb{C}); \mathbb{C}^d)$ acted on by the representation (\ref{UphiOnSxG}) and by the Dirac operator $D_\mathfrak{J}$
of the spectral tuple (\ref{1stSpectralTupleForSxG})
or equivalently in (\ref{2ndSpectralTupleForSxG}) corresponding to $V$. 

Using the identification (\ref{Fnl=HoHoHV}) of the Hilbert space of 
``square trace summable'' $\mathbb{C}^d$ -valued matrix 
functions $F: \widehat{n}\cdot \widehat{l} \mapsto F(\widehat{n} \cdot \widehat{l})$
which are concentrated on single point $\widehat{n} \cdot \widehat{l}$, with the tensor product space
of the subrepresentation 
$[ \, \widehat{n} \cdot \widehat{l} \, ] \, \times \, [ \, \overline{\widehat{l}} \otimes V \, ]$,
we can now easily see that the Fourier transform 
$\mathcal{F} D_{\mathfrak{J}}\mathcal{F}^{-1} = \widetilde{D_{\mathfrak{J}}}$
in action on the space of 
``square trace summable'' $\mathbb{C}^d$ -valued matrix 
functions $F: \widehat{n}\cdot \widehat{l} \mapsto F(\widehat{n} \cdot \widehat{l})$
is equal to the operator of pointwise (with the points $\widehat{n} \cdot \widehat{l}
\in \widehat{\widetilde{\mathbb{S}^1} \times SU(2, \mathbb{C})}$)
matrix multiplication by the following matrix function
\begin{equation}\label{Tilde(DJ)nl}
\widehat{n} \cdot \widehat{l} \mapsto
\widetilde{D_{\mathfrak{J}}}(\widehat{n} \cdot \widehat{l})
= 
-{\textstyle\frac{n}{2}} 
\boldsymbol{1}_{\widehat{l}} \otimes \boldsymbol{1}_{\overline{\widehat{l}}} \otimes \Gamma^0 -
\sum \limits_{k=1,2,3}\boldsymbol{1}_{\widehat{l}} \otimes 
\overline{i\mathbb{A}^{k}_{{}_{\widetilde{l}}}} \otimes \Gamma^k
\end{equation}
or equivalently
\[
\mathcal{F} \big(D_{\mathfrak{J}} \phi)^{a}_{{}_{ij}}(\widehat{n}\cdot \widehat{l})
= 
\sum \limits_{b,m}\Bigg[
-{\textstyle\frac{n}{2}} 
\delta_{{}_{jm}} \big(\Gamma^0\big)_{b}^{a} -
\sum \limits_{k=1,2,3}
\overline{i\mathbb{A}^{k}_{{}_{\widetilde{l}, \,\,\, im}}} \big(\Gamma^k\big)_{b}^{a} \Bigg]
\mathcal{F} \big(D_{\mathfrak{J}} \phi^b)_{{}_{mj}}(\widehat{n}\cdot \widehat{l})
\]

Because $D_\mathfrak{J} = D_{\mathfrak{J}, V}$ commutes with
$\Delta = D_{\mathfrak{J}}^{2}$ and therefore it is decomposable with respect to the commutative 
$C^*$-algebra generated by $\Delta = D_{\mathfrak{J}}^{2}$. Therefore the Fourier transform 
$\widetilde{D_{\mathfrak{J}}}$ of
$D_\mathfrak{J}$ acts on each direct summand space of the representation
\begin{equation}\label{[n.l]x[loV]}
[ \, \widehat{n} \cdot \widehat{l} \, ] \, \times \, [ \, \overline{\widehat{l}} \otimes V \, ]
\end{equation}
as a self-adjoint operator (multiplication by self-adjoint matrix)
$\widetilde{D_{\mathfrak{J}}}(\widehat{n}\cdot \widehat{l})$ 
whose square is equal 
\[
\big({\textstyle\frac{n^2}{4}} + l(l+1) + {\textstyle\frac{1}{4}} \big) \,\,\,\, \boldsymbol{1}
\]
and thus as an operator 
\[
\widetilde{D_{\mathfrak{J}}}(\widehat{n}\cdot \widehat{l}) = 
\sqrt{{\textstyle\frac{n^2}{4}} + l(l+1) + {\textstyle\frac{1}{4}}} \,\,\,\, \mathbb{F}_{{}_{\mathfrak{J}}}
\]
where
\[
\mathbb{F}_{{}_{\mathfrak{J}}}^{*} = \mathbb{F}_{{}_{\mathfrak{J}}}, \,\,\,\, 
\mathbb{F}_{{}_{\mathfrak{J}}}^{2} = \boldsymbol{1}.
\]
Thus, the only eigenvalues of the operator $\mathbb{F}_{{}_{\mathfrak{J}}}$ are $+1$ and $-1$ and the invariant subspace
\[
\mathcal{H}_{{}_{[ \, \widehat{n} \cdot \widehat{l} \, ] \, \times \, [ \, \overline{\widehat{l}} \otimes V \, ]}} = \mathcal{H}_{\widehat{l}} \otimes \mathcal{H}_{\overline{\widehat{l}}} \otimes \mathcal{H}_V
=
\mathcal{H}_{{}_{[ \, \widehat{n} \cdot \widehat{l} \, ] \, \times \, [ \, \overline{\widehat{l}} \otimes V \, ]}}^{+}
\oplus 
\mathcal{H}_{{}_{[ \, \widehat{n} \cdot \widehat{l} \, ] \, \times \, [ \, \overline{\widehat{l}} \otimes V \, ]}}^{-}
\]
of the representation (\ref{[n.l]x[loV]}) is a direct sum of two invariant subspaces on which
the eigenvalue of $\mathbb{F}_{{}_{\mathfrak{J}}}$ is $+1$ and $-1$, respectively. Thus, the Dirac operator
$\widetilde{D_{\mathfrak{J}}}$ is equal to
\[
\widetilde{D_{\mathfrak{J}}} = 
+ \sqrt{{\textstyle\frac{n^2}{4}} + l(l+1) + {\textstyle\frac{1}{4}}} \,\,\,\,\, \boldsymbol{1}
\,\,\,\,\, \textrm{on} \,\,\
\mathcal{H}_{{}_{[ \, \widehat{n} \cdot \widehat{l} \, ] \, \times \, [ \, \overline{\widehat{l}} \otimes V \, ]}}^{+}
\]
and 
\[
\widetilde{D_{\mathfrak{J}}} = 
- \sqrt{{\textstyle\frac{n^2}{4}} + l(l+1) + {\textstyle\frac{1}{4}}} \,\,\,\,\, \boldsymbol{1}
\,\,\,\,\,\, \textrm{on} \,\,\
\mathcal{H}_{{}_{[ \, \widehat{n} \cdot \widehat{l} \, ] \, \times \, [ \, \overline{\widehat{l}} \otimes V \, ]}}^{-};
\]
or equivalently
\[
\widetilde{D_{\mathfrak{J}}}(\widehat{n} \cdot \widehat{l}) = 
+ \sqrt{{\textstyle\frac{n^2}{4}} + l(l+1) + {\textstyle\frac{1}{4}}} \,\,\,\,\, \boldsymbol{1}
\,\,\,\,\, \textrm{on} \,\,\
\mathcal{H}_{{}_{[ \, \widehat{n} \cdot \widehat{l} \, ] \, \times \, [ \, \overline{\widehat{l}} \otimes V \, ]}}^{+}
\]
and 
\[
\widetilde{D_{\mathfrak{J}}}(\widehat{n} \cdot \widehat{l}) = 
- \sqrt{{\textstyle\frac{n^2}{4}} + l(l+1) + {\textstyle\frac{1}{4}}} \,\,\,\,\, \boldsymbol{1}
\,\,\,\,\,\, \textrm{on} \,\,\
\mathcal{H}_{{}_{[ \, \widehat{n} \cdot \widehat{l} \, ] \, \times \, [ \, \overline{\widehat{l}} \otimes V \, ]}}^{-}.
\]
Because our manifold $\widetilde{\mathbb{S}^1} \times SU(2, \mathbb{C})$ has even dimension
we have the grading self-adjoint operator $\mathbb{G}$ of multiplication by the Clifford algebra matrix
$\Gamma^0 \Gamma^1 \Gamma^2 \Gamma^3$ anticommuting with $D_\mathfrak{J}$
and commuting with $\Delta = D_{\mathfrak{J}}^{2}$. Thus the Fourier transform 
$\mathcal{F} \mathbb{G}\mathcal{F}^{-1} = \widetilde{\mathbb{G}}$ of the grading $\mathbb{G}$ 
transforms the invariant space 
\[
\mathcal{H}_{{}_{[ \, \widehat{n} \cdot \widehat{l} \, ] \, \times \, [ \, \overline{\widehat{l}} \otimes V \, ]}} = \mathcal{H}_{\widehat{l}} \otimes \mathcal{H}_{\overline{\widehat{l}}} \otimes \mathcal{H}_V
\]
into itself, and moreover for any (Fourier transformed) eigenstate $\mathcal{F}\phi_\lambda 
= \widetilde{\phi_\lambda}$:
\[
\widetilde{D_\mathfrak{J}} \, \widetilde{\phi_\lambda}
 \,\,\, = \,\,\, \lambda \, \widetilde{\phi_\lambda}
\]
of the Fourier transformed Dirac operator
$\widetilde{D_\mathfrak{J}}$ belonging to this subspace we have
\[
\widetilde{D_\mathfrak{J}} \, \widetilde{\mathbb{G}} \, \widetilde{\phi_\lambda}
= - \widetilde{\mathbb{G}} \,  \widetilde{D_\mathfrak{J}} \, 
 \widetilde{\phi_\lambda}  = - \lambda  \, \widetilde{\phi_\lambda}
\]
and thus the Fourier transform of the grading transforms unitarily 
\[
\mathcal{H}_{{}_{[ \, \widehat{n} \cdot \widehat{l} \, ] \, \times \, [ \, \overline{\widehat{l}} \otimes V \, ]}}^{+}
\,\,\,\,\,\,\,\,\,
\textrm{onto}
\,\,\,\,\,\,\,\,\,
\mathcal{H}_{{}_{[ \, \widehat{n} \cdot \widehat{l} \, ] \, \times \, [ \, \overline{\widehat{l}} \otimes V \, ]}}^{-},
\]
and \emph{vice versa}. Therefore, the last two subspaces have equal dimension:
\begin{multline*}
\textrm{dim} \,\, \mathcal{H}_{{}_{[ \, \widehat{n} \cdot \widehat{l} \, ] \, \times \, [ \, \overline{\widehat{l}} \otimes V \, ]}}^{+}
=
\textrm{dim} \,\, \mathcal{H}_{{}_{[ \, \widehat{n} \cdot \widehat{l} \, ] \, \times \, [ \, \overline{\widehat{l}} \otimes V \, ]}}^{-} =
{\textstyle\frac{1}{2}} \textrm{dim} \,\,
\Big[ \mathcal{H}_{{}_{[ \, \widehat{n} \cdot \widehat{l} \, ] \, \times \, [ \, \overline{\widehat{l}} \otimes V \, ]}} \Big]  \\ = 
{\textstyle\frac{1}{2}} \textrm{dim} \,\, \big[\mathcal{H}_{\widehat{l}} \otimes \mathcal{H}_{\overline{\widehat{l}}} \otimes \mathcal{H}_V \Big]
=
{\textstyle\frac{1}{2}}  (2l+1)^2 \textrm{dim} \,\, \mathcal{H}_V = 
{\textstyle\frac{1}{2}} \, (2l+1)^2 \, d
\end{multline*}
with
\[
\textrm{dim} \,\,
\Big[ \mathcal{H}_{{}_{[ \, \widehat{n} \cdot \widehat{l} \, ] \, \times \, [ \, \overline{\widehat{l}} \otimes V \, ]}} \Big]
= \textrm{dim} \,\, \big[\mathcal{H}_{\widehat{l}} \otimes \mathcal{H}_{\overline{\widehat{l}}} \otimes \mathcal{H}_V \Big]
=
\textrm{dim} \,\,\mathcal{H}_{\widehat{l}} \,\,
\textrm{dim} \,\,  \mathcal{H}_{\overline{\widehat{l}}} \,\,
\textrm{dim} \,\, \mathcal{H}_V
=
(2l+1)^2 \, d
\]
necessary even, which is assured by the very existence of the grading operator $\mathbb{G}$.

Summing up, we have found explicit formula for the following self-adjoint projection operator
\[
{P}_{{}_{\mathfrak{J} \,\,\, +|\lambda_{{}{nl}}|}}(\widehat{n} \cdot \widehat{l}) = 
\frac{\boldsymbol{1} + \mathbb{F}_{{}_{\mathfrak{J}}}}{2} = {\textstyle\frac{1}{2}} \boldsymbol{1} +
{\textstyle\frac{1}{2|\lambda_{{}{nl}}|}}  \,
\widetilde{D_{\mathfrak{J}}}(\widehat{n} \cdot \widehat{l})
\]
which projects on the eigenspace
\[
\mathcal{H}_{{}_{[ \, \widehat{n} \cdot \widehat{l} \, ] \, \times \, [ \, \overline{\widehat{l}} \otimes V \, ]}}^{+} \subset 
\mathcal{H}_{{}_{[ \, \widehat{n} \cdot \widehat{l} \, ] \, \times \, [ \, \overline{\widehat{l}} \otimes V \, ]}}
= \mathcal{H}_{\widehat{l}} \otimes \mathcal{H}_{\overline{\widehat{l}}} \otimes \mathcal{H}_V
\]
of all eigenstates of the operator $\widetilde{D_{\mathfrak{J}}}$ corresponding to the eigenvalue
\[
|\lambda_{{}{nl}}| = \sqrt{{\textstyle\frac{n^2}{4}} + l(l+1) + {\textstyle\frac{1}{4}}}.
\]
And similarly 
\[
{P}_{{}_{\mathfrak{J} \,\,\, -|\lambda_{{}{nl}}|}}(\widehat{n} \cdot \widehat{l}) =
\frac{\boldsymbol{1} - \mathbb{F}_{{}_{\mathfrak{J}}}}{2} = {\textstyle\frac{1}{2}} \boldsymbol{1} -
{\textstyle\frac{1}{2|\lambda_{{}{nl}}|}}   \,
\widetilde{D_{\mathfrak{J}}}(\widehat{n} \cdot \widehat{l})
\]
is the self adjoint projector in $\mathcal{H}_{{}_{[ \, \widehat{n} \cdot \widehat{l} \, ] \, \times \, [ \, \overline{\widehat{l}} \otimes V \, ]}}
= \mathcal{H}_{\widehat{l}} \otimes \mathcal{H}_{\overline{\widehat{l}}} \otimes \mathcal{H}_V$,
which projects on the eigenspace
\[
\mathcal{H}_{{}_{[ \, \widehat{n} \cdot \widehat{l} \, ] \, \times \, [ \, \overline{\widehat{l}} \otimes V \, ]}}^{-} \subset 
\mathcal{H}_{{}_{[ \, \widehat{n} \cdot \widehat{l} \, ] \, \times \, [ \, \overline{\widehat{l}} \otimes V \, ]}}
= \mathcal{H}_{\widehat{l}} \otimes \mathcal{H}_{\overline{\widehat{l}}} \otimes \mathcal{H}_V
\]
of all eigenstates corresponding to the eigenvalue
\[
-|\lambda_{{}{nl}}|= -\sqrt{{\textstyle\frac{n^2}{4}} + l(l+1) + {\textstyle\frac{1}{4}}}
\]
of the self adjoint operator $\widetilde{D_\mathfrak{J}}$. 

Note that 
\[
\frac{\boldsymbol{1} + \mathbb{F}_{{}_{\mathfrak{J}}}}{2}
\]
is the self-adjoint projection which projects on the eigen-space of the operator $\mathbb{F}$
corresponding to the eigen-value $+1$ and 
\[
\frac{\boldsymbol{1} - \mathbb{F}_{{}_{\mathfrak{J}}}}{2}
\]
is the self adjoint projection which projects on the eigenspace of the operator $\mathbb{F}_{{}_{\mathfrak{J}}}$ corresponding to the eigenvalue $-1$.  
Note also that by the anti-commutation rule
$D_\mathfrak{J} \,\mathbb{G} = - \mathbb{G} \, D_\mathfrak{J}$ of the operators
$D_\mathfrak{J}$ and $\mathbb{G}$, and by $\mathbb{G}^2 = \boldsymbol{1}$, the unitary operator 
$\widetilde{\mathbb{G}}$ transforms the projector
\[
{\textstyle\frac{1}{2}} ( \boldsymbol{1} + \mathbb{F}_{{}_{\mathfrak{J}}} )
\,\,\,\,
\textrm{into}
\,\,\,\,
{\textstyle\frac{1}{2}} ( \boldsymbol{1} - \mathbb{F}_{{}_{\mathfrak{J}}}):
\,\,\,\,
\widetilde{\mathbb{G}} \,\, {\textstyle\frac{1}{2}} ( \boldsymbol{1} + \mathbb{F}_{{}_{\mathfrak{J}}} ) \,\, \widetilde{\mathbb{G}}
= {\textstyle\frac{1}{2}} ( \boldsymbol{1} - \mathbb{F}_{{}_{\mathfrak{J}}} )
\]
and \emph{vice versa}
\[
\widetilde{\mathbb{G}} \,\, {\textstyle\frac{1}{2}} ( \boldsymbol{1} - \mathbb{F}_{{}_{\mathfrak{J}}} ) \,\, \widetilde{\mathbb{G}}
= {\textstyle\frac{1}{2}} ( \boldsymbol{1} + \mathbb{F}_{{}_{\mathfrak{J}}} ),
\]
in accordance with what we have already shown, but perhaps using the projectors and the commutation rules
makes the argument more transparent. We have the analogue situation for the unitary involutive 
$\mathbb{F}_{{}_{\mathfrak{J}}}$ and the projectors
\[
\frac{\boldsymbol{1} + \widetilde{\mathbb{G}}}{2}
\,\,\
\textrm{and}
\,\,\,
\frac{\boldsymbol{1} - \widetilde{\mathbb{G}}}{2}
\]
which project, respectively, on the eigenspaces of the opertator $\widetilde{\mathbb{G}}$
corresponding to the eigenvalues $+1$ and $-1$. Namely, $\mathbb{F}_{{}_{\mathfrak{J}}}$ gives the unitary equivalence of the last 
two projectors by the same commutation rules 
$\mathbb{F}_{{}_{\mathfrak{J}}} \widetilde{\mathbb{G}} = 
- \widetilde{\mathbb{G}} \mathbb{F}_{{}_{\mathfrak{J}}}$
and $\mathbb{F}_{{}_{\mathfrak{J}}}^{2} = \boldsymbol{1}$. 

Thus, the operator of pointwise matrix multiplication 
by the matrix functions
\begin{multline*}
\widehat{n} \cdot \widehat{l} \,\,\, \mapsto \,\,\,
{\textstyle\frac{1}{2}} \boldsymbol{1} +
{\textstyle\frac{1}{2|\lambda_{{}{nl}}|}}
\widetilde{D_{\mathfrak{J}}}(\widehat{n} \cdot \widehat{l})
= {P}_{{}_{\mathfrak{J} \,\,\, +|\lambda_{{}{nl}}|}}(\widehat{n} \cdot \widehat{l}) 
\\
\,\,\,\,\,\,\,\,\,\,\,
\textrm{and}
\,\,\,\,\,\,\,\,\,\,\,\,\,\,
\\
\widehat{n} \cdot \widehat{l} \,\,\, \mapsto \,\,\,
{\textstyle\frac{1}{2}} \boldsymbol{1} -
{\textstyle\frac{1}{2|\lambda_{{}{nl}}|}}
\widetilde{D_{\mathfrak{J}}}(\widehat{n} \cdot \widehat{l})
= {P}_{{}_{\mathfrak{J} \,\,\,-|\lambda_{{}{nl}}|}}(\widehat{n} \cdot \widehat{l}) 
\end{multline*}
which are concentrated on the single point $\widehat{n}\cdot \widehat{l} \in 
\widehat{\widetilde{\mathbb{S}^1} \times SU(2, \mathbb{C})}$ are the projection operators
in the Hilbert space $\mathcal{F} \mathscr{H}$ of all``square trace summable'' $\mathbb{C}^d$ -valued matrix functions $F: \widehat{n}\cdot \widehat{l} \mapsto F(\widehat{n} \cdot \widehat{l})$,
which project respectively on the eigenspaces of the Dirac operator
$\widetilde{D_\mathfrak{J}}$ corresponding to the eigenvalues
\[
|\lambda_{{}{nl}}| = \sqrt{{\textstyle\frac{n^2}{4}} + l(l+1) + {\textstyle\frac{1}{4}}}
 \,\,\, \textrm{and}
 \,\,\,
-|\lambda_{{}{nl}}| = -\sqrt{{\textstyle\frac{n^2}{4}} + l(l+1) + {\textstyle\frac{1}{4}}}.
\]
Because we have the matrices 
$\widetilde{D_{\mathfrak{J}}}(\widehat{n} \cdot \widehat{l})$
in explicit form (\ref{Tilde(DJ)nl}), the projectors 
${P}_{{}_{\mathfrak{J} \,\,\,\pm|\lambda_{{}{nl}}|}}(\widehat{n} \cdot \widehat{l})$ are also explicitly given.

However, the projectors 
${P}_{{}_{\mathfrak{J} \,\,\, \pm|\lambda_{{}{nl}}|}}(\widehat{n} \cdot \widehat{l})$ 
on the eigenspaces of the operator
$\widetilde{D_{\mathfrak{J}}}$ in the invariant subspaces $\mathcal{H}_{{}_{[ \, \widehat{n} \cdot \widehat{l} \, ] \, \times \, [ \, \overline{\widehat{l}} \otimes V \, ]}}$ of the subrepresentations (\ref{[n.l]x[loV]}),
corresponding to the eigen-values $\pm|\lambda_{{}{nl}}|$ are insufficient for our purposes. We need 
to construct explicitly all 
\[
{\textstyle\frac{1}{2}} (2l+1)^2 \, d \, = {\textstyle\frac{1}{2}} \textrm{dim} \,\,
\Big[ \mathcal{H}_{{}_{[ \, \widehat{n} \cdot \widehat{l} \, ] \, \times \, [ \, \overline{\widehat{l}} \otimes V \, ]}} \Big] 
\]
Fourier transformed mutually orthogonal eigenstates (for any fixed $\widehat{n} \cdot \widehat{l}$) 
\[
\begin{array}{c}  
 \mathcal{F}\big(\phi^{a}_{{}_{{V},1,+ |\lambda_{{}_{n,l}}|}}\big)_{{}_{ji}}(\widehat{n} \cdot \widehat{l}), \\
\mathcal{F}\big(\phi^{a}_{{}_{{V},2,+ |\lambda_{{}_{n,l}}|}}\big)_{{}_{ji}}(\widehat{n} \cdot \widehat{l}), \\
\vdots \\
\mathcal{F}\big(\phi^{a}_{{}_{{V},s,+ |\lambda_{{}_{n,l}}|}}\big)_{{}_{ji}}(\widehat{n} \cdot \widehat{l}), \\
\vdots \\

                     \end{array}   
                     \,\,\,\,\,\,\,\,\,\,\,\,
\textrm{or respectively}
\,\,\,\,\,\,\,\,\,\,\,\,
\begin{array}{c}  
 \mathcal{F}\big(\phi^{a}_{{}_{{V},1,- |\lambda_{{}_{n,l}}|}}\big)_{{}_{ji}}(\widehat{n} \cdot \widehat{l}), \\
\mathcal{F}\big(\phi^{a}_{{}_{{V},2,- |\lambda_{{}_{n,l}}|}}\big)_{{}_{ji}}(\widehat{n} \cdot \widehat{l}), \\
\vdots \\
\mathcal{F}\big(\phi^{a}_{{}_{{V},s,- |\lambda_{{}_{n,l}}|}}\big)_{{}_{ji}}(\widehat{n} \cdot \widehat{l}), \\
\vdots \\
 \end{array}   
\]
\[
1 \leq s \leq {\textstyle\frac{1}{2}} (2l+1)^2 \, d 
\]
corresponding to the eigenvalue  $+ |\lambda_{{}_{n,l}}|$ or respectively $- |\lambda_{{}_{n,l}}|$,
which span
\[
\mathcal{H}_{{}_{[ \, \widehat{n} \cdot \widehat{l} \, ] \, \times \, [ \, \overline{\widehat{l}} \otimes V \, ]}}^{+}
\,\,\,\,
\textrm{or respectively}
\,\,\,\,
\mathcal{H}_{{}_{[ \, \widehat{n} \cdot \widehat{l} \, ] \, \times \, [ \, \overline{\widehat{l}} \otimes V \, ]}}^{-}.
\]
Note that for each fixed $\widehat{n} \cdot \widehat{l}$ and $s$, the matrix
\[
\mathcal{F}\big(\phi^{a}_{{}_{{V},s,\pm |\lambda_{{}_{n,l}}|}}\big)_{{}_{ji}}(\widehat{n} \cdot \widehat{l})
\,\,\,\,\,
1 \leq a \leq d, \,\, 
-l \leq i,j \leq l,
\]
is in a canonical manner an element of the tensor product Hilbert space
\[
\mathcal{H}_{{}_{[ \, \widehat{n} \cdot \widehat{l} \, ] \, \times \, [ \, \overline{\widehat{l}} \otimes V \, ]}} =
\mathcal{H}_{\widehat{l}} \otimes \mathcal{H}_{\overline{\widehat{l}}} \otimes \mathcal{H}_V.
\] 

Now each eigenspace 
\[
\mathcal{H}_{{}_{[ \, \widehat{n} \cdot \widehat{l} \, ] \, \times \, [ \, \overline{\widehat{l}} \otimes V \, ]}}^{\pm} \subset 
\mathcal{H}_{{}_{[ \, \widehat{n} \cdot \widehat{l} \, ] \, \times \, [ \, \overline{\widehat{l}} \otimes V \, ]}}
\]
corresponding to the eigenvalue  $\pm |\lambda_{{}_{n,l}}|$, respectively, is invariant under the
(Fourier transformed) representation (\ref{UphiOnSxG}), and thus the subrepresentation 
(\ref{[n.l]x[loV]}) acting on 
\[
\mathcal{H}_{{}_{[ \, \widehat{n} \cdot \widehat{l} \, ] \, \times \, [ \, \overline{\widehat{l}} \otimes V \, ]}} = \mathcal{H}_{{}_{[ \, \widehat{n} \cdot \widehat{l} \, ] \, \times \, [ \, \overline{\widehat{l}} \otimes V \, ]}}^{+} \oplus 
\mathcal{H}_{{}_{[ \, \widehat{n} \cdot \widehat{l} \, ] \, \times \, [ \, \overline{\widehat{l}} \otimes V \, ]}}^{-}
\]  
is a direct sum 
\[
[ \, \widehat{n} \cdot \widehat{l} \, ] \, \times \, [ \, \overline{\widehat{l}} \otimes V \, ]
\cong_U
[ \, \widehat{n} \cdot \widehat{l} \, ] \, \times \, [ \, \overline{\widehat{l}} \otimes V \, ]_{+}
\,\, \bigoplus \,\,
[ \, \widehat{n} \cdot \widehat{l} \, ] \, \times \, [ \, \overline{\widehat{l}} \otimes V \, ]_{-}
\]
of two unitarily equivalent subsrepresentations 
\[
[ \, \widehat{n} \cdot \widehat{l} \, ] \, \times \, [ \, \overline{\widehat{l}} \otimes V \, ]_{\pm}
\]
acting, respectively, on
\[
\mathcal{H}_{{}_{[ \, \widehat{n} \cdot \widehat{l} \, ] \, \times \, [ \, \overline{\widehat{l}} \otimes V \, ]}}^{\pm},
\]
with the (Fourier transform of the) grading operator $\widetilde{\mathbb{G}}$ defining equivalence
between them. Thus, it is sufficient to decompose the representation (\ref{[n.l]x[loV]})
of $\widetilde{\mathbb{S}^1} \times SU(2, \mathbb{C}) \times SU(2, \mathbb{C})$ into irreducible components, 
which in turn is equivalent to the task of decomposition of the factor 
representation $\overline{\widehat{l}} \otimes V$ of $SU(2, \mathbb{C})$ into irreducible components,
by the fact that the outer Kronecker product of the product group is irreducible iff the factor
representations are irreducible. Then we pick up the canonical basis in each of the direct summands
and look for linear combinations of vectors which span representations of the same weight, which are
nonzero under the action of the respective projector 
$\mathbb{P}_{{}_{\mathfrak{J} \,\,\, \pm|\lambda_{{}{nl}}|}}$.
This procedure would have required a considerable further analysis for general allowable $V$. We will now 
use the fact that explicit determination of the eigenstates corresponding to the fundamental allowable $V = \tfrac{1}{2} \oplus \tfrac{1}{2}$ is sufficient for the explicit determination of the eigenstates
corresponding to all allowable $V$. Therefore, in the determination of the concrete 
matrices (with arbitrary but fixed $\widehat{n} \cdot \widehat{l}$)
\[
\mathcal{F}\big(\phi^{a}_{{}_{{V},s,\pm |\lambda_{{}_{n,l}}|}}\big)_{{}_{ji}}(\widehat{n} \cdot \widehat{l})
\,\,\,\,\,
1 \leq a \leq d, \,\, 
-l \leq i,j \leq l,
\,\,\,
\widehat{n} \cdot \widehat{l} \,\, \textrm{fixed},
\]
we confine ourselves to the fundamental allowable $V=\tfrac{1}{2} \oplus \tfrac{1}{2}$ with $d=4$.
In this case determination of all $(2l+1)^2 2$ (Fourier transformed) eigenstates 
$\mathcal{F}\big(\phi^{a}_{{}_{{V},s,\pm |\lambda_{{}_{n,l}}|}}\big)_{{}_{ji}}$ concentrated on 
$\widehat{n} \cdot \widehat{l}$, \emph{i. e.} such that
\[
\mathcal{F}\big(\phi^{a}_{{}_{{V},s,\pm |\lambda_{{}_{n,l}}|}}\big)_{{}_{ji}}(\widehat{n'} \cdot \widehat{l'}) = \delta_{nn'}\delta_{ll'}
\mathcal{F}\big(\phi^{a}_{{}_{{V},s,\pm |\lambda_{{}_{n,l}}|}}\big)_{{}_{ji}}(\widehat{n} \cdot \widehat{l}),
\] 
reduces to the determination of four complex numbers
$a^{\pm}_{l}$, $b^{\pm}_{l}$, $c^{\pm}_{l}$, $d^{\pm}_{l}$ depending on $l$, which can be easily
computed.

Thus, we return to the fundamental \emph{allowed} $V$ equal (\ref{spinorV}),
and to the Hilbert space $\mathscr{H} = L^2(\widetilde{\mathbb{S}^1} \times SU(2, \mathbb{C}); \mathbb{C}^4)$ acted on by the representation (\ref{UphiOnSxG}) and by the Dirac operator $D_\mathfrak{J}$
of the spectral tuple (\ref{1stSpectralTupleForSxG})
or equivalently in (\ref{2ndSpectralTupleForSxG}) corresponding to 
\[
V = \widehat{{\textstyle\frac{1}{2}}} \oplus \widehat{{\textstyle\frac{1}{2}}}.
\]
We distinguish the two copies of the representation $\widehat{\tfrac{1}{2}}$ by the superscript
$\pm$, respectively, and write
\[
V = \widehat{{\textstyle\frac{1}{2}}}^{+} \oplus \widehat{{\textstyle\frac{1}{2}}}^{-}
\]
with the canonical basis vectors 
\[
\varepsilon^{+}_{{}_{i=1/2}} = 
\left( \begin{array}{c}  
1 \\
0\\
0 \\
0
 \end{array} \right), \,\,
 \varepsilon^{+}_{{}_{i=-1/2}} = 
\left( \begin{array}{c}  
0 \\
1\\
0 \\
0
 \end{array} \right) 
\]
in the representation subspace of the first copy $\widehat{{\textstyle\frac{1}{2}}}^{+}$
and with the canonical basis vectors 
\[
\varepsilon^{-}_{{}_{i=1/2}} = 
\left( \begin{array}{c}  
0 \\
0\\
1 \\
0
 \end{array} \right), \,\,
 \varepsilon^{-}_{{}_{i=-1/2}} = 
\left( \begin{array}{c}  
0 \\
0\\
0 \\
1
 \end{array} \right) 
\]
in the representation subspace of the second copy $\widehat{{\textstyle\frac{1}{2}}}^{-}$
with both copies acting in 
\[
\mathbb{C}^4 =  \mathcal{H}_V = \mathcal{H}_{\widehat{{\textstyle\frac{1}{2}}}^{+}} \oplus
\mathcal{H}_{\widehat{{\textstyle\frac{1}{2}}}^{-}}.
\]
Thus the representation (\ref{[n.l]x[loV]}) is unitarily equivalent to 
\begin{multline*}
[ \, \widehat{n} \cdot \widehat{l} \, ] \, \times \, [ \, \overline{\widehat{l}} \otimes V \, ]
\cong_U
[ \, \widehat{n} \cdot \widehat{l} \, ] \, \times \, [ \, \overline{\widehat{l}} \otimes \widehat{{\textstyle\frac{1}{2}}}^{+} \, ]
\,\, \bigoplus \,\,
[ \, \widehat{n} \cdot \widehat{l} \, ] \, \times \, [ \, \overline{\widehat{l}} \otimes \widehat{{\textstyle\frac{1}{2}}}^{-} \, ] \\
\cong_U
[ \, \widehat{n} \cdot \widehat{l} \, ] \, \times \, [ \, \widehat{l+{\textstyle\frac{1}{2}}} \, ]_+
\,\, \bigoplus \,\,
[ \, \widehat{n} \cdot \widehat{l} \, ] \, \times \, [ \, \widehat{l-{\textstyle\frac{1}{2}}} \, ]_+
\\
\,\,\,\,\, \bigoplus \,\,
[ \, \widehat{n} \cdot \widehat{l} \, ] \, \times \, [ \, \widehat{l+{\textstyle\frac{1}{2}}} \, ]_-
\,\, \bigoplus \,\,
[ \, \widehat{n} \cdot \widehat{l} \, ] \, \times \, [ \, \widehat{l-{\textstyle\frac{1}{2}}} \, ]_-
\end{multline*}
Let us introduce the canonical basis vectors $e_j$, $-l \leq j \leq l$,
and similarly $\overline{e}_k$, $-l \leq k \leq l$, 
in the Hilbert spaces (compare \cite{Geland-Minlos-Shapiro} for definition of the canonical basis
of irreducible representations $\widehat{l}$ of $SU(2, \mathbb{C})$\footnote{Recall that our first generator has opposite sign in comparison to that used in \cite{Geland-Minlos-Shapiro}.}) 
\[
\mathcal{H}_{\widehat{l}} \,\,\, \textrm{and} \,\,\,
\mathcal{H}_{\overline{\,\widehat{l}\,}},
\]
respectively.
Using the Clebsch-Gordan coefficients we can now express the canonical basis elements 
$e^{(l+1/2) \,\, +}_{m}$, $-(l+1/2) \leq m \leq l+1/2$, which span the direct summand
$\widehat{l + \tfrac{1}{2}}_+$ in the tensor product representation
\[
\overline{\widehat{l}} \otimes \widehat{{\textstyle\frac{1}{2}}}^{+} \cong_U
\widehat{l+{\textstyle\frac{1}{2}}}_+ \oplus \widehat{l-{\textstyle\frac{1}{2}}}_+
\]
in terms of the basis vectors $\overline{e}_k \otimes \varepsilon^{+}_{i}$:
\[
e^{(l+1/2) \,\, +}_{m} = 
\sqrt{{\textstyle\frac{l+m+{\textstyle\frac{1}{2}}}{2l+1}}}
 \,\,
\overline{e}_{-(m-1/2)} \otimes \varepsilon^{+}_{1/2}
+
\sqrt{{\textstyle\frac{l-m+{\textstyle\frac{1}{2}}}{2l+1}}} 
 \,\, 
\overline{e}_{-(m+1/2)} \otimes \varepsilon^{+}_{-1/2},
\]
\[
-(l+1/2) \leq m \leq l+1/2,
\]
compare  \cite{Geland-Minlos-Shapiro}, Part I, Sect. 10, \S 1, \S2.
Similarly, we can express the canonical basis elements
 $e^{(l-1/2) \,\, +}_{m}$, $-(l-1/2) \leq m \leq l-1/2$,
which span the direct summand
$\widehat{l - \tfrac{1}{2}}_+$ in the tensor product representation
\[
\overline{\widehat{l}} \otimes \widehat{{\textstyle\frac{1}{2}}}^{+} \cong_U
\widehat{l+{\textstyle\frac{1}{2}}}_+ \oplus \widehat{l-{\textstyle\frac{1}{2}}}_+
\]
in terms of the basis vectors $\overline{e}_k \otimes \varepsilon^{+}_{i}$:
\[
e^{(l-1/2) \,\, +}_{m} = -
\sqrt{{\textstyle\frac{l-m+{\textstyle\frac{1}{2}}}{2l+1}}}
 \,\,
\overline{e}_{-(m-1/2)} \otimes \varepsilon^{+}_{1/2}
+ 
\sqrt{{\textstyle\frac{l-m+{\textstyle\frac{1}{2}}}{2l+1}}}
 \,\, 
\overline{e}_{-(m+1/2)} \otimes \varepsilon^{+}_{-1/2},
\]
\[
-(l-1/2) \leq m \leq l-1/2.
\] 
Analogously  we can express the canonical basis elements
 $e^{(l+1/2) \,\, -}_{m}$, $-(l+1/2) \leq m \leq l+1/2$,
which span the direct summand
$\widehat{l + \tfrac{1}{2}}_-$ in the tensor product representation
\[
\overline{\widehat{l}} \otimes \widehat{{\textstyle\frac{1}{2}}}^{-} \cong_U
\widehat{l+{\textstyle\frac{1}{2}}}_- \oplus \widehat{l-{\textstyle\frac{1}{2}}}_-
\]
in terms of the basis vectors $\overline{e}_k \otimes \varepsilon^{-}_{i}$:
\[
e^{(l+1/2) \,\, -}_{m} = 
\sqrt{{\textstyle\frac{l+m+{\textstyle\frac{1}{2}}}{2l+1}}}
 \,\,
\overline{e}_{-(m-1/2)} \otimes \varepsilon^{-}_{1/2}
+ 
\sqrt{{\textstyle\frac{l-m+{\textstyle\frac{1}{2}}}{2l+1}}}
 \,\, 
\overline{e}_{-(m+1/2)} \otimes \varepsilon^{-}_{-1/2},
\]
\[
-(l+1/2) \leq m \leq l+1/2.
\]
Similarly,  we can express the canonical basis elements
 $e^{(l-1/2) \,\, -}_{m}$, $-(l-1/2) \leq m \leq l-1/2$,
which span the direct summand
$\widehat{l - \tfrac{1}{2}}_-$ in the tensor product representation
\[
\overline{\widehat{l}} \otimes \widehat{{\textstyle\frac{1}{2}}}^{-} \cong_U
\widehat{l+{\textstyle\frac{1}{2}}}_- \oplus \widehat{l-{\textstyle\frac{1}{2}}}_-
\]
in terms of the basis vectors $\overline{e}_k \otimes \varepsilon^{-}_{i}$:
\[
e^{(l-1/2) \,\, -}_{m} = 
-\sqrt{{\textstyle\frac{l-m+{\textstyle\frac{1}{2}}}{2l+1}}}
 \,\,
\overline{e}_{-m+1/2} \otimes \varepsilon^{-}_{1/2}
+ 
\sqrt{{\textstyle\frac{l-m+{\textstyle\frac{1}{2}}}{2l+1}}}
 \,\, 
\overline{e}_{-m-1/2} \otimes \varepsilon^{-}_{-1/2},
\]
\[
-(l-1/2) \leq m \leq l-1/2.
\] 
Note that the Fourier transformed representation  (\ref{UphiOnSxG}), and thus also the representation
(\ref{[n.l]x[loV]}) commutes with the projectors 
${P}_{{}_{\mathfrak{J} \,\,\, \pm|\lambda_{{}{nl}}|}}(\widehat{n} \cdot \widehat{l})$. 
Because on the other hand the representations 
\[
[ \, \widehat{n} \cdot \widehat{l} \, ] \, \times \, [ \, \overline{\widehat{l}} \otimes V \, ]_{\pm}
\]
have to be equivalent to the representation
\[
[ \, \widehat{n} \cdot \widehat{l} \, ] \, \times \, [ \, \widehat{l+{\textstyle\frac{1}{2}}} \, ]
\,\, \bigoplus \,\,
[ \, \widehat{n} \cdot \widehat{l} \, ] \, \times \, [ \, \widehat{l-{\textstyle\frac{1}{2}}} \, ]
\]
then it follows that the subspace 
\[
\mathcal{H}_{{}_{[ \, \widehat{n} \cdot \widehat{l} \, ] \, \times \, [ \, \overline{\widehat{l}} \otimes V \, ]}}^{\pm} \subset 
\mathcal{H}_{{}_{[ \, \widehat{n} \cdot \widehat{l} \, ] \, \times \, [ \, \overline{\widehat{l}} \otimes V \, ]}}
\]
of the eigenspace which span the irreducible subrepresentation equivalent to
\[
[ \, \widehat{n} \cdot \widehat{l} \, ] \, \times \, [ \, \widehat{l+{\textstyle\frac{1}{2}}} \, ]
\]
must be spanned by the vectors
\[
{P}_{{}_{\mathfrak{J} \,\,\, \pm |\lambda_{{}{nl}}|}}(\widehat{n} \cdot \widehat{l}) e_j \otimes e^{(l+1/2) \,\, +}_{m} \,\,\, \textrm{and} \,\,\, 
{P}_{{}_{\mathfrak{J} \,\,\, \pm |\lambda_{{}{nl}}|}}(\widehat{n} \cdot \widehat{l}) e_j \otimes e^{(l+1/2) \,\, -}_{m}
\]
\[
-(l+1/2) \leq m \leq l+1/2.
\]
Similarly  it follows that the eigen-subspace of 
\[
\mathcal{H}_{{}_{[ \, \widehat{n} \cdot \widehat{l} \, ] \, \times \, [ \, \overline{\widehat{l}} \otimes V \, ]}}^{\pm} \subset 
\mathcal{H}_{{}_{[ \, \widehat{n} \cdot \widehat{l} \, ] \, \times \, [ \, \overline{\widehat{l}} \otimes V \, ]}}
\]
which span the irreducible subrepresentation equivalent to
\[
[ \, \widehat{n} \cdot \widehat{l} \, ] \, \times \, [ \, \widehat{l-{\textstyle\frac{1}{2}}} \, ]
\]
must be spanned by the vectors
\[
{P}_{{}_{\mathfrak{J} \,\,\, \pm |\lambda_{{}{nl}}|}}(\widehat{n} \cdot \widehat{l}) e_j \otimes e^{(l-1/2) \,\, +}_{m} \,\,\, \textrm{and} \,\,\, 
{P}_{{}_{\mathfrak{J} \,\,\, \pm |\lambda_{{}{nl}}|}}(\widehat{n} \cdot \widehat{l}) e_j \otimes e^{(l-1/2) \,\, -}_{m}
\]
\[
-(l-1/2) \leq m \leq l-1/2.
\]
It then follows that there exist such numbers $a^{+}_{l}$, $b^{+}_{l}$, $c^{+}_{l}$,
$d^{+}_{l}$, that 
\begin{multline*}
a^{+}_{l} {P}_{{}_{\mathfrak{J} \,\,\, +|\lambda_{{}{nl}}|}}(\widehat{n} \cdot \widehat{l}) e_j \otimes e^{(l+1/2) \,\, +}_{m} 
+
b^{+}_{l} {P}_{{}_{\mathfrak{J} \,\,\, +|\lambda_{{}{nl}}|}}(\widehat{n} \cdot \widehat{l}) e_j \otimes e^{(l+1/2) \,\, -}_{m},
\\
c^{+}_{l} {P}_{{}_{\mathfrak{J} \,\,\, +|\lambda_{{}{nl}}|}}(\widehat{n} \cdot \widehat{l}) e_j \otimes e^{(l-1/2) \,\, +}_{m'}
+ d^{+}_{l} {P}_{{}_{\mathfrak{J} \,\,\, +|\lambda_{{}{nl}}|}}(\widehat{n} \cdot \widehat{l}) e_j \otimes e^{(l-1/2) \,\, -}_{m'}
\end{multline*}
\[
-(l+1/2) \leq m \leq l+1/2, \,\,\,\,
-(l-1/2) \leq m' \leq l-1/2,
\]
\[
-l \leq j \leq l
\]
span
\[
\mathcal{H}_{{}_{[ \, \widehat{n} \cdot \widehat{l} \, ] \, \times \, [ \, \overline{\widehat{l}} \otimes V \, ]}}^{+}.
\]
We choose them in such a manner that 
\[
a^{+}_{l} {P}_{{}_{\mathfrak{J} \,\,\, +|\lambda_{{}{nl}}|}}(\widehat{n} \cdot \widehat{l}) e_j \otimes e^{(l+1/2) \,\, +}_{m} 
+
b^{+}_{l} {P}_{{}_{\mathfrak{J} \,\,\, +|\lambda_{{}{nl}}|}}(\widehat{n} \cdot \widehat{l}) e_j \otimes e^{(l+1/2) \,\, -}_{m} \neq 0
\]
and 
\[
c^{+}_{l} {P}_{{}_{\mathfrak{J} \,\,\, -|\lambda_{{}{nl}}|}}(\widehat{n} \cdot \widehat{l}) e_j \otimes e^{(l-1/2) \,\, +}_{m'}
+ d^{+}_{l} {P}_{{}_{\mathfrak{J} \,\,\, -|\lambda_{{}{nl}}|}}(\widehat{n} \cdot \widehat{l}) e_j \otimes e^{(l-1/2) \,\, -}_{m'} \neq 0,
\]
are vectors of unit norm (which by construction is possible), and they will automatically be orthonormal. 
Analogously we can choose the numbers $a^{-}_{l}$, $b^{-}_{l}$, $c^{-}_{l}$,
$d^{-}_{l}$, such that 
\begin{multline*}
a^{-}_{l} {P}_{{}_{\mathfrak{J} \,\,\, -|\lambda_{{}{nl}}|}}(\widehat{n} \cdot \widehat{l}) e_j \otimes e^{(l+1/2) \,\, +}_{m} 
+
b^{-}_{l} {P}_{{}_{\mathfrak{J} \,\,\, -|\lambda_{{}{nl}}|}}(\widehat{n} \cdot \widehat{l}) e_j \otimes e^{(l+1/2) \,\, -}_{m},
\\
c^{-}_{l} {P}_{{}_{\mathfrak{J} \,\,\, -|\lambda_{{}{nl}}|}}(\widehat{n} \cdot \widehat{l}) e_j \otimes e^{(l-1/2) \,\, +}_{m'}
+ d^{-}_{l} {P}_{{}_{\mathfrak{J} \,\,\, -|\lambda_{{}{nl}}|}}(\widehat{n} \cdot \widehat{l}) e_j \otimes e^{(l-1/2) \,\, -}_{m'}
\end{multline*}
\[
-(l+1/2) \leq m \leq l+1/2, \,\,\,\,
-(l-1/2) \leq m' \leq l-1/2,
\]
\[
-l \leq j \leq  l,
\]
span
\[
\mathcal{H}_{{}_{[ \, \widehat{n} \cdot \widehat{l} \, ] \, \times \, [ \, \overline{\widehat{l}} \otimes V \, ]}}^{-},
\]
and are orthonormal. By construction, we have in each case $(2l+1)^2 \, 2$ orthonormal vectors
which span the eigenspace corresponding to the eigenvalue $\pm |\lambda_{{}{nl}}|$.

Summing up, the complete system of 
orthonormal ``eigen-matrices'' 
\[
\mathcal{F}\big(\phi_{{}_{{V},s,\pm |\lambda_{{}_{n,l}}|}}\big)(\widehat{n}\cdot \widehat{l}),
\,\,\,\,\,
\widehat{n} \cdot \widehat{l} \,\, \textrm{fixed},
\]
corresponding to the fundamental allowable $V=\widehat{\tfrac{1}{2}} \oplus 
\widehat{\tfrac{1}{2}}$ with $d=4$, is
numbered by the index $s \in \{ j \times m , -(l+1/2) \leq m \leq l+1/2, -l \leq j \leq l \} 
\sqcup \{j\times m',
-(l-1/2) \leq m' \leq l-1/2, -l \leq j \leq l\}$ , and is equal
\begin{multline}\label{F(phi^a_V,lambda)}
\mathcal{F}\big(\phi_{{}_{{V},j \times m,\pm |\lambda_{{}_{n,l}}|}}\big)(\widehat{n}\cdot \widehat{l}) = \\
a^{\pm}_{l} 
\sqrt{{\textstyle\frac{l+m+{\textstyle\frac{1}{2}}}{2l+1}}}
 \,\,
{P}_{{}_{\mathfrak{J} \,\,\, \pm|\lambda_{{}{nl}}|}}(\widehat{n} \cdot \widehat{l}) \, e_j \otimes \overline{e}_{-(m-1/2)} \otimes \varepsilon^{+}_{1/2}
\\
+ 
a^{\pm}_{l} 
\sqrt{{\textstyle\frac{l-m+{\textstyle\frac{1}{2}}}{2l+1}}}
 \,\, 
{P}_{{}_{\mathfrak{J} \,\,\, \pm|\lambda_{{}{nl}}|}}(\widehat{n} \cdot \widehat{l}) \, e_j \otimes \overline{e}_{-(m+1/2)} \otimes \varepsilon^{+}_{-1/2}
 \\
+
b^{\pm}_{l} 
\sqrt{{\textstyle\frac{l+m+{\textstyle\frac{1}{2}}}{2l+1}}}
 \,\,
{P}_{{}_{\mathfrak{J} \,\,\, \pm|\lambda_{{}{nl}}|}}(\widehat{n} \cdot \widehat{l}) \, e_j \otimes \overline{e}_{-(m-1/2)} \otimes \varepsilon^{-}_{1/2}
\\
+ 
b^{\pm}_{l} 
\sqrt{{\textstyle\frac{l-m+{\textstyle\frac{1}{2}}}{2l+1}}}
 \,\, 
{P}_{{}_{\mathfrak{J} \,\,\, \pm|\lambda_{{}{nl}}|}}(\widehat{n} \cdot \widehat{l}) \, e_j \otimes \overline{e}_{-(m+1/2)} \otimes \varepsilon^{-}_{-1/2}, 
\\
\mathcal{F}\big(\phi_{{}_{{V},j \times m',\pm |\lambda_{{}_{n,l}}|}}\big)(\widehat{n}\cdot \widehat{l}) =  \,\,\,\,\,\,\,\,\,\,\,\,\,\,\,\,\,\,\,\,\,\,\,\,\,\,\,\,\,\,\,\,\,\,\,
\,\,\,\,\,\,\,\,\,\,\,\,\,\,\,\,\,\,\,\,\,\,\,\,\,\,\,\,\,\,\,\,\,\,\,\,\,\,\,\,\,\,\,\,\,\,
\,\,\,\,\,\,\,\,\,\,\,\,\,\,\,\,\,\,\,\,\,\,\,\,\,\,\,\,\,\,\,\,\,\,\,\,\,\,\,\,\,\,\,\,\,\,\,
\,\,\,\,\,\,\,  \\ 
-c^{\pm}_{l} 
\sqrt{{\textstyle\frac{l-m+{\textstyle\frac{1}{2}}}{2l+1}}}
 \,\,
{P}_{{}_{\mathfrak{J} \,\,\, \pm|\lambda_{{}{nl}}|}}(\widehat{n} \cdot \widehat{l}) \, e_j \otimes \overline{e}_{-(m-1/2)} \otimes \varepsilon^{+}_{1/2}
\\
+ 
c^{\pm}_{l} 
\sqrt{{\textstyle\frac{l-m+{\textstyle\frac{1}{2}}}{2l+1}}}
 \,\, 
{P}_{{}_{\mathfrak{J} \,\,\, \pm|\lambda_{{}{nl}}|}}(\widehat{n} \cdot \widehat{l}) \, e_j \otimes \overline{e}_{-(m+1/2)} \otimes \varepsilon^{+}_{-1/2} 
\\
- d^{\pm}_{l} 
\sqrt{{\textstyle\frac{l-m+{\textstyle\frac{1}{2}}}{2l+1}}}
 \,\,
{P}_{{}_{\mathfrak{J} \,\,\, \pm|\lambda_{{}{nl}}|}}(\widehat{n} \cdot \widehat{l}) \, e_j \otimes \overline{e}_{-m+1/2} \otimes \varepsilon^{-}_{1/2}
\\
+ 
d^{\pm}_{l} 
\sqrt{{\textstyle\frac{l-m+{\textstyle\frac{1}{2}}}{2l+1}}}
 \,\, 
{P}_{{}_{\mathfrak{J} \,\,\, \pm|\lambda_{{}{nl}}|}}(\widehat{n} \cdot \widehat{l}) \, e_j \otimes \overline{e}_{-m-1/2} \otimes \varepsilon^{-}_{-1/2}, 
\end{multline}
\[
V=\widehat{{\textstyle\frac{1}{2}}} \oplus \widehat{{\textstyle\frac{1}{2}}}, 
\]
\[
-(l+1/2) \leq m \leq l+1/2, \,\,\,\,\,
-(l-1/2) \leq m' \leq l-1/2, \,\,\,\,\, 
-l \leq j \leq l.
\]
Note here that for fixed indices $i,j,a$ and for the fixed point $\widehat{n} \cdot \widehat{l}$
of the dual group, the matrix 
\[
F(\widehat{n}\cdot \widehat{l}) = \mathcal{F}\big(\phi\big)(\widehat{n}\cdot \widehat{l}),
\,\,\,\,\,
\widehat{n} \cdot \widehat{l} \,\, \textrm{fixed},
\]
which has all coefficients equal zero except the coefficient
\[
F^{a}_{{}_{ji}}(\widehat{n}\cdot \widehat{l}) =
\mathcal{F}\big(\phi^a\big)_{{}_{ji}}(\widehat{n}\cdot \widehat{l}) = 
1
\]
is represented by the following simple tensor
\[
F(\widehat{n}\cdot \widehat{l}) = \mathcal{F}\big(\phi\big)(\widehat{n}\cdot \widehat{l})
= e_j \otimes \overline{e}_{-i} \otimes \varepsilon^{+}_{1/2}, 
\,\,\,\,
\textrm{iff}
\,\,\,\, a=1,
\]
\[
F(\widehat{n}\cdot \widehat{l}) = \mathcal{F}\big(\phi\big)(\widehat{n}\cdot \widehat{l})
= e_j \otimes \overline{e}_{-i} \otimes \varepsilon^{+}_{-1/2}, 
\,\,\,\,
\textrm{iff}
\,\,\,\, a=2,
\]
\[
F(\widehat{n}\cdot \widehat{l}) = \mathcal{F}\big(\phi\big)(\widehat{n}\cdot \widehat{l})
= e_j \otimes \overline{e}_{-i} \otimes \varepsilon^{-}_{1/2}, 
\,\,\,\,
\textrm{iff}
\,\,\,\, a=3,
\]
\[
(\widehat{n}\cdot \widehat{l}) = \mathcal{F}\big(\phi\big)(\widehat{n}\cdot \widehat{l})
= e_j \otimes \overline{e}_{-i} \otimes \varepsilon^{-}_{-1/2}, 
\,\,\,\,
\textrm{iff}
\,\,\,\, a=4;
\]
or, more generally,
\[
F(\widehat{n}\cdot \widehat{l}) = \mathcal{F}\big(\phi\big)(\widehat{n}\cdot \widehat{l})
= e_j \otimes \overline{e}_{-i} \otimes \overline{\overline{e}}_a, 
\]
where $\overline{\overline{e}}_1, \ldots, \overline{\overline{e}}_a, \ldots, \overline{\overline{e}}_d$
is the canonical basis in the Hilbert space $\mathcal{H}_V$, and where $e_j$, $-l \leq j \leq l$ 
and $\overline{e}_i$, $-l \leq i \leq l$ are the canonical basis elements in the Hilbert spaces
$\mathcal{H}_{\widehat{l}}$ and $\mathcal{H}_{\overline{\, \widehat{l}\,}}$, respectively.
Of course in the formula (\ref{F(phi^a_V,lambda)}) the projector 
${P}_{{}_{\mathfrak{J} \,\,\, \pm|\lambda_{{}{nl}}|}}(\widehat{n} \cdot \widehat{l})$
acts on the whole simple tensor $e_j \otimes \overline{e}_{i} \otimes \varepsilon^{-}_{k}$
standing after it.

Before passing to QFT on the Einstein Universe we should emphasize that the formulas for 
$\widehat{\Gamma}$ and $\Gamma$ in the Dirac operators
$D$, $D_\mathfrak{J}$ in the space-time spectral tuple (\ref{1stSpectralTupleForSxG})
or equivalently in (\ref{2ndSpectralTupleForSxG}), together with the commutation rules
(\ref{[X,X]}) are valid in the units in which $\hbar = c = R = 1$, where $R$ is the radius
of the Cauchy surface $t=\textrm{const.}:$ $SU(2, \mathbb{C}) = \mathbb{S}^3$ of the Einstein Universe. In other units
the operators $\widehat{\Gamma}$ and $\Gamma$ respectively in $D$, $D_\mathfrak{J}$  will get the
additional factor $\tfrac{1}{R}$, the invariant vector fields $X^1, X^2, X^3$, regarded as 
differential operators, gain the factor $\tfrac{1}{R}$, the operators $P^1, P^2, P^3$ gain the additional 
factor $\tfrac{\hbar}{R}$, and finally the operator $P^0$ gains the factor $\tfrac{\hbar c}{R}$,
so that:
\[
\widehat{\Gamma} = - {\textstyle\frac{i}{2R}}
\widehat{\Gamma}^{{}^{1}} \widehat{\Gamma}^{{}^{2}} \widehat{\Gamma}^{{}^{3}},
\,\,\,\,
\Gamma = i \widehat{\Gamma} = {\textstyle\frac{i}{2R}} \Gamma^1 \Gamma^2 \Gamma^3,
\]
and 
\[
D = {\textstyle\frac{1}{\hbar c}} \widehat{\Gamma}^{{}^{0}}P^0 + 
{\textstyle\frac{1}{\hbar}}\widehat{\Gamma}^{{}^{1}}P^1 +
{\textstyle\frac{1}{\hbar}} \widehat{\Gamma}^{{}^{2}}P^2 + 
{\textstyle\frac{1}{\hbar}} \widehat{\Gamma}^{{}^{3}}P^3
-{\textstyle\frac{i}{2R}}
\widehat{\Gamma}^{{}^{1}} \widehat{\Gamma}^{{}^{2}} \widehat{\Gamma}^{{}^{3}},
\]
\[
D_\mathfrak{J} = {\textstyle\frac{1}{\hbar c}} \Gamma^{{}^{0}}P^0 + 
{\textstyle\frac{1}{\hbar}} \Gamma^{{}^{1}}P^1 +
{\textstyle\frac{1}{\hbar}} \Gamma^{{}^{2}}P^2 + 
{\textstyle\frac{1}{\hbar}} \Gamma^{{}^{3}}P^3
+{\textstyle\frac{i}{2R}} \Gamma^1 \Gamma^2 \Gamma^3,
\]
\[
\Delta =  {\textstyle\frac{1}{\hbar^2 c^2}} \big(P^0\big)^2 + 
{\textstyle\frac{1}{\hbar^2}} \big(P^1\big)^2 + {\textstyle\frac{1}{\hbar^2}} \big(P^2\big)^2 + {\textstyle\frac{1}{\hbar^2}} \big(P^3\big)^2 
+ {\textstyle\frac{1}{4R^2}} \boldsymbol{1},
\]
\[
\square = {\textstyle\frac{1}{\hbar^2 c^2}}\big(P^0\big)^2 - 
{\textstyle\frac{1}{\hbar^2}} \big(P^1\big)^2 - 
{\textstyle\frac{1}{\hbar^2}} \big(P^2\big)^2 - {\textstyle\frac{1}{\hbar^2}} \big(P^3\big)^2
-  {\textstyle\frac{1}{4R^2}}\boldsymbol{1}, 
\]
and similarly the commutation rules (\ref{[X,X]}) will get additional factor $\tfrac{1}{R}$ or 
$\tfrac{\hbar}{R}$ on the right-hand sides:
\begin{align*}\label{[X,X]*}
[X^2,X^1] =  {\textstyle\frac{1}{R}} \, X^3  
& 
\,\,\,\,\,\,\,\,\,\,\,\,\,\,\,\,\,\,\,\,\,\,\,\, \textrm{or}
&
[P^2,P^1] = {\textstyle\frac{i\hbar}{R}} \, P^3 
\\
[X^3,X^1] = - {\textstyle\frac{1}{R}} \, X^2  
&
\,\,\,\,\,\,\,\,\,\,\,\,\,\,\,\,\,\,\,\,\,\,\,\, \textrm{or}
&
[P^3,P^1] = - {\textstyle\frac{i\hbar}{R}} \, P^2 
\\
[X^3,X^2] =  {\textstyle\frac{1}{R}}\, X^1  
& 
\,\,\,\,\,\,\,\,\,\,\,\,\,\,\,\,\,\,\,\,\,\,\,\, \textrm{or}
&
[P^3,P^2] =  {\textstyle\frac{i\hbar}{R}}\, P^1
\\
[X^0, X^k] = 0, 
&
\,\,\,\,\,\,\,\,\,\,\,\,\,\,\,\,\,\,\,\,\,\,\,\, \textrm{or}
&
[P^0, P^k] = 0, 
\\
P^0 =  i\hbar X^0  
&
\,\,\,\,\,\,\,\,\,\,\,\,\,\,\,\,\,\,\,\,\,\,\,\, \textrm{and}
&
P^k = i\hbar X^k,
\end{align*}
on $L^2(\widetilde{\mathbb{S}^1}  \times SU(2, \mathbb{C}))$.
Thus, in the limit $R \rightarrow \infty$ when the Einstein Universe becomes (locally at least)
indistinguishable from the flat Minkowski space-time the vector fields $X^1, X^2, X^3$
and thus the operators $P^1, P^2, P^3$ become
commutative, and pass (locally at least) into the ordinary translation operators on the 
flat Minkowski space-time.
Similarly, the operators $D$, $D_\mathfrak{J}$, $\Delta$, $\square$ pass (at least locally)
into their flat counterparts $D$, $D_\mathfrak{J}$,$\Delta$, $\square$ when $R \rightarrow \infty$.
For a deeper discussion on the causal relationship of the flat Minkowski space-time
to the Einstein Universe we refer to the following works of Segal and his co-workers:
\cite{SegalPNAS}, \cite{PaneitzSegalI}-\cite{PaneitzSegalIII}, \cite{SegalZhouQED},
\cite{SegalEU}. Note however that the normalizations used in \cite{SegalPNAS}, 
\cite{PaneitzSegalI}-\cite{PaneitzSegalIII}, \cite{SegalZhouQED} are slightly different from ours. 
The right invariant vector fields $Y_1,Y_2, Y_3$ (and similarly the left invariant
vector fields $X_0,X_1, X_2, X_3$) of \cite{SegalPNAS}, \cite{PaneitzSegalI}-\cite{PaneitzSegalIII}, 
\cite{SegalZhouQED} have additional factor $2$ in comparison to our right invariant
vector fields $X^0,X^1, X^2, X^3$ on $\mathbb{R} \times SU(2, \mathbb{C})$. Correspondingly the Dirac operator used 
\cite{SegalPNAS}, \cite{PaneitzSegalI}-\cite{PaneitzSegalIII}, \cite{SegalZhouQED} get the additional
factor $2$ in comparison to our $D$, and similarly the invariant wave operator on the Einstein Universe
$\mathbb{R} \times SU(2, \mathbb{C}) = \mathbb{R} \times \mathbb{S}^3$ used in  
\cite{SegalPNAS}, \cite{PaneitzSegalI}-\cite{PaneitzSegalIII}, \cite{SegalZhouQED}
(and denoted there $\square_c$) is equal $4\square$ in terms of the invariant wave operator (say generalized d'Alembertian) $\square$ used by us. 
Besides this in our constructions the roles of right and left invariant vector fields and 
correspondingly the right anf left regular representations are reversed in comparison
to \cite{SegalPNAS}, \cite{PaneitzSegalI}-\cite{PaneitzSegalIII}, \cite{SegalZhouQED}.
This differences are of course completely inessential for the analysis of representations 
acting in single particle Hilbert spaces and in Fock spaces of local fields on the Einstein
Universe presented in \cite{PaneitzSegalI}-\cite{PaneitzSegalIII},  which will be used here.

0

\subsection{White noise construction of free local quantum fields on the Einstein Universe}\label{WhiteNoiseFreeFieldsonEU}

Before continuing our discussion of the extension of the space-time tuple
(\ref{SpacetimeTupleFields}) of Subsection \ref{G}, over to the quantum fields on the Einstein Universe,  
we will construct the free fields usig Hida operators as creation-annihilation operators, 
and then the interacting fields, using the causal 
perturbative method. 
We will use the results
of \cite{PaneitzSegalI}-\cite{PaneitzSegalIII}, \cite{SegalZhouQED}, 
especially concerning the single particle Hilbert spaces of free local fields on the Einstein Universe,
but we introduce several modifications. Namely, we will construct the quantum free fields with the help of Hida operators as the creation-annilhilataion operators of free fields in the ``momentum picture''
(\emph{i.e.} using Fourier transformed states). 
Further, we will use explicitly the Fourier transform as the central technique, in the form adequate
for the implementation of the white noise Hida operators along the lines perfectly analogues to the white noise construction of free fields on the Minkowski space-time, which we have presented in
Sections \ref{e+e-}, \ref{white-noise-proofs}. The Fourier transform was also used in 
\cite{PaneitzSegalI}-\cite{PaneitzSegalIII}, but only implicitly through the harmonic analysis on the (compactified) isometry group of the Einstein Universe, but in the form not very much adequate
for the white noise construction of free quantum fields. Our white noise construction of free fields
on the Einstein Universe will be given in a form which allows immediatte application of the theorems on integral kernel generalized operators, which we have already presented in Subsections \ref{psiBerezin-Hida}, \ref{OperationsOnXi} and Section \ref{A(1)psi(1)}, with the convolution on the Minkowski space-time replaced with the convolution on the Einstein Universe which, as a group, also allows the correponding and natural convolution operation on the function and distribution spaces
over the Einstein Universe. This allows to avoid unnecessary repeatitions and, which is crucial,
will allow us to prove that the massive free fields on the Einstein Universe (and higher order contributions to massive or massless fields)
even evaluated at single space-time point
are ordinary operators on the Fock space, which continuously transform the Hida space into itself.
The methods used in \cite{SegalZhouQED}, \cite{SegalZhouPhi4} are not effective in
showing these result for the massless fields (in fact this result was also proved in
\cite{SegalZhouQED}, \cite{PaneitzSegalI}-\cite{PaneitzSegalIII}, but only for massive free fields
on the Einstein Universe). Also our method is more effective in the proof of the 
Noether theorem involved into treatment of Wick polynomials (conserved currents) of free fields 
and their integrals over Cauchy surfaces.
Finally, we will use the causal perturbative
construction of interacting fields (method not used in \cite{SegalZhouQED}, \cite{SegalZhouPhi4}).
Again the causal method, with the free fields understood as generalized integral kernel
operators with vector-valued kernels (in the sense of \cite{obataJFA}), will allow us to prove
that each higher order contribution to interacting fields on the Einstein Universe, evaluated at space-time point, are ordinary operators on the Fock space transforming continuously the Hida space into itself. Similarly, using the white noise methods and causal perturbative approach we are able to prove
that the scattering generalized operator $S(g=1)$ is an ordinary operator on the Fock space transforming continuously the Hida space into itself.  
It is difficult to see how (if possible) one could achieve these results using the hamiltonian method of \cite{SegalZhouQED}, \cite{PaneitzSegalI}-\cite{PaneitzSegalIII}. 

First we construct the free fields (in the white noise sense
as vector valued integral kernel operators in the sense of Obata \cite{obataJFA}) on the Einstein Universe 
$\mathbb{R} \times SU(2, \mathbb{C}) = \mathbb{R} \times  \mathbb{S}^3$ using the white noise Hida operators as the creation-annihilation operators. We need to associate in a natural way the Gelfand triple corresponding to each single particle space of each free field underlying the considered QFT 
on the Einstein Universe, say QED on the Einstein Universe.  To this end we need to construct
the single particle Hilbert spaces of generalized distributional eigensolutons $\phi$ 
of the differential (say wave) operator, with the wave operator being determined by the equations of motion of the corresponding classical free field. 
The single particle states are distributional solutions corresponding to the generalized eigenvalue of that operator associated to the free field in question and are constructed as continuous functionals over the appropriate nuclear countably Hilbert test spaces $\mathcal{S}_{A}$ 
densely embedded into the appropriate Hilbert spaces $\mathscr{H}$.
In fact, they compose solutions of the corresponding classical free field equations.
Because the tangent bundle $T\big(\mathbb{R} \times SU(2, \mathbb{C}) \big) = T\big(\mathbb{R} \times  \mathbb{S}^3\big)$ is trivial, the distributional solutions (being always regular in our context) -- 
\emph{i.e.} continuous functionals on the nuclear test function space, can be identified with 
ordinary $\mathbb{C}^d$-valued functions $\phi$ on $\mathbb{R} \times SU(2, \mathbb{C})$. 
Depending on the free field in question we consider the Hilbert space 
\[
\mathscr{H} = L^2(\widetilde{\mathbb{S}^1} \times SU(2, \mathbb{C}); \mathbb{C}^d)
= \oplus_{1}^{d} L^2(\widetilde{\mathbb{S}^1} \times SU(2, \mathbb{C}); \mathbb{C})
\]
of $\mathbb{C}^d$-valued (equivalence classes of) square summable functions on the Einstein Universe
$\mathbb{R} \times SU(2, \mathbb{C})$, acted on by the representation
(\ref{UphiOnRxG}), with the finite dimensional representation $V$ of $G= SU(2, \mathbb{C})$ 
in the transformation formula (\ref{UphiOnRxG}) depending on the field in question. 

For these distributions we use the nuclear test space
\begin{multline}\label{TestSpaceOnRxS^3}
\bigoplus \limits_{1}^{d} \mathcal{A} =
\mathcal{S}(\mathbb{R} \times \mathbb{S}^3; \mathbb{C}^d)  =
\mathcal{S}(\mathbb{R} \times SU(2, \mathbb{C}); \mathbb{C}^d) \\ = 
\mathcal{S}(\mathbb{R}) \otimes \mathscr{C}^\infty(SU(2, \mathbb{C}); \mathbb{C}^d)
= \mathcal{S}_A(\mathbb{R} \times SU(2, \mathbb{C}); \mathbb{C}^d)
= \mathcal{S}_{\oplus A'}(\mathbb{R} \times SU(2, \mathbb{C}); \mathbb{C}^d)
\end{multline} 
as the space of space-time test $\mathbb{C}^d$-valued functions, with $d$ depending on the particular
field. Here $\mathcal{A}$ is the algebra entering the spectral tuple 
(\ref{1stSpectralTupleForRxG}) or (\ref{2ndSpectralTupleForRxG}), if we identify the operator
of pointwise  multiplication by a function $f \in \mathcal{A}$ on 
\[
\mathscr{H} = \bigoplus \limits_{1}^{d} L^2(\mathbb{R} \times SU(2, \mathbb{C}); \mathbb{C})
= L^2(\mathbb{R} \times SU(2, \mathbb{C}); \mathbb{C}^d),
\]
with the function $f$ itself. 

This nuclear spacetime test space $\oplus \mathcal{A}$ can be constructed as the standard countably Hilbert 
nuclear space $\mathcal{S}_A(\mathbb{R} \times SU(2, \mathbb{C}))$ (in the sense of Gelfand \cite{GelfandIV}, \cite{obataJFA}, \cite{obata-book} and in notation of \cite{obataJFA},
\cite{obata-book}) based on the standard operator $A$ equal to the direct sum
\begin{multline}\label{StandardAofTestSpaceOnRxS^3}
A = 
\bigoplus \Big[ \big(-{\textstyle\frac{d^2}{dt^2}} +t^2 + 1 \big)\otimes \boldsymbol{1} +
\boldsymbol{1} \otimes \Delta_{{}_{SU(2, \mathbb{C})}} \Big] \\
=\bigoplus \Big[ \Delta_{{}_{\mathbb{R}}} \otimes \boldsymbol{1} +
\boldsymbol{1} \otimes \Delta_{{}_{SU(2, \mathbb{C})}} + V'_{A'} \Big]
=\oplus \big[ \Delta +  V'_{A'} \big] = \oplus A'
\end{multline}
of $d$ copies of
\begin{multline*}
 \Delta_{{}_{\mathbb{R}}} \otimes \boldsymbol{1} +
\boldsymbol{1} \otimes \Delta_{{}_{SU(2, \mathbb{C})}}  + V'_{A'} \\
= \big(- \big( X^0\big)^2 \big) \otimes \boldsymbol{1} +
\boldsymbol{1} \otimes \Delta_{{}_{SU(2, \mathbb{C})}} + V'_{A'} \\
= - \big( X^0\big)^2 -\big(X^1\big)^2 - \big(X^2\big)^2 - \big(X^3\big)^2 
+ {\textstyle\frac{1}{4}} \boldsymbol{1} + V'_{A'}
=\Delta +  V'_{A}
\end{multline*}
on
\[
L^2(\mathbb{R} \times SU(2, \mathbb{C})) = L^2(\mathbb{R}) \otimes L^2(SU(2, \mathbb{C})),
\]
where
\[
\Delta_{{}_{SU(2, \mathbb{C})}} = -\big(X^1\big)^2 - \big(X^2\big)^2 - \big(X^3\big)^2 
+ {\textstyle\frac{1}{4}} \boldsymbol{1},
\,\,\,
\Delta_{{}_{\mathbb{R}}} = - \big( X^0\big)^2,
\]
and $V'_{A'}$ is the operator of pointwise multiplication by the function $t\times w \mapsto 1 + t^2$
on $L^2(\mathbb{R} \times SU(2, \mathbb{C});\mathbb{C})$.

The distributional solutions $\phi$ of the single particle space have a local transformation
law (\ref{UphiOnRxG}) with $V$ being a $d$-dimansional representation $V$ of
the group $G= SU(2, \mathbb{C})$, not necessary of the class (\ref{multispinorV}),
with $V$ being determined by the kind of local field in question and its local transformation
law. In fact the single particle space of distributional eigensolutions composes a representation of the symmetry group $\mathbb{R} \times G \times G 
= \mathbb{R} \times SU(2, \mathbb{C}) \times SU(2, \mathbb{C})$ of the Einstein Universe
$\mathbb{R} \times SU(2, \mathbb{C})$, determined by the transformation formula (\ref{UphiOnRxG})
and by the whole space of distributional eigensolutions of the wave operator corresponding 
to the field in question, which in general is assumed to be invariant under (\ref{UphiOnRxG}). 

Here the natural relation between the causal compactification $\overline{\mathcal{M}_0}^{\, c}$ of the Minkowski space-time $\mathcal{M}_0$ and the 
Einstein Universe is essential, and which essentially can be expressed by the causal periodic
embedding of  $\overline{\mathcal{M}_0}$
into the Einstein Universe $\mathbb{R} \times SU(2, \mathbb{C})$. The point 
is that all eigensolutions of the wave operators which serve as single particle Hilbert spaces
of local free fields on the Einstein Universe are necessarily periodic in time 
\emph{with a fixed common period for all solutions and all local fields}. This holds for 
all higher spin massive and massless fields as well as for the gauge fields such as the electromagnetic
potential field, and for the Dirac spinor field. In fact it holds for  all free fields underlying
the Standard Model. This result has been proved in the series of papers  
\cite{PaneitzSegalI}-\cite{PaneitzSegalIII}, and this is in fact the main result 
of \cite{PaneitzSegalI}-\cite{PaneitzSegalIII} which we will use here. 

Speaking more precisely, the Einstein Universe $\mathbb{R} \times SU(2, \mathbb{C})$ composes in a canonical fashion a causal (``conformal'' for space-time metric) covering of the Minkowski space-time, or rather smooth causality preserving covering of $\overline{\mathcal{M}_0}^{\, c}$ (with appropriate periodic boundary identifications) with the associated Deck discrete covering group generated by the following 
isometry $\zeta$
\[
\zeta: \, t\times w \mapsto (t + \pi) \times (\textrm{An} \, w), \,\,\,\,\,
t \times w \in \mathbb{R} \times SU(2, \mathbb{R}) = \mathbb{R} \times \mathbb{S}^3
\]
of the Einstein Universe $\mathbb{R} \times SU(2, \mathbb{C})$, where ``$\textrm{An} \, w$''
is the antipode of $w$ in $SU(2, \mathbb{C}) = \mathbb{S}^3$. This particular
form of $\zeta$ is valid in units in which $c=R=1$ where $R$ is the radius of 
the Cauchy surface $SU(2, \mathbb{C}) = \mathbb{S}^3$. For an elegant  analysis of this covering 
we refer to \cite{SegalEU}, \cite{SegalZhouQED}, but the reader may also consult 
\cite{HawkingEllis}, \S 5.1, Figure 14. In other words the Einstein 
Universe is the universal cover of the causal (conformal) compactification 
$\overline{\mathcal{M}_0}^{\,c}$ (with appropriately identified boundary points) of the Minkowski space-time $\mathcal{M}_0$.

The local fields on Einstein Universe fall into two classes: the eigensolutions of the corresponding  single-particle spaces  
\begin{enumerate}
\item[I)]
are invariant with respect to $\zeta$ (this is for the first class including the free electromagnetic 
potential field)
\item[II)]
or are invariant with respect to $\zeta^4$ (for the second class including the free Dirac spinor 
field).
\end{enumerate}   
In either case the eigen solutions composing the single particle states are functions on  
$\mathbb{R} \times SU(2, \mathbb{C})$, which are always invariant under $\zeta^4$ 
and are periodic in time with period $4\pi$ (in units in which $c=R=1$), and thus they effectively live on 
$\widetilde{\mathbb{S}^1} \times SU(2, \mathbb{C})$, irrespectively of the kind
of the free field in question. 

This theorem has been proved in \cite{PaneitzSegalI}-\cite{PaneitzSegalIII} (compare also
\cite{SegalZhouQED}) under the natural assumption that only those plane wave packet solutions are
allowed on the Einstein Universe, which arise as extensions of the Minkowski plane wave mass packets solutions on the Minkowski space-time $\mathcal{M}_0$, regarded as a causal submanifold of the Einstein
Universe which is equal to the universal periodic cover of $\overline{\mathcal{M}_0}^{\,c}$ with the
period $\zeta$. This assumption is not arbitrary. All the remaining plane wave  solutions 
on the Einstein Universe $\mathbb{R} \times SU(2, \mathbb{C})$ are unphysical,
and when restricted to the Minkowski manifold $\mathcal{M}_0$, regarded as a submanifold 
$\mathcal{M}_0 \subset \mathbb{R} \times SU(2, \mathbb{C})$, would give plane waves with 
four-momenta lying outside the light cone (tachyonic) and corresponding to scattered
waves (with the asymptotic \emph{in} and \emph{out} four-momenta differing from each other).
It turns out that in fact all plane wave mass packets on the Minkowski manifold 
$\mathcal{M}_0$, regarded as a submanifold 
$\mathcal{M}_0 \subset \mathbb{R} \times SU(2, \mathbb{C})$, can be uniquely and covariantly
(with respect to the Weyl group\footnote{Generated by the Poincar\'e group and the scaling transformations.}) extended over the whole Einstein Universe, 
and all these extensions turn out to be periodic, \emph{i.e.}
invariant under $\zeta$ or $\zeta^4$, depending on the kind of local field or type of wave, 
compare \cite{PaneitzSegalI}-\cite{PaneitzSegalIII}
(in particular Thm. 7.8 of \cite{PaneitzSegalII} for spinor plane waves, and 
\cite{PaneitzSegalIII} for the electromagnetic potential field). In fact the canonical relationship
between Minkowski plane wave packets and Einstein plane wave packets given by the Einstein Universe as
the canonical periodic cover of the causal compactification of the Minkowski spacetime (\cite{PaneitzSegalI}-\cite{PaneitzSegalIII}) in fact allows to compare the scattering phenomenna
involving the many particle plane wave generalized states on the Minkowski space-time, with the corresponding scattering phenomena on the Einstein Universe. We insist on the assumption that the effective cross sections for high energy scattering between the many particle plane wave states on the Minkowski space-time computed from the assumptions 1) -- 3) of  Introduction (confirmed experimentally) should be preserved in passing to the corresponding scattering phenomena between the corresponding states on the Einstein Universe. This would be impossible if we had ignored the natural relationship between the Minkowski plane wave packets and the corresponding plane wave Einstein packets, accounted for in the works \cite{SegalZhouQED}, \cite{PaneitzSegalI}-\cite{PaneitzSegalIII}.

Thus the single particle state functions and thus free quantum fields on the Einstein Universe
$\mathbb{R} \times SU(2, \mathbb{C})$ live effectively on the compactified 
version $\widetilde{\mathbb{S}^1} \times SU(2, \mathbb{C})$ of the Einstein Universe.
Thus in particular when analysing and constructing quantum fields on the Einstein Universe 
the space-time can effectively be replaced with $\widetilde{\mathbb{S}^1} \times SU(2, \mathbb{C})$,
the Hilbert space
\[
\mathscr{H} = \bigoplus \limits_{1}^{d} L^2(\mathbb{R} \times SU(2, \mathbb{C}); \mathbb{C})
= L^2(\mathbb{R} \times SU(2, \mathbb{C}); \mathbb{C}^d),
\]
can be replaced with
\[
\mathscr{H} = \bigoplus \limits_{1}^{d} L^2(\widetilde{\mathbb{S}^1} \times SU(2, \mathbb{C}); \mathbb{C})
= L^2(\widetilde{\mathbb{S}^1} \times SU(2, \mathbb{C}); \mathbb{C}^d),
\]
and the space-time test function spaces (\ref{TestSpaceOnRxS^3}) 
$\oplus\mathcal{A} = \mathcal{S}_{\oplus A'}(\mathbb{R} \times SU(2, \mathbb{C}); \mathbb{C}^d)$,
with $d$ depending on the kind of field, and determined by the standard
operator $A = \oplus A'$ (direct sum of $d$ copies), defined by 
(\ref{StandardAofTestSpaceOnRxS^3}), can be replaced with the following space-time test 
spaces
\begin{multline}\label{PeriodicSpacetimeTestSpace}
\mathscr{E} = \oplus\mathcal{A} = \mathcal{S}_{ A}(\widetilde{\mathbb{S}^1} \times SU(2, \mathbb{C}); \mathbb{C}^d)
=\mathcal{S}_{\oplus A'}(\widetilde{\mathbb{S}^1} \times SU(2, \mathbb{C}); \mathbb{C}^d) \\
= \oplus \mathcal{S}_{A'}(\widetilde{\mathbb{S}^1} \times SU(2, \mathbb{C}); \mathbb{C})
\end{multline}
where $\mathcal{A}=\mathcal{S}_{A'}(\widetilde{\mathbb{S}^1} \times SU(2, \mathbb{C}); \mathbb{C})$ is the algebra entering the spectral triple (\ref{1stSpectralTupleForSxG}) or equivalently (\ref{2ndSpectralTupleForSxG}), if we identify the operator of multiplication by a function $f$, which belongs to the algebra of operators in $\mathcal{A}$ defining the triple (\ref{2ndSpectralTupleForSxG}), with the corresponding function $f$. Thus the algebra $\mathcal{A}=\mathcal{S}_{A'}(\widetilde{\mathbb{S}^1} \times SU(2, \mathbb{C}); \mathbb{C})$ is equal to the standard countably Hilbert nuclear space
\begin{multline*}
\mathcal{A}=\mathcal{S}_{A'}(\widetilde{\mathbb{S}^1} \times SU(2, \mathbb{C}); \mathbb{C})= 
\mathcal{S}_{\Delta}(\widetilde{\mathbb{S}^1} \times SU(2,\mathbb{C})) = 
\mathcal{S}_{{}_{\Delta_{\widetilde{\mathbb{S}^1}} \otimes \boldsymbol{1}
+ \boldsymbol{1} \otimes \Delta_{SU(2,\mathbb{C})}}}(\widetilde{\mathbb{S}^1} \times SU(2,\mathbb{C})) \\ =
\mathcal{S}_{\Delta_{\widetilde{\mathbb{S}^1}}}(\widetilde{\mathbb{S}^1}) 
\otimes \mathcal{S}_{\Delta_{SU(2,\mathbb{C})}}(SU(2,\mathbb{C}))
= \mathscr{C}^\infty(\widetilde{\mathbb{S}^1}) \otimes \mathscr{C}^\infty(SU(2,\mathbb{C}))
= \mathscr{C}^\infty (\widetilde{\mathbb{S}^1} \times SU(2,\mathbb{C})),
\end{multline*} 
determined by the standard operator
\begin{equation}\label{StandardA'forPeriodicSpacetimeTestSpace}
A' = \Delta = \Delta_{\widetilde{\mathbb{S}^1}} \otimes \boldsymbol{1}
+ \boldsymbol{1} \otimes \Delta_{SU(2,\mathbb{C})}
= - \big( X^0\big)^2 -\big(X^1\big)^2 - \big(X^2\big)^2 - \big(X^3\big)^2 
+ {\textstyle\frac{1}{4}} \boldsymbol{1}
\end{equation}
(plus eventually the unit operator in order to achieve a standard operator) on
\[
L^2(\widetilde{\mathbb{S}^1}) \otimes L^2(SU(2,\mathbb{C})) 
= L^2(\widetilde{\mathbb{S}^1} \times SU(2, \mathbb{C})).
\]
The action (\ref{UphiOnRxG}) of the Einstein isometry group on $\mathscr{H}$ degenerates to the
compact action (\ref{UphiOnSxG}) on the Hilbert space $\mathscr{H}$ of time periodic 
$\mathbb{C}^d$-valued functions. It moreover follows that all distributional solutions
which define single particle states belong to the Hilbert space 
\[
\mathscr{H} = \bigoplus \limits_{1}^{d} L^2(\widetilde{\mathbb{S}^1} \times SU(2, \mathbb{C}); \mathbb{C})
= L^2(\widetilde{\mathbb{S}^1} \times SU(2, \mathbb{C}); \mathbb{C}^d),
\]
being regular and periodic, as $\widetilde{\mathbb{S}^1} \times SU(2, \mathbb{C})$ is compact.

Now a single particle Hilbert space of a quantum field on the Einstein Universe, which corresponds
to the specific local transformation formula (\ref{UphiOnSxG}) on the Hilbert space
$\mathscr{H}$ of time periodic $\mathbb{C}^d$-valued functions, is constructed from the invariant subspace
of $\mathscr{H}$ of all distributional eigen solutions of the corresponding invariant 
linear hyperbolic differetial (system of) equation(s), symbolically  $D_{{}_{iff}} \, \phi=0$. 
Then the construction of the single particle space depends on the kind of field, and for 
essentially neutral fields (corresponding to real classical fields) is slightly simpler in comparison to
non-neutral fields (corresponding to complex classical fields). But construction of both kinds of fields
use at the first stage the invariat Hilbert subspace $\mathscr{H}_{{}_{\textrm{sol}}}
\subset \mathscr{H}$ of all solutions of $D_{{}_{iff}} \, \phi=0$. 
In order to construct this invariant subspace $\mathscr{H}_{{}_{\textrm{sol}}}$ of solutions of $D_{{}_{iff}} \, \phi=0$  
we use the Fourier transform, and compute the action 
of the Fourier transformed $\mathcal{F} D_{{}_{iff}} \mathcal{F}^{-1} 
= \widetilde{D_{{}_{iff}}}$ differential operator $D_{{}_{iff}}$ on the Fourier transform 
$\mathcal{F} \phi = \widetilde{\phi}$, $\phi \in \mathscr{H}$. 
Because $D_{{}_{iff}} = U \, D_{{}_{iff}} \, U^{-1}$ is invariant under the representation 
$U$ of the Einstein isometry given by (\ref{UphiOnSxG}), then the  space $\mathscr{H}_{{}_{\textrm{sol}}}$ 
of solutions  $\phi$ of $D_{{}_{iff}} \, \phi = 0$, 
composes a subrepresentation of (\ref{UphiOnSxG}). This method is effective because 
the space of all values $\mathcal{F} \phi(\widehat{n}\cdot \widehat{l}) = 
\widetilde{\phi}(\widehat{n}\cdot \widehat{l})$,
$\phi \in \mathscr{H}$, for fixed point $\widehat{n}\cdot \widehat{l}
\in \widehat{\mathbb{S}^1 \times G}$, $G = SU(2, \mathbb{C})$, is equal 
to the invariant Hilbert subspace of the subrepresentation (\ref{[n.l]x[loV]}):
\[
[\,\widehat{n}\cdot \widehat{l} \,]\, \times \,[\, \overline{\,\widehat{l}\,} \otimes V \,]
\]
of (the Fourier transformed) representation (\ref{UphiOnSxG}). Thus all those 
$\phi \in \mathscr{H}$ whose Fourier transforms are concentrated on the single point $\widehat{n}\cdot \widehat{l}$, compose the invariant subspace of the subrepresentation of (\ref{UphiOnSxG}) 
equivalent to (\ref{[n.l]x[loV]}) with the unitary equivalence given by the Fourier transform.
The representation (\ref{[n.l]x[loV]}) is reducible (in case in which both $\overline{\,\widehat{l}\,}$
and $V$ are non-trivial representations of $SU(2, \mathbb{C})$), and can be easily further decomposed,
by applying decomposition of the factor $\overline{\,\widehat{l}\,} \otimes V$, utilizing the
Clebsch-Gordan decomposition for unitary tensor product representations of the $SU(2, \mathbb{C})$ group. 
In general only for some exceptional (say unitary characters
$\widehat{n}\cdot \widehat{l}$) 
\[ 
\widetilde{\mathbb{S}^1} \times SU(2, \mathbb{C}) \ni t \times w 
\longmapsto \widehat{n}\cdot \widehat{l}\, (t \times w) = \widehat{n}(n) \, \widehat{l}(w)
= e^{-i{\textstyle\frac{n}{2}}t} \, \widehat{l}(w)
\]
of the group $\widetilde{\mathbb{S}^1} \times SU(2, \mathbb{C})$,
the proper invariant subspace 
\begin{equation}\label{H[n.l]x[loV]}
\mathcal{H}_{{}_{[\,\widehat{n}\cdot \widehat{l} \,]\, \times \,[\, \overline{\,\widehat{l}\,} \otimes V \,]}}
= \big \{ \widetilde{\phi}, \,\, \phi \in \mathscr{H}, \textrm{with $\widetilde{\phi}$ concetrated on $\widehat{n}\cdot \widehat{l}$} \big \} 
\end{equation}
of the subrepresentation $[\,\widehat{n}\cdot \widehat{l} \,]\, \times \,[\, \overline{\,\widehat{l}\,} \otimes V \,]$ contains Fourier transformed solutions 
$\widetilde{\phi} \in \widetilde{\mathscr{H}_{{}_{\textrm{sol}}}}$ of the 
equation $D_{{}_{iff}} \, \phi = 0$. The set of all irreducible unitary characters  $\widehat{n}\cdot \widehat{l}$ for which
\begin{equation}\label{singleHintersectionWithH[n.l]x[loV]}
\widetilde{\mathscr{H}_{{}_{\textrm{sol}}}} \cap \mathcal{H}_{{}_{[\,\widehat{n}\cdot \widehat{l} \,]\, \times \,[\, \overline{\,\widehat{l}\,} \otimes V \,]}} = 
\widetilde{\mathscr{H}_{{}_{\textrm{sol}}}} \cap \big \{ \widetilde{\phi}, \,\, \phi \in \mathscr{H}, \textrm{with $\widetilde{\phi}$ concetrated on $\widehat{n}\cdot \widehat{l}$} \big \} \neq \emptyset
\end{equation}
composes (for positive and negative $n \in \mathbb{Z}$) the analogue of the (set theoretical sum of positive and negative energy) orbit\footnote{Orbits of the corresponding points $(\pm,0,0,0)$
or $(\pm1,0,0,1)$ under the hyperbolic Lorentz rotations.} 
$\mathscr{O}_{m,0,0,0} \sqcup \mathscr{O}_{-m,0,0,0}$ (or $\mathscr{O}_{1,0,0,1} \sqcup \mathscr{O}_{-1,0,0,1}$) of the single particle space on the Minkowski space-time.
Namely they are the analogues of the characters 
\[
T_4 \ni x \mapsto e^{-ip\cdot x} \in \mathbb{C}
\] 
of the translation group $T_4$, with
$p = (p_0(\boldsymbol{\p}), \boldsymbol{\p})$ ranging over the orbits in the four-momentum 
space.\footnote{The various directions of 
$\boldsymbol{\p}$ which exhaust all characters with fixed $p_0$ and  $|\boldsymbol{\p}|$,
are the analogues af the unitary irreducible representation (say character) 
\[
\mathbb{R} \times SU(2, \mathbb{C}) \ni t \times w \mapsto \widehat{n}\cdot \widehat{l}\,(t \times w) 
= \widehat{n}(t)\widehat{l}(w) = e^{-i{\textstyle\frac{n}{2}}t} \, \widehat{l}(w)
\]
of $\widetilde{\mathbb{S}^1} \times SU(2, \mathbb{C})$, with the matrix indices $ij$ 
of the character $t \times w \mapsto  \widehat{n}(t)\widehat{l}_{{}_{ij}}(w)$  corresponding
to the various directions of $\boldsymbol{\p}$ in the character
\[ 
\mathbb{R}^4 \ni x=t\times \boldsymbol{\x} \mapsto e^{-ip_0 t}e^{i\boldsymbol{\p} \cdot \boldsymbol{\x}}
= e^{ip\cdot x}.
\]
This is reflected by the analogue form of the inverses of the Fourier transform
formulas (and the Fourier transform formulas themselves):
\begin{multline*}
\phi^a(t,w)
=
{\textstyle\frac{1}{\sqrt{4\pi}}}
\sum \limits_{n\in \mathbb{Z}, \widehat{l} \in \widehat{SU(2, \mathbb{C})}} 
l(l+1)\,\,\,
\textrm{Tr}\,\Big[\, \widetilde{\phi^a}(\widehat{n} \cdot \widehat{l}) \,\, 
\widehat{n} \cdot \widehat{l}(t,w) \,\Big] \\
 =
{\textstyle\frac{1}{\sqrt{4\pi}}}
\sum \limits_{n\in \mathbb{Z}, \widehat{l} \in \widehat{SU(2, \mathbb{C})}} 
l(l+1)\,\,\,
\textrm{Tr}\,\Big[\, \widetilde{\phi^a}(\widehat{n} \cdot \widehat{l}) \,\, \widehat{l}(w) \,\Big] \,\, e^{-i{\textstyle\frac{n}{2}}t} \\
 = 
{\textstyle\frac{1}{\sqrt{4\pi}}}
\sum \limits_{n\in \mathbb{Z}, \widehat{l} \in \widehat{SU(2, \mathbb{C})}, -l \leq i,j \leq l} 
l(l+1)\,\,\,
 \widetilde{\phi^{a}}_{{}_{ji}}(\widehat{n} \cdot \widehat{l}) \,\, 
e^{-i{\textstyle\frac{n}{2}}t} \,\, 
 \widehat{l}_{ij}(w) 
\end{multline*}
and
\[
\phi^a(t, \boldsymbol{\x}) = {\textstyle\frac{1}{(2\pi)^2}} \int 
\widetilde{\phi^a}(p_0, \boldsymbol{\p}) \, e^{-ip_0 t}e^{i\boldsymbol{\p}\cdot \boldsymbol{\x}}
\, \ud p_0 \, \ud^3 \boldsymbol{\p}.
\]
} 
Therefore the set of all irreducible unitary representations (say characters) 
$\widehat{n}\cdot \widehat{l}$ for which the intersection (\ref{singleHintersectionWithH[n.l]x[loV]})
is nonempty, we analogously will denote by $\mathscr{O}_{{}_{+}} \sqcup \mathscr{O}_{{}_{-}}$ and call the orbit corresponding to the Hilbert space $\mathscr{H}_{{}_{\textrm{sol}}} \subset \mathscr{H}$ of the solutions of the differential equation
$D_{{}_{iff}} \phi = 0$ and corresponding to the free field in question (although in this case 
$\mathscr{O}_{{}_{+}} \sqcup \mathscr{O}_{{}_{-}}$ is not equal to any (sum of) actual orbit(s) of any fixed point(s) under any natural action of a group). We have written $\mathscr{O}_{{}_{+}}$
for the set of all characters $\widehat{n}\cdot \widehat{l}$ with nonempty (\ref{singleHintersectionWithH[n.l]x[loV]})
 and $n \geq 0$, and  $\mathscr{O}_{{}_{-}}$ for the set of characters
$\widehat{n}\cdot \widehat{l}$  with nonempty (\ref{singleHintersectionWithH[n.l]x[loV]})
and $n<0$.

It will be convenient to introduce the notation 
\begin{multline}\label{H^oplus_nl}
\widetilde{\mathscr{H}_{{}_{\textrm{sol}}}} \cap \mathcal{H}_{{}_{[\,\widehat{n}\cdot \widehat{l} \,]\, \times \,[\, \overline{\,\widehat{l}\,} \otimes V \,]}}  \\ = 
\widetilde{\mathscr{H}_{{}_{\textrm{sol}}}} \cap \big \{ \widetilde{\phi}, \,\, \phi \in \mathscr{H}, \textrm{with $\widetilde{\phi}$ concentrated on $\widehat{n}\cdot \widehat{l}$} \big \} =
\mathcal{H}^{\oplus}_{{}_{\widehat{n}\cdot \widehat{l}}}, \\
\,\,\, \textrm{for} \, n \geq 0
\end{multline}
\begin{multline}\label{H^ominus_nl}
\widetilde{\mathscr{H}_{{}_{\textrm{sol}}}} \cap \mathcal{H}_{{}_{[\,\widehat{n}\cdot \widehat{l} \,]\, \times \,[\, \overline{\,\widehat{l}\,} \otimes V \,]}} \\ = 
\widetilde{\mathscr{H}_{{}_{\textrm{sol}}}} \cap \big \{ \widetilde{\phi}, \,\, \phi \in \mathscr{H}, \textrm{with $\widetilde{\phi}$ concentrated on $\widehat{n}\cdot \widehat{l}$} \big \} =
\mathcal{H}^{\ominus}_{{}_{\widehat{n}\cdot \widehat{l}}},\\
\,\,\, \textrm{for} \, n < 0
\end{multline}
for the intersection Hilbert subspace (\ref{singleHintersectionWithH[n.l]x[loV]})
of Fourier transforms of positive, respectively, negative, energy solutions, whose Fourier
transforms are concentrated on the single point $\widehat{n}\cdot \widehat{l} \in
\mathscr{O}_{{}_{\pm}}$ of the dual $\widehat{\widetilde{S}^1 \times G}$, $G=SU(2, \mathbb{C})$
(with $\widetilde{S}^1 \times G$ plying the role of translations).

This terminology and notation will hep the comparison with the construction of free fields
on the Minkowski space-time. 

But even if $\widehat{n}\cdot \widehat{l} \in \mathscr{O}_{{}_{+}} \sqcup \mathscr{O}_{{}_{-}}$, so that (\ref{H[n.l]x[loV]})
has nonempty intersection with the (Fourier transformed)  Hilbert subspace 
$\mathscr{H}_{{}_{\textrm{sol}}}$, in general
only a proper subspace of (\ref{H[n.l]x[loV]}) belongs to the (Fourier transformed)
 Hilbert space $\mathscr{H}_{{}_{\textrm{sol}}}$. Therefore, for $\widehat{n}\cdot \widehat{l} \in \mathscr{O}_{{}_{+}} \sqcup \mathscr{O}_{{}_{-}}$ in order to determine the closed subspace of  (\ref{H[n.l]x[loV]}) with the Fourier transformed  Hilbert space $\mathscr{H}_{{}_{\textrm{sol}}}$, equal to the intersection (\ref{singleHintersectionWithH[n.l]x[loV]}),
we decompose the representation (\ref{[n.l]x[loV]}) further (using the Clebsch-Gordan decomposition of the factor $\overline{\, \widehat{l} \,} \otimes V$). Now we choose a particular direct summand in the decomposition  of (\ref{[n.l]x[loV]})  into irreducible components. 
If in the decomposition of (\ref{[n.l]x[loV]}) 
into irreducible components all other direct summands are not equivalent to the choosen one, 
then we have for the chosen direct summand only two possibilities: it enters as a whole into the Hilbert subspace $\mathscr{H}_{{}_{\textrm{sol}}}$ or its intersection with $\mathscr{H}_{{}_{\textrm{sol}}}$ is empty. If among the direct summands there are several equivalent
to the chosen direct summand, then the decomposition can be realized in various unitarily equivalent manners, and we need to consider linear combinations of vectors coming from the equivalent direct summands
as the possible elements of the intersection (\ref{singleHintersectionWithH[n.l]x[loV]}). In any case using the decomposition into irreducible components of the subrepesentation (\ref{[n.l]x[loV]}) we can rather effectively find the complete orthonormal
sets $\{\widetilde{\phi_s}\}$, $1 \leq s \leq \textrm{dim} \, \mathcal{H}_{{}_{[\,\widehat{n}\cdot \widehat{l} \,]\, \times \,[\, \overline{\,\widehat{l}\,} \otimes V \,]}}$, of Fourier transforms  of solutions of $D_{{}_{iff}} \, \phi = 0$, which span the intersections (\ref{singleHintersectionWithH[n.l]x[loV]}),
\emph{i.e.} all Fourier transformed solutions of $D_{{}_{iff}} \, \phi = 0$
whose Fourier transform lie in (\ref{H[n.l]x[loV]}) and are concentrated at 
$\widehat{n}\cdot \widehat{l}$, especially
if the representation $V$ in (\ref{[n.l]x[loV]}) and in (\ref{UphiOnSxG}), is of relatively
low dimension. Because elements $\widetilde{\phi_s}$ of the complete set 
\[
\{\widetilde{\phi_s}\} \subset 
\mathcal{H}_{{}_{[\,\widehat{n}\cdot \widehat{l} \,]\, \times \,[\, \overline{\,\widehat{l}\,} \otimes V \,]}}
\]
are all concentrated on the single point $\widehat{n}\cdot \widehat{l}$:
\[
\widetilde{\phi_s}(\widehat{n'}\cdot \widehat{l'}) = 0,
\,\,\,
\widehat{n'}\cdot \widehat{l'} \neq \widehat{n}\cdot \widehat{l}
\]
then we will write for them $ \widetilde{\phi_s}(\widehat{n}\cdot \widehat{l})$, indicating
explicitly the only element $\widehat{n}\cdot \widehat{l}$ in their support and write
instead  $\{\widetilde{\phi_s}(\widehat{n}\cdot \widehat{l})\}$ for the complete orthonormal
set of solutions which span the intersection (\ref{singleHintersectionWithH[n.l]x[loV]}).
Having the complete set 
\[
\{\widetilde{\phi_s}\}(\widehat{n}\cdot \widehat{l}) \subset 
\mathcal{H}_{{}_{[\,\widehat{n}\cdot \widehat{l} \,]\, \times \,[\, \overline{\,\widehat{l}\,} \otimes V \,]}}
\]
of Fourier transformed solutions whose Fourier transforms are concentrated on the single point 
$\widehat{n}\cdot \widehat{l} \in \mathscr{O}_{{}_{+}} \sqcup \mathscr{O}_{{}_{-}}$, we can
therefore compute explicitly the projector $P(\widehat{n}\cdot \widehat{l})$ which acts
in (\ref{H[n.l]x[loV]}) and projects on the subspace of Fourier transformed solutions
whose Fourier transforms are concentrated on the single point 
$\widehat{n}\cdot \widehat{l} \in \mathscr{O}_{{}_{+}} \sqcup \mathscr{O}_{{}_{-}} \subset \widehat{\widetilde{\mathbb{S}^1}\times SU(2, \mathbb{C})}$. Then the projector operator, equal to the operator of pointwise multiplication
by the function $\widehat{n}\cdot \widehat{l} \mapsto P(\widehat{n}\cdot \widehat{l})$,
projects the Fourier transform $\widetilde{\phi}$ of any 
$\phi \in \mathscr{H}$, onto the Hilbert subspace
$\widetilde{\mathscr{H}_{{}_{\textrm{sol}}}}$  of Fourier transformed solutions.
Note in particular that $P(\widehat{n}\cdot \widehat{l}) = 0$ if $\widehat{n}\cdot \widehat{l} \notin
\mathscr{O}_{{}_{+}} \sqcup \mathscr{O}_{{}_{-}}$.
The projection function 
which restricted to the positive energy orbit $\mathscr{O}_{{}_{+}}$ coincides
with the projector $P$ and which is zero outside $\mathscr{O}_{{}_{+}}$ will be denoted
by\footnote{Here $G = SU(2, \mathbb{C})$.} 
\[
P^\oplus: \, \widehat{\widetilde{S}^1 \times G} \ni \widehat{n}\cdot \widehat{l} \mapsto P^\oplus(\widehat{n}\cdot \widehat{l})
\]
and the projection which restricted to the positive energy orbit $\mathscr{O}_{{}_{-}}$ coincides
with the projector $P$ and which is zero outside $\mathscr{O}_{{}_{-}}$ will be denoted
by\footnote{Here $G = SU(2, \mathbb{C})$.} 
\[
P^\ominus: \, \widehat{\widetilde{S}^1 \times G} \ni \widehat{n}\cdot \widehat{l} 
\mapsto P^\ominus(\widehat{n}\cdot \widehat{l})
\]
These $P^\oplus, P^\ominus$ are the immediate analogues of the projectors $P^\oplus, P^\ominus$ 
used in the white noise construction
of the free Dirac field on the Minkowski space-time in Subsections \ref{psiWightman}, \ref{psiBerezin-Hida}, 
and in Appendix \ref{fundamental,u,v}.  
Let us denote by $\{\widetilde{\phi_{s}^{+}}(\widehat{n}\cdot \widehat{l})\}$
elements of the complete set $\{\widetilde{\phi_s}(\widehat{n}\cdot \widehat{l})\}$ of solutions 
with $n \geq 0$, and by $\{\widetilde{\phi_{s}^{-}}(\widehat{n}\cdot \widehat{l})\}$
those with $n<0$. In order to make the relationship with the white noise construction of the free quantum fields 
on the Minkowski space-time presented in Sections \ref{e+e-} and \ref{white-noise-proofs}, still more transparent, 
we will denote the complete sets of solutions
$\{\widetilde{\phi_{s}^{\pm}}(\widehat{\pm|n|}\cdot \widehat{l})\}$, by
\[
\widetilde{\phi_{s}^{+}}(\widehat{n}\cdot \widehat{l}) = u_s(\widehat{n}\cdot \widehat{l}), \,\,\, 
\widehat{n}\cdot \widehat{l} \in \mathscr{O}_{{}_{+}}
\,\,\,\,\, 
\widetilde{\phi_{s}^{-}}(\widehat{n}\cdot \widehat{l}) = v_s(\widehat{n}\cdot \widehat{l}), \,\,\, 
\widehat{n}\cdot \widehat{l} \in \mathscr{O}_{{}_{-}},
\]
as they are the immediate analogues of the Fourier transforms
\[
\widetilde{\phi_{s}^{+}}(p) = u_s(p), \,\,\, 
p \in \mathscr{O}_{{}_{+m,0,0,0}}
\,\,\,\,\, 
\widetilde{\phi_{s}^{-}}(p) = v_s(p), \,\,\, 
p \in \mathscr{O}_{{}_{-m,0,0,0}},
\]
of the complete systems of solutions of the Dirac equation (in case of the construction of the free Dirac field
on the Minkowski space-time) evaluated at the point $p$ of the corresponding orbit
and determining the corresponding character
\[
\widehat{p}:\, x \mapsto e^{-ip\cdot x}
\]
of the translation group $T_4$. 
Note however that on the Einstein Universe the Fourier transforms
$\widetilde{\phi_{s}^{\pm}}(\widehat{n}\cdot \widehat{l}) 
= \mathcal{F} (\phi_{s}^{\pm})(\widehat{n}\cdot \widehat{l})$ denoted respectively by  $u_{s}(\widehat{n}\cdot \widehat{l})$
or by $v_s(\widehat{n}\cdot \widehat{l})$ (with $n = |n|$ or resp. $n = - |n|$), are for each $s$ equal to matrices, which we interpret as the columns of $d = \textrm{dim} \, V$ matrices 
\begin{multline*}
\left( \begin{array}{c}  
\mathcal{F}(\phi^{1 +}_{s})_{{}_{ji}}(\widehat{n}\cdot \widehat{l}) \\
 \vdots \\
\mathcal{F}(\phi^{d +}_{s})_{{}_{ji}}(\widehat{n}\cdot \widehat{l})
                     \end{array}\right)
                     =
                     \left( \begin{array}{c}  
u^{1}_{s \, {}_{ji}}(\widehat{n}\cdot \widehat{l}) \\
 \vdots \\
u^{d}_{s \, {}_{ji}}(\widehat{n}\cdot \widehat{l})
                     \end{array}\right), 
                     \\
                      \textrm{and respectively} \,\,\,\, 
\left( \begin{array}{c}  
\mathcal{F}(\phi^{1 -}_{s})_{{}_{ji}}(\widehat{n}\cdot \widehat{l}) \\
 \vdots \\
\mathcal{F}(\phi^{d -}_{s})_{{}_{ji}}(\widehat{n}\cdot \widehat{l})
                     \end{array}\right)
                     =
                     \left( \begin{array}{c}  
v^{1}_{s \, {}_{ji}}(\widehat{n}\cdot \widehat{l}) \\
 \vdots \\
v^{d}_{s \, {}_{ji}}(\widehat{n}\cdot \widehat{l})
                     \end{array}\right), 
\end{multline*}
\[
1 \leq i,j \leq  \textrm{dim} \, \widehat{l} = 2l +1
\]
with each matrix 
$\mathcal{F}(\phi^{a +}_{s})_{{}_{ji}}(\widehat{n}\cdot \widehat{l})=
u^{a}_{s \, {}_{ji}}(\widehat{n}\cdot \widehat{l})$,
and $\mathcal{F}(\phi^{a -}_{s})_{{}_{ji}}(\widehat{n}\cdot \widehat{l})=
v^{a}_{s \, {}_{ji}}(\widehat{n}\cdot \widehat{l})$, $1 \leq a \leq d$, 
being a $\textrm{dim} \, \widehat{l} \times \textrm{dim} \, \widehat{l}$ matrix in the indices
$i,j$. Equivalently 
$\mathcal{F}\phi^{\pm}_{s}(\widehat{n}\cdot \widehat{l})$ for each fixed $s$ and the sign
$\pm$, is regarded as a matrix
\[
\Big[F^{\pm}_{s}\Big]_{{}_{a \,\,\,\,\,ji}} = \mathcal{F}(\phi^{a \pm}_{s})_{{}_{ji}}(\widehat{x})
\]
with matrix indices $a$ and $ij$, of a linear operator from the Hilbert space of $\textrm{dim} \, \widehat{l} \times \textrm{dim} \, \widehat{l}$ matrices (regarded as 
the linear space of Hilbert-Schmidt operators with the Hilbert-Schmidt norm multiplied by $\textrm{dim} \, \widehat{l} = 2l +1$) to the conjugation of the Hilbert space $\mathbb{C}^d$.

Of course by definition
\[
\{\widetilde{\phi_{s}^{+}}(\widehat{n}\cdot \widehat{l})\} = \{u^{a}_{s}(\widehat{n}\cdot \widehat{l}) \}
\,\,\,
\textrm{and resp.}
\,\,\,
\{\widetilde{\phi_{s}^{-}}(\widehat{n}\cdot \widehat{l})\} = \{v^{a}_{s}(\widehat{n}\cdot \widehat{l}) \}
\]
compose complete orthonormal systems, respectively, in $\mathcal{H}^{\oplus}_{{}_{\widehat{n}\cdot \widehat{l}}}$ and $\mathcal{H}^{\ominus}_{{}_{\widehat{n}\cdot \widehat{l}}}$.

 Now we define the Hilbert
space subspace $\mathscr{H}_{{}_{\textrm{sol}}}^{{}^{+}} \subset \mathscr{H}_{{}_{\textrm{sol}}}$  of 
$\mathscr{H}$ of all positive energy solutions of $\mathscr{H}_{{}_{\textrm{sol}}}$.
Namely, if we again write 
\[
P^\oplus: \, \widehat{\widetilde{\mathbb{S}^1} \times G}
\ni \widehat{n}\cdot \widehat{l} \longmapsto
P^\oplus(\widehat{n}\cdot \widehat{l}), \,\,\,\,\,\,\,\,\,\,\,\,\,\,\,\,
G = SU(2, \mathbb{C}) 
\]
for the projection function which acts by pointwise multiplication on the Fourier transform
$\widetilde{\mathscr{H}}$ of the Hilbert space $\mathscr{H}$,
for which $P^\oplus(\widehat{n}\cdot \widehat{l}) =0$ if $\widehat{n}\cdot \widehat{l}
\notin \mathscr{O}_{{}_{+}}$ and coincides with $P^\oplus(\widehat{n}\cdot \widehat{l})$
defined above if $\widehat{n}\cdot \widehat{l} \in \mathscr{O}_{{}_{+}}$  
then $\widetilde{\mathscr{H}_{{}_{\textrm{sol}}}^{{}^{+}}} \subset \widetilde{\mathscr{H}}$ is precisely equal to the range of the 
last projector $P^\oplus$. Equivalently $\phi \in \mathscr{H}_{{}_{\textrm{sol}}}^{{}^{+}}$ iff its Fourier transform $\widetilde{\phi}$
is supported in $\mathscr{O}_{{}_{+}}$ and each value 
$\widetilde{\phi}(\widehat{n}\cdot \widehat{l})$ at 
$\widehat{n}\cdot \widehat{l} \in \mathscr{O}_{{}_{+}}$ lies in the range
of $P^\oplus(\widehat{n}\cdot \widehat{l})$, or is spanned by   
$\{\widetilde{\phi_s}^{+}(\widehat{n}\cdot \widehat{l})\} = \{u_{s}(\widehat{n}\cdot \widehat{l})\}$.

Similarly, we define the Hilbert
space subspace $\mathscr{H}_{{}_{\textrm{sol}}}^{{}^{-}} \subset \mathscr{H}_{{}_{\textrm{sol}}}$  of 
$\mathscr{H}$ of all negative energy solutions of $\mathscr{H}_{{}_{\textrm{sol}}}$.
Namely, if we again write 
\[
P^\ominus: \, \widehat{\widetilde{\mathbb{S}^1} \times G}
\ni \widehat{n}\cdot \widehat{l} \longmapsto
P^\ominus(\widehat{n}\cdot \widehat{l}), \,\,\,\,\,\,\,\,\,\,\,\,\,\,\,\,
G = SU(2, \mathbb{C}) 
\]
for the projection function which acts by pointwise multiplication on the Fourier transform
$\widetilde{\mathscr{H}}$ of the Hilbert space $\mathscr{H}$,
for which $P^\ominus(\widehat{n}\cdot \widehat{l}) =0$ if $\widehat{n}\cdot \widehat{l}
\notin \mathscr{O}_{{}_{-}}$ and coincides with $P^\ominus(\widehat{n}\cdot \widehat{l})$
defined above if $\widehat{n}\cdot \widehat{l} \in \mathscr{O}_{{}_{-}}$,  
then $\widetilde{\mathscr{H}_{{}_{\textrm{sol}}}^{{}^{-}}} \subset \widetilde{\mathscr{H}}$ is precisely equal to 
the range of the last projector $P^\ominus$. Equivalently $\phi \in \mathscr{H}_{{}_{\textrm{sol}}}^{{}^{-}}$ iff its Fourier transform  $\widetilde{\phi}$
is supported in $\mathscr{O}_{{}_{-}}$ and each value 
$\widetilde{\phi}(\widehat{n}\cdot \widehat{l})$ at 
$\widehat{n}\cdot \widehat{l} \in \mathscr{O}_{{}_{-}}$ lies in the range
of $P^\ominus(\widehat{n}\cdot \widehat{l})$, or is spanned by   
$\{\widetilde{\phi_s}^{-}(\widehat{n}\cdot \widehat{l})\} = \{v_{s}(\widehat{n}\cdot \widehat{l})\}$.

The Hilbert subspaces 
\[
\mathcal{H}^\oplus = \widetilde{\mathscr{H}_{{}_{\textrm{sol}}}^{{}^{+}}} \subset \widetilde{\mathscr{H}}
\,\,\,
\textrm{and}
\,\,\,
\mathcal{H}^\ominus = \widetilde{\mathscr{H}_{{}_{\textrm{sol}}}^{{}^{-}}} \subset \widetilde{\mathscr{H}}
\]
are the immediate analogues of the Hilbert spaces 
\[
\mathcal{H}^\oplus_{m,0} 
\,\,\,
\textrm{and}
\,\,\,
\mathcal{H}^\ominus_{-m,0} 
\]
used in Subsections \ref{FirstStepH}-\ref{electron+positron},
\ref{psiBerezin-Hida}, \ref{StandardDiracPsiField}, in
the construction of the free Dirac field on the Minkowski space-time.

Of course, we have, by the very construction, the following direct sum decomposition
\[
\mathcal{H}^{\oplus} = \bigoplus \limits_{\widehat{n}\cdot \widehat{l} \in \mathscr{O}_{+}} 
\mathcal{H}^{\oplus}_{{}_{\widehat{n}\cdot \widehat{l}}}, \,\,\,\,
\mathcal{H}^{\ominus} = \bigoplus \limits_{\widehat{n}\cdot \widehat{l} \in \mathscr{O}_{-}} 
\mathcal{H}^{\ominus}_{{}_{\widehat{n}\cdot \widehat{l}}},
\]
which, in order to make the analogue with the construction of the field on the non-compact 
Minkowski space-time still more transparent, can also be written
\[
\mathcal{H}^{\oplus} =  \int \limits_{\mathscr{O}_{+}} 
\mathcal{H}^{\oplus}_{{}_{\widehat{n}\cdot \widehat{l}}} \,\,\,\, 
\ud \mu_{{}_{\mathscr{O}_{+}}}(\widehat{n}\cdot \widehat{l}), \,\,\,\,\,\,\,\,\,
\mathcal{H}^{\ominus} = \int \limits_{\mathscr{O}_{-}} 
\mathcal{H}^{\ominus}_{{}_{\widehat{n}\cdot \widehat{l}}} \,\,\, 
\ud \mu_{{}_{\mathscr{O}_{-}}}(\widehat{n}\cdot \widehat{l}),
\]
with the discrete measure $\mu_{{}_{\mathscr{O}_{\pm}}}$ on the orbit $\mathscr{O}_{\pm}$, 
which ascribes the measure $1$ to each single point. The (finite dimensional)
Hilbert space 
$\mathcal{H}^{\oplus}_{{}_{\widehat{n}\cdot \widehat{l}}}$ of values  (or resp.
 $\mathcal{H}^{\ominus}_{{}_{\widehat{n}\cdot \widehat{l}}}$) of 
values of the Fourier transforms $\widetilde{\phi}$, $\phi \in \mathscr{H}_{{}_{\textrm{sol}}}^{{}^{\pm}}$,
at the single point $\widehat{n}\cdot \widehat{l} \in \mathscr{O}_{\pm}$, plays the role
of the (finite dimensional) Hilbert space $\mathcal{H}^{\oplus}_{{}_{p}}$
(or resp. $\mathcal{H}^{\ominus}_{{}_{p}}$) of values $\widetilde{\phi}(p)$ at fixed\footnote{Or
$p \in \mathscr{O}_{{}_{\pm 1, 0,0,1}}$.}
$p \in \mathscr{O}_{{}_{\pm m, 0,0,0}}$, of the Fourier transforms
of distributional (positive, respectively, negative energy) solutions $\phi$. Note that
in case of the Minkowski space-time, the Hilbert spaces of Fourier transformed positive 
and negative energy solutions $\mathcal{H}^\oplus_{m,0} $ (resp. $\mathcal{H}^\ominus_{-m,0}$) 
can be naturally represented as the direct integrals
\[
\mathcal{H}^{\oplus}_{m,0} =  \int \limits_{\mathscr{O}_{{}_{m,0,0,0}}} 
\mathcal{H}^{\oplus}_{{}_{p}} \,\,\,\, 
\ud \mu_{{}_{\mathscr{O}_{m,0,0,0}}}(p), \,\,\,\,\,\,\,\,\,
\mathcal{H}^{\ominus}_{-m,0} = \int \limits_{\mathscr{O}_{{}_{-m,0,0,0}}} 
\mathcal{H}^{\ominus}_{{}_{p}} \,\,\, 
\ud \mu_{{}_{\mathscr{O}_{-m,0,0,0}}}(p),
\]
or, for massless fields,
\[
\mathcal{H}^{\oplus}_{1,0,0,1} =  \int \limits_{\mathscr{O}_{{}_{1,0,0,1}}} 
\mathcal{H}^{\oplus}_{{}_{p}} \,\,\,\, 
\ud \mu_{{}_{\mathscr{O}_{1,0,0,1}}}(p), \,\,\,\,\,\,\,\,\,
\mathcal{H}^{\ominus}_{-1,0,0,1} = \int \limits_{\mathscr{O}_{{}_{-1,0,0,1}}} 
\mathcal{H}^{\ominus}_{{}_{p}} \,\,\, 
\ud \mu_{{}_{\mathscr{O}_{-1,0,0,1}}}(p).
\]

After this introduction we can define the single particle Hilbert space $\mathcal{H}'$ in case of an essentially neutral field, 
say $\mathbb{A}$, which corresponds to a real classical field.
This is defined as 
\[
\mathcal{H}' = \mathcal{H}^\oplus
\]
in case of positive energy free quantum field or 
\[
\mathcal{H}' = \mathcal{H}^\ominus
\]
in case of negative energy free quantum field (we will need both for technical reasons).

Next, in order to construct the free field $\mathbb{A}$ with the help of Hida operators, we need
to have a countably Hilbert nuclear dense subspace $E$ of restrictions of Fourier
transforms $P^\oplus\widetilde{f}|_{{}_{\mathscr{O}_{+}}}$
(resp. $P^\ominus\widetilde{f}|_{{}_{\mathscr{O}_{-}}}$) of space-time test functions  
\begin{multline*}
f \in \mathcal{S}_{ A}(\widetilde{\mathbb{S}^1} \times SU(2, \mathbb{C}); \mathbb{C}^d)
=\mathcal{S}_{\oplus A'}(\widetilde{\mathbb{S}^1} \times SU(2, \mathbb{C}); \mathbb{C}^d) 
= \oplus \mathcal{S}_{A'}(\widetilde{\mathbb{S}^1} \times SU(2, \mathbb{C}); \mathbb{C}) \\
= \oplus \mathscr{C}^{\infty}(\widetilde{\mathbb{S}^1} \times SU(2, \mathbb{C}); \mathbb{C})
= \mathscr{C}^{\infty}(\widetilde{\mathbb{S}^1} \times SU(2, \mathbb{C}); \mathbb{C}^d)
\end{multline*}
(recall that the standard nuclear spacetime test function space is defined
by (\ref{PeriodicSpacetimeTestSpace})), projected on positive (resp. negative)
energy solution space $\mathcal{H}^\oplus = \widetilde{\mathscr{H}_{{}_{\textrm{sol}}}^{{}^{+}}}$ 
(resp. $\mathcal{H}^\ominus = \widetilde{\mathscr{H}_{{}_{\textrm{sol}}}^{{}^{-}}}$).
Note that the space-time test space (\ref{PeriodicSpacetimeTestSpace}) is defined by the standard operator $A = \oplus_{1}^{d}\Delta = \Delta \boldsymbol{1}_{{}_{d}}$ (plus eventually the unit operator) in the sense of \cite{GelfandIV}
or \cite{hida}, \cite{obata-book}. We will simply write $\Delta$ for the scalar operator
$\Delta \boldsymbol{1}_{{}_{d}}$ on $\mathscr{H}$. 
Here $P^\oplus$ and $P^\ominus$ are the projection operators defied above as the respective
projection functions $\widehat{n}\cdot \widehat{l} \mapsto P^\oplus(\widehat{n}\cdot \widehat{l})$
and $\widehat{n}\cdot \widehat{l} \mapsto P^\ominus(\widehat{n}\cdot \widehat{l})$
on\footnote{$G= SU(2, \mathbb{C})$.} $\widehat{\widetilde{S}^1 \times G}$, with the action
on $\widetilde{\mathscr{H}}$ defined by pointwise matrix multiplication.
This nuclear space $E$, together with its strong dual $E^*$ and the single particle Hilbert 
space $\mathcal{H}'$ composes the so called Gelfand triple:
\begin{equation}\label{SingleGelfandTripleNeutralEU}
\left. \begin{array}{ccccc}              E         & \subset &  \mathcal{H'} & \subset & E^*        \\
                                                   &         & \parallel      &         &  \\
                        &   & \mathcal{H}^{\oplus} \, \textrm{or resp.} \mathcal{H}^{\ominus} \, &  &  \end{array}\right..
\end{equation}  
Because the spacetime test space is equal to the standard countably Hilbert nuclear space 
$\mathscr{E} = \mathcal{S}_{\Delta}(\widetilde{\mathbb{S}^1} \times G); \mathbb{C}^d)$,
and because the Fourier transform is unitary, then
\[
\mathcal{F}\mathcal{S}_{\Delta}(\widetilde{\mathbb{S}^1} \times G; \mathbb{C}^d)
= \mathcal{S}_{\widetilde{\Delta}}(\widehat{\widetilde{\mathbb{S}^1} \times G} \times \mathbb{Z} \times \mathbb{Z}; \mathbb{C}^d)
\]
and 
\[
\mathcal{S}_{\widetilde{\Delta}}(\widehat{\widetilde{\mathbb{S}^1} \times G} \times \mathbb{Z} \times \mathbb{Z}; \mathbb{C}^d)
\]
is equal to a standard countably Hilbert nuclear space determined (in the sense
of \cite{GelfandIV}, \cite{hida}, \cite{obata-book}) by the standard operator
\[
\widetilde{\Delta} \,\,\,
\textrm{on}
\,\,\, \widetilde{\mathscr{H}}.
\]
Note, please, that the subspaces 
$\mathcal{H}^\oplus = P^\oplus \widetilde{\mathscr{H}}, \mathcal{H}^\ominus
= P^\ominus \widetilde{\mathscr{H}} \subset \widetilde{\mathscr{H}}$
are invariant for $\widetilde{\Delta}$, and that 
\begin{multline*}
E = P^\oplus \mathcal{S}_{\widetilde{\Delta}}(\widehat{\widetilde{\mathbb{S}^1} \times G} \times \mathbb{Z} \times \mathbb{Z}; \mathbb{C}^d) \subset \mathcal{H}' = \mathcal{H}^\oplus \\
\textrm{resp.} \,\,\,\,
E = P^\ominus \mathcal{S}_{\widetilde{\Delta}}(\widehat{\widetilde{\mathbb{S}^1} \times G} \times \mathbb{Z} \times \mathbb{Z}; \mathbb{C}^d) \subset \mathcal{H}' = \mathcal{H}^\ominus
\end{multline*}
is equal to that abstract countably Hilbert nuclear space, which is determined
by the standard operator
\[
P^\oplus \widetilde{\Delta} P^\oplus = P^\oplus \widetilde{\Delta} = 
\widetilde{\Delta} P^\oplus \,\,\, \textrm{on} \,\,\, \mathcal{H}^\oplus, 
\]
or respectively by the standard operator
\[
P^\ominus \widetilde{\Delta} P^\ominus = P^\ominus \widetilde{\Delta} = 
\widetilde{\Delta} P^\ominus \,\,\, \textrm{on} \,\,\, \mathcal{H}^\ominus, 
\] 
because for projectors $P^\oplus, P^\ominus$ commuting with $\widetilde{\Delta}$
the space $P^\oplus \mathcal{S}_{\widetilde{\Delta}}$ is determined by
the standard operator $P^\oplus\widetilde{\Delta}P^\oplus$  
if $\mathcal{S}_{\widetilde{\Delta}}$ is determined by $\widetilde{\Delta}$;
and identically we have for
$P^\ominus \mathcal{S}_{\widetilde{\Delta}}$ which is
determined by the standard operator $P^\ominus\widetilde{\Delta}P^\ominus$.

Now using the observation by Hida we conclude that the constructed Gelfand triple 
(\ref{SingleGelfandTripleNeutralEU}) over the single particle Hilbert space $\mathcal{H}'$ can be lifted to the second quantized level:
\begin{equation}\label{FockGelfandTripleNeutralEU}
\left. \begin{array}{ccccc}              (E)         & \subset &  \Gamma(\mathcal{H'}) & \subset & (E)^*        \\
                                                   &         & \parallel      &         &  \\
                        &   & \Gamma\big(\mathcal{H}^{\oplus}) \, \textrm{or resp.} \, \Gamma(\mathcal{H}^{\ominus}\big) &  &  \end{array}\right.,
\end{equation}  
which is determined by the (likewise) standard operator
\[
\Gamma\big(P^\oplus \widetilde{\Delta} P^\oplus\big),
\,\,\, \textrm{or respectively} \,\, \Gamma\big(P^\ominus \widetilde{\Delta} P^\ominus\big),
\,\,\,
\textrm{on} \,\,\, \Gamma(\mathcal{H}')
\]
(here we need $\textrm{Spec} \, \widetilde{\Delta}>1$ which we can achieve by addition of the unit operator to $\Delta$).

However, in order to construct the creation-annihilation operators as the Hida operators, we 
need to have the Gelfand triple (\ref{SingleGelfandTripleNeutralEU}) over the single particle Hilbert space
in the so-called standard form, which respect the so called Kubo-Takenaka conditions,
compare \cite{obata-book}, or Subsection \ref{white-setup}. Namely, suppose 
$\mathscr{O}, \mu_{{}_{\mathscr{O}}}$ be a topological space $\mathscr{O}$ with a Baire or 
Borel measure $\mu_{{}_{\mathscr{O}}}$. We then assume that $\mathcal{H}' = L^2(\mathscr{O},
\ud \mu_{{}_{\mathscr{O}}}; \mathbb{C})$, and that $E= \mathcal{S}_{A}(\mathscr{O}; \mathbb{C})$
for a standard operator $A$ on $L^2(\mathscr{O}, \ud \mu_{{}_{\mathscr{O}}}; \mathbb{C})$ and that this  
$\mathcal{S}_{A}(\mathscr{O}; \mathbb{C})$
respects Kubo-Takenaka conditions, \cite{obata-book}. We need these requirements because 
under these requirements the Dirac delta function $\delta_{p}$, $\ p \in \mathscr{O}$
belongs to the strong dual $E^* = \mathcal{S}_{A}(\mathscr{O}; \mathbb{C})^*$ and has some further continuity properties; and on the other hand the Dirac delta function enters into the definition
of  the Hida operators, compare \cite{obata-book}, Subsection \ref{psiBerezin-Hida}, \ref{white-setup},
\ref{WhiteNoiseA}. 

However, the Gelfand triple (\ref{SingleGelfandTripleNeutralEU}) does not have 
in general the standard form, because in general  
$\mathcal{H}' \neq L^2(\mathscr{O}, \ud \mu_{{}_{\mathscr{O}}}; \mathbb{C})$. The Gelfand triple (\ref{SingleGelfandTripleNeutralEU}) would have the standard form e.g. in 
the special case when the projectors $P^\oplus(\widehat{n}\cdot \widehat{l})$, for each point 
$\widehat{n}\cdot \widehat{l}$ of the dual group are equal $0$ or $\boldsymbol{1}$. In that case
$\mathcal{H}'$ (equal $\mathcal{H}^\oplus$ or $\mathcal{H}^\ominus$) is equal to all trace square 
summable matrices over $\mathscr{O}_{{}_{+}}$ or respectively over $\mathscr{O}_{{}_{-}}$. Thus, by adjoining
$(2l+1) \times (2l+1) \times d$ copies of each point $\widehat{n} \cdot \widehat{l} \in 
\mathscr{O}_{{}_{\pm}}$, corresponding to each matrix component of the matrix at the point 
$\widehat{n} \cdot \widehat{l}$, we can present equivalently $\mathcal{H}'$ as the Hilbert space
of square summable functions over the appropriately weighted discrete set 
\[
\mathscr{O} = \Big\{ \widehat{n} \cdot \widehat{l} \times a
\times j \times i \in \mathscr{O}_{{}_{\pm}} \times \{1, \ldots, d \} \times \{-l, \ldots, l \} \times \{-l, \ldots, l \} \Big\}.
\]
But this is not the case in general. 

Nonetheless, this lack of the standard form is not essential, because
we have the following general situation, which allows us to solve this problem. In general using the Fourier transforms, $u_s$ or $v_s$, of the complete sets of positive or negative energy solutions, respectively, we can construct a natural unitary isomorphism $U$ which defines a natural equivalence of the single particle Hilbert space $\mathcal{H}'$ with a said $L^2(\mathscr{O}, \ud \mu_{{}_{\mathscr{O}}}; \mathbb{C})$. This unitary isomorphism $U$ defines at the same time isomorphism of the countably Hilbert nuclear spaces $E$ and $\mathcal{S}_{A}(\mathscr{O}; \mathbb{C})$, and its linear transpose $U^*$ (between the corresponding strong dual spaces) defines the isomorphism between the strong dual spaces
$E^*$ and $\mathcal{S}_{A}(\mathscr{O}; \mathbb{C})^*$, thus giving us isomorphism between the
initial Gelfand triple (\ref{SingleGelfandTripleNeutralEU}) and a canonical standard Gelfand
triple which preserves the Kubo-Takenaka conditions:
\begin{equation}\label{StandardGelfandTripleNeutralEU}
\left. \begin{array}{ccccc}              \mathcal{S}_{A}(\mathscr{O};\mathbb{C})         & \subset &  L^2(\mathscr{O};\mathbb{C}) & \subset & \mathcal{S}_{A}(\mathscr{O}; \mathbb{C})^*        \\
                               \downarrow \uparrow &         & \downarrow \uparrow      &         & \downarrow \uparrow  \\
                                         E         & \subset &  \mathcal{H'} & \subset & E^*        \\
                                                    &         & \parallel     &         &  \\
                         &  & \mathcal{H}^{\oplus} \, \textrm{or resp.} \, \mathcal{H}^{\ominus} &  & 
\end{array}\right.,
\end{equation}  
with the vertical arrows indicating the unitary operator (and its inverse) $U: \mathcal{H}' \rightarrow  L^2(\mathscr{O};\mathbb{C})$ whose restriction to $E$ defines an isomorphism 
$U: E \rightarrow \mathcal{S}_{A}(\mathscr{O}; \mathbb{C})$ of nuclear spaces
and whose linear transposition $U^*$ defines isomorphism 
$\mathcal{S}_{A}(\mathscr{O}; \mathbb{C})^* \rightarrow E^*$. 
The standard nuclear space $\mathcal{S}_{A}(\mathscr{O}; \mathbb{C})$  
corresponds to the standard operator $A=UP^\oplus \widetilde{\Delta} P^\oplus U^{-1}$
(or resp. $A = UP^\ominus \widetilde{\Delta} P^\ominus U^{-1}$) on $L^2(\mathscr{O};\mathbb{C})$, and is determined by the standard operator $A$ through the standard construction (compare  \cite{obata-book}, Subsection \ref{white-setup}, \ref{psiBerezin-Hida}).

Let $d'(\widehat{n}\cdot \widehat{l}) 
= \textrm{dim} \mathcal{H}^\oplus_{{}_{\widehat{n}\cdot \widehat{l}}}$
and let $d''(\widehat{n}\cdot \widehat{l}) = 
\textrm{dim} \mathcal{H}^\ominus_{{}_{\widehat{n}\cdot \widehat{l}}}$. Of course by construction
$d'(\widehat{n}\cdot \widehat{l})$ is equal to the number of elements of the complete
orthonormal system
\[
\{ u_{s}(\widehat{n}\cdot \widehat{l}) \} \,\,\,
\textrm{in}
\,\,\, \mathcal{H}^\oplus_{{}_{\widehat{n}\cdot \widehat{l}}}, \,\,\, n \geq 0,
\]
and $d''(\widehat{n}\cdot \widehat{l})$ is equal to the number of elements of the complete
orthonormal system
\[
\{ v_{s}(\widehat{n}\cdot \widehat{l}) \} \,\,\,
\textrm{in}
\,\,\, \mathcal{H}^\ominus_{{}_{\widehat{n}\cdot \widehat{l}}} \,\,\, 
\textrm{here} \, n<0.
\]
The said unitary isomorphism $U$ acts pointwisely as an operator
$U(\widehat{n}\cdot \widehat{l})$ on the Fourier transforms concentrated at $\widehat{n}\cdot \widehat{l}$. 
In case of $\mathcal{H}' = \mathcal{H}^\oplus$, the operator $U$ sends each
$\widetilde{\phi} \in \mathcal{\mathcal{H}'}$ into the $d'(\widehat{n}\cdot \widehat{l})$-component 
$\mathbb{C}$-valued function\footnote{We have written shortly $d'$ for $d'(\widehat{n}\cdot \widehat{l})$
in this formula.} 
\[
\mathscr{O}_{+} \ni \widehat{n} \cdot \widehat{l}
\longmapsto 
\Big((\widetilde{\phi})_1 \oplus (\widetilde{\phi})_2 \oplus \cdots \oplus (\widetilde{\phi})_{d'}\Big)(\widehat{n}\cdot \widehat{l}) 
=
(\widetilde{\phi})_1(\widehat{n}\cdot \widehat{l}) \oplus 
(\widetilde{\phi})_2(\widehat{n}\cdot \widehat{l}) \oplus 
\cdots \oplus 
(\widetilde{\phi})_{d'}(\widehat{n}\cdot \widehat{l})
\]
with each value $(\widetilde{\phi})_{s}(\widehat{n}\cdot \widehat{l})$
of the $s$-th component $(\widetilde{\phi})_{s}$ at 
the point $\widehat{n}\cdot \widehat{l} \in \mathscr{O}_{+}$ equal to the $s$-th
coefficient in the decomposition of the vaule $\widetilde{\phi}(\widehat{n}\cdot \widehat{l})$
with respect to the orthonormal system $\{ u_{s}(\widehat{n}\cdot \widehat{l}) \}$,
$1 \leq s \leq d'(\widehat{n}\cdot \widehat{l})$. Thus, the $s$-th component
of the value $\big(U\widetilde{\phi}\big)(\widehat{n}\cdot \widehat{l}) =
U(\widehat{n}\cdot \widehat{l}) \widetilde{\phi}(\widehat{n}\cdot \widehat{l})$ 
is equal to the projection on the $s$-th element $u_s(\widehat{n}\cdot \widehat{l})$
of the complete system of Fourier transformed solutions.

Similarly, the said unitary isomorphism $U$ acts pointwisely through an 
operator $U(\widehat{n}\cdot \widehat{l})$, and in case of $\mathcal{H}' = \mathcal{H}^\ominus$, 
$U$ sends each
$\widetilde{\phi} \in \mathcal{\mathcal{H}'}$ into the $d''(\widehat{n}\cdot \widehat{l})$-component 
$\mathbb{C}$-valued function
\[
\mathscr{O}_{-} \ni \widehat{n} \cdot \widehat{l}
\longmapsto 
\Big((\widetilde{\phi})_1 \oplus (\widetilde{\phi})_2 \oplus \cdots \oplus (\widetilde{\phi})_{d''}\Big)(\widehat{n}\cdot \widehat{l}) 
=
(\widetilde{\phi})_1(\widehat{n}\cdot \widehat{l}) \oplus 
(\widetilde{\phi})_2(\widehat{n}\cdot \widehat{l}) \oplus 
\cdots \oplus 
(\widetilde{\phi})_{d''}(\widehat{n}\cdot \widehat{l})
\]
with each value $(\widetilde{\phi})_{s}(\widehat{n}\cdot \widehat{l})$
of the $s$-th component $(\widetilde{\phi})_{s}$ at 
the point $\widehat{n}\cdot \widehat{l} \in \mathscr{O}_{+}$ equal to  the $s$-th
coefficient in the decomposition of the vaule $\widetilde{\phi}(\widehat{n}\cdot \widehat{l})$
with respect to the orthonormal system $\{v_{s}(\widehat{n}\cdot \widehat{l})\}$,
$1 \leq s \leq d''(\widehat{n}\cdot \widehat{l})$.
Therefore,  the $s$-th component
of the value $\big(U\widetilde{\phi}\big)(\widehat{n}\cdot \widehat{l})=
U(\widehat{n}\cdot \widehat{l}) \widetilde{\phi}(\widehat{n}\cdot \widehat{l})$ 
is equal to the projection on the $s$-th element $v_s(\widehat{n}\cdot \widehat{l})$
of the complete system of Fourier transformed solutions.

Of course the projections in the definition of $U(\widehat{n}\cdot \widehat{l})$
are with respect to the Hilbert space inner product 
(\ref{<tildephi(n.l),tildephi'(n.l)>}) of the Fourier transforms
$\widetilde{\mathscr{H}}$ concentrated on the single point $\widehat{n}\cdot \widehat{l}$
and determined by the Plancherel formula: 
\begin{multline}\label{<tildephi(n.l),tildephi'(n.l)>}
\big\langle \widetilde{\phi}(\widehat{n}\cdot \widehat{l}), \, 
\widetilde{\phi}'(\widehat{n}\cdot \widehat{l}) \big\rangle \\ =
(2l+1) \textrm{Tr} [\widetilde{\phi}(\widehat{n}\cdot \widehat{l}) \,\, 
\widetilde{\phi}'(\widehat{n}\cdot \widehat{l})]
= (2l+1) \sum \limits_{a,j,i} \overline{\widetilde{\phi}^{a}_{{}_{ji}}(\widehat{n}\cdot \widehat{l})} \, \widetilde{\phi}^{' \, a}_{{}_{ji}}(\widehat{n}\cdot \widehat{l}),
\end{multline}
so that
\begin{multline*}
\langle \phi, \phi' \rangle = 
\sum \limits_{a} \int \overline{\phi^a(t,w)} \, \phi^{' \,a}(t,w) \,\ud t \ud w
\\
= \sum_{\widehat{n}\cdot \widehat{l} \in \widehat{\widetilde{S}^1}\times G}
\big\langle \widetilde{\phi}(\widehat{n}\cdot \widehat{l}), \, 
\widetilde{\phi}'(\widehat{n}\cdot \widehat{l}) \big\rangle
= \langle \widetilde{\phi},\widetilde{\phi}' \rangle.
\end{multline*}

In all practical situations the dimensions $d'(\widehat{n}\cdot \widehat{l})$
and $d''(\widehat{n}\cdot \widehat{l})$
of $\mathcal{H}^\oplus_{{}_{\widehat{n}\cdot \widehat{l}}}$
and $\mathcal{H}^\ominus_{{}_{\widehat{n}\cdot \widehat{l}}}$ are equal:
$d'(\widehat{n}\cdot \widehat{l}) = d''(\widehat{n}\cdot \widehat{l})$, but here we 
keep the presentation at the level of greater generality and \emph{a priori} will
distinguish between $d'$ and $d''$. In order to simplify notation we will
simply write $d'$ and $d''$ for $d'(\widehat{n}\cdot \widehat{l})$
and $d''(\widehat{n}\cdot \widehat{l})$ because it will be clear from the context
which particular point $\widehat{n}\cdot \widehat{l}$ is meant as the argument
in the functions $d'$ and $d''$.

We thus have the unitary isomorphism $U$ in each case with the said properties,
which defines equivalence of the single particle Hilbert space $\mathcal{H}' = \mathcal{H}^\oplus$
(respectively $\mathcal{H}' = \mathcal{H}^\ominus$)
with a standard Hilbert space $L^2(\mathscr{O}; \ud \mu_{{}_{\mathscr{O}}})$. The 
set $\mathscr{O}$ is equal to the enlargement of the set $\mathscr{O}_{{}_{+}}$
(respectively of the set $\mathscr{O}_{{}_{-}}$) of points 
$\widehat{n}\cdot \widehat{l} \in \widehat{\widetilde{\mathbb{S}^1}\times G}$ in which we have exactly $d'(\widehat{n}\cdot \widehat{l})$ copies (respectively $d''(\widehat{n}\cdot \widehat{l})$ copies) of each point $\widehat{n}\cdot \widehat{l} \in \mathscr{O}_{{}_{\pm}}$. 
Therefore, $\mathscr{O}$ is in each case equal to a discrete topological space in which each single element
set is open, closed  and compact, with the Baire measure equal to the discrete measure assigning
to each single point set the measure equal $1$.

The unitary operator $U$, say in case  of $\mathcal{H}' = \mathcal{H}^\oplus$,
is equal to the operator function 
\[
\mathscr{O}_{+} \ni \widehat{n} \cdot \widehat{l}
\longmapsto U(\widehat{n} \cdot \widehat{l})
\]
with the pointwise action on $\mathcal{H}' \subset \widetilde{\mathscr{H}}$, and with each
operator $U(\widehat{n} \cdot \widehat{l})$ acting on 
$\mathcal{H}^{\oplus}_{{}_{\widehat{n}\cdot \widehat{l}}}$. Now the $s$-th
component $(\widetilde{\phi})_{s}(\widehat{n}\cdot \widehat{l})$ of the image
$U(\widehat{n} \cdot \widehat{l}) \, \widetilde{\phi}(\widehat{n} \cdot \widehat{l})$  
is by definition equal to the projection of the matrix 
$\widetilde{\phi}(\widehat{n} \cdot \widehat{l}) \in 
\mathcal{H}^{\oplus}_{{}_{\widehat{n}\cdot \widehat{l}}}$
on the one dimensional subspace spanned by the $s$-th element
$u^{a}_{s}(\widehat{n}\cdot \widehat{l})$  of the complete system
$\{u^{a}_{s}(\widehat{n}\cdot \widehat{l}) \}$ in 
$\mathcal{H}^{\oplus}_{{}_{\widehat{n}\cdot \widehat{l}}}$
with respect to the natural inner product (\ref{<tildephi(n.l),tildephi'(n.l)>}).
Of course we have the analogous formula for the isomorphism $U$ in case 
$\mathcal{H}' = \mathcal{H}^\ominus$, with the complete system 
$\{u^{a}_{s}(\widehat{n}\cdot \widehat{l}) \}$ replaced by
$\{v^{a}_{s}(\widehat{n}\cdot \widehat{l}) \}$. 

We may represent the action of the isomorphism $U$ as the pointwise matrix multiplication
operation, if we only fix the manner in which each value 
$\widetilde{\phi}(\widehat{n} \cdot \widehat{l}) \in 
\mathcal{H}^{\oplus}_{{}_{\widehat{n}\cdot \widehat{l}}}$ 
or ${} \in \mathcal{H}^{\ominus}_{{}_{\widehat{n}\cdot \widehat{l}}}$
we treat as a column vector. In particular let us regard it as the following column
vector
\begingroup\makeatletter\def\f@size{5}\check@mathfonts
\def\maketag@@@#1{\hbox{\m@th\large\normalfont#1}}%
\[
\left( \begin{array}{c} 
\widetilde{\phi^1}_{{}_{ji}}(\widehat{n}\cdot \widehat{l}) \\
\vdots \\
\vdots \\
\widetilde{\phi^a}_{{}_{ji}}(\widehat{n}\cdot \widehat{l}) \\
\vdots \\
\vdots \\
\widetilde{\phi^d}_{{}_{ji}}(\widehat{n}\cdot \widehat{l}) \\
\vdots 
\end{array}\right).
\]
\endgroup
in which  values of the multi-index ${}^a_{{}_{\,\,\, j\,\,i}}$ are in the lexicographical order, 
inherited form the ordinary orderings of the separate indices $a,j,i$ with respect to their
values, and with the lexicographical ordering proviso that the index $a$ comes as the first, the index $j$ as the second, and finally $i$ as the last, ``letter'' 
of the word ``${}^a_{{}_{\,\,\, j \,\,i}}$''. Thus, we have the ordering
\[ 
{}^1_{{}_{\,\,\, -l \,\,\, -l}} < {}^1_{{}_{\,\,\,-l \,\,\, -l+1}} < \ldots < {}^1_{{}_{\,\,\, -l \,\,\, -l+2}} < \ldots
\]
with the fixed index $a=1$ coming first and with the indices $j\times i$ in the lexicographical order,
next comes the fixed value $a=2$ with the indices  $j\times i$ in the lexicographical order,
and so on. This means that each vertical ``three-dots'' immediately below $\widetilde{\phi^a}_{{}_{ji}}(\widehat{n}\cdot \widehat{l})$ with fixed $a \in (1, \ldots, d)$  represent the matrix components 
$\widetilde{\phi^a}_{{}_{ji}}(\widehat{n}\cdot \widehat{l})$ with
fixed $a \in (1, \ldots, d)$ and with 
$j\times i \in (-l, \ldots, l)\times (-l, \dots, l)$, arranged in the lexicographical order.

Then, for example in case of $\mathcal{H}' = \mathcal{H}^\oplus$, the isomorphism
$U: \widehat{n} \cdot \widehat{l}
\longmapsto U(\widehat{n} \cdot \widehat{l})$ can be represented as the pointwise matrix multiplication with the matrix 
\begingroup\makeatletter\def\f@size{5}\check@mathfonts
\def\maketag@@@#1{\hbox{\m@th\large\normalfont#1}}%
\[
U(\widehat{n}\cdot \widehat{l}) = (2l+1)
\left( \begin{array}{cccc}   \overline{{u_{1}^{1}}_{{}_{ji}}(\widehat{n}\cdot \widehat{l})} \cdots &  \overline{{u_{1}^{2}}_{{}_{ji}}(\widehat{n}\cdot \widehat{l})} \cdots &
\cdots & \overline{{u_{1}^{d}}_{{}_{ji}}(\widehat{n}\cdot \widehat{l})}\cdots   \\
&\cdots&& \\
\overline{{u_{s}^{1}}_{{}_{ji}}(\widehat{n}\cdot \widehat{l})} \cdots &  \overline{{u_{s}^{2}}_{{}_{ji}}(\widehat{n}\cdot \widehat{l})} \cdots &
\cdots & \overline{{u_{s}^{d}}_{{}_{ji}}(\widehat{n}\cdot \widehat{l})}\cdots   \\
&\cdots&& \\
\overline{{u_{d'}^{1}}_{{}_{ji}}(\widehat{n}\cdot \widehat{l})}\cdots &  \overline{{u_{d'}^{2}}_{{}_{ji}}(\widehat{n}\cdot \widehat{l})}\cdots &
\cdots & \overline{{u_{d'}^{4}}_{{}_{ji}}(\widehat{n}\cdot \widehat{l})}\cdots   \\
  \end{array}\right).
\]
\endgroup
Here we have the matrix which inherits the assumed lexicographical ordering of the multi-indices
${}^{a}_{{}_{\,\,\,j\,\, i}}$, \emph{i.e.} each horizontal ``three-dots'' lying immediately
after the element $\overline{{u_{s}^{a}}_{{}_{ji}}(\widehat{n}\cdot \widehat{l})}$
with fixed $a$ and $s$ represents all matrix elements  $\overline{{u_{s}^{a}}_{{}_{ji}}(\widehat{n}\cdot \widehat{l})}$ with fixed $a$ and $s$, and with the matrix indices $ji \in (-l, \ldots, l)
\times (-l, \ldots, l)$ arranged in the lexicographical order. Of course, we have identical
formula for $U$ in case of $\mathcal{H}' = \mathcal{H}^\ominus$, with 
$u_s$ replaced by $v_s$.

The inverse $U^{-1}$ of the isomorphism\footnote{Compare the formula of the Appendix \ref{fundamental,u,v}for the analogous isomorphism (\ref{isomorphismU}), Subsection \ref{psiBerezin-Hida}, in the construction
of the free Dirac field on the Minkowski space-time.} $U$
can be regarded as the operator of pointwise multiplication by the matrix
\begingroup\makeatletter\def\f@size{5}\check@mathfonts
\def\maketag@@@#1{\hbox{\m@th\large\normalfont#1}}%
\[
U^{-1}(\widehat{n}\cdot \widehat{l}) =
\left( \begin{array}{ccc}   
{u_{1}^{1}}_{{}_{ji}}(\widehat{n}\cdots \widehat{l}) & \cdots & u_{d'}^{1}(\widehat{n}\cdot \widehat{l}) \\
\vdots& &\vdots \\
{u_{1}^{2}}_{{}_{ji}}(\widehat{n}\cdot \widehat{l}) & \cdots&  {u_{d'}^{2}}_{{}_{ji}}(\widehat{n}\cdot \widehat{l}) \\
\vdots& &\vdots \\
\vdots& &\vdots\\
{u_{1}^{d}}_{{}_{ji}}(\widehat{n}\cdot \widehat{l}) &  \cdots & {u_{d'}^{d}}_{{}_{ji}}(\widehat{n}\cdot \widehat{l}) \\
\vdots& &\vdots\\
  \end{array}\right).
\]
\endgroup
Here each vertical ``three-dots'' immediately below 
${u_{s}^{a}}_{{}_{ji}}(\widehat{n}\cdot \widehat{l})$ with fixed $a,s$, represent all the matrix
elements ${u_{s}^{a}}_{{}_{ji}}(\widehat{n}\cdot \widehat{l})$ with fixed $a,s$ and 
with $ji$ arranged in the lexicographical order.

Here the matrix $U^{-1}(\widehat{n}\cdot \widehat{l})$ is acting by matrix multiplication
on the value 
$\Big((\widetilde{\phi})_{{}_{1}} \oplus (\widetilde{\phi})_{{}_{2}} \oplus \cdots 
\oplus (\widetilde{\phi})_{{}_{d'(\widehat{n}\cdot \widehat{l})}}\Big)(\widehat{n}\cdot \widehat{l}) $ of the element 
\[
(\widetilde{\phi})_{{}_{1}} \oplus \cdots \oplus (\widetilde{\phi})_{{}_{d'}} 
\in  L^{2}(\mathscr{O}; \mathbb{C})
\]
regarded as a column
\[
\left( \begin{array}{c} 
(\widetilde{\phi})_1(\widehat{n}\cdot \widehat{l}) \\
\vdots \\
(\widetilde{\phi})_{d'}(\widehat{n}\cdot \widehat{l})
 \end{array}\right).  
\]
Note that 
\[
U(\widehat{n}\cdot \widehat{l})U^{-1}(\widehat{n}\cdot \widehat{l}) = \boldsymbol{1}_{d'}, \,\,\,\,
U^{-1}(\widehat{n}\cdot \widehat{l}) U(\widehat{n}\cdot \widehat{l}) = 
\boldsymbol{1}_{{}_{\mathcal{H}^{\oplus}_{{}_{\widehat{n}\cdot\widehat{l}}}}}. 
\]
The first identity follows from the orthonormality relations of the
Fourier transforms $u_s$ of the fundamental solutions. The second identity follows from the fact 
that the actions of the projection operators 
$P^{\oplus}(\widehat{n}\cdot \widehat{l})$  
in the Hilbert space (\ref{H[n.l]x[loV]}) which project, respectively, on the subspaces
$\mathcal{H}^{\oplus}_{{}_{\widehat{n}\cdot\widehat{l}}}$,  are equal
\[
P^{\oplus}(\widehat{n}\cdot \widehat{l})w =
\sum \limits_{s=1}^{d'} 
u_s(\widehat{n}\cdot \widehat{l}) \, \langle u_s(\widehat{n}\cdot \widehat{l}), w \rangle, 
\]
for all
\[
 w \in 
\mathcal{H}_{{}_{[\,\widehat{n}\cdot \widehat{l} \,]\, \times \,[\, \overline{\,\widehat{l}\,} \otimes V \,]}} 
= \big \{ \widetilde{\phi}, \,\, \phi \in \mathscr{H}, \textrm{with $\widetilde{\phi}$ concentrated on $\widehat{n}\cdot \widehat{l}$} \big \}
\]
with the natural inner product $\langle \cdot, \cdot \rangle$ defined by the Plancherel formula 
and equal (\ref{<tildephi(n.l),tildephi'(n.l)>}). Thus, $P^{\oplus}(\widehat{n}\cdot \widehat{l})$ has the following matrix elements
\[
\big[P^{\oplus}(\widehat{n}\cdot \widehat{l})\big]^{aji \,\,\,\, a'j'i'} =
\sum \limits_{s=1}^{d'} 
(2l+1) {u^{a}_{s}}_{{}_{ji}}(\widehat{n}\cdot \widehat{l}) \, 
\overline{{u^{a'}_{s}}_{{}_{j'i'}}(\widehat{n}\cdot \widehat{l})},
\]
which indeed immediately confirms that $U(\widehat{n}\cdot \widehat{l})$
and $U^{-1}(\widehat{n}\cdot \widehat{l})$ are mutually inverse.
Of course, we have the analogous formulas for $\mathcal{H}' = \mathcal{H}^\ominus$,
with the operator function $P^{\oplus}$, the dimension function $d'$ and the complete system $\{u_{s}(\widehat{n}\cdot \widehat{l})\}$, $1 \leq s \leq d'(\widehat{n}\cdot \widehat{l})$ replaced
with $P^{\ominus}$, $d''$ and $\{v_{s}(\widehat{n}\cdot \widehat{l})\}$, $1 \leq s \leq d''(\widehat{n}\cdot \widehat{l})$. 

We should note that for the essentially neutral quantum free fields 
(e. g.  the electromagnetic potential field) we encounter in 
construction of realistically interacting fields by using the causal perturbative method, 
we have not only the equality 
$d'(\widehat{n}\cdot \widehat{l}) =
d''(\widehat{-n}\cdot \widehat{l})$. For such fields which we encounter in practice, whenever we replace $n$ 
and the matrix indices $j,i$ in the Fourier transform of a positive energy solution by $-n$, $-j,-i$, 
we obtain complex conjugation of the Fourier transform of positive energy solution equal to the Fourier transform 
of a negative energy solution, and we can put
\[
{v_{s}^{a}}_{{}_{ji}}(\widehat{n}\cdot \widehat{l}) = \overline{{u_{s}^{a}}_{{}_{ji}}(\widehat{n}\cdot \widehat{l})}
= {u_{s}^{a}}_{{}_{-j-i}}(\widehat{-n}\cdot \widehat{l}),
\,\,\, \widehat{n}\cdot \widehat{l} \in \mathscr{O}_{{}_{-}},
\,\,\,\,\,\,\, \textrm{\small (for neutral fields)}
\]
For the analysis of the commutation and the pairing generalized functions
of such neutral free fields $\mathbb{A}$ it is convenient to introduce 
a concise notation for the function $w_s$
\[
w_s(\widehat{n}\cdot \widehat{l}) = u_s(\widehat{n}\cdot \widehat{l}), \,\,\,
\widehat{n}\cdot \widehat{l} \in \mathscr{O}_{{}_{+}}, \,\,\,
w_s(\widehat{n}\cdot \widehat{l}) = \overline{u_s(\widehat{n}\cdot \widehat{l})}
= v_s(\widehat{n}\cdot \widehat{l}), 
\,\,\, \widehat{n}\cdot \widehat{l} \in \mathscr{O}_{{}_{-}},
\]
embracing all, positive and negative energy solutions, defined on $\mathscr{O}_{{}_{+}} \sqcup 
\mathscr{O}_{{}_{+}}$ with the property
\begin{equation}\label{n->-nReflectionNeutralSolutions}
{w_{s}^{a}}_{{}_{j \,\, i}}(\widehat{n}\cdot \widehat{l}) = {w_{s}^{a}}_{{}_{-j \,\, -i}}(\widehat{-n}\cdot \widehat{l}).
\end{equation}

Prepared with the single particle Gelfand triple and its standard form (\ref{StandardGelfandTripleNeutralEU}) joined by the unitary isomorphism $U$
we can now repeat the construction of the essentially neutral (\emph{i.e.} real,
positive or negative energy) quantum free field $\mathbb{A}$, as an integral kernel
operator with vector-valued distributional kernel in the sense of Obata
\cite{obataJFA}, repeating exactly the construction
of Subsection \ref{psiBerezin-Hida}. 

Before we give explicit general formula 
for an essentially neutral (say ``real'') free field $\mathbb{A}$ and its commutation ``function'', 
we pass now to 
the analogue preparation
of the single particle Gelfand triple and its standard form in case of a``complex''
quantum free field $\boldsymbol{\psi}$, \emph{i.e.} field which is not essentially neutral
(with antiparticle not coinciding with the particle). In that case both Hilbert spaces
$\mathcal{H}^\oplus$ and $\mathcal{H}^\ominus$ participate in the construction of the single particle
Hilbert space of the positive energy (and separately also of the negative energy)
field $\boldsymbol{\psi}$, compare Subsections \ref{FirstStepH}-\ref{electron+positron}. 
But we treat the elements $\widetilde{\phi}$ of $\mathcal{H}^\ominus$ not as the Fourier transforms
of negative energy solutions $\phi$ of the
original system of differential equations $D_{{}_{iff}} \phi = 0$ defining the field, but instead as 
conjugations $\phi^\flat$ of positive
energy solutions of a conjugated system ${D_{{}_{iff}}}^\flat \phi^\flat = \big(D_{{}_{iff}} \phi \big)^\flat=0$,
of differential equations. The conjugation $\phi^\flat$ of $\phi \in \mathscr{H}$ is defined in the ordinary manner:
\[
\phi^\flat (t,w) = \phi(t,w)^{+} = \overline{\phi(t,w)}^{T},
\]
where the bar stands for ordinary complex conjugation and $T$ for the ordinary transposition,
which converts our column-vectors $\phi$ into the row-vectors with $d$ components
($d = \textrm{dim} \, V$ of the representation $V$ in the transformation formula
(\ref{UphiOnSxG}) acting on $\mathscr{H}$ associated with the considered field). To this conjugation corresponds via the Fourier transform, the conjugation operation (denoted likewise by the superscript $\flat$) 
$(\widetilde{\phi})^\flat = \widetilde{\phi^{\flat}}$ of the Fourier transform images 
$\widetilde{\phi} \in \widetilde{\mathscr{H}}$ of 
$\phi \in \mathscr{H}$. It is easily seen that each value $(\widetilde{\phi})^\flat(\widehat{n}\cdot \widehat{l})$ of $(\widetilde{\phi})^\flat$
at $\widehat{n}\cdot \widehat{l}$ is equal to the row
\[
(\widetilde{\phi})^\flat(\widehat{n}\cdot \widehat{l}) = 
\overline{\widetilde{\phi^{{}^{T}}}\Big(\overline{\widehat{n}}\cdot \overline{\,\widehat{l}\,}\Big)}
=
\overline{\widetilde{\phi}{\overset{{}}{{}^{{}^{}}}}^{{}^{T}}\Big(\overline{\widehat{n}}\cdot \overline{\,\widehat{l}\,}\Big)}
=
\overline{\widetilde{\phi}\Big(\overline{\widehat{n}}\cdot \overline{\,\widehat{l}\,}\Big)}^{T}
\]
of $d$ matrices, each being equal to a $(2l+1) \times (2l+1)$ matrix.
Because we have several possible transpositions in the linear spaces of values 
$\widetilde{\phi}(\widehat{n}\cdot \widehat{l})$, we need to specify the last two transpositions 
in the last formula. 
Recall that for  $\mathbb{C}^d$-valued functions $\phi$ on the group $\widetilde{S}^1 \times G$ 
we regard $\mathcal{F}\phi(\widehat{n}\cdot \widehat{l}) =
\widetilde{\phi}(\widehat{n}\cdot \widehat{l})$ as a matrix  
\[
\Phi_{{}_{a \,\,\,\,\,ji}} = \widetilde{\phi^{a}}_{{}_{ji}}(\widehat{n}\cdot \widehat{l})
\]
of a linear operator from the Hilbert space of $\textrm{dim} \, \widehat{n}\cdot \widehat{l} \times \textrm{dim} \, \widehat{n}\cdot \widehat{l}$ matrices (regarded as 
the linear space of Hilbert-Schmidt operators with the Hilbert-Schmidt norm multiplied by $\textrm{dim} \, \widehat{n}\cdot \widehat{l} = 2l+1$) to the conjugation of the Hilbert space $\mathbb{C}^d$, so that
\[
\widetilde{\phi^T}(\widehat{n}\cdot \widehat{l}) 
= \big(\widetilde{\phi} \big)^T(\widehat{n}\cdot \widehat{l}),
\]
compare (\ref{TranspositionConventionForFourierTr}). In more explicit form
we ragard  $\widetilde{\phi}(\widehat{n}\cdot \widehat{l})$ as a column 
\[
\left( \begin{array}{c}  
\widetilde{\phi^{1}}(\widehat{n}\cdot \widehat{l}) \\
 \vdots \\
\widetilde{\phi^{d}}(\widehat{n}\cdot \widehat{l})
                     \end{array}\right) 
\]
of $d$ matrices  $\widetilde{\phi^{a}}_{{}_{ji}}(\widehat{n}\cdot \widehat{l})$, $a=1, \ldots, d$,
with the matrix indices $-l \leq j,i\leq l$. The 
value $\widetilde{\phi^T}(\widehat{n}\cdot \widehat{l})
= \big(\widetilde{\phi}\big)^T(\widehat{n}\cdot \widehat{l})$ is equal to the row
\[
\left( \begin{array}{ccc}  
\widetilde{\phi^{1}}(\widehat{n}\cdot \widehat{l}), & 
 \ldots, &
\widetilde{\phi^{d}}(\widehat{n}\cdot \widehat{l})
                     \end{array}\right)
\]
of $d$ matrices  $\widetilde{\phi^{a}}_{{}_{ji}}(\widehat{n}\cdot \widehat{l})$,
$1 \leq a \leq d$, with the matrix indices $-l \leq j,i\leq l$.
Thus, for the chosen interpretation of the matrix $\widetilde{\phi}(\widehat{n}\cdot \widehat{l})$
and the transposition, we have the equality $\widetilde{\phi^T}(\widehat{n}\cdot \widehat{l}) = 
\big(\widetilde{\phi}\big)^T(\widehat{n}\cdot \widehat{l}) = 
\big(\widetilde{\phi}(\widehat{n}\cdot \widehat{l})\big)^T$.
But note that in general for the other interpretations of $\widetilde{\phi}(\widehat{n}\cdot \widehat{l})$,
and thus for the other possible transpositions in the space of values
$\widetilde{\phi}(\widehat{n}\cdot \widehat{l})$ of the Fourier transform,
we will in general have $\widetilde{\phi^T}(\widehat{n}\cdot \widehat{l}) \neq \big(\widetilde{\phi}\big)^T(\widehat{n}\cdot \widehat{l}) =\big(\widetilde{\phi}(\widehat{n}\cdot \widehat{l})\big)^T$.

Note that
\[
\widetilde{(\phi)^{\flat\,\,\,a}}_{{}_{\,\,\,\,\,\,\,\, j \,\,\, i}}(\widehat{n}\cdot \widehat{l})=
(\widetilde{\phi})^{\flat\,\,\, a}_{{}_{\,\,\,\,\,\,\,\, j \,\,\, i}}(\widehat{n}\cdot \widehat{l}) 
= \overline{\widetilde{(\phi^T)^a}_{{}_{j \,\,\, i}}\Big(\overline{\widehat{n}}\cdot \overline{\,\widehat{l}\,}\Big)}
= \overline{\widetilde{(\phi^T)^a}_{{}_{-j \,\,\, -i}}(\widehat{-n}\cdot \widehat{l})}.
\]

In the sequel the notation $\widetilde{\phi^{a}}(\widehat{n}\cdot \widehat{l})^T$
with fixed index $a$ will mean the ordinary transposition of the matrix 
$\widetilde{\phi^{a}}_{{}_{ji}}(\widehat{n}\cdot \widehat{l})$, $-l \leq j,i \leq l$.

Next we define the linear space $\mathcal{H}^{\ominus \,\flat}$ of conjugations
$(\widetilde{\phi})^{\flat}$ of Fourier transforms $\widetilde{\phi} \in \mathcal{H}{^\ominus}$ 
of all negative energy solutions $\phi$. It is naturally a Hilbert space with the 
inner product
\[
\big( (\widetilde{\phi})^{\flat}, (\widetilde{\phi}')^{\flat} \big)_{\mathcal{H}^{\ominus \,\flat}} = \big( \widetilde{\phi}', \widetilde{\phi} \big)_{\mathcal{H}^{\ominus}}
\]
(note the reversed position of the primed and unprimed elements). The linear operations
of multiplication by the complex number $\alpha$ and additions are naturally defined in 
 $\mathcal{H}^{\ominus \,\flat}$ by the formulas
\[
 \alpha \cdot (\widetilde{\phi})^{\flat} = (\overline{\alpha} \widetilde{\phi})^{\flat}
 = \alpha (\widetilde{\phi})^{\flat}, \,\,
 (\widetilde{\phi})^{\flat} 
 +
 (\widetilde{\phi}')^{\flat}
 =
 (\widetilde{\phi} +\widetilde{\phi}')^{\flat}, 
 \,\,\,
 \widetilde{\phi}, \widetilde{\phi}' \in \mathcal{H}^{\ominus},
\]
compare Subsection \ref{positron}. 

In the Hilbert space $\mathcal{H}^{\ominus \,\flat}$ act the conjugation
$U^{\flat}$ of the representation (\ref{UphiOnSxG}) with the general definition of the 
conjugation of the representation
\[
U^{\flat} \phi^{\flat} = (U\phi)^{\flat},
\,\,\, \phi \in \mathcal{H}^\ominus \subset \mathscr{H}
\,\,\, \textrm{and $U$ of the form (\ref{UphiOnSxG})}.
\]

The single particle Hilbert space of the ``complex'' quantum free field $\boldsymbol{\psi}$
is defined as the Hilbert space direct sum 
$\mathcal{H}^{\oplus} \oplus \mathcal{H}^{\ominus \,\flat}$
acted on by the representation 
\[
U \oplus U^{\flat},
\,\,\, \textrm{with $U$ of the form (\ref{UphiOnSxG})}.
\]
The Fock-Hilbert space of the ``complex'' field $\boldsymbol{\psi}$ is defined as the 
Fock-Hilbert space 
\[
\Gamma\big(\mathcal{H}^{\oplus} \oplus \mathcal{H}^{\ominus \,\flat}
\big) = \Gamma\big(\mathcal{H}^{\oplus}\big) \otimes 
\Gamma\big(\mathcal{H}^{\ominus \,\flat} \big)
\]  
over the direct sum $\mathcal{H}^{\oplus} \oplus \mathcal{H}^{\ominus \,\flat}$,
with the creation-annihilation operators $a_{\oplus}(g)^+, a_{\oplus}(g)$ in 
$\Gamma\big(\mathcal{H}^{\oplus}\big)$, the creation-annihilation operators
$a_{\ominus}(h)^+, a_{\ominus}(h)$ in 
$\Gamma\big(\mathcal{H}^{\ominus \,\flat}\big)$, the creation-annihilation
operators $a'(g \oplus h)^+, a'(g\oplus h)$ in 
$\Gamma\big(\mathcal{H}^{\oplus} \oplus \mathcal{H}^{\ominus \,\flat}\big)$,
and the parity number operators $\textrm{In}_\oplus$ in $\Gamma\big(\mathcal{H}^{\oplus}\big)$,
and $\textrm{In}_\oplus$ in $\Gamma\big(\mathcal{H}^{\ominus \, \flat}\big)$,
defined exactly as in Subsections \ref{electron} -- \ref{electron+positron}.

Next we introduce the nuclear space $E$, similarly as above, which composes
with the single particle Hilbert space 
$\mathcal{H}'= \mathcal{H}^{\oplus} \oplus \mathcal{H}^{\ominus \, \flat}$
of the field $\boldsymbol{\psi}$, a Gelfand triple
\begin{equation}\label{SinlgeGelfand3pleGeneralpsi}
\left. \begin{array}{ccccc}              E         & \subset &  \mathcal{H'} & \subset & E^*        \\
                                                   &         & \parallel      &         &  \\
                        &   & \mathcal{H}^{\oplus} \oplus \mathcal{H}^{\ominus \, \flat} &  &  \end{array}\right..
\end{equation}
The required countably Hilbert nuclear space $E$ has the direct sum structure
$E = E^\oplus \oplus E^{\oplus \, \flat}$ of the following two nuclear spaces
\[
E^\oplus = \Big\{ g = P^\oplus\widetilde{f}|_{{}_{\mathscr{O}_{+}}}, \,\,\
f \in \mathcal{S}_{\Delta}(\widetilde{\mathbb{S}^1}\times SU(2, \mathbb{C}) ; \mathbb{C}^d)
\Big\}
\]
and 
\[
E^\ominus = \Big\{ h = \big(P^\ominus\widetilde{f}|_{{}_{\mathscr{O}_{-}}}\big)^\flat, \,\,\
f \in \mathcal{S}_{\Delta}(\widetilde{\mathbb{S}^1}\times SU(2, \mathbb{C}) ; \mathbb{C}^d)
\Big\}.
\]
Note that the nuclear space $E^\oplus$ (respectively $E^\ominus$) coincides with the nuclear space 
$E\subset \mathcal{H}^\oplus$ (respectively $E\subset \mathcal{H}^\ominus$)
used above in the construction of the Gelfand triple over the single particle Hilbert space of the positive (respectively, negative) energy neutral field and is canonically associated to the standard operator 
$A = P^\oplus \widetilde{\Delta}P^\oplus$ (respectively $A = P^\ominus \widetilde{\Delta}P^\ominus$)
on $\mathcal{H}^\oplus$ (respectively $\mathcal{H}^\ominus$). $E^{\ominus \, \flat}$
is equal to the conjugation of the nuclear space $E^\ominus$, and therefore is determined canonically
(in the sense of \cite{GelfandIV} or \cite{obata-book}) by the standard operator
$A = \big(P^\ominus \widetilde{\Delta}P^\ominus \big)^{\flat}$
(\emph{i.e.} with $r$-th power of its inverse being nuclear for $r>2$). Therefore, 
the nuclear space $E = E^\oplus \oplus E^{\oplus \, \flat}$ is likewise equal to the abstract countably Hilbert nuclear space determined by the standard operator
\[
A' = P^\oplus \widetilde{\Delta}P^\oplus \oplus  \big(P^\ominus \widetilde{\Delta}P^\ominus \big)^{\flat}
\,\,\,
\textrm{on}
\,\,\, 
\mathcal{H}' = \mathcal{H}^{\oplus} \oplus \mathcal{H}^{\ominus \,\flat}.
\]
 Therefore, it has the form which allows lifting of this construction to the second quantized level
with the corresponding Gelfand triple (if we add to $A'$ the unit operator):
\[
\left. \begin{array}{ccccc}              (E)         & \subset &  \Gamma(\mathcal{H'}) & \subset & (E)^*        \\
                                                   &         & \parallel      &         &  \\
                        &   & \Gamma\big(\mathcal{H}^{\oplus} \oplus \mathcal{H}^{\ominus \, \flat}\big) &  &  \end{array}\right.,
\]  

Again in general the Gelfand triple (\ref{SinlgeGelfand3pleGeneralpsi}) does not have the required standard form. But using the Fourier transform of the complete systems of positive and negative energy solutions of the system $D_{{}_{iff}} \phi = 0$ of differential equations associated to the field we construct the unitary isomorphism of the Gelfand triple (\ref{SinlgeGelfand3pleGeneralpsi})
with a standard Gelfand triple 
\begin{equation}\label{StandardSinlgeGelfand3pleGeneralpsi}
\left. \begin{array}{ccccc}              \mathcal{S}_{A}(\mathscr{O};\mathbb{C})         & \subset &  L^2(\mathscr{O};\mathbb{C}) & \subset & \mathcal{S}_{A}(\mathscr{O}; \mathbb{C})^*        \\
                               \downarrow \uparrow &         & \downarrow \uparrow      &         & \downarrow \uparrow  \\
                                         E         & \subset &  \mathcal{H'} & \subset & E^*        \\
                                                    &         & \parallel     &         &  \\
                         &  & \mathcal{H}^{\oplus} \oplus \mathcal{H}^{\ominus \, \flat} &  & 
\end{array}\right.,
\end{equation} 
with the vertical arrows indicating the unitary operator (and its inverse) $U: \mathcal{H}' \rightarrow  L^2(\mathscr{O};\mathbb{C})$ whose restriction to $E$ defines an isomorphism 
$U: E \rightarrow \mathcal{S}_{A}(\mathscr{O}; \mathbb{C})$ of nuclear spaces
and whose linear transposition $U^*$ defines isomorphism 
$\mathcal{S}_{A}(\mathscr{O}; \mathbb{C})^* \rightarrow E^*$. 
The nuclear space $\mathcal{S}_{A}(\mathscr{O}; \mathbb{C}) \subset 
L^2(\mathscr{O};\mathbb{C})$ then corresponds to the standard operator
$A= U^{-1}A'U^{-1}$ on $L^2(\mathscr{O};\mathbb{C})$, and can be constructed from it 
(compare  \cite{obata-book}, or Subsection \ref{white-setup}). 

The unitary isomorphism\footnote{Comapre the formula of Appendix \ref{fundamental,u,v} for the analogue operator $U$ defined by (\ref{isomorphismU})
in Subsection \ref{psiBerezin-Hida}, in the construction of the free Dirac field on the Minkowski space-time.} operator $U$ in (\ref{StandardSinlgeGelfand3pleGeneralpsi}), 
can be regarded as the operator of pointwise multiplication by the matrix (if we assume the lexicographical order of the three-index ``word'' ${}^{a}_{{}_{\,\,\,\,\,ji}}$)
\begingroup\makeatletter\def\f@size{5}\check@mathfonts
\def\maketag@@@#1{\hbox{\m@th\large\normalfont#1}}%
\begin{multline*}
U(\widehat{n}\cdot \widehat{l}) = \\ (2l+1)
\left( \begin{array}{cccccccc}   \overline{{u_{1}^{1}}_{{}_{ji}}(\widehat{n}\cdot \widehat{l})} \cdots &  \overline{{u_{1}^{2}}_{{}_{ji}}(\widehat{n}\cdot \widehat{l})} \cdots &
\cdots & \overline{{u_{1}^{d}}_{{}_{ji}}(\widehat{n}\cdot \widehat{l})}\cdots & & & &   \\
&\cdots&&&&&& \\
\overline{{u_{s}^{1}}_{{}_{ji}}(\widehat{n}\cdot \widehat{l})} \cdots &  \overline{{u_{s}^{2}}_{{}_{ji}}(\widehat{n}\cdot \widehat{l})} \cdots &
\cdots & \overline{{u_{s}^{d}}_{{}_{ji}}(\widehat{n}\cdot \widehat{l})}\cdots & & 0 & &   \\
&\cdots&&&&&& \\
\overline{{u_{d'}^{1}}_{{}_{ji}}(\widehat{n}\cdot \widehat{l})}\cdots &  \overline{{u_{d'}^{2}}_{{}_{ji}}(\widehat{n}\cdot \widehat{l})}\cdots &
\cdots & \overline{{u_{d'}^{4}}_{{}_{ji}}(\widehat{n}\cdot \widehat{l})}\cdots & & & &   \\
& & \cdots & & {v_{1}^{1}}_{{}_{ji}}(\widehat{n}\cdot \widehat{l})\cdots & {v_{1}^{2}}_{{}_{ji}}(\widehat{n}\cdot \widehat{l})\cdots & 
\cdots & {v_{1}^{d}}_{{}_{ji}}(\widehat{n}\cdot \widehat{l}) \\
&&&&&\cdots&& \\
& 0& \cdots & & {v_{s}^{1}}_{{}_{ji}}(\widehat{n}\cdot \widehat{l})\cdots & {v_{s}^{2}}_{{}_{ji}}(\widehat{n}\cdot \widehat{l})\cdots & 
\cdots & {v_{s}^{d}}_{{}_{ji}}(\widehat{n}\cdot \widehat{l}) \\
&&&&&\cdots && \\
& & \cdots & & {v_{d''}^{1}}_{{}_{ji}}(\widehat{n}\cdot \widehat{l}) & {v_{d''}^{2}}_{{}_{ji}}(\widehat{n}\cdot \widehat{l}) & 
\cdots & {v_{d''}^{d}}_{{}_{ji}}(\widehat{n}\cdot \widehat{l})  \end{array}\right)
\end{multline*}
\endgroup
acting on the element  
$\widetilde{\phi} \oplus (\widetilde{\phi}')^\flat \in \mathcal{H}^{\oplus} \oplus \mathcal{H}^{\ominus \,\flat}$;
where the value $\big(\widetilde{\phi} \oplus (\widetilde{\phi}')^\flat \big)(\widehat{n}\cdot \widehat{l})$ 
at $\widehat{n}\cdot \widehat{l} \in \mathscr{O}_{+}$ of 
$\widetilde{\phi} \oplus (\widetilde{\phi}')^\flat$ is
written as a column vector
\begingroup\makeatletter\def\f@size{5}\check@mathfonts
\def\maketag@@@#1{\hbox{\m@th\large\normalfont#1}}%
\[
\left( \begin{array}{c} 
\widetilde{\phi^1}_{{}_{ji}}(\widehat{n}\cdot \widehat{l}) \\
\vdots \\
\widetilde{\phi^a}_{{}_{ji}}(\widehat{n}\cdot \widehat{l}) \\
\vdots \\
\widetilde{\phi^d}_{{}_{ji}}(\widehat{n}\cdot \widehat{l}) \\
\vdots \\
(\widetilde{\phi}')^{\flat \,\,\, 1}_{{}_{\,\,\,\,\,\,\,\,\, ji}}(\widehat{n}\cdot \widehat{l}) \\
\vdots \\
(\widetilde{\phi}')^{\flat \,\,\, a}_{{}_{\,\,\,\,\,\,\,\,\, ji}}(\widehat{n}\cdot \widehat{l}) \\
\vdots \\
(\widetilde{\phi}')^{\flat \,\,\, d}_{{}_{\,\,\,\,\,\,\,\,\, ji}}(\widehat{n}\cdot \widehat{l}) \\
\vdots
\end{array}\right), \,\,\,\,\,\, \widehat{n}\cdot \widehat{l} \in \mathscr{O}_{{}_{+}}.
\]
\endgroup
Similarly, the inverse $U^{-1}$ of the isomorphism\footnote{Compare the formula of Appendix \ref{fundamental,u,v} of the analogue isomorphism $U$ operator  (\ref{isomorphismU}), Subsection \ref{psiBerezin-Hida}, in the construction of the free Dirac field on the Minkowski space-time.} $U$   
can be regarded as the operator
of pointwise multiplication by the matrix
\begingroup\makeatletter\def\f@size{5}\check@mathfonts
\def\maketag@@@#1{\hbox{\m@th\large\normalfont#1}}%
\[
U^{-1}(\widehat{n}\cdot \widehat{l}) =
\left( \begin{array}{cccccc}   
{u_{1}^{1}}_{{}_{ji}}(\widehat{n}\cdots \widehat{l}) & \cdots & u_{d'}^{1}(\widehat{n}\cdot \widehat{l}) & 0& \cdots & 0   \\
\vdots& &\vdots &&& \\
{u_{1}^{2}}_{{}_{ji}}(\widehat{n}\cdot \widehat{l}) & \cdots&  {u_{d'}^{2}}_{{}_{ji}}(\widehat{n}\cdot \widehat{l}) & 0 & \cdots & 0   \\
\vdots& &\vdots &&& \\
\vdots& &\vdots &&& \\
{u_{1}^{d}}_{{}_{ji}}(\widehat{n}\cdot \widehat{l}) &  \cdots & {u_{d'}^{d}}_{{}_{ji}}(\widehat{n}\cdot \widehat{l}) & 0 & \cdots & 0   \\
\vdots& &\vdots &&& \\
0 &  & 0 & \overline{{v_{1}^{1}}_{{}_{ji}}(\widehat{n}\cdot \widehat{l})} & \cdots & 
\overline{{v_{d''}^{1}}_{{}_{ji}}(\widehat{n}\cdot \widehat{l})} \\
& & &\vdots& &\vdots \\
0 & & 0 & \overline{{v_{1}^{2}}_{{}_{ji}}(\widehat{n}\cdot \widehat{l})} & \cdots & 
\overline{{v_{d''}^{2}}_{{}_{ji}}(\widehat{n}\cdot \widehat{l})} \\
& & &\vdots& &\vdots \\
& & &\vdots& &\vdots \\
0 & & 0 & \overline{{v_{1}^{d}}_{{}_{ji}}(\widehat{n}\cdot \widehat{l})}\cdots & \cdots & 
\overline{{v_{d''}^{d}}_{{}_{ji}}(\widehat{n}\cdot \widehat{l})} \\
& & &\vdots& &\vdots \\
  \end{array}\right)
\]
\endgroup
with the value 
$\Big((\widetilde{\phi})_{{}_{1}} \oplus (\widetilde{\phi})_{{}_{2}} \oplus \cdots 
\oplus (\widetilde{\phi})_{{}_{d'(\widehat{n}\cdot \widehat{l})+d''(\widehat{n}\cdot \widehat{l})}}\Big)(\widehat{n}\cdot \widehat{l})$ 
of the element 
\[
(\widetilde{\phi})_{{}_{1}} \oplus \cdots \oplus (\widetilde{\phi})_{{}_{d'+d''}} 
\in  L^{2}(\mathscr{O}; \mathbb{C})
\]
regarded as a column
\[
\left( \begin{array}{c} 
(\widetilde{\phi})_{{}_{1}}(\widehat{n}\cdot \widehat{l}) \\
\vdots \\
(\widetilde{\phi})_{{}_{d'+d''}}(\widehat{n}\cdot \widehat{l})
 \end{array}\right).  
\]
Note that 
\[
U(\widehat{n}\cdot \widehat{l})U^{-1}(\widehat{n}\cdot \widehat{l}) = \boldsymbol{1}_{d'+d''}, \,\,\,\,
U^{-1}(\widehat{n}\cdot \widehat{l}) U(\widehat{n}\cdot \widehat{l}) = 
\boldsymbol{1}_{{}_{\mathcal{H}^{\oplus}_{{}_{\widehat{n}\cdot\widehat{l}}} \oplus \mathcal{H}^{\ominus \, \flat}_{{}_{\widehat{n}\cdot\widehat{l}}}}}, 
\]
which follows from the fact that the actions of the projection operators 
$P^{\oplus}(\widehat{n}\cdot \widehat{l})$ and $P^{\ominus}(\widehat{n}\cdot \widehat{l})$ 
in the Hilbert space (\ref{H[n.l]x[loV]}) which project, respectively, on the subspaces
$\mathcal{H}^{\oplus}_{{}_{\widehat{n}\cdot\widehat{l}}}$ and $\mathcal{H}^{\ominus}_{{}_{\widehat{n}\cdot\widehat{l}}}$,  are equal
\begin{multline*}
P^{\oplus}(\widehat{n}\cdot \widehat{l})w =
\sum \limits_{s=1}^{d'} 
u_s(\widehat{n}\cdot \widehat{l}) \, \langle u_s(\widehat{n}\cdot \widehat{l}), w \rangle, 
\\
\textrm{or respectively}
\,\,\,
P^{\ominus}(\widehat{n}\cdot \widehat{l}) w =
\sum \limits_{s=1}^{d'} 
v_s(\widehat{n}\cdot \widehat{l}) \, \langle v_s(\widehat{n}\cdot \widehat{l}), w \rangle,
\end{multline*}
for all
\[
 w \in 
\mathcal{H}_{{}_{[\,\widehat{n}\cdot \widehat{l} \,]\, \times \,[\, \overline{\,\widehat{l}\,} \otimes V \,]}} 
= \big \{ \widetilde{\phi}, \,\, \phi \in \mathscr{H}, \textrm{with $\widetilde{\phi}$ concentrated on $\widehat{n}\cdot \widehat{l}$} \big \}
\]
with the natural inner product $\langle \cdot, \cdot \rangle$ defined by the Plancherel formula 
and equal (\ref{<tildephi(n.l),tildephi'(n.l)>}). Thus, $P^{\oplus}(\widehat{n}\cdot \widehat{l})$ and $P^{\ominus}(\widehat{n}\cdot \widehat{l})$ have the following matrix elements
\begin{multline*}
\big[P^{\oplus}(\widehat{n}\cdot \widehat{l})\big]^{aji \,\,\,\, a'j'i'} =
\sum \limits_{s=1}^{d'} 
(2l+1) {u^{a}_{s}}_{{}_{ji}}(\widehat{n}\cdot \widehat{l}) \, 
\overline{{u^{a'}_{s}}_{{}_{j'i'}}(\widehat{n}\cdot \widehat{l})},
\\
\textrm{or respectively}
\,\,\,
\big[P^{\ominus}(\widehat{n}\cdot \widehat{l})\big]^{aji \,\,\,\, a'j'i'} =
\sum \limits_{s=1}^{d'} 
(2l+1) {v^{a}_{s}}_{{}_{ji}}(\widehat{n}\cdot \widehat{l}) \, 
\overline{{v^{a'}_{s}}_{{}_{j'i'}}(\widehat{n}\cdot \widehat{l})},
\end{multline*}
which indeed immediately confirms that $U(\widehat{n}\cdot \widehat{l})$
and $U^{-1}(\widehat{n}\cdot \widehat{l})$ are mutually inverse.

This is the case of the positive energy ``complex'' field $\boldsymbol{\psi}$.
Of course, we have the analogous construction of the Gelfand triple
$E=E^{\oplus \, \flat} \oplus E^\ominus \subset \mathcal{H}' \subset E^*$
over the single particle Hilbert space $\mathcal{H}' = \mathcal{H}^{\oplus \, \flat} \oplus \mathcal{H}^\ominus$ of the negative energy complex field $\boldsymbol{\psi}$ and its standard form (\ref{StandardSinlgeGelfand3pleGeneralpsi}) with the analogous isomorphism $U$ between them. The replacements are obvious and can be omitted.

Thus, in case of the ``complex'' positive (respectively negative) energy field $\boldsymbol{\psi}$ 
we also have the unitary isomorphism $U$  
with the said properties,
which defines equivalence of the single particle Hilbert space $\mathcal{H}' = \mathcal{H}^\oplus
\oplus \mathcal{H}^{\ominus \, \flat}$
(respectively $\mathcal{H}' = \mathcal{H}^{\oplus \, \flat} \oplus \mathcal{H}^\ominus$)
with a standard Hilbert space $L^2(\mathscr{O}; \ud \mu_{{}_{\mathscr{O}}})$. The 
set $\mathscr{O}$ is equal to the enlargement of the set $\mathscr{O}_{{}_{+}}$
(respectively of the set $\mathscr{O}_{{}_{-}}$) of points 
$\widehat{n}\cdot \widehat{l} \in \widehat{\widetilde{\mathbb{S}^1}\times G}$ in which we have exactly $d'(\widehat{n}\cdot \widehat{l}) + d''(\widehat{n}\cdot \widehat{l})$ copies of each point $\widehat{n}\cdot \widehat{l} \in \mathscr{O}_{{}_{\pm}}$. 
Therefore, $\mathscr{O}$ is in each case equal to a discrete topological space in which each single element
set is open, closed  and compact, with the Baire measure equal to the discrete measure assigning
to each single point set the measure equal $1$.

In practice, we encounter only ``complex'' fields $\boldsymbol{\psi}$ in which the associated 
differential equation $D_{{}_{iff}} \phi = 0$ has the property that $\widehat{n}\cdot \widehat{l}
\in \mathscr{O}_{{}_{-}}$ iff $\widehat{-n}\cdot \widehat{l}
\in \mathscr{O}_{{}_{+}}$, and $d'(\widehat{n}\cdot \widehat{l}) 
= d''(\widehat{-n}\cdot \widehat{l})$ ($n \geq 0$). 

Moreover, we consider only the free fields with the corresponding  invariant\footnote{Under (\ref{UphiOnSxG}) acting in the corresponding Hilbert space $\mathscr{H}$ defined above.} 
 differential equations (say of motion) $D_{{}_{iff}} \phi = 0$ which are determined by the extremum action principle with the invariant Lagrange density function. For a general survey
of such Lagrange density functions corresponding to the various local fields
on the Einstein Universe including the Dirac and the electromagnetic potential field, compare \cite{PaneitzSegalI}-\cite{PaneitzSegalIII}. 

In order to construct concrete free fields, e.g. the free Dirac field or the electromagnetic
potential field on the Einstein Universe
according to the general prescription given above it is sufficient to construct the sets of Fourier transforms $u,v$
of the fundamental solutions in the single particle Hilbert spaces. Accordingly to the general procedure mentioned above also this task can be achieved 
relatively easy after the application of the decomposition of the representation associated to  the field in question.
In addition another method has been presented in \cite{PaneitzSegalI}-\cite{PaneitzSegalIII} and the particular cases 
of Dirac and electromagnetic fields have been discussed in detail in
\cite{PaneitzSegalI}-\cite{PaneitzSegalIII}, so we will not present explicit formulas for $u,v$ in this case 
-- the particular case of the general construction of $u,v$ given above. 

Here we give general Lemmas and Theorems concerning general
local free fields on the Einstein Universe together with some general properties of their commutation
and pairing generalized functions and their splitting into retarded and advanced parts -- task not undertaken
by \cite{PaneitzSegalI}-\cite{PaneitzSegalIII}.

First let us note existence of the important class of the fields each solution $\phi$ of the associated
equation $D_{{}_{iff}} \phi = 0$ has the property that $\square \phi = m^2\phi$, for a fixed parameter
$m$ (which we assume non negative and call mass). This is for example the case for all 
higher spin fields associated to the representation (\ref{UphiOnSxG}) with the representation $V$
of $SU(2, \mathbb{C})$  in the formula (\ref{UphiOnSxG}) equal to the bispinor representation 
(\ref{spinorV}), or to the direct sum of tensor products of (\ref{spinorV}) (each direct summand being a tensor product of only even or only odd number of terms (\ref{spinorV})). For the 
associated system of differential equations $D_{{}_{iff}} \phi = 0$  we put $D\phi = m\phi$,
where $D$ is the generalized Dirac operator associated to this representation, which we have
constructed in the previous Subsection. Because $D^2 = \square$, then each solution $\phi$
of  $D\phi = m\phi$ necessary respects the equation $\square \phi = m^2\phi$.
Note that the Fourier transform $\widetilde{\square}$ of the operator $\square$
in action on the element $\widetilde{\phi} \in \widetilde{\mathscr{H}}$ is equal to the operator
of pointwise multiplication by the function
\[
\widehat{\widetilde{S}^1 \times G} \ni \widehat{n}\cdot \widehat{l}
\longmapsto \widetilde{\square} (\widehat{n}\cdot \widehat{l}) = 
{\textstyle\frac{n^2}{4}} - l(l+1) - {\textstyle\frac{1}{4}}.
\]
Therefore, the Fourier transform $\widetilde{\phi}$ of each solution $\phi$
of $D\phi = m\phi$, which necessary respects the equation 
$\square \phi = m^2\phi$, must be concentrated on the set 
\[
\mathscr{O}_{{}_{+}} \sqcup \mathscr{O}_{{}_{+}} = 
\big\{\widehat{n}\cdot \widehat{l}, \,  {\textstyle\frac{n^2}{4}} - l(l+1) - {\textstyle\frac{1}{4}}
= m^2 \big\}.
\]
Therefore, in this case 
\[
\mathscr{O}_{{}_{+}}  = 
\big\{\widehat{n}\cdot \widehat{l}, \,  {\textstyle\frac{n^2}{4}} - l(l+1) - {\textstyle\frac{1}{4}}
= m^2, n\geq 0 \big\}, 
\,\,\,\,\,
\mathscr{O}_{{}_{-}}  = 
\big\{\widehat{n}\cdot \widehat{l}, \,  {\textstyle\frac{n^2}{4}} - l(l+1) - {\textstyle\frac{1}{4}}
= m^2, n < 0 \big\}, 
\]
for the massive field ($m \neq 0$) with the property that each solution fulfils $\square \phi  =
m^2 \phi$; and
\[
\mathscr{O}_{{}_{+}}  = 
\big\{\widehat{n}\cdot \widehat{l}, \, {\textstyle\frac{n^2}{4}} - l(l+1) - {\textstyle\frac{1}{4}}
= 0, n\geq 0 \big\}, 
\,\,\,\,\,
\mathscr{O}_{{}_{-}}  = 
\big\{\widehat{n}\cdot \widehat{l}, \, {\textstyle\frac{n^2}{4}} - l(l+1) - {\textstyle\frac{1}{4}}
= 0, n < 0 \big\}, 
\]
for massless field, with the said property that each solution $\phi$ fulfills
$\square \phi = 0$.

\begin{center}
\begin{tikzpicture}[yscale=1]
    \draw[thick, ->] (2,0) -- (2.5,0);
    \draw[thick, ->] (0,2.5) -- (0,3);
    
     \draw[ultra thin] (-1.2,-2.275) -- (-0.03125,0.0625);
     \draw[ultra thin] (-1.2,2.275) -- (-0.03125,-0.0625);
     
      \draw[ultra thin] (1.25,2.625) -- (1.75,3.625);
      \draw[ultra thin] (1.25,-2.625) -- (1.75,-3.625);

\fill[color=gray, opacity=0.05] (0,0.125) -- (1.125,2.375) -- (0,2.375) -- (0,0.125) -- cycle;
\fill[color=gray, opacity=0.05] (0,-0.125) -- (1.125,-2.375) -- (0,-2.375) -- (0,-0.125) -- cycle;
    
\node [right] at (2.2,-0.3) {$l$};
\node [left] at (0,3) {$n$};
\node [left] at (-1.2,-2.22) {\textrm{\tiny $n=2l+1$}};
\node [left] at (-1.2,2.22) {\textrm{\tiny $n=-(2l+1)$}};
\node [right] at (1.8,3.625) {\textrm{\tiny $\tfrac{n^2}{4}-l(l+1) -\tfrac{1}{4} = 0$}};
\node [right] at (1.8,3.2) {\textrm{\tiny or \,\, $n^2 = (2l+1)^2$}};
\node [right] at (1.8,-3.2) {\textrm{\tiny $n^2 = (2l+1)^2$ \,\, or}};
\node [left] at (-1.8,-3.2) {\textrm{\tiny $n^2 - (2l+1)^2 = (2m)^2$ \,\, or}};
\node [right] at (1.8,-3.625) {\textrm{\tiny $\tfrac{n^2}{4}-l(l+1) -\tfrac{1}{4} = 0$}};
\node [left] at (-1.8,-3.625) {\textrm{\tiny $\tfrac{n^2}{4}-l(l+1) -\tfrac{1}{4} = m^2$}};
\node [left] at (-1.8,3.625) {\textrm{\tiny $\tfrac{n^2}{4}-l(l+1) -\tfrac{1}{4} = m^2$}};

\foreach \x in {0,0.0625,0.125,0.1875,0.25,0.3125,0.375,0.4375,0.5,0.5625,0.625,0.6875,0.75,0.8125,0.875,0.9375,1,
1.0625,1.125}
    \foreach \y in {-2.3125,-2.25,-2.1875,-2.125,-2.0625,-2,-1.9375,-1.875,-1.8125,-1.75,-1.6875,-1.625,-1.5625,-1.5,-1.4375,
    -1.375,-1.3125,-1.25,-1.1875,-1.125,-1.0625,-1,-0.9375,-0.875,-0.8125,-0.75,-0.6875,-0.625,-0.5625,
    -0.5,-0.4375,-0.375,-0.3125,-0.25,-0.1875,-0.125,-0.0625,
  0,0.0625,0.125,0.1875,0.25,0.3125,0.375,0.4375,0.5,0.5625,0.625,0.6875,0.75,0.8125,0.875,0.9375,1,
    1.0625,1.125,1.1875,1.25,1.3125,1.375,1.4375,1.5,1.5625,1.625,1.6875,1.75,1.8125,1.875,1.9375,2,
    2.0625,2.125,2.1875,2.25,2.3125}
    {
    \fill (\x,\y) circle (0.25pt);
    }  
    \fill (0,0.125) circle (0.5pt);
    \fill (0.0625,0.25) circle (0.5pt);
    \fill (0.125,0.375) circle (0.5pt);
    \fill (0.1875,0.5) circle (0.5pt);
    \fill (0.25,0.625) circle (0.5pt);
    \fill (0.3125,0.75) circle (0.5pt);
    \fill (0.375,0.875) circle (0.5pt);
    \fill (0.4375,1) circle (0.5pt);
    \fill (0.5,1.125) circle (0.5pt);
    \fill (0.5625,1.25) circle (0.5pt);
    \fill (0.625,1.375) circle (0.5pt);
    \fill (0.6875,1.5) circle (0.5pt);
    \fill (0.75,1.625) circle (0.5pt);
    \fill (0.8125,1.75) circle (0.5pt);
    \fill (0.875,1.875) circle (0.5pt);
    \fill (0.9375,2) circle (0.5pt);
    \fill (1,2.125) circle (0.5pt);
    \fill (1.0625,2.25) circle (0.5pt);
    \fill (1.125,2.375) circle (0.5pt);
    \fill (1.1875,2.5) circle (0.5pt);
       
    \fill (0,-0.125) circle (0.5pt);
    \fill (0.0625,-0.25) circle (0.5pt);
    \fill (0.125,-0.375) circle (0.5pt);
    \fill (0.1875,-0.5) circle (0.5pt);
    \fill (0.25,-0.625) circle (0.5pt);
    \fill (0.3125,-0.75) circle (0.5pt);
    \fill (0.375,-0.875) circle (0.5pt);
    \fill (0.4375,-1) circle (0.5pt);
    \fill (0.5,-1.125) circle (0.5pt);
    \fill (0.5625,-1.25) circle (0.5pt);
    \fill (0.625,-1.375) circle (0.5pt);
    \fill (0.6875,-1.5) circle (0.5pt);
    \fill (0.75,-1.625) circle (0.5pt);
    \fill (0.8125,-1.75) circle (0.5pt);
    \fill (0.875,-1.875) circle (0.5pt);
    \fill (0.9375,-2) circle (0.5pt);
    \fill (1,-2.125) circle (0.5pt);
    \fill (1.0625,-2.25) circle (0.5pt);
    \fill (1.125,-2.375) circle (0.5pt);
    \fill (1.1875,-2.5) circle (0.5pt);
    
    \fill (0.125,0.8125) circle (0.6pt);
    \fill (0.25,0.9375) circle (0.6pt);
    \fill (0.46875,1.25) circle (0.6pt);
    \fill (1.0625,2.3125) circle (0.6pt);
    
    \fill (0.125,-0.8125) circle (0.6pt);
    \fill (0.25,-0.9375) circle (0.6pt);
    \fill (0.46875,-1.25) circle (0.6pt);
    \fill (1.0625,-2.3125) circle (0.6pt);
    
    \draw[domain=-2.2:-0.1,smooth,variable=\t,line width=0.125pt] plot ({0.3656*sinh(\t)},{-0.7525*cosh(\t)});
    \draw[domain= 1.85:2,smooth,variable=\t,line width=0.125pt] plot ({0.3656*sinh(\t)},{-0.7525*cosh(\t)});
    
    \draw[domain=-2.2:-0.1,smooth,variable=\t,line width=0.125pt] plot ({0.3656*sinh(\t)},{0.7525*cosh(\t)});
    \draw[domain= 1.85:2,smooth,variable=\t,line width=0.125pt] plot ({0.3656*sinh(\t)},{0.7525*cosh(\t)});
\end{tikzpicture}
\end{center}

\begin{center}
\begin{tikzpicture}[yscale=1]
\draw[thick, ->] (2,0) -- (2.5,0);
    \draw[thick, ->] (0,1.5) -- (0,2);
    
    \node [right] at (2.2,-0.3) {$l$};
\node [left] at (0,2) {$n$};
    
\fill[color=gray, opacity=0.05] (0,0.125) -- (1.125,2.375) -- (0,2.375) -- (0,0.125) -- cycle;
\fill[color=gray, opacity=0.05] (0,-0.125) -- (1.125,-2.375) -- (0,-2.375) -- (0,-0.125) -- cycle;

\foreach \x in {0,0.0625,0.125,0.1875,0.25,0.3125,0.375,0.4375,0.5,0.5625,0.625,0.6875,0.75,0.8125,0.875,0.9375,1,
1.0625,1.125}
    \foreach \y in {-2.3125,-2.25,-2.1875,-2.125,-2.0625,-2,-1.9375,-1.875,-1.8125,-1.75,-1.6875,-1.625,-1.5625,-1.5,-1.4375,
    -1.375,-1.3125,-1.25,-1.1875,-1.125,-1.0625,-1,-0.9375,-0.875,-0.8125,-0.75,-0.6875,-0.625,-0.5625,
    -0.5,-0.4375,-0.375,-0.3125,-0.25,-0.1875,-0.125,-0.0625,
  0,0.0625,0.125,0.1875,0.25,0.3125,0.375,0.4375,0.5,0.5625,0.625,0.6875,0.75,0.8125,0.875,0.9375,1,
    1.0625,1.125,1.1875,1.25,1.3125,1.375,1.4375,1.5,1.5625,1.625,1.6875,1.75,1.8125,1.875,1.9375,2,
    2.0625,2.125,2.1875,2.25,2.3125}
    {
    \fill (\x,\y) circle (0.25pt);
    }  
    \fill (0,0.125) circle (0.5pt);
    \fill (0.0625,0.25) circle (0.5pt);
    \fill (0.125,0.375) circle (0.5pt);
    \fill (0.1875,0.5) circle (0.5pt);
    \fill (0.25,0.625) circle (0.5pt);
    \fill (0.3125,0.75) circle (0.5pt);
    \fill (0.375,0.875) circle (0.5pt);
    \fill (0.4375,1) circle (0.5pt);
    \fill (0.5,1.125) circle (0.5pt);
    \fill (0.5625,1.25) circle (0.5pt);
    \fill (0.625,1.375) circle (0.5pt);
    \fill (0.6875,1.5) circle (0.5pt);
    \fill (0.75,1.625) circle (0.5pt);
    \fill (0.8125,1.75) circle (0.5pt);
    \fill (0.875,1.875) circle (0.5pt);
    \fill (0.9375,2) circle (0.5pt);
    \fill (1,2.125) circle (0.5pt);
    \fill (1.0625,2.25) circle (0.5pt);
    \fill (1.125,2.375) circle (0.5pt);
    \fill (1.1875,2.5) circle (0.5pt);
       
    \fill (0,-0.125) circle (0.5pt);
    \fill (0.0625,-0.25) circle (0.5pt);
    \fill (0.125,-0.375) circle (0.5pt);
    \fill (0.1875,-0.5) circle (0.5pt);
    \fill (0.25,-0.625) circle (0.5pt);
    \fill (0.3125,-0.75) circle (0.5pt);
    \fill (0.375,-0.875) circle (0.5pt);
    \fill (0.4375,-1) circle (0.5pt);
    \fill (0.5,-1.125) circle (0.5pt);
    \fill (0.5625,-1.25) circle (0.5pt);
    \fill (0.625,-1.375) circle (0.5pt);
    \fill (0.6875,-1.5) circle (0.5pt);
    \fill (0.75,-1.625) circle (0.5pt);
    \fill (0.8125,-1.75) circle (0.5pt);
    \fill (0.875,-1.875) circle (0.5pt);
    \fill (0.9375,-2) circle (0.5pt);
    \fill (1,-2.125) circle (0.5pt);
    \fill (1.0625,-2.25) circle (0.5pt);
    \fill (1.125,-2.375) circle (0.5pt);
    \fill (1.1875,-2.5) circle (0.5pt);
    
    \fill (0.125,0.8125) circle (0.8pt);
    \fill (0.25,0.9375) circle (0.8pt);
    \fill (0.46875,1.25) circle (0.8pt);
    \fill (1.0625,2.3125) circle (0.8pt);
    
    \fill (0.125,-0.8125) circle (0.8pt);
    \fill (0.25,-0.9375) circle (0.8pt);
    \fill (0.46875,-1.25) circle (0.8pt);
    \fill (1.0625,-2.3125) circle (0.8pt);
    
    \draw [->,very thin] (-0.7,-1.2) to [out=-70,in=130] (0.25, -0.9375);
    \draw [->,very thin] (-0.7,0.7) to [out=-70,in=130] (0.25, 0.9375);
    
    \draw [<-,very thin] (0.25,-0.625) to [out=-45,in=135] (1.5, -0.625);
    \draw [<-,very thin] (0.25,0.625) to [out=-45,in=135] (1.5, 0.625);
    
    \node [left] at (-0.7,-1.2) {\textrm{\tiny points of $\mathscr{O}_{-}$ for massive field}};
    \node [left] at (-0.7,0.7) {\textrm{\tiny points of $\mathscr{O}_{+}$ for massive field}};
    
    \node [right] at (1.5,-0.9) {\textrm{\tiny points of $\mathscr{O}_{-}$ for massless field}};
    \node [right] at (1.5,0.65) {\textrm{\tiny points of $\mathscr{O}_{+}$ for massless field}};
    
    \draw[dashed,domain=-1.8:0.1,smooth,variable=\t,line width=0.07pt] plot ({0.3656*sinh(\t)},{-0.7525*cosh(\t)});
    
    \draw[dashed,domain=-1.8:0.1,smooth,variable=\t,line width=0.07pt] plot ({0.3656*sinh(\t)},{0.7525*cosh(\t)});
\end{tikzpicture}
\end{center}

But this is only accidentally that the orbits $\mathscr{O}_{{}_{\pm}}$ lie, respectively,
exactly on the positive or negative energy sheets of the massive hyperboloid or the ``light'' cone
in the momentum space, and is immediately related to the specific
property of the Dirac operator $D$ whose square is the scalar $\square$. 

There are also other natural local fields on the Einstein Universe, associated e.g.
to the representation (\ref{UphiOnSxG}) with $V$ equal (\ref{V^(1)forms}), or with
$V$ equal to the direct sum of tensor products of any numbers 
of factors (\ref{V^(1)forms}) in each summand. We then consider the signature\footnote{Of 
course corresponding to the space-time pseudo-Riemannian metric, and not to the associated invariant 
Riemannian metric $g_\mathfrak{J}$.}
operator $D = d - \star d \star$ (or its generalization constructed in the previous Subsection).
We then consider the field associated with the invariant system of differential
equations $D_{{}_{iff}} \phi = 0$ of the analogous form $D\phi = m\phi$, but here
with the signature operator $D$, whose square is not a scalar. Only the principal
highest order (here order two) part of $D^2$ is equal $4 \square$, but there are 
also lower order terms present in $D^2$. In this situation the orbits
$\mathscr{O}_{{}_{\pm}}$ lie not exactly on the sheets of the cone or hyperboloid,
but only approximately. In particular in case of the free electromagnetic potential
field each four-vector solution $\phi$ of the Maxwell equations (in the Lorentz gauge)
necessary respects the equation
\begin{multline*}
D^2 \phi = P_{{}_{\textrm{1-form}}} D^2 P_{{}_{\textrm{1-form}}} \phi = 0,
\,\,\,\,\,\, 
D = d - \star d \star
\\
\textrm{on} 
\,\,\,
\mathscr{H} = L^2\Big(\widetilde{S}^1 \times G; {\bigwedge}^{*}_{\mathbb{C}} \Big)
\cong L^2(\widetilde{S}^1 \times G; \mathbb{C}^{16}),
\,\, G= SU(2, \mathbb{C}),
\end{multline*}
where $P_{{}_{\textrm{1-form}}}$ is the projection (commuting with $D^2$) on $1$-forms
in $\mathscr{H}$. The orbits $\mathscr{O}_{{}_{\pm}}$, in case of the electromagnetic
potential field in Lorentz gauge, are the sets of all points $\widehat{n} \cdot \widehat{l}$,
with $n = \pm(2l+1)$, $n = \pm (2l+2)$, $n= \pm 2l$ and with $l$ arbitrary non-negative integer or half 
an odd integer, compare \cite{PaneitzSegalIII}. Below (and above) we have pictured only the points of 
$\mathscr{O}_{}{}_{\pm}$ with integer, non-negative of course, $l$:
\begin{center}
\begin{tikzpicture}[yscale=1]
\draw[thick, ->] (2,0) -- (2.5,0);
    \draw[thick, ->] (0,1.5) -- (0,2);
    
    \node [right] at (2.2,-0.3) {$l$};
\node [left] at (0,2) {$n$};
    
\fill[color=gray, opacity=0.05] (0,0.125) -- (1.125,2.375) -- (0,2.375) -- (0,0.125) -- cycle;
\fill[color=gray, opacity=0.05] (0,-0.125) -- (1.125,-2.375) -- (0,-2.375) -- (0,-0.125) -- cycle;

\foreach \x in {0,0.0625,0.125,0.1875,0.25,0.3125,0.375,0.4375,0.5,0.5625,0.625,0.6875,0.75,0.8125,0.875,0.9375,1,
1.0625,1.125}
    \foreach \y in {-2.3125,-2.25,-2.1875,-2.125,-2.0625,-2,-1.9375,-1.875,-1.8125,-1.75,-1.6875,-1.625,-1.5625,-1.5,-1.4375,
    -1.375,-1.3125,-1.25,-1.1875,-1.125,-1.0625,-1,-0.9375,-0.875,-0.8125,-0.75,-0.6875,-0.625,-0.5625,
    -0.5,-0.4375,-0.375,-0.3125,-0.25,-0.1875,-0.125,-0.0625,
  0,0.0625,0.125,0.1875,0.25,0.3125,0.375,0.4375,0.5,0.5625,0.625,0.6875,0.75,0.8125,0.875,0.9375,1,
    1.0625,1.125,1.1875,1.25,1.3125,1.375,1.4375,1.5,1.5625,1.625,1.6875,1.75,1.8125,1.875,1.9375,2,
    2.0625,2.125,2.1875,2.25,2.3125}
    {
    \fill (\x,\y) circle (0.25pt);
    }  
    \fill (0,0.125) circle (0.5pt);
    \fill (0.0625,0.25) circle (0.5pt);
    \fill (0.125,0.375) circle (0.5pt);
    \fill (0.1875,0.5) circle (0.5pt);
    \fill (0.25,0.625) circle (0.5pt);
    \fill (0.3125,0.75) circle (0.5pt);
    \fill (0.375,0.875) circle (0.5pt);
    \fill (0.4375,1) circle (0.5pt);
    \fill (0.5,1.125) circle (0.5pt);
    \fill (0.5625,1.25) circle (0.5pt);
    \fill (0.625,1.375) circle (0.5pt);
    \fill (0.6875,1.5) circle (0.5pt);
    \fill (0.75,1.625) circle (0.5pt);
    \fill (0.8125,1.75) circle (0.5pt);
    \fill (0.875,1.875) circle (0.5pt);
    \fill (0.9375,2) circle (0.5pt);
    \fill (1,2.125) circle (0.5pt);
    \fill (1.0625,2.25) circle (0.5pt);
    \fill (1.125,2.375) circle (0.5pt);
    \fill (1.1875,2.5) circle (0.5pt);
    
   \fill (0,0.1875) circle (0.5pt);
    \fill (0.0625,0.3125) circle (0.5pt);
    \fill (0.125,0.4375) circle (0.5pt);
    \fill (0.1875,0.5625) circle (0.5pt);
    \fill (0.25,0.6875) circle (0.5pt);
    \fill (0.3125,0.8125) circle (0.5pt);
    \fill (0.375,0.9375) circle (0.5pt);    
    \fill (0.4375,1.0625) circle (0.5pt);
    \fill (0.5,1.1875) circle (0.5pt);  
    \fill (0.5625,1.3125) circle (0.5pt);
    \fill (0.625,1.4375) circle (0.5pt);
    \fill (0.6875,1.5625) circle (0.5pt);
    \fill (0.75,1.6875) circle (0.5pt);
    \fill (0.8125,1.8125) circle (0.5pt);
    \fill (0.875,1.9375) circle (0.5pt);
    \fill (0.9375,2.0625) circle (0.5pt);
    \fill (1,2.1875) circle (0.5pt);
    \fill (1.0625,2.3125) circle (0.5pt);
    \fill (1.125,2.4375) circle (0.5pt);
    \fill (1.1875,2.5625) circle (0.5pt); 
    
    \fill (0,0.0625) circle (0.5pt);
    \fill (0.0625,0.1875) circle (0.5pt);
    \fill (0.125,0.3125) circle (0.5pt);
    \fill (0.1875,0.4375) circle (0.5pt);
    \fill (0.25,0.5625) circle (0.5pt);
    \fill (0.3125,0.6875) circle (0.5pt);
    \fill (0.375,0.8125) circle (0.5pt);
    \fill (0.4375,0.9375) circle (0.5pt);
    \fill (0.5,1.0625) circle (0.5pt);
    \fill (0.5625,1.1875) circle (0.5pt);
    \fill (0.625,1.3125) circle (0.5pt);
    \fill (0.6875,1.4375) circle (0.5pt);
    \fill (0.75,1.5625) circle (0.5pt);
    \fill (0.8125,1.6875) circle (0.5pt);
    \fill (0.875,1.8125) circle (0.5pt);
    \fill (0.9375,1.9375) circle (0.5pt);
    \fill (1,2.0625) circle (0.5pt);
    \fill (1.0625,2.1875) circle (0.5pt);
    \fill (1.125,2.3125) circle (0.5pt);
    \fill (1.1875,2.4375) circle (0.5pt);
   
 \fill (0,0) circle (0.5pt);   
         
    \fill (0,-0.125) circle (0.5pt);
    \fill (0.0625,-0.25) circle (0.5pt);
    \fill (0.125,-0.375) circle (0.5pt);
    \fill (0.1875,-0.5) circle (0.5pt);
    \fill (0.25,-0.625) circle (0.5pt);
    \fill (0.3125,-0.75) circle (0.5pt);
    \fill (0.375,-0.875) circle (0.5pt);
    \fill (0.4375,-1) circle (0.5pt);
    \fill (0.5,-1.125) circle (0.5pt);
    \fill (0.5625,-1.25) circle (0.5pt);
    \fill (0.625,-1.375) circle (0.5pt);
    \fill (0.6875,-1.5) circle (0.5pt);
    \fill (0.75,-1.625) circle (0.5pt);
    \fill (0.8125,-1.75) circle (0.5pt);
    \fill (0.875,-1.875) circle (0.5pt);
    \fill (0.9375,-2) circle (0.5pt);
    \fill (1,-2.125) circle (0.5pt);
    \fill (1.0625,-2.25) circle (0.5pt);
    \fill (1.125,-2.375) circle (0.5pt);
    \fill (1.1875,-2.5) circle (0.5pt);
    
     \fill (0,-0.1875) circle (0.5pt);
    \fill (0.0625,-0.3125) circle (0.5pt);
    \fill (0.125,-0.4375) circle (0.5pt);
    \fill (0.1875,-0.5625) circle (0.5pt);
    \fill (0.25,-0.6875) circle (0.5pt);
    \fill (0.3125,-0.8125) circle (0.5pt);
    \fill (0.375,-0.9375) circle (0.5pt);    
    \fill (0.4375,-1.0625) circle (0.5pt);
    \fill (0.5,-1.1875) circle (0.5pt);  
    \fill (0.5625,-1.3125) circle (0.5pt);
    \fill (0.625,-1.4375) circle (0.5pt);
    \fill (0.6875,-1.5625) circle (0.5pt);
    \fill (0.75,-1.6875) circle (0.5pt);
    \fill (0.8125,-1.8125) circle (0.5pt);
    \fill (0.875,-1.9375) circle (0.5pt);
    \fill (0.9375,-2.0625) circle (0.5pt);
    \fill (1,-2.1875) circle (0.5pt);
    \fill (1.0625,-2.3125) circle (0.5pt);
    \fill (1.125,-2.4375) circle (0.5pt);
    \fill (1.1875,-2.5625) circle (0.5pt); 
    
    \fill (0,-0.0625) circle (0.5pt);
    \fill (0.0625,-0.1875) circle (0.5pt);
    \fill (0.125,-0.3125) circle (0.5pt);
    \fill (0.1875,-0.4375) circle (0.5pt);
    \fill (0.25,-0.5625) circle (0.5pt);
    \fill (0.3125,-0.6875) circle (0.5pt);
    \fill (0.375,-0.8125) circle (0.5pt);
    \fill (0.4375,-0.9375) circle (0.5pt);
    \fill (0.5,-1.0625) circle (0.5pt);
    \fill (0.5625,-1.1875) circle (0.5pt);
    \fill (0.625,-1.3125) circle (0.5pt);
    \fill (0.6875,-1.4375) circle (0.5pt);
    \fill (0.75,-1.5625) circle (0.5pt);
    \fill (0.8125,-1.6875) circle (0.5pt);
    \fill (0.875,-1.8125) circle (0.5pt);
    \fill (0.9375,-1.9375) circle (0.5pt);
    \fill (1,-2.0625) circle (0.5pt);
    \fill (1.0625,-2.1875) circle (0.5pt);
    \fill (1.125,-2.3125) circle (0.5pt);
    \fill (1.1875,-2.4375) circle (0.5pt);

    
    
    
    \draw [<-,very thin] (0.25,-0.625) to [out=-45,in=135] (1.5, -0.625);
    \draw [<-,very thin] (0.25,0.625) to [out=-45,in=135] (1.5, 0.625);
    
    
    \node [right] at (1.5,-0.9) {\textrm{\tiny points of $\mathscr{O}_{-}$ for the el.m. potential field $A$}};
    \node [right] at (1.5,0.65) {\textrm{\tiny points of $\mathscr{O}_{+}$ for the el.m. potential field $A$}};
    

 
\end{tikzpicture}
\end{center}

This form of $\mathscr{O}_{{}_{\pm}}$ takes place in case in which $R=\hslash=c=1$. For the more realistic
value of the radius $R$ of the equal-time Cauchy surface, and for the realistic values of light's velocity $c$, of particle masses $m$, and of the elementary Planck's action $\hslash$, the quantum numbers $n, l$ which enter already the single particle states and define $\mathscr{O}_{{}_{\pm}}$ are enormously large.
In case of all the massive fields $\boldsymbol{\psi}$ we will encounter in practice only such 
whose each component respects the equation 
$\square \boldsymbol{\psi} = m^2\boldsymbol{\psi}$ (for which each solution $\phi$ of the associated equation $D_{{}_{iff}}\phi$ fulfills 
$\square \phi = m^2 \phi$). For such massive fields the orbits $\mathscr{O}_{{}_{\pm}}$ 
consist of all points $\widehat{n}\cdot \widehat{l}$ with $n\in \mathbb{Z}$ and $l$ positive integer or positive half an odd integer, which fulfill
\begin{equation}\label{FiniteOrbit1}
n^2 - (2l+1)^2 = \Big({\textstyle\frac{2mcR}{\hslash}}\Big)^2, \,\,\,\,\,\,\,\, n \geq 0 \,\,
\textrm{for} \,\, \mathscr{O}_{{}_{+}}, \,\,\,\,\,\,\,\,\,\,
n < 0 \,\,
\textrm{for} \,\, \mathscr{O}_{{}_{-}}.
\end{equation}
In particular for the realistic value $R = 8 \cdot 10^{26} [\textrm{meter}]$, and for 
$m$ of the order of the electron mass $m_e$ we obtain enormous number
of the order 
\begin{equation}\label{FiniteOrbit2}
{\textstyle\frac{2m_e cR}{\hslash}} = N   \sim 10^{38}.
\end{equation}

But moreover whatever constant number $N$ (depending on the particle's mass $m$
of the considered massive field and the radius $R$) we assume, only integer values
of $N$ are allowed, because $n$ and $2l+1$ are necessary integer quantum numbers. But
what is of most importance, whatever integer value $N$ we assume for
$\tfrac{2mcR}{\hslash}$, there will exist only a finite 
number of solutions\footnote{We have finite number of rectangles with integer sides
and with fixed integer area $N^2$.} of the 
equation (with fixed particle mass and fixed $R$, and thus with fixed $N$)
\[
n^2 - (2l+1)^2 = [n + (2l+1)][n-(2l+1)] = N^2
\]
in each case $n>0$ or $n<0$. Of course, we have always $N < |n|$, 
and the greatest possible value of $|n|$ is of the order of $N^2$. Therefore, the sets $\mathscr{O}_{{}_{\pm}}$
have finite number of elements for all massive free fields $\boldsymbol{\psi}$
which fulfill  $\square \boldsymbol{\psi} = m^2\boldsymbol{\psi}$, and this is the case for all
free massive fields we encounter in practice.
 Using the classical Euclid's formula which gives all Pythagorean
triples, compare e.g. \cite{Sierpinski}, p. 232, we can give all elements 
$\widehat{n} \cdot \widehat{l}$ 
of $\mathscr{O}_{{}_{+}}$
determined by a fixed value $N$ of $\tfrac{2mcR}{\hslash}$ in the algorithmic form which
may be of use for large numbers $N$. We consider the set of three strictly positive 
integers $n_1, n_2$ and $k$, such that $n_1 < n_2$ with 
$n_1$ and $n_2$ co-prime and not both odd. Then we consider two sets of factorizations
of the number $N$ with the corresponding solution to each factorization. 
For each factorization of the first class 1) $N = k(n_2-n_1)(n_2 +n_1)$,
there corresponds bi-uniquely the solution $n=k\big((n_1)^2 + (n_2)^2\big)$, $l = kn_1n_2 - \tfrac{1}{2}$.
For each factorization of the second class 2) $N= k2n_1n_2$, there corresponds
bi-uniquely the solution  $n=k\big((n_1)^2 + (n_2)^2\big)$, $l = \tfrac{1}{2}k\big((n_2)^2 - (n_1)^2\big)
-\tfrac{1}{2}$. These exhaust all solutions. There may appear repetition of the solution 
common for the two classes only in case both the respective factorizations are possible 
$N =  k(n_2-n_1)(n_2 +n_1) = k'2n'_{1}n'_{2}$. In particular for odd $N$ the second 
factorization is impossible and 
the first factorization exhaust all solutions in bi-unique manner.

That the orbits $\mathscr{O}_{{}_{\pm}}$ of massive free quantum fields on the Einstein
Universe are finite (this is in particular the case for the free Dirac field)
has been noted in \cite{SegalZhouQED}. This fact is used in \cite{SegalZhouQED} in the proof that the
massive free quantum fields evaluated at space-time point are well defined operators in the Fock space
of free fields, and where it also serves (together with the orthogonality relations
for the character functions $u \mapsto \widehat{l}_{{}_{ij}}(u)$) as the basis of the proof
that the interaction hamiltonian of QED is a well defined essentially self-adjoint
operator on the Fock space of the free Dirac and electromagnetic potential fields
on the Einstein Universe. 

In  the remaining part of this Subsection we extend this result, concerned with free fields,
their commutation and pairing functions,
and using white noise construction of free fields we show that each local free massive
field evaluated at space-time point is a well defined operator on the Fock space,
which transforms continuously the Hida space into itself. In Subsection
\ref{CausalSonEU} we extend this result, using the causal method, where we will show its validity
for each higher order contribution
to interacting (massive and massless) fields on the Einstein Universe. 
Finally and still in this Subsection we give general formulas for commutation and pairing
functions as well as for their splitting into the advanced and retarded parts.

Having given the standard Gelfand triples (\ref{StandardGelfandTripleNeutralEU})
in the single particle Hilbert space of a neutral field $\mathbb{A}$
and respectively (\ref{StandardSinlgeGelfand3pleGeneralpsi})
in the single particle Hilbert space of ``complex'' quantum field
$\boldsymbol{\psi}$, we can now
construct, exactly as in Subsection \ref{psiBerezin-Hida}, the 
fields $\mathbb{A}$ and $\boldsymbol{\psi}$ as integral kernel operators
in the sense of \cite{obataJFA}, using the Hida operators as the creation-annihilation
operators. Using the Hida annihilation-creation operators $a,a^+$ in the Fock spaces
over the standard Hilbert spaces $L^2(\mathscr{O; \mathbb{C}})$ and the Hida annihiliation-creation operators $a', a'^{+}$ over the Fock spaces $\Gamma(\mathcal{H}')$, defined as in Subsection \ref{psiBerezin-Hida}, it follows (by the same proof using the second quantized  versions $\Gamma(U)$ of the isomorphisms $U$ in (\ref{StandardGelfandTripleNeutralEU}) and (\ref{StandardSinlgeGelfand3pleGeneralpsi}) joining the triples over $\Gamma(\mathcal{H}')$ 
with the standard triples) that\footnote{This is the case for the positive energy neutral field $\mathbb{A}$, for the negative-energy version
of the neutral field $\mathbb{A}$ the Fourier transforms $u_s$ of positive energy solutions are replaced with the Fourier transforms of the negative energy 
solutions and the complex conjugation
of the positive energy solutions by the conjugation of the negative energy solutions. In order to obtain negative energy versions
of the non-neutral fields $\boldsymbol{\psi}$ from the formulas for the positive energy fields the replacements are analogous: the roles of the  
positive and the negative energy fundamental solutions are interchanged.}
\begin{multline}\label{IntKerOpForNeutralAonEU}
\mathbb{A}^b(x) = \mathbb{A}^{(-) \, b}(x) 
+ \mathbb{A}^{(+) \, b}(x) = \Xi\big(\kappa_{0,1}(b,x)\big) 
+ \Xi\big(\kappa_{1,0}(b,x)\big) \\
= \sum \limits_{s\in \{1,\ldots, d'(\widehat{l}\cdot \widehat{n})\}, \widehat{n}\cdot \widehat{l} \in \mathscr{O}_+} 
\kappa_{0,1}(s, \widehat{n}\cdot \widehat{l}; b,x) \,\, a_{{}_{s}}(\widehat{n}\cdot \widehat{l})
\\
+
\sum \limits_{s\in \{1,\ldots,d'(\widehat{l}\cdot \widehat{n}) \}, \widehat{n}\cdot \widehat{l} \in \mathscr{O}_+} 
\kappa_{1,0}(s, \widehat{n}\cdot \widehat{l}; b ,x) \,\, a_{{}_{s}}(\widehat{n}\cdot \widehat{l})^{+} \\ =
\sum \limits_{(s, \widehat{l}\cdot \widehat{n}) \in \mathscr{O}} 
\kappa_{0,1}(s, \widehat{n}\cdot \widehat{l}; b,x) \,\, a_{{}_{s}}(\widehat{n}\cdot \widehat{l})
\\
+
\sum \limits_{(s, \widehat{l}\cdot \widehat{n}) \in \mathscr{O}} 
\kappa_{1,0}(s, \widehat{n}\cdot \widehat{l}; b ,x) \,\, a_{{}_{s}}(\widehat{n}\cdot \widehat{l})^{+}
\\
b \in \{1, \ldots, d = \textrm{dim} \, V \},
\end{multline}
with the corresponding plane-wave kernels $\kappa_{0,1}(b,x), \kappa_{1,0}(b,x)$ determined by ordinary functions
\begin{multline*}
(s, \widehat{n}\cdot \widehat{l}) \mapsto \kappa_{0,1}(s, \widehat{n}\cdot \widehat{l}; b,x) = 
\sum \limits_{i,j \in \{-l, \ldots, l \}} \sqrt{2l+1}
{u^{b}_{s}}_{{}_{ji}}(\widehat{n}\cdot \widehat{l}) 
\widehat{n}(t) \widehat{l}_{{}_{ij}}(\boldsymbol{w}),
\\
(s, \widehat{n}\cdot \widehat{l}) \mapsto \kappa_{1,0}(s, \widehat{n}\cdot \widehat{l}; b,x) = 
\sum \limits_{i,j \in \{-l, \ldots, l \}} \sqrt{2l+1}
\overline{{u^{b}_{s}}_{{}_{ji}}(\widehat{n}\cdot \widehat{l})} 
\overline{\widehat{n}(t)} \overline{\widehat{l}_{{}_{ij}}(\boldsymbol{w})},\\
\textrm{here}
\,\,\,
x = t\times \boldsymbol{w} \in \mathbb{R} \times SU(2, \mathbb{C}).
\end{multline*}
determining vector-valued kernels $\kappa_{0,1}, \kappa_{1,0}$, defined by the smooth functions
$(s, \widehat{n}\cdot \widehat{l}, \mu,x) \mapsto \kappa_{0,1}(s, \widehat{n}\cdot \widehat{l}; b,x)$,
$(s, \widehat{n}\cdot \widehat{l}, \mu,x) \mapsto \kappa_{1,0}(s, \widehat{n}\cdot \widehat{l}; b,x)$,
as functions on 
$\mathscr{O}\times \{1, \ldots, d \} \times \widetilde{\mathbb{S}^1} \times SU(2, \mathbb{C})$ with
in general infinite set $\mathscr{O}$, which can be regarded as 
$\mathscr{O}_+$ including $d'(\widehat{n}\cdot \widehat{l})$ copies of each point $\widehat{n}\cdot \widehat{l} \in \mathscr{O}_+$. Note that for zero mass neutral field $\mathbb{A}$ the orbit 
$\mathscr{O}_+$ is infinite, so that the sum 
(\ref{IntKerOpForNeutralAonEU}) is infinite and includes infinte number of creation and annihilation
Hida operators $a_{{}_{s}}(\widehat{n}\cdot \widehat{l})^{+}, a_{{}_{s}}(\widehat{n}\cdot \widehat{l})$ 
transforming continuously the Hida space into itself, but still, as we will see, the sum
(\ref{IntKerOpForNeutralAonEU}) represents a well defined generalized integral kernel operator in the sense of Hida-Obata-Sait\^o in the Fock space
transforming continuously the Hida space $(E)$ into $(E)^*$.

Similarly we have for the quantum field $\boldsymbol{\psi}$, which corresponds to 
a complex classical field, with antiparticles not coinciding with particles, where
$\boldsymbol{\psi}$ evaluated at space-time point $x$ is equal to the following
integral kernel operator
\begin{multline}\label{IntKerOpForGeneralPsionEU}
\boldsymbol{\psi}^a(x) = \boldsymbol{\psi}^{(-) \, a}(x) 
+ \boldsymbol{\psi}^{(+) \, a}(x) = \Xi\big(\kappa_{0,1}(a,x)\big) 
+ \Xi\big(\kappa_{1,0}(a,x)\big) \\
= \sum \limits_{s\in \{1,\ldots,d'(\widehat{n}\cdot \widehat{l})\}, \widehat{n}\cdot \widehat{l} \in \mathscr{O}_+} 
\kappa_{01}(s, \widehat{n}\cdot \widehat{l}; a,x) \,\, b_{{}_{s}}(\widehat{n}\cdot \widehat{l})
\\
+
\sum \limits_{s\in \{1,\ldots,d''(\widehat{n}\cdot \widehat{l})\}, \widehat{n}\cdot \widehat{l} \in \mathscr{O}_-} 
\kappa_{1,0}(s, \widehat{n}\cdot \widehat{l}; a,x) \,\, d_{{}_{s}}(\widehat{n}\cdot \widehat{l})^{+} \\ =
\sum \limits_{(s, \widehat{n}\cdot \widehat{l}) \in \mathscr{O}} 
\kappa_{0,1}(s, \widehat{n}\cdot \widehat{l}; a,x) \,\, b_{{}_{s}}(\widehat{n}\cdot \widehat{l})
\\
+
\sum \limits_{(s, \widehat{n}\cdot \widehat{l}) \in \mathscr{O}} 
\kappa_{1,0}(s, \widehat{n}\cdot \widehat{l}; a,x) \,\, d_{{}_{s}}(\widehat{n}\cdot \widehat{l})^{+}
\end{multline}
with Hida creation and annihilation operators $b_{{}_{s}}(\widehat{n}\cdot \widehat{l})^+,
 d_{{}_{s}}(\widehat{n}\cdot \widehat{l})^+$, $b_{{}_{s}}(\widehat{n}\cdot \widehat{l}),
 d_{{}_{s}}(\widehat{n}\cdot \widehat{l})$,
each transforming continuously the Hida space into itself (which here in the discrete
case also the creation Hida operators $b_{{}_{s}}(\widehat{n}\cdot \widehat{l})^+,
 d_{{}_{s}}(\widehat{n}\cdot \widehat{l})^+$ transform continuously
 the Hida space $\big(\mathcal{S}_{A}(\mathscr{O};\mathbb{C})\big)$), so that 
 each summand in (\ref{IntKerOpForGeneralPsionEU}) is an ordinary operator 
transforming continuously the Hida space $(E)$
into itself (as we will soon see), with the corresponding plane-wave 
kernels $\kappa_{01}(a,x), \kappa_{10}(a,x)$ determined by ordinary functions
\begin{multline*}
(s, \widehat{n}\cdot \widehat{l}) \mapsto \kappa_{0,1}(s, \widehat{n}\cdot \widehat{l}; a,x) = \\
=\left\{ \begin{array}{ll}
\sum \limits_{i,j \in \{-l, \ldots, l \}} \sqrt{2l+1}
{u^{a}_{s}}_{{}_{ji}}(\widehat{n}\cdot \widehat{l}) 
\widehat{n}(t) \widehat{l}_{{}_{ij}}(\boldsymbol{w}) & \textrm{if $s\in \{1,\dots d'(\widehat{n}\cdot \widehat{l})\}, \widehat{n}\cdot \widehat{l}
\in \mathscr{O}_+$}
\\
0 & \textrm{if $\widehat{n}\cdot \widehat{l}
\notin \mathscr{O}_+$}
\end{array} \right.
\end{multline*}

\begin{multline*}
(s, \widehat{n}\cdot \widehat{l}) \mapsto \kappa_{1,0}(s, \widehat{n}\cdot \widehat{l}; a,x) = \\
=\left\{ \begin{array}{ll}
0 & \textrm{if $\widehat{n}\cdot \widehat{l}
\notin \mathscr{O}_-$}
\\
\sum \limits_{i,j \in \{-l, \ldots, l \}} \sqrt{2l+1}
{v^{a}_{s}}_{{}_{ji}}(\widehat{n}\cdot \widehat{l}) 
\overline{\widehat{n}(t)} \overline{\widehat{l}_{{}_{ij}}(\boldsymbol{w})} & \textrm{if $s\in \{1,\dots d''(\widehat{n}\cdot \widehat{l})\}, \widehat{n}\cdot \widehat{l}
\in \mathscr{O}_-$}
\end{array} \right.
\end{multline*}
\[
\textrm{here}
\,\,\,
x = t\times \boldsymbol{w} \in \mathbb{R} \times SU(2, \mathbb{C}).
\]
They, in turn, determine vector-valued kernels $\kappa_{01}, \kappa_{10}$
defined by ordinary functions
$(s, \widehat{n}\cdot \widehat{l},a,x) \mapsto \kappa_{01}(s, \widehat{n}\cdot \widehat{l}; a,x)$,
$(s, \widehat{n}\cdot \widehat{l},a,x) \mapsto \kappa_{10}(s, \widehat{n}\cdot \widehat{l}; a,x)$, on 
$\mathscr{O}\times \{1, \ldots,4 \} \times \widetilde{\mathbb{S}^1} \times SU(2, \mathbb{C})$ with the finite set $\mathscr{O}$, which can be regarded as disjoint sum of
$\mathscr{O}_+$ including $d'(\widehat{n}\cdot \widehat{l})$ copies of each point and of
$\mathscr{O}_-$ including $d''(\widehat{n}\cdot \widehat{l})$ copies of each point $\widehat{n}\cdot \widehat{l} \in \mathscr{O}_-$. But as we will show $\boldsymbol{\psi}^a(x)$ is in fact
an ordinary operator in the Fock space transforming continuously the Hida space 
$(E)$ into $(E)$.

Recall that here $b_{{}_{s}}(\widehat{n}\cdot \widehat{l}),
d_{{}_{s}}(\widehat{n}\cdot \widehat{l}),
b_{{}_{s}}(\widehat{n}\cdot \widehat{l})^+, d_{{}_{s}}(\widehat{n}\cdot \widehat{l})^+$
are the Hida operators, which respect the canonical commutation/anti-commutation relations:
\begin{equation}\label{General[b,b^+],[d,d^+]}
\boxed{
\begin{split}
\big[b_{{}_{s}}(\widehat{n}\cdot \widehat{l}), 
b_{{}_{s'}}(\widehat{n'}\cdot \widehat{l'})^+ \big]_\pm = 
\delta_{ss'} \delta_{nn'} \delta_{ll'}, \\
\big[ b_{{}_{s}}(\widehat{n}\cdot \widehat{l}), 
b_{{}_{s'}}(\widehat{n'}\cdot \widehat{l'}) \big]_\pm = 
\big[ b_{{}_{s}}(\widehat{n}\cdot \widehat{l})^+, 
b_{{}_{s'}}(\widehat{n'}\cdot \widehat{l'})^+ \big]_\pm =0, \\
\big[ d_{{}_{s}}(\widehat{n}\cdot \widehat{l}), 
d_{{}_{s'}}(\widehat{n;}\cdot \widehat{l'})^+ \big]_\pm = 
\delta_{ss'} \delta_{nn'} \delta_{ll'}, \\
\big[ d_{{}_{s}}(\widehat{n}\cdot \widehat{l}), 
d_{{}_{s'}}(\widehat{n'}\cdot \widehat{l'}) \big]_\pm = 
\big[ d_{{}_{s}}(\widehat{n}\cdot \widehat{l})^+, 
d_{{}_{s'}}(\widehat{n'}\cdot \widehat{l'})^+ \big]_\pm =0,\\
\big[ b_{{}_{s}}(\widehat{n}\cdot \widehat{l}), 
d_{{}_{s'}}(\widehat{n'}\cdot \widehat{l'})^+ \big]_\pm = 0, 
\end{split}
}
\end{equation}
and $b_{{}_{s}}(\widehat{n}\cdot \widehat{l})$ are defined through the Hida
annihilation operator $a$ (compare Subsection \ref{psiBerezin-Hida}) as 
\[
b_{{}_{s}}(\widehat{n}\cdot \widehat{l})
\overset{\textrm{df}}{=}
a(\delta_{s, \widehat{n}\cdot \widehat{l}})
\]
for $s, \widehat{n}\cdot \widehat{l} \in \mathscr{O}$ with the index $s$
representing the $s$-th fundamental positive energy solution, and 
\[
d_{{}_{s}}(\widehat{n}\cdot \widehat{l})
\overset{\textrm{df}}{=}
a(\delta_{s, \widehat{n}\cdot \widehat{l}})
\]
for $s, \widehat{n}\cdot \widehat{l} \in \mathscr{O}$ with the index $s$
representing the $s$-th fundamental negative energy solution, compare
Subsection \ref{psiBerezin-Hida}. Here in the definition of Hida operators
$b_{{}_{s}}(\widehat{n}\cdot \widehat{l})$, $d_{{}_{s}}(\widehat{n}\cdot \widehat{l})$,
the symbol  $\delta_{s, \widehat{n}\cdot \widehat{l}}$
denotes the Dirac delta functional understood as an element of the space $\mathcal{S}_{A}(\mathscr{O};\mathbb{C})^*$ strongly dual to the standard nuclear space 
$\mathcal{S}_{A}(\mathscr{O};\mathbb{C})$ in (\ref{StandardSinlgeGelfand3pleGeneralpsi}).

The operators $a_{{}_{s}}(\widehat{n}\cdot \widehat{l})$
in the formula (\ref{IntKerOpForNeutralAonEU}) are the Hida operators
\[
a_{{}_{s}}(\widehat{n}\cdot \widehat{l})
\overset{\textrm{df}}{=}
a(\delta_{s, \widehat{n}\cdot \widehat{l}})
\]
in the standard Fock space corresponding to the Fock space of the neutral
field $\mathbb{A}$. Here in the definition of Hida operators
$a_{{}_{s}}(\widehat{n}\cdot \widehat{l})$,
the symbol  $\delta_{s, \widehat{n}\cdot \widehat{l}}$
denotes the Dirac delta functional understood as an element of the space $\mathcal{S}_{A}(\mathscr{O};\mathbb{C})^*$ strongly dual to the standard nuclear space 
$\mathcal{S}_{A}(\mathscr{O};\mathbb{C})$ in (\ref{StandardGelfandTripleNeutralEU}).
Of course the Hida operators $a_{{}_{s}}(\widehat{n}\cdot \widehat{l}), a_{{}_{s}}(\widehat{n}\cdot \widehat{l})^+$ respect the canonical commutation/anticommutation relations
\begin{equation}\label{General[a,a^+]}
\boxed{
\begin{split}
\big[a_{{}_{s}}(\widehat{n}\cdot \widehat{l}),
a_{{}_{s'}}(\widehat{n'}\cdot \widehat{l'})^+ \big]_\pm = \delta_{ss'} \delta_{nn'} \delta_{ll'},
\\
\big[a_{{}_{s}}(\widehat{n}\cdot \widehat{l}), 
a_{{}_{s'}}(\widehat{n'}\cdot \widehat{l'}) \big]_\pm  
=  \big[a_{{}_{s}}(\widehat{n}\cdot \widehat{l})^+,
a_{{}_{s'}}(\widehat{n'}\cdot \widehat{l'})^+ \big]_\pm  = 0.
\end{split}
}
\end{equation}
Note that the Fock spaces are bose or fermi Fock spaces with the corresponding commutation or, respectively, anticommutation
relations of the Hida operators. The bose Fock spaces together with canonical commutation relations of the Hida creation-annihilation 
operators are applied for the fields $\mathbb{A}$ or $\boldsymbol{\psi}$ if in the decomposition of the representation $V$ of $SU(2, \mathbb{C})$, 
in the formula for the representation (\ref{UphiOnSxG}) associated to the field in question, into irreducible components there are present
only the representations $\widehat{l}$ with integer weight $l$. This is the case for example for the representations $V$
of the class (\ref{multispinorV}) with even $n$ or for the class (\ref{V^(n)forms}) of representations $V$. 
If in the decomposition of $V$ into irreducible components $\widehat{l}$ there are present only summands
with half an odd integer weight $l$, then the free field is constructed with the help of Fermi Fock space with canonical
anti-commutation relations of the Hida creation-annihilation operators. This relation between spin and statistics is nontrivial and has been discovered
by Pauli for fields on the Minkowski space-time, and goes back to the expression of Emmy Noether energy integral, which
is positive definite for half an odd integer spin fields only under the canonical anti-commutation relation of the
creation-annihilation operators, and under canonical commutation relations for integer spin fields. The same theorem holds true for free
fields on the Einstein Universe. For the classification of the invariant Lagrange density functions (and thus the corresponding
Noether energy integrals) of local fields with arbitrarily high spins on the Einstein Universe, compare \cite{PaneitzSegalI}-\cite{PaneitzSegalIII}.

Strictly speaking we have the following equalities
\begin{multline}\label{neutralA(phi)}
\mathbb{A}(\phi) = a'(\overline{\check{\widetilde{\phi}}}\big|_{{}_{\mathscr{O}_+}})
+a'(\widetilde{\phi}\big|_{{}_{\mathscr{O_+}}})^+ \\
= 
\sum \limits_{s, \widehat{n}\cdot \widehat{l}\in \mathscr{O}_+}
\kappa_{0,1}(\phi)(s, \widehat{n}\cdot \widehat{l})
\, a_{{}_{s}}(\widehat{n}\cdot \widehat{l})
+
\sum \limits_{s, \widehat{n}\cdot \widehat{l}\in \mathscr{O}_+}
\kappa_{1,0}(\phi)(s, \widehat{n}\cdot \widehat{l})
\, a_{{}_{s}}(\widehat{n}\cdot \widehat{l})^+
\\
= \Xi\big(\kappa_{0,1}(\phi)\big) + \Xi\big(\kappa_{1,0}(\phi)\big),
\,\,\,\,\, \phi \in \mathscr{E}
\end{multline}
and
\begin{multline}\label{generalPsi(phi)}
\boldsymbol{\psi}(\phi) = a'\big(P^\oplus \widetilde{\phi}\big|_{{}_{\mathscr{O}_+}} \oplus 0 \big)
+a'\Big(0 \oplus \big(P^\ominus \widetilde{\phi}\big|_{{}_{\mathscr{O}_-}}\big)^\flat \Big)^+ \\
= 
\sum \limits_{s, \widehat{n}\cdot \widehat{l}\in \mathscr{O}_+}
\kappa_{0,1}(\overline{\phi})(s, \widehat{n}\cdot \widehat{l})
\, a_{{}_{s}}(\widehat{n}\cdot \widehat{l})
+
\sum \limits_{s, \widehat{n}\cdot \widehat{l}\in \mathscr{O}_+}
\kappa_{1,0}(\overline{\phi})(s, \widehat{n}\cdot \widehat{l})
\, a_{{}_{s}}(\widehat{n}\cdot \widehat{l})^+
\\
= \Xi\big(\kappa_{0,1}(\phi)\big) + \Xi\big(\kappa_{1,0}(\phi)\big),
\,\,\,\,\, \phi \in \mathscr{E}.
\end{multline}
Here
\[
\kappa_{0,1}(\phi)(s, \widehat{n}\cdot \widehat{l}) = \sum \limits_{b=1}^{d} \int \limits_{\widetilde{\mathbb{S}^1}\times SU(2, \mathbb{C})}
\kappa_{0,1}(s, \widehat{n}\cdot \widehat{l}; b,x) \, \phi^{b}(x) \ud^4 x,
\]
with $\phi$ ranging over the corresponding space-time standard nuclear test space (\ref{PeriodicSpacetimeTestSpace}):
\[
\phi \in \mathscr{E} = \mathcal{S}_{\Delta}(\widetilde{\mathbb{S}^1} \times G); \mathbb{C}^d)
\]
as usual (compare Subsection \ref{psiBerezin-Hida}). Note also that for $\phi \in \mathscr{E}$, we define
\[
\check{\widetilde{\phi}}_{{}_{j \, i}}(\widehat{n} \cdot \widehat{l}) \overset{\textrm{df}}{=}
\widetilde{\phi}_{{}_{-j \,\,\, -i}}(\widehat{-n} \cdot \widehat{l}),
\]
so that for real $\phi$
\[
\overline{\check{\widetilde{\phi}}} = \widetilde{\phi}.
\]

Strictly speaking the right-hand sides, say $R$, of (\ref{IntKerOpForNeutralAonEU})
and (\ref{IntKerOpForGeneralPsionEU}) should be replaced with $\Gamma(U)^{-1} R \Gamma(U)$
with the corresponding unitary isomorphisms $U$ between the Gelfand triples
(\ref{StandardGelfandTripleNeutralEU}) and respectively 
(\ref{StandardSinlgeGelfand3pleGeneralpsi}). We have omitted the isomorphism
$\Gamma(U)^{-1} (\cdot) \Gamma(U)$ in order to simplify notation, as we did in Subsection \ref{psiBerezin-Hida}. Also in the second and third line of the equalities
(\ref{neutralA(phi)}) and (\ref{generalPsi(phi)}) the isomorphism 
$\Gamma(U)^{-1} (\cdot) \Gamma(U)$ should be present, and was omitted for simplicity of notation.
This omission is inessential and can be summarized in the following proviso: all Hida operators
$a_{{}_{s}}(\widehat{n}\cdot \widehat{l}), b_{{}_{s}}(\widehat{n}\cdot \widehat{l}), d_{{}_{s}}(\widehat{n}\cdot \widehat{l}), \ldots$ which act in the Fock space over the standard Hilbet space should be replaced with $\Gamma(U)^{-1}a_{{}_{s}}(\widehat{n}\cdot \widehat{l})\Gamma(U), 
\Gamma(U)^{-1}b_{{}_{s}}(\widehat{n}\cdot \widehat{l})\Gamma(U), \Gamma(U)^{-1}d_{{}_{s}}(\widehat{n}\cdot \widehat{l})\Gamma(U), \ldots$, whenever they appear in the formulas for field
operators. Of course because $a_{{}_{s}}(\widehat{n}\cdot \widehat{l}), b_{{}_{s}}(\widehat{n}\cdot \widehat{l}), d_{{}_{s}}(\widehat{n}\cdot \widehat{l})$ 
transform continuously the Hida space $\big(\mathcal{S}_{A}(\mathscr{O};\mathbb{C})\big)$ into
itself, then  $\Gamma(U)^{-1}a_{{}_{s}}(\widehat{n}\cdot \widehat{l})\Gamma(U), 
\Gamma(U)^{-1}b_{{}_{s}}(\widehat{n}\cdot \widehat{l})\Gamma(U), \Gamma(U)^{-1}d_{{}_{s}}(\widehat{n}\cdot \widehat{l})\Gamma(U)$ 
are the corresponding Hida operators which transform
the Hida space $(E)$ into itself. Similarly we have for their Hermitian transpositions
(linear transpose preceded and followed with complex conjugation operation, compare Subsection \ref{psiBerezin-Hida}), 
\emph{i.e.} the corresponding Hida creation operators: because
$a_{{}_{s}}(\widehat{n}\cdot \widehat{l})^+, b_{{}_{s}}(\widehat{n}\cdot \widehat{l})^+, d_{{}_{s}}(\widehat{n}\cdot \widehat{l})^+$ 
transform continuously the Hida space $\big(\mathcal{S}_{A}(\mathscr{O};\mathbb{C})\big)$ into
it's strong dual $\big(\mathcal{S}_{A}(\mathscr{O};\mathbb{C})\big)^*$, then  $\Gamma(U)^{-1}a_{{}_{s}}(\widehat{n}\cdot \widehat{l})^+\Gamma(U), 
\Gamma(U)^{-1}b_{{}_{s}}(\widehat{n}\cdot \widehat{l})^+\Gamma(U), \Gamma(U)^{-1}d_{{}_{s}}(\widehat{n}\cdot \widehat{l})^+\Gamma(U)$ are the corresponding Hida operators which transform
the Hida space $(E)$ into the strong dual $(E)^*$. But in practice we can ignore
the isomorphism, and consider literally the expressions which are written explicitly
in the formulas (\ref{IntKerOpForNeutralAonEU})
and (\ref{IntKerOpForGeneralPsionEU}) expressed through the Hida operators 
$a_{{}_{s}}(\widehat{n}\cdot \widehat{l}), b_{{}_{s}}(\widehat{n}\cdot \widehat{l}), d_{{}_{s}}(\widehat{n}\cdot \widehat{l}), \ldots$ 
acting in the Fock space over the standard space.
In particular a theorem proved for such operators, saying in particular that they transform
the standard Hida space $\big(\mathcal{S}_{A}(\mathscr{O};\mathbb{C})\big)$ continuously 
into itself, mean that the corresponding operators obtained by the application 
of the isomorphism $\Gamma(U)^{-1} (\cdot) \Gamma(U)$, transform continuously
the corresponding (non-standard) Hida space $(E)$ into itself $(E)$. And analogous statements
we have for the other theorems in which we replace only the standard Hida space and its
strong dual by the corresponding non-standard Hida space $(E)$ and its strong dual.

Now we are ready to apply the general theory of Fock expansions of integral kernel operators
\cite{hida}, \cite{obata-book}, \cite{obataJFA} (and its extension to the Fermi Fock space
of Subsection \ref{psiBerezin-Hida}, and further extension of the calculus of integral kernel 
operators including convolution and Wick product operations of Subsection \ref{OperationsOnXi})
to the special case of integral kernel operators in the Fock spaces of free
fields on the Einstein Universe. The crucial point is that the orbits
$\mathscr{O}_\pm$ corresponding to all free fields are discrete, and in particular
the measure spaces $(\mathscr{O}, \mu_{{}_{\mathscr{O}}})$ in the construction of the
standard Hida spaces $\big(\mathcal{S}_{A}(\mathscr{O};\mathbb{C})\big)$ are discrete.
Moreover, each delta function $\delta_{s, \widehat{n}\cdot \widehat{l}}$ (characteristic
function of the single-point set $\{ (s, \widehat{n}\cdot \widehat{l})\} \subset \mathscr{O}$)
on the topological discrete Borel space $(\mathscr{O}, \mu_{{}_{\mathscr{O}}})$
belongs to the standard nuclear space:
\[
\delta_{s, \widehat{n}\cdot \widehat{l}} \in \mathcal{S}_{A}(\mathscr{O};\mathbb{C}).
\]
This statement is equivalent to the fact that the Fourier transform
$u_s(\widehat{n}\cdot \widehat{l})$ or $v_s(\widehat{n}\cdot \widehat{l})$ (concentrated at
single point $\widehat{n}\cdot \widehat{l}$) of each
$s$-th positive or negative energy fundamental solution belongs to the domain of the 
standard operator
\begin{align*}
\big(P^\oplus \widetilde{\Delta}P^\oplus\big)^r
\,\,\,
\textrm{or}
\,\,\,
\big(P^\ominus \widetilde{\Delta}P^\ominus\big)^r
\,\,\,
\textrm{or}
\\
\Big(P^\oplus \widetilde{\Delta}P^\oplus \oplus \big(P\ominus \widetilde{\Delta}P^\ominus \big)^\flat \Big)^r
\,\,\,
\textrm{or}
\,\,\,
\Big(\big(P^\oplus \widetilde{\Delta}P^\oplus\big)^\flat \oplus P\ominus \widetilde{\Delta}P^\ominus  \Big)^r 
\\
r \in \mathbb{N},
\end{align*}
respectively (depending on we are dealing with neutral or non-neutral, positive or negative
energy free field). This in turn follows immediately from the fact that Fourier transform 
$u_s(\widehat{n}\cdot \widehat{l})$ or $v_s(\widehat{n}\cdot \widehat{l})$ (concentrated at
single point $\widehat{n}\cdot \widehat{l}$) of each 
$s$-th positive or negative energy fundamental solution belongs to the domain
of the operator
\[
\widetilde{\Delta}^r, \,\,\,\,\, r \in \mathbb{N}
\]
(recall here that the projectors $P^\oplus, P^\ominus$ commute with $\widetilde{\Delta}$).
The last statement, in turn, follows trivially by definition,
because
\[
\widetilde{\Delta} u_s(\widehat{n}\cdot \widehat{l}) 
= \big({\textstyle \frac{n^2}{4}} +l(l+1) + {\textstyle \frac{1}{4}}\big)
u_s(\widehat{n}\cdot \widehat{l}),
\,\,\,\,
\widetilde{\Delta} v_s(\widehat{n}\cdot \widehat{l}) 
= \big({\textstyle \frac{n^2}{4}} +l(l+1) + {\textstyle \frac{1}{4}}\big)
v_s(\widehat{n}\cdot \widehat{l}). 
\]
On the other hand the Dirac
delta functional $\delta_{s, \widehat{n}\cdot \widehat{l}}$  can be identified and represented by the
characteristic function $\delta_{s, \widehat{n}\cdot \widehat{l}}$:
\begin{multline*}
\delta_{s, \widehat{n}\cdot \widehat{l}} \,(f)
= \int \limits_{\mathscr{O}} f(s', \widehat{n'}\cdot \widehat{l'}) \,\,
\delta_{s, \widehat{n}\cdot \widehat{l}} \, (s', \widehat{n'}\cdot \widehat{l'})
\, \ud \mu_{{}_{\mathscr{O}}}(s', \widehat{n'}\cdot \widehat{l'})
\\
= \sum \limits_{s', \widehat{n'}\cdot \widehat{l'} \in \mathscr{O}}
f(s', \widehat{n'}\cdot \widehat{l'}) \,\,
\delta_{s, \widehat{n}\cdot \widehat{l}}(s', \widehat{n'}\cdot \widehat{l'})
= 
f(s, \widehat{n}\cdot \widehat{l}),
\,\,\,\,
f \in  \mathcal{S}_{A}(\mathscr{O};\mathbb{C})
\end{multline*}
Therefore the Dirac delta function belongs to the nuclear test space
$\mathcal{S}_{A}(\mathscr{O};\mathbb{C})$
\begin{equation}\label{deltaInE}
\textrm{Dirac} \, \delta_{s, \widehat{n}\cdot \widehat{l}} \,\, \textrm{function}
\,\, \in \,\,
\mathcal{S}_{A}(\mathscr{O};\mathbb{C}).
\end{equation}
This fact has capital analytical consequences. 

From this fact (\ref{deltaInE}) and from Theorem 2.6 of  \cite{hida}
and its Fermi analogue Thm. \ref{Xi_l,m:Hida->Hida} of Subsection \ref{psiBerezin-Hida}
(compare also Prop. 4.3.10 of \cite{obata-book} and its Fermi analogue
 Corollary \ref{D_xi=int(xiPartial)} of Subsection \ref{psiBerezin-Hida}) we infer
the conclusion that all Hida operators encountered here
\begin{equation}\label{Hidaa^+: (E)->(E)}
a_{{}_{s}}(\widehat{n}\cdot \widehat{l})^+, b_{{}_{s}}(\widehat{n}\cdot \widehat{l})^+, d_{{}_{s}}(\widehat{n}\cdot \widehat{l})^+,
a_{{}_{s}}(\widehat{n}\cdot \widehat{l}), b_{{}_{s}}(\widehat{n}\cdot \widehat{l}), d_{{}_{s}}(\widehat{n}\cdot \widehat{l})  
\in 
\mathscr{L} \Big(\big(\mathcal{S}_{A}(\mathscr{O};\mathbb{C})\big),
\big(\mathcal{S}_{A}(\mathscr{O};\mathbb{C})\big) \Big),
\end{equation}
including the creation Hida operators evaluated at single
points of $\mathscr{O}$, transform continuously the standard Hida space
into itself. Correspondingly the Hida operators 
\begin{equation}\label{HidaGamma(U)^-1a^+Gamma(U): (E)->(E)}
\Gamma(U)^{-1}a_{{}_{s}}(\widehat{n}\cdot \widehat{l})^+\Gamma(U), \,\,
\Gamma(U)^{-1}b_{{}_{s}}(\widehat{n}\cdot \widehat{l})^+\Gamma(U), \,\, \ldots
\in \mathscr{L} \big((E), (E) \big)
\end{equation}
and transform the Hida space $(E)$ continuously into itself, and thus are ordinary
operators. 

Now we give an important consequence of (\ref{Hidaa^+: (E)->(E)})
and correspondingly (\ref{HidaGamma(U)^-1a^+Gamma(U): (E)->(E)}) and thus of (\ref{deltaInE}):
\begin{prop*}
On the Einstein Universe the negative as well as the positive frequency parts 
\[
\mathbb{A}^{(-) b}(x), \boldsymbol{\psi}^{(-) a}(x),
\mathbb{A}^{(+) b}(x), \boldsymbol{\psi}^{(+) a}(x),
\]
of free local and  \emph{massive} fields $\mathbb{A}(x)$, $\boldsymbol{\psi}(x)$,
as well as their space-time derivatives, evaluated at arbitrary but fixed space-time point
$x$, are ordinary operators on the Fock space transforming continuously the test Hida
space into itself. In particular all free local and  \emph{massive} fields $\mathbb{A}(x)$, $\boldsymbol{\psi}(x)$
on the Einstein Universe, evaluated at space-time point, represent ordinary operators
on the Fock space transforming continuously the test Hida space into itself.
\end{prop*}
\qedsymbol \, 
 In this particular case of massive fields matters substantially simplify
because, as we have already seen, for massive fields the discrete space 
$\mathscr{O}$ is finite and thus the sums respectively in (\ref{IntKerOpForNeutralAonEU}) or in (\ref{IntKerOpForGeneralPsionEU}) 
become finite. Because, as we have already shown, all Hida creation-annihilation operators in (\ref{IntKerOpForNeutralAonEU}) 
or in (\ref{IntKerOpForGeneralPsionEU}) transform continuously the Hida space into itself,
then the Proposition follows trivially in this case.
\qed

A weakened form of this Proposition for massive fields 
has been discovered in the series of papers \cite{PaneitzSegalI}-\cite{PaneitzSegalI},
\cite{SegalZhouQED}.

Now we look more closely at the vector-valued distributional kernels $\kappa_{0,1}, \kappa_{1,0}$ defining the general
local (neutral or non-neutral) quantum free fields on the Einstein Universe. Recall that the above constructed free field 
operators $\mathbb{A}, \boldsymbol{\psi}$ are well defined generalized integral kernel operators with vector-valued kernels 
 $\kappa_{0,1}, \kappa_{1,0}$, in the sense of \cite{obataJFA}, for any elements
\begin{align*}
\kappa_{0,1}, \kappa_{1,0} \in \mathscr{L}(\mathscr{E}, E^*)
\cong \mathscr{L}(E, \mathscr{E}^*) \,\,\, \textrm{or resp.} \\
\kappa_{0,1}, \kappa_{1,0} \in \mathscr{L}(\mathscr{E}, \mathcal{S}_{A}(\mathscr{O};\mathbb{C})^*)
\cong \mathscr{L}(\mathcal{S}_{A}(\mathscr{O};\mathbb{C}), \mathscr{E}^*). 
\end{align*}
In this general case the free field operators 
\begin{multline*}
\mathbb{A}, \boldsymbol{\psi} \in
\mathscr{L}\big(\mathscr{E}, \mathscr{L}((E), (E)^* \big) \,\,\, \textrm{or resp.} \\
\mathbb{A}, \boldsymbol{\psi} \in
\mathscr{L}\Big(\mathscr{E}, \mathscr{L}\big((\mathcal{S}_{A}(\mathscr{O};\mathbb{C})), (\mathcal{S}_{A}(\mathscr{O};\mathbb{C}))^* \Big).
\end{multline*}
Recall that the pairings $\langle \cdot, \cdot \rangle$ of the vector-valued distributions  $\kappa_{0,1}, \kappa_{1,0}$ are defined by ordinary integrations, respectively, over space-time and summation over the field components index $a$ or integration over the orbit $\mathscr{O}$ (which degenerates to ordinary summation):
\[
\sum \limits_{a} \int \limits_{\widetilde{\mathbb{S}^1}\times SU(2, \mathbb{C})} \ldots \ud^4 x,
\,\,\,\,\textrm{or resp.} \,\, \int \limits_{\mathscr{O}} \ldots \ud \mu_{{}_{\mathscr{O}}} = \sum \limits_{\mathscr{O}} \ldots,
\]
compare Subsection \ref{psiBerezin-Hida}. In particular for the space-time test function $\phi \in \mathscr{E}$ the distribution
$\kappa_{0,1}(\phi) \in \mathcal{S}_{A}(\mathscr{O};\mathbb{C})^*$ or resp. $\in E^*$ is defined through the kernel
\[
\kappa_{0,1}(\phi)(s, \widehat{n}\cdot \widehat{l}) = 
\sum \limits_{a \in \{1, \ldots, d\}} \,\, \int \limits_{\widetilde{\mathbb{S}^1}\times SU(2, \mathbb{C})} 
\kappa_{0,1}(s, \widehat{n}\cdot \widehat{l}; a, x) \, \phi^a(x)
\, \ud^4 x
\]
and for $\xi \in \mathcal{S}_{A}(\mathscr{O};\mathbb{C})$ or $\in E$, we have the pairings 
\begin{multline*}
\langle \kappa_{0,1}(\phi), \xi \rangle \\
=  \sum \limits_{a \in \{1, \ldots, d\}} \,\, \int \limits_{\mathscr{O}} \,\, \int \limits_{\widetilde{\mathbb{S}^1}\times SU(2, \mathbb{C})} 
\kappa_{0,1}(s, \widehat{n}\cdot \widehat{l}; a, x) \, \phi^a(x) \,  \xi(s, \widehat{n}\cdot \widehat{l}) \,\,
\, \ud^4 x \, \ud \mu_{{}_{\mathscr{O}}}(s, \widehat{n}\cdot \widehat{l}) \\
= 
\sum \limits_{(s, \widehat{n}\cdot \widehat{l}) \in \mathscr{O}} \,\, \sum \limits_{a \in \{1, \ldots, d\}} \,\, \int \limits_{\widetilde{\mathbb{S}^1}\times SU(2, \mathbb{C})} 
\kappa_{0,1}(s, \widehat{n}\cdot \widehat{l}; a, x) \, \phi^a(x) \, \xi(s, \widehat{n}\cdot \widehat{l}) \,\,
\, \ud^4 x,
\end{multline*}
compare Subsection \ref{psiBerezin-Hida} or \cite{obataJFA}. 

However the free fields $\mathbb{A}, \boldsymbol{\psi}$ on the Einstein Universe are much more regular integral kernel operators.
Namely we have the following
\begin{prop*}
The vector-valued kernel distributions  $\kappa_{0,1}, \kappa_{1,0}$ defining the free fields
$\mathbb{A}, \boldsymbol{\psi}$ on the Einstein Universe as integral kernel operators with vector valued kernels,
actually transform continuously the space-time test space $\mathscr{E}$ into the nuclear space $E$ (or respectively
into the standard nuclear space $\mathcal{S}_{A}(\mathscr{O};\mathbb{C})$), \emph{i.e.}
\begin{multline*}
\kappa_{0,1}, \kappa_{1,0} \in \mathscr{L}(\mathscr{E}, E)
\cong \mathscr{L}(E^*, \mathscr{E}^*) \,\,\, \textrm{or resp.} \\
\kappa_{0,1}, \kappa_{1,0} \in \mathscr{L}(\mathscr{E}, \mathcal{S}_{A}(\mathscr{O};\mathbb{C}))
\cong \mathscr{L}(\mathcal{S}_{A}(\mathscr{O};\mathbb{C})^*, \mathscr{E}^*). 
\end{multline*}

From this it follows by the general theory of vector-valued integral kernel operators due to Obata, that
the free field operators
\begin{multline*}
\mathbb{A}, \boldsymbol{\psi} \in
\mathscr{L}\big(\mathscr{E}, \mathscr{L}((E), (E)) \big) \,\,\, \textrm{or resp.} \\
\mathbb{A}, \boldsymbol{\psi} \in
\mathscr{L}\Big(\mathscr{E}, \mathscr{L}\big((\mathcal{S}_{A}(\mathscr{O};\mathbb{C})), (\mathcal{S}_{A}(\mathscr{O};\mathbb{C}) \big) \Big).
\end{multline*}
\end{prop*}
\qedsymbol \,
Let $\kappa_{1,0}$ be the positive frequency kernel defining a neutral field $\mathbb{A}$ on the Einstein Universe.
Let the space-time points $x$ be written as $x = t \times \boldsymbol{w} \in \mathbb{R} \times SU(2, \mathbb{C})$.
Note that 
\begin{multline*}
\kappa_{1,0}(\phi)(s, \widehat{n}\cdot \widehat{l}) = \\
\sum \limits_{b \in \{1, \ldots, d\}} \,\, \sum \limits_{-l \leq i,j \leq l} \, 
\sqrt{2l+1} \,
\overline{{u^{b}_{s}}_{{}_{ji}}(\widehat{n}\cdot \widehat{l})}
\int \limits_{\widetilde{\mathbb{S}^1}\times SU(2, \mathbb{C})} 
 \, \phi^b(t, \boldsymbol{w}) \, \overline{\widehat{n}(t)} \, \overline{\widehat{l}_{{}_{ij}}(\boldsymbol{w})} \, dt \, d\boldsymbol{w}
 \\
= \sum \limits_{b \in \{1, \ldots, d\}} \,\, \sum \limits_{-l \leq i,j \leq l} \, 
\sqrt{2l+1} \,
\overline{{u^{b}_{s}}_{{}_{ji}}(\widehat{n}\cdot \widehat{l})}
\widetilde{\phi^b}_{{}_{ji}}(\widehat{n}\cdot \widehat{l}),
\,\,\, (s, \widehat{n}\cdot \widehat{l}) \in \mathscr{O}.
\end{multline*}
The formula for the other kernels $\kappa_{0,1}$, $\kappa_{1,0}$ of the free fields is analogous and expressed through 
the Fourier transform of (or of the conjugation of) the space-time test function $\phi$. 

Let us estimate the the $q$-th norm $| A^q \cdot| = | \cdot |_{{}_{q}}$ of the function $\kappa_{1,0}(\phi)$
on $\mathscr{O}$, where $| \cdot |$ is the ordinary $L^2$ norm on $L^2(\mathscr{O}; \mu_{{}_{\mathscr{O}}})$
on the discrete measure space $(\mathscr{O}; \mu_{{}_{\mathscr{O}}})$. 
Recall that the eigenvalues of the standard operator $A$ defining the Gelfand triple in the single particle
space of free fields has the same eigenvalues 
\[
\lambda_{nl} = {\textstyle\frac{n^2}{4}} + l(l+1) + {\textstyle\frac{1}{4}} +1
\]
on the invariant subspaces spanned by the functions whose Fourier transforms
are concentrated at $\widehat{n}\cdot \widehat{l}$ as does the standard operator $\Delta +1$
defining the standard nuclear space-time test space $\mathscr{E}$.
We have
\begin{multline*}
\big| A^q \kappa_{1,0}(\phi) \big|^2 = \big| \kappa_{1,0}(\phi) \big|_{{}_{q}}^{2} \\ =
\sum \limits_{(s, \widehat{n}\cdot \widehat{l}) \in \mathscr{O}}
\, \lambda_{nl}^{2q} \, |\kappa_{1,0}(\phi)(s, \widehat{n}\cdot \widehat{l})|^2 \\ \leq
\sum \limits_{(s, \widehat{n}\cdot \widehat{l}) \in \mathscr{O}} \,
\sum \limits_{b \in \{1, \ldots, d\}} \,\, \sum \limits_{-l \leq i,j \leq l} \, 2 \lambda_{nl}^{2q} \, 
(2l+1)^3 |\overline{{u^{b}_{s}}_{{}_{ji}}(\widehat{n}\cdot \widehat{l})}
\widetilde{\phi^b}_{{}_{ji}}(\widehat{n}\cdot \widehat{l})|^2 \\ \leq
\sum \limits_{(s, \widehat{n}\cdot \widehat{l}) \in \mathscr{O}} \,
\sum \limits_{b \in \{1, \ldots, d\}} \,\, \sum \limits_{-l \leq i,j \leq l} \, 2(2l+1)^2 \lambda_{nl}^{2q} \, 
|\widetilde{\phi^b}_{{}_{ji}}(\widehat{n}\cdot \widehat{l})|^2 \\ =
\sum \limits_{(s, \widehat{n}\cdot \widehat{l}) \in \mathscr{O}} \, 
\sum \limits_{b \in \{1, \ldots, d\}} \,\, \sum \limits_{-l \leq i,j \leq l} \, 
2 (2l+1) \, (2l+1) \lambda_{nl}^{2q}
|\widetilde{\phi^b}_{{}_{ji}}(\widehat{n}\cdot \widehat{l})|^2 \\ \leq
\sum \limits_{(s, \widehat{n}\cdot \widehat{l}) \in \mathscr{O}} \, 
\sum \limits_{b \in \{1, \ldots, d\}} \,\, \sum \limits_{-l \leq i,j \leq l} \, 
2 (2l+1) \, \lambda_{nl}^{2(q +1)}
|\widetilde{\phi^b}_{{}_{ji}}(\widehat{n}\cdot \widehat{l})|^2
\leq 2 \big| \phi \big|_{{}_{q+1}}^{2},
\end{multline*}
where the second inequality follows by the fact that the Fourier transforms $u^{b}_{s}(\widehat{n}\cdot \widehat{l})$ of the fundamental solutions are normalized,
whence
\[
\big|{u^{b}_{s}}_{{}_{ji}}(\widehat{n}\cdot \widehat{l})\big| \leq {\textstyle\frac{1}{\sqrt{2l+1}}}.
\]
But this is equivalent to the continuity of the map
\[
\mathscr{E} \ni \phi \longrightarrow \kappa_{1,0}(\phi) \in 
\mathcal{S}_{A}(\mathscr{O};\mathbb{C}) 
\,\,\, \textrm{or respectively} \,\,\,
\mathscr{E} \ni \phi \longrightarrow \kappa_{1,0}(\phi) \in E.
\]
This means that 
\begin{multline*}
\kappa_{1,0} \in \mathscr{L}(\mathscr{E}, E)
\cong \mathscr{L}(E^*, \mathscr{E}^*) \,\,\, \textrm{or resp.} \\
\kappa_{1,0} \in \mathscr{L}(\mathscr{E}, \mathcal{S}_{A}(\mathscr{O};\mathbb{C}))
\cong \mathscr{L}(\mathcal{S}_{A}(\mathscr{O};\mathbb{C})^*, \mathscr{E}^*). 
\end{multline*}
Identical proof works for all remaining vector-valued kernels $\kappa_{0,1}, \kappa_{1,0}$ defining
free fields on the Einstein Universe. 

From this result it follows, by the Theorem 3.13 of \cite{obataJFA} (in Bose case) or by its Fermi analogue -- Theorem \ref{obataJFA.Thm.3.13} 
of Subsection \ref{psiBerezin-Hida} (in Fermi case), that 
\begin{multline*}
\mathbb{A}, \boldsymbol{\psi} \in
\mathscr{L}\big(\mathscr{E}, \mathscr{L}((E), (E)) \big) \,\,\, \textrm{or resp.} \\
\mathbb{A}, \boldsymbol{\psi} \in
\mathscr{L}\Big(\mathscr{E}, \mathscr{L}\big((\mathcal{S}_{A}(\mathscr{O};\mathbb{C})), (\mathcal{S}_{A}(\mathscr{O};\mathbb{C}) \big) \Big).
\end{multline*}
\qed

\begin{prop*}
Let $x = t \times \boldsymbol{u} \in \mathbb{R} \times SU(2, \mathbb{C})$ be a fixed, but arbitrary, space-time point and let
$a$ be a fixed, but again arbitrary, field component index. Let $(\ell, m) = (0,1)$ or $(\ell, m) = (1,0)$ and let
$\kappa_{\ell, m}$ be the vector-valued kernels defining free fields on the Einstein Universe.
Then the kernels $\kappa_{\ell, m}(s, \widehat{n}\cdot \widehat{l}; a, x)$ with fixed $(a,x)$
(and treated as a fixed parameter) represent distributions (continuous functionals) on the corresponding standard nuclear test space
$\mathcal{S}_{A}(\mathscr{O};\mathbb{C})$ (or respectively on $E$), \emph{i.e.}
\[
\kappa_{\ell, m}(s, \widehat{n}\cdot \widehat{l}; a, x) \in \mathcal{S}_{A}(\mathscr{O};\mathbb{C})^*
\,\,\,\,\,\, \textrm{or resp.} \,\,\,\,\,\,
\kappa_{\ell,m}(s, \widehat{n}\cdot \widehat{l}; a, x) \in E^*
\,\,\, \textrm{for each fixed} \, ( a, x), 
\]
\[
(\ell, m) = (0,1) \,\,\,\,\,
\textrm{or} \,\,\,\,\,
(\ell, m) = (1,0).
\]
\end{prop*}
\qedsymbol \,
Let $x = t \times \boldsymbol{w} \in \mathbb{R} \times SU(2, \mathbb{C})$ be a fixed, but arbitrary, space-time point and let
$a$ be a fixed, but again arbitrary, field component index. Let, for example, $\kappa_{0,1}$ be the negative-frequency vector-valued kernel of a
non-neutral free field $\boldsymbol{\psi}$ or of a neutral free field $\mathbb{A}$. Let $\xi \in \mathcal{S}_{A}(\mathscr{O};\mathbb{C})$. 
Recall that the pairing $\langle \kappa_{0,1}(a, x), \xi \rangle$, \emph{i.e.} the value of the distribution (functional) 
$\kappa_{0,1}(a, x)$ (with fixed parameter $(a, x) \in \{1, \ldots, \textrm{dim} \, V \} \times \mathbb{R} \times SU(2, \mathbb{C})$)
on the function $\xi$ is given by integration (summation) over $\mathscr{O}$ of the function (sequence) $\xi$ with the kernel
$\kappa_{0,1}(s, \widehat{n}\cdot \widehat{l}; a, x)$:
\begin{multline}\label{PAIRING<kappa,xi>}
\langle \kappa_{0,1}(a, x), \xi \rangle = 
\sum \limits_{(s, \widehat{n}\cdot \widehat{l}) \in \mathscr{O}} 
\kappa_{0,1}(s, \widehat{n}\cdot \widehat{l}; a, x) \, \xi(s, \widehat{n}\cdot \widehat{l}) 
\\ 
=\sum \limits_{1\leq s\leq d'(\widehat{n}\cdot \widehat{l}), \widehat{n}\cdot \widehat{l} \in \mathscr{O}_+} 
\kappa_{0,1}(s, \widehat{n}\cdot \widehat{l}; a, x) \, \xi(s, \widehat{n}\cdot \widehat{l}) 
\\
= \sum \limits_{1\leq s\leq d'(\widehat{n}\cdot \widehat{l}), \widehat{n}\cdot \widehat{l} \in \mathscr{O}_+} \,\,
\sum \limits_{-l \leq i,j \leq l} \,\,
\sqrt{2l+1} {u^{a}_{s}}_{{}_{ji}}(\widehat{n}\cdot \widehat{l}) \, \widehat{l}_{{}_{ij}}(\boldsymbol{w}) \, \widehat{n}(t)  \, \xi(s, \widehat{n}\cdot \widehat{l}).
\end{multline}
We represent this expression (\ref{PAIRING<kappa,xi>}) for the paring as inner product $(\cdot, \cdot)$ on $L^2(\mathscr{O})$ of auxiliary functions 
belonging to $L^2(\mathscr{O})$. 

Let us define the auxiliary function $\zeta$ by the formula
\[
\zeta(s, \widehat{n}\cdot \widehat{l}) = \sum \limits_{-l \leq i,j \leq l} \,\,
\sqrt{2l+1} \overline{{u^{a}_{s}}_{{}_{ji}}(\widehat{n}\cdot \widehat{l}) \, \widehat{l}_{{}_{ij}}(\boldsymbol{w}) \, \widehat{n}(t)}
\,\,\,\,\,\, \textrm{for} \,\, s \in \{1, \ldots, d'(\widehat{n}\cdot \widehat{l}) \}, \widehat{n}\cdot \widehat{l} \in \mathscr{O}_+ 
\]
and $\zeta(s, \widehat{n}\cdot \widehat{l})$ is defined to be zero for $\widehat{n}\cdot \widehat{l} \notin \mathscr{O}_+$
or $s \notin \{1, \ldots, d'(\widehat{n}\cdot \widehat{l}) \}$. 

Recall that the Fourier transforms $u_s(\widehat{n} \cdot \widehat{l})$ of the fundamental solutions, concentrated respectively
on the single points $\widehat{n} \cdot \widehat{l}$, are normalized:
\[
\sum \limits_{-l \leq i,j \leq l, 1 \leq a \leq \textrm{dim} \, V} \,\,
(2l+1) \big|{u^{a}_{s}}_{{}_{ji}}(\widehat{n}\cdot \widehat{l})\big|^2 = 1,
\]
whence 
\begin{equation}\label{|u^a_s(n.l)|<(sqrt{2l+1})^-1}
\big|{u^{a}_{s}}_{{}_{ji}}(\widehat{n}\cdot \widehat{l})\big| \leq
{\textstyle\frac{1}{\sqrt{2l+1}}}.
\end{equation}
Because $\widehat{l}_{{}_{ij}}(\boldsymbol{w})$ are the matrices of the unitary representation of $SU(2, \mathbb{C})$,
then
\begin{multline*}
\sum \limits_{-l \leq i,j \leq l} \big| \widehat{l}_{{}_{ij}}(\boldsymbol{w}) \big|^2
= \sum \limits_{-l \leq i,j \leq l}  \widehat{l}_{{}_{ij}}(\boldsymbol{w}) \overline{\widehat{l}_{{}_{ij}}(\boldsymbol{w})} 
\\
= \sum \limits_{-l \leq i,j \leq l}  \widehat{l}_{{}_{ij}}(\boldsymbol{w}) \widehat{l}_{{}_{ji}}(\boldsymbol{w}^{-1}) 
= \textrm{Tr} \, \boldsymbol{1}_{{}_{2l+1}} = 2l+1,
\end{multline*}
whence
\[
\big|  \widehat{l}_{{}_{ij}}(\boldsymbol{w}) \big| \leq \sqrt{2l+1}.
\]
Therefore
\[
|\zeta(s, \widehat{n}\cdot \widehat{l})|^2 \leq (2l+1)^2.
\]
Because the range of the index $s$ depends on the point $\widehat{n}\cdot \widehat{l}$,
and is confined to the range $1 \leq s \leq d'(\widehat{n}\cdot \widehat{l}) \leq (2l+1)^2 \, \textrm{dim} \, V
= (2l+1)^2 \, d$, then there exists finite positive integer $r$ such that the function $A^{-r} \zeta$: 
\[
A^{-r} \zeta(s, \widehat{n}\cdot \widehat{l}) = \lambda_{nl}^{-r} \zeta(s, \widehat{n}\cdot \widehat{l})
\]
is square summable, \emph{i.e.} belongs to $L^2(\mathscr{O})$. Here $A$ is the standard operator on $L^2(\mathscr{O})$ definining
the Gelfand triple in the standard version of the single particle Hilbert space of the field $\boldsymbol{\psi}$ in question.

On the other hand, because $\xi \in \mathcal{S}_{A}(\mathscr{O};\mathbb{C})$ and  because $A$ transforms continuously the Hida space
$\mathcal{S}_{A}(\mathscr{O};\mathbb{C})$ into itself, then the function $A^{r} \xi$: 
\[
A^{r} \xi(s, \widehat{n}\cdot \widehat{l}) = \lambda_{nl}^{r} \zeta(s, \widehat{n}\cdot \widehat{l})
\] 
belongs to $\mathcal{S}_{A}(\mathscr{O};\mathbb{C}) \subset L^2(\mathscr{O})$, and moreover there exists positive integer
$q$ (depending on $r$) and a finite constant $C_r$, such that
\[
\sum \limits_{(s, \widehat{n}\cdot \widehat{l}) \in \mathscr{O}} 
\lambda_{nl}^{2r} | \zeta(s, \widehat{n}\cdot \widehat{l})|^2 = \big|A^{r} \xi \big|^2 = \big|\xi \big|_{{}_{r}}^{2}
\leq C_{r}^{2} \, \big|\xi \big|_{{}_{q}}^{2},
\]
where $| \cdot |$ used for functions denotes the $L^2$ norm on $L^2(\mathscr{O})$.

Thus, the pairing (\ref{PAIRING<kappa,xi>}) can be represented as the $L^2$-inner product on 
$L^2(\mathscr{O})$ of the auxiliary functions $A^{-r} \zeta, A^{r} \xi \in L^2(\mathscr{O})$, and by the Schwartz inequality
and by above norm estimations we obtain the inequality
\[
|\langle \kappa_{0,1}(a, x), \xi \rangle| \leq | (A^{-r} \zeta, A^{r} \xi )|\leq 
\big|A^{-r} \zeta \big| \, \big|A^{r} \xi \big| = \big|A^{-r} \zeta \big| \, \big|\xi \big|_{{}_{r}}
\leq C_r \big|A^{-r} \zeta \big| \, \big|\xi \big|_{{}_{q}},
\]
which is equivalent to the continuity of the functional $\kappa_{0,1}(a, x)$, with fixed $(a, x)$, defined by the kernel
$\kappa_{0,1}(s, \widehat{n}\cdot \widehat{l}; a, x)$ with fixed $(a, x)$.

Proof of the continuity of the functionals defined by all the remaining kernels 
$\kappa_{1,0}(s, \widehat{n}\cdot \widehat{l}; a, x)$ and $\kappa_{0,1}(s, \widehat{n}\cdot \widehat{l}; a, x)$
corresponding to non-neutral or neutral fields is the same.
\qed

From the last Proposition it follows (by Thm. 2.6 \cite{hida} in bose case or by its Fermi analog -- Thm. \ref{Xi_l,m:Hida->Hida} 
of Subsection \ref{psiBerezin-Hida}) the following
\begin{prop*}
\begin{multline*}
\mathbb{A}(x), \mathbb{A}^{(+)}(x) , \boldsymbol{\psi}(x), \boldsymbol{\psi}^{(+)}(x) \in
\mathscr{L}\big(\mathscr{E}, \mathscr{L}((E), (E)^*) \big) \,\,\, \textrm{or resp.} \\
\mathbb{A}(x), \mathbb{A}^{(+)}(x) , \boldsymbol{\psi}(x), \boldsymbol{\psi}^{(+)}(x) \in
\mathscr{L}\Big(\mathscr{E}, \mathscr{L}\big((\mathcal{S}_{A}(\mathscr{O};\mathbb{C})), (\mathcal{S}_{A}(\mathscr{O};\mathbb{C}))^* \big) \Big), \\
\mathbb{A}^{(-)}(x), \boldsymbol{\psi}^{(-)}(x) \in
\mathscr{L}\big(\mathscr{E}, \mathscr{L}((E), (E)) \big) \,\,\, \textrm{or resp.} \\
\mathbb{A}^{(-)}(x) , \boldsymbol{\psi}^{(-)}(x) \in
\mathscr{L}\Big(\mathscr{E}, \mathscr{L}\big((\mathcal{S}_{A}(\mathscr{O};\mathbb{C})), (\mathcal{S}_{A}(\mathscr{O};\mathbb{C})) \big) \Big).
\end{multline*}
\end{prop*}

Note that only the first negative frequency summands 
\begin{align*}
\mathbb{A}^{(-) b}(x) \Phi = 
\sum \limits_{s\in \{1,\ldots, d'(\widehat{l}\cdot \widehat{n})\}, \widehat{n}\cdot \widehat{l} \in \mathscr{O}_+} 
\kappa_{01}(s, \widehat{n}\cdot \widehat{l}; b,x) \, a_{{}_{s}}(\widehat{n}\cdot \widehat{l}) \Phi
\\
\boldsymbol{\psi}^{(-) a}(x) \Phi =
\sum \limits_{s\in \{1,\ldots,d'(\widehat{n}\cdot \widehat{l})\}, \widehat{n}\cdot \widehat{l} \in \mathscr{O}_+} 
\kappa_{01}(s, \widehat{n}\cdot \widehat{l}; a,x) b_{{}_{s}}(\widehat{n}\cdot \widehat{l})\Phi
\end{align*}
respectively in (\ref{IntKerOpForNeutralAonEU}) and (\ref{IntKerOpForGeneralPsionEU})
are Bochner summable for each element $\Phi$ of the Hida space $(E)= \cap_{k>0} (E)_{k}$ (respectively of the standard Hida space
$\big(\mathcal{S}_{A}(\mathscr{O};\mathbb{C})\big) = \cap_{k>0} \big(\mathcal{S}_{A}(\mathscr{O};\mathbb{C})\big)_k$), if the sums
\[
\sum \limits_{\mathscr{O}} \ldots
\]
are understood as discrete integrals 
\[
\int \limits_{\mathscr{O}} \ldots \ud \mu_{{}_{\mathscr{O}}},
\]
as sums (integrals) in the Hilbert space $(E)_{k}$ (respectively in the Hilbert space
$\big(\mathcal{S}_{A}(\mathscr{O};\mathbb{C})\big)_k$) for each fixed, but arbitrary, natural $k$.
Thus $\mathbb{A}^{(-) b}(x) \Phi, \boldsymbol{\psi}^{(-) a}(x) \Phi$ belong to 
$(E)_{k}$ (respectively in the Hilbert space
$\big(\mathcal{S}_{A}(\mathscr{O};\mathbb{C})\big)_k$) for each positive $k$, an thus
belong to the Hida space $(E) = \cap_{k>0} (E)_{k}$ (respectively to the standard Hida space). 
Indeed, recall that by Lemma 2.1 of \cite{hida} (or its Fermi analogue -- Lemma \ref{etaPhiPsi} of Subsection \ref{psiBerezin-Hida}) 
the functions 
\[
\begin{split}
\mathscr{O} \ni (s,\widehat{n}\cdot \widehat{l}) \longrightarrow
\| b_{{}_{s}}(\widehat{n}\cdot \widehat{l}) \Phi\|, \\
\mathscr{O} \ni (s,\widehat{n}\cdot \widehat{l}) \longrightarrow
\| d_{{}_{s}}(\widehat{n}\cdot \widehat{l}) \Phi\|, \\
\mathscr{O} \ni (s,\widehat{n}\cdot \widehat{l}) \longrightarrow
\| a_{{}_{s}}(\widehat{n}\cdot \widehat{l}) \Phi\|,
\end{split}
\]
as well as their squares and higher order powers, are summable, because (by the said Lemmas) the functions
\begin{multline*}
\mathscr{O} \times \mathscr{O}  \ni (s,\widehat{n}\cdot \widehat{l}) \times (s',\widehat{n'}\cdot \widehat{l'}) \longrightarrow
\big\langle \big\langle b_{{}_{s}}(\widehat{n}\cdot \widehat{l}) \Phi, b_{{}_{s'}}(\widehat{n'}\cdot \widehat{l'}) \Phi \big\rangle \big\rangle \\
= \big\langle \big\langle  \Phi, b_{{}_{s}}(\widehat{n}\cdot \widehat{l})^+ b_{{}_{s'}}(\widehat{n'}\cdot \widehat{l'}) \Phi \big\rangle \big\rangle
= \eta_{{}_{\Phi, \Phi}}\big(s,\widehat{n}\cdot \widehat{l}); s',\widehat{n'}\cdot \widehat{l'}\big), 
\end{multline*}
\begin{multline*}
\mathscr{O} \times \mathscr{O}  \ni (s,\widehat{n}\cdot \widehat{l}) \times (s',\widehat{n'}\cdot \widehat{l'}) \longrightarrow
\big\langle \big\langle d_{{}_{s}}(\widehat{n}\cdot \widehat{l}) \Phi, d_{{}_{s'}}(\widehat{n'}\cdot \widehat{l'}) \Phi \big\rangle \big\rangle \\ 
= \big\langle \big\langle  \Phi, d_{{}_{s}}(\widehat{n}\cdot \widehat{l})^+ d_{{}_{s'}}(\widehat{n'}\cdot \widehat{l'}) \Phi \big\rangle \big\rangle
= \eta_{{}_{\Phi, \Phi}}\big(s,\widehat{n}\cdot \widehat{l}; s',\widehat{n'}\cdot \widehat{l'}\big),
\end{multline*}
\begin{multline*}
\mathscr{O} \times \mathscr{O}  \ni (s,\widehat{n}\cdot \widehat{l}) \times (s',\widehat{n'}\cdot \widehat{l'}) \longrightarrow
\big\langle \big\langle a_{{}_{s}}(\widehat{n}\cdot \widehat{l}) \Phi, a_{{}_{s'}}(\widehat{n'}\cdot \widehat{l'}) \Phi \big\rangle \big\rangle \\ 
= \big\langle \big\langle  \Phi, a_{{}_{s}}(\widehat{n}\cdot \widehat{l})^+ a_{{}_{s'}}(\widehat{n'}\cdot \widehat{l'}) \Phi \big\rangle \big\rangle
= \eta_{{}_{\Phi, \Phi}}\big(s,\widehat{n}\cdot \widehat{l}; s',\widehat{n'}\cdot \widehat{l'}\big),
\end{multline*}
belong to the respective nuclear spaces $E^{\otimes 2} = E \otimes E$ (or their standard versions).

But note that for massless fields (or for fields with infinite $\mathscr{O}$) this is not the case for 
$\mathbb{A}^{(+) b}(x) \Phi, \boldsymbol{\psi}^{(+) a}(x) \Phi$, 
which do not belong to $(E)_k$, for any $k>0$ (respectively do not belong to any 
$\big(\mathcal{S}_{A}(\mathscr{O};\mathbb{C})\big)_k$, for any $k>0$) and thus 
$\mathbb{A}^{(+) b}(x) \Phi, \boldsymbol{\psi}^{(+) a}(x) \Phi$ \emph{in general do not belong} to the 
Hida space for $\Phi$ being an element of the Hida space, compare Subsection \ref{BSH}.
This is so irrespectively of (\ref{Hidaa^+: (E)->(E)}) or (\ref{HidaGamma(U)^-1a^+Gamma(U): (E)->(E)}). Indeed, recall that
by the canonical commutation or anticommutation relations we have
\begin{multline*}
\big\langle \big\langle a_{{}_{s}}(\widehat{n}\cdot \widehat{l})^+ \Phi, a_{{}_{s'}}(\widehat{n'}\cdot \widehat{l'})^+ \Phi \big\rangle \big\rangle  
= \mp \big\langle \big\langle  \Phi, a_{{}_{s}}(\widehat{n}\cdot \widehat{l})^+ a_{{}_{s'}}(\widehat{n'}\cdot \widehat{l'}) \Phi \big\rangle \big\rangle \\
\pm \delta_{ss'}\delta_{nn'}\delta_{ll'} \langle \langle \Phi, \Phi \rangle \rangle \\
= \mp \eta_{{}_{\Phi, \Phi}}\big(s,\widehat{n}\cdot \widehat{l}; s',\widehat{n'}\cdot \widehat{l'}\big)
\pm \delta_{ss'}\delta_{nn'}\delta_{ll'} \langle \langle \Phi, \Phi \rangle \rangle
\end{multline*}
and similarly for the other Hida creation operators $b_{{}_{s}}(\widehat{n}\cdot \widehat{l})^+$ and 
$b_{{}_{s}}(\widehat{n}\cdot \widehat{l})^+$, so that restriction to the diagonal of the functions
\begin{multline*}
\mathscr{O} \times \mathscr{O}  \ni (s,\widehat{n}\cdot \widehat{l}) \times (s',\widehat{n'}\cdot \widehat{l'}) \longrightarrow
\big\langle \big\langle b_{{}_{s}}(\widehat{n}\cdot \widehat{l})^+ \Phi, b_{{}_{s'}}(\widehat{n'}\cdot \widehat{l'})^+ \Phi \big\rangle \big\rangle \\
= \big\langle \big\langle  \Phi, b_{{}_{s}}(\widehat{n}\cdot \widehat{l}) b_{{}_{s'}}(\widehat{n'}\cdot \widehat{l'})^+ \Phi \big\rangle \big\rangle
= \mp \eta_{{}_{\Phi, \Phi}}\big(s,\widehat{n}\cdot \widehat{l}; s',\widehat{n'}\cdot \widehat{l'}\big)
\pm \delta_{ss'}\delta_{nn'}\delta_{ll'} \langle \langle \Phi, \Phi \rangle \rangle, 
\end{multline*}
\begin{multline*}
\mathscr{O} \times \mathscr{O}  \ni (s,\widehat{n}\cdot \widehat{l}) \times (s',\widehat{n'}\cdot \widehat{l'}) \longrightarrow
\big\langle \big\langle d_{{}_{s}}(\widehat{n}\cdot \widehat{l})^+ \Phi, d_{{}_{s'}}(\widehat{n'}\cdot \widehat{l'})^+ \Phi \big\rangle \big\rangle \\ 
= \big\langle \big\langle  \Phi, d_{{}_{s}}(\widehat{n}\cdot \widehat{l}) d_{{}_{s'}}(\widehat{n'}\cdot \widehat{l'})^+ \Phi \big\rangle \big\rangle
= \mp \eta_{{}_{\Phi, \Phi}}\big(s,\widehat{n}\cdot \widehat{l}; s',\widehat{n'}\cdot \widehat{l'}\big)
\pm \delta_{ss'}\delta_{nn'}\delta_{ll'} \langle \langle \Phi, \Phi \rangle \rangle,
\end{multline*}
cannot be summable as functions  on $\mathscr{O}$. In particular the functions
\[
\begin{split}
\mathscr{O} \ni (s,\widehat{n}\cdot \widehat{l}) \longrightarrow
\| b_{{}_{s}}(\widehat{n}\cdot \widehat{l})^+ \Phi\|, \\
\mathscr{O} \ni (s,\widehat{n}\cdot \widehat{l}) \longrightarrow
\| d_{{}_{s}}(\widehat{n}\cdot \widehat{l})^+ \Phi\|, \\
\mathscr{O} \ni (s,\widehat{n}\cdot \widehat{l}) \longrightarrow
\| a_{{}_{s}}(\widehat{n}\cdot \widehat{l})^+ \Phi\|,
\end{split}
\]
as well as their squares and higher order powers, are \emph{not summable}. In particular  
for massless fields (or for fields with infinite $\mathscr{O}$) 
$\mathbb{A}^{(+) b}(x) \Phi, \boldsymbol{\psi}^{(+) a}(x) \Phi$, 
do not belong to $(E)_k$, for any $k>0$ (respectively do not belong to any 
$\big(\mathcal{S}_{A}(\mathscr{O};\mathbb{C})\big)_k$, for any $k>0$) and thus 
$\mathbb{A}^{(+) b}(x) \Phi, \boldsymbol{\psi}^{(+) a}(x) \Phi$ \emph{in general do not belong} to the 
Hida space, for $\Phi$ being an element of the Hida space. But 
for each element $\Phi$ of the Hida space, 
$\mathbb{A}^{(+) b}(x) \Phi, \boldsymbol{\psi}^{(+) a}(x) \Phi$ \emph{in general do belong} to the 
space strong dual to the Hida space. Moreover, $\mathbb{A}^{(+) b}(x), \boldsymbol{\psi}^{(+) a}(x)$
are generalized operators, particular cases of the integral kernel operators, which transform continuously the Hida space into its strong
dual. In proving this assertion, equivalent to the last Proposition, we could repeat the proof of Subsection \ref{BSH} using the symbols of Berezin-Obata.
However, the proof, presented above, of the last Proposition is simpler.
Nonetheless we present now  still another proof of the last Proposition, also based on the inequaity (2-2) of Lemma 2.1 of \cite{hida} (or respectively
its Fermi analogue -- Lemma \ref{etaPhiPsi} of Subsection \ref{psiBerezin-Hida}). Namely, we show first (using the method of the last Proposition
of Subsection \ref{BSH}) that the negative-frequency operators $\mathbb{A}^{(-) b}(x), \boldsymbol{\psi}^{(-) a}(x)$ and the Hermitian transposition (conjugation) 
$\boldsymbol{\psi}^{(+) a}(x)^+$ transform continuously the Hida space into itself, independently of the argument used in the Proof of the last Proposition. 
Our assertion will then follow by duality.

Therefore, we can repeat the proof of the last Proposition of Subsection \ref{BSH}
utilizing the inequality (2-2) of Lemma 2.1 of \cite{hida} (respectively Lemma \ref{etaPhiPsi} of Subsection \ref{psiBerezin-Hida}), and show that all negative frequency parts of the free local fields and their space-time 
derivatives evaluated at space-time point $x$ are ordinary operators in the Fock space which transform
continuously the Hida space into itself. Summing up we have the following
\begin{prop*}
On the Einstein Universe the negative frequency parts 
\[
\mathbb{A}^{(-) b}(x), \boldsymbol{\psi}^{(-) a}(x),
\]
as well as the hermitian transpositions 
\[
\boldsymbol{\psi}^{(+) a}(x)^+,
\]
of the positive energy parts of free local fields $\mathbb{A}(x)$, $\boldsymbol{\psi}(x)$,
as well as their space-time derivatives, evaluated at arbitrary but fixed space-time point
$x$ are generalized integral kernel operators transforming continuously the test Hida
space into itself, and thus are ordinary operators on the Fock space.

On the Einstein Universe the negative as well as the positive frequency parts 
\[
\mathbb{A}^{(-) b}(x), \boldsymbol{\psi}^{(-) a}(x),
\mathbb{A}^{(+) b}(x), \boldsymbol{\psi}^{(+) a}(x),
\]
of free local fields $\mathbb{A}(x)$, $\boldsymbol{\psi}(x)$,
as well as their space-time derivatives, evaluated at arbitrary but fixed space-time point
$x$ are generalized integral kernel operators transforming continuously the test Hida
space into its strong dual. In particular all local fields $\mathbb{A}(x)$, $\boldsymbol{\psi}(x)$
on the Einstein Universe, evaluated at space-time point $x$, represent integral kernel operators
transforming continuously the test Hida space into its strong dual.
\end{prop*}
\qedsymbol \,
Because for each $\Phi \in (E)$, $\mathbb{A}^{(-) \, b}(x)\Phi \in (E) \subset (E)^*$,
then the canonical pairing $\langle \langle \mathbb{A}^{(-) \, b}(x)\Phi, \mathbb{A}^{(-) \, b}(x)\Phi
\rangle \rangle$ coincides with the ordinary inner product. Recall that the second quantized 
$\Gamma(A)$ version of the standard operator 
$A$ defininig the single particle Gelfand triple of the field $\mathbb{A}$, transforms continuously
the Hida test space $(E)$ into itself. In particular, $\Gamma(A)^r\Phi \in (E)$, $r>0$.
Therefore, we can form the canonical pairing  $\langle \langle \mathbb{A}^{(-) \, b}(x)
\Gamma(A)^r\Phi, \mathbb{A}^{(-) \, b}(x)\Gamma(A)^r\Phi \rangle \rangle$, which coincides with the inner product and is equal 
$\| \mathbb{A}^{(-) \, b}(x)\Phi\|_{r}^{2}$. Estimation of this
$r$-th squared norm with the help of the inequality (2-2) of Lemma 2.1 of \cite{hida} (for bose field) or Lemma \ref{etaPhiPsi} of 
Subsection \ref{psiBerezin-Hida} (for fermi field), will give us continuity of the operator $\mathbb{A}^{(-) \, b}(x)$. Namely, we have 
\begingroup\makeatletter\def\f@size{5}\check@mathfonts
\def\maketag@@@#1{\hbox{\m@th\large\normalfont#1}}%
\begin{multline*}
\| \mathbb{A}^{(-) \, b}(x)\Phi\|_{r}^{2} 
= \langle \langle \mathbb{A}^{(-) \, b}(x)
\Gamma(A)^r\Phi, \mathbb{A}^{(-) \, b}(x)\Gamma(A)^r\Phi \rangle \rangle 
\\
= \sum \limits_{(s, \widehat{n}\cdot \widehat{l}) \in \mathscr{O}} \, \sum \limits_{(s', \widehat{n'}\cdot \widehat{l'}) \in \mathscr{O}} \,
\sum \limits_{-l \leq i,j \leq l} \, \sum \limits_{-l' \leq i',j' \leq l'} \,
\sqrt{2l+1} \sqrt{2l'+1} {u^{b}_{s}}_{{}_{ji}}(\widehat{n}\cdot \widehat{l}) \overline{{u^{b}_{s'}}_{{}_{j'i'}}(\widehat{n'}\cdot \widehat{l'})}
\, \widehat{l}_{{}_{ij}}(\boldsymbol{u}) \overline{\widehat{l'}_{{}_{i'j'}}(\boldsymbol{u})}
\widehat{n-n'}(t) \, \times \\ \,\,\,\,\,\,\,\,\,\,\,\,\,\,\,\,\,\,\,\,\,\,\,\, \,\,\,\,\,\,\,\,\,\,\,\,\,\,\,\,\,\,\,\,\,\,\,\, 
\,\,\,\,\,\,\,\,\,\,\,\,\,\,\,\,\,\,\,\,\,\,\,\, \,\,\,\,\,\,\,\,\,\,\,\,\,\,\,\,\,\,\,\,\,\,\,\, 
\,\,\,\,\,\,\,\,\,\,\,\,\,\,\,\,\,\,\,\,\,\,\,\, \,\,\,\,\,\,\,\,\,\,\,\,\,\,\,\,\,\,\,\,\,\,\,\, \times
\, \langle \langle a_{s'} (\widehat{n'}\cdot \widehat{l'})^+ a_{s} (\widehat{n}\cdot \widehat{l})^+ \Gamma(A)^r \Phi, \Gamma(A)^r \Phi  \rangle \rangle \\
= 
\sum \limits_{(s, \widehat{n}\cdot \widehat{l}) \in \mathscr{O}} \, \sum \limits_{(s', \widehat{n'}\cdot \widehat{l'}) \in \mathscr{O}} \,
\sum \limits_{-l \leq i,j \leq l} \, \sum \limits_{-l' \leq i',j' \leq l'} \,
\sqrt{2l+1} \sqrt{2l'+1} {u^{b}_{s}}_{{}_{ji}}(\widehat{n}\cdot \widehat{l}) \overline{{u^{b}_{s'}}_{{}_{j'i'}}(\widehat{n'}\cdot \widehat{l'})}
\, \widehat{l}_{{}_{ij}}(\boldsymbol{u}) \overline{\widehat{l'}_{{}_{i'j'}}(\boldsymbol{u})}
\widehat{n-n'}(t) \, \times \\ \,\,\,\,\,\,\,\,\,\,\,\,\,\,\,\,\,\,\,\,\,\,\,\,\,\,\,\,\,\,\,\,\,\,\,\,\,\,\,\,\,\,\,\,\,\,\,\,
\,\,\,\,\,\,\,\,\,\,\,\,\,\,\,\,\,\,\,\,\,\,\,\, \,\,\,\,\,\,\,\,\,\,\,\,\,\,\,\,\,\,\,\,\,\,\,\, 
\,\,\,\,\,\,\,\,\,\,\,\,\,\,\,\,\,\,\,\,\,\,\,\, \,\,\,\,\,\,\,\,\,\,\,\,\,\,\,\,\,\,\,\,\,\,\,\, \times
\eta_{{}_{\Gamma(A)^r \Phi, \Gamma(A)^r \Phi}}(s', \widehat{n'}\cdot \widehat{l'}; \, s, \widehat{n}\cdot \widehat{l}),
\\
\textrm{for} \,\,
x = t \times \boldsymbol{u} \in \mathbb{R} \times SU(2, \mathbb{C}).
\end{multline*}
\endgroup
Recall, please, that by the Lemma 2.1 of  \cite{hida} (for bose case) or respectively Lemma \ref{etaPhiPsi}
of Subsection \ref{psiBerezin-Hida} (for the fermi case) the function $\eta_{{}_{\Gamma(A)^r \Phi, \Gamma(A)^r \Phi}}$:
\[
\eta_{{}_{\Gamma(A)^r \Phi, \Gamma(A)^r \Phi}}(s', \widehat{n'}\cdot \widehat{l'}; \, s, \widehat{n}\cdot \widehat{l})
\overset{\textrm{df}}{=}
\langle \langle a_{s'} (\widehat{n'}\cdot \widehat{l'})^+ a_{s} (\widehat{n}\cdot \widehat{l})^+ \Gamma(A)^r \Phi, \Gamma(A)^r \Phi  \rangle \rangle
\]
belongs to the nuclear tensor product space $E^{\otimes 2}$ (or respectively to $\mathcal{S}_{A}(\mathscr{O};\mathbb{C})^{\otimes 2}$), and that the the standard operator $A$ in action on the characteristic function $\delta_{\widehat{n}\cdot \widehat{l}}$ od the single element set $\{(s,\widehat{n}\cdot \widehat{l}) \} \subset \mathscr{O}$ is equal
\[
A \delta_{\widehat{n}\cdot \widehat{l}} = \lambda_{nl} \delta_{\widehat{n}\cdot \widehat{l}}
= \big({\textstyle\frac{n^2}{4}} + l(l+1) + {\textstyle\frac{1}{4}} +1 \big) \delta_{\widehat{n}\cdot \widehat{l}}.
\]
Finally, note that the function (sequence in fact) $f$:
\begin{multline*}
f(s, \widehat{n}\cdot \widehat{l}; \, s', \widehat{n'}\cdot \widehat{l'}) \overset{\textrm{df}}{=} \\
\sum \limits_{-l \leq i,j \leq l} \, \sum \limits_{-l' \leq i',j' \leq l'} \,
\sqrt{2l+1} \sqrt{2l'+1} {u^{b}_{s}}_{{}_{ji}}(\widehat{n}\cdot \widehat{l}) \overline{{u^{b}_{s'}}_{{}_{j'i'}}(\widehat{n'}\cdot \widehat{l'})}
\, \widehat{l}_{{}_{ij}}(\boldsymbol{u}) \overline{\widehat{l'}_{{}_{i'j'}}(\boldsymbol{u})}
\widehat{n-n'}(t), \\
(s,\widehat{n}\cdot \widehat{l}; \, s', \widehat{n'}\cdot \widehat{l'})  \in \mathscr{O} \times \mathscr{O}, \,\,\,
t \times \boldsymbol{u} = x \in \mathbb{R} \times SU(2, \mathbb{C})
\end{multline*}
is bounded by the function $(s, \widehat{n}\cdot \widehat{l}) \rightarrow 4 \lambda_{nl} \lambda_{n'l'}$
\[
|f(s, \widehat{n}\cdot \widehat{l}; \, s, \widehat{n'}\cdot \widehat{l'}) | \leq (2l+1)(2l'+1) \leq 4  \lambda_{nl}\lambda_{n'l'}
\]
so that for some finite positive integer $n_0$ the function 
\[
(s,\widehat{n}\cdot \widehat{l}; \, s', \widehat{n'}\cdot \widehat{l'}) \rightarrow 
4 \lambda_{nl}^{-n_0}\lambda_{n'l'}^{-n_0}f(s, \widehat{n}\cdot \widehat{l}; \, s', \widehat{n'}\cdot \widehat{l'})
= 4 \big((A^{-n_0})^{\otimes 2} f \big) (s, \widehat{n}\cdot \widehat{l}; \, s', \widehat{n'}\cdot \widehat{l'})
\]
is square summable on $\mathscr{O} \times \mathscr{O}$. On the other hand $(A^{n_0})^{\otimes 2} \eta_{{}_{\Gamma(A)^r \Phi, \Gamma(A)^r \Phi}}
\in E^{\otimes 2}$ (or respectively $\in \mathcal{S}_{A}(\mathscr{O};\mathbb{C})^{\otimes 2}$), where
\begin{multline*}
\big((A^{n_0})^{\otimes 2} \eta_{{}_{\Gamma(A)^r \Phi, \Gamma(A)^r \Phi}} \big)(s, \widehat{n}\cdot \widehat{l}; \, s', \widehat{n'}\cdot \widehat{l'})
\\
= \lambda_{nl}^{-n_0}\lambda_{n'l'}^{-n_0} \eta_{{}_{\Gamma(A)^r \Phi, \Gamma(A)^r \Phi}} (s, \widehat{n}\cdot \widehat{l}; \, s', \widehat{n'}\cdot \widehat{l'})
\end{multline*}
belongs to the nuclear space $E^{\otimes 2}$ (or its standard counterpart)  and it must be square summable on
$\mathscr{O} \times \mathscr{O}$. Therefore, by the Schwartz inequality we have the estimation
\begin{multline*}
\| \mathbb{A}^{(-) \, b}(x)\Phi\|_{r}^{2} 
\leq C \big| (A^{n_0})^{\otimes 2} \eta_{{}_{\Gamma(A)^r \Phi, \Gamma(A)^r \Phi}}\big|
= C \big|\eta_{{}_{\Gamma(A)^r \Phi, \Gamma(A)^r \Phi}} \big|_{{}_{n_0}},
\end{multline*}
with the constant $C$ depending on the $L^2$-norm $| \cdot |$ of the function $(A^{-n_0})^{\otimes 2} f$ on $\mathscr{O} \times \mathscr{O}$.

Now, again by the inequality (2-2) of the Lemma 2.1 of \cite{hida} (for bose case) or by the inequality
of Lemma \ref{etaPhiPsi}
of Subsection \ref{psiBerezin-Hida} (for the fermi case) we have the estimation
\begin{multline*}
\| \mathbb{A}^{(-) \, b}(x)\Phi\|_{r}^{2} 
\leq C \big| \eta_{{}_{\Gamma(A)^r \Phi, \Gamma(A)^r \Phi}} \big|_{n_0} \\
\leq C 
{\textstyle\frac{\rho^{-2q}}{-2n_0\, e \, \textrm{ln}\, \rho}}
\|\Gamma(A)^r \Phi\|_{n_0} \, \| \Gamma(A)^r \Phi\|_{n_0}
\\
\leq C C_{n_0 r}
{\textstyle\frac{\rho^{-2q}}{-2q\, e \, \textrm{ln}\, \rho}}
\|\Phi\|_{k(n_0,r)}^{2},
\end{multline*}
which is equivalent to the continuity of the operator $\mathbb{A}^{(-) \, b}(x): (E) \rightarrow (E)$.
The existence of such positive $C_{n_0 r}$ and natural $k(n_0,r)$ that the last inequality is preserved follows from the
continuity of the operator $\Gamma(A)^r: (E) \rightarrow (E)$.
The proof of the continuity of space-time derivatives $X^{\alpha}A^{(-) \, \mu}(x): (E) \rightarrow (E)$ is identical. Here $X^\alpha
= X_{0}^{\alpha_0} \cdots X_{3}^{\alpha_3}$ is the Schwartz' notation for higher order space-time differential operators
with multiidex $\alpha$. Similarly, we show the continuity of the operator 
$\boldsymbol{\psi}^{(-) \, b}(x): (E) \rightarrow (E)$ and of the Hermitian transpose $\boldsymbol{\psi}^{(+) \, b}(x)^+: (E) \rightarrow (E)$
of the operator $\boldsymbol{\psi}^{(+) \, b}(x): (E)^* \rightarrow (E)^*$,
as well as for their space-time derivatives. 

Because the operators  $\mathbb{A}^{(-) \, b}(x): (E) \rightarrow (E)$ and $\boldsymbol{\psi}^{(+) \, b}(x)^+: (E) \rightarrow (E)$
are continuous, then their Hermitian linear transpositions (ordinary linear transposition preceded and followed by the complex conjugation
operation, compare Subsection \ref{psiBerezin-Hida}) $\mathbb{A}^{(+) \, b}(x): (E)^* \rightarrow (E)^*$ 
and respectively $\boldsymbol{\psi}^{(+) \, b}(x): (E)^* \rightarrow (E)^*$, are continuous (compare \cite{treves}). 
Taking into account the natural topological inclusion $(E) \subset (E)^*$, we see that the operators 
$\mathbb{A}^{(+) \, b}(x): (E)^* \rightarrow (E)^*$ 
and respectively $\boldsymbol{\psi}^{(+) \, b}(x): (E)^* \rightarrow (E)^*$ are naturally identified with continuous
operators $(E) \rightarrow (E)^*$. Similarly, the continuous operators $\mathbb{A}^{(-) \, b}(x): (E) \rightarrow (E)$ and 
$\boldsymbol{\psi}^{(-) \, b}(x): (E) \rightarrow (E)$ are naturally identified with continuous operators
$(E) \rightarrow (E)^*$.

 \qed

In particular for the \emph{massive} free quantum essentially neutral field $\mathbb{A}(x)$ corresponding to real classical field, the operator
is symmetric on the Hida space invariant for $\mathbb{A}(x)$. Because the Hida space is perfect (being nuclear) then again by the
Riesz-Sz\"okefalvy-Nagy criterion, the operator $\mathbb{A}(x)$ possesses self adjoint extension. It is less easy (in Bose case) to show that the one-parameter
group generated by the operator $\mathbb{A}(x)$ leaves invariant the Hida space, from which essential self adjointness
of the operator $\mathbb{A}(x)$ follows. The fermi massive case is analytically trivial as in this case the Fock space becomes finite dimensional,
as we have already emphasized (finite $\mathscr{O}$ means finite dimensional single particle space, and thus finite dimensional Fermi Fock space over it).
 Similar methods and the above Propositions will be used to the analysis of the higher order contributions
to interacting fields, expressed in terms of free fields (their Wick product and convolutions with the advanced and retarded parts
of the pairing functions of free fields) in Subsection \ref{CausalSonEU}.

We should also note that our presentation is general and subsumes construction of a free field -- a single object -- whose corresponding quanta may have 
in principle not only various energy (the quantum number $n$) or various squares of the momentum (the quantum number $l(l+1)$) but likewise varios spins
(although with fixed parity of the spin). Only in simplest cases, when in the decomposition of the representation $V$ defining the representation
(\ref{UphiOnSxG}), Subsect. \ref{GeneralizedSchrodinger-VonNeumannPairs}, associated to the field, there are present direct irreducible
summands with the same weight $l$, spins of all quanta are the same and equal $l$. This is in particular the case for the
scalar field associated to the representation  (\ref{UphiOnSxG}) with $V = \widehat{0}$, and with all quanta of the field
having spin zero. The same situation we have for the Dirac field, associated to the representation (\ref{UphiOnSxG}) with
$V = \widehat{\tfrac{1}{2}} \oplus \widehat{\tfrac{1}{2}}$ with all the quanta (positrons and electrons)
having spin $1/2$. But in more involved case, e.g. of the representation $V$ of the the form (\ref{multispinorV}) 
(of Subsect. \ref{GeneralizedSchrodinger-VonNeumannPairs}), the representation $V$ is equal to direct sum with various weights and with the 
various spins of the corresponding quanta. It is true that in this case the field can be decomposed into the corresponding
independent constituent each corresponding to the transformation (\ref{UphiOnSxG})  with $V$ which decomposes into irreducible
components with the same weight. The Fock space of the initial field becomes equal to the tensor product of the Fock spaces
of the constituent fields each with fixed spin. In this case it would be desirable to pass to the constituent fields, 
and to the explicit construction of the tensor product of Fock spaces each associated with definite spin, which means that the representation
$V$ should be first decomposed into direct summands each equivalent to direct sum of several copes of representation with the same weight.
Next we should construct fields corresponding to each summand. In practice however this is not always possible. In particular
the free quantum electromagnetic potential field is  associated with the representation
(\ref{UphiOnSxG})  with $V$ equal to (\ref{LopRepV}). Now the representation $V$ of the form 
(\ref{LopRepV}) is equivalent to the direct sum $\widehat{0} \oplus \widehat{1}$, with the spin zero component
which on the Einstein Universe correspond precisely to the zero-component of the electromagnetic potential describing unphysical
scalar spin zero quanta (``scalar photons''), and the direct summand $\widehat{1}$ of the representation $V$ is associated to the 
spatial components of the potential. We could in principle pass to the Coulomb gauge and start with the representation 
$V = \widehat{1}$ and with the representation (\ref{UphiOnSxG}) acting on the three-component spatial potential 
which respects the transversality condition (which reduces the spin projection on fixed direction to just only two 
values $\pm 1$). This would be satisfactory only at the free field level. But, when passing to interactions, this would force us to 
account for separately the constraint global part of Maxwell equations
and separately the radiation part. In causal approach we need the whole four-potential which subsumes
the whole local interaction Lagrange density, as the construction of the scattering matrix is based on causality principle.
We therefore are forced to use local gauge which does not separate the radiation part seating in the spatial components from the zero component
and uses potential with the help of which the total local Lagrange interaction density can be expressed and which in turn uses the whole 
four-component electromagnetic potential and not only its spatial components.  In such local gauge, the Lorentz gauge, we are thus forced
to use redundant components (redundant, because the spin of real photons is $1$ and its projection assumes only two
values $\pm1$ because of zero rest mass). In this case we pick up the subclass of so-called physical states $\Phi$ which respect a condition
$X_\mu \mathbb{A}^{(-) \, \mu} \Phi = 0$
(negative frequency part $X_\mu \mathbb{A}^{(-) \, \mu}$ of the quantum counterpart of the classical Lorentz condition operator is zero when operating on the physical states) with the property that the Lorentz condition (at the quantum level) is fulfilled only for the averages in the states of the physical subclass. 
This forces us to use the indefinite inner product space of Krein (Gupta-Bleuler formalism) with two states being equivalent whenever differing
by a Krein-norm zero vector, and reduces the redundant unphysical freedom. 

A simpler situation we encounter when we try to adopt the \emph{massive} free quantum four-vector field, associated to the representation 
(\ref{UphiOnSxG}) with $V$ equal (\ref{LopRepV}), as a description of free quanta all having spin $1$. If for some deeper reasons
we need a four-vector field (e.g. the covariant local Lagrange density utilizes all four components of the field) then because the representation
$V$ equal (\ref{LopRepV}) is equivalent to the direct sum $\widehat{0} \oplus \widehat{1}$, we again have the redundant spin zero
component $\widehat{0}$ of V. This time however the subsidiary ``Lorentz condition''  (in operator form) reduces the redundant components
and moreover is compatible with equations of motion. This is what we could expect from  the fact that spin $1$ particle has three independent values of the projection
and on the other hand the subsidiary ```Lorentz condition'' reduces the four spin degrees of freedom to the required three, as expected for spin $1$. 
For zero mass four-vector field we cannot preserve four component character of the field just by imposing subsidiary operator condition (say ``Lorentz condition''),
because this condition reduces by one the four spin degrees of freedom, so we remain with three spin degrees of freedom. But the zero mass particle of spin $1$ possesses only two spin projections $\pm1$, so the obtained three spin degrees of freedom do not fit with zero mass spin $1$ particle. 
Thus, we construct the single particle space of the massive four-vector field as consisting of solutions $\phi$ whose each component 
$\phi^\mu$ respects the massive Klein-Gordon equation $[\square -m^2]\phi^\mu = 0$ and the subsidiary condition $X_\mu \phi^\mu = 0$
and regard the system of differential equations
\begin{align*}
[\square -m^2]\phi^\mu = 0, \\
X_\mu \phi^\mu = 0,
\end{align*} 
as the system $D_{{}_{iff}} \phi = 0$ defining the field, which we have used above in the general construction of a free field.

Now we pass to the analysis of the commutation (generalized) functions and of the pairing (generalized) functions
of the free fields, as well as of their splitting into retarded and advanced parts.

It is true that at the general
level of free fields, we cannot exactly specify the relevant commutators of free fields which are relevant. In order to learn out
what commutators are relevant we should know in what linear combinations the specified free fields enter into the important 
operators of the theory, primary to the Wick polynomial expression for the Lagrange interaction density,
and to the Wick polynomial expression of the Noether conserved currents. Nonetheless we can form general useful theorems
concerning the most important properties of the commutation and paring functions which are fundamental for the 
causal construction of the scattering operator and the interacting fields, pertinent to the realistic interactions
(like QED or generally pertinent to the Standard Model). 

Thus, the essentially neutral fields $\mathbb{A}$ in general enter the mentioned important operators simply as factors
of Wick monomials. The non-neutral (corresponding to complex classical) fields $\boldsymbol{\psi}$ enter into the relevant 
operators (which, as observables, are sesquilinear forms at the classical level in the complex unobservable field) together 
with their conjugations $\boldsymbol{\psi}^\sharp$ as factors of the Wick monomials corresponding to the 
classical sesquilinear forms. The concrete form of the conjugation operation $\boldsymbol{\psi} \rightarrow \boldsymbol{\psi}^\sharp$ 
depends on the specific type of the field $\boldsymbol{\psi}$. For the field corresponding
to the classical scalar complex field the conjugation at the classical level is just equal to the ordinary complex conjugation, to which correspond the 
Hermitian linear transposition $\boldsymbol{\psi}^+$ at the quantum field theory level. For the field $\boldsymbol{\psi}$ corresponding to a many 
component complex classical field, the conjugation $\boldsymbol{\psi}^\sharp$  at the classical level is equal to the complex conjugation and transposition followed in general by a linear transformation expressed through multiplication by a constant matrix in field components. At the quantum level $\boldsymbol{\psi}^\sharp$ correspoding to this classical conjugation is equal to the Hermitian linear transposition of each component of the field followed by the transposition and the linear transformation using exactly the same constant matrix in field components as does the classical transposition. In particular for the Dirac
field $\boldsymbol{\psi}$ the Noether conserved currents and the interaction Lagrange density uses besides the field $\boldsymbol{\psi}$
also the Dirac-conjugated field $\boldsymbol{\psi}^{\sharp} = \big[\boldsymbol{\psi}^{+}\big]^\textrm{Tr} \, \gamma_0$. 
For other kinds another matrix $\widehat{\Gamma}_0$ in field components can serve the role of $\gamma_0$ in the conjugation operation. 
 We thus will need the following commutation functions $D^{ab}$, $S^{ab}$, and the following pairings $D^{(-) \, ab}$, $S^{(-) \, ab}$:
\[
-i D^{ab}(x,y) \,\, \boldsymbol{1} \, = [\mathbb{A}^{a}(x), \mathbb{A}^{b}(y)]_{\mp},
\]
\[
-i D^{(-) \, ab}(x,y) \,\, \boldsymbol{1} \, = [\mathbb{A}^{(-) \, a}(x), \mathbb{A}^{(+) \, b}(y)]_{\mp} =
\quad \underbracket{\mathbb{A}^{a}(x) \mathbb{A}^{b}}(y)
\]
\[
-iD^{(+) \, ab}(x,y) \,\, \boldsymbol{1} \, = [\mathbb{A}^{(+) \, a}(x), \mathbb{A}^{(-) \, b}(y)]_{\mp} =
\mp (-i) D^{(-) \, ab}(x,y),
\]
\[
-i S^{ab}(x,y) \,\, \boldsymbol{1} \, = [\boldsymbol{\psi}^{a}(x), \boldsymbol{\psi}^{\sharp \, b}(y)]_{\mp},
\]
\[
-i S^{(-) \, ab}(x,y) \,\, \boldsymbol{1} \, = [\boldsymbol{\psi}^{(-) \, a}(x), \boldsymbol{\psi}^{\sharp \, (+) \, b}(y)]_{\mp} =
\quad \underbracket{\boldsymbol{\psi}^{a}(x), \boldsymbol{\psi}^{\sharp \, b}}(y),
\]
\[
-i S^{(+) \, ab}(x,y) \,\, \boldsymbol{1} \, = [\boldsymbol{\psi}^{(+) \, a}(x), \boldsymbol{\psi}^{\sharp \, (-) \, b}(y)]_{\mp} =
\mp (-i) S^{(-) \, ab}(x,y).
\]

Recall that the spacetime points $y = t \times \boldsymbol{w}$ can be identified with the elements of the subgroup $\mathbb{R}\times SU(2, \mathbb{C})
\cong \mathbb{R}\times SU(2, \mathbb{C}) \times \{1\}$ 
of the Einstein isometry group $\mathbb{R}\times SU(2, \mathbb{C}) \times SU(2, \mathbb{C})$. In particular 
$y^{-1} = (-t) \times \boldsymbol{w}^{-1}$, where $-t$ is the inversion in the Abelian factor of time translations, and $\boldsymbol{w}^{-1}$
is the group inversion in the $SU(2, \mathbb{C})$ group. Let us denote the unit of the $\mathbb{R}\times SU(2, \mathbb{C})$ just by $e
= 0 \times 1$, with $1$ denoting the unit in $SU(2, \mathbb{C})$.
Note that because the left hand sides of the above commutation relations are $c$-numbers proportional to the unit operator, then
applying to the commutation relations the ``second quantized'' operator $\Gamma(U_{{}_{y\times 1}}) ( \, \cdot \, ) \Gamma(U_{{}_{y \times 1}})^{-1}$ of the
representation operator $U_{{}_{y\times 1}}$ given by (\ref{UphiOnRxG}) associated to the free field in question and restricted to the subgroup
$\mathbb{R}\times SU(2, \mathbb{C}) \times \{1\}$, we arrive at the conclusion that
\begin{align*}
D^{ab}(x,y) = D^{ab}(xy^{-1},e) \eqqcolon D^{ab}(xy^{-1}), \\ D^{(\pm) \, ab}(x,y) = D^{(\pm) \, ab}(xy^{-1}, e) \eqqcolon D^{(\pm) \, ab}(xy^{-1}), \\
S^{ab}(x,y) = S^{ab}(xy^{-1}, e) \eqqcolon S^{ab}(xy^{-1}), \\ S^{(\pm) \, ab}(x,y) = S^{(\pm) \, ab}(xy^{-1}, e) \eqqcolon  S^{(\pm) \, ab}(xy^{-1}),
\end{align*}
for the corresponding generalized commutation and pairing functions $D^{ab}$, $S^{ab}$, $D^{(-) \, ab}$, $S^{(-) \, ab}$
in one space-time variable, which we have likewise denoted by the same symbol as the two-variable
commutation and pairing generalized functions. 

It can be easily seen that
\begin{multline}\label{D(x,y)}
D^{ab}(t\times \boldsymbol{w}, t' \times \boldsymbol{w}') = \\ =
\sum\limits_{\substack{\widehat{n}\cdot \widehat{l} \in \mathscr{O}_+ \\ 1\leq s \leq d'(\widehat{n}\cdot \widehat{l}) \\ -l \leq i,j \leq l 
\\ -l' \leq i',j' \leq l'}} 
i (2l+1) {u^{a}_{s}}_{{}_{ji}}(\widehat{n}\cdot \widehat{l}) \widehat{l}_{{}_{ij}}(\boldsymbol{w})
\widehat{n}(t) \,
\overline{{u^{b}_{s}}_{{}_{j' i'}}(\widehat{n}\cdot \widehat{l})} \, \overline{\widehat{l}_{{}_{i' j'}}(\boldsymbol{w}')} \overline{\widehat{n}(t')}
\\
\mp 
\sum\limits_{\substack{\widehat{n}\cdot \widehat{l} \in \mathscr{O}_+ \\ 1\leq s \leq d'(\widehat{n}\cdot \widehat{l}) \\ -l \leq i,j \leq l 
\\ -l' \leq i',j' \leq l'}} 
i (2l+1) \overline{{u^{a}_{s}}_{{}_{ji}}(\widehat{n}\cdot \widehat{l})} \, \overline{\widehat{l}_{{}_{ij}}(\boldsymbol{w})}
 \overline{\widehat{n}(t)}
{u^{b}_{s}}_{{}_{j' i'}}(\widehat{n}\cdot \widehat{l}) \widehat{l}_{{}_{i' j'}}(\boldsymbol{w}') \widehat{n}(t'),
\end{multline}
\begin{multline}\label{S(x,y)}
S^{ab}(t\times \boldsymbol{w}, t' \times \boldsymbol{w}') = \\ =
\sum\limits_{\substack{1 \leq c \leq d \\ \widehat{n}\cdot \widehat{l} \in \mathscr{O}_+ \\ 1\leq s \leq d'(\widehat{n}\cdot \widehat{l}) \\ -l \leq i,j \leq l 
\\ -l' \leq i',j' \leq l'}} 
i (2l+1) {u^{a}_{s}}_{{}_{ji}}(\widehat{n}\cdot \widehat{l}) \widehat{l}_{{}_{ij}}(\boldsymbol{w})
\widehat{n}(t) \, \widehat{\Gamma}_{0}^{bc} \,
\overline{{u^{c}_{s}}_{{}_{j' i'}}(\widehat{n}\cdot \widehat{l})} \, \overline{\widehat{l}_{{}_{i' j'}}(\boldsymbol{w}')} \overline{\widehat{n}(t')}
\\
\mp 
\sum\limits_{\substack{1 \leq c \leq d \\ \widehat{n}\cdot \widehat{l} \in \mathscr{O}_- \\ 1\leq s \leq d''(\widehat{n}\cdot \widehat{l}) \\ -l \leq i,j \leq l 
\\ -l' \leq i',j' \leq l'}} 
i (2l+1) \overline{{v^{a}_{s}}_{{}_{ji}}(\widehat{n}\cdot \widehat{l})} \,  \overline{\widehat{l}_{{}_{ij}}(\boldsymbol{w})}
 \overline{\widehat{n}(t)} \, \widehat{\Gamma}_{0}^{bc} \,
{v^{b}_{s}}_{{}_{j' i'}}(\widehat{n}\cdot \widehat{l}) \widehat{l}_{{}_{i' j'}}(\boldsymbol{w}') \widehat{n}(t'),
\end{multline}
so that
\begin{multline}\label{D(x)}
D^{ab}(t\times \boldsymbol{w}) = 
\sum\limits_{\substack{\widehat{n}\cdot \widehat{l} \in \mathscr{O}_+ \\ 1\leq s \leq d'(\widehat{n}\cdot \widehat{l}) \\ -l \leq i,j,i' \leq l}} 
i (2l+1) {u^{a}_{s}}_{{}_{ji}}(\widehat{n}\cdot \widehat{l}) \widehat{l}_{{}_{ij}}(\boldsymbol{w})
\widehat{n}(t) \,
\overline{{u^{b}_{s}}_{{}_{i' i'}}(\widehat{n}\cdot \widehat{l})} 
\\
\mp 
\sum\limits_{\substack{\widehat{n}\cdot \widehat{l} \in \mathscr{O}_+ \\ 1\leq s \leq d'(\widehat{n}\cdot \widehat{l}) \\ -l \leq i,j,i' \leq l}} 
i (2l+1) \overline{{u^{a}_{s}}_{{}_{ji}}(\widehat{n}\cdot \widehat{l})} \, \overline{\widehat{l}_{{}_{ij}}(\boldsymbol{w})}
 \overline{\widehat{n}(t)}
{u^{b}_{s}}_{{}_{i' i'}}(\widehat{n}\cdot \widehat{l}),
\end{multline}
\begin{multline}\label{S(x)}
S^{ab}(t\times \boldsymbol{w}) = 
\sum\limits_{\substack{1 \leq c \leq d \\ \widehat{n}\cdot \widehat{l} \in \mathscr{O}_+ \\ 1\leq s \leq d'(\widehat{n}\cdot \widehat{l}) \\ -l \leq i,j,i' \leq l}} 
i (2l+1) {u^{a}_{s}}_{{}_{ji}}(\widehat{n}\cdot \widehat{l}) \widehat{l}_{{}_{ij}}(\boldsymbol{w})
\widehat{n}(t) \, \widehat{\Gamma}_{0}^{bc} \,
\overline{{u^{c}_{s}}_{{}_{i' i'}}(\widehat{n}\cdot \widehat{l})} 
\\
\mp 
\sum\limits_{\substack{1 \leq c \leq d \\ \widehat{n}\cdot \widehat{l} \in \mathscr{O}_- \\ 1\leq s \leq d''(\widehat{n}\cdot \widehat{l}) \\ -l \leq i,j,i' \leq l}} 
i (2l+1) \overline{{v^{a}_{s}}_{{}_{ji}}(\widehat{n}\cdot \widehat{l})} \,  \overline{\widehat{l}_{{}_{ij}}(\boldsymbol{w})}
 \overline{\widehat{n}(t)} \, \widehat{\Gamma}_{0}^{bc} \,
{v^{b}_{s}}_{{}_{i' i'}}(\widehat{n}\cdot \widehat{l}),
\end{multline}
\begin{multline}\label{D^(-)(x,y)}
D^{(-) \,ab}(t\times \boldsymbol{w}, t' \times \boldsymbol{w}') = \\ =
\sum\limits_{\substack{\widehat{n}\cdot \widehat{l} \in \mathscr{O}_+ \\ 1\leq s \leq d'(\widehat{n}\cdot \widehat{l}) \\ -l \leq i,j \leq l 
\\ -l' \leq i',j' \leq l'}} 
i (2l+1) {u^{a}_{s}}_{{}_{ji}}(\widehat{n}\cdot \widehat{l}) \widehat{l}_{{}_{ij}}(\boldsymbol{w})
\widehat{n}(t) \,
\overline{{u^{b}_{s}}_{{}_{j' i'}}(\widehat{n}\cdot \widehat{l})} \, \overline{\widehat{l}_{{}_{i' j'}}(\boldsymbol{w}')} \overline{\widehat{n}(t')},
\end{multline}
\begin{multline}\label{S^(-)(x,y)}
S^{(-) \, ab}(t\times \boldsymbol{w}, t' \times \boldsymbol{w}') = \\ =
\sum\limits_{\substack{1 \leq c \leq d \\ \widehat{n}\cdot \widehat{l} \in \mathscr{O}_+ \\ 1\leq s \leq d'(\widehat{n}\cdot \widehat{l}) \\ -l \leq i,j \leq l 
\\ -l' \leq i',j' \leq l'}} 
i (2l+1) {u^{a}_{s}}_{{}_{ji}}(\widehat{n}\cdot \widehat{l}) \widehat{l}_{{}_{ij}}(\boldsymbol{w})
\widehat{n}(t) \, \widehat{\Gamma}_{0}^{bc} \,
\overline{{u^{c}_{s}}_{{}_{j' i'}}(\widehat{n}\cdot \widehat{l})} \, \overline{\widehat{l}_{{}_{i' j'}}(\boldsymbol{w}')} \overline{\widehat{n}(t')},
\end{multline}
\begin{equation}\label{D^(-)(x)}
D^{(-) \,ab}(t\times \boldsymbol{w}) = 
\sum\limits_{\substack{\widehat{n}\cdot \widehat{l} \in \mathscr{O}_+ \\ 1\leq s \leq d'(\widehat{n}\cdot \widehat{l}) \\ -l \leq i,j,i' \leq l}} 
i (2l+1) {u^{a}_{s}}_{{}_{ji}}(\widehat{n}\cdot \widehat{l}) \widehat{l}_{{}_{ij}}(\boldsymbol{w})
\widehat{n}(t) \,
\overline{{u^{b}_{s}}_{{}_{i' i'}}(\widehat{n}\cdot \widehat{l})},
\end{equation}
\begin{equation}\label{S^(-)(x)}
S^{(-) \, ab}(t\times \boldsymbol{w}) = 
\sum\limits_{\substack{1 \leq c \leq d \\ \widehat{n}\cdot \widehat{l} \in \mathscr{O}_+ \\ 1\leq s \leq d'(\widehat{n}\cdot \widehat{l}) \\ -l \leq i,j,i' \leq l}} 
i (2l+1) {u^{a}_{s}}_{{}_{ji}}(\widehat{n}\cdot \widehat{l}) \widehat{l}_{{}_{ij}}(\boldsymbol{w})
\widehat{n}(t) \, \widehat{\Gamma}_{0}^{bc} \,
\overline{{u^{c}_{s}}_{{}_{i' i'}}(\widehat{n}\cdot \widehat{l})},
\end{equation}

As we have already explained, the fundamental solutions have common period in time, and so do the commutation and pairing functions
whose time dependence is expressed through the periodic characters $\widehat{n}(t)$, $n \in \mathbb{Z}$, with common period in time. 
Thus the commutation and pairing functions effectively live on the compactified Einstein Univerese $\widetilde{\mathbb{S}^1} \times SU(2, \mathbb{C})$, 
and the pairing functions
define distributions on the spacetime test periodic function space
\[
\mathscr{E} = \mathcal{S}_{{}_{\Delta +1}}(\widetilde{\mathbb{S}^1} \times SU(2, \mathbb{C}); \mathbb{C}^d)
\]
where $d = \textrm{dim} \, V$ is equal to the dimension of the representation $V$ of $SU(2, \mathbb{C})$ entering the transformation formula 
(\ref{UphiOnSxG}) associated to the field in question. Note that we have used abbreviated notation $\Delta + 1$ for the standard operator
$\Delta \boldsymbol{1}_{{}_{d}} + \boldsymbol{1}_{{}_{d}}$ acting on the Hilbert space 
$L^2(\widetilde{\mathbb{S}^1} \times SU(2, \mathbb{C}); \mathbb{C}^d)$. 

Each finite subsummand in (\ref{D(x,y)})-(\ref{S^(-)(x)}) represents a well defined and even smooth function belonging respectively
to $\mathscr{E}$ or $\mathscr{E}^{\otimes 2}$, but of course in general, when the orbits $\mathscr{O}_{\pm}$
are infinite, the whole series in (\ref{D(x,y)})-(\ref{D^(-)(x)}) are not convergent in $\mathscr{E}$ or $\mathscr{E}^{\otimes 2}$
and even do not represent any ordinary functions. Nonetheless, the series (\ref{D(x,y)})-(\ref{D^(-)(x)}) are convergent in 
distribution spaces $\mathscr{E}^*$ or $\mathscr{E}^{* \otimes 2} = \big(\mathscr{E}^{\otimes 2}\big)^*$, respectively. 
Namely, each finite subsummand in (\ref{D(x,y)})-(\ref{D^(-)(x)}), regarded as an element of $\mathscr{E}^*$ or of $\mathscr{E}^{*\otimes 2}$, respectively,
evaluated at any $\phi \in \mathscr{E}$ or respectively at $\phi \otimes \varphi \in \mathscr{E}^{\otimes 2}$, with evaluation given by integration with
test function $\phi$ or $\phi \otimes \varphi$, gives a convergent numerical series. Because $\mathscr{E}$ and $\mathscr{E}^{\otimes 2}$
are complete Fr\'echet spaces then, by the Banach-Steinhaus theorem, the series (\ref{D(x,y)})-(\ref{D^(-)(x)}) represent continuous
functionals on $\mathscr{E}$ or, respectively, on $\mathscr{E}^{\otimes 2}$. Therefore, we have the following
\begin{lem}
The series (\ref{D(x,y)})-(\ref{S^(-)(x)}) converge, respectively, in $\mathscr{E}^*$ or in $\mathscr{E}^{* \otimes 2}$,
and represent continuous functionals, respectively, on $\mathscr{E}$ or $\mathscr{E}^{\otimes 2}$.
\label{Pairings}
\end{lem}
\qedsymbol \,
Let, for example, $D$ be the two-variable commutation function (\ref{D(x,y)}) of the essentially neutral free field $\mathbb{A}$,
associated to the representation (\ref{UphiOnSxG}) with a $d$-dimensional representation $V$ of $SU(2, \mathbb{C})$ in (\ref{UphiOnSxG}), and let
$\phi \otimes \varphi \in \mathscr{E}^{\otimes 2}$. 
It is easily seen that the canonical pairing $\langle D, \phi \otimes \varphi \rangle$ is equal
\begin{multline}\label{<D,phixvarphi>}
\langle D, \phi \otimes \varphi \rangle = \\ = 
\sum\limits_{a,b =1}^d \,\, \int\limits_{\big[\widetilde{\mathbb{S}^1} \times SU(2, \mathbb{C})\big]^{\times 2}}
D^{ab}(t\times \boldsymbol{w}, t' \times \boldsymbol{w}') \, \phi^a(t\times \boldsymbol{w}) \,
\varphi^b(t'\times \boldsymbol{w}') \,
dt d\boldsymbol{w}dt' d\boldsymbol{w}' \\ =
\sum\limits_{\substack{\widehat{n}\cdot \widehat{l} \in \mathscr{O}_+ \\ 1\leq s \leq d'(\widehat{n}\cdot \widehat{l})}} 
{\textstyle\frac{i}{2l+1}}
\overline{\big\langle u_s(\widehat{n}\cdot \widehat{l}), \widetilde{\overline{\phi}}(\widehat{n}\cdot \widehat{l}) \big\rangle}
\big\langle u_s(\widehat{n}\cdot \widehat{l}), \widetilde{\varphi}(\widehat{n}\cdot \widehat{l}) \big\rangle \\
\mp
\sum\limits_{\substack{\widehat{n}\cdot \widehat{l} \in \mathscr{O}_+ \\ 1\leq s \leq d'(\widehat{n}\cdot \widehat{l})}} 
{\textstyle\frac{i}{2l+1}}
\big\langle u_s(\widehat{n}\cdot \widehat{l}), \widetilde{\phi}(\widehat{n}\cdot \widehat{l}) \big\rangle
\overline{\big\langle u_s(\widehat{n}\cdot \widehat{l}), \widetilde{\overline{\varphi}}(\widehat{n}\cdot \widehat{l}) \big\rangle} 
\end{multline}
where the symbol $\langle \, \cdot \, ,  \, \cdot \, \rangle$ on the right-hand side denotes the inner product 
(\ref{<tildephi(n.l),tildephi'(n.l)>}) determined by the Plancherel formula. Now, repeating essentially the argument already applied in the 
proof of the previous Propositions, we represent each of the two summands in the expression (\ref{<D,phixvarphi>}) for the paring as the 
inner product\footnote{This specific formula holds for the first summad in the third line of (\ref{<D,phixvarphi>}). For the second summand --
that in the last line of (\ref{<D,phixvarphi>}) -- we have analogous expression 
$\overline{\langle P_{{}_{\xi}} \widetilde{A^r \phi} , \widetilde{A^r \overline{\varphi'}}\rangle}$ with the complex conjugation moved from $\phi$
and put on $\varphi'$.} 
$\langle P_{{}_{\xi}} \widetilde{A^r \overline{\phi}} , \widetilde{A^r \varphi'}\rangle$ of an auxiliary function $\widetilde{A^r \varphi'}$
belonging to the Fourier transform image $\widetilde{\mathscr{H}}$ of $\mathscr{H} = L^2(\widetilde{\mathbb{S}^1} \times SU(2, \mathbb{C}); \mathbb{C}^d)$
with the orthogonal projection $P_{{}_{\xi}} \widetilde{A^r \overline{\phi}} \in \widetilde{\mathscr{H}}$ of the function 
$\widetilde{A^r \phi} \in \widetilde{\mathscr{H}}$ on another auxiliary matrix-valued function $\xi \in \widetilde{\mathscr{H}}$.

Consider, for example, the first main summand in (\ref{<D,phixvarphi>}) -- \emph{i.e.} that in the third line of the formula
(\ref{<D,phixvarphi>}).

Next, consider the auxiliary matrix valued functions $\zeta_{{}_{s,\widehat{n}\cdot \widehat{l}}}$, 
numbered by the parameters $(s,\widehat{n}\cdot \widehat{l}) \in \{1, \ldots, d'(\widehat{n}\cdot \widehat{l})\} \times \mathscr{O}_+$, 
and defined on the dual $\widehat{\widetilde{\mathbb{S}^1}\times G}$, $G= SU(2, \mathbb{C})$, by the following formula
\[
\big[\zeta_{{}_{s,\widehat{n}\cdot \widehat{l}}}\big]^{a}_{{}_{ji}}(\widehat{n'}\cdot \widehat{l'}) = 
\delta_{nn'} \delta_{ll'} \,\, {u_{s}^{a}}_{{}_{ji}}(\widehat{n}\cdot \widehat{l}). 
\]
In particular each $\zeta_{{}_{s,\widehat{n}\cdot \widehat{l}}}$ is concentrated on single point set $\{\widehat{n}\cdot \widehat{l}\}
\subset \mathscr{O}_+$ and differs from zero only for $s \in \{1, \ldots, d'(\widehat{n}\cdot \widehat{l})\}$. 

It is clear that for the standard operator $A = \Delta +1$ defining the nuclear space $\mathscr{E}$ we have
\[
\widetilde{A} \zeta_{{}_{s,\widehat{n}\cdot \widehat{l}}} = \lambda_{nl} \zeta_{{}_{s,\widehat{n}\cdot \widehat{l}}}
\]
and that the system $\{\zeta_{{}_{s,\widehat{n}\cdot \widehat{l}}}\}$ is by construction orthonormal in $\widetilde{\mathscr{H}}$
by the orthonomal property of the system $\{u_{s}(\widehat{n}\cdot \widehat{l})\}$ of Fourier transforms 
$u_{s}$ of fundamental solutions, concentrated on single point $\widehat{n}\cdot \widehat{l}$. 
It is likewise clear that for sufficiently large positive integer $r$ the series
\[
\sum\limits_{\substack{\widehat{n}\cdot \widehat{l} \in \mathscr{O}_+ \\ 1\leq s \leq d'(\widehat{n}\cdot \widehat{l})}} 
\lambda_{nl}^{-r} \zeta_{{}_{s,\widehat{n}\cdot \widehat{l}}} = 
\sum\limits_{\substack{\widehat{n}\cdot \widehat{l} \in \mathscr{O}_+ \\ 1\leq s \leq d'(\widehat{n}\cdot \widehat{l})}} 
\widetilde{A}^{-r} \zeta_{{}_{s,\widehat{n}\cdot \widehat{l}}} = \xi
\]
converges in  $\widetilde{\mathscr{H}}$ to an element $\xi \in \widetilde{\mathscr{H}}$. 
Note finally that if $\varphi \in \mathscr{E}$, then all the more the inverse Fourier transform $\varphi'$
of the function
\[
\widehat{n}\cdot \widehat{l} \longrightarrow  {\textstyle\frac{i}{2l+1}}\widetilde{\varphi}(\widehat{n}\cdot \widehat{l})
\]
belongs to $\mathscr{E}$, and because
\[
\big\langle {\textstyle\frac{i}{2l+1}}\widetilde{\varphi}(\widehat{n}\cdot \widehat{l}),
 {\textstyle\frac{i}{2l+1}}\widetilde{\varphi}(\widehat{n}\cdot \widehat{l})
\big\rangle <
\big\langle \widetilde{\varphi}(\widehat{n}\cdot \widehat{l}),
 \widetilde{\varphi}(\widehat{n}\cdot \widehat{l})
\big\rangle
\]
then $|A^r \varphi'| < |A^r \varphi|$, for all positive $r$. 

Now we can see that the first summand in (\ref{<D,phixvarphi>}) -- \emph{i.e.} that in the third line of (\ref{<D,phixvarphi>}) --
can be written
\begin{multline*}
\sum\limits_{\substack{\widehat{n}\cdot \widehat{l} \in \mathscr{O}_+ \\ 1\leq s \leq d'(\widehat{n}\cdot \widehat{l})}} 
{\textstyle\frac{i}{2l+1}}
\overline{\big\langle u_s(\widehat{n}\cdot \widehat{l}), \widetilde{\overline{\phi}}(\widehat{n}\cdot \widehat{l}) \big\rangle}
\big\langle u_s(\widehat{n}\cdot \widehat{l}), \widetilde{\varphi}(\widehat{n}\cdot \widehat{l}) \big\rangle
\\
\sum\limits_{\substack{\widehat{n}\cdot \widehat{l} \in \mathscr{O}_+ \\ 1\leq s \leq d'(\widehat{n}\cdot \widehat{l})}} 
\overline{\big\langle \lambda_{nl} ^{-r} u_s(\widehat{n}\cdot \widehat{l}), \lambda_{nl}^{r} \widetilde{\overline{\phi}}(\widehat{n}\cdot \widehat{l}) \big\rangle}
\big\langle \lambda_{nl} ^{-r} u_s(\widehat{n}\cdot \widehat{l}), \lambda_{nl} ^{r} \widetilde{\varphi'}(\widehat{n}\cdot \widehat{l}) \big\rangle
= \langle P_{{}_{\xi}} \widetilde{A^r \overline{\phi}} , \widetilde{A^r \varphi'}\rangle.
\end{multline*}
Thus, again by Schwartz inequality, we have the following estimation
\[
\big|\langle P_{{}_{\xi}} \widetilde{A^r \overline{\phi}} , \widetilde{A^r \varphi'}\rangle  \big|
\leq |\xi|^2 \, |A^r \phi| \, |A^r \varphi' | < |\xi|^2 \, |A^r \phi| \, |A^r \varphi |
= |\xi|^2 \, |\phi|_{{}_{r}} \, |\varphi |_{{}_{r}}.
\]
Similar estimation we have for the second summand in (\ref{<D,phixvarphi>}) -- \emph{i.e.} that in the last line of (\ref{<D,phixvarphi>})
-- which is equal $\mp \overline{\langle P_{{}_{\xi}} \widetilde{A^r \phi} , \widetilde{A^r \overline{\varphi'}}\rangle}$.
Therefore convergence of the series (\ref{D(x,y)}), when evaluated at any fixed element $\phi \otimes \varphi \in \mathscr{E}^{\otimes \, 2}$
follows, and the continuity of the functional on $\mathscr{E}^{\otimes 2}$, defined by kernel $D^{ab}(t\times \boldsymbol{w}, t' \times \boldsymbol{w}')$. 

Let now, for example, $D$ be the single-variable commutation function (\ref{D(x)}) of the essentially neutral free field $\mathbb{A}$,
associated to the representation (\ref{UphiOnSxG}) with a $d$-dimensional representation $V$ of $SU(2, \mathbb{C})$ in (\ref{UphiOnSxG}), and let
$\phi \in \mathscr{E} = \mathcal{S}_{{}_{\Delta +1}}(\widetilde{\mathbb{S}^1} \times SU(2, \mathbb{C}); \mathbb{C}^d)$. 
It is easily seen that the canonical pairing $\langle D, \phi \rangle$ is equal (the index $b \in \{1, \ldots, d\}$ is fixed)
\begin{multline}\label{<D,phi>}
\langle D, \phi \rangle  = 
\sum\limits_{a =1}^d \,\, \int\limits_{\widetilde{\mathbb{S}^1} \times SU(2, \mathbb{C})}
D^{ab}(t\times \boldsymbol{w}) \, \phi^a(t\times \boldsymbol{w}) \,
dt d\boldsymbol{w} \\ =
\sum\limits_{\substack{\widehat{n}\cdot \widehat{l} \in \mathscr{O}_+ \\ 1\leq s \leq d'(\widehat{n}\cdot \widehat{l}) \\ -l \leq j \leq l}} 
i \overline{\big\langle u_s(\widehat{n}\cdot \widehat{l}), \widetilde{\overline{\phi}}(\widehat{n}\cdot \widehat{l}) \big\rangle
\, {u_{s}^{b}}_{{}_{jj}}(\widehat{n}\cdot \widehat{l})} \\
\mp
\sum\limits_{\substack{\widehat{n}\cdot \widehat{l} \in \mathscr{O}_+ \\ 1\leq s \leq d'(\widehat{n}\cdot \widehat{l}) \\ -l \leq j \leq l}} 
i \big\langle u_s(\widehat{n}\cdot \widehat{l}), \widetilde{\phi}(\widehat{n}\cdot \widehat{l}) \big\rangle \,
 {u_{s}^{b}}_{{}_{jj}}(\widehat{n}\cdot \widehat{l})
\end{multline}
where again the symbol $\langle \, \cdot \, ,  \, \cdot \, \rangle$ on the right-hand side denotes the inner product 
(\ref{<tildephi(n.l),tildephi'(n.l)>}) determined by the Plancherel formula. 

We will proceed similarly and, in addition to the above orthonormal system $\zeta_{{}_{\widehat{n}\cdot \widehat{l}}}$, we define here another
system of auxiliary matrix valued functions $\chi_{{}_{s,\widehat{n}\cdot \widehat{l}}}$, 
numbered by the parameters $(s,\widehat{n}\cdot \widehat{l}) \in \{1, \ldots, d'(\widehat{n}\cdot \widehat{l})\} \times \mathscr{O}_+$, 
and defined on the dual $\widehat{\widetilde{\mathbb{S}^1}\times G}$, $G= SU(2, \mathbb{C})$, by the following formula
\[
\big[\chi_{{}_{s,\widehat{n}\cdot \widehat{l}}}\big]^{a}_{{}_{ji}}(\widehat{n'}\cdot \widehat{l'}) = 
\delta^{ab} \delta_{nn'} \delta_{ll'} \,\delta_{ji}. 
\] 
Again it is obvious that for some positive integer $r$ the series
\[
\sum\limits_{\substack{\widehat{n}\cdot \widehat{l} \in \mathscr{O}_+ \\ 1\leq s \leq d'(\widehat{n}\cdot \widehat{l})}} 
\lambda_{nl}^{-r} \chi_{{}_{s,\widehat{n}\cdot \widehat{l}}} = 
\sum\limits_{\substack{\widehat{n}\cdot \widehat{l} \in \mathscr{O}_+ \\ 1\leq s \leq d'(\widehat{n}\cdot \widehat{l})}} 
\widetilde{A}^{-r} \chi_{{}_{s,\widehat{n}\cdot \widehat{l}}} = \eta
\]
converges in  $\widetilde{\mathscr{H}}$ to an element $\eta \in \widetilde{\mathscr{H}}$. 

Now it is easily seen that the pairing (\ref{<D,phi>}) can be written as
\begin{multline*}
\langle D, \phi \rangle = 
\sum\limits_{\substack{\widehat{n}\cdot \widehat{l} \in \mathscr{O}_+ \\ 1\leq s \leq d'(\widehat{n}\cdot \widehat{l}) \\ -l \leq i \leq l}} 
i \overline{\big\langle \lambda_{nl}^{-r} u_s(\widehat{n}\cdot \widehat{l}), \lambda_{nl}^{2r} \widetilde{\overline{\phi}}(\widehat{n}\cdot \widehat{l}) \big\rangle
\, \lambda_{nl}^{-r} {u_{s}^{b}}_{{}_{ii}}(\widehat{n}\cdot \widehat{l})} \\
\mp
\sum\limits_{\substack{\widehat{n}\cdot \widehat{l} \in \mathscr{O}_+ \\ 1\leq s \leq d'(\widehat{n}\cdot \widehat{l}) \\ -l \leq i \leq l}} 
i \big\langle \lambda_{nl}^{-r} u_s(\widehat{n}\cdot \widehat{l}), \lambda_{nl}^{2r} \widetilde{\phi}(\widehat{n}\cdot \widehat{l}) \,
\lambda_{nl}^{-r} {u_{s}^{b}}_{{}_{ii}}(\widehat{n}\cdot \widehat{l}) \\ =
i \overline{\big\langle \langle \xi, \widetilde{A^{2r} \overline{\phi}} \rangle \xi, \, \chi \big\rangle}
\mp 
i \big\langle \langle \xi, \widetilde{A^{2r} \phi} \rangle \xi, \, \chi \big\rangle.
\end{multline*}
Therefore, again by Schwartz inequality, 
\begin{multline*}
|\langle D, \phi \rangle| \leq \Big| \big\langle \langle \xi, \widetilde{A^{2r} \overline{\phi}} \rangle \xi, \, \chi \big\rangle \Big|
\\ \leq 
2 \, \big| \big\langle \langle \xi, \widetilde{A^{2r} \phi} \rangle \xi, \, \chi \big\rangle \big|  =
2 \, |\langle \xi, \widetilde{A^{2r} \phi} \rangle \langle \xi, \chi \rangle | 
\\
2 \, |\xi|^2 \, |\chi| \, |\widetilde{A^{2r} \phi} | = 2 \, |\xi|^2 \, |\chi| \, |\phi |_{{}_{2r}},
\end{multline*}
and the convergence of the series (\ref{D(x)}), when evaluated at any fixed element $\phi$ of $\mathscr{E}$, follows, 
as well as the continuity of the functional, defined on 
$\mathscr{E}$ by the kernel $D^{ab}(t\times \boldsymbol{w})$ with any fixed index $b$.

Convergence of the series (\ref{D(x,y)})-(\ref{S^(-)(x)}), when evaluated at any fixed elements of $\mathscr{E}, \mathscr{E}^{\otimes \, 2}$, for all the remaining
kernels (\ref{D(x,y)})-(\ref{S^(-)(x)}), is shown in the same manner.
\qed

It follows by construction, that the commuation functions (\ref{D(x,y)})-(\ref{S^(-)(x)}) respect the corresponding differential equations,
analogously as do the commutation functions on the Minkowski space-time. Nonetheless, the supports of the single-variable 
commutation functions (\ref{D(x,y)})-(\ref{S^(-)(x)}) are not confined to the closure of the interior
of the foreward and backward light cones. It is easily seen for massive fields, because in this case the sums in 
(\ref{D(x,y)})-(\ref{S^(-)(x)}) become finite, and the commutation functions are in this case not only ordinary smooth functios,
but even real analytic (recall that the Lie group $\widetilde{\mathbb{S}^1} \times SU(2, \mathbb{C})$ posses natural real analytic structure, as do each 
Lie group). For such functions it is impossible to vanish identically on an open set, except in the trivial case when they are identically zero. 

We shoud emphasize that this is in agreement with causality for the scattering operator $S$, contrary to what one could expect by a naive comparison to the Minkowski space-time, where the commutation and pairing functions are supported on the closures of the interiors of future and past light cones. But this support property  holds for free fields on the Minkowski space-time only accidentally and is compatible with causality due to the particular geometry of the Minkowski space-time which admits the hyperbolic Loretnz rotations as the symmetry transformations. 
In more general case, e.g. for the Einstein Universe, causality for the scattering operator does not require this particular support property,
essentially because the hyperbolic Lorentz rotations are not included as symmetry transformations, compare Subsection \ref{CausalSonEU}. As we will explain in 
Subsection \ref{CausalSonEU} the causality condition 
\[
(\textrm{causality}) \,\,\,\,\,\,\,\,\,\,\,\,
S\big((g_1 + g_2)\mathcal{L}\big) = S(g_2\mathcal{L}) S(g_1\mathcal{L})
\,\,\,\,\,\,
\textrm{for}
\,\, G_1 \prec G_2.
\]
is sufficient for the construction of the scattering operator $S$, where $\mathcal{L}$ is the interaction Lagrange density,  
$g_{i}$ are the auxiliary ``intensity-of-interaction'' space-time functions supported on such regions $G_i$ 
that each point of $G_1$ precede all points of $G_2$ (in short $ G_1 \prec G_2$). Because we do not have at our disposal
the hyperbolic Lorentz rotations, then this causality condition cannot be extended over regions $G_i$ which are
mutually space-like separated, compare Subsection \ref{CausalSonEU} or \cite{Bogoliubov_Shirkov} or \cite{Bogoliubov-Shirkov}.

Nonetheless, the above form of causality is strong enough for the construction of the scattering operator, as we will see in
Subsection \ref{CausalSonEU}. 
In order to make practical use of this causality condiction we observe that the single-variable commutation and pairing functions
 (\ref{D(x,y)})-(\ref{S^(-)(x)}) of the free fields of the theory can be naturally split into the retarded and advanced 
parts:
\begin{align*}
\mathbb{S}= \mathbb{S}_{\textrm{ret}} - \mathbb{S}_{\textrm{av}}, \\
\textrm{for} \\
\mathbb{S}(x)= D^{ab}(x), S^{ab}(x), D^{(\pm) \, ab}(x), S^{(\pm) \, ab}(x),  
\end{align*}
where the support of the retarded parts is included ($\textrm{mod} \, 4\pi$) in the closed future half-space $V_+$ -- the future of the 
Cachy surface $t=0$, and the advanced parts are supported on the past half-space $V_-$ --
the set of points which lie on or which precede the Cauchy surface $t=0$. Because the pairing functions are in fact periodic, 
then this condition should be expressed in terms of
the periodic step function 
\[
\theta(t+m4\pi)=
\left\{ \begin{array}{ll}
1, &  0 \leq t < 2\pi, \\
0, & -2\pi \leq t <0
\end{array} \right.,
m \in \mathbb{Z}
\]
on the reals and the corresponding step function (denoted likewise $\theta$)
\[
\theta(x) = \theta(t),
\,\,\,\, 
\textrm{for}
\,\,\,
x= t \times \boldsymbol{w} \in \mathbb{R} \times SU(2, \mathbb{C})
\]
on the Einstein Universe $\mathbb{R} \times SU(2, \mathbb{C})$. Then introducing $\theta^{-1}(x) = \theta(x^{-1})$
we have 
\[
V_+ \overset{\textrm{df}}{=} \textrm{supp} \, \theta,
\,\,\,\,\,
V_- \overset{\textrm{df}}{=} \textrm{supp} \, \theta^{-1}.
\]

In general a (periodic) distribution on the Einstein Universe has infinitely many divisions into retarded and advanced parts with the indicated support properties. 
Indeed, having a particular splitting into retarded and advanced parts, we can add to the advanced and to the retarded part any distribution concentrated on the Cauchy surfaces $t=0$ and $t=2\pi$. We have however a natural splitting which is determined by the periodic theta function $\theta$
within the physically relevant class of free fields and some of their Wick products. Indeed
the operation of multiplication of the generalized integral kernel operators $\boldsymbol{\psi}, \mathbb{A}
\in \mathscr{L}(\mathscr{E}^{*}, \,  \mathscr{L}((\boldsymbol{E}), (\boldsymbol{E}))$ by theta function $\theta$ is well defined
provided $\boldsymbol{\psi}, \mathbb{A}$ are free fields or equal to the Wick products of free fields which contain at most one massless field (or
one field with infinite orbit $\mathscr{O}_\pm$). 
This is mainly because these are well defined generalized operators $\mathscr{L}((\boldsymbol{E}), (\boldsymbol{E}))$, when evaluated at space-time point $x$ 
and there is no need in treating them as distributions -- in this lies the usefulness of 
the white noise analysis. From the white nose point of view it cannot \emph{a priori} be excluded that the pointwise multiplication
by theta function
acts within the some subclass of the operators $\mathscr{L}(\mathscr{E}, \mathscr{L}((\boldsymbol{E}), (\boldsymbol{E}))$ (although not obvious). 
The said multiplication operation has the interpretation of the multiplication of a well defined operator
$(\boldsymbol{E}) \rightarrow (\boldsymbol{E})^*$ by an ordinary number $\theta(x)$, as $x$ is fixed here.
Moreover, as we will see in this Subsection, on the Einstein Universe Wick polynomials in free fields evaluated at $x$, also belong to 
$\mathscr{L}((\boldsymbol{E}), (\boldsymbol{E})^*)$ and are well defined generalized operators which represent operators in 
\[
\mathscr{L}(\mathscr{E}, \mathscr{L}((\boldsymbol{E}), (\boldsymbol{E}))),
\]
and moreover, when multiplied pointwisely by the theta function,
they represent well defined generalized operators in 
\[
\mathscr{L}(\mathscr{E}, \mathscr{L}((\boldsymbol{E}), (\boldsymbol{E})))
\]
provided each of the Wick monomials contains at most one massless (or infinite orbit) field.
This is contrary to what we have on the 
Minkowski space-time\footnote{Recall that on the Minkowski space-time the operation of pointwise multiplication by the step theta function
$\theta$ applied to the kernels of an operator in $\mathscr{L}\big( \mathscr{E}, \mathscr{L}((\boldsymbol{E}), (\boldsymbol{E})) \big)$,
gives kernels of a more singular operator, which belongs only to $\mathscr{L}\big( \mathscr{E}, \mathscr{L}((\boldsymbol{E}), (\boldsymbol{E})^*) \big)$.
Recall also that on the Minkowski space-time the Wick product also behaves more singularly if among the factor fields there are massless fields,
and in this case the  Wick product belongs only to $\mathscr{L}\big( \mathscr{E}, \mathscr{L}((\boldsymbol{E}), (\boldsymbol{E})^*) \big)$,
but not to $\mathscr{L}\big( \mathscr{E}, \mathscr{L}((\boldsymbol{E}), (\boldsymbol{E})) \big)$. 
Compare Subsection \ref{WickForChronological} and Subsection \ref{OperationsOnXi}.}, where the operation of pointwise multiplication by the step theta function behaves
more singularly (compare Subsection \ref{WickForChronological}). On the Einstein Universe the Wick product of free fields is an operator
which belongs to $\mathscr{L}(\mathscr{E}, \mathscr{L}((\boldsymbol{E}), (\boldsymbol{E})))$.
Therefore, on the Einstein Universe the operation of multiplication
of the Wick polynomial in free fields, evaluated at space-time point $x$, by the number $\theta(x)$, is a well defined operation 
and represents operation acting within the space $\mathscr{L}(\mathscr{E}, \mathscr{L}((\boldsymbol{E}), (\boldsymbol{E})))$, \emph{i.e.} whenever applied to the Wick product operator in $\mathscr{L}(\mathscr{E}, \mathscr{L}((\boldsymbol{E}), (\boldsymbol{E})))$ which contains at most one mess less (or infinite orbit)
field, as for example the Lagrange interaction density $\mathcal{L}$ in QED.  
Therefore we have at our disposal the natural chronological product in QED on the Einstein Universe, or any other QFT on the Einstein Universe,
which can be naturally constructed with the help of the step theta function. 
Chronological product for the other interactions, containing more than two massless free (or infinite orbit) fields is also possible on the 
Einstein Universe but, as we will see, it is more elaborate with more complicated Epstein-Glaser-type distribution splitting.

All the more we also  have on the Einstein Universe the natural construction of
the retarded part $\mathbb{S}_{\textrm{ret}}$ for any single-variable commutation or pairing function $\mathbb{S}(x)$ represented
by any one of the series (\ref{D(x,y)})-(\ref{S^(-)(x)}). We take any finite subsummand in (\ref{D(x,y)})-(\ref{S^(-)(x)})
representing the single variable commutation or pairing function $\mathbb{S}(x)$, and multiply it by the number $\theta(x)$. Finally we pass to the limit
in the series obtained in this manner, with respect to the topology on $\mathscr{E}^*$. By replacing $\theta(x)$ with 
$\big[- \theta(x^{-1}) \big]$ in this construction we obtain the advanced
part $\mathbb{S}_{\textrm{av}}$. Therefore we have the following formulas
\begin{multline}\label{Dret(x)}
D^{\textrm{ret} \, ab}(t\times \boldsymbol{w}) = 
\sum\limits_{\substack{\widehat{n}\cdot \widehat{l} \in \mathscr{O}_+ \\ 1\leq s \leq d'(\widehat{n}\cdot \widehat{l}) \\ -l \leq i,j,i' \leq l}} 
i (2l+1) {u^{a}_{s}}_{{}_{ji}}(\widehat{n}\cdot \widehat{l}) \widehat{l}_{{}_{ij}}(\boldsymbol{w})
\widehat{n}(t) \, \theta(t) \,
\overline{{u^{b}_{s}}_{{}_{i' i'}}(\widehat{n}\cdot \widehat{l})} 
\\
\mp 
\sum\limits_{\substack{\widehat{n}\cdot \widehat{l} \in \mathscr{O}_+ \\ 1\leq s \leq d'(\widehat{n}\cdot \widehat{l}) \\ -l \leq i,j,i' \leq l}} 
i (2l+1) \overline{{u^{a}_{s}}_{{}_{ji}}(\widehat{n}\cdot \widehat{l})} \, \overline{\widehat{l}_{{}_{ij}}(\boldsymbol{w})}
 \overline{\widehat{n}(t)} \, \theta(t)
{u^{b}_{s}}_{{}_{i' i'}}(\widehat{n}\cdot \widehat{l}),
\end{multline}
\begin{multline}\label{Sret(x)}
S^{ab}_{\textrm{ret}}(t\times \boldsymbol{w}) = 
\sum\limits_{\substack{1 \leq c \leq d \\ \widehat{n}\cdot \widehat{l} \in \mathscr{O}_+ \\ 1\leq s \leq d'(\widehat{n}\cdot \widehat{l}) \\ -l \leq i,j,i' \leq l}} 
i (2l+1) {u^{a}_{s}}_{{}_{ji}}(\widehat{n}\cdot \widehat{l}) \widehat{l}_{{}_{ij}}(\boldsymbol{w})
\widehat{n}(t) \, \theta(t) \, \widehat{\Gamma}_{0}^{bc} \,
\overline{{u^{c}_{s}}_{{}_{i' i'}}(\widehat{n}\cdot \widehat{l})} 
\\
\mp 
\sum\limits_{\substack{1 \leq c \leq d \\ \widehat{n}\cdot \widehat{l} \in \mathscr{O}_- \\ 1\leq s \leq d''(\widehat{n}\cdot \widehat{l}) \\ -l \leq i,j,i' \leq l}} 
i (2l+1) \overline{{v^{a}_{s}}_{{}_{ji}}(\widehat{n}\cdot \widehat{l})} \,  \overline{\widehat{l}_{{}_{ij}}(\boldsymbol{w})}
 \overline{\widehat{n}(t)} \, \theta(t) \widehat{\Gamma}_{0}^{bc} \,
{v^{b}_{s}}_{{}_{i' i'}}(\widehat{n}\cdot \widehat{l}),
\end{multline}
\begin{equation}\label{D^(-)ret(x)}
D^{(-) \, \textrm{ret} \, ab}(t\times \boldsymbol{w}) = 
\sum\limits_{\substack{\widehat{n}\cdot \widehat{l} \in \mathscr{O}_+ \\ 1\leq s \leq d'(\widehat{n}\cdot \widehat{l}) \\ -l \leq i,j,i' \leq l}} 
i (2l+1) {u^{a}_{s}}_{{}_{ji}}(\widehat{n}\cdot \widehat{l}) \widehat{l}_{{}_{ij}}(\boldsymbol{w})
\widehat{n}(t) \, \theta(t)
\overline{{u^{b}_{s}}_{{}_{i' i'}}(\widehat{n}\cdot \widehat{l})},
\end{equation}
\begin{equation}\label{S^(-)ret(x)}
S^{(-) \, ab}_{\textrm{ret}}(t\times \boldsymbol{w}) = 
\sum\limits_{\substack{1 \leq c \leq d \\ \widehat{n}\cdot \widehat{l} \in \mathscr{O}_+ \\ 1\leq s \leq d'(\widehat{n}\cdot \widehat{l}) \\ -l \leq i,j,i' \leq l}} 
i (2l+1) {u^{a}_{s}}_{{}_{ji}}(\widehat{n}\cdot \widehat{l}) \widehat{l}_{{}_{ij}}(\boldsymbol{w})
\widehat{n}(t) \, \theta(t) \, \widehat{\Gamma}_{0}^{bc} \,
\overline{{u^{c}_{s}}_{{}_{i' i'}}(\widehat{n}\cdot \widehat{l})},
\end{equation}

\begin{multline}\label{Dav(x)}
D^{\textrm{av} \, ab}(t\times \boldsymbol{w}) = 
\sum\limits_{\substack{\widehat{n}\cdot \widehat{l} \in \mathscr{O}_+ \\ 1\leq s \leq d'(\widehat{n}\cdot \widehat{l}) \\ -l \leq i,j,i' \leq l}} 
i (2l+1) {u^{a}_{s}}_{{}_{ji}}(\widehat{n}\cdot \widehat{l}) \widehat{l}_{{}_{ij}}(\boldsymbol{w})
\widehat{n}(t) \, \big[-\theta(-t)\big] \,
\overline{{u^{b}_{s}}_{{}_{i' i'}}(\widehat{n}\cdot \widehat{l})} 
\\
\mp 
\sum\limits_{\substack{\widehat{n}\cdot \widehat{l} \in \mathscr{O}_+ \\ 1\leq s \leq d'(\widehat{n}\cdot \widehat{l}) \\ -l \leq i,j,i' \leq l}} 
i (2l+1) \overline{{u^{a}_{s}}_{{}_{ji}}(\widehat{n}\cdot \widehat{l})} \, \overline{\widehat{l}_{{}_{ij}}(\boldsymbol{w})}
 \overline{\widehat{n}(t)} \, \big[-\theta(-t)\big]
{u^{b}_{s}}_{{}_{i' i'}}(\widehat{n}\cdot \widehat{l}),
\end{multline}
\begin{multline}\label{Sav(x)}
S^{ab}_{\textrm{av}}(t\times \boldsymbol{w}) = 
\sum\limits_{\substack{1 \leq c \leq d \\ \widehat{n}\cdot \widehat{l} \in \mathscr{O}_+ \\ 1\leq s \leq d'(\widehat{n}\cdot \widehat{l}) \\ -l \leq i,j,i' \leq l}} 
i (2l+1) {u^{a}_{s}}_{{}_{ji}}(\widehat{n}\cdot \widehat{l}) \widehat{l}_{{}_{ij}}(\boldsymbol{w})
\widehat{n}(t) \, \big[-\theta(-t)\big] \, \widehat{\Gamma}_{0}^{bc} \,
\overline{{u^{c}_{s}}_{{}_{i' i'}}(\widehat{n}\cdot \widehat{l})} 
\\
\mp 
\sum\limits_{\substack{1 \leq c \leq d \\ \widehat{n}\cdot \widehat{l} \in \mathscr{O}_- \\ 1\leq s \leq d''(\widehat{n}\cdot \widehat{l}) \\ -l \leq i,j,i' \leq l}} 
i (2l+1) \overline{{v^{a}_{s}}_{{}_{ji}}(\widehat{n}\cdot \widehat{l})} \,  \overline{\widehat{l}_{{}_{ij}}(\boldsymbol{w})}
 \overline{\widehat{n}(t)} \, \big[-\theta(-t)\big] \widehat{\Gamma}_{0}^{bc} \,
{v^{b}_{s}}_{{}_{i' i'}}(\widehat{n}\cdot \widehat{l}),
\end{multline}
\begin{equation}\label{D^(-)av(x)}
D^{(-) \, \textrm{av} \, ab}(t\times \boldsymbol{w}) = 
\sum\limits_{\substack{\widehat{n}\cdot \widehat{l} \in \mathscr{O}_+ \\ 1\leq s \leq d'(\widehat{n}\cdot \widehat{l}) \\ -l \leq i,j,i' \leq l}} 
i (2l+1) {u^{a}_{s}}_{{}_{ji}}(\widehat{n}\cdot \widehat{l}) \widehat{l}_{{}_{ij}}(\boldsymbol{w})
\widehat{n}(t) \, \big[-\theta(-t)\big]
\overline{{u^{b}_{s}}_{{}_{i' i'}}(\widehat{n}\cdot \widehat{l})},
\end{equation}
\begin{equation}\label{S^(-)av(x)}
S^{(-)}_{\textrm{av}}(t\times \boldsymbol{w}) = 
\sum\limits_{\substack{1 \leq c \leq d \\ \widehat{n}\cdot \widehat{l} \in \mathscr{O}_+ \\ 1\leq s \leq d'(\widehat{n}\cdot \widehat{l}) \\ -l \leq i,j,i' \leq l}} 
i (2l+1) {u^{a}_{s}}_{{}_{ji}}(\widehat{n}\cdot \widehat{l}) \widehat{l}_{{}_{ij}}(\boldsymbol{w})
\widehat{n}(t) \, \big[-\theta(-t)\big] \, \widehat{\Gamma}_{0}^{bc} \,
\overline{{u^{c}_{s}}_{{}_{i' i'}}(\widehat{n}\cdot \widehat{l})}.
\end{equation}

Of course in case of infinite $\mathscr{O}_\pm$, the series (\ref{Dret(x)})-(\ref{S^(-)av(x)}) do not converge in $\mathscr{E}$ as functions,
but similarly as for (\ref{D(x,y)})-(\ref{S^(-)(x)}), the series (\ref{Dret(x)})-(\ref{S^(-)av(x)}) all converge in 
$\mathscr{E}^*$.  Namely for any fixed space-time test function $\phi \in \mathscr{E}$ each finite subsummand in (\ref{Dret(x)})-(\ref{S^(-)av(x)}), 
understood as an element of $\mathscr{E}^*$ and evaluated at any fixed $\phi \in \mathscr{E}$, converges absolutely as a numerical series and, by the Banach-Steinhaus
theorem, represents continuous functional on $\mathscr{E}$. We have so prepared the proof of the Lemma \ref{Pairings} that it can also be adopted here
for the proof of the following

\begin{lem}
The series (\ref{Dret(x)})-(\ref{S^(-)av(x)}) converge in $\mathscr{E}^*$,
and represent continuous functionals on $\mathscr{E}$.
\label{retOfPairings}
\end{lem}
\qedsymbol \, 
Before we apply the proof of the last Proposition, we need the following expression
\[
\int\limits_{\widetilde{\mathbb{S}^1} \times SU(2, \mathbb{C})} 
\theta(t) \, \phi^a(t, \boldsymbol{w}) \, \overline{\widehat{n}(t)} \overline{\widehat{l}_{{}_{ij}}(\boldsymbol{w})}
\, dt \, d \boldsymbol{w},
\,\,\,\,
\phi \in \mathscr{E} = \mathcal{S}_{{}_{\Delta +1}}(\widetilde{\mathbb{S}^1} \times SU(2, \mathbb{C}); \mathbb{C}^d),
\]
to be written in terms of the Fourier transform $\widetilde{\phi}$. We can do this by using separately the Fourier transform
on $\widetilde{\mathbb{S}^1}$ in time variable. Namely we have
\begin{align*}
\widehat{\theta}_{n} \coloneqq
{\textstyle\frac{1}{\sqrt{4\pi}}} \int\limits_{\widetilde{\mathbb{S}^1}} \theta(t) \, \overline{\widehat{n}(t)} \, dt, \\
\theta(t) = {\textstyle\frac{1}{\sqrt{4\pi}}} \sum\limits_{n \in \mathbb{Z}} \widehat{\theta}_{n} \, \widehat{n}(t), \\
\widehat{\phi}_{n}(\boldsymbol{w}) \coloneqq
{\textstyle\frac{1}{\sqrt{4\pi}}} \int\limits_{\widetilde{\mathbb{S}^1}} \phi(t, \boldsymbol{w}) \, \overline{\widehat{n}(t)} \, dt, \\
\phi(t, \boldsymbol{w}) = {\textstyle\frac{1}{\sqrt{4\pi}}} \sum\limits_{n \in \mathbb{Z}} \widehat{\phi}_{n}(\boldsymbol{w}) \, \widehat{n}(t), \\
\theta(t)\phi(t, \boldsymbol{w}) = {\textstyle\frac{1}{4\pi}} \sum\limits_{n', n'' \in \mathbb{Z}} \widehat{\theta}_{n'} \widehat{\phi}_{n''-n'}(\boldsymbol{w}) \, \widehat{n''}(t).
\end{align*}
Note that
\[
\widehat{\theta}_{n}=
\left\{ \begin{array}{ll}
{\textstyle\frac{i}{\sqrt{\pi} n}}, &  n \in 2 \mathbb{Z} +1, \\
0, & n \in 2 \mathbb{Z} \,\,\, \textrm{and} \,\,\, n\neq 0, \\
\sqrt{\pi}, & n=0
\end{array} \right..
\]

In the sequel we will need to choose (arbitrary) one of the two possible square  roots $\pm \sqrt{\widehat{\theta}_{n}}$
for each $n \in \mathbb{Z}$, and we choose
\[
\sqrt{\widehat{\theta}_{n}}=
\left\{ \begin{array}{ll}
{\textstyle\frac{e^{i\pi/4}}{\sqrt{\pi} \sqrt{n}}}, &  n \in 2 \mathbb{Z} +1, \\
0, & n \in 2 \mathbb{Z} \,\,\, \textrm{and} \,\,\, n\neq 0, \\
\pi^{1/4}, & n=0.
\end{array} \right..
\]
Finally because
\[
\int\limits_{\widetilde{\mathbb{S}^1}} \widehat{n''-n}(t) dt = 4\pi \delta_{n'' \, n},
\]
then we have
\begin{multline*}
\int\limits_{\widetilde{\mathbb{S}^1} \times SU(2, \mathbb{C})} 
\theta(t) \, \phi^a(t, \boldsymbol{w}) \, \overline{\widehat{n}(t)} \overline{\widehat{l}_{{}_{ij}}(\boldsymbol{w})}
\, dt \, d \boldsymbol{w} = \\ =
{\textstyle\frac{1}{4\pi}} \sum\limits_{n', n'' \in \mathbb{Z}} \int\limits_{\widetilde{\mathbb{S}^1} \times SU(2, \mathbb{C})} 
\widehat{\theta}_{n'} \widehat{\phi}_{n''-n'}(\boldsymbol{w}) \, \widehat{n''}(t) \, \overline{\widehat{n}(t)} \overline{\widehat{l}_{{}_{ij}}(\boldsymbol{w})}
\, dt \, d \boldsymbol{w} = \\ =
{\textstyle\frac{1}{4\pi}} \sum\limits_{n', n'' \in \mathbb{Z}}
\int\limits_{SU(2, \mathbb{C})} 
\widehat{\theta}_{n'} \widehat{\phi}_{n''-n'}(\boldsymbol{w}) \, 
\Big[\int\limits_{\widetilde{\mathbb{S}^1}} 
\widehat{n''-n}(t) dt
\Big]
 \, \overline{\widehat{l}_{{}_{ij}}(\boldsymbol{w})}
 \, d \boldsymbol{w} = \\ =
\sum\limits_{n' \in \mathbb{Z}}
\int\limits_{SU(2, \mathbb{C})} 
\widehat{\theta}_{n'} \widehat{\phi}_{n-n'}(\boldsymbol{w}) \, 
 \, \overline{\widehat{l}_{{}_{ij}}(\boldsymbol{w})}
 \, d \boldsymbol{w} = \\ =
\sum\limits_{n'\in \mathbb{Z}} \widehat{\theta}_{n'}
{\textstyle\frac{1}{\sqrt{4\pi}}} \int\limits_{\widetilde{\mathbb{S}^1} \times SU(2, \mathbb{C})}
\phi^a(t, \boldsymbol{w}) \, \overline{\widehat{n-n'}(t)} \overline{\widehat{l}_{{}_{ij}}(\boldsymbol{w})} \, dt \, d \boldsymbol{w}
\\
= \sum\limits_{n'\in \mathbb{Z}} \widehat{\theta}_{n'} \widetilde{\phi^a}_{{}_{ji}}(\widehat{n-n'}\cdot \widehat{l}), \\
\,\,\,\,
\phi \in \mathscr{E} = \mathcal{S}_{{}_{\Delta +1}}(\widetilde{\mathbb{S}^1} \times SU(2, \mathbb{C}); \mathbb{C}^d).
\end{multline*}
Summing up we have the formula
\begin{equation}\label{int(theta)x(phi)n^l^}
\int\limits_{\widetilde{\mathbb{S}^1} \times SU(2, \mathbb{C})} 
\theta(t) \, \phi^a(t, \boldsymbol{w}) \, \overline{\widehat{n}(t)} \overline{\widehat{l}_{{}_{ij}}(\boldsymbol{w})}
\, dt \, d \boldsymbol{w}  =
\sum\limits_{n'\in \mathbb{Z}} \widehat{\theta}_{n'} \widetilde{\phi^a}_{{}_{ji}}(\widehat{n-n'}\cdot \widehat{l}), \\
\end{equation}
\[
\textrm{for} \,\,\,
\phi \in \mathscr{E} = \mathcal{S}_{{}_{\Delta +1}}(\widetilde{\mathbb{S}^1} \times SU(2, \mathbb{C}); \mathbb{C}^d).
\]

Now we can repeat the proof of the last Lemma using the formula (\ref{int(theta)x(phi)n^l^}) in it.
Namely, let, for example, $D$ be the single-variable retarded function (\ref{Dret(x)}) corresponding to a neutral free
field $\mathbb{A}$. Let $\phi \in \mathscr{E}$, and let $b \in \{1, \ldots, d = \textrm{dim} \, V \}$ be a fixed value
of the field index. Then for the pairing $\langle D^{\textrm{ret}}, \phi \rangle$ of the distribution defined by the kernel 
(\ref{Dret(x)}) with fixed index $b$, with the test function $\phi$, we obtain the following value, on using the
formula (\ref{int(theta)x(phi)n^l^}) for the expression in the square bracket, as below: 
\begin{multline*}
\langle D^{\textrm{ret}}, \phi \rangle  = 
\sum\limits_{a =1}^d \,\, \int\limits_{\widetilde{\mathbb{S}^1} \times SU(2, \mathbb{C})}
D^{\textrm{ret} \, ab}(t\times \boldsymbol{w}) \, \phi^a(t\times \boldsymbol{w}) \,
dt d\boldsymbol{w} 
\\ 
= \sum\limits_{\substack{\widehat{n}\cdot \widehat{l} \in \mathscr{O}_+ \\ 1\leq s \leq d'(\widehat{n}\cdot \widehat{l}) \\ -l \leq i,j,i' \leq l \\
1 \leq a \leq \textrm{dim} \, V}} 
(2l+1) {u^{a}_{s}}_{{}_{ji}}(\widehat{n}\cdot \widehat{l}) 
\Bigg[
\,\,
\int\limits_{\widetilde{\mathbb{S}^1} \times SU(2, \mathbb{C})} 
\theta(t) \, \phi^a(t, \boldsymbol{w}) \, \widehat{n}(t) \,  \widehat{l}_{{}_{ij}}(\boldsymbol{w})
\, dt \, d \boldsymbol{w}
\Bigg]
\, {u^{b}_{s}}_{{}_{i'i'}}(\widehat{n}\cdot \widehat{l})
\\
= \sum\limits_{\substack{\widehat{n}\cdot \widehat{l} \in \mathscr{O}_+ \\ 1\leq s \leq d'(\widehat{n}\cdot \widehat{l}) \\ -l \leq i,j,i' \leq l \\
1 \leq a \leq \textrm{dim} \, V}} 
(2l+1) {u^{a}_{s}}_{{}_{ji}}(\widehat{n}\cdot \widehat{l}) 
\overline{\Bigg[
\,\,
\int\limits_{\widetilde{\mathbb{S}^1} \times SU(2, \mathbb{C})} 
\theta(t) \, \overline{\phi^a(t, \boldsymbol{w})} \, \overline{\widehat{n}(t)} \overline{\widehat{l}_{{}_{ij}}(\boldsymbol{w})}
\, dt \, d \boldsymbol{w}
\Bigg]}
\, {u^{b}_{s}}_{{}_{i'i'}}(\widehat{n}\cdot \widehat{l})
\\
= \sum\limits_{\substack{\widehat{n}\cdot \widehat{l} \in \mathscr{O}_+ \\ 1\leq s \leq d'(\widehat{n}\cdot \widehat{l}) \\ n' \in \mathbb{Z} 
\\ -l \leq i,j,i' \leq l \\ 1 \leq a \leq \textrm{dim} \, V}} 
(2l+1) 
\overline{
\overline{{u^{a}_{s}}_{{}_{ji}}(\widehat{n}\cdot \widehat{l}) \widehat{\theta}_{n'}} \, \, \widetilde{\overline{\phi^{a}}}_{{}_{ji}}(\widehat{n-n'}\cdot \widehat{l})
}
\, {u^{b}_{s}}_{{}_{i'i'}}(\widehat{n}\cdot \widehat{l})
\\
= \sum\limits_{\substack{\widehat{n}\cdot \widehat{l} \in \mathscr{O}_+ \\ 1\leq s \leq d'(\widehat{n}\cdot \widehat{l}) \\ n' \in \mathbb{Z} 
\\ -l \leq i,j,i' \leq l \\ 1 \leq a \leq \textrm{dim} \, V}} 
(2l+1) 
\overline{
\overline{{u^{a}_{s}}_{{}_{ji}}(\widehat{n}\cdot \widehat{l}) \sqrt{\widehat{\theta}_{n'}}} \, \, \widetilde{\overline{\phi^{a}}}_{{}_{ji}}(\widehat{n-n'}\cdot \widehat{l})
}
\, {u^{b}_{s}}_{{}_{i'i'}}(\widehat{n}\cdot \widehat{l})\sqrt{\widehat{\theta}_{n'}}
\end{multline*}
Summing up we have the following expression for the pairing
\begin{multline}\label{<Dret,phi>}
\langle D^{\textrm{ret}}, \phi \rangle  = 
\sum\limits_{\substack{\widehat{n}\cdot \widehat{l} \in \mathscr{O}_+ \\ 1\leq s \leq d'(\widehat{n}\cdot \widehat{l}) \\ n' \in \mathbb{Z} 
\\ -l \leq i,j,i' \leq l \\ 1 \leq a \leq \textrm{dim} \, V}} 
(2l+1) 
\overline{
\overline{{u^{a}_{s}}_{{}_{ji}}(\widehat{n}\cdot \widehat{l}) \sqrt{\widehat{\theta}_{n'}}} \, \, \widetilde{\overline{\phi^{a}}}_{{}_{ji}}(\widehat{n-n'}\cdot \widehat{l})
}
\, {u^{b}_{s}}_{{}_{i'i'}}(\widehat{n}\cdot \widehat{l})\sqrt{\widehat{\theta}_{n'}}.
\end{multline}

We stop here and define the family of auxiliary matrix valued functions $\zeta_{{}_{s,\widehat{n-n'}\cdot \widehat{l}}}$, 
numbered by the parameters $s,n,l,n'$ such that 
\[
(s,\widehat{n}\cdot \widehat{l}) \in \{1, \ldots, d'(\widehat{n}\cdot \widehat{l})\} \times \mathscr{O}_+,
\,\, n' \in \mathbb{Z}
\] 
and defined on the dual $\widehat{\widetilde{\mathbb{S}^1}\times G}$, with $G= SU(2, \mathbb{C})$, by the following formula
\begin{equation}\label{zeta(n''-n')general}
\big[\zeta_{{}_{s,\widehat{n-n'}\cdot \widehat{l}}}\big]^{a}_{{}_{ji}}(\widehat{n''-n'''}\cdot \widehat{l''}) = 
\delta_{n \, n''} \delta_{ll''} \delta_{n' \, n'''} \,\, {u_{s}^{a}}_{{}_{ji}}(\widehat{n}\cdot \widehat{l}) \, \sqrt{\widehat{\theta}_{n'}}, 
\end{equation}
each concentrated on the single point set $\widehat{n-n'}\cdot \widehat{l} \in \widehat{\widetilde{\mathbb{S}^1}\times G}$.
However now in case we have several $\widehat{n}\cdot \widehat{l} \in \mathscr{O}_+$, say $\widehat{n_{1}(l)}\cdot \widehat{l}, \ldots,
\widehat{n_{k_l}(l)}\cdot \widehat{l} \in \mathscr{O}_+$, for each fixed $l$,
 we have to consider several such families of functions, correspondingly to the number of points $\widehat{n}\cdot \widehat{l} \in \mathscr{O}_+$
with fixed $l$. Let us consider the simplest case where all $k_l$ are equal say $n_{{}_{\mathscr{O}_+}}$, as for the electromagnetic potential
field, where all $k_l = 3 = n_{{}_{\mathscr{O}_+}}$, compare the last picture. In fact let us start with a still simpler case
of massless field $\mathbb{A}$,  whose each component respects the massless wave equation 
$\square \mathbb{A}=0$, and thus which has the corresponding orbit 
\[
\mathscr{O}_{{}_{+}}  = 
\big\{\widehat{n}\cdot \widehat{l}, \, {\textstyle\frac{n^2}{4}} - l(l+1) - {\textstyle\frac{1}{4}}
= 0, n\geq 0 \big\},
\] 
lying exactly on the ``light cone'' in the momentum space. In this case for each fixed non-negative integer or half-an-odd  positive integer $l$ 
there exist at most one
positive integer $n$, such that $\widehat{n} \cdot \widehat{l} \in \mathscr{O}_+$. Therefore the set of points
$\widehat{n-n'}\cdot \widehat{l} \in \widehat{\widetilde{\mathbb{S}^1} \times G}$ covers each point of the dual $\widehat{\widetilde{\mathbb{S}^1} \times G}$
at most once when $\widehat{n} \cdot \widehat{l}$ runs over the orbit $\mathscr{O}_+$ and the integer $n'$ runs independently 
over $\mathbb{Z}$. In this case we define just one family of auxiliary functions $\zeta_{{}_{s,\widehat{n}\cdot \widehat{l}}}$, given by the 
formula (\ref{zeta(n''-n')general}), which is well defined in this case. When we have the situation in which all $k_l$ are equal and for each fixed
$\widehat{n} \cdot \widehat{l} \in \mathscr{O}_+$ we have exactly $k_l = n_{{}_{\mathscr{O}_+}}$ points $\widehat{n} \cdot \widehat{l} \in \mathscr{O}_+$
with the same $l$, then in the formula  (\ref{zeta(n''-n')general}) the choice of the specific value of $n$ for each fixed $l$
in $\widehat{n}\cdot \widehat{l} \in \mathscr{O}_+$ need to be made. The various specifications are understood to be different when they have no common choice
of $n$ for any common $l$, for each $\widehat{n} \cdot \widehat{l} \in \mathscr{O}_+$.  Choice of different specifications,
(and we have exactly $n_{{}_{\mathscr{O}_+}}$ different specifications) gives the corresponding $n_{{}_{\mathscr{O}_+}}$ families of auxiliary functions 
$\zeta_{{}_{s,\widehat{n-n'}\cdot \widehat{l}}}$. Correspondingly to each specification we define another family of 
auxiliary matrix valued functions $\chi_{{}_{s,\widehat{n-n'}\cdot \widehat{l}}}$, 
numbered by the same parameters $s,\widehat{n-n'}\cdot \widehat{l}$ as the family $\zeta_{{}_{s,\widehat{n-n'}\cdot \widehat{l}}}$, and defined
by the following formula
\[
\big[\chi_{{}_{s,\widehat{n-n'}\cdot \widehat{l}}}\big]^{a}_{{}_{ji}}(\widehat{n''-n'''}\cdot \widehat{l''}) = 
\delta_{n \, n''} \delta_{ll''} \delta_{n' \, n'''} \,\, \delta^{ab} \delta_{ji}.
\]
Correspondinly to each specification we divide the total sum (\ref{<Dret,phi>}) into $n_{{}_{\mathscr{O}_+}}$ subsums
each containing at most one representant $\widehat{n}\cdot \widehat{l}$ of the orbit $\mathscr{O}_+$ for each fixed value of $l$.

Next we estimate each of the $n_{{}_{\mathscr{O}_+}}$ subsums separately using the corresponding families of auxiliary functions
$\zeta_{{}_{s,\widehat{n-n'}\cdot \widehat{l}}}$ and $\chi_{{}_{s,\widehat{n-n'}\cdot \widehat{l}}}$, exactly as we did in the proof of the last Proposition. 
Namely, we observe that for sufficiently large positive integer $r$ the series
\begin{align*}
\sum\limits_{\substack{\widehat{n}\cdot \widehat{l} \in \mathscr{O}_+ \\ 1\leq s \leq d'(\widehat{n}\cdot \widehat{l}) \\ n' \in \mathbb{Z}}} 
\lambda_{n-n', l}^{-r} \zeta_{{}_{s,\widehat{n-n'}\cdot \widehat{l}}} = 
\sum\limits_{\substack{\widehat{n}\cdot \widehat{l} \in \mathscr{O}_+ \\ 1\leq s \leq d'(\widehat{n}\cdot \widehat{l}) \\ n' \in \mathbb{Z}}} 
\widetilde{A}^{-r} \zeta_{{}_{s,\widehat{n-n'}\cdot \widehat{l}}} = \xi, \\
\sum\limits_{\substack{\widehat{n}\cdot \widehat{l} \in \mathscr{O}_+ \\ 1\leq s \leq d'(\widehat{n}\cdot \widehat{l}) \\ n' \in \mathbb{Z}}} 
\lambda_{n-n', l}^{-r} \chi_{{}_{s,\widehat{n-n'}\cdot \widehat{l}}} = 
\sum\limits_{\substack{\widehat{n}\cdot \widehat{l} \in \mathscr{O}_+ \\ 1\leq s \leq d'(\widehat{n}\cdot \widehat{l}) \\ n' \in \mathbb{Z}}} 
\widetilde{A}^{-r} \chi_{{}_{s,\widehat{n-n'}\cdot \widehat{l}}} = \eta
\end{align*}
converge in  $\widetilde{\mathscr{H}}$ to elements $\xi, \eta \in \widetilde{\mathscr{H}}$.  Here the summations run effectively 
over the corresponding subrange which contains at most one representant $\widehat{n}\cdot \widehat{l}$ of the orbit $\mathscr{O}_+$ 
for each fixed value of $l$, as the families of auxiliary functions $\zeta_{{}_{s,\widehat{n-n'}\cdot \widehat{l}}}$
and $\chi_{{}_{s,\widehat{n-n'}\cdot \widehat{l}}}$ used here correspond to one of the $n_{{}_{\mathscr{O}_+}}$ specifications of this range and  
all functions of these families of auxiliary functions  are supported within this range.

Now we can estimate each one of the $n_{{}_{\mathscr{O}_+}}$ subsums in the expression (\ref{<Dret,phi>}) for the pairing, exactly
as we did in the proof of the last Lemma (the prime over the sum means that the range of the sum (\ref{<Dret,phi>})
is restricted to one of the $n_{{}_{\mathscr{O}_+}}$ possible specifications and is confined to the corresponding subrange)
\begin{multline*}
\sum'\limits_{\substack{\widehat{n}\cdot \widehat{l} \in \mathscr{O}_+ \\ 1\leq s \leq d'(\widehat{n}\cdot \widehat{l}) \\ n' \in \mathbb{Z} 
\\ -l \leq i,j,i' \leq l \\ 1 \leq a \leq \textrm{dim} \, V}} 
(2l+1) 
\overline{
\overline{{u^{a}_{s}}_{{}_{ji}}(\widehat{n}\cdot \widehat{l}) \sqrt{\widehat{\theta}_{n'}}} \, \, \widetilde{\overline{\phi^{a}}}_{{}_{ji}}(\widehat{n-n'}\cdot \widehat{l})
}
\, {u^{b}_{s}}_{{}_{i'i'}}(\widehat{n}\cdot \widehat{l})\sqrt{\widehat{\theta}_{n'}}
\\
=
\sum\limits_{\substack{\widehat{n}\cdot \widehat{l} \in \mathscr{O}_+ \\ 1\leq s \leq d'(\widehat{n}\cdot \widehat{l}) \\ n' \in \mathbb{Z} 
\\ -l \leq i,j,i' \leq l \\ 1 \leq a \leq \textrm{dim} \, V}} 
(2l+1) 
\overline{
\overline{\big[\zeta_{{}_{s,\widehat{n-n'}\cdot \widehat{l}}}\big]^{a}_{{}_{ji}}(\widehat{n-n'}\cdot \widehat{l})} \,\,\,\,\,
 \widetilde{\overline{\phi^{a}}}_{{}_{ji}}(\widehat{n-n'}\cdot \widehat{l})
}
\, \big[\zeta_{{}_{s,\widehat{n-n'}\cdot \widehat{l}}}\big]^{b}_{{}_{i'i'}}(\widehat{n-n'}\cdot \widehat{l}) 
\end{multline*}
\begingroup\makeatletter\def\f@size{5}\check@mathfonts
\def\maketag@@@#1{\hbox{\m@th\large\normalfont#1}}%
\begin{multline*}
=
\sum\limits_{\substack{\widehat{n}\cdot \widehat{l} \in \mathscr{O}_+ \\ 1\leq s \leq d'(\widehat{n}\cdot \widehat{l}) \\ n' \in \mathbb{Z} 
\\ -l \leq i,j,i' \leq l \\ 1 \leq a \leq \textrm{dim} \, V}} 
(2l+1) 
\overline{
\overline{\lambda_{n-n', l}^{-r}\big[\zeta_{{}_{s,\widehat{n-n'}\cdot \widehat{l}}}\big]^{a}_{{}_{ji}}(\widehat{n-n'}\cdot \widehat{l})} \,\,\,\,\,
 \lambda_{n-n', l}^{2r} \widetilde{\overline{\phi^{a}}}_{{}_{ji}}(\widehat{n-n'}\cdot \widehat{l})
}
\, \lambda_{n-n', l}^{-r} \big[\zeta_{{}_{s,\widehat{n-n'}\cdot \widehat{l}}}\big]^{b}_{{}_{i'i'}}(\widehat{n-n'}\cdot \widehat{l})
\\
=
\sum\limits_{\substack{\widehat{n}\cdot \widehat{l} \in \mathscr{O}_+ \\ 1\leq s \leq d'(\widehat{n}\cdot \widehat{l}) \\ n' \in \mathbb{Z} 
\\ -l \leq i' \leq l}} 
\overline{
\Bigg\langle \lambda_{n-n', l}^{-r} \zeta_{{}_{s,\widehat{n-n'}\cdot \widehat{l}}}(\widehat{n-n'}\cdot \widehat{l}) \, ,
 \lambda_{n-n', l}^{2r} \widetilde{\overline{\phi}}(\widehat{n-n'}\cdot \widehat{l}) \Bigg\rangle
}
\, \lambda_{n-n', l}^{-r} \big[\zeta_{{}_{s,\widehat{n-n'}\cdot \widehat{l}}}\big]^{b}_{{}_{i'i'}}(\widehat{n-n'}\cdot \widehat{l})
\end{multline*}
\endgroup
\[
\,\,\,\,\,\,\,\,\,\,\,\,\,\,\,\,\,\,\,\,\,\,\,\,\,\,\,\,\,\,\,\,\,\,\,\,\,\,\,\,\,\,\,\,\,\,\,\,\,\,\,\,\,\,\,\,\,\,\,\,\,\,\,
\,\,\,\,\,\,\,\,\,\,\,\,\,\,\,\,\,\,\,\,\,\,\,\,\,\,\,\,\,\,\,\,\,\,\,\,\,\,\,\,\,\,\,\,\,\,\,\,\,\,\,\,\,\,\,\,\,\,\,\,\,\,\,
= \overline{\Big\langle \xi, \widetilde{A^{2r} \overline{\phi}} \Big\rangle} \, \langle \xi, \eta \rangle,
\]
where $\langle \, \cdot \, , \, \cdot \, \rangle$ in the last but one line is again the inner product (\ref{<tildephi(n.l),tildephi'(n.l)>}) 
so that $\langle \, \cdot \, , \, \cdot \, \rangle$ in the last line is the inner product in $\widetilde{\mathscr{H}}$.
Therefore, for each one of the $n_{{}_{\mathscr{O}_+}}$ subsums in the expression (\ref{<Dret,phi>}) we obtain the following estimation for its
absolute value
\begin{multline*}
\Bigg|\sum'\limits_{\substack{\widehat{n}\cdot \widehat{l} \in \mathscr{O}_+ \\ 1\leq s \leq d'(\widehat{n}\cdot \widehat{l}) \\ n' \in \mathbb{Z} 
\\ -l \leq i,j,i' \leq l \\ 1 \leq a \leq \textrm{dim} \, V}} 
(2l+1) 
\overline{
\overline{{u^{a}_{s}}_{{}_{ji}}(\widehat{n}\cdot \widehat{l}) \sqrt{\widehat{\theta}_{n'}}} \, \, \widetilde{\overline{\phi^{a}}}_{{}_{ji}}(\widehat{n-n'}\cdot \widehat{l})
}
\, {u^{b}_{s}}_{{}_{i'i'}}(\widehat{n}\cdot \widehat{l})\sqrt{\widehat{\theta}_{n'}} \Bigg|
\\
= \Big|\overline{\Big\langle \xi, \widetilde{A^{2r} \overline{\phi}} \Big\rangle} \Big| \, |\langle \xi, \eta \rangle |
\leq |\xi|^2 \, |\eta| \, |\phi|_{{}_{2r}},
\end{multline*}
with the auxiliary functions $\xi, \eta \in \widetilde{\mathscr{H}}$ corresponding to each of the $n_{{}_{\mathscr{O}_+}}$ subsums $\sum'$. 
Therefore, for each $\phi \in \mathscr{E}$, the value $\langle \textrm{finite subsummand of} \, D^\textrm{ret}, \,\, \phi\rangle$
of each finite subsummand in (\ref{Dret(x)}) evaluated at $\phi$ is absolutely convergent, so that (\ref{Dret(x)})
converges in $\mathscr{E}^*$ and $D^\textrm{ret}$ given by (\ref{Dret(x)}) represents a continuous functional on $\mathscr{E}$.

Proof of the convergence of the remaining series (\ref{Dret(x)})-(\ref{S^(-)av(x)}) is identical.

Note that the same proof works if the number $k_l$ of points $\widehat{n} \cdot \widehat{l} \in \mathscr{O}_\pm$ with the same $l$
is not necessary equal for all $\widehat{n} \cdot \widehat{l} \in \mathscr{O}_\pm$ and independent of $l$, but it works also if
$k_l$ varies with $l$, but is uniformly bounded, with a bound equal $n_{{}_{\mathscr{O}_+}}$. 
In fact a modification of this proof also allows to include situations in which $k_l$
grows at most polynomially with $l$. However, we do not enter into the analysis of such situations, because the presented proof embraces
all relevant free fields underlying QFT with realistic interactions.
\qed

{\bf REMARK}. In order to make the analogue between the constructions and particular formulas concerning free fields on the 
Minkowski space-time $\mathbb{R}^4$, which can also be regarded as the additive Lie group $\mathbb{R}^4$,
and on the Einstein Universe, which can also be regarded as the Lie group $\mathbb{R}\times SL(2, \mathbb{C})$,
we should introduce the convolution between the Fourier transforms of  smooth functions
on the space-time. Because the fields effectively live on the compactification
$\widetilde{\mathbb{S}^1}\times SL(2, \mathbb{C})$ of $\mathbb{R}\times SL(2, \mathbb{C})$, which also is a Lie group,
then we should introduce the convolution between Fourier transforms $\widetilde{\varphi}, \widetilde{\phi}$ 
of functions $\varphi, \phi \in \mathscr{E} = \mathcal{S}_{\Delta =1}\big(\widetilde{\mathbb{S}^1}\times SL(2, \mathbb{C})\big)$.
Namely, we define the convolution through the standard formula
\[
\widetilde{\varphi} \ast \widetilde{\phi} \overset{\textrm{df}}{=} \widetilde{\varphi.\phi},
\]
as the Fourier transform of the pointwise product $\varphi.\phi$. By definition this convolution is commutative. It should be distinguished
from the convolution $\varphi \ast \phi$ naturally arising from the group structure of the (compactified) space-time
Lie group $\widetilde{\mathbb{S}^1}\times SL(2, \mathbb{C})$, 
which in case of the Einstein Universe is non-commutative due to the non-Abelian character of the group
$\widetilde{\mathbb{S}^1}\times SL(2, \mathbb{C})$. We hope that the context
and the explicit usage of the Fourier transform sign will indicate clearly which convolution is meant in each particular
case. The convolution of the Fourier transforms of more general square integrable functions
$\phi, \varphi \in L^2\big(\widetilde{\mathbb{S}^1}\times SL(2, \mathbb{C})$ of $\mathbb{R}\times SL(2, \mathbb{C}) \big)$ (with respect to
the normalized invariant measure) can be written more explicitly in the form
\begin{multline*}
{\textstyle\frac{1}{\sqrt{4\pi}}} 
\int\limits_{\widetilde{\mathbb{S}^1} \times SU(2, \mathbb{C})} 
\varphi(t, \boldsymbol{w}) \, \phi(t, \boldsymbol{w}) \, \overline{\widehat{n}(t)} \overline{\widehat{l}_{{}_{ij}}(\boldsymbol{w})}
\, dt \, d \boldsymbol{w} 
\\
=
{\textstyle\frac{1}{\sqrt{4\pi}}} 
\sum\limits_{\substack{ n',l',j',i'\\ l'',j'',i''}} 
{\textstyle\frac{(2l'+1)(2l''+1)}{2l +1}} 
\widetilde{\varphi}_{{}_{\,\, j' i'}} \big(\widehat{n'} \cdot \widehat{l'}\big)
\,
\widetilde{\phi}_{{}_{\,\, j'' i''}} \big(\widehat{n-n'} \cdot \widehat{l''}\big) \,
\underset{l' \,\, i'j'}{M}_{{}_{l'' \,\, i''j''}}^{{}^{l \,\, ij}}
\\
= \Big[ \widetilde{\varphi} \ast \widetilde{\phi} \Big]_{{}_{ji}}\big(\widehat{n} \cdot \widehat{l}\big),
\end{multline*}
where 
\[
\widehat{l'}_{{}_{i'j'}}(\boldsymbol{w})\widehat{l''}_{{}_{i''j''}}(\boldsymbol{w}) = 
\sum\limits_{\substack{|l' -l''| \leq l \leq l'+l'' \\ -l \leq i,j \leq l}} 
\underset{l' \,\, i'j'}{M}_{{}_{l'' \,\, i''j''}}^{{}^{l \,\, ij}}
\widehat{l}_{{}_{ij}}(\boldsymbol{w}), \,\,\,\,\,
\boldsymbol{w} \in SU(2, \mathbb{C})
\]
is the formula determined by the tensor product decomposition formula and the Clebsh-Gordan coefficients.
Because
\[
\underset{l'=0 \,\, i'=0 j'=0}{M}_{{}_{l'' \,\, i''j''}}^{{}^{l \,\, ij}} = \delta_{l''}^{l} \delta_{i''}^{i} \delta_{j''}^{j}
\]
then in particular inserting $\varphi = \theta$ into the convolution formula we obtain (\ref{int(theta)x(phi)n^l^}) easily, 
because the invariant measure $d \boldsymbol{w}$
on $SU(2, \mathbb{C}) \cong \mathbb{S}^3$ is so normalized that
\[
\int\limits_{SU(2, \mathbb{C})}  d \boldsymbol{w} = 1.
\]
\qed

Let us introduce the general class $\mathfrak{K}_0$ of kernels and the corresponding integral kernel operators (fields),
the immediate analogue of the class $\mathfrak{K}_0$ of kernels on the Minkowski space-time introduced in Subsection \ref{OperationsOnXi},
Def. \ref{K_0}: 

\begin{defin}
Let the class $\mathfrak{K}_0$ consists of the plane wave kernels $\kappa_{1,0},\kappa_{0,1}$ defining the free fields of the theory and 
of their derivatives 
\[
(X)^\alpha\kappa_{\ell,m} = (X^{0})^{\alpha_0}\ldots (X^{3})^{\alpha_3} \kappa_{\ell,m}, \,\, \ell,m=0,1,
\]
where $X^{0}, \ldots, X^{3}$ are the right invariant vector fields-derivations on the Einstein Universe 
introduced in the previous Subsection, and with $X^\mu$, $\mu=0,1,2,3$, plying the role analogues to the role which translation derivations
\[
{\textstyle\frac{\partial}{\partial x_0}}, \ldots, {\textstyle\frac{\partial}{\partial x_3}},
\]
do on the Minkowski space-time.

We assume that the orbits $\mathscr{O}_\pm$, associated to the considered class of free fields, have the property that the number 
$k_l$ of points $\widehat{n} \cdot \widehat{l} \in \mathscr{O}_\pm$ with the same $l$
is not necessary equal for all $\widehat{n} \cdot \widehat{l} \in \mathscr{O}_\pm$, in general
$k_l$ may depend on $l$, but $k_l$ is uniformly bounded for $\widehat{n} \cdot \widehat{l} \in \mathscr{O}_\pm$, with the bound 
which we denote by $n_{{}_{\mathscr{O}_\pm}}$.
\label{K_0onEU}
\end{defin}

Note that having given any $\kappa_{1,0}, \kappa_{1,0}$ of the class $\mathfrak{K}_0$, corresponding to a free field
with the single particle Gelfand triple $E \subset \mathcal{H}' \subset {E}^*$, it can be easily seen
that for any $\xi \in E$ the function $(X)^\alpha \kappa_{1,0}(\xi) \in \mathscr{E}$ and $(X)^\alpha \kappa_{0,1}(\xi) \in \mathscr{E}$, 
i.e both functions $(X)^\alpha \kappa_{1,0}(\xi)$ and $(X)^\alpha \kappa_{0,1}(\xi)$ belong to the space-time standard nuclear test space
$\mathscr{E}$, and that in this compact case the multiplier algebra $\mathcal{O}_{M}(\widetilde{\mathbb{S}^1} \times SU(2, \mathbb{C}))$
of $\mathscr{E} = \mathcal{S}_{A}(\widetilde{\mathbb{S}^1} \times SU(2, \mathbb{C}))$ coincides with $\mathscr{E}$ (in case of $\mathbb{C}$-valued functions). 
Because the convolution $F \ast \phi$ on the compact group $\widetilde{\mathbb{S}^1} \times SU(2, \mathbb{C})$ of a distribution
$F \in \mathscr{E}^*$ with $\phi \in \mathscr{E}$, defines a continuous map $\mathscr{E} \rightarrow \mathscr{E}$ (below we give a proof of it),
then the convolutor algebra\footnote{In this Schwartz notation for $\mathcal{O}'_{C}(\widetilde{\mathbb{S}^1} \times SU(2, \mathbb{C}))$
we should also write the strong dual with prime sign insted of the asterisk, in particular using the Schwartz notation more consequently
we should write $\mathcal{S}_{A}'(\widetilde{\mathbb{S}^1} \times SU(2, \mathbb{C}))$ instead of our 
$\mathcal{S}_{A}(\widetilde{\mathbb{S}^1} \times SU(2, \mathbb{C}))^*$,
but the asterisk notation  for strong dual space is more frequently used in the Japanese school of Hida to which we frequently refer,
compare \cite{Schwartz} or Appendix \ref{convolutorsO'_C}.} 
$\mathcal{O}'_{C}(\widetilde{\mathbb{S}^1} \times SU(2, \mathbb{C}))$ of 
$\mathscr{E} = \mathcal{S}_{A}(\widetilde{\mathbb{S}^1} \times SU(2, \mathbb{C}))$, $A = \Delta +1$, coincides with the distribution
space $\mathscr{E}^*$ and thus the predual $\mathcal{O}_{C}(\widetilde{\mathbb{S}^1} \times SU(2, \mathbb{C}))$ of 
$\mathcal{O}'_{C}(\widetilde{\mathbb{S}^1} \times SU(2, \mathbb{C})) = \mathscr{E}^*$ coincides with 
$\mathscr{E} = \mathcal{S}_{A}(\widetilde{\mathbb{S}^1} \times SU(2, \mathbb{C}))$ itself. Thus
$\mathcal{O}_{C}(\widetilde{\mathbb{S}^1} \times SU(2, \mathbb{C})) = \mathcal{O}_{M}(\widetilde{\mathbb{S}^1} \times SU(2, \mathbb{C}))$ and we obtain 
 $(X)^\alpha \kappa_{1,0}(\xi) \in \mathcal{O}_{C}(\widetilde{\mathbb{S}^1} \times SU(2, \mathbb{C}))$ and $(X)^\alpha \kappa_{0,1}(\xi) \in \mathcal{O}_{C}(\widetilde{\mathbb{S}^1} \times SU(2, \mathbb{C}))$ trivially in this compact case. In particular, we can repeat the general method
of Subsection \ref{OperationsOnXi} and show that the Wick product of any two free fields on the Einstein Universe
or of their derivatives are integral kernel operators in $\mathscr{L}(\mathscr{E}, \mathscr{L}((\boldsymbol{E}), (\boldsymbol{E})))$.
Of course, we do it by investigation of the continuity analysis of the linear maps (distributions) defined by the pointwise products
\[
(X)^\alpha \kappa'_{1,0} \overset{\cdot}{\otimes} (X)^\beta \kappa''_{1,0}, \,\,\,
(X)^\alpha \kappa'_{1,0} \overset{\cdot}{\otimes} (X)^\beta \kappa''_{0,1},
\]
with $\kappa'_{1,0},\kappa''_{0,1} \in \mathfrak{K}_0$ corresponding to the free fields with the corresponding single particle Gelfand triples
$E_{{}_{'}} \subset \mathcal{H}' \subset E_{{}_{'}}^*$ and $E_{{}_{''}} \subset \mathcal{H}'' \subset E_{{}_{''}}^*$, compare Subsection \ref{OperationsOnXi}. 
Because the differential operators $(X)^\alpha = (X^{0})^{\alpha_0}\ldots (X^{3})^{\alpha_3}$
are continuous differential operators on $\mathscr{E}$, it is sufficient to investigate continuity of the distributions defined by the pointwise
product kernels
\[
\kappa'_{1,0} \overset{\cdot}{\otimes} \kappa''_{1,0}, \,\,\,
\kappa'_{1,0} \overset{\cdot}{\otimes}  \kappa''_{0,1},
\]
of the kernels $\kappa'_{1,0}, \kappa''_{0,1} \in \mathfrak{K}_0$.
Recall that $\kappa'_{1,0} \overset{\cdot}{\otimes} \kappa''_{0,1}$ is the vector-valued distribution defined by the 
kernel equal to the ordinary pointwise product
\[
\kappa_{1,0} \overset{\cdot}{\otimes} \kappa_{0,1}(s', \widehat{n'}\cdot \widehat{l'}, s'', \widehat{n''}\cdot \widehat{l''}; a, b, x)
= \kappa_{1,0} (s', \widehat{n'}\cdot \widehat{l'}; a,x) \kappa_{0,1}(s'', \widehat{n''}\cdot \widehat{l''}; b, x),
\]
\[
(s', \widehat{n'}\cdot \widehat{l'}) \in \mathscr{O}', \,\,\,\,
(s'', \widehat{n''}\cdot \widehat{l''}) \in \mathscr{O}'',
\]
where $\widehat{n'}\cdot \widehat{l'} \in \mathscr{O}_{\pm}'$, $\widehat{n''}\cdot \widehat{l''} \in \mathscr{O}_{\pm}''$
range over the orbits $\mathscr{O}_{\pm}'$, $\mathscr{O}_{\pm}''$ corresponding to the fields defined by the kernels
$\kappa'_{\ell,m}, \kappa''_{\ell,m} \in \mathfrak{K}_0$, $\ell,m = 0,1$.

On the Einstein Universe, where the distribution kernels live (concerning the space-time variable $x$) on the compact manifold 
$\widetilde{\mathbb{S}^1} \times SU(2, \mathbb{C})$
the Lemma \ref{Hypocont.Ofkappa.kappa} as well as the Lemma \ref{kappaBarDotOtimeskappa} (including the point
4)) of Subsection \ref{OperationsOnXi}, are valid for the kernels $\kappa'_{1,0}, \kappa''_{0,1} \in \mathfrak{K}_0$
corresponding to massive as well as for massless fields. 

We encourage the reader to go through the explicit verification of these Lemmas in case of free fields on the Einstein Universe. 
Here we present another, perhaps more immediate, proof of the said Lemmas for the kernels defininig free fields
(or their derivatives) on the Einstein Universe. 

\begin{lem}
Let $\kappa_{\ell, m}, \dots, \kappa'_{\ell, m}, \kappa''_{\ell, m}, \ldots, \kappa'''_{\ell, m}   \in \mathfrak{K}_0$, $\ell,m = 1,0$, be the kernels 
of free fields on the Einstein Universe (all positive or all negative-energy fields)
with the corresponding single particle Gelfand triples 
\[
\begin{array}{ccccc} E & \subset & \mathcal{H} & \subset & E^* 
\\ \vdots&&\vdots&& \vdots \\ 
 E_{{}_{'}} & \subset & \mathcal{H}' & \subset & E_{{}_{'}}^*
 \\  
E_{{}_{''}} & \subset & \mathcal{H}'' & \subset & E_{{}_{''}}^*
\\\vdots &&\vdots&&\vdots  \\  
E_{{}_{'''}} & \subset & \mathcal{H}''' & \subset & E_{{}_{'''}}^*.
\end{array} 
\]
Let $\mathscr{E}=\mathcal{S}_{{}_{\Delta +1}}(\widetilde{\mathbb{S}^1} \times SU(2, \mathbb{C}); \mathbb{C})$.
Then 
\begin{enumerate}
\item[1)]
\[
\kappa'_{1, 0} \overset{\cdot}{\otimes} \kappa''_{1, 0} \in  \mathscr{L}(\mathscr{E}, E_{{}_{'}} \otimes E_{{}_{''}}) \cong
\mathscr{L}(E_{{}_{'}}^* \otimes E_{{}_{''}}^*, \mathscr{E}^*).
\]
\item[2)]
\[
\kappa'_{1, 0} \overset{\cdot}{\otimes} \kappa''_{0, 1} \in  \mathscr{L}(\mathscr{E}, E_{{}_{'}} \otimes E_{{}_{''}}^*) \cong
\mathscr{L}(E_{{}_{'}}^* \otimes E_{{}_{''}}, \mathscr{E}^*).
\]
\item[3)]
\begin{multline*}
\kappa_{1, 0} \overset{\cdot}{\otimes} \cdots \overset{\cdot}{\otimes} \kappa'_{1, 0} \overset{\cdot}{\otimes} 
\kappa''_{0, 1}\overset{\cdot}{\otimes} \cdots \overset{\cdot}{\otimes} \kappa'''_{0, 1} 
\in  \mathscr{L}(\mathscr{E}, E_{{}_{}} \otimes \cdots \otimes E_{{}_{'}} \otimes E_{{}_{''}}^* \otimes \cdots \otimes E_{{}_{'''}}^*) 
\\
\cong
\mathscr{L}(E_{{}_{}}^* \otimes \cdots \otimes E_{{}_{'}}^* \otimes E_{{}_{''}} \otimes \cdots \otimes E_{{}_{'''}}, \mathscr{E}^*).
\end{multline*}
\item[4)] The above statements 1)-3) remain valid if we replace the kernels $\kappa_{1,0}, \ldots$ with their derivations
$(X)^\alpha \kappa_{1,0}, \ldots$.
\end{enumerate}
\label{kappaBarDotOtimeskappaOnEU}
\end{lem}
\qedsymbol \,
Ad 1). Let, for example, $\kappa'_{1,0}, \kappa''_{1,0}$ be the negative frequency kernels of two neutral fields. 
Let $\phi \in \mathscr{E}=\mathcal{S}_{{}_{\Delta +1}}(\widetilde{\mathbb{S}^1} \times SU(2, \mathbb{C}); \mathbb{C})
=\mathscr{C}^\infty(\widetilde{\mathbb{S}^1} \times SU(2, \mathbb{C}); \mathbb{C})$ 
(of $\mathbb{C}$-valued space-time test functions) and let $x=t\times \boldsymbol{w} \in \mathbb{R} \times SU(2, \mathbb{C})$. 
By definition
\begin{multline*}
\kappa'_{1,0} \overset{\cdot}{\otimes} \kappa''_{1,0}(\phi)(s', \widehat{n'}\cdot \widehat{l'}, s'', \widehat{n''}\cdot \widehat{l''}; a, b) 
\\
=
\int \,
\kappa'_{1,0} (s', \widehat{n'}\cdot \widehat{l'}; a,x) \kappa''_{1,0}(s'', \widehat{n''}\cdot \widehat{l''}; b, x) \phi(x) \ud^4 x
\end{multline*}
\begin{multline*}
=
\sum \limits_{-l' \leq i',j' \leq l'} \,\,\,\,\, \sum \limits_{-l'' \leq i'',j'' \leq l''} \,\,\,\,\,
\int \,
\sqrt{2l'+1} \sqrt{2l''+1}  \,\,\, \times
\\
\times \,\,\,
\overline{{u^{a}_{s'}}_{{}_{\,\,j'i'}}(\widehat{n'}\cdot \widehat{l'})} \,\,
\overline{{u^{b}_{s''}}_{{}_{\,\,j''i''}}(\widehat{n''}\cdot \widehat{l''})} \,\, 
\overline{\widehat{n'}(t)} \,\,
\overline{\widehat{l'}_{{}_{i'j'}}(\boldsymbol{w})} \,\,
\overline{\widehat{n''}(t)} \,\,
\overline{\widehat{l''}_{{}_{i''j''}}(\boldsymbol{w})}
\,
\phi(t, \boldsymbol{w}) \,\, \ud t \ud \boldsymbol{w}
\end{multline*}
\begin{multline}\label{kappa10.kappa10}
=
\sum \limits_{-l' \leq i',j' \leq l'} \,\,\,\,\, \sum \limits_{-l'' \leq i'',j'' \leq l''} \,\,\,\,\,
\sum\limits_{\substack{|l' -l''| \leq l \leq l'+l'' \\ -l \leq i,j \leq l}} \,\,\,\,
\sqrt{2l'+1} \sqrt{2l''+1}  \,\,\, \times
\\
\times \,\,\,
\overline{{u^{a}_{s'}}_{{}_{\,\,j'i'}}(\widehat{n'}\cdot \widehat{l'})} \,
\overline{{u^{b}_{s''}}_{{}_{\,\,j''i''}}(\widehat{n''}\cdot \widehat{l''})} \, 
\overline{\underset{\widehat{l'} \,\, i'j'}{M}_{{}_{\widehat{l''} \,\, i''j''}}^{{}^{\widehat{l} \,\, ij}}} 
\widetilde{\phi}_{{}_{ji}}(\widehat{n'+n''}\cdot \widehat{l}),
\end{multline}
where
\[
\overline{\widehat{n'}(t)} \,\,\overline{\widehat{n''}(t)} = \overline{\widehat{n'+n''}(t)}
\]
and where the pointwise product of the matrix elements of the irreducible representations $\widehat{l'}, \widehat{l''}$ of $SU(2, \mathbb{C})$  
is given by the known formula
\[
\widehat{l'}_{{}_{i'j'}}(\boldsymbol{w})\widehat{l''}_{{}_{i''j''}}(\boldsymbol{w}) = 
\sum\limits_{\substack{|l' -l''| \leq l \leq l'+l'' \\ -l \leq i,j \leq l}} 
\underset{\widehat{l'} \,\, i'j'}{M}_{{}_{\widehat{l''} \,\, i''j''}}^{{}^{\widehat{l} \,\, ij}}
\widehat{l}_{{}_{ij}}(\boldsymbol{w}), \,\,\,\,\,
\boldsymbol{w} \in SU(2, \mathbb{C}).
\]
Recall that the matrix element $\widehat{l'}\otimes \widehat{l''}_{{}_{i'i'' \,\,\,j'j''}}$ of the tensor product reperesentation 
$\widehat{l'}\otimes \widehat{l''}$ is equal to the above pointwise product
\[
\widehat{l'}\otimes \widehat{l''}_{{}_{i'i'' \,\,\,j'j''}}(\boldsymbol{w})
= \widehat{l'}_{{}_{i'j'}}(\boldsymbol{w})\widehat{l''}_{{}_{i''j''}}(\boldsymbol{w}).
\]
Therefore, each coefficient (with fixed indices $l',i',j',l'',i'',j'',l,i,j$)
\[
\underset{\widehat{l'} \,\, i'j'}{M}_{{}_{\widehat{l''} \,\, i''j''}}^{{}^{\widehat{l} \,\, ij}}
\]
is equal to the product of a matrix element of a unitary matrix by a matrix element of its inverse, where the unitary matrix is the one
which gives the unitary equivalence between
\[
\widehat{l'}\otimes \widehat{l''} \,\,\,
\textrm{and}
\,\,\,
\widehat{l'+l''} \oplus \widehat{l'+l''-1} \oplus \cdots \oplus \widehat{|l'-l''|},
\]
and the last direct sum is represented by the matrix with the matrix blocks $\widehat{l}_{{}_{ij}}$ arranged along the diagonal.
Therefore 
\begin{equation}\label{|M|<1}
\Bigg| \underset{\widehat{l'} \,\, i'j'}{M}_{{}_{\widehat{l''} \,\, i''j''}}^{{}^{\widehat{l} \,\, ij}} \Bigg| \leq 1.
\end{equation}

In order to prove continuity of the map
\[
\phi \longmapsto \kappa'_{1,0} \overset{\cdot}{\otimes} \kappa''_{1,0}(\phi)
\]
we estimate the the $q$-norm or more precisely the $q,q$-norm of the function $\kappa_{1,0} \overset{\cdot}{\otimes} \kappa_{1,0}(\phi)$
on $\mathscr{O}' \times \mathscr{O}''$.
Let $| \cdot |$ be the $L^2$ norm on $L^2(\mathscr{O}' \times \mathscr{O}''; \mathbb{C})$, with $A', A''$ the standard operators defining
the standard nuclear spaces $E_{{}_{'}}$ and $E_{{}_{''}}$. Recall that $A'$ is equal to the restriction of $\widetilde{A}= \widetilde{\Delta}+1$
to $\mathscr{O}'$ and respectively $A''$ is equal to the restriction of $\widetilde{A}= \widetilde{\Delta}+1$ to the orbit $\mathscr{O}''$.
Recall that $\widetilde{A}= \widetilde{\Delta}+1$ is an ordinary function on $\widehat{\widetilde{\mathbb{S}^1}\times G}$, $G= SU(2, \mathbb{C})$,
and that $\widetilde{A}= \widetilde{\Delta}+1$ is an ordinray function on $\mathscr{O}'$ or resp. on $\mathscr{O}''$:
\begin{multline*}
\widetilde{A}\xi(s', \widehat{n'}\cdot \widehat{l'}) = \big[\widetilde{\Delta}+1 \big]\xi(s', \widehat{n'}\cdot \widehat{l'})
= \lambda_{{}_{n',l'}} \xi(s', \widehat{n'}\cdot \widehat{l'}) 
\\
 \textrm{or resp.}
\\
\,\,\,
\widetilde{A}\xi(s', \widehat{n''}\cdot \widehat{l''}) = \big[\widetilde{\Delta}+1 \big]\xi(s'', \widehat{n''}\cdot \widehat{l''})
= \lambda_{{}_{n'',l''}}\xi(s'', \widehat{n''}\cdot \widehat{l''}),
\end{multline*}
where
\[
\lambda_{n,l} = {\textstyle\frac{n^2}{4}} + l(l+1) + {\textstyle\frac{1}{4}} +1,
\]
and where $\xi$ is a function on $\mathscr{O}'$ or resp. on $\mathscr{O}''$.
Thus by (\ref{kappa10.kappa10}) we have the following norm estimation
\begin{multline*}
\big|(A'^q \otimes A''^q ) \,\,  \kappa'_{1,0} \overset{\cdot}{\otimes} \kappa''_{1,0}(\phi) \big|^{2}
\\
=\big|(\widetilde{A}^q \otimes \widetilde{A}^q ) \,\,  \kappa'_{1,0} \overset{\cdot}{\otimes} \kappa''_{1,0}(\phi) \big|^{2}
= \big|\kappa_{1,0} \overset{\cdot}{\otimes} \kappa_{1,0}(\phi) \big|_{q}^{2}
\\
= \sum \limits_{\substack{(s', \widehat{n'}\cdot \widehat{l'}) \in  \mathscr{O}'\\ (s'', \widehat{n''}\cdot \widehat{l''}) \in  \mathscr{O}'' }}
 \Big[ \lambda_{{}_{n',l'}}^q \lambda_{{}_{n'',l''}}^q
\big|\kappa_{1,0} \overset{\cdot}{\otimes} \kappa_{1,0}(\phi) (s', \widehat{n'}\cdot \widehat{l'}, s'', \widehat{n''}\cdot \widehat{l''}) \big|
\Big]^2
\end{multline*}
\begin{multline*}
\leq
\sum \limits_{\substack{(s', \widehat{n'}\cdot \widehat{l'}) \in  \mathscr{O}'\\ (s'', \widehat{n''}\cdot \widehat{l''}) \in  \mathscr{O}'' }}
\Bigg[
\sum \limits_{\substack{-l' \leq i',j' \leq l' \\ -l'' \leq i'',j'' \leq l''}} \,\,\,\,\,
\sum\limits_{\substack{|l' -l''| \leq l \leq l'+l'' \\ -l \leq i,j \leq l}} \,\,\,\,
\sqrt{2l'+1} \sqrt{2l''+1}  \,\,\, \times
\\
\times \,\,\,
\lambda_{{}_{n',l'}}^q \lambda_{{}_{n'',l''}}^q
\Big|
\overline{{u^{a}_{s'}}_{{}_{\,\,j'i'}}(\widehat{n'}\cdot \widehat{l'})} \,
\overline{{u^{b}_{s''}}_{{}_{\,\,j''i''}}(\widehat{n''}\cdot \widehat{l''})} \, 
\overline{\underset{\widehat{l'} \,\, i'j'}{M}_{{}_{\widehat{l''} \,\, i''j''}}^{{}^{\widehat{l} \,\, ij}}} 
\widetilde{\phi}_{{}_{ji}}(\widehat{n'+n''}\cdot \widehat{l})
\Big|
\Bigg]^2
\end{multline*}
\begin{multline*}
\leq
\Bigg[
\sum \limits_{\substack{(s', \widehat{n'}\cdot \widehat{l'}) \in  \mathscr{O}'\\ (s'', \widehat{n''}\cdot \widehat{l''}) \in  \mathscr{O}'' }}
\sum \limits_{\substack{-l' \leq i',j' \leq l' \\ -l'' \leq i'',j'' \leq l''}} \,\,\,\,\,
\sum\limits_{\substack{|l' -l''| \leq l \leq l'+l'' \\ -l \leq i,j \leq l}} \,\,\,\,
\sqrt{2l'+1} \sqrt{2l''+1}  \,\,\, \times
\\
\times \,\,\,
\lambda_{{}_{n',l'}}^q \lambda_{{}_{n'',l''}}^q
\Big|
\overline{{u^{a}_{s'}}_{{}_{\,\,j'i'}}(\widehat{n'}\cdot \widehat{l'})} \,
\overline{{u^{b}_{s''}}_{{}_{\,\,j''i''}}(\widehat{n''}\cdot \widehat{l''})} \, 
\overline{\underset{\widehat{l'} \,\, i'j'}{M}_{{}_{\widehat{l''} \,\, i''j''}}^{{}^{\widehat{l} \,\, ij}}} 
\widetilde{\phi}_{{}_{ji}}(\widehat{n'+n''}\cdot \widehat{l})
\Big|
\Bigg]^2
\end{multline*}
\begin{multline*}
=
\Bigg[
\sum \limits_{\substack{(s', \widehat{n'}\cdot \widehat{l'}) \in  \mathscr{O}'\\ (s'', \widehat{n''}\cdot \widehat{l''}) \in  \mathscr{O}'' }}
\sum \limits_{\substack{-l' \leq i',j' \leq l' \\ -l'' \leq i'',j'' \leq l''}} \,\,\,\,\,
\sum\limits_{\substack{|l' -l''| \leq l \leq l'+l'' \\ -l \leq i,j \leq l}} \,\,\,\,
\sqrt{2l'+1} \sqrt{2l''+1}  \,\,\, \times
\\
\times \,\,\,
{\textstyle\frac{\lambda_{{}_{n',l'}}^q}{\lambda_{{}_{n'+n'',l}}^r}}
{\textstyle\frac{\lambda_{{}_{n'',l''}}^q}{\lambda_{{}_{n'+n'',l}}^r}}
\Big|
\overline{{u^{a}_{s'}}_{{}_{\,\,j'i'}}(\widehat{n'}\cdot \widehat{l'})} \,
\overline{{u^{b}_{s''}}_{{}_{\,\,j''i''}}(\widehat{n''}\cdot \widehat{l''})} \, 
\overline{\underset{\widehat{l'} \,\, i'j'}{M}_{{}_{\widehat{l''} \,\, i''j''}}^{{}^{\widehat{l} \,\, ij}}} 
\lambda_{{}_{n'+n'',l}}^{2r}
\widetilde{\phi}_{{}_{ji}}(\widehat{n'+n''}\cdot \widehat{l})
\Big|
\Bigg]^2
\end{multline*}
valid for any $r\in \mathbb{N}$.
Now the summation 
\[
\sum \limits_{\substack{(s', \widehat{n'}\cdot \widehat{l'}) \in  \mathscr{O}'\\ (s'', \widehat{n''}\cdot \widehat{l''}) \in  \mathscr{O}'' }}
\]
in the last formula we perform in two stages. First we perform the subsum with respect to all $(s', \widehat{n'}\cdot \widehat{l'}) \in  \mathscr{O}'$
and $(s'', \widehat{n''}\cdot \widehat{l''}) \in  \mathscr{O}''$ for which $n'+n''$ has fixed value $n= n'+n''$, and then perform summation with respect
to all $n$ (positive in case the fields are positive energy fields, or negative, in case both considered fields are negative energy fields). 
Note that if the covering numbers $n_{{}_{\mathscr{O'}_\pm}}$, $n_{{}_{\mathscr{O''}_\pm}}$ of the orbits $\mathscr{O'}_\pm$, $\mathscr{O''}_\pm$
(compare Definition \ref{K_0onEU}) are both equal $1$, \emph{i.e.} the energy quantum numbers $n'$ and respectively $n''$ \footnote{In perfect analogy to the Minkowski space-time case, where on the orbit $\mathscr{O}_{\pm m,0}$ or $\mathscr{O}_{\pm 1,0,0,1}$, $p_{0}$ is a function $p_{0}(|\boldsymbol{\p}|)$ 
of the absolute value $|\boldsymbol{\p}|$ of spatial momentum.} on the orbits $\mathscr{O'}_{\pm}$ and $\mathscr{O''}_{\pm}$ are functions
of the total momentum quantum number $l'(l'+1)$ resp. $l''(l''+1)$ (or simply of $l'$ or $l''$), then the sum
\[
\sum \limits_{\substack{(s', \widehat{n'}\cdot \widehat{l'}) \in  \mathscr{O}'\\ (s'', \widehat{n''}\cdot \widehat{l''}) \in  \mathscr{O}'' 
\\ n'+n''=n}} \ldots =
\sum \limits_{\substack{\widehat{n'}\cdot \widehat{l'} \in  \mathscr{O}'_\pm\\ \widehat{n''}\cdot \widehat{l''} \in  \mathscr{O}''_\pm 
\\ n'+n''=n}} \,\,\,\,
\sum \limits_{\substack{1 \leq s' \leq d^{(','')}(\widehat{n'}\cdot \widehat{l'}) \\ 1 \leq s'' \leq d^{(','')}(\widehat{n''}\cdot \widehat{l''})}} \ldots
\]
contains at most $|n|= |n'+n''|$ terms for each fixed pair $(s',s'')$ of indices $s',s''$. This maximum value is reached if both fields
are massless with orbits $\mathscr{O'}_{\pm}$ and $\mathscr{O''}_{\pm}$ lying precisely on the ``light cones'' in the momentum space,
compare the pictures placed above in this Subsection. Of course in case the covering numbers $n_{{}_{\mathscr{O'}_\pm}}$, $n_{{}_{\mathscr{O''}_\pm}}$
are not equal $1$ we have at most $n_{{}_{\mathscr{O'}_\pm}} n_{{}_{\mathscr{O''}_\pm}}|n|= n_{{}_{\mathscr{O'}_\pm}} n_{{}_{\mathscr{O''}_\pm}}|n'+n''|$
terms in this sum
\[
\sum \limits_{\substack{(s', \widehat{n'}\cdot \widehat{l'}) \in  \mathscr{O}'\\ (s'', \widehat{n''}\cdot \widehat{l''}) \in  \mathscr{O}'' 
\\ n'+n''=n}},
\]
for each fixed pair $(s',s'')$ of indices $s',s''$. Recall that
\begin{eqnarray*}
1 \leq s' \leq d^{(','')}(\widehat{n'}\cdot \widehat{l'}) \leq (2l'+1)^2 \textrm{dim} \, V', 
\\
1 \leq s'' \leq d^{(','')}(\widehat{n''}\cdot \widehat{l''}) \leq (2l''+1)^2 \textrm{dim} \, V''
\end{eqnarray*}
where $V'$ and $V''$ are finite dimensional representations of $SU(2, \mathbb{C})$ associated to the considered fields
and defined by the kernels, $\kappa'_{\ell,m}, \kappa''_{\ell,m}$, $\ell,m = 0,1$, and that $d^{(','')}(\widehat{n'}\cdot \widehat{l'})$
is equal to the dimension function $d'(\widehat{n''}\cdot \widehat{l''})$ for $\mathscr{O}'_+$ or $d''(\widehat{n'}\cdot \widehat{l'})$
for $\mathscr{O}'_-$ and similarly for $d^{(','')}(\widehat{n''}\cdot \widehat{l''})$, introduced above.
Therefore taking onto account the estimations
(\ref{|u^a_s(n.l)|<(sqrt{2l+1})^-1}) and (\ref{|M|<1}) 
we arrive at the following inequality for each fixed $l,i,j$:
\begin{multline*}
\sum \limits_{\substack{(s', \widehat{n'}\cdot \widehat{l'}) \in  \mathscr{O}'\\ (s'', \widehat{n''}\cdot \widehat{l''}) \in  \mathscr{O}''
\\ n'+n''=n }}
\sum \limits_{\substack{-l' \leq i',j' \leq l' \\ -l'' \leq i'',j'' \leq l''}} \,\,
\sqrt{2l'+1} \sqrt{2l''+1}  \,\,\, \times
\\
\times \,\,\,
{\textstyle\frac{\lambda_{{}_{n',l'}}^q}{\lambda_{{}_{n'+n'',l}}^r}}
{\textstyle\frac{\lambda_{{}_{n'',l''}}^q}{\lambda_{{}_{n'+n'',l}}^r}}
\Big|
\overline{{u^{a}_{s'}}_{{}_{\,\,j'i'}}(\widehat{n'}\cdot \widehat{l'})} \,
\overline{{u^{b}_{s''}}_{{}_{\,\,j''i''}}(\widehat{n''}\cdot \widehat{l''})} \, 
\overline{\underset{\widehat{l'} \,\, i'j'}{M}_{{}_{\widehat{l''} \,\, i''j''}}^{{}^{\widehat{l} \,\, ij}}} 
\Big|
\end{multline*}
\begin{multline*}
=
\sum \limits_{\substack{\widehat{n'}\cdot \widehat{l'} \in  \mathscr{O}'_\pm\\ \widehat{n''}\cdot \widehat{l''} \in  \mathscr{O}''_\pm 
\\ n'+n''=n}} \,\,\,\,
\sum \limits_{\substack{1 \leq s' \leq (2l'+1)^2 \textrm{dim} \, V' \\ 1 \leq s'' \leq (2l''+1)^2 \textrm{dim} \, V''}}
\sum \limits_{\substack{-l' \leq i',j' \leq l' \\ -l'' \leq i'',j'' \leq l''}} \,\,
\sqrt{2l'+1} \sqrt{2l''+1}  \,\,\, \times
\\
\times \,\,\,
{\textstyle\frac{\lambda_{{}_{n',l'}}^q}{\lambda_{{}_{n'+n'',l}}^r}}
{\textstyle\frac{\lambda_{{}_{n'',l''}}^q}{\lambda_{{}_{n'+n'',l}}^r}}
\Big|
\overline{{u^{a}_{s'}}_{{}_{\,\,j'i'}}(\widehat{n'}\cdot \widehat{l'})} \,
\overline{{u^{b}_{s''}}_{{}_{\,\,j''i''}}(\widehat{n''}\cdot \widehat{l''})} \, 
\overline{\underset{\widehat{l'} \,\, i'j'}{M}_{{}_{\widehat{l''} \,\, i''j''}}^{{}^{\widehat{l} \,\, ij}}} 
\Big|
\end{multline*}
\begin{multline*}
\leq
\textrm{dim} \, V' \textrm{dim} \, V'' \,
\sum \limits_{\substack{\widehat{n'}\cdot \widehat{l'} \in  \mathscr{O}'_\pm\\ \widehat{n''}\cdot \widehat{l''} \in  \mathscr{O}''_\pm \\ n'+n''=n }}
{\textstyle\frac{\lambda_{{}_{n',l'}}^q(2l'+1)^4}{\lambda_{{}_{n'+n'',l}}^r}}
{\textstyle\frac{\lambda_{{}_{n'',l''}}^q(2l''+1)^4}{\lambda_{{}_{n'+n'',l}}^r}}
\\
\leq
\textrm{dim} \, V' \textrm{dim} \, V'' \,
\sum \limits_{\substack{\widehat{n'}\cdot \widehat{l'} \in  \mathscr{O}'_\pm\\ \widehat{n''}\cdot \widehat{l''} \in  \mathscr{O}''_\pm \\ n'+n''=n }}
{\textstyle\frac{\lambda_{{}_{n',l'}}^{q+4}}{\lambda_{{}_{n'+n'',l}}^r}}
{\textstyle\frac{\lambda_{{}_{n'',l''}}^{q+4}}{\lambda_{{}_{n'+n'',l}}^r}}
\\
=
\textrm{dim} \, V' \textrm{dim} \, V'' \,
\sum \limits_{\substack{\widehat{n'}\cdot \widehat{l'} \in  \mathscr{O}'_\pm\\ \widehat{n''}\cdot \widehat{l''} \in  \mathscr{O}''_\pm \\ n'+n''=n }}
{\textstyle\frac{\lambda_{{}_{n',l'}}^{q+4}}{\lambda_{{}_{n,l}}^r}}
{\textstyle\frac{\lambda_{{}_{n'',l''}}^{q+4}}{\lambda_{{}_{n,l}}^r}}.
\end{multline*}
Because each orbit $\mathscr{O'}_{\pm}$ or $\mathscr{O''}_\pm$ lies (at least asymtotically) within the positive (resp. negative)
energy cone in the momentum space (compare the pictures placed above in this Subsection), then for $r$ sufficiently greater than $q$
\[
{\textstyle\frac{\lambda_{{}_{n',l'}}^{q+4}}{\lambda_{{}_{n'+n'',l}}^r}}= {\textstyle\frac{\lambda_{{}_{n',l'}}^{q+4}}{\lambda_{{}_{n,l}}^r}} \leq 1,
\,\,\,
{\textstyle\frac{\lambda_{{}_{n'',l''}}^{q+4}}{\lambda_{{}_{n'+n'',l}}^r}} = {\textstyle\frac{\lambda_{{}_{n'',l''}}^{q+4}}{\lambda_{{}_{n,l}}^r}}  \leq 1,
\]
and because the last sum contains at most $n_{{}_{\mathscr{O'}_\pm}} n_{{}_{\mathscr{O''}_\pm}}|n|= n_{{}_{\mathscr{O'}_\pm}} n_{{}_{\mathscr{O''}_\pm}}|n'+n''|$
terms, then we can alaways choose such a natural $r$ (e.g. $r=2q+5$), that 
\[
\textrm{dim} \, V' \textrm{dim} \, V'' 
\sum \limits_{\substack{\widehat{n'}\cdot \widehat{l'} \in  \mathscr{O}'_\pm\\ \widehat{n''}\cdot \widehat{l''} \in  \mathscr{O}''_\pm \\ n'+n''=n }}
{\textstyle\frac{\lambda_{{}_{n',l'}}^{q+4}}{\lambda_{{}_{n,l}}^r}}
{\textstyle\frac{\lambda_{{}_{n'',l''}}^{q+4}}{\lambda_{{}_{n,l}}^r}}
\leq
\textrm{dim} \, V' \textrm{dim} \, V' \, n_{{}_{\mathscr{O'}_\pm}} n_{{}_{\mathscr{O''}_\pm}}.
\]

Thus for sufficiently large $r$ and depending only on $q$ we obtain the following upper bound estimation for each fixed $l,i,j$ and $n$
\begin{multline}\label{sum(n'+n''=n)}
\sum \limits_{\substack{(s', \widehat{n'}\cdot \widehat{l'}) \in  \mathscr{O}'\\ (s'', \widehat{n''}\cdot \widehat{l''}) \in  \mathscr{O}''
\\ n'+n''=n }}
\sum \limits_{\substack{-l' \leq i',j' \leq l' \\ -l'' \leq i'',j'' \leq l''}} \,\,
\sqrt{2l'+1} \sqrt{2l''+1}  \,\,\, \times
\\
\times \,\,\,
{\textstyle\frac{\lambda_{{}_{n',l'}}^q}{\lambda_{{}_{n'+n'',l}}^r}}
{\textstyle\frac{\lambda_{{}_{n'',l''}}^q}{\lambda_{{}_{n'+n'',l}}^r}}
\Big|
\overline{{u^{a}_{s'}}_{{}_{\,\,j'i'}}(\widehat{n'}\cdot \widehat{l'})} \,
\overline{{u^{b}_{s''}}_{{}_{\,\,j''i''}}(\widehat{n''}\cdot \widehat{l''})} \, 
\overline{\underset{\widehat{l'} \,\, i'j'}{M}_{{}_{\widehat{l''} \,\, i''j''}}^{{}^{\widehat{l} \,\, ij}}} 
\Big|
\\
\leq
\textrm{dim} \, V' \textrm{dim} \, V' \, n_{{}_{\mathscr{O'}_\pm}} n_{{}_{\mathscr{O''}_\pm}},
\end{multline}
independent of  $l,i,j$ and $n$.

Therefore for sufficiently large $r$, e.g. $r>2q+5$
\begin{multline*}
\big|(A'^q \otimes A''^q ) \,\,  \kappa'_{1,0} \overset{\cdot}{\otimes} \kappa''_{1,0}(\phi) \big|^{2}
\\
=\big|\kappa'_{1,0} \overset{\cdot}{\otimes} \kappa''_{1,0}(\phi) \big|_{q}^{2}
\\
\leq
\big[\textrm{dim} \, V' \textrm{dim} \, V'' \, n_{{}_{\mathscr{O'}_\pm}} n_{{}_{\mathscr{O''}_\pm}}\big]^2
\Big[ 
\sum \limits_{\substack{n\in \mathbb{Z} \\ l \in \mathbb{N} + {\textstyle\frac{1}{2}} \mathbb{N}
\\ -l \leq i,j \leq l}} 
\big|\lambda_{{}_{n,l}}^{2r}\widetilde{\phi}_{{}_{ji}}(\widehat{n}\cdot \widehat{l})\big|
\Big]^2
\end{multline*}
\begin{multline*}
=
\big[\textrm{dim} \, V' \textrm{dim} \, V'' \, n_{{}_{\mathscr{O'}_\pm}} n_{{}_{\mathscr{O''}_\pm}}\big]^2
\Big[ 
\sum \limits_{\substack{n\in \mathbb{Z} \\ l \in \mathbb{N} + {\textstyle\frac{1}{2}} \mathbb{N}
\\ -l \leq i,j \leq l}} 
(2l+1)
\big|{\textstyle\frac{\overline{\lambda_{{}_{n,l}}^{-r'}}}{2l+1}} \lambda_{{}_{n,l}}^{2r+r'}\widetilde{\phi}_{{}_{ji}}(\widehat{n}\cdot \widehat{l})\big|
\Big]^2
\end{multline*}
\begin{multline}\label{|kappa10.kappa10|q<|phi|2r+r'}
= 
\big[\textrm{dim} \, V' \textrm{dim} \, V'' \, n_{{}_{\mathscr{O'}_\pm}} n_{{}_{\mathscr{O''}_\pm}}\big]^2
\Big|\Big\langle \widetilde{\xi}, \widetilde{A^{2r+r'}\phi}\Big\rangle \Big|^2 \leq 
\\
\big[\textrm{dim} \, V' \textrm{dim} \, V'' \, n_{{}_{\mathscr{O'}_\pm}} n_{{}_{\mathscr{O''}_\pm}}\big]^2
|\xi|^2 |\phi|_{2r+r'}^{2},
\end{multline}
where $\xi \in L^2(\widetilde{\mathbb{S}^1}\times SU(2, \mathbb{C}); \mathbb{C})$
whose Fourier transform matrix coefficients at the point $\widehat{n}\cdot \widehat{l} \in \widehat{\widetilde{\mathbb{S}^1}\times G}$,
$G= SU(2, \mathbb{C})$, are equal
\[
\widetilde{\xi}_{{}_{ij}}(\widehat{n}\cdot \widehat{l}) = {\textstyle\frac{\lambda_{{}_{n,l}}^{-r'}}{2l+1}},
\,\,\, -l \leq i,j \leq l,
\]
which for sufficiently large natural $r'$ indeed $\widetilde{\xi} \in \mathscr{F} L^2(\widetilde{\mathbb{S}^1}\times G; \mathbb{C})
= L^2(\widehat{\widetilde{\mathbb{S}^1}\times G}; \mathbb{C})$. In the last line of the formulae
(\ref{|kappa10.kappa10|q<|phi|2r+r'}) $|\cdot |$ and $|\cdot|_{p}$ is the $L^2$-norm 
in $L^2(\widetilde{\mathbb{S}^1}\times G; \mathbb{C})$ and the $|\cdot|_{p}$-norm in 
$\mathscr{E} \subset L^2(\widetilde{\mathbb{S}^1}\times G; \mathbb{C})$. Recall that the sign $|\cdot|$
in the first line of (\ref{|kappa10.kappa10|q<|phi|2r+r'}) denotes the $L^2$-norm in 
\[
L^2(\mathscr{O}'\times\mathscr{O}';\mathbb{C})
\supset E_{{}_{'}} \otimes E_{{}_{''}}.
\]
Recall that we have identified the standard nuclear spaces with their non-standard counterparts:
\begin{eqnarray*}
L^2(\mathscr{O}';\mathbb{C}) \supset \mathcal{S}_{A'}(\mathscr{O}';\mathbb{C}) \cong E_{{}_{'}}, \,\,\,\,
L^2(\mathscr{O}'';\mathbb{C}) \supset \mathcal{S}_{A''}(\mathscr{O}'';\mathbb{C}) \cong E_{{}_{''}},
\end{eqnarray*}
\begin{multline*}
L^2(\mathscr{O}';\mathbb{C}) \otimes L^2(\mathscr{O}'';\mathbb{C}) \cong_{U} 
L^2(\mathscr{O}'\times \mathscr{O}'';\mathbb{C}) \supset \mathcal{S}_{A'\otimes A''}(\mathscr{O}' \times \mathscr{O}';\mathbb{C}) 
\\
\cong \mathcal{S}_{A'}(\mathscr{O}';\mathbb{C}) \otimes  \mathcal{S}_{A''}(\mathscr{O}'';\mathbb{C}) \cong E_{{}_{'}} \otimes E_{{}_{''}}, 
\end{multline*}
and here the tensor product $\otimes$ applied to Hilbert spaces denotes the Hilbert space tensor product and, when applied to the nuclear spaces,
$\otimes$ denotes the projective (equivalent here to the equicotinuous) tensor product, compare \cite{obataJFA} or Subsection \ref{white-setup}.

The inequality (\ref{|kappa10.kappa10|q<|phi|2r+r'}) proves continuity of the map
\[
\mathscr{E} \ni \phi \longmapsto \kappa'_{1,0} \overset{\cdot}{\otimes} \kappa''_{1,0}(\phi) \in E_{{}_{'}} \otimes E_{{}_{''}}
\]
and thus assertion 1).

Ad 2). Let, for example, $\kappa'_{1,0}$ and $\kappa''_{0,1}$ be the negative frequency kernel of a neutral field and 
the positive frequency kernel of another neutral field, and let $\phi$ be as in 1). Analogously we have the formula
\begin{multline*}
\kappa_{1,0} \overset{\cdot}{\otimes} \kappa_{0,1}(\phi)(s', \widehat{n'}\cdot \widehat{l'}, s'', \widehat{n''}\cdot \widehat{l''}; a, b) 
\\
=
\int 
\kappa'_{1,0} (s', \widehat{n'}\cdot \widehat{l'}; a,x) \kappa''_{0,1}(s'', \widehat{n''}\cdot \widehat{l''}; b, x) \phi(x) \ud^4 x
\end{multline*}
\begin{multline*}
=
\sum \limits_{-l' \leq i',j' \leq l'} \,\,\,\,\, \sum \limits_{-l'' \leq i'',j'' \leq l''} \,\,\,\,\,
\sum\limits_{\substack{|l' -l''| \leq l \leq l'+l'' \\ -l \leq i,j \leq l}} \,\,\,\,
\sqrt{2l'+1} \sqrt{2l''+1}  \,\,\, \times
\\
\times \,\,\,
\overline{{u^{a}_{s'}}_{{}_{\,\,j'i'}}(\widehat{n'}\cdot \widehat{l'})} \,
{u^{b}_{s''}}_{{}_{\,\,j''i''}}(\widehat{n''}\cdot \widehat{l''}) \, 
\overline{\underset{\widehat{l'}, \, i', j'}{M}_{{}_{\widehat{l''} \,\, -i''-j''}}^{{}^{\widehat{l} \,\, ij}}} 
\widetilde{\phi}_{{}_{ji}}(\widehat{n'-n''}\cdot \widehat{l}).
\end{multline*}

In order to show the continuity of the map
\[
\mathscr{E} \ni \phi \longmapsto \kappa'_{1,0} \overset{\cdot}{\otimes} \kappa''_{0,1}(\phi) \in  E_{{}_{'}} \otimes E_{{}_{''}}^{*}
\]
we estimate the $(q,-p)$-norm of the function $\kappa'_{1,0} \overset{\cdot}{\otimes} \kappa''_{0,1}(\phi)$
on $\mathscr{O}' \times \mathscr{O}''$, with $p,q \in \mathbb{N}$. We have the following analogous inequality
\begin{multline*}
\big|(A'^q \otimes A''^{-p} ) \,\,  \kappa'_{1,0} \overset{\cdot}{\otimes} \kappa''_{0,1}(\phi) \big|^{2}
\\
=\big|(\widetilde{A}^q \otimes \widetilde{A}^{-p} ) \,\,  \kappa'_{1,0} \overset{\cdot}{\otimes} \kappa''_{0,1}(\phi) \big|^{2}
= \big|\kappa_{1,0} \overset{\cdot}{\otimes} \kappa_{0,1}(\phi) \big|_{q,-p}^{2}
\end{multline*}
\begin{multline}\label{|kappa10.kappa01(phi)|q,-p|}
\leq 
\Bigg[
\sum \limits_{\substack{(s', \widehat{n'}\cdot \widehat{l'}) \in  \mathscr{O}'\\ (s'', \widehat{n''}\cdot \widehat{l''}) \in  \mathscr{O}'' }} \,\,\,
\sum \limits_{\substack{-l' \leq i',j' \leq l' \\ -l'' \leq i'',j'' \leq l''}} \,\,\,\,\,
\sum\limits_{\substack{|l' -l''| \leq l \leq l'+l'' \\ -l \leq i,j \leq l}} \,\,\,\,
\sqrt{2l'+1} \sqrt{2l''+1}  \,\,\, \times
\\
\times \,\,\,'
{\textstyle\frac{\lambda_{{}_{n',l'}}^q}{\lambda_{{}_{n'-n'',l}}^r}}
{\textstyle\frac{1}{\lambda_{{}_{n'',l''}}^p}}
\Big|
\overline{{u^{a}_{s'}}_{{}_{\,\,j'i'}}(\widehat{n'}\cdot \widehat{l'})} \,
{u^{b}_{s''}}_{{}_{\,\,j''i''}}(\widehat{n''}\cdot \widehat{l''}) \, 
\overline{\underset{\widehat{l'} \,\, i'j'}{M}_{{}_{\widehat{l''} \,\, -i''-j''}}^{{}^{\widehat{l} \,\, ij}}} 
\lambda_{{}_{n'-n'',l}}^{r}
\widetilde{\phi}_{{}_{ji}}(\widehat{n'-n''}\cdot \widehat{l})
\Big|
\Bigg]^2
\end{multline}
Analogously the summation 
\[
\sum \limits_{\substack{(s', \widehat{n'}\cdot \widehat{l'}) \in  \mathscr{O}'\\ (s'', \widehat{n''}\cdot \widehat{l''}) \in  \mathscr{O}'' }}
\]
in the last formula we perform in two stages. First we perform the subsum with respect to all $(s', \widehat{n'}\cdot \widehat{l'}) \in  \mathscr{O}'$
and $(s'', \widehat{n''}\cdot \widehat{l''}) \in  \mathscr{O}''$ for which $n'-n''$ has fixed value $n= n'-n''$, and then perform summation with respect
to $n$. That is we perform first
\[
\sum \limits_{\substack{(s', \widehat{n'}\cdot \widehat{l'}) \in  \mathscr{O}'\\ (s'', \widehat{n''}\cdot \widehat{l''}) \in  \mathscr{O}'' 
\\ n'-n''=n}} \ldots
=
\sum \limits_{\substack{\widehat{n'}\cdot \widehat{l'} \in  \mathscr{O}'_\pm\\ \widehat{n''}\cdot \widehat{l''} \in  \mathscr{O}''_\pm 
\\ n'-n''=n}} \,\,\,\,
\sum \limits_{\substack{1 \leq s' \leq d^{(','')}(\widehat{n'}\cdot \widehat{l'}) \\ 1 \leq s'' \leq d^{(','')}(\widehat{n''}\cdot \widehat{l''})}} \ldots 
\,\, .
\]

Analogously as in 1) we have the following inequalty for each fixed $l,i,j$:
\begin{multline*}
\sum \limits_{\substack{(s', \widehat{n'}\cdot \widehat{l'}) \in  \mathscr{O}'\\ (s'', \widehat{n''}\cdot \widehat{l''}) \in  \mathscr{O}''
\\ n'-n''=n }}
\sum \limits_{\substack{-l' \leq i',j' \leq l' \\ -l'' \leq i'',j'' \leq l''}} \,\,
\sqrt{2l'+1} \sqrt{2l''+1}  \,\,\, \times
\\
\times \,\,\,
{\textstyle\frac{\lambda_{{}_{n',l'}}^q}{\lambda_{{}_{n'-n'',l}}^r}}
{\textstyle\frac{1}{\lambda_{{}_{n'',l''}}^p}}
\Big|
\overline{{u^{a}_{s'}}_{{}_{\,\,j'i'}}(\widehat{n'}\cdot \widehat{l'})} \,
{u^{b}_{s''}}_{{}_{\,\,j''i''}}(\widehat{n''}\cdot \widehat{l''}) \, 
\overline{\underset{\widehat{l'} \,\, i'j'}{M}_{{}_{\widehat{l''} \,\, -i''-j''}}^{{}^{\widehat{l} \,\, ij}}} 
\Big|
\end{multline*}
\begin{multline*}
=
\sum \limits_{\substack{\widehat{n'}\cdot \widehat{l'} \in  \mathscr{O}'_\pm\\ \widehat{n''}\cdot \widehat{l''} \in  \mathscr{O}''_\pm 
\\ n'-n''=n}} \,\,\,\,
\sum \limits_{\substack{1 \leq s' \leq (2l'+1)^2 \textrm{dim} \, V' \\ 1 \leq s'' \leq (2l''+1)^2 \textrm{dim} \, V''}}
\sum \limits_{\substack{-l' \leq i',j' \leq l' \\ -l'' \leq i'',j'' \leq l''}} \,\,
\sqrt{2l'+1} \sqrt{2l''+1}  \,\,\, \times
\\
\times \,\,\,
{\textstyle\frac{\lambda_{{}_{n',l'}}^q}{\lambda_{{}_{n'-n'',l}}^r}}
{\textstyle\frac{1}{\lambda_{{}_{n'',l''}}^p}}
\Big|
\overline{{u^{a}_{s'}}_{{}_{\,\,j'i'}}(\widehat{n'}\cdot \widehat{l'})} \,
{u^{b}_{s''}}_{{}_{\,\,j''i''}}(\widehat{n''}\cdot \widehat{l''}) \, 
\overline{\underset{\widehat{l'} \,\, i'j'}{M}_{{}_{\widehat{l''} \,\, -i''-j''}}^{{}^{\widehat{l} \,\, ij}}} 
\Big|
\end{multline*}
\[
\leq
\textrm{dim} \, V' \textrm{dim} \, V'' \,
\sum \limits_{\substack{\widehat{n'}\cdot \widehat{l'} \in  \mathscr{O}'_\pm\\ \widehat{n''}\cdot \widehat{l''} \in  \mathscr{O}''_\pm \\ n'-n''=n }}
{\textstyle\frac{\lambda_{{}_{n',l'}}^q(2l'+1)^4}{\lambda_{{}_{n'-n'',l}}^r}}
{\textstyle\frac{(2l''+1)^4}{\lambda_{{}_{n'',l''}}^p}}
\]
\begin{multline*}
\leq
\textrm{dim} \, V' \textrm{dim} \, V'' \,
\sum \limits_{\substack{\widehat{n'}\cdot \widehat{l'} \in  \mathscr{O}'_\pm\\ \widehat{n''}\cdot \widehat{l''} \in  \mathscr{O}''_\pm \\ n'-n''=n }}
{\textstyle\frac{\lambda_{{}_{n',l'}}^{q+4}}{\lambda_{{}_{n'-n'',l}}^r}}
{\textstyle\frac{\lambda_{{}_{n'',l''}}^{4}}{\lambda_{{}_{n'',l''}}^p}}
\\
=
\textrm{dim} \, V' \textrm{dim} \, V'' \,
\sum \limits_{\substack{\widehat{n'}\cdot \widehat{l'} \in  \mathscr{O}'_\pm\\ \widehat{n''}\cdot \widehat{l''} \in  \mathscr{O}''_\pm \\ n'-n''=n }}
{\textstyle\frac{\lambda_{{}_{n',l'}}^{q+4}}{\lambda_{{}_{n'-n'',l}}^r}}
{\textstyle\frac{1}{\lambda_{{}_{n'',l''}}^{p-4}}}.
\end{multline*}

Now it is not difficult to see that in case both covering numbers\footnote{In this case numbers $n'$ and $n''$ are well defined functions of $l'$ resp. $l''$ on the respective orbits $\mathscr{O}'_\pm$, $\mathscr{O}''_\pm$.} $n_{{}_{\mathscr{O'}_\pm}}, n_{{}_{\mathscr{O''}_\pm}}$ are equal one, for each $q \in \mathbb{N}$
we can choose such $p,r \in \mathbb{N}$ that the sum 
\[
\sum \limits_{\substack{\widehat{n'}\cdot \widehat{l'} \in  \mathscr{O}'_\pm\\ \widehat{n''}\cdot \widehat{l''} \in  \mathscr{O}''_\pm \\ n'-n''=n }}
{\textstyle\frac{\lambda_{{}_{n',l'}}^{q+4}}{\lambda_{{}_{n'-n'',l}}^r}}
{\textstyle\frac{1}{\lambda_{{}_{n'',l''}}^{p-4}}} \leq C
\]
is convergent and bounded by a constant $C$ independent of $n$. In case $n_{{}_{\mathscr{O'}_\pm}} \neq 1, n_{{}_{\mathscr{O''}_\pm}} \neq 1$
we have
\[
\sum \limits_{\substack{\widehat{n'}\cdot \widehat{l'} \in  \mathscr{O}'_\pm \\ \widehat{n''}\cdot \widehat{l''} \in  \mathscr{O}''_\pm \\ n'-n''=n }}
{\textstyle\frac{\lambda_{{}_{n',l'}}^{q+4}}{\lambda_{{}_{n'-n'',l}}^r}}
{\textstyle\frac{1}{\lambda_{{}_{n'',l''}}^{p-4}}} \leq C n_{{}_{\mathscr{O'}_\pm}} n_{{}_{\mathscr{O''}_\pm}}.
\]

Therefore  for each $q \in \mathbb{N}$ there exist $r,p \in \mathbb{N}$ such that
\begin{multline}\label{sum(n'-n''=n)}
\sum \limits_{\substack{(s', \widehat{n'}\cdot \widehat{l'}) \in  \mathscr{O}'\\ (s'', \widehat{n''}\cdot \widehat{l''}) \in  \mathscr{O}''
\\ n'-n''=n }}
\sum \limits_{\substack{-l' \leq i',j' \leq l' \\ -l'' \leq i'',j'' \leq l''}} \,\,
\sqrt{2l'+1} \sqrt{2l''+1}  \,\,\, \times
\\
\times \,\,\,
{\textstyle\frac{\lambda_{{}_{n',l'}}^q}{\lambda_{{}_{n'-n'',l}}^r}}
{\textstyle\frac{1}{\lambda_{{}_{n'',l''}}^p}}
\Big|
\overline{{u^{a}_{s'}}_{{}_{\,\,j'i'}}(\widehat{n'}\cdot \widehat{l'})} \,
{u^{b}_{s''}}_{{}_{\,\,j''i''}}(\widehat{n''}\cdot \widehat{l''}) \, 
\overline{\underset{\widehat{l'} \,\, i'j'}{M}_{{}_{\widehat{l''} \,\, -i''-j''}}^{{}^{\widehat{l} \,\, ij}}} 
\Big|
\\
\leq
C \, \textrm{dim} \, V' \textrm{dim} \, V'' \, n_{{}_{\mathscr{O'}_\pm}} n_{{}_{\mathscr{O''}_\pm}}
\end{multline}
for all $l,i,j,$ and $n$.
Thus from (\ref{|kappa10.kappa01(phi)|q,-p|}) and (\ref{sum(n'-n''=n)}) it follows 
that for each $q\in \mathbb{N}$ there exist $p,r \in \mathbb{N}$ such that the inequality holds
\begin{multline}\label{q,-p-normOfkappa10.kappa01(phi)<C|phi|r}
\big|(A'^q \otimes A''^{-p} ) \,\,  \kappa'_{1,0} \overset{\cdot}{\otimes} \kappa''_{0,1}(\phi) \big|^{2}
\\
=\big|(\widetilde{A}^q \otimes \widetilde{A}^{-p} ) \,\,  \kappa'_{1,0} \overset{\cdot}{\otimes} \kappa''_{0,1}(\phi) \big|^{2}
= \big|\kappa_{1,0} \overset{\cdot}{\otimes} \kappa_{0,1}(\phi) \big|_{q,-p}^{2}
\\
\leq
\big[C \, \textrm{dim} \, V' \textrm{dim} \, V'' \, n_{{}_{\mathscr{O'}_\pm}} n_{{}_{\mathscr{O''}_\pm}}\big]^2
\Big[ 
\sum \limits_{\substack{n\in \mathbb{Z} \\ l \in \mathbb{N} + {\textstyle\frac{1}{2}} \mathbb{N}
\\ -l \leq i,j \leq l}} 
\big|\lambda_{{}_{n,l}}^{r}\widetilde{\phi}_{{}_{ji}}(\widehat{n}\cdot \widehat{l})\big|
\Big]^2
\\
=
 \big[C \, \textrm{dim} \, V' \textrm{dim} \, V'' \, n_{{}_{\mathscr{O'}_\pm}} n_{{}_{\mathscr{O''}_\pm}}\big]^2
\Big[ 
\sum \limits_{\substack{n\in \mathbb{Z} \\ l \in \mathbb{N} + {\textstyle\frac{1}{2}} \mathbb{N}
\\ -l \leq i,j \leq l}} 
(2l+1)
\big|{\textstyle\frac{\overline{\lambda_{{}_{n,l}}^{-r'}}}{2l+1}} \lambda_{{}_{n,l}}^{r+r'}\widetilde{\phi}_{{}_{ji}}(\widehat{n}\cdot \widehat{l})\big|
\Big]^2
\\
= 
\big[C \, \textrm{dim} \, V' \textrm{dim} \, V'' \, n_{{}_{\mathscr{O'}_\pm}} n_{{}_{\mathscr{O''}_\pm}}\big]^2
\Big|\Big\langle \widetilde{\xi}, \widetilde{A^{r+r'}\phi}\Big\rangle \Big|^2 \leq 
\\
\big[C \, \textrm{dim} \, V' \textrm{dim} \, V'' \, n_{{}_{\mathscr{O'}_\pm}} n_{{}_{\mathscr{O''}_\pm}}\big]^2
|\xi|^2 |\phi|_{r+r'}^{2},
\end{multline}
for all sufficiently large $r'\in\mathbb{N}$.
Here $\xi \in L^2(\widetilde{\mathbb{S}^1}\times SU(2, \mathbb{C}); \mathbb{C})$ is the function defined in the proof of 1) and
whose Fourier transform matrix coefficients at the point $\widehat{n}\cdot \widehat{l} \in \widehat{\widetilde{\mathbb{S}^1}\times G}$,
$G= SU(2, \mathbb{C})$, are equal
\[
\widetilde{\xi}_{{}_{ij}}(\widehat{n}\cdot \widehat{l}) = {\textstyle\frac{\lambda_{{}_{n,l}}^{-r'}}{2l+1}},
\,\,\, -l \leq i,j \leq l,
\]
and which for sufficiently large natural $r'$ indeed  $\widetilde{\xi} \in \mathscr{F} L^2(\widetilde{\mathbb{S}^1}\times G; \mathbb{C})
= L^2(\widehat{\widetilde{\mathbb{S}^1}\times G}; \mathbb{C})$.

Recall, please, that we have identified the standard nuclear spaces with their non-standard counterparts:
\begin{eqnarray*}
L^2(\mathscr{O}';\mathbb{C}) \supset \mathcal{S}_{A'}(\mathscr{O}';\mathbb{C}) \cong E_{{}_{'}}, \,\,\,\,
L^2(\mathscr{O}'';\mathbb{C}) \supset \mathcal{S}_{A''}(\mathscr{O}'';\mathbb{C}) \cong E_{{}_{''}},
\end{eqnarray*}
and thus we have made also the identification
\[
\mathcal{S}_{A'}(\mathscr{O}';\mathbb{C}) \otimes  \mathcal{S}_{A''}(\mathscr{O}'';\mathbb{C})^* \cong E_{{}_{'}} \otimes E_{{}_{''}}^{*}. 
\]

Note that the inequality (\ref{q,-p-normOfkappa10.kappa01(phi)<C|phi|r}) proves the continuity of the map
\[
\mathscr{E} \ni \phi \longmapsto \kappa'_{1,0} \overset{\cdot}{\otimes} \kappa''_{0,1}(\phi) \in  E_{{}_{'}} \otimes E_{{}_{''}}^{*}.
\]
Indeed, recall that the 0-neighborhood basis in the inductive limit $E_{{}_{''}}^{*}$:
\[
E_{{}_{'' \,-1}} \subset E_{{}_{'' \,-2}} \subset \ldots \subset E_{{}_{'' \,-p}} \subset \ldots \subset E_{{}_{''}}^{*} 
\]
is given by the convex and ballanced hulls of the sums \cite{Schaefer}:
\[
\bigcup_{p \in \mathbb{N}} K''_{-p}
\]
where each $K''_{-p}$ is any non-empty open ball in the Hilbert space $E_{{}_{'' \,-p}} \subset L^2(\mathscr{O}'';\mathbb{C})$ centered at zero.
On the other hand the 0-neighborhood basis in the projective (coinciding with equicontinuous) tensor product
$E_{{}_{'}} \otimes E_{{}_{''}}^{*}$ of the nuclear spaces $E_{{}_{'}}, E_{{}_{''}}^{*}$ is given by the convex ballanced hulls of
\cite{treves}:
\[
U_{\alpha} \otimes W_{\beta}
\]
where $U_{\alpha}$ ranges over the 0-neighborhood basis in $E_{{}_{'}}$ and $W_\beta$ ranges over the 0-neighborhood basis
in $E_{{}_{''}}^{*}$. From this and from the inequality (\ref{q,-p-normOfkappa10.kappa01(phi)<C|phi|r}) the said continuity follows.

Ad 3). The case of more than just two factors can be proved in the the same manner as 1) and 2).

Ad 4). Replacement of the kernels $\kappa_{1,0}, \ldots$ with their derivations $(X)^\alpha \kappa_{1,0}= (X^0)^{\alpha_0} \ldots (X^3)^{\alpha_3}
\kappa_{1,0}, \ldots$ brings no essentially new difficulties in the proof presented for the kernels themselves. This is because the operators
$X^1, X^2, X^3$ act on the matrix of the irreducible representation $\widehat{l}$ as multiplication by the matrix,
respectively, equal (\ref{tildeX_1}), (\ref{tildeX_2}), (\ref{tildeX_3}) of preceding Subsection, with non-zero matrix coefficients
given by 
\[
\alpha_m = \sqrt{(l+m)(l-m+1)}, \,\,\, -l \leq m \leq l,
\]
and 
\[
iX^0 \, \widehat{n} = {\textstyle\frac{n}{2}} \, \widehat{n}.
\]

\qed

Now repeating the method presented in the proof of the last Proposition and Lemma \ref{kappaBarDotOtimeskappaOnEU} 
we will prove a Lemma which, among other things, will allow us to show that
on the Einstein Universe the operation of pointwise multiplication by the step theta function applied to the kernel, is a well defined operation 
and represents operation acting within the space of integral kernel operators of a subclass of the class $\mathscr{L}(\mathscr{E}, \mathscr{L}((\boldsymbol{E}), (\boldsymbol{E})))$. In fact we will show this for the free field operators and for the 
Wick products of free fields and their derivations which contain at most one massless, or more generally,
at most one free field with infinite orbit $\mathscr{O}_\pm$ as a factor in the Wick monomial, 
to which we apply the pointwise operation of multiplication by the theta function. 

Let us denote the operation of pointwise multiplication (with resect to space-time variable) by the step-theta function $\theta$ by a kernel $\kappa_{\ell,m}$
simply by 
\[
\theta \kappa_{\ell,m}.
\]
Namely if the kernel $\kappa_{\ell, m}$ is defined by a function
\[
\kappa_{\ell,m}(s, \widehat{n}\cdot\widehat{l}, \ldots, s', \widehat{n'}\cdot\widehat{l'}; a,x)
= \kappa_{\ell,m}(s, \widehat{n}\cdot\widehat{l}, \ldots, s', \widehat{n'}\cdot\widehat{l'}; a,t,\boldsymbol{w}), \,\,\,
x = (t, \boldsymbol{w}),
\]
with $\ell+m$ pairs $s, \widehat{n}\cdot\widehat{l}, \ldots, s', \widehat{n'}\cdot\widehat{l'}$ of variables, and one space-time variable
$x = (t, \boldsymbol{w})$, then $\theta \kappa_{\ell,m}$ is defined by 
\[
\theta(x) \kappa_{\ell,m}(s, \widehat{n}\cdot\widehat{l}, \ldots, s', \widehat{n'}\cdot\widehat{l'}; a,x)
= \theta(t)\kappa_{\ell,m}(s, \widehat{n}\cdot\widehat{l}, \ldots, s', \widehat{n'}\cdot\widehat{l'}; a,t,\boldsymbol{w}), \,\,\,
x = (t, \boldsymbol{w}).
\]
Recall that the class of kernels $\kappa_{\ell,m}$ we are considering embraces the plane-wave kernels $\kappa_{0,1}, \kappa_{1,0} \in \mathfrak{K}_0$
defining free fields, or their derivations $(X)^\alpha\kappa_{0,1}, (X)^\alpha \kappa_{1,0}$, and their pointwise products (with respect to the space-time variable)
\[
\kappa_{1,0} \overset{\cdot}{\otimes} \kappa_{0,1}, \ldots
\]
(eventually symmetrized/antisymmetrized, compare Subsection \ref{OperationsOnXi}, which correspond to the Wick products of the free fields
or their derivations), which indeed are regular and defined by ordinary functions, so that the operation of pointwise multiplication of the kernel
of this class by the step theta function $\theta$ is well defined as the operation performed upon functions representing these kernels.
In the following Lemma we show that the operation of pointwise multiplication by $\theta$ do not lead us out of the class of distributions
defined in the last Lemma \ref{kappaBarDotOtimeskappaOnEU}. 

\begin{lem}
Let $\kappa_{\ell, m}, \dots, \kappa'_{\ell, m}, \kappa''_{\ell, m}, \ldots, \kappa'''_{\ell, m}   \in \mathfrak{K}_0$, $\ell,m = 1,0$, be the kernels 
of free fields on the Einstein Universe (all positive or all negative-energy fields)
with the corresponding single particle Gelfand triples 
\[
\begin{array}{ccccc} E & \subset & \mathcal{H} & \subset & E^* 
\\ \vdots&&\vdots&& \vdots \\ 
 E_{{}_{'}} & \subset & \mathcal{H}' & \subset & E_{{}_{'}}^*
 \\  
E_{{}_{''}} & \subset & \mathcal{H}'' & \subset & E_{{}_{''}}^*
\\\vdots &&\vdots&&\vdots  \\  
E_{{}_{'''}} & \subset & \mathcal{H}''' & \subset & E_{{}_{'''}}^*.
\end{array} 
\]
Let $\mathscr{E}=\mathcal{S}_{{}_{\Delta +1}}(\widetilde{\mathbb{S}^1} \times SU(2, \mathbb{C}); \mathbb{C})$
or respectively  $\mathscr{E}=\mathcal{S}_{{}_{\Delta +1}}(\widetilde{\mathbb{S}^1} \times SU(2, \mathbb{C}); \mathbb{C}^d)$.
Then 
\begin{enumerate}
\item[1)]
\[
\theta \kappa'_{1,0}, \theta \kappa'_{0,1} \in  \mathscr{L}(\mathscr{E}, E_{{}_{'}}) \cong
\mathscr{L}(E_{{}_{'}}^*, \mathscr{E}^*) \subset \mathscr{L}(E_{{}_{'}}, \mathscr{E}^*) .
\]
\item[2)]
\[
\theta \kappa'_{1, 0} \overset{\cdot}{\otimes} \kappa''_{1, 0} \in  \mathscr{L}(\mathscr{E}, E_{{}_{'}}^{*} \otimes E_{{}_{''}}^{*}) \cong
\mathscr{L}(E_{{}_{'}} \otimes E_{{}_{''}}, \mathscr{E}^*).
\]
\item[3)]
\[
\theta \kappa'_{1, 0} \overset{\cdot}{\otimes} \kappa''_{0, 1} \in  \mathscr{L}(\mathscr{E}, E_{{}_{'}}^{*} \otimes E_{{}_{''}}^*) \cong
\mathscr{L}(E_{{}_{'}} \otimes E_{{}_{''}}, \mathscr{E}^*).
\]
\item[4)]
\begin{multline*}
\theta \kappa_{1, 0} \overset{\cdot}{\otimes} \cdots \overset{\cdot}{\otimes} \kappa'_{1, 0} \overset{\cdot}{\otimes} 
\kappa''_{0, 1}\overset{\cdot}{\otimes} \cdots \overset{\cdot}{\otimes} \kappa'''_{0, 1} 
\in  \mathscr{L}(\mathscr{E}, E_{{}_{}}^{*} \otimes \cdots \otimes E_{{}_{'}}^{*} \otimes E_{{}_{''}}^* \otimes \cdots \otimes E_{{}_{'''}}^*) 
\\
\cong
\mathscr{L}(E_{{}_{}} \otimes \cdots \otimes E_{{}_{'}} \otimes E_{{}_{''}} \otimes \cdots \otimes E_{{}_{'''}}, \mathscr{E}^*).
\end{multline*}
\item[5)]
\[
\theta \kappa'_{1,0}(\phi), \theta \kappa'_{0,1}(\phi)
\]
are uniformly bounded in $E_{{}_{'}}$ and
\begin{align*}
\theta \kappa'_{1, 0} \overset{\cdot}{\otimes} \kappa''_{1, 0}(\phi), 
\,\, \theta \kappa'_{1, 0} \overset{\cdot}{\otimes} \kappa''_{0, 1}(\phi),
\\
\theta \kappa_{1, 0} \overset{\cdot}{\otimes} \cdots \overset{\cdot}{\otimes} \kappa'_{1, 0} \overset{\cdot}{\otimes} 
\kappa''_{0, 1}\overset{\cdot}{\otimes} \cdots \overset{\cdot}{\otimes} \kappa'''_{0, 1} (\phi)
\end{align*}
are uniformly bounded, respectively in $E_{{}_{'}}^{*} \otimes E_{{}_{''}}^{*}$, $E_{{}_{'}}^{*} \otimes E_{{}_{''}}^*$,
or in 
\[
E_{{}_{}}^{*} \otimes \cdots \otimes E_{{}_{'}}^{*} \otimes E_{{}_{''}}^* \otimes \cdots \otimes E_{{}_{'''}}^*,
\]
whenever $\phi$ ranges over any bounded set in $\mathscr{E}$.
\item[6)] The above statements 1)-5) remain valid if we replace the kernels $\kappa_{1,0}, \ldots$ with their derivations
$(X)^\alpha \kappa_{1,0}, \ldots$.
\end{enumerate}
\label{theta.kappaBarDotOtimeskappaOnEU}
\end{lem}

\qedsymbol \, 
Ad 1). Let, for example, $\kappa'_{1,0}$ be the negative frequency kernel of a neutral field.
Let $\phi \in \mathscr{E}=\mathcal{S}_{{}_{\Delta +1}}(\widetilde{\mathbb{S}^1} \times SU(2, \mathbb{C}); \mathbb{C}^d)
=\mathscr{C}^\infty(\widetilde{\mathbb{S}^1} \times SU(2, \mathbb{C}); \mathbb{C}^d)$ 
 and let $x=t\times \boldsymbol{w} \in \mathbb{R} \times SU(2, \mathbb{C})$. By definition and by the formula 
(\ref{int(theta)x(phi)n^l^})
\begin{multline*}
\theta \kappa'_{1,0}(\phi)(s', \widehat{n'}\cdot \widehat{l'})
= \sum \limits_{1 \leq a\leq d} \,\, \int \limits_{\widetilde{\mathbb{S}^1}\times SU(2, \mathbb{C})} 
\theta(t) \kappa'_{1,0}(s', \widehat{n'}\cdot \widehat{l'}; a, t, \boldsymbol{w})
\, \phi(t, \boldsymbol{w}) \, \ud t \ud \boldsymbol{w}
\\
= 
\sum \limits_{\substack{1 \leq a \leq d \\ n \in \mathbb{Z} \\ -l' \leq i',j' \leq l'}} 
\sqrt{2l'+1} \overline{{u_{s'}^{a}}_{{}_{ \, j'i'}}(\widehat{n'}\cdot \widehat{l'})} \widehat{\theta}_{n} 
\widetilde{\phi^{a}}_{{}_{j'i'}}(\widehat{n'-n} \cdot \widehat{l'}).
\end{multline*}
Let $A'$ be the standard operator (equal to $\widetilde{\Delta} +1 = \widetilde{A}$ restricted to $\mathscr{O}'$) on
\[
L^2(\mathscr{O}'; \mathbb{C}) \supset \mathcal{S}_{A'}(\mathscr{O}'; \mathbb{C}) \cong E_{{}_{'}}
\]
defining the nuclear space $E_{{}_{'}}$, which we identify with its standard counterpart $\mathcal{S}_{A'}(\mathscr{O}'; \mathbb{C})$.
We have the following estimation for the $|\cdot|_{q}$-norm of the function $\theta \kappa'_{1,0}(\phi)$
on $\mathscr{O}'$:
\begin{multline*}
\big|\theta \kappa'_{1,0}(\phi) \big|_{q}^{2} = \big|A'^q \theta \kappa'_{1,0}(\phi) \big|^2
\leq
\sum \limits_{\substack{\widehat{n'}\cdot \widehat{l'} \in \mathscr{O}'_\pm \\ 1 \leq s' \leq d^{(','')}(\widehat{n'}\cdot \widehat{l'} )}} 
\lambda_{n',l'}^{2q} \big|\theta \kappa_{1,0}(\phi)(s', \widehat{n'}\cdot \widehat{l'}) \Big|^2
\\
\leq
\sum \limits_{\substack{\widehat{n'}\cdot \widehat{l'} \in \mathscr{O}'_\pm \\ 1 \leq s' \leq d^{(','')}(\widehat{n'}\cdot \widehat{l'} )}} 
\Bigg[
\sum \limits_{\substack{1 \leq a \leq d \\ n \in \mathbb{Z} \\ -l' \leq i',j' \leq l'}} 
\lambda_{n',l'}^{q} \,
\sqrt{2l'+1} \big| \overline{{u_{s'}^{a}}_{{}_{ \, j'i'}}(\widehat{n'}\cdot \widehat{l'})} \widehat{\theta}_{n} 
\widetilde{\phi^{a}}_{{}_{j'i'}}(\widehat{n'-n} \cdot \widehat{l'}) \big|
\Bigg]^2
\\
\leq 
\Bigg[
\sum \limits_{\substack{\widehat{n'}\cdot \widehat{l'} \in \mathscr{O}'_\pm \\ 1 \leq s' \leq d^{(','')}(\widehat{n'}\cdot \widehat{l'} )}} 
\sum \limits_{\substack{1 \leq a \leq d \\ n \in \mathbb{Z} \\ -l' \leq i',j' \leq l'}} 
\lambda_{n',l'}^{q} \,
\sqrt{2l'+1} \big| \overline{{u_{s'}^{a}}_{{}_{ \, j'i'}}(\widehat{n'}\cdot \widehat{l'})} \widehat{\theta}_{n} 
\widetilde{\phi^{a}}_{{}_{j'i'}}(\widehat{n'-n} \cdot \widehat{l'}) \big|
\Bigg]^2
\\
= 
\Bigg[
\sum \limits_{\substack{\widehat{n'}\cdot \widehat{l'} \in \mathscr{O}'_\pm}} 
\sum \limits_{\substack{1 \leq s' \leq d^{(','')}(\widehat{n'}\cdot \widehat{l'}) \\ 1 \leq a \leq d \\ n \in \mathbb{Z} \\ -l' \leq i',j' \leq l'}} 
\lambda_{n',l'}^{q} \,
\sqrt{2l'+1} \big| \overline{{u_{s'}^{a}}_{{}_{ \, j'i'}}(\widehat{n}\cdot \widehat{l'})} \widehat{\theta}_{n} 
\widetilde{\phi^{a}}_{{}_{j'i'}}(\widehat{n'-n} \cdot \widehat{l'}) \big|
\Bigg]^2
\end{multline*}

Next, exactly as in the proof of Lemma \ref{retOfPairings}, we divide the orbit $\mathscr{O}'_\pm$ of the considered
(positive or negative energy) field into (at most) $n_{{}_{\mathscr{O}'_\pm}}$ disjoint subsets
\[
\mathscr{O}'_\pm = \mathscr{O}_{\pm \, 1} \sqcup \mathscr{O}'_{\pm \, 2} \sqcup \ldots \sqcup \mathscr{O}'_{\pm \, n_{{}_{\mathscr{O}'_\pm}}}, 
\]
on each of which there exist at most one $\widehat{n'} \cdot \widehat{l'} \in \mathscr{O}'_{\pm \, k}$ for each $l'$.
Correspondingly to each subset
\[
\mathscr{O}'_{\pm \, 1}, \mathscr{O}'_{\pm \, 2}, \ldots, \mathscr{O}'_{\pm \, n_{{}_{\mathscr{O}'_\pm}}}
\]
we divide the sum 
\[
\sum \limits_{\substack{\widehat{n'}\cdot \widehat{l'} \in \mathscr{O}'_\pm}} 
\]
into $n_{{}_{\mathscr{O}_\pm}}$ subsums
\[
\sum' \limits_{\substack{\widehat{n}\cdot \widehat{l} \in \mathscr{O}_\pm}} = 
\sum \limits_{\substack{\widehat{n'}\cdot \widehat{l'} \in \mathscr{O}'_{\pm \, k} }} 
\]
for $k = 1, \ldots, n_{{}_{\mathscr{O}'_\pm}}$, and on each of the suborbits $\mathscr{O}'_{\pm \, k}$, $k = 1,2, \ldots,  n_{{}_{\mathscr{O}'_\pm}}$
we define the auxiliary functions $\zeta$ and $\xi$ exactly as in the proof of the last Proposition,
multiplied by 
\[
{\textstyle\frac{1}{\sqrt{2l'+1}}}
\]
at each point $\widehat{n'}\cdot \widehat{l'} \in \widehat{\widetilde{\mathbb{S}^1}\times G}$, $G = SU(2, \mathbb{C})$.
Therefore, for each of the $n_{{}_{\mathscr{O}'_\pm}}$ subsums we obtain the following estimation
\begin{multline*}
\Bigg[
\sum' \limits_{\substack{\widehat{n'}\cdot \widehat{l'} \in \mathscr{O}'_\pm}} \,\,\,
\sum \limits_{\substack{1 \leq s' \leq d^{(','')}(\widehat{n'}\cdot
 \widehat{l'}) \\ 1 \leq a \leq d \\ n \in \mathbb{Z} \\ -l' \leq i',j' \leq l'}} 
\lambda_{n',l'}^{q} \,
\sqrt{2l'+1} \big| \overline{{u_{s'}^{a}}_{{}_{ \, j'i'}}(\widehat{n'}\cdot \widehat{l'})} \widehat{\theta}_{n} 
\widetilde{\phi^{a}}_{{}_{j'i'}}(\widehat{n'-n} \cdot \widehat{l'}) \big|
\Bigg]^2 
\\
\leq 
\Big| \Big\langle \xi, \widetilde{A^{q+r} \phi} \Big\rangle \Big|^2
\leq |\xi|^2 \Big| \widetilde{A^{q+r} \phi} \Big|^2
=|\xi|^2 \Big| \widetilde{A^{q+r} \phi} \Big|^2
=|\xi|^2 \big| A^{q+r} \phi \big|^{2} 
=|\xi|^2 \big| \phi \big|_{q+r}^{2}. 
\end{multline*}
Denoting the auxiliary functions $\zeta, \xi$ corresponding to the suborbit $\mathscr{O}_{\pm \, k}$
by $\zeta_k, \xi_k$, we obtain the estimation
\[
\big|\theta \kappa_{1,0}(\phi) \big|_{q}^{2} = \big|A^q \theta \kappa_{1,0}(\phi) \big|^2
\leq 
\big[|\xi_1|^2 + \ldots + |\xi_{n_{{}_{\mathscr{O}'_\pm}}}|^2 \big] \big| \phi \big|_{q+r}^{2}
\]
because we have only finite number $n_{{}_{\mathscr{O}'_\pm}}$  of such suborits and corresponding subsums. From the last inequality
continuity of the map
\[
\mathscr{E} \ni \phi \longmapsto \kappa_{1,0}(\phi) \in E_{{}_{'}}
\]
and assertion 1) follows.

Ad 2). We proceed as in the proof of Lemma \ref{kappaBarDotOtimeskappaOnEU}, 1), 2). 
Let, for example, $\kappa'_{1,0}, \kappa''_{1,0}$ be the negative frequency kernels of two neutral fields. 
Let $\phi \in \mathscr{E}=\mathcal{S}_{{}_{\Delta +1}}(\widetilde{\mathbb{S}^1} \times SU(2, \mathbb{C}); \mathbb{C})
=\mathscr{C}^\infty(\widetilde{\mathbb{S}^1} \times SU(2, \mathbb{C}); \mathbb{C})$ 
(of $\mathbb{C}$-valued space-time test functions) and let $x=t\times \boldsymbol{w} \in \mathbb{R} \times SU(2, \mathbb{C})$.
By definition and by the formula (\ref{int(theta)x(phi)n^l^})
\begin{multline}\label{theta.kappa10.kappa10}
\theta \kappa'_{1,0} \overset{\cdot}{\otimes} \kappa''_{1,0}(\phi)(s', \widehat{n'}\cdot \widehat{l'}, s'', \widehat{n''}\cdot \widehat{l''}; a, b) 
\\
=
\int \,
\theta(x) \kappa'_{1,0} (s', \widehat{n'}\cdot \widehat{l'}; a,x) \kappa''_{1,0}(s'', \widehat{n''}\cdot \widehat{l''}; b, x) \phi(x) \ud^4 x
\\
=
\sum \limits_{-l' \leq i',j' \leq l'} \,\,\,\,\, \sum \limits_{-l'' \leq i'',j'' \leq l''} \,\,\,\,\,
\int \,
\sqrt{2l'+1} \sqrt{2l''+1}  \,\,\, \times
\\
\times \,\,\,
\overline{{u^{a}_{s'}}_{{}_{\,\,j'i'}}(\widehat{n'}\cdot \widehat{l'})} \,\,
\overline{{u^{b}_{s''}}_{{}_{\,\,j''i''}}(\widehat{n''}\cdot \widehat{l''})} \,\, 
\overline{\widehat{n'}(t)} \,\,
\overline{\widehat{l'}_{{}_{i'j'}}(\boldsymbol{w})} \,\,
\overline{\widehat{n''}(t)} \,\,
\overline{\widehat{l''}_{{}_{i''j''}}(\boldsymbol{w})}
\, \theta(t) \,
\phi(t, \boldsymbol{w}) \,\, \ud t \ud \boldsymbol{w}
\\
=
\sum \limits_{n''' \in \mathbb{Z}} \,\,
\sum \limits_{\substack{-l' \leq i',j' \leq l' \\ -l'' \leq i'',j'' \leq l''}} 
\,\,\,\,\,
\sum\limits_{\substack{|l' -l''| \leq l \leq l'+l'' \\ -l \leq i,j \leq l}} \,\,\,\,
\sqrt{2l'+1} \sqrt{2l''+1}  \,\,\, \times
\\
\times \,\,\,
\overline{{u^{a}_{s'}}_{{}_{\,\,j'i'}}(\widehat{n'}\cdot \widehat{l'})} \,
\overline{{u^{b}_{s''}}_{{}_{\,\,j''i''}}(\widehat{n''}\cdot \widehat{l''})} \, 
\overline{\underset{\widehat{l'} \,\, i'j'}{M}_{{}_{\widehat{l''} \,\, i''j''}}^{{}^{\widehat{l} \,\, ij}}} 
\widehat{\theta}_{n'''}
\widetilde{\phi}_{{}_{ji}}(\widehat{n'+n''-n'''}\cdot \widehat{l}),
\end{multline}
Thus by (\ref{theta.kappa10.kappa10}) and for $q \in \mathbb{N}$ we have the following $|\cdot|_{-q}$-norm estimation
\begin{multline*}
\big|\theta \kappa_{1,0} \overset{\cdot}{\otimes} \kappa_{1,0}(\phi) \big|_{-q}^{2} =
\big|(A'^{-q} \otimes A''^{-q} ) \,\,  \kappa'_{1,0} \overset{\cdot}{\otimes} \kappa''_{1,0}(\phi) \big|^{2}
\\
=\big|(\widetilde{A}^{-q} \otimes \widetilde{A}^{-q} ) \,\, \theta \kappa'_{1,0} \overset{\cdot}{\otimes} \kappa''_{1,0}(\phi) \big|^{2}
\\
= \sum \limits_{\substack{(s', \widehat{n'}\cdot \widehat{l'}) \in  \mathscr{O}'\\ (s'', \widehat{n''}\cdot \widehat{l''}) \in  \mathscr{O}'' }}
 \Big[ \lambda_{{}_{n',l'}}^{-q} \lambda_{{}_{n'',l''}}^{-q}
\big|\theta \kappa_{1,0} \overset{\cdot}{\otimes} \kappa_{1,0}(\phi) (s', \widehat{n'}\cdot \widehat{l'}, s'', \widehat{n''}\cdot \widehat{l''}) \big|
\Big]^2
\end{multline*}
\begin{multline*}
\leq
\sum \limits_{\substack{(s', \widehat{n'}\cdot \widehat{l'}) \in  \mathscr{O}'\\ (s'', \widehat{n''}\cdot \widehat{l''}) \in  \mathscr{O}'' }}
\Bigg[
\sum \limits_{n''' \in \mathbb{Z}} \,\,
\sum \limits_{\substack{-l' \leq i',j' \leq l' \\ -l'' \leq i'',j'' \leq l''}} \,\,\,\,\,
\sum\limits_{\substack{|l' -l''| \leq l \leq l'+l'' \\ -l \leq i,j \leq l}} \,\,\,\,
\sqrt{2l'+1} \sqrt{2l''+1}  \,\,\, \times
\\
\times \,\,\,
\lambda_{{}_{n',l'}}^{-q} \lambda_{{}_{n'',l''}}^{-q}
\Big|
\overline{{u^{a}_{s'}}_{{}_{\,\,j'i'}}(\widehat{n'}\cdot \widehat{l'})} \,
\overline{{u^{b}_{s''}}_{{}_{\,\,j''i''}}(\widehat{n''}\cdot \widehat{l''})} \, 
\overline{\underset{\widehat{l'} \,\, i'j'}{M}_{{}_{\widehat{l''} \,\, i''j''}}^{{}^{\widehat{l} \,\, ij}}} 
\widehat{\theta}_{n'''} \widetilde{\phi}_{{}_{ji}}(\widehat{n'+n''-n'''}\cdot \widehat{l})
\Big|
\Bigg]^2
\end{multline*}
\begin{multline*}
\leq
\Bigg[
\sum \limits_{\substack{(s', \widehat{n'}\cdot \widehat{l'}) \in  \mathscr{O}'\\ (s'', \widehat{n''}\cdot \widehat{l''}) \in  \mathscr{O}'' }} \,\,
\sum \limits_{n''' \in \mathbb{Z}} \,\,
\sum \limits_{\substack{-l' \leq i',j' \leq l' \\ -l'' \leq i'',j'' \leq l''}} \,\,\,\,\,
\sum\limits_{\substack{|l' -l''| \leq l \leq l'+l'' \\ -l \leq i,j \leq l}} \,\,\,\,
\sqrt{2l'+1} \sqrt{2l''+1}  \,\,\, \times
\\
\times \,\,\,
\lambda_{{}_{n',l'}}^{-q} \lambda_{{}_{n'',l''}}^{-q}
\Big|
\overline{{u^{a}_{s'}}_{{}_{\,\,j'i'}}(\widehat{n'}\cdot \widehat{l'})} \,
\overline{{u^{b}_{s''}}_{{}_{\,\,j''i''}}(\widehat{n''}\cdot \widehat{l''})} \, 
\overline{\underset{\widehat{l'} \,\, i'j'}{M}_{{}_{\widehat{l''} \,\, i''j''}}^{{}^{\widehat{l} \,\, ij}}} 
\widehat{\theta}_{n'''} \widetilde{\phi}_{{}_{ji}}(\widehat{n'+n''-n'''}\cdot \widehat{l})
\Big|
\Bigg]^2
\end{multline*}
\begin{multline*}
=
\Bigg[
\sum \limits_{\substack{(s', \widehat{n'}\cdot \widehat{l'}) \in  \mathscr{O}'\\ (s'', \widehat{n''}\cdot \widehat{l''}) \in  \mathscr{O}''
\\ n''' \in \mathbb{Z} }} \,\,
\sum \limits_{\substack{-l' \leq i',j' \leq l' \\ -l'' \leq i'',j'' \leq l''}} \,\,\,\,\,
\sum\limits_{\substack{|l' -l''| \leq l \leq l'+l'' \\ -l \leq i,j \leq l}} 
{\textstyle\frac{\sqrt{2l'+1}}{\lambda_{{}_{n',l'}}^q \lambda_{{}_{n'+n''-n,l}}^r}}
{\textstyle\frac{\sqrt{2l''+1}}{\lambda_{{}_{n'',l''}}^q \lambda_{{}_{n'+n''-n,l}}^r}} \,\, \times
\\
\times \,\,\,
\Big|
\overline{{u^{a}_{s'}}_{{}_{\,\,j'i'}}(\widehat{n'}\cdot \widehat{l'})} \,
\overline{{u^{b}_{s''}}_{{}_{\,\,j''i''}}(\widehat{n''}\cdot \widehat{l''})} \, 
\overline{\underset{\widehat{l'} \,\, i'j'}{M}_{{}_{\widehat{l''} \,\, i''j''}}^{{}^{\widehat{l} \,\, ij}}} 
\lambda_{{}_{n'+n''-n,l}}^{2r}
\widehat{\theta}_{n'''} \widetilde{\phi}_{{}_{ji}}(\widehat{n'+n''-n'''}\cdot \widehat{l})
\Big|
\Bigg]^2
\end{multline*}

For fixed $n'+n''-n'''=n$ and $l,i,j$, the series (smilarly as in the proof of Lemma \ref{kappaBarDotOtimeskappaOnEU}, 1), 2)):
\begin{multline*}
\sum \limits_{\substack{(s', \widehat{n'}\cdot \widehat{l'}) \in  \mathscr{O}'\\ (s'', \widehat{n''}\cdot \widehat{l''}) \in  \mathscr{O}''
\\ n''' \in \mathbb{Z} \\ n'+n''-n'''=n }} \,\,
\sum \limits_{\substack{-l' \leq i',j' \leq l' \\ -l'' \leq i'',j'' \leq l''}} 
{\textstyle\frac{\sqrt{2l'+1}}{\lambda_{{}_{n',l'}}^q \lambda_{{}_{n'+n''-n''',l}}^r}}
{\textstyle\frac{\sqrt{2l''+1}}{\lambda_{{}_{n'',l''}}^q \lambda_{{}_{n'+n''-n''',l}}^r}} \,\, \times
\\
\times \,\,\,
\Big|
\overline{{u^{a}_{s'}}_{{}_{\,\,j'i'}}(\widehat{n'}\cdot \widehat{l'})} \,
\overline{{u^{b}_{s''}}_{{}_{\,\,j''i''}}(\widehat{n''}\cdot \widehat{l''})} \, 
\overline{\underset{\widehat{l'} \,\, i'j'}{M}_{{}_{\widehat{l''} \,\, i''j''}}^{{}^{\widehat{l} \,\, ij}}} 
\Big|
\end{multline*}
\begin{multline*}
=
\sum \limits_{\substack{ \widehat{n'}\cdot \widehat{l'} \in  \mathscr{O}'_\pm \\ \widehat{n''}\cdot \widehat{l''} \in  \mathscr{O}''_\pm 
\\ n''' \in \mathbb{Z} \\ n'+n''-n'''=n }} \,\,
\sum \limits_{\substack{1 \leq s' \leq d^{(','')}(\widehat{n'}\cdot \widehat{l'}) \\ 1 \leq s'' \leq d^{(','')}(\widehat{n''}\cdot \widehat{l''}) }} \,\,
\sum \limits_{\substack{-l' \leq i',j' \leq l' \\ -l'' \leq i'',j'' \leq l''}} 
{\textstyle\frac{\sqrt{2l'+1}}{\lambda_{{}_{n',l'}}^q \lambda_{{}_{n'+n''-n''',l}}^r}}
{\textstyle\frac{\sqrt{2l''+1}}{\lambda_{{}_{n'',l''}}^q \lambda_{{}_{n'+n''-n''',l}}^r}} \,\, \times
\\
\times \,\,\,
\Big|
\overline{{u^{a}_{s'}}_{{}_{\,\,j'i'}}(\widehat{n'}\cdot \widehat{l'})} \,
\overline{{u^{b}_{s''}}_{{}_{\,\,j''i''}}(\widehat{n''}\cdot \widehat{l''})} \, 
\overline{\underset{\widehat{l'} \,\, i'j'}{M}_{{}_{\widehat{l''} \,\, i''j''}}^{{}^{\widehat{l} \,\, ij}}} 
\Big|
\end{multline*}
\begin{multline*}
\leq 
\sum \limits_{\substack{ \widehat{n'}\cdot \widehat{l'} \in  \mathscr{O}'_\pm \\ \widehat{n''}\cdot \widehat{l''} \in  \mathscr{O}''_\pm 
\\ n''' \in \mathbb{Z} \\ n'+n''-n'''=n }} \,\,
{\textstyle\frac{(2l'+1)^4}{\lambda_{{}_{n',l'}}^q \lambda_{{}_{n'+n''-n''',l}}^r}}
{\textstyle\frac{ (2l''+1)^4}{\lambda_{{}_{n'',l''}}^q \lambda_{{}_{n'+n''-n''',l}}^r}}
\\
\leq 
\sum \limits_{\substack{ \widehat{n'}\cdot \widehat{l'} \in  \mathscr{O}'_\pm \\ \widehat{n''}\cdot \widehat{l''} \in  \mathscr{O}''_\pm 
\\ n''' \in \mathbb{Z} \\ n'+n''-n'''=n}} \,\,
{\textstyle\frac{1}{\lambda_{{}_{n',l'}}^{q-4} \lambda_{{}_{n'+n''-n''',l}}^r}}
{\textstyle\frac{1}{\lambda_{{}_{n'',l''}}^{q-4} \lambda_{{}_{n'+n''-n''',l}}^r}} \leq C
\end{multline*}
is evidently convergent and bounded by a number $C$, for sufficiently large $q$ and $r \in \mathbb{C}$,
with the bound $C$ independent of the value of $n=n'+n''+n'''$ and independent of the value of the indices $l,i,j$.
From this, similarly as in proof of 1), Lemma \ref{kappaBarDotOtimeskappaOnEU}, it follows existence of such $q\in \mathbb{N}$ that
the inequality
\begin{multline*}
\big|\theta \kappa_{1,0} \overset{\cdot}{\otimes} \kappa_{1,0}(\phi) \big|_{-q}^{2} =
\big|(A'^{-q} \otimes A''^{-q} ) \,\, \theta  \kappa'_{1,0} \overset{\cdot}{\otimes} \kappa''_{1,0}(\phi) \big|^{2}
\\
=\big|(\widetilde{A}^{-q} \otimes \widetilde{A}^{-q} ) \,\, \theta \kappa'_{1,0} \overset{\cdot}{\otimes} \kappa''_{1,0}(\phi) \big|^{2}
\leq
\Big[  
C
\sum \limits_{\substack{n \in \mathbb{Z} \\ l \in \mathbb{N} \cup {\textstyle\frac{1}{2}} \mathbb{N} \\ -l \leq i,j \leq l }}
\big|
\lambda_{{}_{n,l}}^{2r} 
\widetilde{\phi}_{{}_{ji}}(\widehat{n}\cdot \widehat{l}) \big|
\Big]^2
\\
=
 C^2
\Big[ 
\sum \limits_{\substack{n\in \mathbb{Z} \\ l \in \mathbb{N} \cup {\textstyle\frac{1}{2}} \mathbb{N}
\\ -l \leq i,j \leq l}} 
(2l+1)
\big|{\textstyle\frac{\overline{\lambda_{{}_{n,l}}^{-r'}}}{2l+1}} \lambda_{{}_{n,l}}^{2r+r'}\widetilde{\phi}_{{}_{ji}}(\widehat{n}\cdot \widehat{l})\big|
\Big]^2
\\
= 
C^2
\Big|\Big\langle \widetilde{\xi}, \widetilde{A^{2r+r'}\phi}\Big\rangle \Big|^2 \leq 
C^2
|\xi|^2 |\phi|_{2r+r'}^{2},
\end{multline*}
holds with sufficiently large $r',r$ and the function $\xi$ defined as in the proof of Lemma \ref{kappaBarDotOtimeskappaOnEU}.
The last inequality proves continuity of the map
\[
\mathscr{E} \ni \phi \longmapsto \theta \kappa'_{1,0} \overset{\cdot}{\otimes} \kappa''_{1,0}(\phi) \in E_{{}_{'}}^* \otimes E_{{}_{''}}^*
\]
and thus assertion 2).

Ad 3), 4). Proof of these statements is similar as that presented above for 2). 

Ad 5). This point immediately follows from the explicit norm estimation given in the proof 1)-4).

Ad 6). Its proof follows similarly as 4) of Lemma \ref{kappaBarDotOtimeskappaOnEU}.
\qed

Let us pass now to the more interesting case, in which the multipliction of the regular kernel by the step theta function gives again a regular kernel
with the corresponding operator transforming the test Hida space continuously again into the Hida space.
Namely we have

\begin{lem}
Let $\kappa_{\ell, m}, \dots, \kappa'_{\ell, m}, \kappa''_{\ell, m}, \ldots, \kappa'''_{\ell, m}   \in \mathfrak{K}_0$, $\ell,m = 1,0$, be the kernels 
of free fields $\mathbb{A}, \mathbb{A}', \ldots$ on the Einstein Universe (all positive or all negative-energy fields)
with the corresponding single particle Gelfand triples 
\[
\begin{array}{ccccc} E & \subset & \mathcal{H} & \subset & E^* 
\\ \vdots&&\vdots&& \vdots \\ 
 E_{{}_{'}} & \subset & \mathcal{H}' & \subset & E_{{}_{'}}^*
 \\  
E_{{}_{''}} & \subset & \mathcal{H}'' & \subset & E_{{}_{''}}^*
\\\vdots &&\vdots&&\vdots  \\  
E_{{}_{'''}} & \subset & \mathcal{H}''' & \subset & E_{{}_{'''}}^*.
\end{array} 
\]
Let $\mathscr{E}=\mathcal{S}_{{}_{\Delta +1}}(\widetilde{\mathbb{S}^1} \times SU(2, \mathbb{C}); \mathbb{C})$
or respectively  $\mathscr{E}=\mathcal{S}_{{}_{\Delta +1}}(\widetilde{\mathbb{S}^1} \times SU(2, \mathbb{C}); \mathbb{C}^d)$.
Suppose that at most one of the fields  $\mathbb{A}, \mathbb{A}', \ldots$ has the corresponding orbit $\mathscr{O}_\pm$
infinite and all remaning fields being massive and thus havig finite orbits $\ldots, \mathscr{O}'_{\pm}, \ldots$ with the corresponding
nuclear spaces $\ldots E_{{}_{'}}, E_{{}_{''}}, \ldots $ finite dimensional.
Then 
\begin{enumerate}
\item[1)]
\begin{multline*}
\theta \kappa_{1, 0} \overset{\cdot}{\otimes} \cdots \overset{\cdot}{\otimes} \kappa'_{1, 0} \overset{\cdot}{\otimes} 
\kappa''_{0, 1}\overset{\cdot}{\otimes} \cdots \overset{\cdot}{\otimes} \kappa'''_{0, 1} 
\in  \mathscr{L}(\mathscr{E}, E_{{}_{}} \otimes \cdots \otimes E_{{}_{'}} \otimes E_{{}_{''}}^* \otimes \cdots \otimes E_{{}_{'''}}^*) 
\\
\cong
\mathscr{L}(E_{{}_{}}^{*} \otimes \cdots \otimes E_{{}_{'}}^{*} \otimes E_{{}_{''}} \otimes \cdots \otimes E_{{}_{'''}}, \mathscr{E}^*).
\end{multline*}
\item[2)]
\[
\theta \kappa_{1, 0} \overset{\cdot}{\otimes} \cdots \overset{\cdot}{\otimes} \kappa'_{1, 0} \overset{\cdot}{\otimes} 
\kappa''_{0, 1}\overset{\cdot}{\otimes} \cdots \overset{\cdot}{\otimes} \kappa'''_{0, 1} (\phi)
\]
ranges over a bounded set in
\[
E_{{}_{}} \otimes \cdots \otimes E_{{}_{'}} \otimes E_{{}_{''}}^* \otimes \cdots \otimes E_{{}_{'''}}^*
\]
whenever $\phi$ ranges over a bounded set in $\mathscr{E}$.
\item[3)] 
The above statements 1) and 2) remain valid if we replace the kernels $\kappa_{1,0}, \ldots$ with their derivations
$(X)^\alpha \kappa_{1,0}, \ldots$.
\end{enumerate}
\label{theta.kappaBarDotOtimeskappaWithOneMasslessOnEU}
\end{lem}

\qedsymbol \,
Because all, except at most one, of the orbits $\mathscr{O}_\pm, \ldots, \mathscr{O}'_\pm, \mathscr{O}''_\pm, \ldots\mathscr{O}'''_\pm$
are finite, then we can repeat the proof of Lemma \ref{retOfPairings} 
or of points 1), 5) and 6) of Lemma \ref{theta.kappaBarDotOtimeskappaOnEU}.
\qed

For the further analysis of the contraction of the tensor product of kernels being
equal to dot products of plane wave kernels we need the following
\begin{lem}
Let
\[
\kappa^{(1)}_{0,1},\kappa^{(1)}_{1,0}, \,\,\, \ldots, \,\,\, \kappa^{(q)}_{0,1}, \kappa^{(q)}_{1,0}
\] 
be the kernels of free fields $\mathbb{A}^{(1)}, \ldots, \mathbb{A}^{(q')}$ of the general
class of Definition \ref{K_0onEU}. Let 
\[
E_{{}_{(1)}} \subset \mathcal{H}_{{}_{(1)}} \subset E_{{}_{(1)}}^{*},
\,\,\,\,
\ldots,
\,\,\,\,
E_{{}_{(q)}} \subset \mathcal{H}_{{}_{(q)}} \subset E_{{}_{(q)}}^{*},
\]
be, respectively, their single particle Gelfand triples. 
Then for
\[
\kappa^{(1)}_{0,1} \dot{\otimes} \dots \dot{\otimes} \kappa^{(q)}_{0,1}, \,\,\,\,
\kappa^{(1)}_{1,0} \dot{\otimes} \dots \dot{\otimes}  \kappa^{(q)}_{1,0}
\]
\[
\kappa^{(1)}_{0,1} \dot{\otimes} \dots \dot{\otimes} \kappa^{(m)}_{0,1} \dot{\otimes}
\kappa^{(1)}_{1,0} \dot{\otimes} \dots \dot{\otimes}  \kappa^{(\ell)}_{1,0}, \,\,\,\, \ell,m \leq q
\]
understood as elements of
\[
\mathscr{L}\big(E_{{}_{(1)}} \otimes \ldots \otimes E_{{}_{(q)}}, \mathscr{E}^* \big)
\]
or respectively
\[
\mathscr{L}\big(E_{{}_{(1)}} \otimes \ldots \otimes E_{{}_{(m+\ell)}}, \mathscr{E}^* \big)
\]
and for 
\[
\xi \in E_{{}_{(1)}} \otimes \ldots \otimes E_{{}_{(\ell+m)}}
\] 
or, respectively, for
\[
\xi \in E_{{}_{(1)}} \otimes \ldots \otimes E_{{}_{(q)}}
\]
the values
\[
\kappa^{(1)}_{0,1} \dot{\otimes} \dots \dot{\otimes} \kappa^{(m)}_{0,1} \dot{\otimes}
\kappa^{(1)}_{1,0} \dot{\otimes} \dots \dot{\otimes}  \kappa^{(\ell)}_{1,0}(\xi),
\]
\begin{multline*}
\kappa^{(1)}_{0,1} \dot{\otimes} \dots \dot{\otimes} \kappa^{(q)}_{0,1}(\xi), \,\,\,\,
\kappa^{(1)}_{1,0} \dot{\otimes} \dots \dot{\otimes}  \kappa^{(q)}_{1,0}(\xi) \in \mathscr{E} 
= \mathcal{S}_{A}\big(\widetilde{\mathbb{S}^1}\times SU(2, \mathbb{C}) \big) 
\\
= \mathcal{S}_{\Delta +1}\big(\widetilde{\mathbb{S}^1}\times SU(2, \mathbb{C}) \big) 
\subset \mathcal{S}_{\Delta +1}\big(\widetilde{\mathbb{S}^1}\times SU(2, \mathbb{C}) \big)^* = \mathscr{E}^*.
\end{multline*}
Equivalently 
\[
\kappa^{(1)}_{0,1} \dot{\otimes} \dots \dot{\otimes} \kappa^{(m)}_{0,1} \dot{\otimes}
\kappa^{(1)}_{1,0} \dot{\otimes} \dots \dot{\otimes}  \kappa^{(\ell)}_{1,0}(\xi),
\]
\[
\kappa^{(1)}_{0,1} \dot{\otimes} \dots \dot{\otimes} \kappa^{(q)}_{0,1}(\xi), \,\,\,\,
\kappa^{(1)}_{1,0} \dot{\otimes} \dots \dot{\otimes}  \kappa^{(q)}_{1,0}(\xi) \in 
\textrm{Dom} \, [\Delta+1]^m, \,\,\,  m \in \mathbb{N}.
\]
Moreover, for each natural $m$ there exist a natural $n$, depending on $m$, and a constant $c_m$, such that the inequalities
\[
\Big|\kappa^{(1)}_{0,1} \dot{\otimes} \dots \dot{\otimes} \kappa^{(m)}_{0,1} \dot{\otimes}
\kappa^{(1)}_{1,0} \dot{\otimes} \dots \dot{\otimes}  \kappa^{(\ell)}_{1,0}(\xi)\Big|_{m} \leq c_m \, |\xi|_{n}^{2},
\]
\begin{align*}
\Big|\kappa^{(1)}_{0,1} \dot{\otimes} \dots \dot{\otimes} \kappa^{(q)}_{0,1}(\xi) \Big|_{m} \leq c_m \, |\xi|_{n}^{2},
\\
\Big|\kappa^{(1)}_{0,1} \dot{\otimes} \dots \dot{\otimes} \kappa^{(q)}_{0,1}(\xi) \Big|_{m} \leq c_m \, |\xi|_{n}^{2}
\end{align*} 
are fulfilled. Here $|\cdot|_m$, $|\cdot|_{n}$ are the Hilbert systems of norms defining the toplogy of the standard
 nuclear countably Hilbert spaces, respectively, in $\mathscr{E}$ and 
in
\[
E_{{}_{(1)}} \otimes \ldots \otimes E_{{}_{(\ell +m)}}
\]
or, respectively, in
\[
E_{{}_{(1)}} \otimes \ldots \otimes E_{{}_{(q)}}.
\]
Therefore it follows that
\[
\kappa^{(1)}_{0,1} \dot{\otimes} \dots \dot{\otimes} \kappa^{(q)}_{0,1}, \,\,\,\,
\kappa^{(1)}_{1,0} \dot{\otimes} \dots \dot{\otimes}  \kappa^{(q)}_{1,0} \in 
\mathscr{L}\big(E_{{}_{(1)}} \otimes \ldots \otimes E_{{}_{(q)}}, \mathscr{E} \big),
\]
\[
\kappa^{(1)}_{0,1} \dot{\otimes} \dots \dot{\otimes} \kappa^{(m)}_{0,1} \dot{\otimes}
\kappa^{(1)}_{1,0} \dot{\otimes} \dots \dot{\otimes}  \kappa^{(\ell)}_{1,0}
\in
\mathscr{L}\big(E_{{}_{(1)}} \otimes \ldots \otimes E_{{}_{(\ell+m)}}, \mathscr{E} \big)
\]
\label{(kappa...kappa)(xi)inSA}
\end{lem}
Recall, please, that
\begin{multline*}
\kappa^{(1)}_{0,1} \dot{\otimes} \dots \dot{\otimes} \kappa^{(q)}_{0,1}(\xi)(a, \ldots, b, x)
=
\sum \limits_{\mathscr{O}^{(1)} } 
\cdots
\sum \limits_{ \mathscr{O}^{(q)} } \,\,\,
\Bigg\{
\kappa^{(1)}_{0,1}\big(s^{(1)},\widehat{n^{(1)}}\cdot\widehat{l^{(1)}}; a,x\big) \ldots \,\,\, \times
\\
\times \,\,\, \ldots
\kappa^{(q)}_{0,1}\big(s^{(q)},\widehat{n^{(q)}}\cdot\widehat{l^{(q)}}; b,x\big) \,\, 
\xi\big(s^{(1)},\widehat{n^{(1)}}\cdot\widehat{l^{(1)}}, \ldots, s^{(q)},\widehat{n^{(q)}}\cdot\widehat{l^{(q)}}\big)
\Bigg\}
\end{multline*}
with the sum ranging over all momentum variables 
$s^{(1)},\widehat{n^{(1)}}\cdot\widehat{l^{(1)}}, \ldots, s^{(q)},\widehat{n^{(q)}}\cdot\widehat{l^{(q)}}$.
We have the analogous formulas for
\[
\kappa^{(1)}_{1,0} \dot{\otimes} \dots \dot{\otimes} \kappa^{(q)}_{1,0}(\xi)(a, \ldots, b, x),
\,\,\,\,
\kappa^{(1)}_{0,1} \dot{\otimes} \dots \dot{\otimes} \kappa^{(m)}_{0,1} \dot{\otimes}
\kappa^{(1)}_{1,0} \dot{\otimes} \dots \dot{\otimes} \kappa^{(\ell)}_{1,0}(\xi)(a, \ldots, b', x)
\]
\qedsymbol \,\,
1) First way (an outline).
Poof that
\[
\kappa^{(1)}_{0,1} \dot{\otimes} \dots \dot{\otimes} \kappa^{(q)}_{0,1}(\xi), \,\,\,
\kappa^{(1)}_{1,0} \dot{\otimes} \dots \dot{\otimes}  \kappa^{(q)}_{1,0}(\xi) \in 
\textrm{Dom} \, [\Delta+1]^m, \,\,\,  m \in \mathbb{N}
\]
follows easily by looking at the Fourier transforms 
\[
\mathscr{F}\Big[\kappa^{(1)}_{0,1} \dot{\otimes} \dots \dot{\otimes} \kappa^{(q)}_{0,1}(\xi) \Big], \,\,\,\,
\mathscr{F}\Big[\kappa^{(1)}_{1,0} \dot{\otimes} \dots \dot{\otimes}  \kappa^{(q)}_{1,0}(\xi)  \Big],
\]
of the functions
\[
\kappa^{(1)}_{0,1} \dot{\otimes} \dots \dot{\otimes} \kappa^{(q)}_{0,1}(\xi), \,\,\,\,
\kappa^{(1)}_{1,0} \dot{\otimes} \dots \dot{\otimes}  \kappa^{(q)}_{1,0}(\xi),
\]
which we obtain without calculation by the simple form of the plane wave kernels being equal, as functions
of space-time coordinate, to the characters.
Then we use the fact that
\[
\xi \in  \textrm{Dom} \, \big[\otimes (\widetilde{\Delta}+1)\big]^m, \,\,\, m \in \mathbb{N}.
\]
Finally we are using the simple formula 
\[
[\Delta +1]^m (\widehat{n} \cdot\widehat{l})_{{}_{ij}}(t, \boldsymbol{w}) = 
\lambda_{{}_{n,l}}^m .
(\widehat{n} \cdot\widehat{l})_{{}_{ij}}(t, \boldsymbol{w})
\]
for the action of $[\Delta +1]^m$ on the matrix elements of the chatacter $\widehat{n} \cdot\widehat{l}$
and the simple action formula of the tensor product of Fourier transformed $\Delta +1$ on $\xi$  
\begin{multline*}
\big[\otimes (\widetilde{\Delta}+1)\big]^m \xi
\big(s^{(1)},\widehat{n^{(1)}} \cdot\widehat{l^{(1)}}, \ldots, s^{(q)},\widehat{n^{(q)}} \cdot\widehat{l^{(q)}} \big)
\\
=
\lambda_{{}_{n^{(1)},l^{(1)}}}^m \ldots \lambda_{{}_{n^{(q)},l^{(q)}}}^m \,\,
\xi
\big(s^{(1)},\widehat{n^{(1)}} \cdot\widehat{l^{(1)}}, \ldots, s^{(q)},\widehat{n^{(q)}} \cdot\widehat{l^{(q)}} \big).
\end{multline*}

2) Second way (a complete proof).
We start with the simplest case of $q=1$.  Let
\begin{multline*}
\kappa_{0,1}\big(s, \widehat{n}\cdot \widehat{l}; a,x\big) 
= \sum \limits_{-l \leq i,j \leq l} \sqrt{2l+1} {u^{a}_{s}}_{{}_{ji}}(\widehat{n}\cdot \widehat{l})
\widehat{n}(t) \widehat{l}_{{}_{i \, j}}(\boldsymbol{w}), 
\\ 
\,\,\,\,\,\,\,\,\,\,\,\,\,\,\,\,\,\,\,\,\,\,\,\,\,\,\,\,\,\,\,\,\,\,\,\,\,\,\,\,\,\,\,\,\,\,\,\,\,\,\,\,\,\,\,\,  
x = (t, \boldsymbol{w}), \big(s, \widehat{n}\cdot \widehat{l}\big) \in \mathscr{O},
\end{multline*} 
be the plane wave kernel function of a positive energy free field, which belongs to the general class
of Definition \ref{K_0onEU} with the single particle Gelfand triple 
$E \subset \mathcal{H} \subset E^*$, $\mathcal{H} \cong L^2(\mathscr{O}, \mathbb{C})$, and is
associated to a representation $V$ of $SL(2, \mathbb{C})$. 
Let $\xi \in E = \mathcal{S}_{\widetilde{A}}(\mathscr{O})$.
It is immediately seen that
\[
\widetilde{\kappa_{0,1}(\xi)}_{{}_{ji}}(\widehat{n}\cdot \widehat{l}) =  \sum\limits_{1 \leq s \leq d'(\widehat{n}\cdot \widehat{l})}
{\textstyle\frac{\sqrt{4\pi}}{2l+1}}
{u^{a}_{s}}_{{}_{ji}}(\widehat{n}\cdot \widehat{l}) \xi\big(s, \widehat{n}\cdot \widehat{l}\big)
\,\,\,\, \textrm{for} \,\, \widehat{n}\cdot \widehat{l} \in \mathscr{O}_+
\]
and $\widetilde{\kappa_{0,1}(\xi)}_{{}_{ji}}(\widehat{n}\cdot \widehat{l}) =0$ for $\widehat{n}\cdot \widehat{l} \notin \mathscr{O}_+$, so that 
$\widetilde{\kappa_{0,1}(\xi)}$ is concentrated on the orbit $\mathscr{O}_+$. 
Therefore 
\begin{multline}\label{|Amkappa01(xi)|2Estimation}
\big|A^m \kappa_{0,1}(\xi) \big|_{{}_{L^2}}^{2} = \Big|\big(\widetilde{A}\big)^m \widetilde{\kappa_{0,1}(\xi)} \Big|_{{}_{L^2}}^{2}
=
{\textstyle\frac{1}{\sqrt{4\pi}}}
\sum \limits_{\substack{ \widehat{n}\cdot \widehat{l} \in \mathscr{O}_+ \\ -l \leq i,j \leq l }} 
(2l+1) \Big|
\lambda_{{}_{n,l}}^{m} \widetilde{\kappa_{0,1}(\xi)}_{{}_{ji}}(\widehat{n}\cdot \widehat{l}) 
\Big|^2
\\
\leq
\sum \limits_{\substack{ (s,\widehat{n}\cdot \widehat{l}) \in \mathscr{O} }} 
 \big|\sqrt{4\pi} \textrm{dim} \, V
\lambda_{{}_{n,l}}^{m} \xi\big(s,\widehat{n}\cdot \widehat{l}\big)
\big|^2,
\end{multline}
where we have used the inequality
\[
\big|{u^{a}_{s}}_{{}_{ji}}(\widehat{n}\cdot \widehat{l}) \big| \leq {\textstyle\frac{1}{\sqrt{2l+1}}},
\]
and the fact that $d'(\widehat{n}\cdot \widehat{l}) < (2l+1)^2 \textrm{dim} \, V$, so the range of the
index $s$ being equal $1 \leq s \leq d'(\widehat{n}\cdot \widehat{l})$ is not larger than $1 \leq s  \leq (2l+1)^2 \textrm{dim} \, V$.
Now we use the fact that $\xi \in \textrm{Dom} \big(\widetilde{A}\big)^n$ for all $n \in \mathbb{N}$, which allows us to write the
inequality  (\ref{|Amkappa01(xi)|2Estimation}) in the following form
\begin{multline*}
\big| A^m \kappa_{0,1}(\xi) \big|_{{}_{L^2}}^{2} \leq
\sum \limits_{\substack{ (s,\widehat{n}\cdot \widehat{l}) \in \mathscr{O} }} 
 \Big|
{\textstyle\frac{\sqrt{4\pi} \textrm{dim} \, V \lambda_{{}_{n,l}}^{m}}{\lambda_{{}_{n,l}}^{n}}}
 \,\, \lambda_{{}_{n,l}}^{n} \xi\big(s,\widehat{n}\cdot \widehat{l}\big).
\Big|^2
\\
\leq 
\underset{(s, \widehat{n}\cdot \widehat{l}) \in \mathscr{O}}{\textrm{sup}} \big|\eta\big(s, \widehat{n}\cdot \widehat{l}\big)\big|
\sum \limits_{\substack{ (s,\widehat{n}\cdot \widehat{l}) \in \mathscr{O} }} 
 \big| \lambda_{{}_{n,l}}^{n} \xi\big(s,\widehat{n}\cdot \widehat{l}\big).
\big|^2 
\\
\leq
|\eta|_{{}_{L^2}} |\big(\widetilde{A}\big)^n \xi|_{{}_{L^2}}^{2} = |\eta|_{0} |\xi|_{n}^{2},
\end{multline*}
for the auxiliary function 
\[
\eta\big(s,\widehat{n}\cdot \widehat{l}\big) = {\textstyle\frac{\sqrt{4\pi}\textrm{dim} \, V \lambda_{{}_{n,l}}^{2m}}{\lambda_{{}_{n,l}}^{2n}}},
\,\,\,\,\,\, \widehat{n}\cdot \widehat{l} \in \mathscr{O}_+, \,\, 1 \leq s \leq d'(\widehat{n}\cdot \widehat{l}), 
\]
which, of course, belongs to $L^2(\mathscr{O})$, if $n$ (depending on $m$) is taken sufficiently large.
Therefore, in case $q=1$ and for each natural $m$ there exists a natural $n$, depending on $m$, such that the inequality
\[
\big|\kappa_{0,1}(\xi) \big|_{m} \, \leq \, \sqrt{|\eta|_{0}} \,\,\, |\xi|_{n}
\] 
is fulfilled.

The case $q>1$ follows from the case $q=1$
by the obvious product formula
\begin{multline*}
\big(\kappa^{(1)}_{0,1} \dot{\otimes} \ldots  \dot{\otimes} \kappa^{(m)}_{0,1} \dot{\otimes}
\kappa^{(1)}_{1,0} \dot{\otimes} \ldots \dot{\otimes} \ldots \dot{\otimes} \kappa^{(\ell)}_{1,0}\big)
(\xi_{{}_{(1)}} \otimes \ldots \otimes \xi_{{}_{(m)}} \otimes \xi_{{}_{(m+1)}} \otimes \ldots \otimes  \xi_{{}_{(\ell+m)}}) 
\\
= 
\kappa^{(1)}_{0,1}(\xi_{{}_{(1)}})\cdot \ldots \cdot \kappa^{(m)}_{0,1}(\xi_{{}_{(m)}}) \cdot
\kappa^{(1)}_{1,0}(\xi_{{}_{(m+1)}}) \cdot \ldots \cdot \kappa^{(\ell)}_{1,0}(\xi_{{}_{(m+\ell)}})
\end{multline*}
and the fact that each factor
\[
\kappa^{(i)}_{0,1}(\xi_{{}_{(i)}})  \,\,\,\,\, \textrm{or} \,\,\,\,\, \kappa^{(i)}_{1,0}(\xi_{{}_{(m+i)}})
\]
is an element of $\mathscr{E}$ continuously depending on $\xi_{{}_{(i)}}$, or, respectively, on $\xi_{{}_{(m+i)}}$.  
\qed

Note, please, that the Lemma \ref{(kappa...kappa)(xi)inSA} is the immediate analogue of the 
continuity of the maps
\[
\xi \longmapsto 
\kappa^{(1)}_{0,1} \dot{\otimes} \dots \dot{\otimes} \kappa^{(q)}_{0,1}(\xi), \,\,\,\,
\kappa^{(1)}_{1,0} \dot{\otimes} \dots \dot{\otimes}  \kappa^{(q)}_{1,0}(\xi) 
\in \mathcal{O}_M,
\]
\[
\xi \longmapsto 
\kappa^{(1)}_{0,1} \dot{\otimes} \dots \dot{\otimes} \kappa^{(m)}_{0,1} \dot{\otimes}
\kappa^{(1)}_{1,0} \dot{\otimes} \dots \dot{\otimes}  \kappa^{(\ell)}_{1,0}(\xi) 
\in  \mathcal{O}_M 
\]
for the dot products of plane wave kernels correpsonding to free fields on the Minkowski
space-time. The case $q=1$ of this assertion for Minkowski space-time is given 
in Lemma \ref {kappa0,1,kappa1,0psi}, Subsection \ref{psiBerezin-Hida} -- for massive fields, and Lemma \ref{kappa0,1,kappa1,0ForA},
Subsection \ref{A=Xi0,1+Xi1,0} for massless and electromagnetic potential field kernels. 
The case $q>1$ on the Minkowski space-time follows by the same product formula, exactly as in the proof presented above 
on the Einstein Universe. Here, instead of the non-compact $\mathbb{R}^4$, we have the compact 
group $\widetilde{\mathbb{S}^1} \times SL(2, \mathbb{C})$
and the multipliers of $\mathscr{E}^*$ coincide with $\mathscr{E}$. Therefore, Lemma \ref{(kappa...kappa)(xi)inSA}
asserts exactly the same for the Einstein Universe, that $\xi \mapsto \kappa_{\ell,m}(\xi)$, $\xi \in E^{\otimes \, (\ell+m)}$,
is continuous into the algebra of multipliers of $\mathscr{E}^*$, provided $\kappa_{\ell,m}$ are equal to dot products of the 
plane wave kernels $\kappa_{1,0}, \kappa_{0,1}$ of free fields with single particle Gelfand triples $E \subset \mathcal{H} \subset E^*$.

\begin{lem}
Let $\kappa^{(1)}_{\ell, m}, \dots, \kappa^{(q)}_{\ell, m}   \in \mathfrak{K}_0$, $\ell,m = 1,0$, be the kernels 
of $q$ free fields $\mathbb{A}^{(1)}, \ldots, \mathbb{A}^{(q)}$, respectively, 
on the Einstein Universe (all positive or all negative-energy fields)
with the corresponding single particle Gelfand triples 
\[
\begin{array}{ccccc} 
 E_{{}_{(1)}} & \subset & \mathcal{H}_{{}_{(1)}} & \subset & E_{{}_{(1)}}^*
\\\vdots &&\vdots&&\vdots  \\  
E_{{}_{(q)}} & \subset & \mathcal{H}_{{}_{(q)}} & \subset & E_{{}_{(q)}}^*.
\end{array} 
\]
Let $\mathscr{E}=\mathcal{S}_{{}_{\Delta +1}}(\widetilde{\mathbb{S}^1} \times SU(2, \mathbb{C}); \mathbb{C})$
or respectively  $\mathscr{E}=\mathcal{S}_{{}_{\Delta +1}}(\widetilde{\mathbb{S}^1} \times SU(2, \mathbb{C}); \mathbb{C}^d)$.
Suppose that at most one, say $\mathbb{A}^{(i)}$, of the $q$ free fields $\mathbb{A}^{(1)}, \ldots, \mathbb{A}^{(q)}$ has the corresponding orbit 
$\mathscr{O}^{(i)}_\pm$ infinite and all remaining fields being massive and thus having finite orbits $\mathscr{O}^{(j)}_{\pm}$, $j\neq i$, 
with the corresponding
nuclear spaces $E_{{}_{(j)}}$, $j\neq i$, finite dimensional. 
Let the common space-time variable
of the kernels 
\[
\kappa^{(1)}_{0,1}, \ldots, \kappa^{(q)}_{0, 1}
\]
be denoted $x$ and of the kernels 
\[
\kappa^{(1)}_{1,0}, \ldots, \kappa^{(q)}_{1,0}
\]
be denoted $y$, and let 
\begin{multline*}
\theta.\big(\kappa^{(q)}_{1, 0} \overset{\cdot}{\otimes} \cdots \overset{\cdot}{\otimes} \kappa^{(q)}_{1, 0}\big) \otimes_{q}
\big(\kappa^{(1)}_{0, 1}\overset{\cdot}{\otimes} \cdots \overset{\cdot}{\otimes} \kappa^{(q)}_{0, 1}\big)(x,y) 
\\
\overset{\textrm{df}}{=}
\theta\big(xy^{-1}\big)\big(\kappa^{(1)}_{1, 0} \overset{\cdot}{\otimes} \cdots \overset{\cdot}{\otimes} \kappa^{(q)}_{1, 0}\big) \otimes_{q}
\big(\kappa^{(q)}_{0, 1}\overset{\cdot}{\otimes} \cdots \overset{\cdot}{\otimes} \kappa^{(q)}_{0, 1}\big)(x,y)
\\
= 
\big(\theta_y\kappa^{(1)}_{1, 0} \overset{\cdot}{\otimes} \cdots \overset{\cdot}{\otimes} \kappa^{(q)}_{1, 0}\big) \otimes_{q}
\big(\kappa^{(q)}_{0, 1}\overset{\cdot}{\otimes} \cdots \overset{\cdot}{\otimes} \kappa^{(q)}_{0, 1}\big)(x,y) 
\end{multline*}
where $\theta_y(x) = \theta\big(xy^{-1}\big)$. 

Then for the $q$-contraction $\otimes_{q}$, in which all momentum variables are contracted, the following assertions hold:
\begin{enumerate}
\item[1)]
The $q$-contraction 
\[
\theta.\big(\kappa^{(1)}_{0,1} \overset{\cdot}{\otimes} \cdots \overset{\cdot}{\otimes} \kappa^{(q)}_{0,1}\big) \otimes_{q}
\big(\kappa^{(1)}_{1,0}\overset{\cdot}{\otimes} \cdots \overset{\cdot}{\otimes} \kappa^{(q)}_{1,0}\big)
\in  \mathscr{L}(\mathscr{E}^{\otimes \, 2}, \mathbb{C}) 
\cong \mathscr{E}^{* \, \otimes \, 2}.
\]
\item[2)] 
The above statement 1) remains valid if we replace the kernels $\kappa^{(1)}_{1,0}, \ldots$ with their derivations
$(X)^\alpha \kappa_{1,0}, \ldots$.
\end{enumerate}
\label{(q'=q)-contractionkappaBarDotOtimeskappaWithOneMasslessOnEU}
\end{lem}

\qedsymbol \,
Because all, except at most one, of the orbits $\mathscr{O}_\pm, \ldots, \mathscr{O}'_\pm, \mathscr{O}''_\pm, \ldots\mathscr{O}'''_\pm$
are finite, then we can repeat the proof of Lemma \ref{retOfPairings} 
or of points 1), 5) and 6) of Lemma \ref{theta.kappaBarDotOtimeskappaOnEU}.
\qed

\begin{lem}
Let $\kappa^{(1)}_{\ell, m}, \dots, \kappa^{(q)}_{\ell, m}   \in \mathfrak{K}_0$, $\ell,m = 1,0$, be the kernels 
of $q$ free fields $\mathbb{A}^{(1)}, \ldots, \mathbb{A}^{(q)}$, respectively, 
on the Einstein Universe (all positive or all negative-energy fields)
with the corresponding single particle Gelfand triples 
\[
\begin{array}{ccccc} 
 E_{{}_{(1)}} & \subset & \mathcal{H}_{{}_{(1)}} & \subset & E_{{}_{(1)}}^*
\\\vdots &&\vdots&&\vdots  \\  
E_{{}_{(q)}} & \subset & \mathcal{H}_{{}_{(q)}} & \subset & E_{{}_{(q)}}^*.
\end{array} 
\]
Let $\mathscr{E}=\mathcal{S}_{{}_{\Delta +1}}(\widetilde{\mathbb{S}^1} \times SU(2, \mathbb{C}); \mathbb{C})$
or respectively  $\mathscr{E}=\mathcal{S}_{{}_{\Delta +1}}(\widetilde{\mathbb{S}^1} \times SU(2, \mathbb{C}); \mathbb{C}^d)$.
Suppose that at most one, say $\mathbb{A}^{(i)}$, of the $q$ free fields $\mathbb{A}^{(1)}, \ldots, \mathbb{A}^{(q)}$ has the corresponding orbit 
$\mathscr{O}^{(i)}_\pm$ infinite and all remaining fields being massive and thus having finite orbits $\mathscr{O}^{(j)}_{\pm}$, $j\neq i$, 
with the corresponding
nuclear spaces $E_{{}_{(j)}}$, $j\neq i$, finite dimensional. 
Let the common space-time variable
of the kernels 
\[
\kappa^{(1)}_{0,1}, \ldots, \kappa^{(q)}_{0, 1}
\]
be denoted $x$ and of the kernels 
\[
\kappa^{(1)}_{1,0}, \ldots, \kappa^{(q)}_{1,0}
\]
be denoted $y$, and let 
\begin{multline*}
\theta.\big(\kappa^{(q)}_{1, 0} \overset{\cdot}{\otimes} \cdots \overset{\cdot}{\otimes} \kappa^{(q)}_{1, 0}\big) \otimes_{q'}
\big(\kappa^{(1)}_{0, 1}\overset{\cdot}{\otimes} \cdots \overset{\cdot}{\otimes} \kappa^{(q)}_{0, 1}\big)(x,y) 
\\
\overset{\textrm{df}}{=}
\theta\big(xy^{-1}\big)\big(\kappa^{(1)}_{1, 0} \overset{\cdot}{\otimes} \cdots \overset{\cdot}{\otimes} \kappa^{(q)}_{1, 0}\big) \otimes_{q'}
\big(\kappa^{(1)}_{0, 1}\overset{\cdot}{\otimes} \cdots \overset{\cdot}{\otimes} \kappa^{(q)}_{0, 1}\big)(x,y)
\\
= 
\big(\theta_y\kappa^{(1)}_{1, 0} \overset{\cdot}{\otimes} \cdots \overset{\cdot}{\otimes} \kappa^{(q)}_{1, 0}\big) \otimes_{q'}
\big(\kappa^{(q)}_{0, 1}\overset{\cdot}{\otimes} \cdots \overset{\cdot}{\otimes} \kappa^{(q)}_{0, 1}\big)(x,y) 
\end{multline*}
where $\theta_y(x) = \theta\big(xy^{-1}\big)$. Assume, finally, that  in the $q'$-contraction 
$\otimes_{q'}$, $q' < q$, the infinite orbit momentum variables are not contracted.
Then the following assertions hold for such $q'$-contraction $\otimes_{q'}$:
\begin{enumerate}
\item[1)]
The $q'$-contraction ($q' < q$)
\begin{multline*}
\theta.\big(\kappa^{(1)}_{0,1} \overset{\cdot}{\otimes} \cdots \overset{\cdot}{\otimes} \kappa^{(q)}_{0,1}\big) \otimes_{q'}
\big(\kappa^{(1)}_{1,0}\overset{\cdot}{\otimes} \cdots \overset{\cdot}{\otimes} \kappa^{(q)}_{1,0}\big)
\\ 
\in  \mathscr{L}(\mathscr{E}^{\otimes \, 2}, E_{{}_{(q'+1)}}^{*} \otimes \cdots \otimes E_{{}_{q}}^{*} \otimes E_{{}_{q'+1}}
\otimes \cdots \otimes E_{{}_{(q)}}) 
\\
\cong
\mathscr{L}(E_{{}_{(q'+1)}} \otimes \cdots \otimes E_{{}_{(q)}} \otimes E_{{}_{q'+1}}^{*} \otimes \cdots
\otimes E_{{}_{(q)}}^{*}, \mathscr{E}^{* \, \otimes \, 2}).
\end{multline*}
\item[2)] 
For each natural $m$ there exist  natural $n,k$ and a constant $c_{-n,m}$, 
depending only on the kernels $\kappa^{(1)}_{0,1}, \kappa^{(1)}_{1,0}, \ldots \kappa^{(q)}_{0,1},
\kappa^{(q)}_{1,0}$ and on $m,n$, such that
\[
\Big| 
\Big(\theta.\big(\kappa^{(1)}_{0,1} \dot{\otimes} \ldots \dot{\otimes} \kappa^{(q)}_{0,1} \big) 
\otimes_{q'} \big(\kappa^{(1)}_{1,0} \dot{\otimes} \ldots \dot{\otimes} \kappa^{(q)}_{1,0} \big) \Big)(\phi \otimes \varphi)
 \Big|_{-n,m}
\leq 
c_{-n,m} \, 
\big|\phi \big|_{k}\big|\phi \big|_{k},
\]
for
\[
\chi= \phi\otimes \varphi \in \mathscr{E}\otimes \mathscr{E}.
\]
In particular
\[
\theta.\big(\kappa^{(1)}_{0,1} \dot{\otimes} \cdots \cdot{\otimes} \kappa^{(q)}_{0,1}\big) \otimes_{q'} 
\big(\kappa^{(1)}_{1,0}\cdot{\otimes} \cdots \cdot{\otimes} \kappa^{(q)}_{1,0}\big) (\chi)
\]
ranges over a bounded set in
\[
E_{{}_{q'+1}}^{*} \otimes \cdots \otimes E_{{}_{q}}^{*} \otimes E_{{}_{q'+1}} \otimes \cdots \otimes E_{{}_{(q)}}
\]
whenever $\chi$ ranges over a bounded set in $\mathscr{E}$.
\item[3)] 
The above statements 1), 2), 3) remain valid if we replace the kernels $\kappa^{(1)}_{1,0}, \ldots$ with their derivations
$(X)^\alpha \kappa_{1,0}, \ldots$.
\end{enumerate}
\label{q-contractionkappaBarDotOtimeskappaWithOneMasslessNotContractedOnEU}
\end{lem}
Here the symbol $\otimes_{q'}$ stands for $q'$-contraction: $q'$ momentum variables
of the  kernel 
\[
\kappa^{(1)}_{0,1} \dot{\otimes} \cdots \cdot{\otimes} \kappa^{(q)}_{0,1}
\]
are equated with $q'$ momentum variables of the kernel 
\[
\kappa^{(1)}_{1,0}\cdot{\otimes} \cdots \cdot{\otimes} \kappa^{(q)}_{1,0}
\]
and then summation is performed with respect to the equated momentum variables. By assumption the contracted variables
do not include the only pair of momentum variables which ranges over the infinite orbit $\mathscr{O}^{(i)}$. 
In order to simplify notation we have assumed that the first $q'$ momentum variables of the kernel
\[
\kappa^{(1)}_{0,1} \dot{\otimes} \cdots \cdot{\otimes} \kappa^{(q)}_{0,1}
\]
are contracted with the first $q'$ momentum variables of the kernel 
\[
\kappa^{(1)}_{1,0}\cdot{\otimes} \cdots \cdot{\otimes} \kappa^{(q)}_{1,0},
\]
so the massless (or infinite orbit) kernels are not placed among the first $q'$ ``annihilation''
and ``creation'' plane wave  kernels.
 
This contraction makes sense because
\[
\big(\kappa^{(1)}_{0,1} \dot{\otimes} \cdots \cdot{\otimes} \kappa^{(q)}_{0,1}\big) \otimes 
\big(\kappa^{(1)}_{1,0}\cdot{\otimes} \cdots \cdot{\otimes} \kappa^{(q)}_{1,0}\big) \in
E_{{}_{(q'+1)}}^{*} \otimes \cdots \otimes E_{{}_{(q)}}^{*} \otimes E_{{}_{(q'+1)}} \otimes 
\cdots \otimes E_{{}_{(q)}} \otimes \mathscr{E}^{*\, \otimes \, 2}.
\]
Finally $|\xi|_{m,-n}$ for 
\[
\xi = \xi_{(q'+1)} \otimes \ldots \xi_{(q)}\otimes \zeta_{(q'+1)} \otimes \zeta_{(q)}
 \in E_{{}_{(q'+1)}}^{*} \otimes \cdots \otimes E_{{}_{q}}^{*} \otimes E_{{}_{q'+1}} 
\otimes \cdots \otimes E_{{}_{(q)}}
\]
stand for the norms
\[
\Big| \big[\widetilde{A}^{\otimes(q-q')}\big]^{-n} \big[\widetilde{A}^{\otimes(q-q')}\big]^{m} \xi \Big|_{L^2}
= 
\Big| 
\widetilde{A} ^{-n}\xi_{(q'+1)} \otimes \widetilde{A}^{-n} \ldots \xi_{(q)}\otimes \widetilde{A}^{m}\zeta_{(q'+1)} \otimes \widetilde{A}^{m}\zeta_{(q)}
\Big|_{L^2}
\]
definining the topology on the nuclear space
\[
E_{{}_{(q'+1)}}^{*} \otimes \cdots \otimes E_{{}_{q}}^{*} \otimes E_{{}_{q'+1}} 
\otimes \cdots \otimes E_{{}_{(q)}},
\]
and $| \cdot|_{{}_{L2}}$ is the $L^2$ norm on the Hilbert space tensor product 
\begin{multline*}
L^2(\mathscr{O}^{(q'+1)}) \otimes \ldots \otimes  L^2(\mathscr{O}^{(q)}) \otimes L^2(\mathscr{O}^{(q'+1)}) \otimes \ldots
\otimes L^2(\mathscr{O}^{(q)})
\\
= L^2\big(\mathscr{O}^{(q'+1)} \times \mathscr{O}^{(q)} \times \mathscr{O}^{(q'+1)} \times \ldots \times \mathscr{O}^{(q)}\big).
\end{multline*}

\qedsymbol \,
Because all, except at most one, of the orbits $\mathscr{O}_\pm, \ldots, \mathscr{O}'_\pm, \mathscr{O}''_\pm, \ldots\mathscr{O}'''_\pm$
are finite, then again we can repeat the proof of Lemma \ref{retOfPairings} 
or of points 1), 5), 6) of Lemma \ref{theta.kappaBarDotOtimeskappaOnEU}.
\qed

\begin{lem}
Let $\kappa^{(1)}_{\ell, m}, \dots, \kappa^{(q)}_{\ell, m}   \in \mathfrak{K}_0$, $\ell,m = 1,0$, be the kernels 
of $q$ free fields $\mathbb{A}^{(1)}, \ldots, \mathbb{A}^{(q)}$, respectively, 
on the Einstein Universe (all positive or all negative-energy fields)
with the corresponding single particle Gelfand triples 
\[
\begin{array}{ccccc} 
 E_{{}_{(1)}} & \subset & \mathcal{H}_{{}_{(1)}} & \subset & E_{{}_{(1)}}^*
\\\vdots &&\vdots&&\vdots  \\  
E_{{}_{(q)}} & \subset & \mathcal{H}_{{}_{(q)}} & \subset & E_{{}_{(q)}}^*.
\end{array} 
\]
Let $\mathscr{E}=\mathcal{S}_{{}_{\Delta +1}}(\widetilde{\mathbb{S}^1} \times SU(2, \mathbb{C}); \mathbb{C})$
or respectively  $\mathscr{E}=\mathcal{S}_{{}_{\Delta +1}}(\widetilde{\mathbb{S}^1} \times SU(2, \mathbb{C}); \mathbb{C}^d)$.
Suppose that at most one, say $\mathbb{A}^{(i)}$, of the $q$ free fields $\mathbb{A}^{(1)}, \ldots, \mathbb{A}^{(q)}$ has the corresponding orbit 
$\mathscr{O}^{(i)}_\pm$ infinite and all remaining fields being massive and thus having finite orbits $\mathscr{O}^{(j)}_{\pm}$, $j\neq i$, 
with the corresponding
nuclear spaces $E_{{}_{(j)}}$, $j\neq i$, finite dimensional. 
Let the common space-time variable
of the kernels 
\[
\kappa^{(1)}_{0,1}, \ldots, \kappa^{(q)}_{0, 1}
\]
be denoted $x$ and of the kernels 
\[
\kappa^{(1)}_{1,0}, \ldots, \kappa^{(q)}_{1,0}
\]
be denoted $y$, and let 
\begin{multline*}
\theta.\big(\kappa^{(q)}_{1, 0} \overset{\cdot}{\otimes} \cdots \overset{\cdot}{\otimes} \kappa^{(q)}_{1, 0}\big) \otimes_{q'}
\big(\kappa(1)_{0, 1}\overset{\cdot}{\otimes} \cdots \overset{\cdot}{\otimes} \kappa^{(q)}_{0, 1}\big)(x,y) 
\\
\overset{\textrm{df}}{=}
\theta\big(xy^{-1}\big)\big(\kappa^{(1)}_{1, 0} \overset{\cdot}{\otimes} \cdots \overset{\cdot}{\otimes} \kappa^{(q)}_{1, 0}\big) \otimes_{q'}
\big(\kappa^{(1)}_{0, 1}\overset{\cdot}{\otimes} \cdots \overset{\cdot}{\otimes} \kappa^{(q)}_{0, 1}\big)(x,y)
\\
= 
\big(\theta_y\kappa^{(1)}_{1, 0} \overset{\cdot}{\otimes} \cdots \overset{\cdot}{\otimes} \kappa^{(q)}_{1, 0}\big) \otimes_{q'}
\big(\kappa^{(q)}_{0, 1}\overset{\cdot}{\otimes} \cdots \overset{\cdot}{\otimes} \kappa^{(q)}_{0, 1}\big)(x,y) 
\end{multline*}
where $\theta_y(x) = \theta\big(xy^{-1}\big)$. 

Then the following assertions hold for the $q'$-contraction $\otimes_{q'}$ (here we assume that the 
first $q'$ annihilation kernels are contracted with the first $q'$ creation kernels):
\begin{enumerate}
\item[1)]
The $q'$-contraction ($q' \leq q$)
\begin{multline*}
\theta.\big(\kappa^{(1)}_{0,1} \overset{\cdot}{\otimes} \cdots \overset{\cdot}{\otimes} \kappa^{(q)}_{0,1}\big) \otimes_{q'}
\big(\kappa^{(1)}_{1,0}\overset{\cdot}{\otimes} \cdots \overset{\cdot}{\otimes} \kappa^{(q)}_{1,0}\big)
\\ 
\in  \mathscr{L}(\mathscr{E}^{\otimes \, 2}, E_{{}_{(q'+1)}}^{*} \otimes \cdots \otimes E_{{}_{q}}^{*} \otimes E_{{}_{q'+1}}
\otimes \cdots \otimes E_{{}_{(q)}}) 
\\
\cong
\mathscr{L}(E_{{}_{(q'+1)}} \otimes \cdots \otimes E_{{}_{(q)}} \otimes E_{{}_{q'+1}}^{*} \otimes \cdots
\otimes E_{{}_{(q)}}^{*}, \mathscr{E}^{* \, \otimes \, 2}).
\end{multline*}
\item[2)] 
The above statements 1), 2) remain valid if we replace the kernels $\kappa^{(1)}_{1,0}, \ldots$ with their derivations
$(X)^\alpha \kappa_{1,0}, \ldots$.
\end{enumerate}
\label{q-contractionkappaBarDotOtimeskappaWithOneMasslessOnEU}
\end{lem}

\qedsymbol \,
The case in which the infinite orbit momentum variables are not contracted follows from 
Lemma \ref{q-contractionkappaBarDotOtimeskappaWithOneMasslessNotContractedOnEU}. Therefore,
we can assume that the only pair of infinite orbit momentum variables are among the contracted 
momentum variables. 

Note, please, that the contraction kenrel is equal to the absolutely convergent discrete integral/sum
\begin{equation}\label{KernelOf(q'<q)Contraction}
 \Big(\theta.\big(\kappa^{(1)}_{0,1} \dot{\otimes} \ldots  \kappa^{(q)}_{0,1} \big) 
\otimes_{q'} \big(\kappa^{(1)}_{1,0} \dot{\otimes} \ldots \dot{\otimes} \kappa^{(q)}_{1,0} \big)\Big)
\big(\xi\otimes \zeta \big)(a, \ldots, b, c, \ldots, d, x,y)
\end{equation}
\begin{multline*}
=
\sum \limits_{\mathscr{O}^{(1)} } 
\cdots
\sum \limits_{ \mathscr{O}^{(q')} } 
\,\,\,\,
\sum \limits_{[\mathscr{O}^{(q'+1)}]^{\times \, 2} } 
\cdots
\sum \limits_{ [\mathscr{O}^{(q)}]^{\times \, 2} }
\,\,\,
\Bigg\{
\\
\theta\big(xy^{-1}\big)
\kappa^{(1)}_{0,1}\big(s^{(1)},\widehat{n^{(1)}}\cdot\widehat{l^{(1)}}; a,x\big) \ldots 
\kappa^{(q')}_{0,1}\big(s^{(q')},\widehat{n^{(q')}}\cdot\widehat{l^{(q')}}; b',x\big) \,\, \times
\\
\times \,\, 
\kappa^{(q'+1)}_{0,1}\big(s^{(q'+1)},\widehat{n^{(q'+1)}}\cdot\widehat{l^{(q'+1)}}; b'',x\big) 
\ldots
\kappa^{(q)}_{0,1}\big(s^{(q)},\widehat{n^{(q)}}\cdot\widehat{l^{(q)}}; b,x\big) 
\,\, \times
\\
\times \,\,
\kappa^{(1)}_{1,0}\big(s^{(1)},\widehat{n^{(1)}}\cdot\widehat{l^{(1)}}; c,y\big) \ldots
\kappa^{(q')}_{1,0}\big(s^{(q')},\widehat{n^{(q')}}\cdot\widehat{l^{(q')}}; d',y\big)
\,\, \times
\\
\times \,\,
\kappa^{(q'+1)}_{1,0}\big(s'^{(q'+1)},\widehat{n'^{(q'+1)}}\cdot\widehat{l'^{(q'+1)}}; d'',y\big) 
\ldots
\kappa^{(q)}_{1,0}\big(s'^{(q)},\widehat{n'^{(q)}}\cdot\widehat{l'^{(q)}}; d,y\big) 
\,\, \times
\\
\xi_{(q'+1)}\big(s^{(q'+1)},\widehat{n^{(q'+1)}}\cdot\widehat{l^{(q'+1)}}\big) \ldots 
\xi_{(q)}\big(s^{(q)},\widehat{n^{(q)}}\cdot\widehat{l^{(q)}} \big)
\\
\zeta_{(q'+1)}\big( s'^{(q'+1)},\widehat{n'^{(q'+1)}}\cdot\widehat{l'^{(q'+1)}}\big) \dots 
\zeta_{(q)}\big(s'^{(q)},\widehat{n'^{(q)}}\cdot\widehat{l'^{(q)}} \big)
\Bigg\},
\end{multline*}
for
\[
\xi \otimes \zeta = \xi_{(q'+1)} \otimes \ldots \xi_{(q)}\otimes \zeta_{(q'+1)} \otimes \zeta_{(q)}
 \in E_{{}_{(q'+1)}} \otimes \cdots \otimes E_{{}_{q}} \otimes E_{{}_{q'+1}} 
\otimes \cdots \otimes E_{{}_{(q)}}.
\]
From Lemma \ref{(kappa...kappa)(xi)inSA} it follows that
\begin{equation}\label{KernelOf(q'0,1Dotproduct)}
 \big(\kappa^{(q'+1)}_{0,1} \dot{\otimes} \ldots  \kappa^{(q)}_{0,1} \big) 
\big(\xi\big)(a, \ldots, b, x)
\end{equation}
\begin{multline*}
=
\sum \limits_{\mathscr{O}^{(q'+1)} } 
\cdots
\sum \limits_{ \mathscr{O}^{(q)} } \,\,\,
\Bigg\{
\\
\kappa^{(q'+1)}_{0,1}\big(s^{(q'+1)},\widehat{n^{(q'+1)}}\cdot\widehat{l^{(q'+1)}}; b'',x\big) 
\ldots
\kappa^{(q)}_{0,1}\big(s^{(q)},\widehat{n^{(q)}}\cdot\widehat{l^{(q)}}; b,x\big) \,\, \times
\\
\times \,\,
\xi_{(q'+1)}\big(s^{(q'+1)},\widehat{n^{(q'+1)}}\cdot\widehat{l^{(q'+1)}}\big) \ldots 
\xi_{(q)}\big(s^{(q)},\widehat{n^{(q)}}\cdot\widehat{l^{(q)}} \big)
\Bigg\},
\end{multline*}
regarded as a function of $(b'', \dots, b, x)$, belongs to $\mathscr{E}$. Here the sum ranges, of course, over the 
momentum variables 
\[
\big(s^{(q'+1)},\widehat{n^{(q'+1)}}\cdot\widehat{l^{(q'+1)}}, \ldots, s^{(q)},\widehat{n^{(q)}}\cdot\widehat{l^{(q)}} \big)
\]
of the function $\xi$. By the same Lemma \ref{(kappa...kappa)(xi)inSA} the map 
\begin{equation}\label{ContinuityOfq'0,1Dotproduct:E->E}
E_{{}_{(q'+1)}} \otimes \ldots \otimes E_{{}_{q}} \ni \xi \longrightarrow
\big(\kappa^{(q'+1)}_{0,1} \dot{\otimes} \ldots  \kappa^{(q)}_{0,1} \big) 
\big(\xi\big) \in \mathscr{E}
\end{equation}

Similarly, it follows from Lemma 
\ref{(kappa...kappa)(xi)inSA} that 
\begin{equation}\label{KernelOf(q'1,0Dotproduct)}
 \big(\kappa^{(q'+1)}_{1,0} \dot{\otimes} \ldots  \kappa^{(q)}_{1,0} \big) 
\big(\xi\big)(d'', \ldots, d, y)
\end{equation}
\begin{multline*}
=
\sum \limits_{\mathscr{O}^{(q'+1)} } 
\cdots
\sum \limits_{ \mathscr{O}^{(q)} } \,\,\,
\Bigg\{
\\
\kappa^{(q'+1)}_{1,0}\big(s'^{(q'+1)},\widehat{n'^{(q'+1)}}\cdot\widehat{l'^{(q'+1)}}; d'',y\big) 
\ldots
\kappa^{(q)}_{1,0}\big(s'^{(q)},\widehat{n'^{(q)}}\cdot\widehat{l'^{(q)}}; d,y\big) 
\,\, \times
\\
\zeta_{(q'+1)}\big( s'^{(q'+1)},\widehat{n'^{(q'+1)}}\cdot\widehat{l'^{(q'+1)}}\big) \dots 
\zeta_{(q)}\big(s'^{(q)},\widehat{n'^{(q)}}\cdot\widehat{l'^{(q)}} \big)
\Bigg\},
\end{multline*}
regarded as a function of $(d'', \dots, d, y)$, belongs to $\mathscr{E}$. Here the sum ranges, of course, over the 
momentum variables 
\[
\big(s'^{(q'+1)},\widehat{n'^{(q'+1)}}\cdot\widehat{l'^{(q'+1)}}, \ldots, s'^{(q)},\widehat{n'^{(q)}}\cdot\widehat{l'^{(q)}} \big)
\]
of the function $\zeta$. Similarly, by the same Lemma \ref{(kappa...kappa)(xi)inSA}, the map 
\[
E_{{}_{(q'+1)}} \otimes \ldots \otimes E_{{}_{q}} \ni \zeta \longrightarrow
\big(\kappa^{(q'+1)}_{1,0} \dot{\otimes} \ldots  \kappa^{(q)}_{1,0} \big) 
\big(\zeta\big) \in \mathscr{E}
\]
is continuous.

Because by assumption the non-contracted momentum variables, corresponding to the orbits
\[
\mathscr{O}^{(q'+1)}, \ldots \mathscr{O}^{(q)},
\]
are finite, and thus the nuclear spaces
\[
E_{{}_{(q'+1)}}, \ldots, E_{{}_{q}} 
\]
are finite dimensional, then the continuous map
\[
E_{{}_{(q'+1)}} \otimes \ldots \otimes E_{{}_{q}} \ni \zeta \longrightarrow
\big(\kappa^{(q'+1)}_{1,0} \dot{\otimes} \ldots  \kappa^{(q)}_{1,0} \big) 
\big(\zeta\big) \in \mathscr{E}
\]
can obviously be extended to (even canonically coincides with) a continuous map
\begin{equation}\label{ContinuityOfq'1,0Dotproduct:E*->E}
E_{{}_{(q'+1)}}^{*} \otimes \ldots \otimes E_{{}_{q}}^{*} \ni \zeta \longrightarrow
\big(\kappa^{(q'+1)}_{1,0} \dot{\otimes} \ldots  \kappa^{(q)}_{0,1} \big) 
\big(\zeta\big) \in \mathscr{E}.
\end{equation}
Next, recall please, that by Lemma \ref{(q'=q)-contractionkappaBarDotOtimeskappaWithOneMasslessOnEU}
the contraction
\begin{equation}\label{KernelOf(q'=q)Contraction}
 \Big(\theta.\big(\kappa^{(1)}_{0,1} \dot{\otimes} \ldots  \kappa^{(q')}_{0,1} \big) 
\otimes_{q'} \big(\kappa^{(1)}_{1,0} \dot{\otimes} \ldots \dot{\otimes} \kappa^{(q')}_{1,0} \big)\Big)
(a, \ldots, b', c, \ldots, d', x,y)
\end{equation}
\begin{multline*}
=
\sum \limits_{\mathscr{O}^{(1)} } 
\cdots
\sum \limits_{ \mathscr{O}^{(q')} } 
\,\,\,
\Bigg\{
\\
\theta\big(xy^{-1}\big)
\kappa^{(1)}_{0,1}\big(s^{(1)},\widehat{n^{(1)}}\cdot\widehat{l^{(1)}}; a,x\big) \ldots 
\kappa^{(q)}_{0,1}\big(s^{(q')},\widehat{n^{(q')}}\cdot\widehat{l^{(q')}}; b',x\big) \,\, \times
\\
\times \,\, 
\kappa^{(1)}_{1,0}\big(s^{(1)},\widehat{n^{(1)}}\cdot\widehat{l^{(1)}}; c,y\big) \ldots
\kappa^{(q')}_{1,0}\big(s^{(q')},\widehat{n^{(q')}}\cdot\widehat{l^{(q')}}; d',y\big)
\Bigg\},
\end{multline*}
is a well defined element of $\mathscr{E}^*\otimes \mathscr{E}^*$.
Therefore for each
\[
\xi \otimes \zeta = \xi_{(q'+1)} \otimes \ldots \xi_{(q)}\otimes \zeta_{(q'+1)} \otimes \zeta_{(q)}
 \in E_{{}_{(q'+1)}}^{*} \otimes \cdots \otimes E_{{}_{q}}^{*} \otimes E_{{}_{q'+1}} 
\otimes \cdots \otimes E_{{}_{(q)}}.
\]
the contraction (\ref{KernelOf(q'<q)Contraction}) has the product form
\begin{multline}\label{ProductFormOfContractionOneMassless}
\Big(\theta. \big(\kappa^{(1)}_{0,1} \dot{\otimes} \ldots  \kappa^{(q)}_{0,1} \big) 
\otimes_{q'} \big(\kappa^{(1)}_{1,0} \dot{\otimes} \ldots \dot{\otimes} \kappa^{(q)}_{1,0} \big)\Big)
\big(\xi\otimes \zeta \big) 
\\
=
\Big(\theta.\big(\kappa^{(1)}_{0,1} \dot{\otimes} \ldots  \kappa^{(q')}_{0,1} \big) 
\otimes_{q'} \big(\kappa^{(1)}_{1,0} \dot{\otimes} \ldots \dot{\otimes} \kappa^{(q')}_{1,0} \big)\Big) 
\big(\kappa^{(q'+1)}_{0,1} \dot{\otimes} \ldots  \kappa^{(q)}_{0,1} \big) \big(\xi\big) 
\,\, \times \,\,
\\
\times \,\,
\big(\kappa^{(q'+1)}_{1,0} \dot{\otimes} \ldots  \kappa^{(q)}_{1,0} \big) 
\big(\zeta\big),
\end{multline}
which evidently is continuous as a map
\[
E_{{}_{(q'+1)}}^{*} \otimes \cdots \otimes E_{{}_{q}}^{*} \otimes E_{{}_{q'+1}} 
\otimes \cdots \otimes E_{{}_{(q)}} \longrightarrow \mathscr{E}^*\otimes\mathscr{E}^*
\]
because of the continuity of the maps (\ref{ContinuityOfq'0,1Dotproduct:E->E})
and (\ref{ContinuityOfq'1,0Dotproduct:E*->E}) into the algebra $\mathscr{E}$ of multipliers of $\mathscr{E}^*$,
\emph{i.e.} pointwise product of elements of $\mathscr{E}^*$ by any element of $\mathscr{E}$ gives a continuous
map  $\mathscr{E}^* \longrightarrow \mathscr{E}^*$. 
\qed

{\bf REMARK}. Note that, correspondingly to the formula (\ref{ProductFormOfContractionOneMassless}), the contraction kernel function 
has the form of product
\begin{multline}\label{ProductFormOfContractionOneMassless(x,y)}
\Big(\theta. \big(\kappa^{(1)}_{0,1} \dot{\otimes} \ldots  \kappa^{(q)}_{0,1} \big) 
\otimes_{q'} \big(\kappa^{(1)}_{1,0} \dot{\otimes} \ldots \dot{\otimes} \kappa^{(q)}_{1,0} \big)\Big)
\big(\, (\ldots)\, , \, (\ldots)' \, ; \, a, \ldots, b, c, \ldots, d, x,y\big) 
\\
=
\Big(\theta.\big(\kappa^{(1)}_{0,1} \dot{\otimes} \ldots  \kappa^{(q')}_{0,1} \big) 
\otimes_{q'} \big(\kappa^{(1)}_{1,0} \dot{\otimes} \ldots \dot{\otimes} \kappa^{(q')}_{1,0} \big)\Big)(a, \ldots, b', c, \ldots, d', x, y) 
\,\, \times 
\\
\times \,\, 
\big(\kappa^{(q'+1)}_{0,1} \dot{\otimes} \ldots  \kappa^{(q)}_{0,1} \otimes
\kappa^{(q'+1)}_{1,0} \dot{\otimes} \ldots  \kappa^{(q)}_{1,0} \big) \big(\, (\ldots) \, , \, (\ldots)' \, , \, b'', \ldots, b, d'', \ldots, d, x, y\big),
\end{multline}
of scalar valued $\widetilde{\mathbb{S}^1}\times SL(2, \mathbb{C})$-invariant contraction kernel (distribution)
\[
\Big(\theta.\big(\kappa^{(1)}_{0,1} \dot{\otimes} \ldots  \kappa^{(q')}_{0,1} \big) 
\otimes_{q'} \big(\kappa^{(1)}_{1,0} \dot{\otimes} \ldots \dot{\otimes} \kappa^{(q')}_{1,0} \big)\Big)(a, \ldots, b', c, \ldots, d',x,y) 
\]
and of a vector-valued kernel (distribution)
\[
\big(\kappa^{(q'+1)}_{0,1} \dot{\otimes} \ldots  \kappa^{(q)}_{0,1} \otimes
\kappa^{(q'+1)}_{1,0} \dot{\otimes} \ldots  \kappa^{(q)}_{1,0} \big) \big(\, (\ldots) \, , \, (\ldots)' \, , \, b'', \ldots, b, d'', \ldots, d, x, y\big).
\]

Here, in order to simplify notation, the momentum variables of the dot product of the ``annihilation'' plane wave kernels 
\[
\kappa^{(q'+1)}_{0,1} \dot{\otimes} \ldots  \kappa^{(q)}_{0,1}
\]
are written as dots $(\ldots)$ in parenthesis, similarly for the momentum variables of the dot product
\[
\kappa^{(q'+1)}_{1,0} \dot{\otimes} \ldots  \kappa^{(q)}_{1,0}
\]
of the ``creation'' plane wave kernels written as $(\ldots)'$ with prime sign.

Omitting the obvious space-time variables $x,y$, the fields component indices $a, \ldots, d$ and momentum variables 
$s^{(i)}, \widehat{n^{(i)}}\cdot\widehat{l^{(i)}}$, we can write the product formula shortly
\begin{multline}\label{ProductFormOfContractionOneMasslessShortForm}
\theta. \big(\kappa^{(1)}_{0,1} \dot{\otimes} \ldots  \kappa^{(q)}_{0,1} \big) 
\otimes_{q'} \big(\kappa^{(1)}_{1,0} \dot{\otimes} \ldots \dot{\otimes} \kappa^{(q)}_{1,0} \big)
\\
=
\Big(\theta.\big(\kappa^{(1)}_{0,1} \dot{\otimes} \ldots  \kappa^{(q')}_{0,1} \big) 
\otimes_{q'} \big(\kappa^{(1)}_{1,0} \dot{\otimes} \ldots \dot{\otimes} \kappa^{(q')}_{1,0} \big)\Big)
\,\, \times 
\\
\times \,\, 
\big(\kappa^{(q'+1)}_{0,1} \dot{\otimes} \ldots  \kappa^{(q)}_{0,1} \otimes
\kappa^{(q'+1)}_{1,0} \dot{\otimes} \ldots  \kappa^{(q)}_{1,0} \big).
\end{multline}

\qed

From Lemma \ref{theta.kappaBarDotOtimeskappaWithOneMasslessOnEU} and Thm. \ref{obataJFA.Thm.3.13} of Subsection
\ref{psiBerezin-Hida} it follows  the following
\begin{cor*}
If we multiply a Wick product field 
\[
\mathcal{L} \in \mathscr{L}\big(\mathscr{E}, \,\, \mathscr{L}((\boldsymbol{E}),(\boldsymbol{E})) \big),
\]
which contains at most one massless field and the remaining being massive, by the step $\theta$-function, then we obtain again a regular operator
\[
\theta \mathcal{L} \in \mathscr{L}\big(\mathscr{E}, \,\, \mathscr{L}((\boldsymbol{E}),(\boldsymbol{E})) \big).
\]
This is in particular the case for the Lagrange QED interaction density operator $\mathcal{L}$.
\end{cor*}

We will make use of this Corollary  and of the Lemmas \ref{theta.kappaBarDotOtimeskappaWithOneMasslessOnEU}, 
\ref{q-contractionkappaBarDotOtimeskappaWithOneMasslessOnEU}, 
in the natural construction of the scattering operator for QED on the Einstein Universe
in the following Sections. 

Suppose we have a free field $\mathbb{A}$, constructed as the integral kernel operator
\[
\mathbb{A} = \Xi(\kappa_{0,1}) + \Xi(\kappa_{1,0}) = \mathbb{A}^{(-)} + \mathbb{A}^{(+)},
\]
with the corresponding orbit $\mathscr{O}$, single particle Gelfand triple $E \subset \mathcal{H} \subset E^*$ and 
with the corresponding ``plane wave'' kernels defined by the ``plane wave'' functions
\[
(s, \widehat{n}\cdot \widehat{l}; a,x\big) \mapsto \kappa_{0,1}\big(s, \widehat{n}\cdot \widehat{l}; a,x\big), 
\,\,\,\,
(s, \widehat{n}\cdot \widehat{l}; a,x\big) \mapsto \kappa_{1,0}\big(s, \widehat{n}\cdot \widehat{l}; a,x\big).
\]
Recall also that, according to the above discussion, for any free field on the Einstein Universe
\begin{align*}
\kappa_{0,1} \in \mathscr{L}(E^*, \mathscr{E}^*) \cong E \otimes \mathscr{E}^* \subset E^* \otimes \mathscr{E}^*,
\\
\kappa_{1,0} \in \mathscr{L}(E^*, \mathscr{E}^*) \cong E \otimes \mathscr{E}^* \subset E^* \otimes \mathscr{E}^*.
\end{align*}
Therefore
\[
\kappa_{0,1} \otimes \kappa_{1,0} \in E \otimes E \otimes \mathscr{E}^* \otimes \mathscr{E}^*
\subset E^* \otimes E \otimes \mathscr{E}^* \otimes \mathscr{E}^*,
\]
and is represented by the function
\begin{multline*}
\big(s, \widehat{n}\cdot \widehat{l}, s', \widehat{n'}\cdot \widehat{l'}; a,x, b,y\big)
\longmapsto \kappa_{0,1} \otimes \kappa_{1,0} \big(s, \widehat{n}\cdot \widehat{l}, s', \widehat{n'}\cdot \widehat{l'}; a,x, b,y\big)
\\
=
\kappa_{0,1}\big(s, \widehat{n}\cdot \widehat{l}; a,x\big)\kappa_{1,0}\big(s', \widehat{n'}\cdot \widehat{l'};b,y\big).
\end{multline*}

Thus, the last (discrete) momentum variable of the kernel $\kappa_{0,1}$ (designating the points $(s, \widehat{n}\cdot \widehat{l}\big)$ of the orbit 
$\mathscr{O}$) can be contracted with the first (discrete) momentum variable of the kernel $\kappa_{1,0}$ 
(designating the points $(s', \widehat{n'}\cdot \widehat{l'}\big)$ of the orbit 
$\mathscr{O}$)
and gives a well defined element of $\mathscr{E}^* \otimes \mathscr{E}^*$.
Namely
\begin{multline*}
\kappa_{0,1} \otimes_1 \kappa_{1,0}(a,x,b,y) = 
\sum \limits_{\substack{(s,\widehat{n}\cdot\widehat{l}) \in \mathscr{O} \\ (s',\widehat{n;}\cdot\widehat{l'})  \in \mathscr{O} }}
\delta_{s \, s'}\delta_{n \, n'} \delta_{l \, l'}
\kappa_{0,1}\big(s, \widehat{n}\cdot \widehat{l}; a,x\big)\kappa_{1,0}\big(s', \widehat{n'}\cdot \widehat{l'};b,y\big)
\\
=
\sum\limits_{(s,\widehat{n}\cdot\widehat{l}) \in \mathscr{O}}
\kappa_{0,1}\big(s, \widehat{n}\cdot \widehat{l}; a,x\big)\kappa_{1,0}\big(s, \widehat{n}\cdot \widehat{l};b,y\big),
\end{multline*}
with
\[
\kappa_{0,1} \otimes_1 \kappa_{1,0} \in \mathscr{E}^* \otimes \mathscr{E}^*.
\]

The pairing distribution is equal to the dual pairing ($1$-contraction)
\begin{multline*}
-i D^{(-) \, ab}(x,y) \,\, \boldsymbol{1} \, = [\mathbb{A}^{(-) \, a}(x), \mathbb{A}^{(+) \, b}(y)]_{\mp} 
\\
=
\quad \underbracket{\mathbb{A}^{a}(x) \mathbb{A}^{b}}(y)
= \kappa_{0,1} \otimes_1 \kappa_{1,0}(a,x,b,y), 
\end{multline*}
which follows from our 
general analysis of Subsections \ref{psiBerezin-Hida} and \ref{WickForChronological}, and
which remains generally valid also for discrete orbit $\mathscr{O}$ free fields. 
Then it follows immediately the assertion of Lemma \ref{Pairings} as the immediate consequence of the fact that
\begin{align*}
\kappa_{0,1} \in \mathscr{L}(E^*, \mathscr{E}^*) \cong E \otimes \mathscr{E}^* \subset E^* \otimes \mathscr{E}^*,
\\
\kappa_{1,0} \in \mathscr{L}(E^*, \mathscr{E}^*) \cong E \otimes \mathscr{E}^* \subset E^* \otimes \mathscr{E}^*
\end{align*}

We nonetheless have proved the Lemma \ref{Pairings} immediately, because the method of the proof
of Lemma \ref{Pairings} can be adopted to the proof of the point
1) of Lemma \ref{theta.kappaBarDotOtimeskappaOnEU}.  Speaking otherwise we need Lemma \ref{retOfPairings} or
1) of Lemma \ref{theta.kappaBarDotOtimeskappaOnEU}, which cannot be inferred from
\begin{align*}
\kappa_{0,1} \in \mathscr{L}(E^*, \mathscr{E}^*) \cong E \otimes \mathscr{E}^* \subset E^* \otimes \mathscr{E}^*,
\\
\kappa_{1,0} \in \mathscr{L}(E^*, \mathscr{E}^*) \cong E \otimes \mathscr{E}^* \subset E^* \otimes \mathscr{E}^*,
\end{align*}
because the properties of pointwise multiplication 
operation by the step theta function are not obvious and cannot be established without any use of an explicit norm estimation. 
Of course having given 1) of Lemma \ref{theta.kappaBarDotOtimeskappaOnEU}, or 
equivalently
\begin{align*}
\theta_y \kappa_{0,1} \in \mathscr{L}(E^*, \mathscr{E}^*) \cong E \otimes \mathscr{E}^* \subset E^* \otimes \mathscr{E}^*,
\\
\kappa_{1,0} \in \mathscr{L}(E^*, \mathscr{E}^*) \cong E \otimes \mathscr{E}^* \subset E^* \otimes \mathscr{E}^*,
\end{align*}
we obtain immediately
\[
\theta_y \kappa_{0,1} \otimes_1 \kappa_{1,0} \in \mathscr{E}^* \otimes \mathscr{E}^*,
\]
where $\theta_y(x) = \theta\big(xy^{-1}\big)$.
From this we immediately obtain the assertion of Lemma \ref{retOfPairings} because 
\[
-i \textrm{ret} D^{(-) \, ab}(x,y) = -i D^{(-)\, \textrm{ret} \, ab}(x,y) = \theta_y \kappa_{0,1} \otimes_1 \kappa_{1,0}(x,y),
\]
but the explicit proof of Lemma \ref{retOfPairings} or of 1) of Lemma \ref{theta.kappaBarDotOtimeskappaOnEU} cannot be avoided.
Indeed, Lemmas \ref{theta.kappaBarDotOtimeskappaWithOneMasslessOnEU} and
\ref{q-contractionkappaBarDotOtimeskappaWithOneMasslessOnEU} are no longer true if we have more than one massless kernel in them
or more than one kernel associated to infinite orbit, which illustrate non-trivial character of the pointwise multiplication operation 
by the step theta function. Below in this Subsection we analyse this more subtle situation.

Note that in the general case, \emph{i.e.} with arbitrary number of
massless (or infinite orbit) plane wave kernels in the dot product $\dot{\otimes}$
we have
\begin{align*}
\big(\kappa^{(1)}_{1, 0} \overset{\cdot}{\otimes} \cdots \overset{\cdot}{\otimes} \kappa^{(q)}_{1, 0}\big) & \in
E_{{}_{(1)}} \otimes \cdots \otimes E_{{}_{(q)}} \otimes \mathscr{E}^*
\\
\big(\kappa^{(1)}_{0, 1}\overset{\cdot}{\otimes} \cdots \overset{\cdot}{\otimes} \kappa^{(q)}_{0, 1}\big) & \in
E_{{}_{(1)}}^* \otimes \cdots \otimes E_{{}_{(q)}}^* \otimes \mathscr{E}^*
\end{align*}
by the general kernel theorem and by Lemma \ref{kappaBarDotOtimeskappaOnEU}, so
\begin{multline*}
\big(\kappa^{(1)}_{1, 0} \overset{\cdot}{\otimes} \cdots \overset{\cdot}{\otimes} \kappa^{(q)}_{1, 0}\big) 
\otimes
\big(\kappa^{(1)}_{0, 1}\overset{\cdot}{\otimes} \cdots \overset{\cdot}{\otimes} \kappa^{(q)}_{0, 1}\big)
\\
\in
E_{{}_{(1)}} \otimes \cdots \otimes E_{{}_{(q)}} 
\otimes E_{{}_{(1)}}^* \otimes \cdots \otimes E_{{}_{(q)}}^* \otimes \mathscr{E}^* \otimes \mathscr{E}^*,
\end{multline*}
and the $q'$-contraction 
\begin{multline*}
\big(\kappa^{(1)}_{1, 0} \overset{\cdot}{\otimes} \cdots \overset{\cdot}{\otimes} \kappa^{(q)}_{1, 0}\big) 
\otimes_{q'}
\big(\kappa^{(1)}_{0, 1}\overset{\cdot}{\otimes} \cdots \overset{\cdot}{\otimes} \kappa^{(q)}_{0, 1}\big)
\\
\in 
E_{{}_{(1)}} \otimes \widehat{\cdots} \otimes E_{{}_{(q)}} 
\otimes E_{{}_{(1)}}^* \otimes \widehat{\cdots} \otimes E_{{}_{(q)}}^* \otimes \mathscr{E}^* \otimes \mathscr{E}^*
\end{multline*}
makes sense. Here the hat $\widehat{\cdots}$ means that $q'$ pairs $E_{{}_{(i)}},E_{{}_{(i)}}^*$ of the contracted 
momentum nuclear spaces are removed from the tensor product 
\[
E_{{}_{(1)}} \otimes \cdots \otimes E_{{}_{(q)}} 
\otimes E_{{}_{(1)}}^* \otimes \cdots \otimes E_{{}_{(q)}}^* \otimes \mathscr{E}^* \otimes \mathscr{E}^*.
\]
As usual $\otimes_{q'}$ stands always for the contraction of $q'$ pairs of momentum variables and the contraction
in our work is never  applied to the space-time variables, compare also Subsections \ref{WickForProduct}
and \ref{WickForChronological}. In particular in case $q'=q$
\[
\big(\kappa^{(1)}_{1, 0} \overset{\cdot}{\otimes} \cdots \overset{\cdot}{\otimes} \kappa^{(q)}_{1, 0}\big) 
\otimes_{q}
\big(\kappa^{(1)}_{0, 1}\overset{\cdot}{\otimes} \cdots \overset{\cdot}{\otimes} \kappa^{(q)}_{0, 1}\big)
\in  \mathscr{E}^* \otimes \mathscr{E}^*
\]
and all pairs of the corresponding momentum variables are paired so that 
\[
\big(\kappa^{(1)}_{1, 0} \overset{\cdot}{\otimes} \cdots \overset{\cdot}{\otimes} \kappa^{(q)}_{1, 0}\big) 
\otimes_{q}
\big(\kappa^{(1)}_{0, 1}\overset{\cdot}{\otimes} \cdots \overset{\cdot}{\otimes} \kappa^{(q)}_{0, 1}\big)
\]
is a scalar-valued distribution in two space-time variables.

We pass now to the more subtle situation: the construction of the retarded part 
\begin{equation}\label{thetay(kk'...k''')q-contraction(k,k',...k''')}
\big(\theta_y\kappa^{(1)}_{1, 0} \overset{\cdot}{\otimes} \cdots \overset{\cdot}{\otimes} \kappa^{(q)}_{1, 0}\big) \otimes_{q'}
\big(\kappa^{(1)}_{0, 1}\overset{\cdot}{\otimes} \cdots \overset{\cdot}{\otimes} \kappa^{(q)}_{0, 1}\big),
\,\,\,\,\,\, q' \leq q 
\end{equation}
of the contraction
\begin{equation}\label{(kk'...k''')q-contraction(k,k',...k''')}
\big(\kappa^{(1)}_{1, 0} \overset{\cdot}{\otimes} \cdots \overset{\cdot}{\otimes} \kappa^{(q)}_{1, 0}\big) \otimes_{q'}
\big(\kappa^{(1)}_{0, 1}\overset{\cdot}{\otimes} \cdots \overset{\cdot}{\otimes} \kappa^{(q)}_{0, 1}\big) 
\end{equation}
of Lemma \ref{q-contractionkappaBarDotOtimeskappaWithOneMasslessOnEU}, in case we have more than just one 
massless (or infinite orbit) plane wave kernel in ech of the two dot product $\dot{\otimes}$
factors. This retarded (and advanced) part is necessary for the construction of the scattering matrix operator 
in Quantum Field Theory (QFT) on the Einstein Universe with interaction
$\mathcal{L}$ whose Wick monomials contain more than one massless (or infinite orbit) field. If each of the dot factors of the
contraction includes more than one zero mass fields
(or infinite orbit fields) then the contraction discrete integral/sum (\ref{thetay(kk'...k''')q-contraction(k,k',...k''')}) 
is not convergent in the whole test space $\mathscr{E} \otimes \mathscr{E}$, but only in a closed
subspace of it, similarly as for the retarded part 
(\ref{theta(x-y)masslesskappa01.masslesskappa10contractionmasslesskappa10.masslesskappa10q}), Subsection \ref{WickForChronological}, 
of the contracted kernels on the Minkowski space-time, Subsection \ref{WickForChronological}.

Note that whenever $q'=q$ in (\ref{(kk'...k''')q-contraction(k,k',...k''')}), then
(\ref{(kk'...k''')q-contraction(k,k',...k''')}) is $\widetilde{\mathbb{S}^1}\times SU(2, \mathbb{C})$-invariant.

Using the $\widetilde{\mathbb{S}^1}\times SU(2, \mathbb{C})$-invariance of the contraction kernel 
(\ref{(kk'...k''')q-contraction(k,k',...k''')}) with $q'=q$, and of 
(\ref{thetay(kk'...k''')q-contraction(k,k',...k''')}) with $q'=q$, we can
write 
\begin{align*}
\big(\theta_y\kappa^{(1)}_{1, 0} \overset{\cdot}{\otimes} \cdots \overset{\cdot}{\otimes} \kappa^{(q)}_{1, 0}\big) \otimes_q
\big(\kappa^{(1)}_{0, 1}\overset{\cdot}{\otimes} \cdots \overset{\cdot}{\otimes} \kappa^{(q)}_{0, 1}\big)(x,y)
=
\textrm{ret}\kappa_q\big(xy^{-1}\big)
\\
\big(\kappa^{(1)}_{1, 0} \overset{\cdot}{\otimes} \cdots \overset{\cdot}{\otimes} \kappa^{(q)}_{1, 0}\big) \otimes_q
\big(\kappa^{(1)}_{0, 1}\overset{\cdot}{\otimes} \cdots \overset{\cdot}{\otimes} \kappa^{(q)}_{0, 1}\big)(x,y) 
= \kappa_q\big(xy^{-1}\big),
\end{align*}
with the corresponding kernels $\kappa_q, \textrm{ret}\kappa_q$ in one space-time variable,
analogously as on the Minkowski space-time, Subsection \ref{WickForChronological}. It turns out that if $q>1$, then the integral
\[
\textrm{ret}\kappa_q(\phi) = \int \limits_{\widetilde{\mathbb{S}^1}\times SU(2, \mathbb{C})} \textrm{ret}\kappa_q(x) \, \phi(x) dx,
\]
giving the evaluation of $\textrm{ret}\kappa_q$ at $\phi$,
is not convergent for each $\phi \in \mathscr{E}$, similarly as on the Minkowski space-time. But this integral becomes absolutely
convergent and even uniformly bounded whenever $\phi$ ranges over a bounded set in $\mathscr{E}$ of all $\phi$ whose time
derivatives on the two Cauchy surfaces $t=0$ and $t=2\pi$ vanish up to some finite order $\omega$, depending
on the distribution (\ref{(kk'...k''')q-contraction(k,k',...k''')}) (or, more precisely, on the concrete plane wave kernels contracted in it 
and on the number of contracted kernels). Here, on the (compactified) Einstein Universe the retarded part of the $q$-contraction 
constructed with 
the $\theta$ function  is convergent on test functions $\phi$ which have vanishing time derivatives on 
the two Cauchy surfaces, $t=0$ and $t=2\pi$,
because these are the two Cauchy surfaces ($\textrm{mod}4\pi$)
\[
\mathbb{S}^3 \sqcup \mathbb{S}^3 = SU(2, \mathbb{C}) \sqcup SU(2, \mathbb{C}) \subset \widetilde{\mathbb{S}^1}\times SU(2, \mathbb{C})
\]
at which the periodic step theta function
$\theta$ has a jump. The other difference, in comparison to the flat Minkowski space-time, is that here
on the Einstein Universe, where the fields live effectively on its compactification
$\widetilde{\mathbb{S}^1}\times SU(2, \mathbb{C})$, the contraction distributions (\ref{(kk'...k''')q-contraction(k,k',...k''')})  
do not have in general their support confined to the closure of the interior of the past and future light cone, 
and here only time derivatives must vanish on the whole
Cauchy surfaces $t=0$ and $t=2\pi$, and not only at single point of one Cauchy surface, as was the case on the Minkowski space-time. 
Therefore, the abstract splitting of the contraction kernel $\kappa_q$
into retarded and advanced part is much less unique in comparison to the Minkowski space-time. This arbitrariness can however be eliminated,
similarly as on the Minkowski space time, as we explain the following Subsections.
 
Here $\kappa_q$ can be split
\begin{align*}
\kappa_q =  \textrm{ret}\kappa_q - \textrm{av} \kappa_q, & & \check{\theta}(x) = \theta(x^{-1}) 
\\
\textrm{supp} \, \textrm{ret}\kappa_q =  \textrm{supp} \, \theta, 
&&\textrm{supp} \, \textrm{av}\kappa_q =  \textrm{supp} \,  \check{\theta},
\\
\textrm{for} \, x= (t, \boldsymbol{w}) & & x^{-1} = (-t, \boldsymbol{w}^{-1}),
\end{align*}
with the splitting determined up to a common additive term in the retarded
and advanced parts which is a distribution concentrated on the two Cauchy surfaces 
$t=0$ and $t=2\pi$. Accordingly, the retarded and advanced parts of the contraction kernel 
(\ref{(kk'...k''')q-contraction(k,k',...k''')}) are determined up to a kernel concentrated on
\[
\big[SU(2, \mathbb{C}) \sqcup SU(2, \mathbb{C})\big] \times \widetilde{\mathbb{S}^1}\times SU(2, \mathbb{C})
\subset \big[ \widetilde{\mathbb{S}^1}\times SU(2, \mathbb{C}) \big]^{\times \, 2}.
\]
In fact the construction runs completely analogously as for the Minkowski space-time $\mathbb{R}^4$, Subsection \ref{WickForChronological}.
We replace the Abelian additive Lie group $\mathbb{R}^4$ and the additive group multiplication and its inverse $x+y, x-y$ in it by the non-Abelian
Lie group $\widetilde{\mathbb{S}^1}\times SU(2, \mathbb{C})$ with the non-Abelian group muliplication and its inverse $xy, xy^{-1}$ in 
$\widetilde{\mathbb{S}^1}\times SU(2, \mathbb{C})$. Analogously the direct product Lie group
\[
\mathbb{R}^{4n} = \big[\mathbb{R}^4 \big]^{\times \, n}
\]
we replace with the direct product Lie group
\[
\big[ \widetilde{\mathbb{S}^1}\times SU(2, \mathbb{C}) \big]^{\times \, n}.
\]
In particular, the invertible homogeneous and additive group maps
\begin{multline*}
L: \big[\mathbb{R}^4 \big]^{\times \, n} \ni (x_1, \ldots, x_n)
\\
\longmapsto (x_1 + \ldots + x_n, x_2+ x_n, \ldots, x_{n-1} +x_n, x_n ) \in  \big[\mathbb{R}^4 \big]^{\times \, n}
\end{multline*}
and
\begin{multline*}
L^{-1}: \big[\mathbb{R}^4 \big]^{\times \, n} \ni (x_1, \ldots, x_n)
\\
\longmapsto (x_1 - x_n, x_2 - x_n, \ldots, x_{n-1} - x_n, x_n ) \in  \big[\mathbb{R}^4 \big]^{\times \, n}
\end{multline*}
of Subsection \ref{WickForChronological} we replace here on the Einstein Universe by the invertible group maps
\begin{multline*}
L: \big[ \widetilde{\mathbb{S}^1}\times SU(2, \mathbb{C}) \big]^{\times \, n}
\ni (x_1, \ldots, x_n)
\\
\longmapsto (x_1x_n, x_2 x_n, \ldots, x_{n-1} x_n, x_n ) \in
\big[ \widetilde{\mathbb{S}^1}\times SU(2, \mathbb{C}) \big]^{\times \, n}
\end{multline*}
and
\begin{multline*}
L: \big[ \widetilde{\mathbb{S}^1}\times SU(2, \mathbb{C}) \big]^{\times \, n}
\ni (x_1, \ldots, x_n)
\\
\longmapsto (x_1x_{n}^{-1}, x_2x_{n}^{-1}, \ldots, x_{n-1} x_{n}^{-1}, x_n ) \in
\big[ \widetilde{\mathbb{S}^1}\times SU(2, \mathbb{C}) \big]^{\times \, n}
\end{multline*}
acting on the Lie group
\[
\big[ \widetilde{\mathbb{S}^1}\times SU(2, \mathbb{C}) \big]^{\times \, n}.
\]

The only difference is that we replace the auxiliary functions $\omega_{{}_{0 \,\, \alpha}}$ and the operator
$\Omega'$ of Subsection \ref{WickForChronological} with the corresponding two similar sets of the analogous functions
$\{\omega_{{}_{0 \,\, I \,\, \alpha}}, \omega_{{}_{0 \,\, II \,\, \alpha}} \}$ and the operator $\Omega'$ projecting on the subspace 
of the test space $\mathscr{E}^{\otimes \, (k-1)} = \mathcal{S}_{\Delta+1}\big( \widetilde{\mathbb{S}^1}\times SU(2, \mathbb{C})\big)^{\otimes \, (k-1)}$
consisting of functions whose time derivatives   
vanish on the two Cauchy surfaces $t=0$ and $t= 2\pi$ up to a finite order $\omega$ in each space-time variable.

Now we explain this in detail on the particular example of (\ref{(kk'...k''')q-contraction(k,k',...k''')})
in which we have just two massless (or any infinite orbit) plane wave kernels in each of the two factors
of (\ref{(kk'...k''')q-contraction(k,k',...k''')}), \emph{i.e} with $q'=q$ and with $q=2$. Compare the Minkowski analogue
(\ref{theta(x-y)masslesskappa01.masslesskappa10contractionmasslesskappa10.masslesskappa10}) of 
(\ref{thetay(kk'...k''')q-contraction(k,k',...k''')}) with $q'=q$ and $q=2$.
The analysis of the general case of 
(\ref{thetay(kk'...k''')q-contraction(k,k',...k''')}) is the same as 
(\ref{thetay(kk'...k''')q-contraction(k,k',...k''')}) with $q'=q$ and with $q=2$. 

Let 
\begin{align*}
\mathbb{A}' = \Xi(\kappa'_{0,1}) +\Xi(\kappa'_{1,0})  & = \mathbb{A}'^{(-)} +\mathbb{A}'^{(+)}, 
\\
\mathbb{A}''  = \Xi(\kappa''_{0,1}) +\Xi(\kappa''_{1,0}) & = \mathbb{A}''^{(-)} +\mathbb{A}''^{(+)}
\end{align*}
be two massless free fields, or more generally, free fields with infinite orbits $\mathscr{O}', \mathscr{O}''$
of the general class of Definition \ref{K_0onEU}.  Let both $\mathbb{A}', \mathbb{A}''$ be positive energy fields.
Recall that $\mathscr{O}', \mathscr{O}''$ are equal to  $\mathscr{O}'_+,\mathscr{O}''_+ $ with each of the 
points of the orbit $\mathscr{O}'_+,\mathscr{O}''_+ $ counting several times, depending on the complete set of fundamental solutions in the single particle Hilbert spaces, compare discussion above. Accordingly to the notation explained above in this Subsection
we write $\big(s', \widehat{n'}\cdot \widehat{l'}\big) \in \mathscr{O}'$ for the points 
of $\mathscr{O}'$ with  $\widehat{n'}\cdot \widehat{l'} \in \mathscr{O}'_+$. Analogously we denote
the points  $\big(s'', \widehat{n''}\cdot \widehat{l''}\big)$ of $\mathscr{O}''$ with $\widehat{n''}\cdot \widehat{l''} \in \mathscr{O}''_+$.

Let 
\begin{align*}
E_{{}_{'}} & \,\,\,\,\,\,\,\,\,\,\,\,\,\,\,\,\,\,\,\,\,\,\,\,\,\,\,\,\,  \subset & \mathscr{H}_{{}_{'}} 
& \,\,\,\,\,\,\,\,\,\,\,\,\,\,\,\,\,\,\,\,\,\,\,\,\,\,\,\,\,  \subset & E_{{}_{'}}^{*}, 
\\
E_{{}_{''}} & \,\,\,\,\,\,\,\,\,\,\,\,\,\,\,\,\,\,\,\,\,\,\,\,\,\,\,\,\,  \subset & \mathscr{H}_{{}_{''}} 
& \,\,\,\,\,\,\,\,\,\,\,\,\,\,\,\,\,\,\,\,\,\,\,\,\,\,\,\,\,  \subset & E_{{}_{''}}^{*},
\end{align*}
be the single particle Gelfand triples, respectively, of the fields $\mathbb{A}', \mathbb{A}''$ with the spaces
\begin{align*}
E_{{}_{'}} = S_{{}_{\widetilde{A}}}(\mathscr{O}') = S_{{}_{\widetilde{\Delta +1}}}(\mathscr{O}')&  & \mathscr{H}_{{}_{'}} = L^2(\mathscr{O}')&  & E_{{}_{'}}^{*} 
= S_{{}_{\widetilde{A}}}(\mathscr{O}')^*= S_{{}_{\widetilde{\Delta +1}}}(\mathscr{O}')^*, 
\\
E_{{}_{''}} = S_{{}_{\widetilde{A}}}(\mathscr{O}'') = S_{{}_{\widetilde{\Delta +1}}}(\mathscr{O}'') &  & \mathscr{H}_{{}_{''}} = L^2(\mathscr{O}'')&  & E_{{}_{''}}^{*} 
= S_{{}_{\widetilde{A}}}(\mathscr{O}'')^*  = S_{{}_{\widetilde{\Delta +1}}}(\mathscr{O}')^*,
\end{align*}
of $\mathbb{C}^{d_{{}_{'}}}$ or $\mathbb{C}^{d_{{}_{''}}}$-valued functions, depending on the particular type of the free fields $\mathbb{A}', \mathbb{A}''$.
Recall that 
\[
d_{{}_{'}} = \textrm{dim} \, V', \,\,\,\,\,\, d_{{}_{''}} = \textrm{dim} \, V'',
\]
where $V', V''$ are the finite dimensional representations of $SL(2, \mathbb{C})$ associated to the fields $\mathbb{A}', \mathbb{A}''$,
as we have explained in details above in this Subsection. 
Let the (discrete) momentum variables designating respectively the points of the orbits $\mathscr{O}', \mathscr{O}''$
be denoted, respectively, by
\begin{align*}
\big(s', \widehat{n'}\cdot \widehat{l'}\big) \in \mathscr{O}' 
& \,\,\,\,\,\,\,\,\,\,\,\,\,\,\,\,\,\,\,\,\,\,\,\,\,\,\,\,\,  \textrm{or} & \big(s''', \widehat{n'}\cdot \widehat{l'''}\big) \in \mathscr{O}', 
\\
\big(s'', \widehat{n''}\cdot \widehat{l''}\big) \in \mathscr{O}'' 
& \,\,\,\,\,\,\,\,\,\,\,\,\,\,\,\,\,\,\,\,\,\,\,\,\,\,\,\,\,  \textrm{or} & \big(s'''', \widehat{n'}\cdot \widehat{l''''}\big) \in \mathscr{O}''.
\end{align*}
Let the corresponding plane wave kernel functions be
\begin{align*}
&\kappa'_{0,1}\big(s', \widehat{n'}\cdot \widehat{l'}; a,x\big) 
= \sum \limits_{-l' \leq i',j' \leq l'} \sqrt{2l'+1} {u^{a}_{s'}}_{{}_{j'i'}}(\widehat{n'}\cdot \widehat{l'})
\widehat{n'}(t) \widehat{l'}_{{}_{i' \, j'}}(\boldsymbol{w}), 
\\ 
& \,\,\,\,\,\,\,\,\,\,\,\,\,\,\,\,\,\,\,\,\,\,\,\,\,\,\,\,\,\,\,\,\,\,\,\,\,\,\,\,\,\,\,\,\,\,\,\,\,\,\,\,\,\,\,\,  
x = (t, \boldsymbol{w}), \big(s', \widehat{n'}\cdot \widehat{l'}\big) \in \mathscr{O}',
\\
&\kappa'_{1,0}\big(s''', \widehat{n'''}\cdot \widehat{l'''}; c,y\big) 
\\
&= \sum \limits_{-l''' \leq i''',j' \leq l'''} \sqrt{2l'''+1} 
\overline{{u^{a}_{s'''}}_{{}_{j'''i'''}}(\widehat{n'''}\cdot \widehat{l'''})} \,
\overline{\widehat{n'''}(t)} \, \overline{\widehat{l'''}_{{}_{i''' \, j'''}}(\boldsymbol{v})},
\\
& \,\,\,\,\,\,\,\,\,\,\,\,\,\,\,\,\,\,\,\,\,\,\,\,\,\,\,\,\,\,\,\,\,\,\,\,\,\,\,\,\,\,\,\,\,\,\,\,\,\,\,\,\,\,\,\,  
y = (\tau, \boldsymbol{v}), \big(s''', \widehat{n'''}\cdot \widehat{l'''}\big) \in \mathscr{O}'
\\
&\kappa''_{0,1}\big(s'', \widehat{n''}\cdot \widehat{l''}; b,x\big) 
= \sum \limits_{-l'' \leq i'',j'' \leq l''} \sqrt{2l''+1} {u^{b}_{s''}}_{{}_{j''i''}}(\widehat{n''}\cdot \widehat{l''})
\widehat{n''}(t) \widehat{l'}_{{}_{i' \, j'}}(\boldsymbol{w}), 
\\
& \,\,\,\,\,\,\,\,\,\,\,\,\,\,\,\,\,\,\,\,\,\,\,\,\,\,\,\,\,\,\,\,\,\,\,\,\,\,\,\,\,\,\,\,\,\,\,\,\,\,\,\,\,\,\,\, 
x = (t, \boldsymbol{w}), \big(s'', \widehat{n''}\cdot \widehat{l''}\big) \in \mathscr{O}''
\\
&\kappa''_{1,0}\big(s'''', \widehat{n''''}\cdot \widehat{l''''}; d,y\big) 
\\
&= \sum \limits_{-l'''' \leq i'''',j'''' \leq l''''} \sqrt{2l''''+1} 
\overline{{u^{a}_{s''''}}_{{}_{j'''i'''}}(\widehat{n''''}\cdot \widehat{l''''})} \,
\overline{\widehat{n''''}(t)} \, \overline{\widehat{l''''}_{{}_{i'''' \, j''''}}(\boldsymbol{v})}, 
\\ 
& \,\,\,\,\,\,\,\,\,\,\,\,\,\,\,\,\,\,\,\,\,\,\,\,\,\,\,\,\,\,\,\,\,\,\,\,\,\,\,\,\,\,\,\,\,\,\,\,\,\,\,\,\,\,\,\, 
y = (\tau, \boldsymbol{v}), \big(s'''', \widehat{n''''}\cdot \widehat{l''''}\big) \in \mathscr{O}''.
\end{align*}
\[
1 \leq a,c \leq d_{{}_{'}}, \,\,\,\, 1 \leq b,d \leq d_{{}_{''}}.
\]
Here $u$ stand for the Fourier transforms of the fundamental solutions, introduced above. Here if  the variables
in $u$ have the odd number of primes, then $u$ is understood as the Fourier transform of the fundamental
solution in the single particle Hilbert space of the field $\mathbb{A}'$. If the number of primes in the variables
of $u$ is even, then $u$ is understood as the Fourier transform of the fundamental solution in the single particle
Hilbert space of the field $\mathbb{A}''$. We introduce this notation in order to reduce the number of indices.
In order to compute the contraction kernel function
\[
(a,b,c,d,x,y) \longmapsto \big(\kappa'_{0,1} \dot{\otimes} \kappa''_{0,1} \big) 
\otimes_2 \big(\kappa'_{1,0} \dot{\otimes} \kappa''_{1,0} \big)(a,b,c,d,x,y )
\]
at
\[
(a,b,c,d,x,y)
\]
we equate the momentum variables
\begin{align*}
\big(s', \widehat{n'}\cdot \widehat{l'}\big) = \big(s''', \widehat{n'''}\cdot \widehat{l'''}\big) \in \mathscr{O}',
\\
\big(s'', \widehat{n''}\cdot \widehat{l''}\big) = \big(s'''', \widehat{n''''}\cdot \widehat{l''''}\big) \in \mathscr{O}''
\end{align*}
in
\begin{multline*}
\big(\kappa'_{0,1} \dot{\otimes} \kappa''_{0,1} \big) 
\otimes \big(\kappa'_{1,0} \dot{\otimes} \kappa''_{1,0} \big) 
\big(s', \widehat{n'}\cdot \widehat{l'}, s'', \widehat{n''}\cdot \widehat{l''}, s''', 
\widehat{n'''}\cdot \widehat{l'''}, s'''', \widehat{n''''}\cdot \widehat{l''''}; a,b,c,d,x,y  \big)
\end{multline*}
\begin{multline*}
=
\kappa'_{0,1}\big(s', \widehat{n'}\cdot \widehat{l'}; a,x\big) 
\kappa''_{0,1}\big(s'', \widehat{n''}\cdot \widehat{l''}; b,x\big) 
\kappa'_{1,0}\big(s''', \widehat{n'''}\cdot \widehat{l'''}; c,y\big) 
\kappa''_{1,0}\big(s'''', \widehat{n''''}\cdot \widehat{l''''}; d,y\big) 
\end{multline*}
and perform summation with respect to
\[
\big(s', \widehat{n'}\cdot \widehat{l'}\big) \in \mathscr{O}'
\,\,\,\, \textrm{and} \,\,\,\,
\big(s'', \widehat{n''}\cdot \widehat{l''}\big) \in \mathscr{O}''.
\]
Next we compute the kernel function
\begin{multline*}
\theta_y \big(\kappa'_{0,1} \dot{\otimes} \kappa''_{0,1} \big) \otimes_2 \big(\kappa'_{1,0} \dot{\otimes} \kappa''_{1,0} \big)(a,b,c,d,x,y)
\\
=
\theta_y(x) \big(\kappa'_{0,1} \dot{\otimes} \kappa''_{0,1} \big) \otimes_2 \big(\kappa'_{1,0} \dot{\otimes} \kappa''_{1,0} \big)(a,b,c,d,x,y)
\\
=
\theta\big(xy^{-1}\big) \big(\kappa'_{0,1} \dot{\otimes} \kappa''_{0,1} \big) \otimes_2 \big(\kappa'_{1,0} \dot{\otimes} \kappa''_{1,0} \big)(a,b,c,d,x,y),
\end{multline*}
by the ordinary pointwise (in space-time variables $(x,y)$) multiplication by $\theta_y(x) = \theta\big(xy^{-1}\big)$.
Then the evaluation of the contraction integral kernel
\[
\theta_y \big(\kappa'_{0,1} \dot{\otimes} \kappa''_{0,1} \big) \otimes_2 \big(\kappa'_{1,0} \dot{\otimes} \kappa''_{1,0} \big),
\]
at a general simpe tensor 
\begin{multline*}
\chi = \phi \otimes \varphi \in \mathscr{E} \otimes \mathscr{E} 
= S_{A}\big(\widetilde{\mathbb{S}^1}\times SU(2, \mathbb{C}); \, \mathbb{C}^{d_{{}_{'}}d_{{}_{''}}}\big) \otimes 
S_{A}\big(\widetilde{\mathbb{S}^1}\times SU(2, \mathbb{C}); \mathbb{C}^{d_{{}_{'}}d_{{}_{''}}}\big) 
\\
= 
S_{A\otimes A}\big(\big[ \widetilde{\mathbb{S}^1}\times SU(2, \mathbb{C}) \big]^{\times \, 2}; \, \mathbb{C}^{d_{{}_{'}}d_{{}_{'}}d_{{}_{''}}d_{{}_{''}}}\big),
\,\,\,\,\,\,\,\,\,\, A = \Delta +1,
\end{multline*}
is equal
\begin{multline*}
\Big(\theta_y \big(\kappa'_{0,1} \dot{\otimes} \kappa''_{0,1} \big) 
\otimes_2 \big(\kappa'_{1,0} \dot{\otimes} \kappa''_{1,0} \big)\Big)(\phi \otimes \varphi)
=  \Big\langle \theta_y \big(\kappa'_{0,1} \dot{\otimes} \kappa''_{0,1} \big) 
\otimes_2 \big(\kappa'_{1,0} \dot{\otimes} \kappa''_{1,0} \big),  \, \phi \otimes \varphi \Big\rangle
\end{multline*}
\begin{multline}\label{<thetaykappa_2,phixvarphi>'}
=\sum
\int\limits_{\big[ \widetilde{\mathbb{S}^1}\times SU(2, \mathbb{C}) \big]^{\times \, 2}}
(2l' +1)(2l''+1) {u^{a}_{s'}}_{{}_{j'i'}}(\widehat{n'}\cdot \widehat{l'}) \, {u^{b}_{s''}}_{{}_{j''i''}}(\widehat{n''}\cdot \widehat{l''}) \,\, \times
\\
\times \,\,
\underset{l' \,\, i'j'}{M}_{{}_{l'' \,\, i''j''}}^{{}^{l^0 \,\, i^0j^0}}
\, \overline{\widehat{l^0}_{{}_{-i^0 \,\, -j^0}}(\boldsymbol{w})} \,
\overline{\widehat{-n'-n''}(t-\tau)} \, 
 \theta(t-\tau) \phi^{ab}(t, \boldsymbol{w}) \,\, \times 
\\
\times \,\,
\overline{{u^{c}_{s'}}_{{}_{j'''i'''}}(\widehat{n'}\cdot \widehat{l'})} \, 
\overline{{u^{d}_{s''}}_{{}_{j''''i''''}}(\widehat{n''}\cdot \widehat{l''})} \, 
\overline{\underset{l' \,\, i'''j'''}{M}_{{}_{l'' \,\, i''''j''''}}^{{}^{l \,\, ij}}} \, 
\overline{\widehat{l}_{{}_{ij}}(\boldsymbol{v})} \, 
\varphi^{cd}(\tau, \boldsymbol{v})
\, \ud t \, \ud \boldsymbol{w} \, \ud \tau \, \ud \boldsymbol{v} 
\end{multline}
\begin{multline}\label{<thetaykappa_2,phixvarphi>}
=
\sum\limits_{n\in \mathbb{Z}}
\sum 
(2l' +1)(2l''+1) {u^{a}_{s'}}_{{}_{j'i'}}(\widehat{n'}\cdot \widehat{l'}) \, {u^{b}_{s''}}_{{}_{j''i''}}(\widehat{n''}\cdot \widehat{l''}) \,\, \times
\\
\times \,\,
\underset{l' \,\, i'j'}{M}_{{}_{l'' \,\, i''j''}}^{{}^{l^0 \,\, i^0j^0}} \,
 \widehat{\theta}_n \widetilde{\phi^{ab}}_{{}_{j^0i^0}}\big((\widehat{-n'-n''-n})\cdot \overline{\widehat{l^0}} \big) \,\, \times 
\\
\times \,\,
\overline{{u^{c}_{s'}}_{{}_{j'''i'''}}(\widehat{n'}\cdot \widehat{l'})} \, 
\overline{{u^{d}_{s''}}_{{}_{j''''i''''}}(\widehat{n''}\cdot \widehat{l''})} \, 
\overline{\underset{l' \,\, i'''j'''}{M}_{{}_{l'' \,\, i''''j''''}}^{{}^{l \,\, ij}}} \, 
\widetilde{\varphi^{cd}}_{{}_{ji}}\big((\widehat{n'+n''+n})\cdot \overline{\widehat{l}} \big) 
\end{multline}
Here the sum $\Sigma$, written in explicit form, reads
\begin{equation}\label{RangeOfSummation}
\sum  = 
\sum \limits_{\substack{(s',\widehat{n'}\cdot\widehat{l'}) \in \mathscr{O}' \\ (s'',\widehat{n'}\cdot\widehat{l''})  \in \mathscr{O}'' }} 
\,\,\,
\sum \limits_{\substack{-l' \leq i',j',i'',j'' \leq l' \\ -l'' \leq i'',j'',i'''',j'''' \leq l''}} 
\,\,\,
\sum \limits_{\substack{|l'-l''| \leq l,l^{0} \leq l'+l'' \\ -l^0 \leq i^0,j^0 \leq l^0 }} 
\,\,\,
\sum \limits_{\substack{-l \leq i,j \leq l \\ a,b,c,d}},
\end{equation}
and $\overline{\underset{l' \,\, i'''j'''}{M}_{{}_{l'' \,\, i''''j''''}}^{{}^{l \,\, ij}}}, \, 
\underset{l' \,\, i'j'}{M}_{{}_{l'' \,\, i''j''}}^{{}^{l^0 \,\, i^0j^0}}$ come from the product
formula 
\begin{align*}
\widehat{l'}_{{}_{i' \, j'}}(\boldsymbol{w})\widehat{l''}_{{}_{i'' \, j''}}(\boldsymbol{w}) &
=
\sum \limits_{\substack{|l'-l''| \leq l^{0} \leq l'+l'' \\ -l^0 \leq i^0,j^0 \leq l^0 }} 
\underset{l' \,\, i'j'}{M}_{{}_{l'' \,\, i''j''}}^{{}^{l^0 \,\, i^0j^0}} \, \widehat{l^0}_{{}_{i^0,j^0}}(\boldsymbol{w})
\\
\overline{\widehat{l'}_{{}_{i''' \,\, j'''}}(\boldsymbol{v})}  \,\,\,\, \overline{\widehat{l''}_{{}_{i'''' \, j''''}}(\boldsymbol{v})} &
=
\sum \limits_{\substack{|l'-l''| \leq l \leq l'+l'' \\ -l \leq i,j \leq l }} 
\overline{\underset{l' \,\, i'''j'''}{M}_{{}_{l'' \,\, i''''j''''}}^{{}^{l \,\, ij}}}
 \, \overline{\widehat{l}_{{}_{ij}}(\boldsymbol{v})},
\end{align*}
for the matrix elements of the irreducible representations $\widehat{l'},\widehat{l''}$.

It is easily seen that for general $\chi = \phi \otimes \varphi$ the sum (\ref{<thetaykappa_2,phixvarphi>}) is 
logarithmically divergent  in $n$ because 
\[
\widehat{\theta}_{n} \coloneqq
{\textstyle\frac{1}{\sqrt{4\pi}}} \int\limits_{\widetilde{\mathbb{S}^1}} \theta(t) \, \overline{\widehat{n}(t)} \, dt
=
\left\{ \begin{array}{ll}
{\textstyle\frac{i}{\sqrt{\pi} n}}, &  n \in 2 \mathbb{Z} +1, \\
0, & n \in 2 \mathbb{Z} \,\,\, \textrm{and} \,\,\, n\neq 0, \\
\sqrt{\pi}, & n=0.
\end{array} \right.
\]
Note here that the sum (\ref{<thetaykappa_2,phixvarphi>}), or discrete integral (\ref{<thetaykappa_2,phixvarphi>}), 
is the immediate analogue of the integral (\ref{theta(x-y)masslesskappa01.masslesskappa10contractionmasslesskappa10.masslesskappa10})
of Subsection \ref{WickForChronological},
with the first integral of (\ref{theta(x-y)masslesskappa01.masslesskappa10contractionmasslesskappa10.masslesskappa10})
corresponding to that contribution in (\ref{<thetaykappa_2,phixvarphi>}) which has $n\neq 0$, and with the second integral
in (\ref{theta(x-y)masslesskappa01.masslesskappa10contractionmasslesskappa10.masslesskappa10}) corresponding to the
contribution of (\ref{<thetaykappa_2,phixvarphi>}) with $n=0$, and with
\begin{align*}
\sum\limits_{n\neq 0, n\in \mathbb{Z}} \widehat{\theta}_{n} 
 \ldots & \,\,\,\,\,\,\,\,\,\,\,\, \textrm{corresponding to} & 
\textrm{reg} \, \int {\textstyle\frac{i}{p_0}} \ud p_0 \ldots, 
\\
\sum \limits_{\substack{(s',\widehat{n'}\cdot\widehat{l'}) \in \mathscr{O}' \\ -l' \leq i',j',i''',j''' \leq l' \\ }} 
\ldots & \,\,\,\,\,\,\,\,\,\,\,\, \textrm{corresponding to} & 
 \int  \ud^3 \boldsymbol{p}' \ldots,
\\
\sum \limits_{\substack{(s'',\widehat{n''}\cdot\widehat{l''}) \in \mathscr{O}'' \\ -l'' \leq i'',j'',i'''',j'''' \leq l' \\ }} 
\ldots & \,\,\,\,\,\,\,\,\,\,\,\, \textrm{corresponding to} & 
 \int  \ud^3 \boldsymbol{p}'' \ldots,
\end{align*}
and with the summation 
\[
\sum \limits_{\substack{|l'-l''| \leq l,l^{0} \leq l'+l'' \\ -l^0 \leq i^0,j^0 \leq l^0 }} 
\,\,\,
\sum \limits_{-l \leq i,j \leq l },
\]
in (\ref{<thetaykappa_2,phixvarphi>}) coming from the formulas for the products 
\[
\widehat{l'}_{{}_{i' \, j'}}(\boldsymbol{w})\widehat{l''}_{{}_{i'' \, j''}}(\boldsymbol{w})
\,\,\,\,\,\,\,
\textrm{and} \,\,\,\,\,\,\,
\overline{\widehat{l'}_{{}_{i''' \,\, j'''}}(\boldsymbol{v})}  \, \overline{\widehat{l''}_{{}_{i'''' \, j''''}}(\boldsymbol{v})}
= \widehat{l'}_{{}_{-i''' \,\, -j'''}}(\boldsymbol{v})  \,\, \widehat{l''}_{{}_{-i'''' \, -j''''}}(\boldsymbol{v})
\]
of the the matrix elements of the irreducible representations
$\widehat{l'},\widehat{l''}$, and corresponding to the simpler product formula of plane wave exponents 
in the integrals  (\ref{theta(x-y)masslesskappa01.masslesskappa10contractionmasslesskappa10.masslesskappa10})
on the flat Minkowski space-time $\mathbb{R}^4$. The formulas for the product of the irreducible characters (plane waves) of the Abelian
additive group $\mathbb{R}^{4}$ is of course simpler in comparison to the product formula for the matrix elements
of the characters $\widehat{n}\cdot \widehat{l}$ -- irreducible represebtations of the non-Abelian group $\widetilde{\mathbb{S}^1}\times SU(2, \mathbb{C})$.

But we consider now, similarly as in Subsection \ref{WickForChronological}, a closed subspace 
$\mathscr{E}_{{}_{\omega}}$
of 
\[
\mathscr{E} = S_{{}_{\Delta +1}}(\widetilde{\mathbb{S}^1}\times SU(2, \mathbb{C}); \mathbb{C}^{d_{{}_{'}}d_{{}_{''}}}) = 
S_{{}_{A}}(\widetilde{\mathbb{S}^1}\times SU(2, \mathbb{C}); \mathbb{C}^{d_{{}_{'}}d_{{}_{''}}})
\]
consisting of such elements $\phi \in \mathscr{E}$, whose time derivatives at both Cauchy surfaces $t=0$
and $t=2\pi$, at which $\theta$ has the jump, vanish up to some finite order $\omega$. 
Each such function $\phi \in \mathscr{E}_{{}_{\omega}}$ has the property that the pointwise product 
\[
\phi_{{}_{\theta}} \overset{\textrm{df}}{=}  \theta.\phi \in \textrm{Dom} \, A^{\omega/2} = \textrm{Dom} \, [\Delta +1]^{\omega/2},
\]
and in particular 
\begin{equation}\label{|A^omegatheta.phi|L2<infty}
\big|A^{\omega/2} \phi_{{}_{\theta}} \big|_{{}_{L^2}} = \big|[\Delta +1]^{\omega/2} \phi_{{}_{\theta}}  \big|_{{}_{L^2}} < \infty.
\end{equation}
Moreover, because
\begin{multline*}
\big|A^{\omega/2} \phi_{{}_{\theta}} \big|_{{}_{L^2}}^{2} 
=
\int\limits_{\widetilde{\mathbb{S}^1}\times SU(2, \mathbb{C})}
\big|A^{\omega/2} \phi_{{}_{\theta}}(x) \big|^2
\, \ud x 
=
\int\limits_{\textrm{supp} \, \theta}
\big|A^{\omega/2} \phi(x) \big|^2
\, \ud x 
\\
\leq 
\int\limits_{\widetilde{\mathbb{S}^1}\times SU(2, \mathbb{C})}
\big|A^{\omega/2} \phi(x) \big|^2
\, \ud x 
=
\big|A^{\omega/2} \phi \big|_{{}_{L^2}}^{2}, 
\end{multline*}
then (\ref{|A^omegatheta.phi|L2<infty}) holds uniformly if 
$\phi$ ranges over any intersection of a bounded set in $\mathscr{E}$ with $\mathscr{E}_{{}_{\omega}} \subset \mathscr{E}$. Indeed on 
bounded subset of $\mathscr{E}$ each of the norms
\[
| \cdot |_m = \big|A^m \cdot \big|_{{}{L^2}},
\]
defininig the nuclear topology of $\mathscr{E} = \mathcal{S}_A(\widetilde{\mathbb{S}^1}\times SU(2, \mathbb{C}); \mathbb{C}^{d_{{}_{'}}d_{{}_{''}}})$,
taken separately, is bounded.
Because the Fourier transform is unitary, then
\begin{equation}\label{|tildeA^omegatildetheta.phi|L2<infty}
\big|\widetilde{A}^{\omega/2} \widetilde{\phi_{{}_{\theta}}} \big|_{{}_{L^2}} 
= \Big|\Big[\widetilde{\Delta +1}\Big]^{\omega/2} \widetilde{\phi_{{}_{\theta}}}  \Big|_{{}_{L^2}} < \infty,
\end{equation} 
holds uniformly whenever $\phi \in \mathscr{E}$ ranges over bounded set in $\mathscr{E}$ 
of all those elements $\phi$ whose time derivatives vanish at the Cauchy surfaces $t=0$ and 
$t= 2\pi$ up to order $\omega$, including, of course, the zero order, \emph{i.e.} the value of the function itself at 
the whole Cauchy surfaces  $t=0$ and  $t=2\pi$). 

For any $\chi \in \mathscr{E} \otimes \mathscr{E}$ of the form
\begin{equation}\label{FirstSpecialchi}
\chi(x,y) = \phi(xy^{-1}) \varphi(y), \,\,\, \phi \in \mathscr{E}_{{}_{\omega}}, \varphi \in \mathscr{E},
\end{equation}  
the integral (\ref{<thetaykappa_2,phixvarphi>'}) becomes convergent and uniformly bounded,
whenever $\phi \in \mathscr{E}_{{}_{\omega}}$, and $\varphi \in \mathscr{E}$ range over any bounded sets
in $\mathscr{E}$ (or whenever $\chi$ of the above form ranges over bounded set in $\mathscr{E}\otimes \mathscr{E}$).
Indeed, for $\chi$ of the above form (\ref{FirstSpecialchi}) and by the invariance of the measure $\ud x = \ud t \, \ud \boldsymbol{w}$, 
we have 
\[
\Big(\theta_y \big(\kappa'_{0,1} \dot{\otimes} \kappa''_{0,1} \big) 
\otimes_2 \big(\kappa'_{1,0} \dot{\otimes} \kappa''_{1,0} \big)\Big)(\chi) 
=  \Big\langle \theta_y \big(\kappa'_{0,1} \dot{\otimes} \kappa''_{0,1} \big) 
\otimes_2 \big(\kappa'_{1,0} \dot{\otimes} \kappa''_{1,0} \big),  \, \chi \Big\rangle
\]
\begin{multline*}
=\sum 
\int\limits_{\big[ \widetilde{\mathbb{S}^1}\times SU(2, \mathbb{C}) \big]^{\times \, 2}}
(2l' +1)(2l''+1) {u^{a}_{s'}}_{{}_{j'i'}}(\widehat{n'}\cdot \widehat{l'}) \, {u^{a}_{s''}}_{{}_{j''i''}}(\widehat{n''}\cdot \widehat{l''}) \,\, \times
\\
\times \,\,
\underset{l' \,\, i'j'}{M}_{{}_{l'' \,\, i''j''}}^{{}^{l^0 \,\, i^0j^0}}
\, \overline{\widehat{l^0}_{{}_{-i^0 \,\, -j^0}}(\boldsymbol{w})} \,
\overline{\widehat{-n'-n''}(t-\tau)} \, 
 \theta(t-\tau) \phi^{ab}(t-\tau, \boldsymbol{w}\boldsymbol{v}^{-1}) \,\, \times 
\\
\times \,\,
\overline{{u^{c}_{s'}}_{{}_{j'''i'''}}(\widehat{n'}\cdot \widehat{l'})} \, 
\overline{{u^{d}_{s''}}_{{}_{j''''i''''}}(\widehat{n''}\cdot \widehat{l''})} \, 
\overline{\underset{l' \,\, i'''j'''}{M}_{{}_{l'' \,\, i''''j''''}}^{{}^{l \,\, ij}}} \, 
\overline{\widehat{l}_{{}_{ij}}(\boldsymbol{v})} \, 
\varphi^{cd}(\tau, \boldsymbol{v})
\, \ud t \, \ud \boldsymbol{w} \, \ud \tau \, \ud \boldsymbol{v} 
\end{multline*}
\begin{multline*}
=\sum 
\int\limits_{\big[ \widetilde{\mathbb{S}^1}\times SU(2, \mathbb{C}) \big]^{\times \, 2}}
(2l' +1)(2l''+1) {u^{a}_{s'}}_{{}_{j'i'}}(\widehat{n'}\cdot \widehat{l'}) \, {u^{a}_{s''}}_{{}_{j''i''}}(\widehat{n''}\cdot \widehat{l''}) \,\, \times
\\
\times \,\,
\underset{l' \,\, i'j'}{M}_{{}_{l'' \,\, i''j''}}^{{}^{l^0 \,\, i^0j^0}}
\, \overline{\widehat{l^0}_{{}_{-i^0 \,\, -j^0}}(\boldsymbol{w})} \,
\overline{\widehat{-n'-n''}(t)} \, 
 \theta(t) \phi^{ab}(t, \boldsymbol{w}) \,\, \times 
\\
\times \,\,
\overline{{u^{c}_{s'}}_{{}_{j'''i'''}}(\widehat{n'}\cdot \widehat{l'})} \, 
\overline{{u^{d}_{s''}}_{{}_{j''''i''''}}(\widehat{n''}\cdot \widehat{l''})} \, 
\overline{\underset{l' \,\, i'''j'''}{M}_{{}_{l'' \,\, i''''j''''}}^{{}^{l \,\, ij}}} \, 
\overline{\widehat{l}_{{}_{ij}}(\boldsymbol{v})} \, 
\varphi^{cd}(\tau, \boldsymbol{v})
\, \ud t \, \ud \boldsymbol{w} \, \ud \tau \, \ud \boldsymbol{v} 
\end{multline*}
\begin{multline*}
=\sum 
\int\limits_{\big[ \widetilde{\mathbb{S}^1}\times SU(2, \mathbb{C}) \big]^{\times \, 2}}
(2l' +1)(2l''+1) {u^{a}_{s'}}_{{}_{j'i'}}(\widehat{n'}\cdot \widehat{l'}) \, {u^{a}_{s''}}_{{}_{j''i''}}(\widehat{n''}\cdot \widehat{l''}) \,\, \times
\\
\times \,\,
\underset{l' \,\, i'j'}{M}_{{}_{l'' \,\, i''j''}}^{{}^{l^0 \,\, i^0j^0}}
\, \overline{\widehat{l^0}_{{}_{-i^0 \,\, -j^0}}(\boldsymbol{w})} \,
\overline{\widehat{-n'-n''}(t)} \, 
\phi_{{}_{\theta}}^{ab}(t, \boldsymbol{w}) \,\, \times 
\\
\times \,\,
\overline{{u^{c}_{s'}}_{{}_{j'''i'''}}(\widehat{n'}\cdot \widehat{l'})} \, 
\overline{{u^{d}_{s''}}_{{}_{j''''i''''}}(\widehat{n''}\cdot \widehat{l''})} \, 
\overline{\underset{l' \,\, i'''j'''}{M}_{{}_{l'' \,\, i''''j''''}}^{{}^{l \,\, ij}}} \, 
\overline{\widehat{l}_{{}_{ij}}(\boldsymbol{v})} \, 
\varphi^{cd}(\tau, \boldsymbol{v})
\, \ud t \, \ud \boldsymbol{w} \, \ud \tau \, \ud \boldsymbol{v} 
\end{multline*}
\begin{multline}\label{<thetaykappa_2,phixvarphiomega>}
=
4\pi
\sum 
(2l' +1)(2l''+1) {u^{a}_{s'}}_{{}_{j'i'}}(\widehat{n'}\cdot \widehat{l'}) \, {u^{a}_{s''}}_{{}_{j''i''}}(\widehat{n''}\cdot \widehat{l''}) \,\, \times
\\
\times \,\,
\underset{l' \,\, i'j'}{M}_{{}_{l'' \,\, i''j''}}^{{}^{l^0 \,\, i^0j^0}}
\widetilde{\phi^{ab}_{{}_{\theta}}}_{{}_{-j^0 \,\, -i^0}}\big(\widehat{-n'-n''} \cdot \widehat{l^0} \big) 
 \,\, \times 
\\
\times \,\,
\overline{{u^{c}_{s'}}_{{}_{j'''i'''}}(\widehat{n'}\cdot \widehat{l'})} \,\,\, 
\overline{{u^{d}_{s''}}_{{}_{j''''i''''}}(\widehat{n''}\cdot \widehat{l''})} \,\,\, 
\overline{\underset{l' \,\, i'''j'''}{M}_{{}_{l'' \,\, i''''j''''}}^{{}^{l \,\, ij}}} \,\,\, 
\widetilde{\varphi^{cd}}_{{}_{ji}}\big(\widehat{0}\cdot \widehat{l} \big) 
\end{multline}
The range of the sum $\Sigma$ is equal (\ref{RangeOfSummation}). 

Because (\ref{|A^omegatheta.phi|L2<infty}) or, equivalently, (\ref{|tildeA^omegatildetheta.phi|L2<infty}) holds,
and $\varphi \in \mathscr{E}$,
then (\ref{<thetaykappa_2,phixvarphiomega>}) can be written in the following form
\begin{multline*}
4\pi
\sum 
{\textstyle\frac{(2l' +1)(2l''+1)}{\lambda_{{}_{-n'-n'',l^0}}^{\omega/2} }}
{u^{a}_{s'}}_{{}_{j'i'}}(\widehat{n'}\cdot \widehat{l'}) \, {u^{b}_{s''}}_{{}_{j''i''}}(\widehat{n''}\cdot \widehat{l''}) \,\, \times
\\
\times \,\,
\underset{l' \,\, i'j'}{M}_{{}_{l'' \,\, i''j''}}^{{}^{l^0 \,\, i^0j^0}}
\widetilde{A}^{\omega/2}\widetilde{\phi^{ab}_{{}_{\theta}}}_{{}_{-j^0 \,\, -i^0}}\big(\widehat{-n'-n''} \cdot \widehat{l^0} \big) 
 \,\, \times 
\\
\times \,\,
\overline{{u^{c}_{s'}}_{{}_{j'''i'''}}(\widehat{n'}\cdot \widehat{l'})} \,\,\, 
\overline{{u^{d}_{s''}}_{{}_{j''''i''''}}(\widehat{n''}\cdot \widehat{l''})} \,\,\, 
\overline{\underset{l' \,\, i'''j'''}{M}_{{}_{l'' \,\, i''''j''''}}^{{}^{l \,\, ij}}} \,\,\, 
\widetilde{\varphi^{cd}}_{{}_{ji}}\big(\widehat{0}\cdot \widehat{l} \big) 
\end{multline*}
for $\omega \in \mathbb{N}$ as above, which can be chosen to be any
natural number and with the range of the sum equal (\ref{RangeOfSummation}), because the Fourier transform
$\widetilde{A} = \widetilde{\Delta}+1$ of the operator $A=\Delta+1$ acts through the pointwise multiplication by the
function
\[
\widehat{n}\cdot \widehat{l} \mapsto \lambda_{{}_{n,l}},
\]
\emph{i.e.}
\[
\Big(\widetilde{A}\widetilde{\phi}\Big)_{{}_{j \,\, i}}\big(\widehat{n} \cdot \widehat{l} \big) 
=  \lambda_{{}_{n,l}} \widetilde{\phi}_{{}_{j \,\, i}}\big(\widehat{n} \cdot \widehat{l} \big).
\]
Therefore
\begin{multline*}
\Big|
\Big(\theta_y \big(\kappa'_{0,1} \dot{\otimes} \kappa'_{0,1} \big) 
\otimes_2 \big(\kappa'_{1,0} \dot{\otimes} \kappa'_{1,0} \big)\Big)(\chi) \Big|
=  \Big|\Big\langle \theta_y \big(\kappa'_{0,1} \dot{\otimes} \kappa'_{0,1} \big) 
\otimes_2 \big(\kappa'_{1,0} \dot{\otimes} \kappa'_{1,0} \big),  \, \chi \Big\rangle
 \Big|
\\
\leq
4\pi
\sum 
{\textstyle\frac{(2l' +1)(2l''+1)}{\lambda_{{}_{-n'-n'',l^0}}^{\omega/2} }}
\big|{u^{a}_{s'}}_{{}_{j'i'}}(\widehat{n'}\cdot \widehat{l'}) \, {u^{b}_{s''}}_{{}_{j''i''}}(\widehat{n''}\cdot \widehat{l''})\big| \,\, \times
\\
\times \,\,
\big|\underset{l' \,\, i'j'}{M}_{{}_{l'' \,\, i''j''}}^{{}^{l^0 \,\, i^0j^0}}\big|
\big|\widetilde{A}^{\omega/2}\widetilde{\phi^{ab}_{{}_{\theta}}}_{{}_{-j^0 \,\, -i^0}}\big(\widehat{-n'-n''} \cdot \widehat{l^0} \big) \big|
 \,\, \times 
\\
\times \,\,
\big|{u^{c}_{s'}}_{{}_{j'''i'''}}(\widehat{n'}\cdot \widehat{l'}) \big| \,\,\, 
\big|{u^{d}_{s''}}_{{}_{j''''i''''}}(\widehat{n''}\cdot \widehat{l''}) \big| \,\,\, 
\big|\underset{l' \,\, i'''j'''}{M}_{{}_{l'' \,\, i''''j''''}}^{{}^{l \,\, ij}} \big| \,\,\, 
\big|\widetilde{\varphi^{cd}}_{{}_{ji}}\big(\widehat{0}\cdot \widehat{l} \big)\big|. 
\end{multline*}
Now we apply the inequalities
\begin{align*}
\big| \underset{l' \,\, i'''j'''}{M}_{{}_{l'' \,\, i''''j''''}}^{{}^{l \,\, ij}} \big| \leq 1,
& \,\,\,\,\,\,\,\,\,\, \big| {u^{a}_{s'}}_{{}_{j'i'}}(\widehat{n'}\cdot \widehat{l'}) \big| \leq {\textstyle\frac{1}{\sqrt{2l'+1}}},
& \big| {u^{b}_{s''}}_{{}_{j''i''}}(\widehat{n'}\cdot \widehat{l''}) \big| \leq {\textstyle\frac{1}{\sqrt{2l''+1}}},
\\
\big|\underset{l' \,\, i'j'}{M}_{{}_{l'' \,\, i''j''}}^{{}^{l^0 \,\, i^0j^0}} \big| \leq 1, 
& \,\,\,\,\,\,\,\,\,\, \big| {u^{c}_{s'}}_{{}_{j'''i'''}}(\widehat{n'}\cdot \widehat{l'}) \big| \leq {\textstyle\frac{1}{\sqrt{2l'+1}}},
& \big| {u^{d}_{s''}}_{{}_{j''''i''''}}(\widehat{n''}\cdot \widehat{l''}) \big| \leq {\textstyle\frac{1}{\sqrt{2l''+1}}},
\end{align*}
which we have already proved, and which give here
\begin{multline*}
=  \Big|\Big\langle \theta_y \big(\kappa'_{0,1} \dot{\otimes} \kappa'_{0,1} \big) 
\otimes_2 \big(\kappa'_{1,0} \dot{\otimes} \kappa'_{1,0} \big),  \, \chi \Big\rangle
 \Big|
\\
\leq 
4\pi
\sum 
{\textstyle\frac{1}{\lambda_{{}_{-n'-n'',l^0}}^{\omega/2} }}
  \,
\big|\widetilde{A}^{\omega/2}\widetilde{\phi^{ab}_{{}_{\theta}}}_{{}_{-j^0 \,\, -i^0}}\big(\widehat{-n'-n''} \cdot \widehat{l^0} \big) \big|
 \,\,\,\,
\big|\widetilde{\varphi^{cd}}_{{}_{ji}}\big(\widehat{0}\cdot \widehat{l} \big)\big| 
\end{multline*}
with the range of the sum equal (\ref{RangeOfSummation}).
Keeping fixed $\widehat{n'}\cdot \widehat{l'} \in \mathscr{O}'_{+}$ 
and $\widehat{n''}\cdot \widehat{l''} \in \mathscr{O}''_{+}$ in $\Sigma$ and the other summation indices, we perform first the summation
with respect to $s',s''$ which, as we know, have the range lying within the following intervals
\[
1 \leq s' \leq (2l'+1)^2 \, \textrm{dim} \, V', \,\,\, 1 \leq s'' \leq (2l''+1)^2 \, \textrm{dim} \, V''.
\]
Therefore the summation with respect to $s',s''$ is equivalent to putting the additional factor 
$(2l'+1)^2 \, \textrm{dim} \, V' \, \leq (2l''+1)^2 \, \textrm{dim} \, V''$ in front of the summand.
Then, $(\widehat{n'}\cdot \widehat{l'}) \in \mathscr{O}'_{+}$
and $(\widehat{n''}\cdot \widehat{l''}) \in \mathscr{O}''_{+}$ and the remaining summation indices still keeping fixed , 
we perform the summation with respect to
$i',j'$, $i''',j'''$, $i'', j''$, $i'''', j''''$, which can be replaced by the further additional factor $(2l'+1)^4(2l''+1)^4$
in front of the summand, in view of the range of this summation. Next we pull 
\[
\underset{c,d,l,i,j}{\textrm{sup}} \big| \widetilde{\varphi^{cd}}_{{}_{ji}}\big(\widehat{0}\cdot \widehat{l} \big) \big|
\leq \big| \widetilde{\varphi} \big|_{{}_{L^2}} 
=  \big| \varphi \big|_{{}_{L^2}},
\]
out of the sum, and perform the summation with respect to $l$, which operation allows us to majorize further the sum
with respect to $l$ by the additional factor replacing summation in $l$, and equal $l'+l''$, because $l$ ranges between $|l'-l''|$ and $l'+l''$,
and cannot assume the number of values greater than $l'+l''$. Next we perform the summation with respect to $c,d$,
which can be replaced by the factor $\textrm{dim} \, V' \, \textrm{dim} \, V''$ in the summand. Thus we arrive at the estimation
\begin{multline*}
\Big|\Big\langle \theta_y \big(\kappa'_{0,1} \dot{\otimes} \kappa'_{0,1} \big) 
\otimes_2 \big(\kappa'_{1,0} \dot{\otimes} \kappa'_{1,0} \big),  \, \chi \Big\rangle
 \Big|
\leq 
(\textrm{dim} \, V')^2 \, (\textrm{dim} \, V'')^2 \, \times
\\
\times \,\,
\big| \varphi \big|_{{}_{L^2}}
\sum \limits_{\substack{\widehat{n'}\cdot\widehat{l'} \in \mathscr{O}'_{+} \\ \widehat{n'}\cdot\widehat{l''}  \in \mathscr{O}''_{+} 
\\ |l'-l''| \leq l^{0} \leq l'+l'' \\ -l^0 \leq i^0,j^0 \leq l^0 \\a,b }} 
\Big|
{\textstyle\frac{(2l'+1)^2(2l''+1)^2 (l'+l'')}{\lambda_{{}_{-n'-n'',l^0}}^{\omega/2} }} 
\widetilde{A}^{\omega/2}\widetilde{\phi^{ab}_{{}_{\theta}}}_{{}_{-j^0 \,\, -i^0}}\big(\widehat{-n'-n''} \cdot \widehat{l^0} \big)
\Big|.
\end{multline*}
Now recall that the points $\widehat{n}\cdot \widehat{l}$ of the orbit
 $\mathscr{O}_{+}$ of positive energy field lie (at least asymptotically) within
the ``positive energy cone in momentum space'' 
\[
n^2 \geq (2l+1)^2, \,\,\,\, n \geq 0
\]
and the points of the orbit $\mathscr{O}_{-}$ of negative energy field lie (at least asymptotically) 
within the ``negative energy cone in momentum space''
\[
n^2 \geq (2l+1)^2, \,\,\,\, n \leq 0.
\]
In each case $(2l+1) \leq |n|$ (at least assymptotically in more general case) for all points $\widehat{n}\cdot \widehat{l}$ of the orbit.
Therefore we have the estimation
\begin{multline*}
\Big|\Big\langle \theta_y \big(\kappa'_{0,1} \dot{\otimes} \kappa''_{0,1} \big) 
\otimes_2 \big(\kappa'_{1,0} \dot{\otimes} \kappa''_{1,0} \big),  \, \chi \Big\rangle
 \Big|
\leq 
(\textrm{dim} \, V')^2 \, (\textrm{dim} \, V'')^2 \, \times
\\
\times \,\,
\big| \varphi \big|_{{}_{L^2}}
\sum \limits_{\substack{\widehat{n'}\cdot\widehat{l'} \in \mathscr{O}'_{+} \\ \widehat{n'}\cdot\widehat{l''}  \in \mathscr{O}''_{+} 
\\ |l'-l''| \leq l^{0} \leq l'+l'' \\ -l^0 \leq i^0,j^0 \leq l^0 \\ a,b }} 
\Big|
{\textstyle\frac{(n'+n'')^2(n'+n'')^2 (n'+n'')}{\lambda_{{}_{-n'-n'',l^0}}^{\omega/2} }} 
\widetilde{A}^{\omega/2}\widetilde{\phi^{ab}_{{}_{\theta}}}_{{}_{-j^0 \,\, -i^0}}\big(\widehat{-n'-n''} \cdot \widehat{l^0} \big)
\Big|.
\end{multline*}

Suppose for a while that the orbits, $\mathscr{O}'_{+}, \mathscr{O}''_{+} $ of our fields $\mathbb{A}', \mathbb{A}''$
of the general class defined as in Definition \ref{K_0onEU}, have the covering numbers
$n_{{}_{\mathscr{O}'_+}}, n_{{}_{\mathscr{O}''_+}} \leq 1$.  Thus, the number 
of points $\widehat{n'} \cdot \widehat{l'} \in \mathscr{O}'_+$ with the same $l'$ is at most one
and on the orbit $\mathscr{O}'_+$ the energy quantum number $n'$ is a well defined function of the total
momentum quantum number $l'$. The same we have, by our temporary assumption, for the orbit 
$\mathscr{O}''_+$. In this situation $\widehat{-n'-n''} \cdot \widehat{l^0}$ in the last summation 
ranges (asymptotically at least) within the``negative energy cone in the momentum space'' 
and each point of this cone is reached at most $n'+n''$ times
whenever $n',l',n'', l'', l^0,i^0,j^0$ range over the whole summation range in this expression. 
Therefore
\begin{multline*}
\Big|\Big\langle \theta_y \big(\kappa'_{0,1} \dot{\otimes} \kappa''_{0,1} \big) 
\otimes_2 \big(\kappa'_{1,0} \dot{\otimes} \kappa''_{1,0} \big),  \, \chi \Big\rangle
 \Big|
\leq 
(\textrm{dim} \, V')^2 \, (\textrm{dim} \, V'')^2 \, \times
\\
\times \,\,
\big| \varphi \big|_{{}_{L^2}}
\sum \limits_{\substack{\widehat{n}\cdot\widehat{l} \in \widehat{\widetilde{\mathbb{S}^1}\times G}  
 \\ -l \leq i,j \leq l \\ a,b }} 
\Big|
{\textstyle\frac{n^2n^2n^2}{\lambda_{{}_{-n,l}}^{\omega/2} }} 
\widetilde{A}^{\omega/2}\widetilde{\phi^{ab}_{{}_{\theta}}}_{{}_{-j \,\, -i}}\big(\widehat{-n} \cdot \widehat{l} \big)
\Big|,
\end{multline*}
with $G = SL(2, \mathbb{C})$. Thus changing the sign of the summation indices $n,i,j$
\begin{multline*}
\Big|\Big\langle \theta_y \big(\kappa'_{0,1} \dot{\otimes} \kappa'_{0,1} \big) 
\otimes_2 \big(\kappa'_{1,0} \dot{\otimes} \kappa'_{1,0} \big),  \, \chi \Big\rangle
 \Big|
\leq 
(\textrm{dim} \, V')^2 \, (\textrm{dim} \, V'')^2 \, \times
\\
\times \,\,
\big| \varphi \big|_{{}_{L^2}}
\sum \limits_{\substack{\widehat{n}\cdot\widehat{l} \in \widehat{\widetilde{\mathbb{S}^1}\times G}  
 \\ -l \leq i,j \leq l \\ a,b }} 
\Big|
{\textstyle\frac{n^6}{\lambda_{{}_{n,l}}^{\omega/2} }} 
\widetilde{A}^{\omega/2}\widetilde{\phi^{ab}_{{}_{\theta}}}_{{}_{j \,\, i}}\big(\widehat{n} \cdot \widehat{l} \big)
\Big|.
\end{multline*}
It is easily seen that $\omega \in \mathbb{N}$ can be so chosen that the matrix-valued function 
$\widetilde{\eta}$ on $\widehat{\widetilde{\mathbb{S}^1}\times G} $, equal by definition 
\[
\widetilde{\eta^{ab}}_{{}_{j \,\, i}}\big(\widehat{n} \cdot \widehat{l} \big)
=
{\textstyle\frac{n^6}{\lambda_{{}_{n,l}}^{\omega/2} \, (2l+1)}}, \,\,\,\, - l \leq  i,j \leq l,
\,\,\, 1 \leq a \leq \textrm{dim} \, V', \,\,\, 1 \leq b \leq \textrm{dim} \, V''
\]
is square summable, \emph{i.e.} has the Plancherel inner product norm finite
\[
\big| \widetilde{\eta} \big|_{{}_{L^2}} < \infty.
\]
Let now $\zeta \in L^2\big(\widehat{\widetilde{\mathbb{S}^1}\times G}, \mathbb{C}^{d'd''} \big)$, $G = SL(2, \mathbb{C})$, 
be the function whose Fourier transform is by definition equal
\[
\widetilde{\zeta^{ab}}_{{}_{j \,\, i}}\big(\widehat{n} \cdot \widehat{l} \big)
=
\big| \widetilde{A}^{\omega/2}\widetilde{\phi^{ab}_{{}_{\theta}}}_{{}_{j \,\, i}}\big(\widehat{n} \cdot \widehat{l} \big) \big|
\]
Of course the Plancherel inner product norms of $\widetilde{\zeta}$ and $\widetilde{A}^{\omega/2}\widetilde{\phi_{{}_{\theta}}}$
coincide
\[
\big| \widetilde{\zeta} \big|_{{}_{L^2}} = \big| \widetilde{A}^{\omega/2}\widetilde{\phi_{{}_{\theta}}} \big|_{{}_{L^2}}
= \big| A^{\omega/2}\phi_{{}_{\theta}} \big|_{{}_{L^2}}.
\]
By the Schwartz inequality
\begin{multline*}
\Big|\Big\langle \theta_y \big(\kappa'_{0,1} \dot{\otimes} \kappa''_{0,1} \big) 
\otimes_2 \big(\kappa'_{1,0} \dot{\otimes} \kappa''_{1,0} \big),  \, \chi \Big\rangle
 \Big|
\\
\leq 
(\textrm{dim} \, V')^2 \, (\textrm{dim} \, V'')^2 \, \big| \varphi \big|_{{}_{L^2}}
\langle \widetilde{\eta}, \widetilde{\zeta} \rangle
\leq 
(\textrm{dim} \, V')^2 \, (\textrm{dim} \, V'')^2 \, \big| \varphi \big|_{{}_{L^2}}
\big| \widetilde{\eta} \big|_{{}_{L^2}}  \big| \widetilde{\zeta} \big|_{{}_{L^2}} 
\\
=
(\textrm{dim} \, V')^2 \, (\textrm{dim} \, V'')^2 \, \big| \widetilde{\eta}\big|_{{}_{L^2}} \, 
\big| \varphi \big|_{{}_{L^2}} \, \big| A^{\omega/2}\phi_{{}_{\theta}} \big|_{{}_{L^2}}.
\end{multline*}
Of course in case of more general fields $\mathbb{A}', \mathbb{A}''$ of Definition 
\ref{K_0onEU} for which the covering numbers
$n_{{}_{\mathscr{O}'_+}}, n_{{}_{\mathscr{O}''_+}} > 1$, we will have the inequality
\begin{multline}\label{<thetaykappa(2-contraction)kappa,chi><infty}
\Big|\Big\langle \theta_y \big(\kappa'_{0,1} \dot{\otimes} \kappa''_{0,1} \big) 
\otimes_2 \big(\kappa'_{1,0} \dot{\otimes} \kappa''_{1,0} \big),  \, \chi \Big\rangle
 \Big|
\\
\leq 
n_{{}_{\mathscr{O}'_+}} n_{{}_{\mathscr{O}''_+}}
(\textrm{dim} \, V')^2 \, (\textrm{dim} \, V'')^2 \, \big| \widetilde{\eta}\big|_{{}_{L^2}} \, 
\big| \varphi \big|_{{}_{L^2}} \, \big| A^{\omega/2}\phi_{{}_{\theta}} \big|_{{}_{L^2}}
\\
\leq 
n_{{}_{\mathscr{O}'_+}} n_{{}_{\mathscr{O}''_+}}
(\textrm{dim} \, V')^2 \, (\textrm{dim} \, V'')^2 \, \big| \widetilde{\eta}\big|_{{}_{L^2}} \, 
\big| \varphi \big|_{{}_{L^2}} \, \big| A^{\omega/2}\phi \big|_{{}_{L^2}}.
\end{multline}

We have thus proved that the contraction discrete integral (\ref{<thetaykappa_2,phixvarphiomega>}), or sum
(\ref{<thetaykappa_2,phixvarphiomega>}), is absolutely convergent
on the closed subspace of functions $\chi$ of the form
(\ref{FirstSpecialchi}). Moreover, because the inequality (\ref{|A^omegatheta.phi|L2<infty}) holds uniformly
for $\phi \in \mathscr{E}_{{}_{\omega}}$ ranging over any bounded set in $\mathscr{E}$, and
\[
\big| \varphi \big|_{{}_{L^2}} < \infty
\]
holds uniformly for $\varphi$ ranging over bounded set in $\mathscr{E}$,
then it follows from (\ref{<thetaykappa(2-contraction)kappa,chi><infty}) that the discrete integral/sum (\ref{<thetaykappa_2,phixvarphiomega>})
is uniformly bounded whenever $\chi$ is of the form (\ref{FirstSpecialchi}) and ranges over any bounded set 
in $\mathscr{E} \otimes \mathscr{E}$ or whenever $\phi \in \mathscr{E}_{{}_{\omega}}$, $\varphi \in \mathscr{E}$
range over bounded sets in $\mathscr{E}$.

We can repeat exactly the same method of estimation of the contraction integral/sum (\ref{thetay(kk'...k''')q-contraction(k,k',...k''')}) with $q'=q$
and $q>2$. Of course in this case we will have $q$ factors in each of the two products 
\[
\widehat{l^{(1)}}_{{}_{i' \, j'}}(\boldsymbol{w}) \,\,\, \ldots \,\,\, \widehat{l^{(q)}}_{{}_{i'' \, j''}}(\boldsymbol{w})
\,\,\,\,\,\,\,
\textrm{and} \,\,\,\,\,\,\,
\overline{\widehat{l^{(1)}}_{{}_{i''' \,\, j'''}}(\boldsymbol{v})}  \,\,\, \ldots \,\,\, \overline{\widehat{l^{(q)}}_{{}_{i'''' \, j''''}}(\boldsymbol{v})}
\]
of the matrix components of the irreducible representations $\widehat{l^{(1)}}, \ldots, \widehat{l^{(q)}}$. Correspondingly
we will have more (at most $q/2$) factors in each of the two products of the ``Clebsh coefficients''
\[
\underset{l^{(1)} \,\, i_{11}j_{11}}{M}_{{}_{l^{(2)} \,\, i_{21}j_{21}}}^{{}^{l_1 \,\, i_1 j_1}} \, 
\underset{l^{(3)} \,\, i_{31}j_{31}}{M}_{{}_{l^{(4)} \,\, i_{41}j_{41}}}^{{}^{l_2 \,\, i_2 j_2}}
\ldots
\,\,\,\,\,\, \textrm{and} \,\,\,\,\,\, 
\overline{\underset{l^{(1)} \,\, i'_{11}j'_{11}}{M}_{{}_{l^{(2)} \,\, i'_{21}j_{21}}}^{{}^{l'_{1} \,\, i'_{1} j'_{1}}}} \, 
\overline{\underset{l^{(3)} \,\, i'_{31}j'_{31}}{M}_{{}_{l^{(4)} \,\, i'_{41}j_{41}}}^{{}^{l'_{2} \,\, i'_{2} j'_{2}}}} 
\ldots
\]
corresponding to each of the above two products of matrix elements of irreducible representations
$\widehat{l^{(1)}}, \ldots, \widehat{l^{(q)}}$.  We will also have $q$ factors 
\[
(2l^{(1)}+1) \ldots (2l^{(q)}+1) \,\,\, \textrm{and} \,\,\,\,  u \ldots u
\]
in the summand with the summation over all $q$ orbits $\mathscr{O}^{(1)}, \ldots, \mathscr{O}^{(q)}$. 
It is easily seen that enlarging respectively the number $\omega$ we can likewise achieve the absolute convergence 
of the discrete contraction integral (\ref{thetay(kk'...k''')q-contraction(k,k',...k''')}) with $q'=q$, and the analogue
 inequality
\begin{multline}\label{<thetaykappa(q-contraction)kappa,chi><infty}
\Big|\Big(\theta_y \big(\kappa^{(1)}_{0,1} \dot{\otimes} \ldots \dot{\otimes} \kappa^{(q)}_{0,1} \big) 
\otimes_q \big(\kappa^{(1)}_{1,0} \dot{\otimes} \ldots \dot{\otimes} \kappa^{(q)}_{1,0} \big) \Big)(\chi)
 \Big|
\\
\leq 
n_{{}_{\mathscr{O}^{(1)}_+}} \ldots n_{{}_{\mathscr{O}^{(q)}_+}}
(\textrm{dim} \, V^{(1)})^2 \, \ldots \, (\textrm{dim} \, V^{(q)})^2 \, \big| \widetilde{\eta}\big|_{{}_{L^2}} \, 
\big| \varphi \big|_{{}_{L^2}} \, \big| A^{\omega/2}\phi_{{}_{\theta}} \big|_{{}_{L^2}}
\\
\leq 
n_{{}_{\mathscr{O}^{(1)}_+}} \ldots n_{{}_{\mathscr{O}^{(q)}_+}}
(\textrm{dim} \, V^{(1)})^2 \, \ldots \, (\textrm{dim} \, V^{(q)})^2 \, \big| \widetilde{\eta}\big|_{{}_{L^2}} \, 
\big| \varphi \big|_{{}_{L^2}} \, \big| A^{\omega/2}\phi \big|_{{}_{L^2}},
\end{multline}
with the analogue function 
\[
\eta \in L^2\big( \widetilde{\mathbb{S}^1}\times SU(2, \mathbb{C})\big),
\]
and for $\chi(x,y) = \phi\big(xy^{-1}\big)\varphi(y)$ of the form (\ref{FirstSpecialchi}), \emph{i.e.}
with $\phi \in \mathscr{E}_{{}_{\omega}} \subset \mathscr{E}$ and $\varphi \in \mathscr{E}$. 
By the invarince of the space-time measure $\ud x$ and the operator $A= \Delta +1$ under left (and right) actions of the group
$\widetilde{\mathbb{S}^1}\times SU(2, \mathbb{C})$, we have 
\[
|\chi|_{{}_{m}} = |A^m \otimes A^m \chi|_{{}_{L^2}} = |A^m \phi|_{{}_{L^2}}|A^m \varphi|_{{}_{L^2}}
= |\phi|_{{}_{m}} |\varphi|_{{}_{m}}, 
\]
for $\chi \in \mathscr{E^{\otimes \, 2}}$
of the form $\chi(x,y) = \phi\big(xy^{-1}\big)\varphi(y)$.
Because $|\cdot|_n  < |\cdot|_m$, $n<m$, then the inequality (\ref{<thetaykappa(q-contraction)kappa,chi><infty})
can be rewritten in the following form
\begin{multline}\label{|thetaykappa(q-contraction)kappa(chi)|<C|chi|omega}
\Big|\Big(\theta_y \big(\kappa^{(1)}_{0,1} \dot{\otimes} \ldots \dot{\otimes} \kappa^{(q)}_{0,1} \big) 
\otimes_q \big(\kappa^{(1)}_{1,0} \dot{\otimes} \ldots \dot{\otimes} \kappa^{(q)}_{1,0} \big) \Big)(\chi)
 \Big|
\\
\leq 
n_{{}_{\mathscr{O}^{(1)}_+}} \ldots n_{{}_{\mathscr{O}^{(q)}_+}}
(\textrm{dim} \, V^{(1)})^2 \, \ldots \, (\textrm{dim} \, V^{(q)})^2 \, \big| \widetilde{\eta}\big|_{{}_{L^2}} \, 
\big| \theta.\chi \big|_{{}_{\omega/2}}
\\
\leq 
n_{{}_{\mathscr{O}^{(1)}_+}} \ldots n_{{}_{\mathscr{O}^{(q)}_+}}
(\textrm{dim} \, V^{(1)})^2 \, \ldots \, (\textrm{dim} \, V^{(q)})^2 \, \big| \widetilde{\eta}\big|_{{}_{L^2}} \, 
\big| \chi \big|_{{}_{\omega/2}},
\end{multline}
for
\[
\chi(x,y) = \phi\big(xy^{-1}\big)\varphi(y), \,\,\, \phi \in \mathscr{E}_{{}_{\omega}}, \,\,\,\,  \varphi \in \mathscr{E},
\]
\[
\theta.\chi(x,y) = \theta\big(xy^{-1}\big) \phi\big(xy^{-1}\big)\varphi(y),
\]
with a finite $\omega \in \mathbb{N}$ depending on $q$. 

Similar estimation and absolute convergence for (\ref{thetay(kk'...k''')q-contraction(k,k',...k''')}) we obtain, 
repeating the same method, for the general case $q'<q$:
\begin{multline}\label{|<thetaykappa(q'-contraction)kappa(chi),xi>|<C|chi|omega|xi|}
\Big| \Big\langle \Big(\theta_y \big(\kappa^{(1)}_{0,1} \dot{\otimes} \ldots \dot{\otimes} \kappa^{(q)}_{0,1} \big) 
\otimes_{q'} \big(\kappa^{(1)}_{1,0} \dot{\otimes} \ldots \dot{\otimes} \kappa^{(q)}_{1,0} \big) \Big)(\chi), \,\,
\xi \Big\rangle
 \Big|
\\
\leq 
n_{{}_{\mathscr{O}^{(1)}_+}} \ldots n_{{}_{\mathscr{O}^{(q)}_+}}
(\textrm{dim} \, V^{(1)})^2 \, \ldots \, (\textrm{dim} \, V^{(q)})^2 \, \big| \widetilde{\eta}\big|_{{}_{L^2}} \, 
\big| \theta.\chi \big|_{{}_{\omega/2}} \big|\xi\big|_{{}_{m}}
\\
\leq 
n_{{}_{\mathscr{O}^{(1)}_+}} \ldots n_{{}_{\mathscr{O}^{(q)}_+}}
(\textrm{dim} \, V^{(1)})^2 \, \ldots \, (\textrm{dim} \, V^{(q)})^2 \, \big| \widetilde{\eta}\big|_{{}_{L^2}} \, 
\big| \chi \big|_{{}_{\omega/2}} \big|\xi\big|_{{}_{m}}, 
\end{multline}
for
\[
\chi(x,y) = \phi\big(xy^{-1}\big)\varphi(y), \,\,\, \phi \in \mathscr{E}_{{}_{\omega}}, \,\,\,\,  \varphi \in \mathscr{E},
\,\,\,\, m > m_0 > 0
\]
with a finite $\omega \in \mathbb{N}$ depending on $q',q$, and for
\[
\xi \in E_{{}_{(q'+1)}} \otimes \cdots \otimes E_{{}_{(q)}}.
\]
In this contraction $\otimes_{q'}$ we have contracted the first $q'$ pairs of momentum variables, so the the momentum variables 
of the first $q'$ plane wave ``annihilation'' kernels $\kappa^{(1)}_{0,1}, \ldots, \kappa^{(q')}_{0,1}$ coincide respectively
with the momentum variables of the first $q'$ plane wave ``creation'' kernels $\kappa^{(1)}_{1,0}, \ldots, \kappa^{(q')}_{1,0}$.
Of course we can contract any other $q'$ pairs of the corresponding plane wave kernels, remembering only that 
a plane wave ``annihilation'' kernel can be contracted only with ``creation'' plane wave kernel which is associated 
to the same free field and thus to the same orbit. 
Recall that $|\xi|_0 = |\xi|_{{}_{L^2}}$ is the  Hilbert space norm of $\xi$ in the Hilbert space tensor product 
\[
L^2\big(\mathscr{O}^{(q'+1)}; \mathbb{C}\big) \otimes \ldots \otimes L^2\big(\mathscr{O}^{(q)}; \mathbb{C}\big) 
= L^2\big(\mathscr{O}^{(q'+1)} \times \mathscr{O}^{(q)} ; \mathbb{C}\big), 
\]
which we sometimes denote simply by $|\xi| = |\xi|_0$ as the $0$-norm of the countable system of norms
\[
|\cdot|_{m} = \big|\big[\widetilde{A}^{\otimes \, (q-q')}\big]^m \cdot \big|_{{}_{L^2}}, \,\,\,\, A = \Delta +1,
\,\,\, m = 0, 1, 2, \ldots 
\]
defining the standard nuclear countably Hilbert space 
\[
E_{{}_{(q'+1)}} \otimes \cdots \otimes E_{{}_{(q)}} = \mathcal{S}_{\widetilde{A}}(\mathscr{O}^{(q'+1)}) \otimes 
\ldots \otimes \mathcal{S}_{\widetilde{A}}(\mathscr{O}^{(q)}) 
= \mathcal{S}_{\otimes_{q'+1}^{q} \widetilde{A}}(\mathscr{O}^{(q'+1)}\times \ldots \times \mathscr{O}^{(q)}). 
\]
The function $\widetilde{\eta}$ in (\ref{|<thetaykappa(q'-contraction)kappa(chi),xi>|<C|chi|omega|xi|}), 
defined on the dual of $\widetilde{\mathbb{S}^1}\times SL(2, \mathbb{C})$,
with finite Plancherel Hilbert space norm, depends only on the contracted plane wave kernels in
(\ref{|<thetaykappa(q'-contraction)kappa(chi),xi>|<C|chi|omega|xi|})
and on the numbers $q',q$ in (\ref{|<thetaykappa(q'-contraction)kappa(chi),xi>|<C|chi|omega|xi|}), 
but does not dependent on $\chi, \xi$. The value of the number $\omega$ is chosen to be large enough
in order to ensure $\big|\widetilde{\eta}\big|_{{}_{L^2}} < \infty$.  

Here we should also emphasize that (\ref{|<thetaykappa(q'-contraction)kappa(chi),xi>|<C|chi|omega|xi|}) is the estimation
not merely of the absolute value of the contraction sum/integral 
(\ref{thetay(kk'...k''')q-contraction(k,k',...k''')}) but of this sum/integral in which the summand/integrand
is replaced with its absolute value. Whence the absolute and uniform convergence of the contraction kernel
(\ref{thetay(kk'...k''')q-contraction(k,k',...k''')}) follows.

Recall that for $0 \leq q' \leq 1$ 
the inequality (\ref{|<thetaykappa(q'-contraction)kappa(chi),xi>|<C|chi|omega|xi|}) 
holds for all $\phi, \varphi \in \mathscr{E}$, which we formally indicate by writing
$\omega<0$ for this case, \emph{i.e.} we have no restrictions on the time derivatives of $\phi$
at the Cauchy surfaces $t=0$ and $t=2\pi$, including the zero order time derivative.
Indeed, in case we have no contractions at all, when $q'=0$, or just only one contraction, when $q=1$,
 it is easily seen that instead 
of the inequality (\ref{|<thetaykappa(q'-contraction)kappa(chi),xi>|<C|chi|omega|xi|})
we have 
\begin{multline}\label{|<thetaykappa(1-contraction)kappa(chi),xi>|<C|chi|omega|xi|}
\Big| \Big\langle \Big(\theta_y \big(\kappa^{(1)}_{0,1} \dot{\otimes} \ldots \dot{\otimes} \kappa^{(q)}_{0,1} \big) 
\otimes_{q'} \big(\kappa^{(1)}_{1,0} \dot{\otimes} \ldots \dot{\otimes} \kappa^{(q)}_{1,0} \big) \Big)(\chi), \,\,
\xi \Big\rangle
 \Big|
\leq 
c \, 
\big| \theta.\chi \big|_{{}_{L^2}} \big|\xi\big|_{{}_{m}}, 
\\
 \textrm{for} \,\, q' = 0, 1,
\end{multline}
for
\[
\chi(x,y) = \phi\big(xy^{-1}\big)\varphi(y), \,\,\, \phi \in \mathscr{E}, \,\,\,\,  \varphi \in \mathscr{E},
\,\,\,\, m > m_0 > 0,
\]
and where we have put
\[
\theta. \chi(x,y) = \theta\big(xy^{-1}\big)\phi\big(xy^{-1}\big)\varphi(y).
\]
Recall, please, that
\[
\big|\xi\big|_{{}_{m}} = \big| \big[\otimes_{q'+1}^{q} \widetilde{A}\big]^{m} \xi \big|_{L^2}, \,\,\, m= 0, 1, 2 \ldots
\]
are the Hilbert norms defining the standard nuclear topology on the standard nuclear space
\[
E_{{}_{(q'+1)}} \otimes \cdots \otimes E_{{}_{(q)}}. 
\]
The constant $c$ is independent of $\chi, \xi$ and depends only on the kernels
\[
\kappa^{(1)}_{0,1}, \kappa^{(1)}_{1,0}, \ldots \kappa^{(q)}_{0,1}, \kappa^{(q)}_{1,0}.
\]

To the  $q'$-contractions $\otimes_{q'}$ with $q'<2$ we ascribe $\omega<0$,
in analogy to the contractions of negative singularity degree on the Minkowski space-time,
whose retarded and advanced parts can be computed through ordinary pointwise multiplication by the step
theta function. In particular, and similarly as on the Minkowski space-time, the pairing and commutation functions
on the Einstein Universe have negative $\omega$ and can be computed through pointwise multiplication
by theta function, in accordance to what we have already proved. But on the Einstein Universe we have 
an extraordinarily regular behavior in case each of the two contracted dot product factors includes at most
one massless (or infinite orbit) plane wave kernel. In this case 
(\ref{thetay(kk'...k''')q-contraction(k,k',...k''')}) with $q'>2$ converges absolutely for any $q'\leq q$
and for all $\chi \in \mathscr{E}^{\otimes \, 2}$
and the retarded and advanced part of the contraction $\otimes_q$ can be computed in this case 
by the ordinary pointwise multiplication by the step theta function. 
This would be impossible on the Minkowski space-time even if we had at most one massless plane-wave
kernel $\kappa_{0,1}$ and $\kappa_{1,0}$ in each 
dot product factor.

Now, similarly as in Subsection \ref{WickForChronological} we construct the continuous operator on $\mathscr{E}$
which projects on the subspace $\mathscr{E}_{{}_{\omega}} = \Omega'\mathscr{E}$.

Let $\omega_{{}_{o \,\, I \,\, \alpha}}, \omega_{{}_{o \,\, II \,\, \alpha}} \in \mathscr{E}$ be smooth functions on 
$\widetilde{\mathbb{S}^1}\times SU(2, \mathbb{C})$, defined for each index $\alpha$ such that $0 \leq \alpha \leq \omega$, 
which fulfill
\begin{align*}
& \partial_{t}^{\beta} \omega_{{}_{o \,\,I \,\, \alpha}} (t=0) = \delta^{\beta}_{\alpha},&  & \,\,\,
\partial_{t}^{\beta} \omega_{{}_{o \,\, II \,\,\alpha}} (t=0) = 0,
\\
&\partial_{t}^{\beta} \omega_{{}_{o \,\, II \,\,\alpha}} (t=2\pi) = \delta^{\beta}_{\alpha}, & & \,\,\,
\partial_{t}^{\beta} \omega_{{}_{o \,\,I \,\, \alpha}}(t=2\pi) = 0, 
\\
&\,\,\,\, & 0 \leq \alpha, \beta \leq \omega. &
\end{align*}

Such functions $\omega_{{}_{o \,\, I \,\, \alpha}}, \omega_{{}_{o \,\, II \,\, \alpha}}  \in \mathscr{E}$, $0 \leq \alpha, \beta \leq \omega$, do exist
for each fixed $\omega \in \mathbb{N}$.
Indeed, let
\[
f_{{}_{I \,\, \alpha}}(x) = 
{\textstyle\frac{t^\alpha}{\alpha!}}, 
\,\,\,
f_{{}_{II \,\, \alpha}}(x) = 
{\textstyle\frac{(t-2\pi)^\alpha}{\alpha!}},
\,\,\,\,\, x = (t, \boldsymbol{w}) \in \widetilde{\mathbb{S}^1}\times SU(2, \mathbb{C}), 
\,\,\,\, 0 \leq \alpha \leq \omega,
\]
Let $w_{I},w_{II}$ be smooth, \emph{i.e.} 
\[
w_{I},w_{II} \in \mathscr{E} = S_{\Delta+1}\big(\widetilde{\mathbb{S}^1}\times SU(2, \mathbb{C}); \mathbb{C}^{d_{{}_{'}}d_{{}_{''}}}\big)
= \mathscr{C}^\infty\big( \widetilde{\mathbb{S}^1}\times SU(2, \mathbb{C}); \mathbb{C}^{d_{{}_{'}}d_{{}_{''}}}\big),
\]
which depend only on time. Let further $w_{I}$ be equal $1$ in a neighborhood of $t=0$ and equal zero in a neighborhood
of $t= 2\pi$ and \emph{vice versa} for $w_{II}$, being equal zero in a neighborhood of $t=0$ and $1$ in a neighborhood
of $t= 2\pi$. Then we can put
\[
\omega_{{}_{o \,\, I \,\, \alpha}} = f_{{}_{I \,\, \alpha}}.w_I, \,\,\, \omega_{{}_{o \,\,II \,\, \alpha}} = f_{{}_{II \,\, \alpha}}.w_{II}.
\]

Let for any $\phi \in \mathscr{E}$  the projection operator $\Omega'$ be defined, analogously as in Subsection \ref{WickForChronological}
\begin{equation}\label{OmegaphiEU}
\Omega' \phi = \phi - \sum \limits_{0 \leq \alpha \leq \omega} \partial_{t}^{\alpha} \phi(t=0) \,\,  \omega_{{}_{o \,\, I \,\, \alpha}}
- \sum \limits_{0 \leq \alpha \leq \omega} \partial_{t}^{\alpha} \phi(t=2\pi) \,\,  \omega_{{}_{o \,\, II \,\, \alpha}}.
\end{equation}
Writing the evalutaion at  $x= (t, \boldsymbol{w}) \in \widetilde{\mathbb{S}^1}\times SU(2, \mathbb{C})$ explicitly
we have
\begin{multline*}
\Omega' \phi(t, \boldsymbol{w}) = \phi(t, \boldsymbol{w}) 
- \sum \limits_{0 \leq \alpha \leq \omega} \partial_{t}^{\alpha} \phi(t=0, \boldsymbol{w}) \,\,  
\omega_{{}_{o \,\, I \,\, \alpha}}(t)
\\
- \sum \limits_{0 \leq \alpha \leq \omega} \partial_{t}^{\alpha} \phi(t=2\pi, \boldsymbol{w}) \,\,  
\omega_{{}_{o \,\, II \,\, \alpha}}(t).
\end{multline*}

It is easily seen that $\Omega'$ is a continuous idempotent which projects on the subspace $\mathscr{E}_{{}_{\omega}} \subset \mathscr{E}$ of all smooth
functions whose time derivatives vanish on the two Cauchy surfaces, $t=0$ and $t=2\pi$, up to order $\omega$. 

We extend this operator on functions in $\mathscr{E}^{\otimes \, 2}$ according to the general formula sated in Subsection \ref{WickForChronological}.
Namely, let for any $\chi \in \mathscr{E}^{\otimes \, 2}$ the function $\chi^\natural \in \mathscr{E}^{\otimes \, 2}$ be defined by the rule
\[
\chi^\natural(x,y) = \chi(xy,y) = \chi \circ L (x,y),
\]
with the group map
\[
L: \,\, \big[ \widetilde{\mathbb{S}^1}\times SU(2, \mathbb{C}) \big]^{\times \, 2} \ni (x,y)
\longmapsto (xy,y)
\in
\big[ \widetilde{\mathbb{S}^1}\times SU(2, \mathbb{C}) \big]^{\times \, 2}
\]
which induces linear operator on functions on
\[
\big[ \widetilde{\mathbb{S}^1}\times SU(2, \mathbb{C}) \big]^{\times \, 2},
\]
which commutes with the time differentiation $\partial_t$ with respect to the time coordinate $t$ of the first space-time variable $x= (t, \boldsymbol{w})$ in 
\[
(x,y) \in \big[ \widetilde{\mathbb{S}^1}\times SU(2, \mathbb{C}) \big]^{\times \, 2}.
\]
It is obvious that $\chi^\natural = \chi \circ L \in \mathscr{E}^{\otimes \, 2}$.
Let 
\[
\chi^\natural = \chi \circ L = \sum \limits_{j} \phi_j \otimes \varphi_j, \,\,\,\,\,\,\, \chi = \sum \limits_{j} (\phi_j \otimes \varphi_j) \circ L^{-1},
\]
be its expansion into simple tensors $\phi_j \otimes \varphi_j$ in $\mathscr{E}^{\otimes \, 2}$, so that
\[
\chi^\natural(x,y) = \sum \limits_{j} \phi_j(x)\varphi_j(y), \,\,\,\, \chi(x,y) = \sum \limits_{j} \phi_j\big(xy^{-1}\big) \varphi_j(y),
\]
where
\[
L^{-1}(x,y) = \big(xy^{-1},y\big)
\]
is the inverse of $L$.
Then for $\chi \in \mathscr{E}\otimes \mathscr{E}$, $\Omega \chi$ be defined by the formula
\begin{equation}\label{SumOfFirstSpecialchi-s}
\Omega \chi(x,y) = \sum \limits_{j} \big(\Omega' \phi_j\big)\big(xy^{-1}\big)\varphi_j(y)
=
\sum \limits_{j} \big[\big(\Omega' \phi_j\big) \otimes \varphi_j\big] \circ L^{-1}(x,y),
\end{equation}
\[
\Omega (\chi) \circ L(x,y) = \sum \limits_{j} \big(\Omega' \phi_j\big)(x)\varphi_j(y).
\]
with $\Omega' \phi_j$ given by (\ref{OmegaphiEU}).
Equivalently
\begin{multline*}
\Omega (\chi) \circ L (x,y) = \chi \circ L(x,y) -  \, \sum \limits_{\beta=0}^{\omega}  \omega_{{}_{o \,\, I \,\, \beta}} (x) 
\,\, \partial_{{}_{t}}^{\beta} \big(\chi \circ L\big) (t=0, \boldsymbol{w},y) 
\\
-  \, \sum \limits_{\beta=0}^{\omega}  \omega_{{}_{o \,\, II \,\, \beta}} (x) 
\,\, \partial_{{}_{t}}^{\beta} \big(\chi \circ L\big) (t=2\pi, \boldsymbol{w},y) 
\end{multline*}
\[
 \big(x= (t, \boldsymbol{w}), \, y \big) \in \big[ \widetilde{\mathbb{S}^1}\times SU(2, \mathbb{C}) \big]^{\times \, 2}.
\]
From this the explicit formula for $\Omega \chi (x,y)$ immediately  follows 
for any $\chi\in \mathscr{E}$:
\begin{multline}\label{EpsteinGlaserOmegaEU}
\Omega \chi (x,y) = \chi(x,y) 
- \, \sum \limits_{\beta=0}^{\omega} \omega_{{}_{o \,\,I \,\, \beta}} \big(xy^{-1}\big)
 \,\,\, \partial_{{}_{t}}^{\beta}\chi (t=\tau , \boldsymbol{w},y)
\\
- \, \sum \limits_{\beta=0}^{\omega} \omega_{{}_{o \,\, II \,\, \beta}} \big(xy^{-1}\big)
 \,\,\, \partial_{{}_{t}}^{\beta}\chi (t=\tau + 2\pi, \boldsymbol{w},y). 
\end{multline}

Because the contraction discrete integral (\ref{<thetaykappa_2,phixvarphiomega>}), or sum
(\ref{<thetaykappa_2,phixvarphiomega>}), is absolutely convergent
on the closed subspace of functions $\chi$ of the form
(\ref{FirstSpecialchi}) and uniformly absolutely bounded for $\chi$ of the form
(\ref{FirstSpecialchi}) and ranging over bounded set, then (compare (\ref{FirstSpecialchi}) and (\ref{SumOfFirstSpecialchi-s}))
the contraction sum/integral 
\[
\Big(\theta_y \big(\kappa'_{0,1} \dot{\otimes} \kappa''_{0,1} \big) 
\otimes_2 \big(\kappa'_{1,0} \dot{\otimes} \kappa''_{1,0} \big)\Big)(\Omega\chi) 
\]
\begin{multline*}
=
\sum \limits_{\substack{(s',\widehat{n'}\cdot\widehat{l'}) \in \mathscr{O}' \\ (s'',\widehat{n'}\cdot\widehat{l''})  \in \mathscr{O}'' 
\\ a,b,c,d }} 
\int\limits_{\big[ \widetilde{\mathbb{S}^1}\times SU(2, \mathbb{C}) \big]^{\times \, 2}}
\kappa'_{0,1}\big(s', \widehat{n'}\cdot \widehat{l'}; a,x\big) 
\kappa''_{0,1}\big(s'', \widehat{n''}\cdot \widehat{l''}; b,x\big) \,\, \times
\\
\times \,
\kappa'_{1,0}\big(s', \widehat{n'}\cdot \widehat{l'}; c,y\big) 
\kappa''_{1,0}\big(s'', \widehat{n''}\cdot \widehat{l''}; d,y\big) 
\, \Omega \chi^{abcd}(x,y)
\, \ud x \, \ud y
\end{multline*}
is absolutely convergent, and uniformly and absolutely bounded whenever $\chi$ ranges over any bounded set in $\mathscr{E} \otimes \mathscr{E}$.
More generally
\begin{equation}\label{AbsoluteUniformConvergenceOf(q'<q)Contraction}
\Bigg\langle \Big(\theta_y \big(\kappa^{(1)}_{0,1} \dot{\otimes} \ldots  \kappa^{(q)}_{0,1} \big) 
\otimes_{q'} \big(\kappa^{(1)}_{1,0} \dot{\otimes} \ldots \dot{\otimes} \kappa^{(q)}_{1,0} \big)\Big)(\Omega\chi), \, \xi \Bigg\rangle 
\end{equation}
\begin{multline*}
=
\sum \limits_{\mathscr{O}^{(1)} } 
\cdots
\sum \limits_{ \mathscr{O}^{(q)} } \,\,\,
\sum \limits_{\substack{ a, \ldots, b \\ c, \ldots, d }} \,\,\,
\int\limits_{\big[ \widetilde{\mathbb{S}^1}\times SU(2, \mathbb{C}) \big]^{\times \, 2}} 
\Bigg\{
\\
\theta\big(xy^{-1}\big)
\kappa^{(1)}_{0,1}\big(s^{(1)},\widehat{n^{(1)}}\cdot\widehat{l^{(1)}}; a,x\big) \ldots 
\kappa^{(q')}_{0,1}\big(s^{(q')},\widehat{n^{(q')}}\cdot\widehat{l^{(q')}}; b',x\big) \,\, \times
\\
\times \,\, 
\kappa^{(q'+1)}_{0,1}\big(s^{(q'+1)},\widehat{n^{(q'+1)}}\cdot\widehat{l^{(q'+1)}}; b'',x\big) 
\ldots
\kappa^{(q)}_{0,1}\big(s^{(q)},\widehat{n^{(q)}}\cdot\widehat{l^{(q)}}; b,x\big) 
\,\, \times
\\
\times \,\,
\kappa^{(1)}_{1,0}\big(s^{(1)},\widehat{n^{(1)}}\cdot\widehat{l^{(1)}}; c,y\big) \ldots
\kappa^{(q')}_{1,0}\big(s^{(q')},\widehat{n^{(q')}}\cdot\widehat{l^{(q')}}; d',y\big)
\,\, \times
\\
\times \,\,
\kappa^{(q'+1)}_{1,0}\big(s'^{(q'+1)},\widehat{n'^{(q'+1)}}\cdot\widehat{l'^{(q'+1)}}; d'',y\big) 
\ldots
\kappa^{(q)}_{1,0}\big(s'^{(q)},\widehat{n'^{(q)}}\cdot\widehat{l'^{(q)}}; d,y\big) 
\,\, \times
\\
\xi\big(s^{(q'+1)},\widehat{n^{(q'+1)}}\cdot\widehat{l^{(q'+1)}}, \ldots, s^{(q)},\widehat{n^{(q)}}\cdot\widehat{l^{(q)}}, 
s'^{(q'+1)},\widehat{n'^{(q'+1)}}\cdot\widehat{l'^{(q'+1)}}, \ldots, s'^{(q)},\widehat{n'^{(q)}}\cdot\widehat{l'^{(q)}}\big)
\\
\, \Omega \chi^{a \ldots bc\ldots d}(x,y)
\Bigg\}
\, \ud x \, \ud y
\end{multline*}
is absolutely convergent, and uniformly and absolutely bounded whenever $\chi$ ranges over any bounded set in 
\begin{multline*}
\chi \in \mathscr{E} \otimes \mathscr{E} 
= S_{A}\big(\widetilde{\mathbb{S}^1}\times SU(2, \mathbb{C}); \, \mathbb{C}^{d_{{}_{(1)}} \cdots d_{{}_{q}}} \big) \otimes 
S_{A}\big(\widetilde{\mathbb{S}^1}\times SU(2, \mathbb{C}); \mathbb{C}^{d_{{}_{(1)}} \cdots d_{{}_{q}}} \big) 
\\
= 
S_{A\otimes A}\big(\big[ \widetilde{\mathbb{S}^1}\times SU(2, \mathbb{C}) \big]^{\times \, 2}; \, 
\mathbb{C}^{d_{{}_{(1)}} \cdots d_{{}_{q}}d_{{}_{(1)}} \cdots d_{{}_{q}}}\big),
\,\,\,\,\,\,\,\,\,\, A = \Delta +1.
\end{multline*}
and for $\xi$ ranging over any bounded set
\[
E_{{}_{(q'+1)}} \otimes \cdots \otimes E_{{}_{(q)}}.
\]
In this contraction we contracted the first $q'$ pairs of momentum variables, so the momentum variables 
of the first $q'$ plane wave ``annihilation'' kernels $\kappa^{(1)}_{0,1}, \ldots, \kappa^{(q')}_{0,1}$ coincide respectively
with the momentum variables of the first $q'$ plane wave ``creation'' kernels $\kappa^{(1)}_{1,0}, \ldots, \kappa^{(q')}_{1,0}$.
The absolute uniform convergence follows from the inequalities
(\ref{<thetaykappa(q-contraction)kappa,chi><infty}) and (\ref{|<thetaykappa(q'-contraction)kappa(chi),xi>|<C|chi|omega|xi|})
with $\chi$ replaced by $\Omega \chi$:
\begin{multline}\label{|<thetaykappa(q'-contraction)kappa(Omegachi),xi>|<C|chi|omega|xi|}
\Big| \Big\langle \Big(\theta_y \big(\kappa^{(1)}_{0,1} \dot{\otimes} \ldots \dot{\otimes} \kappa^{(q)}_{0,1} \big) 
\otimes_{q'} \big(\kappa^{(1)}_{1,0} \dot{\otimes} \ldots \dot{\otimes} \kappa^{(q)}_{1,0} \big) \Big)(\Omega\chi), \,\,
\xi \Big\rangle
 \Big|
\\
\leq 
n_{{}_{\mathscr{O}^{(1)}_+}} \ldots n_{{}_{\mathscr{O}^{(q)}_+}}
(\textrm{dim} \, V^{(1)})^2 \, \ldots \, (\textrm{dim} \, V^{(q)})^2 \, \big| \widetilde{\eta}\big|_{{}_{L^2}} \, 
\big| \theta.\Omega \chi \big|_{{}_{\omega/2}} \big|\xi\big|_{{}_{0}}
\\
\leq 
n_{{}_{\mathscr{O}^{(1)}_+}} \ldots n_{{}_{\mathscr{O}^{(q)}_+}}
(\textrm{dim} \, V^{(1)})^2 \, \ldots \, (\textrm{dim} \, V^{(q)})^2 \, \big| \widetilde{\eta}\big|_{{}_{L^2}} \, 
\big| \Omega \chi \big|_{{}_{\omega/2}} \big|\xi\big|_{{}_{0}},
\end{multline}
and continuity of the projection operator $\Omega$, because the inequality 
(\ref{|<thetaykappa(q'-contraction)kappa(Omegachi),xi>|<C|chi|omega|xi|}) gives us not only  the estimation for 
the absolute value of the contraction sum/integral (\ref{AbsoluteUniformConvergenceOf(q'<q)Contraction}), but in fact for this sum in which the 
summand/integrand is replaced with its absolute value. Indeed, recall that 
(\ref{|<thetaykappa(q'-contraction)kappa(chi),xi>|<C|chi|omega|xi|}) gives an estimation not merely
of the absolute value of the sum/integral (\ref{thetay(kk'...k''')q-contraction(k,k',...k''')}) but of the 
sum/integral with the summand/integrand replaced by its absolute value. 

Here 
\[
\theta.\Omega \chi(x,y) =  \theta\big(xy^{-1}\big) \big(\Omega \chi\big)\big(xy^{-1}\big). 
\]
In particular (compare (\ref{SumOfFirstSpecialchi-s}))
\[
\theta.\Omega \chi(x,y) 
= \sum \limits_{j} \theta\big(xy^{-1}\big)\big(\Omega'\phi_j\big)\big(xy^{-1}\big)\varphi_j(y),
\]
\[
(\theta.\Omega)(\chi) \circ L (x,y) 
= \sum \limits_{j} \theta(x)\big(\Omega'\phi_j\big)(x)\varphi_j(y).
\]

\begin{defin}
Let $\theta_{\varepsilon}$ be a one-parameter family of smooth functions which, 
when $\theta, \theta_{\varepsilon}$
are regarded as elements of $\mathscr{E}^*$, converges 
to $\theta$ in $\mathscr{E}^*$ when $\varepsilon \rightarrow 0$. We need to specify more precisely the
convergence $\theta_{\varepsilon} \overset{\varepsilon \rightarrow 0}{\longrightarrow} \theta$.
Namely we require that
\begin{multline}\label{thetavarepsilon->theta}
\big|A^{\omega/2} (\phi_{{}_{\theta}}-\phi_{{}_{\theta_{\varepsilon}}}) \big|_{{}_{L^2}}^{2} 
=
\int\limits_{\widetilde{\mathbb{S}^1}\times SU(2, \mathbb{C})}
\big|A^{\omega/2} (\phi_{{}_{\theta}}-\phi_{{}_{\theta_{\varepsilon}}})(x)\big|^2
\, \ud x 
\\
=
\int\limits_{\widetilde{\mathbb{S}^1}\times SU(2, \mathbb{C})}
\Big|A^{\omega/2} \big((\theta- \theta_{\varepsilon}).\phi\big)(x) \Big|^2
\, \ud x 
\overset{\varepsilon \rightarrow 0}{\longrightarrow} 0, \,\,\,\,\, A= \Delta +1,
\end{multline}
uniformly for $\phi \in \mathscr{E}_{{}_{\omega}}$ varying over any bounded 
set in $\mathscr{E}_{{}_{\omega}}$. Equivalently we require
\begin{equation}\label{Omegathetavarepsilon->theta}
\big|A^{\omega/2} (\theta-\theta_{\varepsilon})\Omega\phi \big|_{{}_{L^2}}^{2} 
=
\int\limits_{\widetilde{\mathbb{S}^1}\times SU(2, \mathbb{C})}
\Big|A^{\omega/2} \big((\theta- \theta_{\epsilon}).\phi\big)(x) \Big|^2
\, \ud x 
\overset{\varepsilon \rightarrow 0}{\longrightarrow} 0,
\end{equation}
uniformly for $\phi$ ranging over any bounded set in $\mathscr{E}$. 
\label{ConvergenceOfthetavarepsilon}
\end{defin}

Note, please, that this Definition of convergence is not empty and such families $\theta_\varepsilon$
do exist. To see this we confine ourselves to the sequence $\varepsilon_j = 1/(2j+1)$, $j = 2,3,4, \ldots$ 
and to the corresponding sequence $\theta_{\varepsilon_j}$ of functions, which is to be convergent to $\theta$,
which is sufficient for us. In our case we can restrict consideration to functions $\theta_{\varepsilon_j}$ which,
similarly as $\theta$ itself, depend only on time. In the time interval $-1 \leq t \leq 1$
around the jump of $\theta$ at $t=0$ we can define:
\[
\theta_{\varepsilon_j}(t) = {\textstyle\frac{1}{2}}\big(f_j(t)\big)^{{}^{1/(2j+1)}} 
+ {\textstyle\frac{1}{2}},  \,\,\,\,\,\,\,\,  -1 \leq t \leq 1
\]
and similarly in the inerval  $-1 + 2\pi \leq t \leq 1+ 2\pi$
around the jump of $\theta$ at $t=2\pi$:
\[
\theta_{\varepsilon_j}(t) = {\textstyle\frac{1}{2}}\big(f_j(2\pi-t)\big)^{{}^{1/(2j+1)}} + {\textstyle\frac{1}{2}}, 
\,\,\,\,\,\,\,\, -1 + 2\pi \leq t \leq 1+ 2\pi,   
\]
with
\[
f_j(t)= {\textstyle\frac{1}{1-j}}t^2 +t + {\textstyle\frac{1}{j-1}}.
\]
One can easily see that
\begin{align}
\int\limits_{[-1,1] \times SU(2, \mathbb{C})}
\Big|A^{\omega/2} \big((\theta- \theta_{\varepsilon_j}).\phi\big)(x) \Big|^2
\, \ud x  \,\,\, & \overset{j \rightarrow \infty}{\longrightarrow} 0,
\label{ConvergenceOn[-1,1]xSL(2,C)}
\\
\int\limits_{[-1+2\pi,1+2\pi] \times SU(2, \mathbb{C})}
\Big|A^{\omega/2} \big((\theta- \theta_{\varepsilon_j}).\phi\big)(x) \Big|^2
\, \ud x \,\,\, & \overset{j \rightarrow \infty}{\longrightarrow} 0,
\label{ConvergenceOn[2pi-1,2pi+1]xSL(2,C)}
\end{align}
uniformly for $\phi \in \mathscr{E}_{{}_{\omega}}$ varying over any bounded 
set in $\mathscr{E}_{{}_{\omega}}$. Note that
\begin{align*}
\partial_{{}_{t}}\theta_{\varepsilon_j}(t) & =  {\textstyle\frac{1}{2}}{\textstyle\frac{1}{2j+1}} 
\big(f_j(t)\big)^{{}^{-1 + 1/(2j+1)}} 
+ {\textstyle\frac{1}{1-j}}{\textstyle\frac{1}{2j+1}} 
\big(f_j(t)\big)^{{}^{1/(2j+1)}}, 
\\
\partial_{{}_{t}}^{2}\theta_{\varepsilon_j}(t) & =  {\textstyle\frac{1}{2}}{\textstyle\frac{1}{2j+1}}
\big(-1 + {\textstyle\frac{1}{2j+1}} \big) 
\big(f_j(t)\big)^{{}^{-2+1/(2j+1)}} 
\\
&+{\textstyle\frac{2}{1-j}}{\textstyle\frac{1}{2j+1}}
\big(-1 + {\textstyle\frac{1}{2j+1}} \big) 
\big(f_j(t)\big)^{{}^{-1+1/(2j+1)}} 
\\
&+ {\textstyle\frac{1}{(1-j)^2}}{\textstyle\frac{1}{(2j+1)^2}} 
\big(f_j(t)\big)^{{}^{1/(2j+1)}}
+ {\textstyle\frac{1}{1-j}}{\textstyle\frac{1}{(2j+1)^2}} 
\big(f_j(t)\big)^{{}^{-1 +1/(2j+1)}}, 
\\
\partial_{{}_{t}}^{3}\theta_{\varepsilon_j}(t) & =  \ldots, 
\\
\vdots \,\,\,\,\ & 
\end{align*}
on the time interval $[-1,1]$, and similarly on the time interval $[-1+2\pi, 1+2\pi]$.
We have the dominated pointwise convergence (except possibly at $t=0$ or $t=2\pi$)
\[
\big|\big(A^{\omega}(\theta - \theta_{\varepsilon_j}).\phi\big)(x)\big|  \overset{j \rightarrow \infty}{\longrightarrow} 0, 
\]
\[ 
x \in [-1, 1] \times SL(2, \mathbb{C}) \,\,\, \textrm{and} \,\,\, x \in [-1+2\pi, 1+2\pi] \times SL(2, \mathbb{C}).
\]
We see that all the singularities
of the limits (for $j \rightarrow \infty$) of the time derivatives  $\partial_{{}_{t}}^{m}\theta_{\varepsilon_j}$ at $t=0$ and at $t= 2\pi$,
up to order $m = \omega$, are cancelled by the vanishing time derivatives of $\phi$ at $t=0$ and at $t=2\pi$,
and contributions to the Taylor expansion of $\phi \in \mathscr{E}_{{}_{\omega}}$ around $t=0$ and around $t=2\pi$ vanish up to order $\omega$
in the time variable. The required uniform convergence (\ref{ConvergenceOn[-1,1]xSL(2,C)}) 
and (\ref{ConvergenceOn[2pi-1,2pi+1]xSL(2,C)})
follows easily from the ``Dominated Convergence Theorem'' (e.g. Corollary 3.4.5 in \cite{Segal_Kunze}) and the assumed uniform 
boundedness of the derivatives of $\phi$ and the vanishing of time derivatives up to order $\omega$ at $t=0$ and $t=2\pi$.  
Of course, it remains to prolong the functions $\theta_{\varepsilon_j}$ smoothly outside the two time intervals
$[-1,1]$ and $[-1+2\pi, 1+2\pi]$, with the preservation of the inequality of Definition 
\ref{ConvergenceOfthetavarepsilon}. But this is a rather simple and elementary task, so we do not present
details of this construction. 

Note also that for $\omega=0$ the convergence $\theta_\varepsilon \overset{\varepsilon \rightarrow 0}{\longrightarrow}   \theta$ of Definition \ref{ConvergenceOfthetavarepsilon} follows from the convergence in $\mathscr{E}^*$, and is even weaker than 
convergence in $\mathscr{E}^*$, because requires the pointwise convergence only on the elements of $\mathscr{E}_{{}_{\omega}} \subset \mathscr{E}$.
In case of $\omega>0$ it is not clear if the convergence of  Definition \ref{ConvergenceOfthetavarepsilon} follows from 
the convergence in $\mathscr{E}^*$. 

Because of the assumed convergence $\theta_\varepsilon \overset{\varepsilon \rightarrow 0}{\longrightarrow}   \theta$, stated in Definition \ref{ConvergenceOfthetavarepsilon}, 
\[
\big| (\theta-\theta_\varepsilon).\Omega \chi \big|_{{}_{\omega/2}} \overset{\varepsilon \rightarrow 0}{\longrightarrow} 0
\]
uniformly whenever $\chi \in \mathscr{E}^{\otimes\, 2}$ ranges over any bounded set in $\mathscr{E}^{\otimes\, 2}$
(with the convention that $\big((\theta-\theta_\varepsilon).\Omega \chi\big)(x,y) = (\theta-\theta_\varepsilon)\big(xy^{-1}\big)\Omega \chi(x,y)$).
Assuming $\omega$ in this convergence the same as $\omega$ in the inequality 
(\ref{|<thetaykappa(q'-contraction)kappa(Omegachi),xi>|<C|chi|omega|xi|}), it follows from
the inequality (\ref{|<thetaykappa(q'-contraction)kappa(Omegachi),xi>|<C|chi|omega|xi|}) that 
\begin{multline*}
\Bigg|\Big\langle\Big(\theta_{y}\big(\kappa^{(1)}_{0,1} \dot{\otimes} \ldots  \kappa^{(q)}_{0,1} \big) 
\otimes_{q'} \big(\kappa^{(1)}_{1,0} \dot{\otimes} \kappa^{(q)}_{1,0} \big)\Big)(\Omega \chi)
\\
- 
\Big(\theta_{\varepsilon \, y}\big(\kappa^{(1)}_{0,1} \dot{\otimes} \ldots  \kappa^{(q)}_{0,1} \big) 
\otimes_{q'} \big(\kappa^{(1)}_{1,0} \dot{\otimes} \kappa^{(q)}_{1,0} \big)\Big)(\Omega \chi), \,\, \xi \Big\rangle \Bigg|
\end{multline*}
holds
uniformly whenever 
\[
\xi \in E_{{}_{(q'+1)}} \otimes \cdots \otimes E_{{}_{(q)}} \,\,\, \textrm{and} \,\,\, \chi \in \mathscr{E}^{\otimes \, 2}
\]
 range over bounded sets.
Then 
\begin{multline}\label{ConvergencethetavarepsilonOfContractionKernelInL(E,E*)}
\theta.\Big(\big(\kappa^{(1)}_{0,1} \dot{\otimes} \ldots  \kappa^{(q)}_{0,1} \big) 
\otimes_{q'} \big(\kappa^{(1)}_{1,0} \dot{\otimes} \kappa^{(q)}_{1,0} \big)\Big) \circ \Omega
\\
=
\Big(\theta_{\varepsilon y}\big(\kappa^{(1)}_{0,1} \dot{\otimes} \ldots  \kappa^{(q)}_{0,1} \big) 
\otimes_{q'} \big(\kappa^{(1)}_{1,0} \dot{\otimes} \kappa^{(q)}_{1,0} \big)\Big) \circ \Omega
\\
\overset{\varepsilon \rightarrow 0}{\longrightarrow}
\,\,\,
\Big(\theta_y\big(\kappa^{(1)}_{0,1} \dot{\otimes} \ldots  \kappa^{(q)}_{0,1} \big) 
\otimes_{q'} \big(\kappa^{(1)}_{1,0} \dot{\otimes} \kappa^{(q)}_{1,0} \big)\Big) \circ \Omega
\end{multline}
converge uniformly on each bounded set, when regarded as elements of
\[
\mathscr{L}\big(\mathscr{E}^{\otimes \, 2}, \,\, E_{{}_{(q'+1)}}^{*} \otimes \cdots \otimes E_{{}_{(q)}}^{*}  \big)
\cong E_{{}_{(q'+1)}}^{*} \otimes \cdots \otimes E_{{}_{(q)}}^{*} \otimes \mathscr{E}^{* \, \otimes \, 2}.
\] 
and thus in the ordinary topology of uniform convergence on bounded sets in
\[
\mathscr{L}\big(\mathscr{E}^{\otimes \, 2}, \,\, E_{{}_{(q'+1)}}^{*} \otimes \cdots \otimes E_{{}_{(q)}}^{*}  \big),
\]
and thus simply in 
\[
\mathscr{L}\big(\mathscr{E}^{\otimes \, 2}, \,\, E_{{}_{(q'+1)}}^{*} \otimes \cdots \otimes E_{{}_{(q)}}^{*}  \big)
\cong 
\mathscr{L}\big( \, E_{{}_{(q'+1)}} \otimes \cdots \otimes E_{{}_{(q)}}, \, \mathscr{E}^{* \, \otimes \, 2}  \big).
\]

For the further analysis of the contraction (\ref{AbsoluteUniformConvergenceOf(q'<q)Contraction}) we need 
Lemma \ref{(kappa...kappa)(xi)inSA}. From this Lemma we immediately see
that the contraction kernel (\ref{AbsoluteUniformConvergenceOf(q'<q)Contraction}) can be written 
canonically as the product 
\begin{multline}\label{ProductFormOfContraction}
\Big(\theta. \big(\kappa^{(1)}_{0,1} \dot{\otimes} \ldots  \kappa^{(q)}_{0,1} \big) 
\otimes_{q'} \big(\kappa^{(1)}_{1,0} \dot{\otimes} \ldots \dot{\otimes} \kappa^{(q)}_{1,0} \big)\Big)
\big(\xi\otimes \zeta \big) 
\\
=
\Big(\theta.\big(\kappa^{(1)}_{0,1} \dot{\otimes} \ldots  \kappa^{(q')}_{0,1} \big) 
\otimes_{q'} \big(\kappa^{(1)}_{1,0} \dot{\otimes} \ldots \dot{\otimes} \kappa^{(q')}_{1,0} \big)\Big) 
\big(\kappa^{(q'+1)}_{0,1} \dot{\otimes} \ldots  \kappa^{(q)}_{0,1} \big) \big(\xi\big) 
\,\, \times \,\,
\\
\times \,\,
\big(\kappa^{(q'+1)}_{1,0} \dot{\otimes} \ldots  \kappa^{(q)}_{1,0} \big) \big(\zeta\big),
\end{multline}
of scalar valued $\widetilde{\mathbb{S}^1}\times SL(2, \mathbb{C})$-invariant contraction kernel (distribution)
which captures all $q'$ contractions
\[
\Big(\theta.\big(\kappa^{(1)}_{0,1} \dot{\otimes} \ldots  \kappa^{(q')}_{0,1} \big) 
\otimes_{q'} \big(\kappa^{(1)}_{1,0} \dot{\otimes} \ldots \dot{\otimes} \kappa^{(q')}_{1,0} \big)\Big) 
\]
and of the kernel without any contractions
\[
\big(\kappa^{(q'+1)}_{0,1} \dot{\otimes} \ldots  \kappa^{(q)}_{0,1} \big) \big(\xi\big) 
\big(\kappa^{(q'+1)}_{1,0} \dot{\otimes} \ldots  \kappa^{(q)}_{1,0} \big) \big(\zeta\big).
\]
From this product formula it evidently follows the continuity of the contraction kernel, regarded as a map
\[
E_{{}_{(q'+1)}} \otimes \cdots \otimes E_{{}_{q}} \otimes E_{{}_{q'+1}} 
\otimes \cdots \otimes E_{{}_{(q)}} \longrightarrow \mathscr{E}^* \otimes\mathscr{E}^*,
\]
because of the continuity of the maps 
\begin{align*}
E_{{}_{(q'+1)}} \otimes \cdots \otimes E_{{}_{q}} \otimes E_{{}_{q'+1}} \ni \,\,\,\,\, \xi & \longmapsto
\,\,\,\,\,\,\,\,\,\, \big(\kappa^{(q'+1)}_{0,1} \dot{\otimes} \ldots  \kappa^{(q)}_{0,1} \big) \big(\xi\big) \in & \mathscr{E}
\\
E_{{}_{(q'+1)}} \otimes \cdots \otimes E_{{}_{q}} \otimes E_{{}_{q'+1}} \ni \,\,\,\,\, \zeta & \longmapsto
\,\,\,\,\,\,\,\,\,\, \big(\kappa^{(q'+1)}_{1,0} \dot{\otimes} \ldots  \kappa^{(q)}_{1,0} \big) \big(\zeta\big) \in & \mathscr{E}
\end{align*}
into the algebra $\mathscr{E}$ of multipliers of $\mathscr{E}^*$,
\emph{i.e.} pointwise product of elements of $\mathscr{E}^*$ by any element of $\mathscr{E}$ gives a continuous
map  $\mathscr{E}^* \longrightarrow \mathscr{E}^*$. 

This continuity also follows from the uniform convergence
(\ref{ConvergencethetavarepsilonOfContractionKernelInL(E,E*)}), because  the nuclear space
\[
\mathscr{L}\big(\mathscr{E}^{\otimes \, 2}, \,\, E_{{}_{(q'+1)}}^{*} \otimes \cdots \otimes E_{{}_{(q)}}^{*}  \big)
\cong E_{{}_{(q'+1)}}^{*} \otimes \cdots \otimes E_{{}_{(q)}}^{*} \otimes \mathscr{E}^{* \, \otimes \, 2}
\]  
is complete.

The kernel function can thus be written in the product form. Correspondingly to the formula
 (\ref{ProductFormOfContraction}) the contraction kernel function 
has the form of product
\begin{multline}\label{ProductFormOfContraction(x,y)}
\bigg[\Big(\theta. \big(\kappa^{(1)}_{0,1} \dot{\otimes} \ldots  \kappa^{(q)}_{0,1} \big) 
\otimes_{q'} \big(\kappa^{(1)}_{1,0} \dot{\otimes} \ldots \dot{\otimes} \kappa^{(q)}_{1,0} \big) \Big) \circ \Omega \bigg]
\big(\, (\ldots)\, , \, (\ldots)' \, ; \, a, \ldots, b, c, \ldots, d,x,y\big) 
\\
=
\bigg[\Big(\theta.\big(\kappa^{(1)}_{0,1} \dot{\otimes} \ldots  \kappa^{(q')}_{0,1} \big) 
\otimes_{q'} \big(\kappa^{(1)}_{1,0} \dot{\otimes} \ldots \dot{\otimes} \kappa^{(q')}_{1,0} \big)\Big) \circ \Omega \bigg](a, \ldots, b', c, \ldots, d', x, y) 
\,\, \times 
\\
\times \,\, 
\big(\kappa^{(q'+1)}_{0,1} \dot{\otimes} \ldots  \kappa^{(q)}_{0,1} \otimes
\kappa^{(q'+1)}_{1,0} \dot{\otimes} \ldots  \kappa^{(q)}_{1,0} \big) \big(\, (\ldots) \, , \, (\ldots)' \, , \, b'', \ldots, b, d'', \ldots, d, x, y\big),
\end{multline}
of scalar valued $\widetilde{\mathbb{S}^1}\times SL(2, \mathbb{C})$-invariant contraction kernel (distribution) including
all $q'$ contractions of all $q'$ contracted pairs of plane wave kernels
\[
\bigg[\Big(\theta.\big(\kappa^{(1)}_{0,1} \dot{\otimes} \ldots  \kappa^{(q')}_{0,1} \big) 
\otimes_{q'} \big(\kappa^{(1)}_{1,0} \dot{\otimes} \ldots \dot{\otimes} \kappa^{(q')}_{1,0} \big)\Big) \circ \Omega \bigg](a, \ldots, b', c, \ldots, d', x, y) 
\]
and of a vector-valued kernel (distribution) without any contractions
\[
\big(\kappa^{(q'+1)}_{0,1} \dot{\otimes} \ldots  \kappa^{(q)}_{0,1} \otimes
\kappa^{(q'+1)}_{1,0} \dot{\otimes} \ldots  \kappa^{(q)}_{1,0} \big) \big(\, (\ldots) \, , \, (\ldots)' \, , \, b'', \ldots, b, d'', \ldots, d, x, y\big).
\]

In order to simplify notation, the momentum variables of the dot product of the ``annihilation'' plane wave kernels 
\[
\kappa^{(q'+1)}_{0,1} \dot{\otimes} \ldots  \kappa^{(q)}_{0,1}
\]
are written as dots $(\ldots)$ in parenthesis, similarly for the momentum variables of the dot product
\[
\kappa^{(q'+1)}_{1,0} \dot{\otimes} \ldots  \kappa^{(q)}_{1,0}
\]
of the ``creation'' plane wave kernels written as $(\ldots)'$ with prime.

We have removed the projection operator $\Omega$ from the second vector valued factor in (\ref{ProductFormOfContraction(x,y)}) 
which does not involve any contractions, because it is well defined on the whole test space $\mathscr{E}\otimes \mathscr{E}$, and 
$\Omega$ is not needed there.

Also the product formula (\ref{ProductFormOfContraction}) can be written shortly as the pointwise product of kernel functions
\begin{multline}\label{GeneralProductFormOfContractionShortForm}
\Big(\theta. \big(\kappa^{(1)}_{0,1} \dot{\otimes} \ldots  \kappa^{(q)}_{0,1} \big) 
\otimes_{q'} \big(\kappa^{(1)}_{1,0} \dot{\otimes} \ldots \dot{\otimes} \kappa^{(q)}_{1,0} \big)\Big)
\\
=
\Big(\theta.\big(\kappa^{(1)}_{0,1} \dot{\otimes} \ldots  \kappa^{(q')}_{0,1} \big) 
\otimes_{q'} \big(\kappa^{(1)}_{1,0} \dot{\otimes} \ldots \dot{\otimes} \kappa^{(q')}_{1,0} \big)\Big) 
\,\, \times \,\,
\\
\times \,\,
\big(\kappa^{(q'+1)}_{0,1} \dot{\otimes} \ldots  \kappa^{(q)}_{0,1} \big) \otimes
\big(\kappa^{(q'+1)}_{1,0} \dot{\otimes} \ldots  \kappa^{(q)}_{1,0} \big),
\end{multline}
where the first factor depends only on space-time variables $x,y$ and field component indices,
and the second factor depends in addition on momentum variables and where $\times$
stands for pointwise multiplication, well defined even with the first factor being 
a proper distribution, because the second factor is a multiplier of the distribution
space $\mathscr{E}^{* \, \otimes \, 2}$.

As on the Minkowski space-time it is the scalar factor
\[
\big(\kappa^{(1)}_{0,1} \dot{\otimes} \ldots  \kappa^{(q')}_{0,1} \big) 
\otimes_{q'} \big(\kappa^{(1)}_{1,0} \dot{\otimes} \ldots \dot{\otimes} \kappa^{(q')}_{1,0} \big)
\]
 which captures all contractions and
which brings in the non-uniqueness in the construction of the retarded and advanced part of it if $q' \geq 2$.
Indeed, the image $\textrm{Im} \, \Omega$ of $\Omega$ is spanned by elements $\chi(x,y) = \phi\big(xy^{-1}\big)\varphi(y)$, 
$\phi \in \mathscr{E}_{{}_{\omega}} = \textrm{Im} \, \Omega'$,
$\varphi \in \mathscr{E}$. The kernel $\textrm{Ker} \, \Omega$ of $\Omega$ is spanned by elements 
$\chi(x,y) = \phi\big(xy^{-1}\big)\varphi(y)$, $\phi \in \textrm{Ker} \, \Omega'$,
$\varphi \in \mathscr{E}$.
It immediately follows that the kernel $\textrm{Ker} \, \Omega'$ is spanned by elements of the form
\[
\sum\limits_{\alpha=0}^{\omega} \omega_{{}_{0 \,\, I \, \alpha}} \otimes f_\alpha + 
\sum\limits_{\beta=0}^{\omega} \omega_{{}_{0 \,\, II \, \beta}} \otimes g_\beta
\]
where 
\[
f_\alpha = \eta_\alpha\big|_{t=0} , \,\, g_\beta = \zeta_\beta\big|_{t=2\pi}, \,\,\,
\eta_\alpha, \zeta_\beta \in \mathscr{E} = \mathcal{S}_{\Delta+1}(\widetilde{\mathbb{S}^1}\times SU(2, \mathbb{C})),
\]
and $\omega_{{}_{0 \,\, I \, \alpha}}, \omega_{{}_{0 \,\, II \, \alpha}} \in \mathscr{C}^\infty(\widetilde{\mathbb{S}^1})$ are the functions
constructed above in the definition of the idempotent $\Omega'$. Therefore,   $f_\alpha \in \mathscr{C}^\infty(SU(2, \mathbb{C})_{{}_{t=0}})
= \mathcal{S}_{\Delta_{SU(2, \mathbb{C})}}(SU(2, \mathbb{C}))$ on the Cauchy surface $SU(2, \mathbb{C})_{{}_{t=0}} = \mathbb{S}^{3}_{{}_{t=0}}$ at $t=0$
and $g_\beta \in \mathscr{C}^\infty(SU(2, \mathbb{C})_{{}_{t=2\pi}})
= \mathcal{S}_{\Delta_{SU(2, \mathbb{C})}}(SU(2, \mathbb{C}))$ on the Cauchy surface $SU(2, \mathbb{C})_{{}_{t=2\p}} = \mathbb{S}^{3}_{{}_{t=2\pi}}$ at $t=2\pi$.
We see that $\textrm{Ker} \, \Omega$ is spanned by elements of the form
\[
\chi(x,y) = \sum\limits_{\alpha=0}^{\omega} \omega_{{}_{0 \,\, I \, \alpha}}(t-\tau)  f_\alpha\big(\boldsymbol{w}\boldsymbol{v}^{-1}\big)
\varphi(\tau, \boldsymbol{v}) + 
\sum\limits_{\beta=0}^{\omega} \omega_{{}_{0 \,\, II \, \beta}}(t-\tau)  g_\beta\big(\boldsymbol{w}\boldsymbol{v}^{-1}\big)\varphi(\tau, \boldsymbol{v}),
\]
\[
\varphi \in \mathscr{E}, f_\alpha \in \mathscr{C}^\infty(SU(2, \mathbb{C})_{{}_{t=0}}), g_\beta \in \mathscr{C}^\infty(SU(2, \mathbb{C})_{{}_{t=2\pi}})
\] 
\[
x = (t, \boldsymbol{w}), \, y = (\tau, \boldsymbol{v}) \in \widetilde{\mathbb{S}^1}\times SU(2, \mathbb{C}).
\] 
Therefore, the most general functional on the kernel $\textrm{Ker} \, \Omega$ is of the form
\[
G(x,y) = \delta^{\omega}_{{}_{1;2}}(x,y)F(y), \,\,\,\, F \in \mathscr{E}^*,
\]
where
\[
\delta^{\omega}_{{}_{1;2}}(x,y) \overset{\textrm{df}}{=}
\sum\limits_{\alpha = 0}^{\omega} C_{\alpha}^{I}\big(\boldsymbol{w}\boldsymbol{v}^{-1}\big) \delta^{(\alpha)}(t-\tau) 
+ \sum\limits_{\beta = 0}^{\omega} C_{\beta}^{II}\big(\boldsymbol{w}\boldsymbol{v}^{-1}\big) \delta^{(\beta)}(t-\tau - 2\pi),
\]
\[
x = (t, \boldsymbol{w}), \, y = (\tau, \boldsymbol{v}) \in \widetilde{\mathbb{S}^1}\times SU(2, \mathbb{C}).
\] 
This functional $G$ moreover vanishes identically on the image $\textrm{Im} \, \Omega$. Thus the most general
$\widetilde{\mathbb{S}^1}\times SU(2, \mathbb{C})$-invariant functional on the kernel 
$\textrm{Ker} \, \Omega$  is of the form
\[
\delta^{\omega}_{{}_{1;2}}(x,y) \overset{\textrm{df}}{=}
\sum\limits_{\alpha = 0}^{\omega} C_{\alpha}^{I}\big(\boldsymbol{w}\boldsymbol{v}^{-1}\big) \delta^{(\alpha)}(t-\tau) 
+ \sum\limits_{\beta = 0}^{\omega} C_{\beta}^{II}\big(\boldsymbol{w}\boldsymbol{v}^{-1}\big) \delta^{(\beta)}(t-\tau - 2\pi).
\]
Moreover, this functional $\delta^{\omega}_{{}_{1;2}}$ vanishes identically on the image $\textrm{Im} \, \Omega$. Because the image
$\textrm{Im} \, \Omega = \textrm{Ker} \, [\boldsymbol{1} - \Omega]$, then $\textrm{Im} \, \Omega$
is a closed subspace and 
\[
\mathscr{E}^{\otimes \, 2} = \textrm{Ker} \, \Omega \oplus \textrm{Im} \, \Omega.
\]
Therefore in order to construct the most general extension of the retarded 
\[
\theta.\big(\kappa^{(1)}_{0,1} \dot{\otimes} \ldots  \kappa^{(q')}_{0,1} \big) 
\otimes_{q'} \big(\kappa^{(1)}_{1,0} \dot{\otimes} \ldots \dot{\otimes} \kappa^{(q')}_{1,0} \big) 
\]
and the advanced part 
\[
-\check{\theta}.\big(\kappa^{(1)}_{0,1} \dot{\otimes} \ldots  \kappa^{(q')}_{0,1} \big) 
\otimes_{q'} \big(\kappa^{(1)}_{1,0} \dot{\otimes} \ldots \dot{\otimes} \kappa^{(q')}_{1,0} \big) 
\]
well defined on the image $\textrm{Im} \, \Omega$, we should add to 
\[
\theta.\big(\kappa^{(1)}_{0,1} \dot{\otimes} \ldots  \kappa^{(q')}_{0,1} \big) 
\otimes_{q'} \big(\kappa^{(1)}_{1,0} \dot{\otimes} \ldots \dot{\otimes} \kappa^{(q')}_{1,0} \big) \circ \Omega
\]
and, respectively, to
\[
-\check{\theta}.\big(\kappa^{(1)}_{0,1} \dot{\otimes} \ldots  \kappa^{(q')}_{0,1} \big) 
\otimes_{q'} \big(\kappa^{(1)}_{1,0} \dot{\otimes} \ldots \dot{\otimes} \kappa^{(q')}_{1,0} \big) \circ \Omega
\]
the $\widetilde{\mathbb{S}^1}\times SU(2, \mathbb{C})$-invariant
scalar distribution of the general form
\[
\delta^{\omega}_{{}_{1;2}}(x,y) \overset{\textrm{df}}{=}
\sum\limits_{\alpha = 0}^{\omega} C_{\alpha}^{I}\big(\boldsymbol{w}\boldsymbol{v}^{-1}\big) \delta^{(\alpha)}(t-\tau) 
+ \sum\limits_{\beta = 0}^{\omega} C_{\beta}^{II}\big(\boldsymbol{w}\boldsymbol{v}^{-1}\big) \delta^{(\beta)}(t-\tau - 2\pi),
\]
\[
x = (t, \boldsymbol{w}), \, y = (\tau, \boldsymbol{v}) \in \widetilde{\mathbb{S}^1}\times SL(2, \mathbb{C}),
\] 
concentrated on the cartesian product 
\[
\big[SU(2, \mathbb{C}) \sqcup SU(2, \mathbb{C}) \big] \times \big[ \widetilde{\mathbb{S}^1}\times SU(2, \mathbb{C})\big]
\]
of the disjoint sum of the two Cauchy surfaces $t=0$ and $t= 2\pi$,
with the whole $\widetilde{\mathbb{S}^1}\times SU(2, \mathbb{C})$. Here $C_{\alpha}^{I},  C_{\beta}^{II}$
are arbitrary smooth functions, or even distributions, on the Cauchy surfaces, respectively, $t=0$ and $t=2\pi$. Writing
in more symmetric notation
\begin{equation}\label{delta^omega_1;2}
\delta^{\omega}_{{}_{1;2}}(x_1,x_2) \overset{\textrm{df}}{=}
\sum\limits_{\alpha = 0}^{\omega} C_{\alpha}^{I}\big(\boldsymbol{w}_1\boldsymbol{w}_{2}^{-1}\big) \delta^{(\alpha)}(t_1-t_2) 
+ \sum\limits_{\beta = 0}^{\omega} C_{\beta}^{II}\big(\boldsymbol{w}_1\boldsymbol{w}_{2}^{-1}\big) \delta^{(\beta)}(t_1-t_2 - 2\pi),
\end{equation}
\[
x_1 = (t_1, \boldsymbol{w}_2), \, x_2 = (t_2, \boldsymbol{w}_2) \in \widetilde{\mathbb{S}^1}\times SU(2, \mathbb{C}).
\] 

This can be also seen by another method, which we have used in Subsection \ref{WickForChronological}. Using the 
$\widetilde{\mathbb{S}^1}\times SU(2, \mathbb{C})$-invariance of the scalar contraction kernel we see that
\[
\big(\kappa^{(1)}_{0,1} \dot{\otimes} \ldots  \kappa^{(q')}_{0,1} \big) 
\otimes_{q'} \big(\kappa^{(1)}_{1,0} \dot{\otimes} \ldots \dot{\otimes} \kappa^{(q')}_{1,0} \big)(x,y) = \kappa_{q'}\big(xy^{-1}\big)
\]
and
\[
\big(\theta_y \kappa^{(1)}_{0,1} \dot{\otimes} \ldots  \kappa^{(q')}_{0,1} \big) 
\otimes_{q'} \big(\kappa^{(1)}_{1,0} \dot{\otimes} \ldots \dot{\otimes} \kappa^{(q')}_{1,0} \big)(x,y) = \theta\big(xy^{-1}\big)\kappa_{q'}\big(xy^{-1}\big)
\]
for an element $\kappa_{q'}$ of $\mathscr{E}^*$. We have shown that $\theta_{\varepsilon}\kappa_{q'}$
converges to  $\theta \kappa_{q'}$ in $[\textrm{Im} \, \mathscr{E}]^* \subset \mathscr{E}^*$, provided $\theta_\varepsilon$
is converging to $\theta$ in the sense of Definition \ref{ConvergenceOfthetavarepsilon}.
Note, please, that the idempotent $\Omega' = \Omega'^{2}$ is a continuous operator on $\mathscr{E}$, its image
$\textrm{Im} \, \Omega' = \textrm{Ker} \, [\boldsymbol{1} - \Omega]$ and its kernel $\textrm{Ker} \, \Omega'$
are thus closed and
\[
\mathscr{E} = \textrm{Ker} \, \Omega' \oplus \textrm{Im} \, \Omega'.
\]
It is immediately seen that the kernel $\textrm{Ker} \, \Omega'$ is spanned by elements of the form
\[
\sum\limits_{\alpha=0}^{\omega} \omega_{{}_{0 \,\, I \, \alpha}} \otimes f_\alpha + 
\sum\limits_{\beta=0}^{\omega} \omega_{{}_{0 \,\, II \, \beta}} \otimes g_\beta
\]
where 
\[
f_\alpha \in \mathscr{C}^\infty(SU(2, \mathbb{C})_{{}_{t=0}}), \,\, g_\beta \in \mathscr{C}^\infty(SU(2, \mathbb{C})_{{}_{t=2\pi}}).
\] 
It is therefore easily seen that the functional 
\[
\delta^{\omega}_{{}_{1;2}}(x) \overset{\textrm{df}}{=}
\sum\limits_{\alpha = 0}^{\omega} C_{\alpha}^{I}\big(\boldsymbol{w}\big) \delta^{(\alpha)}(t) 
+ \sum\limits_{\beta = 0}^{\omega} C_{\beta}^{II}\big(\boldsymbol{w}\big) \delta^{(\beta)}(t- 2\pi),
\]
\[
x = (t, \boldsymbol{w}) \in \widetilde{\mathbb{S}^1}\times SU(2, \mathbb{C}),
\] 
is the most general functional on $\textrm{Ker} \, \Omega'$, concentrated on the disjoint sum 
\[
\{t=0\} \times SU(2, \mathbb{C}) \sqcup \{t=0\} \times SU(2, \mathbb{C}) \subset \widetilde{\mathbb{S}^1}\times SU(2, \mathbb{C})
\]
of the two Cauchy surfaces $t=0$ and $t=2\pi$. This functional vanishes on the image
$\textrm{Im} \, \Omega'$. Therefore, the most general retarded part of $\kappa_{q'}$ is of the form
\[
\textrm{ret} \, \kappa_{q'} = \theta \kappa_{q'} \circ \Omega' + \delta^{\omega}_{{}_{1;2}}.
\]

We thus again arrive at the conclusion that the retarded part of the contraction kernel (\ref{ProductFormOfContraction(x,y)})
is determined up to the corresponding additive reminder kernel
\begin{multline*}
\delta^{\omega}_{{}{1;2}}(a, \ldots, b', c, \ldots, d', x, y) 
\,\, \times 
\\
\times \,\, 
\big(\kappa^{(q'+1)}_{0,1} \dot{\otimes} \ldots  \kappa^{(q)}_{0,1} \otimes
\kappa^{(q'+1)}_{1,0} \dot{\otimes} \ldots  \kappa^{(q)}_{1,0} \big) \big(\, (\ldots) \, , \, (\ldots)' \, , \, b'', \ldots, b, d'', \ldots, d, x, y\big),
\end{multline*}
where
\begin{multline*}
\delta^{\omega}_{{}_{1;2}}(a, \ldots, b', c, \ldots, d',x,y) \overset{\textrm{df}}{=}
\sum\limits_{\alpha = 0}^{\omega} C_{\alpha}^{I\, a, \ldots, b', c, \ldots, d'}\big(\boldsymbol{w}\boldsymbol{v}^{-1}\big) \delta^{(\alpha)}(t-\tau) 
\\
+ \sum\limits_{\beta = 0}^{\omega} C_{\beta}^{II \, a, \ldots, b', c, \ldots, d'}\big(\boldsymbol{w}\boldsymbol{v}^{-1}\big) \delta^{(\beta)}(t-\tau - 2\pi),
\end{multline*}
with $C_{\alpha}^{I}, C_{\beta}^{II}$ being now distributions on the two Cauchy surfaces $t=0$ and $t=2\pi$ regarded as functionals on the test 
spaces of smooth multi-component functions on these Cauchy surfaces.

Similarly as on the Minkowski space-time, we introduce the \emph{double limit} contraction 
$\otimes||_{{}_{q'}}$
\[
\bigg[\Big(\big(\theta_y \kappa^{(1)}_{0,1} \dot{\otimes} \ldots  \kappa^{(q)}_{0,1} \big) 
\otimes||_{{}_{q'}} \big(\kappa^{(1)}_{1,0} \dot{\otimes} \ldots \dot{\otimes} \kappa^{(q)}_{1,0} \big) \Big) \bigg]
\big(\, (\ldots)\, , \, (\ldots)' \, ; \, a, \ldots, b, c, \ldots, d,x,y\big) 
\overset{\textrm{df}}{=}
\]
\begin{multline*}
=
\bigg[\Big(\theta.\big(\kappa^{(1)}_{0,1} \dot{\otimes} \ldots  \kappa^{(q')}_{0,1} \big) 
\otimes_{q'} \big(\kappa^{(1)}_{1,0} \dot{\otimes} \ldots \dot{\otimes} \kappa^{(q')}_{1,0} \big)\Big) \circ \Omega \bigg](a, \ldots, b', c, \ldots, d', x, y) 
\,\, \times 
\\
\times \,\, 
\big(\kappa^{(q'+1)}_{0,1} \dot{\otimes} \ldots  \kappa^{(q)}_{0,1} \otimes
\kappa^{(q'+1)}_{1,0} \dot{\otimes} \ldots  \kappa^{(q)}_{1,0} \big) \big(\, (\ldots) \, , \, (\ldots)' \, , \, b'', \ldots, b, d'', \ldots, d, x, y\big)
\\
+
\delta^{\omega}_{{}{1;2}}(a, \ldots, b', c, \ldots, d', x, y) 
\,\, \times 
\\
\times \,\, 
\big(\kappa^{(q'+1)}_{0,1} \dot{\otimes} \ldots  \kappa^{(q)}_{0,1} \otimes
\kappa^{(q'+1)}_{1,0} \dot{\otimes} \ldots  \kappa^{(q)}_{1,0} \big) \big(\, (\ldots) \, , \, (\ldots)' \, , \, b'', \ldots, b, d'', \ldots, d, x, y\big).
\end{multline*}
\begin{multline*}
=
\bigg[\Big(\big(\theta_y\kappa^{(1)}_{0,1} \dot{\otimes} \ldots  \kappa^{(q')}_{0,1} \big) 
\otimes_{q'} \big(\kappa^{(1)}_{1,0} \dot{\otimes} \ldots \dot{\otimes} \kappa^{(q')}_{1,0} \big)\Big) \circ \Omega \bigg](a, \ldots, b', c, \ldots, d', x, y) 
\,\, \times 
\\
\times \,\, 
\big(\kappa^{(q'+1)}_{0,1} \dot{\otimes} \ldots  \kappa^{(q)}_{0,1} \otimes
\kappa^{(q'+1)}_{1,0} \dot{\otimes} \ldots  \kappa^{(q)}_{1,0} \big) \big(\, (\ldots) \, , \, (\ldots)' \, , \, b'', \ldots, b, d'', \ldots, d, x, y\big)
\\
+
\delta^{\omega}_{{}{1;2}}(a, \ldots, b', c, \ldots, d', x, y) 
\,\, \times 
\\
\times \,\, 
\big(\kappa^{(q'+1)}_{0,1} \dot{\otimes} \ldots  \kappa^{(q)}_{0,1} \otimes
\kappa^{(q'+1)}_{1,0} \dot{\otimes} \ldots  \kappa^{(q)}_{1,0} \big) \big(\, (\ldots) \, , \, (\ldots)' \, , \, b'', \ldots, b, d'', \ldots, d, x, y\big).
\end{multline*}
In short notation
\begin{multline}\label{ProductFormOfContraction(x,y)}
\Big(\big(\theta_y \kappa^{(1)}_{0,1} \dot{\otimes} \ldots  \kappa^{(q)}_{0,1} \big) 
\otimes||_{{}_{q'}} \big(\kappa^{(1)}_{1,0} \dot{\otimes} \ldots \dot{\otimes} \kappa^{(q)}_{1,0} \big) \Big)
\\
=
\Big(\big(\theta_y\kappa^{(1)}_{0,1} \dot{\otimes} \ldots  \kappa^{(q')}_{0,1} \big) 
\otimes_{q'} \big(\kappa^{(1)}_{1,0} \dot{\otimes} \ldots \dot{\otimes} \kappa^{(q')}_{1,0} \big)\Big) \circ \Omega 
\,\, \times 
\\
\times \,\, 
\big(\kappa^{(q'+1)}_{0,1} \dot{\otimes} \ldots  \kappa^{(q)}_{0,1} \otimes
\kappa^{(q'+1)}_{1,0} \dot{\otimes} \ldots  \kappa^{(q)}_{1,0} \big) 
\\
+
\delta^{\omega}_{{}{1;2}} \,\, . \,\,
\big(\kappa^{(q'+1)}_{0,1} \dot{\otimes} \ldots  \kappa^{(q)}_{0,1} \otimes
\kappa^{(q'+1)}_{1,0} \dot{\otimes} \ldots  \kappa^{(q)}_{1,0} \big).
\end{multline}

We can summarize our analysis of the pointwise multiplication by the step theta function of a contraction
kernel involving dot products of arbitrary number of massless or infinite orbit plane wave kernels 
in the following 
\begin{lem}
Let $\kappa^{(1)}_{\ell, m}, \dots, \kappa^{(q)}_{\ell, m}   \in \mathfrak{K}_0$, $\ell,m = 1,0$, be the kernels 
of $q$ free fields $\mathbb{A}^{(1)}, \ldots, \mathbb{A}^{(q)}$, respectively, 
on the Einstein Universe (all positive or all negative-energy fields)
with the corresponding single particle Gelfand triples 
\[
\begin{array}{ccccc} 
 E_{{}_{(1)}} & \subset & \mathcal{H}_{{}_{(1)}} & \subset & E_{{}_{(1)}}^*
\\\vdots &&\vdots&&\vdots  \\  
E_{{}_{(q)}} & \subset & \mathcal{H}_{{}_{(q)}} & \subset & E_{{}_{(q)}}^*.
\end{array} 
\]
Let $\mathscr{E}=\mathcal{S}_{{}_{\Delta +1}}(\widetilde{\mathbb{S}^1} \times SU(2, \mathbb{C}); \mathbb{C})$
or respectively  $\mathscr{E}=\mathcal{S}_{{}_{\Delta +1}}(\widetilde{\mathbb{S}^1} \times SU(2, \mathbb{C}); \mathbb{C}^d)$.

Let the common space-time variable
of the kernels 
\[
\kappa^{(1)}_{0,1}, \ldots, \kappa^{(q)}_{0, 1}
\]
be denoted $x$ and of the kernels 
\[
\kappa^{(1)}_{1,0}, \ldots, \kappa^{(q)}_{1,0}
\]
be denoted $y$, and let 
\begin{multline*}
\theta.\big(\kappa^{(q)}_{0,1} \overset{\cdot}{\otimes} \cdots \overset{\cdot}{\otimes} \kappa^{(q)}_{0,1}\big) \otimes_{q'}
\big(\kappa^{(1)}_{1,0}\overset{\cdot}{\otimes} \cdots \overset{\cdot}{\otimes} \kappa^{(q)}_{1,0}\big)(x,y) 
\\
\overset{\textrm{df}}{=}
\theta\big(xy^{-1}\big)\big(\kappa^{(1)}_{0,1} \overset{\cdot}{\otimes} \cdots \overset{\cdot}{\otimes} \kappa^{(q)}_{0,1}\big) \otimes_{q'}
\big(\kappa^{(1)}_{1,0}\overset{\cdot}{\otimes} \cdots \overset{\cdot}{\otimes} \kappa^{(q)}_{1,0}\big)(x,y)
\\
= 
\big(\theta_y\kappa^{(1)}_{0,1} \overset{\cdot}{\otimes} \cdots \overset{\cdot}{\otimes} \kappa^{(q)}_{0,1}\big) \otimes_{q'}
\big(\kappa^{(q)}_{1,0}\overset{\cdot}{\otimes} \cdots \overset{\cdot}{\otimes} \kappa^{(q)}_{1,0}\big)(x,y) 
\end{multline*}
where $\theta_y(x) = \theta\big(xy^{-1}\big)$. Let as assume that these are the first $q'$ momentum variables
of the kernel 
\[
\kappa^{(1)}_{0,1} \overset{\cdot}{\otimes} \cdots \overset{\cdot}{\otimes} \kappa^{(q)}_{0,1}
\]
which are contracted with the the first momentum variables of the kernel
\[
\kappa^{(1)}_{1,0}\overset{\cdot}{\otimes} \cdots \overset{\cdot}{\otimes} \kappa^{(q)}_{1,0}.
\]

Let $\theta_\varepsilon$ converges to $\theta$ when $\varepsilon \rightarrow 0$ in the sense of Definition 
\ref{ConvergenceOfthetavarepsilon} with some integer $\omega$ in Definition 
\ref{ConvergenceOfthetavarepsilon}.

Then there exists a finite integer $\omega$ in the convergence condition of Definition 
\ref{ConvergenceOfthetavarepsilon} for the convergence $\theta_\varepsilon \rightarrow \theta$,
depending only on the contracted plane wave kernels
\[
\kappa^{(1)}_{0,1}, \kappa^{(1)}_{0, 1}, \ldots, \kappa^{(q')}_{0,1}, \kappa^{(q')}_{1,0}, 
\]
such that:
\begin{enumerate}
\item[1)]
\begin{multline*}
\theta_{\varepsilon}.\big(\kappa^{(1)}_{0,1} \overset{\cdot}{\otimes} \cdots \overset{\cdot}{\otimes} \kappa^{(q)}_{0,1}\big) \otimes_{q'}
\big(\kappa^{(1)}_{1,0}\overset{\cdot}{\otimes} \cdots \overset{\cdot}{\otimes} \kappa^{(q)}_{1,0}\big) \circ \Omega
\\
\overset{\varepsilon \rightarrow 0}{\longrightarrow}
\theta.\big(\kappa^{(1)}_{0,1} \overset{\cdot}{\otimes} \cdots \overset{\cdot}{\otimes} \kappa^{(q)}_{0,1}\big) \otimes_{q'}
\big(\kappa^{(1)}_{1,0}\overset{\cdot}{\otimes} \cdots \overset{\cdot}{\otimes} \kappa^{(q)}_{1,0}\big) \circ \Omega
\\ 
\,\,\, \textrm{in} \,\,\,  \mathscr{L}(\mathscr{E}^{\otimes \, 2}, 
E_{{}_{(q'+1)}}^{*} \otimes \cdots \otimes E_{{}_{q}}^{*} \otimes E_{{}_{q'+1}}^{*}
\otimes \cdots \otimes E_{{}_{(q)}}^{*}) 
\\
\cong
\mathscr{L}(E_{{}_{(q'+1)}} \otimes \cdots \otimes E_{{}_{(q)}} \otimes E_{{}_{q'+1}} \otimes \cdots
\otimes E_{{}_{(q)}}, \mathscr{E}^{* \, \otimes \, 2}).
\end{multline*}
\item[2)]
The most general retarded and, respectively, advanced part of 
\[
\big(\kappa^{(1)}_{0,1} \overset{\cdot}{\otimes} \cdots \overset{\cdot}{\otimes} \kappa^{(q)}_{0,1}\big) \otimes_{q'}
\big(\kappa^{(1)}_{1,0}\overset{\cdot}{\otimes} \cdots \overset{\cdot}{\otimes} \kappa^{(q)}_{1,0}\big)
\]
has the form
\begin{multline*}
\big(\theta_y \kappa^{(1)}_{0,1} \dot{\otimes} \ldots  \kappa^{(q)}_{0,1} \big) 
\otimes||_{{}_{q'}} \big(\kappa^{(1)}_{1,0} \dot{\otimes} \ldots \dot{\otimes} \kappa^{(q)}_{1,0} \big) 
\\
=
\Big(\big(\theta_y\kappa^{(1)}_{0,1} \dot{\otimes} \ldots  \kappa^{(q)}_{0,1} \big) 
\otimes_{q'} \big(\kappa^{(1)}_{1,0} \dot{\otimes} \ldots \dot{\otimes} \kappa^{(q)}_{1,0} \big)\Big) \circ \Omega 
\\
+
\delta^{\omega}_{{}{1;2}} \,\, . \,\,
\big(\kappa^{(q'+1)}_{0,1} \dot{\otimes} \ldots  \kappa^{(q)}_{0,1} \otimes
\kappa^{(q'+1)}_{1,0} \dot{\otimes} \ldots  \kappa^{(q)}_{1,0} \big)
\\
=
\Big(\big(\theta_y\kappa^{(1)}_{0,1} \dot{\otimes} \ldots  \kappa^{(q')}_{0,1} \big) 
\otimes_{q'} \big(\kappa^{(1)}_{1,0} \dot{\otimes} \ldots \dot{\otimes} \kappa^{(q')}_{1,0} \big)\Big) \circ \Omega 
\,\, \times 
\\
\times \,\, 
\big(\kappa^{(q'+1)}_{0,1} \dot{\otimes} \ldots  \kappa^{(q)}_{0,1} \otimes
\kappa^{(q'+1)}_{1,0} \dot{\otimes} \ldots  \kappa^{(q)}_{1,0} \big) 
\\
+
\delta^{\omega}_{{}{1;2}} \,\, . \,\,
\big(\kappa^{(q'+1)}_{0,1} \dot{\otimes} \ldots  \kappa^{(q)}_{0,1} \otimes
\kappa^{(q'+1)}_{1,0} \dot{\otimes} \ldots  \kappa^{(q)}_{1,0} \big)
\end{multline*}
and
\begin{multline*}
\big(-\check{\theta}_y \kappa^{(1)}_{0,1} \dot{\otimes} \ldots  \kappa^{(q)}_{0,1} \big) 
\otimes||_{{}_{q'}} \big(\kappa^{(1)}_{1,0} \dot{\otimes} \ldots \dot{\otimes} \kappa^{(q)}_{1,0} \big) 
\\
=
\Big(\big(-\check{\theta}_y\kappa^{(1)}_{0,1} \dot{\otimes} \ldots  \kappa^{(q)}_{0,1} \big) 
\otimes_{q'} \big(\kappa^{(1)}_{1,0} \dot{\otimes} \ldots \dot{\otimes} \kappa^{(q)}_{1,0} \big)\Big) \circ \Omega 
\\
+
\delta^{\omega}_{{}{1;2}} \,\, . \,\,
\big(\kappa^{(q'+1)}_{0,1} \dot{\otimes} \ldots  \kappa^{(q)}_{0,1} \otimes
\kappa^{(q'+1)}_{1,0} \dot{\otimes} \ldots  \kappa^{(q)}_{1,0} \big)
\\
=
\Big(\big(-\check{\theta}_y\kappa^{(1)}_{0,1} \dot{\otimes} \ldots  \kappa^{(q')}_{0,1} \big) 
\otimes_{q'} \big(\kappa^{(1)}_{1,0} \dot{\otimes} \ldots \dot{\otimes} \kappa^{(q')}_{1,0} \big)\Big) \circ \Omega 
\,\, \times 
\\
\times \,\, 
\big(\kappa^{(q'+1)}_{0,1} \dot{\otimes} \ldots  \kappa^{(q)}_{0,1} \otimes
\kappa^{(q'+1)}_{1,0} \dot{\otimes} \ldots  \kappa^{(q)}_{1,0} \big) 
\\
+
\delta^{\omega}_{{}{1;2}} \,\, . \,\,
\big(\kappa^{(q'+1)}_{0,1} \dot{\otimes} \ldots  \kappa^{(q)}_{0,1} \otimes
\kappa^{(q'+1)}_{1,0} \dot{\otimes} \ldots  \kappa^{(q)}_{1,0} \big).
\end{multline*}
Here
\begin{multline*}
\Omega \chi (x,y) = \chi(x,y) 
- \, \sum \limits_{\beta=0}^{\omega} \omega_{{}_{o \,\,I \,\, \beta}} \big(xy^{-1}\big)
 \,\,\, \partial_{{}_{t}}^{\beta}\chi (t=\tau , \boldsymbol{w},y)
\\
- \, \sum \limits_{\beta=0}^{\omega} \omega_{{}_{o \,\, II \,\, \beta}} \big(xy^{-1}\big)
 \,\,\, \partial_{{}_{t}}^{\beta}\chi (t=\tau + 2\pi, \boldsymbol{w},y). 
\end{multline*}
with the set $\{ \omega_{{}_{o \,\, I \,\, \alpha}} , \omega_{{}_{o \,\, II \,\, \beta}}\}$  of functions defined 
as above; and where
\[
\delta^{\omega}_{{}_{1;2}}(x,y) \overset{\textrm{df}}{=}
\sum\limits_{\alpha = 0}^{\omega} C_{\alpha}^{I}\big(\boldsymbol{w}\boldsymbol{v}^{-1}\big) \delta^{(\alpha)}(t-\tau) 
+ \sum\limits_{\beta = 0}^{\omega} C_{\beta}^{II}\big(\boldsymbol{w}\boldsymbol{v}^{-1}\big) \delta^{(\beta)}(t-\tau - 2\pi),
\]
\[
x = (t, \boldsymbol{w}), \, y = (\tau, \boldsymbol{v}) \in \widetilde{\mathbb{S}^1}\times SU(2, \mathbb{C}),
\] 
with arbitrary distributions $C^{I}_\alpha, C^{II}_{\beta}$ on the two Cauchy surfaces
$t=0$ and $t=2\pi$, common for the retarded and advanced part.
\item[3)] 
The above statements, 1) and 2), remain valid if we replace the kernels $\kappa^{(1)}_{1,0}, \ldots$ with their derivations
$(X)^\alpha \kappa_{1,0}, \ldots$.
\end{enumerate}
\label{q-contractionkappaBarDotOtimeskappaWithManyMasslessOnEU}
\end{lem}

We have also proved the following
\begin{lem}
Let $\kappa^{(1)}_{\ell, m}, \dots, \kappa^{(q)}_{\ell, m}   \in \mathfrak{K}_0$, $\ell,m = 1,0$, be the kernels 
of $q$ free fields $\mathbb{A}^{(1)}, \ldots, \mathbb{A}^{(q)}$, respectively, 
on the Einstein Universe (all positive or all negative-energy fields)
with the corresponding single particle Gelfand triples 
\[
\begin{array}{ccccc} 
 E_{{}_{(1)}} & \subset & \mathcal{H}_{{}_{(1)}} & \subset & E_{{}_{(1)}}^*
\\\vdots &&\vdots&&\vdots  \\  
E_{{}_{(q)}} & \subset & \mathcal{H}_{{}_{(q)}} & \subset & E_{{}_{(q)}}^*.
\end{array} 
\]
Let $\mathscr{E}=\mathcal{S}_{{}_{\Delta +1}}(\widetilde{\mathbb{S}^1} \times SU(2, \mathbb{C}); \mathbb{C})$
or respectively  $\mathscr{E}=\mathcal{S}_{{}_{\Delta +1}}(\widetilde{\mathbb{S}^1} \times SU(2, \mathbb{C}); \mathbb{C}^d)$.

Let the common space-time variable
of the kernels 
\[
\kappa^{(1)}_{0,1}, \ldots, \kappa^{(q)}_{0, 1}
\]
be denoted $x$ and of the kernels 
\[
\kappa^{(1)}_{1,0}, \ldots, \kappa^{(q)}_{1,0}
\]
be denoted $y$, and let 
\begin{multline*}
\theta.\big(\kappa^{(q)}_{0,1} \overset{\cdot}{\otimes} \cdots \overset{\cdot}{\otimes} \kappa^{(q)}_{0,1}\big) \otimes_{q'}
\big(\kappa^{(1)}_{1,0}\overset{\cdot}{\otimes} \cdots \overset{\cdot}{\otimes} \kappa^{(q)}_{1,0}\big)(x,y) 
\\
\overset{\textrm{df}}{=}
\theta\big(xy^{-1}\big)\big(\kappa^{(1)}_{0,1} \overset{\cdot}{\otimes} \cdots \overset{\cdot}{\otimes} \kappa^{(q)}_{0,1}\big) \otimes_{q'}
\big(\kappa^{(1)}_{1,0}\overset{\cdot}{\otimes} \cdots \overset{\cdot}{\otimes} \kappa^{(q)}_{1,0}\big)(x,y)
\\
= 
\big(\theta_y\kappa^{(1)}_{0,1} \overset{\cdot}{\otimes} \cdots \overset{\cdot}{\otimes} \kappa^{(q)}_{0,1}\big) \otimes_{q'}
\big(\kappa^{(q)}_{1,0}\overset{\cdot}{\otimes} \cdots \overset{\cdot}{\otimes} \kappa^{(q)}_{1,0}\big)(x,y) 
\end{multline*}
where $\theta_y(x) = \theta\big(xy^{-1}\big)$. Let us assume that these are the first $q'$ momentum variables
of the kernel 
\[
\kappa^{(1)}_{0,1} \overset{\cdot}{\otimes} \cdots \overset{\cdot}{\otimes} \kappa^{(q)}_{0,1}
\]
which are contracted with the the first momentum variables of the kernel
\[
\kappa^{(1)}_{1,0}\overset{\cdot}{\otimes} \cdots \overset{\cdot}{\otimes} \kappa^{(q)}_{1,0}.
\]

Let $0 \leq q' <2$.

Let $\theta_\varepsilon$ converges to $\theta$ when $\varepsilon \rightarrow 0$ in the sense of Definition 
\ref{ConvergenceOfthetavarepsilon} with  $\omega =0$ in Definition 
\ref{ConvergenceOfthetavarepsilon}.

Then 
\begin{enumerate}
\item[1)]
\begin{multline*}
\theta_{\varepsilon}.\big(\kappa^{(1)}_{0,1} \overset{\cdot}{\otimes} \cdots \overset{\cdot}{\otimes} \kappa^{(q)}_{0,1}\big) \otimes_{q'}
\big(\kappa^{(1)}_{1,0}\overset{\cdot}{\otimes} \cdots \overset{\cdot}{\otimes} \kappa^{(q)}_{1,0}\big) 
\\
\overset{\varepsilon \rightarrow 0}{\longrightarrow}
\theta.\big(\kappa^{(1)}_{0,1} \overset{\cdot}{\otimes} \cdots \overset{\cdot}{\otimes} \kappa^{(q)}_{0,1}\big) \otimes_{q'}
\big(\kappa^{(1)}_{1,0}\overset{\cdot}{\otimes} \cdots \overset{\cdot}{\otimes} \kappa^{(q)}_{1,0}\big) 
\\ 
\,\,\, \textrm{in} \,\,\,  \mathscr{L}(\mathscr{E}^{\otimes \, 2}, 
E_{{}_{(q'+1)}}^{*} \otimes \cdots \otimes E_{{}_{q}}^{*} \otimes E_{{}_{q'+1}}^{*}
\otimes \cdots \otimes E_{{}_{(q)}}^{*}) 
\\
\cong
\mathscr{L}(E_{{}_{(q'+1)}} \otimes \cdots \otimes E_{{}_{(q)}} \otimes E_{{}_{q'+1}} \otimes \cdots
\otimes E_{{}_{(q)}}, \mathscr{E}^{* \, \otimes \, 2}), \,\,\,\,\,\, 0 \leq q' <2.
\end{multline*}
\item[2)]
The most general retarded and, respectively, advanced part of 
\[
\big(\kappa^{(1)}_{0,1} \overset{\cdot}{\otimes} \cdots \overset{\cdot}{\otimes} \kappa^{(q)}_{0,1}\big) \otimes_{q'}
\big(\kappa^{(1)}_{1,0}\overset{\cdot}{\otimes} \cdots \overset{\cdot}{\otimes} \kappa^{(q)}_{1,0}\big), \,\,\, 0 \leq q' <2
\]
has the form
\begin{multline*}
\big(\theta_y \kappa^{(1)}_{0,1} \overset{\cdot}{\otimes} \cdots \overset{\cdot}{\otimes} \kappa^{(q)}_{0,1}\big) \otimes_{q'}
\big(\kappa^{(1)}_{1,0}\overset{\cdot}{\otimes} \cdots \overset{\cdot}{\otimes} \kappa^{(q)}_{1,0}\big)
\\
=
\big(\theta_y\kappa^{(1)}_{0,1} \dot{\otimes} \ldots  \kappa^{(q')}_{0,1} \big) 
\otimes_{q'} \big(\kappa^{(1)}_{1,0} \dot{\otimes} \ldots \dot{\otimes} \kappa^{(q')}_{1,0} \big)\Big) 
\,\, \times 
\\
\times \,\, 
\big(\kappa^{(q'+1)}_{0,1} \dot{\otimes} \ldots  \kappa^{(q)}_{0,1} \otimes
\kappa^{(q'+1)}_{1,0} \dot{\otimes} \ldots  \kappa^{(q)}_{1,0} \big) 
\end{multline*}
and
\begin{multline*}
\big(-\check{\theta}_y \kappa^{(1)}_{0,1} \overset{\cdot}{\otimes} \cdots \overset{\cdot}{\otimes} \kappa^{(q)}_{0,1}\big) \otimes_{q'}
\big(\kappa^{(1)}_{1,0}\overset{\cdot}{\otimes} \cdots \overset{\cdot}{\otimes} \kappa^{(q)}_{1,0}\big)
\\
=
\big(-\check{\theta}_y\kappa^{(1)}_{0,1} \dot{\otimes} \ldots  \kappa^{(q')}_{0,1} \big) 
\otimes_{q'} \big(\kappa^{(1)}_{1,0} \dot{\otimes} \ldots \dot{\otimes} \kappa^{(q')}_{1,0} \big)
\,\, \times 
\\
\times \,\, 
\big(\kappa^{(q'+1)}_{0,1} \dot{\otimes} \ldots  \kappa^{(q)}_{0,1} \otimes
\kappa^{(q'+1)}_{1,0} \dot{\otimes} \ldots  \kappa^{(q)}_{1,0} \big).
\end{multline*}
\item[3)] 
The above statements, 1) and 2), remain valid if we replace the kernels $\kappa^{(1)}_{1,0}, \ldots$ with their derivations
$(X)^\alpha \kappa_{1,0}, \ldots$.
\end{enumerate}
\label{1-contractionkappaBarDotOtimeskappaWithManyMasslessOnEU}
\end{lem}
\qedsymbol \, 
Immediate consequence of the inequality (\ref{|<thetaykappa(1-contraction)kappa(chi),xi>|<C|chi|omega|xi|}).
\qed

\begin{defin}
The minimal value of the integer $\omega$ such that the assertion of Lemma \ref{q-contractionkappaBarDotOtimeskappaWithManyMasslessOnEU}
is true, we call \emph{singularity degree} of the vector valued kernel
\[
\big(\kappa^{(1)}_{0,1} \overset{\cdot}{\otimes} \cdots \overset{\cdot}{\otimes} \kappa^{(q)}_{0,1}\big) \otimes_{q'}
\big(\kappa^{(1)}_{1,0}\overset{\cdot}{\otimes} \cdots \overset{\cdot}{\otimes} \kappa^{(q)}_{1,0}\big),
\]
equal to the singularity degree of the scalar factor (which captures all contractions) of this kernel, coming from its canonical
factorization. 
\label{SingularityDegree}
\end{defin}

Concerning the analogue of Lemma \ref{S*Xi} of Subsection \ref{OperationsOnXi} on the Einstein Universe, we recall
that the convolution of a distribution with a smooth function of compact support on the Abelian additive Lie group $\mathbb{R}^n$ is again a smooth
function on $\mathbb{R}^n$, compare e.g. \cite{Rudin}, Thm. 6.30 (b).
This suggests that the convolution of a functional $F \in \mathscr{E}^* = \mathcal{S}_A(\widetilde{\mathbb{S}^1} \times G; \mathbb{C})^*$
with any 
\[
\phi \in \mathscr{E} = \mathcal{S}_A(\widetilde{\mathbb{S}^1} \times G; \mathbb{C}) = \mathscr{C}^\infty(\widetilde{\mathbb{S}^1} \times G; \mathbb{C}),
\,\,\,\,\, G = SU(2, \mathbb{C})
\]
should define a continuous map $\mathscr{E} \rightarrow \mathscr{E}$ in case of our compact Lie group $\widetilde{\mathbb{S}^1} \times G$.

Indeed, let $F, \phi$ be functions on $\widetilde{\mathbb{S}^1} \times G$, $G = SU(2, \mathbb{C})$, let $A= \Delta + 1$
be the standard operator on $L^2(\widetilde{\mathbb{S}^1} \times G; \mathbb{C})$  defining 
$\mathscr{E} = \mathcal{S}_A(\widetilde{\mathbb{S}^1} \times G; \mathbb{C})$.
Let $\phi \in \mathscr{E}$. Finally let $F \in L^2(\widetilde{\mathbb{S}^1} \times G; \mathbb{C})$ or more generally
let $F \in \mathscr{E}^* = \mathcal{S}_A(\widetilde{\mathbb{S}^1} \times G; \mathbb{C})^*$, \emph{i.e.} for some $p \in \mathbb{N}$, 
$|A^{-p}F| = |F|_{-p} < \infty$, where $| \cdot |$ is the $L^2$-norm on $L^2(\widetilde{\mathbb{S}^1} \times G; \mathbb{C})$.
It is easily seen that the Fourier transform $\widetilde{F \ast \phi}$ of the convolution
\[
F \ast \phi(x) = \int \limits_{\widetilde{\mathbb{S}^1} \times G} 
F(y)\phi(y^{-1}x) dy,
\,\,\,
x=(t, \boldsymbol{w}), y = (\tau, \boldsymbol{u}) \in \widetilde{\mathbb{S}^1} \times G,
\]
is equal
\begin{multline*}
\widetilde{F \ast \phi}_{{}_{ji}}(\widehat{n}\cdot \widehat{l}) = 
\Big(\widetilde{\phi}(\widehat{n}\cdot \widehat{l}) \widetilde{F}(\widehat{n}\cdot \widehat{l}) \Big)_{{}_{ji}}
= \sum \limits_{-l \leq m \leq l} \widetilde{\phi}_{{}_{jm}}(\widehat{n}\cdot \widehat{l}) \, \widetilde{F}_{{}_{mi}}(\widehat{n}\cdot \widehat{l}).
\end{multline*}
We have the following norm estimation
\begin{multline*}
\big| F \ast \phi \big|_{q}^{2} = \big|A^q F \ast \phi \big|^{2} = \big|\widetilde{A^q F \ast \phi} \big|^{2}
= \big|\widetilde{A^q}  \widetilde{F \ast \phi} \big|^{2}
\\
\leq 
\Bigg[ 
\sum \limits_{\substack{n\in \mathbb{Z} \\ l \in \mathbb{N}+ {\textstyle\frac{1}{2}} \mathbb{N}  \\-l \leq m,i,j \leq l}} 
\big|
\lambda_{{}_{n,l}}^{q}
\widetilde{\phi}_{{}_{jm}}(\widehat{n}\cdot \widehat{l}) \, \widetilde{F}_{{}_{mi}}(\widehat{n}\cdot \widehat{l})
\big|
\Bigg]^2
\end{multline*}
\begin{multline*}
= 
\Bigg[ 
\sum \limits_{\substack{n\in \mathbb{Z} \\ l \in \mathbb{N}+ {\textstyle\frac{1}{2}} \mathbb{N}  \\-l \leq m,i,j \leq l}} 
\big|
\lambda_{{}_{n,l}}^{q+p}
\widetilde{\phi}_{{}_{jm}}(\widehat{n}\cdot \widehat{l}) \, 
\lambda_{{}_{n,l}}^{-p}
\widetilde{F}_{{}_{mi}}(\widehat{n}\cdot \widehat{l})
\big|
\Bigg]^2
\\
=
\Bigg[ 
\sum \limits_{\substack{n\in \mathbb{Z} \\ l \in \mathbb{N}+ {\textstyle\frac{1}{2}} \mathbb{N}  \\-l \leq m,i,j \leq l}} 
\big|
(2l+1)(2l+1)
\lambda_{{}_{n,l}}^{q+p}
\widetilde{\phi}_{{}_{jm}}(\widehat{n}\cdot \widehat{l}) \, 
{\textstyle\frac{\lambda_{{}_{n,l}}^{-p}}{(2l+1)^2}}
\widetilde{F}_{{}_{mi}}(\widehat{n}\cdot \widehat{l})
\big|
\Bigg]^2
\end{multline*}
\begin{multline*}
= 
\Bigg[ 
\sum \limits_{\substack{n\in \mathbb{Z} \\ l \in \mathbb{N}+ {\textstyle\frac{1}{2}} \mathbb{N}  \\-l \leq m,i,j \leq l}} 
\Big|
(2l+1)(2l+1)
\lambda_{{}_{n,l}}^{q+p}
\widetilde{\phi}_{{}_{jm}}(\widehat{n}\cdot \widehat{l}) \, 
\Big[
\sum \limits_{-l \leq i \leq l}
{\textstyle\frac{\lambda_{{}_{n,l}}^{-p}}{(2l+1)^2}}
\widetilde{F}_{{}_{mi}}(\widehat{n}\cdot \widehat{l})
\Big]
\Big|
\Bigg]^2
\\
\leq 
\Bigg[ 
\sum \limits_{\substack{n\in \mathbb{Z} \\ l \in \mathbb{N}+ {\textstyle\frac{1}{2}} \mathbb{N}  \\-l \leq m,i,j \leq l}} 
(2l+1)
\Big|
\lambda_{{}_{n,l}}^{q+p+2}
\widetilde{\phi}_{{}_{jm}}(\widehat{n}\cdot \widehat{l}) \, 
\overline{
\Big[
\sum \limits_{-l \leq i \leq l}
{\textstyle\frac{\lambda_{{}_{n,l}}^{-p}}{(2l+1)^2}}
\overline{\widetilde{F}_{{}_{mi}}(\widehat{n}\cdot \widehat{l})}
\Big]
}
\Big|
\Bigg]^2
\end{multline*}
\begin{equation}\label{|F*phi|_q <|phi|_q+p+2}
= 
\Big|
\Big\langle \widetilde{A^{-p} G} , \widetilde{A^{q+p+2}\phi} \Big\rangle 
\Big|^2
\leq 
|G|_{-p}^{2} \, |\phi|_{q+p+2}^{2}.
\end{equation}
Here
\[
\widetilde{G}_{{}_{jm}}(\widehat{n}\cdot \widehat{l}) \overset{\textrm{df}}{=}
\sum \limits_{-l \leq i \leq l}
{\textstyle\frac{1}{(2l+1)^2}}
\overline{\widetilde{F}_{{}_{mi}}(\widehat{n}\cdot \widehat{l})},
\,\,\, j = -l, -l+1, \ldots, l.
\]
It is easily seen that
\begin{multline*}
(2l+1) \sum \limits_{-l \leq m \leq l} \sum \limits_{-l \leq j \leq l}
\Big| \widetilde{G}_{{}_{jm}}(\widehat{n}\cdot \widehat{l}) \Big|^2
\leq 
\sum \limits_{-l \leq m \leq l} \, 
\Bigg[ 
\sum \limits_{-l \leq i \leq l}
\Big| \widetilde{F}_{{}_{mi}}(\widehat{n}\cdot \widehat{l})\Big|
\Bigg]^2
\\
\leq
(2l+1) \sum \limits_{-l \leq m \leq l} \sum \limits_{-l \leq j \leq l}
\Big| \widetilde{F}_{{}_{mi}}(\widehat{n}\cdot \widehat{l}) \Big|^2, 
\end{multline*}
so that 
\[
(2l+1) \lambda_{{}_{n,l}}^{-2p} \sum \limits_{-l \leq m \leq l} \sum \limits_{-l \leq j \leq l}
\Big| \widetilde{G}_{{}_{jm}}(\widehat{n}\cdot \widehat{l}) \Big|^2
\leq
(2l+1) \lambda_{{}_{n,l}}^{-2p} \sum \limits_{-l \leq m \leq l} \sum \limits_{-l \leq j \leq l}
\Big| \widetilde{F}_{{}_{mi}}(\widehat{n}\cdot \widehat{l}) \Big|^2, 
\]
and thus
\[
|G|_{-p}^{2} = |A^{-p}G|^2 \leq  |A^{-p}F|^2 = |F|_{-p}^{2} < \infty.
\]

From the inequlity (\ref{|F*phi|_q <|phi|_q+p+2})  it follows that $F \ast \phi \in \mathscr{E}$ as well as continuity of the map
\[
\mathscr{E} \ni \phi \longrightarrow F \ast \phi \in \mathscr{E}.
\]
Because for any $F \in \mathscr{E}^*$ the convolution $F \ast \phi$ defines a continuous map
\[
\mathscr{E} \ni \phi \longrightarrow F \ast \phi \in \mathscr{E},
\]
then the analogue of Lemma  \ref{S*Xi} of Subsection \ref{OperationsOnXi} and of its strengthened form
follow immediately on the Einstein Universe. Namely we have the following
\begin{lem}
Let $\kappa_{\ell, m}, \dots, \kappa'_{\ell, m}, \kappa''_{\ell, m}, \ldots, \kappa'''_{\ell, m}   \in \mathfrak{K}_0$, $\ell,m = 1,0$, be the kernels 
of free fields on the Einstein Universe (all positive or all negative-energy fields)
with the corresponding single particle Gelfand triples 
\[
\begin{array}{ccccc} E & \subset & \mathcal{H} & \subset & E^* 
\\ \vdots&&\vdots&& \vdots \\ 
 E_{{}_{'}} & \subset & \mathcal{H}' & \subset & E_{{}_{'}}^*
 \\  
E_{{}_{''}} & \subset & \mathcal{H}'' & \subset & E_{{}_{''}}^*
\\\vdots &&\vdots&&\vdots  \\  
E_{{}_{'''}} & \subset & \mathcal{H}''' & \subset & E_{{}_{'''}}^*.
\end{array} 
\]
Let $\mathscr{E}=\mathcal{S}_{{}_{\Delta +1}}(\widetilde{\mathbb{S}^1} \times SU(2, \mathbb{C}); \mathbb{C})$,
and let $F \in \mathscr{E}^*$.
Then 
\begin{enumerate}
\item[1)]
\[
F \ast \kappa'_{1,0}, F \ast \theta \kappa'_{0,1} \in  \mathscr{L}(\mathscr{E}, E_{{}_{'}}) \cong
\mathscr{L}(E_{{}_{'}}^*, \mathscr{E}^*) \subset \mathscr{L}(E_{{}_{'}}, \mathscr{E}^*) .
\]
\item[2)]
\[
F \ast \kappa'_{1, 0} \overset{\cdot}{\otimes} \kappa''_{1, 0} \in  \mathscr{L}(\mathscr{E}, E_{{}_{'}}^{*} \otimes E_{{}_{''}}^{*}) \cong
\mathscr{L}(E_{{}_{'}} \otimes E_{{}_{''}}, \mathscr{E}^*).
\]
\item[3)]
\[
F \ast \kappa'_{1, 0} \overset{\cdot}{\otimes} \kappa''_{0, 1} \in  \mathscr{L}(\mathscr{E}, E_{{}_{'}}^{*} \otimes E_{{}_{''}}^*) \cong
\mathscr{L}(E_{{}_{'}} \otimes E_{{}_{''}}, \mathscr{E}^*).
\]
\item[4)]
\begin{multline*}
F \ast \kappa_{1, 0} \overset{\cdot}{\otimes} \cdots \overset{\cdot}{\otimes} \kappa'_{1, 0} \overset{\cdot}{\otimes} 
\kappa''_{0, 1}\overset{\cdot}{\otimes} \cdots \overset{\cdot}{\otimes} \kappa'''_{0, 1} 
\in  \mathscr{L}(\mathscr{E}, E_{{}_{}}^{*} \otimes \cdots \otimes E_{{}_{'}}^{*} \otimes E_{{}_{''}}^* \otimes \cdots \otimes E_{{}_{'''}}^*) 
\\
\cong
\mathscr{L}(E_{{}_{}} \otimes \cdots \otimes E_{{}_{'}} \otimes E_{{}_{''}} \otimes \cdots \otimes E_{{}_{'''}}, \mathscr{E}^*).
\end{multline*}
\item[5)] The above statements 1)-4) remain valid if we replace the kernels $\kappa_{1,0}, \ldots$ with their derivations
$(X)^\alpha \kappa_{1,0}, \ldots$.
\end{enumerate}
\label{F*kappaBarDotOtimeskappaSingularOnEU}
\end{lem}

\begin{lem}
Let $\kappa_{\ell, m}, \dots, \kappa'_{\ell, m}, \kappa''_{\ell, m}, \ldots, \kappa'''_{\ell, m}   \in \mathfrak{K}_0$, $\ell,m = 1,0$, be the kernels 
of free fields on the Einstein Universe (all positive or all negative-energy fields)
with the corresponding single particle Gelfand triples 
\[
\begin{array}{ccccc} E & \subset & \mathcal{H} & \subset & E^* 
\\ \vdots&&\vdots&& \vdots \\ 
 E_{{}_{'}} & \subset & \mathcal{H}' & \subset & E_{{}_{'}}^*
 \\  
E_{{}_{''}} & \subset & \mathcal{H}'' & \subset & E_{{}_{''}}^*
\\\vdots &&\vdots&&\vdots  \\  
E_{{}_{'''}} & \subset & \mathcal{H}''' & \subset & E_{{}_{'''}}^*.
\end{array} 
\]
Let $\mathscr{E}=\mathcal{S}_{{}_{\Delta +1}}(\widetilde{\mathbb{S}^1} \times SU(2, \mathbb{C}); \mathbb{C})$,
and let $F \in \mathscr{E}^*$.
Then 
\begin{enumerate}
\item[1)]
\[
F \ast \kappa'_{1,0}, F \ast \theta \kappa'_{0,1} \in  \mathscr{L}(\mathscr{E}, E_{{}_{'}}) \cong
\mathscr{L}(E_{{}_{'}}^*, \mathscr{E}^*) \subset \mathscr{L}(E_{{}_{'}}, \mathscr{E}^*) .
\]
\item[2)]
\[
F \ast \kappa'_{1, 0} \overset{\cdot}{\otimes} \kappa''_{1, 0} \in  \mathscr{L}(\mathscr{E}, E_{{}_{'}} \otimes E_{{}_{''}}) \cong
\mathscr{L}(E_{{}_{'}}^* \otimes E_{{}_{''}}^*, \mathscr{E}^*).
\]
\item[3)]
\[
F \ast \kappa'_{1, 0} \overset{\cdot}{\otimes} \kappa''_{0, 1} \in  \mathscr{L}(\mathscr{E}, E_{{}_{'}} \otimes E_{{}_{''}}^*) \cong
\mathscr{L}(E_{{}_{'}}^* \otimes E_{{}_{''}}, \mathscr{E}^*).
\]
\item[4)]
\begin{multline*}
F \ast \kappa_{1, 0} \overset{\cdot}{\otimes} \cdots \overset{\cdot}{\otimes} \kappa'_{1, 0} \overset{\cdot}{\otimes} 
\kappa''_{0, 1}\overset{\cdot}{\otimes} \cdots \overset{\cdot}{\otimes} \kappa'''_{0, 1} 
\in  \mathscr{L}(\mathscr{E}, E_{{}_{}} \otimes \cdots \otimes E_{{}_{'}} \otimes E_{{}_{''}}^* \otimes \cdots \otimes E_{{}_{'''}}^*) 
\\
\cong
\mathscr{L}(E_{{}_{}}^* \otimes \cdots \otimes E_{{}_{'}}^* \otimes E_{{}_{''}} \otimes \cdots \otimes E_{{}_{'''}}, \mathscr{E}^*).
\end{multline*}
\item[5)] The above statements 1)-3) remain valid if we replace the kernels $\kappa_{1,0}, \ldots$ with their derivations
$(X)^\alpha \kappa_{1,0}, \ldots$.
\end{enumerate}
\label{F*kappaBarDotOtimeskappaOnEU}
\end{lem}
\qedsymbol \,
Let $\mathscr{E}=\mathcal{S}_{{}_{\Delta +1}}(\widetilde{\mathbb{S}^1} \times SU(2, \mathbb{C}); \mathbb{C})$,
and let $F \in \mathscr{E}^*$, $\phi \in \mathscr{E}$.
Let for example
\begin{multline*}
\kappa_{\ell, m} = \kappa_{1, 0} \overset{\cdot}{\otimes} \cdots \overset{\cdot}{\otimes} \kappa'_{1, 0} \overset{\cdot}{\otimes} 
\kappa''_{0, 1}\overset{\cdot}{\otimes} \cdots \overset{\cdot}{\otimes} \kappa'''_{0, 1} 
\in  \mathscr{L}(\mathscr{E}, E_{{}_{}} \otimes \cdots \otimes E_{{}_{'}} \otimes E_{{}_{''}}^* \otimes \cdots \otimes E_{{}_{'''}}^*) 
\\
\cong
\mathscr{L}(E_{{}_{}}^* \otimes \cdots \otimes E_{{}_{'}}^* \otimes E_{{}_{''}} \otimes \cdots \otimes E_{{}_{'''}}, \mathscr{E}^*),
\end{multline*}
with $\kappa_{1,0}, \ldots , \kappa'_{1,0}, \ldots \in \mathfrak{K}_0$ equal to the plane-wave kernels of free fields (all positive or all negative-energy fields)
on the Einstein Universe.
Because $F \ast \phi \in \mathscr{E}$,
\[
F \ast \kappa_{\ell, m}(\phi) = \kappa_{\ell, m}(F \ast \phi).
\]
Because the map
\[
\mathscr{E} \ni \phi \longrightarrow \kappa_{\ell, m}(\phi) \in
E_{{}_{}} \otimes \cdots \otimes E_{{}_{'}} \otimes E_{{}_{''}}^* \otimes \cdots \otimes E_{{}_{'''}}^*
\]
is by assumption continuous, and because the map 
\[
\mathscr{E} \ni \phi \longrightarrow F \ast \phi \in \mathscr{E},
\]
is continuous, then we immediately see that the map
\[
\mathscr{E} \ni \phi \longrightarrow  F \ast \kappa_{\ell, m}(\phi) = \kappa_{\ell, m}(F \ast \phi)
\in E_{{}_{}} \otimes \cdots \otimes E_{{}_{'}} \otimes E_{{}_{''}}^* \otimes \cdots \otimes E_{{}_{'''}}^*
\]
is continuous, which proves assertion 4) of Lemma \ref{F*kappaBarDotOtimeskappaOnEU}. Proof of assertion
4) of  Lemma \ref{F*kappaBarDotOtimeskappaSingularOnEU}, as well as of the remaining assertions of Lemmas 
\ref{F*kappaBarDotOtimeskappaOnEU} and \ref{F*kappaBarDotOtimeskappaSingularOnEU} is identical.
\qed

We give here another proof of Lemmas \ref{F*kappaBarDotOtimeskappaOnEU} and \ref{F*kappaBarDotOtimeskappaSingularOnEU},
fully analogous to the proof of Lemma \ref{S*Xi} of Subsection \ref{OperationsOnXi}.

Because for each $F \in  \mathcal{S}_{A}'(\widetilde{\mathbb{S}^1} \times SU(2, \mathbb{C}))
= \mathcal{S}_{A}(\widetilde{\mathbb{S}^1} \times SU(2, \mathbb{C}))^* = \mathscr{E}^*$, the map 
\[
\mathcal{S}_{A}(\widetilde{\mathbb{S}^1} \times SU(2, \mathbb{C})) = \mathscr{E} \ni \phi \longrightarrow F \ast \phi \in \mathscr{E}
=\mathcal{S}_{A}(\widetilde{\mathbb{S}^1} \times SU(2, \mathbb{C})),
\]
is continuous, then the Schwartz convolutor algebra 
$\mathcal{O}_{C}'(\widetilde{\mathbb{S}^1} \times SU(2, \mathbb{C}))$ of the algebra 
$\mathcal{S}_{A}(\widetilde{\mathbb{S}^1} \times SU(2, \mathbb{C})) = \mathscr{E}$ coincides with 
\[
\mathcal{S}_{A}'(\widetilde{\mathbb{S}^1} \times SU(2, \mathbb{C}))
\mathcal{S}_{A}(\widetilde{\mathbb{S}^1} \times SU(2, \mathbb{C}))^* \mathscr{E}^*,
\]
with its predual equal to 
\[
\mathcal{O}_{C}(\widetilde{\mathbb{S}^1} \times SU(2, \mathbb{C})) = \mathcal{S}_{A}(\widetilde{\mathbb{S}^1} \times SU(2, \mathbb{C})) = \mathscr{E}
\]
itself, as we have already remarked. 

It turns out that if $F \in \mathcal{S}_{A}' = \mathscr{E}^*$ then the operator 
\[
C_F: \phi \mapsto F \ast \phi = C_F(\phi)
\] 
of convolution with $F \in \mathcal{S}_{A}' = \mathscr{E}^*$,
 corresponding to $F$, maps continuously 
$\mathcal{S}_{A} = \mathscr{E} \rightarrow \mathcal{S}_{A} = \mathscr{E}$, i.e. $C_F \in \mathscr{L}(\mathscr{E}, \mathscr{E})$.

Similarly, in case of the compact Lie group $\widetilde{\mathbb{S}^1} \times SU(2, \mathbb{C})$,
te multiplier algebra $\mathcal{O}_M(\widetilde{\mathbb{S}^1} \times SU(2, \mathbb{C}))$ 
of the algebra $\mathcal{S}_{A}(\widetilde{\mathbb{S}^1} \times SU(2, \mathbb{C})) = \mathscr{E}$ coincides with
the algebra $\mathcal{S}_{A}(\widetilde{\mathbb{S}^1} \times SU(2, \mathbb{C})) = \mathscr{E}$ itself.
We can introduce after Schwartz the operator $M_{F} \in \mathscr{L}(\mathscr{E}, \mathscr{E})$ of pointwise multiplication 
by $F$, corresponding to each 
$F \in \mathcal{O}_M(\widetilde{\mathbb{S}^1} \times SU(2, \mathbb{C})) = \mathscr{E}$.

It is easily seen that the composition of convolution operators
\[
C_{F} \circ C_{H}, \,\,\, F, H \in \mathscr{E}^*,
\]
is again equal to a convolution operator
\[
C_{F} \circ C_{H} = C_{F \ast H}
\]
for a functional $F\ast H \in \mathscr{E}^*$, moreover explicit Fourier decomposition shows that
\[
\widetilde{F\ast H}_{{}_{ji}}(\widehat{n} \cdot \widehat{l}) =
\Bigg( \widetilde{H}(\widehat{n} \cdot \widehat{l}) \widetilde{F}(\widehat{n} \cdot \widehat{l}) \Bigg)_{{}_{ji}}
= \sum \limits_{m=-l}^{l} \widetilde{H}_{{}_{jm}}(\widehat{n} \cdot \widehat{l}) \widetilde{F}_{{}_{mi}}(\widehat{n} \cdot \widehat{l}),
\]
with a norm estimation similar to that presented in (\ref{|F*phi|_q <|phi|_q+p+2}) proving that
\[
| F\ast H |_{-(p+q)} < \infty, \,\,\,\, |F|_{-p}<\infty, |H|_{-q} <\infty.
\]

Therefore, we can, again after Schwartz \cite{Schwartz}, introduce the topology on
$\mathcal{O}'_{C}= \mathscr{E}^*$ and on $\mathcal{O}_{M}= \mathscr{E}$, 
induced from the topology of uniform convergence on bounded sets 
on $\mathscr{L}(\mathscr{E}, \mathscr{E})$. 

These are the Schwartz operator topologies on 
$\mathcal{O}_M$ and $\mathcal{O}'_{C}$. These spaces become nuclear with these topologies,
(quasi-) complete and barreled. 

Let us introduce, in addition to the convolution operator $C_F$, a closely
related operator $C_{F}^{+}$, defined similarly by
\begin{multline*}
C_{F}^{+}(\phi)(x) \overset{\textrm{df}}{=} 
\int \limits_{\widetilde{\mathbb{S}^1} \times G} 
F(x^{-1}y)\phi(y) dy 
\\
= 
\int \limits_{\widetilde{\mathbb{S}^1} \times G} 
F(y)\phi(xy) dy 
\,\,\,
x=(t, \boldsymbol{w}), y = (\tau, \boldsymbol{u}) \in \widetilde{\mathbb{S}^1} \times G,
\end{multline*}
Similarly as the operator $C_F$, also the operator  $C_{F}^{+} \in \mathscr{L}(\mathscr{E}, \mathscr{E})$   if and only if 
$F \in \mathcal{O}'_{C}(\widetilde{\mathbb{S}^1} \times SU(2, \mathbb{C}))= \mathscr{E}^*$.
That indeed  $C_{F}^{+}$ is a continuous operator $\mathscr{E}\rightarrow \mathscr{E}$ we show exactly 
as continuity of the operator $C_{F}$, by Fourier transforming the function
$C_{F}^{+}(\phi)$, and using norm estimation identical as that in (\ref{|F*phi|_q <|phi|_q+p+2}).

\begin{twr*}
Let $\mathcal{S}_{A}'(\widetilde{\mathbb{S}^1} \times SU(2, \mathbb{C})) = \mathscr{E}^*$ be endowed with its 
(ordinary nuclear) strong dual topology, and $\mathcal{O}_M(\widetilde{\mathbb{S}^1} \times SU(2, \mathbb{C}))
= \mathscr{E}$, 
$\mathcal{O}'_{C}(\widetilde{\mathbb{S}^1} \times SU(2, \mathbb{C})) = \mathscr{E}^*$
with the Schwartz' operator topologies defined as above. 
On the space $\mathscr{E}^*$ we can define the operation of multiplication by $F\in \mathcal{O}_M$ 
through the linear transpose of the map $M_F$, which maps continuously 
$\mathscr{E}^* \rightarrow \mathscr{E}^*$ and defines a bilinear hypocontinuous multiplication map 
$\mathscr{E}^* \times \mathcal{O}_{M} \rightarrow \mathscr{E}^*$.
Similarly on the space $\mathscr{E}^*$ we can define the operation of convolution by $F\in \mathcal{O}'_{C}$ 
through the linear transpose of the map $C_{F}^{+}$, which maps continuously 
$\mathscr{E}^* \rightarrow \mathscr{E}^*$ and defines a bilinear hypocontinuous convolution map 
$\mathcal{O}'_{C} \times \mathscr{E}^* \rightarrow \mathscr{E}^*$.  
\end{twr*}
It is easily checked that the multiplication $F.H$ and convolution operations $F\ast H$ coincide with their
operator definitions due to Schwartz:
\[
M_F \circ M_H = M_{F.H}, \,\,\,\,\,\,
C_F \circ C_H = C_{F \ast H}.
\]
This is the compact Lie group $\widetilde{\mathbb{S}^1} \times SU(2, \mathbb{C})$ version of the corresponding theorem
for the Abelian non-compact additive Lie group $\mathbb{R}^n$, \cite{Schwartz}, Thm. X and Thm. XI, Chap. VII, \S 5, pp. 245-248,
and Appendix \ref{convolutorsO'_C}. 

In the sequel we need only separate continuity of the convolution operation $\mathcal{O}'_{C} \times \mathscr{E}^*
\rightarrow \mathscr{E}^*$, which in our case of compact Lie group defines an operation 
$\mathscr{E}^* \times \mathscr{E}^* \rightarrow \mathscr{E}^*$, with the Schwartz operator topology on the first factor and with
the ordinary strong dual topology on the second factor of $\mathscr{E}^* \times \mathscr{E}^*$, and with the ordinary strong dual
topology on the image $\mathscr{E}^*$. In fact we need only the continuity of the convolution with the the first factor fixed,
as a map $\mathscr{E}^* \rightarrow \mathscr{E}^*$ with respect to ordinary strong dual topology. This continuity
follows by duality already from the continuity of the operator $C_{F}^{+}: \mathscr{E} \rightarrow \mathscr{E}$ with respect to ordinary
countably Hilbert nuclear Fr\'echet topology on $\mathscr{E}$.

\qedsymbol \, 
{\bf (Another proof of  Lemmas \ref{F*kappaBarDotOtimeskappaOnEU} and \ref{F*kappaBarDotOtimeskappaSingularOnEU})}

Let 
\begin{multline*}
\kappa_{\ell, m} = \kappa_{1, 0} \overset{\cdot}{\otimes} \cdots \overset{\cdot}{\otimes} \kappa'_{1, 0} \overset{\cdot}{\otimes} 
\kappa''_{0, 1}\overset{\cdot}{\otimes} \cdots \overset{\cdot}{\otimes} \kappa'''_{0, 1} 
\in  \mathscr{L}(\mathscr{E}, E_{{}_{}} \otimes \cdots \otimes E_{{}_{'}} \otimes E_{{}_{''}}^* \otimes \cdots \otimes E_{{}_{'''}}^*) 
\\
\cong
\mathscr{L}(E_{{}_{}}^* \otimes \cdots \otimes E_{{}_{'}}^* \otimes E_{{}_{''}} \otimes \cdots \otimes E_{{}_{'''}}, \mathscr{E}^*),
\end{multline*}
be regarded as element 
\[
\kappa_{\ell, m} = \kappa_{1, 0} \overset{\cdot}{\otimes} \cdots \overset{\cdot}{\otimes} \kappa'_{1, 0} \overset{\cdot}{\otimes} 
\kappa''_{0, 1}\overset{\cdot}{\otimes} \cdots \overset{\cdot}{\otimes} \kappa'''_{0, 1} 
\in 
\mathscr{L}(E_{{}_{}}^* \otimes \cdots \otimes E_{{}_{'}}^* \otimes E_{{}_{''}} \otimes \cdots \otimes E_{{}_{'''}}, \mathscr{E}^*).
\]
Let 
\[
\xi \in E_{{}_{}}^* \otimes \cdots \otimes E_{{}_{'}}^* \otimes E_{{}_{''}} \otimes \cdots \otimes E_{{}_{'''}}.
\]
Then the map 
\[
E_{{}_{}}^* \otimes \cdots \otimes E_{{}_{'}}^* \otimes E_{{}_{''}} \otimes \cdots \otimes E_{{}_{'''}} \ni \xi 
\longrightarrow 
\kappa_{\ell, m}(\xi)
\in \mathscr{E}^*
\]
is continuous. Because $\mathscr{E}^* = \mathcal{O}'_{C}(\widetilde{\mathbb{S}^1} \times SU(2, \mathbb{C}))$,
then for any fixed $F \in \mathscr{E}^*$ the map
\[
E_{{}_{}}^* \otimes \cdots \otimes E_{{}_{'}}^* \otimes E_{{}_{''}} \otimes \cdots \otimes E_{{}_{'''}} \ni \xi 
\longrightarrow 
F \ast \kappa_{\ell, m}(\xi)
\in \mathscr{E}^*
\]
is continuous, which proves assertion
4) of  Lemma \ref{F*kappaBarDotOtimeskappaOnEU}. Proof of the remaining assertions of Lemmas 
\ref{F*kappaBarDotOtimeskappaOnEU} and \ref{F*kappaBarDotOtimeskappaSingularOnEU} is identical.
\qed

\begin{twr}
Let $\kappa_{\ell, m}, \dots, \kappa'_{\ell, m}, \kappa''_{\ell, m}, \ldots, \kappa'''_{\ell, m}   \in \mathfrak{K}_0$, $\ell,m = 1,0$, be the kernels 
of free fields on the Einstein Universe (all positive or all negative-energy fields)
\begin{eqnarray*}
\Xi_{0,1}(\kappa_{0, 1}) + \Xi_{1,0}(\kappa_{1,0}),
\\
\vdots  
\\
\Xi_{0,1}(\kappa'_{0, 1}) + \Xi_{1,0}(\kappa'_{1,0}),
\\
\Xi_{0,1}(\kappa''_{0, 1}) + \Xi_{1,0}(\kappa''_{1,0}),
\\
\vdots
\\
\Xi_{0,1}(\kappa'''_{0, 1}) + \Xi_{1,0}(\kappa'''_{1,0}),
\end{eqnarray*}
with the corresponding single particle Gelfand triples 
\[
\begin{array}{ccccc} E & \subset & \mathcal{H} & \subset & E^* 
\\ \vdots&&\vdots&& \vdots \\ 
 E_{{}_{'}} & \subset & \mathcal{H}' & \subset & E_{{}_{'}}^*
 \\  
E_{{}_{''}} & \subset & \mathcal{H}'' & \subset & E_{{}_{''}}^*
\\\vdots &&\vdots&&\vdots  \\  
E_{{}_{'''}} & \subset & \mathcal{H}''' & \subset & E_{{}_{'''}}^*.
\end{array} 
\]
Let $\mathscr{E}=\mathcal{S}_{{}_{\Delta +1}}(\widetilde{\mathbb{S}^1} \times SU(2, \mathbb{C}); \mathbb{C})$.
Then 
\begin{enumerate}
\item[1)]
The Wick product operator
\[
\Xi_{2,0}(\kappa_{2,0}(x)) = \boldsymbol{{:}} \Xi_{1,0}(\kappa'_{1,0}(x)) \Xi_{1,0}(\kappa''_{1,0}(x))
\boldsymbol{{:}}, \,\,\, x= (t, \boldsymbol{w}) \in \mathbb{R} \times SU(2, \mathbb{C}),
\]
defines an integral kernel operator
\[
\Xi_{2,0}(\kappa_{2,0})
\in 
\mathscr{L}((\boldsymbol{E}) \otimes \mathscr{E}, \, (\boldsymbol{E}))
\cong 
\mathscr{L}\big(\mathscr{E}, \,\, \mathscr{L}(\boldsymbol{E}), \, (\boldsymbol{E})) \big)
\]
with the vector valued kernel
\[
\kappa_{2,0}=\kappa'_{1, 0} \overline{\dot{\otimes}} \kappa''_{1, 0} \in  \mathscr{L}(\mathscr{E}, E_{{}_{'}} \widehat{\otimes} E_{{}_{''}}) \cong
\mathscr{L}(E_{{}_{'}}^* \widehat{\otimes} E_{{}_{''}}^*, \mathscr{E}^*).
\]
\item[2)]
Let $(\boldsymbol{E}) = (E_{{}_{'}}) \otimes (E_{{}_{''}})$. The Wick product integral kernel operator
\[
\Xi_{1,1}(\kappa_{1,1}(x)) = \boldsymbol{{:}} \Xi_{1,0}(\kappa'_{1,0}(x)) \Xi_{0,1}(\kappa''_{0,1}(x))
\boldsymbol{{:}} 
\]
defines an integral kernel operator
\[
\Xi_{1,1}(\kappa_{1,1})
\in 
\mathscr{L}((\boldsymbol{E}) \otimes \mathscr{E}, \, (\boldsymbol{E}))
\cong 
\mathscr{L}\big(\mathscr{E}, \,\, \mathscr{L}(\boldsymbol{E}), \, (\boldsymbol{E})) \big)
\]
with the vector valued kernel
\[
\kappa_{1,1} = \kappa'_{1, 0} \overset{\cdot}{\otimes} \kappa''_{0, 1} \in  \mathscr{L}(\mathscr{E}, E_{{}_{'}} \otimes E_{{}_{''}}^*) \cong
\mathscr{L}(E_{{}_{'}}^* \otimes E_{{}_{''}}, \mathscr{E}^*).
\]
\item[3)]
Let $(\boldsymbol{E}) = (E) \otimes \ldots \otimes (E_{{}_{'}}) \otimes \ldots \otimes (E_{{}_{''}})$ with $\ell+m$ factors. 
The Wick product integral kernel operator
\[
\Xi_{\ell,m}(\kappa_{\ell,m}(x)) = \boldsymbol{{:}} \Xi_{1,0}(\kappa_{1,0}(x)) \cdots \Xi_{1,0}(\kappa'_{1,0}(x))
\Xi_{0,1}(\kappa''_{0,1}(x)) \cdots \Xi_{0,1}(\kappa'''_{0,1}(x))
\boldsymbol{{:}} 
\]
with $\ell$ ``first'' factors 
\[
\Xi_{1,0}(\kappa_{1,0}(x)) \cdots \Xi_{1,0}(\kappa'_{1,0}(x))
\]
and $m$ ``last'' factors 
\[
\Xi_{0,1}(\kappa''_{0,1}(x)) \cdots \Xi_{0,1}(\kappa'''_{0,1}(x)),
\]
defines an integral kernel operator
\[
\Xi_{\ell,m}(\kappa_{\ell,m})
\in 
\mathscr{L}((\boldsymbol{E}) \otimes \mathscr{E}, \, (\boldsymbol{E}))
\cong 
\mathscr{L}\big(\mathscr{E}, \,\, \mathscr{L}(\boldsymbol{E}), \, (\boldsymbol{E})) \big)
\]
with the vector valued kernel
\begin{multline*}
\kappa_{\ell,m} = \kappa_{1, 0} \overline{\dot{\otimes}} \cdots \overline{\dot{\otimes}} \kappa'_{1, 0} \overline{\dot{\otimes}}
\kappa''_{0, 1} \overline{\dot{\otimes}} \cdots \overline{\dot{\otimes}} \kappa'''_{0, 1} 
\in  \mathscr{L}(\mathscr{E}, E_{{}_{}} \widehat{\otimes} \cdots \widehat{\otimes} E_{{}_{'}} \widehat{\otimes} E_{{}_{''}}^* \widehat{\otimes} 
\cdots \widehat{\otimes} E_{{}_{'''}}^*) 
\\
\cong
\mathscr{L}(E_{{}_{}}^* \widehat{\otimes} \cdots \widehat{\otimes} E_{{}_{'}}^* \widehat{\otimes} E_{{}_{''}} \widehat{\otimes} 
\cdots \widehat{\otimes} E_{{}_{'''}}, \mathscr{E}^*).
\end{multline*}
\item[4)] The above statements 1)-3) remain valid if we replace the kernels $\kappa_{1,0}, \ldots$ with their derivations
$(X)^\alpha \kappa_{1,0}, \ldots$.
\end{enumerate}
\label{WickProdFreeFieldsOnEU}
\end{twr}

\begin{twr}
Let $\kappa_{\ell, m}, \dots, \kappa'_{\ell, m}, \kappa''_{\ell, m}, \ldots, \kappa'''_{\ell, m}   \in \mathfrak{K}_0$, $\ell,m = 1,0$, be the kernels 
of free fields on the Einstein Universe (all positive or all negative-energy fields)
\begin{eqnarray*}
\Xi_{0,1}(\kappa_{0, 1}) + \Xi_{1,0}(\kappa_{1,0}),
\\
\vdots  
\\
\Xi_{0,1}(\kappa'_{0, 1}) + \Xi_{1,0}(\kappa'_{1,0}),
\\
\Xi_{0,1}(\kappa''_{0, 1}) + \Xi_{1,0}(\kappa''_{1,0}),
\\
\vdots
\\
\Xi_{0,1}(\kappa'''_{0, 1}) + \Xi_{1,0}(\kappa'''_{1,0}),
\end{eqnarray*}
with the corresponding single particle Gelfand triples 
\[
\begin{array}{ccccc} E & \subset & \mathcal{H} & \subset & E^* 
\\ \vdots&&\vdots&& \vdots \\ 
 E_{{}_{'}} & \subset & \mathcal{H}' & \subset & E_{{}_{'}}^*
 \\  
E_{{}_{''}} & \subset & \mathcal{H}'' & \subset & E_{{}_{''}}^*
\\\vdots &&\vdots&&\vdots  \\  
E_{{}_{'''}} & \subset & \mathcal{H}''' & \subset & E_{{}_{'''}}^*.
\end{array} 
\]
Let $\mathscr{E}=\mathcal{S}_{{}_{\Delta +1}}(\widetilde{\mathbb{S}^1} \times SU(2, \mathbb{C}); \mathbb{C}))$ 
and $F \in \mathscr{E}^*=\mathcal{S}_{{}_{\Delta +1}}(\widetilde{\mathbb{S}^1} \times SU(2, \mathbb{C}); \mathbb{C}))^*$. 
Let $(\boldsymbol{E}) = (E) \otimes \ldots \otimes (E_{{}_{'}}) \otimes \ldots \otimes (E_{{}_{''}})$. 
Finally let 
\[
\Xi_{\ell,m}(\kappa_{\ell,m}(x)) = \boldsymbol{{:}} \Xi_{1,0}(\kappa_{1,0}(x)) \cdots \Xi_{1,0}(\kappa'_{1,0}(x))
\Xi_{0,1}(\kappa''_{0,1}(x)) \cdots \Xi_{0,1}(\kappa'''_{0,1}(x))
\boldsymbol{{:}}
\]
\[
x= (t, \boldsymbol{w}) \in \mathbb{R} \times SU(2, \mathbb{C}),
\]
be the Wick product integral kernel operator determining an integral kernel operator
\[
\Xi_{\ell,m}(\kappa_{\ell,m})
\in 
\mathscr{L}((\boldsymbol{E}) \otimes \mathscr{E}, \, (\boldsymbol{E}))
\cong 
\mathscr{L}\big(\mathscr{E}, \,\, \mathscr{L}(\boldsymbol{E}), \, (\boldsymbol{E})) \big)
\]
with the vector valued kernel
\begin{multline*}
\kappa_{\ell,m} = \kappa_{1, 0} \overline{\dot{\otimes}} \cdots \overline{\dot{\otimes}} \kappa'_{1, 0} \overline{\dot{\otimes}}
\kappa''_{0, 1} \overline{\dot{\otimes}} \cdots \overline{\dot{\otimes}} \kappa'''_{0, 1} 
\in  \mathscr{L}(\mathscr{E}, E_{{}_{}} \widehat{\otimes} \cdots \widehat{\otimes} E_{{}_{'}} \widehat{\otimes} E_{{}_{''}}^* \widehat{\otimes} 
\cdots \widehat{\otimes} E_{{}_{'''}}^*) 
\\
\cong
\mathscr{L}(E_{{}_{}}^* \widehat{\otimes} \cdots \widehat{\otimes} E_{{}_{'}}^* \widehat{\otimes} E_{{}_{''}} \widehat{\otimes} 
\cdots \widehat{\otimes} E_{{}_{'''}}, \mathscr{E}^*).
\end{multline*}

Then
\begin{enumerate}
\item[1)]
The operator, given by the convolution of $F$ with $\Xi_{\ell,m}(\kappa_{\ell,m})$,
\begin{multline*}
F \ast \Xi_{\ell,m}(\kappa_{\ell,m})(x)
= \int \limits_{\widetilde{\mathbb{S}^1} \times SU(2, \mathbb{C})} F(xy^{-1})
\Xi_{\ell,m}(\kappa_{\ell,m}(y)) \, dy
\\
=
\Xi_{\ell,m} \Bigg( \,\, \int \limits_{\widetilde{\mathbb{S}^1} \times SU(2, \mathbb{C})} F(xy^{-1})
\kappa_{\ell,m}(y) \, dy  \Bigg) \\
= 
\Xi_{\ell,m} \big(F \ast \kappa_{\ell,m}(x) \big),
\end{multline*}
defines an integral kernel operator
\[
\Xi\big(F  \ast \kappa_{\ell,m}\big)
\in 
\mathscr{L}((\boldsymbol{E}) \otimes \mathscr{E}, \, (\boldsymbol{E}))
\cong 
\mathscr{L}\big(\mathscr{E}, \,\, \mathscr{L}(\boldsymbol{E}), \, (\boldsymbol{E})) \big)
\]
with the vector valued kernel
\begin{multline*}
F \ast \kappa_{\ell,m} 
\\
= F \ast \big(\kappa_{1, 0} \overline{\dot{\otimes}} \cdots \overline{\dot{\otimes}} \kappa'_{1, 0} \overline{\dot{\otimes}}
\kappa''_{0, 1} \overline{\dot{\otimes}} \cdots \overline{\dot{\otimes}} \kappa'''_{0, 1}\big) 
\in  \mathscr{L}(\mathscr{E}, E_{{}_{}} \widehat{\otimes} \cdots \widehat{\otimes} E_{{}_{'}} \widehat{\otimes} E_{{}_{''}}^* \widehat{\otimes} 
\cdots \widehat{\otimes} E_{{}_{'''}}^*) 
\\
\cong
\mathscr{L}(E_{{}_{}}^* \widehat{\otimes} \cdots \widehat{\otimes} E_{{}_{'}}^* \widehat{\otimes} E_{{}_{''}} \widehat{\otimes} 
\cdots \widehat{\otimes} E_{{}_{'''}}, \mathscr{E}^*).
\end{multline*}
\item[2)] The above statement 1) remains valid if we replace the kernels $\kappa_{1,0}, \ldots$ with their derivations
$(X)^\alpha \kappa_{1,0}, \ldots$.
\end{enumerate}
\label{F*WickProdFreeFieldsOnEU}
\end{twr}
\qedsymbol \, 
Theorems \ref{WickProdFreeFieldsOnEU} and \ref{F*WickProdFreeFieldsOnEU} follow from the Rules of Subsection \ref{OperationsOnXi}
joining the kernels with the corresponding integral kernel operators and from the correspondence between the operations performed upon the kernels
and the corresponding operations on the corresponding integral kernel operators, and from Theorem \ref{obataJFA.Thm.3.13} of Subsection \ref{psiBerezin-Hida}. 
Indeed, Theorem \ref{WickProdFreeFieldsOnEU} follows from Lemma \ref{kappaBarDotOtimeskappaOnEU} and Theorem \ref{obataJFA.Thm.3.13} of Subsection \ref{psiBerezin-Hida} (note that symmetrization/antisymmetrization
$\overline{\dot{\otimes}}$ of the pointwise product $\dot{\otimes}$ of the kernels does not change continuity of the kernels stated in Lemma 
\ref{kappaBarDotOtimeskappaOnEU}). Theorem \ref{F*WickProdFreeFieldsOnEU} follows from Theorem \ref{obataJFA.Thm.3.13} of Subsection \ref{psiBerezin-Hida}
and from Lemma \ref{F*kappaBarDotOtimeskappaOnEU}. 
\qed

\subsection{Wick's theorem for ``products'' and the scattering operator on the Einstein Universe}\label{WickForProductOnEU}

Let us consider the Wick product theorem for free fields on Einstein Universe, along the lines presented in Subsection \ref{WickForProduct}. 
It is used in the intermediate stage of the computations
of the scattering operator and interacting fields.  The so called ``Wick theorem'' (compare \cite{Bogoliubov_Shirkov}, \S 17.2) is used for 
decomposition of the ``product'' 
\begin{equation}\label{Wick(x)Wick(y)EU}
\boldsymbol{{:}} \mathbb{A}_{{}_{1}}^{a}(x) \ldots \mathbb{A}_{{}_{N}}^{a}(x) \boldsymbol{{:}} \, 
\boldsymbol{{:}} \mathbb{A}_{{}_{N+1}}^{b}(y) \ldots \mathbb{A}_{{}_{M}}^{b}(y)  \boldsymbol{{:}}
\end{equation}
of Wick product monomials 
\begin{equation}\label{WickMonomialsInGeneralFreeFieldsEU}
\boldsymbol{{:}} \mathbb{A}_{{}_{1}}^{a}(x) \ldots \mathbb{A}_{{}_{N}}^{a}(x) \boldsymbol{{:}} 
\,\,\,
\textrm{and}
\,\,\,
\boldsymbol{{:}} \mathbb{A}_{{}_{N+1}}^{b}(y) \ldots \mathbb{A}_{{}_{M}}^{b}(y)  \boldsymbol{{:}}
\end{equation}
in fixed components $\mathbb{A}_{{}_{k}}^{a}$ of free fields $\mathbb{A}_{{}_{k}}$, each separately evaluated at the same space-time point $x$ or respectively, $y$,
into the sum of Wick monomials (each in the so-called ``normal order'').

The point lies, as we have said in Subsection \ref{WickForProduct}, in the correct definition of such ``product'', 
because each factor evaluated respectively at $x$ or $y$,
represents a generalized integral kernel operator transforming continuously the Hida space $(\boldsymbol{E})$ into its strong dual
$(\boldsymbol{E})^*$, so that the product cannot be understood as ordinary operator composition, and therefore a correct definition is required.
Recall that  $(\boldsymbol{E}) = (E_1) \otimes \ldots \otimes (E_k) \otimes \ldots \otimes (E_M)$ is the Hida test space in the total Fock space of the free fields involved in the product (in fact in the total Fock space of all free fields underlying the QFT in question).

The crucial point is that the free fields and their Wick products define (finite sums of) integral kernel operators with vector-valued
kernels in the sense of \cite{obataJFA}, as we have explained above, and the ``product'' can be given as a distributional
kernel operator. Indeed, from Theorem \ref{WickProdFreeFieldsOnEU} of Subsection \ref{WhiteNoiseFreeFieldsonEU}, it follows that each factor (\ref{WickMonomialsInGeneralFreeFieldsEU}) separately
represents an integral kernel operator which belongs to 
\[
\mathscr{L}\big(\mathscr{E}, \, \mathscr{L}((\boldsymbol{E}), \, (\boldsymbol{E})) \big),
\]
irrespectively if among the factors $\mathbb{A}_{{}_{k}}$, $1 \leq k \leq M$ there are or not massless fields (or their derivatives).
This means that the first factor in (\ref{WickMonomialsInGeneralFreeFieldsEU}) defines the corresponding continuous map
\[
\mathscr{E} \ni \phi \longmapsto \Xi'(\phi) =
\sum\limits_{\substack{\ell,m \\ \ell'+m'=N}} 
\Xi_{\ell',m'}\big(\kappa'_{\ell',m'}(\phi)\big)
\in \mathscr{L}\big( (\boldsymbol{E}), (\boldsymbol{E})\big),  
\]
and similarly the second factor in (\ref{WickMonomialsInGeneralFreeFieldsEU}) defines continuous map
\[
\mathscr{E} \ni \varphi \longmapsto \Xi''(\varphi) =
\sum\limits_{\substack{\ell'',m'' \\ \ell''+m''=M-N}} 
\Xi_{\ell'',m''}\big({\kappa''}_{\ell'',m''}(\varphi)\big)
\in \mathscr{L}\big( (\boldsymbol{E}), (\boldsymbol{E})\big),  
\]
where $\mathscr{E} = \mathcal{S}_{A}(\widetilde{\mathbb{S}^1} \times SU(2, \mathbb{C}); \mathbb{C})$, 
$A = \Delta+1$, and $E_j = \mathcal{S}_{{}_{\widetilde{A}}}(\mathscr{O}_j; \mathbb{C})$,
$i,j = 1,2, \ldots, M$,  (compare Subsection \ref{WhiteNoiseFreeFieldsonEU} for the definition of the standard operators
determining $E_j$ as equal to $\widetilde{A}$ restricted to the corresponding orbit $\mathscr{O}_j$ on the correspondng standard Hilbert space
$L^2(\mathscr{O}_j; \mathbb{C})$).  Both factors $\Xi'$ and $ \Xi''$
are equal to finite sums of integral kernel operators with $\mathscr{E}^{*}$-valued distributional kernels
$\kappa'_{\ell',m'}, \kappa''_{\ell'',m''}$. In this case both factors $\Xi'(\phi)$ and $ \Xi''(\varphi)$, when evaluated at the test functions 
$\phi, \varphi \in \mathscr{E}$, are ordinary operators on the Fock space transforming continuously the Hida space $(\boldsymbol{E})$ into itself,
and thus can be composed $\Xi'(\phi) \circ  \Xi''(\varphi)$ as operators, giving the composition operator
\[
\Xi'(\phi) \circ  \Xi''(\varphi) \in \mathscr{L}\big( (\boldsymbol{E}), (\boldsymbol{E})\big),
\]
defining the map
\[
\mathscr{E} \otimes \mathscr{E} \ni \phi \otimes \varphi \longmapsto \Xi'(\phi) \circ  \Xi''(\varphi) \in \mathscr{L}\big( (\boldsymbol{E}), (\boldsymbol{E})\big),
\]
which by construction is separately continuous in the arguments $\phi \in \mathscr{E}$ and $\varphi \in \mathscr{E}$. Because
$\mathscr{E}$ is a complete Fr\'echet space, then by Proposition 1.3.11 of \cite{obataJFA} there exists
the corresponding operator-valued continuous map (say operator-valued distribution) 
\begin{multline*}
\phi \otimes \varphi \longmapsto 
 \Xi(\phi \otimes \varphi) \overset{\textrm{df}}{=} \\ \overset{\textrm{df}}{=}
\sum\limits_{a,b}
\int\limits_{\big[\widetilde{\mathbb{S}^1} \times SU(2, \mathbb{C})\big]^{\times \, 2}} 
\boldsymbol{{:}} \mathbb{A}_{{}_{1}}^{a}(x) \ldots \mathbb{A}_{{}_{N}}^{a}(x) \boldsymbol{{:}} \, 
\boldsymbol{{:}} \mathbb{A}_{{}_{N+1}}^{b}(y) \ldots \mathbb{A}_{{}_{M}}^{b}(y)  \boldsymbol{{:}} \,
\phi^a \otimes \varphi^b(x,y) \, \ud^4x \, \ud^4 y
\\
= \Xi'(\phi) \circ  \Xi''(\varphi). 
\end{multline*}
In particular the operator map $\phi \otimes \varphi \mapsto \Xi(\phi \otimes \varphi )$ defines a 
generalized operator 
\[
\Xi \in \mathscr{L}\big(\mathscr{E} \otimes \mathscr{E}, \, \mathscr{L}((\boldsymbol{E}), \, (\boldsymbol{E}) \big)
\cong \mathscr{L}((\boldsymbol{E}) \otimes \mathscr{E} \otimes \mathscr{E}, \, (\boldsymbol{E})) 
\]
which by Theorem 4.8 of \cite{obataJFA} possesses unique Fock expansion
\[
\Xi = \sum\limits_{\ell, m} \Xi_{\ell, m}({\kappa}_{\ell, m}), 
\]
into integral kernel operators
with $\mathscr{E}^{*} \otimes \mathscr{E}^{*} = \mathscr{L}(\mathscr{E} \otimes \mathscr{E}, \mathbb{C})$-valued
kernels ${\kappa}_{\ell', m}$. This Fock expansion becomes finite, with all $\ell+ m \leq M$ with one pair of indices $(\ell,m) = (0,0)$
in it, and its computation can be easily reduced\footnote{Compare Subsection \ref{WickForProduct} and our remarks in Subsection 
\ref{psiBerezin-Hida}.} to the canonical commutation/anticommutation relations for Hida creation-annihilation operators, 
if among the integral kernel operators $\Xi_{\ell, m}({\kappa}_{\ell, m})$ we count for also the scalar
integral kernel operator
\[
\Xi_{0, 0}(\kappa_{0,0}(\phi \otimes \varphi)) = \kappa_{0,0}(\phi \otimes \varphi)  \, \boldsymbol{1}  
=  \langle \kappa_{0,0}, \phi \otimes \varphi \rangle \, \boldsymbol{1}, 
\]
with the kernel $\kappa_{0, 0} \in \mathscr{E}^{*} \otimes \mathscr{E}^{*}$, which together with the remaining $\kappa_{\ell,m}$ in the Fock expansion,
is determined by the kernels $\kappa_{0,1}, \kappa_{1,0}$ defining the free fields $\mathbb{A}_{{}_{k}}$ involved into the Wick product. 

This gives us the Wick theorem for ``product'' (\ref{Wick(x)Wick(y)EU}) of the Wick product factors of free fields
$\mathbb{A}_{{}_{k}}$ or their derivatives on the Einstein Universe, which gives a strict mathematical sense to the 
``Wick theorem'' of \cite{Bogoliubov_Shirkov}, \S 17.2, on the Einstein Universe space-time, compare Subsection \ref{WickForProduct}.

Thus we can summarize the results in the Wick theorem for the tensor product (\ref{Wick(x)Wick(y)EU}) of normally ordered free field factors 
on the Einstein Universe, compare Subsection \ref{WickForProduct}. 
Namely, if $\mathbb{A}_{{}_{k}}$, $k=1, \ldots, M$  
in (\ref{Wick(x)Wick(y)}) are free massless or massive fields on the Einstein Universe, then the operators
\begin{eqnarray*}
\Xi'(\phi) = \sum \limits_{a} \int\limits_{\widetilde{\mathbb{S}^1} \times SU(2, \mathbb{C})}  
\boldsymbol{{:}} \mathbb{A}_{{}_{1}}^{a}(x) \ldots \mathbb{A}_{{}_{N}}^{a}(x) \boldsymbol{{:}} \, \phi^a(x) \,\,\,\, \ud^4x
\,\,\,\,
\in \mathscr{L}((\boldsymbol{E}),(\boldsymbol{E})),
\\
\Xi''(\varphi) = \sum \limits_{b} \int\limits_{\widetilde{\mathbb{S}^1} \times SU(2, \mathbb{C})}  
\boldsymbol{{:}} \mathbb{A}_{{}_{N+1}}^{b}(y) \ldots \mathbb{A}_{{}_{M}}^{b}(y)  \boldsymbol{{:}}  \,\,\,\, \varphi^b(y) \, \ud^4y
\,\,\,\,
\in \mathscr{L}((\boldsymbol{E}),(\boldsymbol{E})),
\\
\Xi(\phi \otimes \varphi) = 
\sum \limits_{a,b}
\int\limits_{\big[\widetilde{\mathbb{S}^1} \times SU(2, \mathbb{C})\big]^{\times \, 2}} 
\boldsymbol{{:}} \mathbb{A}_{{}_{1}}^{a}(x) \ldots \mathbb{A}_{{}_{N}}^{a}(x) \boldsymbol{{:}} \,\,
\boldsymbol{{:}} \mathbb{A}_{{}_{N+1}}^{b}(y) \ldots \mathbb{A}_{{}_{M}}^{b}(y)  \boldsymbol{{:}} \,\, \times
\\ \times \,\,\,\,\,\,\,
\phi^a \otimes \varphi^b(x,y) \, \ud^4x \ud^4y \,\,\,\,\,\, \in \mathscr{L}((\boldsymbol{E}),(\boldsymbol{E})),
\end{eqnarray*} 
and moreover
\begin{eqnarray*}
\Xi' \in \mathscr{L}(\mathscr{E}, \mathscr{L}((\boldsymbol{E}),(\boldsymbol{E}))\big),
\\
\Xi''\in \mathscr{L}(\mathscr{E}, \mathscr{L}((\boldsymbol{E}),(\boldsymbol{E}))\big),
\\
\Xi\in \mathscr{L}(\mathscr{E} \otimes \mathscr{E}, \mathscr{L}((\boldsymbol{E}),(\boldsymbol{E}))\big),
\end{eqnarray*}
and the following Wick theorem for the generalized operator $\Xi$ with kernel (\ref{Wick(x)Wick(y)EU}) holds
\begin{twr}
\[
\Xi =
\sum_{\substack{\kappa'_{\ell',m'} \\ \kappa''_{\ell'',m''}}} \sum \limits_{0\leq q \leq \textrm{min} \{m',\ell''\}} (-1)^{c(q)} \,
\Xi_{\ell'+\ell''-q,m'+m''-q}\big( \kappa'_{\ell',m'}\overline{\otimes_{q}} \, {\kappa''}_{\ell'',m''}\big),
\]
where the kernels $\kappa'_{\ell',m'}, \kappa''_{\ell'',m''}$ range, respectively, over the  kernels of finite Fock expansions 
of the operators $\Xi', \Xi''$. The (symmetrized/antisymmetrized) $q$-contractions $\overline{\otimes_{q}}$ are performed
upon the pairs of variables in which the first element of the contracted pair lies among the last $m''$ variables of the kernel 
$\kappa'_{\ell',m'}$ and the second variable
of the contracted pair lies among the first $l''$ variables of the kernel $\kappa''_{\ell'',m''}$, and to both variables
of the contracted pair correspond respectively annihilation and creation operator of one and the same free field. The number 
$c(q)$ is equal to the number of fermi commutations performed in the contraction $\overline{\otimes_{q}}$. 
\label{WickEU}
\end{twr}

Note that on the Einstein Universe the ``formal'' or ``symbolic'' version of the Wick theorem (\cite{Bogoliubov_Shirkov}, \S 17.2,
\cite{Bogoliubov-Shirkov}, \S 17.2, the ``first Wick theorem''),
becomes strictly true without any need for ``regularization'' if among the  free field factors 
\[
\mathbb{A}_{{}_{1}}^{a}(x) \ldots \mathbb{A}_{{}_{N}}^{a}(x) 
\]
or among the free field factors
\[
 \mathbb{A}_{{}_{N+1}}^{b}(y) \ldots \mathbb{A}_{{}_{M}}^{b}(y)
\]
in (\ref{Wick(x)Wick(y)EU}) there is at most one massless field with the corresponding orbit $\mathscr{O}_\pm$ which is an infinite set, and with
all remaining free fields $\mathbb{A}_{{}_{i}}$ which are massive and respecting
\[
\big[\square -m_{i}^2 \big] \mathbb{A}_{{}_{i}} = 0, \,\,\,\, m_i \neq 0.
\]
Thus all, eventually all but one, of the factor fields $\mathbb{A}_{{}_{i}}$ of the first normal factor in  (\ref{Wick(x)Wick(y)EU}) 
have corresponding orbits
$\mathscr{O}_\pm$ finite, \emph{i.e.} are finite sets with the operators evaluated at spacetime point
\[
\mathbb{A}_{{}_{i}}(x) \in \mathscr{L}\big((\boldsymbol{E}), (\boldsymbol{E})\big)
\]
which are ordinary operators in the Fock space transforming continuously the Hida space $(\boldsymbol{E})$ into itself, compare Subsection
\ref{WhiteNoiseFreeFieldsonEU}. The same situation we have by assumption for the second normal factor in (\ref{Wick(x)Wick(y)EU}).

Recall that having given a free field $\mathbb{A}'$ with single particle Gelfand triple $E \subset \mathcal{H} \subset E^*$ 
and defining kernels
\[ 
\kappa_{0,1}, \kappa_{1,0} \in \mathscr{L}(\mathscr{E}, E) \cong E \otimes \mathscr{E}^*,
\]
\emph{i.e.}
\[
\mathbb{A}' = \Xi_{0,1}(\kappa_{0,1}) + \Xi_{1,0}(\kappa_{1,0}) = \mathbb{A}'^{(-)} + \mathbb{A}'^{(+)},
\]
the pairing distribution
\[
\quad \underbracket{\mathbb{A}'(x) \mathbb{A}'}(y) = [\mathbb{A}'^{(-)}(x), \mathbb{A}'^{(+)}(y)]_\pm
= -i D'^{(-)}(xy^{-1}) \, \boldsymbol{1}
\]
can be computed as the contraction
\[
\langle \kappa'_{0,1}, \kappa'_{1,0} \rangle(x,y) = \kappa'_{0,1} \otimes_1 \kappa'_{1,0}(x,y) 
\]
and represents a distribution
\[
\langle \kappa'_{0,1}, \kappa'_{1,0} \rangle = \kappa'_{0,1} \otimes_1 \kappa'_{1,0} \in \mathscr{E}^* \otimes \mathscr{E}^*.
\]
If $\mathbb{A}''$ is another such field with defining kernels $\kappa''_{0,1}, \kappa''_{1,0}$, and at least one of the
fields  $\mathbb{A}', \mathbb{A}''$ has finite orbit $\mathscr{O}'_{\pm}$ or $\mathscr{O}''_{\pm}$ (which is the case when one of them is massive
and respects $\square \mathbb{A}' = m'^{2} \mathbb{A}'$ or $\square \mathbb{A}'' = m''^{2} \mathbb{A}''$ with $m',m''\neq 0$)
then in this particular case the contraction
\[
\big(\kappa'_{0,1} \dot{\otimes} \kappa''_{0,1}\big) \otimes_2 
\big(\kappa'_{1,0} \dot{\otimes} \kappa''_{1,0}\big)(x,y) 
\] 
is equal to the pointwise product (with respect to space-time variable)
\begin{multline*}
\big(\kappa'_{0,1} \dot{\otimes} \kappa''_{0,1}\big) \otimes_2 
\big(\kappa'_{1,0} \dot{\otimes} \kappa''_{1,0}\big)(x,y) = \kappa'_{0,1} \otimes_1 \kappa'_{1,0}(x,y) \kappa''_{0,1} \otimes_1 \kappa''_{1,0}(x,y)
\\
= (-i) D'^{(-)}(xy^{-1}) (-i) D''^{(-)}(xy^{-1})
\end{multline*}
of the pairing functions $D'^{(-)}, D''^{(-)}$, well defined because the orbit of the massive field is finite and the pairing of the massive field is an 
ordinary smooth function on the compactified Einstein Universe. Therefore, from the Wick Theorem \ref{WickEU} we obtain the following
\begin{cor*}
When all, or all except one, factor fields $\mathbb{A}_{{}_{i}}$ are massive in each of the normal factors
of (\ref{Wick(x)Wick(y)EU}), then the ``formal'' Wick theorem with pairings (compare \cite{Bogoliubov_Shirkov}, \S 17.2):
\begingroup\makeatletter\def\f@size{5}\check@mathfonts
\def\maketag@@@#1{\hbox{\m@th\large\normalfont#1}}%
\begin{multline*}
\boldsymbol{{:}} \mathbb{A}_{{}_{1}}^{a}(x) \ldots \mathbb{A}_{{}_{N}}^{a}(x) \boldsymbol{{:}} \,\,
\boldsymbol{{:}} \mathbb{A}_{{}_{N+1}}^{b}(y) \ldots \mathbb{A}_{{}_{M}}^{b}(y)  \boldsymbol{{:}}
\\
= \,\,\,\,\,\,\,\,
\boldsymbol{{:}} \mathbb{A}_{{}_{1}}^{a}(x)  \ldots \mathbb{A}_{{}_{N}}^{a}(x) 
 \mathbb{A}_{{}_{N+1}}^{b}(y) \ldots \mathbb{A}_{{}_{M}}^{b}(y)  \boldsymbol{{:}}
\\
+
\sum_{\substack{1\leq i \leq N \\ 1 \leq j \leq M-N}}
\boldsymbol{{:}} \mathbb{A}_{{}_{1}}^{a}(x) \ldots \quad \underbracket{\mathbb{A}_{{}_{i}}^{a}(x) \ldots \mathbb{A}_{{}_{N}}^{a}(x) 
 \mathbb{A}_{{}_{N+1}}^{b}(y) \ldots \mathbb{A}_{{}_{N+j}}^{b}}(y) \ldots \mathbb{A}_{{}_{M}}^{b}(y)  \boldsymbol{{:}}
\\
+
\sum_{\substack{1\leq i \leq N \\ 1 \leq j \leq M-N \\ 1 \leq k \leq N \\ 1 \leq n \leq M-N}} 
\boldsymbol{{:}} \mathbb{A}_{{}_{1}}^{a}(x) \ldots \quad \underbracket{\mathbb{A}_{{}_{i}}^{a}(x) \ldots 
\quad \underbracket{\mathbb{A}_{{}_{k}}^{a}(x) \ldots  \mathbb{A}_{{}_{N}}^{a}(x) 
 \mathbb{A}_{{}_{N+1}}^{b}(y) \ldots \mathbb{A}_{{}_{N+j}}^{b}}(y) \ldots \mathbb{A}_{{}_{N+n}}^{b}}(y)  \ldots \mathbb{A}_{{}_{M}}^{b}(y)  \boldsymbol{{:}}
\\
+ \ldots
\\
= \,\,\,\,\,\,\,\,
\boldsymbol{{:}} \mathbb{A}_{{}_{1}}^{a}(x)  \ldots \mathbb{A}_{{}_{N}}^{a}(x) 
 \mathbb{A}_{{}_{N+1}}^{b}(y) \ldots \mathbb{A}_{{}_{M}}^{b}(y)  \boldsymbol{{:}}
\\
+
\sum_{\substack{1\leq i \leq N \\ 1 \leq j \leq M-N}}
(-1)^{p_{ij}} \, \quad \underbracket{\mathbb{A}_{{}_{i}}^{a}(x) \mathbb{A}_{{}_{N+j}}^{b}}(y) \,
\boldsymbol{{:}} \mathbb{A}_{{}_{1}}^{a}(x)  \ldots \overbrace{\mathbb{A}_{{}_{i}}^{a}(x)}^{\textrm{deleted}} \ldots  
 \ldots \overbrace{\mathbb{A}_{{}_{N+j}}^{b}(y)}^{\textrm{deleted}} \ldots  \mathbb{A}_{{}_{M}}^{b}(y)  \boldsymbol{{:}}
\\
+
\sum_{\substack{1\leq i \leq N \\ 1 \leq j \leq M-N \\ 1 \leq k \leq N \\ 1 \leq n \leq M-N}} 
(-1)^{p_{in}} \, (-1)^{p_{kj}} \,  \quad \underbracket{\mathbb{A}_{{}_{i}}^{a}(x) \mathbb{A}_{{}_{N+n}}^{b}}(y) \,
\quad \underbracket{\mathbb{A}_{{}_{k}}^{a}(x) \mathbb{A}_{{}_{N+j}}^{b}}(y) \,\,\, \times
\\
\times \,\,\,
\boldsymbol{{:}} \mathbb{A}_{{}_{1}}^{a}(x)  \ldots \overbrace{\mathbb{A}_{{}_{i}}^{a}(x)}^{\textrm{deleted}} \ldots  
\overbrace{\mathbb{A}_{{}_{k}}^{a}(x)}^{\textrm{deleted}} \ldots
 \ldots \overbrace{\mathbb{A}_{{}_{N+j}}^{b}(y)}^{\textrm{deleted}} \ldots
\ldots \overbrace{\mathbb{A}_{{}_{N+n}}^{b}(y)}^{\textrm{deleted}}
\ldots  \mathbb{A}_{{}_{M}}^{b}(y)  \boldsymbol{{:}}
\\
+ \ldots
\end{multline*}
\endgroup
is rigorously true, without the need for regularization. 
\end{cor*}
\qedsymbol \,
Indeed, the pairing functions of the massive free field factors become ordinary smooth functions in $\mathscr{E}^{\otimes \, 2}$,
because their orbits $\mathscr{O}_\pm$ are finite and thus such pairings are multipliers of $\mathscr{E}^{* \otimes \, 2}$, 
compare Subsection \ref{WhiteNoiseFreeFieldsonEU}. Therefore, 
if a term,
\begingroup\makeatletter\def\f@size{5}\check@mathfonts
\def\maketag@@@#1{\hbox{\m@th\large\normalfont#1}}%
\[
\quad \underbracket{\mathbb{A}_{{}_{i}}^{a}(x) \mathbb{A}_{{}_{N+n}}^{b}}(y) \ldots
\quad \underbracket{\mathbb{A}_{{}_{k}}^{a}(x) \mathbb{A}_{{}_{N+j}}^{b}}(y) \,\,\,\,\,
\boldsymbol{{:}} \mathbb{A}_{{}_{1}}^{a}(x)  \ldots \overbrace{\mathbb{A}_{{}_{i}}^{a}(x)}^{\textrm{deleted}} \ldots  
\overbrace{\mathbb{A}_{{}_{k}}^{a}(x)}^{\textrm{deleted}} \ldots
 \ldots \overbrace{\mathbb{A}_{{}_{N+j}}^{b}(y)}^{\textrm{deleted}} \ldots
\ldots \overbrace{\mathbb{A}_{{}_{N+n}}^{b}(y)}^{\textrm{deleted}}
\ldots  \mathbb{A}_{{}_{M}}^{b}(y)  \boldsymbol{{:}}
\]
\endgroup
in the above normal order expansion, has among the front pairings
\[
\quad \underbracket{\mathbb{A}_{{}_{i}}^{a}(x) \mathbb{A}_{{}_{N+n}}^{b}}(y) \ldots
\quad \underbracket{\mathbb{A}_{{}_{k}}^{a}(x) \mathbb{A}_{{}_{N+j}}^{b}}(y)
\]
the pairing of massless fields (and by assumption there is at most one such pairing in each term, if any at all), 
then the remaining factor pairings are smooth multipliers, and thus the whole 
product of pairings is a well defined distribution in $\mathscr{E}^{* \, \otimes \, 2}$ which is multiplied by the ordinary operator
\[
\boldsymbol{{:}} \mathbb{A}_{{}_{1}}^{a}(x)  \ldots \overbrace{\mathbb{A}_{{}_{i}}^{a}(x)}^{\textrm{deleted}} \ldots  
\overbrace{\mathbb{A}_{{}_{k}}^{a}(x)}^{\textrm{deleted}} \ldots
 \ldots \overbrace{\mathbb{A}_{{}_{N+j}}^{b}(y)}^{\textrm{deleted}} \ldots
\ldots \overbrace{\mathbb{A}_{{}_{N+n}}^{b}(y)}^{\textrm{deleted}}
\ldots  \mathbb{A}_{{}_{M}}^{b}(y)  \boldsymbol{{:}} 
\]
which belongs to
\[
\mathscr{L}\big((\boldsymbol{E}), (\boldsymbol{E})\big),
\]
so that the whole term 
\begingroup\makeatletter\def\f@size{5}\check@mathfonts
\def\maketag@@@#1{\hbox{\m@th\large\normalfont#1}}%
\[
\quad \underbracket{\mathbb{A}_{{}_{i}}^{a}(x) \mathbb{A}_{{}_{N+n}}^{b}}(y) \ldots
\quad \underbracket{\mathbb{A}_{{}_{k}}^{a}(x) \mathbb{A}_{{}_{N+j}}^{b}}(y) \,\,\,\,\,
\boldsymbol{{:}} \mathbb{A}_{{}_{1}}^{a}(x)  \ldots \overbrace{\mathbb{A}_{{}_{i}}^{a}(x)}^{\textrm{deleted}} \ldots  
\overbrace{\mathbb{A}_{{}_{k}}^{a}(x)}^{\textrm{deleted}} \ldots
 \ldots \overbrace{\mathbb{A}_{{}_{N+j}}^{b}(y)}^{\textrm{deleted}} \ldots
\ldots \overbrace{\mathbb{A}_{{}_{N+n}}^{b}(y)}^{\textrm{deleted}}
\ldots  \mathbb{A}_{{}_{M}}^{b}(y)  \boldsymbol{{:}}
\]
\endgroup
represents a well defined operator in 
\[
\mathscr{L}\big(\mathscr{E}, \,  \mathscr{L}\big((\boldsymbol{E}), (\boldsymbol{E})\big)\big).
\]
Similar situation we have for the term 
\begingroup\makeatletter\def\f@size{5}\check@mathfonts
\def\maketag@@@#1{\hbox{\m@th\large\normalfont#1}}%
\[
\quad \underbracket{\mathbb{A}_{{}_{i}}^{a}(x) \mathbb{A}_{{}_{N+n}}^{b}}(y) \ldots
\quad \underbracket{\mathbb{A}_{{}_{k}}^{a}(x) \mathbb{A}_{{}_{N+j}}^{b}}(y) \,\,\,\,
\boldsymbol{{:}} \mathbb{A}_{{}_{1}}^{a}(x)  \ldots \overbrace{\mathbb{A}_{{}_{i}}^{a}(x)}^{\textrm{deleted}} \ldots  
\overbrace{\mathbb{A}_{{}_{k}}^{a}(x)}^{\textrm{deleted}} \ldots
 \ldots \overbrace{\mathbb{A}_{{}_{N+j}}^{b}(y)}^{\textrm{deleted}} \ldots
\ldots \overbrace{\mathbb{A}_{{}_{N+n}}^{b}(y)}^{\textrm{deleted}}
\ldots  \mathbb{A}_{{}_{M}}^{b}(y)  \boldsymbol{{:}}
\]
\endgroup
in front of which there are no pairings of massless fields, and thus all the pairings in it are smooth. Because the 
operator 
\[
\boldsymbol{{:}} \mathbb{A}_{{}_{1}}^{a}(x)  \ldots \overbrace{\mathbb{A}_{{}_{i}}^{a}(x)}^{\textrm{deleted}} \ldots  
\overbrace{\mathbb{A}_{{}_{k}}^{a}(x)}^{\textrm{deleted}} \ldots
 \ldots \overbrace{\mathbb{A}_{{}_{N+j}}^{b}(y)}^{\textrm{deleted}} \ldots
\ldots \overbrace{\mathbb{A}_{{}_{N+n}}^{b}(y)}^{\textrm{deleted}}
\ldots  \mathbb{A}_{{}_{M}}^{b}(y)  \boldsymbol{{:}} 
\]
represents a generalized operator in
\[
\mathscr{L}\big(\mathscr{E}^{\otimes \, 2}, \,  \mathscr{L}\big((\boldsymbol{E}), (\boldsymbol{E})\big)\big)
\] 
then multiplied by a smooth element of $\mathscr{E}^{\otimes \, 2}$ will stay in
\[
\mathscr{L}\big(\mathscr{E}^{\otimes \, 2}, \,  \mathscr{L}\big((\boldsymbol{E}), (\boldsymbol{E})\big)\big).
\]
Thus all terms of the formal Wick theorem become, in the considered special case, well defined operators of
\[
\mathscr{L}\big(\mathscr{E}^{\otimes \, 2}, \,  \mathscr{L}\big((\boldsymbol{E}), (\boldsymbol{E})\big)\big).
\]
It can be easily checked that indeed they represent Fock expansion of the operator $\Xi$ with the kernel (\ref{Wick(x)Wick(y)EU})
by explicit computation of the contractions
\[
\kappa'_{\ell',m'}\overline{\otimes_{q}} \, {\kappa''}_{\ell'',m''}
\]
in the Theorem \ref{WickEU}.
\qed

Summing up, using the rigorous version of the Wick Theorem \ref{WickEU} or its Corollary
for the tensor product (\ref{Wick(x)Wick(y)EU}) in the situation, where we have at most one massless free field $\mathbb{A}_{{}_{k}}$ in 
each of the normal factors of (\ref{Wick(x)Wick(y)EU}),
we can write the kernel (\ref{Wick(x)Wick(y)EU}) in the following normal decomposition:
\begingroup\makeatletter\def\f@size{5}\check@mathfonts
\def\maketag@@@#1{\hbox{\m@th\large\normalfont#1}}%
\begin{multline}\label{Wick(x)Wick(y)NormalKernelDecompositionEU}
\boldsymbol{{:}} \mathbb{A}_{{}_{1}}^{a}(x) \ldots \mathbb{A}_{{}_{N}}^{a}(x) \boldsymbol{{:}} \,\,
\boldsymbol{{:}} \mathbb{A}_{{}_{N+1}}^{b}(y) \ldots \mathbb{A}_{{}_{M}}^{b}(y)  \boldsymbol{{:}}
\\
= \,\,\,\,\,\,\,\,
\boldsymbol{{:}} \mathbb{A}_{{}_{1}}^{a}(x)  \ldots \mathbb{A}_{{}_{N}}^{a}(x) 
 \mathbb{A}_{{}_{N+1}}^{b}(y) \ldots \mathbb{A}_{{}_{M}}^{b}(y)  \boldsymbol{{:}}
\\
+
\sum_{\substack{1\leq i \leq N \\ 1 \leq j \leq M-N}}
t^{ij}(xy^{-1}) \,
\boldsymbol{{:}} \mathbb{A}_{{}_{1}}^{a}(x)  \ldots \overbrace{\mathbb{A}_{{}_{i}}^{a}(x)}^{\textrm{deleted}} \ldots  
 \ldots \overbrace{\mathbb{A}_{{}_{N+j}}^{b}(y)}^{\textrm{deleted}} \ldots  \mathbb{A}_{{}_{M}}^{b}(y)  \boldsymbol{{:}}
\\
+
\sum_{\substack{1\leq i \leq N \\ 1 \leq j \leq M-N \\ 1 \leq k \leq N \\ 1 \leq n \leq M-N}} 
t^{inkj}(xy^{-1}) \,\,\, \times
\\
\times \,\,\,
\boldsymbol{{:}} \mathbb{A}_{{}_{1}}^{a}(x)  \ldots \overbrace{\mathbb{A}_{{}_{i}}^{a}(x)}^{\textrm{deleted}} \ldots  
\overbrace{\mathbb{A}_{{}_{k}}^{a}(x)}^{\textrm{deleted}} \ldots
 \ldots \overbrace{\mathbb{A}_{{}_{N+j}}^{b}(y)}^{\textrm{deleted}} \ldots
\ldots \overbrace{\mathbb{A}_{{}_{N+n}}^{b}(y)}^{\textrm{deleted}}
\ldots  \mathbb{A}_{{}_{M}}^{b}(y)  \boldsymbol{{:}}
\\
+ \ldots + \overbrace{t^{in\ldots kj}(xy^{-1})}^{\kappa_{0,0}(x-y)} \,\, \boldsymbol{1},
\end{multline}
\endgroup
with the well defined and $\widetilde{\mathbb{S}^1} \times SU(2, \mathbb{C})$-invariant distributions
\[
t^{ij}, t^{inkj}, \ldots, t^{in\ldots kj} \in \mathscr{E}^*
\]
defining the kernels 
\[
t^{ij}(xy^{-1}), t^{inkj}(xy^{-1}), \ldots, t^{in\ldots kj}(xy^{-1})
\]
of $\widetilde{\mathbb{S}^1} \times SU(2, \mathbb{C})$-invariant distributions in $\mathscr{E}^*\otimes \mathscr{E}^*$. Note that the distributions
\[
t^{ij}, t^{inkj}, \ldots, t^{in\ldots kj} \in \mathscr{E}^*
\]
can be easily read-off from the rigorous version of the Wick Theorem \ref{WickEU}, on application of the product formula
(\ref{ProductFormOfContractionOneMassless(x,y)}) of Subsection \ref{WhiteNoiseFreeFieldsonEU},
by the repeated application of the pointed product $\dot{\otimes}$, contraction $\otimes_q$ 
and finally by symmetrization and antisymmetrization. 

Note that the retarded and advanced part of each scalar distribution $t^{in\ldots j}$ in (\ref{Wick(x)Wick(y)NormalKernelDecompositionEU})
can be computed through pointwise multiplication by the periodic step theta function, in accordance with the results of the
previous Subsection, because each  $t^{in\ldots j}$ contains at most one pairing of massless fields as a factor.

Moreover, in case we have at most one mas less field in each normal factor
(\ref{WickMonomialsInGeneralFreeFieldsEU}) of (\ref{Wick(x)Wick(y)EU}), each of the normal contributions
to (\ref{Wick(x)Wick(y)NormalKernelDecompositionEU}) has an extraordinary property. Namely, consider a contribution
\begingroup\makeatletter\def\f@size{5}\check@mathfonts
\def\maketag@@@#1{\hbox{\m@th\large\normalfont#1}}%
\begin{equation}\label{GeneralContributionToS2EU}
t^{ink\ldots j}(xy^{-1}) \,\,\, 
\boldsymbol{{:}} \mathbb{A}_{{}_{1}}^{a}(x)  \ldots \overbrace{\mathbb{A}_{{}_{i}}^{a}(x)}^{\textrm{deleted}} \ldots  
\overbrace{\mathbb{A}_{{}_{k}}^{a}(x)}^{\textrm{deleted}} \ldots
 \ldots \overbrace{\mathbb{A}_{{}_{N+j}}^{b}(y)}^{\textrm{deleted}} \ldots
\ldots \overbrace{\mathbb{A}_{{}_{N+n}}^{b}(y)}^{\textrm{deleted}}
\ldots  \mathbb{A}_{{}_{M}}^{b}(y)  \boldsymbol{{:}}
\end{equation}
\endgroup
with the paired free fields  
\[
\mathbb{A}_{{}_{i}}^{a}, \mathbb{A}_{{}_{k}}^{a}, \mathbb{A}_{{}_{N+n}}, \ldots\mathbb{A}_{{}_{N+j}}
\]
removed from the normal product in (\ref{GeneralContributionToS2EU}).
Each such contribution (\ref{GeneralContributionToS2EU}) has the following important property. 
Namely let 
\[
\sum\limits_{\ell',m'} \Xi'_{\ell',m'}(\kappa'_{\ell',m'}(x)) = \,\,\,\,\,
\boldsymbol{{:}} \mathbb{A}_{{}_{1}}^{a}(x)  \ldots \overbrace{\mathbb{A}_{{}_{i}}^{a}(x)}^{\textrm{deleted}} \ldots   \mathbb{A}_{{}_{N}}^{a}(x) \boldsymbol{{:}}
\]
and 
\[
 \sum\limits_{\ell'',m''} \Xi''_{\ell'',m''}(\kappa''_{\ell'',m''}(y))  = \,\,\,\,\,
\boldsymbol{{:}} \mathbb{A}_{{}_{N+1}}^{b}(y) \ldots \overbrace{\mathbb{A}_{{}_{N+j}}^{b}(y)}^{\textrm{deleted}} \ldots  \mathbb{A}_{{}_{M}}^{b}(y)  \boldsymbol{{:}}.
\]
be the normal products of the free fields which are present in (\ref{GeneralContributionToS2EU}).
Let
\[
\kappa'''_{\ell,m}(x,y) = \kappa'_{\ell',m'} \overline{\otimes} \kappa''_{\ell'',m''}(x,y) = \kappa'_{\ell',m'}(x)  \kappa''_{\ell'',m''}(y),
\,\,\, \ell = \ell'+ \ell'', \,\, m = m' +m'',
\]
be the kernels of the total normal product 
\begingroup\makeatletter\def\f@size{5}\check@mathfonts
\def\maketag@@@#1{\hbox{\m@th\large\normalfont#1}}%
\begin{equation}\label{NormalFactorOfTheGeneralContributionToS2EU}
\boldsymbol{{:}} \mathbb{A}_{{}_{1}}^{a}(x)  \ldots \overbrace{\mathbb{A}_{{}_{i}}^{a}(x)}^{\textrm{deleted}} \ldots  
\overbrace{\mathbb{A}_{{}_{k}}^{a}(x)}^{\textrm{deleted}} \ldots
 \ldots \overbrace{\mathbb{A}_{{}_{N+j}}^{b}(y)}^{\textrm{deleted}} \ldots
\ldots \overbrace{\mathbb{A}_{{}_{N+n}}^{b}(y)}^{\textrm{deleted}}
\ldots  \mathbb{A}_{{}_{M}}^{b}(y)  \boldsymbol{{:}}
\,\,\,\,\,
=
\sum \limits_{\ell,m}
\Xi(\kappa'''_{\ell,m}(x,y)) 
\end{equation}
\endgroup
in (\ref{GeneralContributionToS2EU}). The bar over $\otimes$ means that the kernel should be symmetrized in (discrete) Bose momentum and spin
variables and antisymmetrized in Fermi momentum and spin variables -- numbering the discrete points of the corresponding orbits.
Then we have the following alternative:
\begin{enumerate}
\item[1)]
The causal distribution $t^{ink\ldots j}(xy^{-1})$ in (\ref{GeneralContributionToS2EU}) is proper and not equal to any smooth function but 
the kernel $\kappa'''_{\ell,m}$ of the normal product (\ref{NormalFactorOfTheGeneralContributionToS2EU}) in
(\ref{GeneralContributionToS2EU}) is defined on the finite dimensional nuclear spaces
$E_1, \ldots E_{\ell+m}$, where 
\begin{eqnarray*}
E_{1} \subset \mathcal{H}_{1} \subset E_{1}^{*},
\\
\vdots
\\
E_{\ell+m} \subset \mathcal{H}_{\ell+m} \subset E_{\ell+m}^{*}
\end{eqnarray*}
are single particle and \emph{finite dimensional} Gelfand triples of the massive fields 
$\mathbb{A}_{{}_{1}}^{a}, \ldots, \mathbb{A}_{{}_{\ell+m}}^{a} $ in the normal product
(\ref{NormalFactorOfTheGeneralContributionToS2EU}). This is the case if the only pair of massless fields
is paired in (\ref{GeneralContributionToS2EU}).
\item[]
or
\item[2)]
The causal distribution $t^{ink\ldots j}(xy^{-1})$ in (\ref{GeneralContributionToS2EU}) is a smooth function 
and the kernel $\kappa'''_{\ell,m}$ of the normal product (\ref{NormalFactorOfTheGeneralContributionToS2EU})
can be extended to an element 
\[
\kappa'''_{\ell,m} \in  \mathscr{L}\big(E_{1}^{*} \otimes \ldots \otimes E_{\ell}^{*} \otimes E_{\ell+1} \otimes \ldots \otimes E_{\ell+m}, 
\,\, \mathscr{E}^{* \, \otimes \,2}  \big).
\]
This is the case if the pair of massless fields is not paired in (\ref{GeneralContributionToS2EU}).
\end{enumerate}
It is evident that in both cases, 1) and 2), of the alternative, the kernels
\[
\kappa_{\ell,m}(x,y) =  t^{ink\ldots j}(xy^{-1})\kappa_{\ell,m}(x,y) = t^{ink\ldots j}(xy^{-1})\kappa'_{\ell',m'}(x)\kappa''_{\ell'',m''}(y)
\]
of the contribution (\ref{GeneralContributionToS2EU}) and the kernels (discrete momentum and spin variables -- designating the discrete points of the corresponding orbits -- are not written explicitly)
\begin{align}
\textrm{ret} \kappa_{\ell,m}(x,y) & \overset{\textrm{df}}{=} \big(\textrm{ret} t^{ink\ldots j}\big)(xy^{-1})\kappa'_{\ell',m'}(x)\kappa''_{\ell'',m''}(y)
\label{retkappal,m(x,y)}
\\
\textrm{av} \kappa_{\ell,m}(x,y) & \overset{\textrm{df}}{=} \big(\textrm{av} t^{ink\ldots j}\big)(xy^{-1})\kappa'_{\ell',m'}(x)\kappa''_{\ell'',m''}(y)
\label{avkappal,m(x,y)}
\end{align}
can  be extended to elements 
\begin{equation}\label{extensionretkappaavkappa}
\kappa_{\ell,m}, \,\, \textrm{ret}\kappa_{\ell,m}, \,\, 
\textrm{av}\kappa_{\ell,m} 
\in  \mathscr{L}\big(E_{1}^{*} \otimes \ldots \otimes E_{\ell}^{*} \otimes E_{\ell+1} \otimes \ldots \otimes E_{\ell+m}, 
\,\, \mathscr{E}^{* \, \otimes \,2}  \big).
\end{equation}
Note also that in the first case 1) of the above alternative the kernels
\[
\kappa'_{\ell',m'}(\xi) \,\,\, \textrm{and} \,\,\, \kappa''_{\ell'',m''}(\eta) \,\,\, \in \mathscr{E} \subset \mathscr{E}^*
\]
are smooth, as functions of the space-time variable, 
for (with the appropriate permutation $\pi$ of $1, \ldots, \ell'+m'$ and a permutation
$\rho$ of $1, \ldots, \ell''+m''$)
\[
\xi \in E_{\pi(1)} \otimes \ldots E_{\pi(\ell'+m')} \,\,\,\,\,\, \eta \in E_{N+\rho(1)} \otimes \ldots E_{N+ \rho(\ell''+m'')} 
\]
and 
\[
\kappa'_{\ell',m'}(\xi) \,\,\, \textrm{and} \,\,\, \kappa''_{\ell'',m''}(\xi) 
\]
remin to be smooth for 
\begin{multline*}
\xi \in E_{\pi(1)}^{*} \otimes \ldots E_{\pi(\ell')}^{*} \otimes E_{\pi(\ell'+1)} \otimes \ldots \otimes E_{\pi(\ell'+m')} 
\\ 
\textrm{or} \,\,\, \eta \in E_{N+\rho(1)}^{*} \otimes \ldots E_{N+\rho(\ell'')}^{*} 
\otimes E_{N+ \rho(\ell''+ 1)} \otimes \ldots \otimes E_{N+\rho(\ell''+m'')}, 
\end{multline*}
as the nuclear spaces $E_\pi(1), \ldots $ are here, in case 1), finite dimensional.
Because  
\[
\kappa'_{\ell',m'} \in \mathscr{L}\big(E_{\pi(1)}^{*} \otimes \ldots E_{\pi(\ell')}^{*} \otimes E_{\pi(\ell'+1)} \otimes \ldots \otimes E_{\pi(\ell'+m')},
\,\, \mathscr{E}^* \big)
\]
and 
\[
\kappa''_{\ell'',m''} \in \mathscr{L}\big(E_{N+\rho(1)}^{*} \otimes \ldots E_{N+ \rho(\ell'')}^{*} 
\otimes E_{N+ \rho(\ell''+1)} \otimes \ldots \otimes E_{N+\rho(\ell''+m'')},
\,\, \mathscr{E}^* \big)
\]
then the convolution $t^{ink\ldots j} \ast \kappa'_{\ell',m'}$ and the dot products 
\begin{align*}
\big(t^{ink\ldots j} \ast \kappa'_{\ell',m'}\big) \dot{\otimes} \kappa''_{\ell'',m''}(y) 
& = (t^{ink\ldots j} \ast \kappa'_{\ell',m'}\big)(y) \kappa''_{\ell'',m''}(y),
\\
\big(\textrm{ret}t^{ink\ldots j} \ast \kappa'_{\ell',m'}\big) \dot{\otimes} \kappa''_{\ell'',m''}(y) 
& = \big(\textrm{ret}t^{ink\ldots j} \ast \kappa'_{\ell',m'}\big)(y) \kappa''_{\ell'',m''}(y),
\\
\big(\textrm{av}t^{ink\ldots j} \ast \kappa'_{\ell',m'}\big) \dot{\otimes} \kappa''_{\ell'',m''}(y) 
& = \big(\textrm{av}t^{ink\ldots j} \ast \kappa'_{\ell',m'}\big)(y) \kappa''_{\ell'',m''}(y),
\end{align*}
make sense and the maps
\begin{align*}
\xi \otimes \eta & \longrightarrow \kappa_{\ell,m}(\xi\otimes\eta)(x,y) & = t^{ink\ldots j}(xy^{-1})
\kappa'_{\ell',m'}(\xi)(x)\kappa''_{\ell'',m''}(\eta)(y),
\\
\xi \otimes \eta &  \longrightarrow \textrm{ret}  \kappa_{\ell,m}(\xi\otimes\eta)(x,y) & = \textrm{ret} t^{ink\ldots j}(xy^{-1})
\kappa'_{\ell',m'}(\xi)(x)\kappa''_{\ell'',m''}(\eta)(y),
\\
\xi \otimes \eta & \longrightarrow \textrm{av} \kappa_{\ell,m}(\xi\otimes\eta)(x,y) & = \textrm{av} t^{ink\ldots j}(xy^{-1})
\kappa'_{\ell',m'}(\xi)(x)\kappa''_{\ell'',m''}(\eta)(y),
\end{align*}
are continuous, when regarded as maps
\[
E_{\pi(1)}^{*} \otimes \ldots E_{\pi(\ell')}^{*} \otimes E_{\pi(\ell'+1)} \otimes \ldots \otimes E_{\pi(\ell'+m')}
\longrightarrow \mathscr{E}^{* \, \otimes \, 2},
\]
and the maps
\begin{align*}
\xi \otimes \eta \longrightarrow \big(t^{ink\ldots j} \ast \kappa'_{\ell',m'}\big) \dot{\otimes} \kappa''_{\ell'',m''}(y) 
= & \big(t^{ink\ldots j} \ast \kappa'_{\ell',m'}\big)(y) \kappa''_{\ell'',m''}(y),
\\
\xi \otimes \eta \longrightarrow \big(\textrm{ret}t^{ink\ldots j} \ast \kappa'_{\ell',m'}\big) \dot{\otimes} \kappa''_{\ell'',m''}(y) 
= & \big(t^{ink\ldots j} \ast \kappa'_{\ell',m'}\big)(y) \kappa''_{\ell'',m''}(y),
\\
\xi \otimes \eta \longrightarrow \big(\textrm{av}t^{ink\ldots j} \ast \kappa'_{\ell',m'}\big) \dot{\otimes} \kappa''_{\ell'',m''}(y) 
= &  \big(t^{ink\ldots j} \ast \kappa'_{\ell',m'}\big)(y) \kappa''_{\ell'',m''}(y),
\end{align*}
are continuous, when regarded as maps
\[
E_{\pi(1)}^{*} \otimes \ldots E_{\pi(\ell')}^{*} \otimes E_{\pi(\ell'+1)} \otimes \ldots \otimes E_{\pi(\ell'+m')}
\longrightarrow \mathscr{E}^{*}.
\]
Therefore the extendibility property (\ref{extensionretkappaavkappa}) of the kernels (\ref{retkappal,m(x,y)}) and
(\ref{retkappal,m(x,y)}) 
together with
\begin{multline}\label{t*kappa'.kappa''}
t^{ink\ldots j} \ast \kappa'_{\ell',m'} \dot{\otimes} \kappa''_{\ell'',m''}, \,\,
\big(\textrm{ret}t^{ink\ldots j} \ast \kappa'_{\ell',m'}\big) \dot{\otimes} \kappa''_{\ell'',m''}, \,\,
\big(\textrm{av}t^{ink\ldots j} \ast \kappa'_{\ell',m'}\big) \dot{\otimes} \kappa''_{\ell'',m''}
\\
\in 
\mathscr{L}\big(E_{1}^{*} \otimes \ldots \otimes E_{\ell}^{*} \otimes E_{\ell+1} \otimes \ldots \otimes E_{\ell+m}, 
\,\, \mathscr{E}^{* \, \otimes \,2}  \big), 
\end{multline}
\[
\ell = \ell'+\ell'', \,\, m = m' +m'',
\]
follows in the case 1) of the above alternative.
The same assertion holds, \emph{i.e.} the continuous extendibility property 
(\ref{extensionretkappaavkappa}) of the kernels (\ref{retkappal,m(x,y)}) and
(\ref{retkappal,m(x,y)})  and (\ref{t*kappa'.kappa''}) also
in the second case, 2), of the above alternative, because in the case 2) the scalar
distribution $t^{ink\ldots j}$ is a smooth function. By the general theory of
integral kernel operators (compare Subsection \ref{psiBerezin-Hida}, \ref{WhiteNoiseFreeFieldsonEU})
it follows that the integral kernel operators corresponding to the kernels
\begin{align*}
\kappa_{\ell,m}, \,\,\,\, \textrm{ret}\kappa_{\ell,m}, \,\,\,\,\, 
\textrm{av}\kappa_{\ell,m}, \,\,\,\,\,  \big(t^{ink\ldots j} \ast \kappa'_{\ell',m'}\big) \dot{\otimes} \kappa''_{\ell'',m''}, 
\\
\big(\textrm{ret}t^{ink\ldots j} \ast \kappa'_{\ell',m'}\big) \dot{\otimes} \kappa''_{\ell'',m''}, \,\,\,\,\,
\big(\textrm{av}t^{ink\ldots j} \ast \kappa'_{\ell',m'}\big) \dot{\otimes} \kappa''_{\ell'',m''}
\end{align*}
belong, respectively, to
\[
\mathscr{L}\big(\mathscr{E}^{\otimes \,2}, \,\, \mathscr{L}((\boldsymbol{E}), (\boldsymbol{E}) \big)
\,\,\, \textrm{or} \,\,\,
\mathscr{L}\big(\mathscr{E}, \,\, \mathscr{L}((\boldsymbol{E}), (\boldsymbol{E}) \big).
\]
This, as we will see below in this Subsection, implies that the $2$-nd order, and more generally, $n$-th order contribution $S_n$ 
to the scattering operator $S$ belongs to
\[
\mathscr{L}\big(\mathscr{E}^{\otimes \, n}, \,\, \mathscr{L}((\boldsymbol{E}), (\boldsymbol{E}) \big)
\]
only if the interaction Lagrange density generalized operator $\mathscr{L}$ contains at most one massless field
(as e.g. in QED). 

Of course, we can arrive at the conclusion (\ref{t*kappa'.kappa''}) immediately if we apply 
Theorem \ref{F*WickProdFreeFieldsOnEU} of Subsection \ref{WhiteNoiseFreeFieldsonEU}, from which it follows that
(\ref{t*kappa'.kappa''}) holds generally,
independently of the number of massless fields in (\ref{Wick(x)Wick(y)EU}).
However, the presented proof of the continuous extendibility property (\ref{extensionretkappaavkappa}) is applicable only 
in case we have at most one masless field in each normal factor
(\ref{WickMonomialsInGeneralFreeFieldsEU}) of (\ref{Wick(x)Wick(y)EU}).

Let us also emphasize once again  that according to the results of the previous Subsection, 
in this case of at most one massless free field in each normal factor, the retarded and advanced part of each scalar distribution  
$t^{ink\ldots j}$ can be realized naturally and uniquely by the pointwise multiplication by the (periodic) step
theta function $\theta$, because each  $t^{ink\ldots j}$ contains at most one pairing of massless fields
as a factor. This will subsantially simplify the algorithmic computation of the scattering matrix through the natural
definition of the chronological product and the Wick theorem in case when the interaction
$\mathcal{L}$ contains at most one mass less field. We will present this Wick theorem for this special case 
in the next Subsection.  For this reason we prefer to distinguish this special case.

Let, in particular, $M=2N$ and $\mathbb{A}_{{}_{k}} = \mathbb{A}_{{}_{N+k}}$, $k=1, \ldots, N$, with
\[
i\mathcal{L}(x) = \boldsymbol{{:}} \mathbb{A}_{{}_{1}}^{a}(x) \ldots \mathbb{A}_{{}_{N}}^{a}(x) \boldsymbol{{:}}
\] 
and 
\[
i\mathcal{L}(y) = \boldsymbol{{:}} \mathbb{A}_{{}_{N+1}}^{b}(y) \ldots \mathbb{A}_{{}_{N+N=M}}^{b}(y)  \boldsymbol{{:}}
\]
in (\ref{Wick(x)Wick(y)EU}) with $\mathcal{L}$ equal to the Lagrange interaction density of a QFT theory with at most
one massless free field in $\mathcal{L}$. 
The formula (\ref{Wick(x)Wick(y)NormalKernelDecompositionEU}) gives us the normal decomposition of the operator valued distribution
\[
A'_{(2)}(x,y) = -i^2 \mathcal{L}(x) \mathcal{L}(y)= \overline{S_1}(x)S_1(y) = -S_1(x)S_1(y),
\]
used in the construction of the kernel $S_2(x,y)$ of the second order contribution to the scattering matrix, compare
\ref{MotivationForHida}. Replacing $x$ and $y$ in the formula (\ref{Wick(x)Wick(y)NormalKernelDecompositionEU})
computed in this special case, we obtain from (\ref{Wick(x)Wick(y)NormalKernelDecompositionEU}) the normal decomposition
of the operator valued distribution 
\[
R'_{(2)}(x,y) = -i^2 \mathcal{L}(y) \mathcal{L}(x) = S_1(y)\overline{S_1}(x) = -S_1(y)S_1(x)
\]
needed in the computation of the second order term $S_2(x,y)$, compare
\ref{MotivationForHida}. Thus for the distribution kernel (compare
\ref{MotivationForHida})
\[
D_{(2)} = R'_{(2)} - A'_{(2)},
\]
we again obtain the formula similar to (\ref{Wick(x)Wick(y)NormalKernelDecompositionEU}), in fact the formula
(\ref{Wick(x)Wick(y)NormalKernelDecompositionEU}) plus the same expression (\ref{Wick(x)Wick(y)NormalKernelDecompositionEU}) with $x$ and $y$ interchanged. 
Thus, for the retarded part $R_{(2)}$ of $D_{(2)}$ we obtain again the formula (\ref{Wick(x)Wick(y)NormalKernelDecompositionEU})
minus the same expression with $x$ and $y$ interchanged, and with the scalar distributions 
\[
t^{ij}, t^{inkj}, \ldots, t^{in\ldots kj} 
\]
replaced by their retarded parts, computed according to the Epstein-Glaser-type splitting prescription, 
which reduces here to the ordinary pointwise multiplication by the step theta function, working
for these distributions in case $\mathcal{L}$ contains at most one massless (or infinte orbit) field. 
Similar expression (\ref{Wick(x)Wick(y)NormalKernelDecompositionEU})
minus the expression (\ref{Wick(x)Wick(y)NormalKernelDecompositionEU}) with interchanged $x$ and $y$
and with the scalar distributions 
\[
t^{ij}, t^{inkj}, \ldots, t^{in\ldots kj} 
\]
replaced with their advanced parts, we obtain for the advanced part $A_{(2)}$ of the distribution $D_{(2)}$.

Therefore the second order contribution
\[
S_2(x,y) = A_{(2)}(x,y) - A'_{(2)}(x,y)
\]
or equivalently
\[
S_2(x,y) = R_{(2)}(x,y) - R'_{(2)}(x,y),
\]
again has the general form (\ref{Wick(x)Wick(y)NormalKernelDecompositionEU}) with the scalar distributions
\[
t^{ij}, t^{inkj}, \ldots, t^{in\ldots kj} 
\]
replaced with the respective advanced and retarded parts, in any case with well defined 
$\widetilde{\mathbb{S}^1} \times SU(2, \mathbb{C})$-invariant distributions. Let for example (\ref{GeneralContributionToS2EU})
be one of the finite  contributions to $S_2(x,y)$. 
Let 
\[
 \Xi'_{\ell',m'}(\kappa'_{\ell',m'}(x)) = 
\boldsymbol{{:}} \mathbb{A}_{{}_{1}}^{a}(x)  \ldots \overbrace{\mathbb{A}_{{}_{i}}^{a}(x)}^{\textrm{deleted}} \ldots   
\mathbb{A}_{{}_{N}}^{a}(x) \boldsymbol{{:}}
\]
and 
\[
 \Xi''_{\ell'',m''}(\kappa''_{\ell'',m''}(y))  =
\boldsymbol{{:}} \mathbb{A}_{{}_{N+1}}^{b}(y) \ldots \overbrace{\mathbb{A}_{{}_{N+j}}^{b}(y)}^{\textrm{deleted}} \ldots  
\mathbb{A}_{{}_{2N}}^{b}(y)  \boldsymbol{{:}}.
\]
Therefore the value of the contribution 
to 
\[
S_2(g)=S_2(1 \otimes g_2) = \int S_2(x,y) \big(1_{{}_{\mathbb{R}^4}} \otimes g_2\big) (x, y) \, \ud^4 x \, \ud^4 y 
= \int S_2(x,y) g_2(y) \, \ud^4 x \, \ud^4 y, 
\]
\[
g(x,y) = \big(1_{{}_{\mathbb{R}^4}} \otimes g_2\big) (x, y) = 1_{{}_{\mathbb{R}^4}}(x)  g_2(y) = g_2(y), \,\, g_2 \in \mathscr{E},
\]
and coming from (\ref{GeneralContributionToS2EU}) is equal to the finite sum
of integral kernel operators
\begin{multline*}
\sum\limits_{\ell,m}\Xi_{\ell,m}(\kappa_{\ell,m}(g_n)) 
\\
= 
\int
\textrm{ret} \, t^{in \ldots j}(xy^{-1}) \,
\boldsymbol{{:}} \mathbb{A}_{{}_{1}}^{a}(x)  \ldots \overbrace{\mathbb{A}_{{}_{i}}^{a}(x)}^{\textrm{deleted}} \ldots  
 \ldots \overbrace{\mathbb{A}_{{}_{N+j}}^{b}(y)}^{\textrm{deleted}} \ldots  \mathbb{A}_{{}_{M}}^{b}(y)  \boldsymbol{{:}}
\, g_2(y)
\, 
\ud^4 x \, \ud^4 y
\end{multline*}
with 
\[
\kappa_{\ell,m} = \big(\textrm{ret} t^{in \ldots j} \ast \kappa'_{\ell',m'}\big) \dot{\otimes} \kappa''_{\ell'',m''},
\,\,\, \ell = \ell'+\ell'', m = m' + m''.
\]

Thus we see fom Theorem \ref{F*WickProdFreeFieldsOnEU} of Subsection \ref{WhiteNoiseFreeFieldsonEU} that
\[
\sum\limits_{\ell,m}\Xi_{\ell,m}(\kappa_{\ell,m}(g_n)) \in  \mathscr{L}(\boldsymbol{E}), (\boldsymbol{E}))
\]
and 
\[
\mathscr{E} \ni g_2 \longrightarrow \sum\limits_{\ell,m}\Xi_{\ell,m}(\kappa_{\ell,m}(g_2)) \in \mathscr{L}(\boldsymbol{E}), (\boldsymbol{E}) ),
\]
is continuous. The same conclusion aso follows from (\ref{t*kappa'.kappa''}). Note also that 
\[
\kappa'_{\ell',m'}(\xi), \kappa''_{\ell'',m''}(\eta) \in \mathcal{O}_C(\mathbb{R}^4) \subset \mathcal{O}_M(\mathbb{R}^4)
= \mathscr{E} \subset \mathscr{E}^*
\]
for 
\[
\xi \in E_{\pi(1)} \otimes \ldots \otimes E_{\pi(\ell'+m')},
\,\,\, \textrm{respectively} \,\,\,
\eta \in E_{N+\rho(1)} \otimes \ldots \otimes E_{N+ \rho(\ell''+m'')},
\]
\emph{i.e.} for elements of tensor products of the nuclear spaces $E_{\pi(k)}$ 
composing the single particle Gelfand triples $E_{\pi(k)} \subset \mathcal{H}_{\pi(k)} \subset E_{\pi(k)}^{*}$
of the fields
\[
\mathbb{A}_{{}_{1}}^{a}(x)  \ldots \overbrace{\mathbb{A}_{{}_{i}}^{a}(x)}^{\textrm{deleted}} \ldots \mathbb{A}_{{}_{N}}^{b}(x)
\]
or, respectively, composing the Gelfand triples  $E_{N+\rho(k)} \subset \mathcal{H}_{N+\rho(k)} \subset E_{N+\rho(k)}^{*}$ 
in the single particle Hilbert spaces  
of the fields
\[
\mathbb{A}_{{}_{N+1}}^{b}(y) \ldots \overbrace{\mathbb{A}_{{}_{N+j}}^{b}(y)}^{\textrm{deleted}} \ldots  \mathbb{A}_{{}_{2N}}^{b}(y),
\]
compare Subsection \ref{WhiteNoiseFreeFieldsonEU}. Moreover $\kappa'_{\ell',m'}(\xi), \kappa''_{\ell'',m''}(\eta)$ 
depend continuously, respectively, on $\xi$ and $\eta$, in the strong dual topology of $\mathscr{E}^*$ in the image. 
Therefore, the kernel $\kappa_{\ell, m}(\xi \otimes \eta)$ of the operator 
$\Xi_{\ell,m}(\kappa_{\ell, m}(\xi \otimes \eta))$, being equal
\[
\kappa_{\ell, m}(\xi \otimes \eta)(x,y) = \textrm{ret} t^{ij}(xy^{-1}) \kappa'_{\ell',m'}(\xi)(x) \kappa''_{\ell'',m''}(\eta)(y)
\]
represents a continuous map
\[
\xi \otimes \eta \longmapsto \kappa_{\ell, m}(\xi \otimes \eta) \in \mathscr{E}^{* \otimes \, 2}
\]
which by (\ref{extensionretkappaavkappa}) can be extended to
\[
\kappa_{\ell, m} \in \mathscr{L}\big(E_{1}^{*} \otimes \ldots \otimes E_{\ell}^{*} \otimes E_{\ell+1} \otimes E_{\ell+m}, \,\, \mathscr{E}^{* \otimes \, 2} \big)
\]
and by the general theory of integral kernel operators due to Hida, Obata and Sait\^o 
(Subsection \ref{psiBerezin-Hida}, Thm. \ref{obataJFA.Thm.3.13},
compare also Subsection \ref{WhiteNoiseFreeFieldsonEU}) 
\[
\sum\limits_{\ell,m}\Xi_{\ell,m}(\kappa_{\ell,m}(g_2)) \in  \mathscr{L} \big( \mathscr{E}^{\otimes \, 2}, \, 
\mathscr{L}(\boldsymbol{E}), (\boldsymbol{E})) \big).
\]

Presented analysis is the same for all remaining contributions to $S_2(1\otimes g_2)$
and by the same Theorem \ref{F*WickProdFreeFieldsOnEU} of Subsection \ref{WhiteNoiseFreeFieldsonEU} 
\[
S_2(1 \otimes g_2) = \int S_2(x,y) g_2(y) \, \ud^4 x \, \ud^4 y \in \mathscr{L}(\boldsymbol{E}), (\boldsymbol{E}))
\]
and 
\[
 \mathscr{E} \ni g_2 \longrightarrow S(1 \otimes g_2) \in \mathscr{L}(\boldsymbol{E}), (\boldsymbol{E}))
\]
is continuous, including the case with one massless field $\mathbb{A}_{{}_{k}}$ in the interaction lagrange density
\[
i\mathcal{L}(x) = \boldsymbol{{:}} \mathbb{A}_{{}_{1}}^{a}(x) \ldots \mathbb{A}_{{}_{N}}^{a}(x) \boldsymbol{{:}},
\] 
as e.g. in QED, and moreover, by (\ref{extensionretkappaavkappa})
\[
S_2 \in  \mathscr{L} \big( \mathscr{E}^{\otimes \, 2}, \, 
\mathscr{L}(\boldsymbol{E}), (\boldsymbol{E})) \big).
\]

Of course the Wick Theorem \ref{WickEU} can be applied in the computation of the higher order 
terms $S_n$ of the scattering operator (compare Subsection \ref{MotivationForHida})
and now it should be clear that the repeated application of Theorem \ref{F*WickProdFreeFieldsOnEU} of Subsection \ref{WhiteNoiseFreeFieldsonEU} 
and (\ref{extensionretkappaavkappa}) gives 
\[
S_n(1 \otimes 1 \ldots \otimes 1 \otimes g_n) \in \mathscr{L}(\boldsymbol{E}), (\boldsymbol{E})), \,\,\,\, g_n \in \mathscr{E},
\]
in case with at most one massless factor $\mathbb{A}_{{}_{k}}$ in the interaction density $\mathcal{L}$.

Moreover 
\[
\mathscr{E} \ni g_n \longrightarrow S_n(1 \otimes 1 \ldots \otimes 1 \otimes g_n) \in \mathscr{L}(\boldsymbol{E}), (\boldsymbol{E})), 
\]
is continuous, also in case there are massless factors $\mathbb{A}_{{}_{k}}$ in the interaction density $\mathcal{L}$, and moreover
\[
S_n \in \mathscr{L} \big(\mathscr{E}^{\otimes \, n}, \, \mathscr{L}(\boldsymbol{E}), (\boldsymbol{E})) \big), 
\]
if there is at most one massless field in the interaction density $\mathcal{L}$.

Therefore we can summarize aur analysis with 
\begin{twr}
Let $S_n$ be the $n$-th order contribution to the scattering integral kernel operator. Then
\begin{enumerate}
\item[1)]
\[
\mathscr{E} \ni g_n \longrightarrow S_n(1 \otimes 1 \ldots \otimes 1 \otimes g_n) \in \mathscr{L}(\boldsymbol{E}), (\boldsymbol{E})), 
\]
is continuous,
including the case with at most one massless factor  $\mathbb{A}_{{}_{k}}$ in the interaction density $\mathcal{L}$.
In particular
\[
S_n(1 \otimes 1 \ldots \otimes 1 \otimes 1) = S_n(1) \in \mathscr{L}(\boldsymbol{E}), (\boldsymbol{E}))
\]
because the constant function $1 \in \mathscr{E} = \mathcal{S}_{\Delta+1}(\widetilde{\mathbb{S}^1}\times SU(2, \mathbb{C}))$.
 \item[2)]
\[
S_n \in \mathscr{L} \big(\mathscr{E}^{\otimes \, n}, \, \mathscr{L}(\boldsymbol{E}), (\boldsymbol{E})) \big), 
\]
including the case with at most one massless factor $\mathbb{A}_{{}_{k}}$ in the interaction density $\mathcal{L}$.
\end{enumerate}
\label{Sin(SE)xn->((E)->(E))EU}
\end{twr}

The above theorem can be extended on the case in which we have more than one masless field, or more generally,
free field with infinite orbit in the interaction  Lagrange density $\mathcal{L}$. But in this case the the scattering operator
$S$ obtained is far not unique, due to the considerable freedom in the choice of the splitting in the computation
of the retarded and advanced parts of the scalar products of pairings (contractions).

We can proceed identically in general situation, where we have arbitrary number of massless free field factors $\mathbb{A}_{{}_{k}}$ in (\ref{Wick(x)Wick(y)EU}). 
We can use Wick Theorem \ref{WickEU} 
for the tensor product (\ref{Wick(x)Wick(y)EU}) of generalized operators,
and we can write the kernel (\ref{Wick(x)Wick(y)EU}) in the normal decomposition (\ref{Wick(x)Wick(y)NormalKernelDecompositionEU}) as above,
with the well defined $\widetilde{\mathbb{S}^1}\times SL(2, \mathbb{C})$-invariant distributions
\[
t^{ij}, t^{inkj}, \ldots, t^{in\ldots kj} \in \mathscr{E}^*
\]
and defining the kernels 
\[
t^{ij}(xy^{-1}), t^{inkj}(xy^{-1}), \ldots, t^{in\ldots kj}(xy^{-1})
\]
of $\widetilde{\mathbb{S}^1} \times SU(2, \mathbb{C})$-invariant distributions in $\mathscr{E}^*\otimes \mathscr{E}^*$. 
The scalar invariant distributions $t^{ij}, \ldots$ can be easily computed by application of the product
formula (\ref{ProductFormOfContraction(x,y)}) of Subsection \ref{WhiteNoiseFreeFieldsonEU} for the contraction kernels,
and are equal to the canonical scalar factors of the contraction vector-valued kernels.

Then we apply Theorem \ref{F*WickProdFreeFieldsOnEU} of Subsection \ref{WhiteNoiseFreeFieldsonEU}. 

As before we put $M=2N$ and $\mathbb{A}_{{}_{k}} = \mathbb{A}_{{}_{N+k}}$, $k=1, \ldots, N$, with
\[
i\mathcal{L}(x) = \boldsymbol{{:}} \mathbb{A}_{{}_{1}}^{a}(x) \ldots \mathbb{A}_{{}_{N}}^{a}(x) \boldsymbol{{:}}
\] 
and 
\[
i\mathcal{L}(y) = \boldsymbol{{:}} \mathbb{A}_{{}_{N+1}}^{b}(y) \ldots \mathbb{A}_{{}_{N+N=M}}^{b}(y)  \boldsymbol{{:}}
\]
in (\ref{Wick(x)Wick(y)NormalKernelDecompositionEU}) with $\mathcal{L}$ equal to the Lagrange interaction density of a QFT theory.
In this case the formula (\ref{Wick(x)Wick(y)NormalKernelDecompositionEU}) gives us, as before, the normal decomposition of the operator valued distribution
\[
A'_{(2)}(x,y) = -i^2 \mathcal{L}(x) \mathcal{L}(y),
\]
used in the construction of the kernel $S_2(x,y)$ of the second order contribution to the scattering matrix. 
Replacing $x$ and $y$ in the formula (\ref{Wick(x)Wick(y)NormalKernelDecompositionEU})
computed in this special case, we obtain from (\ref{Wick(x)Wick(y)NormalKernelDecompositionEU}) the normal decomposition
of the operator valued distribution 
\[
R'_{(2)}(x,y) = -i^2 \mathcal{L}(y) \mathcal{L}(x)
\]
needed in the computation of the second order term $S_2(x,y)$, compare
\ref{CausalSonEU} and \ref{MotivationForHida}. Thus, for the distribution kernel (compare
Subsections \ref{CausalSonEU} and \ref{MotivationForHida})
\[
D_{(2)} = R'_{(2)} - A'_{(2)},
\]
we again obtain the formula similar to (\ref{Wick(x)Wick(y)NormalKernelDecompositionEU}), in fact the formula
(\ref{Wick(x)Wick(y)NormalKernelDecompositionEU}) plus the same expression (\ref{Wick(x)Wick(y)NormalKernelDecompositionEU}) with $x$ and $y$ interchanged. 
Thus, for the retarded part $R_{(2)}$ of $D_{(2)}$ we obtain again the formula (\ref{Wick(x)Wick(y)NormalKernelDecompositionEU})
minus the same expression with $x$ and $y$ interchanged, and with the scalar distributions 
\[
t^{ij}, t^{inkj}, \ldots, t^{in\ldots kj} 
\]
replaced by their retarded parts, computed according to the Epstein-Glaser-type splitting prescription of Subsection \ref{WhiteNoiseFreeFieldsonEU},
(which now if $t^{in\ldots kj}$ is equal to a kernel which includes two, or more, contractions of massless plane wave kernels, is not unique),
however an operation which is well defined on the Einstein Universe by the results of Subsection \ref{WhiteNoiseFreeFieldsonEU}. 
Similar expression (\ref{Wick(x)Wick(y)NormalKernelDecompositionEU})
minus the expression (\ref{Wick(x)Wick(y)NormalKernelDecompositionEU}) with interchanged $x$ and $y$
and with the scalar distributions 
\[
t^{ij}, t^{inkj}, \ldots, t^{in\ldots kj} 
\]
replaced with their advanced parts, we obtain for the advanced part $A_{(2)}$ of the distribution $D_{(2)}$.

Therefore, the second order contribution
\[
S_2(x,y) = A_{(2)}(x,y) - A'_{(2)}(x,y)
\]
or equivalently
\[
S_2(x,y) = R_{(2)}(x,y) - R'_{(2)}(x,y),
\]
again has the general form (\ref{Wick(x)Wick(y)NormalKernelDecompositionEU}) with the scalar distributions
\[
t^{ij}, t^{inkj}, \ldots, t^{in\ldots kj} 
\]
replaced with the respective advanced and retarded parts, in any case with well defined 
$\widetilde{\mathbb{S}^1}\times SU(2, \mathbb{C})$-invariant distributions. 

Proceeding as above we see that $S_2(1\otimes g_2)$, $g_2 \in \mathscr{E}$, is equal to the finite sum
\[
\sum\limits_{\ell,m}\Xi_{\ell,m}(\kappa_{\ell,m}(g_2)) 
\]
of integral kernel operators with the kernels of the form
\[
\kappa_{\ell,m} = \big(\textrm{ret} t^{in \ldots j} \ast \kappa'_{\ell',m'}\big) \dot{\otimes} \kappa''_{\ell'',m''},
\,\,\, \ell = \ell'+\ell'', m = m' + m''.
\]
Thus by Theorem \ref{F*WickProdFreeFieldsOnEU} of Subsection \ref{WhiteNoiseFreeFieldsonEU} as above we see
that $S_2(x,y)$ is a kernel of a generalized integral kernel operator $S_2$ such that
\[
\mathscr{E} \ni g_2 \longrightarrow S_2(1\otimes g_2) \in \mathscr{L} \big((\boldsymbol{E}), (\boldsymbol{E}) \big)
\]
is continuos.

Also in the general case with more than one massless field in $\mathcal{L}$
the kernels $\kappa_{\ell, m}$ (discrete momentum variables are not written explicitly) 
\[
\kappa_{\ell, m}(x,y) = \textrm{ret} t^{ij}(xy^{-1}) \kappa'_{\ell',m'}(x) \kappa''_{\ell'',m''}(y)
\]
of the operator 
\[
S_2(x,y) = \Xi_{\ell,m}(\kappa_{\ell, m}(x,y))
\]
represent  continuous maps
\[
\xi \otimes \eta \longmapsto \kappa_{\ell, m}(\xi \otimes \eta) \in \mathscr{E}^{* \otimes \, 2}
\]
when regarded as maps of
\[
E_{1} \otimes \ldots \otimes E_{\ell} \otimes E_{\ell+1} \otimes E_{\ell+m} \ni \xi\otimes\eta
\]
into 
\[
\mathscr{E}^{* \otimes \, 2},
\]
because 
\[
\kappa'_{\ell',m'}(\xi), \,\, \kappa''_{\ell'',m''}(\eta) \,\, \in \mathscr{E}
\]
and the maps
\[
\xi \mapsto \kappa'_{\ell',m'}(\xi), \,\, \eta \mapsto \kappa''_{\ell'',m''}(\eta)
\]
are continuous. Therefore, again by the general theory of integral kernel operators due to Hida, Obata and Sait\^o 
(Subsection \ref{psiBerezin-Hida}, Thm. \ref{obataJFA.Thm.3.13},
compare also Subsection \ref{WhiteNoiseFreeFieldsonEU}) 
\[
S_2 \in  \mathscr{L} \big( \mathscr{E}^{\otimes \, 2}, \, 
\mathscr{L}(\boldsymbol{E}), (\boldsymbol{E})^*) \big).
\]
If we have more than one massless free field in $\mathcal{L}$ the property (\ref{extensionretkappaavkappa})
is also true and $\kappa_{\ell, m}$ can be extended to elements of
\[
\mathscr{L}\big(E_{1}^{*} \otimes \ldots \otimes E_{\ell}^{*} 
\otimes E_{\ell+1} \otimes E_{\ell+m}, \,\, \mathscr{E}^{* \otimes \, 2} \big)
\]
but now, in the kerel
\[
\kappa_{\ell, m}(x,y) = \textrm{ret} \, t^{in \ldots j}(xy^{-1}) \kappa'_{\ell',m'}(x) \kappa''_{\ell'',m''}(y),
\]
is far not unique, and the kernel $\textrm{ret} \, t^{in \ldots j}$ is defined on test functions $\chi(x,y)=\phi(xy^{-1})\varphi$
with $\phi$ lying in the image of the idempotent opertor $\Omega'$ of subsection. We can add to $\textrm{ret} \, t^{in \ldots j}$,
defined in the previous Subsection, any distribution which is zero on $\textrm{Im} \, \Omega'$ in order to obtain any other
possible retarded part of $t^{in \ldots j}$. Because now the complementary space $\textrm{ker} \, \Omega'$ is nonzero (even of infinite dimension)
the arbitrarines is enormous, at least in principle.

Therefore we arrive at the following
\begin{twr}
Let $S_n$ be the $n$-th order contribution to the scattering integral kernel operator. Suppose that some Wick monomials
of the Wick polynomial interaction density operator $\mathcal{L}$
contain more than one massless factor $\mathbb{A}_{{}_{k}}$. Then
\begin{enumerate}
\item[1)]
\[
\mathscr{E} \ni g_n \longrightarrow S_n(1 \otimes 1 \ldots \otimes 1 \otimes g_n) \in \mathscr{L}(\boldsymbol{E}), (\boldsymbol{E})), 
\]
is continuous. In particular
\[
S_n(1 \otimes 1 \ldots \otimes 1 \otimes 1) = S_n(1) \in \mathscr{L}(\boldsymbol{E}), (\boldsymbol{E}))
\]
because the constant function $1 \in \mathscr{E} = \mathcal{S}_{\Delta+1}(\widetilde{\mathbb{S}^1}\times SU(2, \mathbb{C}))$.
 \item[2)]
\[
S_n \in \mathscr{L} \big(\mathscr{E}^{\otimes \, n}, \, \mathscr{L}(\boldsymbol{E}), (\boldsymbol{E})) \big). 
\]
\end{enumerate}
\label{GeneralSin(SE)xn->((E)->(E))EU}
\end{twr}

However, now in Theorem \ref{GeneralSin(SE)xn->((E)->(E))EU}, if we have monomials with more than one massless free field 
factor in $\mathcal{L}$ we cannot apply the 
simple splitting into retarded and advanced parts of the scalar distributions $t^{in \ldots j}$ using ordinary multiplication by $\theta$.
A more general Epstein-Glaser-type splitting must be performed. 
For the construction of the retarded and advanced parts we use the ``natural'' formula of multiplication by the step 
$\theta$ function, but in general not on the whole test space, but only on the image of the idempotent
operator $\Omega$, or respectively $\Omega'$. We have illustrated it in the previous Subsection 
using the operator $\Omega'$ equal (\ref{OmegaphiEU}). This operator served rather as an illustration of the general method only, 
as its image $\textrm{Im} \, \Omega'$ is relatively small leaving a tremendous ambiguity in the choice
of possible retarded and advanced parts, determined up to  distributions concentrated at the whole Cauchy surfaces
$t=0$ and $t=2\pi$. In Subsection \ref{splittingEU} we are using a projection $\Omega'$ with finite codimension
and construct the retarded part up to a distribution of singularity order less than or equal the order 
of the splitted distribution, and constentrated at the single point $(0,e)$, 
and eventally also at $(2\pi,e)$ in case of periodic $\theta$.

As we have seen in the previous Subsection, using (\ref{OmegaphiEU}),
in case the distribution  $t^{in \ldots j}$ includes more than one pair of contracted plane wave kernels
(with the corresponding orbits being infinite) 
 $\kappa^{(1)}_{0,1}(x_i), \kappa^{(1)}_{1,0}(x_j)$ with the same pair of space-time coordinates $x_i, x_j$
then  splitting of $t^{in \ldots j}$  into advanced and retarded part is non-unique. If there is $k$ such space-time coordinates, 
then the splitting of  $t^{in \ldots j}$ is determined up to a distribution 
supported on the following submanifold of lower dimension
\[
\big[SU(2, \mathbb{C}) \sqcup SU(2, \mathbb{C}) \big]^{\times \, (k-1)} \times \big[\widetilde{\mathbb{S}^1}\times SU(2, \mathbb{C}) \big]^{\times \, (n-k+1)}
\subset \big[\widetilde{\mathbb{S}^1}\times SU(2, \mathbb{C}) \big]^{\times \, n}
\] 
of $k-1$ copies of  disjoint sums $SU(2, \mathbb{C}) \sqcup SU(2, \mathbb{C}) = \mathbb{S}^{3} \sqcup \mathbb{S}^{3}$ of two Cauchy surfaces
$t=0$ and $t=2\pi$ and $n-k+1$ copies of the compactified Einstein Universe $\widetilde{\mathbb{S}^1}\times SU(2, \mathbb{C})$.

\subsection{Splitting of distributions on the Einstein Universe}\label{splittingEU}

In this Subsection we use another $\Omega'$ (instead of $\Omega$), which replaces (\ref{OmegaphiEU}), Subsection \ref{WhiteNoiseFreeFieldsonEU}.
This new $\Omega'$ has the image of finite codimension, consisting of test functions whose derivatives vanish up to a finite order $\omega$
at only two points, $(0,e)$ and $(2\pi,e)$, with the order $\omega$ equal to the order of splitted distribution. 
This new $\Omega'$ allows  construction of the advanced and retarded parts 
of the products $\kappa_q$ of massless pairings on the compactified 
Einstein Universe (with periodic $\theta$) up to a finite number of constants, depending on the 
UV singularity order $\omega$ of the product $\kappa_q$ 
of massless pairings which is to be splitted.  In fact each particular QFT is determined
by the products $\kappa_q$ (and their retarded and advanced parts), which are  present in the Wick decomposition
of the operator $\mathcal{L}(x)\mathcal{L}(y)$, compare Subsection \ref{WickForChronological}.

We give here a general deiniton for $\widetilde{\textrm{ret} \, \kappa_q}$ such that its inverse Fourier transform $\textrm{ret} \, \kappa_q$
is equal $\theta\kappa_q$ on the image of the operator $\Omega'$. But also we give a practical method for computation of 
$[\widetilde{\textrm{ret}]''' \, \kappa_q}$ which coincides with $\widetilde{\textrm{ret} \, \kappa_q}$ up to a distribution of order 
$\leq \omega$, concentrated at the single point $(0,e)$ (and eventualy also at $(2\pi, e)$, in case of the 
compactified EU with periodic $\theta$).

We also present here the analogous formula for the ordinary (non-compactified) Einstein Universe with ordinary (non-periodic)
$\theta$-function, with the new $\Omega'$ having the image of finite codimension, consisting of test functions whose derivatives 
vanish up to a finite order $\omega$ at only one point $(0,e)$. Also in this case, we construct
explicitly $\textrm{ret} \, \kappa_q$ up to a distribution of degree $\leq \omega$, concentrated at the single
point $(0,e)$ -- the unit of the group $G$ with which EU can be identified.

We need the construction of the retarded and advanced parts of the products $\kappa_q$ of more than $q=1$ 
massless pairings whenever there are Wick monomial in the interaction Lagrangian $\mathcal{L}(x)$ that contain 
more than one massless free field. If all Wick monomials are massive in $\mathcal{L}(x)$, the 
$\textrm{ret}, \textrm{av}$ parts are given by the ordinary multiplication by the
step $\theta$ function, with products of pairings being of the order $\omega = - \infty$.

We have assumed the ordinary causal axioms (I)-(IV) plus (V), which can be adopted
to the Einstein Universe, both compactified with periodic $\theta$, and noncompactified with ordinary non-periodic
$\theta$. For the formulation of (I)-(V) on the Einstein Universe, in both versions -- the compactified and ordinary non compactified --
compare Subsection \ref{CausalSonEU}. Here we look more carefully at the axiom (V), or at the splitting problem of products of pairings
into retarded and advanced part in both cases, the compactified and non compactified.

We have assumed covariance of the scattering operator $S$ under the full
symmetry group of the Einstein Universe (or, respectively, its compactification, in case we are using periodic $\theta$).
Let us consider the kernels of the integral kernel operators whose finite sum is equal to the product 
$S_1(x)S_1(y) =i^2\mathcal{L}(x)\mathcal{L}(y)$ or to $S_2(x,y)$. By construction the scalar factors, in their canonical
product form, are equal to 
with scalar contractions, which capture all pairings, or product 
$\kappa_q(x,y)= \kappa_q(xy^{-1},e)$ of $q$ pairings, which can be represented as distributions of single space-time variable
equal to the product $xy^{-1}$, which we denote by $d_q(xy^{-1})$. 
If we assume covariance of $S$
under the full symmetry group, then $d_q(x)$, regarded as distributions of one space-time variable, should still be
covariant under the symmetry subgroup $1 \times 1 \times SU(2,\mathbb{C})$, or even invariant under this 
subgroup, if  $d_q(xy^{-1})=\kappa_q(x,y)= \kappa_q(xy^{-1},e)$ represents the scalar contribution
\[
\kappa_q(x,y) \boldsymbol{1}
\]
to $S_2(x,y)$. Now let us consider the retarded part 
\[
\langle \textrm{ret} \, d_q, \phi \rangle
= \langle d_q, \theta \Omega'\phi\rangle, 
\]
but with the operator $\Omega'$ not equal to (\ref{OmegaphiEU}), Subsection \ref{WhiteNoiseFreeFieldsonEU},
but with some other $\Omega'$ with image of finite codimension. Let for example, the subtraction terms
of $\Omega'$ contain only time derivatives $\partial_{t}^\alpha\phi(0,\boldsymbol{w}_1)$ and   
$\partial_{t}^\alpha\phi(2\pi,\boldsymbol{w}_1)$ at a fixed and just one point $\boldsymbol{w}_1$ of the Cauchy surfaces
$t=0$ (and evenuall also $t=2\pi$ in case with periodic $\theta$).
Otherwise, we consider $\textrm{ret} \, d_q$ as defined by the multiplication by the step theta function
on the subspace $\textrm{Im} \Omega'$ of all those test functions whose time derivatives up to order $\omega$ vanish at
just one point $(0,\boldsymbol{w}_1)$ of the Cauchy surfaces $t=0$, and, in case of periodic $\theta$, in addition at the point $(2\pi,\boldsymbol{w}_1)$
of the Cauchy surface $t=2\pi$. But such definition of $\textrm{ret} \, d_q$ would not be $1 \times 1 \times SU(2,\mathbb{C})$-covariant,
because such subspace  $\textrm{Im} \Omega'$ is not $1 \times 1 \times SU(2,\mathbb{C})$-invariant. Indeed, if we look
at the action (\ref{UphiOnRxG}), Subsetion \ref{GeneralizedSchrodinger-VonNeumannPairs}, 
of the symmetry group, than we see that the subgroup $1 \times 1 \times SU(2,\mathbb{C})$ 
acts on vector valued $d_q$ in the following manner
\[
\textrm{Ad}_{{}_{\boldsymbol{v}}}d_q(t, \boldsymbol{w}) = V(\boldsymbol{v}) d_q(t, \boldsymbol{v}^{-1}\boldsymbol{w}\boldsymbol{v})
\] 
and 
\[
\textrm{Ad}_{{}_{\boldsymbol{v}}}d_q(t, \boldsymbol{w}) =d_q(t, \boldsymbol{v}^{-1}\boldsymbol{w}\boldsymbol{v})
\] 
for scalar valued $d_q$. We, thus, see that
the definition of $\textrm{ret} \, d_q = \theta d_q \circ \Omega'$ is covariant if  $\textrm{Im} \Omega'$ 
is invariant under the $\textrm{Ad}$-action of  $1 \times 1 \times \boldsymbol{v}$ $\in 1 \times 1\times SU(2,\mathbb{C})$ given by the rule
\[
\textrm{Ad}_{{}_{\boldsymbol{v}}}\phi(t, \boldsymbol{w}) = V(\boldsymbol{v}) \phi(t, \boldsymbol{v}^{-1}\boldsymbol{w}\boldsymbol{v}).
\] 
Thus, we see that
 the space of functions with (space-) time derivatives  vanishing at the two points
$(0,\boldsymbol{w}_1)$ and $(2\pi,\boldsymbol{w}_1)$ up to order $\omega$ is transformed,
by an element $1\times 1\times \boldsymbol{v}$ of the subgroup $1 \times 1 \times SU(2,\mathbb{C})$,
into the space of functions with (space-) time derivatives  vanishing at the two points
$(0,\boldsymbol{v}^{-1}\boldsymbol{w}_1\boldsymbol{v})$ and $(0,\boldsymbol{v}^{-1}\boldsymbol{w}_1 \boldsymbol{v})$ up to order $\omega$. 
Thus the covariance under the full symmetry group would be broken if $\boldsymbol{w}_1 \neq e$, where $e$ is the unit in $SU(2,\mathbb{C})$.
In order to retain covariance under the full symmetry group, we should enlarge our starting space $\textrm{Im} \Omega'$ and
consider the smallest subspace containing it which is invariant under $\textrm{Ad}$-action of $1\times 1\times \boldsymbol{v}$ which, as is easily seen,
is equal to the sum of all  $\textrm{Ad}$-images of  $\textrm{Im} \Omega'$ under $1\times 1\times \boldsymbol{v}$ in  $1\times 1\times SU(2,\mathbb{C})$ 
and consists of all those functions
whose time derivatives vanish up to order $\omega$ in all points of the Cauchy surfaces $t=0$ and, in case of periodic $\theta$, 
also at $t=2\pi$ if $\boldsymbol{w_1} \neq e$. 
Therefore, if $\boldsymbol{w_1} \neq e$, then using the covariance of $S$ under the full symmetry group  we obtain the image of the operator
$\Omega'$ equal to (\ref{OmegaphiEU}) of Subsection \ref{WhiteNoiseFreeFieldsonEU}, in case with periodic $\theta$, or the operator
$\Omega'$ equal to (\ref{OmegaphiEU}) of Subsection \ref{WhiteNoiseFreeFieldsonEU} without the term with the auxiliary function $w_{II}$,
in case of non periodic $\theta$. This case we have already discussed in Subsection \ref{WhiteNoiseFreeFieldsonEU}
for the periodic $\theta$. The case with non periodic $\theta$ and ordinary non compactified EU is even simpler.

It remains to consider the case in which $\boldsymbol{w}_1=e$.
Let us consider the case with $\boldsymbol{w}_1=e$. In this case $\textrm{Im} \, \Omega'$ is likewise
invariant under the $\textrm{Ad}$-action of $1\times 1\times \boldsymbol{v}$, so that
the distribution $d_q$ is still covariant under the $\textrm{Ad}$-action of $1\times 1\times \boldsymbol{v}$.
Therefore we should try to use the new operator $\Omega'$ with subtraction of time derivatives at just $(0,e)$, and
eventually at $(2\pi,e)$ in case with periodic $\theta$,
without going into any conflict with the covariace of $S$ under the full symmetry
group  $\widetilde{\mathbb{S}^1}\times SU(2,\mathbb{C}) \times SU(2,\mathbb{C})$.

We start with periodic $\theta$ on compactified EU.
We check now if we can replace the operator $\Omega'$ equal to (\ref{OmegaphiEU}) of Subsection \ref{WhiteNoiseFreeFieldsonEU}
with some another $\Omega'$ containing the subtraction of derivatives at only two points, $(0,e)$ and $(2\pi,e)$, 
with larger $\textrm{Im} \, \Omega'$  of finite codimension and still converging $\langle d_q, \theta \Omega' \phi\rangle$,
$\phi \in \mathscr{E}$. 

Below $d_q$ and $\kappa_q$ will be denoted by the same symbol $\kappa_q$.

It is quite nontrivial if the splitting 
with $\Omega'$ equal to (\ref{OmegaphiEU}) of Subsection \ref{WhiteNoiseFreeFieldsonEU}
into retarded and advanced part fulfils axiom (V) (compare Subsection \ref{WickForProduct}) of ``maximal naturality'' of the splitting, 
which accordingly to this axiom,
should coincide with the natural formula given by the multiplication by the periodic theta function, whenever this natural formula makes sense
and can be applied to a test function. 

We would like also the Fourier transform $\widetilde{\textrm{ret} \, \kappa_q}$ of the 
retarded part $\textrm{ret} \, \kappa_q$ of a product $\kappa_q$ of $q$ (massless) pairings to have
the least possible order(on the Minkowski space-time it was equal to the aorder of $\widetilde{\kappa_q}$). 
Let us remind that the standard countably Hilbert nuclear space 
\[
\mathscr{E}= \mathcal{S}_{\Delta+1}(G)= \mathscr{C}^\infty(G), \,\,\,\,\, G = \widetilde{\mathbb{S}^1}\times SU(2, \mathbb{C}),
\] 
is a projective limit $\cap \mathscr{E}_k$ of $\mathscr{E}_{k}$,
and also its Fourier transformed space $\widetilde{\mathscr{E}}$ is an isomorphic projective limit $\cap \widetilde{\mathscr{E}}_k$,
$k=1,2,3 \ldots$ of Hilbert spaces $\widetilde{\mathscr{E}}_{k} = \widetilde{\mathscr{E}_{k}}$. The strong dual space
$\mathscr{E}^*$ is the inductive limit $\cup \mathscr{E}_k$  of Hilbert spaces  $\mathscr{E}_{-k}$, $k=0,1,2,3$, and 
also its Fourier transformed space $\widetilde{\mathscr{E}^*}$ is an isomorphic inductive limit $\cup \widetilde{\mathscr{E}}_{-k}$,
$k=1,2,3 \ldots$ of Hilbert spaces $\widetilde{\mathscr{E}}_{-k} = \widetilde{\mathscr{E}_{-k}}$. The axiom (V) says
that 
\[
\textrm{ret} \, \kappa_q \in \widetilde{\mathscr{E}}_{-k} \,\,\, \textrm{if}
\,\,\, \widetilde{\kappa_q} \in \widetilde{\mathscr{E}}_{-k}.
\] 
The least number $k$ (we can also consider real $k$, in this case we use infimum) such that
\[
\widetilde{d} \in \widetilde{\mathscr{E}}_{-k}.
\] 
In other words, we assume that the singularity order $\omega$ is the same for $\kappa_q$
and $\textrm{ret} \, \kappa_q$ (compare axiom (V), Subsection \ref{axiomsS}).
Let us recall, that the UV singularity order $\omega$ is the singularity order of the quasiasymptotic of the splitted
distribution $\kappa_q$ at zero (in space-time, i.e., in our case, at the unit element of the group $G$), and being a local
quantity, it is well-defined on any (Lorentz) manifold, and not only on the flat Minkowski space-time. Compare the definition
of the UV quasiasymptotic singularity degree $\omega$ in Section \ref{WickForChronological}, or \cite{Vladimirov1}.

It is, in principle at least, possible that the above condition 
put on $\phi \in \textrm{Im} \Omega'$ is too strong, or that the operator $\Omega'$, equal (\ref{OmegaphiEU}), Subsection \ref{WhiteNoiseFreeFieldsonEU},
 has too small image (at least in some particular cases, or for particular orders of products $\kappa_q$ of pairings),
and that putting the condition of vanishing derivatives of $\phi \in \textrm{Im} \Omega'$ at only two points, $(0,e)$ and $(2\pi,e)$, 
with the corresponding idempotent operator $\Omega'$, the analog of the formula (\ref{OmegaphiEU}) 
of Subsection \ref{WhiteNoiseFreeFieldsonEU}, replaced with another $\Omega'$ with the image $\textrm{Im} \, \Omega'$ of finite codimension,
would be sufficient.
Below in this Subsection we will show, that in general for products $\kappa_q$ of $q>1$ massless pairings (thus, having arbitrary orders) 
the condition of vanishing of derivatives of $\phi$ up to any finite order at 
$(0,e)$ and at $(2\pi,e)$ in $\widetilde{\mathbb{S}^1}\times SU(2,\mathbb{C})$, with the corresponding $\Omega'$
is, in general, sufficient for the convergence of $\langle \theta \kappa_q, \Omega'\phi\rangle$, for $\phi$ in $\mathscr{E}$
with periodic $\theta$. 

In general, for any product $\kappa_q$ of pairings, which we encounter in the Wick product
decomposition of $\mathcal{L}(x)\mathcal{L}(y)$, with  $\kappa_q$ being of the orders $\omega'$, possibly higher than or equal to zero 
(which is the case with more than one massless field in $\mathcal{L}$), the 
condition of vanishing of time (and also all space-time) derivatives up to the finite order $\omega=\omega'$ at the``subtraction points''
$(0, e)$, $(2\pi, e)$, on the
Cauchy surfaces $t=0$ and, respectively, $t=2\pi$, with the corresponding
\[
\Omega'\phi(x,\boldsymbol{w}) = \phi(t,\boldsymbol{w}) 
- \sum\limits_{a=0}^{\omega}\partial_{t}^{a}\phi(0,e) \,\, {\textstyle\frac{t^\alpha}{a!}} \, w_{{}_{I}}(t) 
- \sum\limits_{a=0}^{\omega}\partial_{t}^{a}\phi(2\pi,e) \,\, (-1)^a{\textstyle\frac{(t-2\pi)^a}{a!}} \, w_{{}_{II}}(t)
\] 
will be sufficient, and still multiplication by the step periodic $\theta$
will give convergent $\langle \kappa_q, \theta \phi\rangle$ on $\phi$ in $\textrm{Im} \, \Omega'$. 
We give a proof of this theorem below in this Subsection. Thus for such QFT and for periodic $\theta$
defining the time evolution, with more than one massless field in $\mathcal{L}$ and periodic $\theta$,
with orders of some of $\kappa_q$ in the Wick decomposition of  $\mathcal{L}(x)\mathcal{L}(y)$ greater than or equal to zero,
the splitting of $\kappa_q$ (or $t^{ij}$) into retarded and advanced parts can, in general, be determined up to
a finite number of arbitrary constants.

As we can see, in case of QFT in which we have more than one massless field in $\mathcal{L}$ and periodic $\theta$, 
the following problem
is crucial: to investigate for what products  $\kappa_q$, $q>1$, of massless pairings 
the operator $\Omega'$, given by the formula (\ref{OmegaphiEU}) 
of Subsection \ref{WhiteNoiseFreeFieldsonEU}, can be replaced with another with larger $\textrm{Im} \, \Omega'$
on which multiplication $\theta \kappa_q$  by the periodic theta function $\theta$ gives well-defined distribution on $\textrm{Im} \, \Omega'$.
We undertake this problem now. We do it gradually by imposing
the condition of vanishing of derivatives, always at a single point -- the unit element $(0,e)$ of the group
$\widetilde{\mathbb{S}^1}\times SU(2, \mathbb{C})$, or at $(2\pi,e)$, using exactly the same method
for the computation of $\textrm{ret} \, \kappa_q$ and investigation of convergence of
\[
\langle \textrm{ret} \, \kappa_q, \phi\rangle \overset{\textrm{df}}{=}\langle \theta \kappa_q, \Omega'\phi\rangle, \,\,\, \phi \in \mathscr{E},
\]
as in \cite{Scharf} on the Minkowski space-time, pp. 178-181 (reported in Subsection \ref{WickForChronological}),
always with just one subtraction point, $(0,e)$ or $(2\pi,e)$, in $\Omega'$.

The  said method of \cite{Scharf} can naturally be
extended over the Einstein Universe, and more generally, on space-times where we have natural Fourier transform construction
associated to the standard operator $\Delta$ defining the space-time test space as a standard countably Hilbert nuclear space,
associated to the Dirac operators $D, D_{{}_{\mathfrak{J}}}$ corresponding, respectively, to the Riemann and the Lorentz metrics 
associated in the way explained
in Subsection \ref{GeneralizedSchrodinger-VonNeumannPairs}, with the 
Fourier transform given through the spectral decomposition of the standard operator and the Dirac operators $D, D_{{}_{\mathfrak{J}}}$. 
In case of highly symmetric space-times,
where the complete systems of solutions naturally arise from representation theory
(e.g. Einstein Universe), explicit calculations can be performed relatively easily. In this case explicit
formula for the Fourier transforms $\widetilde{\kappa_q}$ of the products of pairings, or scalar $\otimes_q$-contractions, 
can be explicitly computed in this case, on using completeness
relations for the complete systems of solutions underlying the free fields in a way quite analogous to that
presented in \cite{Scharf} (compare derivation of the formulas (2.7.31), (2.8.33) and (3.6.8) in \cite{Scharf}), 
again using representation theory.
 
In particular, depending on the order $\omega$ of contraction $\kappa_q$ which is to be splitted, we put the condition of vanishing of all derivatives of 
test functions $\phi$ up to order at least $\omega$ at the  space-time ``subtraction point'' $(0,e)$, or at ``subtraction point''
$(2\pi,e)$, which gives us
a subspace of finite codimension in $\mathscr{E}$, with the corresponding operator $\Omega'$ whose image is a subspace of finite codimension. 
Then we investigate convergence of $\langle \theta \kappa_q, \Omega'\phi\rangle$, for $\phi$ in $\mathscr{E}$, and determine
the linear subspace on which this formula is convergent.

We consider the condition of vanishing derivatives up to finite order at $(0,e)$ or, respectively, at  $(2\pi,e)$
in $\widetilde{\mathbb{S}^1}\times SU(2,\mathbb{C})$  
\begin{equation}\label{x+-}
(t=0,e) \,\,\,\,\,\, (t=2\pi, \boldsymbol{w}_1=e),
\end{equation} 
with the corresponding $\Omega'$ with, respectively, only one ``subtraction point'' $(0,e)$, or $(2\pi,e)$:
\begin{equation}\label{Omega'phiEU}
\Omega'\phi(x,\boldsymbol{w}) = \phi(t,\boldsymbol{w}) 
- \sum\limits_{a=0}^{\omega}\big(X_0\big)^{a}\phi(0,e) \,\, {\textstyle\frac{(x_0)^a}{a!}} \, w(t),
\end{equation}
or
\begin{equation}\label{Omega'phiEUseveral}
\Omega'\phi(x,\boldsymbol{w}) = \phi(t,\boldsymbol{w}) 
- \sum\limits_{a=0}^{\omega} \big(X_0\big)^{a}\phi(2\pi,e) \,\, {\textstyle\frac{(x_0)^a}{a!}} \, w(t),
\end{equation}
or still more generally
\begin{gather}
\Omega'\phi(x,\boldsymbol{w}) = \phi(t,\boldsymbol{w}) 
- \sum\limits_{|\alpha|=0}^{\omega} X^{\alpha}\phi(0,e) \,\, {\textstyle\frac{x^\alpha}{\alpha!}} \, w(t) 
\label{Omega'phiEUseveralMultiindex}
\\
\Omega'\phi(x,\boldsymbol{w}) = \phi(t,\boldsymbol{w}) 
- \sum\limits_{|\alpha|=0}^{\omega} X^{\alpha}\phi(2\pi,e) \,\,{\textstyle\frac{x^\alpha}{\alpha!}} \, w(t),
\nonumber
\end{gather}
\[
X^\alpha \phi = \big(X_0\big)^{\alpha_0}\big(X_1\big)^{\alpha_1}\big(X_2\big)^{\alpha_2}\big(X_3\big)^{\alpha_3} \phi,
\,\,\,\, x^\alpha = \big(x_0\big)^{\alpha_0}\big(x_1\big)^{\alpha_1}\big(x_2\big)^{\alpha_2}\big(x_3\big)^{\alpha_3}.
\]
Each subtraction term
at the subtraction point $(0,e)$, respectively $(2\pi,e)$, contains the system of functions $x_\mu$ with differentials
dual to the right invariant vector fields $X_\mu$ up to order $\omega$, at  $(0,e)$ and $(2\pi,e)$. This means that
\begin{gather*}\label{dualityXmu-xmu}
X_\mu(x_\nu)(0,e) = X_\mu(x_\nu)(2\pi,e) = \delta_{\mu \nu}, 
\\
X^\alpha (x^\beta)(0,e) = X^\alpha (x^\beta)(2\pi,e) =0, \,\, \beta <\alpha \leq \omega
\end{gather*}
The functions $x_\mu$ can be easily chosen the same for the points
$(0,e)$ and  $(2\pi,e)$. 
The auxiliary function $w$ is equal one on a neighborhood of $t=0$ and zero on a neighborhood of $t=2\pi$, and,
respectively, \emph{vice versa}
for subtraction at $(0,e)$ or $(2\pi,e)$.  

As an intermediate step between (\ref{Omega'phiEU}) and (\ref{Omega'phiEUseveralMultiindex}) we consider
(\ref{Omega'phiEUseveral}).

In order to verify convergence of $\langle \theta \kappa_q, \Omega'\phi\rangle$, for $\phi$ in $\mathscr{E}$,
we need to construct a simple system of smooth functions $x_0, x_1,x_2,x_3$ vanishing at the``subtraction points'', whose differentials compose
a basis dual up to order $\omega$ to the right invariant vector fields $X_0=\partial_t, X_1,X_2,X_3$, 
constructed in Subsection \ref{GeneralizedSchrodinger-VonNeumannPairs}, 
at the ``subtraction points'' $(0,e)$ and $(2\pi,e)$. 
The functions $x_\mu$ should be chosen naturally, and should be simple combinations of the characters $\widehat{n}(t)$ 
and $\widehat{l}_{{}_{ij}}(\boldsymbol{w})$,
with low values of $n$ and $l$, such that their Fourier transform should be easily computable. In particular, for the  order $\omega=2$, we can put
\begin{gather}
x_0(t,\boldsymbol{w}) = \sin t = {\textstyle\frac{-i}{2}}\big[\widehat{2}(t)-\widehat{-2}(t)\big] = {\textstyle\frac{1}{2i}}\big[e^{it}-e^{-it}\big],
\label{x0}
\\
x_1(t,\boldsymbol{w}) = {\textstyle\frac{-i}{\sqrt{2}}}\big[\widehat{1}_{{}_{0 \, 1}}(\boldsymbol{w})+\widehat{1}_{{}_{1 \, 0}}(\boldsymbol{w})\big],
\label{xmu}
\\
x_2(t,\boldsymbol{w}) = {\textstyle\frac{1}{\sqrt{2}}}\big[\widehat{1}_{{}_{0 \, 1}}(\boldsymbol{w})-\widehat{1}_{{}_{1 \, 0}}(\boldsymbol{w})\big], 
\nonumber
\\
x_3(t,\boldsymbol{w}) = {\textstyle\frac{-i}{2}} \big[\widehat{1}_{{}_{1 \,\, 1}}(\boldsymbol{w})-\widehat{1}_{{}_{-1 \,\, -1}}(\boldsymbol{w})\big].
\nonumber
\end{gather}
In order to obtain the corresponding $x_\mu$ dual to $X_\mu$ at $(0,e)$ and $(2\pi,e)$ up to
higher order $\omega$ we need to consider the corresponding linear combinations of the powers of the
functions $x_\mu$, which are dual up to order $\omega=2$. In particular $x_0(t) = \sin t + (\sin^3 t)/3!$
will be the function $x_0$ of the system $x_\mu$ dual to $X_\mu$ at $(0,e)$ and $(2\pi,e)$ up order $\omega=4$.
Analogously by addition of the higher powers $\sin^5 t, \sin^7 t, \ldots$, with the appropriate coefficients, we obtain
the functions $x_0$ of the system $x_\mu$ dual to $X_\mu$ at $(0,e)$ and $(2\pi,e)$ up order $\omega=6,8, \ldots$.      
In order to obtain the corresponding functions $x_1, x_2, x_3$, which together with the above $x_0$,
have differentials dual to the right invariant $X_0, X_1, X_2, X_3$ at $(0,e)$ and $(2\pi, e)$ up to higher
orders $\omega$ we similarly we compose linear combinations of odd powers of (\ref{xmu}) with appropriate coefficients
we obtain functions $x_1,x_2,x_3$ of the system $x_0$, $x_1$, $x_2$, $x_3$ dual to $X_0, \ldots, X_3$ at $(0,e)$, $(2\pi, e)$
of higher order $\omega$. 
Because the functions (\ref{xmu}) are simple combinations of the matrix elements of the irreducible representation
$\widehat{l}$, here with $l=1$, the computation of  their pointwise products $\widehat{1}_{{}_{m_1 \, k_1}}(\boldsymbol{w})
\widehat{1}_{{}_{m_2 \, k_2}}(\boldsymbol{w})$ in terms of the matrix elements $\widehat{l}_{{}_{ji}}(\boldsymbol{w})$,
for $\leq 0 \leq l \leq 2$ is immediate by the pointwise product
rule for the matrix elements of the irreducible representations.

Fourier transforms of (\ref{x0}) and (\ref{xmu}) are equal
\begin{gather}
\big(\widetilde{x_0}\big)_{{}_{ji}}(\widehat{n}\cdot \widehat{l}) = 
4 \sqrt{\pi} \,\,
{\textstyle\frac{-i}{4}} \big[\delta_{{}_{n\, 2}}-\delta_{{}_{n\, -2}}\big]
\delta_{{}_{l \, 0}} \delta_{{}_{i \, 0}} \delta_{{}_{j \, 0}},
\label{FT(x0)}
\\
\big(\widetilde{x_1}\big)_{{}_{ji}}(\widehat{n}\cdot \widehat{l}) =
{\textstyle\frac{-i}{\sqrt{2}}} \delta_{{}_{n\, 0}} \, \delta_{{}_{l \, 1}} \, \big[\delta_{{}_{j \, 1}}\delta_{{}_{i \, 0}} 
+\delta_{{}_{j \, 0}}\delta_{{}_{i \, 1}} \big]
\label{FT(xmu)}
\\
\big(\widetilde{x_2}\big)_{{}_{ji}}(\widehat{n}\cdot \widehat{l}) =
{\textstyle\frac{1}{\sqrt{2}}} \delta_{{}_{n\, 0}} \, \delta_{{}_{l \, 1}} \, \big[\delta_{{}_{j \, 1}}\delta_{{}_{i \, 0}} 
-\delta_{{}_{j \, 0}}\delta_{{}_{i \, 1}} \big],
\nonumber
\\
\big(\widetilde{x_3}\big)_{{}_{ji}}(\widehat{n}\cdot \widehat{l}) =
{\textstyle\frac{i}{2}} \,  \delta_{{}_{n\, 0}} \, \delta_{{}_{l \, 1}} \, \big[\delta_{{}_{j \,\, 1}}\delta_{{}_{i \,\, 1}}
-\delta_{{}_{j \,\, -1}}\delta_{{}_{i \,\, -1}}\big].
\nonumber
\end{gather}
The Fourier transorm image of the multiplication by $x_\mu$ action plays the role of ``differentiation'' in ``momentum space''
here on the discrete dual of the (compactified) space-time group $\widetilde{\mathbb{S}^1}\times SU(2,\mathbb{C})$, 
and has the following form in case of (\ref{x0}) and (\ref{xmu}) and order $\omega=2$: 
\begin{equation}\label{FT(x0.phi)}
\big(\widetilde{x_0 \phi}\big)_{{}_{ji}}(\widehat{n}\cdot \widehat{l})
= \Big(\widetilde{x_0} \,\, \ast \,\, \widetilde{\phi}\Big)_{{}_{ji}}(\widehat{n}\cdot \widehat{l})
= 
2 \,\,\,
i\, {\textstyle\frac{\widetilde{\phi}_{{}_{\,\, j i}} \big(\widehat{n+2} \cdot \widehat{l}\big) 
-\widetilde{\phi}_{{}_{\,\, j i}} \big(\widehat{n-2} \cdot \widehat{l}\big)}{(n+2) - (n-2)}},
\end{equation}
\begin{multline*}
\big(\widetilde{x_1 \phi}\big)_{{}_{ji}}(\widehat{n}\cdot \widehat{l})
= \Big(\widetilde{x_1} \,\, \ast \,\, \widetilde{\phi}\Big)_{{}_{ji}}(\widehat{n}\cdot \widehat{l})
\\
=
{\textstyle\frac{-3i}{\sqrt{2}4\pi}} 
\sum\limits_{\substack{ l'',j'',i''}} 
{\textstyle\frac{2l''+1}{2l +1}} 
\,
\widetilde{\phi}_{{}_{\,\, j'' i''}} \big(\widehat{n} \cdot \widehat{l''}\big) \,
\Big[
\underset{1 \,\, 01}{M}_{{}_{l'' \,\, i''j''}}^{{}^{l \,\, ij}} 
+\underset{1 \,\, 10}{M}_{{}_{l'' \,\, i''j''}}^{{}^{l \,\, ij}}
\Big],
\end{multline*}
\begin{multline}\label{FT(xmu.phi)}
\big(\widetilde{x_2 \phi}\big)_{{}_{ji}}(\widehat{n}\cdot \widehat{l})
= \Big(\widetilde{x_2} \,\, \ast \,\, \widetilde{\phi}\Big)_{{}_{ji}}(\widehat{n}\cdot \widehat{l})
\\
=
{\textstyle\frac{3}{\sqrt{2}4\pi}} 
\sum\limits_{\substack{ l'',j'',i''}} 
{\textstyle\frac{2l''+1}{2l +1}} 
\,
\widetilde{\phi}_{{}_{\,\, j'' i''}} \big(\widehat{n} \cdot \widehat{l''}\big) \,
\Big[
\underset{1 \,\, 01}{M}_{{}_{l'' \,\, i''j''}}^{{}^{l \,\, ij}} 
-\underset{1 \,\, 10}{M}_{{}_{l'' \,\, i''j''}}^{{}^{l \,\, ij}}
\Big],
\end{multline}
\begin{multline*}
\big(\widetilde{x_3 \phi}\big)_{{}_{ji}}(\widehat{n}\cdot \widehat{l})
= \Big(\widetilde{x_3} \,\, \ast \,\, \widetilde{\phi}\Big)_{{}_{ji}}(\widehat{n}\cdot \widehat{l})
\\
=
{\textstyle\frac{i}{8\pi}} 
\sum\limits_{\substack{ l'',j'',i''}} 
{\textstyle\frac{2l''+1}{2l +1}} 
\,
\widetilde{\phi}_{{}_{\,\, j'' i''}} \big(\widehat{n} \cdot \widehat{l''}\big) \,
\Big[
\underset{1 \,\, 1 \,\,1}{M}_{{}_{l'' \,\, i''j''}}^{{}^{l \,\, ij}}
-\underset{1 \,\, -1 \,\, -1}{M}_{{}_{l'' \,\, i''j''}}^{{}^{l \,\, ij}}
\Big], 
\end{multline*}
where 
\begin{equation}\label{l'(w)l''(w)}
\widehat{l'}_{{}_{i'j'}}(\boldsymbol{w})\widehat{l''}_{{}_{i''j''}}(\boldsymbol{w}) = 
\sum\limits_{\substack{|l' -l''| \leq l \leq l'+l'' \\ -l \leq i,j \leq l}} 
\underset{l' \,\, i'j'}{M}_{{}_{l'' \,\, i''j''}}^{{}^{l \,\, ij}}
\widehat{l}_{{}_{ij}}(\boldsymbol{w}), \,\,\,\,\,
\boldsymbol{w} \in SU(2, \mathbb{C})
\end{equation}
is the product formula determined by the tensor product decomposition and the Clebsh-Gordan coefficients. 
Let us recall that we can use the Clebsch-Gordan coefficients 
$C^{{}^{\widehat{y},s}}_{{}_{\widehat{z},m \,\,\, \widehat{x},i}}$ 
of the group $G = SU(2,\mathbb{C})$ to compute the multiplication matrix $M$:
\[
\underset{\widehat{z}, m,k}{M}_{{}_{\widehat{x} \,\, ji}}^{{}^{\widehat{y} \,\, sq}} =
C^{{}^{\widehat{y},s}}_{{}_{\widehat{z},m \,\,\, \widehat{x},-i}}C^{{}^{\widehat{y},q}}_{{}_{\widehat{z},k \,\,\, \widehat{x},-j}}.
\] 
For this particular group the Clebsch-Gordan coefficients respect the identity
\[
C^{{}^{\widehat{y},s}}_{{}_{\widehat{z},0 \,\,\, \widehat{x},i}} = \delta_{{}_{i}}^{{}^{s}} C^{{}^{\widehat{y},s}}_{{}_{\widehat{z},0 \,\,\, \widehat{x},i}}.
\]
Thus, it follows
that in the formulas for the values of $\widetilde{x_\mu \phi}$ at $\widehat{n}\cdot\widehat{l}$, the number $l''$
ranges at most over the following three possible values $l''=l-1,l,l+1$ and with the finite range of $j'',i''$,
similarly restricted by the value of $j,i$, and differing from $j,i$ by a quantity of the order of magnitude comparable to unity. 
For large values of $n,l$, the above formulas contain differences of the Fourier transform $\widetilde{\phi}_{ji}(\widehat{n}\cdot\widehat{l})$
taken at $n,l,j,i$ differing be a tiny value of the order of magnitude comparable to unity, so in passing to 
very large $n,l$ (possible for the large radius $R$ of the Cauchy surface), we indeed obtain a ``differentiation'' operation.   
This is in general the case only up to constants $c_\mu$, respectively, for $\widetilde{x_\mu} \, \ast \, \widetilde{\phi}$,
which, as we see, is equal $c_0 = 4\sqrt{\pi}$ for (\ref{FT(x0.phi)}) and for our particular choice of $x_0$. These constants $c_\mu$
may, of course, be converted into $1$ after multiplication of $x_\mu$ by the respective $1/c_\mu$.

We should emphasize that the convolution $\ast$ considered here is the \emph{commutative} convolution
$\ast$, being defined through
the Fourier transform of the commutative pointwise multiplication of functions on 
$\widetilde{\mathbb{S}^1}\times SU(2,\mathbb{C})$. In the Fourier transform image of
$\mathscr{E}$, or more generally, in (a proper subspace of the) Fourier transform image of the dual space 
$\mathscr{E}^*$ of distributions, there is also another \emph{noncommutative} product, which is the Fourier transform image
of the noncommutative convolution product of functions on the noncommutative group $\widetilde{\mathbb{S}^1}\times SU(2,\mathbb{C})$,
and in the Fourier image space, consisting of matrix valued functions on the discrete space of characters $\widehat{n}\cdot \widehat{l}$
it is given by the point-$\widehat{n}\cdot \widehat{l}$-vise matrix multiplication, which is, of course, noncommutative,
compare Subsection \ref{GeneralizedSchrodinger-VonNeumannPairs}.

Using the formulas (\ref{FT(X_0)})-(\ref{tildeX_3}) of Subsection \ref{GeneralizedSchrodinger-VonNeumannPairs}, 
we can easily compute Fourier transforms of the Dirac
delta function and all its derivatives $X^\alpha \delta$, where $\alpha$ is a multiindex sequence $(\alpha_0,\alpha_1, \alpha_2,\alpha_3)$
and $X^\alpha\delta = (X_{0})^{\alpha_0}(X_{1})^{\alpha_1}(X_{2})^{\alpha_2}(X_{3})^{\alpha_3}\delta$, although we have to be carefull,
as now the operators $X_\mu, X_\nu$, $\nu \neq 0$ do not commute, so that the order of the $\alpha_1,\alpha_2,\alpha_3$ is important. We have:
\begin{multline}\label{FT(Xalphadelta)}
\widetilde{X^\alpha \delta}(\widehat{n}\cdot \widehat{l})
\\
 =  
\big(\mathbb{A}_{{}_{0}}(\widehat{n}\cdot \widehat{l})\big)^{\alpha_0} 
 \big(\mathbb{A}_{{}_{1}}(\widehat{n}\cdot \widehat{l}) \big)^{\alpha_1}  
\big(\mathbb{A}_{{}_{2}}(\widehat{n}\cdot \widehat{l}) \big)^{\alpha_2}
 \big(\mathbb{A}_{{}_{3}}(\widehat{n}\cdot \widehat{l}) \big)^{\alpha_3} 
= \mathbb{A}^\alpha(\widehat{n}\cdot \widehat{l}), 
\\
\alpha = (\alpha_0,\alpha_1, \alpha_2,\alpha_3),
\end{multline}
\begin{equation}\label{A0(n.l)}
\mathbb{A}_{{}_{0}}(\widehat{n}\cdot \widehat{l}) = \mathbb{A}_{{}_{0 \, \widehat{n}}} =  {\textstyle\frac{-in}{2}} \, \boldsymbol{1},
\end{equation}
and with the matrix valued functions 
\[
\widehat{n} \cdot \widehat{l} \longrightarrow \mathbb{A}_{{}_{i}}(\widehat{n}\cdot \widehat{l}) =  {\overline{\mathbb{A}_{{}_{i \, \widehat{l}}}}}^T, 
\,\,\,\,\, i=1,2,3,
\]
where $\mathbb{A}_{{}_{i \, \widehat{l}}}$ are equal to the standard generators of the irreducible representation $\widehat{l}$ of $SU(2,\mathbb{C})$,
and are given by the formulas (\ref{tildeX_1})-(\ref{tildeX_1}) of Subsection \ref{GeneralizedSchrodinger-VonNeumannPairs}.
In particular
\begin{equation}\label{A_0^a(n.l)}
\widetilde{(X_{0})^a \delta}(\widehat{n}\cdot \widehat{l}) = 
\big(\mathbb{A}_{{}_{0}}\big)^{a}(\widehat{n}\cdot \widehat{l}) = \big(\mathbb{A}_{{}_{0 \, \widehat{n}}}\big)^a =
\big({\textstyle\frac{-in}{2}}\big)^a \, \boldsymbol{1}.
\end{equation}
This is the case for the standard Dirac delta function $\delta$, centered at $(t,\boldsymbol{w}) = (0,e)$. For the general Dirac delta function
$\delta_{{}_{t_1,\boldsymbol{w}_1}}$, centered at $(t_1,\boldsymbol{w}_1)$, we have
\[
\widetilde{X^\alpha \delta_{{}_{t_1,\boldsymbol{w}_1}}}(\widehat{n}\cdot \widehat{l}) = 
e^{int_1/2} \mathbb{A}^\alpha(\widehat{n}\cdot \widehat{l})\widehat{l}(\boldsymbol{w}_1), 
\]
in particular
\begin{equation}\label{FT(Xalphadelta2pie)}
\widetilde{X^\alpha \delta_{{}_{2\pi,e}}}(\widehat{n}\cdot \widehat{l}) = 
e^{in\pi} \mathbb{A}^\alpha(\widehat{n}\cdot \widehat{l}) = (-1)^n \mathbb{A}^\alpha(\widehat{n}\cdot \widehat{l}). 
\end{equation}

We can remove the additional
constant $2$ from (\ref{FT(x0.phi)}) by redefining the function $x_0$, (\ref{x0}), and multiplying it by $1/2$, 
and get for such function:
\begin{equation}\label{x0canonical}
x_0(t,w) = {\textstyle\frac{1}{2}} \sin t = 
 {\textstyle\frac{-i}{4}}\big[\widehat{2}(t)-\widehat{-2}(t)\big] = 
{\textstyle\frac{1}{4i}}\big[e^{it}-e^{-it}\big].
\end{equation}
After this redefinition of $x_0$ we get
\begin{equation}\label{FT(x0.phi)Canonical}
\big(\widetilde{x_0 \phi}\big)_{{}_{ji}}(\widehat{n}\cdot \widehat{l})
= \Big(\widetilde{x_0} \,\, \ast \,\, \widetilde{\phi}\Big)_{{}_{ji}}(\widehat{n}\cdot \widehat{l})
=
i\, {\textstyle\frac{\widetilde{\phi}_{{}_{\,\, j i}} \big(\widehat{n+2} \cdot \widehat{l}\big) 
-\widetilde{\phi}_{{}_{\,\, j i}} \big(\widehat{n-2} \cdot \widehat{l}\big)}{(n+2) - (n-2)}}.
\end{equation}
But the vector field $X_0 = \partial_t$,
or differential  operator $X_0 = \partial_{t}$, will have to be replaced with $X_0 = 2 \, \partial_t$,
in order to keep the mutually dual character of $X_0$ and $x_0$ at the ``subtraction points'' $(0,e),(2\pi,e)$
\begin{equation}\label{duality=X_0(x_0)=1}
X_0\big(x_0\big)(0,e) = 1, \,\,\,\, X_0\big(x_0\big)(2\pi,e) =1.
\end{equation}
For the new $X_0$, dual to (\ref{x0canonical}), we have
\begin{equation}\label{A_0(n.l)2}
\widetilde{X_{0} \delta}(\widehat{n}\cdot \widehat{l}) = 
\big(\mathbb{A}_{{}_{0}}\big)(\widehat{n}\cdot\widehat{l}) = \mathbb{A}_{{}_{0 \, \widehat{n}}} =
2 \,\, \big({\textstyle\frac{-in}{2}}\big) \, \boldsymbol{1}
=  \big(-i \, n\big) \,\, \boldsymbol{1},
\end{equation}
and
\[
\widetilde{(X_{0})^a \delta}(\widehat{n}\cdot \widehat{l}) = 
\big(\mathbb{A}_{{}_{0}}\big)^{a}(\widehat{n}\cdot\widehat{l}) = \big(\mathbb{A}_{{}_{0 \, \widehat{n}}}\big)^a 
=  \big(-i \, n\big)^a \,\, \boldsymbol{1}.
\]

Still, we need explicit formulas for $\widetilde{\theta}$ and $\widetilde{x^\alpha \theta} = \widetilde{x^\alpha} \, \ast \, \widetilde{\theta}$, 
for the calculations that we are going to do soon. In fact, on the Einstein Universe, which is static, with the time translation symmetry, and
the corresponding distinguished time like Killing vector field $\partial_t$ and with no \emph{global} Lorentz group symmetry, the case in which the subtraction terms
contain only time derivatives is crucial, with $\Omega'$ equal (\ref{Omega'phiEU}), and containing terms with time derivatives only.
Therefore, first of all, we will need   
\[
\widetilde{(x_0)^a \theta} = \underbrace{\widetilde{x_0} \, \ast \ldots \, \ast \widetilde{x_0}}_{\textrm{$a$ times}} \, \ast \, \widetilde{\theta},
\,\,\,\,\,\,\, n=0,1,2, \ldots.
\]
In fact, $\widetilde{\theta}$ we have already computed in Subsection \ref{WhiteNoiseFreeFieldsonEU}, and it is equal
\[
  \widetilde{\theta}_{{}_{ji}}(\widehat{n}\cdot\widehat{l})
=\begin{cases}
    {\textstyle\frac{1}{\sqrt{\pi}}}{\textstyle\frac{i}{n}} \delta_{{}_{l \,0}} \delta_{{}_{j \,0}}\delta_{{}_{i \,0}}, 
& \text{if $n \in 2\mathbb{Z} +1$},\\
    0, & \text{if $n \in 2\mathbb{Z}\setminus \{0\}$}, \\
    \sqrt{\pi} \, \delta_{{}_{l \,0}} \delta_{{}_{j \,0}}\delta_{{}_{i \,0}}, & \text{if $n=0$}.
  \end{cases}
\]
For $x_0$ of order $\omega=2$, normalized as in (\ref{x0canonical}), for which (\ref{FT(x0.phi)Canonical}) holds, repeated application of
the formula (\ref{FT(x0.phi)Canonical}) to $\widetilde{\phi} = \widetilde{\theta}$ gives the following result
\begin{multline}\label{FT((x^a)theta)}
  \big(\widetilde{(x_0)^a \theta}\big)_{{}_{ji}}(\widehat{n}\cdot\widehat{l}) 
= \big(\widetilde{(x_0)^a} \, \ast  \, \widetilde{\theta}\big)_{{}_{ji}}(\widehat{n}\cdot\widehat{l})
= \big(\widetilde{x_0} \, \ast \ldots \, \ast \widetilde{x_0} \, \ast \widetilde{\theta}\big)_{{}_{ji}}(\widehat{n}\cdot\widehat{l})
\\
=\begin{cases}
    {\textstyle\frac{1}{\sqrt{\pi}}}\, 
{\textstyle\frac{i^{a+1}(-1)^a a!}{(n+2a)(n+2a-4)(n+2a-2\cdot4)(n+2a-3\cdot 4) \ldots (n-2a)}}
\delta_{{}_{l \,0}} \delta_{{}_{j \,0}}\delta_{{}_{i \,0}}, 
& \text{if $n \in 2\mathbb{Z} +1$},\\
  \sqrt{\pi} (-1)^a {a \choose 0} {\textstyle\frac{i^a}{4^a}}\, \delta_{{}_{l \,0}} \delta_{{}_{j \,0}}\delta_{{}_{i \,0}}, & \text{if $n=2a$}, \\
  \sqrt{\pi} (-1)^{a+1} {a \choose 1} {\textstyle\frac{i^a}{4^a}}\, \delta_{{}_{l \,0}} \delta_{{}_{j \,0}}\delta_{{}_{i \,0}}, & \text{if $n=2a-4$}, \\
  \sqrt{\pi} (-1)^{a+2} {a \choose 2} {\textstyle\frac{i^a}{4^a}}\, \delta_{{}_{l \,0}} \delta_{{}_{j \,0}}\delta_{{}_{i \,0}}, & \text{if $n=2a-2\cdot 4$}, \\
  \sqrt{\pi} (-1)^{a+3} {a \choose 3} {\textstyle\frac{i^a}{4^a}}\, \delta_{{}_{l \,0}} \delta_{{}_{j \,0}}\delta_{{}_{i \,0}}, & \text{if $n=2a-3\cdot 4$}, \\
  \vdots & \vdots \\
  \sqrt{\pi} (-1)^{a+a} {a \choose a} {\textstyle\frac{i^a}{4^a}}\, \delta_{{}_{l \,0}} \delta_{{}_{j \,0}}\delta_{{}_{i \,0}}, & \text{if $n=-2a$}, \\
 0, & \text{otherwise}. \\
  \end{cases}
\end{multline}
Note that indeed for $n$ large (and odd, where $\widetilde{\theta}\neq 0$) 
in comparison to $a$, this ``finite difference'' formula $\widetilde{(x_0)^a} \, \ast \, \widetilde{\theta}$
tends to the ordinary differential 
\[
\big(i\partial_{n}\big)^{a}\big({\textstyle\frac{1}{\sqrt{\pi}}}{\textstyle\frac{i}{n}}\big)\delta_{{}_{l \,0}} \delta_{{}_{j \,0}}\delta_{{}_{i \,0}},
\]
as expected.   

In order to obtain the formula analogue to (\ref{FT((x^a)theta)}) for $x_0$ of the system $x_0,x_1,x_2,x_3$, dual to $X_0 = 2\, \partial_t$, $X_1$,
$X_2$, $X_3$,
up to any finite order $\omega$,
at $(0,e)$ and at $(2\pi,e)$, we just use (\ref{FT((x^a)theta)}), as these $x_0$ are linear combinantions of powers of $x_0$
equal (\ref{x0canonical}), and the formula (\ref{FT((x^a)theta)}) gives convolution of the Fourier transform of any natuar power of this $x_0$ with
$\widetilde{\theta}$. Similarly, repeated application of (\ref{FT(xmu.phi)}) will give the formula for 
\[
\Big(x_{1}^{\alpha_1}  x_{2}^{\alpha^2} x_{3}^{\alpha_3}\theta\Big)^{\widetilde{\,\,\,\,\,\,\,\,\,\,\,\,\,}} 
= \widetilde{x_{1}^{\alpha_1}} \, \ast \,  \widetilde{x_{2}^{\alpha_2}} \, \ast \,  \widetilde{x_{3}^{\alpha_3}} \, \ast \, \widetilde{\theta}, 
\]
for the functions $x_1,x_2,x_3$ of the system $x_0,x_1,x_2,x_3$, dual to $X_0 = 4\sqrt{\pi}$, $X_1$, $X_2$, $X_3$,
up to any finite order $\omega$, at $(0,e)$ and at $(2\pi,e)$. Thus, the explicit
formula for $\widetilde{x^\alpha \theta} = \widetilde{x^\alpha} \ast  \widetilde{\theta}$ can be easily obtained
for the system $x_0$, $x_1$, $x_2$, $x_3$, which is dual to $X_0 = 2\, \partial_t$, $X_1$, $X_2$, $X_3$,
up to finite order $\omega$, at $(0,e)$ and at $(2\pi,e)$, for any natural $\omega$.

With this at hand, and with the natural Fourier transform at our disposal, we are ready to implement the method of Scharf, reported
in Subsection \ref{WickForChronological}, eqs. (\ref{FT(retkappaq)})-(\ref{FT(retCausalkappaq)}),
and (\ref{DispersionFormulaForRetarded}), (\ref{InvariantDispersionFormulaForRetarded}), for the computation
of the Fourier transform of the retarded part of $\kappa_q$ and for the investigation  of convergence of 
\begin{equation}\label{<thetakappa,Omega'phi>}
\big\langle \textrm{ret} \, \kappa_q, \phi\big\rangle \overset{\textrm{df}}{=} \big\langle \theta\kappa_q, \Omega'\phi\big\rangle, \,\,\,\,\, \phi \in \mathscr{E}.
\end{equation} 

We start with the simplest situation 
with only one ``subtraction point'' $(0,e)$ and with the operator $\Omega'$ equal (\ref{Omega'phiEU}).
We are using $x_0$ equal (\ref{x0canonical}) in $\Omega'$, and with $X_0 = 2 \, \partial_t$, dual to $x_0$
for which (\ref{FT(x0.phi)Canonical}) and (\ref{A_0(n.l)2}) hold.

We assume that $\omega'$ is equal to the order of $\kappa_q$. More precisely, we assume that for each $l$ and each matrix  $ji$-element 
of the Fourier transform $\widetilde{\kappa_q}$ at $\widehat{n}\cdot\widehat{l}$, the function 
\[
n \longmapsto \widetilde{\kappa_q}_{{}_{ji}}(\widehat{n}\cdot\widehat{l})
\] 
behaves asymtotically as a polynomial in $n$ of degree $\omega'$, or that $\omega'$
is the infimum of numbers $\nu$ such that for a polynomial $P(n)$ of degree $\nu$ the function
\[
n \longmapsto 
{\textstyle\frac{\widetilde{\kappa_q}_{{}_{ji}}(\widehat{n}\cdot\widehat{l})}{P(n)}}
\]
is bounded. 

The functions $x_0, x_1,x_2,x_3$, play the role around $(0,e)$ (and around $(2\pi,e)$),
analogue to the Cartesian coordinates $x_0, x_1,x_2,x_3$ on $\mathbb{R}^4$ around zero. Next we verify convergence of
(\ref{<thetakappa,Omega'phi>}).

Since
\begin{gather*}
X^\alpha \phi(0,e) = (-1)^{|\alpha|} \big\langle X^\alpha \delta, \phi \big\rangle
= (-1)^{|\alpha|} \big\langle \widetilde{X^\alpha \delta}, \widetilde{\phi} \big\rangle =
(-1)^{|\alpha|} \big\langle \mathbb{A}^{\alpha}, \widetilde{\phi} \big\rangle,
\\
\big(X_0\big)^{a} \phi (0,e) = (-1)^a \big\langle \big(X_0\big)^{a} \delta, \phi \big\rangle
= (-1)^a \Big\langle \widetilde{\big(X_0\big)^{a} \delta}, \,\, \widetilde{\phi} \Big\rangle =
(-1)^a \big\langle \big(\mathbb{A}_{{}_{0}}\big)^{a}, \widetilde{\phi} \big\rangle,
\\
\widetilde{x^\alpha w} = \widetilde{x^\alpha} \,  \ast \, \widetilde{w},
\,\,\, \widetilde{(x_{0})^a w} = \widetilde{(x_0)^a} \,\, \ast \,\, \widetilde{w}
= \underbrace{\widetilde{x_0} \, \ast \, \ldots \, \ast \,  \widetilde{x_0}}_{\textrm{$a$ times}} \,\, \ast \,\, \widetilde{w},
\\
\mathbb{A}_{{}_{0}}(\widehat{n}\cdot\widehat{l}) 
=   \big(- i \, n \big) \,\, \boldsymbol{1},
\end{gather*}
then from (\ref{<thetakappa,Omega'phi>}) and (\ref{Omega'phiEU}) we get 
\begin{multline*}
\langle  \widetilde{\textrm{ret} \, \kappa_q}, \widetilde{\phi} \rangle
= \big\langle \kappa_q, \theta \Omega'\phi\big\rangle = 
\big\langle \widetilde{\kappa_q}, \widetilde{\theta} \ast \widetilde{\Omega'\phi}\big\rangle
\\
=\Bigg\langle \widetilde{\kappa_q}, \widetilde{\theta} \ast \Bigg[\widetilde{\phi} 
- \sum\limits_{a=0}^{\omega} {\textstyle\frac{(-1)^a}{a!}} \widetilde{(x_0)^a} \,\, \ast \,\, \widetilde{w} 
\,\,\,
\Big\langle \big(\mathbb{A}_{{}_{0}}\big)^{a}, \widetilde{\phi} \Big\rangle \Bigg] \Bigg\rangle
\\
=\Bigg\langle \widetilde{\theta} \ast \widetilde{\kappa_q}, \,\,\, \widetilde{\phi} 
- \sum\limits_{a=0}^{\omega} {\textstyle\frac{(-1)^a}{a!}} \widetilde{(x_0)^a} \,\, \ast \,\, \widetilde{w} 
\,\,\,
\Big\langle \big(\mathbb{A}_{{}_{0}}\big)^{a}, \widetilde{\phi} \Big\rangle \Bigg\rangle
\end{multline*}
\begin{multline*}
=
\sum\limits_{n,l} (2l+1) 
\textrm{Tr} \Big[ \widetilde{\theta} \ast \widetilde{\kappa_q}(\widehat{n}\cdot \widehat{l}) \widetilde{\phi}(\widehat{n}\cdot \widehat{l})  \Big]
\\
-
\sum\limits_{a=0}^{\omega}
{\textstyle\frac{(-1)^a}{a!}}
 \sum_{\substack{n,l \\ n',l'}}
 (2l+1) 
\textrm{Tr} \Big[ \widetilde{\theta} \ast \widetilde{\kappa_q}(\widehat{n}\cdot \widehat{l}) \widetilde{(x_0)^a} \, \ast \, \widetilde{w}(\widehat{n}\cdot \widehat{l})  \Big]
(2l'+1)
\textrm{Tr} \Big[ \big(\mathbb{A}_{{}_{0}}\big)^{a}(\widehat{n'}\cdot \widehat{l'}) \widetilde{\phi}(\widehat{n'}\cdot \widehat{l'})  \Big]
\end{multline*}
We exchange the summation indices $n,l$ and $n',l'$
in both discrete pairing integrals or sums $\langle \cdot , \langle \cdot, \cdot\rangle \rangle$ 
(given by the Plancherel inner product induced by the Fourier transform,
compare Subsection \ref{GeneralizedSchrodinger-VonNeumannPairs}) in the subtraction term, and get
\begin{multline*}
\langle  \widetilde{\textrm{ret} \, \kappa_q}, \widetilde{\phi} \rangle
=
\sum\limits_{n,l} (2l+1) 
\textrm{Tr} \Big[ \widetilde{\theta} \ast \widetilde{\kappa_q}(\widehat{n}\cdot \widehat{l}) \widetilde{\phi}(\widehat{n}\cdot \widehat{l})  \Big]
\\
-
\sum\limits_{a=0}^{\omega}
{\textstyle\frac{(-1)^a}{a!}}
 \sum_{\substack{n,l \\ n',l'}}
 (2l'+1) 
\textrm{Tr} \Big[ \widetilde{\theta} \ast \widetilde{\kappa_q}(\widehat{n'}\cdot \widehat{l'}) \widetilde{(x_0)^a} \, \ast \, \widetilde{w}(\widehat{n'}\cdot \widehat{l'})  \Big]
(2l+1)
\textrm{Tr} \Big[ \big(\mathbb{A}_{{}_{0}}\big)^{a}(\widehat{n}\cdot \widehat{l}) \widetilde{\phi}(\widehat{n}\cdot \widehat{l})  \Big]
\end{multline*}
\begin{multline*}
=
\sum\limits_{n,l} (2l+1) 
\textrm{Tr} \Big[ \widetilde{\theta} \ast \widetilde{\kappa_q}(\widehat{n}\cdot \widehat{l}) \widetilde{\phi}(\widehat{n}\cdot \widehat{l})  \Big]
\\
-
\sum\limits_{a=0}^{\omega}
{\textstyle\frac{(-1)^a}{a!}}
 \sum_{\substack{n,l}}
 \Big\langle \widetilde{\theta} \ast \widetilde{\kappa_q}, \,\, \widetilde{(x_0)^a} \, \ast \, \widetilde{w} \Big\rangle
(2l+1)
\textrm{Tr} \Big[ \big(\mathbb{A}_{{}_{0}}\big)^{a}(\widehat{n}\cdot \widehat{l}) \widetilde{\phi}(\widehat{n}\cdot \widehat{l})  \Big]
\end{multline*}
\begin{multline*}
=
\sum\limits_{n,l} (2l+1) 
\textrm{Tr} \Bigg[ \Bigg( \widetilde{\theta} \ast \widetilde{\kappa_q}(\widehat{n}\cdot \widehat{l}) 
-
\sum\limits_{a=0}^{\omega}
{\textstyle\frac{(-1)^a (\mathbb{A}_{{}_{0}})^{a}(\widehat{n}\cdot \widehat{l})}{a!}}
\,\,
 \Big\langle \widetilde{\theta} \ast \widetilde{\kappa_q},  \,\, \widetilde{(x_0)^a} \, \ast \, \widetilde{w} \Big\rangle
\Bigg)
 \widetilde{\phi}(\widehat{n}\cdot \widehat{l}) 
 \Bigg]
\\
=
\sum\limits_{n,l} (2l+1) 
\textrm{Tr} \Bigg[ \Bigg( \widetilde{\theta} \ast \widetilde{\kappa_q}(\widehat{n}\cdot \widehat{l}) 
-
\sum\limits_{a=0}^{\omega}
{\textstyle\frac{(-1)^a (\mathbb{A}_{{}_{0}})^{a}(\widehat{n} \cdot \widehat{l})}{a!}}
\,\,
 \Big\langle \widetilde{\theta} \ast \, \widetilde{(x_0)^a} \,  \ast \, \widetilde{\kappa_q},  \,\, \widetilde{w} \Big\rangle
\Bigg)
 \widetilde{\phi}(\widehat{n}\cdot \widehat{l}) 
 \Bigg]
\end{multline*}
Thus it follows that
\[
\widetilde{\textrm{ret} \, \kappa_q}(\widehat{n}\cdot \widehat{l})
=
\widetilde{\theta} \ast \widetilde{\kappa_q}(\widehat{n}\cdot \widehat{l}) 
-
\sum\limits_{a=0}^{\omega}
{\textstyle\frac{(-1)^a (\mathbb{A}_{{}_{0}})^{a}(\widehat{n} \cdot \widehat{l})}{a!}}
\,\,
\Big\langle \widetilde{\theta} \ast \, \widetilde{(x_0)^a} \,  \ast \, \widetilde{\kappa_q},  \,\, \widetilde{w} \Big\rangle.
\]
Using the convolution formula of Subsection \ref{WhiteNoiseFreeFieldsonEU} 
and  the formula for $\widetilde{\theta}$ we have
\begin{multline*}
\big(\widetilde{\theta} \ast \widetilde{f}\big)_{{}{ji}}(\widehat{n}\cdot \widehat{l}) = 
\big(\widetilde{\theta f}\big)_{{}{ji}}(\widehat{n}\cdot \widehat{l})
\\
=
{\textstyle\frac{1}{\sqrt{4\pi}}} \int\limits_{{}_{\widetilde{\mathbb{S}^1}\times SU(2,\mathbb{C})}} 
\theta(t) f(t, \boldsymbol{w}) \, \overline{\widehat{n}(t)} \, \overline{\widehat{l}_{{}_{ij}}(\boldsymbol{w})} \,
dt \, d \boldsymbol{w}
=
{\textstyle\frac{1}{\sqrt{4\pi}}}
\sum\limits_{n''} \widehat{\theta}_{n''} \widetilde{f}_{{}_{ji}}(\widehat{n-n''}
\cdot \widehat{l})
\end{multline*}
for $f \in \mathscr{E}^*$ (whenever convergent), with 
\[
  \widehat{\theta}_{{}_{n}}
=\begin{cases}
    {\textstyle\frac{1}{\sqrt{\pi}}}{\textstyle\frac{i}{n}}, 
& \text{if $n \in 2\mathbb{Z} +1$},\\
    0, & \text{if $n \in 2\mathbb{Z}\setminus \{0\}$}, \\
    \sqrt{\pi}, & \text{if $n=0$}.
  \end{cases}
\]
and substituting into the last formula for $\widetilde{\textrm{ret} \, \kappa_q}$ we get
\begin{multline*}
\widetilde{\textrm{ret} \, \kappa_q}(\widehat{n}\cdot \widehat{l}) =
{\textstyle\frac{1}{\sqrt{4\pi}}}
\sum\limits_{n''} \widehat{\theta}_{n''} \widetilde{\kappa_q}(\widehat{n-n''} \cdot \widehat{l})
\\
-
\sum\limits_{a=0}^{\omega}
{\textstyle\frac{(-1)^a (\mathbb{A}_{{}_{0}})^{a}(\widehat{n} \cdot \widehat{l})}{a!}}
\sum\limits_{n',l'} (2l'+1) \textrm{Tr} \Bigg[\Big(\widetilde{\theta} \ast \, \widetilde{(x_0)^a} \,  \ast \, \widetilde{\kappa_q}\Big)(\widehat{n'}\cdot \widehat{l'})
\widetilde{w}(\widehat{n'}\cdot \widehat{l'}) \Bigg]  
\\
=
{\textstyle\frac{1}{\sqrt{4\pi}}}\sum\limits_{n''} \widehat{\theta}_{{}_{n''}} \widetilde{\kappa_q}(\widehat{n-n''} \cdot \widehat{l})
\\
-
{\textstyle\frac{1}{\sqrt{4\pi}}}
\sum\limits_{a=0}^{\omega}
{\textstyle\frac{(-1)^a (\mathbb{A}_{{}_{0}})^{a}(\widehat{n} \cdot \widehat{l})}{a!}}
\sum\limits_{n'',n',l'} (2l'+1) 
\textrm{Tr} \Bigg[\widehat{\theta}_{{}_{n''}} \, \Big(\widetilde{(x_0)^a} \,  \ast \, \widetilde{\kappa_q}\Big)(\widehat{n'-n''}\cdot \widehat{l'})
\widetilde{w}(\widehat{n'}\cdot \widehat{l'}) \Bigg].  
\end{multline*}
Therefore we arrive at the formula
\begin{multline}\label{FT(retkappaq)1EU}
\widetilde{\textrm{ret} \, \kappa_q}(\widehat{n}\cdot \widehat{l}) 
={\textstyle\frac{1}{\sqrt{4\pi}}}
\sum\limits_{n''} \widehat{\theta}_{{}_{n''}} \Bigg\{ \widetilde{\kappa_q}(\widehat{n-n''} \cdot \widehat{l})
\\
-
\sum\limits_{a=0}^{\omega}
{\textstyle\frac{(-1)^a (\mathbb{A}_{{}_{0}})^{a}(\widehat{n} \cdot \widehat{l})}{a!}}
\sum\limits_{n',l'} (2l'+1) \textrm{Tr} \Big[\Big(\widetilde{(x_0)^a} \,  \ast \, \widetilde{\kappa_q}\Big)(\widehat{n'-n''}\cdot \widehat{l'})
\widetilde{w}(\widehat{n'}\cdot \widehat{l'}) \Big]
\Bigg\}.  
\end{multline}
Derivation of this formula should be compared with the derivation of the analogue formula (\ref{FT(retkappaq)1}) of
Subsection \ref{WickForChronological} on the Minkowski space-time. Here, similarly as in Subsection
\ref{WickForChronological}, both terms in (\ref{FT(retkappaq)1EU}), if taken separately, are not expected to be convergent.

Now, on the compact $\widetilde{\mathbb{S}^1}\times SU(2, \mathbb{C})$ the Fourier transform $\widetilde{\textrm{ret} \, \kappa_q}$ 
of any well-defined distribution, say of $\textrm{ret} \, \kappa_q$ in $\mathscr{E}^*$, is a well-defined function
of the charcters $\widehat{n}\cdot\widehat{l}$, \emph{i.e.} on the dual group. Moreover, because the function $x_0$ is smooth, then it is a multiplier
of $\mathscr{E}$ and, by duality, a mulitplier of $\mathscr{E}^*$. Therefore $\widetilde{(x_0)}$ is a convolutor of $\widetilde{\mathscr{E}^*}$. It means that
for each natural $a$, $\widetilde{(x_0)^a} \, \ast \, \widetilde{\textrm{ret} \, \kappa_q}$ is in $\mathscr{E}^*$
whenever $\widetilde{\textrm{ret} \, \kappa_q}$ is in $\mathscr{E}^*$, or that for any natural $a$ 
\[
\widehat{n}\cdot\widehat{l} \longmapsto \Big(\widetilde{(x_0)^a} \, \ast \, \widetilde{\textrm{ret} \, \kappa_q}\Big)(\widehat{n}\cdot\widehat{l})
\]
is a well-defined function on the dual of the group  $\widetilde{\mathbb{S}^1}\times SU(2, \mathbb{C})$, 
whenever $\widetilde{\textrm{ret} \, \kappa_q}$ is in $\mathscr{E}^*$. This means that if
$\widetilde{\textrm{ret} \, \kappa_q}$ is in $\mathscr{E}^*$, then each character $\widehat{n}\cdot\widehat{l}$
can be chosen as the ``normalization point'' at which all time like ``derivatives'' $\widetilde{(x_0)^a} \, \ast \, \widetilde{\textrm{ret} \, \kappa_q}$
in the discrete ``momentum space'' exist, up to any order $a$. 
Thus for each  $\widehat{n}\cdot\widehat{l}$ in the argument of $\widetilde{\textrm{ret} \, \kappa_q}$ we can choose
a ``normalization point'' $\widehat{n'''}\cdot\widehat{l}$, depending on  $\widehat{n}\cdot\widehat{l}$,
and subtract the first $\omega$ time-like-terms of the ``Taylor expansion''
around the ``normalization point'',
and define Fourier transform $\big[\widetilde{\textrm{ret} \, \kappa_q}\big]'''$  of another possible retarded part 
of $\kappa_q$, putting
\begin{equation}\label{TaylorSubtracionEU}
\big[\widetilde{\textrm{ret} \, \kappa_q}\big]'''(\widehat{n}\cdot\widehat{l})
\overset{\textrm{df}}{=}
\widetilde{\textrm{ret} \, \kappa_q}(\widehat{n}\cdot\widehat{l}) - \sum\limits_{b=0}^{\omega} \,\,
{\textstyle\frac{\big(\mathbb{A}_{{}_{0}}(\widehat{n} \cdot \widehat{l})-\mathbb{A}_{{}_{0}}(\widehat{n'''} \cdot \widehat{l})\big)^b}{b!}}
\,\,\, \Big(\widetilde{(x_0)^b} \,\, \ast  \,\, \widetilde{\textrm{ret} \, \kappa_q}\Big)(\widehat{n'''}\cdot\widehat{l}).
\end{equation}
This is so because in this subtraction of the first ``Taylor expansion coefficients'' we have subtracted distribution of the form
(here with $\delta$ in single time variable $t$ and with $n'''$ understood as a constant)
\begin{equation}\label{FT(EpsteinGlaserReminderEU)}
\sum\limits_{b=0}^{\omega} \widetilde{C_{b,n'''}}(\widehat{l}) \,\,  \widetilde{\big(X_0\big)^b\delta}(\widehat{n}),
\,\,\,\, \widetilde{C_{b,n'''}}(\widehat{l}) = \sum\limits_{a=b}^{\omega} {\textstyle\frac{i^{a+b}(n''')^{a-b}}{b!(a-b)!}}\Big(\widetilde{(x_0)^a} \,\, \ast  \,\, \widetilde{\textrm{ret} \, \kappa_q}\Big)(\widehat{n'''}\cdot\widehat{l}),
\end{equation}
from $\widetilde{\textrm{ret} \, \kappa_q}$, equal to the Fourier transform of a distribution of the form
\begin{equation}\label{sumC_b(w)delta^(b)}
\sum\limits_{b=0}^{\omega} C_{b,n'''}(\boldsymbol{w}) \,\, \big(X_0\big)^b\delta(t),
\end{equation}
which is concentrated at the Cauchy surface $t=0$. But in general it is not equal zero on $\textrm{Im} \, \Omega'$. Therefore,
we get the inverse Fourier transform of
$\big[\widetilde{\textrm{ret} \, \kappa_q}\big]'''$
which, in principle, may be different from the ``natural''
$\textrm{ret} \, \kappa_q$ on $\textrm{Im} \, \Omega'$,
disturbing the ``natural'' formula of multiplication by $\theta$ on the subspace $\textrm{Im} \, \Omega'$.

Now we show that all time-like ``derivatives'' up to order $\omega$ of (\ref{TaylorSubtracionEU}) vanish at the ``normalization point'' 
$\widehat{n'''}\cdot\widehat{l}$, \emph{i.e.}
\[
\Big(\widetilde{(x_0)^a} \, \ast \, \big[\widetilde{\textrm{ret} \, \kappa_q}\big]'''\Big)(\widehat{n'''}\cdot\widehat{l}) = 0,
\,\,\,\, a=0,1,2, \ldots, \omega, \,\,\, \widehat{l} \in \widehat{SU(2,\mathbb{C})}
\]  
or, equivalently, that all terms containing the auxiliary function $\widetilde{w}$ drop out in  (\ref{TaylorSubtracionEU}).

We compute $\widetilde{(x_0)^b} \,\, \ast  \,\, \widetilde{\textrm{ret} \, \kappa_q}$ by application of $\widetilde{(x_0)^b} \,\, \ast \ldots$
to the formula (\ref{FT(retkappaq)1EU}), and substitute into (\ref{TaylorSubtracionEU}).
Since (\ref{dualityXmu-xmu}) holds, and $x_0$ and $X_0$ are mutually dual up to order $\omega$ at $(0,e)$, or 
\begin{gather*}
X_0(x_0)(0,e) = X_0(x_0)(2\pi,e) =1, 
\\
\Big(\big(X_0)^{a}\big(x_0\big)^{b}\Big)(0,e) = \Big(\big(X_0)^{a}\big(x_0\big)^{b}\Big)(2\pi,e) = 0, b<a \leq \omega
\end{gather*}
and thus 
\[
\Big(\big(X_0)^{a}\big(x_0\big)^{b})\Big)(0,e) = a! \delta_{ab}, \,\,\, a,b \leq \omega,
\]
(for example $x_0$ equal (\ref{x0canonical}) and $X_0 = 4\sqrt{\pi}\partial_t$ works for $\omega\leq 2$),
then 
\[
  \Big(\widetilde{(x_0)^b} \, \ast \, (\mathbb{A}_{{}_{0}})^a \Big)(\widehat{n}\cdot\widehat{l}) 
=\begin{cases}
    {\textstyle\frac{a!}{(a-b)!}} (\mathbb{A}_{{}_{0}})^{a-b}(\widehat{n}\cdot\widehat{l}) 
& \text{if $a\geq b$},\\
    0, & \text{if $a<b\leq \omega$},
  \end{cases}
\]
because 
\[
\widetilde{(x_0)^b} \ast (\mathbb{A}_{{}_{0}})^a =  \widetilde{(x_0)^b} \ast \widetilde{(X_0)^a\delta} 
= \Big((x_0)^b (X_0)^a\delta\Big)^{\widetilde{\,\,\,\,\,\,\,}}.
\]
Therefore, we get
\begin{multline*}
\Big(\widetilde{(x_0)^b} \,\, \ast  \,\, \widetilde{\textrm{ret} \, \kappa_q}\Big)(\widehat{n'''}\cdot \widehat{l}) 
=
{\textstyle\frac{1}{\sqrt{4\pi}}}
\sum\limits_{n''} \widehat{\theta}_{{}_{n''}} \Bigg\{ \Big(\widetilde{(x_0)^b} \, \ast \, \widetilde{\kappa_q}\Big)(\widehat{n'''-n''} \cdot \widehat{l})
\\
-
\sum\limits_{b\leq a}
{\textstyle\frac{(-1)^a (\mathbb{A}_{{}_{0}})^{a-b}(\widehat{n'''} \cdot \widehat{l})}{(a-b)!}}
\sum\limits_{n',l'} (2l'+1) \textrm{Tr} \Big[\Big(\widetilde{(x_0)^a} \,  \ast \, \widetilde{\kappa_q}\Big)(\widehat{n'-n''}\cdot \widehat{l'})
\widetilde{w}(\widehat{n'}\cdot \widehat{l'}) \Big]
\Bigg\}.  
\end{multline*}
Now we substitute this formula into (\ref{TaylorSubtracionEU}).
Since
\begin{multline}\label{BinomialFormulaA0}
\sum\limits_{b\leq a} \,\,
{\textstyle\frac{\big(\mathbb{A}_{{}_{0}}(\widehat{n} \cdot \widehat{l})-\mathbb{A}_{{}_{0}}(\widehat{n'''} \cdot \widehat{l})\big)^b}{b!}}
{\textstyle\frac{\big(\mathbb{A}_{{}_{0}}(\widehat{n'''} \cdot \widehat{l})\big)^{a-b}}{(a-b)!}}
\\
=
{\textstyle\frac{1}{a!}}
\sum\limits_{b\leq a}
{a \choose b}
\big(\mathbb{A}_{{}_{0}}(\widehat{n} \cdot \widehat{l})-\mathbb{A}_{{}_{0}}(\widehat{n'''} \cdot \widehat{l})\big)^b\big(\mathbb{A}_{{}_{0}}(\widehat{n'''} \cdot \widehat{l})\big)^{a-b}
=
{\textstyle\frac{\big(\mathbb{A}_{{}_{0}}(\widehat{n} \cdot \widehat{l})\big)^a}{a!}},
\end{multline}
we get
\begin{multline}\label{FT(retkappaq)3EU}
\big[\widetilde{\textrm{ret} \, \kappa_q}\big]'''(\widehat{n}\cdot\widehat{l}) 
=
{\textstyle\frac{1}{\sqrt{4\pi}}}
\sum\limits_{n''} \widehat{\theta}_{{}_{n''}} \Bigg\{ \widetilde{\kappa_q}(\widehat{n-n''} \cdot \widehat{l})
\\
-
\sum\limits_{b=0}^{\omega} \,\,
{\textstyle\frac{\big(\mathbb{A}_{{}_{0}}(\widehat{n} \cdot \widehat{l})-\mathbb{A}_{{}_{0}}(\widehat{n'''} \cdot \widehat{l})\big)^b}{b!}}
\Big(\widetilde{(x_0)^b} \,  \ast \, \widetilde{\kappa_q}\Big)(\widehat{n'''-n''}\cdot \widehat{l})
\Bigg\}
\end{multline}
The last term 
\[
\sum\limits_{n''} \widehat{\theta}_{{}_{n''}} \Bigg\{ 
\sum\limits_{b=0}^{\omega} \,\,
{\textstyle\frac{\big(\mathbb{A}_{{}_{0}}(\widehat{n} \cdot \widehat{l})-\mathbb{A}_{{}_{0}}(\widehat{n'''} \cdot \widehat{l})\big)^b}{b!}}
\Big(\widetilde{(x_0)^a} \,  \ast \, \widetilde{\kappa_q}\Big)(\widehat{n'''-n''}\cdot \widehat{l})
\Bigg\}
\]
can be weritten as
\[
\sum\limits_{b=0}^{\omega} \,\,
{\textstyle\frac{\big(\mathbb{A}_{{}_{0}}(\widehat{n} \cdot \widehat{l})-\mathbb{A}_{{}_{0}}(\widehat{n'''} \cdot \widehat{l})\big)^b}{b!}}
\Big(\widetilde{\theta} \, \ast \, \widetilde{(x_0)^b} \,  \ast \, \widetilde{\kappa_q}\Big)(\widehat{n'''}\cdot \widehat{l})
\]
Now we use the commutativity of the convolution $\ast$ used here, we commute $\widetilde{\theta}$ and $\widetilde{(x_0)^b}$, 
which correspods to the ``integration by parts'' operation in the $p_0$-component of the momentum in the corresponding derivation 
of Subsection \ref{WickForChronological} on the Minkowski space-time,  
in which the discrete quantum number $n$ ranges over the continuous zero component of momentum). This allows to write
the last term in the form
\begin{equation}\label{SubtractionTermFT(x0b)*FT(theta*FT(kappa))}
\sum\limits_{b=0}^{\omega} \,\,
{\textstyle\frac{\big(\mathbb{A}_{{}_{0}}(\widehat{n} \cdot \widehat{l})-\mathbb{A}_{{}_{0}}(\widehat{n'''} \cdot \widehat{l})\big)^b}{b!}}
\Big(\widetilde{(x_0)^b} \,  \ast \, \widetilde{\theta} \, \ast \, \widetilde{\kappa_q}\Big)(\widehat{n'''}\cdot \widehat{l}).
\end{equation}

Up to this point we have proceeded generally with any function $x_0$ dual to $X_0$ up to order $\omega$ at $(0,e)$.  
Let us, for definiteness, use the function $x_0$ equal (\ref{x0canonical}), which can be used for $\omega\leq 2$, 
and for which we have (\ref{FT((x^a)theta)}). 
Substituting (\ref{FT((x^a)theta)}) into (\ref{SubtractionTermFT(x0b)*FT(theta*FT(kappa))}) we write 
(\ref{SubtractionTermFT(x0b)*FT(theta*FT(kappa))}) as follows
\begin{multline*}
\sum\limits_{b=0}^{\omega} \,\,
{\textstyle\frac{\big(\mathbb{A}_{{}_{0}}(\widehat{n} \cdot \widehat{l})-\mathbb{A}_{{}_{0}}(\widehat{n'''} \cdot \widehat{l})\big)^b}{b!}}
\, \times
\\
\times \,
\sum\limits_{n''\in 2\mathbb{Z}+1}
    {\textstyle\frac{1}{\sqrt{\pi}}}\, 
{\textstyle\frac{i^{b+1}(-1)^b b!}{(n''+2b)(n''+2b-4)(n''+2b-2\cdot4)(n''+2b-3\cdot 4) \ldots (n''-2b)}}
\,\,
\widetilde{\kappa_q}(\widehat{n'''-n''}\cdot \widehat{l})
\\
+
\sum\limits_{b=0}^{\omega} \,\,
{\textstyle\frac{\big(\mathbb{A}_{{}_{0}}(\widehat{n} \cdot \widehat{l})-\mathbb{A}_{{}_{0}}(\widehat{n'''} \cdot \widehat{l})\big)^b}{b!}}
\sum\limits_{n''\in \mathbb{Z}_b}
\sqrt{\pi} (-1)^{b+(2b-n'')/4} {b \choose (2b-n'')/4} {\textstyle\frac{i^b}{4^b}}
\,\,
\widetilde{\kappa_q}(\widehat{n'''-n''} \cdot \widehat{l})  
\end{multline*}
where $\mathbb{Z}_b$ is the finite set $\{2b, 2b-4, 2b-2\cdot 4, \ldots, -2b\}$. Substituting
(\ref{A_0(n.l)2}) we can write this term as
\begin{multline*}
\sum\limits_{b=0}^{\omega} \,\,
\sum\limits_{n''\in 2\mathbb{Z}+1}
    {\textstyle\frac{i}{\sqrt{\pi}}}\, 
{\textstyle\frac{(-1)^b(n-n''')^b}{(n''+2b)(n''+2b-4)(n''+2b-2\cdot4)(n''+2b-3\cdot 4) \ldots (n''-2b)}}
\,\,
\widetilde{\kappa_q}(\widehat{n'''-n''}\cdot \widehat{l})
\\
+
\sum\limits_{b=0}^{\omega} \,\,
{\textstyle\frac{(\sqrt{\pi})^{b+1}(n-n''')^b}{2^b b!}}
\sum\limits_{n''\in \mathbb{Z}_b}
(-1)^{b+(2b-n'')/4} {b \choose (2b-n'')/4} 
\,\,
\widetilde{\kappa_q}(\widehat{n'''-n''} \cdot \widehat{l})  
\end{multline*}
Thus, (\ref{FT(retkappaq)3EU}) can be written as
\begin{multline}\label{FT(retkappaq)4EU}
\big[\widetilde{\textrm{ret} \, \kappa_q}\big]'''(\widehat{n}\cdot\widehat{l}) 
\\
=\sum\limits_{n''\in 2\mathbb{Z}+1} {\textstyle\frac{i}{2\pi}} {\textstyle\frac{1}{n''}}\widetilde{\kappa_q}(\widehat{n-n''} \cdot \widehat{l})
+ {\textstyle\frac{1}{2}} \, \widetilde{\kappa_q}(\widehat{n} \cdot \widehat{l})
\\
-
\sum\limits_{b=0}^{\omega} \,\,
\sum\limits_{n''\in 2\mathbb{Z}+1}
    {\textstyle\frac{i}{2\pi}}\, 
{\textstyle\frac{(-1)^b(n-n''')^b}{(n''+2b)(n''+2b-4)(n''+2b-2\cdot4)(n''+2b-3\cdot 4) \ldots (n''-2b)}}
\,\,
\widetilde{\kappa_q}(\widehat{n'''-n''}\cdot \widehat{l})
\\
-
\sum\limits_{b=0}^{\omega} \,\,
{\textstyle\frac{(\sqrt{\pi})^{b}(n-n''')^b}{2^{b+1} b!}}
\sum\limits_{n''\in \mathbb{Z}_b}
(-1)^{b+(2b-n'')/4} {b \choose (2b-n'')/4} 
\,\,
\widetilde{\kappa_q}(\widehat{n'''-n''} \cdot \widehat{l})  
\end{multline}
Now we arrive at the final formula, which depends on the parity of the number $n'''$.
Let us assume first that $n'''$ is even. Then we introduce the new summation variable
$n_1=n-n''$ in the first term in (\ref{FT(retkappaq)4EU}), and new summation variable 
$n_1= n'''-n''$ in the last two subtraction terms. 
Then for $n\in 2\mathbb{Z}$
\begin{multline}\label{DispersionFormulasn'''EvennEven}
\big[\widetilde{\textrm{ret} \, \kappa_q}\big]'''(\widehat{n}\cdot\widehat{l}) 
\\
=
\sum\limits_{n_1\in 2\mathbb{Z}+1}
    {\textstyle\frac{i}{2\pi}}\, 
\Bigg[
{\textstyle\frac{1}{n-n_1}}\,
- 
\sum\limits_{b=0}^{\omega} \,\,
{\textstyle\frac{(-1)^b(n-n''')^b}{(n'''-n_1+2b)(n'''-n_1+2b-4)(n'''-n_1+2b-2\cdot 4) \ldots (n'''-n_1-2b)}}
\\
\Bigg]
\,\,
\widetilde{\kappa_q}(\widehat{n_1} \cdot \widehat{l})  + {\textstyle\frac{1}{2}} \, \widetilde{\kappa_q}(\widehat{n} \cdot \widehat{l})
\end{multline}
For $n\in 2\mathbb{Z}+1$
\begin{multline}\label{DispersionFormulasn'''EvennOdd}
\big[\widetilde{\textrm{ret} \, \kappa_q}\big]'''(\widehat{n}\cdot\widehat{l}) 
\\
=
\sum\limits_{n_1\in 2\mathbb{Z}}
{\textstyle\frac{i}{2\pi}}\, 
{\textstyle\frac{1}{n-n_1}}\,
\,\,
\widetilde{\kappa_q}(\widehat{n_1} \cdot \widehat{l})  + {\textstyle\frac{1}{2}}\, \widetilde{\kappa_q}(\widehat{n} \cdot \widehat{l})
\\
-
\sum\limits_{b=0}^{\omega} \,\,
{\textstyle\frac{(\sqrt{\pi})^{b}(n-n''')^b}{2^{b+1} b!}}
\sum\limits_{n_1\in \mathbb{Z}_b+n'''}
(-1)^{b+(2b-n'''+n_1)/4} {b \choose (2b-n'''+n_1)/4} 
\,
\widetilde{\kappa_q}(\widehat{n_1} \cdot \widehat{l})  
\end{multline}
If $n'''$ is odd, then the parity of the summation ranges in the above formulas is exachanged, so that
for $n\in 2\mathbb{Z}+1$
\begin{multline}\label{DispersionFormulasn'''OddnOdd}
\big[\widetilde{\textrm{ret} \, \kappa_q}\big]'''(\widehat{n}\cdot\widehat{l}) 
\\
=
\sum\limits_{n_1\in 2\mathbb{Z}}
    {\textstyle\frac{i}{2\pi}}\, 
\Bigg[
{\textstyle\frac{1}{n-n_1}}\,
- 
\sum\limits_{b=0}^{\omega} \,\,
{\textstyle\frac{(-1)^b(n-n''')^b}{(n'''-n_1+2b)(n'''-n_1+2b-4)(n'''-n_1+2b-2\cdot 4) \ldots (n'''-n_1-2b)}}
\\
\Bigg]
\,\,
\widetilde{\kappa_q}(\widehat{n_1} \cdot \widehat{l}) + {\textstyle\frac{1}{2}} \, \widetilde{\kappa_q}(\widehat{n} \cdot \widehat{l}),  
\end{multline}
and for $n\in 2\mathbb{Z}$
\begin{multline}\label{DispersionFormulasn'''OddnEven}
\big[\widetilde{\textrm{ret} \, \kappa_q}\big]'''(\widehat{n}\cdot\widehat{l}) 
\\
=
\sum\limits_{n_1\in 2\mathbb{Z}+1}
{\textstyle\frac{i}{2\pi}}\, 
{\textstyle\frac{1}{n-n_1}}\,
\,\,
\widetilde{\kappa_q}(\widehat{n_1} \cdot \widehat{l})  + {\textstyle\frac{1}{2}} \, \widetilde{\kappa_q}(\widehat{n} \cdot \widehat{l})
\\
-
\sum\limits_{b=0}^{\omega} \,\,
{\textstyle\frac{(\sqrt{\pi})^{b}(n-n''')^b}{2^{b+1} b!}}
\sum\limits_{n_1\in \mathbb{Z}_b+n'''}
(-1)^{b+(2b-n'''+n_1)/4} {b \choose (2b-n'''+n_1)/4} 
\,
\widetilde{\kappa_q}(\widehat{n_1} \cdot \widehat{l})  
\end{multline}

Now we observe that the same method 
applied to the ``subtraction point'' $(2\pi,e)$
and the corresponding  operator $\Omega'$ equal (\ref{Omega'phiEUseveral}), 
gives exactly the same formulas (\ref{DispersionFormulasn'''EvennEven})
and (\ref{DispersionFormulasn'''EvennOdd}) for even $n''',n$. The formulas 
(\ref{DispersionFormulasn'''OddnOdd}) and (\ref{DispersionFormulasn'''OddnEven}) for odd $n''',n$ will also remain
the same in this case, except that $n$ and $n'''$ and, thus, the difference $n-n'''$ will get the minus sign
in the numerators of the subtracted terms, so that the numerators of the subtracted terms
will get the additional factor $(-1)^b$, \emph{i.e.} become equal $(-1)^b(-1)^b(n-n''')^b=(n-n''')^b$, in 
(\ref{DispersionFormulasn'''OddnOdd}) and (\ref{DispersionFormulasn'''OddnEven}), due to
the formula (\ref{FT(Xalphadelta2pie)}), valid for $\delta_{{}_{2\pi,e}}$ concentrated at $(2\pi,e)$, 
as both $n$ and $n'''$ are odd in (\ref{DispersionFormulasn'''OddnOdd}) and (\ref{DispersionFormulasn'''OddnEven}). 
This is the case because the function $x_0$ (in fact all functions $x_\mu$, $0,1,2,3$) is identical
for both points $(0,e)$ and $(2\pi,e)$.

For the purpose, which will become clear later, we note here that we can remove the additional $(-1)^b$ 
from the numerators of and get identical formulas
(\ref{DispersionFormulasn'''OddnOdd}) and (\ref{DispersionFormulasn'''OddnEven}) also for odd $n,n'''$ and for the 
subtraction point $(2\pi,e)$ if we replace the operator $\Omega'$ equal (\ref{Omega'phiEUseveral}) with a slightly different:
\begin{equation}\label{CheckOmega'phiEUseveral}
\check{\Omega'}\phi(x,\boldsymbol{w}) = \phi(t,\boldsymbol{w}) 
- \sum\limits_{a=0}^{\omega} \big(X_0\big)^{a}\phi(2\pi,e) \,\, (-1)^a{\textstyle\frac{(x_0)^a}{a!}} \, w(t),
\end{equation}
which defines $\check{\textrm{ret}} \, \kappa_q = \theta \kappa_q \circ \check{\Omega'}$
coinciding with $\textrm{ret} \, \kappa_q = \theta \kappa_q \circ \Omega'$ on thest functions
$\phi$, whose time derivatives vanish up to order $\omega$ at $(2\pi,e)$, because for such 
functions $\Omega'\phi = \check{\Omega'}\phi=\phi$. In this case we apply in the subtraction terms of the first ``Taylor 
expansion terms'' at the ``normalization point $\widehat{n'''}\cdot\widehat{l}$, the corresponding additional facor $(-1)^b$.

Before we pass to the investigation of convergence of the final formulas  (\ref{DispersionFormulasn'''EvennEven})-(\ref{DispersionFormulasn'''OddnEven}), 
let us note that introduction of the spatial derivations $X_1,X_2,X_3$, in addition to $X_0$, into the operator $\Omega'$ or $\check{\Omega}$,
brings no addintional effect. Still exactly the same method of subtraction 
\begin{multline*}
\big[\widetilde{\textrm{ret} \, \kappa_q}\big]'''(\widehat{n}\cdot\widehat{l})
\overset{\textrm{df}}{=}
\widetilde{\textrm{ret} \, \kappa_q}(\widehat{n}\cdot\widehat{l}) 
\\
- \sum\limits_{|\beta|=0}^{\omega} \,\,
{\textstyle\frac{\big(\mathbb{A}_{{}_{0}}(\widehat{n} \cdot \widehat{l})-\mathbb{A}_{{}_{0}}(\widehat{n'''} \cdot \widehat{l})\big)^{\beta_0}
\big(\mathbb{A}_{{}_{1}}(\widehat{n} \cdot \widehat{l})-\mathbb{A}_{{}_{1}}(\widehat{n'''} \cdot \widehat{l})\big)^{\beta_1}
\ldots \big(\mathbb{A}_{{}_{3}}(\widehat{n} \cdot \widehat{l})-\mathbb{A}_{{}_{3}}(\widehat{n'''} \cdot \widehat{l})\big)^{\beta_3}}{\beta!}}
\, \Big(\widetilde{x^\beta} \,\, \ast  \,\, \widetilde{\textrm{ret} \, \kappa_q}\Big)(\widehat{n'''}\cdot\widehat{l})
\end{multline*}
\begin{equation}\label{TaylorSubtracionEUX0X1X2X3}
=
\widetilde{\textrm{ret} \, \kappa_q}(\widehat{n}\cdot\widehat{l}) 
- \sum\limits_{|\beta|=0}^{\omega} \,\,
{\textstyle\frac{\big(\mathbb{A}(\widehat{n} \cdot \widehat{l})-\mathbb{A}(\widehat{n'''} \cdot \widehat{l})\big)^\beta}{\beta!}}
\, \Big(\widetilde{x^\beta} \,\, \ast  \,\, \widetilde{\textrm{ret} \, \kappa_q}\Big)(\widehat{n'''}\cdot\widehat{l}).
\end{equation}
of the first terms up to order $\omega$ of the ``Taylor expansion''
at the normalization point $\widehat{n'''}\cdot \widehat{l}$ works  
when there are also spatial derivations $X_k$, $k=1,2,3$ in the subtraction terms.
This method of complete elimination of the terms containing the auxiliary function $w$,
or functions $w$, remains the the same, for three reasons 1)-3). 

First, 1) because
the functions $x_\mu$
are common for the two points $(0,e)$ and $(2\pi,e)$, and depend solely on the order $\omega$. 

Next, 2) since (\ref{dualityXmu-xmu}) holds, and $x_\mu$ and $X_\mu$ are mutually dual up to order $\omega$ at $(0,e)$
and $(2\pi,e)$,
then 
\[
  \Big(\widetilde{x^\beta} \, \ast \, \mathbb{A}^\alpha \Big)(\widehat{n}\cdot\widehat{l}) 
=\begin{cases}
    {\textstyle\frac{\alpha!}{(\alpha-\beta)!}} \mathbb{A}^{\alpha-\beta}(\widehat{n}\cdot\widehat{l}) 
& \text{if $\beta \leq \alpha$, $|\alpha|\leq \omega$},\\
    0, & \text{if $\alpha<\beta \leq \omega$},
  \end{cases}
\]
and identically for the Fourier transform 
\[
e^{in\pi} \mathbb{A}^\alpha(\widehat{n}\cdot \widehat{l})=
\widetilde{X^\alpha \delta_{{}_{2\pi,e}}}(\widehat{n}\cdot \widehat{l})
\overset{\textrm{df}}{=} 
\hat{\hat{2\pi}}(\widehat{n}) \,\, \mathbb{A}^\alpha(\widehat{n}\cdot \widehat{l}), \,\,\, 
\hat{\hat{2\pi}}(\widehat{n}) \overset{\textrm{df}}{=} \widehat{n}(2\pi) = e^{in\pi}, 
\]
of $X^\alpha\delta_{{}_{2\pi,e}}$ (compare (\ref{FT(Xalphadelta2pie)})), because 
\[
\widetilde{x^\beta} \ast \mathbb{A}^\alpha =  \widetilde{x^\beta} \ast \widetilde{X^\alpha\delta} 
= \Big(x^\beta X^\alpha\delta\Big)^{\widetilde{\,\,\,\,\,\,\,}} \,\,\, 
\textrm{and} \,\,\,
\widetilde{x^\beta} \ast \,\, \big(\hat{\hat{2\pi}} \, \mathbb{A}^\alpha\big) =  \widetilde{x^\beta} \ast \widetilde{X^\alpha\delta_{{}_{2\pi,e}}} 
= \Big(x^\beta X^\alpha\delta_{{}_{2\pi,e}}\Big)^{\widetilde{\,\,\,\,\,\,\,}}.
\]
Finally, 3) we have the binomial formula    
\begin{multline}\label{BinomaialFormulaAmu}
\sum\limits_{\beta\leq \alpha} \,\,
{\textstyle\frac{\big(\mathbb{A}_{{}_{0}}(\widehat{n} \cdot \widehat{l})-\mathbb{A}_{{}_{0}}(\widehat{n'''} \cdot \widehat{l})\big)^{\beta_0}
\big(\mathbb{A}_{{}_{1}}(\widehat{n} \cdot \widehat{l})-\mathbb{A}_{{}_{1}}(\widehat{n'''} \cdot \widehat{l})\big)^{\beta_1}
\ldots \big(\mathbb{A}_{{}_{3}}(\widehat{n} \cdot \widehat{l})-\mathbb{A}_{{}_{3}}(\widehat{n'''} \cdot \widehat{l})\big)^{\beta_3}}{\beta!}}
{\textstyle\frac{\big(\mathbb{A}(\widehat{n'''} \cdot \widehat{l})\big)^{\alpha-\beta}}{(\alpha-\beta)!}}
\\
=
{\textstyle\frac{1}{\alpha!}}
\sum\limits_{\beta\leq \alpha}
{\alpha \choose \beta}
\big(\mathbb{A}_{{}_{0}}(\widehat{n} \cdot \widehat{l})-\mathbb{A}_{{}_{0}}(\widehat{n'''} \cdot \widehat{l})\big)^{\beta_0}
\ldots 
\big(\mathbb{A}_{{}_{3}}(\widehat{n} \cdot \widehat{l})-\mathbb{A}_{{}_{3}}(\widehat{n'''} \cdot \widehat{l})\big)^{\beta_3}
\big(\mathbb{A}(\widehat{n'''} \cdot \widehat{l})\big)^{\alpha-\beta}
\\
=
{\textstyle\frac{\big(\mathbb{A}(\widehat{n} \cdot \widehat{l})\big)^\alpha}{\alpha!}},
\end{multline}
which still holds for 
\begin{gather*}
\mathbb{A}_{{}_{0}}(\widehat{n}\cdot\widehat{l}) 
=  \big(- i \, n \big) \,\, \boldsymbol{1},
\\
\mathbb{A}_{{}_{i}}(\widehat{n}\cdot \widehat{l}) =  {\overline{\mathbb{A}_{{}_{i \, \widehat{l}}}}}^T, 
\,\,\,\,\, i=1,2,3.
\end{gather*}
It follows immediately from the fact that $\mathbb{A}_{{}_{0}}$ commutes with all other
$\mathbb{A}_{{}_{\mu}}$, and from the fact that $\mathbb{A}_{{}_{k}}$, with $k=1,2,3$, is independent of $\widehat{n}$, 
so that
\begin{equation}\label{partial_n(Ai)=0}
\mathbb{A}_{{}_{k}}(\widehat{n} \cdot \widehat{l}) - \mathbb{A}_{{}_{k}}(\widehat{n'''} \cdot \widehat{l}) = 0,
\,\,\, k=1,2,3,
\end{equation}
\emph{i.e.} the ``differential''  of the ``spatial components'' vanishes. This means that   
\[
\big(\mathbb{A}_{{}_{k}}(\widehat{n} \cdot \widehat{l}) - \mathbb{A}_{{}_{k}}(\widehat{n'''} \cdot \widehat{l})\big)^{\beta_k} = 
\begin{cases}
\boldsymbol{1}, &, \beta_k = 0,\\
0, & \beta_k >0
\end{cases}
\]
in (\ref{BinomaialFormulaAmu}), which immediately gives (\ref{BinomaialFormulaAmu}).

Using 1)-3) we see that also in the general case with two``subtraction points'' (\ref{x+-})
and for the corresponding operator $\Omega'$, equal (\ref{Omega'phiEUseveralMultiindex}), and containing 
all space-time derivatives, all terms containng the auxiliary function
$w$ can be eliminated by the same method, and we arrive at the formula
\begin{multline}\label{FT(retkappaq)3EUX0X1X2X3}
\big[\widetilde{\textrm{ret} \, \kappa_q}\big]'''(\widehat{n}\cdot\widehat{l}) 
=
{\textstyle\frac{1}{\sqrt{4\pi}}}\sum\limits_{n''} \widehat{\theta}_{{}_{n''}} \Bigg\{ \widetilde{\kappa_q}(\widehat{n-n''} \cdot \widehat{l})
\\
-
\sum\limits_{|\beta|=0}^{\omega} \,\,
{\textstyle\frac{\big(\mathbb{A}(\widehat{n} \cdot \widehat{l})-\mathbb{A}(\widehat{n'''} \cdot \widehat{l})\big)^\beta}{\beta!}}
\Big(\widetilde{x^\beta} \,  \ast \, \widetilde{\kappa_q}\Big)(\widehat{n'''-n''}\cdot \widehat{l})
\Bigg\}
\end{multline}
analogue to (\ref{FT(retkappaq)3EU}), with the last subtracted term, which can be written as
\begin{equation}\label{SubtractionTermFT(x0b)*FT(theta*FT(kappa))X0X1X2X3}
\sum\limits_{|\beta|=0}^{\omega} \,\,
{\textstyle\frac{\big(\mathbb{A}(\widehat{n} \cdot \widehat{l})-\mathbb{A}(\widehat{n'''} \cdot \widehat{l})\big)^\beta}{\beta!}}
\Big(\widetilde{x^\beta} \,  \ast \, \widetilde{\theta} \, \ast \, \widetilde{\kappa_q}\Big)(\widehat{n'''}\cdot \widehat{l}),
\end{equation}
and which is the analogue of (\ref{SubtractionTermFT(x0b)*FT(theta*FT(kappa))}).
Because of the property (\ref{partial_n(Ai)=0}), the formulas (\ref{FT(retkappaq)3EUX0X1X2X3})
(\ref{SubtractionTermFT(x0b)*FT(theta*FT(kappa))X0X1X2X3}) reduce, respectively,
to the formulas  (\ref{FT(retkappaq)3EU}) and (\ref{SubtractionTermFT(x0b)*FT(theta*FT(kappa))}) with the summation reduced
to the summation over $\beta_0=b$.
Thus we arrive at the final formulas identical to
(\ref{DispersionFormulasn'''EvennEven})-(\ref{DispersionFormulasn'''OddnEven}) also in this case.

Let us investigate now convergence of the final formulas.
We can restrict our consideration to the case in which we have only time derivatives of $\phi$ in $\Omega'$.
Let the subtraction point be equal $(0,e)$. Let us start with the analysis for all cases with  $\omega\leq 2$.
In this case we may use the final formulas 
(\ref{DispersionFormulasn'''EvennEven})-(\ref{DispersionFormulasn'''OddnEven}).

Let, for definiteness, $n'''$ be even.
We see from the  formulas (\ref{DispersionFormulasn'''EvennEven}) and (\ref{DispersionFormulasn'''EvennOdd}), that 
whatever the number $\omega\leq 2$ we choose, we cannot improve the convergence of 
$\widetilde{\textrm{ret} \, \kappa_q}(\widetilde{n}\cdot\widetilde{l})$ for odd $n$, as from the formula 
(\ref{DispersionFormulasn'''EvennOdd}) it follows that the soubtraction has always the effect of adding
a finite number of terms there, \emph{i.e.} for $\widetilde{\textrm{ret} \, \kappa_q}(\widetilde{n}\cdot\widetilde{l})$ 
with odd $n$. But situation is the same for any natural $\omega$. Indeed, this is evident for $\omega\leq 2$, as in this case we can use
(\ref{DispersionFormulasn'''EvennEven})-(\ref{DispersionFormulasn'''EvennOdd}). 
In order to enlarge $\omega$ by $1$, $2, \ldots$, we need to modify $x_0$, used in 
(\ref{DispersionFormulasn'''EvennEven})-(\ref{DispersionFormulasn'''OddnEven}) by addition to it the powers of it,
staring with the third power.
 From the formula (\ref{FT((x^a)theta)}) it follows that we obtain
the same formulas (\ref{DispersionFormulasn'''EvennEven}) and (\ref{DispersionFormulasn'''OddnOdd})
with additional subtraction terms of order $\sim a_4/(n_1)^4 + a_5/(n_1)^5 + \ldots$ at infinity,
and the same formulas (\ref{DispersionFormulasn'''EvennOdd}) and (\ref{DispersionFormulasn'''OddnEven})
with additional subtraction terms which are likewise equal to finite sums in 
(\ref{DispersionFormulasn'''EvennOdd}) and (\ref{DispersionFormulasn'''OddnEven}). 
This cannot repair divergence with any assumed value of $\omega$.  
The situation with odd $n'''$ is the same, as follows from   (\ref{DispersionFormulasn'''OddnOdd}) 
and (\ref{DispersionFormulasn'''OddnEven}), with the role of the parities of $n$ reversed.

Subtraction can have any influence on the convergence only under the assumption that $\widetilde{\kappa_q}$
is non zero only for $\widehat{n}\cdot \widehat{l}$ with $n$ of fixed parity. Let, for definitness,
$\widetilde{\kappa_q}(\widehat{n}\cdot \widehat{l})$ be nonzero only for odd $n$. In this case
we choose even $n'''$. In this case contribution from (\ref{DispersionFormulasn'''EvennOdd}) vanishes identically,
\emph{i.e.} $[\widetilde{\textrm{ret} \, \kappa_q}]'''(\widehat{n}\cdot\widehat{l})=0$ for odd $n$, and we are left
with the only non trivial contribution for $\widetilde{\textrm{ret} \, \kappa_q}(\widehat{n}\cdot\widehat{l})$
with even $n$, given by the formula (\ref{DispersionFormulasn'''EvennEven}). 
If $\widetilde{\kappa_q}(\widehat{n}\cdot \widehat{l})$ is nonzero only for even $n$,
then we choose odd $n'''$. In this case contribution from (\ref{DispersionFormulasn'''OddnEven}) vanishes identically, 
\emph{i.e.} $[\widetilde{\textrm{ret} \, \kappa_q}]'''(\widehat{n}\cdot\widehat{l})=0$ for even $n$,
with the only non trivial contribution for $\widetilde{\textrm{ret} \, \kappa_q}(\widehat{n}\cdot\widehat{l})$
with odd $n$, given by the formula (\ref{DispersionFormulasn'''OddnOdd}).

We therefore proceed in the following way. Having given $\kappa_q$, we see that it can always be uniquely
written as the sum
\[
\kappa_q = \kappa_{q \,\textrm{odd}}  + \kappa_{q \,\textrm{even}},
\]
of the odd $\kappa_{q \,\textrm{odd}}$ and the even part $\kappa_{q \,\textrm{even}}$. The Fourier transform of the odd part is equal to  
$\widetilde{\kappa_q}(\widehat{n}\cdot\widehat{l})$ if $n$ is odd and zero if $n$ is even.
And \emph{vice versa} fot the even part, whose Fourier transform is equal to  
$\widetilde{\kappa_q}(\widehat{n}\cdot\widehat{l})$ if $n$ is even and zero if $n$ is odd.
 
Simply speaking we divide the Fourier transform $\widetilde{\kappa_q}$ into the part $\widetilde{\kappa_{q \,\textrm{odd}}}$ 
supported on $\widehat{n}\cdot\widehat{l}$
with odd $n$ and the part $\widetilde{\kappa_{q \,\textrm{even}}}$ supported on $\widehat{n}\cdot\widehat{l}$ with even $n$, 
and compute $\widetilde{\textrm{ret} \,\kappa_{q \,\textrm{odd}}}$ and
$\widetilde{\textrm{ret} \,\kappa_{q \,\textrm{even}}}$ and the corresponding
$\big[\widetilde{\textrm{ret} \,\kappa_{q \,\textrm{odd}}}\big]'''$ 
and $\big[\widetilde{\textrm{ret} \,\kappa_{q \,\textrm{even}}}\big]'''$, separately for the even and odd supported parts.
In fact we define the Fourer transform of the retarded part separately for the odd characters $\widehat{n}\cdot \widehat{l}$
and separately for the even characters $\widehat{n}\cdot \widehat{l}$.

It is sufficient to analyze only the Fourier transform of the retarded part of the odd part
of $\kappa_q$, \emph{i.e.} for the part of $\widetilde{\kappa_q}$ concentrated on the
characters with odd energy number, \emph{i.e.} restricting consideration to the formula 
(\ref{DispersionFormulasn'''EvennEven}). This is obvious, because the expressions in the square bracket
in (\ref{DispersionFormulasn'''EvennEven}) and (\ref{DispersionFormulasn'''OddnOdd}) are identical. 
Indeed, we just only reverse the parity of $n,n'''$ in the final formulas for the retarded part of the odd part
in order to obtain the result valid for the retarded part of the even part.

So, let us analyze the Fourier transform of the retarded part of the odd part
of $\kappa_q$ only, \emph{i.e.} of the part of $\widetilde{\kappa_q}$ concetrated on $\widehat{n}\cdot \widehat{l}$ with odd $n$.
In this situation the subtraction terms can repair the convergence, in principle at least, just
by improving the summability with respect to $n_1$ only in the formula (\ref{DispersionFormulasn'''EvennEven})
for even $n$. Indeed, let us consider the expression
\begin{equation}\label{[omega]}
{\textstyle\frac{1}{n-n_1}}\,
- 
\sum\limits_{b=0}^{\omega} \,\,
{\textstyle\frac{(-1)^b(n-n''')^b}{(n'''-n_1+2b)(n'''-n_1+2b-4)(n'''-n_1+2b-2\cdot 4) \ldots (n'''-n_1-2b)}}
\end{equation}
in the square bracket in (\ref{DispersionFormulasn'''EvennEven}) for various $\omega$.   
For $\omega =0$ (\ref{[omega]}) is equal    
\[
{\textstyle\frac{1}{n-n_1}}-{\textstyle\frac{1}{n'''-n_1}}={\textstyle\frac{n'''-n}{(n-n_1)(n'''-n_1)}}.
\]
Therefore if in addition for each $l$, $\widetilde{\kappa_q}(\widehat{n_1}\cdot\widehat{l})$ is a bounded function 
of $n_1$ , \emph{i.e.} if $\kappa_q$ is of zero order, then we see that (\ref{DispersionFormulasn'''EvennEven}):
\[
\big[\widetilde{\textrm{ret} \, \kappa_{q \,\textrm{odd}}}\big]'''(\widehat{n}\cdot\widehat{l}) 
=
\sum\limits_{n_1\in 2\mathbb{Z}+1}
    {\textstyle\frac{i}{2\pi}}\, 
\Big[
{\textstyle\frac{n'''-n}{(n-n_1)(n'''-n_1)}}
\Big]
\,\,
\widetilde{\kappa_{q \,\textrm{odd}}}(\widehat{n_1} \cdot \widehat{l})  
+ {\textstyle\frac{1}{2}} \, \widetilde{\kappa_{q \,\textrm{odd}}}(\widehat{n} \cdot \widehat{l}),
\]
\[
n,n''' \in 2 \mathbb{Z}, \,\, \omega=\omega'=0,
\]
indeed becomes convergent
if we put $\omega=0$. We have, of course, analoge result for the even part, so that analogously
\[
\big[\widetilde{\textrm{ret} \, \kappa_{q \,\textrm{even}}}\big]'''(\widehat{n}\cdot\widehat{l}) 
=
\sum\limits_{n_1\in 2\mathbb{Z}}
    {\textstyle\frac{i}{2\pi}}\, 
\Big[
{\textstyle\frac{n'''-n}{(n-n_1)(n'''-n_1)}}
\Big]
\,\,
\widetilde{\kappa_{q \,\textrm{even}}}(\widehat{n_1} \cdot \widehat{l})  
+ {\textstyle\frac{1}{2}} \, \widetilde{\kappa_{q \,\textrm{even}}}(\widehat{n} \cdot \widehat{l}),
\]
\[
n,n''' \in 2 \mathbb{Z}+1, \,\, \omega=\omega'=0,
\]
becomes convergent if we put $\omega=0$, and we obtan now the full formula for all, even and odd $n$. Nonetheless we have to remember
that $n'''$ is odd for $n$ odd and $n'''$ is even for $n$ even and we cannot use the same $n'''$ for all $n$, requiring
the normalization points with at least two different energy component $n'''$ in order to reconstruct 
$\big[\widetilde{\textrm{ret} \, \kappa_q}\big]'''$ for all $\widehat{n}\cdot\widehat{l}$.

Let us look now at (\ref{[omega]}) for $\omega=1$:
\begin{multline*}
{\textstyle\frac{1}{n-n_1}} -{\textstyle\frac{1}{n'''-n_1}}
+{\textstyle\frac{n-n'''}{(n'''-n_1+2)(n'''-n_1-2)}}
=
{\textstyle\frac{n-n'''}{n-n_1}} \,\,
{\textstyle\frac{(nn'''-{n'''}^2+4) +(n'''-n)n_1}{(n'''-n_1+2)(n'''-n_1)(n'''-n_1-2)}}
\end{multline*}
behaving like $\sim 1/(n_1)^3$ for $n_1$ tending to infinity. Therefore,  if
$\widetilde{\kappa_q}$ is nonzero only for $\widehat{n}\cdot\widehat{l}$ with odd $n$
(which, by definition, is the case for the odd part), the only nonzero
values of $\widetilde{\textrm{ret} \, \kappa_q}(\widehat{n}\cdot\widehat{l})$ are that with even $n$, and
are given by the formula (\ref{DispersionFormulasn'''EvennEven}) which is convergent if the order $\omega'$ of $\kappa_q$
is less than or equal one, \emph{i.e.}
\[
{\textstyle\frac{|\widetilde{\kappa_q}(\widetilde{n}\cdot\widetilde{l})|}{|n|^\nu}}  
\]
is bounded for $\nu\geq 1$. Speaking more suggestively, if for each $l$,
$\widetilde{\kappa_q}(\widehat{n}\cdot\widehat{l})$, regarded as a function of $n$, grows at infinity not faster than 
$\sim (\textrm{const}) \cdot n$ (in fact not faster than $\sim (\textrm{const}) \cdot n^\nu$, $\nu<2$, but in QFT we encounter only integer $\omega'$), 
then the sum (\ref{DispersionFormulasn'''EvennEven}):
\begin{multline*}
\big[\widetilde{\textrm{ret} \, \kappa_{q \,\textrm{odd}}}\big]'''(\widehat{n}\cdot\widehat{l}) 
\\
=
\sum\limits_{n_1\in 2\mathbb{Z}+1}
    {\textstyle\frac{i}{2\pi}}\, 
\Big[
{\textstyle\frac{n-n'''}{n-n_1}} \,\,
{\textstyle\frac{(nn'''-{n'''}^2+4) +(n'''-n)n_1}{(n'''-n_1+2)(n'''-n_1)(n'''-n_1-2)}}
\Big]
\,\,
\widetilde{\kappa_{q \,\textrm{odd}}}(\widehat{n_1} \cdot \widehat{l})  
+ {\textstyle\frac{1}{2}} \, \widetilde{\kappa_{q \,\textrm{odd}}}(\widehat{n} \cdot \widehat{l}),
\end{multline*}
\[
n,n''' \in 2 \mathbb{Z}, \,\, \omega=\omega'=1,
\]
is convergent. We have, of course, analoge result for the even part, so that analogously
\begin{multline*}
\big[\widetilde{\textrm{ret} \, \kappa_{q \,\textrm{even}}}\big]'''(\widehat{n}\cdot\widehat{l}) 
\\
=
\sum\limits_{n_1\in 2\mathbb{Z}}
    {\textstyle\frac{i}{2\pi}}\, 
\Big[
{\textstyle\frac{n-n'''}{n-n_1}} \,\,
{\textstyle\frac{(nn'''-{n'''}^2+4) +(n'''-n)n_1}{(n'''-n_1+2)(n'''-n_1)(n'''-n_1-2)}}
\Big]
\,\,
\widetilde{\kappa_{q \,\textrm{even}}}(\widehat{n_1} \cdot \widehat{l})  
+ {\textstyle\frac{1}{2}} \, \widetilde{\kappa_{q \,\textrm{even}}}(\widehat{n} \cdot \widehat{l}),
\end{multline*}
\[
n,n''' \in 2 \mathbb{Z}+1, \,\, \omega=\omega'=1,
\]
is convergent if we put $\omega=1$, obtaning the full formula for all, even and odd $n$. However
$n'''$ is odd for $n$ odd and $n'''$ is even for $n$ even, so we need at least two differnt $n'''$ in order
to reconstruct $\big[\widetilde{\textrm{ret} \, \kappa_q}\big]'''$ for all $\widehat{n}\cdot\widehat{l}$.

Consider now (\ref{[omega]}) for $\omega=2$:
\begin{multline*}
{\textstyle\frac{1}{n-n_1}} -{\textstyle\frac{1}{n'''-n_1}}
+{\textstyle\frac{n-n'''}{(n'''-n_1+2)(n'''-n_1-2)}}
- {\textstyle\frac{(n-n''')^2}{(n'''-n_1+4)(n'''-n_1)(n'''-n_1-4)}}
\\
=
{\textstyle\frac{n'''-n}{n-n_1}}
{\textstyle\frac{C + B n_1
+A n_{1}^{2}}{(n'''-n_1+4)(n'''-n_1+2)(n'''-n_1)(n'''-n_1-2)(n'''-n_1-4)}},
\end{multline*}
where
\begin{gather*}
A = (n-n''')^2-4,
\\
 B= -2n^2n'''+4n(n''')^2-12n-2(n''')^3+20n''',
\\
C = n^2(n''')^2-4n^2-2n(n''')^3+20nn''' +(n''')^4 - 20(n''')^2+64.
\end{gather*}
It behaves like $\sim 1/(n_1)^4$ for $n_1$ tending to infinity. Therefore,  if
$\widetilde{\kappa_q}$ is nonzero only for $\widehat{n}\cdot\widehat{l}$ with odd $n$ (which, by definition, is the case for the odd part), the only nonzero
values of $\widetilde{\textrm{ret} \, \kappa_q}(\widehat{n}\cdot\widehat{l})$ are that with even $n$, and
are given by the formula (\ref{DispersionFormulasn'''EvennEven}):
\begin{multline*}
\big[\widetilde{\textrm{ret} \, \kappa_{q \,\textrm{odd}}}\big]'''(\widehat{n}\cdot\widehat{l}) 
\\
=
\sum\limits_{n_1\in 2\mathbb{Z}+1}
    {\textstyle\frac{i}{2\pi}}\, 
\Big[
{\textstyle\frac{n'''-n}{n-n_1}}
{\textstyle\frac{C + B n_1
+A n_{1}^{2}}{(n'''-n_1+4)(n'''-n_1+2)(n'''-n_1)(n'''-n_1-2)(n'''-n_1-4)}}
\\
\Big]
\,\,
\widetilde{\kappa_{q \,\textrm{odd}}}(\widehat{n_1} \cdot \widehat{l})  
+ {\textstyle\frac{1}{2}} \, \widetilde{\kappa_{q \,\textrm{odd}}}(\widehat{n} \cdot \widehat{l}),
\end{multline*}
\[
n,n''' \in 2 \mathbb{Z}, \,\, \omega=\omega'=2,
\]
which is convergent if the order $\omega'$ of $\kappa_q$
is less than or equal two. Analogously we have for the even part with
\begin{multline*}
\big[\widetilde{\textrm{ret} \, \kappa_{q \,\textrm{even}}}\big]'''(\widehat{n}\cdot\widehat{l}) 
\\
=
\sum\limits_{n_1\in 2\mathbb{Z}}
    {\textstyle\frac{i}{2\pi}}\, 
\Big[
{\textstyle\frac{n'''-n}{n-n_1}}
{\textstyle\frac{C + B n_1
+A n_{1}^{2}}{(n'''-n_1+4)(n'''-n_1+2)(n'''-n_1)(n'''-n_1-2)(n'''-n_1-4)}}
\\
\Big]
\,\,
\widetilde{\kappa_{q \,\textrm{even}}}(\widehat{n_1} \cdot \widehat{l})  
+ {\textstyle\frac{1}{2}} \, \widetilde{\kappa_{q \,\textrm{even}}}(\widehat{n} \cdot \widehat{l}),
\end{multline*}
\[
n,n''' \in 2 \mathbb{Z}+1, \,\, \omega=\omega'=2,
\]
which is convergent if the order $\omega'$ of $\kappa_q$
is less than or equal two, but here with odd $n'''$ for odd $n$. Thus we have recontructed 
$\big[\widetilde{\textrm{ret} \, \kappa_q}\big]'''$ fully.

So far we can go with $x_0(t) =(1/2)\sin t$, used in 
(\ref{DispersionFormulasn'''EvennEven})- (\ref{DispersionFormulasn'''OddnEven}), and cannot further enlarge
$\omega$ in these formulas, because  $x_0(t) ={\tfrac{1}{2}}\sin t$ is dual to $X_0=2\, \partial_t$ at $t=0$ up to order $\omega=2$
in the sense explainded above, or because the third order term in the Taylor expansion 
\[
{\textstyle\frac{\sin t}{2}} \,\,\,\,\,\,\, = \,\,\,\,\,\,\, {\textstyle\frac{1}{2}} t \,\,\,\,\,\,  -{\textstyle\frac{1}{2}} {\textstyle\frac{1}{3!}} t^3
\,\,\,\,\, + \,\,\,\, \ldots
\]
is nonzero, so that this $x_0$ cannot be used in our operator $\Omega'$ with $\omega \geq 3$. 
In order to get a function $x_0$ dual to $X_0=2\, \partial_t$ up to order $\omega=4$, which can be used
in $\Omega'$ with $\omega \leq 4$, we need to add to the previous $x_0={\tfrac{1}{2}}\sin $ its power $\big({\tfrac{1}{2}}\sin \big)^3$
multiplied by $2/3$:
\[
x_0(t) = {\textstyle\frac{\sin t }{2}} + {\textstyle\frac{2}{3}} \left({\textstyle\frac{\sin t}{2}}\right)^3
\] 
in order to obtain new function $x_0$ with Taylor expansion
\[
x_0(t) \,\,\,\,\,\,\, = \,\,\,\,\,\,\,  {\textstyle\frac{1}{2}} t \,\,\,\,\,\,  -{\textstyle\frac{9}{2}}{\textstyle\frac{1}{5!}} t^5 
\,\,\,\,\, + \,\,\,\, \ldots.
\]
Because we add to $x_0(t) ={\tfrac{1}{2}}\sin t$ only its power, we need not make any new computations,
and still use the formula (\ref{FT((x^a)theta)}) valid for any natural power of $x_0(t) ={\tfrac{1}{2}}\sin t$, 
and proceed with the new $x_0$ along the lines presented above and get the final formulas valid with $\omega \leq 4$.
The analoge of the expression (\ref{[omega]}) for $\omega=3$ computed with the new $x_0$ (which is dual to $X_0$
up to order $\omega= 4$) 
has the following form 
\begin{multline*}
{\textstyle\frac{1}{n-n_1}} -{\textstyle\frac{1}{n'''-n_1}} +{\textstyle\frac{n-n'''}{(n'''-n_1+2)(n'''-n_1-2)}}
-{\textstyle\frac{(n-n''')^2}{(n'''-n_1+4)(n'''-n_1)(n'''-n_1-4)}}
\\
+{\textstyle\frac{(n-n''')^3-2^2 \cdot (n-n''')}{(n'''-n_1+6)(n'''-n_1+2)(n'''-n_1-2)(n'''-n_1-6)}}
\\
+ {\textstyle\frac{2^4 \cdot (n-n''')^2}{(n'''-n_1+8)(n'''-n_1+4)(n'''-n_1)(n'''-n_1-4)(n'''-n_1-8)}}
+ \left(\sim{\textstyle\frac{a_6}{n_{1}^{6}}}\right) 
\end{multline*}
\begin{multline*}
=
{\textstyle\frac{n'''-n}{n-n_1}}
{\textstyle\frac{D \cdot n_{1}^{3}+ E \cdot n_{1}^2 + F\cdot n_{1}+G}{(n'''-n_1+6)(n'''-n_1+4)(n'''-n_1+2)(n'''-n_1)(n'''-n_1-2)(n'''-n_1-4)(n'''-n_1+-6)}}
\\
+ {\textstyle\frac{2^4 \cdot (n-n''')^2}{(n'''-n_1+8)(n'''-n_1+4)(n'''-n_1)(n'''-n_1-4)(n'''-n_1-8)}}
+ \left(\sim{\textstyle\frac{a_6}{n_{1}^{6}}}\right) 
\end{multline*}
Here
\[
D= \big[(n-n''')^2-4^2 \big](n-n'''),
\]
\begin{multline*}
E= -3n^3n'''+5n^2(n''')^2+52n^2n'''-24n^2-n(n''')^3-104n(n''')^2
\\
+72nn'''+312n-(n''')^4+52(n''')^3-32(n''')^2-520n'''+144
\end{multline*}
\begin{multline*}
F= 3n^3(n''')^2-16n^3-7n^2(n''')^3-26n^2(n''')^2+88n^2n'''+104n^2
\\
+5n(n''')^4+52a(n''')^3
-152n(n''')^2-520nn'''+496n-(n''')^5-26(n''')^4
\\
+72(n''')^3+520(n''')^2-656n'''-1664,
\end{multline*}
\begin{multline*}
G= -n^3(n''')^3+16n^3n'''+3n^2(n''')^4-72n^2(n''')^2+144n^2-3n(n''')^5
\\
+112n(n''')^3-784nn'''+(n''')^6-56(n''')^4+784(n''')^2-2304,
\end{multline*}
and
\[
\left(\sim{\textstyle\frac{a_6}{n_{1}^{6}}}\right) 
\]
denotes finite number of rational terms in $n_1,n,n'''$ which decrease at infinity like $\sim \tfrac{1}{n_{1}^{6}}$
or faster.
Writing the analogue of the expression (\ref{[omega]}) for $\omega =3$ and $x_0(t) = \tfrac{\sin t}{2} + \tfrac{2}{3}(\tfrac{\sin t}{2})^3$
fully we have
\begin{multline*}
{\textstyle\frac{n'''-n}{n-n_1}}
{\textstyle\frac{D \cdot n_{1}^{3}+ E \cdot n_{1}^2 + F \cdot n_{1}+G}{(n'''-n_1+6)(n'''-n_1+4)(n'''-n_1+2)(n'''-n_1)(n'''-n_1-2)(n'''-n_1-4)(n'''-n_1-6)}}
\\
+ {\textstyle\frac{2^4 (n-n''')^2}{(n'''-n_1+8)(n'''-n_1+4)(n'''-n_1)(n'''-n_1-4)(n'''-n_1-8)}}
\\
+ {\textstyle\frac{-2^3 \cdot 5 \cdot (n-n''')^3}{(n'''-n_1+10)(n'''-n_1+6) \ldots (n'''-n_1-10)}}
+ {\textstyle\frac{-2^5 \cdot 5 \cdot (n-n''')^2}{(n'''-n_1+12)(n'''-n_1+8) \ldots (n'''-n_1-12)}}
\\
+ {\textstyle\frac{2^6 \cdot 3 \cdot 5 \cdot 7 \cdot (n-n''')^3}{(n'''-n_1+14)(n'''-n_1+10) \ldots (n'''-n_1-14)}}
+ {\textstyle\frac{-2^9\cdot 5 \cdot 7 \cdot (n-n''')^3}{(n'''-n_1+18)(n'''-n_1+14) \ldots (n'''-n_1-18)}}
\end{multline*}
so that for the odd part
\begin{multline*}
\big[\widetilde{\textrm{ret} \, \kappa_{q \,\textrm{odd}}}\big]'''(\widehat{n}\cdot\widehat{l}) 
\\
=
\sum\limits_{n_1\in 2\mathbb{Z}+1}
    {\textstyle\frac{i}{2\pi}}\, 
\Bigg[
{\textstyle\frac{n'''-n}{n-n_1}}
{\textstyle\frac{D \cdot n_{1}^{3}+ E \cdot n_{1}^2 + F \cdot n_{1}+G}{(n'''-n_1+6)(n'''-n_1+4)(n'''-n_1+2)\ldots(n'''-n_1-6)}}
\\
+ {\textstyle\frac{2^4 (n-n''')^2}{(n'''-n_1+8)(n'''-n_1+4)(n'''-n_1)(n'''-n_1-4)(n'''-n_1-8)}}
\\
+ {\textstyle\frac{-2^3 \cdot 5 \cdot (n-n''')^3}{(n'''-n_1+10)(n'''-n_1+6) \ldots (n'''-n_1-10)}}
+ {\textstyle\frac{-2^5 \cdot 5 \cdot (n-n''')^2}{(n'''-n_1+12)(n'''-n_1+8) \ldots (n'''-n_1-12)}}
\\
+ {\textstyle\frac{2^6 \cdot 3 \cdot 5 \cdot 7 \cdot (n-n''')^3}{(n'''-n_1+14)(n'''-n_1+10) \ldots (n'''-n_1-14)}}
+ {\textstyle\frac{-2^9\cdot 5 \cdot 7 \cdot (n-n''')^3}{(n'''-n_1+18)(n'''-n_1+14) \ldots (n'''-n_1-18)}}
\\
\Bigg]
\,\,
\widetilde{\kappa_{q \,\textrm{odd}}}(\widehat{n_1} \cdot \widehat{l})  
+ {\textstyle\frac{1}{2}} \, \widetilde{\kappa_{q \,\textrm{odd}}}(\widehat{n} \cdot \widehat{l}).
\end{multline*}
\[
n,n'''\in 2\mathbb{Z}, \,\,\, \omega=3,
\]
is convergent whenever the degree $\omega'$ of $\kappa_q\leq 3$. Analogously for the even part
\begin{multline*}
\big[\widetilde{\textrm{ret} \, \kappa_{q \,\textrm{even}}}\big]'''(\widehat{n}\cdot\widehat{l}) 
\\
=
\sum\limits_{n_1\in 2\mathbb{Z}}
    {\textstyle\frac{i}{2\pi}}\, 
\Bigg[
{\textstyle\frac{n'''-n}{n-n_1}}
{\textstyle\frac{D \cdot n_{1}^{3}+ E \cdot n_{1}^2 + F \cdot n_{1}+G}{(n'''-n_1+6)(n'''-n_1+4)(n'''-n_1+2)\ldots(n'''-n_1-6)}}
\\
+ {\textstyle\frac{2^4 (n-n''')^2}{(n'''-n_1+8)(n'''-n_1+4)(n'''-n_1)(n'''-n_1-4)(n'''-n_1-8)}}
\\
+ {\textstyle\frac{-2^3 \cdot 5 \cdot (n-n''')^3}{(n'''-n_1+10)(n'''-n_1+6) \ldots (n'''-n_1-10)}}
+ {\textstyle\frac{-2^5 \cdot 5 \cdot (n-n''')^2}{(n'''-n_1+12)(n'''-n_1+8) \ldots (n'''-n_1-12)}}
\\
+ {\textstyle\frac{2^6 \cdot 3 \cdot 5 \cdot 7 \cdot (n-n''')^3}{(n'''-n_1+14)(n'''-n_1+10) \ldots (n'''-n_1-14)}}
+ {\textstyle\frac{-2^9\cdot 5 \cdot 7 \cdot (n-n''')^3}{(n'''-n_1+18)(n'''-n_1+14) \ldots (n'''-n_1-18)}}
\\
\Bigg]
\,\,
\widetilde{\kappa_{q \,\textrm{even}}}(\widehat{n_1} \cdot \widehat{l})  
+ {\textstyle\frac{1}{2}} \, \widetilde{\kappa_{q \,\textrm{even}}}(\widehat{n} \cdot \widehat{l}).
\end{multline*}
\[
n,n'''\in 2\mathbb{Z}+1, \,\,\, \omega=3,
\]
convergent whenever the degree $\omega'$ of $\kappa_q\leq 3$.

For the  expression,  analogue to (\ref{[omega]}), for $\omega =4$ and $x_0(t) = \tfrac{\sin t}{2} + \tfrac{2}{3}(\tfrac{\sin t}{2})^3$
we have
\begin{multline*}
{\textstyle\frac{n'''-n}{n-n_1}}
{\textstyle\frac{D \cdot n_{1}^{3}+ E \cdot n_{1}^2 + F \cdot n_{1}+G}{(n'''-n_1+6)(n'''-n_1+4)(n'''-n_1+2)(n'''-n_1)(n'''-n_1-2)(n'''-n_1-4)(n'''-n_1-6)}}
\\
+ {\textstyle\frac{-(n-n''')^4 +2^4 (n-n''')^2}{(n'''-n_1+8)(n'''-n_1+4)\ldots (n'''-n_1-8)}}
\\
+ {\textstyle\frac{-2^3 \cdot 5 \cdot (n-n''')^3}{(n'''-n_1+10)(n'''-n_1+6) \ldots (n'''-n_1-10)}}
+ {\textstyle\frac{-2^5 \cdot 5 \cdot (n-n''')^2}{(n'''-n_1+12)(n'''-n_1+8) \ldots (n'''-n_1-12)}}
\\
+ {\textstyle\frac{2^6 \cdot 3 \cdot 5 \cdot 7 \cdot (n-n''')^3}{(n'''-n_1+14)(n'''-n_1+10) \ldots (n'''-n_1-14)}}
+ {\textstyle\frac{-2^9\cdot 5 \cdot 7 \cdot (n-n''')^3}{(n'''-n_1+18)(n'''-n_1+14) \ldots (n'''-n_1-18)}}
\\
+{\textstyle\frac{2^3\cdot 5\cdot 3 \cdot (n-n''')^4}{(n'''-n_1+12)(n'''-n_1+8)\ldots (n'''-n_1-12)}}
\\
+{\textstyle\frac{(-1)2^7\cdot 5\cdot 7 \cdot (n-n''')^4}{(n'''-n_1+16)(n'''-n_1+12)\ldots (n'''-n_1-16)}}
+{\textstyle\frac{2^{10}\cdot 5^2 \cdot 7 \cdot (n-n''')^4}{(n'''-n_1+20)(n'''-n_1+16)\ldots (n'''-n_1-20)}}
\end{multline*}
By reducing the first two fractions to the common denominator form, 
we get the fraction whose numerator is a polynomial of degree $4$ in $n_1$ and whose denominator 
is a polynomial of degree $10$ in $n_1$. The numerators of subsequent fractions do not depend on $n_1$, 
and their denominators are of $6$-th, or greater than the $6$-th, order in $n_1$. Thus, the whole expression
behaves like $\sim 1/n_{1}^{6}$ for $n_1$ going to infinity, and is equal: 
\begin{multline*}
{\textstyle\frac{n'''-n}{n-n_1}}
{\textstyle\frac{\big[((n-n''')^2-16)(n^2+nn''-2n''') +E\big]\cdot n_{1}^{4}+ \ldots}
{(n'''-n_1+8)(n'''-n_1+6)(n'''-n_1+4)(n'''-n_1+2)\ldots (n'''-n_1-8)}}
\\
+ {\textstyle\frac{-2^3 \cdot 5 \cdot (n-n''')^3}{(n'''-n_1+10)(n'''-n_1+6) \ldots (n'''-n_1-10)}}
+ {\textstyle\frac{-2^5 \cdot 5 \cdot (n-n''')^2}{(n'''-n_1+12)(n'''-n_1+8) \ldots (n'''-n_1-12)}}
\\
+ {\textstyle\frac{2^6 \cdot 3 \cdot 5 \cdot 7 \cdot (n-n''')^3}{(n'''-n_1+14)(n'''-n_1+10) \ldots (n'''-n_1-14)}}
+ {\textstyle\frac{-2^9\cdot 5 \cdot 7 \cdot (n-n''')^3}{(n'''-n_1+18)(n'''-n_1+14) \ldots (n'''-n_1-18)}}
\\
+{\textstyle\frac{2^3\cdot 5\cdot 3 \cdot (n-n''')^4}{(n'''-n_1+12)(n'''-n_1+8)\ldots (n'''-n_1-12)}}
\\
+{\textstyle\frac{(-1)2^7\cdot 5\cdot 7 \cdot (n-n''')^4}{(n'''-n_1+16)(n'''-n_1+12)\ldots (n'''-n_1-16)}}
+{\textstyle\frac{2^{10}\cdot 5^2 \cdot 7 \cdot (n-n''')^4}{(n'''-n_1+20)(n'''-n_1+16)\ldots (n'''-n_1-20)}}
\end{multline*}
so that for the odd part we have
\begin{multline*}
\big[\widetilde{\textrm{ret} \, \kappa_{q \,\textrm{odd}}}\big]'''(\widehat{n}\cdot\widehat{l}) 
\\
=
\sum\limits_{n_1\in 2\mathbb{Z}+1}
    {\textstyle\frac{i}{2\pi}}\, 
\Bigg[
{\textstyle\frac{n'''-n}{n-n_1}}
{\textstyle\frac{\big[((n-n''')^2-16)(n^2+nn''-2n''') +E\big]\cdot n_{1}^{4}+ \ldots}
{(n'''-n_1+8)(n'''-n_1+6)(n'''-n_1+4)(n'''-n_1+2)\ldots (n'''-n_1-8)}}
\\
+ {\textstyle\frac{-2^3 \cdot 5 \cdot (n-n''')^3}{(n'''-n_1+10)(n'''-n_1+6) \ldots (n'''-n_1-10)}}
+ {\textstyle\frac{-2^5 \cdot 5 \cdot (n-n''')^2}{(n'''-n_1+12)(n'''-n_1+8) \ldots (n'''-n_1-12)}}
\\
+ {\textstyle\frac{2^6 \cdot 3 \cdot 5 \cdot 7 \cdot (n-n''')^3}{(n'''-n_1+14)(n'''-n_1+10) \ldots (n'''-n_1-14)}}
+ {\textstyle\frac{-2^9\cdot 5 \cdot 7 \cdot (n-n''')^3}{(n'''-n_1+18)(n'''-n_1+14) \ldots (n'''-n_1-18)}}
\\
+{\textstyle\frac{2^3\cdot 5\cdot 3 \cdot (n-n''')^4}{(n'''-n_1+12)(n'''-n_1+8)\ldots (n'''-n_1-12)}}
\\
+{\textstyle\frac{(-1)2^7\cdot 5\cdot 7 \cdot (n-n''')^4}{(n'''-n_1+16)(n'''-n_1+12)\ldots (n'''-n_1-16)}}
+{\textstyle\frac{2^{10}\cdot 5^2 \cdot 7 \cdot (n-n''')^4}{(n'''-n_1+20)(n'''-n_1+16)\ldots (n'''-n_1-20)}}
\\
\Bigg]
\,\,
\widetilde{\kappa_{q \,\textrm{odd}}}(\widehat{n_1} \cdot \widehat{l})  
+ {\textstyle\frac{1}{2}} \, \widetilde{\kappa_{q \,\textrm{odd}}}(\widehat{n} \cdot \widehat{l}),
\end{multline*}
\[
n,n'''\in 2\mathbb{Z}, \,\,\, \omega=4,
\]
convergent whenever the degree $\omega'$ of $\kappa_q\leq 4$. Analogously for the even part
\begin{multline*}
\big[\widetilde{\textrm{ret} \, \kappa_{q \,\textrm{even}}}\big]'''(\widehat{n}\cdot\widehat{l}) 
\\
=
\sum\limits_{n_1\in 2\mathbb{Z}}
    {\textstyle\frac{i}{2\pi}}\, 
\Bigg[
{\textstyle\frac{n'''-n}{n-n_1}}
{\textstyle\frac{\big[((n-n''')^2-16)(n^2+nn''-2n''') +E\big]\cdot n_{1}^{4}+ \ldots}
{(n'''-n_1+8)(n'''-n_1+6)(n'''-n_1+4)(n'''-n_1+2)\ldots (n'''-n_1-8)}}
\\
+ {\textstyle\frac{-2^3 \cdot 5 \cdot (n-n''')^3}{(n'''-n_1+10)(n'''-n_1+6) \ldots (n'''-n_1-10)}}
+ {\textstyle\frac{-2^5 \cdot 5 \cdot (n-n''')^2}{(n'''-n_1+12)(n'''-n_1+8) \ldots (n'''-n_1-12)}}
\\
+ {\textstyle\frac{2^6 \cdot 3 \cdot 5 \cdot 7 \cdot (n-n''')^3}{(n'''-n_1+14)(n'''-n_1+10) \ldots (n'''-n_1-14)}}
+ {\textstyle\frac{-2^9\cdot 5 \cdot 7 \cdot (n-n''')^3}{(n'''-n_1+18)(n'''-n_1+14) \ldots (n'''-n_1-18)}}
\\
+{\textstyle\frac{2^3\cdot 5\cdot 3 \cdot (n-n''')^4}{(n'''-n_1+12)(n'''-n_1+8)\ldots (n'''-n_1-12)}}
\\
+{\textstyle\frac{(-1)2^7\cdot 5\cdot 7 \cdot (n-n''')^4}{(n'''-n_1+16)(n'''-n_1+12)\ldots (n'''-n_1-16)}}
+{\textstyle\frac{2^{10}\cdot 5^2 \cdot 7 \cdot (n-n''')^4}{(n'''-n_1+20)(n'''-n_1+16)\ldots (n'''-n_1-20)}}
\\
\Bigg]
\,\,
\widetilde{\kappa_{q \,\textrm{even}}}(\widehat{n_1} \cdot \widehat{l}) 
+ {\textstyle\frac{1}{2}} \, \widetilde{\kappa_{q \,\textrm{even}}}(\widehat{n} \cdot \widehat{l}),
\end{multline*}
\[
n,n'''\in 2\mathbb{Z}+1, \,\,\, \omega=4,
\]
convergent whenever the degree $\omega'$ of $\kappa_q\leq 4$ (or even for $\omega'<5$, but in QFT we encounter
only integer $\omega'$).

We cannot apply the same $x_0(t) = \tfrac{\sin t}{2} + \tfrac{2}{3}(\tfrac{\sin t}{2})^3$ (which is dual of order $4$)
for $\omega \geq 5$.
 
In order to get $x_0$ dual to $X_0= 2\, \partial_t$ at $t=0$ up to order $\omega=6$,
we need to add to the previous $x_0$, which was of order $\omega=4$, the power   $\big({\tfrac{1}{2}}\sin \big)^5$
multiplied by $2\cdot 3/5$:
\[
x_0(t) = {\textstyle\frac{\sin t }{2}} + {\textstyle\frac{2}{3}} \left({\textstyle\frac{\sin t}{2}}\right)^3
+ {\textstyle\frac{2\cdot 3}{5}}\left({\textstyle\frac{\sin t}{2}}\right)^5
\] 
in order to obtain new function $x_0$ dual of order $\omega=6$, with Taylor expansion
\[
x_0(t) \,\,\,\,\,\,\, = \,\,\,\,\,\,\,  {\textstyle\frac{1}{2}} t \,\,\,\,\,\, -{\textstyle\frac{9\cdot 5^2}{2}}{\textstyle\frac{1}{7!}} t^7 
\,\,\,\,\, + \,\,\,\, \ldots,
\]
and similarly
\begin{align*}
x_0(t) = {\textstyle\frac{\sin t }{2}} + {\textstyle\frac{2}{3}} \left({\textstyle\frac{\sin t}{2}}\right)^3
+ {\textstyle\frac{2\cdot 3}{5}}\left({\textstyle\frac{\sin t}{2}}\right)^5 
+ {\textstyle\frac{4\cdot 5}{7}}\left({\textstyle\frac{\sin t}{2}}\right)^7, 
\,\,\,\,\,\,\,\,\,\,\,\,\,\,\,\,\,\,\,\,\,\,\,\,\,\,\,\,\,\,\,\,\,\,\,\,\,\,\,\, & \,\,\,\,\,\,\,\,\,\,  \textrm{for} \,\,\, \omega \leq 8 
\\
\\
x_0(t) = {\textstyle\frac{\sin t }{2}} + {\textstyle\frac{2}{3}} \left({\textstyle\frac{\sin t}{2}}\right)^3
+ {\textstyle\frac{2\cdot 3}{5}}\left({\textstyle\frac{\sin t}{2}}\right)^5 
+ {\textstyle\frac{4\cdot 5}{7}}\left({\textstyle\frac{\sin t}{2}}\right)^7
+ {\textstyle\frac{2\cdot 5\cdot 7}{3^2}}\left({\textstyle\frac{\sin t}{2}}\right)^9,  & \,\,\,\,\,\,\,\,\,\,  \textrm{for} \,\,\, \omega \leq 10
\\
  \vdots \,\,\,\,\,\,\,\,\,\,\,\,\,\,\,\,\,\,\,\,\,\,\,\,\,\,\,\,\,\,\,\,\,\,\,\,\,\,\,\,  
\,\,\,\,\,\,\,\,\,\,\,\,\,\,\,\,\,\,\,\,\,\,\,\,\,\,\,\,\,\,\,\,\,\,\,\,\,\,\,\,   &   \,\,\,\,\,\,\,\,\,\,\,\,\,\,\,\,\,\,\,\,   \vdots
\end{align*}
always adding powers of  ${\tfrac{1}{2}}\sin $, and using repeatedly the formula (\ref{FT((x^a)theta)}) 
for the powers of $x_0= {\tfrac{1}{2}}\sin$ we can compute 
$\big[\widetilde{\textrm{ret} \, \kappa_q}\big]'''(\widehat{n}\cdot\widehat{l})$ for any
 $\widetilde{\kappa_q}$ with any order $\omega'$,
computing it separately for the characters $\widehat{n}\cdot\widehat{l}$ with odd and separately with even $n$
using  in each case different $n'''$, which is, respectively, odd or even.

This conclusion concerns convergence of $\big[\widetilde{\textrm{ret} \, \kappa_q}\big]'''(\widehat{n}\cdot\widehat{l})$.
But $\big[\widetilde{\textrm{ret} \, \kappa_q}\big]'''(\widehat{n}\cdot\widehat{l})$ and
$\widetilde{\textrm{ret} \, \kappa_q}(\widehat{n}\cdot\widehat{l})$ 
differ by (\ref{FT(EpsteinGlaserReminderEU)}). 
Therefore, by the very construction we are sure only that 
$\big[\widetilde{\textrm{ret} \, \kappa_q}\big]'''(\widehat{n}\cdot\widehat{l})$
and $\widetilde{\textrm{ret} \, \kappa_q}(\widehat{n}\cdot\widehat{l})$ coincide on the 
relatively small subspace of test functions $\widetilde{\phi}$, with $\phi$ whose all time derivatives up to order $\omega=\omega'$ 
vanish at all points of the Cauchy surface $t=0$.

Summing up, and taking into account
\[
\widetilde{\kappa_{q \,\textrm{odd}}}(\widehat{n}\cdot\widehat{l}) = 0, \,\, n \in 2\mathbb{Z},
\,\,\,\,\,\,\,\,\,\,
\widetilde{\kappa_{q \,\textrm{even}}}(\widehat{n}\cdot\widehat{l}) = 0, \,\, n \in 2\mathbb{Z}+1,
\]
we have proved the following
\begin{twr} 
 Let $\textrm{ret} \, \kappa_q = \theta \kappa_q \circ \Omega'$, for periodic step theta function $\theta$
on the compactified Einstein Universe. Here 
\begin{multline}\label{Omega'phiEU(0,e)-(2pi,e)}
\Omega'\phi(t,\boldsymbol{w}) = \phi(t,\boldsymbol{w}) - 
\sum\limits_{|\alpha| \leq \omega} (X)^\alpha\phi(0,e)  {\textstyle\frac{(x_0)^a}{a!}} w_{I}(t)
\\
-
\sum\limits_{|\alpha| \leq \omega}  (X)^\alpha\phi(2\pi,e) {\textstyle\frac{(x_0)^a}{a!}} w_{II}(t), \,\,\, \phi \in \mathscr{E}, 
\end{multline}
with the function $x_0$ coming from the system of functions $x_0$,$x_1$,$x_2$,$x_3$, depending on $\omega$ and defined above, 
with differentials dual to $X_0=2\partial_t$,$X_1$,$X_2$,$X_3$ up to order $\omega$. $w_{II}$, $w_{I}$ are smooth functions
of $t\in \widetilde{\mathbb{S}^1}$, with $w_{I}$ equal $1$ on a neighborhood of $t=0$ and equal $0$ on a neighborhood of $t=2\pi$,
and \emph{vice versa} for $w_{II}$. Let the polynomial functions $A, \dots, G$ of the three integers $n,n_1,n'''$, be defined as above.  
Let $\omega'$ be the growing order of $\widetilde{\kappa_q}$ at infinity. Define
\[
\big[\widetilde{\textrm{ret} \, \kappa_{q \,\textrm{odd}}}\big]'''(\widehat{n}\cdot\widehat{l}) 
=
\sum\limits_{n_1\in 2\mathbb{Z}+1}
    {\textstyle\frac{i}{2\pi}}\, 
\Big[
{\textstyle\frac{n'''-n}{(n-n_1)(n'''-n_1)}}
\Big]
\,\,
\widetilde{\kappa_{q \,\textrm{odd}}}(\widehat{n_1} \cdot \widehat{l}),
\]
\[
\textrm{for} \,\,\, n,n''' \in 2 \mathbb{Z}, \,\, \omega=0,
\]
and
\[
\big[\widetilde{\textrm{ret} \, \kappa_{q \,\textrm{even}}}\big]'''(\widehat{n}\cdot\widehat{l}) 
=
\sum\limits_{n_1\in 2\mathbb{Z}}
    {\textstyle\frac{i}{2\pi}}\, 
\Big[
{\textstyle\frac{n'''-n}{(n-n_1)(n'''-n_1)}}
\Big]
\,\,
\widetilde{\kappa_{q \,\textrm{even}}}(\widehat{n_1} \cdot \widehat{l}),
\]
\[
\textrm{for} \,\,\, n,n''' \in 2 \mathbb{Z}+1, \,\, \omega=0,
\] 
are convergent if $\omega' \leq \omega = 0$.
\begin{multline*}
\big[\widetilde{\textrm{ret} \, \kappa_{q \,\textrm{odd}}}\big]'''(\widehat{n}\cdot\widehat{l}) 
\\
=
\sum\limits_{n_1\in 2\mathbb{Z}+1}
    {\textstyle\frac{i}{2\pi}}\, 
\Big[
{\textstyle\frac{n-n'''}{n-n_1}} \,\,
{\textstyle\frac{(nn'''-{n'''}^2+4) +(n'''-n)n_1}{(n'''-n_1+2)(n'''-n_1)(n'''-n_1-2)}}
\Big]
\,\,
\widetilde{\kappa_{q \,\textrm{odd}}}(\widehat{n_1} \cdot \widehat{l}),
\end{multline*}
\[
\textrm{for} \,\,\, n,n''' \in 2 \mathbb{Z}, \,\, \omega=1,
\]
and
\begin{multline*}
\big[\widetilde{\textrm{ret} \, \kappa_{q \,\textrm{even}}}\big]'''(\widehat{n}\cdot\widehat{l}) 
\\
=
\sum\limits_{n_1\in 2\mathbb{Z}}
    {\textstyle\frac{i}{2\pi}}\, 
\Big[
{\textstyle\frac{n-n'''}{n-n_1}} \,\,
{\textstyle\frac{(nn'''-{n'''}^2+4) +(n'''-n)n_1}{(n'''-n_1+2)(n'''-n_1)(n'''-n_1-2)}}
\Big]
\,\,
\widetilde{\kappa_{q \,\textrm{even}}}(\widehat{n_1} \cdot \widehat{l}),
\end{multline*}
\[
\textrm{for} \,\,\,  n,n''' \in 2 \mathbb{Z}+1, \,\, \omega=1,
\]
are convergent if $\omega' \leq 1$.
\begin{multline*}
\widetilde{\textrm{ret} \, \kappa_{q \,\textrm{odd}}}(\widehat{n}\cdot\widehat{l}) 
\\
=
\sum\limits_{n_1\in 2\mathbb{Z}+1}
    {\textstyle\frac{i}{2\pi}}\, 
\Big[
{\textstyle\frac{n'''-n}{n-n_1}}
{\textstyle\frac{C + B n_1
+A n_{1}^{2}}{(n'''-n_1+4)(n'''-n_1+2)(n'''-n_1)(n'''-n_1-2)(n'''-n_1-4)}}
\Big]
\,\,
\widetilde{\kappa_{q \,\textrm{odd}}}(\widehat{n_1} \cdot \widehat{l}),
\end{multline*}
\[
\textrm{for} \,\,\, n,n''' \in 2 \mathbb{Z}, \,\, \omega=2,
\]
and
\begin{multline*}
\big[\widetilde{\textrm{ret} \, \kappa_{q \,\textrm{even}}}\big]'''(\widehat{n}\cdot\widehat{l}) 
\\
=
\sum\limits_{n_1\in 2\mathbb{Z}}
    {\textstyle\frac{i}{2\pi}}\, 
\Big[
{\textstyle\frac{n'''-n}{n-n_1}}
{\textstyle\frac{C + B n_1
+A n_{1}^{2}}{(n'''-n_1+4)(n'''-n_1+2)(n'''-n_1)(n'''-n_1-2)(n'''-n_1-4)}}
\Big]
\,\,
\widetilde{\kappa_{q \,\textrm{even}}}(\widehat{n_1} \cdot \widehat{l}),
\end{multline*}
\[
\textrm{for} \,\,\, n,n''' \in 2 \mathbb{Z}+1, \,\, \omega=2,
\]
are convergent if $\omega'\leq 2$. 
\begin{multline*}
\big[\widetilde{\textrm{ret} \, \kappa_{q \,\textrm{odd}}}\big]'''(\widehat{n}\cdot\widehat{l}) 
\\
=
\sum\limits_{n_1\in 2\mathbb{Z}+1}
    {\textstyle\frac{i}{2\pi}}\, 
\Bigg[
{\textstyle\frac{n'''-n}{n-n_1}}
{\textstyle\frac{D \cdot n_{1}^{3}+ E \cdot n_{1}^2 + F \cdot n_{1}+G}{(n'''-n_1+6)(n'''-n_1+4)(n'''-n_1+2)\ldots(n'''-n_1-6)}}
\\
+ {\textstyle\frac{2^4 (n-n''')^2}{(n'''-n_1+8)(n'''-n_1+4)(n'''-n_1)(n'''-n_1-4)(n'''-n_1-8)}}
\\
+ {\textstyle\frac{-2^3 \cdot 5 \cdot (n-n''')^3}{(n'''-n_1+10)(n'''-n_1+6) \ldots (n'''-n_1-10)}}
+ {\textstyle\frac{-2^5 \cdot 5 \cdot (n-n''')^2}{(n'''-n_1+12)(n'''-n_1+8) \ldots (n'''-n_1-12)}}
\\
+ {\textstyle\frac{2^6 \cdot 3 \cdot 5 \cdot 7 \cdot (n-n''')^3}{(n'''-n_1+14)(n'''-n_1+10) \ldots (n'''-n_1-14)}}
+ {\textstyle\frac{-2^9\cdot 5 \cdot 7 \cdot (n-n''')^3}{(n'''-n_1+18)(n'''-n_1+14) \ldots (n'''-n_1-18)}}
\\
\Bigg]
\,\,
\widetilde{\kappa_{q \,\textrm{odd}}}(\widehat{n_1} \cdot \widehat{l}),
\end{multline*}
\[
\textrm{for} \,\, n,n'''\in 2\mathbb{Z}, \,\,\, \omega=3,
\]
and
\begin{multline*}
\big[\widetilde{\textrm{ret} \, \kappa_{q \,\textrm{even}}}\big]'''(\widehat{n}\cdot\widehat{l}) 
\\
=
\sum\limits_{n_1\in 2\mathbb{Z}}
    {\textstyle\frac{i}{2\pi}}\, 
\Bigg[
{\textstyle\frac{n'''-n}{n-n_1}}
{\textstyle\frac{D \cdot n_{1}^{3}+ E \cdot n_{1}^2 + F \cdot n_{1}+G}{(n'''-n_1+6)(n'''-n_1+4)(n'''-n_1+2)\ldots(n'''-n_1-6)}}
\\
+ {\textstyle\frac{2^4 (n-n''')^2}{(n'''-n_1+8)(n'''-n_1+4)(n'''-n_1)(n'''-n_1-4)(n'''-n_1-8)}}
\\
+ {\textstyle\frac{-2^3 \cdot 5 \cdot (n-n''')^3}{(n'''-n_1+10)(n'''-n_1+6) \ldots (n'''-n_1-10)}}
+ {\textstyle\frac{-2^5 \cdot 5 \cdot (n-n''')^2}{(n'''-n_1+12)(n'''-n_1+8) \ldots (n'''-n_1-12)}}
\\
+ {\textstyle\frac{2^6 \cdot 3 \cdot 5 \cdot 7 \cdot (n-n''')^3}{(n'''-n_1+14)(n'''-n_1+10) \ldots (n'''-n_1-14)}}
+ {\textstyle\frac{-2^9\cdot 5 \cdot 7 \cdot (n-n''')^3}{(n'''-n_1+18)(n'''-n_1+14) \ldots (n'''-n_1-18)}}
\\
\Bigg]
\,\,
\widetilde{\kappa_{q \,\textrm{even}}}(\widehat{n_1} \cdot \widehat{l}),
\end{multline*}
\[
\textrm{for} \,\,\, n,n'''\in 2\mathbb{Z}+1, \,\,\, \omega=3,
\]
are convergent if $\omega' \leq 3$.
\begin{multline*}
\big[\widetilde{\textrm{ret} \, \kappa_{q \,\textrm{odd}}}\big]'''(\widehat{n}\cdot\widehat{l}) 
\\
=
\sum\limits_{n_1\in 2\mathbb{Z}+1}
    {\textstyle\frac{i}{2\pi}}\, 
\Bigg[
{\textstyle\frac{n'''-n}{n-n_1}}
{\textstyle\frac{\big[((n-n''')^2-16)(n^2+nn''-2n''') +E\big]\cdot n_{1}^{4}+ \ldots}
{(n'''-n_1+8)(n'''-n_1+6)(n'''-n_1+4)(n'''-n_1+2)\ldots (n'''-n_1-8)}}
\\
+ {\textstyle\frac{-2^3 \cdot 5 \cdot (n-n''')^3}{(n'''-n_1+10)(n'''-n_1+6) \ldots (n'''-n_1-10)}}
+ {\textstyle\frac{-2^5 \cdot 5 \cdot (n-n''')^2}{(n'''-n_1+12)(n'''-n_1+8) \ldots (n'''-n_1-12)}}
\\
+ {\textstyle\frac{2^6 \cdot 3 \cdot 5 \cdot 7 \cdot (n-n''')^3}{(n'''-n_1+14)(n'''-n_1+10) \ldots (n'''-n_1-14)}}
+ {\textstyle\frac{-2^9\cdot 5 \cdot 7 \cdot (n-n''')^3}{(n'''-n_1+18)(n'''-n_1+14) \ldots (n'''-n_1-18)}}
\\
+{\textstyle\frac{2^3\cdot 5\cdot 3 \cdot (n-n''')^4}{(n'''-n_1+12)(n'''-n_1+8)\ldots (n'''-n_1-12)}}
\\
+{\textstyle\frac{(-1)2^7\cdot 5\cdot 7 \cdot (n-n''')^4}{(n'''-n_1+16)(n'''-n_1+12)\ldots (n'''-n_1-16)}}
+{\textstyle\frac{2^{10}\cdot 5^2 \cdot 7 \cdot (n-n''')^4}{(n'''-n_1+20)(n'''-n_1+16)\ldots (n'''-n_1-20)}}
\\
\Bigg]
\,\,
\widetilde{\kappa_{q \,\textrm{odd}}}(\widehat{n_1} \cdot \widehat{l}),
\end{multline*}
\[
\textrm{for} \,\,\, n,n'''\in 2\mathbb{Z}, \,\,\, \omega=4,
\]
and
\begin{multline*}
\big[\widetilde{\textrm{ret} \, \kappa_{q \,\textrm{even}}}\big]'''(\widehat{n}\cdot\widehat{l}) 
\\
=
\sum\limits_{n_1\in 2\mathbb{Z}}
    {\textstyle\frac{i}{2\pi}}\, 
\Bigg[
{\textstyle\frac{n'''-n}{n-n_1}}
{\textstyle\frac{\big[((n-n''')^2-16)(n^2+nn''-2n''') +E\big]\cdot n_{1}^{4}+ \ldots}
{(n'''-n_1+8)(n'''-n_1+6)(n'''-n_1+4)(n'''-n_1+2)\ldots (n'''-n_1-8)}}
\\
+ {\textstyle\frac{-2^3 \cdot 5 \cdot (n-n''')^3}{(n'''-n_1+10)(n'''-n_1+6) \ldots (n'''-n_1-10)}}
+ {\textstyle\frac{-2^5 \cdot 5 \cdot (n-n''')^2}{(n'''-n_1+12)(n'''-n_1+8) \ldots (n'''-n_1-12)}}
\\
+ {\textstyle\frac{2^6 \cdot 3 \cdot 5 \cdot 7 \cdot (n-n''')^3}{(n'''-n_1+14)(n'''-n_1+10) \ldots (n'''-n_1-14)}}
+ {\textstyle\frac{-2^9\cdot 5 \cdot 7 \cdot (n-n''')^3}{(n'''-n_1+18)(n'''-n_1+14) \ldots (n'''-n_1-18)}}
\\
+{\textstyle\frac{2^3\cdot 5\cdot 3 \cdot (n-n''')^4}{(n'''-n_1+12)(n'''-n_1+8)\ldots (n'''-n_1-12)}}
\\
+{\textstyle\frac{(-1)2^7\cdot 5\cdot 7 \cdot (n-n''')^4}{(n'''-n_1+16)(n'''-n_1+12)\ldots (n'''-n_1-16)}}
+{\textstyle\frac{2^{10}\cdot 5^2 \cdot 7 \cdot (n-n''')^4}{(n'''-n_1+20)(n'''-n_1+16)\ldots (n'''-n_1-20)}}
\\
\Bigg]
\,\,
\widetilde{\kappa_{q \,\textrm{even}}}(\widehat{n_1} \cdot \widehat{l}),
\end{multline*}
\[
\textrm{for} \,\,\, n,n'''\in 2\mathbb{Z}+1, \,\,\, \omega=4,
\]
are convergent if  $\omega'\leq 4$. Dots in the numerators of the last two formulas denote a polynomial in $n_1,n,n'''$, which is of degree $3$
in $n_1$, and which we have constructed above. 

Then the distribution  $\big[\widetilde{\textrm{ret} \, \kappa_q}\big]'''$ is equal to 
$\widetilde{\textrm{ret} \, \kappa_q}$ up to a distribution concentrated on the Cauchy surfaces $t=0$ and $t= 2\pi$:
\[
\textrm{ret} \, \kappa_q -
\big[\textrm{ret} \, \kappa_q\big]'''
=
\sum\limits_{b=0}^{\omega} C_{I, b, n'''}(w) \cdot 
(X_0)^{b}\delta(t)
+
\sum\limits_{b=0}^{\omega} C_{II, b, n'''}(w) \cdot
(X_0)^{b}\delta(t-2\pi)
\]
with the functions $C_{I, b, n'''} = C_{b,n'''}$, where $C_{b, n''''}$ are given in the formulas
(\ref{FT(EpsteinGlaserReminderEU)}), (\ref{sumC_b(w)delta^(b)}). We have the analogous formula for $C_{II, b, n'''}$.

We have analogous convergent formulas for $\omega\geq 5$ using $x_0, \ldots x_3$ dual to $X_0, \ldots, X_3$ at $(0,e)$ and $(2\pi,e)$, 
up to order not less than $\omega$, with the general method of their construction presented above. 
\qed
\label{retkappa_qEUcompact}
\end{twr}

As is easily seen the orders of the above given distributions 
$\big[\widetilde{\textrm{ret} \, \kappa_q}]'''$, $\widetilde{\textrm{ret} \, \kappa_q}$, $\kappa_q$ are equal $\omega'=\omega$,
whenever the order of $\widetilde{\kappa_q}$ is equal $\omega'=\omega$. 
Contrary to the Minkowski space-time, we have no
well-defined scaling on the compactified Einstein Universe. It is easily seen that whenever the 
scaling makes sense, then it is preserved by the ``natural'' retarded part. In particular on the ordinary (non compactified)
Einstein Universe, with ordinary step theta function in definition of the natural retarded part of $\kappa_q$ (whenever exists), 
the time scaling degree indeed is preserved, \emph{i.e.} equal to the scaling degree of $\kappa_q$, or respectivly, the scaling degree in the zero momentum
component of the Fourier transform of the natural retarded part of $\kappa_q$ coincides with the scaling degree 
in the zero component of momentum of the Fourier transform of $\kappa_q$.

It is easily seen why we have arrived at the distribution  $\big[\textrm{ret} \, \kappa_q\big]'''$
for which
\[
\textrm{ret} \, \kappa_q \,\, \overset{\textrm{df}}{=} \,\, \theta\kappa_q \circ \Omega' \,\, = \,\, 
\big[\textrm{ret} \, \kappa_q\big]''',
\,\,\,\,\,\,\, \textrm{not on the whole} \,\,\, \textrm{Im} \, \Omega',
\]
with $\Omega'$ equal (\ref{Omega'phiEU(0,e)-(2pi,e)}), but which agrees with $\textrm{ret} \, \kappa_q$, 
only up to a distribution of order $\leq \omega$, concentrated on the whole Cauchy surfaces $t=0$ and $t=2\pi$.  
This is because we have used separately for each  $\widehat{n}\cdot\widehat{l}$ in the argument of $\widetilde{\textrm{ret} \, \kappa_q}$ 
a ``subtraction point'' $\widehat{n'''}\cdot\widehat{l}$, depending on  $\widehat{l}$,
and have subtracted the first $\omega$ time-like-terms of the ``Taylor expansion''
around the ``subtraction point'',
and have defined Fourier transform $\big[\widetilde{\textrm{ret} \, \kappa_q}\big]'''$  of another possible retarded part 
of $\kappa_q$, putting (\ref{TaylorSubtracionEUX0X1X2X3}). But in order to achieve a retarded part, which agrees with
the natural $\textrm{ret} \, \kappa_q$ on the whole image of $\Omega'$, i.e. up to a distribution of order $\omega$,
concentrated at just two points $(0,e)$ and $(2\pi, e)$, we need use in (\ref{TaylorSubtracionEUX0X1X2X3}) 
a fixed ``subtraction point'' $\widehat{n'''}\cdot\widehat{l'''}$,
independent of $\widehat{n}\cdot\widehat{l}$, and should replace the formula (\ref{TaylorSubtracionEUX0X1X2X3})
with 
\begin{equation}\label{TaylorSubtracionEUX0X1X2X3n'''l'''}
\big[\widetilde{\textrm{ret} \, \kappa_q}\big]'''(\widehat{n}\cdot\widehat{l})
\overset{\textrm{df}}{=}
\widetilde{\textrm{ret} \, \kappa_q}(\widehat{n}\cdot\widehat{l}) 
- \sum\limits_{|\beta|=0}^{\omega} \,\,
{\textstyle\frac{\big(\mathbb{A}(\widehat{n} \cdot \widehat{l})-\mathbb{A}(\widehat{n'''} \cdot \widehat{l'''})\big)^\beta}{\beta!}}
\, \Big(\widetilde{x^\beta} \,\, \ast  \,\, \widetilde{\textrm{ret} \, \kappa_q}\Big)(\widehat{n'''}\cdot\widehat{l'''}).
\end{equation}
But for general $\widehat{n'''}\cdot\widehat{l'''}$ this formula is meaningless, because the matrices 
$\mathbb{A}_{{}_{\mu}}(\widehat{n} \cdot \widehat{l})$, $\mu = 0, \ldots, 3$, do not live in general in the same matrix space as the matrices
$\mathbb{A}_{{}_{\mu}}(\widehat{n'''} \cdot \widehat{l'''})$, $\mu = 0, \ldots, 3$, and 
$\widetilde{\textrm{ret} \, \kappa_q}(\widehat{n'''}\cdot\widehat{l'''})$,
so that the product and subtraction operations in (\ref{TaylorSubtracionEUX0X1X2X3n'''l'''}) would not be well-defined.
But the ``subtraction points''  $\widehat{n'''}\cdot\widehat{l'''}$ with $l'''=0$ are exceptional and for any such ``subtration point''
the matrix $\widetilde{\textrm{ret} \, \kappa_q}(\widehat{n'''}\cdot\widehat{l'''})$ degenerates to a scalar, which can be regarded 
as proportional to a unit matrix. Similarly 
the matrices $\mathbb{A}_{{}_{k}}(\widehat{n'''} \cdot \widehat{l'''})$, $k=1,2,3$, degenrate to the zero scalars, and
the matrix  $\mathbb{A}_{{}_{0}}(\widehat{n'''} \cdot \widehat{l'''})$ degenerates to a scalar. Thus, all the matrices
 $\mathbb{A}_{{}_{\mu}}(\widehat{n'''} \cdot \widehat{l'''})$  can, in this case, be regarded
as proportional to the unit matrix in the matrix space of the actual $\mathbb{A}_{{}_{\mu}}(\widehat{n} \cdot \widehat{l})$.  
Therefore, the formula  (\ref{TaylorSubtracionEUX0X1X2X3n'''l'''}) makes sense and, moreover, the computation method, presented above,
can be applied without any changes also for the fixed ``subtraction point'' $\widehat{n'''}\cdot\widehat{l'''=0}$. 
Moreover, the subtracted ``Taylor expansion coefficients'' in (\ref{TaylorSubtracionEUX0X1X2X3n'''l'''})
are equal to the Fourier transform of a distribution of degree $\leq \omega$ and concentrated at the single point $(0,e)$ -- the unit
of our group $G$ if $\widehat{n'''}\cdot\widehat{l'''=0}$. Therefore, our new 
$\big[\widetilde{\textrm{ret} \, \kappa_q}\big]'''$, computed with the method presented above, agrees with the ``natural''
$\textrm{ret} \, \kappa_q$ on $\textrm{Im} \, \Omega'$,
not disturbing the ``natural'' formula of multiplication by $\theta$ on the subspace $\textrm{Im} \, \Omega'$.
Repeating the construction given above for a fixed ``subtraction 
point'' $\widehat{n'''}\cdot\widehat{l'''} = \widehat{n'''}\cdot\widehat{0}$, we can quite easily compute
explicit formula for $\big[\widetilde{\textrm{ret} \, \kappa_q}\big]'''$, which respects the condition
\begin{multline*}
\textrm{ret} \, \kappa_q \,\, - \,\,
\big[\textrm{ret} \, \kappa_q\big]''' 
\\
= 
P_I(X_0, \ldots, X_3) \delta(x_0)\dots \delta(x_3)
+ P_{II}(X_0, \ldots, X_3) \delta(x_0-x_{00})\dots \delta(x_3 - x_{30}), 
\\
 (0,e) = (x_0=0, \ldots, x_3=0),
\,\,\, 
(2\pi,e) = (x_{00}, \ldots, x_{30}),
\end{multline*}
with polynomials 
\[
P_I(X_0, \ldots, X_3) = \sum\limits_{|\alpha| \leq \omega} a_{I,\alpha} X^\alpha,
\,\,\,
P_{II}(X_0, \ldots, X_3) = \sum\limits_{|\alpha| \leq \omega} a_{II,\alpha} X^\alpha
\]
of degree $\textrm{max} |\alpha| \leq \omega$, and smooth coefficients $a_{I,\alpha}, a_{II,\alpha}$, 
when expressed in a local coordinate system $(x_0, \ldots,  x_3)$
which is dual to $X_0, \ldots, X_3$ at $(0,e)$ and at $(2\pi,e)$, up to order $\omega$ at least. We therefore 
denote such $\big[\textrm{ret} \, \kappa_q\big]'''$ simply by $\textrm{ret} \, \kappa_q$.

Now we pass to the case with ordinary non-periodic $\theta$ on the ordinary (non compactified) Einstein
Universe $\mathbb{R}\times SU(2\mathbb{C})$. We start with
\begin{equation}\label{Omega'phiEUnonCompact}
\Omega'\phi(x,\boldsymbol{w}) = \phi(t,\boldsymbol{w}) 
- \sum\limits_{a=0}^{\omega}\partial_{t}^{a}\phi(0,e) \,\, {\textstyle\frac{t^a}{a!}} \, w(t), 
\end{equation}
with time derivations only $X_0= \partial_t$, and one ``subtraction point'' $(0,e)$ -- 
the unit of $\mathbb{R}\times SU(2\mathbb{C})$. The method of subtraction of the first Taylor
expansion terms at the ``normalization point'', given above, leading from the formula
\[
\langle  \widetilde{\textrm{ret} \, \kappa_q}, \widetilde{\phi} \rangle
= \big\langle \kappa_q, \theta \Omega'\phi\big\rangle = 
\big\langle \widetilde{\kappa_q}, \widetilde{\theta} \ast \widetilde{\Omega'\phi}\big\rangle
\]
to the analogue final formula, through the analogues of the intermediate formulas  
(\ref{FT(retkappaq)1EU})-(\ref{FT(retkappaq)3EU}), is much simpler in this non compact case, and even more,
is closely related to the original method of Scharf reported in Subsection \ref{WickForChronological}.  
Here we have the additive noncompact Abelian group $\mathbb{R}$ instead of the compact Abelian group
$\widetilde{\mathbb{S}^1}$.  The discrete ``energy quantum number'' variable $n\in \mathbb{Z}$ in the characters 
$\widehat{n}(t) = e^{int/2}$ of the group $\widetilde{\mathbb{S}^1}$, 
is replaced with the continuous energy variable
$p_0 \in \mathbb{R}$ in the characters $\widehat{p_0}(t)= e^{ip_0t}$ of the additive group $\mathbb{R}$, and the purely discrete
characters  $\widehat{n}\cdot\widehat{l}$ of the compact group $\widetilde{\mathbb{S}^1}\times SU(2,\mathbb{C})$ 
are replaced with  $\widehat{p_0}\cdot \widehat{l}$,
$\widehat{n'''}\cdot\widehat{l}$, $\widehat{n'''}\cdot\widehat{l'''=0}$ are replaced with  the continousy-discrete characters $\widehat{p'''_{0}}\cdot\widehat{l}$
$\widehat{p'''_0}\cdot\widehat{l'''=0}$ of the non compact group $\mathbb{R}\times SU(2,\mathbb{C})$. Correspondingly
the discrete integrals (sums) 
\[
\sum\limits_{n,n'\in \mathbb{Z}} \ldots, \,\,\,\,\,\,\,\,\,
\sum\limits_{n''\in \mathbb{Z}} \ldots,
\]
are replaced with continuous integrals
\[
\int\limits_{\mathbb{R}\times\mathbb{R}} dp_0 \, dp'_{0} \ldots, \,\,\,\,\,
\int\limits_{\mathbb{R}} dp''_{o} \ldots, \,\,\,\, 
\textrm{or} \,\,\,\,\,\,\,\,\,\,\,\,\,\,\,
{\textstyle\frac{1}{\sqrt{2\pi}}}{\textstyle\frac{1}{\sqrt{2\pi}}}\int\limits_{\mathbb{R}\times\mathbb{R}} dp_0 \, dp'_{0} \ldots, \,\,\,\,\,\
{\textstyle\frac{1}{\sqrt{2\pi}}}\int\limits_{\mathbb{R}} dp''_{o} \ldots. 
\]
This is of considerable importance. The first consequence is that we can use $x_0(t)=t$ as the analytic function dual to $X_0=\partial_t$
up to infinite order at $t=0$, which works for all natural $\omega$. In the compact case we have no such analytic function $x_0$
of $t$ at our disposal (smooth would be impractical). The simple function $x_0(t) = t$ cannot be used in the compact case because this function 
has a jump, if regarded as a function on the group $\widetilde{\mathbb{S}^1}$, and multiplication
by this function, or convolution with $\widetilde{x_0}$, would be as singular as multiplication by $\theta$, or convolution with $\widetilde{\theta}$
itself. In particular multiplcation by such discontinuous $x_0$ does not act within the test space and, by duallity, 
within the distribution space. Construction of the remaining analytic functions, $x_1, x_2,x_3$, which togetehr with 
$x_0$ are dual to $X_0, \ldots, X_3$ at the unit $(0, e)$ up to order $\omega$, remain the same.  
In the non compact case, with the ordinary, non-periodic step $\theta$-function, the final formula is, by this, slightly simpler.

We repeat the same computations, but with the following replacements.
\begin{gather*}
\Omega'\phi(x,\boldsymbol{w}) = \phi(t,\boldsymbol{w}) 
- \sum\limits_{a=0}^{\omega}\partial_{t}^{a}\phi(0,e) \,\, {\textstyle\frac{t^a}{a!}} \, w(t) 
\\
x_0(t) = t, \,\,\,\,\,X_0 = \partial_t, \,\,\,\,  \widetilde{x_0} \ast \widetilde{\phi} = i \partial_{{}_{p_0}} \widetilde{\phi}, \,\,\,\,
\\
\widetilde{X_{0} \delta}(\widehat{p_0}\cdot \widehat{l}) = \mathbb{A}_{{}_{0}}(\widehat{p_0}\cdot \widehat{l}) = -ip_0 \, \boldsymbol{1}.
\\
\widetilde{\theta}_{{}_{ji}}(\widehat{p_0}\cdot\widehat{l})
= {\textstyle\frac{1}{\sqrt{2\pi}}}{\textstyle\frac{i}{p_0+ i0}} \delta_{{}_{l \,\, 0}}\delta_{{}_{j \,\, 0}}\delta_{{}_{i \,\, 0}}
\\
\big(X_0\big)^{a} \phi (0,e) = (-1)^a \big\langle \big(X_0\big)^{a} \delta, \phi \big\rangle
= (-1)^a \Big\langle \widetilde{\big(X_0\big)^{a} \delta}, \,\, \widetilde{\phi} \Big\rangle =
(-1)^a \big\langle \big(-ip_0)^{a}, \widetilde{\phi} \big\rangle,
\end{gather*}

Because $\mathbb{R}$ or $\mathbb{R}\times SU(2\mathbb{C})$, is noncompact,
existence of a ``normalization point'' $\widehat{p'''_{0}}\cdot\widehat{l}$, for each $\widehat{l}\in \widehat{SU(2,\mathbb{C})}$,
and with common energy component $p'''_{0}$ for all characters $\widehat{l}$ is now slightly less trivial, and in fact should
be proved. Let us recall that the ``normalization point'' $\widehat{p'''_{0}}\cdot\widehat{l}$ is a point such that 
$\widetilde{\textrm{ret} \, \kappa_q}(\widehat{p_{0}}\cdot\widehat{l})$,
regarded as a distribution in the variable $p_0$, is a regular distribution in a neighborhood of $p'''_{0}$ 
and possess derivatives in $p_0$ at $p'''_{0}$, in the ordinary
function sense, up to order $\omega$ at least. Existence of $\widehat{p'''_{0}}\cdot\widehat{l}$ is non trivial in this non compact case. 
On the Minkowski space-time we have computed
explicitly $\widetilde{\kappa_q}$, where $\widetilde{\kappa_q}$ are regular everywhere and analytic except the finite jump at the characteristic
Lorentz invariant submanifold, so we can prove existence of the ``normalization point'' by explicit inspection. 
But in general the Fourier transform of distributions $\kappa_q$ with cone-shaped supports, is regular and a boundary value of an analytic function,
compare Thm. IX.16, \cite{Reed_SimonII}. We have not computed explicitly $\widetilde{\kappa_q}$ for the ordinary (non compact) 
Einstein Universe (although it should be emphasized here that it can be done effectively and explicitly on the basis of representation theory), 
but the analogue property is valid for the Fourier transform $\widetilde{\textrm{ret} \, \kappa_q}$ 
of a distribution $\textrm{ret} \, \kappa_q$ on $\mathbb{R}\times SU(2\mathbb{C})$, because $\textrm{ret} \, \kappa_q$ is 
supported on the half-space 
\[
V^+ = \{(t,\boldsymbol{w})\in \mathbb{R}\times SU(2,\mathbb{C}): \,\,  t \geq 0\}.
\]   
Indded, it is sufficient to apply  Thm. IX.16, \cite{Reed_SimonII}, to distributions on $\mathbb{R}$, supported at the closed half real line. 
Therefore, our distribution $\widetilde{\textrm{ret} \, \kappa_q}(\widehat{p_{0}}\cdot\widehat{l})$, whenever exists,
has to be regular in $\mathbb{R}+i\mathbb{R}_{+}$ 
for each $\widehat{l}\in \widehat{SU(2,\mathbb{C})}$, when regarded as a distribution in the real variable $p_0$, and moreover
is equal to a boundary value of an analytic function. Anyway, existence of the ``normalization points'' 
seems to be justified, similarly as on the Minkowski space-time. We proceed as on the Minkowski space-time. We assume 
existence of the ``normalization points'' and get the final formula. At the very end we check if the final formula 
indeed gives a retarded part of $\kappa_q$.  

Having given the ``normalization points'' $\widehat{p'''_{0}}\cdot\widehat{l}$, $\widehat{l}\in \widehat{SU(2\mathbb{C})}$,
at our disposal, we apply, after Scharf, the subtraction 
\begin{multline}\label{TaylorSubtracionEUnonCompact}
\big[\widetilde{\textrm{ret} \, \kappa_q}\big]'''(\widehat{p_0}\cdot\widehat{l}) \,
\overset{\textrm{df}}{=} \,
\widetilde{\textrm{ret} \, \kappa_q}(\widehat{p_0}\cdot\widehat{l}) - \sum\limits_{b=0}^{\omega} \,\,
{\textstyle\frac{\big(\mathbb{A}_{{}_{0}}(\widehat{p_0} \cdot \widehat{l})-\mathbb{A}_{{}_{0}}(\widehat{p'''_{0}} \cdot \widehat{l})\big)^b}{b!}}
\,\,\, \Big(\widetilde{(ip_0)^b} \,\, \ast  \,\, \widetilde{\textrm{ret} \, \kappa_q}\Big)(\widehat{p'''_{0}}\cdot\widehat{l})
\\
=
\widetilde{\textrm{ret} \, \kappa_q}(\widehat{p_0}\cdot\widehat{l}) - \sum\limits_{b=0}^{\omega} \,\,
{\textstyle\frac{\big(p_{0} -p'''_{0}\big)^b}{b!}}
\,\,\, \Big(\partial_{{}_{p_0}}^{b}\widetilde{\textrm{ret} \, \kappa_q}\Big)(\widehat{p'''_{0}}\cdot\widehat{l}).
\end{multline}
obtaining Fourier transform of another possible retarded part of $\kappa_q$.
This is so because in this subtraction of the first Taylor expansion terms we have subtracted distribution 
(here with $\delta$ in single time variable $t$ and with $p'''_{0}$ understood as a constant)
 equal to the Fourier transform 
\begin{equation}\label{FT(EpsteinGlaserReminderEUnonCompact)}
\sum\limits_{b=0}^{\omega} \widetilde{C_{b,p'''_{0}}}(\widehat{l}) \,\,  \widetilde{\big(X_0\big)^b\delta}(\widehat{p_0}),
\,\,\,\, \widetilde{C_{b,p'''_{0}}}(\widehat{l}) = \sum\limits_{a=b}^{\omega} {\textstyle\frac{i^{a+b}(p'''_{0})^{a-b}}{b!(a-b)!}}
\,\,
i^a\Big(\partial_{{}_{p_0}}^{a}\widetilde{\textrm{ret} \, \kappa_q}\Big)(\widehat{p'''_{0}}\cdot\widehat{l}),
\end{equation}
of a distribution of the form
\begin{equation}\label{EpsteinGlaserReminderEUnonCompact}
\sum\limits_{b=0}^{\omega} C_{b,p'''_{0}}(\boldsymbol{w}) \,\, \big(X_0\big)^b\delta(t),
\end{equation}
which is concentrated at the Cauchy surface $t=0$. But the inverse Fourier transform $\big[\textrm{ret} \, \kappa_q\big]'''$
of $\big[\widetilde{\textrm{ret} \, \kappa_q}\big]'''$ does not coincide with the ``natural retarded part'' on the whole image of $\Omega'$, 
because (\ref{EpsteinGlaserReminderEUnonCompact}) can in general
be nonzero on $\textrm{Im} \, \Omega'$.  Namely, the difference (\ref{EpsteinGlaserReminderEUnonCompact}) between 
the natural $\textrm{ret} \, \kappa_q$ and the inverse Fourier transform of $\big[\widetilde{\textrm{ret} \, \kappa_q}\big]'''$
may, in general, be nonzero on $\textrm{Im} \, \Omega'$. 
We see that the inverse Fourier transform of $\big[\widetilde{\textrm{ret} \, \kappa_q}\big]'''$
can affect the natural formula of multiplication by $\theta$ on $\textrm{Im} \, \Omega'$. We can be sure only, on general grounds,
that 
\[
\big[\textrm{ret} \, \kappa_q\big]'''=\theta \kappa_q,
\]
on the test functions $\phi$ whose time derivatives vanish up to order $\omega$ at all points of the Cauchy surface $t=0$.

Applying the subtraction (\ref{TaylorSubtracionEUnonCompact}) we arrive at
\begin{multline*}
\big[\widetilde{\textrm{ret} \, \kappa_q}\big]'''(\widehat{p_0}\cdot\widehat{l}) 
={\textstyle\frac{1}{2\pi}}\int dp''_{0} {\textstyle\frac{i}{p''_{0}+i0}} \Bigg\{ \widetilde{\kappa_q}(\widehat{p_0-p''_{0}} \cdot \widehat{l})
\\
-
\sum\limits_{b=0}^{\omega} \,\,
{\textstyle\frac{\big(\mathbb{A}_{{}_{0}}(\widehat{p_0} \cdot \widehat{l})-\mathbb{A}_{{}_{0}}(\widehat{p'''_{0}} \cdot \widehat{l})\big)^b}{b!}}
\Big(\widetilde{t^b} \,  \ast \, \widetilde{\kappa_q}\Big)(\widehat{p'''_{0}-p''_{0}}\cdot \widehat{l})
\Bigg\}
\end{multline*}
\begin{multline}\label{FT(retkappaq)3NonCompactEU}
={\textstyle\frac{1}{2\pi}}\int dp''_{0} {\textstyle\frac{i}{p''_{0}+i0}} \Bigg\{ \widetilde{\kappa_q}(\widehat{p_0-p''_{0}} \cdot \widehat{l})
\\
-
\sum\limits_{b=0}^{\omega} \,\,
{\textstyle\frac{\big(p_0 -p'''_{0}\big)^b}{b!}}
\Big(\partial_{{}_{p_0}}^{b}\widetilde{\kappa_q}\Big)(\widehat{p'''_{0}-p''_{0}}\cdot \widehat{l})
\Bigg\}
\end{multline}
analogous to (\ref{FT(retkappaq)3EU}) or, transferring the differentiation in the variable $p'''_{0}-p''_{0}$ into the differetiation
in the variable $p''_{0}$,
\begin{multline*}
\big[\widetilde{\textrm{ret} \, \kappa_q}\big]'''(\widehat{p_0}\cdot\widehat{l}) 
={\textstyle\frac{1}{2\pi}}\int dp''_{0} {\textstyle\frac{i}{p''_{0}+i0}} \Bigg\{ \widetilde{\kappa_q}(\widehat{p_0-p''_{0}} \cdot \widehat{l})
\\
-
\sum\limits_{b=0}^{\omega} \,\,
{\textstyle\frac{\big(p_0 -p'''_{0}\big)^b}{b!}}
(-1)^b\partial_{{}_{p''_{0}}}^{b}\big[\widetilde{\kappa_q}(\widehat{p'''_{0}-p''_{0}}\cdot \widehat{l})\big]
\Bigg\}.
\end{multline*} 
We apply  the integration by parts in the subtraction terms and transfer the derivativation operator $\partial_{{}_{p''_{0}}}^{b}$
on the factor $i/(p''_{0}+i0)$. Next, we introduce the new integration variable $p_{10} = p_0-p''_{0}$ in the first term
and the new integration variable $p_{10}=p'''_{0}-p''_{0}$ in the subtraction term, and arrive at
\[
\big[\widetilde{\textrm{ret} \, \kappa_q}\big]'''(\widehat{p_0}\cdot\widehat{l}) 
={\textstyle\frac{i}{2\pi}}\int dp_{10} \,
\Bigg[
{\textstyle\frac{1}{p_0-p_{10}+i0}} 
-
\sum\limits_{b=0}^{\omega} \,\,
{\textstyle\frac{(-1)^b\big(p_0 -p'''_{0}\big)^b}{b!}}
{\textstyle\frac{1}{(p'''_{0}-p_{10}+i0)^{b+1}}} 
\Bigg] 
\widetilde{\kappa_q}(\widehat{p_{10}}\cdot\widehat{l}). 
\]
The expression in the square bracket in the integrand is equal
\[
{\textstyle\frac{1}{p_0-p_{10}+i0}} 
-
\sum\limits_{b=0}^{\omega} \,\,
{\textstyle\frac{(-1)^b\big(p_0 -p'''_{0}\big)^b}{b!}}
{\textstyle\frac{1}{(p'''_{0}-p_{10}+i0)^{b+1}}} 
\,\,
=
\,\,
\left( -
{\textstyle\frac{p_0-p'''_{0}}{p'''_{0}-p_{10}+i0}}
\right)^{\omega+1}
{\textstyle\frac{1}{p_0-p_{10}+i0}}. 
\]
Substituting this expression into the integral we get the final formula
\begin{gather}\label{DisparsionFormulaRetkappaqEUnonCompactIntInp10}
\big[\widetilde{\textrm{ret} \, \kappa_q}\big]'''(\widehat{p_0}\cdot \widehat{l}) = {\textstyle\frac{i}{2\pi}}
\int\limits_{-\infty}^{+\infty}
{\textstyle\frac{\left(p'''_{0}-p_0\right)^{\omega +1} \widetilde{\kappa_q}(\widehat{p_{10}}\cdot\widehat{l}) \,\, dp_{10}}{\left(p'''_{0} -p_{10}
+ i\, 0 \right)^{\omega+1}\left(p_0-p_{10} +i \, 0 \right)}}.
\end{gather}
Introducing the new integration variable $t = p_{10}/p_{0}$,
(which we can apply for each $p_0\neq 0$) we obtain the formula in the form analogue to 
(\ref{dispesionFm=0}), Subsection \ref{OperationsOnXi}, and (\ref{FT(retCausalkappaq)}),
(\ref{InvariantDispersionFormulaForRetarded}) of
Subsection \ref{WickForChronological}:
\begin{gather}\label{DisparsionFormulaRetkappaqEUnonCompact}
\big[\widetilde{\textrm{ret} \, \kappa_q}\big]'''(\widehat{p_0}\cdot \widehat{l}) = {\textstyle\frac{i}{2\pi}}
\left(p'''_{0}-p_0\right)^{\omega +1} \int\limits_{-\infty}^{+\infty}
{\textstyle\frac{\widetilde{\kappa_q}(\widehat{tp_0}\cdot\widehat{l}) \,\, dt}{\left(p'''_{0} -p_0t
+ i\, 0 \right)^{\omega+1}\left(1+t +i \, 0 \, \textrm{sgn}(p_0)\right)}},
\end{gather}
which, in general, makes sense only for $p_0 \neq 0$, 
and, in this case it can also be written as
\begin{gather}\label{DisparsionFormulaRetkappaqEUnonCompact'}
\big[\widetilde{\textrm{ret} \, \kappa_q}\big]'''(\widehat{p_0}\cdot \widehat{l}) = {\textstyle\frac{i}{2\pi}}
\left({\textstyle\frac{p'''_{0}-p_0}{p_0}}\right)^{\omega +1} \int\limits_{-\infty}^{+\infty}
{\textstyle\frac{\widetilde{\kappa_q}(\widehat{tp_0}\cdot\widehat{l}) \,\, dt}{\left(t-{\textstyle\frac{p'''_{0}}{p_0}} - i \textrm{sgn}(p_0)
\, 0 \right)^{\omega+1}\left(1+t +i \textrm{sgn}(p_0)
\, 0 \right)}}.
\end{gather}
In fact (\ref{DisparsionFormulaRetkappaqEUnonCompact}) resembles most the formula (\ref{dispesionFm=0}) for the product $d$
of massless pairings on the Minkowski space-time, in case of $\omega=2$, of course, because (\ref{dispesionFm=0}) is valid for $\omega=2$,
and for the product $d$ of massless kernels of degree $2$. This could have been expected, as on the EU there are the massless fields
which have infinite orbits, and there are only the products of $q$ massless pairings ($q$-contractions, $q>1$, of massless kernels)
which cannot be splitted on EU into the retarded and advanced parts by the simple multiplication by the theta function,
without the subtraction terms, and the formula like  (\ref{DisparsionFormulaRetkappaqEUnonCompact})  with non-trivial
subtraction terms has to be used.

Let us investigate now convergence of (\ref{DisparsionFormulaRetkappaqEUnonCompactIntInp10}).
Let the order of $\widetilde{\kappa_q}$ be equal $\omega'$. This means that for large $(p_{10},l)$,  
$\widetilde{\kappa_q}$ behaves, as a function of $\widehat{p_{10}}\cdot\widehat{l}$, as a power function 
$\big(\sqrt{(p_{10})^2+l(l+1)}\big)^{\omega'}$.
In particular, for each fixed $l$, $\widetilde{\kappa_q}(\widehat{p_{10}}\cdot\widehat{l})$ behaves as 
the power function $p_{10}^{\omega'}$ for $p_{10} \rightarrow \pm \infty$. 
We see from the formula (\ref{DisparsionFormulaRetkappaqEUnonCompactIntInp10}) that for any value of the order $\omega'$,
(\ref{DisparsionFormulaRetkappaqEUnonCompactIntInp10}) will be convergent if we put $\omega=\omega'$, in case $\omega'$ is natural,
or if we put $\omega$ equal to the least natural, which is greater than $\omega'$. 

But $\big[\textrm{ret} \, \kappa_q\big]'''$ and $\textrm{ret} \, \kappa_q$ differ by the distribution
(\ref{EpsteinGlaserReminderEUnonCompact}), and, as we have already seen, 
the formula (\ref{DisparsionFormulaRetkappaqEUnonCompactIntInp10})
does not have to be equal to the Fourier transform of a distribution 
which coincides with multiplication by $\theta$ on the whole $\textrm{Im} \, \Omega'$:
\[
\theta\kappa_q \circ \Omega' \,\, = \,\, 
\big[\textrm{ret} \, \kappa_q\big]''',
\,\,\,\,\,\,\, \textrm{not on the whole} \,\,\, \textrm{Im} \, \Omega'
\]
because the added term (\ref{EpsteinGlaserReminderEUnonCompact}) may, in principle, be nonzero 
on $\textrm{Im} \, \Omega'$. Using the same method as above on the compactified EU with periodic $\theta$
and a single ``subtraction point'' $\widehat{p'''_{0}}\cdot\widehat{l'''}= \widehat{p'''_{0}}\cdot\widehat{0}$
independent of $\widehat{l}$,
we can obtain a formula for a different $\big[\textrm{ret} \, \kappa_q\big]'''$, 
not equal to (\ref{DisparsionFormulaRetkappaqEUnonCompact'}),
and if we start with the operator
$\Omega'$ equal (\ref{Omega'phiEUseveralMultiindex}), which respects
\[
\theta\kappa_q \circ \Omega' \,\, = \,\, 
\big[\textrm{ret} \, \kappa_q\big]''',
\,\,\,\,\,\,\, \textrm{on the whole} \,\,\, \textrm{Im} \, \Omega'
\]
and thus with the difference $\textrm{ret} \, \kappa_q \,\, - \,\, 
\big[\textrm{ret} \, \kappa_q\big]'''$ equal to a distribution of order $\leq \omega$ and concentrated
at the single point $(0,e)$ -- the unit of $G$, in accordance with axiom (V). Therefore, such  $\big[\textrm{ret} \, \kappa_q\big]'''$
is natural, and we simply denote it likewise by $\textrm{ret} \, \kappa_q$. 

By the axiom (V), the natural $\textrm{ret} \, \kappa_q = \theta\kappa_q \circ \Omega'$, with the idempotent
operator $\Omega'$ equal (\ref{Omega'phiEUseveralMultiindex}), 
is determined up to a distribution of degree $\leq \omega$:
\[
\Delta(x) = P(X_0, \ldots, X_3) \delta(x_0)\dots \delta(x_3), \,\,\, (0,e) = (x_0=0, \ldots, x_3=0)
\]
concentrated at the point $(0,e)$, with the polynomial 
\[
P(X_0, \ldots, X_3) = \sum\limits_{|\alpha| \leq \omega} a_\alpha X^\alpha
\]
of degree $\textrm{max} |\alpha| \leq \omega$, with smooth coefficients, when expressed in a local coordinate system $(x_0, \ldots,  x_3)$
which is dual to $X_0, \ldots, X_3$ at $(0,e)$ up to order $\omega$ at least. 
Here $\omega$ is the singularity degree
of $\kappa_q$, and $a_\alpha$ are smooth functions.
The assumption of invariance allow us to use
\[
\Delta(x) = \sum\limits_{|\alpha| \leq \omega} c_\alpha \nabla^\alpha \delta(x),
\]     
where in $ \nabla^\alpha = \nabla_0^{\alpha_0}\nabla_1^{\alpha_1}\nabla_2^{\alpha_2}\nabla_3^{\alpha_3}$,
we have the covariant derivatives $\nabla_i$, and smooth functions $c_\alpha$ such that the linear operator in $\Delta$ is a covariant operator. 

We have analogous formulas for the retarded and advanced part of a distribution of several space-time variables on EU, provided it has finite, 
in case it is non-negative, singularity order at zero (or rather at the unit of the group $G$ identified with the EU space-time, or its Cartesian product with itself, 
identified with a product of the group $G$ with itself),  which is $G$-invariant (a generalization of translation invariance to EU), 
similarly as on the Minkowski space-time.

\subsection{Wick Theorem for the natural chronological product 
\\ on the Einstein Universe}\label{WickForChronologicalEU}

The main motivation of this Subsection is to give a simple recurrence formula for the scattering operator,
or for the chronological product -- \emph{The Wick Theorem for the Natural Chronological Product}. Correspondingly
to the two cases of the previous Subsections, namely Thm. \ref{Sin(SE)xn->((E)->(E))EU}
and \ref{GeneralSin(SE)xn->((E)->(E))EU}, we will have two cases of the Wick Theorem for the Chronological Product. 
In fact we give a simple systematic method for practical computation of the scattering operator, 
whose existence and general properties have been established
in Thm. \ref{Sin(SE)xn->((E)->(E))EU}
and \ref{GeneralSin(SE)xn->((E)->(E))EU}, Subsection \ref{WickForProductOnEU}, correspondingly to the two kinds of interaction. 
Exactly as in Subsection 
\ref{WickForProduct}  the heuristic definition of the chronological product using the step
theta function is the main idea which helps to organize the calculation.

Proceeding as in Subsection \ref{WickForChronological} we give a rigorous meaning to the 
chronological product formula
\begin{multline}\label{thetaS_nEU}
S_n(x_1, \ldots, x_n) = i^n \, T\big(\mathcal{L}(x_1) \ldots \mathcal{L}(x_n) \big) \\ = i^{n}
\sum\limits_{\pi} \theta(t_{\pi(1)} -t_{\pi(2)}) \theta(t_{\pi(2)} -t_{\pi(3)}) \ldots \theta(t_{\pi(n-1)} -t_{\pi(n)})  \,
\mathcal{L}(x_{\pi(1)}) \ldots \mathcal{L}(x_{\pi(n)}), \\ 
\,\,\,\,\,\,\,\,\,\,\,\,\,\,\,\,\,\,\,\,\,\,\,\,\,\,\,\,\,\,\,\,
x_{\pi(k)} = (t_{\pi(k)}, \boldsymbol{w}_{\pi(k)}), \,\,\, \pi \in \textrm{Permutations of} \,\{1, \ldots, n\}.
\end{multline}
on the Einstein Universe. 
We consider as the Lagrange density interaction $\mathcal{L}$ a general Wick polynomial in free fields. Each monomial
may contain any number of massless (or infinite orbit) free fields. But in accordance to the 
results of Subsections \ref{WhiteNoiseFreeFieldsonEU} and \ref{WickForProductOnEU}, there is a fundamental
difference between the following  two situations. 1)  Each monomial of $\mathcal{L}$ contains at most
one massless (or infinite orbit) free field, and the remaining fields in the monomial being massive (or finite orbit fields).
2) Some monomials of $\mathcal{L}$ contain more than one massless (or infinite orbit) free field.

In the first case, which we have called 1), the multiplication of the Wick product operator, understood as a finite sum of integral kernel operators,
 by the step theta function $\theta$ is well-behaved  and transforms the initial operator into a finite sum of integral
kernel operators of the same regular class which, when evaluated at space-time test function, transform continuously
the test Hida space into itself  and become ordinary operators on the Fock space. In particular the formula (\ref{thetaS_nEU}), with the operator 
$\mathcal{L}$ understood as the finite sum of integral kernel operators, becomes mathematically meaningful and represents
a well defined generalized operator, which after being evaluated at space-time test function, represents
an ordinary operator transforming continuously the Hida test space $(\boldsymbol{E})$ into itself. Recall that 
$(\boldsymbol{E})$ is the Hida test space in the total Fock space (tensor product of all Fock spaces of all free fields of the theory)
and equal to the projective tensor product of the Hida spaces of all free fields, compare Subsection \ref{psiBerezin-Hida}.

In the second case, which we have named 2), situation is more subtle. The chronological product is more singular, and 
we have to proceed exactly as in Subsection \ref{WickForChronological} in  order to  give a rigorous sense
to the expression (\ref{thetaS_nEU}) which in this case, when taken literally, is meaningless. In this Subsection we give the rigorous
sense to the chronological product (\ref{thetaS_nEU}) in both cases: 1) and 2).

\vspace*{1cm}

\begin{center}
\small{CASE 1)}
\end{center}

Let us start with the first case 1), in which each monomial of the Lagrange density $\mathcal{L}$ of interaction 
contains at most one massless free field or infinite orbit free field, and all remaining free fields in each monomial
are massive or finite orbit free fields.

Let for example
\[
\mathcal{L}(x) = \boldsymbol{{:}} \boldsymbol{\psi}(x)^{+}\gamma_{0} \gamma^\mu \boldsymbol{\psi}(x) A_\mu(x) \boldsymbol{{:}},
\]
as in spinor QED, but our analysis is general and we can replace $\mathcal{L}$ with any Wick polynomial of free fields
with each of the Wick monomials possibly containing at most one massless (or infinite orbit) free field factors.

From Lemma \ref{kappaBarDotOtimeskappaOnEU} of Subsection \ref{WhiteNoiseFreeFieldsonEU} and 
Theorem \ref{obataJFA.Thm.3.13} of Subsection \ref{psiBerezin-Hida} it follows that $\mathcal{L}$ is a finite sum 
\[
\mathcal{L} =
\sum\limits_{\substack{\ell,m \\ \ell'+m'\leq N}} 
\Xi_{\ell',m'}\big(\kappa'_{\ell',m'}\big)
\in \mathscr{L}( (\boldsymbol{E}) \otimes \mathscr{E} , (\boldsymbol{E}) ).
\]
of well defined integral kernel operators. In case of QED $\ell'+m'=3$ in the range of the above sum, but in general case $\ell'+m'\leq N$
in the summation range, where $N$ is equal to the degree of the Wick polynomial $\mathcal{L}$.

Because each Wick monomial of $\mathcal{L}$ contains at most one massless (or infinte orbit) free field,
then by Lemma \ref{theta.kappaBarDotOtimeskappaWithOneMasslessOnEU}
\begin{multline*}
\theta \kappa'_{\ell',m'} 
\,\,\, \in \,\,\, \mathscr{L}\big(E_{i_1}^{*} \otimes \ldots  \otimes E_{i_{{\ell'}}}^{*} \otimes 
E_{i_{\ell'+1}} \otimes \ldots E_{i_{\ell'+m'}}, \mathscr{E}^*\big) 
\\
\cong \mathscr{L}\big(\mathscr{E}, E_{i_1} \otimes \ldots  \otimes E_{i_{{\ell'}}} \otimes E_{i_{\ell'+1}}^{*} \otimes \ldots \otimes E_{i_{\ell'+m'}}^{*}\big),
\end{multline*}
where
\[
E_{i_1}, \ldots, E_{i_{\ell'}},
\]
correspond to the first $\ell$ momentum variables\footnote{Corresponding to the number $\ell'$ of creation operators in the integral kernel operators $\Xi_{\ell',m'}(\theta\kappa'_{\ell',m'})$, $\Xi(\kappa'_{\ell',m'})$ corresponding to the kernels $\theta\kappa'_{\ell',m'}$, $\kappa'_{\ell',m'}$.} 
of the kernels $\theta\kappa'_{\ell',m'}$, $\kappa'_{\ell',m'}$
and 
\[
E_{i_{\ell'+1}}, \ldots, E_{i_{\ell'+m'}},
\]
correspond to the last momentum variables\footnote{Corresponding to the number $m'$ of annihilation operators in the integral 
kernel operators $\Xi_{\ell',m'}(\theta\kappa'_{\ell',m'})$,
$\Xi(\kappa'_{\ell',m'})$ corresponding to the kernels $\theta\kappa'_{\ell',m'}$, $\kappa'_{\ell',m'}$.}, 
of the kernels $\theta\kappa'_{\ell',m'}$, $\kappa'_{\ell',m'}$.
Thus the kernel $\theta\kappa'_{\ell',m'}$ has the same extendibility property as
the kernel $\kappa'_{\ell',m'}$, \emph{i.e.} we have in the domain
of these kernels $\theta\kappa'_{\ell',m'}$,$\kappa'_{\ell',m'}$
exactly $\ell'$ dual spaces $E^*, \ldots$, corresponding to the first $\ell'$ momentum variables of the kernels 
$\theta\kappa'_{\ell',m'}$, $\kappa'_{\epsilon \,\, \ell',m'}$ and $m'$ initial spaces  $E, \ldots$,
corresponding to the last $m'$ momentum variables of the kernels $\theta\kappa'_{\ell',m'}$, 
$\kappa'_{\epsilon \,\, \ell',m'}$.
From this it follows, by Theorem \ref{obataJFA.Thm.3.13}, Subsection \ref{psiBerezin-Hida}, that
the operator
\[
\theta\mathcal{L} =
\sum\limits_{\substack{\ell,m \\ \ell'+m'=3}} 
\Xi_{\ell',m'}\big(\theta\kappa'_{\ell',m'}\big)
\in \mathscr{L}( (\boldsymbol{E}) \otimes \mathscr{E} , (\boldsymbol{E}) )
\]
belongs to the same class 
\[
\mathscr{L}( (\boldsymbol{E}) \otimes \mathscr{E} , (\boldsymbol{E}) )
\]
as the initial operator $\mathcal{L}$ does, which was impossible on the Minkowski space-time,
even if we had at most one massless field in each monomial in $\mathcal{L}$.

Therefore we can (compare Subsections \ref{WickForProduct}, \ref{WickForProductOnEU}) form the following continuous map,
and the corresponding generalized operator
\begin{multline*}
\mathscr{E}^{\otimes \, n}\ni \phi_1 \otimes \ldots \phi_n  \longmapsto 
\Xi(\phi_1 \otimes \ldots \otimes \phi_n) \overset{\textrm{df}}{=} \\ \overset{\textrm{df}}{=} 
\sum\limits_{\pi}
\theta_{{}_{ \pi(2)}} \mathcal{L}(\phi_{\pi(1)}) \circ 
\theta_{{}_{\pi(3)}} \mathcal{L}(\phi_{\pi(2)}) \circ \ldots \circ
\theta_{{}_{\pi(n)}} \mathcal{L}(\phi_{\pi(n-1)}) \circ
\mathcal{L}(\phi_{\pi(n)}) 
\\
\in \mathscr{L}\big( (\boldsymbol{E}), (\boldsymbol{E})\big),
\end{multline*}
defined by ordinary compositon $\circ$ of operators transforming continuously the Hida space into itself.
This composition exists for each permutation $\pi$, with the kernels of the composed operators
sufficiently regular, in case each monomial of $\mathcal{L}$ contains at most one massless (or infinite orbit) free field, 
by Theorem \ref{obataJFA.Thm.3.13} of Subsection \ref{psiBerezin-Hida}. 
By this Theorem the composed operators belong to 
\[
\mathscr{L}\big(\mathscr{E}, \, \mathscr{L} ((\boldsymbol{E}), (\boldsymbol{E})) \big)
\]
and, when evaluated at $\phi_{\pi(i)} \in \mathscr{E}$, can be composed and thus in general the above composition
is well defined and moreover defines
\[
\Xi
\in \mathscr{L}( (\boldsymbol{E}) \otimes \mathscr{E}^{\otimes \, n} , (\boldsymbol{E}) )
\cong  \mathscr{L}\big(\mathscr{E}^{\otimes \, n}, \, \mathscr{L}( (\boldsymbol{E}), (\boldsymbol{E}) ) \, \big)
\]
(compare Subsections \ref{WickForProduct}, \ref{WickForProductOnEU}).

We can apply the Wick Theorem \ref{WickEU} of Subsection \ref{WickForProductOnEU} to this product operator $\Xi$, which immediately gives the following finite
Fock expansion
\begin{multline*}
\Xi =
i^n \sum \limits_{\pi}\sum_{\substack{\kappa^{\pi(i)}_{\ell_{\pi(i)},m_{\pi(i)}}}} \,\,\,
\sum \limits_{0\leq q_i \leq \textrm{min} \, \{m_{\pi(i), \ell_{\pi(i+1)} \}}} \,\,\, (-1)^{c(q_1) + \ldots c(q_{n-1})} 
\,\, \times 
\\
\times \,\,
\Xi_{{}_{\ell_{\pi(1)}+ \ldots +\ell_{\pi(n)} -q_1 \ldots - q_{n-1}, \,\, m_{\pi(1)}+ \ldots +m_{\pi(n)} -q_1 \ldots - q_{n-1}}},
\end{multline*}
where
\begin{multline*}
\Xi_{{}_{\ell_{\pi(1)}+ \ldots +\ell_{\pi(n)} -q_1 \ldots - q_{n-1}, \,\, m_{\pi(1)}+ \ldots +m_{\pi(n)} -q_1 \ldots - q_{n-1}}}
\\
=
\Xi\big( \theta_{{}_{\pi(2)}} \kappa^{\pi(1)}_{\ell_{\pi(1)},m_{\pi(1)}}\otimes_{{}_{q_1}} \, 
\theta_{{}_{\pi(3)}} \kappa^{\pi(2)}_{\ell_{\pi(2)},m_{\pi(2)}} \,\,
\otimes_{{}_{q_2}}
\ldots  \,\,\, \times 
\\
\times \,\, \ldots
\otimes_{{}_{q_{n-2}}} \, 
\theta_{{}_{\pi(n)}} \kappa^{\pi(n-1)}_{\ell_{\pi(n-1)},m_{\pi(n-1)}}\otimes_{{}_{q_{n-1}}} \,  
\kappa^{\pi(n)}_{\ell_{\pi(n)},m_{\pi(n)}}\big),
\end{multline*}
with $\kappa^{\pi(i)}_{\ell'_{i}, m'_{i}}$ ranging independently over all kernels taken from the finite Fock
expansion of the Wick polynomial operator $\mathcal{L}(x_{\pi(i)})$.

For each pair of the kernels $\kappa^{1}_{\ell_{1}, m_{1}}$ and $\kappa^{2}_{\ell_{2}, m_{2}}$, which are taken from the finite Fock
expansion, respectively, of the Wick polynomial operator $\mathcal{L}(x_1)$ and $\mathcal{L}(x_2)$, it follows by Lemma \ref{q-contractionkappaBarDotOtimeskappaWithOneMasslessOnEU} of Subsection
\ref{WhiteNoiseFreeFieldsonEU} that
\[
\kappa_{\ell,m}(x_1,x_2) = \theta_{{}_{2}}\kappa^{1}_{\ell_{1}, m_{1}} \otimes_q \kappa^{2}_{\ell_{2}, m_{2}}(x_1,x_2)
=
\theta\big(x_1x_{2}^{-1}\big)\kappa^{1}_{\ell_{1}, m_{1}} \otimes_q \kappa^{2}_{\ell_{2}, m_{2}}(x_1,x_2),
\]
\[
\ell = \ell_{1} + \ell_{2} - q, \,\,\, m = m_{1}+m_{2} - q
\] 
represents a kernel
\[
\theta_{{}_{2}}\kappa^{1}_{\ell_{1}, m_{1}} \otimes_q \kappa^{2}_{\ell_{2}, m_{2}} \in \mathscr{L}\big(E^{* \otimes \, \ell} \otimes E^{\otimes \, m},
\mathscr{E}^{* \, \otimes \, 2}\big),
\]
where, for simplicity of notation, all single particle nuclear spaces, corresponding to the momentum variables of the kernels $\kappa^{1}_{\ell_{1}, m_{1}}$, 
$\kappa^{2}_{\ell_{2}, m_{2}}$, we have denoted by the same symbol $E$. 
By the repeated application of the Lemma \ref{q-contractionkappaBarDotOtimeskappaWithOneMasslessOnEU} of Subsection
\ref{WhiteNoiseFreeFieldsonEU} we obtain more generally 
\begin{multline*}
\kappa_{\ell, m} =
\theta_{{}_{\pi(2)}} \kappa^{\pi(1)}_{\ell_{\pi(1)},m_{\pi(1)}}\otimes_{{}_{q_1}} \, 
\theta_{{}_{\pi(3)}} \kappa^{\pi(2)}_{\ell_{\pi(2)},m_{\pi(2)}} \,\,
\otimes_{{}_{q_2}}
\ldots  \,\,\, \times 
\\
\times \,\, \ldots
\otimes_{{}_{q_{n-2}}} \, 
\theta_{{}_{\pi(n)}} \kappa^{\pi(n-1)}_{\ell_{\pi(n-1)},m_{\pi(n-1)}}\otimes_{{}_{q_{n-1}}} \,  \kappa^{\pi(n)}_{\ell_{\pi(n)},m_{\pi(n)}}
\\
\in \mathscr{L}\big(E^{* \otimes \, \ell} \otimes E^{\otimes \, m}, \,\, \mathscr{E}^{* \, \otimes \, n}\big),
\\
\ell = \ell_{\pi(1)}+ \ldots +\ell_{\pi(n)} -q_1 \ldots - q_{n-1},
\,\,\,\, m= m_{\pi(1)}+ \ldots +m_{\pi(n)} -q_1 \ldots - q_{n-1}
\end{multline*}
for contraction of $n$ kernels $\kappa^{i}_{\ell_{i}, m_{i}}$ which are, rspectively, taken from the finite Fock
expansion of the Wick polynomial operator $\mathcal{L}(x_{\pi(i)})$, respectively multiplied by the appropriately ``translated'' theta functions
$\theta_{{}_{\pi(i+1)}}\big(x_{\pi(i)}\big) = \theta\big(x_{\pi(i))}x_{\pi(i+1)}^{-1}\big)$.
Here also we have used simplified notation in which all single particle nuclear spaces corresponding to the momentum variables 
we have denoted by the same symbol $E$.  Therefore we see again, this time by looking at the kernels
of finite Fock expansion of the product operator $\Xi$, that indeed 
\[
\Xi
\in \mathscr{L}( (\boldsymbol{E}) \otimes \mathscr{E}^{\otimes \, n} , (\boldsymbol{E}) )
\cong  \mathscr{L}\big(\mathscr{E}^{\otimes \, n}, \, \mathscr{L}( (\boldsymbol{E}), (\boldsymbol{E}) ) \, \big),
\]
but moreover we can compute the kernels $\kappa_{\ell,m}$ of the expansion of the chronological product $\Xi$
by using very simple operations: multiplication of the plane-wave kernels of free fields
\emph{i.e.} the operation $\otimes$, with some multiplications
of the plane wave kernels taken at the same space-time point, \emph{i.e.} the dot product operation $\dot{\otimes}$,
then symmetrization in the first and separately last bose momentum variables, antisymmetrization 
in the first and separately in the last momentum fermi variables, and finally summation (discrete integration)
with respect to the contracted momentum variables. 
We therefore arrive at the following

\begin{twr}
Let each of the Wick monomials
of the Wick polynomial $\mathcal{L}$ contain at most one massless (or infinite orbit) free field  
and be of degree at most $N$, and with each of the remaining free fields in each monomial of the Lagrange interaction density
operator $\mathcal{L}$
being massive (or finite orbit) field. This is the case in particular for QED.
In particular $\mathcal{L}$ for spinor QED interaction has degree $N=3$. 
Then for $\phi, \phi_k \in \mathscr{E}$
the operators
\[
\Xi'(\phi) =  \int\limits_{\widetilde{\mathbb{S}^1}\times SU(2, \mathbb{C})}  \mathcal{L}(x) \, \phi(x) \,\,\,\, \ud^4x
\in \mathscr{L}((\boldsymbol{E}),(\boldsymbol{E})),
\]
\begin{multline*}
\Xi(\phi_1 \otimes \ldots \phi_n)
\\
=
\sum\limits_{\pi}
\int\limits_{\big[\widetilde{\mathbb{S}^1}\times SU(2, \mathbb{C})\big]^{\times \, n}} 
\theta_{{}_{\pi(2)}} \mathcal{L}(x_{\pi(1)}) 
\theta_{{}_{\pi(3)}} \mathcal{L}(x_{\pi(2)})  \ldots 
\theta_{{}_{\pi(n)}} \mathcal{L}(x_{\pi(n-1)}) 
\mathcal{L}(\phi_{\pi(n)}) 
\,\, \times
\\
\times \,\,\,
\phi_{1} \otimes \ldots \otimes \phi_{n}(x_1, \ldots, x_n) \, \ud^4x_1 \ldots \ud^4x_n 
\\
=
\sum \limits_{\ell +m \leq Nn} \Xi_{\ell,m}\big( \kappa_{\ell,m}(\phi_1 \otimes \ldots \phi_n)\big)
\in \mathscr{L}((\boldsymbol{E}),(\boldsymbol{E}))
\end{multline*}
\begin{multline*}
\Xi(\phi_1 \otimes \ldots \otimes \phi_n) =
\\
= 
\int\limits_{\big[\widetilde{\mathbb{S}^1}\times SU(2, \mathbb{C})\big]^{\times \, n}} 
i^n \, T\big(\mathcal{L}(x_1) \ldots \mathcal{L}(x_n) \big) \,
\phi_{1} \otimes \ldots \otimes \phi_{n}(x_1, \ldots, x_n) \, \ud^4x_1 \ldots \ud^4x_n 
\\
\overset{\textrm{df}}{=} 
\Xi(\phi_1 \otimes \ldots \phi_n)
\,\,\, \in \mathscr{L}((\boldsymbol{E}),(\boldsymbol{E})),
\end{multline*} 
and moreover
\begin{eqnarray*}
\Xi' \in \mathscr{L}(\mathscr{E}, \mathscr{L}((\boldsymbol{E}),(\boldsymbol{E}))\big),
\\
\Xi \in \mathscr{L}(\mathscr{E}^{\otimes \, n}, \mathscr{L}((\boldsymbol{E}),(\boldsymbol{E}))\big),
\end{eqnarray*}
and the following Wick theorem holds
\begin{multline*}
\Xi =
i^n \sum \limits_{\pi}\sum_{\substack{\kappa^{\pi(i)}_{\ell_{\pi(i)},m_{\pi(i)}}}} \,\,\,
\sum \limits_{0\leq q_i \leq \textrm{min} \, \{m_{\pi(i), \ell_{\pi(i+1)} \}}} \,\,\, (-1)^{c(q_1) + \ldots c(q_{n-1})} 
\,\, \times 
\\
\times \,\,
\Xi_{{}_{\ell_{\pi(1)}+ \ldots +\ell_{\pi(n)} -q_1 \ldots - q_{n-1}, \,\, m_{\pi(1)}+ \ldots +m_{\pi(n)} -q_1 \ldots - q_{n-1}}},
\end{multline*}
where
\begin{multline*}
\Xi_{{}_{\ell_{\pi(1)}+ \ldots +\ell_{\pi(n)} -q_1 \ldots - q_{n-1}, \,\, m_{\pi(1)}+ \ldots +m_{\pi(n)} -q_1 \ldots - q_{n-1}}}
\\
=
\Xi\big( \theta_{{}_{\pi(2)}} \kappa^{\pi(1)}_{\ell_{\pi(1)},m_{\pi(1)}}\otimes_{{}_{q_1}} \, 
\theta_{{}_{\pi(3)}} \kappa^{\pi(2)}_{\ell_{\pi(2)},m_{\pi(2)}} \,\,
\otimes_{{}_{q_2}}
\ldots  \,\,\, \times 
\\
\times \,\, \ldots
\otimes_{{}_{q_{n-2}}} \, 
\theta_{{}_{\pi(n)}} \kappa^{\pi(n-1)}_{\ell_{\pi(n-1)},m_{\pi(n-1)}}\otimes_{{}_{q_{n-1}}} \,  \kappa^{\pi(n)}_{\ell_{\pi(n)},m_{\pi(n)}}\big).
\end{multline*}
Here the kernels $\kappa^{\pi(i)}_{\ell_{\pi(i)},m_{\pi(i)}}$ range independently over the  kernels of the finite Fock expansion
of the operator $\Xi' = \mathcal{L}(x_{\pi(i)})$. The (symmetrized/antisymmetrized) $q_i$-contractions $\otimes_{{}_{q_{i}}}$ are performed
upon the pairs of variables in which the first element of the contracted pair lies among the last $m_{\pi(i)}$ variables of the kernel 
$\theta_{{}_{\pi(i+1)}}\kappa^{{\pi(i)}}_{\ell_{\pi(i)},m_{\pi(i)}}$ and the second variable
of the contracted pair lies among the first $l_{\pi(i+1)}$ variables of the kernel 
$\theta_{{}_{\pi(i+2)}}\kappa^{\pi(i+1)}_{\ell_{\pi(i+1)},m_{\pi(i+1)}}$, 
and to both variables of the contracted pair correspond respectively annihilation and creation operator of the free fields 
with nonzero mutual contraction
(pairing). The number 
$c(q_{i})$ is equal to the number of fermi commutations performed in the contraction $\otimes_{{}_{q_{i}}}$. 
\label{WickThmForChronologicalOneMassLessEU}
\end{twr}

\vspace*{1cm}

\begin{center}
\small{CASE 2)}
\end{center}

Suppose that some Wick monomials of the Lagrange interaction density $\mathcal{L}$ contain more than
one massless (or infinite orbit) free field. Without any loss of generality we can assume
$\mathcal{L}$ to be a single such Wick monomial of degree $N$.

From Lemma \ref{kappaBarDotOtimeskappaOnEU} of Subsection \ref{WhiteNoiseFreeFieldsonEU} and 
Theorem \ref{obataJFA.Thm.3.13} of Subsection \ref{psiBerezin-Hida} it follows that also in this case $\mathcal{L}$ is a finite sum 
\[
\mathcal{L} = \Xi' =
\sum\limits_{\substack{\ell,m \\ \ell'+m'=N}} 
\Xi_{\ell',m'}\big(\kappa'_{\ell',m'}\big)
\in \mathscr{L}( (\boldsymbol{E}) \otimes \mathscr{E} , (\boldsymbol{E}) ).
\]
of well defined integral kernel operators.

The difference in comparison to case 1) is that in this case 2) the pointwise multiplication of $\mathcal{L}$
by the step theta function $\theta$ is more singular when we have more than one massless (or infinite orbit)
free field in $\mathcal{L}$ in accordance to our general results of Subsection \ref{WhiteNoiseFreeFieldsonEU}.

Proceeding exactly as in Subsection \ref{WickForChronological} we replace the step $\theta$ function
 by a one-parameter family of smooth functions $\theta_\varepsilon$, which converges
to $\theta$ in the sense of
Definition \ref{ConvergenceOfthetavarepsilon} of Subsection \ref{WhiteNoiseFreeFieldsonEU}. The number
$\omega$ in Definition \ref{ConvergenceOfthetavarepsilon} of Subsection \ref{WhiteNoiseFreeFieldsonEU} we will specify later.

Let us consider in particular the contraction
\[
\theta_{{}_{ \varepsilon \,\, 2}} \kappa^{1}_{\ell_1, m_1} \otimes_q \kappa^{2}_{\ell_2, m_2}, \,\,\,\,\,\,
q>1,
\]
where the space-time variable of the kerel $\kappa^{1}_{\ell_1, m_1}$ is $x_1$, and the space-time variable
of the kernel $\kappa^{2}_{\ell_2, m_2}$ is $x_2$, and, as usual $\theta_{{}_{ \varepsilon \,\, 2}}(x_1) = \theta_\varepsilon\big(x_1x_{2}^{-1}\big)$.
Let $\Omega$ be equal to the projection $\Omega'$ equal \ref{Omega'phiEU(0,e)-(2pi,e)}) for periodic $\theta$ and, respectively,
(\ref{Omega'phiEUseveralMultiindex}) for ordinary non-periodic $\theta$, with $\omega$ in these projections equal to the singulrity order 
of this particular contraction at $(0,e)$.
Because each of the kernels $ \kappa^{1}_{\ell_1, m_1}$, $\kappa^{2}_{\ell_2, m_2}$ of the Fock expansion of the Wick
product (monomial) $\Xi' = \mathcal{L}(x_1)$ or, respectively, $\Xi' = \mathcal{L}(x_2)$, is equal to a finite sum of simple dot products of the form 
\begin{align*}
\kappa^{1}_{\ell_1, m_1} =
\kappa^{(1)}_{0,1} \overset{\cdot}{\otimes} \cdots \overset{\cdot}{\otimes} \kappa^{(m_1)}_{0,1} \dot{\otimes}
\kappa'^{(1)}_{1,0}\overset{\cdot}{\otimes} \cdots \overset{\cdot}{\otimes} \kappa'^{(\ell_1)}_{1,0}, 
\\
\kappa^{2}_{\ell_2, m_2} =
\kappa''^{(1)}_{0,1} \overset{\cdot}{\otimes} \cdots \overset{\cdot}{\otimes} \kappa''^{(m_2)}_{0,1} \dot{\otimes}
\kappa^{(1)}_{1,0}\overset{\cdot}{\otimes} \cdots \overset{\cdot}{\otimes} \kappa^{(\ell_2)}_{1,0}, 
\end{align*}
where $\kappa^{(i)}_{0,1}, \kappa^{(i)}_{1,0}$, $\kappa''^{(i)}_{0,1}, \kappa'^{(i)}_{1,0}$ are the plane wave kernels of the free fields in the Wick
product $\Xi' = \mathcal{L}(x_1)$ and, resectively, $\Xi' = \mathcal{L}(x_2)$,  then by Lemma \ref{q-contractionkappaBarDotOtimeskappaWithManyMasslessOnEU} 
of Subsection \ref{WhiteNoiseFreeFieldsonEU} and Subsection \ref{splittingEU}
\[
\theta_{{}_{ \varepsilon \,\, 2}} \kappa^{1}_{\ell_1, m_1} \otimes_q \kappa^{2}_{\ell_2, m_2} \circ \Omega 
\,\,\,\,\,
\overset{\varepsilon \rightarrow 0}{\longrightarrow}
\,\,\,\,\,
\theta_{{}_{2}} \kappa^{1}_{\ell_1, m_1} \otimes_q \kappa^{2}_{\ell_2, m_2} \circ \Omega
\]
in 
\begin{multline*}
\mathscr{L}\big(\mathscr{E}^{\otimes \, 2}, 
E_{{}_{(1')}}^{*} \otimes  \ldots \otimes E_{{}_{\ell'_{1}}}^{*} \otimes 
E_{{}_{q+1}}^{*} \otimes \ldots \otimes E_{{}_{\ell_2}}^{*} \otimes  
E_{{}_{q+1}}^{*} \otimes \ldots \otimes E_{{}_{m_1}}^{*} \otimes
E_{{}_{1''}}^{*} \otimes \ldots \otimes  E_{{}_{(\ell''_{2})}}^{*}\big) 
\\
\cong
\mathscr{L}\big(E_{{}_{(1')}} \otimes  \ldots \otimes E_{{}_{\ell'_{1}}} \otimes 
E_{{}_{q+1}} \otimes \ldots \otimes E_{{}_{\ell_2}} \otimes  
E_{{}_{q+1}} \otimes \ldots \otimes E_{{}_{m_1}} \otimes
E_{{}_{1''}} \otimes \ldots \otimes  E_{{}_{(\ell''_{2})}}
, \mathscr{E}^{* \, \otimes \, 2}\big),
\end{multline*}
with $\theta_\varepsilon$ converging to $\theta$
in the sense of Definition \ref{ConvergenceOfthetavarepsilon} of Subsection \ref{WhiteNoiseFreeFieldsonEU}.
Here, for simplicity of notation, we have assumed that the first $q$ momentum variables of the ``annihilation'' kernels
\[
\kappa^{(1)}_{0,1} \overset{\cdot}{\otimes} \cdots \overset{\cdot}{\otimes} \kappa^{(m_1)}_{0,1}
\]
have been contracted with the first $q$ momentum variables of the ``creation'' kernels
\[
\kappa^{(1)}_{1,0}\overset{\cdot}{\otimes} \cdots \overset{\cdot}{\otimes} \kappa^{(\ell_2)}_{1,0}.
\]

From Lemma \ref{q-contractionkappaBarDotOtimeskappaWithManyMasslessOnEU} 
of Subsection \ref{WhiteNoiseFreeFieldsonEU} and the results of Subsection \ref{splittingEU}, 
it follows that, for the kernels $\kappa^{1}_{\ell_1, m_1}$, $\kappa^{2}_{\ell_2, m_2}$
 equal to simple dot
products of plane wave kernels, the most general $G$-invariant extension of the limit
\[
\underset{\varepsilon \rightarrow 0}{\textrm{lim}}
\theta_{{}_{ \varepsilon \,\, 2}} \kappa^{1}_{\ell_1, m_1} \otimes_q \kappa^{2}_{\ell_2, m_2} 
= \theta_{{}_{2}} \kappa^{1}_{\ell_1, m_1} \otimes_q \kappa^{2}_{\ell_2, m_2},
\]
well defined on the closed subspace $\textrm{Im} \, \Omega \subset \mathscr{E}$, over the whole
test space 
\[
\mathscr{E} = \textrm{Ker} \, \Omega \oplus \textrm{Im} \, \Omega
\]
 is equal
\begingroup\makeatletter\def\f@size{5}\check@mathfonts
\def\maketag@@@#1{\hbox{\m@th\large\normalfont#1}}%
\begin{multline*}
\big(\theta_2 \kappa^{(1)}_{0,1} \overset{\cdot}{\otimes} \cdots \overset{\cdot}{\otimes} \kappa^{(m_1)}_{0,1} \dot{\otimes}
\kappa'^{(1)}_{1,0}\overset{\cdot}{\otimes} \cdots \overset{\cdot}{\otimes} \kappa'^{(\ell_1)}_{1,0} \big) 
\otimes||_{{}_{q}} 
\big(\kappa''^{(1)}_{0,1} \overset{\cdot}{\otimes} \cdots \overset{\cdot}{\otimes} \kappa''^{(m_2)}_{0,1} \dot{\otimes}
\kappa^{(1)}_{1,0}\overset{\cdot}{\otimes} \cdots \overset{\cdot}{\otimes} \kappa^{(\ell_2)}_{1,0} \big) 
\\
=\theta_{{}_{2}} \kappa'_{\ell_1, m_1} \otimes_q \kappa'_{\ell_2, m_2} \circ \Omega
\,\,\,\,\,\,\,\,\,\,\,\,\,\,\,\,\,\,\,\,\,\,\,\,\,\,\,\,\,\,\,\,\,\,\,\,\,\,\,\,\,\,\,\,\,\,\,\,\,\,\,\,\,\,\,\,\,\,\,\,
\,\,\,\,\,\,\,\,\,\,\,\,\,\,\,\,\,\,\,\,\,\,\,\,\,\,\,\,\,\,\,\,\,\,\,\,\,\,\,\,\,\,\,\,\,\,\,\,\,\,\,\,\,\,\,\,\,\,\,\,
\,\,\,\,\,\,\,\,\,\,\,\,\,\,\,\,\,\,\,\,\,\,\,\,\,\,\,\,\,\,\,\,\,\,\,\,\,\,\,\,\,\,\,\,\,\,\,\,\,\,\,\,\,\,\,\,\,\,\,\,
\\
\,\,\,\,\,\,\,\,\,\,\,\,\,\,\,\,\,\,\,\,\,\,\,\,\,\,\,\,\,\,
+
\delta^{\omega}_{{}{1;2}} \,\, . \,\,
\big(\kappa^{(q+1)}_{0,1} \overset{\cdot}{\otimes} \cdots \overset{\cdot}{\otimes} \kappa^{(m_1)}_{0,1} \dot{\otimes}
\kappa'^{(1)}_{1,0}\overset{\cdot}{\otimes} \cdots \overset{\cdot}{\otimes} \kappa'^{(\ell_1)}_{1,0} \big) 
\otimes 
\big(\kappa''^{(1)}_{0,1} \overset{\cdot}{\otimes} \cdots \overset{\cdot}{\otimes} \kappa''^{(m_2)}_{0,1} \dot{\otimes}
\kappa^{(q+1)}_{1,0}\overset{\cdot}{\otimes} \cdots \overset{\cdot}{\otimes} \kappa^{(\ell_2)}_{1,0} \big), 
\end{multline*}
\endgroup
\[
=
\theta_{{}_{2}} \kappa^{1}_{\ell_1, m_1} \otimes_q \kappa^{2}_{\ell_2, m_2} \circ \Omega + \textrm{Remainder kernel},
\]
where
\begin{multline*}
\delta^{\omega}_{{}_{1;2}}(x_1x_2^{-1}) \overset{\textrm{df}}{=}
\sum\limits_{|\alpha| \leq \omega} C_{I,\alpha} \,\, \delta^{(\alpha_0)}(t_{1}-t_{2}) \delta^{(\alpha_1)}({x}_{11}{x}_{12}^{-1})
\delta^{(\alpha_2)}({x}_{21}{x}_{22}^{-1})\delta^{(\alpha_3)}({x}_{31}{x}_{32}^{-1}) 
\\
+
\sum\limits_{|\alpha| \leq \omega} C_{II, \alpha} \,\, \delta^{(\alpha_0)}(t_{1}-t_{2}-2\pi) \delta^{(\alpha_1)}({x}_{11}{x}_{12}^{-1})
\delta^{(\alpha_2)}({x}_{21}{x}_{22}^{-1})\delta^{(\alpha_3)}({x}_{31}{x}_{32}^{-1}), 
\end{multline*}
\[
x_1 = (t_1, {x}_{11}, {x}_{21}, {x}_{31}), \, x_2 = (t_2, {x}_{12},{x}_{22}, {x}_{32}) \in G,
\,\,\, C_{I,\alpha}, C_{II,\alpha_1} \in \mathbb{C},
\] 
with $G =\widetilde{\mathbb{S}^1}\times SU(2, \mathbb{C})$ for periodic $\theta$ and $G = \mathbb{R}\times SU(2, \mathbb{C})$
for ordinary non-periodic $\theta$, and with the analytic fuctions $x_0=t,x_1,x_2,x_2$ on $G$ which can serve as a local coordinate
system in a neighborhood of the unit $(0,e)$, which are dual to the invariant vector fields $X_0, \ldots, X_3$ on $G$ at $(0,e)$ (and
at $(2\pi, e)$ for periodic $\theta$) up to order $\omega$ at least (compare Subsection \ref{splittingEU}). 
For the case with ordinary non-peridic $\theta$, $C_{II,\alpha} = 0$.

Expressing the same otherwise the above formula gives the most general form of the retarded part of the vector-valued kernel
\begingroup\makeatletter\def\f@size{5}\check@mathfonts
\def\maketag@@@#1{\hbox{\m@th\large\normalfont#1}}%
\[
\kappa^{1}_{\ell_1, m_1} \otimes_q \kappa^{2}_{\ell_2, m_2} =
\big(\kappa^{(1)}_{0,1} \overset{\cdot}{\otimes} \cdots \overset{\cdot}{\otimes} \kappa^{(m_1)}_{0,1} \dot{\otimes}
\kappa'^{(1)}_{1,0}\overset{\cdot}{\otimes} \cdots \overset{\cdot}{\otimes} \kappa'^{(\ell_1)}_{1,0} \big) 
\otimes_{{}_{q}} 
\big(\kappa''^{(1)}_{0,1} \overset{\cdot}{\otimes} \cdots \overset{\cdot}{\otimes} \kappa''^{(m_2)}_{0,1} \dot{\otimes}
\kappa^{(1)}_{1,0}\overset{\cdot}{\otimes} \cdots \overset{\cdot}{\otimes} \kappa^{(\ell_2)}_{1,0} \big). 
\]
\endgroup

The same distribution  $\delta^{\omega}_{{}{1;2}}$ and the same \emph{remainder kernel} has to be used in the construction of the most general
advanced part of this kernel, which has to be equal
\begingroup\makeatletter\def\f@size{5}\check@mathfonts
\def\maketag@@@#1{\hbox{\m@th\large\normalfont#1}}%
\begin{multline*}
\big(-\check{\theta}_2 \kappa^{(1)}_{0,1} \overset{\cdot}{\otimes} \cdots \overset{\cdot}{\otimes} \kappa^{(m_1)}_{0,1} \dot{\otimes}
\kappa'^{(1)}_{1,0}\overset{\cdot}{\otimes} \cdots \overset{\cdot}{\otimes} \kappa'^{(\ell_1)}_{1,0} \big) 
\otimes||_{{}_{q}} 
\big(\kappa''^{(1)}_{0,1} \overset{\cdot}{\otimes} \cdots \overset{\cdot}{\otimes} \kappa''^{(m_2)}_{0,1} \dot{\otimes}
\kappa^{(1)}_{1,0}\overset{\cdot}{\otimes} \cdots \overset{\cdot}{\otimes} \kappa^{(\ell_2)}_{1,0} \big) 
=- \check{\theta}_{{}_{2}} \kappa^{1}_{\ell_1, m_1} \otimes_q \kappa^{2}_{\ell_2, m_2} \circ \Omega
+
\delta^{\omega}_{{}{1;2}}, 
\end{multline*}
\endgroup

In case the contraction number $q \leq 1$ in 
\[
\theta_{{}_{ \varepsilon \,\, 2}} \kappa^{1}_{\ell_1, m_1} \otimes_q \kappa^{2}_{\ell_2, m_2}, \,\,\,\,\,\,
q \leq 1,
\]
 the limit
\[
\theta_{{}_{ \varepsilon \,\, 2}} \kappa^{1}_{\ell_1, m_1} \otimes_q \kappa^{2}_{\ell_2, m_2} 
\,\,\,\,\,
\overset{\varepsilon \rightarrow 0}{\longrightarrow}
\,\,\,\,\,
\theta_{{}_{2}} \kappa^{1}_{\ell_1, m_1} \otimes_q \kappa^{2}_{\ell_2, m_2} 
\]
exists in 
\begin{multline*}
\mathscr{L}\big(\mathscr{E}^{\otimes \, 2}, 
E_{{}_{(1')}}^{*} \otimes  \ldots \otimes E_{{}_{\ell'_{1}}}^{*} \otimes 
E_{{}_{q+1}}^{*} \otimes \ldots \otimes E_{{}_{\ell_2}}^{*} \otimes  
E_{{}_{q+1}}^{*} \otimes \ldots \otimes E_{{}_{m_1}}^{*} \otimes
E_{{}_{1''}}^{*} \otimes \ldots \otimes  E_{{}_{(\ell''_{2})}}^{*}\big) 
\\
\cong
\mathscr{L}\big(E_{{}_{(1')}} \otimes  \ldots \otimes E_{{}_{\ell'_{1}}} \otimes 
E_{{}_{q+1}} \otimes \ldots \otimes E_{{}_{\ell_2}} \otimes  
E_{{}_{q+1}} \otimes \ldots \otimes E_{{}_{m_1}} \otimes
E_{{}_{1''}} \otimes \ldots \otimes  E_{{}_{(\ell''_{2})}}
, \mathscr{E}^{* \, \otimes \, 2}\big),
\end{multline*}
by Lemma \ref{1-contractionkappaBarDotOtimeskappaWithManyMasslessOnEU} 
of Subsection \ref{WhiteNoiseFreeFieldsonEU}.
Therefore, in case of zero-contraction $\otimes_0 = \otimes$ or $1$-contraction $\otimes_1$ we can put $\Omega= \boldsymbol{1}$
and the contraction kernel 
\[
\theta_{{}_{ \varepsilon \,\, 2}} \kappa^{1}_{\ell_1, m_1} \otimes_q \kappa^{2}_{\ell_2, m_2}
, \,\,\,\,\,\,
q \leq 1,
\]
converges to the contraction
\[
\theta_{{}_{2}} \kappa^{1}_{\ell_1, m_1} \otimes_q \kappa^{2}_{\ell_2, m_2}
, \,\,\,\,\,\,
q \leq 1,
\]
on the whole test space and with the \emph{remainder kernel} equal zero.

Recall, please, that $\otimes_q$ in
\[
\theta_{{}_{ \varepsilon \,\, 2}} \kappa^{1}_{\ell_1, m_1}\otimes_q \kappa^{2}_{\ell_2, m_2},
\,\,\,\,\,
\theta_{{}_{ 2}} \kappa^{1}_{\ell_1, m_1}\otimes_q \kappa^{2}_{\ell_2, m_2},
\]
means that, after performing the contraction summations, the kernels
\[
\theta_{{}_{ \varepsilon \,\, 2}} \kappa^{1}_{\ell_1, m_1} \otimes \kappa^{2}_{\ell_2, m_2},
\,\,\,\,\,
\theta_{{}_{ 2}} \kappa^{1}_{\ell_1, m_1} \otimes \kappa^{2}_{\ell_2, m_2}
\]
should be symmetrized, separately in the Bose variables momentum variables, and antisymmetrized in fermi
momentum variables. This is because we want to have one-to-one correspondence between kernels
$\kappa_{\ell,m}$ and the integral kernel operators $\Xi(\kappa_{\ell,m})$, compare Subsection \ref{psiBerezin-Hida}.

The general principle used in passing from the formulas of Subsection \ref{WickForChronological} on the Minkowski space-time
to the formulas on the Einstein Universe for the kernels of higer order contributions $S_n$,  $n>2$, to the scattering operator
is that we replace the additive Lie group $\mathbb{R}^4$ of Subsection 
\ref{WickForChronological} by the non-Abelian Lie group $G$ (with $G=\widetilde{\mathbb{S}^1}\times SU(2, \mathbb{C})$ for periodic $\theta$ and, respectively,
$G=\mathbb{R}\times SU(2, \mathbb{C})$ for non-periodic $\theta$). 
The additive group action $x_1+ x_2$ in $\mathbb{R}^4$ we replace with the group action
$x_1x_2$ in  $G$. The direct product additive group
$\big[\mathbb{R}^4 \big]^{\times \, n}$ we replace with the direct product
group $G^{\times \, n}$. 

Instead of the uniform and additive (linear) map $L$ and its inverse $L^{-1}$ of Subsection \ref{WickForChronological} we put
\[
L(x_1, \ldots, x_n) = (x_1x_n, \, x_2x_n, \ldots, \, x_{n-1}x_{n}, \, x_n)
\]
\[
L^{-1}(x_1, \ldots, x_n) = \big(x_{1}x_{n}^{-1}, x_{2}x_{n}^{-1}, x_{3}x_{n}^{-1},\,  \ldots, \, x_{n-1}x_{n}^{-1}, \, x_{n}\big)
\]
as group maps acting on the direct product group 
\[
\big[\widetilde{\mathbb{S}^1}\times SU(2, \mathbb{C}) \big]^{\times \, n}.
\]

For $\chi \in  \mathscr{E}^{\otimes \, n}$ of the form
\[
\chi = (\phi \otimes \varphi) \circ L^{-1}, \,\,\,\, \phi \in \mathscr{E}^{\otimes \, (n-1)}, \,\, \varphi \in \mathscr{E},
\]
we put (instead of the operator $\Omega$ given by (\ref{SimpleFormulaForOmega}) of Subsection \ref{WickForChronological})
\[
\Omega \chi = \big[ (\Omega' \phi) \otimes \varphi \big] \circ L^{-1}
\]
where for
\begin{multline*}
\phi(x_{1}, \ldots, x_{n-1}) = \phi_{{}_{1}} \otimes \ldots \otimes \phi_{{}_{n-1}}(x_{1}, \ldots, x_{n-1}) 
\\
=
\phi_{{}_{1}}(x_{1})\phi_{{}_{2}}(x_{2}) \ldots \phi_{{}_{n-1}}(x_{n-1}),
\,\,\,\, \phi_{{}_{i}} \in \mathscr{E},
\end{multline*}
we define the continuous idempotent operator analogously as the operator $\Omega'$ given by (\ref{OmegaphiR(k-1)} in Subsection  \ref{WickForChronological})
on the Minkowski space-time, adding similar suntraction also at the point $(2\pi,e)$ for the case with periodic $\theta$.

Therefore, we can summarize our recurrence rule for the scattering operator or chronological product, with the interaction $\mathcal{L}$
containing monomials with more than one massless or inifinite orbit free field, in the following

\begin{twr}
Let some Wick monomials
of the Wick interaction polynomial $\mathcal{L}$ contain more than one massless (or infinite orbit) free field  
and let $\mathcal{L}$ be of degree at most $N$. 
Then for $\phi, \phi_k \in \mathscr{E}$ 
the operators
\[
\Xi'(\phi) =  \int\limits_{\widetilde{\mathbb{S}^1}\times SU(2,\mathbb{C})}  \mathcal{L}(x) \, \phi(x) \,\,\,\, \ud^4x
\in \mathscr{L}((\boldsymbol{E}),(\boldsymbol{E})),
\]
\begin{multline*}
\Xi_{\varepsilon,\epsilon}(\phi_1 \otimes \phi_2)
\\
=
\sum\limits_{\pi}
\int\limits_{\big[\widetilde{\mathbb{S}^1}\times SU(2,\mathbb{C})\big]^{\times \, 2}} 
\theta_{{}_{\varepsilon \,\, \pi(2)}} \mathcal{L}_\epsilon(x_{\pi(1)}) 
\mathcal{L}_\epsilon(x_{\pi(2)}) \,\,
\phi_{1} \otimes \phi_{2}(x_1, x_2) \, \ud^4x_1 \ud^4x_2 
\\
=
\sum \limits_{\ell +m \leq Nn} \Xi_{\ell,m}\big( \kappa_{\varepsilon, \epsilon \,\, \ell,m}(\phi_1 \otimes \phi_2)\big)
\in \mathscr{L}((\boldsymbol{E}),(\boldsymbol{E}))
\end{multline*}
or, respectively, for $n>2$,
\begin{multline*}
\Xi_{\varepsilon,\epsilon}(\phi_1 \otimes \ldots \phi_n)
\\
=
\int\limits_{\big[\widetilde{\mathbb{S}^1}\times SU(2,\mathbb{C})]^{\times \, n}} 
\big[
\theta_{{}_{\varepsilon}}(x_1x_{n}^{-1}) \ldots \theta_{{}_{\varepsilon}}(x_{n-1}x_{n}^{-1}) \, D_{(n)}(x_1, \ldots, x_n)
-R'_{(n)}(x_1, \ldots, x_n)
\big]
\,\, \times
\\
\times \,\,\,
\phi_{1} \otimes \ldots \otimes \phi_{n}(x_1, \ldots, x_n) \, \ud^4x_1 \ldots \ud^4x_n 
\\
=
\sum \limits_{\ell +m \leq Nn} \Xi_{\ell,m}\big( \kappa_{\varepsilon, \epsilon \,\, \ell,m}(\phi_1 \otimes \ldots \phi_n)\big)
\in \mathscr{L}((\boldsymbol{E}),(\boldsymbol{E}))
\end{multline*}

\begin{multline*}
\Xi^\Omega(\phi_1 \otimes \ldots \otimes \phi_n) =
\\
= 
\int\limits_{\big[\widetilde{\mathbb{S}^1}\times SU(2,\mathbb{C})\big]^{\times \, n}} 
i^n \, T\big(\mathcal{L}(x_1) \ldots \mathcal{L}(x_n) \big) \,
\Omega \phi_{1} \otimes \ldots \otimes \phi_{n}(x_1, \ldots, x_n) \, \ud^4x_1 \ldots \ud^4x_n 
\\
\overset{\textrm{df}}{=} 
\underset{\varepsilon,\epsilon \rightarrow 0}{\textrm{lim}}
\Xi_{\varepsilon,\epsilon}^{\Omega}(\phi_1 \otimes \ldots \phi_n)
\,\,\, \in \mathscr{L}((\boldsymbol{E}),(\boldsymbol{E})^*),
\end{multline*} 
and moreover
\begin{eqnarray*}
\Xi' \in \mathscr{L}(\mathscr{E}, \mathscr{L}((\boldsymbol{E}),(\boldsymbol{E}))\big),
\\
\Xi_{\varepsilon,\epsilon}, \Xi_{\varepsilon,\epsilon}^{\Omega} \in \mathscr{L}(\mathscr{E}^{\otimes \, n}, \mathscr{L}((\boldsymbol{E}),(\boldsymbol{E}))\big)
\\
\Xi^\Omega \in \mathscr{L}(\mathscr{E}^{\otimes \, n}, \mathscr{L}((\boldsymbol{E}),(\boldsymbol{E}))\big).
\end{eqnarray*}
\[
 \Xi_{\varepsilon,\epsilon}^{\Omega} = \sum \limits_{\ell +m \leq Nn} 
\Xi_{\ell,m}\big( \kappa_{\varepsilon, \epsilon \,\, \ell,m} \circ \Omega_{{}_{\ell,m}}\big);
\]
(in QED $N=3$). The following Wick theorem holds. 
\[
\textrm{In case} \, n=2:
\]
\[
\Xi^\Omega =
i^2 \sum \limits_{\pi}\sum_{\substack{\kappa^{\pi(i)}_{\ell_{\pi(i)},m_{\pi(i)}}}} \,\,\,
\sum \limits_{q} \,\,\, (-1)^{c(q)} 
\,\, 
\Xi_{{}_{\ell_{\pi(1)}+\ell_{\pi(2)} -q, \,\, m_{\pi(1)}+m_{\pi(2)} -q}},
\]
where
\[
\Xi_{{}_{\ell_{\pi(1)}+ \ell_{\pi(2)} -q, \,\, m_{\pi(1)}+m_{\pi(2)} -q}}
=
\Xi\big( \theta_{{}_{\pi(2)}} \kappa^{\pi(1)}_{\ell_{\pi(1)},m_{\pi(1)}}\otimes||_{{}_{q}} \, 
\kappa^{\pi(2)}_{\ell_{\pi(2)},m_{\pi(2)}} \big).
\]
Here the kernels $\kappa^{\pi(i)}_{\ell_{\pi(i)},m_{\pi(i)}}$ range independently over the  kernels of the finite Fock expansion
of the operator $\Xi' = \mathcal{L}(x_{\pi(i)})$. The (symmetrized/antisymmetrized) double limit $q_i$-contractions $\otimes||_{{}_{q_{i}}}$ are performed
upon the pairs of variables in which the first element of the contracted pair lies among the last $m_{\pi(i)}$ variables of the kernel 
$\theta_{{}_{\pi(2)}}\kappa^{\pi(1)}_{\ell_{\pi(1)},m_{\pi(1)}}$ and the second variable
of the contracted pair lies among the first $l_{\pi(2)}$ variables of the kernel 
$\kappa^{\pi(2)}_{\ell_{\pi(2)},m_{\pi(2)}}$, and to both variables, respectively,
correspond to the annihilation and creation operator of the 
free fields with non-zero pairing. The number 
$c(q)$ is equal to the number of Fermi commutations performed 
in the double limit contraction $\otimes||_{{}_{q}}$. 
\[
\textrm{In case} \, n \geq 2:
\]
\begin{multline*}
\Xi^\Omega =
\sum_{\substack{\kappa^{i}_{\ell_{i},m_{i}} \\ i\in\{1,2\}, \, q}} \,\,\,
(-1)^{c(q)}
\Xi\Big(\textrm{ret} \, \big( \kappa^{1}_{\ell_{1},m_{1}}\otimes_{{}_{q}} \, 
\kappa^{2}_{\ell_{2},m_{2}} \big) \Big)
-
\sum_{\substack{\kappa^{i}_{\ell_{i},m_{i}} \\ i\in\{1,2\}, \, q}} \,\,\,
(-1)^{c(q)}
\Xi\Big(\textrm{ret} \, \big( \kappa^{2}_{\ell_{2},m_{2}}\otimes_{{}_{q}} \, 
\kappa^{1}_{\ell_{1},m_{1}} \big) \Big)
\\
- 
\sum_{\substack{\kappa^{i}_{\ell_{i},m_{i}} \\ i\in\{3,4\}, \, q}} \,\,\,
(-1)^{c(q)}
\Xi\Big(\kappa^{1}_{\ell_{1},m_{1}}\otimes_{{}_{q}} \, 
\kappa^{2}_{\ell_{2},m_{2}} \Big).
\end{multline*}
The kernels
$\kappa^{1}_{\ell_{1},m_{1}}$  range over the kernels of $S(Y,x_n)$ and
$\kappa^{2}_{\ell_{2},m_{2}}$ range over the kernels of $\overline{S}(X)$, with $X \sqcup Y= \{x_1, \ldots, x_{n-1}\}$,
$X \neq \emptyset$.

\label{WickThmForChronologicalEU}
\end{twr}

{\bf REMARK}. In Subsection \ref{splittingEU}, Thm. \ref{retkappa_qEUcompact}, we have given explicit formula for the computation
of
\[
\theta \kappa^{1}_{\ell_{1},m_{1}}\otimes||_{{}_{q}} \, 
\kappa^{2}_{\ell_{2},m_{2}} =
\theta \kappa^{1}_{\ell_{1},m_{1}}\otimes_{{}_{q}} \, 
\kappa^{2}_{\ell_{2},m_{2}} \circ \Omega',
\]
for the kernels $\kappa^{i}_{\ell_{i},m_{i}}$ ranging over the products of kernels of free fields (e.g. products kernels of $\mathcal{L}(x)$
and $\mathcal{L}(y)$), in case of scalar contractions
\[
\textrm{ret} \, \kappa_q = \kappa^{1}_{\ell_{1},m_{1}}\otimes||_{{}_{q}} \, 
\kappa^{2}_{\ell_{2},m_{2}}, \,\,\,\, q=\ell_{1}+m_{1}=\ell_{2}+ m_{2},
\]
in which all (discrete) momenta variables are contracted, i.e. for products of $q$ pairings, provided we are given
the Fourier transform $\widetilde{\kappa_q}$ explicitly. This assumption is not arbitrary, because for the (compactified) Einstein Universe
with (periodic) $\theta$-function, the complete system of solutions underlying single particle states of free fields
can be computed explicitly, and the completeness relations allow to compute also $\widetilde{\kappa_q}$ explicitly.   
In fact knowledge of $\textrm{ret} \, \kappa_q$ for the pairings entering the Wick decomposition of $\mathcal{L}(x)\mathcal{L}(y)$
is sufficient to reconstruct all higher order contributions $S_n$ to the scattering operator, compare Subsection \ref{WickForChronological}. 
\qed

In the same way we construct the Fock decomposition, or the Wick Theorem, for the anti-chronological
product, by replacing the step theta function $\theta$ in the above Theorem with the step function
$-\check{\theta}$, where $\check{\theta}(x) = \theta\big(x^{-1}\big)$.

The first difference between the above stated two cases, 1) and 2), is that only in the case 1) the scattering operator
$S$ is determined uniquely. In case 2) we have arbitrary constatnts $C_{\alpha}^{I}$ and $C_{\beta}^{II}$ of the reminder kernels,
in the construction of $\textrm{ret} \, \kappa_q$ for the basic distributions $\kappa_q$ of the theory -- 
the products of pairings in the Wick decomposition of $\mathcal{L}(x)\mathcal{L}(y)$, 
which remain to be established. Some further important differences between the two 
cases are given in this Subsection and in the next Subsection.

Having given the chronological $S_n$ and anti-chronological product $\overline{S_n}$, we can compute the interacting field
$\big(\mathbb{A} \big)_{{}_{\textrm{int}}}$ for each Wick polynomial $\mathbb{A}$ in free fields, accordingly to the
general Bogoliubov formula, compare Subsection \ref{CausalSonEU}. It is important that each higher order contribution
to interacting field   $\big(\mathbb{A} \big)_{{}_{\textrm{int}}}$ is equal to finite sum of integral kernel operators.
Their Wick product makes sense and can immediately be computed by the ordinary pointwise product of the 
corresponding kernels (taken in the same space-time coordinate, \emph{i.e.} the dot product operation $\dot{\otimes}$, eventually
symmetrized and antisymmetrized), compare Subsection \ref{psiBerezin-Hida}. We can use this opportunity to impose the natural
condition 
\[
\big({:}\mathbb{A}_1 \mathbb{A}_2{:}\big)_{{}_{\textrm{int}}} \,\,\, = \,\,\,\ 
{:}\big(\mathbb{A}_1 \big)_{{}_{\textrm{int}}}\big(\mathbb{A}_2 \big)_{{}_{\textrm{int}}}{:},
\]
well-defined, because each higher order contribution to $\big(\mathbb{A}_i \big)_{{}_{\textrm{int}}}$ 
has the form of finite sum of integral kernel operators, and, as we know, Wick product for such operators
can be very easily computed (of course we order the terms as in the formal power series formula with respect to the coupling
constant or constants). It is an open problem if, in case 2), this condition can eliminate (partially at least) 
the arbitrariness involved in the constatnts $C_{\alpha}^{I}$ and $C_{\beta}^{II}$ of the reminder kernels
in the construction of $\textrm{ret} \, \kappa_q$. On the Minkowski space-time the arbitrariness in $\textrm{ret} \, \kappa_q$ 
could have been eliminated
by imposing the natural equations of motion for the interacting fields. It is an open problem, if the same holds, in case 2),
for the Einstein Universe (compactified or not, with the $\theta$-function periodic, or, respectively, ordinary non periodic
on the ordinary non compactified EU).

Moreover, having given the Wick theorem for the chronological and anti-chronological product, now determined uniquely, in case 1) at least, 
we can compute the Schwinger's complete Green functions,
compare Subsection \ref{Green}. But now 
on the Einstein Universe, where the Noether integral operators 
$\boldsymbol{P}^\mu$ and $\boldsymbol{P}^\mu\boldsymbol{P}_\mu$ 
on the Fock space (corresponding to the one parameter subgroups of space-time symmetry subgroup 
generated by $\boldsymbol{X}^\mu$ -- the formal analogue of the translation subgroup on the Minkowski space-time) have purely
discrete spectra, by the results of Subsection \ref{GeneralizedSchrodinger-VonNeumannPairs}, and \ref{WhiteNoiseFreeFieldsonEU}. 
Therefore, the Schwinger-Bogoliubov method for the construction of the well defined state in the Fock space
corresponding to Schwinger's many particle wave function is justified, in accordance to Subsection \ref{Green}.
Possibly this is so also in case 2), but we have the non uniqueness problem not solved yet in this case. 

In case 2) we have the nonuniqueness problem of the scattering operator unsolved yet but, even assuming
that it can be positively solved (in analogy to the Minkowski space-time) we have another problem in case 2):
the higher order contributions to the
interacting $\boldsymbol{P}^\mu$ and $\boldsymbol{P}^\mu\boldsymbol{P}_\mu$ (the Noether integrals in which we replace 
the products of classical fields with the interacting counterparts of the Wick products of the free fields), 
are too singular and transform the Hida space $(\boldsymbol{E})$ into its strong dual 
$(\boldsymbol{E})^*$ in case 2). 

The QFT with Lagrange density of interaction operator $\mathcal{L}$, with monomials containing more than one massless 
(or infinite orbit) free field, seem to have very restricted physical status. Such QFT can be applied to the scattering phenomena,
but not to the bound state problem.  Also in case 2) the Emmy Noether integrals 
computed for interacting fields
(for space-times with symmetry, such as the Einstein Universe)
are not ordinary self-adjoint operators on the Fock space although they should be. They again belong to 
\[
\mathscr{L}((\boldsymbol{E}),(\boldsymbol{E})^*),
\]
but not to 
\[
\mathscr{L}((\boldsymbol{E}),(\boldsymbol{E})).
\]
In particular the spectral theory cannot be applied to these Noether integrals, which, from the physical point of view,
is a serious flow, and suggests that QFT with the interactions $\mathcal{L}$, which fall into the case 2), should be considered as not
fully legitimate Quantum Field Theories.

To the contrary the QFT with interactions $\mathcal{L}$ of case 1), whose each Wick monomial contains at most one massless,
or infinite orbit, free field, should be regarded as fully legitimate Quantum Field Theory. For example QED falls into this class,
and is a fully legitimate QFT on the Einstein Universe (contrary to the case of Minkowski space-time, where even QED has very restricted meaning,
and can be applied only to the scattering processes with the multi-particle gneralized plane waves as the \emph{in} and \emph{out} states). 
In such theories 
\[
\big(\mathbb{A} \big)_{{}_{\textrm{int}}} \in
\mathscr{L}(\mathscr{E}, \mathscr{L}((\boldsymbol{E}),(\boldsymbol{E}))\big),
\]
\[
\big(\mathbb{A} \big)_{{}_{\textrm{int}}}(\phi) \in 
\mathscr{L}((\boldsymbol{E}),(\boldsymbol{E})), \,\,\,\,\,\, \textrm{for} \,\,\,\,\ \phi \in \mathscr{E},
\]
the local fluctuations of the observable fields make sense and the Noether integrals, computed for interacting fields
(at least this is the case for each higher order contribution) belong to 
\[
\mathscr{L}((\boldsymbol{E}),(\boldsymbol{E})), 
\]
and are essentially self-adjoint operators on the Fock space if they are observable local fields. 
Therefore, in our closer investigation of the next Subsection we restrict our
attention to interacting fields in QFT theories with interaction $\mathcal{L}$ whose each Wick monomial contains 
at most one massless (or infinite orbit) free field, e.g. such as QED.

\subsection{Causal perturbative construction of interacting fields \\ 
on the Einstein Universe}\label{CausalSonEU}

In this Subsection we show that for any Wick polynomial $\mathbb{A}$ in free fields, and in theory whose each monomial of 
the Lagrange density $\mathcal{L}$ of interaction contains at most one massless, or infinite orbit, field -- the case 1) 
of Subsection \ref{WickForChronologicalEU} (as e.g. QED) the higher order contributions to the interacting field, $\big(\mathbb{A} \big)_{{}_{\textrm{int}}}$, 
defined through the chronological product or through the scattering operator,
due to Bogoliubov formula \cite{Bogoliubov_Shirkov}, \S 40.5, belong to
\[
\mathscr{L}(\mathscr{E}, \mathscr{L}((\boldsymbol{E}),(\boldsymbol{E}))\big),
\]
\emph{i.e.} define continuous maps
\[
\mathscr{E} \ni \phi \longmapsto
\big(\mathbb{A} \big)_{{}_{\textrm{int}}}(\phi) \in
\mathscr{L}((\boldsymbol{E}),(\boldsymbol{E})).
\] 
Moreover, we will show in this Subsection that QFT of case 1), including QED, are so regular that higher order contributions to $\big(\mathbb{A} \big)_{{}_{\textrm{int}}}(x)$, evaluated at single space-time point, transform the Hida space $(\boldsymbol{E})$ continuously into itself $(\boldsymbol{E})$, and become
ordinary operators on the Fock space:
\[
\textrm{all contributions of order $>0$ to} \,\,
\big(\mathbb{A} \big)_{{}_{\textrm{int}}}(x) \in \mathscr{L}((\boldsymbol{E}),(\boldsymbol{E})),
\]
whenever the Wick polynomial $\mathbb{A} $ is of degree one. Correspondingly,  the higher order contributions to the conserved Noether integrals
are ordinary self-adjoint operators acting within the Fock space in case 1). 
The higher order contributions to the interacting
electromagnetic potential field and the interacting Dirac field, evaluated at single space-time point $x$,
are ordinary operators on the Fock space transforming continuously the Hida space into itself. Also zero order term,
\emph{i.e.} the free Dirac field, being massive, evaluated at single space-time point is an ordinary operator on the Fock space, 
transforming continuously the Hida space into itself, by the results of Subsection \ref{WhiteNoiseFreeFieldsonEU}. Of course the massless free
fields, and also the free electromagnetic potential field, are more singular and, when evaluated at single space-time
point $x$, are not equal to ordinary operators on the Fock space, but belong to
\[
\mathscr{L}((\boldsymbol{E}),(\boldsymbol{E})^*),
\] 
in accordance to the general results of  Subsection \ref{WhiteNoiseFreeFieldsonEU}. Therefore, in case of mass less
(or infinte orbit) interacting fields, all higher order, except the zero order, contributions evaluated at single
space-time point are ordinary operators on the Fock space and belong to 
\[
\mathscr{L}((\boldsymbol{E}),(\boldsymbol{E})),
\] 
in theories with the Lagrange density of interaction $\mathcal{L}$ which falls into the case 1) of Subsection \ref{WickForChronologicalEU}. 

The above stated result can be inferred, even together with the recurrence rule for the higher order
contributions to the interacting fields, using the recurrence rule for the scattering operator, or the Wick Theorem for the natural 
chronological product, of the previous Subsection. We nonetheless present here a ``slowly'' analysis
by introducing the scattering operator with an explicit adaptation of the causality axioms (I)-(V) to the Einstein Universe.
For the formulation of the axioms (I)-(V) in the particular case of Minkowski space-time, compare Subsections \ref{WickForProduct},
\ref{MotivationForHida}. 
Although our method is general, we concentrate attention on the particular example of QFT with interaction $\mathcal{L}$
of the first case 1) of the previous Subsection. Namely, we consider the spinor QED interaction density $\mathcal{L}$
(but the reader may substitute any other example of case 1) of the interaction $\mathcal{L}$). 
Necessary repetitions of the arguments used in the proof of Theorem \ref{Sin(SE)xn->((E)->(E))EU} of Subsection \ref{WickForProductOnEU}, 
will appear here.

For interactions $\mathcal{L}$ of case 2) the interacting fields are more singular. In particular, when 
evaluated at single space-time point, transform the Hida space $(\boldsymbol{E})$ continuously into $(\boldsymbol{E})^*$, and become
generalized operators:
\[
\textrm{all contributions of order $>0$ to} \,\,
\big(\mathbb{A} \big)_{{}_{\textrm{int}}}(x) \in \mathscr{L}((\boldsymbol{E}),(\boldsymbol{E})^*),
\]
not acting within the Fock space. Correspondingly,  the higher order contributions to the conserved Noether integrals
are no longer ordinary operators acting within the Fock space in case 2). But still
\[
\textrm{all higher order contributions to} \,\,
\big(\mathbb{A} \big)_{{}_{\textrm{int}}} \in 
\mathscr{L}(\mathscr{E}, \mathscr{L}((\boldsymbol{E}),(\boldsymbol{E}))\big),
\]
and interacting fields are well-defined operator valued distributions, also in case 2). Explicit formulas in case 2) are more complicated, due to the
more complicated formula for distribution splitting, which must be applied in case 2), 
namely the formula of Thm. \ref{retkappa_qEUcompact} of Subsection \ref{splittingEU}, for the compactified Einstein Universe with 
periodic $\theta$. The splitting formula (\ref{DisparsionFormulaRetkappaqEUnonCompactIntInp10}), Subsection \ref{splittingEU}}, 
in case 2) for the ordinary non compactified Einstein Universe with ordinary non periodic $\theta$
is slightly simpler, but we have not finished its investigation yet. Also the problem of non uniqueness of the scattering operator, due to non uniqueness
of the splitting in case 2), is still unsolved yet on the Einstein Universe.

Thus, in this Subsection we construct the scattering functional $S(g)$ or $S(g\mathcal{L})$
on the Einstein Universe $\mathbb{R}\times SU(2,\mathbb{C})= \mathbb{R}\times \mathbb{S}^3$, 
using precisely the causal perturbative  method of St\"uckelberg-Bogoliubov-Epstein-Glaser,
compare \cite{Bogoliubov_Shirkov}, \cite{Epstein-Glaser} or Subsection \ref{MotivationForHida}.
The interacting fields on the Einstein universe are defined due to Bogoliubov-Shirkov
from the scattering operator by a functional derivation of the scattering functional,
compare \cite{Bogoliubov_Shirkov} or Subsection \ref{A(1)psi(1)}. In fact Theorem \ref{Sin(SE)xn->((E)->(E))EU}
of Subsection \ref{WickForProductOnEU} already subsumes the construction of the scattering operator on the Einstein Universe
in case 1) (and Theorem \ref{GeneralSin(SE)xn->((E)->(E))EU}
of Subsection \ref{WickForProductOnEU} in case 2)) but, as we have already said, here we base this construction
on the causality axioms (I)-(IV) in more details. 

As we have seen the isometry group of the Einstein Universe 
$\mathbb{R}\times SU(2,\mathbb{C})$ is $\mathbb{R}\times SU(2,\mathbb{C})\times SU(2,\mathbb{C})$
with the Einstein Universe (our space-time) being the group 
$\mathbb{R}\times SU(2,\mathbb{C})$ naturally identified with the subgroup 
$\mathbb{R}\times SU(2,\mathbb{C}) \times \{ \boldsymbol{1} \}$ of the isometry group.
The essential point is that the causal compactification of the Minkowski spacetime is naturally periodically
and causally covered by the Einstein Universe, which among other things determines a natural relationship 
between the Minkowski and Einstein wave packets, which allows comparison of the 
scattering phenomena on the Minkowski and Einstein Universe, \cite{SegalZhouQED}, \cite{PaneitzSegalI}
-- \cite{PaneitzSegalIII}. Most important fact concerning free fields is that all single particle
states are periodic with common period $4\pi$ for all fields.
All commutation $[\mathbb{A}^\mu(x), \mathbb{A}^{\nu}(y)]_{\pm}$,
$[\boldsymbol{\psi}^a(x), \boldsymbol{\psi}^{b}(y)]_{\pm}$, $\ldots$
 and pairing functions $[\mathbb{A}^{(-) \, \mu}(x), \mathbb{A}^{(+) \, \nu}(y)]_{\pm}$,
$[\boldsymbol{\psi}^{(-) \, a}(x), \boldsymbol{\psi}^{(+) \, b}(y)]_{\pm}$
 of free fields $\mathbb{A}^a(x), \boldsymbol{\psi}^a(x)$ on the Einstein Universe have the general form
\[
\begin{split}
[\mathbb{A}^\mu(x), \mathbb{A}^{\nu}(y)]_{\pm} = -i D^{\mu \nu}(xy^{-1}), \\
[\boldsymbol{\psi}^a(x), \boldsymbol{\psi}^{\sharp b}(y)]_{\pm} = -i S^{ab}(xy^{-1}), \\
\vdots
 \\
[\mathbb{A}^{(-) \, \mu}(x), \mathbb{A}^{(+) \, \nu}(y)]_{\pm} = -i D^{(-) \, \mu \nu}(xy^{-1}), \\ 
[\boldsymbol{\psi}^{(-) \, a}(x), \boldsymbol{\psi}^{(+) \, \sharp b}(y)]_{\pm} = -i S^{(-) \, ab}(xy^{-1}), 
\\
\vdots
\end{split}
\]
with
\[
D^{\mu \nu},S^{ab}, \ldots,  D^{(-) \, \mu \nu}, S^{(-) \, ab}, \ldots \in \mathscr{E}^* = \mathcal{S}_{\Delta}\big(\widetilde{\mathbb{S}^1}\times SU(2,\mathbb{C}); \mathbb{C}\big)^*
\]
where the group multiplication operation
\[
xx' = (t+t')\times w w',
\,\,\,\, x=t \times w, x'= t'\times w' \in \mathbb{R}\times SU(2,\mathbb{C})
\] 
in $\mathbb{R}\times SU(2,\mathbb{C})$ we write multiplicatively. Because all single and many-particle states are periodic with common period $4\pi$ then all said distributions 
\[
D^{\mu \nu},S^{ab}, \ldots,  D^{(-) \, \mu \nu}, S^{(-) \, ab}, \ldots
\] 
are periodic and effectively belong to $\mathscr{E}^* = \mathcal{S}_{\Delta}\big(\widetilde{\mathbb{S}^1}\times SU(2,\mathbb{C}); \mathbb{C}\big)^*$. Thus in constructing the fee fields, the commutation and pairing functions
as well as the scattering operator functional we can replace the Einstein Universe space-time
by its compatification $\widetilde{\mathbb{S}^1}\times SU(2,\mathbb{C})$. This we will done in the sequel.

The said commutation and  pairing functions are the immediate analogues of 
the commutation and pairing functions 
\[
\begin{split}
[\mathbb{A}^\mu(x), \mathbb{A}^{\nu}(y)]_{-} = -i D^{\mu \nu}(x-y), \\
[\boldsymbol{\psi}^a(x), \boldsymbol{\psi}^{\sharp b}(y)]_{+} = -i S^{ab}(x-y), \\
\vdots
 \\
[\mathbb{A}^{(-) \, \mu}(x), \mathbb{A}^{(+) \, \nu}(y)]_{-} = -i D^{(-) \, \mu \nu}(x-y), \\ 
[\boldsymbol{\psi}^{(-) \, a}(x), \boldsymbol{\psi}^{(+) \, \sharp b}(y)]_{+} = -i S^{(-) \, ab}(x-y), 
\\
\vdots
\end{split}
\]
on the Minkowski space-time, isomorphic to the Abelian group of translations 
$T_4 \subset T_4 \circledS SU(2, \mathbb{C})$, where the group action in $T_4$ was written
additively. The (compactified Einstein Universe)  group $\widetilde{\mathbb{S}^1}\times SU(2,\mathbb{C})$ plays in construction of free fields and the scattering operator on the Einstein Universe exactly the same role which the group of translations $T_4$ played in construction of free fields and the scattering 
matrix on the  Minkowski space-time. In particular  $\widetilde{\mathbb{S}^1}\times SU(2,\mathbb{C})$
being a group allows a natural construction of convolution in the function spaces on it, in particular in the standard nuclear test function space 
$\mathscr{E} = \mathcal{S}_{\Delta}\big(\widetilde{\mathbb{S}^1}\times SU(2,\mathbb{C}); \mathbb{C}\big)$ and in its strong dual $\mathscr{E}^* = \mathcal{S}_{\Delta}\big(\widetilde{\mathbb{S}^1}\times SU(2,\mathbb{C}); \mathbb{C}\big)^*$.

In particular the higher order contributions to the scattering operator, and in particular to the interacting
fields, on the Einstein Universe have the same general form as on the Minkowski space-time,
with the nuclear test function spaces $E_1$ $E_2$, $\ldots$ in the corresponding single particle Hilbert 
spaces of the corresponding free fields of the theory (constructed in the previous Subsections) and
with the corresponding nuclear space-time test function space  $\mathscr{E}$ on the Einstein Universe
in the formulas. In the general formulas for the higher order terms the convolution on the Minkowski space-time (identifilable with the translation group
$T_4$), is replaced with the convolution on the (compactified) Einstein Universe group
$\widetilde{\mathbb{S}^1}\times SU(2,\mathbb{C})$.
In particular the general theorems of 
Subsection \ref{OperationsOnXi}, are likwise applicable to higher order contributions on the Einstein Universe with the convolution on the translation group replaced with the convolution on the compactified
Einstein Universe group.  

Although the replacements are rather obvious we present them explicitly (even if some repetitions will necessary appear) in order to emphasize those circumstances thanks to which the higher order contributions to interacting fields on the Einstein Universe are not only integral kernel operators transforming
continuously the Hida space into itself, but moreover the higher order contributions to interacting fields even when evaluated at single space-time point are well defined operators transforming continuously the Hida space into itself, and  thus represent ordinary (unbounded) operators
in the Fock space. Similarly, the higher order contributions to the scattering operator
on the Einstein Universe are equal to  ordinary operators in the Fock space (as we already know form Theorem \ref{Sin(SE)xn->((E)->(E))EU}
of Subsection \ref{WickForProductOnEU}). 

Although our presentation is in principle general we will concentrate here on the 
QED example with the free electromagnetic field $A^\mu(x)$ and the free Dirac field
$\boldsymbol{\psi}^a(x)$ as the only free fields of the theory
with the commutation and pairing functions
\[
\begin{split}
[A^\mu(x), A^{\nu}(y)]_{-} = -i D^{\mu \nu}_{0}(xy^{-1}), \\
[\boldsymbol{\psi}^a(x), \boldsymbol{\psi}^{\sharp b}(y)]_{+} = -i S^{ab}(xy^{-1}), \\
[A^{(-) \, \mu}(x), A^{(+) \, \nu}(y)]_{-} = -i D^{(-) \, \mu \nu}_{0}(xy^{-1}), \\ 
[\boldsymbol{\psi}^{(-) \, a}(x), \boldsymbol{\psi}^{(+) \, \sharp b}(y)]_{+} = -i S^{(-) \, ab}(xy^{-1}), 
\\
\end{split}
\]
computed in the previous Subsections, and with the
interaction Lagrangian density 
\[
\mathcal{L}(x) = e \, \boldsymbol{{:}} \psi^\sharp(x) \gamma_\nu A^\nu(x) \psi(x_1) \boldsymbol{{:}},
\]
and with $\sharp$ denoting the Dirac conjugation.

The free fields, understood as integral kernel operators in the sense of Obata,
we insert into the formulas for the causal perturbative series for interacting fields. The necessary
operations of Wick product, splitting, integrations, have a rigorous meaning as operations
performed upon integral kernel operations explained in  Subsection \ref{OperationsOnXi}. 
Their adaptation to the Einstein Universe is given in Subsections \ref{WhiteNoiseFreeFieldsonEU}
and \ref{splittingEU}. In case of QED with just one massless field in $\mathcal{L}$ the splitting
simplifies substantially, and reduces to the ordinary multiplication by $\theta$-function (periodic
in case of compactified Einstein Universe), and this is the case for the splitting of products
of pairing which contains at most one massless (or infinite orbit) pairing, for the proof compare
Subsection \ref{WhiteNoiseFreeFieldsonEU}.

The formulas for the higher order contributions are exactly the same as in the standard perturbative 
causal spinor QED, compare e.g. \cite{DKS1} or \cite{Scharf}, but with the Wick product and integration
in these formulas rigorously understood as performed upon integral kernel operators and expressed
by the Rules of  Subsection \ref{OperationsOnXi}. 

Recall once again that in order to define (the higher order contributions to) the local interacting 
field $\mathbb{A}_{{}_{\textrm{int}}}(x)$ corresponding to a free field $\mathbb{A}(x)$ we 
are using the Bogoliubov-Shirkov \cite{Bogoliubov_Shirkov} definition of the interacting field as the variational derivative 
\[
\mathbb{A}_{{}_{\textrm{int}}}(x) = \mathbb{A}_{{}_{\textrm{int}}}(g=1,x)
\]
where
\begin{multline*}
\mathbb{A}_{{}_{\textrm{int}}}(g,x) =  S^{-1}(g\mathcal{L})
\frac{\delta S(g\mathcal{L}+h\mathbb{A})}{\delta h(x)}\Bigg{|}_{{}_{h=0}}
\\
= \mathbb{A}(x) + \sum \limits_{n=1}^{\infty} {\textstyle\frac{1}{n!}} 
\int \ud^4 x_1 \cdots \ud^4 x_n \mathbb{A}^{(n)}(x_1, \ldots, x_n,x) \,
g(x_1) \cdots g(x_n)
\end{multline*}
has the form of the formal functional power series. Here we are using the scattering
operator $S(g\mathcal{L})$, denoted in Subsection \ref{MotivationForHida}
shortly by $S(g)$, and with $g$ equal to the ``switching-of-interaction-function'' $g$. 
Here on the Einstein Universe the free fields, the pairing and commutation functions,
the``intensity-of-interaction'' function $g$ are all periodic and the integrals run over the 
compactified Einstein Universe or over its $n$-fold Cartesian product, with the time variable
confined to the interval $(-2\pi, 2\pi]$ (in units in which $R=c=\hslash = 1$, in other units
the time interval will have to be rescaled as we have already indicated in Subsection \ref{GeneralizedSchrodinger-VonNeumannPairs}). 

The operator functional $S(g)=S(g\mathcal{L})$ with the Lagrange density
interaction of the theory (here the Lagrange density interaction $\mathcal{L}$ of QED)
is constructed as the formal functional power series
\[
\begin{split}
S(g\mathcal{L}) = \boldsymbol{1}+
\sum \limits_{n=1}^{\infty} {\textstyle\frac{1}{n!}} 
\int \ud^4 x_1 \cdots \ud^4 x_n S(x_1, \ldots, x_n) \,
g(x_1) \cdots g(x_n),
\\
S^{-1}(g\mathcal{L}) = \boldsymbol{1} +
\sum \limits_{n=1}^{\infty} {\textstyle\frac{1}{n!}} 
\int \ud^4 x_1 \cdots \ud^4 x_n \overline{S}(x_1, \ldots, x_n) \,
g(x_1) \cdots g(x_n),
\end{split}
\]
inductively within the causal method  explained in Subsection
\ref{MotivationForHida}, based on the axioms (I)-(IV) of Subsection \ref{MotivationForHida}.
These axioms can be naturally formulated and applied to the scattering operator functional on the Einstein Universe, compare \cite{Bogoliubov_Shirkov}, \cite{Bogoliubov-Shirkov}.
Let us start by going through shortly the motivation for the causality axiom (I) and its formulation in terms of operator kernels $S_n(x_1, \ldots, x_n)$.
For this purpose let $G_1$ and $G_2$ be two regions in space-time. By the periodicity,
we may confine ourselves without any loss of generality to the case in which both, $G_1$ and $G_2$, lie
between two equal time Cauchy surfaces separated by the time interval not exceeding
$2\pi$ (in units $R=c=\hslash=1$). Let moreover all points of $G_1$ precede an equal-time Cauchy surface $t=\tau$  and all points of $G_2$ have time coordinate greater than $\tau$,
which we denote $G_1 \prec G_2$. 
\begin{center}
\begin{tikzpicture}[yscale=1]
    \draw[thin, ->] (0,1.8) -- (0,3.5);
    \draw[ultra thick] (0.99,0.1) -- (0.99,-2.9);
    \draw[ultra thick] (-0.99,-0.1) -- (-0.99,-3.25);


\draw[ultra thick] (0.99,0.1) -- (0.99,2.1);
 \draw[ultra thick] (-0.99,-0.1) -- (-0.99,1.9);

\draw[ultra thick] (0.99,1) -- (0.99,2.1);
 \draw[ultra thick] (-0.99,0.7) -- (-0.99,1.9);





\draw[ultra thin,dashed, rotate=10] (0.9,-0.2) arc(10:-160:1cm and 0.5cm);

\draw[ultra thick, rotate=10] (0.45,-2.85) arc(16:-175:1cm and 0.5cm);
\draw[ultra thick, rotate around = {10:(0,2)}] (0,2) ellipse (1cm and 0.5cm);



\draw [<-,very thin] (1,1.7) to [out=-70,in=130] (2.2,1.5);

\draw [<-,very thin] (0.7,-0.6) to [out=-70,in=130] (1.6,-0.8);
\draw [->,very thin] (-1.5,-3.5) to [out=10,in=230] (-0.75,-3.5);
\draw [->,very thin] (-2,2.2) to [out=-70,in=130] (-1,2);

\fill[color=gray!, fill opacity=0.1,rotate around = {20:(0,0.15)}] (0,0.4) ellipse (0.5 and 0.3);
\fill[color=gray!, fill opacity=0.1] (0,-2.2) ellipse (0.3 and 0.8);

\node [right] at (2.2, 1.5) {$\textrm{Einstein Universe}$};

\node [right] at (1.6, -0.8) {$t = \textrm{const}$};
\node [left] at (0,3.5) {$t$};
\node [left] at (-3,0) {$G_1 \prec G_2$};
\node [left] at (-1.5,-3.5) {$t=-2\pi$};
\node [left] at (-2,2.2) {$t=2\pi$};

\node [above] at (0,0.13) {$G_2$};
\node [below] at (0,-1.8) {$G_1$};
\end{tikzpicture}
\end{center}
Let $g_1$ and $g_2$ be two smooth ``intensity of interaction'' functions with the property
$\textrm{supp} \, g_1 \subset G_1$ and $\textrm{supp} \, g_2 \subset G_2$.
Then consider the ``intensity of interaction'' 
$g = g_1 + g_2$. Then the interaction $g\mathcal{L}$ is switched off completely
outside $G_1 \cup G_2$, and the state of the system $\Phi$ in times preceding 
both regions $G_i$ evolves as a state of a free system, and  it is a constant state in
the ``interaction picture'' in the times preceding both $G_i$. Similarly the state of the system in the ``interaction picture'' is constant in time for all times later than both $G_i$. The scattering operator $S(g) = S(g\mathcal{L})$ transforms the initial state $\Phi$ into the final  $S(g)\Phi
= S(g\mathcal{L})\Phi$. This final state can be obtained from the
intermediate state $\Phi_{{}_{\tau}}$ of the system which it assumes at time $t=\tau$.
Now the causality requirement says that the state  $\Phi_{{}_{\tau}}$ should not depend
on the interaction in the region $G_2$ lying in the future of $t=\tau$, \emph{i.e.}
\[
\Phi_{{}_{\tau}} = S(g_1)\Phi = S(g_1\mathcal{L})\Phi,
\]
where $S(g_1\mathcal{L})$ is the scattering operator in which the interaction is switched on
with the intensity $g_1$; and moreover the final state $S(g\mathcal{L})\Phi$
should be obtainable from the intermediate one $\Phi_{{}_{\tau}}$
by the application of the scattering operator $S(g_2\mathcal{L})$
with the intesity of interaction equal $g_2$:
\[
S(g\mathcal{L})\Phi = S(g_2\mathcal{L})\Phi_{{}_{\tau}}.
\]
We therefore arrive at the causality condition
\[
(\textrm{I}) \,\,\,\,\,\,\,\,\,\,\,\,
S\big((g_1 + g_2)\mathcal{L}\big) = S(g_2\mathcal{L}) S(g_1\mathcal{L})
\,\,\,\,\,\,
\textrm{if}
\,\, G_1 \prec G_2.
\]
Here we do not have at our disposal the Lorentz transformations, and we cannot prolong
the said requirement to the case the regions $G_1$ and $G_2$ are mutually space-like,
as we did in Subsection \ref{A(1)psi(1)}, compare also \cite{Bogoliubov-Shirkov},
Chap. IV.\S 15.3 or \cite{Bogoliubov_Shirkov}, Chap. IV \S 20.5. Going after
\cite{Bogoliubov_Shirkov}, Chap. IV \S 20 (compare also \cite{Bogoliubov-Shirkov}, Chap. III \S\S15,
16 or \cite{Scharf}, Chap. 3.1) we can express the causality condition in terms
of the kernels $S_n(x_1, \ldots, x_n)$. Because of the periodicity, we identify the space-time points
$x,x'$ whenever difference $|x_0 - x'_{0}| = m4\pi$ of their time coordinates $x_0$ and $x'_{0}$ is equal to integer $m$ multiple of $4\pi$, and  we can confine ourselves
to the case where time coordinates $x_{i0}$ of the space-time points $x_1, \ldots, x_n$
are separated by the time distance not exceeding $4\pi$:  $|x_{i0} - x_{k0}| < 4\pi$
for $1 \leq i,k \leq n$. Then using
the periodic step function 
\[
\theta(t+m4\pi)=
\left\{ \begin{array}{ll}
1, &  0 \leq t < 2\pi, \\
0, & -2\pi \leq t <0
\end{array} \right.,
m \in \mathbb{Z}
\]
on the reals and the corresponding step function (denoted likewise $\theta$)
\[
\theta(x) = \theta(x_0),
\,\,\,\, 
\textrm{for}
\,\,\,
x= x_0 \times w \in \mathbb{R} \times SU(2, \mathbb{C})
\]
on the Einstein Universe $\mathbb{R} \times SU(2, \mathbb{C})$, we can easily see
that, expressed in terms of the kernels $S_n(x_1, \ldots, x_n)$, the causality
condition (I) reads\footnote{Recall that 
\[
x_1(x_2)^{-1} = (x_{10}-x_{20}) \times w_1 w_{2}^{-1}
\,\,\, \textrm{for}
\,\,\, x_1 = x_{10} \times w_1, x_2 = x_{20} \times w_2 \in \mathbb{R} \times SU(2, \mathbb{C}).
\]}
\begin{multline*}
S_n(x_1, \ldots, x_n)= S_k(x_1, \ldots, x_k)S_{n-k}(x_{k+1}, \ldots, x_n)
\\
\textrm{if} \,\, 
\{x_{k+1}, \ldots, x_n\} \prec \{x_1, \ldots, x_k \}
\end{multline*}
or equivalently
\begin{multline*}
S_n(x_1, \ldots, x_n)= S_k(x_1, \ldots, x_k)S_{n-k}(x_{k+1}, \ldots, x_n)
\\
\textrm{if} \,\, \textrm{not} \,
x_{i0}  > x_{j0},
\,\,
i\in \{1, \ldots, k\}, \, j \in \{k+1, \ldots, n\}, 
\end{multline*}
or
\begin{multline*}
S_n(x_1, \ldots, x_n)= S_k(x_1, \ldots, x_k)S_{n-k}(x_{k+1}, \ldots, x_n)
\\
\textrm{if} \,\, 
x_{j0}  \leq x_{i0},
\,\,
i\in \{1, \ldots, k\}, \, j \in \{k+1, \ldots, n\}, 
\end{multline*}
Let us introduce the right translation $R_{x^{-1}}\theta$:
\[
y \longmapsto R_{x^{-1}}\theta(y) = \theta(yx^{-1}),
\]
of the periodic function $\theta$ by $x^{-1} \in \mathbb{R}\times SU(2, \mathbb{C})$.
Then we can see that the immediate analogue of the closed future light cone 
$x+\overline{V_+}$ with the apex at $x$ in the Minkowski spacetime in the causal 
construction of the 
scattering operator of Subsection \ref{MotivationForHida} and Section \ref{A(1)psi(1)} is played 
by the following closed set, say ``forward half-space'':
\[
V_{x}^{+} = \textrm{supp} \, R_{x^{-1}}\theta 
= \{y+ m4\pi: 2\pi \geq y_0 - x_0 \geq 0, m\in \mathbb{Z}  \}
\,\,\,
\textrm{analogue of}
\,\,
x+\overline{V_+},
\]
on the Einstein Universe.
Here we identify the space-time points separated by integer multiple of 
$4\pi$ in time and we have chosen the representative time interval
$(-2\pi < x_0, y_0 \leq 2\pi]$ for the space-time points $x,y$. 

Analogously we have the following closed set, say ``backward half-space''
\[
V_{x}^{-} = \textrm{supp} \, R_{x^{-1}}\theta^{-1} 
= \{y+ m4\pi: -2\pi \leq y_0 - x_0 \leq 0, m\in \mathbb{Z}  \}
\,\,\,
\textrm{analogue of}
\,\,
x+\overline{V_-},
\]
where 
\[
\theta^{-1}(x) = \theta(x^{-1}).
\]

The remaining axioms: (II) covariance under the representation of the Einstein isometry group
$\mathbb{R}\times SU(2, \mathbb{C}) \times SU(2, \mathbb{C})$ representation $\boldsymbol{U} =
\Gamma(U_1 \oplus U_2) = \Gamma(U_1) \otimes \Gamma(U_2)$ acting in the total Fock space
of the free fields $\boldsymbol{\psi}$, $A$ of the theory, (III) unitarity (Krein-isometricity
in case of QED or the Standard Model with the Higgs field with the gauge fields) and (IV)
correspondence principle identifying the first order kernel $S_1(x)$
with $i \mathcal{L}(x)$ (compare \cite{Bogoliubov_Shirkov}), where $\mathcal{L}$ is the Lagrange density of interaction of the theory
in question, remain unchanged. Here, as on Minkowski space-time, we have two obtions.
First, and more natural, assumes covariance of $S$ under the full symmetry group 
$\mathbb{R}\times SU(2, \mathbb{C}) \times SU(2, \mathbb{C})$ of the Einstein universe (in case with ordinary 
step theta function $\theta$).  We have used periodic $\theta$ function and compactified Einstein Universe 
$\widetilde{\mathbb{S}^1}\times SU(2, \mathbb{C})$ with the 
covariance of $S$ under the full symmetry group 
$\widetilde{\mathbb{S}^1}\times SU(2, \mathbb{C}) \times SU(2, \mathbb{C})$ of the compactified
Einstein Universe $\widetilde{\mathbb{S}^1}\times SU(2, \mathbb{C})$. 
Thus, we have the following axioms
\begin{align*}
(\textrm{I}) & \,\,\,\,\,\,\,\,\,\,\,\, &
S_{n}(x_1, \ldots, x_n) = S_{k}(x_1, \ldots, x_k)S_{n-k}(x_{k+1}, \ldots, x_n), \\ 
& & \,\,\,\,\,\,\, 
\textrm{if} \,\, \textrm{not} \,
x_{j0}  > x_{i0},
 \,\, i \in\{1, \ldots, k\}, j \in \{k+1, \ldots, n\}
\\
(\textrm{II} )& \,\,\,\,\,\,\,\,\,\,\,\, &
\boldsymbol{U}_{{}_{s\times u \times v}} S_n(x_1, \ldots, x_n){\boldsymbol{U}_{{}_{s \times u \times v}}}^{-1} = S_n(U_{{}_{s\times u \times v}} x_1, \ldots, U_{{}_{s\times u \times v}} x_n),
\\ 
(\textrm{III} )& \,\,\,\,\,\,\,\,\,\,\,\, &
\eta S_n(x_1, \ldots, x_n)^{+} \eta = \overline{S_n}(x_1, \ldots, x_n), \\
(\textrm{IV}) & \,\,\,\,\,\,\,\,\,\,\,\, &
S_1(x) = i \mathcal{L}(x),
\end{align*}
\begin{enumerate}
\item[(V)] \,\,\,\,\,\,\,\,\,\,\,\,
The value of the retarded part of a vector valued kernel on a test function should coincide with the natural formula given by the multiplication by the step theta 
function on a space-time test function, whenever the natural formula is meaningful for this test function.
\,\,\,\,\,\,\,\,\,\,\,\,\,\,\,\,\,\,\,\,\,\,\,\,\,\,\,\,\,\,\,\,\,\,\,\,\,\,\,\,\,\,\,\,\,\,\,\,\,\,\,\,\,\,\,\,\,\,\,\,
\end{enumerate}
It is not clear if the axiom (V) is equivalent to the following assertion:
\begin{enumerate}
\item[] \,\,\,\,\,\,\,\,\,\,\,\,
The orders of each vector valued kernel and of its retadred part for the integral kernel operators representing causal combinanions of products of $S_n$
with $S_k$, should coincide.
\,\,\,\,\,\,\,\,\,\,\,\,\,\,\,\,\,\,\,\,\,\,\,\,\,\,\,\,\,\,\,\,\,\,\,\,\,\,\,\,\,\,\,\,\,\,\,\,\,\,\,\,\,\,\,\,\,\,\,\,
\end{enumerate}
For some partial results on the relation between this asserion and axiom (V), compare Subsection \ref{splittingEU}.

Recall that kernels $\kappa_{l,m}$ of our generalized operators are elements of the (projective) tensor products
of nuclear spaces
$\mathscr{E}^{*\otimes \, n} \otimes E^*\otimes \ldots$,
being inductive limits
\[
\mathscr{E}_{-1} \subset \mathscr{E}_{-2} \subset \mathscr{E}_{-3} \subset   \ldots \subset \mathscr{E}_{-k} \subset  \ldots \subset \mathscr{E}^*
\]
 of Hilbert spaces $\mathscr{E}_{-k}$ , where $\mathscr{E}$ is the space-time test space. Similarly for the single particle nuclear spaces
$E$ and their duals $E^*$, and similarly for $\mathscr{E}^{*\otimes \, n}$. The minimal number $k =k_{1} + \ldots k_{n}$ such that
\[
\kappa_{l,m} \in \mathscr{E}_{-k_{1}} \otimes \ldots \otimes \mathscr{E}_{-k_{n}} \otimes  E^*\otimes \ldots 
\]
is called the order of $\kappa_{l,m}$. 

\vspace*{0.5cm}

Here $U_{{}_{s\times u \times v}}$ is given by the formula (\ref{UphiOnRxG}) of Subsection \ref{GeneralizedSchrodinger-VonNeumannPairs}, namely 
\[
U_{{}_{s\times u \times v}} \phi(t \times x) = 
V(v) \phi\big((t+s) \times (v^{-1}xu) \big),
\]
with
\[
s\times u \times v \in \widetilde{\mathbb{S}^1} \times G \times G, \,\, \textrm{or respectivey},
s\times u \times v \in \mathbb{R} \times G \times G, \,\,\, D-SU(2, \mathbb{C}).
\]

Applying the axiom (II) to the the subgroup 
$\widetilde{\mathbb{S}^1}\times SU(2, \mathbb{C}) \cong \widetilde{\mathbb{S}^1}\times SU(2, \mathbb{C}) \times \{\boldsymbol{1}\} \subset \widetilde{\mathbb{S}^1}\times SU(2, \mathbb{C}) \times SU(2, \mathbb{C})$ only, with periodic $\theta$ or, respectively, to the subgroup
$\mathbb{R}\times SU(2, \mathbb{C}) \cong \mathbb{R}\times SU(2, \mathbb{C}) \times \{\boldsymbol{1}\} \subset \mathbb{R}\times SU(2, \mathbb{C}) \times SU(2, \mathbb{C})$ with the ordinary non periodic $\theta$ (correspondingly to the restriction to covariance under translations on the Minkowski space-time)
we immediately coclude from the axiom (II): 
\begin{align*}
& \,\,\,\,\,\,\,\,\,\,\,\, &
\boldsymbol{U}_{s \times u} S_n(x_1, \ldots, x_n)\boldsymbol{U}_{s \times u}^{-1} = S_n(U_{{}_{s\times u}}x_1, \ldots, U_{{}_{s\times u}}x_n).
\end{align*}

Here
\[
(s,u) \in \widetilde{\mathbb{S}^{1}}\times SU(2, \mathbb{C}) \cong \mathbb{R}\times SU(2, \mathbb{C}) \times \{\boldsymbol{1}\} \subset \mathbb{R}\times SU(2, \mathbb{C}) \times SU(2, \mathbb{C}),
\]
or, respectively,
\[
(s,u) \in \mathbb{R}\times SU(2, \mathbb{C}) \cong \mathbb{R}\times SU(2, \mathbb{C}) \times \{\boldsymbol{1}\} \subset \mathbb{R}\times SU(2, \mathbb{C}) \times SU(2, \mathbb{C}),
\]
and the right action
\[
U_{{}_{s\times u}}x = U_{{}_{s\times u \times 1}}x= (s+t) \times wu,
\,\,\,
\textrm{where}
\,\,\, x=t \times w \in \widetilde{\mathbb{S}^1} \times SU(2, \mathbb{C}) 
\]
or, respectively,
\[
x=t \times w \in \mathbb{R} \times SU(2, \mathbb{C})
\]
of $(s,u)$ on the space-time point $x= t \times w \in \widetilde{\mathbb{S}^{1}} \times SU(2, \mathbb{C})$ or, respectively, 
$x= t \times w \in \mathbb{R} \times SU(2, \mathbb{C})$.

\vspace*{0.5cm}

Using this form of causality (I), covariance under the symmetry group 
$\widetilde{\mathbb{S}^1} \times SU(2, \mathbb{C}) \times SU(2, \mathbb{C})$, \emph{i.e.} (II), 
unitarity (or Krein-isometricity) (III),
correspondence principle (IV) identifying the kernel of the first order contribution
with the Lagrange density of interaction $\mathcal{L}$ multiplied by $i$, and naturality of the splitting, \emph{i.e.} axiom (V),  
we are able to construct all higher order kernels as determined by the axioms (I)--(V). 

Although the method is essentially the same as that introduced by Bogoliubov Shirkov, Epstein and Glaser, compare Subsection \ref{MotivationForHida} and Section \ref{A(1)psi(1)}, let us indicate shortly how this can be achieved in case of the Einstein Universe space-time, because there appear essential simplifications of the causal method when applied to the Einstein Universe space-time. We are using the periodic step $\theta$-fuction and, thus, compactified Einstein Universe.

So let us assume we have all $S_k(x_1, \ldots, x_k)$ for $k=1, \ldots, n-1$. After Epstein and Glaser let us introduce the following operator-valued distributions 
\[
\begin{split}
A'_{(n)}(x_1, \ldots, x_{n-1}, x_n) = \sum \limits_{P_2} \overline{S}(X)S(Y,x_n), \\
R'_{(n)}(x_1, \ldots, x_{n-1}, x_n) = \sum \limits_{P_2} S(Y,x_n)\overline{S}(X), 
\end{split}
\]
where the sums run over all divisions $P_2$ of the set $\{x_1, \ldots, x_{n-1} \}$
into two disjoint subsets $X$ and $Y$:
\[
\{x_1, \ldots, x_{n-1} \} = X \sqcup Y,
\,\,\, \textrm{with} \,\,\,
X \neq \emptyset.
\] 
Thus by assumption $A'_{(n)}$ and $R'_{(n)}$ are known. Next after Epstein and Glaser 
we introduce the following operator-valued distributions
\[
\begin{split}
A_{(n)}(x_1, \ldots, x_{n-1}, x_n) = \sum \limits_{P_{2}^{0}} \overline{S}(X)S(Y,x_n)
= \sum \limits_{P_2} \overline{S}(X)S(Y,x_n) + S(x_1, \ldots, x_n), \\
R_{(n)}(x_1, \ldots, x_{n-1}, x_n) = \sum \limits_{P_{2}^{0}} S(Y,x_n)\overline{S}(X)
= \sum \limits_{P_2} S(Y,x_n)\overline{S}(X) + S(x_1, \ldots, x_n), 
\end{split}
\]
where now summation is extended over all divisions $P_{2}^{0}$ of the set 
 $\{x_1, \ldots, x_{n-1} \}$
into two disjoint subsets $X$ and $Y$, which include the empty set $X= \emptyset$. Note that
\[
D_{(n)} = R'_{(n)} - A'_{(n)} = R_{(n)} - A_{(n)}.
\]
The point is that the construction of the (analogue of the Epstein-Glaser's) \emph{splitting} of a scalar 
$\widetilde{\mathbb{S}^1} \times SU(2, \mathbb{C})$-invariant distribution into the retarded and advanced part
 can be extended over the Einstein Universe, compare Subsection \ref{WhiteNoiseFreeFieldsonEU}. 
Moreover in case 1) of interaction $\mathcal{L}$, e.g. QED interaction, containing at most one massless 
field and with the remaining fields being massive (Dirac in QED, and thus finite orbit fields, 
compare Subsetion \ref{WhiteNoiseFreeFieldsonEU}) the splitting of the scalar 
$\widetilde{\mathbb{S}^1} \times SU(2, \mathbb{C})$-invariant  distributions which are to be splitted 
becomes unique and can be reduced to the multiplication by the the step theta function, 
in accordance to our results proved in Subsections \ref{WhiteNoiseFreeFieldsonEU} and \ref{WickForProductOnEU}.

Namely, let us introduce after \cite{Epstein-Glaser} higher 
dimensional generalization of the backward and forward ``half-spaces'':
\[
\Gamma_{\pm}^{(n)}(y) = \big\{X \in \mathcal{M}^n: x_j \in V_{y}^{\pm}  \big\}, \,\,\,\,\,
X = \{x_1, \ldots, x_n \},
\]
\[
\mathcal{M} = \widetilde{\mathbb{S}^1} \times SU(2, \mathbb{C}).
\]
Then it is shown by repeating the argument of \cite{Epstein-Glaser}, or \cite{Scharf}, Chap. 3.1
(compare also \cite{DKS1}) that 
\[
\begin{split}
\textrm{supp} \, R_{(n)}(x_1, \ldots, x_{n-1}, x_n) \subseteq \Gamma_{+}^{(n-1)}(x_n),
\\
\textrm{supp} \, A_{(n)}(x_1, \ldots, x_{n-1}, x_n) \subseteq \Gamma_{-}^{(n-1)}(x_n),
\\
\textrm{supp} \, D_{(n)}(x_1, \ldots, x_{n-1}, x_n) \subseteq \Gamma_{+}^{(n-1)}(x_n)
\sqcup \Gamma_{-}^{(n-1)}(x_n).
\\
\end{split}
\]
But the point is that each $D_{(n)}$ can be (in case of Einstein Universe space-time) uniquely 
splitted into a sum of operator distributions each having the support, respectively, in
$\Gamma_{+}^{(n-1)}(x_n)$ or in $\Gamma_{-}^{(n-1)}(x_n)$ and that this splitting can be made explicitly and independently of the conditions (I) -- (IV). The essential point is that $R_{(n)}$ and 
$A_{(n)}$ can be separately computed as the spitting of $D_{(n)}$ into the advanced $A_{(n)}$ and retarded $R_{(n)}$ parts, so that 
\[
\begin{split}
S_n(x_1, \ldots, x_{n-1}, x_n) = A_{(n)}(x_1, \ldots, x_{n-1}, x_n) - A'_{(n)}(x_1, \ldots, x_{n-1}, x_n)
\\
\,\,\,
\textrm{or equivalently}
\,\,\,
\\
S_n(x_1, \ldots, x_{n-1}, x_n) = R_{(n)}(x_1, \ldots, x_{n-1}, x_n) - R'_{(n)}(x_1, \ldots, x_{n-1}, x_n)
\end{split}
\]
and the inductive step from $n-1$ to $n$ can be computed. But we should emphasize that now
$V_{x}^{\pm}$ are equal to the foreward and backward ``half spaces''and not to the forward
and backward light cones emerging from $x$ and the splitting is unique only in case 1) when each Wick monomial
of $\mathcal{L}$ contains at most one massless (or infinite orbit field), as is the case e.g. for QED.
In case 2), as we have seen in Subsection \ref{WhiteNoiseFreeFieldsonEU}, the operator valued distributions are 
splitted only up to terms supported on the Cauchy
surfaces $t=0$ and $t=2\pi$ in each space-time variable, which complicates slightly the computation
of the splitting and of the scattering operator and, contrary to the Minkowski case, 
is far not determined solely by the single point supported distributions. Fortunately in case 1) we can forget 
about this complication in the splitting.  Here on the Einstein Universe and with interaction of 
case 1), e.g. QED, situation is much better than on the Minkowski space-time,
the kernels  $D_{(n)}$ of generalized operators (say operator valued distributions) can be uniquely and naturally splitted,
and have the same general form with the scalar $\mathbb{R}\times SU(2, \mathbb{C})$-invariant scalar distributions
depending on the variables $x_1, \ldots, x_n$ joined in pairs $x_ix_{k}^{-1}$ as their separate variables, 
which in fact follows by the said invariance. The splitting is reduced to the 
splitting of these scalar 
$\mathbb{R}\times SU(2, \mathbb{C})$-invariant distributions, which in case 1) are regular enough to have unique splitting
 defined through the pointwise multiplication by the step theta function.\footnote{This we have
already proved in Theorem \ref{Sin(SE)xn->((E)->(E))EU}
of Subsection \ref{WickForProductOnEU}, but here we look more closely at the relation of the construction of the scattering operator
to the causality axioms, and extend our analysis over the interacting fields.} 
  Let us look at this point more closely. 

Recall that the commutation functions and the pairing functions
\[
(x,y) \mapsto S^{ab}(xy^{-1}), S^{(-) \, ab}(xy^{-1}), S^{(-) \, ab}(xy^{-1})
\]
of the Dirac field (and in general for local massive field on the Einstein Universe) 
are special kinds of generalized functions 
in $\mathscr{E}^* = \mathcal{S}_{\Delta}\big(\widetilde{\mathbb{S}^1}\times SU(2, \mathbb{C})\big)^*$,
and are in fact ordinary smooth functions, \emph{i.e.} elements of 
$\mathscr{E} = \mathcal{S}_{\Delta}\big(\widetilde{\mathbb{S}^1}\times SU(2, \mathbb{C})\big)
\mathscr{C}^\infty\big(\widetilde{\mathbb{S}^1}\times SU(2, \mathbb{C})\big)$.
For them we can realize the splitting into  retarded and advanced part
with the help of ordinary multiplication by the (periodic) theta function $\theta$:
\begin{align*}
S^{ab}_{\textrm{ret}}(xy^{-1}) = \theta(x_0-y_0)S^{ab}(xy^{-1}), \\
S^{ab}_{\textrm{av}}(xy^{-1}) = -\theta(y_0-x_0)S^{ab}(xy^{-1}), \\
S^{(-) \, ab}_{\textrm{ret}}(xy^{-1}) = \theta(x_0-y_0)S^{(-) \, ab}(xy^{-1}), \\ 
S^{(-) \, ab}_{\textrm{av}}(xy^{-1}) = -\theta(y_0-x_0) S^{(-) \, ab}(xy^{-1}), \\
S^{(+) \, ab}_{\textrm{ret}}(xy^{-1}) = \theta(x_0-y_0)S^{(+) \, ab}(xy^{-1}), \\
S^{(+) \, ab}_{\textrm{av}}(xy^{-1}) = -\theta(y_0-x_0) S^{(+) \, ab}(xy^{-1}), \\
S^{ab}(xy^{-1}) = S^{ab}_{\textrm{ret}}(xy^{-1}) - S^{ab}_{\textrm{av}}(xy^{-1}), \\
S^{(-) \, ab}(xy^{-1}) =S^{(-) \, ab}_{\textrm{ret}}(xy^{-1}) -
S^{(-) \, ab}_{\textrm{av}}(xy^{-1}), \\
S^{(+) \, ab}(xy^{-1}) = S^{(+) \, ab}_{\textrm{ret}}(xy^{-1}) -
S^{(+) \, ab}_{\textrm{av}}(xy^{-1}).
\end{align*}

The commutation and paring functions 
\[
(x,y) \mapsto  D^{\mu \nu}_{0}(xy^{-1}), D^{(-) \, \mu \nu}_{0}(xy^{-1}),  D^{(-) \, \mu \nu}_{0}(xy^{-1})
\]
of a free massless field, in our case
of the electromagnetic potential field, are distributions
in $\mathscr{E}^* = \mathcal{S}_{\Delta}\big(\widetilde{\mathbb{S}^1}\times SU(2, \mathbb{C})\big)^*$
and are not representable by ordinary smooth functions. Nonetheless
the splitting can likewise be uniquely realized for them
with the help of the multiplication by the  (periodic) step function 
$\theta$ and a limit process, as we have shown in the previous Subsections, and which is 
convenient to  be denoted likwise with the help of a multiplication by $\theta$:
\begin{align*}
D^{\textrm{ret} \, \mu \nu}_{0}(xy^{-1}) = \theta(x_0-y_0)D^{\mu \nu}_{0}(xy^{-1}), \\
D^{\textrm{av} \, \mu \nu}_{0}(xy^{-1}) = -\theta(y_0-x_0)D^{\mu \nu}_{0}(xy^{-1}), \\
D^{(-) \, \textrm{ret} \, \mu \nu}_{0}(xy^{-1}) = \theta(x_0-y_0)D^{(-) \, \mu \nu}_{0}(xy^{-1}), \\
D^{(-) \, \textrm{av} \, \mu \nu}_{0}(xy^{-1}) = -\theta(y_0-x_0) D^{(-) \mu \nu}_{0}(xy^{-1}),\\
D^{(+) \, \textrm{ret} \, \mu \nu}_{0}(xy^{-1}) = \theta(x_0-y_0)D^{(+) \, \mu \nu}_{0}(xy^{-1}),\\
D^{(+) \, \textrm{av} \, \mu \nu}_{0}(xy^{-1}) = -\theta(y_0-x_0) D^{(+) \, \mu \nu}_{0}(xy^{-1}), \\
D^{\mu \nu}_{0}(xy^{-1}) = D^{\textrm{ret} \, \mu \nu}_{0}(xy^{-1}) - D^{\textrm{av} \, \mu \nu}_{0}(xy^{-1}), \\
D^{(-) \, \mu \nu}_{0}(xy^{-1}) =D^{(-) \, \textrm{ret} \, \mu \nu}_{0}(xy^{-1}) -
D^{(-) \, \textrm{av} \, \mu \nu}_{0}(xy^{-1}), \\
D^{(+) \, \mu \nu}_{0}(xy^{-1}) = D^{(+) \, \textrm{ret} \, \mu \nu}_{0}(xy^{-1}) -
D^{(+) \, \textrm{av} \, \mu \nu}_{0}(xy^{-1}).
\end{align*}
For a proof of this assertion, compare Subsection \ref{WhiteNoiseFreeFieldsonEU}.

Of course the operation of taking the retarded part is idempotential, \emph{i.e.} repeated
twice or $k$ times is equal to the operation of taking the retarded part taken just once.
This is well reflected by the notation of retarded part by (formal) multiplication by the the periodic
theta function $\theta$ which is likewise idempotential. Similarly the operation of 
taking the advanced part is idempotential up to the factor $(-1)^{k-1}$, \emph{i.e.}
repeated $k$ times is equal to the operation of taking the retarded part taken just once
and multiplied by $(-1)^{k-1}$. Again this is well reflected by the notation of the
advanced part by the (formal) multiplication by $-\theta$, having the same 
idempotential property up to the factor $(-1)^{k-1}$.

Easy analysis will show that the generalized operator 
$D_{(n)}$ has the form of sum of components of the form
\begin{equation}\label{GeneralD_nOnEU}
\cdots \theta(x_{i0}-x_{k0})D^{(\pm)}_{0}(x_ix_{k}^{-1}) \cdots 
\theta(x_{j0}-x_{q0}) S^{(\pm)}_{0}(x_jx_{q}^{-1}) \ldots
 \,\,\,\, \boldsymbol{:} \cdots \boldsymbol{:}
\end{equation}
with a Wick product $\boldsymbol{:} \cdots \boldsymbol{:}$ of free field operators
in each component. Moreover, the number of scalar factors in each such component
includes scalar factors in which there are always present $n-1$ different pairs of variables  
$(x_i,x_k)$ in the form $x_ix_{k}^{-1}$. Possibly one and the same pair $(x_i,x_k)$ 
will appear more than just once in the form of a variable $x_ix_{k}^{-1}$ in the scalar factors.
But the factor 
\[
\theta(x_{i0}-x_{k0})D^{(\pm)}_{0}(x_ix_{k}^{-1})
\]
representing the (retarded or minus advanced part of the) pairing function of the massless
electromagnetic potential field with the same pair $(x_i,x_{k})$ of variables appear at most once.
Even more, if the paring function of the massless electromagnetic potential field
appears more than once, say
\[
\cdots \theta(x_{i0}-x_{k0})D^{(\pm)}_{0}(x_ix_{k}^{-1}) \cdots 
\theta(x_{j0}-x_{q0}) D^{(\pm)}_{0}(x_jx_{q}^{-1}) \ldots
 \,\,\,\, \boldsymbol{:} \cdots \boldsymbol{:}
\]
then the two-element sets of variables $\{x_i, x_{k}\}$ and $\{x_j,x_{q} \}$
are disjoint, because of the form of the Lagrange interaction 
density $\mathcal{L}$ for QED, into which the electromagnetic potential field enters once as a factor
and the Dirac field twice all (as the Wick product) at the same space-time point. 

Suppose that the said component
\[
\cdots \theta(x_{i0}-x_{k0})D^{(\pm)}_{0}(x_ix_{k}^{-1}) \cdots 
\theta(x_{j0}-x_{q0}) S^{(\pm)}_{0}(x_jx_{q}^{-1}) \ldots
 \,\,\,\, \boldsymbol{:} \cdots \boldsymbol{:}
\] 
of $D_{(n)}$ has $N\geq n$ scalar factors
\[
\cdots \theta(x_{i0}-x_{k0})D^{(\pm)}_{0}(x_ix_{k}^{-1}) \cdots 
\theta(x_{j0}-x_{q0}) S^{(\pm)}_{0}(x_jx_{q}^{-1}) \ldots.
\]
Each such factor we will write as the sum of two terms
\[
\theta(x_{i0}-x_{k0})D^{(\pm)}_{0}(x_ix_{k}^{-1})
= \big[\theta(x_{i0}-x_{k0}) + \theta(x_{k0}-x_{i0}) \big] \theta(x_{i0}-x_{k0})D^{(\pm)}_{0}(x_ix_{k}^{-1})
\]
or respectively
\[
\theta(x_{j0}-x_{q0}) S^{(\pm)}_{0}(x_jx_{q}^{-1})=
\big[\theta(x_{j0}-x_{q0}) + \theta(x_{q0}-x_{j0}) \big]\theta(x_{j0}-x_{q0}) S^{(\pm)}_{0}(x_jx_{q}^{-1}),
\]
and insert into the expression 
\[
\cdots \theta(x_{i0}-x_{k0})D^{(\pm)}_{0}(x_ix_{k}^{-1}) \cdots 
\theta(x_{j0}-x_{q0}) S^{(\pm)}_{0}(x_jx_{q}^{-1}) \ldots
 \,\,\,\, \boldsymbol{:} \cdots \boldsymbol{:}
\] 
for the said component of $D_{(n)}$, obtaining $2^N$ terms.
Part of them are immediately seen to be zero (those containing
the factors $\theta(x_{i0}-x_{k0})\theta(x_{k0}-x_{i0})$ with opposite order).
It is easily seen that the the remaining terms are divided into the two
disjoint groups, those having the support in 
\[
\Gamma_{+}^{(n-1)}(x_n)
\]
and composing the retarded part  $R_{(n)}$
and all those having the support in
\[
\Gamma_{-}^{(n-1)}(x_n)
\]
which, taken with the minus sign, compose the advanced part $A_{(n)}$. 

Here we should emphasize that due to Theorem \ref{WickProdFreeFieldsOnEU}, Subsection \ref{WhiteNoiseFreeFieldsonEU}, 
the Wick product factor $\boldsymbol{:} \cdots \boldsymbol{:}$
in (\ref{GeneralD_nOnEU}) is well defined finite sum of integral kernel operators
$\Xi(\kappa_{\ell,m})$ each with vector valued kernel $\kappa_{\ell,m}$ (with $\ell+m\leq n$ equal to the total number of the creation and annihilation operators as factors in the cofficient of the Wick product $\boldsymbol{:} \cdots \boldsymbol{:}$ represented by the integral kernel operator $\Xi(\kappa_{\ell,m})$ corresponding to the kernel $\kappa_{\ell, m}$). Each such kernel $\kappa_{\ell,m}$
transforms continuously the nuclear space $E_{i_{1}} \otimes \cdots \otimes E_{i_{\ell+m}}$
into $\mathscr{E}^{*\otimes N'}$, where $N'$ is the number of factors in $\boldsymbol{:} \cdots \boldsymbol{:}$ with $N'$ different space-time coordinate variables.  By Theorem \ref{WickProdFreeFieldsOnEU}, Subsection \ref{WhiteNoiseFreeFieldsonEU} (the analogue of Lemma \ref{kappaBarDotOtimeskappa} of Subsection \ref{OperationsOnXi}), 
each such kernel $\kappa_{\ell,m}$ can be extended to a continuous map from 
\[
\big(E_{i_{1}} \otimes \cdots \otimes E_{i_{\ell}} \big)^* \otimes E_{i_{\ell+1}} \otimes \ldots \otimes E_{i_{\ell+m}}=
E_{i_{1}}^{*} \otimes \cdots \otimes E_{i_{\ell}}^* \otimes E_{i_{\ell+1}} \otimes \ldots \otimes E_{i_{\ell+m}}
\]
into $\mathscr{E}^{* \otimes N'}$. Therefore, for the Wick product $\Xi(\kappa_{\ell,m})$ of the integral kernel operators with the kernels of the class 
of Definition \ref{K_0onEU}, Subsection  \ref{WhiteNoiseFreeFieldsonEU} (analogue to the class\footnote{With the differential operators $\partial_\mu$ replaced with $X^\mu$, although differentiation operation is not used in case of spinor QED lagrange interaction and correspondigly differentiation may be discarded in preparation of the class $\mathfrak{K_0}$ in this case.} $\mathfrak{K}_0$, Def. \ref{K_0}, compare Subsection \ref{WhiteNoiseFreeFieldsonEU}) 
we can extend the kernel $\kappa_{\ell,m}$ 
to a continuous map from 
\[
\big(E_{i_{1}} \otimes \cdots \otimes E_{i_{\ell}} \big)^* \otimes E_{i_{\ell+1}} \otimes \ldots \otimes E_{i_{\ell+m}}=
E_{i_{1}}^{*} \otimes \cdots \otimes E_{i_{\ell}}^* \otimes E_{i_{\ell+1}} \otimes \ldots \otimes E_{i_{\ell+m}}
\]
into $\mathscr{E}^{* \otimes N'}$. Note also that by Theorem \ref{F*WickProdFreeFieldsOnEU}, Subsection \ref{WhiteNoiseFreeFieldsonEU} the analogue of the Lemma
\ref{S*Xi} of Subsect. \ref{OperationsOnXi} (in the strengthened form with the Minkowski space-time replaced with the compactified Einstein Universe 
$\widetilde{\mathbb{S}^1}\times SU(2, \mathbb{C})$, with the convolutor algebra
$\mathcal{O}_C( \mathbb{R}^4; \mathbb{C})$ in its formulation
replaced with $\mathscr{E} = \mathcal{S}_{\Delta}(\widetilde{\mathbb{S}^1}\times SU(2, \mathbb{C}); \mathbb{C}))= \mathscr{C}^\infty(\widetilde{\mathbb{S}^1}\times SU(2, \mathbb{C}); \mathbb{C})$), the kernel $S\ast\kappa_{\ell,m}$, with $S\in \mathscr{E}^*$
 is extendible to a continuous map from  
\[
\big(E_{i_{1}} \otimes \cdots \otimes E_{i_{\ell}} \big)^* \otimes E_{i_{\ell+1}} \otimes \ldots \otimes E_{i_{\ell+m}}=
E_{i_{1}}^{*} \otimes \cdots \otimes E_{i_{\ell}}^* \otimes E_{i_{\ell+1}} \otimes \ldots \otimes \otimes E_{i_{\ell+m}}
\]
into $\mathscr{E}^{* \otimes N'}$. Therefore, by the repeated application of Theorem \ref{F*WickProdFreeFieldsOnEU} of 
Subsection \ref{WhiteNoiseFreeFieldsonEU} (analogue of Lemma
\ref{S*Xi}, Subsect. \ref{OperationsOnXi}), we infer that each component  
(\ref{GeneralD_nOnEU}) is represented by a finite sum of integral kerel operators $\Xi(\kappa_{\ell,m})$, with each kernel $\kappa_{\ell,m}$ belonging to 
\[
\mathscr{L}(E_{i_{1}} \otimes \cdots \otimes E_{i_{\ell+m}}, \mathscr{E}^{* \otimes n})
\]
and being (uniquely) extendible to an element of 
\[
\mathscr{L}(E_{i_{1}}^{*} \otimes \cdots \otimes E_{i_{\ell}}^* \otimes E_{i_{\ell+1}} \otimes \ldots \otimes E_{i_{\ell+m}}; \mathscr{E}^{* \otimes n}).
\]
Therefore (by Thm. 3.13 of  \cite{obataJFA} or Thm \ref{obataJFA.Thm.3.13})
each such $\Xi(\kappa_{\ell,m})$ represents an integral kernel operator
belongng to
\[
\mathscr{L}\big(\mathscr{E}^{\otimes n}, \mathscr{L}((\boldsymbol{E}), (\boldsymbol{E})) \big).
\]
In particular each higher order contribution $S_n\big((g=1)\mathcal{L}\big)$ represents 
a well defined operator in the Fock space, transforming continuously the Hida
test space $(\boldsymbol{E})$ into itself. This however is what we have expected by the very
compactness of the group  $\widetilde{\mathbb{S}^1}\times SU(2, \mathbb{C})$ where
adiabatic-limit-type problems disappear completely.

Similarly, using the causal perturbative 
method, we construct inductively 
the scattering operator functional
\begingroup\makeatletter\def\f@size{5}\check@mathfonts
\def\maketag@@@#1{\hbox{\m@th\large\normalfont#1}}%
\begin{multline*}
S(g\mathcal{L} + h\mathbb{A}) = \boldsymbol{1}+
\\
\sum \limits_{n=1}^{\infty} {\textstyle\frac{1}{(n+m)!}} 
\int \ud^4 x_1 \cdots \ud^4 x_n y_1 \cdots \ud^4 y_m S(x_1, \ldots, x_n,y_1, \ldots, y_m) \,
g(x_1) \cdots g(x_n) h(y_1) \cdots h(y_m),
\end{multline*}
\begin{multline*}
S^{-1}(g\mathcal{L}+h\mathbb{A}) = \boldsymbol{1} + \\
\sum \limits_{n=1}^{\infty} {\textstyle\frac{1}{n!}} 
\int \ud^4 x_1 \cdots \ud^4 x_n\ud^4 y_1 \cdots \ud^4 y_m \overline{S}(x_1, \ldots, x_n,y_1, \ldots, y_m) \,
g(x_1) \cdots g(x_n)h(y_1) \cdots h(y_m),
\end{multline*}
\endgroup
on using the interaction Lagrange density $g\mathcal{L} + h\mathbb{A}$ with two-component
$(g,h)$ ``intensity-of-interaction'' function $(g,h)$ serving as a tool for implementing the causality condition
\begin{multline*}
S\Big((g_1+g_2)\mathcal{L} + (h_1+h_2)\mathbb{A}\Big)
= S\big(g_2\mathcal{L} + h_2\mathbb{A}\big)S\big(g_1\mathcal{L} + h_1\mathbb{A}\big)
\\
\textrm{\tiny whenever $\textrm{supp} \, (g_1,h_1) \prec \textrm{supp} \, (g_2,h_2)$},
\end{multline*}
compare Subsection \ref{MotivationForHida} or \cite{DKS1}, \cite{Scharf}, \cite{DutFred}.
The kernel $S(x_1, \ldots, x_n,y)$ of the $n+1$-order term
\[ 
{\textstyle\frac{1}{(n+1)!}} 
\int \ud^4 x_1 \cdots \ud^4 x_n\ud^4 y S(x_1, \ldots, x_n,y) \,
g(x_1) \cdots g(x_n)h(y) 
\]
contribution to $S(g\mathcal{L} + h\mathbb{A})$ into which $h$ enters linearly,
we denote shortly by $S(Z,y)$, using the abbreviated notation of Epstein-Glaser for the set $\{x_1, \ldots, x_n\}$ of space-time variables. The kernels of the $n$-th order contributions 
\begin{multline*}
{\textstyle\frac{1}{(n)!}} 
\int \ud^4 x_1 \cdots \ud^4 x_n S(x_1, \ldots, x_n) \,
g(x_1) \cdots g(x_n)
\\
\textrm{and} \,\,
{\textstyle\frac{1}{(n)!}} 
\int \ud^4 x_1 \cdots \ud^4 x_n \overline{S}(x_1, \ldots, x_n) \,
g(x_1) \cdots g(x_n)
\end{multline*}
to $S(g\mathcal{L})$, and respectively to $S^{-1}(g\mathcal{L})$,
we denote simply by $S(x_1, \ldots, x_n) = S(Z)$ and $\overline{S}(x_1, \ldots, x_n)
= \overline{S}(Z)$.

The variational derivative at $h=0$ in definition of interacting fields is understood,
at least initially, only formally at the present stage of the theory and
means nothing else but taking the sum 
\[
\int \Sigma(g,x)h(x) dx
\]
of all terms which are linear in $h$ in the formal expansion for 
$S(g\mathcal{L} +h\mathbb{A})$ and putting the kernel $\Sigma(g,x)$ of this term as the derivative. 
For a rigorous definition of the higher order contributions to $S(g,h) = S(g\mathcal{L}+h\mathbb{A})$,
with the creation-annihilation operators understood as the Hida operators, including the case of the Grassmann-valued
test functions $h$, compare Subsection \ref{WickForProduct}. For a rigorous treatement of the variational derivatives with 
respect to Grassmann-valued test functions $h$, evaluated at $h\neq 0$, where we have to distinguish the left-hand-side
and right-hand-side derivatives, compare \cite{Berezin}.

Note that $\mathbb{A}$
need not be equal to one of the elementary free fields of the theory, but it can be equal to any 
local field expressed as a Wick polynomial of free fields.  

Performing the formal functional variation we can easily see that
\[
\mathbb{A}^{(n)}(x_1, \ldots, x_n,x) = \mathbb{A}^{(n)}(Z,x)
\]
is eqaual
\begin{multline*}
\mathbb{A}^{(n)}(Z,x) =  {\textstyle\frac{1}{i}} \sum \limits_{Z = X \sqcup Y}
\overline{S}(X)S(Y,x),
\\ 
\textrm{\tiny sum over all partitions $X \sqcup Y$ of $Z$ including $X = \emptyset$}
\\
= {\textstyle\frac{1}{i}} \sum' \limits_{Z = X \sqcup Y}
\overline{S}(X)S(Y,x) + S(Z,x),
\\
\textrm{\tiny sum over all partitions $X \sqcup Y$ of $Z$ with $X \neq \emptyset$},
\end{multline*}
compare Section \ref{A(1)psi(1)} and Subsection \ref{WickForProduct} as well as \cite{Epstein-Glaser}, \cite{DKS1}, \cite{Scharf}.
This is precisely the kernel $A_{(n+1)}(Z,x)$ of Epstein-Glaser introduced 
above (compare also Subsection \ref{MotivationForHida}, Section \ref{A(1)psi(1)}),
but computed for the scattering 
operator $S(g,h) = S(g\mathcal{L}+h\mathbb{A})$, with the support properties 
summarized above (compare Section \ref{A(1)psi(1)}  and Subsections \ref{MotivationForHida} and \ref{WickForProduct}).

In particular, it follows
that $\mathbb{A}^{(n)}(Z,x)$ is equal to the advanced part of the kernel 
$D_{(n+1)}(Z,x)$ in the inductive constrution the $n+1$-order contribution (defined above)
to the scattering operator $S(g\mathcal{L}+h\mathbb{A})$ , multiplied
by $-i$. More precisely let us introduce, once again, after Epstein and Glaser, besides
\[
A'_{(n+1)}(Z,x) = \sum' \limits_{Z = X \sqcup Y}
\overline{S}(X)S(Y,x)
\]
the kernel
\[
R'_{(n+1)}(Z,x) = \sum' \limits_{Z = X \sqcup Y} S(Y,x) \overline{S}(X)
\]
(note that the primed sums run over those partitions which do not include the 
empty set $X=\emptyset$).
Introducing further after Epstein and Glaser the kernel
\[
D_{(n+1)}(Z,x) = R'_{(n+1)}(Z,x) - A'_{(n+1)}(Z,x)
\]
we see (compare Subsection \ref{MotivationForHida}) that
\[
\mathbb{A}^{(n)}(Z,x) = {\textstyle\frac{1}{i}} A_{(n+1)}(Z,x)
\]
where
\[
A_{(n+1)}(Z,x) = \textrm{advanced part} \big[D_{(n+1)}(Z,x)  \big]
\]
in the decomposition 
\[
D_{(n+1)}(Z,x) = R_{(n+1)}(Z,x)-A_{(n+1)}(Z,x)
\]
of $D_{(n+1)}$ into the advanced $A_{(n+1)}$ and retarded part $R_{(n+1)}(Z,x)$.

In fact the operator kernel  $D_{(n+1)}$ has the form of sum of components
each having the general form (\ref{GeneralD_nOnEU}) with 
the scalar terms in pairs of variables $(x_i,x_k)$ entering in the form $x_ix_{k}^{-1}$,
and each such component contains the scalar factors for $n$ different pairs
of variables. In general situation, even for the Wick product $\mathbb{A}(x)$
of free fields containing more than just one electromagnetic free field factor,
the number of factors $\theta(x_{i0}-x_{k0})D^{(\pm)}_{0}(x_ix_{k}^{-1})$
in each term (\ref{GeneralD_nOnEU})
with the paring function of the massless electromagnetic potential field
and the same pair $(x_i,x_{k})$ of variables is equal at most one.
Of course in general situation, when we have more than just one electromagnetic free field factor
in $\mathbb{A}(x)$, we can encounter two factors
\[
\theta(x_{i0}-x_{k0})D^{(\pm)}_{0}(x_ix_{k}^{-1})
\theta(x_{k0}-x_{j0})D^{(\pm)}_{0}(x_kx_{j}^{-1})
\]
in (\ref{GeneralD_nOnEU}) with one common variable $x_k$, but never
\[
\theta(x_{i0}-x_{k0})D^{(\pm)}_{0}(x_ix_{k}^{-1})
\theta(x_{i0}-x_{k0})D^{(\pm)}_{0}(x_ix_{k}^{-1})
\]
with both common variables $(x_i,x_k)$, by the form of interaction Lagrange density, and  which allows the simple splitting method summarized above.

By the analogues of the Lemmas \ref{kappaBarDotOtimeskappa} and \ref{S*Xi}
of Subsection \ref{OperationsOnXi}, mentioned above in the analysis of (\ref{GeneralD_nOnEU}),
each higher order contribution 
\[
\mathbb{A}^{(n)}(g=1)
\]
with the operator kernel
\[
\mathbb{A}^{(n)}(g=1,x) =  {\textstyle\frac{1}{n!}} 
\int \ud^4 x_1 \cdots \ud^4 x_n \mathbb{A}^{(n)}(x_1, \ldots, x_n,x),
\]
to the interacting field
\[
\mathbb{A}_{{}_{\textrm{int}}} = \mathbb{A}_{{}_{\textrm{int}}}(g=1),
\]
is a finite sum of integral kernel operators $\Xi(\kappa_{\ell,m})$,
each belonging to 
\[
\mathscr{L}\big(\mathscr{E}, \, \mathscr{L}((\boldsymbol{E}), (\boldsymbol{E}))\big),
\]
so that 
\[
\mathbb{A}^{(n)}(g=1)
\in \mathscr{L}\big(\mathscr{E}, \, \mathscr{L}((\boldsymbol{E}), (\boldsymbol{E}))\big),
\]
and thus with $\mathbb{A}^{(n)}(g=1)$ being a well defined operator-valued 
distribution transforming continuously the space-time test space
$\mathscr{E}$ into $\mathscr{L}((\boldsymbol{E}), (\boldsymbol{E}))$.

In particular for $\mathbb{A}(x) = A^\mu(x)$ equal to the $\mu$-th component 
of electromagnetic potential field we have the following formula
\[
A^{\mu \, (1)}(x_1,x) = {\textstyle\frac{1}{i}} \textrm{advanced part} \big[D_{(2)}(x_1,x)  \big]
\]
for the kernel of the first order contribution to the interacting
field $A_{{}_{\textrm{int}}}^{\mu}(x)$,
where ($\eta$ is the Gupta-Bleuler operator in the total Fock space of the free
$A$ and $\boldsymbol{\psi}$ fields acting trivially in the factor Fock space
of the field $\boldsymbol{\psi}$)
\begin{multline*}
D_{(2)}(x_1,x) = \eta \big(i\mathcal{L}(x_1)\big)^{+} \eta \,\, iA^\mu(x) 
\,\,\,
-
\,\,\,
iA^\mu(x)   \,\,\, \eta \big(i\mathcal{L}(x_1)\big)^{+} \eta 
\\
=
eA^\mu(x) \, \boldsymbol{{:}} \psi^\sharp(x_1) \gamma_\nu A^\nu(x_1) \psi(x_1) \boldsymbol{{:}}
\,\,\,\,
- 
\,\,\,\,
e \, \boldsymbol{{:}} \psi^\sharp(x_1) \gamma_\nu  A^\nu(x_1) \psi(x_1) \boldsymbol{{:}} \, A^\mu(x) 
\\
=
-e[A^\nu(x_1), A^\mu(x) ] \,
\boldsymbol{{:}} \psi^\sharp(x_1) \gamma_\nu \psi(x_1) \boldsymbol{{:}} 
\,\,\,\,\,\,
=
-ie D_{0}^{\mu \nu}(x_1x^{-1}) \,
\boldsymbol{{:}} \psi^\sharp(x_1) \gamma_\nu \psi(x_1) \boldsymbol{{:}}. 
\end{multline*}

For $\mathbb{A}(x) = \boldsymbol{\psi}^a(x)$ equal to the $a$-th component 
of Dirac's spinor field we have the following formula
\[
\boldsymbol{\psi}^{a \, (1)}(x_1,x) = {\textstyle\frac{1}{i}} \textrm{advanced part} \big[D_{(2)}(x_1,x)  \big]
\]
for the kernel of the first order contribution to the interacting
field $\boldsymbol{\psi}_{{}_{\textrm{int}}}^{a}(x)$, with
\begin{multline*}
D_{(2)}(x_1,x) = \eta \big(i\mathcal{L}(x_1)\big)^{+} \eta \,\, i\boldsymbol{\psi}^a(x) 
\,\,\,
-
\,\,\,
i\boldsymbol{\psi}^a(x)  \,\,\, \eta \big(i\mathcal{L}(x_1)\big)^{+} \eta 
\\
=
ie S\big(xx_{1}^{-1}\big)\gamma_\nu \boldsymbol{\psi}(x_1)A^\nu(x_1).
\end{multline*}

The computation of the higher order $D_{(n)}$ and their advanced parts being essentially simple can therefore be omitted. We give only the explicit formulas for the first order contributions to interacting fields (with summation convention and Dirac adjoined bispinor $\boldsymbol{\psi}^\sharp(x) = \boldsymbol{\psi}(x)^\sharp = \boldsymbol{\psi}(x)^+\gamma_0$, 
compare \cite{DKS1}, \cite{Scharf}, \cite{DutFred} where $\boldsymbol{\psi}^\sharp(x)$
is denoted by $\overline{\boldsymbol{\psi}}(x)$) 
\begin{multline*}
\boldsymbol{\psi}_{{}_{\textrm{int}}}^{a}(g, x) = \\ =
\boldsymbol{\psi}^{a}(x) + \sum \limits_{n=1}^{\infty} \frac{1}{n!}
\int \limits_{\big[\widetilde{\mathbb{S}^1}\times SU(2, \mathbb{C})\big]^{\times \, n}} \ud^4x_1 \cdots \ud^4 x_n \boldsymbol{\psi}^{a \, (n)}(x_1, \ldots, x_n; x)
g(x_1) \cdots g(x_n), 
\end{multline*}
with for example 
\[
\boldsymbol{\psi}^{a \, (1)}(x_1; x) = 
-e S_{{}_{\textrm{ret}}}^{aa_1}\big(xx_{1}^{-1}\big) \gamma^{\nu_1 \, a_1a_2} \boldsymbol{\psi}^{a_2}(x_1)A_{\nu_1}(x_1), 
\]
\[
\textrm{e. t. c.}
\]
and  
\begin{multline*}
{A_{{}_{\textrm{int}}}}^{\mu}(g, x) = \\ =
A_{\mu}(x) + \sum \limits_{n=1}^{\infty} \frac{1}{n!}
\int \limits_{\big[\widetilde{\mathbb{S}^1}\times SU(2, \mathbb{C})\big]^{\times \, n}} \ud^4x_1 \cdots \ud^4 x_n A^{\mu \, (n)}(x_1, \ldots, x_n; x)
g(x_1) \cdots g(x_n),
\end{multline*}
with for example
\[
A^{\mu \, (1)}(x_1;x) = -e D^{\textrm{av} \, \mu \nu}_{0}\big(x_1x^{-1}\big) \,
{:}\boldsymbol{\psi}^{\sharp \, a_1}(x_1) \gamma_{\nu}^{a_1a_2} \boldsymbol{\psi}^{a_2}(x_1){:},
\]
\[
\textrm{e. t. c.}
\]

But the point is that on the Einstein Universe not only
\begin{align*}
\boldsymbol{\psi}^{a \, (n)}(g=1) \in \mathscr{L}\big(\mathscr{E}, \, \mathscr{L}((\boldsymbol{E}), (\boldsymbol{E}))\big),
\\
A^{\mu \, (n)}(g=1) \in \mathscr{L}\big(\mathscr{E}, \, \mathscr{L}((\boldsymbol{E}), (\boldsymbol{E}))\big),
\\
\textrm{and generally}
\\
\mathbb{A}^{(n)}(g=1)
\in \mathscr{L}\big(\mathscr{E}, \, \mathscr{L}((\boldsymbol{E}), (\boldsymbol{E}))\big),
\end{align*}
for a Wick polynomial $\mathbb{A}(x)$ in free fields, with the kernel operators
\begin{align*}
\boldsymbol{\psi}^{a \, (n)}(g=1, x) = 
\frac{1}{n!}
\int \limits_{\big[\widetilde{\mathbb{S}^1}\times SU(2, \mathbb{C})\big]^{\times \, n}} \ud^4x_1 \cdots \ud^4 x_n \boldsymbol{\psi}^{a \, (n)}(x_1, \ldots, x_n; x) \\
{A_{{}_{\textrm{int}}}}^{\mu \, (n)}(g=1, x) =  
\frac{1}{n!}
\int \limits_{\big[\widetilde{\mathbb{S}^1}\times SU(2, \mathbb{C})\big]^{\times \, n}} \ud^4x_1 \cdots \ud^4 x_n A^{\mu \, (n)}(x_1, \ldots, x_n; x) \\
\mathbb{A}^{(n)}(g=1, x) = \frac{1}{n!}
\int \limits_{\big[\widetilde{\mathbb{S}^1}\times SU(2, \mathbb{C})\big]^{\times \, n}} \ud^4x_1 \cdots \ud^4 x_n \mathbb{A}^{(n)}(x_1, \ldots, x_n; x).
\end{align*}
We have much more regular behavior of all higher order contributions to interacting fields.
Namely, on the Einstein Universe the higher order contributions $\boldsymbol{\psi}^{a \, (n)}(g=1), A^{\mu \, (n)}(g=1)$ 
to interacting fields $\boldsymbol{\psi}_{{}_{\textrm{int}}}^{a}(g=1), {A_{{}_{\textrm{int}}}}^{\mu}(g=1)$, 
evaluated at single space-time point $x$, \emph{i.e}
\[
\boldsymbol{\psi}^{a \, (n)}(g=1, x), \,\,\,
A^{\mu \, (n)}(g=1;x),
\]
are well defined operators on the Fock space transforming continuously the Hida test
space $(\boldsymbol{E})$ into itself. Recall that the free fields are understood as the 
integral kernel operators determined by the corresponding vector-valued distributional kernels. 
The Wick products of free fields, according to Subsection \ref{OperationsOnXi}, are likewise 
integral kernel operators with vector-valued kernels obtained by the operation of (symmetrized and 
respectively anti-symmetrized) pointwise product of the kernels corresponding to free fields.
The general form of higher order contributions has the form of finite sums of repeated convolutions
of retarded and advanced parts of the pairings with the corresponding Wick products. 
These operations are understood as performed on integral kernel operators and by the Rules of  Subsection \ref{OperationsOnXi} are realized through the corresponding repeated operation on the kernels of convolution of the retarded and advanced pairing functions with the kernels corresponding to the Wick products. Here the fact that the orbit $\mathscr{O}_+$ (and $\mathscr{O}_-$)  is a finite set for the free Dirac field (and for general massive field on the Einstein Universe) intervenes decisively. Indeed, it is easily seen by the general formula for the higher order contributions
\[
\boldsymbol{\psi}^{a \, (n)}(g=1, x), \,\,\,
A^{\mu \, (n)}(g=1;x),
\]
that each such contribution is a finite sum of the Hida creation and annihilation operator products,
and thus is equal to ordinary operator in the Fock space transforming the Hida test space continuously
into itself. Indeed, this easily follows by the said general formulas for the higher order contributions
expressed through operations of successive convolutions performed on the corresponding kernels, by taking into account 
the orthogonality of the functions $t \mapsto \widehat{n}(t)$ and $\boldsymbol{w} \mapsto 
\widehat{l}_{{}_{ij}}(\boldsymbol{w})$ defining the character 
$\widehat{n}\cdot \widehat{l}$ of the group $\widetilde{\mathbb{S}^1} \times SU(2, \mathbb{C})$
(\emph{i.e.} Peter-Weyl theorem for the group $\widetilde{\mathbb{S}^1} \times SU(2, \mathbb{C})$)
and the formula for the pointwise product $\widehat{l'}_{{}_{i'j'}}(\boldsymbol{w})\widehat{l''}_{{}_{i''j''}}(\boldsymbol{w})$ expressing this product as finite linear combinations of the functions
$\widehat{l}_{{}_{ij}}(\boldsymbol{w})$ with $|l'-l''| \leq l \leq l'+l''$. The essential point
is that each massive free field factor (Dirac free field in case of spinor QED), enters with 
kernels which contain only a finite number of functions $\boldsymbol{w} \mapsto 
\widehat{l}_{{}_{ij}}(\boldsymbol{w})$, \emph{i.e.} with finite number of weights $l$,
and the same holds for the commutation functions of the massive fields (here the Dirac field),
 pairings of massive fields, and their retarded and advanced parts. Thanks to this opportunity all but finite terms will be cancelled out by the orthogonality of the functions $\boldsymbol{w} \mapsto \widehat{l}_{{}_{ij}}(\boldsymbol{w})$.

Let us look at this computation more closely on the example of the first order contribution
in order to understand better the general mechanism of cancelling all but finite number of terms.
The commutation $S^{a'a}(x'x^{-1})$ and the pairing functions
$S^{(\pm) \, a'a}(x'x^{-1})$ of the free Dirac field, as well as their retarded and advanced parts, are equal to the finite sums (with respect to $\widehat{n}\cdot \widehat{l} \in \mathscr{O}_+$,
$s \in \{1, \ldots, d'(\widehat{n}\cdot \widehat{l})=d''(\widehat{n}\cdot \widehat{l}) = (2l+1)^2 2\}$,
$i,j,i',j' \in \{-l, \ldots, l \}$
and $b,c' \in \{1, \ldots, 4\}$)
of products of the form (we discard here the irrelevant constant factors)
\begin{multline}\label{FactorFormOfSonEU}
{u^{b}_{s}}_{{}_{ji}}(\widehat{n}\cdot \widehat{l}) 
\widehat{n}(t) \widehat{l}_{{}_{ij}}(\boldsymbol{w})
{v^{c'}_{s}}_{{}_{j'i'}}(\widehat{n}\cdot \widehat{l})
\overline{\widehat{n}(t')} \overline{\widehat{l}_{{}_{i'j'}}(\boldsymbol{w}')} \\
= {u^{b}_{s}}_{{}_{ji}}(\widehat{n}\cdot \widehat{l}) 
\widehat{n}(t) \widehat{l}_{{}_{ij}}(\boldsymbol{w})
{v^{c'}_{s}}_{{}_{j'i'}}(\widehat{n}\cdot \widehat{l})
\widehat{n}(-t') \widehat{l}_{{}_{-i \,\,\, -j}}(\boldsymbol{w}'), \\
\textrm{here}
\,\,\,
x = t\times \boldsymbol{w}, x' = t' \times \boldsymbol{w}' \in \mathbb{R} \times SU(2, \mathbb{C}).
\end{multline}
Each component $\boldsymbol{\psi}^a(x)$ of the free Dirac field $\boldsymbol{\psi}$ evaluated at
space-time point $x$ is equal to the following finite sum
\begin{multline}\label{IntKerOpForPsionEU}
\boldsymbol{\psi}^a(x) = \boldsymbol{\psi}^{(-) \, a}(x) 
+ \boldsymbol{\psi}^{(+) \, a}(x) = \Xi\big(\kappa_{01}(a,x)\big) 
+ \Xi\big(\kappa_{10}(a,x)\big) \\
= \sum \limits_{s\in \{1,\ldots,(2l+1)^2 2\}, \widehat{n}\cdot \widehat{l} \in \mathscr{O}_+} 
\kappa_{01}(s, \widehat{n}\cdot \widehat{l}; a,x) b_{{}_{s}}(\widehat{n}\cdot \widehat{l})
\\
+
\sum \limits_{s\in \{1,\ldots,(2l+1)^2 2\}, \widehat{n}\cdot \widehat{l} \in \mathscr{O}_-} 
\kappa_{10}(s, \widehat{n}\cdot \widehat{l}; a,x) d_{{}_{s}}(\widehat{n}\cdot \widehat{l})^{+}
\end{multline}
 of Hida creation and annihilation operators $b_{{}_{s}}(\widehat{n}\cdot \widehat{l}),
 d_{{}_{s}}(\widehat{n}\cdot \widehat{l})^{+}$
(because  $\mathscr{O}_+$ is finite for the Dirac field),
each transforming continuously the Hida space into itself, so that 
$\boldsymbol{\psi}^a(x)$ is an ordinary operator 
transforming continuously the Hida space
into itself, with the corresponding plane-wave kernels $\kappa_{01}(a,x), \kappa_{10}(a,x)$
determined by ordinary functions
\begin{multline*}
(s, \widehat{n}\cdot \widehat{l}) \mapsto \kappa_{0,1}(s, \widehat{n}\cdot \widehat{l}; a,x) = 
\sum \limits_{i,j \in \{-l, \ldots, l \}} \sqrt{2l+1}
{u^{a}_{s}}_{{}_{ji}}(\widehat{n}\cdot \widehat{l}) 
\widehat{n}(t) \widehat{l}_{{}_{ij}}(\boldsymbol{w}),
\\
(s, \widehat{n}\cdot \widehat{l}) \mapsto \kappa_{1,0}(s, \widehat{n}\cdot \widehat{l}; a,x) = 
\sum \limits_{i,j \in \{-l, \ldots, l \}} \sqrt{2l+1}
{v^{a}_{s}}_{{}_{ji}}(\widehat{n}\cdot \widehat{l}) 
\overline{\widehat{n}(t)} \overline{\widehat{l}_{{}_{ij}}(\boldsymbol{w})},\\
\textrm{here}
\,\,\,
x = t\times \boldsymbol{w} \in \mathbb{R} \times SU(2, \mathbb{C}).
\end{multline*}
Recall that the above value of $\kappa_{0,1}(s, \widehat{n}\cdot \widehat{l}; a,x)$ is assumed for
$\widehat{n}\cdot \widehat{l} \in \mathscr{O}_+$, for $\widehat{n}\cdot \widehat{l} \notin \mathscr{O}_+$
it is equal zero. Similarly the above value of $\kappa_{1,0}(s, \widehat{n}\cdot \widehat{l}; a,x)$ is assumed for
$\widehat{n}\cdot \widehat{l} \in \mathscr{O}_-$, for $\widehat{n}\cdot \widehat{l} \notin \mathscr{O}_-$
it is equal zero.
The kernel functions $\kappa_{0,1}(s, \widehat{n}\cdot \widehat{l}; a,x)$, $\kappa_{1,0}(s, \widehat{n}\cdot \widehat{l}; a,x)$, 
in turn, determine vector-valued kernels $\kappa_{01}, \kappa_{10}$
defined by ordinary functions
$(s, \widehat{n}\cdot \widehat{l},a,x) \mapsto \kappa_{01}(s, \widehat{n}\cdot \widehat{l}; a,x)$,
$(s, \widehat{n}\cdot \widehat{l},a,x) \mapsto \kappa_{10}(s, \widehat{n}\cdot \widehat{l}; a,x)$, on 
$\mathscr{O}\times \{1, \ldots,4 \} \times \widetilde{\mathbb{S}^1} \times SU(2, \mathbb{C})$ with the finite set $\mathscr{O}$
introduced in Subsetion \ref{WhiteNoiseFreeFieldsonEU}, which can be regarded as 
$\mathscr{O}_+$ including $d'(\widehat{n}\cdot \widehat{l}) = d''(\widehat{n}\cdot \widehat{l})
= (2l+1)^2 2$ copies of each point $\widehat{n}\cdot \widehat{l} \in \mathscr{O}_+$.
Thus each vector-valued plane wave kernel $\kappa_{01}$ and $\kappa_{10}$ of the free Dirac field can be regarded as finite system of smooth functions living effectively on the compacitfied Einstein Universe $\widetilde{\mathbb{S}^1} \times SU(2, \mathbb{C})$. 
Similarly the free electromagnetic potential field is equal to the
integral kernel operator
\begin{multline}\label{IntKerOpForAonEU}
A^\mu(x) = A^{(-) \, \mu}(x) 
+ A^{(+) \, \mu}(x) = \Xi\big(\kappa_{01}(\mu,x)\big) 
+ \Xi\big(\kappa_{10}(\mu,x)\big) \\
= \sum \limits_{s\in \{1,\ldots, d'(\widehat{l}\cdot \widehat{n})\}, \widehat{n}\cdot \widehat{l} \in \mathscr{O}_+} 
\kappa_{01}(s, \widehat{n}\cdot \widehat{l}; \mu,x) a_{{}_{s}}(\widehat{n}\cdot \widehat{l})
\\
+
\sum \limits_{s\in \{1,\ldots,d'(\widehat{l}\cdot \widehat{n}) \}, \widehat{n}\cdot \widehat{l} \in \mathscr{O}_+} 
\kappa_{10}(s, \widehat{n}\cdot \widehat{l}; \mu,x) a_{{}_{s}}(\widehat{n}\cdot \widehat{l})^{+}
\end{multline}
with the corresponding plane-wave kernels $\kappa_{01}(\mu,x), \kappa_{10}(\mu,x)$ determined by ordinary functions
\begin{multline*}
(s, \widehat{n}\cdot \widehat{l}) \mapsto \kappa_{01}(s, \widehat{n}\cdot \widehat{l}; \mu,x) = 
\sum \limits_{i,j \in \{-l, \ldots, l \}} \sqrt{2l+1}
{w^{\mu}_{s}}_{{}_{ji}}(\widehat{n}\cdot \widehat{l}) 
\widehat{n}(t) \widehat{l}_{{}_{ij}}(\boldsymbol{w}),
\\
(s, \widehat{n}\cdot \widehat{l}) \mapsto \kappa_{10}(s, \widehat{n}\cdot \widehat{l}; \mu,x) = 
\sum \limits_{i,j \in \{-l, \ldots, l \}} \sqrt{2l+1}
\overline{{w^{\mu}_{s}}_{{}_{ji}}(\widehat{n}\cdot \widehat{l})} 
\overline{\widehat{n}(t)} \overline{\widehat{l}_{{}_{ij}}(\boldsymbol{w})},\\
\textrm{here}
\,\,\,
x = t\times \boldsymbol{w} \in \mathbb{R} \times SU(2, \mathbb{C}).
\end{multline*}
determining vector-valued kernels $\kappa_{01}, \kappa_{10}$, defined by the smooth functions
$(s, \widehat{n}\cdot \widehat{l}, \mu,x) \mapsto \kappa_{01}(s, \widehat{n}\cdot \widehat{l}; \mu,x)$,
$(s, \widehat{n}\cdot \widehat{l}, \mu,x) \mapsto \kappa_{10}(s, \widehat{n}\cdot \widehat{l}; \mu,x)$,
as functions on 
$\mathscr{O}\times \{0, \ldots, 3 \} \times \widetilde{\mathbb{S}^1} \times SU(2, \mathbb{C})$ with the infinite set $\mathscr{O}$
introduced in Subsection \ref{WhiteNoiseFreeFieldsonEU}, which can be regarded as 
$\mathscr{O}_+$ including $d'(\widehat{n}\cdot \widehat{l})$ copies of each point $\widehat{n}\cdot \widehat{l} \in \mathscr{O}_+$. But here $\mathscr{O}_+$  is the orbit corresponding to the electromagnetic potential field, and as we know it is infinite, so that the sum 
(\ref{IntKerOpForAonEU}) is infinite and includes infinte number of creation and annihilation
Hida operators $a_{{}_{s}}(\widehat{n}\cdot \widehat{l})^{+}, a_{{}_{s}}(\widehat{n}\cdot \widehat{l})$ 
transforming continuously the Hida space into itself. 

Now utilizing the orthonormality of the character functions 
$\boldsymbol{w} \mapsto \widehat{l}_{{}_{ij}}(\boldsymbol{w})$, the law for their pointwise product and the general form of the advanced part $D^{\textrm{av} \, \mu' \mu}_{0}\big(x'x^{-1}\big) $ of commutation function $D^{\mu' \mu}_{0}\big(x'x^{-1}\big)$
and of the commutation function $S_{{}_{\textrm{ret}}}^{a'a}(x'x^{-1})$ which has the form of finite sum of terms (\ref{IntKerOpForPsionEU}) and the fact that the fields $\boldsymbol{\psi}^a(x), \boldsymbol{\psi}^{\sharp \, a}(x)$ include the kernels $\kappa_{01}(s, \widehat{n}\cdot \widehat{l}; \mu,x)$,
$\kappa_{10}(s, \widehat{n}\cdot \widehat{l}; \mu,x)$ with only finite number
of values for  $s$ and for $\widehat{n}\cdot \widehat{l}$, we can easily see that for example 
the first order contributions
\begin{align*}
\boldsymbol{\psi}^{a \, (1)}(g=1; x) = 
-e \int \limits_{\widetilde{\mathbb{S}^1} \times SU(2, \mathbb{C})}
S_{{}_{\textrm{ret}}}^{aa_1}\big(xx_{1}^{-1}\big) \gamma^{\nu_1 \, a_1a_2} \boldsymbol{\psi}^{a_2}(x_1)A_{\nu_1}(x_1) \, \ud^4 x_1, \\
A^{\mu \, (1)}(g=1;x) = -e\int \limits_{\widetilde{\mathbb{S}^1} \times SU(2, \mathbb{C})} D^{\textrm{av} \, \mu \nu}_{0}\big(x_1x^{-1}\big) \,
{:}\boldsymbol{\psi}^{\sharp \, a_1}(x_1) \gamma_{\nu}^{a_1a_2} \boldsymbol{\psi}^{a_2}(x_1){:}
\, \ud^4 x_1, 
\end{align*}
are equal to sums with at most finite number of products of Hida creation and annihilation operators.

The same is easily seen to be true for the higher order contributions
$\boldsymbol{\psi}^{a \, (n)}(g=1; x), A^{\mu \, (n)}(g=1;x)$.
In particular for each fixed space-time point $x$ all $\boldsymbol{\psi}^{a \, (n)}(g=1; x), A^{\mu \, (n)}(g=1;x)$ are ordinary operators in the Fock space transforming continuously the Hida space $(\boldsymbol{E})$ into itself, \emph{i.e.} all the said higher order contributions evaluated at $x$ belong to $\mathscr{L}((\boldsymbol{E}), (\boldsymbol{E}))$.

Thus, we can summarize our analysis with the following
\begin{twr}
Let each of the  Wick monomials of the Lagrange density interaction operator $\mathcal{L}$ contains at most one massless (or infinite orbit)
free field with all remaining free fields in the monomial being massive (or finte orbit) fields. This is e.g. the case for QED.
Then for each Wick polynomial field $\mathbb{A}$  
\[
\Big[\textrm{each contribution of order $>0$ to} \,\,
\big(\mathbb{A} \big)_{{}_{\textrm{int}}}\Big] \in
\mathscr{L}(\mathscr{E}, \mathscr{L}((\boldsymbol{E}),(\boldsymbol{E}))\big),
\]
\emph{i.e.} each higher order contribution to $\big(\mathbb{A} \big)_{{}_{\textrm{int}}}$ defines a continuous map
\[
\mathscr{E} \ni \phi \longmapsto
\Big[\textrm{higher order contribution to} \,\,
\big(\mathbb{A} \big)_{{}_{\textrm{int}}}\Big](\phi) \in
\mathscr{L}((\boldsymbol{E}),(\boldsymbol{E})).
\] 
Moreover, the higher order contributions to $\big(\mathbb{A} \big)_{{}_{\textrm{int}}}$,
evaluated at single space-time point $x$, transform the Hida space $(\boldsymbol{E})$ continuously into itself $(\boldsymbol{E})$, and become
ordinary operators on the Fock space:
\[
\textrm{each} \,  \Big[\textrm{contribution of order $>0$ to} \, \big(\mathbb{A} \big)_{{}_{\textrm{int}}}\Big](x) \in \mathscr{L}((\boldsymbol{E}),(\boldsymbol{E})),
\]
if $\mathbb{A}$ is of degree $1$ in the free massless (or infinite orbit) fields. In paricular in case of spinor QED and for each space-time point $x$
\begin{align*}
\boldsymbol{\psi}^{a \, (n)}(g=1; x) \in \mathscr{L}((\boldsymbol{E}),(\boldsymbol{E})), \, n=0,1, \dots, 
\\
A^{\mu \, (n)}(g=1;x) \in \mathscr{L}((\boldsymbol{E}),(\boldsymbol{E})), \, n=1,2, \ldots.
\end{align*}
\label{InteractingFieldsAtxOnEU}
\end{twr}

\begin{cor*}
If a field $\boldsymbol{\psi}$ is (minimally) coupled to the electromagnetic potential field, and all higher order contributions
to $\boldsymbol{\psi}_{{}_{\textrm{}int}}(x)$, evaluated at space-time point $x$ belong to
\[
\mathscr{L}((\boldsymbol{E}),(\boldsymbol{E}))
\]
\emph{i.e.} are well defined operators with (all higher order contributions to) the Noether integrals of conserved currents
computed for interacting fields (e.g. the operator of the electric charge) being well defined essentially self-adjoint operators on the Fock space,
then the field $\boldsymbol{\psi}$ is massive (or more generally finite-orbit). 
\end{cor*}

The property of $\boldsymbol{\psi}_{{}_{\textrm{}int}}$ that the value $\boldsymbol{\psi}_{{}_{\textrm{}int}}(x)$  of the interacting charged field 
at space-time point $x$ is a well-defined operator is not arbitrary, compare Subsection \ref{ASTheoryextended}. Also, the property that the
higher order contributions to the conserved Noether integrals should be well-defined operators, is not arbitrary. It should be expected
in connection to the possible particle interpretation of the theory, associated with the spectra of these conserved charges. 
In particular (compare Subsection \ref{ASTheoryextended})
the Noether integral
\[
Q = \int\limits_{{}_{\textrm{Cauchy surface}}} \ud^3 x \, j^{0}_{{}_{\textrm{int}}}(x), 
\]
with the conserved current $j^{0}_{{}_{\textrm{int}}}(x)$, associated to the global gauge group $U(1)$ is the total electric 
charge $Q$, and it is expected to be an ordinary self-adjoint operator 
canonically conjugated to the phase (say, the ``second quantized unitary gauge operator'' $e^{iS(x)}$) 
\[
S_0 = \int\limits_{{}_{\textrm{Cauchy surface}}} \ud^3 x \, S(x)
\]
acting by composition on the operator $\boldsymbol{\psi}_{{}_{\textrm{}int}}(x)$. Phase in principle makes sense
for example for any closable non-singular $\boldsymbol{\psi}_{{}_{\textrm{}int}}(x)$ through its polar decomposition 
(phase makes sense e.g. for the free massive half-integer spin field $\boldsymbol{\psi}(x)$ on the Einstein Universe, 
although it is not unique whenever $\boldsymbol{\psi}(x)$ is singular). Note that the polar decomposition of 
 $\boldsymbol{\psi}_{{}_{\textrm{}int}}(x)$ is well-defined (at least at each order separately) on the Einstein Universe, 
and the action of $e^{iS(x)}$ on the partial isometry factor of $\boldsymbol{\psi}_{{}_{\textrm{}int}}(x)$ 
through the operator composition.

Schwinger was one of the first physicists who noticed the non-trivial experimental fact that only massive particles are coupled
to the electromagnetic potential field. Now we have much larger experimental evidence which still confirms it.
Therefore, it is remarkable that we are able to prove it as a consequence of simple causal axioms
for the scattering operator and from the natural existence requirement put on the higher order contributions
to the conserved Noether integrals  (generalized total charges), as ordinary self-adjoint operators.

Moreover, the interacting fields $\mathbb{A}_{{}_{\textrm{int}}}$ and the scattering operator $S(g=1)$ are given through Fock expansions
\begin{align}
\mathbb{A}_{{}_{\textrm{int}}} = \sum \limits_{\ell,m} \Xi(\kappa_{\ell,m}), \,\,\,
\Xi(\kappa_{\ell,m})
\in \mathscr{L}(\mathscr{E}, \mathscr{L}((\boldsymbol{E}),(\boldsymbol{E}))\big) \cong 
\mathscr{L}((\boldsymbol{E}) \otimes \mathscr{E}, \, (\boldsymbol{E})\big),
\label{generalAintEU}
\\
S(g=1) = \sum \limits_{\ell,m} \Xi(\kappa_{\ell,m}),
\,\,\,\, \Xi(\kappa_{\ell,m}) \in \mathscr{L}((\boldsymbol{E}),(\boldsymbol{E})),
\label{generalSonEU}
\end{align}
into integral kernel operators in the sense of \cite{obataJFA} (with vector valued kernels $\kappa_{\ell,m}$ for $\mathbb{A}_{{}_{\textrm{int}}}$)
or in the sense of \cite{hida} for $S(g=1)$ (with scalar valued kernels $\kappa_{\ell,m}$) which can be subject to a precise and computable
convergence criteria, which utilize the symbol calculus of Obata, compare \cite{obataJFA},
\cite{obata}, \cite{obata-book}. Indeed Theorem \ref{WickThmForChronologicalOneMassLessEU} of Subsection  \ref{WickForChronologicalEU} 
(\emph{Wick Theorem for the Natural Chronological Product})  
 gives the Fock expansion of the scattering operator $S(g=1)$ or, more generally, for the natural chronological product,
in case 1) of interaction $\mathcal{L}$. The same Wick theorem
can also be applied for the simple recurrent construction of the Fock expansion of $\mathbb{A}_{{}_{\textrm{int}}}$
in terms of the product operation applied to the plane wave kernels of the free fields of the theory (of course
with some products taken at the same space-time point, then symmetrizations and antisymmetrizations and contractions). 

Therefore, the symbol calculus of Hida-Obata-Sait\^o can easily be applied.

 This allows us to verify the convergence of the perturbative series 
with the tools which were beyond our reach before. 

Writing the total Fock space as the tensor product of Fermi Fock spaces by the tensor product
of Bose Fock spaces, the symbol calculus is, in accordance to our Remarks 
\ref{TretmentFermiIntegKerOp=TretmentBoseIntegKerOp} and \ref{TretmentFermiIntegKerOp=TretmentBoseIntegKerOpII}
of Subsection \ref{psiBerezin-Hida}, based on the tensor product of the finite particle and the exponential states
\[
\Phi_N \otimes \Phi_\xi \in (E_1) \otimes (E_2) = (\boldsymbol{E}),
\]
with the state 
\[
\Phi_N = \Phi_0 \oplus \xi_1 \oplus \big(\Sigma_{{}_{i_1,i_2}} \xi_{i_1}  \widehat{\otimes} \xi_{i_2} \big)
\oplus \ldots \oplus \big(\Sigma_{{}_{j_1,\ldots j_N}} \xi_{j_1}  \widehat{\otimes} \ldots  \widehat{\otimes} \xi_{j_N} \big),
\,\,\,\,\, \xi_i \in E_1,
\] 
in the Fermi Fock space equal to any finite direct sum of simple (antisymmetric) tensors,
with at most finite number of particles, say equal N, and with the ordinary
exponential state 
\[
\Phi_\xi = \sum\limits_{n=0}^{\infty} {\textstyle\frac{1}{n!}} \xi^{\widehat{\otimes} \, n},
\,\,\,\, \xi \in E_2
\] 
in the Bose Fock space. Recall that in order to apply Thm 4.8 of \cite{obataJFA}, we need to investigate convergence of the series of ``symbols'',
 or simply the of averages
\[
\sum\limits_{\ell,m}
\big\langle \big\langle \overline{\Phi_{N'} \otimes \Phi_{\eta}}, \Xi(\kappa_{\ell,m}) \Phi_N \otimes \Phi_\xi \big\rangle \big\rangle
= 
\sum\limits_{\ell,m}
\big( \big( \Phi_{N'} \otimes \Phi_{\eta}, \Xi(\kappa_{\ell,m}) \Phi_N \otimes \Phi_\xi \big) \big)_{{}_{0}}
\]
of the integral kernel operators $\Xi(\kappa_{\ell,m})$. The notation for the dual pairing 
$\langle\langle \cdot, \cdot \rangle\rangle$ and for the inner product $(( \cdot, \cdot ))_{{}_{0}}$ in the Fock space
is taken from Subsection \ref{psiBerezin-Hida}.
Convergence of the series of ``symbols'' guarantees, by Thm 4.8 of \cite{obataJFA},
convergence of the Fock expansion
\[
\sum\limits_{\ell,m} \Xi(\kappa_{\ell,m}).
\]
The symbols 
\[
\big\langle \big\langle \overline{\Phi_{\eta}},  a_s(\widehat{n}\cdot \widehat{l}) \Phi_\xi \big\rangle \big\rangle,
\,\,\,\,\,\,
\big\langle \big\langle \overline{\Phi_{\eta}},  a_s(\widehat{n}\cdot \widehat{l})^+ \Phi_\xi \big\rangle \big\rangle
\]
of the Bose Hida operators and of their normal products 
can be easily computed from the property
\[
a_s(\widehat{n}\cdot \widehat{l}) \Phi_\xi = \xi(s, \widehat{n}\cdot \widehat{l}) \, \Phi_\xi, \,\,\,\,\,\, \xi \in E_1.
\]
The estimation of the averages in the states $\Phi_N$ of the Fermi operators (integral kernel operators)
can be computed as in the said Remarks, by the trivial replacement of the continuous orbits of momentum variables
and integrations of momenta replaced with discrete orbits of discrete momentum variables and summations, on the Einstein Universe. 
Suppose that we have a zero mass (or infinite orbit) free Fermi fields among the free fields underlying the theory in question.
This is, by what we know from experiment, rather unexpected. Even in this infinite dimensional case the operator norms 
\[
\| b_s(\widehat{n}\cdot \widehat{l}) \|_{{}_{\textrm{\emph{Op}}}} = \| b_{s}(\widehat{n}\cdot \widehat{l})^{+}\|_{{}_{\textrm{\emph{Op}}}} = 
1,
\]
of the Fermi Hida operators $b_s(\widehat{n}\cdot \widehat{l}), b_s(\widehat{n}\cdot \widehat{l})^+$ are finite. More generally the operator norms of the operators 
\[
\| b(\xi) \|_{{}_{\textrm{\emph{Op}}}} = \| b^{+}(\xi)\|_{{}_{\textrm{\emph{Op}}}} = \|\xi\|_{{}_{L^{2}}},
\,\,\,\,\, \xi \in E_1,
\]
are finite and the operators 
\begin{align*}
b(\xi) = \sum \limits_{(s, \widehat{n}\cdot \widehat{l}) \in \mathscr{O}} \xi(s, \widehat{n}\cdot \widehat{l}) \, b_{s}(\widehat{n}\cdot \widehat{l}),
\\
b^{+}(\xi) =
\sum \limits_{(s, \widehat{n}\cdot \widehat{l}) \in \mathscr{O}} \xi(s, \widehat{n}\cdot \widehat{l}) \, b_{s}(\widehat{n}\cdot \widehat{l})^{+},
\end{align*}
are bounded. Therefore the averages of the Fermi Hida operators and their normal products, and thus of the integral kernel Fermi 
contributions, or the averages of the Fermi contributions to the 
Fock expansion in the states of the form $\Phi_N$, can immediatly be obtained with the help of the above formulas 
for operator norm of the Fermi Hida operators and from the Schwarz inequality.

Recall please that the symbol calculus of Fermi integral kernel operators (fields) based on the even exponential states $\Phi_{\zeta}^{+}$
in the even part of Fermi Fock space is not effective here
if the subspace of coherent even Fermi exponential states with multiplicative  symbols is not dense
in the even part $(E_1)_+$ of the Fermi Hida space $(E_1)$. This is hardly possible even for the infinie dimensional Fermi Fock spaces 
in accordance with Remarks \ref{TretmentFermiIntegKerOp=TretmentBoseIntegKerOp} and \ref{TretmentFermiIntegKerOp=TretmentBoseIntegKerOpII}
of Subsection \ref{psiBerezin-Hida}. Therefore we do not use Fermi even exponential states, but instead
we are using the states $\Phi_N$ in the Fermi Fock space. 

Moreover in case when the Fermi Fock spaces are finite the explicit form of averages in the states 
$\Phi_N$ of the Fermi contributions are not relevant, because in this finite dimensional
case the contribution to the convergence coming from fields acting on finite-dimensional spaces trivializes. 

In particular, the Fermi integral kernel contributions to perturbative series (\ref{generalAintEU}) and (\ref{generalSonEU}) can be
trivially estimated in case of spinor QED on the Einstein Universe. Indeed, recall please, that the single particle Hilbert space
of the Dirac field, being massive, is finite dimensional, compare Subsection \ref{WhiteNoiseFreeFieldsonEU}. 
Because the Dirac field is half an odd integer spin field,
it is a Fermi field. Its Fock space, equal to the Fermi Fock space over a finite dimensional single particle Hilbert space,
is also finite dimensional. This means that each integral kernel operator $\Xi(\kappa_{\ell,m})$ in the expansion of  
(\ref{generalAintEU}) and (\ref{generalSonEU}) contains finite number of Fermi Hida creation and annihilation operators
(by the anticommutation relations for the Fermi creation-annihilation operators) and this number is bounded by a fixed
number depending on the dimension of the Fermi Fock space of the Dirac spinor field. Therefore the convergence of 
spinor QED, \emph{i.e.} convergence of the Fock expansions (\ref{generalAintEU}) and (\ref{generalSonEU}) in spinor QED
is governed only by the Bose integral kernel contributions in (\ref{generalAintEU}) and (\ref{generalSonEU}) 
(coming from the Bose electromagnetic potential field).
Therefore, the Bose-Berezin-Obata symbol calculus is sufficient for the analysis of the convergence  
of the Fock expansions (\ref{generalAintEU}) and (\ref{generalSonEU}), respectively, in 
\[
\mathscr{L}(\mathscr{E}, \mathscr{L}((\boldsymbol{E}),(\boldsymbol{E}))\big)
\,\,\,\,\textrm{and} \,\,\,\,
\mathscr{L}\big((\boldsymbol{E}),(\boldsymbol{E}) \big),
\]
for spinor QED on the Einstein Universe, and Thm 4.8 of \cite{obataJFA} can be applied for investigation of the convergence.

We have of course similar situation for any other massive Fermi half an odd integer spin field, 
coupled minimally to the electromagnetic potential field.

But we also immediately see that in QFT on the Einstein Universe of purely massive Fermi fields (no Bose fields) with the interaction
$\mathcal{L}$ defined through
a Wick polynomial in free Fermi fields, the the Fock expansions (\ref{generalAintEU}) and (\ref{generalSonEU}) are finite
and trivially convergent in this case. Therefore we have
\begin{twr}
The perturbative Fock expansions (\ref{generalAintEU}) and (\ref{generalSonEU}) are finite and thus convergent, respectively, in 
\[
\mathscr{L}(\mathscr{E}, \mathscr{L}((\boldsymbol{E}),(\boldsymbol{E}))\big)
\,\,\,\,\textrm{and} \,\,\,\,
\mathscr{L}\big((\boldsymbol{E}),(\boldsymbol{E}) \big),
\]
for QFT of massive Fermi fields with interaction $\mathcal{L}$ equal to a Wick polynomial in Fermi fields on the Einstein Universe.
\label{ConvergenceOfMassiveFermiQFTonEU}
\end{twr}

In particular this is the case for the old Fermi model -- the first model for electroweak interactions of the neutron, proton, electron,
positron and neutrino (provided we assume the neutrino to be massive).

This however does not finish the investigation of QFT with exclusively massive Fermi fields, as a successful theory.
There remain an important point to be investigated. 

Although the perturbative series (\ref{generalAintEU}) and (\ref{generalSonEU}) are finite in massive Fermi
QFT on the Einstein Universe, we  cannot yet be fully content with such QFT theories. Although we are 
content with the finiteness of the perturbative series (\ref{generalAintEU}) for interacting fields representing
well defined interacting fields, we cannot be fully content yet with the finite series (\ref{generalSonEU})
for the scattering operator $S(g=1) \in \mathscr{L}\big((\boldsymbol{E}),(\boldsymbol{E}) \big)$. 

This is because we are 
not fully content with $S(g=1)$ regarded as an operator transforming continuously the Hida space into itself (here simply bounded operator). 
In case of Einstein Universe and massive Fermi fields the finite expansion (\ref{generalSonEU}) represents
a well defined and bounded operator, because the Hida operators are bounded, the orbits of the fields are finite
as well as the total Fock space  is finite dimensional.  
We expect however of the fully successful QFT that the scattering operator $S(g=1)$ is not only a bounded operator
in the Fock space, but it should be a unitary operator in the Fock space. Thus, the unitarity is to be investigated
(and expected by the formal unitarity to be preserved by assumption in the causal construction of the statternig operator).

Although any bounded operator
admits a Fock expansion (\ref{generalSonEU}) into integral kernel operators $\Xi(\kappa_{\ell,m})$
with scalar valued distributional kernels $\kappa_{\ell,m}$ (\cite{Berezin}, or \cite{obata} Cor. 6.2)
it is possible only for infinite Fock expansion (\ref{generalSonEU}), except the trivial case
of the scalar operator $\Xi(\kappa_{0,0})$ proportional to the unit operator, or except the case the expansion  (\ref{generalSonEU})
includes purely Fermi integral kernel operators.  Therefore, in theories including Bose fields, e.g. QED, with the Bose integral kernel operators
necessary included in  (\ref{generalSonEU}) through the Bose fields contributions (electromgnetic potential field in QED)
the expansion (\ref{generalSonEU}) has to be infinite in case it represents a bounded scattering operator $S(g=1)$. 
Indeed, by Proposition
6.3 of \cite{obata}, if the integral kernel operator $\Xi(\kappa_{\ell,m})$ in the Bose Fock space admits extension to a bounded
operator in the Bose Fock space, then $\Xi(\kappa_{\ell,m})=0$ or $\ell=m=0$. Therefore the Fock expansions of bounded operators
(including Bose integral kernel operators) are necessary infinite. Recall, please, that such expansion cannot be investigated
within the Hilbert space structure, and has nothing to do e.g. with convergence in Hilbert space operator topology,
or strong or weak operator topology. In particular each integral kernel operator $\Xi(\kappa_{\ell,m})$, with non trivial Bose contribution,
is unbounded and cannot be treated purely within the Hilbert space structure, but rather within the infinite dimensional analogue of the Schwartz
test nucler space -- the test Hida space. The bounded operator in its Fock expansion is regarded as an element of    
\[
\mathscr{L}\big((\boldsymbol{E}),(\boldsymbol{E})^* \big),
\]
and regarded as a generalized operator, which, as such, admits Fock expansion,
compare Thm. 6.1 of \cite{obata}. 

The situation for scalar QED, where we have the electromagnetic potential field 
coupled minimally to a (necessary massive) spin zero
field, is not so trivial. In this case the spin-zero field, although being massive
and thus with finite dimensional single particle Hilbert space (Subsection \ref{WhiteNoiseFreeFieldsonEU}), still has the Bose Fock space
of infinite dimension. Similar less trivial situation we have for any other massive and integer
spin field minimally coupled to the electromagnetic potential field. Because the massive field
is now a Bose field, then it has the Bose Fock space infinite dimensional, although it has the Bose
Fock space over a finite dimensional single particle Hilbert space. Fortunately in this case, of 
Bose fields, we have the powerful symbol calculus of Berezin-Obata, and Thm 4.8 of \cite{obataJFA} 
can immediately be applied to such cases. More generally because in the ``Standard Model'', in its version with
the Higgs field, we do not encounter massless (or infinte orbit) Fermi fields, then the convergence of the perturbative
series for this ``Model'' is expected by the effect of couplings of the Bose fields (most of which are massive) between themselves
and their couplings to the massive fermi quark fields, with finite dimensional Fermi Fock spaces, which cancels out most of the terms
of the expansion. Therefore Thm 4.8 of \cite{obataJFA} can be applied to the investigation of the convergence
of this ``Standard Model''. 
Of course investigation of the convergence of the modified version of the ``Standard Model'' including also
massless Fermi fields (which seem to be rather less interesting from the physical point of view)  
also can be reduced to the application of Thm 4.8 of \cite{obataJFA} by the method which we have explained above.

Application of Thm 4.8 of \cite{obataJFA} will allow us to classify all Wick polynomial interactions $\mathcal{L}$, or QFT, 
on the Einstein Universe
for which the perturbative Fock expansions (\ref{generalAintEU}) and (\ref{generalSonEU}) 
for $\mathbb{A}_{{}_{\textrm{int}}}$ and $S(g=1)$ are convergent, when regarded as perturbative Quantum Field Theories. 
Contrary to the Minkowski
space-time, situation on the Einstein Universe is much better also regarding convergence of QFT with interactions
which fall into the case 1) of Subsection \ref{WickForChronologicalEU}, including QED. 
We hope to present results concerning
convergence of the perturbative series, including spinor and scalar QED, in a separate work.

In particular the series (\ref{generalAintEU}) for interacting field 
$\mathbb{A}_{{}_{\textrm{int}}}$ converges as the series of integral kernel operators with vector-valued kernels $\kappa_{l,m}$ in the sense of Thm 4.8 of \cite{obataJFA} if and only if it represents an element $\mathbb{A}_{{}_{\textrm{int}}}$ of
\[
\mathscr{L}\big( (\boldsymbol{E}) \otimes \mathscr{E}, \, (\boldsymbol{E}) \big) \cong
\mathscr{L}\Big( \mathscr{E}, \,\, \mathscr{L}\big( (\boldsymbol{E}), (\boldsymbol{E})\big) \, \Big),
\]
with continuous map
\[
\mathscr{E} \ni \phi \longmapsto
\mathbb{A}_{{}_{\textrm{int}}}(\phi) = \sum \limits_{l,m} \Xi\big(\kappa_{l,m}(\phi)\big)
\in \mathscr{L}\big( (\boldsymbol{E}), (\boldsymbol{E})\big).
\]
In this case the variational derivative in the Bogoliubov definition of the interacting
field $\mathbb{A}_{{}_{\textrm{int}}}$ is not only formal but can be given a sense
of the limit
\[
\mathbb{A}_{{}_{\textrm{int}}}(\phi)\Phi
= \lim \limits_{\epsilon \rightarrow 0}
S^{-1}(\mathcal{L})\frac{S\big(\mathcal{L} +\epsilon\phi \mathbb{A}\big) - S(\mathcal{L})}{\epsilon}
\Phi, \,\,\,\, \Phi \in (\boldsymbol{E}), \,\,\,
\phi \in \mathscr{E},
\]
in the nuclear topology of the Hida space $(\boldsymbol{E})$.

Of course one of the most important applications of Theorem \ref{InteractingFieldsAtxOnEU} is the computation of the interacting
field $\mathbb{A}_{{}_{\textrm{int}}}(x)$ which corresponds to the conserved current
(generalization of the energy-momentum density) $\mathbb{A}(x)$ which is equal to the Wick
product of quantum free fields corresponding to the classical conserved free fields current 
in the Emmy Noether integral corresponding to the one-parameter group of symmetries generated by one of the vector fields
$X^\mu$ of Subsection \ref{GeneralizedSchrodinger-VonNeumannPairs}. In particular not only the higher order 
contribution to the Emmy Noether conserved integral are well defined essentially self-adjoint operators in the Fock space 
transforming continuously the Hida space into itself. But moreover if the Noether integral corresponds to the global gauge transformation
and represents the conserved total electric charge for interacting fields, each higher order contribution to the 
conserved current (the density whose integration over the equal-time Cauchy surface gives the Noether generator) 
evaluated at specified space-time point is a well defined operator $\mathbb{A}_{{}_{\textrm{int}}}(x)$ in the 
Fock space transforming continuously the Hida space into itself. This is because the density $\mathbb{A}_{{}_{\textrm{int}}}$ in this case is equal to the interacting
field which corresponds to the Wick product $\mathbb{A}$ of free charge carrying fields. According to the last Corollary these fields
must be massive. Thus by Theorem \ref{InteractingFieldsAtxOnEU}, all higher order (including zero order) contributions to these densities, 
even evaluated at single space-time point $x$,
give ordinary operators on the Fock space continuously transforming the Hida space into itself. 

This gives the positive
solution to the Problem  I of Subsection \ref{G} of Introduction.
In particular let  $\boldsymbol{P}^{{}^{0}}$, $\mathbb{P}^{{}^{k}}$, $k=1,2,3$,
be the Noether integral generators in the Fock space of free fields, 
corresponding to the one-parameter subgroups of symmetries generated by $X^0$, $X^k$,
respectively. Note that the generators $\boldsymbol{P}^{{}^{0}}$, $\mathbb{P}^{{}^{k}}$, $k=1,2,3$,
restricted to the respective invariant subspaces of the Fock space, described in Subsection
\ref{GeneralizedSchrodinger-VonNeumannPairs}, can be identified with the generators
$\boldsymbol{P}^{{}^{0}}$, $\mathbb{P}^{{}^{k}}$, of the space-time spectral tuple
(\ref{3rdSpectralTupleForSxG}) of Subsection \ref{GeneralizedSchrodinger-VonNeumannPairs}.
In particular all higher order contributions to 
$\boldsymbol{P}^{{}^{0}}_{{}_{\textrm{int}}}, \mathbb{P}^{{}^{k}}_{{}_{\textrm{int}}}$
are, by what we have just shown, well defined operators in the Fock space, even transforming
continuously the Hida space into itself. In particular easy application  of the 
Riesz-Sz\"okefalvy-Nagy  criterion (\cite{Riesz-Szokefalvy}, p. 120) together with the nuclearity 
of the Hida space
$(\boldsymbol{E})$, Hermicity of the higher order contributions and invariance of $(\boldsymbol{E})$ under the higher order contributions
to $\boldsymbol{P}^{{}^{0}}_{{}_{\textrm{int}}}, \mathbb{P}^{{}^{k}}_{{}_{\textrm{int}}}$,
will show (\cite{Segal_Kunze}, Chap. 10.3) that $\boldsymbol{P}^{{}^{0}}_{{}_{\textrm{int}}}, \mathbb{P}^{{}^{k}}_{{}_{\textrm{int}}}$,
up to any arbitrarily high order, are essentially self-adjoint operators. We thus have solved partially
the Problem II of Subsection \ref{G}. Further investigation of this Problem will be continued 
in the next Subsection.

\subsection{The key points used in the proofs}

We have obtained results which distinguish two cases of QFT on EU, compactified and ordinary EU. In case 1)
all Wick monomials in the interaction Lagrangian $\mathcal{L}$ contain at most one massless field. In case 
2) some Wick monomials in $\mathcal{L}$ contain at least two massless factors.

The difference between the two cases 1) and 2) on EU, compactified or not, which are 
distinguished in theorems of Subsections 
\ref{WickForChronologicalEU} and \ref{CausalSonEU}, arises only because
we are using the natural free fields, constructed in \cite{PaneitzSegalI}-\cite{PaneitzSegalIII}, \cite{SegalZhouQED}, 
on the ordinary EU (non compactified).  These fields are all invariant under the discrete
symmetry $\zeta^4$, where $\zeta: (t, \boldsymbol{w}) \rightarrow (t+\pi, \textrm{Antipode of} \, \boldsymbol{w})$ in units 
in which $\hslash=c=R=1$ and in the realization of EU as the product 
$\mathbb{R} \times \mathbb{S}^3$ of $\mathbb{R}$ and the unit 3-sphere.
The symmetry $\zeta$ lies in the center of the group of all smooth causal automorphisms 
of the natural casual structure of EU. This restriction, put on the free fields, is not arbitrary,
as noted in \cite{SegalZhouQED}, in accordance with the justification we have already 
referred to at the beginning of Subsection \ref{WhiteNoiseFreeFieldsonEU}. It is best understood for the 
massless fields, which are causally invariant. The Minkowski space-time is causally periodically
embedded into the EU, with the period $\zeta$. Minkowski mass packets extend uniquely to wave functions defined on the whole
ambient EU, regarded as the causal manifold. Natural fields are those and only those fields which can
arise from such extension. They turn out to be periodic with period $\zeta$ for the e.m. potential field,
and periodic with period $\zeta^4$ for the spinor field. In each case, they are invariant under $\zeta^4$. 
In principle, there are many other solutions of the corresponding
wave equations on EU, which are not periodic, but they are pathological and, when extended far enough on EU, 
lead to tachyonic behavior, \emph{i.e.} with the Fourier transform (when regarded as functions on the 
embedded causal Minkowski space-time) with the support lying outside the cone $p^2>0$ in the momentum space. 
The above restriction put on free fields is supported  also by the local scattering phenomena to which we successfully
apply the Minkowski wave packets. We can consistently assume that space-time is not globally flat but instead, say, the EU, 
without disturbing the successful use of the Minkowski free fields in the local scattering phenomena, only if we eliminate
the superfluous amount of non-extendible waves, which coincide locally with the plane Minkowski waves. Otherwise, we would
obtain enormous excess of quantum numbers, which is not observed. Only taking into account the natural
relation of the free fields on EU with the free fields on the Minkowski space-time, we arrive at the conclusion
that the realistic wave functions on EU are $\zeta^4$-invariant. This periodicity, together with the nonzero curvature of EU
leads to the conclusion that the Fourier support (with the Fourier transform on EU) of massive fields on EU
is finite, compare \cite{PaneitzSegalI}-\cite{PaneitzSegalIII}, \cite{SegalZhouQED}, and thus the orbits
$\mathscr{O}$,$\mathscr{O}^+$,$\mathscr{O}^-$ of massive free fields on EU are finite. This is the key point in the proofs
of Subsections \ref{WhiteNoiseFreeFieldsonEU}, \ref{WickForChronologicalEU} and \ref{CausalSonEU}.

The singularity order $\omega$ of distributions, which we need to split into retarded and advanced part to compute the kernels of 
the higher order contributions to the scattering operator, is a local quantity. $\omega$ makes sense on any space-time 
(provided we have constructed division of the waves into positive and negative frequency solutions, 
using \emph{e.g.} the Hamiltonian formulation of the classical theory, which in general is not easy, and is accompanied by
the problem of constrains, and which trivializes in case of EU).  Here, in case of EU, we have two choices: either (A) to put the 
condition on the free fields, which eliminates
the superfluous excess of waves on EU, in order to keep the natural relation between the free fields on EU and on the Minkowski space-time, 
or (B) to put the requirement that the Green functions on EU should have the same singularity order $\omega$
as the corresponding Green functions on the Minkowski space-time. The problem is that both (A) and (B) are contradictory
and both cannot be preserved. In order to preserve the singularity degree $\omega$ we need to allow all solutions
of the invariant wave equations on EU, which correspond to the wave equations on the Minkowski space-time, including the
non-periodic solutions. We agree with \cite{PaneitzSegalI}-\cite{PaneitzSegalIII}, \cite{SegalZhouQED}, and reject
(B) as nonphysical. The results on EU, based on (A), and given in 
Subsections \ref{WickForChronologicalEU} and \ref{CausalSonEU}, 
we have extended over the Minkowski space-time, in  a weakened version. This supports (A).
The use of the white-noise calculus in proving the difference between the two cases, 1) and 2), mentioned above,
is not essential on EU. This result could have been achieved using the method of 
\cite{PaneitzSegalI}-\cite{PaneitzSegalIII}, \cite{SegalZhouQED}, where it was proved for spinor QED on EU. But the white
noise calculus is indispensable in the proof of a weakened extension of these results on the Minkowski space-time,
compare Subsection \ref{OperationsOnXiIF}, Theorems \ref{ExistenceIntFields.g=1.m>0}--\ref{NonExistenceIntFields.g=1.m=0QFT}.  

That the results are weakened in passing to Minkowski spacetime 
means that on this space-time the higher-order corrections to interacting fields 
are singular, being continuous mappings from the test space to the space 
$\mathcal{L}\big((\boldsymbol{E}), (\boldsymbol{E})^*\big)$. 
On flat space-time, it is impossible to obtain these corrections as operator distributions, \emph{i.e.}
continuous maps of the test space into $\mathcal{L}\big((\boldsymbol{E}), (\boldsymbol{E})\big)$.

But if we had assumed (B) and rejected (A), we would get QFT on EU which is as singular in case 1) as in case 2), and equally singular
as on the Minkowski space-time, with the higher order contributions to interacting fields that are not operator-valued
distributions, but more singular generalized operators, transforming continuously the test space into the space of generalized
operators $ \mathscr{L}\big( (\boldsymbol{E}), (\boldsymbol{E})^*\big)$.

\subsection{Experimental consequences}

We have distinguished two substantially different QFT's: 1) with each of the Wick monomials in the 
interaction Lagrangian $\mathcal{L}$ containing at most one massless field, or 2) with some of the Wick
monomials in $\mathcal{L}$ containing more than one massless field.
The case 2) is singular by the theorems of Subsections \ref{WickForChronologicalEU} and \ref{CausalSonEU}, 
and we classify the QFT's of 2)-class as nonphysical.

The substantial difference between the above two cases, 1) and 2), of  QFT's,
has a profound relation to the so-called
``Higgs mechanism'' and nonzero neutrino mass. In principle, the Yang-Mills fields with more complicated gauge group, as, e.g.,
Yang-Mills fields, which are used in the ``Standard Model'', even when not coupled to the other Fermi charged fields, compose 
a non-linear system of self-interacting massless fields. Taken literally, such a theory would have a very restricted meaning,
in accordance with the mentioned results, because some of the Wick monomials of the interaction Lagrangian contain 
products of massless gauge fields. 
But, using the Higgs ``mechanism'', we arrive at a system with massive vector Bose fields
and a massive Higgs field, interacting with the Fermi charged fields, with the first order interaction,
containing no Wick monomials with more than one massless Bose field. But in case of the Glashow-Weinberg-Salam system,
we still have Wick monomials in the (first order) interaction Lagrangian, which contain
products of neutrino fields. Rejecting the singular case 2) as nonphysical QFT's, we arrive at the conclusion
that neutrino should be massive. 
Only with the spontaneously broken symmetry and nonzero neutrino mass 
for the Glashow-Weinberg-Salam system of interacting fields 
we arrive with QFT with the (first order) interaction Lagrangian, that no longer contains
Wick monomials with more than one massless field.

Taking into account the substantial difference between the two cases, 1) and 2), this ``mechanism'' 
seems to have a more profound meaning, allowing us to avoid the singularities of case 2). 
Assuming that QFT's of type 2) are nonphysical, we expect spontaneous symmetry breaking 
also in QCD with non-zero gluon masses. Whether the gluon mass is non-zero 
or not remains experimentally open for now. But based on the obtained results, we expect non-zero 
gluon masses and the existence of additional Higgs bosons associated with spontaneous symmetry breaking in QCD.

Another corollary is that, e.g., the scalar QED on EU is a singular theory on EU, with the interaction 
Lagrangian containing the Wick monomial of second 
order in the massless e.m. potential. Rejecting QFT's of the 2)-class as nonphysical, we arrive at the conclusion that 
the scalar QED is a nonphysical theory. This is in agreement with experiment, as we do not observe elementary 
spin-zero electrically charged particles, but only composite and unstable. This is again in agreement with what we obtain 
on the Minkowski space-time using the white-noise calculus. Let us recall
that for the scalar QED we have
\[
S_1(x_1) = i {:} \mathcal{L}(x_1){:},
\]     
with
\begin{equation}\label{LscalarQED}
\mathcal{L}(x_1) = i e \left( \nabla_\mu \boldsymbol{\varphi}^{+}(x_1) \boldsymbol{\varphi}(x_1) 
-\boldsymbol{\varphi}^{+}(x_1) \nabla_\mu  \boldsymbol{\varphi}(x_1) \right) A^\mu(x_1) 
+ e^2 \boldsymbol{\varphi}^{+}(x_1) \boldsymbol{\varphi}(x_1)    A_\mu(x_1) A^\mu(x_1),
\end{equation}
containing the term with the product of two massless fields. Some of the the fourth-, 
and higher-order contributions to interacting fields will be singular.
We cannot avoid singularities of this theory by simply using a more consequent perturbative approach, in which
we consequently use only the first order term of the above interaction Lagrangian in the first order
contribution $S_1(x)$, and recover the second order term of (\ref{LscalarQED}) when computing $S_2(x_1,x_2)$, 
using the charge conjugation invariance, gauge invariance, and the freedom in the splitting of the corresponding distribution, multiplying
\[
e^2 {:} \boldsymbol{\varphi}^{+}(x_1) \boldsymbol{\varphi}(x_2)    A_\mu(x_1) A^\mu(x_2){:}
\]
(which was of singularity order $\omega=0$ on Minkowski space-time). If it was of order $\omega=0$ on EU 
(as on the Minkowski space-time) we would have a free choice of the Epstein-Glaser remainder term containing arbitrary
constant $C$: 
\begin{equation}\label{EUremainderScalarQED}
-i e^2 C \delta(x_1x_2^{-1}) {:}\boldsymbol{\varphi}^{+}(x_1) \boldsymbol{\varphi}(x_2)    A_\mu(x_1) A^\mu(x_2){:}
+ i e^2 C \delta(x_1x_2^{-1}) {:}\boldsymbol{\varphi}^{+}(x) \boldsymbol{\varphi}(x)    A_\mu(x) A^\mu(x){:},
\end{equation}
fixed by the charge conjugation invariance and the gauge invariance, and equal to the 
second order term in (\ref{LscalarQED}), after integration over $x_2$ and division by $2!$, in accordance with the definition 
of the second order contribution to the Lagrangian arising from the free choice terms in computation of the splitting, 
\cite{Bogoliubov_Shirkov}. But, with the physical assumption (A) and natural free fields on EU, the massive field 
$\boldsymbol{\varphi}$ is a field with finite orbit $\mathscr{O}= \mathscr{O}^+ \sqcup \mathscr{O}^-$, 
and the said distribution is of negative order $\omega$ on EU, so that
the splitting is unique, and the retarded and advanced part is unique and given by ordinary multiplication by $\theta$. 
Thus, if we use only the first order term of (\ref{LscalarQED}) in the first order contribution $S_1(x_1)$ on EU, 
we cannot preserve gauge invariance. Even if we ignore the axiom (V) of preservation of the singularity degree (which will make
the QFT undetermined), and add the terms (\ref{EUremainderScalarQED}) to the retarded part of the said 
distribution in order to restore gauge invariance, then we recover the second order part of (\ref{LscalarQED}), 
but then some fourth- and higher-even-order contributions to interacting fields
become singular, as we already said. In each case, scalar QED is singular on EU. Analogous result we have
on the Minkowski space-time, compare the end of Subsection \ref{OperationsOnXiIF}.

\subsection{What are the Hida operators and the compactified EU for? Feynman integral}

The white noise calculus and Hida operators are essential for the results we have achieved 
for perturbative QFT on the Minkowski space-time. The analog results for perturbative
QFT on EU could have been achieved without the use of white noise analysis \cite{SegalZhouQED}.
But in going beyond the perturbative methods to the computation of the full Green functions
\[
\Delta(x,y) = i {\textstyle\frac{\left\langle T \boldsymbol{\varphi}(x)\boldsymbol{\varphi}(y) S \right\rangle_{{}_{0}}}{\langle S \rangle_{{}_{0}}}},
\,\,\,\,\,
\textrm{$\langle \cdot \rangle_{{}_{0}}$ -- vacuum average},
\] 
which are based on the functional averaging, we found the white noise Hida test space and the Hida 
generalized functions indispensable on EU.

We present here a rigorous construction of the integral averaging, based on the method outlined in
\cite{Bogoliubov_Shirkov}, \S\S 43.1 -- 43.4, and performed within the momentum representation.
For the compactified EU $\simeq G = [\mathbb{R} \, \textrm{mod} \, 4\pi] \times SU(2,\mathbb{C})$, 
and periodic $\theta$ (in ret and av), the Fourier transform has purely discrete 
range, which allows us to avoid passing to the continuum
limit of the original idea of \cite{Bogoliubov_Shirkov}, \S\S 43.1. We construct first the averaging
for the compactified EU and then eventually pass to the case in which $\theta$ has larger, 
but always an integral multiple of the period of the  free fields on EU, but arbitrarily
large period. Here we recall the idea of \cite{Bogoliubov_Shirkov}, \S 43,
but applied to the compactified EU, and propose spaces of functions $\nu,\varphi$ 
to ensure the convergence of the method -- the problem posed there.

For simplicity, we consider after \cite{Bogoliubov_Shirkov} the real (essentially neutral)
scalar field $\boldsymbol{\varphi}(x)$ on $G$, and whenever we are using bold letter, we mean
the free field operator, and not ordinary functions. Ordinary functions are written with non-bold letters. 

We are using the standard Fourier transform $\varphi \rightarrow \widetilde{\varphi}$ on $G$, as in the previous Subsections,
and the ordinary $L^2$-inner product $\langle \cdot, \cdot \rangle$ for (here real) functions on
$G$ with the ordinary invariant measure on $G$, normalized to $4\pi$ (our $G$ is compact).

We start at once with the definition of the nuclear spaces within which we will be seeking
the proper domains for $\nu,\varphi$. First we use the real space-time Gelfand triple
\[
E = \mathcal{S}_{A}(G;\mathbb{R}) \subset L^2(G;\mathbb{R}) = H \subset  \mathcal{S}_{A}(G;\mathbb{R})^* = E^*
\]  
with the standard operator $A = -4\partial_{t}^2 \otimes 1 - 1 \otimes \Delta_{{}_{G}} +1$ on $L^2(G;\mathbb{R})$,
and its Bose-Fock lifting to the Gelfand triple over the Bose-Fock space $\Gamma(H)$:
\[
(E) = \mathcal{S}_{\Gamma(A)}(E^*;\mathbb{R}) \subset L^2(E^*, d\mu) = \Gamma(H) \subset  \mathcal{S}_{\Gamma(A)}(E^*;\mathbb{R})^* = (E)^*.
\]  
Here $\mu$ is the standard Gaussian probability measure on $E^*$, with the canonical isomorphism
$L^2(E^*, d\mu) = \Gamma(H)$ given by the Wiener-It\^o-Segal chaos decomposition, compare Subsection \ref{white-setup}
or \cite{obata-book}, \cite{GelfandIV}. Below we will be referring also to theorems
involving complexifications $E_{\mathbb{C}}, H_\mathbb{C}, E_\mathbb{C}^*$ of the spaces $E,H,E^*$
and their Fock liftings, but we should emphasize that the real versions of these spaces play a distinguished
role not only for the real field, and the existence of the measure $\mu$ is associated with the reality of the
considered spaces. In what follows we identify the spaces  $E,H,E^*$ with their Fourier transform
isomorphic images $\widetilde{E},\widetilde{H},\widetilde{E}^*$, and correspondingly
we identify their Fock liftings, with the inner product canonically 
inducing the dual pairings $\langle \nu, \varphi \rangle = \langle \widetilde{\nu}, \widetilde{\varphi}\rangle$, 
given by the Plancherel formula, with the standard operator $\widetilde{A}$ equal to the Fourier transform of $A$.
The dual paring in the Fock lifted Gelfand triple is induced by the inner product $\langle\langle \cdot, \cdot \rangle\rangle$
in $\Gamma(H)$, which in the function representation, via the Wiener-It\^o-Segal isomorphism, is equal to the
$L^2$ norm on  $L^2(E^*, d\mu)$, compare Subsection \ref{white-setup}.  In the sequel, we will use the Hilbert
spaces $E_k$, $k \in \mathbb{Z}$ -- the closures of the domains of $A^k$ in $H$, with respect to the Hilbertian norms 
$| \cdot |_{{}_{k}} \overset{def}{=} |A^k \cdot|_{{}_{0}}$, where $|\cdot |_{{}_{0}}$ is the Hilbert space norm of $H$.
The space $E$ becomes then a projective limit $\cap_{{}_{k}} E_k$ of $E_k$, with its dual equal to the inductive limit 
$E^* = \cup_{{}_{k}} E_k$. We again identify these spaces with their Fourier transform images, $\widetilde{E}_k$,
with the standard operator equal $\widetilde{A}$. We have the natural continuous inclusions 
\[
E \subset \ldots E_{k} \subset \ldots E_{k-1} \subset \ldots \subset E_{1} \subset E_0 = H 
\subset E_{-1} \subset E_{-k +1} \subset \ldots E_{-k} \subset \ldots \subset E^* 
\] 
with $E^*_{k}$ canonically isomorphic to $E_{-k}$ for $k\in \mathbb{Z}$ (in fact, we can use any real numbers for $k$).
We have analogous Hilbert $(E)_k$ spaces and continuous inclusions, and the realizations as the projective and inductive limits
for the Hida nuclear spaces $(E), (E)^*$, 
with the standard operator $A$ replaced with the standard operator $\Gamma(A)$. For a more detailed description of this construction
of Hida test space and its dual, we recommend \cite{obata-book}. In what follows, we use the fact that it has already been exploited by us in the construction
of the free fields on EU, that we have at our disposal a natural way of constructing closed subspaces of the Gelfand triple
$E\subset H\subset E^*$, which again compose Gelfand triples naturally embedded in the original Gelfand triple $E\subset H\subset E^*$. 
Indeed, we can do it just by passing to the Fourier images $\widetilde{E}\subset \widetilde{H}\subset \widetilde{E}^*$ 
and by selecting any subset $\mathscr{R}$ of the characters $\widehat{n}\cdot\widehat{l}$, 
and restrict the elements of the spaces to these and only these
elements with the Fourier transform, supported at this subset. The corresponding standard operator becomes equal to 
the restriction of $\widetilde{A}$ to the elemets of the initial $\widetilde{H}$, with the support at this fixed subset $\mathscr{R}$.
The corresponding closed subspaces $E_{{}_{\mathscr{R} \,\, k}} \subset E_{{}_{k}}$, 
and their productive and inductive limits, composing the corresponding Gelfand triples 
$E_{{}_{\mathscr{R}}} \subset H_{{}_{\mathscr{R}}} \subset E_{{}_{\mathscr{R}}}^*$, 
will frequently be denoted without the subscript $\mathscr{R}$,
in order to simplify notation, whenever it is clear whether we mean an element $\widetilde{\xi} \in \widetilde{E}_{{}_{k}}$
supported at $\mathscr{R}$, and belonging to $\widetilde{E}_{{}_{\mathscr{R} \,\, k}}$, or generally whenever it is clear that
we mean $\widetilde{\xi}$ supported at $\mathscr{R}$. In particular the Gelfand triples
corresponding to the real field with the positive energy orbit $\mathscr{O}^+$, are constructed from the Gelfand
triple $\widetilde{E}\subset \widetilde{H}\subset \widetilde{E}^*$, just by restriction of the Fourier transform components 
to the subset $\mathscr{R} = \mathscr{O}^+$.

Note here, that also each  element of $\widetilde{E}^*$ can be regarded as a Fourier series, but the condition
$\widetilde{\varphi} \in \widetilde{E}^*$ is equivalent with existence of $k$, such that  $\widetilde{\varphi} \in \widetilde{E}_{-k}$,
which, in turn, is equivalent to
\[
\sum (2l+1) \textstyle{\frac{1}{(n^2 + 1 + l(l+1))^{2k}}}
\widetilde{\varphi}(\widehat{n}\cdot\widehat{l})_{{}_{i \, j}}\overline{\widetilde{\varphi}(\widehat{n}\cdot\widehat{l})_{{}_{i \, j}}}
< +\infty, 
\]
with the summation range $n\in \mathbb{Z}, l \in \mathbb{N}, -l \leq i,j \leq l$ for the initial Gelfand triple
$\widetilde{E}\subset \widetilde{H}\subset \widetilde{E}^*$ or, respectively, with the
summation range $\widehat{n}\cdot\widehat{l} \in \mathscr{R}$, $ -l \leq i,j \leq l$, for 
$\widetilde{\varphi} \in E_{{}_{\mathscr{R}}}^*$ defined by the restriction of the support to  $\mathscr{R}$. Below, we consider real
$\nu, \varphi$, so that 
\[
\overline{\widetilde{\varphi}(\widehat{n}\cdot\widehat{l})_{{}_{i \, j}}} = \widetilde{\varphi}(\widehat{-n}\cdot\widehat{l})_{{}_{-i \, -j}},
\,\,\,
\overline{\widetilde{\nu}(\widehat{n}\cdot\widehat{l})_{{}_{i \, j}}} = \widetilde{\nu}(\widehat{-n}\cdot\widehat{l})_{{}_{-i \, -j}} 
\]
and for real $\varphi$
\begin{multline*}
\langle \widetilde{\varphi}, \widetilde{\varphi} \rangle
=
\sum (2l+1)
\widetilde{\varphi}(\widehat{n}\cdot\widehat{l})_{{}_{i \, j}}\overline{\widetilde{\varphi}(\widehat{n}\cdot\widehat{l})_{{}_{i \, j}}}
\\
= 
\sum (2l+1)
\widetilde{\varphi}(\widehat{n}\cdot\widehat{l})_{{}_{i \, j}} \widetilde{\varphi}(\widehat{-n}\cdot\widehat{l})_{{}_{-i \, -j}}
\\
=
\sum (2l+1) \left[
x_{{}_{\widehat{n}, \widehat{l} \, i \, j}}^2  + y_{{}_{\widehat{n}, \widehat{l} \, i \, j}}^2
\right]
\end{multline*}
if 
\begin{multline*}
\widetilde{\varphi}(\widehat{n}\cdot\widehat{l})_{{}_{i \, j}} =
x_{{}_{\widehat{n}, \widehat{l}, \, i \, j}} + i y_{{}_{\widehat{n}, \widehat{l}, \, i \, j}}^2,
\\
\textrm{Re}  \, \left[ \widetilde{\varphi}(\widehat{n}\cdot\widehat{l})_{{}_{i \, j}} \right] = 
x_{{}_{\widehat{n}, \widehat{l}, \, i \, j}},
\,\,\,
\textrm{Im}  \, \left[ \widetilde{\varphi}(\widehat{n}\cdot\widehat{l})_{{}_{i \, j}} \right] = 
y_{{}_{\widehat{n}, \widehat{l}, \, i \, j}},
\end{multline*}
because the real part is symmetric, and the imaginary part is skew-symmetric:
\[
x_{{}_{\widehat{-n}, \widehat{l}, \, -i \, -j}} = x_{{}_{\widehat{n}, \widehat{l} \, i \, j}},
\,\,\,
y_{{}_{\widehat{-n}, \widehat{l}, \, -i \, -j}} = - y_{{}_{\widehat{n}, \widehat{l}, \, i \, j}}.
\]

We will do computations in the momentum representation (after the Fourier transformation). Recall that the free field has the 
form of the Fourier transform, which, in the case of compactified EU $=G$ (and also for ordinary non-compactified EU), 
has the form of the Fourier series with well-defined Fourier-operator coefficients 
$\widetilde{\boldsymbol{\varphi}}(\widehat{n}\cdot\widehat{l})_{{}_{i \,\, j}}$.
This is the consequence of the $\zeta^4$invariance \cite{PaneitzSegalI}-\cite{PaneitzSegalIII}.   
This series is moreover finite, if the field $\boldsymbol{\varphi}$ is massive. For the real scalar field, we have
\[
\widetilde{\boldsymbol{\varphi}}(\widehat{n}\cdot\widehat{l})_{{}_{i \,\, j}} =
\begin{cases}
{\textstyle\frac{\sqrt{4\pi}}{2l+1}}\sum\limits_{s=1}^{(2l+1)^2}\left[
u_{{}_{s}}(\widehat{n}\cdot\widehat{l})_{{}_{i \,\, j}} a_{{}_{s}}(\widehat{n}\cdot\widehat{l})
+
\overline{u_{{}_{s}}(\widehat{n}\cdot\widehat{l})_{{}_{-j \,\, -j}}} a_{{}_{s}}(\widehat{n}\cdot\widehat{l})^+\right], & 
\textrm{if} \,\,\, 
\widehat{n}^+\cdot\widehat{l} \in \mathscr{O}^+, \\
0, &  \textrm{if} \,\,\, 
\widehat{n}^+\cdot\widehat{l} \notin \mathscr{O}^+,
\end{cases}
\]
where $\mathscr{O}^+$ is the positive energy orbit
corresponding to the real scalar field ($\mathscr{O}^+, \mathscr{O}^-$ are finite sets if the field is massive, compare Subsection \ref{WhiteNoiseFreeFieldsonEU}).
The orbit is determined by the invariant equation respected by the free field, compare Subsection \ref{WhiteNoiseFreeFieldsonEU}, and the finite range
matrices $u_{{}_{s}}(\widehat{n}\cdot\widehat{l})$ are the Fourier transforms of fundamental solutions, concetrated on the single point character
set $\{ \widehat{n}\cdot\widehat{l}\}$ (non-scalar fields will have $u$ with one more index $a$ counting the number of components
of the field $\boldsymbol{\varphi}$, compare Subsection \ref{WhiteNoiseFreeFieldsonEU}). Aso the finite range of the discrete index $s$ depends 
on the actual character $\widehat{n}\cdot\widehat{l}$ and on the kind of field. For the scalar field $s$ ranges among all positive integers
from $1$ to $(2l+1)^2$, and below in the sums, we will not write it explicitly in order to simplify notation.   
$a_{{}_{s}}(\widehat{n}\cdot\widehat{l}), a_{{}_{s}}(\widehat{n}\cdot\widehat{l})^+$ are 
the Hida annihilation-creation operators, which for real scalar field in addition fulfill
\[
a_{{}_{s}}(\widehat{n}\cdot\widehat{l}) = a_{{}_{s}}(\widehat{-n}\cdot\widehat{l}),
\,\,\,
\left[a_{{}_{s}}(\widehat{n}\cdot\widehat{l}), a_{{}_{s'}}(\widehat{n'}\cdot\widehat{l'})^+ \right] = 
\delta_{{}_{s \, s'}} \delta_{{}_{n \, n'}} \delta_{{}_{l \, l'}}.
\]
For non-scalar fields, we have an analogous generalization (Subsection \ref{WhiteNoiseFreeFieldsonEU})
with various roles of the positive and negative energy parts of the corresponding orbit, depending on whether the field is neutral (real)
or charged (complex).  The method of computation of the matrices $u$ is given in Subsection \ref{WhiteNoiseFreeFieldsonEU}. 
In the computation of $u$ we include only these
$u$ which leads to the field invariant under $\zeta$ or $\zeta^4$, where $\zeta(t,w)= (t+\pi, \textrm{Antipode of} \, w)$.
For the real scalar field, $\zeta\boldsymbol{\varphi} = \boldsymbol{\varphi}$. 
$u$ can also be computed by comparison with the classification
of the fields on EU given in \cite{PaneitzSegalI}-\cite{PaneitzSegalIII}.

For the compactified EU, with the $\theta$-function of the same period as the common 
period of all free fields on EU, also such periodic $\theta$ can be Fourier transformed and decomposed into the same complete
system of characters $\widehat{n}\cdot\widehat{l}$, as the free fields themselves.
In particular, using the Fourier transform isomorphism, 
the vacuum average of the chronological product (chronological pairing) 
$\left\langle T(\boldsymbol{\varphi}(x_1)\boldsymbol{\varphi}(x_2))\right\rangle_{{}_{0}} $ can be written
in the momentum picture in the form
\begin{equation}\label{chronologicalProd}
\left\langle T\left(\widetilde{\boldsymbol{\varphi}}(\widehat{n_1}\cdot\widehat{l_1})_{{}_{i_1 \,\, j_1}} 
\widetilde{\boldsymbol{\varphi}}(\widehat{n_2}\cdot\widehat{l_2})_{{}_{i_2 \,\, j_2}}\right)\right\rangle_{{}_{0}} 
=
-i{\textstyle\frac{\sqrt{4\pi}}{2l_1+1}}
\widetilde{\Delta_{c}}(\widehat{n_1}\cdot\widehat{l_1})_{{}_{j_1 \,\, j_1}}
\delta_{{}_{n_1 \, -n_2}} \delta_{{}_{i_1 \, -i_2}} \delta_{{}_{j_1 \, -j_2}} \delta_{{}_{l_1 \, l_2}},
\end{equation}
where
\begin{multline*}
{\textstyle\frac{1}{\sqrt{4\pi}^2}} \int\limits_{G\times G} dt_1dw_1dt_2dw_2 \Delta_c(t_1-t_2,w_1w_2^{-1})  
\overline{\widehat{n_1}(t_1)} \, \overline{\widehat{l_1}(w_1)_{{}_{j_1 \, i_1}}} \,
\overline{\widehat{n_2}(t_2)} \, \overline{\widehat{l_2}(w_2)_{{}_{j_2 \, i_2}}} 
\\
=
{\textstyle\frac{\sqrt{4\pi}}{2l_1+1}}
\widetilde{\Delta_{c}}(\widehat{n_1}\cdot\widehat{l_1})_{{}_{j_1 \,\, j_1}}
\delta_{{}_{n_1 \, -n_2}} \delta_{{}_{i_1 \, -i_2}} \delta_{{}_{j_1 \, -j_2}} \delta_{{}_{l_1 \, l_2}}   
\end{multline*}
is the Fourier transform of the function $(x_1,x_2) \rightarrow \Delta_c(x_1x_2^{-1})$, treated as the function of two independent space-time 
variabes, $x_1=(t_1,w_1),x_2=(t_2,w_2)$, and such that 
\[
\left\langle T(\boldsymbol{\varphi}(x_1)\boldsymbol{\varphi}(x_2))\right\rangle_{{}_{0}}  = D_c(x_1x_2^{-1}) = -i\Delta_c(x_1x_2^{-1})
\]
is the chronological pairing of the free real field,
and where
\[
\widetilde{\Delta_{c}}(\widehat{n}\cdot\widehat{l})_{{}_{j \,\, i}}
=
{\textstyle\frac{1}{\sqrt{4\pi}}} \int\limits_{G} dtdw \,\, \Delta_c(t,w)  
\overline{\widehat{n}(t)} \, \overline{\widehat{l}(w)_{{}_{i \, j}}}
\] 
is the Fourier transform of $\Delta_c$, treated as a function of one space-time variable $x=(t,w)$. Recall
that for massive field on EU, the causal distribution $D_c$ is equal to ordinary function
and its Fourier transform represents an $L^2$-summable series. We use the reality and parity of $\Delta_c$,
which together imply
\[
\widetilde{\Delta_{c}}(\widehat{n}\cdot\widehat{l})_{{}_{j \,\, i}} = \widetilde{\Delta_{c}}(\widehat{-n}\cdot\widehat{l})_{{}_{-i \,\, -j}},
\,\,\,\,
\widetilde{\Delta_{c}}(\widehat{n}\cdot\widehat{l})_{{}_{j \,\, j}} = \widetilde{\Delta_{c}}(\widehat{-n}\cdot\widehat{l})_{{}_{-j \,\, -j}}.
= \overline{\widetilde{\Delta_{c}}(\widehat{n}\cdot\widehat{l})_{{}_{j \,\, j}} }
\]

Let us first briefly give the general idea of \cite{Bogoliubov_Shirkov}, \S 43, and only then will we pass to a more detailed
discussion. It is suggested there to compute first the vacuum averaging of the chronological product of the simple 
functional $e^{i\langle \nu,\boldsymbol{\varphi}\rangle}$ 
$=e^{i\langle \widetilde{\nu},\widetilde{\boldsymbol{\varphi}}\rangle}$ -- the quantum counterpart of the
character functional $F(\widetilde{\varphi}) = e^{i\langle \nu,\varphi\rangle} = e^{i\langle \widetilde{\nu},\widetilde{\varphi}\rangle}$ 
of the nuclear group $E$, with addition of vectors in $E$ as the group action,
and with $\varphi$ presumably coming from a closed subspace of $E^*$, and $\nu$ representing a functional from $E^*$,
which can sensibly be restricted to this subspace. The closer identification of these subspaces will come only 
after a more detailed analysis of the method of \cite{Bogoliubov_Shirkov}, \S 43. The determination of these subspaces 
will be reduced to the determination of the proper support $\mathscr{R}$, mentioned above. In the case of the chracter
functional, the computation of  
\begin{equation}\label{I(nu)}
\mathcal{T}\mu_T\left(\widetilde{\nu}\right)  = \left\langle T\left( e^{i\langle \nu,\boldsymbol{\varphi}\rangle}\right)\right\rangle_{{}_{0}} 
= \left\langle T\left( e^{i\langle \widetilde{\nu},\widetilde{\boldsymbol{\varphi}}\rangle}\right)\right\rangle_{{}_{0}}
\end{equation}
in terms of a  ``functional averaging'' can accordingly be simplified, \cite{Bogoliubov_Shirkov}, \S 43.1-43.2,
when passing to the momentum representation:
\begin{equation}\label{BSstep1}
\mathcal{T}\mu_T\left(\widetilde{\nu}\right) = \int\limits_{E^*} e^{i\langle \widetilde{\nu},\widetilde{\varphi}\rangle} d\mu_T(\widetilde{\varphi}),
\end{equation}
in which we have an operation closely related to integration over $E^*$ (we will identify it later).
The next essential point of the idea of Bogoliubov and Shirkov, consists in passing from the 
simlpe character functional to more involved fuctionals $F(\widetilde{\varphi})$ which, hopefully, can be
subjected to the $d\mu_T$-``integration''. First of all they ask for the functionals $F$ (in general, nonlinear)
which are equal to Fourier transforms $\mathcal{T} \Lambda$ of ``something'' $\Lambda$ (let say here, measures, or more generally,
Hida infinite dimensional distributions $\Lambda$ lying in $(E)^* = \mathcal{S}_{\Gamma(A)}(E^*;\mathbb{R})^*$):
\begin{equation}\label{BSstep2}
F(\widetilde{\varphi}) = \int\limits_{E^*} e^{i\langle\widetilde{\nu},\widetilde{\varphi}\rangle} \Lambda(\widetilde{\nu})
d\mu(\widetilde{\nu}),
\end{equation}
here with the standard Gaussian probability measure $\mu$ on $E^*$. 
Having a functional $F$ as the Fourier transform image of a distribution or a measure, we define
after \cite{Bogoliubov_Shirkov}, \S 43.1-43.2: 
\begin{multline}\label{BSstep3}
\left\langle T\left(F(\boldsymbol{\varphi})  \right)\right\rangle_{{}_{0}} 
= \int\limits_{E^*} \left\langle T\left( e^{i\langle \nu,\boldsymbol{\varphi}\rangle}\right)\right\rangle_{{}_{0}}
 \Lambda(\widetilde{\nu})
d\mu(\widetilde{\nu})
\\
=
\int\limits_{E^*} \int\limits_{E^*} e^{i\langle \widetilde{\nu},\widetilde{\varphi}\rangle} d\mu_T(\widetilde{\varphi}) 
\Lambda(\widetilde{\nu})
d\mu(\widetilde{\nu})
\\
=
\int\limits_{E^*} \int\limits_{E^*} e^{i\langle \widetilde{\nu},\widetilde{\varphi}\rangle} \Lambda(\widetilde{\nu})
d\mu(\widetilde{\nu})
d\mu_T(\widetilde{\varphi}) 
= 
\int\limits_{E^*} F(\widetilde{\varphi})
d\mu_T(\widetilde{\varphi}).
\end{multline}
Here for the functionals $F,G$ for which $F(\varphi) = G(\widetilde{\varphi})$, we use the same 
symbol $F$, i.e. we denote $G$ also by $F$, in order to simplify notation, 
and use $\mathcal{T}$ for the infinite dimensional Fourier transform, which we define below.

Let us now take a closer look at each of the three steps (\ref{BSstep1})-(\ref{BSstep3}), and try to choose
the spaces of allowed $\nu$ and $\varphi$ more closely, which make the definition of $d\mu_T$ rigorous,
and will give meaning and secure the convergence of limit transitions used in the computations.

The computation of (\ref{BSstep1}) is proposed to be performed in two stages in \cite{Bogoliubov_Shirkov}. In the first
stage, we compute explicitly
\[
T\left( e^{i\langle \nu,\boldsymbol{\varphi}\rangle}\right), 
\]
as an operator series, in terms of the components of $\widetilde{\nu}$ and the operator components of
$\widetilde{\boldsymbol{\varphi}}$ and then compute
\[
\left\langle T\left( e^{i\langle \nu,\boldsymbol{\varphi}\rangle}\right)\right\rangle_{{}_{0}} 
\]
in terms of an exponent of a series involving $\widetilde{\nu}, \widetilde{\varphi}$, regarded as 
elements of $E^*$, without any operators. 

First we compute
\[
T\left(F(\boldsymbol{\varphi})  \right)
\]
for the functionals of the following monomial form
\[
\widetilde{\boldsymbol{\varphi}} \mapsto \langle \widetilde{\nu},\widetilde{\boldsymbol{\varphi}}\rangle^m, 
\,\,\, 
\widetilde{\boldsymbol{\varphi}} \mapsto \langle \widetilde{\nu_1},\widetilde{\boldsymbol{\varphi}}\rangle \ldots 
\langle \widetilde{\nu_m},\widetilde{\boldsymbol{\varphi}}\rangle
\]  
which correspond to the classical functionals $F$, which are symmetric monomials
\begin{multline*}
\widetilde{E}^* \ni \widetilde{\varphi} \mapsto \langle \widetilde{\nu},\widetilde{\varphi}\rangle^m 
= \langle \widetilde{\nu}^{\otimes \, m}, \widetilde{\varphi}^{\otimes \, m} \rangle,
\\
\widetilde{E}^* \ni \widetilde{\varphi} \mapsto \langle \widetilde{\nu_1},\widetilde{\varphi}\rangle \ldots \langle \widetilde{\nu_m},\widetilde{\varphi}\rangle
= \left\langle \widetilde{\nu}_1 \otimes \ldots \otimes \widetilde{\nu_m}, \widetilde{\varphi}^{\widehat{\otimes} m} \right\rangle,
\,\,\, \widetilde{\nu}, \widetilde{\nu_i} \in \widetilde{E}
\end{multline*}
on $\widetilde{E}^*$ or, equivalently, on $E^*$. Note that the role of $\nu$ and $\phi$ is symmetric, 
as we can exchange here $E$ and $E^*$ with each other, or if $\nu, \nu_i \in E_k$, then $\varphi \in E_{-k}$, 
in order to make the pairings well-defined. We will specify 
the domains of admissible $\nu, \nu_i$ and $\varphi$ soon. $\widehat{\otimes}$ denotes symmetrized tensor roduct 
(which for nuclear spaces is uniquely defined, as the projective and injective tensor products coincide in this case). 
When we use the tensor product of Hilbert spaces, we mean the Hilbert-space tensor product.

For example, by the Wick theorem and the formula (\ref{chronologicalProd}) for the chronological pairing,
we obtain
\begin{multline*}
T\left( \langle \widetilde{\nu},\widetilde{\boldsymbol{\varphi}}\rangle^2\right) 
= T\left( \sum\limits_{\widehat{l_1},\widehat{n_1}} (2l_1+1)
\textrm{Tr} \, \left[\widetilde{\boldsymbol{\varphi}}(\widehat{n_1}\cdot\widehat{l})
\widetilde{\nu}(\widehat{n_1}\cdot\widehat{l_1})\right] 
\sum\limits_{\widehat{l_2},\widehat{n_2}} (2l_2+1)\textrm{Tr} \, 
\left[\widetilde{\boldsymbol{\varphi}}(\widehat{n_2}\cdot\widehat{l_2})
\widetilde{\nu}(\widehat{n_1}\cdot\widehat{l_2})\right]
\right) 
\\
= \,\,\,\,
{:}
\left( \sum\limits_{\widehat{l_1},\widehat{n_1}} (2l_1+1)
\textrm{Tr} \, \left[\widetilde{\boldsymbol{\varphi}}(\widehat{n_1}\cdot\widehat{l_1})
\widetilde{\nu}(\widehat{n_1}\cdot\widehat{l_1})\right] 
\sum\limits_{\widehat{l_2},\widehat{n_2}} (2l_2+1)
\textrm{Tr} \, \left[\widetilde{\boldsymbol{\varphi}}(\widehat{n_2}\cdot\widehat{l_2})
\widetilde{\nu}(\widehat{n_2}\cdot\widehat{l_2})\right]
\right) 
{:}
\\
-i \sum\limits_{\widehat{l},\widehat{n}, i,j} (2l+1) 
\widetilde{\nu}(\widehat{n}\cdot\widehat{l})_{{}_{j \, i}} \,\,
\sqrt{4\pi} \widetilde{\Delta_{c}}(\widehat{n}\cdot\widehat{l})_{{}_{jj}}
\,\,
\widetilde{\nu}(\widehat{-n}\cdot\widehat{l})_{{}_{-j \, -i}},
\end{multline*}
\begin{multline*}
=
{:}
\left( \sum\limits_{\widehat{l_1}\cdot\widehat{n_1} \in \mathscr{O}^+} (2l_1+1)
\textrm{Tr} \, \left[\widetilde{\boldsymbol{\varphi}}(\widehat{n_1}\cdot\widehat{l_1})
\widetilde{\nu}(\widehat{n_1}\cdot\widehat{l_1})\right] 
\sum\limits_{\widehat{l_2}\cdot\widehat{n_2} \in \mathscr{O}^+} (2l_2+1)
\textrm{Tr} \, \left[\widetilde{\boldsymbol{\varphi}}(\widehat{n_2}\cdot\widehat{l_2})
\widetilde{\nu}(\widehat{n_2}\cdot\widehat{l_2})\right]
\right) 
{:}
\\
-i \left\langle \widetilde{\nu}, \widetilde{T_c} \widetilde{\nu}^*\right\rangle \,\,\,
= \,\,\,\, {:} \langle \widetilde{\nu},\widetilde{\boldsymbol{\varphi}}\rangle^2 {:} 
-i \left\langle \widetilde{\nu}, \widetilde{T_c} \widetilde{\nu}^*\right\rangle,
\end{multline*}
so that
\begin{multline}\label{<T(<v,phi>)>_0}
\left\langle T\left( \langle \widetilde{\nu},\widetilde{\boldsymbol{\varphi}}\rangle^2\right)\right\rangle_{{}_{0}}
=
-i \sum\limits_{\widehat{l},\widehat{n}, i,j} (2l+1) 
\widetilde{\nu}(\widehat{n}\cdot\widehat{l})_{{}_{j \, i}} \,\,
\sqrt{4\pi} \widetilde{\Delta_{c}}(\widehat{n}\cdot\widehat{l})_{{}_{jj}}
\,\,
\widetilde{\nu}(\widehat{-n}\cdot\widehat{l})_{{}_{-j \, -i}}
\\
=
-i \sum\limits_{\widehat{l},\widehat{n}} (2l+1) 
\textrm{Tr} \, \left[
\widetilde{\nu}(\widehat{n}\cdot\widehat{l}) \,\,
\widetilde{T_{c}}(\widehat{n}\cdot\widehat{l})
\,\,
\widetilde{\nu}(\widehat{n}\cdot\widehat{l})^*
\right] 
= -i \left\langle \widetilde{\nu}, \widetilde{T_c} \widetilde{\nu}^*\right\rangle.
\end{multline} 
Here ${:} \ldots {:}$ denotes the Wick ordered product, and we have introduced 
the following multiplication operator $\widetilde{T_c}$ with the following (matrix) multiplication 
components
\[
\widetilde{T_{c}}(\widehat{n}\cdot\widehat{l})_{{}_{i \, j}} = \delta_{{}_{i \, j}}
\sqrt{4\pi} \,\, \widetilde{\Delta_{c}}(\widehat{n}\cdot\widehat{l})_{{}_{jj}},
\]
and where $M^*$  denotes the Hermitian conjugation of the matrix $M$.   
Note that in the pairings 
$\langle \widetilde{\nu},\widetilde{\boldsymbol{\varphi}}\rangle, \langle \widetilde{\nu},
\widetilde{\boldsymbol{\varphi}}\rangle^2, \ldots$ involving free field operator $\widetilde{\boldsymbol{\varphi}}$,
we have the summation range of the characters $\widehat{n}\cdot\widehat{l}$ restricted to the positive energy orbit $\mathscr{O}^+$, 
in accordance with the support of the free field operator components of $\widetilde{\boldsymbol{\varphi}}$, while in the pairing 
$-i \left\langle \widetilde{\nu}, \widetilde{T_c} \widetilde{\nu}^*\right\rangle$ equal to the chronological
pairing, we have, in principle, the full summation range over \emph{all} characters $\widehat{n}\cdot\widehat{l}$ of the group $G=$ EU. 
But the expression (\ref{<T(<v,phi>)>_0}) is meaningfull if $\left\langle \widetilde{\nu}, \widetilde{T_c} \widetilde{\nu}^*\right\rangle$
is convergent. This will be the case for all $\widetilde{\nu} \in E_1$ which are supported on the set $\mathscr{R}$ of characters 
$\widehat{n}\cdot\widehat{l}$ for which 
\begin{equation}\label{R}
{\textstyle\frac{1}{(n^2 +1 + l(l+1))^4}} \leq \widetilde{\Delta_{c}}(\widehat{n}\cdot\widehat{l})_{{}_{jj}} \leq (n^2 +1 + l(l+1))^2. 
\end{equation}
This is so, because for $\widetilde{\nu} \in \widetilde{E}$ supported at $\mathscr{R}$ we have the inequality
\begin{equation}\label{|<nuTcnu>|<|nu|_1^2}
\left| \left\langle \widetilde{\nu}, \widetilde{T_c} \widetilde{\nu}^*\right\rangle \right| 
\leq \, \sqrt{4\pi} \, \left| \widetilde{A} \widetilde{\nu}\right|_{{}_{0}}^{2} = \sqrt{4\pi} \, \left|\widetilde{\nu} \right|_{{}_{1}}^{2}.
\end{equation}
Here, we only use the right-hand part of the inequality (\ref{R}). The role of the left-hand inequality will be explained later.
The character range of the summation in the operator pairing  $\langle \widetilde{\nu},\widetilde{\boldsymbol{\varphi}}\rangle$
is automatically restricted to the corresponding orbit $\mathscr{O}^+$, which for the massive field on EU
is automatically a finite subset of the  set of all characters. In passing on to the massless case, we generalize this
property and will assume that the allowed range $\mathscr{R}$ intersects $\mathscr{O}^+$ and $\mathscr{O}^-$ at a finite set
\begin{equation}\label{RnO+-=Finite}
\mathscr{R} \cap \mathscr{O}^+ = \mathscr{O}^{+}_{0} = \textrm{finite set},
\,\,\,\,\,     \mathscr{R} \cap \mathscr{O}^- = \mathscr{O}^{-}_{0} = \textrm{finite set}.
\end{equation}
More precisely, we assume existence
of $R>0$ such that all $\widehat{n}\cdot\widehat{l} \in \mathscr{O}^{\pm}$ with $n^2 +1 + l(l+1) < R$ 
are contained in $\mathscr{R}$ and all characters of $\mathscr{O}^{\pm}$ with  $n^2 +1 + l(l+1) > R$ are not contained in $\mathscr{R}$.
Also it will be sufficient to fulfill the condition  (\ref{R}) asymptotically: we assume the existence of $R'>0$,
such that (\ref{R}) is fulfilled for all $\widehat{n}\cdot\widehat{l} \in \mathscr{R}$ with $n^2 +1 + l(l+1) > R'$. 

In order to compute the chronological product $T\left( \langle \widetilde{\nu},\widetilde{\boldsymbol{\varphi}}\rangle^n\right) $
of the higher order monomial $\langle \widetilde{\nu},\widetilde{\boldsymbol{\varphi}}\rangle^n$, let us consider the product
\[
\widetilde{\boldsymbol{\varphi}}(\widehat{n_1}\cdot\widehat{l_1})_{{}_{i_1 \,\, j_1}} \ldots
\widetilde{\boldsymbol{\varphi}}(\widehat{n_n}\cdot\widehat{l_n})_{{}_{i_n \,\, j_n}}
\]
of $n$-order. It is well-defined as an operator transforming continuously the Hida space $(E)$ again
into the Hida space, because this is so for the Hida annihilation-creation operators on EU (compare Subsection
\ref{WhiteNoiseFreeFieldsonEU}). Next, let us consider such $n$-order product containing exactly $k$ pairings. Let us denote 
the number of all $k$ pairings in this $n$-order product by $\left[\begin{smallmatrix} n \\ k \end{smallmatrix} \right]$.
We observe that
\begin{equation}\label{[n,k]}
\left[\begin{smallmatrix} n \\ k \end{smallmatrix} \right] = 
\begin{cases}
{\textstyle\frac{n!}{k!(n-2k)!2^k}}, & n \geq 2k, \\
0, & n < 2k,
\end{cases}
\end{equation}
which can be proven by induction using the following identities
\[
\left[\begin{smallmatrix} n +1 \\ k \end{smallmatrix} \right] = \left[\begin{smallmatrix} n  \\ k \end{smallmatrix} \right]
+ n \left[\begin{smallmatrix} n -1 \\ k-1 \end{smallmatrix} \right],
\,\,\,
\left[\begin{smallmatrix} n  \\ k +1 \end{smallmatrix} \right] 
= \left[\begin{smallmatrix} n -2k  \\ 1 \end{smallmatrix} \right] {\textstyle\frac{1}{k+1}},
\,\,\,
\left[\begin{smallmatrix} n  \\ 0 \end{smallmatrix} \right] = 1, \left[\begin{smallmatrix} n  \\ 1 \end{smallmatrix} \right]
= \textstyle{\binom{n}{2}}.
\]
Again using induction with respect to the order, and the formula (\ref{[n,k]}), we show that for each natural $n$
\begin{equation}\label{InductionCP}
\sum\limits_{k=0}^{n} T\left({\textstyle\frac{i^k\langle \widetilde{\nu},\widetilde{\boldsymbol{\varphi}}\rangle^k}{k!}}\right)
=
\textrm{all terms of} \,\, \sum\limits_{k=0}^{n} {:}{\textstyle\frac{i^k\left(\langle \widetilde{\nu},\widetilde{\boldsymbol{\varphi}}\rangle
+ \left\langle \widetilde{\nu}, \widetilde{T_c} \widetilde{\nu}^*\right\rangle/2 \right)^k}{k!}}{:} \,\,
\textrm{up to order} \, n,
\end{equation}
where each $\widetilde{\boldsymbol{\varphi}}$ is counted as a quantity of $1$-order and each pairing  
$\left\langle \widetilde{\nu}, \widetilde{T_c} \widetilde{\nu}^*\right\rangle/2$ is counted as a quantity of $2$-order. Therefore
if the series on the r.h.s. of (\ref{InductionCP}) is convergent, then we obtain
\begin{equation}\label{T(expi<nu,phi>)}
T\left( e^{i\langle \nu,\boldsymbol{\varphi}\rangle}\right)  \,\, = \,\,\,\,  {:}e^{i\left[\langle \widetilde{\nu},\widetilde{\boldsymbol{\varphi}}\rangle
+ \left\langle \widetilde{\nu}, \widetilde{T_c} \widetilde{\nu}^*\right\rangle/2 \right]}{:} \,\,\,\, = \,\,\,\,
{:}e^{i \langle \widetilde{\nu},\widetilde{\boldsymbol{\varphi}}\rangle}{:} \,\,\,\,
e^{\frac{i}{2}\left\langle \widetilde{\nu}, \widetilde{T_c} \widetilde{\nu}^*\right\rangle}.
\end{equation}

Let us investigate now the convergence of the operator series
\begin{equation}\label{SeriesXi}
\Xi = \,\,\, {:}e^{i\left[\langle \widetilde{\nu},\widetilde{\boldsymbol{\varphi}}\rangle
+ \left\langle \widetilde{\nu}, \widetilde{T_c} \widetilde{\nu}^*\right\rangle/2 \right]}{:} \,\,\, =
\sum\limits_{k=0}^{+\infty} {:}{\textstyle\frac{i^k\left(\langle \widetilde{\nu},\widetilde{\boldsymbol{\varphi}}\rangle
+ \left\langle \widetilde{\nu}, \widetilde{T_c} \widetilde{\nu}^*\right\rangle/2 \right)^k}{k!}}{:},
\end{equation}
regarded as an operator defined on the Hida test space $(\widetilde{E})$, with the space of elements in $\widetilde{E}$ supported at the orbit $\mathscr{O}^+$ of the
free field $\boldsymbol{\varphi}$, i.e. with $\Xi$ acting in the dense subspace $(\widetilde{E})$ 
of the Fock space of the free field $\boldsymbol{\varphi}$. To this end, let us note that the operator
\[
\langle \widetilde{\nu},\widetilde{\boldsymbol{\varphi}}\rangle 
=
\sum\limits_{\widehat{n}\cdot\widehat{l} \in \mathscr{O}^+} (2l+1)
\textrm{Tr} \, \left[
\widetilde{\nu}(\widehat{n}\cdot\widehat{l})
\widetilde{\boldsymbol{\varphi}}(\widehat{n}\cdot\widehat{l}) 
\right] = 
\Xi(\kappa_{0,1}) + \Xi(\kappa_{1,0})
\]
is equal to the sum of two well-defined integral kernel operators
\[
\Xi(\kappa_{0,1}) = \sum\limits_{\widehat{n}\cdot\widehat{l} \in \mathscr{O}^+,s} \kappa_{0,1}(s,\widehat{n}\cdot\widehat{l}) \, a_{{}_{s}}(\widehat{n}\cdot\widehat{l}),
\,\,\, 
\Xi(\kappa_{1,0}) = \sum\limits_{\widehat{n}\cdot\widehat{l} \in \mathscr{O}^+,s} \kappa_{1,0}(s,\widehat{n}\cdot\widehat{l}) \, a_{{}_{s}}(\widehat{n}\cdot\widehat{l})^+
\]
acting on the Hida space $(\widetilde{E})$, again with the space of elements of $\widetilde{E}$ supported at the orbit $\mathscr{O}^+$ of the
free field $\boldsymbol{\varphi}$, i.e. in the Fock space of the free field $\boldsymbol{\varphi}$. Here
\begin{multline*}
\kappa_{0,1}(s,\widehat{n}\cdot\widehat{l}) = \sqrt{4\pi}
\sum\limits_{ -l \leq i,j \leq l} 
u_{{}_{s}}(\widehat{n}\cdot\widehat{l})_{{}_{i j}} \widetilde{\nu}(\widehat{n}\cdot\widehat{l})_{{}_{j i}},
\\
\kappa_{1,0}(s,\widehat{n}\cdot\widehat{l}) = \sqrt{4\pi}
\sum\limits_{ -l \leq i,j \leq l} 
\overline{u_{{}_{s}}(\widehat{-n}\cdot\widehat{l})_{{}_{-j \, -i}}} \widetilde{\nu}(\widehat{n}\cdot\widehat{l})_{{}_{j i}}. 
\end{multline*}
In case of a massive scalar field, the kernels $\kappa_{0,1},\kappa_{1,0}$ 
have finite support $\mathscr{O}_0 =\{(\widehat{n}\cdot\widehat{l},s):
\widehat{n}\cdot\widehat{l} \in \mathscr{O}^+, 1 \leq s \leq (2l+1)^2\}$, and by the assumption
(\ref{RnO+-=Finite}), put on the allowable $\widetilde{\nu}$, these kernels have finite support
$\mathscr{O}_0 =\{(\widehat{n}\cdot\widehat{l},s):
\widehat{n}\cdot\widehat{l} \in \mathscr{O}^{+}_{0}, 1 \leq s \leq (2l+1)^2\}$ in general,
where $\mathscr{R} \cap\mathscr{O}^{+} = \mathscr{O}^{+}_{0}$. From this finite support property
of $\kappa_{0,1},\kappa_{1,0}$ and Theorem 2.6 of \cite{hida}, it follows that the integral kernel
operators $\Xi(\kappa_{0,1}),\Xi(\kappa_{1,0})$ transform continuously the test Hida space $(\widetilde{E})$ into itself, where
$(\widetilde{E})$ is the Hida test space $(\widetilde{E}_{{}_{\mathscr{O}^+}})$ in the Fock space of the field $\boldsymbol{\varphi}$, \emph{i.e.}
constructed upon the elements with Fourier transforms supported at $\mathscr{O}^+$.   

Recall that we have the pairings
\begin{multline*}
\langle \kappa_{0,1}, \widetilde{\xi}\rangle = \sum\limits_{\widehat{n}\cdot\widehat{l} 
\in \mathscr{O}^+,s} \kappa_{0,1}(s,\widehat{n}\cdot\widehat{l}) \, \widetilde{\xi}(s,\widehat{n}\cdot\widehat{l}),
\\
\langle \kappa_{1,0}, \widetilde{\xi}\rangle = \sum\limits_{\widehat{n}\cdot\widehat{l} 
\in \mathscr{O}^+,s} \kappa_{1,0}(s,\widehat{n}\cdot\widehat{l}) \, \widetilde{\xi}(s,\widehat{n}\cdot\widehat{l}),
\,\,\,\,
\widetilde{\xi}\in S_{\widetilde{A}}(\mathscr{O}),
\end{multline*}
and that $\widetilde{\xi}\in S_{\widetilde{A}}(\mathscr{O}) \simeq \widetilde{E}_{{}_{\mathscr{O}^+}}$ iff
\[
\widetilde{\xi}(s,\widehat{n}\cdot\widehat{l}) = U\widetilde{\phi}(s,\widehat{n}\cdot\widehat{l}) = 
(2l+1) \textrm{Tr} \, [u_{{}_{s}}(\widehat{n}\cdot\widehat{l})^*\widetilde{\phi}(\widehat{n}\cdot\widehat{l})] 
\]
for $\widetilde{\phi} \in \widetilde{E}$, supported at $\mathscr{O}^+$, with the unitary isomorphism $U$ between the standard 
realization $S_{\widetilde{A}}(\mathscr{O})$ and $\widetilde{E}_{{}_{\mathscr{O}^+}}$, induced by the matrices $u$
as in Subsection \ref{WhiteNoiseFreeFieldsonEU}. Recall that the discrete set $\mathscr{O}$ is an enlargement of the orbit
$\mathscr{O}^+$, with each character $\widehat{n}\cdot\widehat{l}\in \mathscr{O}^+$ having $(2l+1)^2$ 
copies in $\mathscr{O}$, compare Subsection \ref{WhiteNoiseFreeFieldsonEU}.
After indetifying the isomorphic spaces we see that
$\widetilde{\xi} \in \widetilde{E}_k$, iff 
\[
\sum\limits_{\widehat{n}\cdot\widehat{l} \in \mathscr{O}^+,s}  
\left(n^2+1 +l(l+1)\right)^{2k}\left|\widetilde{\xi}(s,\widehat{n}\cdot\widehat{l})\right|^2 < +\infty.
\] 
It is important that in each case, massive or massless field $\boldsymbol{\varphi}$, the intersetion of the support 
$\mathscr{R}$ of $\widetilde{\nu}$ with the orbit $\mathscr{O}^+$, is a finite set $\mathscr{O}^{+}_{0}$ (in massive case this 
is authomatic, becuse the whole $\mathscr{O}^+$ is finite itself in this case). Thus, by assumption, the support of $\kappa_{0,1}, \kappa_{1,0}$,
is in each case finite. Therefore the pairings 
\[
\widetilde{\xi} \longmapsto \langle \kappa_{0,1}, \widetilde{\xi}\rangle, \,\,\,\,\,
\widetilde{\xi} \longmapsto \langle \kappa_{1,0}, \widetilde{\xi}\rangle
\]
can be extended to continuous functionals on the spaces $E_{k}$ for all negative $k$,
with finite respective norms $| \cdot |_{k}$, $k \in \mathbb{Z}$, or $k \in \mathbb{R}$.

For the investigation of the convergence of the series $\Xi$ we use the symbol calculus.
Recall that the symbol of a generalized operator $\Xi$ transforming continuously $(E)$ into $(E)^*$ 
is defined in the following manner
\[
\textrm{Symbol} \, \left[ \Xi \right] (\widetilde{\xi}, \widetilde{\eta}) = \left\langle\left\langle \,\,\, \Xi \, \epsilon_{{}_{\widetilde{\xi}}} \,\, ,  
\,\,\, \epsilon_{{}_{\widetilde{\eta}}} \,\, \right\rangle\right\rangle,
\,\,\,\,\,\, \widetilde{\xi}, \widetilde{\eta} \in \widetilde{E},
\]
where $\epsilon_{{}_{\widetilde{\eta}}} \in (\widetilde{E})_\mathbb{C}$ is the standard coherent state, \emph{i.e.}
the corresponding function on $E^*$ defining $\epsilon_{{}_{\widetilde{\eta}}}$ is equal
\[
\epsilon_{{}_{\widetilde{\eta}}}(x) = \textrm{exp} \left( \left\langle x, \widetilde{\eta}\right\rangle 
- {\textstyle\frac{1}{2}} \left\langle \widetilde{\eta}, \widetilde{\eta}\right\rangle  \right),
\,\,\,\, x \in E^*, \, \widetilde{\eta} \in E_\mathbb{C}.
\]
For example the symbol of an integral kernel operator
\[
\Xi(\kappa_{l,m}), \,\,\, \kappa_{l,m} \in E^{*\widehat{\otimes} l}\otimes E^{*\widehat{\otimes} l} 
\]
is equal
\[
\textrm{Symbol} \, \left[ \Xi(\kappa_{l,m})\right](\widetilde{\xi}, \widetilde{\eta}) = e^{\langle \widetilde{\xi}, \widetilde{\eta}\rangle}
\left\langle\kappa_{l,m}, \,\, \widetilde{\eta}^{\otimes l} \otimes \widetilde{\xi}^{\otimes m}\right\rangle.
\]
In this place we are using the Hida spaces constructed
with the spaces of elemets supported at the orbit $\mathscr{O}^+$. 

By the fact that the kernel $\kappa$ of the Wick product 
\[
{:}\Xi(\kappa_{l,m})\Xi(\kappa'_{l',m'}){:}
\]
of integral kernel operators 
\[
\Xi(\kappa_{l,m}),\Xi(\kappa'_{l',m'})
\]
is equal to the symmetrized tensor product $\kappa = \kappa_{l,m}\widehat{\otimes}\kappa'_{l',m'}$ 
of the kernels $\kappa_{l,m},\kappa'_{l',m'}$, we obtain
\begin{multline*}
e^{-\langle \widetilde{\xi}, \widetilde{\eta}\rangle} \,\, 
\textrm{Symbol} \, \left[{:}{\textstyle\frac{i^n\left(\langle \widetilde{\nu},\widetilde{\boldsymbol{\varphi}}\rangle
+ \left\langle \widetilde{\nu}, \widetilde{T_c} \widetilde{\nu}^*\right\rangle/2 \right)^n}{n!}}{:}  \right](\widetilde{\xi}, \widetilde{\eta})
\\
=
{\textstyle\frac{i^n}{n!}} \left(  
\langle \kappa_{0,1}, \widetilde{\xi}\rangle + \langle \kappa_{1,0}, \widetilde{\eta}\rangle 
+ \textstyle{\frac{1}{2}} \left\langle \widetilde{\nu}, \widetilde{T_c} \widetilde{\nu}^*\right\rangle 
\right)^n
\end{multline*}
and for the series $\Xi$, (\ref{SeriesXi}),  we obtain 
\[
\textrm{Symbol} \, \left[ \Xi \right] (\widetilde{\xi}, \widetilde{\eta}) = 
e^{\langle \widetilde{\xi}, \widetilde{\eta}\rangle} 
\exp i \left[\langle \kappa_{0,1}, \widetilde{\xi}\rangle + \langle \kappa_{1,0}, \widetilde{\eta}\rangle 
+ \textstyle{\frac{1}{2}} \left\langle \widetilde{\nu}, \widetilde{T_c} \widetilde{\nu}^*\right\rangle \right].
\]

\begin{lem}
For any $p\geq 0$ and $\varepsilon >0$, there exist $C \geq 0$ and $q \geq 0$ such that
\begin{multline*}
\left| 
\exp i\left[\langle \kappa_{0,1}, \widetilde{\xi}\rangle + \langle \kappa_{1,0}, \widetilde{\eta}\rangle 
+ \textstyle{\frac{1}{2}} \left\langle \widetilde{\nu}, \widetilde{T_c} \widetilde{\nu}^*\right\rangle \right] 
\right|
\\
\leq 
\,\,\,\,\,\,\,\,\, 
C  \exp \varepsilon \left(
\left| \widetilde{\xi} \right|_{{}_{p+q}}^{2}
+
\left| \widetilde{\eta} \right|_{{}_{-p}}^{2}
\right),
\,\,\,\,\,\,\,\,\,\,\,\,\,\,\,\,\,\,\,\,\,\,\,\,\,\,\,\,\,\,\,\,\,\,\,\,\, 
\widetilde{\xi}, \widetilde{\eta} \in \widetilde{E}_\mathbb{C}.
\end{multline*}
Here $\widetilde{E}_\mathbb{C}$ is the complexification of the space 
of elements of $\widetilde{E}_{{}_{\mathscr{O}^+}}$, supported at $\mathscr{O}^+$. 
$\widetilde{\xi}, \widetilde{\eta} \in \widetilde{E}_\mathbb{C}$
means that $\widetilde{\xi}, \widetilde{\eta}$ are representants of the elements of $\widetilde{E}_{{}_{\mathscr{O}^+}}$
in the standard realization $\mathcal{S}_{\widetilde{A}}(\mathscr{O})$ of 
$\widetilde{E}_{{}_{\mathscr{O}^+}}$.
\label{WickPartSymboInequality}
\end{lem}
\qedsymbol \,
Let $p\geq 0$ and $\varepsilon >0$.
By the assumption put on $\mathscr{R}$, the support $\mathscr{O}_{0}$ of $\kappa_{0,1}, \kappa_{1,0}$
is finite. Therefore, $\kappa_{0,1}, \kappa_{1,0}$,
regarded as linear maps on each fixed $E_k$, are finite dimensional, and supported at the elements 
with the series components, which are non-zero only on the finite subset $\mathscr{O}_{0}$.
It follows that for any $q \geq 0$, there exist $c_{0,1}(p+q),c_{1,0}(-p)\geq 0$, such that 
\[
\left| \langle \kappa_{0,1}, \widetilde{\xi}\rangle\right| \leq c_{0,1}(p+q) \left|\widetilde{\xi} \right|_{{}_{p+q}},
\,\,\,
\left| \langle \kappa_{1,0}, \widetilde{\eta}\rangle\right| \leq c_{1,0}(-p) \left| \widetilde{\eta}  \right|_{{}_{-p}},
\,\,\,
\widetilde{\eta}, \widetilde{\xi} \in E_\mathbb{C}.
\]
It is sufficient to take 
\begin{multline*}
c_{0,1}(p+q) = \underset{\mathscr{O}_0}{\textrm{sup}} 
\left|\kappa_{0,1}(s,\widehat{n}\cdot\widehat{l})(n^2+1+l(l+1))^{-p-q}\right|,
\\
c_{1,0}(-p) = \underset{\mathscr{O}_0}{\textrm{sup}} 
\left|\kappa_{1,0}(s,\widehat{n}\cdot\widehat{l})(\widehat{n}\cdot\widehat{l})(n^2+1+l(l+1))^{p}\right|.
\end{multline*}
For all $\widetilde{\eta}, \widetilde{\xi} \in E_\mathbb{C}$ such that
\begin{equation}\label{xi,eta-BoundedDomain}
{\textstyle\frac{c_{1,0}(p+q) }{\varepsilon}} \,\,\,\,\,\,\, <
\,\,\,\,\,\, \left| \widetilde{\xi}  \right|_{{}_{p+q}}, 
\,\,\,\,\,\,\,\,\,\,\,\,\,\,\,\,\,\,\,\,\,\,\,\,\,\,\,\,\,\,\,\,\,\,
{\textstyle\frac{c_{1,0}(-p) }{\varepsilon}} \,\,\,\,\,\,\,\,
< \,\,\,\,\,\,\,\, \left| \widetilde{\eta}  \right|_{{}_{-p}} 
\end{equation}
we have
\begin{multline*}
\left| \langle \kappa_{0,1}, \widetilde{\xi}\rangle\right|
+
\left| \langle \kappa_{1,0}, \widetilde{\eta}\rangle\right|
\\
\,\,\,\,\,\,\,\,\,\,\,\,\,\,\,
\leq
\,\,\,\,\,\,\,\,\,\,\,\,\,\,\,
c_{0,1}(p+q) \left|\widetilde{\xi} \right|_{{}_{p+q}}
+
c_{1,0}(-p) \left| \widetilde{\eta}  \right|_{{}_{-p}}
\,\,\,\,\,
<
\,\,\,\,\,\,\,\,\,\,\,\,\,\,\,
\varepsilon \left( 
\left| \widetilde{\xi} \right|_{{}_{p+q}}^{2}
+
\left| \widetilde{\eta} \right|_{{}_{-p}}^{2}
\right).
\end{multline*}
Therefore, for each $p\geq 0$ and $\varepsilon >0$, we can find $C, q\geq 0$, such that
the inequality of the lemma holds in the domain (\ref{xi,eta-BoundedDomain}). It is immediate
that for fixed $p,\varepsilon$ we can find appropriate $C,q$, on the bounded completion 
\[
{\textstyle\frac{c_{1,0}(p+q) }{\varepsilon}} \,\,\,\,\,\,\,
\geq \,\,\,\,\,\, \left| \widetilde{\xi}  \right|_{{}_{p+q}}, 
\,\,\,\,\,\,\,\,\,\,\,\,\,\,\,\,\,\,\,\,\,\,\,\,\,\,\,\,\,\,\,\,\,\,
{\textstyle\frac{c_{1,0}(-p) }{\varepsilon}} \,\,\,\,\,\,\
\geq \,\,\,\,\,\,\, \left| \widetilde{\eta}  \right|_{{}_{-p}} 
\]
of the domain (\ref{xi,eta-BoundedDomain}), with the function on the l.h.s. of the inequality of the lemma
being continuous on the bounded closed subset of a finite dimensional linear space. 
Choosing the maximal values of $C,q$, which fit the inequality
of lemma separately in these two domains, we construct $C,q$, required by the lemma.
\qed

\begin{lem}
Let $\widetilde{E}_\mathbb{C}$ be the complexification of the space of elements of $\widetilde{E}$, 
supported at $\mathscr{O}^+$. Let $\widetilde{\xi}, \widetilde{\eta}$ be representants of the elements of $\widetilde{E}_{{}_{\mathscr{O}^+}}$ in the
standard realization $\mathcal{S}_{\widetilde{A}}(\mathscr{O})$ of 
$\widetilde{E}_{{}_{\mathscr{O}^+}}$.  

For any $p\geq 0$ and $\varepsilon >0$, there exist $C \geq 0$ and $q \geq 0$ such that
\[
\left| 
e^{\langle \widetilde{\xi}, \widetilde{\eta}\rangle} 
\right|
\,\,\,\,\, 
\leq 
\,\,\,\,\, 
C  \exp \varepsilon \left(
\left| \widetilde{\xi} \right|_{{}_{p+q}}^{2}
+
\left| \widetilde{\eta} \right|_{{}_{-p}}^{2}
\right),
\]
for all $\widetilde{\xi}, \widetilde{\eta} \in \mathcal{S}_{\widetilde{A}}(\mathscr{O})$,
with the support contained in the finite set $\mathscr{O}_0$.
\label{ScalarPartSymboInequality}
\end{lem}
\qedsymbol \,
For each point $j$ of the discrete support set $\mathscr{O}_0$ we introduce an auxiliary integral kernel
operator $\Xi(\kappa_{0,1})$, with the kernel $\kappa_{0,1}$ supported at this point $j$ and having the value $1$
at this point. By Theorem 2.6, \cite{hida}, each such $\Xi(\kappa_{0,1})$ transforms continuously
$(\widetilde{E})$ into $(\widetilde{E})$, where we mean Hida space  $(\widetilde{E}_{{}_{\mathscr{O}^+}})$,
and use the standard realization $\mathcal{S}_{\widetilde{A}}(\mathscr{O})$ of $\widetilde{E}_{{}_{\mathscr{O}^+}}$.
Therefore the symbol of each such auxiliary operator respects the following condition of \cite{obataJFA}, \S 4.2.
For any $p\geq 0$ and $\varepsilon >0$, there exist $C_j \geq 0$ and $q_j \geq 0$ such that (\cite{obataJFA}, \S 4.2)
\begin{multline*}
\left|
\textrm{Symbol} \, \left[ \Xi(\kappa_{0,1}) \right] (\widetilde{\xi}, \widetilde{\eta}) 
\right|
=
\left| 
e^{\langle \widetilde{\xi}, \widetilde{\eta}\rangle}
\langle \kappa_{1,0}, \widetilde{\xi}\rangle 
\right|
\\
\leq 
\,\,\,\,\,\,\,\,\, 
C_j  \exp \varepsilon \left(
\left| \widetilde{\xi} \right|_{{}_{p+q_j}}^{2}
+
\left| \widetilde{\eta} \right|_{{}_{-p}}^{2}
\right),
\,\,\,\,\,\,\,\,\,\,\,\,\,\,\,\,\,\,\,\,\,\,\,\,\,\,\,\,\,\,\,\,\,\,\,\,\, 
\widetilde{\xi}, \widetilde{\eta} \in \widetilde{E}_\mathbb{C}.
\end{multline*}
Now we put
\[
C = \underset{j \in \mathscr{O}_0}{\textrm{max}} C_j,
\,\,\,
q = \underset{j \in \mathscr{O}_0}{\textrm{max}} q_j.
\]
Thus for any $p\geq 0$ and $\varepsilon >0$, we thus constructed $C \geq 0$ and $q \geq 0$ such that
\[
\left| 
e^{\langle \widetilde{\xi}, \widetilde{\eta}\rangle}
\langle \kappa_{1,0}, \widetilde{\xi}\rangle 
\right|
\,\,\,\,\,
\leq 
\,\,\,\,\,\,\,\,\, 
C  \exp \varepsilon \left(
\left| \widetilde{\xi} \right|_{{}_{p+q}}^{2}
+
\left| \widetilde{\eta} \right|_{{}_{-p}}^{2}
\right),
\]
for all $\widetilde{\xi}, \widetilde{\eta} \in \mathcal{S}_{\widetilde{A}}(\mathscr{O})$,
with the support contained in the finite set $\mathscr{O}_0$,
and for each of the auxiliary integral kernel operators. In particular the last inequality
holds for all elements $\widetilde{\xi}, \widetilde{\eta}$ in the assertion of the lemma 
which moreover respect
\begin{equation}\label{Boundedxi}
\left| \widetilde{\xi}\right|_{{}_{0}}^{2} > \sharp \mathscr{O}_0,
\,\,\,\,
\left| \widetilde{\eta}\right|_{{}_{0}}^{2} > \sharp \mathscr{O}_0,
\end{equation}
where $\sharp \mathscr{O}_0$ is the number of elements of $\mathscr{O}_0$. But each $\widetilde{\xi}$,
supported in $\mathscr{O}_0$ and fulfiling (\ref{Boundedxi}), 
must have at least one component, say $j$-th, with the absolute value greather than one,
with $|\langle \kappa_{1,0}, \widetilde{\xi}\rangle | >1$, where $\kappa_{1,0}$ is the kernel supported at $j$.
Therefore,  for any $p\geq 0$ and $\varepsilon >0$, we thus constructed $C \geq 0$ and $q \geq 0$ such that
\[
\left| 
e^{\langle \widetilde{\xi}, \widetilde{\eta}\rangle}
\right|
\,\,\,\,\,
\leq 
\,\,\,\,\,\,\,\,\, 
C  \exp \varepsilon \left(
\left| \widetilde{\xi} \right|_{{}_{p+q}}^{2}
+
\left| \widetilde{\eta} \right|_{{}_{-p}}^{2}
\right),
\]
for all $\widetilde{\xi}, \widetilde{\eta} \in \mathcal{S}_{\widetilde{A}}(\mathscr{O})$,
with the support contained in the finite set $\mathscr{O}_0$, which moreover respect (\ref{Boundedxi}).
On the domain complementary to (\ref{Boundedxi}), the l.h.s.and the r.h.s of the inequality of lemma are
equal to absolute values of continuous functions
on a compact subset of a finite-dimensional linear space, and for each $p\geq 0$, $\varepsilon >0$, 
we can find appropriate $C,q$ for this domain. Taking again the maximal $C,q$, corresponding to these domains,
we construct $C,q$, required by the lemma.  
\qed

\begin{lem}
For any $p\geq 0$ and $\varepsilon >0$, there exist $C \geq 0$ and $q \geq 0$ such that
\begin{multline*}
\left| 
\textrm{Symbol} \, \left[ \Xi \right] (\widetilde{\xi}, \widetilde{\eta})
\right|
 =
\left| 
e^{\langle \widetilde{\xi}, \widetilde{\eta}\rangle} 
\exp i \left[\langle \kappa_{0,1}, \widetilde{\xi}\rangle + \langle \kappa_{1,0}, \widetilde{\eta}\rangle 
+ \textstyle{\frac{1}{2}} \left\langle \widetilde{\nu}, \widetilde{T_c} \widetilde{\nu}^*\right\rangle \right]
\right|
\\
\leq 
\,\,\,\,\,\,\,\,\, 
C  \exp \varepsilon \left(
\left| \widetilde{\xi} \right|_{{}_{p+q}}^{2}
+
\left| \widetilde{\eta} \right|_{{}_{-p}}^{2}
\right),
\,\,\,\,\,\,\,\,\,\,\,\,\,\,\,\,\,\,\,\,\,\,\,\,\,\,\,\,\,\,\,\,\,\,\,\,\, 
\widetilde{\xi}, \widetilde{\eta} \in \widetilde{E}_\mathbb{C}.
\end{multline*}
Here $\widetilde{E}_\mathbb{C}$ is the complexification of the space 
of elements of $\widetilde{E}_{{}_{\mathscr{O}^+}}$, supported at $\mathscr{O}^+$. 
$\widetilde{\xi}, \widetilde{\eta} \in \widetilde{E}_\mathbb{C}$
means that $\widetilde{\xi}, \widetilde{\eta}$ are representants of the elements of $\widetilde{E}_{{}_{\mathscr{O}^+}}$
in the standard realization $\mathcal{S}_{\widetilde{A}}(\mathscr{O})$ of 
$\widetilde{E}_{{}_{\mathscr{O}^+}}$.
\label{SymboInequality}
\end{lem}
\qedsymbol \,
This lemma follows from lemmas \ref{WickPartSymboInequality} and \ref{ScalarPartSymboInequality}.
\qed

\begin{twr}
Let the support $\mathscr{R}$ of $\widetilde{\nu}$ respects the assumptions (\ref{R}) and (\ref{RnO+-=Finite}).
Then the series representing the following vacuum averages of the following chronological products 
\[
\left\langle T\left( \langle \widetilde{\nu},\widetilde{\boldsymbol{\varphi}}\rangle^2\right)\right\rangle_{{}_{0}}
= -i \left\langle \widetilde{\nu}, \widetilde{T_c} \widetilde{\nu}^*\right\rangle,
\]
\[
\left\langle
T\left( e^{i\langle \nu,\boldsymbol{\varphi}\rangle}\right)
\right\rangle_{{}_{0}}  \,\, = \,\,\,\,  
e^{\frac{i}{2}\left\langle \widetilde{\nu}, \widetilde{T_c} \widetilde{\nu}^*\right\rangle},
\]
are convergent for all $\widetilde{\nu} \in \widetilde{E}_1$, supported at $\mathscr{R}$.

Let, moreover, $(\widetilde{E})$ be the Hida space in the Fock space of the free field $\boldsymbol{\varphi}$, with 
the space of elements of $\widetilde{E}$ supported at $\mathscr{O}^+$.
For each $\Phi \in (\widetilde{E})$ the series  
\[
T\left( e^{i\langle \nu,\boldsymbol{\varphi}\rangle}\right)\Phi = \,\,\, {:}e^{i\left[\langle \widetilde{\nu},\widetilde{\boldsymbol{\varphi}}\rangle
+ \left\langle \widetilde{\nu}, \widetilde{T_c} \widetilde{\nu}^*\right\rangle/2 \right]}{:} \Phi \,\,\, =
\sum\limits_{k=0}^{+\infty} {:}{\textstyle\frac{i^k\left(\langle \widetilde{\nu},\widetilde{\boldsymbol{\varphi}}\rangle
+ \left\langle \widetilde{\nu}, \widetilde{T_c} \widetilde{\nu}^*\right\rangle/2 \right)^k}{k!}}{:}\Phi
\]
converges in $(\widetilde{E})$, and the operator 
\[
T\left( e^{i\langle \nu,\boldsymbol{\varphi}\rangle}\right)
\]
maps continuously the Hida space $(\widetilde{E})$ into 
$(\widetilde{E})$. 
\label{TProdConvergence}
\end{twr}
\qedsymbol \, 
The proof of the part concerning vacuum averages we have already given above. The proof concerning convergence
of the chronological product follows from theorem 4.4 of \cite{obataJFA}, and the property of the symbol
of the operator
\[
\Xi = T\left( e^{i\langle \nu,\boldsymbol{\varphi}\rangle}\right),
\]
given in lemma \ref{SymboInequality}.
\qed

Before we pass to the functional integral form of (\ref{T(expi<nu,phi>)}), let us
remind the guessing method of the validity of the formulas of theorem \ref{TProdConvergence}, which is given in \cite{Bogoliubov_Shirkov}, \S 42.3.
For the functionals $F(\boldsymbol{\varphi})$  of the above monomial type $F(\boldsymbol{\varphi}) = \langle \nu,\boldsymbol{\varphi}\rangle^n$ it is observed 
in \cite{Bogoliubov_Shirkov}, \S 43.2, validity of the following
rule. To compute  
\[
 T\left(F(\widetilde{\boldsymbol{\varphi}})  \right)
\]
we insert into $F(\widetilde{\boldsymbol{\varphi}})$ an ordinary Fourier series components $\widetilde{\varphi}$ for the
operator components $\widetilde{\boldsymbol{\varphi}}$   (with $\widetilde{\varphi}$ coming from a subspace of $\widetilde{E}^*$). Next, 
in the classical $F(\widetilde{\varphi})$, obtained in this way, we replace each of the Fourier series components
\[
\widetilde{\varphi}(\widehat{n}\cdot\widehat{l})_{{}_{i \, j}}
\]  
with the following variational operator
\[
\widetilde{\varphi'}(\widehat{n}\cdot\widehat{l})_{{}_{i \, j}} =
\widetilde{\varphi}(\widehat{n}\cdot\widehat{l})_{{}_{i \, j}}  
-i{\textstyle\frac{\sqrt{4\pi}}{2l+1}}
\widetilde{\Delta_{c}}(\widehat{n}\cdot\widehat{l})_{{}_{j \,\, j}}
{\textstyle\frac{\delta}{\delta \widetilde{\varphi}(\widehat{-n}\cdot\widehat{l})_{{}_{-i \, -j}} }}
\]
and form the following variational operator
\[
U(1|\widetilde{\varphi}) = F\left(\widetilde{\varphi}(\widehat{n}\cdot\widehat{l})_{{}_{i \, j}}  \,\,
- \,\,\, i{\textstyle\frac{\sqrt{4\pi}}{2l+1}} \,\,
\widetilde{\Delta_{c}}(\widehat{n}\cdot\widehat{l})_{{}_{j \, j}} \,\,
{\textstyle\frac{\delta}{\delta \widetilde{\varphi}(\widehat{-n}\cdot\widehat{l})_{{}_{-i \, -j}} }}  
\right) \boldsymbol{1}
\]
understood as acting on the constant unit function $\boldsymbol{1}$, equal 
$\boldsymbol{1}(\widetilde{\varphi}) = 1$ for each $\widetilde{\varphi}$.
Finally, we compose the required time ordered operator, by replacing in the final formula, 
in which all variational differentiations are done, 
the $c$-number components $\widetilde{\varphi}$
with the operator Fourier components $\boldsymbol{\widetilde{\varphi}}$, regarding the products in 
such obtained expression as the Wick products. We get in this way 
\[
T\left(F(\widetilde{\boldsymbol{\varphi}})  \right).
\]
It is observed next that this rule works for $F$ equal to any sum of such monomials, \emph{i.e.} polynomials,
and for  $F$ in the form of convergent series of monomials. We apply this rule to the functional
$F(\widetilde{\varphi}) = e^{i\langle \widetilde{\nu},\widetilde{\varphi}\rangle}$.
Let us introduce the following notation
\[
\left[
\widetilde{\varphi} -
i{\textstyle\frac{\widetilde{T_{c}}}{2l+1}} \,\,
\,\,
{\textstyle\frac{\delta}{\delta \overline{\widetilde{\varphi}}}}
\right]_{{}_{i \,j}}(\widehat{n}\cdot\widehat{l}) 
=
\widetilde{\varphi}(\widehat{n}\cdot\widehat{l})_{{}_{i \, j}}  \,\,
- \,\,\, i{\textstyle\frac{\sqrt{4\pi}}{2l+1}} \,\,
\widetilde{\Delta_{c}}(\widehat{n}\cdot\widehat{l})_{{}_{j \, j}} \,\,
{\textstyle\frac{\delta}{\delta \widetilde{\varphi}(\widehat{-n}\cdot\widehat{l})_{{}_{-i \, -j}} }}.  
\]
With this notation 
\begin{multline*}
U(1|\widetilde{\varphi}) = F\left(\widetilde{\varphi} \,\,
- \,\,\, i{\textstyle\frac{\widetilde{T_{c}}}{2l+1}} \,\,
 \,\,
{\textstyle\frac{\delta}{\delta \overline{\widetilde{\varphi}}}}
\right) \boldsymbol{1}
\\
=
\exp i \left[
\sum\limits_{\widehat{n}\cdot\widehat{l}}
(2l+1)
\textrm{Tr} \,
\left[
\widetilde{\nu}(\widehat{n}\cdot\widehat{l})
\left(\widetilde{\varphi} \,\,
- \,\,\, i{\textstyle\frac{\widetilde{T_{c}}}{2l+1}} \,\,
 \,\,
{\textstyle\frac{\delta}{\delta \overline{\widetilde{\varphi}}}}
\right)(\widehat{n}\cdot\widehat{l})
\right] 
\right]\boldsymbol{1}.
\end{multline*}
To compute $U(1|\widetilde{\varphi})$, it is introduced the auxiliary quantity
\[
U(\lambda|\widetilde{\varphi}) =
\exp i \lambda \left[
\sum\limits_{\widehat{n}\cdot\widehat{l}}
(2l+1)
\textrm{Tr} \,
\left[
\widetilde{\nu}(\widehat{n}\cdot\widehat{l})
\left(\widetilde{\varphi} \,\,
- \,\,\, i{\textstyle\frac{\widetilde{T_{c}}}{2l+1}} \,\,
 \,\,
{\textstyle\frac{\delta}{\delta \overline{\widetilde{\varphi}}}}
\right) (\widehat{n}\cdot\widehat{l})
\right]
\right]\boldsymbol{1},
\]
which respects the equation
\[
{\textstyle\frac{\partial U(\lambda|\widetilde{\varphi})}{\partial \lambda}}
=
\exp i \lambda \left[
\sum\limits_{\widehat{n}\cdot\widehat{l}}
(2l+1)
\textrm{Tr} \,
\left[
\widetilde{\nu}(\widehat{n}\cdot\widehat{l})
\left(\widetilde{\varphi} \,\,
- \,\,\, i{\textstyle\frac{\widetilde{T_{c}}}{2l+1}} \,\,
 \,\,
{\textstyle\frac{\delta}{\delta \overline{\widetilde{\varphi}}}}
\right)(\widehat{n}\cdot\widehat{l})
\right]
\right] U(\lambda|\widetilde{\varphi}).
\]
Solution of this equation gives our $U(1|\widetilde{\varphi})$ for $\lambda=1$ if it
respects the initial condition
\[
U(\lambda=0|\widetilde{\varphi}) =1.
\]
We search the solution in this form
\[
U(\lambda|\widetilde{\varphi}) = e^{s(\lambda|\widetilde{\varphi})}
\]
with $s(\lambda|\widetilde{\varphi})$ fulfilling 
\[
{\textstyle\frac{\partial s}{\partial \lambda}}
=
\sum\limits_{\widehat{n}\cdot\widehat{l}}
(2l+1)
\textrm{Tr} \,
\left[
\widetilde{\nu}(\widehat{n}\cdot\widehat{l})
\left(i\widetilde{\varphi} \,\,
+ \,\,\, {\textstyle\frac{\widetilde{T_{c}}}{2l+1}} \,\,
 \,\,
{\textstyle\frac{\delta s}{\delta \overline{\widetilde{\varphi}}}}
\right)(\widehat{n}\cdot\widehat{l})
\right]. 
\]
We search $s$ in the form
\[
s(\lambda|\widetilde{\varphi}) = i \lambda 
\sum\limits_{\widehat{n}\cdot\widehat{l}}
(2l+1)
\textrm{Tr} \,
\left[
\widetilde{\nu}(\widehat{n}\cdot\widehat{l})
\widetilde{\varphi}(\widehat{n}\cdot\widehat{l})
\right]
+ 
r(\lambda)
=
i\lambda\langle \widetilde{\nu},\widetilde{\varphi}\rangle
+ 
r(\lambda),
\]
with $r$ independent of $\widetilde{\varphi}$. 
Because with such $s$ we have
\begin{multline*}
{\textstyle\frac{\partial s}{\partial \lambda}}
=
 i 
\sum\limits_{\widehat{n}\cdot\widehat{l}}
(2l+1)
\textrm{Tr} \,
\left[
\widetilde{\nu}(\widehat{n}\cdot\widehat{l})
\widetilde{\varphi}(\widehat{n}\cdot\widehat{l})
\right]
+ 
{\textstyle\frac{\partial r}{\partial \lambda}}
\\
=
\sum\limits_{\widehat{n}\cdot\widehat{l}}
(2l+1)
\textrm{Tr} \,
\left[
\widetilde{\nu}(\widehat{n}\cdot\widehat{l})
\left(i\widetilde{\varphi} \,\,
+ \,\,\, {\textstyle\frac{\widetilde{T_{c}}}{2l+1}} \,\,
 \,\,
{\textstyle\frac{\delta s}{\delta \overline{\widetilde{\varphi}}}}
\right)(\widehat{n}\cdot\widehat{l})
\right]
\end{multline*}
\[
=
i 
\sum\limits_{\widehat{n}\cdot\widehat{l}}
(2l+1)
\textrm{Tr} \,
\left[
\widetilde{\nu}(\widehat{n}\cdot\widehat{l})
\widetilde{\varphi}(\widehat{n}\cdot\widehat{l})
\right]
+
i\lambda
\left\langle \widetilde{\nu}, \widetilde{T_c} \widetilde{\nu}^*\right\rangle,
\]
then
\[
{\textstyle\frac{\partial r}{\partial \lambda}} =
i\lambda
\left\langle \widetilde{\nu}, \widetilde{T_c} \widetilde{\nu}^*\right\rangle.
\]
Integrating this equation, with the initial condition $r(\lambda=0) = 0$, determined by the initial condition
for $U(\lambda|\widetilde{\varphi})$, we obtain
\[
r(\lambda) = {\textstyle\frac{i\lambda^2}{2}} \left\langle \widetilde{\nu}, \widetilde{T_c} \widetilde{\nu}^*\right\rangle,
\,\,\,\,
U(\lambda|\widetilde{\varphi}) =
e^{i\left[\lambda\langle \widetilde{\nu},\widetilde{\varphi}\rangle
+ \lambda^2 \left\langle \widetilde{\nu}, \widetilde{T_c} \widetilde{\nu}^*\right\rangle/2 \right]},
\]
and 
\begin{multline*}
U(1|\widetilde{\varphi}) =
e^{i\left[\langle \widetilde{\nu},\widetilde{\varphi}\rangle
+ \left\langle \widetilde{\nu}, \widetilde{T_c} \widetilde{\nu}^*\right\rangle/2 \right]},
\,\,\,\,\,\,\,\,\,\,\,\,\,\,\,\,
T\left( e^{i\langle \nu,\boldsymbol{\varphi}\rangle}\right) = \,\,\, {:}e^{i\left[\langle \widetilde{\nu},\widetilde{\boldsymbol{\varphi}}\rangle
+ \left\langle \widetilde{\nu}, \widetilde{T_c} \widetilde{\nu}^*\right\rangle/2 \right]}{:},
\\
U(1|0) =
\left\langle
T\left( e^{i\langle \widetilde{\nu},\widetilde{\boldsymbol{\varphi}}\rangle}\right)
\right\rangle_{{}_{0}}  \,\, = \,\,\,\,  
e^{\frac{i}{2}\left\langle \widetilde{\nu}, \widetilde{T_c} \widetilde{\nu}^*\right\rangle}.
\end{multline*}
  
Now we pass to the construction of the functional integral form of (\ref{T(expi<nu,phi>)}), following
the idea of \cite{Bogoliubov_Shirkov}, \S 43.3. Using reality and parity of $\Delta_c$,
 (\ref{T(expi<nu,phi>)}) can be written in the following form
\begin{multline}\label{Product}
\left\langle
T\left( e^{i\langle \widetilde{\nu},\widetilde{\boldsymbol{\varphi}}\rangle}\right)
\right\rangle_{{}_{0}}  \,\, = \,\,\,\,  
e^{\frac{i}{2}\left\langle \widetilde{\nu}, \widetilde{T_c} \widetilde{\nu}^*\right\rangle}
\\
= \exp \left[
{\textstyle\frac{i}{2}}
\sum\limits_{\widehat{l},\widehat{n}, i,j} (2l+1) 
\widetilde{\nu}(\widehat{n}\cdot\widehat{l})_{{}_{i \, j}} \,\,
\sqrt{4\pi} \widetilde{\Delta_{c}}(\widehat{n}\cdot\widehat{l})_{{}_{jj}}
\,\,
\widetilde{\nu}(\widehat{-n}\cdot\widehat{l})_{{}_{-i \, -j}}
\right]
\\
=
\underset{\widehat{l},\widehat{n}, i,j}{\prod} \exp 
\left[ 
{\textstyle\frac{i}{2}}
(2l+1) 
\widetilde{\nu}(\widehat{n}\cdot\widehat{l})_{{}_{i \, j}} \,\,
\sqrt{4\pi} \widetilde{\Delta_{c}}(\widehat{n}\cdot\widehat{l})_{{}_{jj}}
\,\,
\widetilde{\nu}(\widehat{-n}\cdot\widehat{l})_{{}_{-i \, -j}}
\right]
\\
=
\underset{\widehat{l},\widehat{n \geq 0}, i,j}{\prod} \exp 
\left[ 
i
(2l+1) 
\widetilde{\nu}(\widehat{n}\cdot\widehat{l})_{{}_{i \, j}} \,\,
\sqrt{4\pi} \widetilde{\Delta_{c}}(\widehat{n}\cdot\widehat{l})_{{}_{jj}}
\,\,
\widetilde{\nu}(\widehat{-n}\cdot\widehat{l})_{{}_{-i \, -j}}
\right].
\end{multline}
We consider now a fixed factor of the last product, with fixed $n,l,j,i$ and express its value in terms of a two-dimensional
Gaussian-like integral over two real variables $x_{{}_{\widehat{n},\widehat{l},j,i}}, y_{{}_{\widehat{n},\widehat{l},j,i}}$,
which we interpret, after \cite{Bogoliubov_Shirkov}, as, respectively, the real and imaginary part
of the Fourier components
\[
\widetilde{\varphi}(\widehat{n}\cdot\widehat{l})_{{}_{j \, i}}
= x_{{}_{\widehat{n},\widehat{l},j,i}} + i y_{{}_{\widehat{n},\widehat{l},j,i}}
\]
of a real function $\varphi$, therefore, respecting
\[
x_{{}_{\widehat{-n},\widehat{l},-j,-i}} = x_{{}_{\widehat{n},\widehat{l},j,i}},
\,\,\,
y_{{}_{\widehat{-n},\widehat{l},-j,-i}} = -y_{{}_{\widehat{n},\widehat{l},j,i}}.
\]
To this end,  we introduce the following parameters $\lambda,\mu,a$ of the Gaussian exponent in the Gaussian-like
integral expression of our factor 
\[
\lambda = \left[
\widetilde{\nu}(\widehat{n}\cdot\widehat{l})_{{}_{i \, j}}
+
\widetilde{\nu}(\widehat{-n}\cdot\widehat{l})_{{}_{-i \, -j}}
\right](2l+1),
\]
\[
\mu = \left[
\widetilde{\nu}(\widehat{n}\cdot\widehat{l})_{{}_{i \, j}}
-
\widetilde{\nu}(\widehat{-n}\cdot\widehat{l})_{{}_{-i \, -j}}
\right](2l+1),
\,\,\,
a = {\textstyle\frac{2l+1}{\sqrt{4\pi} \widetilde{\Delta_{c}}(\widehat{n}\cdot\widehat{l})_{{}_{jj}}}},
\]
so that
\[
\lambda^2-\mu^2 = (2l+1)^2 4 \widetilde{\nu}(\widehat{n}\cdot\widehat{l})_{{}_{i \, j}}
\widetilde{\nu}(\widehat{-n}\cdot\widehat{l})_{{}_{-i \, -j}},
\]
\[
\widetilde{\varphi}(\widehat{n}\cdot\widehat{l})_{{}_{j \, i}}\widetilde{\varphi}(\widehat{-n}\cdot\widehat{l})_{{}_{-j \, -i}}
= x_{{}_{\widehat{n},\widehat{l},j,i}}^{2} + x_{{}_{\widehat{n},\widehat{l},j,i}}^{2},
\]
\[
\lambda x_{{}_{\widehat{n},\widehat{l},j,i}} 
+ i\mu y_{{}_{\widehat{n},\widehat{l},j,i}} 
=
(2l+1)\left[
\widetilde{\nu}(\widehat{n}\cdot\widehat{l})_{{}_{i \, j}}\widetilde{\varphi}(\widehat{n}\cdot\widehat{l})_{{}_{j \, i}}
+
\widetilde{\nu}(\widehat{-n}\cdot\widehat{l})_{{}_{-i \, -j}}\widetilde{\varphi}(\widehat{-n}\cdot\widehat{l})_{{}_{-j \, -i}}
\right]
\]
Using the standard Gaussian-like quadrature, we obtain
\begin{multline*}
\exp \left[ 
i
\widetilde{\nu}(\widehat{n}\cdot\widehat{l})_{{}_{i \, j}} \,\,
\sqrt{4\pi} \widetilde{\Delta_{c}}(\widehat{n}\cdot\widehat{l})_{{}_{jj}}
\,\,
\widetilde{\nu}(\widehat{-n}\cdot\widehat{l})_{{}_{-i \, -j}}
\right] = e^{i\frac{\lambda^2-\mu^2}{4a}}
\\
=
\underset{\epsilon \rightarrow 0}{\textrm{lim}}
{\textstyle\frac{ia}{\pi}}
\int\limits_{-\infty}^{+\infty} dx_{{}_{\widehat{n},\widehat{l},j,i}} \int\limits_{-\infty}^{+\infty}
dy_{{}_{\widehat{n},\widehat{l},j,i}}
\exp
i
\left[
(-2a+i\epsilon)
{\textstyle\frac{1}{2}}
\left( x_{{}_{\widehat{n},\widehat{l},j,i}}^{2} + y_{{}_{\widehat{n},\widehat{l},j,i}}^{2} \right)
+\lambda x_{{}_{\widehat{n},\widehat{l},j,i}} 
+ i\mu y_{{}_{\widehat{n},\widehat{l},j,i}} 
\right]
\end{multline*}
\begin{multline*}
=
{\textstyle\frac{i(2l+1)}{\sqrt{4\pi^{3}}\widetilde{\Delta_{c}}(\widehat{n}\cdot\widehat{l})_{{}_{jj}}}}
\int\limits_{-\infty}^{+\infty} dx_{{}_{\widehat{n},\widehat{l},j,i}} \int\limits_{-\infty}^{+\infty}
dy_{{}_{\widehat{n},\widehat{l},j,i}}
\Bigg\lbrace
\exp
{\textstyle\frac{i}{2}}
\left[
(-{\textstyle\frac{2l+1}{\sqrt{\pi} \widetilde{\Delta_{c}}(\widehat{n}\cdot\widehat{l})_{{}_{jj}}}})
\left( x_{{}_{\widehat{n},\widehat{l},j,i}}^{2} + y_{{}_{\widehat{n},\widehat{l},j,i}}^{2} \right)
\right] 
\,\, \times
\\
\times \,\,
\exp i \left[
(2l+1)\widetilde{\nu}(\widehat{n}\cdot\widehat{l})_{{}_{i \, j}}\widetilde{\varphi}(\widehat{n}\cdot\widehat{l})_{{}_{j \, i}}
+
(2l+1)\widetilde{\nu}(\widehat{-n}\cdot\widehat{l})_{{}_{-i \, -j}}\widetilde{\varphi}(\widehat{-n}\cdot\widehat{l})_{{}_{-j \, -i}}
\right]
\Bigg\rbrace
\end{multline*}

\begin{multline*}
=
{\textstyle\frac{i(2l+1)}{\sqrt{4\pi^{3}}\widetilde{\Delta_{c}}(\widehat{n}\cdot\widehat{l})_{{}_{jj}}}}
\int\limits_{-\infty}^{+\infty} dx_{{}_{\widehat{n},\widehat{l},j,i}} \int\limits_{-\infty}^{+\infty}
dy_{{}_{\widehat{n},\widehat{l},j,i}}
\Bigg\lbrace
\exp
{\textstyle\frac{i}{2}}
\left[
\widetilde{\varphi}(\widehat{n}\cdot\widehat{l})_{{}_{j \, i}}
(-{\textstyle\frac{2l+1}{\sqrt{\pi} \widetilde{\Delta_{c}}(\widehat{n}\cdot\widehat{l})_{{}_{jj}}}})
\widetilde{\varphi}(\widehat{-n}\cdot\widehat{l})_{{}_{-j \, -i}}
\right] 
\,\, \times
\\
\times \,\,
\exp i \left[
(2l+1)\widetilde{\nu}(\widehat{n}\cdot\widehat{l})_{{}_{i \, j}}\widetilde{\varphi}(\widehat{n}\cdot\widehat{l})_{{}_{j \, i}}
+
(2l+1)\widetilde{\nu}(\widehat{-n}\cdot\widehat{l})_{{}_{-i \, -j}}\widetilde{\varphi}(\widehat{-n}\cdot\widehat{l})_{{}_{-j \, -i}}
\right]
\Bigg\rbrace
\end{multline*}
where in the last two integrals, $\epsilon$ and the limit $\epsilon \rightarrow 0$, is not explicitly written. 

Let us introduce, for each $n\geq 0$, $l$, $-l \leq j,i \leq -l$, the following
two-dimensional normalized ``complex measure''
\begin{multline*}
\delta\widetilde{\varphi}(\widehat{n}\cdot\widehat{l})_{{}_{ji}}=
{\textstyle\frac{i(2l+1)}{\sqrt{4\pi^{3}}\widetilde{\Delta_{c}}(\widehat{n}\cdot\widehat{l})_{{}_{jj}}}}
\exp
{\textstyle\frac{i}{2}}
\left[
\widetilde{\varphi}(\widehat{n}\cdot\widehat{l})_{{}_{j \, i}}
(-{\textstyle\frac{2l+1}{\sqrt{\pi} \widetilde{\Delta_{c}}(\widehat{n}\cdot\widehat{l})_{{}_{jj}}}})
\widetilde{\varphi}(\widehat{-n}\cdot\widehat{l})_{{}_{-j \, -i}}
\right] dx_{{}_{\widehat{n},\widehat{l},j,i}}dy_{{}_{\widehat{n},\widehat{l},j,i}}
\\
=
{\textstyle\frac{i(2l+1)}{\sqrt{4\pi^{3}}\widetilde{\Delta_{c}}(\widehat{n}\cdot\widehat{l})_{{}_{jj}}}}
\exp
{\textstyle\frac{i}{2}}
\left[
(-{\textstyle\frac{2l+1}{\sqrt{\pi} \widetilde{\Delta_{c}}(\widehat{n}\cdot\widehat{l})_{{}_{jj}}}})
\left( x_{{}_{\widehat{n},\widehat{l},j,i}}^{2} + y_{{}_{\widehat{n},\widehat{l},j,i}}^{2} \right)
\right] dx_{{}_{\widehat{n},\widehat{l},j,i}}dy_{{}_{\widehat{n},\widehat{l},j,i}}
\\
=
{\textstyle\frac{\exp
{\textstyle\frac{i}{2}}
\left[
(-{\textstyle\frac{2l+1}{\sqrt{\pi} \widetilde{\Delta_{c}}(\widehat{n}\cdot\widehat{l})_{{}_{jj}}}})
\left( x_{{}_{\widehat{n},\widehat{l},j,i}}^{2} + y_{{}_{\widehat{n},\widehat{l},j,i}}^{2} \right)
\right] dx_{{}_{\widehat{n},\widehat{l},j,i}}dy_{{}_{\widehat{n},\widehat{l},j,i}}}
{\int\int\exp
{\textstyle\frac{i}{2}}
\left[
(-{\textstyle\frac{2l+1}{\sqrt{\pi} \widetilde{\Delta_{c}}(\widehat{n}\cdot\widehat{l})_{{}_{jj}}}})
\left( x_{{}_{\widehat{n},\widehat{l},j,i}}^{2} + y_{{}_{\widehat{n},\widehat{l},j,i}}^{2} \right)
\right] dx_{{}_{\widehat{n},\widehat{l},j,i}}dy_{{}_{\widehat{n},\widehat{l},j,i}}}}
\end{multline*}
where, again, $\epsilon$ and the limit $\epsilon \rightarrow 0$ is not explicitly written.

Returning to the product (\ref{Product}) we arrive, still at the heuristic level, with the following formula:
\begin{equation}\label{BSInt}
\left\langle
T\left( e^{i\langle \widetilde{\nu},\widetilde{\boldsymbol{\varphi}}\rangle}\right)
\right\rangle_{{}_{0}}  \,\, = \,\,\,\,  
e^{\frac{i}{2}\left\langle \widetilde{\nu}, \widetilde{T_c} \widetilde{\nu}^*\right\rangle}
=
\int
e^{i\langle \widetilde{\nu},\widetilde{\varphi}\rangle}
\prod_{n\geq 0, l,j,i} \delta\widetilde{\varphi}(\widehat{n}\cdot\widehat{l})_{{}_{ji}}
\end{equation}
It is therefore tempting to search for a normalized complex measure
\[
d\mu_T(\widetilde{\varphi}) = \prod_{n\geq 0, l,j,i} \delta\widetilde{\varphi}(\widehat{n}\cdot\widehat{l})_{{}_{ji}}
\]
in the form of a product measure, first on the big space $\mathbb{R}^\infty$ of all real sequences (say, of the real and imaginary parts of 
a ``Fourier series components''), hoping that it will have support in the interesting for us linear subspace, which moreover allows us to compute
the r.h.s. of (\ref{BSInt}), and respects the condition (\ref{BSInt}):
\[
e^{\frac{i}{2}\left\langle \widetilde{\nu}, \widetilde{T_c} \widetilde{\nu}^*\right\rangle}
=
\int
e^{i\langle \widetilde{\nu},\widetilde{\varphi}\rangle}
\, d \mu_T(\widetilde{\varphi}).
\]
Although this approach may seem natural at the very first sight, we abandon it. 
We do it, because the very existence of such a product ``complex measure'', already on the huge product space
$\mathbb{R}^\infty$ is a delicate problem. Already, its infinite variation (if it woud exist at all)
introduces additional difficulties in its construction, if we would like to get the required measure as the product measure on
$\mathbb{R}^\infty$. For example, using the $\epsilon$-limit trick, we have normalized each 
finite-dimensional projection of such a hypothetical 
product measure to 1, \emph{i.e.} with the projected measure of the whole finite-dimensional subspace on which we project,
equal $1$. Initially, it may thus seem that this normalization, which for probability measures on $\mathbb{R}^\infty$
saves the consistency, and by Kolmogorov's theorem, the existence of the product measure on $\mathbb{R}^\infty$, is not yet
sufficient in our case, for the existence of the product measure.  In particular, it is immediately
seen that the sequence of measures of balls (in the finite dimensional subspace on which we project) with the radii going to infinity,
is not convergent. For a $\sigma$-measure of finite variation, it should converge to $1$. Therefore, we need to check, e.g.,
if the $\sigma$-additivity holds at least in cases in which the corresponding limits of the sums (series)
of the measures are convergent, and check if the measure is independent of countable decomposition into disjoint
measurable subsets, with the series of measures of the subsets convergent. 

In order to avoid subtleties of this construction, we are searching for
a measure on $E^* \simeq \widetilde{E}^*$, or even more generally, we are searching for a Hida distribution $P \in (E)^* \simeq(\widetilde{E})$
whose Fourier transform is equal to the l.h.s. of (\ref{BSInt}), equivalently
\[
e^{\frac{i}{2}\left\langle \widetilde{\nu}, \widetilde{T_c} \widetilde{\nu}^*\right\rangle} =
\mathcal{T}P(\widetilde{\nu}) = \int\limits_{\widetilde{E}^*} 
e^{i\langle \widetilde{\nu},\widetilde{\varphi}\rangle} P(\widetilde{\varphi})
\, d \mu(\widetilde{\varphi})
=
\left\langle\left\langle e_{{}_{\widetilde{\nu}}}, P \right\rangle\right\rangle, 
\] 
possibly with the elements $\widetilde{\varphi}, \widetilde{\nu}$ appropriately supported.
Here we have the standard Gaussian probability measure $\mu$ on $E^* \simeq \widetilde{E}^*$, which
arises in the natural Wigner-It\^o-Segal isomorphism $\Gamma(H) \simeq L^2(E^*, d\mu; \mathbb{R})$,
mentioned at the beginning of this Subsection and where we have introduced the character function
\[
\widetilde{E}^* \ni \widetilde{\varphi} \longmapsto 
e_{{}_{\widetilde{\nu}}}(\widetilde{\varphi}) = e^{i\langle \widetilde{\nu},\widetilde{\varphi}\rangle}
\] 
of the nuclear group $\widetilde{E}^*$, well-defined for all $\widetilde{\varphi} \in \widetilde{E}^*$
if $\widetilde{\nu} \in \widetilde{E}$. Note however, that the standard Gaussian measure $\mu$
is supported at $E_{-1} \simeq \widetilde{E}_{-1} \subset \widetilde{E}^*$. In particular, the support of our distribution
$P$ (if it exists at all) is contained in $\widetilde{E}_{-1}$. Below, we investigate it more closely.
Therefore the character function $e_{{}_{\widetilde{\nu}}}$ makes sense for all $\widetilde{\varphi}$ in the support of $\mu$ 
if $\widetilde{\nu} \in \widetilde{E}_{1}$. This is also the condition for existence of the l.h.s. of (\ref{BSInt}), as we know,
if the Fourier transform support $\mathscr{R}$ of $\widetilde{\nu}$ respects the assumptions
(\ref{R}) and (\ref{RnO+-=Finite}).
Let us remind that for $\widetilde{\eta} \in \widetilde{E}_\mathbb{C}$, 
$e_{{}_{\widetilde{\eta}}} \in (\widetilde{E})_\mathbb{C}$. Therefore, we have a well-defined Fourier transform \cite{obata-book}
\[
\mathcal{T}P(\widetilde{\eta}) =
\left\langle\left\langle e_{{}_{\widetilde{\eta}}}, P \right\rangle\right\rangle
=\int\limits_{\widetilde{E}^*} 
e^{i\langle \widetilde{\eta},\widetilde{\varphi}\rangle} P(\widetilde{\varphi})
\, d \mu(\widetilde{\varphi}), \,\,\,  \eta \in \widetilde{E}_\mathbb{C}, \,\,\,
\]
of any Hida distribution $P \in (E)^* \simeq(\widetilde{E})^*$. 
The integral form of the pairing $\langle\langle \cdot , \cdot \rangle\rangle$ 
can be used here, not only purely formally,
because, similarly to finite-dimensional tempered distributions, also the infinite-dimensional
distributions $P$ can be approximated by the infinite-dimensional test functions $P_n \in (\widetilde{E})$:
\[
\mathcal{T}P(\widetilde{\eta}) =
\left\langle\left\langle e_{{}_{\widetilde{\eta}}}, P \right\rangle\right\rangle
= \underset{n \rightarrow \infty}{\textrm{lim}}
\int\limits_{\widetilde{E}^*} 
e^{i\langle \widetilde{\eta},\widetilde{\varphi}\rangle} P_n(\widetilde{\varphi})
\, d \mu(\widetilde{\varphi}).
\]

To formulate our next theorem, we need to characterize the finite dimensional restrictions
of the distribution $P$, which should coincide with the finite dimensional Gaussian distribution
of our heuristic construction. To this end, note the following. The system of functions 
$\eta_{{}_{\widehat{n},\widehat{l},j,i}}, \zeta_{{}_{\widehat{n},\widehat{l},j,i}}$, respectively,
with the real and imaginary parts of the Fourier transform components all equal zero
except, respectively, the real component equal $x_{{}_{\widehat{n},\widehat{l},j,i}}=1$, or, respectively, except the imaginary part component
equal $y_{{}_{\widehat{n},\widehat{l},j,i}}=1$, for a fixed $n,l,j,i$, compose a complete orthogonal system,
with the square of norms equal to $\tfrac{1}{2l+1}$. Let us denote the Fourier transform of any such system of functions
by
\[
\widetilde{\xi}_{{}_{\widehat{n},\widehat{l},j,i}} = \widetilde{\eta}_{{}_{\widehat{n},\widehat{l},j,i}} 
+\overset{\cdot}{\i} \,\, \widetilde{\zeta}_{{}_{\widehat{n},\widehat{l},j,i}}.
\]

Therefore, there arise two problems. First, we need to check if the distribution $d\mu_T(\widetilde{\varphi}) = P((\widetilde{\varphi})) d\mu(\widetilde{\varphi})$
exists. Second, if it exists, then its image under the map
\begin{multline*}
\widetilde{E}^* \ni \widetilde{\varphi} \longmapsto 
\left(
\langle \widetilde{\varphi},\eta_{{}_{\widehat{n_1},\widehat{l_1},j_1,i_1}}\rangle, 
\langle \widetilde{\varphi},\overset{\cdot}{\i} \, \zeta_{{}_{\widehat{n_1},\widehat{l_1},j_1,i_1}}\rangle, \ldots,
\langle \widetilde{\varphi},\eta_{{}_{\widehat{n_k},\widehat{l_k},j_k,i_k}}\rangle, 
\langle \widetilde{\varphi},\overset{\cdot}{\i} \, \zeta_{{}_{\widehat{n_k},\widehat{l_k},j_k,i_k}}\rangle
\right)
\\
=
(x_{{}_{\widehat{n_1},\widehat{l_1},j_1,i_1}}, y_{{}_{\widehat{n_1},\widehat{l_1},j_1,i_1}}, \ldots
x_{{}_{\widehat{n_k},\widehat{l_k},j_k,i_k}}, y_{{}_{\widehat{n_k},\widehat{l_k},j_k,i_k}})
\in \mathbb{R}^{2k}
\end{multline*}
should be equal to the following distribution:
\[
\prod\limits_{i=1}^{k}
{\textstyle\frac{\overset{\cdot}{\i} \, (2l_i+1)}{\sqrt{4\pi^{3}}\widetilde{\Delta_{c}}(\widehat{n_i}\cdot\widehat{l_i})_{{}_{j_ij_i}}}}
\exp
{\textstyle\frac{\overset{\cdot}{\i}}{2}}
\left[
(-{\textstyle\frac{2l+1}{\sqrt{\pi} \widetilde{\Delta_{c}}(\widehat{n_i}\cdot\widehat{l_i})_{{}_{j_ij_i}}}})
\left( x_{{}_{\widehat{n_i},\widehat{l_i},j_i,i_i}}^{2} + y_{{}_{\widehat{n_i},\widehat{l_i},j_i,i_i}}^{2} \right)
\right].
\]
We have
\begin{twr}
If $\widetilde{\nu},\widetilde{\varphi}$ are supported at $\mathscr{R}$, which respects (\ref{R}) and 
(\ref{RnO+-=Finite}), then the distribution $P$ fulfills the condition
\begin{equation}\label{FTmuT}
e^{\frac{\overset{\cdot}{\i}}{2}\left\langle \widetilde{\nu}, \widetilde{T_c} \widetilde{\nu}^*\right\rangle} =
\mathcal{T}P(\widetilde{\nu}) = \int\limits_{\widetilde{E}^*} 
e^{\overset{\cdot}{\i} \, \langle \widetilde{\nu},\widetilde{\varphi}\rangle} P(\widetilde{\varphi})
\, d \mu(\widetilde{\varphi})
\end{equation}
exists and is uniquely determined by this condition. 

Let 
\[
\widetilde{\xi}_{{}_{\widehat{n},\widehat{l},j,i}} = \widetilde{\eta}_{{}_{\widehat{n},\widehat{l},j,i}} 
+
\overset{\cdot}{\i} \,\, 
\widetilde{\zeta}_{{}_{\widehat{n},\widehat{l},j,i}},
\]
for $\widehat{n}\cdot\widehat{l} \in \mathscr{R}$ and with real $\widetilde{\eta},\widetilde{\zeta}$-s, 
compose an orthogonal system of functions in $\widetilde{E}$, supported at $\mathscr{R}$,
which respect the orthogonality conditions
\begin{multline*}
\langle\widetilde{\eta}_{{}_{\widehat{n},\widehat{l},j,i}}, \widetilde{\eta}_{{}_{\widehat{n'},\widehat{l'},j',i'}} \rangle
= {\textstyle\frac{1}{2l+1}}\delta_{{}_{n \, n'}}\delta_{{}_{l \, l'}}\delta_{{}_{j \, j'}}\delta_{{}_{i \, i'}},
\\
\langle\widetilde{\zeta}_{{}_{\widehat{n},\widehat{l},j,i}}, \widetilde{\zeta}_{{}_{\widehat{n'},\widehat{l'},j',i'}} \rangle
= {\textstyle\frac{1}{2l+1}} \delta_{{}_{n \, n'}}\delta_{{}_{l \, l'}}\delta_{{}_{j \, j'}}\delta_{{}_{i \, i'}},
\,\,\,
\langle\widetilde{\eta}_{{}_{\widehat{n},\widehat{l},j,i}}, \widetilde{\zeta}_{{}_{\widehat{n'},\widehat{l'},j',i'}} \rangle = 0.
\end{multline*}
Then the image of the distribution $d\mu_T(\widetilde{\varphi}) = P(\widetilde{\varphi}) d\mu(\widetilde{\varphi})$ under the map
\begin{multline*}
\widetilde{E}^* \ni \widetilde{\varphi} \longmapsto 
\left(
\langle \widetilde{\varphi},\eta_{{}_{\widehat{n_1},\widehat{l_1},j_1,i_1}}\rangle, 
\langle \widetilde{\varphi},\overset{\cdot}{\i} \, \zeta_{{}_{\widehat{n_1},\widehat{l_1},j_1,i_1}}\rangle, \ldots,
\langle \widetilde{\varphi},\eta_{{}_{\widehat{n_k},\widehat{l_k},j_k,i_k}}\rangle, 
\langle \widetilde{\varphi},\overset{\cdot}{\i} \, \zeta_{{}_{\widehat{n_k},\widehat{l_k},j_k,i_k}}\rangle
\right)
\\
=
(x_{{}_{\widehat{n_1},\widehat{l_1},j_1,i_1}}, y_{{}_{\widehat{n_1},\widehat{l_1},j_1,i_1}}, \ldots
x_{{}_{\widehat{n_k},\widehat{l_k},j_k,i_k}}, y_{{}_{\widehat{n_k},\widehat{l_k},j_k,i_k}})
\in \mathbb{R}^{2k}
\end{multline*}
is equal to the following distribution
\[
\prod\limits_{i=1}^{k}
\left[
{\textstyle\frac{\overset{\cdot}{\i} \, (2l_i+1)}{\sqrt{4\pi^{3}}\widetilde{\Delta_{c}}(\widehat{n_i}\cdot\widehat{l_i})_{{}_{j_ij_i}}}}
\exp
{\textstyle\frac{\overset{\cdot}{\i}}{2}}
\left[
(-{\textstyle\frac{2l+1}{\sqrt{\pi} \widetilde{\Delta_{c}}(\widehat{n_i}\cdot\widehat{l_i})_{{}_{j_ij_i}}}})
\left( x_{{}_{\widehat{n_i},\widehat{l_i},j_i,i_i}}^{2} + y_{{}_{\widehat{n_i},\widehat{l_i},j_i,i_i}}^{2} \right)
\right]
dx_{{}_{\widehat{n_i},\widehat{l_i},j_i,i_i}} y_{{}_{\widehat{n_i},\widehat{l_i},j_i,i_i}}
\right].
\]
\label{muTexistence}
\end{twr}

\qedsymbol \,
From the inequality (\ref{|<nuTcnu>|<|nu|_1^2}) we get the following inequalities
\[
\left| e^{ \frac{\overset{\cdot}{\i}}{2}\left\langle \widetilde{\nu}, \widetilde{T_c} \widetilde{\nu}^*\right\rangle} \right|
\leq 
e^{\frac{1}{2}\left| \left\langle \widetilde{\nu}, \widetilde{T_c} \widetilde{\nu}^*\right\rangle \right|}
\leq
e^{ \sqrt{\pi} \, \left|\widetilde{\nu} \right|_{{}_{1}}^{2}},
\]
for all complex valued $\widetilde{\nu} \in \widetilde{E}_\mathbb{C}$, supported at $\mathscr{R}$. From this and theorem 5.6.12 of \cite{obata-book}
we obtain existence of $P$. Uniqueness follows from the fact that the kernel of the Fourier transform is zero,
\emph{i.e.} the Fourier transform $\mathcal{T}$ is injective, by Corollary 5.6.11 of \cite{obata-book}.

Let us, for simplicity of notation, introduce the following abbreviation:
\[
\widetilde{\xi}_{{}_{i}} = 
\widetilde{\xi}_{{}_{\widehat{n_i},\widehat{l_i},j_i,i_i}} = \widetilde{\eta}_{{}_{\widehat{n_i},\widehat{l_i},j_i,i_i}} 
+\overset{\cdot}{\i} \,\, \widetilde{\zeta}_{{}_{\widehat{n_i},\widehat{l_i},j_i,i_i}} = \widetilde{\eta}_{{}_{i}} 
+ \, \overset{\cdot}{\i} \, \widetilde{\zeta}_{{}_{i}}  
\]
\[
x_{{}_{i}} = x_{{}_{\widehat{n_i},\widehat{l_i},j_i,i_i}}, 
\,\,\,\,
y_{{}_{i}} = y_{{}_{\widehat{n_i},\widehat{l_i},j_i,i_i}}.
\]
Let
\[
\upsilon(x_{{}_{1}}, y_{{}_{1}}, \ldots
x_{{}_{k}}, y_{{}_{k}})
\]
be the tempered distribution on $\mathbb{R}^{2k}$, which is the image
of the distribution $P$. The finite dimensional tempered
distribution $\upsilon$ is uniquely determined by its finite-dimensional Fourier transform $\widehat{\upsilon}$
on $\mathbb{R}^{2k}$.
Let us compute $\widehat{\upsilon}$ first, and then recover $\upsilon$ by the inverse Fourier transform.
By definition, we have
\begin{multline*}
\widehat{\upsilon}(u_{{}_{1}}, w_{{}_{1}}, \ldots, u_{{}_{k}}, w_{{}_{k}}) 
\\
= 
\int\limits_{-\infty}^{+\infty} \ldots \int\limits_{-\infty}^{+\infty} \exp i \left[
\sum\limits_{i=1}^{k} u_i x_i +  w_i y_i 
\right]
\,
\upsilon(x_{{}_{1}},y_{{}_{1}}, \ldots, x_{{}_{k}}, y_{{}_{k}}) \, dx_{{}_{1}} dy_{{}_{1}} \ldots dx_{{}_{k}}dy_{{}_{k}}
\\
=
\int\limits_{\widetilde{E}^*} 
\exp \overset{\cdot}{\i}
\sum\limits_{i=1}^{k} 
\left[
\langle \widetilde{\varphi},u_i \widetilde{\eta}_{{}_{i}} \rangle + 
\langle \widetilde{\varphi},w_i \, \overset{\cdot}{\i} \, \widetilde{\zeta}_{{}_{i}} \rangle
\right]
P(\widetilde{\varphi})
\, d\mu(\widetilde{\varphi})
\\
=
\int\limits_{\widetilde{E}^*} 
\exp \overset{\cdot}{\i} \, \left\langle \widetilde{\varphi}, \sum\limits_{i=1}^{k} u_i \widetilde{\eta}_{{}_{i}} +\sum\limits_{i=1}^{k} w_i \, \overset{\cdot}{\i} \, \widetilde{\zeta}_{{}_{i}} \right\rangle
P(\widetilde{\varphi})
\, d\mu(\widetilde{\varphi}),
\end{multline*}
which by (\ref{FTmuT}) is equal to 
\begin{equation}\label{FTPojmuT}
\exp \textstyle{\frac{\overset{\cdot}{\i}}{2}}
\left[
\left\langle \sum\limits_{i=1}^{k} u_i \widetilde{\eta}_{{}_{i}} +\sum\limits_{i=1}^{k} w_i \, \overset{\cdot}{\i} \, \widetilde{\zeta}_{{}_{i}}, \widetilde{T_c} \left(\sum\limits_{i=1}^{k} u_i \widetilde{\eta}_{{}_{i}} +\sum\limits_{i=1}^{k} w_i \, \overset{\cdot}{\i} \, \widetilde{\zeta}_{{}_{i}}\right)^*\right\rangle
\right],
\end{equation}
and by orthogonality is equal
\[
\prod\limits_{i=1}^{k}
\exp \textstyle{\frac{\overset{\cdot}{\i}}{2}}
\left[
\textstyle{\frac{\sqrt{4\pi}\widetilde{\Delta_{c}}(\widehat{n_i}\cdot\widehat{l_i})_{{}_{j_ij_i}}}{2l_i+1}}
\left(u_{{}_{i}}^2 + w_{{}_{i}}^2\right)
\right]
= \widehat{\upsilon}(u_{{}_{1}}, w_{{}_{1}}, \ldots, u_{{}_{k}}, w_{{}_{k}}) 
\]
with the inverse Fourier transform 
\[
\upsilon(x_{{}_{1}},y_{{}_{1}}, \ldots, x_{{}_{k}},y_{{}_{k}})
\] 
equal to what is asserted by theorem.
\qed      

Theorems \ref{TProdConvergence} and \ref{muTexistence} give a complete analysis of the
first step (\ref{BSstep1}) -- definition and existence of $\mu_T$. We pass now to the
second step (\ref{BSstep2}) -- investigation which physically interesting functionals $F$
are equal to Fourier transform of Hida distributions.
We note first that we have a complete
characterization of the class of functionals $F$ which
are images of Hida distributions $\Lambda \in (\widetilde{E})^* \simeq (E)^*$. 
It is given in theorem 5.6.12 of \cite{obata-book}. Roughly speaking, it follows from this classification,
that the functional of the type
\[
F(\varphi) = e^{i\int \mathcal{L}(\varphi(x)) dx}
\]
with interaction Lagrangian density $\mathcal{L}(x)$, is in the Fourier transform image of a Hida distribution
if $\mathcal{L}$ contains natural powers of the field $\varphi(x)$ and its derivatives,
which are of second order at most in $\varphi$. This may seem to be a quite narrow class
from the physical point of view. But in the system of several interacting
fields, we can perform averaging with respect to each kind of field separately. Therefore, if each
interaction monomial in $\mathcal{L}(x)$ contains one and the same field in degree not greater than two,
then we can reach the functional as a Fourier transform, separately at each step corresponding 
to the averaging with respect to each kind
of field separately (a generalization of the Fubini theorem). In this case we can execute the second and third step (\ref{BSstep2}),
(\ref{BSstep3}), of the idea of \cite{Bogoliubov_Shirkov}, without essential modifications. 
In particular, for the Weinberg-Salam model of electroweak interactions, exponentiation of the only term of  
$\mathcal{L}(x)$, which cannot be reached as a Fourier transform
comes from the Higgs field $\varphi$ and is of the following third-order form
\begin{equation}\label{e^intphi^3}
F(\varphi) = e^{i \int \varphi(x)^3 dx}.
\end{equation}

In order to treat functionals (\ref{e^intphi^3}) of third or four order in $\varphi$ in the exponent, we need
to have the support of the distribution $P$ appropriately narrow, and need to know the distribution
$P$ more explicitly. By theorem \ref{muTexistence} and on heuristic grounds 
we expect $P$ to have the function-like, Gaussian-type form close to
\[
P(\widetilde{\varphi}) = e^{i \mathscr{A}_{{}_{0}}( \widetilde{\varphi})}  = e^{i \mathscr{A}_{{}_{0}}( \varphi)} 
\]   
up to a Gaussian factor ``flattening'' the standard probability measure $\mu$,
where $\mathscr{A}_{{}_{0}}$ is the action of the free classical (here scalar real) field $\varphi$, 
and which in the momentum representation
should have the form, up to a normalization,  close to
\[
e^{-\frac{\overset{\cdot}{\i}}{2}\left\langle \widetilde{\varphi}, \widetilde{T_c}^{-1} \widetilde{\varphi}^*\right\rangle},
\]
up to a Gaussian factor ``flattening'' $\mu$.
For the nonrelativistic systems with one degree of freedom (and potentials equal to Fourier transforms of measures),
a similar rigorous construction of the Feynman integral has been achieved, which is based on the
standard Gaussian probability measure $\mu$. In case of potentials for which propagators are \emph{a priori} known,
it gives the formulas for propagators which agree with the standard formulas, \cite{Hida}. 
For such systems,  with the trajectories $y(t)$ replacing our $\varphi(x)$,  $P(y)$
has the Gaussian-type form $e^{i \mathscr{A}_{{}_{0}}(y)}$, but with the additional Brownian factor 
of Gaussian-type, ``flattening'' the probability measure $\mu$, \cite{Hida}, theorem 10.2. We expect similar
``flattening'' factor also in our case.

For this to be possible at all, the support of $P$ should be appropriately narrow, and concentrated on 
functions subjectable to a unique and natural extension of differentiation up to the second order, 
as we do have a second-order differential operator in $\mathscr{A}_{{}_{0}}$ acting on $\varphi$. 
$\textrm{supp} \, P \subset \widetilde{E}_1 \simeq E_1$
would be sufficient, \emph{i.e.} $\textrm{Domain} \, A \simeq \textrm{Domain} \, \widetilde{A}$
of the self-adjoint extension of the standard operator $A$ introduced at the beginning of this Subsection. 
That the support of $P$ must be concentrated on some proper subspace $E_{k} \simeq \widetilde{E}_{k} \subset \widetilde{E}_{-1}$, 
with $k\geq 1$, can also be seen in the following way. Except for the functionals (\ref{e^intphi^3}) we need to have also
\begin{equation}\label{phi(x)phi(y)e^intphi^3}
F(\varphi) =  \varphi(x)\varphi(y)e^{i \int \varphi(z)^3 dz},
\end{equation}
for example, when computing the Green function of the Higgs boson. But the valuation
functional $\varphi \mapsto \varphi(x)$, cannot sensibly be extended (with the uniqueness of the extension achieved 
by the preservation of continuity) on the whole $E_{-1}$, but on $E_{1}$, or $E_{k}$, $k\geq 1$. 
Otherwise, we would have (\ref{phi(x)phi(y)e^intphi^3}) which is not defined on the whole
support $E_{-1}$ of $\mu$, and could not sensibly be integrated. Up to now, we have used only the right-hand inequality
of (\ref{R}). The left-hand inequality of (\ref{R}) influences the support of $P$, and we are going to exploit it now.

\begin{twr}
If the support $\mathscr{R}$ of $\widetilde{\nu}, \widetilde{\varphi}$ in (\ref{FTmuT})
 respects (\ref{R}) then $\textrm{supp} \, P \subset \widetilde{E}_1 \simeq E_1$
\end{twr}
\qedsymbol \,
From the left-hand inequality of (\ref{R}) it follows the inequality
\begin{equation}\label{TPcontinuity}
\sqrt{4\pi} \, \left|\widetilde{\nu} \right|_{{}_{-2}}^{2}
=
\sqrt{4\pi} \, \left| \widetilde{A}^2 \widetilde{\nu}\right|_{{}_{-2}}^{2} 
\leq
\left| \left\langle \widetilde{\nu}, \widetilde{T_c} \widetilde{\nu}^*\right\rangle \right|.
\end{equation}
It follows that the Fourier transform
\[
\mathcal{T} P (\widetilde{\nu}) = e^{\frac{\overset{\cdot}{\i}}{2}\left\langle \widetilde{\nu}, \widetilde{T_c} \widetilde{\nu}^*\right\rangle} 
\]
of $P$ has a continuous extension on $\widetilde{E}_{-2} \simeq E_{-2}$. Now, we observe that
the proof of the Minlos theorem 17.1 (or thm. 1.5.3 of \cite{obata-book}), as presented in \cite{Yamasaki} , 
can be extended from probability measures on the Hida distribution $P(\widetilde{\varphi}) d\mu(\widetilde{\varphi})$, 
with the finite dimensional projections equal to the standard complex Gaussian distribution of theorem \ref{muTexistence}. 
Using this generalization of Minlos' theorem and the continuous extendibility of $\mathcal{T} P$ on 
$\widetilde{E}_{-2} \simeq E_{-2}$,  we conclude that $\textrm{supp} \, P \subset \widetilde{E}_1 \simeq E_1$,
because $A^{-1} \simeq \widetilde{A}^{-1}$ is Hilbert-Schmidt, and the inclusion $\widetilde{E}_{-1} \rightarrow \widetilde{E}_{-2}$
is Hilbert-Schmidt with the dual $E_{-1}^{*}$ of the Hilbert space $E_{-1}$ equal $E_{1}$. 

\qed

Theorem 1.5.3 is formulated in \cite{obata-book} only for probability measures on $E^* \simeq \widetilde{E}^*$. 
It is immediately seen to have a natural extension, at least on a wide class of Hida distributions. 
It is the case, e.g. for distribution representable by a test Hida function, or more generally, 
by absolutely summable function on $\widetilde{E}^*$, because the pairing is given by integration with the standard
Gaussian probability measure $\mu$.

To treat the functional (\ref{e^intphi^3}) or (\ref{phi(x)phi(y)e^intphi^3}) we expand them, using the exponent series
into the functionals
\[
F_{n}(\widetilde{\varphi})
= \sum\limits_{k}^{n} {\textstyle\frac{{\overset{\cdot}{\i}}^k}{k!}}
\left(
\int \varphi(z)^3 dz
\right)^k
\,\,\,
\textrm{or}
\,\,\,
F_{n}(\widetilde{\varphi})
= \varphi(x)\varphi(y)\sum\limits_{k}^{n} {\textstyle\frac{{\overset{\cdot}{\i}}^k}{k!}}
\left(
\int \varphi(z)^3 dz
\right)^k
\]
which, by theorem 5.6.12 of \cite{obata-book}, are Fourier transforms of Hida distributions $\Lambda_n$:
\[
\mathcal{T}\Lambda_n(\widetilde{\varphi}) = F_n(\widetilde{\varphi}),
\]  
with $F_{n}$ converging pointwisely to $F$. Here we interrupt our analysis of convergence implicitly 
used in (\ref{BSstep3}). The idea is that the Hida distribution
calculus is largely reduced to integration technics with respect to the standard measure $\mu$. These integrals
may be approximated by ordinary integrals with high multiplicity, on using density and orthogonality 
of the (ordinary and Wick ordered) polynomials on the Gaussian space $E^* \simeq \widetilde{E}^*$.

This opens us to a method which, from the point of view of practical computations, 
may constitute an improvement of the method based on replacing space-time by a periodic finite lattice
\cite{Creutz}. As is known, the main problems associated with the method based on finite 
periodic lattices are related to the so-called artifacts. Checking whether the choice of 
UV- and IR-renormalization, given by the mere restriction to a finite periodic lattice, 
corresponds in the limit to ordinary renormalization, is problematic. 
The second main problem is related to the ambiguity in averaging over 
Fermionic variables.

Therefore, to make our solution to the problem posed in \S 43.3, \cite{Bogoliubov_Shirkov} 
interesting, we should have to give an analogous construction of $d\mu_T$ for averaging 
over Fermi fields. This can be done through the analogous refinement of the method given 
in \S 43.4, \cite{Bogoliubov_Shirkov}. This problem we have, to a large extent, already solved, because (following
Berezin \cite{Berezin}) we have given the analogue of the Gelfand space-time triple $E \subset H \subset E^*$ (given at the beginning of this Section)
with the nuclear space $E$ of Grassmann valued space-time test functions, and its nuclear dual $E^*$ of Grassmann 
valued Hida distributions, compare \cite{wawrzyckiInfinite} or Subsection \ref{axiomsS}. 
We need to extend the isomorphism of the Wiener-It\^o-Segal from the Bose over the Fermi Fock space. Concerning
the Wick ordered polynomial functions on $E^*$ of the Bose case, we replace in them the pointwise (symmetric) products of ordinary 
functions with the Grassmann products of Grassmann valued functions. The lacking part is the 
analogue of the standard cylindrical measure $\mu$ on $E^*$. But this analogue can be constructed largely along 
the standard lines, given in \cite{Gelfand}, using the Berezin integration over Grassmann variables and the
transformation properties of the Gaussian integral over Grassmann variables \cite{Berezin}.

Recall that the fact that an assumption of the type (\ref{R}) should be 
imposed on the domain of the functional averaging, has been clear 
to physicists from the very beginning of the application of the 
averaging method to the calculation of Green's functions,
compare \S 2.5 (\S 2.6 of Russ. Ed.) of \cite{SlavnovFaddeev}.
Possibly the condition may be to some 
extent relaxed (e.g. with slightly different power 
of the denominator on the left), but the above analysis show that
condition of the type (\ref{R}) goes in the right direction.

\subsection{Space-time geometry and quantum fields. Quasi-classical limit of quantum fields}\label{CurvatureQFTandG}

In the previous Subsections we have proved that the simple causality axioms (I)-(V)
for the scattering operator of some perturbative Quantum Field Theories (QFT), lead to well behaved interacting fields,
if the globally causal space-time has compact Cauchy surfaces and non-zero curvature. This is the case only for some classes of QFT
with some particulalr interaction Lagrangians.  For example this is the case for spinor QED with minimal $U(1)$ coupling, 
in which moreover the charged fields are necessary massive. In particular, we obtained a theoretical proof 
of the experimentally confirmed fact that charged fields are massive.  
Only on globally causal space-times with compact Cauchy surfaces and non-zero curvature QFT may exist with
the smeared out interacting fields being ordinary operators 
with well-defined fluctuations, and with the Noether integrals (if symmetries exist)
for interacting fields being well-defined self-adjoint operators on the Fock space. 
This is the case e.g. for spinor QED with massive charges. 
On the flat Minkowski space-time this is impossible. Moreover, on globally causal space-times with 
compact Cauchy surfaces and non-zero curvature
several theorems concerning QFT interactions can be proved (we have proved 
them explicitly in case of the Einstein Universe), which are confirmed by experiment, 
and which have no other theoretical justification, e.g. that only
massive fields couple to the electromagnetic potential field. On the flat space-time
the QFT with the same general causality axioms (I)-(V) gives much more singular 
interacting fields, which do not allow the basic quantum mechanical
interpretation of the theory. Although me may prove the analogous result: that charged fields are massive, 
using the Hida operators, but only for interacting fields which are quite singular (Subsection \ref{OperationsOnXi}). 
In particular on the flat Minkowski space-time, or more generally
on space-time with non-compact or compact Cauchy surfaces but with zero curvature, the interacting fields, 
after smearing out with test functions
are not ordinary operators on the Fock space and the fluctuations of the smeared out interacting fields
are meaningless. QFT, determined by the causality axioms (I)-(V) on such space-times, e.g. on the Minkowski
space-time, have very restricted meaning and can be applied only to the scattering processes
involving nonnormalizable generalized plane wave states as the \emph{in} and \emph{out} states. 
The Schwinger Green functions can be computed also in the flat Minkowski case, but the associated so called
many particle wave function tied to the Green function through the Huyghens principle, cannot sensibly be
associated with any state of the Fock space acted on by the interacting fields. Only passing to globally causal space-time
with compact Cauchy surfaces and non-zero curvature we can associate a well defined
Fock state to the Schwinger many-particle wave function (in the general way we have learned from Bogoliubov).

It therefore seems (and the above cited results can in principle be interpreted as a proof) that already from 
the causality axioms (I)-(V) of the scattering operator it follows that interacting quantum fields which admit quasiclassical limit
have nonzero (gravitational) weight and cannot live on the flat space-time. 

Essentially no back reaction of the quantum field to the space-time geometry is explicitly included 
into the causality axioms (I)-(V) of such
QFT. Nonetheless the cited results suggest that, at least some important aspects concerning relation 
between quantum fields and space-time geometry are inherited in the causality axioms (I)-(V) of the scattering operator,
which obviously depend on the space-time geometry.
 Because in addition we expect of any 
successful QFT to have correspondence to the initial classical theory, then using this correspondence
for the specific quasi-classical states, and the classical Einstein equations relating the space-time geometry to the Hilbert-Einstein energy-momentum
tensor, we try to reconstruct the relation of quantum states of quantum fields to space-time geometry, which is valid also for 
states with large particle numbers where the back reaction to the space-time geometry is expected to be non-negligible.

At the begining let us remind the general and universal division of the general laws of  a physical (classical) 
system (having finite or, as in case of fields, infinite number of degrees of freedom) into the \emph{dynamical} laws and \emph{constrains}.
Presence of the constrains means that (at the level of Lagrangian formulation) the momenta of the system are not independent
functions of the velocities. In that case generalized momenta $p$ and positions $q$ of the system are relatad
by some equations $F(p,q)=0$. We do not enter now into details of the possible types of such constrains, but mention only
that presence of constrains is quite universally pertinent to the most important systems 
(including QED, more complicated and quite realistic theories with gauge freedom,
or theory of the massive fourvector field, and many other realistic systems).  Constrains have tremendous consequece in passing 
to the quantum counterpart of the theory. As we have lerned so far, the quantum mechanical counterparts of the 
constrains should be treated in substantially different manner
in comparison to the equations of motion, exactly as at the classical level.
There are various types of constrains which may have operator counterparts at quantum level.  A large class of constrains,
which have been divided into some subclasses (at the classical evel) in \cite{DiracLectures}, pass after quantization
into subsidiary conditons at the operator level, i.e. into operator equations. This is e.g. the case for the quantum field of 
the massive fourvector field, where the constrain has the form of an operator equation, compatible with the operator equations which incorporate dynamics. But 
there are also systems with constrains (not investigated in \cite{DiracLectures}, e.g. free electromagnetic potential field with the Lorentz condition, 
or QED with the Lorentz condition), for which, the constrains in passing to the quantized system,  do not have any operator counterparts,
but only the dynamical equations do have. These constrains do not admit operator form not only because of the inconsitency with the dynamical
equations but also for other reasons, e.g. inconsistency with the definition of the vacuum (as is the case for the Lorentz condition
for the free e. m. potential or in QED). In particular consider the free Maxwell equations 
divided into the Lorentz constrain $\partial A=0$ and the dynaical wave equation $\square A=0$, 
or Maxwell equations with currents divided into the Lorenz constrain $\partial A=0$ and the dynamical equation $\square A = j$. In passing to the 
quantized system the constrains can only be satisfied in the sense of averages $\big\langle \partial A \big\rangle = 0$ 
and only $\square A = 0$ in operator form. Similarly the constrain $\big\langle \partial A_{{}_{\textrm{int}}} \big\rangle=0$ is fulfiled for the averaged values  
and only $\square A_{{}_{\textrm{int}}} =  j_{{}_{\textrm{int}}}$ as operator equation, \emph{i.e.} 
if computed in the physical states $\Phi$ selected by the condition $\partial A^{-}\Phi = 0$ (or $\partial A^{-}_{{}_{\textrm{int}}}\Phi = 0$) 
plying the role of the constrain at the quantum level. Thus the full Maxwell equations at the quantum level are fulfilled in the sense 
of averages in physical states, on which the indefinite Krein inner product is positive. 
In this last type of constrains (which are very important and seem indispensable) 
no operator form of the constrains is compatible with the operator equations of motion. 
In this situation when passing to quantum theory we are working with dynamical equations of motion ignoring altogether the constrains,
and only at the very end select the physical states in which the quantum analogs of the classical constrains are fulfilled 
in the sense of averaging in physical states. Having this in mind let us have a look at the classical
system of matter field equations coupled to the gravitational field in accordance to the classic version of Einstein's theory 
of gravity. First, we select the Lagrangian $\mathcal{L}_{{}_{M}}$ for the matter fields
$\psi$. It can be for example the Lagrangian $\mathcal{L}_{{}_{M}}$ of the classical system underlying scalar or spinor QED, \emph{i.e.}
e.m. potential field coupled to the classical de Broglie-Dirac field (understood, of course, not as a Schr\"odinger 
probability state, but as a classical de Broglie matter field with polarization, compare \cite{TomonagaII}). $\mathcal{L}_{{}_{M}}$ 
together with the Hilbert-Einstein Lagrangian density $\mathcal{L}_{{}_{G}}$ 
(for the vacuous Einstein equations of the gravitational field) compose the total Lagrangian 
\[
\mathcal{L} = \mathcal{L}_{{}_{G}} + \alpha_{{}_{M}}\mathcal{L}_{{}_{M}}
\]
of the system, with the constant $\alpha_{{}_{M}}$ depending on the concrete type of the matter fields $\psi$. Since $\mathcal{L}_{{}_{G}}$
does not depend on the matter fields $\psi$, variation $\tfrac{\delta S}{\delta \psi}$ of the total action  
\[
S =  \int \big[ \mathcal{L}_{{}_{G}} + \alpha_{{}_{M}}\mathcal{L}_{{}_{M}} \big] \, d^4 x = S_{{}_{G}} + \alpha_{{}_{M}} S_{{}_{M}}
\]
with respect to matter fields will yeld the same matter equations
\begin{equation}\label{ME}
\frac{\delta S_{{}_{M}}}{\delta \psi} = 0
\end{equation}
of motion as the variation of $S_{{}_{M}}$ alone. 
Variation of the total action $S$ with respect to the metric $g_{\mu\nu}$ yelds the Einstein equations
\begin{equation}\label{EE}
G_{\mu \nu} = - \kappa \mathbb{T}_{\mu\nu},
\end{equation}
where
\[
\mathbb{T}_{\mu\nu} = {\textstyle{\frac{\alpha_{{}_{M}}}{\kappa}}} {\textstyle{\frac{1}{\sqrt{-g}}}} \,\,
\frac{\delta S_{{}_{M}}}{\delta g_{\mu\nu}},
\]
is the Hilbert-Einstein energy momentum tensor of the matter fields.
There are various possible ways of looking at this system. Ultimately, however, it is all about 
determinding 1) the space-time $\mathcal{M}$ (with a specific $g_{\mu\nu}$) and 2) fields $\psi$ on $\mathcal{M}$ 
that satisfy the equations (\ref{ME}) and (\ref{EE}). In any case we are seeking for $\mathcal{M},g$
and $\psi$ which are consistent, \emph{i.e.} veryfying both (\ref{ME}) and (\ref{EE}). 
In principle, in searching for a consistent $\mathcal{M},g, \psi$, 
we can always put things in the follownig order. First, we take any (say globally hyperbolic) 
but fixed space-time $\mathcal{M},g$. Next, we take a generally covariant Lagrangian $\mathcal{L}_{{}_{M}}$ on $\mathcal{M}$
for the matter fields $\psi$ of required type on $\mathcal{M}$, and investigate all possible solutions of matter equations (\ref{ME}). 
The system of matter equations (on fixed  $\mathcal{M},g$) can be treated completely independently 
of (\ref{EE}), as only the Lagrangian $\mathcal{L}_{{}_{M}}$  of matter fields contributes into (\ref{ME}). 
As such the system (\ref{ME}) of matter equations can be divided into constrains and dynamical equations and treated by themselves.
Finally, the whole system is consistent if and only if among the solutions of (\ref{ME}) (on the chosen and fixed $\mathcal{M},g$),
there are ones for which equations (\ref{EE}) are fulfilled. As we see the choice of a fixed space-time $\mathcal{M},g$
and the Einstein equations (\ref{EE})  play the role of constrains for the system of matter fields governed by the
equations (\ref{ME}). 

In passing to quantum theory we keep the same order and ordinary laws of quantization. First, 
we fix a (globally hyperbolic) space-time, and according to ordinary rules keep the metric 
as a purely classical object (otherwise the quantization of fields would not have any sensible meaning). 
Applying ordinary laws of quantization we cannot expect to have the counterpart of (\ref{EE}) in any operator form as (\ref{EE}) 
are the constrains for the matter system on the fixed space-time. 
Similarly the part of equations (\ref{ME}) which is subsumed by constrains cannot, in general 
(e.g. for QED with Lorentz constrains) be expected to have any quantum  
operator analogue, because they are also the constrains of the system. Only the part of equations (\ref{ME})  
which composes the dynamical part of the system (\ref{ME}) is expected to have sensible quantum operator analogue.
Concerning the constrains (\ref{EE}) they obviously cannot have any operator analogue at the quantum level, as the left hand side 
of (\ref{EE}) is a $c$-number after quantization, or just the operator proportional to the identity (operator-valued distribution with 
values proportional to the identity operator), which is impossible for the right hand side of (\ref{EE}) at the quantum level. 
In this the constrains (\ref{EE}) are similar to the Lorentz constrains in QED.  
We propose to treat the constrains (\ref{EE}) similarly as the Lorentz constrains in QED, \emph{i.e.}
ignore  (\ref{EE}) in the proces of quantization of the matter fields on the assumed and fixed space-time 
using the causality axioms (I)-(IV) (with the respective treatment of the constrain part and dynamical part of (\ref{ME})) 
and only regain the analogue of (\ref{EE}) at the quantum level as a condition selecting physical states. 
The mode of selection of the physical states cannot be fully analouge to the Lorentz constrains 
as we do not expect to have any exact equation for averages because the operator corresponding to the left hand side of  (\ref{EE}) is an ordinary
$c$-number and the operator corresponding to the right hand side is not a $c$-number. Nonetheless if the QFT
for matter fields is successful then we can compute the averages $\big\langle \mathbb{T}_{{}_{\mu \nu}}(\phi)\big\rangle$ 
and fluctuation 
\[
\Delta\mathbb{T}_{{}_{\mu \nu}}(\phi) = \sqrt{\langle \mathbb{T}_{{}_{\mu \nu}}(\phi)^2\rangle - \langle \mathbb{T}_{{}_{\mu \nu}}(\phi) \rangle^2}
\]
of the operator corresponding to the Hilbert-Einstein energy-momentum tensor 
$\mathbb{T}$ smeared out with space time test functions $\phi$, for a dense subset of the states $\Phi$
in its domain. The condition $\big\langle G(\phi)\big\rangle = G(\phi) = \big\langle -\kappa \mathbb{T}(\phi)\big\rangle$
selecting physical states would be analogous to the selection of physical states imposed by the Lorentz condition,
but it seems to be rather unrealistic. For the space of states in which the said equality of averages holds seem to be
too restrictive in rejecting too much possible superpositions. Success and everyday practice of quantum theory shows that 
a physical state $\Phi$ admits a considerable class of superpositions in which some states of microscopic subsystems 
are varied. Thus a wide class of superpositions has to be admitted.
Let $\Phi, \Phi'$ be any two different states in the domain of the self-adjoint operator 
$\mathbb{T}_{{}_{\mu \nu}}(\phi)$ in which the average $\big\langle \mathbb{T}_{{}_{\mu \nu}}(\phi)\big\rangle$ 
is \emph{exactly} the same. It is rather exceptional that their normalized superposition 
will keep \emph{exactly} the same average value of $\mathbb{T}_{{}_{\mu \nu}}(\phi)$.
An example of rather unrealistic exception we have when zero is a discre point of the 
spectrum of $\mathbb{T}_{{}_{\mu \nu}}(\phi)$ with $\Phi$ and $\Phi'$ differing by a proper vector corresponding to the zero proper value
of the self-adjoint $\mathbb{T}_{{}_{\mu \nu}}(\phi)$. More realistic exceptions can be constructed if we recall that 
self-adjoint $\mathbb{T}_{{}_{\mu \nu}}(\phi)$ is unitarily equivalent to a multiplication operator by a measurable function 
on a measure space. If under this unitary equivalence $\Phi$ and $\Phi'$ are transformed into square integrable functions with disjoint supports
then the average in their normalized superposition will stay the same as the common average in the states $\Phi,\Phi'$. 
It is however fairly not clear if the exceptions are reach enough to explain the success of the ordinary quantum mechanisc 
applied to microscopic systems.  But we expect that for physical quasi-classical states for which 
$\Delta \mathbb{T}(\phi)$ is small in comparison to  the average $\big\langle \mathbb{T}(\phi)\big\rangle$ 
the average $\big\langle -\kappa\mathbb{T}(\phi)\big\rangle$ agrees with the left hand side of (\ref{EE}) 
at least up to the uncertainty $\kappa\Delta \mathbb{T}(\phi)$. 
We propose to include all states verifying this condition, to the class of physical states, irrespectively 
of the value of the uncertainty $\Delta \mathbb{T}(\phi)$.  Thus, it is meaningfull and seems reasonable 
to put the constrain selecting the physical states $\Phi$ for which
the average $\big\langle -\kappa \mathbb{T}(\phi)\big\rangle$ agrees with the smeared out
\[
G(\phi) \overset{\textrm{df}}{=} \int G(x)\phi(x) \, d^4x
\]
Einstein geometric tensor components up to the fluctuation $\kappa\Delta \mathbb{T}(\phi)$ in the state $\Phi$:
\begin{equation}\label{QEE}
-\kappa\Delta \mathbb{T}(\phi) \leq G(\phi) - \big\langle -\kappa\mathbb{T}(\phi)\big\rangle \leq \kappa\Delta \mathbb{T}(\phi).
\end{equation}     
and put (\ref{QEE}) as the analogue for the classical constrains (\ref{EE}) at quantum level.

On the space-time with nonzero curvarute and compact Cauchy surface, with causal QFT (with axioms (I)-(IV)) upon it,
with well-behaved interacting fields, such as the  scalar or spinor QED on the Einstein Universe, the condition
(\ref{QEE}) is nonempty in the sense that whenever classical fields allow solutions of the classical equations
of matter (\ref{ME}) which verify also (\ref{EE}), then so is at the quantum level: there should exist states
in the Fock space of (interacting) fields, which verify (\ref{QEE}). It is most easily to see it for
the systems consisting exclusively of Bose fields (eg. for scalar QED), first without interaction. At the quantum level for free quantum fields
$\mathbb{T}$ is equal to the Wick poduct of the free fields corresponding to the classical fields in the classical $\mathbb{T}$.
We can use the exponential (coherent) states $\Phi_{\xi} \in (E)$ for $\xi$ in the full single particle test space $E$.
We have the property (say, on the Einstein Universe)
\[
a(\widehat{n}\cdot\widehat{l}) \Phi_\xi = \xi(\widehat{n}\cdot\widehat{l}) \Phi_\xi, 
\]  
for the (Bose) annihilation operators $a(\widehat{n}\cdot\widehat{l})$,  from which it follows that the (Berezin, or normalized) symbol 
\[
S_\xi[A(x)] = \frac{\left\langle\left\langle\Phi_{\xi}, A\Phi_{\xi} \right\rangle\right\rangle}{\left\langle\left\langle \Phi_\xi, \Phi_\xi \right\rangle\right\rangle}
\]
of a generalized integral kernel operator (such as free fields and their Wick products) behaves multiplicatively 
\[
S_\xi[{:}A(x){:} {:} B(x){:}] = S_\xi[{:}A(x){:}] S_\xi[ {:}B(x){:}] 
\]
under the Wick product. For each smooth classical solution of the matter fields we can find $\xi \in E$ such 
that the symbols of the free field operators coincide with the classical corresponding fields.
Choosing $\xi$ to be the element corresponding to the classical free field which verifies (\ref{EE}) we get
\[
G(x) = -\kappa S_\xi[\mathbb{T}(x)]
\,\,\,\,\,\,\,\,\, \textrm{or}
\,\,\,\,\,\,\,\,\,
G(\phi) = - \kappa \int S_\xi[\mathbb{T}(x)] \, \phi(x) \, d^4x,
\]
for each space-time test function $\phi$.
The Wick product fields behave regularly on EU. In particular they map continuously the space-time test space into the space of continuous
operators on the test Hida space $(E)$ in the total Fock space, \emph{i.e.} $\phi \mapsto \mathbb{T}(\phi)$ is a continuous map to the space
of linear continuous operators on $(E)$. It follows that the last equality can be rewritten as equality for average 
in the normalized coherent state $\Phi_\xi$
\[
G(\phi) = - \kappa \frac{\left\langle \Phi_\xi, \mathbb{T}(\phi) \Phi_\xi \right\rangle}{\| \Phi_\xi \|^2}.
\]
Note here, that in the definition of the symbol we have dual pairing $\langle\langle \cdot, \cdot \rangle\rangle$,
which becomes the ordinary inner product if both paired elements are elements of finite Hilbert space norm,
which is in particular the case for $\Phi_\xi$ and  $\mathbb{T}(\phi) \Phi_\xi$ in our case. On Minkowski space time, for example, 
even the Wick products of free fields behave more singularily, if the products contain more than one massless field factor, 
so that the smeared out Wick product will in general transform the Hida space into its strong dual, and the pairing in the symbol for such operators cannot
be written as an actual average in the inner product sense for a well-defined state in the Fock space. In passing to the quantum counterpart of the system with interaction with perturbative QFT with regular interacting fields
(as on EU) still for the system with exclusivly Bose fields (say, scalar QED on EU), the symbols of the higher order contributions to the perturbed operators 
are in correspondence with the classical perturbation contributions. Therefore, if the corresponding classical system 
of interacting fields is consistent and admitts solutions of both (\ref{ME}) and (\ref{EE}), we expect the same at the quantum level.

We finish this Subsection with a remark
that the system of free fields on the Einstein Universe defines in a canonical manner the space-time geometry in an 
operator form, in which the smooth functions are represented by commuting operators.
In fact this operator-spectral construction is based on the representors of the symmetry group
associated to the system of free fields.

This is possible because the system of free quantum fields, together with the Emmy Noether
conserved integrals $\boldsymbol{P}^{{}^{0}}, \mathbb{P}^{{}^{k}}$  (corresponding to the
one-parameter groups of symmetries generated by the derivatives $X^0, X^k$, $k=1,2,3$, of Subsection \ref{GeneralizedSchrodinger-VonNeumannPairs})
provide in the Fock space an infinite system of spectral triples in the total Fock space in the sense of Connes,
which include the whole information of the space-time manifold together with its metric structure, in the operator format
and has been constructed in Subsection \ref{GeneralizedSchrodinger-VonNeumannPairs}. 

Here we only mention that the construction of the invariant subspaces 
corresponding to the spectral tuple (\ref{3rdSpectralTupleForSxG}) of Subsection \ref{GeneralizedSchrodinger-VonNeumannPairs}
in the Fock space of purely positive energy fields would be impossible. But if we add to each kind of positive energy field
a negative energy counterpart, and extend the Fock space by tensoring it with the Fock space of purely negative energy fields,
then the construction of the invariant subspaces of such extended Fock space, which can be identified
with the Hilbert space of the spectral tuple (\ref{3rdSpectralTupleForSxG}) of Subsection \ref{GeneralizedSchrodinger-VonNeumannPairs}
becomes possible. The action of the Noether integrals can then be identified with the 
operators $X^0, X^k$, $k=1,2,3$ acting on the smooth functions on the space-time, and the triple 
(\ref{3rdSpectralTupleForSxG}) of Subsection \ref{GeneralizedSchrodinger-VonNeumannPairs} can be identified
with 
\begin{multline}\label{specral-space-time-free-fields-EU}
\Bigg( \,\,\,\,\,\,\,\, \mathcal{A} = 
\{f (\boldsymbol{Q}^{0}, \boldsymbol{Q}^{1}, \ldots, \boldsymbol{Q}^{m}), 
f \in \mathscr{C}^\infty (\mathbb{R}^m)\} \,\,,
\,\,\,\,\,\,\,\,
\mathcal{H}_{{}_{\textrm{inv}}} \subset \Gamma(\mathcal{H}_{{}_{\textrm{total}}}) \,, \\
D_{{}_{\textrm{ell}}} =  D_{{}_{\mathfrak{J}}} \, {} = \Gamma^{{}^{0}} \boldsymbol{P}^{{}^{0}} + \Gamma^{{}^{1}} \mathbb{P}^{{}^{1}} + 
\ldots + \Gamma^{{}^{3}} \mathbb{P}^{{}^{3}}  + \Gamma 
- i\Gamma^{{}^{1}} A^{{}^{1}} - 
\ldots -i \Gamma^{{}^{3}} A^{{}^{3}} \,\,, 
\,\,\,\,\,\,\,\,\, \\
D \, {} = \widehat{\Gamma}^{{}^{0}} \boldsymbol{P}^{{}^{0}} + \widehat{\Gamma}^{{}^{1}} \mathbb{P}^{{}^{1}} + 
\ldots + \widehat{\Gamma}^{{}^{3}} \mathbb{P}^{{}^{3}} + \widehat{\Gamma}
- i\widehat{\Gamma}^{{}^{1}} A^{{}^{1}} - 
\ldots -i \widehat{\Gamma}^{{}^{3}} A^{{}^{3}} \,,
\,\,\,\,\,\,\,\,\,
\mathfrak{J}  \,\,\,\, \Bigg),
\end{multline}
where $\boldsymbol{P}^{{}^{0}}, \mathbb{P}^{{}^{k}}$ are conserved Noether integrals, being equal to the generators
of the symmetry subgroup $\widetilde{\mathbb{S}^1}\times SU(2, \mathbb{C}) \subset \widetilde{\mathbb{S}^1}\times SU(2, \mathbb{C}) \times SU(2, \mathbb{C})$
of the symmetry group of the Einstein Universe, and $\mathcal{H}_{{}_{\textrm{inv}}}$ is a subspace of the
(extended) Fock space, invariant for $\boldsymbol{P}^{{}^{0}}, \mathbb{P}^{{}^{k}}$. 
Recall that here $\boldsymbol{Q}^0, \ldots, \boldsymbol{Q}^k$, are the operators on  $\mathcal{H}_{{}_{\textrm{inv}}}$, which together
with $\boldsymbol{P}^0 = \mathbb{P}^0 +iA^0$,  $\boldsymbol{P}^k = \mathbb{P}^k - iA^k$, compose the generalized Schr\"odinger-Von Neumann
pairs on $\mathcal{H}_{{}_{\textrm{inv}}}$, and where the operators $A^\mu$ are equal to the operators of multiplication
by constant matrices -- the generators of the finite dimensional representation $V$ of $SU(2, \mathbb{C})$ associated to 
the invariant subspace $\mathcal{H}_{{}_{\textrm{inv}}}$ and thus to the triple in the manner explained in Subsection 
\ref{GeneralizedSchrodinger-VonNeumannPairs} (compare the formula (\ref{3rdSpectralTupleForSxG}) of Subsection \ref{GeneralizedSchrodinger-VonNeumannPairs}).

There are infinitely many such invariant subspaces $\mathcal{H}_{{}_{\textrm{inv}}}$ in the (extended Fock space) 
with arbitrary large and uniform multiplicity of $\boldsymbol{P}^{{}^{0}}, \mathbb{P}^{{}^{k}}$ acting on $\mathcal{H}_{{}_{\textrm{inv}}}$,
compare Subsection \ref{GeneralizedSchrodinger-VonNeumannPairs}. 

Finally we note that this spectral-operator construction of space-time geometry is in principle also
possible on the space-times without any symmetries. This is in particular possible for space-times
which are images of the Einstein Universe under causal (``conformal'') isomorphisms. 
As manifolds, these space-times are Lie groups $\mathbb{R}\times SL(2, \mathbb{C})$, but of course the left and right invariant
vector fields do not compose actual space-time symmetries.
The additional complication follows from the fact that no one-parameter groups of diffeomorphisms, corresponding to the right invariant vector fields
$X^\mu$, and the corresponding derivations $X^\mu$, 
compose any symmetries of the differential equations corresponding to the free fields. In particular
the action of any such one-parameter group on the elements of the single particle Hilbert space
of any fixed free field leads us out of the single particle Hilbert space. Therefore, the whole system of free fields
must be sufficiently large in order that the direct sum of the single particle Hilbert saces of all
free fields of the system composes an invariant space for the action of $X^\mu$. Then the group of diffeomorphisms
can be regarded as a symmetry of the whole system of free fields, considered as a generalized free field
in the sense of Greenberg, \cite{Greenberg}, \cite{SegalZhouQED}, \cite{SegalZhouPhi4}, whose action in the Fock space can be 
constructed as the application of the functor of second quantization.

The spectral relationship (\ref{specral-space-time-free-fields-EU}) of spacetime geometry to free fields can be prolonged
on the interacting fields defined perturbatively, by switching on the interaction separately in the subsystem of positive energy fields
and separately in the subsystem of negative energy fields, and by replacing the conserved Noether integrals 
$\boldsymbol{P}^{{}^{0}}, \mathbb{P}^{{}^{k}}$  in (\ref{specral-space-time-free-fields-EU}) by the interacting
counterparts (with the conserved densities replaced by the interacting counterparts) 
$\boldsymbol{P}^{{}^{0}}_{{}_{\textrm{int}}}, \mathbb{P}^{{}^{k}}_{{}_{\textrm{int}}}$:
\begin{multline}\label{specral-space-time-interacting-fields-EU}
\Bigg( \,\,\,\,\,\,\,\, \mathcal{A} = 
\{f (\boldsymbol{Q}^{0}_{{}_{\textrm{int}}}, \boldsymbol{Q}^{1}_{{}_{\textrm{int}}}, \ldots, \boldsymbol{Q}^{m}_{{}_{\textrm{int}}}), 
f \in \mathscr{C}^\infty (\mathbb{R}^m)\} \,\,,
\,\,\,\,\,\,\,\,
\mathcal{H}_{{}_{\textrm{inv} \, \textrm{int}}} \subset \Gamma(\mathcal{H}_{{}_{\textrm{total}}}) \,, \\
D_{{}_{\textrm{ell}\, \textrm{int}}} =  D_{{}_{\mathfrak{J} \, \textrm{int}}} \, {} 
= \Gamma^{{}^{0}} \boldsymbol{P}^{{}^{0}}_{{}_{\textrm{int}}} + \Gamma^{{}^{1}} \mathbb{P}^{{}^{1}} + 
\ldots + \Gamma^{{}^{3}} \mathbb{P}^{{}^{3}}_{{}_{\textrm{int}}}  + \Gamma 
- i\Gamma^{{}^{1}} A^{{}^{1}} - 
\ldots -i \Gamma^{{}^{3}} A^{{}^{3}} \,\,, 
\,\,\,\,\,\,\,\,\, \\
D_{{}_{\textrm{int}}} \, {} = \widehat{\Gamma}^{{}^{0}} \boldsymbol{P}^{{}^{0}}_{{}_{\textrm{int}}} 
+ \widehat{\Gamma}^{{}^{1}} \mathbb{P}^{{}^{1}}_{{}_{\textrm{int}}} + 
\ldots + \widehat{\Gamma}^{{}^{3}} \mathbb{P}^{{}^{3}}_{{}_{\textrm{int}}} 
+ \widehat{\Gamma} - i\widehat{\Gamma}^{{}^{1}} A^{{}^{1}} - 
\ldots -i \widehat{\Gamma}^{{}^{3}} A^{{}^{3}} \,,
\,\,\,\,\,\,\,\,\,
\mathfrak{J}  \,\,\,\, \Bigg),
\end{multline}

In accordance to our results (compare the last Subsection) each higher order contribution to
$\boldsymbol{P}^{{}^{0}}_{{}_{\textrm{int}}}, \mathbb{P}^{{}^{k}}_{{}_{\textrm{int}}}$ is a well defined 
self-adjoint operator on the total Fock space (which was impossible on the flat Minkowski space-time). 

We believe that the spectral relation of the space-time geometry to quantum fields, namely
(\ref{specral-space-time-free-fields-EU}) in case of free fields and (\ref{specral-space-time-interacting-fields-EU})
in case of interacting fields, gives us a deeper insight. First of all we see from 
(\ref{specral-space-time-free-fields-EU}) or (\ref{specral-space-time-interacting-fields-EU}) that the space-time geometry
spectrally described by the Dirac operators $D,D_{{}_{\mathfrak{J}}}$ or, respectively, 
$D_{{}_{\textrm{int}}}, D_{{}_{\mathfrak{J} \, \textrm{int}}}$, is intimately related to the Noether energy-momentum
operator, as these operators are expressed through the conserved Noether energy-momentum components or, respectively,
through their interacting counterparts. Moreover it seems important to emphasize that (\ref{specral-space-time-free-fields-EU})
is a reformulation of the Quantization Postulate for free fields (compare \cite{Bogoliubov_Shirkov}, \S 9.4), 
without any extra arbitrary assumptions. The spectral tuple (\ref{specral-space-time-interacting-fields-EU}) is nothing more but 
the standard perturbation of (\ref{specral-space-time-free-fields-EU}) and can be look upon as an extension of the Quantization
Postulate over the interacting fields defined perturbatively.

\section{APPENDIX: On the spectral characterization of non compact manifolds}\label{AppendixNonCompMani}

In this Appendix consisting of four Subsections we give a spectral characterization of paracompact open non compact oriented complete 
Riemannian manifolds $(\mathcal{M}, g)$. We gradually -- in three steps -- reduce the task to the problem of spectral 
characterization of compact manifolds, as resolved in \cite{Connes_spectral}. As is well known on every orientable 
open non compact (paracompact) complete Riemannian manifold there exists the natural self-adjoint 
Dirac operator $D$,
in the Hilbert space of square integrable spinors $\mathcal{H}$ (or resp. sections of  bundles of de Rham forms) \cite{Roe} 
and the pointwise multiplication representation in $\mathcal{H}$ of an ideal of smooth functions -- a nonunital nuclear algebra 
of operators $\mathcal{A}$ in $\mathcal{H}$ -- such that $(\mathcal{A}, \mathcal{H}, D)$ is a (nonunital) spectral triple \cite{Gay}. 
Recall that it is the pseudo-Riemannian (Lorentzian) globally causal metric which plays the role in QFT, together with corresponding 
Krein fundamental operator $\mathfrak{J}$ in the module of square integrable sections of multi-spinors
or other higher integral or half an odd integral spin sections. But we start here with analysis of the naturally 
associated ordinary Riemannian metric $g_\mathfrak{J}$ and the elliptic Dirac operator 
$D_\mathfrak{J}$. In order to simplify notation we will denote them in this Appendix simply
by $g$ and $D$.  Thus the spectral triple $(\mathcal{A}, \mathcal{H}, D)$ of this Appendix
should be compared to the spectral triple which
in the Introduction and in Subsection \ref{DirectIntRepVF}
was denoted by $(\mathcal{A}, \mathcal{H}_{\textrm{inv}}, D_\mathfrak{J})$.

The algebra $\mathcal{A}$ 
is canonically a nuclear algebra $K\{ M_p \}$  of functions on $\mathcal{M}$ of Gelfand and Shilov \cite{GelfandII} with the 
functions $M_p $ on $\mathcal{M}$ defined by a Morse function associated to a non-focal point and a natural Nash closed and isometric embedding of the manifold 
$\mathcal{M}$ into the Euclidean space $\mathbb{R}^L$ with appropriately large $L$,
as we have already mentioned at the end of Subsection \ref{DirectIntRepVF}, compare also discussion of this Appendix. 
But likewise the nuclear algebra $\mathcal{A} = K\{ M_p \}$ can be naturally constructed as a
standard nuclear countably Hilbert space $\mathcal{A} = \mathcal{S}_{A'}(\mathcal{M}; \mathbb{C})$ 
(or counting with multiplicity $\pi(\mathcal{A}) = \mu_\pi \, \mathcal{S}_{A'}(\mathcal{M}; \mathbb{C})$ 
with the operator in $\pi(\mathcal{A})$, defined  
on $L^2(\mathcal{M}; \mathbb{C}^{\mu_\pi}) = \oplus L^2(\mathcal{M}; \mathbb{C})
= L^2(\mathcal{M} \sqcup \ldots \mathcal{M}; \mathbb{C})$
as operator of multiplication by a function on $\mathcal{M} \sqcup \ldots \mathcal{M}$ (disjoint sum of $\mu_\pi$ copies
of $\mathcal{M}$)
 identified with that function (identical on each copy),
in the notation of Obata \cite{obata-book}, \cite{hida} used in Subsection \ref{white-setup},
defined by a standard operator $A'$ on $L^2(\mathcal{M}; \mathbb{C})$, such that
\[
D^2 + V_A = \oplus \Delta + V_A = \oplus A'
\]
on
\[
L^2(\mathcal{M}; \mathbb{C}^{\mu_\pi}) = \oplus L^2(\mathcal{M}; \mathbb{C}) =  L^2(\mathcal{M} \sqcup \ldots \mathcal{M}; \mathbb{C}).
\]
Here $\Delta$ is the Laplace operator on $\mathcal{M}$ corresponding to the Riemannan metric $g$, and
also the volume form in $L^2(\mathcal{M})$ is that determined by the Riemannian metric $g$, finally 
$V_A$ is the operator of multiplication by the function $\big(M_1\big)^2$ on $\mathcal{M}$, 
where $M_1$ is the first of the sequence of functions  $M_1 \in \{M_1, M_2, \ldots\}$ defining
$\mathcal{A} = K\{ M_p \}$, 
again determined by the 
Nash isometric closed embedding of $\mathcal{M}$ and (associated to this embedding and to a non-focal point
$x_0$, compare discussion below) Morse function, as was explained at the end of Subsection \ref{DirectIntRepVF}.
Namely for $x \in \mathcal{M}$, $M_1(x)$ is the Euclidean distance of $x$ to a fixed non-focal
$x_0 \in \mathbb{R}^L$ (where $\mathcal{M}$ is identified here with its closed Nash embedding into the 
Euclidean space $\mathbb{R}^L$). Here we have direct sum of ${\mu_\pi}$ copies of the indicated operators
with ${\mu_\pi}$ equal to the uniform multiplicity of the action $\pi$ of the algebra
$\mathcal{A}$ equal to ${\mu_\pi}$-fold uniform copy of $\mathcal{A}$ acting by pointwise multiplication on 
$L^2(\mathcal{M}; \mathbb{C})$.

The standard operator 
\[
A = D^2 + V_A = \oplus \Delta + V_A = \oplus A'
\]
defines the nuclear test space
\[
\bigoplus \limits_{1}^{d} \mathcal{A} = \oplus \mathcal{S}_{A'}(\mathcal{M}; \mathbb{C}) = 
\mathcal{S}_{\oplus A'}(\mathcal{M}; \mathbb{C}^d) = 
\mathcal{S}_{A}(\mathcal{M}; \mathbb{C}^d), \,\,\,\,
d = \mu_\pi,
\]
which is important for QFT on $\mathcal{M}$ whenever a globally causal Lorentz metric exists on 
$\mathcal{M}$ and $\mathcal{M}$ serves as a space-time.
As we have already seen, it is important for the white
noise construction of free and interacting quantum fields on $\mathcal{M}$ to have 
$\oplus \mathcal{A}$ in the standard form
\[
\oplus \mathcal{A} = \mathcal{S}_{A}(\mathcal{M}; \mathbb{C}^d),
\]
where we have direct sum of $d= \mu_{\pi}$ copies of $\mathcal{A}$. 
We could also write
\[
A = D^2 + V_A = \oplus \Delta + \oplus V'_{A} = \oplus A'
\]
with
\[
\oplus V'_{A} = V_A, \,\,\,\,\,  A' = \Delta + V'_{A} 
\]
and with the operator $V'_{A}$ of pointwise  multiplication by $\big(M_1\big)^2$, 
regarded as the operator on $L^2(\mathcal{M}; \mathbb{C})$.
 
We start with the simplest possible situation (Subsection \ref{AppRn}) 
of the standard euclidean manifold $\mathbb{R}^n$ whose canonical (stereographic projection)
leads to the close interplay of the standard manifold and metric structure of $\mathbb{R}^n$ with the standard
manifold and metric structure of the $n$-sphere 
$\mathbb{S}^n$. We observe that this interplay has general properties which are common for any pairs of riemannian manifolds
$(\mathcal{M}, \mathcal{M}_+)$ of the following type: an open complete noncompact Riemannian manifold 
$(\mathcal{M}, g)$ is embedded conformally as open dense submanifold
into a compact Riemannian manifold $(\mathcal{M}_+, g_+)$. In this case we can always
construct a ``scaling operator'' $Q$ and a ''binding potential operator'' $V$ affiliated with the algebra of operators
$\mathcal{A}''$, where $\mathcal{A}''$ is the  double commutant (or weak closure) of $\mathcal{A}$, 
such that $QD + V = U^{-1}D_+U$ is unitary equivalent to the Dirac operator
$D_+$ of the compact manifold $(\mathcal{M}_+, g_+)$, and such that
$\mathcal{A} = U^{-1} \mathcal{A}_{0}^{+} U$
is unitary equivalent to an essential ideal $\mathcal{A}_{0}^{+}$ of the representation of the algebra 
$\mathcal{A}^+$ of the spectral triple $(\mathcal{A}^+, \mathcal{H}^+, D_+)$ characterizing spectrally
the compact manifold $(\mathcal{M}_+, g_+)$. The ``scaling'' operator $Q$ being canonically determined 
by the conformal factor and the ``binding potential'' operator $V$ control the behavior of the regular 
functions of $\mathcal{A}_{0}^{+}$  at infinity and are both uniquely determined by the
conformal embedding $\mathcal{M} \rightarrow \mathcal{M}_+$.
We can thus reduce the spectral characterization of the standard Riemannian manifold $\mathbb{R}^n$
to the compact case by the canonical dense open conformal embedding $\mathbb{R}^n \rightarrow \mathbb{S}^n$ 
(stereographic projection in this particular case). We should emphasize here that the algebra
\[
\oplus \mathcal{A} = \mathcal{S}_{A}(\mathcal{M}; \mathbb{C}^d)
\]
being standard countably Hilbert and nuclear turns to be fully characterized spectrally
by the initial Dirac operator $D$, by the ``scaling''  operator $Q$, by the ``binding potential operator
$V$ and by the algebra of operators $\mathcal{A}^+$. Indeed it is the unique standard
countably Hilbert space $\mathcal{S}_{A}(\mathcal{M})$ of functions, determined by the standard 
(respecting the (A1), (A2), (A3)
conditions of Subsection \ref{white-setup}) operator $A$, and uniquely determined by
\[
D^2 + V_A = A = \oplus A'
\]
where the number of copies of $A'$ is equal to the uniform multiplicity of the algebra
$\big(\mathcal{A}^+\big)''$ and where $V_A$ is affiliated to $\mathcal{A}^+$ and such that
the operator $A$ in the last formula is indeed standard, \emph{i. e.} respecting the conditions
(A1), (A2) and (A3) of Subsect. \ref{white-setup}.
   
Next we observe (Subsect. \ref{AppRn}) that the method of reduction of spectral characterization of the standard $\mathbb{R}^n$ to the 
compact case $\mathbb{S}^n$ can be applied to any
open non compact geodesically complete manifold $(\mathcal{M}, g)$ provided $\mathcal{M}$ is diffeomorphic
to the interior of a compact manifold $W$ with boundary $\partial W$, and provided there exists a smooth Riemannian 
metric $h$ on the whole $\mathcal{W}$ (including boundary $\partial \mathcal{W}$) which is conformally equivalent to $g$
on $\intt \mathcal{W}$. In this case we can construct a diffeomorphic copy $\mathcal{W}'$ of $\mathcal{W}$ with a complete metric $g'$ on $\mathcal{W}'$  and glue along the diffeomorphic common boundary $\partial \mathcal{W}$ of 
$\mathcal{W}$ and $\mathcal{W}'$ obtaining a compact manifold $\mathcal{W} \bigcup_{\partial \mathcal{W}} \mathcal{W}'$ with a metric  $\widetilde{g}$ plying the role of the standard
$n$-sphere $\mathbb{S}^n$ with the metric $\widetilde{g}$ such that 
$\big( \mathcal{W} \bigcup_{\partial \mathcal{W}} \mathcal{W}' - \partial \mathcal{W}, 
\widetilde{g}|_{{}_{\mathcal{W} \bigcup_{\partial W} W' - \partial W}} \big)$ 
is conformally equivalent to  $(\intt \mathcal{W} \bigcup \intt \mathcal{W}', g \sqcup g')$ (plying the role of $\mathbb{R}^n$). We can therefore construct the ``scaling'' operator corresponding to the conformal factor and the ``binding potential'' 
operator exactly as in the preceding Subsection \ref{AppRn} for $\mathbb{R}^n$ embedded conformally in $\mathbb{S}^n$. 

Finally (Subsection \ref{AppOpenNonCompactComplete}) we reduce the general case of (paracompact) open non compact
complete Riemannian manifold $(\mathcal{M}, g)$ to the case described in Subsection \ref{AppRn} by decomposing $\mathcal{M}$
into closed compact submanifolds 
$\mathcal{W}_i$, $i \in \mathbb{Z}$  with compact boundaries $\partial \mathcal{W}_i$ 
with $\intt \mathcal{W}_i \cap \intt \mathcal{W}_j = \emptyset$, $i \neq j$ and with $\partial \mathcal{W}_i 
= \partial_i \mathcal{W}_i \bigsqcup \partial_{i+1} \mathcal{W}_{i+1}$, where $\partial_i \mathcal{W}_i 
= \mathcal{W}_{i-1} \bigcap \mathcal{W}_i$. We achieve this decomposition using a nondegenerate Morse function $f$ on $(\mathcal{M}, g)$. We construct the nondegenerate function $f$ exactly as Morse replacing the Whitney embedding by the closed version of isometric Nash embedding 
of a complete Riemannian manifold $(\mathcal{M}, g)$ into the Euclidean manifold $\mathbb{R}^L$ with sufficiently
large $L$. For each $i \in \mathbb{Z}$ we define a diffeomorphic copy $\mathcal{W}'_i$ of 
$\mathcal{W}_i$ and the same Morse function $f$ will serve to construct complete Riemannian manifold
$(\intt \mathcal{W}_i \bigsqcup \intt \mathcal{W}'_i, g_i \sqcup g'_i$ conformally equivalent to the open submanifold 
$(\mathcal{W}_i \bigcup_{\partial \mathcal{W}_i} \mathcal{W}'_i- \partial \mathcal{W}_i, \widetilde{g}_i )$ of a compact manifold
$(\mathcal{W}_i \bigcup_{\partial \mathcal{W}_i} \mathcal{W}'_i, \widetilde{g}_i)$ with 
$\widetilde{g}_i|_{\mathcal{W}_i} = g|_{\mathcal{W}_i}$, thus obtaining for each triple of submanifolds
$\mathcal{W}_i , \mathcal{W}'_i, \mathcal{W}_i \bigcup_{\partial \mathcal{W}_i} \mathcal{W}'_i$  situation exactly the same as that for $\mathcal{W}, \mathcal{W}', \mathcal{W} \bigcup_{\partial \mathcal{W}} \mathcal{W}'$
in the Subsection \ref{App-conformal}.

Coming back to the globally causal Lorentzian structure we add  Subsection \ref{App-lorentz} where 
we give a short remark on the situations in which the presented above conformal compactifications of the ordinary Riemannian manifolds 
$(\mathcal{M}, g_\mathfrak{J})$ arise from the causal compactifications 
$\overline{\mathcal{M}}^{\, c}$ of the corresponding  Lorentzian manifolds $\mathcal{M}, g$, 
where the compact comapctification  $\overline{\mathcal{M}}^{\, c}$ is periodically embedded 
into a Lorentzian globally causal manifold $\widetilde{\mathcal{M}}, \widetilde{g}$, with the period
being generated by a diffeomorphism preserving the pseudo-Riemannian (lorentz) $\widetilde{g}$  and the corresponding
Riemannian metric  $\widetilde{g}_{\mathfrak{J}}$ on $\widetilde{\mathcal{M}}$.

\subsection{APPENDIX: standard $\mathbb{R}^n$ with its natural 
compactification -- the standard $\mathbb{S}^n$}\label{AppRn}

For the sake of simplicity we restrict attention to the case of $\dim = n =2$ in all computations
of this Subsection,
although all the formulas and operators have their immediate counterparts in higher dimensions.

We consider the unit $2$-sphere as isometrically embedded submanifold in $\mathbb{R}^3$
of all those points $(X,Y,Z)$ for which $X^2 + Y^2 + (Z -1)^2 = 1$, i.e unit sphere with the center
$(0,0,1)$, and denote it by $\mathbb{S}^2((0,0,1), 1)$, and let $(0,0,2)$ be ``the point at infinity $\infty$''. Then we consider the sphere $\mathbb{S}^2((0,0,1/2), 1/2)$
of radius $1/2$ centered at $(0,0,1/2)$, and the embedding $s^+:$ 
$\mathbb{R}^2 \xrightarrow{s} \mathbb{S}^2((0,0,1), 1) - \{ \infty \}$ being given by the composition
\[
\mathbb{R}^2 \xrightarrow{\textrm{stereographic projection}} \mathbb{S}^2((0,0,1/2), 1/2) 
\xrightarrow{\textrm{isotropic scaling}}
\mathbb{S}^2((0,0,1), 1):
\]
\[
(x,y) \xrightarrow{s^+} (X(x,y), Y(x,y), Z(x,y)) = (2x q^{-1}, 2y q^{-1}, 2 - 2q^{-1}),
\]
(where $q(x,y) = 1 + x^2 + y^2$) with the first map being the inverse of the stereographic 
projection $\mathbb{S}^2((0,0,1/2), 1/2) - \{(0,0,1/2)\} \rightarrow \mathbb{R}^2$ 
from the ``North Pole'' $(0,0,1)$ of the sphere $\mathbb{S}^2((0,0,1/2),1/2)$ on the plane tangent to 
the sphere $\mathbb{S}^2((0,0,1/2),1/2)$ at the ``South Pole'' $(0,0,0)$, and the second map is the isotropic 
scaling with factor $2$: $(X,Y,Z) \mapsto (2X, 2Y, 2Z)$.
The conformal embedding (projection from the ``North Pole'') $\mathbb{R}^2 \xrightarrow{s^+} \mathbb{S}^2 - \{ \infty \}$
generates two metrics on $\mathbb{R}^2$ (regarded as the manifold with the standard manifold structure
in case $\mathbb{R}^4$ when $\dim = 4$).
Namely, the standard Euclidean metric
\[
g_{{}_{\mathbb{R}^2}} = dz \otimes \overline{dz} = dx \otimes dx + dy \otimes dy,
\]
coming from the euclidean structure and giving the standard open noncompact complete Riemannian manifold
$(\mathbb{R}^2, g_{{}_{\mathbb{R}^2}})$; and the one induced from (the standard in case $\mathbb{S}^n$ for $n \geq 4$) 
$\mathbb{S}^2$ by the open dense 
conformal embedding $\mathbb{R}^2 \xrightarrow{s^+} \mathbb{S}^2$:
\[
g_{{}_{\mathbb{S}^2}} = d \theta^2 + \sin^2 \theta \, d \phi^2 
=  4 q(z)^{-2} dz \otimes \overline{dz} = 4 q(x,y)^{-2} g_{{}_{\mathbb{R}^2}},
\]
where $q(z) = 1 + z \overline{z} = 1 + x^2 + y^2 = q (x,y)$.
The two metrics give rise to the two versions of each structure induced naturally by the metric:
the two volume forms
\[
{\dvol}_{{}_{\mathbb{R}^2}} = \frac{i}{2}dz \wedge \overline{dz} = dx \wedge dy, \,\,\,\,\,
{\dvol}_{{}_{\mathbb{S}^2}} = 2i q^{-2} \, dz \wedge \overline{dz} = 4q^{-2} \, dx \wedge dy;
\] 
and the two Hilbert spaces of square integrable spinors
\[
L^2 (\mathbb{R}^2, S; {\dvol}_{{}_{\mathbb{R}^2}}) \,\,\, \textrm{and} \,\,\,
L^2 (\mathbb{S}^2, S; {\dvol}_{{}_{\mathbb{S}^2}}),
\]
on $\mathbb{R}^2$ with the inner products equal
\[
(\psi, \psi)_{\mathbb{R}^2} = \int \limits_{\mathbb{R}^2} \, (|\psi_1|^2 + |\psi_2|^2) \, dx \wedge dy,
\]
\[
(\phi, \phi)_{\mathbb{S}^2} = \int \limits_{\mathbb{R}^2} \, (|\phi_1|^2 + |\phi_2|^2) \, 4q^{-2} dx \wedge dy,
\]
respecively for $ \psi \in L^2 (\mathbb{R}^2, S; {\dvol}_{{}_{\mathbb{R}^2}})$ and 
$\phi \in L^2 (\mathbb{S}^2, S; {\dvol}_{{}_{\mathbb{S}^2}})$
and the two corresponding Dirac operators
\[
D_{\mathbb{R}^2} = \gamma^1 (-i \partial_x) + \gamma^2 (-i \partial_y) \,\,\, \textrm{and} \,\,\,
D_{\mathbb{S}^2} = - i \gamma(dx_j) \nabla^{\mathbb{S}^2}_{\partial_j} 
= \gamma^1 (-i q \partial_x +ix) + \gamma^2 (-i q \partial_y +iy) 
\]
acting respectively in $L^2 (\mathbb{R}^2, S; {\dvol}_{{}_{\mathbb{R}^2}})$ and 
$L^2 (\mathbb{S}^2, S; {\dvol}_{{}_{\mathbb{S}^2}})$,
where $\gamma^1 = \sigma^1$, $\gamma^2 = \sigma^2$, with $\sigma^i$, $i = 1,2$ being the Pauli matrices. In fact
$D_{\mathbb{S}^2}$ is nothing but the ordinary Dirac operator $\slashed{D}_{\mathbb{S}^2}$ on $\mathbb{S}^2$ in the coordinate chart given 
by the projection $s^+$ from the ``North Pole''.

The conformal embedding $\mathbb{R}^2 \xrightarrow{s^+} \mathbb{S}^2$ induces a unitary
map 
\[
L^2 (\mathbb{R}^2, S; {\dvol}_{{}_{\mathbb{R}^2}}) \xrightarrow{U}
L^2 (\mathbb{S}^2, S; {\dvol}_{{}_{\mathbb{S}^2}}),
\]
given by the formula
\[
U\psi(x,y) = \frac{1}{2} q(x,y) \psi(x,y)
\]
with the unitary inverse
\[
U^{-1} \phi (x,y) = 2 q^{-1}(x,y) \phi(x,y),
\]
where
\[
\frac{1}{2} q
\]
may be though of as a square root of the Radon-Nikodym derivative
\[
\frac{{\dvol}_{{}_{\mathbb{R}^2}}}{{\dvol}_{{}_{\mathbb{S}^2}}}.
\]

\begin{defin*}
Let us define the ``scaling'' operator $Q$ of pointwise multiplication by the number
\[
q(x,y)
\]
at the point $(x,y)$, and the operator $V$ of pointwise multiplication by
the matrix
\[
\gamma^1 V_1(x,y) + \gamma^2 V_2(x,y) \,\,\,\,\,\,\, (V_i \in C^\infty (\mathbb{R}^2))
\]
at the point $(x,y)$ in the Hilbert space $L^2 (\mathbb{R}^2, S; \dvol_{\mathbb{R}^2})$. The operators 
$Q$ and $V$ are by construction self adjoint and are affiliated with the double commutant (weak closure) of the
pointwise multiplication representation $\pi_{\mathbb{R}^2}$ 
of the nuclear algebra of Schwarz functions $\mathcal{S}(\mathbb{R}^2)$ in $L^2 (\mathbb{R}^2, S; \dvol_{\mathbb{R}^2})$.
Moreover, the core $\mathcal{S}(\mathbb{R}^2) \oplus \mathcal{S}(\mathbb{R}^2) \subset 
L^2 (\mathbb{R}^2, S; \dvol_{{}_{\mathbb{R}^2}})$ of $Q$ is contained in the core $\bigcap_k \Dom (D_{\mathbb{R}^2})^k$ of $D_{\mathbb{R}^2}$. 
If in addition the core of $V$ is contained in $\bigcap_k \Dom (D_{\mathbb{R}^2})^k$
then we call $V$ the ``binding potential'' operator. 

\end{defin*}

We have the following simple

\begin{lem*}
There exists the binding potential operator $V =  x \, \gamma^1 + y \, \gamma^2$ on 
$L^2 (\mathbb{R}^2, S; \dvol_{{}_{\mathbb{R}^2}})$ such that 
\[
U(QD_{\mathbb{R}^2} + V)U^{-1} = D_{\mathbb{S}^2}.
\]
\end{lem*}
\qed

Now consider the ordinary spectral triple $(C^\infty(\mathbb{S}^2), \mathcal{H}_{\mathbb{S}^2}, \slashed{D}_{\mathbb{S}^2})$
of $\mathbb{S}^2$, where $\mathcal{H}_{\mathbb{S}^2}$ is the Hilbert space of square integrable sections of the 
spinor bundle over $\mathbb{S}^2$, with the representation $\pi$ of the algebra $C^\infty (\mathbb{S}^2)$
in $\mathcal{H}_{\mathbb{S}^2}$ given by the ordinary pointwise multiplication.  
Because the image of the conformal embedding $s^+$ is equal to the whole $\mathbb{S}^2$
except a set of measure zero -- the one point ``$\infty$'' -- then the unitary map $U$ can be regarded as a unitary map between the Hilbert spaces of square integrable spinors on $\mathbb{R}^2$ and $\mathbb{S}^2$ respectively.
Every element $\varphi \in C^\infty (\mathbb{S}^2)$ can be represented as the restriction 
of a smooth function $f \in C^\infty (\mathbb{R}^3)$ to $\mathbb{S}^2$ regarded as isometrically 
embedded in $\mathbb{R}^3$. Moreover, every such 
$\varphi \in C^\infty (\mathbb{S}^2)$ can be uniquely represented by a smooth function 
$\varphi_{\mathbb{S}^2} = f \circ s^+$ on 
$\mathbb{R}^2$ (with all its derivatives of all orders greater than zero vanishing at infinity).
Let us denote the algebra of all smooth functions 
$\varphi_{\mathbb{S}^2}, \varphi \in C^\infty(\mathbb{S}^2)$ on $\mathbb{R}^2$ by $\mathcal{A}^+$. Similarly every (square integrable) section 
of  $\mathcal{H}_{\mathbb{S}^2}$ can be naturally identified with a unique element of 
$L^2 (\mathbb{S}^2, S; \dvol_{{}_{\mathbb{S}^2}})$, 
and the representation $\pi$ of pointwise multiplication by $\varphi \in  C^\infty (\mathbb{S}^2)$ in  $\mathcal{H}_{\mathbb{S}^2}$ can be identified with the representation $\pi_{\mathbb{S}^2}$ of pointwise multiplication by 
$\varphi_{\mathbb{S}^2}$ in $L^2 (\mathbb{S}^2, S; \dvol_{{}_{\mathbb{S}^2}})$, and similarly the action $(\slashed{D}_{\mathbb{S}^2}, \mathcal{H}_{\mathbb{S}^2})$ of the Dirac operator $\slashed{D}_{\mathbb{S}^2}$ in $\mathcal{H}_{\mathbb{S}^2}$
can be identified with the action $(D_{\mathbb{S}^2}, L^2 (\mathbb{S}^2, S; \dvol_{{}_{\mathbb{S}^2}}))$ of $D_{\mathbb{S}^2}$ in $L^2 (\mathbb{S}^2, S; \dvol_{{}_{\mathbb{S}^2}})$ with all interrelations between $\pi$ and $(\slashed{D}_{\mathbb{S}^2}, \mathcal{H}_{\mathbb{S}^2})$ being preserved by $\pi_{\mathbb{S}^2}$ and 
$(D_{\mathbb{S}^2}, L^2 (\mathbb{S}^2, S; \dvol_{{}_{\mathbb{S}^2}}))$. In short the triple 
$(C^\infty(\mathbb{S}^2), \mathcal{H}_{\mathbb{S}^2}, \slashed{D}_{\mathbb{S}^2})$ with the action $\pi$
of $C^\infty (\mathbb{S}^2)$ in $\mathcal{H}_{\mathbb{S}^2}$ can be naturally identified with 
$(\mathcal{A}^+, L^2 (\mathbb{S}^2, S; \dvol_{\mathbb{S}^2}),  D_{\mathbb{S}^2})$ 
with the action of $\mathcal{A}^+$ given by $\pi_{\mathbb{S}^2}$. 

Using the Lemma and the spectral triple of the $2$-sphere
$(\mathcal{A}^+, L^2 (\mathbb{S}^2, S; \dvol_{\mathbb{S}^2}), D_{\mathbb{S}^2})$  with the action of 
$\mathcal{A}^+$ in $L^2 (\mathbb{S}^2, S; \dvol_{{}_{\mathbb{S}^2}})$ 
given by $\pi_{\mathbb{S}^2}$,
which respects all conditions of Connes necessary and sufficient for  $\mathcal{A}^+$
to be isomorphic to the algebra of all smooth functions on a compact manifold we obtain the following
\begin{twr*}
For the spectral triple $(\mathcal{A} = \mathcal{S}(\mathbb{R}^2), 
L^2 (\mathbb{R}^2, S; \dvol_{{}_{\mathbb{R}^2}}), D_{\mathbb{R}^2})$ with the action $\pi_{\mathbb{R}^2}$ of $\mathcal{A}$ 
in $L^2 (\mathbb{R}^2, S; \dvol_{{}_{\mathbb{R}^2}})$ given by pointwise multiplication, there exist a self-adjoint ``binding
potential''  operator $V$ and the ``scalling'' self-adjoint operator $Q$ in 
$L^2 (\mathbb{R}^2, S; \dvol_{{}_{\mathbb{R}^2}})$ both affiliated with $\pi_{\mathbb{R}^2}(\mathcal{A})''$ and a commutative
algebra $\mathcal{A}^+$ of operators in $L^2 (\mathbb{R}^2, S; \dvol_{{}_{\mathbb{R}^2}})$ 
containing $\pi_{\mathbb{R}^2}(\mathcal{A})$ as an essential ideal such that
\begin{equation}\label{(A^+,H^+,D_+)sphere}
(\mathcal{A}^+, \mathcal{H}^+ = L^2 (\mathbb{R}^2, S; \dvol_{{}_{\mathbb{R}^2}}), QD_{\mathbb{R}^2}+ V)
\end{equation} 
is a spectral triple fulfilling all conditions of Connes sufficient and neccesary
for $\mathcal{A}^+$ to be identifiable with the algebra of all smooth functions on a compact
manifold.
\end{twr*}
\begin{rem*}
Thus the general strategy is to reconstruct first $\mathcal{A}^+$ as an algebra of smooth
functions, using the triple (\ref{(A^+,H^+,D_+)sphere}).
Recall please that
\[
\mathcal{A} = \mathcal{S}(\mathbb{R}^2) = \mathcal{S}_{A'}(\mathbb{R}^2;\mathbb{C})
\]
for the standard operator
\[
A' = \Delta_{\mathbb{R}^2} + r^2 + \boldsymbol{1} 
= -\partial_{x_1}^{2} -\partial_{x_2}^{2} + \big(x_1\big)^2 + \big(x_2\big)^2
+ \boldsymbol{1} 
\,\,\,\,\,\,\,\,\,
\textrm{on}
\,\,\,
L^2(\mathbb{R}^2;\mathbb{C}).
\]
Having the operators in $\mathcal{A}^+$ identified with
operators of multiplication by ordinary $\mathbb{C}$-valued functions on the Hilbert space
$L^2 (\mathbb{R}^2; S; \dvol_{{}_{\mathbb{R}^2}}) = L^2 (\mathbb{R}^2; \mathbb{C}^2)$ as well as
the Hilbert space $\mathcal{H}^+$ in the form of square integrable (equivalence classes) 
of ordinary functions we can proceed then to reconstruction of $\mathcal{A}$ as the standard 
countably Hilbert nuclear space, using these identifications. 

Recall please that 
\begin{multline*}
f \in S_{A}(\mathbb{R}^2; \mathbb{C}^2) = S_{A'\oplus A'}(\mathbb{R}^2; \mathbb{C}^2) 
= \bigoplus \limits_{1}^{2} \mathcal{S}_{A'}(\mathbb{R}^2; \mathbb{C})
\subset \\ \subset \bigoplus \limits_{1}^{2} L^2(\mathbb{R}^2; \mathbb{C}) = 
L^2(\mathbb{R}^2; \mathbb{C}^2) =
L^2 (\mathbb{R}^2; S; \dvol_{{}_{\mathbb{R}^2}}) = \mathcal{H}^+
\end{multline*}
if and only if 
\[
f \in \bigcap \limits_{k\in \mathbb{N}} \textrm{Dom} \, A^k
\]
where $A= A' \oplus A'$ is the standard operator
\[
A = \oplus A' = \big(D_{\mathbb{R}^2} \big)^2 + V_A =  \oplus \Delta + V_A
\]
on 
\[
\oplus L^2(\mathbb{R}^2;\mathbb{C}) = L^2(\mathbb{R}^2;\mathbb{C}^2) = L^2 (\mathbb{R}^2; S; \dvol_{{}_{\mathbb{R}^2}})  = \mathcal{H}^+.
\]
Here we have direct sum of two copies of the indicated operators because the uniform multiplicity
of $\big( \mathcal{A}^+\big)''$ is equal two. The operator $V_A$ is by construction affiliated
to $\mathcal{A}^+$ and  thus is equal to an operator of multiplication by a function. 
Now using the \emph{continuous version Theorem} of Obata \cite{obata.Cont.Version.Thm}, 
we can choose the ordinary functions (ad unique) representants of the elements 
$f \in S_{A}(\mathbb{R}^2; \mathbb{C}^2)$ to be smooth and with both components of $f$ belonging to $\mathcal{A}^+$.

Now we define $\pi(\mathcal{A}) \subset \mathcal{A}^+$ as the operators of multiplication by
any of the two components of $f \in \mathcal{A}^+$, with $f$ representing any element
of $S_{A}(\mathbb{R}^2; \mathbb{C}^2)$.
 
There are many operators $V_A$ of multiplication by a function, \emph{i. e} affiliated
to $\mathcal{A}^+$, which give standard $A  = \oplus \Delta_{\mathbb{R}^2} + V_A = \oplus A'$ 
with the same\footnote{And the same $\oplus \mathcal{A} = \mathcal{S}(\mathbb{R}^2;\mathbb{C}^2) = \mathcal{S}_A(\mathbb{R}^2; \mathbb{C}) = \mathcal{S}_{\oplus A'}(\mathbb{R}^2; \mathbb{C}^2)$}
$\mathcal{A}  =
\mathcal{S}_{A'}(\mathbb{R}^2;\mathbb{C})$. 
Let us emphasize that because $\Delta_{\mathbb{R}^2}$ is determined by 
$D_{\mathbb{R}^2}$ we can characterize\footnote{And $\mathcal{A} = \mathcal{S}(\mathbb{R}^2) 
= \mathcal{S}_A(\mathbb{R}^2)$.} $\oplus \mathcal{A} =
\mathcal{S}_{A}(\mathbb{R}^2; \mathbb{C}^2)$ and $V_A$ spectrally by the requirement that the operator
\[
A = \big(D_{\mathbb{R}^2}\big)^{2} +  V_A = \oplus \Delta_{\mathbb{R}^2} + V_A = \oplus A'
\]
determining it is standard, \emph{i. e.} respects the conditions (A1), (A2) and (A3) of Subsection \ref{white-setup}.
Here we have the direct sum of two copies of the indicated operators and algebras, because the uniform mltiplicity of $\big(\mathcal{A}^+\big)''$ is two.

Also in order to fix the spectral characterization of $\mathcal{A} = \mathcal{S}(\mathbb{R}^2) 
= \mathcal{S}_{A'}(\mathbb{R}^2)$ (and \emph{a fortiori} of the allowed operators $V_A$) we should fix the asymptotics of the spectra of the allowed operators $A=\oplus A'$
by requiring the spectrum of $A'$ to be the same (counting with multiplicity)
as the spectrum of the hamiltonian operator $H_{(2)}$ of the two dimensional oscillator (compare
Appendix \ref{asymptotics}), and thus that $A'$
and $H_{(2)}$ are unitarily equivalent. In case of the standard manifold $\mathbb{R}^{n}$
(replacing our $\mathbb{R}^2$) we assume that the operator $A'$ in $A = \oplus A'$ is unitarily equivalent
to the hamiltonian operator $H_{(n)}$ of the $n$ dimensional  oscillator (compare Appendix 
\ref{asymptotics}). In general for other manifolds $\mathcal{M} \neq \mathbb{R}^{n}$ the appropriate spectral 
characterization of the operators $A=\oplus A'$ defining 
$\mathcal{A} = \mathcal{S}_{A'}(\mathcal{M}; \mathbb{C})$ (and
$\oplus \mathcal{A} = \mathcal{S}_{A}(\mathcal{M}; \mathbb{C}^d)$) will of course be different. 
They can be read off from the explicit geometric construction of the operators $A= \oplus A'$ and $V_A$ 
which we give in Appendix \ref{AppOpenNonCompactComplete}. The operators $A = \oplus A'$, constructed by geometric means, 
should serve for $\mathcal{M}$ as the reference operator replacing  $H_{(n)}$.
\end{rem*}
\qedsymbol \, 
Indeed if $(\mathcal{A}^+, \mathcal{H}_2, D_2, c_2, \gamma_2)$, with a faithful
representation $\pi_2$ of $\mathcal{A}^+$  in $\mathcal{H}_2$,
is a spectral triple respecting the afore mentioned Connes conditions \cite{Connes_spectral} (we use the notation 
$\mathcal{A}^+, \mathcal{H}_i, D_i, c_i, \gamma_i$, $i= 1,2$, for $\mathcal{A}, \mathcal{H}, D, c, \gamma$ of \cite{Connes_spectral}) 
necessary and sufficent for $\mathcal{A}^+$ to be identifiable with the algebra of all smooth functions on a compact manifold,
then for any unitary operator $U : \mathcal{H}_1 \rightarrow \mathcal{H}_2$ the spectral
triple $(\mathcal{A}^+, \mathcal{H}_1, U^{-1}D_2U, U^{-1}c_2U, U^{-1}\gamma_2 U)$
with the representation $\pi_1 = U^{-1} \pi_2 U$,
is a spectral triple respecting all afore mentioned Connes conditions. It is sufficient
to apply this observation to the spectral triple $(\mathcal{A}^+, L^2 (\mathbb{S}^2, S; \dvol_{{}_{\mathbb{S}^2}}), D_{\mathbb{S}^2})$ 
of the standard $2$-sphere constructed above,  with the action $\pi_{\mathbb{S}^2}$ of $\mathcal{A}^+$ in 
$L^2 (\mathbb{S}^2, S; \dvol_{{}_{\mathbb{S}^2}})$  and with the unitary and ``scalling'' operators $U, Q$ defined 
by the conformal factor and the ``binding potential'' operator $V$ of the Lemma of the first Subsection of this Appendix. 
\qed

\begin{rem*}
\[
[Q D_{\mathbb{R}^2} + V, a] = Q[D_{\mathbb{R}^2}, a] \,\,\, \textrm{for all $a \in \pi_{\mathbb{R}^2}(\mathcal{A})
= \pi_{\mathbb{R}^2}(\mathcal{S}(\mathbb{R}^2))$}
\]
because $Q$ and $V$ are affiliated with $\pi_{\mathbb{R}^2}(\mathcal{A})''$ and in particular
\[
[V, a] = [Q, a] = 0, \,\,\,\,\,\, a \in \pi_{\mathbb{R}^2}(\mathcal{A}).
\]
Therefore the Connes spectral formula
\[
\dist (p,p') = \sup \{ |f(p) - f(p')|; f \in \mathcal{A}, \| [D_{\mathbb{R}^2}, f] \| \leq 1 \}
\]
for the geodesic distance between any two points $p, p' \in \mathbb{R}^2$ determined by the Dirac 
operator $D_{\mathbb{R}^2}$ (and coinciding
with the background euclidean distance on $\mathbb{R}^2$) coincides with the Connes spectral distance formula
determined by the Dirac operator $Q D_{\mathbb{R}^2} + V$ except for the conformal factor $q$. The conformal factor
$q$ in turn may be regarded to be equal twice the square root
\[
\frac{1}{2} q = \sqrt{\frac{\dvol_{\mathbb{R}^2}}{\dvol_{\mathbb{S}^2}}}
\]
of the Radon-Nikodym derivative
\[
\frac{\dvol_{\mathbb{R}^2}}{\dvol_{\mathbb{S}^2}},
\]
and comes from the operator $Q$, and \emph{vice versa} the operator $Q$ is uniquely determined by the
conformal factor equal to the square root of the Radon-Nikodym derivative mentioned to above. This is why we have
called $Q$ the ``scaling operator''.
\end{rem*}

\subsection{APPENDIX: orientable open complete non compact manifolds with compactly conformally 
fittable boundary}\label{App-conformal}

Here we extend the method of the preceding Subsection of the Appendix, and we construct the corresponding ``scaling''
and ``binding potential'' operators $Q$ and $V$ for orientable manifolds behaving sufficiently
``regularly'' at infinity, leaving over the most general case of an open complete Riemannian manifold to the last Subsection of the Appendix.  

\begin{defin*}
We say that (paracompact) open (non-compact) Riemannian manifold $(\mathcal{M}, g)$ has compactly conformally
fittable boundary iff there exists a smooth function $\sigma$ on $\mathcal{M}$ and a compact manifold $\mathcal{W}$
with boundary $\partial \mathcal{W}$ such that
\begin{enumerate}
\item[1)]
$\mathcal{M}$ is diffeomorphic to $\intt \mathcal{W}$, thus there exists an embedding $i:$ 
\[
\mathcal{M} \xrightarrow{i} \intt \mathcal{W} \subset \mathcal{W}
\]
such that the image of $i$ equals $\intt \mathcal{W}$.
\item[2)]
The (metric induced by the) metric
\[
h = e^{-\sigma}g
\]
(through the embedding $i: \mathcal{M} \rightarrow \mathcal{W}$) extends to a smooth Riemannian metric on $\mathcal{W}$.
\end{enumerate}
\end{defin*}

We assume that the open (without boundary) complete orientable Riemannian manifold $(\mathcal{M}, g)$ considered in 
this Subsection of the Appendix has compactly conformally fittable boundary. The point is that in this case
we can use a diffeomorphic copy $\mathcal{W}'$ of $\mathcal{W}$ (with $\mathcal{M}$ diffeomorphic to 
$\intt \mathcal{W}$ and to $\intt \mathcal{W}'$)
and using any diffeomorphism between $\partial \mathcal{W}$ and $\partial \mathcal{W}'$ we can glue 
$\mathcal{W}$ and $\mathcal{W}'$ along the
common boundary $\partial \mathcal{W}$ in order to obtain a compact closed manifold 
$\mathcal{W} \bigcup_{\partial \mathcal{W}} \mathcal{W}'$.
Moreover we can construct a Riemannian metric $g''$ on $\mathcal{W} \bigcup_{\partial \mathcal{W}} \mathcal{W}'$ in such a way that 
$(\mathcal{W} \bigcup_{\partial \mathcal{W}} \mathcal{W}' - \partial \mathcal{W}, 
g''|_{{}_{\mathcal{W} \bigcup_{\partial \mathcal{W}} \mathcal{W}' - \partial \mathcal{W}}})$ is conformally
equivalent to the disjoint sum complete Riemannian manifold $(\intt W \bigsqcup \intt W', g \sqcup g')$ with the metric on 
$\intt \mathcal{W} \cong_{\textrm{diff}} \mathcal{M}$ equal to the original metric $g$ (under the diffeomorphic 
identification of $\mathcal{M}$ with $\intt \mathcal{W}$). In this case we can apply the method of the preceding Subsection
of the Appendix
with the conformal embedding $(\intt \mathcal{W} \bigsqcup \intt \mathcal{W}', g \sqcup g') 
\rightarrow (\mathcal{W} \bigcup_{\partial \mathcal{W}} \mathcal{W}', g'')$
instead of the conformal embedding $\mathbb{R}^2 \xrightarrow{s} \mathbb{S}^2$ of the preceding Subsection.
Using canonical spectral triples $(\mathcal{A}_1 , \mathcal{H}_1 , D_1)$ and $(\mathcal{A}_2 , \mathcal{H}_2 , D_2)$ 
corresponding to the open complete Riemannian manifolds $(\intt \mathcal{W} \cong_{\textrm{diff}} \mathcal{M}, g)$
and $(\intt \mathcal{W}', g')$ we can prove the following
\begin{twr*}
For a spectral triple $(\mathcal{A}_1 , \mathcal{H}_1 , D_1)$ with a faithful representation $\pi_1$ of $\mathcal{A}_1$
in $\mathcal{H}_1$ the algebra $\mathcal{A}_1$ can be idetifiable with an essential ideal of the algebra of all smooth functions on an open (boundary-less) complete orientable manifold $(\mathcal{M}, g)$ 
with compactly conformally fittable boundary  
iff there exist another spectral triple $(\mathcal{A}_2 , \mathcal{H}_2 , D_2)$ with the action $\pi_2$ of $\mathcal{A}_2$
in $\mathcal{H}_2$, and self adjoint ``scaling'' and ``binding potential'' operators $Q_i,V_i$, $i = 1,2$ resp. in 
$\mathcal{H}_i$ affiliated resp. with $\pi_i (\mathcal{A}_i)''$, and a unital commutative algebra 
$\mathcal{A}^+$
of operators in $\mathcal{H}_1 \oplus \mathcal{H}_2$ containing $\pi_1(\mathcal{A}_1) \oplus \pi_2 (\mathcal{A}_2)$
as an essential ideal, such that 
\[
\big( \mathcal{A}^+, \mathcal{H}_1 \oplus \mathcal{H}_2, (Q_1 D_1 + V_1) \oplus (Q_2 D_2 + V_2) \big)
\]
respects all Connes conditions sufficient and necessary for $\mathcal{A}^+$ being identifiable with the
algebra of operators of multiplication by all smooth functions on a compact manifold, acting with uniform multiplicity $d$ on the Hilbert space of square summable functions on the manifold and with the Hilbert space  $\mathcal{H}_1 \oplus \mathcal{H}_2$ identifiable to the direct sum of $d$ copies of the Hilbert space of square summable functions on the manifold; and thus   
\[
\big( \mathcal{A}^+, \mathcal{H}_1 \oplus \mathcal{H}_2, (Q_1 D_1 + V_1) \oplus (Q_2 D_2 + V_2) \big)
\]
respects conditions necessary and sufficient to
being identifiable with the spectral triple of a compact manifold. 
\end{twr*}

\subsection{APPENDIX: open noncompact complete orientable Riemannian manifold}\label{AppOpenNonCompactComplete}

In this Subsection of the Appendix, concerned with spectral characterization of 
orientable Riemannian geodesically complete manifolds $(\mathcal{M}, g)$ we use the Morse function construction 
and a ``closed version'' of Nash embedding theorem for complete Riemannian manifold in reducing the 
situation to the one described in the preceding Subsection of the Appendix. Thus we construct
the ``scaling'' and ``binding potential'' operators for general geodesically complete Riemannian 
manifold $(\mathcal{M}, g)$. 

In the first step we construct a decomposition of $\mathcal{M}$ into a a countable family of compact
submanifolds $\mathcal{W}_i \subset \mathcal{M}$ with boundaries $\partial \mathcal{W}_i$, $i \in \mathbb{Z}$ 
such that $\mathcal{M} \subset \bigcup \mathcal{W}_i$, $\intt \mathcal{W}_i \bigcap \intt \mathcal{W}_j = \emptyset$,
$i \neq j$, and $\partial \mathcal{W}_i = \partial_i \mathcal{W}_i \bigsqcup \partial_{i +1} (-\mathcal{W}_{i+1})$,
where $\partial_i \mathcal{M}_i = \mathcal{M}_{i-1} \bigcap \mathcal{M}_i$, and such that 
\[
\mathcal{M} = \ldots \bigcup_{\partial_i \mathcal{W}_i} \mathcal{W}_i \bigcup_{\partial_{i+1} \mathcal{M}_{i+1}} 
\mathcal{W}_{i+1} \bigcup_{\partial_{i+2} \mathcal{W}_{i+2}} \ldots \,\,\,\,\, i \in \mathbb{Z}.
\] 
We achieve this decomposition by constructing a non-degenerate Morse function $f$ on $\mathcal{M}$. 
Construction of the Morse function $f$ is exactly the same as the one performed by Morse with 
the only difference that instead of the Whitney embedding we use a closed isometric Nash 
embedding of a complete manifold $\mathcal{M}$ into the Euclidean space of appropriately
high dimension. Let us remind that utilizing this function $f$ as at the end of Subsect. \ref{DirectIntRepVF},
we construct the nuclear algebra $\mathcal{A} = K\{ M_p \} =\mathcal{S}_{A'}(\mathcal{M})$
(and the algebra $\mathcal{A} = \mathcal{S}_{A'}(\mathcal{M}; \mathbb{C})$),
with the requirement that the operator
\[
A = \oplus A' = D^2 + V_A= \oplus \Delta + V_A, \,\,\, V_A \in \big( \mathcal{A}^+\big)''
\]
is standard (and thus characterized spectrally) and where $\mathcal{A}^+$ is also 
characterized spectrally, see discussion below.
Recall, please, that $\Delta$ is the Laplace operator on $\mathcal{M}$ determined by the metric
$g$ and $\oplus \Delta = D^2$. $V_A$ can be chosen to be the operator of multiplication by the
function $\big(M_1\big)^2$, determined by the Nash embedding used below. For $M_p$
we put $f^p$. The function $M_1$ can be chosen to be equal to the Morse function $f$, which we construct below (in THE FIRST STEP). 
The indicated direct sums are equal to $d$ copies of indicated operators acting on 
$L^2(\mathcal{M}; \mathbb{C}^d) = \oplus L^2(\mathcal{M}; \mathbb{C})$, with $d$ equal to the uniform
multiplicity of the algebras $\pi_{\mathcal{M}}(\mathcal{A})''$ and $\big(\mathcal{A}^+\big)''$.

Let us emphasize that the Gelfand-Shilov nuclear space $\mathcal{A} = K\{ M_p \}$ contains essentially
all functions of compact support (provided the support does not contain points of a measure zero set whose complement is open). 
This is of interest for QFT on $\mathcal{M}$ understood as space-time manifold
admitting appropriate lorenz globally causal structure.

In the second step we show that for each Riemannian manifold $(\mathcal{W}_i, g|_{{}_{\mathcal{W}_i}})$
there exists $\mathcal{W}'_i \cong_{\textrm{diff}} \mathcal{W}_i$ and a metric $g'_i$ on $\mathcal{W}'_i$
such that 
\[
(\mathcal{W}_i \bigcup_{\partial \mathcal{W}_i} \mathcal{W}'_i, g_i \cup_{{}_{\partial \mathcal{W}_i}} g'_i = g''_i)
\]
is a smooth Riemannian closed (compact) manifold with $g|_{{}_{\mathcal{W}_i}} = g''_i|_{{}_{\mathcal{W}_i}}$.

In the third step we show that there exist complete metrics $h_i, h'_i$ on $\intt W_i$ and $\intt W'_i$
such that
\[
\big(\mathcal{W}_i \bigcup_{\partial \mathcal{W}_i} \mathcal{W}'_i - \partial_i W, 
{g''_i}|_{{}_{\mathcal{W}_i \bigcup_{\partial \mathcal{W}_i} \mathcal{W}'_i - \partial_i W}} \big)
\]
is conformally equivalent to
\[
\big( \intt W_i \bigsqcup \intt W'_i , h_i \sqcup h'_i = h''_i  \big).
\]
Thus the metric $h''_i$ on the open dense subset 
$\mathcal{W}_i \bigcup_{\partial \mathcal{W}_i} \mathcal{W}'_i - \partial_i \mathcal{W} 
= \intt \mathcal{W}_i \sqcup \intt \mathcal{W}'_i$
of the closed manifold $\mathcal{W}_i \bigcup_{\partial \mathcal{W}_i} \mathcal{W}'_i$ is
conformally equivalent to a metric $g''_i|_{{}_{\mathcal{W}_i \cup_{{}_{\partial \mathcal{W}}}\mathcal{W}'_i - \partial \mathcal{W}_i}}$ 
which has a smooth extension $g''_i$ to the whole compact manifold
\[
\big( \mathcal{W}_i \bigcup_{\partial \mathcal{W}_i} \mathcal{W}'_i, g''_i  \big).
\]
We arrive thus at the situation for 
$\mathcal{W}_i, \mathcal{W}'_i, \mathcal{W}_i \bigcup_{\partial \mathcal{W}_i} \mathcal{W}'_i$ the same as for 
$\mathcal{W}, \mathcal{W}', \mathcal{W} \bigcup_{\partial \mathcal{W}} \mathcal{W}'$ in the preceding Subsection.

\begin{center}
\small THE FIRST STEP
\end{center}

We in order to achieve the decomposition 
\[
\mathcal{M} = \ldots \bigcup_{\partial_i \mathcal{W}_i} \mathcal{W}_i \bigcup_{\partial_{i+1} \mathcal{M}_{i+1}} 
\mathcal{W}_{i+1} \bigcup_{\partial_{i+2} \mathcal{W}_{i+2}} \ldots \,\,\,\,\, i \in \mathbb{Z}.
\] 
we construct a nondegenerate Morse function $f$ on $(\mathcal{M}, g)$.  
Let $\mathcal{M} \rightarrow \mathbb{R}^L$ be the isometric Nash embedding of $(\mathcal{M}, g)$
into the euclidean manifold $\mathbb{R}^L$. Because $(\mathcal{M}, g)$ is geodesically complete
we can improve the embedding in such a way that it will be not only isometric but also
closed \cite{Muller}, i.e. with the the closed image in $\mathbb{R}^L$. Enlarging eventually the 
dimension $L$ of the euclidean space $\mathbb{R}^L$ we can construct a closed isometric
embedding $\mathcal{M} \rightarrow {\mathbb{R}^N}_+$ into the half space 
${\mathbb{R}^N}_+ = \{ (x_1, \ldots x_N); x_N \geq 0 \}$ of the euclidean space $\mathbb{R}^N$. 
Now we choose a point $p_0 \in {\mathbb{R}^N}_- 
= \mathbb{R}^N - {\mathbb{R}^N}_+= \{ (x_1, \ldots x_N); x_N < 0 \}$, which is not a focal
point for the embedded $\mathcal{M}$. Note that the euclidean distance of $p_0$ from the hyperplane
$x_N = 0$ is strictly positive, and thus its distance from the embedded $\mathcal{M}$
is strictly positive. We define the function $f$ on $\mathcal{M} \subset \mathbb{R}^N$
\[
f: \mathcal{M} \ni p \longmapsto \textrm{euclidean distance of $p$ from $p_0$}.
\] 
It is well known that all critical points of $f$ are non-degenerate because $p_0$ is not focal
and that $f > \epsilon$ for some fixed $\epsilon>0$. 
Let
\[
\mathcal{M}^a =^{\textrm{def}} f^{-1}((-\infty, a]) = f^{-1} ([0,a]).
\]
Because $f$ is non-degenerate then for all values $a$, except for at most the denumerable
subset of critical values of $f$, the subset $f^{-1}(\{a\})$ is a submanifold of $\mathcal{M}$ and $\mathcal{M}^a$
is a submanifold of $\mathcal{M}$ with boundary $\partial \mathcal{M}^a = f^{-1}(\{a\})$.
Moreover $\mathcal{M}^a \subset \mathcal{M} \subset \mathbb{R}^N$ is compact 
as the intersection of the closed ball $D^N(a, p_0)$ of radius $a$ centered at $p_0$ (which is compact
in $\mathbb{R}^N$) with the closed subset $\mathcal{M}$ of $\mathbb{R}^N$ (as the embedding 
$\mathcal{M} \rightarrow \mathbb{R}^N$ is closed). 
Suppose $0 < a_1 < a_2 < \ldots$ is an unbounded increasing sequence of non-critical values of $f$ 
(there exists such a sequence because the set of critical values of $f$ is at most denumerable)
Therefore  
\[
\mathcal{W}_i = \mathcal{M}^{a_{i+1}} - \intt \mathcal{M}^{a_i}
= \big(D^N(a_{i+1}, p_0) - \intt D^N(a_i, p_0) \big) \cap \mathcal{M}
\]
is a compact submanifold of $\mathcal{M}$ for all $i \in \mathbb{N}$ with boundaries
\[
\partial \mathcal{W}_i = \partial_i \mathcal{W}_i \sqcup \partial_{i+1} \mathcal{W}_{i+1},
\]
where 
\[
\partial_i \mathcal{W}_i = f^{-1}(\{ a_i \}).
\]
In this way we obtain the desired decomposition
\[
\mathcal{M} = \ldots \bigcup_{\partial_i \mathcal{W}_i} \mathcal{W}_i \bigcup_{\partial_{i+1} \mathcal{M}_{i+1}} 
\mathcal{W}_{i+1} \bigcup_{\partial_{i+2} \mathcal{W}_{i+2}} \ldots \,\,\,\,\, i \in \mathbb{Z}.
\]

\begin{center}
\small THE SECOND STEP
\end{center}

For each submanifold $\mathcal{W}_i \subset \mathcal{M}$ we consider a difeomorphic copy
$\mathcal{W}'_i$ and glue $\mathcal{W}_i$ with $\mathcal{W}'_i$ using a diffeomorphism
$\partial \mathcal{W}_i \rightarrow \partial \mathcal{W}'_i$, obtaining a closed (compact)
manifold $\mathcal{W}_i \bigcup_{\partial \mathcal{W}_i} \mathcal{W}'_i$.

By the collar neighborhood theorem there exists an open set in 
$\mathcal{W}_i \bigcup_{\partial \mathcal{W}_i} \mathcal{W}'_i$
containing $\mathcal{W}_i$ on which a smooth metric is defined coinciding on $\mathcal{W}_i$
with the metric $g$ on $\mathcal{M}$. Using the partition--of--unity--construction
we may extend smoothly this metric obtaining a smooth metric $g''_i$
on the closed manifold $\mathcal{W}_i \bigcup_{\partial \mathcal{W}_i} \mathcal{W}'_i$.

\begin{center}
\small THE THIRD STEP
\end{center}

The non-degenerate Morse function $f$ on $\mathcal{M}$ defines smooth and non-degenerate functions $f_i$
on $\mathcal{W}_i \subset \mathcal{M}$ and, respectively, smooth non-degenerate functions $f'_i$  
on $\mathcal{W}'_i$. We define the metrics 
\[
h_i = \Big(\frac{1}{f_i - a_i} + \frac{1}{a_{i+1} - f_i} \Big) g''_i, \,\,\,
h'_i = \Big(\frac{1}{f'_i - a_i} + \frac{1}{a_{i+1} - f'_i} \Big) g''_i
\]
which are smooth on $\intt \mathcal{W}_i$ and $\intt \mathcal{W}'_i$ respectively and the metric
$h_i \sqcup h'_i$ is smooth on $\mathcal{W}_i \bigcup_{\partial \mathcal{W}_i} \mathcal{W}'_i - \partial \mathcal{W}_i
= \intt \mathcal{W}_i \sqcup \intt \mathcal{W}'_i$ and any smooth curve joining any 
point outside $\partial \mathcal{W}_i$ with any point of $\partial \mathcal{W}_i$ has infinite $h_i \sqcup h'_i$-length 
so that $h_i$ and $h'_i$ are complete on $\intt \mathcal{W}_i$ and resp. $\intt \mathcal{W}'_i$.
We thus arrive at the situation for 
$\mathcal{W}_i, \mathcal{W}'_i, \mathcal{W}_i \bigcup_{\partial \mathcal{W}_i} \mathcal{W}'_i$ the same as for 
$\mathcal{W}, \mathcal{W}', \mathcal{W} \bigcup_{\partial \mathcal{W}} \mathcal{W}'$ in the preceding Subsection
of the Appendix. 

Therefore, using the spectral triples $\mathcal{A}_i, \mathcal{H}_i, D_i$ and 
$\mathcal{A}'_i, \mathcal{H}'_i, D'_i$ for the  oriented complete Riemannian manifolds 
$(\intt \mathcal{W}_i, h_i)$ and  $(\intt \mathcal{W}'_i, h'_i)$ we can prove the following
\begin{twr*}
Let $(\mathcal{A}, \mathcal{H}, D)$ be a spectral triple with an ivolutive faithful representation 
$\pi = (\mathcal{A}, \mathcal{H})$ of an ivolutive (nonunital) algebra $\mathcal{A}$ in a separable Hilbert 
space $\mathcal{H}$. Then $\mathcal{A}$ is identifiable with an essential ideal of the algebra of all smooth functions
on a complete Riemannian manifold iff 
\begin{enumerate}
\item[1)]
There exists a Hilbert space $\mathcal{H}' = \oplus_i \mathcal{H}_i$ and self adjoint operator 
$D' = \oplus_i D'_i$, $D'_i = D'|_{{}_{\mathcal{H}'_i}}$ and an involutive representation $\pi' = (\mathcal{A}', \mathcal{H}) = \oplus_i (\mathcal{A}_i,
\mathcal{H}_i)$ of an involutive (nonunital) algebra $\mathcal{A}' = \oplus_i \mathcal{A}'_i$ such that 
each $(\mathcal{A}'_i, \mathcal{H}'_i, D'_i)$ with the representation $\pi_i$ of $\mathcal{A}'_i$ is a spectral triple;
\item[2)]
$\pi = \oplus_i \pi_i =  \oplus_i (\mathcal{A}_i, \mathcal{H}_i)$, $D = \oplus_i D_i$, $D_i = D|_{{}_{\mathcal{H}_i}}$,
each $(\mathcal{A}_i, \mathcal{H}_i, D_i)$ with the representation $\pi_i$ of $\mathcal{A}_i$ is a spectral
triple;
\item[3)] For every $i, k \in \mathbb{N}$ there exists a unital algebra ${\mathcal{A}_{ik}}^+$ of operators
in 
\[
\mathcal{H}_i \oplus \mathcal{H}_{i+1} \oplus \ldots \oplus \mathcal{H}_{i+k} 
\oplus \mathcal{H}'_{i}  \oplus \mathcal{H}'_{i+1} \oplus \ldots \oplus \mathcal{H}'_{i+k}
\]
containing $(\mathcal{A}_i, \mathcal{H}_i) \oplus \ldots \oplus (\mathcal{A}_{i+k}, \mathcal{H}_{i+k})
\oplus (\mathcal{A}'_i, \mathcal{H}'_i) \oplus \ldots \oplus (\mathcal{A}'_{i+k}, \mathcal{H}'_{i+k})$ as an essential ideal
and selfadjoint ``scaling'' and ``binding potential'' operators $Q_i, V_i$ and $Q'_i, V'_i$ affiliated
respectively with $(\mathcal{A}_i, \mathcal{H}_i)''$ and $(\mathcal{A}'_i, \mathcal{H}'_i)''$ such that  
\begin{multline*}
\Big({\mathcal{A}_{ik}}^+, \,\, \mathcal{H}_i \oplus \ldots \oplus \mathcal{H}_{i+k} \oplus  \mathcal{H}'_i \oplus \ldots \oplus  \mathcal{H}'_{i+k}, \\
(Q_iD_i + V_i) \oplus \ldots \oplus (Q_{i+k}D_{i+k} + V_{i+k}) \oplus (Q'_iD'_i + V'_i)
\oplus \ldots \oplus (Q'_{i+k}D'_{i+k} + V'_{i+k})  \Big)
\end{multline*}
is a spectral triple which respects all conditions of Connes sufficient 
and necessary for ${\mathcal{A}_{ik}}^+$
to be identifiable with the algebra of all smooth functions on a compact manifold and thus being a spectral triple
of a closed compact manifold. 
\end{enumerate}
\end{twr*}

\subsection{APPENDIX: periodic causal embeddings}
\label{App-lorentz}

Of course, we are interesting in construction of the Riemannian compact spectral triples
\begin{equation}\label{compactifiedMinkowski(A,H,D)}
\big(\mathcal{A}, \mathcal{H}_{\textrm{inv}}, QD_\mathfrak{J} +V\big)
\end{equation}
of Subsection \ref{DirectIntRepVF},
as arising from the Noether integrals corresponding to the free quantum fields on a 
compactified Minkowski space-time $\mathcal{M}_0$, say on a causal compactification 
$\overline{\mathcal{M}_0}^{\, c}$
with periodic boundary conditions. Thus wee need a periodic causal inclusion 
of the compactification $\overline{\mathcal{M}_0}^{\, c}$ into a globally causal
Lorentzian manifold $\widetilde{\mathcal{M}}, \widetilde{g}, \widetilde{g}_\mathfrak{J}$, with the period
generated by a map $\zeta$ on $\widetilde{\mathcal{M}}, \widetilde{g}$ with $\zeta$
preserving the lorentz metric $\widetilde{g}$ as well as the corresponding Riemannian metric
$\widetilde{g}_{\mathfrak{J}}$. The free quantum fields on the compactified space-time should be
naturally induced by the embedding and by the fields on the initial space-time manifold
$\mathcal{M}_0$. 

We use the following general principle when constructing free fields on the spacetime  
$\widetilde{\mathcal{M}}, \widetilde{g}$ which causally periodically includes the  causal compactification 
$\overline{\mathcal{M}_0}^{\, c}$ of the
Minkowski space-time $\mathcal{M}_0$: the only allowed plane waves on $\widetilde{\mathcal{M}}, \widetilde{g}$ are equal to the extensions of the plane waves on $\mathcal{M}_0$ regarded as a submanifold
of $\widetilde{\mathcal{M}}, \widetilde{g}$ given by the periodic causal embedding. 

For the physically interesting space-times, such as Minkowski space-time, $\mathcal{M}_0, g_0$, de Sitter space-time, or for the whole family of Robertson-Walker-Lema\^itre space-times, there exists
such periodic causal embedding of their causal compactifications into the Einstein Universe
space-timie, plying the role of $\widetilde{\mathcal{M}}, \widetilde{g}$, compare e.g. \cite{HawkingEllis}.
 
The periodic embedding of causal compactification $\overline{\mathcal{M}_0}^{\, c}$ 
of the Minkowski space-time $\mathcal{M}_0, g_0$ into the Einstein Universe 
is worked out in details, together with the construction of the free fields induced by this embedding,
in the series of works \cite{PaneitzSegalI}\cite{PaneitzSegalIII}. In Subsection \ref{GeneralizedSchrodinger-VonNeumannPairs} 
we  analyze in details the  relationship of the Noether integral generators of the free quantum fields
on the Einstein Universe to the space-time spectral tuples 
\[
\big(\mathcal{A}, \mathcal{H}_{\textrm{inv}}, D_\mathfrak{J}, D\big)
\]
on the invariant subspaces $\mathcal{H}_{\textrm{inv}}$ of these Noether generators. 
Using the unitary  map $U$ of \cite{PaneitzSegalI}\cite{PaneitzSegalIII} relating the 
free plane waves (serving as single particle states, and $\Gamma(U)$ relating the corresponding fields)
with the corresponding on the Einstein Universe, which are periodic and live essentially 
on the compactified Einstein Universe $\widetilde{\mathbb{S}^1} \times \mathbb{S}^3$, we can regard
the $\mathcal{H}_{\textrm{inv}}$ as a subspace in the Fock space of free fields on the Minkowski space-time
and the Noether integral operators of fields on the Einstein Universe we can regard as operators 
on the Fock space of free fields on the Minkowski space-time. We can thus compute the last
spectral tuple (which can be regarded as that determining the compact 
spectral tuple (\ref{compactifiedMinkowski(A,H,D)}) in terms of operators and operator fields
of the free fields living on the Minkowski space-time.

\section{APPENDIX: Comparison of the asymptotics of the spectra of $A^{(n)}$ and $H_{(n)}$}\label{asymptotics}

In this Subsection we investigate the spectra of the operators which are equal to $A^{(n_{{}_{0}})}$ and $H_{(n_{{}_{0}})}$
respectively modulo irrelevant additive constant (in order to simplify notation and keep closer to the existing conventions). 
Namely, in this Subsection we define
\[
\begin{split}
H_{(1)} = - \frac{d^2}{dp^2} + p^2,\\
H_{(n_{{}_{0}})} = - \Delta_{\mathbb{R}^{n_{{}_{0}}}} + r^2 = - \frac{\partial^2}{{\partial p_1}^2} \ldots 
- \frac{\partial^2}{{\partial p_{n_{{}_{0}}}}^2} +(p_1)^2 + \ldots (p_{n_{{}_{0}}})^2 = d\Gamma_{n_{{}_{0}}}(H_{(1)}), \\
\end{split}
\] 

We have the following known facts
\begin{enumerate}
\item[Fact.1]
\[
\{ \lambda = l(l+n_{{}_{0}}-2), l= 0,1,2, \ldots\} = \Sp \Delta_{\mathbb{S}^{n_{{}_{0}}-1}}
\]
with the multiplicity of each $\lambda = l(l+n_{{}_{0}}-2)$ equal to
\[
{l+n_{{}_{0}}-1 \choose n_{{}_{0}}-1} 
- {l+n_{{}_{0}}-3 \choose n_{{}_{0}}-1},
\]
compare e.g. \cite{Shubin}, Ch. III. \S 22.
\item[Fact.2]
\[
\{ \lambda = 2(n_1 + \ldots + n_{n_{{}_{0}}}) + n_{{}_{0}}, n_1, \ldots n_{n_{{}_{0}}} \in \mathbb{N} \cup\{0\} \} = \Sp H_{(n_{{}_{0}})}
\]
with the multiplicity of each $\lambda = 2(n_1 + \ldots + n_{n_{{}_{0}}}) + n_{{}_{0}}$ equal to the number of ordered partitions
of 
\[
\frac{\lambda - n_{{}_{0}}}{2} = n_1 + \ldots + n_{n_{{}_{0}}} 
\]
into a sum of $n_{{}_{0}}$ non-negative integers $n_1, \ldots n_{n_{{}_{0}}}$.
\item[Fact.3]
The number of ordered partitions
of $k = n_1 + \ldots + n_{n_{{}_{0}}}$ into a sum of $n_{{}_{0}}$ non-negative integers $n_1, \ldots n_{n_{{}_{0}}}$
is equal to 
\[
{n_{{}_{0}} + k -1\choose n_{{}_{0}} -1}.
\]
\end{enumerate}

Joining these Facts together we obtain after not very complicated analysis the following

\begin{lem*}
${}$
\begin{enumerate}
\item[(Sp.1)]
For odd dimension $n_{{}_{0}}$: $\Sp H_{(n_{{}_{0}})}^{2} \subset \Sp (A^{(n_{{}_{0}})} + n_{{}_{0}} -1)$ with the multiplicity of each 
$\lambda \in \Sp H_{(n_{{}_{0}})}^{2}$
less then the multiplicity of that $\lambda \in \Sp (A^{(n_{{}_{0}})} + n_{{}_{0}} -1)$; 
and $\Sp (A^{(n_{{}_{0}})} + n_{{}_{0}} -1) \subset \Sp H_{(n_{{}_{0}})}$ with the multiplicity of 
each $\lambda \in \Sp (A^{(n_{{}_{0}})} + n_{{}_{0}} -1)$
less than the multiplicity of that $\lambda \in \Sp H_{(n_{{}_{0}})}$.
\item[(Sp.2)]
For even dimension $n_{{}_{0}}$: $\Sp H_{(n_{{}_{0}})}^{2} \subset \Sp (A^{(n_{{}_{0}})} + n_{{}_{0}} -1)$ with the multiplicity of each $\lambda \in \Sp H_{(n_{{}_{0}})}^{2}$
less then the multiplicity of that $\lambda \in \Sp (A^{(n_{{}_{0}})} + n_{{}_{0}} -1)$;
and $\Sp 2(A^{(n_{{}_{0}})} + n_{{}_{0}} -1) \subset \Sp H_{(n_{{}_{0}})}$ with the multiplicity of each 
$\lambda \in \Sp 2(A^{(n_{{}_{0}})} + n_{{}_{0}} -1)$
less than the multiplicity of that $\lambda \in \Sp H_{(n_{{}_{0}})}$; 
\end{enumerate}
where the inequalities for multiplicites hold true asymptotically, i.e. for all eigenvalues $\lambda$ greather than
a fixed constant depending only on the dimension $n_{{}_{0}}$.
\end{lem*}

From this Lemma we obtain the following 
\begin{cor}\label{AsymptoticsA^(n0)=AsymptoticsH_(n0)}
If $\{\lambda_{m}^{0}\}_{m \in \mathbb{N}}
= \Sp H_{(n_{{}_{0}})}$ and  $\{\lambda_{n}\}_{n \in \mathbb{N}}
= \Sp A^{(n_{{}_{0}})} = \Sp A^{(n_{{}_{0}})}$, counted with multiplicities, 
then a sequense $\{C_n\}_{n \in \mathbb{N}}$ of numbers 
fulfills
\[
\sum \limits_{m \in \mathbb{N}} {(\lambda_{m}^{0})}^{N} |C_m|^2 < +\infty, \,\,\, 
N = 2,3, \ldots 
\]

if and only if 
\[
\sum \limits_{m \in \mathbb{N}} {(\lambda_{m})}^{N} |C_m|^2 < +\infty, \,\,\, 
N= 2, 3, \ldots. 
\]
\end{cor}

\vspace*{0.5cm}

\begin{center}
\small EXAMPLE: ASYMPTOTICS OF  $\Sp A^{(3)}$ AND $\Sp H_{(3)}$ 
\end{center}
Let us make a closer look at the three dimensional case. It is easily seen that
\begin{equation}\label{sp.3.1}
\begin{split}
\Sp H_{(3)}^{2} \subset \Sp (A^{(3)} + 2) \,\,\,\textrm{if the multiplicity is ignored}.\\
\Sp (A^{(3)} + 2) = \Sp H_{(3)} \,\,\,\textrm{if the multiplicity is ignored}.
\end{split}
\end{equation}
Moreover, we have in this case ($k$ is any natural number greater than a fixed constant) 
\begin{multline}\label{sp.3.2}
\textrm{multiplicity} \Big[ (2k + 3)^2 \in \Sp H_{(3)}^{2} \Big] = {\textstyle\frac{1}{2}}(k+1)(k+2) \\
< (2k +1)^2 \leq \textrm{multiplicity}\Big[(2k + 3)^2 \in \Sp(A^{(3)} + 2)\Big];
\end{multline}
and 
\begin{multline}\label{sp.3.3}
\textrm{multiplicity}\Big[(2k + 3) \in \Sp(A^{(3)} + 2)\Big] < {\textstyle\frac{1}{2}}(7k+5) \\
< {\textstyle\frac{1}{2}}(k+1)(k+2) = \textrm{multiplicity} \Big[ (2k + 3) \in \Sp H_{(3)} \Big].
\end{multline}

Further we may use the spherical coordinates in which the radial and angular variables may be separated.
Then the complete system of eigenfunctions of the operator $H_{(3)}$ is equal to
\[
e_{n,l}^{m}(t,\theta, \phi) = h_n \otimes Y_{l}^{m}(t,\theta, \phi) = h_n(t)Y_{l}^{m}(\theta, \phi), \,\,\, {-l \leq m \leq l}, n,l = 1,2, \ldots
\]
with the Hermite functions $h_n$ and the spherical functions $Y_{l}^{m}$; and with the corresponding eigenvalues
\[
{\lambda^{0}}_{n,l}^{m} = 4n +2l +3 
\]
composing the $\Sp H_{(3)}$. Then if $U = U_2 U_1$ is the unitary operator constructed in Subsection
\ref{dim=n} then
\[
U e_{n,l}^{m}
\]
gives the complete orthonormal system of $A^{(3)} = U\big(H_{(3)} \otimes \boldsymbol{1} + \boldsymbol{1} \otimes
A^{(3)}\big)U^{-1}$ corresponding to the eigenvalues
\[
\lambda_{n,l}^{m} = 2n + l(l+1) +3, \,\, n,l = 0,1,2, \ldots \,\,\, -l \leq m \leq l.
\]
In this situation, i.e. using (\ref{sp.3.1}) - (\ref{sp.3.3}), one can prove the following special case of the last Corollary 
\begin{cor}
A sequence $\{C_{n,l}^{m}\}_{n,l=0,1,\ldots, \, -l \leq m \leq l}$ of numbers 
fulfills
\[
\sum \limits_{n,l\in \mathbb{N}, -l \leq m \leq l} {({\lambda^{0}}_{n,l}^{m})}^{N} |C_{n,l}^{m}|^2 < +\infty, \,\,\, 
N = 2,3, \ldots 
\]
if and only if 

\[
\sum \limits_{n,l\in \mathbb{N}, -l \leq m \leq l} {(\lambda_{n,l}^{m})}^{N} |C_{n,l}^{m}|^2 < +\infty, \,\,\, 
N= 2, 3, \ldots. 
\]
\end{cor}

The primary use of the Corollaries of this Appendix is their application to the problem of determination
of multipliers of $\mathcal{S}_{A^{(n)}}(\mathbb{R}^n) = \mathcal{S}^0(\mathbb{R}^n)$ (and convolutors of 
$\mathscr{F} \mathcal{S}_{A^{(n)}}(\mathbb{R}^n) 
= \mathcal{S}_{\mathscr{F}A^{(n)}\mathscr{F}^{-1}}(\mathbb{R}^n)
= \mathcal{S}^{00}(\mathbb{R}^n)$). It allows us to reduce this problem to the problem  (solved by L. Schwartz) 
of determination of multipliers and convolutors
of $\mathcal{S}(\mathbb{R}^n) = \mathcal{S}_{H_{(n)}}(\mathbb{R}^n)$.

But we give here a caution joining these Corollaries with the problem of Gelfand's abstract
characterization of the standard countably Hilbert nuclear algebras $\mathcal{S}_{A}(\mathcal{M})$
by standard operators $A$ on $L^2(\mathcal{M})$.
Namely, one could think that the similarity of asymptotics  of two standard operators 
$A_1$ and $A_2$ should guarantee a close relationship of the two associated standard
nuclear algebras $\mathcal{S}_{A_1}(\mathcal{M})$ and $\mathcal{S}_{A_2}(\mathcal{M})$. In particular 
if $\mathcal{S}_{A_1}(\mathcal{M})$ contains essentially all smooth functions of compact
support we expect the same of $\mathcal{S}_{A_2}(\mathcal{M})$. In fact this is true only after
additional requirements which cannot be expressed solely in terms of the spectra, and 
concern in particular commutation relations of the operators, and other requirements which are under the scope of the spectral geometry.
It is true that the algebra $\mathcal{S}^0(\mathbb{R}^n) = \mathcal{S}_{A^{(n)}}(\mathbb{R}^n)$
indeed posses all smooth functions of compact support, provided the set which serves as the
support does not contain the zero point (thus essentially all functions of compact support).
This perhaps one could expect because the Schwartz algebra $\mathcal{S}(\mathbb{R}^n) 
= \mathcal{S}_{H_{(n)}}(\mathbb{R}^n)$ contains all functions of compact support
and the operators $A^{(n)}$ and $H_{(n)}$ have similar asymptotics, expressed by the Corollary
stated above. But in general the condition of the above Corollary \ref{AsymptoticsA^(n0)=AsymptoticsH_(n0)}
is insufficient to guarantee this. In fact, we can give two standard operators $A_1$ and
$A_2$ on $L^2(\mathbb{R}^n)$ with exactly the same spectra (counting with multiplicity) and even 
unitary equivalent, 
such that $\mathcal{S}_{A_1}(\mathbb{R}^n)$
contains essentially\footnote{For all supports if only the support does not contain a fixed point.} 
all smooth functions of compact support and $\mathcal{S}_{A_2}(\mathbb{R}^n)$
contains no elements of compact support. Indeed, we can define
\[
A_1 = A^{(n)}, \,\,\, A_2 = \mathscr{F} A^{(n)} \mathscr{F}^{-1},
\] 
where $\mathscr{F}$ is the Fourier transform. Indeed, 
by the first Proposition of Subsection \ref{splitting}
\[
\mathcal{S}_{\mathscr{F}A^{(n)}\mathscr{F}^{-1}}(\mathbb{R}^n)
= \mathcal{S}^{00}(\mathbb{R}^n)
\]
contains no elements of compact support; although the algebra of functions
\[
\mathcal{S}_{A^{(n)}}(\mathbb{R}^n)
= \mathcal{S}^{0}(\mathbb{R}^n)
\]
contains essentially all smooth functions of compact support, \emph{i. e.} except the supports 
containing the zero point. 

This is because the Gelfand's abstract characterization of the nuclear space
$\mathcal{S}_A(\mathcal{M})$ through the standard $A$ on $L^2(\mathcal{M})$ ignores the algebraic
multiplicative structure of $\mathcal{S}_A(\mathcal{M})$, and concerns only the linear-toplogy part. 
Therefore, questions related to the multiplicative structure need to reflect the relation between 
the chosen multiplicative structure  and the linear-toplogy structure involved into the abstract characterization
of $\mathcal{S}_A(\mathcal{M})$. In particular the question concerning support of
$f \in \mathcal{S}_A(\mathcal{M})$ involves (implicitly) the multiplicative structure, because 
each point of $p\in \mathcal{M}$ determines the corresponding character $\chi_p$ of
$\mathcal{S}_A(\mathcal{M})$ regarded as the  commutative algebra under pointwise multiplication.
$\chi_p \in \textrm{supp} \, f$ if and only if $\chi_p(f) \neq 0$. Preserving the relationship of the
unitary equivalence $U$ between $A_1$ and $A_2 = UA_1U^{-1}$ and the multiplicative structures we recover the equality of the corresponding
algebras $\mathcal{S}_{A_1}(\mathcal{M}_1)$ and $\mathcal{S}_{A_2}(\mathcal{M}_2)$ (up to equivalence induced by a diffeomorphism $\textrm{Spec} \, \mathcal{S}_{A_2}(\mathcal{M}_2) \rightarrow 
\textrm{Spec} \, \mathcal{S}_{A_1}(\mathcal{M}_1)$ of the spectra). 
Namely, we choose $A_1$ and the multiplication in $\mathcal{S}_{A_1}(\mathcal{M}_1)$ in such a manner
that it is continuous in the nuclear topology and $\mathcal{S}_{A_1}(\mathcal{M}_1)$ becomes a nuclear algebra. Next we assume
that the equivalence $U$ preserves multiplications $\cdot$ and $\ast$ (not necessarily pointwise
but then in general $\textrm{Spec} \, \mathcal{S}_{A_2}(\mathcal{M}_2) \neq \mathcal{M}_2$,
 $\textrm{Spec} \, \mathcal{S}_{A_1}(\mathcal{M}_1) \neq \mathcal{M}_1$)
\[
U(f\cdot g) = Uf \ast Ug
\]
in the respective nuclear algebras 
$\mathcal{S}_{A_1}(\mathcal{M}_1)$ and $\mathcal{S}_{A_2}(\mathcal{M}_2)$. Then the 
algebras become essentially equal, up to equivalence induced a  map between the spectra
\[
\textrm{Spec} \, \mathcal{S}_{A_2}(\mathcal{M}_2) 
\rightarrow \textrm{Spec} \, \mathcal{S}_{A_1}(\mathcal{M}_1),
\]
naturally induced by $U$.
 This will remain to be true
if $U$ preserves the multiplication only up to an invertible multiplier $M$ of the image
algebra $\mathcal{S}_{A_2}(\mathcal{M}_2)$:
\[
U(f\cdot g) = M \ast Uf \ast Ug,
\]
as one can see for the particular case of the examples worked out in Subsections \ref{dim=1}, \ref{dim=n}.

Indeed, the algebras $\mathcal{S}^{0}$ and $\mathcal{S}^{00}$ become essentially equivalent
and even equal under the Fourier transform, but we have to remember that this is the case if 
the commutative multiplication
$\ast$ in $\mathcal{S}^{00}$ is defined through convolution and not by pointwise multiplication and in particular that 
$\chi_x \in \textrm{supp} \, g$ provided $\chi_x(g) \neq 0$, and $\chi_x$ is a character of
$\mathcal{S}^{00}$ for the multiplication defined by convolution $\ast$.

In Appendix \ref{AppendixNonCompMani} we use this observations for the special case of pointwise multiplications $\cdot$
and $\ast$ in the function algebras $\mathcal{S}_{A_1}(\mathcal{M}_1)$ and 
$\mathcal{S}_{A_2}(\mathcal{M}_2)$, 
with
\[
\textrm{Spec} \, \mathcal{S}_{A_2}(\mathcal{M}_2) = \mathcal{M}_2, \,\,\,\,\, 
 \textrm{Spec} \, \mathcal{S}_{A_1}(\mathcal{M}_1) = \mathcal{M}_1,
\]
and with the multiplier $M$ equal to the Radon-Nikodym derivation induced by the diffeomorphism
\[
\mathcal{M}_1 \rightarrow \mathcal{M}_2 
\]
associated to the unitary equivalence $U$.

\section{APPENDIX: Fourier transforms $u_s(\boldsymbol{\p})$ and $v_s(-\boldsymbol{\p})$
of a complete system of distributional solutions
of the homogeneous Dirac equation}\label{fundamental,u,v}

As we have seen the Hilbert spaces $\mathcal{H}_{m,0}^{\oplus}$ and $\mathcal{H}_{-m,0}^{\ominus}$ of Fourier transforms of bispinor solutions of the Dirac equation, concentrated respectively on the orbit $\mathscr{O}_{m,0,0,0}$
and $\mathscr{O}_{-m,0,0,0}$, are equal to the images of the corresponding projection operators
$P^\oplus$ and $P^\ominus$ -- the multiplication operators by the corresponding
orthogonal projections $P^\oplus(p)$, $p \in \mathscr{O}_{m,0,0,0}$ and $P^\ominus(p), p \in \mathscr{O}_{-m,0,0,0}$
-- compare Subsection \ref{e1}. Recall that 
\[
\textrm{rank} P^\oplus(p) = 2, p \in \mathscr{O}_{m,0,0,0}, \,\,\,\,\,\,\,\,\,
\textrm{rank} P^\ominus(p) = 2, p \in \mathscr{O}_{-m,0,0,0}.
\]
It is therefore possible to choose at each point $p = (\boldsymbol{\p}, p_0(\boldsymbol{\p}))
= (\boldsymbol{\p}, E(\boldsymbol{\p}) = \sqrt{|\boldsymbol{\p}|^2 +m^2})$ of the orbit $\mathscr{O}_{m,0,0,0}$
(specified uniquely by $\boldsymbol{\p} \in \mathbb{R}^3$) 
a pair of vectors $u_s(\boldsymbol{\p})$, $s=1,2$, which span the image $\textrm{Im} \, P^\oplus(\boldsymbol{\p}, p_0(\boldsymbol{\p})) = \textrm{Im} \, P^\oplus(\boldsymbol{\p}, E(\boldsymbol{\p}))$
of $P^\oplus(p) = P^\oplus(\boldsymbol{\p}, p_0(\boldsymbol{\p}))$.
Similarly, for each point $p = (\boldsymbol{\p}, p_0(\boldsymbol{\p}))
= (\boldsymbol{\p}, -E(\boldsymbol{\p}) = -\sqrt{|\boldsymbol{\p}|^2 +m^2})$ of the orbit $\mathscr{O}_{-m,0,0,0}$
(specified by $\boldsymbol{\p} \in \mathbb{R}^3$) we can find a pair of two vectors
$v_s(\boldsymbol{\p})$, $s=1,2$, which span the image $\textrm{Im} \, P^\ominus(\boldsymbol{\p}, p_0(\boldsymbol{\p})) = \textrm{Im} \, P^\ominus(\boldsymbol{\p}, -E(\boldsymbol{\p}))$, 
$E(\boldsymbol{\p}) = \sqrt{|\boldsymbol{\p}|^2 +m^2}$ for 
$p = \big(\boldsymbol{\p}, -E(\boldsymbol{\p})\big) = (\boldsymbol{\p}, -\sqrt{|\boldsymbol{\p}|^2 +m^2})
\in \mathscr{O}_{-m,0,0,0}$. We choose these vectors in such a manner that their components
depend smoothly on $\boldsymbol{\p}$ and are multipliers and even convolutors of the Schwartz nuclear
algebra $\mathcal{S}(\mathbb{R}^3; \mathbb{C})$. Moreover, we choose them in such a manner that 
$\boldsymbol{\p} \mapsto u_s(\boldsymbol{\p})$ and $\boldsymbol{\p} \mapsto v_s(-\boldsymbol{\p})$
represent Fourier transforms of certain solutions of the free Dirac equation concentrated respectively on the orbits
$\mathscr{O}_{m,0,0,0}$ and $\mathscr{O}_{-m,0,0,0}$. That $\boldsymbol{\p} \mapsto v_s(-\boldsymbol{\p})$, $s=1,2$, represent the Fourier transforms of solutions of 
the Dirac equation and not simply $\boldsymbol{\p} \mapsto v_s(\boldsymbol{\p})$, $s=1,2$, is a matter
of tradition and does not have any deeper justification. Of course there is a whole infinity of different choices
for $u_s(\boldsymbol{\p})$ and  $v_s(\boldsymbol{\p})$, giving unitary equivalent constructions
of the Dirac field.

In this Appendix we construct one useful example of $u_s(\boldsymbol{\p})$ and  $v_s(\boldsymbol{\p})$, $s=1,2$
for the chiral representation of the Clifford algebra generators (Dirac matrices)
\begin{equation}\label{chiralgamma}
\gamma^0 = \left( \begin{array}{cc}   0 &  \bold{1}_2  \\
                                           
                                                   \bold{1}_2              & 0 \end{array}\right), \,\,\,\,
\gamma^k = \left( \begin{array}{cc}   0 &  -\sigma_k  \\
                                           
                                                   \sigma_k             & 0 \end{array}\right),
\end{equation}
which we have used in Subsection \ref{e1} as well as for the so called standard representation
\begin{multline}\label{standardgamma}
\gamma^0 = C \, \left( \begin{array}{cc}   0 &  \bold{1}_2  \\                    
                                                   \bold{1}_2              & 0 \end{array}\right)C^{-1}
= \left( \begin{array}{cc}  \bold{1}_2  &  0  \\
                                                          0       & -\bold{1}_2 \end{array}\right),
\\
\gamma^k = C \, \left( \begin{array}{cc}   0 &  -\sigma_k  \\
                                                   \sigma_k             & 0 \end{array}\right) C^{-1}
= \left( \begin{array}{cc}   0 &  \sigma_k  \\
                                                   -\sigma_k             & 0 \end{array}\right),
\end{multline}
of the Dirac matrices, where 
\[
C = \frac{1}{\sqrt{2}}\left( \begin{array}{cc}   \bold{1}_2 &  \bold{1}_2  \\
                                                   \bold{1}_2             & -\bold{1}_2 \end{array}\right)
= C^{*} = C^{-1}
\]
is unitary involutive $4 \times 4$ matrix.

\begin{center}
{\small THE SOLUTIONS  $u_s(\boldsymbol{\p})$ AND $v_s(\boldsymbol{\p})$ IN THE CHIRAL REPRESENTATION
(\ref{chiralgamma})}
\end{center}

Let us start with the chiral representation (used in Subsection \ref{e1}). Recall that 
\[
P^\oplus(p) = \frac{1}{2}
\left( \begin{array}{cc} 1 & \beta(p)^{-2}  \\
                                                  \beta(p)^2 &  1 \end{array}\right), \,\,\,\,
p \in \mathscr{O}_{m,0,0,0}
\]
with $\beta(p)$ (chosen correspondingly to the chiral representation, as there is infinitum of other possible choices
of $\beta(p)$, compare Subsect. \ref{e1}) corresponding to the orbit $ \mathscr{O}_{m,0,0,0}$, i.e.
\begin{equation}\label{betaO_m,0,0,0}
\begin{split}
\beta(p)^{-2} = \frac{1}{m} \big(p^0 \bold{1} + \vec{p} \cdot \vec{\sigma} \big), \,\,\,
p^0(\vec{p}) = \sqrt{\vec{p} \cdot \vec{p} + m^2} = E(\vec{p}), \\
\beta(p)^{2} = \frac{1}{m} \big(p^0 \bold{1} - \vec{p} \cdot \vec{\sigma} \big), \,\,\,
p^0(\vec{p}) = \sqrt{\vec{p} \cdot \vec{p} + m^2} = E(\vec{p}).
\end{split}
\end{equation}
Similarly recall that here
\[
P^\ominus(p) = \frac{1}{2}
\left( \begin{array}{cc} 1 & -\beta(p)^{-2}  \\
                                                  -\beta(p)^2 &  1 \end{array}\right), \,\,\,\,
p \in \mathscr{O}_{-m,0,0,0}
\]
with $\beta(p)$ corresponding to the orbit $ \mathscr{O}_{-m,0,0,0}$, i.e.
\begin{equation}\label{betaO_-m,0,0,0}
\begin{split}
\beta(p)^{-2} = \frac{1}{m} \big(-p^0 \bold{1} - \vec{p} \cdot \vec{\sigma} \big), \,\,\,
p^0(\vec{p}) = - \sqrt{\vec{p} \cdot \vec{p} + m^2} = -E(\vec{p}), \\
\beta(p)^{2} = \frac{1}{m} \big(-p^0 \bold{1} + \vec{p} \cdot \vec{\sigma} \big), \,\,\,
p^0(\vec{p}) = - \sqrt{\vec{p} \cdot \vec{p} + m^2} = - E(\vec{p}),
\end{split}
\end{equation}
compare Subsection \ref{e1}. In this case (of chiral representation (\ref{chiralgamma})) 
one can put 
\begin{multline}\label{chiral,u,v}
u_s(\boldsymbol{\p}) =  \frac{1}{\sqrt{2}} \sqrt{\frac{E(\boldsymbol{\p}) + m}{2 E(\boldsymbol{\p})}}
\left( \begin{array}{c}   \chi_s + \frac{\boldsymbol{\p} \cdot \boldsymbol{\sigma}}{E(\boldsymbol{\p}) + m} \chi_s
\\                                           
              \chi_s - \frac{\boldsymbol{\p} \cdot \boldsymbol{\sigma}}{E(\boldsymbol{\p}) + m} \chi_s                         \end{array}\right) \\ =
\frac{1}{\sqrt{2}} \sqrt{\frac{E(\boldsymbol{\p}) + m}{2 E(\boldsymbol{\p})}}
\left( \begin{array}{c}   \chi_s + \frac{\boldsymbol{\p} \cdot \boldsymbol{\sigma}}{E(\boldsymbol{\p}) + m} \chi_s
\\                                           
              \beta\big(p_0(\boldsymbol{\p}), \boldsymbol{\p}\big)^2 \, \big(\chi_s + \frac{\boldsymbol{\p} \cdot \boldsymbol{\sigma}}{E(\boldsymbol{\p}) + m} \chi_s\big)                         \end{array}\right),
\\
v_s(\boldsymbol{\p}) =  \frac{1}{\sqrt{2}} \sqrt{\frac{E(\boldsymbol{\p}) + m}{2 E(\boldsymbol{\p})}}
\left( \begin{array}{c}   \chi_s + \frac{\boldsymbol{\p} \cdot \boldsymbol{\sigma}}{E(\boldsymbol{\p}) + m} \chi_s
\\                                           
              -\big(\chi_s - \frac{\boldsymbol{\p} \cdot \boldsymbol{\sigma}}{E(\boldsymbol{\p}) + m}\chi_s \big)                          \end{array}\right) = \\
\frac{1}{\sqrt{2}} \sqrt{\frac{E(\boldsymbol{\p}) + m}{2 E(\boldsymbol{\p})}}
\left( \begin{array}{c}   \chi_s + \frac{\boldsymbol{\p} \cdot \boldsymbol{\sigma}}{E(\boldsymbol{\p}) + m} \chi_s
\\                                           
              -\beta(p_0\big(\boldsymbol{\p}), -\boldsymbol{\p}\big)^{2} \, \big(\chi_s + \frac{\boldsymbol{\p} \cdot \boldsymbol{\sigma}}{E(\boldsymbol{\p}) + m}\chi_s \big)                          \end{array}\right)
\end{multline}
where
\[
\chi_1 = \left( \begin{array}{c} 1  \\
                                                  0 \end{array}\right), \,\,\,\,\,
\chi_2 = \left( \begin{array}{c} 0  \\
                                                  1 \end{array}\right).
\]
Here $\beta(p)$ in the formula for $u_{s}(\boldsymbol{\p})$ is that (\ref{betaO_m,0,0,0}) corresponding
to the orbit $\mathscr{O}_{m,0,0,0}$ and in the formula for $v_{s}(\boldsymbol{\p})$ the matrix function
$\beta(p)$ equals 
(\ref{betaO_-m,0,0,0}) correspondingly
to the orbit $\mathscr{O}_{-m,0,0,0}$, so that by construction  the solutions $u_{s}(\boldsymbol{\p}), v_{s}(-\boldsymbol{\p})$ have the general form (with the respective $\beta(p)$ corresponding
to the respective orbit $\mathscr{O}_{\pm m,0,0,0}$)
\[
\begin{split}
u_{s}(\boldsymbol{\p}) \overset{\textrm{df}}{=}
u_s(p_0(\boldsymbol{\p}), \boldsymbol{\p}) =  \left( \begin{array}{c} \widetilde{\varphi}_{s +}(p)  \\
                                                  \beta(p)^2 \widetilde{\varphi}_{s+}(p) \end{array}\right), \,\,\,
p = (p_0(\boldsymbol{\p}), \boldsymbol{\p}) \in \mathscr{O}_{m,0,0,0}, \\
v_{s}(\boldsymbol{-\p}) \overset{\textrm{df}}{=}
v_s(p_0(\boldsymbol{\p}), -\boldsymbol{\p}) =  \left( \begin{array}{c} \widetilde{\varphi}_{s-}(p)  \\
                                                  -\beta(p)^2 \widetilde{\varphi}_{s-}(p) \end{array}\right), \,\,\,
p = (p_0(\boldsymbol{\p}), \boldsymbol{\p}) \in \mathscr{O}_{-m,0,0,0},
\end{split}
\]
with
\[
\begin{split}
\widetilde{\varphi}_{s+}\big(p = (p_0(\boldsymbol{\p}), \boldsymbol{\p}) \big) = 
\chi_s + \frac{\boldsymbol{\p} \cdot \boldsymbol{\sigma}}{E(\boldsymbol{\p}) + m} \chi_s, \,\,\,
p = (p_0(\boldsymbol{\p}), \boldsymbol{\p}) \in \mathscr{O}_{m,0,0,0}, \\
\widetilde{\varphi}_{s-}\big(p = (p_0(\boldsymbol{\p}), \boldsymbol{\p}) \big) = 
\chi_s - \frac{\boldsymbol{\p} \cdot \boldsymbol{\sigma}}{E(\boldsymbol{\p}) + m} \chi_s, \,\,\,
p = (p_0(\boldsymbol{\p}), \boldsymbol{\p}) \in \mathscr{O}_{-m,0,0,0}. 
\end{split}
\]
as expected by construction of $\mathcal{H}_{m,0}^{\oplus}$ and $\mathcal{H}_{-m,0}^{\ominus}$ in Subsection \ref{e1}.

The vectors $u_s(\boldsymbol{\p})$ and  $v_s(\boldsymbol{\p})$, $s=1,2$,
respect the following orthonormality relations:
\begin{equation}\label{u^+u=delta}
\begin{split}
u_s(\boldsymbol{\p})^+ u_{s'}(\boldsymbol{\p}) = \delta_{ss'}, \,\,\,
v_s(\boldsymbol{\p})^+ v_{s'}(\boldsymbol{\p}) = \delta_{ss'}, \,\,\,
u_s(\boldsymbol{\p})^+ v_{s'}(-\boldsymbol{\p}) = 0.
\end{split}
\end{equation}
By construction we have 
\begin{equation}\label{E+-(p)}
\begin{split}
E_+(\boldsymbol{\p}) 
= \sum_{s=1,2} u_{s}(\boldsymbol{\p}) u_s(\boldsymbol{\p})^+ =
\frac{1}{2E(\boldsymbol{\p})} \big( E(\boldsymbol{\p}) \boldsymbol{1} 
+\boldsymbol{\p} \cdot \boldsymbol{\alpha} + \beta m \big), \,\,\,\,
E(\boldsymbol{\p}) = \sqrt{|\boldsymbol{\p}|^2 + m^2} \\
E_-(\boldsymbol{\p}) 
= \sum_{s=1,2} v_{s}(\boldsymbol{\p}) v_s(\boldsymbol{\p})^+ =
\frac{1}{2E(\boldsymbol{\p})} \big( E(\boldsymbol{\p}) \boldsymbol{1} 
+\boldsymbol{\p} \cdot \boldsymbol{\alpha} - \beta m \big), \,\,\,\,
E(\boldsymbol{\p}) = \sqrt{|\boldsymbol{\p}|^2 + m^2}.
\end{split}
\end{equation}
Here
\[
\begin{split}
\boldsymbol{\sigma} = (\sigma_1, \sigma_2, \sigma_3), \,\,\,
\boldsymbol{\alpha} = (\alpha^1, \alpha^2, \alpha^3), \,\,\,\\
\boldsymbol{\p} \cdot \boldsymbol{\sigma} =
\sum_{i=1}^{3} p_i \sigma_i, \,\,\,
\boldsymbol{\p} \cdot \boldsymbol{\alpha} =
\sum_{i=1}^{3} p_i \alpha^i, \\
\alpha^i = \gamma^0 \gamma^i, \,\, \beta = \gamma^0.
\end{split}
\]
Note that $E_{+}(\boldsymbol{p})$ and $E_{-}(-\boldsymbol{p})$ are mutually orthogonal
projectors on $\mathbb{C}^4$ such that $E_{+}(\boldsymbol{p}) + E_{-}(-\boldsymbol{p}) = \boldsymbol{1}$ and
such that the operators $E_+$ and $E_-$ of Subsection \ref{FirstStepH} are equal to the operators of pointwise multiplications by the matrices
$E_{\pm}(\pm \boldsymbol{p})$ on the Hilbert spaces  $\mathcal{H}_{m,0}^{\oplus}$ and $\mathcal{H}_{-m,0}^{\ominus}$
of bispinors concentrated respectively on $\mathscr{O}_{m,0,0,0}$
and $\mathscr{O}_{-m,0,0,0}$ (with the point $p= (p_0(\boldsymbol{\p}), \boldsymbol{\p})$ of the respective orbit identified with its Cartesian coordinates $\boldsymbol{\p}$).

Moreover, recall that for any element $\widetilde{\phi} \in \mathcal{H}_{m,0}^{\oplus}$ 
the following algebraic relation holds (summation with respect to $i=1,2,3$)
\[
p_0 \gamma^0\widetilde{\phi}(p) =  \big[ p^i \gamma^i + m \boldsymbol{1} \big]\widetilde{\phi}(p), \,\,\,
p \in \mathscr{O}_{m,0,0,0},  
\]
compare Subsection \ref{e1}, so that
\[
E(\boldsymbol{\p}) \widetilde{\phi}(p) = \big[\boldsymbol{\p} \cdot \boldsymbol{\alpha} 
+ m \beta \big] \widetilde{\phi}(p), \,\,\, p = (p_0(\boldsymbol{\p}), \boldsymbol{\p}) \in \mathscr{O}_{m,0,0,0},
\]
for all $\widetilde{\phi} \in \mathcal{H}_{m,0}^{\oplus}$ and thus 
\begin{multline}\label{E_+Phi=Phi}
E_+(\boldsymbol{\p}) \widetilde{\phi}(p)
= \Bigg(\sum_{s=1,2} u_{s}(\boldsymbol{\p}) u_s(\boldsymbol{\p})^+\Bigg) \widetilde{\phi}(p)  \\ =
\frac{1}{2E(\boldsymbol{\p})} \big( E(\boldsymbol{\p}) \boldsymbol{1} 
+\boldsymbol{\p} \cdot \boldsymbol{\alpha} + \beta m \big) \widetilde{\phi}(p)
= \widetilde{\phi}(p), \\
p = (p_0(\boldsymbol{\p}), \boldsymbol{\p}) \in \mathscr{O}_{m,0,0,0},
\end{multline}
for each $\widetilde{\phi} \in \mathcal{H}_{m,0}^{\oplus}$.

Similarly for any element $\widetilde{\phi} \in \mathcal{H}_{-m,0}^{\ominus}$ 
the following algebraic relation holds (summation with respect to $i=1,2,3$)
\begin{multline*}
p_0 \gamma^0\widetilde{\phi}(p) =  \big[ p^i \gamma^i + m \boldsymbol{1} \big]\widetilde{\phi}(p), \\
p= (p_0(\boldsymbol{\p}, \boldsymbol{\p})) = (-E(\boldsymbol{\p}), \boldsymbol{\p}) \in \mathscr{O}_{-m,0,0,0},  
\end{multline*}
compare Subsection \ref{e1}, so that
\begin{multline*}
-E(\boldsymbol{\p}) \widetilde{\phi}(-E(\boldsymbol{\p}), \boldsymbol{\p}) = \big[\boldsymbol{\p} \cdot \boldsymbol{\alpha} 
+ m \beta \big] \widetilde{\phi}(-E(\boldsymbol{\p}), \boldsymbol{\p}), \\
p = (p_0(\boldsymbol{\p}), \boldsymbol{\p}) = (-E(\boldsymbol{\p}), \boldsymbol{\p}) \in \mathscr{O}_{-m,0,0,0},
\end{multline*}
for all $\widetilde{\phi} \in \mathcal{H}_{-m,0}^{\ominus}$ and thus 
\[
E(\boldsymbol{\p}) \widetilde{\phi}(-E(\boldsymbol{\p}), -\boldsymbol{\p}) =
\big(\boldsymbol{\p} \cdot \boldsymbol{\alpha} - \beta m \big)\widetilde{\phi}(-E(\boldsymbol{\p}), -\boldsymbol{\p}),
\,\,\, \widetilde{\phi} \in \mathcal{H}_{-m,0}^{\ominus}. 
\]
Therefore we have
\begin{multline}\label{E_-Phi=Phi}
E_-(\boldsymbol{\p}) \widetilde{\phi}(-E(\boldsymbol{\p}), -\boldsymbol{\p})
= \Bigg(\sum_{s=1,2} v_{s}(\boldsymbol{\p}) v_s(\boldsymbol{\p})^+\Bigg) 
\widetilde{\phi}(-E(\boldsymbol{\p}), -\boldsymbol{\p})  \\ =
\frac{1}{2E(\boldsymbol{\p})} \big( E(\boldsymbol{\p}) \boldsymbol{1} 
+\boldsymbol{\p} \cdot \boldsymbol{\alpha} - \beta m \big) \widetilde{\phi}(-E(\boldsymbol{\p}), -\boldsymbol{\p})
= \widetilde{\phi}(-E(\boldsymbol{\p}), -\boldsymbol{\p}), \\
p = (p_0(\boldsymbol{\p}), \boldsymbol{\p}) \in \mathscr{O}_{-m,0,0,0},
\end{multline}
for each $\widetilde{\phi} \in \mathcal{H}_{-m,0}^{\ominus}$. 

By construction we have
\begin{equation}\label{P^plusu=u,P^minusv=v}
P^\oplus \big(E(\boldsymbol{\p}), \boldsymbol{\p} \big) \, u_s(\boldsymbol{\p}) = u_s(\boldsymbol{\p}),
\,\,\,\,\,
P^\ominus \big(-E(\boldsymbol{\p}), \boldsymbol{\p} \big) \, v_s(-\boldsymbol{\p}) = v_s(-\boldsymbol{\p})
\end{equation}
or
\begin{equation}\label{P^minusv=v}
P^\ominus \big(-E(\boldsymbol{\p}), -\boldsymbol{\p} \big) \, v_s(\boldsymbol{\p}) = v_s(\boldsymbol{\p}),
\end{equation}
and 
\begin{equation}\label{P^plusPhi=Phi,P^minusPhi=Phi}
\begin{split}
P^\oplus \big(E(\boldsymbol{\p}), \boldsymbol{\p} \big) \, \widetilde{\phi}((E(\boldsymbol{\p}),\boldsymbol{\p}) 
= \widetilde{\phi}((E(\boldsymbol{\p}),\boldsymbol{\p}), \,\,\,\,\,\,
\widetilde{\phi} \in \mathcal{H}_{m,0}^{\oplus}, \\
P^\ominus \big(-E(\boldsymbol{\p}), \boldsymbol{\p} \big) \, \widetilde{\phi}(-E(\boldsymbol{\p}),\boldsymbol{\p}) 
= \widetilde{\phi}(-E(\boldsymbol{\p}),\boldsymbol{\p}), \,\,\,\,\,\,
\widetilde{\phi} \in \mathcal{H}_{-m,0}^{\ominus}.
\end{split}
\end{equation}

In particular for any function $\widetilde{\phi}$ there do exist the following decompositions
\begin{equation}\label{P^plusTilde(phi),P^minusTilde(phi)}
\begin{split}
P^\oplus \big(|p_0(\boldsymbol{\p})|, \boldsymbol{\p} \big) \, \widetilde{\phi}(|p_0(\boldsymbol{\p})|,\boldsymbol{\p}) 
=f_1(\boldsymbol{\p})u_1(\boldsymbol{\p}) + f_2(\boldsymbol{\p})u_2(\boldsymbol{\p}),
\\
P^\oplus \big(|p_0(\boldsymbol{\p})|, \boldsymbol{\p} \big)^* \, \widetilde{\phi}(|p_0(\boldsymbol{\p})|,\boldsymbol{\p}) 
=f_1(\boldsymbol{\p})\gamma^0 u_1(\boldsymbol{\p}) + f_2(\boldsymbol{\p})\gamma^0 u_2(\boldsymbol{\p}),
\\
P^\ominus \big(-|p_0(\boldsymbol{\p})|, \boldsymbol{\p} \big) \, \widetilde{\phi}(-|p_0(\boldsymbol{\p})|,\boldsymbol{\p}) 
=f_1(\boldsymbol{\p})v_1(-\boldsymbol{\p}) + f_2(\boldsymbol{\p})v_2(-\boldsymbol{\p}),
\\
P^\ominus \big(-|p_0(\boldsymbol{\p})|, -\boldsymbol{\p} \big) \, \widetilde{\phi}(-|p_0(\boldsymbol{\p})|,-\boldsymbol{\p}) 
=f_1(-\boldsymbol{\p})v_1(\boldsymbol{\p}) + f_2(-\boldsymbol{\p})v_2(\boldsymbol{\p}),
\\
P^\ominus \big(-|p_0(\boldsymbol{\p})|, -\boldsymbol{\p} \big)^* \, \widetilde{\phi}(-|p_0(\boldsymbol{\p})|,-\boldsymbol{\p}) 
=f_1(-\boldsymbol{\p})\gamma^0v_1(\boldsymbol{\p}) + f_2(-\boldsymbol{\p})\gamma^0v_2(\boldsymbol{\p}),
\end{split}
\end{equation}
with the corresponding component functions $f_s$, because 
\[
P^\oplus \big(|p_0(\boldsymbol{\p})|, \boldsymbol{\p} \big)^* = 
\gamma^0 P^\oplus \big(|p_0(\boldsymbol{\p})|, \boldsymbol{\p} \big)\gamma^0,
\,\,\,
P^\ominus \big(-|p_0(\boldsymbol{\p})|, \boldsymbol{\p} \big)^* = 
\gamma^0 P^\ominus \big(-|p_0(\boldsymbol{\p})|, \boldsymbol{\p} \big)\gamma^0,
\]
correspond to the projections determined by the fundamental solutions $\gamma^0u, \gamma^0v$.

From the formulas (\ref{P^plusu=u,P^minusv=v}) or (\ref{P^minusv=v}) it follows in particular that
\begin{multline}\label{u^+P^plusPhi=u^+Phi}
u_s(\boldsymbol{\p})^+\widetilde{\phi}(E(\boldsymbol{\p}), \boldsymbol{\p}) =
\sum_{a=1}^{4} \overline{u_{s}^{a}(\boldsymbol{\p})}\widetilde{\phi}^{a}(E(\boldsymbol{\p}), \boldsymbol{\p}) =
\Big(u_{s}(\boldsymbol{\p}), \, \widetilde{\phi}(E(\boldsymbol{\p}), \boldsymbol{\p}) \Big)_{\mathbb{C}^4} \\
= \Big(P^\oplus(E(\boldsymbol{\p}), \,  \boldsymbol{\p}) u_{s}(\boldsymbol{\p}), \widetilde{\phi}(E(\boldsymbol{\p}), \boldsymbol{\p}) \Big)_{\mathbb{C}^4} =
\Big( u_{s}(\boldsymbol{\p}), \, P^\oplus(E(\boldsymbol{\p}), \boldsymbol{\p})^*\widetilde{\phi}(E(\boldsymbol{\p}), \boldsymbol{\p}) \Big)_{\mathbb{C}^4} \\ =
u_s(\boldsymbol{\p})^+ \big(P^\oplus(E(\boldsymbol{\p}), \boldsymbol{\p})^*\widetilde{\phi}(E(\boldsymbol{\p}), \boldsymbol{\p})\big) = u_s(\boldsymbol{\p})^+ \big(P^{\oplus *}\widetilde{\phi}\big)(E(\boldsymbol{\p}), \boldsymbol{\p}),
\\
\,\,\, \textrm{for any smooth} \, \widetilde{\phi},
\end{multline}
\begin{equation}\label{u^+P^plusPhi=u^+Phi'}
u_s(\boldsymbol{\p})^+ \gamma^0 u_r(\boldsymbol{\p})= \big(u_s(\boldsymbol{\p}), \gamma^0 u_r(\boldsymbol{\p}) \big)_{{}_{\mathbb{C}^4}}
= {\textstyle\frac{m}{|p_0(\boldsymbol{\p})|}} \delta_{sr}
\end{equation}
and
\begin{multline}\label{v^+P^minusPhi=v^+Phi}
v_s(\boldsymbol{\p})^+\widetilde{\phi}(-E(\boldsymbol{\p}), -\boldsymbol{\p}) =
\sum_{a=1}^{4} \overline{v_{s}^{a}(\boldsymbol{\p})}\widetilde{\phi}^{a}(-E(\boldsymbol{\p}), -\boldsymbol{\p}) \\
\Big(v_{s}(\boldsymbol{\p}), \, \widetilde{\phi}(-E(\boldsymbol{\p}), -\boldsymbol{\p}) \Big)_{\mathbb{C}^4}
= \Big(P^\ominus(-E(\boldsymbol{\p}), \,  -\boldsymbol{\p}) v_{s}(\boldsymbol{\p}), \widetilde{\phi}(-E(\boldsymbol{\p}), -\boldsymbol{\p}) \Big)_{\mathbb{C}^4} = \\
\Big( v_{s}(\boldsymbol{\p}), \, P^\ominus(-E(\boldsymbol{\p}), -\boldsymbol{\p})^*\widetilde{\phi}(-E(\boldsymbol{\p}), -\boldsymbol{\p}) \Big)_{\mathbb{C}^4} \\ =
v_s(\boldsymbol{\p})^+ \big(P^\ominus(-E(\boldsymbol{\p}), -\boldsymbol{\p})^*\widetilde{\phi}(-E(\boldsymbol{\p}), -\boldsymbol{\p})\big) = v_s(\boldsymbol{\p})^+ \big(P^{\ominus *}\widetilde{\phi}\big)(-E(\boldsymbol{\p}), -\boldsymbol{\p}),
\\
\,\,\, \textrm{for any smooth} \, \widetilde{\phi},
\end{multline}
\begin{equation}\label{v^+P^minusPhi=v^+Phi'}
v_s(\boldsymbol{\p})^+ \gamma^0 v_r(\boldsymbol{\p})= \big(v_s(\boldsymbol{\p}), \gamma^0 v_r(\boldsymbol{\p}) \big)_{{}_{\mathbb{C}^4}}
= -{\textstyle\frac{m}{|p_0(\boldsymbol{\p})|}} \delta_{sr}.
\end{equation}
Joining (\ref{P^plusTilde(phi),P^minusTilde(phi)}), (\ref{u^+P^plusPhi=u^+Phi}) and (\ref{u^+P^plusPhi=u^+Phi'}) we get
\begin{multline*}
\big(u_s(\boldsymbol{\p}), \widetilde{\phi}(|p_0(\boldsymbol{\p})|, \boldsymbol{\p}) \big)_{{}_{\mathbb{C}^4}}
=
\big(P^{\oplus}(|p_0(\boldsymbol{\p}), \boldsymbol{\p})u_s(\boldsymbol{\p}), \widetilde{\phi}(|p_0(\boldsymbol{\p})|, \boldsymbol{\p}) \big)_{{}_{\mathbb{C}^4}}
\\
=
\big(u_s(\boldsymbol{\p}), P^\oplus(|p_0(\boldsymbol{\p}), \boldsymbol{\p})^* \widetilde{\phi}(|p_0(\boldsymbol{\p})|, \boldsymbol{\p}) \big)_{{}_{\mathbb{C}^4}}
=
\big(u_s(\boldsymbol{\p}), \gamma^0u_1(\boldsymbol{\p}) f_1(\boldsymbol{\p})+ \gamma^0u_2(\boldsymbol{\p}) f_2(\boldsymbol{\p})\big)_{{}_{\mathbb{C}^4}}
\end{multline*}
\begin{multline*}
= {\textstyle\frac{m}{|p_0(\boldsymbol{\p})|}} 
\big(\gamma^0 u_s(\boldsymbol{\p}), \gamma^0u_1(\boldsymbol{\p}) f_1(\boldsymbol{\p})+ \gamma^0u_2(\boldsymbol{\p}) f_2(\boldsymbol{\p})\big)_{{}_{\mathbb{C}^4}}
\\
= {\textstyle\frac{m}{|p_0(\boldsymbol{\p})|}} 
\big(\gamma^0 u_s(\boldsymbol{\p}), P^{\oplus *}\widetilde{\phi}(|p_0(\boldsymbol{\p})|, \boldsymbol{\p}) \big)_{{}_{\mathbb{C}^4}}.
\end{multline*}
Similarly, joining (\ref{P^plusTilde(phi),P^minusTilde(phi)}), (\ref{v^+P^minusPhi=v^+Phi}) and (\ref{v^+P^minusPhi=v^+Phi'}) we get
\begin{multline*}
\big(v_s(\boldsymbol{\p}), \widetilde{\phi}(-|p_0(\boldsymbol{\p})|, -\boldsymbol{\p}) \big)_{{}_{\mathbb{C}^4}}
=
\big(P^{\ominus}(-|p_0(\boldsymbol{\p}), -\boldsymbol{\p})v_s(\boldsymbol{\p}), \widetilde{\phi}(-|p_0(\boldsymbol{\p})|, -\boldsymbol{\p}) \big)_{{}_{\mathbb{C}^4}}
\\
=
\big(v_s(\boldsymbol{\p}), P^\ominus(-|p_0(\boldsymbol{\p}), -\boldsymbol{\p})^* \widetilde{\phi}(-|p_0(\boldsymbol{\p})|, -\boldsymbol{\p}) \big)_{{}_{\mathbb{C}^4}}
\\
=
\big(v_s(\boldsymbol{\p}), \gamma^0v_1(\boldsymbol{\p}) f_1(-\boldsymbol{\p})+ \gamma^0v_2(\boldsymbol{\p}) f_2(-\boldsymbol{\p})\big)_{{}_{\mathbb{C}^4}}
\end{multline*}
\begin{multline*}
= - {\textstyle\frac{m}{|p_0(\boldsymbol{\p})|}} 
\big(\gamma^0 v_s(\boldsymbol{\p}), \gamma^0v_1(\boldsymbol{\p}) f_1(-\boldsymbol{\p})+ \gamma^0u_2(\boldsymbol{\p}) f_2(-\boldsymbol{\p})\big)_{{}_{\mathbb{C}^4}}
\\
= - {\textstyle\frac{m}{|p_0(\boldsymbol{\p})|}} 
\big(\gamma^0 u_s(\boldsymbol{\p}), P^{\oplus *}\widetilde{\phi}(-|p_0(\boldsymbol{\p})|, -\boldsymbol{\p}) \big)_{{}_{\mathbb{C}^4}}.
\end{multline*}

Summing up, we have the following identities valid for any funtion $\widetilde{\phi}$
\begin{equation}\label{u^+P^plusPhi=u^+Phi''}
\big(u_s(\boldsymbol{\p}), \widetilde{\phi}(|p_0(\boldsymbol{\p})|, \boldsymbol{\p}) \big)_{{}_{\mathbb{C}^4}}
= {\textstyle\frac{m}{|p_0(\boldsymbol{\p})|}} 
\big(\gamma^0 u_s(\boldsymbol{\p}), P^{\oplus *}\widetilde{\phi}(|p_0(\boldsymbol{\p})|, \boldsymbol{\p}) \big)_{{}_{\mathbb{C}^4}}
\end{equation}
and
\begin{equation}\label{v^+P^minusPhi=v^+Phi''}
\big(v_s(\boldsymbol{\p}), \widetilde{\phi}(-|p_0(\boldsymbol{\p})|, -\boldsymbol{\p}) \big)_{{}_{\mathbb{C}^4}}
= -{\textstyle\frac{m}{|p_0(\boldsymbol{\p})|}} 
\big(\gamma^0 v_s(\boldsymbol{\p}), P^{\ominus *}\widetilde{\phi}(-|p_0(\boldsymbol{\p})|, -\boldsymbol{\p}) \big)_{{}_{\mathbb{C}^4}}
\end{equation}

It should be stressed that the formulas (\ref{u^+P^plusPhi=u^+Phi}), (\ref{u^+P^plusPhi=u^+Phi''}) and (\ref{v^+P^minusPhi=v^+Phi}), (\ref{v^+P^minusPhi=v^+Phi''})
are valid for any $\widetilde{\phi}$ not necessary belonging to $\mathcal{H}_{m,0}^{\oplus}$ 
or $\mathcal{H}_{-m,0}^{\ominus}$. 

It is obvious that the projectors $P^\oplus(p)$, $p \in \mathscr{O}_{m,0,0,0}$ 
and $P^\ominus(p)$, $p \in \mathscr{O}_{-m,0,0,0}$, an be expressed in the following
manifestly covariant form
\begin{equation}\label{covariantPplusPminus}
\begin{split}
P^\oplus(p) = \frac{1}{2m} \big[g_{\nu \mu} p^\nu \gamma^\mu + m \boldsymbol{1}_{{}_{4}} \big]
=  \frac{1}{2m} \big[\slashed{p} + m  \big], \,\,\,
p \in \mathscr{O}_{m,0,0,0}, \\
P^\ominus(p) = \frac{1}{2m} \big[g_{\nu \mu} p^\nu \gamma^\mu + m \boldsymbol{1}_{{}_{4}} \big]
=  \frac{1}{2m} \big[\slashed{p} + m  \big], \,\,\,
p \in \mathscr{O}_{-m,0,0,0}.
\end{split}
\end{equation} 

Finally  let us give the formulas useful in computation of the commutation functions 
and pairing functions for the Dirac field and its Dirac adjoined field. To this end 
let us recall that for a bispinor $u(\boldsymbol{\p})$ the Dirac adjoint
$u^\sharp(\boldsymbol{\p}) = u(\boldsymbol{\p})^\sharp$ 
is defined to be equal $u(\boldsymbol{\p})^+ \gamma^0$. 
There is frequently used (common) notation  $\overline{u}(\boldsymbol{\p})$ for the Dirac adjoint
$u^\sharp(\boldsymbol{\p})$. 
This common notation is somewhat unfortunate, because the Dirac adjoint may be mislead with the ordinary complex conjugaton, 
particularly so in expressions with explicitly used component indices
of bispinor, which we have already agreed to be denoted by overset bar (which also is a traditional notation for complex conjugation). It must be explicitly stated 
what is meant in each case in working with common notation of Dirac adjoint of bispinors. When working with quantum Dirac field 
$\boldsymbol{\psi}(x)$, the superscript $\sharp$ in $\boldsymbol{\psi}^\sharp(x)$ will
always mean the Dirac adjoint. Therefore, we use the superscript $\sharp$ for Dirac-adjoning operation. 
Denoting here $u^{\sharp}_{s}(\boldsymbol{\p}), v^{\sharp}_{s}(-\boldsymbol{\p})$ the
Dirac adjoints of the complete system of solutions $u_s(\boldsymbol{\p}), v_s(-\boldsymbol{\p})$, 
we get (summation with respect to $i= 1,2,3$)
\[
\begin{split}
\sum_{s=1,2} u_{s}(\boldsymbol{\p}) u^{\sharp}_{s}(\boldsymbol{\p}) =
\frac{1}{2E(\boldsymbol{\p})} \big( E(\boldsymbol{\p}) \gamma^0 
-p^i \gamma^i + \boldsymbol{1} m \big), \,\,\,\,
E(\boldsymbol{\p}) = \sqrt{|\boldsymbol{\p}|^2 + m^2} \\
 \sum_{s=1,2} v_{s}(\boldsymbol{\p}) v^{\sharp}_{s}(\boldsymbol{\p}) =
\frac{1}{2E(\boldsymbol{\p})} \big( E(\boldsymbol{\p}) \gamma^0 
-p^i \gamma^i - \boldsymbol{1} m \big), \,\,\,\,
E(\boldsymbol{\p}) = \sqrt{|\boldsymbol{\p}|^2 + m^2},
\end{split}
\] 
on multiplying the formulas (\ref{E+-(p)}) for $E_{\pm}(\boldsymbol{\p})$ by $\gamma^0$
on the right, and which is frequently written as
\begin{equation}\label{covariantProj}
\begin{split}
\sum_{s=1,2} u_{s}(\boldsymbol{\p}) u^{\sharp}_{s}(\boldsymbol{\p}) =
\frac{\slashed{p} +m}{2E(\boldsymbol{\p})} = 
\frac{p_\mu \gamma^\mu +m}{2E(\boldsymbol{\p})}, \,\,\,\,
E(\boldsymbol{\p}) = \sqrt{|\boldsymbol{\p}|^2 + m^2} \\
 \sum_{s=1,2} v_{s}(\boldsymbol{\p}) v^{\sharp}_{s}(\boldsymbol{\p}) =
\frac{\slashed{p} -m}{2E(\boldsymbol{\p})} =
\frac{p_\mu \gamma^\mu - m}{2E(\boldsymbol{\p})}, \,\,\,\,
E(\boldsymbol{\p}) = \sqrt{|\boldsymbol{\p}|^2 + m^2}.
\end{split}
\end{equation}

\begin{center}
{\small THE SOLUTIONS  $u_s(\boldsymbol{\p})$ AND $v_s(\boldsymbol{\p})$ IN THE STANDARD REPRESENTATION
(\ref{standardgamma})}
\end{center}

Now let us give the formulas for the fundamental solutions $u_s(\boldsymbol{\p}), v_s(-\boldsymbol{\p})$, 
$s=1,2$, and projections $P^\oplus, P^\ominus$
$E_+, E_-$, in the so called standard representation (\ref{standardgamma}) of the Dirac gamma
matrices. It is not necessary to start the whole analysis with unitary Mackey's induced representations 
using the other choice of the functions $\beta(p)$ corresponding
to the orbits $\mathscr{O}_{m,0,0,0}$ and $\mathscr{O}_{-m,0,0,0}$, which determines
the Hilbert spaces of solutions of the Dirac equation with the standard Dirac matrices
(\ref{standardgamma}). Indeed, in order to determine the corresponding projectors it is sufficient
to apply the adjoint homomorphism $C^{-1} (\cdot)C$, and in order to determine the corresponding solutions
$u_s(\boldsymbol{\p}), v_s(-\boldsymbol{\p})$ it is sufficient to apply the unitary operator of multiplication by $C$
\begin{multline}\label{standard,u,v}
u_s(\boldsymbol{\p}) = C \frac{1}{\sqrt{2}} \sqrt{\frac{E(\boldsymbol{\p}) + m}{2 E(\boldsymbol{\p})}}
\left( \begin{array}{c}   \chi_s + \frac{\boldsymbol{\p} \cdot \boldsymbol{\sigma}}{E(\boldsymbol{\p}) + m}\chi_s 
\\                                           
              \chi_s - \frac{\boldsymbol{\p} \cdot \boldsymbol{\sigma}}{E(\boldsymbol{\p}) + m}\chi_s                          \end{array}\right) \\ 
=  \sqrt{\frac{E(\boldsymbol{\p}) + m}{2 E(\boldsymbol{\p})}}
\left( \begin{array}{c}   \chi_s 
\\                                           
              \frac{\boldsymbol{\p} \cdot \boldsymbol{\sigma}}{E(\boldsymbol{\p}) + m} \chi_s                         \end{array}\right),
\\
v_s(\boldsymbol{\p}) =  C \frac{1}{\sqrt{2}} \sqrt{\frac{E(\boldsymbol{\p}) + m}{2 E(\boldsymbol{\p})}}
\left( \begin{array}{c}   \chi_s + \frac{\boldsymbol{\p} \cdot \boldsymbol{\sigma}}{E(\boldsymbol{\p}) + m} \chi_s
\\                                           
              -\big(\chi_s - \frac{\boldsymbol{\p} \cdot \boldsymbol{\sigma}}{E(\boldsymbol{\p}) + m}\chi_s \big)                          \end{array}\right)  \\ =
\sqrt{\frac{E(\boldsymbol{\p}) + m}{2 E(\boldsymbol{\p})}}
\left( \begin{array}{c}   \frac{\boldsymbol{\p} \cdot \boldsymbol{\sigma}}{E(\boldsymbol{\p}) + m} \chi_s
\\                                           
            \chi_s                            \end{array}\right)
\end{multline}
to the complete system of solutions in the chiral representation. For the corresponding
projectors in the standard representation (\ref{standardgamma}) we thus have
\begin{multline*}
P^\oplus(p) = C^{-1}\frac{1}{2}
\left( \begin{array}{cc} \boldsymbol{1}_2 & \beta(p)^{-2}  \\
                                                  \beta(p)^2 &  \boldsymbol{1}_2 \end{array}\right) C  \\ =
\frac{1}{2}
\left( \begin{array}{cc} \frac{m+ E(\boldsymbol{\p})}{m} \boldsymbol{1}_2 & 
-\frac{\boldsymbol{\p} \cdot \boldsymbol{\sigma}}{m}  \\
                           \frac{\boldsymbol{\p} \cdot \boldsymbol{\sigma}}{m} &  \frac{m- E(\boldsymbol{\p})}{m} \boldsymbol{1}_2 \end{array}\right), \,\,\,\,
p = (E(\boldsymbol{\p}), \boldsymbol{\p}) \in \mathscr{O}_{m,0,0,0},
\end{multline*}
(here with $\beta(p)$ equal (\ref{betaO_m,0,0,0})) and similarly for $P^\ominus(-E(\boldsymbol{\p}), \boldsymbol{\p})$
(with $\beta(p)$ equal (\ref{betaO_-m,0,0,0}) in the formula below)
\begin{multline*}
P^\ominus(p) = C^{-1}\frac{1}{2}
\left( \begin{array}{cc} \boldsymbol{1}_2 & -\beta(p)^{-2}  \\
                                                  -\beta(p)^2 &  \boldsymbol{1}_2 \end{array}\right) C  \\ =
\frac{1}{2}
\left( \begin{array}{cc} \frac{m - E(\boldsymbol{\p})}{m} \boldsymbol{1}_2 & 
-\frac{\boldsymbol{\p} \cdot \boldsymbol{\sigma}}{m}  \\
                           \frac{\boldsymbol{\p} \cdot \boldsymbol{\sigma}}{m} &  \frac{m+E(\boldsymbol{\p})}{m} \boldsymbol{1}_2 \end{array}\right), \,\,\,\,\,
p = (-E(\boldsymbol{\p}), \boldsymbol{\p}) \in \mathscr{O}_{-m,0,0,0}.
\end{multline*}
Of course, we have the analogous formulas for $E_{\pm}(\boldsymbol{p})$ but we have to remember that 
with the corresponding matrices $\alpha^i = \gamma^0\gamma^i$ in the standard representation (\ref{standardgamma}).
By construction the (Fourier transforms) $u_s(\boldsymbol{\p}), v_s(-\boldsymbol{\p})$ of solutions 
in the standard representation (\ref{standardgamma}) respect the analogous relations 
(\ref{u^+u=delta})-(\ref{covariantProj}).

\begin{center}
{\small ON THE UNITARY ISOMORPHISM $U$ OF SUBSECTION \ref{psiBerezin-Hida} FOR THE DIRAC FIELD}
\end{center}

Note that the unitary isomorphism operator $U$, defined by (\ref{isomorphismU})
in Subsection \ref{psiBerezin-Hida},
can be regarded as the operator of pointwise multiplication by the matrix
\begingroup\makeatletter\def\f@size{5}\check@mathfonts
\def\maketag@@@#1{\hbox{\m@th\large\normalfont#1}}%
\[
U(\boldsymbol{\p}) =
\frac{1}{2|p_0(\boldsymbol{\p})|}
\left( \begin{array}{cccccccc}   \overline{u_{1}^{1}(\boldsymbol{\p})} &  \overline{u_{1}^{2}(\boldsymbol{\p})} &
\overline{u_{1}^{3}(\boldsymbol{\p})} & \overline{u_{1}^{4}(\boldsymbol{\p})} & & 0 & &   \\
\overline{u_{2}^{1}(\boldsymbol{\p})} &  \overline{u_{2}^{2}(\boldsymbol{\p})} &
\overline{u_{2}^{3}(\boldsymbol{\p})} & \overline{u_{2}^{4}(\boldsymbol{\p})} & & & &   \\
& & & & v_{1}^{1}(\boldsymbol{\p}) & v_{1}^{2}(\boldsymbol{\p}) & 
v_{1}^{3}(\boldsymbol{\p}) & v_{1}^{4}(\boldsymbol{\p}) \\
& & 0 & & v_{2}^{1}(\boldsymbol{\p}) & v_{2}^{2}(\boldsymbol{\p}) & 
v_{2}^{3}(\boldsymbol{\p}) & v_{2}^{4}(\boldsymbol{\p})  \end{array}\right)
\]
\endgroup
acting on the element  
$\widetilde{\phi} \oplus (\widetilde{\phi}')^\flat \in \mathcal{H}_{m,0}^{\oplus} \oplus \mathcal{H}_{-m,0}^{\ominus \, \flat}$;
where the value $\big(\widetilde{\phi} \oplus (\widetilde{\phi}')^\flat \big)(|p_0(\boldsymbol{\p})|, \boldsymbol{\p})$ 
at $p = ((|p_0(\boldsymbol{\p})|, \boldsymbol{\p})) \in \mathscr{O}_{m,0,0,0}$ of 
$\widetilde{\phi} \oplus (\widetilde{\phi}')^\flat$ is
written as a column vector
\[
\left( \begin{array}{c} 
\widetilde{\phi}(|p_0(\boldsymbol{\p})|, \boldsymbol{\p}) \\
\big[(\widetilde{\phi}')^\flat(|p_0(\boldsymbol{\p})|, \boldsymbol{\p})\big]^T
\end{array}\right).
\]
Similarly, the inverse $U^{-1}$ of the isomorphism (\ref{isomorphismU}), Subsection \ref{psiBerezin-Hida}, 
can be regarded as the operator
of pointwise multiplication by the matrix
\begingroup\makeatletter\def\f@size{5}\check@mathfonts
\def\maketag@@@#1{\hbox{\m@th\large\normalfont#1}}%
\[
U^{-1}(\boldsymbol{\p}) =
2|p_0(\boldsymbol{\p})|
\left( \begin{array}{cccc}   
u_{1}^{1}(\boldsymbol{\p}) &  u_{2}^{1}(\boldsymbol{\p}) & 0 & 0   \\
u_{1}^{2}(\boldsymbol{\p}) &  u_{2}^{2}(\boldsymbol{\p}) & 0 & 0   \\
u_{1}^{3}(\boldsymbol{\p}) &  u_{2}^{3}(\boldsymbol{\p}) & 0 & 0   \\
u_{1}^{4}(\boldsymbol{\p}) &  u_{2}^{4}(\boldsymbol{\p}) & 0 & 0   \\
0 & 0 & \overline{v_{1}^{1}(\boldsymbol{\p})} & \overline{v_{2}^{1}(\boldsymbol{\p})} \\
0 & 0 & \overline{v_{1}^{2}(\boldsymbol{\p})} & \overline{v_{2}^{2}(\boldsymbol{\p})} \\
0 & 0 & \overline{v_{1}^{3}(\boldsymbol{\p})} & \overline{v_{2}^{3}(\boldsymbol{\p})} \\
0 & 0 & \overline{v_{1}^{4}(\boldsymbol{\p})} & \overline{v_{2}^{4}(\boldsymbol{\p})} \\
  \end{array}\right)
\]
\endgroup
with the value 
$\Big((\widetilde{\phi})_1 \oplus (\widetilde{\phi})_2 \oplus (\widetilde{\phi})_3 \oplus (\widetilde{\phi})_4\Big)(\boldsymbol{\p}) $ 
of the element 
\[
(\widetilde{\phi})_1 \oplus (\widetilde{\phi})_2 \oplus (\widetilde{\phi})_3 \oplus (\widetilde{\phi})_4 
\in \oplus L^{2}(\mathbb{R}^3; \mathbb{C}) =  L^{2}(\mathbb{R}^3; \mathbb{C}^4)
\]
regarded as a column
\[
\left( \begin{array}{c} 
(\widetilde{\phi})_1(\boldsymbol{\p}) \\
(\widetilde{\phi})_2(\boldsymbol{\p}) \\
(\widetilde{\phi})_3(\boldsymbol{\p}) \\
(\widetilde{\phi})_4(\boldsymbol{\p})
 \end{array}\right).  
\]
Note that 
\[
U(\boldsymbol{\p})U^{-1}(\boldsymbol{\p}) = \boldsymbol{1}_4, \,\,\,\,
U^{-1}(\boldsymbol{\p}) U(\boldsymbol{\p}) = 
\left( \begin{array}{cc} 
E_+(\boldsymbol{\p}) & \boldsymbol{0}_4 \\
\boldsymbol{0}_4 & E_-(\boldsymbol{\p})^T
 \end{array}\right).
\]
Note also that 
\[
\left( \begin{array}{cc} 
E_+(\boldsymbol{\p}) & 0 \\
0 & E_-(\boldsymbol{\p})^T
 \end{array}\right) \left( \begin{array}{c} 
\widetilde{\phi}(|p_0(\boldsymbol{\p})|, \boldsymbol{\p}) \\
\big[(\widetilde{\phi}')^\flat(|p_0(\boldsymbol{\p})|, \boldsymbol{\p})\big]^T
\end{array}\right) = \left( \begin{array}{c} 
\widetilde{\phi}(|p_0(\boldsymbol{\p})|, \boldsymbol{\p}) \\
\big[(\widetilde{\phi}')^\flat(|p_0(\boldsymbol{\p})|, \boldsymbol{\p})\big]^T
\end{array}\right)
\]
for $\widetilde{\phi} \oplus 
(\widetilde{\phi}')^\flat \in \mathcal{H}' = \mathcal{H}_{m,0}^{\oplus} \oplus \mathcal{H}_{-m,0}^{\ominus \, \flat}$,
which follows from (\ref{E_+Phi=Phi}) and (\ref{E_-Phi=Phi}).

\section{APPENDIX: Schwartz' spaces of convolutors $\mathcal{O}'_{C}$ 
and multipliers $\mathcal{O}_M$ of $\mathcal{S}$}\label{convolutorsO'_C}

Schwartz \cite{Schwartz} introduced the following linear function spaces 
(in this Appendix we use notation of Schwartz including his notation $\mathcal{E}$
for $\mathscr{C}^\infty(\mathbb{R}^n,; \mathbb{C})$ and its strong dual space $\mathcal{E}'$ of distributions with compact support,
which should not be misled with our notation $\mathscr{E}$ for a class of countably-Hilbert nuclear space-time test spaces
$\mathcal{S}(\mathbb{R}^4; \mathbb{C}^m)$ or $\mathcal{S}^{00}(\mathbb{R}^4; \mathbb{C}^m)$)
\begin{enumerate}
\item[]
$\mathcal{D} = \{\varphi \in \mathscr{C}^\infty(\mathbb{R}^n; \mathbb{C}), \,
\textrm{supp} \varphi \, \textrm{compact}  \}$,
\item[]
$\mathcal{S} = \mathcal{S}_{H_{(n)}}(\mathbb{R}^n; \mathbb{C}) = \mathcal{S}(\mathbb{R}^n; \mathbb{C}) 
= \{\varphi \in \mathscr{C}^\infty(\mathbb{R}^n; \mathbb{C}),
\forall \alpha, \beta \in \mathbb{N}_{0}^{n}: x^\alpha \partial^\beta \varphi \in \mathscr{C}_0  \}$,
\item[]
$\mathcal{D}_{L^p} = \{\varphi \in \mathscr{C}^\infty(\mathbb{R}^n; \mathbb{C}),
\forall \alpha \in \mathbb{N}_{0}^{n}: \partial^\alpha \varphi \in L^p  \}$ (Sobolev space $W^{\infty, p}$)
$1 \leq p < \infty$,
\item[]
$\mathcal{B} = \mathcal{D}_{L^{\infty}} = \{\varphi \in \mathscr{C}^\infty(\mathbb{R}^n; \mathbb{C}),
\forall \alpha \in \mathbb{N}_{0}^{n}: \partial^\alpha \varphi \in L^{\infty}  \}$,
\item[]
$\overset{\cdot}{\mathcal{B}} = \{\varphi \in \mathscr{C}^\infty(\mathbb{R}^n; \mathbb{C}),
\forall \alpha \in \mathbb{N}_{0}^{n}: \partial^\alpha \varphi \in \mathscr{C}_0  \}$,
\item[]
$\mathcal{O}_{C} = \{\varphi \in \mathscr{C}^\infty(\mathbb{R}^n; \mathbb{C}),
\exists k \in \mathbb{N}_0 \forall \alpha \in \mathbb{N}_{0}^{n}: \,  (1 + |x|^2)^{-k} \partial^\alpha 
\varphi \in \mathscr{C}_0 \}$ (very slowly increasing functions),
\item[]
$\mathcal{O}_{M} = \{\varphi \in \mathscr{C}^\infty(\mathbb{R}^n; \mathbb{C}),
\forall \alpha \in \mathbb{N}_{0}^{n} \exists k \in \mathbb{N}_0: \,  (1 + |x|^2)^{-k} \partial^\alpha 
\varphi \in \mathscr{C}_0 \}$ (slowly increasing functions),
\item[]
$\mathcal{E} = \mathscr{C}^\infty(\mathbb{R}^n; \mathbb{C})$;
\end{enumerate}
and their strong duals, which we will denote in this Appendix (after Schwartz \cite{Schwartz}) with the prime sign
$(\cdot)'$ 
\begin{enumerate}
\item[]
$\mathcal{D}'$ (distributions),
\item[]
$\mathcal{S}'$ (tempered distributions, denoted by us $\mathcal{S}(\mathbb{R}^n; \mathbb{C})^*$),
\item[]
$\mathcal{D}'_{L^p} = \{T \in \mathcal{D}',
\exists m \in \mathbb{N}_{0}: \,  T = \sum_{|\alpha| \leq m} \partial^\alpha f_\alpha \, \textrm{with} 
\, f_\alpha \in L^p  \}$,
\item[]
$\mathcal{O}'_{C} = \{T \in \mathcal{D}',
\forall k \in \mathbb{N}_{0} \exists m \in \mathbb{N}_{0}^{n}: \, 
(1 + |x|^2)^{k} T = \sum_{|\alpha| \leq m } \partial^\alpha f_{\alpha} \, \textrm{with} \, 
f_\alpha \in L^{\infty}  \}$ (rapidly decreasing distributions),
\item[]
$\mathcal{O}'_{M} = \{T \in \mathcal{D}',
\exists m \in \mathbb{N}_{0}^{n} \forall k \in \mathbb{N}_{0}: \,  
(1 + |x|^2)^{k} T = \sum_{|\alpha| \leq m } \partial^\alpha f_{\alpha} \, \textrm{with} \, 
f_\alpha \in L^{\infty}  \}$ (very rapidly decreasing distributions),
\item[]
$\mathcal{E}'$ (distributions with compact support).
\end{enumerate}
Here $\mathscr{C}_0$ is the space of continuous $\mathbb{C}$-valued functions on $\mathbb{R}^n$,
tending to zero at infinity.  

All these linear topological spaces together with the topology were constructed in \cite{Schwartz}, except the
space $\mathcal{O}_{C}$ -- the predual of the Schwartz convolutor algebra $\mathcal{O}'_{C}$
of rapidly decreasing distributions. 
The function space $\mathcal{O}_{C}$ together with its inductive limit topology such that 
 $\mathcal{O}'_{C}$ with the Schwartz operator topology of uniform convergence on bounded sets, 
becomes the strong dual of $\mathcal{O}_{C}$, has been determined by Horv\'ath. Namely, 
$\mathcal{O}'_{C} = \{T\in \mathcal{S}': T \, \textrm{extends uniquely to a continuous linear functional $\tilde{T}$
on  $\mathcal{O}_{C}$}  \}$, with the operator Schwartz topology of uniform convergence on bounded sets on
 $\mathcal{O}'_{C}$ coinciding with the strong dual topology on the space dual to  $\mathcal{O}_{C}$.

We have the following topological inclusions 
(with $E \subset F$ meaning that the topology of $E$ is finer than that of $F$):
\[
\left. \begin{array}{ccccccccccccccc}   & & 1 & \leq & p & \leq  & q & & & & & & & &  \\
 & & & & & & & & & & & & & &  \\
\mathcal{D} & \subset & \mathcal{S} & \subset & \mathcal{D}_{L^p} & \subset &  \mathcal{D}_{L^q} & \subset & \overset{\cdot}{\mathcal{B}} & \subset & \mathcal{B} & \subset & \mathcal{O}_{M} & \subset & \mathcal{E}     \\
\cap & \cap & \cap & \cap & \cap & \cap & \cap & \cap & \cap & \cap & \cap & \cap & \cap & \cap & \cap   \\
\mathcal{E}'& \subset & \mathcal{O}'_{C} & \subset & \mathcal{D}'_{L^p} & \subset & \mathcal{D}'_{L^q} & \subset & \overset{\cdot}{\mathcal{B}}'  & \subset & \mathcal{B}' & \subset & \mathcal{S}' & \subset & \mathcal{D}' 
\end{array}\right.,
\]
\[
\left. \begin{array}{ccccccccccccc}   
\mathcal{D} & \subset & \mathcal{S} & \subset & \mathcal{D}_{L^p} & \subset & \overset{\cdot}{\mathcal{B}} & \subset & \mathcal{O}_{C} & \subset & \mathcal{O}_{M} & \subset & \mathcal{E}     \\
\cap &  &  &  &  &  &  &  &  &  &  &  &   \cap   \\
\mathcal{E}'& \subset & \mathcal{O}'_{M} & \subset & \mathcal{O}'_{C} & \subset & \mathcal{D}'_{L^p} & \subset & \mathcal{D}'_{L^q} &  \subset & \mathcal{S}' & \subset & \mathcal{D}' \\
 & & & & & & & & & & & & \\
& & & & 1 & \leq & p & \leq  & q & & & & 
\end{array}\right.,
\]
compare \cite{Schwartz}, p. 420, or \cite{Horvath}, \cite{Larcher}, \cite{Kisynski}.

Therefore, elements of all indicated spaces (except the whole of $\mathcal{E} = \mathscr{C}^\infty$ and $\mathcal{D}'$)
\[
\mathcal{D}, \mathcal{S}, \mathcal{D}_{L^p}, 
\mathcal{E}',
\mathcal{D}'_{L^p}, \mathcal{O}'_{M}, 
\mathcal{O}'_{C},
\]  
can be naturally regarded as tempered distributions, i.e. as elements of $\mathcal{S}'$. 
But we should emphasize that the 
topology of each individual space is strictly stronger than the topology induced from the topology 
of the strong dual space $\mathcal{S}'$ of tempered distributions. 

Let us recall that the Fourier transform $\mathscr{F}$ maps isomorphically $\mathcal{S}$
onto $\mathcal{S}$. The Fourier transform is defined on the space of tempered distributions $\mathcal{S}'$
through the linear transpose (dual) of the Fourier transform on $\mathcal{S}$, which by the general
properties of the linear transpose \cite{treves} defines a continuous linear isomorphism 
$\mathcal{S}' \rightarrow \mathcal{S}'$ for the strong dual topology on $\mathcal{S}'$, and 
denoted by the same symbol $\mathscr{F}$. 

Because the elements of the linear spaces 
\[
\mathcal{D}, \mathcal{S}, \mathcal{D}_{L^p}, 
\mathcal{E}',
\mathcal{D}'_{L^p}, \mathcal{O}'_{M}, 
\mathcal{O}'_{C},
\]  
are naturally identified with elements of $\mathcal{S}'$ then
in particular the Fourier transform is a well defined
liner map on these spaces (although in general it leads us out of the particular space in question). 

Recall further that the operator $M_S$ of multiplication by any element $S$ of $\mathcal{O}_M$
maps isomorphically $\mathcal{S} \rightarrow \mathcal{S}$. Thus elements $S$ of $\mathcal{O}_M$
are naturally idetified with continuous multiplication operators $M_S$ mapping continuously 
$\mathcal{S}$ into $\mathcal{S}$,
i.e. with elements of $\mathscr{L}(\mathcal{S}, \mathcal{S})$. Therefore, we can introduce 
on $\mathcal{O}_M$ after 
Schwartz \cite{Schwartz} the topology of uniform convergence on bounded sets induced from 
$\mathscr{L}(\mathcal{S}, \mathcal{S})$. 

Further, recall that translation 
\[
T_b: \varphi \rightarrow T_b\varphi, \,\,\,\,\,\,\,
T_b\varphi(x) \overset{\textrm{df}}{=} \varphi(x+b)
\]
maps isomorphically $\mathcal{S} \rightarrow \mathcal{S}$. Again by duality we define 
\[
S \ast \varphi(x) \overset{\textrm{df}}{=} \langle S, T_{-x} \varphi \rangle = S(T_{-x} \varphi),
\]
where $\langle \cdot, \cdot \rangle$ stands for the canonical bilinear form on 
$\mathcal{S}' \times \mathcal{S} = \mathcal{S}^* \times \mathcal{S}$, i.e. the pairing
defined by taking the value of the functional. 
It turns out that if $S \in \mathcal{S}'$ then the operator 
\[
C_S: \varphi \mapsto S \ast \varphi = C_S(\varphi)
\] 
of convolution with $S \in \mathcal{S}'$ corresponding to $S$ maps continuously 
$\mathcal{S} \rightarrow \mathcal{O}_C$, i.e. $C_S \in \mathscr{L}(\mathcal{S}, \mathcal{O}_C)$. 
Moreover, $S \in \mathcal{O}'_{C}$ if and only if
the corresponding convloution operator $C_S \in \mathscr{L}(\mathcal{S}, \mathcal{S})$,
i.e. if and only if $C_S$ maps (continuously) the Schwartz space $\mathcal{S}$
into itself. Moreover, if $S \in \mathcal{O}'_{C}$ then 
$C_{\tilde{S}} \in \mathscr{L}(\mathcal{O}_{C}, \mathcal{O}_{C})$, where $\tilde{S}$
is the unique extension of the functional $S$ on $\mathcal{S}$ over $\mathcal{O}_C$. 

Therefore, we can, again after Schwartz \cite{Schwartz}, introduce the topology on
$\mathcal{O}'_{C}$ induced from the topology of uniform convergence on bounded sets 
on $\mathscr{L}(\mathcal{S}, \mathcal{S})$. 

These are the Schwartz operator topologies on 
$\mathcal{O}_M$ and $\mathcal{O}'_{C}$. These spaces become nuclear with these topologies,
(quasi-) complete and barreled. For their definitions as induced by systems of semi-norms we refer
the reader to the classic work \cite{Schwartz} or \cite{Horvath}, \cite{Larcher}, \cite{Kisynski}. 
In fact all indicated spaces are barreled,
although all of them are endowed with topology strictly stronger than the topology induced by the strong dual topology of
$\mathcal{S}'$ (for all of them except the whole of the space $\mathcal{E}$ and $\mathcal{D}'$
which cannot be naturally included into $\mathcal{S}'$).  

Let us introduce, in addition to the convolution operator $C_S$, a closely
related operator $C_{S}^{+}$, defined similarly by
\[
C_{S}^{+}(\varphi)(x) \overset{\textrm{df}}{=} \langle S, T_{x} \varphi \rangle = S(T_{x} \varphi).
\]
Similarly, as the operator $C_S$, also the operator  $C_{S}^{+} \in \mathscr{L}(\mathcal{S}, \mathcal{S})$   if and only if 
$S \in \mathcal{O}'_{C}$.

\begin{twr*}
Let $\mathcal{S}'$ be endowed with the strong dual topology, and $\mathcal{O}_{M}$, $\mathcal{O}'_{C}$
with the Schwartz' operator topologies defined as above. 
On the space $\mathcal{S}'$ we can define the operation of multiplication by $S\in \mathcal{O}_M$ 
through the linear transpose of the map $M_S$, which maps continuously 
$\mathcal{S}' \rightarrow \mathcal{S}'$ and defines a bilinear hypocontinuous multiplication map 
$\mathcal{S}' \times \mathcal{O}_{M} \rightarrow \mathcal{S}'$.
Similarly, on the space $\mathcal{S}'$ we can define the operation of convolution by $S\in \mathcal{O}'_{C}$ 
through the linear transpose of the map $C_{S}^{+}$, which maps continuously 
$\mathcal{S}' \rightarrow \mathcal{S}'$ and defines a bilinear hypocontinuous convolution map 
$\mathcal{S}' \times \mathcal{O}'_{C} \rightarrow \mathcal{S}'$.  
\end{twr*}
Compare \cite{Schwartz}, Thm. X and Thm. XI, Chap. VII, \S 5, pp. 245-248.

On the space $\mathcal{O}_M$ we can define the commutative  multiplication operation $S_1 \cdot S_2$:
\[
\mathcal{O}_{M}  \times  \mathcal{O}_{M} \ni S_1 \times S_2 \mapsto  S_1 \cdot S_2 \in \mathcal{O}_M
\]
through the composition
of the corresponding multiplication operators $M_{S_1} \circ M_{S_2} = M_{S_2} \circ M_{S_1} = M_{S_1 \cdot S_2}$,
which corresponds to the ordinary pointwise multiplication of functions $f_1, f_2 \in \mathcal{O}_M$
representing the corresponding tempered distributions $S_1, S_2 \in \mathcal{O}_M \subset \mathcal{S}'$.
Similarly, we can define commutative convolution operation  $S_1 \ast S_2$:
\[
\mathcal{O}'_{C}  \times  \mathcal{O}'_{C} \ni S_1 \times S_2 \mapsto  S_1 \ast S_2 \in \mathcal{O}'_{C}
\]
through the composition
of the corresponding convolution operators $C_{S_1} \circ C_{S_2} = C_{S_2} \circ C_{S_1} = C_{S_1 \ast S_2}$,
which coincides with the ordinary convolution $f_1 \ast f_2$ of functions $f_1, f_2$
if the tempered distributions $S_1, S_2, S_1 \ast S_2  \in \mathcal{O}_M \subset \mathcal{S}'$
can be represented by ordinary functions $f_1, f_2, f_1 \ast f_2$.

\begin{twr*}
\begin{enumerate}
\item[1)]
The multiplication $S_1 \cdot S_2$ operation is not only hypocontinuous as a map 
$\mathcal{O}_{M}  \times  \mathcal{O}_{M} \rightarrow \mathcal{O}_{M}$,
but likewise (jointly) continuous.
\item[2)]
The convolution $S_1 \ast S_2$ operation is not only hypocontinuous as a map 
$\mathcal{O}'_{C}  \times  \mathcal{O}'_{C} \rightarrow \mathcal{O}'_{C}$,
but likewise (jointly) continuous.
\end{enumerate}
\end{twr*}
Compare \cite{Schwartz}, Remark on page 248, or \cite{Larcher}, Proposition 5. 

Similarly, we define a function to be a multiplier (convolutor) of the indicated function space if the corresponding 
multiplication (convolution) operator maps the space continuously into itself. Similarly, we define by duality the
multipliers (convolutors) of the strong dual of the indicated function space.  

Recall the Schwartz' \emph{Fourier exchange Theorem} (\cite{Schwartz}, Chap. VII.8, Thm. XV)  
\begin{twr*}
If the linear topological spaces $\mathcal{O}_{M}$ and $\mathcal{O}'_{C}$ are endowed with the 
Schwartz' operator topologies, defined as above, then the 
Fourier transform $\mathscr{F}$, regarded as a map on $\mathcal{S}'$ restricted to $\mathcal{O}'_{C}$, 
transforms isomorphically $\mathcal{O}'_{C}$ onto $\mathcal{O}_{M}$, and the following formula
\[
\mathscr{F}(S \ast T) = \mathscr{F}S \cdot \mathscr{F}T,
\] 
is valid for any $S \in \mathcal{O}'_{C}$ and $T \in \mathcal{S}'$. 
\end{twr*}

All cited results in this Appendix are essentially contained in the classic work \cite{Schwartz}
of L. Schwartz. Some stated above results are only remarked there or sometimes formulated without (detailed) proofs,
but the reader will find all details in the subsequent literature on distribution theory. 
In particular a topological supplement to the proof of the Fourier exchange Theorem XV (Chap. VII.8 \cite{Schwartz}) 
can be found e.g. in \cite{Kakita}, but a full and systematic treatment of this theorem can be found
in \cite{Kisynski}, where a detailed construction of the predual $\mathcal{O}_{C}$ of $\mathcal{O}'_{C}$
is also given. For further details on the indicated spaces and their multipliers and convolutors
compare \cite{Schwartz}, \cite{Yosida}, \cite{Larcher}, \cite{Larcher-Wengenroth}, \cite{Horvath}. 

\begin{rem*}
Note that the multiplication $\cdot$ map 
$\mathcal{O}_{M}  \times  \mathcal{O}_{M} \rightarrow \mathcal{O}_{M}$ (as well as the convolution $\ast$ map: 
$\mathcal{O}'_{C}  \times  \mathcal{O}'_{C} \rightarrow \mathcal{O}'_{C}$)
is not hypocontinuous with respect to the topology on $\mathcal{O}_{M}$ (resp. on $\mathcal{O}'_{C}$)
induced from the strong dual topology on $\mathcal{S}'$. Indeed if it was hypocontinuous then by the well known
extension theorem, compare  the Proposition of Chap. III, \S 5.4, p.90 in \cite{Schaefer}, a hypocontinuous extension
of the multiplication to a product $\mathcal{S}' \times \mathcal{S}' \rightarrow \mathcal{S}'$ 
(resp. extension of the convolution) 
could have been constructed, which coincides with the ordinary function pointwise multiplication 
(resp. convolution) product if the distributions can be represented by functions. Because $\mathcal{S'}$ is the strong dual of 
a reflexive Fr\'echet space $\mathcal{S}$, then by Thm. 41.1
of \cite{treves}, we could have obtained in this way a continuous extension of the product of distributions
respecting the natural algebraic laws under multiplication and differentiation and coinciding with the ordinary 
pointwise multiplication (resp. convolution) product of functions whenever the distributions coincide with ordinary functions. 
But this would be in contradiction to 
the classic result of Schwartz, which says that such extension is impossible, compare \cite{Schwart-mult.impossible}
or \cite{Schwartz}, Chap. V.1. 
Similarly, we can show that the extension of the convolution product on the convolution algebra of 
$\mathcal{S}^{0}(\mathbb{R}^n; \mathbb{C})$
is not hypocontinuous with respect to the topology inherited from the strong dual 
$\mathcal{S}^{0}(\mathbb{R}^n; \mathbb{C})^*$, because of the topological inclusions
$\mathcal{S}^{0}(\mathbb{R}^n; \mathbb{C}) \subset \mathcal{S}(\mathbb{R}^n; \mathbb{C})$ and
$\mathcal{S}(\mathbb{R}^n; \mathbb{C})^* \subset \mathcal{S}^{0}(\mathbb{R}^n; \mathbb{C})^*$,
with the topology on $\mathcal{S}^{0}(\mathbb{R}^n; \mathbb{C})$ coinciding with that inherited fram
$\mathcal{S}(\mathbb{R}^n; \mathbb{C})$, compare Subsection \ref{SA=S0}. Equivalently: 
the pointwise multiplication product defined on the multiplier algebra of 
$\mathcal{S}^{00}(\mathbb{R}^n; \mathbb{C})$
is not hypocontinuous with respect to the topology inherited from the strong dual 
$\mathcal{S}^{00}(\mathbb{R}^n; \mathbb{C})^*$.
\end{rem*}

\section{APPENDIX: A Generalization of Mackey's theory}\label{PartIIMackey}

\subsection{Preliminaries}\label{pre}

It should be stressed that the analysis we give here is inapplicable for general linear spaces with indefinite inner product. We are concerned with non-degenerate, decomposable and complete inner product spaces in the terminology of 
\cite{Bog}, which have been called Krein spaces in \cite{DutFred}, \cite{Stro}, \cite{Bog} and \cite{Wawr} for the reasons we explain below. They emerged naturally in solving physical problems concerned with quantum mechanics (\cite{Dirac}, \cite{Pauli}) and quantum field theory (\cite{Gupta}, \cite{Bleuler}) in quantization of electromagnetic field and turned up generally to be very important (and even seem indispensable) in construction of quantum fields with non-trivial gauge freedom. Similarly we have to emphasize that we are not dealing
with general unitary (i. e. preserving the indefinite inner product in Krein space) representations of the double cover 
$\mathfrak{G} = T_{4} \circledS SL(2, \mathbb{C})$ of the Poincar\'e group, but only with the exceptional representations
of {\L}opusza\'nski-type, which naturally emerge in construction of the free photon field, which have a rather exceptional structure of induced representations, and allow non-trivial analytic constructions of tensoring and decomposing, which 
is truly exceptional among Krein-unitary (preserving the indefinite product) representations in Krein spaces.

The non-degenerate, decomposable and complete indefinite inner product space $\mathcal{H}$, hereafter called \emph{Krein space}, may equivalently be described as an ordinary Hilbert space $\mathcal{H}$ with an ordinary strictly positive inner product $(\cdot, \cdot)$, together with a distinguished self-adjoint (in the ordinary Hilbert space sense) fundamental symmetry (Gupta-Bleuler operator) $\mathfrak{J} = P_+ - P_- $, where $P_+$ and $P_-$ are ordinary self-adjoint (with respect to the Hilbert space inner product $(\cdot, \cdot)$)) projections such that their sum is the identity operator: $P_+ + P_- = I$. The indefinite inner product is given by $(\cdot, \cdot)_{\mathfrak{J}} = (\mathfrak{J} \cdot, \cdot) = (\cdot, \mathfrak{J} \cdot)$. 
Recall that in our previous paper \cite{Wawr} the indefinite product was designated by $(\cdot, \cdot)$ and the ordinary Hilbert space inner product associated with the fundamental symmetry $\mathfrak{J}$ was designated by $(\cdot, \cdot)_{\mathfrak{J}}$. The indefinite and the associated definite inner product play symmetric roles in the sense that one may start with a fixed indefinite inner product in the Krein space and construct the Hilbert space associated with an admissible fundamental symmetry, or vice versa: one can start with a fixed Hilbert space and for every fundamental symmetry construct the indefinite inner product in it, both approaches are completely equivalent provided the fundamental symmetry being admissible (in the sense of \cite{Stro}) and fixed. We hope the slight change of notation will not cause any serious misunderstandings and is introduced because our analytical arguments will be based on the ordinary Hilbert space properties, so will frequently refer to the standard literature on the subject, so we designated the ordinary strictly definite inner product by $(\cdot, \cdot)$ which is customary.

Let an operator $A$ in $\mathcal{H}$ be given. The operator $A^{\dag}$ in $\mathcal{H}$ is called Krein-adjoint
of the operator $A$ in $\mathcal{H}$ in case it is adjoint in the sense of the indefinite inner product:   
$(Ax, y)_{\mathfrak{J}} = (\mathfrak{J} Ax, y) = (\mathfrak{J} x, A^{\dag}y) = (x, A^{\dag}y)_{\mathfrak{J}}$
for all $x, y \in \mathcal{H}$, or equivalently $A^{\dag} = \mathfrak{J} A^{*}\mathfrak{J}$, where $A^{*}$ is the ordinary adjoint operator with respect to the definite inner product $(\cdot, \cdot)$. The operator $U$ and its inverse $U^{-1}$ isometric with respect to the indefinite product $(\cdot, \cdot)_{\mathfrak{J}}$, e. g. 
$(Ux, Uy)_{\mathfrak{J}} = (x, y)_{\mathfrak{J}}$ for all $x, y \in \mathcal{H}$ (same for $U^{-1}$),
equivalently $UU^{\dag} = U^{\dag}U = I$, will also be called 
unitary (sometimes $\mathfrak{J}$-unitary or Krein-unitary) trusting to the context or explanatory remarks to make clear what is meant in each instance: unitarity for the indefinite inner product or the ordinary unitarity for the strictly definite Hilbert space inner product.

In particular we may consider $\mathfrak{J}$-\emph{symmetric} representations $x \mapsto A_x$ of involutive algebras,
i. e. such that $x^{*} \mapsto {A_x}^{\dag}$, where $(\cdot) \mapsto (\cdot)^{*}$ is the involution in the algebra in question. A fundamental role for the spectral analysis in Krein spaces is likewise played by commutative (Krein) self-adjoint, or $\mathfrak{J}$-symmetric weakly closed subalgebras. However their structure is far from being completely described, with the exception of the special case when the rank of $P_+$ or $P_-$ is finite dimensional (here the analysis is complete and was done by Neumark). Even in this particular case a unitary representation of a separable locally compact group in the Krein space, although reducible, may not in general be decomposable, compare \cite{Neumark1, Neumark2, Neumark3, Neumark4}.

In case the dimension of the rank 
$\mathcal{H}_+ = P_{+} \mathcal{H}$ or $\mathcal{H}_{-} = P_{-} \mathcal{H}$ of $P_+$ or $P_-$ is finite we get the spaces analysed by Pontrjagin, Krein and Neumark, compare e. g. \cite{Pontrjagin}, \cite{Krein} and the literature in \cite{Bog}.

The circumstance that the Krein space may be defined as an ordinary Hilbert $\mathcal{H}$ space with a distinguished non-degenerate fundamental symmetry (or Gupta-Bleuler operator) $\mathfrak{J} = \mathfrak{J}^{*}$, $\mathfrak{J}^2 = I$ in it , say a pair $(\mathcal{H}, \mathfrak{J})$, allows us to extend the fundamental
analytical constructions on a wide class of induced Krein-isometric representations of $\mathfrak{G} = T_{4} \circledS SL(2, \mathbb{C})$  in Krein spaces. In particular we may define a Krein-isometric representation of $T_{4} \circledS SL(2, \mathbb{C})$ induced by a Krein-unitary representation of a subgrup $H$ corresponding to a particular class of 
$SL2,\mathbb{C})$-orbits  on the dual group $\widehat{T_4}$ of $T_4$ (in our case we consider the class corresponding to the representation with the spectrum of the four-momenta concentrated on the ``light cone'') 
word for word as in the ordinary Hilbert space by replacing the representation of the subgroup $H$ by a Krein-unitary representation $L$ in a Krein space $(\mathcal{H}_L, \mathfrak{J}_L)$. This leads to a Krein-isometric representation 
$U^L$ in a Krein space $(\mathcal{H}^L , \mathfrak{J}^L)$ (see Sect. \ref{def_ind_krein}). Application of  
Lemma \ref{lop_ind_1}, Section \ref{lop_ind}, leads to the ordinary direct integral $\mathcal{H} = \int_{\mathfrak{G}/H} \, \mathcal{H}_q \, d\mu_{\mathfrak{G}/H}(q)$ of Hilbert spaces $\mathcal{H}_q = \mathcal{H}_L$ over the coset measure space $\mathfrak{G}/H$ with the measure induced by the Haar measure on $\mathfrak{G}$. One obtains in this manner the Krein space $(\mathcal{H}, U^{-1}\mathfrak{J}^L U)$ given by the ordinary Hilbert space $\mathcal{H}$ equal to the  above mentioned direct integral of the ordinary Hilbert spaces 
$\mathcal{H}_q$ all of them equal to $\mathcal{H}_L$ together with the fundamental symmetry $ \mathfrak{J} = U^{-1}\mathfrak{J}^L U$ equal to the ordinary direct integral $\int_{\mathfrak{G}/H} \, \mathfrak{J}_q \, d\mu_{\mathfrak{G}/H}(q)$ of fundamental symmetries $\mathfrak{J}_q = \mathfrak{J}_L$ as operators in $\mathcal{H}_q = \mathcal{H}_L$ and with the  representation $U^{-1}U^LU$ of $\mathfrak{G}$ in the Krein space $(\mathcal{H}, \mathfrak{J})$ (and  $U$ given by a completely analogous formula as that in Lemma \ref{lop_ind_1} of Section \ref{lop_ind}) of Wigner's form  
\cite{Wigner_Poincare} (imprimitivity system). 

This is the case for the indecomposable (although reducible) representation of 
$\mathfrak{G} = T_{4} \circledS SL(2, \mathbb{C})$ constructed by 
{\L}opusza\'nski with $H = T_4 \cdot G_\chi$, with the "small" subgroup $G_\chi \cong \tilde{E_2}$ of $SL(2, \mathbb{C})$ corresponding to the ``light-cone'' orbit in the spectrum of four-momenta operators. One may give to it the form of  representation $U^{-1}U^L U$ equivalent to an induced representation $U^L$, because the representors of the normal factor (that is of the translation subgroup $T_4$) of the semidirect product $T_{4} \circledS SL(2, \mathbb{C})$ as well as their generators, i. e. four-momentum operators $P_0, \ldots , P_3$, commute with the fundamental symmetry $\mathfrak{J} = \int_{\mathfrak{G}/H} \, \mathfrak{J}_q \, d\mu_{\mathfrak{G}/H}(q)$, so that all of them are not only $\mathfrak{J}$-unitary but unitary with respect to the ordinary Hilbert space inner product (and their generators $P_0, \ldots, P_3$ are not only Krein-self-adjoint but also self-adjoint in the ordinary sense with respect to the ordinary definite inner product of the Hilbert space $\mathcal{H}$), so the algebra generated by $P_0, \ldots , P_3$ leads to the ordinary direct integral decomposition with the decomposition corresponding to the ordinary spectral measure, contrary to what happens for general Krein-selfadjoint commuting operators in Krein space $(\mathcal{H}, \mathfrak{J})$ (for details see Sect. \ref{lop_ind}). This in case of 
$\mathfrak{G} = T_{4} \circledS SL(2, \mathbb{C})$, gives to the representation 
$U^{-1}U^L U$ of $\mathfrak{G}$ the form of Wigner \cite{Wigner_Poincare} (viz. a
\emph{system of imprimitivity} in mathematicians' parlance) with the only difference that $L$ is not unitary but 
Krein-unitary in $(\mathcal{H}_L , \mathfrak{J}_L)$.   

Another gain we have thanks to the above mentioned circumstance is that we can construct tensor product 
$(\mathcal{H}_1, \mathfrak{J}_1) \otimes (\mathcal{H}_2, \mathfrak{J}_2)$ of Krein spaces
$(\mathcal{H}_1, \mathfrak{J}_1)$ and $(\mathcal{H}_2, \mathfrak{J}_2)$ as $(\mathcal{H}_1 \otimes \mathcal{H}_2, \mathfrak{J}_1 \otimes \mathfrak{J}_2)$ where in the last expression we have the ordinary tensor products of Hilbert spaces and operators in Hilbert spaces (compare Sect. \ref{kronecker}).

Similarly having any two such ($\mathfrak{J}_1$- and  $\mathfrak{J}_2$-)isometric representations $U^L$ and $U^M$ induced by ($\mathfrak{J}_L$ and $\mathfrak{J}_M$-unitary) representations $L$ and $M$ of subgroups $G_1$ and $G_2$  in Krein spaces $(\mathcal{H}_L, \mathfrak{J}_L)$ and $(\mathcal{H}_M, \mathfrak{J}_M)$ respectively we may construct the tensor product $U^L \otimes U^M$ of Krein-isometric representations in the tensor product Krein space $(\mathcal{H}_1, \mathfrak{J}_1) \otimes (\mathcal{H}_2, \mathfrak{J}_2)$, which is likewise ($\mathfrak{J}_1 \otimes \mathfrak{J}_2$-)isometric. It turns out that the Kronecker product  $U^L \times U^M$ and $U^{L \times M}$ are (Krein-)unitary equivalent (see Sect. \ref{kronecker})
as representations of $\mathfrak{G} \times \mathfrak{G}$. Because the tensor product $U^L \otimes U^M$ as a representation of $\mathfrak{G}$ is the restriction of the Kronecker product $U^L \times U^M$ to the diagonal subgroup of 
$\mathfrak{G} \times \mathfrak{G}$ we may analyse the representation $U^L \otimes U^M$ by analysing the restriction of the induced representation $U^{L \times M}$ to the diagonal subgroup exactly as in the Mackey theory of induced representations in Hilbert spaces.  Although in general for $\mathfrak{J}_1 \otimes \mathfrak{J}_2$-unitary representations in Krein space $(\mathcal{H}_1, \mathfrak{J}_1) \otimes (\mathcal{H}_2, \mathfrak{J}_2)$ ordinary decomposability breaks down, we can nonetheless still decompose the representation $U^{L \times M}$ restricted to the diagonal into induced representations which, by the above mentioned  Krein-unitary equivalence, gives us a decomposition of the tensor product representation 
$U^L \otimes U^M$  of $\mathfrak{G}$.  Indeed, it turns out that the whole argument of Mackey \cite{Mackey} preserves its validity and effectiveness in the construction of  decomposition of tensor product of induced representations for the case in which the representations $L$ and $M$ of the subgrups $G_1$ and $G_2$ are replaced with (specific) unitary (or $\mathfrak{J}_L$- and $\mathfrak{J}_M$-unitary) representations in Krein spaces $(\mathcal{H}_L, \mathfrak{J}_L )$ and $(\mathcal{H}_M, \mathfrak{J}_M )$ respectively. We give details on the subject below in Section \ref{subgroup}.  Because the 
{\L}opusza\'nski representation is (Krein-unitary equivalent to) an induced representation in a Krein space (Sect. \ref{lop_ind}), we can decompose the tensor product of {\L}opusza\'nski representations. The specific property
of the group $T_{4} \circledS SL(2, \mathbb{C})$ is that  this decomposition  may be performed explicitly into indecomposable sub-representations.  

The Krein-isometric induced representations of $T_4 \circledS SL(2, \mathbb{C})$ 
which we describe here cover all representations important
for QFT. All the representations which act on single particle states of local fields (including zero mass gauge fields) have three important properties: 1) They are strongly continuous on a common dense invariant 
subdomain. 2) Translations commute with the fundamental symmetry
$\mathfrak{J}$, so that translations are unitary with respect to the Hilbert space inner product as well as are Krein-unitary, and thus compose ordinary (strongly continuous) unitary representation of the translation subgroup. 3) The representations are ``locally'' bounded with respect to the joint spectrum of translation generators in the sense (\ref{circumstance2}) (see the beginning of Section \ref{constr-of-VF}).

Nonetheless the relevant representations, or the associated imprimitivity systems (e.g. {\L}opusza\'nski representation) are unbounded, and require a special care in the correct definition of the Kronecker product and moreover contain analytic subtleties which could have been omitted in the original Mackey theory.
The other difference in comparison to the original Mackey theory is that we exploit (and prove)
a decomposition/disintegration theorem for measures which are not finite, which makes the proof much longer
in comparison to Mackey's proof. In principle we could have confine ourselves after Mackey
to decomposition of finite measures (much easier). However the representations encountered in QFT are naturally related to Poincar\'e invariant measures which are not finite. Avoiding them by utilizing finite measures would not be very economical for a physicist, because in further computations he had to recover then the 
``Clebsch-Gordan'' coefficients relating obtained decompositions to the original representations naturally connected with infinite invariant measures.

\begin{rem}
Let us emphasize that here ``continuity'', ``density'', ``boundedness'', and other standard analytic notions,
as the ``closure of a densely defined operator'' or ``weak''or ``strong'' topologies in the algebra of bounded operators,  
refer to the ordinary Hilbert space norm and definite Hilbert space inner product in $\mathcal{H}$ of the Krein space 
$(\mathcal{H}, \mathfrak{J})$ in question. We are mainly concerned with the Lie group 
$\mathfrak{G} = T_{4} \circledS SL(2, \mathbb{C})$ but the general theory of induced representations in Krein spaces presented here is valid for general separable locally compact topological groups $\mathfrak{G}$. Thus separability and local compactness of $\mathfrak{G}$ is assumed to be valid throughout the whole paper whenever the identification
$\mathfrak{G} = T_{4} \circledS SL(2, \mathbb{C})$ is not explicitly stated.
\label{pre.1}
\end{rem}

\subsection{Definition of the induced representation $U^L$ in Krein 
space $(\mathcal{H}^L , \mathfrak{J}^L )$}\label{def_ind_krein}

Here by a Krein-unitary and strongly continuous representation $L: G \ni x \mapsto L_x$ of a separable 
locally compact group $G$ we shall mean a homomorphism of $G$ into the group of all (Krein-)unitary transformations of some separable Krein space $(\mathcal{H}_L, \mathfrak{J}_L)$ (i. e. with separable Hilbert space $\mathcal{H}_L$) onto itself which is:

\begin{enumerate}

\item[(a)]

Strongly continuous: for each $\upsilon \in \mathcal{H}_L$ the function $x \mapsto L_x \upsilon$ is continuous 
with respect to the ordinary strictly definite Hilbert space norm $\| \upsilon \| = \sqrt{(\upsilon, \upsilon)}$
in $\mathcal{H}_L$.

\item[(b)]
Almost uniformly bounded: there exist a compact neighbourhood $V$ of unity $e \in G$ such that the set
$\| L_{x} \|$, $x \in V \subset G$ is bounded or, what is the same thing, that
the set $\| L_{x} \|$ with $x$ ranging over a compact set $K$ is bounded for every compact 
subset $K$ of $G$. 

\end{enumerate}
Because the strong operator topology in $\mathcal{B}(\mathcal{H}_L)$ is stronger than the weak operator topology then for each $\upsilon , \varphi \in \mathcal{H}_L$ the function $x \mapsto (L_x \upsilon , \varphi )$ is continuous on $G$. One point has to be noted: because the range and domain of each $L_x$ equals 
$\mathcal{H}_L$, which as a Krein space $(\mathcal{H}_U, \mathfrak{J}_U)$ is closed and non-degenerate, then by Theorem 3.10 of \cite{Bog} each $L_x$ is continuous i. e. bounded with respect to the Hilbert space norm $\| \cdot \|$ in 
$\mathcal{H}_L$, and each $L_x$ indeed belongs to the algebra $\mathcal{B}(\mathcal{H}_L)$
of bounded  operators in the Hilbert space $\mathcal{H}_L$ (which is non-trivial as an $\mathfrak{J}_L$-isometric densely defined operator in the Krein space $(\mathcal{H}_L, \mathfrak{J}_L)$ may be discontinuous, as we will see in this Section, compare also \cite{Bog}). We also could immediately refer to a theorem which says that Krein-unitary operator is continuous, i. e. Hilbert-space-norm bounded (compare Theorem 4.1 in \cite{Bog}).

Besides in this paper will be considered a very specific class of Krein-isometric representations $U$ of $\mathfrak{G}$ in Krein spaces, to which the induced representations of $\mathfrak{G}$ in Krein spaces, hereby defined, belong. Namely
here by a Krein-isometric and strongly continuous representation  of a separable 
locally compact group $\mathfrak{G}$ we shall mean a homomorphism $U: \mathfrak{G} \ni x \mapsto U_x$ of $\mathfrak{G}$ into a group of Krein-isometric and closable operators of some separable Krein space $(\mathcal{H}, \mathfrak{J})$
with dense common domain $\mathfrak{D}$ equal to their common range in $\mathcal{H}$ and such that  
\begin{enumerate}

\item[]

$U$ is strongly continuous on the common domain $\mathfrak{D}$: for each $f \in \mathfrak{D} \subset \mathcal{H}$ the function $x \mapsto U_x f$ is continuous with respect to the ordinary strictly definite Hilbert space norm 
$\| f \| = \sqrt{(f, f)}$ in $\mathcal{H}$.  

\end{enumerate}

Let $H$ be a closed subgroup of a separable locally compact group $\mathfrak{G}$. In the applications
we have in view\footnote{E. g. in decomposing tensor products of the representations of the double cover 
$\mathfrak{G}$ of the Poincar\'e group in Krein spaces encountered in QFT} 
the right $H$-cosets, i. e. elements of $\mathfrak{G}/H$, are exceptionally regular,
and have a ``measure product property''. 
Namely every element (with a possible exception of a subset of $\mathfrak{G}$ of 
Haar measure zero) $\mathfrak{g} \in \mathfrak{G}$ can be uniquely represented as a product
$\mathfrak{g} = h \cdot q$, where $h \in H$ and $q \in Q \cong \mathfrak{G}/H$ with a subset $Q$ 
of $\mathfrak{G}$ which is not only measurable but, outside a null set, is a sub-manifold of 
$\mathfrak{G}$, such that $\mathfrak{G}$ is the product 
$H \times \mathfrak{G}/H$ measure space, with the regular Baire measure space structure on $\mathfrak{G}/H$ 
associated to the canonical locally compact topology on $\mathfrak{G}/H$ induced by the natural
projection $\pi: \mathfrak{G} \mapsto \mathfrak{G}/H$ and with the ordinary right Haar measure space
structure $(H, \mathscr{R}_H, \mu_H)$ on $H$, which is known to be regular with the ring 
$\mathscr{R}_H$ of Baire sets\footnote{We will need the complete measure spaces on $\mathfrak{G}, \mathfrak{G}/H$ 
but the Baire measures are sufficient to generate them by the Carath\'eodory method, 
because we have assumed the topology on $\mathfrak{G}$ to fulfil the second axiom
of countability.}. In short   
$(\mathfrak{G}, \mathscr{R}_{\mathfrak{G}}, \mu) = 
(H \times \mathfrak{G}/H,\mathscr{R}_{H \times Q}, \mu_H \times \mu_{\mathfrak{G}/H} )$.   
In our applications we are dealing with pairs $H \subset \mathfrak{G} $ of Lie subgroups of 
the double cover $T_{4} \circledS SL(2, \mathbb{C})$ of the Poincar\'e group $\mathfrak{P}$ including the group 
$T_{4} \circledS SL(2, \mathbb{C})$ itself, with a sub-manifold structure of $H$ and $Q \cong \mathfrak{G}/H$.
This  opportunities allow us to reduce the analysis  of the induced representation $U^L$ in the Krein space
defined in this Section to an application of the Fubini theorem and to the von Neumann analysis of the direct integral
of ordinary Hilbert spaces. (The same assumption together with its analogue for the double cosets in $\mathfrak{G}$ simplifies also the  problem of decomposition of tensor products of induced representations of $\mathfrak{G}$ 
and reduces it mostly to an application of the Fubini theorem and harmonic analysis on the "small" subgroups: 
namely  at the initial stage we reduce the problem to the geometry of right cosets and 
double cosets with the observation that Mackey's theorem on Kronecker product and subgroup theorem of induced representations likewise work for the induced representations in Krein spaces defined here, and then apply the Fubini theorem and harmonic analysis on the "small" subgroups.). Driving by the physical examples we assume for a while 
that the ``measure product property'' is fulfilled by the right $H$-cosets in $\mathfrak{G}$. 
(We abandon soon this assumption so that our results, namely 
\emph{the subgroup theorem} and \emph{the Kronecker product theorem}, hold true for induced representations 
in Krein spaces, without this assumption.)    

Let $L$ be any ($\mathfrak{J}_L$-)unitary strongly continuous and almost uniformly bounded representation of $H$ 
in a Krein space $(\mathcal{H}_L , \mathfrak{J}_L )$. Let $\mu_H$ and $\mu_{\mathfrak{G}/H}$ be (quasi) invariant measures on $H$ and on the homogeneous space $\mathfrak{G}/H$ of right $H$-cosets in $\mathfrak{G}$ induced by the (right) Haar measure $\mu$ on $\mathfrak{G}$ by the ``unique factorization''. Let us denote\footnote{$L$ in superscript! The $\mathcal{H}_L$ with the lower case of the index $L$ is reserved for the space of the representation $L$ of the subgroup $H \subset \mathfrak{G}$.} by 
$\mathcal{H}^L$ the set of all functions $f: \mathfrak{G} \ni x \mapsto f_x$ from $\mathfrak{G}$ to $\mathcal{H}_L$ such that   

\begin{enumerate}

\item[(i)]
            $(f_x, \upsilon)$ is measurable function of $x \in \mathfrak{G}$ 
            for all $\upsilon \in \mathcal{H}_L$.    
               
\item[(ii)]
            $f_{hx} = L_h (f_{x})$ for all $h \in H$ and $x \in \mathfrak{G}$.

\item[(iii)] Into the linear space of functions $f$ fulfilling (i) and (ii) let us introduce the operator
             $\mathfrak{J}^L$ by the formula $(\mathfrak{J}^L f)_{x} = L_h \mathfrak{J}_L L_{h^{-1}} (f_{x})$, 
             where $x = h\cdot q$ is the unique decomposition 
             of $x \in  \mathfrak{G}$. Besides (i) and (ii) we require 
\[
\int \, (\, \mathfrak{J}_L ((\mathfrak{J}^L f)_x ), f_x \,) \, d\mu_{\mathfrak{G}/H} < \infty,
\]
             where the meaning of the integral is to be found in the fact that the integrand is constant on the right 
             $H$-cosets and hence defines a function on the coset space $\mathfrak{G}/H$.

\end{enumerate}

Because every $x \in \mathfrak{G}$ has a unique factorization $x = h \cdot q$ with $h \in  H$ and 
$q \in Q \cong \mathfrak{G}/H$, then by ``unique factorization'' the functions $f \in \mathcal{H}^L$ as well as the functions $x \mapsto (f_x , \upsilon)$
with $\upsilon \in \mathcal{H}_L$, on $\mathfrak{G}$, may be treated as functions on the Cartesian product $H \times Q
\cong H \times \mathfrak{G}/H \cong \mathfrak{G}$. The axiom (i) means that the functions 
$(h,q) \mapsto (f_{h\cdot q}, \upsilon)$ for $\upsilon \in \mathcal{H}_L$ are measurable on the product measure space 
$(H \times \mathfrak{G}/H,\mathscr{R}_{H \times \mathfrak{G}/H}, \mu_H \times \mu_{\mathfrak{G}/H}) 
\cong (H \times Q,\mathscr{R}_{H \times Q}, \mu_H \times \mu_{\mathfrak{G}/H} )$. In particular let $W : q \mapsto W_q \in \mathcal{H}_L$ be a function on $Q$ such that $q \mapsto (W_q , \upsilon)$ is measurable with respect to the standard measure space $(Q, \mathscr{R}_Q , dq)$ for all $\upsilon \in \mathcal{H}_L$, and such that 
$\int \, (W_q, W_q ) \, d\mu_{\mathfrak{G}/H}(q) < \infty$. Then by the  analysis of \cite{von_neumann_dec} (compare also \cite{Neumark_dec}, \S 26.5) which is by now standard, the set of such functions $W$ (when functions equal almost everywhere are identified) compose the direct integral $\int \, \mathcal{H}_L \, d\mu_{\mathfrak{G}/H}(q)$ Hilbert space with the inner product $ (W, F) = \int \, (W_q, F_q ) \, d\mu_{\mathfrak{G}/H}(q)$. 
For every such $W \in \int \, \mathcal{H}_L \, d\mu_{\mathfrak{G}/H}(q)$ the function 
$(h,q) \mapsto f_{h\cdot q} = L_h W_q$ fulfils (i) and
(ii). (ii) is trivial. For each $\upsilon \in \mathcal{H}_L$ the function $(h,q) \mapsto (f_{h\cdot q}, \upsilon ) 
= (L_h W_q , \upsilon )$ is measurable on the product measure space $(H \times Q,\mathscr{R}_{H \times Q}, \mu_H \times \mu_{\mathfrak{G}/H} ) \cong (\mathfrak{G}, \mathscr{R}_{\mathfrak{G}}, \mu)$ because for any orthonormal basis $\{ e_n \}_{n \in \mathbb{N}}$ of the Hilbert space $\mathcal{H}_L$ we have:
\[
\begin{split}
(f_{h\cdot q}, \upsilon) = (f_{h\cdot q}, \mathfrak{J}_L \mathfrak{J}_L  \upsilon) 
= (L_h W_q, \mathfrak{J}_L \mathfrak{J}_L  \upsilon)
= (\mathfrak{J}_L L_h W_q,  \mathfrak{J}_L  \upsilon) \\
= (\mathfrak{J}_L W_q , L_{h^{-1}} \mathfrak{J}_L  \upsilon)
= \sum_{n \in \mathbb{N}} (\mathfrak{J}_L W_q ,e_n)(e_n , L_{h^{-1}} \mathfrak{J}_L  \upsilon)
\end{split}
\]
where each summand gives a measurable function 
$(h,q)$ ${}\mapsto{}$ $(\mathfrak{J}_L W_q ,e_n)(e_n , L_h \mathfrak{J}_L  \upsilon)$ on the product measure space 
$(H \times Q,\mathscr{R}_{H \times Q}, \mu_H \times \mu_{\mathfrak{G}/H} )$ by Scholium 3.9 of \cite{Segal_Kunze}. On the other hand for every function $(h,q) \mapsto (f_{h \cdot q} ,\upsilon)$ measurable on the product measure space the restricted functions $q \mapsto (f_{h \cdot q} ,\upsilon)$ and $h \mapsto (f_{h \cdot q} ,\upsilon)$, i. e. with one of the arguments $h$ and $q$ fixed, are measurable, which follows from the Fubini theorem (compare e. g. \cite{Segal_Kunze}, Theorem 3.4) and thus $q \mapsto (f_q , \upsilon)$ is measurable (i.e. with the argument $h$ fixed and equal $e$ in $(h,q) \mapsto (f_{h \cdot q} ,\upsilon)$). Because a simple computation shows that  
\[
\begin{split}
\int \, (\, \mathfrak{J}_L ((\mathfrak{J}^L f)_x ), f_x \,) \, d\mu_{\mathfrak{G}/H} 
= \int \, (\, \mathfrak{J}_L ((\mathfrak{J}^L f)_{h\cdot q} ), f_{h\cdot q} \,) \, d\mu_{\mathfrak{G}/H}(q) \\
=\int \, (f_q , f_q \,) \, d\mu_{\mathfrak{G}/H}(q),
\end{split}
\]
one can see that when functions equal almost everywhere are identified 
$\mathcal{H}^L$ becomes a Hilbert space with the inner product 
\begin{equation}\label{inn_ind_def}
(f, g) = \int \, (\, \mathfrak{J}_L ((\mathfrak{J}^L f)_x ), g_x \,) \, d\mu_{\mathfrak{G}/H}. 
\end{equation}
(In fact because the values of $f \in \mathcal{H}^L$ are in the fixed Hilbert space $\mathcal{H}_L$ we do not
have to tangle into the the whole machinery of direct integral Hilbert spaces of von Neumann. It suffices
to make obvious modifications in the corresponding proof that $L^2(\mathfrak{G}/H)$ is a Hilbert space, compare
\cite{Neumark_dec}, , \S 26.5.).

A simple verification shows that $\mathfrak{J}^L$ is a bounded self-adjoint operator in the Hilbert space $\mathcal{H}^L$
with respect to the definite inner product (\ref{inn_ind_def}) and that $(\mathfrak{J}^L)^2 = I$. Therefore
$(\mathcal{H}^L , \mathfrak{J}^L )$ is a Krein space with the indefinite product
\begin{equation}\label{ind_ind}
\big(f, g \big)_{\mathfrak{J}^L} = (\mathfrak{J}^L f, g) = \int \, (\, \mathfrak{J}_L (f_x ), g_x \,) \, d\mu_{\mathfrak{G}/H}
\end{equation}
which is meaningful because the integrand is constant on the right $H$-cosets, i. e. it is a function of $q \in Q \cong \mathfrak{G}/H$. 

Let the function $[x] \mapsto \lambda ([x], g)$ on $\mathfrak{G}/H$ be the Radon-Nikodym derivative 
$\lambda( \cdot ,g) = \frac{\ud (R_g \mu)}{\ud \mu}(\cdot)$, where $[x]$ stands for the right 
$H$-coset $Hx$ of $x \in \mathfrak{G}$
($\mu$ stands for the (quasi) invariant measure $\mu_{\mathfrak{G}/H}$ on $\mathfrak{G}/H$ induced by the assumed ``factorization'' property from the Haar measure $\mu$ on $\mathfrak{G}$ and $R_g \mu$
stands for the right translation of the measure $\mu$: $R_g \mu (E) = \mu(Eg)$). 

For every $g_0 \in \mathfrak{G}$ let us consider a densely defined operator $U^L_{g_0}$.
Its domain $\mathfrak{D}(U^L_{g_0})$  is equal to the set of all those  $f \in  \mathcal{H}^L$ 
for which the function
\[
x \mapsto   f'_x = \sqrt{\lambda ([x], g_0)} \, f_{xg_0}
\]
has finite Hilbert space norm (i. e. ordinary norm with respect to the ordinary definite 
inner product (\ref{inn_ind_def})) in $\mathcal{H}^L$:
\[
\begin{split}
 \big( f' , f' \big) = \int \, (\, \mathfrak{J}_L ((\mathfrak{J}^L f')_x ), f'_x \,) \, d\mu_{\mathfrak{G}/H} \\
= \int \, (\, \mathfrak{J}_L L_{h(x)} \mathfrak{J}_L L_{h(x)^{-1}} \sqrt{\lambda ([x], g_0)} \, f_{xg_0}  , 
\sqrt{\lambda ([x], g_0)} \, f_{xg_0} \,) \, 
d\mu_{\mathfrak{G}/H}(x) < \infty,
\end{split}
\]
where $h(x) \in H$ is the unique element corresponding to $x$ such that $h(x)^{-1}x \in Q$; and whenever $ f \in \mathfrak{D} (U^{L}_{g_0})$ we put 
\[
(U^L_{g_0} f)_x = \sqrt{\lambda ([x], g_0)} \, f_{xg_0}.
\]
$U^L$, after restriction to a suitable sub-domain, becomes a group homomorphism of 
$\mathfrak{G}$ into a group of densely defined $\mathfrak{J}^L$-isometries of the Krein space 
$(\mathcal{H}^L , \mathfrak{J}^L )$. Let us formulate this statement more precisely in a form of a Theorem:

\begin{twr}
The operators $U^L_{g_0}$, $g_0 \in \mathfrak{G}$, are closed and $\mathfrak{J}^L$-isometric with dense domains
$\mathfrak{D}(U^L_{g_0})$, dense ranges $\mathfrak{R}(U^L_{g_0})$ and dense intersection $\bigcap_{g_0 \in \mathfrak{G}} \mathfrak{D}(U^L_{g_0}) = \bigcap_{g_0 \in \mathfrak{G}} \mathfrak{R}(U^L_{g_0})$. $U^{L}_{g_0 k_0}$ is equal to the closure of the composition $\widetilde{U^{L}_{g_0}} \, \widetilde{U^{L}_{k_0}} = 
\widetilde{U^{L}_{g_0 k_0}}$
of the restrictions $\widetilde{U^{L}_{g_0}}$ and $\widetilde{U^{L}_{k_0}}$ of $U^{L}_{g_0}$ and $U^{L}_{k_0}$
to the domain $\bigcap_{g_0 \in \mathfrak{G}} \mathfrak{D}(U^L_{g_0})$, i. e.  the map $g_0 \mapsto \widetilde{U^L_{g_0}}$ is a Krein-isometric representation of 
$\mathfrak{G}$. There exists a dense sub-domain $\mathfrak{D} \subset \bigcap_{g_0 \in \mathfrak{G}} \mathfrak{D}(U^L_{g_0})$ such that $U^L_{g_0} \mathfrak{D} = \mathfrak{D}$, $U^L_{g_0}$ is the closure of the restriction 
$\widetilde{\widetilde{U^{L}_{g_0}}}$ of $U^L_{g_0}$ to the sub-domain $\mathfrak{D}$, and  $g_0 \mapsto \widetilde{\widetilde{U^L_{g_0}}}$ is strongly continuous Krein-isometric representation of $\mathfrak{G}$ on its domain 
$\mathfrak{D}$.

\label{def_ind_krein:twr.1}
\end{twr}

\qedsymbol \, 
Let us introduce the class $C^{L}_{00} \subset \mathcal{H}^L$ of functions 
$h\cdot q \mapsto f_{h \cdot q} = L_h W_q$ with 
$q \mapsto W_q \in \mathcal{H}_L$ continuous and compact support on $Q \cong \mathfrak{G}/H$. Of course 
each such function $W$ is an element of the direct integral Hilbert space 
$\int \, \mathcal{H}_L \, d\mu_{\mathfrak{G}/H}$. One easily verifies that all the conditions of Lemma \ref{lem:dense.4}
of (the next) Sect. \ref{dense} are true for the class $C^{L}_{00}$. Therefore $C^{L}_{00}$ is 
dense in $\mathcal{H}^L$. Let $h\cdot q \mapsto f_{h \cdot q} = L_h W_q$ be an element of $C^{L}_{00}$ and let 
$K$ be the compact support of the function $W$.  Using the ``unique factorization'' let us introduce the functions
$(q,h_0 , q_0 ) \mapsto h'_{{}_{q,h_0 , q_0}} \in H$ 
and $(q,h_0 , q_0 ) \mapsto q'_{{}_{q,h_0 , q_0}} \in Q \cong \mathfrak{G}/H$
in the following way. Let $g_0 = q_0 \cdot h_0$. We define $ h'_{{}_{q,h_0 , q_0}} \in H$ 
and $q'_{{}_{q,h_0 , q_0}} \in Q \subset \mathfrak{G}$ 
to be the elements, uniquely corresponding to $(q,h_0 , q_0 )$, such that 
\begin{equation}\label{q'_h'}
q\cdot h_0 \cdot q_0 = h'_{{}_{q,h_0 , q_0}} \cdot q'_{{}_{q,h_0 , q_0}}.
\end{equation} 
Finally let $c_{K, g_0} = \sup_{q \in K} \big\| L_{h'_{{}_{q,h_0 , q_0}}} \big\|$, which is finite
outside a null set, on account of the almost uniform boundedness of the representation $L$,
and because  $q \mapsto h'_{{}_{q,h_0 , q_0}}$ is continuous outside a $\mu_{\mathfrak{G}/H}$-null set
(``measure product property'')\footnote{It holds true even if the ``measure product property is not assumed''
-- compare the comments below in this Section.}. 

\begin{equation}\label{def_ind_krein:ineq}
\begin{split}
\| U^{L}_{h_0 \cdot q_0} f \|^2 =  \big(U^{L}_{h_0 \cdot q_0} f , U^{L}_{h_0 \cdot q_0} f \big) \\
= \int \, (\, \mathfrak{J}_L ((\mathfrak{J}^L U^{L}_{h_0 \cdot q_0}f)_{h\cdot q} ), (U^{L}_{h_0 \cdot q_0}f)_{h\cdot q} \,) \, 
d\mu_{\mathfrak{G}/H}(q) \\
= \int \, (\, L_{h'_{q, h_0 , q_0}}f_{q'_{q, h_0 , q_0}} , L_{h'_{q, h_0 , q_0}}f_{q'_{q, h_0 , q_0}} \,) \, 
d\mu_{\mathfrak{G}/H}(q'_{q, h_0 , q_0}) \\
\leq  c_{K,g_0}^{2} \, \int \, (\, f_{q'_{q, h_0 , q_0}} , f_{q'_{q, h_0 , q_0}} \,) \, 
d\mu_{\mathfrak{G}/H}(q'_{q, h_0 , q_0}) \\
= c_{K,g_0}^{2} \, \| f \|^2 , \,\,\,\,\,\,\, 
g_0 = h_0 \cdot q_0 \in \mathfrak{G}.  
\end{split}
\end{equation}
Thus it follows that $C^{L}_{00} \subset \mathfrak{D}(U^{L}_{g_0})$ for every $g_0 \in \mathfrak{G}$. Similarly it is easily verifiable that $C^{L}_{00} \subset \mathfrak{R}(U^{L}_{g_0})$ whenever the Radon-Nikodym derivative 
$\lambda ([x], g_0)$ is continuous in $[x]$. It follows from definition
that for $f \in \mathcal{H}^L$ being a member of $\bigcap_{g_0 \in \mathfrak{G}} \mathfrak{D}(U^L_{g_0})$ is equivalent to being a member of $\bigcap_{g_0 \in \mathfrak{G}} \mathfrak{R}(U^L_{g_0})$.

We shall show that $\big( \, U^{L}_{g_0} \, \big)^\dagger = U^{L}_{{g_0}^{-1}}$, 
where $T^\dagger$ stands for the adjoint of the operator $T$ in the sense of Krein \cite{Bog}, page 121: 
for any linear operator $T$ with dense domain $\mathfrak{D}(T)$ the vector 
$g \in \mathcal{H}^L$ belongs to $\mathfrak{D}(T^\dagger)$ if and only if there exists a $k \in \mathcal{H}^L$ 
such that 
\[
(\mathfrak{J}^L Tf , g) = (\mathfrak{J}^L f , k), \,\,\, \textrm{for all} \,\, f \in \mathfrak{D}(T),
\] 
and in this case we put $T^\dagger g = k$, with the unique $k$ as $\mathfrak{D}(T)$ is dense
(i. e. same definition as for the 
ordinary adjoint with the definite Hilbert space inner product $(\cdot, \cdot)$ given by (\ref{inn_ind_def})
replaced with the indefinite one $(\mathfrak{J}^L \cdot, \cdot)$, given by (\ref{ind_ind})). 

Now let $g$ be arbitrary in $\mathfrak{D}\big( \, \big( \,U^{L}_{g_0} \, \big)^\dagger \, \big)$, 
and let $\big(U^{L}_{g_0} \big)^\dagger g = k$. 

The inclusion $\big( \, U^{L}_{g_0} \, \big)^\dagger \subset U^{L}_{{g_0}^{-1}}$ is equivalent
to the equation $U^{L}_{{g_0}^{-1}} g = k$. By the definition of the Krein adjoint of an operator, for any
$f \in \mathfrak{D}\big(U^{L}_{g_0}\big)$ we have 
\[
\begin{split}
\big( \, \mathfrak{J}^L \, U^{L}_{g_0} f , g \,\big) = (\mathfrak{J}^L \, f, k), \\
\,\,\, \textrm{i. \, e.}
\int \, \Big( \mathfrak{J}_L \, \big( \, U^{L}_{g_0} \, f \big)_x , g_x  \, \Big) \, d\mu_{\mathfrak{G}/H}(x)
= \int \, (\mathfrak{J}_L \, f_x , k_x ) \, d\mu_{\mathfrak{G}/H}(x);
\end{split}
\]
which by the definition of $U^{L}_{g_0}$ and quasi invariance of the measure $\mu_{\mathfrak{G}/H}$
means that
\[
\begin{split}
\int \, \Big( \mathfrak{J}_L \, f_x , \sqrt{\frac{d\mu_{\mathfrak{G}/H}(x{g_0}^{-1})}{d\mu_{\mathfrak{G}/H}(x)}} 
g_{x{g_0}^{-1}}  \, \Big) \, d\mu_{\mathfrak{G}/H}(x) \\
= \int \, (\mathfrak{J}_L \, f_x , k_x ) \, d\mu_{\mathfrak{G}/H}(x) \,\,\, \textrm{for all} \,\, f \in 
\mathfrak{D}\big(U^{L}_{g_0} \big);
\end{split}
\]
i. e. the function $u$
\[
x \mapsto u_x =  \sqrt{\frac{d\mu_{\mathfrak{G}/H}(x{g_0}^{-1})}{d\mu_{\mathfrak{G}/H}(x)}} 
g_{x{g_0}^{-1}} - k_x
\]
is $\mathfrak{J}^L$-orthogonal to all elements of $\mathfrak{D}\big( \big( \,U^{L}_{g_0} \big)$: $(\mathfrak{J}^L f , u) = 0$ for all $f \in \mathfrak{D}\big(U^{L}_{g_0} \big)$.
Because $\mathfrak{D}\big(U^{L}_{g_0}\big)$ is dense in $\mathcal{H}^L$, and $\mathfrak{J}^L$ is unitary with respect to the ordinary Hilbert space inner product (\ref{inn_ind_def}) in $\mathcal{H}^L$ it follows that $\mathfrak{J}^L \mathfrak{D}\big(U^{L}_{g_0}\big)$ is dense in 
$\mathcal{H}^L$. Therefore $u$ must be zero as a vector orthogonal to 
$\mathfrak{J}^L \mathfrak{D}\big(U^{L}_{g_0} \big)$ in the sense of the Hilbert space inner product (\ref{inn_ind_def}). Thus
\[
\sqrt{\frac{d\mu_{\mathfrak{G}/H}(x{g_0}^{-1})}{d\mu_{\mathfrak{G}/H}(x)}} 
g_{x{g_0}^{-1}} = k_x
\] 
almost everywhere, and because by definition $(k, k) < \infty$, we have shown that $U^{L}_{{g_0}^{-1}} g = k$.

Next we show that $\big( \, U^{L}_{g_0} \, \big)^\dagger \supset U^{L}_{{g_0}^{-1}}$. Let $g$ be arbitrary
in $\mathfrak{D}(U^L_{{g_0}^{-1}})$ and let $U^{L}_{{g_0}^{-1}} g = k$. It must be shown that for any 
$f \in \mathfrak{D}\big(U^{L}_{g_0} \big)$,Section \ref{PartIIMackey}
$(\mathfrak{J}^L \, U^{L}_{g_0} f  , g) = (\mathfrak{J}^L \, f, k)$. This is the same as showing that 
\[
\int \, \Big( \mathfrak{J}_L \, \big( \, U^{L}_{g_0} \, f \big)_x , g_x  \, \Big) \, d\mu_{\mathfrak{G}/H}(x)
= \int \, (\mathfrak{J}_L \, f_x , k_x ) \, d\mu_{\mathfrak{G}/H}(x),
\] 
which again easily follows from definition of $U^{L}_{g_0}$ and quasi invariance of the measure 
$d\mu_{\mathfrak{G}/H}(x)$:
\[
\begin{split}
\int \, \Big( \mathfrak{J}_L \, \big( \, U^{L}_{g_0} \, f \big)_x , g_x  \, \Big) \, d\mu_{\mathfrak{G}/H}(x) 
= \int \, \sqrt{\frac{d\mu_{\mathfrak{G}/H}(xg_0)}{d\mu_{\mathfrak{G}/H}(x)}} \big(\mathfrak{J}_L \, f_{xg_0} , g_x \big) \,
d\mu_{\mathfrak{G}/H}(x) \\
=  \int \, \sqrt{\frac{d\mu_{\mathfrak{G}/H}(x{g_0}^{-1}g_0)}{d\mu_{\mathfrak{G}/H}(x{g_0}^{-1})}} \Big(\mathfrak{J}_L \, f_{x{g_0}^{-1}g_0} , g_{x{g_0}^{-1}} \Big) \,
\, \frac{d\mu_{\mathfrak{G}/H}(x{g_0}^{-1})}{d\mu_{\mathfrak{G}/H}(x)} \, d\mu_{\mathfrak{G}/H}(x) \\
= \int \,  \Big( \mathfrak{J}_L \, f_{x} , \sqrt{\frac{d\mu_{\mathfrak{G}/H}(x{g_0}^{-1}}{d\mu_{\mathfrak{G}/H}(x)}} g_{x{g_0}^{-1}} \Big)  \, d\mu_{\mathfrak{G}/H}(x) \\
=  \int \,  \Big( \mathfrak{J}_L \, f_{x} , \big(U^{L}_{{g_0}^{-1}} g\big)_{x} \Big)  \, 
d\mu_{\mathfrak{G}/H}(x)  
= \int \, (\mathfrak{J}_L \, f_x , k_x ) \, d\mu_{\mathfrak{G}/H}(x).
\end{split}
\] 
Thus we have shown that $\big(U^{L}_{g_0} \big)^\dagger = U^{L}_{{g_0}^{-1}}$. 

Because $C^{L}_{00} \subset \mathfrak{D}(U^L_{{g_0}^{-1}})$ then $\mathfrak{D}(U^L_{{g_0}^{-1}})$
is dense, thus  $U^{L}_{{g_0}^{-1}}$, equal to $\big(U^{L}_{g_0} \big)^\dagger$, is closed by Theorem 2.2 of \cite{Bog} (Krein adjoint $T^\dagger$ is always closed,
as it is equal $\mathfrak{J}^L  T^* \mathfrak{J}^L$ with the ordinary adjoint $T^*$ operator, 
and because the fundamental symmetry $\mathfrak{J}^L$
is unitary in the associated Hilbert space $\mathcal{H}^L$, compare Lemma 2.1 in \cite{Bog}).

In order to prove the second statement it will be sufficient to show that 
$\big( \, \widetilde{U^{L}_{g_0}} \, \big)^\dagger = U^{L}_{{g_0}^{-1}}$
because the homomorphism property of the map $g_0 \mapsto U^{L}_{g_0}$ restricted to $\bigcap_{g \in \mathfrak{G}}
\mathfrak{D}(U^L_{g})$
is a simple consequence of the definition of $U^{L}_{g_0}$. But the proof of the equality 
$\big( \, \widetilde{U^{L}_{g_0}} \, \big)^\dagger = U^{L}_{{g_0}^{-1}}$ runs exactly the same way as the 
proof of the equality $\big( \, U^{L}_{g_0} \, \big)^\dagger = U^{L}_{{g_0}^{-1}}$, with the trivial 
replacement of $\mathfrak{D}\big(U^{L}_{g_0}\big)$ by $\mathfrak{D}$, as it is valid for any dense sub-domain 
$\mathfrak{D}$ contained in $\mathfrak{D}\big(U^{L}_{g_0}\big)$ instead of $\mathfrak{D}\big(U^{L}_{g_0}\big)$.  
Then by Theorem 2.5 of \cite{Bog}
it follows that $\big( \, \widetilde{U^{L}_{g_0}} \, \big)^{\dagger \dagger} 
= \big( \, U^{L}_{{g_0}^{-1}} \, \big)^\dagger$ is equal to the closure $\overline{\widetilde{U^{L}_{g_0}}}$
of the operator $\widetilde{U^{L}_{g_0}}$. Because $\big( \, U^{L}_{{g_0}^{-1}} \, \big)^\dagger
= U^{L}_{g_0}$, we get $U^{L}_{g_0} = \overline{\widetilde{U^{L}_{g_0}}}$. 

By the above remark we also have $U^{L}_{g_0} = \overline{\widetilde{\widetilde{U^{L}_{g_0}}}}$ for any
restriction $\widetilde{\widetilde{U^{L}_{g_0}}}$ of $U^{L}_{g_0}$ to a dense sub-domain $\mathfrak{D}
\subset \mathfrak{D}(U^L_{g_0})$

In order to prove the third statement, let us introduce a dense sub-domain $C^{L}_{0} \subset C^{L}_{00}$
of continuous functions with compact support on $\mathfrak{G}/H$. Its full definition and properties are given in the 
next Section. In particular $U^{L}_{g_0} C^{L}_{0} = C^{L}_{0}$ whenever the Radon-Nikodym derivative 
$\lambda ([x], g_0)$ is continuous in $[x]$. For each element $f^0$ of $C^{L}_{0}$ we have the inequality
shown to be valid in the course of proof of Lemma \ref{lem:dense.1}, Sect. \ref{dense}: 
\[
\| f^{0}_{x_1} - f^{0}_{x_2} \|^2 \leq \sup_{h \in H} \big\|  f^{L, V}_{(h,e) \cdot (e,x_1 )} - f^{L,V}_{(h,e)\cdot (e,x_2 )} \big\|^2 \, 2 \,\sup_{x \in \mathfrak{G}} \mu_{H}(Kx^{-1} \cap H)
\]
where $f^{L, V}$ is a function depending on $f^0$, continuous on the direct product group $H \times \mathfrak{G}$
and with compact support $K_H \times V$ with $V$ being a compact neighbourhood of the two points $x_1$ and $x_2$.
Because any such function $f^{L, V}$ must be uniformly continuous, the strong continuity of $U^{L}$ on 
the sub-domain $C^{L}_{0}$ follows. Because $U^{L}_{g_0} C^{L}_{0} = C^{L}_{0}$, the third statement is proved
with $\mathfrak{D} = C^{L}_{0}$ (In case the Radon-Nikodym derivative was not continuous and ``measure product property'' not satisfied it would be sufficient to use all finite sums 
$U^{L}_{g_1} f^1 + \ldots U^{L}_{g_n} f^n , f^k \in C^{L}_{0}$ as the common sub-domain 
$\mathfrak{D}$ instead of $C^{L}_{0}$).
\qed

\begin{rem}
By definition of the Krein-adjoint operator and the properties: 
1) $U^L_{g} \mathfrak{D} = \mathfrak{D}$, $g \in \mathfrak{G}$,  2)
$\big( \, U^{L}_{g} \, \big)^\dagger = U^{L}_{g^{-1}}$, $g \in \mathfrak{G}$, it easily follows that for each $g \in \mathfrak{G}$ 
\begin{equation}\label{UD=D}
\big( \, U^{L}_{g} \, \big)^\dagger \, U^{L}_{g} = I \,\,\, \textrm{and} \,\,\,
U^{L}_{g} \, \big( \, U^{L}_{g} \, \big)^\dagger = I
\end{equation}
on the domain $\mathfrak{D}$. We may easily modify the common domain $\mathfrak{D}$ so as to achieve
the additional property: 3) $\mathfrak{J}^L \mathfrak{D} = \mathfrak{D}$ together with 1) and 2)
and thus with (\ref{UD=D}). 
Indeed, to achieve this one may define
$\mathfrak{D}$ to be the linear span of the set $\Big\{ \, \Big( (\mathfrak{J}^L)^{m_1} U^{L}_{g_1}   \ldots 
U^{L}_{g_n} (\mathfrak{J}^L)^{m_{n+1}} \Big) f \, \Big\}$: with $g_k$ ranging over $\mathfrak{G}$, $f \in  C^{L}_{0}$, 
$n \in \mathbb{N}$ and $k \mapsto m_{k}$ over the sequences with $m_k$ equal 0 or 1. 
In case the Radon-Nikodym derivative $\lambda$ is continuous and the 
``measure product property'' fulfilled, $\mathfrak{D} = C^{L}_{0}$ meets all the requirements. 

\label{rem:def_ind_krein.1}
\end{rem}

\begin{cor}
For every $U^{L}_{g_0}$ there exists a unique unitary (with respect to the definite inner product 
(\ref{inn_ind_def})) operator $U_{g_0}$ in $\mathcal{H}^L$ and unique selfadjoint (with respect to
(\ref{inn_ind_def})) positive operator $H_{g_0}$, with dense domain $\mathfrak{D}(U^{L}_{g_0})$ and dense range such that $U^{L}_{g_0} = U_{g_0} H_{g_0}$.
\label{def_ind_krein:cor.1}
\end{cor}
\qedsymbol \,
Immediate consequence of the von Neumann polar decomposition theorem and closedness of $U^{L}_{g_0}$.
\qed

\vspace*{0.2cm}

Of course the ordinary unitary operators $U_{g_0}$ of the Corollary do not compose any representation in general as the 
operators $U_{g_0}$ and $H_{g_0}$ of the polar decomposition do not commute if $U_{g_0}$ is non normal.

\begin{twr}
$L$ and $\mathfrak{J}_L$ commute  if and only if $U^L$ and $\mathfrak{J}^L$ commute.
If $U^L$ and $\mathfrak{J}^L$ commute, then $L$ is not only $\mathfrak{J}_L$-unitary but also unitary in the ordinary sense for the definite inner product in the Hilbert space $\mathcal{H}_L$.
If $U^L$ and $\mathfrak{J}^L$ commute then $U^L$ is not only $\mathfrak{J}^L$-isometric
but unitary with respect to the ordinary Hilbert space inner product (\ref{inn_ind_def}) in $\mathcal{H}^L$, 
i . e. the operators $U^{L}_{g_0}$ are bounded and unitary with respect to (\ref{inn_ind_def}).
The representation $L$ is uniformly bounded if and only if the induced representation $U^{L}$ is Krein-unitary 
(with each $U^{L}_{g_0}$ bounded) and uniformly bounded.
\label{def_ind_krein:twr.2}
\end{twr}

\qedsymbol \,
Using the functions $(q,h_0 , q_0 ) \mapsto h'_{{}_{q,h_0 , q_0}} \in H$ and 
$(q,h_0 , q_0 ) \mapsto q'_{{}_{q,h_0 , q_0}} \in Q \cong \mathfrak{G}/H$ defined by (\ref{q'_h'}),
one easily verifies that $L$ and $\mathfrak{J}_L$ commute (and thus $L$ is not only $\mathfrak{J}_L$-unitary but also unitary in the ordinary sense for the definite inner product in the Hilbert space 
$\mathcal{H}_L$) if and only if $U^L$ and $\mathfrak{J}^L$ commute (i. e. when $U^L$ is not only $\mathfrak{J}^L$-isometric
but unitary with respect to the ordinary Hilbert space norm (\ref{inn_ind_def}) in $\mathcal{H}^L$). To this end we utilize 
the fact that for each fixed $x$, $f_x$ with $f$ ranging over $C^{L}_{00}$ has $\mathcal{H}^L$ as their 
closed linear span. We leave details to the reader.  

\qed

\vspace*{0.2cm}

\begin{cor}
If $N \subset H \subset \mathfrak{G}$ is a normal subgroup of $\mathfrak{G}$ such that the restriction of $L$
to $N$ is uniformly bounded (or commutes with $\mathfrak{J}_L$) then the restriction of $U^L$ to the subgroup
$N$ is a Krein-unitary representation of the subgroup with each $U^{L}_{n}, n \in N$ bounded uniformly in $n$ 
(or $U^{L}$ restricted to $N$ commutes with $\mathfrak{J}^L$ and is an ordinary unitary representation of $N$ in the Hilbert space $\mathcal{H}^L$). 

\label{def_ind_krein:cor.2}
\end{cor}
\qed

In the proof of the strong continuity of $U^L$ on the dense domain $\mathfrak{D}$ we have used a specific dense subspace 
$C^{L}_{0}$ of $\mathcal{H}^L$. In the next Subsection we give its precise definition and provide the remaining relevant analytic underpinnings which we introduce after Mackey. In the proof of strong continuity we did as in the classical proof of strong continuity of the right regular representation of $\mathfrak{G}$ in $L^2 (\mathfrak{G})$ or in $L^2 (\mathfrak{G}/H)$ (of course with the obvious Radon-Nikodym factor in the latter case), with the necessary modifications required for the Krein space. In our  proof of strong continuity on $\mathfrak{D}$ the strong continuity of the representation $L$ plays a much more profound role in comparison to the original Mackey's theory.

The additional assumption posed on right $H$-cosets, i. e. ``measure product property'' is unnecessary. In order to give to this paper a more independent character we point out that the above construction of the induced representation in Krein space is possible without this assumption which may be of use for spectral analysis for (unnecessary elliptic) operators on manifolds uniform for more general semi-direct product Lie groups preserving indefinite pseudo-Riemann structures. Namely for any closed subgroup $H \subset \mathfrak{G}$ (with the ``measure product property'' unnecessary fulfilled)
the right action of $H$ on $\mathfrak{G}$ is proper and both $\mathfrak{G}$ and $\mathfrak{G}/H$ are metrizable so that a theorem of Federer and Morse \cite{Federer_Morse} can be applied (with the regular Baire (or Borel) Haar measure space structure $(\mathfrak{G}, \mathscr{R}_{\mathfrak{G}}, \mu)$ on $\mathfrak{G}$) 
in proving that  there exists a Borel subset $B \subset \mathfrak{G}$ such that: 
(a) $B$ intersects each right $H$-coset in exactly one point and (b) for each compact subset $K$ of $\mathfrak{G}$,
$\pi^{-1}(\pi(K))\cap B$ has a compact closure (compare Lemma 1.1 of \cite{Mackey}). 
In short $B$ is a ``regular Borel section of $\mathfrak{G}$ with respect to $H$''. In particular it follows that any 
$\mathfrak{g} \in \mathfrak{G}$ has unique factorization $\mathfrak{g} = h \cdot b$, $h \in H, b \in B$.  Using the Lemma and extending a technique of A. Weil used in studying relatively invariant measures Mackey gave in \cite{Mackey} a general construction of quasi invariant measures in $\mathfrak{G}/H$ (all being equivalent).

\vspace*{0.2cm}

\begin{scriptsize} 

The general construction of quasi invariant (standard) Baire (or Borel) measures on the locally compact homogeneous space $\mathfrak{G}/H$ was proposed in a somewhat shortened form in \S 1 of \cite{Mackey}, where the technique of A. Weil was adopted and developed into a $\rho$- and $\lambda$-functions construction. Today it is known as a standard construction of \emph{the quotient of a measure space by a group}, detailed exposition can be found e.g. in \cite{Bourbaki}. Only for sake of completeness let us remind the main Lemmas and Theorem of \S 1 of  \cite{Mackey} (details omitted in the exposition of \cite{Mackey} are to be found e. g. in \cite{Bourbaki} with the trivial interchanging of left and right). Let 
$L_g \mu$ and $R_g \mu$ be the left and right translations of a measure $\mu$ on $\mathfrak{G}/H$: $L_g \mu (E) = \mu (gE)$ and $R_g \mu(E) = \mu (Eg)$. Let $\mu$ be the right Haar measure on $\mathfrak{G}$. Denoting the the constant Radon-Nikodym derivative of the right Haar measure $L_g \mu$ with respect to $\mu$ by $\Delta_{\mathfrak{G}}(g)$, and similarly defined constant Radon-Nikodym derivative  for the closed subgroup $H$ by $\Delta_{H}(g)$ we have the the following Lemmas and Theorems. 
\begin{enumerate}

\item[] LEMMA. \emph{Let $\mu$ be a non-zero measure on $\mathfrak{G}/H$ and $\mu_0 = \mu_\mathfrak{G}$ be the right Haar measure on $\mathfrak{G}$. The following conditions are equivalent}:

\item[a)] \emph{$\mu$ is quasi invariant with respect to $\mathfrak{G}$;}

\item[b)] \emph{a set $E \subset \mathfrak{G}/H$ is of $\mu$-measure zero if and only if $\pi^{-1}(E)$ is
of $\mu_0$-measure zero;}

\item[c)] \emph{the ``pseudo-counter-image'' measure $\mu^{\sharp}$ is equivalent to $\mu_0$.} 

\item[] \emph{Assume one (and thus all) of the conditions to be fulfilled and thus let $\mu^{\sharp} = 
\rho \cdot \mu_0$, where $\rho$ is a Baire (or Borel) $\mu$-measurable function non zero everywhere on $\mathfrak{G}$.
Then for every $s \in \mathfrak{G}$ the Radon-Nikodym derivative $\lambda( \cdot, s)$ of the measure 
$R_s \mu$ with respect to the measure $\mu$ is equal to
\[
\lambda( \pi(x), s) = \frac{\ud (R_s \mu)}{\ud \mu}(\pi(x)) = \rho(xs)/\rho(x)
\]
almost $\mu$-everywhere on $\mathfrak{G}$.}

\end{enumerate}

\begin{enumerate}

\item[] THEOREM. a) \emph{Any two non zero quasi invariant measures on $\mathfrak{G}/H$ are equivalent.}

\item[b)] \emph{If $\mu$ and $\mu'$ are two non zero quasi invariant measures on $\mathfrak{G}/H$
and $\ud (R_s \mu)/ \ud \mu = d(R_s \mu')/ \ud \mu'$ 
almost $\mu$-everywhere (and thus almost $\mu'$-everywhere), 
then $\mu' = c \cdot \mu$, where $c$ is a positive number.}

\end{enumerate}

\begin{enumerate}

\item[] LEMMA. \emph{Measure $\rho \cdot \mu_0$ has the form $\mu^{\sharp}$ if and only if for each $h \in H$
the equality 
\[
\rho (hx) = \frac{\Delta_{H}(h)}{\Delta_{\mathfrak{G}}(h)} \rho(x)
\]
is fulfilled almost $\mu_0$-everywhere on $\mathfrak{G}$.}

\end{enumerate}

\begin{enumerate}

\item[] THEOREM. a) \emph{There exist functions $\rho$ fulfilling the conditions of the preceding Lemma, for example}
\[
\rho(x) = \frac{\Delta_H (h(x))}{\Delta_{\mathfrak{G}}(h(x))}, 
\]
\emph{where $h(x) \in H$ is the only element of $H$ corresponding to 
$x \in \mathfrak{G}$ such that $h(x)^{-1}x \in B$.}

\item[b)] \emph{$\rho$ can be chosen to be continuous.}

\item[c)] \emph{One may chose the regular section $B$ to be continuous outside a discrete countable set in $\mathfrak{G}/H$
whenever $\mathfrak{G}$ is a topological manifold with $H$ as closed topological sub-manifold; thus $x \mapsto h(x)$
becomes continuous outside a set of measure zero in $\mathfrak{G}$.}

\item[d)] \emph{Given such a function $\rho$ one can construct a quasi invariant measure $\mu$ on $\mathfrak{G}/H$
such that $\mu^{\sharp} = \rho \cdot \mu_0$.}

\item[e)] \emph{$\rho(xs)/\rho(x)$ with $s,x \in \mathfrak{G}$ does not depend on $x$ within the class $\pi(x)$ and determinates a function $(\pi(x), s) \mapsto \lambda(\pi(x), s)$
on $\mathfrak{G}/H \times \mathfrak{G}$ equal to the Radon-Nikodym derivative 
$\ud (R_s \mu)/ \ud \mu (\pi(x))$.}

\item[f)] \emph{Given any Baire (or Borel) function $\lambda (\cdot, \cdot)$ on $\mathfrak{G}/H \times \mathfrak{G}$
fulfilling the general properties of Radon-Nikodym derivative: (i) for all $x, s, z \in \mathfrak{G}$, 
$\lambda(\pi(z), xs) = \lambda(\pi(zx), s) \lambda(\pi(z), x)$, (ii) for all $h \in H$, $\lambda(\pi(e), h)
= \Delta_H (h)/\Delta_{\mathfrak{G}}(h)$, (iii) $\lambda(\pi(e), s)$ is bounded on compact sets as a function of $s$,
one can construct a quasi invariant measure $\mu$ on $\mathfrak{G}/H$ such that 
$\ud (R_s \mu)/ \ud \mu (\pi(x)) = \lambda(\pi(x), s)$, 
almost $\mu$-everywhere with respect to $s,x$ on $\mathfrak{G}$.}

\end{enumerate}

Thus every non zero quasi invariant measure $\mu$ on $\mathfrak{G}/H$ gives rise to a $\rho$-function
and $\lambda$-function and vice versa every ``abstract Radon-Nikodym derivative'' i.e. $\lambda$-function 
(or equivalently every $\rho$-function)
gives rise to a quasi invariant measure $\mu$ on $\mathfrak{G}/H$ determined up to a non zero constant factor. 
Every quasi invariant measure $\mu$ on $\mathfrak{G}/H$ is thus a pseudo-image of the right Haar measure $\mu$
on $\mathfrak{G}$ under the canonical projection $\pi$ in the terminology of \cite{Bourbaki}. In particular if 
the groups $\mathfrak{G}$ and $H$ are unimodular (i. e. $\Delta_{\mathfrak{G}} = 1_{\mathfrak{G}}$ and $\Delta_H = 1_H$) then among quasi invariant measures on $\mathfrak{G}/H$ there exists a strictly invariant measure. 

\end{scriptsize} 

\vspace*{0.2cm}

The measure space structure of 
$\mathfrak{G}/H$ uniform for the group $\mathfrak{G}$ may be transferred to $B$ together with the uniform structure, such that $(\mathfrak{G}/H,\mathscr{R}_{\mathfrak{G}/H}, \mu_{\mathfrak{G}/H}) \cong (B,\mathscr{R}_{B}, \mu_B )$. The set $B$ plays the role of the sub-manifold $Q$ in the ``measure product property''. This however would be insufficient, and we have to prove  a kind of regularity of right $H$-cosets instead of ``measure product property''. Namely let us 
define $h(x) \in H$,
which corresponds uniquely to $x \in \mathfrak{G}$, such that $h(x)^{-1}x \in B$. We have to prove that the functions 
$x \mapsto h(x)$ and  $x \mapsto h(x)^{-1}x$ are Borel (thus in particular measurable), which however was carried through in the proof of Lemma 1.4 of \cite{Mackey}. Now the only point which has to be changed is the definition of the fundamental symmetry operator $\mathfrak{J}^L$ in $\mathcal{H}^L$. We put
\[
(\mathfrak{J}^L f)_x = L_{h(x)} \mathfrak{J}_{L} L_{h(x)^{-1}} \, f_x .
\] 
We define $\mathcal{H}^L$ as 
the set of functions $\mathfrak{G} \mapsto \mathcal{H}_L$ fulfilling the conditions (i), (ii)  and such that 
\[
\int \limits_{B} \, (\, \mathfrak{J}_L ((\mathfrak{J}^L f)_x ), f_x \,) \, d\mu < \infty.
\]
The proof that $\mathcal{H}^L$ is a Hilbert space with the inner product 
\[
(f, g) = \int \, (\, \mathfrak{J}_L ((\mathfrak{J}^L f)_x ), g_x \,) \, d\mu
=\int \limits_{B} \, (f_b , g_b \,) \, d\mu_{B}(b), \,\,\, \textrm{where} \,\, b \in B, 
\]
is the same in this case with the only difference that the regularity of $H$-cosets is used instead of the Fubini theorem in reducing the problem to the von Neumann's direct integral Hilbert space construction. Namely we define a unitary map $V:
f \mapsto W^{f} = f\vert_B$
from the space $\mathcal{H}^L$ to the direct integral Hilbert space $\int \, \mathcal{H}_L \, d\mu_B$  of functions $b \mapsto W_b \in \mathcal{H}_L$ by a simple restriction to $B$ which is ``onto'' in consequence of the regularity of $H$-cosets. Its isometric character is trivial. $V$ has the inverse $W \mapsto f^W$ with $\big(f^{W}\big)_{x} = 
L_{h(x)}W_{h(x)^{-1}x}$. In particular $f^W$ is measurable on $\mathfrak{G}$ as for an orthonormal basis $\{ e_n \}_{n \in \mathbb{N}}$ of the Hilbert space $\mathcal{H}_L$ and any $\upsilon \in \mathcal{H}_L$ we have:
\[
\begin{split}
(f^{W}_{x}, \upsilon) = (f^{W}_{x}, \mathfrak{J}_L \mathfrak{J}_L  \upsilon) 
= (L_{h(x)} W_{h(x)^{-1}x}, \mathfrak{J}_L \mathfrak{J}_L  \upsilon)
= (\mathfrak{J}_L L_{h(x)} W_{h(x)^{-1}x},  \mathfrak{J}_L  \upsilon) \\
= (\mathfrak{J}_L W_{h(x)^{-1}x} , L_{h(x)^{-1}} \mathfrak{J}_L  \upsilon)
= \sum_{n \in \mathbb{N}} (\mathfrak{J}_L W_{h(x)^{-1}x} ,e_n)(e_n , L_{h(x)^{-1}} \mathfrak{J}_L  \upsilon)
\end{split}
\]
which, as a pointwise convergent series of measurable (again by Scholium 3.9 of \cite{Segal_Kunze}) functions in $x$ 
is measurable in $x$. 
We have to prove in addition that the induced representations $U^L$ in Krein spaces $(\mathcal{H}^L , \mathfrak{J}^L )$ corresponding to different choices of regular Borel sections $B$ are (Krein-)unitary equivalent. Namely let $B_1$ and $B_2$ be the two Borel sections in question. The Krein-unitary operator 
$U_{12}: (U_{12}f)_x = L_{h_{12}(x)} f_x $, where $h_{12}(x) \in H$ transforms the intersection point of the right $H$-coset $Hx$ with the section $B_1$ into the intersection point of the same coset $Hx$ with the Borel section $B_2$,
gives the Krein-unitary equivalence. The proof is similar to the proof of Lemma \ref{lop_ind_1} of Sect. \ref{lop_ind}. 

Therefore, from now on everything which concerns induced representations in Krein spaces, with the group $\mathfrak{G}$
not explicitly assumed to be equal $T_{4} \circledS SL(2, \mathbb{C})$, does not assume 
``measure product property''. Also Theorems \ref{def_ind_krein:twr.1} and \ref{def_ind_krein:twr.2} and Corollaries 
\ref{def_ind_krein:cor.1} and \ref{def_ind_krein:cor.2}
remain true without the ``measure product property'' for any locally compact and separable
$\mathfrak{G}$ and its closed subgroup $H$. Indeed using the regular Borel section $B$
of $\mathfrak{G}$ the functions (\ref{q'_h'}):
$(q,h_0 , q_0 ) \mapsto h'_{{}_{q,h_0 , q_0}}$ and $(q,h_0 , q_0 ) \mapsto q'_{{}_{q,h_0 , q_0}}$ 
may likewise be defined in this more general situation. Moreover, by Lemma 1.1 and the proof 
of Lemma 1.4 of \cite{Mackey}, $h'_{{}_{q,h_0 , q_0}}$ ranges within a compact subset of $H$, 
whenever $q$ ranges within in a compact subset of $\mathfrak{G}$, so that the proofs remain unchanged.

The construction of the induced representation in Krein space has also another invariance property: it does not depend
on the choice of a quasi invariant measure $\mu$ on $\mathfrak{G}/H$ in the unique equivalence class. Let 
$\frac{\ud \mu'}{\ud \mu}$ be the Radon-Nikodym derivative corresponding to measures 
$\mu'$ and $\mu$. Introducing the left-handed-superscript $\mu$ in ${}^{\mu}\mathcal{H}^L$ and ${}^{\mu}U^L$ 
for indicating the measure used in the construction of $\mathcal{H}^L$ and $U^L$, we may formulate a Theorem:

\begin{twr}
Let $\mu'$ and $\mu$ be quasi invariant measures in $\mathfrak{G}/H$ with Radon-Nikodym
derivative $\psi = \frac{\ud \mu'}{\ud \mu}$.
Then there exists a unitary and Krein-unitary transformation $V$ from ${}^{\mu}\mathcal{H}^L$ onto 
${}^{\mu'}\mathcal{H}^L$ such that $V \big( {}^{\mu}U^{L}_y \big) V^{-1} = {}^{\mu'}U^{L}_y$
for all $y \in \mathfrak{G}$; that is the representations ${}^{\mu}U^L$ and ${}^{\mu'}U^L$
are Krein-unitary equivalent.
\label{def_ind_krein:twr.3}
\end{twr}

\qedsymbol \,
Let $f$ be any element of ${}^{\mu}\mathcal{H}^L$ and let $\pi$ be the canonical map 
$\mathfrak{G} \mapsto \mathfrak{G}/H$. This ensures 
$(\sqrt{\psi \circ \pi} \,f , \sqrt{\psi \circ \pi} \,f )$ to be finite in ${}^{\mu'}\mathcal{H}^L$
and equal $(f , f )$ in ${}^{\mu}\mathcal{H}^L$, i. e. ensures
$\sqrt{\psi \circ \pi} \,f$  to be a member of ${}^{\mu'}\mathcal{H}^L$ as $\sqrt{\psi \circ \pi}$
is measurable with the same norm as $f$; and moreover the Krein-square-inner product
$(\sqrt{\psi \circ \pi} \,f , \sqrt{\psi \circ \pi} \,f)_{\mathfrak{J}^L}$ in ${}^{\mu'}\mathcal{H}^L$
is equal to that $(f , f)_{\mathfrak{J}^L}$ in ${}^{\mu}\mathcal{H}^L$. Moreover every $g$ in ${}^{\mu'}\mathcal{H}^L$ is evidently of the form $\sqrt{\psi \circ \pi} \,f$
for some $f \in {}^{\mu}\mathcal{H}^L$. Let $V$ be the operator of multiplication by $\sqrt{\psi \circ \pi}$.
Then $V$ defines a unitary and Krein-unitary map of ${}^{\mu}\mathcal{H}^L$ onto ${}^{\mu'}\mathcal{H}^L$. 
The verification that $V \big( {}^{\mu}U^{L} \big) V^{-1} = {}^{\mu'}U^{L}$ is immediate. 
\qed

Finally we mention the following easy but useful 
\begin{twr}
Let $L$ and $L'$ be Krein-unitary representations in $(\mathcal{H}_L , \mathfrak{J}_L)$,
which are Krein-unitary and unitary equivalent, then the induced representations $U^{L}$ and $U^{L'}$ are 
Krein-unitary equivalent.  
\label{def_ind_krein:twr.4}
\end{twr}

\subsection{Certain dense subspaces of $\mathcal{H}^L$}\label{dense}

We present here some lemmas of analytic character which we shall need later and which we have used in the proof
of Thm. \ref{def_ind_krein:twr.1} of Sect. \ref{def_ind_krein}.
Let $\mu_H$ be the right invariant Haar measure on $H$. Let $C^L$ denote the set of all functions
$f: \mathfrak{G} \ni x \mapsto f_x \in \mathcal{H}_L$, which are continuous with respect to the Hilbert space 
norm $\| \cdot \| = \sqrt{(\cdot , \cdot)}$ in the Hilbert space $\mathcal{H}_L$, and with compact support.
Let us denote the support of $f$ by $K_f$.
\begin{lem}
For each $f \in C^L$ there is a unique function $f^0$ from $\mathfrak{G}$ to $\mathcal{H}_L$ such that 
$\int \, ( \mathfrak{J}_L L_{h^{-1}} f_{hx} , \upsilon) \, d\mu_H (h) = (\mathfrak{J}_L f^{0}_{x}, \upsilon )$ for all
$x \in \mathfrak{G}$ and all $\upsilon \in \mathcal{H}_L$. This function is continuous and it is a member
of $\mathcal{H}^L$. The function $\mathfrak{G}/H \ni [x] \mapsto \big(\mathfrak{J}_L 
(\mathfrak{J}^Lf^{0})_{x}, f^{0}_{x} \big)$ as well as the function $\mathfrak{G}/H \ni [x] \mapsto f^{0}_x$ 
has a compact support. Finally $\sup_{x \in \mathfrak{G}} 
\big(\mathfrak{J}_L (\mathfrak{J}^Lf^{0})_{x}, f^{0}_{x} \big) = \sup_{b \in B} (f^{0}_{b} , f^{0}_{b}) < \infty$, 
where $B$ is a regular Borel section 
of $\mathfrak{G}$ with respect to $H$ of Sect. \ref{def_ind_krein}. 
\label{lem:dense.1}
\end{lem}
\qedsymbol \,
 Let $f \in C^L$. For each fixed $x \in \mathfrak{G}$ consider the anti-linear functional
\[ 
\upsilon \mapsto F_{x}(\upsilon) =  \int \, ( \mathfrak{J}_L L_{h^{-1}} f_{hx} , \upsilon) \, d\mu_H (h)
\] 
on $\mathcal{H}_L$.  From the Cauchy-Schwarz inequality for the Hilbert space inner product $(\cdot , \cdot )$ in the Hilbert space $\mathcal{H}_L$ and unitarity of $\mathfrak{J}_L$  with respect to the inner product 
$(\cdot , \cdot )$ in $\mathcal{H}_L$, one gets 
\[
\begin{split} 
|F_{x}(\upsilon)| \leq  \int \, 
|( \mathfrak{J}_L L_{h^{-1}} f_{hx} , \upsilon)| \, d\mu_H (h) 
\leq   \int \, \| \mathfrak{J}_L  L_{h^{-1}}f_{hx} \| \|  \upsilon \| \, d\mu_H (h) \\
=  \Big( \int \, \| \mathfrak{J}_L  L_{h^{-1}}f_{hx} \|  \, d\mu_H (h) \Big) \,\| \upsilon \|
=  \Big( \int \, \| L_{h^{-1}}f_{hx} \|  \, d\mu_H (h) \Big) \,\| \upsilon \|;
\end{split}
\] 
where the integrand in the last expression is a compactly supported continuous function of $h$ as a consequence of the strong continuity of the representation $L$ and because $f$ is compactly supported norm continuous. Therefore the integral in the last expression is finite, so that the functional $F_x$ is continuous. Thus by Riesz's theorem (in the conjugate version) there exists a unique element $g_x$ of $\mathcal{H}_L$  (depending of course on $x$) such that for all $\upsilon \in \mathcal{H}_L : F_x (\upsilon) = (g_x , \upsilon)$. We put $f^{0}_{x} = \mathfrak{J}_L g_x$, so that 
$F_x (\upsilon) = (\mathfrak{J}_L f^{0}_{x} , \upsilon)$, $\upsilon \in \mathcal{H}_L$. We have to show that 
$f^{0}: x \mapsto f^{0}_{x}$ has the desired properties. 

That $f^{0}_{h'x} = L_{h'} f^{0}_{x}$ for all $h' \in H$ and $x \in \mathfrak{G}$  follows from right invariance of 
the Haar measure $\mu_H$ on $H$:
\[
\begin{split}
(\mathfrak{J}_L L_{h'}f^{0}_{x} , \upsilon) = (\mathfrak{J}_L f^{0}_{x} , L_{h'^{-1}} \upsilon)
= \int \, ( \mathfrak{J}_L L_{h^{-1}} f_{hx} , L_{h'^{-1}} \upsilon) \, d\mu_H (h) \\
= \int \, ( \mathfrak{J}_L L_{h'} L_{h^{-1}} f_{hx} ,  \upsilon) \, d\mu_H (h) 
= \int \, ( \mathfrak{J}_L L_{(hh'^{-1})^{-1}} f_{hx} ,  \upsilon) \, d\mu_H (h) \\
= \int \, ( \mathfrak{J}_L L_{(hh'h'^{-1})^{-1}} f_{hh'x} ,  \upsilon) \, d\mu_H (hh')
= \int \, ( \mathfrak{J}_L L_{(hh'h'^{-1})^{-1}} f_{hh'x} ,  \upsilon) \, d\mu_H (h) \\
= \int \, ( \mathfrak{J}_L L_{(h)^{-1}} f_{hh'x} ,  \upsilon) \, d\mu_H (h)
= (\mathfrak{J}_L f^{0}_{h'x} , \upsilon), 
\end{split}
\]
for all $\upsilon \in \mathcal{H}_L$, $ h' \in H$, $x \in \mathfrak{G}$.

Denote the compact support of $f$ by $K$. From the strong continuity of the representation $L$ it follows immediately that the function
\[
(h, x) \mapsto f^{L}_{(h,x)} = L_{h^{-1}}f_{hx}
\]
is a norm continuous function on the direct product group $H \times \mathfrak{G}$
and compactly supported with respect to the first variable, i. e. for every $x \in \mathfrak{G}$ the 
function $h \mapsto f^{L}_{(h,x)}$
has compact support equal $K x^{-1} \cap H$. It is therefore 
uniformly norm continuous on the direct product group $H \times \mathfrak{G}$ with respect to the first variable.
For any compact subset $V$ of $\mathfrak{G}$ let $\phi_V$ be a real continuous function on $\mathfrak{G}$ with compact support equal 1 everywhere on $V$ (there exists such a function because $\mathfrak{G}$ as a topological 
space is normal). For $f \in C^L$ and any compact $V \subset \mathfrak{G}$ we introduce a 
norm continuous function on the direct product group $H\times \mathfrak{G}$ as a product $f^L \, \phi_V$:
\[
(h, x) \mapsto f^{L, V}_{(h,x)} = f^{L}_{(h,x)}  \phi_V (x),
\] 
which in addition is compactly supported and has the property that 
\[
f^{L, V}_{(h,x)} = L_{h^{-1}}f_{hx}
\]
for $(h,x) \in H \times V \subset H \times \mathfrak{G}$. In particular $f^{L,V}$ as compactly supported is not
only norm continuous but uniformly continuous on the direct product group $H \times \mathfrak{G}$ (i. e. uniformly 
in both variables jointly). 
Let $\{e_n\}_{n \in \mathbb{N}}$ be an orthonormal basis in the Hilbert space 
$\mathcal{H}_L$ and let $\mathcal{O} \subset \mathfrak{G}$ be any open set containing $x_1 , x_2 \in \mathfrak{G}$ 
with compact closure $V$. From the definition of $f^{0}$ it follows that
\[
\begin{split}
\| f^{0}_{x_1} - f^{0}_{x_2} \|^2 = \| \mathfrak{J}_L (f^{0}_{x_1} - f^{0}_{x_2}) \|^2 
= \sum_{n \in \mathbb{N}} \big| (\mathfrak{J}_L (f^{0}_{x_1} - f^{0}_{x_2}) , e_n ) \big|^2 \\
= \sum_{n \in \mathbb{N}}  \Big| \int \, ( \mathfrak{J}_L L_{h^{-1}} (f_{hx_1} - f_{hx_2}) , e_n) \, d\mu_H (h) \Big|^2 \\
\leq \sum_{n \in \mathbb{N}}   \int \, 
\big|( \mathfrak{J}_L L_{h^{-1}} (f_{hx_1} - f_{hx_2}) , e_n) \big|^2 \, d\mu_H (h) \\
=    \int \, \sum_{n \in \mathbb{N}}
\big|( \mathfrak{J}_L L_{h^{-1}} (f_{hx_1} - f_{hx_2}) , e_n) \big|^2 \, d\mu_H (h) \\
=  \int \, \big\| \mathfrak{J}_L  L_{h^{-1}}(f_{hx_1} - f_{hx_2}) \big\|^2 \, d\mu_H (h)
=  \int \, \big\|  L_{h^{-1}}(f_{hx_1} - f_{hx_2}) \big\|^2 \, d\mu_H (h) \\
\leq \sup_{h \in H} \big\|  L_{h^{-1}}f_{hx_1} - L_{h^{-1}}f_{hx_2} \big\|^2 \, 
\mu_{H}\big((Kx_{1}^{-1} \cap H) \cup (Kx_{2}^{-1} \cap H) \big) \\G_2 \cap \, ({x_0}^{-1}G_1 x_0)
\leq \sup_{h \in H} \big\|  f^{L, V}_{(h,e) \cdot (e,x_1 )} - f^{L,V}_{(h,e)\cdot (e,x_2 )} \big\|^2 \, 2 \,\sup_{x \in \mathfrak{G}} \mu_{H}(Kx^{-1} \cap H).
\end{split}
\]
Because the function $f^{L, V}$ is norm continuous on $H \times \mathfrak{G}$ and the continuity is uniform and $\sup_{x \in \mathfrak{G}} \mu_{H}(Kx^{-1} \cap H)< \infty$
(\cite{Mackey}, proof of Lemma 3.1) the norm continuity of $f^0$ is proved.
 
Similarly we get
\[
\begin{split}
 \| f^{0}_{x} \|^2 \leq  \sup_{h \in H} \big\|  L_{h^{-1}}f_{hx} \big\|^2 \,  \mu_{H}(Kx^{-1} \cap H) \\ 
= \sup_{h \in H} \big\|  f^{L, V^x }_{(h,e) \cdot (e,x )} \big\|^2 \,  \mu_{H}(Kx^{-1} \cap H) < \infty,
\end{split}
\]
because $K$ is compact and $f^{L, V^x}$ is norm continuous on $H \times \mathfrak{G}$ and compactly supported,
where $V^x$ is a compact neighbourhood of $x \in \mathfrak{G}$. Therefore $\| f^{0}_{x} \| = 0$ for all $x \notin HK$.
Thus as a function on $\mathfrak{G}/H$: $[x] \mapsto \big((\mathfrak{J}^Lf^{0})_{x}, f^{0}_{x} \big)$ and \emph{a fortiori} the function $[x] \mapsto \big(\mathfrak{J}_L (\mathfrak{J}^Lf^{0})_{x}, f^{0}_{x} \big)$ vanishes outside the compact canonical image of $HK$ in $\mathfrak{G}/H$. 

Finally let us note that if $h(x)$ is the element of $H$ defined in Sect. \ref{def_ind_krein} 
corresponding to $x \in \mathfrak{G}$, then 
\[
\big(\mathfrak{J}_L (\mathfrak{J}^Lf^{0})_{x}, f^{0}_{x} \big) =  (f^{0}_{b} , f^{0}_{b})
\] 
with $b = h(x)^{-1}x$ -- the unique intersection point of the coset $Hx$ with the Borel section $B$.
Because $f$ is continuous with compact support, then the last assertion of the Lemma follows from 
Lemma 1.1 of \cite{Mackey}.
\qedsymbol \,

\vspace*{0.5cm}

We shall denote the class of functions $f^0$ for $f \in C^L$ of Lemma \ref{lem:dense.1} by $C^{L}_{0}$.

\begin{lem}
For each fixed $x \in \mathfrak{G}$ the vectors $f^{0}_{x}$ for $f^{0} \in C^{L}_{0}$ form a dense linear subspace
of $\mathcal{H}_L$.
\label{lem:dense.2}
\end{lem}

\qedsymbol \,
Note that if $f^{0} \in C^{L}_{0}$ and $R_s f$ is defined by the equation $(R_s f)_x = f_{xs}$
for all $x$ and $s$ in $\mathfrak{G}$ then $R_s f^{0} = (R_s f)^{0}$ so that for all $f \in C^L$ and $s \in \mathfrak{G}$, 
$R_s f^0 \in C^{L}_{0}$. Therefore the set $\mathcal{H}_{L}''$ of vectors $f^{0}_x$ for $f^0 \in C^{L}_{0}$ and $x$ fixed is independent of $x$. Let $\mathcal{H}_{L}'$ be the $\mathfrak{J}_L$-orthogonal complement of $\mathcal{H}_{L}''$,
i. e. the set of all $\upsilon \in \mathcal{H}_L$ such that $(\mathfrak{J}_L g, \upsilon) = 0$ for all 
$g \in \mathcal{H}_{L}''$. Then if $\upsilon \in \mathcal{H}_{L}'$ we have $(f^{0}_{x}, \upsilon) = 0$ for all
$f^0 \in C^{L}_{0}$ and all $x \in \mathfrak{G}$. Therefore $(\mathfrak{J}_L f^{0}_{hx}, \upsilon) 
= (\mathfrak{J}_L f^{0}_{x}, L_{h^{-1}}\upsilon) = 0$ for all $f^0$ in $C^{L}_{0}$, all $x$ in $\mathfrak{G}$ and  
all $h \in H$. Hence $\mathcal{H}_{L}'$ is invariant under the representation, as $L$ is $\mathfrak{J}_L$-unitary.
Let $L'$ be the restriction of $L$ to $\mathcal{H}_{L}'$. Suppose that there exists a non zero member $f^0$ of
$C^{L'}_{0}$. Thus \emph{a fortiori} $f^0 \in C^{L}_{0}$ and we have a contradiction since the values of $f^0$
are all in $\mathcal{H}_{L}'$, so that we would have in $( \mathfrak{J}^L , \mathcal{H}^L )$: 
\[
\begin{split}
(f^0 , g)_{\mathfrak{J}^L} = ( \mathfrak{J}^L f^0, g ) \\
= \int \, (\mathfrak{J}_L f^{0}_{x} , g_x \, d\mu_{\mathfrak{G}/H} = 0
\end{split}
\]   
for all $g \in \mathcal{H}^L$, which would give us $f^0 = 0$, because the Krein space 
$( \mathfrak{J}^L , \mathcal{H}^L )$ of the induced representation $U^L$ is non degenerate (or $\mathfrak{J}^L$ invertible). Thus in order to show that 
$\mathcal{H}_{L}' = 0$ we need only show that when $\mathcal{H}_{L}' \neq 0$ there exists a non zero member 
$f^0$ of $C^{L'}_{0}$. But if none existed then 
\[
\int \, (\mathfrak{J}_L L'_{h^{-1}} f_{hx}, \upsilon) \, d\mu_H (h)
\] 
would be zero for all $x$, all $\upsilon$ in $\mathcal{H}_L$ and all $f$ in $C^{L'}$. In particular the integral would be 
zero for $f = u \upsilon'$, for all continuous complex functions  $u$ on $\mathfrak{G}$ of compact support and all 
$\upsilon' \in \mathcal{H}_{L}'$, i .e  
\[
\int \, u(hx) (\mathfrak{J}_L L'_{h^{-1}} \upsilon' , \upsilon ) \, d\mu_H (h)
\] 
would be zero for all $x$, all $\upsilon$ in $\mathcal{H}_L$, all $\upsilon'$ in $\mathcal{H}_{L}'$ and all 
complex continuous $u$ of compact support on $\mathfrak{G}$, which, because $L$ (and thus $L'$) 
is strongly continuous, would imply that 
\[
(\mathfrak{J}_L L'_{h^{-1}} \upsilon' , \upsilon ) = 0
\]
for all $\upsilon$ in $\mathcal{H}_L$, all $\upsilon'$ in $\mathcal{H}_{L}'$ and all $h \in H$.
This is impossible because the Krein space $( \mathfrak{J}_L , \mathcal{H}_L )$ of the representation $L$ is non degenerate and $L'_{h^{-1}}$ non-singular as a Krein-unitary operator. Thus we have proved that $\mathcal{H}_{L}' = 0$. This means that
$\mathfrak{J}_L \mathcal{H}_{L}''$ is dense in the Hilbert space $\mathcal{H}_L$, and because $\mathfrak{J}_L$ is
unitary in $\mathcal{H}_L$ with respect to the ordinary definite inner product $(\cdot , \cdot)$, this means that 
$\mathcal{H}_{L}''$ is dense in the Hilbert space $\mathcal{H}_L$.
\qed

\vspace*{0.5cm}

\begin{lem}
Let $C$ be any family of functions from $\mathfrak{G}$ to $\mathcal{H}_L$ such that:
\begin{enumerate}

\item[(a)]
$C \subset \mathcal{H}^L$.

\item[(b)]
For each $s \in \mathfrak{G}$ there exists a positive Borel function $\rho_s$ such that for all $f \in C$,
$\rho_s R_s f \in C$ where $(R_s f)_x = f_{xs}$.

\item[(c)]
If $f \in C$ then $gf \in C$ for all bounded continuous complex valued functions $g$ on $\mathfrak{G}$ which
are constant on the right $H$-cosets.

\item[(d)]

There exists a sequence $f^1 , f^2 , \ldots$ of members of $C$ and a subset $P$ of $\mathfrak{G}$ of positive 
Haar measure such that for each $x \in P$ the members $f^{1}_{x} , f^{2}_{x} , \ldots$ of $\mathcal{H}_L$ 
have $\mathcal{H}_L$ as their closed linear span.

\end{enumerate}

Then the members of $C$ have $\mathcal{H}^L$ as their closed linear span.
\label{lem:dense.3}
\end{lem}

\qedsymbol \,
 Choose $f^1 , f^2 , \ldots$ as in the condition (d). Let $u$ be any member of $\mathcal{H}^L$
which is $\mathfrak{J}^L$-orthogonal to all members of $C$:  
\[
\big(f, u \big)_{\mathfrak{J}^L} = (\mathfrak{J}^L f, u) = \int \, (\, \mathfrak{J}_L (f_x ), u_x \,) \, d\mu_{\mathfrak{G}/H} = 0
\]
for all $f \in C$. Then
\[
\begin{split}
(\mathfrak{J}^L (\rho_s g)(R_s f^j), u) = \int \, (\, \mathfrak{J}_L ((\rho_s g)(x)(R_s f^j)_x ), u_x \,) \, d\mu_{\mathfrak{G}/H} = 0
\end{split}
\] 
for every $j \in \mathbb{N}$, all $s$ and every bounded continuous $g$ on $\mathfrak{G}$ which is constant on the right $H$-cosets. It follows at once that for all $s$ and all $j \in \mathbb{N}$ $(\mathfrak{J}_L f^{j}_{xs}, u_x) = 0$ for almost all $x \in \mathfrak{G}$. Since $x \mapsto (\mathfrak{J}_L f^{j}_{x}, u_x)$ is a Borel function on $\mathfrak{G}$ the
function 
\[
(x, s) \mapsto (\mathfrak{J}_L f^{j}_{xs}, u_x) = \sum_{n \in \mathbb{N}}(\mathfrak{J}_L f^{j}_{xs}, e_n ) (e_n , u_x) 
\]
is Borel on the product measure space $\mathfrak{G} \times \mathfrak{G}$ on repeating the argument of Sect. \ref{def_ind_krein} 
(Scholium 3.9 of \cite{Segal_Kunze}) and joining it with the fact that composition of a measurable (Borel) function on 
$\mathfrak{G}$ with the
continuous function $\mathfrak{G} \times \mathfrak{G} \ni (x, s) \mapsto xs \in \mathfrak{G}$ is measurable (Borel)
on the product measure space $\mathfrak{G} \times \mathfrak{G}$ (compare e. g. \cite{Segal_Kunze}). Thus we may apply 
the Fubini theorem (Thm. 3.4 in \cite{Segal_Kunze}) and conclude that for almost all $x$, $(\mathfrak{J}_L f^{j}_{xs}, u_x)$ is zero for almost all $s$. Since $j$ runs over a countable class 
we may select a single null set $N \subset \mathfrak{G}$ such that for each $x \notin N$,  
$(\mathfrak{J}_L f^{j}_{xs}, u_x)$ is, for almost all $s$, zero for all $j \in \mathbb{N}$. It follows that for each 
$x \notin N$ there exists $s \in x^{-1}P$ such that $(\mathfrak{J}_L f^{j}_{xs}, u_x) = 0$ for $j \in \mathbb{N}$
and hence that $u_x = 0$ because $\mathfrak{J}_L$ is unitary with respect to the ordinary definite Hilbert space inner product in the Hilbert space $\mathcal{H}_L$. Thus $u$ is almost everywhere zero and $\mathfrak{J}^L C$ must be 
dense in $\mathcal{H}^L$. 
Because $\mathfrak{J}^L$ is unitary in the ordinary sense with respect to the definite inner product 
(eq. (\ref{inn_ind_def}) of Sect. \ref{def_ind_krein}) in $\mathcal{H}^L$, $C$ must be dense in $\mathcal{H}^L$.  
\qed

\vspace*{0.5cm}

\begin{lem}
Let $C^1$ be any family of functions from $\mathfrak{G}$ to $\mathcal{H}_L$ such that:
\begin{enumerate}

\item[(a)]
For each $f \in C^1$ there exists a positive Borel function $\rho$ on $\mathfrak{G}$ such that
\[
\Big(\mathfrak{J}_L \frac{1}{\rho (x)}f_x , \upsilon \Big) = \Big(\frac{1}{\rho (x)}\mathfrak{J}_L f_x , \upsilon \Big)
= \frac{1}{\rho (x)} \Big(\mathfrak{J}_L f_x , \upsilon \Big)
\]
is continuous as a function of $x$ for all $\upsilon \in \mathcal{H}_L$.

\item[(b)]
$C^1 \subset \mathcal{H}^L$.

\item[(c)]
For each $s \in \mathfrak{G}$ there exists a positive Borel function $\rho_s$ such that for all $f \in C^1$,
$\rho_s R_s f \in C^1$ where $(R_s f)_x = f_{xs}$.

\item[(d)]
If $f \in C^1$ then $gf \in C^1$ for all bounded continuous complex valued functions $g$ on $\mathfrak{G}$ which
are constant on the right $H$-cosets and vanish outside of $\pi^{-1}(K)$ for some compact subset $K$ of 
$\mathfrak{G}/H$.

\item[(e)]
For some (and hence all) $x \in \mathfrak{G}$ the members $f_{x}$ of $\mathcal{H}_L$  for $f \in C^1$
have $\mathcal{H}_L$ as their closed linear span.

\end{enumerate}

Then the members of $C^1$ have $\mathcal{H}^L$ as their closed linear span.

\label{lem:dense.4}
\end{lem}

\qedsymbol \,
 Choose $f^1 , f^2 , \ldots$ in $C^1$ so that $f^{1}_{e} , f^{2}_{e} , \ldots$ 
have $\mathcal{H}_L$ as their closed linear span; $e$ being the identity of $\mathfrak{G}$. 
Let $u$ be any member of $\mathcal{H}^L$
which is $\mathfrak{J}^L$-orthogonal to all members of $C^1$. Then
\[
\begin{split}
(\mathfrak{J}^L (\rho_s g)(R_s f^j), u) = \int \, (\, \mathfrak{J}_L ((\rho_s g)(x)(R_s f^j)_x ), u_x \,) \, d\mu_{\mathfrak{G}/H} = 0
\end{split}
\] 
for every $j \in \mathbb{N}$, all $s$ and every bounded continuous $g$ on $\mathfrak{G}$ which is constant on the 
right $H$-cosets. It follows at once that for all $s$ and all $j \in \mathbb{N}$ $(\mathfrak{J}_L f^{j}_{xs}, u_x) = 0$ for almost all $x \in \mathfrak{G}$. Since $(x, s) \mapsto (\mathfrak{J}_L f^{j}_{xs}, u_x)$ is a Borel function on 
the product measure space $\mathfrak{G} \times \mathfrak{G}$ (compare the proof of Lemma \ref{lem:dense.3}) 
we may apply the Fubini theorem as in the preceding Lemma and conclude that for almost all $x$, 
$(\mathfrak{J}_L f^{j}_{xs}, u_x)$ is zero for almost all $s$. Since $j$ runs over a countable class we may select a 
single null set $N$ in $\mathfrak{G}$ such that for each $x \notin N$, $(\mathfrak{J}_L f^{j}_{xs}, u_x)$ is for
almost all $s$ zero for all $j$. Suppose that $u_{x_1} \neq 0$ for some $x_1 \notin N$. Then 
$(\mathfrak{J}_L f^{j}_{e}, u_{x_1}) \neq 0$ for some $j$ as $\mathfrak{J}_L$ is unitary with respect to the ordinary
Hilbert space inner product $(\cdot, \cdot)$ in $\mathcal{H}_L$ (as in the proof of the preceding Lemma). But for some positive Borel function $\rho$, $(\mathfrak{J}_L f^{j}_{x}, u_{x_1})\big/\rho(x)$ is continuous in $x$. 
Hence $(\mathfrak{J}_L f^{j}_{x_1 s}, u_{x_1})\big/\rho(x_1 s) \neq 0$ for $s$ in some neighbourhood of $x_{1}^{-1}$.
Thus $(\mathfrak{J}_L f^{j}_{x_1 s}, u_{x_1}) \neq 0$ for $s$ in some neighbourhood of $x_{1}^{-1}$. But this contradicts 
the fact that $(\mathfrak{J}_L f^{j}_{x_1 s}, u_{x_1})$ is zero for almost all $s \in \mathfrak{G}$. Therefore
$u_x$ is zero almost everywhere. Thus only the zero element is orthogonal (in the ordinary positive inner product space
in $\mathcal{H}^L$) to all members of $\mathfrak{J}^L C^1$ and it follows that $\mathfrak{J}^L C^1$ must be dense
in $\mathcal{H}^L$. Because $\mathfrak{J}^L$ is unitary with respect to the ordinary  definite inner product 
$(\cdot, \cdot)$ in $\mathcal{H}^L$, it follows that $C^1$ is dense in $\mathcal{H}^L$.
\qed

\vspace*{0.5cm}

\begin{lem}
$C^{L}_{0}$ is dense in $\mathcal{H}^L$.
\label{lem:dense.5}
\end{lem}

\qedsymbol \,
The Lemma is an immediate consequence of Lemmas \ref{lem:dense.2} and \ref{lem:dense.4}. 
\qed

\vspace*{0.5cm}

\begin{lem}
There exists a sequence $f^1 , f^2 , \ldots$ of elements $C^{L}_{0} \subset \mathcal{H}^L$ such that for each fixed $x \in \mathfrak{G}$ the vectors $f^{k}_{x}$, $k = 1, 2, \ldots$ form a dense linear subspace of $\mathcal{H}_L$.
\label{lem:dense.6}
\end{lem}

\qedsymbol \,
 We have seen in the previous Sect. that as a Hilbert space $\mathcal{H}^L$  is unitary equivalent to the direct integral Hilbert space $\int \, \mathcal{H}_L \, d\mu_{\mathfrak{G}/H}$ over the $\sigma$-finite and regular Baire  (or Borel) measure space $(\mathfrak{G}/H , \mathscr{R}_{\mathfrak{G}/H}, \mu_{\mathfrak{G}/H})$ with 
separable $\mathcal{H}_L$. Because $\mathfrak{G}/H = \mathfrak{X}$ is locally compact metrizable and fulfils the second axiom of countability its minimal (one point or Alexandroff) compactification $\mathfrak{X}_+$ is likewise metrizable
(compare e. g. \cite{Engelking}, Corollary 7.5.43). Thus the Banach algebra $C(\mathfrak{X}_+)$ is separable,
compare e. g. \cite{Krein}, Thm. 2 or \cite{Gelfand_Silov}). Because $C(\mathfrak{X}_+)$ is equal to the minimal unitization
$C_0 (\mathfrak{X})^+$ of the Banach algebra $C_0 (\mathfrak{X})$ of continuous functions on $\mathfrak{X}$ 
vanishing at infinity (compare \cite{Neumark_dec}), thus by the construction of minimal unitization
it follows that $C_0 (\mathfrak{X})$ is separable (of course with respect to the supremum norm in 
$C_0 (\mathfrak{X})$) as a closed ideal
in $C_0 (\mathfrak{X})^+$ of codimension one. Because the  
measure space $(\mathfrak{G}/H , \mathscr{R}_{\mathfrak{G}/H}, \mu_{\mathfrak{G}/H})$ is the regular Baire measure space, 
induced by the integration lattice $C_\mathcal{K}(\mathfrak{X}) \subset C_0 (\mathfrak{X})$ of continuous functions with compact support (compare \cite{Segal_Kunze}), it follows
from Corollary 4.4.2 of \cite{Segal_Kunze} that the Hilbert space 
$L^2 (\mathfrak{G}/H , \mu_{\mathfrak{G}/H})$ of square summable functions over $\mathfrak{X} = \mathfrak{G}/H$
is separable\footnote{For the reasons explained in Sect. \ref{decomposition} we are interesting 
in complete measure spaces on $\mathfrak{G}/H$ and on all other quotient spaces encountered later in this paper. 
But the Baire or Borel measure is pretty sufficient in the investigation of the associated Hilbert spaces 
$L^2 (\mathfrak{G}/H , \mu_{\mathfrak{G}/H})$ or $\mathcal{H}^L$ as all measurable sets differ from the Borel sets just by 
null sets, and the space of equivalence classes of Borel square summable functions in $L^2 (\mathfrak{G}/H , \mu_{\mathfrak{G}/H})$ is the same as the space of equivalence classes of square summable measurable functions. Recall that the Baire measure space may be completed to a Lebesgue-type measure space, e. g. using the 
Carath\'eodory method. In other words the Baire or Borel (the same in this case) measure space may be completed such that any subset of measurable null set will be measurable.}. Let $\{e_n \}_{n \in  \mathbb{N}}$ be an orthonormal basis in $\mathcal{H}_L$. Using standard -- by now -- Hilbert space (\cite{Neumark_dec}) and measure space 
(e. g. Fubini theorem\footnote{Compare eq. (\ref{dir_int_L^2:decompositions}) of Sect.\ref{decomposition}. }) techniques  and the results of \cite{von_neumann_dec} one can prove that 
\[
\begin{split}
\int \, \mathcal{H}_L \, d\mu_{\mathfrak{G}/H} 
= \bigoplus \limits_{n \in \mathbb{N}} \, \int \, \mathbb{C} e_n \, d\mu_{\mathfrak{G}/H} \\
= \bigoplus \limits_{n \in \mathbb{N}} \, L^2 (\mathfrak{G}/H , \mu_{\mathfrak{G}/H}) .
\end{split}
\]
Thus  
$\int \, \mathcal{H}_L \, d\mu_{\mathfrak{G}/H}$ itself must be separable and therefore 
$\mathcal{H}^L$ is separable. Thus we may choose a sequence $f^1 , f^2 , \ldots$ of elements 
$C^{L}_{0} \subset \mathcal{H}^L$ such that for each $f \in C^{L}_{0}$ there exists a subsequence 
$f^{n_1} , f^{n_2} , \ldots$ which converges in norm $\| \cdot \|$ of $\mathcal{H}^L$ to $f$. Then a slight and obvious
modification of the standard proof of the Riesz-Fischer theorem (e. g. \cite{Segal_Kunze}, Thm. 4.2) gives a sub-subsequence $f^{n_{m_1}} , f^{n_{m_2}} , \ldots$ which, after restriction to the regular Borel section $B \cong \mathfrak{G}/H$ converges almost uniformly to the restriction of $f$ to $B$ (where $B \cong \mathfrak{G}/H$ is locally compact with the natural topology induced by the canonical projection $\pi$, with the Baire measure space structure 
$(\mathfrak{G}/H,\mathscr{R}_{\mathfrak{G}/H}, \mu_{\mathfrak{G}/H}) \cong (B,\mathscr{R}_{B}, \mu_B )$)
obtained by Mackey's technique of quotiening the measure space $\mathfrak{G}$ by the group $H$
recapitulated shortly in Sect. \ref{def_ind_krein}. As $f^k , f$ are continuous and compactly supported as functions on 
$B \cong \mathfrak{G}/H$, the convergence is uniform on $B$. The Lemma now, for $x \in B$, is an immediate consequence of Lemma \ref{lem:dense.2}. Because for each $x \in \mathfrak{G}$ we have $f^{k}_{x} = L_{h(x)} f^{k}_{h(x)^{-1}x}, f_{x} = L_{h(x)} f_{h(x)^{-1}x}$ with $h(x)^{-1}x \in B$ and because $L_h$ is invertible (and bounded) for every $h \in H$, the Lemma is proved.   
\qed

\vspace*{0.5cm}














\subsection{{\L}opusza\'nski representation as an induced representation}\label{lop_ind}

Let $\mathfrak{G}$ be a separable locally compact group and $H$ its closed subgroup.
In this section we shall need Lemma \ref{lop_ind_1} (below), which we prove assuming 
the ``measure product property'',
because it is sufficient for the analysis of the {\L}opusza\'nski representation
of the double covering of the Poincar\'e group. However it can be proved without this assumption,
as the reader will easily see by recalling the respective remarks of Sect. \ref{def_ind_krein}.

Thus we assume (for simplicity)
that the right Haar measure space $\Big( \, \mathfrak{G}\, , \,\, \mathscr{R}_{{}_{\mathfrak{G}}} \, , \,\,
\mu_{{}_{\mathfrak{G}}} \, \Big)$ 
be equal to the product measure space  $\Big( \, H \times \mathfrak{G}/H, \,\, 
\mathscr{R}_{{}_{H \times \mathfrak{G}/H}}\, , \,\, \mu_{{}_{H}} \times \mu_{{}_{\mathfrak{G}/H}} \, \Big)$
with $\Big( \, H , \,\, 
\mathscr{R}_{{}_{H}}\, , \,\, \mu_{{}_{H}} \, \Big)$ equal to the right Haar measure space on $H$
and with the Mackey quotient measure space $\Big( \, \mathfrak{G}/H \, , \,\, \mathscr{R}_{{}_{\mathfrak{G}/H}} \, , \,\,
\mu_{{}_{\mathfrak{G}/H}} \, \Big)$ on $\mathfrak{G}/H$ (described briefly in Sect \ref{def_ind_krein}).
In most cases of physical applications both $\mathfrak{G}$ and $H$ are unimodular.  
Let $g = h \cdot q$ be the corresponding unique factorization of $g \in \mathfrak{G}$
with $h \in H$ and $q \in Q \subset \mathfrak{G}$ representing the class $[g] \in \mathfrak{G}/H$. Uniqueness of the factorization allows us to introduce the following functions ((already mentioned in Sect. \ref{def_ind_krein}) 
$(q,h_0 , q_0 ) \mapsto h'_{{}_{q,h_0 , q_0}} \in H$ 
and $(q,h_0 , q_0 ) \mapsto q'_{{}_{q,h_0 , q_0}} \in Q \cong \mathfrak{G}/H$, where for any
$g_0 = q_0 \cdot h_0 \in \mathfrak{G}$ we define $ h'_{{}_{q,h_0 , q_0}} \in H$ 
and $q'_{{}_{q,h_0 , q_0}} \in Q \subset \mathfrak{G}$ 
to be the elements, uniquely corresponding to $(q,h_0 , q_0 )$, such that 
\[
q\cdot h_0 \cdot q_0 = h'_{{}_{q,h_0 , q_0}} \cdot q'_{{}_{q,h_0 , q_0}}.
\] 
In particular if  $g = h q$, then $q$ represents $[g] \in \mathfrak{G}/H$, and  
$q'_{{}_{q,h_0 , q_0}}$ represents $[gg_0]$, i.e. the right action of $\mathfrak{G}$ on $\mathfrak{G}/H$.
It is easily verifiable that $(q,h_0 , q_0 ) \mapsto h'_{{}_{q,h_0 , q_0}}$ behaves like a multiplier,
i.e. denoting $h'_{{}_{q,h_0 , q_0}}$ and $q'_{{}_{q,h_0 , q_0}}$ just by $h'_{{}_{q,g_{{}_0}}}$
and $q'_{{}_{q,g_{{}_0}}}$ we have
\[
\boxed{h'_{{}_{q,\, g_{{}_0}}} \cdot h'_{{}_{q'_{{}_{q,g_{{}_0}}}, \, g_{{}_1}}} = h'_{{}_{q , \, g_{{}_0} g_{{}_1}}}.}
\] 

Let $U^L$ be the Krein isometric representation of $\mathfrak{G}$ induced by an almost uniformly bounded Krein-unitary
representation of $H$ in the Krein space $(\mathcal{H}_L , \mathfrak{J}_L)$, defined as in Sect. \ref{def_ind_krein}. 
Let us introduce the Hilbert space
\begin{equation}\label{hilbert_system_prim}
\mathcal{H} = \int \limits_{\mathfrak{G}/H} \, \mathcal{H}_L \, \ud \mu_{{}_{\mathfrak{G}/H}}
\end{equation}
and the fundamental symmetry $\mathfrak{J}$
\begin{equation}\label{fund_sym_dec}
\mathfrak{J} = \int \limits_{\mathfrak{G}/H} \, \mathfrak{J}_L \, \ud \mu_{{}_{\mathfrak{G}/H}}
\end{equation}
in $\mathcal{H}$, i.e. operator decomposable with respect to the decomposition (\ref{hilbert_system_prim}) 
whose all components in its decomposition are equal $\mathfrak{J}_L$. Because
 $\mathfrak{J}_{L}^{*} = \mathfrak{J}_L$
and $\mathfrak{J}_{L}^{*} \mathfrak{J}_L = \mathfrak{J}_L \mathfrak{J}_{L}^{*} = I$, then by \cite{von_neumann_dec}
the same holds true of the operator $\mathfrak{J}$, i.e. it is unitary and selfadjoint, i.e. $\mathfrak{J}^* = \mathfrak{J}$
and $\mathfrak{J}^* \mathfrak{J} = \mathfrak{J} \mathfrak{J}^* = I$, so that  $\mathfrak{J}^2 = I$ and 
$\mathfrak{J}$ is a fundamental symmetry. We may therefore introduce the Krein space $(\mathcal{H}, \mathfrak{J})$.

\begin{lem}\label{InducedFormToImprimitivity}
Let $\mathfrak{G}$ be a separable locally compact group and $H$ its closed subgroup. Assume (for simplicity)
that the ''measure product property'' is fulfilled by $\mathfrak{G}$ and $H$. Then the operators
\[
U : \mathcal{H} \mapsto \mathcal{H}^L, \,\,\, \textrm{and} \,\,\,
S : \mathcal{H}^L \mapsto \mathcal{H},
\]
defined as follows 
\[
\Big( U W \Big)_{h\cdot q} = L_h W_{{}_q}, \,\,\, \textrm{and} \,\,\,
\Big( S f \Big)_{q} = L_{{}_{h^{-1}}} f_{{}_{h\cdot q}},
\] 
for all $W \in \mathcal{H}$ and $f \in \mathcal{H}^L$, are well defined operators, both are isometric
and Krein-isometric between $(\mathcal{H}, \mathfrak{J})$ and $(\mathcal{H}^L , \mathfrak{J}^L)$ and 
moreover $US = I$ and $SU = I$ and moreover 
\[
U^{-1} \mathfrak{J}^L U = \mathfrak{J},
\]
so that $U$ and $S$ are unitary and Krein-uinitary. We have 
\[
\Big( V_{g_{{}_0}} W \Big)_{q} = \Big( U^{-1} U^{L}_{g_{{}_0}} U W \Big)_{q} \\
= \sqrt{\lambda(q, g_{{}_0})} L_{{}_{h'_{{}_{q,g_{{}_0}}}}} W_{{}_{q'_{{}_{q,g_{{}_0}}}}};
\] 
or equivalently
\[
\Big( V_{g_{{}_0}} W \Big)_{[g]} = \Big( U^{-1} U^{L}_{g_{{}_0}} U W \Big)_{[g]} \\
= \sqrt{\lambda([g], g_{{}_0})} L_{{}_{h'_{{}_{[g],g_{{}_0}}}}} W_{{}_{[g \cdot g_{{}_0}]}}.
\]  
In short: $U^L$ is unitary and Krein unitary equivalent to the Krein-isometric representation
$V$ of $\mathfrak{G}$ in $(\mathcal{H}, \mathfrak{J})$.  

\label{lop_ind_1}
\end{lem}

\qedsymbol \,
That the functions $UW$, $W \in \mathcal{H}$, and $U^{-1}f$, $f \in \mathcal{H}^L$ fulfil the required measurability conditions has been already shown in Sect. \ref{def_ind_krein}). Verification of the isometric and Krein-isometric
character of both $U$ and $S$ is easy, and we leave it to the reader. Checking $US = I$ and $SU = I$ as well as the
last equality is likewise simple.   
\qed

Now let us turn our attention to the construction of semi-direct product groups and their specific class of 
Krein-isometric representations to which the {\L}opusza\'nski representation belong
together with the related systems of imprimitivity in the Krein space $(\mathcal{H}, \mathfrak{J})$,
say of Lemma \ref{lop_ind_1}.
Let $G_1$ and $G_2$ be separable locally compact groups and let $G_1$ be Abelian 
($G_1$ plays the role of four translations subgroup $T_4$ and $G_2$ plays the role of the
$SL(2, \mathbb{C})$ subgroup of the double covering $\mathfrak{G} = T_4 \circledS SL(2,\mathbb{C})$
of the Poincar\'e group). Let there be given a homomorphism
of $G_2$ into the group of automorphisms of $G_1$ and let $y [x] \in G_1$ be the action of the automorphism
corresponding to $y$ on $x \in G_1$. We assume that $(x,y) \mapsto y [x]$ is jointly continuous in both variables.
We define the semi-direct product $\mathfrak{G} = G_1 \circledS G_2$ as the topological product
$G_1 \times G_2$ with the multiplication rule $(x_1 , y_1)(x_2 , y_2) = (x_1 y_{{}_1} [x_2], y_1 y_2)$. 
$\mathfrak{G} = G_1 \circledS G_2$ under this operation is a separable locally compact group. Recall that the subset of elements $(x,e)$ with $x \in G_1$ and $e$ being the identity is a closed subgroup of the semi direct product $\mathfrak{G}$ naturally isomorphic to $G_1$ and similarly the set of elements
$(e,y)$, $y \in G_2$ is a closed subgroup of $\mathfrak{G} = G_1 \circledS G_2$ naturally isomorphic to
$G_2$. Let us identify those subgroups with $G_1$ and $G_2$ respectively. Since $(x,e)(e,y) = (x,y)$
it follows at once that any Krein-isometric representation $(x,y) \mapsto V_{(x,y)}$ of 
$\mathfrak{G} = G_1 \circledS G_2$ in the Krein space  $(\mathcal{H}, \mathfrak{J})$ is determined by its 
restrictions $N$ and $U$ to the subgroups $G_1$ and $G_2$ respectively:
$V_{(x,y)} = N_x U_y$. Conversely if $N$ and $U$ are Krein-isometric representations of $G_1$
and $G_2$ which act in the same Krein space $(\mathcal{H}, \mathfrak{J})$
and with the same core invariant domain $\mathfrak{D}$, and moreover if the representation $N$
commutes with the fundamental symmetry $\mathfrak{J}$ and is therefore unitary, then one easily checks that
$(x,y) \mapsto N_x U_y$ defines a Krein-isometric representation if and only if $U_y N_x U_{y^{-1}} = N_{ y[x]}$.
Indeed the ``if'' part is easy. Assume then that $ V_{(x,y)} = N_x U_y$ is a representation. Then for any 
$(x,y), (x',y') \in G_1 \circledS G_2$ one has $N_x U_y N_{x'} U_{y^{-1}} U_{yy'} = N_x N_{y[x']}U_{yy'}$
on the core dense set $\mathfrak{D}$.
Because $N_x$ is unitary it follows that $U_y N_{x'} U_{y^{-1}} U_{yy'} = N_{y[x']}U_{yy'}$ on $\mathfrak{D}$.
Because $U_y \mathfrak{D} = \mathfrak{D}$ for all $y \in y \in G_2$ and $U_y U_{y^{-1}} = I$ on $\mathfrak{D}$,
then it follows that $U_y N_{x'}U_{y^{-1}} = N_{y[x']}$ on $\mathfrak{D}$ for all $x,x' \in G_1$ and all $y \in G_2$.
Because the right hand side is unitary, then $U_y N_{x'} U_{y^{-1}}$ can be extended to a unitary operator,
although $U$ is in general unbounded.  
 Now assume (which is the case for representations of translations acting in one particle states in QFT,
for example this is the case for the restriction of the {\L}opusza\'nski representation to the translation subgroup) 
that the representation $N$ of the Abelian subgroup $G_1$ commutes with the fundamental symmetry 
$\mathfrak{J}$ in $\mathcal{H}$, and thus it is not only Krein-isometric but unitary in $\mathcal{H}$ 
in the usual sense. Moreover, the restrictions  $N$ of representations acting in one particle states are in 
fact of uniform (even finite) multiplicity. Because $N$ is a unitary representation 
of a separable locally compact Abelian group $G_1$ in the Hilbert space the Neumak's theorem is applicable,
which says that $N$ is determined by a projection valued (spectral) measure $S \mapsto E_S$ 
(which as we will see may be associated with the direct integral decomposition 
(\ref{hilbert_system_prim}) with the appropriate subgroup $H$), defined on the Borel (or Baire)
sets $S$ of the character group $\widehat{G_1}$ of $G_1$:  
\[
N_x = \int \limits_{\widehat{G_1}} \chi(x) \, dE(\chi).
\]

It is readily verified that 
$N$ and $U$ satisfy the above identity if and only if the spectral measure $E$ and the 
representation $U$ satisfy $U_y E_S U_{y^{-1}} = E_{[S]y}$, for all $y \in G_2$ and
all Borel sets $S \subset \widehat{G_1}$; where the action $[\chi]y$ of $y \in G_2$
on $\chi \in \widehat{G_1}$ is defined
by the equation $\langle [\chi]y, x \rangle = \langle \chi, y^{-1}[x] \rangle$ (with $\langle \chi, x \rangle$
denoting the value of the character $\chi \in \widehat{G_1}$ on the element $x \in G_1$). 
Indeed:
\begin{multline}\label{ueu}
U_y N_x U_{y^{-1}} = \int \limits_{\widehat{G_1}} \chi(x) \, d(U_yE(\chi)U_{y^{-1}}) 
= N_{y[x]} = \int \limits_{\widehat{G_1}} \chi(y[x]) \, dE(\chi) \\ 
=\int \limits_{\widehat{G_1}} \big([\chi]y^{-1}\big)(x) \, dE(\chi) = \int \limits_{\widehat{G_1}} \chi(x) \, dE([\chi]y) .
\end{multline}
We call such $E$, $N$, and $U$
a \emph{system of imprimitivity in the Krein space} $(\mathcal{H}, \mathfrak{J})$, after Mackey \cite{Mackey_imprimitivity}
who defined the structure for representations $N$ and $U$ in Hilbert space $\mathcal{H}$
which are both unitary in the ordinary sense.

Consider now the action of $G_2$ on $\widehat{G_1}$. If the spectral measure $E$ is concentrated in one
of the orbits of $\widehat{G_1}$ under $G_2$ let $\chi_0$ be any member of this orbit $\mathscr{O}_{\chi_0}$ and
let $G_{\chi_0}$ be the subgroup of all $y \in G_2$ for which $[\chi_0]y = \chi_0$. 
Then $y \mapsto [\chi_0]y$ defines a one-to-one Borel set preserving map between the points
of this orbit $\mathscr{O}_{\chi_0}$ and the points of the homogeneous space $G_2 / G_{\chi_0} = \mathfrak{G}/H$,
where $H = G_1 \cdot G_{\chi_0}$. In this way 
$E$, $N$, $U$, becomes a system of imprimitivity based on the homogeneous space $\mathfrak{G} /H$.
Now when $E$ is concentrated on a single orbit the assumption of uniform multiplicity of $N$ would be unnecessary,
but instead we may require $U$ to be ``locally bounded'': $||U_y f || < c_{\Delta} ||f ||$ for all $f \in \mathcal{H}$ 
whose spectral support (in their decomposition with respect to $E$) is contained within compact subset $\Delta \subset G_2 / G_{\chi_0} = \mathfrak{G}/H$, with a positive constant $c_{\Delta}$ depending on $\Delta$. 
(In fact we have implicitly used the ``local boundedness'' in the first equality of (\ref{ueu}).) Then using ergodicity
of the action of $G_2$ (resp. $\mathfrak{G}$) on $G_2 / G_{\chi_0}$ (resp. $\mathfrak{G}/H$) one can prove
uniform multiplicity of the spectral measure $E$. 
A computation similar to that performed by Mackey in \cite{Mackey_imprimitivity} (compare also
\cite{Mackey1}, \S 6 or \cite{Mackey2}, \S 3.7) shows that the representation $V_{(x,y)} = N_x U_y$ defined by the system is just equal to the Krein-isometric 
representation $V$ of $\mathfrak{G} = G_1 \circledS G_2$ in the 
Krein space $(\mathcal{H}, \mathfrak{J})$ of the Lemma (\ref{lop_ind_1}) with a representation $L$ of the
subgroup $H$, which is easily checked to be Krein-unitary in case the multiplicity of $N$ is
assumed to be finite. Thus it follows the following theorem

\begin{twr}
Let $E$, $N$, $U$ be a system of imprimitivity giving a Krein-isometric representation 
$V_{(x,y)} = N_x U_y$ of a semi direct product $\mathfrak{G} = G_1 \circledS G_2$ 
of separable locally compact groups $G_1$ and $G_2$ with $G_1$ Abelian in a Krein space
$(\mathcal{H}, \mathfrak{J})$ and with the representation
$N$ commuting with $\mathfrak{J}$ and thus being unitary in $\mathcal{H}$, 
for which the following assumptions are satisfied:
\begin{enumerate}

\item[1)]
The spectral measure is concentrated on a single orbit $\mathscr{O}_{\chi_0}$ in $\widehat{G_1}$ under $G_2$.

\item[2)]
The representation $U$ (equivalently the representation $V$)  is ``locally bounded''
with respect to $E$. 

\end{enumerate}

Then the representation $N$ (and equivalently the spectral measure $E$) is of uniform multiplicity.
The fundamental symmetry $\mathfrak{J}$ is decomposable with respect to the decomposition
of $\mathcal{H}$ associated (in the sense of \cite{von_neumann_dec}) to the spectral measure $E$
of the system, and has a decomposition of the form (\ref{fund_sym_dec}).

Assume moreover that:

\begin{enumerate}

\item[3)]
The representation $N$ has finite multiplicity.

\end{enumerate}

Then $V$ is unitary and Krein-unitary equivalent to a Krein-isometric representation $U^L$ 
induced by a Krein unitary representation $L$ of the subgroup $H = G_1 \cdot G_{\chi_0}$ 
associated to the orbit. 
\label{lop_ind:twr.1}
\end{twr}
\qed

This theorem may be given a more general form by discarding 3), but the given version is sufficient 
for the representations acting in one particle states
of free fields with non trivial gauge freedom, and thus acting in Krein spaces (with the 
fundamental symmetry operator $\mathfrak{J}$ called Gupta-Bleuler operator in physicists parlance),
where the representations $L$ act in Krein spaces $(\mathcal{H}_L , \mathfrak{J}_L)$
of finite dimension.

Consider for example the double covering $\mathfrak{G} = T_4 \circledS SL(2,\mathbb{C})$
of the Poincar\'e group with the semi direct product structure defined
by the following homomorphism: $\alpha [t_x]= \alpha x \alpha^*$,
where the translation $t_x:  (a_0 , a_1 , a_2 , a_3) \mapsto (a_0 , a_1 , a_2 , a_3) + (x_0 , x_1 , x_2 , x_3)$
is written as a Hermitian matrix 
\[
x = \left( \begin{array}{cc} x_0 + x_3 & x_1 - i x_2 \\ 
                            x_1 + i x_2 & x_0 - x_3  \end{array}\right)
\] 
in the formula $\alpha x \alpha^*$ giving $\alpha [t_x]$ and $\alpha^*$ is the Hermitian 
adjoint of $\alpha \in SL(2, \mathbb{C})$. 

Characters $\chi_p \in \widehat{T_4}$ of the group $T_4$ have the following form
\[
\chi _p (t_x) = e^{i( - p_0a_0 + p_1 x_1 + p_2 x_2 + p_3 x_3)},
\]
for $p = (p_0 , p_1 , p_2 , p_3)$ ranging over $\mathbb{R}^4$. For each character $\chi_p \in \widehat{T_4}$
let us consider the orbit $\mathscr{O}_{\chi_p}$ passing through $\chi_p$,
under the action $\chi_p \mapsto [\chi_p]\alpha$, $\alpha \in SL(2, \mathbb{C})$,
where $[\chi_p]\alpha$ is the character given by the formula
\[
T_4 \ni t_x \xrightarrow{[\chi_p]\alpha} \big( [\chi_p]\alpha \big)(t_x)
= \chi_p (\alpha^{-1} [t_x])  = \chi_p (\alpha^{-1} x {\alpha^{*}}^{-1}) = \chi_{\alpha p \alpha^{*}}(x)
= \chi_{\alpha p\alpha^{*}}(t_x),
\]
where in the formulas $\alpha p \alpha^*$ and $\alpha^{-1}x{\alpha^{*}}^{-1}$, $x$ and $p$
are regarded as Hermitian $2 \times 2$ matrices:
\[
x = \left( \begin{array}{cc} x_0 + x_3 & x_1 - i x_2 \\ 
                            x_1 + i x_2 & x_0 - x_3  \end{array}\right) \,\,\, \textrm{and} \,\,\,
p = \left( \begin{array}{cc} p_0 + p_3 & p_1 - i p_2 \\ 
                            p_1 + i p_2 & p_0 - p_3  \end{array}\right).
\] 
Let $G_{\chi_p}$ be the stationary subgroup of the point $\chi_p \in \widehat{T_4}$. 
Let $H = H_{\chi_p} = T_4 \cdot G_{\chi_p}$, and let $L'$ be a Krein-unitary representation 
of the stationary group $G_{\chi_p}$. Then $L$ given by 
\[
L_{t_x \cdot g} = \chi_p (t_x) L'_{g}, \,\,\, t_x \in T_4 , g \in G_{\chi_p}
\]
is a well defined Krein-unitary representation of $H_{\chi_p} = T_4 \cdot G_{\chi_p}$ because $G_{\chi_p}$
is the stationary subgroup for the point $\chi_p$. 
The functions $(q,h_0 , q_0 ) \mapsto h'_{{}_{q,h_0 , q_0}} \in H$ 
and $(q,h_0 , q_0 ) \mapsto q'_{{}_{q,h_0 , q_0}} \in Q \cong \mathfrak{G}/H_{\chi_p}$ 
corresponding to the respective $H= H_{\chi_p}$ or the respective orbits $\mathscr{O}_{\chi_p}$ are known
for all orbits in $\widehat{T_4}$ under $SL(2, \mathbb{C})$ and may be explicitly computed.

For example for $p = (1, 0, 0, 1)$ lying on the light cone in the joint spectrum sp$(P_0 , \ldots P_3)$
of the canonical generators of one parameter subgroups of translations, the stationary
subgroup $G_{\chi_p} = G_{\chi_{{}_{(1, 0, 0, 1)}}}$ is equal to the group of matrices 
\[
\left( \begin{array}{cc} e^{i\phi/2} & e^{i \phi/2}z \\ 
                                     0 & e^{-i\phi/2}  \end{array}\right), \,\,\,
 0 \leq \phi < 4\pi , \,\,\, z \in \mathbb{C} 
\]
isomorphic to (the double covering of) the 
symmetry group $E_2$ of the Euclidean plane and with the orbit $\mathscr{O}_{\chi_{{}_{(1, 0, 0, 1)}}}$
equal to the forward cone with the apex removed.

Consider then the Hilbert space $\mathcal{H}_L$ to be equal $\mathbb{C}^4$
with the standard inner product and with the fundamental symmetry equal
\[
\mathfrak{J}_L
 = \left( \begin{array}{cccc} -1 & 0 & 0 & 0 \\ 
                                           0 & 1 & 0 & 0 \\
                                           0 & 0 & 1 & 0 \\
                                           0 & 0 & 0 & 1  \end{array}\right).
\] 

Finally let $L'$ be the following Krein-unitary representation
\begin{equation}\label{L'}
\left(\begin{array}{cc} e^{i\phi/2} & e^{i \phi/2}z \\ 
                                     0 & e^{-i\phi/2}  \end{array}\right) \xrightarrow{L'_{(z, \phi)}}
\left( \begin{array}{cccc} 
1 + \frac{1}{2}|z|^2                     & \frac{1}{\sqrt{2}}z & \frac{1}{\sqrt{2}}\overline{z} & -\frac{1}{2}|z|^2 \\ 
\frac{1}{\sqrt{2}} e^{-i\phi}\overline{z} & e^{-i\phi} & 0                    & \frac{1}{\sqrt{2}} e^{-i\phi}\overline{z} \\
\frac{1}{\sqrt{2}} e^{i\phi}z            & 0         & e^{i\phi}             & -\frac{1}{\sqrt{2}} e^{i\phi}z  \\
\frac{1}{2}|z|^2                          & \frac{1}{\sqrt{2}}z & \frac{1}{\sqrt{2}}\overline{z} & 1 
- \frac{1}{2}|z|^2  \end{array}\right)
\end{equation}
of $G_{\chi_{{}_{(1, 0, 0, 1)}}} \cong \widetilde{E_2}$ in the Krein space $(\mathcal{H}_L , \mathfrak{J}_L)$ and 
define the Krein-unitary representation $L$: 
$H = T_4 \cdot G_{\chi_{{}_{(1, 0, 0, 1)}}} \ni t_x \cdot (z,\phi) \xrightarrow{L_{t_x \cdot (z, \phi)}}
\chi_{{}_{(1, 0, 0, 1)}}(t_x) L'_{(z, \phi)}$ corresponding to the Krein-unitary representation $L'$
of $G_{\chi_{{}_{(1, 0, 0, 1)}}}$. 
Then one obtains in this way the system of imprimitivity with the representation
$V$ of the Lemma \ref{lop_ind_1} equal to the {\L}opusza\'nski representation acting in the 
one particle states of the free photon field in the momentum representation, having exactly Wigner's form
\cite{Wigner_Poincare} with the only difference that $L$ is not unitary but Krein-unitary.

\vspace*{1cm}

Several remarks are in order.

1) In case of $\mathfrak{G} = T_4 \circledS SL(2,\mathbb{C})$,
$\widehat{T_4} = \mathbb{R}^4$ with the natural smooth action of $SL(2,\mathbb{C})$ 
giving it the Lorentz structure. The possible orbits  
$\mathscr{O}_{\chi_p} \subset \widehat{T_4} = \mathbb{R}^4$ 
are: the single point $(0,0,0,0,)$ -- the apex of ``the light-cone'', the upper/lower half
of the light cone (without the apex), the upper/lower sheet of the two-sheeted hyperboloid, and
the one-sheet hyperboloid. Thus all of them are smooth manifolds (with the exclusion of the apex, of course). 
Joining this with the
Mackey analysis of quasi invariant measures on  homogeneous $\mathfrak{G}/H$ spaces one can see that
the spectral measures of the translation generators (for representations with the joint spectrum
sp$(P_0 , \ldots P_3)$ concentrated on single orbits) are equivalent to measures  induced by the 
Lebesgue measure on $\mathbb{R}^4 = \widehat{T_4}$ (of course with the exclusion of the representations corresponding
to the apex -- the single point orbit, with the zero $(0,0,0,0)$
as the only value of the joint spectrum sp$(P_0 , \ldots P_3)$.

2) Note that for the system of imprimitivity $E$, $N$, $U$ in the Krein space the condition:
\begin{multline*}
V_{(x,y)} E_S V_{(x,y)^{-1}} = N_x U_y E_S U_{y^{-1}} N_{x^{-1}} \\
= N_x E_{[S]y} N_{x^{-1}}
= E_{[S]y} \,\,\, \textrm{for all} \,\,\, (x,y) \in G_1 \circledS G_2 \,\,\, \textrm{and all Borel sets}
\,\,\, S \subset \widehat{G_1}
\end{multline*}
holds, and is essentially equivalent to the condition: 
\[
U_y E_S U_{y^{-1}} = E_{[S]y}, \,\,\, \textrm{for all} \,\,\, y \in G_2 , 
 \,\,\, \textrm{and all Borel sets} \,\,\, S \subset \widehat{G_1}.
\]
We may write it as 
$V_{(x,y)} E_S V_{(x,y)^{-1}} = E_{[S](x,y)}$, with the trivial action 
$[\chi](x,e) = \chi$, $x \in G_1$ and $[\chi](e,y) = [\chi]y$.
It is more convenient to relate the system of imprimitivity immediately to $V$ and inspired by Mackey put
the following more general definition.

Let $V$ be a Krein-isometric representation of a separable locally compact group $\mathfrak{G}$
in a Krein space $(\mathcal{H}, \mathfrak{J})$. By a system of imprimitivity for $V$, we 
mean the system $E$, $B$, $\varphi$ consisting of
\begin{enumerate}

\item[a)]
an analytic Borel set $B$;

\item[b)]
an anti-homomorphism $\varphi$ of $\mathfrak{G}$ into the group of all Borel automorphisms of
$B$ such that $(y,b) \mapsto (y,[b]y)$ is a Borel automorphism of $\mathfrak{G} \times B$;

here we have written $[b]y$ for the action of the automorphism $\varphi(y)$ on $b \in B$.

\item[c)]
The spectral measure $E$ consists of selfadjont and Krein selfadjoint projections commuting 
with $\mathfrak{J}$ in $(\mathcal{H}, \mathfrak{J})$, and is such that 
$V_y E_S {V_{y}}^{-1} = E_{[S]y^{-1}}$.

\item[d)]  
The representation $V$ is ``locally bounded'' with respect to $E$.

\end{enumerate} 
   
Any induced Krein-isometric representation ${}^{\mu}U^L$  possesses a canonical system of
imprimitivity in $(\mathcal{H}^L , \mathfrak{J}^L)$ related to it. Namely
let $S$ be a Borel set on $\mathfrak{G}/H$, and let $S'$ be its inverse under the 
quotient map $\mathfrak{G} \rightarrow \mathfrak{G}/H$. Let $1_{S'}$ be the characteristic
function of $S'$. Then $f \xrightarrow{E_S}  1_{S'} f$, $f \in \mathcal{H}^L$
is a self adjoint and Krein self adjoint projection, which commutes with $\mathfrak{J}^L$.
Thus $S \mapsto E_S$ is a spectral measure based on the analytic Borel space $\mathfrak{G}/H$.
By the inequality (\ref{def_ind_krein:ineq}) in the proof of Theorem \ref{def_ind_krein:twr.1}
the representation ${}^{\mu}U^L$ is ``locally bounded'', i.e. fulfils condition 3) 
of Theorem \ref{lop_ind:twr.1} or condition d).

The representation $V$ of Lemma \ref{lop_ind_1} in the Krein space $(\mathcal{H},\mathfrak{J})$
together with the spectral measure $E'$ on $B = \mathfrak{G}/H$ associated with the decomposition 
(\ref{hilbert_system_prim}) is a system of imprimitivity in Krein space which by Lemma \ref{lop_ind_1}
is Krein-unitary and unitary equivalent to the canonical system of imprimitivity $U^L$, $E$, $\varphi$ 
defined above.  That $V, E'$ of Lemma \ref{lop_ind_1} with $\varphi_{g_{{}_0}}(q) = q'_{{}_{q,g_{{}_0}}}$ 
composes a system of imprimitivity can be checked directly using the multiplier property of the function 
$(q,g_{{}_0}) \mapsto h'_{{}_{q,g_{{}_0}}}$.

3) The plan for further computations is the following. First we start with the systems of imprimitivity fulfilling the conditions 1)-3) of Theorem \ref{lop_ind:twr.1} sufficient for accounting for the representations acting in one particle states of free fields. Then we prove the ``subgroup'' and ``Kronecker product theorems'' for the induced representations
in order to achieve decompositions of tensor products of these representations into direct integrals of representations
connected with imprimitivity systems concentrated on single orbits (using Mackey double-coset-type technics).
The component representations of the decomposition will not in general have the standard form of induced 
representations (contrary to what happens for tensor products of induced representations of Mackey which are unitary in ordinary sense). But then we back to Theorem  \ref{lop_ind:twr.1} applied again to each of the component representations
in order to restore the standard form of induced representation in Krein space to each of them separately. 
In this way we may repeat
the procedure of decomposing tensor product of the component representations (now in the standard form)
and continue it potentially in infinitum.
It turns out that the condition 3) of finite multiplicity will have to be abandoned and replaced with infinite uniform multiplicity in further stages of this process, 
but we have all the grounds for the condition 2) of ``local boundedness'' to be preserved in all
cases at all levels of the decomposition.
Indeed recall that the spectral values $(p_0, \ldots p_3 )$ of the translation generators (four-momentum operators) in the tensor product of representations corresponding to imprimitivity systems concentrated on single orbits
$\mathscr{O}',\mathscr{O}'' \subset \widehat{T_4}$, are the sums
$(p'_0, \ldots p'_3 ) + (p''_0, \ldots p''_3 )$, with the spectral values $(p'_0, \ldots p'_3 )$ and 
$(p''_0, \ldots p''_3 ) $ ranging over $\mathscr{O}'$ and $\mathscr{O}''$ respectively. Now the geometry 
of the orbits in case of $\mathfrak{G}
= T_4 \circledS SL(2,\mathbb{C})$ is such that the sets of all values  $(p'_0, \ldots p'_3 )$ and 
$(p''_0, \ldots p''_3 )$ for which $(p_0, \ldots p_3 )$ ranges over a compact set, are compact 
(discarding irrelevant null sets of $(p_0, \ldots p_3 )$ not belonging to the joint spectrum of momentum operators of the tensor product representation -- the light cones -- in the only case of tensoring representation corresponding to the positive energy light cone orbit with the representation corresponding to the negative energy light cone).

4) In fact the representation of one particle states in the Fock space (with the Gupta-Bleuler or fundamental
symmetry operator) is induced by the following representation $L''$ in the
above defined Krein space $(\mathcal{H}_L , \mathfrak{J}_L)$ of the double covering
of the symmetry group of the Euclidean plane:
\begin{equation}\label{L''}
L''_{(z, \phi)} =
\left( \begin{array}{cccc} 
1 + \frac{1}{2}|z|^2                     & \frac{1}{2}(\overline{z}+z) & \frac{i}{2}(z-\overline{z}) & -\frac{1}{2}|z|^2 \\ 
\frac{1}{2} (e^{-i\phi}\overline{z} +e^{i\phi}z) & \cos \phi & \sin \phi & -\frac{1}{2}(e^{-i\phi}\overline{z} +e^{i\phi}z \\
\frac{i}{2}(e^{i\phi}z-e^{-i\phi}\overline{z}) & -\sin \phi & \cos \phi & -\frac{1}{2}(e^{i\phi}z- e^{-\phi}\overline{z}) \\
\frac{1}{2}|z|^2  & \frac{1}{2}(\overline{z} + z) & \frac{i}{2}(z-\overline{z}) & 1- \frac{1}{2}|z|^2  \end{array}\right),
\end{equation}
compare e.g. \cite{Weinberg1, Weinberg2}, or \cite{lop1,lop2}. But the operator
\[
\left( \begin{array}{cccc} 1 &    0       &          0  & 0 \\ 
                           0 & 1/\sqrt{2} & i/\sqrt{2}  & 0 \\
                           0 & 1/\sqrt{2} & -i/\sqrt{2} & 0 \\
                           0 &    0       &    0        & 1  \end{array}\right).
\] 
which is Krein-unitary and unitary in $(\mathcal{H}_L , \mathfrak{J}_L)$ sets up 
Krein-unitary and unitary equivalence between the representation $L'$ of (\ref{L'})
and the representation $L''$ of (\ref{L''}) as well as between the associated representations $L$.
By Theorem \ref{def_ind_krein:twr.4} it makes no difference which one we use, but for some technical reasons we
prefer the representation $L$ associated with (\ref{L'}).

5) The representation which we have called by the name of {\L}opusza\'nski has appeared in physics
rather very early, compare \cite{Wigner}, and then in relation to the Gupta-Bleuler quantization
of the free photon field:  \cite{Weinberg1, Weinberg2}, \cite{Halpern}, \cite{Kupersztych}. 
But it was {\L}opusza\'nski \cite{lop1,lop2} who initiated a systematic study of the relation
of the representation with the Gupta-Bleuler 
formalism. That's why we call the representation after him.

\subsection{Kronecker product of induced representations in Krein spaces}\label{kronecker}

In this Section we define the outer Kronecker product and inner Kronecker product of Krein
isometric (and Krein unitary) representations and give an important theorem concerning 
Krein isometric representation induced by a Kronecker product of Krein-unitary representations.

The whole construction is based on the ordinary tensor product of the associated Hilbert spaces 
and operators in the Hilbert spaces. We recapitulate shortly a specific realization of the 
tensor product of Hilbert spaces as trace class conjugate-linear operators, in short we realize it by the  Hilbert-Schmidt 
class of conjugate-linear operators\footnote{Alternatively one may consider linear Hilbert-Schmidt class operators, but
replace one of the Hilbert spaces in question by its conjugate space, compare \cite{Mackey}, \S 5.} with the standard 
operator $L^2$-norm, for details we refer the reader to the original
paper by Murray and von Neumann \cite{Murray_von_Neumann}.

Let $\mathcal{H}_1$ and $\mathcal{H}_2$ be two
separable Hilbert spaces over $\mathbb{C}$ (recall that by the proof of Lemma \ref{lem:dense.6}
the Hilbert space $\mathcal{H}^L$ of the Krein-isometric representation $U^L$ of a separable locally compact group 
$\mathfrak{G}$ induced by a Krein-unitary representation $L$ of a closed subgroup $G_1 \subset \mathfrak{G}$
is separable). A mapping $T$ of $\mathcal{H}_2$ to $\mathcal{H}_1$
is conjugate-linear iff $T(\alpha f + \beta g) = \overline{\alpha} \, T(f) + \overline{\beta} \, T(g)$
for all $f,g \in \mathcal{H}_2$ and all complex numbers $\alpha$ and $\beta$, with the ``over-line'' sign standing
for complex conjugation. For any such conjugate-linear operator $T$ we define the conjugate 
version of its adjoint $T^{{}^\circledast}$, namely this is the operator fulfilling 
$(Tg, f) = (T^{{}^\circledast} f, g)$ for all $f \in \mathcal{H}_1$ and all $g \in \mathcal{H}_2$.
In particular if $T$ is bounded, conjugate-linear, finite-rank operator so is its conjugate
adjoint $T^{{}^\circledast}$. If $U_1$ and $U_2$ are bounded operators in $\mathcal{H}_1$
and $\mathcal{H}_2$ respectively then $U_1 T U_2$ is a finite rank operator from $\mathcal{H}_2$
into $\mathcal{H}_1$. One easily verifies that $(A T B)^{{}^\circledast} = B^* T^{{}^\circledast} A^*$,
where $A$ and $B$ are linear operators in $\mathcal{H}_1$ and $\mathcal{H}_2$ with $A^*$ and
$B^*$ equal to their ordinary adjoint operators. If $U_1$ and $U_2$ are densely defined operators in $\mathcal{H}_1$
and $\mathcal{H}_2$ respectively on linear domains $\mathfrak{D}_1 \subset \mathcal{H}_1$
and $\mathfrak{D}_2 \subset \mathcal{H}_2$ and $T$ is finite rank operator with the rank contained
in $\mathfrak{D}_1$ and supported in $\mathfrak{D}_2$, then $U_1 T U_2$ is a well defined finite rank operator.
Let $\mathcal{H}' = \mathcal{H}_1 \otimes' \mathcal{H}_2$ be the linear space of finite rank 
conjugate-linear operators $T$ of $\mathcal{H}_2$ into $\mathcal{H}_1$. For any two such operators
$T$ and $S$ the operator $T S^{{}^\circledast}$ is linear from $\mathcal{H}_1$
into $\mathcal{H}_1$ and of finite rank (similarly $T^{{}^\circledast} S$ is linear and finite rank
from $\mathcal{H}_2$ into $\mathcal{H}_2$). We may therefore introduce the following 
inner product in $\mathcal{H}'$:  
\begin{multline*}
\langle T, S \rangle = \Tr [\, T S^{{}^\circledast} \,] = \sum \limits_{n} (T \, S^{{}^\circledast} e_n, e_n) \\
= \sum \limits_{n} (T^{{}^\circledast} e_n , S^{{}^\circledast} e_n) 
= \sum \limits_{m} (T \varepsilon_m , S \varepsilon_m)    \\
\sum \limits_{m} (T^{{}^\circledast} S \varepsilon_m , \varepsilon_m)
= \Tr [\, T^{{}^\circledast} S \,], 
\end{multline*}
where $\{ e_n \}_{n \in \mathbb{N}}$ and $\{ \varepsilon_m \}_{m \in \mathbb{N}}$ are orthonormal bases
in $\mathcal{H}_1$ and $\mathcal{H}_2$ respectively. The completion of $\mathcal{H}'$ with respect to
this inner product composes the tensor product $\mathcal{H}  = \mathcal{H}_1 \otimes \mathcal{H}_2$.

Let $A$ and $B$ be bounded operators in $\mathcal{H}_1$
and $\mathcal{H}_2$. Their tensor product $A \otimes B$ acting in $\mathcal{H}_1 \otimes \mathcal{H}_2$
is defined as the operator $T \mapsto ATB^*$, for $T \in \mathcal{H}_1 \otimes \mathcal{H}_2$. 
In particular if for any $f \in \mathcal{H}_1$ and $g \in \mathcal{H}_2$ we define the finite rank 
conjugate-linear operator $T_{{}_{f, g}}: w \mapsto f \, \cdot \, (g, w)$ supported on the 
linear subspace generated by $g$ with the range generated by $f$, then 
$T_{{}_{f, g}} \in \mathcal{H}_1 \otimes \mathcal{H}_2$ is written as $f \otimes g$ and we have
$(f_1 \otimes g_1 , f_2 \otimes g_2) = \Tr \big[ T_{{}_{f_1, g_1}} \, 
\big(T_{{}_{f_2, g_2}}\big)^{{}^\circledast} \big]
= \Tr \big[ T_{{}_{f_1, g_1}} \, T_{{}_{g_2, f_2}} \big] = (f_1 , f_2)\cdot (g_1 , g_2)$ 
because $\big(T_{{}_{f_2, g_2}}\big)^{{}^\circledast} = T_{{}_{g_2, f_2}}$.

If $(\mathcal{H}_1 , \mathfrak{J}_1)$ and $(\mathcal{H}_2 , \mathfrak{J}_2)$ are two Krein spaces,
then we define their tensor product as the Krein space 
$(\mathcal{H}_1 \otimes \mathcal{H}_2 , \mathfrak{J}_1 \otimes \mathfrak{J}_2)$; verification
of the self-adjointness of $\mathfrak{J}_1 \otimes \mathfrak{J}_2$ and the property 
$\big( \mathfrak{J}_1 \otimes \mathfrak{J}_2 \big)^2 = I$ is immediate.

We say an operator $T$ from $\mathcal{H}_2$ into $\mathcal{H}_1$ is supported by finite dimensional (or 
more generally: closed) 
linear subspace $\mathfrak{M} \subset \mathcal{H}_2$ or by the projection $P_{{}_\mathfrak{M}}$, in case 
$T = TP_{{}_\mathfrak{M}}$, where $P_{{}_\mathfrak{M}}$ is the self adjoint projection with range $\mathfrak{M}$.
Similarly we say an operator $T$ from $\mathcal{H}_2$ into $\mathcal{H}_1$ has range in a finite dimensional
(or more generally: closed) linear subspace $\mathfrak{N} \subset \mathfrak{H}_1$, in case 
$T = P_{{}_\mathfrak{N}} T$, where $P_{{}_\mathfrak{N}}$ is the self adjoint projection with range $\mathfrak{N}$. 
One easily verifies the following tracial property. Let $B$ be any finite rank and \emph{linear} operator
from $\mathcal{H}_1$ into $\mathcal{H}_1$ supported on a finite dimensional linear subspace of the domain $\mathfrak{D}_1$ and with the range also finite dimensional and lying in $\mathfrak{D}_1$. Then for any linear
operator defined on the dense domain $\mathfrak{D}_1 \subset \mathcal{H}_1$ and preserving 
it, i. e.  with $\mathfrak{D}_1$ contained in the common domain of $A$ and its 
adjoint $A^*$, we have the tracial property
\[
\Tr [B A] = \Tr [A B]. 
\]  
Indeed any such linear $B$ is a finite linear combination of the operators $\mathbb{T}_{{}_{f, f'}}$
defined as follows: $\mathbb{T}_{{}_{f, f'}} (w) = (w, f) \cdot f'$. 
By linearity it will be sufficient to establish the tracial property for the linear operator
$B$ of the form $B = \mathbb{T}_{{}_{f_{{}_1}, f_{{}_2}}} + \mathbb{T}_{{}_{f_{{}_3}, f_{{}_4}}}$
with $f_i \in \mathfrak{D}_1$, $i = 1,2,3,4$. Using the Gram-Schmidt orthogonalization 
we construct an orthonormal basis $\{ e_n \}_{n \in \mathbb{N}}$ of $\mathcal{H}_1$ with
$e_n \in \mathfrak{D}_1$. We have in this case
\begin{multline}\label{tr_BA}
\Tr [B A] = \Tr \Big[ \big( \mathbb{T}_{{}_{f_{{}_1}, f_{{}_2}}} 
+ \mathbb{T}_{{}_{f_{{}_3}, f_{{}_4}}} \big) A \Big] 
= \Tr \Big[ \mathbb{T}_{{}_{f_{{}_1}, f_{{}_2}}} A \big]
+ \Tr \Big[ \mathbb{T}_{{}_{f_{{}_3}, f_{{}_4}}}  A \Big]  \\
= \sum \limits_n \big( \mathbb{T}_{{}_{f_{{}_1}, f_{{}_2}}} A e_n , e_n \big) 
+ \sum \limits_n \big( \mathbb{T}_{{}_{f_{{}_3}, f_{{}_4}}}  A e_n , e_n \big)  \\
= \sum \limits_n (A e_n , f_1) \cdot (f_2 , e_n) + \sum \limits_n (A e_n , f_3) \cdot (f_4 , e_n) \\
= \sum \limits_n ( e_n , A^* f_1) \cdot (f_2 , e_n) + \sum \limits_n ( e_n , A^* f_3) \cdot (f_4 , e_n)
= (f_2 , A^* f_1) + (f_4 , A^* f_3)   \\
= (A f_2 , f_1) + (A f_4 , f_3) < \infty,
\end{multline}
because by the assumed properties of the operator $A$ the vectors $f_1 , f_3 \in \mathfrak{D}_1$ are
contained in the domain of $A^*$ and likewise the vectors $f_2 , f_4 \in \mathfrak{D}_1$ lie in the domain of $A$. 
Similarly we have:

\begin{multline}\label{tr_AB}
\Tr [A B] = \Tr \Big[ \big( A \mathbb{T}_{{}_{f_{{}_1}, f_{{}_2}}} 
+  \mathbb{T}_{{}_{f_{{}_3}, f_{{}_4}}} \big) \Big] 
= \Tr \Big[ A \mathbb{T}_{{}_{f_{{}_1}, f_{{}_2}}} \big]
+ \Tr \Big[ A \mathbb{T}_{{}_{f_{{}_3}, f_{{}_4}}}  \Big]  \\
= \sum \limits_n \big(  A \mathbb{T}_{{}_{f_{{}_1}, f_{{}_2}}} e_n , e_n \big) 
+ \sum \limits_n \big( A \mathbb{T}_{{}_{f_{{}_3}, f_{{}_4}}}  e_n , e_n \big)  \\
= \sum \limits_n ( e_n , f_1) \cdot ( A f_2 , e_n) + \sum \limits_n ( e_n , f_3) \cdot ( A f_4 , e_n)
= (A f_2 , f_1) + (A f_4 , f_3) < \infty.
\end{multline}
Comparing (\ref{tr_BA}) and (\ref{tr_AB}) we obtain the tracial property.

Now let  $U_1 = U^{L}_{x}$ and $U_2 = U^{M}_{y}$ be densely defined and closable Krein isometric 
operators of the respective Krein isometric induced representations of the groups $\mathfrak{G}_1$ and $\mathfrak{G}_2$ 
in $\mathcal{H}_1 = \mathcal{H}^L$
and $\mathcal{H}_2 = \mathcal{H}^M$ respectively with linear domains $\mathfrak{D}_i \subset \mathcal{H}_i$,
$i = 1,2$, equal to the corresponding
domains $\mathfrak{D}$ of Theorem \ref{def_ind_krein:twr.1} and Remark \ref{rem:def_ind_krein.1}
and with the respective fundamental symmetries $\mathfrak{J}_1 = \mathfrak{J}^L$, $\mathfrak{J}_2 = \mathfrak{J}^M$. 
Therefore by Theorem \ref{def_ind_krein:twr.1} and Remark \ref{rem:def_ind_krein.1} 
$U^i (\mathfrak{D}_i) = \mathfrak{D}_i$ and 
$\mathfrak{J}_i (\mathfrak{D}_i) = \mathfrak{D}_i$, $i = 1, 2$, so that $\mathfrak{D}_i$ is contained in the domain
of ${U_i}^*$ and ${U_i}^* (\mathfrak{D}_1 ) = \mathfrak{D}_i$.
Finally let $T,S$ be any finite rank operators in the linear subspace $\mathfrak{D}_{12}
= \textrm{linear\,span}\{ T_{{}_{f, g}}, f \in \mathfrak{D}_1 , g \in \mathfrak{D}_2 \}$ of finite rank 
operators supported in $\mathfrak{D}_2$ and with ranges in $\mathfrak{D}_1$. In particular for each 
$S \in \mathfrak{D}_{12}$, $S^{{}^\circledast}$ is supported in $\mathfrak{D}_1$ and has rank in $\mathfrak{D}_2$.
By the known property of Hilbert Schmidt operators $\mathfrak{D}_1 \otimes \mathfrak{D}_2 = \mathfrak{D}_{12}$
is dense in $\mathcal{H}_1 \otimes \mathcal{H}_2$.
We claim that $U_1 \otimes U_2$ is well defined on $\mathfrak{D}_1 \otimes \mathfrak{D}_2 = \mathfrak{D}_{12}$.

Indeed, by the Gram-Schmidt orthonormalization we may construct an orthonormal base $\{e_n\}_{n \in \mathbb{N}}$
of $\mathcal{H}_1$ with each $e_n$ being an element of the linear dense domain $\mathfrak{D}_1$. 
For any $f_1 , f_2 \in \mathfrak{D}_1$ and $g_1 , g_2 \in \mathfrak{D}_2$ we have 
\begin{multline*}
\Big\| \big( U_1 \otimes U_2 \big) \big( f_1 \otimes g_1 + f_2 \otimes g_2 \big) \Big\|^2  \\
= \Big\langle \,\, U_1 \big( T_{{}_{f_{{}_1}, g_{{}_1}}} + T_{{}_{f_{{}_2}, g_{{}_2}}} \big){U_2}^* \,\, , \,\,\,\, 
U_1 \big( T_{{}_{f_{{}_1}, g_{{}_1}}} + T_{{}_{f_{{}_2}, g_{{}_2}}} \big)  {U_2}^* \,\, \Big\rangle  \\
= \Tr \Big[ \, U_1 \big( T_{{}_{f_{{}_1}, g_{{}_1}}} \big) {U_2}^* 
\big(  U_1 \big( T_{{}_{f_{{}_1}, g_{{}_1}}}  \big) {U_2}^* \big)^{{}^\circledast} \, \Big]
+ \Tr \Big[ \, U_1 \big( T_{{}_{f_{{}_1}, g_{{}_1}}} \big) {U_2}^* 
\big(  U_1 \big( T_{{}_{f_{{}_2}, g_{{}_2}}}  \big) {U_2}^* \big)^{{}^\circledast} \, \Big] \\
+ \Tr \Big[ \, U_1 \big( T_{{}_{f_{{}_2}, g_{{}_2}}} \big) {U_2}^* 
\big(  U_1 \big( T_{{}_{f_{{}_1}, g_{{}_1}}}  \big) {U_2}^* \big)^{{}^\circledast} \, \Big]
+ \Tr \Big[ \, U_1 \big( T_{{}_{f_{{}_2}, g_{{}_2}}} \big) {U_2}^* 
\big(  U_1 \big( T_{{}_{f_{{}_2}, g_{{}_2}}} \big) {U_2}^* \big)^{{}^\circledast} \, \Big] \\
= \sum \limits_n \big( U_1 f_1, e_n  \big) \cdot \big( g_1, {U_2}^* U_2 g_1  \big) 
\cdot \big( {U_1}^* e_n, f_1  \big) \\
+ \sum \limits_n \big( U_1 f_1, e_n  \big) \cdot \big( g_1, {U_2}^* U_2 g_2  \big) 
\cdot \big( {U_1}^* e_n, f_2  \big) \\
+ \sum \limits_n \big( U_1 f_2, e_n  \big) \cdot \big( g_2, {U_2}^* U_2 g_1  \big) 
\cdot \big( {U_1}^* e_n, f_1  \big) \\
+ \sum \limits_n \big( U_1 f_2, e_n  \big) \cdot \big( g_2, {U_2}^* U_2 g_2  \big) 
\cdot \big( {U_1}^* e_n, f_2  \big).
\end{multline*}  
Because $\mathfrak{D}_1$ is in the domain of ${U_1}^*$ and ${U_1}^* (\mathfrak{D}_1) = U_1 (\mathfrak{D}_1)
= \mathfrak{D}_1$ and similarly for $U_2$, the last expression is equal to 
\begin{multline*}
\sum \limits_n \big( U_1 f_1, e_n  \big) \cdot \big( U_2 g_1,  U_2 g_1  \big) 
\cdot \big(  e_n, U_1 f_1  \big) \\
+ \sum \limits_n \big( U_1 f_1, e_n  \big) \cdot \big( U_2 g_1,  U_2 g_2  \big) 
\cdot \big(  e_n, U_1 f_2  \big) \\
+ \sum \limits_n \big( U_1 f_2, e_n  \big) \cdot \big( U_2 g_2,  U_2 g_1  \big) 
\cdot \big(  e_n, U_1 f_1  \big) \\
+ \sum \limits_n \big( U_1 f_2, e_n  \big) \cdot \big( U_2 g_2,  U_2 g_2  \big) 
\cdot \big(  e_n, U_1 f_2  \big) \\
= \big( U_1 f_1,   U_1 f_1 \big) \cdot \big( U_2 g_1,  U_2 g_1  \big) 
+ \big( U_1 f_1,  U_1 f_2 \big) \cdot \big( {U_2} g_1,  U_2 g_2  \big) \\ 
+ \big( U_1 f_2, U_1 f_1  \big) \cdot \big( U_2 g_2,  U_2 g_1  \big) 
+ \big( U_1 f_2, U_1 f_2  \big) \cdot \big( U_2 g_2,  U_2 g_2  \big) < \infty,
\end{multline*}
so that 
\begin{multline*}
\Big\| \big( U_1 \otimes U_2 \big) \big( f_1 \otimes g_1 + f_2 \otimes g_2 \big) \Big\|^2  \\
= \Big\langle \,\, U_1 \big( T_{{}_{f_{{}_1}, g_{{}_1}}} + T_{{}_{f_{{}_2}, g_{{}_2}}} \big){U_2}^* \,\, , \,\,\,\, 
U_1 \big( T_{{}_{f_{{}_1}, g_{{}_1}}} + T_{{}_{f_{{}_2}, g_{{}_2}}} \big)  {U_2}^* \,\, \Big\rangle < \infty 
\end{multline*}  
and $\big( U_1 \otimes U_2 \big) \big( f_1 \otimes g_1 + f_2 \otimes g_2 \big)$ is well defined. By induction
for each $T \in \mathfrak{D}_{12}$,  
$\big( U_1 \otimes U_2 \big) \big( T \big)= U_1 T { U_2}^*$ is well defined conjugate-linear operator of 

Hilbert-Schimdt class, so that $U_1 \otimes U_2$ is well defined on the linear domain $\mathfrak{D}_{12}$
dense in $\mathcal{H}_1 \otimes \mathcal{H}_2$.  By the Proposition of Chap. VIII.10, page 298 of 
\cite{Reed_Simon} it follows that $U_1 \otimes U_2$ is closable.  
Next, let $T, S \in \mathfrak{D}_{12}$, then by Theorem \ref{def_ind_krein:twr.1} 
and Remark \ref{rem:def_ind_krein.1}
\begin{multline*}
\mathfrak{J}_i (\mathfrak{D}_i) = \mathfrak{D}_i \,\,\, \textrm{and} \,\,\, 
U_i (\mathfrak{D}_i) = \mathfrak{D}_i  \,\,\, \textrm{and} \,\,\, \textrm{and} \,\,\,
{U_i}^* (\mathfrak{D}_i) = \mathfrak{D}_i  \\
\big( \, U_{i} \, \big)^\dagger \, U_{i} = \mathfrak{J}_i {U_i}^* \mathfrak{J}_i U_{i} 
= I  \,\,\, \textrm{and} \,\,\,
U_{i} \, \mathfrak{J}_i {U_i}^* \mathfrak{J}_i = I \,\,\, \textrm{on} \,\,\, \mathfrak{D}_i .
\end{multline*}
Thus for each $T, S \in \mathfrak{D}_{12}$ the following expressions are well defined and 
(e. g. for $T = T_{{}_{f_1, g_1}}$ and $S = T_{{}_{f_2, g_2}}$) 
\begin{multline*}
\Big( \,\, (\mathfrak{J}_1 \otimes \mathfrak{J}_2) \, (U_1 \otimes U_2)  \, (f_1 \otimes g_1) \,\, , 
\,\,\,\, (U_1 \otimes U_2) \,  (f_2 \otimes g_2) \,\,  \Big) \\
= \Big\langle \,\, \mathfrak{J}_1 U_1 T {U_2}^*\mathfrak{J}_2 \,\, , \,\,\,\,  U_1 S {U_2}^*  \,\, \Big\rangle  
= \Tr \Big[ \, \mathfrak{J}_1 U_1 T {U_2}^* \mathfrak{J}_2 \big(  U_1 S {U_2}^*  \big)^{{}^\circledast} \, \Big]  \\
= \Tr \Big[ \, \mathfrak{J}_1 U_1 T {U_2}^* \mathfrak{J}_2  U_2 S^{{}^\circledast} {U_1}^*   \, \Big] 
= \Tr \Big[ \, \mathfrak{J}_1 U_1 T \mathfrak{J}_2 \mathfrak{J}_2 {U_2}^* \mathfrak{J}_2  U_2 S^{{}^\circledast} {U_1}^*   \, \Big] \\ 
= \Tr \Big[ \,  \mathfrak{J}_1 U_1 T \mathfrak{J}_2 \{ \mathfrak{J}_2 {U_2}^* \mathfrak{J}_2  U_2 \} 
S^{{}^\circledast} {U_1}^*  \, \Big] 
= \Tr \Big[ \, \{ \mathfrak{J}_1 U_1 \} T \mathfrak{J}_2 S^{{}^\circledast} {U_1}^* \, \Big] \\
= \Tr \Big[ \,  T \mathfrak{J}_2 S^{{}^\circledast} {U_1}^* \{ \mathfrak{J}_1 U_1 \} \, \Big] 
= \Tr \Big[ \,  T \mathfrak{J}_2 S^{{}^\circledast} \mathfrak{J}_1 \{ \mathfrak{J}_1  {U_1}^* \mathfrak{J}_1 U_1 \} \, \Big]  \\ 
= \Tr \Big[ \,  T \mathfrak{J}_2 S^{{}^\circledast} \mathfrak{J}_1 \, \Big] 
= \Tr \Big[ \, \mathfrak{J}_1 T \mathfrak{J}_2 S^{{}^\circledast}  \, \Big] \\
= \Big( \,\, (\mathfrak{J}_1 \otimes \mathfrak{J}_2)  \, (f_1 \otimes g_1) \,\, , \,\,\,\,   f_2 \otimes g_2  \Big), 
\end{multline*}  
because the tracial property is applicable to the pair of operators 
\[
B =  T \mathfrak{J}_2 S^{{}^\circledast} {U_1}^*
\,\,\, \textrm{and} \,\,\,
A = \mathfrak{J}_1 U_1 
\]
as well as to the pair of operators
\[
B =  T \mathfrak{J}_2 S^{{}^\circledast}
\,\,\, \textrm{and} \,\,\,
A = \mathfrak{J}_1, 
\]
as both the operators $B$ are linear finite rank operators supported on finite dimensional subspaces contained 
in $\mathfrak{D}_1$ and with finite dimensional ranges contained in $\mathfrak{D}_1$
and for the operators $A$ indicated to above the linear domain $\mathfrak{D}_1$ is contained in the common domain of $A$
and $A^*$;  and moreover $\mathfrak{J}_1 (U_1)^* \mathfrak{J}_1 U_1$ and 
$ \mathfrak{J}_2 {U_2}^* \mathfrak{J}_2  U_2$ are well defined unit operators on the domains $\mathfrak{D}_1$
and $\mathfrak{D}_2$ respectively. Therefore $U_1 \otimes U_2 = U^L \otimes U^M$ is Krein-isometric on its
domain $\mathfrak{D}_{12}$ which holds by continuity for its closure.

We may therefore define the outer Kronecker product Krein-isometric representation 
$U^L \times U^M : \mathfrak{G}_1 \times \mathfrak{G}_2 \ni (x,y) \mapsto U^{L}_{x} \otimes U^{M}_{y}$ 
of the product group $\mathfrak{G}_1 \times \mathfrak{G}_2$, which is Krein isometric in the
Krein space $\big(\mathcal{H}^L \otimes \mathcal{H}^M  , \,\, \mathfrak{J}^L \otimes \mathfrak{J}^M \big)$.
All the more, if $U_1$ and $U_2$ are Krein-unitary representations of $G_1$ and $G_2$, 
respectively in $(\mathcal{H}_1 , \mathfrak{J}_1)$ and 
$(\mathcal{H}_2 , \mathfrak{J}_2)$, so is $U_1 \times U_2$ in the Krein space
$\big(\mathcal{H}_1 \otimes \mathcal{H}_2  , \,\, \mathfrak{J}_1 \otimes \mathfrak{J}_2 \big)$. Similarly
one easily verifies that $U_1 \times U_2$ is almost uniformly bounded whenever  $U_1$ and $U_2$ are.
In particular if $G_1$ and $G_2$ are two closed subgroups of the separable locally compact groups 
$\mathfrak{G}_1$ and $\mathfrak{G}_2$ respectively and $L$ and $M$ their Krein unitary and uniformly
bounded representations, then we may define the outer Kronecker
product representation $L \times M$ of the product group $G_1 \times G_2$ by the ordinary formula 
$\mathfrak{G}_1 \times \mathfrak{G}_2 \ni (\xi , \eta) \mapsto  L_\xi \otimes M_\eta$, which is 
Krein unitary and almost uniformly bounded in the Krein space
$\big(\mathcal{H}_L \otimes \mathcal{H}_M  , \,\, \mathfrak{J}_L \otimes \mathfrak{J}_M \big)$
whenever $L$ and $M$ are in the respective Krein spaces $(\mathcal{H}_L, \mathfrak{J}_L)$
and $(\mathcal{H}_M, \mathfrak{J}_M)$. We may therefore define the Krein-isometric 
representation ${}^{\mu_1 \times \mu_2}U^{L \times M}$ of the group $\mathfrak{G}_1 \times \mathfrak{G}_2$
in the Krein space $\mathcal{H}^{L \times M}$ induced by the representation $L \times M$ of the 
closed subgroup $G_1 \times G_2$, where $\mu_i$ are the respective quasi invariant measures in 
$\mathfrak{G}_i /G_i$.

Let us make an observation used in the proof of the Theorem of this Section.
Let $B_1$ be a Borel section of $\mathfrak{G}_1$ with respect to 
$G_1$ and respectively $B_2$ a Borel section of $\mathfrak{G}_2$ with respect to $G_2$ 
defined as in Section \ref{def_ind_krein} with the associated Borel functions
$h_1 : \mathfrak{G}_1 \ni x \mapsto h_1 (x) \in G_1$ such that ${h_1 (x)}^{-1}x \in B_1$ and
$h_2: \mathfrak{G}_2 \ni y \mapsto h_2 (y) \in G_2$ such that ${h_2 (y)}^{-1}y \in B_2$.
Then $B_1 \times B_2$ is a Borel section of $\mathfrak{G}_1 \times \mathfrak{G}_2$ with respect
to the closed subgroup $G_1 \times G_2$ with the associated Borel function
$h: (x, y) \mapsto h(x,y) \in G_1 \times G_2$ such that ${h(x, y)}^{-1} (x,y) \in B_1 \times B_2$, equal to
$h(x, y) = \big( \, h_{{}_1} (x) \, , \,\, h_{{}_2} (y) \, \big) \, = \, h_{{}_1} (x) \times h_{{}_2} (y)$. 
Let $w \in \mathcal{H}^{L \times M}$. Thus the corresponding operator $\mathfrak{J}^{L \times M}$
acts as follows 
\begin{multline*}
\big(\mathfrak{J}^{L \times M} w \big)_{{}_{(x, y)}} = 
(L \times M)_{{}_{h_1 (x) \times h_2 (y)}} \circ \big( \mathfrak{J}_{L \times M} \big) \circ
(L \times M)_{{}_{{h_{{}_1} (x)}^{-1} \times {h_{{}_2} (y)}^{-1}}} w_{{}_{(x,y)}} \\
= \big( L_{{}_{h_{{}_1} (x)}} \otimes M_{{}_{h_{{}_2} (y)}} \big) \circ
\big( \mathfrak{J}_L \otimes \mathfrak{J}_M  \big) \circ
\big( L_{{}_{{h_{{}_1} (x)}^{-1}}} \otimes M_{{}_{{h_{{}_2} (y)}^{-1}}} \big) w_{{}_{(x, y)}} \\
= \big( L_{{}_{h_{{}_1} (x)}} \mathfrak{J}_L   L_{{}_{{h_{{}_1} (x)}^{-1}}} \big) \otimes 
\big( M_{{}_{h_{{}_2} (y)}} \mathfrak{J}_M  M_{{}_{{h_{{}_2} (y)}^{-1}}} \big) w_{{}_{(x, y)}}.      
\end{multline*}
Thus the vector $\mathfrak{J}_{L \times M} \big( \mathfrak{J}^{L \times M}  w \big)_{{}_{(x, y)}}$ in the integrand in the formula for the inner product in $\mathcal{H}^{L \times M}$
\[
(w, u)
= \int \limits_{\mathfrak{G}_1 \times \mathfrak{G}_2} 
\Big( \mathfrak{J}_{L \times M} \big( \mathfrak{J}^{L \times M}  w \big)_{{}_{(x, y)}} , 
\big(u \big)_{{}_{(x, y)}}  \Big)  \,\, \ud (\mu_1 \times \mu_2 ) ([(x,y)])
\]
may be written as follows
\begin{multline*}
\mathfrak{J}_{L \times M} \big( \mathfrak{J}^{L \times M}  w \big)_{{}_{(x,y)}}
= \big( \mathfrak{J}_L \otimes \mathfrak{J}_M \big) \circ \big( \mathfrak{J}^{L \times M}  w \big)_{{}_{(x, y)}} \\
=  \big( \mathfrak{J}_L L_{{}_{h_{{}_1} (x)}} \mathfrak{J}_L   L_{{}_{{h_{{}_1} (x)}^{-1}}} \big) \otimes 
\big( \mathfrak{J}_M  M_{{}_{h_{{}_2} (y)}} \mathfrak{J}_M  M_{{}_{{h_{{}_2} (y)}^{-1}}} \big) w_{{}_{(x, y)}}
= \big( {}_{{}_x}\mathfrak{J}^L \otimes {}_{{}_y}\mathfrak{J}^M  \big) w_{{}_{(x, y)}},
\end{multline*}
where we have introduced the following self-adjoint operators 
\[
{}_{{}_x}\mathfrak{J}^L = \mathfrak{J}_L L_{{}_{h_{{}_1} (x)}} \mathfrak{J}_L   L_{{}_{{h_{{}_1} (x)}^{-1}}}
\,\,\, \textrm{and} \,\,\,
{}_{{}_y}\mathfrak{J}^M = \mathfrak{J}_M  M_{{}_{h_{{}_2} (y)}} \mathfrak{J}_M  M_{{}_{{h_{{}_2} (y)}^{-1}}}
\]
acting in $\mathcal{H}_L$ and $\mathcal{H}_M$, respectively, with the ordinary tensor product 
operator ${}_{{}_x}\mathfrak{J}^L \otimes {}_{{}_y}\mathfrak{J}^M $ acting in the tensor product
$\mathcal{H}_L \otimes \mathcal{H}_M$ Hilbert space.

Checking their self-adjointness is immediate. Indeed, because $L$ is Krein unitary in $(\mathcal{H}_L , \mathfrak{J}_L)$
we have (and similarly for the rep. $M$):
\[
L_{{}_{{h_{{}_1} (x)}^{-1}}} = \big( L_{{}_{h_{{}_1} (x)}} \big)^\dagger 
= \mathfrak{J}_L \big( L_{{}_{h_{{}_1} (x)}} \big)^* \mathfrak{J}_L .
\]
Therefore 
\[
{}_{{}_x}\mathfrak{J}^L = \mathfrak{J}_L L_{{}_{h_{{}_1} (x)}} \big( L_{{}_{h_{{}_1} (x)}} \big)^* \mathfrak{J}_L ,  
\]
because $\big( \mathfrak{J}_L \big)^2 = I$. Because $\mathfrak{J}_L$ is self-adjoint, self-adjointness
of ${}_{{}_x}\mathfrak{J}^L$ is now immediate (self-adjointness of ${}_{{}_y}\mathfrak{J}^M$
follows similarly).   

We are ready now to formulate the main goal of this Section:

\begin{twr}
Let $L$ and $M$ be Krein-unitary strongly continuous and almost uniformly bounded representations of the
closed subgroups $G_1$ and $G_2$ of the separable locally compact groups $\mathfrak{G}_1$ and $\mathfrak{G}_2$,
respectively, in the Krein spaces $(\mathcal{H}_L , \mathfrak{J}_L)$ and $(\mathcal{H}_M , \mathfrak{J}_M)$.
Then the Krein isometric representation ${}^{\mu_1 \times \mu_2}U^{L \times M}$ of the group
$\mathfrak{G}_1 \times \mathfrak{G}_2$ with the representation space equal to the Krein space 
$(\mathcal{H}^{L \times M} , \mathfrak{J}^{L \times M})$ is unitary and Krein-unitary equivalent to the 
Krein-isometric representation ${}^{\mu_1}U^L \, \times \, {}^{\mu_2}U^M$ of the group 
$\mathfrak{G}_1 \times \mathfrak{G}_2$
with the representation space equal to the Krein space 
$(\mathcal{H}^L \otimes \mathcal{H}^M , \mathfrak{J}^L \otimes \mathfrak{J}^M)$. More precisely:
there exists a map $V: \mathcal{H}^L \otimes \mathcal{H}^M \mapsto \mathcal{H}^{L \times M}$
which is unitary between the indicated Hilbert spaces and Krein-unitary between the Krein spaces 
$(\mathcal{H}_L \otimes \mathcal{H}_M , \mathfrak{J}_L \otimes \mathfrak{J}_M)$ and
$(\mathcal{H}^{L \times M} , \mathfrak{J}^{L \times M})$ and such that 
\begin{equation}\label{ulxm=ulxum}
\boxed{V^{-1} \, \Big( \, {}^{\mu_1 \times \mu_2}U^{L \times M} \, \Big) \, V \,\, 
= \,\, {}^{\mu_1}U^L \, \times \, {}^{\mu_2}U^M .}  
\end{equation}  

\label{twr.1:kronecker}
\end{twr}
\qedsymbol \,
Let $T$ be any member of $\mathcal{H}^L \otimes \mathcal{H}^M$, regarded as a \emph{conjugate-linear} operator
from ${}^{\mu_2}\mathcal{H}^M$ into ${}^{\mu_1}\mathcal{H}^L$, with the corresponding \emph{linear} operator 
$T \, T^{{}^\circledast}$ on  ${}^{\mu_1}\mathcal{H}^L$ having finite trace. Let moreover $T$ be a finite rank
operator. Then there exist $f_{{}_1}, f_{{}_2}, \ldots , f_{{}_n} \in {}^{\mu_1}\mathcal{H}^L$ and 
$g_{{}_1}, g_{{}_2}, \ldots , g_{{}_n} \in {}^{\mu_2}\mathcal{H}^M$ such that
$T(g) = T_{{}_{f_{{}_1}, g_{{}_1}}} (g) + \ldots + T_{{}_{f_{{}_n}, g_{{}_n}}} (g)
= f_{{}_1} \cdot \big( g_{{}_1} , w \big) + \ldots + f_{{}_n} \cdot \big( g_{{}_n} , g \big)$. For each
$(x, y) \in \mathfrak{G}_1 \times \mathfrak{G}_2$ we may define a conjugate-linear finite rank operator 
$\big(V(T)\big)_{{}_{(x, y)}}$ from $\mathcal{H}_M$ into $\mathcal{H}_L$ as follows. Let $\upsilon \in \mathcal{H}_M$,
then we put $\big(V(T)\big)_{{}_{(x, y)}}(\upsilon) = f_{{}_1} \cdot \big( g_{{}_1} , \upsilon \big) + \ldots + 
f_{{}_n} \cdot \big( g_{{}_n} , \upsilon \big)$. Note, please, that $\big(V(T)\big)_{{}_{(\xi x, \eta y)}} \, = \,
L_\xi \big(V(T)\big)_{{}_{(x, y)}} (M_\eta)^*$ for all $(x, y) \in \mathfrak{G}_1 \times \mathfrak{G}_2$
and all $(\xi, \eta) \in G_1 \times G_2$, so that the function $V(T): 
\mathfrak{G}_1 \times \mathfrak{G}_2 \ni (x, y) \mapsto \big(V(T)\big)_{{}_{(x, y)}} \in \mathcal{H}_L 
\otimes \mathcal{H}_M$ fulfils $\big(V(T)\big)_{{}_{(\xi x, \eta y)}} = \big( L_\xi \otimes M_\eta \big) 
\big(V(T)\big)_{{}_{(x, y)}}$ for all $(x, y) \in \mathfrak{G}_1 \times \mathfrak{G}_2$
and all $(\xi, \eta) \in G_1 \times G_2$.  

We shall show that the function $V(T)$ is a member of $\mathcal{H}^{L\times M}$ and moreover, that $V$ is unitary. To this
end we observe first, that $V$ is isometric (for the ordinary definite inner products), i. e. 
$\| V(T) \| = \| T \|$. Indeed, let $\{e_k\}_{k \in \mathbb{N}}$ be an orthonormal basis in $\mathcal{H}_L$. 
Using the observation we have made just before the formulation of the Theorem, self-adjointness
of the operators ${}_{{}_x}\mathfrak{J}^L$ and ${}_{{}_y}\mathfrak{J}^M$ and Scholium 3.9 and 5.3 of
\cite{Segal_Kunze}, we obtain:
\begin{multline*}
\| T \|^2 =  \big( \, f_{{}_1}\otimes g_{{}_1} + \ldots +  f_{{}_n} \otimes g_{{}_n} \, , \,\, 
f_{{}_1}\otimes g_{{}_1} + \ldots +  f_{{}_n} \otimes g_{{}_n} \, \big)   \\
= \Tr \Big[ \big( T_{{}_{f_{{}_1}, g_{{}_1}}} + \ldots + T_{{}_{f_{{}_n}, g_{{}_n}}} \big) 
\big( T_{{}_{g_{{}_1}, f_{{}_1}}} + \ldots + T_{{}_{g_{{}_n}, f_{{}_n}}} \big) \Big] 
= \sum \limits_{i,j = 1}^{n} (f_i , f_j) \cdot (g_i , g_j)    \\
= \sum \limits_{i,j = 1}^{n} 
\bigg( 
\int \limits_{\mathfrak{G}_1} 
\Big( \mathfrak{J}_L  \big( \mathfrak{J}^L  f_i \big)_{{}_{x}} , 
\big(f_j \big)_{{}_{x}}  \Big)  \,\, \ud \mu_1 ([x]) \,\, \bigg) \cdot
\bigg( 
\int \limits_{\mathfrak{G}_2} 
\Big( \mathfrak{J}_M  \big( \mathfrak{J}^M  g_i \big)_{{}_{y}} , 
\big(g_j \big)_{{}_{y}}  \Big)  \,\, \ud \mu_2 ([y]) \,\, 
\bigg)      \\
= \int \limits_{\mathfrak{G}_1 \times \mathfrak{G}_2} \, \bigg( \sum \limits_{i,j = 1}^{n}  
\Big( \mathfrak{J}_L  \big( \mathfrak{J}^L  f_i \big)_{{}_{x}} , 
\big(f_j \big)_{{}_{x}}  \Big) \cdot
\Big( \mathfrak{J}_M  \big( \mathfrak{J}^M  g_i \big)_{{}_{y}} , 
\big(g_j \big)_{{}_{y}}  \Big) 
\bigg)  
\,\,\,\, \ud (\mu_1 \times \mu_2 ) ([(x,y)]) \\
= \int \limits_{\mathfrak{G}_1 \times \mathfrak{G}_2} 
\bigg( \sum \limits_{i,j = 1}^{n}  
\Big( {}_{{}_x}\mathfrak{J}^L \big( f_i \big)_{{}_{x}} , 
\big(f_j \big)_{{}_{x}}  \Big) \cdot
\Big( {}_{{}_y}\mathfrak{J}^M \big( g_i \big)_{{}_{y}} , 
\big(g_j \big)_{{}_{y}}  \Big) 
\bigg)  
\,\,\,\, \ud (\mu_1 \times \mu_2 ) ([(x,y)])   \\
= \int \, \bigg( \sum \limits_{i,j = 1}^{n} \sum \limits_{k \in \mathbb{N}}  
\Big( {}_{{}_x}\mathfrak{J}^L \big( f_i \big)_{{}_{x}} , \, e_k \, \Big) \cdot \Big( \, e_k  , \,
\big(f_j \big)_{{}_{x}}  \Big) \cdot
\Big( {}_{{}_y}\mathfrak{J}^M \big( g_i \big)_{{}_{y}} , 
\big(g_j \big)_{{}_{y}}  \Big) 
\bigg)  
\,\,\,\, \ud (\mu_1 \times \mu_2) ([(x,y)])  \\
= \int \, \bigg( \sum \limits_{i,j = 1}^{n} \sum \limits_{k \in \mathbb{N}}  
\Big( \, e_k  , \, \big(f_j \big)_{{}_{x}}  \Big) \cdot  
\Big( \big( g_i \big)_{{}_{y}} , {}_{{}_y}\mathfrak{J}^M  \big(g_j \big)_{{}_{y}}  \Big) \cdot 
\Big( {}_{{}_x}\mathfrak{J}^L \big( f_i \big)_{{}_{x}} , \, e_k \, \Big)
 \bigg)  
\,\,\,\, \ud (\mu_1 \times \mu_2) ([(x,y)])  \\
= \int \, \bigg( \sum \limits_{i,j = 1}^{n} \sum \limits_{k \in \mathbb{N}}  
\Big( \, e_k  , \, \big(f_j \big)_{{}_{x}}  \Big) \cdot   
\Big( {}_{{}_x}\mathfrak{J}^L \Big( 
\big( f_i \big)_{{}_{x}} \cdot \Big( \big( g_i \big)_{{}_{y}} , {}_{{}_y}\mathfrak{J}^M  \big(g_j \big)_{{}_{y}}  \Big)
\Big) 
\, , \, e_k  \, \Big)
 \bigg)  
\,\,\,\, \ud (\mu_1 \times \mu_2 ) ([(x,y)])  \\
= \int \, \bigg( \sum \limits_{i,j = 1}^{n} \sum \limits_{k \in \mathbb{N}}  
\Big( \, e_k  , \, \big(f_j \big)_{{}_{x}}  \Big) \cdot   
\Big( \, {}_{{}_x}\mathfrak{J}^L \, \circ \, T_{{}_{(f_{{}_i})_{{}_x}, (g_{{}_i})_{{}_y} }} \, \circ \,
{}_{{}_y}\mathfrak{J}^M \big( \big(g_j \big)_{{}_{y}} \big) \, ,  \, e_k \, \Big) 
 \bigg)  
\,\,\,\, \ud (\mu_1 \times \mu_2 ) ([(x,y)])  \\
\end{multline*}
\begin{multline*}
= \int \, \bigg( \sum \limits_{i,j = 1}^{n} \sum \limits_{k \in \mathbb{N}}   
\Big( \, {}_{{}_x}\mathfrak{J}^L \, \circ \, T_{{}_{(f_{{}_i})_{{}_x}, (g_{{}_i})_{{}_y} }} \, \circ \,
{}_{{}_y}\mathfrak{J}^M \, \circ \, T_{{}_{(g_{{}_j})_{{}_x}, (f_{{}_j})_{{}_y} }}
\big( e_k \big) \, ,  \, e_k \, \Big)
\bigg)  
\,\,\,\, \ud (\mu_1 \times \mu_2 ) ([(x,y)])  \\
= \int \, \bigg( \sum \limits_{i,j = 1}^{n}  
\Tr \Big[ \, {}_{{}_x}\mathfrak{J}^L \, \circ \, T_{{}_{(f_{{}_i})_{{}_x}, (g_{{}_i})_{{}_y} }} \, \circ \,
{}_{{}_y}\mathfrak{J}^M \, \circ \, T_{{}_{(g_{{}_j})_{{}_x}, (f_{{}_j})_{{}_y} }} \, \Big]
\bigg)  
\,\,\,\, \ud (\mu_1 \times \mu_2 ) ([(x,y)])  \\
= \int \, \bigg( \sum \limits_{i,j = 1}^{n}  
\Tr \Big[ \, {}_{{}_x}\mathfrak{J}^L \, \circ \, T_{{}_{(f_{{}_i})_{{}_x}, (g_{{}_i})_{{}_y} }} \, \circ \,
{}_{{}_y}\mathfrak{J}^M \, \circ \, \big( T_{{}_{(f_{{}_j})_{{}_x}, (g_{{}_j})_{{}_y} }} \big)^{{}^\circledast} \, \Big]
\bigg)  
\,\,\,\, \ud (\mu_1 \times \mu_2 ) ([(x,y)])  \\
=\int \limits_{\mathfrak{G}_1 \times \mathfrak{G}_2} 
\Bigg( \, \Big( {}_{{}_x}\mathfrak{J}^L \otimes {}_{{}_y}\mathfrak{J}^M \Big) \Big( V(T) \Big)_{{}_{(x, y)}}  \, , \,
\Big( V(T) \Big)_{{}_{(x, y)}} \,  \Bigg)  \,\, \ud (\mu_1 \times \mu_2 ) ([(x,y)]) \\
=\int \limits_{\mathfrak{G}_1 \times \mathfrak{G}_2} 
\Bigg( \, \mathfrak{J}_{L \times M} \Big( \mathfrak{J}^{L \times M}  V(T) \Big)_{{}_{(x, y)}} \, , \,
\Big( V(T) \Big)_{{}_{(x, y)}} \,  \Bigg)  \,\, \ud (\mu_1 \times \mu_2 ) ([(x,y)]) = \| V(T) \|^2.
\end{multline*}
(The unspecified domain of integration in the above formulas is of course equal $\mathfrak{G}_1 \times \mathfrak{G}_2$.)

Therefore $V$ is isometric and $V(T) \in \mathcal{H}^{L\times M}$ for the indicated $T$, as the 
required measurability conditions
again easily follow from Scholium 3.9 of \cite{Segal_Kunze}. 
Now by the properties of Hilbert-Schmidt operators, the finite rank conjugate-linear operators 
$T: {}^{\mu_2}\mathcal{H}^M \mapsto {}^{\mu_1}\mathcal{H}^L$ are dense 
in ${}^{\mu_1}\mathcal{H}^L \otimes {}^{\mu_2}\mathcal{H}^M$  (compare e. g. \cite{Murray_von_Neumann}, Chap. II 
or \cite{Segal_Kunze}, Chap. 14.2 or \cite{Segal}). Thus the domain of the operator $V$ is dense. 

In order to show that the range of $V$ is likewise dense, consider the closure $C^1$ under the norm in 
$\mathcal{H}^{L\times M}$ of the linear set  of all functions $V(T)$,
where $T = T_{{}_{f_{{}_1}, g_{{}_1}}} + \ldots + T_{{}_{f_{{}_n}, g_{{}_n}}}$ with $f_i$ ranging over  $C^{L}_{0}
\subset \mathcal{H}^L$ and  $g_j$ over the corresponding set $C^{M}_{0} \subset \mathcal{H}^M$. Because $V$
is isometric it can be uniquely extended so that $C^1$ lies in the range of this unique extension. Let us denote the 
extension likewise by $V$. (For a densely defined Krein-isometric map this
would in general be impossible because $V$ could be discontinuous, this is the reason why we need to know if $V$ is continuous, i. e.  bounded for the ordinary positive definite inner products.)

Now by the property of Hilbert-Schmidt operators (mentioned above) the linear span of operators 
$T_{\upsilon, \textrm{v}} : \mathcal{H}_M \mapsto \mathcal{H}_L$
with $\upsilon$ and $\textrm{v}$ ranging over dense subsets of $\mathcal{H}_L$ and $\mathcal{H}_M$, respectively,
is dense in $\mathcal{H}_L \otimes \mathcal{H}_M$. This property of Hilbert-Schmidt operators together with
a repeated application of Lemma \ref{lem:dense.2} and \ref{lem:dense.5} of Sect. \ref{dense} and Scholium 3.9 and 5.3 of  \cite{Segal_Kunze} will show that all the conditions, (a)-(e), of Lemma \ref{lem:dense.4} are satisfied for 
$C^1 \subset \mathcal{H}^{L \times M}$. 

In particular if $\psi$ is a complex valued continuous function on $\mathfrak{G}_1 \times \mathfrak{G}_2$
which is constant on the right $G_1 \times G_2$ cosets and vanish outside of\footnote{$\pi$ denotes 
here the canonical quotient map $\mathfrak{G}_1 \times \mathfrak{G}_2 \mapsto 
(\mathfrak{G}_1 \times \mathfrak{G}_2 )/(G_1 \times G_2)$.} $\pi^{-1}(K)$
for some compact subset $K$ of $(\mathfrak{G}_1 \times \mathfrak{G}_2 )/(G_1 \times G_2)$, then it is measurable
and $\psi \in L^2 ((\mathfrak{G}_1 \times \mathfrak{G}_2 )/(G_1 \times G_2), \mu_1 \times \mu_2)$ 
and by Scholim 3.9 and 5.3 of  \cite{Segal_Kunze} it is an $L^2$-limit  of continuous such functions of ``product
form'' $\phi \cdot \varphi : \mathfrak{G}_1 \times \mathfrak{G}_2  \ni (x,y) \mapsto
\phi(x) \cdot \varphi(y)$. Thus the condition (d) of Lemma \ref{lem:dense.4} follows. The above mentioned
property of Hilbert-Schmidt operators and Lemma \ref{lem:dense.2} applied to $C^{L}_{0}
\subset \mathcal{H}^L$ and to $C^{M}_{0} \subset \mathcal{H}^M$, proves condition (e) of 
Lemma \ref{lem:dense.4}. Condition (b) follows from the the fact that $V(T) \in \mathcal{H}^{L\times M}$
for finite rank operators $T$, proved in the first part of the proof. An application of
the Lusin Theorem (Corollary 5.2.2 of \cite{Segal_Kunze}, together with an obvious adaptation
of the the standard proof of the Riesz-Fischer theorem already used in the proof of Lemma \ref{lem:dense.5}) 
proves condition (a) of Lemma \ref{lem:dense.4}. 
By the remark opening the proof of Lemma \ref{lem:dense.2} the linear sets $ C^{L}_{0}$
and $C^{M}_{0}$ of functions are closed with respect to right $\mathfrak{G}_1$ and $\mathfrak{G}_2$-translations,
respectively. Thus it easily follows that the linear 
set of functions $V\big( T_{{}_{f_{{}_1}, g_{{}_1}}} + \ldots + T_{{}_{f_{{}_n}, g_{{}_n}}} \big)$
with $f_{{}_i} \in C^{L}_{0}$, $g_{{}_j} \in C^{M}_{0}$ is closed under the right  
$\mathfrak{G}_1 \times \mathfrak{G}_2$-translations. Then, a simple continuity argument shows that 
$C^1$ is closed under right $\mathfrak{G}_1 \times \mathfrak{G}_2$-translations. Thus condition

(c) of Lemma \ref{lem:dense.4} is satisfied with trivial functions $\rho_s$ all equal identically to
the constant unit function. 
 
Thus  Lemma \ref{lem:dense.4} may be applied to $C^1$ lying in the range of $V$, so that the range
is dense in $\mathcal{H}^{L \times M}$. Therefore $C^1 = \mathcal{H}^{L \times M}$ and $V$ is unitary.  

We shall show that $V$ is Krein-unitary.  By the unitarity of $V$, it will be sufficient by continuity
to show that $V$ is Krein-isometric on finite rank operators $T \in \mathcal{H}^L \otimes \mathcal{H}^M$.
By self-adjointness of $\mathfrak{J}^L$ and $\mathfrak{J}^M$ we have  the following equalities for $T$ of the form indicated to above: 
\begin{multline*}
\big( \, \| T \|_{{}_{\mathfrak{J}^L \otimes \mathfrak{J}^M}} \,\big)^2 
=  \Big( \big(\mathfrak{J}^L \otimes \mathfrak{J}^M \big) \, \big( f_{{}_1}\otimes g_{{}_1} + \ldots +  
f_{{}_n} \otimes g_{{}_n} \big) \, , 
\,\, f_{{}_1}\otimes g_{{}_1} + \ldots +  f_{{}_n} \otimes g_{{}_n} \, \Big)   \\
= \Tr \Big[ \mathfrak{J}^L \big( T_{{}_{f_{{}_1}, g_{{}_1}}} + \ldots + T_{{}_{f_{{}_n}, g_{{}_n}}} \big) \mathfrak{J}^M
\big( T_{{}_{f_{{}_1}, g_{{}_1}}} + \ldots + T_{{}_{f_{{}_n}, g_{{}_n}}}  \big)^{{}^\circledast}  \Big]   \\
= \sum \limits_{i,j = 1}^{n} (\mathfrak{J}^L f_i , f_j) \cdot (\mathfrak{J}^M g_i , g_j) 
\end{multline*}   
\begin{multline*}
= \sum \limits_{i,j = 1}^{n} 
\bigg( 
\int \limits_{\mathfrak{G}_1} 
\Big( \mathfrak{J}_L  \big(f_i \big)_{{}_{x}} , 
\big(f_j \big)_{{}_{x}}  \Big)  \,\, \ud \mu_1 ([x]) \,\, \bigg) \cdot
\bigg( 
\int \limits_{\mathfrak{G}_2} 
\Big( \mathfrak{J}_M  \big( g_i \big)_{{}_{y}} , 
\big(g_j \big)_{{}_{y}}  \Big)  \,\, \ud \mu_2 ([y]) \,\, 
\bigg)      \\
= \int \limits_{\mathfrak{G}_1 \times \mathfrak{G}_2} \, \bigg( \sum \limits_{i,j = 1}^{n}  
\Big( \mathfrak{J}_L  \big( f_i \big)_{{}_{x}} , 
\big(f_j \big)_{{}_{x}}  \Big) \cdot
\Big( \mathfrak{J}_M  \big( g_i \big)_{{}_{y}} , 
\big(g_j \big)_{{}_{y}}  \Big) 
\bigg)  
\,\,\,\, \ud (\mu_1 \times \mu_2 ) ([(x,y)]) 
\end{multline*}
\begin{multline*}
=\int \limits_{\mathfrak{G}_1 \times \mathfrak{G}_2} 
\Tr \Big[ \mathfrak{J}_L \big( T_{{}_{(f_{{}_1})_{{}_x}, (g_{{}_1})_{{}_y} }}
+ \ldots + 
T_{{}_{(f_{{}_n})_{{}_x}, (g_{{}_n})_{{}_y} }}
\big) \mathfrak{J}_M
\big( T_{{}_{(f_{{}_1})_{{}_x}, (g_{{}_1})_{{}_y} }}
+ \ldots       \\
\ldots + T_{{}_{(f_{{}_n})_{{}_x}, (g_{{}_n})_{{}_y} }}
\big)^{{}^\circledast} \Big]
  \,\, \ud (\mu_1 \times \mu_2 ) ([(x,y)])     \\
=\int \limits_{\mathfrak{G}_1 \times \mathfrak{G}_2} 
\Bigg( \, \Big( \mathfrak{J}_L \otimes \mathfrak{J}_M \Big) \Big( V(T) \Big)_{{}_{(x, y)}} \, , \,
\Big( V(T) \Big)_{{}_{(x, y)}} \,  \Bigg)  \,\, \ud (\mu_1 \times \mu_2 ) ([(x,y)])     \\
=\int \limits_{\mathfrak{G}_1 \times \mathfrak{G}_2} 
\Bigg( \, \mathfrak{J}_{L \times M} \Big( V(T) \Big)_{{}_{(x, y)}} \, , \,
\Big( V(T) \Big)_{{}_{(x, y)}} \,  \Bigg)  \,\, \ud (\mu_1 \times \mu_2 ) ([(x,y)])   \\
= \big( \, \| V(T) \|_{{}_{\mathfrak{J}^{L \times M}}} \,\big)^2 .
\end{multline*}

Recall that  the domain $\mathfrak{D}_{12}$ (common for all $(x,y) \in \mathfrak{G}_1 \times \mathfrak{G}_2$) 
of the operators $U_1 \otimes U_2 = {}^{\mu_1}U^{L}_{x} \, \otimes \, {}^{\mu_2}U^{M}_{y}
= \big( {}^{\mu_1}U^{L} \, \times \, {}^{\mu_2}U^{M} \big)_{(x,y)}$  representing $(x,y) \in \mathfrak{G}_1 \times \mathfrak{G}_2$, is invariant for the operators $U_1 \otimes U_2 = {}^{\mu_1}U^{L}_{x} \, \otimes \, {}^{\mu_2}U^{M}_{y}
= \big( {}^{\mu_1}U^L \, \times \, {}^{\mu_2}U^M \big)_{(x, y)}$. For each
$(x,y)$ let us denote the closure of ${}^{\mu_1}U^L \, \times \, {}^{\mu_2}U^M
=  {}^{\mu_1}U^{L}_{x} \, \otimes \, {}^{\mu_2}U^{M}_{y}$ likewise by ${}^{\mu_1}U^L \, \times \, {}^{\mu_2}U^M$.
Note that $V(T)$, $T \in \mathfrak{D}_{12}$ compose an invariant domain of the representation
${}^{\mu_1 \times \mu_2}U^{L \times M}$. Denote the closures of the operators 
${}^{\mu_1 \times \mu_2}U^{L \times M}_{(x, y)}$ with the common invariant domain $V(\mathfrak{D}_{12})$ likewise by 
${}^{\mu_1 \times \mu_2}U^{L \times M}_{(x, y)}$.

The equality (\ref{ulxm=ulxum}) is regarded as equality for the closures of the operators
${}^{\mu_1 \times \mu_2}U^{L \times M}$ and ${}^{\mu_1}U^L \, \times \, {}^{\mu_2}U^M$. 

By Theorem \ref{def_ind_krein:twr.1} and its proof the closures of ${}^{\mu_1 \times \mu_2}U^{L \times M}$ do not depend
on the choice of the dense common invariant domain. Therefore in order to show the equality 
(\ref{ulxm=ulxum}) it is sufficient that the respective closed operators in (\ref{ulxm=ulxum}) coincide
on the domain of all finite rank operators $T \in \mathfrak{D}_{12}$. This however is immediate.
Indeed, let $T = T_{{}_{f_{{}_1}, g_{{}_1}}} + \ldots + T_{{}_{f_{{}_n}, g_{{}_n}}}$
with $f_i \in \mathfrak{D}_1$ and $g_j \in \mathfrak{D}_2$. Then
\begin{multline}\label{ultum*}
\big( {}^{\mu_1}U^{L} \, \times \, {}^{\mu_2}U^{M} \big)_{(x_0 ,y_0 )} (T) 
= \big( {}^{\mu_1}U^{L}_{x_0} \, \otimes \, {}^{\mu_2}U^{M}_{y_0} \big) (T) \,\,
= \,\, {}^{\mu_1}U^{L}_{x_0} \,\,\, T \,\,\, \big( {}^{\mu_2}U^{M}_{y_0} \big)^* \\
= \sqrt{\lambda_1(\cdot, x_0)} \sqrt{\lambda_2 (\cdot, y_0)} \,
\big(T_{{}_{R_{x_0}f_{{}_1}, R_{y_0}g_{{}_1}}} + \ldots 
+  T_{{}_{R_{x_0}f_{{}_n}, R_{y_0}g_{{}_n}}} \big).
\end{multline}
On the other hand we have:
\begin{multline*}
 \Big( {}^{\mu_1 \times \mu_2}U^{L \times M}_{(x_0 ,y_0 )} V(T) \Big)_{{}_{(x, y)}} 
=\sqrt{\lambda_1([x], x_0)} \sqrt{\lambda_2 ([y], y_0)} \,
\Big( V(T) \Big)_{{}_{(x \cdot x_0, y \cdot y_0)}} \\
= \sqrt{\lambda_1([x], x_0)} \sqrt{\lambda_2 ([y], y_0)} \,
\Big(
T_{{}_{(R_{x_0}f_{{}_1})_{{}_x}, (R_{y_0}g_{{}_1})_{{}_y} }} + \ldots
\ldots + T_{{}_{(R_{x_0}f_{{}_n})_{{}_x}, (R_{y_0}g_{{}_n})_{{}_y} }}
\Big),
\end{multline*}
so that 
\begin{multline*}
\Big( V^{-1} \, \big( \, {}^{\mu_1 \times \mu_2}U^{L \times M} \, \big) \, V \Big) \big( T \big) \\
= \sqrt{\lambda_1(\cdot, x_0)} \sqrt{\lambda_2 (\cdot, y_0)} \,
\big(T_{{}_{R_{x_0}f_{{}_1}, R_{y_0}g_{{}_1}}} + \ldots 
+  T_{{}_{R_{x_0}f_{{}_n}, R_{y_0}g_{{}_n}}} \big).
\end{multline*}
Comparing it with (\ref{ultum*}) one can see that (\ref{ulxm=ulxum}) holds
on $\mathfrak{D}_{12}$. Thus the proof of (\ref{ulxm=ulxum}) is complete now. The Theorem
is hereby proved completely.
\qed

Presented proof of Theorem \ref{twr.1:kronecker} is an extended and modified version of the
Mackey's proof of Theorem 5.2 in \cite{Mackey}.

Note, please, that the equality (\ref{ulxm=ulxum}) for the closures of the operators 
${}^{\mu_1 \times \mu_2}U^{L \times M}$ and ${}^{\mu_1}U^L \, \times \, {}^{\mu_2}U^M$ 
is non trivial. Indeed, recall that in general almost all kinds of pathology not excluded 
by general theorems can be shown to exist for unbounded operators. In particular two \emph{distinct} 
and closed operators may still coincide on a dense domain. This is why we need to be careful 
in proving (\ref{ulxm=ulxum}).  This in particular shows that the fundamental theorems of the original
Mackey theory by no means are automatic for the induced Krein-isometric representations, where the
representors are in general densely defined and unbounded. Here we saw it for the Theorem 
\ref{twr.1:kronecker}. But differences in the proofs arise likewise in the latter part of the theory. 
In particular if we 
want to prove the \emph{subgroup theorem} and the so called \emph{Kronecker product theorem}
for the induced Krein-isometric representations with precisely the same assumptions
posed on the group as in Mackey's theory, then some additional analysis will have to be made 
in treating decompositions of non finite quasi invariant measures. Compare Sect. \ref{decomposition}.

\subsection{Subgroup theorem in Krein spaces. Preliminaries}\label{subgroup.preliminaries}

This Section is a word for word repetition of the argument of \S 6 of \cite{Mackey}. That 
the general Mackey's argument may be applied to induced representations in Krein spaces is the whole point.  
Although  it is rather clear that his general argumentation is applicable in the Krein space, we  
restate it here because it lies at the very heart of the presented method of decomposition
of tensor product of induced representations, and will make the paper self contained. It should be noted however
that it requires some additional analysis in decomposing non finite quasi invariant measures,
which makes a difference in proving the existence of the corresponding direct integral decompositions.

The circumstance that the {\L}opusza\'nski representation of $\mathfrak{G}$ is equivalent to an induced 
representation in a Krein space greatly simplifies the problem of decomposing tensor product of {\L}opusza\'nski 
representations and reduces it largely to the geometry of right cosets and double cosets in the group $\mathfrak{G}$ and to 
a ``Fubini-like'' theorem, just like for the ordinary induced representations of Mackey. Similar decomposition
method of quotiening by a subgroup in construction of complete sets of unitary representations of semi simple 
Lie groups was applied by Gelfand and Neumark, and by several authors in constructing harmonic analysis on classical Lie groups. The main gain is that the subtle analytic properties of the {\L}opusza\'nski representation (unboundedness)  
does not intervene dramatically after this reduction to geometry of cosets and double cosets.

Our main theorem asserts the existence of a certain useful direct integral decomposition of
the tensor product $U^L \otimes U^M$ of two induced representations of a group $\mathfrak{G}$
in a Krein space, whose construction is completely analogous to that of Mackey for ordinary
unitary representations, compare \cite{Mackey}. By definition $U^L \otimes U^M$ is obtained 
from the outer Kronecker product representation $U^L \times U^M$ of $\mathfrak{G} \times \mathfrak{G}$
by restricting $U^L \times U^M$ to the diagonal subgroup $\overline{\mathfrak{G}} \cong \mathfrak{G}$
of all $(x,y) \in \mathfrak{G}\times \mathfrak{G}$ with $x = y$. By the Theorem of Sect. \ref{kronecker},
$U^L \times U^M$ is Krein-unitary equivalent to $U^{L \times M}$. Thus $U^L \otimes U^M$ can be analysed by analysing
the restriction of $U^{L \times M}$ to the diagonal subgroup $\overline{\mathfrak{G}} \cong \mathfrak{G}$. 
Our theorem on tensor product decomposition follows (just as in \cite{Mackey}) from these remarks and
a theorem on restriction to a subgroup of an induced representation in a Krein space, say a \emph{subgroup theorem}.
\emph{Subgroup theorem} gives a decomposition of the restriction of an induced representation (in a Krein space)
to a closed subgroup, with the component representations in the decomposition themselves Krein-unitary equivalent to induced representations. Namely, let $L$ be strongly continuous almost uniformly bounded Krein-unitary representation of the
closed subgroup $G_1$ of $\mathfrak{G}$ and consider the restriction ${}_{G_2}U^L$ of $U^L$ to a second 
closed subgroup $G_2$. While $\mathfrak{G}$ acts transitively on the homogeneous space $\mathfrak{G}/G_1$ of
right $G_1$-cosets this will not be true in general of $G_2$. Moreover, and this is the main advantage of
induced representations, any division of $\mathfrak{G}/G_1$ into two parts $S_1$ and $S_2$, each a Baire (or Borel)
set which is not a null set (with respect to any, and hence every quasi invariant measure on $\mathfrak{G}/G_1$),
and each invariant under $G_2$ leads to a corresponding direct sum decomposition of ${}_{G_2}U^L$. Indeed
the closed subspaces $\mathcal{H}^{L}_{S_1}$ and $\mathcal{H}^{L}_{S_2}$ of all $f \in \mathcal{H}^L$ which vanish
respectively outside of $\pi^{-1}(S_1)$ and $\pi^{-1}(S_2)$ are invariant and are orthogonal complements of each other
with respect to the ordinary (as well as the Krein) inner product on $\mathcal{H}^L$.

Assume for a while, just for illustrative purposes, that there is a null set $N$ in $\mathfrak{G}/G_1$ whose complement
is the union of countably many non null orbits $C_1 , C_2 , \ldots$ of $\mathfrak{G}/G_1$ under $G_2$.  
Then by the above remarks we obtain a direct sum decomposition of ${}_{G_2}U^L$ into as many parts as there are non null orbits. Our analysis reaches its goal after analysing the nature of these parts. Analysis of these parts is our goal of the rest of this Section.

In our paper we shall consider a more general case in which all of the orbits can be null sets and the sum becomes an integral and we have to use the von Neumann theory of direct integral Hilbert spaces \cite{von_neumann_dec}.
Of course according to the definition given above (with $S_1$ or $S_2$ equal to 
a $G_2$ orbit $C$ in $\mathfrak{G}/G_1$), $\mathcal{H}^{L}_{C}$ will be zero dimensional whenever the orbit 
$C$ is a null set. However it is possible to reword the definition so that it always gives a non zero Hilbert
space (with the respective Krein structure) and so that when $C$ is not a null set this definition is essentially the same as that already given, compare \cite{Mackey}, \S 6. Indeed note that when $C$ is a non null set then $\mathcal{H}^{L}_{C}$
may be equivalently defined as follows. Let $x_c$ be any member of $\mathfrak{G}$ such that $\pi(x_c) \in C$
and consider the set ${\mathcal{H}^{L}_{C}}'$ of all functions $f$ from the double coset $G_1 x_c G_2$ to $\mathcal{H}_L$
such that: (i) $x \mapsto (f_x ,\upsilon)$ is a Borel function for all $\upsilon  \in \mathcal{H}_L$, 
(ii) $f_{\xi x} = L_\xi (f_x )$ for all $\xi \in G_1$ and all $x \in G_1 x_c G_2$ and 
(iii): 
\[
\| f \|_C = \int \limits_{C} \, (\, \mathfrak{J}_L ((\mathfrak{J}^L f)_x ), f_x \,) \, d\mu_{\mathfrak{G}/G_1}
= \int \limits_{(G_1 x_c G_2 ) \, \cap B} \, (f_b , f_b \,) \, d\mu_{B}(b) < \infty,
\]  
where $B$ is the regular Borel section of $\mathfrak{G}$ with respect to $G_1$ of Sect. \ref{def_ind_krein}
(we could use as well the sub-manifold $Q$ of Sect. \ref{def_ind_krein} but we prefer to proceed generally and independently of the ``factorization'' assumption). The operator $\mathfrak{J}^L$
in $\mathcal{H}^{L}_{C}$ is given by simple restriction, and its definition on 
${\mathcal{H}^{L}_{C}}'$ is obvious:
\[
(\mathfrak{J}^{L,C} f)_x = L_{h(x)} \mathfrak{J}_{L} L_{h(x)^{-1}} \, f_x ;
\] 
with the obvious definition of the Krein inner product in ${\mathcal{H}^{L}_{C}}'$
\[
\big(f, g \big)_{\mathfrak{J}^{L,C}} = (\mathfrak{J}^{L,C} f, g) 
= \int \limits_{C} \, (\, \mathfrak{J}_L (f_x ), g_x \,) \, d\mu_{\mathfrak{G}/G_1}, \,\,\, f,g \in {\mathcal{H}^{L}_{C}}'.
\]
Similarly we define the operator $U^{L, C}_{\xi}$ in $\mathcal{H}^{L}_{C}$ for $\xi \in G_2$ as the restriction of $U^{L}_{\xi}$ to $\mathcal{H}^{L}_{C}$, i. e. to the functions supported by the orbit $C$, and its definition giving an equivalent representation on ${\mathcal{H}^{L}_{C}}'$ is likewise obvious:
\[
(U^{L, C}_{\xi} f)_x = \sqrt{\lambda ([x], \xi)} \, f_{x \xi},
\]
with the $\lambda$-function of the quasi invariant measure $\mu$ restricted to $C \times G_2$. 

Moreover, and this is the whole point, the measure in $C$
need not be defined by restricting $\mu = \mu_{\mathfrak{G}/G_1}$ to $C$. There exists a non zero measure $\mu_C$ on $C$ quasi invariant with respect to $G_2$ determined up to a constant factor, whose
Radon-Nikodym function $\ud(R_\eta \mu_C )/\ud \mu_C$, $\eta \in G_2$ 
(i. e. the associated $\lambda_C$-function) is equal to the restriction to the subspace $C \times G_2$ of the 
$\lambda$-function, i. e. Radon-Nikodym derivative $\ud (R_\eta \mu)/\ud \mu$ , associated with $\mu = \mu_{\mathfrak{G}/G_1}$. Indeed, although $C$ does not have the form of a quotient of a group by its closed subgroup, it follows from Theorem 3, page 253 of \cite{Kuratowski} that the map $x \mapsto \pi(x_c x)$ induces a Borel isomorphism\footnote{With the Borel structure on $C$ induced from the surrounding space $\mathfrak{G}/G_1$:
we define $E \subset C$ to be Borel iff $E = E' \cap C$ for a Borel set $E'$ in $\mathfrak{G}/G_1$.
However our assumptions concerning the group $\mathfrak{G}$ and the subgroups $G_1$ and $G_2$ are exactly the same as those of Mackey, and they do not even guarantee the local compactness of the orbits $C$, compare Sect. \ref{decomposition}.} 
$\psi$ of the quotient space $G_2 /G_{x_c}$ onto $C$, where $G_{x_c} = G_2 \cap ({x_c}^{-1}G_1 x_c)$ is the closed subgroup of all $x \in G_2$ such that $\pi(x_c x) = \pi(x_c)$. Thus $C \times G_2 \cong G_2 /G_{x_c} \times G_2$
as Borel spaces under the indicated isomorphism and moreover if $[x] \in G_2 /G_{x_c}$ and $[z] = \pi(x_c x)$ correspond
under this isomorphism and $\eta \in G_2$ then $[x]\eta$ and $[z]\eta$ do also, where $[x]\eta = [x\eta]$ and 
$[z]\eta = [z\eta]$ denote the action of $\eta \in G_2$ on $[x] \in G_2 / G_{x_c}$ and $[z] \in C$ respectively. 
Thus the existence of the quasi invariant measure  
$\mu_C$ on $C$ follows from the general Mackey classification of quasi invariant 
measures on the quotient of a locally compact group by a closed subgroup, compare the respective Theorem of 
Sect. \ref{def_ind_krein}. Using the quasi invariant measure $\mu_C$ on $C$ gives a non trivial space 
${\mathcal{H}^{L}_{C}}'$ for every orbit $C$, which in case of a non null
orbit $C$ is trivially equivalent to $\mathcal{H}^{L}_{C}$.

We are now in a position to formulate the main goal 
of this Section:

\begin{lem}
Let $C$ be any orbit in $\mathfrak{G}/G_1$ under $G_2$ and let $x_c$ be such that $\pi(x_c) \in C$. Let 
${\mathcal{H}^{L}_{C}}'$ be defined as above. Let ${^{\mu^{x_c}}}U^{L^{x_c}}$ be the representation of $G_2$ induced by the
strongly continuous almost uniformly bounded Krein-unitary representation $L^{x_c}: \eta \mapsto 
L_{x_c \eta {x_c}^{-1}}$ of $G_2 \cap ({x_c}^{-1}G_1 x_c )$ with 
the representation space of $L^{x_c}$ equal to $\mathcal{H}_{L^{x_c}} = \mathcal{H}_L$ and the fundamental symmetry 
$\mathfrak{J}_{L^{x_c}} = \mathfrak{J}_L$; and with the quasi invariant measure $\mu^{x_c}$ in the homogeneous space
$G_2 / (G_2 \cap \, ({x_c}^{-1}G_1 x_c))$ equal to the transfer of the measure $\mu_C$ in $C$ over to the homogeneous space by the map $\psi$.
Let ${}^{\mu^{x_c}}\mathcal{H}^{L^{x_c}}$ be the Krein space of the induced 
representation ${}^{\mu^{x_c}}U^{L^{x_c}}$. We assume the fundamental symmetry $\mathfrak{J}_{x_c}$ in 
${}^{\mu^{x_c}}\mathcal{H}^{L^{x_c}}$ to be defined by the equation $(\mathfrak{J}_{x_c} g)_t = 
L_{h(x_c t)} \mathfrak{J}_L L_{h(x_c t)^{-1}} g_t$ and the Krein inner product given by the ordinary formula
\[
\int \limits_{G_2 \big{/} \big(G_2 \cap \, ({x_c}^{-1}G_1 x_c)\big)} \, 
(\mathfrak{J}_{L} \tilde{f}_{t} , \tilde{f}_{t} ) \,\, \ud \mu^{x_c}([t]), \,\,\,t \in G_2 .
\]
Then there is a Krein-unitary map $V_{x_c}$ of ${\mathcal{H}^{L}_{C}}'$ onto ${}^{\mu^{x_c}}\mathcal{H}^{L^{x_c}}$
such that if $g \in {}^{\mu^{x_c}}\mathcal{H}^{L^{x_c}}$ corresponds to 
$f \in {\mathcal{H}^{L}_{C}}'$ then ${}^{\mu^{x_c}}U^{L^{x_c}}_s g$ corresponds to $U^{L, C}_{s} f$ where 
$(U^{L, C}_{s} f)_x = f_{xs} \sqrt{\lambda_C ([x],s)}$
for all $x \in C$ and all $s \in G_2$.

\label{lem:subgroup.preliminaries.1}
\end{lem}

\qedsymbol \,
 For each function $f$ on $G_1 x_c G_2$ satisfying the conditions (i) and (ii) of 
the definition of ${\mathcal{H}^{L}_{C}}'$ let $\tilde{f}$ be defined by $\tilde{f}_t = f_{x_c t}$
for all $t \in G_2$. Then $(\tilde{f}_t , \upsilon )$ is a Borel function of $t$ on $G_2$ for all 
$\upsilon \in \mathcal{H}_L$. If $\eta \in G_{x_c} = G_2 \cap \, ({x_c}^{-1}G_1 x_c)$ then if $\xi = x_c \eta {x_c}^{-1}$
we have $\tilde{f}_{\eta t} = \tilde{f}_{{x_c}^{-1}\xi x_c t} = f_{\xi x_c t} = L_\xi \tilde{f}_t
= L_{x_c \eta {x_c}^{-1}}(\tilde{f}_t)$; that is 
\begin{equation}\label{g_2_induced}
\tilde{f}_{\eta t} = L_{x_c \eta {x_c}^{-1}}(\tilde{f}_t)
\end{equation}
for all $t \in G_2$ and all $\eta \in G_2 \cap \, ({x_c}^{-1} G_1 x_c)$. Conversely let $g$ be any function from $G_2$ to 
$\mathcal{H}_L$ which is Borel in the sense that $x \mapsto (g_x ,\upsilon)$ is a Borel function on $G_2$ for all 
$\upsilon  \in \mathcal{H}_L$ and which satisfies (\ref{g_2_induced}). We define the corresponding function $f$ by the equation $f_{\xi x_c t} = L_\xi (g_t)$ for all $\xi \in G_1$ and $t \in G_2$. 
If $\xi_1 x_c t_1 = \xi_2 x_c t_2$ then ${\xi_2}^{-1}\xi_1 = x_c t_2 {t_1}^{-1}{x_c}^{-1}$ so that 
$g_{t_2{t_1}^{-1}t} = L_{{\xi_2}^{-1}\xi_1}(g_t)$. Therefore $L_{\xi_2}(g_{t_2}) = L_{\xi_1}(g_{t_1})$
and $f$ is well defined. Next we show that $(f_x , \upsilon)$ is Borel function of $x$ on $G_1 x_c G_2$ 
for all $\upsilon \in \mathcal{H}_L$. Let $f'$ be the function on $G_1
 \times G_2$ defined by 
$f'(\xi , \eta) = L_\xi (g_\eta)$ for all $(\xi , \eta) \in G_1 \times G_2$. Choose now an orthonormal
basis $\{ \varphi_i \}_{i \in \mathbb{N}}$ in $\mathcal{H}_L$. Then we have 
$(f'(\xi, \eta), \upsilon) = (f'(\xi, \eta), \mathfrak{J}_L \mathfrak{J}_L \upsilon) = 
(\mathfrak{J}_L f'(\xi, \eta),  \mathfrak{J}_L \upsilon) 
= (\mathfrak{J}_L L_\xi (g_\eta) ,  \mathfrak{J}_L \upsilon) =  
(\mathfrak{J}_L  g_\eta , L_{\xi^{-1}} \mathfrak{J}_L \upsilon)
= \sum_{i = 1}^{\infty} = ( \mathfrak{J}_L  g_\eta , \varphi_i ) 
( \varphi_i , L_{\xi^{-1}} \mathfrak{J}_L \upsilon )$. By Scholium 3.9 of \cite{Segal_Kunze} 
$(f'(\xi, \eta), \upsilon)$ is a Borel function of $(\xi , \eta)$ on $G_1 \times G_2$ regarded as
the product measure space, for all $\upsilon \in \mathcal{H}_L$. Let us introduce after Mackey a new
group operation in $G_1 \times G_2$ putting $(\xi_1 , \eta_1) (\xi_2 , \eta_2) = (\xi _1 \xi_2 , \eta_2 \eta_1)$
and call the resulting group $G_3$. Then $\xi_1 x_c \eta_1 = \xi_2 x_c \eta_2$ if and only if 
$(\xi_2 , \eta_2)^{-1} (\xi_1 , \eta_1) = ({\xi_2}^{-1}\xi_1 , \eta_1 {\eta_2}^{-1})$ has the form 
$(\xi , {x_c}^{-1}\xi^{-1}x_c)$. The set of all  $(\xi , {x_c}^{-1}\xi^{-1}x_c)$, $\xi \in G_1$ is a subgroup
$G_4$ of $G_3$. Thus the map $(\xi ,\eta ) \mapsto \xi x_c \eta$ sets up a one-to-one correspondence 
between the points of the homogeneous space $G_3 /G_4$ of left $G_4$-cosets and the points of the double
coset $G_1 x_c G_2$. The map is continuous and on account of the assumed separability it follows again from
Theorem 3, page 253 of \cite{Kuratowski} that the map sets up a Borel isomorphism. Moreover the function
$(\xi , \eta) \mapsto (f'(\xi , \eta) , \upsilon)$ is constant on left $G_4$-cosets in $G_3$, as an easy
computation shows that $(f'((\xi , \eta)\omega_0) , \upsilon)
= (f'(\xi , \eta) , \upsilon)$ for all $\omega_0 = (\xi_0 ,{x_c}^{-1}{\xi_0}^{-1}x_c ) \in G_4$. 
Therefore $(\xi , \eta) \mapsto (f'(\xi , \eta) , \upsilon)$ defines a function on $G_3 / G_4$
which by Lemma 1.2 of \cite{Mackey} must be Borel because $(\xi , \eta) \mapsto (f'(\xi , \eta) , \upsilon)$ itself is Borel on $G_3$. That $(f_x , \upsilon)$ is a Borel function of $x \in G_1 x_c G_2$ now follows from the fact 
that the mapping of $G_3 / G_4$ onto $G_1 x_c G_2$ is a Borel isomorphism and preserves Borel sets. Finally observe that
$\tilde{f} = g$. Therefore $f \mapsto \tilde{f}$ is a one-to-one map of functions satisfying 
(i) and (ii) of the definition of ${\mathcal{H}^{L}_{C}}'$ onto Borel functions satisfying (\ref{g_2_induced}).
Consider the mapping $t \mapsto \pi(x_c t)$ of $G_2$ onto $C$. It defines one-to-one and Borel set preserving map $\psi$ from $G_2 / (G_2 \cap \, ({x_c}^{-1}G_1 x_c))$ onto $C$ and such that if $[t] = \pi'(t)$ and $[z] = \pi(z)$ 
correspond under the map $\psi$ and 
$\eta \in G_2$ then $[x]\eta$ and $[z]\eta$ do also ($\pi'$ stands for the canonical projection 
$G_2 / (G_2 \cap \, ({x_c}^{-1}G_1 x_c)) \mapsto G_2$). Finally $z \mapsto (\mathfrak{J}_L f_z , f_z)$ and 
$t \mapsto (\mathfrak{J}_{L} \tilde{f}_t , \tilde{f}_t)$ define functions
$\pi(z) \mapsto (\mathfrak{J}_L f_{\pi(z)} , f_{\pi(z)})$ and 
$\pi'(t) \mapsto (\mathfrak{J}_{L} \tilde{f}_{\pi'(t)} , \tilde{f}_{\pi'(t)})$ on $C$ and 
$G_2 / (G_2 \cap \, ({x_c}^{-1}G_1 x_c))$ respectively which correspond under the same map $\psi$:
$(\mathfrak{J}_L f_{\psi(\pi'(t))} , f_{\psi(\pi'(t))}) = (\mathfrak{J}_L f_{\pi(x_c t)} , f_{\pi(x_c t)})
= (\mathfrak{J}_L f_{x_c t} , f_{x_c t}) = (\mathfrak{J}_{L} \tilde{f}_{\pi'(t)} , \tilde{f}_{\pi'(t)})$.  

If we use this same map $\psi$ to transfer the measure $\mu_C$ on $C$ over to the homogeneous space
$G_2 / (G_2 \cap \, ({x_c}^{-1}G_1 x_c))$ we will get a quasi invariant measure $\mu^{x_c}$ there such that 
\[
\begin{split}
\int \limits_{C} \, (\mathfrak{J}_L f_z , f_z) \,\, \ud \mu_C ([z]) = 
\int \limits_{C} \, (\mathfrak{J}_L f_{[z]} , f_{[z]}) \,\, \ud \mu_C ([z]) \\
\int \limits_{C} \, (\mathfrak{J}_L f_{\psi([t])} , f_{\psi([t])}) \,\, \ud \mu_C (\psi([t])) 
= \int \limits_{G_2 / (G_2 \cap \, ({x_c}^{-1}G_1 x_c))} \, 
(\mathfrak{J}_{L} \tilde{f}_{[t]} , \tilde{f}_{[t]}) \,\, \ud \mu^{x_c}([t]) \\
= \int \limits_{G_2 / (G_2 \cap \, ({x_c}^{-1}G_1 x_c))} \, 
(\mathfrak{J}_{L} \tilde{f}_{t} , \tilde{f}_{t}) \,\, \ud \mu^{x_c}([t]).
\end{split} 
\]
Thus by the polarization identity (compare e. g. \cite{Segal_Kunze}, \S 8.3, page 222 or \cite{Bog}, page 4) the map $f \mapsto \tilde{f}$ sets up the Krein-unitary transformation $V_	{x_c}$ demanded by the Lemma as the verification 
of $V_{x_c} U^{L, C}_{s} V_{x_c}^{-1} = {}^{\mu^{x_c}}U^{L^{x_c}}_s$, $s \in G_2$, and 
$V_{x_c}\mathfrak{J}^{L,C} V_{x_c}^{-1}  = \mathfrak{J}_{x_c}$ is almost immediate
as $V_{x_c}$ is bounded, which we show below in Lemma \ref{lem:subgroup.preliminaries.2}. 
Similarly verification that $\mathfrak{J}_{x_c} \mathfrak{J}_{x_c} = I$ and that $\mathfrak{J}_{x_c}$ is self adjoint
with respect to the definite inner product    
\begin{equation}\label{g_2'_def_inn}
(\tilde{f}, \tilde{g})_{x_0} =  \int \limits_{G_2 / (G_2 \cap \, ({x_c}^{-1}G_1 x_c))} \, 
\Big(\mathfrak{J}_L (\mathfrak{J}_{x_0} \tilde{f}_t), \tilde{g}_t \Big) \,\, \ud \mu^{x_c}([t])
\end{equation}
in the Hilbert space ${}^{\mu^{x_c}}\mathcal{H}^{L^{x_c}}$, is likewise immediate. 
\qed

\vspace*{0.5cm}

Note that in general the norm and topology induced by the inner product (\ref{g_2'_def_inn}) defined by 
$\mathfrak{J}_{x_c}$ is not equivalent to the norm 
\[
\| \tilde{f} \|^2 = (\tilde{f}, \tilde{f}) =  \int \limits_{G_2 \big{/} \big(G_2 \cap \, ({x_c}^{-1}G_1 x_c) \big)} \, 
\Big(\mathfrak{J}_L (\mathfrak{J}^{L^{x_c}} \tilde{f}_t), \tilde{f}_t \Big) \,\, \ud \mu^{x_c}([t])
\]
and topology defined by the ordinary fundamental symmetry $\mathfrak{J}^{L^{x_c}}$
of Sect. \ref{def_ind_krein} (of course with $\mathfrak{G}$ and $H$ replaced with $G_2$ and 
$G_2 \cap \, ({x_c}^{-1}G_1 x_c)$): 
\[
\mathfrak{J}^{L^{x_c}} \tilde{f}_t = L^{x_c}_{h_{x_c}(t)} \mathfrak{J}_{L} L^{x_c}_{h_{x_c}(t)^{-1}} \, \tilde{f}_t ,
\] 
where $h_{x_c}(t) \in G_2 \cap \, ({x_c}^{-1}G_1 x_c)$ is defined as in Remark \ref{rem:def_ind_krein.1}
by a regular Borel section $B_{x_c}$ of $G_2$ with respect to the subgroup $G_2 \cap \, ({x_c}^{-1}G_1 x_c)$.
However if for each $t \in G_2$, $h(x_c t) \in G_{x_c}$, then the two topologies coincide. Similarly whenever the homogeneous space $G_2 \big{/} \big(G_2 \cap \, ({x_c}^{-1}G_1 x_c) \big)$
is compact then the two topologies coincide (but this case is not interesting). 

\vspace*{0.5cm}

\begin{lem}
The operators $V_{x_c}$ of the preceding Lemma are also isometric  with respect to the norms 
$\| \cdot \|_C$ in ${\mathcal{H}^{L}_{C}}'$ and $\| \cdot \|_{x_c} = \sqrt{(\cdot , \cdot)_{x_c}}$ in 
${}^{\mu^{x_c}}\mathcal{H}^{L^{x_c}}$, where $(\cdot, \cdot)_{x_c}$ is defined as by (\ref{g_2'_def_inn}), 
giving the norm in ${}^{\mu_{x_c}}\mathcal{H}^{L^{x_c}}$ 
induced by $\mathfrak{J}_{x_c}$. In particular we have $\| V_{x_c} \| = 1$ for all $x_c$.
\label{lem:subgroup.preliminaries.2}
\end{lem}

\qedsymbol \,
Denote the subgroup $G_2 \cap \, ({x_c}^{-1}G_1 x_c)$ by $G_{x_c}$. 
The Lemma is an immediate consequence of definitions of $\| \cdot \|_C$, $V_{x_c}$ 
and (\ref{g_2'_def_inn}) giving the norm $\| \cdot \|$ in ${}^{\mu^{x_c}}\mathcal{H}^{L^{x_c}}$:
\begin{equation}\label{norm_tilde}
\begin{split}
{\| V_{x_c}f \|_{{x_c}}}^2 = (\tilde{f}, \tilde{f})_{x_c} =  \int \limits_{G_2 / G_{x_c}} \, 
\big( \mathfrak{J}_L (\mathfrak{J}_{x_c} \tilde{f})_t , \tilde{f}_t \big) \,\, \ud \mu^{x_c}([t]) \\ 
= \int \limits_{G_2 / G_{x_c}} \, \big( \mathfrak{J}_L (V_{x_c}^{-1} \mathfrak{J}^{L,C} 
V_{x_c} V_{x_c}^{-1} f)_t , (V_{x_c}^{-1}f)_t \big) \,\, \ud \mu^{x_c}([t])  \\
=  \int \limits_{G_2 / G_{x_c}} \, \big( \mathfrak{J}_L (V_{x_c}^{-1} \mathfrak{J}^{L,C} 
 f)_t , (V_{x_c}^{-1}f)_t \big) \,\, \ud \mu^{x_c}([t])  
\end{split}
\end{equation}
and because $V_{x_c}$ is Krein-unitary, i. e. isometric for the Krein inner products
\[
\int \limits_{C} \, \big( \mathfrak{J}_L (\cdot)_z , (\cdot)_z \big) \,\, \ud \mu_C ([z])
\,\,\,\textrm{and} \,\,\,
\int \limits_{G_2 / G_{x_c}} \, \big( \mathfrak{J}_L (\cdot)_t , (\cdot)_t \big) \,\, \ud \mu^{x_c}([t]), 
\]
the last integral in (\ref{norm_tilde}) is equal to
\[
\int \limits_{C} \, \big( \mathfrak{J}_L (\mathfrak{J}^{L,C} f)_z , f_z \big) \,\, \ud \mu_C ([z]) 
= \| f \|_{C}^{2}.
\]
\qed

\vspace*{0.5cm}

Note, please, that the Lemmas of Sect. \ref{dense}, i. e. Lemmas \ref{lem:dense.1} -- \ref{lem:dense.6},
are equally applicable to the Krein space $(\mathcal{H}^{L^{x_c}}, \mathfrak{J}_{x_c})$,
with $\mathfrak{J}^{L^{x_c}}$ replaced by $\mathfrak{J}_{x_c}$, and with the section $B_{x_c}$
replaced with the image of $G_2 \big{/} \big(G_2 \cap \, ({x_c}^{-1}G_1 x_c)\big)$ under the inverse of the map
$t \mapsto x_c t$. We formulate this remark as a separate 

\begin{lem}
The Lemmas \ref{lem:dense.1} -- \ref{lem:dense.6} are true for the Hilbert space $\mathcal{H}^{L^{x_c}}$
of the Krein space  $(\mathcal{H}^{L^{x_c}}, \mathfrak{J}_{x_c})$, i.  e. with $L$ replaced by  $L^{x_c}$,
$\mathfrak{J}_L$ replaced by $\mathfrak{J}_{L^{x_c}}= \mathfrak{J}_L$,
$\mathcal{H}_L$replaced with $\mathcal{H}_{L^{x_c}} = \mathcal{H}_L$, $\mathfrak{J}^{L} = \mathfrak{J}^{L^{x_c}}$ replaced by $\mathfrak{J}_{x_c}$ and finally with the section $B_{x_c}$ replaced with the image of 
$G_2 \big{/} \big(G_2 \cap \, ({x_c}^{-1}G_1 x_c)\big)$ under the inverse of the map
$t \mapsto x_c t$.   
\label{lem:subgroup.preliminaries.3}
\end{lem} 
\qedsymbol \, The proofs remain unchanged.
\qed

\vspace*{0.5cm}

In Subsection \ref{subgroup} we explain why we are using $\mathfrak{J}_{x_c}$
in ${^{\mu^{x_c}}}\mathcal{H}^{L^{x_c}}$  instead of $\mathfrak{J}^{L^{x_c}}$.

\subsection{Decomposition (disintegration) of measures}\label{decomposition}

In this section we present a decomposition theorem for non finite measures. Although by Thm. 
\ref{def_ind_krein:twr.3} we could, after Mackey, restrict ourselves to finite measures in
the analysis of tensor products of induced representations, we insist to stay with
induced representations connected with natural infinite measures encountered in physics,
in order to avoid computation of the Clebsh-Gordan coefficients in latter stages of computations.

Let $\mathfrak{G}$, $G_1$ and $G_2$ be such as in Sect. \ref{subgroup.preliminaries}. 
Because the base of the system of neighbourhoods of unity in $\mathfrak{G}$ is countable, the uniform space 
$\mathfrak{X} = \mathfrak{G}/G_1$ is metrizable (compare e. g. \cite{Weil}, \S 2) for any closed subgroup 
$G_1 \subset \mathfrak{G}$. The right action of $G_1$ on $\mathfrak{G}$ is proper and the quotient map
$\pi: \mathfrak{G} \mapsto \mathfrak{G}/G_1$ is open, so that the space $\mathfrak{X} = \mathfrak{G}/G_1$ of 
right $G_1$ orbits ($G_1$ cosets) automatically has the required regularity: measurability of the equivalence
relation defined by the $G_1$ orbits. In particular the quotient 
space $\mathfrak{X}$ is Hausdorff, separable and locally compact and the measure $\rho \cdot \mu_0$ 
(with the $\rho$-function of Sect. \ref{def_ind_krein} and right Haar measure $\mu_0$ on $\mathfrak{G}$)
is decomposable into a direct integral of measures $\rho \cdot \mu_0 = \int \limits_{\mathfrak{G}/G_1} \, \beta_{[x]} 
\, \ud \mu ([x])$ with the component measures 
$\beta_{[x]}$ of the decomposition concentrated in the $G_1$ orbit (right coset) $[x]$
and with Radon-Nikodym derivative associated with the action of the subgroup $G_1$ (i. e. $\lambda_{[x]}$-function) corresponding to $\beta_{[x]}$ equal to the
restriction to the orbit $[x]$ and to the subgroup $G_1$ of the Radon-Nikodym (i. e. $\lambda$-function) corresponding to the measure  $\rho \cdot \mu_0$. This in particular gives us the quasi invariant regular Baire (or Borel) measure 
$\mu = \mu_{\mathfrak{G}/G_1}$ on the uniform space $\mathfrak{X}$ corresponding to $\rho$,
i. e. the factor measure of $\rho \cdot \mu_0$ (Mackey's method of constructing 
general regular quasi invariant measure on the quotient space $\mathfrak{X} = \mathfrak{G}/G_1$).  

This is not the case if we replace $\mathfrak{G}$ with $\mathfrak{X} = \mathfrak{G}/G_1$ acted on by a second closed subgroup $G_2 \subset \mathfrak{G}$. The quotient space $\mathfrak{X}/G_2$ 
is in general a badly behaved non Hausdorff space with non measurable equivalence relation defined in $\mathfrak{X}$
with the $G_2$ orbits as equivalence classes. We require a regularity condition in order to achieve an effective
tool for constructing effectively a dual of the group
$\mathfrak{G}$ in question with the help of decomposition of tensor product of induced representations.  

Let $\mathfrak{X}$, for example $\mathfrak{X} = \mathfrak{G}/G_1$, be any separable locally compact metrizable space
with an equivalence relation $R$ in $\mathfrak{X}$, for example with the equivalence classes given by right $G_2$-orbits in
$\mathfrak{X} = \mathfrak{G}/G_1$ under the right action of a second closed subgroup $G_2 \subset \mathfrak{G}$.  
Let the equivalence classes form a set $\mathfrak{C}$ and for each

$\mathfrak{x} \in \mathfrak{X}$ let $\pi_{\mathfrak{X}}(\mathfrak{x}) \in \mathfrak{C}$ denote the equivalence class of 
$\mathfrak{x}$. Let $\mathfrak{X}$ be endowed with a regular measure $\mu$ (quasi invariant in case 
$\mathfrak{X} = \mathfrak{G}/G_1$).
We define following \cite{Rohlin} the relation $R$ to be measurable\footnote{Strictly speaking in Rohlin's definition
of measurability of $R$, accepted by Mackey in \cite{Mackey}, the set $E_0$ is empty and $\pi_{\mathfrak{X}}^{-1}(E_k )$, $k\geq 1$, are just $\mu$-measurable
and not necessary Borel. But the difference is unessential as we explain below in this Sect..} 
if there exists a countable family 
$E_0 , E_1 , E_2 , \ldots$ of subsets of $\mathfrak{C}$ such that $\pi_{\mathfrak{X}}^{-1}(E_i)$ is a Baire (or Borel) set for each $i$ and such that $\mu (\pi_{\mathfrak{X}}^{-1}(E_0 )) = 0$, and such that each point $C$ of $\mathfrak{C}$ not belonging to $E_0$ is the intersection of the $E_i$ which contain it. Under this assumption of measurability $\mu$ may be decomposed (disintegrated) as an integral $\mu = \int \limits_{\mathfrak{C}} \, \mu_{C} 
\, d\nu(C)$ over $\mathfrak{C}$ of measures $\mu_C$, with each $\mu_C$ concentrated on the corresponding equivalence class $C$, i .e $G_2$ orbit in case $\mathfrak{X}= \mathfrak{G}/G_1$, with a regular measure $\nu = \mu_{\mathfrak{X}/G_2}$ on 
$\mathfrak{C} = \mathfrak{X}/G_2$ i. e. the factor measure of $\mu$, which we may call the ``double factor measure'' $\mu_{(\mathfrak{G}/G_1)/G_2}$ of $\mu_0 = \mu_\mathfrak{G}$ in case  $\mathfrak{X}= \mathfrak{G}/G_1$; 
and moreover in this case  when $\mathfrak{X}= \mathfrak{G}/G_1$ the Radon-Nikodym derivative (i. e. $\lambda_C$-function) corresponding to $\mu_C$ and associated with action of the subgroup $G_2$ is equal to the restriction to the orbit $C$
and to the subgroup $G_2$ of the Radon Nikodym derivative ($\lambda$-function) corresponding to $\mu$.
In this case we say after Mackey that the subgroups $G_1$ and $G_2$ are \emph{regularly related}. In short:
the orbits in $\mathfrak{G}/G_1$ under the right action of $G_2$  form the equivalence classes of a measurable equivalence relation\footnote{Using literally Rohlin's definition of measurability: almost all of the orbits in $\mathfrak{G}/G_1$ under the right action of $G_2$  form the equivalence classes of a Rohlin-measurable equivalence relation.}.    

Let us explain the meaning of the regularity condition. Even if $G_1$ and $G_2$ were not regularly related we could
of course find a countable set $E_1 , E_2 , \ldots$ of Borel unions of orbits which generate the $\sigma$-ring
of all measurable unions of orbits. The unique equivalence relation $R$ such that 
$\mathfrak{x} \in \mathfrak{G}/G_1$ and $\mathfrak{y} \in \mathfrak{G}/G_1$ are in the relation 
whenever $\mathfrak{x}$ and $\mathfrak{y}$ are in the same sets $E_j$ will be measurable. This equivalence relation
gives us a decomposition of the quasi invariant measure $\mu$ into quasi invariant component measures 
$\mu_P$ concentrated on subsets $P \subset \mathfrak{C}$, but in this general non regular situation
the subsets $P$ are unions of many orbits $C \in \mathfrak{C}$. This would give us decomposition of $U^L$ 
restricted to $G_2$, but in this decomposition the component representations will not be associated with single 
orbits, i. e. with single double cosets $G_1 x_0 G_2$ and will not be identifiable as 
``induced representations''\footnote{In fact the representations $U^{L^{x_c}}$ of Lemma 
\ref{lem:subgroup.preliminaries.1} of Sect. \ref{subgroup.preliminaries} do not have the standard form of induced representations defined in Sect. \ref{def_ind_krein} as $\mathfrak{J}_{x_c} \neq \mathfrak{J}^{L^{x_c}}$, but in relevant cases of representations encountered in QFT they may be shown to be Krein-unitary equivalent to standard induced representations (in the sense of Sect. \ref{def_ind_krein}).
Anyway they are concentrated in single orbits.}
$U^{L^{c_c}}$ of 
$G_2$ of Lemma \ref{lem:subgroup.preliminaries.1} of Sect. \ref{subgroup.preliminaries}. Little or nothing is 
known of such component representations related to non transitive systems of imprimitivity. In fact the regularity
of the $G_2$-orbits in $\mathfrak{G}/G_1$ is essentially equivalent\footnote{One may characterise the space of orbits by considering the respective group algebra or the associated universal enveloping $C^*$algebra. Connes developed
a general theory of cross-product $C^*$-algebras and von Neumann algebras associated with foliations,
strongly motivated by the Mackey theory of induced representations, compare \cite{Connes} and references
there in.} for the group $\mathfrak{G}$
to be of type I. Because of the bi-unique correspondence between $G_2$ orbits in $\mathfrak{G}/G_1$ 
and double cosets $G_1 x G_2$ in $\mathfrak{G}$, and because of the relation between Borel structures
on $\mathfrak{X} = \mathfrak{G}/G_1$ and on $\mathfrak{X}/G_2$, we may reformulate the regularity condition as follows.
We assume that there exists a sequence $E_0 , E_1 , E_2 , \ldots$ of measurable subsets of $\mathfrak{G}$
each of which is a union of double cosets such that $E_0$ has Haar measure zero and each double coset not 
in $E_0$ is the intersection of the $E_j$ which contain it (compare Lemma \ref{lem:decomposition.10}).

\vspace*{0.5cm}

\begin{ex}

The equivalence relation on the two-torus $\mathfrak{X} = \mathbb{R}^2 / \mathbb{Z}^2$ given by the 
leaves of the Kronecker foliation associated to an irrational number $\theta$, i. e. given by the differential
equation
\[
dy = \theta dx,
\] 
is not measurable. The leaves, i. e. equivalence classes, can be viewed as orbits of the additive group $\mathbb{R}$ on the two-torus
$\mathfrak{X} = \mathbb{R}^2 / \mathbb{Z}^2$.  
\label{ex:decomposition.1}
\end{ex}

\vspace*{0.5cm}

In the original Mackey's theory the induced representations ${}^{\mu}U^L$ and ${}^{\mu'}U^L$
are unitary equivalent (in our case unitary and Krein-unitary equivalent) whenever the quasi 
invariant measures $\mu$ and $\mu'$ on 
$\mathfrak{G}/G_1$ are equivalent, which is always the case, as all such measures are equivalent. 
We could assume all measures $\mu$ in the induced representations ${}^{\mu}U^L$ to be finite without
any lost of generality. In particular (and this simplifies matter if we are interested in
computing decompositions of tensor products up to unitary and Krein-unitary equivalence) we may restrict ourself to finite 
measures $\mu$ on $\mathfrak{G}/G_1$, as Mackey did in \cite{Mackey},  in constructing decomposition (disintegration) $\mu = \int \limits_{\mathfrak{C}} \, \mu_{C} \, \ud \nu(C)$ with each of the measures 
$\mu_{C}$  concentrated on the corresponding orbit $C$ and the corresponding Radon-Nikodym derivative associated with $\mu_{C}$ under the action of $G_2$ equal to the restriction to the orbit $C$ and to the subgroup $G_2$ 
of the Radon Nikodym derivative associated with $\mu$ (this is proved in \S 11 of \cite{Mackey}). 
because our aim is to reduce computations and because we are interesting in tensor product decompositions themselves 
(not only up to unitary and Krein unitary equivalence) we insist in stayng with the original infinite measures  
$\mu$ in construction of the decomposition $\mu = \int \limits_{\mathfrak{C}} \, \mu_{C} \, \ud \nu(C)$
with the above mentioned properties. Because Mackey's construction of decomposition of finite measure 
$\mu$ is sufficient for the theory of unitary group representations (as well as 
for the extension of the construction of induced representation to representations of $C^*$-algebras 
along the lines proposed by Rieffel) decomposition having the above mentioned properties of a quasi invariant measure 
$\mu$ which is not finite has not been constructed explicitly in the classical mathematical literature, 
at least the author was not able to find it
(in the Bourbaki's course on integration \cite{Bourbaki}, Chap. 7.2.1-7.2.3  
decomposition of this type is constructed but under stronger assumption than measurability of the equivalence relation given by right $G_2$ action on $\mathfrak{X} = \mathfrak{G}/G_1$
where it is assumed instead that the action is proper and moreover where it is assumed that the measure 
$\mu$ is relatively invariant and not merely quasi invariant -- assumptions too strong for us).
Because the required decomposition of not necessary finite quasi invariant measure $\mu$ on 
$\mathfrak{G}/G_1$ is important for the decomposition of the restriction of the induced representation 
${}^{\mu}U^L$ in a Krein space to a closed subgroup (and \emph{a fortiori} to a decomposition
of tensor product of induced representations
${}^{\mu}U^L$ and ${}^{\mu}U^M$ in Krein spaces) we present here its construction explicitly
only for the sake of completeness. 
The construction presented here uses a localization procedure in reducing the problem 
of decomposition to the Mackey-Godement decomposition (\cite{Mackey}, \S 11) of a finite quasi invariant measure.

\vspace*{0.2cm}

\begin{scriptsize}

Whenever the action of $G_2$ on $\mathfrak{X} = 
\mathfrak{G}/G_1$ is proper one can just replace the continuous homomorphism $\chi: G_2 \mapsto \mathbb{R}_+$ in \cite{Bourbaki}, Chap. 7.2.1-7.2.3, 
by the Radon-Nikodym derivative associated with the measure $\mu$ on $\mathfrak{X} = \mathfrak{G}/G_1$ in this case. Using the Federer and Morse theorem \cite{Federer_Morse} one constructs a regular Borel section of $\mathfrak{X}$ with respect to $G_2$ which enables the construction of the factor measure $\nu$ on the quotient $\mathfrak{C} = \mathfrak{X}/G_2$ of the space $\mathfrak{X}$ by the group $G_2$ with the method of \cite{Bourbaki} changed in  minor points only.

\end{scriptsize}

\vspace*{0.2cm}

Let $\mathfrak{X}$ be the separable locally compact metrizable (in fact complete metric) space 
$\mathfrak{G}/G_1$ equipped with a regular 
quasi invariant measure $\mu$. Let $R$ be the equivalence relation 
in $\mathfrak{X}$ given by the right action of a second closed subgroup $G_2$ with the associated quotient map 
$\pi_\mathfrak{X} : \mathfrak{X} \mapsto \mathfrak{X}/R = \mathfrak{X}/G_2$, 
and let $K$ be a compact subset of $\mathfrak{X}$. There is canonically defined equivalence 
relation $R_K$ on $K$ induced by $R$ on $K$ with the associated quotient map 
$\pi_K : K \mapsto K/R_K$ equal to the restriction of $\pi_\mathfrak{X}$ to the subset $K$.

Note please that for an equivalence relation $R$ in the separable locally compact and metrizable space 
$\mathfrak{X} = \mathfrak{G}/G_1$ the above mentioned (Rohlin's \cite{Rohlin}) condition of measurability of $R$ is equivalent to the following condition: the family $\mathfrak{K}$ of those compact sets $K \subset \mathfrak{X}$ 
for which the quotient space $K/R_K$ is Hausdorff is $\mu$-dense, i. e. one of the following and equivalent conditions is fulfilled:  
\begin{enumerate}

\item[(I)]  For a subset $A \subset \mathfrak{X}$ to be locally $\mu$-negligible it is necessary and 

            sufficient that $\mu(A \cap K) = 0$ for all $K \in \mathfrak{K}$.

\item[(II)]  For any compact subset $K_0$ of $\mathfrak{X}$ and for any $\epsilon > 0$ there exists a subset
           $K \in \mathfrak{K}$ contained in $K_0$ and such that $\mu(K_0 - K) \leq \epsilon$.

\item[(III)] For each compact subset $A$ of $\mathfrak{X}$ there exists a partition of $A$ into 
             a $\mu$-negligible subset $N$ and a sequence $\{ K_n \}_{n \in \mathbb{N}}$ of compact subsets 
             belonging to $\mathfrak{K}$.

\item[(IV)]  For each compact subset $K$ of $\mathfrak{X}$ there exists an increasing 
             $H_1 \subseteq H_2 \subseteq \ldots$ sequence $\{ H_n \}_{n \in \mathbb{N}}$ of compact 
             sets belonging to $\mathfrak{K}$ contained in $K$ and such that the set 
             $Z = K - \bigcup \limits_{n \in \mathbb{N}} H_n$ is $\mu$-negligible.

\end{enumerate}

Indeed, because the the system of neighbourhoods of unity in $\mathfrak{G}$ is countable, the uniform space 
$\mathfrak{X} = \mathfrak{G}/G_1$ is completely metrizable and locally compact (compare e. g. \cite{Weil}, \S 2) for any closed subgroup $G_1 \subset \mathfrak{G}$. Therefore Proposition 3 of \cite{Bourbaki_i}, Chap. VI, \S 3.4, is
applicable. By this Proposition we need only show that using the family $\mathfrak{K}$ one can
construct the sets $E_0, E_1 , \ldots$ of the Rohlin's measurability condition of $R$, for which 
$\pi_{\mathfrak{X}}^{-1}(E_k )$, $k\geq 1$, are not only $\mu$-measurable but moreover Borel. This however follows
from the fact that $\mathfrak{X}$ is countable at infinity: there exists a sequence of compact subsets 
$K_1 \subset K_2 \subset \ldots$ of $\mathfrak{X}$ such that $\mathfrak{X} = \cup_i K_i$ and moreover we may assume that 
they are regular closed sets: $\cl \intt K_m = K_m$.   

Indeed, let $\{ \mathcal{O}_k \}_{k \in \mathbb{N}}$ be a countable base of the topology in $\mathfrak{X}$,
such that the closure $\overline{\mathcal{O}_k }$ of each $\mathcal{O}_k$ is compact (there exists such a base because
$\mathfrak{X}$ is second countable and locally compact). For each $\overline{\mathcal{O}_k }$ choose
a sequence $\{K_{kl}\}_{l \in \mathbb{N}}$ of compact sets belonging to $\mathfrak{K}$ and a $\mu$-negligible
subset $M_k$ giving the partition $\overline{\mathcal{O}_k } = M_k \dot{\cup} K_{k1} \dot{\cup} K_{k2} \dot{\cup} \ldots$
of $\overline{\mathcal{O}_k }$, existence of which is assured by the condition (III). 
Define the $\mu$-negligible set $M = \cup_k M_k$ and a maximal subset $M_0$ of $M$ invariant under the action of $G_2$
on $\mathfrak{X}$. 

By the condition (IV) we can construct for each $K_{m}$ 
a sequence $H_{m1} \subset H_{m2} \subset H_{m3} \subset \ldots $ of compact subsets of $K_m$ belonging to
$\mathfrak{K}$ and a $\mu$-negligible subset $Z_m$ such that $K_{m} = Z_m 
\dot{\cup} \big( \cup_n K_{mn}\big)$. Define the $\mu$-negligible set $Z = \bigcup \limits_{n \in \mathbb{N}} Z_m$
and the maximal subset $Z_0$ of $Z$ invariant under the action of $G_2$.

 Let us define a countable family of sets $E_0 = \pi_\mathfrak{X}(Z_0 \cup M_0)$, $E_{mn} = 
\pi_\mathfrak{X}(K_{mn}) = \pi_{K_{mn}}(K_{mn})
= K_{mn} / R_{K_{mn}}$, $m,n \in \mathbb{N}$ in $\mathfrak{X}/G_2 = \mathfrak{X}/R$, where 
$K_{mn} / R_{K_{mn}}$ is Hausdorff by assumption. 

Now let $\mathfrak{x}_1$ and $\mathfrak{x}_2$
be two elements of $\mathfrak{X}$ not in $N_0 = Z_0 \cup M_0$ such that  $\pi_\mathfrak{X}(\mathfrak{x}_1 ) \neq 
\pi_\mathfrak{X}(\mathfrak{x}_2 )$. Then by construction there exists $H_{mn} \in \mathfrak{K}$ containing  
$\mathfrak{x'}_1$ and $\mathfrak{x'}_2$ with $\pi_\mathfrak{X}(\mathfrak{x'}_1 ) = \pi_\mathfrak{X}(\mathfrak{x}_1 )$ and 
$\pi_\mathfrak{X}(\mathfrak{x'}_2 ) = \pi_\mathfrak{X}(\mathfrak{x}_2 )$. 

$H_{mn} / R_{H_{mn}}$ containing $\pi_\mathfrak{X}(\mathfrak{x}_1 ) = \pi_\mathfrak{X}(\mathfrak{x'}_1 )$ and $\pi_\mathfrak{X}(\mathfrak{x}_2 ) = \pi_\mathfrak{X}(\mathfrak{x'}_2 )$ is Hausdorff by construction. Thus there exist two compact non intersecting neighbourhoods $\overline{\mathcal{O}}_{\mathfrak{x'}_1}$ 
and $\overline{\mathcal{O}}_{\mathfrak{x'}_2}$ of $\mathfrak{x'}_1$ and $\mathfrak{x'}_2$ respectively
such that for $K_{\mathfrak{x'}_1} = \overline{\mathcal{O}}_{\mathfrak{x'}_1} \cap H_{mn}$
and $K_{\mathfrak{x'}_2} = \overline{\mathcal{O}}_{\mathfrak{x'}_2} \cap H_{mn}$
we have $\pi_\mathfrak{X}^{-1} (K_{\mathfrak{x'}_1}) \cap \pi_\mathfrak{X}^{-1} (K_{\mathfrak{x'}_2}) = \emptyset$.
By construction we may choose  $K_{{m_1}{n_1}} \subset K_{\mathfrak{x}_1}$ and $K_{{m_2}{n_2}} \subset K_{\mathfrak{x}_2}$ in $\mathfrak{K}$ such that 
$\mathfrak{x'}_1 \in K_{{m_1}{n_1}}$  and $\mathfrak{x'}_2 \in K_{{m_2}{n_2}}$. Of course we have 
$E_{{m_1}{n_1}} \cap E_{{m_2}{n_2}} = \pi_\mathfrak{X}^{-1} (K_{{m_1}{n_1}}) \cap 
\pi_\mathfrak{X}^{-1} (K_{{m_2}{n_2}}) = \emptyset$. Thus the intersection of all $E_{mn} \in \mathfrak{K}$ containing
$\pi_\mathfrak{X}(\mathfrak{x}_1 ) \in \mathfrak{X}/G_2$ is equal $\{ \pi_\mathfrak{X}(\mathfrak{x}_1 ) \}$.
We have to show that $\pi_\mathfrak{X}^{-1}(E_{mn}) = \pi_\mathfrak{X}^{-1}(\pi_\mathfrak{X}(K_{mn}))$
are Baire (or Borel) sets. To this end observe please that $\pi_\mathfrak{X}^{-1}(\pi_\mathfrak{X}(K_{mn}))$
is equal to the saturation of $K_{mn}$, i. e. $\pi_\mathfrak{X}^{-1}(\pi_\mathfrak{X}(K_{mn}))
= K_{mn} \cdot G_2$. Choose a compact neighbourhood $V$ of the unit in $G_2$ such that $V = V^{-1}$. Then
if $G_2$ is connected then $G_2 = \bigcup \limits_{n \in \mathbb{N}} V^n$; if $G_2$ is not connected
then it is still a countable sum of connected components of the form $\bigcup \limits_{n \in \mathbb{N}} V^n \eta_m$, 
with $\eta_m \in G_2$  chosen from $m$-th connected component $G_{2m}$ of $G_2$. Thus in each case $G_2$ is a countable sum
$\bigcup \limits_{k,l \in \mathbb{N}} V_{kl}$ of compact sets $V_{kl}$. Therefore
$\pi_\mathfrak{X}^{-1}(E_{mn}) = K_{mn} \cdot G_2 = \bigcup \limits_{k,l \in \mathbb{N}} K_{mn} \cdot V_{kl}$      
being a countable sum of compact sets is contained in the $\sigma$-ring generated by the compact sets and all the more 
it is a Borel set contained in the $\sigma$-ring generated by the closed sets. Thus both definitions of measurability
of the equivalence relation $R$ on $\mathfrak{X}$ are equivalent.

\vspace*{0.5cm}

\begin{lem}
There exists a Borel set $B_0$ in $\mathfrak{X} = \mathfrak{G}/G_1$ 
and a $\mu$-negligible subset $N_0 \subset \mathfrak{X}$ consisting of $G_2$ orbits in 
$\mathfrak{X} = \mathfrak{G}/G_1$
such that $B_0$ intersects each $G_2$ orbit not contained in $N_0$ in exactly one point.

\label{lem:decomposition.1}
\end{lem}

\qedsymbol \,

For the proof compare e. g. \cite{Bourbaki_i}, Chap. VI, \S 3.4, Thm. 3.
\qed

\vspace*{0.5cm}

Adding to $B_0$ any section of the $\mu$-negligible set $N_0$ we obtain a measurable section $B_{00}$
for the whole space $\mathfrak{X}$. For equivalence relations $R$ on smooth manifold $\mathfrak{X}$
defined by foliations on $\mathfrak{X}$ (i. e. smooth and integrable sub-bundles of $T\mathfrak{X}$)
existence of a measurable section is equivalent for the foliation to be of type I: i. e. the von Neumann algebra associated
to the foliation is of type I iff the foliation admits a Lebesgue measurable section, compare \cite{Connes}, 
Chap. I.4.$\gamma$, Proposition 5.

Because the Borel space $\mathfrak{X} = \mathfrak{G}/G_1$ is standard it follows by the second Theorem on page 74 of
\cite{Mackey2} that the quotient Borel structure on $\mathfrak{X}/G_2 - N_0 /G_2$ is likewise standard;
i. e. there exits a Borel isomorphism $\psi_0$: $(\mathfrak{X} - N_0)/G_2 \rightarrow S_0 \subset$ onto a 
Borel subset $S_0$  of a complete separable metric space $S$.

The space $(\mathfrak{X} - N_0 )/G_2$ however  need not be locally compact and it is not if the action of
$G_2$ on $\mathfrak{X} = \mathfrak{G}/G_1$ is not proper but only measurable, i.e. with measurable equivalence relation determined by the action of $G_2$. Similarly $G_2$-orbit $C$ in $\mathfrak{X}$ as a subset of a locally compact space
$\mathfrak{X}$ need not be closed if the action of $G_2$ is not proper and thus need not be locally compact with the 
topology induced from the surrounding space $\mathfrak{X}$.

\vspace*{0.5cm}

\begin{lem}

Let $N_0$ be as in Lemma \ref{lem:decomposition.1}. A necessary and sufficient condition that a 
subset $E$ of $\mathfrak{X}/G_2 - N_0 /G_2$ be a Borel set is that $\pi_{\mathfrak{X}}^{-1}(E)$
be a Borel set in $\mathfrak{X} - N_0$. A necessary and sufficient condition that a function 
$f$ on $\mathfrak{X}/G_2 - N_0 /G_2$ be a Borel function is that $f \circ \pi_{\mathfrak{X}}$
be a Borel function on $\mathfrak{X} - N_0$.  

\label{lem:decomposition.3}
\end{lem}

\qedsymbol \, Let $p_0$ be the Borel function $\psi_0 \circ  \pi_{\mathfrak{X}} : \mathfrak{X} - N_0
\rightarrow S_0$. Let $E'$ be any subset of $S_0$ such that $p_{0}^{-1}(E')$ is a Borel set. Let 
$B_0$ be the Borel section of $\mathfrak{X} - N_0$ with respect to $G_2$, existence of which has been
proved in Lemma \ref{lem:decomposition.1}. Then $p_0 (p_{0}^{-1}(E') \cap B_0) = E'$, and thus $E'$ is a Borel set
by Theorem 3, page 253 of \cite{Kuratowski}, compare likewise the Theorems on pages 72-73 of \cite{Mackey2}, 
because $p_0$ is one-to-one Borel function on $B_0$. 
Conversely: if $E'$ is Borel in $S_0$ then because $p_0$ is a Borel function, so is the set $p_{0}^{-1}(E')$. The first part of the Lemma  follows now from this and from definition 
of the Borel structure induced on $\psi_0 ((\mathfrak{X} - N_0 )/G_2)$ and \emph{a fortiori} on $\mathfrak{X}/G_2 - N_0 /G_2$. The remaining part of the Lemma is an immediate consequence of the first part.   
\qed

\vspace*{0.5cm}

We have the following disintegration theorem for the (not necessarily finite) measure $\mu$
and any of its pseudo image measures $\nu$ on $\mathfrak{X}/G_2$ (for definition of pseudo image measure $\nu$
compare e. g. \cite{Bourbaki_i}, Chap. VI.3.2):

\vspace*{0.5cm}

\begin{lem}
For each orbit $C = \pi_{\mathfrak{X}}^{-1}(d_0) \subset \mathfrak{X}$ with $d_0 \in \mathfrak{X}/G_2$ there exists a 
Borel measure $\mu_C$ in $\mathfrak{X}$ concentrated on the orbit $C$, i. e. $\mu_C (\mathfrak{X} - C)
= \mu_C (\mathfrak{X} - \pi_{\mathfrak{X}}^{-1}(d_0)) = 0$.
For any $g \in L^1 (\mathfrak{X} , \mu)$ 
the set of all those $G_2$ orbits $C$ for which $g$ is not 
$\mu_C$-integrable is $\nu$-negligible and the function 
\[
C \mapsto \int \, g(x) d\mu_C(x)
\]
is $\nu$-summable and $\nu$-measurable, and
\begin{equation}\label{dec_m}
\int \, d\nu(C) \, \int \, g(x) \, d\mu_C (x)  = \int \, g(x) \, d \mu(x).
\end{equation}
In short
\[
\mu = \int \, \mu_C (x) \, d\nu(C).
\]

\label{lem:decomposition.4}
\end{lem}

\begin{rem}
For each orbit $C$ the measure $\mu_C$ may also be naturally viewed as a measure on the 
 $\sigma$-ring $\mathscr{R}_C$ of measurable subsets of $C$ induced from the surrounding space $\mathfrak{X}$:
$E \in \mathscr{R}_C$ iff $E = E' \cap C$ for some $E' \in \mathscr{R}_{\mathfrak{X}}$, i. e. with the 
subspace Borel structure. 
\label{rem:decomposition.1}
\end{rem}

\qedsymbol \,
 For the proof we refer the reader e. g.  to \cite{Bourbaki_i}, Chap. VI, \S 3.5. 
\qed

\vspace*{0.5cm}

We shall show that for each $C$ the measure $\mu_C$ is quasi invariant and that for all $\eta \in G_2$ 
the Radon-Nikodym derivative $\lambda_C (\cdot , \eta) = 
\frac{\ud (R_\eta \mu_C)}{\ud \mu_C}(\cdot)$ is equal to the restriction of the Radon-Nikodym 
derivative  $\lambda (\cdot , \eta) = \frac{\ud (R_\eta \mu)}{\ud \mu}(\cdot)$
to the orbit $C$. In doing so we prefer reducing the problem to the
Mackey-Godement decomposition  of a finite measure (\cite{Mackey}, \S 11) 
using a localization of the measure space 
$(\mathfrak{X}, \mathscr{R}_{\mathfrak{X}}, \mu)$ and its disintegration.  
Toward this end we need some further Lemmas.   

\vspace*{0.5cm}

\begin{lem}
Let $\mu$, $\mu_C$ and $\nu$ be as in the preceding Lemma. Let $K$ be a compact subset of $\mathfrak{X}$.
Then $\pi_{\mathfrak{X}} (K)$ is measurable on $\mathfrak{X}/G_2$.

\label{lem:decomposition.5'}
\end{lem}

\qedsymbol \,
Let $K$ be any compact subset of $\mathfrak{X}$ and let $Z , K_{n}$ be the subsets of condition
(IV), i. e. $K_n \in \mathfrak{K}$ is an increasing sequence of compact subsets of $K$, and $Z$ is $\mu$-negligible subset of $K$ such that $K = Z \dot{\cup} \big( K_1 \cup K_2 \cup \ldots  \big)$. Let us define the subset (if any) $Z_0 \subset Z$
consisting of intersections of full  $G_2$-orbits with $K$, i. e.  the maximal subset of $Z$ invariant under the action of $G_2$ on $\mathfrak{X}$. 
\begin{center}
\begin{tikzpicture}

\path[fill=gray] (2,4) to [out=45,in=45] (4,2)

to [out=-135,in=-135] (2,4);

\draw[thin] (1,1.25) to [out=47.5,in=-92.5] (2.25,5);



\draw[thin] (1.5,1) to [out=40,in=-85] (3,5);


\draw[thin] (2,1) to [out=35,in=-80] (3.5,5);


\draw[thin] (2.5,1) to [out=30,in=-75] (4,5);


\draw[ultra thick] (2.19,4) to [out=-70,in=170] (2.91,3.25);

\draw[ultra thick] (3.445,1.82) to [out=50,in=-109] (4.05,2.99);

\draw[ultra thick] (2.99,1.99) to [out=50,in=-98] (3.545,3.6);

\draw [->,very thin] (1,4) to [out=-70,in=130] (2, 4);

\draw [->,very thin] (1,2) to [out=-70,in=130] (1.44, 1.8);

\draw [<-,very thin] (3.48,3) to [out=0,in=180] (5,4);

\draw [<-,very thin] (2.5,3.5) to [out=45,in=135] (5,4);

\draw [<-,very thin] (3.645,2) to [out=-45,in=-95] (5,4);

\draw [<-,very thin] (3.1,2.2) to [out=135,in=90] (2.5,2) to [out=-90,in=-135] (4,1);

\draw [<-,very thin] (3.5,1.95) to [out=135,in=90] (3,1.5) to [out=-90,in=190] (4,1);

\node [right] at (4,1) {$Z_{0}$};

\node [right] at (5,4) {$Z$};

\node [left] at (1,2) {$G_2$-orbits in $\mathfrak{X} = \mathfrak{G}/G_1$};

\node [left] at (1,4) {$K$};

\end{tikzpicture} 
\end{center}
Then $\pi_\mathfrak{X} (K - Z) = \pi_\mathfrak{X} (K - Z_{0})$. We shall show that
$\mu(Z_{0} \cdot G_2) = \mu( \pi_{\mathfrak{X}}^{-1} (\pi_\mathfrak{X} ( Z_{0}) ) = 0$. Toward this end observe
that because $\mathfrak{X}$ is metrizable and separable we may assume the elements $\mathcal{O}_m$, $m \in \mathbb{N}$,
of basis of topology to be the balls with compact closure $\overline{\mathcal{O}_m}$; and the $\sigma$-ring of Borel sets on $\mathfrak{X}$
generated by the open $\mathcal{O}_m$ or closed $\overline{\mathcal{O}_m}$ balls.
\begin{center}
\begin{tikzpicture}

\path [fill=gray] (5,2) circle (0.5);

\draw[thick] (5,2) ellipse (0.7 and 0.7);

\draw[thick] (4.293,4) ellipse (0.6 and 0.5);

\draw[thick] (3.368,5) ellipse (0.575 and 0.4);

\draw[thin] (5,0) to [out=60.75,in=-90] (5.7,2)
to [out=90,in=-45] (3,6);

\draw[thin] (4.7,0) to [out=60.75,in=-90] (5.4,2)
to [out=90,in=-45] (2.7,6);

\draw[thin] (4.6,0) to [out=60.75,in=-90] (5.3,2)
to [out=90,in=-45] (2.6,6);

\draw[thin] (3.6,0) to [out=60.75,in=-90] (4.3,2)

to [out=90,in=-45] (1.6,6);

\draw[ultra thick] (5.4,1.7) to [out=89,in=-89] (5.4,2.3);

\draw[ultra thick] (5.3,1.6) to [out=88,in=-88] (5.3,2.4);


\draw [->,very thin] (3,2) to [out=-70,in=130] (4.505, 2.495);

\draw [->,very thin] (3,4) to [out=-20,in=135] (3.8, 4.3);

\draw [->,very thin] (3,1) to [out=45,in=-135] (4.6465, 1.6465);

\draw [<-,very thin] (4.7,0) to [out=-45,in=135] (6, 0);


\node [right] at (6,0) {$Z_{0} \cdot G_2$};

\node [left] at (3,4) {$\mathcal{O}_\epsilon \cdot \eta$,};

\node [left] at (3.5,3.6) {$\eta \in G_2$};

\node [left] at (3,2) {$\mathcal{O}_\epsilon$};

\node [left] at (3,1) {$K$};

\node [right] at (-2.75,4) {For each $\epsilon > 0$ };

\node [right] at (-2.75,3.5) {there exists open $\mathcal{O}_\epsilon 
\supset K$ };

\node [left] at (0.75,3) {with: $\mu(\mathcal{O}_\epsilon 
- K) < \epsilon$};

\node [right] at (-2.75,2.5) {by regularity of $\mu$};

\end{tikzpicture} 
\end{center}
By the regularity and quasi invariance of the measure $\mu$ it easily follows that the $\mu$-measure of the intersection of 
$Z_{0} \cdot G_2$ with any open set in $\mathfrak{X}$ is equal zero, and thus again by the regularity
of $\mu$ and second countability of $\mathfrak{X}$ it easily follows that 
$\mu(Z_{0} \cdot G_2) = \mu ({\pi_\mathfrak{X}}^{-1} (\pi_\mathfrak{X} ( Z_{0}) )) = 0$. Thus 
$\pi_\mathfrak{X} ( Z_{0})$ is a subset of a measurable null set, and so must be a measurable set with
$\nu(\pi_\mathfrak{X} ( Z_{0})) = 0$, because $\nu$ is a pseudo-image measure of $\mu$ under $\pi_\mathfrak{X}$. Moreover, we have: 
\[
\pi_\mathfrak{X}(K-Z) 
= \pi_\mathfrak{X}(K - Z_0) 
=  \pi_\mathfrak{X}(K) - \pi_\mathfrak{X}(Z_0), 
\]
because $Z_0$ consists of intersections of $G_2$-orbits with 
$K$. 

On the other hand
\[
\psi_0 \circ \pi_\mathfrak{X} (K -Z) 
\]
is a Borel set in $S$, and thus $\pi_\mathfrak{X} (K -Z)$ is a Borel set in $\mathfrak{X}/G_2$ as $\psi_0$
is a Borel isomorphism. Indeed, because images preserve the set theoretic sum operation we have
\[
\psi_0 \circ \pi_\mathfrak{X} (K -Z) =  \bigcup \limits_{n \in \mathbb{N}} \psi_0 \circ \pi_\mathfrak{X} (K_i). 
\]
Because $K_j \in \mathfrak{K}$ then $K_{j} / R_{K_{j}}$ is Hausdorff and the quotient map $\pi_{K_j}$ is closed and thus the quotient space $K_{j} / R_{K_{j}}$ is homeomorphic to the compact space $\pi_{K_j}(K_j)$, and moreover because  $K_j$ is compact and metrizable (as a subspace of the metrizable space $\mathfrak{X}$) the quotient space  $K_{j} / R_{K_{j}}$ is likewise metrizable (\cite{Engelking}, Thm. 7.5.22). We can therefore apply the Federer and Morse Theorem 5.1 
of \cite{Federer_Morse} in order to prove the existence for each $j$ of a Borel subset $B_{j} \subset K_j$ such that $\pi_{K_j}(B_j) = \pi_{K_j}(K_j) (= \pi_{\mathfrak{X}}(K_j))$ 
and such that $\pi_{K_j}$ is one-to-one on $B_j$. Therefore $\psi_0 \circ \pi_{\mathfrak{X}}$ is one-to-one Borel
function on a Borel subset $B_j$ of the complete separable metric space  $\mathfrak{X}$ to a complete separable metric
space $S$. Therefore again by the Theorem on page 253 of \cite{Kuratowski} (compare likewise the Theorem on page 72 
of \cite{Mackey2}), it follows that $\psi_0 \circ \pi_{\mathfrak{X}}(B_j) = \psi_0 \circ \pi_{\mathfrak{X}}(K_j)$ 
is a Borel set. Because $\psi_0$ is a Borel isomorphism it follows that $\pi_{\mathfrak{X}}(K_j)$ is a Borel set
in $\mathfrak{X}/G_2$. 

Thus $\pi_\mathfrak{X}(K)$ differs from a Borel set
$\pi_\mathfrak{X}(K- Z)$ by a measurable $\nu$-negligible subset $\pi_\mathfrak{X}(Z_{0})
\subset \pi_\mathfrak{X}(K)$; so we have shown that  $\pi_\mathfrak{X}(K)$ is measurable. 
\qed

\vspace*{0.5cm}

Note that the Lemma \ref{lem:decomposition.5'} is non trivial. 
By the well known theorem of Suslin -- continuous image of a Borel set is not always Borel,
but it is always measurable, compare e. g. \cite{Jech}, Lemm. 11.6, page 142 and Thm. 11.18, page 150,
where the references to the original literature are provided. However this argument would be insufficient for 
$\pi_{\mathfrak{X}}(K)$ to be measurable in $\mathfrak{X}/G_2$
for any compact set $K \subset \mathfrak{X}$. Indeed it would in addition require to be shown that the quotient Borel structure on $\mathfrak{X}/G_2$ is equal to the $\sigma$-ring of Borel sets generated by the closed (open)
sets of the quotient topology on $\mathfrak{X}/G_2$.

\vspace*{0.5cm}

\begin{lem}
Let $\mu, \mu_C , \nu$ be as in Lemma \ref{lem:decomposition.4} and let $K$ be a compact subset of $\mathfrak{X}$.
Let $\eta \in G_2$ and let $\mathscr{R}_K$ be the $\sigma$-ring  of Borel\footnote{The $\sigma$-ring of Borel sets with a regular measure on this ring is sufficient to recover all measurable subsets and their measures obtained by the standard completion of the Borel measure space.} subsets of $K$ induced form the surrounding measure space $\mathfrak{X}$. Let $(\mu)'_K$ and $(\mu_C)'_K$ denote the restrictions of 
$\mu$ and $\mu_C$ to $K$ defined on the $\sigma$-ring $\mathscr{R}_K$ respectively, and let $R_\eta \mu , R_\eta \mu_C$ denote their right translations; and similarly let $(\nu)'_{\pi_\mathfrak{X}(K)}$ be the restriction of the measure $\nu$ to the subset $\pi_{\mathfrak{X}}(K)$. 
Then

\begin{enumerate}

\item[(a)] 
\[
(\mu)'_K = \int \, (\mu_C )'_K  \, d (\nu)'_{\pi_\mathfrak{X}(K)}(C)
\]
with each $(\mu_C)'_K$ concentrated on $C \cap K$.

\item[(b)]
\[
R_\eta \mu = \int \, R_\eta \mu_C  \, d\nu(C).
\]

\end{enumerate}

\label{lem:decomposition.5}
\end{lem}

\qedsymbol \,
  Part (a) of the Lemma is an immediate consequence 
of Lemmas  \ref{lem:decomposition.4} and  \ref{lem:decomposition.5'} with $1_K \cdot g$ 
inserted for $g$ in the formula (\ref{dec_m}), where $1_K$ is the characteristic
function of the compact set $K$. The only non-trivial part of the proof lies in showing that 
$\pi_\mathfrak{X}(K)$ is measurable, which was proved in Lemma \ref{lem:decomposition.5'}. 

For (b) observe that if $R_{\eta^{-1}}g \in L^1 (\mathfrak{X}, \mu)
\Leftrightarrow g \in L^1 (\mathfrak{X}, R_\eta \mu)$, then by Lemma \ref{lem:decomposition.4}:
\[
\begin{split}
\int g(x) \, d ( R_\eta \mu) = \int g(x \cdot \eta^{-1}) \, d \mu 
= \int d \nu (C) \, \int g(x \cdot \eta^{-1}) \, d \mu_C (x) \\
= \int d \nu (C) \, \int g(x) \, d (R_\eta \mu_C ) (x), 
\end{split}
\] 
thus 
\[
R_\eta \mu = \int R_\eta \mu_C \, d \nu (C).
\]
\qed

\vspace*{0.5cm}

Note that the operations of restriction $(\cdot)'_{K}$
to $K$ and right translation $R_\eta (\cdot)$ do not commute. Indeed if we write $R_\eta \circ (\cdot)'_{K}$
for $ R_\eta((\cdot)'_K )$, then $R_\eta \circ (\cdot)'_{K} 
= (\cdot)'_{K \cdot \eta^{-1}} \circ R_\eta = (R_\eta (\cdot))'_{K \cdot \eta^{-1}}$
i. e. first restrict to $K$ and then translate $R_\eta$ is the same as first translate $R_\eta$ and then
restrict to $K \cdot \eta^{-1}$ (and not to $K$).

\vspace*{0.5cm}

\begin{rem}

Let $Op(\mu)$ denote a repeated application of several restrictions to compact sets and translations:
$(\cdot)'_{K_1}, R_{\eta_1}(\cdot), \ldots$ performed on the measure $\mu$. Then the repeated application of 
Lemma \ref{lem:decomposition.5} (a) and (b) gives

\[
Op( \mu) = \int  Op( \mu_C ) \,\, \ud \widetilde{Op}(\nu)(C),
\]
where $\widetilde{Op}(\nu)$ denotes the restriction $()'_{\pi_\mathfrak{X}(K)}$ with the compact set
$K \subset \mathfrak{X}$ which arises in the following way: $(\cdot)'_K$ is the restriction which arises from 
$Op$ by commuting all translations to the right (so as to be performed first) and all restrictions to the left
(so as to be performed after all translations): $Op = (\cdot)'_K \circ R_\eta (\cdot)$ or 
$Op (\cdot) = (R_\eta (\cdot))'_K$.
\label{rem:decomposition.2}
\end{rem}

\vspace*{0.5cm}

\begin{lem}
Let $K, (\mu)'_K, (\mu_C)'_K, (\nu)'_{\pi_\mathfrak{X}(K)}$ be as in the preceding Lemma. 
For any bounded and $(\mu)'_K$-measurable function  $g$ 
and for any $f \in L^1 (\pi_{\mathfrak{X}}^{-1}(K) , (\nu)'_{\pi_\mathfrak{X}(K)})$
the set of all those $G_2$ orbits $C$ having non empty intersection $C \cap K$ for which $g$ is not 
$\mu_C$-integrable is $\nu$-negligible and the the function 
\[
C \mapsto \int \, g(x) d (\mu_C)'_K (x)
\]
on this set of orbits $C$ is $(\nu)'_{\pi_\mathfrak{X}(K)}$-summable and $(\nu)'_{\pi_\mathfrak{X}(K)}$-measurable, and
\begin{equation}\label{dec_m_1}
\int f(C) \int \, g(x) \, d (\mu_C)'_K (x) \, d (\nu)'_{\pi_\mathfrak{X}(K)} (C)   
= \int \, f(\pi_{\mathfrak{X}}(x) )g(x) \, d (\mu)'_K (x).
\end{equation}

\label{lem:decomposition.6}

\end{lem}

\qedsymbol \,
 The Lemma is an immediate consequence of the preceding Lemma. The only non-trivial part of the proof is
is to show that $f$ is measurable on $\mathfrak{X}/G_2$ if and only if $f\circ \pi_{\mathfrak{X}}$ 
is measurable on $\mathfrak{X}$. But this is an immediate consequence of Lemma \ref{lem:decomposition.3}.
\qed

\vspace*{0.5cm}

In order to simplify notation  let us denote the operation of 
restriction $(\cdot)'_K$ to $K$ just by $(\cdot)'$ in the next Lemma and its proof.
In all other restrictions $(\cdot)'_D$ the sets $D$ will be specified explicitly. 

\begin{lem}
Let $\mu, \mu_C$ be as in Lemma \ref{lem:decomposition.4} and let $K$ be a compact subset
of $\mathfrak{X}$. Let $\eta \in G_2$
and let $C$ be any $G_2$-orbit having non empty intersection $C \cap K \cdot \eta^{-1} \cap K$.
Then for the respective measures obtained by right translations and restrictions  
performed on $\mu$ and $\mu_C$ respectively we have:
\begin{enumerate}

\item[(a)] The measures $((\mu_C)')'_{K \cdot \eta^{-1}}$ and $(R_\eta (\mu_C)')'$, 
defined on measurable subsets of $C \cap K \cap  K \cdot \eta^{-1}$, are equivalent.

\item[(b)]
\[
\begin{split}
\lambda_C (\cdot , \eta) = 
\frac{d(R_\eta \mu_C)}{d\mu_C}(\cdot) \\
= \frac{\ud \,(R_\eta (\mu_C)')'}{\ud \, ((\mu_C)')'_{K \cdot \eta^{-1}}}(\cdot) 
= \frac{\ud \, (R_\eta \mu')'}{d (\mu')'_{K \cdot \eta^{-1}}}(\cdot) \\
= \frac{\ud \, R_\eta \mu}{\ud \mu}(\cdot)
= \lambda (\cdot , \eta)
\end{split}
\]
on $C \cap K \cap  K \cdot \eta^{-1}$.
\end{enumerate}

\label{lem:decomposition.7}
\end{lem}

\qedsymbol \,
 In addition to the operations of translation and restriction let us introduce after Mackey, \cite{Mackey}, \S 11, one more operation $\widetilde{\cdot}$ defined on finite measures $\mu$ on $\mathfrak{X}$, giving measures
$\widetilde{\mu}$ on $\mathfrak{X}/G_2$. Namely we put $\widetilde{\mu}(E) = \mu(\pi_\mathfrak{X} ^{-1} (E))$. 
$\widetilde{\mu'}$ is well defined for any quasi invariant measure $\mu$ on $\mathfrak{X}/G_2$ because $\mu'$ is finite.  More precisely $\widetilde{\mu'}$ is defined on the 
$\sigma$-ring of measurable subsets $E$ of $\pi_\mathfrak{X}(K)$ by the formula: 
$\widetilde{\mu'}(E) = \mu'(\pi_{\mathfrak{X}}^{-1}(E)) = \mu(K \cap  \pi_{\mathfrak{X}}^{-1}(E))$. 
A simple verification of definitions shows that $\widetilde{\mu'}$ is a pseudo image measure
of the measure $\mu'$ under $\pi_\mathfrak{X}$, so that 
\[
\mu' = \int \,  \mu'_C \, d\widetilde{\mu'}(C), 
\]
on measurable subsets of $K$ and 
where the integral is over the orbits $C$ having non void intersection with $K$
and with $\mu'_C$ concentrated on $C \cap K$. 
Similarly we have for the pairs of measures 
\begin{equation}\label{pairs}
\Big( \,\, (\mu')'_{K  \eta^{-1}} \,\, , \,\,\,
\widetilde{(\mu')'_{K  \eta^{-1}} } \,\, \Big)
 \,\, \textrm{and} \,\, 
\Big( \,\, (R_\eta \mu')' \,\, , \,\,\, \widetilde{(R_\eta \mu')'} \,\, \Big):
\end{equation}

\[
(\mu')'_{K  \eta^{-1}} 
= \int \,  \Big( (\mu')'_{K \eta^{-1}} \Big)_{{}_C} 
\,\,\,\, \ud \, \widetilde{(\mu')'_{K \eta^{-1}}} \,\, (C)
\]
and
\[
(R_\eta \mu')' = \int \Big( R_\eta \mu')' \Big)_{{}_C} \,\,\,\, \ud \, \widetilde{(R_\eta \mu')'} \,\, (C),
\]
both $(\mu')'_{K  \eta^{-1}}$ and $(R_\eta \mu')'$ defined on measurable subsets of $K \cdot \eta^{-1} \cap K$ 
(instead of $K$): with the measure $(R_\eta \mu')'$ equal to the measure $R_\eta \mu$
restricted to $K \cdot \eta^{-1} \cap K$, and $(\mu')'_{K  \eta^{-1}}
= (\mu)'_{K  \eta^{-1} \cap K}$ equal to the measure $\mu$ restricted to 
the same compact subset $K \cdot \eta^{-1} \cap K$; and with the corresponding tilde measures 
both defined on measurable subsets of the measurable (Lemma \ref{lem:decomposition.5'}) set 
$\pi_\mathfrak{X}(K \cdot \eta^{-1} \cap K)$; namely 
\[
\begin{split}
\widetilde{(R_\eta \mu')'} \, (E) = (R_\eta \mu')'(\pi_{\mathfrak{X}}^{-1}(E))
= R_\eta \mu'(K \cap \pi_{\mathfrak{X}}^{-1}(E)) \\ 
=  (R_\eta \mu )'_{K \eta^{-1}}(K \cap \pi_{\mathfrak{X}}^{-1}(E)) 
= R_\eta \mu (K \eta^{-1} \cap K \cap \pi_{\mathfrak{X}}^{-1}(E))
\end{split}
\]
and 

\[
\widetilde{(\mu')'_{K  \eta^{-1}} } \, (E) = \widetilde{(\mu)'_{K  \eta^{-1} \cap K} } \, (E)
= \mu (K \eta^{-1} \cap K \cap \pi_{\mathfrak{X}}^{-1}(E)).  
\] 

Note please that our Lemma \ref{lem:decomposition.6}  holds true for any pseudo-image
measure $\nu$ of $\mu$. By Lemma \ref{lem:decomposition.5'}, any pseudo-image measure
of the restriction $\mu'$ is a restriction $(\nu)'_{\pi_\mathfrak{X}(K)}$ of a pseudo-image measure of $\mu$.
It follows that the Lemma  \ref{lem:decomposition.6} is applicable to the pairs of measures (\ref{pairs}).
Indeed it is sufficient to insert $K \cdot \eta^{-1} \cap K$ instead of $K$ in the Lemma \ref{lem:decomposition.6}
and apply it to $(\mu')'_{K  \eta^{-1}} = (\mu)'_{K  \eta^{-1} \cap K}$ (or to $(R_\eta \mu')'
= (R_\eta \mu )'_{K \eta^{-1} \cap K}$)
instead of $\mu'$, because for an appropriate $\nu$,  $(\nu)'_{\pi_\mathfrak{X}(K \cdot \eta^{-1} \cap K)}$ gives
the pseudo-image measure $\widetilde{(\mu')'_{K  \eta^{-1}} }$ (or respectively $\widetilde{(R_\eta \mu')'}$) of 
$(\mu')'_{K  \eta^{-1}}$ (or respectively of $(R_\eta \mu' )'$). 
We may thus apply Lemma 11.4 of \cite{Mackey}, \S 11, to the pairs of measures (\ref{pairs}). 
Because $\mu$ is quasi invariant, the measures $(\mu')'_{K  \eta^{-1}} = (\mu)'_{K  \eta^{-1} \cap K}$ and 
$(R_\eta \mu')' = (R_\eta \mu )'_{K \eta^{-1} \cap K}$ are equivalent as measures 
on $K \cdot \eta^{-1} \cap K$, and thus by Lemma 11.4 of \cite{Mackey} 
it follows that $\widetilde{(\mu')'_{K  \eta^{-1}}}$ and $\widetilde{(R_\eta \mu')'}$ are equivalent as measures
on $\pi_\mathfrak{X}(K \cdot \eta^{-1} \cap K)$. Introducing the corresponding measurable weight function 
$f_1$ on $\mathfrak{X}/G_2$ which is non zero on $\pi_\mathfrak{X}(K \cdot \eta^{-1} \cap K)$, 
we have
\[
f_1 \cdot \ud \, \widetilde{(\mu')'_{K  \eta^{-1}}} = \ud \, \widetilde{(R_\eta \mu')'}
\]
and 
\begin{equation}\label{pair2}
(R_\eta \mu')' = \int f_1 (C) \Big(  R_\eta \mu')' \Big)_{{}_C} \,\,\,\, 
\ud \, \widetilde{(\mu')'_{K \eta^{-1}}} \,\, (C),
\end{equation}
\begin{equation}\label{pair1}
(\mu')'_{K  \eta^{-1}} 
= \int \,  \Big( (\mu')'_{K \eta^{-1}} \Big)_{{}_C} 
\,\,\,\, \ud \, \widetilde{(\mu')'_{K \eta^{-1}}} \,\, (C).
\end{equation}
Now applying again the Lemma 11.4 of \cite{Mackey} to the pairs of measures:
\[ 
\Big( \,\, (\mu')'_{K  \eta^{-1}} \,\, , \,\,\,
\widetilde{(\mu')'_{K  \eta^{-1}} } \,\, \Big)
 \,\, \textrm{and} \,\, 
\Big( \,\, (R_\eta \mu')' \,\, , \,\,\, \widetilde{(\mu')'_{K  \eta^{-1}} }  \,\, \Big)
\]
with the respective decompositions (\ref{pair1}) and (\ref{pair2}) we prove that the measures 
$\Big( (\mu')'_{K \eta^{-1}} \Big)_{{}_C}$ and $\Big( R_\eta \mu')' \Big)_{{}_C}$ are equivalent and
\[
\begin{split}
f_1 (C) \cdot \frac{\ud \, \Big( R_\eta \mu')' \Big)_{{}_C} } { \ud \, \Big( (\mu')'_{K \eta^{-1}} \Big)_{{}_C}  }(\cdot) 
=  \frac{\ud \, (R_\eta \mu')'}{\ud \, (\mu')'_{K \cdot \eta^{-1}}}(\cdot) \\
= \frac{\ud \, (R_\eta \mu)}{\ud \, \mu}(\cdot) = \lambda( \cdot , \eta),
\end{split}
\]
on $C \cap K \cdot \eta^{-1} \cap K$, where the last two equalities follow from definitions and where
\[
f_1 = \frac{\ud \, \widetilde{(R_\eta \mu')'}}{\ud \, \widetilde{(\mu')'_{K  \eta^{-1}}}}.
\]

On the other hand it follows from Lemma \ref{lem:decomposition.5} and Remark \ref{rem:decomposition.2}
that 
\[
(R_\eta \mu')' = \int  (R_\eta (\mu_{{}_C})')'  \,\,\,\, \ud \, (\nu)'_{\pi_\mathfrak{X}(K \eta^{-1} \cap K)} (C)
\]
and 
\[
(\mu')'_{K \eta^{-1}} = \int  ((\mu_{{}_C})')'_{K \eta^{-1}}  \,\,\,\, 
\ud \, (\nu)'_{\pi_\mathfrak{X}(K \eta^{-1} \cap K)} (C).
\]
Thus both $\widetilde{(R_\eta \mu')'}$ and $(\nu)'_{\pi_\mathfrak{X}(K \eta^{-1} \cap K)}$ being 
pseudo-image measures of the measure $(R_\eta \mu')'$ under $\pi_\mathfrak{X}$ (of course restricted to
$K \eta^{-1} \cap K)$) are equivalent. Introducing the respective measurable, non zero on 
$\pi_\mathfrak{X}(K \eta^{-1} \cap K)$, weight function $f_2$ we have 
\[
f_2 \cdot \ud \, (\nu)'_{\pi_\mathfrak{X}(K \eta^{-1} \cap K)}  = \ud \, \widetilde{(R_\eta \mu')'},
\]
so that 
\[
\ud \, (R_\eta (\mu_{{}_C})')' = f_2 (C) \cdot \ud \,  \Big( R_\eta \mu')' \Big)_{{}_C}.  
\]
Similarly because $\mu$ is quasi invariant, the measures  $(\mu')'_{K  \eta^{-1}}$ and $(R_\eta \mu')'$ are 
equivalent, and thus again by Lemma 11.4 of \cite{Mackey} the measures $\widetilde{(\mu')'_{K \eta^{-1}}}$
and $(\nu)'_{\pi_\mathfrak{X}(K \eta^{-1} \cap K)}$ are likewise equivalent. Introducing the respective non zero
on $\pi_\mathfrak{X}(K \eta^{-1} \cap K)$ and measurable weight function $f_3$ we have 
\[
f_3 \cdot \ud \, (\nu)'_{\pi_\mathfrak{X}(K \eta^{-1} \cap K)}  = \ud \, \widetilde{(\mu')'_{K  \eta^{-1}} },
\]
so that 
\[
\ud \, ((\mu_{{}_C})')'_{K \eta^{-1}} = f_3 (C) \cdot \ud \,  \Big( (\mu')'_{K \eta^{-1}} \Big)_{{}_C}.  
\]
Joining the above equalities we obtain (the last two equalities follows from definition of $\lambda_C$ and 
from definition of Radon-Nikodym derivative, i. e. its local character)
\[
\begin{split}
\lambda( \cdot , \eta) 
= f_1 (C) \cdot \frac{\ud \, \Big( R_\eta \mu')' \Big)_{{}_C} } { \ud \, \Big( (\mu')'_{K \eta^{-1}} \Big)_{{}_C}  }(\cdot) 
= f_1 (C) \cdot \frac{1}{f_2 (C)} \cdot f_3 (C) \cdot \frac{ \ud \, (R_\eta (\mu_C)')' }{ \ud \, ((\mu_C)')'_{K \eta^{-1}} } \\
= \frac{ \ud \, (R_\eta (\mu_C)')' }{ \ud \, ((\mu_C)')'_{K \eta^{-1}} } 
= \frac{\ud \, (R_\eta \mu_C)}{\ud \, \mu_C}(\cdot) = \lambda_C (\cdot , \eta)
\end{split}
\]
on $C \cap K \cap  K \cdot \eta^{-1}$, because by the known property of Radon-Nikodym derivatives (compare e. g. Scholium 4.5 of \cite{Segal_Kunze})
\[
f_1  \cdot \frac{1}{f_2} \cdot f_3  
= \frac{\ud \, \widetilde{(R_\eta \mu')'} }{ \ud \, \widetilde{(\mu')'_{K  \eta^{-1}} } } 
\cdot \frac{ \ud \, (\nu)'_{\pi_\mathfrak{X}(K \eta^{-1} \cap K)} }{ \ud \, \widetilde{(R_\eta \mu')'} } 
\cdot \frac{ \ud \, \widetilde{(\mu')'_{K  \eta^{-1}} } }{ \ud \, (\nu)'_{\pi_\mathfrak{X}(K \eta^{-1} \cap K)} } = 1,
\]
on all orbits $C$ with non void intersection $C \cap K \cap  K \cdot \eta^{-1}$. 
\qed

\vspace*{0.5cm}

We are are now in a position to formulate the main goal of this Section.

\vspace*{0.5cm}

\begin{lem}
Let $\mu$ be any quasi invariant measure on $\mathfrak{X}$ and
let $\nu$ be any pseudo image measure of $\mu$. Then the measures $\mu_C$
in the decomposition 
\[
\mu = \int \, \mu_C (x) \, d\nu(C) 
\]
of Lemma \ref{lem:decomposition.4} are also quasi invariant and for each $\eta \in G_2$ the Radon-Nikodym 
derivative $\lambda_C (\cdot , \eta) = 
\frac{\ud (R_\eta \mu_C)}{\ud \mu_C}(\cdot)$ is equal to the restriction of the Radon-Nikodym 
derivative  $\lambda (\cdot , \eta) = \frac{\ud (R_\eta \mu)}{\ud \mu}(\cdot)$ to the orbit $C$.

\label{lem:decomposition.8}
\end{lem}

\qedsymbol \,
 Indeed, let $x$ be any point in $\mathfrak{X}$ and $\eta$ any element of $G_2$. We show that on a neighbourhood of $x$ the statement of the Theorem holds true. To this end let $\mathcal{O}_m$ be a neighbourhood of $x$ chosen from the basis of topology constructed above. Then $\mathcal{O}_m \cdot \eta$ is a neighbourhood of $x \cdot \eta$. Therefore the compact set
$K = \overline{\mathcal{O}_m} \cup (\overline{\mathcal{O}_m} \cdot \eta )$ has the property that 
$K \cap ( K \cdot \eta^{-1} )$ contains an open neighbourhood of $x$. 
Now it is sufficient to apply Lemma \ref{lem:decomposition.7} with this $K$ in order to show
that the equality of the Theorem holds true on some open neighbourhood of $x$.    
\qedsymbol \,

\begin{rem}
It has been proved in Sect. \ref{subgroup.preliminaries} that for each orbit $C$  there exists a measure $\mu_C$, 
concentrated on $C$, with the associated Radon-Nikodym derivative equal to the restriction to the orbit 
$C$ of the Radon-Nikodym derivative associated with $\mu$. This however would be insufficient because we need to know that
the measures $\mu_C$ conspire together so as to compose a decomposition of the measure $\mu$. This is why we need Lemma 
\ref{lem:decomposition.8}. Although the Lemma was not explicitly formulated in \cite{Mackey}, it easily follows
for the case of finite $\mu$ from the Lemmas of \cite{Mackey}, \S 11.

\label{rem:decompositions.3}
\end{rem}

\vspace*{0.5cm}

Using Lemma \ref{lem:decomposition.4} and the general properties of the integral and the algebra of measurable functions one can prove a slightly strengthened version of 
Lemma \ref{lem:decomposition.4} which 
may be called a skew version of the Fubini theorem, because it extends the Fubini theorem to the case 
where we have a skew product measure $\mu$ with only one projection, i.e. the quotient map $\pi_\mathfrak{X}$:
\begin{lem}[Skew Version of the Fubini Theorem]
Let $\mu$, $\mu_C$ and $\nu$ be such as in Lemma \ref{lem:decomposition.4}. Let $g$ 
be a positive complex valued and measurable function on $\mathfrak{X}$. Then
 \begin{equation}\label{skew_fubini.1:decompositions}
C \mapsto \int \, g(x) d\mu_C(x)
\end{equation}
is measurable, and if any one of the following two integrals:
\[
\int \, d\nu(C) \, \int \, g(x) \, d\mu_C (x)  \,\,\, \textrm{and} \,\,\,  \int \, g(x) \, d \mu(x),
\]
does exist, then there exists the other and both are equal in this case. 

In particular it follows that if $g$ is integrable on $(\mathfrak{X} , \mathscr{R}_\mathfrak{X}, \mu)$
then  
\begin{equation}\label{skew_fubini.2:decompositions}
\int \, d\nu(C) \, \int \, g(x) \, d\mu_C (x)  = \int \, g(x) \, d \mu(x).
\end{equation}
\label{lem:decomposition.9}
\end{lem}

\qedsymbol \,

 For the proof compare \cite{Bourbaki_i}, Chap. VI, Remark of \S 3.4. 
Here we give only few comments:
The Lemma holds for positive and continuous $g$ with compact support as a consequence of
Lemma \ref{lem:decomposition.4}. Next we note that the class of functions which
satisfy (\ref{skew_fubini.1:decompositions}) and (\ref{skew_fubini.2:decompositions}) is closed 
under sequential convergence of increasing sequences.

The Lemma follows by repeated application of the sequential continuity of the integral for increasing sequences;
compare, please, the proof of Thm. 3.4 and Corollary 3.6.2 of \cite{Segal_Kunze}. 
\qed

\vspace*{0.5cm}

Note that the integral 
\[
\int \, g(x) d\mu_C(x)
\]
in (\ref{skew_fubini.1:decompositions}) and (\ref{skew_fubini.2:decompositions})
may be replaced with
\[
\int \limits_{C} \, g^C (x) d\mu_C(x),
\]
where $g^C$ is the restriction of $g$ to the orbit $C$, because $\mu_C$ is concentrated
on $C$. However just like in the ordinary Fubini theorem the whole difficulty in application
of the skew version of the Fubini Theorem lies in proving the measurability
of $g$ on the ``skew product''$\mathfrak{X} \xrightarrow{\pi_\mathfrak{X}}  \mathfrak{X}/G_2$ measure 
space $(\mathfrak{X}, \mathscr{R}_\mathfrak{X} , \mu)$. Indeed even if  the orbits $C$ were nice closed 
subsets and $g^C$ measurable on $C$ (with respect to the  measure structure induced from the 
surrounding space $\mathfrak{X}$)   
the function $g$ still could be non measurable on $(\mathfrak{X}, \mathscr{R}_\mathfrak{X} , \mu)$; 
for simple examples we refer e. g. to \cite{Segal_Kunze} or to any other book on measure theory.
More restrictive constrains are to be put on the separate $g^C$ as functions on the orbits $C$ 
in order to guarantee the measurability of $g$ on the measure space $\mathfrak{X}$. 

We face the same problem with the ordinary Fubini theorem. If in addition
$g^C \in L^2 (C, \mu_C)$ for each $C$ (or $\nu$-almost all orbits $C$), the required additional requirement
is just the von Neumann direct integral structure put on $C \mapsto g^C$ which
is the necessary and sufficient condition for 
the existence of a function $f \in  L^2 (\mathfrak{X}, \mu)$ such that
$f^C = g^C$ for $\nu$-almost all orbits $C$. Namely, consider the space of functions
$C \mapsto g^C \in L^2 (C, \mu_C)$, which composes 
\begin{equation}\label{dir_int_skew:decompositions}
\int \limits_{\mathfrak{X}/G_2} L^2 (C, \mu_C) \,\, \ud \nu (C),
\end{equation}  
then for every element $C \mapsto g^C$ of direct integral (\ref{dir_int_skew:decompositions}) 
there exists a function $f \in  L^2 (\mathfrak{X}, \mu)$ such that
$f^C = g^C$ for $\nu$-almost all orbits $C$. In short 
\begin{equation}\label{dir_int_skew_L^2:decompositions}
\boxed{\int \limits_{\mathfrak{X}/G_2} L^2 (C, \mu_C) \,\, \ud \nu (C) = L^2(\mathfrak{X} , \mu).}
\end{equation}
We skip proving the equality (\ref{dir_int_skew_L^2:decompositions}) because in the next Section we 
prove a more general version of (\ref{dir_int_skew_L^2:decompositions}) for vector valued functions 
$g \in \mathcal{H}^L$ on $\mathfrak{X} = \mathfrak{G}/G_1$, compare Lemma \ref{lem:subgroup.1} (a).
This strengthened version (\ref{dir_int_skew_L^2:decompositions}) of the skew Fubini theorem 
lies behind harmonic analysis on classical Lie groups and provides also an effective tool for 
tensor product decompositions of induced representations in Krein spaces. In practice 
the classical groups with the harmonic analysis relatively complete on them, have the structure of cosets and double 
cosets (corresponding to the orbits $C$) much more nice in comparison to what we have actually assumed, 
so that a vector valued version of the strengthened version of the ordinary Fubini theorem: 
\begin{equation}\label{dir_int_L^2:decompositions}
\boxed{\int \limits_{X} L^2 (Y, \mu_Y) \,\, \ud \mu_X  = L^2(X \times Y , \mu_X \times \mu_Y)}
\end{equation}
would be sufficient for our applications. Namely the  
``measure product property'' holds also in our practical applications for the double coset
space:
\begin{align*} 
\big( \mathfrak{G}, & \mathscr{R}_\mathfrak{G}, \mu_\mathfrak{G} \big) \\
= & \Big(\, G_1 \times \mathfrak{G}/G_1 \times ( \mathfrak{G}/G_1 )/G_2 \, , \,\, 
\mathscr{R}_{{}_{{}_{G_1 \times \mathfrak{G}/G_1 \times ( \mathfrak{G}/G_1 )/G_2} }} \, , \,\, 
\mu_{{}_{{}_{G_1}}} \times \mu_{{}_{{}_{\mathfrak{G}/G_1}}} \times \mu_{{}_{{}_{(\mathfrak{G}/G_1 )/G_2}}} \, \Big)
\end{align*} 
with the analogous functions (\ref{q'_h'}), measure $\mu = \mu_{{}_{{}_{\mathfrak{G}/G_1}}} $ and the pseudo image measure 
$\nu = \mu_{{}_{{}_{(\mathfrak{G}/G_1 )/G_2}}}$ effectively computable. 

Note that (\ref{dir_int_skew_L^2:decompositions}) and (\ref{dir_int_L^2:decompositions}) 
may be proved for more general measure spaces\footnote{Our proof of (\ref{dir_int_skew_L^2:decompositions}) may be easily adopted to general non-separable case, provided that the assertion of Lemma \ref{lem:decomposition.9} holds
true for the measures $\mu$ and $\nu$.}. 

Here the measure spaces are not ``too big'', so that
the associated Hilbert spaces of square summable functions are separable.

\vspace*{0.5cm}

At the end of this Section we transfer the measure structure on $\mathfrak{X}/G_2$ over
to the the set  $G_1 : G_2$ of all double cosets $G_1 x G_2$, using the natural 
bi-unique correspondence $C \mapsto D_{{}_C} = \pi^{-1}(C)$ between the orbits $C$
and double cosets $D$. Next we transfer it again to a measurable section $\mathfrak{B}$
of $\mathfrak{G}$ cutting every double coset at exactly one point and give measurability
criterion for a function on $\mathfrak{B}$ with this measure structure inherited from 
$\mathfrak{X}/G_2$. We shall use it in Sections \ref{subgroup} and \ref{Kronecker_product}.

\begin{defin}
We put $\ud \nu _0 (D) = \ud \nu(C_{{}_D})$ 
for the measure $\nu_0$ transferred over to measurable subsets of the set of all double cosets, where $C_{{}_D}$ 
is the orbit corresponding to the double coset, i. e. $D = \pi^{-1}(C)$. 
Let $B_0$ be a measurable section of $\mathfrak{X}$ with respect to $G_2$, existence of which has been proved
in Lemma \ref{lem:decomposition.1}. Let $B$ be a measurable (even Borel) 
section of $\mathfrak{G}$ with respect to $G_1$ (which exists by Lemma 1.1 of \cite{Mackey}). 
Next we define the set $\mathfrak{B} = \pi^{-1}(B_0) \cap B$. We call $\mathfrak{B}$ the section
of $\mathfrak{G}$ with respect to double cosets.

\label{def:decomposition.1}
\end{defin}

$\mathfrak{B}$ is measurable by Lemma 1.1 of \cite{Mackey} and by Lemmas 

\ref{lem:decomposition.1}, \ref{lem:decomposition.3}  of this Section. 
It has the property that every double coset intersects $\mathfrak{B}$ at exactly one point.
We may transfer the measure space structure $(\mathfrak{X}/G_2, \mathscr{R}_{\mathfrak{X}/G_2},
\nu)$ over to get $(\mathfrak{B}, \mathscr{R}_\mathfrak{B}, \nu_{{}_\mathfrak{B}})$. 

\begin{defin}

For each double coset $D$ 
there exists exactly one element $x_{{}_{D}} \in \mathfrak{B} \cap D$.
We define $\ud \nu_{{}_\mathfrak{B}}(x_{{}_D}) = \ud \nu_0(D)$. 
The same holds for orbits $C$: to each orbit $C$ there exists exactly one element 
$x_c \in \mathfrak{B} \cap \pi^{-1}(C)$. We put respectively
$\ud \nu_{{}_\mathfrak{B}} (x_c) = \ud \nu(C)$. Note that $x_c = x_{{}_D}$ 
iff $C$ and $D$ correspond.

\label{def:decomposition.2}
\end{defin}

\begin{lem}
A set $E$ of orbits $C$ is measurable iff the sum of
the corresponding double cosets, regarded as subsets of $\mathfrak{G}$, is measurable in $\mathfrak{G}$.
Thus in particular a function $g$ on $\mathfrak{B}$ is measurable iff there exists a function $f$ measurable on 
$\mathfrak{G}$ and constant along each double coset, such that the restriction of $f$ to $\mathfrak{B}$ 
is equal to $g$.  

\label{lem:decomposition.10}
\end{lem} 

\qedsymbol \, 
By Lemma 1.2 of \cite{Mackey} a set $F \subset \mathfrak{X} = \mathfrak{G}/G_1$
is measurable iff $A = \pi^{-1}(F)$ is measurable in $\mathfrak{G}$ and by Lemma \ref{lem:decomposition.3}
a subset $E \subset \mathfrak{X}/G_2$ is measurable iff $F = \pi_\mathfrak{X}^{-1}(E)$
is measurable on $\mathfrak{X}$. Thus a set $E$ of orbits $C$ is measurable iff the sum of
the corresponding double cosets, regarded as subsets of $\mathfrak{G}$, is measurable in $\mathfrak{G}$,
(as already claimed at the beginning of this Section). This proves the Lemma.
\qed

In particular if we define $s(x)$ to be the double coset containing
$x$, then we transfer the measure $\nu$ over to the subsets of double cosets correctly
if we define the set $E$ of double orbits to be measurable if and only if $s^{-1}(E)$ 
is measurable on $\mathfrak{G}$. 

Writing $x$ for the variable with values in $\mathfrak{G}$, and writing $[x]$
for $\pi(x)$ varying over $\mathfrak{X} = \mathfrak{G}/G_1$ we have

\begin{lem}
Let $\mu$, $\mu_C$ and $\nu$ be such as in Lemma \ref{lem:decomposition.4}. Let $g$ 
be a positive complex valued and measurable function on $\mathfrak{X}$.
Let $\mu_{D} = \mu_{x_{{}_D}} = \mu_{C_{{}_D}}$ be the measure concentrated on the orbit
$C_{{}_D}$ corresponding to the double coset $D$.Then: 

\begin{equation}\label{skew_fubini.11:decompositions}
D \mapsto \int \, g([x]) \,\, \ud \mu_{D} ([x]) \,\,\, \textrm{and} \,\,\,
\mathfrak{B} \ni x_{{}_D} \mapsto \int \, g([x]) \,\, \ud \mu_{x_{{}_D}} ([x])
\end{equation}
are measurable, and
\begin{enumerate}

\item[1)]
 if any one of the following two integrals:
\[
\int \, d\nu_0 (D) \, \int \, g([x]) \, \ud \mu_{D} ([x])  \,\,\, \textrm{and} \,\,\,  \int \, g([x]) \, \ud \mu([x]),
\]
does exist, then there exists the other and both are equal in this case. 

In particular it follows that if $g$ is integrable on $(\mathfrak{X} , \mathscr{R}_\mathfrak{X}, \mu)$
then  
\begin{equation}\label{skew_fubini.22:decompositions}
\int \, \ud \nu_0 (D) \, \int \, g([x]) \, \ud \mu_{D} ([x])  = \int \, g([x]) \, \ud \mu([x]).
\end{equation}

\item[2)]
Similarly if any one of the following two integrals:
\[
\int \, \ud \nu_{{}_\mathfrak{B}} (x_{{}_D}) \, \int \, g([x]) \, \ud \mu_{x_{{}_D}}  \,\,\, \textrm{and} \,\,\,  
\int \, g([x]) \, \ud \mu([x]),
\]
does exist, then there exists the other and both are equal in this case. 

In particular it follows that if $g$ is integrable on $(\mathfrak{X} , \mathscr{R}_\mathfrak{X}, \mu)$
then  
\begin{equation}\label{skew_fubini.222:decompositions}
\int \, \ud \nu_{{}_\mathfrak{B}} (x_{{}_D}) \, \int \, g([x]) \, \ud \mu_{x_{{}_D}} ([x])  = \int \, g([x]) \, 
\ud \mu([x]).
\end{equation}

\end{enumerate}

\label{lem:decomposition.11}
\end{lem} 

\qedsymbol \,
Because by definition (with $\mathfrak{x} \in \mathfrak{X} = \mathfrak{G}/G_1$ and $x \in \mathfrak{G}$)
\[
\int \limits_{C_{{}_D}} \, g(\mathfrak{x}) \,\, \ud \mu_{C} (\mathfrak{x}) 
=  \int \limits_{D} \, g([x]) \,\, \ud \mu_{D} ([x]),
\]
the Lemma is an immediate consequence of definitions Def \ref{def:decomposition.1} 
and \ref{def:decomposition.2} and Lemma \ref{lem:decomposition.9}. 
\qed

\vspace*{0.5cm}

\subsection{Subgroup theorem in Krein spaces}\label{subgroup}

Let $G_1$ and $G_2$ be regularly related closed subgroups of $\mathfrak{G}$ (for definition
compare Sect. \ref{decomposition}).
Consider the restriction ${}_{{}_{G_2}}U^L$ to the subgroup $G_2 \subset \mathfrak{G}$ of the representation 
${}^{\mu}U^L$ of $\mathfrak{G}$ in the Krein 
space ${}^{\mu}\mathcal{H}^L$, induced from a representation $L$ of the subgroup $H = G_1$, defined as in 
Sect \ref{def_ind_krein}.  
For each $G_2$-orbit $C$ in $\mathfrak{X} = \mathfrak{G}/G_1$ let us introduce the Krein-isometric 
representation $U^{L, C}$, 
defined in Sect. \ref{subgroup.preliminaries},
and acting in the Krein space $({\mathcal{H}^{L}_{C}}' , \mathfrak{J}^{L,C})$. 
Let $\nu$ be any pseudo image measure of 
$\mu$ on $\mathfrak{X}/G_2$, for its definition compare \cite{Bourbaki_i}, Chap. VI.3.2. For simplicity 
we drop the $\mu$ superscript in  ${}^{\mu}U^L$
and ${}^{\mu}\mathcal{H}^L$ and just write $U^L$ and $\mathcal{H}^L$. 

Let us remind the definition of the direct integral of Hilbert spaces after \cite{Segal_dec_I}, 
but compare also \cite{von_neumann_dec}:
 
\begin{defin}[Direct integral of Hilbert spaces]

Let $(\mathfrak{X}/G_2, \mathscr{R}_{\mathfrak{X}/G_2}, \nu)$ be a measure space $M$, and

suppose that for each point $C$ of $\mathfrak{X}/G_2$ there is a Hilbert space ${\mathcal{H}^{L}_{C}}'$.
A Hilbert space $\mathcal{H}^L$ is called a \emph{direct integral} of the ${\mathcal{H}^{L}_{C}}'$
over $M$, symbolically
\begin{equation}\label{dec_H^L}
\mathcal{H}^L = \int {\mathcal{H}^{L}_{C}}' \,\, \ud \nu (C), 
\end{equation}
if for each $g \in \mathcal{H}^L$ there is a function $C \mapsto g^C$ on $\mathfrak{X}/G_2$ to the disjoint union 
$\coprod \limits_{C \in \mathfrak{X}/G_2} {\mathcal{H}^{L}_{C}}'$, such that 
$g^C \in {\mathcal{H}^{L}_{C}}'$ for all $C$, and with the following properties 1) and 2):
\begin{enumerate}

\item[1)]

If $g$ and $k$ are in $\mathcal{H}^L$ and if $u = \alpha g + \beta k$, and if $\big( \cdot , \cdot \big)_C$
is the inner product in ${\mathcal{H}^{L}_{C}}'$ then 
$C \mapsto \Big( g^C , k^C \Big)_C$ is integrable on $M$, and the inner product $(g,k)$
on $\mathcal{H}^L$ is equal to 
\[
( g, k) = \int \limits_{\mathfrak{X}/G_2} \big( g^C , k^C \big)_C \,\, \ud \nu (C),  
\]
and $u^C = \alpha g^C + \beta k^C$ for almost all $C \in \mathfrak{X}/G_2$, and all $\alpha, \beta \in \mathbb{C}$.

\item[2)]

If $C \mapsto u^C$ is a function with $u^C \in {\mathcal{H}^{L}_{C}}'$ for all $C$, if $C \mapsto \big( g^C , u^C \big)_C$
is measurable for all $g \in \mathcal{H}^L$, and if $C \mapsto \big( u^C , u^C \big)_C$ is integrable
on $M$, then there exists an element $u'$ of $\mathcal{H}^L$ such that
\[
u'^C = u^C \,\,\,\textrm{almost everywhere on} \,\, M.
\]
\end{enumerate} 
The function $C \mapsto g^C$ is called the decomposition of $g$ and is symbolized by 
\[
g = \int \limits_{\mathfrak{X}/G_2} g^C \,\, \ud \nu (C).
\]

A linear operator $U$ on $\mathcal{H}^L$ is said to be decomposable with respect to the direct integral
Hilbert space decomposition (\ref{dec_H^L}) if there is a function $C \mapsto U^C$ on $\mathfrak{X}/G_2$
with $U^C$ being a linear operator in ${\mathcal{H}^{L}_{C}}'$ for each $C$, and 
\begin{enumerate}

\item[3)]
the property that for each $g$ in its domain and all $k$ in $\mathcal{H}^L$,
$(Ug)^C = U^C g^C$ almost everywhere on $M$ and the function 
$C \mapsto \big( U^C g^C , k^C \big)_C$ is integrable on $M$.

\end{enumerate}

If $U$ is densely defined the property 3) is equivalent to the following: 

\begin{enumerate}

\item[3')]
for all $g, k$ in $\mathcal{H}^L$ in the domain of $U$, $C \mapsto \big( U^C g^C , k^C \big)_C$
is integrable on $M$ and 
\[
\int \limits_{\mathfrak{X}/G_2} \big( U^C g^C , k^C \big)_C \,\, \ud \nu (C) = (Ug,k).
\]
\end{enumerate}

The function $C \mapsto U^C$ is then called the decomposition of $U$ with respect to 
(\ref{dec_H^L}) and symbolized by
\[
U = \int \limits_{\mathfrak{X}/G_2} U^C \,\, \ud \nu (C).
\]
If $C \mapsto U^C$ is almost everywhere a scalar operator, $U$ is called diagonalizable
with respect to (\ref{dec_H^L}). The totality of all bounded operators diagonalizable with respect
to (\ref{dec_H^L}) composes the commutative von Neumann algebra $\mathfrak{A}_{\mathfrak{G}/G_2}$ associated with the decomposition (\ref{dec_H^L}), compare \cite{von_neumann_dec}. A bounded operator $U$
in $\mathcal{H}^L$ is decomposable with respect to (\ref{dec_H^L}) if and only if 
it commutes with all elements of $\mathfrak{A}_{\mathfrak{G}/G_2}$ $\Leftrightarrow$
$U \in \big( \mathfrak{A}_{\mathfrak{G}/G_2} \big)'$. This condition may easily be extended on unbounded operators:
e. g. closable $U$ is decomposable with respect to (\ref{dec_H^L}) if the spectral projectors of  both the
factors in its polar decomposition commute with all elements of $\mathfrak{A}_{\mathfrak{G}/G_2}$;
or still more generally: $U$ is decomposable with respect to (\ref{dec_H^L}) $\Leftrightarrow$  $U$ is affiliated with 
the commutor $\big( \mathfrak{A}_{\mathfrak{G}/G_2} \big)'$ of $\mathfrak{A}_{\mathfrak{G}/G_2}$, 
i.e. iff it commutes with every unitary operator in the commutor
$\big( \mathfrak{A}_{\mathfrak{G}/G_2} \big)'' = \mathfrak{A}_{\mathfrak{G}/G_2}$ 
of $\big( \mathfrak{A}_{\mathfrak{G}/G_2} \big)'$.

\label{direct_int:subgroup}
\end{defin}

Note that the map $T$ which transforms $g$ into its decomposition $C \mapsto g^C$ may be viewed as a unitary
operator decomposing $U$:  
\[
T U T^{-1} = \int \limits_{\mathfrak{X}/G_2} U^C \,\, \ud \nu (C). 
\]

There are many possible realizations $T: f \mapsto T(f)$ of the Hilbert space 
$\mathcal{H}^L$ as the direct integral (\ref{dec_H^L}) all corresponding to the same commutative 
decomposition algebra $\mathfrak{A}_{\mathfrak{G}/G_2}$. However the difference between 
any two $T: f \mapsto T(f) = \Big( C \mapsto f^C \Big) $ and $T': f \mapsto T'(f) = \Big( C \mapsto \big(f^C \big)' \Big)$ 
of them is irrelevant: there exists for them a map $C \mapsto U^C$ with each $U^C$ unitary in ${\mathcal{H}^{L}_{C}}'$
and such that:

\begin{enumerate}

\item[1)]
$U^C f^C = \big(f^C \big)'$ for almost all $C$.  

\item[2)]
$C \mapsto \big(f^C , g^C \big)_C$ is measurable in realization $T$ $\Leftrightarrow$ 
$C \mapsto \Big( \, U^C f^C \, , \,\, U^C f^C \, \Big)_C$ is measurable in realization $T'$.

\end{enumerate}
(Compare \cite{von_neumann_dec}).

For the reasons explained in the footnote to Lemma \ref{lem:dense.6} it is sufficient to consider
the $\sigma$-rings $\mathscr{R}_{\mathfrak{X}/G_2}$ and $\mathscr{R}_{\mathfrak{X}}$ of Borel sets, with the 
Borel structure on $\mathfrak{X}/G_2$ defined as in Sect. \ref{decomposition}, in the investigation of the respective
Hilbert and Krein spaces.
    
We shall need a 

\begin{lem}  

\begin{enumerate}

\item[(a)] 

\[
\mathcal{H}^L \cong \int \limits_{\mathfrak{X}/G_2} {\mathcal{H}^{L}_{C}}' \,\, \ud \nu (C).
\]

\item[(b)]

\[
{}_{{}_{G_2}}U^L \cong \int \limits_{\mathfrak{X}/G_2} U^{L, C} \,\, \ud \nu (C).
\]

\item[(c)]

\[
\mathfrak{J}^L \cong \int \limits_{\mathfrak{X}/G_2} \mathfrak{J}^{L,C} \,\, \ud \nu (C).
\]

The equivalences $\cong$ are all under the same map (or realization) 
$T: \mathcal{H}^L \mapsto \int \limits_{\mathfrak{X}/G_2} {\mathcal{H}^{L}_{C}}' \,\, \ud \nu (C)$
giving the corresponding decomposition $T(f): C \mapsto f^C$
for each $f \in \mathcal{H}^L$, in which $f^C$ is the restriction of $f$ to the double 
coset $D_{{}_C} = G_1 x_c G_2 = \pi^{-1} (C)$ corresponding
\footnote{I. e. we chose $x_c \in \mathfrak{B} \subset \mathfrak{G}$ for which $\pi(x_c) \in C$, 
compare Def. \ref{def:decomposition.1} and \ref{def:decomposition.2}.} to $C$.

In particular $T$ is unitary and Krein-unitary map between the Krein spaces 
\[
(\mathcal{H}^L , \mathfrak{J}^L) \,\,\, \textrm{and} \,\,\, 
\Big( \int \limits_{\mathfrak{X}/G_2} {\mathcal{H}^{L}_{C}}' \,\,
 \ud \nu (C) , \int \limits_{\mathfrak{X}/G_2} \mathfrak{J}^{L,C} \,\, \ud \nu (C) \Big).
\]
\end{enumerate}
\label{lem:subgroup.1}
\end{lem}

\begin{rem}
The equivalences $\cong$ may be read in fact as ordinary equalities.
\label{rem:subgroup.1}
\end{rem}

\qedsymbol \,
Let 
\[
( \cdot , \cdot)_C = {\| \cdot \|_C}^2
= \int \limits_{C}
\big( \mathfrak{J}_L (\mathfrak{J}^{L,C} \, \cdot \,)_x , (\, \cdot \,)_x \big) \,\, \ud \mu_C (x)
\] 
be defined on ${\mathcal{H}^{L}_{C}}'$ as in Sect. \ref{subgroup.preliminaries}. 
Recall that for any element $g$ of $\int \limits_{\mathfrak{X}/G_2} {\mathcal{H}^{L}_{C}}' \,\, \ud \nu (C)$ i. e. a function $C \mapsto g^C$ from the set of $G_2$-orbits 
$\mathfrak{X}/G_2$ to the disjoint union 
$\coprod \limits_{C \in \mathfrak{X}/G_2} {\mathcal{H}^{L}_{C}}'$ such that 
$g^C \in {\mathcal{H}^{L}_{C}}'$ for all $C$, the function $C \mapsto {\| g^C \|_C}^2 =
(g^C , g^C)_C$ is $\nu$-summable and $\nu$-measurable and defines inner product 
for any $g, k \in \int \limits_{\mathfrak{X}/G_2} {\mathcal{H}^{L}_{C}}' \,\, \ud \nu (C)$ by the formula 
\begin{equation}\label{inn_dir_int_1:subgroup}
( g, k) = \int \limits_{\mathfrak{X}/G_2} \,\, \ud \nu (C) \, \int \limits_{C}
\big( \mathfrak{J}_L (\mathfrak{J}^{L,C} g^C)_x , k^{C}_{x} \big) \,\, \ud \mu_C (x)
= \int \limits_{\mathfrak{X}/G_2} ( g^C , k^C)_C \,\, \ud \nu (C).  
\end{equation}

We shall exhibit a natural unitary map $T$ from $\mathcal{H}^L$ onto 
$\int \limits_{\mathfrak{X}/G_2} {\mathcal{H}^{L}_{C}}' \,\, \ud \nu (C)$
or, what is equivalent, we shall show that the decomposition $T(f) = \big( C \mapsto f^C \big)$  
corresponding to each $f \in \mathcal{H}^L$, with $f^C$ equal to the restriction of $f$
to the double coset $D_{{}_C} = G_1 x_c G_2 = \pi^{-1}(C)$ corresponding to $C$, has all 
the properties required in Definition \ref{direct_int:subgroup}.

Let $f$ and $k$ be any functions in $\mathcal{H}^L$. Then by Lemma \ref{lem:decomposition.4} we have 
\[
\int \limits_{\mathfrak{X}/G_2} \,\, \ud \nu (C) \, \int \limits_{C}
\big( \mathfrak{J}_L (\mathfrak{J}^L f)_x , k_x \big) \,\, \ud \mu_C (x)
= \int \limits_{\mathfrak{X}} \big( \mathfrak{J}_L (\mathfrak{J}^L f)_x , k_x \big) \,\, \ud \mu(x)
= \| f \|^2 < \infty,
\] 
with the set of all $G_2$ orbits $C$ for which $x \mapsto \big( \mathfrak{J}_L (\mathfrak{J}^L f)_x , k_x \big)$ 
is not $\mu_C$-integrable being $\nu$-negligible and the function 
\[
C \mapsto \int \limits_{C}
\big( \mathfrak{J}_L (\mathfrak{J}^L f)_x , k_x \big) \,\, \ud \mu_C (x)
\]
being $\nu$-summable and $\nu$-measurable. Moreover, because for each orbit  $C$ the measure $\mu_C$
is concentrated on $C$ (Lemma \ref{lem:decomposition.4}), the integral 
\[
\int \limits_{C}
\big( \mathfrak{J}_L (\mathfrak{J}^L f)_x , k_x \big) \,\, \ud \mu_C (x)
\] 
is equal 
\[
\int \limits_{C}
\big( \mathfrak{J}_L ((\mathfrak{J}^L f)^C)_x , (k^C)_x \big) \,\, \ud \mu_C (x)
= \int \limits_{C}
\big( \mathfrak{J}_L (\mathfrak{J}^L f^C)_x , (k^C)_x \big) \,\, \ud \mu_C (x)
\]
where $f^C$ (and similarly for $k^C$) is the restriction of $f$ to the double coset 
$D_{{}_C} = G_1 x G_2 = \pi^{-1}(C)$ corresponding to $C$. i.e. with 
any $x$ for which\footnote{We have chosen $x = x_c$ to belong to the measurable section $\mathfrak{B}$ of double cosets 
in $\mathfrak{G}$ constructed in Sect. \ref{decomposition}, but this is unnecessary here.}
$\pi(x) \in C$, say $x = x_c$, with $C \mapsto x_c \in \mathfrak{B}$ of Sect. \ref{decomposition}. 
Because $f^C \in {\mathcal{H}^{L}_{C}}'$ and likewise $\mathfrak{J}^{L,C}$ are defined 
as the ordinary restrictions, $(\mathfrak{J}^L f)^C = \mathfrak{J}^L f^C = \mathfrak{J}^{L,C} f$ is the restriction
of $\mathfrak{J}^L f$ to the double coset 
$D_{{}_C} = G_1 x_c G_2$ corresponding to $C$. We thus obtain
\[
\int \limits_{C}
\big( \mathfrak{J}_L (\mathfrak{J}^L f)_x , k_x \big) \,\, \ud \mu_C (x) 
= \int \limits_{C}
\big( \mathfrak{J}_L (\mathfrak{J}^{L,C} f^C)_x , (k^C)_x \big) \,\, \ud \mu_C (x).
\]
Therefore it follows that the map $T: f \mapsto \big( C \mapsto f^C \big)$, where $f^C$
is the restriction of $f$ to the double coset corresponding to the orbit $C$,
fulfils the requirements of Part 1) of Definition \ref{direct_int:subgroup};
in particular $\| T(f) \| = \| f \|$ and the range $T(\mathcal{H}^L)$ is a Hilbert space
with the inner product (\ref{inn_dir_int_1:subgroup}).

We shall verify Part 2) of the Definition
\ref{direct_int:subgroup}: i. e. that the decomposition map $T(f) = \big( C \mapsto f^C \big)$
defined as above has the properties indicated in 2) of Definition
\ref{direct_int:subgroup} on its whole range 
$T(\mathcal{H}^L)$.  Toward this end let $C \mapsto u^C$ fulfil the conditions required
in 2) of Def. \ref{direct_int:subgroup}: 
\begin{equation}\label{condition1}
C \mapsto \int \limits_{C}
\Big( \mathfrak{J}_L (\mathfrak{J}^{L,C} u^C)_x , (k^{C})_{x} \Big) \,\, \ud \mu_C (x)
= \big( u^C , k^C \big)_C 
\end{equation}
is measurable for each $k \in \mathcal{H}^L$ and 
\begin{equation}\label{condition2}
C \mapsto \int \limits_{C}
\Big( \mathfrak{J}_L (\mathfrak{J}^{L,C} u^C)_x , (u^{C})_{x} \Big) \,\, \ud \mu_C (x)
=  \big( u^C , u^C \big)_C 
\end{equation}
is measurable and integrable. Consider the space $\mathfrak{F}$ of all functions $C \mapsto k^C \in {\mathcal{H}^{L}_{C}}'$ fulfilling the following conditions:
\[
C \mapsto \int \limits_{C}
\Big( \mathfrak{J}_L (\mathfrak{J}^{L,C} g^C)_x , g^{C}_{x} \Big) \,\, \ud \mu_C (x)
=  \big( k^C , k^C \big)_C 
\]
is measurable and integrable. Let $X$ be the maximal \emph{linear}  
subspace of $\mathfrak{F}$, where a subspace of $\mathfrak{F}$ we have called linear, 
whenever it is closed under formation of finite linear combinations over $\mathbb{C}$.
$X$ is not empty as it contains the subspace $T(\mathcal{H}^L)$ itself, which is a Hilbert space. Moreover
if $C \mapsto k^C, C \mapsto r^C$ are any two functions belonging to $X$ the formula
\[
\begin{split}
h\Big(\, C \mapsto k^C \, , \,\, C \mapsto r^C \, \Big) 
=  \int \limits_{\mathfrak{X}/G_2}  ( k^C , r^C)_C  \,\, \ud \nu (C)   \\
= \int \limits_{\mathfrak{X}/G_2} \Big( \int \limits_{C}
\big( \mathfrak{J}_L (\mathfrak{J}^{L,C} k^C)_x , ( r^C )_x \big) \,\, \ud \mu_C (x) \Big) \, \ud \nu(C)
\end{split}
\] 
defines a Hermitian form on $X$. Thus by the Cauchy-Schwarz inequality we have:
\begin{equation}\label{Cauchy-Schwartz:subgroup}
\Big{|} \int \limits_{\mathfrak{X}/G_2}  ( k^C , r^C)_C  \,\, \ud \nu (C) \Big{|}^2
\leq \Big( \int \limits_{\mathfrak{X}/G_2}  ( k^C , k^C)_C  \,\, \ud \nu (C)  \Big)
\cdot \Big( \int \limits_{\mathfrak{X}/G_2}  ( r^C , r^C)_C  \,\, \ud \nu (C)  \Big).
\end{equation}
Now by the first part of the proof, $T(\mathcal{H}^L)$ is a Hilbert space with the inner product
(\ref{inn_dir_int_1:subgroup}) and in particular a linear subspace of $\mathfrak{F}$.
We may thus insert for $C \mapsto k^C$ in (\ref{Cauchy-Schwartz:subgroup}) any
decomposition $C \mapsto f^C$ of $f \in \mathcal{H}^{L}$, with $f^C$ equal to the restriction 
of $f$ to the double coset $D_{{}_C} = \pi^{-1} (C)$ corresponding to 
$C$. Similarly we may insert the function $C \mapsto u^C$ for the function $C \mapsto r^C$ in (\ref{Cauchy-Schwartz:subgroup}). 
Indeed, because of the conditions (\ref{condition1}) and (\ref{condition2}), fulfilled by 
the function $C \mapsto u^C$, the function
\[
C \mapsto \big( f^C + u^C , f^C + u^C \big)_C  =
\big( f^C , f^C \big)_C 
+ \big( f^C ,  u^C \big)_C 
+ \big( u^C , f^C  \big)_C 
+ \big(  u^C , u^C \big)_C 
\]
is measurable and by the Cauchy-Schwarz inequality integrable, for all $f \in \mathcal{H}^L$.
Therefore $C \mapsto u^C$ and $T(\mathcal{H}^L)$ are both contained in one linear subspace of 
$\mathfrak{F}$, and thus by the maximality of $X$ they are contained in $X$, so that we can insert
$C \mapsto u^C$ for $C \mapsto r^C$ in (\ref{Cauchy-Schwartz:subgroup}).
Thus the indicated insertions in the inequality (\ref{Cauchy-Schwartz:subgroup}) lead us to the inequality
\[
\Big{|} \int \limits_{\mathfrak{X}/G_2} ( f^C , u^C)_C  \,\, \ud \nu (C) \Big{|}^2
\leq \Big( \int \limits_{\mathfrak{X}/G_2}  ( f^C , f^C)_C  \,\, \ud \nu (C)  \Big)
\cdot \Big( \int \limits_{\mathfrak{X}/G_2}  ( u^C , u^C)_C  \,\, \ud \nu (C)  \Big)
\]
for all $C \mapsto f^C$ in $T(\mathcal{H}^L)$. 
Therefore the linear functional 
\[
T(f) \mapsto L \big( \, T(f)  \, \big) = L \Big( \,  C \mapsto f^C  \, \Big)
= h\Big(\, C \mapsto f^C  \, , \,\,  C \mapsto u^C  \, \Big), 
\]
on $T(\mathcal{H}^L)$ is bounded by the last inequality. Because the range $T(\mathcal{H}^L)$ of $T$
is a Hilbert space it follows by the Riesz theorem ((e. g. Corollary 8.3.2. of \cite{Segal_Kunze}) applied 
to the linear functional $L$ that there exists exactly one element $T(f')$ in the range of 
$T$ such that 
\[
\begin{split}
(f,f') = \big( \, T(f) \, , \,\, T(f') \, \big) \,\,\,\,\,\,\,\,\,\,\,\,\,\,\,\,\,\,\,\,\,\,\,\,\,\,\,
\,\,\,\,\,\,\,\,\,\,\,\,\,\,\,\,\,\,\,\,\,\,\,\,\,\,\,\,\,\,\,\,\,\,\,\,\,\,\,\,\,\,\,\,\,\,\,\,\,
\,\,\,\,\,\,\,\,\,\,\,\,\,\,\,\,\,\,\,\,\,\,\,\,\,\,\,\,\,\,\,\,\,\,\,\,  \\ 
= \int \limits_{\mathfrak{X}/G_2} ( f^C , f'^C)_C  \,\, \ud \nu (C) 
= h\Big(\, C \mapsto f^C \, , \,\, C \mapsto u^C \, \Big)
\end{split}
\] 
for all $f \in \mathcal{H}^L$. Therefore
\[
\int \limits_{\mathfrak{X}/G_2}  \big( f^C , f'^C \big)_C  \,\, \ud \nu (C) 
= \int \limits_{\mathfrak{X}/G_2}  \big( f^C , u^C \big)_C  \,\, \ud \nu (C) 
\] 
for all $f \in \mathcal{H}^L$ and for a fixed $f' \in \mathcal{H}^L$, or equivalently 
\[
\int \limits_{\mathfrak{X}/G_2}  \big( f^C , f'^C - u^C \big)_C  \,\, \ud \nu (C) = 0,
\] 
for all $f \in \mathcal{H}^L$. Inserting the definition of $\big( f^C , f'^C - u^C \big)_C $
we get:
\begin{equation}\label{f_j_dense:subgroup}
\begin{split}
\int \limits_{\mathfrak{X}/G_2} \,\, \int \limits_{C}
\Big( \mathfrak{J}_L (\mathfrak{J}^{L,C} f^C)_x , (f'^C - u^C)_{x} \Big) \,\, \ud \mu_C (x)  \,\, \ud \nu (C) 
\,\,\,\,\,\,\,\,\,\,\,\,\,\,\,\,\,\,\,\,\,\,\,\,\,\,\,\,\,\,\,\,\,\,\,\,\,\,\,\,\,\,\,\,\,\,\,\,
\,\,\,\,\,\,\,\,\,\,\,\,\,\,  \\
=\int \limits_{\mathfrak{X}/G_2} \,\, \int \limits_{C}
\Big( \mathfrak{J}_L (\mathfrak{J}^L f)_x , (f'^C - u^C)_{x} \Big) \,\, \ud \mu_C (x)  \,\, \ud \nu (C) = 0,
\end{split}
\end{equation}
for all $f \in \mathcal{H}^L$. By Lemma \ref{lem:dense.6} there exists a sequence $f^1 , f^2 , \ldots$ of elements $C^{L}_{0} \subset \mathcal{H}^L$ such that for each fixed $x \in \mathfrak{G}$ the vectors $f^{k}_{x}$, $k = 1, 2, \ldots$ form a dense linear subspace of $\mathcal{H}_L$. By the proof of the same Lemma \ref{lem:dense.6} there exists 
a sequence $g_1 , g_2 , \ldots$ of continuous complex valued functions on $\mathfrak{X} = \mathfrak{G}/G_1$ with compact 
supports, dense in $L^2 (\mathfrak{X},\mu)$ with respect to the $L^2$ norm 
$\| \cdot \|_{L^2}$. For each $g_j$ define the
corresponding function $g'_j$ on $\mathfrak{G}$ by the formula $g'_j (x) = g_j (\pi (x))$, where
$\pi$ is the canonical quotient map $\mathfrak{G} \mapsto \mathfrak{G}/G_1 = \mathfrak{X}$. Note, please, that
$\Big( \mathfrak{J}_L (\mathfrak{J}^L g'_j \cdot f)_{{}_x} , (f'^C - u^C)_{{}_x} \Big)
= (g'_j)_{{}_x} \cdot \Big( \mathfrak{J}_L (\mathfrak{J}^L \cdot f)_{{}_x} , (f'^C - u^C)_{{}_x} \Big)$ for all 
$j \in \mathbb{N}$ and all $f \in \mathcal{H}^L$. Inserting now $g'_j \cdot f^i$ for $f$ in 
(\ref{f_j_dense:subgroup}) we get
\[
\int \limits_{\mathfrak{X}/G_2} \,\, g_j (C) \cdot \int \limits_{C} 
\Big( \mathfrak{J}_L (\mathfrak{J}^L f^i)_{{}_x} , (f'^C - u^C)_{{}_x} \Big) \,\, \ud \mu_C (x)  \,\, \ud \nu (C) = 0,
\]  
for all $i,j \in \mathbb{N}$. Because $\{g_j\}_{j \in \mathbb{N}}$ is dense in $L^2 (\mathfrak{X},\mu)$
and the function
\[
C \mapsto \int \limits_{C} 
\Big( \mathfrak{J}_L (\mathfrak{J}^L f^i)_{{}_x} , (f'^C - u^C)_{{}_x} \Big) \,\, \ud \mu_C (x)
\]
by construction belongs to $L^2 (\mathfrak{X},\mu)$, it follows that outside a $\nu$-negligible
subset $N$ of orbits $C$
\[
\int \limits_{C} 
\Big( \mathfrak{J}_L (\mathfrak{J}^L f^i)_{{}_x} , (f'^C - u^C)_{{}_x} \Big) \,\, \ud \mu_C (x)  = 0,
\]  
for all $i \in \mathbb{N}$. Thus if $C \notin N$, then
\begin{equation}\label{*f_j_dense:subgroup}
\int \limits_{C} 
\Big( \mathfrak{J}_L (\mathfrak{J}^L f^i)_{{}_x} , (f'^C - u^C)_{{}_x} \Big) \,\, \ud \mu_C (x)  = 0,
\end{equation} 
for all $i \in \mathbb{N}$. Applying Lemma \ref{lem:subgroup.preliminaries.1} to this orbit $C$ 
and the associated ${\mathcal{H}^{L}_{C}}'$ we get an isomorphism of it with a Krein space 
$\mathcal{H}^{L^{x_c}}$  of an induced representation (recall that $x_c \in \mathfrak{B} \subset \mathfrak{G}$
with $\pi(x_c) \in C$, compare Def. \ref{def:decomposition.2}). 
Then (\ref{*f_j_dense:subgroup}) together with Lemma \ref{lem:subgroup.preliminaries.3} and 
Lemma \ref{lem:dense.3} or \ref{lem:dense.4} applied to $\mathcal{H}^{L^{x_c}}$
gives $f'^C - u^C = 0$. This shows that the decomposition $T: f \mapsto \big( C \mapsto f^C \big)$
fulfils Part 2) of Definition \ref{direct_int:subgroup}. We have thus proved 
Part (a) of the Lemma.

Then we have to prove that the operators $T \, {}_{{}_{G_2}}U^L \, T^{-1}$ and $T \, \mathfrak{J}^L \, T^{-1}$
are decomposable with respect to (\ref{dec_H^L}) and $C \mapsto U^{L, C}$ and
$C \mapsto \mathfrak{J}^{L,C}$ are their respective decompositions. Let  $\eta \in G_2$.
Writing $\lambda(\eta)$ for the $\lambda$-function $[x] \mapsto \lambda([x], \eta)$ 
corresponding to the measure $\mu$ and analogously writing $\lambda_C (\eta)$  for the $\lambda_C$ function 
$[x] \mapsto \lambda_C ([x], \eta)$ corresponding to $\mu_C$ we have: 
\[
\begin{split}
\Big( T \, {}_{{}_{G_2}}U^{L}_{\eta} \, T^{-1} \Big) \Big( C \mapsto f^C \Big)
 = \big( T \, {}_{{}_{G_2}}U^{L}_{\eta} \big) \big(f\big)  \\
 =  T \big( \sqrt{\lambda(\eta)}  R_\eta f \big) = \Big( \, C \mapsto \sqrt{\lambda(\eta)|_{{}_C}}  R_\eta f^C \, \Big),
\end{split}
\]
where $\lambda(\eta)|_{{}_C}$ denotes the restriction of $\lambda(\eta)$ to the orbit $C$. By 
Lemma \ref{lem:decomposition.8} the restriction $\lambda(\eta)|_{{}_C}$ of $\lambda(\eta)$ to the orbit $C$
is equal to $\lambda_C (\eta)$, so that
\[
\Big( T \, {}_{{}_{G_2}}U^{L}_{\eta} \, T^{-1} \Big) \Big( C \mapsto f^C \Big)
= \Big( C \mapsto \sqrt{\lambda_C (\eta)}  R_\eta f^C \Big)
= \Big( C \mapsto U^{L, C}_{\eta} f^C \Big),
\]
which means that
\[
{}_{{}_{G_2}}U^L \cong \int \limits_{\mathfrak{X}/G_2} U^{L, C} \,\, \ud \nu (C),
\]
and proves (b). Similarly for the operator $\mathfrak{J}^L$:
\[
\begin{split}
\Big( T \, \mathfrak{J}^L \, T^{-1} \Big) \Big( C \mapsto f^C \Big)
 = \big(T \, \mathfrak{J}^L \big) \big(f\big) \\
 =  T \big( \mathfrak{J}^L f \big) = \Big( \, C \mapsto \big(\mathfrak{J}^L f\big)^C \, \Big).
\end{split}
\]
By definition of the operator $\mathfrak{J}^{L,C}$ we have $\big(\mathfrak{J}^L f\big)^C
= \mathfrak{J}^L f^C = \mathfrak{J}^{L,C} f^C$. Therefore
\[
\Big( T \, \mathfrak{J}^L \, T^{-1} \Big) \Big( C \mapsto f^C \Big)
= \Big( \, C \mapsto \mathfrak{J}^{L,C} f^C \, \Big),
\]
which means that
\[
\mathfrak{J}^L \cong \int \limits_{\mathfrak{X}/G_2} \mathfrak{J}^{L,C} \,\, \ud \nu (C),
\]
and proves (c).

Because for each $C$, $\mathfrak{J}^{L,C}$ is unitary and self adjoint in ${\mathcal{H}^{L}_{C}}'$
and $\big( \mathfrak{J}^{L,C} \big)^2 = I$,
then by \cite{von_neumann_dec}, \S 14, the same holds true for the operator
\[
\int \limits_{\mathfrak{X}/G_2} \mathfrak{J}^{L,C} \,\, \ud \nu (C) \,\,\, \textrm{in} \,\,\, 
\int \limits_{\mathfrak{X}/G_2} {\mathcal{H}^{L}_{C}}' \,\, \ud \nu (C), 
\]
so that 

\[
\Big( \int \limits_{\mathfrak{X}/G_2} {\mathcal{H}^{L}_{C}}' \,\,
 \ud \nu (C) , \int \limits_{\mathfrak{X}/G_2} \mathfrak{J}^{L,C} \,\, \ud \nu (C) \Big),
\]
is a Krein space.

Finally we have to show that $T$ is Krein unitary. To this end observe that for each $f,g \in \mathcal{H}^L$
\[
\begin{split}
\Big( T(f), T(g) \Big)_{\int \mathfrak{J}^{L,C} \, \ud \nu (C)}
= \int \limits_{\mathfrak{X}/G_1}  \Big( \mathfrak{J}^{L,C} f^C , g^C \Big)_C \,\, \ud \nu (C) \\
= \int \limits_{\mathfrak{X}/G_1} \int \limits_{C} \Big( \mathfrak{J}_L \big( (\mathfrak{J}^{L,C})^2 f^C\big)_{{}_x} , 
\big( g^C \big)_{{}_x} \Big)_C \,\, \ud \mu_C (x) \,\, \ud \nu (C) \\ 
= \int \limits_{\mathfrak{X}/G_1} \int \limits_{C} \Big( \mathfrak{J}_L \big( f^C\big)_{{}_x} , 
\big( g^C \big)_{{}_x} \Big) \,\, \ud \mu_C (x) \,\, \ud \nu (C).
\end{split}
\] 
Because $f^C$ and $g^C$ are the ordinary restrictions of $f$ and $g$ to $G_1 x_c G_2$ and the measure
$\mu_C$ is concentrated on $C$ (Lemma \ref{lem:decomposition.4}), the integrand in the last integral may be 
replaced with $\Big( \mathfrak{J}_L \big( f \big)_{{}_x} , 
\big( g \big)_{{}_x} \Big)$. Because $f,g \in \mathcal{H}^L$, the function 
$x \mapsto \Big( \mathfrak{J}_L \big( f \big)_{{}_x} \big( g \big)_{{}_x} \Big)$ is constant 
on the right $G_1$-cosets and measurable and integrable on $\mathfrak{X} = \mathfrak{G}/G_1$
as a function of right $G_1$-cosets. Thus by Lemma \ref{lem:decomposition.9} the last integral is equal to
\[
\begin{split}
\int \limits_{\mathfrak{X}/G_1} \int \limits_{C} \Big( \mathfrak{J}_L \big( f \big)_{{}_x} , 
\big( g \big)_{{}_x} \Big) \,\, \ud \mu_C (x) \,\, \ud \nu (C) 
= \int \limits_{\mathfrak{X}} \Big( \mathfrak{J}_L \big( f \big)_{{}_x} , 
\big( g \big)_{{}_x} \Big) \,\, \ud \mu (x)
= \big( f , g \big)_{\mathfrak{J}^L},
\end{split}
\] 
so that 
\[
\Big( T(f), T(g) \Big)_{\int \mathfrak{J}^{L,C} \, \ud \nu(C)} 
= \big( f , g \big)_{\mathfrak{J}^L}.
\]
\qed

Actually we could merely use all  $g' \cdot f$, with $g \in C_{\mathcal{K}}(\mathfrak{X})$ and 
$f \in C^{L}_{0}$ instead of its denumerable subset $g'_j \cdot f^i$ , $i,j \in \mathbb{N}$
in the proof of Lemma \ref{lem:subgroup.1}. Its denumerability
shows that $\mathcal{H}^L$ is separable as the direct integral (a). This however is superfluous because 
separability of $\mathcal{H}^L$ has been already shown within the proof of Lemma \ref{lem:dense.6}.

\vspace*{0.5cm}

\begin{lem}

Let $\mathfrak{B}$ be the section of $\mathfrak{G}$ with respect to double cosets
of Def. \ref{def:decomposition.1} and let $C \mapsto x_c \in \mathfrak{B}$, $D \mapsto x_{{}_D} \in \mathfrak{B}$
be the bi-unique maps of Def. \ref{def:decomposition.2}.

Let $\nu_0$ be the measure on the subsets of the set $G_1 : G_2$ of all double cosets $D$
equal to the transfer of the measure $\nu$ on $\mathfrak{X}/G_2$ over
to the set of double cosets by the natural bi-unique map $C \mapsto D_{{}_C} = \pi^{-1}(C)$. 
Let $\nu_{\mathfrak{B}}$ be the measure on the section $\mathfrak{B}$ equal to the transfer
of $\nu$ over to the section $\mathfrak{B}$ by the map $C \mapsto x_c$ (or equivalently
equal to the transfer of $\nu_0$ by the map $D \mapsto x_{{}_D}$). 
Let $\mu_D = \mu_{C_{{}_D}}$, where $C_{{}_D}$ is the orbit corresponding to the double coset $D$,
be the measure concentrated on $C_{{}_D}$, where $\mu_C$ is the measure of Lemma \ref{lem:decomposition.4}.
Let us denote the space of functions
${\mathcal{H}^{L}_{C}}'$ of Sect. \ref{subgroup.preliminaries}, defined on the double coset $D$
corresponding to $C$ just by $\mathcal{H}^{L}_{D}$ and similarly if $U^{L, C}$ and 
$\mathfrak{J}^{L,C}$ is the representation and the operator
of Sect. \ref{subgroup.preliminaries}, then we put $U^{L, D} = U^{L, C_{{}_D}}$
and $\mathfrak{J}^{L,D} = \mathfrak{J}^{L,C_{{}_D}}$;
analogously we define  $U^{L, x_{{}_D}} = U^{L, C_{{}_D}}$
and $\mathfrak{J}^{L, x_{{}_D}} = \mathfrak{J}^{L, C_{{}_D}}$. Then we have

\begin{enumerate}

\item[(a)] 

\[
\mathcal{H}^L \cong \int \limits_{\mathfrak{X}/G_2} {\mathcal{H}^{L}_{C}}' \,\, \ud \nu (C)
= \int \limits_{G_1 : G_2} \mathcal{H}^{L}_{D} \,\, \ud \nu_0 (D)
= \int \limits_{\mathfrak{B}} \mathcal{H}^{L}_{x_{{}_D}} \,\, \ud \nu_{{}_\mathfrak{B}} (x_{{}_D}).
\]

\item[(b)]

\[
{}_{{}_{G_2}}U^L \cong \int \limits_{\mathfrak{X}/G_2} U^{L, C} \,\, \ud \nu (C)
= \int \limits_{G_1 : G_2} U^{L, D} \,\, \ud \nu_0 (D)
= \int \limits_{\mathfrak{B}} U^{L, x_{{}_D}} \,\, 
\ud \nu_{{}_\mathfrak{B}} (x_{{}_D}).
\]

\item[(c)]

\[
\mathfrak{J}^L \cong \int \limits_{\mathfrak{X}/G_2} \mathfrak{J}^{L,C} \,\, \ud \nu (C)
= \int \limits_{G_1 : G_2} \mathfrak{J}^{L,D} \,\, \ud \nu_0 (D)
= \int \limits_{\mathfrak{B}} \mathfrak{J}^{L,x_{{}_D}} \,\, 
\ud \nu_{{}_\mathfrak{B}} (x_{{}_D}).
\]

The equivalences $\cong$ are all under the same map 
$T: \mathcal{H}^L \mapsto \int \limits_{G_1 : G_2} \mathcal{H}^{L}_{D} \,\, \ud \nu (C)$
giving the corresponding decomposition $T(f): D \mapsto f^{C_{{}_D}}$ (or respectively
$T(f): x_{{}_D}  \mapsto f^{C_{{}_D}}$)
for each $f \in \mathcal{H}^L$, in which $f^{C_{{}_D}}$ is the restriction of $f$ to the double 
coset $D = D_{{}_{C_{{}_D}}} = G_1 x_{{}_D} G_2 = \pi^{-1} (C_{{}_D})$ corresponding to $C_{{}_D}$.
In particular $T$ is unitary and Krein-unitary map between the Krein spaces 
\[
(\mathcal{H}^L , \mathfrak{J}^L)
\]
and
\[
\Big( \int \limits_{G_1 : G_2} \mathcal{H}^{L}_{D} \,\,
 \ud \nu (C) , \int \limits_{G_1 : G_2} \mathfrak{J}^{L,D} \,\, \ud \nu_0 (D) \Big)
\]
or respectively
\[
\Big( \int \limits_\mathfrak{B} \mathcal{H}^{L}_{x_{{}_D}} \,\,
 \ud \nu_{{}_\mathfrak{B}} (x_{{}_D}) , \int \limits_{\mathfrak{B}} \mathfrak{J}^{L,x_{{}_D}} \,\, 
\ud \nu_{{}_\mathfrak{B}} (x_{{}_D}) \Big).
\]

\end{enumerate}

\label{lem:subgroup.2}
\end{lem}

\qedsymbol \,
The Lemma follows from Lemma \ref{lem:subgroup.1} by a mere renaming of the 
points of the measure space $\mathfrak{X}/G_2$ of $G_2$-orbits $C$ in $\mathfrak{X}$, with the preservation
of the measure structure under the indicated renaming, which is guaranteed by Def. \ref{def:decomposition.1}
and \ref{def:decomposition.2}. 
\qed

\vspace*{0.5cm}

\begin{lem}
Let $\big( \, {}^{\mu^{x_c}}\mathcal{H}^{L^{x_c}} \, , \,\, \mathfrak{J}_{x_c} \big)$
be the Krein space of the representation ${}^{\mu^{x_c}}U^{L^{x_c}}$ of the subgroup $G_2$ defined
in Lemma \ref{lem:subgroup.preliminaries.1} with the inner product $(\cdot,\cdot)_{x_c}$ in 
${}^{\mu^{x_c}}\mathcal{H}^{L^{x_c}}$ defined by eq. (\ref{g_2'_def_inn}) in the proof of 
Lemma \ref{lem:subgroup.preliminaries.1}. For each $x_{{}_D} \in \mathfrak{B}$ we put 
${}^{\mu^{x_{{}_D}}}\mathcal{H}^{L^{x_{{}_D}}} = {}^{\mu^{x_c}}\mathcal{H}^{L^{x_c}}$,
$\mathfrak{J}_{x_{{}_D}} = \mathfrak{J}_{x_c}$, $G_{{}_{x_{{}_D}}} = G_{x_c}$
and $\big(\cdot , \cdot \big)_{x_{{}_D}} = (\cdot,\cdot)_{x_c}$
with the orbit $C$ corresponding to $D$. For each fixed element $f \in \mathcal{H}^L$
consider the following function 
\[
\mathfrak{B} \ni x_{{}_D} \mapsto {\widetilde{f}}^{{}^{{}^{x_{{}_D}}}} \in \,\,\, {}^{\mu^{x_{{}_D}}}\mathcal{H}^{L^{x_{{}_D}}}
\]
where for each $x_{{}_D}$, ${\widetilde{f}}^{{}^{{}^{x_{{}_D}}}}$ is defined as the function
\[
G_2  \ni t \mapsto 
\big( {\widetilde{f}}^{{}^{{}^{x_{{}_D}}}} \big)_{{}_t} = 
\big( f^D \big)_{{}_{x_{{}_D} \cdot t }},
\]   
with $f^D$ equal to the restriction of $f$ to $D$. The linear set $\mathcal{H}$ of all such functions 
$x_{{}_D} \mapsto {\widetilde{f}}^{{}^{{}^{x_{{}_D}}}}$ with $f$ ranging over the whole space $\mathcal{H}^L$ and with the inner product 
\begin{equation}\label{inn_prod_D:subgroup}
(\widetilde{f}, \widetilde{g}) = \int \limits_\mathfrak{B} 
\big( {\widetilde{f}}^{{}^{{}^{x_{{}_D}}}} , {\widetilde{g}}^{{}^{{}^{x_{{}_D}}}}  \big)_{x_{{}_D}} \,\,
 \ud \nu_{{}_\mathfrak{B}}(x_{{}_D}),
\end{equation}  
is equal to
\[
\int \limits_\mathfrak{B}  \,\,\,\, {}^{\mu^{x_{{}_D}}}\mathcal{H}^{L^{x_{{}_D}}} \,\,\,\,\,\, \ud \nu_{{}_\mathfrak{B}}(x_{{}_D}).
\]

\label{lem:subgroup.3}
\end{lem}
\qedsymbol \, Note, please, that by definition of the measures $\mu^{x_{{}_D}}$ and 
the operators $\mathfrak{J}_{x_{{}_D}}$ 
\begin{multline*}
\big( {\widetilde{f}}^{{}^{{}^{x_{{}_D}}}} , {\widetilde{g}}^{{}^{{}^{x_{{}_D}}}}  \big)_{x_{{}_D}} 
= \int \limits_{G_2 / G_{{}_{x_{{}_D}}}} 
\Big( \mathfrak{J}_L \big( \mathfrak{J}_{x_{{}_D}} {\widetilde{f}}^{{}^{{}^{x_{{}_D}}}} \big)_{{}_t} , 
\big( {\widetilde{g}}^{{}^{{}^{x_{{}_D}}}} \big)_{{}_t}  \Big)  \,\, \ud \mu^{x_{{}_D}} ([t]) \\
= \int \limits_{G_2 / G_{{}_{x_{{}_D}}}} 
\Big( \mathfrak{J}_L L_{h(x_{{}_D} \cdot t)}  \mathfrak{J}_L  L_{h(x_{{}_D} \cdot t)^{-1}} 
\big( f^D \big)_{{}_{x_{{}_D} \cdot t}} , 
\big( g^D \big)_{{}_{x_{{}_D} \cdot t}}  \Big) \,\, \ud \mu^{x_{{}_D}} ([t])     \\
=  \int \limits_{D}
\Big( \mathfrak{J}_L \big( \mathfrak{J}^L  f^D \big)_{{}_x} , 
\big(g^D \big)_{{}_x}  \Big)  \,\, \ud \mu_D ([x])
=\int \limits_{D}
\Big( \mathfrak{J}_L \big( \mathfrak{J}^L f\big)_{{}_x} , g_{{}_x}  \Big)  \,\, \ud \mu_D ([x])
\end{multline*}
and because 
\[
\mathfrak{G}/G_1 \ni [x] \mapsto  \Big( \mathfrak{J}_L  \big( \mathfrak{J}^L  f \big)_{{}_{[x]}} , 
 g_{{}_{[x]}}  \Big)
=  \Big( \mathfrak{J}_L \big( \mathfrak{J}^L  f \big)_{{}_x} , g_{{}_x}  \Big)
\]
is measurable it follows from (\ref{skew_fubini.1:decompositions}) of  
Lemma \ref{lem:decomposition.11} that the function
\[
x_{{}_D} \mapsto  \big( {\widetilde{f}}^{{}^{{}^{x_{{}_D}}}} , {\widetilde{g}}^{{}^{{}^{x_{{}_D}}}}  \big)_{x_{{}_D}} 
\]
is measurable for all $f,g \in \mathcal{H}^L$. Similarly by (\ref{skew_fubini.2:decompositions}) 
of part 2) of  Lemma  \ref{lem:decomposition.11}

\begin{multline*}
(\widetilde{f}, \widetilde{g}) = \int \limits_\mathfrak{B}  
\big( {\widetilde{f}}^{{}^{{}^{x_{{}_D}}}} , {\widetilde{g}}^{{}^{{}^{x_{{}_D}}}}  \big)_{x_{{}_D}} \,\,
 \ud \nu_{{}_\mathfrak{B}}(x_{{}_D}) \\
 = \int \limits_\mathfrak{B} \,\,  \int \limits_{G_2 / G_{{}_{x_{{}_D}}}} 
\Big( \mathfrak{J}_L \big( \mathfrak{J}_{x_{{}_D}} {\widetilde{f}}^{{}^{{}^{x_{{}_D}}}} \big)_{{}_t} , 
\big( {\widetilde{g}}^{{}^{{}^{x_{{}_D}}}} \big)_{{}_t}  \Big)_{x_{{}_D}}  \,\, \ud \mu^{x_{{}_D}} ([t])
 \,\,\, \ud \nu_{{}_\mathfrak{B}}(x_{{}_D})   \\
\int \limits_\mathfrak{B} \,\,  \int \limits_{D}
\Big( \mathfrak{J}_L \big( \mathfrak{J}^L f\big)_{{}_x} , g_{{}_x}  \Big)  \,\, \ud \mu_D ([x])
\,\,\, \ud \nu_{{}_\mathfrak{B}}(x_{{}_D}) 
= \int \limits_{\mathfrak{G}/G_1}
\Big( \mathfrak{J}_L \big( \mathfrak{J}^L f\big)_{{}_x} , g_{{}_x}  \Big)  \,\, \ud \mu ([x]) \\
= (f,g).
\end{multline*}
Therefore $\mathcal{H}$ is a Hilbert space with the inner product (\ref{inn_prod_D:subgroup}) as the 
isometric image of the Hilbert space 
$\mathcal{H}^L$. We need only show Part 2) of Def. \ref{direct_int:subgroup} to be fulfilled.
Toward this end let $x_{{}_D} \mapsto {u}^{{}^{{}^{x_{{}_D}}}} \in {}^{\mu^{x_{{}_D}}}\mathcal{H}^{L^{x_{{}_D}}}$ be a function fulfilling the conditions of Part 2) of Def. \ref{direct_int:subgroup} (of course with the obvious replacements 
of $C$ with $D$ and ${\mathcal{H}^{L}_{C}}'$ with ${}^{\mu^{x_{{}_D}}}\mathcal{H}^{L^{x_{{}_D}}}$). We have to show
existence of a function $f' \in \mathcal{H}^L$ such that the function 
$x_{{}_D} \mapsto {\widetilde{f'}}^{{}^{{}^{x_{{}_D}}}}$ is equal almost everywhere
to the function $x_{{}_D} \mapsto {u}^{{}^{{}^{x_{{}_D}}}}$. We proceed exactly as in the proof
of Part (a) of Lemma \ref{lem:subgroup.1} by formation of the analogous maximal linear subspace $X$ in the
space $\mathfrak{F}$ of all functions $x_{{}_D} \mapsto {k}^{{}^{{}^{x_{{}_D}}}}$ for which
\[
x_{{}_D} \mapsto \big( {k}^{{}^{{}^{x_{{}_D}}}} , 
{k}^{{}^{{}^{x_{{}_D}}}}  \big)_{x_{{}_D}} 
\] 
is measurable and integrable and then using Riesz theorem and Lemma \ref{lem:dense.3} or 
\ref{lem:dense.4} in proving the existence of $f'$ (in this case the proof is even simpler because 
the Lemma \ref{lem:subgroup.preliminaries.1} is not necessary in proving 
${\widetilde{f'}}^{{}^{{}^{x_{{}_D}}}} - {u}^{{}^{{}^{x_{{}_D}}}} = 0$ from the analogue of (\ref{*f_j_dense:subgroup});
indeed it is sufficient to apply Lemma \ref{lem:subgroup.preliminaries.3} and Lemma \ref{lem:dense.3} or 
\ref{lem:dense.4}).  
\qed

From now on we identify the Hilbert space $\mathcal{H}^L$ with the direct integral:
\[
\mathcal{H}^L = \int \limits_{\mathfrak{B}} \mathcal{H}^{L}_{x_{{}_D}} \,\, \ud \nu_{{}_\mathfrak{B}} (x_{{}_D}).
\]
with the realization $T \mapsto T(f)$ of the direct integral equal to $T(f): x_{{}_D} \mapsto f^D$, where $f^D$
is the ordinary restriction of $f \in \mathcal{H}^L$ to the double coset $D$. Similarly by
\[
\int \limits_\mathfrak{B}  \,\,\,\, {}^{\mu^{x_{{}_D}}}\mathcal{H}^{L^{x_{{}_D}}} \,\,\,\,\,\, \ud \nu_{{}_\mathfrak{B}}(x_{{}_D}),
\]
we understand the direct integral with the realization of Lemma \ref{lem:subgroup.3}.

\begin{lem}
For each orbit $C$ let $V_{x_c}$ be the Krein-unitary map defined in Lemma \ref{lem:subgroup.preliminaries.1}.
For each $x_{{}_D} \in \mathfrak{B}$ (equivalently: each double coset $D$) let us put $V_{x_{{}_D}} = V_{x_c}$ 
with $C$ corresponding to $D$. Then $x_{{}_D} \mapsto V_{x_{{}_D}}$ is a decomposition of a well defined operator
\begin{multline*}
\mathcal{H}^L = \int \limits_{\mathfrak{B}} \mathcal{H}^{L}_{x_{{}_D}} \,\, \ud \nu_{{}_\mathfrak{B}} (x_{{}_D}) 
\xrightarrow{V}
\int \limits_\mathfrak{B}  \,\,\,\, {}^{\mu^{x_{{}_D}}}\mathcal{H}^{L^{x_{{}_D}}} \,\,\,\,\,\, 
\ud \nu_{{}_\mathfrak{B}}(x_{{}_D}) :   \\
\Big( x_{{}_D} \mapsto {f}^{{}^{{}^{x_{{}_D}}}} \Big) \mapsto 
\Big( x_{{}_D} \mapsto  V_{x_{{}_D}}  {f}^{{}^{{}^{x_{{}_D}}}}  \Big).
\end{multline*}
In short
\[
V = \int \limits_{\mathfrak{B}} V_{x_{{}_D}} \,\, 
\ud \nu_{{}_\mathfrak{B}} (x_{{}_D}).
\]
The operator $V$ is unitary and Krein-unitary between the Krein spaces
\begin{multline*}
\Big( \int \limits_\mathfrak{B}  \,\,\,\, {}^{\mu^{x_{{}_D}}}\mathcal{H}^{L^{x_{{}_D}}} \,\,\,\,\,\, 
\ud \nu_{{}_\mathfrak{B}}(x_{{}_D}) \,\,
 , \,\,\,\, \int \limits_\mathfrak{B}  \,\,\,\, \mathfrak{J}_{x_{{}_D}} \,\,\,\,\,\, 
\ud \nu_{{}_\mathfrak{B}}(x_{{}_D}) \,\, \Big)   \\       
\,\,\, \textrm{and} \,\,\, 
\Big( \int \limits_\mathfrak{B} \mathcal{H}^{L}_{x_{{}_D}} \,\,
 \ud \nu_{{}_\mathfrak{B}} (x_{{}_D}) \,\, , \,\,\,\, \int \limits_{\mathfrak{B}} \mathfrak{J}^{L,x_{{}_D}} \,\, 
\ud \nu_{{}_\mathfrak{B}} (x_{{}_D}) \,\, \Big) =(\mathcal{H}^L , \mathfrak{J}^L);
\end{multline*}
and moreover:
\[
V \Big( \, {}_{{}_{G_2}}U^L \Big) V^{-1}
= V \Big( \, \int \limits_{\mathfrak{B}} U^{L, x_{{}_D}} \,\, 
\ud \nu_{{}_\mathfrak{B}} (x_{{}_D}) \, \Big) V^{-1}
=  \int \limits_{\mathfrak{B}} \,\,\, {}^{\mu^{x_{{}_D}}}U^{L^{x_{{}_D}}} \,\,\,\,\,\, 
\ud \nu_{{}_\mathfrak{B}} (x_{{}_D})  
\]
and 
\[
V \Big( \, \mathfrak{J}^L  \Big) V^{-1}
= V \Big( \, \int \limits_{\mathfrak{B}} \mathfrak{J}^{L,x_{{}_D}} \,\, 
\ud \nu_{{}_\mathfrak{B}} (x_{{}_D}) \, \Big) V^{-1}
=  \int \limits_{\mathfrak{B}} \mathfrak{J}_{x_{{}_D}} \,\, 
\ud \nu_{{}_\mathfrak{B}} (x_{{}_D}). 
\]

\label{lem:subgroup.4}
\end{lem}
\qedsymbol \, 
Let $f$ be any element of $\mathcal{H}^L$ and $t \in G_2$. By definition we have
\[
\Big( V_{x_{{}_D}}  {f}^{{}^{{}^{x_{{}_D}}}}  \Big)_{{}_t} 
= \big(f^D \big)_{{}_{x_{{}_D} \cdot t}}  
=  \Big( {\widetilde{f}}^{{}^{{}^{x_{{}_D}}}} \Big)_{{}_t}, 
\]
with ${\widetilde{f}}^{{}^{{}^{x_{{}_D}}}}$ defined in Lemma \ref{lem:subgroup.3}. Thus by the realization of 
\[
\int \limits_\mathfrak{B}  \,\,\,\, {}^{\mu^{x_{{}_D}}}\mathcal{H}^{L^{x_{{}_D}}} \,\,\,\,\,\, \ud \nu_{{}_\mathfrak{B}}(x_{{}_D})
\]
given in Lemma \ref{lem:subgroup.3}, $V$ is onto. Moreover, by the proof of Lemma \ref{lem:subgroup.3}
\[
x_{{}_D} \mapsto \big( V_{x_{{}_D}}  {f}^{{}^{{}^{x_{{}_D}}}} , {\widetilde{g}}^{{}^{{}^{x_{{}_D}}}}  \big)_{x_{{}_D}}
= \big( {\widetilde{f}}^{{}^{{}^{x_{{}_D}}}} , {\widetilde{g}}^{{}^{{}^{x_{{}_D}}}}  \big)_{x_{{}_D}} \,\,
\]
is measurable for all $g \in \mathcal{H}^L$, and thus for all 
\[
\big( x_{{}_D} \mapsto  {\widetilde{g}}^{{}^{{}^{x_{{}_D}}}} \big)  \in
\int \limits_\mathfrak{B}  \,\,\,\, {}^{\mu^{x_{{}_D}}}\mathcal{H}^{L^{x_{{}_D}}} \,\,\,\,\,\, 
\ud \nu_{{}_\mathfrak{B}}(x_{{}_D});
\]
therefore $V$ is a well defined operator. Moreover, by the proof of Lemma \ref{lem:subgroup.3} 
\begin{multline*}
(Vf,Vg)
= \int \limits_\mathfrak{B} 
\big( V_{x_{{}_D}}  {f}^{{}^{{}^{x_{{}_D}}}} , \, V_{x_{{}_D}} {g}^{{}^{{}^{x_{{}_D}}}}  \big)_{x_{{}_D}} \,\,
 \ud \nu_{{}_\mathfrak{B}}(x_{{}_D})  \\
= \int \limits_\mathfrak{B} 
\big( {\widetilde{f}}^{{}^{{}^{x_{{}_D}}}} , {\widetilde{g}}^{{}^{{}^{x_{{}_D}}}}  \big)_{x_{{}_D}} \,\,
 \ud \nu_{{}_\mathfrak{B}}(x_{{}_D}) = (f,g),
\end{multline*}
so that $V$ is unitary (it likewise follows from Lemma \ref{lem:subgroup.preliminaries.2}). 

Again by Lemma \ref{lem:decomposition.11} we have:

\begin{multline*}
(Vf, Vg)_{{}_{\int  \mathfrak{J}_{x_{{}_D}} \,\, 
\ud \nu_{{}_\mathfrak{B}} (x_{{}_D})}} 
= \int \limits_\mathfrak{B} 
\big( \mathfrak{J}_{x_{{}_D}} {\widetilde{f}}^{{}^{{}^{x_{{}_D}}}} , 
{\widetilde{g}}^{{}^{{}^{x_{{}_D}}}}  \big)_{x_{{}_D}} \,\,
 \ud \nu_{{}_\mathfrak{B}}(x_{{}_D})   \\
= \int \limits_\mathfrak{B} 
\int \limits_{D}
\Big( \mathfrak{J}_L  \big( f^D \big)_{{}_x} , 
\big(g^D \big)_{{}_x}  \Big)  \,\, \ud \mu_D ([x]) \,\,
 \ud \nu_{{}_\mathfrak{B}}(x_{{}_D})                      \\
=  \int \limits_{\mathfrak{G}/G_1}
\Big( \mathfrak{J}_L  f_{{}_x} , g_{{}_x}  \Big)  \,\, \ud \mu_D ([x]) \,\,
 \ud \nu_{{}_\mathfrak{B}}(x_{{}_D}) 
= (f,g)_{{}_{\mathfrak{J}^L}}
\end{multline*}
which shows that $V$ is Krein unitary.

Because by Lemma \ref{lem:subgroup.preliminaries.1}
\[
V_{x_{{}_D}} \,\, U^{L, x_{{}_D}} \,\, {V_{x_{{}_D}}}^{-1}
=  \,\, {}^{\mu^{x_{{}_D}}}U^{L^{x_{{}_D}}} \,\,\, \textrm{and}  \,\,\, 
V_{x_{{}_D}} \,\, \mathfrak{J}^{L,x_{{}_D}} \,\, {V_{x_{{}_D}}}^{-1}
= \mathfrak{J}_{x_{{}_D}} ,
\] 
the rest of the Lemma is thereby proved.  
\qed

\begin{rem}
By a mere renaming of points associated to the isomorphisms
 $\mathfrak{B} \cong G_1 : G_2 \cong \mathfrak{X}/G_2$ of measure spaces, e.g introducing
$V_{{}_{D}} = V_{x_{{}_D}}$, $\mu^{D} = \mu^{x_{{}_D}}$ and the measure $\nu_0$ as in 
Def. \ref{def:decomposition.2} we may rephrase Lemma \ref{lem:subgroup.4} as follows.
$D \mapsto V_{{}_{D}}$ is a decomposition of a well defined operator

\[
V = \int \limits_{G_1 : G_2} V_{{}_{D}} \,\, 
\ud \nu_0 (D) :
\]
\begin{multline*}
\mathcal{H}^L = \int \limits_{G_1 : G_2} \mathcal{H}^{L}_{D} \,\, \ud \nu_0 (D) 
\xrightarrow{V}
\int \limits_{G_1 : G_2}  \,\,\,\, {}^{\mu^{{}^D}}\mathcal{H}^{L^{{}^D}} \,\,\,\,\,\, 
\ud \nu_0 (D) :   \\
\Big(D \mapsto f^D \Big) \mapsto 
\Big( D \mapsto  V_{{}_{D}}  \, f^D  \Big).
\end{multline*}
The operator $V$ is unitary and Krein-unitary between the Krein spaces
\begin{multline*}
\Big( \int \limits_{G_1 : G_2}  \,\,\,\, {}^{\mu^{{}^D}}\mathcal{H}^{L^{{}^D}} \,\,\,\,\,\, 
\ud \nu_0(D) \,\,
 , \,\,\,\, \int \limits_{G_1 : G_2}  \,\,\,\, \mathfrak{J}_{{}_D} \,\,\,\,\,\, 
\ud \nu_0(D) \,\, \Big)   \\       
\,\,\, \textrm{and} \,\,\, 
\Big( \int \limits_{G_1 : G_2} \mathcal{H}^{L}_{D} \,\,
 \ud \nu_0 (D) \,\, , \,\,\,\, \int \limits_{G_1 : G_2} \mathfrak{J}^{L,D} \,\, 
\ud \nu_0 (D) \,\, \Big) = (\mathcal{H}^L , \mathfrak{J}^L);
\end{multline*}
and moreover: 
\[
V \Big( \, {}_{{}_{G_2}}U^L \Big) V^{-1}
= V \Big( \, \int \limits_{G_1 : G_2} U^{L, D} \,\, 
\ud \nu_0 (D) \, \Big) V^{-1}
=  \int \limits_{G_1 : G_2} \,\,\, {}^{\mu^{{}^D}}U^{L^{{}^D}} \,\,\,\,\,\, 
\ud \nu_0 (D)  
\]
and 
\[
V \Big( \, \mathfrak{J}^L  \Big) V^{-1}
= V \Big( \, \int \limits_{G_1 : G_2} \mathfrak{J}^{L,D} \,\, 
\ud \nu_{{}_\mathfrak{B}} (x_{{}_D}) \, \Big) V^{-1}
=  \int \limits_{G_1 : G_2} \,\,\, \mathfrak{J}_{{}_D} \,\,\,\,\,\, 
\ud \nu_0 (D).  
\]

\label{rem:subgroup.2}
\end{rem}

\begin{defin}
Let $G_1$ and $G_2$ be two closed subgroups of a separable locally compact 
group $\mathfrak{G}$. Let $B$ be any Borel section of $\mathfrak{G}$ with respect
to $G_1$ and for each $x \in \mathfrak{G}$  let $h(x)$ be the unique element of $G_1$
such that $h(x)^{-1} \cdot x \in B$. 
Let $\mu$ be any quasi invariant measure $\mu$
on $\mathfrak{G}/G_1$ and let $\nu$ be any pseudo-image measure on $(\mathfrak{G}/G_1 )/G_2$ of the measure
$\mu$ under the quotient map  $\pi_{\mathfrak{G}/G_1}: \mathfrak{G}/G_1 \mapsto (\mathfrak{G}/G_1 )/G_2$;
so that:
\[ 
\mu =  \int \limits_{(\mathfrak{G}/G_1 )/G_2} \mu_C \, \ud \nu(C).
\]
Let us call any measure $\nu_0$ on measurable subsets of the set $G_1 : G_2$   of all double cosets 
\emph{admissible} iff it is equal to the transfer of $\nu$ over to $G_1 : G_2$ by the natural map 
$(\mathfrak{G}/G_1 )/G_2 \ni C \mapsto \pi^{-1}(C) \in G_1 : G_2$. Finally let $x$ be any
element of $\mathfrak{G}$ with $\pi(x) \in C$. We put $\mu^x$ for the measure on 
$G_2 / G_x$ equal to the transfer of the measure $\mu_C$ over to
$G_2 / G_x$ by the map $G_2 / G_x \ni [y] \mapsto [xy] \in C \subset \mathfrak{G}/G_1$, 
where $G_x = G_2 \cap (x^{-1} G_1 x)$
and where $[\cdot]$ denotes the respective equivalence classes. 
  
\label{defin:subgroup.2}
\end{defin}

Summing up we have just proved the following

\begin{twr}[Subgroup Theorem]
Let $U^L$ be the isometric representation of the separable locally compact group $\mathfrak{G}$
in the Krein space $(\mathcal{H}^L , \mathfrak{J}^L)$, induced by the Krein-unitary 
representation $L$ of the closed subgroup
$G_1$ of $\mathfrak{G}$ and the quasi invariant measure $\mu$ on 
$\mathfrak{G}/G_1$ and the Borel section $B$ of $\mathfrak{G}$
with respect to $G_1$. Then $U^L$ is independent to within Krein-unitary 
equivalence of the choice of $B$. Let $G_2$ be a second closed subgroup of $\mathfrak{G}$ and suppose
that $G_1$ and $G_2$ are regularly related. For each $x \in \mathfrak{G}$ consider the closed subgroup
$G_x = G_2 \cap (x^{-1} G_1 x)$ and let $U^{L^{x}}$ denote the representation of $G_2$
in the Krein space $(\mathcal{H}^{L^{x}}, \mathfrak{J}_{{}_x})$ induced by the Krein-unitary representation $L^{x}:
\eta \mapsto L_{x\eta x^{-1}}$ of the subgroup $G_x$ in the Krein space 
$(\mathcal{H}_L , \mathfrak{J}_L )$, where $\big( \mathfrak{J}_{{}_x} g \big)_{{}_t}
= L_{h(x\cdot t)} \mathfrak{J}_L L_{{h(x \cdot t)}^{-1}} \big( g \big)_{{}_t}$
and with the inner product in $\mathcal{H}^{L^{x}}$ and Krein-inner product 
in $(\mathcal{H}^{L^{x}}, \mathfrak{J}_{{}_x})$ defined respectively by the formulas
\[
(f,g)_{{}_x} = \int \limits_{G_2 / G_x} 
\Big( \mathfrak{J}_L \big( \mathfrak{J}_{{}_x} f \big)_{{}_t} , 
\big( g \big)_{{}_t}  \Big)  \,\, \ud \mu^x ([t])
\]
and
\[
(f,g)_{{}_{\mathfrak{J}_{{}_x}}} = (\mathfrak{J}_{{}_x} f , g)_{{}_x} = 
 \int \limits_{G_2 / G_x} 
\Big( \mathfrak{J}_L \big(f \big)_{{}_t} , 
\big( g \big)_{{}_t}  \Big)  \,\, \ud \mu^x ([t]);
\]
and with the quasi invariant measure $\mu^x$ on $G_2 /G_x$ given by Def. \ref{defin:subgroup.2}. 
Then $U^{L^{x}}$ is determined to 
within Krein-unitary and unitary equivalence by the double coset $G_1 x G_2 = s(x)$ to which 
$x$ belongs and we may write $U^{L^{{}^D}} = U^{L^{x}}$, where $D = s(x)$. Finally 
$U^L$ restricted to $G_2$ is a direct integral over $G_1 : G_2$ with respect to any admissible 
(Def. \ref{defin:subgroup.2}) measure in $G_1 : G_2$, of the representations $U^{L^{{}^D}}$.

\label{twr:subgroup.1}
\end{twr}

It may happen that all the component representations ${}^{\mu^{x_{{}_D}}}U^{L^{x_{{}_D}}}$
are bounded and thus Krein-unitary, although $U^L$ is unbounded. In this case the norms
$\big{\|} {}^{\mu^{x_{{}_D}}}U^{L^{x_{{}_D}}} \big{\|}_{x_{{}_D}}$ are unbounded functions
of $x_{{}_D}$ (resp. $D$). Unfortunately instead of $\mathfrak{J}_{x_{{}_D}}$ we cannot use any standard 
fundamental symmetry in 
${^{\mu^{x_{{}_D}}}}\mathcal{H}^{L^{x_{{}_D}}}$:
\[
\Big( \mathfrak{J}^{L^{x_{{}_D}}} 
{\widetilde{f}}^{{}^{{}^{x_{{}_D}}}} \Big)_{{}_t} = L^{x_{{}_D}}_{h_{x_{{}_D}}(t)} 
\mathfrak{J}_{L} L^{x_{{}_D}}_{h_{x_{{}_D}}(t)^{-1}} \, 
\Big({\widetilde{f}}^{{}^{{}^{x_{{}_D}}}}\Big)_{{}_t} ,
\] 
where $h_{x_{{}_D}}(t) \in G_{x_{{}_D}}$ is defined as in Sect. \ref{def_ind_krein}
by a regular Borel section $B_{x_{{}_D}}$ of $G_2$ with respect to the subgroup 
$G_{x_{{}_D}} = G_2 \cap \, ({x_{{}_D}}^{-1}G_1 x_{{}_D})$. A difficulty will arise with this 
$\mathfrak{J}^{L^{x_{{}_D}}}$.
Namely in general the norms $\big{\|} {}^{\mu^{x_{{}_D}}}U^{L^{x_{{}_D}}} \big{\|}_{x_{{}_D}}$  
are such that the operator $V$ would be unbounded with the standard fundamental symmetries
in ${^{\mu^{x_{{}_D}}}}\mathcal{H}^{L^{x_{{}_D}}}$.

It is important that in practical computations, e.g. with $\mathfrak{G}$ equal to the double covering 
of the Pouncar\'e group, much stronger regularity is preserved, e. g. the ``measure product property''
(see the end of Sect. \ref{decomposition}), with the measurable sections $B$ and $\mathfrak{B}$ as differential
sub-manifolds (if we discard unimportant null subset), so that the function $x \mapsto h(x)$ and
all the remaining functions -- analogue of (\ref{q'_h'}) -- associated 
to the measure product structure are effectively computable together with the measures
$\mu$ and $\nu_0$. This is important because together with the theorem of the next Section
give an effective tool for decomposing tensor product of induced representations of the double cover 
of the Poincar\'e group in Krein spaces.
Moreover the operator $V$ of Lemma \ref{lem:subgroup.4} and 
Remark \ref{rem:subgroup.2} is likewise effectively computable in this case.

\subsection{Kronecker product theorem in Krein spaces}\label{Kronecker_product}

Let ${}^{\mu_1}U^L$ and ${}^{\mu_2}U^M$ be Krein-isometric representations
of the separable locally compact group $\mathfrak{G}$ induced from Krein-unitary
representations of the closed subgroups $G_1 \subset \mathfrak{G}$ and $G_2 \subset \mathfrak{G}$
respectively. 
The Krein-isometric representation ${}^{\mu_1}U^L \, \otimes \, {}^{\mu_2}U^M$ of $\mathfrak{G}$ is obtained from the 
Krein-isometric representation ${}^{\mu_1}U^L \, \times \, {}^{\mu_2}U^M$ of $\mathfrak{G} \times \mathfrak{G}$ by restriction to the 
diagonal subgroup $\overline{\mathfrak{G}}$ of all those 
$(x,y) \in \mathfrak{G} \times \mathfrak{G}$ for which $x = y$, which is naturally isomorphic
to $\mathfrak{G}$ itself: $\overline{\mathfrak{G}} \cong \mathfrak{G}$, with
the natural isomorphism $(x,x) \mapsto x$. Thus by the natural isomorphism 
the representation ${}^{\mu_1}U^L \, \otimes \, {}^{\mu_2}U^M$ of $\mathfrak{G}$ may be identified
with the restriction of the representation ${}^{\mu_1}U^L \, \times \, {}^{\mu_2}U^M$ of 
 the group $\mathfrak{G} \times \mathfrak{G}$ to the diagonal subgroup $\overline{\mathfrak{G}}$. By 
Theorem \ref{twr.1:kronecker}, ${}^{\mu_1}U^L \, \times \, {}^{\mu_2}U^M$ is Krein-unitary and unitary equivalent to the 
Krein-isometric representation ${}^{\mu_1 \times \mu_2}U^{L \times M}$ of $\mathfrak{G} \times \mathfrak{G}$
induced by the Krein-unitary representation $L \times M$ of the closed subgroup $G_1 \times G_2$. Thus the
Krein-isometric representation $U^L \otimes U^M$ of $\mathfrak{G}$ is naturally 
equivalent to the restriction of the Krein-isometric representation ${}^{\mu_1 \times \mu_2}U^{L \times M}$ of 
$\mathfrak{G} \times \mathfrak{G}$ to the closed diagonal subgroup $\overline{\mathfrak{G}}$. Thus we are trying
to apply the \emph{Subgroup Theorem} \ref{twr:subgroup.1} inserting 
$\mathfrak{G} \times \mathfrak{G}$ for $\mathfrak{G}$, $\overline{\mathfrak{G}}$ for  $G_2$,
and the subgroup $G_1 \times G_2 \subset \mathfrak{G} \times \mathfrak{G}$ for
$G_1$ in the Subgroup Theorem. But the Subgroup Theorem 
is applicable in that way if the subgroups $G_1 \times G_2$ and 
$\overline{\mathfrak{G}}$ are regularly related.
Mackey recognized that they are indeed regularly related in 
$\mathfrak{G} \times \mathfrak{G}$ if and only if $G_1$ and $G_2$ are in $\mathfrak{G}$, 
pointing out a natural measure isomorphism between the
measure spaces $(G_1 \times G_2 ) : \overline{\mathfrak{G}}$ and $G_1 : G_2$ of double cosets
respectively in $\mathfrak{G} \times \mathfrak{G}$ and $\mathfrak{G}$.   
The isomorphism is induced by the map 
$\mathfrak{G} \times \mathfrak{G} \ni (x,y) \mapsto xy^{-1} \in \mathfrak{G}$. However his 
argumentation strongly depends on the finiteness of the quasi invariant measures in the homogeneous spaces
$(\mathfrak{G} \times \mathfrak{G}) /(G_1 \times G_2)$ and $\mathfrak{G}/G_1$ which slightly 
simplifies the construction of the $\sigma$-rings of measurable subsets in the corresponding
spaces of double cosets. Our proof that the map $(x,y) \mapsto xy^{-1}$ induces isomorphism
of the respective spaces of double cosets must have been slightly changed at this point
by addition of Lemma \ref{lem:decomposition.10}. The rest of the proof of 
Theorem \ref{twr:Kronecker_product} of this 
Section follows from the Subgroup Theorem \ref{twr:subgroup.1} in the same way as Theorem
7.2 from Theorem 7.1 in \cite{Mackey}.

By the above remarks we shall show that the measure spaces
$(G_1 \times G_2 ) : \overline{\mathfrak{G}}$ and $G_1 : G_2$ of double cosets constructed as in
Sect. \ref{decomposition} are isomorphic, with the isomorphism induced by the map 
$\mathfrak{G} \times \mathfrak{G} \ni (x,y) \mapsto xy^{-1} \in \mathfrak{G}$. Note first of all
that the indicated map sets up a one-to-one correspondence between the double cosets
in  $(G_1 \times G_2 ) : \overline{\mathfrak{G}}$ and double cosets in $G_1 : G_2$, in which the double coset
$(G_1 \times G_2 ) (x,y) \overline{\mathfrak{G}}$ corresponds to the double coset 
$G_1 xy^{-1} G_2$. Moreover in this mapping a set is measurable if and only if its 
image is measurable and \emph{vice versa}, a set is measurable if and only if its inverse image is measurable.
Thus it is an isomorphism of measure spaces. Indeed $(x_1 , x_2)$ and $(x_2 , y_2)$ 
go into the same point of $\mathfrak{G}$ under the indicated map
if and only if they belong to the same left $\overline{\mathfrak{G}}$ coset in 
$\mathfrak{G} \times \mathfrak{G}$. Now by Lemma \ref{lem:decomposition.10} of sect. \ref{decomposition} and 
by Lemma 1.2 of \cite{Mackey} (equally applicable to left coset spaces) the indicated one-to-one map
of double coset spaces is an isomorphism of measure spaces. Thus the Subgrop Theorem \ref{twr:subgroup.1} is 
applicable to ${}^{\mu_1 \times \mu_2}U^{L \times M}$ with $L$ replaced by $L \times M$, $\mathfrak{G}$
replaced by $\mathfrak{G} \times \mathfrak{G}$, $G_1$ replaced by $G_1 \times G_2$ 
and $G_2$ replaced by $\overline{\mathfrak{G}}$
and the function $\mathfrak{G} \ni x \mapsto h(x) \in G_1$
replaced by the function $\mathfrak{G} \times \mathfrak{G}
\ni (x,y) \mapsto h(x,y) = \big( h_{{}_1}(x) , h_{{}_2}(y) \big) \in G_1 \times G_2$,
where the functions $\mathfrak{G} \ni x \mapsto  h_{{}_1}(x) \in G_1$
and $\mathfrak{G} \ni y \mapsto  h_{{}_2}(y) \in G_2$
correspond to the respective Borel sections of $\mathfrak{G}$
with respect to $G_1$ and $G_2$ respectively used in the construction
of the representations ${}^{\mu_1}U^L$ and ${}^{\mu_2}U^M$ 
(compare Sect. \ref{def_ind_krein} and \ref{kronecker}).

In order to simplify formulation of the upcoming theorem let us give the following

\begin{defin}
Let $\nu_{0}^{12}$ be the admissible measure on the set of double cosets 
$\big(G_1 \times G_2 ):\overline{\mathfrak{G}}$ in $\mathfrak{G} \times \mathfrak{G}$

given by Def. \ref{defin:subgroup.2}, where we have used the product quasi invariant measure 
$\mu = \mu_1 \times \mu_2$ on the homogeneous space $\big( \mathfrak{G} \times \mathfrak{G} \big)
\big{/} \big( G_1 \times G_2 \big)$. Let us define the measure 
$\nu_{12}$ on the space $G_1 : G_2$ of double cosets in $\mathfrak{G}$ to be equal to the transfer
of $\nu_{0}^{12}$ by the map induced by 
$\mathfrak{G} \times \mathfrak{G} \ni (x,y) \mapsto xy^{-1} \in \mathfrak{G}$.
If $(\mu_1 \times \mu_2)^{(x,y)}$ is the quasi invariant measure  on
$\overline{\mathfrak{G}} / G_{(x,y)}$ given by Def. \ref{defin:subgroup.2}
with $G_{(x,y)} = \overline{\mathfrak{G}} \cap \big( (x,y)^{-1} (G_1 \times G_2) (x,y) \big)$
then we define $\mu^{x,y}$ to be the transfer of the measure $(\mu_1 \times \mu_2)^{(x,y)}$
over to the homogeneous space $\mathfrak{G} \big{/} \big( x^{-1}G_1 x \, \, \cap \, y^{-1} G_2 y \big)$
by the map $(x,x) \mapsto x$. 

\label{def:Kronecker_product.1}
\end{defin}   

Now we are ready to formulate the main goal of this paper:

\begin{twr}[Kronecker Product Theorem]
Let $G_1$ and $G_2$ be regularly related closed subgroups of the separable locally compact group 
$\mathfrak{G}$. Let $L$ and $M$ be Krein-unitary representations of $G_1$ and $G_2$ respectively
in the Krein spaces $(\mathcal{H}_L , \mathfrak{J}_L)$ and $(\mathcal{H}_M , \mathfrak{J}_M)$.
For each $(x, y) \in \mathfrak{G} \times \mathfrak{G}$ consider the Krein-unitary 
representations $L^x : s \mapsto L_{xsx^{-1}}$ and $M^y : s \mapsto M_{ysy^{-1}}$
of the subgroup $(x^{-1}G_1 x) \cap (y^{-1}G_2 y)$
in the Krein spaces $(\mathcal{H}_L , \mathfrak{J}_L)$ and $(\mathcal{H}_M , \mathfrak{J}_M)$
respectively.
Let us denote the tensor product $L^x \otimes M^y$ Krein-unitary representation
acting in the Krein space $(\mathcal{H}_L \otimes \mathfrak{J}_M , \mathfrak{J}_L \otimes \mathfrak{J}_M)$,
by $N^{x,y}$. 
Let $U^{N^{x,y}}$ be the Krein-isometric representation of $\mathfrak{G}$ induced
by $N^{x,y}$ acting in the Krein space $\big(\mathcal{H}^{N^{x,y}}, \mathfrak{J}{{}_{x,y}}\big)$, where
for each $w \in \mathcal{H}^{N^{x,y}}$
\[
\big( \mathfrak{J}_{{}_{x,y}} w \big)_{{}_{s}}
= \big( L_{h_{{}_1}(xs)} \,  \mathfrak{J}_L \,  L_{h_{{}_1}(xs)^{-1}} \big) \otimes 
\big( M_{h_{{}_2}(ys)} \, \mathfrak{J}_M \,  M_{h_{{}_2}(ys)^{-1}} \big) \big( w \big)_{{}_{s}};
\]
and with the inner product in Hilbert space $\mathcal{H}^{N^{x,y}}$ and the Krein-inner product
in the Krein space $\big(\mathcal{H}^{N^{x,y}}, \mathfrak{J}{{}_{x,y}}\big)$
given by the formulas
\[
(w,g)_{{}_{x,y}} 
= \int \limits_{\mathfrak{G} \big{/} \big( x^{-1}G_1 x \, \, \cap \, y^{-1} G_2 y \big)} 
\Big( \mathfrak{J}_L \otimes \mathfrak{J}_M \big( \mathfrak{J}_{{}_{x,y}}  w \big)_{{}_s} , 
\big( g \big)_{{}_s}  \Big)  \,\, \ud \mu^{x,y} ([s])
\]
and 
\begin{multline*}
(w,g)_{{}_{\mathfrak{J}_{{}_{x,y}}}} = (\mathfrak{J}_{{}_{x,y}} w , g)_{{}_{x,y}}   \\
= \int \limits_{\mathfrak{G} \big{/} \big( x^{-1}G_1 x \, \, \cap \, y^{-1} G_2 y \big)} 
\Big( \mathfrak{J}_L \otimes \mathfrak{J}_M \big( w \big)_{{}_s} , 
\big( g \big)_{{}_s}  \Big)  \,\, \ud \mu^{x,y} ([s]),
\end{multline*}
with the quasi invariant measure $\mu^{x,y}$ given by Def. \ref{def:Kronecker_product.1}.
Then $U^{N^{x,y}}$ is determined to within Krein-unitary equivalence by the double coset
$D =  G_1 xy^{-1}G_2$ to which $xy^{-1}$ belongs and we may write $U^{N^{x,y}} = U^D$. Finally
$U^L \otimes U^M$ is Krein-unitary equivalent to the direct integral of $U^D$ with respect to the
measure $\nu_{12}$ (Def. \ref{def:Kronecker_product.1}) on $G_1 : G_2$.

\label{twr:Kronecker_product}
\end{twr}
\qedsymbol \, 
By the above remarks the Subgroup Theorem \ref{twr:subgroup.1} is applicable to the restriction
of the representation ${}^{\mu_1 \times \mu_2}U^{L \times M}$ of $\mathfrak{G} \times \mathfrak{G}$
to the subgroup $\overline{\mathfrak{G}}$. 
By this theorem, ${}^{\mu_1 \times \mu_2}U^{L \times M}$ restricted to 
$\overline{\mathfrak{G}}$ is a direct integral over the space of  double cosets\footnote{I. e. with $(x,y)$ ranging 
over $\mathfrak{B}_1 \times \mathfrak{B}_2$ -- the corresponding section 
of $\mathfrak{G} \times \mathfrak{G}$ with respect to double cosets 
$(G_1 \times G_2 ) : \overline{\mathfrak{G}}$.}  $(G_1 \times G_2 ) (x,y) \overline{\mathfrak{G}}$
with exactly one representant $(x,y)$ for each
double coset, of the representations $U^{(L \times M)^{(x,y)}}$ of 
the subgroup $\overline{\mathfrak{G}}$. 
Each of the representations
$U^{(L \times M)^{(x,y)}}$ of $\overline{\mathfrak{G}}$ is induced by the Krein-unitary representation 
$(L \times M)^{(x,y)}: (s,s) \mapsto (L \times M)_{{}_{(x,y)(s,s)(x,y)^{-1}}}
= L_{xsy^{-1}} \otimes M_{xsy^{-1}}$
of the subgroup $G_{(x,y)} = \overline{\mathfrak{G}} \cap \big( (x,y)^{-1} (G_1 \times G_2) (x,y) \big) 
\subset \overline{\mathfrak{G}}$
in the Krein space $(\mathcal{H}_L \otimes \mathcal{H}_M , \mathfrak{J}_L \otimes \mathfrak{J}_M)$. 
Moreover $U^{(L \times M)^{(x,y)}}$ acts in the Krein space 
$\big( \mathcal{H}^{(L \times M)^{(x,y)}} , \mathfrak{J}_{{}_{(x,y)}} \big)$ where for each function $w \in \mathcal{H}^{(L \times M)^{(x,y)}}$ we have
\begin{multline*}
\big( \mathfrak{J}_{{}_{(x,y)}} w \big)_{{}_{(s,s)}}
= (L \times M)_{h((x,y)\cdot (s,s))} \mathfrak{J}_{L\times M} (L \times M)_{{h((x,y) \cdot (s,s))}^{-1}} \big( w \big)_{{}_{(s,s)}}   \\
= L \times M)_{(h_{{}_1}(xs) , h_{{}_2}(ys))} \mathfrak{J}_{L\times M} 
(L \times M)_{{(h_{{}_1}(xs)^{-1} , h_{{}_2}(ys)^{-1})}} \big( w \big)_{{}_{(s,s)}} \\
= \big( L_{h_{{}_1}(xs)} \,  \mathfrak{J}_L \,  L_{h_{{}_1}(xs)^{-1}} \big) \otimes 
\big( M_{h_{{}_2}(ys)} \, \mathfrak{J}_M \,  M_{h_{{}_2}(xs)^{-1}} \big) \big( w \big)_{{}_{(s,s)}}.
\end{multline*}
The inner product in $\mathcal{H}^{(L \times M)^{(x,y)}}$ and Krein-inner product in
$\big( \mathcal{H}^{(L \times M)^{(x,y)}} , \mathfrak{J}_{{}_{(x,y)}} \big)$ are defined by 
\begin{multline*}
(w,g)_{{}_{(x,y)}} = \int \limits_{\overline{\mathfrak{G}} / G_{(x,y)}} 
\Big( \mathfrak{J}_{L \times M} \big( \mathfrak{J}_{{}_{(x,y)}}  w \big)_{{}_{(s,s)}} , 
\big( g \big)_{{}_{(s,s)}}  \Big)  \,\, \ud (\mu_1 \times \mu_2)^{(x,y)} ([(s,s)])  \\
= \int \limits_{\overline{\mathfrak{G}} / G_{(x,y)}} 
\Big( \mathfrak{J}_L \otimes \mathfrak{J}_M \big( \mathfrak{J}_{{}_{(x,y)}}  w \big)_{{}_{(s,s)}} , 
\big( g \big)_{{}_{(s,s)}}  \Big)  \,\, \ud (\mu_1 \times \mu_2)^{(x,y)} ([(s,s)])
\end{multline*}
and
\begin{multline*}
(w,g)_{{}_{\mathfrak{J}_{{}_{(x,y)}}}} = (\mathfrak{J}_{{}_{(x,y)}} w , g)_{{}_{(x,y)}}   \\
=\int \limits_{\overline{\mathfrak{G}} / G_{(x,y)}} 
\Big( \mathfrak{J}_{L \times M} \big(w \big)_{{}_{(s,s)}} , 
\big( g \big)_{{}_{(s,s)}}  \Big)  \,\, \ud (\mu_1 \times \mu_2)^{(x,y)} ([(s,s)])  \\
= \int \limits_{\overline{\mathfrak{G}} / G_{(x,y)}} 
\Big( \mathfrak{J}_L \otimes \mathfrak{J}_M \big(w \big)_{{}_{(s,s)}} , 
\big( g \big)_{{}_{(s,s)}}  \Big)  \,\, \ud (\mu_1 \times \mu_2)^{(x,y)} ([(s,s)]);
\end{multline*}
with the quasi invariant measure $(\mu_1 \times \mu_2)^{(x,y)} $ on
$\overline{\mathfrak{G}} / G_{(x,y)}$ given by Def. \ref{defin:subgroup.2}.

Now under the natural isomorphism $(x,x) \mapsto x$ transferring $\overline{\mathfrak{G}}$ onto $\mathfrak{G}$ 
the group $G_{(x,y)} = \overline{\mathfrak{G}} \cap \big( (x,y)^{-1} (G_1 \times G_2) (x,y) \big)$
is transferred onto the subgroup $x^{-1}G_1 x \, \, \cap \, y^{-1} G_2 y$ of $\mathfrak{G}$ 
and the homogeneous space $\overline{\mathfrak{G}} / G_{(x,y)}$ with the quasi invariant
measure $(\mu_1 \times \mu_2)^{(x,y)}$ is transferred over to the homogeneous space 
$\mathfrak{G} \big{/} \big( x^{-1}G_1 x \, \, \cap \, y^{-1} G_2 y \big)$ with the quasi
invariant measure, which we denote by $\mu^{x,y}$. 
\qed

\subsection{Krein-isometric representations induced by decomposable Krein-unitary
representations}\label{decomposable_L}

We say a family $\mathfrak{S}$ of operators in a Hilbert space $\mathcal{H}$ is \emph{reducible}
by an idempotent $P$ (i. e. a bounded operator $P$ which satisfies the identity $P^2 = P$),
or equivalently by a closed subspace equal to the range $P\mathcal{H}$ of $P$, in case 
$PUP = UP$ for all $U \in \mathfrak{S}$. We say the family $\mathfrak{S}$ is \emph{decomposable} 
in case $PU = UP$ for all $U \in \mathfrak{S}$. In this case the Hilbert space $\mathcal{H}$ 
is the direct sum of closed subspaces $\mathcal{H}_1 = P \mathcal{H}$ and 
$\mathcal{H}_2 = (I - P)\mathcal{H}$ and every operator in $\mathfrak{S}$
is a direct sum of operators $U_1$ and $U_2$ with $U_i$ acting in $\mathcal{H}_i$, $i = 1,2$.
The closed subspaces $\mathcal{H}_i$, $i = 1,2$, are orthogonal iff $P$ is self adjoint.
Moreover if $(\mathcal{H}, \mathfrak{J})$ is a Krein space, the closed subspaces
$\mathcal{H}_i$, $i = 1,2$, are Krein-orthogonal iff the idempotent $P$ is Krein-self-adjoint:
$P^\dagger = P$.
Now the Krein-isometric representations $U^L$  inherit
decomposability from decomposability of $L$. Namely for each idempotent $P_L$ acting in the Krein space of the representation $L$ we may define a natural idempotent $P^L$
by the formula $(P^L f)_{{}_x} = P_L f_{{}_x}$ for $f \in \mathcal{H}^L$ provided $P_L$ commutes with the
representation $L$. Checking that $P^L$ is well defined (with measurable $x \mapsto \Big(\, \big(P^L f\big)_{{}_x} \, , \,\, \upsilon \, \Big)$ for each $\upsilon \in \mathcal{H}_L$ and $\big(P^L f\big)_{{}_{hx}} = L_h \big(P^L f\big)_{{}_{x}}$) 
and that $P^L$ is a bounded idempotent is immediate.
Moreover $P^L$ likewise commutes with $U^L$ and is self-adjoint whenever $P_L$ is.

Thus in particular for the standard Krein-isometric representation we have the following

\begin{twr}
Let $H$ be a closed subgroup of the separable locally compact group $\mathfrak{G}$. Let 
$U^L$ be the Krein-isometric representation of $\mathfrak{G}$ acting in the
Krein space $(\mathcal{H}^L , \mathfrak{J}^L)$, induced by the Krein-unitary
representation $L$ of the subgroup $H$, acting in the Krein space $(\mathcal{H}_L , \mathfrak{J}_L)$. 
Let for a measure space $(\mathbb{R}, \mathscr{R}_{\mathbb{R}}, m)$ 
the operators of the representation $L$ and the fundamental symmetry 
$\mathfrak{J}_L$ be decomposable:
\[
L = \int \limits_\mathbb{R} \,\, {}_{{}_\lambda}L \,\,\,\, \ud m (\lambda)
\,\,\,\,\,\,\,\, \textrm{and} \,\,\,\,\,\,\,\,
\mathfrak{J}_L = \int \limits_\mathbb{R} \,\, \mathfrak{J}_{{}_{{}_{{}_\lambda}L}} \,\,\,\, \ud m (\lambda)
\]
with respect to a direct integral decomposition
\begin{equation}\label{hl_dec_h_lambda}
\mathcal{H}_L = \int \limits_\mathbb{R} \,\, \mathcal{H}_{{}_{{}_\lambda}L} \,\,\,\, \ud m (\lambda),
\end{equation}
of the Hilbert space $\mathcal{H}_L$.  Then 
\begin{equation}\label{decomosable_L_dec_H^L}
\mathcal{H}^L = \int \limits_\mathbb{R} \,\, \mathcal{H}^{{}^{{}_{{}_\lambda}L}}  \,\,\,\, \ud m (\lambda)
\end{equation}
and all operators of the representation $U^L$ and the fundamental symmetry $\mathfrak{J}^L$ are
decomposable with respect to (\ref{decomosable_L_dec_H^L}), i. e.

\[
U^L \, = \, \int \limits_\mathbb{R} \,\, U^{{}^{{}_{{}_\lambda}L}}  \,\,\,\, \ud m (\lambda)
\,\,\,\,\,\,\,\, \textrm{and} \,\,\,\,\,\,\,\,
\mathfrak{J}^L \, = \, \int \limits_\mathbb{R} \,\, \mathfrak{J}^{{}^{{}_{{}_\lambda}L}}  \,\,\,\, \ud m (\lambda);
\] 
where $U^{{}^{{}_{{}_\lambda}L}}$ is the Krein-isometric representation in the Krein space 
$\big( \mathcal{H}^{{}^{{}_{{}_\lambda}L}} , \mathfrak{J}^{{}^{{}_{{}_\lambda}L}} \big)$ induced by
the Krein-unitary representation ${}_{{}_\lambda}L$ of the subgroup $H$, acting in the Krein space  
$\big( \mathcal{H}_{{}_{{}_\lambda}L} , \mathfrak{J}_{{}_{{}_{{}_\lambda}L}} \big)$. 

\label{twr:decomposable_L.1}
\end{twr}   
\qedsymbol \, [Outline of the proof.]
Let $\lambda \mapsto E (\lambda)_L$ be the spectral measure associated with the decomposition
(\ref{hl_dec_h_lambda}). Consider the direct integral decompositions 
\begin{multline*}
\mathcal{H}^L = \int \limits_\mathbb{R} \,\, \mathcal{H}^{L}(\lambda)  \,\,\,\, \ud m (\lambda), \\
\mathfrak{J}^L \, = \, \int \limits_\mathbb{R} \,\, \mathfrak{J}^{L}(\lambda)  \,\,\,\, \ud m (\lambda)
\,\,\,\,\,\,\,\, \textrm{and} \,\,\,\,\,\,\,\,
U^L \, = \, \int \limits_\mathbb{R} \,\, U^{L}(\lambda)  \,\,\,\, \ud m (\lambda),\label{decomposable_L:uniform_mult}
\end{multline*}
of $\mathcal{H}^L$, $\mathfrak{J}^L$ and $U^L$,
associated with the corresponding spectral measure $\lambda \mapsto E (\lambda)^L$ 
and the same measure $m$. Using the vector-valued version of (\ref{dir_int_L^2:decompositions})
and the Fubini theorem one shows that 
$\mathcal{H}^{L}(\lambda) = \mathcal{H}_{{}_{{}_\lambda}L}$ and the equalities of the Radon-Nikodym
derivatives
\begin{multline*}
\frac{\ud \, \Big( \, E(\lambda)^L \mathfrak{J}^L f \, , \,\, E(\lambda)^L g \, \Big)}{\ud m (\lambda)} 
= \frac{\ud \, \Big( \, E(\lambda)^L \mathfrak{J}^{{}^{{}_{{}_\lambda}L}} f \, , \,\, E(\lambda)^L g \, \Big)}
{\ud m (\lambda)},  \\
\frac{\ud \, \Big( \, E(\lambda)^L U^L f \, , \,\, E(\lambda)^L g \, \Big)}{\ud m (\lambda)} 
= \frac{\ud \, \Big( \, E(\lambda)^L U^{{}^{{}_{{}_\lambda}L}} f \, , \,\, E(\lambda)^L g \, \Big)}
{\ud m (\lambda)}, 
\end{multline*}
for all $f, g \in \mathcal{H}^L$ in the domain of $U^L$, which means that 
$\mathfrak{J}^L (\lambda) = \mathfrak{J}^{{}^{{}_{{}_\lambda}L}}$ and $U^L (\lambda) = U^{{}^{{}_{{}_\lambda}L}}$.

\qed

Using the Dunford-Gelfand-Mackey \cite{Dunford, Gelfand_spectral} (or more general \cite{Foias,Lanze}) spectral measures and corresponding decompositions, we could generalize the last theorem
keeping Krein self adjointness of the idempotents of the decomposition of $L$ just using Dunford 
or more general spectral measures), 
but abandoning their commutativity with $\mathfrak{J}_L$, and thus discarding their self-adjointness.  

\vspace*{0.5cm}

But in decomposition of the Krein-isometric induced representation restricted to a closed subgroup as
in the Subgroup Theorem (or respectively in decomposition of the tensor product of Krein-isometric 
induced representations as in the Kronecker Product Theorem) we have encountered Krein-isometric induced representations 
$U^{L^{x}}$ in the Krein space  $(\mathcal{H}^{L^{x}}, \mathfrak{J}_{{}_x})$ 
with the non-standard fundamental symmetry $\mathfrak{J}_{{}_x}$ instead of the standard one
$\mathfrak{J}^{L^x}$ 
(respectively their tensor product $U^{N^{x,y}}$ acting in the tensor product Krein space 
$\big(\mathcal{H}^{N^{x,y}}, \mathfrak{J}{{}_{x,y}}\big) = 
\big(\mathcal{H}^{L^{x}} \otimes \mathcal{H}^{M^{y}}, \mathfrak{J}_{{}_x} \otimes \mathfrak{J}_{{}_y} \big)$). 
In this case for each idempotent $P_L$ acting in the representation space $(\mathcal{H}_L , \mathfrak{J}_L)$
and commuting with $L^x$ we could similarly define the corresponding operator $P^L$:
$(P^L f)_{{}_x} = P_L f_{{}_x}$ for $f \in \mathcal{H}^{L^{x}}$. (Similarly we can define $P^{{}^{N^{x,y}}}$
for each idempotent $P_{{}_{N^{x,y}}}$ commuting with $N^{x,y}$.) However in this case with non-standard fundamental
symmetry $\big( \mathfrak{J}_{{}_{x}} g \big)_{{}_{s}} \, = \, L_{h_{{}_1}(xs)} \,  \mathfrak{J}_L \,  L_{h_{{}_1}(xs)^{-1}} \big( g \big)_{{}_{s}}$, $g \in \mathcal{H}^{L^x}$ (resp. $\big( \mathfrak{J}_{{}_{x,y}} w \big)_{{}_{s}}
= \big( L_{h_{{}_1}(xs)} \,  \mathfrak{J}_L \,  L_{h_{{}_1}(xs)^{-1}} \big) \otimes 
\big( M_{h_{{}_2}(ys)} \, \mathfrak{J}_M \,  M_{h_{{}_2}(ys)^{-1}} \big) \big( w \big)_{{}_{s}}$, $w \in 
\mathcal{H}^{N^{x,y}}$) the operator $P^L$ (or $P^{{}^{N^{x,y}}}$) is in general
\emph{unbounded}. Moreover $P^L$ (resp. $P^{{}^{N^{x,y}}}$) is non 
self-adjoint in this case even if $P_L$ (resp. $P_{{}_{N^{x,y}}}$) is self-adjoint.
We hope the slightly misleading (unjustified) notation $U^{N^{x,y}}$ will cause no serious troubles. 

Thus in particular the Theorem \ref{twr:decomposable_L.1} (and its generalizations with Dunford-Gelfand-Mackey
spectral measure decompositions) cannot in general be immediately applied to the representations
$U^{N^{x,y}}$ standing in the Kronecker Product Theorem for the tensor product of {\L}opusza\'nski
representations of the double covering $\mathfrak{G} = T_4 \circledS SL(2,\mathbb{C})$
of the Poincar\'e group. But $U^{N^{x,y}}$ as a Krein-isometric representation
of the semi-direct product $\mathfrak{G} = T_4 \circledS SL(2,\mathbb{C})$ defines an imprimitivity
system in Krein space (in the sense of Sect \ref{lop_ind}) which is concentrated on a single orbit.
We then restore the form of the ordinary induced representation to $U^{N^{x,y}}$
by applying Theorem \ref{lop_ind:twr.1} of Sect. \ref{lop_ind}  but we need the generalized version 
of this theorem with the finite multiplicity condition 3) discarded and replaced with infinite uniform
multiplicity. It follows that $U^{N^{x,y}}$ is equivalent to a standard Krein-isometric induced 
representation with the equivalence given by a Krein-isometric operator which is nonsingular (unbo
unded) in the sense that its domain and image are dense core domains of the equivalent representations. 
Then  using the vector valued Fubini like theorem (eq. \ref{dir_int_skew_L^2:decompositions} of Sect. \ref{decomposition})
we find explicit form of the standard induced representation.  We hope to present in a subsequent 
paper the full analysis of the component representations $U^{N^{x,y}}$ in the decomposition of tensor 
product of {\L}opusza\'nski representations.

The necessity of restoring the standard form of induced Krein-isometric representation to the component 
Krein-isometric representations in the decomposition
of tensor product of standard induced Krein-isometric representations is the main difference in comparison
to the Mackey theory of unitary induced representations. In case of the double covering $\mathfrak{G}$ 
of the Poincar\'e group this ``restoring'' is quite elaborate, but effectively computable. 
The case of tensor products of ordinary unitary induced representations may be rather effectively reduced to the harmonic
analysis on ``small groups''\footnote{For the harmonic analysis on $SL(2,\mathbb{R})$ compare 
\cite{Harish-Chandra,Ehrenpreis_Mautner1,Ehrenpreis_Mautner2,Ehrenpreis_Mautner3}.} $G_{\chi_p} = SU(2,\mathbb{C})$ or $SL(2,\mathbb{R})$ (see Sect. \ref{lop_ind}) 
and to the tensor products\footnote{Computed in \cite{nai1,nai2,nai3}.} 
of Gelfand-Neumark representations of $SL(2,\mathbb{C})$, with the help of the original Mackey's
Subgroup and Kronecker Product Theorems for unitary induced representations. Indeed for tensor products
of integer spin representations (for both versions of the energy sign) these decompositions
have indeed been computed by Tatsuuma \cite{tat}. Unfortunately the paper \cite{tat} presents only 
the results without proofs, and some of the results presented there are not correct, namely those
under X).  

\begin{rem}
Because the representation of the translation subgroup $T_4 \subset T_4 \circledS SL(2, \mathbb{C})$
in  {\L}opusza\'nski-type representation is equivalent to the representation of $T_4$ in 
direct sum of several 
(four in case of the  {\L}opusza\'nski representation) representations of, say helicity zero, ordinary unitary
induced representations of $T_4 \circledS SL(2, \mathbb{C})$, corresponding to the 
``light-cone orbit'' in the momentum space, and the representation of $T_4 \circledS SL(2, \mathbb{C})$ in 
the Fock space is the direct sum of symmetrized/anitisymmetrized
tensor products of one-particle representations, then investigation of the multiplicity of the representation of
$T_4$ in the Fock space is reduced to the decomposition of tensor products of ordinary unitary induced 
representations of $T_4 \circledS SL(2, \mathbb{C})$. 
\label{decomposable_L:uniform_mult}
\end{rem}

\pagebreak
    
\vspace*{5cm}

\section{ACKNOWLEDGEMENTS}

\vspace*{0.3cm}

The author is indebted for helpful discussions to Professor A. Staruszkiewicz.  
The author would especially like to thank Professor A. Staruszkiewicz and 
Professor M. Je\.zabek  
for the warm encouragement. He would like to thank Professor M. Je\.zabek 
for the excellent conditions for work at INP PAS in Krak\'ow, Poland, 
where the  $40\%$ of this work has been written. 

The $60\%$ of this work has been written in The
Bogoliubov Laboratory of Theoretical Physics of the Joint Institute for Nuclear Research
in Dubna, Russia. More precisely: 
Subsections \ref{OperationsOnXiIF}, \ref{WickForProduct}, \ref{WickForChronological}, \ref{psichi}, \ref{equivalentA-s},
\ref{AS}, \ref{A(1)chi}, \ref{ASTheoryextended}, \ref{splittingEU}, Section \ref{A(1)psi(1)}, 
and Section \ref{EUandG} have been written in JINR, Dubna,
as well as some of the remaining Subsections have been enlarged or corrected in JINR, Dubna.
The author would like to express his deep gratitude to Professor D. Kazakov for the very helpful discussions.
The author would like to thank for discussions with Professor I. Volovich.
I would like to thank for the excellent conditions for work at JINR, Dubna.

\pagebreak

\vspace*{5cm}

\printindex

\end{document}
